\documentclass[11pt,a4paper,twocolumn,times]{elsarticle} 
\makeatletter
\def\ps@pprintTitle{%
\let\@oddhead\@empty
\let\@evenhead\@empty
\def\@oddfoot{}%
\let\@evenfoot\@oddfoot}
\makeatother

\usepackage{multibib}
\newcites{intro}{Introduction}
\usepackage{color}
\usepackage{amsmath}
\usepackage{bm}
\usepackage{calligra}
\usepackage{caption}
\usepackage{cprotect}
\usepackage{enumerate}
\usepackage[T1]{fontenc}
\usepackage[normalem]{ulem}
\usepackage{lineno}
\usepackage{multicol}
\usepackage{tabularx}
\usepackage[margin=1.0cm,includefoot]{geometry} 
\usepackage{bibentry}
\usepackage[colorlinks]{hyperref}
\nobibliography*
\makeatletter
\providecommand{\doi}[1]{%
\begingroup
\let\bibinfo\@secondoftwo
\urlstyle{rm}%
\href{http://dx.doi.org/#1}{%
doi:\discretionary{}{}{}%
\nolinkurl{#1}%
}%
\endgroup
}
\makeatother
\usepackage{environ}

\newenvironment{table_col}
{\par\medskip\noindent\minipage{\linewidth}}
{\endminipage\par\medskip}

\NewEnviron{equationCol}{
\begin{equation}
\BODY
\end{equation}
}
\usepackage{tikz} 
\usetikzlibrary{scopes} 
\usetikzlibrary{decorations.markings,patterns} 
\usepackage{tabu} 
\usepackage{longtable} 
\usepackage{mhchem} 
\usepackage{makecell} 
\usepackage{mwe}
\usepackage{float}
\usepackage{subcaption}
\usepackage[]{titlesec}
\usepackage[shortlabels]{enumitem}
\setlist[enumerate]{label*=\arabic*.}
\usepackage{fancyvrb}
\usepackage{fancyhdr} 
\usepackage{graphicx}
\usepackage{listings}
\lstdefinelanguage{mylang}{
alsoletter=0123456789,
alsodigit={.-}
}

\makeatletter
\newcommand\footnoteref[1]{\protected@xdef\@thefnmark{\ref{#1}}\@footnotemark}
\makeatother

\definecolor{pranab_green}{rgb}{0.31,0.53,0.10}
\definecolor{pranab_red}{rgb}{0.85,0.23,0.11}

\def\titlename{The AFLOW Library of Crystallographic Prototypes: Part 2}

\def\NUMPROTOSPARTONE{288}
\def\NUMPROTOSPARTTWO{302}
\def\NUMPROTOSTOTAL{590}

\def\AFLOW{{\small AFLOW}}
\def\QUANTUMESPRESSO{\textsc{Quantum {\small ESPRESSO}}}
\def\FHIAIMS{{\small FHI-AIMS}}
\def\ABINIT{{\small ABINIT}}

\def\VASP{{\small VASP}}

\def\CIF{{\small CIF}}
\def\POSCAR{{\small POSCAR}}

\def\citeAFLOW{\citeintro{aflowPAPER,curtarolo:art110,curtarolo:art58,curtarolo:art57,monsterPGM,curtarolo:art121,aflowPI}} 

\begin{document}

\newgeometry{left=2.25cm,right=2.25cm,top=3.0cm,bottom=3.0cm,footskip=1.5cm}

\onecolumn
\begin{frontmatter}
\title{\titlename{}}


\author[dukeGenomics,dukeMEMS]{David Hicks}
\author[nrl]{Michael J. Mehl} \ead{mmehl@.usna.edu}
\author[dukeGenomics,dukeMEMS]{Eric Gossett}
\author[dukeGenomics,dukeMEMS]{Cormac Toher}
\author[dukeGenomics,dukeMEMS,nrcn]{Ohad Levy}
\author[sto]{\\Robert M. Hanson}
\author[byu]{Gus Hart}
\author[dukeGenomics,dukeCurtarolo,FHICurtarolo]{Stefano Curtarolo}\ead{stefano@duke.edu}

\address[dukeGenomics]{ Center for Materials Genomics, Duke University, Durham, NC 27708, USA}
\address[dukeMEMS]{Department of Mechanical Engineering and Materials Science, Duke University, Durham NC 27708}
\address[nrl]{Center for Materials Physics and Technology, Code 6390, U.S. Naval Research Laboratory, Washington DC 20375}
\address[nrcn]{Department of Physics, NRCN, P.O. Box 9001, Beer-Sheva 84190, Israel}
\address[sto]{Department of Chemistry, St. Olaf College, Northfield, Minnesota 55057}
\address[byu]{Department of Physics and Astronomy, Brigham Young University, Provo UT 84602}
\address[dukeCurtarolo]{Materials Science, Electrical Engineering, Physics and Chemistry, Duke University, Durham, North Carolina 27708}
\address[FHICurtarolo]{Fritz-Haber-Institut der Max-Planck-Gesellschaft, 14195 Berlin-Dahlem, Germany}

\begin{abstract}
Materials discovery via high-throughput methods relies on the availability of structural prototypes,
which are generally decorated with varying combinations of elements to produce potential new materials.
To facilitate the automatic generation of these materials, we developed
{\it The \AFLOW\ Library of Crystallographic Prototypes} --- a collection of crystal prototypes
that can be rapidly decorated using the \AFLOW\ software.
Part 2 of this work introduces an additional \NUMPROTOSPARTTWO\ crystal
structure prototypes, including at least one from each of the 138 space groups not included in Part 1.
Combined with Part 1, the entire library consists of \NUMPROTOSTOTAL\ unique crystallographic prototypes
covering all 230 space groups.
We also present discussions of enantiomorphic space groups, Wigner-Seitz cells,
the two-dimensional plane groups, and the various different space group notations
used throughout crystallography.
All structures --- from both Part 1 and Part 2 --- are listed in the web version of the library available at
\url{http://www.aflow.org/CrystalDatabase}.
\end{abstract}

\end{frontmatter}

\twocolumn
\renewcommand{\thefootnote}{\fnsymbol{footnote}} 
\newgeometry{left=1.0cm,right=1.0cm,top=1.5cm,bottom=2.0cm,footskip=1.5cm}
{\large\textbf{Table of Contents}}
\begin{enumerate}
\vspace{-0.25cm} \item Introduction \dotfill {\hyperref[sec:intro]{\pageref{sec:intro}}} \\
\vspace{-0.75cm} \item Enantiomorphic Space Groups \dotfill {\hyperref[sec:chiral]{\pageref{sec:chiral}}} \\
\vspace{-0.75cm} \item The Wigner-Seitz Primitive Cell \dotfill {\hyperref[sec:wigner]{\pageref{sec:wsc}}} \\
\vspace{-0.75cm} \item Plane Groups (Two-Dimensional Space Groups) \dotfill {\hyperref[sec:plane]{\pageref{sec:plane}}} \\
\begin{enumerate}
\vspace{-0.65cm} \item The Parallelogram Crystal System \dotfill {\hyperref[parallel]{\pageref{parallel}}} \\
\begin{enumerate}
\vspace{-0.5cm} \item Plane Group \#1: $p1$ \dotfill {\hyperref[p1]{\pageref{p1}}} \\
\vspace{-0.5cm} \item Plane Group \#2: $p2$ \dotfill {\hyperref[p2]{\pageref{p2}}} \\
\end{enumerate}
\vspace{-0.65cm} \item The Rectangular Crystal System \dotfill {\hyperref[rect]{\pageref{rect}}} \\
\begin{enumerate}
\vspace{-0.5cm} \item Plane Group \#3: $p1m1$ \dotfill {\hyperref[p1m1]{\pageref{p1m1}}} \\
\vspace{-0.5cm} \item Plane Group \#4: $p1g1$ \dotfill {\hyperref[p1g1]{\pageref{p1g1}}} \\
\vspace{-0.5cm} \item Plane Group \#5: $c1m1$ \dotfill {\hyperref[c1m1]{\pageref{c1m1}}} \\
\vspace{-0.5cm} \item Plane Group \#6: $p2mm$ \dotfill {\hyperref[p2mm]{\pageref{p2mm}}} \\
\vspace{-0.5cm} \item Plane Group \#7: $p2mg$ \dotfill {\hyperref[p2mg]{\pageref{p2mg}}} \\
\vspace{-0.5cm} \item Plane Group \#8: $p2gg$ \dotfill {\hyperref[p2gg]{\pageref{p2gg}}} \\
\vspace{-0.5cm} \item Plane Group \#9: $c2mm$ \dotfill {\hyperref[c2mm]{\pageref{c2mm}}} \\
\end{enumerate}
\vspace{-0.65cm} \item The Square Crystal System \dotfill {\hyperref[square]{\pageref{square}}} \\
\begin{enumerate}
\vspace{-0.65cm} \item Plane Group \#10: $p4$ \dotfill {\hyperref[p4]{\pageref{p4}}} \\
\vspace{-0.5cm} \item Plane Group \#11: $p4mm$ \dotfill {\hyperref[p4mm]{\pageref{p4mm}}} \\
\vspace{-0.5cm} \item Plane Group \#12: $p4gm$ \dotfill {\hyperref[p4gm]{\pageref{p4gm}}} \\
\end{enumerate}
\vspace{-0.65cm} \item The Trigonal Crystal System \dotfill {\hyperref[trigonal]{\pageref{trigonal}}} \\
\begin{enumerate}
\vspace{-0.5cm} \item Plane Group \#13: $p3$ \dotfill {\hyperref[p3]{\pageref{p3}}} \\
\vspace{-0.5cm} \item Plane Group \#14: $p3m1$ \dotfill {\hyperref[p3m1]{\pageref{p3m1}}} \\
\vspace{-0.5cm} \item Plane Group \#15: $p31m$ \dotfill {\hyperref[p31m]{\pageref{p31m}}} \\
\end{enumerate}
\vspace{-0.65cm} \item The Hexagonal Crystal System \dotfill {\hyperref[trigonal]{\pageref{trigonal}}} \\
\begin{enumerate}
\vspace{-0.5cm} \item Plane Group \#16: $p6$ \dotfill {\hyperref[p6]{\pageref{p6}}} \\
\vspace{-0.5cm} \item Plane Group \#17: $p6mm$ \dotfill {\hyperref[p6mm]{\pageref{p6mm}}} \\
\end{enumerate}
\end{enumerate}
\vspace{-0.75cm} \item Space Group Notation \dotfill {\hyperref[sec:space_group_notation]{\pageref{sec:space_group_notation}}} \\
\vspace{-0.75cm} \item Conclusion \dotfill {\hyperref[sec:conclusion]{\pageref{sec:conclusion}}} \\
\vspace{-0.75cm} \item Acknowledgments \dotfill {\hyperref[sec:Acknowledgements]{\pageref{sec:Acknowledgements}}} \\
\vspace{-0.75cm} \item References \dotfill {\hyperref[sec:Ref]{\pageref{sec:Ref}}} \\
\end{enumerate}
\vspace{-0.5cm}
\textbf{Prototypes} \\
\noindent
\textbf{$\bm{P\bar{1}}$ (2) \dotfill} \\
\begin{enumerate}
\vspace{-0.75cm} \item H$_{2}$S: {\small A2B\_aP6\_2\_aei\_i} \dotfill {\hyperref[A2B_aP6_2_aei_i]{\pageref{A2B_aP6_2_aei_i}}} \\
\end{enumerate} \vspace{-0.75cm}
\textbf{$\bm{Pm}$ (6) \dotfill} \\
\begin{enumerate}
\vspace{-0.75cm} \item Mo$_{8}$P$_{5}$: {\small A8B5\_mP13\_6\_a7b\_3a2b} \dotfill {\hyperref[A8B5_mP13_6_a7b_3a2b]{\pageref{A8B5_mP13_6_a7b_3a2b}}} \\
\vspace{-0.75cm} \item FeNi: {\small AB\_mP4\_6\_2b\_2a} \dotfill {\hyperref[AB_mP4_6_2b_2a]{\pageref{AB_mP4_6_2b_2a}}} \\
\end{enumerate} \vspace{-0.75cm}
\textbf{$\bm{Pc}$ (7) \dotfill} \\
\begin{enumerate}
\vspace{-0.75cm} \item H$_{2}$S IV: {\small A2B\_mP12\_7\_4a\_2a} \dotfill {\hyperref[A2B_mP12_7_4a_2a]{\pageref{A2B_mP12_7_4a_2a}}} \\
\vspace{-0.75cm} \item As$_{2}$Ba: {\small A2B\_mP18\_7\_6a\_3a} \dotfill {\hyperref[A2B_mP18_7_6a_3a]{\pageref{A2B_mP18_7_6a_3a}}} \\
\vspace{-0.75cm} \item $\epsilon$-WO$_{3}$: {\small A3B\_mP16\_7\_6a\_2a} \dotfill {\hyperref[A3B_mP16_7_6a_2a]{\pageref{A3B_mP16_7_6a_2a}}} \\
\vspace{-0.75cm} \item Rh$_{2}$Ga$_{9}$: {\small A9B2\_mP22\_7\_9a\_2a} \dotfill {\hyperref[A9B2_mP22_7_9a_2a]{\pageref{A9B2_mP22_7_9a_2a}}} \\
\end{enumerate} \vspace{-0.75cm}
\textbf{$\bm{Cc}$ (9) \dotfill} \\
\begin{enumerate}
\vspace{-0.75cm} \item $\alpha$-P$_3$N$_5$: {\small A5B3\_mC32\_9\_5a\_3a} \dotfill {\hyperref[A5B3_mC32_9_5a_3a]{\pageref{A5B3_mC32_9_5a_3a}}} \\
\vspace{-0.75cm} \item H$_{3}$Cl: {\small AB3\_mC16\_9\_a\_3a} \dotfill {\hyperref[AB3_mC16_9_a_3a]{\pageref{AB3_mC16_9_a_3a}}} \\
\end{enumerate} \vspace{-0.75cm}
\textbf{$\bm{P2/m}$ (10) \dotfill} \\
\begin{enumerate}
\vspace{-0.75cm} \item $\delta$-PdCl$_{2}$: {\small A2B\_mP6\_10\_mn\_bg} \dotfill {\hyperref[A2B_mP6_10_mn_bg]{\pageref{A2B_mP6_10_mn_bg}}} \\
\vspace{-0.75cm} \item H$_{3}$Cl: {\small AB3\_mP16\_10\_mn\_3m3n} \dotfill {\hyperref[AB3_mP16_10_mn_3m3n]{\pageref{AB3_mP16_10_mn_3m3n}}} \\
\vspace{-0.75cm} \item Muthmannite: {\small ABC2\_mP8\_10\_ac\_eh\_mn} \dotfill {\hyperref[ABC2_mP8_10_ac_eh_mn]{\pageref{ABC2_mP8_10_ac_eh_mn}}} \\
\vspace{-0.75cm} \item LiSn: {\small AB\_mP6\_10\_en\_am} \dotfill {\hyperref[AB_mP6_10_en_am]{\pageref{AB_mP6_10_en_am}}} \\
\vspace{-0.75cm} \item S-carbon: {\small A\_mP8\_10\_2m2n} \dotfill {\hyperref[A_mP8_10_2m2n]{\pageref{A_mP8_10_2m2n}}} \\
\end{enumerate} \vspace{-0.75cm}
\textbf{$\bm{C2/m}$ (12) \dotfill} \\
\begin{enumerate}
\vspace{-0.75cm} \item Thortveitite: {\small A7B2C2\_mC22\_12\_aij\_h\_i} \dotfill {\hyperref[A7B2C2_mC22_12_aij_h_i]{\pageref{A7B2C2_mC22_12_aij_h_i}}} \\
\vspace{-0.75cm} \item M-carbon: {\small A\_mC16\_12\_4i} \dotfill {\hyperref[A_mC16_12_4i]{\pageref{A_mC16_12_4i}}} \\
\end{enumerate} \vspace{-0.75cm}
\textbf{$\bm{P2/c}$ (13) \dotfill} \\
\begin{enumerate}
\vspace{-0.75cm} \item H$_{2}$S: {\small A2B\_mP12\_13\_2g\_ef} \dotfill {\hyperref[A2B_mP12_13_2g_ef]{\pageref{A2B_mP12_13_2g_ef}}} \\
\end{enumerate} \vspace{-0.75cm}
\textbf{$\bm{P2_{1}/c}$ (14) \dotfill} \\
\begin{enumerate}
\vspace{-0.75cm} \item $\gamma$-PdCl$_{2}$: {\small A2B\_mP6\_14\_e\_a} \dotfill {\hyperref[A2B_mP6_14_e_a]{\pageref{A2B_mP6_14_e_a}}} \\
\vspace{-0.75cm} \item $\alpha$-Toluene: {\small A7B8\_mP120\_14\_14e\_16e} \dotfill {\hyperref[A7B8_mP120_14_14e_16e]{\pageref{A7B8_mP120_14_14e_16e}}} \\
\end{enumerate} \vspace{-0.75cm}
\textbf{$\bm{C2/c}$ (15) \dotfill} \\
\begin{enumerate}
\vspace{-0.75cm} \item H$_{3}$Cl: {\small AB3\_mC16\_15\_e\_cf} \dotfill {\hyperref[AB3_mC16_15_e_cf]{\pageref{AB3_mC16_15_e_cf}}} \\
\vspace{-0.75cm} \item H-III: {\small A\_mC24\_15\_2e2f} \dotfill {\hyperref[A_mC24_15_2e2f]{\pageref{A_mC24_15_2e2f}}} \\
\end{enumerate} \vspace{-0.75cm}
\textbf{$\bm{P222_{1}}$ (17) \dotfill} \\
\begin{enumerate}
\vspace{-0.75cm} \item $\alpha$-Naumannite: {\small A2B\_oP12\_17\_abe\_e} \dotfill {\hyperref[A2B_oP12_17_abe_e]{\pageref{A2B_oP12_17_abe_e}}} \\
\end{enumerate} \vspace{-0.75cm}
\textbf{$\bm{P2_{1}2_{1}2_{1}}$ (19) \dotfill} \\
\begin{enumerate}
\vspace{-0.75cm} \item H$_{3}$Cl: {\small AB3\_oP16\_19\_a\_3a} \dotfill {\hyperref[AB3_oP16_19_a_3a]{\pageref{AB3_oP16_19_a_3a}}} \\
\end{enumerate} \vspace{-0.75cm}
\textbf{$\bm{C222}$ (21) \dotfill} \\
\begin{enumerate}
\vspace{-0.75cm} \item Ta$_{2}$H: {\small AB2\_oC6\_21\_a\_k} \dotfill {\hyperref[AB2_oC6_21_a_k]{\pageref{AB2_oC6_21_a_k}}} \\
\end{enumerate} \vspace{-0.75cm}
\textbf{$\bm{F222}$ (22) \dotfill} \\
\begin{enumerate}
\vspace{-0.75cm} \item CeRu$_{2}$B$_{2}$: {\small A2BC2\_oF40\_22\_fi\_ad\_gh} \dotfill {\hyperref[A2BC2_oF40_22_fi_ad_gh]{\pageref{A2BC2_oF40_22_fi_ad_gh}}} \\
\vspace{-0.75cm} \item FeS: {\small AB\_oF8\_22\_a\_c} \dotfill {\hyperref[AB_oF8_22_a_c]{\pageref{AB_oF8_22_a_c}}} \\
\end{enumerate} \vspace{-0.75cm}
\textbf{$\bm{I222}$ (23) \dotfill} \\
\begin{enumerate}
\vspace{-0.75cm} \item H$_{3}$S: {\small A3B\_oI32\_23\_ij2k\_k} \dotfill {\hyperref[A3B_oI32_23_ij2k_k]{\pageref{A3B_oI32_23_ij2k_k}}} \\
\vspace{-0.75cm} \item \begin{raggedleft}Stannoidite: \end{raggedleft} \\ {\small A8B2C12D2E\_oI50\_23\_bcfk\_i\_3k\_j\_a} \dotfill {\hyperref[A8B2C12D2E_oI50_23_bcfk_i_3k_j_a]{\pageref{A8B2C12D2E_oI50_23_bcfk_i_3k_j_a}}} \\
\vspace{-0.75cm} \item NaFeS$_{2}$: {\small ABC2\_oI16\_23\_ab\_i\_k} \dotfill {\hyperref[ABC2_oI16_23_ab_i_k]{\pageref{ABC2_oI16_23_ab_i_k}}} \\
\vspace{-0.75cm} \item BPS$_{4}$: {\small ABC4\_oI12\_23\_a\_b\_k} \dotfill {\hyperref[ABC4_oI12_23_a_b_k]{\pageref{ABC4_oI12_23_a_b_k}}} \\
\end{enumerate} \vspace{-0.75cm}
\textbf{$\bm{I2_{1}2_{1}2_{1}}$ (24) \dotfill} \\
\begin{enumerate}
\vspace{-0.75cm} \item Weberite: {\small AB7CD2\_oI44\_24\_a\_b3d\_c\_ac} \dotfill {\hyperref[AB7CD2_oI44_24_a_b3d_c_ac]{\pageref{AB7CD2_oI44_24_a_b3d_c_ac}}} \\
\end{enumerate} \vspace{-0.75cm}
\textbf{$\bm{Pmc2_{1}}$ (26) \dotfill} \\
\begin{enumerate}
\vspace{-0.75cm} \item H$_{2}$S\footnote[1]{\label{note:A2B_oP12_26_abc_ab-sg}H$_{2}$S and $\beta$-SeO$_{2}$ have the same \AFLOW\ prototype label. They are generated by the same symmetry operations with different sets of parameters.}: {\small A2B\_oP12\_26\_abc\_ab} \dotfill {\hyperref[A2B_oP12_26_abc_ab-H2S]{\pageref{A2B_oP12_26_abc_ab-H2S}}} \\
\vspace{-0.75cm} \item $\beta$-SeO$_{2}$\footnoteref{note:A2B_oP12_26_abc_ab-sg}: {\small A2B\_oP12\_26\_abc\_ab} \dotfill {\hyperref[A2B_oP12_26_abc_ab-SeO2]{\pageref{A2B_oP12_26_abc_ab-SeO2}}} \\
\vspace{-0.75cm} \item TlP$_{5}$: {\small A5B\_oP24\_26\_3a3b2c\_ab} \dotfill {\hyperref[A5B_oP24_26_3a3b2c_ab]{\pageref{A5B_oP24_26_3a3b2c_ab}}} \\
\end{enumerate} \vspace{-0.75cm}
\textbf{$\bm{Pcc2}$ (27) \dotfill} \\
\begin{enumerate}
\vspace{-0.75cm} \item \begin{raggedleft}Ca$_{4}$Al$_{6}$O$_{16}$S: \end{raggedleft} \\ {\small A6B4C16D\_oP108\_27\_abcd4e\_4e\_16e\_e} \dotfill {\hyperref[A6B4C16D_oP108_27_abcd4e_4e_16e_e]{\pageref{A6B4C16D_oP108_27_abcd4e_4e_16e_e}}} \\
\end{enumerate} \vspace{-0.75cm}
\textbf{$\bm{Pca2_{1}}$ (29) \dotfill} \\
\begin{enumerate}
\vspace{-0.75cm} \item ZrO$_{2}$\footnote[3]{\label{note:AB2_oP12_29_a_2a-sg}ZrO$_{2}$ and Pyrite have similar \AFLOW\ prototype labels ({\it{i.e.}}, same symmetry and set of Wyckoff positions with different stoichiometry labels due to alphabetic ordering of atomic species). They are generated by the same symmetry operations with different sets of parameters.}: {\small A2B\_oP12\_29\_2a\_a} \dotfill {\hyperref[A2B_oP12_29_2a_a]{\pageref{A2B_oP12_29_2a_a}}} \\
\vspace{-0.75cm} \item Pyrite\footnoteref{note:AB2_oP12_29_a_2a-sg}: {\small AB2\_oP12\_29\_a\_2a} \dotfill {\hyperref[AB2_oP12_29_a_2a]{\pageref{AB2_oP12_29_a_2a}}} \\
\vspace{-0.75cm} \item Cobaltite: {\small ABC\_oP12\_29\_a\_a\_a} \dotfill {\hyperref[ABC_oP12_29_a_a_a]{\pageref{ABC_oP12_29_a_a_a}}} \\
\end{enumerate} \vspace{-0.75cm}
\textbf{$\bm{Pnc2}$ (30) \dotfill} \\
\begin{enumerate}
\vspace{-0.75cm} \item Bi$_{5}$Nb$_{3}$O$_{15}$: {\small A5B3C15\_oP46\_30\_a2c\_bc\_a7c} \dotfill {\hyperref[A5B3C15_oP46_30_a2c_bc_a7c]{\pageref{A5B3C15_oP46_30_a2c_bc_a7c}}} \\
\vspace{-0.75cm} \item CuBrSe$_{3}$: {\small ABC3\_oP20\_30\_2a\_c\_3c} \dotfill {\hyperref[ABC3_oP20_30_2a_c_3c]{\pageref{ABC3_oP20_30_2a_c_3c}}} \\
\end{enumerate} \vspace{-0.75cm}
\textbf{$\bm{Pba2}$ (32) \dotfill} \\
\begin{enumerate}
\vspace{-0.75cm} \item Re$_{2}$O$_{5}$[SO$_{4}$]$_{2}$: {\small A13B2C2\_oP34\_32\_a6c\_c\_c} \dotfill {\hyperref[A13B2C2_oP34_32_a6c_c_c]{\pageref{A13B2C2_oP34_32_a6c_c_c}}} \\
\end{enumerate} \vspace{-0.75cm}
\textbf{$\bm{Pna2_{1}}$ (33) \dotfill} \\
\begin{enumerate}
\vspace{-0.75cm} \item $\kappa$-alumina: {\small A2B3\_oP40\_33\_4a\_6a} \dotfill {\hyperref[A2B3_oP40_33_4a_6a]{\pageref{A2B3_oP40_33_4a_6a}}} \\
\end{enumerate} \vspace{-0.75cm}
\textbf{$\bm{Pnn2}$ (34) \dotfill} \\
\begin{enumerate}
\vspace{-0.75cm} \item TiAl$_{2}$Br$_{8}$: {\small A2B8C\_oP22\_34\_c\_4c\_a} \dotfill {\hyperref[A2B8C_oP22_34_c_4c_a]{\pageref{A2B8C_oP22_34_c_4c_a}}} \\
\vspace{-0.75cm} \item FeSb$_{2}$: {\small AB2\_oP6\_34\_a\_c} \dotfill {\hyperref[AB2_oP6_34_a_c]{\pageref{AB2_oP6_34_a_c}}} \\
\end{enumerate} \vspace{-0.75cm}
\textbf{$\bm{Cmm2}$ (35) \dotfill} \\
\begin{enumerate}
\vspace{-0.75cm} \item V$_{2}$MoO$_{8}$: {\small AB8C2\_oC22\_35\_a\_ab3e\_e} \dotfill {\hyperref[AB8C2_oC22_35_a_ab3e_e]{\pageref{AB8C2_oC22_35_a_ab3e_e}}} \\
\end{enumerate} \vspace{-0.75cm}
\textbf{$\bm{Cmc2_{1}}$ (36) \dotfill} \\
\begin{enumerate}
\vspace{-0.75cm} \item HCl: {\small AB\_oC8\_36\_a\_a} \dotfill {\hyperref[AB_oC8_36_a_a]{\pageref{AB_oC8_36_a_a}}} \\
\end{enumerate} \vspace{-0.75cm}
\textbf{$\bm{Ccc2}$ (37) \dotfill} \\
\begin{enumerate}
\vspace{-0.75cm} \item Li$_{2}$Si$_{2}$O$_{5}$: {\small A2B5C2\_oC36\_37\_d\_c2d\_d} \dotfill {\hyperref[A2B5C2_oC36_37_d_c2d_d]{\pageref{A2B5C2_oC36_37_d_c2d_d}}} \\
\end{enumerate} \vspace{-0.75cm}
\textbf{$\bm{Abm2}$ (39) \dotfill} \\
\begin{enumerate}
\vspace{-0.75cm} \item Ta$_{3}$S$_{2}$: {\small A2B3\_oC40\_39\_2d\_2c2d} \dotfill {\hyperref[A2B3_oC40_39_2d_2c2d]{\pageref{A2B3_oC40_39_2d_2c2d}}} \\
\vspace{-0.75cm} \item VPCl$_{9}$: {\small A9BC\_oC44\_39\_3c3d\_a\_c} \dotfill {\hyperref[A9BC_oC44_39_3c3d_a_c]{\pageref{A9BC_oC44_39_3c3d_a_c}}} \\
\end{enumerate} \vspace{-0.75cm}
\textbf{$\bm{Ama2}$ (40) \dotfill} \\
\begin{enumerate}
\vspace{-0.75cm} \item K$_{2}$CdPb: {\small AB2C\_oC16\_40\_a\_2b\_b} \dotfill {\hyperref[AB2C_oC16_40_a_2b_b]{\pageref{AB2C_oC16_40_a_2b_b}}} \\
\vspace{-0.75cm} \item CeTe$_{3}$: {\small AB3\_oC16\_40\_b\_3b} \dotfill {\hyperref[AB3_oC16_40_b_3b]{\pageref{AB3_oC16_40_b_3b}}} \\
\end{enumerate} \vspace{-0.75cm}
\textbf{$\bm{Fmm2}$ (42) \dotfill} \\
\begin{enumerate}
\vspace{-0.75cm} \item W$_{3}$O$_{10}$: {\small A10B3\_oF52\_42\_2abce\_ab} \dotfill {\hyperref[A10B3_oF52_42_2abce_ab]{\pageref{A10B3_oF52_42_2abce_ab}}} \\
\vspace{-0.75cm} \item BN: {\small AB\_oF8\_42\_a\_a} \dotfill {\hyperref[AB_oF8_42_a_a]{\pageref{AB_oF8_42_a_a}}} \\
\end{enumerate} \vspace{-0.75cm}
\textbf{$\bm{Iba2}$ (45) \dotfill} \\
\begin{enumerate}
\vspace{-0.75cm} \item MnGa$_{2}$Sb$_{2}$: {\small A2BC2\_oI20\_45\_c\_b\_c} \dotfill {\hyperref[A2BC2_oI20_45_c_b_c]{\pageref{A2BC2_oI20_45_c_b_c}}} \\
\end{enumerate} \vspace{-0.75cm}
\textbf{$\bm{Ima2}$ (46) \dotfill} \\
\begin{enumerate}
\vspace{-0.75cm} \item TiFeSi: {\small ABC\_oI36\_46\_ac\_bc\_3b} \dotfill {\hyperref[ABC_oI36_46_ac_bc_3b]{\pageref{ABC_oI36_46_ac_bc_3b}}} \\
\end{enumerate} \vspace{-0.75cm}
\textbf{$\bm{Pnnn}$ (48) \dotfill} \\
\begin{enumerate}
\vspace{-0.75cm} \item \begin{raggedleft}$\alpha$-RbPr[MoO$_{4}$]$_{2}$: \end{raggedleft} \\ {\small A2B8CD\_oP24\_48\_k\_2m\_d\_b} \dotfill {\hyperref[A2B8CD_oP24_48_k_2m_d_b]{\pageref{A2B8CD_oP24_48_k_2m_d_b}}} \\
\end{enumerate} \vspace{-0.75cm}
\textbf{$\bm{Pccm}$ (49) \dotfill} \\
\begin{enumerate}
\vspace{-0.75cm} \item $\beta$-Ta$_{2}$O$_{5}$: {\small A5B2\_oP14\_49\_dehq\_ab} \dotfill {\hyperref[A5B2_oP14_49_dehq_ab]{\pageref{A5B2_oP14_49_dehq_ab}}} \\
\vspace{-0.75cm} \item CsPr[MoO$_{4}$]$_{2}$: {\small AB2C8D\_oP24\_49\_g\_q\_2qr\_e} \dotfill {\hyperref[AB2C8D_oP24_49_g_q_2qr_e]{\pageref{AB2C8D_oP24_49_g_q_2qr_e}}} \\
\end{enumerate} \vspace{-0.75cm}
\textbf{$\bm{Pban}$ (50) \dotfill} \\
\begin{enumerate}
\vspace{-0.75cm} \item La$_{2}$NiO$_{4}$: {\small A2BC4\_oP28\_50\_ij\_ac\_ijm} \dotfill {\hyperref[A2BC4_oP28_50_ij_ac_ijm]{\pageref{A2BC4_oP28_50_ij_ac_ijm}}} \\
\vspace{-0.75cm} \item $\alpha$-Tl$_{2}$TeO$_{3}$: {\small A3BC2\_oP48\_50\_3m\_m\_2m} \dotfill {\hyperref[A3BC2_oP48_50_3m_m_2m]{\pageref{A3BC2_oP48_50_3m_m_2m}}} \\
\end{enumerate} \vspace{-0.75cm}
\textbf{$\bm{Pnna}$ (52) \dotfill} \\
\begin{enumerate}
\vspace{-0.75cm} \item GaCl$_{2}$: {\small A2B\_oP24\_52\_2e\_cd} \dotfill {\hyperref[A2B_oP24_52_2e_cd]{\pageref{A2B_oP24_52_2e_cd}}} \\
\vspace{-0.75cm} \item Sr$_{2}$Bi$_{3}$: {\small A3B2\_oP20\_52\_de\_cd} \dotfill {\hyperref[A3B2_oP20_52_de_cd]{\pageref{A3B2_oP20_52_de_cd}}} \\
\end{enumerate} \vspace{-0.75cm}
\textbf{$\bm{Pmna}$ (53) \dotfill} \\
\begin{enumerate}
\vspace{-0.75cm} \item TaNiTe$_{2}$: {\small ABC2\_oP16\_53\_h\_e\_gh} \dotfill {\hyperref[ABC2_oP16_53_h_e_gh]{\pageref{ABC2_oP16_53_h_e_gh}}} \\
\vspace{-0.75cm} \item CuBrSe$_{3}$: {\small ABC3\_oP20\_53\_e\_g\_hi} \dotfill {\hyperref[ABC3_oP20_53_e_g_hi]{\pageref{ABC3_oP20_53_e_g_hi}}} \\
\end{enumerate} \vspace{-0.75cm}
\textbf{$\bm{Pcca}$ (54) \dotfill} \\
\begin{enumerate}
\vspace{-0.75cm} \item BiGaO$_{3}$: {\small ABC3\_oP20\_54\_e\_d\_cf} \dotfill {\hyperref[ABC3_oP20_54_e_d_cf]{\pageref{ABC3_oP20_54_e_d_cf}}} \\
\end{enumerate} \vspace{-0.75cm}
\textbf{$\bm{Pbam}$ (55) \dotfill} \\
\begin{enumerate}
\vspace{-0.75cm} \item GeAs$_{2}$: {\small A2B\_oP24\_55\_2g2h\_gh} \dotfill {\hyperref[A2B_oP24_55_2g2h_gh]{\pageref{A2B_oP24_55_2g2h_gh}}} \\
\vspace{-0.75cm} \item Rh$_{5}$Ge$_{3}$: {\small A3B5\_oP16\_55\_ch\_agh} \dotfill {\hyperref[A3B5_oP16_55_ch_agh]{\pageref{A3B5_oP16_55_ch_agh}}} \\
\vspace{-0.75cm} \item R-carbon: {\small A\_oP16\_55\_2g2h} \dotfill {\hyperref[A_oP16_55_2g2h]{\pageref{A_oP16_55_2g2h}}} \\
\end{enumerate} \vspace{-0.75cm}
\textbf{$\bm{Pnnm}$ (58) \dotfill} \\
\begin{enumerate}
\vspace{-0.75cm} \item $\alpha$-PdCl$_{2}$: {\small A2B\_oP6\_58\_g\_a} \dotfill {\hyperref[A2B_oP6_58_g_a]{\pageref{A2B_oP6_58_g_a}}} \\
\end{enumerate} \vspace{-0.75cm}
\textbf{$\bm{Pmmn}$ (59) \dotfill} \\
\begin{enumerate}
\vspace{-0.75cm} \item FeOCl: {\small ABC\_oP6\_59\_a\_b\_a} \dotfill {\hyperref[ABC_oP6_59_a_b_a]{\pageref{ABC_oP6_59_a_b_a}}} \\
\end{enumerate} \vspace{-0.75cm}
\textbf{$\bm{Pbcn}$ (60) \dotfill} \\
\begin{enumerate}
\vspace{-0.75cm} \item Rh$_{2}$S$_{3}$: {\small A2B3\_oP20\_60\_d\_cd} \dotfill {\hyperref[A2B3_oP20_60_d_cd]{\pageref{A2B3_oP20_60_d_cd}}} \\
\vspace{-0.75cm} \item WO$_{3}$: {\small A3B\_oP32\_60\_3d\_d} \dotfill {\hyperref[A3B_oP32_60_3d_d]{\pageref{A3B_oP32_60_3d_d}}} \\
\vspace{-0.75cm} \item $\beta$-Toluene: {\small A7B8\_oP120\_60\_7d\_8d} \dotfill {\hyperref[A7B8_oP120_60_7d_8d]{\pageref{A7B8_oP120_60_7d_8d}}} \\
\end{enumerate} \vspace{-0.75cm}
\textbf{$\bm{Pbca}$ (61) \dotfill} \\
\begin{enumerate}
\vspace{-0.75cm} \item Benzene: {\small AB\_oP48\_61\_3c\_3c} \dotfill {\hyperref[AB_oP48_61_3c_3c]{\pageref{AB_oP48_61_3c_3c}}} \\
\end{enumerate} \vspace{-0.75cm}
\textbf{$\bm{Pnma}$ (62) \dotfill} \\
\begin{enumerate}
\vspace{-0.75cm} \item Tongbaite: {\small A2B3\_oP20\_62\_2c\_3c} \dotfill {\hyperref[A2B3_oP20_62_2c_3c]{\pageref{A2B3_oP20_62_2c_3c}}} \\
\vspace{-0.75cm} \item Forsterite: {\small A2B4C\_oP28\_62\_ac\_2cd\_c} \dotfill {\hyperref[A2B4C_oP28_62_ac_2cd_c]{\pageref{A2B4C_oP28_62_ac_2cd_c}}} \\
\vspace{-0.75cm} \item SrH$_{2}$: {\small A2B\_oP12\_62\_2c\_c} \dotfill {\hyperref[A2B_oP12_62_2c_c]{\pageref{A2B_oP12_62_2c_c}}} \\
\vspace{-0.75cm} \item $\epsilon$-NiAl$_{3}$: {\small A3B\_oP16\_62\_cd\_c} \dotfill {\hyperref[A3B_oP16_62_cd_c]{\pageref{A3B_oP16_62_cd_c}}} \\
\vspace{-0.75cm} \item Cubanite: {\small AB2C3\_oP24\_62\_c\_d\_cd} \dotfill {\hyperref[AB2C3_oP24_62_c_d_cd]{\pageref{AB2C3_oP24_62_c_d_cd}}} \\
\vspace{-0.75cm} \item Molybdite: {\small AB3\_oP16\_62\_c\_3c} \dotfill {\hyperref[AB3_oP16_62_c_3c]{\pageref{AB3_oP16_62_c_3c}}} \\
\vspace{-0.75cm} \item Barite: {\small AB4C\_oP24\_62\_c\_2cd\_c} \dotfill {\hyperref[AB4C_oP24_62_c_2cd_c]{\pageref{AB4C_oP24_62_c_2cd_c}}} \\
\vspace{-0.75cm} \item Westerveldite: {\small AB\_oP8\_62\_c\_c} \dotfill {\hyperref[AB_oP8_62_c_c]{\pageref{AB_oP8_62_c_c}}} \\
\end{enumerate} \vspace{-0.75cm}
\textbf{$\bm{Cmcm}$ (63) \dotfill} \\
\begin{enumerate}
\vspace{-0.75cm} \item Rasvumite: {\small A2BC3\_oC24\_63\_e\_c\_cg} \dotfill {\hyperref[A2BC3_oC24_63_e_c_cg]{\pageref{A2BC3_oC24_63_e_c_cg}}} \\
\vspace{-0.75cm} \item \begin{raggedleft}La$_{43}$Ni$_{17}$Mg$_{5}$: \end{raggedleft} \\ {\small A43B5C17\_oC260\_63\_c8fg6h\_cfg\_ce3f2h} \dotfill {\hyperref[A43B5C17_oC260_63_c8fg6h_cfg_ce3f2h]{\pageref{A43B5C17_oC260_63_c8fg6h_cfg_ce3f2h}}} \\
\vspace{-0.75cm} \item MnAl$_{6}$: {\small A6B\_oC28\_63\_efg\_c} \dotfill {\hyperref[A6B_oC28_63_efg_c]{\pageref{A6B_oC28_63_efg_c}}} \\
\vspace{-0.75cm} \item Post-perovskite: {\small AB3C\_oC20\_63\_a\_cf\_c} \dotfill {\hyperref[AB3C_oC20_63_a_cf_c]{\pageref{AB3C_oC20_63_a_cf_c}}} \\
\vspace{-0.75cm} \item MgSO$_{4}$: {\small AB4C\_oC24\_63\_a\_fg\_c} \dotfill {\hyperref[AB4C_oC24_63_a_fg_c]{\pageref{AB4C_oC24_63_a_fg_c}}} \\
\vspace{-0.75cm} \item Anhydrite: {\small AB4C\_oC24\_63\_c\_fg\_c} \dotfill {\hyperref[AB4C_oC24_63_c_fg_c]{\pageref{AB4C_oC24_63_c_fg_c}}} \\
\end{enumerate} \vspace{-0.75cm}
\textbf{$\bm{Cmca}$ (64) \dotfill} \\
\begin{enumerate}
\vspace{-0.75cm} \item H$_{2}$S: {\small A2B\_oC24\_64\_2f\_f} \dotfill {\hyperref[A2B_oC24_64_2f_f]{\pageref{A2B_oC24_64_2f_f}}} \\
\end{enumerate} \vspace{-0.75cm}
\textbf{$\bm{Cccm}$ (66) \dotfill} \\
\begin{enumerate}
\vspace{-0.75cm} \item SrAl$_{2}$Se$_{4}$: {\small A2B4C\_oC28\_66\_l\_kl\_a} \dotfill {\hyperref[A2B4C_oC28_66_l_kl_a]{\pageref{A2B4C_oC28_66_l_kl_a}}} \\
\vspace{-0.75cm} \item H$_{3}$S: {\small A3B\_oC64\_66\_gi2lm\_2l} \dotfill {\hyperref[A3B_oC64_66_gi2lm_2l]{\pageref{A3B_oC64_66_gi2lm_2l}}} \\
\vspace{-0.75cm} \item $\beta$-ThI$_{3}$: {\small A3B\_oC64\_66\_kl2m\_bdl} \dotfill {\hyperref[A3B_oC64_66_kl2m_bdl]{\pageref{A3B_oC64_66_kl2m_bdl}}} \\
\end{enumerate} \vspace{-0.75cm}
\textbf{$\bm{Cmma}$ (67) \dotfill} \\
\begin{enumerate}
\vspace{-0.75cm} \item Al$_{2}$CuIr\footnote[4]{\label{note:ABC2_oC16_67_b_g_ag-sg}Al$_{2}$CuIr and HoCuP$_{2}$ have similar \AFLOW\ prototype labels ({\it{i.e.}}, same symmetry and set of Wyckoff positions with different stoichiometry labels due to alphabetic ordering of atomic species). They are generated by the same symmetry operations with different sets of parameters.}: {\small A2BC\_oC16\_67\_ag\_b\_g} \dotfill {\hyperref[A2BC_oC16_67_ag_b_g]{\pageref{A2BC_oC16_67_ag_b_g}}} \\
\vspace{-0.75cm} \item HoCuP$_{2}$\footnoteref{note:ABC2_oC16_67_b_g_ag-sg}: {\small ABC2\_oC16\_67\_b\_g\_ag} \dotfill {\hyperref[ABC2_oC16_67_b_g_ag]{\pageref{ABC2_oC16_67_b_g_ag}}} \\
\vspace{-0.75cm} \item $\alpha$-FeSe\footnote[2]{\label{note:AB_oC8_67_a_g-sg}$\alpha$-FeSe and $\alpha$-PbO have the same \AFLOW\ prototype label. They are generated by the same symmetry operations with different sets of parameters.}: {\small AB\_oC8\_67\_a\_g} \dotfill {\hyperref[AB_oC8_67_a_g-FeSe]{\pageref{AB_oC8_67_a_g-FeSe}}} \\
\vspace{-0.75cm} \item $\alpha$-PbO\footnoteref{note:AB_oC8_67_a_g-sg}: {\small AB\_oC8\_67\_a\_g} \dotfill {\hyperref[AB_oC8_67_a_g-PbO]{\pageref{AB_oC8_67_a_g-PbO}}} \\
\end{enumerate} \vspace{-0.75cm}
\textbf{$\bm{Ccca}$ (68) \dotfill} \\
\begin{enumerate}
\vspace{-0.75cm} \item PdSn$_{4}$: {\small AB4\_oC20\_68\_a\_i} \dotfill {\hyperref[AB4_oC20_68_a_i]{\pageref{AB4_oC20_68_a_i}}} \\
\end{enumerate} \vspace{-0.75cm}
\textbf{$\bm{Fddd}$ (70) \dotfill} \\
\begin{enumerate}
\vspace{-0.75cm} \item Mn$_{2}$B: {\small AB2\_oF48\_70\_f\_fg} \dotfill {\hyperref[AB2_oF48_70_f_fg]{\pageref{AB2_oF48_70_f_fg}}} \\
\end{enumerate} \vspace{-0.75cm}
\textbf{$\bm{Immm}$ (71) \dotfill} \\
\begin{enumerate}
\vspace{-0.75cm} \item Ta$_{3}$B$_{4}$: {\small A4B3\_oI14\_71\_gh\_cg} \dotfill {\hyperref[A4B3_oI14_71_gh_cg]{\pageref{A4B3_oI14_71_gh_cg}}} \\
\vspace{-0.75cm} \item NbPS: {\small ABC\_oI12\_71\_h\_j\_g} \dotfill {\hyperref[ABC_oI12_71_h_j_g]{\pageref{ABC_oI12_71_h_j_g}}} \\
\end{enumerate} \vspace{-0.75cm}
\textbf{$\bm{Ibca}$ (73) \dotfill} \\
\begin{enumerate}
\vspace{-0.75cm} \item KAg[CO$_{3}$]: {\small ABCD3\_oI48\_73\_d\_e\_e\_ef} \dotfill {\hyperref[ABCD3_oI48_73_d_e_e_ef]{\pageref{ABCD3_oI48_73_d_e_e_ef}}} \\
\end{enumerate} \vspace{-0.75cm}
\textbf{$\bm{Imma}$ (74) \dotfill} \\
\begin{enumerate}
\vspace{-0.75cm} \item KHg$_{2}$: {\small A2B\_oI12\_74\_h\_e} \dotfill {\hyperref[A2B_oI12_74_h_e]{\pageref{A2B_oI12_74_h_e}}} \\
\vspace{-0.75cm} \item Al$_{4}$U: {\small A4B\_oI20\_74\_beh\_e} \dotfill {\hyperref[A4B_oI20_74_beh_e]{\pageref{A4B_oI20_74_beh_e}}} \\
\end{enumerate} \vspace{-0.75cm}
\textbf{$\bm{P4}$ (75) \dotfill} \\
\begin{enumerate}
\vspace{-0.75cm} \item \begin{raggedleft}BaCr$_{2}$Ru$_{4}$O$_{12}$: \end{raggedleft} \\ {\small AB2C12D4\_tP76\_75\_2a2b\_2d\_12d\_4d} \dotfill {\hyperref[AB2C12D4_tP76_75_2a2b_2d_12d_4d]{\pageref{AB2C12D4_tP76_75_2a2b_2d_12d_4d}}} \\
\end{enumerate} \vspace{-0.75cm}
\textbf{$\bm{P4_{1}}$ (76) \dotfill} \\
\begin{enumerate}
\vspace{-0.75cm} \item LaRhC$_{2}$: {\small A2BC\_tP16\_76\_2a\_a\_a} \dotfill {\hyperref[A2BC_tP16_76_2a_a_a]{\pageref{A2BC_tP16_76_2a_a_a}}} \\
\vspace{-0.75cm} \item Cs$_{3}$P$_{7}$: {\small A3B7\_tP40\_76\_3a\_7a} \dotfill {\hyperref[A3B7_tP40_76_3a_7a]{\pageref{A3B7_tP40_76_3a_7a}}} \\
\end{enumerate} \vspace{-0.75cm}
\textbf{$\bm{P4_{2}}$ (77) \dotfill} \\
\begin{enumerate}
\vspace{-0.75cm} \item Pinnoite: {\small A2B6CD7\_tP64\_77\_2d\_6d\_d\_ab6d} \dotfill {\hyperref[A2B6CD7_tP64_77_2d_6d_d_ab6d]{\pageref{A2B6CD7_tP64_77_2d_6d_d_ab6d}}} \\
\vspace{-0.75cm} \item H$_{2}$S III: {\small A2B\_tP48\_77\_8d\_4d} \dotfill {\hyperref[A2B_tP48_77_8d_4d]{\pageref{A2B_tP48_77_8d_4d}}} \\
\end{enumerate} \vspace{-0.75cm}
\textbf{$\bm{P4_{3}}$ (78) \dotfill} \\
\begin{enumerate}
\vspace{-0.75cm} \item Sr$_{2}$As$_{2}$O$_{7}$: {\small A2B7C2\_tP88\_78\_4a\_14a\_4a} \dotfill {\hyperref[A2B7C2_tP88_78_4a_14a_4a]{\pageref{A2B7C2_tP88_78_4a_14a_4a}}} \\
\end{enumerate} \vspace{-0.75cm}
\textbf{$\bm{I4}$ (79) \dotfill} \\
\begin{enumerate}
\vspace{-0.75cm} \item TlZn$_{2}$Sb$_{2}$: {\small A2BC2\_tI20\_79\_c\_2a\_c} \dotfill {\hyperref[A2BC2_tI20_79_c_2a_c]{\pageref{A2BC2_tI20_79_c_2a_c}}} \\
\end{enumerate} \vspace{-0.75cm}
\textbf{$\bm{I4_{1}}$ (80) \dotfill} \\
\begin{enumerate}
\vspace{-0.75cm} \item $\beta$-NbO$_{2}$: {\small AB2\_tI48\_80\_2b\_4b} \dotfill {\hyperref[AB2_tI48_80_2b_4b]{\pageref{AB2_tI48_80_2b_4b}}} \\
\end{enumerate} \vspace{-0.75cm}
\textbf{$\bm{P\bar{4}}$ (81) \dotfill} \\
\begin{enumerate}
\vspace{-0.75cm} \item GeSe$_{2}$: {\small AB2\_tP12\_81\_adg\_2h} \dotfill {\hyperref[AB2_tP12_81_adg_2h]{\pageref{AB2_tP12_81_adg_2h}}} \\
\end{enumerate} \vspace{-0.75cm}
\textbf{$\bm{I\bar{4}}$ (82) \dotfill} \\
\begin{enumerate}
\vspace{-0.75cm} \item Ni$_{3}$P: {\small A3B\_tI32\_82\_3g\_g} \dotfill {\hyperref[A3B_tI32_82_3g_g]{\pageref{A3B_tI32_82_3g_g}}} \\
\end{enumerate} \vspace{-0.75cm}
\textbf{$\bm{P4/m}$ (83) \dotfill} \\
\begin{enumerate}
\vspace{-0.75cm} \item Ti$_{2}$Ge$_{3}$: {\small A3B2\_tP10\_83\_adk\_j} \dotfill {\hyperref[A3B2_tP10_83_adk_j]{\pageref{A3B2_tP10_83_adk_j}}} \\
\end{enumerate} \vspace{-0.75cm}
\textbf{$\bm{P4/n}$ (85) \dotfill} \\
\begin{enumerate}
\vspace{-0.75cm} \item SrBr$_{2}$: {\small A2B\_tP30\_85\_ab2g\_cg} \dotfill {\hyperref[A2B_tP30_85_ab2g_cg]{\pageref{A2B_tP30_85_ab2g_cg}}} \\
\end{enumerate} \vspace{-0.75cm}
\textbf{$\bm{P4_{2}/n}$ (86) \dotfill} \\
\begin{enumerate}
\vspace{-0.75cm} \item Ti$_{3}$P: {\small AB3\_tP32\_86\_g\_3g} \dotfill {\hyperref[AB3_tP32_86_g_3g]{\pageref{AB3_tP32_86_g_3g}}} \\
\end{enumerate} \vspace{-0.75cm}
\textbf{$\bm{I4_{1}/a}$ (88) \dotfill} \\
\begin{enumerate}
\vspace{-0.75cm} \item ThCl$_{4}$: {\small A4B\_tI20\_88\_f\_a} \dotfill {\hyperref[A4B_tI20_88_f_a]{\pageref{A4B_tI20_88_f_a}}} \\
\vspace{-0.75cm} \item $\alpha$-NbO$_{2}$: {\small AB2\_tI96\_88\_2f\_4f} \dotfill {\hyperref[AB2_tI96_88_2f_4f]{\pageref{AB2_tI96_88_2f_4f}}} \\
\end{enumerate} \vspace{-0.75cm}
\textbf{$\bm{P422}$ (89) \dotfill} \\
\begin{enumerate}
\vspace{-0.75cm} \item C$_{17}$FeO$_{4}$Pt: {\small A17BC4D\_tP184\_89\_17p\_p\_4p\_io} \dotfill {\hyperref[A17BC4D_tP184_89_17p_p_4p_io]{\pageref{A17BC4D_tP184_89_17p_p_4p_io}}} \\
\end{enumerate} \vspace{-0.75cm}
\textbf{$\bm{P42_{1}2}$ (90) \dotfill} \\
\begin{enumerate}
\vspace{-0.75cm} \item \begin{raggedleft}Na$_{4}$Ti$_{2}$Si$_{8}$O$_{22}$[H$_{2}$O]$_{4}$: \end{raggedleft} \\ {\small A4B2C13D\_tP40\_90\_g\_d\_cef2g\_c} \dotfill {\hyperref[A4B2C13D_tP40_90_g_d_cef2g_c]{\pageref{A4B2C13D_tP40_90_g_d_cef2g_c}}} \\
\vspace{-0.75cm} \item \begin{raggedleft}BaCu$_{4}$[VO][PO$_{4}$]$_{4}$: \end{raggedleft} \\ {\small AB4C17D4E\_tP54\_90\_a\_g\_c4g\_g\_c} \dotfill {\hyperref[AB4C17D4E_tP54_90_a_g_c4g_g_c]{\pageref{AB4C17D4E_tP54_90_a_g_c4g_g_c}}} \\
\end{enumerate} \vspace{-0.75cm}
\textbf{$\bm{P4_{1}22}$ (91) \dotfill} \\
\begin{enumerate}
\vspace{-0.75cm} \item ThBC: {\small ABC\_tP24\_91\_d\_d\_d} \dotfill {\hyperref[ABC_tP24_91_d_d_d]{\pageref{ABC_tP24_91_d_d_d}}} \\
\end{enumerate} \vspace{-0.75cm}
\textbf{$\bm{P4_{2}22}$ (93) \dotfill} \\
\begin{enumerate}
\vspace{-0.75cm} \item \begin{raggedleft}AsPh$_{4}$CeS$_{8}$P$_{4}$Me$_{8}$: \end{raggedleft} \\ {\small AB32CD4E8\_tP184\_93\_i\_16p\_af\_2p\_4p} \dotfill {\hyperref[AB32CD4E8_tP184_93_i_16p_af_2p_4p]{\pageref{AB32CD4E8_tP184_93_i_16p_af_2p_4p}}} \\
\end{enumerate} \vspace{-0.75cm}
\textbf{$\bm{P4_{2}2_{1}2}$ (94) \dotfill} \\
\begin{enumerate}
\vspace{-0.75cm} \item Na$_{5}$Fe$_{3}$F$_{14}$: {\small A14B3C5\_tP44\_94\_c3g\_ad\_bg} \dotfill {\hyperref[A14B3C5_tP44_94_c3g_ad_bg]{\pageref{A14B3C5_tP44_94_c3g_ad_bg}}} \\
\vspace{-0.75cm} \item Li$_{2}$MoF$_{6}$: {\small A6B2C\_tP18\_94\_eg\_c\_a} \dotfill {\hyperref[A6B2C_tP18_94_eg_c_a]{\pageref{A6B2C_tP18_94_eg_c_a}}} \\
\end{enumerate} \vspace{-0.75cm}
\textbf{$\bm{P4_{3}22}$ (95) \dotfill} \\
\begin{enumerate}
\vspace{-0.75cm} \item ThBC: {\small ABC\_tP24\_95\_d\_d\_d} \dotfill {\hyperref[ABC_tP24_95_d_d_d]{\pageref{ABC_tP24_95_d_d_d}}} \\
\end{enumerate} \vspace{-0.75cm}
\textbf{$\bm{I422}$ (97) \dotfill} \\
\begin{enumerate}
\vspace{-0.75cm} \item NaGdCu$_{2}$F$_{8}$: {\small A2B8CD\_tI24\_97\_d\_k\_a\_b} \dotfill {\hyperref[A2B8CD_tI24_97_d_k_a_b]{\pageref{A2B8CD_tI24_97_d_k_a_b}}} \\
\vspace{-0.75cm} \item Ta$_{2}$Se$_{8}$I: {\small AB8C2\_tI44\_97\_e\_2k\_cd} \dotfill {\hyperref[AB8C2_tI44_97_e_2k_cd]{\pageref{AB8C2_tI44_97_e_2k_cd}}} \\
\end{enumerate} \vspace{-0.75cm}
\textbf{$\bm{I4_{1}22}$ (98) \dotfill} \\
\begin{enumerate}
\vspace{-0.75cm} \item CdAs$_{2}$: {\small A2B\_tI12\_98\_f\_a} \dotfill {\hyperref[A2B_tI12_98_f_a]{\pageref{A2B_tI12_98_f_a}}} \\
\end{enumerate} \vspace{-0.75cm}
\textbf{$\bm{P4bm}$ (100) \dotfill} \\
\begin{enumerate}
\vspace{-0.75cm} \item Fresnoite: {\small A2B8C2D\_tP26\_100\_c\_abcd\_c\_a} \dotfill {\hyperref[A2B8C2D_tP26_100_c_abcd_c_a]{\pageref{A2B8C2D_tP26_100_c_abcd_c_a}}} \\
\vspace{-0.75cm} \item Ce$_{3}$Si$_{6}$N$_{11}$: {\small A3B11C6\_tP40\_100\_ac\_bc2d\_cd} \dotfill {\hyperref[A3B11C6_tP40_100_ac_bc2d_cd]{\pageref{A3B11C6_tP40_100_ac_bc2d_cd}}} \\
\end{enumerate} \vspace{-0.75cm}
\textbf{$\bm{P4_{2}cm}$ (101) \dotfill} \\
\begin{enumerate}
\vspace{-0.75cm} \item $\gamma$-MgNiSn: {\small A7B7C2\_tP32\_101\_bde\_ade\_d} \dotfill {\hyperref[A7B7C2_tP32_101_bde_ade_d]{\pageref{A7B7C2_tP32_101_bde_ade_d}}} \\
\end{enumerate} \vspace{-0.75cm}
\textbf{$\bm{P4_{2}nm}$ (102) \dotfill} \\
\begin{enumerate}
\vspace{-0.75cm} \item Gd$_{3}$Al$_{2}$: {\small A2B3\_tP20\_102\_2c\_b2c} \dotfill {\hyperref[A2B3_tP20_102_2c_b2c]{\pageref{A2B3_tP20_102_2c_b2c}}} \\
\end{enumerate} \vspace{-0.75cm}
\textbf{$\bm{P4cc}$ (103) \dotfill} \\
\begin{enumerate}
\vspace{-0.75cm} \item NbTe$_{4}$: {\small AB4\_tP10\_103\_a\_d} \dotfill {\hyperref[AB4_tP10_103_a_d]{\pageref{AB4_tP10_103_a_d}}} \\
\end{enumerate} \vspace{-0.75cm}
\textbf{$\bm{P4nc}$ (104) \dotfill} \\
\begin{enumerate}
\vspace{-0.75cm} \item Ba$_{5}$In$_{4}$Bi$_{5}$: {\small A5B5C4\_tP28\_104\_ac\_ac\_c} \dotfill {\hyperref[A5B5C4_tP28_104_ac_ac_c]{\pageref{A5B5C4_tP28_104_ac_ac_c}}} \\
\vspace{-0.75cm} \item Tl$_{4}$HgI$_{6}$: {\small AB6C4\_tP22\_104\_a\_2ac\_c} \dotfill {\hyperref[AB6C4_tP22_104_a_2ac_c]{\pageref{AB6C4_tP22_104_a_2ac_c}}} \\
\end{enumerate} \vspace{-0.75cm}
\textbf{$\bm{P4_{2}mc}$ (105) \dotfill} \\
\begin{enumerate}
\vspace{-0.75cm} \item BaGe$_{2}$As$_{2}$: {\small A2BC2\_tP20\_105\_f\_ac\_2e} \dotfill {\hyperref[A2BC2_tP20_105_f_ac_2e]{\pageref{A2BC2_tP20_105_f_ac_2e}}} \\
\end{enumerate} \vspace{-0.75cm}
\textbf{$\bm{P4_{2}bc}$ (106) \dotfill} \\
\begin{enumerate}
\vspace{-0.75cm} \item NaZn[OH]$_{3}$: {\small A3BC3D\_tP64\_106\_3c\_c\_3c\_c} \dotfill {\hyperref[A3BC3D_tP64_106_3c_c_3c_c]{\pageref{A3BC3D_tP64_106_3c_c_3c_c}}} \\
\end{enumerate} \vspace{-0.75cm}
\textbf{$\bm{I4mm}$ (107) \dotfill} \\
\begin{enumerate}
\vspace{-0.75cm} \item Co$_{5}$Ge$_{7}$: {\small A5B7\_tI24\_107\_ac\_abd} \dotfill {\hyperref[A5B7_tI24_107_ac_abd]{\pageref{A5B7_tI24_107_ac_abd}}} \\
\vspace{-0.75cm} \item GeP: {\small AB\_tI4\_107\_a\_a} \dotfill {\hyperref[AB_tI4_107_a_a]{\pageref{AB_tI4_107_a_a}}} \\
\end{enumerate} \vspace{-0.75cm}
\textbf{$\bm{I4cm}$ (108) \dotfill} \\
\begin{enumerate}
\vspace{-0.75cm} \item Sr$_{5}$Si$_{3}$: {\small A3B5\_tI32\_108\_ac\_a2c} \dotfill {\hyperref[A3B5_tI32_108_ac_a2c]{\pageref{A3B5_tI32_108_ac_a2c}}} \\
\end{enumerate} \vspace{-0.75cm}
\textbf{$\bm{I4_{1}md}$ (109) \dotfill} \\
\begin{enumerate}
\vspace{-0.75cm} \item LaPtSi: {\small ABC\_tI12\_109\_a\_a\_a} \dotfill {\hyperref[ABC_tI12_109_a_a_a]{\pageref{ABC_tI12_109_a_a_a}}} \\
\vspace{-0.75cm} \item NbAs: {\small AB\_tI8\_109\_a\_a} \dotfill {\hyperref[AB_tI8_109_a_a]{\pageref{AB_tI8_109_a_a}}} \\
\end{enumerate} \vspace{-0.75cm}
\textbf{$\bm{I4_{1}cd}$ (110) \dotfill} \\
\begin{enumerate}
\vspace{-0.75cm} \item Be[BH$_{4}$]$_{2}$: {\small A2BC8\_tI176\_110\_2b\_b\_8b} \dotfill {\hyperref[A2BC8_tI176_110_2b_b_8b]{\pageref{A2BC8_tI176_110_2b_b_8b}}} \\
\end{enumerate} \vspace{-0.75cm}
\textbf{$\bm{P\bar{4}2m}$ (111) \dotfill} \\
\begin{enumerate}
\vspace{-0.75cm} \item MnF$_{2}$: {\small A2B\_tP12\_111\_2n\_adf} \dotfill {\hyperref[A2B_tP12_111_2n_adf]{\pageref{A2B_tP12_111_2n_adf}}} \\
\vspace{-0.75cm} \item NV: {\small AB\_tP8\_111\_n\_n} \dotfill {\hyperref[AB_tP8_111_n_n]{\pageref{AB_tP8_111_n_n}}} \\
\end{enumerate} \vspace{-0.75cm}
\textbf{$\bm{P\bar{4}2c}$ (112) \dotfill} \\
\begin{enumerate}
\vspace{-0.75cm} \item $\alpha$-CuAlCl$_{4}$: {\small AB4C\_tP12\_112\_b\_n\_e} \dotfill {\hyperref[AB4C_tP12_112_b_n_e]{\pageref{AB4C_tP12_112_b_n_e}}} \\
\end{enumerate} \vspace{-0.75cm}
\textbf{$\bm{P\bar{4}2_{1}m}$ (113) \dotfill} \\
\begin{enumerate}
\vspace{-0.75cm} \item Akermanite: {\small A2BC7D2\_tP24\_113\_e\_a\_cef\_e} \dotfill {\hyperref[A2BC7D2_tP24_113_e_a_cef_e]{\pageref{A2BC7D2_tP24_113_e_a_cef_e}}} \\
\end{enumerate} \vspace{-0.75cm}
\textbf{$\bm{P\bar{4}2_{1}c}$ (114) \dotfill} \\
\begin{enumerate}
\vspace{-0.75cm} \item SeO$_{3}$: {\small A3B\_tP32\_114\_3e\_e} \dotfill {\hyperref[A3B_tP32_114_3e_e]{\pageref{A3B_tP32_114_3e_e}}} \\
\vspace{-0.75cm} \item Pd$_{4}$Se: {\small A4B\_tP10\_114\_e\_a} \dotfill {\hyperref[A4B_tP10_114_e_a]{\pageref{A4B_tP10_114_e_a}}} \\
\end{enumerate} \vspace{-0.75cm}
\textbf{$\bm{P\bar{4}m2}$ (115) \dotfill} \\
\begin{enumerate}
\vspace{-0.75cm} \item Rh$_{3}$P$_{2}$: {\small A2B3\_tP5\_115\_g\_ag} \dotfill {\hyperref[A2B3_tP5_115_g_ag]{\pageref{A2B3_tP5_115_g_ag}}} \\
\vspace{-0.75cm} \item HgI$_{2}$: {\small AB2\_tP12\_115\_j\_egi} \dotfill {\hyperref[AB2_tP12_115_j_egi]{\pageref{AB2_tP12_115_j_egi}}} \\
\end{enumerate} \vspace{-0.75cm}
\textbf{$\bm{P\bar{4}c2}$ (116) \dotfill} \\
\begin{enumerate}
\vspace{-0.75cm} \item Ru$_{2}$Sn$_{3}$: {\small A2B3\_tP20\_116\_bci\_fj} \dotfill {\hyperref[A2B3_tP20_116_bci_fj]{\pageref{A2B3_tP20_116_bci_fj}}} \\
\end{enumerate} \vspace{-0.75cm}
\textbf{$\bm{P\bar{4}b2}$ (117) \dotfill} \\
\begin{enumerate}
\vspace{-0.75cm} \item $\beta$-Bi$_{2}$O$_{3}$: {\small A2B3\_tP20\_117\_i\_adgh} \dotfill {\hyperref[A2B3_tP20_117_i_adgh]{\pageref{A2B3_tP20_117_i_adgh}}} \\
\end{enumerate} \vspace{-0.75cm}
\textbf{$\bm{P\bar{4}n2}$ (118) \dotfill} \\
\begin{enumerate}
\vspace{-0.75cm} \item RuIn$_{3}$: {\small A3B\_tP16\_118\_ei\_f} \dotfill {\hyperref[A3B_tP16_118_ei_f]{\pageref{A3B_tP16_118_ei_f}}} \\
\vspace{-0.75cm} \item Ir$_{3}$Ga$_{5}$: {\small A5B3\_tP32\_118\_g2i\_aceh} \dotfill {\hyperref[A5B3_tP32_118_g2i_aceh]{\pageref{A5B3_tP32_118_g2i_aceh}}} \\
\end{enumerate} \vspace{-0.75cm}
\textbf{$\bm{I\bar{4}m2}$ (119) \dotfill} \\
\begin{enumerate}
\vspace{-0.75cm} \item RbGa$_{3}$: {\small A3B\_tI24\_119\_b2i\_af} \dotfill {\hyperref[A3B_tI24_119_b2i_af]{\pageref{A3B_tI24_119_b2i_af}}} \\
\vspace{-0.75cm} \item GaSb: {\small AB\_tI4\_119\_c\_a} \dotfill {\hyperref[AB_tI4_119_c_a]{\pageref{AB_tI4_119_c_a}}} \\
\end{enumerate} \vspace{-0.75cm}
\textbf{$\bm{I\bar{4}c2}$ (120) \dotfill} \\
\begin{enumerate}
\vspace{-0.75cm} \item KAu$_{4}$Sn$_{2}$: {\small A4BC2\_tI28\_120\_i\_d\_e} \dotfill {\hyperref[A4BC2_tI28_120_i_d_e]{\pageref{A4BC2_tI28_120_i_d_e}}} \\
\end{enumerate} \vspace{-0.75cm}
\textbf{$\bm{P4/mmm}$ (123) \dotfill} \\
\begin{enumerate}
\vspace{-0.75cm} \item CaRbFe$_{4}$As$_{4}$: {\small A4BC4D\_tP10\_123\_gh\_a\_i\_d} \dotfill {\hyperref[A4BC4D_tP10_123_gh_a_i_d]{\pageref{A4BC4D_tP10_123_gh_a_i_d}}} \\
\end{enumerate} \vspace{-0.75cm}
\textbf{$\bm{P4/mcc}$ (124) \dotfill} \\
\begin{enumerate}
\vspace{-0.75cm} \item Nb$_{4}$CoSi: {\small AB4C\_tP12\_124\_a\_m\_c} \dotfill {\hyperref[AB4C_tP12_124_a_m_c]{\pageref{AB4C_tP12_124_a_m_c}}} \\
\vspace{-0.75cm} \item NbTe$_{4}$: {\small AB4\_tP10\_124\_a\_m} \dotfill {\hyperref[AB4_tP10_124_a_m]{\pageref{AB4_tP10_124_a_m}}} \\
\end{enumerate} \vspace{-0.75cm}
\textbf{$\bm{P4/nbm}$ (125) \dotfill} \\
\begin{enumerate}
\vspace{-0.75cm} \item PtPb$_{4}$: {\small A4B\_tP10\_125\_m\_a} \dotfill {\hyperref[A4B_tP10_125_m_a]{\pageref{A4B_tP10_125_m_a}}} \\
\vspace{-0.75cm} \item KCeSe$_{4}$: {\small ABC4\_tP12\_125\_a\_b\_m} \dotfill {\hyperref[ABC4_tP12_125_a_b_m]{\pageref{ABC4_tP12_125_a_b_m}}} \\
\end{enumerate} \vspace{-0.75cm}
\textbf{$\bm{P4/nnc}$ (126) \dotfill} \\
\begin{enumerate}
\vspace{-0.75cm} \item BiAl$_{2}$S$_{4}$: {\small A2BC4\_tP28\_126\_cd\_e\_k} \dotfill {\hyperref[A2BC4_tP28_126_cd_e_k]{\pageref{A2BC4_tP28_126_cd_e_k}}} \\
\end{enumerate} \vspace{-0.75cm}
\textbf{$\bm{P4/mbm}$ (127) \dotfill} \\
\begin{enumerate}
\vspace{-0.75cm} \item ThB$_{4}$: {\small A4B\_tP20\_127\_ehj\_g} \dotfill {\hyperref[A4B_tP20_127_ehj_g]{\pageref{A4B_tP20_127_ehj_g}}} \\
\end{enumerate} \vspace{-0.75cm}
\textbf{$\bm{P4/mnc}$ (128) \dotfill} \\
\begin{enumerate}
\vspace{-0.75cm} \item K$_{2}$SnCl$_{6}$: {\small A6B2C\_tP18\_128\_eh\_d\_b} \dotfill {\hyperref[A6B2C_tP18_128_eh_d_b]{\pageref{A6B2C_tP18_128_eh_d_b}}} \\
\vspace{-0.75cm} \item FeCu$_{2}$Al$_{7}$: {\small A7B2C\_tP40\_128\_egi\_h\_e} \dotfill {\hyperref[A7B2C_tP40_128_egi_h_e]{\pageref{A7B2C_tP40_128_egi_h_e}}} \\
\end{enumerate} \vspace{-0.75cm}
\textbf{$\bm{P4/ncc}$ (130) \dotfill} \\
\begin{enumerate}
\vspace{-0.75cm} \item CuBi$_{2}$O$_{4}$: {\small A2BC4\_tP28\_130\_f\_c\_g} \dotfill {\hyperref[A2BC4_tP28_130_f_c_g]{\pageref{A2BC4_tP28_130_f_c_g}}} \\
\vspace{-0.75cm} \item Ba$_{5}$Si$_{3}$: {\small A5B3\_tP32\_130\_cg\_cf} \dotfill {\hyperref[A5B3_tP32_130_cg_cf]{\pageref{A5B3_tP32_130_cg_cf}}} \\
\end{enumerate} \vspace{-0.75cm}
\textbf{$\bm{P4_{2}/mcm}$ (132) \dotfill} \\
\begin{enumerate}
\vspace{-0.75cm} \item Rb$_{2}$TiCu$_{2}$S$_{4}$: {\small A2B2C4D\_tP18\_132\_e\_i\_o\_d} \dotfill {\hyperref[A2B2C4D_tP18_132_e_i_o_d]{\pageref{A2B2C4D_tP18_132_e_i_o_d}}} \\
\vspace{-0.75cm} \item AgUF$_{6}$: {\small AB6C\_tP16\_132\_d\_io\_a} \dotfill {\hyperref[AB6C_tP16_132_d_io_a]{\pageref{AB6C_tP16_132_d_io_a}}} \\
\end{enumerate} \vspace{-0.75cm}
\textbf{$\bm{P4_{2}/nbc}$ (133) \dotfill} \\
\begin{enumerate}
\vspace{-0.75cm} \item $\beta$-V$_{3}$S: {\small AB3\_tP32\_133\_h\_i2j} \dotfill {\hyperref[AB3_tP32_133_h_i2j]{\pageref{AB3_tP32_133_h_i2j}}} \\
\end{enumerate} \vspace{-0.75cm}
\textbf{$\bm{P4_{2}/mbc}$ (135) \dotfill} \\
\begin{enumerate}
\vspace{-0.75cm} \item Downeyite: {\small A2B\_tP24\_135\_gh\_h} \dotfill {\hyperref[A2B_tP24_135_gh_h]{\pageref{A2B_tP24_135_gh_h}}} \\
\vspace{-0.75cm} \item ZnSb$_{2}$O$_{4}$: {\small A4B2C\_tP28\_135\_gh\_h\_d} \dotfill {\hyperref[A4B2C_tP28_135_gh_h_d]{\pageref{A4B2C_tP28_135_gh_h_d}}} \\
\end{enumerate} \vspace{-0.75cm}
\textbf{$\bm{P4_{2}/nmc}$ (137) \dotfill} \\
\begin{enumerate}
\vspace{-0.75cm} \item Zn$_{3}$P$_{2}$: {\small A2B3\_tP40\_137\_cdf\_3g} \dotfill {\hyperref[A2B3_tP40_137_cdf_3g]{\pageref{A2B3_tP40_137_cdf_3g}}} \\
\vspace{-0.75cm} \item ZrO$_{2}$\footnote[5]{\label{note:AB2_tP6_137_a_d-sg}ZrO$_{2}$ and HgI$_{2}$ have similar \AFLOW\ prototype labels ({\it{i.e.}}, same symmetry and set of Wyckoff positions with different stoichiometry labels due to alphabetic ordering of atomic species). They are generated by the same symmetry operations with different sets of parameters.}: {\small A2B\_tP6\_137\_d\_a} \dotfill {\hyperref[A2B_tP6_137_d_a]{\pageref{A2B_tP6_137_d_a}}} \\
\vspace{-0.75cm} \item CeCo$_{4}$B$_{4}$: {\small A4BC4\_tP18\_137\_g\_b\_g} \dotfill {\hyperref[A4BC4_tP18_137_g_b_g]{\pageref{A4BC4_tP18_137_g_b_g}}} \\
\vspace{-0.75cm} \item HgI$_{2}$\footnoteref{note:AB2_tP6_137_a_d-sg}: {\small AB2\_tP6\_137\_a\_d} \dotfill {\hyperref[AB2_tP6_137_a_d]{\pageref{AB2_tP6_137_a_d}}} \\
\end{enumerate} \vspace{-0.75cm}
\textbf{$\bm{P4_{2}/ncm}$ (138) \dotfill} \\
\begin{enumerate}
\vspace{-0.75cm} \item C: {\small A\_tP12\_138\_bi} \dotfill {\hyperref[A_tP12_138_bi]{\pageref{A_tP12_138_bi}}} \\
\end{enumerate} \vspace{-0.75cm}
\textbf{$\bm{I4/mmm}$ (139) \dotfill} \\
\begin{enumerate}
\vspace{-0.75cm} \item Calomel: {\small AB\_tI8\_139\_e\_e} \dotfill {\hyperref[AB_tI8_139_e_e]{\pageref{AB_tI8_139_e_e}}} \\
\end{enumerate} \vspace{-0.75cm}
\textbf{$\bm{I4/mcm}$ (140) \dotfill} \\
\begin{enumerate}
\vspace{-0.75cm} \item W$_{5}$Si$_{3}$: {\small A3B5\_tI32\_140\_ah\_bk} \dotfill {\hyperref[A3B5_tI32_140_ah_bk]{\pageref{A3B5_tI32_140_ah_bk}}} \\
\vspace{-0.75cm} \item Cr$_{5}$B$_{3}$: {\small A3B5\_tI32\_140\_ah\_cl} \dotfill {\hyperref[A3B5_tI32_140_ah_cl]{\pageref{A3B5_tI32_140_ah_cl}}} \\
\end{enumerate} \vspace{-0.75cm}
\textbf{$\bm{I4_{1}/amd}$ (141) \dotfill} \\
\begin{enumerate}
\vspace{-0.75cm} \item $\alpha$-ThSi$_{2}$: {\small A2B\_tI12\_141\_e\_a} \dotfill {\hyperref[A2B_tI12_141_e_a]{\pageref{A2B_tI12_141_e_a}}} \\
\end{enumerate} \vspace{-0.75cm}
\textbf{$\bm{I4_{1}/acd}$ (142) \dotfill} \\
\begin{enumerate}
\vspace{-0.75cm} \item S-III: {\small A\_tI16\_142\_f} \dotfill {\hyperref[A_tI16_142_f]{\pageref{A_tI16_142_f}}} \\
\end{enumerate} \vspace{-0.75cm}
\textbf{$\bm{P3}$ (143) \dotfill} \\
\begin{enumerate}
\vspace{-0.75cm} \item Simpsonite: {\small A4B14C3\_hP21\_143\_bd\_ac4d\_d} \dotfill {\hyperref[A4B14C3_hP21_143_bd_ac4d_d]{\pageref{A4B14C3_hP21_143_bd_ac4d_d}}} \\
\vspace{-0.75cm} \item ScRh$_{6}$P$_{4}$: {\small A4B6C\_hP11\_143\_bd\_2d\_a} \dotfill {\hyperref[A4B6C_hP11_143_bd_2d_a]{\pageref{A4B6C_hP11_143_bd_2d_a}}} \\
\vspace{-0.75cm} \item MoS$_{2}$: {\small AB2\_hP12\_143\_cd\_ab2d} \dotfill {\hyperref[AB2_hP12_143_cd_ab2d]{\pageref{AB2_hP12_143_cd_ab2d}}} \\
\end{enumerate} \vspace{-0.75cm}
\textbf{$\bm{P3_{1}}$ (144) \dotfill} \\
\begin{enumerate}
\vspace{-0.75cm} \item IrGe$_{4}$: {\small A4B\_hP15\_144\_4a\_a} \dotfill {\hyperref[A4B_hP15_144_4a_a]{\pageref{A4B_hP15_144_4a_a}}} \\
\vspace{-0.75cm} \item TeZn: {\small AB\_hP6\_144\_a\_a} \dotfill {\hyperref[AB_hP6_144_a_a]{\pageref{AB_hP6_144_a_a}}} \\
\end{enumerate} \vspace{-0.75cm}
\textbf{$\bm{P3_{2}}$ (145) \dotfill} \\
\begin{enumerate}
\vspace{-0.75cm} \item \begin{raggedleft}Sheldrickite: \end{raggedleft} \\ {\small A2B3C3DE7\_hP48\_145\_2a\_3a\_3a\_a\_7a} \dotfill {\hyperref[A2B3C3DE7_hP48_145_2a_3a_3a_a_7a]{\pageref{A2B3C3DE7_hP48_145_2a_3a_3a_a_7a}}} \\
\end{enumerate} \vspace{-0.75cm}
\textbf{$\bm{R3}$ (146) \dotfill} \\
\begin{enumerate}
\vspace{-0.75cm} \item $\gamma$-Ag$_{3}$SI: {\small A3BC\_hR5\_146\_b\_a\_a} \dotfill {\hyperref[A3BC_hR5_146_b_a_a]{\pageref{A3BC_hR5_146_b_a_a}}} \\
\vspace{-0.75cm} \item FePSe$_{3}$: {\small ABC3\_hR10\_146\_2a\_2a\_2b} \dotfill {\hyperref[ABC3_hR10_146_2a_2a_2b]{\pageref{ABC3_hR10_146_2a_2a_2b}}} \\
\end{enumerate} \vspace{-0.75cm}
\textbf{$\bm{R\bar{3}}$ (148) \dotfill} \\
\begin{enumerate}
\vspace{-0.75cm} \item Phenakite: {\small A2B4C\_hR42\_148\_2f\_4f\_f} \dotfill {\hyperref[A2B4C_hR42_148_2f_4f_f]{\pageref{A2B4C_hR42_148_2f_4f_f}}} \\
\vspace{-0.75cm} \item $\beta$-PdCl$_2$: {\small A2B\_hR18\_148\_2f\_f} \dotfill {\hyperref[A2B_hR18_148_2f_f]{\pageref{A2B_hR18_148_2f_f}}} \\
\end{enumerate} \vspace{-0.75cm}
\textbf{$\bm{P312}$ (149) \dotfill} \\
\begin{enumerate}
\vspace{-0.75cm} \item Ti$_{3}$O: {\small AB3\_hP24\_149\_acgi\_3l} \dotfill {\hyperref[AB3_hP24_149_acgi_3l]{\pageref{AB3_hP24_149_acgi_3l}}} \\
\end{enumerate} \vspace{-0.75cm}
\textbf{$\bm{P3_{2}12}$ (153) \dotfill} \\
\begin{enumerate}
\vspace{-0.75cm} \item CrCl$_{3}$: {\small A3B\_hP24\_153\_3c\_2b} \dotfill {\hyperref[A3B_hP24_153_3c_2b]{\pageref{A3B_hP24_153_3c_2b}}} \\
\end{enumerate} \vspace{-0.75cm}
\textbf{$\bm{P3_{2}21}$ (154) \dotfill} \\
\begin{enumerate}
\vspace{-0.75cm} \item S-II: {\small A\_hP9\_154\_bc} \dotfill {\hyperref[A_hP9_154_bc]{\pageref{A_hP9_154_bc}}} \\
\end{enumerate} \vspace{-0.75cm}
\textbf{$\bm{P3m1}$ (156) \dotfill} \\
\begin{enumerate}
\vspace{-0.75cm} \item CdI$_{2}$: {\small AB2\_hP9\_156\_b2c\_3a2bc} \dotfill {\hyperref[AB2_hP9_156_b2c_3a2bc]{\pageref{AB2_hP9_156_b2c_3a2bc}}} \\
\vspace{-0.75cm} \item CuI: {\small AB\_hP12\_156\_2ab3c\_2ab3c} \dotfill {\hyperref[AB_hP12_156_2ab3c_2ab3c]{\pageref{AB_hP12_156_2ab3c_2ab3c}}} \\
\vspace{-0.75cm} \item $\beta$-CuI: {\small AB\_hP4\_156\_ac\_ac} \dotfill {\hyperref[AB_hP4_156_ac_ac]{\pageref{AB_hP4_156_ac_ac}}} \\
\end{enumerate} \vspace{-0.75cm}
\textbf{$\bm{P31m}$ (157) \dotfill} \\
\begin{enumerate}
\vspace{-0.75cm} \item Ag$_{5}$Pb$_{2}$O$_{6}$: {\small A5B6C2\_hP13\_157\_2ac\_2c\_b} \dotfill {\hyperref[A5B6C2_hP13_157_2ac_2c_b]{\pageref{A5B6C2_hP13_157_2ac_2c_b}}} \\
\end{enumerate} \vspace{-0.75cm}
\textbf{$\bm{P3c1}$ (158) \dotfill} \\
\begin{enumerate}
\vspace{-0.75cm} \item $\beta$-RuCl$_{3}$: {\small A3B\_hP8\_158\_d\_a} \dotfill {\hyperref[A3B_hP8_158_d_a]{\pageref{A3B_hP8_158_d_a}}} \\
\end{enumerate} \vspace{-0.75cm}
\textbf{$\bm{P31c}$ (159) \dotfill} \\
\begin{enumerate}
\vspace{-0.75cm} \item Bi$_{2}$O$_{3}$: {\small A2B3\_hP20\_159\_bc\_2c} \dotfill {\hyperref[A2B3_hP20_159_bc_2c]{\pageref{A2B3_hP20_159_bc_2c}}} \\
\vspace{-0.75cm} \item Nierite: {\small A4B3\_hP28\_159\_ab2c\_2c} \dotfill {\hyperref[A4B3_hP28_159_ab2c_2c]{\pageref{A4B3_hP28_159_ab2c_2c}}} \\
\vspace{-0.75cm} \item YbBaCo$_{4}$O$_{7}$: {\small AB4C7D\_hP26\_159\_b\_ac\_a2c\_b} \dotfill {\hyperref[AB4C7D_hP26_159_b_ac_a2c_b]{\pageref{AB4C7D_hP26_159_b_ac_a2c_b}}} \\
\end{enumerate} \vspace{-0.75cm}
\textbf{$\bm{R3m}$ (160) \dotfill} \\
\begin{enumerate}
\vspace{-0.75cm} \item H$_{3}$S: {\small A3B\_hR4\_160\_b\_a} \dotfill {\hyperref[A3B_hR4_160_b_a]{\pageref{A3B_hR4_160_b_a}}} \\
\vspace{-0.75cm} \item Al$_{8}$Cr$_{5}$: {\small A8B5\_hR26\_160\_a3bc\_a3b} \dotfill {\hyperref[A8B5_hR26_160_a3bc_a3b]{\pageref{A8B5_hR26_160_a3bc_a3b}}} \\
\vspace{-0.75cm} \item Carbonyl Sulphide: {\small ABC\_hR3\_160\_a\_a\_a} \dotfill {\hyperref[ABC_hR3_160_a_a_a]{\pageref{ABC_hR3_160_a_a_a}}} \\
\vspace{-0.75cm} \item Moissanite-15R: {\small AB\_hR10\_160\_5a\_5a} \dotfill {\hyperref[AB_hR10_160_5a_5a]{\pageref{AB_hR10_160_5a_5a}}} \\
\end{enumerate} \vspace{-0.75cm}
\textbf{$\bm{P\bar{3}m1}$ (164) \dotfill} \\
\begin{enumerate}
\vspace{-0.75cm} \item La$_{2}$O$_{3}$: {\small A2B3\_hP5\_164\_d\_ad} \dotfill {\hyperref[A2B3_hP5_164_d_ad]{\pageref{A2B3_hP5_164_d_ad}}} \\
\vspace{-0.75cm} \item $\delta_{H}^{II}$-NW$_2$: {\small AB2\_hP9\_164\_bd\_c2d} \dotfill {\hyperref[AB2_hP9_164_bd_c2d]{\pageref{AB2_hP9_164_bd_c2d}}} \\
\vspace{-0.75cm} \item CuNiSb$_{2}$: {\small ABC2\_hP4\_164\_a\_b\_d} \dotfill {\hyperref[ABC2_hP4_164_a_b_d]{\pageref{ABC2_hP4_164_a_b_d}}} \\
\end{enumerate} \vspace{-0.75cm}
\textbf{$\bm{P\bar{3}c1}$ (165) \dotfill} \\
\begin{enumerate}
\vspace{-0.75cm} \item Cu$_{3}$P: {\small A3B\_hP24\_165\_bdg\_f} \dotfill {\hyperref[A3B_hP24_165_bdg_f]{\pageref{A3B_hP24_165_bdg_f}}} \\
\end{enumerate} \vspace{-0.75cm}
\textbf{$\bm{R\bar{3}m}$ (166) \dotfill} \\
\begin{enumerate}
\vspace{-0.75cm} \item Al$_{4}$C$_{3}$: {\small A4B3\_hR7\_166\_2c\_ac} \dotfill {\hyperref[A4B3_hR7_166_2c_ac]{\pageref{A4B3_hR7_166_2c_ac}}} \\
\vspace{-0.75cm} \item SmSI: {\small ABC\_hR6\_166\_c\_c\_c} \dotfill {\hyperref[ABC_hR6_166_c_c_c]{\pageref{ABC_hR6_166_c_c_c}}} \\
\end{enumerate} \vspace{-0.75cm}
\textbf{$\bm{R\bar{3}c}$ (167) \dotfill} \\
\begin{enumerate}
\vspace{-0.75cm} \item PrNiO$_{3}$: {\small AB3C\_hR10\_167\_b\_e\_a} \dotfill {\hyperref[AB3C_hR10_167_b_e_a]{\pageref{AB3C_hR10_167_b_e_a}}} \\
\vspace{-0.75cm} \item KBO$_{2}$: {\small ABC2\_hR24\_167\_e\_e\_2e} \dotfill {\hyperref[ABC2_hR24_167_e_e_2e]{\pageref{ABC2_hR24_167_e_e_2e}}} \\
\end{enumerate} \vspace{-0.75cm}
\textbf{$\bm{P6}$ (168) \dotfill} \\
\begin{enumerate}
\vspace{-0.75cm} \item K$_{2}$Ta$_{4}$O$_{9}$F$_{4}$: {\small A2B13C4\_hP57\_168\_d\_c6d\_2d} \dotfill {\hyperref[A2B13C4_hP57_168_d_c6d_2d]{\pageref{A2B13C4_hP57_168_d_c6d_2d}}} \\
\vspace{-0.75cm} \item Al[PO$_{4}$]: {\small AB4C\_hP72\_168\_2d\_8d\_2d} \dotfill {\hyperref[AB4C_hP72_168_2d_8d_2d]{\pageref{AB4C_hP72_168_2d_8d_2d}}} \\
\end{enumerate} \vspace{-0.75cm}
\textbf{$\bm{P6_{1}}$ (169) \dotfill} \\
\begin{enumerate}
\vspace{-0.75cm} \item $\alpha$-Al$_{2}$S$_{3}$: {\small A2B3\_hP30\_169\_2a\_3a} \dotfill {\hyperref[A2B3_hP30_169_2a_3a]{\pageref{A2B3_hP30_169_2a_3a}}} \\
\end{enumerate} \vspace{-0.75cm}
\textbf{$\bm{P6_{5}}$ (170) \dotfill} \\
\begin{enumerate}
\vspace{-0.75cm} \item Al$_{2}$S$_{3}$: {\small A2B3\_hP30\_170\_2a\_3a} \dotfill {\hyperref[A2B3_hP30_170_2a_3a]{\pageref{A2B3_hP30_170_2a_3a}}} \\
\end{enumerate} \vspace{-0.75cm}
\textbf{$\bm{P6_{2}}$ (171) \dotfill} \\
\begin{enumerate}
\vspace{-0.75cm} \item Sr[S$_{2}$O$_{6}$][H$_{2}$O]$_{4}$: {\small A10B2C\_hP39\_171\_5c\_c\_a} \dotfill {\hyperref[A10B2C_hP39_171_5c_c_a]{\pageref{A10B2C_hP39_171_5c_c_a}}} \\
\end{enumerate} \vspace{-0.75cm}
\textbf{$\bm{P6_{4}}$ (172) \dotfill} \\
\begin{enumerate}
\vspace{-0.75cm} \item Sr[S$_{2}$O$_{6}$][H$_{2}$O]$_{4}$: {\small A10B2C\_hP39\_172\_5c\_c\_a} \dotfill {\hyperref[A10B2C_hP39_172_5c_c_a]{\pageref{A10B2C_hP39_172_5c_c_a}}} \\
\end{enumerate} \vspace{-0.75cm}
\textbf{$\bm{P6_{3}}$ (173) \dotfill} \\
\begin{enumerate}
\vspace{-0.75cm} \item PI$_{3}$: {\small A3B\_hP8\_173\_c\_b} \dotfill {\hyperref[A3B_hP8_173_c_b]{\pageref{A3B_hP8_173_c_b}}} \\
\vspace{-0.75cm} \item $\beta$-Si$_{3}$N$_{4}$: {\small A4B3\_hP14\_173\_bc\_c} \dotfill {\hyperref[A4B3_hP14_173_bc_c]{\pageref{A4B3_hP14_173_bc_c}}} \\
\end{enumerate} \vspace{-0.75cm}
\textbf{$\bm{P\bar{6}}$ (174) \dotfill} \\
\begin{enumerate}
\vspace{-0.75cm} \item Fe$_{12}$Zr$_{2}$P$_{7}$: {\small A12B7C2\_hP21\_174\_2j2k\_ajk\_cf} \dotfill {\hyperref[A12B7C2_hP21_174_2j2k_ajk_cf]{\pageref{A12B7C2_hP21_174_2j2k_ajk_cf}}} \\
\vspace{-0.75cm} \item GdSI: {\small ABC\_hP12\_174\_cj\_fk\_aj} \dotfill {\hyperref[ABC_hP12_174_cj_fk_aj]{\pageref{ABC_hP12_174_cj_fk_aj}}} \\
\end{enumerate} \vspace{-0.75cm}
\textbf{$\bm{P6/m}$ (175) \dotfill} \\
\begin{enumerate}
\vspace{-0.75cm} \item Nb$_{7}$Ru$_{6}$B$_{8}$: {\small A8B7C6\_hP21\_175\_ck\_aj\_k} \dotfill {\hyperref[A8B7C6_hP21_175_ck_aj_k]{\pageref{A8B7C6_hP21_175_ck_aj_k}}} \\
\vspace{-0.75cm} \item Mg[NH]: {\small ABC\_hP36\_175\_jk\_jk\_jk} \dotfill {\hyperref[ABC_hP36_175_jk_jk_jk]{\pageref{ABC_hP36_175_jk_jk_jk}}} \\
\end{enumerate} \vspace{-0.75cm}
\textbf{$\bm{P6_{3}/m}$ (176) \dotfill} \\
\begin{enumerate}
\vspace{-0.75cm} \item Er$_{3}$Ru$_{2}$: {\small A3B2\_hP10\_176\_h\_bd} \dotfill {\hyperref[A3B2_hP10_176_h_bd]{\pageref{A3B2_hP10_176_h_bd}}} \\
\vspace{-0.75cm} \item Fe$_{3}$Te$_{3}$Tl: {\small A3B3C\_hP14\_176\_h\_h\_d} \dotfill {\hyperref[A3B3C_hP14_176_h_h_d]{\pageref{A3B3C_hP14_176_h_h_d}}} \\
\vspace{-0.75cm} \item UCl$_{3}$: {\small A3B\_hP8\_176\_h\_d} \dotfill {\hyperref[A3B_hP8_176_h_d]{\pageref{A3B_hP8_176_h_d}}} \\
\end{enumerate} \vspace{-0.75cm}
\textbf{$\bm{P622}$ (177) \dotfill} \\
\begin{enumerate}
\vspace{-0.75cm} \item SiO$_{2}$: {\small A2B\_hP36\_177\_j2lm\_n} \dotfill {\hyperref[A2B_hP36_177_j2lm_n]{\pageref{A2B_hP36_177_j2lm_n}}} \\
\end{enumerate} \vspace{-0.75cm}
\textbf{$\bm{P6_{1}22}$ (178) \dotfill} \\
\begin{enumerate}
\vspace{-0.75cm} \item AuF$_{3}$: {\small AB3\_hP24\_178\_b\_ac} \dotfill {\hyperref[AB3_hP24_178_b_ac]{\pageref{AB3_hP24_178_b_ac}}} \\
\vspace{-0.75cm} \item Sc-V: {\small A\_hP6\_178\_a} \dotfill {\hyperref[A_hP6_178_a]{\pageref{A_hP6_178_a}}} \\
\end{enumerate} \vspace{-0.75cm}
\textbf{$\bm{P6_{5}22}$ (179) \dotfill} \\
\begin{enumerate}
\vspace{-0.75cm} \item AuF$_{3}$: {\small AB3\_hP24\_179\_b\_ac} \dotfill {\hyperref[AB3_hP24_179_b_ac]{\pageref{AB3_hP24_179_b_ac}}} \\
\end{enumerate} \vspace{-0.75cm}
\textbf{$\bm{P6_{4}22}$ (181) \dotfill} \\
\begin{enumerate}
\vspace{-0.75cm} \item $\beta$-SiO$_{2}$: {\small A2B\_hP9\_181\_j\_c} \dotfill {\hyperref[A2B_hP9_181_j_c]{\pageref{A2B_hP9_181_j_c}}} \\
\end{enumerate} \vspace{-0.75cm}
\textbf{$\bm{P6mm}$ (183) \dotfill} \\
\begin{enumerate}
\vspace{-0.75cm} \item AuCN: {\small ABC\_hP3\_183\_a\_a\_a} \dotfill {\hyperref[ABC_hP3_183_a_a_a]{\pageref{ABC_hP3_183_a_a_a}}} \\
\vspace{-0.75cm} \item CrFe$_{3}$NiSn$_{5}$: {\small AB\_hP6\_183\_c\_ab} \dotfill {\hyperref[AB_hP6_183_c_ab]{\pageref{AB_hP6_183_c_ab}}} \\
\end{enumerate} \vspace{-0.75cm}
\textbf{$\bm{P6cc}$ (184) \dotfill} \\
\begin{enumerate}
\vspace{-0.75cm} \item Al[PO$_{4}$]: {\small AB4C\_hP72\_184\_d\_4d\_d} \dotfill {\hyperref[AB4C_hP72_184_d_4d_d]{\pageref{AB4C_hP72_184_d_4d_d}}} \\
\end{enumerate} \vspace{-0.75cm}
\textbf{$\bm{P6_{3}cm}$ (185) \dotfill} \\
\begin{enumerate}
\vspace{-0.75cm} \item KNiCl$_{3}$: {\small A3BC\_hP30\_185\_cd\_c\_ab} \dotfill {\hyperref[A3BC_hP30_185_cd_c_ab]{\pageref{A3BC_hP30_185_cd_c_ab}}} \\
\vspace{-0.75cm} \item Cu$_{3}$P\footnote[6]{\label{note:AB3_hP24_185_c_ab2c-sg}Cu$_{3}$P and Na$_{3}$As have similar \AFLOW\ prototype labels ({\it{i.e.}}, same symmetry and set of Wyckoff positions with different stoichiometry labels due to alphabetic ordering of atomic species). They are generated by the same symmetry operations with different sets of parameters.}: {\small A3B\_hP24\_185\_ab2c\_c} \dotfill {\hyperref[A3B_hP24_185_ab2c_c]{\pageref{A3B_hP24_185_ab2c_c}}} \\
\vspace{-0.75cm} \item $\beta$-RuCl$_{3}$: {\small A3B\_hP8\_185\_c\_a} \dotfill {\hyperref[A3B_hP8_185_c_a]{\pageref{A3B_hP8_185_c_a}}} \\
\vspace{-0.75cm} \item Na$_{3}$As\footnoteref{note:AB3_hP24_185_c_ab2c-sg}: {\small AB3\_hP24\_185\_c\_ab2c} \dotfill {\hyperref[AB3_hP24_185_c_ab2c]{\pageref{AB3_hP24_185_c_ab2c}}} \\
\end{enumerate} \vspace{-0.75cm}
\textbf{$\bm{P6_{3}mc}$ (186) \dotfill} \\
\begin{enumerate}
\vspace{-0.75cm} \item Fe$_{3}$Th$_{7}$: {\small A3B7\_hP20\_186\_c\_b2c} \dotfill {\hyperref[A3B7_hP20_186_c_b2c]{\pageref{A3B7_hP20_186_c_b2c}}} \\
\end{enumerate} \vspace{-0.75cm}
\textbf{$\bm{P\bar{6}m2}$ (187) \dotfill} \\
\begin{enumerate}
\vspace{-0.75cm} \item Re$_{3}$N: {\small AB3\_hP4\_187\_e\_fh} \dotfill {\hyperref[AB3_hP4_187_e_fh]{\pageref{AB3_hP4_187_e_fh}}} \\
\end{enumerate} \vspace{-0.75cm}
\textbf{$\bm{P\bar{6}c2}$ (188) \dotfill} \\
\begin{enumerate}
\vspace{-0.75cm} \item LiScI$_{3}$: {\small A3BC\_hP10\_188\_k\_a\_e} \dotfill {\hyperref[A3BC_hP10_188_k_a_e]{\pageref{A3BC_hP10_188_k_a_e}}} \\
\vspace{-0.75cm} \item BaSi$_{4}$O$_{9}$: {\small AB9C4\_hP28\_188\_e\_kl\_ak} \dotfill {\hyperref[AB9C4_hP28_188_e_kl_ak]{\pageref{AB9C4_hP28_188_e_kl_ak}}} \\
\end{enumerate} \vspace{-0.75cm}
\textbf{$\bm{P\bar{6}2m}$ (189) \dotfill} \\
\begin{enumerate}
\vspace{-0.75cm} \item \begin{raggedleft}$\pi$-FeMg$_{3}$Al$_{8}$Si$_{6}$: \end{raggedleft} \\ {\small A8BC3D6\_hP18\_189\_bfh\_a\_g\_i} \dotfill {\hyperref[A8BC3D6_hP18_189_bfh_a_g_i]{\pageref{A8BC3D6_hP18_189_bfh_a_g_i}}} \\
\vspace{-0.75cm} \item \begin{raggedleft}$\pi$-FeMg$_{3}$Al$_{9}$Si$_{5}$: \end{raggedleft} \\ {\small A9BC3D5\_hP18\_189\_fi\_a\_g\_bh} \dotfill {\hyperref[A9BC3D5_hP18_189_fi_a_g_bh]{\pageref{A9BC3D5_hP18_189_fi_a_g_bh}}} \\
\end{enumerate} \vspace{-0.75cm}
\textbf{$\bm{P\bar{6}2c}$ (190) \dotfill} \\
\begin{enumerate}
\vspace{-0.75cm} \item Li$_{2}$Sb: {\small A2B\_hP18\_190\_gh\_bf} \dotfill {\hyperref[A2B_hP18_190_gh_bf]{\pageref{A2B_hP18_190_gh_bf}}} \\
\vspace{-0.75cm} \item $\alpha$-Sm$_{3}$Ge$_{5}$: {\small A5B3\_hP16\_190\_bdh\_g} \dotfill {\hyperref[A5B3_hP16_190_bdh_g]{\pageref{A5B3_hP16_190_bdh_g}}} \\
\vspace{-0.75cm} \item Troilite: {\small AB\_hP24\_190\_i\_afh} \dotfill {\hyperref[AB_hP24_190_i_afh]{\pageref{AB_hP24_190_i_afh}}} \\
\end{enumerate} \vspace{-0.75cm}
\textbf{$\bm{P6/mcc}$ (192) \dotfill} \\
\begin{enumerate}
\vspace{-0.75cm} \item Beryl: {\small A2B3C18D6\_hP58\_192\_c\_f\_lm\_l} \dotfill {\hyperref[A2B3C18D6_hP58_192_c_f_lm_l]{\pageref{A2B3C18D6_hP58_192_c_f_lm_l}}} \\
\vspace{-0.75cm} \item AlPO$_{4}$: {\small AB2\_hP72\_192\_m\_j2kl} \dotfill {\hyperref[AB2_hP72_192_m_j2kl]{\pageref{AB2_hP72_192_m_j2kl}}} \\
\end{enumerate} \vspace{-0.75cm}
\textbf{$\bm{P6_{3}/mcm}$ (193) \dotfill} \\
\begin{enumerate}
\vspace{-0.75cm} \item Mavlyanovite: {\small A5B3\_hP16\_193\_dg\_g} \dotfill {\hyperref[A5B3_hP16_193_dg_g]{\pageref{A5B3_hP16_193_dg_g}}} \\
\end{enumerate} \vspace{-0.75cm}
\textbf{$\bm{P6_{3}/mmc}$ (194) \dotfill} \\
\begin{enumerate}
\vspace{-0.75cm} \item Ni$_{3}$Ti: {\small A3B\_hP16\_194\_gh\_ac} \dotfill {\hyperref[A3B_hP16_194_gh_ac]{\pageref{A3B_hP16_194_gh_ac}}} \\
\vspace{-0.75cm} \item Co$_{2}$Al$_{5}$: {\small A5B2\_hP28\_194\_ahk\_ch} \dotfill {\hyperref[A5B2_hP28_194_ahk_ch]{\pageref{A5B2_hP28_194_ahk_ch}}} \\
\vspace{-0.75cm} \item Al$_{9}$Mn$_{3}$Si: {\small A9B3C\_hP26\_194\_hk\_h\_a} \dotfill {\hyperref[A9B3C_hP26_194_hk_h_a]{\pageref{A9B3C_hP26_194_hk_h_a}}} \\
\end{enumerate} \vspace{-0.75cm}
\textbf{$\bm{P23}$ (195) \dotfill} \\
\begin{enumerate}
\vspace{-0.75cm} \item PrRu$_{4}$P$_{12}$: {\small A12BC4\_cP34\_195\_2j\_ab\_2e} \dotfill {\hyperref[A12BC4_cP34_195_2j_ab_2e]{\pageref{A12BC4_cP34_195_2j_ab_2e}}} \\
\end{enumerate} \vspace{-0.75cm}
\textbf{$\bm{F23}$ (196) \dotfill} \\
\begin{enumerate}
\vspace{-0.75cm} \item Cu$_{2}$Fe[CN]$_{6}$: {\small A12B2C\_cF60\_196\_h\_bc\_a} \dotfill {\hyperref[A12B2C_cF60_196_h_bc_a]{\pageref{A12B2C_cF60_196_h_bc_a}}} \\
\vspace{-0.75cm} \item \begin{raggedleft}MgB$_{12}$H$_{12}$[H$_{2}$O]$_{12}$: \end{raggedleft} \\ {\small A12B36CD12\_cF488\_196\_2h\_6h\_ac\_fgh} \dotfill {\hyperref[A12B36CD12_cF488_196_2h_6h_ac_fgh]{\pageref{A12B36CD12_cF488_196_2h_6h_ac_fgh}}} \\
\end{enumerate} \vspace{-0.75cm}
\textbf{$\bm{P2_{1}3}$ (198) \dotfill} \\
\begin{enumerate}
\vspace{-0.75cm} \item Sodium Chlorate: {\small ABC3\_cP20\_198\_a\_a\_b} \dotfill {\hyperref[ABC3_cP20_198_a_a_b]{\pageref{ABC3_cP20_198_a_a_b}}} \\
\end{enumerate} \vspace{-0.75cm}
\textbf{$\bm{Pm\bar{3}}$ (200) \dotfill} \\
\begin{enumerate}
\vspace{-0.75cm} \item Mg$_{2}$Zn$_{11}$: {\small A2B11\_cP39\_200\_f\_aghij} \dotfill {\hyperref[A2B11_cP39_200_f_aghij]{\pageref{A2B11_cP39_200_f_aghij}}} \\
\end{enumerate} \vspace{-0.75cm}
\textbf{$\bm{Pn\bar{3}}$ (201) \dotfill} \\
\begin{enumerate}
\vspace{-0.75cm} \item KSbO$_{3}$: {\small AB3C\_cP60\_201\_ce\_fh\_g} \dotfill {\hyperref[AB3C_cP60_201_ce_fh_g]{\pageref{AB3C_cP60_201_ce_fh_g}}} \\
\end{enumerate} \vspace{-0.75cm}
\textbf{$\bm{Fm\bar{3}}$ (202) \dotfill} \\
\begin{enumerate}
\vspace{-0.75cm} \item KB$_{6}$H$_{6}$: {\small A6B6C\_cF104\_202\_h\_h\_c} \dotfill {\hyperref[A6B6C_cF104_202_h_h_c]{\pageref{A6B6C_cF104_202_h_h_c}}} \\
\vspace{-0.75cm} \item \begin{raggedleft}FCC C$_{60}$ Buckminsterfullerine: \end{raggedleft} \\ {\small A\_cF240\_202\_h2i} \dotfill {\hyperref[A_cF240_202_h2i]{\pageref{A_cF240_202_h2i}}} \\
\end{enumerate} \vspace{-0.75cm}
\textbf{$\bm{Fd\bar{3}}$ (203) \dotfill} \\
\begin{enumerate}
\vspace{-0.75cm} \item Pyrochlore: {\small A2BCD3E6\_cF208\_203\_e\_c\_d\_f\_g} \dotfill {\hyperref[A2BCD3E6_cF208_203_e_c_d_f_g]{\pageref{A2BCD3E6_cF208_203_e_c_d_f_g}}} \\
\vspace{-0.75cm} \item Tychite: {\small A4B2C6D16E\_cF232\_203\_e\_d\_f\_eg\_a} \dotfill {\hyperref[A4B2C6D16E_cF232_203_e_d_f_eg_a]{\pageref{A4B2C6D16E_cF232_203_e_d_f_eg_a}}} \\
\vspace{-0.75cm} \item Rb$_{3}$AsSe$_{16}$: {\small AB3C16\_cF160\_203\_b\_ad\_eg} \dotfill {\hyperref[AB3C16_cF160_203_b_ad_eg]{\pageref{AB3C16_cF160_203_b_ad_eg}}} \\
\end{enumerate} \vspace{-0.75cm}
\textbf{$\bm{Pa\bar{3}}$ (205) \dotfill} \\
\begin{enumerate}
\vspace{-0.75cm} \item Ca$_{3}$Al$_{2}$O$_{6}$: {\small A2B3C6\_cP264\_205\_2d\_ab2c2d\_6d} \dotfill {\hyperref[A2B3C6_cP264_205_2d_ab2c2d_6d]{\pageref{A2B3C6_cP264_205_2d_ab2c2d_6d}}} \\
\vspace{-0.75cm} \item \begin{raggedleft}Simple Cubic C$_{60}$ Buckminsterfullerine: \end{raggedleft} \\ {\small A\_cP240\_205\_10d} \dotfill {\hyperref[A_cP240_205_10d]{\pageref{A_cP240_205_10d}}} \\
\end{enumerate} \vspace{-0.75cm}
\textbf{$\bm{Ia\bar{3}}$ (206) \dotfill} \\
\begin{enumerate}
\vspace{-0.75cm} \item AlLi$_{3}$N$_{2}$: {\small AB3C2\_cI96\_206\_c\_e\_ad} \dotfill {\hyperref[AB3C2_cI96_206_c_e_ad]{\pageref{AB3C2_cI96_206_c_e_ad}}} \\
\end{enumerate} \vspace{-0.75cm}
\textbf{$\bm{P432}$ (207) \dotfill} \\
\begin{enumerate}
\vspace{-0.75cm} \item Pd$_{17}$Se$_{15}$: {\small A17B15\_cP64\_207\_acfk\_eij} \dotfill {\hyperref[A17B15_cP64_207_acfk_eij]{\pageref{A17B15_cP64_207_acfk_eij}}} \\
\end{enumerate} \vspace{-0.75cm}
\textbf{$\bm{P4_{2}32}$ (208) \dotfill} \\
\begin{enumerate}
\vspace{-0.75cm} \item PH$_{3}$: {\small A3B\_cP16\_208\_j\_b} \dotfill {\hyperref[A3B_cP16_208_j_b]{\pageref{A3B_cP16_208_j_b}}} \\
\vspace{-0.75cm} \item \begin{raggedleft}Cs$_{2}$ZnFe[CN]$_{6}$: \end{raggedleft} \\ {\small A6B2CD6E\_cP64\_208\_m\_ad\_b\_m\_c} \dotfill {\hyperref[A6B2CD6E_cP64_208_m_ad_b_m_c]{\pageref{A6B2CD6E_cP64_208_m_ad_b_m_c}}} \\
\end{enumerate} \vspace{-0.75cm}
\textbf{$\bm{F432}$ (209) \dotfill} \\
\begin{enumerate}
\vspace{-0.75cm} \item F$_{6}$KP: {\small A24BC\_cF104\_209\_j\_a\_b} \dotfill {\hyperref[A24BC_cF104_209_j_a_b]{\pageref{A24BC_cF104_209_j_a_b}}} \\
\end{enumerate} \vspace{-0.75cm}
\textbf{$\bm{F4_{1}32}$ (210) \dotfill} \\
\begin{enumerate}
\vspace{-0.75cm} \item Te[OH]$_{6}$: {\small A12B6C\_cF608\_210\_4h\_2h\_e} \dotfill {\hyperref[A12B6C_cF608_210_4h_2h_e]{\pageref{A12B6C_cF608_210_4h_2h_e}}} \\
\end{enumerate} \vspace{-0.75cm}
\textbf{$\bm{I432}$ (211) \dotfill} \\
\begin{enumerate}
\vspace{-0.75cm} \item SiO$_{2}$: {\small A2B\_cI72\_211\_hi\_i} \dotfill {\hyperref[A2B_cI72_211_hi_i]{\pageref{A2B_cI72_211_hi_i}}} \\
\end{enumerate} \vspace{-0.75cm}
\textbf{$\bm{P4_{3}32}$ (212) \dotfill} \\
\begin{enumerate}
\vspace{-0.75cm} \item SrSi$_{2}$: {\small A2B\_cP12\_212\_c\_a} \dotfill {\hyperref[A2B_cP12_212_c_a]{\pageref{A2B_cP12_212_c_a}}} \\
\end{enumerate} \vspace{-0.75cm}
\textbf{$\bm{I4_{1}32}$ (214) \dotfill} \\
\begin{enumerate}
\vspace{-0.75cm} \item Ca$_{3}$PI$_{3}$: {\small A3B3C\_cI56\_214\_g\_h\_a} \dotfill {\hyperref[A3B3C_cI56_214_g_h_a]{\pageref{A3B3C_cI56_214_g_h_a}}} \\
\vspace{-0.75cm} \item Petzite: {\small A3BC2\_cI48\_214\_f\_a\_e} \dotfill {\hyperref[A3BC2_cI48_214_f_a_e]{\pageref{A3BC2_cI48_214_f_a_e}}} \\
\end{enumerate} \vspace{-0.75cm}
\textbf{$\bm{P\bar{4}3m}$ (215) \dotfill} \\
\begin{enumerate}
\vspace{-0.75cm} \item $\gamma$-brass: {\small A4B9\_cP52\_215\_ei\_3efgi} \dotfill {\hyperref[A4B9_cP52_215_ei_3efgi]{\pageref{A4B9_cP52_215_ei_3efgi}}} \\
\end{enumerate} \vspace{-0.75cm}
\textbf{$\bm{F\bar{4}3m}$ (216) \dotfill} \\
\begin{enumerate}
\vspace{-0.75cm} \item \begin{raggedleft}Quartenary Heusler: \end{raggedleft} \\ {\small ABCD\_cF16\_216\_c\_d\_b\_a} \dotfill {\hyperref[ABCD_cF16_216_c_d_b_a]{\pageref{ABCD_cF16_216_c_d_b_a}}} \\
\end{enumerate} \vspace{-0.75cm}
\textbf{$\bm{P\bar{4}3n}$ (218) \dotfill} \\
\begin{enumerate}
\vspace{-0.75cm} \item Ag$_{3}$[PO$_{4}$]: {\small A3B4C\_cP16\_218\_c\_e\_a} \dotfill {\hyperref[A3B4C_cP16_218_c_e_a]{\pageref{A3B4C_cP16_218_c_e_a}}} \\
\end{enumerate} \vspace{-0.75cm}
\textbf{$\bm{F\bar{4}3c}$ (219) \dotfill} \\
\begin{enumerate}
\vspace{-0.75cm} \item Boracite: {\small A7BC3D13\_cF192\_219\_de\_b\_c\_ah} \dotfill {\hyperref[A7BC3D13_cF192_219_de_b_c_ah]{\pageref{A7BC3D13_cF192_219_de_b_c_ah}}} \\
\end{enumerate} \vspace{-0.75cm}
\textbf{$\bm{I\bar{4}3d}$ (220) \dotfill} \\
\begin{enumerate}
\vspace{-0.75cm} \item Cu$_{15}$Si$_{4}$: {\small A15B4\_cI76\_220\_ae\_c} \dotfill {\hyperref[A15B4_cI76_220_ae_c]{\pageref{A15B4_cI76_220_ae_c}}} \\
\vspace{-0.75cm} \item Th$_{3}$P$_{4}$: {\small A4B3\_cI28\_220\_c\_a} \dotfill {\hyperref[A4B3_cI28_220_c_a]{\pageref{A4B3_cI28_220_c_a}}} \\
\end{enumerate} \vspace{-0.75cm}
\textbf{$\bm{Pm\bar{3}m}$ (221) \dotfill} \\
\begin{enumerate}
\vspace{-0.75cm} \item Ca$_{3}$Al$_{2}$O$_{6}$: {\small A2B3C6\_cP33\_221\_cd\_ag\_fh} \dotfill {\hyperref[A2B3C6_cP33_221_cd_ag_fh]{\pageref{A2B3C6_cP33_221_cd_ag_fh}}} \\
\end{enumerate} \vspace{-0.75cm}
\textbf{$\bm{Pn\bar{3}n}$ (222) \dotfill} \\
\begin{enumerate}
\vspace{-0.75cm} \item Ce$_{5}$Mo$_{3}$O$_{16}$: {\small A5B3C16\_cP96\_222\_ce\_d\_fi} \dotfill {\hyperref[A5B3C16_cP96_222_ce_d_fi]{\pageref{A5B3C16_cP96_222_ce_d_fi}}} \\
\end{enumerate} \vspace{-0.75cm}
\textbf{$\bm{Fm\bar{3}m}$ (225) \dotfill} \\
\begin{enumerate}
\vspace{-0.75cm} \item Th$_{6}$Mn$_{23}$: {\small A23B6\_cF116\_225\_bd2f\_e} \dotfill {\hyperref[A23B6_cF116_225_bd2f_e]{\pageref{A23B6_cF116_225_bd2f_e}}} \\
\vspace{-0.75cm} \item K$_{2}$PtCl$_{6}$: {\small A6B2C\_cF36\_225\_e\_c\_a} \dotfill {\hyperref[A6B2C_cF36_225_e_c_a]{\pageref{A6B2C_cF36_225_e_c_a}}} \\
\end{enumerate} \vspace{-0.75cm}
\textbf{$\bm{Fm\bar{3}c}$ (226) \dotfill} \\
\begin{enumerate}
\vspace{-0.75cm} \item NaZn$_{13}$: {\small AB13\_cF112\_226\_a\_bi} \dotfill {\hyperref[AB13_cF112_226_a_bi]{\pageref{AB13_cF112_226_a_bi}}} \\
\end{enumerate} \vspace{-0.75cm}
\textbf{$\bm{Fd\bar{3}m}$ (227) \dotfill} \\
\begin{enumerate}
\vspace{-0.75cm} \item \begin{raggedleft}Pyrochlore Iridate: \end{raggedleft} \\ {\small A2B2C7\_cF88\_227\_c\_d\_af} \dotfill {\hyperref[A2B2C7_cF88_227_c_d_af]{\pageref{A2B2C7_cF88_227_c_d_af}}} \\
\vspace{-0.75cm} \item Spinel: {\small A3B4\_cF56\_227\_ad\_e} \dotfill {\hyperref[A3B4_cF56_227_ad_e]{\pageref{A3B4_cF56_227_ad_e}}} \\
\end{enumerate} \vspace{-0.75cm}
\textbf{$\bm{Fd\bar{3}c}$ (228) \dotfill} \\
\begin{enumerate}
\vspace{-0.75cm} \item CuCrCl$_{5}$[NH$_{3}$]$_{6}$: {\small A5BCD6\_cF416\_228\_eg\_c\_b\_h} \dotfill {\hyperref[A5BCD6_cF416_228_eg_c_b_h]{\pageref{A5BCD6_cF416_228_eg_c_b_h}}} \\
\vspace{-0.75cm} \item TeO$_{6}$H$_{6}$: {\small A6B\_cF224\_228\_h\_c} \dotfill {\hyperref[A6B_cF224_228_h_c]{\pageref{A6B_cF224_228_h_c}}} \\
\end{enumerate} \vspace{-0.75cm}
\textbf{$\bm{Im\bar{3}m}$ (229) \dotfill} \\
\begin{enumerate}
\vspace{-0.75cm} \item $\gamma$-brass: {\small A3B10\_cI52\_229\_e\_fh} \dotfill {\hyperref[A3B10_cI52_229_e_fh]{\pageref{A3B10_cI52_229_e_fh}}} \\
\vspace{-0.75cm} \item $\beta$-Hg$_{4}$Pt: {\small A4B\_cI10\_229\_c\_a} \dotfill {\hyperref[A4B_cI10_229_c_a]{\pageref{A4B_cI10_229_c_a}}} \\
\vspace{-0.75cm} \item Ir$_{3}$Ge$_{7}$: {\small A7B3\_cI40\_229\_df\_e} \dotfill {\hyperref[A7B3_cI40_229_df_e]{\pageref{A7B3_cI40_229_df_e}}} \\
\end{enumerate} \vspace{-0.75cm}
\textbf{$\bm{Ia\bar{3}d}$ (230) \dotfill} \\
\begin{enumerate}
\vspace{-0.75cm} \item Garnet: {\small A2B3C12D3\_cI160\_230\_a\_c\_h\_d} \dotfill {\hyperref[A2B3C12D3_cI160_230_a_c_h_d]{\pageref{A2B3C12D3_cI160_230_a_c_h_d}}} \\
\end{enumerate}
\vspace{-0.5cm}
\textbf{Index}
\vspace{-0.3cm}
\begin{enumerate}
\item Prototype Index \dotfill {\hyperref[sec:protoInd]{\pageref{sec:protoInd}}} \\
\vspace{-0.75cm} \item Pearson Symbol Index \dotfill {\hyperref[sec:pearsonInd]{\pageref{sec:pearsonInd}}} \\
\vspace{-0.75cm} \item Strukturbericht Designation Index \dotfill {\hyperref[sec:strukInd]{\pageref{sec:strukInd}}} \\
\vspace{-0.75cm} \item Duplicate AFLOW Label \dotfill {\hyperref[sec:dupInd]{\pageref{sec:dupInd}}} \\
\vspace{-0.75cm} \item Similar AFLOW Label \dotfill {\hyperref[sec:simInd]{\pageref{sec:simInd}}} \\
\vspace{-0.75cm} \item \CIF\ Index \dotfill {\hyperref[sec:cifInd]{\pageref{sec:cifInd}}} \\
\vspace{-0.75cm} \item \POSCAR\ Index \dotfill {\hyperref[sec:poscarInd]{\pageref{sec:poscarInd}}} \\
\end{enumerate} \vspace{-0.5cm}
\renewcommand{\thefootnote}{\arabic{footnote}}

\renewcommand{\arraystretch}{1.25}
\section{\label{sec:intro} Introduction}

The advent of high-throughput computing has paved the way for
efficient and rapid exploration of materials space.
Exploiting such computational frameworks has facilitated the
predictions of novel materials, including
metallic glasses~\citeintro{curtarolo:art112},
high-entropy oxides~\citeintro{curtarolo:art99},
candidate photovoltaic absorbers~\citeintro{Yu_photovoltaics_2012},
rechargeable battery materials~\citeintro{CederMRSB2011},
and magnetic Heuslers~\citeintro{curtarolo:art109}.
Equipped with automated symmetry~\citeintro{Hicks_aflowsym_2018},
mechanical~\citeintro{curtarolo:art115,curtarolo:art96},
and thermal analyses~\citeintro{Nath_QHA_2016,Nath_QHA_thermal_conductivity_2016,curtarolo:art125},
a panoply of material properties can be readily calculated and stored in material-property repositories 
--- such as \AFLOW~\citeAFLOW,
\underline{No}vel \underline{Ma}terials \underline{D}iscovery (NoMaD)~\citeintro{nomad},
Materials Project~\citeintro{materialsproject.org},
and the \underline{O}pen \underline{Q}uantum \underline{M}aterials \underline{D}atabase (OQMD)~\citeintro{oqmd.org}.
Creating new materials in these computational databases generally entails decorating
structural prototypes with various atomic species.
To increase the likelihood of synthesizing the conjectured materials, it is advisable to
explore previously studied/observed crystal structures as the basis for developing prototype structures.

A variety of resources, such as the {\it Strukturbericht} series~\citeintro{ewald43:struk},
{\it The Structure of Crystals}~\citeintro{wyckoff63:structures},
{\it Pearson's Handbook}~\citeintro{pearson58:handbook},
and {\it The American Mineralogist Crystal Database}~\citeintro{downs03:amsdb},
have cataloged crystal structures over the past century
(for a succinct historical recount of crystal structure information, see Ref.~\citeintro{curtarolo:art121}).
These collections are invaluable for finding crystals of certain symmetries, compositions, {\it Strukturbericht}
designations, and other structural classifications.
Furthermore, they are prime sources for structural prototypes in computational frameworks.
However, until recently, much of the data was not easily accessible to the computational materials science community,
hindering automatic generation of these prototypes.

A combined effort involving the U.S. Naval Research Laboratory's {\it Crystal Lattice Structures} web page
and the \AFLOW\ consortium yielded the construction of a new online crystal structure database:
{\it The \AFLOW\ Library of Crystallographic Prototypes}.
The library is the result of a synergistic effort to gather crystal structure prototypes from literature
and integrate them into the \AFLOW\ computational framework for automatic generation.
The first stage of this venture, Part 1, introduces \NUMPROTOSPARTONE\ crystallographic prototypes from 92 different space groups~\citeintro{curtarolo:art121}.
Each prototype entry contains its symmetry descriptions,
lattice and atomic basis vector equations, elements/compounds exhibiting the structure,
and citations to the original references.
With the \AFLOW\ software, geometry files of the structure can be created in
common {\it ab-initio} code formats, {\it i.e.},
\VASP~\citeintro{kresse_vasp,VASP4_2,vasp_cms1996,vasp_prb1996},
\QUANTUMESPRESSO~\citeintro{quantum_espresso_2009},
\FHIAIMS~\citeintro{Blum_CPC2009_AIM},
and \ABINIT~\citeintro{gonze:abinit}.
Additionally, the infrastructure allows users to tune the internal degree(s) of freedom (lattice
and Wyckoff parameters) and alter the atomic species of the structure, providing a robust crystal prototyping tool.
The content associated with each prototype entry is detailed in Part 1~\citeintro{curtarolo:art121}.

The prototype information is also available online at the following URL: \url{http://www.aflow.org/CrystalDatabase}.
Along with the structural information listed in the article, the website features additional functionality.
An interactive Jmol applet is shown on each prototype entry page, allowing multiple viewing
perspectives and differing cell representations (conventional, primitive, supercell, and Wigner-Seitz).
Each page is also accompanied by a prototype generator that interfaces with the \AFLOW\
software.

This article presents the continued work and second installment of the \AFLOW\ Library of
Crystallographic Prototypes.
In Part 2, the crystallographic library is extended by \NUMPROTOSPARTTWO\ structure prototypes with
representatives from the remaining 138 space groups not included in Part 1.
The online version of the library contains all of the prototypes from Part 1 and Part 2.

The outline of this article is as follows:
Section 2 highlights the enantiomorphic space groups.
Section 3 discusses the Wigner-Seitz cell and showcases the Jmol functionality.
Section 4 introduces the two-dimensional plane groups (or ``wallpaper'' groups).
Section 5 describes the different space group symbols, including the Hermann-Mauguin, Hall, International, and Sch{\"o}nflies
notations, along with the origin/setting choices used throughout the library.

\section{\label{sec:chiral} Enantiomorphic Space Groups}

If we look at space group $P4_1 (\#76)$, we see that it has one
Wyckoff position (4a), with
operations~\citeintro{Aroyo_Bulgar_Chem_Comm_43_183_2011}
\begin{displaymath}
\left(x, y, z\right) ~ \left( - x, - y, z + \frac{1}{2}\right) ~ \left( - y, x, z + \frac{1}{4} \right) ~
\left(y, - x, z + \frac{3}{4} \right) ~ .
\end{displaymath}
If we then look at space group $P4_3 (\#78)$, we find it also has one
(4a) Wyckoff position, with operations
\begin{displaymath}
\left(x, y, z\right) ~ \left( - x, - y, z + \frac{1}{2} \right) ~ \left( - y, x, z + \frac{3}{4} \right) ~
\left(y, - x, z + \frac{1}{4} \right) ~ ,
\end{displaymath}
where the only difference is that the $1/4$ and $3/4$ fractions have
swapped positions. We can easily show that space group \#78 is a
mirror reflection of \#76 in the $z = 0$ plane.

\begin{figure}
\begin{center}
\includegraphics[width=\linewidth]{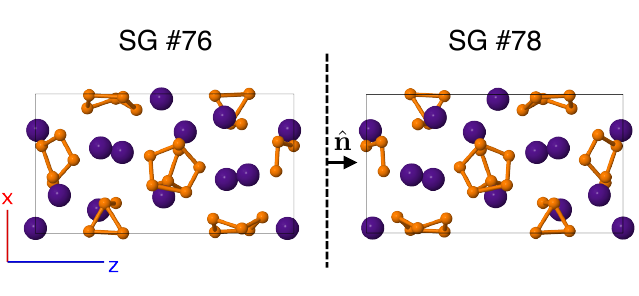}
\end{center}
\caption{\label{fig:Cs3P7}\textbf{Illustration of enantiomorphic structures.}
Cs$_3$P$_7$ in space group $P4_1$ (\#76)
(left), and reflected through the $z = 0$ plane into space group
$P4_3$ (\#78) (right). The positive $z$ direction is to the right
in both figures, with the mirror plane perpendicular to the page.
The figures were produced by
Jmol~\protect\citeintro{Jmol}.
}
\end{figure}

To see this more clearly, consider the Cs$_3$P$_7$ structure
(A3B7\_tP40\_76\_3a\_7a\footnote{This structure can be found in the
Library of Crystallographic Prototypes at
\href{http://aflow.org/CrystalDatabase/A3B7\_tP40\_76\_3a\_7a.html}{http://aflow.org/CrystalDatabase/A3B7\_tP40\_76\_3a\_7a.html.}}).
This structure was found in space group \#76, but if we reflect all of the
coordinates through the $z = 0$ plane, it transforms into a structure
in space group \#78, as shown in the Jmol~\citeintro{Jmol} rendering in
Figure~\ref{fig:Cs3P7}.

The distance between any pair of atoms is the same in the $P4_3$
structure as it is in the $P4_1$ structure, and the angle between any
three atoms is the same in both structures. It follows that the
structures are degenerate, there is no difference in energy between
them, and they should be equally likely to form.

Any structure in space group $P4_1$ can be transformed into $P4_3$ by
this method. Pairs of space groups which allow these transformations
are said to be enantiomorphic~\citeintro{cdict,Han_Int_Table_A_3}, or
chiral.\footnote{Formally, any object has {\em chirality} if it is not
superposable on its mirror image~\citeintro{cdict}.
Chirality is a fundamental aspect of life on Earth.
All amino acids found in living organisms are
left-handed~\citeintro{Sedbrook_Smithsonian_2016}.}

There are eleven pairs of enantiomorphic space
groups~\citeintro{cdict,Han_Int_Table_A_3}:
\begin{itemize}
\item $P4_1$ (\#76) and $P4_3$ (\#78),
\item $P4_122$ (\#91) and $P4_322$ (\#95),
\item $P4_12_12$ (\#92) and $P4_32_12$ (\#96),
\item $P3_1$ (\#144) and $P3_2$ (\#145),
\item $P3_112$ (\#151) and $P3_212$ (\#153),
\item $P3_121$ (\#152) and $P3_221$ (\#154),
\item $P6_1$ (\#169) and $P6_5$ (\#170),
\item $P6_2$ (\#171) and $P6_4$ (\#172),
\item $P6_122$ (\#178) and $P6_522$ (\#179),
\item $P6_222$ (\#180) and $P6_422$ (\#181), and
\item $P4_132$ (\#213) and $P4_332$ (\#212).
\end{itemize}

In addition, forty-three other space groups allow chiral crystal structures.
The complete set of sixty-five space groups are known as the Sohncke groups~\citeintro{cdict}.

\section{\label{sec:wsc}The Wigner-Seitz Primitive Cell}

Given a lattice described by a set of primitive lattice vectors,
$\mathbf{a}_1$, $\mathbf{a}_2$, and $\mathbf{a}_3$, we can define a
{\em unit cell} as a volume which, when translated through every
vector of the form $N_1 \, \mathbf{a}_1 + N_2 \, \mathbf{a}_2 + N_3 \,
\mathbf{a}_3$, completely fills space without overlap. Even if we
require that this cell have the minimum volume,
\begin{equation}
\label{equ:cellvolume}
V = \mathbf{a}_1 \cdot (\mathbf{a}_2 \times \mathbf{a}_3),
\end{equation}
this is not a unique definition, and in fact there are an infinite
number of choices for the primitive vectors.

As an example, consider the two-dimensional hexagonal lattice with
primitive vectors
\begin{eqnarray}
\label{equ:hex1}
\mathbf{a}_1 & = & \frac12 \, a \, \mathbf{\hat{x}} - \frac{\sqrt{3}}2
a \, \mathbf{\hat{y}} \nonumber \\
\mathbf{a}_2 & = & \frac12 \, a \, \mathbf{\hat{x}} + \frac{\sqrt{3}}2
a \, \mathbf{\hat{y}} ~ .
\end{eqnarray}
We show this lattice in Figure~\ref{fig:lat1}, along with a two-site
basis (the gray squares). Perhaps the simplest unit cell we can
construct here is a parallelogram, as seen in Figure~\ref{fig:lat1}.
This is a unit cell, as it has the area of the primitive cell,
\begin{eqnarray}
\label{equ:hexarea}
A = \frac{\sqrt3}2 \, a^2,
\end{eqnarray}
each of its replicas contains two basis points, and the cell with its
replicas tile the space.

\begin{figure}
\begin{center}
\begin{tikzpicture}
\filldraw[draw=black,
fill=gray!30](2,0)--(3,-1.732050808)--(4,0)--(3,1.732050808);
\foreach \x in{0,2,...,6}
\draw[fill=black](\x,0)circle(.1cm);
\foreach \x in{0,2,...,6}
\draw[fill=black](\x,3.464101615)circle(.1cm);
\foreach \x in{0,2,...,6}
\draw[fill=black](\x,-3.464101615)circle(.1cm);
\foreach \x in{1,3,...,5}
\draw[fill=black](\x,1.732050808)circle(.1cm);
\foreach \x in{1,3,...,5}
\draw[fill=black](\x,-1.732050808)circle(.1cm);
\draw[ultra thick,->](2,0)--(3,1.732050808)node[above]{$\mathbf{a}_2$};
\draw[ultra
thick,->](2,0)--(3,-1.732050808)node[below]{$\mathbf{a}_1$};
\foreach \x in{0.60,2.60,...,4.60}
\draw[fill=gray=!70](\x,0.4) rectangle (\x+0.1,0.5);
\foreach \x in{0.60,2.60,...,4.60}
\draw[fill=gray=!70](\x,-3.064101615) rectangle (\x+0.1,-2.964101615);
\foreach \x in{1.60,3.60,...,5.60}
\draw[fill=gray=!70](\x,2.132050808) rectangle (\x+0.1,2.232050808);
\foreach \x in{1.60,3.60,...,5.60}
\draw[fill=gray=!70](\x,-1.332050808) rectangle (\x+0.1,-1.232050808);
\foreach \x in{1.30,3.30,...,5.30}
\draw[fill=gray=!70](\x,-0.7) rectangle (\x+0.1,-0.6);
\foreach \x in{1.30,3.30,...,5.30}
\draw[fill=gray=!70](\x,3.464101615-0.7) rectangle (\x+0.1,3.464101615-0.6);
\foreach \x in{0.30,2.30,...,4.30}
\draw[fill=gray=!70](\x,1.732050808-0.7) rectangle (\x+0.1,1.732050808-0.6);
\foreach \x in{0.30,2.30,...,4.30}
\draw[fill=gray=!70](\x,-1.732050808-0.7) rectangle (\x+0.1,-1.732050808-0.6);
\end{tikzpicture}
\end{center}
\caption{\label{fig:lat1}
\textbf{A two-dimensional hexagonal lattice with basis.}
The black circles show the positions of the lattice
vectors, $N_1 \, \mathbf{a}_1 + N_2 \, \mathbf{a}_2$, where $N_1$ and
$N_2$ are integers and $\mathbf{a}_1$ and $\mathbf{a}_2$ are given
by Equation~(\ref{equ:hex1}). The gray squares are a two-site basis for
this lattice.}
\end{figure}

The choice of unit cell is not unique. Consider, for example, an
equivalent set of primitive vectors,
\begin{eqnarray}
\label{equ:hex2}
\mathbf{a}'_1 & = & \frac32 \, a \, \mathbf{\hat{x}} + \frac{\sqrt{3}}2
a \, \mathbf{\hat{y}} \nonumber \\
\mathbf{a}'_2 & = & \frac12 \, a \, \mathbf{\hat{x}} + \frac{\sqrt{3}}2
a \, \mathbf{\hat{y}} ~ ,
\end{eqnarray}
and the accompanying unit cell shown in Figure~\ref{fig:lat2}. This
choice of unit cell has the proper area given by Equation~(\ref{equ:hexarea}), contains
both basis points, and tiles the space. Both Figures~\ref{fig:lat1}
and \ref{fig:lat2} describe the same lattice plus basis, and so are
both unit cells for the lattice.

\begin{figure}
\begin{center}
\begin{tikzpicture}
\filldraw[draw=black,
fill=gray!30](1,-1.732050808)--(2,0)--(5,1.732050808)--(4,0);
\foreach \x in{0,2,...,6}
\draw[fill=black](\x,0)circle(.1cm);
\foreach \x in{0,2,...,6}
\draw[fill=black](\x,3.464101615)circle(.1cm);
\foreach \x in{0,2,...,6}
\draw[fill=black](\x,-3.464101615)circle(.1cm);
\foreach \x in{1,3,...,5}
\draw[fill=black](\x,1.732050808)circle(.1cm);
\foreach \x in{1,3,...,5}
\draw[fill=black](\x,-1.732050808)circle(.1cm);
\draw[ultra thick,->](1,-1.732050808)--(2,0)node[above]{$\mathbf{a}'_2$};
\draw[ultra
thick,->](1,-1.732050808)--(4,0)node[right]{$\mathbf{a}'_1$};
\foreach \x in{0.60,2.60,...,4.60}
\draw[fill=gray=!70](\x,0.4) rectangle (\x+0.1,0.5);
\foreach \x in{0.60,2.60,...,4.60}
\draw[fill=gray=!70](\x,-3.064101615) rectangle (\x+0.1,-2.964101615);
\foreach \x in{1.60,3.60,...,5.60}
\draw[fill=gray=!70](\x,2.132050808) rectangle (\x+0.1,2.232050808);
\foreach \x in{1.60,3.60,...,5.60}
\draw[fill=gray=!70](\x,-1.332050808) rectangle (\x+0.1,-1.232050808);
\foreach \x in{1.30,3.30,...,5.30}
\draw[fill=gray=!70](\x,-0.7) rectangle (\x+0.1,-0.6);
\foreach \x in{1.30,3.30,...,5.30}
\draw[fill=gray=!70](\x,3.464101615-0.7) rectangle (\x+0.1,3.464101615-0.6);
\foreach \x in{0.30,2.30,...,4.30}
\draw[fill=gray=!70](\x,1.732050808-0.7) rectangle (\x+0.1,1.732050808-0.6);
\foreach \x in{0.30,2.30,...,4.30}
\draw[fill=gray=!70](\x,-1.732050808-0.7) rectangle (\x+0.1,-1.732050808-0.6);
\end{tikzpicture}
\end{center}
\caption{\label{fig:lat2}
\textbf{A two-dimensional hexagonal lattice with non-standard primitive vectors.}
This is the same hexagonal lattice from Figure~\ref{fig:lat1} with primitive vectors
given by Equation~(\ref{equ:hex2}).
Each unit cell still contains images of the two sites in the basis.}
\end{figure}

Though not required, it is frequently useful to have a primitive cell
which is uniquely defined and exhibits the symmetry of the lattice.
Such a cell, known as the Wigner-Seitz cell~\citeintro{Ashcroft_p73_1976},
exists for every lattice. The Wigner-Seitz cell is defined as the
locus of all points closer to a given lattice point than to any other
lattice point. We have constructed the Wigner-Seitz cell for our
two-dimensional hexagonal lattice as shown in Figure~\ref{fig:lat3}.
Like all primitive cells, the Wigner-Seitz cell tiles its space, but
unlike other cells it manifests the symmetry of the underlying
hexagonal lattice,\footnote{Note that we say that the Wigner-Seitz
cell has the symmetry of the lattice, not the crystal structure.
Taken without a basis, the Wigner-Seitz cell in
Figure~\ref{fig:lat3} has a six-fold rotation axis about the origin,
in agreement with the symmetry of the hexagonal lattice. With the
basis, however, the cell shown in Figure~\ref{fig:lat3} does {\em
not} have full hexagonal symmetry, even though its shape is a
hexagon.} in this case exhibiting the six-fold rotation symmetry
characteristic of a hexagonal lattice.

\begin{figure}
\begin{center}
\begin{tikzpicture}
\filldraw[draw=black,
fill=gray!30](3,-0.5773502692) -- (3,0.5773502692) --
(2,1.154700538) -- (1,0.5773502692) -- (1,-0.5773502692) --
(2,-1.1547003538) -- (3,-0.5773502692);
\foreach \x in{0,2,...,6}
\draw[fill=black](\x,0)circle(.1cm);
\foreach \x in{0,2,...,6}
\draw[fill=black](\x,3.464101615)circle(.1cm);
\foreach \x in{0,2,...,6}
\draw[fill=black](\x,-3.464101615)circle(.1cm);
\foreach \x in{1,3,...,5}
\draw[fill=black](\x,1.732050808)circle(.1cm);
\foreach \x in{1,3,...,5}
\draw[fill=black](\x,-1.732050808)circle(.1cm);
\draw[ultra thick,->](2,0)--(3,1.732050808)node[above]{$\mathbf{a}_2$};
\draw[ultra
thick,->](2,0)--(3,-1.732050808)node[below]{$\mathbf{a}_1$};
\foreach \x in{0.60,2.60,...,4.60}
\draw[fill=gray=!70](\x,0.4) rectangle (\x+0.1,0.5);
\foreach \x in{0.60,2.60,...,4.60}
\draw[fill=gray=!70](\x,-3.064101615) rectangle (\x+0.1,-2.964101615);
\foreach \x in{1.60,3.60,...,5.60}
\draw[fill=gray=!70](\x,2.132050808) rectangle (\x+0.1,2.232050808);
\foreach \x in{1.60,3.60,...,5.60}
\draw[fill=gray=!70](\x,-1.332050808) rectangle (\x+0.1,-1.232050808);
\foreach \x in{1.30,3.30,...,5.30}
\draw[fill=gray=!70](\x,-0.7) rectangle (\x+0.1,-0.6);
\foreach \x in{1.30,3.30,...,5.30}
\draw[fill=gray=!70](\x,3.464101615-0.7) rectangle (\x+0.1,3.464101615-0.6);
\foreach \x in{0.30,2.30,...,4.30}
\draw[fill=gray=!70](\x,1.732050808-0.7) rectangle (\x+0.1,1.732050808-0.6);
\foreach \x in{0.30,2.30,...,4.30}
\draw[fill=gray=!70](\x,-1.732050808-0.7) rectangle (\x+0.1,-1.732050808-0.6);
\end{tikzpicture}
\end{center}
\caption{\label{fig:lat3}
\textbf{A two-dimensional hexagonal lattice with its corresponding Wigner-Seitz cell centered on a lattice point.}}
\end{figure}

By definition, all points in the Wigner-Seitz cell are closer to its
origin than to any other lattice point. Since we can center the cell
on a point other than the lattice points defined by
Equation~(\ref{equ:cellvolume}), we can use this fact to determine the nearest
neighbors of a given Wyckoff position. Consider the Wigner-Seitz cell
in Figure~\ref{fig:lat4}, where we center the cell on one of the
points in the basis. Only one of the images of the other point in the
basis exists in this unit cell, so these two points are nearest
neighbors. The only ambiguity in this definition is when a basis
point is on the boundary of the Wigner-Seitz zone. In this case there
will be multiple copies of the basis point, all of them the same
distance from the origin.

\begin{figure}
\begin{center}
\begin{tikzpicture}
\filldraw[draw=black,
fill=gray!30](3+0.6,-0.5773502692+0.5) -- (3+0.6,0.5773502692+0.5) --
(2+0.6,1.154700538+0.5) -- (1+0.6,0.5773502692+0.5) -- (1+0.6,-0.5773502692+0.5) --
(2+0.6,-1.1547003538+0.5) -- (3+0.6,-0.5773502692+0.5);
\foreach \x in{0,2,...,6}
\draw[fill=black](\x,0)circle(.1cm);
\foreach \x in{0,2,...,6}
\draw[fill=black](\x,3.464101615)circle(.1cm);
\foreach \x in{0,2,...,6}
\draw[fill=black](\x,-3.464101615)circle(.1cm);
\foreach \x in{1,3,...,5}
\draw[fill=black](\x,1.732050808)circle(.1cm);
\foreach \x in{1,3,...,5}
\draw[fill=black](\x,-1.732050808)circle(.1cm);
\draw[ultra thick,->](2,0)--(3,1.732050808)node[above]{$\mathbf{a}_2$};
\draw[ultra
thick,->](2,0)--(3,-1.732050808)node[below]{$\mathbf{a}_1$};
\foreach \x in{0.60,2.60,...,4.60}
\draw[fill=gray=!70](\x,0.4) rectangle (\x+0.1,0.5);
\foreach \x in{0.60,2.60,...,4.60}
\draw[fill=gray=!70](\x,-3.064101615) rectangle (\x+0.1,-2.964101615);
\foreach \x in{1.60,3.60,...,5.60}
\draw[fill=gray=!70](\x,2.132050808) rectangle (\x+0.1,2.232050808);
\foreach \x in{1.60,3.60,...,5.60}
\draw[fill=gray=!70](\x,-1.332050808) rectangle (\x+0.1,-1.232050808);
\foreach \x in{1.30,3.30,...,5.30}
\draw[fill=gray=!70](\x,-0.7) rectangle (\x+0.1,-0.6);
\foreach \x in{1.30,3.30,...,5.30}
\draw[fill=gray=!70](\x,3.464101615-0.7) rectangle (\x+0.1,3.464101615-0.6);
\foreach \x in{0.30,2.30,...,4.30}
\draw[fill=gray=!70](\x,1.732050808-0.7) rectangle (\x+0.1,1.732050808-0.6);
\foreach \x in{0.30,2.30,...,4.30}
\draw[fill=gray=!70](\x,-1.732050808-0.7) rectangle (\x+0.1,-1.732050808-0.6);
\end{tikzpicture}
\end{center}
\caption{\label{fig:lat4}
\textbf{A two-dimensional hexagonal lattice with its corresponding Wigner-Seitz cell centered on one of the atoms.}}
\end{figure}

\begin{figure*}
\begin{center}
\includegraphics[width=0.995\textwidth]{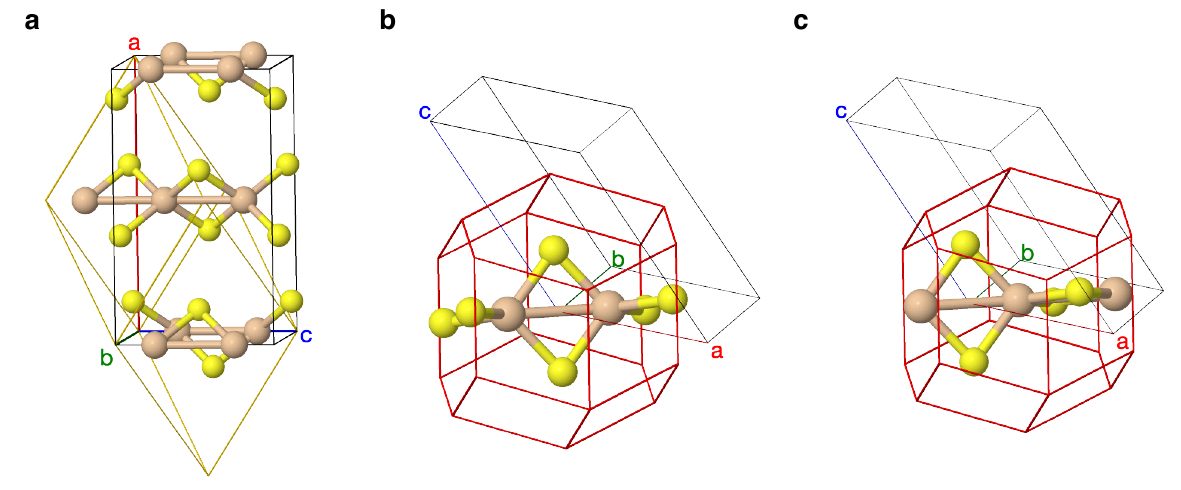}
\end{center}
\caption{\label{fig:cellreps}
\textbf{Unit cell representations for body-centered orthorhombic SiS$_{2}$.}
The silicon and sulfur atoms are beige and yellow, respectively.
(\textbf{a}) The primitive (yellow) and conventional (black) unit
cell representations, where the axes are shown with respect to the conventional cell.
(\textbf{b}) The Wigner-Seitz cell (red polyhedron) centered at the origin between
two silicon atoms.
The axes are shown with respect to the body-centered orthorhombic primitive unit cell (black)
defined in Part 1 of the library~\protect\citeintro{curtarolo:art121}.
(\textbf{c}) The Wigner-Seitz cell (red polyhedron) centered on one of the silicon atoms
in the primitive cell.
The axes are shown with respect to the body-centered orthorhombic primitive unit cell (black)
defined in Part 1 of the library~\protect\citeintro{curtarolo:art121}.}
\end{figure*}

Three dimensions complicate the visualization of the Wigner-Seitz
cell. We have added the ability to view the Wigner-Seitz cell to
every structure in the \AFLOW\ Library of Crystallographic Prototypes, using
JSmol~\citeintro{Hanson:JSmol} to render the cells. As an example, consider
the SiS$_2$ structure A2B\_oI12\_72\_j\_a.\footnote{This structure can
be found in the Library of Crystallographic Prototypes at
\href{http://aflow.org/CrystalDatabase/A2B_oI12_72_j_a.html}{http://aflow.org/CrystalDatabase/A2B\_oI12\_72\_j\_a.html.}}
Figure~\ref{fig:cellreps}(a) shows the primitive and conventional unit
cells of this system, as well as the positions of the silicon and
sulfur atoms.

Figure~\ref{fig:cellreps}(b) shows the Wigner-Seitz cell, centered on the
origin of the lattice. One can clearly see that the unit cell
contains two silicon atoms, with two sulfur atoms in the interior of the zone
and two on the boundary.

We can also center the Wigner-Seitz cell on the a particular atom, as
shown with an silicon atom in Figure~\ref{fig:cellreps}(c). This construction
unambiguously identifies atoms which are nearest neighbors to the
target silicon atom. This procedure can of course be extended to any atom
in the primitive cell.

The Wigner-Seitz cell has its analog in reciprocal space. If
$\{\mathbf{R}\}$ is the set of all lattice points in three dimensional
space, the reciprocal lattice is defined as the set of all vectors
$\{\mathbf{K}\}$ such that~\citeintro{Ashcroft_c5_1976}
\begin{equation}
\label{equ:recip}
e^{i \mathbf{R} \cdot \mathbf{K}} = 1.
\end{equation}
If the real-space lattice is defined by primitive vectors
$\mathbf{a}_1$, $\mathbf{a}_2$, and $\mathbf{a}_3$, then the
reciprocal lattice is defined by vectors
\begin{equation}
\label{equ:bdef}
\mathbf{b}_i = \left(\frac{2\pi}{V}\right) \mathbf{a}_j \times
\mathbf{a}_k,
\end{equation}
where $(i,j,k)$ are in cyclic order. The reciprocal lattice is dual
to the real-space lattice, with
\begin{equation}
\label{equ:adob}
\mathbf{a}_i \cdot \mathbf{b}_j = 2\pi \delta_{ij} ~ .
\end{equation}
In matrix notation, we can write this as
\begin{equation}
\label{equ:bdefmat}
2 \pi \mathbf{B} = \left( \mathbf{A}^T \right)^{-1} ~ ,
\end{equation}
where $\mathbf{A}$, $\mathbf{B}$ are matrices whose columns are the
lattice vectors $\mathbf{a}$ and reciprocal lattice vectors $\mathbf{b}$,
respectively.

This reciprocal lattice has its own Wigner-Seitz cell, the {\em first
Brillouin zone}. The relationship between real space lattices and
their reciprocal space Brillouin zones are discussed in
Ref.~\citeintro{curtarolo:art58}.

\section{\label{sec:plane} Plane Groups (Two-Dimensional Space Groups)}

Even though we live in a world with three physical dimensions, we can
often achieve insight into that world by considering corresponding
situations in two dimensions~\citeintro{Abbott_Flatland}. Two-dimensional
physics is also important when we consider such materials as
graphene~\citeintro{Allen_Chem_Rev_110_132_2010},
MoS$_2$~\citeintro{Li_J_Materio_1_33_2015}, and similar materials. These
systems still form periodic structures, and have symmetry operations
associated with translations, rotations and reflections in two
dimensions. These form the seventeen ``plane groups'', often called
``wallpaper groups'' for obvious reasons.

Since two-dimensional figures readily translate to the page, there are
numerous discussions of group properties readily
available~\citeintro{Joyce_Wallpaper_1997,Morandi_Wallpaper_2003}, along
with some beautiful examples of wallpaper
symmetries~\citeintro{Eck_Wallpaper}.

At the atomic level, plane groups, like space groups, each possess a
set of Wyckoff positions that determine allowed atomic positions in
keeping with the symmetry of the group. These have been tabulated in
the {\em International Tables}~\citeintro{tables_crystallography_A} and
online~\citeintro{Aroyo_Bulgar_Chem_Comm_43_183_2011}. Here, we present a
graphical illustration of the Wyckoff positions of each of the
seventeen space groups, using the notation of the {\em International
Tables} for each group.

Much as with three-dimensional space groups, the plane groups can be
divided into crystal systems, each member of a given system having the
same holohedry (rotational symmetry of the point group of the
lattice).

There are five crystal systems in two dimensions:

\begin{itemize}
\item Simple parallelograms (plane groups \#1-2): the analog to the
triclinic system in three dimensions. The primitive vectors for
these lattices are given by
\begin{eqnarray}
\label{lat:parallel}
\mathbf{a}_1 & = & a \, \hat{\mathbf{x}} \nonumber \\
\mathbf{a}_2 & = & b \, \cos\theta \, \hat{\mathbf{x}} + b \,
\sin\theta \, \hat{\mathbf{y}}.
\end{eqnarray}
The choice of $a$, $b$, and $\theta$ is completely arbitrary,
provided that the choices do not fall into another of the crystal
systems.

\item Rectangles (plane groups \#3-9): the system is defined by a
rectangular lattice
\begin{eqnarray}
\label{lat:rectangle}
\mathbf{a}_1 & = & a \, \hat{\mathbf{x}} \nonumber \\
\mathbf{a}_2 & = & b \, \hat{\mathbf{y}}.
\end{eqnarray}
with perpendicular lattice vectors [$\theta = 90^{\circ}$ in
Equation~(\ref{lat:parallel})]. The system contains both simple
rectangular lattices, where the lattice is defined by
Equation~(\ref{lat:rectangle}), and centered rectangular lattices, analogous
to base-centered lattices in three dimensions. For the centered
lattices (plane groups \#5 and \#9), the conventional cell is defined by
Equation~(\ref{lat:rectangle}), and the primitive unit cell is defined by
\begin{eqnarray}
\label{lat:centered}
\mathbf{a}_1 & = & \frac12 a \, \hat{\mathbf{x}} - \frac12 b \,
\hat{\mathbf{y}} \nonumber \\
\mathbf{a}_2 & = & \frac12 a \, \hat{\mathbf{x}} + \frac12 b \,
\hat{\mathbf{y}}.
\end{eqnarray}

\item Squares (plane groups \#10-12): special cases of the
rectangular system given by Equation~(\ref{lat:rectangle}) with $b = a$, so that the
primitive lattice is specified by the vectors
\begin{eqnarray}
\label{lat:square}
\mathbf{a}_1 & = & a \, \hat{\mathbf{x}} \nonumber \\
\mathbf{a}_2 & = & a \, \hat{\mathbf{y}}.
\end{eqnarray}
The centered rectangular lattice Equation~(\ref{lat:centered}) also becomes a
square lattice when $a = b$.

\item Trigonal (plane groups \#13-15): In this system the primitive
vectors are of equal length and separated by an angle of
120$^\circ$, with a three-fold rotation axis about the origin [$b =
a$ and $\theta = 60^\circ$ is given by Equation~(\ref{lat:parallel}), or $b = \sqrt3
\, a$ in Equation~(\ref{lat:centered})]. The lattice vectors are
\begin{eqnarray}
\label{lat:hex}
\mathbf{a}_1 & = & \frac12 a \, \hat{\mathbf{x}} - \frac{\sqrt3}2 a \,
\hat{\mathbf{y}} \nonumber \\
\mathbf{a}_2 & = & \frac12 a \, \hat{\mathbf{x}} + \frac{\sqrt3}2 a \,
\hat{\mathbf{y}}.
\end{eqnarray}
This can be regarded as a special case of the parallelogram
in Equation~(\ref{lat:parallel}) with $b = a$ and $\theta = 120^\circ$, or of
the centered rectangular lattice with $b = \sqrt3 \, a$.

\item Hexagonal (plane groups \#16-17): As in the three-dimensional
case, a distinction is made between the trigonal system, which has a
three-fold rotational axis about the origin, and the hexagonal
system, which has a six-fold rotational axis. In either case, the
primitive vectors of the unit cell are given by Equation~(\ref{lat:hex}).

\end{itemize}

Each section below lists the primitive vectors that describe the
lattice associated with the plane group, a table of Wyckoff positions
allowed for the space group, and a figure showing possible Wyckoff
positions, along with the Wigner-Seitz cell for the lattice.

\subsection{\label{parallel} The Parallelogram Crystal System}

\subsubsection{\label{p1} Plane Group \#1: $p1$}

This plane group has the lowest symmetry possible for a periodic
lattice. The primitive lattice vectors are given by
Equation~(\ref{lat:parallel}), and the single Wyckoff position is completely
general:
\begin{table_col}
\begin{center}
\begin{tabular}{||c|c||}
\hline\hline
{\bf Label} & {\bf Lattice Coordinates} \\
\hline
(1a) & $(x,y)$ \\
\hline\hline
\end{tabular}
\end{center}
\end{table_col}

\begin{figure}[h]
\begin{center}
\includegraphics[width=\linewidth]{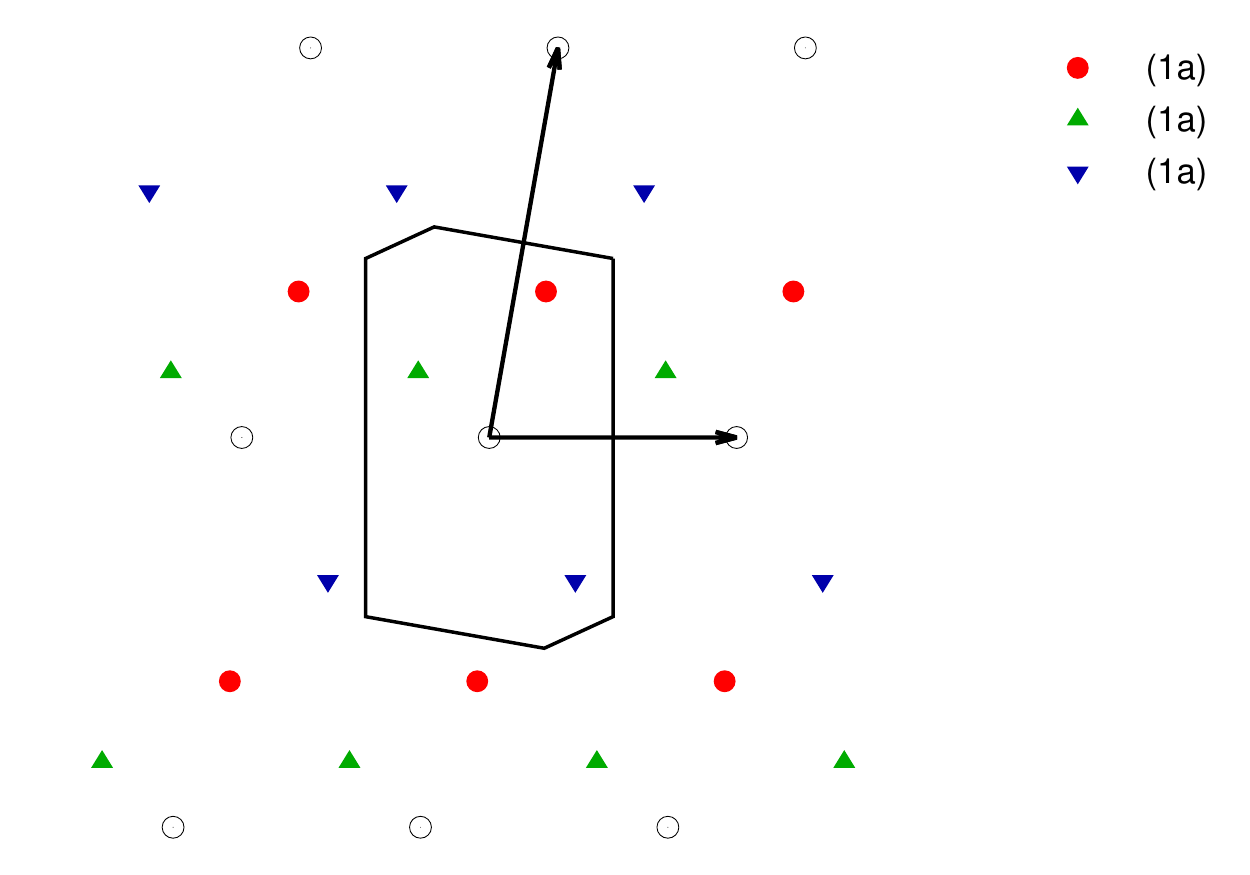}
\end{center}
\caption{\label{fig:p1}
\textbf{Possible Wyckoff positions for plane group
\#1, $\bm{p1}$.} We show multiple possible Wyckoff positions, as if
only one is occupied it can be placed at the origin, and the
plane group achieves the higher symmetry of plane group \#2, $p
2$. The black outline represents the boundary of the Wigner-Seitz
cell for the lattice in Equation~(\ref{lat:parallel}). The open circles
indicate the lattice points.}
\end{figure}

Figure~\ref{fig:p1} shows a selection of Wyckoff positions for plane
group $p1$. It is worthwhile to note that if there is only one
atom in the plane group defined by Equation~(\ref{lat:parallel}), then we can place
it at the origin, which then becomes an inversion site. This
immediately promotes the structure to the higher symmetry of plane
group \#2, $p2$. If the structure contains only two identical
atoms, then there is an inversion site between them, and it is once
again appropriate to place the structure in space group $p2$.

\subsubsection{\label{p2} Plane Group \#2: $p2$}

Plane group \#2, $p2$, has the same type of primitive lattice as
$p1$, in Equation~(\ref{lat:parallel}), but now inversion through the origin
produces an unchanged structure. This means that any function
operating in this crystal system must have the property that
\begin{equation}
\label{equ:inverse}
f(x,y) = f(-x,-y).
\end{equation}
This provides several relatively high-symmetry Wyckoff positions for
single atoms, in addition to the general Wyckoff position.
\begin{table_col}
\begin{center}
\begin{tabular}{||c|c||}
\hline\hline
{\bf Label} & {\bf Lattice Coordinates} \\
\hline
(2e) & $(x,y) ~ (-x,-y)$ \\
\hline
(1d) & $(1/2, 1/2)$ \\
\hline
(1c) & $(1/2, 0)$ \\
\hline
(1b) & $(0, 1/2)$ \\
\hline
(1a) & $(0,0)$ \\
\hline\hline
\end{tabular}
\end{center}
\end{table_col}

\begin{figure}[h]
\begin{center}
\includegraphics[width=\linewidth]{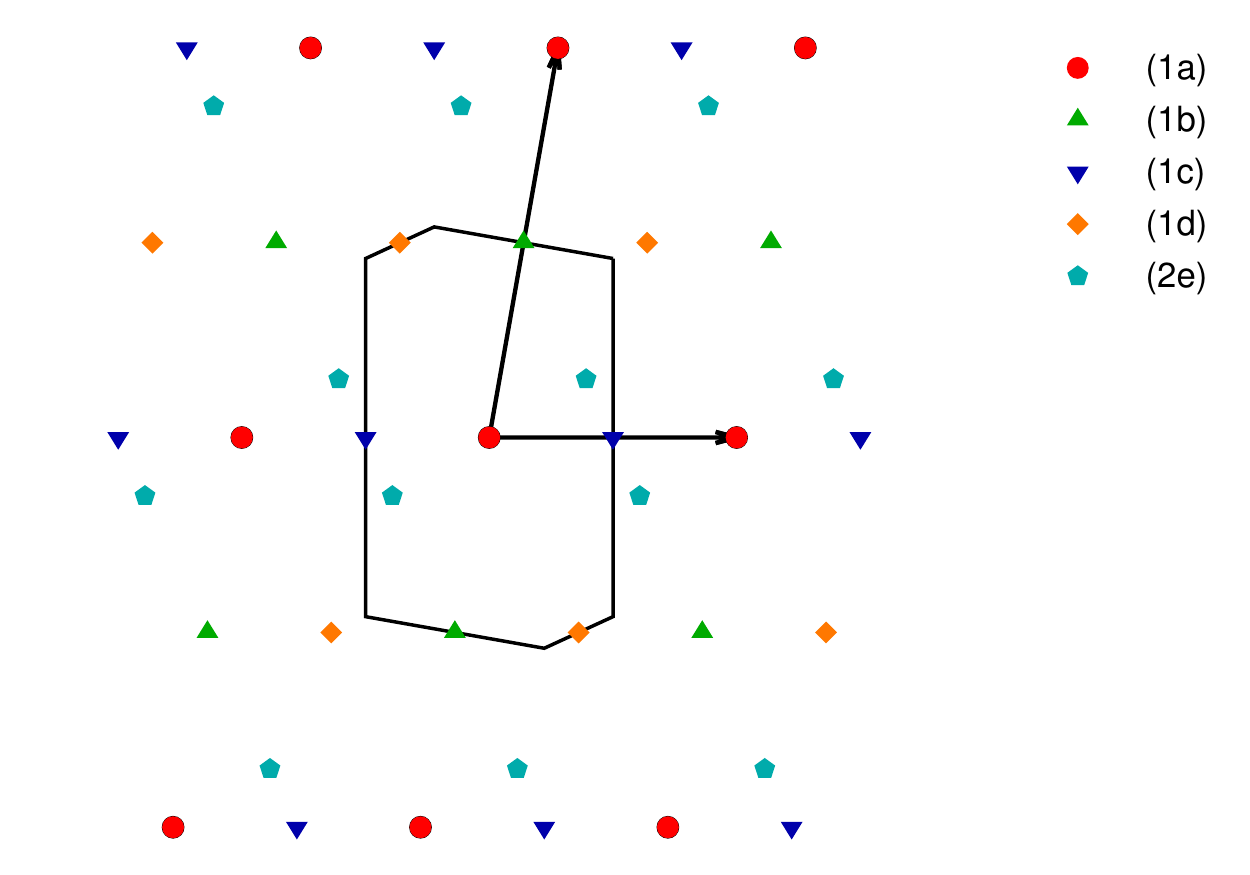}
\end{center}
\caption{\label{fig:p2}
\textbf{Possible Wyckoff positions for plane group
\#2, $\bm{p2}$.} The positions of the (1a)-(1d) sites are fixed,
while the (2e) sites are arbitrary.
The black outline represents the boundary of the Wigner-Seitz
cell for the lattice in Equation~(\ref{lat:parallel}).}
\end{figure}

\subsection{\label{rect} The Rectangular Crystal System}

All of the rectangular plane groups have a conventional cell defined
by the primitive vectors in Equation~(\ref{lat:rectangle}). This is also the
primitive cell for space groups \#3, \#4, \#6, \#7 and \#8.

\subsubsection{\label{p1m1} Plane Group \#3: $p1m1$}

This rectangular space group has primitive vectors
in Equation~(\ref{lat:rectangle}) and a reflection around $x = 0$; that is, any
function operating in this space group must have the property that
\begin{equation}
\label{equ:xmirror}
f(-x,y) = f(x,y)
\end{equation}
as well as the periodic properties
\begin{equation}
f(x + a,y) = f(x, y + b) = f(x,y).
\end{equation}
The Wyckoff positions for space group \#3 are given by
\begin{table_col}
\begin{center}
\begin{tabular}{||c|c||}
\hline\hline
{\bf Label} & {\bf Lattice Coordinates} \\
\hline
(2c) & $(x,y) ~ (-x,y)$ \\
\hline
(1b) & $(1/2, y)$ \\
\hline
(1a) & $(0,y)$ \\
\hline\hline
\end{tabular}
\end{center}
\end{table_col}
\noindent The plane group does not contain the inversion.

\begin{figure}[h]
\begin{center}
\includegraphics[width=\linewidth]{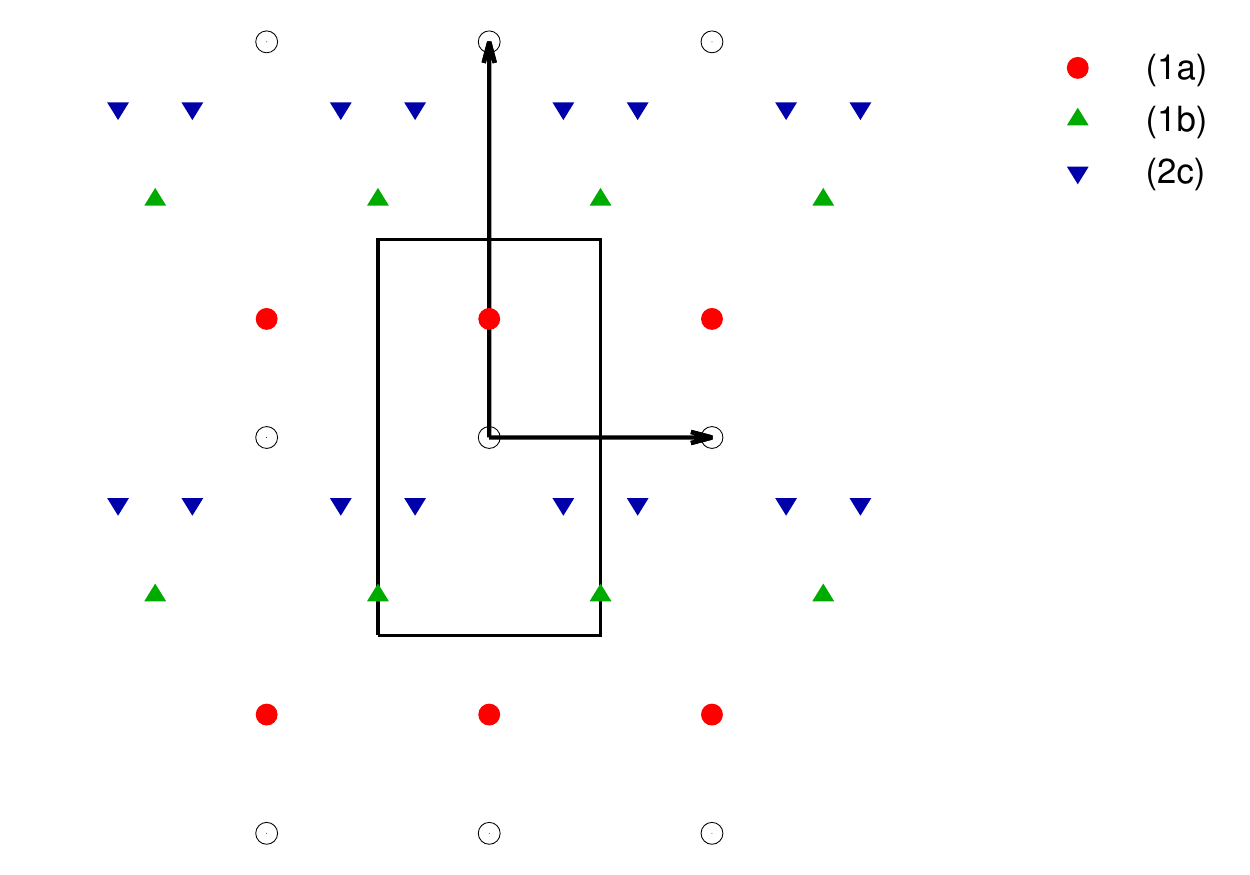}
\end{center}
\caption{\label{fig:p1m1}
\textbf{Possible Wyckoff positions for plane group
\#3, $\bm{p1m1}$.}
The black outline represents the boundary of the Wigner-Seitz cell
for the lattice in Equation~(\ref{lat:rectangle}). The open circles indicate
the lattice points.}
\end{figure}

A graphical representation of the Wyckoff positions and the
Wigner-Seitz cell for plane group \#3 are given in
Figure~\ref{fig:p1m1}.

\subsubsection{\label{p1g1} Plane Group \#4: $p1g1$}

This rectangular space group has a glide
reflection~\citeintro{Joyce_Wallpaper_1997}, with a reflection about $x = 0$
combined with a translation of $1/2 \, b$ along the $y$ direction.
That is, any function showing this symmetry must have the property
that
\begin{equation}
\label{equ:glidex}
f(x,y) = f(-x,y+1/2)
\end{equation}
There is only one Wyckoff position:
\begin{center}
\begin{tabular}{||c|c||}
\hline\hline
{\bf Label} & {\bf Lattice Coordinates} \\
\hline
(2a) & $(x,y) ~ (-x,y + 1/2)$ \\
\hline\hline
\end{tabular}
\end{center}
There is no inversion in this crystal structure.

\begin{figure}[h]
\begin{center}
\includegraphics[width=\linewidth]{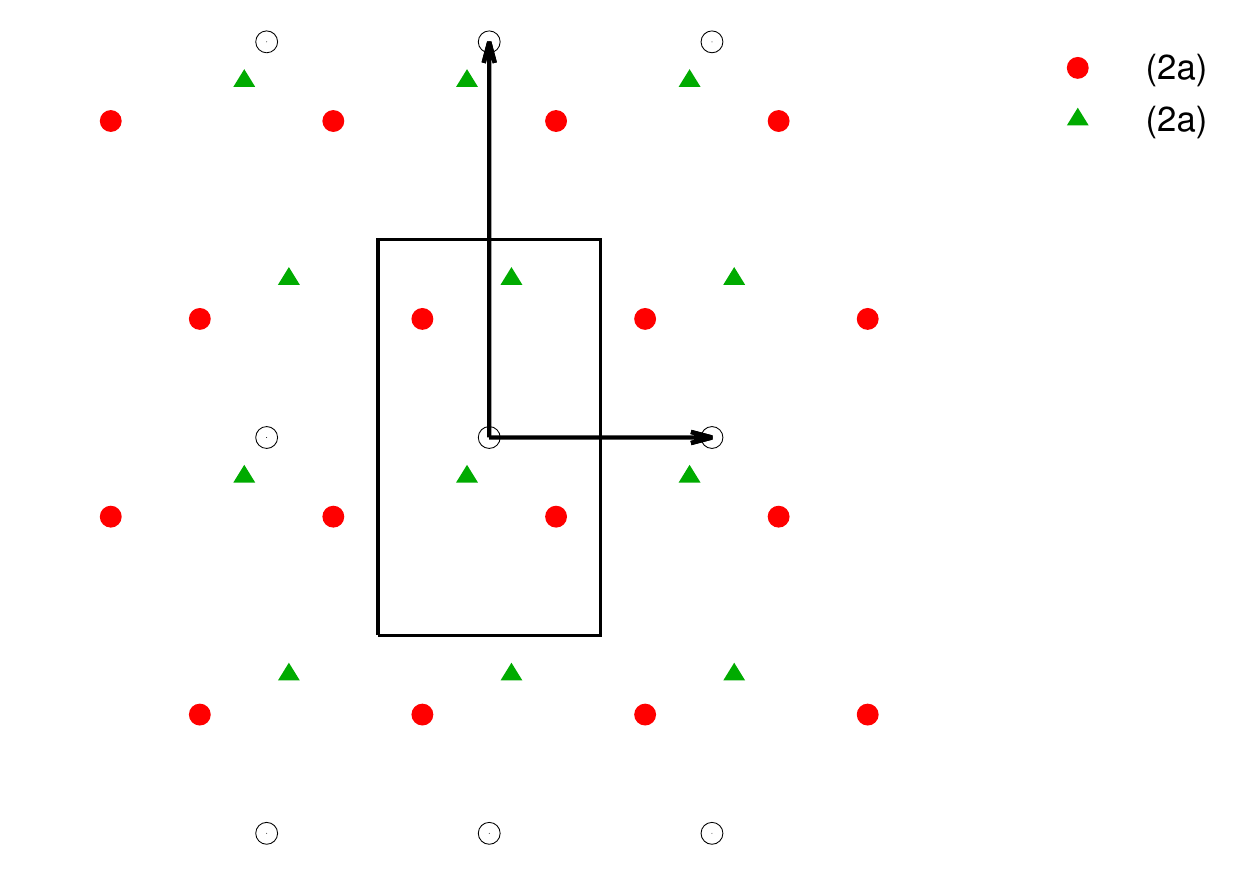}
\end{center}
\caption{\label{fig:p1g1}
\textbf{Possible Wyckoff positions for plane group
\#4, $\bm{p1g1}$.}
The black outline represents the boundary of the Wigner-Seitz cell
for the lattice in Equation~(\ref{lat:rectangle}). We show two possible
occupations of the (2a) Wyckoff position to show that this is not in
the higher symmetry plane group $p2mg$ (\#7). The open
circles indicate the lattice points.}
\end{figure}

Figure~\ref{fig:p1g1} shows possible occupations of this Wyckoff
position. We show two possible occupations. If only one (2a) site is
occupied in this crystal system, there is an inversion between the two
atoms, and we can place the origin there. In that case $y = 1/4$, and
the system actually has the higher symmetry $p2mg$ (\#7),
with atoms on the (2c) sites.

\subsubsection{\label{c1m1} Plane Group \#5: $c1m1$}

This is a centered rectangular space group, with the conventional cell
given by Equation~(\ref{lat:rectangle}) and the primitive vectors given by
Equation~(\ref{lat:centered}). Just as in space group \#4, there is a glide
reflection in Equation~(\ref{equ:glidex}). The Wyckoff positions are given by
\begin{table_col}
\begin{center}
\begin{tabular}{||c|c||}
\hline\hline
{\bf Label} & {\bf Conventional Lattice Coordinates} \\
\hline
& $(x,y) ~ (-x,y)$ \\
(4b) & $- - - - - - - - - - - - - - - - -$ \\
& $(x + 1/2, y + 1/2) ~ (-x + 1/2, y + 1/2)$ \\
\hline
& $(0,y)$ \\
(2a) & $- - - - - - - - - - - - - - - - -$ \\
& $(1/2, y + 1/2)$ \\
\hline\hline
\end{tabular}
\end{center}
\end{table_col}
\noindent where, following convention, we give the coordinates of the positions
in terms of the conventional lattice in Equation~(\ref{lat:rectangle}), but we
also explicitly show the translations due to the centered primitive
cell below the dashed line. The plane group does not contain the
inversion.

\begin{figure}[h]
\begin{center}
\includegraphics[width=\linewidth]{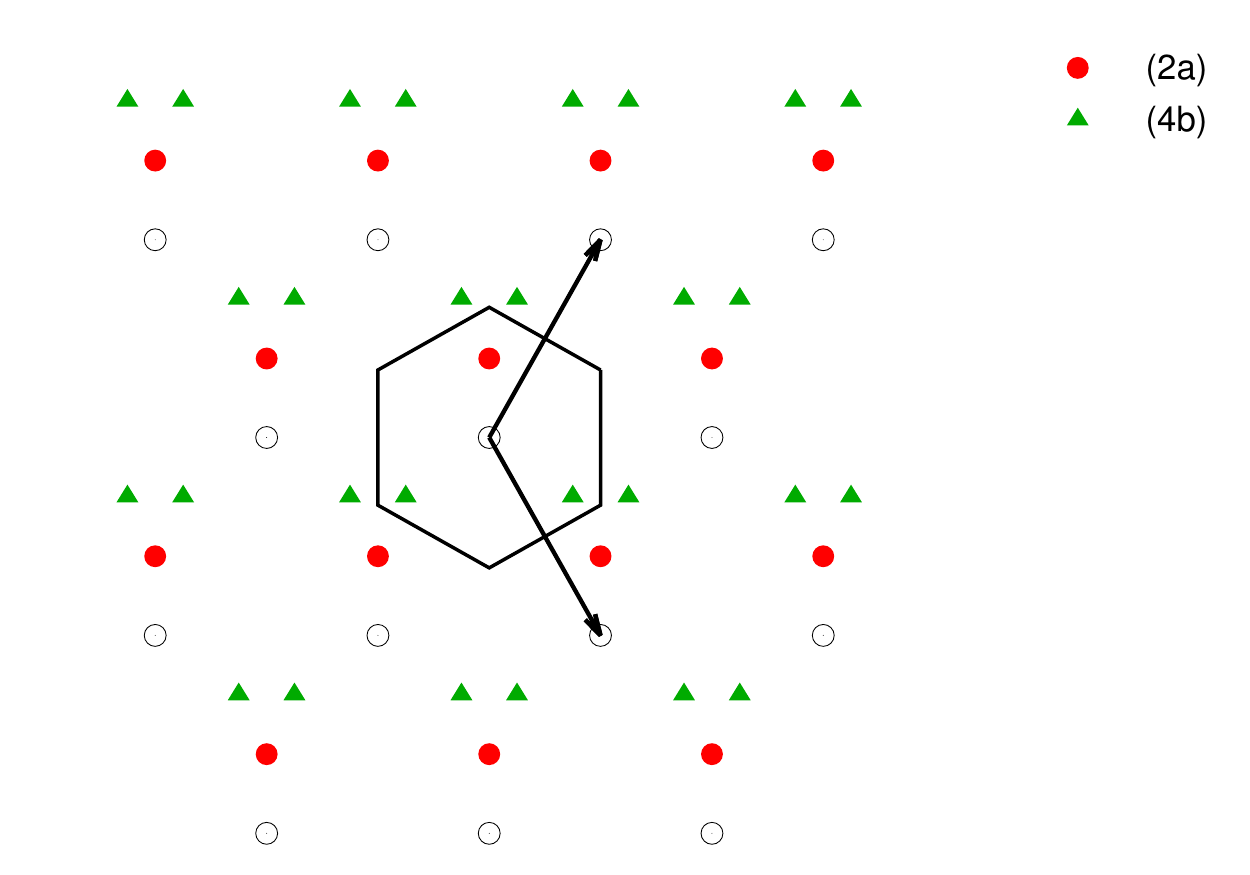}
\end{center}
\caption{\label{fig:c1m1}
\textbf{Possible Wyckoff positions for plane group
\#5, $\bm{c1m1}$.}
The black outline represents the boundary of the Wigner-Seitz cell
for the primitive cell in Equation~(\ref{lat:centered}).
The open circles indicate the lattice points.}
\end{figure}

The Wyckoff positions for this lattice are sketched in
Figure~\ref{fig:c1m1}.

\subsubsection{\label{p2mm} Plane Group \#6: $p2mm$}

\begin{figure}[h]
\begin{center}
\includegraphics[width=\linewidth]{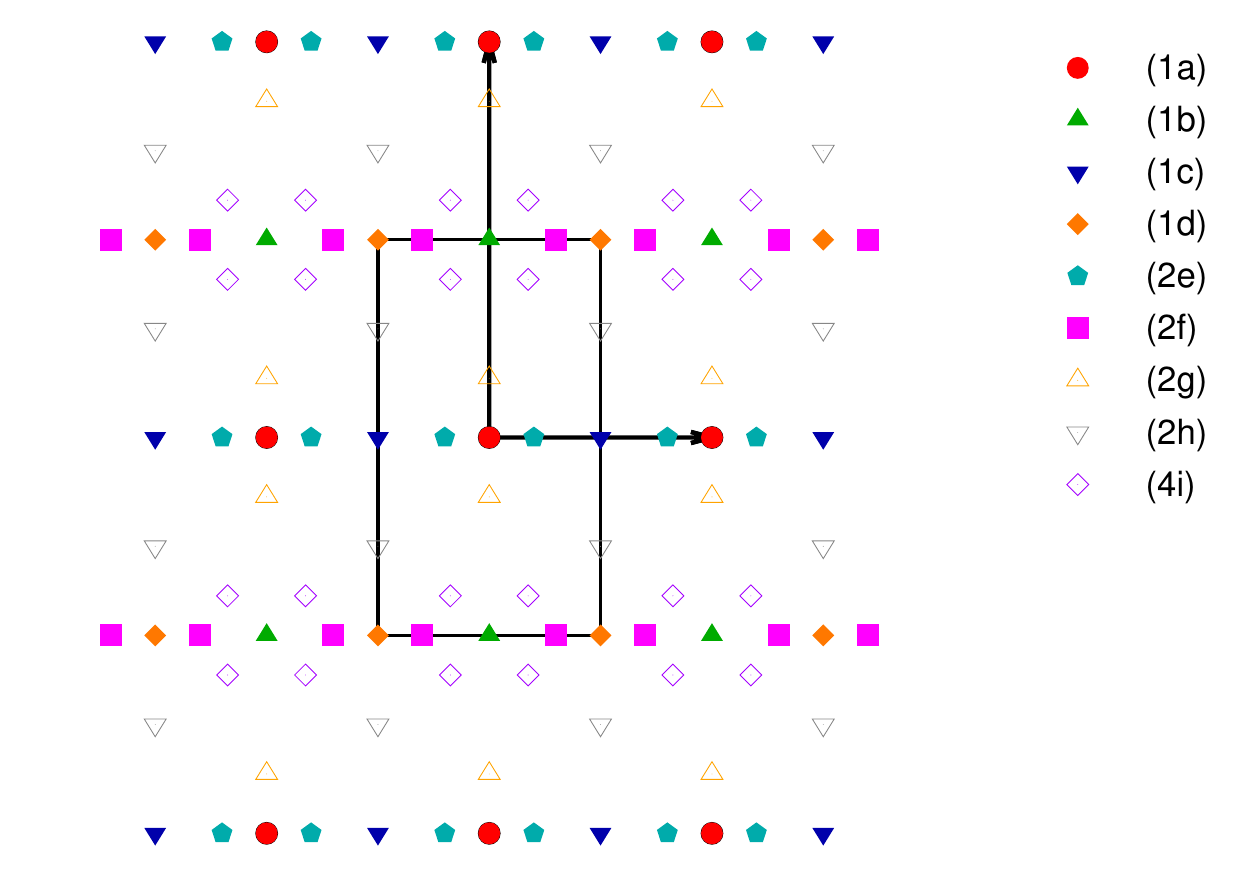}
\end{center}
\caption{\label{fig:p2mm}
\textbf{Possible Wyckoff positions for plane group
\#6, $\bm{p2mm}$.}
The black outline represents the boundary of the Wigner-Seitz cell
for the lattice in Equation~(\ref{lat:rectangle}). Note the mirroring about $x
= 0$ and $y = 0$.}
\end{figure}

The remaining rectangular plane groups all contain the inversion
given by Equation~(\ref{equ:inverse}). Group \#6 also includes a mirror reflection
about $x = 0$ 
in Equation~(\ref{equ:xmirror}). Combining both of these operations
show us that there must also be a mirror reflection about $y = 0$, as
well:
\begin{equation}
\label{equ:ymirror}
f(x,y) = f(x,-y).
\end{equation}
This leads to several Wyckoff positions, all of which are outlined in
Figure~\ref{fig:p2mm}.
\begin{table_col}
\begin{center}
\begin{tabular}{||c|c||}
\hline\hline
{\bf Label} & {\bf Lattice Coordinates} \\
\hline
(4i) & $(x,y) ~ (-x, -y) ~ (-x,y) ~ (x, -y)$ \\
\hline
(2h) & $(1/2,y) ~ (1/2, -y)$ \\
\hline
(2g) & $(0,y) ~ (0,-y)$ \\
\hline
(2f) & $(x,1/2) ~ (-x,1/2)$ \\
\hline
(2e) & $(x,0) ~ (-x,0)$ \\
\hline
(1d) & $(1/2,1/2)$ \\
\hline
(1c) & $(1/2,0)$ \\
\hline
(1b) & $(0, 1/2)$ \\
\hline
(1a) & $(0,0)$ \\
\hline\hline
\end{tabular}
\end{center}
\end{table_col}

\subsubsection{\label{p2mg} Plane Group \#7: $p2mg$}

\begin{figure}[h]
\begin{center}
\includegraphics[width=\linewidth]{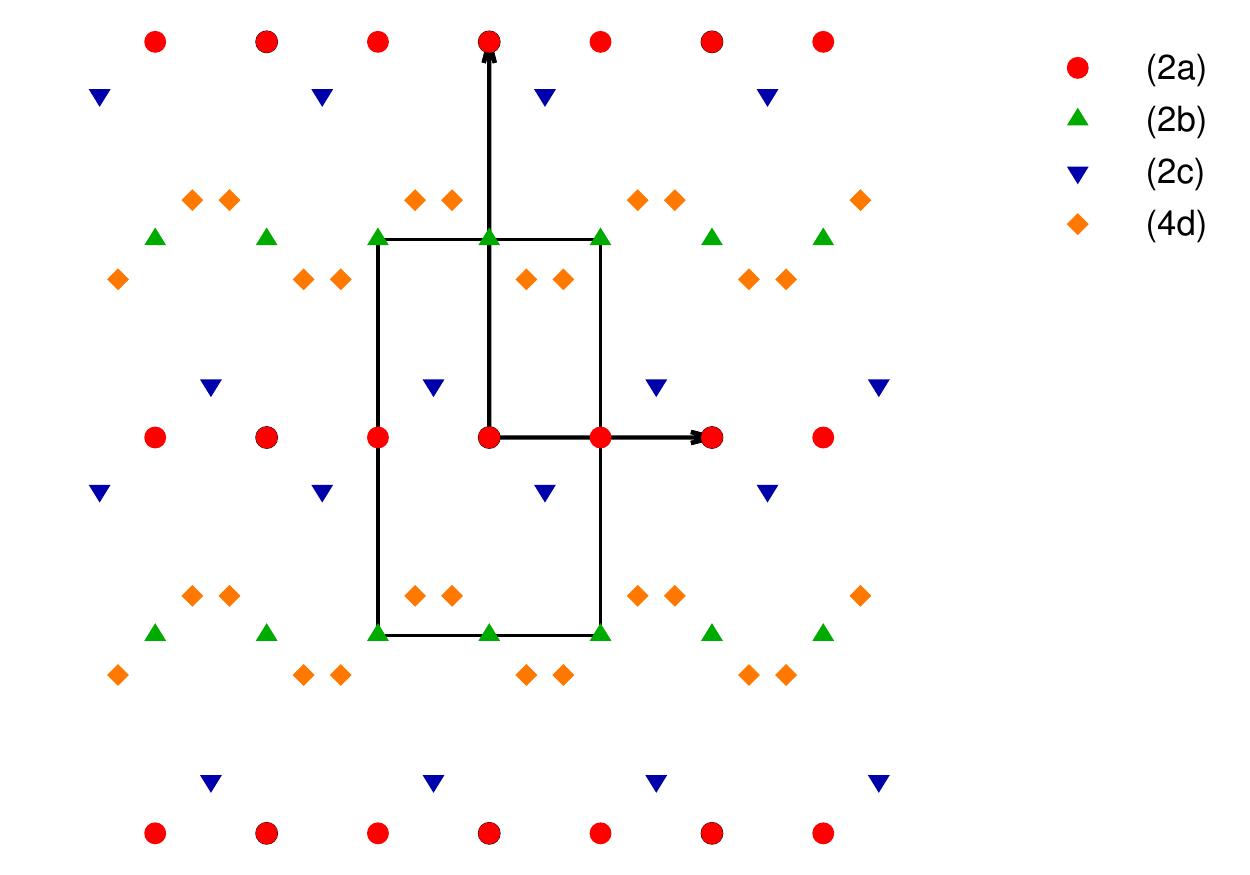}
\end{center}
\caption{\label{fig:p2mg}
\textbf{Possible Wyckoff positions for plane group
\#7, $\bm{p2mg}$.}
The black outline represents the boundary of the Wigner-Seitz cell
for the lattice in Equation~(\ref{lat:rectangle}). Note the mirroring about $x
= 0$ and $y = 0$.}
\end{figure}

In addition to the inversion, this group has a glide reflection
about $y = 0$ combined with a translation of $1/2
\, a$ along the $x$ direction:
\begin{equation}
\label{equ:glidey}
f(x,y) = f(-x + 1/2,y)
\end{equation}
The Wyckoff positions are sketched in
Figure~\ref{fig:p2mg} and are listed below.
\begin{table_col}
\begin{center}
\begin{tabular}{||c|c||}
\hline\hline
{\bf Label} & {\bf Lattice Coordinates} \\
\hline
(4d) & $(x,y) ~ (-x, -y) ~ (-x + 1/2,y) ~ (x + 1/2, -y)$ \\
\hline
(2c) & $(1/4,y) ~ (3/4, -y)$ \\
\hline
(2b) & $(0,1/2) ~ (1/2,1/2)$ \\
\hline
(2a) & $(0,0) ~ (1/2,0)$ \\
\hline\hline
\end{tabular}
\end{center}
\end{table_col}

\subsubsection{\label{p2gg} Plane Group \#8: $p2gg$}

\begin{figure}[h]
\begin{center}
\includegraphics[width=\linewidth]{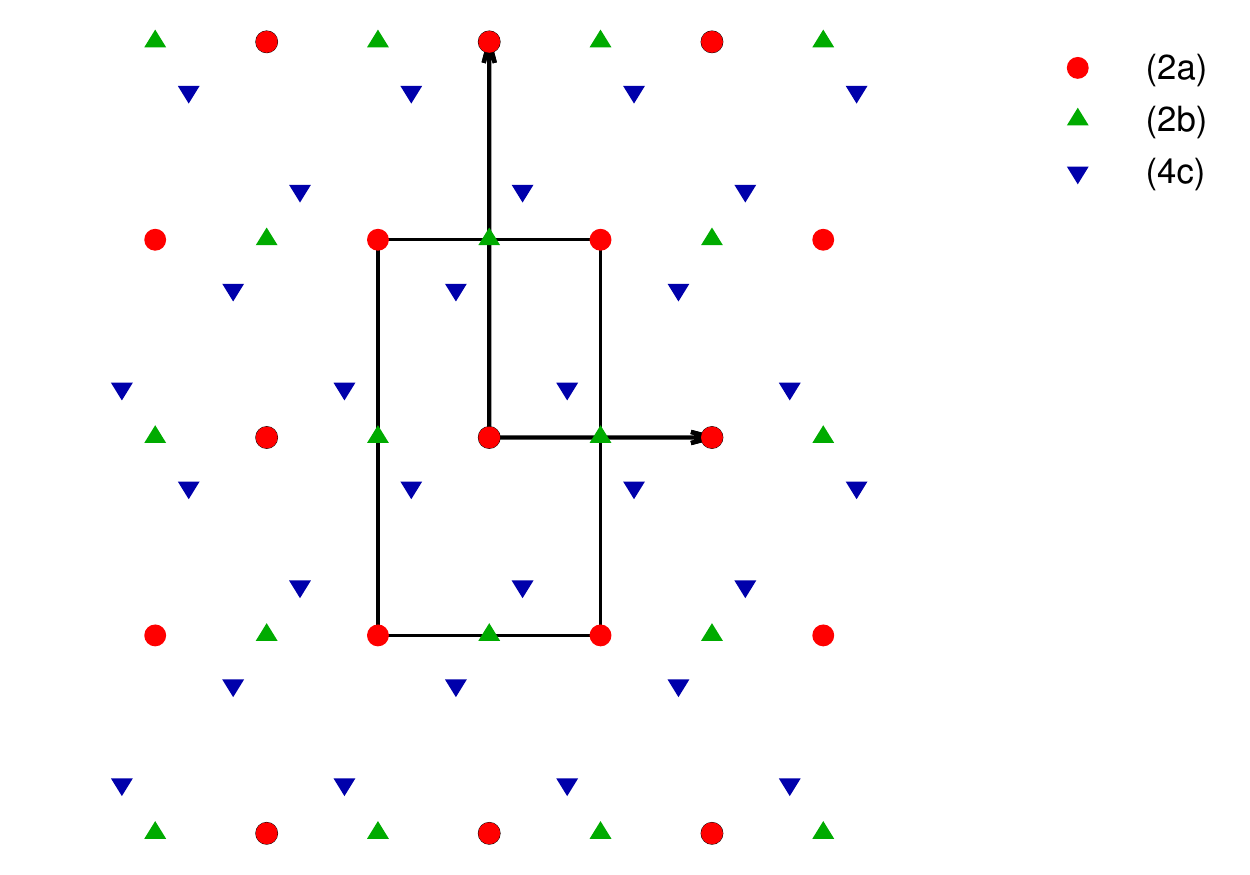}
\end{center}
\caption{\label{fig:p2gg}
\textbf{Possible Wyckoff positions for plane group
\#8, $\bm{p2gg}$.}
The black outline represents the boundary of the Wigner-Seitz cell
for the lattice in Equation~(\ref{lat:rectangle}).}
\end{figure}

This group has two glide reflections as well as the inversion. Its
Wyckoff positions are shown in Figure~\ref{fig:p2gg} and are listed
below.
\begin{table_col}
\begin{center}
\begin{tabular}{||c|c||}
\hline\hline
{\bf Label} & {\bf Lattice Coordinates} \\
\hline
(4c) & \makecell[c]{$(x,y) ~ (-x, -y)$ \\ $(-x + 1/2,y + 1/2) ~ (x + 1/2, -y + 1/2)$} \\
\hline
(2b) & $(1/2, 0) ~ (0,1/2)$ \\
\hline
(2a) & $(0,0) ~ (1/2,1/2)$ \\
\hline\hline
\end{tabular}
\end{center}
\end{table_col}

\subsubsection{\label{c2mm} Plane Group \#9: $c2mm$}

\begin{figure}[h]
\begin{center}
\includegraphics[width=\linewidth]{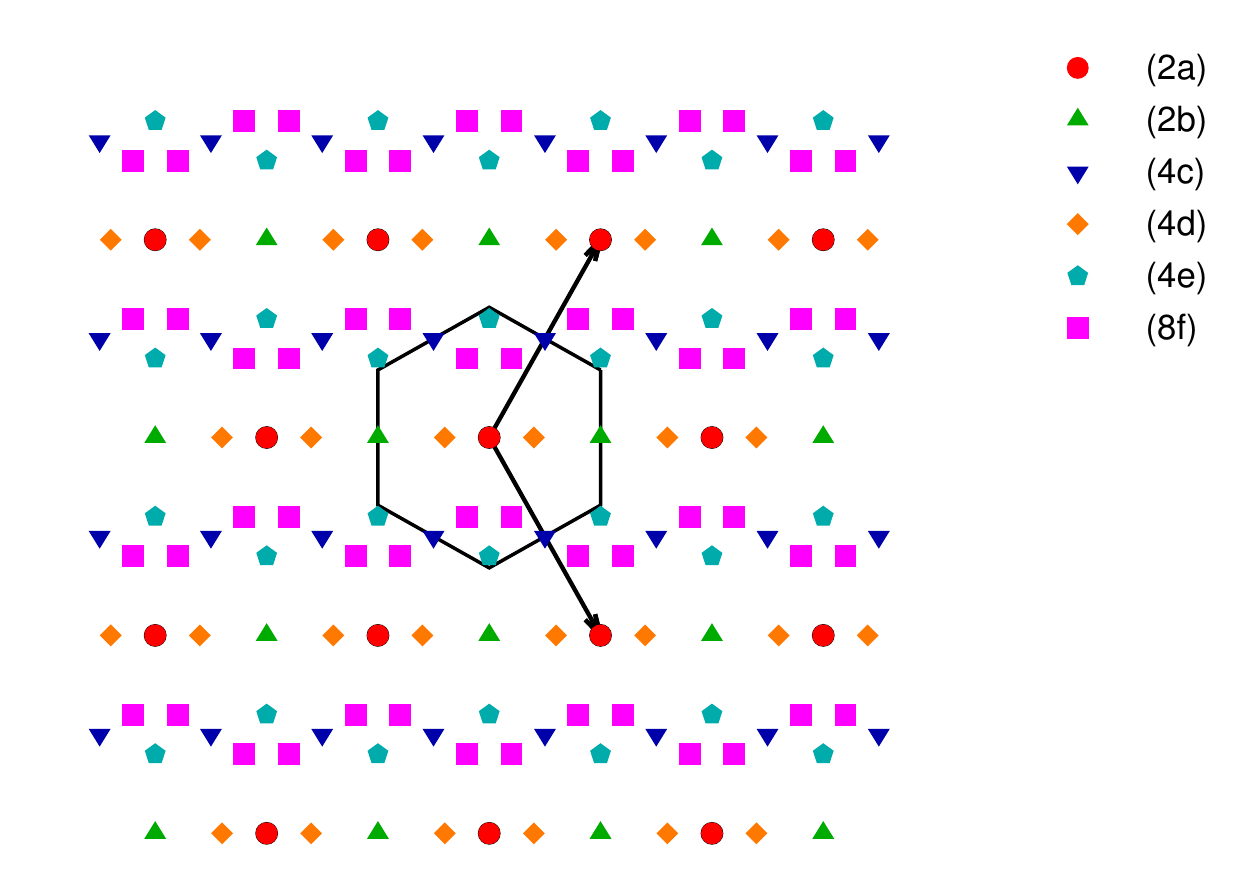}
\end{center}
\caption{\label{fig:c2mm}
\textbf{Possible Wyckoff positions for plane group
\#9, $\bm{c2mm}$.}
The black outline represents the boundary of the Wigner-Seitz cell
for the primitive cell in Equation~(\ref{lat:centered}). The lattice vectors
and Wigner-Seitz cell for the conventional cell
in Equation~(\ref{lat:rectangle}) are identical to those shown in
Figure~\ref{fig:p2mm}.}
\end{figure}

Like plane group \#6, $p2mm$, this group has mirror reflections
around both $x = 0$ and $y = 0$. However, it is a centered lattice,
with the primitive cell defined by Equation~(\ref{lat:centered}) and the
conventional cell defined by Equation~(\ref{lat:rectangle}). The Wyckoff
positions are sketched in Figure~\ref{fig:c2mm} and listed below.
Note that as with group $c1m1$, \#5, we have listed the
coordinates in terms of the conventional unit cell, and included the
translations associated with the centered cell in the list.
\begin{table_col}
\begin{center}
\begin{tabular}{||c|c||}
\hline\hline
{\bf Label} & {\bf Conventional Lattice Coordinates} \\
\hline
(8f) & \makecell[c]{$(x,y) ~ (-x, -y) ~ (-x, y) ~ (x, -y)$ \\
$- - - - - - - - - - - - - - - - - - - - -$ \\
$(x+ 1/2, y + 1/2) ~ (-x+ 1/2, -y + 1/2)$ \\
$(-x+ 1/2, y + 1/2) ~ (x+ 1/2, -y + 1/2)$} \\
\hline
(4e) & \makecell[c]{$(0,y) ~ (0,-y)$ \\
$- - - - - - - - - - - - - - - - - - - - -$ \\
$(1/2,y + 1/2) ~ (1/2,- y + 1/2)$} \\
\hline
(4d) & \makecell[c]{$(x,0) ~ (-x,0)$ \\
$- - - - - - - - - - - - - - - - - - - - -$ \\
$(x + 1/2,1/2) ~ (-x + 1/2,1/2)$} \\
\hline
(4c) & \makecell[c]{$(1/4,1/4) ~ (3/4,1/4)$ \\
$- - - - - - - - - - - - - - - - - - - - -$ \\
$(3/4,3/4) ~ (1/4,3/4)$} \\
\hline
(2b) & \makecell[c]{$(0,1/2)$ \\
$- - - - - - - - - - - - - - - - - - - - -$ \\
$(1/2, 0)$} \\
\hline
(2a) & \makecell[c]{$(0,0)$ \\
$- - - - - - - - - - - - - - - - - - - - -$ \\
$(1/2, 1/2)$} \\
\hline\hline
\end{tabular}
\end{center}
\end{table_col}

\subsection{\label{square} The Square Crystal System}

All of the lattices in this system remain unchanged when rotated by
$90^\circ$ about the origin, so that the crystal obeys the relationship
\begin{equation}
f(x,y) = f(-y,x) = f(-x,-y) = f(y,-x) ~ ,
\end{equation}
automatically including the inversion from Equation~(\ref{equ:inverse}).
They thus form perfect squares, with the
primitive cell given by Equation~(\ref{lat:square}).

\subsubsection{\label{p4} Plane Group \#10: $p4$}

\begin{figure}[h]
\begin{center}
\includegraphics[width=\linewidth]{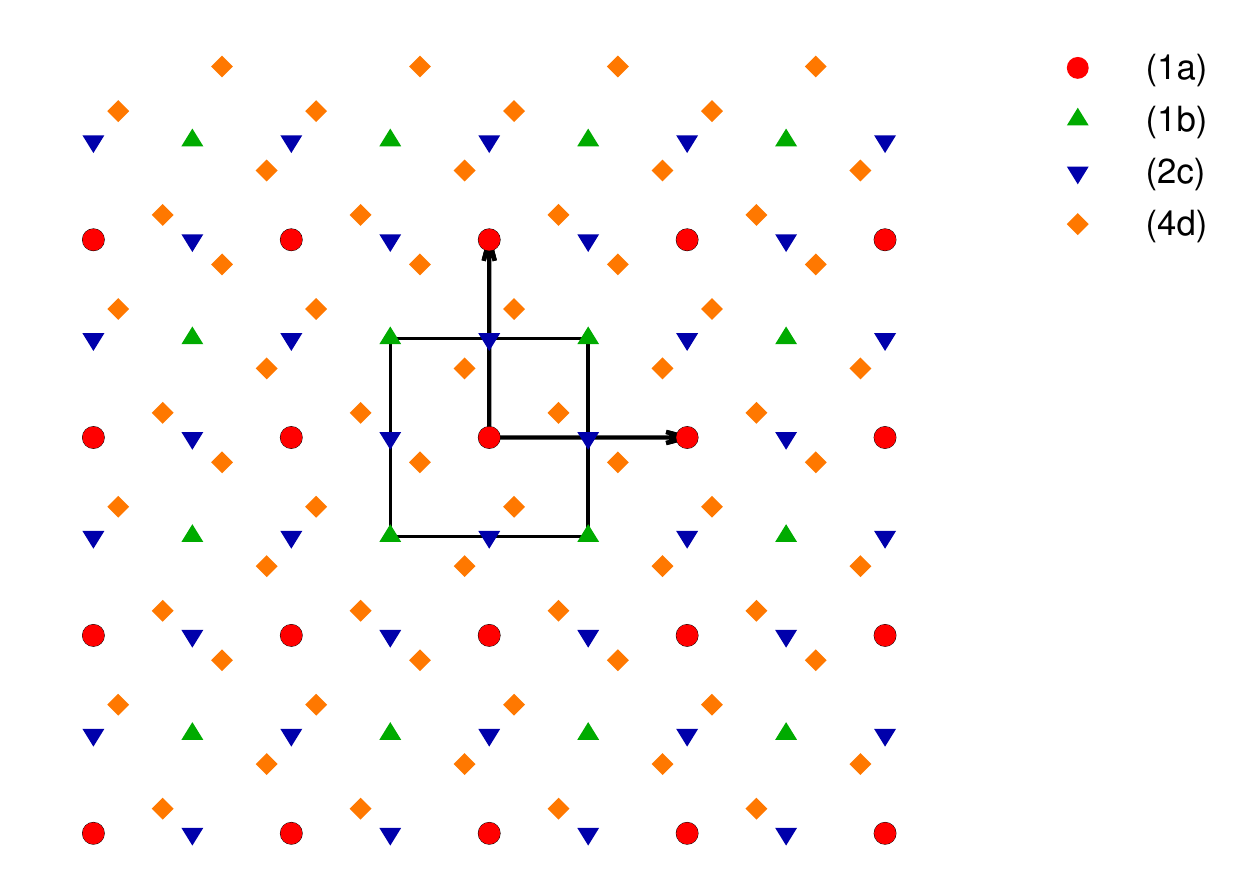}
\end{center}
\caption{\label{fig:p4}
\textbf{Possible Wyckoff positions for plane group
\#10, $\bm{p4}$.}
The black outline represents the boundary of the Wigner-Seitz cell for the
lattice in Equation~(\ref{lat:square}). As with all of the square plane
groups, this group is invariant with respect to a rotation of
$90^\circ$ about the origin and contains the inversion.}
\end{figure}

This is the simplest plane group, with no reflections or glide
reflections. The Wyckoff positions are sketched in
Figure~\ref{fig:p4} and listed below.

\begin{table_col}
\begin{center}
\begin{tabular}{||c|c||}
\hline\hline
{\bf Label} & {\bf Lattice Coordinates} \\
\hline
(4d) & $(x,y) ~ (-x, -y) ~ (-y,x) ~ (y,-x)$ \\
\hline
(2c) & $(1/2,0) ~ (0,1/2)$ \\
\hline
(1b) & $(1/2,1/2)$ \\
\hline
(1a) & $(0,0)$ \\
\hline\hline
\end{tabular}
\end{center}
\end{table_col}

\subsubsection{\label{p4mm} Plane Group \#11: $p4mm$}

\begin{figure}[h]
\begin{center}
\includegraphics[width=\linewidth]{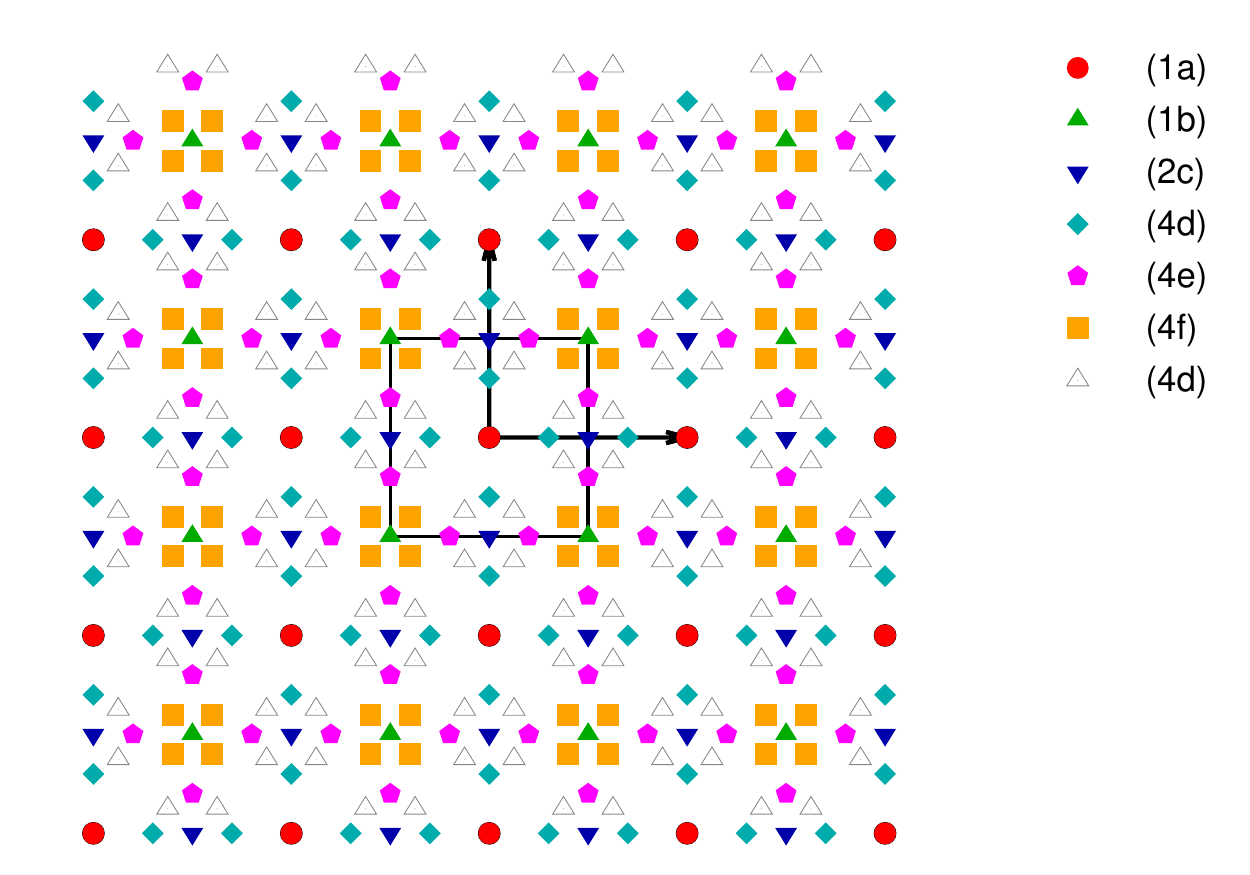}
\end{center}
\caption{\label{fig:p4mm}
\textbf{Possible Wyckoff positions for plane group
\#11, $\bm{p4mm}$.}
The black outline represents the boundary of the Wigner-Seitz cell for the
lattice in Equation~(\ref{lat:square}).}
\end{figure}

This square plane group includes reflections about $x = 0$ and $y =0$.
When the $90^\circ$ rotations are included, this also generates mirror
reflections around the lines $y = \pm x$, so this group admits
operations of the form
\begin{equation}
\label{equ:4mirror}
f(x,y) = f(-x,y) = f(x,-y) = f(y,x) = \cdots.
\end{equation}
The Wyckoff positions are
sketched in Figure~\ref{fig:p4mm} and listed below.
\begin{table_col}
\begin{center}
\begin{tabular}{||c|c||}
\hline\hline
{\bf Label} & {\bf Lattice Coordinates} \\
\hline
(8g) & \makecell[c]{$(x,y) ~ (-x, -y) ~ (-y,x) ~ (y,-x)$ \\
$(-x,y) ~ (x,-y) ~ (y,x) ~ (-y,-x)$} \\
\hline
(4f) & $(x,x) ~ (-x,-x) ~ (-x,x) ~ (x,-x)$ \\
\hline
(4e) & $(x,1/2) ~ (-x,1/2) ~ (1/2,x) ~ (1/2,-x)$ \\
\hline
(4d) & $(x,0) ~ (-x,0) ~ (0,x) ~ (0,-x)$ \\
\hline
(2c) & $(1/2,0) ~ (0,1/2)$ \\
\hline
(1b) & $(1/2,1/2)$ \\
\hline
(1a) & $(0,0)$ \\
\hline\hline
\end{tabular}
\end{center}
\end{table_col}

\subsubsection{\label{p4gm} Plane Group \#12: $p4gm$}

\begin{figure}[h]
\begin{center}
\includegraphics[width=\linewidth]{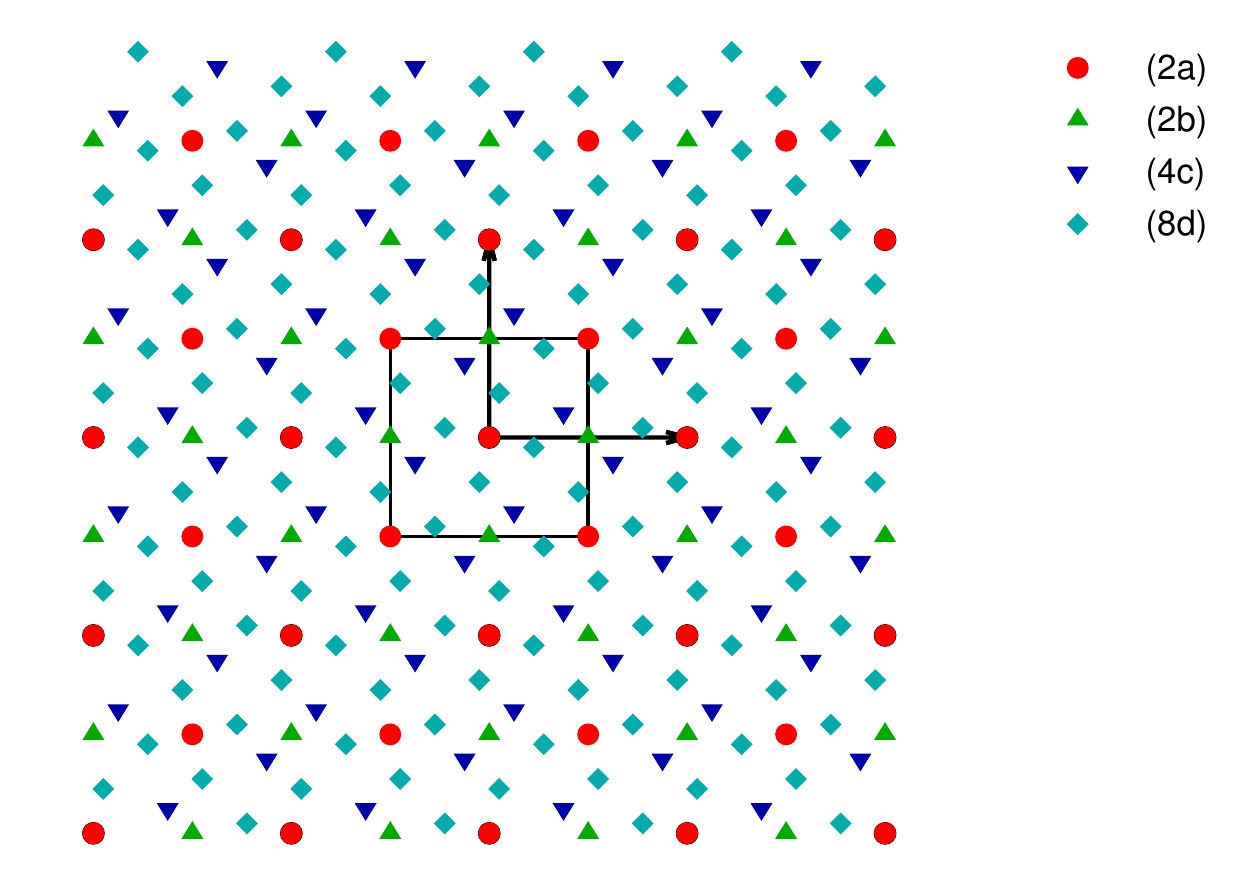}
\end{center}
\caption{\label{fig:p4gm}
\textbf{Possible Wyckoff positions for plane group
\#12, $\bm{p4gm}$.}
The black outline represents the boundary of the Wigner-Seitz cell for the
lattice in Equation~(\ref{lat:square}).}
\end{figure}

This group contains reflections, but they are not centered on the
center of rotation, as they are with plane group \#11. The Wyckoff
positions are listed below and sketched in Figure~\ref{fig:p4gm}.
\begin{center}
\begin{tabular}{||c|c||}
\hline\hline
{\bf Label} & {\bf Lattice Coordinates} \\
\hline
(8d) & \makecell[c]{$(x,y) ~ (-x, -y) ~ (-y,x) ~ (y,-x)$ \\
$(-x+1/2,y+1/2) ~ (x+1/2,-y+1/2)$ \\
$(y+1/2,x+1/2) ~ (-y+1/2,-x+1/2)$} \\
\hline
(4c) & \makecell[c]{$(x,x+1/2) ~ (-x,-x+1/2)$ \\
$(-x+1/2,x) ~ (x+1/2,-x)$} \\
\hline
(2b) & $(1/2,0) ~ (0,1/2)$ \\
\hline
(2a) & $(0,0) ~ (1/2,1/2)$ \\
\hline\hline
\end{tabular}
\end{center}

\subsection{\label{trigonal} The Trigonal Crystal System}

All of the plane groups in the trigonal crystal system are described
by the primitive vectors in Equation~(\ref{lat:hex}) and are invariant under
$120^\circ$ rotations about the origin. This system explicitly
excludes groups which are invariant with respect to $60^\circ$
rotations, and so none of these groups contains the inversion
operation in Equation~(\ref{equ:inverse}).

\subsubsection{\label{p3} Plane Group \#13: $p3$}

\begin{figure}[h]
\begin{center}
\includegraphics[width=\linewidth]{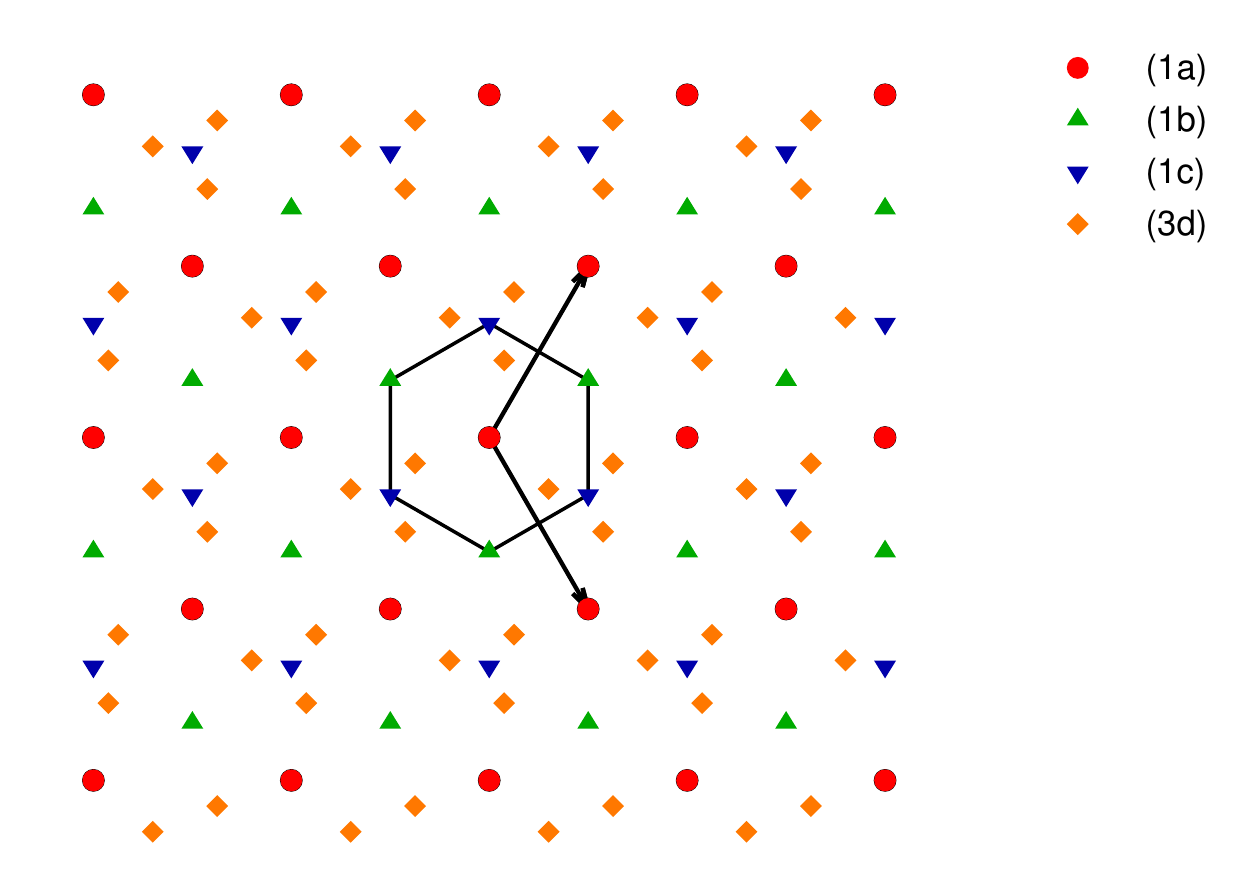}
\end{center}
\caption{\label{fig:p3}
\textbf{Possible Wyckoff positions for plane group
\#13, $\bm{p3}$.}
The black outline represents the boundary of the Wigner-Seitz cell for the
lattice in Equation~(\ref{lat:hex}). Note the 3-fold rotation axis about the
origin.}
\end{figure}

This is the simplest of the trigonal groups, with only the $120^\circ$
symmetry operation. The Wyckoff positions are listed below, and
sketched in Figure~\ref{fig:p3}.
\begin{table_col}
\begin{center}
\begin{tabular}{||c|c||}
\hline\hline
{\bf Label} & {\bf Lattice Coordinates} \\
\hline
(3d) & $(x,y) ~ (- y, x-y) ~ (-x+y,-x)$ \\
\hline
(1c) & $(2/3,1/3)$ \\
\hline
(1b) & $(1/3,2/3)$ \\
\hline
(1a) & $(0,0)$ \\
\hline\hline
\end{tabular}
\end{center}
\end{table_col}

\subsubsection{\label{p3m1} Plane Group \#14: $p3m1$}

\begin{figure}[h]
\begin{center}
\includegraphics[width=\linewidth]{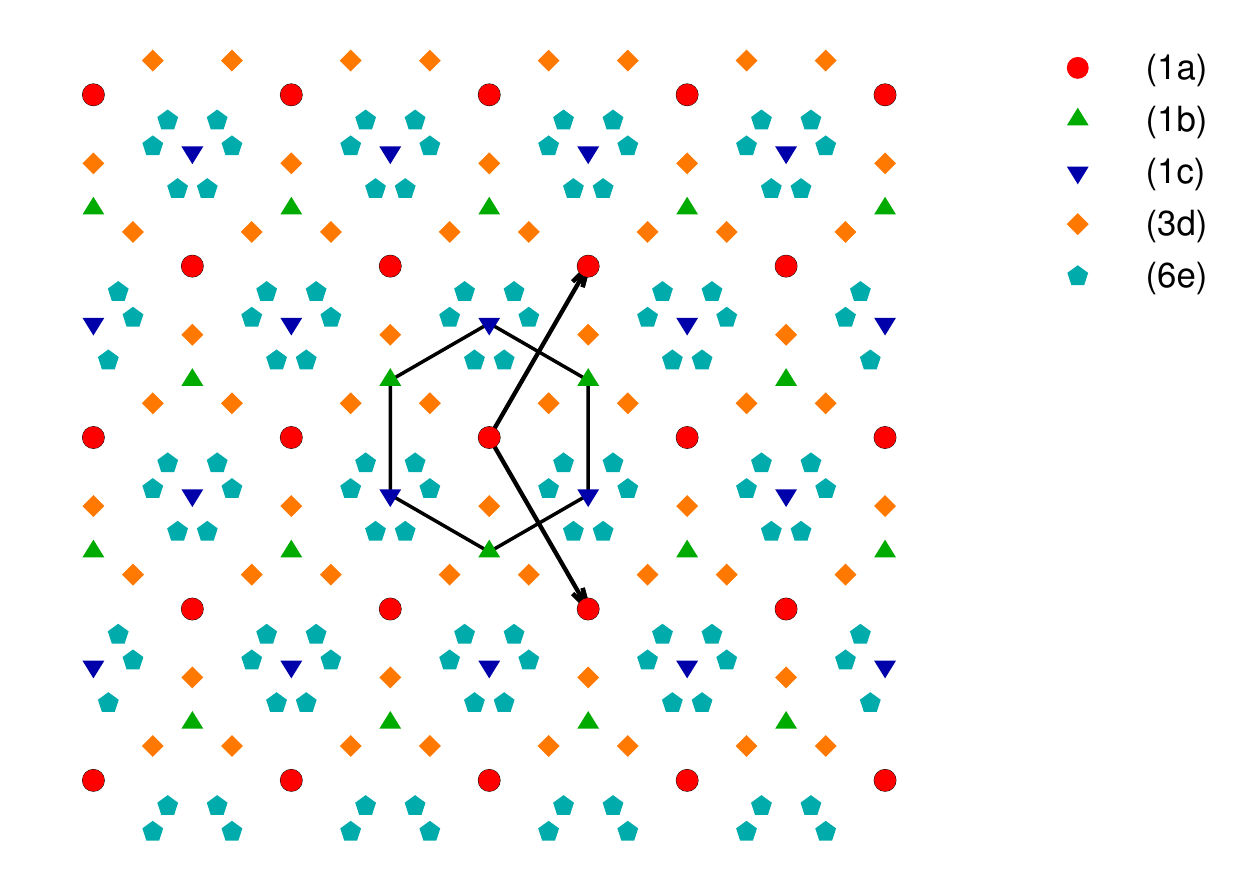}
\end{center}
\caption{\label{fig:p3m1}
\textbf{Possible Wyckoff positions for plane group
\#14, $\bm{p3m1}$.}
The black outline represents the boundary of the Wigner-Seitz
cell for the lattice in Equation~(\ref{lat:hex}). Note the reflection about
the Cartesian $y$ axis.}
\end{figure}

This group includes a reflection about the Cartesian $y$-axis ({\em
not} the $x$ or $y$ {\em lattice} coordinates). The Wyckoff
positions are listed below, and sketched in Figure~\ref{fig:p3m1}.
\begin{table_col}
\begin{center}
\begin{tabular}{||c|c||}
\hline\hline
{\bf Label} & {\bf Lattice Coordinates} \\
\hline
(6e) & \makecell[c]{$(x,y) ~ (- y, x-y) ~ (-x+y,-x)$ \\
$(-y,-x) ~ (-x + y, y) ~ (x, x-y)$} \\
\hline
(3d) & $(x,-x) ~ (x, 2x) ~ (-2x,-x)$ \\
\hline
(1c) & $(2/3,1/3)$ \\
\hline
(1b) & $(1/3,2/3)$ \\
\hline
(1a) & $(0,0)$ \\
\hline\hline
\end{tabular}
\end{center}
\end{table_col}

\subsubsection{\label{p31m} Plane Group \#15: $p31m$}

\begin{figure}[h]
\begin{center}
\includegraphics[width=\linewidth]{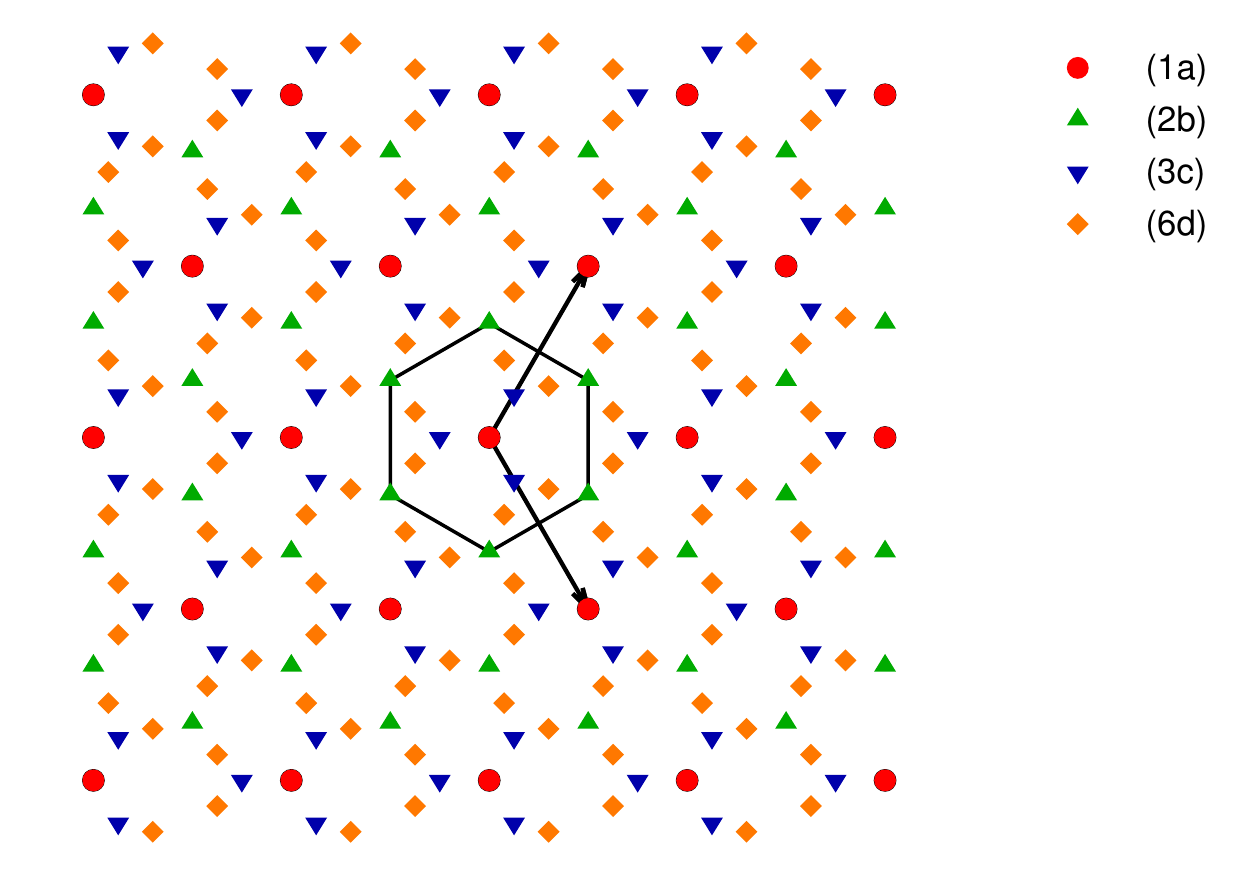}
\end{center}
\caption{\label{fig:p31m}
\textbf{Possible Wyckoff positions for plane group
\#15, $\bm{p31m}$.}
The black outline represents the boundary of the Wigner-Seitz
cell for the lattice in Equation~(\ref{lat:hex}). Note the reflection about
the Cartesian $x$ axis.}
\end{figure}

This is similar to plane group \#14, but now, instead of a reflection about the $y$-axis,
the group includes a reflection about the Cartesian $x$-axis.
The Wyckoff positions are listed below, and sketched in Figure~\ref{fig:p31m}.
\begin{table_col}
\begin{center}
\begin{tabular}{||c|c||}
\hline\hline
{\bf Label} & {\bf Lattice Coordinates} \\
\hline
(6d) & \makecell[c]{$(x,y) ~ (- y, x-y) ~ (-x+y,-x)$ \\
$(y,x) ~ (x - y,-y) ~ (-x, -x+y)$} \\
\hline
(3c) & $(x,0) ~ (0,x) ~ (-x,-x)$ \\
\hline
(2b) & $(1/3,2/3) ~ (2/3,1/3)$ \\
\hline
(1a) & $(0,0)$ \\
\hline\hline
\end{tabular}
\end{center}
\end{table_col}

\subsection{\label{hexagonal} The Hexagonal Crystal System}

Like the trigonal crystal system, the lattice associated with the
hexagonal crystal system is described by the primitive vectors
in Equation~(\ref{lat:hex}), but here both groups are invariant under a $60^\circ$
degree rotation, and so include the inversion operation
in Equation~(\ref{equ:inverse}).

\subsubsection{\label{p6} Plane Group \#16: $p6$}

This group includes the $60^\circ$ rotation operation, but no
reflections. The Wyckoff positions are given below, and described
graphically in Figure~\ref{fig:p6}.
\begin{table_col}
\begin{center}
\begin{tabular}{||c|c||}
\hline\hline
{\bf Label} & {\bf Lattice Coordinates} \\
\hline
(6d) & \makecell[c]{$(x,y) ~ (-y, x-y) ~ (-x+y,-x)$ \\
$(-x,-y) ~ (y,-x+y) ~ (x-y,x)$} \\
\hline
(3c) & $(1/2,0) ~ (0,1/2) ~ (1/2,1/2)$ \\
\hline
(2b) & $(1/3,2/3) ~ (2/3,1/3)$ \\
\hline
(1a) & $(0,0)$ \\
\hline\hline
\end{tabular}
\end{center}
\end{table_col}

\begin{figure}[h]
\begin{center}
\includegraphics[width=\linewidth]{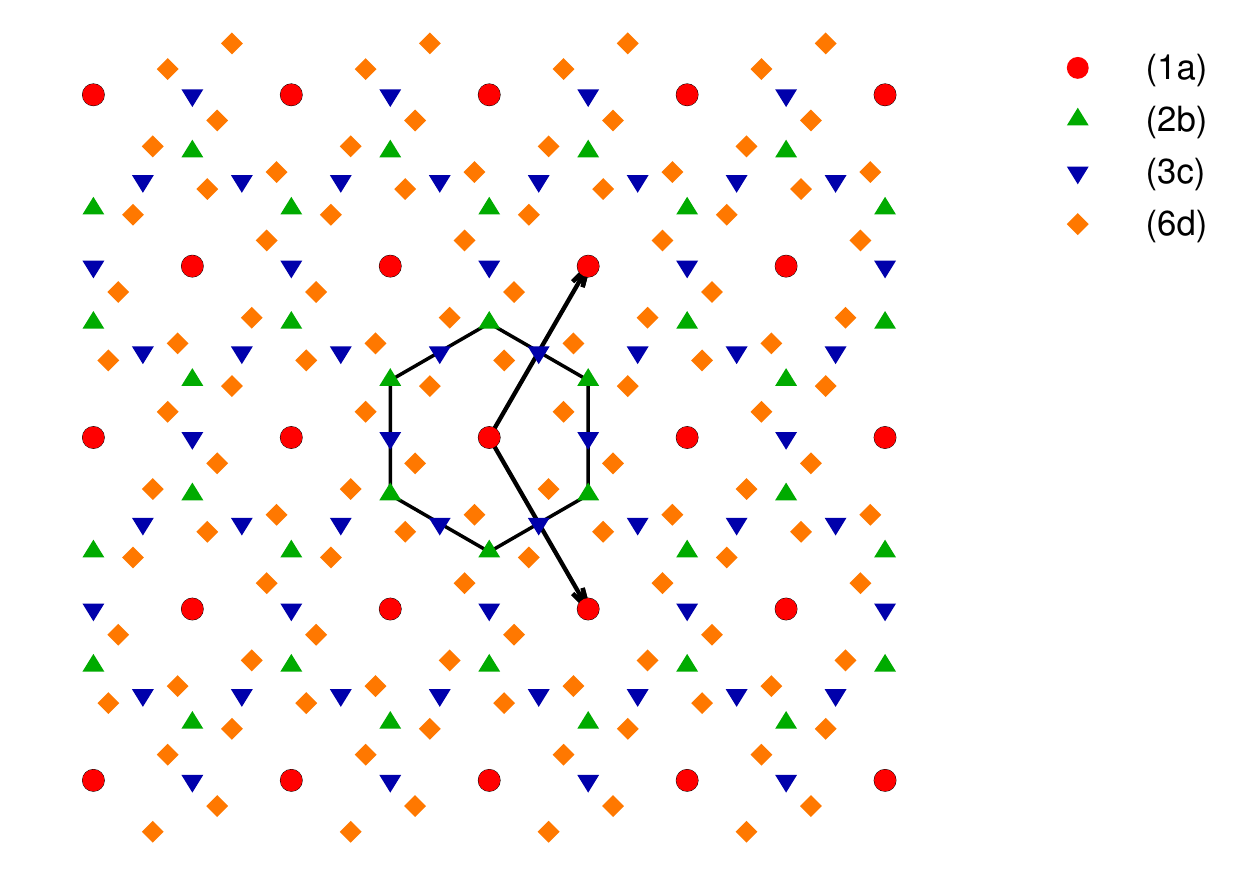}
\end{center}
\caption{\label{fig:p6}
\textbf{Possible Wyckoff positions for plane group
\#16, $\bm{p6}$.}
The black outline represents the boundary of the Wigner-Seitz
cell for the lattice in Equation~(\ref{lat:hex}). Note the $60^\circ$
rotation symmetry about the origin.}
\end{figure}

\subsubsection{\label{p6mm} Plane Group \#17: $p6mm$}

The final plane group includes the $60^\circ$ rotation operation, as
well as reflections along both the Cartesian axes. The Wyckoff
positions are given below, and described graphically in
Figure~\ref{fig:p6mm}.
\begin{table_col}
\begin{center}
\begin{tabular}{||c|c||}
\hline\hline
{\bf Label} & {\bf Lattice Coordinates} \\
\hline
(12f) & \makecell[c]{$(x,y) ~ (-y, x-y) ~ (-x+y,-x) ~ (-x,-y)$ \\
$(y,-x+y) ~ (x-y,x) ~ (-y,-x) ~ (-x+y,y)$ \\
$(x,x-y) ~ (y,x) ~ (x-y,-y) ~ (-x,-x+y)$} \\
\hline
(6e) & \makecell[c]{$(x,-x) ~ (x,2x) ~ (-2x,-x)$ \\
$(-x,x) ~ (-x,-2x) ~ (2x,x)$} \\
\hline
(6d) & \makecell[c]{$(x,0) ~ (0,x) ~ (-x,-x)$ \\
$(-x,0) ~ (0,-x) ~ (x,x)$} \\
\hline
(3c) & $(1/2,0) ~ (0,1/2) ~ (1/2,1/2)$ \\
\hline
(2b) & $(1/3,2/3) ~ (2/3,1/3)$ \\
\hline
(1a) & $(0,0)$ \\
\hline\hline
\end{tabular}
\end{center}
\end{table_col}

\begin{figure}[h]
\begin{center}
\includegraphics[width=\linewidth]{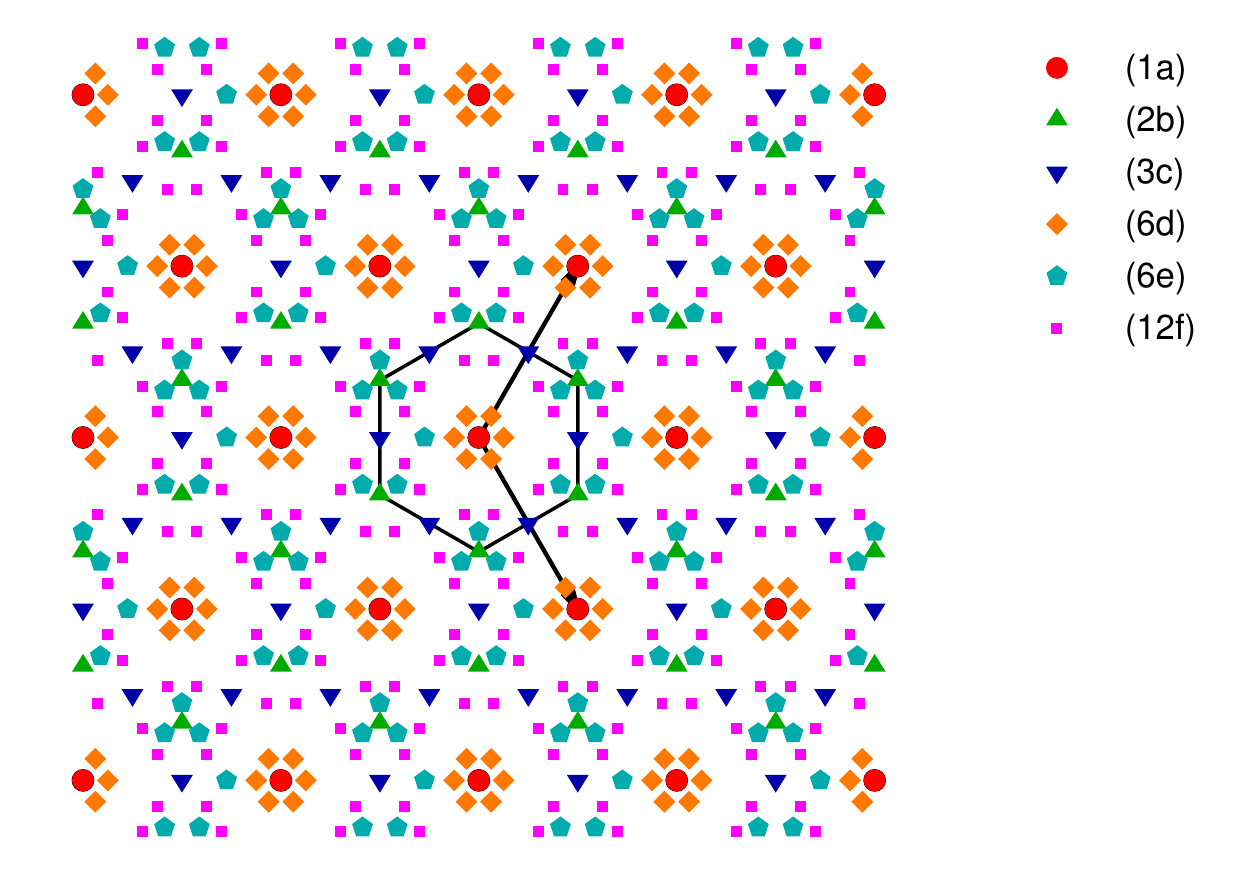}
\end{center}
\caption{\label{fig:p6mm}
\textbf{Possible Wyckoff positions for plane group
\#17, $\bm{p6mm}$.}
The black outline represents the boundary of the Wigner-Seitz
cell for the lattice in Equation~(\ref{lat:hex}). Note the $60^\circ$
rotation symmetry about the origin as well as the reflections
about the Cartesian axes.}
\end{figure}

\section{\label{sec:space_group_notation} Space Group Notation}
There are a variety of space group notation methods.
Each structural prototype page includes
the space group number and International symbol.
The Crystallographic Information Files (\CIF)~\citeintro{hall06:CIF} used
in the library indicates the number and the Hermann-Mauguin and Hall symbols.
Lastly, the \VASP\ \POSCAR\ lists the number and the Hermann-Mauguin and Sch{\"o}nflies
symbols.

The notations we use are
\begin{itemize}
\item Hermann-Mauguin~\citeintro{Hermann_1928,Mauguin_1931} (in \CIF\ and \POSCAR)
\item Hall~\citeintro{Hall_1981} (in \CIF)
\item ``International'', a compact form of the Hermann-Mauguin
notation used in the International Tables of
Crystallography~\citeintro{Brock_2016} (in entry page).
\item Sch\"{o}nflies~\citeintro{Shoenflies_1891,Shoenflies_1923} (in \POSCAR)
\end{itemize}

Note that most notations provide different values for each orientation
of a crystal. A complete list of orientations can be found in Hall
and Grosse-Kunstleve~\citeintro{hall15:sgtable}. In general we use the first
space-group orientation listed for a space group on that page. There
are two exceptions:

\begin{itemize}
\item For some space groups, {\em e.g.} \#227, there are two settings,
reflecting a choice of origin. We always take the second setting,
which places the origin of the real-space coordinate system at an
inversion site.
\item For rhombohedral unit cells we always use the label for the full
hexagonal unit cell (H) rather than the primitive rhombohedral cell
(R), which follows our choice to describe the unit cell in terms of
the hexagonal lattice parameters $a$ and $c$, rather than the
rhombohedral lattice parameters $a$ and $\alpha$.
\end{itemize}

Complete listings of all space group orientations can be found at

\begin{itemize}
\item \begin{flushleft}Concise Space Group Symbols,\end{flushleft} http://cci.lbl.gov/sginfo/hall\_symbols.html
\item \begin{flushleft}Elk Spacegroup Manual,\end{flushleft} http://elk.sourceforge.net/spacegroup.pdf
\end{itemize}

\section{\label{sec:conclusion} Conclusion}
Herein, we present the second part of {\it The \AFLOW\ Library of Crystallographic Prototypes}.
The article includes \NUMPROTOSPARTTWO\ crystallographic prototypes and provides the corresponding structural
information for each entry.
The geometry file for each structure can be generated via \AFLOW\ to facilitate high-throughput
computation of material properties.
This information is also available online at \url{http://www.aflow.org/CrystalDatabase},
where it is combined with the prototypes listed in Part 1.

\onecolumn
{
\centering
{\small{
\captionsetup{width=13cm}
\begin{longtabu}{cccc>$c<$>$c<$}
\caption{\textbf{A list of the various space group notations.} The space group number, orientation, Hermann-Mauguin symbol,
Hall symbol, International symbol, and Sch\"{o}nflies symbol are listed.\label{tab:sg_notation}} \\
\hline
\mbox{Number} & \mbox{Orientation} & \mbox{Hermann-Mauguin} & \mbox{Hall} & \mbox{International} & \mbox{Sch\"{o}nflies} \\
\hline
\endfirsthead
\caption*{Table~\ref*{tab:sg_notation} (continued): \textbf{A list of the various space group notations.} The space group number, orientation, Hermann-Mauguin symbol,
Hall symbol, International symbol, and Sch\"{o}nflies symbol are listed.} \\
\hline
\mbox{Number} & \mbox{Orientation} & \mbox{Hermann-Mauguin} & \mbox{Hall} & \mbox{International} & \mbox{Sch\"{o}nflies} \\
\hline
\endhead
1 & & P 1 & P 1 & P1 & C_{1}^{1} \\
2 & & P -1 & -P 1 & P\overline{1} & C_{i}^{1} \\
3 & b & P 1 2 1 & P 2y & P2 & C_{2}^{1} \\
4 & b & P 1 21 1 & P 2yb & P2_{1} & C_{2}^{2} \\
5 & b1 & C 1 2 1 & C 2y & C2 & C_{2}^{3} \\
6 & b & P 1 m 1 & P -2y & Pm & C_{s}^{1} \\
7 & b1 & P 1 c 1 & P -2yc & Pc & C_{s}^{2} \\
8 & b1 & C 1 m 1 & C -2y & Cm & C_{s}^{3} \\
9 & b1 & C 1 c 1 & C -2yc & Cc & C_{s}^{4} \\
10 & b & P 1 2/m 1 & -P 2y & P2/m & C_{2h}^{1} \\
11 & b & P 1 21/m 1 & -P 2yb & P2_{1}/m & C_{2h}^{2} \\
12 & b1 & C 1 2/m 1 & -C 2y & C2/m & C_{2h}^{3} \\
13 & b1 & P 1 2/c 1 & -P 2yc & P2/c & C_{2h}^{4} \\
14 & b1 & P 1 21/c 1 & -P 2ybc & P2_{1}/c & C_{2h}^{5} \\
15 & b1 & C 1 2/c 1 & -C 2yc & C2/c & C_{2h}^{6} \\
16 & & P 2 2 2 & P 2 2 & P222 & D_{2}^{1} \\
17 & & P 2 2 21 & P 2c 2 & P222_{1} & D_{2}^{2} \\
18 & & P 21 21 2 & P 2 2ab & P2_{1}2_{1}2 & D_{2}^{3} \\
19 & & P 21 21 21 & P 2ac 2ab & P2_{1}2_{1}2_{1} & D_{2}^{4} \\
20 & & C 2 2 21 & C 2c 2 & C222_{1} & D_{2}^{5} \\
21 & & C 2 2 2 & C 2 2 & C222 & D_{2}^{6} \\
22 & & F 2 2 2 & F 2 2 & F222 & D_{2}^{7} \\
23 & & I 2 2 2 & I 2 2 & I222 & D_{2}^{8} \\
24 & & I 21 21 21 & I 2b 2c & I2_{1}2_{1}2_{1} & D_{2}^{9} \\
25 & & P m m 2 & P 2 -2 & Pmm2 & C_{2v}^{1} \\
26 & & P m c 21 & P 2c -2 & Pmc2_{1} & C_{2v}^{2} \\
27 & & P c c 2 & P 2 -2c & Pcc2 & C_{2v}^{3} \\
28 & & P m a 2 & P 2 -2a & Pma2 & C_{2v}^{4} \\
29 & & P c a 21 & P 2c -2ac & Pca2_{1} & C_{2v}^{5} \\
30 & & P n c 2 & P 2 -2bc & Pnc2 & C_{2v}^{6} \\
31 & & P m n 21 & P 2ac -2 & Pmn2_{1} & C_{2v}^{7} \\
32 & & P b a 2 & P 2 -2ab & Pba2 & C_{2v}^{8} \\
33 & & P n a 21 & P 2c -2n & Pna2_{1} & C_{2v}^{9} \\
34 & & P n n 2 & P 2 -2n & Pnn2 & C_{2v}^{10} \\
35 & & C m m 2 & C 2 -2 & Cmm2 & C_{2v}^{11} \\
36 & & C m c 21 & C 2c -2 & Cmc2_{1} & C_{2v}^{12} \\
37 & & C c c 2 & C 2 -2c & Ccc2 & C_{2v}^{13} \\
38 & & A m m 2 & A 2 -2 & Amm2 & C_{2v}^{14} \\
39 & & A b m 2 & A 2 -2c & Aem2 & C_{2v}^{15} \\
40 & & A m a 2 & A 2 -2a & Ama2 & C_{2v}^{16} \\
41 & & A b a 2 & A 2 -2ac & Aea2 & C_{2v}^{17} \\
42 & & F m m 2 & F 2 -2 & Fmm2 & C_{2v}^{18} \\
43 & & F d d 2 & F 2 -2d & Fdd2 & C_{2v}^{19} \\
44 & & I m m 2 & I 2 -2 & Imm2 & C_{2v}^{20} \\
45 & & I b a 2 & I 2 -2c & Iba2 & C_{2v}^{21} \\
46 & & I m a 2 & I 2 -2a & Ima2 & C_{2v}^{22} \\
47 & & P 2/m 2/m 2/m & -P 2 2 & Pmmm & D_{2h}^{1} \\
48 & 2 & P 2/n 2/n 2/n:2 & -P 2ab 2bc & Pnnn & D_{2h}^{2} \\
49 & & P 2/c 2/c 2/m & -P 2 2c & Pccm & D_{2h}^{3} \\
50 & 2 & P 2/b 2/a 2/n:2 & -P 2ab 2b & Pban & D_{2h}^{4} \\
51 & & P 21/m 2/m 2/a & -P 2a 2a & Pmma & D_{2h}^{5} \\
52 & & P 2/n 21/n 2/a & -P 2a 2bc & Pnna & D_{2h}^{6} \\
53 & & P 2/m 2/n 21/a & -P 2ac 2 & Pmna & D_{2h}^{7} \\
54 & & P 21/c 2/c 2/a & -P 2a 2ac & Pcca & D_{2h}^{8} \\
55 & & P 21/b 21/a 2/m & -P 2 2ab & Pbam & D_{2h}^{9} \\
56 & & P 21/c 21/c 2/n & -P 2ab 2ac & Pccn & D_{2h}^{10} \\
57 & & P 2/b 21/c 21/m & -P 2c 2b & Pbcm & D_{2h}^{11} \\
58 & & P 21/n 21/n 2/m & -P 2 2n & Pnnm & D_{2h}^{12} \\
59 & 2 & P 21/m 21/m 2/n:2 & -P 2ab 2a & Pmmn & D_{2h}^{13} \\
60 & & P 21/b 2/c 21/n & -P 2n 2ab & Pbcn & D_{2h}^{14} \\
61 & & P 21/b 21/c 21/a & -P 2ac 2ab & Pbca & D_{2h}^{15} \\
62 & & P 21/n 21/m 21/a & -P 2ac 2n & Pnma & D_{2h}^{16} \\
63 & & C 2/m 2/c 21/m & -C 2c 2 & Cmcm & D_{2h}^{17} \\
64 & & C 2/m 2/c 21/a & -C 2bc 2 & Cmca & D_{2h}^{18} \\ 
65 & & C 2/m 2/m 2/m & -C 2 2 & Cmmm & D_{2h}^{19} \\
66 & & C 2/c 2/c 2/m & -C 2 2c & Cccm & D_{2h}^{20} \\
67 & & C 2/m 2/m 2/a & -C 2b 2 & Cmma & D_{2h}^{21} \\ 
68 & 2 & C 2/c 2/c 2/a:2 & -C 2b 2bc & Ccca & D_{2h}^{22} \\ 
69 & & F 2/m 2/m 2/m & -F 2 2 & Fmmm & D_{2h}^{23} \\
70 & 2 & F 2/d 2/d 2/d:2 & -F 2uv 2vw & Fddd & D_{2h}^{24} \\
71 & & I 2/m 2/m 2/m & -I 2 2 & Immm & D_{2h}^{25} \\
72 & & I 2/b 2/a 2/m & -I 2 2c & Ibam & D_{2h}^{26} \\
73 & & I 2/b 2/c 2/a & -I 2b 2c & Ibca & D_{2h}^{27} \\
74 & & I 2/m 2/m 2/a & -I 2b 2 & Imma & D_{2h}^{28} \\
75 & & P 4 & P 4 & P4 & C_{4}^{1} \\
76 & & P 41 & P 4w & P4_{1} & C_{4}^{2} \\
77 & & P 42 & P 4c & P4_{2} & C_{4}^{3} \\
78 & & P 43 & P 4cw & P4_{3} & C_{4}^{4} \\
79 & & I 4 & I 4 & I4 & C_{4}^{5} \\
80 & & I 41 & I 4bw & I4_{1} & C_{4}^{6} \\
81 & & P -4 & P -4 & P\overline{4}& S_{4}^{1} \\
82 & & I -4 & I -4 & I\overline{4}& S_{4}^{2} \\
83 & & P 4/m & -P 4 & P4/m & C_{4h}^{1} \\
84 & & P 42/m & -P 4c & P4_{2}/m & C_{4h}^{2} \\
85 & 2 & P 4/n:2 & -P 4a & P4/n & C_{4h}^{3} \\
86 & 2 & P 42/n:2 & -P 4bc & P4_{2}/n & C_{4h}^{4} \\ 
87 & & I 4/m & -I 4 & I4/m & C_{4h}^{5} \\
88 & 2 & I 41/a:2 & -I 4ad & I4_{1}/a & C_{4h}^{6} \\
89 & & P 4 2 2 & P 4 2 & P422 & D_{4}^{1} \\
90 & & P 4 21 2 & P 4ab 2ab & P42_{1}2 & D_{4}^{2} \\ 
91 & & P 41 2 2 & P 4w 2c & P4_{1}22 & D_{4}^{3} \\
92 & & P 41 21 2 & P 4abw 2nw & P4_{1}2_{1}2 & D_{4}^{4} \\
93 & & P 42 2 2 & P 4c 2 & P4_{2}22 & D_{4}^{5} \\
94 & & P 42 21 2 & P 4n 2n & P4_{2}2_{1}2 & D_{4}^{6} \\
95 & & P 43 2 2 & P 4cw 2c & P4_{3}22 & D_{4}^{7} \\
96 & & P 43 21 2 & P 4nw 2abw & P4_{3}2_{1}2 & D_{4}^{8} \\
97 & & I 4 2 2 & I 4 2 & I422 & D_{4}^{9} \\
98 & & I 41 2 2 & I 4bw 2bw & I4_{1}22 & D_{4}^{10} \\
99 & & P 4 m m & P 4 -2 & P4mm & C_{4v}^{1} \\
100 & & P 4 b m & P 4 -2ab & P4bm & C_{4v}^{2} \\
101 & & P 42 c m & P 4c -2c & P4_{2}cm & C_{4v}^{3} \\
102 & & P 42 n m & P 4n -2n & P4_{2}nm & C_{4v}^{4} \\
103 & & P 4 c c & P 4 -2c & P4cc & C_{4v}^{5} \\
104 & & P 4 n c & P 4 -2n & P4nc & C_{4v}^{6} \\
105 & & P 42 m c & P 4c -2 & P4_{2}mc & C_{4v}^{7} \\
106 & & P 42 b c & P 4c -2ab & P_{2}bc & C_{4v}^{8} \\
107 & & I 4 m m & I 4 -2 & I4mm & C_{4v}^{9} \\
108 & & I 4 c m & I 4 -2c & I4cm & C_{4v}^{10} \\
109 & & I 41 m d & I 4bw -2 & I4_{1}md & C_{4v}^{11} \\
110 & & I 41 c d & I 4bw -2c & I4_{1}cd & C_{4v}^{12} \\
111 & & P -4 2 m & P -4 2 & P\overline{4}2m & D_{2d}^{1} \\
112 & & P -4 2 c & P -4 2c & P\overline{4}2c & D_{2d}^{2} \\
113 & & P -4 21 m & P -4 2ab & P\overline{4}2_{1}m & D_{2d}^{3} \\
114 & & P -4 21 c & P -4 2n & P\overline{4}2_{1}c & D_{2d}^{4} \\
115 & & P -4 m 2 & P -4 -2 & P\overline{4}m2 & D_{2d}^{5} \\
116 & & P -4 c 2 & P -4 -2c & P\overline{4}c2 & D_{2d}^{6} \\
117 & & P -4 b 2 & P -4 -2ab & P\overline{4}b2 & D_{2d}^{7} \\
118 & & P -4 n 2 & P -4 -2n & P\overline{4}n2 & D_{2d}^{8} \\
119 & & I -4 m 2 & I -4 -2 & I\overline{4}m2 & D_{2d}^{9} \\
120 & & I -4 c 2 & I -4 -2c & I\overline{4}c2 & D_{2d}^{10} \\
121 & & I -4 2 m & I -4 2 & I\overline{4}2m & D_{2d}^{11} \\
122 & & I -4 2 d & I -4 2bw & I\overline{4}2d & D_{2d}^{12} \\
123 & & P 4/m 2/m 2/m & -P 4 2 & P4/mmm & D_{4h}^{1} \\
124 & & P 4/m 2/c 2/c & -P 4 2c & P4/mcc & D_{4h}^{2} \\
125 & 2 & P 4/n 2/b 2/m:2 & -P 4a 2b & P4/nbm & D_{4h}^{3} \\
126 & 2 & P 4/n 2/n 2/c:2 & -P 4a 2bc & P4/nnc & D_{4h}^{4} \\
127 & & P 4/m 21/b 2/m & -P 4 2ab & P4/mbm & D_{4h}^{5} \\
128 & & P 4/m 21/n 2/c & -P 4 2n & P4/mnc & D_{4h}^{6} \\
129 & 2 & P 4/n 21/m 2/m:2 & -P 4a 2a & P4/nmm & D_{4h}^{7} \\
130 & 2 & P 4/n 21/c 2/c:2 & -P 4a 2ac & P4/ncc & D_{4h}^{8} \\
131 & & P 42/m 2/m 2/c & -P 4c 2 & P4_{2}/mmc & D_{4h}^{9} \\
132 & & P 42/m 2/c 2/m & -P 4c 2c & P4_{2}/mcm & D_{4h}^{10} \\
133 & 2 & P 42/n 2/b 2/c:2 & -P 4ac 2b & P4_{2}/nbc & D_{4h}^{11} \\
134 & 2 & P 42/n 2/n 2/m:2 & -P 4ac 2bc & P4_{2}/nnm & D_{4h}^{12} \\
135 & & P 42/m 21/b 2/c & -P 4c 2ab & P4_{2}/mbc & D_{4h}^{13} \\
136 & & P 42/m 21/n 2/m & -P 4n 2n & P4_{2}/mnm & D_{4h}^{14} \\
137 & 2 & P 42/n 21/m 2/c:2 & -P 4ac 2a & P4_{2}/nmc & D_{4h}^{15} \\
138 & 2 & P 42/n 21/c 2/m:2 & -P 4ac 2ac & P4_{2}/ncm & D_{4h}^{16} \\
139 & & I 4/m 2/m 2/m & -I 4 2 & I4/mmm & D_{4h}^{17} \\
140 & & I 4/m 2/c 2/m & -I 4 2c & I4/mcm & D_{4h}^{18} \\
141 & 2 & I 41/a 2/m 2/d:2 & -I 4bd 2 & I4_{1}/amd & D_{4h}^{19} \\
142 & 2 & I 41/a 2/c 2/d:2 & -I 4bd 2c & I4_{1}/acd & D_{4h}^{20} \\
143 & & P 3 & P 3 & P3 & C_{3}^{1} \\
144 & & P 31 & P 31 & P3_{1} & C_{3}^{2} \\
145 & & P 32 & P 32 & P3_{2} & C_{3}^{3} \\
146 & H & R 3:H & R 3 & R3 & C_{3}^{4} \\ 
147 & & P -3 & -P 3 & P\overline{3} & C_{3i}^{1} \\
148 & H & R -3:H & -R 3 & R\overline{3} & C_{3i}^{2} \\ 
149 & & P 3 1 2 & P 3 2 & P312 & D_{3}^{1} \\
150 & & P 3 2 1 & P 3 2'' & P321 & D_{3}^{2} \\
151 & & P 31 1 2 & P 31 2c (0 0 1) & P3_{1}12 & D_{3}^{3} \\
152 & & P 31 2 1 & P 31 2'' & P3_{1}21 & D_{3}^{4} \\
153 & & P 32 1 2 & P 32 2c (0 0 -1) & P3_{2}12 & D_{3}^{5} \\
154 & & P 32 2 1 & P 32 2'' & P3_{2}21 & D_{3}^{6} \\
155 & H & R 32:H & R 3 2'' & R32 & D_{3}^{7} \\ 
156 & & P 3 m 1 & P 3 -2'' & P3m1 & C_{3v}^{1} \\
157 & & P 3 1 m & P 3 -2 & P31m & C_{3v}^{2} \\
158 & & P 3 c 1 & P 3 -2''c & P3c1 & C_{3v}^{3} \\
159 & & P 3 1 c & P 3 -2c & P31c & C_{3v}^{4} \\
160 & H & R 3 m:H & R 3 -2'' & R3m & C_{3v}^{5} \\
161 & H & R 3 c:H & R 3 -2''c & R3c & C_{3v}^{6} \\ 
162 & & P -3 1 2/m & -P 3 2 & P\overline{3}1m & D_{3d}^{1} \\
163 & & P -3 1 2/c & -P 3 2c & P\overline{3}1c & D_{3d}^{2} \\
164 & & P -3 2/m 1 & -P 3 2'' & P\overline{3}m1 & D_{3d}^{3} \\
165 & & P -3 2/c 1 & -P 3 2''c & P\overline{3}c1 & D_{3d}^{4} \\
166 & H & R -3 2/m:H & -R 3 2'' & R\overline{3}m & D_{3d}^{5} \\
167 & H & R -3 2/c:H & -R 3 2''c & R\overline{3}c & D_{3d}^{6} \\
168 & & P 6 & P 6 & P6 & C_{6}^{1} \\
169 & & P 61 & P 61 & P6_{1} & C_{6}^{2} \\
170 & & P 65 & P 65 & P6_{5} & C_{6}^{3} \\
171 & & P 62 & P 62 & P6_{2} & C_{6}^{4} \\
172 & & P 64 & P 64 & P6_{4} & C_{6}^{5} \\
173 & & P 63 & P 6c & P6_{3} & C_{6}^{6} \\
174 & & P -6 & P -6 & P\overline{6} & C_{3h}^{1} \\
175 & & P 6/m & -P 6 & P6/m & C_{6h}^{1} \\
176 & & P 63/m & -P 6c & P6_{3}/m & C_{6h}^{2} \\
177 & & P 6 2 2 & P 6 2 & P622 & D_{6}^{1} \\
178 & & P 61 2 2 & P 61 2 (0 0 -1) & P6_{1}22 & D_{6}^{2} \\
179 & & P 65 2 2 & P 65 2 (0 0 1) & P6_{5}22 & D_{6}^{3} \\
180 & & P 62 2 2 & P 62 2c (0 0 1) & P6_{2}22 & D_{6}^{4} \\
181 & & P 64 2 2 & P 64 2c (0 0 -1) & P6_{4}22 & D_{6}^{5} \\
182 & & P 63 2 2 & P 6c 2c & P6_{3}22 & D_{6}^{6} \\
183 & & P 6 m m & P 6 -2 & P6mm & C_{6v}^{1} \\
184 & & P 6 c c & P 6 -2c & P6cc & C_{6v}^{2} \\
185 & & P 63 c m & P 6c -2 & P6_{3}cm & C_{6v}^{3} \\
186 & & P 63 m c & P 6c -2c & P6_{3}mc & C_{6v}^{4} \\
187 & & P -6 m 2 & P -6 2 & P\overline{6}m2 & D_{3h}^{1} \\
188 & & P -6 c 2 & P -6c 2 & P\overline{6}c2 & D_{3h}^{2} \\
189 & & P -6 2 m & P -6 -2 & P\overline{6}2m & D_{3h}^{3} \\
190 & & P -6 2 c & P -6c -2c & P\overline{6}2c & D_{3h}^{4} \\
191 & & P 6/m 2/m 2/m & -P 6 2 & P6/mmm & D_{6h}^{1} \\
192 & & P 6/m 2/c 2/c & -P 6 2c & P6/mcc & D_{6h}^{2} \\
193 & & P 63/m 2/c 2/m & -P 6c 2 & P6_{3}/mcm & D_{6h}^{3} \\
194 & & P 63/m 2/m 2/c & -P 6c 2c & P6_{3}/mmc & D_{6h}^{4} \\
195 & & P 2 3 & P 2 2 3 & P23 & T_{}^{1} \\
196 & & F 2 3 & F 2 2 3 & F23 & T_{}^{2} \\
197 & & I 2 3 & I 2 2 3 & I23 & T_{}^{3} \\
198 & & P 21 3 & P 2ac 2ab 3 & P2_{1}3 & T_{}^{4} \\
199 & & I 21 3 & I 2b 2c 3 & I2_{1}3 & T_{}^{5} \\
200 & & P 2/m -3 & -P 2 2 3 & Pm\overline{3} & T_{h}^{1} \\
201 & 2 & P 2/n -3:2 & -P 2ab 2bc 3 & Pn\overline{3} & T_{h}^{2} \\
202 & & F 2/m -3 & -F 2 2 3 & Fm\overline{3} & T_{h}^{3} \\
203 & 2 & F 2/d -3:2 & -F 2uv 2vw 3 & Fd\overline{3} & T_{h}^{4} \\
204 & & I 2/m -3 & -I 2 2 3 & Im\overline{3} & T_{h}^{5} \\
205 & & P 21/a -3 & -P 2ac 2ab 3 & Pa\overline{3} & T_{h}^{6} \\
206 & & I 21/a -3 & -I 2b 2c 3 & Ia\overline{3} & T_{h}^{7} \\
207 & & P 4 3 2 & P 4 2 3 & P432 & O^{1} \\
208 & & P 42 3 2 & P 4n 2 3 & P4_{2}32 & O^{2} \\ 
209 & & F 4 3 2 & F 4 2 3 & F432 & O^{3} \\
210 & & F 41 3 2 & F 4d 2 3 & F4_{1}32 & O^{4} \\
211 & & I 4 3 2 & I 4 2 3 & I432 & O^{5} \\
212 & & P 43 3 2 & P 4acd 2ab 3 & P4_{3}32 & O^{6} \\
213 & & P 41 3 2 & P 4bd 2ab 3 & P4_{1}32 & O^{7} \\
214 & & I 41 3 2 & I 4bd 2c 3 & I4_{1}32 & O^{8} \\
215 & & P -4 3 m & P -4 2 3 & P\overline{4}3m & T_{d}^{1} \\
216 & & F -4 3 m & F -4 2 3 & F\overline{4}3m & T_{d}^{2} \\
217 & & I -4 3 m & I -4 2 3 & I\overline{4}3m & T_{d}^{3} \\
218 & & P -4 3 n & P -4n 2 3 & P\overline{4}3n & T_{d}^{4} \\
219 & & F -4 3 c & F -4c 2 3 & F\overline{4}3c & T_{d}^{5} \\
220 & & I -4 3 d & I -4bd 2c 3 & I\overline{4}3d & T_{d}^{6} \\
221 & & P 4/m -3 2/m & -P 4 2 3 & Pm\overline{3}m & O_{h}^{1} \\
222 & 2 & P 4/n -3 2/n:2 & -P 4a 2bc 3 & Pn\overline{3}n & O_{h}^{2} \\
223 & & P 42/m -3 2/n & -P 4n 2 3 & Pm\overline{3}n & O_{h}^{3} \\
224 & 2 & P 42/n -3 2/m:2 & -P 4bc 2bc 3 & Pn\overline{3}m & O_{h}^{4} \\
225 & & F 4/m -3 2/m & -F 4 2 3 & Fm\overline{3}m & O_{h}^{5} \\
226 & & F 4/m -3 2/c & -F 4c 2 3 & Fm\overline{3}c & O_{h}^{6} \\
227 & 2 & F 41/d -3 2/m:2 & -F 4vw 2vw 3 & Fd\overline{3}m & O_{h}^{7} \\
228 & 2 & F 41/d -3 2/c:2 & -F 4cvw 2vw 3 & Fd\overline{3}c & O_{h}^{8} \\
229 & & I 4/m -3 2/m & -I 4 2 3 & Im\overline{3}m & O_{h}^{9} \\
230 & & I 41/a -3 2/d & -I 4bd 2c 3 & Ia\overline{3}d & O_{h}^{10}
\end{longtabu}
}}
}
\twocolumn

\section{\label{sec:Acknowledgements}Acknowledgments}

The authors would like to thank D. A. Papaconstantopoulos, who first proposed the
\textit{Crystal Lattice Structures} database, and R. Benjamin Young, who help set
up the original web site in the summer of 1995. Special thanks are due to the H.
Stokes, for providing updates to the {\tt FINDSYM} code, and to the librarians at
the U.S. Naval Research Laboratory Ruth H. Hooker Research Library, who tracked
down many dozens of research articles which are not yet available online.
M. J. Mehl is supported by the Kinnear Foundation and under contract from Duke University.
We also acknowledge support by the by DOD-ONR
(N00014-17-1-2090, 
N00014-16-1-2326, 
N00014-15-1-2863, 
and N00014-16-1-2781). 
DH acknowledges support from the Department of Defense through the National
Defense Science and Engineering Graduate (NDSEG) Fellowship Program. 
The \AFLOW\ consortium would like to acknowledge the Duke University Center for
Materials Genomics.


\onecolumn
\begin{multicols*}{2}
\bibliographyintro{bib-intro,xstefano-20180531}
\end{multicols*}
\twocolumn

\bibliographystyleintro{PhysRevwithTitles_DOI_v1b}
\vfill

\onecolumn
{\phantomsection\label{A2B_aP6_2_aei_i}}
\subsection*{\huge \textbf{{\normalfont H$_{2}$S (90~GPa) Structure: A2B\_aP6\_2\_aei\_i}}}
\noindent \hrulefill
\vspace*{0.25cm}
\begin{figure}[htp]
  \centering
  \vspace{-1em}
  {\includegraphics[width=1\textwidth]{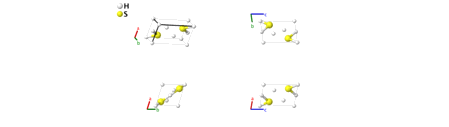}}
\end{figure}
\vspace*{-0.5cm}
\renewcommand{\arraystretch}{1.5}
\begin{equation*}
  \begin{array}{>{$\hspace{-0.15cm}}l<{$}>{$}p{0.5cm}<{$}>{$}p{18.5cm}<{$}}
    \mbox{\large \textbf{Prototype}} &\colon & \ce{H2S} \\
    \mbox{\large \textbf{\AFLOW\ prototype label}} &\colon & \mbox{A2B\_aP6\_2\_aei\_i} \\
    \mbox{\large \textbf{\textit{Strukturbericht} designation}} &\colon & \mbox{None} \\
    \mbox{\large \textbf{Pearson symbol}} &\colon & \mbox{aP6} \\
    \mbox{\large \textbf{Space group number}} &\colon & 2 \\
    \mbox{\large \textbf{Space group symbol}} &\colon & P\bar{1} \\
    \mbox{\large \textbf{\AFLOW\ prototype command}} &\colon &  \texttt{aflow} \,  \, \texttt{-{}-proto=A2B\_aP6\_2\_aei\_i } \, \newline \texttt{-{}-params=}{a,b/a,c/a,\alpha,\beta,\gamma,x_{3},y_{3},z_{3},x_{4},y_{4},z_{4} }
  \end{array}
\end{equation*}
\renewcommand{\arraystretch}{1.0}

\vspace*{-0.25cm}
\noindent \hrulefill
\begin{itemize}
  \item{This structure was found by first-principles electronic structure
calculations and is predicted to be the stable structure of H$_{2}$S
in the range $80 - 140$~GPa.}
  \item{The data presented here was computed at 90~GPa.}
  \item{The original reference places H atoms on (1g), (1f) and (2i) sites,
with S atoms on (2i) sites.  We have changed the origin so that the H
atoms are now on (1a), (1e) and (2i) sites.
}
\end{itemize}

\noindent \parbox{1 \linewidth}{
\noindent \hrulefill
\\
\textbf{Triclinic primitive vectors:} \\
\vspace*{-0.25cm}
\begin{tabular}{cc}
  \begin{tabular}{c}
    \parbox{0.6 \linewidth}{
      \renewcommand{\arraystretch}{1.5}
      \begin{equation*}
        \centering
        \begin{array}{ccc}
              \mathbf{a}_1 & = & a \mathbf{\hat{x}} \\
    \mathbf{a}_2 & = & b \cos\gamma \, \mathbf{\hat{x}} + b \sin\gamma \,\mathbf{\hat{y}} \\
    \mathbf{a}_3 & = & c_x \mathbf{\hat{x}} + c_y \, \mathbf{\hat{y}} + c_z \, \mathbf{\hat{z}}\\\\
     c_x & = & c \, \cos\beta \\
    c_y & = & c \, (\cos\alpha -\cos\beta \cos\gamma)/\sin\gamma \\
    c_z & = & \sqrt{c^2-c_x^2-c_y^2}

        \end{array}
      \end{equation*}
    }
    \renewcommand{\arraystretch}{1.0}
  \end{tabular}
  \begin{tabular}{c}
    \includegraphics[width=0.3\linewidth]{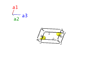} \\
  \end{tabular}
\end{tabular}

}
\vspace*{-0.25cm}

\noindent \hrulefill
\\
\textbf{Basis vectors:}
\vspace*{-0.25cm}
\renewcommand{\arraystretch}{1.5}
\begin{longtabu} to \textwidth{>{\centering $}X[-1,c,c]<{$}>{\centering $}X[-1,c,c]<{$}>{\centering $}X[-1,c,c]<{$}>{\centering $}X[-1,c,c]<{$}>{\centering $}X[-1,c,c]<{$}>{\centering $}X[-1,c,c]<{$}>{\centering $}X[-1,c,c]<{$}}
  & & \mbox{Lattice Coordinates} & & \mbox{Cartesian Coordinates} &\mbox{Wyckoff Position} & \mbox{Atom Type} \\  
  \mathbf{B}_{1} & = & 0 \, \mathbf{a}_{1} + 0 \, \mathbf{a}_{2} + 0 \, \mathbf{a}_{3} & = & 0 \, \mathbf{\hat{x}} + 0 \, \mathbf{\hat{y}} + 0 \, \mathbf{\hat{z}} & \left(1a\right) & \mbox{H I} \\ 
\mathbf{B}_{2} & = & \frac{1}{2} \, \mathbf{a}_{1} + \frac{1}{2} \, \mathbf{a}_{2} & = & \frac{1}{2}\left(a+b\cos\gamma\right) \, \mathbf{\hat{x}} + \frac{1}{2}b\sin\gamma \, \mathbf{\hat{y}} & \left(1e\right) & \mbox{H II} \\ 
\mathbf{B}_{3} & = & x_{3} \, \mathbf{a}_{1} + y_{3} \, \mathbf{a}_{2} + z_{3} \, \mathbf{a}_{3} & = & \left(x_{3}a+y_{3}b\cos\gamma+z_{3}c_{x}\right) \, \mathbf{\hat{x}} + \left(y_{3}b\sin\gamma+z_{3}c_{y}\right) \, \mathbf{\hat{y}} + z_{3}c_{z} \, \mathbf{\hat{z}} & \left(2i\right) & \mbox{H III} \\ 
\mathbf{B}_{4} & = & -x_{3} \, \mathbf{a}_{1}-y_{3} \, \mathbf{a}_{2}-z_{3} \, \mathbf{a}_{3} & = & \left(-x_{3}a-y_{3}b\cos\gamma-z_{3}c_{x}\right) \, \mathbf{\hat{x}} + \left(-y_{3}b\sin\gamma-z_{3}c_{y}\right) \, \mathbf{\hat{y}}-z_{3}c_{z} \, \mathbf{\hat{z}} & \left(2i\right) & \mbox{H III} \\ 
\mathbf{B}_{5} & = & x_{4} \, \mathbf{a}_{1} + y_{4} \, \mathbf{a}_{2} + z_{4} \, \mathbf{a}_{3} & = & \left(x_{4}a+y_{4}b\cos\gamma+z_{4}c_{x}\right) \, \mathbf{\hat{x}} + \left(y_{4}b\sin\gamma+z_{4}c_{y}\right) \, \mathbf{\hat{y}} + z_{4}c_{z} \, \mathbf{\hat{z}} & \left(2i\right) & \mbox{S} \\ 
\mathbf{B}_{6} & = & -x_{4} \, \mathbf{a}_{1}-y_{4} \, \mathbf{a}_{2}-z_{4} \, \mathbf{a}_{3} & = & \left(-x_{4}a-y_{4}b\cos\gamma-z_{4}c_{x}\right) \, \mathbf{\hat{x}} + \left(-y_{4}b\sin\gamma-z_{4}c_{y}\right) \, \mathbf{\hat{y}}-z_{4}c_{z} \, \mathbf{\hat{z}} & \left(2i\right) & \mbox{S} \\ 
\end{longtabu}
\renewcommand{\arraystretch}{1.0}
\noindent \hrulefill
\\
\textbf{References:}
\vspace*{-0.25cm}
\begin{flushleft}
  - \bibentry{Li_JCP_140_2014}. \\
\end{flushleft}
\noindent \hrulefill
\\
\textbf{Geometry files:}
\\
\noindent  - CIF: pp. {\hyperref[A2B_aP6_2_aei_i_cif]{\pageref{A2B_aP6_2_aei_i_cif}}} \\
\noindent  - POSCAR: pp. {\hyperref[A2B_aP6_2_aei_i_poscar]{\pageref{A2B_aP6_2_aei_i_poscar}}} \\
\onecolumn
{\phantomsection\label{A8B5_mP13_6_a7b_3a2b}}
\subsection*{\huge \textbf{{\normalfont \begin{raggedleft}Mo$_{8}$P$_{5}$ (High-temperature) Structure: \end{raggedleft} \\ A8B5\_mP13\_6\_a7b\_3a2b}}}
\noindent \hrulefill
\vspace*{0.25cm}
\begin{figure}[htp]
  \centering
  \vspace{-1em}
  {\includegraphics[width=1\textwidth]{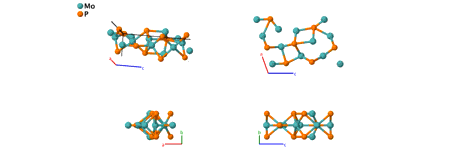}}
\end{figure}
\vspace*{-0.5cm}
\renewcommand{\arraystretch}{1.5}
\begin{equation*}
  \begin{array}{>{$\hspace{-0.15cm}}l<{$}>{$}p{0.5cm}<{$}>{$}p{18.5cm}<{$}}
    \mbox{\large \textbf{Prototype}} &\colon & \ce{Mo8P5} \\
    \mbox{\large \textbf{\AFLOW\ prototype label}} &\colon & \mbox{A8B5\_mP13\_6\_a7b\_3a2b} \\
    \mbox{\large \textbf{\textit{Strukturbericht} designation}} &\colon & \mbox{None} \\
    \mbox{\large \textbf{Pearson symbol}} &\colon & \mbox{mP13} \\
    \mbox{\large \textbf{Space group number}} &\colon & 6 \\
    \mbox{\large \textbf{Space group symbol}} &\colon & Pm \\
    \mbox{\large \textbf{\AFLOW\ prototype command}} &\colon &  \texttt{aflow} \,  \, \texttt{-{}-proto=A8B5\_mP13\_6\_a7b\_3a2b } \, \newline \texttt{-{}-params=}{a,b/a,c/a,\beta,x_{1},z_{1},x_{2},z_{2},x_{3},z_{3},x_{4},z_{4},x_{5},z_{5},x_{6},z_{6},x_{7},z_{7},x_{8},z_{8},} \newline {x_{9},z_{9},x_{10},z_{10},x_{11},z_{11},x_{12},z_{12},x_{13},z_{13} }
  \end{array}
\end{equation*}
\renewcommand{\arraystretch}{1.0}

\vspace*{-0.25cm}
\noindent \hrulefill
\begin{itemize}
  \item{This high-temperature phase of the Mo-P system is observed between $1580^{\circ}\mathrm{C} - 1680^{\circ}\mathrm{C}$ (Johnsson, 1972).
}
\end{itemize}

\noindent \parbox{1 \linewidth}{
\noindent \hrulefill
\\
\textbf{Simple Monoclinic primitive vectors:} \\
\vspace*{-0.25cm}
\begin{tabular}{cc}
  \begin{tabular}{c}
    \parbox{0.6 \linewidth}{
      \renewcommand{\arraystretch}{1.5}
      \begin{equation*}
        \centering
        \begin{array}{ccc}
              \mathbf{a}_1 & = & a \, \mathbf{\hat{x}} \\
    \mathbf{a}_2 & = & b \, \mathbf{\hat{y}} \\
    \mathbf{a}_3 & = & c \cos\beta \, \mathbf{\hat{x}} + c \sin\beta \, \mathbf{\hat{z}} \\

        \end{array}
      \end{equation*}
    }
    \renewcommand{\arraystretch}{1.0}
  \end{tabular}
  \begin{tabular}{c}
    \includegraphics[width=0.3\linewidth]{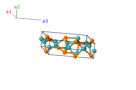} \\
  \end{tabular}
\end{tabular}

}
\vspace*{-0.25cm}

\noindent \hrulefill
\\
\textbf{Basis vectors:}
\vspace*{-0.25cm}
\renewcommand{\arraystretch}{1.5}
\begin{longtabu} to \textwidth{>{\centering $}X[-1,c,c]<{$}>{\centering $}X[-1,c,c]<{$}>{\centering $}X[-1,c,c]<{$}>{\centering $}X[-1,c,c]<{$}>{\centering $}X[-1,c,c]<{$}>{\centering $}X[-1,c,c]<{$}>{\centering $}X[-1,c,c]<{$}}
  & & \mbox{Lattice Coordinates} & & \mbox{Cartesian Coordinates} &\mbox{Wyckoff Position} & \mbox{Atom Type} \\  
  \mathbf{B}_{1} & = & x_{1} \, \mathbf{a}_{1} + z_{1} \, \mathbf{a}_{3} & = & \left(x_{1}a+z_{1}c\cos\beta\right) \, \mathbf{\hat{x}} + z_{1}c\sin\beta \, \mathbf{\hat{z}} & \left(1a\right) & \mbox{Mo I} \\ 
\mathbf{B}_{2} & = & x_{2} \, \mathbf{a}_{1} + z_{2} \, \mathbf{a}_{3} & = & \left(x_{2}a+z_{2}c\cos\beta\right) \, \mathbf{\hat{x}} + z_{2}c\sin\beta \, \mathbf{\hat{z}} & \left(1a\right) & \mbox{P I} \\ 
\mathbf{B}_{3} & = & x_{3} \, \mathbf{a}_{1} + z_{3} \, \mathbf{a}_{3} & = & \left(x_{3}a+z_{3}c\cos\beta\right) \, \mathbf{\hat{x}} + z_{3}c\sin\beta \, \mathbf{\hat{z}} & \left(1a\right) & \mbox{P II} \\ 
\mathbf{B}_{4} & = & x_{4} \, \mathbf{a}_{1} + z_{4} \, \mathbf{a}_{3} & = & \left(x_{4}a+z_{4}c\cos\beta\right) \, \mathbf{\hat{x}} + z_{4}c\sin\beta \, \mathbf{\hat{z}} & \left(1a\right) & \mbox{P III} \\ 
\mathbf{B}_{5} & = & x_{5} \, \mathbf{a}_{1} + \frac{1}{2} \, \mathbf{a}_{2} + z_{5} \, \mathbf{a}_{3} & = & \left(x_{5}a+z_{5}c\cos\beta\right) \, \mathbf{\hat{x}} + \frac{1}{2}b \, \mathbf{\hat{y}} + z_{5}c\sin\beta \, \mathbf{\hat{z}} & \left(1b\right) & \mbox{Mo II} \\ 
\mathbf{B}_{6} & = & x_{6} \, \mathbf{a}_{1} + \frac{1}{2} \, \mathbf{a}_{2} + z_{6} \, \mathbf{a}_{3} & = & \left(x_{6}a+z_{6}c\cos\beta\right) \, \mathbf{\hat{x}} + \frac{1}{2}b \, \mathbf{\hat{y}} + z_{6}c\sin\beta \, \mathbf{\hat{z}} & \left(1b\right) & \mbox{Mo III} \\ 
\mathbf{B}_{7} & = & x_{7} \, \mathbf{a}_{1} + \frac{1}{2} \, \mathbf{a}_{2} + z_{7} \, \mathbf{a}_{3} & = & \left(x_{7}a+z_{7}c\cos\beta\right) \, \mathbf{\hat{x}} + \frac{1}{2}b \, \mathbf{\hat{y}} + z_{7}c\sin\beta \, \mathbf{\hat{z}} & \left(1b\right) & \mbox{Mo IV} \\ 
\mathbf{B}_{8} & = & x_{8} \, \mathbf{a}_{1} + \frac{1}{2} \, \mathbf{a}_{2} + z_{8} \, \mathbf{a}_{3} & = & \left(x_{8}a+z_{8}c\cos\beta\right) \, \mathbf{\hat{x}} + \frac{1}{2}b \, \mathbf{\hat{y}} + z_{8}c\sin\beta \, \mathbf{\hat{z}} & \left(1b\right) & \mbox{Mo V} \\ 
\mathbf{B}_{9} & = & x_{9} \, \mathbf{a}_{1} + \frac{1}{2} \, \mathbf{a}_{2} + z_{9} \, \mathbf{a}_{3} & = & \left(x_{9}a+z_{9}c\cos\beta\right) \, \mathbf{\hat{x}} + \frac{1}{2}b \, \mathbf{\hat{y}} + z_{9}c\sin\beta \, \mathbf{\hat{z}} & \left(1b\right) & \mbox{Mo VI} \\ 
\mathbf{B}_{10} & = & x_{10} \, \mathbf{a}_{1} + \frac{1}{2} \, \mathbf{a}_{2} + z_{10} \, \mathbf{a}_{3} & = & \left(x_{10}a+z_{10}c\cos\beta\right) \, \mathbf{\hat{x}} + \frac{1}{2}b \, \mathbf{\hat{y}} + z_{10}c\sin\beta \, \mathbf{\hat{z}} & \left(1b\right) & \mbox{Mo VII} \\ 
\mathbf{B}_{11} & = & x_{11} \, \mathbf{a}_{1} + \frac{1}{2} \, \mathbf{a}_{2} + z_{11} \, \mathbf{a}_{3} & = & \left(x_{11}a+z_{11}c\cos\beta\right) \, \mathbf{\hat{x}} + \frac{1}{2}b \, \mathbf{\hat{y}} + z_{11}c\sin\beta \, \mathbf{\hat{z}} & \left(1b\right) & \mbox{Mo VIII} \\ 
\mathbf{B}_{12} & = & x_{12} \, \mathbf{a}_{1} + \frac{1}{2} \, \mathbf{a}_{2} + z_{12} \, \mathbf{a}_{3} & = & \left(x_{12}a+z_{12}c\cos\beta\right) \, \mathbf{\hat{x}} + \frac{1}{2}b \, \mathbf{\hat{y}} + z_{12}c\sin\beta \, \mathbf{\hat{z}} & \left(1b\right) & \mbox{P IV} \\ 
\mathbf{B}_{13} & = & x_{13} \, \mathbf{a}_{1} + \frac{1}{2} \, \mathbf{a}_{2} + z_{13} \, \mathbf{a}_{3} & = & \left(x_{13}a+z_{13}c\cos\beta\right) \, \mathbf{\hat{x}} + \frac{1}{2}b \, \mathbf{\hat{y}} + z_{13}c\sin\beta \, \mathbf{\hat{z}} & \left(1b\right) & \mbox{P V} \\ 
\end{longtabu}
\renewcommand{\arraystretch}{1.0}
\noindent \hrulefill
\\
\textbf{References:}
\vspace*{-0.25cm}
\begin{flushleft}
  - \bibentry{Johnsson_Mo8P5_ActChemScand_1972}. \\
\end{flushleft}
\textbf{Found in:}
\vspace*{-0.25cm}
\begin{flushleft}
  - \bibentry{Villars_PearsonsCrystalData_2013}. \\
\end{flushleft}
\noindent \hrulefill
\\
\textbf{Geometry files:}
\\
\noindent  - CIF: pp. {\hyperref[A8B5_mP13_6_a7b_3a2b_cif]{\pageref{A8B5_mP13_6_a7b_3a2b_cif}}} \\
\noindent  - POSCAR: pp. {\hyperref[A8B5_mP13_6_a7b_3a2b_poscar]{\pageref{A8B5_mP13_6_a7b_3a2b_poscar}}} \\
\onecolumn
{\phantomsection\label{AB_mP4_6_2b_2a}}
\subsection*{\huge \textbf{{\normalfont FeNi Structure: AB\_mP4\_6\_2b\_2a}}}
\noindent \hrulefill
\vspace*{0.25cm}
\begin{figure}[htp]
  \centering
  \vspace{-1em}
  {\includegraphics[width=1\textwidth]{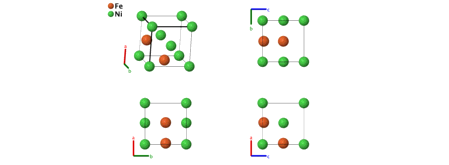}}
\end{figure}
\vspace*{-0.5cm}
\renewcommand{\arraystretch}{1.5}
\begin{equation*}
  \begin{array}{>{$\hspace{-0.15cm}}l<{$}>{$}p{0.5cm}<{$}>{$}p{18.5cm}<{$}}
    \mbox{\large \textbf{Prototype}} &\colon & \ce{FeNi} \\
    \mbox{\large \textbf{\AFLOW\ prototype label}} &\colon & \mbox{AB\_mP4\_6\_2b\_2a} \\
    \mbox{\large \textbf{\textit{Strukturbericht} designation}} &\colon & \mbox{None} \\
    \mbox{\large \textbf{Pearson symbol}} &\colon & \mbox{mP4} \\
    \mbox{\large \textbf{Space group number}} &\colon & 6 \\
    \mbox{\large \textbf{Space group symbol}} &\colon & Pm \\
    \mbox{\large \textbf{\AFLOW\ prototype command}} &\colon &  \texttt{aflow} \,  \, \texttt{-{}-proto=AB\_mP4\_6\_2b\_2a } \, \newline \texttt{-{}-params=}{a,b/a,c/a,\beta,x_{1},z_{1},x_{2},z_{2},x_{3},z_{3},x_{4},z_{4} }
  \end{array}
\end{equation*}
\renewcommand{\arraystretch}{1.0}

\vspace*{-0.25cm}
\noindent \hrulefill
\begin{itemize}
  \item{In the original the site occupations are mixed with Ni majority (0.85) on sites 1a and Fe majority on 1b.}
\end{itemize}

\noindent \parbox{1 \linewidth}{
\noindent \hrulefill
\\
\textbf{Simple Monoclinic primitive vectors:} \\
\vspace*{-0.25cm}
\begin{tabular}{cc}
  \begin{tabular}{c}
    \parbox{0.6 \linewidth}{
      \renewcommand{\arraystretch}{1.5}
      \begin{equation*}
        \centering
        \begin{array}{ccc}
              \mathbf{a}_1 & = & a \, \mathbf{\hat{x}} \\
    \mathbf{a}_2 & = & b \, \mathbf{\hat{y}} \\
    \mathbf{a}_3 & = & c \cos\beta \, \mathbf{\hat{x}} + c \sin\beta \, \mathbf{\hat{z}} \\

        \end{array}
      \end{equation*}
    }
    \renewcommand{\arraystretch}{1.0}
  \end{tabular}
  \begin{tabular}{c}
    \includegraphics[width=0.3\linewidth]{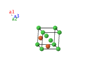} \\
  \end{tabular}
\end{tabular}

}
\vspace*{-0.25cm}

\noindent \hrulefill
\\
\textbf{Basis vectors:}
\vspace*{-0.25cm}
\renewcommand{\arraystretch}{1.5}
\begin{longtabu} to \textwidth{>{\centering $}X[-1,c,c]<{$}>{\centering $}X[-1,c,c]<{$}>{\centering $}X[-1,c,c]<{$}>{\centering $}X[-1,c,c]<{$}>{\centering $}X[-1,c,c]<{$}>{\centering $}X[-1,c,c]<{$}>{\centering $}X[-1,c,c]<{$}}
  & & \mbox{Lattice Coordinates} & & \mbox{Cartesian Coordinates} &\mbox{Wyckoff Position} & \mbox{Atom Type} \\  
  \mathbf{B}_{1} & = & x_{1} \, \mathbf{a}_{1} + z_{1} \, \mathbf{a}_{3} & = & \left(x_{1}a+z_{1}c\cos\beta\right) \, \mathbf{\hat{x}} + z_{1}c\sin\beta \, \mathbf{\hat{z}} & \left(1a\right) & \mbox{Ni I} \\ 
\mathbf{B}_{2} & = & x_{2} \, \mathbf{a}_{1} + z_{2} \, \mathbf{a}_{3} & = & \left(x_{2}a+z_{2}c\cos\beta\right) \, \mathbf{\hat{x}} + z_{2}c\sin\beta \, \mathbf{\hat{z}} & \left(1a\right) & \mbox{Ni II} \\ 
\mathbf{B}_{3} & = & x_{3} \, \mathbf{a}_{1} + \frac{1}{2} \, \mathbf{a}_{2} + z_{3} \, \mathbf{a}_{3} & = & \left(x_{3}a+z_{3}c\cos\beta\right) \, \mathbf{\hat{x}} + \frac{1}{2}b \, \mathbf{\hat{y}} + z_{3}c\sin\beta \, \mathbf{\hat{z}} & \left(1b\right) & \mbox{Fe I} \\ 
\mathbf{B}_{4} & = & x_{4} \, \mathbf{a}_{1} + \frac{1}{2} \, \mathbf{a}_{2} + z_{4} \, \mathbf{a}_{3} & = & \left(x_{4}a+z_{4}c\cos\beta\right) \, \mathbf{\hat{x}} + \frac{1}{2}b \, \mathbf{\hat{y}} + z_{4}c\sin\beta \, \mathbf{\hat{z}} & \left(1b\right) & \mbox{Fe II} \\ 
\end{longtabu}
\renewcommand{\arraystretch}{1.0}
\noindent \hrulefill
\\
\textbf{References:}
\vspace*{-0.25cm}
\begin{flushleft}
  - \bibentry{Tagai_FeNi_ZKristallogr_1995}. \\
\end{flushleft}
\textbf{Found in:}
\vspace*{-0.25cm}
\begin{flushleft}
  - \bibentry{Villars_PearsonsCrystalData_2013}. \\
\end{flushleft}
\noindent \hrulefill
\\
\textbf{Geometry files:}
\\
\noindent  - CIF: pp. {\hyperref[AB_mP4_6_2b_2a_cif]{\pageref{AB_mP4_6_2b_2a_cif}}} \\
\noindent  - POSCAR: pp. {\hyperref[AB_mP4_6_2b_2a_poscar]{\pageref{AB_mP4_6_2b_2a_poscar}}} \\
\onecolumn
{\phantomsection\label{A2B_mP12_7_4a_2a}}
\subsection*{\huge \textbf{{\normalfont H$_{2}$S IV Structure: A2B\_mP12\_7\_4a\_2a}}}
\noindent \hrulefill
\vspace*{0.25cm}
\begin{figure}[htp]
  \centering
  \vspace{-1em}
  {\includegraphics[width=1\textwidth]{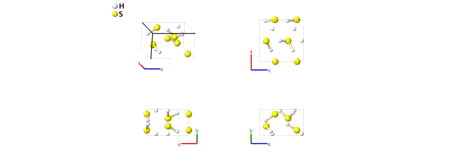}}
\end{figure}
\vspace*{-0.5cm}
\renewcommand{\arraystretch}{1.5}
\begin{equation*}
  \begin{array}{>{$\hspace{-0.15cm}}l<{$}>{$}p{0.5cm}<{$}>{$}p{18.5cm}<{$}}
    \mbox{\large \textbf{Prototype}} &\colon & \ce{H2S} \\
    \mbox{\large \textbf{\AFLOW\ prototype label}} &\colon & \mbox{A2B\_mP12\_7\_4a\_2a} \\
    \mbox{\large \textbf{\textit{Strukturbericht} designation}} &\colon & \mbox{None} \\
    \mbox{\large \textbf{Pearson symbol}} &\colon & \mbox{mP12} \\
    \mbox{\large \textbf{Space group number}} &\colon & 7 \\
    \mbox{\large \textbf{Space group symbol}} &\colon & Pc \\
    \mbox{\large \textbf{\AFLOW\ prototype command}} &\colon &  \texttt{aflow} \,  \, \texttt{-{}-proto=A2B\_mP12\_7\_4a\_2a } \, \newline \texttt{-{}-params=}{a,b/a,c/a,\beta,x_{1},y_{1},z_{1},x_{2},y_{2},z_{2},x_{3},y_{3},z_{3},x_{4},y_{4},z_{4},x_{5},y_{5},z_{5},x_{6},} \newline {y_{6},z_{6} }
  \end{array}
\end{equation*}
\renewcommand{\arraystretch}{1.0}

\vspace*{-0.25cm}
\noindent \hrulefill
\begin{itemize}
  \item{The H$_{2}$S-IV phase is stable at pressures greater than 5~GPa and
temperatures below 150~K, and is stable at higher pressures for higher
temperatures. Shimizu {\em et. al.} (Shimizu, 1995) determined
the phase diagram. Endo {\em et al.} (Endo, 998) found the
crystal structure at 11.4~GPa, but could not determine the positions
of the hydrogen atoms.  Here we use the predicted structure from Li
{\em et al.} at 30~GPa.
}
\end{itemize}

\noindent \parbox{1 \linewidth}{
\noindent \hrulefill
\\
\textbf{Simple Monoclinic primitive vectors:} \\
\vspace*{-0.25cm}
\begin{tabular}{cc}
  \begin{tabular}{c}
    \parbox{0.6 \linewidth}{
      \renewcommand{\arraystretch}{1.5}
      \begin{equation*}
        \centering
        \begin{array}{ccc}
              \mathbf{a}_1 & = & a \, \mathbf{\hat{x}} \\
    \mathbf{a}_2 & = & b \, \mathbf{\hat{y}} \\
    \mathbf{a}_3 & = & c \cos\beta \, \mathbf{\hat{x}} + c \sin\beta \, \mathbf{\hat{z}} \\

        \end{array}
      \end{equation*}
    }
    \renewcommand{\arraystretch}{1.0}
  \end{tabular}
  \begin{tabular}{c}
    \includegraphics[width=0.3\linewidth]{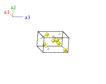} \\
  \end{tabular}
\end{tabular}

}
\vspace*{-0.25cm}

\noindent \hrulefill
\\
\textbf{Basis vectors:}
\vspace*{-0.25cm}
\renewcommand{\arraystretch}{1.5}
\begin{longtabu} to \textwidth{>{\centering $}X[-1,c,c]<{$}>{\centering $}X[-1,c,c]<{$}>{\centering $}X[-1,c,c]<{$}>{\centering $}X[-1,c,c]<{$}>{\centering $}X[-1,c,c]<{$}>{\centering $}X[-1,c,c]<{$}>{\centering $}X[-1,c,c]<{$}}
  & & \mbox{Lattice Coordinates} & & \mbox{Cartesian Coordinates} &\mbox{Wyckoff Position} & \mbox{Atom Type} \\  
  \mathbf{B}_{1} & = & x_{1} \, \mathbf{a}_{1} + y_{1} \, \mathbf{a}_{2} + z_{1} \, \mathbf{a}_{3} & = & \left(x_{1}a+z_{1}c\cos\beta\right) \, \mathbf{\hat{x}} + y_{1}b \, \mathbf{\hat{y}} + z_{1}c\sin\beta \, \mathbf{\hat{z}} & \left(2a\right) & \mbox{H I} \\ 
\mathbf{B}_{2} & = & x_{1} \, \mathbf{a}_{1}-y_{1} \, \mathbf{a}_{2} + \left(\frac{1}{2} +z_{1}\right) \, \mathbf{a}_{3} & = & \left(\frac{1}{2}c\cos\beta +x_{1}a + z_{1}c\cos\beta\right) \, \mathbf{\hat{x}}-y_{1}b \, \mathbf{\hat{y}} + \left(\frac{1}{2} +z_{1}\right)c\sin\beta \, \mathbf{\hat{z}} & \left(2a\right) & \mbox{H I} \\ 
\mathbf{B}_{3} & = & x_{2} \, \mathbf{a}_{1} + y_{2} \, \mathbf{a}_{2} + z_{2} \, \mathbf{a}_{3} & = & \left(x_{2}a+z_{2}c\cos\beta\right) \, \mathbf{\hat{x}} + y_{2}b \, \mathbf{\hat{y}} + z_{2}c\sin\beta \, \mathbf{\hat{z}} & \left(2a\right) & \mbox{H II} \\ 
\mathbf{B}_{4} & = & x_{2} \, \mathbf{a}_{1}-y_{2} \, \mathbf{a}_{2} + \left(\frac{1}{2} +z_{2}\right) \, \mathbf{a}_{3} & = & \left(\frac{1}{2}c\cos\beta +x_{2}a + z_{2}c\cos\beta\right) \, \mathbf{\hat{x}}-y_{2}b \, \mathbf{\hat{y}} + \left(\frac{1}{2} +z_{2}\right)c\sin\beta \, \mathbf{\hat{z}} & \left(2a\right) & \mbox{H II} \\ 
\mathbf{B}_{5} & = & x_{3} \, \mathbf{a}_{1} + y_{3} \, \mathbf{a}_{2} + z_{3} \, \mathbf{a}_{3} & = & \left(x_{3}a+z_{3}c\cos\beta\right) \, \mathbf{\hat{x}} + y_{3}b \, \mathbf{\hat{y}} + z_{3}c\sin\beta \, \mathbf{\hat{z}} & \left(2a\right) & \mbox{H III} \\ 
\mathbf{B}_{6} & = & x_{3} \, \mathbf{a}_{1}-y_{3} \, \mathbf{a}_{2} + \left(\frac{1}{2} +z_{3}\right) \, \mathbf{a}_{3} & = & \left(\frac{1}{2}c\cos\beta +x_{3}a + z_{3}c\cos\beta\right) \, \mathbf{\hat{x}}-y_{3}b \, \mathbf{\hat{y}} + \left(\frac{1}{2} +z_{3}\right)c\sin\beta \, \mathbf{\hat{z}} & \left(2a\right) & \mbox{H III} \\ 
\mathbf{B}_{7} & = & x_{4} \, \mathbf{a}_{1} + y_{4} \, \mathbf{a}_{2} + z_{4} \, \mathbf{a}_{3} & = & \left(x_{4}a+z_{4}c\cos\beta\right) \, \mathbf{\hat{x}} + y_{4}b \, \mathbf{\hat{y}} + z_{4}c\sin\beta \, \mathbf{\hat{z}} & \left(2a\right) & \mbox{H IV} \\ 
\mathbf{B}_{8} & = & x_{4} \, \mathbf{a}_{1}-y_{4} \, \mathbf{a}_{2} + \left(\frac{1}{2} +z_{4}\right) \, \mathbf{a}_{3} & = & \left(\frac{1}{2}c\cos\beta +x_{4}a + z_{4}c\cos\beta\right) \, \mathbf{\hat{x}}-y_{4}b \, \mathbf{\hat{y}} + \left(\frac{1}{2} +z_{4}\right)c\sin\beta \, \mathbf{\hat{z}} & \left(2a\right) & \mbox{H IV} \\ 
\mathbf{B}_{9} & = & x_{5} \, \mathbf{a}_{1} + y_{5} \, \mathbf{a}_{2} + z_{5} \, \mathbf{a}_{3} & = & \left(x_{5}a+z_{5}c\cos\beta\right) \, \mathbf{\hat{x}} + y_{5}b \, \mathbf{\hat{y}} + z_{5}c\sin\beta \, \mathbf{\hat{z}} & \left(2a\right) & \mbox{S I} \\ 
\mathbf{B}_{10} & = & x_{5} \, \mathbf{a}_{1}-y_{5} \, \mathbf{a}_{2} + \left(\frac{1}{2} +z_{5}\right) \, \mathbf{a}_{3} & = & \left(\frac{1}{2}c\cos\beta +x_{5}a + z_{5}c\cos\beta\right) \, \mathbf{\hat{x}}-y_{5}b \, \mathbf{\hat{y}} + \left(\frac{1}{2} +z_{5}\right)c\sin\beta \, \mathbf{\hat{z}} & \left(2a\right) & \mbox{S I} \\ 
\mathbf{B}_{11} & = & x_{6} \, \mathbf{a}_{1} + y_{6} \, \mathbf{a}_{2} + z_{6} \, \mathbf{a}_{3} & = & \left(x_{6}a+z_{6}c\cos\beta\right) \, \mathbf{\hat{x}} + y_{6}b \, \mathbf{\hat{y}} + z_{6}c\sin\beta \, \mathbf{\hat{z}} & \left(2a\right) & \mbox{S II} \\ 
\mathbf{B}_{12} & = & x_{6} \, \mathbf{a}_{1}-y_{6} \, \mathbf{a}_{2} + \left(\frac{1}{2} +z_{6}\right) \, \mathbf{a}_{3} & = & \left(\frac{1}{2}c\cos\beta +x_{6}a + z_{6}c\cos\beta\right) \, \mathbf{\hat{x}}-y_{6}b \, \mathbf{\hat{y}} + \left(\frac{1}{2} +z_{6}\right)c\sin\beta \, \mathbf{\hat{z}} & \left(2a\right) & \mbox{S II} \\ 
\end{longtabu}
\renewcommand{\arraystretch}{1.0}
\noindent \hrulefill
\\
\textbf{References:}
\vspace*{-0.25cm}
\begin{flushleft}
  - \bibentry{Li_JCP_140_2014}. \\
  - \bibentry{Shimizu_PRB_51_1995}. \\
  - \bibentry{Endo_PRB_57_1998}. \\
\end{flushleft}
\noindent \hrulefill
\\
\textbf{Geometry files:}
\\
\noindent  - CIF: pp. {\hyperref[A2B_mP12_7_4a_2a_cif]{\pageref{A2B_mP12_7_4a_2a_cif}}} \\
\noindent  - POSCAR: pp. {\hyperref[A2B_mP12_7_4a_2a_poscar]{\pageref{A2B_mP12_7_4a_2a_poscar}}} \\
\onecolumn
{\phantomsection\label{A2B_mP18_7_6a_3a}}
\subsection*{\huge \textbf{{\normalfont As$_{2}$Ba Structure: A2B\_mP18\_7\_6a\_3a}}}
\noindent \hrulefill
\vspace*{0.25cm}
\begin{figure}[htp]
  \centering
  \vspace{-1em}
  {\includegraphics[width=1\textwidth]{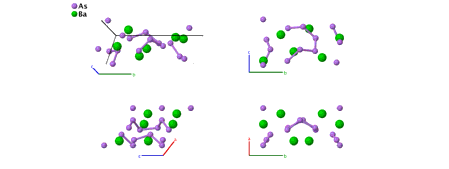}}
\end{figure}
\vspace*{-0.5cm}
\renewcommand{\arraystretch}{1.5}
\begin{equation*}
  \begin{array}{>{$\hspace{-0.15cm}}l<{$}>{$}p{0.5cm}<{$}>{$}p{18.5cm}<{$}}
    \mbox{\large \textbf{Prototype}} &\colon & \ce{As2Ba} \\
    \mbox{\large \textbf{\AFLOW\ prototype label}} &\colon & \mbox{A2B\_mP18\_7\_6a\_3a} \\
    \mbox{\large \textbf{\textit{Strukturbericht} designation}} &\colon & \mbox{None} \\
    \mbox{\large \textbf{Pearson symbol}} &\colon & \mbox{mP18} \\
    \mbox{\large \textbf{Space group number}} &\colon & 7 \\
    \mbox{\large \textbf{Space group symbol}} &\colon & Pc \\
    \mbox{\large \textbf{\AFLOW\ prototype command}} &\colon &  \texttt{aflow} \,  \, \texttt{-{}-proto=A2B\_mP18\_7\_6a\_3a } \, \newline \texttt{-{}-params=}{a,b/a,c/a,\beta,x_{1},y_{1},z_{1},x_{2},y_{2},z_{2},x_{3},y_{3},z_{3},x_{4},y_{4},z_{4},x_{5},y_{5},z_{5},x_{6},} \newline {y_{6},z_{6},x_{7},y_{7},z_{7},x_{8},y_{8},z_{8},x_{9},y_{9},z_{9} }
  \end{array}
\end{equation*}
\renewcommand{\arraystretch}{1.0}

\noindent \parbox{1 \linewidth}{
\noindent \hrulefill
\\
\textbf{Simple Monoclinic primitive vectors:} \\
\vspace*{-0.25cm}
\begin{tabular}{cc}
  \begin{tabular}{c}
    \parbox{0.6 \linewidth}{
      \renewcommand{\arraystretch}{1.5}
      \begin{equation*}
        \centering
        \begin{array}{ccc}
              \mathbf{a}_1 & = & a \, \mathbf{\hat{x}} \\
    \mathbf{a}_2 & = & b \, \mathbf{\hat{y}} \\
    \mathbf{a}_3 & = & c \cos\beta \, \mathbf{\hat{x}} + c \sin\beta \, \mathbf{\hat{z}} \\

        \end{array}
      \end{equation*}
    }
    \renewcommand{\arraystretch}{1.0}
  \end{tabular}
  \begin{tabular}{c}
    \includegraphics[width=0.3\linewidth]{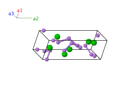} \\
  \end{tabular}
\end{tabular}

}
\vspace*{-0.25cm}

\noindent \hrulefill
\\
\textbf{Basis vectors:}
\vspace*{-0.25cm}
\renewcommand{\arraystretch}{1.5}
\begin{longtabu} to \textwidth{>{\centering $}X[-1,c,c]<{$}>{\centering $}X[-1,c,c]<{$}>{\centering $}X[-1,c,c]<{$}>{\centering $}X[-1,c,c]<{$}>{\centering $}X[-1,c,c]<{$}>{\centering $}X[-1,c,c]<{$}>{\centering $}X[-1,c,c]<{$}}
  & & \mbox{Lattice Coordinates} & & \mbox{Cartesian Coordinates} &\mbox{Wyckoff Position} & \mbox{Atom Type} \\  
  \mathbf{B}_{1} & = & x_{1} \, \mathbf{a}_{1} + y_{1} \, \mathbf{a}_{2} + z_{1} \, \mathbf{a}_{3} & = & \left(x_{1}a+z_{1}c\cos\beta\right) \, \mathbf{\hat{x}} + y_{1}b \, \mathbf{\hat{y}} + z_{1}c\sin\beta \, \mathbf{\hat{z}} & \left(2a\right) & \mbox{As I} \\ 
\mathbf{B}_{2} & = & x_{1} \, \mathbf{a}_{1}-y_{1} \, \mathbf{a}_{2} + \left(\frac{1}{2} +z_{1}\right) \, \mathbf{a}_{3} & = & \left(\frac{1}{2}c\cos\beta +x_{1}a + z_{1}c\cos\beta\right) \, \mathbf{\hat{x}}-y_{1}b \, \mathbf{\hat{y}} + \left(\frac{1}{2} +z_{1}\right)c\sin\beta \, \mathbf{\hat{z}} & \left(2a\right) & \mbox{As I} \\ 
\mathbf{B}_{3} & = & x_{2} \, \mathbf{a}_{1} + y_{2} \, \mathbf{a}_{2} + z_{2} \, \mathbf{a}_{3} & = & \left(x_{2}a+z_{2}c\cos\beta\right) \, \mathbf{\hat{x}} + y_{2}b \, \mathbf{\hat{y}} + z_{2}c\sin\beta \, \mathbf{\hat{z}} & \left(2a\right) & \mbox{As II} \\ 
\mathbf{B}_{4} & = & x_{2} \, \mathbf{a}_{1}-y_{2} \, \mathbf{a}_{2} + \left(\frac{1}{2} +z_{2}\right) \, \mathbf{a}_{3} & = & \left(\frac{1}{2}c\cos\beta +x_{2}a + z_{2}c\cos\beta\right) \, \mathbf{\hat{x}}-y_{2}b \, \mathbf{\hat{y}} + \left(\frac{1}{2} +z_{2}\right)c\sin\beta \, \mathbf{\hat{z}} & \left(2a\right) & \mbox{As II} \\ 
\mathbf{B}_{5} & = & x_{3} \, \mathbf{a}_{1} + y_{3} \, \mathbf{a}_{2} + z_{3} \, \mathbf{a}_{3} & = & \left(x_{3}a+z_{3}c\cos\beta\right) \, \mathbf{\hat{x}} + y_{3}b \, \mathbf{\hat{y}} + z_{3}c\sin\beta \, \mathbf{\hat{z}} & \left(2a\right) & \mbox{As III} \\ 
\mathbf{B}_{6} & = & x_{3} \, \mathbf{a}_{1}-y_{3} \, \mathbf{a}_{2} + \left(\frac{1}{2} +z_{3}\right) \, \mathbf{a}_{3} & = & \left(\frac{1}{2}c\cos\beta +x_{3}a + z_{3}c\cos\beta\right) \, \mathbf{\hat{x}}-y_{3}b \, \mathbf{\hat{y}} + \left(\frac{1}{2} +z_{3}\right)c\sin\beta \, \mathbf{\hat{z}} & \left(2a\right) & \mbox{As III} \\ 
\mathbf{B}_{7} & = & x_{4} \, \mathbf{a}_{1} + y_{4} \, \mathbf{a}_{2} + z_{4} \, \mathbf{a}_{3} & = & \left(x_{4}a+z_{4}c\cos\beta\right) \, \mathbf{\hat{x}} + y_{4}b \, \mathbf{\hat{y}} + z_{4}c\sin\beta \, \mathbf{\hat{z}} & \left(2a\right) & \mbox{As IV} \\ 
\mathbf{B}_{8} & = & x_{4} \, \mathbf{a}_{1}-y_{4} \, \mathbf{a}_{2} + \left(\frac{1}{2} +z_{4}\right) \, \mathbf{a}_{3} & = & \left(\frac{1}{2}c\cos\beta +x_{4}a + z_{4}c\cos\beta\right) \, \mathbf{\hat{x}}-y_{4}b \, \mathbf{\hat{y}} + \left(\frac{1}{2} +z_{4}\right)c\sin\beta \, \mathbf{\hat{z}} & \left(2a\right) & \mbox{As IV} \\ 
\mathbf{B}_{9} & = & x_{5} \, \mathbf{a}_{1} + y_{5} \, \mathbf{a}_{2} + z_{5} \, \mathbf{a}_{3} & = & \left(x_{5}a+z_{5}c\cos\beta\right) \, \mathbf{\hat{x}} + y_{5}b \, \mathbf{\hat{y}} + z_{5}c\sin\beta \, \mathbf{\hat{z}} & \left(2a\right) & \mbox{As V} \\ 
\mathbf{B}_{10} & = & x_{5} \, \mathbf{a}_{1}-y_{5} \, \mathbf{a}_{2} + \left(\frac{1}{2} +z_{5}\right) \, \mathbf{a}_{3} & = & \left(\frac{1}{2}c\cos\beta +x_{5}a + z_{5}c\cos\beta\right) \, \mathbf{\hat{x}}-y_{5}b \, \mathbf{\hat{y}} + \left(\frac{1}{2} +z_{5}\right)c\sin\beta \, \mathbf{\hat{z}} & \left(2a\right) & \mbox{As V} \\ 
\mathbf{B}_{11} & = & x_{6} \, \mathbf{a}_{1} + y_{6} \, \mathbf{a}_{2} + z_{6} \, \mathbf{a}_{3} & = & \left(x_{6}a+z_{6}c\cos\beta\right) \, \mathbf{\hat{x}} + y_{6}b \, \mathbf{\hat{y}} + z_{6}c\sin\beta \, \mathbf{\hat{z}} & \left(2a\right) & \mbox{As VI} \\ 
\mathbf{B}_{12} & = & x_{6} \, \mathbf{a}_{1}-y_{6} \, \mathbf{a}_{2} + \left(\frac{1}{2} +z_{6}\right) \, \mathbf{a}_{3} & = & \left(\frac{1}{2}c\cos\beta +x_{6}a + z_{6}c\cos\beta\right) \, \mathbf{\hat{x}}-y_{6}b \, \mathbf{\hat{y}} + \left(\frac{1}{2} +z_{6}\right)c\sin\beta \, \mathbf{\hat{z}} & \left(2a\right) & \mbox{As VI} \\ 
\mathbf{B}_{13} & = & x_{7} \, \mathbf{a}_{1} + y_{7} \, \mathbf{a}_{2} + z_{7} \, \mathbf{a}_{3} & = & \left(x_{7}a+z_{7}c\cos\beta\right) \, \mathbf{\hat{x}} + y_{7}b \, \mathbf{\hat{y}} + z_{7}c\sin\beta \, \mathbf{\hat{z}} & \left(2a\right) & \mbox{Ba I} \\ 
\mathbf{B}_{14} & = & x_{7} \, \mathbf{a}_{1}-y_{7} \, \mathbf{a}_{2} + \left(\frac{1}{2} +z_{7}\right) \, \mathbf{a}_{3} & = & \left(\frac{1}{2}c\cos\beta +x_{7}a + z_{7}c\cos\beta\right) \, \mathbf{\hat{x}}-y_{7}b \, \mathbf{\hat{y}} + \left(\frac{1}{2} +z_{7}\right)c\sin\beta \, \mathbf{\hat{z}} & \left(2a\right) & \mbox{Ba I} \\ 
\mathbf{B}_{15} & = & x_{8} \, \mathbf{a}_{1} + y_{8} \, \mathbf{a}_{2} + z_{8} \, \mathbf{a}_{3} & = & \left(x_{8}a+z_{8}c\cos\beta\right) \, \mathbf{\hat{x}} + y_{8}b \, \mathbf{\hat{y}} + z_{8}c\sin\beta \, \mathbf{\hat{z}} & \left(2a\right) & \mbox{Ba II} \\ 
\mathbf{B}_{16} & = & x_{8} \, \mathbf{a}_{1}-y_{8} \, \mathbf{a}_{2} + \left(\frac{1}{2} +z_{8}\right) \, \mathbf{a}_{3} & = & \left(\frac{1}{2}c\cos\beta +x_{8}a + z_{8}c\cos\beta\right) \, \mathbf{\hat{x}}-y_{8}b \, \mathbf{\hat{y}} + \left(\frac{1}{2} +z_{8}\right)c\sin\beta \, \mathbf{\hat{z}} & \left(2a\right) & \mbox{Ba II} \\ 
\mathbf{B}_{17} & = & x_{9} \, \mathbf{a}_{1} + y_{9} \, \mathbf{a}_{2} + z_{9} \, \mathbf{a}_{3} & = & \left(x_{9}a+z_{9}c\cos\beta\right) \, \mathbf{\hat{x}} + y_{9}b \, \mathbf{\hat{y}} + z_{9}c\sin\beta \, \mathbf{\hat{z}} & \left(2a\right) & \mbox{Ba III} \\ 
\mathbf{B}_{18} & = & x_{9} \, \mathbf{a}_{1}-y_{9} \, \mathbf{a}_{2} + \left(\frac{1}{2} +z_{9}\right) \, \mathbf{a}_{3} & = & \left(\frac{1}{2}c\cos\beta +x_{9}a + z_{9}c\cos\beta\right) \, \mathbf{\hat{x}}-y_{9}b \, \mathbf{\hat{y}} + \left(\frac{1}{2} +z_{9}\right)c\sin\beta \, \mathbf{\hat{z}} & \left(2a\right) & \mbox{Ba III} \\ 
\end{longtabu}
\renewcommand{\arraystretch}{1.0}
\noindent \hrulefill
\\
\textbf{References:}
\vspace*{-0.25cm}
\begin{flushleft}
  - \bibentry{Emmerling_As2Ba_ZeitAnorgAllgChem_2004}. \\
\end{flushleft}
\textbf{Found in:}
\vspace*{-0.25cm}
\begin{flushleft}
  - \bibentry{Villars_PearsonsCrystalData_2013}. \\
\end{flushleft}
\noindent \hrulefill
\\
\textbf{Geometry files:}
\\
\noindent  - CIF: pp. {\hyperref[A2B_mP18_7_6a_3a_cif]{\pageref{A2B_mP18_7_6a_3a_cif}}} \\
\noindent  - POSCAR: pp. {\hyperref[A2B_mP18_7_6a_3a_poscar]{\pageref{A2B_mP18_7_6a_3a_poscar}}} \\
\onecolumn
{\phantomsection\label{A3B_mP16_7_6a_2a}}
\subsection*{\huge \textbf{{\normalfont $\epsilon$-WO$_{3}$ (Low-temperature) Structure: A3B\_mP16\_7\_6a\_2a}}}
\noindent \hrulefill
\vspace*{0.25cm}
\begin{figure}[htp]
  \centering
  \vspace{-1em}
  {\includegraphics[width=1\textwidth]{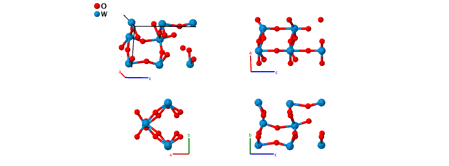}}
\end{figure}
\vspace*{-0.5cm}
\renewcommand{\arraystretch}{1.5}
\begin{equation*}
  \begin{array}{>{$\hspace{-0.15cm}}l<{$}>{$}p{0.5cm}<{$}>{$}p{18.5cm}<{$}}
    \mbox{\large \textbf{Prototype}} &\colon & \ce{$\epsilon$-WO3} \\
    \mbox{\large \textbf{\AFLOW\ prototype label}} &\colon & \mbox{A3B\_mP16\_7\_6a\_2a} \\
    \mbox{\large \textbf{\textit{Strukturbericht} designation}} &\colon & \mbox{None} \\
    \mbox{\large \textbf{Pearson symbol}} &\colon & \mbox{mP16} \\
    \mbox{\large \textbf{Space group number}} &\colon & 7 \\
    \mbox{\large \textbf{Space group symbol}} &\colon & Pc \\
    \mbox{\large \textbf{\AFLOW\ prototype command}} &\colon &  \texttt{aflow} \,  \, \texttt{-{}-proto=A3B\_mP16\_7\_6a\_2a } \, \newline \texttt{-{}-params=}{a,b/a,c/a,\beta,x_{1},y_{1},z_{1},x_{2},y_{2},z_{2},x_{3},y_{3},z_{3},x_{4},y_{4},z_{4},x_{5},y_{5},z_{5},x_{6},} \newline {y_{6},z_{6},x_{7},y_{7},z_{7},x_{8},y_{8},z_{8} }
  \end{array}
\end{equation*}
\renewcommand{\arraystretch}{1.0}

\vspace*{-0.25cm}
\noindent \hrulefill
\begin{itemize}
  \item{This is a low-temperature phase of tungsten trioxide (WO$_{3}$), which was observed at 15~K (Woodward, 1997).
}
\end{itemize}

\noindent \parbox{1 \linewidth}{
\noindent \hrulefill
\\
\textbf{Simple Monoclinic primitive vectors:} \\
\vspace*{-0.25cm}
\begin{tabular}{cc}
  \begin{tabular}{c}
    \parbox{0.6 \linewidth}{
      \renewcommand{\arraystretch}{1.5}
      \begin{equation*}
        \centering
        \begin{array}{ccc}
              \mathbf{a}_1 & = & a \, \mathbf{\hat{x}} \\
    \mathbf{a}_2 & = & b \, \mathbf{\hat{y}} \\
    \mathbf{a}_3 & = & c \cos\beta \, \mathbf{\hat{x}} + c \sin\beta \, \mathbf{\hat{z}} \\

        \end{array}
      \end{equation*}
    }
    \renewcommand{\arraystretch}{1.0}
  \end{tabular}
  \begin{tabular}{c}
    \includegraphics[width=0.3\linewidth]{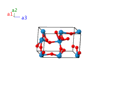} \\
  \end{tabular}
\end{tabular}

}
\vspace*{-0.25cm}

\noindent \hrulefill
\\
\textbf{Basis vectors:}
\vspace*{-0.25cm}
\renewcommand{\arraystretch}{1.5}
\begin{longtabu} to \textwidth{>{\centering $}X[-1,c,c]<{$}>{\centering $}X[-1,c,c]<{$}>{\centering $}X[-1,c,c]<{$}>{\centering $}X[-1,c,c]<{$}>{\centering $}X[-1,c,c]<{$}>{\centering $}X[-1,c,c]<{$}>{\centering $}X[-1,c,c]<{$}}
  & & \mbox{Lattice Coordinates} & & \mbox{Cartesian Coordinates} &\mbox{Wyckoff Position} & \mbox{Atom Type} \\  
  \mathbf{B}_{1} & = & x_{1} \, \mathbf{a}_{1} + y_{1} \, \mathbf{a}_{2} + z_{1} \, \mathbf{a}_{3} & = & \left(x_{1}a+z_{1}c\cos\beta\right) \, \mathbf{\hat{x}} + y_{1}b \, \mathbf{\hat{y}} + z_{1}c\sin\beta \, \mathbf{\hat{z}} & \left(2a\right) & \mbox{O I} \\ 
\mathbf{B}_{2} & = & x_{1} \, \mathbf{a}_{1}-y_{1} \, \mathbf{a}_{2} + \left(\frac{1}{2} +z_{1}\right) \, \mathbf{a}_{3} & = & \left(\frac{1}{2}c\cos\beta +x_{1}a + z_{1}c\cos\beta\right) \, \mathbf{\hat{x}}-y_{1}b \, \mathbf{\hat{y}} + \left(\frac{1}{2} +z_{1}\right)c\sin\beta \, \mathbf{\hat{z}} & \left(2a\right) & \mbox{O I} \\ 
\mathbf{B}_{3} & = & x_{2} \, \mathbf{a}_{1} + y_{2} \, \mathbf{a}_{2} + z_{2} \, \mathbf{a}_{3} & = & \left(x_{2}a+z_{2}c\cos\beta\right) \, \mathbf{\hat{x}} + y_{2}b \, \mathbf{\hat{y}} + z_{2}c\sin\beta \, \mathbf{\hat{z}} & \left(2a\right) & \mbox{O II} \\ 
\mathbf{B}_{4} & = & x_{2} \, \mathbf{a}_{1}-y_{2} \, \mathbf{a}_{2} + \left(\frac{1}{2} +z_{2}\right) \, \mathbf{a}_{3} & = & \left(\frac{1}{2}c\cos\beta +x_{2}a + z_{2}c\cos\beta\right) \, \mathbf{\hat{x}}-y_{2}b \, \mathbf{\hat{y}} + \left(\frac{1}{2} +z_{2}\right)c\sin\beta \, \mathbf{\hat{z}} & \left(2a\right) & \mbox{O II} \\ 
\mathbf{B}_{5} & = & x_{3} \, \mathbf{a}_{1} + y_{3} \, \mathbf{a}_{2} + z_{3} \, \mathbf{a}_{3} & = & \left(x_{3}a+z_{3}c\cos\beta\right) \, \mathbf{\hat{x}} + y_{3}b \, \mathbf{\hat{y}} + z_{3}c\sin\beta \, \mathbf{\hat{z}} & \left(2a\right) & \mbox{O III} \\ 
\mathbf{B}_{6} & = & x_{3} \, \mathbf{a}_{1}-y_{3} \, \mathbf{a}_{2} + \left(\frac{1}{2} +z_{3}\right) \, \mathbf{a}_{3} & = & \left(\frac{1}{2}c\cos\beta +x_{3}a + z_{3}c\cos\beta\right) \, \mathbf{\hat{x}}-y_{3}b \, \mathbf{\hat{y}} + \left(\frac{1}{2} +z_{3}\right)c\sin\beta \, \mathbf{\hat{z}} & \left(2a\right) & \mbox{O III} \\ 
\mathbf{B}_{7} & = & x_{4} \, \mathbf{a}_{1} + y_{4} \, \mathbf{a}_{2} + z_{4} \, \mathbf{a}_{3} & = & \left(x_{4}a+z_{4}c\cos\beta\right) \, \mathbf{\hat{x}} + y_{4}b \, \mathbf{\hat{y}} + z_{4}c\sin\beta \, \mathbf{\hat{z}} & \left(2a\right) & \mbox{O IV} \\ 
\mathbf{B}_{8} & = & x_{4} \, \mathbf{a}_{1}-y_{4} \, \mathbf{a}_{2} + \left(\frac{1}{2} +z_{4}\right) \, \mathbf{a}_{3} & = & \left(\frac{1}{2}c\cos\beta +x_{4}a + z_{4}c\cos\beta\right) \, \mathbf{\hat{x}}-y_{4}b \, \mathbf{\hat{y}} + \left(\frac{1}{2} +z_{4}\right)c\sin\beta \, \mathbf{\hat{z}} & \left(2a\right) & \mbox{O IV} \\ 
\mathbf{B}_{9} & = & x_{5} \, \mathbf{a}_{1} + y_{5} \, \mathbf{a}_{2} + z_{5} \, \mathbf{a}_{3} & = & \left(x_{5}a+z_{5}c\cos\beta\right) \, \mathbf{\hat{x}} + y_{5}b \, \mathbf{\hat{y}} + z_{5}c\sin\beta \, \mathbf{\hat{z}} & \left(2a\right) & \mbox{O V} \\ 
\mathbf{B}_{10} & = & x_{5} \, \mathbf{a}_{1}-y_{5} \, \mathbf{a}_{2} + \left(\frac{1}{2} +z_{5}\right) \, \mathbf{a}_{3} & = & \left(\frac{1}{2}c\cos\beta +x_{5}a + z_{5}c\cos\beta\right) \, \mathbf{\hat{x}}-y_{5}b \, \mathbf{\hat{y}} + \left(\frac{1}{2} +z_{5}\right)c\sin\beta \, \mathbf{\hat{z}} & \left(2a\right) & \mbox{O V} \\ 
\mathbf{B}_{11} & = & x_{6} \, \mathbf{a}_{1} + y_{6} \, \mathbf{a}_{2} + z_{6} \, \mathbf{a}_{3} & = & \left(x_{6}a+z_{6}c\cos\beta\right) \, \mathbf{\hat{x}} + y_{6}b \, \mathbf{\hat{y}} + z_{6}c\sin\beta \, \mathbf{\hat{z}} & \left(2a\right) & \mbox{O VI} \\ 
\mathbf{B}_{12} & = & x_{6} \, \mathbf{a}_{1}-y_{6} \, \mathbf{a}_{2} + \left(\frac{1}{2} +z_{6}\right) \, \mathbf{a}_{3} & = & \left(\frac{1}{2}c\cos\beta +x_{6}a + z_{6}c\cos\beta\right) \, \mathbf{\hat{x}}-y_{6}b \, \mathbf{\hat{y}} + \left(\frac{1}{2} +z_{6}\right)c\sin\beta \, \mathbf{\hat{z}} & \left(2a\right) & \mbox{O VI} \\ 
\mathbf{B}_{13} & = & x_{7} \, \mathbf{a}_{1} + y_{7} \, \mathbf{a}_{2} + z_{7} \, \mathbf{a}_{3} & = & \left(x_{7}a+z_{7}c\cos\beta\right) \, \mathbf{\hat{x}} + y_{7}b \, \mathbf{\hat{y}} + z_{7}c\sin\beta \, \mathbf{\hat{z}} & \left(2a\right) & \mbox{W I} \\ 
\mathbf{B}_{14} & = & x_{7} \, \mathbf{a}_{1}-y_{7} \, \mathbf{a}_{2} + \left(\frac{1}{2} +z_{7}\right) \, \mathbf{a}_{3} & = & \left(\frac{1}{2}c\cos\beta +x_{7}a + z_{7}c\cos\beta\right) \, \mathbf{\hat{x}}-y_{7}b \, \mathbf{\hat{y}} + \left(\frac{1}{2} +z_{7}\right)c\sin\beta \, \mathbf{\hat{z}} & \left(2a\right) & \mbox{W I} \\ 
\mathbf{B}_{15} & = & x_{8} \, \mathbf{a}_{1} + y_{8} \, \mathbf{a}_{2} + z_{8} \, \mathbf{a}_{3} & = & \left(x_{8}a+z_{8}c\cos\beta\right) \, \mathbf{\hat{x}} + y_{8}b \, \mathbf{\hat{y}} + z_{8}c\sin\beta \, \mathbf{\hat{z}} & \left(2a\right) & \mbox{W II} \\ 
\mathbf{B}_{16} & = & x_{8} \, \mathbf{a}_{1}-y_{8} \, \mathbf{a}_{2} + \left(\frac{1}{2} +z_{8}\right) \, \mathbf{a}_{3} & = & \left(\frac{1}{2}c\cos\beta +x_{8}a + z_{8}c\cos\beta\right) \, \mathbf{\hat{x}}-y_{8}b \, \mathbf{\hat{y}} + \left(\frac{1}{2} +z_{8}\right)c\sin\beta \, \mathbf{\hat{z}} & \left(2a\right) & \mbox{W II} \\ 
\end{longtabu}
\renewcommand{\arraystretch}{1.0}
\noindent \hrulefill
\\
\textbf{References:}
\vspace*{-0.25cm}
\begin{flushleft}
  - \bibentry{Woodward_WO3_JSolStateChem_1997}. \\
\end{flushleft}
\textbf{Found in:}
\vspace*{-0.25cm}
\begin{flushleft}
  - \bibentry{Villars_PearsonsCrystalData_2013}. \\
\end{flushleft}
\noindent \hrulefill
\\
\textbf{Geometry files:}
\\
\noindent  - CIF: pp. {\hyperref[A3B_mP16_7_6a_2a_cif]{\pageref{A3B_mP16_7_6a_2a_cif}}} \\
\noindent  - POSCAR: pp. {\hyperref[A3B_mP16_7_6a_2a_poscar]{\pageref{A3B_mP16_7_6a_2a_poscar}}} \\
\onecolumn
{\phantomsection\label{A9B2_mP22_7_9a_2a}}
\subsection*{\huge \textbf{{\normalfont Rh$_{2}$Ga$_{9}$ Structure: A9B2\_mP22\_7\_9a\_2a}}}
\noindent \hrulefill
\vspace*{0.25cm}
\begin{figure}[htp]
  \centering
  \vspace{-1em}
  {\includegraphics[width=1\textwidth]{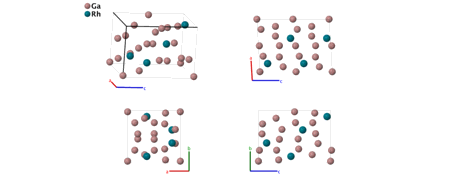}}
\end{figure}
\vspace*{-0.5cm}
\renewcommand{\arraystretch}{1.5}
\begin{equation*}
  \begin{array}{>{$\hspace{-0.15cm}}l<{$}>{$}p{0.5cm}<{$}>{$}p{18.5cm}<{$}}
    \mbox{\large \textbf{Prototype}} &\colon & \ce{Rh2Ga9} \\
    \mbox{\large \textbf{\AFLOW\ prototype label}} &\colon & \mbox{A9B2\_mP22\_7\_9a\_2a} \\
    \mbox{\large \textbf{\textit{Strukturbericht} designation}} &\colon & \mbox{None} \\
    \mbox{\large \textbf{Pearson symbol}} &\colon & \mbox{mP22} \\
    \mbox{\large \textbf{Space group number}} &\colon & 7 \\
    \mbox{\large \textbf{Space group symbol}} &\colon & Pc \\
    \mbox{\large \textbf{\AFLOW\ prototype command}} &\colon &  \texttt{aflow} \,  \, \texttt{-{}-proto=A9B2\_mP22\_7\_9a\_2a } \, \newline \texttt{-{}-params=}{a,b/a,c/a,\beta,x_{1},y_{1},z_{1},x_{2},y_{2},z_{2},x_{3},y_{3},z_{3},x_{4},y_{4},z_{4},x_{5},y_{5},z_{5},x_{6},} \newline {y_{6},z_{6},x_{7},y_{7},z_{7},x_{8},y_{8},z_{8},x_{9},y_{9},z_{9},x_{10},y_{10},z_{10},x_{11},y_{11},z_{11} }
  \end{array}
\end{equation*}
\renewcommand{\arraystretch}{1.0}

\noindent \parbox{1 \linewidth}{
\noindent \hrulefill
\\
\textbf{Simple Monoclinic primitive vectors:} \\
\vspace*{-0.25cm}
\begin{tabular}{cc}
  \begin{tabular}{c}
    \parbox{0.6 \linewidth}{
      \renewcommand{\arraystretch}{1.5}
      \begin{equation*}
        \centering
        \begin{array}{ccc}
              \mathbf{a}_1 & = & a \, \mathbf{\hat{x}} \\
    \mathbf{a}_2 & = & b \, \mathbf{\hat{y}} \\
    \mathbf{a}_3 & = & c \cos\beta \, \mathbf{\hat{x}} + c \sin\beta \, \mathbf{\hat{z}} \\

        \end{array}
      \end{equation*}
    }
    \renewcommand{\arraystretch}{1.0}
  \end{tabular}
  \begin{tabular}{c}
    \includegraphics[width=0.3\linewidth]{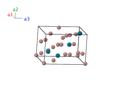} \\
  \end{tabular}
\end{tabular}

}
\vspace*{-0.25cm}

\noindent \hrulefill
\\
\textbf{Basis vectors:}
\vspace*{-0.25cm}
\renewcommand{\arraystretch}{1.5}
\begin{longtabu} to \textwidth{>{\centering $}X[-1,c,c]<{$}>{\centering $}X[-1,c,c]<{$}>{\centering $}X[-1,c,c]<{$}>{\centering $}X[-1,c,c]<{$}>{\centering $}X[-1,c,c]<{$}>{\centering $}X[-1,c,c]<{$}>{\centering $}X[-1,c,c]<{$}}
  & & \mbox{Lattice Coordinates} & & \mbox{Cartesian Coordinates} &\mbox{Wyckoff Position} & \mbox{Atom Type} \\  
  \mathbf{B}_{1} & = & x_{1} \, \mathbf{a}_{1} + y_{1} \, \mathbf{a}_{2} + z_{1} \, \mathbf{a}_{3} & = & \left(x_{1}a+z_{1}c\cos\beta\right) \, \mathbf{\hat{x}} + y_{1}b \, \mathbf{\hat{y}} + z_{1}c\sin\beta \, \mathbf{\hat{z}} & \left(2a\right) & \mbox{Ga I} \\ 
\mathbf{B}_{2} & = & x_{1} \, \mathbf{a}_{1}-y_{1} \, \mathbf{a}_{2} + \left(\frac{1}{2} +z_{1}\right) \, \mathbf{a}_{3} & = & \left(\frac{1}{2}c\cos\beta +x_{1}a + z_{1}c\cos\beta\right) \, \mathbf{\hat{x}}-y_{1}b \, \mathbf{\hat{y}} + \left(\frac{1}{2} +z_{1}\right)c\sin\beta \, \mathbf{\hat{z}} & \left(2a\right) & \mbox{Ga I} \\ 
\mathbf{B}_{3} & = & x_{2} \, \mathbf{a}_{1} + y_{2} \, \mathbf{a}_{2} + z_{2} \, \mathbf{a}_{3} & = & \left(x_{2}a+z_{2}c\cos\beta\right) \, \mathbf{\hat{x}} + y_{2}b \, \mathbf{\hat{y}} + z_{2}c\sin\beta \, \mathbf{\hat{z}} & \left(2a\right) & \mbox{Ga II} \\ 
\mathbf{B}_{4} & = & x_{2} \, \mathbf{a}_{1}-y_{2} \, \mathbf{a}_{2} + \left(\frac{1}{2} +z_{2}\right) \, \mathbf{a}_{3} & = & \left(\frac{1}{2}c\cos\beta +x_{2}a + z_{2}c\cos\beta\right) \, \mathbf{\hat{x}}-y_{2}b \, \mathbf{\hat{y}} + \left(\frac{1}{2} +z_{2}\right)c\sin\beta \, \mathbf{\hat{z}} & \left(2a\right) & \mbox{Ga II} \\ 
\mathbf{B}_{5} & = & x_{3} \, \mathbf{a}_{1} + y_{3} \, \mathbf{a}_{2} + z_{3} \, \mathbf{a}_{3} & = & \left(x_{3}a+z_{3}c\cos\beta\right) \, \mathbf{\hat{x}} + y_{3}b \, \mathbf{\hat{y}} + z_{3}c\sin\beta \, \mathbf{\hat{z}} & \left(2a\right) & \mbox{Ga III} \\ 
\mathbf{B}_{6} & = & x_{3} \, \mathbf{a}_{1}-y_{3} \, \mathbf{a}_{2} + \left(\frac{1}{2} +z_{3}\right) \, \mathbf{a}_{3} & = & \left(\frac{1}{2}c\cos\beta +x_{3}a + z_{3}c\cos\beta\right) \, \mathbf{\hat{x}}-y_{3}b \, \mathbf{\hat{y}} + \left(\frac{1}{2} +z_{3}\right)c\sin\beta \, \mathbf{\hat{z}} & \left(2a\right) & \mbox{Ga III} \\ 
\mathbf{B}_{7} & = & x_{4} \, \mathbf{a}_{1} + y_{4} \, \mathbf{a}_{2} + z_{4} \, \mathbf{a}_{3} & = & \left(x_{4}a+z_{4}c\cos\beta\right) \, \mathbf{\hat{x}} + y_{4}b \, \mathbf{\hat{y}} + z_{4}c\sin\beta \, \mathbf{\hat{z}} & \left(2a\right) & \mbox{Ga IV} \\ 
\mathbf{B}_{8} & = & x_{4} \, \mathbf{a}_{1}-y_{4} \, \mathbf{a}_{2} + \left(\frac{1}{2} +z_{4}\right) \, \mathbf{a}_{3} & = & \left(\frac{1}{2}c\cos\beta +x_{4}a + z_{4}c\cos\beta\right) \, \mathbf{\hat{x}}-y_{4}b \, \mathbf{\hat{y}} + \left(\frac{1}{2} +z_{4}\right)c\sin\beta \, \mathbf{\hat{z}} & \left(2a\right) & \mbox{Ga IV} \\ 
\mathbf{B}_{9} & = & x_{5} \, \mathbf{a}_{1} + y_{5} \, \mathbf{a}_{2} + z_{5} \, \mathbf{a}_{3} & = & \left(x_{5}a+z_{5}c\cos\beta\right) \, \mathbf{\hat{x}} + y_{5}b \, \mathbf{\hat{y}} + z_{5}c\sin\beta \, \mathbf{\hat{z}} & \left(2a\right) & \mbox{Ga V} \\ 
\mathbf{B}_{10} & = & x_{5} \, \mathbf{a}_{1}-y_{5} \, \mathbf{a}_{2} + \left(\frac{1}{2} +z_{5}\right) \, \mathbf{a}_{3} & = & \left(\frac{1}{2}c\cos\beta +x_{5}a + z_{5}c\cos\beta\right) \, \mathbf{\hat{x}}-y_{5}b \, \mathbf{\hat{y}} + \left(\frac{1}{2} +z_{5}\right)c\sin\beta \, \mathbf{\hat{z}} & \left(2a\right) & \mbox{Ga V} \\ 
\mathbf{B}_{11} & = & x_{6} \, \mathbf{a}_{1} + y_{6} \, \mathbf{a}_{2} + z_{6} \, \mathbf{a}_{3} & = & \left(x_{6}a+z_{6}c\cos\beta\right) \, \mathbf{\hat{x}} + y_{6}b \, \mathbf{\hat{y}} + z_{6}c\sin\beta \, \mathbf{\hat{z}} & \left(2a\right) & \mbox{Ga VI} \\ 
\mathbf{B}_{12} & = & x_{6} \, \mathbf{a}_{1}-y_{6} \, \mathbf{a}_{2} + \left(\frac{1}{2} +z_{6}\right) \, \mathbf{a}_{3} & = & \left(\frac{1}{2}c\cos\beta +x_{6}a + z_{6}c\cos\beta\right) \, \mathbf{\hat{x}}-y_{6}b \, \mathbf{\hat{y}} + \left(\frac{1}{2} +z_{6}\right)c\sin\beta \, \mathbf{\hat{z}} & \left(2a\right) & \mbox{Ga VI} \\ 
\mathbf{B}_{13} & = & x_{7} \, \mathbf{a}_{1} + y_{7} \, \mathbf{a}_{2} + z_{7} \, \mathbf{a}_{3} & = & \left(x_{7}a+z_{7}c\cos\beta\right) \, \mathbf{\hat{x}} + y_{7}b \, \mathbf{\hat{y}} + z_{7}c\sin\beta \, \mathbf{\hat{z}} & \left(2a\right) & \mbox{Ga VII} \\ 
\mathbf{B}_{14} & = & x_{7} \, \mathbf{a}_{1}-y_{7} \, \mathbf{a}_{2} + \left(\frac{1}{2} +z_{7}\right) \, \mathbf{a}_{3} & = & \left(\frac{1}{2}c\cos\beta +x_{7}a + z_{7}c\cos\beta\right) \, \mathbf{\hat{x}}-y_{7}b \, \mathbf{\hat{y}} + \left(\frac{1}{2} +z_{7}\right)c\sin\beta \, \mathbf{\hat{z}} & \left(2a\right) & \mbox{Ga VII} \\ 
\mathbf{B}_{15} & = & x_{8} \, \mathbf{a}_{1} + y_{8} \, \mathbf{a}_{2} + z_{8} \, \mathbf{a}_{3} & = & \left(x_{8}a+z_{8}c\cos\beta\right) \, \mathbf{\hat{x}} + y_{8}b \, \mathbf{\hat{y}} + z_{8}c\sin\beta \, \mathbf{\hat{z}} & \left(2a\right) & \mbox{Ga VIII} \\ 
\mathbf{B}_{16} & = & x_{8} \, \mathbf{a}_{1}-y_{8} \, \mathbf{a}_{2} + \left(\frac{1}{2} +z_{8}\right) \, \mathbf{a}_{3} & = & \left(\frac{1}{2}c\cos\beta +x_{8}a + z_{8}c\cos\beta\right) \, \mathbf{\hat{x}}-y_{8}b \, \mathbf{\hat{y}} + \left(\frac{1}{2} +z_{8}\right)c\sin\beta \, \mathbf{\hat{z}} & \left(2a\right) & \mbox{Ga VIII} \\ 
\mathbf{B}_{17} & = & x_{9} \, \mathbf{a}_{1} + y_{9} \, \mathbf{a}_{2} + z_{9} \, \mathbf{a}_{3} & = & \left(x_{9}a+z_{9}c\cos\beta\right) \, \mathbf{\hat{x}} + y_{9}b \, \mathbf{\hat{y}} + z_{9}c\sin\beta \, \mathbf{\hat{z}} & \left(2a\right) & \mbox{Ga IX} \\ 
\mathbf{B}_{18} & = & x_{9} \, \mathbf{a}_{1}-y_{9} \, \mathbf{a}_{2} + \left(\frac{1}{2} +z_{9}\right) \, \mathbf{a}_{3} & = & \left(\frac{1}{2}c\cos\beta +x_{9}a + z_{9}c\cos\beta\right) \, \mathbf{\hat{x}}-y_{9}b \, \mathbf{\hat{y}} + \left(\frac{1}{2} +z_{9}\right)c\sin\beta \, \mathbf{\hat{z}} & \left(2a\right) & \mbox{Ga IX} \\ 
\mathbf{B}_{19} & = & x_{10} \, \mathbf{a}_{1} + y_{10} \, \mathbf{a}_{2} + z_{10} \, \mathbf{a}_{3} & = & \left(x_{10}a+z_{10}c\cos\beta\right) \, \mathbf{\hat{x}} + y_{10}b \, \mathbf{\hat{y}} + z_{10}c\sin\beta \, \mathbf{\hat{z}} & \left(2a\right) & \mbox{Rh I} \\ 
\mathbf{B}_{20} & = & x_{10} \, \mathbf{a}_{1}-y_{10} \, \mathbf{a}_{2} + \left(\frac{1}{2} +z_{10}\right) \, \mathbf{a}_{3} & = & \left(\frac{1}{2}c\cos\beta +x_{10}a + z_{10}c\cos\beta\right) \, \mathbf{\hat{x}}-y_{10}b \, \mathbf{\hat{y}} + \left(\frac{1}{2} +z_{10}\right)c\sin\beta \, \mathbf{\hat{z}} & \left(2a\right) & \mbox{Rh I} \\ 
\mathbf{B}_{21} & = & x_{11} \, \mathbf{a}_{1} + y_{11} \, \mathbf{a}_{2} + z_{11} \, \mathbf{a}_{3} & = & \left(x_{11}a+z_{11}c\cos\beta\right) \, \mathbf{\hat{x}} + y_{11}b \, \mathbf{\hat{y}} + z_{11}c\sin\beta \, \mathbf{\hat{z}} & \left(2a\right) & \mbox{Rh II} \\ 
\mathbf{B}_{22} & = & x_{11} \, \mathbf{a}_{1}-y_{11} \, \mathbf{a}_{2} + \left(\frac{1}{2} +z_{11}\right) \, \mathbf{a}_{3} & = & \left(\frac{1}{2}c\cos\beta +x_{11}a + z_{11}c\cos\beta\right) \, \mathbf{\hat{x}}-y_{11}b \, \mathbf{\hat{y}} + \left(\frac{1}{2} +z_{11}\right)c\sin\beta \, \mathbf{\hat{z}} & \left(2a\right) & \mbox{Rh II} \\ 
\end{longtabu}
\renewcommand{\arraystretch}{1.0}
\noindent \hrulefill
\\
\textbf{References:}
\vspace*{-0.25cm}
\begin{flushleft}
  - \bibentry{Bostrom_Rh2Ga9_ZAnorgAllgeChem_2005}. \\
\end{flushleft}
\textbf{Found in:}
\vspace*{-0.25cm}
\begin{flushleft}
  - \bibentry{Villars_PearsonsCrystalData_2013}. \\
\end{flushleft}
\noindent \hrulefill
\\
\textbf{Geometry files:}
\\
\noindent  - CIF: pp. {\hyperref[A9B2_mP22_7_9a_2a_cif]{\pageref{A9B2_mP22_7_9a_2a_cif}}} \\
\noindent  - POSCAR: pp. {\hyperref[A9B2_mP22_7_9a_2a_poscar]{\pageref{A9B2_mP22_7_9a_2a_poscar}}} \\
\onecolumn
{\phantomsection\label{A5B3_mC32_9_5a_3a}}
\subsection*{\huge \textbf{{\normalfont $\alpha$-P$_3$N$_5$ Structure: A5B3\_mC32\_9\_5a\_3a}}}
\noindent \hrulefill
\vspace*{0.25cm}
\begin{figure}[htp]
  \centering
  \vspace{-1em}
  {\includegraphics[width=1\textwidth]{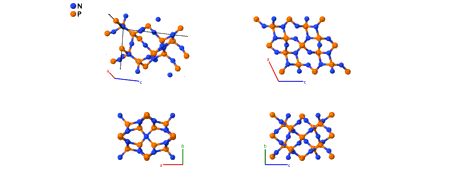}}
\end{figure}
\vspace*{-0.5cm}
\renewcommand{\arraystretch}{1.5}
\begin{equation*}
  \begin{array}{>{$\hspace{-0.15cm}}l<{$}>{$}p{0.5cm}<{$}>{$}p{18.5cm}<{$}}
    \mbox{\large \textbf{Prototype}} &\colon & \ce{$\alpha$-P3N5} \\
    \mbox{\large \textbf{\AFLOW\ prototype label}} &\colon & \mbox{A5B3\_mC32\_9\_5a\_3a} \\
    \mbox{\large \textbf{\textit{Strukturbericht} designation}} &\colon & \mbox{None} \\
    \mbox{\large \textbf{Pearson symbol}} &\colon & \mbox{mC32} \\
    \mbox{\large \textbf{Space group number}} &\colon & 9 \\
    \mbox{\large \textbf{Space group symbol}} &\colon & Cc \\
    \mbox{\large \textbf{\AFLOW\ prototype command}} &\colon &  \texttt{aflow} \,  \, \texttt{-{}-proto=A5B3\_mC32\_9\_5a\_3a } \, \newline \texttt{-{}-params=}{a,b/a,c/a,\beta,x_{1},y_{1},z_{1},x_{2},y_{2},z_{2},x_{3},y_{3},z_{3},x_{4},y_{4},z_{4},x_{5},y_{5},z_{5},x_{6},} \newline {y_{6},z_{6},x_{7},y_{7},z_{7},x_{8},y_{8},z_{8} }
  \end{array}
\end{equation*}
\renewcommand{\arraystretch}{1.0}

\noindent \parbox{1 \linewidth}{
\noindent \hrulefill
\\
\textbf{Base-centered Monoclinic primitive vectors:} \\
\vspace*{-0.25cm}
\begin{tabular}{cc}
  \begin{tabular}{c}
    \parbox{0.6 \linewidth}{
      \renewcommand{\arraystretch}{1.5}
      \begin{equation*}
        \centering
        \begin{array}{ccc}
              \mathbf{a}_1 & = & \frac12 \, a \, \mathbf{\hat{x}} - \frac12 \, b \, \mathbf{\hat{y}} \\
    \mathbf{a}_2 & = & \frac12 \, a \, \mathbf{\hat{x}} + \frac12 \, b \, \mathbf{\hat{y}} \\
    \mathbf{a}_3 & = & c \cos\beta \, \mathbf{\hat{x}} + c \sin\beta \, \mathbf{\hat{z}} \\

        \end{array}
      \end{equation*}
    }
    \renewcommand{\arraystretch}{1.0}
  \end{tabular}
  \begin{tabular}{c}
    \includegraphics[width=0.3\linewidth]{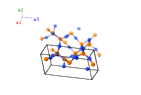} \\
  \end{tabular}
\end{tabular}

}
\vspace*{-0.25cm}

\noindent \hrulefill
\\
\textbf{Basis vectors:}
\vspace*{-0.25cm}
\renewcommand{\arraystretch}{1.5}
\begin{longtabu} to \textwidth{>{\centering $}X[-1,c,c]<{$}>{\centering $}X[-1,c,c]<{$}>{\centering $}X[-1,c,c]<{$}>{\centering $}X[-1,c,c]<{$}>{\centering $}X[-1,c,c]<{$}>{\centering $}X[-1,c,c]<{$}>{\centering $}X[-1,c,c]<{$}}
  & & \mbox{Lattice Coordinates} & & \mbox{Cartesian Coordinates} &\mbox{Wyckoff Position} & \mbox{Atom Type} \\  
  \mathbf{B}_{1} & = & \left(x_{1}-y_{1}\right) \, \mathbf{a}_{1} + \left(x_{1}+y_{1}\right) \, \mathbf{a}_{2} + z_{1} \, \mathbf{a}_{3} & = & \left(x_{1}a+z_{1}c\cos\beta\right) \, \mathbf{\hat{x}} + y_{1}b \, \mathbf{\hat{y}} + z_{1}c\sin\beta \, \mathbf{\hat{z}} & \left(4a\right) & \mbox{N I} \\ 
\mathbf{B}_{2} & = & \left(x_{1}+y_{1}\right) \, \mathbf{a}_{1} + \left(x_{1}-y_{1}\right) \, \mathbf{a}_{2} + \left(\frac{1}{2} +z_{1}\right) \, \mathbf{a}_{3} & = & \left(\frac{1}{2}c\cos\beta +x_{1}a + z_{1}c\cos\beta\right) \, \mathbf{\hat{x}}-y_{1}b \, \mathbf{\hat{y}} + \left(\frac{1}{2} +z_{1}\right)c\sin\beta \, \mathbf{\hat{z}} & \left(4a\right) & \mbox{N I} \\ 
\mathbf{B}_{3} & = & \left(x_{2}-y_{2}\right) \, \mathbf{a}_{1} + \left(x_{2}+y_{2}\right) \, \mathbf{a}_{2} + z_{2} \, \mathbf{a}_{3} & = & \left(x_{2}a+z_{2}c\cos\beta\right) \, \mathbf{\hat{x}} + y_{2}b \, \mathbf{\hat{y}} + z_{2}c\sin\beta \, \mathbf{\hat{z}} & \left(4a\right) & \mbox{N II} \\ 
\mathbf{B}_{4} & = & \left(x_{2}+y_{2}\right) \, \mathbf{a}_{1} + \left(x_{2}-y_{2}\right) \, \mathbf{a}_{2} + \left(\frac{1}{2} +z_{2}\right) \, \mathbf{a}_{3} & = & \left(\frac{1}{2}c\cos\beta +x_{2}a + z_{2}c\cos\beta\right) \, \mathbf{\hat{x}}-y_{2}b \, \mathbf{\hat{y}} + \left(\frac{1}{2} +z_{2}\right)c\sin\beta \, \mathbf{\hat{z}} & \left(4a\right) & \mbox{N II} \\ 
\mathbf{B}_{5} & = & \left(x_{3}-y_{3}\right) \, \mathbf{a}_{1} + \left(x_{3}+y_{3}\right) \, \mathbf{a}_{2} + z_{3} \, \mathbf{a}_{3} & = & \left(x_{3}a+z_{3}c\cos\beta\right) \, \mathbf{\hat{x}} + y_{3}b \, \mathbf{\hat{y}} + z_{3}c\sin\beta \, \mathbf{\hat{z}} & \left(4a\right) & \mbox{N III} \\ 
\mathbf{B}_{6} & = & \left(x_{3}+y_{3}\right) \, \mathbf{a}_{1} + \left(x_{3}-y_{3}\right) \, \mathbf{a}_{2} + \left(\frac{1}{2} +z_{3}\right) \, \mathbf{a}_{3} & = & \left(\frac{1}{2}c\cos\beta +x_{3}a + z_{3}c\cos\beta\right) \, \mathbf{\hat{x}}-y_{3}b \, \mathbf{\hat{y}} + \left(\frac{1}{2} +z_{3}\right)c\sin\beta \, \mathbf{\hat{z}} & \left(4a\right) & \mbox{N III} \\ 
\mathbf{B}_{7} & = & \left(x_{4}-y_{4}\right) \, \mathbf{a}_{1} + \left(x_{4}+y_{4}\right) \, \mathbf{a}_{2} + z_{4} \, \mathbf{a}_{3} & = & \left(x_{4}a+z_{4}c\cos\beta\right) \, \mathbf{\hat{x}} + y_{4}b \, \mathbf{\hat{y}} + z_{4}c\sin\beta \, \mathbf{\hat{z}} & \left(4a\right) & \mbox{N IV} \\ 
\mathbf{B}_{8} & = & \left(x_{4}+y_{4}\right) \, \mathbf{a}_{1} + \left(x_{4}-y_{4}\right) \, \mathbf{a}_{2} + \left(\frac{1}{2} +z_{4}\right) \, \mathbf{a}_{3} & = & \left(\frac{1}{2}c\cos\beta +x_{4}a + z_{4}c\cos\beta\right) \, \mathbf{\hat{x}}-y_{4}b \, \mathbf{\hat{y}} + \left(\frac{1}{2} +z_{4}\right)c\sin\beta \, \mathbf{\hat{z}} & \left(4a\right) & \mbox{N IV} \\ 
\mathbf{B}_{9} & = & \left(x_{5}-y_{5}\right) \, \mathbf{a}_{1} + \left(x_{5}+y_{5}\right) \, \mathbf{a}_{2} + z_{5} \, \mathbf{a}_{3} & = & \left(x_{5}a+z_{5}c\cos\beta\right) \, \mathbf{\hat{x}} + y_{5}b \, \mathbf{\hat{y}} + z_{5}c\sin\beta \, \mathbf{\hat{z}} & \left(4a\right) & \mbox{N V} \\ 
\mathbf{B}_{10} & = & \left(x_{5}+y_{5}\right) \, \mathbf{a}_{1} + \left(x_{5}-y_{5}\right) \, \mathbf{a}_{2} + \left(\frac{1}{2} +z_{5}\right) \, \mathbf{a}_{3} & = & \left(\frac{1}{2}c\cos\beta +x_{5}a + z_{5}c\cos\beta\right) \, \mathbf{\hat{x}}-y_{5}b \, \mathbf{\hat{y}} + \left(\frac{1}{2} +z_{5}\right)c\sin\beta \, \mathbf{\hat{z}} & \left(4a\right) & \mbox{N V} \\ 
\mathbf{B}_{11} & = & \left(x_{6}-y_{6}\right) \, \mathbf{a}_{1} + \left(x_{6}+y_{6}\right) \, \mathbf{a}_{2} + z_{6} \, \mathbf{a}_{3} & = & \left(x_{6}a+z_{6}c\cos\beta\right) \, \mathbf{\hat{x}} + y_{6}b \, \mathbf{\hat{y}} + z_{6}c\sin\beta \, \mathbf{\hat{z}} & \left(4a\right) & \mbox{P I} \\ 
\mathbf{B}_{12} & = & \left(x_{6}+y_{6}\right) \, \mathbf{a}_{1} + \left(x_{6}-y_{6}\right) \, \mathbf{a}_{2} + \left(\frac{1}{2} +z_{6}\right) \, \mathbf{a}_{3} & = & \left(\frac{1}{2}c\cos\beta +x_{6}a + z_{6}c\cos\beta\right) \, \mathbf{\hat{x}}-y_{6}b \, \mathbf{\hat{y}} + \left(\frac{1}{2} +z_{6}\right)c\sin\beta \, \mathbf{\hat{z}} & \left(4a\right) & \mbox{P I} \\ 
\mathbf{B}_{13} & = & \left(x_{7}-y_{7}\right) \, \mathbf{a}_{1} + \left(x_{7}+y_{7}\right) \, \mathbf{a}_{2} + z_{7} \, \mathbf{a}_{3} & = & \left(x_{7}a+z_{7}c\cos\beta\right) \, \mathbf{\hat{x}} + y_{7}b \, \mathbf{\hat{y}} + z_{7}c\sin\beta \, \mathbf{\hat{z}} & \left(4a\right) & \mbox{P II} \\ 
\mathbf{B}_{14} & = & \left(x_{7}+y_{7}\right) \, \mathbf{a}_{1} + \left(x_{7}-y_{7}\right) \, \mathbf{a}_{2} + \left(\frac{1}{2} +z_{7}\right) \, \mathbf{a}_{3} & = & \left(\frac{1}{2}c\cos\beta +x_{7}a + z_{7}c\cos\beta\right) \, \mathbf{\hat{x}}-y_{7}b \, \mathbf{\hat{y}} + \left(\frac{1}{2} +z_{7}\right)c\sin\beta \, \mathbf{\hat{z}} & \left(4a\right) & \mbox{P II} \\ 
\mathbf{B}_{15} & = & \left(x_{8}-y_{8}\right) \, \mathbf{a}_{1} + \left(x_{8}+y_{8}\right) \, \mathbf{a}_{2} + z_{8} \, \mathbf{a}_{3} & = & \left(x_{8}a+z_{8}c\cos\beta\right) \, \mathbf{\hat{x}} + y_{8}b \, \mathbf{\hat{y}} + z_{8}c\sin\beta \, \mathbf{\hat{z}} & \left(4a\right) & \mbox{P III} \\ 
\mathbf{B}_{16} & = & \left(x_{8}+y_{8}\right) \, \mathbf{a}_{1} + \left(x_{8}-y_{8}\right) \, \mathbf{a}_{2} + \left(\frac{1}{2} +z_{8}\right) \, \mathbf{a}_{3} & = & \left(\frac{1}{2}c\cos\beta +x_{8}a + z_{8}c\cos\beta\right) \, \mathbf{\hat{x}}-y_{8}b \, \mathbf{\hat{y}} + \left(\frac{1}{2} +z_{8}\right)c\sin\beta \, \mathbf{\hat{z}} & \left(4a\right) & \mbox{P III} \\ 
\end{longtabu}
\renewcommand{\arraystretch}{1.0}
\noindent \hrulefill
\\
\textbf{References:}
\vspace*{-0.25cm}
\begin{flushleft}
  - \bibentry{Horstmann_AngChemIntEd_36_1873_1997}. \\
\end{flushleft}
\noindent \hrulefill
\\
\textbf{Geometry files:}
\\
\noindent  - CIF: pp. {\hyperref[A5B3_mC32_9_5a_3a_cif]{\pageref{A5B3_mC32_9_5a_3a_cif}}} \\
\noindent  - POSCAR: pp. {\hyperref[A5B3_mC32_9_5a_3a_poscar]{\pageref{A5B3_mC32_9_5a_3a_poscar}}} \\
\onecolumn
{\phantomsection\label{AB3_mC16_9_a_3a}}
\subsection*{\huge \textbf{{\normalfont H$_{3}$Cl (20~GPa) Structure: AB3\_mC16\_9\_a\_3a}}}
\noindent \hrulefill
\vspace*{0.25cm}
\begin{figure}[htp]
  \centering
  \vspace{-1em}
  {\includegraphics[width=1\textwidth]{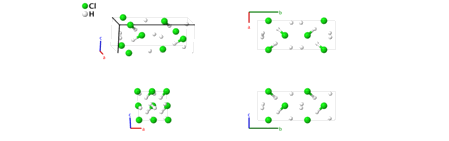}}
\end{figure}
\vspace*{-0.5cm}
\renewcommand{\arraystretch}{1.5}
\begin{equation*}
  \begin{array}{>{$\hspace{-0.15cm}}l<{$}>{$}p{0.5cm}<{$}>{$}p{18.5cm}<{$}}
    \mbox{\large \textbf{Prototype}} &\colon & \ce{H3Cl} \\
    \mbox{\large \textbf{\AFLOW\ prototype label}} &\colon & \mbox{AB3\_mC16\_9\_a\_3a} \\
    \mbox{\large \textbf{\textit{Strukturbericht} designation}} &\colon & \mbox{None} \\
    \mbox{\large \textbf{Pearson symbol}} &\colon & \mbox{mC16} \\
    \mbox{\large \textbf{Space group number}} &\colon & 9 \\
    \mbox{\large \textbf{Space group symbol}} &\colon & Cc \\
    \mbox{\large \textbf{\AFLOW\ prototype command}} &\colon &  \texttt{aflow} \,  \, \texttt{-{}-proto=AB3\_mC16\_9\_a\_3a } \, \newline \texttt{-{}-params=}{a,b/a,c/a,\beta,x_{1},y_{1},z_{1},x_{2},y_{2},z_{2},x_{3},y_{3},z_{3},x_{4},y_{4},z_{4} }
  \end{array}
\end{equation*}
\renewcommand{\arraystretch}{1.0}

\vspace*{-0.25cm}
\noindent \hrulefill
\begin{itemize}
  \item{This structure was found via first-principles calculations.  The data
presented here was computed at a pressure of 20~GPa.
}
\end{itemize}

\noindent \parbox{1 \linewidth}{
\noindent \hrulefill
\\
\textbf{Base-centered Monoclinic primitive vectors:} \\
\vspace*{-0.25cm}
\begin{tabular}{cc}
  \begin{tabular}{c}
    \parbox{0.6 \linewidth}{
      \renewcommand{\arraystretch}{1.5}
      \begin{equation*}
        \centering
        \begin{array}{ccc}
              \mathbf{a}_1 & = & \frac12 \, a \, \mathbf{\hat{x}} - \frac12 \, b \, \mathbf{\hat{y}} \\
    \mathbf{a}_2 & = & \frac12 \, a \, \mathbf{\hat{x}} + \frac12 \, b \, \mathbf{\hat{y}} \\
    \mathbf{a}_3 & = & c \cos\beta \, \mathbf{\hat{x}} + c \sin\beta \, \mathbf{\hat{z}} \\

        \end{array}
      \end{equation*}
    }
    \renewcommand{\arraystretch}{1.0}
  \end{tabular}
  \begin{tabular}{c}
    \includegraphics[width=0.3\linewidth]{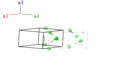} \\
  \end{tabular}
\end{tabular}

}
\vspace*{-0.25cm}

\noindent \hrulefill
\\
\textbf{Basis vectors:}
\vspace*{-0.25cm}
\renewcommand{\arraystretch}{1.5}
\begin{longtabu} to \textwidth{>{\centering $}X[-1,c,c]<{$}>{\centering $}X[-1,c,c]<{$}>{\centering $}X[-1,c,c]<{$}>{\centering $}X[-1,c,c]<{$}>{\centering $}X[-1,c,c]<{$}>{\centering $}X[-1,c,c]<{$}>{\centering $}X[-1,c,c]<{$}}
  & & \mbox{Lattice Coordinates} & & \mbox{Cartesian Coordinates} &\mbox{Wyckoff Position} & \mbox{Atom Type} \\  
  \mathbf{B}_{1} & = & \left(x_{1}-y_{1}\right) \, \mathbf{a}_{1} + \left(x_{1}+y_{1}\right) \, \mathbf{a}_{2} + z_{1} \, \mathbf{a}_{3} & = & \left(x_{1}a+z_{1}c\cos\beta\right) \, \mathbf{\hat{x}} + y_{1}b \, \mathbf{\hat{y}} + z_{1}c\sin\beta \, \mathbf{\hat{z}} & \left(4a\right) & \mbox{Cl} \\ 
\mathbf{B}_{2} & = & \left(x_{1}+y_{1}\right) \, \mathbf{a}_{1} + \left(x_{1}-y_{1}\right) \, \mathbf{a}_{2} + \left(\frac{1}{2} +z_{1}\right) \, \mathbf{a}_{3} & = & \left(\frac{1}{2}c\cos\beta +x_{1}a + z_{1}c\cos\beta\right) \, \mathbf{\hat{x}}-y_{1}b \, \mathbf{\hat{y}} + \left(\frac{1}{2} +z_{1}\right)c\sin\beta \, \mathbf{\hat{z}} & \left(4a\right) & \mbox{Cl} \\ 
\mathbf{B}_{3} & = & \left(x_{2}-y_{2}\right) \, \mathbf{a}_{1} + \left(x_{2}+y_{2}\right) \, \mathbf{a}_{2} + z_{2} \, \mathbf{a}_{3} & = & \left(x_{2}a+z_{2}c\cos\beta\right) \, \mathbf{\hat{x}} + y_{2}b \, \mathbf{\hat{y}} + z_{2}c\sin\beta \, \mathbf{\hat{z}} & \left(4a\right) & \mbox{H I} \\ 
\mathbf{B}_{4} & = & \left(x_{2}+y_{2}\right) \, \mathbf{a}_{1} + \left(x_{2}-y_{2}\right) \, \mathbf{a}_{2} + \left(\frac{1}{2} +z_{2}\right) \, \mathbf{a}_{3} & = & \left(\frac{1}{2}c\cos\beta +x_{2}a + z_{2}c\cos\beta\right) \, \mathbf{\hat{x}}-y_{2}b \, \mathbf{\hat{y}} + \left(\frac{1}{2} +z_{2}\right)c\sin\beta \, \mathbf{\hat{z}} & \left(4a\right) & \mbox{H I} \\ 
\mathbf{B}_{5} & = & \left(x_{3}-y_{3}\right) \, \mathbf{a}_{1} + \left(x_{3}+y_{3}\right) \, \mathbf{a}_{2} + z_{3} \, \mathbf{a}_{3} & = & \left(x_{3}a+z_{3}c\cos\beta\right) \, \mathbf{\hat{x}} + y_{3}b \, \mathbf{\hat{y}} + z_{3}c\sin\beta \, \mathbf{\hat{z}} & \left(4a\right) & \mbox{H II} \\ 
\mathbf{B}_{6} & = & \left(x_{3}+y_{3}\right) \, \mathbf{a}_{1} + \left(x_{3}-y_{3}\right) \, \mathbf{a}_{2} + \left(\frac{1}{2} +z_{3}\right) \, \mathbf{a}_{3} & = & \left(\frac{1}{2}c\cos\beta +x_{3}a + z_{3}c\cos\beta\right) \, \mathbf{\hat{x}}-y_{3}b \, \mathbf{\hat{y}} + \left(\frac{1}{2} +z_{3}\right)c\sin\beta \, \mathbf{\hat{z}} & \left(4a\right) & \mbox{H II} \\ 
\mathbf{B}_{7} & = & \left(x_{4}-y_{4}\right) \, \mathbf{a}_{1} + \left(x_{4}+y_{4}\right) \, \mathbf{a}_{2} + z_{4} \, \mathbf{a}_{3} & = & \left(x_{4}a+z_{4}c\cos\beta\right) \, \mathbf{\hat{x}} + y_{4}b \, \mathbf{\hat{y}} + z_{4}c\sin\beta \, \mathbf{\hat{z}} & \left(4a\right) & \mbox{H III} \\ 
\mathbf{B}_{8} & = & \left(x_{4}+y_{4}\right) \, \mathbf{a}_{1} + \left(x_{4}-y_{4}\right) \, \mathbf{a}_{2} + \left(\frac{1}{2} +z_{4}\right) \, \mathbf{a}_{3} & = & \left(\frac{1}{2}c\cos\beta +x_{4}a + z_{4}c\cos\beta\right) \, \mathbf{\hat{x}}-y_{4}b \, \mathbf{\hat{y}} + \left(\frac{1}{2} +z_{4}\right)c\sin\beta \, \mathbf{\hat{z}} & \left(4a\right) & \mbox{H III} \\ 
\end{longtabu}
\renewcommand{\arraystretch}{1.0}
\noindent \hrulefill
\\
\textbf{References:}
\vspace*{-0.25cm}
\begin{flushleft}
  - \bibentry{Duan_2015}. \\
\end{flushleft}
\noindent \hrulefill
\\
\textbf{Geometry files:}
\\
\noindent  - CIF: pp. {\hyperref[AB3_mC16_9_a_3a_cif]{\pageref{AB3_mC16_9_a_3a_cif}}} \\
\noindent  - POSCAR: pp. {\hyperref[AB3_mC16_9_a_3a_poscar]{\pageref{AB3_mC16_9_a_3a_poscar}}} \\
\onecolumn
{\phantomsection\label{A2B_mP6_10_mn_bg}}
\subsection*{\huge \textbf{{\normalfont $\delta$-PdCl$_{2}$ Structure: A2B\_mP6\_10\_mn\_bg}}}
\noindent \hrulefill
\vspace*{0.25cm}
\begin{figure}[htp]
  \centering
  \vspace{-1em}
  {\includegraphics[width=1\textwidth]{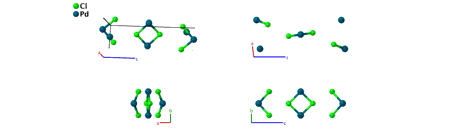}}
\end{figure}
\vspace*{-0.5cm}
\renewcommand{\arraystretch}{1.5}
\begin{equation*}
  \begin{array}{>{$\hspace{-0.15cm}}l<{$}>{$}p{0.5cm}<{$}>{$}p{18.5cm}<{$}}
    \mbox{\large \textbf{Prototype}} &\colon & \ce{$\delta$-PdCl2} \\
    \mbox{\large \textbf{\AFLOW\ prototype label}} &\colon & \mbox{A2B\_mP6\_10\_mn\_bg} \\
    \mbox{\large \textbf{\textit{Strukturbericht} designation}} &\colon & \mbox{None} \\
    \mbox{\large \textbf{Pearson symbol}} &\colon & \mbox{mP6} \\
    \mbox{\large \textbf{Space group number}} &\colon & 10 \\
    \mbox{\large \textbf{Space group symbol}} &\colon & P2/m \\
    \mbox{\large \textbf{\AFLOW\ prototype command}} &\colon &  \texttt{aflow} \,  \, \texttt{-{}-proto=A2B\_mP6\_10\_mn\_bg } \, \newline \texttt{-{}-params=}{a,b/a,c/a,\beta,x_{3},z_{3},x_{4},z_{4} }
  \end{array}
\end{equation*}
\renewcommand{\arraystretch}{1.0}

\vspace*{-0.25cm}
\noindent \hrulefill
\begin{itemize}
  \item{(Evers, 2010) use the unique-axis $c$ setting of space group $P2/m$.  
We have switched this to our standard unique-axis $b$ setting.
The data was taken at 793~K.
}
\end{itemize}

\noindent \parbox{1 \linewidth}{
\noindent \hrulefill
\\
\textbf{Simple Monoclinic primitive vectors:} \\
\vspace*{-0.25cm}
\begin{tabular}{cc}
  \begin{tabular}{c}
    \parbox{0.6 \linewidth}{
      \renewcommand{\arraystretch}{1.5}
      \begin{equation*}
        \centering
        \begin{array}{ccc}
              \mathbf{a}_1 & = & a \, \mathbf{\hat{x}} \\
    \mathbf{a}_2 & = & b \, \mathbf{\hat{y}} \\
    \mathbf{a}_3 & = & c \cos\beta \, \mathbf{\hat{x}} + c \sin\beta \, \mathbf{\hat{z}} \\

        \end{array}
      \end{equation*}
    }
    \renewcommand{\arraystretch}{1.0}
  \end{tabular}
  \begin{tabular}{c}
    \includegraphics[width=0.3\linewidth]{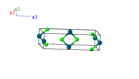} \\
  \end{tabular}
\end{tabular}

}
\vspace*{-0.25cm}

\noindent \hrulefill
\\
\textbf{Basis vectors:}
\vspace*{-0.25cm}
\renewcommand{\arraystretch}{1.5}
\begin{longtabu} to \textwidth{>{\centering $}X[-1,c,c]<{$}>{\centering $}X[-1,c,c]<{$}>{\centering $}X[-1,c,c]<{$}>{\centering $}X[-1,c,c]<{$}>{\centering $}X[-1,c,c]<{$}>{\centering $}X[-1,c,c]<{$}>{\centering $}X[-1,c,c]<{$}}
  & & \mbox{Lattice Coordinates} & & \mbox{Cartesian Coordinates} &\mbox{Wyckoff Position} & \mbox{Atom Type} \\  
  \mathbf{B}_{1} & = & \frac{1}{2} \, \mathbf{a}_{2} & = & \frac{1}{2}b \, \mathbf{\hat{y}} & \left(1b\right) & \mbox{Pd I} \\ 
\mathbf{B}_{2} & = & \frac{1}{2} \, \mathbf{a}_{1} + \frac{1}{2} \, \mathbf{a}_{3} & = & \frac{1}{2}\left(a+c\cos\beta\right) \, \mathbf{\hat{x}} + \frac{1}{2}c\sin\beta \, \mathbf{\hat{z}} & \left(1g\right) & \mbox{Pd II} \\ 
\mathbf{B}_{3} & = & x_{3} \, \mathbf{a}_{1} + z_{3} \, \mathbf{a}_{3} & = & \left(x_{3}a+z_{3}c\cos\beta\right) \, \mathbf{\hat{x}} + z_{3}c\sin\beta \, \mathbf{\hat{z}} & \left(2m\right) & \mbox{Cl I} \\ 
\mathbf{B}_{4} & = & -x_{3} \, \mathbf{a}_{1} + -z_{3} \, \mathbf{a}_{3} & = & \left(-x_{3}a-z_{3}c\cos\beta\right) \, \mathbf{\hat{x}} + -z_{3}c\sin\beta \, \mathbf{\hat{z}} & \left(2m\right) & \mbox{Cl I} \\ 
\mathbf{B}_{5} & = & x_{4} \, \mathbf{a}_{1} + \frac{1}{2} \, \mathbf{a}_{2} + z_{4} \, \mathbf{a}_{3} & = & \left(x_{4}a+z_{4}c\cos\beta\right) \, \mathbf{\hat{x}} + \frac{1}{2}b \, \mathbf{\hat{y}} + z_{4}c\sin\beta \, \mathbf{\hat{z}} & \left(2n\right) & \mbox{Cl II} \\ 
\mathbf{B}_{6} & = & -x_{4} \, \mathbf{a}_{1} + \frac{1}{2} \, \mathbf{a}_{2}-z_{4} \, \mathbf{a}_{3} & = & \left(-x_{4}a-z_{4}c\cos\beta\right) \, \mathbf{\hat{x}} + \frac{1}{2}b \, \mathbf{\hat{y}}-z_{4}c\sin\beta \, \mathbf{\hat{z}} & \left(2n\right) & \mbox{Cl II} \\ 
\end{longtabu}
\renewcommand{\arraystretch}{1.0}
\noindent \hrulefill
\\
\textbf{References:}
\vspace*{-0.25cm}
\begin{flushleft}
  - \bibentry{Evers_AngewChemIntEd_49_5677_2010}. \\
\end{flushleft}
\noindent \hrulefill
\\
\textbf{Geometry files:}
\\
\noindent  - CIF: pp. {\hyperref[A2B_mP6_10_mn_bg_cif]{\pageref{A2B_mP6_10_mn_bg_cif}}} \\
\noindent  - POSCAR: pp. {\hyperref[A2B_mP6_10_mn_bg_poscar]{\pageref{A2B_mP6_10_mn_bg_poscar}}} \\
\onecolumn
{\phantomsection\label{AB3_mP16_10_mn_3m3n}}
\subsection*{\huge \textbf{{\normalfont H$_{3}$Cl (400~GPa) Structure: AB3\_mP16\_10\_mn\_3m3n}}}
\noindent \hrulefill
\vspace*{0.25cm}
\begin{figure}[htp]
  \centering
  \vspace{-1em}
  {\includegraphics[width=1\textwidth]{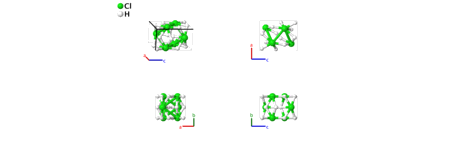}}
\end{figure}
\vspace*{-0.5cm}
\renewcommand{\arraystretch}{1.5}
\begin{equation*}
  \begin{array}{>{$\hspace{-0.15cm}}l<{$}>{$}p{0.5cm}<{$}>{$}p{18.5cm}<{$}}
    \mbox{\large \textbf{Prototype}} &\colon & \ce{H3Cl} \\
    \mbox{\large \textbf{\AFLOW\ prototype label}} &\colon & \mbox{AB3\_mP16\_10\_mn\_3m3n} \\
    \mbox{\large \textbf{\textit{Strukturbericht} designation}} &\colon & \mbox{None} \\
    \mbox{\large \textbf{Pearson symbol}} &\colon & \mbox{mP16} \\
    \mbox{\large \textbf{Space group number}} &\colon & 10 \\
    \mbox{\large \textbf{Space group symbol}} &\colon & P2/m \\
    \mbox{\large \textbf{\AFLOW\ prototype command}} &\colon &  \texttt{aflow} \,  \, \texttt{-{}-proto=AB3\_mP16\_10\_mn\_3m3n } \, \newline \texttt{-{}-params=}{a,b/a,c/a,\beta,x_{1},z_{1},x_{2},z_{2},x_{3},z_{3},x_{4},z_{4},x_{5},z_{5},x_{6},z_{6},x_{7},z_{7},x_{8},z_{8} }
  \end{array}
\end{equation*}
\renewcommand{\arraystretch}{1.0}

\vspace*{-0.25cm}
\noindent \hrulefill
\begin{itemize}
  \item{This structure was found via first-principles calculations.  The data
presented here was computed at a pressure of 400~GPa.
}
\end{itemize}

\noindent \parbox{1 \linewidth}{
\noindent \hrulefill
\\
\textbf{Simple Monoclinic primitive vectors:} \\
\vspace*{-0.25cm}
\begin{tabular}{cc}
  \begin{tabular}{c}
    \parbox{0.6 \linewidth}{
      \renewcommand{\arraystretch}{1.5}
      \begin{equation*}
        \centering
        \begin{array}{ccc}
              \mathbf{a}_1 & = & a \, \mathbf{\hat{x}} \\
    \mathbf{a}_2 & = & b \, \mathbf{\hat{y}} \\
    \mathbf{a}_3 & = & c \cos\beta \, \mathbf{\hat{x}} + c \sin\beta \, \mathbf{\hat{z}} \\

        \end{array}
      \end{equation*}
    }
    \renewcommand{\arraystretch}{1.0}
  \end{tabular}
  \begin{tabular}{c}
    \includegraphics[width=0.3\linewidth]{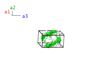} \\
  \end{tabular}
\end{tabular}

}
\vspace*{-0.25cm}

\noindent \hrulefill
\\
\textbf{Basis vectors:}
\vspace*{-0.25cm}
\renewcommand{\arraystretch}{1.5}
\begin{longtabu} to \textwidth{>{\centering $}X[-1,c,c]<{$}>{\centering $}X[-1,c,c]<{$}>{\centering $}X[-1,c,c]<{$}>{\centering $}X[-1,c,c]<{$}>{\centering $}X[-1,c,c]<{$}>{\centering $}X[-1,c,c]<{$}>{\centering $}X[-1,c,c]<{$}}
  & & \mbox{Lattice Coordinates} & & \mbox{Cartesian Coordinates} &\mbox{Wyckoff Position} & \mbox{Atom Type} \\  
  \mathbf{B}_{1} & = & x_{1} \, \mathbf{a}_{1} + z_{1} \, \mathbf{a}_{3} & = & \left(x_{1}a+z_{1}c\cos\beta\right) \, \mathbf{\hat{x}} + z_{1}c\sin\beta \, \mathbf{\hat{z}} & \left(2m\right) & \mbox{Cl I} \\ 
\mathbf{B}_{2} & = & -x_{1} \, \mathbf{a}_{1} + -z_{1} \, \mathbf{a}_{3} & = & \left(-x_{1}a-z_{1}c\cos\beta\right) \, \mathbf{\hat{x}} + -z_{1}c\sin\beta \, \mathbf{\hat{z}} & \left(2m\right) & \mbox{Cl I} \\ 
\mathbf{B}_{3} & = & x_{2} \, \mathbf{a}_{1} + z_{2} \, \mathbf{a}_{3} & = & \left(x_{2}a+z_{2}c\cos\beta\right) \, \mathbf{\hat{x}} + z_{2}c\sin\beta \, \mathbf{\hat{z}} & \left(2m\right) & \mbox{H I} \\ 
\mathbf{B}_{4} & = & -x_{2} \, \mathbf{a}_{1} + -z_{2} \, \mathbf{a}_{3} & = & \left(-x_{2}a-z_{2}c\cos\beta\right) \, \mathbf{\hat{x}} + -z_{2}c\sin\beta \, \mathbf{\hat{z}} & \left(2m\right) & \mbox{H I} \\ 
\mathbf{B}_{5} & = & x_{3} \, \mathbf{a}_{1} + z_{3} \, \mathbf{a}_{3} & = & \left(x_{3}a+z_{3}c\cos\beta\right) \, \mathbf{\hat{x}} + z_{3}c\sin\beta \, \mathbf{\hat{z}} & \left(2m\right) & \mbox{H II} \\ 
\mathbf{B}_{6} & = & -x_{3} \, \mathbf{a}_{1} + -z_{3} \, \mathbf{a}_{3} & = & \left(-x_{3}a-z_{3}c\cos\beta\right) \, \mathbf{\hat{x}} + -z_{3}c\sin\beta \, \mathbf{\hat{z}} & \left(2m\right) & \mbox{H II} \\ 
\mathbf{B}_{7} & = & x_{4} \, \mathbf{a}_{1} + z_{4} \, \mathbf{a}_{3} & = & \left(x_{4}a+z_{4}c\cos\beta\right) \, \mathbf{\hat{x}} + z_{4}c\sin\beta \, \mathbf{\hat{z}} & \left(2m\right) & \mbox{H III} \\ 
\mathbf{B}_{8} & = & -x_{4} \, \mathbf{a}_{1} + -z_{4} \, \mathbf{a}_{3} & = & \left(-x_{4}a-z_{4}c\cos\beta\right) \, \mathbf{\hat{x}} + -z_{4}c\sin\beta \, \mathbf{\hat{z}} & \left(2m\right) & \mbox{H III} \\ 
\mathbf{B}_{9} & = & x_{5} \, \mathbf{a}_{1} + \frac{1}{2} \, \mathbf{a}_{2} + z_{5} \, \mathbf{a}_{3} & = & \left(x_{5}a+z_{5}c\cos\beta\right) \, \mathbf{\hat{x}} + \frac{1}{2}b \, \mathbf{\hat{y}} + z_{5}c\sin\beta \, \mathbf{\hat{z}} & \left(2n\right) & \mbox{Cl II} \\ 
\mathbf{B}_{10} & = & -x_{5} \, \mathbf{a}_{1} + \frac{1}{2} \, \mathbf{a}_{2}-z_{5} \, \mathbf{a}_{3} & = & \left(-x_{5}a-z_{5}c\cos\beta\right) \, \mathbf{\hat{x}} + \frac{1}{2}b \, \mathbf{\hat{y}}-z_{5}c\sin\beta \, \mathbf{\hat{z}} & \left(2n\right) & \mbox{Cl II} \\ 
\mathbf{B}_{11} & = & x_{6} \, \mathbf{a}_{1} + \frac{1}{2} \, \mathbf{a}_{2} + z_{6} \, \mathbf{a}_{3} & = & \left(x_{6}a+z_{6}c\cos\beta\right) \, \mathbf{\hat{x}} + \frac{1}{2}b \, \mathbf{\hat{y}} + z_{6}c\sin\beta \, \mathbf{\hat{z}} & \left(2n\right) & \mbox{H IV} \\ 
\mathbf{B}_{12} & = & -x_{6} \, \mathbf{a}_{1} + \frac{1}{2} \, \mathbf{a}_{2}-z_{6} \, \mathbf{a}_{3} & = & \left(-x_{6}a-z_{6}c\cos\beta\right) \, \mathbf{\hat{x}} + \frac{1}{2}b \, \mathbf{\hat{y}}-z_{6}c\sin\beta \, \mathbf{\hat{z}} & \left(2n\right) & \mbox{H IV} \\ 
\mathbf{B}_{13} & = & x_{7} \, \mathbf{a}_{1} + \frac{1}{2} \, \mathbf{a}_{2} + z_{7} \, \mathbf{a}_{3} & = & \left(x_{7}a+z_{7}c\cos\beta\right) \, \mathbf{\hat{x}} + \frac{1}{2}b \, \mathbf{\hat{y}} + z_{7}c\sin\beta \, \mathbf{\hat{z}} & \left(2n\right) & \mbox{H V} \\ 
\mathbf{B}_{14} & = & -x_{7} \, \mathbf{a}_{1} + \frac{1}{2} \, \mathbf{a}_{2}-z_{7} \, \mathbf{a}_{3} & = & \left(-x_{7}a-z_{7}c\cos\beta\right) \, \mathbf{\hat{x}} + \frac{1}{2}b \, \mathbf{\hat{y}}-z_{7}c\sin\beta \, \mathbf{\hat{z}} & \left(2n\right) & \mbox{H V} \\ 
\mathbf{B}_{15} & = & x_{8} \, \mathbf{a}_{1} + \frac{1}{2} \, \mathbf{a}_{2} + z_{8} \, \mathbf{a}_{3} & = & \left(x_{8}a+z_{8}c\cos\beta\right) \, \mathbf{\hat{x}} + \frac{1}{2}b \, \mathbf{\hat{y}} + z_{8}c\sin\beta \, \mathbf{\hat{z}} & \left(2n\right) & \mbox{H VI} \\ 
\mathbf{B}_{16} & = & -x_{8} \, \mathbf{a}_{1} + \frac{1}{2} \, \mathbf{a}_{2}-z_{8} \, \mathbf{a}_{3} & = & \left(-x_{8}a-z_{8}c\cos\beta\right) \, \mathbf{\hat{x}} + \frac{1}{2}b \, \mathbf{\hat{y}}-z_{8}c\sin\beta \, \mathbf{\hat{z}} & \left(2n\right) & \mbox{H VI} \\ 
\end{longtabu}
\renewcommand{\arraystretch}{1.0}
\noindent \hrulefill
\\
\textbf{References:}
\vspace*{-0.25cm}
\begin{flushleft}
  - \bibentry{Zeng_2017}. \\
\end{flushleft}
\noindent \hrulefill
\\
\textbf{Geometry files:}
\\
\noindent  - CIF: pp. {\hyperref[AB3_mP16_10_mn_3m3n_cif]{\pageref{AB3_mP16_10_mn_3m3n_cif}}} \\
\noindent  - POSCAR: pp. {\hyperref[AB3_mP16_10_mn_3m3n_poscar]{\pageref{AB3_mP16_10_mn_3m3n_poscar}}} \\
\onecolumn
{\phantomsection\label{ABC2_mP8_10_ac_eh_mn}}
\subsection*{\huge \textbf{{\normalfont \begin{raggedleft}Muthmannite (AuAgTe$_{2}$) Structure: \end{raggedleft} \\ ABC2\_mP8\_10\_ac\_eh\_mn}}}
\noindent \hrulefill
\vspace*{0.25cm}
\begin{figure}[htp]
  \centering
  \vspace{-1em}
  {\includegraphics[width=1\textwidth]{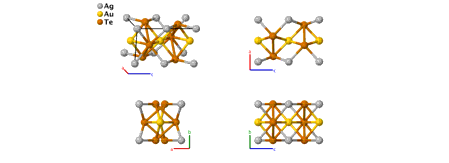}}
\end{figure}
\vspace*{-0.5cm}
\renewcommand{\arraystretch}{1.5}
\begin{equation*}
  \begin{array}{>{$\hspace{-0.15cm}}l<{$}>{$}p{0.5cm}<{$}>{$}p{18.5cm}<{$}}
    \mbox{\large \textbf{Prototype}} &\colon & \ce{AuAgTe2} \\
    \mbox{\large \textbf{\AFLOW\ prototype label}} &\colon & \mbox{ABC2\_mP8\_10\_ac\_eh\_mn} \\
    \mbox{\large \textbf{\textit{Strukturbericht} designation}} &\colon & \mbox{None} \\
    \mbox{\large \textbf{Pearson symbol}} &\colon & \mbox{mP8} \\
    \mbox{\large \textbf{Space group number}} &\colon & 10 \\
    \mbox{\large \textbf{Space group symbol}} &\colon & P2/m \\
    \mbox{\large \textbf{\AFLOW\ prototype command}} &\colon &  \texttt{aflow} \,  \, \texttt{-{}-proto=ABC2\_mP8\_10\_ac\_eh\_mn } \, \newline \texttt{-{}-params=}{a,b/a,c/a,\beta,x_{5},z_{5},x_{6},z_{6} }
  \end{array}
\end{equation*}
\renewcommand{\arraystretch}{1.0}

\noindent \parbox{1 \linewidth}{
\noindent \hrulefill
\\
\textbf{Simple Monoclinic primitive vectors:} \\
\vspace*{-0.25cm}
\begin{tabular}{cc}
  \begin{tabular}{c}
    \parbox{0.6 \linewidth}{
      \renewcommand{\arraystretch}{1.5}
      \begin{equation*}
        \centering
        \begin{array}{ccc}
              \mathbf{a}_1 & = & a \, \mathbf{\hat{x}} \\
    \mathbf{a}_2 & = & b \, \mathbf{\hat{y}} \\
    \mathbf{a}_3 & = & c \cos\beta \, \mathbf{\hat{x}} + c \sin\beta \, \mathbf{\hat{z}} \\

        \end{array}
      \end{equation*}
    }
    \renewcommand{\arraystretch}{1.0}
  \end{tabular}
  \begin{tabular}{c}
    \includegraphics[width=0.3\linewidth]{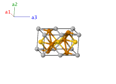} \\
  \end{tabular}
\end{tabular}

}
\vspace*{-0.25cm}

\noindent \hrulefill
\\
\textbf{Basis vectors:}
\vspace*{-0.25cm}
\renewcommand{\arraystretch}{1.5}
\begin{longtabu} to \textwidth{>{\centering $}X[-1,c,c]<{$}>{\centering $}X[-1,c,c]<{$}>{\centering $}X[-1,c,c]<{$}>{\centering $}X[-1,c,c]<{$}>{\centering $}X[-1,c,c]<{$}>{\centering $}X[-1,c,c]<{$}>{\centering $}X[-1,c,c]<{$}}
  & & \mbox{Lattice Coordinates} & & \mbox{Cartesian Coordinates} &\mbox{Wyckoff Position} & \mbox{Atom Type} \\  
  \mathbf{B}_{1} & = & 0 \, \mathbf{a}_{1} + 0 \, \mathbf{a}_{2} + 0 \, \mathbf{a}_{3} & = & 0 \, \mathbf{\hat{x}} + 0 \, \mathbf{\hat{y}} + 0 \, \mathbf{\hat{z}} & \left(1a\right) & \mbox{Ag I} \\ 
\mathbf{B}_{2} & = & \frac{1}{2} \, \mathbf{a}_{3} & = & \frac{1}{2}c\cos\beta \, \mathbf{\hat{x}} + \frac{1}{2}c\sin\beta \, \mathbf{\hat{z}} & \left(1c\right) & \mbox{Ag II} \\ 
\mathbf{B}_{3} & = & \frac{1}{2} \, \mathbf{a}_{1} + \frac{1}{2} \, \mathbf{a}_{2} & = & \frac{1}{2}a \, \mathbf{\hat{x}} + \frac{1}{2}b \, \mathbf{\hat{y}} & \left(1e\right) & \mbox{Au I} \\ 
\mathbf{B}_{4} & = & \frac{1}{2} \, \mathbf{a}_{1} + \frac{1}{2} \, \mathbf{a}_{2} + \frac{1}{2} \, \mathbf{a}_{3} & = & \frac{1}{2}\left(a+c\cos\beta\right) \, \mathbf{\hat{x}} + \frac{1}{2}b \, \mathbf{\hat{y}} + \frac{1}{2}c\sin\beta \, \mathbf{\hat{z}} & \left(1h\right) & \mbox{Au II} \\ 
\mathbf{B}_{5} & = & x_{5} \, \mathbf{a}_{1} + z_{5} \, \mathbf{a}_{3} & = & \left(x_{5}a+z_{5}c\cos\beta\right) \, \mathbf{\hat{x}} + z_{5}c\sin\beta \, \mathbf{\hat{z}} & \left(2m\right) & \mbox{Te I} \\ 
\mathbf{B}_{6} & = & -x_{5} \, \mathbf{a}_{1} + -z_{5} \, \mathbf{a}_{3} & = & \left(-x_{5}a-z_{5}c\cos\beta\right) \, \mathbf{\hat{x}} + -z_{5}c\sin\beta \, \mathbf{\hat{z}} & \left(2m\right) & \mbox{Te I} \\ 
\mathbf{B}_{7} & = & x_{6} \, \mathbf{a}_{1} + \frac{1}{2} \, \mathbf{a}_{2} + z_{6} \, \mathbf{a}_{3} & = & \left(x_{6}a+z_{6}c\cos\beta\right) \, \mathbf{\hat{x}} + \frac{1}{2}b \, \mathbf{\hat{y}} + z_{6}c\sin\beta \, \mathbf{\hat{z}} & \left(2n\right) & \mbox{Te II} \\ 
\mathbf{B}_{8} & = & -x_{6} \, \mathbf{a}_{1} + \frac{1}{2} \, \mathbf{a}_{2}-z_{6} \, \mathbf{a}_{3} & = & \left(-x_{6}a-z_{6}c\cos\beta\right) \, \mathbf{\hat{x}} + \frac{1}{2}b \, \mathbf{\hat{y}}-z_{6}c\sin\beta \, \mathbf{\hat{z}} & \left(2n\right) & \mbox{Te II} \\ 
\end{longtabu}
\renewcommand{\arraystretch}{1.0}
\noindent \hrulefill
\\
\textbf{References:}
\vspace*{-0.25cm}
\begin{flushleft}
  - \bibentry{Bindi_AuAgTe2_PhilosMagLett_2008}. \\
\end{flushleft}
\textbf{Found in:}
\vspace*{-0.25cm}
\begin{flushleft}
  - \bibentry{Villars_PearsonsCrystalData_2013}. \\
\end{flushleft}
\noindent \hrulefill
\\
\textbf{Geometry files:}
\\
\noindent  - CIF: pp. {\hyperref[ABC2_mP8_10_ac_eh_mn_cif]{\pageref{ABC2_mP8_10_ac_eh_mn_cif}}} \\
\noindent  - POSCAR: pp. {\hyperref[ABC2_mP8_10_ac_eh_mn_poscar]{\pageref{ABC2_mP8_10_ac_eh_mn_poscar}}} \\
\onecolumn
{\phantomsection\label{AB_mP6_10_en_am}}
\subsection*{\huge \textbf{{\normalfont LiSn Structure: AB\_mP6\_10\_en\_am}}}
\noindent \hrulefill
\vspace*{0.25cm}
\begin{figure}[htp]
  \centering
  \vspace{-1em}
  {\includegraphics[width=1\textwidth]{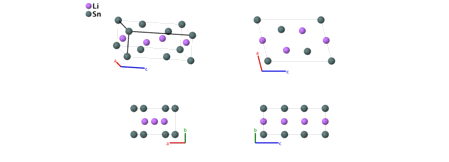}}
\end{figure}
\vspace*{-0.5cm}
\renewcommand{\arraystretch}{1.5}
\begin{equation*}
  \begin{array}{>{$\hspace{-0.15cm}}l<{$}>{$}p{0.5cm}<{$}>{$}p{18.5cm}<{$}}
    \mbox{\large \textbf{Prototype}} &\colon & \ce{LiSn} \\
    \mbox{\large \textbf{\AFLOW\ prototype label}} &\colon & \mbox{AB\_mP6\_10\_en\_am} \\
    \mbox{\large \textbf{\textit{Strukturbericht} designation}} &\colon & \mbox{None} \\
    \mbox{\large \textbf{Pearson symbol}} &\colon & \mbox{mP6} \\
    \mbox{\large \textbf{Space group number}} &\colon & 10 \\
    \mbox{\large \textbf{Space group symbol}} &\colon & P2/m \\
    \mbox{\large \textbf{\AFLOW\ prototype command}} &\colon &  \texttt{aflow} \,  \, \texttt{-{}-proto=AB\_mP6\_10\_en\_am } \, \newline \texttt{-{}-params=}{a,b/a,c/a,\beta,x_{3},z_{3},x_{4},z_{4} }
  \end{array}
\end{equation*}
\renewcommand{\arraystretch}{1.0}

\noindent \parbox{1 \linewidth}{
\noindent \hrulefill
\\
\textbf{Simple Monoclinic primitive vectors:} \\
\vspace*{-0.25cm}
\begin{tabular}{cc}
  \begin{tabular}{c}
    \parbox{0.6 \linewidth}{
      \renewcommand{\arraystretch}{1.5}
      \begin{equation*}
        \centering
        \begin{array}{ccc}
              \mathbf{a}_1 & = & a \, \mathbf{\hat{x}} \\
    \mathbf{a}_2 & = & b \, \mathbf{\hat{y}} \\
    \mathbf{a}_3 & = & c \cos\beta \, \mathbf{\hat{x}} + c \sin\beta \, \mathbf{\hat{z}} \\

        \end{array}
      \end{equation*}
    }
    \renewcommand{\arraystretch}{1.0}
  \end{tabular}
  \begin{tabular}{c}
    \includegraphics[width=0.3\linewidth]{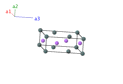} \\
  \end{tabular}
\end{tabular}

}
\vspace*{-0.25cm}

\noindent \hrulefill
\\
\textbf{Basis vectors:}
\vspace*{-0.25cm}
\renewcommand{\arraystretch}{1.5}
\begin{longtabu} to \textwidth{>{\centering $}X[-1,c,c]<{$}>{\centering $}X[-1,c,c]<{$}>{\centering $}X[-1,c,c]<{$}>{\centering $}X[-1,c,c]<{$}>{\centering $}X[-1,c,c]<{$}>{\centering $}X[-1,c,c]<{$}>{\centering $}X[-1,c,c]<{$}}
  & & \mbox{Lattice Coordinates} & & \mbox{Cartesian Coordinates} &\mbox{Wyckoff Position} & \mbox{Atom Type} \\  
  \mathbf{B}_{1} & = & 0 \, \mathbf{a}_{1} + 0 \, \mathbf{a}_{2} + 0 \, \mathbf{a}_{3} & = & 0 \, \mathbf{\hat{x}} + 0 \, \mathbf{\hat{y}} + 0 \, \mathbf{\hat{z}} & \left(1a\right) & \mbox{Sn I} \\ 
\mathbf{B}_{2} & = & \frac{1}{2} \, \mathbf{a}_{1} + \frac{1}{2} \, \mathbf{a}_{2} & = & \frac{1}{2}a \, \mathbf{\hat{x}} + \frac{1}{2}b \, \mathbf{\hat{y}} & \left(1e\right) & \mbox{Li I} \\ 
\mathbf{B}_{3} & = & x_{3} \, \mathbf{a}_{1} + z_{3} \, \mathbf{a}_{3} & = & \left(x_{3}a+z_{3}c\cos\beta\right) \, \mathbf{\hat{x}} + z_{3}c\sin\beta \, \mathbf{\hat{z}} & \left(2m\right) & \mbox{Sn II} \\ 
\mathbf{B}_{4} & = & -x_{3} \, \mathbf{a}_{1} + -z_{3} \, \mathbf{a}_{3} & = & \left(-x_{3}a-z_{3}c\cos\beta\right) \, \mathbf{\hat{x}} + -z_{3}c\sin\beta \, \mathbf{\hat{z}} & \left(2m\right) & \mbox{Sn II} \\ 
\mathbf{B}_{5} & = & x_{4} \, \mathbf{a}_{1} + \frac{1}{2} \, \mathbf{a}_{2} + z_{4} \, \mathbf{a}_{3} & = & \left(x_{4}a+z_{4}c\cos\beta\right) \, \mathbf{\hat{x}} + \frac{1}{2}b \, \mathbf{\hat{y}} + z_{4}c\sin\beta \, \mathbf{\hat{z}} & \left(2n\right) & \mbox{Li II} \\ 
\mathbf{B}_{6} & = & -x_{4} \, \mathbf{a}_{1} + \frac{1}{2} \, \mathbf{a}_{2}-z_{4} \, \mathbf{a}_{3} & = & \left(-x_{4}a-z_{4}c\cos\beta\right) \, \mathbf{\hat{x}} + \frac{1}{2}b \, \mathbf{\hat{y}}-z_{4}c\sin\beta \, \mathbf{\hat{z}} & \left(2n\right) & \mbox{Li II} \\ 
\end{longtabu}
\renewcommand{\arraystretch}{1.0}
\noindent \hrulefill
\\
\textbf{References:}
\vspace*{-0.25cm}
\begin{flushleft}
  - \bibentry{muller_LiSn_ZNaturB_1973}. \\
\end{flushleft}
\textbf{Found in:}
\vspace*{-0.25cm}
\begin{flushleft}
  - \bibentry{Villars_PearsonsCrystalData_2013}. \\
\end{flushleft}
\noindent \hrulefill
\\
\textbf{Geometry files:}
\\
\noindent  - CIF: pp. {\hyperref[AB_mP6_10_en_am_cif]{\pageref{AB_mP6_10_en_am_cif}}} \\
\noindent  - POSCAR: pp. {\hyperref[AB_mP6_10_en_am_poscar]{\pageref{AB_mP6_10_en_am_poscar}}} \\
\onecolumn
{\phantomsection\label{A_mP8_10_2m2n}}
\subsection*{\huge \textbf{{\normalfont S-carbon Structure: A\_mP8\_10\_2m2n}}}
\noindent \hrulefill
\vspace*{0.25cm}
\begin{figure}[htp]
  \centering
  \vspace{-1em}
  {\includegraphics[width=1\textwidth]{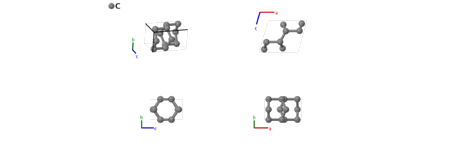}}
\end{figure}
\vspace*{-0.5cm}
\renewcommand{\arraystretch}{1.5}
\begin{equation*}
  \begin{array}{>{$\hspace{-0.15cm}}l<{$}>{$}p{0.5cm}<{$}>{$}p{18.5cm}<{$}}
    \mbox{\large \textbf{Prototype}} &\colon & \ce{C} \\
    \mbox{\large \textbf{\AFLOW\ prototype label}} &\colon & \mbox{A\_mP8\_10\_2m2n} \\
    \mbox{\large \textbf{\textit{Strukturbericht} designation}} &\colon & \mbox{None} \\
    \mbox{\large \textbf{Pearson symbol}} &\colon & \mbox{mP8} \\
    \mbox{\large \textbf{Space group number}} &\colon & 10 \\
    \mbox{\large \textbf{Space group symbol}} &\colon & P2/m \\
    \mbox{\large \textbf{\AFLOW\ prototype command}} &\colon &  \texttt{aflow} \,  \, \texttt{-{}-proto=A\_mP8\_10\_2m2n } \, \newline \texttt{-{}-params=}{a,b/a,c/a,\beta,x_{1},z_{1},x_{2},z_{2},x_{3},z_{3},x_{4},z_{4} }
  \end{array}
\end{equation*}
\renewcommand{\arraystretch}{1.0}

\vspace*{-0.25cm}
\noindent \hrulefill
\begin{itemize}
  \item{This is a predicted ``superhard'' allotrope of carbon.  Shortly after
this paper was published, two other papers predicted similar
structures, differentiated mainly by an origin shift:
\begin{itemize}
\item F-carbon (Tian, 2012): the origin is shifted by $1/2
  \, \mathbf{a}_{3}$.
\item J-carbon (Wang, 2012): the origin is shifted by $1/2
  \left(\mathbf{a}_{1} + \mathbf{a}_{3}\right)$.
\end{itemize}}
  \item{This is {\em not} the orthorhombic phase, also denoted S-carbon, found
by He {\em et al.} (He, 2012)
}
\end{itemize}

\noindent \parbox{1 \linewidth}{
\noindent \hrulefill
\\
\textbf{Simple Monoclinic primitive vectors:} \\
\vspace*{-0.25cm}
\begin{tabular}{cc}
  \begin{tabular}{c}
    \parbox{0.6 \linewidth}{
      \renewcommand{\arraystretch}{1.5}
      \begin{equation*}
        \centering
        \begin{array}{ccc}
              \mathbf{a}_1 & = & a \, \mathbf{\hat{x}} \\
    \mathbf{a}_2 & = & b \, \mathbf{\hat{y}} \\
    \mathbf{a}_3 & = & c \cos\beta \, \mathbf{\hat{x}} + c \sin\beta \, \mathbf{\hat{z}} \\

        \end{array}
      \end{equation*}
    }
    \renewcommand{\arraystretch}{1.0}
  \end{tabular}
  \begin{tabular}{c}
    \includegraphics[width=0.3\linewidth]{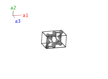} \\
  \end{tabular}
\end{tabular}

}
\vspace*{-0.25cm}

\noindent \hrulefill
\\
\textbf{Basis vectors:}
\vspace*{-0.25cm}
\renewcommand{\arraystretch}{1.5}
\begin{longtabu} to \textwidth{>{\centering $}X[-1,c,c]<{$}>{\centering $}X[-1,c,c]<{$}>{\centering $}X[-1,c,c]<{$}>{\centering $}X[-1,c,c]<{$}>{\centering $}X[-1,c,c]<{$}>{\centering $}X[-1,c,c]<{$}>{\centering $}X[-1,c,c]<{$}}
  & & \mbox{Lattice Coordinates} & & \mbox{Cartesian Coordinates} &\mbox{Wyckoff Position} & \mbox{Atom Type} \\  
  \mathbf{B}_{1} & = & x_{1} \, \mathbf{a}_{1} + z_{1} \, \mathbf{a}_{3} & = & \left(x_{1}a+z_{1}c\cos\beta\right) \, \mathbf{\hat{x}} + z_{1}c\sin\beta \, \mathbf{\hat{z}} & \left(2m\right) & \mbox{C I} \\ 
\mathbf{B}_{2} & = & -x_{1} \, \mathbf{a}_{1} + -z_{1} \, \mathbf{a}_{3} & = & \left(-x_{1}a-z_{1}c\cos\beta\right) \, \mathbf{\hat{x}} + -z_{1}c\sin\beta \, \mathbf{\hat{z}} & \left(2m\right) & \mbox{C I} \\ 
\mathbf{B}_{3} & = & x_{2} \, \mathbf{a}_{1} + z_{2} \, \mathbf{a}_{3} & = & \left(x_{2}a+z_{2}c\cos\beta\right) \, \mathbf{\hat{x}} + z_{2}c\sin\beta \, \mathbf{\hat{z}} & \left(2m\right) & \mbox{C II} \\ 
\mathbf{B}_{4} & = & -x_{2} \, \mathbf{a}_{1} + -z_{2} \, \mathbf{a}_{3} & = & \left(-x_{2}a-z_{2}c\cos\beta\right) \, \mathbf{\hat{x}} + -z_{2}c\sin\beta \, \mathbf{\hat{z}} & \left(2m\right) & \mbox{C II} \\ 
\mathbf{B}_{5} & = & x_{3} \, \mathbf{a}_{1} + \frac{1}{2} \, \mathbf{a}_{2} + z_{3} \, \mathbf{a}_{3} & = & \left(x_{3}a+z_{3}c\cos\beta\right) \, \mathbf{\hat{x}} + \frac{1}{2}b \, \mathbf{\hat{y}} + z_{3}c\sin\beta \, \mathbf{\hat{z}} & \left(2n\right) & \mbox{C III} \\ 
\mathbf{B}_{6} & = & -x_{3} \, \mathbf{a}_{1} + \frac{1}{2} \, \mathbf{a}_{2}-z_{3} \, \mathbf{a}_{3} & = & \left(-x_{3}a-z_{3}c\cos\beta\right) \, \mathbf{\hat{x}} + \frac{1}{2}b \, \mathbf{\hat{y}}-z_{3}c\sin\beta \, \mathbf{\hat{z}} & \left(2n\right) & \mbox{C III} \\ 
\mathbf{B}_{7} & = & x_{4} \, \mathbf{a}_{1} + \frac{1}{2} \, \mathbf{a}_{2} + z_{4} \, \mathbf{a}_{3} & = & \left(x_{4}a+z_{4}c\cos\beta\right) \, \mathbf{\hat{x}} + \frac{1}{2}b \, \mathbf{\hat{y}} + z_{4}c\sin\beta \, \mathbf{\hat{z}} & \left(2n\right) & \mbox{C IV} \\ 
\mathbf{B}_{8} & = & -x_{4} \, \mathbf{a}_{1} + \frac{1}{2} \, \mathbf{a}_{2}-z_{4} \, \mathbf{a}_{3} & = & \left(-x_{4}a-z_{4}c\cos\beta\right) \, \mathbf{\hat{x}} + \frac{1}{2}b \, \mathbf{\hat{y}}-z_{4}c\sin\beta \, \mathbf{\hat{z}} & \left(2n\right) & \mbox{C IV} \\ 
\end{longtabu}
\renewcommand{\arraystretch}{1.0}
\noindent \hrulefill
\\
\textbf{References:}
\vspace*{-0.25cm}
\begin{flushleft}
  - \bibentry{Niu_PRL_108_2012}. \\
  - \bibentry{Tian_JPCM_24_2012}. \\
  - \bibentry{Wang_JCP_137_2012}. \\
  - \bibentry{He_SSC_152_2012}. \\
  - \bibentry{He_JSM_34_2012}. \\
\end{flushleft}
\noindent \hrulefill
\\
\textbf{Geometry files:}
\\
\noindent  - CIF: pp. {\hyperref[A_mP8_10_2m2n_cif]{\pageref{A_mP8_10_2m2n_cif}}} \\
\noindent  - POSCAR: pp. {\hyperref[A_mP8_10_2m2n_poscar]{\pageref{A_mP8_10_2m2n_poscar}}} \\
\onecolumn
{\phantomsection\label{A7B2C2_mC22_12_aij_h_i}}
\subsection*{\huge \textbf{{\normalfont \begin{raggedleft}Thortveitite ([Sc,Y]$_2$Si$_2$O$_7$, $S2_{1}$) Structure: \end{raggedleft} \\ A7B2C2\_mC22\_12\_aij\_h\_i}}}
\noindent \hrulefill
\vspace*{0.25cm}
\begin{figure}[htp]
  \centering
  \vspace{-1em}
  {\includegraphics[width=1\textwidth]{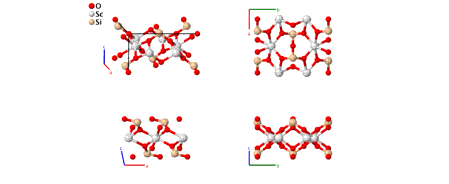}}
\end{figure}
\vspace*{-0.5cm}
\renewcommand{\arraystretch}{1.5}
\begin{equation*}
  \begin{array}{>{$\hspace{-0.15cm}}l<{$}>{$}p{0.5cm}<{$}>{$}p{18.5cm}<{$}}
    \mbox{\large \textbf{Prototype}} &\colon & \ce{[Sc,Y]2Si2O7} \\
    \mbox{\large \textbf{\AFLOW\ prototype label}} &\colon & \mbox{A7B2C2\_mC22\_12\_aij\_h\_i} \\
    \mbox{\large \textbf{\textit{Strukturbericht} designation}} &\colon & \mbox{$S2_{1}$} \\
    \mbox{\large \textbf{Pearson symbol}} &\colon & \mbox{mC22} \\
    \mbox{\large \textbf{Space group number}} &\colon & 12 \\
    \mbox{\large \textbf{Space group symbol}} &\colon & C2/m \\
    \mbox{\large \textbf{\AFLOW\ prototype command}} &\colon &  \texttt{aflow} \,  \, \texttt{-{}-proto=A7B2C2\_mC22\_12\_aij\_h\_i } \, \newline \texttt{-{}-params=}{a,b/a,c/a,\beta,y_{2},x_{3},z_{3},x_{4},z_{4},x_{5},y_{5},z_{5} }
  \end{array}
\end{equation*}
\renewcommand{\arraystretch}{1.0}

\vspace*{-0.25cm}
\noindent \hrulefill
\begin{itemize}
  \item{Thortveitite is the primary source of scandium, and is one of the
simplest sorosilicates, minerals with isolated Si$_{2}$O$_{7}$
groups (Bianchi, 1988).}
  \item{Although the (4h) Wyckoff position is randomly occupied by both Sc and
Y atoms, we use Sc to represent the site.}
  \item{(Bianchi, 1988) gives structural information for several samples of
thortveitite.  We use the data from sample 1, collected in Iveland,
Norway.}
\end{itemize}

\noindent \parbox{1 \linewidth}{
\noindent \hrulefill
\\
\textbf{Base-centered Monoclinic primitive vectors:} \\
\vspace*{-0.25cm}
\begin{tabular}{cc}
  \begin{tabular}{c}
    \parbox{0.6 \linewidth}{
      \renewcommand{\arraystretch}{1.5}
      \begin{equation*}
        \centering
        \begin{array}{ccc}
              \mathbf{a}_1 & = & \frac12 \, a \, \mathbf{\hat{x}} - \frac12 \, b \, \mathbf{\hat{y}} \\
    \mathbf{a}_2 & = & \frac12 \, a \, \mathbf{\hat{x}} + \frac12 \, b \, \mathbf{\hat{y}} \\
    \mathbf{a}_3 & = & c \cos\beta \, \mathbf{\hat{x}} + c \sin\beta \, \mathbf{\hat{z}} \\

        \end{array}
      \end{equation*}
    }
    \renewcommand{\arraystretch}{1.0}
  \end{tabular}
  \begin{tabular}{c}
    \includegraphics[width=0.3\linewidth]{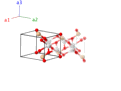} \\
  \end{tabular}
\end{tabular}

}
\vspace*{-0.25cm}

\noindent \hrulefill
\\
\textbf{Basis vectors:}
\vspace*{-0.25cm}
\renewcommand{\arraystretch}{1.5}
\begin{longtabu} to \textwidth{>{\centering $}X[-1,c,c]<{$}>{\centering $}X[-1,c,c]<{$}>{\centering $}X[-1,c,c]<{$}>{\centering $}X[-1,c,c]<{$}>{\centering $}X[-1,c,c]<{$}>{\centering $}X[-1,c,c]<{$}>{\centering $}X[-1,c,c]<{$}}
  & & \mbox{Lattice Coordinates} & & \mbox{Cartesian Coordinates} &\mbox{Wyckoff Position} & \mbox{Atom Type} \\  
  \mathbf{B}_{1} & = & 0 \, \mathbf{a}_{1} + 0 \, \mathbf{a}_{2} + 0 \, \mathbf{a}_{3} & = & 0 \, \mathbf{\hat{x}} + 0 \, \mathbf{\hat{y}} + 0 \, \mathbf{\hat{z}} & \left(2a\right) & \mbox{O I} \\ 
\mathbf{B}_{2} & = & -y_{2} \, \mathbf{a}_{1} + y_{2} \, \mathbf{a}_{2} + \frac{1}{2} \, \mathbf{a}_{3} & = & \frac{1}{2}c\cos\beta \, \mathbf{\hat{x}} + y_{2}b \, \mathbf{\hat{y}} + \frac{1}{2}c\sin\beta \, \mathbf{\hat{z}} & \left(4h\right) & \mbox{Sc} \\ 
\mathbf{B}_{3} & = & y_{2} \, \mathbf{a}_{1}-y_{2} \, \mathbf{a}_{2} + \frac{1}{2} \, \mathbf{a}_{3} & = & \frac{1}{2}c\cos\beta \, \mathbf{\hat{x}}-y_{2}b \, \mathbf{\hat{y}} + \frac{1}{2}c\sin\beta \, \mathbf{\hat{z}} & \left(4h\right) & \mbox{Sc} \\ 
\mathbf{B}_{4} & = & x_{3} \, \mathbf{a}_{1} + x_{3} \, \mathbf{a}_{2} + z_{3} \, \mathbf{a}_{3} & = & \left(x_{3}a+z_{3}c\cos\beta\right) \, \mathbf{\hat{x}} + z_{3}c\sin\beta \, \mathbf{\hat{z}} & \left(4i\right) & \mbox{O II} \\ 
\mathbf{B}_{5} & = & -x_{3} \, \mathbf{a}_{1}-x_{3} \, \mathbf{a}_{2}-z_{3} \, \mathbf{a}_{3} & = & \left(-x_{3}a-z_{3}c\cos\beta\right) \, \mathbf{\hat{x}} + -z_{3}c\sin\beta \, \mathbf{\hat{z}} & \left(4i\right) & \mbox{O II} \\ 
\mathbf{B}_{6} & = & x_{4} \, \mathbf{a}_{1} + x_{4} \, \mathbf{a}_{2} + z_{4} \, \mathbf{a}_{3} & = & \left(x_{4}a+z_{4}c\cos\beta\right) \, \mathbf{\hat{x}} + z_{4}c\sin\beta \, \mathbf{\hat{z}} & \left(4i\right) & \mbox{Si} \\ 
\mathbf{B}_{7} & = & -x_{4} \, \mathbf{a}_{1}-x_{4} \, \mathbf{a}_{2}-z_{4} \, \mathbf{a}_{3} & = & \left(-x_{4}a-z_{4}c\cos\beta\right) \, \mathbf{\hat{x}} + -z_{4}c\sin\beta \, \mathbf{\hat{z}} & \left(4i\right) & \mbox{Si} \\ 
\mathbf{B}_{8} & = & \left(x_{5}-y_{5}\right) \, \mathbf{a}_{1} + \left(x_{5}+y_{5}\right) \, \mathbf{a}_{2} + z_{5} \, \mathbf{a}_{3} & = & \left(x_{5}a+z_{5}c\cos\beta\right) \, \mathbf{\hat{x}} + y_{5}b \, \mathbf{\hat{y}} + z_{5}c\sin\beta \, \mathbf{\hat{z}} & \left(8j\right) & \mbox{O III} \\ 
\mathbf{B}_{9} & = & \left(-x_{5}-y_{5}\right) \, \mathbf{a}_{1} + \left(-x_{5}+y_{5}\right) \, \mathbf{a}_{2}-z_{5} \, \mathbf{a}_{3} & = & \left(-x_{5}a-z_{5}c\cos\beta\right) \, \mathbf{\hat{x}} + y_{5}b \, \mathbf{\hat{y}}-z_{5}c\sin\beta \, \mathbf{\hat{z}} & \left(8j\right) & \mbox{O III} \\ 
\mathbf{B}_{10} & = & \left(-x_{5}+y_{5}\right) \, \mathbf{a}_{1} + \left(-x_{5}-y_{5}\right) \, \mathbf{a}_{2}-z_{5} \, \mathbf{a}_{3} & = & \left(-x_{5}a-z_{5}c\cos\beta\right) \, \mathbf{\hat{x}}-y_{5}b \, \mathbf{\hat{y}}-z_{5}c\sin\beta \, \mathbf{\hat{z}} & \left(8j\right) & \mbox{O III} \\ 
\mathbf{B}_{11} & = & \left(x_{5}+y_{5}\right) \, \mathbf{a}_{1} + \left(x_{5}-y_{5}\right) \, \mathbf{a}_{2} + z_{5} \, \mathbf{a}_{3} & = & \left(x_{5}a+z_{5}c\cos\beta\right) \, \mathbf{\hat{x}}-y_{5}b \, \mathbf{\hat{y}} + z_{5}c\sin\beta \, \mathbf{\hat{z}} & \left(8j\right) & \mbox{O III} \\ 
\end{longtabu}
\renewcommand{\arraystretch}{1.0}
\noindent \hrulefill
\\
\textbf{References:}
\vspace*{-0.25cm}
\begin{flushleft}
  - \bibentry{Bianchi_AmMin_73_601_1988}. \\
\end{flushleft}
\textbf{Found in:}
\vspace*{-0.25cm}
\begin{flushleft}
  - \bibentry{Downs_Am_Min_88_2003}. \\
\end{flushleft}
\noindent \hrulefill
\\
\textbf{Geometry files:}
\\
\noindent  - CIF: pp. {\hyperref[A7B2C2_mC22_12_aij_h_i_cif]{\pageref{A7B2C2_mC22_12_aij_h_i_cif}}} \\
\noindent  - POSCAR: pp. {\hyperref[A7B2C2_mC22_12_aij_h_i_poscar]{\pageref{A7B2C2_mC22_12_aij_h_i_poscar}}} \\
\onecolumn
{\phantomsection\label{A_mC16_12_4i}}
\subsection*{\huge \textbf{{\normalfont M-carbon Structure: A\_mC16\_12\_4i}}}
\noindent \hrulefill
\vspace*{0.25cm}
\begin{figure}[htp]
  \centering
  \vspace{-1em}
  {\includegraphics[width=1\textwidth]{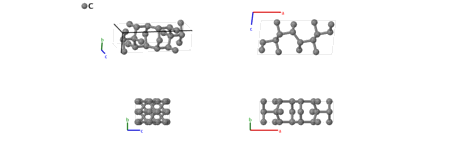}}
\end{figure}
\vspace*{-0.5cm}
\renewcommand{\arraystretch}{1.5}
\begin{equation*}
  \begin{array}{>{$\hspace{-0.15cm}}l<{$}>{$}p{0.5cm}<{$}>{$}p{18.5cm}<{$}}
    \mbox{\large \textbf{Prototype}} &\colon & \ce{C} \\
    \mbox{\large \textbf{\AFLOW\ prototype label}} &\colon & \mbox{A\_mC16\_12\_4i} \\
    \mbox{\large \textbf{\textit{Strukturbericht} designation}} &\colon & \mbox{None} \\
    \mbox{\large \textbf{Pearson symbol}} &\colon & \mbox{mC16} \\
    \mbox{\large \textbf{Space group number}} &\colon & 12 \\
    \mbox{\large \textbf{Space group symbol}} &\colon & C2/m \\
    \mbox{\large \textbf{\AFLOW\ prototype command}} &\colon &  \texttt{aflow} \,  \, \texttt{-{}-proto=A\_mC16\_12\_4i } \, \newline \texttt{-{}-params=}{a,b/a,c/a,\beta,x_{1},z_{1},x_{2},z_{2},x_{3},z_{3},x_{4},z_{4} }
  \end{array}
\end{equation*}
\renewcommand{\arraystretch}{1.0}

\vspace*{-0.25cm}
\noindent \hrulefill
\begin{itemize}
  \item{This structure was originally found by Oganov and
Glass, (Oganov, 2006) and was refined and designated M-Carbon by
Li {\em et al.} (Li, 2009)
}
\end{itemize}

\noindent \parbox{1 \linewidth}{
\noindent \hrulefill
\\
\textbf{Base-centered Monoclinic primitive vectors:} \\
\vspace*{-0.25cm}
\begin{tabular}{cc}
  \begin{tabular}{c}
    \parbox{0.6 \linewidth}{
      \renewcommand{\arraystretch}{1.5}
      \begin{equation*}
        \centering
        \begin{array}{ccc}
              \mathbf{a}_1 & = & \frac12 \, a \, \mathbf{\hat{x}} - \frac12 \, b \, \mathbf{\hat{y}} \\
    \mathbf{a}_2 & = & \frac12 \, a \, \mathbf{\hat{x}} + \frac12 \, b \, \mathbf{\hat{y}} \\
    \mathbf{a}_3 & = & c \cos\beta \, \mathbf{\hat{x}} + c \sin\beta \, \mathbf{\hat{z}} \\

        \end{array}
      \end{equation*}
    }
    \renewcommand{\arraystretch}{1.0}
  \end{tabular}
  \begin{tabular}{c}
    \includegraphics[width=0.3\linewidth]{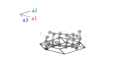} \\
  \end{tabular}
\end{tabular}

}
\vspace*{-0.25cm}

\noindent \hrulefill
\\
\textbf{Basis vectors:}
\vspace*{-0.25cm}
\renewcommand{\arraystretch}{1.5}
\begin{longtabu} to \textwidth{>{\centering $}X[-1,c,c]<{$}>{\centering $}X[-1,c,c]<{$}>{\centering $}X[-1,c,c]<{$}>{\centering $}X[-1,c,c]<{$}>{\centering $}X[-1,c,c]<{$}>{\centering $}X[-1,c,c]<{$}>{\centering $}X[-1,c,c]<{$}}
  & & \mbox{Lattice Coordinates} & & \mbox{Cartesian Coordinates} &\mbox{Wyckoff Position} & \mbox{Atom Type} \\  
  \mathbf{B}_{1} & = & x_{1} \, \mathbf{a}_{1} + x_{1} \, \mathbf{a}_{2} + z_{1} \, \mathbf{a}_{3} & = & \left(x_{1}a+z_{1}c\cos\beta\right) \, \mathbf{\hat{x}} + z_{1}c\sin\beta \, \mathbf{\hat{z}} & \left(4i\right) & \mbox{C I} \\ 
\mathbf{B}_{2} & = & -x_{1} \, \mathbf{a}_{1}-x_{1} \, \mathbf{a}_{2}-z_{1} \, \mathbf{a}_{3} & = & \left(-x_{1}a-z_{1}c\cos\beta\right) \, \mathbf{\hat{x}} + -z_{1}c\sin\beta \, \mathbf{\hat{z}} & \left(4i\right) & \mbox{C I} \\ 
\mathbf{B}_{3} & = & x_{2} \, \mathbf{a}_{1} + x_{2} \, \mathbf{a}_{2} + z_{2} \, \mathbf{a}_{3} & = & \left(x_{2}a+z_{2}c\cos\beta\right) \, \mathbf{\hat{x}} + z_{2}c\sin\beta \, \mathbf{\hat{z}} & \left(4i\right) & \mbox{C II} \\ 
\mathbf{B}_{4} & = & -x_{2} \, \mathbf{a}_{1}-x_{2} \, \mathbf{a}_{2}-z_{2} \, \mathbf{a}_{3} & = & \left(-x_{2}a-z_{2}c\cos\beta\right) \, \mathbf{\hat{x}} + -z_{2}c\sin\beta \, \mathbf{\hat{z}} & \left(4i\right) & \mbox{C II} \\ 
\mathbf{B}_{5} & = & x_{3} \, \mathbf{a}_{1} + x_{3} \, \mathbf{a}_{2} + z_{3} \, \mathbf{a}_{3} & = & \left(x_{3}a+z_{3}c\cos\beta\right) \, \mathbf{\hat{x}} + z_{3}c\sin\beta \, \mathbf{\hat{z}} & \left(4i\right) & \mbox{C III} \\ 
\mathbf{B}_{6} & = & -x_{3} \, \mathbf{a}_{1}-x_{3} \, \mathbf{a}_{2}-z_{3} \, \mathbf{a}_{3} & = & \left(-x_{3}a-z_{3}c\cos\beta\right) \, \mathbf{\hat{x}} + -z_{3}c\sin\beta \, \mathbf{\hat{z}} & \left(4i\right) & \mbox{C III} \\ 
\mathbf{B}_{7} & = & x_{4} \, \mathbf{a}_{1} + x_{4} \, \mathbf{a}_{2} + z_{4} \, \mathbf{a}_{3} & = & \left(x_{4}a+z_{4}c\cos\beta\right) \, \mathbf{\hat{x}} + z_{4}c\sin\beta \, \mathbf{\hat{z}} & \left(4i\right) & \mbox{C IV} \\ 
\mathbf{B}_{8} & = & -x_{4} \, \mathbf{a}_{1}-x_{4} \, \mathbf{a}_{2}-z_{4} \, \mathbf{a}_{3} & = & \left(-x_{4}a-z_{4}c\cos\beta\right) \, \mathbf{\hat{x}} + -z_{4}c\sin\beta \, \mathbf{\hat{z}} & \left(4i\right) & \mbox{C IV} \\ 
\end{longtabu}
\renewcommand{\arraystretch}{1.0}
\noindent \hrulefill
\\
\textbf{References:}
\vspace*{-0.25cm}
\begin{flushleft}
  - \bibentry{Li_PRL_102_2009}. \\
  - \bibentry{Oganov_JCP_124_2006}. \\
\end{flushleft}
\noindent \hrulefill
\\
\textbf{Geometry files:}
\\
\noindent  - CIF: pp. {\hyperref[A_mC16_12_4i_cif]{\pageref{A_mC16_12_4i_cif}}} \\
\noindent  - POSCAR: pp. {\hyperref[A_mC16_12_4i_poscar]{\pageref{A_mC16_12_4i_poscar}}} \\
\onecolumn
{\phantomsection\label{A2B_mP12_13_2g_ef}}
\subsection*{\huge \textbf{{\normalfont H$_{2}$S (15~GPa) Structure: A2B\_mP12\_13\_2g\_ef}}}
\noindent \hrulefill
\vspace*{0.25cm}
\begin{figure}[htp]
  \centering
  \vspace{-1em}
  {\includegraphics[width=1\textwidth]{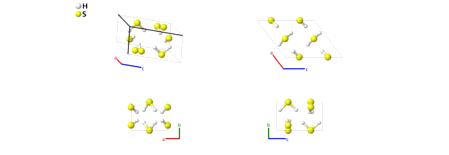}}
\end{figure}
\vspace*{-0.5cm}
\renewcommand{\arraystretch}{1.5}
\begin{equation*}
  \begin{array}{>{$\hspace{-0.15cm}}l<{$}>{$}p{0.5cm}<{$}>{$}p{18.5cm}<{$}}
    \mbox{\large \textbf{Prototype}} &\colon & \ce{H2S} \\
    \mbox{\large \textbf{\AFLOW\ prototype label}} &\colon & \mbox{A2B\_mP12\_13\_2g\_ef} \\
    \mbox{\large \textbf{\textit{Strukturbericht} designation}} &\colon & \mbox{None} \\
    \mbox{\large \textbf{Pearson symbol}} &\colon & \mbox{mP12} \\
    \mbox{\large \textbf{Space group number}} &\colon & 13 \\
    \mbox{\large \textbf{Space group symbol}} &\colon & P2/c \\
    \mbox{\large \textbf{\AFLOW\ prototype command}} &\colon &  \texttt{aflow} \,  \, \texttt{-{}-proto=A2B\_mP12\_13\_2g\_ef } \, \newline \texttt{-{}-params=}{a,b/a,c/a,\beta,y_{1},y_{2},x_{3},y_{3},z_{3},x_{4},y_{4},z_{4} }
  \end{array}
\end{equation*}
\renewcommand{\arraystretch}{1.0}

\vspace*{-0.25cm}
\noindent \hrulefill
\begin{itemize}
  \item{This structure was found by first-principles electronic structure
calculations and is predicted to be the stable structure of H$_{2}$S
in the range $10 - 30$~GPa, which does not agree with the experimental
phase diagram. (Shimizu, 1995)}
  \item{The data presented here was computed at 15~GPa.
}
\end{itemize}

\noindent \parbox{1 \linewidth}{
\noindent \hrulefill
\\
\textbf{Simple Monoclinic primitive vectors:} \\
\vspace*{-0.25cm}
\begin{tabular}{cc}
  \begin{tabular}{c}
    \parbox{0.6 \linewidth}{
      \renewcommand{\arraystretch}{1.5}
      \begin{equation*}
        \centering
        \begin{array}{ccc}
              \mathbf{a}_1 & = & a \, \mathbf{\hat{x}} \\
    \mathbf{a}_2 & = & b \, \mathbf{\hat{y}} \\
    \mathbf{a}_3 & = & c \cos\beta \, \mathbf{\hat{x}} + c \sin\beta \, \mathbf{\hat{z}} \\

        \end{array}
      \end{equation*}
    }
    \renewcommand{\arraystretch}{1.0}
  \end{tabular}
  \begin{tabular}{c}
    \includegraphics[width=0.3\linewidth]{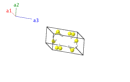} \\
  \end{tabular}
\end{tabular}

}
\vspace*{-0.25cm}

\noindent \hrulefill
\\
\textbf{Basis vectors:}
\vspace*{-0.25cm}
\renewcommand{\arraystretch}{1.5}
\begin{longtabu} to \textwidth{>{\centering $}X[-1,c,c]<{$}>{\centering $}X[-1,c,c]<{$}>{\centering $}X[-1,c,c]<{$}>{\centering $}X[-1,c,c]<{$}>{\centering $}X[-1,c,c]<{$}>{\centering $}X[-1,c,c]<{$}>{\centering $}X[-1,c,c]<{$}}
  & & \mbox{Lattice Coordinates} & & \mbox{Cartesian Coordinates} &\mbox{Wyckoff Position} & \mbox{Atom Type} \\  
  \mathbf{B}_{1} & = & y_{1} \, \mathbf{a}_{2} + \frac{1}{4} \, \mathbf{a}_{3} & = & \frac{1}{4}c\cos\beta \, \mathbf{\hat{x}} + y_{1}b \, \mathbf{\hat{y}} + \frac{1}{4}c\sin\beta \, \mathbf{\hat{z}} & \left(2e\right) & \mbox{S I} \\ 
\mathbf{B}_{2} & = & -y_{1} \, \mathbf{a}_{2} + \frac{3}{4} \, \mathbf{a}_{3} & = & \frac{3}{4}c\cos\beta \, \mathbf{\hat{x}}-y_{1}b \, \mathbf{\hat{y}} + \frac{3}{4}c\sin\beta \, \mathbf{\hat{z}} & \left(2e\right) & \mbox{S I} \\ 
\mathbf{B}_{3} & = & \frac{1}{2} \, \mathbf{a}_{1} + y_{2} \, \mathbf{a}_{2} + \frac{1}{4} \, \mathbf{a}_{3} & = & \left(\frac{1}{2}a+\frac{1}{4}c\cos\beta\right) \, \mathbf{\hat{x}} + y_{2}b \, \mathbf{\hat{y}} + \frac{1}{4}c\sin\beta \, \mathbf{\hat{z}} & \left(2f\right) & \mbox{S II} \\ 
\mathbf{B}_{4} & = & \frac{1}{2} \, \mathbf{a}_{1}-y_{2} \, \mathbf{a}_{2} + \frac{3}{4} \, \mathbf{a}_{3} & = & \left(\frac{1}{2}a+\frac{3}{4}c\cos\beta\right) \, \mathbf{\hat{x}}-y_{2}b \, \mathbf{\hat{y}} + \frac{3}{4}c\sin\beta \, \mathbf{\hat{z}} & \left(2f\right) & \mbox{S II} \\ 
\mathbf{B}_{5} & = & x_{3} \, \mathbf{a}_{1} + y_{3} \, \mathbf{a}_{2} + z_{3} \, \mathbf{a}_{3} & = & \left(x_{3}a+z_{3}c\cos\beta\right) \, \mathbf{\hat{x}} + y_{3}b \, \mathbf{\hat{y}} + z_{3}c\sin\beta \, \mathbf{\hat{z}} & \left(4g\right) & \mbox{H I} \\ 
\mathbf{B}_{6} & = & -x_{3} \, \mathbf{a}_{1} + y_{3} \, \mathbf{a}_{2} + \left(\frac{1}{2} - z_{3}\right) \, \mathbf{a}_{3} & = & \left(\frac{1}{2}c\cos\beta - x_{3}a - z_{3}c\cos\beta\right) \, \mathbf{\hat{x}} + y_{3}b \, \mathbf{\hat{y}} + \left(\frac{1}{2} - z_{3}\right)c\sin\beta \, \mathbf{\hat{z}} & \left(4g\right) & \mbox{H I} \\ 
\mathbf{B}_{7} & = & -x_{3} \, \mathbf{a}_{1}-y_{3} \, \mathbf{a}_{2}-z_{3} \, \mathbf{a}_{3} & = & \left(-x_{3}a-z_{3}c\cos\beta\right) \, \mathbf{\hat{x}}-y_{3}b \, \mathbf{\hat{y}}-z_{3}c\sin\beta \, \mathbf{\hat{z}} & \left(4g\right) & \mbox{H I} \\ 
\mathbf{B}_{8} & = & x_{3} \, \mathbf{a}_{1}-y_{3} \, \mathbf{a}_{2} + \left(\frac{1}{2} +z_{3}\right) \, \mathbf{a}_{3} & = & \left(\frac{1}{2}c\cos\beta +x_{3}a + z_{3}c\cos\beta\right) \, \mathbf{\hat{x}}-y_{3}b \, \mathbf{\hat{y}} + \left(\frac{1}{2} +z_{3}\right)c\sin\beta \, \mathbf{\hat{z}} & \left(4g\right) & \mbox{H I} \\ 
\mathbf{B}_{9} & = & x_{4} \, \mathbf{a}_{1} + y_{4} \, \mathbf{a}_{2} + z_{4} \, \mathbf{a}_{3} & = & \left(x_{4}a+z_{4}c\cos\beta\right) \, \mathbf{\hat{x}} + y_{4}b \, \mathbf{\hat{y}} + z_{4}c\sin\beta \, \mathbf{\hat{z}} & \left(4g\right) & \mbox{H II} \\ 
\mathbf{B}_{10} & = & -x_{4} \, \mathbf{a}_{1} + y_{4} \, \mathbf{a}_{2} + \left(\frac{1}{2} - z_{4}\right) \, \mathbf{a}_{3} & = & \left(\frac{1}{2}c\cos\beta - x_{4}a - z_{4}c\cos\beta\right) \, \mathbf{\hat{x}} + y_{4}b \, \mathbf{\hat{y}} + \left(\frac{1}{2} - z_{4}\right)c\sin\beta \, \mathbf{\hat{z}} & \left(4g\right) & \mbox{H II} \\ 
\mathbf{B}_{11} & = & -x_{4} \, \mathbf{a}_{1}-y_{4} \, \mathbf{a}_{2}-z_{4} \, \mathbf{a}_{3} & = & \left(-x_{4}a-z_{4}c\cos\beta\right) \, \mathbf{\hat{x}}-y_{4}b \, \mathbf{\hat{y}}-z_{4}c\sin\beta \, \mathbf{\hat{z}} & \left(4g\right) & \mbox{H II} \\ 
\mathbf{B}_{12} & = & x_{4} \, \mathbf{a}_{1}-y_{4} \, \mathbf{a}_{2} + \left(\frac{1}{2} +z_{4}\right) \, \mathbf{a}_{3} & = & \left(\frac{1}{2}c\cos\beta +x_{4}a + z_{4}c\cos\beta\right) \, \mathbf{\hat{x}}-y_{4}b \, \mathbf{\hat{y}} + \left(\frac{1}{2} +z_{4}\right)c\sin\beta \, \mathbf{\hat{z}} & \left(4g\right) & \mbox{H II} \\ 
\end{longtabu}
\renewcommand{\arraystretch}{1.0}
\noindent \hrulefill
\\
\textbf{References:}
\vspace*{-0.25cm}
\begin{flushleft}
  - \bibentry{Li_JCP_140_2014}. \\
  - \bibentry{Shimizu_PRB_51_1995}. \\
\end{flushleft}
\noindent \hrulefill
\\
\textbf{Geometry files:}
\\
\noindent  - CIF: pp. {\hyperref[A2B_mP12_13_2g_ef_cif]{\pageref{A2B_mP12_13_2g_ef_cif}}} \\
\noindent  - POSCAR: pp. {\hyperref[A2B_mP12_13_2g_ef_poscar]{\pageref{A2B_mP12_13_2g_ef_poscar}}} \\
\onecolumn
{\phantomsection\label{A2B_mP6_14_e_a}}
\subsection*{\huge \textbf{{\normalfont $\gamma$-PdCl$_{2}$ Structure: A2B\_mP6\_14\_e\_a}}}
\noindent \hrulefill
\vspace*{0.25cm}
\begin{figure}[htp]
  \centering
  \vspace{-1em}
  {\includegraphics[width=1\textwidth]{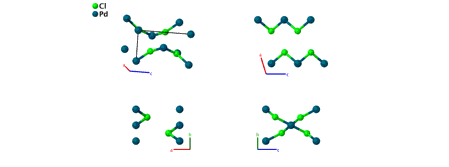}}
\end{figure}
\vspace*{-0.5cm}
\renewcommand{\arraystretch}{1.5}
\begin{equation*}
  \begin{array}{>{$\hspace{-0.15cm}}l<{$}>{$}p{0.5cm}<{$}>{$}p{18.5cm}<{$}}
    \mbox{\large \textbf{Prototype}} &\colon & \ce{$\gamma$-PdCl2} \\
    \mbox{\large \textbf{\AFLOW\ prototype label}} &\colon & \mbox{A2B\_mP6\_14\_e\_a} \\
    \mbox{\large \textbf{\textit{Strukturbericht} designation}} &\colon & \mbox{None} \\
    \mbox{\large \textbf{Pearson symbol}} &\colon & \mbox{mP6} \\
    \mbox{\large \textbf{Space group number}} &\colon & 14 \\
    \mbox{\large \textbf{Space group symbol}} &\colon & P2_{1}/c \\
    \mbox{\large \textbf{\AFLOW\ prototype command}} &\colon &  \texttt{aflow} \,  \, \texttt{-{}-proto=A2B\_mP6\_14\_e\_a } \, \newline \texttt{-{}-params=}{a,b/a,c/a,\beta,x_{2},y_{2},z_{2} }
  \end{array}
\end{equation*}
\renewcommand{\arraystretch}{1.0}

\vspace*{-0.25cm}
\noindent \hrulefill
\begin{itemize}
  \item{(Evers, 2010) place the Pd atoms on the (2c) Wyckoff position.  We have
shifted the origin so that the Pd atoms are at the (2a) position.}
  \item{Data was taken at 300~K.
}
\end{itemize}

\noindent \parbox{1 \linewidth}{
\noindent \hrulefill
\\
\textbf{Simple Monoclinic primitive vectors:} \\
\vspace*{-0.25cm}
\begin{tabular}{cc}
  \begin{tabular}{c}
    \parbox{0.6 \linewidth}{
      \renewcommand{\arraystretch}{1.5}
      \begin{equation*}
        \centering
        \begin{array}{ccc}
              \mathbf{a}_1 & = & a \, \mathbf{\hat{x}} \\
    \mathbf{a}_2 & = & b \, \mathbf{\hat{y}} \\
    \mathbf{a}_3 & = & c \cos\beta \, \mathbf{\hat{x}} + c \sin\beta \, \mathbf{\hat{z}} \\

        \end{array}
      \end{equation*}
    }
    \renewcommand{\arraystretch}{1.0}
  \end{tabular}
  \begin{tabular}{c}
    \includegraphics[width=0.3\linewidth]{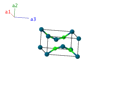} \\
  \end{tabular}
\end{tabular}

}
\vspace*{-0.25cm}

\noindent \hrulefill
\\
\textbf{Basis vectors:}
\vspace*{-0.25cm}
\renewcommand{\arraystretch}{1.5}
\begin{longtabu} to \textwidth{>{\centering $}X[-1,c,c]<{$}>{\centering $}X[-1,c,c]<{$}>{\centering $}X[-1,c,c]<{$}>{\centering $}X[-1,c,c]<{$}>{\centering $}X[-1,c,c]<{$}>{\centering $}X[-1,c,c]<{$}>{\centering $}X[-1,c,c]<{$}}
  & & \mbox{Lattice Coordinates} & & \mbox{Cartesian Coordinates} &\mbox{Wyckoff Position} & \mbox{Atom Type} \\  
  \mathbf{B}_{1} & = & 0 \, \mathbf{a}_{1} + 0 \, \mathbf{a}_{2} + 0 \, \mathbf{a}_{3} & = & 0 \, \mathbf{\hat{x}} + 0 \, \mathbf{\hat{y}} + 0 \, \mathbf{\hat{z}} & \left(2a\right) & \mbox{Pd} \\ 
\mathbf{B}_{2} & = & \frac{1}{2} \, \mathbf{a}_{2} + \frac{1}{2} \, \mathbf{a}_{3} & = & \frac{1}{2}c\cos\beta \, \mathbf{\hat{x}} + \frac{1}{2}b \, \mathbf{\hat{y}} + \frac{1}{2}c\sin\beta \, \mathbf{\hat{z}} & \left(2a\right) & \mbox{Pd} \\ 
\mathbf{B}_{3} & = & x_{2} \, \mathbf{a}_{1} + y_{2} \, \mathbf{a}_{2} + z_{2} \, \mathbf{a}_{3} & = & \left(x_{2}a+z_{2}c\cos\beta\right) \, \mathbf{\hat{x}} + y_{2}b \, \mathbf{\hat{y}} + z_{2}c\sin\beta \, \mathbf{\hat{z}} & \left(4e\right) & \mbox{Cl} \\ 
\mathbf{B}_{4} & = & -x_{2} \, \mathbf{a}_{1} + \left(\frac{1}{2} +y_{2}\right) \, \mathbf{a}_{2} + \left(\frac{1}{2} - z_{2}\right) \, \mathbf{a}_{3} & = & \left(\frac{1}{2}c\cos\beta - x_{2}a - z_{2}c\cos\beta\right) \, \mathbf{\hat{x}} + \left(\frac{1}{2} +y_{2}\right)b \, \mathbf{\hat{y}} + \left(\frac{1}{2} - z_{2}\right)c\sin\beta \, \mathbf{\hat{z}} & \left(4e\right) & \mbox{Cl} \\ 
\mathbf{B}_{5} & = & -x_{2} \, \mathbf{a}_{1}-y_{2} \, \mathbf{a}_{2}-z_{2} \, \mathbf{a}_{3} & = & \left(-x_{2}a-z_{2}c\cos\beta\right) \, \mathbf{\hat{x}}-y_{2}b \, \mathbf{\hat{y}}-z_{2}c\sin\beta \, \mathbf{\hat{z}} & \left(4e\right) & \mbox{Cl} \\ 
\mathbf{B}_{6} & = & x_{2} \, \mathbf{a}_{1} + \left(\frac{1}{2} - y_{2}\right) \, \mathbf{a}_{2} + \left(\frac{1}{2} +z_{2}\right) \, \mathbf{a}_{3} & = & \left(\frac{1}{2}c\cos\beta +x_{2}a + z_{2}c\cos\beta\right) \, \mathbf{\hat{x}} + \left(\frac{1}{2} - y_{2}\right)b \, \mathbf{\hat{y}} + \left(\frac{1}{2} +z_{2}\right)c\sin\beta \, \mathbf{\hat{z}} & \left(4e\right) & \mbox{Cl} \\ 
\end{longtabu}
\renewcommand{\arraystretch}{1.0}
\noindent \hrulefill
\\
\textbf{References:}
\vspace*{-0.25cm}
\begin{flushleft}
  - \bibentry{Evers_AngewChemIntEd_49_5677_2010}. \\
\end{flushleft}
\noindent \hrulefill
\\
\textbf{Geometry files:}
\\
\noindent  - CIF: pp. {\hyperref[A2B_mP6_14_e_a_cif]{\pageref{A2B_mP6_14_e_a_cif}}} \\
\noindent  - POSCAR: pp. {\hyperref[A2B_mP6_14_e_a_poscar]{\pageref{A2B_mP6_14_e_a_poscar}}} \\
\onecolumn
{\phantomsection\label{A7B8_mP120_14_14e_16e}}
\subsection*{\huge \textbf{{\normalfont $\alpha$-Toluene Structure: A7B8\_mP120\_14\_14e\_16e}}}
\noindent \hrulefill
\vspace*{0.25cm}
\begin{figure}[htp]
  \centering
  \vspace{-1em}
  {\includegraphics[width=1\textwidth]{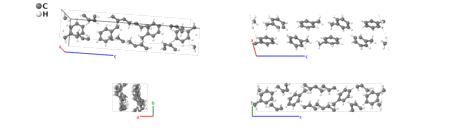}}
\end{figure}
\vspace*{-0.5cm}
\renewcommand{\arraystretch}{1.5}
\begin{equation*}
  \begin{array}{>{$\hspace{-0.15cm}}l<{$}>{$}p{0.5cm}<{$}>{$}p{18.5cm}<{$}}
    \mbox{\large \textbf{Prototype}} &\colon & \ce{$\alpha$-C7H8} \\
    \mbox{\large \textbf{\AFLOW\ prototype label}} &\colon & \mbox{A7B8\_mP120\_14\_14e\_16e} \\
    \mbox{\large \textbf{\textit{Strukturbericht} designation}} &\colon & \mbox{None} \\
    \mbox{\large \textbf{Pearson symbol}} &\colon & \mbox{mP120} \\
    \mbox{\large \textbf{Space group number}} &\colon & 14 \\
    \mbox{\large \textbf{Space group symbol}} &\colon & P2_{1}/c \\
    \mbox{\large \textbf{\AFLOW\ prototype command}} &\colon &  \texttt{aflow} \,  \, \texttt{-{}-proto=A7B8\_mP120\_14\_14e\_16e } \, \newline \texttt{-{}-params=}{a,b/a,c/a,\beta,x_{1},y_{1},z_{1},x_{2},y_{2},z_{2},x_{3},y_{3},z_{3},x_{4},y_{4},z_{4},x_{5},y_{5},z_{5},x_{6},} \newline {y_{6},z_{6},x_{7},y_{7},z_{7},x_{8},y_{8},z_{8},x_{9},y_{9},z_{9},x_{10},y_{10},z_{10},x_{11},y_{11},z_{11},x_{12},y_{12},z_{12},x_{13},y_{13},} \newline {z_{13},x_{14},y_{14},z_{14},x_{15},y_{15},z_{15},x_{16},y_{16},z_{16},x_{17},y_{17},z_{17},x_{18},y_{18},z_{18},x_{19},y_{19},z_{19},x_{20},} \newline {y_{20},z_{20},x_{21},y_{21},z_{21},x_{22},y_{22},z_{22},x_{23},y_{23},z_{23},x_{24},y_{24},z_{24},x_{25},y_{25},z_{25},x_{26},y_{26},z_{26},} \newline {x_{27},y_{27},z_{27},x_{28},y_{28},z_{28},x_{29},y_{29},z_{29},x_{30},y_{30},z_{30} }
  \end{array}
\end{equation*}
\renewcommand{\arraystretch}{1.0}

\vspace*{-0.25cm}
\noindent \hrulefill
\begin{itemize}
  \item{$\alpha$-Toluene is the stable low-temperature crystalline structure of
the toluene molecule, C$_7$H$_8$, which crystallizes below 178~K.
This data was constructed from experiments at 150~K.}
  \item{The hydrogen atomic positions were approximated to agree
with the chemistry of the toluene molecule.
}
\end{itemize}

\noindent \parbox{1 \linewidth}{
\noindent \hrulefill
\\
\textbf{Simple Monoclinic primitive vectors:} \\
\vspace*{-0.25cm}
\begin{tabular}{cc}
  \begin{tabular}{c}
    \parbox{0.6 \linewidth}{
      \renewcommand{\arraystretch}{1.5}
      \begin{equation*}
        \centering
        \begin{array}{ccc}
              \mathbf{a}_1 & = & a \, \mathbf{\hat{x}} \\
    \mathbf{a}_2 & = & b \, \mathbf{\hat{y}} \\
    \mathbf{a}_3 & = & c \cos\beta \, \mathbf{\hat{x}} + c \sin\beta \, \mathbf{\hat{z}} \\

        \end{array}
      \end{equation*}
    }
    \renewcommand{\arraystretch}{1.0}
  \end{tabular}
  \begin{tabular}{c}
    \includegraphics[width=0.3\linewidth]{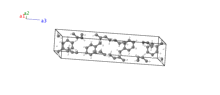} \\
  \end{tabular}
\end{tabular}

}
\vspace*{-0.25cm}

\noindent \hrulefill
\\
\textbf{Basis vectors:}
\vspace*{-0.25cm}
\renewcommand{\arraystretch}{1.5}
\begin{longtabu} to \textwidth{>{\centering $}X[-1,c,c]<{$}>{\centering $}X[-1,c,c]<{$}>{\centering $}X[-1,c,c]<{$}>{\centering $}X[-1,c,c]<{$}>{\centering $}X[-1,c,c]<{$}>{\centering $}X[-1,c,c]<{$}>{\centering $}X[-1,c,c]<{$}}
  & & \mbox{Lattice Coordinates} & & \mbox{Cartesian Coordinates} &\mbox{Wyckoff Position} & \mbox{Atom Type} \\  
  \mathbf{B}_{1} & = & x_{1} \, \mathbf{a}_{1} + y_{1} \, \mathbf{a}_{2} + z_{1} \, \mathbf{a}_{3} & = & \left(x_{1}a+z_{1}c\cos\beta\right) \, \mathbf{\hat{x}} + y_{1}b \, \mathbf{\hat{y}} + z_{1}c\sin\beta \, \mathbf{\hat{z}} & \left(4e\right) & \mbox{C I} \\ 
\mathbf{B}_{2} & = & -x_{1} \, \mathbf{a}_{1} + \left(\frac{1}{2} +y_{1}\right) \, \mathbf{a}_{2} + \left(\frac{1}{2} - z_{1}\right) \, \mathbf{a}_{3} & = & \left(\frac{1}{2}c\cos\beta - x_{1}a - z_{1}c\cos\beta\right) \, \mathbf{\hat{x}} + \left(\frac{1}{2} +y_{1}\right)b \, \mathbf{\hat{y}} + \left(\frac{1}{2} - z_{1}\right)c\sin\beta \, \mathbf{\hat{z}} & \left(4e\right) & \mbox{C I} \\ 
\mathbf{B}_{3} & = & -x_{1} \, \mathbf{a}_{1}-y_{1} \, \mathbf{a}_{2}-z_{1} \, \mathbf{a}_{3} & = & \left(-x_{1}a-z_{1}c\cos\beta\right) \, \mathbf{\hat{x}}-y_{1}b \, \mathbf{\hat{y}}-z_{1}c\sin\beta \, \mathbf{\hat{z}} & \left(4e\right) & \mbox{C I} \\ 
\mathbf{B}_{4} & = & x_{1} \, \mathbf{a}_{1} + \left(\frac{1}{2} - y_{1}\right) \, \mathbf{a}_{2} + \left(\frac{1}{2} +z_{1}\right) \, \mathbf{a}_{3} & = & \left(\frac{1}{2}c\cos\beta +x_{1}a + z_{1}c\cos\beta\right) \, \mathbf{\hat{x}} + \left(\frac{1}{2} - y_{1}\right)b \, \mathbf{\hat{y}} + \left(\frac{1}{2} +z_{1}\right)c\sin\beta \, \mathbf{\hat{z}} & \left(4e\right) & \mbox{C I} \\ 
\mathbf{B}_{5} & = & x_{2} \, \mathbf{a}_{1} + y_{2} \, \mathbf{a}_{2} + z_{2} \, \mathbf{a}_{3} & = & \left(x_{2}a+z_{2}c\cos\beta\right) \, \mathbf{\hat{x}} + y_{2}b \, \mathbf{\hat{y}} + z_{2}c\sin\beta \, \mathbf{\hat{z}} & \left(4e\right) & \mbox{C II} \\ 
\mathbf{B}_{6} & = & -x_{2} \, \mathbf{a}_{1} + \left(\frac{1}{2} +y_{2}\right) \, \mathbf{a}_{2} + \left(\frac{1}{2} - z_{2}\right) \, \mathbf{a}_{3} & = & \left(\frac{1}{2}c\cos\beta - x_{2}a - z_{2}c\cos\beta\right) \, \mathbf{\hat{x}} + \left(\frac{1}{2} +y_{2}\right)b \, \mathbf{\hat{y}} + \left(\frac{1}{2} - z_{2}\right)c\sin\beta \, \mathbf{\hat{z}} & \left(4e\right) & \mbox{C II} \\ 
\mathbf{B}_{7} & = & -x_{2} \, \mathbf{a}_{1}-y_{2} \, \mathbf{a}_{2}-z_{2} \, \mathbf{a}_{3} & = & \left(-x_{2}a-z_{2}c\cos\beta\right) \, \mathbf{\hat{x}}-y_{2}b \, \mathbf{\hat{y}}-z_{2}c\sin\beta \, \mathbf{\hat{z}} & \left(4e\right) & \mbox{C II} \\ 
\mathbf{B}_{8} & = & x_{2} \, \mathbf{a}_{1} + \left(\frac{1}{2} - y_{2}\right) \, \mathbf{a}_{2} + \left(\frac{1}{2} +z_{2}\right) \, \mathbf{a}_{3} & = & \left(\frac{1}{2}c\cos\beta +x_{2}a + z_{2}c\cos\beta\right) \, \mathbf{\hat{x}} + \left(\frac{1}{2} - y_{2}\right)b \, \mathbf{\hat{y}} + \left(\frac{1}{2} +z_{2}\right)c\sin\beta \, \mathbf{\hat{z}} & \left(4e\right) & \mbox{C II} \\ 
\mathbf{B}_{9} & = & x_{3} \, \mathbf{a}_{1} + y_{3} \, \mathbf{a}_{2} + z_{3} \, \mathbf{a}_{3} & = & \left(x_{3}a+z_{3}c\cos\beta\right) \, \mathbf{\hat{x}} + y_{3}b \, \mathbf{\hat{y}} + z_{3}c\sin\beta \, \mathbf{\hat{z}} & \left(4e\right) & \mbox{C III} \\ 
\mathbf{B}_{10} & = & -x_{3} \, \mathbf{a}_{1} + \left(\frac{1}{2} +y_{3}\right) \, \mathbf{a}_{2} + \left(\frac{1}{2} - z_{3}\right) \, \mathbf{a}_{3} & = & \left(\frac{1}{2}c\cos\beta - x_{3}a - z_{3}c\cos\beta\right) \, \mathbf{\hat{x}} + \left(\frac{1}{2} +y_{3}\right)b \, \mathbf{\hat{y}} + \left(\frac{1}{2} - z_{3}\right)c\sin\beta \, \mathbf{\hat{z}} & \left(4e\right) & \mbox{C III} \\ 
\mathbf{B}_{11} & = & -x_{3} \, \mathbf{a}_{1}-y_{3} \, \mathbf{a}_{2}-z_{3} \, \mathbf{a}_{3} & = & \left(-x_{3}a-z_{3}c\cos\beta\right) \, \mathbf{\hat{x}}-y_{3}b \, \mathbf{\hat{y}}-z_{3}c\sin\beta \, \mathbf{\hat{z}} & \left(4e\right) & \mbox{C III} \\ 
\mathbf{B}_{12} & = & x_{3} \, \mathbf{a}_{1} + \left(\frac{1}{2} - y_{3}\right) \, \mathbf{a}_{2} + \left(\frac{1}{2} +z_{3}\right) \, \mathbf{a}_{3} & = & \left(\frac{1}{2}c\cos\beta +x_{3}a + z_{3}c\cos\beta\right) \, \mathbf{\hat{x}} + \left(\frac{1}{2} - y_{3}\right)b \, \mathbf{\hat{y}} + \left(\frac{1}{2} +z_{3}\right)c\sin\beta \, \mathbf{\hat{z}} & \left(4e\right) & \mbox{C III} \\ 
\mathbf{B}_{13} & = & x_{4} \, \mathbf{a}_{1} + y_{4} \, \mathbf{a}_{2} + z_{4} \, \mathbf{a}_{3} & = & \left(x_{4}a+z_{4}c\cos\beta\right) \, \mathbf{\hat{x}} + y_{4}b \, \mathbf{\hat{y}} + z_{4}c\sin\beta \, \mathbf{\hat{z}} & \left(4e\right) & \mbox{C IV} \\ 
\mathbf{B}_{14} & = & -x_{4} \, \mathbf{a}_{1} + \left(\frac{1}{2} +y_{4}\right) \, \mathbf{a}_{2} + \left(\frac{1}{2} - z_{4}\right) \, \mathbf{a}_{3} & = & \left(\frac{1}{2}c\cos\beta - x_{4}a - z_{4}c\cos\beta\right) \, \mathbf{\hat{x}} + \left(\frac{1}{2} +y_{4}\right)b \, \mathbf{\hat{y}} + \left(\frac{1}{2} - z_{4}\right)c\sin\beta \, \mathbf{\hat{z}} & \left(4e\right) & \mbox{C IV} \\ 
\mathbf{B}_{15} & = & -x_{4} \, \mathbf{a}_{1}-y_{4} \, \mathbf{a}_{2}-z_{4} \, \mathbf{a}_{3} & = & \left(-x_{4}a-z_{4}c\cos\beta\right) \, \mathbf{\hat{x}}-y_{4}b \, \mathbf{\hat{y}}-z_{4}c\sin\beta \, \mathbf{\hat{z}} & \left(4e\right) & \mbox{C IV} \\ 
\mathbf{B}_{16} & = & x_{4} \, \mathbf{a}_{1} + \left(\frac{1}{2} - y_{4}\right) \, \mathbf{a}_{2} + \left(\frac{1}{2} +z_{4}\right) \, \mathbf{a}_{3} & = & \left(\frac{1}{2}c\cos\beta +x_{4}a + z_{4}c\cos\beta\right) \, \mathbf{\hat{x}} + \left(\frac{1}{2} - y_{4}\right)b \, \mathbf{\hat{y}} + \left(\frac{1}{2} +z_{4}\right)c\sin\beta \, \mathbf{\hat{z}} & \left(4e\right) & \mbox{C IV} \\ 
\mathbf{B}_{17} & = & x_{5} \, \mathbf{a}_{1} + y_{5} \, \mathbf{a}_{2} + z_{5} \, \mathbf{a}_{3} & = & \left(x_{5}a+z_{5}c\cos\beta\right) \, \mathbf{\hat{x}} + y_{5}b \, \mathbf{\hat{y}} + z_{5}c\sin\beta \, \mathbf{\hat{z}} & \left(4e\right) & \mbox{C V} \\ 
\mathbf{B}_{18} & = & -x_{5} \, \mathbf{a}_{1} + \left(\frac{1}{2} +y_{5}\right) \, \mathbf{a}_{2} + \left(\frac{1}{2} - z_{5}\right) \, \mathbf{a}_{3} & = & \left(\frac{1}{2}c\cos\beta - x_{5}a - z_{5}c\cos\beta\right) \, \mathbf{\hat{x}} + \left(\frac{1}{2} +y_{5}\right)b \, \mathbf{\hat{y}} + \left(\frac{1}{2} - z_{5}\right)c\sin\beta \, \mathbf{\hat{z}} & \left(4e\right) & \mbox{C V} \\ 
\mathbf{B}_{19} & = & -x_{5} \, \mathbf{a}_{1}-y_{5} \, \mathbf{a}_{2}-z_{5} \, \mathbf{a}_{3} & = & \left(-x_{5}a-z_{5}c\cos\beta\right) \, \mathbf{\hat{x}}-y_{5}b \, \mathbf{\hat{y}}-z_{5}c\sin\beta \, \mathbf{\hat{z}} & \left(4e\right) & \mbox{C V} \\ 
\mathbf{B}_{20} & = & x_{5} \, \mathbf{a}_{1} + \left(\frac{1}{2} - y_{5}\right) \, \mathbf{a}_{2} + \left(\frac{1}{2} +z_{5}\right) \, \mathbf{a}_{3} & = & \left(\frac{1}{2}c\cos\beta +x_{5}a + z_{5}c\cos\beta\right) \, \mathbf{\hat{x}} + \left(\frac{1}{2} - y_{5}\right)b \, \mathbf{\hat{y}} + \left(\frac{1}{2} +z_{5}\right)c\sin\beta \, \mathbf{\hat{z}} & \left(4e\right) & \mbox{C V} \\ 
\mathbf{B}_{21} & = & x_{6} \, \mathbf{a}_{1} + y_{6} \, \mathbf{a}_{2} + z_{6} \, \mathbf{a}_{3} & = & \left(x_{6}a+z_{6}c\cos\beta\right) \, \mathbf{\hat{x}} + y_{6}b \, \mathbf{\hat{y}} + z_{6}c\sin\beta \, \mathbf{\hat{z}} & \left(4e\right) & \mbox{C VI} \\ 
\mathbf{B}_{22} & = & -x_{6} \, \mathbf{a}_{1} + \left(\frac{1}{2} +y_{6}\right) \, \mathbf{a}_{2} + \left(\frac{1}{2} - z_{6}\right) \, \mathbf{a}_{3} & = & \left(\frac{1}{2}c\cos\beta - x_{6}a - z_{6}c\cos\beta\right) \, \mathbf{\hat{x}} + \left(\frac{1}{2} +y_{6}\right)b \, \mathbf{\hat{y}} + \left(\frac{1}{2} - z_{6}\right)c\sin\beta \, \mathbf{\hat{z}} & \left(4e\right) & \mbox{C VI} \\ 
\mathbf{B}_{23} & = & -x_{6} \, \mathbf{a}_{1}-y_{6} \, \mathbf{a}_{2}-z_{6} \, \mathbf{a}_{3} & = & \left(-x_{6}a-z_{6}c\cos\beta\right) \, \mathbf{\hat{x}}-y_{6}b \, \mathbf{\hat{y}}-z_{6}c\sin\beta \, \mathbf{\hat{z}} & \left(4e\right) & \mbox{C VI} \\ 
\mathbf{B}_{24} & = & x_{6} \, \mathbf{a}_{1} + \left(\frac{1}{2} - y_{6}\right) \, \mathbf{a}_{2} + \left(\frac{1}{2} +z_{6}\right) \, \mathbf{a}_{3} & = & \left(\frac{1}{2}c\cos\beta +x_{6}a + z_{6}c\cos\beta\right) \, \mathbf{\hat{x}} + \left(\frac{1}{2} - y_{6}\right)b \, \mathbf{\hat{y}} + \left(\frac{1}{2} +z_{6}\right)c\sin\beta \, \mathbf{\hat{z}} & \left(4e\right) & \mbox{C VI} \\ 
\mathbf{B}_{25} & = & x_{7} \, \mathbf{a}_{1} + y_{7} \, \mathbf{a}_{2} + z_{7} \, \mathbf{a}_{3} & = & \left(x_{7}a+z_{7}c\cos\beta\right) \, \mathbf{\hat{x}} + y_{7}b \, \mathbf{\hat{y}} + z_{7}c\sin\beta \, \mathbf{\hat{z}} & \left(4e\right) & \mbox{C VII} \\ 
\mathbf{B}_{26} & = & -x_{7} \, \mathbf{a}_{1} + \left(\frac{1}{2} +y_{7}\right) \, \mathbf{a}_{2} + \left(\frac{1}{2} - z_{7}\right) \, \mathbf{a}_{3} & = & \left(\frac{1}{2}c\cos\beta - x_{7}a - z_{7}c\cos\beta\right) \, \mathbf{\hat{x}} + \left(\frac{1}{2} +y_{7}\right)b \, \mathbf{\hat{y}} + \left(\frac{1}{2} - z_{7}\right)c\sin\beta \, \mathbf{\hat{z}} & \left(4e\right) & \mbox{C VII} \\ 
\mathbf{B}_{27} & = & -x_{7} \, \mathbf{a}_{1}-y_{7} \, \mathbf{a}_{2}-z_{7} \, \mathbf{a}_{3} & = & \left(-x_{7}a-z_{7}c\cos\beta\right) \, \mathbf{\hat{x}}-y_{7}b \, \mathbf{\hat{y}}-z_{7}c\sin\beta \, \mathbf{\hat{z}} & \left(4e\right) & \mbox{C VII} \\ 
\mathbf{B}_{28} & = & x_{7} \, \mathbf{a}_{1} + \left(\frac{1}{2} - y_{7}\right) \, \mathbf{a}_{2} + \left(\frac{1}{2} +z_{7}\right) \, \mathbf{a}_{3} & = & \left(\frac{1}{2}c\cos\beta +x_{7}a + z_{7}c\cos\beta\right) \, \mathbf{\hat{x}} + \left(\frac{1}{2} - y_{7}\right)b \, \mathbf{\hat{y}} + \left(\frac{1}{2} +z_{7}\right)c\sin\beta \, \mathbf{\hat{z}} & \left(4e\right) & \mbox{C VII} \\ 
\mathbf{B}_{29} & = & x_{8} \, \mathbf{a}_{1} + y_{8} \, \mathbf{a}_{2} + z_{8} \, \mathbf{a}_{3} & = & \left(x_{8}a+z_{8}c\cos\beta\right) \, \mathbf{\hat{x}} + y_{8}b \, \mathbf{\hat{y}} + z_{8}c\sin\beta \, \mathbf{\hat{z}} & \left(4e\right) & \mbox{C VIII} \\ 
\mathbf{B}_{30} & = & -x_{8} \, \mathbf{a}_{1} + \left(\frac{1}{2} +y_{8}\right) \, \mathbf{a}_{2} + \left(\frac{1}{2} - z_{8}\right) \, \mathbf{a}_{3} & = & \left(\frac{1}{2}c\cos\beta - x_{8}a - z_{8}c\cos\beta\right) \, \mathbf{\hat{x}} + \left(\frac{1}{2} +y_{8}\right)b \, \mathbf{\hat{y}} + \left(\frac{1}{2} - z_{8}\right)c\sin\beta \, \mathbf{\hat{z}} & \left(4e\right) & \mbox{C VIII} \\ 
\mathbf{B}_{31} & = & -x_{8} \, \mathbf{a}_{1}-y_{8} \, \mathbf{a}_{2}-z_{8} \, \mathbf{a}_{3} & = & \left(-x_{8}a-z_{8}c\cos\beta\right) \, \mathbf{\hat{x}}-y_{8}b \, \mathbf{\hat{y}}-z_{8}c\sin\beta \, \mathbf{\hat{z}} & \left(4e\right) & \mbox{C VIII} \\ 
\mathbf{B}_{32} & = & x_{8} \, \mathbf{a}_{1} + \left(\frac{1}{2} - y_{8}\right) \, \mathbf{a}_{2} + \left(\frac{1}{2} +z_{8}\right) \, \mathbf{a}_{3} & = & \left(\frac{1}{2}c\cos\beta +x_{8}a + z_{8}c\cos\beta\right) \, \mathbf{\hat{x}} + \left(\frac{1}{2} - y_{8}\right)b \, \mathbf{\hat{y}} + \left(\frac{1}{2} +z_{8}\right)c\sin\beta \, \mathbf{\hat{z}} & \left(4e\right) & \mbox{C VIII} \\ 
\mathbf{B}_{33} & = & x_{9} \, \mathbf{a}_{1} + y_{9} \, \mathbf{a}_{2} + z_{9} \, \mathbf{a}_{3} & = & \left(x_{9}a+z_{9}c\cos\beta\right) \, \mathbf{\hat{x}} + y_{9}b \, \mathbf{\hat{y}} + z_{9}c\sin\beta \, \mathbf{\hat{z}} & \left(4e\right) & \mbox{C IX} \\ 
\mathbf{B}_{34} & = & -x_{9} \, \mathbf{a}_{1} + \left(\frac{1}{2} +y_{9}\right) \, \mathbf{a}_{2} + \left(\frac{1}{2} - z_{9}\right) \, \mathbf{a}_{3} & = & \left(\frac{1}{2}c\cos\beta - x_{9}a - z_{9}c\cos\beta\right) \, \mathbf{\hat{x}} + \left(\frac{1}{2} +y_{9}\right)b \, \mathbf{\hat{y}} + \left(\frac{1}{2} - z_{9}\right)c\sin\beta \, \mathbf{\hat{z}} & \left(4e\right) & \mbox{C IX} \\ 
\mathbf{B}_{35} & = & -x_{9} \, \mathbf{a}_{1}-y_{9} \, \mathbf{a}_{2}-z_{9} \, \mathbf{a}_{3} & = & \left(-x_{9}a-z_{9}c\cos\beta\right) \, \mathbf{\hat{x}}-y_{9}b \, \mathbf{\hat{y}}-z_{9}c\sin\beta \, \mathbf{\hat{z}} & \left(4e\right) & \mbox{C IX} \\ 
\mathbf{B}_{36} & = & x_{9} \, \mathbf{a}_{1} + \left(\frac{1}{2} - y_{9}\right) \, \mathbf{a}_{2} + \left(\frac{1}{2} +z_{9}\right) \, \mathbf{a}_{3} & = & \left(\frac{1}{2}c\cos\beta +x_{9}a + z_{9}c\cos\beta\right) \, \mathbf{\hat{x}} + \left(\frac{1}{2} - y_{9}\right)b \, \mathbf{\hat{y}} + \left(\frac{1}{2} +z_{9}\right)c\sin\beta \, \mathbf{\hat{z}} & \left(4e\right) & \mbox{C IX} \\ 
\mathbf{B}_{37} & = & x_{10} \, \mathbf{a}_{1} + y_{10} \, \mathbf{a}_{2} + z_{10} \, \mathbf{a}_{3} & = & \left(x_{10}a+z_{10}c\cos\beta\right) \, \mathbf{\hat{x}} + y_{10}b \, \mathbf{\hat{y}} + z_{10}c\sin\beta \, \mathbf{\hat{z}} & \left(4e\right) & \mbox{C X} \\ 
\mathbf{B}_{38} & = & -x_{10} \, \mathbf{a}_{1} + \left(\frac{1}{2} +y_{10}\right) \, \mathbf{a}_{2} + \left(\frac{1}{2} - z_{10}\right) \, \mathbf{a}_{3} & = & \left(\frac{1}{2}c\cos\beta - x_{10}a - z_{10}c\cos\beta\right) \, \mathbf{\hat{x}} + \left(\frac{1}{2} +y_{10}\right)b \, \mathbf{\hat{y}} + \left(\frac{1}{2} - z_{10}\right)c\sin\beta \, \mathbf{\hat{z}} & \left(4e\right) & \mbox{C X} \\ 
\mathbf{B}_{39} & = & -x_{10} \, \mathbf{a}_{1}-y_{10} \, \mathbf{a}_{2}-z_{10} \, \mathbf{a}_{3} & = & \left(-x_{10}a-z_{10}c\cos\beta\right) \, \mathbf{\hat{x}}-y_{10}b \, \mathbf{\hat{y}}-z_{10}c\sin\beta \, \mathbf{\hat{z}} & \left(4e\right) & \mbox{C X} \\ 
\mathbf{B}_{40} & = & x_{10} \, \mathbf{a}_{1} + \left(\frac{1}{2} - y_{10}\right) \, \mathbf{a}_{2} + \left(\frac{1}{2} +z_{10}\right) \, \mathbf{a}_{3} & = & \left(\frac{1}{2}c\cos\beta +x_{10}a + z_{10}c\cos\beta\right) \, \mathbf{\hat{x}} + \left(\frac{1}{2} - y_{10}\right)b \, \mathbf{\hat{y}} + \left(\frac{1}{2} +z_{10}\right)c\sin\beta \, \mathbf{\hat{z}} & \left(4e\right) & \mbox{C X} \\ 
\mathbf{B}_{41} & = & x_{11} \, \mathbf{a}_{1} + y_{11} \, \mathbf{a}_{2} + z_{11} \, \mathbf{a}_{3} & = & \left(x_{11}a+z_{11}c\cos\beta\right) \, \mathbf{\hat{x}} + y_{11}b \, \mathbf{\hat{y}} + z_{11}c\sin\beta \, \mathbf{\hat{z}} & \left(4e\right) & \mbox{C XI} \\ 
\mathbf{B}_{42} & = & -x_{11} \, \mathbf{a}_{1} + \left(\frac{1}{2} +y_{11}\right) \, \mathbf{a}_{2} + \left(\frac{1}{2} - z_{11}\right) \, \mathbf{a}_{3} & = & \left(\frac{1}{2}c\cos\beta - x_{11}a - z_{11}c\cos\beta\right) \, \mathbf{\hat{x}} + \left(\frac{1}{2} +y_{11}\right)b \, \mathbf{\hat{y}} + \left(\frac{1}{2} - z_{11}\right)c\sin\beta \, \mathbf{\hat{z}} & \left(4e\right) & \mbox{C XI} \\ 
\mathbf{B}_{43} & = & -x_{11} \, \mathbf{a}_{1}-y_{11} \, \mathbf{a}_{2}-z_{11} \, \mathbf{a}_{3} & = & \left(-x_{11}a-z_{11}c\cos\beta\right) \, \mathbf{\hat{x}}-y_{11}b \, \mathbf{\hat{y}}-z_{11}c\sin\beta \, \mathbf{\hat{z}} & \left(4e\right) & \mbox{C XI} \\ 
\mathbf{B}_{44} & = & x_{11} \, \mathbf{a}_{1} + \left(\frac{1}{2} - y_{11}\right) \, \mathbf{a}_{2} + \left(\frac{1}{2} +z_{11}\right) \, \mathbf{a}_{3} & = & \left(\frac{1}{2}c\cos\beta +x_{11}a + z_{11}c\cos\beta\right) \, \mathbf{\hat{x}} + \left(\frac{1}{2} - y_{11}\right)b \, \mathbf{\hat{y}} + \left(\frac{1}{2} +z_{11}\right)c\sin\beta \, \mathbf{\hat{z}} & \left(4e\right) & \mbox{C XI} \\ 
\mathbf{B}_{45} & = & x_{12} \, \mathbf{a}_{1} + y_{12} \, \mathbf{a}_{2} + z_{12} \, \mathbf{a}_{3} & = & \left(x_{12}a+z_{12}c\cos\beta\right) \, \mathbf{\hat{x}} + y_{12}b \, \mathbf{\hat{y}} + z_{12}c\sin\beta \, \mathbf{\hat{z}} & \left(4e\right) & \mbox{C XII} \\ 
\mathbf{B}_{46} & = & -x_{12} \, \mathbf{a}_{1} + \left(\frac{1}{2} +y_{12}\right) \, \mathbf{a}_{2} + \left(\frac{1}{2} - z_{12}\right) \, \mathbf{a}_{3} & = & \left(\frac{1}{2}c\cos\beta - x_{12}a - z_{12}c\cos\beta\right) \, \mathbf{\hat{x}} + \left(\frac{1}{2} +y_{12}\right)b \, \mathbf{\hat{y}} + \left(\frac{1}{2} - z_{12}\right)c\sin\beta \, \mathbf{\hat{z}} & \left(4e\right) & \mbox{C XII} \\ 
\mathbf{B}_{47} & = & -x_{12} \, \mathbf{a}_{1}-y_{12} \, \mathbf{a}_{2}-z_{12} \, \mathbf{a}_{3} & = & \left(-x_{12}a-z_{12}c\cos\beta\right) \, \mathbf{\hat{x}}-y_{12}b \, \mathbf{\hat{y}}-z_{12}c\sin\beta \, \mathbf{\hat{z}} & \left(4e\right) & \mbox{C XII} \\ 
\mathbf{B}_{48} & = & x_{12} \, \mathbf{a}_{1} + \left(\frac{1}{2} - y_{12}\right) \, \mathbf{a}_{2} + \left(\frac{1}{2} +z_{12}\right) \, \mathbf{a}_{3} & = & \left(\frac{1}{2}c\cos\beta +x_{12}a + z_{12}c\cos\beta\right) \, \mathbf{\hat{x}} + \left(\frac{1}{2} - y_{12}\right)b \, \mathbf{\hat{y}} + \left(\frac{1}{2} +z_{12}\right)c\sin\beta \, \mathbf{\hat{z}} & \left(4e\right) & \mbox{C XII} \\ 
\mathbf{B}_{49} & = & x_{13} \, \mathbf{a}_{1} + y_{13} \, \mathbf{a}_{2} + z_{13} \, \mathbf{a}_{3} & = & \left(x_{13}a+z_{13}c\cos\beta\right) \, \mathbf{\hat{x}} + y_{13}b \, \mathbf{\hat{y}} + z_{13}c\sin\beta \, \mathbf{\hat{z}} & \left(4e\right) & \mbox{C XIII} \\ 
\mathbf{B}_{50} & = & -x_{13} \, \mathbf{a}_{1} + \left(\frac{1}{2} +y_{13}\right) \, \mathbf{a}_{2} + \left(\frac{1}{2} - z_{13}\right) \, \mathbf{a}_{3} & = & \left(\frac{1}{2}c\cos\beta - x_{13}a - z_{13}c\cos\beta\right) \, \mathbf{\hat{x}} + \left(\frac{1}{2} +y_{13}\right)b \, \mathbf{\hat{y}} + \left(\frac{1}{2} - z_{13}\right)c\sin\beta \, \mathbf{\hat{z}} & \left(4e\right) & \mbox{C XIII} \\ 
\mathbf{B}_{51} & = & -x_{13} \, \mathbf{a}_{1}-y_{13} \, \mathbf{a}_{2}-z_{13} \, \mathbf{a}_{3} & = & \left(-x_{13}a-z_{13}c\cos\beta\right) \, \mathbf{\hat{x}}-y_{13}b \, \mathbf{\hat{y}}-z_{13}c\sin\beta \, \mathbf{\hat{z}} & \left(4e\right) & \mbox{C XIII} \\ 
\mathbf{B}_{52} & = & x_{13} \, \mathbf{a}_{1} + \left(\frac{1}{2} - y_{13}\right) \, \mathbf{a}_{2} + \left(\frac{1}{2} +z_{13}\right) \, \mathbf{a}_{3} & = & \left(\frac{1}{2}c\cos\beta +x_{13}a + z_{13}c\cos\beta\right) \, \mathbf{\hat{x}} + \left(\frac{1}{2} - y_{13}\right)b \, \mathbf{\hat{y}} + \left(\frac{1}{2} +z_{13}\right)c\sin\beta \, \mathbf{\hat{z}} & \left(4e\right) & \mbox{C XIII} \\ 
\mathbf{B}_{53} & = & x_{14} \, \mathbf{a}_{1} + y_{14} \, \mathbf{a}_{2} + z_{14} \, \mathbf{a}_{3} & = & \left(x_{14}a+z_{14}c\cos\beta\right) \, \mathbf{\hat{x}} + y_{14}b \, \mathbf{\hat{y}} + z_{14}c\sin\beta \, \mathbf{\hat{z}} & \left(4e\right) & \mbox{C XIV} \\ 
\mathbf{B}_{54} & = & -x_{14} \, \mathbf{a}_{1} + \left(\frac{1}{2} +y_{14}\right) \, \mathbf{a}_{2} + \left(\frac{1}{2} - z_{14}\right) \, \mathbf{a}_{3} & = & \left(\frac{1}{2}c\cos\beta - x_{14}a - z_{14}c\cos\beta\right) \, \mathbf{\hat{x}} + \left(\frac{1}{2} +y_{14}\right)b \, \mathbf{\hat{y}} + \left(\frac{1}{2} - z_{14}\right)c\sin\beta \, \mathbf{\hat{z}} & \left(4e\right) & \mbox{C XIV} \\ 
\mathbf{B}_{55} & = & -x_{14} \, \mathbf{a}_{1}-y_{14} \, \mathbf{a}_{2}-z_{14} \, \mathbf{a}_{3} & = & \left(-x_{14}a-z_{14}c\cos\beta\right) \, \mathbf{\hat{x}}-y_{14}b \, \mathbf{\hat{y}}-z_{14}c\sin\beta \, \mathbf{\hat{z}} & \left(4e\right) & \mbox{C XIV} \\ 
\mathbf{B}_{56} & = & x_{14} \, \mathbf{a}_{1} + \left(\frac{1}{2} - y_{14}\right) \, \mathbf{a}_{2} + \left(\frac{1}{2} +z_{14}\right) \, \mathbf{a}_{3} & = & \left(\frac{1}{2}c\cos\beta +x_{14}a + z_{14}c\cos\beta\right) \, \mathbf{\hat{x}} + \left(\frac{1}{2} - y_{14}\right)b \, \mathbf{\hat{y}} + \left(\frac{1}{2} +z_{14}\right)c\sin\beta \, \mathbf{\hat{z}} & \left(4e\right) & \mbox{C XIV} \\ 
\mathbf{B}_{57} & = & x_{15} \, \mathbf{a}_{1} + y_{15} \, \mathbf{a}_{2} + z_{15} \, \mathbf{a}_{3} & = & \left(x_{15}a+z_{15}c\cos\beta\right) \, \mathbf{\hat{x}} + y_{15}b \, \mathbf{\hat{y}} + z_{15}c\sin\beta \, \mathbf{\hat{z}} & \left(4e\right) & \mbox{H I} \\ 
\mathbf{B}_{58} & = & -x_{15} \, \mathbf{a}_{1} + \left(\frac{1}{2} +y_{15}\right) \, \mathbf{a}_{2} + \left(\frac{1}{2} - z_{15}\right) \, \mathbf{a}_{3} & = & \left(\frac{1}{2}c\cos\beta - x_{15}a - z_{15}c\cos\beta\right) \, \mathbf{\hat{x}} + \left(\frac{1}{2} +y_{15}\right)b \, \mathbf{\hat{y}} + \left(\frac{1}{2} - z_{15}\right)c\sin\beta \, \mathbf{\hat{z}} & \left(4e\right) & \mbox{H I} \\ 
\mathbf{B}_{59} & = & -x_{15} \, \mathbf{a}_{1}-y_{15} \, \mathbf{a}_{2}-z_{15} \, \mathbf{a}_{3} & = & \left(-x_{15}a-z_{15}c\cos\beta\right) \, \mathbf{\hat{x}}-y_{15}b \, \mathbf{\hat{y}}-z_{15}c\sin\beta \, \mathbf{\hat{z}} & \left(4e\right) & \mbox{H I} \\ 
\mathbf{B}_{60} & = & x_{15} \, \mathbf{a}_{1} + \left(\frac{1}{2} - y_{15}\right) \, \mathbf{a}_{2} + \left(\frac{1}{2} +z_{15}\right) \, \mathbf{a}_{3} & = & \left(\frac{1}{2}c\cos\beta +x_{15}a + z_{15}c\cos\beta\right) \, \mathbf{\hat{x}} + \left(\frac{1}{2} - y_{15}\right)b \, \mathbf{\hat{y}} + \left(\frac{1}{2} +z_{15}\right)c\sin\beta \, \mathbf{\hat{z}} & \left(4e\right) & \mbox{H I} \\ 
\mathbf{B}_{61} & = & x_{16} \, \mathbf{a}_{1} + y_{16} \, \mathbf{a}_{2} + z_{16} \, \mathbf{a}_{3} & = & \left(x_{16}a+z_{16}c\cos\beta\right) \, \mathbf{\hat{x}} + y_{16}b \, \mathbf{\hat{y}} + z_{16}c\sin\beta \, \mathbf{\hat{z}} & \left(4e\right) & \mbox{H II} \\ 
\mathbf{B}_{62} & = & -x_{16} \, \mathbf{a}_{1} + \left(\frac{1}{2} +y_{16}\right) \, \mathbf{a}_{2} + \left(\frac{1}{2} - z_{16}\right) \, \mathbf{a}_{3} & = & \left(\frac{1}{2}c\cos\beta - x_{16}a - z_{16}c\cos\beta\right) \, \mathbf{\hat{x}} + \left(\frac{1}{2} +y_{16}\right)b \, \mathbf{\hat{y}} + \left(\frac{1}{2} - z_{16}\right)c\sin\beta \, \mathbf{\hat{z}} & \left(4e\right) & \mbox{H II} \\ 
\mathbf{B}_{63} & = & -x_{16} \, \mathbf{a}_{1}-y_{16} \, \mathbf{a}_{2}-z_{16} \, \mathbf{a}_{3} & = & \left(-x_{16}a-z_{16}c\cos\beta\right) \, \mathbf{\hat{x}}-y_{16}b \, \mathbf{\hat{y}}-z_{16}c\sin\beta \, \mathbf{\hat{z}} & \left(4e\right) & \mbox{H II} \\ 
\mathbf{B}_{64} & = & x_{16} \, \mathbf{a}_{1} + \left(\frac{1}{2} - y_{16}\right) \, \mathbf{a}_{2} + \left(\frac{1}{2} +z_{16}\right) \, \mathbf{a}_{3} & = & \left(\frac{1}{2}c\cos\beta +x_{16}a + z_{16}c\cos\beta\right) \, \mathbf{\hat{x}} + \left(\frac{1}{2} - y_{16}\right)b \, \mathbf{\hat{y}} + \left(\frac{1}{2} +z_{16}\right)c\sin\beta \, \mathbf{\hat{z}} & \left(4e\right) & \mbox{H II} \\ 
\mathbf{B}_{65} & = & x_{17} \, \mathbf{a}_{1} + y_{17} \, \mathbf{a}_{2} + z_{17} \, \mathbf{a}_{3} & = & \left(x_{17}a+z_{17}c\cos\beta\right) \, \mathbf{\hat{x}} + y_{17}b \, \mathbf{\hat{y}} + z_{17}c\sin\beta \, \mathbf{\hat{z}} & \left(4e\right) & \mbox{H III} \\ 
\mathbf{B}_{66} & = & -x_{17} \, \mathbf{a}_{1} + \left(\frac{1}{2} +y_{17}\right) \, \mathbf{a}_{2} + \left(\frac{1}{2} - z_{17}\right) \, \mathbf{a}_{3} & = & \left(\frac{1}{2}c\cos\beta - x_{17}a - z_{17}c\cos\beta\right) \, \mathbf{\hat{x}} + \left(\frac{1}{2} +y_{17}\right)b \, \mathbf{\hat{y}} + \left(\frac{1}{2} - z_{17}\right)c\sin\beta \, \mathbf{\hat{z}} & \left(4e\right) & \mbox{H III} \\ 
\mathbf{B}_{67} & = & -x_{17} \, \mathbf{a}_{1}-y_{17} \, \mathbf{a}_{2}-z_{17} \, \mathbf{a}_{3} & = & \left(-x_{17}a-z_{17}c\cos\beta\right) \, \mathbf{\hat{x}}-y_{17}b \, \mathbf{\hat{y}}-z_{17}c\sin\beta \, \mathbf{\hat{z}} & \left(4e\right) & \mbox{H III} \\ 
\mathbf{B}_{68} & = & x_{17} \, \mathbf{a}_{1} + \left(\frac{1}{2} - y_{17}\right) \, \mathbf{a}_{2} + \left(\frac{1}{2} +z_{17}\right) \, \mathbf{a}_{3} & = & \left(\frac{1}{2}c\cos\beta +x_{17}a + z_{17}c\cos\beta\right) \, \mathbf{\hat{x}} + \left(\frac{1}{2} - y_{17}\right)b \, \mathbf{\hat{y}} + \left(\frac{1}{2} +z_{17}\right)c\sin\beta \, \mathbf{\hat{z}} & \left(4e\right) & \mbox{H III} \\ 
\mathbf{B}_{69} & = & x_{18} \, \mathbf{a}_{1} + y_{18} \, \mathbf{a}_{2} + z_{18} \, \mathbf{a}_{3} & = & \left(x_{18}a+z_{18}c\cos\beta\right) \, \mathbf{\hat{x}} + y_{18}b \, \mathbf{\hat{y}} + z_{18}c\sin\beta \, \mathbf{\hat{z}} & \left(4e\right) & \mbox{H IV} \\ 
\mathbf{B}_{70} & = & -x_{18} \, \mathbf{a}_{1} + \left(\frac{1}{2} +y_{18}\right) \, \mathbf{a}_{2} + \left(\frac{1}{2} - z_{18}\right) \, \mathbf{a}_{3} & = & \left(\frac{1}{2}c\cos\beta - x_{18}a - z_{18}c\cos\beta\right) \, \mathbf{\hat{x}} + \left(\frac{1}{2} +y_{18}\right)b \, \mathbf{\hat{y}} + \left(\frac{1}{2} - z_{18}\right)c\sin\beta \, \mathbf{\hat{z}} & \left(4e\right) & \mbox{H IV} \\ 
\mathbf{B}_{71} & = & -x_{18} \, \mathbf{a}_{1}-y_{18} \, \mathbf{a}_{2}-z_{18} \, \mathbf{a}_{3} & = & \left(-x_{18}a-z_{18}c\cos\beta\right) \, \mathbf{\hat{x}}-y_{18}b \, \mathbf{\hat{y}}-z_{18}c\sin\beta \, \mathbf{\hat{z}} & \left(4e\right) & \mbox{H IV} \\ 
\mathbf{B}_{72} & = & x_{18} \, \mathbf{a}_{1} + \left(\frac{1}{2} - y_{18}\right) \, \mathbf{a}_{2} + \left(\frac{1}{2} +z_{18}\right) \, \mathbf{a}_{3} & = & \left(\frac{1}{2}c\cos\beta +x_{18}a + z_{18}c\cos\beta\right) \, \mathbf{\hat{x}} + \left(\frac{1}{2} - y_{18}\right)b \, \mathbf{\hat{y}} + \left(\frac{1}{2} +z_{18}\right)c\sin\beta \, \mathbf{\hat{z}} & \left(4e\right) & \mbox{H IV} \\ 
\mathbf{B}_{73} & = & x_{19} \, \mathbf{a}_{1} + y_{19} \, \mathbf{a}_{2} + z_{19} \, \mathbf{a}_{3} & = & \left(x_{19}a+z_{19}c\cos\beta\right) \, \mathbf{\hat{x}} + y_{19}b \, \mathbf{\hat{y}} + z_{19}c\sin\beta \, \mathbf{\hat{z}} & \left(4e\right) & \mbox{H V} \\ 
\mathbf{B}_{74} & = & -x_{19} \, \mathbf{a}_{1} + \left(\frac{1}{2} +y_{19}\right) \, \mathbf{a}_{2} + \left(\frac{1}{2} - z_{19}\right) \, \mathbf{a}_{3} & = & \left(\frac{1}{2}c\cos\beta - x_{19}a - z_{19}c\cos\beta\right) \, \mathbf{\hat{x}} + \left(\frac{1}{2} +y_{19}\right)b \, \mathbf{\hat{y}} + \left(\frac{1}{2} - z_{19}\right)c\sin\beta \, \mathbf{\hat{z}} & \left(4e\right) & \mbox{H V} \\ 
\mathbf{B}_{75} & = & -x_{19} \, \mathbf{a}_{1}-y_{19} \, \mathbf{a}_{2}-z_{19} \, \mathbf{a}_{3} & = & \left(-x_{19}a-z_{19}c\cos\beta\right) \, \mathbf{\hat{x}}-y_{19}b \, \mathbf{\hat{y}}-z_{19}c\sin\beta \, \mathbf{\hat{z}} & \left(4e\right) & \mbox{H V} \\ 
\mathbf{B}_{76} & = & x_{19} \, \mathbf{a}_{1} + \left(\frac{1}{2} - y_{19}\right) \, \mathbf{a}_{2} + \left(\frac{1}{2} +z_{19}\right) \, \mathbf{a}_{3} & = & \left(\frac{1}{2}c\cos\beta +x_{19}a + z_{19}c\cos\beta\right) \, \mathbf{\hat{x}} + \left(\frac{1}{2} - y_{19}\right)b \, \mathbf{\hat{y}} + \left(\frac{1}{2} +z_{19}\right)c\sin\beta \, \mathbf{\hat{z}} & \left(4e\right) & \mbox{H V} \\ 
\mathbf{B}_{77} & = & x_{20} \, \mathbf{a}_{1} + y_{20} \, \mathbf{a}_{2} + z_{20} \, \mathbf{a}_{3} & = & \left(x_{20}a+z_{20}c\cos\beta\right) \, \mathbf{\hat{x}} + y_{20}b \, \mathbf{\hat{y}} + z_{20}c\sin\beta \, \mathbf{\hat{z}} & \left(4e\right) & \mbox{H VI} \\ 
\mathbf{B}_{78} & = & -x_{20} \, \mathbf{a}_{1} + \left(\frac{1}{2} +y_{20}\right) \, \mathbf{a}_{2} + \left(\frac{1}{2} - z_{20}\right) \, \mathbf{a}_{3} & = & \left(\frac{1}{2}c\cos\beta - x_{20}a - z_{20}c\cos\beta\right) \, \mathbf{\hat{x}} + \left(\frac{1}{2} +y_{20}\right)b \, \mathbf{\hat{y}} + \left(\frac{1}{2} - z_{20}\right)c\sin\beta \, \mathbf{\hat{z}} & \left(4e\right) & \mbox{H VI} \\ 
\mathbf{B}_{79} & = & -x_{20} \, \mathbf{a}_{1}-y_{20} \, \mathbf{a}_{2}-z_{20} \, \mathbf{a}_{3} & = & \left(-x_{20}a-z_{20}c\cos\beta\right) \, \mathbf{\hat{x}}-y_{20}b \, \mathbf{\hat{y}}-z_{20}c\sin\beta \, \mathbf{\hat{z}} & \left(4e\right) & \mbox{H VI} \\ 
\mathbf{B}_{80} & = & x_{20} \, \mathbf{a}_{1} + \left(\frac{1}{2} - y_{20}\right) \, \mathbf{a}_{2} + \left(\frac{1}{2} +z_{20}\right) \, \mathbf{a}_{3} & = & \left(\frac{1}{2}c\cos\beta +x_{20}a + z_{20}c\cos\beta\right) \, \mathbf{\hat{x}} + \left(\frac{1}{2} - y_{20}\right)b \, \mathbf{\hat{y}} + \left(\frac{1}{2} +z_{20}\right)c\sin\beta \, \mathbf{\hat{z}} & \left(4e\right) & \mbox{H VI} \\ 
\mathbf{B}_{81} & = & x_{21} \, \mathbf{a}_{1} + y_{21} \, \mathbf{a}_{2} + z_{21} \, \mathbf{a}_{3} & = & \left(x_{21}a+z_{21}c\cos\beta\right) \, \mathbf{\hat{x}} + y_{21}b \, \mathbf{\hat{y}} + z_{21}c\sin\beta \, \mathbf{\hat{z}} & \left(4e\right) & \mbox{H VII} \\ 
\mathbf{B}_{82} & = & -x_{21} \, \mathbf{a}_{1} + \left(\frac{1}{2} +y_{21}\right) \, \mathbf{a}_{2} + \left(\frac{1}{2} - z_{21}\right) \, \mathbf{a}_{3} & = & \left(\frac{1}{2}c\cos\beta - x_{21}a - z_{21}c\cos\beta\right) \, \mathbf{\hat{x}} + \left(\frac{1}{2} +y_{21}\right)b \, \mathbf{\hat{y}} + \left(\frac{1}{2} - z_{21}\right)c\sin\beta \, \mathbf{\hat{z}} & \left(4e\right) & \mbox{H VII} \\ 
\mathbf{B}_{83} & = & -x_{21} \, \mathbf{a}_{1}-y_{21} \, \mathbf{a}_{2}-z_{21} \, \mathbf{a}_{3} & = & \left(-x_{21}a-z_{21}c\cos\beta\right) \, \mathbf{\hat{x}}-y_{21}b \, \mathbf{\hat{y}}-z_{21}c\sin\beta \, \mathbf{\hat{z}} & \left(4e\right) & \mbox{H VII} \\ 
\mathbf{B}_{84} & = & x_{21} \, \mathbf{a}_{1} + \left(\frac{1}{2} - y_{21}\right) \, \mathbf{a}_{2} + \left(\frac{1}{2} +z_{21}\right) \, \mathbf{a}_{3} & = & \left(\frac{1}{2}c\cos\beta +x_{21}a + z_{21}c\cos\beta\right) \, \mathbf{\hat{x}} + \left(\frac{1}{2} - y_{21}\right)b \, \mathbf{\hat{y}} + \left(\frac{1}{2} +z_{21}\right)c\sin\beta \, \mathbf{\hat{z}} & \left(4e\right) & \mbox{H VII} \\ 
\mathbf{B}_{85} & = & x_{22} \, \mathbf{a}_{1} + y_{22} \, \mathbf{a}_{2} + z_{22} \, \mathbf{a}_{3} & = & \left(x_{22}a+z_{22}c\cos\beta\right) \, \mathbf{\hat{x}} + y_{22}b \, \mathbf{\hat{y}} + z_{22}c\sin\beta \, \mathbf{\hat{z}} & \left(4e\right) & \mbox{H VIII} \\ 
\mathbf{B}_{86} & = & -x_{22} \, \mathbf{a}_{1} + \left(\frac{1}{2} +y_{22}\right) \, \mathbf{a}_{2} + \left(\frac{1}{2} - z_{22}\right) \, \mathbf{a}_{3} & = & \left(\frac{1}{2}c\cos\beta - x_{22}a - z_{22}c\cos\beta\right) \, \mathbf{\hat{x}} + \left(\frac{1}{2} +y_{22}\right)b \, \mathbf{\hat{y}} + \left(\frac{1}{2} - z_{22}\right)c\sin\beta \, \mathbf{\hat{z}} & \left(4e\right) & \mbox{H VIII} \\ 
\mathbf{B}_{87} & = & -x_{22} \, \mathbf{a}_{1}-y_{22} \, \mathbf{a}_{2}-z_{22} \, \mathbf{a}_{3} & = & \left(-x_{22}a-z_{22}c\cos\beta\right) \, \mathbf{\hat{x}}-y_{22}b \, \mathbf{\hat{y}}-z_{22}c\sin\beta \, \mathbf{\hat{z}} & \left(4e\right) & \mbox{H VIII} \\ 
\mathbf{B}_{88} & = & x_{22} \, \mathbf{a}_{1} + \left(\frac{1}{2} - y_{22}\right) \, \mathbf{a}_{2} + \left(\frac{1}{2} +z_{22}\right) \, \mathbf{a}_{3} & = & \left(\frac{1}{2}c\cos\beta +x_{22}a + z_{22}c\cos\beta\right) \, \mathbf{\hat{x}} + \left(\frac{1}{2} - y_{22}\right)b \, \mathbf{\hat{y}} + \left(\frac{1}{2} +z_{22}\right)c\sin\beta \, \mathbf{\hat{z}} & \left(4e\right) & \mbox{H VIII} \\ 
\mathbf{B}_{89} & = & x_{23} \, \mathbf{a}_{1} + y_{23} \, \mathbf{a}_{2} + z_{23} \, \mathbf{a}_{3} & = & \left(x_{23}a+z_{23}c\cos\beta\right) \, \mathbf{\hat{x}} + y_{23}b \, \mathbf{\hat{y}} + z_{23}c\sin\beta \, \mathbf{\hat{z}} & \left(4e\right) & \mbox{H IX} \\ 
\mathbf{B}_{90} & = & -x_{23} \, \mathbf{a}_{1} + \left(\frac{1}{2} +y_{23}\right) \, \mathbf{a}_{2} + \left(\frac{1}{2} - z_{23}\right) \, \mathbf{a}_{3} & = & \left(\frac{1}{2}c\cos\beta - x_{23}a - z_{23}c\cos\beta\right) \, \mathbf{\hat{x}} + \left(\frac{1}{2} +y_{23}\right)b \, \mathbf{\hat{y}} + \left(\frac{1}{2} - z_{23}\right)c\sin\beta \, \mathbf{\hat{z}} & \left(4e\right) & \mbox{H IX} \\ 
\mathbf{B}_{91} & = & -x_{23} \, \mathbf{a}_{1}-y_{23} \, \mathbf{a}_{2}-z_{23} \, \mathbf{a}_{3} & = & \left(-x_{23}a-z_{23}c\cos\beta\right) \, \mathbf{\hat{x}}-y_{23}b \, \mathbf{\hat{y}}-z_{23}c\sin\beta \, \mathbf{\hat{z}} & \left(4e\right) & \mbox{H IX} \\ 
\mathbf{B}_{92} & = & x_{23} \, \mathbf{a}_{1} + \left(\frac{1}{2} - y_{23}\right) \, \mathbf{a}_{2} + \left(\frac{1}{2} +z_{23}\right) \, \mathbf{a}_{3} & = & \left(\frac{1}{2}c\cos\beta +x_{23}a + z_{23}c\cos\beta\right) \, \mathbf{\hat{x}} + \left(\frac{1}{2} - y_{23}\right)b \, \mathbf{\hat{y}} + \left(\frac{1}{2} +z_{23}\right)c\sin\beta \, \mathbf{\hat{z}} & \left(4e\right) & \mbox{H IX} \\ 
\mathbf{B}_{93} & = & x_{24} \, \mathbf{a}_{1} + y_{24} \, \mathbf{a}_{2} + z_{24} \, \mathbf{a}_{3} & = & \left(x_{24}a+z_{24}c\cos\beta\right) \, \mathbf{\hat{x}} + y_{24}b \, \mathbf{\hat{y}} + z_{24}c\sin\beta \, \mathbf{\hat{z}} & \left(4e\right) & \mbox{H X} \\ 
\mathbf{B}_{94} & = & -x_{24} \, \mathbf{a}_{1} + \left(\frac{1}{2} +y_{24}\right) \, \mathbf{a}_{2} + \left(\frac{1}{2} - z_{24}\right) \, \mathbf{a}_{3} & = & \left(\frac{1}{2}c\cos\beta - x_{24}a - z_{24}c\cos\beta\right) \, \mathbf{\hat{x}} + \left(\frac{1}{2} +y_{24}\right)b \, \mathbf{\hat{y}} + \left(\frac{1}{2} - z_{24}\right)c\sin\beta \, \mathbf{\hat{z}} & \left(4e\right) & \mbox{H X} \\ 
\mathbf{B}_{95} & = & -x_{24} \, \mathbf{a}_{1}-y_{24} \, \mathbf{a}_{2}-z_{24} \, \mathbf{a}_{3} & = & \left(-x_{24}a-z_{24}c\cos\beta\right) \, \mathbf{\hat{x}}-y_{24}b \, \mathbf{\hat{y}}-z_{24}c\sin\beta \, \mathbf{\hat{z}} & \left(4e\right) & \mbox{H X} \\ 
\mathbf{B}_{96} & = & x_{24} \, \mathbf{a}_{1} + \left(\frac{1}{2} - y_{24}\right) \, \mathbf{a}_{2} + \left(\frac{1}{2} +z_{24}\right) \, \mathbf{a}_{3} & = & \left(\frac{1}{2}c\cos\beta +x_{24}a + z_{24}c\cos\beta\right) \, \mathbf{\hat{x}} + \left(\frac{1}{2} - y_{24}\right)b \, \mathbf{\hat{y}} + \left(\frac{1}{2} +z_{24}\right)c\sin\beta \, \mathbf{\hat{z}} & \left(4e\right) & \mbox{H X} \\ 
\mathbf{B}_{97} & = & x_{25} \, \mathbf{a}_{1} + y_{25} \, \mathbf{a}_{2} + z_{25} \, \mathbf{a}_{3} & = & \left(x_{25}a+z_{25}c\cos\beta\right) \, \mathbf{\hat{x}} + y_{25}b \, \mathbf{\hat{y}} + z_{25}c\sin\beta \, \mathbf{\hat{z}} & \left(4e\right) & \mbox{H XI} \\ 
\mathbf{B}_{98} & = & -x_{25} \, \mathbf{a}_{1} + \left(\frac{1}{2} +y_{25}\right) \, \mathbf{a}_{2} + \left(\frac{1}{2} - z_{25}\right) \, \mathbf{a}_{3} & = & \left(\frac{1}{2}c\cos\beta - x_{25}a - z_{25}c\cos\beta\right) \, \mathbf{\hat{x}} + \left(\frac{1}{2} +y_{25}\right)b \, \mathbf{\hat{y}} + \left(\frac{1}{2} - z_{25}\right)c\sin\beta \, \mathbf{\hat{z}} & \left(4e\right) & \mbox{H XI} \\ 
\mathbf{B}_{99} & = & -x_{25} \, \mathbf{a}_{1}-y_{25} \, \mathbf{a}_{2}-z_{25} \, \mathbf{a}_{3} & = & \left(-x_{25}a-z_{25}c\cos\beta\right) \, \mathbf{\hat{x}}-y_{25}b \, \mathbf{\hat{y}}-z_{25}c\sin\beta \, \mathbf{\hat{z}} & \left(4e\right) & \mbox{H XI} \\ 
\mathbf{B}_{100} & = & x_{25} \, \mathbf{a}_{1} + \left(\frac{1}{2} - y_{25}\right) \, \mathbf{a}_{2} + \left(\frac{1}{2} +z_{25}\right) \, \mathbf{a}_{3} & = & \left(\frac{1}{2}c\cos\beta +x_{25}a + z_{25}c\cos\beta\right) \, \mathbf{\hat{x}} + \left(\frac{1}{2} - y_{25}\right)b \, \mathbf{\hat{y}} + \left(\frac{1}{2} +z_{25}\right)c\sin\beta \, \mathbf{\hat{z}} & \left(4e\right) & \mbox{H XI} \\ 
\mathbf{B}_{101} & = & x_{26} \, \mathbf{a}_{1} + y_{26} \, \mathbf{a}_{2} + z_{26} \, \mathbf{a}_{3} & = & \left(x_{26}a+z_{26}c\cos\beta\right) \, \mathbf{\hat{x}} + y_{26}b \, \mathbf{\hat{y}} + z_{26}c\sin\beta \, \mathbf{\hat{z}} & \left(4e\right) & \mbox{H XII} \\ 
\mathbf{B}_{102} & = & -x_{26} \, \mathbf{a}_{1} + \left(\frac{1}{2} +y_{26}\right) \, \mathbf{a}_{2} + \left(\frac{1}{2} - z_{26}\right) \, \mathbf{a}_{3} & = & \left(\frac{1}{2}c\cos\beta - x_{26}a - z_{26}c\cos\beta\right) \, \mathbf{\hat{x}} + \left(\frac{1}{2} +y_{26}\right)b \, \mathbf{\hat{y}} + \left(\frac{1}{2} - z_{26}\right)c\sin\beta \, \mathbf{\hat{z}} & \left(4e\right) & \mbox{H XII} \\ 
\mathbf{B}_{103} & = & -x_{26} \, \mathbf{a}_{1}-y_{26} \, \mathbf{a}_{2}-z_{26} \, \mathbf{a}_{3} & = & \left(-x_{26}a-z_{26}c\cos\beta\right) \, \mathbf{\hat{x}}-y_{26}b \, \mathbf{\hat{y}}-z_{26}c\sin\beta \, \mathbf{\hat{z}} & \left(4e\right) & \mbox{H XII} \\ 
\mathbf{B}_{104} & = & x_{26} \, \mathbf{a}_{1} + \left(\frac{1}{2} - y_{26}\right) \, \mathbf{a}_{2} + \left(\frac{1}{2} +z_{26}\right) \, \mathbf{a}_{3} & = & \left(\frac{1}{2}c\cos\beta +x_{26}a + z_{26}c\cos\beta\right) \, \mathbf{\hat{x}} + \left(\frac{1}{2} - y_{26}\right)b \, \mathbf{\hat{y}} + \left(\frac{1}{2} +z_{26}\right)c\sin\beta \, \mathbf{\hat{z}} & \left(4e\right) & \mbox{H XII} \\ 
\mathbf{B}_{105} & = & x_{27} \, \mathbf{a}_{1} + y_{27} \, \mathbf{a}_{2} + z_{27} \, \mathbf{a}_{3} & = & \left(x_{27}a+z_{27}c\cos\beta\right) \, \mathbf{\hat{x}} + y_{27}b \, \mathbf{\hat{y}} + z_{27}c\sin\beta \, \mathbf{\hat{z}} & \left(4e\right) & \mbox{H XIII} \\ 
\mathbf{B}_{106} & = & -x_{27} \, \mathbf{a}_{1} + \left(\frac{1}{2} +y_{27}\right) \, \mathbf{a}_{2} + \left(\frac{1}{2} - z_{27}\right) \, \mathbf{a}_{3} & = & \left(\frac{1}{2}c\cos\beta - x_{27}a - z_{27}c\cos\beta\right) \, \mathbf{\hat{x}} + \left(\frac{1}{2} +y_{27}\right)b \, \mathbf{\hat{y}} + \left(\frac{1}{2} - z_{27}\right)c\sin\beta \, \mathbf{\hat{z}} & \left(4e\right) & \mbox{H XIII} \\ 
\mathbf{B}_{107} & = & -x_{27} \, \mathbf{a}_{1}-y_{27} \, \mathbf{a}_{2}-z_{27} \, \mathbf{a}_{3} & = & \left(-x_{27}a-z_{27}c\cos\beta\right) \, \mathbf{\hat{x}}-y_{27}b \, \mathbf{\hat{y}}-z_{27}c\sin\beta \, \mathbf{\hat{z}} & \left(4e\right) & \mbox{H XIII} \\ 
\mathbf{B}_{108} & = & x_{27} \, \mathbf{a}_{1} + \left(\frac{1}{2} - y_{27}\right) \, \mathbf{a}_{2} + \left(\frac{1}{2} +z_{27}\right) \, \mathbf{a}_{3} & = & \left(\frac{1}{2}c\cos\beta +x_{27}a + z_{27}c\cos\beta\right) \, \mathbf{\hat{x}} + \left(\frac{1}{2} - y_{27}\right)b \, \mathbf{\hat{y}} + \left(\frac{1}{2} +z_{27}\right)c\sin\beta \, \mathbf{\hat{z}} & \left(4e\right) & \mbox{H XIII} \\ 
\mathbf{B}_{109} & = & x_{28} \, \mathbf{a}_{1} + y_{28} \, \mathbf{a}_{2} + z_{28} \, \mathbf{a}_{3} & = & \left(x_{28}a+z_{28}c\cos\beta\right) \, \mathbf{\hat{x}} + y_{28}b \, \mathbf{\hat{y}} + z_{28}c\sin\beta \, \mathbf{\hat{z}} & \left(4e\right) & \mbox{H XIV} \\ 
\mathbf{B}_{110} & = & -x_{28} \, \mathbf{a}_{1} + \left(\frac{1}{2} +y_{28}\right) \, \mathbf{a}_{2} + \left(\frac{1}{2} - z_{28}\right) \, \mathbf{a}_{3} & = & \left(\frac{1}{2}c\cos\beta - x_{28}a - z_{28}c\cos\beta\right) \, \mathbf{\hat{x}} + \left(\frac{1}{2} +y_{28}\right)b \, \mathbf{\hat{y}} + \left(\frac{1}{2} - z_{28}\right)c\sin\beta \, \mathbf{\hat{z}} & \left(4e\right) & \mbox{H XIV} \\ 
\mathbf{B}_{111} & = & -x_{28} \, \mathbf{a}_{1}-y_{28} \, \mathbf{a}_{2}-z_{28} \, \mathbf{a}_{3} & = & \left(-x_{28}a-z_{28}c\cos\beta\right) \, \mathbf{\hat{x}}-y_{28}b \, \mathbf{\hat{y}}-z_{28}c\sin\beta \, \mathbf{\hat{z}} & \left(4e\right) & \mbox{H XIV} \\ 
\mathbf{B}_{112} & = & x_{28} \, \mathbf{a}_{1} + \left(\frac{1}{2} - y_{28}\right) \, \mathbf{a}_{2} + \left(\frac{1}{2} +z_{28}\right) \, \mathbf{a}_{3} & = & \left(\frac{1}{2}c\cos\beta +x_{28}a + z_{28}c\cos\beta\right) \, \mathbf{\hat{x}} + \left(\frac{1}{2} - y_{28}\right)b \, \mathbf{\hat{y}} + \left(\frac{1}{2} +z_{28}\right)c\sin\beta \, \mathbf{\hat{z}} & \left(4e\right) & \mbox{H XIV} \\ 
\mathbf{B}_{113} & = & x_{29} \, \mathbf{a}_{1} + y_{29} \, \mathbf{a}_{2} + z_{29} \, \mathbf{a}_{3} & = & \left(x_{29}a+z_{29}c\cos\beta\right) \, \mathbf{\hat{x}} + y_{29}b \, \mathbf{\hat{y}} + z_{29}c\sin\beta \, \mathbf{\hat{z}} & \left(4e\right) & \mbox{H XV} \\ 
\mathbf{B}_{114} & = & -x_{29} \, \mathbf{a}_{1} + \left(\frac{1}{2} +y_{29}\right) \, \mathbf{a}_{2} + \left(\frac{1}{2} - z_{29}\right) \, \mathbf{a}_{3} & = & \left(\frac{1}{2}c\cos\beta - x_{29}a - z_{29}c\cos\beta\right) \, \mathbf{\hat{x}} + \left(\frac{1}{2} +y_{29}\right)b \, \mathbf{\hat{y}} + \left(\frac{1}{2} - z_{29}\right)c\sin\beta \, \mathbf{\hat{z}} & \left(4e\right) & \mbox{H XV} \\ 
\mathbf{B}_{115} & = & -x_{29} \, \mathbf{a}_{1}-y_{29} \, \mathbf{a}_{2}-z_{29} \, \mathbf{a}_{3} & = & \left(-x_{29}a-z_{29}c\cos\beta\right) \, \mathbf{\hat{x}}-y_{29}b \, \mathbf{\hat{y}}-z_{29}c\sin\beta \, \mathbf{\hat{z}} & \left(4e\right) & \mbox{H XV} \\ 
\mathbf{B}_{116} & = & x_{29} \, \mathbf{a}_{1} + \left(\frac{1}{2} - y_{29}\right) \, \mathbf{a}_{2} + \left(\frac{1}{2} +z_{29}\right) \, \mathbf{a}_{3} & = & \left(\frac{1}{2}c\cos\beta +x_{29}a + z_{29}c\cos\beta\right) \, \mathbf{\hat{x}} + \left(\frac{1}{2} - y_{29}\right)b \, \mathbf{\hat{y}} + \left(\frac{1}{2} +z_{29}\right)c\sin\beta \, \mathbf{\hat{z}} & \left(4e\right) & \mbox{H XV} \\ 
\mathbf{B}_{117} & = & x_{30} \, \mathbf{a}_{1} + y_{30} \, \mathbf{a}_{2} + z_{30} \, \mathbf{a}_{3} & = & \left(x_{30}a+z_{30}c\cos\beta\right) \, \mathbf{\hat{x}} + y_{30}b \, \mathbf{\hat{y}} + z_{30}c\sin\beta \, \mathbf{\hat{z}} & \left(4e\right) & \mbox{H XVI} \\ 
\mathbf{B}_{118} & = & -x_{30} \, \mathbf{a}_{1} + \left(\frac{1}{2} +y_{30}\right) \, \mathbf{a}_{2} + \left(\frac{1}{2} - z_{30}\right) \, \mathbf{a}_{3} & = & \left(\frac{1}{2}c\cos\beta - x_{30}a - z_{30}c\cos\beta\right) \, \mathbf{\hat{x}} + \left(\frac{1}{2} +y_{30}\right)b \, \mathbf{\hat{y}} + \left(\frac{1}{2} - z_{30}\right)c\sin\beta \, \mathbf{\hat{z}} & \left(4e\right) & \mbox{H XVI} \\ 
\mathbf{B}_{119} & = & -x_{30} \, \mathbf{a}_{1}-y_{30} \, \mathbf{a}_{2}-z_{30} \, \mathbf{a}_{3} & = & \left(-x_{30}a-z_{30}c\cos\beta\right) \, \mathbf{\hat{x}}-y_{30}b \, \mathbf{\hat{y}}-z_{30}c\sin\beta \, \mathbf{\hat{z}} & \left(4e\right) & \mbox{H XVI} \\ 
\mathbf{B}_{120} & = & x_{30} \, \mathbf{a}_{1} + \left(\frac{1}{2} - y_{30}\right) \, \mathbf{a}_{2} + \left(\frac{1}{2} +z_{30}\right) \, \mathbf{a}_{3} & = & \left(\frac{1}{2}c\cos\beta +x_{30}a + z_{30}c\cos\beta\right) \, \mathbf{\hat{x}} + \left(\frac{1}{2} - y_{30}\right)b \, \mathbf{\hat{y}} + \left(\frac{1}{2} +z_{30}\right)c\sin\beta \, \mathbf{\hat{z}} & \left(4e\right) & \mbox{H XVI} \\ 
\end{longtabu}
\renewcommand{\arraystretch}{1.0}
\noindent \hrulefill
\\
\textbf{References:}
\vspace*{-0.25cm}
\begin{flushleft}
  - \bibentry{Nayak_2010}. \\
\end{flushleft}
\textbf{Found in:}
\vspace*{-0.25cm}
\begin{flushleft}
  - \bibentry{CSD}. \\
\end{flushleft}
\noindent \hrulefill
\\
\textbf{Geometry files:}
\\
\noindent  - CIF: pp. {\hyperref[A7B8_mP120_14_14e_16e_cif]{\pageref{A7B8_mP120_14_14e_16e_cif}}} \\
\noindent  - POSCAR: pp. {\hyperref[A7B8_mP120_14_14e_16e_poscar]{\pageref{A7B8_mP120_14_14e_16e_poscar}}} \\
\onecolumn
{\phantomsection\label{AB3_mC16_15_e_cf}}
\subsection*{\huge \textbf{{\normalfont H$_{3}$Cl (50~GPa) Structure: AB3\_mC16\_15\_e\_cf}}}
\noindent \hrulefill
\vspace*{0.25cm}
\begin{figure}[htp]
  \centering
  \vspace{-1em}
  {\includegraphics[width=1\textwidth]{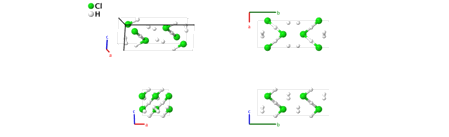}}
\end{figure}
\vspace*{-0.5cm}
\renewcommand{\arraystretch}{1.5}
\begin{equation*}
  \begin{array}{>{$\hspace{-0.15cm}}l<{$}>{$}p{0.5cm}<{$}>{$}p{18.5cm}<{$}}
    \mbox{\large \textbf{Prototype}} &\colon & \ce{H3Cl} \\
    \mbox{\large \textbf{\AFLOW\ prototype label}} &\colon & \mbox{AB3\_mC16\_15\_e\_cf} \\
    \mbox{\large \textbf{\textit{Strukturbericht} designation}} &\colon & \mbox{None} \\
    \mbox{\large \textbf{Pearson symbol}} &\colon & \mbox{mC16} \\
    \mbox{\large \textbf{Space group number}} &\colon & 15 \\
    \mbox{\large \textbf{Space group symbol}} &\colon & C2/c \\
    \mbox{\large \textbf{\AFLOW\ prototype command}} &\colon &  \texttt{aflow} \,  \, \texttt{-{}-proto=AB3\_mC16\_15\_e\_cf } \, \newline \texttt{-{}-params=}{a,b/a,c/a,\beta,y_{2},x_{3},y_{3},z_{3} }
  \end{array}
\end{equation*}
\renewcommand{\arraystretch}{1.0}

\vspace*{-0.25cm}
\noindent \hrulefill
\begin{itemize}
  \item{This structure was found via first-principles calculations.  The data
presented here was computed at a pressure of 50~GPa.
}
\end{itemize}

\noindent \parbox{1 \linewidth}{
\noindent \hrulefill
\\
\textbf{Base-centered Monoclinic primitive vectors:} \\
\vspace*{-0.25cm}
\begin{tabular}{cc}
  \begin{tabular}{c}
    \parbox{0.6 \linewidth}{
      \renewcommand{\arraystretch}{1.5}
      \begin{equation*}
        \centering
        \begin{array}{ccc}
              \mathbf{a}_1 & = & \frac12 \, a \, \mathbf{\hat{x}} - \frac12 \, b \, \mathbf{\hat{y}} \\
    \mathbf{a}_2 & = & \frac12 \, a \, \mathbf{\hat{x}} + \frac12 \, b \, \mathbf{\hat{y}} \\
    \mathbf{a}_3 & = & c \cos\beta \, \mathbf{\hat{x}} + c \sin\beta \, \mathbf{\hat{z}} \\

        \end{array}
      \end{equation*}
    }
    \renewcommand{\arraystretch}{1.0}
  \end{tabular}
  \begin{tabular}{c}
    \includegraphics[width=0.3\linewidth]{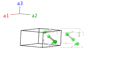} \\
  \end{tabular}
\end{tabular}

}
\vspace*{-0.25cm}

\noindent \hrulefill
\\
\textbf{Basis vectors:}
\vspace*{-0.25cm}
\renewcommand{\arraystretch}{1.5}
\begin{longtabu} to \textwidth{>{\centering $}X[-1,c,c]<{$}>{\centering $}X[-1,c,c]<{$}>{\centering $}X[-1,c,c]<{$}>{\centering $}X[-1,c,c]<{$}>{\centering $}X[-1,c,c]<{$}>{\centering $}X[-1,c,c]<{$}>{\centering $}X[-1,c,c]<{$}}
  & & \mbox{Lattice Coordinates} & & \mbox{Cartesian Coordinates} &\mbox{Wyckoff Position} & \mbox{Atom Type} \\  
  \mathbf{B}_{1} & = & \frac{1}{2} \, \mathbf{a}_{2} & = & \frac{1}{4}a \, \mathbf{\hat{x}} + \frac{1}{4}b \, \mathbf{\hat{y}} & \left(4c\right) & \mbox{H I} \\ 
\mathbf{B}_{2} & = & \frac{1}{2} \, \mathbf{a}_{1} + \frac{1}{2} \, \mathbf{a}_{3} & = & \left(\frac{1}{4}a+\frac{1}{2}c\cos\beta\right) \, \mathbf{\hat{x}}- \frac{1}{4}b  \, \mathbf{\hat{y}} + \frac{1}{2}c\sin\beta \, \mathbf{\hat{z}} & \left(4c\right) & \mbox{H I} \\ 
\mathbf{B}_{3} & = & -y_{2} \, \mathbf{a}_{1} + y_{2} \, \mathbf{a}_{2} + \frac{1}{4} \, \mathbf{a}_{3} & = & \frac{1}{4}c\cos\beta \, \mathbf{\hat{x}} + y_{2}b \, \mathbf{\hat{y}} + \frac{1}{4}c\sin\beta \, \mathbf{\hat{z}} & \left(4e\right) & \mbox{Cl} \\ 
\mathbf{B}_{4} & = & y_{2} \, \mathbf{a}_{1}-y_{2} \, \mathbf{a}_{2} + \frac{3}{4} \, \mathbf{a}_{3} & = & \frac{3}{4}c\cos\beta \, \mathbf{\hat{x}}-y_{2}b \, \mathbf{\hat{y}} + \frac{3}{4}c\sin\beta \, \mathbf{\hat{z}} & \left(4e\right) & \mbox{Cl} \\ 
\mathbf{B}_{5} & = & \left(x_{3}-y_{3}\right) \, \mathbf{a}_{1} + \left(x_{3}+y_{3}\right) \, \mathbf{a}_{2} + z_{3} \, \mathbf{a}_{3} & = & \left(x_{3}a+z_{3}c\cos\beta\right) \, \mathbf{\hat{x}} + y_{3}b \, \mathbf{\hat{y}} + z_{3}c\sin\beta \, \mathbf{\hat{z}} & \left(8f\right) & \mbox{H II} \\ 
\mathbf{B}_{6} & = & \left(-x_{3}-y_{3}\right) \, \mathbf{a}_{1} + \left(-x_{3}+y_{3}\right) \, \mathbf{a}_{2} + \left(\frac{1}{2} - z_{3}\right) \, \mathbf{a}_{3} & = & \left(\frac{1}{2}c\cos\beta - x_{3}a - z_{3}c\cos\beta\right) \, \mathbf{\hat{x}} + y_{3}b \, \mathbf{\hat{y}} + \left(\frac{1}{2} - z_{3}\right)c\sin\beta \, \mathbf{\hat{z}} & \left(8f\right) & \mbox{H II} \\ 
\mathbf{B}_{7} & = & \left(-x_{3}+y_{3}\right) \, \mathbf{a}_{1} + \left(-x_{3}-y_{3}\right) \, \mathbf{a}_{2}-z_{3} \, \mathbf{a}_{3} & = & \left(-x_{3}a-z_{3}c\cos\beta\right) \, \mathbf{\hat{x}}-y_{3}b \, \mathbf{\hat{y}}-z_{3}c\sin\beta \, \mathbf{\hat{z}} & \left(8f\right) & \mbox{H II} \\ 
\mathbf{B}_{8} & = & \left(x_{3}+y_{3}\right) \, \mathbf{a}_{1} + \left(x_{3}-y_{3}\right) \, \mathbf{a}_{2} + \left(\frac{1}{2} +z_{3}\right) \, \mathbf{a}_{3} & = & \left(\frac{1}{2}c\cos\beta +x_{3}a + z_{3}c\cos\beta\right) \, \mathbf{\hat{x}}-y_{3}b \, \mathbf{\hat{y}} + \left(\frac{1}{2} +z_{3}\right)c\sin\beta \, \mathbf{\hat{z}} & \left(8f\right) & \mbox{H II} \\ 
\end{longtabu}
\renewcommand{\arraystretch}{1.0}
\noindent \hrulefill
\\
\textbf{References:}
\vspace*{-0.25cm}
\begin{flushleft}
  - \bibentry{Duan_2015}. \\
\end{flushleft}
\noindent \hrulefill
\\
\textbf{Geometry files:}
\\
\noindent  - CIF: pp. {\hyperref[AB3_mC16_15_e_cf_cif]{\pageref{AB3_mC16_15_e_cf_cif}}} \\
\noindent  - POSCAR: pp. {\hyperref[AB3_mC16_15_e_cf_poscar]{\pageref{AB3_mC16_15_e_cf_poscar}}} \\
\onecolumn
{\phantomsection\label{A_mC24_15_2e2f}}
\subsection*{\huge \textbf{{\normalfont H-III (300~GPa) Structure: A\_mC24\_15\_2e2f}}}
\noindent \hrulefill
\vspace*{0.25cm}
\begin{figure}[htp]
  \centering
  \vspace{-1em}
  {\includegraphics[width=1\textwidth]{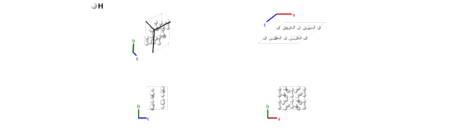}}
\end{figure}
\vspace*{-0.5cm}
\renewcommand{\arraystretch}{1.5}
\begin{equation*}
  \begin{array}{>{$\hspace{-0.15cm}}l<{$}>{$}p{0.5cm}<{$}>{$}p{18.5cm}<{$}}
    \mbox{\large \textbf{Prototype}} &\colon & \ce{H} \\
    \mbox{\large \textbf{\AFLOW\ prototype label}} &\colon & \mbox{A\_mC24\_15\_2e2f} \\
    \mbox{\large \textbf{\textit{Strukturbericht} designation}} &\colon & \mbox{None} \\
    \mbox{\large \textbf{Pearson symbol}} &\colon & \mbox{mC24} \\
    \mbox{\large \textbf{Space group number}} &\colon & 15 \\
    \mbox{\large \textbf{Space group symbol}} &\colon & C2/c \\
    \mbox{\large \textbf{\AFLOW\ prototype command}} &\colon &  \texttt{aflow} \,  \, \texttt{-{}-proto=A\_mC24\_15\_2e2f } \, \newline \texttt{-{}-params=}{a,b/a,c/a,\beta,y_{1},y_{2},x_{3},y_{3},z_{3},x_{4},y_{4},z_{4} }
  \end{array}
\end{equation*}
\renewcommand{\arraystretch}{1.0}

\vspace*{-0.25cm}
\noindent \hrulefill
\begin{itemize}
  \item{This structure was determined by density functional simulations. The
authors claim it is in good agreement with experimental data for
H-III, and is the lowest energy structure at pressures from
approximately 100-250~GPa, including zero-point motion.
The data presented here was computed at 300~GPa.
If we change our description of the unit cell so that
\[ \mathbf{a}_{3} \rightarrow \mathbf{a}_{1} + \mathbf{a}_{2} +
\mathbf{a}_{3}\]
then all of the primitive vectors for the base-centered orthorhombic
structure have approximately equal lengths, and the angles between
them are approximately $60^\circ$.  This structure is very close to
exhibiting a face-centered cubic lattice.
}
\end{itemize}

\noindent \parbox{1 \linewidth}{
\noindent \hrulefill
\\
\textbf{Base-centered Monoclinic primitive vectors:} \\
\vspace*{-0.25cm}
\begin{tabular}{cc}
  \begin{tabular}{c}
    \parbox{0.6 \linewidth}{
      \renewcommand{\arraystretch}{1.5}
      \begin{equation*}
        \centering
        \begin{array}{ccc}
              \mathbf{a}_1 & = & \frac12 \, a \, \mathbf{\hat{x}} - \frac12 \, b \, \mathbf{\hat{y}} \\
    \mathbf{a}_2 & = & \frac12 \, a \, \mathbf{\hat{x}} + \frac12 \, b \, \mathbf{\hat{y}} \\
    \mathbf{a}_3 & = & c \cos\beta \, \mathbf{\hat{x}} + c \sin\beta \, \mathbf{\hat{z}} \\

        \end{array}
      \end{equation*}
    }
    \renewcommand{\arraystretch}{1.0}
  \end{tabular}
  \begin{tabular}{c}
    \includegraphics[width=0.3\linewidth]{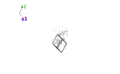} \\
  \end{tabular}
\end{tabular}

}
\vspace*{-0.25cm}

\noindent \hrulefill
\\
\textbf{Basis vectors:}
\vspace*{-0.25cm}
\renewcommand{\arraystretch}{1.5}
\begin{longtabu} to \textwidth{>{\centering $}X[-1,c,c]<{$}>{\centering $}X[-1,c,c]<{$}>{\centering $}X[-1,c,c]<{$}>{\centering $}X[-1,c,c]<{$}>{\centering $}X[-1,c,c]<{$}>{\centering $}X[-1,c,c]<{$}>{\centering $}X[-1,c,c]<{$}}
  & & \mbox{Lattice Coordinates} & & \mbox{Cartesian Coordinates} &\mbox{Wyckoff Position} & \mbox{Atom Type} \\  
  \mathbf{B}_{1} & = & -y_{1} \, \mathbf{a}_{1} + y_{1} \, \mathbf{a}_{2} + \frac{1}{4} \, \mathbf{a}_{3} & = & \frac{1}{4}c\cos\beta \, \mathbf{\hat{x}} + y_{1}b \, \mathbf{\hat{y}} + \frac{1}{4}c\sin\beta \, \mathbf{\hat{z}} & \left(4e\right) & \mbox{H I} \\ 
\mathbf{B}_{2} & = & y_{1} \, \mathbf{a}_{1}-y_{1} \, \mathbf{a}_{2} + \frac{3}{4} \, \mathbf{a}_{3} & = & \frac{3}{4}c\cos\beta \, \mathbf{\hat{x}}-y_{1}b \, \mathbf{\hat{y}} + \frac{3}{4}c\sin\beta \, \mathbf{\hat{z}} & \left(4e\right) & \mbox{H I} \\ 
\mathbf{B}_{3} & = & -y_{2} \, \mathbf{a}_{1} + y_{2} \, \mathbf{a}_{2} + \frac{1}{4} \, \mathbf{a}_{3} & = & \frac{1}{4}c\cos\beta \, \mathbf{\hat{x}} + y_{2}b \, \mathbf{\hat{y}} + \frac{1}{4}c\sin\beta \, \mathbf{\hat{z}} & \left(4e\right) & \mbox{H II} \\ 
\mathbf{B}_{4} & = & y_{2} \, \mathbf{a}_{1}-y_{2} \, \mathbf{a}_{2} + \frac{3}{4} \, \mathbf{a}_{3} & = & \frac{3}{4}c\cos\beta \, \mathbf{\hat{x}}-y_{2}b \, \mathbf{\hat{y}} + \frac{3}{4}c\sin\beta \, \mathbf{\hat{z}} & \left(4e\right) & \mbox{H II} \\ 
\mathbf{B}_{5} & = & \left(x_{3}-y_{3}\right) \, \mathbf{a}_{1} + \left(x_{3}+y_{3}\right) \, \mathbf{a}_{2} + z_{3} \, \mathbf{a}_{3} & = & \left(x_{3}a+z_{3}c\cos\beta\right) \, \mathbf{\hat{x}} + y_{3}b \, \mathbf{\hat{y}} + z_{3}c\sin\beta \, \mathbf{\hat{z}} & \left(8f\right) & \mbox{H III} \\ 
\mathbf{B}_{6} & = & \left(-x_{3}-y_{3}\right) \, \mathbf{a}_{1} + \left(-x_{3}+y_{3}\right) \, \mathbf{a}_{2} + \left(\frac{1}{2} - z_{3}\right) \, \mathbf{a}_{3} & = & \left(\frac{1}{2}c\cos\beta - x_{3}a - z_{3}c\cos\beta\right) \, \mathbf{\hat{x}} + y_{3}b \, \mathbf{\hat{y}} + \left(\frac{1}{2} - z_{3}\right)c\sin\beta \, \mathbf{\hat{z}} & \left(8f\right) & \mbox{H III} \\ 
\mathbf{B}_{7} & = & \left(-x_{3}+y_{3}\right) \, \mathbf{a}_{1} + \left(-x_{3}-y_{3}\right) \, \mathbf{a}_{2}-z_{3} \, \mathbf{a}_{3} & = & \left(-x_{3}a-z_{3}c\cos\beta\right) \, \mathbf{\hat{x}}-y_{3}b \, \mathbf{\hat{y}}-z_{3}c\sin\beta \, \mathbf{\hat{z}} & \left(8f\right) & \mbox{H III} \\ 
\mathbf{B}_{8} & = & \left(x_{3}+y_{3}\right) \, \mathbf{a}_{1} + \left(x_{3}-y_{3}\right) \, \mathbf{a}_{2} + \left(\frac{1}{2} +z_{3}\right) \, \mathbf{a}_{3} & = & \left(\frac{1}{2}c\cos\beta +x_{3}a + z_{3}c\cos\beta\right) \, \mathbf{\hat{x}}-y_{3}b \, \mathbf{\hat{y}} + \left(\frac{1}{2} +z_{3}\right)c\sin\beta \, \mathbf{\hat{z}} & \left(8f\right) & \mbox{H III} \\ 
\mathbf{B}_{9} & = & \left(x_{4}-y_{4}\right) \, \mathbf{a}_{1} + \left(x_{4}+y_{4}\right) \, \mathbf{a}_{2} + z_{4} \, \mathbf{a}_{3} & = & \left(x_{4}a+z_{4}c\cos\beta\right) \, \mathbf{\hat{x}} + y_{4}b \, \mathbf{\hat{y}} + z_{4}c\sin\beta \, \mathbf{\hat{z}} & \left(8f\right) & \mbox{H IV} \\ 
\mathbf{B}_{10} & = & \left(-x_{4}-y_{4}\right) \, \mathbf{a}_{1} + \left(-x_{4}+y_{4}\right) \, \mathbf{a}_{2} + \left(\frac{1}{2} - z_{4}\right) \, \mathbf{a}_{3} & = & \left(\frac{1}{2}c\cos\beta - x_{4}a - z_{4}c\cos\beta\right) \, \mathbf{\hat{x}} + y_{4}b \, \mathbf{\hat{y}} + \left(\frac{1}{2} - z_{4}\right)c\sin\beta \, \mathbf{\hat{z}} & \left(8f\right) & \mbox{H IV} \\ 
\mathbf{B}_{11} & = & \left(-x_{4}+y_{4}\right) \, \mathbf{a}_{1} + \left(-x_{4}-y_{4}\right) \, \mathbf{a}_{2}-z_{4} \, \mathbf{a}_{3} & = & \left(-x_{4}a-z_{4}c\cos\beta\right) \, \mathbf{\hat{x}}-y_{4}b \, \mathbf{\hat{y}}-z_{4}c\sin\beta \, \mathbf{\hat{z}} & \left(8f\right) & \mbox{H IV} \\ 
\mathbf{B}_{12} & = & \left(x_{4}+y_{4}\right) \, \mathbf{a}_{1} + \left(x_{4}-y_{4}\right) \, \mathbf{a}_{2} + \left(\frac{1}{2} +z_{4}\right) \, \mathbf{a}_{3} & = & \left(\frac{1}{2}c\cos\beta +x_{4}a + z_{4}c\cos\beta\right) \, \mathbf{\hat{x}}-y_{4}b \, \mathbf{\hat{y}} + \left(\frac{1}{2} +z_{4}\right)c\sin\beta \, \mathbf{\hat{z}} & \left(8f\right) & \mbox{H IV} \\ 
\end{longtabu}
\renewcommand{\arraystretch}{1.0}
\noindent \hrulefill
\\
\textbf{References:}
\vspace*{-0.25cm}
\begin{flushleft}
  - \bibentry{Pickard_Nature_Physics_3_2007}. \\
\end{flushleft}
\noindent \hrulefill
\\
\textbf{Geometry files:}
\\
\noindent  - CIF: pp. {\hyperref[A_mC24_15_2e2f_cif]{\pageref{A_mC24_15_2e2f_cif}}} \\
\noindent  - POSCAR: pp. {\hyperref[A_mC24_15_2e2f_poscar]{\pageref{A_mC24_15_2e2f_poscar}}} \\
\onecolumn
{\phantomsection\label{A2B_oP12_17_abe_e}}
\subsection*{\huge \textbf{{\normalfont $\alpha$-Naumannite (Ag$_{2}$Se) Structure: A2B\_oP12\_17\_abe\_e}}}
\noindent \hrulefill
\vspace*{0.25cm}
\begin{figure}[htp]
  \centering
  \vspace{-1em}
  {\includegraphics[width=1\textwidth]{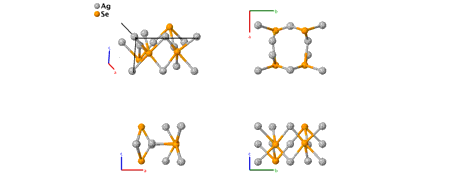}}
\end{figure}
\vspace*{-0.5cm}
\renewcommand{\arraystretch}{1.5}
\begin{equation*}
  \begin{array}{>{$\hspace{-0.15cm}}l<{$}>{$}p{0.5cm}<{$}>{$}p{18.5cm}<{$}}
    \mbox{\large \textbf{Prototype}} &\colon & \ce{$\alpha$-Ag2Se} \\
    \mbox{\large \textbf{\AFLOW\ prototype label}} &\colon & \mbox{A2B\_oP12\_17\_abe\_e} \\
    \mbox{\large \textbf{\textit{Strukturbericht} designation}} &\colon & \mbox{None} \\
    \mbox{\large \textbf{Pearson symbol}} &\colon & \mbox{oP12} \\
    \mbox{\large \textbf{Space group number}} &\colon & 17 \\
    \mbox{\large \textbf{Space group symbol}} &\colon & P222_{1} \\
    \mbox{\large \textbf{\AFLOW\ prototype command}} &\colon &  \texttt{aflow} \,  \, \texttt{-{}-proto=A2B\_oP12\_17\_abe\_e } \, \newline \texttt{-{}-params=}{a,b/a,c/a,x_{1},x_{2},x_{3},y_{3},z_{3},x_{4},y_{4},z_{4} }
  \end{array}
\end{equation*}
\renewcommand{\arraystretch}{1.0}

\noindent \parbox{1 \linewidth}{
\noindent \hrulefill
\\
\textbf{Simple Orthorhombic primitive vectors:} \\
\vspace*{-0.25cm}
\begin{tabular}{cc}
  \begin{tabular}{c}
    \parbox{0.6 \linewidth}{
      \renewcommand{\arraystretch}{1.5}
      \begin{equation*}
        \centering
        \begin{array}{ccc}
              \mathbf{a}_1 & = & a \, \mathbf{\hat{x}} \\
    \mathbf{a}_2 & = & b \, \mathbf{\hat{y}} \\
    \mathbf{a}_3 & = & c \, \mathbf{\hat{z}} \\

        \end{array}
      \end{equation*}
    }
    \renewcommand{\arraystretch}{1.0}
  \end{tabular}
  \begin{tabular}{c}
    \includegraphics[width=0.3\linewidth]{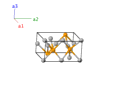} \\
  \end{tabular}
\end{tabular}

}
\vspace*{-0.25cm}

\noindent \hrulefill
\\
\textbf{Basis vectors:}
\vspace*{-0.25cm}
\renewcommand{\arraystretch}{1.5}
\begin{longtabu} to \textwidth{>{\centering $}X[-1,c,c]<{$}>{\centering $}X[-1,c,c]<{$}>{\centering $}X[-1,c,c]<{$}>{\centering $}X[-1,c,c]<{$}>{\centering $}X[-1,c,c]<{$}>{\centering $}X[-1,c,c]<{$}>{\centering $}X[-1,c,c]<{$}}
  & & \mbox{Lattice Coordinates} & & \mbox{Cartesian Coordinates} &\mbox{Wyckoff Position} & \mbox{Atom Type} \\  
  \mathbf{B}_{1} & = & x_{1} \, \mathbf{a}_{1} & = & x_{1}a \, \mathbf{\hat{x}} & \left(2a\right) & \mbox{Ag I} \\ 
\mathbf{B}_{2} & = & -x_{1} \, \mathbf{a}_{1} + \frac{1}{2} \, \mathbf{a}_{3} & = & -x_{1}a \, \mathbf{\hat{x}} + \frac{1}{2}c \, \mathbf{\hat{z}} & \left(2a\right) & \mbox{Ag I} \\ 
\mathbf{B}_{3} & = & x_{2} \, \mathbf{a}_{1} + \frac{1}{2} \, \mathbf{a}_{2} & = & x_{2}a \, \mathbf{\hat{x}} + \frac{1}{2}b \, \mathbf{\hat{y}} & \left(2b\right) & \mbox{Ag II} \\ 
\mathbf{B}_{4} & = & -x_{2} \, \mathbf{a}_{1} + \frac{1}{2} \, \mathbf{a}_{2} + \frac{1}{2} \, \mathbf{a}_{3} & = & -x_{2}a \, \mathbf{\hat{x}} + \frac{1}{2}b \, \mathbf{\hat{y}} + \frac{1}{2}c \, \mathbf{\hat{z}} & \left(2b\right) & \mbox{Ag II} \\ 
\mathbf{B}_{5} & = & x_{3} \, \mathbf{a}_{1} + y_{3} \, \mathbf{a}_{2} + z_{3} \, \mathbf{a}_{3} & = & x_{3}a \, \mathbf{\hat{x}} + y_{3}b \, \mathbf{\hat{y}} + z_{3}c \, \mathbf{\hat{z}} & \left(4e\right) & \mbox{Ag III} \\ 
\mathbf{B}_{6} & = & -x_{3} \, \mathbf{a}_{1}-y_{3} \, \mathbf{a}_{2} + \left(\frac{1}{2} +z_{3}\right) \, \mathbf{a}_{3} & = & -x_{3}a \, \mathbf{\hat{x}}-y_{3}b \, \mathbf{\hat{y}} + \left(\frac{1}{2} +z_{3}\right)c \, \mathbf{\hat{z}} & \left(4e\right) & \mbox{Ag III} \\ 
\mathbf{B}_{7} & = & -x_{3} \, \mathbf{a}_{1} + y_{3} \, \mathbf{a}_{2} + \left(\frac{1}{2} - z_{3}\right) \, \mathbf{a}_{3} & = & -x_{3}a \, \mathbf{\hat{x}} + y_{3}b \, \mathbf{\hat{y}} + \left(\frac{1}{2} - z_{3}\right)c \, \mathbf{\hat{z}} & \left(4e\right) & \mbox{Ag III} \\ 
\mathbf{B}_{8} & = & x_{3} \, \mathbf{a}_{1}-y_{3} \, \mathbf{a}_{2}-z_{3} \, \mathbf{a}_{3} & = & x_{3}a \, \mathbf{\hat{x}}-y_{3}b \, \mathbf{\hat{y}}-z_{3}c \, \mathbf{\hat{z}} & \left(4e\right) & \mbox{Ag III} \\ 
\mathbf{B}_{9} & = & x_{4} \, \mathbf{a}_{1} + y_{4} \, \mathbf{a}_{2} + z_{4} \, \mathbf{a}_{3} & = & x_{4}a \, \mathbf{\hat{x}} + y_{4}b \, \mathbf{\hat{y}} + z_{4}c \, \mathbf{\hat{z}} & \left(4e\right) & \mbox{Se} \\ 
\mathbf{B}_{10} & = & -x_{4} \, \mathbf{a}_{1}-y_{4} \, \mathbf{a}_{2} + \left(\frac{1}{2} +z_{4}\right) \, \mathbf{a}_{3} & = & -x_{4}a \, \mathbf{\hat{x}}-y_{4}b \, \mathbf{\hat{y}} + \left(\frac{1}{2} +z_{4}\right)c \, \mathbf{\hat{z}} & \left(4e\right) & \mbox{Se} \\ 
\mathbf{B}_{11} & = & -x_{4} \, \mathbf{a}_{1} + y_{4} \, \mathbf{a}_{2} + \left(\frac{1}{2} - z_{4}\right) \, \mathbf{a}_{3} & = & -x_{4}a \, \mathbf{\hat{x}} + y_{4}b \, \mathbf{\hat{y}} + \left(\frac{1}{2} - z_{4}\right)c \, \mathbf{\hat{z}} & \left(4e\right) & \mbox{Se} \\ 
\mathbf{B}_{12} & = & x_{4} \, \mathbf{a}_{1}-y_{4} \, \mathbf{a}_{2}-z_{4} \, \mathbf{a}_{3} & = & x_{4}a \, \mathbf{\hat{x}}-y_{4}b \, \mathbf{\hat{y}}-z_{4}c \, \mathbf{\hat{z}} & \left(4e\right) & \mbox{Se} \\ 
\end{longtabu}
\renewcommand{\arraystretch}{1.0}
\noindent \hrulefill
\\
\textbf{References:}
\vspace*{-0.25cm}
\begin{flushleft}
  - \bibentry{Pinsker_Ag2Se_SovPhysCrystallogr_1965}. \\
\end{flushleft}
\textbf{Found in:}
\vspace*{-0.25cm}
\begin{flushleft}
  - \bibentry{Villars_PearsonsCrystalData_2013}. \\
\end{flushleft}
\noindent \hrulefill
\\
\textbf{Geometry files:}
\\
\noindent  - CIF: pp. {\hyperref[A2B_oP12_17_abe_e_cif]{\pageref{A2B_oP12_17_abe_e_cif}}} \\
\noindent  - POSCAR: pp. {\hyperref[A2B_oP12_17_abe_e_poscar]{\pageref{A2B_oP12_17_abe_e_poscar}}} \\
\onecolumn
{\phantomsection\label{AB3_oP16_19_a_3a}}
\subsection*{\huge \textbf{{\normalfont H$_{3}$Cl (100~GPa) Structure: AB3\_oP16\_19\_a\_3a}}}
\noindent \hrulefill
\vspace*{0.25cm}
\begin{figure}[htp]
  \centering
  \vspace{-1em}
  {\includegraphics[width=1\textwidth]{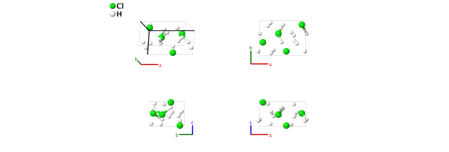}}
\end{figure}
\vspace*{-0.5cm}
\renewcommand{\arraystretch}{1.5}
\begin{equation*}
  \begin{array}{>{$\hspace{-0.15cm}}l<{$}>{$}p{0.5cm}<{$}>{$}p{18.5cm}<{$}}
    \mbox{\large \textbf{Prototype}} &\colon & \ce{H3Cl} \\
    \mbox{\large \textbf{\AFLOW\ prototype label}} &\colon & \mbox{AB3\_oP16\_19\_a\_3a} \\
    \mbox{\large \textbf{\textit{Strukturbericht} designation}} &\colon & \mbox{None} \\
    \mbox{\large \textbf{Pearson symbol}} &\colon & \mbox{oP16} \\
    \mbox{\large \textbf{Space group number}} &\colon & 19 \\
    \mbox{\large \textbf{Space group symbol}} &\colon & P2_{1}2_{1}2_{1} \\
    \mbox{\large \textbf{\AFLOW\ prototype command}} &\colon &  \texttt{aflow} \,  \, \texttt{-{}-proto=AB3\_oP16\_19\_a\_3a } \, \newline \texttt{-{}-params=}{a,b/a,c/a,x_{1},y_{1},z_{1},x_{2},y_{2},z_{2},x_{3},y_{3},z_{3},x_{4},y_{4},z_{4} }
  \end{array}
\end{equation*}
\renewcommand{\arraystretch}{1.0}

\vspace*{-0.25cm}
\noindent \hrulefill
\begin{itemize}
  \item{This structure was found via first-principles calculations.  The data
presented here was computed at a pressure of 100~GPa.
}
\end{itemize}

\noindent \parbox{1 \linewidth}{
\noindent \hrulefill
\\
\textbf{Simple Orthorhombic primitive vectors:} \\
\vspace*{-0.25cm}
\begin{tabular}{cc}
  \begin{tabular}{c}
    \parbox{0.6 \linewidth}{
      \renewcommand{\arraystretch}{1.5}
      \begin{equation*}
        \centering
        \begin{array}{ccc}
              \mathbf{a}_1 & = & a \, \mathbf{\hat{x}} \\
    \mathbf{a}_2 & = & b \, \mathbf{\hat{y}} \\
    \mathbf{a}_3 & = & c \, \mathbf{\hat{z}} \\

        \end{array}
      \end{equation*}
    }
    \renewcommand{\arraystretch}{1.0}
  \end{tabular}
  \begin{tabular}{c}
    \includegraphics[width=0.3\linewidth]{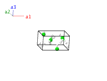} \\
  \end{tabular}
\end{tabular}

}
\vspace*{-0.25cm}

\noindent \hrulefill
\\
\textbf{Basis vectors:}
\vspace*{-0.25cm}
\renewcommand{\arraystretch}{1.5}
\begin{longtabu} to \textwidth{>{\centering $}X[-1,c,c]<{$}>{\centering $}X[-1,c,c]<{$}>{\centering $}X[-1,c,c]<{$}>{\centering $}X[-1,c,c]<{$}>{\centering $}X[-1,c,c]<{$}>{\centering $}X[-1,c,c]<{$}>{\centering $}X[-1,c,c]<{$}}
  & & \mbox{Lattice Coordinates} & & \mbox{Cartesian Coordinates} &\mbox{Wyckoff Position} & \mbox{Atom Type} \\  
  \mathbf{B}_{1} & = & x_{1} \, \mathbf{a}_{1} + y_{1} \, \mathbf{a}_{2} + z_{1} \, \mathbf{a}_{3} & = & x_{1}a \, \mathbf{\hat{x}} + y_{1}b \, \mathbf{\hat{y}} + z_{1}c \, \mathbf{\hat{z}} & \left(4a\right) & \mbox{Cl} \\ 
\mathbf{B}_{2} & = & \left(\frac{1}{2} - x_{1}\right) \, \mathbf{a}_{1}-y_{1} \, \mathbf{a}_{2} + \left(\frac{1}{2} +z_{1}\right) \, \mathbf{a}_{3} & = & \left(\frac{1}{2} - x_{1}\right)a \, \mathbf{\hat{x}}-y_{1}b \, \mathbf{\hat{y}} + \left(\frac{1}{2} +z_{1}\right)c \, \mathbf{\hat{z}} & \left(4a\right) & \mbox{Cl} \\ 
\mathbf{B}_{3} & = & -x_{1} \, \mathbf{a}_{1} + \left(\frac{1}{2} +y_{1}\right) \, \mathbf{a}_{2} + \left(\frac{1}{2} - z_{1}\right) \, \mathbf{a}_{3} & = & -x_{1}a \, \mathbf{\hat{x}} + \left(\frac{1}{2} +y_{1}\right)b \, \mathbf{\hat{y}} + \left(\frac{1}{2} - z_{1}\right)c \, \mathbf{\hat{z}} & \left(4a\right) & \mbox{Cl} \\ 
\mathbf{B}_{4} & = & \left(\frac{1}{2} +x_{1}\right) \, \mathbf{a}_{1} + \left(\frac{1}{2} - y_{1}\right) \, \mathbf{a}_{2}-z_{1} \, \mathbf{a}_{3} & = & \left(\frac{1}{2} +x_{1}\right)a \, \mathbf{\hat{x}} + \left(\frac{1}{2} - y_{1}\right)b \, \mathbf{\hat{y}}-z_{1}c \, \mathbf{\hat{z}} & \left(4a\right) & \mbox{Cl} \\ 
\mathbf{B}_{5} & = & x_{2} \, \mathbf{a}_{1} + y_{2} \, \mathbf{a}_{2} + z_{2} \, \mathbf{a}_{3} & = & x_{2}a \, \mathbf{\hat{x}} + y_{2}b \, \mathbf{\hat{y}} + z_{2}c \, \mathbf{\hat{z}} & \left(4a\right) & \mbox{H I} \\ 
\mathbf{B}_{6} & = & \left(\frac{1}{2} - x_{2}\right) \, \mathbf{a}_{1}-y_{2} \, \mathbf{a}_{2} + \left(\frac{1}{2} +z_{2}\right) \, \mathbf{a}_{3} & = & \left(\frac{1}{2} - x_{2}\right)a \, \mathbf{\hat{x}}-y_{2}b \, \mathbf{\hat{y}} + \left(\frac{1}{2} +z_{2}\right)c \, \mathbf{\hat{z}} & \left(4a\right) & \mbox{H I} \\ 
\mathbf{B}_{7} & = & -x_{2} \, \mathbf{a}_{1} + \left(\frac{1}{2} +y_{2}\right) \, \mathbf{a}_{2} + \left(\frac{1}{2} - z_{2}\right) \, \mathbf{a}_{3} & = & -x_{2}a \, \mathbf{\hat{x}} + \left(\frac{1}{2} +y_{2}\right)b \, \mathbf{\hat{y}} + \left(\frac{1}{2} - z_{2}\right)c \, \mathbf{\hat{z}} & \left(4a\right) & \mbox{H I} \\ 
\mathbf{B}_{8} & = & \left(\frac{1}{2} +x_{2}\right) \, \mathbf{a}_{1} + \left(\frac{1}{2} - y_{2}\right) \, \mathbf{a}_{2}-z_{2} \, \mathbf{a}_{3} & = & \left(\frac{1}{2} +x_{2}\right)a \, \mathbf{\hat{x}} + \left(\frac{1}{2} - y_{2}\right)b \, \mathbf{\hat{y}}-z_{2}c \, \mathbf{\hat{z}} & \left(4a\right) & \mbox{H I} \\ 
\mathbf{B}_{9} & = & x_{3} \, \mathbf{a}_{1} + y_{3} \, \mathbf{a}_{2} + z_{3} \, \mathbf{a}_{3} & = & x_{3}a \, \mathbf{\hat{x}} + y_{3}b \, \mathbf{\hat{y}} + z_{3}c \, \mathbf{\hat{z}} & \left(4a\right) & \mbox{H II} \\ 
\mathbf{B}_{10} & = & \left(\frac{1}{2} - x_{3}\right) \, \mathbf{a}_{1}-y_{3} \, \mathbf{a}_{2} + \left(\frac{1}{2} +z_{3}\right) \, \mathbf{a}_{3} & = & \left(\frac{1}{2} - x_{3}\right)a \, \mathbf{\hat{x}}-y_{3}b \, \mathbf{\hat{y}} + \left(\frac{1}{2} +z_{3}\right)c \, \mathbf{\hat{z}} & \left(4a\right) & \mbox{H II} \\ 
\mathbf{B}_{11} & = & -x_{3} \, \mathbf{a}_{1} + \left(\frac{1}{2} +y_{3}\right) \, \mathbf{a}_{2} + \left(\frac{1}{2} - z_{3}\right) \, \mathbf{a}_{3} & = & -x_{3}a \, \mathbf{\hat{x}} + \left(\frac{1}{2} +y_{3}\right)b \, \mathbf{\hat{y}} + \left(\frac{1}{2} - z_{3}\right)c \, \mathbf{\hat{z}} & \left(4a\right) & \mbox{H II} \\ 
\mathbf{B}_{12} & = & \left(\frac{1}{2} +x_{3}\right) \, \mathbf{a}_{1} + \left(\frac{1}{2} - y_{3}\right) \, \mathbf{a}_{2}-z_{3} \, \mathbf{a}_{3} & = & \left(\frac{1}{2} +x_{3}\right)a \, \mathbf{\hat{x}} + \left(\frac{1}{2} - y_{3}\right)b \, \mathbf{\hat{y}}-z_{3}c \, \mathbf{\hat{z}} & \left(4a\right) & \mbox{H II} \\ 
\mathbf{B}_{13} & = & x_{4} \, \mathbf{a}_{1} + y_{4} \, \mathbf{a}_{2} + z_{4} \, \mathbf{a}_{3} & = & x_{4}a \, \mathbf{\hat{x}} + y_{4}b \, \mathbf{\hat{y}} + z_{4}c \, \mathbf{\hat{z}} & \left(4a\right) & \mbox{H III} \\ 
\mathbf{B}_{14} & = & \left(\frac{1}{2} - x_{4}\right) \, \mathbf{a}_{1}-y_{4} \, \mathbf{a}_{2} + \left(\frac{1}{2} +z_{4}\right) \, \mathbf{a}_{3} & = & \left(\frac{1}{2} - x_{4}\right)a \, \mathbf{\hat{x}}-y_{4}b \, \mathbf{\hat{y}} + \left(\frac{1}{2} +z_{4}\right)c \, \mathbf{\hat{z}} & \left(4a\right) & \mbox{H III} \\ 
\mathbf{B}_{15} & = & -x_{4} \, \mathbf{a}_{1} + \left(\frac{1}{2} +y_{4}\right) \, \mathbf{a}_{2} + \left(\frac{1}{2} - z_{4}\right) \, \mathbf{a}_{3} & = & -x_{4}a \, \mathbf{\hat{x}} + \left(\frac{1}{2} +y_{4}\right)b \, \mathbf{\hat{y}} + \left(\frac{1}{2} - z_{4}\right)c \, \mathbf{\hat{z}} & \left(4a\right) & \mbox{H III} \\ 
\mathbf{B}_{16} & = & \left(\frac{1}{2} +x_{4}\right) \, \mathbf{a}_{1} + \left(\frac{1}{2} - y_{4}\right) \, \mathbf{a}_{2}-z_{4} \, \mathbf{a}_{3} & = & \left(\frac{1}{2} +x_{4}\right)a \, \mathbf{\hat{x}} + \left(\frac{1}{2} - y_{4}\right)b \, \mathbf{\hat{y}}-z_{4}c \, \mathbf{\hat{z}} & \left(4a\right) & \mbox{H III} \\ 
\end{longtabu}
\renewcommand{\arraystretch}{1.0}
\noindent \hrulefill
\\
\textbf{References:}
\vspace*{-0.25cm}
\begin{flushleft}
  - \bibentry{Duan_2015}. \\
\end{flushleft}
\noindent \hrulefill
\\
\textbf{Geometry files:}
\\
\noindent  - CIF: pp. {\hyperref[AB3_oP16_19_a_3a_cif]{\pageref{AB3_oP16_19_a_3a_cif}}} \\
\noindent  - POSCAR: pp. {\hyperref[AB3_oP16_19_a_3a_poscar]{\pageref{AB3_oP16_19_a_3a_poscar}}} \\
\onecolumn
{\phantomsection\label{AB2_oC6_21_a_k}}
\subsection*{\huge \textbf{{\normalfont Ta$_{2}$H Structure: AB2\_oC6\_21\_a\_k}}}
\noindent \hrulefill
\vspace*{0.25cm}
\begin{figure}[htp]
  \centering
  \vspace{-1em}
  {\includegraphics[width=1\textwidth]{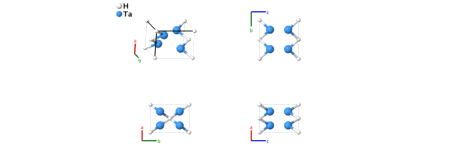}}
\end{figure}
\vspace*{-0.5cm}
\renewcommand{\arraystretch}{1.5}
\begin{equation*}
  \begin{array}{>{$\hspace{-0.15cm}}l<{$}>{$}p{0.5cm}<{$}>{$}p{18.5cm}<{$}}
    \mbox{\large \textbf{Prototype}} &\colon & \ce{Ta2H} \\
    \mbox{\large \textbf{\AFLOW\ prototype label}} &\colon & \mbox{AB2\_oC6\_21\_a\_k} \\
    \mbox{\large \textbf{\textit{Strukturbericht} designation}} &\colon & \mbox{None} \\
    \mbox{\large \textbf{Pearson symbol}} &\colon & \mbox{oC6} \\
    \mbox{\large \textbf{Space group number}} &\colon & 21 \\
    \mbox{\large \textbf{Space group symbol}} &\colon & C222 \\
    \mbox{\large \textbf{\AFLOW\ prototype command}} &\colon &  \texttt{aflow} \,  \, \texttt{-{}-proto=AB2\_oC6\_21\_a\_k } \, \newline \texttt{-{}-params=}{a,b/a,c/a,z_{2} }
  \end{array}
\end{equation*}
\renewcommand{\arraystretch}{1.0}

\noindent \parbox{1 \linewidth}{
\noindent \hrulefill
\\
\textbf{Base-centered Orthorhombic primitive vectors:} \\
\vspace*{-0.25cm}
\begin{tabular}{cc}
  \begin{tabular}{c}
    \parbox{0.6 \linewidth}{
      \renewcommand{\arraystretch}{1.5}
      \begin{equation*}
        \centering
        \begin{array}{ccc}
              \mathbf{a}_1 & = & \frac12 \, a \, \mathbf{\hat{x}} - \frac12 \, b \, \mathbf{\hat{y}} \\
    \mathbf{a}_2 & = & \frac12 \, a \, \mathbf{\hat{x}} + \frac12 \, b \, \mathbf{\hat{y}} \\
    \mathbf{a}_3 & = & c \, \mathbf{\hat{z}} \\

        \end{array}
      \end{equation*}
    }
    \renewcommand{\arraystretch}{1.0}
  \end{tabular}
  \begin{tabular}{c}
    \includegraphics[width=0.3\linewidth]{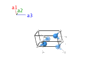} \\
  \end{tabular}
\end{tabular}

}
\vspace*{-0.25cm}

\noindent \hrulefill
\\
\textbf{Basis vectors:}
\vspace*{-0.25cm}
\renewcommand{\arraystretch}{1.5}
\begin{longtabu} to \textwidth{>{\centering $}X[-1,c,c]<{$}>{\centering $}X[-1,c,c]<{$}>{\centering $}X[-1,c,c]<{$}>{\centering $}X[-1,c,c]<{$}>{\centering $}X[-1,c,c]<{$}>{\centering $}X[-1,c,c]<{$}>{\centering $}X[-1,c,c]<{$}}
  & & \mbox{Lattice Coordinates} & & \mbox{Cartesian Coordinates} &\mbox{Wyckoff Position} & \mbox{Atom Type} \\  
  \mathbf{B}_{1} & = & 0 \, \mathbf{a}_{1} + 0 \, \mathbf{a}_{2} + 0 \, \mathbf{a}_{3} & = & 0 \, \mathbf{\hat{x}} + 0 \, \mathbf{\hat{y}} + 0 \, \mathbf{\hat{z}} & \left(2a\right) & \mbox{H} \\ 
\mathbf{B}_{2} & = & \frac{1}{2} \, \mathbf{a}_{2} + z_{2} \, \mathbf{a}_{3} & = & \frac{1}{4}a \, \mathbf{\hat{x}} + \frac{1}{4}b \, \mathbf{\hat{y}} + z_{2}c \, \mathbf{\hat{z}} & \left(4k\right) & \mbox{Ta} \\ 
\mathbf{B}_{3} & = & \frac{1}{2} \, \mathbf{a}_{1} + -z_{2} \, \mathbf{a}_{3} & = & \frac{1}{4}a \, \mathbf{\hat{x}}- \frac{1}{4}b  \, \mathbf{\hat{y}}-z_{2}c \, \mathbf{\hat{z}} & \left(4k\right) & \mbox{Ta} \\ 
\end{longtabu}
\renewcommand{\arraystretch}{1.0}
\noindent \hrulefill
\\
\textbf{References:}
\vspace*{-0.25cm}
\begin{flushleft}
  - \bibentry{Asano_Ta2H_JApplCrystallogr_1978}. \\
\end{flushleft}
\textbf{Found in:}
\vspace*{-0.25cm}
\begin{flushleft}
  - \bibentry{Villars_PearsonsCrystalData_2013}. \\
\end{flushleft}
\noindent \hrulefill
\\
\textbf{Geometry files:}
\\
\noindent  - CIF: pp. {\hyperref[AB2_oC6_21_a_k_cif]{\pageref{AB2_oC6_21_a_k_cif}}} \\
\noindent  - POSCAR: pp. {\hyperref[AB2_oC6_21_a_k_poscar]{\pageref{AB2_oC6_21_a_k_poscar}}} \\
\onecolumn
{\phantomsection\label{A2BC2_oF40_22_fi_ad_gh}}
\subsection*{\huge \textbf{{\normalfont CeRu$_{2}$B$_{2}$ Structure: A2BC2\_oF40\_22\_fi\_ad\_gh}}}
\noindent \hrulefill
\vspace*{0.25cm}
\begin{figure}[htp]
  \centering
  \vspace{-1em}
  {\includegraphics[width=1\textwidth]{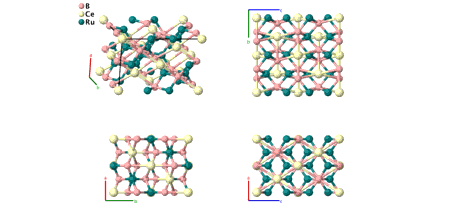}}
\end{figure}
\vspace*{-0.5cm}
\renewcommand{\arraystretch}{1.5}
\begin{equation*}
  \begin{array}{>{$\hspace{-0.15cm}}l<{$}>{$}p{0.5cm}<{$}>{$}p{18.5cm}<{$}}
    \mbox{\large \textbf{Prototype}} &\colon & \ce{CeRu2B2} \\
    \mbox{\large \textbf{\AFLOW\ prototype label}} &\colon & \mbox{A2BC2\_oF40\_22\_fi\_ad\_gh} \\
    \mbox{\large \textbf{\textit{Strukturbericht} designation}} &\colon & \mbox{None} \\
    \mbox{\large \textbf{Pearson symbol}} &\colon & \mbox{oF40} \\
    \mbox{\large \textbf{Space group number}} &\colon & 22 \\
    \mbox{\large \textbf{Space group symbol}} &\colon & F222 \\
    \mbox{\large \textbf{\AFLOW\ prototype command}} &\colon &  \texttt{aflow} \,  \, \texttt{-{}-proto=A2BC2\_oF40\_22\_fi\_ad\_gh } \, \newline \texttt{-{}-params=}{a,b/a,c/a,y_{3},z_{4},z_{5},y_{6} }
  \end{array}
\end{equation*}
\renewcommand{\arraystretch}{1.0}

\noindent \parbox{1 \linewidth}{
\noindent \hrulefill
\\
\textbf{Face-centered Orthorhombic primitive vectors:} \\
\vspace*{-0.25cm}
\begin{tabular}{cc}
  \begin{tabular}{c}
    \parbox{0.6 \linewidth}{
      \renewcommand{\arraystretch}{1.5}
      \begin{equation*}
        \centering
        \begin{array}{ccc}
              \mathbf{a}_1 & = & \frac12 \, b \, \mathbf{\hat{y}} + \frac12 \, c \, \mathbf{\hat{z}} \\
    \mathbf{a}_2 & = & \frac12 \, a \, \mathbf{\hat{x}} + \frac12 \, c \, \mathbf{\hat{z}} \\
    \mathbf{a}_3 & = & \frac12 \, a \, \mathbf{\hat{x}} + \frac12 \, b \, \mathbf{\hat{y}} \\

        \end{array}
      \end{equation*}
    }
    \renewcommand{\arraystretch}{1.0}
  \end{tabular}
  \begin{tabular}{c}
    \includegraphics[width=0.3\linewidth]{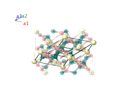} \\
  \end{tabular}
\end{tabular}

}
\vspace*{-0.25cm}

\noindent \hrulefill
\\
\textbf{Basis vectors:}
\vspace*{-0.25cm}
\renewcommand{\arraystretch}{1.5}
\begin{longtabu} to \textwidth{>{\centering $}X[-1,c,c]<{$}>{\centering $}X[-1,c,c]<{$}>{\centering $}X[-1,c,c]<{$}>{\centering $}X[-1,c,c]<{$}>{\centering $}X[-1,c,c]<{$}>{\centering $}X[-1,c,c]<{$}>{\centering $}X[-1,c,c]<{$}}
  & & \mbox{Lattice Coordinates} & & \mbox{Cartesian Coordinates} &\mbox{Wyckoff Position} & \mbox{Atom Type} \\  
  \mathbf{B}_{1} & = & 0 \, \mathbf{a}_{1} + 0 \, \mathbf{a}_{2} + 0 \, \mathbf{a}_{3} & = & 0 \, \mathbf{\hat{x}} + 0 \, \mathbf{\hat{y}} + 0 \, \mathbf{\hat{z}} & \left(4a\right) & \mbox{Ce I} \\ 
\mathbf{B}_{2} & = & \frac{3}{4} \, \mathbf{a}_{1} + \frac{3}{4} \, \mathbf{a}_{2} + \frac{3}{4} \, \mathbf{a}_{3} & = & \frac{3}{4}a \, \mathbf{\hat{x}} + \frac{3}{4}b \, \mathbf{\hat{y}} + \frac{3}{4}c \, \mathbf{\hat{z}} & \left(4d\right) & \mbox{Ce II} \\ 
\mathbf{B}_{3} & = & y_{3} \, \mathbf{a}_{1}-y_{3} \, \mathbf{a}_{2} + y_{3} \, \mathbf{a}_{3} & = & y_{3}b \, \mathbf{\hat{y}} & \left(8f\right) & \mbox{B I} \\ 
\mathbf{B}_{4} & = & -y_{3} \, \mathbf{a}_{1} + y_{3} \, \mathbf{a}_{2}-y_{3} \, \mathbf{a}_{3} & = & -y_{3}b \, \mathbf{\hat{y}} & \left(8f\right) & \mbox{B I} \\ 
\mathbf{B}_{5} & = & z_{4} \, \mathbf{a}_{1} + z_{4} \, \mathbf{a}_{2}-z_{4} \, \mathbf{a}_{3} & = & z_{4}c \, \mathbf{\hat{z}} & \left(8g\right) & \mbox{Ru I} \\ 
\mathbf{B}_{6} & = & -z_{4} \, \mathbf{a}_{1}-z_{4} \, \mathbf{a}_{2} + z_{4} \, \mathbf{a}_{3} & = & -z_{4}c \, \mathbf{\hat{z}} & \left(8g\right) & \mbox{Ru I} \\ 
\mathbf{B}_{7} & = & z_{5} \, \mathbf{a}_{1} + z_{5} \, \mathbf{a}_{2} + \left(\frac{1}{2} - z_{5}\right) \, \mathbf{a}_{3} & = & \frac{1}{4}a \, \mathbf{\hat{x}} + \frac{1}{4}b \, \mathbf{\hat{y}} + z_{5}c \, \mathbf{\hat{z}} & \left(8h\right) & \mbox{Ru II} \\ 
\mathbf{B}_{8} & = & \left(\frac{1}{2} - z_{5}\right) \, \mathbf{a}_{1} + \left(\frac{1}{2} - z_{5}\right) \, \mathbf{a}_{2} + z_{5} \, \mathbf{a}_{3} & = & \frac{1}{4}a \, \mathbf{\hat{x}} + \frac{1}{4}b \, \mathbf{\hat{y}} + \left(\frac{1}{2} - z_{5}\right)c \, \mathbf{\hat{z}} & \left(8h\right) & \mbox{Ru II} \\ 
\mathbf{B}_{9} & = & y_{6} \, \mathbf{a}_{1} + \left(\frac{1}{2} - y_{6}\right) \, \mathbf{a}_{2} + y_{6} \, \mathbf{a}_{3} & = & \frac{1}{4}a \, \mathbf{\hat{x}} + y_{6}b \, \mathbf{\hat{y}} + \frac{1}{4}c \, \mathbf{\hat{z}} & \left(8i\right) & \mbox{B II} \\ 
\mathbf{B}_{10} & = & \left(\frac{1}{2} - y_{6}\right) \, \mathbf{a}_{1} + y_{6} \, \mathbf{a}_{2} + \left(\frac{1}{2} - y_{6}\right) \, \mathbf{a}_{3} & = & \frac{1}{4}a \, \mathbf{\hat{x}} + \left(\frac{1}{2} - y_{6}\right)b \, \mathbf{\hat{y}} + \frac{1}{4}c \, \mathbf{\hat{z}} & \left(8i\right) & \mbox{B II} \\ 
\end{longtabu}
\renewcommand{\arraystretch}{1.0}
\noindent \hrulefill
\\
\textbf{References:}
\vspace*{-0.25cm}
\begin{flushleft}
  - \bibentry{Rogl_CeRu2B2_JLessCommMet_1985}. \\
\end{flushleft}
\textbf{Found in:}
\vspace*{-0.25cm}
\begin{flushleft}
  - \bibentry{Villars_PearsonsCrystalData_2013}. \\
\end{flushleft}
\noindent \hrulefill
\\
\textbf{Geometry files:}
\\
\noindent  - CIF: pp. {\hyperref[A2BC2_oF40_22_fi_ad_gh_cif]{\pageref{A2BC2_oF40_22_fi_ad_gh_cif}}} \\
\noindent  - POSCAR: pp. {\hyperref[A2BC2_oF40_22_fi_ad_gh_poscar]{\pageref{A2BC2_oF40_22_fi_ad_gh_poscar}}} \\
\onecolumn
{\phantomsection\label{AB_oF8_22_a_c}}
\subsection*{\huge \textbf{{\normalfont FeS (Low-temperature) Structure: AB\_oF8\_22\_a\_c}}}
\noindent \hrulefill
\vspace*{0.25cm}
\begin{figure}[htp]
  \centering
  \vspace{-1em}
  {\includegraphics[width=1\textwidth]{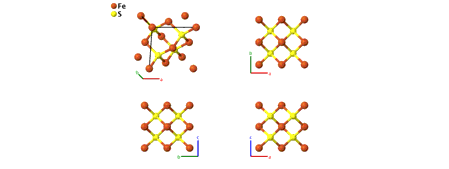}}
\end{figure}
\vspace*{-0.5cm}
\renewcommand{\arraystretch}{1.5}
\begin{equation*}
  \begin{array}{>{$\hspace{-0.15cm}}l<{$}>{$}p{0.5cm}<{$}>{$}p{18.5cm}<{$}}
    \mbox{\large \textbf{Prototype}} &\colon & \ce{FeS} \\
    \mbox{\large \textbf{\AFLOW\ prototype label}} &\colon & \mbox{AB\_oF8\_22\_a\_c} \\
    \mbox{\large \textbf{\textit{Strukturbericht} designation}} &\colon & \mbox{None} \\
    \mbox{\large \textbf{Pearson symbol}} &\colon & \mbox{oF8} \\
    \mbox{\large \textbf{Space group number}} &\colon & 22 \\
    \mbox{\large \textbf{Space group symbol}} &\colon & F222 \\
    \mbox{\large \textbf{\AFLOW\ prototype command}} &\colon &  \texttt{aflow} \,  \, \texttt{-{}-proto=AB\_oF8\_22\_a\_c } \, \newline \texttt{-{}-params=}{a,b/a,c/a }
  \end{array}
\end{equation*}
\renewcommand{\arraystretch}{1.0}

\noindent \parbox{1 \linewidth}{
\noindent \hrulefill
\\
\textbf{Face-centered Orthorhombic primitive vectors:} \\
\vspace*{-0.25cm}
\begin{tabular}{cc}
  \begin{tabular}{c}
    \parbox{0.6 \linewidth}{
      \renewcommand{\arraystretch}{1.5}
      \begin{equation*}
        \centering
        \begin{array}{ccc}
              \mathbf{a}_1 & = & \frac12 \, b \, \mathbf{\hat{y}} + \frac12 \, c \, \mathbf{\hat{z}} \\
    \mathbf{a}_2 & = & \frac12 \, a \, \mathbf{\hat{x}} + \frac12 \, c \, \mathbf{\hat{z}} \\
    \mathbf{a}_3 & = & \frac12 \, a \, \mathbf{\hat{x}} + \frac12 \, b \, \mathbf{\hat{y}} \\

        \end{array}
      \end{equation*}
    }
    \renewcommand{\arraystretch}{1.0}
  \end{tabular}
  \begin{tabular}{c}
    \includegraphics[width=0.3\linewidth]{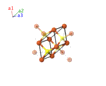} \\
  \end{tabular}
\end{tabular}

}
\vspace*{-0.25cm}

\noindent \hrulefill
\\
\textbf{Basis vectors:}
\vspace*{-0.25cm}
\renewcommand{\arraystretch}{1.5}
\begin{longtabu} to \textwidth{>{\centering $}X[-1,c,c]<{$}>{\centering $}X[-1,c,c]<{$}>{\centering $}X[-1,c,c]<{$}>{\centering $}X[-1,c,c]<{$}>{\centering $}X[-1,c,c]<{$}>{\centering $}X[-1,c,c]<{$}>{\centering $}X[-1,c,c]<{$}}
  & & \mbox{Lattice Coordinates} & & \mbox{Cartesian Coordinates} &\mbox{Wyckoff Position} & \mbox{Atom Type} \\  
  \mathbf{B}_{1} & = & 0 \, \mathbf{a}_{1} + 0 \, \mathbf{a}_{2} + 0 \, \mathbf{a}_{3} & = & 0 \, \mathbf{\hat{x}} + 0 \, \mathbf{\hat{y}} + 0 \, \mathbf{\hat{z}} & \left(4a\right) & \mbox{Fe} \\ 
\mathbf{B}_{2} & = & \frac{1}{4} \, \mathbf{a}_{1} + \frac{1}{4} \, \mathbf{a}_{2} + \frac{1}{4} \, \mathbf{a}_{3} & = & \frac{1}{4}a \, \mathbf{\hat{x}} + \frac{1}{4}b \, \mathbf{\hat{y}} + \frac{1}{4}c \, \mathbf{\hat{z}} & \left(4c\right) & \mbox{S} \\ 
\end{longtabu}
\renewcommand{\arraystretch}{1.0}
\noindent \hrulefill
\\
\textbf{References:}
\vspace*{-0.25cm}
\begin{flushleft}
  - \bibentry{Wintenberger_FeS_ActCrystallogrA_1978}. \\
\end{flushleft}
\textbf{Found in:}
\vspace*{-0.25cm}
\begin{flushleft}
  - \bibentry{Villars_PearsonsCrystalData_2013}. \\
\end{flushleft}
\noindent \hrulefill
\\
\textbf{Geometry files:}
\\
\noindent  - CIF: pp. {\hyperref[AB_oF8_22_a_c_cif]{\pageref{AB_oF8_22_a_c_cif}}} \\
\noindent  - POSCAR: pp. {\hyperref[AB_oF8_22_a_c_poscar]{\pageref{AB_oF8_22_a_c_poscar}}} \\
\onecolumn
{\phantomsection\label{A3B_oI32_23_ij2k_k}}
\subsection*{\huge \textbf{{\normalfont H$_{3}$S (5~GPa) Structure: A3B\_oI32\_23\_ij2k\_k}}}
\noindent \hrulefill
\vspace*{0.25cm}
\begin{figure}[htp]
  \centering
  \vspace{-1em}
  {\includegraphics[width=1\textwidth]{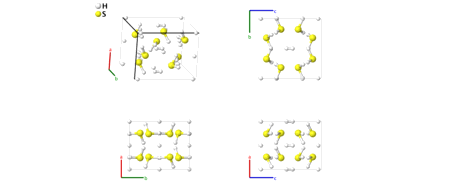}}
\end{figure}
\vspace*{-0.5cm}
\renewcommand{\arraystretch}{1.5}
\begin{equation*}
  \begin{array}{>{$\hspace{-0.15cm}}l<{$}>{$}p{0.5cm}<{$}>{$}p{18.5cm}<{$}}
    \mbox{\large \textbf{Prototype}} &\colon & \ce{H3S} \\
    \mbox{\large \textbf{\AFLOW\ prototype label}} &\colon & \mbox{A3B\_oI32\_23\_ij2k\_k} \\
    \mbox{\large \textbf{\textit{Strukturbericht} designation}} &\colon & \mbox{None} \\
    \mbox{\large \textbf{Pearson symbol}} &\colon & \mbox{oI32} \\
    \mbox{\large \textbf{Space group number}} &\colon & 23 \\
    \mbox{\large \textbf{Space group symbol}} &\colon & I222 \\
    \mbox{\large \textbf{\AFLOW\ prototype command}} &\colon &  \texttt{aflow} \,  \, \texttt{-{}-proto=A3B\_oI32\_23\_ij2k\_k } \, \newline \texttt{-{}-params=}{a,b/a,c/a,z_{1},z_{2},x_{3},y_{3},z_{3},x_{4},y_{4},z_{4},x_{5},y_{5},z_{5} }
  \end{array}
\end{equation*}
\renewcommand{\arraystretch}{1.0}

\vspace*{-0.25cm}
\noindent \hrulefill
\begin{itemize}
  \item{This structure is found in H$_{3}$S in the pressure range 3.5-17~GPa.
The data presented here was taken at 5~GPa.
}
\end{itemize}

\noindent \parbox{1 \linewidth}{
\noindent \hrulefill
\\
\textbf{Body-centered Orthorhombic primitive vectors:} \\
\vspace*{-0.25cm}
\begin{tabular}{cc}
  \begin{tabular}{c}
    \parbox{0.6 \linewidth}{
      \renewcommand{\arraystretch}{1.5}
      \begin{equation*}
        \centering
        \begin{array}{ccc}
              \mathbf{a}_1 & = & - \frac12 \, a \, \mathbf{\hat{x}} + \frac12 \, b \, \mathbf{\hat{y}} + \frac12 \, c \, \mathbf{\hat{z}} \\
    \mathbf{a}_2 & = & ~ \frac12 \, a \, \mathbf{\hat{x}} - \frac12 \, b \, \mathbf{\hat{y}} + \frac12 \, c \, \mathbf{\hat{z}} \\
    \mathbf{a}_3 & = & ~ \frac12 \, a \, \mathbf{\hat{x}} + \frac12 \, b \, \mathbf{\hat{y}} - \frac12 \, c \, \mathbf{\hat{z}} \\

        \end{array}
      \end{equation*}
    }
    \renewcommand{\arraystretch}{1.0}
  \end{tabular}
  \begin{tabular}{c}
    \includegraphics[width=0.3\linewidth]{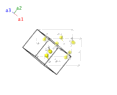} \\
  \end{tabular}
\end{tabular}

}
\vspace*{-0.25cm}

\noindent \hrulefill
\\
\textbf{Basis vectors:}
\vspace*{-0.25cm}
\renewcommand{\arraystretch}{1.5}
\begin{longtabu} to \textwidth{>{\centering $}X[-1,c,c]<{$}>{\centering $}X[-1,c,c]<{$}>{\centering $}X[-1,c,c]<{$}>{\centering $}X[-1,c,c]<{$}>{\centering $}X[-1,c,c]<{$}>{\centering $}X[-1,c,c]<{$}>{\centering $}X[-1,c,c]<{$}}
  & & \mbox{Lattice Coordinates} & & \mbox{Cartesian Coordinates} &\mbox{Wyckoff Position} & \mbox{Atom Type} \\  
  \mathbf{B}_{1} & = & z_{1} \, \mathbf{a}_{1} + z_{1} \, \mathbf{a}_{2} & = & z_{1}c \, \mathbf{\hat{z}} & \left(4i\right) & \mbox{H I} \\ 
\mathbf{B}_{2} & = & -z_{1} \, \mathbf{a}_{1}-z_{1} \, \mathbf{a}_{2} & = & -z_{1}c \, \mathbf{\hat{z}} & \left(4i\right) & \mbox{H I} \\ 
\mathbf{B}_{3} & = & \left(\frac{1}{2} +z_{2}\right) \, \mathbf{a}_{1} + z_{2} \, \mathbf{a}_{2} + \frac{1}{2} \, \mathbf{a}_{3} & = & \frac{1}{2}b \, \mathbf{\hat{y}} + z_{2}c \, \mathbf{\hat{z}} & \left(4j\right) & \mbox{H II} \\ 
\mathbf{B}_{4} & = & \left(\frac{1}{2} - z_{2}\right) \, \mathbf{a}_{1}-z_{2} \, \mathbf{a}_{2} + \frac{1}{2} \, \mathbf{a}_{3} & = & \frac{1}{2}b \, \mathbf{\hat{y}}-z_{2}c \, \mathbf{\hat{z}} & \left(4j\right) & \mbox{H II} \\ 
\mathbf{B}_{5} & = & \left(y_{3}+z_{3}\right) \, \mathbf{a}_{1} + \left(x_{3}+z_{3}\right) \, \mathbf{a}_{2} + \left(x_{3}+y_{3}\right) \, \mathbf{a}_{3} & = & x_{3}a \, \mathbf{\hat{x}} + y_{3}b \, \mathbf{\hat{y}} + z_{3}c \, \mathbf{\hat{z}} & \left(8k\right) & \mbox{H III} \\ 
\mathbf{B}_{6} & = & \left(-y_{3}+z_{3}\right) \, \mathbf{a}_{1} + \left(-x_{3}+z_{3}\right) \, \mathbf{a}_{2} + \left(-x_{3}-y_{3}\right) \, \mathbf{a}_{3} & = & -x_{3}a \, \mathbf{\hat{x}}-y_{3}b \, \mathbf{\hat{y}} + z_{3}c \, \mathbf{\hat{z}} & \left(8k\right) & \mbox{H III} \\ 
\mathbf{B}_{7} & = & \left(y_{3}-z_{3}\right) \, \mathbf{a}_{1} + \left(-x_{3}-z_{3}\right) \, \mathbf{a}_{2} + \left(-x_{3}+y_{3}\right) \, \mathbf{a}_{3} & = & -x_{3}a \, \mathbf{\hat{x}} + y_{3}b \, \mathbf{\hat{y}}-z_{3}c \, \mathbf{\hat{z}} & \left(8k\right) & \mbox{H III} \\ 
\mathbf{B}_{8} & = & \left(-y_{3}-z_{3}\right) \, \mathbf{a}_{1} + \left(x_{3}-z_{3}\right) \, \mathbf{a}_{2} + \left(x_{3}-y_{3}\right) \, \mathbf{a}_{3} & = & x_{3}a \, \mathbf{\hat{x}}-y_{3}b \, \mathbf{\hat{y}}-z_{3}c \, \mathbf{\hat{z}} & \left(8k\right) & \mbox{H III} \\ 
\mathbf{B}_{9} & = & \left(y_{4}+z_{4}\right) \, \mathbf{a}_{1} + \left(x_{4}+z_{4}\right) \, \mathbf{a}_{2} + \left(x_{4}+y_{4}\right) \, \mathbf{a}_{3} & = & x_{4}a \, \mathbf{\hat{x}} + y_{4}b \, \mathbf{\hat{y}} + z_{4}c \, \mathbf{\hat{z}} & \left(8k\right) & \mbox{H IV} \\ 
\mathbf{B}_{10} & = & \left(-y_{4}+z_{4}\right) \, \mathbf{a}_{1} + \left(-x_{4}+z_{4}\right) \, \mathbf{a}_{2} + \left(-x_{4}-y_{4}\right) \, \mathbf{a}_{3} & = & -x_{4}a \, \mathbf{\hat{x}}-y_{4}b \, \mathbf{\hat{y}} + z_{4}c \, \mathbf{\hat{z}} & \left(8k\right) & \mbox{H IV} \\ 
\mathbf{B}_{11} & = & \left(y_{4}-z_{4}\right) \, \mathbf{a}_{1} + \left(-x_{4}-z_{4}\right) \, \mathbf{a}_{2} + \left(-x_{4}+y_{4}\right) \, \mathbf{a}_{3} & = & -x_{4}a \, \mathbf{\hat{x}} + y_{4}b \, \mathbf{\hat{y}}-z_{4}c \, \mathbf{\hat{z}} & \left(8k\right) & \mbox{H IV} \\ 
\mathbf{B}_{12} & = & \left(-y_{4}-z_{4}\right) \, \mathbf{a}_{1} + \left(x_{4}-z_{4}\right) \, \mathbf{a}_{2} + \left(x_{4}-y_{4}\right) \, \mathbf{a}_{3} & = & x_{4}a \, \mathbf{\hat{x}}-y_{4}b \, \mathbf{\hat{y}}-z_{4}c \, \mathbf{\hat{z}} & \left(8k\right) & \mbox{H IV} \\ 
\mathbf{B}_{13} & = & \left(y_{5}+z_{5}\right) \, \mathbf{a}_{1} + \left(x_{5}+z_{5}\right) \, \mathbf{a}_{2} + \left(x_{5}+y_{5}\right) \, \mathbf{a}_{3} & = & x_{5}a \, \mathbf{\hat{x}} + y_{5}b \, \mathbf{\hat{y}} + z_{5}c \, \mathbf{\hat{z}} & \left(8k\right) & \mbox{S} \\ 
\mathbf{B}_{14} & = & \left(-y_{5}+z_{5}\right) \, \mathbf{a}_{1} + \left(-x_{5}+z_{5}\right) \, \mathbf{a}_{2} + \left(-x_{5}-y_{5}\right) \, \mathbf{a}_{3} & = & -x_{5}a \, \mathbf{\hat{x}}-y_{5}b \, \mathbf{\hat{y}} + z_{5}c \, \mathbf{\hat{z}} & \left(8k\right) & \mbox{S} \\ 
\mathbf{B}_{15} & = & \left(y_{5}-z_{5}\right) \, \mathbf{a}_{1} + \left(-x_{5}-z_{5}\right) \, \mathbf{a}_{2} + \left(-x_{5}+y_{5}\right) \, \mathbf{a}_{3} & = & -x_{5}a \, \mathbf{\hat{x}} + y_{5}b \, \mathbf{\hat{y}}-z_{5}c \, \mathbf{\hat{z}} & \left(8k\right) & \mbox{S} \\ 
\mathbf{B}_{16} & = & \left(-y_{5}-z_{5}\right) \, \mathbf{a}_{1} + \left(x_{5}-z_{5}\right) \, \mathbf{a}_{2} + \left(x_{5}-y_{5}\right) \, \mathbf{a}_{3} & = & x_{5}a \, \mathbf{\hat{x}}-y_{5}b \, \mathbf{\hat{y}}-z_{5}c \, \mathbf{\hat{z}} & \left(8k\right) & \mbox{S} \\ 
\end{longtabu}
\renewcommand{\arraystretch}{1.0}
\noindent \hrulefill
\\
\textbf{References:}
\vspace*{-0.25cm}
\begin{flushleft}
  - \bibentry{Strobel_2011}. \\
\end{flushleft}
\noindent \hrulefill
\\
\textbf{Geometry files:}
\\
\noindent  - CIF: pp. {\hyperref[A3B_oI32_23_ij2k_k_cif]{\pageref{A3B_oI32_23_ij2k_k_cif}}} \\
\noindent  - POSCAR: pp. {\hyperref[A3B_oI32_23_ij2k_k_poscar]{\pageref{A3B_oI32_23_ij2k_k_poscar}}} \\
\onecolumn
{\phantomsection\label{A8B2C12D2E_oI50_23_bcfk_i_3k_j_a}}
\subsection*{\huge \textbf{{\normalfont \begin{raggedleft}Stannoidite (Cu$_{8}$(Fe,Zn)$_{3}$Sn$_{2}$S$_{12}$) Structure: \end{raggedleft} \\ A8B2C12D2E\_oI50\_23\_bcfk\_i\_3k\_j\_a}}}
\noindent \hrulefill
\vspace*{0.25cm}
\begin{figure}[htp]
  \centering
  \vspace{-1em}
  {\includegraphics[width=1\textwidth]{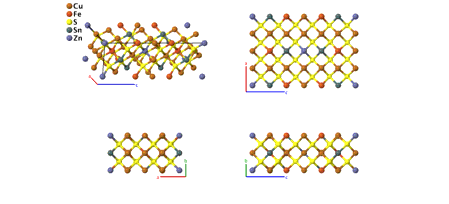}}
\end{figure}
\vspace*{-0.5cm}
\renewcommand{\arraystretch}{1.5}
\begin{equation*}
  \begin{array}{>{$\hspace{-0.15cm}}l<{$}>{$}p{0.5cm}<{$}>{$}p{18.5cm}<{$}}
    \mbox{\large \textbf{Prototype}} &\colon & \ce{Cu8(Fe,Zn)3Sn2S12} \\
    \mbox{\large \textbf{\AFLOW\ prototype label}} &\colon & \mbox{A8B2C12D2E\_oI50\_23\_bcfk\_i\_3k\_j\_a} \\
    \mbox{\large \textbf{\textit{Strukturbericht} designation}} &\colon & \mbox{None} \\
    \mbox{\large \textbf{Pearson symbol}} &\colon & \mbox{oI50} \\
    \mbox{\large \textbf{Space group number}} &\colon & 23 \\
    \mbox{\large \textbf{Space group symbol}} &\colon & I222 \\
    \mbox{\large \textbf{\AFLOW\ prototype command}} &\colon &  \texttt{aflow} \,  \, \texttt{-{}-proto=A8B2C12D2E\_oI50\_23\_bcfk\_i\_3k\_j\_a } \, \newline \texttt{-{}-params=}{a,b/a,c/a,x_{4},z_{5},z_{6},x_{7},y_{7},z_{7},x_{8},y_{8},z_{8},x_{9},y_{9},z_{9},x_{10},y_{10},z_{10} }
  \end{array}
\end{equation*}
\renewcommand{\arraystretch}{1.0}

\vspace*{-0.25cm}
\noindent \hrulefill
\begin{itemize}
  \item{The composition of the Zn (2a) site is actually
Zn$_{0.85}$Fe$_{0.15}$.
}
\end{itemize}

\noindent \parbox{1 \linewidth}{
\noindent \hrulefill
\\
\textbf{Body-centered Orthorhombic primitive vectors:} \\
\vspace*{-0.25cm}
\begin{tabular}{cc}
  \begin{tabular}{c}
    \parbox{0.6 \linewidth}{
      \renewcommand{\arraystretch}{1.5}
      \begin{equation*}
        \centering
        \begin{array}{ccc}
              \mathbf{a}_1 & = & - \frac12 \, a \, \mathbf{\hat{x}} + \frac12 \, b \, \mathbf{\hat{y}} + \frac12 \, c \, \mathbf{\hat{z}} \\
    \mathbf{a}_2 & = & ~ \frac12 \, a \, \mathbf{\hat{x}} - \frac12 \, b \, \mathbf{\hat{y}} + \frac12 \, c \, \mathbf{\hat{z}} \\
    \mathbf{a}_3 & = & ~ \frac12 \, a \, \mathbf{\hat{x}} + \frac12 \, b \, \mathbf{\hat{y}} - \frac12 \, c \, \mathbf{\hat{z}} \\

        \end{array}
      \end{equation*}
    }
    \renewcommand{\arraystretch}{1.0}
  \end{tabular}
  \begin{tabular}{c}
    \includegraphics[width=0.3\linewidth]{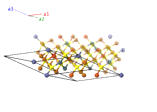} \\
  \end{tabular}
\end{tabular}

}
\vspace*{-0.25cm}

\noindent \hrulefill
\\
\textbf{Basis vectors:}
\vspace*{-0.25cm}
\renewcommand{\arraystretch}{1.5}
\begin{longtabu} to \textwidth{>{\centering $}X[-1,c,c]<{$}>{\centering $}X[-1,c,c]<{$}>{\centering $}X[-1,c,c]<{$}>{\centering $}X[-1,c,c]<{$}>{\centering $}X[-1,c,c]<{$}>{\centering $}X[-1,c,c]<{$}>{\centering $}X[-1,c,c]<{$}}
  & & \mbox{Lattice Coordinates} & & \mbox{Cartesian Coordinates} &\mbox{Wyckoff Position} & \mbox{Atom Type} \\  
  \mathbf{B}_{1} & = & 0 \, \mathbf{a}_{1} + 0 \, \mathbf{a}_{2} + 0 \, \mathbf{a}_{3} & = & 0 \, \mathbf{\hat{x}} + 0 \, \mathbf{\hat{y}} + 0 \, \mathbf{\hat{z}} & \left(2a\right) & \mbox{Zn} \\ 
\mathbf{B}_{2} & = & \frac{1}{2} \, \mathbf{a}_{2} + \frac{1}{2} \, \mathbf{a}_{3} & = & \frac{1}{2}a \, \mathbf{\hat{x}} & \left(2b\right) & \mbox{Cu I} \\ 
\mathbf{B}_{3} & = & \frac{1}{2} \, \mathbf{a}_{1} + \frac{1}{2} \, \mathbf{a}_{2} & = & \frac{1}{2}c \, \mathbf{\hat{z}} & \left(2c\right) & \mbox{Cu II} \\ 
\mathbf{B}_{4} & = & \frac{1}{2} \, \mathbf{a}_{1} + \left(\frac{1}{2} +x_{4}\right) \, \mathbf{a}_{2} + x_{4} \, \mathbf{a}_{3} & = & x_{4}a \, \mathbf{\hat{x}} + \frac{1}{2}c \, \mathbf{\hat{z}} & \left(4f\right) & \mbox{Cu III} \\ 
\mathbf{B}_{5} & = & \frac{1}{2} \, \mathbf{a}_{1} + \left(\frac{1}{2} - x_{4}\right) \, \mathbf{a}_{2}-x_{4} \, \mathbf{a}_{3} & = & -x_{4}a \, \mathbf{\hat{x}} + \frac{1}{2}c \, \mathbf{\hat{z}} & \left(4f\right) & \mbox{Cu III} \\ 
\mathbf{B}_{6} & = & z_{5} \, \mathbf{a}_{1} + z_{5} \, \mathbf{a}_{2} & = & z_{5}c \, \mathbf{\hat{z}} & \left(4i\right) & \mbox{Fe} \\ 
\mathbf{B}_{7} & = & -z_{5} \, \mathbf{a}_{1}-z_{5} \, \mathbf{a}_{2} & = & -z_{5}c \, \mathbf{\hat{z}} & \left(4i\right) & \mbox{Fe} \\ 
\mathbf{B}_{8} & = & \left(\frac{1}{2} +z_{6}\right) \, \mathbf{a}_{1} + z_{6} \, \mathbf{a}_{2} + \frac{1}{2} \, \mathbf{a}_{3} & = & \frac{1}{2}b \, \mathbf{\hat{y}} + z_{6}c \, \mathbf{\hat{z}} & \left(4j\right) & \mbox{Sn} \\ 
\mathbf{B}_{9} & = & \left(\frac{1}{2} - z_{6}\right) \, \mathbf{a}_{1}-z_{6} \, \mathbf{a}_{2} + \frac{1}{2} \, \mathbf{a}_{3} & = & \frac{1}{2}b \, \mathbf{\hat{y}}-z_{6}c \, \mathbf{\hat{z}} & \left(4j\right) & \mbox{Sn} \\ 
\mathbf{B}_{10} & = & \left(y_{7}+z_{7}\right) \, \mathbf{a}_{1} + \left(x_{7}+z_{7}\right) \, \mathbf{a}_{2} + \left(x_{7}+y_{7}\right) \, \mathbf{a}_{3} & = & x_{7}a \, \mathbf{\hat{x}} + y_{7}b \, \mathbf{\hat{y}} + z_{7}c \, \mathbf{\hat{z}} & \left(8k\right) & \mbox{Cu IV} \\ 
\mathbf{B}_{11} & = & \left(-y_{7}+z_{7}\right) \, \mathbf{a}_{1} + \left(-x_{7}+z_{7}\right) \, \mathbf{a}_{2} + \left(-x_{7}-y_{7}\right) \, \mathbf{a}_{3} & = & -x_{7}a \, \mathbf{\hat{x}}-y_{7}b \, \mathbf{\hat{y}} + z_{7}c \, \mathbf{\hat{z}} & \left(8k\right) & \mbox{Cu IV} \\ 
\mathbf{B}_{12} & = & \left(y_{7}-z_{7}\right) \, \mathbf{a}_{1} + \left(-x_{7}-z_{7}\right) \, \mathbf{a}_{2} + \left(-x_{7}+y_{7}\right) \, \mathbf{a}_{3} & = & -x_{7}a \, \mathbf{\hat{x}} + y_{7}b \, \mathbf{\hat{y}}-z_{7}c \, \mathbf{\hat{z}} & \left(8k\right) & \mbox{Cu IV} \\ 
\mathbf{B}_{13} & = & \left(-y_{7}-z_{7}\right) \, \mathbf{a}_{1} + \left(x_{7}-z_{7}\right) \, \mathbf{a}_{2} + \left(x_{7}-y_{7}\right) \, \mathbf{a}_{3} & = & x_{7}a \, \mathbf{\hat{x}}-y_{7}b \, \mathbf{\hat{y}}-z_{7}c \, \mathbf{\hat{z}} & \left(8k\right) & \mbox{Cu IV} \\ 
\mathbf{B}_{14} & = & \left(y_{8}+z_{8}\right) \, \mathbf{a}_{1} + \left(x_{8}+z_{8}\right) \, \mathbf{a}_{2} + \left(x_{8}+y_{8}\right) \, \mathbf{a}_{3} & = & x_{8}a \, \mathbf{\hat{x}} + y_{8}b \, \mathbf{\hat{y}} + z_{8}c \, \mathbf{\hat{z}} & \left(8k\right) & \mbox{S I} \\ 
\mathbf{B}_{15} & = & \left(-y_{8}+z_{8}\right) \, \mathbf{a}_{1} + \left(-x_{8}+z_{8}\right) \, \mathbf{a}_{2} + \left(-x_{8}-y_{8}\right) \, \mathbf{a}_{3} & = & -x_{8}a \, \mathbf{\hat{x}}-y_{8}b \, \mathbf{\hat{y}} + z_{8}c \, \mathbf{\hat{z}} & \left(8k\right) & \mbox{S I} \\ 
\mathbf{B}_{16} & = & \left(y_{8}-z_{8}\right) \, \mathbf{a}_{1} + \left(-x_{8}-z_{8}\right) \, \mathbf{a}_{2} + \left(-x_{8}+y_{8}\right) \, \mathbf{a}_{3} & = & -x_{8}a \, \mathbf{\hat{x}} + y_{8}b \, \mathbf{\hat{y}}-z_{8}c \, \mathbf{\hat{z}} & \left(8k\right) & \mbox{S I} \\ 
\mathbf{B}_{17} & = & \left(-y_{8}-z_{8}\right) \, \mathbf{a}_{1} + \left(x_{8}-z_{8}\right) \, \mathbf{a}_{2} + \left(x_{8}-y_{8}\right) \, \mathbf{a}_{3} & = & x_{8}a \, \mathbf{\hat{x}}-y_{8}b \, \mathbf{\hat{y}}-z_{8}c \, \mathbf{\hat{z}} & \left(8k\right) & \mbox{S I} \\ 
\mathbf{B}_{18} & = & \left(y_{9}+z_{9}\right) \, \mathbf{a}_{1} + \left(x_{9}+z_{9}\right) \, \mathbf{a}_{2} + \left(x_{9}+y_{9}\right) \, \mathbf{a}_{3} & = & x_{9}a \, \mathbf{\hat{x}} + y_{9}b \, \mathbf{\hat{y}} + z_{9}c \, \mathbf{\hat{z}} & \left(8k\right) & \mbox{S II} \\ 
\mathbf{B}_{19} & = & \left(-y_{9}+z_{9}\right) \, \mathbf{a}_{1} + \left(-x_{9}+z_{9}\right) \, \mathbf{a}_{2} + \left(-x_{9}-y_{9}\right) \, \mathbf{a}_{3} & = & -x_{9}a \, \mathbf{\hat{x}}-y_{9}b \, \mathbf{\hat{y}} + z_{9}c \, \mathbf{\hat{z}} & \left(8k\right) & \mbox{S II} \\ 
\mathbf{B}_{20} & = & \left(y_{9}-z_{9}\right) \, \mathbf{a}_{1} + \left(-x_{9}-z_{9}\right) \, \mathbf{a}_{2} + \left(-x_{9}+y_{9}\right) \, \mathbf{a}_{3} & = & -x_{9}a \, \mathbf{\hat{x}} + y_{9}b \, \mathbf{\hat{y}}-z_{9}c \, \mathbf{\hat{z}} & \left(8k\right) & \mbox{S II} \\ 
\mathbf{B}_{21} & = & \left(-y_{9}-z_{9}\right) \, \mathbf{a}_{1} + \left(x_{9}-z_{9}\right) \, \mathbf{a}_{2} + \left(x_{9}-y_{9}\right) \, \mathbf{a}_{3} & = & x_{9}a \, \mathbf{\hat{x}}-y_{9}b \, \mathbf{\hat{y}}-z_{9}c \, \mathbf{\hat{z}} & \left(8k\right) & \mbox{S II} \\ 
\mathbf{B}_{22} & = & \left(y_{10}+z_{10}\right) \, \mathbf{a}_{1} + \left(x_{10}+z_{10}\right) \, \mathbf{a}_{2} + \left(x_{10}+y_{10}\right) \, \mathbf{a}_{3} & = & x_{10}a \, \mathbf{\hat{x}} + y_{10}b \, \mathbf{\hat{y}} + z_{10}c \, \mathbf{\hat{z}} & \left(8k\right) & \mbox{S III} \\ 
\mathbf{B}_{23} & = & \left(-y_{10}+z_{10}\right) \, \mathbf{a}_{1} + \left(-x_{10}+z_{10}\right) \, \mathbf{a}_{2} + \left(-x_{10}-y_{10}\right) \, \mathbf{a}_{3} & = & -x_{10}a \, \mathbf{\hat{x}}-y_{10}b \, \mathbf{\hat{y}} + z_{10}c \, \mathbf{\hat{z}} & \left(8k\right) & \mbox{S III} \\ 
\mathbf{B}_{24} & = & \left(y_{10}-z_{10}\right) \, \mathbf{a}_{1} + \left(-x_{10}-z_{10}\right) \, \mathbf{a}_{2} + \left(-x_{10}+y_{10}\right) \, \mathbf{a}_{3} & = & -x_{10}a \, \mathbf{\hat{x}} + y_{10}b \, \mathbf{\hat{y}}-z_{10}c \, \mathbf{\hat{z}} & \left(8k\right) & \mbox{S III} \\ 
\mathbf{B}_{25} & = & \left(-y_{10}-z_{10}\right) \, \mathbf{a}_{1} + \left(x_{10}-z_{10}\right) \, \mathbf{a}_{2} + \left(x_{10}-y_{10}\right) \, \mathbf{a}_{3} & = & x_{10}a \, \mathbf{\hat{x}}-y_{10}b \, \mathbf{\hat{y}}-z_{10}c \, \mathbf{\hat{z}} & \left(8k\right) & \mbox{S III} \\ 
\end{longtabu}
\renewcommand{\arraystretch}{1.0}
\noindent \hrulefill
\\
\textbf{References:}
\vspace*{-0.25cm}
\begin{flushleft}
  - \bibentry{Kudoh_ZKrist_144_1976}. \\
\end{flushleft}
\noindent \hrulefill
\\
\textbf{Geometry files:}
\\
\noindent  - CIF: pp. {\hyperref[A8B2C12D2E_oI50_23_bcfk_i_3k_j_a_cif]{\pageref{A8B2C12D2E_oI50_23_bcfk_i_3k_j_a_cif}}} \\
\noindent  - POSCAR: pp. {\hyperref[A8B2C12D2E_oI50_23_bcfk_i_3k_j_a_poscar]{\pageref{A8B2C12D2E_oI50_23_bcfk_i_3k_j_a_poscar}}} \\
\onecolumn
{\phantomsection\label{ABC2_oI16_23_ab_i_k}}
\subsection*{\huge \textbf{{\normalfont NaFeS$_{2}$ Structure: ABC2\_oI16\_23\_ab\_i\_k}}}
\noindent \hrulefill
\vspace*{0.25cm}
\begin{figure}[htp]
  \centering
  \vspace{-1em}
  {\includegraphics[width=1\textwidth]{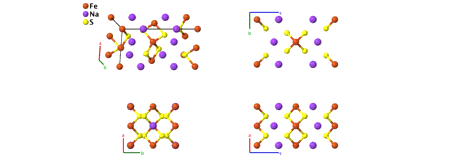}}
\end{figure}
\vspace*{-0.5cm}
\renewcommand{\arraystretch}{1.5}
\begin{equation*}
  \begin{array}{>{$\hspace{-0.15cm}}l<{$}>{$}p{0.5cm}<{$}>{$}p{18.5cm}<{$}}
    \mbox{\large \textbf{Prototype}} &\colon & \ce{NaFeS2} \\
    \mbox{\large \textbf{\AFLOW\ prototype label}} &\colon & \mbox{ABC2\_oI16\_23\_ab\_i\_k} \\
    \mbox{\large \textbf{\textit{Strukturbericht} designation}} &\colon & \mbox{None} \\
    \mbox{\large \textbf{Pearson symbol}} &\colon & \mbox{oI16} \\
    \mbox{\large \textbf{Space group number}} &\colon & 23 \\
    \mbox{\large \textbf{Space group symbol}} &\colon & I222 \\
    \mbox{\large \textbf{\AFLOW\ prototype command}} &\colon &  \texttt{aflow} \,  \, \texttt{-{}-proto=ABC2\_oI16\_23\_ab\_i\_k } \, \newline \texttt{-{}-params=}{a,b/a,c/a,z_{3},x_{4},y_{4},z_{4} }
  \end{array}
\end{equation*}
\renewcommand{\arraystretch}{1.0}

\noindent \parbox{1 \linewidth}{
\noindent \hrulefill
\\
\textbf{Body-centered Orthorhombic primitive vectors:} \\
\vspace*{-0.25cm}
\begin{tabular}{cc}
  \begin{tabular}{c}
    \parbox{0.6 \linewidth}{
      \renewcommand{\arraystretch}{1.5}
      \begin{equation*}
        \centering
        \begin{array}{ccc}
              \mathbf{a}_1 & = & - \frac12 \, a \, \mathbf{\hat{x}} + \frac12 \, b \, \mathbf{\hat{y}} + \frac12 \, c \, \mathbf{\hat{z}} \\
    \mathbf{a}_2 & = & ~ \frac12 \, a \, \mathbf{\hat{x}} - \frac12 \, b \, \mathbf{\hat{y}} + \frac12 \, c \, \mathbf{\hat{z}} \\
    \mathbf{a}_3 & = & ~ \frac12 \, a \, \mathbf{\hat{x}} + \frac12 \, b \, \mathbf{\hat{y}} - \frac12 \, c \, \mathbf{\hat{z}} \\

        \end{array}
      \end{equation*}
    }
    \renewcommand{\arraystretch}{1.0}
  \end{tabular}
  \begin{tabular}{c}
    \includegraphics[width=0.3\linewidth]{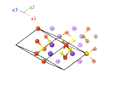} \\
  \end{tabular}
\end{tabular}

}
\vspace*{-0.25cm}

\noindent \hrulefill
\\
\textbf{Basis vectors:}
\vspace*{-0.25cm}
\renewcommand{\arraystretch}{1.5}
\begin{longtabu} to \textwidth{>{\centering $}X[-1,c,c]<{$}>{\centering $}X[-1,c,c]<{$}>{\centering $}X[-1,c,c]<{$}>{\centering $}X[-1,c,c]<{$}>{\centering $}X[-1,c,c]<{$}>{\centering $}X[-1,c,c]<{$}>{\centering $}X[-1,c,c]<{$}}
  & & \mbox{Lattice Coordinates} & & \mbox{Cartesian Coordinates} &\mbox{Wyckoff Position} & \mbox{Atom Type} \\  
  \mathbf{B}_{1} & = & 0 \, \mathbf{a}_{1} + 0 \, \mathbf{a}_{2} + 0 \, \mathbf{a}_{3} & = & 0 \, \mathbf{\hat{x}} + 0 \, \mathbf{\hat{y}} + 0 \, \mathbf{\hat{z}} & \left(2a\right) & \mbox{Fe I} \\ 
\mathbf{B}_{2} & = & \frac{1}{2} \, \mathbf{a}_{2} + \frac{1}{2} \, \mathbf{a}_{3} & = & \frac{1}{2}a \, \mathbf{\hat{x}} & \left(2b\right) & \mbox{Fe II} \\ 
\mathbf{B}_{3} & = & z_{3} \, \mathbf{a}_{1} + z_{3} \, \mathbf{a}_{2} & = & z_{3}c \, \mathbf{\hat{z}} & \left(4i\right) & \mbox{Na} \\ 
\mathbf{B}_{4} & = & -z_{3} \, \mathbf{a}_{1}-z_{3} \, \mathbf{a}_{2} & = & -z_{3}c \, \mathbf{\hat{z}} & \left(4i\right) & \mbox{Na} \\ 
\mathbf{B}_{5} & = & \left(y_{4}+z_{4}\right) \, \mathbf{a}_{1} + \left(x_{4}+z_{4}\right) \, \mathbf{a}_{2} + \left(x_{4}+y_{4}\right) \, \mathbf{a}_{3} & = & x_{4}a \, \mathbf{\hat{x}} + y_{4}b \, \mathbf{\hat{y}} + z_{4}c \, \mathbf{\hat{z}} & \left(8k\right) & \mbox{S} \\ 
\mathbf{B}_{6} & = & \left(-y_{4}+z_{4}\right) \, \mathbf{a}_{1} + \left(-x_{4}+z_{4}\right) \, \mathbf{a}_{2} + \left(-x_{4}-y_{4}\right) \, \mathbf{a}_{3} & = & -x_{4}a \, \mathbf{\hat{x}}-y_{4}b \, \mathbf{\hat{y}} + z_{4}c \, \mathbf{\hat{z}} & \left(8k\right) & \mbox{S} \\ 
\mathbf{B}_{7} & = & \left(y_{4}-z_{4}\right) \, \mathbf{a}_{1} + \left(-x_{4}-z_{4}\right) \, \mathbf{a}_{2} + \left(-x_{4}+y_{4}\right) \, \mathbf{a}_{3} & = & -x_{4}a \, \mathbf{\hat{x}} + y_{4}b \, \mathbf{\hat{y}}-z_{4}c \, \mathbf{\hat{z}} & \left(8k\right) & \mbox{S} \\ 
\mathbf{B}_{8} & = & \left(-y_{4}-z_{4}\right) \, \mathbf{a}_{1} + \left(x_{4}-z_{4}\right) \, \mathbf{a}_{2} + \left(x_{4}-y_{4}\right) \, \mathbf{a}_{3} & = & x_{4}a \, \mathbf{\hat{x}}-y_{4}b \, \mathbf{\hat{y}}-z_{4}c \, \mathbf{\hat{z}} & \left(8k\right) & \mbox{S} \\ 
\end{longtabu}
\renewcommand{\arraystretch}{1.0}
\noindent \hrulefill
\\
\textbf{References:}
\vspace*{-0.25cm}
\begin{flushleft}
  - \bibentry{Boller_NaFeS2_MonatChemMo_1983}. \\
\end{flushleft}
\textbf{Found in:}
\vspace*{-0.25cm}
\begin{flushleft}
  - \bibentry{Villars_PearsonsCrystalData_2013}. \\
\end{flushleft}
\noindent \hrulefill
\\
\textbf{Geometry files:}
\\
\noindent  - CIF: pp. {\hyperref[ABC2_oI16_23_ab_i_k_cif]{\pageref{ABC2_oI16_23_ab_i_k_cif}}} \\
\noindent  - POSCAR: pp. {\hyperref[ABC2_oI16_23_ab_i_k_poscar]{\pageref{ABC2_oI16_23_ab_i_k_poscar}}} \\
\onecolumn
{\phantomsection\label{ABC4_oI12_23_a_b_k}}
\subsection*{\huge \textbf{{\normalfont BPS$_{4}$ Structure: ABC4\_oI12\_23\_a\_b\_k}}}
\noindent \hrulefill
\vspace*{0.25cm}
\begin{figure}[htp]
  \centering
  \vspace{-1em}
  {\includegraphics[width=1\textwidth]{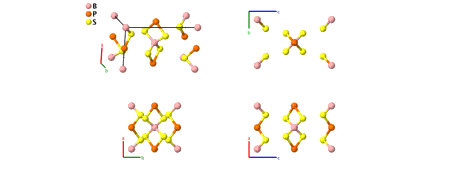}}
\end{figure}
\vspace*{-0.5cm}
\renewcommand{\arraystretch}{1.5}
\begin{equation*}
  \begin{array}{>{$\hspace{-0.15cm}}l<{$}>{$}p{0.5cm}<{$}>{$}p{18.5cm}<{$}}
    \mbox{\large \textbf{Prototype}} &\colon & \ce{BPS4} \\
    \mbox{\large \textbf{\AFLOW\ prototype label}} &\colon & \mbox{ABC4\_oI12\_23\_a\_b\_k} \\
    \mbox{\large \textbf{\textit{Strukturbericht} designation}} &\colon & \mbox{None} \\
    \mbox{\large \textbf{Pearson symbol}} &\colon & \mbox{oI12} \\
    \mbox{\large \textbf{Space group number}} &\colon & 23 \\
    \mbox{\large \textbf{Space group symbol}} &\colon & I222 \\
    \mbox{\large \textbf{\AFLOW\ prototype command}} &\colon &  \texttt{aflow} \,  \, \texttt{-{}-proto=ABC4\_oI12\_23\_a\_b\_k } \, \newline \texttt{-{}-params=}{a,b/a,c/a,x_{3},y_{3},z_{3} }
  \end{array}
\end{equation*}
\renewcommand{\arraystretch}{1.0}

\noindent \parbox{1 \linewidth}{
\noindent \hrulefill
\\
\textbf{Body-centered Orthorhombic primitive vectors:} \\
\vspace*{-0.25cm}
\begin{tabular}{cc}
  \begin{tabular}{c}
    \parbox{0.6 \linewidth}{
      \renewcommand{\arraystretch}{1.5}
      \begin{equation*}
        \centering
        \begin{array}{ccc}
              \mathbf{a}_1 & = & - \frac12 \, a \, \mathbf{\hat{x}} + \frac12 \, b \, \mathbf{\hat{y}} + \frac12 \, c \, \mathbf{\hat{z}} \\
    \mathbf{a}_2 & = & ~ \frac12 \, a \, \mathbf{\hat{x}} - \frac12 \, b \, \mathbf{\hat{y}} + \frac12 \, c \, \mathbf{\hat{z}} \\
    \mathbf{a}_3 & = & ~ \frac12 \, a \, \mathbf{\hat{x}} + \frac12 \, b \, \mathbf{\hat{y}} - \frac12 \, c \, \mathbf{\hat{z}} \\

        \end{array}
      \end{equation*}
    }
    \renewcommand{\arraystretch}{1.0}
  \end{tabular}
  \begin{tabular}{c}
    \includegraphics[width=0.3\linewidth]{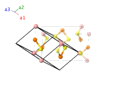} \\
  \end{tabular}
\end{tabular}

}
\vspace*{-0.25cm}

\noindent \hrulefill
\\
\textbf{Basis vectors:}
\vspace*{-0.25cm}
\renewcommand{\arraystretch}{1.5}
\begin{longtabu} to \textwidth{>{\centering $}X[-1,c,c]<{$}>{\centering $}X[-1,c,c]<{$}>{\centering $}X[-1,c,c]<{$}>{\centering $}X[-1,c,c]<{$}>{\centering $}X[-1,c,c]<{$}>{\centering $}X[-1,c,c]<{$}>{\centering $}X[-1,c,c]<{$}}
  & & \mbox{Lattice Coordinates} & & \mbox{Cartesian Coordinates} &\mbox{Wyckoff Position} & \mbox{Atom Type} \\  
  \mathbf{B}_{1} & = & 0 \, \mathbf{a}_{1} + 0 \, \mathbf{a}_{2} + 0 \, \mathbf{a}_{3} & = & 0 \, \mathbf{\hat{x}} + 0 \, \mathbf{\hat{y}} + 0 \, \mathbf{\hat{z}} & \left(2a\right) & \mbox{B} \\ 
\mathbf{B}_{2} & = & \frac{1}{2} \, \mathbf{a}_{2} + \frac{1}{2} \, \mathbf{a}_{3} & = & \frac{1}{2}a \, \mathbf{\hat{x}} & \left(2b\right) & \mbox{P} \\ 
\mathbf{B}_{3} & = & \left(y_{3}+z_{3}\right) \, \mathbf{a}_{1} + \left(x_{3}+z_{3}\right) \, \mathbf{a}_{2} + \left(x_{3}+y_{3}\right) \, \mathbf{a}_{3} & = & x_{3}a \, \mathbf{\hat{x}} + y_{3}b \, \mathbf{\hat{y}} + z_{3}c \, \mathbf{\hat{z}} & \left(8k\right) & \mbox{S} \\ 
\mathbf{B}_{4} & = & \left(-y_{3}+z_{3}\right) \, \mathbf{a}_{1} + \left(-x_{3}+z_{3}\right) \, \mathbf{a}_{2} + \left(-x_{3}-y_{3}\right) \, \mathbf{a}_{3} & = & -x_{3}a \, \mathbf{\hat{x}}-y_{3}b \, \mathbf{\hat{y}} + z_{3}c \, \mathbf{\hat{z}} & \left(8k\right) & \mbox{S} \\ 
\mathbf{B}_{5} & = & \left(y_{3}-z_{3}\right) \, \mathbf{a}_{1} + \left(-x_{3}-z_{3}\right) \, \mathbf{a}_{2} + \left(-x_{3}+y_{3}\right) \, \mathbf{a}_{3} & = & -x_{3}a \, \mathbf{\hat{x}} + y_{3}b \, \mathbf{\hat{y}}-z_{3}c \, \mathbf{\hat{z}} & \left(8k\right) & \mbox{S} \\ 
\mathbf{B}_{6} & = & \left(-y_{3}-z_{3}\right) \, \mathbf{a}_{1} + \left(x_{3}-z_{3}\right) \, \mathbf{a}_{2} + \left(x_{3}-y_{3}\right) \, \mathbf{a}_{3} & = & x_{3}a \, \mathbf{\hat{x}}-y_{3}b \, \mathbf{\hat{y}}-z_{3}c \, \mathbf{\hat{z}} & \left(8k\right) & \mbox{S} \\ 
\end{longtabu}
\renewcommand{\arraystretch}{1.0}
\noindent \hrulefill
\\
\textbf{References:}
\vspace*{-0.25cm}
\begin{flushleft}
  - \bibentry{Weiss_BPS4_ZNaturB_1963}. \\
\end{flushleft}
\textbf{Found in:}
\vspace*{-0.25cm}
\begin{flushleft}
  - \bibentry{Villars_PearsonsCrystalData_2013}. \\
\end{flushleft}
\noindent \hrulefill
\\
\textbf{Geometry files:}
\\
\noindent  - CIF: pp. {\hyperref[ABC4_oI12_23_a_b_k_cif]{\pageref{ABC4_oI12_23_a_b_k_cif}}} \\
\noindent  - POSCAR: pp. {\hyperref[ABC4_oI12_23_a_b_k_poscar]{\pageref{ABC4_oI12_23_a_b_k_poscar}}} \\
\onecolumn
{\phantomsection\label{AB7CD2_oI44_24_a_b3d_c_ac}}
\subsection*{\huge \textbf{{\normalfont \begin{raggedleft}Weberite (Na$_{2}$MgAlF$_{7}$) Structure: \end{raggedleft} \\ AB7CD2\_oI44\_24\_a\_b3d\_c\_ac}}}
\noindent \hrulefill
\vspace*{0.25cm}
\begin{figure}[htp]
  \centering
  \vspace{-1em}
  {\includegraphics[width=1\textwidth]{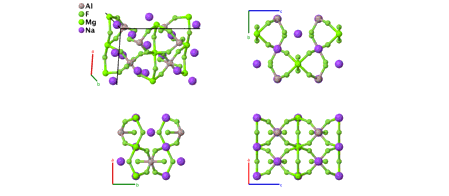}}
\end{figure}
\vspace*{-0.5cm}
\renewcommand{\arraystretch}{1.5}
\begin{equation*}
  \begin{array}{>{$\hspace{-0.15cm}}l<{$}>{$}p{0.5cm}<{$}>{$}p{18.5cm}<{$}}
    \mbox{\large \textbf{Prototype}} &\colon & \ce{Na2MgAlF7} \\
    \mbox{\large \textbf{\AFLOW\ prototype label}} &\colon & \mbox{AB7CD2\_oI44\_24\_a\_b3d\_c\_ac} \\
    \mbox{\large \textbf{\textit{Strukturbericht} designation}} &\colon & \mbox{None} \\
    \mbox{\large \textbf{Pearson symbol}} &\colon & \mbox{oI44} \\
    \mbox{\large \textbf{Space group number}} &\colon & 24 \\
    \mbox{\large \textbf{Space group symbol}} &\colon & I2_{1}2_{1}2_{1} \\
    \mbox{\large \textbf{\AFLOW\ prototype command}} &\colon &  \texttt{aflow} \,  \, \texttt{-{}-proto=AB7CD2\_oI44\_24\_a\_b3d\_c\_ac } \, \newline \texttt{-{}-params=}{a,b/a,c/a,x_{1},x_{2},y_{3},z_{4},z_{5},x_{6},y_{6},z_{6},x_{7},y_{7},z_{7},x_{8},y_{8},z_{8} }
  \end{array}
\end{equation*}
\renewcommand{\arraystretch}{1.0}

\noindent \parbox{1 \linewidth}{
\noindent \hrulefill
\\
\textbf{Body-centered Orthorhombic primitive vectors:} \\
\vspace*{-0.25cm}
\begin{tabular}{cc}
  \begin{tabular}{c}
    \parbox{0.6 \linewidth}{
      \renewcommand{\arraystretch}{1.5}
      \begin{equation*}
        \centering
        \begin{array}{ccc}
              \mathbf{a}_1 & = & - \frac12 \, a \, \mathbf{\hat{x}} + \frac12 \, b \, \mathbf{\hat{y}} + \frac12 \, c \, \mathbf{\hat{z}} \\
    \mathbf{a}_2 & = & ~ \frac12 \, a \, \mathbf{\hat{x}} - \frac12 \, b \, \mathbf{\hat{y}} + \frac12 \, c \, \mathbf{\hat{z}} \\
    \mathbf{a}_3 & = & ~ \frac12 \, a \, \mathbf{\hat{x}} + \frac12 \, b \, \mathbf{\hat{y}} - \frac12 \, c \, \mathbf{\hat{z}} \\

        \end{array}
      \end{equation*}
    }
    \renewcommand{\arraystretch}{1.0}
  \end{tabular}
  \begin{tabular}{c}
    \includegraphics[width=0.3\linewidth]{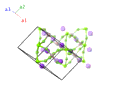} \\
  \end{tabular}
\end{tabular}

}
\vspace*{-0.25cm}

\noindent \hrulefill
\\
\textbf{Basis vectors:}
\vspace*{-0.25cm}
\renewcommand{\arraystretch}{1.5}
\begin{longtabu} to \textwidth{>{\centering $}X[-1,c,c]<{$}>{\centering $}X[-1,c,c]<{$}>{\centering $}X[-1,c,c]<{$}>{\centering $}X[-1,c,c]<{$}>{\centering $}X[-1,c,c]<{$}>{\centering $}X[-1,c,c]<{$}>{\centering $}X[-1,c,c]<{$}}
  & & \mbox{Lattice Coordinates} & & \mbox{Cartesian Coordinates} &\mbox{Wyckoff Position} & \mbox{Atom Type} \\  
  \mathbf{B}_{1} & = & \frac{1}{4} \, \mathbf{a}_{1} + \left(\frac{1}{4} +x_{1}\right) \, \mathbf{a}_{2} + x_{1} \, \mathbf{a}_{3} & = & x_{1}a \, \mathbf{\hat{x}} + \frac{1}{4}c \, \mathbf{\hat{z}} & \left(4a\right) & \mbox{Al} \\ 
\mathbf{B}_{2} & = & \frac{3}{4} \, \mathbf{a}_{1} + \left(\frac{1}{4} - x_{1}\right) \, \mathbf{a}_{2} + \left(\frac{1}{2} - x_{1}\right) \, \mathbf{a}_{3} & = & -x_{1}a \, \mathbf{\hat{x}} + \frac{1}{2}b \, \mathbf{\hat{y}} + \frac{1}{4}c \, \mathbf{\hat{z}} & \left(4a\right) & \mbox{Al} \\ 
\mathbf{B}_{3} & = & \frac{1}{4} \, \mathbf{a}_{1} + \left(\frac{1}{4} +x_{2}\right) \, \mathbf{a}_{2} + x_{2} \, \mathbf{a}_{3} & = & x_{2}a \, \mathbf{\hat{x}} + \frac{1}{4}c \, \mathbf{\hat{z}} & \left(4a\right) & \mbox{Na I} \\ 
\mathbf{B}_{4} & = & \frac{3}{4} \, \mathbf{a}_{1} + \left(\frac{1}{4} - x_{2}\right) \, \mathbf{a}_{2} + \left(\frac{1}{2} - x_{2}\right) \, \mathbf{a}_{3} & = & -x_{2}a \, \mathbf{\hat{x}} + \frac{1}{2}b \, \mathbf{\hat{y}} + \frac{1}{4}c \, \mathbf{\hat{z}} & \left(4a\right) & \mbox{Na I} \\ 
\mathbf{B}_{5} & = & y_{3} \, \mathbf{a}_{1} + \frac{1}{4} \, \mathbf{a}_{2} + \left(\frac{1}{4} +y_{3}\right) \, \mathbf{a}_{3} & = & \frac{1}{4}a \, \mathbf{\hat{x}} + y_{3}b \, \mathbf{\hat{y}} & \left(4b\right) & \mbox{F I} \\ 
\mathbf{B}_{6} & = & \left(\frac{1}{2} - y_{3}\right) \, \mathbf{a}_{1} + \frac{3}{4} \, \mathbf{a}_{2} + \left(\frac{1}{4} - y_{3}\right) \, \mathbf{a}_{3} & = & \frac{1}{4}a \, \mathbf{\hat{x}}-y_{3}b \, \mathbf{\hat{y}} + \frac{1}{2}c \, \mathbf{\hat{z}} & \left(4b\right) & \mbox{F I} \\ 
\mathbf{B}_{7} & = & \left(\frac{1}{4} +z_{4}\right) \, \mathbf{a}_{1} + z_{4} \, \mathbf{a}_{2} + \frac{1}{4} \, \mathbf{a}_{3} & = & \frac{1}{4}b \, \mathbf{\hat{y}} + z_{4}c \, \mathbf{\hat{z}} & \left(4c\right) & \mbox{Mg} \\ 
\mathbf{B}_{8} & = & \left(\frac{1}{4} - z_{4}\right) \, \mathbf{a}_{1} + \left(\frac{1}{2} - z_{4}\right) \, \mathbf{a}_{2} + \frac{3}{4} \, \mathbf{a}_{3} & = & \frac{1}{2}a \, \mathbf{\hat{x}} + \frac{1}{4}b \, \mathbf{\hat{y}}-z_{4}c \, \mathbf{\hat{z}} & \left(4c\right) & \mbox{Mg} \\ 
\mathbf{B}_{9} & = & \left(\frac{1}{4} +z_{5}\right) \, \mathbf{a}_{1} + z_{5} \, \mathbf{a}_{2} + \frac{1}{4} \, \mathbf{a}_{3} & = & \frac{1}{4}b \, \mathbf{\hat{y}} + z_{5}c \, \mathbf{\hat{z}} & \left(4c\right) & \mbox{Na II} \\ 
\mathbf{B}_{10} & = & \left(\frac{1}{4} - z_{5}\right) \, \mathbf{a}_{1} + \left(\frac{1}{2} - z_{5}\right) \, \mathbf{a}_{2} + \frac{3}{4} \, \mathbf{a}_{3} & = & \frac{1}{2}a \, \mathbf{\hat{x}} + \frac{1}{4}b \, \mathbf{\hat{y}}-z_{5}c \, \mathbf{\hat{z}} & \left(4c\right) & \mbox{Na II} \\ 
\mathbf{B}_{11} & = & \left(y_{6}+z_{6}\right) \, \mathbf{a}_{1} + \left(x_{6}+z_{6}\right) \, \mathbf{a}_{2} + \left(x_{6}+y_{6}\right) \, \mathbf{a}_{3} & = & x_{6}a \, \mathbf{\hat{x}} + y_{6}b \, \mathbf{\hat{y}} + z_{6}c \, \mathbf{\hat{z}} & \left(8d\right) & \mbox{F II} \\ 
\mathbf{B}_{12} & = & \left(\frac{1}{2} - y_{6} + z_{6}\right) \, \mathbf{a}_{1} + \left(-x_{6}+z_{6}\right) \, \mathbf{a}_{2} + \left(\frac{1}{2} - x_{6} - y_{6}\right) \, \mathbf{a}_{3} & = & -x_{6}a \, \mathbf{\hat{x}} + \left(\frac{1}{2} - y_{6}\right)b \, \mathbf{\hat{y}} + z_{6}c \, \mathbf{\hat{z}} & \left(8d\right) & \mbox{F II} \\ 
\mathbf{B}_{13} & = & \left(y_{6}-z_{6}\right) \, \mathbf{a}_{1} + \left(\frac{1}{2} - x_{6} - z_{6}\right) \, \mathbf{a}_{2} + \left(\frac{1}{2} - x_{6} + y_{6}\right) \, \mathbf{a}_{3} & = & \left(\frac{1}{2} - x_{6}\right)a \, \mathbf{\hat{x}} + y_{6}b \, \mathbf{\hat{y}}-z_{6}c \, \mathbf{\hat{z}} & \left(8d\right) & \mbox{F II} \\ 
\mathbf{B}_{14} & = & \left(\frac{1}{2} - y_{6} - z_{6}\right) \, \mathbf{a}_{1} + \left(\frac{1}{2} +x_{6} - z_{6}\right) \, \mathbf{a}_{2} + \left(x_{6}-y_{6}\right) \, \mathbf{a}_{3} & = & x_{6}a \, \mathbf{\hat{x}}-y_{6}b \, \mathbf{\hat{y}} + \left(\frac{1}{2} - z_{6}\right)c \, \mathbf{\hat{z}} & \left(8d\right) & \mbox{F II} \\ 
\mathbf{B}_{15} & = & \left(y_{7}+z_{7}\right) \, \mathbf{a}_{1} + \left(x_{7}+z_{7}\right) \, \mathbf{a}_{2} + \left(x_{7}+y_{7}\right) \, \mathbf{a}_{3} & = & x_{7}a \, \mathbf{\hat{x}} + y_{7}b \, \mathbf{\hat{y}} + z_{7}c \, \mathbf{\hat{z}} & \left(8d\right) & \mbox{F III} \\ 
\mathbf{B}_{16} & = & \left(\frac{1}{2} - y_{7} + z_{7}\right) \, \mathbf{a}_{1} + \left(-x_{7}+z_{7}\right) \, \mathbf{a}_{2} + \left(\frac{1}{2} - x_{7} - y_{7}\right) \, \mathbf{a}_{3} & = & -x_{7}a \, \mathbf{\hat{x}} + \left(\frac{1}{2} - y_{7}\right)b \, \mathbf{\hat{y}} + z_{7}c \, \mathbf{\hat{z}} & \left(8d\right) & \mbox{F III} \\ 
\mathbf{B}_{17} & = & \left(y_{7}-z_{7}\right) \, \mathbf{a}_{1} + \left(\frac{1}{2} - x_{7} - z_{7}\right) \, \mathbf{a}_{2} + \left(\frac{1}{2} - x_{7} + y_{7}\right) \, \mathbf{a}_{3} & = & \left(\frac{1}{2} - x_{7}\right)a \, \mathbf{\hat{x}} + y_{7}b \, \mathbf{\hat{y}}-z_{7}c \, \mathbf{\hat{z}} & \left(8d\right) & \mbox{F III} \\ 
\mathbf{B}_{18} & = & \left(\frac{1}{2} - y_{7} - z_{7}\right) \, \mathbf{a}_{1} + \left(\frac{1}{2} +x_{7} - z_{7}\right) \, \mathbf{a}_{2} + \left(x_{7}-y_{7}\right) \, \mathbf{a}_{3} & = & x_{7}a \, \mathbf{\hat{x}}-y_{7}b \, \mathbf{\hat{y}} + \left(\frac{1}{2} - z_{7}\right)c \, \mathbf{\hat{z}} & \left(8d\right) & \mbox{F III} \\ 
\mathbf{B}_{19} & = & \left(y_{8}+z_{8}\right) \, \mathbf{a}_{1} + \left(x_{8}+z_{8}\right) \, \mathbf{a}_{2} + \left(x_{8}+y_{8}\right) \, \mathbf{a}_{3} & = & x_{8}a \, \mathbf{\hat{x}} + y_{8}b \, \mathbf{\hat{y}} + z_{8}c \, \mathbf{\hat{z}} & \left(8d\right) & \mbox{F IV} \\ 
\mathbf{B}_{20} & = & \left(\frac{1}{2} - y_{8} + z_{8}\right) \, \mathbf{a}_{1} + \left(-x_{8}+z_{8}\right) \, \mathbf{a}_{2} + \left(\frac{1}{2} - x_{8} - y_{8}\right) \, \mathbf{a}_{3} & = & -x_{8}a \, \mathbf{\hat{x}} + \left(\frac{1}{2} - y_{8}\right)b \, \mathbf{\hat{y}} + z_{8}c \, \mathbf{\hat{z}} & \left(8d\right) & \mbox{F IV} \\ 
\mathbf{B}_{21} & = & \left(y_{8}-z_{8}\right) \, \mathbf{a}_{1} + \left(\frac{1}{2} - x_{8} - z_{8}\right) \, \mathbf{a}_{2} + \left(\frac{1}{2} - x_{8} + y_{8}\right) \, \mathbf{a}_{3} & = & \left(\frac{1}{2} - x_{8}\right)a \, \mathbf{\hat{x}} + y_{8}b \, \mathbf{\hat{y}}-z_{8}c \, \mathbf{\hat{z}} & \left(8d\right) & \mbox{F IV} \\ 
\mathbf{B}_{22} & = & \left(\frac{1}{2} - y_{8} - z_{8}\right) \, \mathbf{a}_{1} + \left(\frac{1}{2} +x_{8} - z_{8}\right) \, \mathbf{a}_{2} + \left(x_{8}-y_{8}\right) \, \mathbf{a}_{3} & = & x_{8}a \, \mathbf{\hat{x}}-y_{8}b \, \mathbf{\hat{y}} + \left(\frac{1}{2} - z_{8}\right)c \, \mathbf{\hat{z}} & \left(8d\right) & \mbox{F IV} \\ 
\end{longtabu}
\renewcommand{\arraystretch}{1.0}
\noindent \hrulefill
\\
\textbf{References:}
\vspace*{-0.25cm}
\begin{flushleft}
  - \bibentry{Knop_Na2MgAlF7_JSolStateChem_1982}. \\
\end{flushleft}
\textbf{Found in:}
\vspace*{-0.25cm}
\begin{flushleft}
  - \bibentry{Villars_PearsonsCrystalData_2013}. \\
\end{flushleft}
\noindent \hrulefill
\\
\textbf{Geometry files:}
\\
\noindent  - CIF: pp. {\hyperref[AB7CD2_oI44_24_a_b3d_c_ac_cif]{\pageref{AB7CD2_oI44_24_a_b3d_c_ac_cif}}} \\
\noindent  - POSCAR: pp. {\hyperref[AB7CD2_oI44_24_a_b3d_c_ac_poscar]{\pageref{AB7CD2_oI44_24_a_b3d_c_ac_poscar}}} \\
\onecolumn
{\phantomsection\label{A2B_oP12_26_abc_ab-H2S}}
\subsection*{\huge \textbf{{\normalfont H$_{2}$S (70~GPa) Structure: A2B\_oP12\_26\_abc\_ab}}}
\noindent \hrulefill
\vspace*{0.25cm}
\begin{figure}[htp]
  \centering
  \vspace{-1em}
  {\includegraphics[width=1\textwidth]{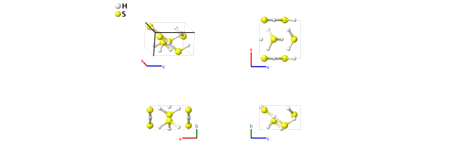}}
\end{figure}
\vspace*{-0.5cm}
\renewcommand{\arraystretch}{1.5}
\begin{equation*}
  \begin{array}{>{$\hspace{-0.15cm}}l<{$}>{$}p{0.5cm}<{$}>{$}p{18.5cm}<{$}}
    \mbox{\large \textbf{Prototype}} &\colon & \ce{H2S} \\
    \mbox{\large \textbf{\AFLOW\ prototype label}} &\colon & \mbox{A2B\_oP12\_26\_abc\_ab} \\
    \mbox{\large \textbf{\textit{Strukturbericht} designation}} &\colon & \mbox{None} \\
    \mbox{\large \textbf{Pearson symbol}} &\colon & \mbox{oP12} \\
    \mbox{\large \textbf{Space group number}} &\colon & 26 \\
    \mbox{\large \textbf{Space group symbol}} &\colon & Pmc2_{1} \\
    \mbox{\large \textbf{\AFLOW\ prototype command}} &\colon &  \texttt{aflow} \,  \, \texttt{-{}-proto=A2B\_oP12\_26\_abc\_ab } \, \newline \texttt{-{}-params=}{a,b/a,c/a,y_{1},z_{1},y_{2},z_{2},y_{3},z_{3},y_{4},z_{4},x_{5},y_{5},z_{5} }
  \end{array}
\end{equation*}
\renewcommand{\arraystretch}{1.0}

\vspace*{-0.25cm}
\noindent \hrulefill
\begin{itemize}
  \item{This structure was found by first-principles electronic structure
calculations and is predicted to be the stable structure of H$_{2}$S
in the range $40 - 80$~GPa.  The data presented here was computed at
70~GPa.
H$_{2}$S (pp. {\hyperref[A2B_oP12_26_abc_ab-H2S]{\pageref{A2B_oP12_26_abc_ab-H2S}}}) 
and $\beta$-SeO$_{2}$ (pp. {\hyperref[A2B_oP12_26_abc_ab-SeO2]{\pageref{A2B_oP12_26_abc_ab-SeO2}}}) 
have the same \AFLOW\ prototype label. 
They are generated by the same symmetry operations with different sets of parameters 
(\texttt{-{}-params}) specified in their corresponding \CIF\ files.
}
\end{itemize}

\noindent \parbox{1 \linewidth}{
\noindent \hrulefill
\\
\textbf{Simple Orthorhombic primitive vectors:} \\
\vspace*{-0.25cm}
\begin{tabular}{cc}
  \begin{tabular}{c}
    \parbox{0.6 \linewidth}{
      \renewcommand{\arraystretch}{1.5}
      \begin{equation*}
        \centering
        \begin{array}{ccc}
              \mathbf{a}_1 & = & a \, \mathbf{\hat{x}} \\
    \mathbf{a}_2 & = & b \, \mathbf{\hat{y}} \\
    \mathbf{a}_3 & = & c \, \mathbf{\hat{z}} \\

        \end{array}
      \end{equation*}
    }
    \renewcommand{\arraystretch}{1.0}
  \end{tabular}
  \begin{tabular}{c}
    \includegraphics[width=0.3\linewidth]{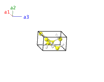} \\
  \end{tabular}
\end{tabular}

}
\vspace*{-0.25cm}

\noindent \hrulefill
\\
\textbf{Basis vectors:}
\vspace*{-0.25cm}
\renewcommand{\arraystretch}{1.5}
\begin{longtabu} to \textwidth{>{\centering $}X[-1,c,c]<{$}>{\centering $}X[-1,c,c]<{$}>{\centering $}X[-1,c,c]<{$}>{\centering $}X[-1,c,c]<{$}>{\centering $}X[-1,c,c]<{$}>{\centering $}X[-1,c,c]<{$}>{\centering $}X[-1,c,c]<{$}}
  & & \mbox{Lattice Coordinates} & & \mbox{Cartesian Coordinates} &\mbox{Wyckoff Position} & \mbox{Atom Type} \\  
  \mathbf{B}_{1} & = & y_{1} \, \mathbf{a}_{2} + z_{1} \, \mathbf{a}_{3} & = & y_{1}b \, \mathbf{\hat{y}} + z_{1}c \, \mathbf{\hat{z}} & \left(2a\right) & \mbox{H I} \\ 
\mathbf{B}_{2} & = & -y_{1} \, \mathbf{a}_{2} + \left(\frac{1}{2} +z_{1}\right) \, \mathbf{a}_{3} & = & -y_{1}b \, \mathbf{\hat{y}} + \left(\frac{1}{2} +z_{1}\right)c \, \mathbf{\hat{z}} & \left(2a\right) & \mbox{H I} \\ 
\mathbf{B}_{3} & = & y_{2} \, \mathbf{a}_{2} + z_{2} \, \mathbf{a}_{3} & = & y_{2}b \, \mathbf{\hat{y}} + z_{2}c \, \mathbf{\hat{z}} & \left(2a\right) & \mbox{S I} \\ 
\mathbf{B}_{4} & = & -y_{2} \, \mathbf{a}_{2} + \left(\frac{1}{2} +z_{2}\right) \, \mathbf{a}_{3} & = & -y_{2}b \, \mathbf{\hat{y}} + \left(\frac{1}{2} +z_{2}\right)c \, \mathbf{\hat{z}} & \left(2a\right) & \mbox{S I} \\ 
\mathbf{B}_{5} & = & \frac{1}{2} \, \mathbf{a}_{1} + y_{3} \, \mathbf{a}_{2} + z_{3} \, \mathbf{a}_{3} & = & \frac{1}{2}a \, \mathbf{\hat{x}} + y_{3}b \, \mathbf{\hat{y}} + z_{3}c \, \mathbf{\hat{z}} & \left(2b\right) & \mbox{H II} \\ 
\mathbf{B}_{6} & = & \frac{1}{2} \, \mathbf{a}_{1}-y_{3} \, \mathbf{a}_{2} + \left(\frac{1}{2} +z_{3}\right) \, \mathbf{a}_{3} & = & \frac{1}{2}a \, \mathbf{\hat{x}}-y_{3}b \, \mathbf{\hat{y}} + \left(\frac{1}{2} +z_{3}\right)c \, \mathbf{\hat{z}} & \left(2b\right) & \mbox{H II} \\ 
\mathbf{B}_{7} & = & \frac{1}{2} \, \mathbf{a}_{1} + y_{4} \, \mathbf{a}_{2} + z_{4} \, \mathbf{a}_{3} & = & \frac{1}{2}a \, \mathbf{\hat{x}} + y_{4}b \, \mathbf{\hat{y}} + z_{4}c \, \mathbf{\hat{z}} & \left(2b\right) & \mbox{S II} \\ 
\mathbf{B}_{8} & = & \frac{1}{2} \, \mathbf{a}_{1}-y_{4} \, \mathbf{a}_{2} + \left(\frac{1}{2} +z_{4}\right) \, \mathbf{a}_{3} & = & \frac{1}{2}a \, \mathbf{\hat{x}}-y_{4}b \, \mathbf{\hat{y}} + \left(\frac{1}{2} +z_{4}\right)c \, \mathbf{\hat{z}} & \left(2b\right) & \mbox{S II} \\ 
\mathbf{B}_{9} & = & x_{5} \, \mathbf{a}_{1} + y_{5} \, \mathbf{a}_{2} + z_{5} \, \mathbf{a}_{3} & = & x_{5}a \, \mathbf{\hat{x}} + y_{5}b \, \mathbf{\hat{y}} + z_{5}c \, \mathbf{\hat{z}} & \left(4c\right) & \mbox{H III} \\ 
\mathbf{B}_{10} & = & -x_{5} \, \mathbf{a}_{1}-y_{5} \, \mathbf{a}_{2} + \left(\frac{1}{2} +z_{5}\right) \, \mathbf{a}_{3} & = & -x_{5}a \, \mathbf{\hat{x}}-y_{5}b \, \mathbf{\hat{y}} + \left(\frac{1}{2} +z_{5}\right)c \, \mathbf{\hat{z}} & \left(4c\right) & \mbox{H III} \\ 
\mathbf{B}_{11} & = & x_{5} \, \mathbf{a}_{1}-y_{5} \, \mathbf{a}_{2} + \left(\frac{1}{2} +z_{5}\right) \, \mathbf{a}_{3} & = & x_{5}a \, \mathbf{\hat{x}}-y_{5}b \, \mathbf{\hat{y}} + \left(\frac{1}{2} +z_{5}\right)c \, \mathbf{\hat{z}} & \left(4c\right) & \mbox{H III} \\ 
\mathbf{B}_{12} & = & -x_{5} \, \mathbf{a}_{1} + y_{5} \, \mathbf{a}_{2} + z_{5} \, \mathbf{a}_{3} & = & -x_{5}a \, \mathbf{\hat{x}} + y_{5}b \, \mathbf{\hat{y}} + z_{5}c \, \mathbf{\hat{z}} & \left(4c\right) & \mbox{H III} \\ 
\end{longtabu}
\renewcommand{\arraystretch}{1.0}
\noindent \hrulefill
\\
\textbf{References:}
\vspace*{-0.25cm}
\begin{flushleft}
  - \bibentry{Li_JCP_140_2014}. \\
\end{flushleft}
\noindent \hrulefill
\\
\textbf{Geometry files:}
\\
\noindent  - CIF: pp. {\hyperref[A2B_oP12_26_abc_ab-H2S_cif]{\pageref{A2B_oP12_26_abc_ab-H2S_cif}}} \\
\noindent  - POSCAR: pp. {\hyperref[A2B_oP12_26_abc_ab-H2S_poscar]{\pageref{A2B_oP12_26_abc_ab-H2S_poscar}}} \\
\onecolumn
{\phantomsection\label{A2B_oP12_26_abc_ab-SeO2}}
\subsection*{\huge \textbf{{\normalfont $\beta$-SeO$_{2}$ Structure: A2B\_oP12\_26\_abc\_ab}}}
\noindent \hrulefill
\vspace*{0.25cm}
\begin{figure}[htp]
  \centering
  \vspace{-1em}
  {\includegraphics[width=1\textwidth]{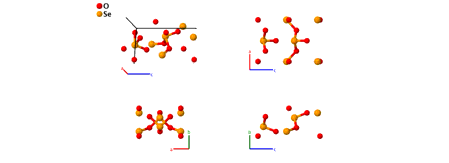}}
\end{figure}
\vspace*{-0.5cm}
\renewcommand{\arraystretch}{1.5}
\begin{equation*}
  \begin{array}{>{$\hspace{-0.15cm}}l<{$}>{$}p{0.5cm}<{$}>{$}p{18.5cm}<{$}}
    \mbox{\large \textbf{Prototype}} &\colon & \ce{$\beta$-SeO2} \\
    \mbox{\large \textbf{\AFLOW\ prototype label}} &\colon & \mbox{A2B\_oP12\_26\_abc\_ab} \\
    \mbox{\large \textbf{\textit{Strukturbericht} designation}} &\colon & \mbox{None} \\
    \mbox{\large \textbf{Pearson symbol}} &\colon & \mbox{oP12} \\
    \mbox{\large \textbf{Space group number}} &\colon & 26 \\
    \mbox{\large \textbf{Space group symbol}} &\colon & Pmc2_{1} \\
    \mbox{\large \textbf{\AFLOW\ prototype command}} &\colon &  \texttt{aflow} \,  \, \texttt{-{}-proto=A2B\_oP12\_26\_abc\_ab } \, \newline \texttt{-{}-params=}{a,b/a,c/a,y_{1},z_{1},y_{2},z_{2},y_{3},z_{3},y_{4},z_{4},x_{5},y_{5},z_{5} }
  \end{array}
\end{equation*}
\renewcommand{\arraystretch}{1.0}

\vspace*{-0.25cm}
\noindent \hrulefill
\begin{itemize}
  \item{H$_{2}$S (pp. {\hyperref[A2B_oP12_26_abc_ab-H2S]{\pageref{A2B_oP12_26_abc_ab-H2S}}})
and $\beta$-SeO$_{2}$ (pp. {\hyperref[A2B_oP12_26_abc_ab-SeO2]{\pageref{A2B_oP12_26_abc_ab-SeO2}}})
have the same \AFLOW\ prototype label.
They are generated by the same symmetry operations with different sets of parameters
(\texttt{-{}-params}) specified in their corresponding \CIF\ files.
}
\end{itemize}

\noindent \parbox{1 \linewidth}{
\noindent \hrulefill
\\
\textbf{Simple Orthorhombic primitive vectors:} \\
\vspace*{-0.25cm}
\begin{tabular}{cc}
  \begin{tabular}{c}
    \parbox{0.6 \linewidth}{
      \renewcommand{\arraystretch}{1.5}
      \begin{equation*}
        \centering
        \begin{array}{ccc}
              \mathbf{a}_1 & = & a \, \mathbf{\hat{x}} \\
    \mathbf{a}_2 & = & b \, \mathbf{\hat{y}} \\
    \mathbf{a}_3 & = & c \, \mathbf{\hat{z}} \\

        \end{array}
      \end{equation*}
    }
    \renewcommand{\arraystretch}{1.0}
  \end{tabular}
  \begin{tabular}{c}
    \includegraphics[width=0.3\linewidth]{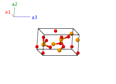} \\
  \end{tabular}
\end{tabular}

}
\vspace*{-0.25cm}

\noindent \hrulefill
\\
\textbf{Basis vectors:}
\vspace*{-0.25cm}
\renewcommand{\arraystretch}{1.5}
\begin{longtabu} to \textwidth{>{\centering $}X[-1,c,c]<{$}>{\centering $}X[-1,c,c]<{$}>{\centering $}X[-1,c,c]<{$}>{\centering $}X[-1,c,c]<{$}>{\centering $}X[-1,c,c]<{$}>{\centering $}X[-1,c,c]<{$}>{\centering $}X[-1,c,c]<{$}}
  & & \mbox{Lattice Coordinates} & & \mbox{Cartesian Coordinates} &\mbox{Wyckoff Position} & \mbox{Atom Type} \\  
  \mathbf{B}_{1} & = & y_{1} \, \mathbf{a}_{2} + z_{1} \, \mathbf{a}_{3} & = & y_{1}b \, \mathbf{\hat{y}} + z_{1}c \, \mathbf{\hat{z}} & \left(2a\right) & \mbox{O I} \\ 
\mathbf{B}_{2} & = & -y_{1} \, \mathbf{a}_{2} + \left(\frac{1}{2} +z_{1}\right) \, \mathbf{a}_{3} & = & -y_{1}b \, \mathbf{\hat{y}} + \left(\frac{1}{2} +z_{1}\right)c \, \mathbf{\hat{z}} & \left(2a\right) & \mbox{O I} \\ 
\mathbf{B}_{3} & = & y_{2} \, \mathbf{a}_{2} + z_{2} \, \mathbf{a}_{3} & = & y_{2}b \, \mathbf{\hat{y}} + z_{2}c \, \mathbf{\hat{z}} & \left(2a\right) & \mbox{Se I} \\ 
\mathbf{B}_{4} & = & -y_{2} \, \mathbf{a}_{2} + \left(\frac{1}{2} +z_{2}\right) \, \mathbf{a}_{3} & = & -y_{2}b \, \mathbf{\hat{y}} + \left(\frac{1}{2} +z_{2}\right)c \, \mathbf{\hat{z}} & \left(2a\right) & \mbox{Se I} \\ 
\mathbf{B}_{5} & = & \frac{1}{2} \, \mathbf{a}_{1} + y_{3} \, \mathbf{a}_{2} + z_{3} \, \mathbf{a}_{3} & = & \frac{1}{2}a \, \mathbf{\hat{x}} + y_{3}b \, \mathbf{\hat{y}} + z_{3}c \, \mathbf{\hat{z}} & \left(2b\right) & \mbox{O II} \\ 
\mathbf{B}_{6} & = & \frac{1}{2} \, \mathbf{a}_{1}-y_{3} \, \mathbf{a}_{2} + \left(\frac{1}{2} +z_{3}\right) \, \mathbf{a}_{3} & = & \frac{1}{2}a \, \mathbf{\hat{x}}-y_{3}b \, \mathbf{\hat{y}} + \left(\frac{1}{2} +z_{3}\right)c \, \mathbf{\hat{z}} & \left(2b\right) & \mbox{O II} \\ 
\mathbf{B}_{7} & = & \frac{1}{2} \, \mathbf{a}_{1} + y_{4} \, \mathbf{a}_{2} + z_{4} \, \mathbf{a}_{3} & = & \frac{1}{2}a \, \mathbf{\hat{x}} + y_{4}b \, \mathbf{\hat{y}} + z_{4}c \, \mathbf{\hat{z}} & \left(2b\right) & \mbox{Se II} \\ 
\mathbf{B}_{8} & = & \frac{1}{2} \, \mathbf{a}_{1}-y_{4} \, \mathbf{a}_{2} + \left(\frac{1}{2} +z_{4}\right) \, \mathbf{a}_{3} & = & \frac{1}{2}a \, \mathbf{\hat{x}}-y_{4}b \, \mathbf{\hat{y}} + \left(\frac{1}{2} +z_{4}\right)c \, \mathbf{\hat{z}} & \left(2b\right) & \mbox{Se II} \\ 
\mathbf{B}_{9} & = & x_{5} \, \mathbf{a}_{1} + y_{5} \, \mathbf{a}_{2} + z_{5} \, \mathbf{a}_{3} & = & x_{5}a \, \mathbf{\hat{x}} + y_{5}b \, \mathbf{\hat{y}} + z_{5}c \, \mathbf{\hat{z}} & \left(4c\right) & \mbox{O III} \\ 
\mathbf{B}_{10} & = & -x_{5} \, \mathbf{a}_{1}-y_{5} \, \mathbf{a}_{2} + \left(\frac{1}{2} +z_{5}\right) \, \mathbf{a}_{3} & = & -x_{5}a \, \mathbf{\hat{x}}-y_{5}b \, \mathbf{\hat{y}} + \left(\frac{1}{2} +z_{5}\right)c \, \mathbf{\hat{z}} & \left(4c\right) & \mbox{O III} \\ 
\mathbf{B}_{11} & = & x_{5} \, \mathbf{a}_{1}-y_{5} \, \mathbf{a}_{2} + \left(\frac{1}{2} +z_{5}\right) \, \mathbf{a}_{3} & = & x_{5}a \, \mathbf{\hat{x}}-y_{5}b \, \mathbf{\hat{y}} + \left(\frac{1}{2} +z_{5}\right)c \, \mathbf{\hat{z}} & \left(4c\right) & \mbox{O III} \\ 
\mathbf{B}_{12} & = & -x_{5} \, \mathbf{a}_{1} + y_{5} \, \mathbf{a}_{2} + z_{5} \, \mathbf{a}_{3} & = & -x_{5}a \, \mathbf{\hat{x}} + y_{5}b \, \mathbf{\hat{y}} + z_{5}c \, \mathbf{\hat{z}} & \left(4c\right) & \mbox{O III} \\ 
\end{longtabu}
\renewcommand{\arraystretch}{1.0}
\noindent \hrulefill
\\
\textbf{References:}
\vspace*{-0.25cm}
\begin{flushleft}
  - \bibentry{Orosel_SeO2_JSolidStateChem_2004}. \\
\end{flushleft}
\textbf{Found in:}
\vspace*{-0.25cm}
\begin{flushleft}
  - \bibentry{Villars_PearsonsCrystalData_2013}. \\
\end{flushleft}
\noindent \hrulefill
\\
\textbf{Geometry files:}
\\
\noindent  - CIF: pp. {\hyperref[A2B_oP12_26_abc_ab-SeO2_cif]{\pageref{A2B_oP12_26_abc_ab-SeO2_cif}}} \\
\noindent  - POSCAR: pp. {\hyperref[A2B_oP12_26_abc_ab-SeO2_poscar]{\pageref{A2B_oP12_26_abc_ab-SeO2_poscar}}} \\
\onecolumn
{\phantomsection\label{A5B_oP24_26_3a3b2c_ab}}
\subsection*{\huge \textbf{{\normalfont TlP$_{5}$ Structure: A5B\_oP24\_26\_3a3b2c\_ab}}}
\noindent \hrulefill
\vspace*{0.25cm}
\begin{figure}[htp]
  \centering
  \vspace{-1em}
  {\includegraphics[width=1\textwidth]{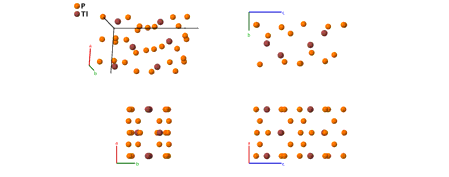}}
\end{figure}
\vspace*{-0.5cm}
\renewcommand{\arraystretch}{1.5}
\begin{equation*}
  \begin{array}{>{$\hspace{-0.15cm}}l<{$}>{$}p{0.5cm}<{$}>{$}p{18.5cm}<{$}}
    \mbox{\large \textbf{Prototype}} &\colon & \ce{TlP5} \\
    \mbox{\large \textbf{\AFLOW\ prototype label}} &\colon & \mbox{A5B\_oP24\_26\_3a3b2c\_ab} \\
    \mbox{\large \textbf{\textit{Strukturbericht} designation}} &\colon & \mbox{None} \\
    \mbox{\large \textbf{Pearson symbol}} &\colon & \mbox{oP24} \\
    \mbox{\large \textbf{Space group number}} &\colon & 26 \\
    \mbox{\large \textbf{Space group symbol}} &\colon & Pmc2_{1} \\
    \mbox{\large \textbf{\AFLOW\ prototype command}} &\colon &  \texttt{aflow} \,  \, \texttt{-{}-proto=A5B\_oP24\_26\_3a3b2c\_ab } \, \newline \texttt{-{}-params=}{a,b/a,c/a,y_{1},z_{1},y_{2},z_{2},y_{3},z_{3},y_{4},z_{4},y_{5},z_{5},y_{6},z_{6},y_{7},z_{7},y_{8},z_{8},x_{9},} \newline {y_{9},z_{9},x_{10},y_{10},z_{10} }
  \end{array}
\end{equation*}
\renewcommand{\arraystretch}{1.0}

\noindent \parbox{1 \linewidth}{
\noindent \hrulefill
\\
\textbf{Simple Orthorhombic primitive vectors:} \\
\vspace*{-0.25cm}
\begin{tabular}{cc}
  \begin{tabular}{c}
    \parbox{0.6 \linewidth}{
      \renewcommand{\arraystretch}{1.5}
      \begin{equation*}
        \centering
        \begin{array}{ccc}
              \mathbf{a}_1 & = & a \, \mathbf{\hat{x}} \\
    \mathbf{a}_2 & = & b \, \mathbf{\hat{y}} \\
    \mathbf{a}_3 & = & c \, \mathbf{\hat{z}} \\

        \end{array}
      \end{equation*}
    }
    \renewcommand{\arraystretch}{1.0}
  \end{tabular}
  \begin{tabular}{c}
    \includegraphics[width=0.3\linewidth]{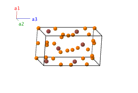} \\
  \end{tabular}
\end{tabular}

}
\vspace*{-0.25cm}

\noindent \hrulefill
\\
\textbf{Basis vectors:}
\vspace*{-0.25cm}
\renewcommand{\arraystretch}{1.5}
\begin{longtabu} to \textwidth{>{\centering $}X[-1,c,c]<{$}>{\centering $}X[-1,c,c]<{$}>{\centering $}X[-1,c,c]<{$}>{\centering $}X[-1,c,c]<{$}>{\centering $}X[-1,c,c]<{$}>{\centering $}X[-1,c,c]<{$}>{\centering $}X[-1,c,c]<{$}}
  & & \mbox{Lattice Coordinates} & & \mbox{Cartesian Coordinates} &\mbox{Wyckoff Position} & \mbox{Atom Type} \\  
  \mathbf{B}_{1} & = & y_{1} \, \mathbf{a}_{2} + z_{1} \, \mathbf{a}_{3} & = & y_{1}b \, \mathbf{\hat{y}} + z_{1}c \, \mathbf{\hat{z}} & \left(2a\right) & \mbox{P I} \\ 
\mathbf{B}_{2} & = & -y_{1} \, \mathbf{a}_{2} + \left(\frac{1}{2} +z_{1}\right) \, \mathbf{a}_{3} & = & -y_{1}b \, \mathbf{\hat{y}} + \left(\frac{1}{2} +z_{1}\right)c \, \mathbf{\hat{z}} & \left(2a\right) & \mbox{P I} \\ 
\mathbf{B}_{3} & = & y_{2} \, \mathbf{a}_{2} + z_{2} \, \mathbf{a}_{3} & = & y_{2}b \, \mathbf{\hat{y}} + z_{2}c \, \mathbf{\hat{z}} & \left(2a\right) & \mbox{P II} \\ 
\mathbf{B}_{4} & = & -y_{2} \, \mathbf{a}_{2} + \left(\frac{1}{2} +z_{2}\right) \, \mathbf{a}_{3} & = & -y_{2}b \, \mathbf{\hat{y}} + \left(\frac{1}{2} +z_{2}\right)c \, \mathbf{\hat{z}} & \left(2a\right) & \mbox{P II} \\ 
\mathbf{B}_{5} & = & y_{3} \, \mathbf{a}_{2} + z_{3} \, \mathbf{a}_{3} & = & y_{3}b \, \mathbf{\hat{y}} + z_{3}c \, \mathbf{\hat{z}} & \left(2a\right) & \mbox{P III} \\ 
\mathbf{B}_{6} & = & -y_{3} \, \mathbf{a}_{2} + \left(\frac{1}{2} +z_{3}\right) \, \mathbf{a}_{3} & = & -y_{3}b \, \mathbf{\hat{y}} + \left(\frac{1}{2} +z_{3}\right)c \, \mathbf{\hat{z}} & \left(2a\right) & \mbox{P III} \\ 
\mathbf{B}_{7} & = & y_{4} \, \mathbf{a}_{2} + z_{4} \, \mathbf{a}_{3} & = & y_{4}b \, \mathbf{\hat{y}} + z_{4}c \, \mathbf{\hat{z}} & \left(2a\right) & \mbox{Tl I} \\ 
\mathbf{B}_{8} & = & -y_{4} \, \mathbf{a}_{2} + \left(\frac{1}{2} +z_{4}\right) \, \mathbf{a}_{3} & = & -y_{4}b \, \mathbf{\hat{y}} + \left(\frac{1}{2} +z_{4}\right)c \, \mathbf{\hat{z}} & \left(2a\right) & \mbox{Tl I} \\ 
\mathbf{B}_{9} & = & \frac{1}{2} \, \mathbf{a}_{1} + y_{5} \, \mathbf{a}_{2} + z_{5} \, \mathbf{a}_{3} & = & \frac{1}{2}a \, \mathbf{\hat{x}} + y_{5}b \, \mathbf{\hat{y}} + z_{5}c \, \mathbf{\hat{z}} & \left(2b\right) & \mbox{P IV} \\ 
\mathbf{B}_{10} & = & \frac{1}{2} \, \mathbf{a}_{1}-y_{5} \, \mathbf{a}_{2} + \left(\frac{1}{2} +z_{5}\right) \, \mathbf{a}_{3} & = & \frac{1}{2}a \, \mathbf{\hat{x}}-y_{5}b \, \mathbf{\hat{y}} + \left(\frac{1}{2} +z_{5}\right)c \, \mathbf{\hat{z}} & \left(2b\right) & \mbox{P IV} \\ 
\mathbf{B}_{11} & = & \frac{1}{2} \, \mathbf{a}_{1} + y_{6} \, \mathbf{a}_{2} + z_{6} \, \mathbf{a}_{3} & = & \frac{1}{2}a \, \mathbf{\hat{x}} + y_{6}b \, \mathbf{\hat{y}} + z_{6}c \, \mathbf{\hat{z}} & \left(2b\right) & \mbox{P V} \\ 
\mathbf{B}_{12} & = & \frac{1}{2} \, \mathbf{a}_{1}-y_{6} \, \mathbf{a}_{2} + \left(\frac{1}{2} +z_{6}\right) \, \mathbf{a}_{3} & = & \frac{1}{2}a \, \mathbf{\hat{x}}-y_{6}b \, \mathbf{\hat{y}} + \left(\frac{1}{2} +z_{6}\right)c \, \mathbf{\hat{z}} & \left(2b\right) & \mbox{P V} \\ 
\mathbf{B}_{13} & = & \frac{1}{2} \, \mathbf{a}_{1} + y_{7} \, \mathbf{a}_{2} + z_{7} \, \mathbf{a}_{3} & = & \frac{1}{2}a \, \mathbf{\hat{x}} + y_{7}b \, \mathbf{\hat{y}} + z_{7}c \, \mathbf{\hat{z}} & \left(2b\right) & \mbox{P VI} \\ 
\mathbf{B}_{14} & = & \frac{1}{2} \, \mathbf{a}_{1}-y_{7} \, \mathbf{a}_{2} + \left(\frac{1}{2} +z_{7}\right) \, \mathbf{a}_{3} & = & \frac{1}{2}a \, \mathbf{\hat{x}}-y_{7}b \, \mathbf{\hat{y}} + \left(\frac{1}{2} +z_{7}\right)c \, \mathbf{\hat{z}} & \left(2b\right) & \mbox{P VI} \\ 
\mathbf{B}_{15} & = & \frac{1}{2} \, \mathbf{a}_{1} + y_{8} \, \mathbf{a}_{2} + z_{8} \, \mathbf{a}_{3} & = & \frac{1}{2}a \, \mathbf{\hat{x}} + y_{8}b \, \mathbf{\hat{y}} + z_{8}c \, \mathbf{\hat{z}} & \left(2b\right) & \mbox{Tl II} \\ 
\mathbf{B}_{16} & = & \frac{1}{2} \, \mathbf{a}_{1}-y_{8} \, \mathbf{a}_{2} + \left(\frac{1}{2} +z_{8}\right) \, \mathbf{a}_{3} & = & \frac{1}{2}a \, \mathbf{\hat{x}}-y_{8}b \, \mathbf{\hat{y}} + \left(\frac{1}{2} +z_{8}\right)c \, \mathbf{\hat{z}} & \left(2b\right) & \mbox{Tl II} \\ 
\mathbf{B}_{17} & = & x_{9} \, \mathbf{a}_{1} + y_{9} \, \mathbf{a}_{2} + z_{9} \, \mathbf{a}_{3} & = & x_{9}a \, \mathbf{\hat{x}} + y_{9}b \, \mathbf{\hat{y}} + z_{9}c \, \mathbf{\hat{z}} & \left(4c\right) & \mbox{P VII} \\ 
\mathbf{B}_{18} & = & -x_{9} \, \mathbf{a}_{1}-y_{9} \, \mathbf{a}_{2} + \left(\frac{1}{2} +z_{9}\right) \, \mathbf{a}_{3} & = & -x_{9}a \, \mathbf{\hat{x}}-y_{9}b \, \mathbf{\hat{y}} + \left(\frac{1}{2} +z_{9}\right)c \, \mathbf{\hat{z}} & \left(4c\right) & \mbox{P VII} \\ 
\mathbf{B}_{19} & = & x_{9} \, \mathbf{a}_{1}-y_{9} \, \mathbf{a}_{2} + \left(\frac{1}{2} +z_{9}\right) \, \mathbf{a}_{3} & = & x_{9}a \, \mathbf{\hat{x}}-y_{9}b \, \mathbf{\hat{y}} + \left(\frac{1}{2} +z_{9}\right)c \, \mathbf{\hat{z}} & \left(4c\right) & \mbox{P VII} \\ 
\mathbf{B}_{20} & = & -x_{9} \, \mathbf{a}_{1} + y_{9} \, \mathbf{a}_{2} + z_{9} \, \mathbf{a}_{3} & = & -x_{9}a \, \mathbf{\hat{x}} + y_{9}b \, \mathbf{\hat{y}} + z_{9}c \, \mathbf{\hat{z}} & \left(4c\right) & \mbox{P VII} \\ 
\mathbf{B}_{21} & = & x_{10} \, \mathbf{a}_{1} + y_{10} \, \mathbf{a}_{2} + z_{10} \, \mathbf{a}_{3} & = & x_{10}a \, \mathbf{\hat{x}} + y_{10}b \, \mathbf{\hat{y}} + z_{10}c \, \mathbf{\hat{z}} & \left(4c\right) & \mbox{P VIII} \\ 
\mathbf{B}_{22} & = & -x_{10} \, \mathbf{a}_{1}-y_{10} \, \mathbf{a}_{2} + \left(\frac{1}{2} +z_{10}\right) \, \mathbf{a}_{3} & = & -x_{10}a \, \mathbf{\hat{x}}-y_{10}b \, \mathbf{\hat{y}} + \left(\frac{1}{2} +z_{10}\right)c \, \mathbf{\hat{z}} & \left(4c\right) & \mbox{P VIII} \\ 
\mathbf{B}_{23} & = & x_{10} \, \mathbf{a}_{1}-y_{10} \, \mathbf{a}_{2} + \left(\frac{1}{2} +z_{10}\right) \, \mathbf{a}_{3} & = & x_{10}a \, \mathbf{\hat{x}}-y_{10}b \, \mathbf{\hat{y}} + \left(\frac{1}{2} +z_{10}\right)c \, \mathbf{\hat{z}} & \left(4c\right) & \mbox{P VIII} \\ 
\mathbf{B}_{24} & = & -x_{10} \, \mathbf{a}_{1} + y_{10} \, \mathbf{a}_{2} + z_{10} \, \mathbf{a}_{3} & = & -x_{10}a \, \mathbf{\hat{x}} + y_{10}b \, \mathbf{\hat{y}} + z_{10}c \, \mathbf{\hat{z}} & \left(4c\right) & \mbox{P VIII} \\ 
\end{longtabu}
\renewcommand{\arraystretch}{1.0}
\noindent \hrulefill
\\
\textbf{References:}
\vspace*{-0.25cm}
\begin{flushleft}
  - \bibentry{Olofsson_TlP5_ActChemScand_1971}. \\
\end{flushleft}
\textbf{Found in:}
\vspace*{-0.25cm}
\begin{flushleft}
  - \bibentry{Villars_PearsonsCrystalData_2013}. \\
\end{flushleft}
\noindent \hrulefill
\\
\textbf{Geometry files:}
\\
\noindent  - CIF: pp. {\hyperref[A5B_oP24_26_3a3b2c_ab_cif]{\pageref{A5B_oP24_26_3a3b2c_ab_cif}}} \\
\noindent  - POSCAR: pp. {\hyperref[A5B_oP24_26_3a3b2c_ab_poscar]{\pageref{A5B_oP24_26_3a3b2c_ab_poscar}}} \\
\onecolumn
{\phantomsection\label{A6B4C16D_oP108_27_abcd4e_4e_16e_e}}
\subsection*{\huge \textbf{{\normalfont \begin{raggedleft}Ca$_{4}$Al$_{6}$O$_{16}$S Structure: \end{raggedleft} \\ A6B4C16D\_oP108\_27\_abcd4e\_4e\_16e\_e}}}
\noindent \hrulefill
\vspace*{0.25cm}
\begin{figure}[htp]
  \centering
  \vspace{-1em}
  {\includegraphics[width=1\textwidth]{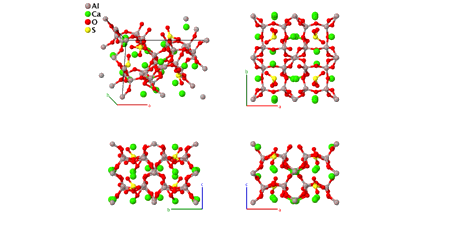}}
\end{figure}
\vspace*{-0.5cm}
\renewcommand{\arraystretch}{1.5}
\begin{equation*}
  \begin{array}{>{$\hspace{-0.15cm}}l<{$}>{$}p{0.5cm}<{$}>{$}p{18.5cm}<{$}}
    \mbox{\large \textbf{Prototype}} &\colon & \ce{Ca4Al6O16S} \\
    \mbox{\large \textbf{\AFLOW\ prototype label}} &\colon & \mbox{A6B4C16D\_oP108\_27\_abcd4e\_4e\_16e\_e} \\
    \mbox{\large \textbf{\textit{Strukturbericht} designation}} &\colon & \mbox{None} \\
    \mbox{\large \textbf{Pearson symbol}} &\colon & \mbox{oP108} \\
    \mbox{\large \textbf{Space group number}} &\colon & 27 \\
    \mbox{\large \textbf{Space group symbol}} &\colon & Pcc2 \\
    \mbox{\large \textbf{\AFLOW\ prototype command}} &\colon &  \texttt{aflow} \,  \, \texttt{-{}-proto=A6B4C16D\_oP108\_27\_abcd4e\_4e\_16e\_e } \, \newline \texttt{-{}-params=}{a,b/a,c/a,z_{1},z_{2},z_{3},z_{4},x_{5},y_{5},z_{5},x_{6},y_{6},z_{6},x_{7},y_{7},z_{7},x_{8},y_{8},z_{8},x_{9},} \newline {y_{9},z_{9},x_{10},y_{10},z_{10},x_{11},y_{11},z_{11},x_{12},y_{12},z_{12},x_{13},y_{13},z_{13},x_{14},y_{14},z_{14},x_{15},y_{15},z_{15},x_{16},} \newline {y_{16},z_{16},x_{17},y_{17},z_{17},x_{18},y_{18},z_{18},x_{19},y_{19},z_{19},x_{20},y_{20},z_{20},x_{21},y_{21},z_{21},x_{22},y_{22},z_{22},} \newline {x_{23},y_{23},z_{23},x_{24},y_{24},z_{24},x_{25},y_{25},z_{25},x_{26},y_{26},z_{26},x_{27},y_{27},z_{27},x_{28},y_{28},z_{28},x_{29},y_{29},} \newline {z_{29} }
  \end{array}
\end{equation*}
\renewcommand{\arraystretch}{1.0}

\noindent \parbox{1 \linewidth}{
\noindent \hrulefill
\\
\textbf{Simple Orthorhombic primitive vectors:} \\
\vspace*{-0.25cm}
\begin{tabular}{cc}
  \begin{tabular}{c}
    \parbox{0.6 \linewidth}{
      \renewcommand{\arraystretch}{1.5}
      \begin{equation*}
        \centering
        \begin{array}{ccc}
              \mathbf{a}_1 & = & a \, \mathbf{\hat{x}} \\
    \mathbf{a}_2 & = & b \, \mathbf{\hat{y}} \\
    \mathbf{a}_3 & = & c \, \mathbf{\hat{z}} \\

        \end{array}
      \end{equation*}
    }
    \renewcommand{\arraystretch}{1.0}
  \end{tabular}
  \begin{tabular}{c}
    \includegraphics[width=0.3\linewidth]{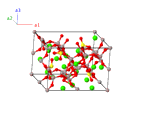} \\
  \end{tabular}
\end{tabular}

}
\vspace*{-0.25cm}

\noindent \hrulefill
\\
\textbf{Basis vectors:}
\vspace*{-0.25cm}
\renewcommand{\arraystretch}{1.5}
\begin{longtabu} to \textwidth{>{\centering $}X[-1,c,c]<{$}>{\centering $}X[-1,c,c]<{$}>{\centering $}X[-1,c,c]<{$}>{\centering $}X[-1,c,c]<{$}>{\centering $}X[-1,c,c]<{$}>{\centering $}X[-1,c,c]<{$}>{\centering $}X[-1,c,c]<{$}}
  & & \mbox{Lattice Coordinates} & & \mbox{Cartesian Coordinates} &\mbox{Wyckoff Position} & \mbox{Atom Type} \\  
  \mathbf{B}_{1} & = & z_{1} \, \mathbf{a}_{3} & = & z_{1}c \, \mathbf{\hat{z}} & \left(2a\right) & \mbox{Al I} \\ 
\mathbf{B}_{2} & = & \left(\frac{1}{2} +z_{1}\right) \, \mathbf{a}_{3} & = & \left(\frac{1}{2} +z_{1}\right)c \, \mathbf{\hat{z}} & \left(2a\right) & \mbox{Al I} \\ 
\mathbf{B}_{3} & = & \frac{1}{2} \, \mathbf{a}_{2} + z_{2} \, \mathbf{a}_{3} & = & \frac{1}{2}b \, \mathbf{\hat{y}} + z_{2}c \, \mathbf{\hat{z}} & \left(2b\right) & \mbox{Al II} \\ 
\mathbf{B}_{4} & = & \frac{1}{2} \, \mathbf{a}_{2} + \left(\frac{1}{2} +z_{2}\right) \, \mathbf{a}_{3} & = & \frac{1}{2}b \, \mathbf{\hat{y}} + \left(\frac{1}{2} +z_{2}\right)c \, \mathbf{\hat{z}} & \left(2b\right) & \mbox{Al II} \\ 
\mathbf{B}_{5} & = & \frac{1}{2} \, \mathbf{a}_{1} + z_{3} \, \mathbf{a}_{3} & = & \frac{1}{2}a \, \mathbf{\hat{x}} + z_{3}c \, \mathbf{\hat{z}} & \left(2c\right) & \mbox{Al III} \\ 
\mathbf{B}_{6} & = & \frac{1}{2} \, \mathbf{a}_{1} + \left(\frac{1}{2} +z_{3}\right) \, \mathbf{a}_{3} & = & \frac{1}{2}a \, \mathbf{\hat{x}} + \left(\frac{1}{2} +z_{3}\right)c \, \mathbf{\hat{z}} & \left(2c\right) & \mbox{Al III} \\ 
\mathbf{B}_{7} & = & \frac{1}{2} \, \mathbf{a}_{1} + \frac{1}{2} \, \mathbf{a}_{2} + z_{4} \, \mathbf{a}_{3} & = & \frac{1}{2}a \, \mathbf{\hat{x}} + \frac{1}{2}b \, \mathbf{\hat{y}} + z_{4}c \, \mathbf{\hat{z}} & \left(2d\right) & \mbox{Al IV} \\ 
\mathbf{B}_{8} & = & \frac{1}{2} \, \mathbf{a}_{1} + \frac{1}{2} \, \mathbf{a}_{2} + \left(\frac{1}{2} +z_{4}\right) \, \mathbf{a}_{3} & = & \frac{1}{2}a \, \mathbf{\hat{x}} + \frac{1}{2}b \, \mathbf{\hat{y}} + \left(\frac{1}{2} +z_{4}\right)c \, \mathbf{\hat{z}} & \left(2d\right) & \mbox{Al IV} \\ 
\mathbf{B}_{9} & = & x_{5} \, \mathbf{a}_{1} + y_{5} \, \mathbf{a}_{2} + z_{5} \, \mathbf{a}_{3} & = & x_{5}a \, \mathbf{\hat{x}} + y_{5}b \, \mathbf{\hat{y}} + z_{5}c \, \mathbf{\hat{z}} & \left(4e\right) & \mbox{Al V} \\ 
\mathbf{B}_{10} & = & -x_{5} \, \mathbf{a}_{1}-y_{5} \, \mathbf{a}_{2} + z_{5} \, \mathbf{a}_{3} & = & -x_{5}a \, \mathbf{\hat{x}}-y_{5}b \, \mathbf{\hat{y}} + z_{5}c \, \mathbf{\hat{z}} & \left(4e\right) & \mbox{Al V} \\ 
\mathbf{B}_{11} & = & x_{5} \, \mathbf{a}_{1}-y_{5} \, \mathbf{a}_{2} + \left(\frac{1}{2} +z_{5}\right) \, \mathbf{a}_{3} & = & x_{5}a \, \mathbf{\hat{x}}-y_{5}b \, \mathbf{\hat{y}} + \left(\frac{1}{2} +z_{5}\right)c \, \mathbf{\hat{z}} & \left(4e\right) & \mbox{Al V} \\ 
\mathbf{B}_{12} & = & -x_{5} \, \mathbf{a}_{1} + y_{5} \, \mathbf{a}_{2} + \left(\frac{1}{2} +z_{5}\right) \, \mathbf{a}_{3} & = & -x_{5}a \, \mathbf{\hat{x}} + y_{5}b \, \mathbf{\hat{y}} + \left(\frac{1}{2} +z_{5}\right)c \, \mathbf{\hat{z}} & \left(4e\right) & \mbox{Al V} \\ 
\mathbf{B}_{13} & = & x_{6} \, \mathbf{a}_{1} + y_{6} \, \mathbf{a}_{2} + z_{6} \, \mathbf{a}_{3} & = & x_{6}a \, \mathbf{\hat{x}} + y_{6}b \, \mathbf{\hat{y}} + z_{6}c \, \mathbf{\hat{z}} & \left(4e\right) & \mbox{Al VI} \\ 
\mathbf{B}_{14} & = & -x_{6} \, \mathbf{a}_{1}-y_{6} \, \mathbf{a}_{2} + z_{6} \, \mathbf{a}_{3} & = & -x_{6}a \, \mathbf{\hat{x}}-y_{6}b \, \mathbf{\hat{y}} + z_{6}c \, \mathbf{\hat{z}} & \left(4e\right) & \mbox{Al VI} \\ 
\mathbf{B}_{15} & = & x_{6} \, \mathbf{a}_{1}-y_{6} \, \mathbf{a}_{2} + \left(\frac{1}{2} +z_{6}\right) \, \mathbf{a}_{3} & = & x_{6}a \, \mathbf{\hat{x}}-y_{6}b \, \mathbf{\hat{y}} + \left(\frac{1}{2} +z_{6}\right)c \, \mathbf{\hat{z}} & \left(4e\right) & \mbox{Al VI} \\ 
\mathbf{B}_{16} & = & -x_{6} \, \mathbf{a}_{1} + y_{6} \, \mathbf{a}_{2} + \left(\frac{1}{2} +z_{6}\right) \, \mathbf{a}_{3} & = & -x_{6}a \, \mathbf{\hat{x}} + y_{6}b \, \mathbf{\hat{y}} + \left(\frac{1}{2} +z_{6}\right)c \, \mathbf{\hat{z}} & \left(4e\right) & \mbox{Al VI} \\ 
\mathbf{B}_{17} & = & x_{7} \, \mathbf{a}_{1} + y_{7} \, \mathbf{a}_{2} + z_{7} \, \mathbf{a}_{3} & = & x_{7}a \, \mathbf{\hat{x}} + y_{7}b \, \mathbf{\hat{y}} + z_{7}c \, \mathbf{\hat{z}} & \left(4e\right) & \mbox{Al VII} \\ 
\mathbf{B}_{18} & = & -x_{7} \, \mathbf{a}_{1}-y_{7} \, \mathbf{a}_{2} + z_{7} \, \mathbf{a}_{3} & = & -x_{7}a \, \mathbf{\hat{x}}-y_{7}b \, \mathbf{\hat{y}} + z_{7}c \, \mathbf{\hat{z}} & \left(4e\right) & \mbox{Al VII} \\ 
\mathbf{B}_{19} & = & x_{7} \, \mathbf{a}_{1}-y_{7} \, \mathbf{a}_{2} + \left(\frac{1}{2} +z_{7}\right) \, \mathbf{a}_{3} & = & x_{7}a \, \mathbf{\hat{x}}-y_{7}b \, \mathbf{\hat{y}} + \left(\frac{1}{2} +z_{7}\right)c \, \mathbf{\hat{z}} & \left(4e\right) & \mbox{Al VII} \\ 
\mathbf{B}_{20} & = & -x_{7} \, \mathbf{a}_{1} + y_{7} \, \mathbf{a}_{2} + \left(\frac{1}{2} +z_{7}\right) \, \mathbf{a}_{3} & = & -x_{7}a \, \mathbf{\hat{x}} + y_{7}b \, \mathbf{\hat{y}} + \left(\frac{1}{2} +z_{7}\right)c \, \mathbf{\hat{z}} & \left(4e\right) & \mbox{Al VII} \\ 
\mathbf{B}_{21} & = & x_{8} \, \mathbf{a}_{1} + y_{8} \, \mathbf{a}_{2} + z_{8} \, \mathbf{a}_{3} & = & x_{8}a \, \mathbf{\hat{x}} + y_{8}b \, \mathbf{\hat{y}} + z_{8}c \, \mathbf{\hat{z}} & \left(4e\right) & \mbox{Al VIII} \\ 
\mathbf{B}_{22} & = & -x_{8} \, \mathbf{a}_{1}-y_{8} \, \mathbf{a}_{2} + z_{8} \, \mathbf{a}_{3} & = & -x_{8}a \, \mathbf{\hat{x}}-y_{8}b \, \mathbf{\hat{y}} + z_{8}c \, \mathbf{\hat{z}} & \left(4e\right) & \mbox{Al VIII} \\ 
\mathbf{B}_{23} & = & x_{8} \, \mathbf{a}_{1}-y_{8} \, \mathbf{a}_{2} + \left(\frac{1}{2} +z_{8}\right) \, \mathbf{a}_{3} & = & x_{8}a \, \mathbf{\hat{x}}-y_{8}b \, \mathbf{\hat{y}} + \left(\frac{1}{2} +z_{8}\right)c \, \mathbf{\hat{z}} & \left(4e\right) & \mbox{Al VIII} \\ 
\mathbf{B}_{24} & = & -x_{8} \, \mathbf{a}_{1} + y_{8} \, \mathbf{a}_{2} + \left(\frac{1}{2} +z_{8}\right) \, \mathbf{a}_{3} & = & -x_{8}a \, \mathbf{\hat{x}} + y_{8}b \, \mathbf{\hat{y}} + \left(\frac{1}{2} +z_{8}\right)c \, \mathbf{\hat{z}} & \left(4e\right) & \mbox{Al VIII} \\ 
\mathbf{B}_{25} & = & x_{9} \, \mathbf{a}_{1} + y_{9} \, \mathbf{a}_{2} + z_{9} \, \mathbf{a}_{3} & = & x_{9}a \, \mathbf{\hat{x}} + y_{9}b \, \mathbf{\hat{y}} + z_{9}c \, \mathbf{\hat{z}} & \left(4e\right) & \mbox{Ca I} \\ 
\mathbf{B}_{26} & = & -x_{9} \, \mathbf{a}_{1}-y_{9} \, \mathbf{a}_{2} + z_{9} \, \mathbf{a}_{3} & = & -x_{9}a \, \mathbf{\hat{x}}-y_{9}b \, \mathbf{\hat{y}} + z_{9}c \, \mathbf{\hat{z}} & \left(4e\right) & \mbox{Ca I} \\ 
\mathbf{B}_{27} & = & x_{9} \, \mathbf{a}_{1}-y_{9} \, \mathbf{a}_{2} + \left(\frac{1}{2} +z_{9}\right) \, \mathbf{a}_{3} & = & x_{9}a \, \mathbf{\hat{x}}-y_{9}b \, \mathbf{\hat{y}} + \left(\frac{1}{2} +z_{9}\right)c \, \mathbf{\hat{z}} & \left(4e\right) & \mbox{Ca I} \\ 
\mathbf{B}_{28} & = & -x_{9} \, \mathbf{a}_{1} + y_{9} \, \mathbf{a}_{2} + \left(\frac{1}{2} +z_{9}\right) \, \mathbf{a}_{3} & = & -x_{9}a \, \mathbf{\hat{x}} + y_{9}b \, \mathbf{\hat{y}} + \left(\frac{1}{2} +z_{9}\right)c \, \mathbf{\hat{z}} & \left(4e\right) & \mbox{Ca I} \\ 
\mathbf{B}_{29} & = & x_{10} \, \mathbf{a}_{1} + y_{10} \, \mathbf{a}_{2} + z_{10} \, \mathbf{a}_{3} & = & x_{10}a \, \mathbf{\hat{x}} + y_{10}b \, \mathbf{\hat{y}} + z_{10}c \, \mathbf{\hat{z}} & \left(4e\right) & \mbox{Ca II} \\ 
\mathbf{B}_{30} & = & -x_{10} \, \mathbf{a}_{1}-y_{10} \, \mathbf{a}_{2} + z_{10} \, \mathbf{a}_{3} & = & -x_{10}a \, \mathbf{\hat{x}}-y_{10}b \, \mathbf{\hat{y}} + z_{10}c \, \mathbf{\hat{z}} & \left(4e\right) & \mbox{Ca II} \\ 
\mathbf{B}_{31} & = & x_{10} \, \mathbf{a}_{1}-y_{10} \, \mathbf{a}_{2} + \left(\frac{1}{2} +z_{10}\right) \, \mathbf{a}_{3} & = & x_{10}a \, \mathbf{\hat{x}}-y_{10}b \, \mathbf{\hat{y}} + \left(\frac{1}{2} +z_{10}\right)c \, \mathbf{\hat{z}} & \left(4e\right) & \mbox{Ca II} \\ 
\mathbf{B}_{32} & = & -x_{10} \, \mathbf{a}_{1} + y_{10} \, \mathbf{a}_{2} + \left(\frac{1}{2} +z_{10}\right) \, \mathbf{a}_{3} & = & -x_{10}a \, \mathbf{\hat{x}} + y_{10}b \, \mathbf{\hat{y}} + \left(\frac{1}{2} +z_{10}\right)c \, \mathbf{\hat{z}} & \left(4e\right) & \mbox{Ca II} \\ 
\mathbf{B}_{33} & = & x_{11} \, \mathbf{a}_{1} + y_{11} \, \mathbf{a}_{2} + z_{11} \, \mathbf{a}_{3} & = & x_{11}a \, \mathbf{\hat{x}} + y_{11}b \, \mathbf{\hat{y}} + z_{11}c \, \mathbf{\hat{z}} & \left(4e\right) & \mbox{Ca III} \\ 
\mathbf{B}_{34} & = & -x_{11} \, \mathbf{a}_{1}-y_{11} \, \mathbf{a}_{2} + z_{11} \, \mathbf{a}_{3} & = & -x_{11}a \, \mathbf{\hat{x}}-y_{11}b \, \mathbf{\hat{y}} + z_{11}c \, \mathbf{\hat{z}} & \left(4e\right) & \mbox{Ca III} \\ 
\mathbf{B}_{35} & = & x_{11} \, \mathbf{a}_{1}-y_{11} \, \mathbf{a}_{2} + \left(\frac{1}{2} +z_{11}\right) \, \mathbf{a}_{3} & = & x_{11}a \, \mathbf{\hat{x}}-y_{11}b \, \mathbf{\hat{y}} + \left(\frac{1}{2} +z_{11}\right)c \, \mathbf{\hat{z}} & \left(4e\right) & \mbox{Ca III} \\ 
\mathbf{B}_{36} & = & -x_{11} \, \mathbf{a}_{1} + y_{11} \, \mathbf{a}_{2} + \left(\frac{1}{2} +z_{11}\right) \, \mathbf{a}_{3} & = & -x_{11}a \, \mathbf{\hat{x}} + y_{11}b \, \mathbf{\hat{y}} + \left(\frac{1}{2} +z_{11}\right)c \, \mathbf{\hat{z}} & \left(4e\right) & \mbox{Ca III} \\ 
\mathbf{B}_{37} & = & x_{12} \, \mathbf{a}_{1} + y_{12} \, \mathbf{a}_{2} + z_{12} \, \mathbf{a}_{3} & = & x_{12}a \, \mathbf{\hat{x}} + y_{12}b \, \mathbf{\hat{y}} + z_{12}c \, \mathbf{\hat{z}} & \left(4e\right) & \mbox{Ca IV} \\ 
\mathbf{B}_{38} & = & -x_{12} \, \mathbf{a}_{1}-y_{12} \, \mathbf{a}_{2} + z_{12} \, \mathbf{a}_{3} & = & -x_{12}a \, \mathbf{\hat{x}}-y_{12}b \, \mathbf{\hat{y}} + z_{12}c \, \mathbf{\hat{z}} & \left(4e\right) & \mbox{Ca IV} \\ 
\mathbf{B}_{39} & = & x_{12} \, \mathbf{a}_{1}-y_{12} \, \mathbf{a}_{2} + \left(\frac{1}{2} +z_{12}\right) \, \mathbf{a}_{3} & = & x_{12}a \, \mathbf{\hat{x}}-y_{12}b \, \mathbf{\hat{y}} + \left(\frac{1}{2} +z_{12}\right)c \, \mathbf{\hat{z}} & \left(4e\right) & \mbox{Ca IV} \\ 
\mathbf{B}_{40} & = & -x_{12} \, \mathbf{a}_{1} + y_{12} \, \mathbf{a}_{2} + \left(\frac{1}{2} +z_{12}\right) \, \mathbf{a}_{3} & = & -x_{12}a \, \mathbf{\hat{x}} + y_{12}b \, \mathbf{\hat{y}} + \left(\frac{1}{2} +z_{12}\right)c \, \mathbf{\hat{z}} & \left(4e\right) & \mbox{Ca IV} \\ 
\mathbf{B}_{41} & = & x_{13} \, \mathbf{a}_{1} + y_{13} \, \mathbf{a}_{2} + z_{13} \, \mathbf{a}_{3} & = & x_{13}a \, \mathbf{\hat{x}} + y_{13}b \, \mathbf{\hat{y}} + z_{13}c \, \mathbf{\hat{z}} & \left(4e\right) & \mbox{O I} \\ 
\mathbf{B}_{42} & = & -x_{13} \, \mathbf{a}_{1}-y_{13} \, \mathbf{a}_{2} + z_{13} \, \mathbf{a}_{3} & = & -x_{13}a \, \mathbf{\hat{x}}-y_{13}b \, \mathbf{\hat{y}} + z_{13}c \, \mathbf{\hat{z}} & \left(4e\right) & \mbox{O I} \\ 
\mathbf{B}_{43} & = & x_{13} \, \mathbf{a}_{1}-y_{13} \, \mathbf{a}_{2} + \left(\frac{1}{2} +z_{13}\right) \, \mathbf{a}_{3} & = & x_{13}a \, \mathbf{\hat{x}}-y_{13}b \, \mathbf{\hat{y}} + \left(\frac{1}{2} +z_{13}\right)c \, \mathbf{\hat{z}} & \left(4e\right) & \mbox{O I} \\ 
\mathbf{B}_{44} & = & -x_{13} \, \mathbf{a}_{1} + y_{13} \, \mathbf{a}_{2} + \left(\frac{1}{2} +z_{13}\right) \, \mathbf{a}_{3} & = & -x_{13}a \, \mathbf{\hat{x}} + y_{13}b \, \mathbf{\hat{y}} + \left(\frac{1}{2} +z_{13}\right)c \, \mathbf{\hat{z}} & \left(4e\right) & \mbox{O I} \\ 
\mathbf{B}_{45} & = & x_{14} \, \mathbf{a}_{1} + y_{14} \, \mathbf{a}_{2} + z_{14} \, \mathbf{a}_{3} & = & x_{14}a \, \mathbf{\hat{x}} + y_{14}b \, \mathbf{\hat{y}} + z_{14}c \, \mathbf{\hat{z}} & \left(4e\right) & \mbox{O II} \\ 
\mathbf{B}_{46} & = & -x_{14} \, \mathbf{a}_{1}-y_{14} \, \mathbf{a}_{2} + z_{14} \, \mathbf{a}_{3} & = & -x_{14}a \, \mathbf{\hat{x}}-y_{14}b \, \mathbf{\hat{y}} + z_{14}c \, \mathbf{\hat{z}} & \left(4e\right) & \mbox{O II} \\ 
\mathbf{B}_{47} & = & x_{14} \, \mathbf{a}_{1}-y_{14} \, \mathbf{a}_{2} + \left(\frac{1}{2} +z_{14}\right) \, \mathbf{a}_{3} & = & x_{14}a \, \mathbf{\hat{x}}-y_{14}b \, \mathbf{\hat{y}} + \left(\frac{1}{2} +z_{14}\right)c \, \mathbf{\hat{z}} & \left(4e\right) & \mbox{O II} \\ 
\mathbf{B}_{48} & = & -x_{14} \, \mathbf{a}_{1} + y_{14} \, \mathbf{a}_{2} + \left(\frac{1}{2} +z_{14}\right) \, \mathbf{a}_{3} & = & -x_{14}a \, \mathbf{\hat{x}} + y_{14}b \, \mathbf{\hat{y}} + \left(\frac{1}{2} +z_{14}\right)c \, \mathbf{\hat{z}} & \left(4e\right) & \mbox{O II} \\ 
\mathbf{B}_{49} & = & x_{15} \, \mathbf{a}_{1} + y_{15} \, \mathbf{a}_{2} + z_{15} \, \mathbf{a}_{3} & = & x_{15}a \, \mathbf{\hat{x}} + y_{15}b \, \mathbf{\hat{y}} + z_{15}c \, \mathbf{\hat{z}} & \left(4e\right) & \mbox{O III} \\ 
\mathbf{B}_{50} & = & -x_{15} \, \mathbf{a}_{1}-y_{15} \, \mathbf{a}_{2} + z_{15} \, \mathbf{a}_{3} & = & -x_{15}a \, \mathbf{\hat{x}}-y_{15}b \, \mathbf{\hat{y}} + z_{15}c \, \mathbf{\hat{z}} & \left(4e\right) & \mbox{O III} \\ 
\mathbf{B}_{51} & = & x_{15} \, \mathbf{a}_{1}-y_{15} \, \mathbf{a}_{2} + \left(\frac{1}{2} +z_{15}\right) \, \mathbf{a}_{3} & = & x_{15}a \, \mathbf{\hat{x}}-y_{15}b \, \mathbf{\hat{y}} + \left(\frac{1}{2} +z_{15}\right)c \, \mathbf{\hat{z}} & \left(4e\right) & \mbox{O III} \\ 
\mathbf{B}_{52} & = & -x_{15} \, \mathbf{a}_{1} + y_{15} \, \mathbf{a}_{2} + \left(\frac{1}{2} +z_{15}\right) \, \mathbf{a}_{3} & = & -x_{15}a \, \mathbf{\hat{x}} + y_{15}b \, \mathbf{\hat{y}} + \left(\frac{1}{2} +z_{15}\right)c \, \mathbf{\hat{z}} & \left(4e\right) & \mbox{O III} \\ 
\mathbf{B}_{53} & = & x_{16} \, \mathbf{a}_{1} + y_{16} \, \mathbf{a}_{2} + z_{16} \, \mathbf{a}_{3} & = & x_{16}a \, \mathbf{\hat{x}} + y_{16}b \, \mathbf{\hat{y}} + z_{16}c \, \mathbf{\hat{z}} & \left(4e\right) & \mbox{O IV} \\ 
\mathbf{B}_{54} & = & -x_{16} \, \mathbf{a}_{1}-y_{16} \, \mathbf{a}_{2} + z_{16} \, \mathbf{a}_{3} & = & -x_{16}a \, \mathbf{\hat{x}}-y_{16}b \, \mathbf{\hat{y}} + z_{16}c \, \mathbf{\hat{z}} & \left(4e\right) & \mbox{O IV} \\ 
\mathbf{B}_{55} & = & x_{16} \, \mathbf{a}_{1}-y_{16} \, \mathbf{a}_{2} + \left(\frac{1}{2} +z_{16}\right) \, \mathbf{a}_{3} & = & x_{16}a \, \mathbf{\hat{x}}-y_{16}b \, \mathbf{\hat{y}} + \left(\frac{1}{2} +z_{16}\right)c \, \mathbf{\hat{z}} & \left(4e\right) & \mbox{O IV} \\ 
\mathbf{B}_{56} & = & -x_{16} \, \mathbf{a}_{1} + y_{16} \, \mathbf{a}_{2} + \left(\frac{1}{2} +z_{16}\right) \, \mathbf{a}_{3} & = & -x_{16}a \, \mathbf{\hat{x}} + y_{16}b \, \mathbf{\hat{y}} + \left(\frac{1}{2} +z_{16}\right)c \, \mathbf{\hat{z}} & \left(4e\right) & \mbox{O IV} \\ 
\mathbf{B}_{57} & = & x_{17} \, \mathbf{a}_{1} + y_{17} \, \mathbf{a}_{2} + z_{17} \, \mathbf{a}_{3} & = & x_{17}a \, \mathbf{\hat{x}} + y_{17}b \, \mathbf{\hat{y}} + z_{17}c \, \mathbf{\hat{z}} & \left(4e\right) & \mbox{O V} \\ 
\mathbf{B}_{58} & = & -x_{17} \, \mathbf{a}_{1}-y_{17} \, \mathbf{a}_{2} + z_{17} \, \mathbf{a}_{3} & = & -x_{17}a \, \mathbf{\hat{x}}-y_{17}b \, \mathbf{\hat{y}} + z_{17}c \, \mathbf{\hat{z}} & \left(4e\right) & \mbox{O V} \\ 
\mathbf{B}_{59} & = & x_{17} \, \mathbf{a}_{1}-y_{17} \, \mathbf{a}_{2} + \left(\frac{1}{2} +z_{17}\right) \, \mathbf{a}_{3} & = & x_{17}a \, \mathbf{\hat{x}}-y_{17}b \, \mathbf{\hat{y}} + \left(\frac{1}{2} +z_{17}\right)c \, \mathbf{\hat{z}} & \left(4e\right) & \mbox{O V} \\ 
\mathbf{B}_{60} & = & -x_{17} \, \mathbf{a}_{1} + y_{17} \, \mathbf{a}_{2} + \left(\frac{1}{2} +z_{17}\right) \, \mathbf{a}_{3} & = & -x_{17}a \, \mathbf{\hat{x}} + y_{17}b \, \mathbf{\hat{y}} + \left(\frac{1}{2} +z_{17}\right)c \, \mathbf{\hat{z}} & \left(4e\right) & \mbox{O V} \\ 
\mathbf{B}_{61} & = & x_{18} \, \mathbf{a}_{1} + y_{18} \, \mathbf{a}_{2} + z_{18} \, \mathbf{a}_{3} & = & x_{18}a \, \mathbf{\hat{x}} + y_{18}b \, \mathbf{\hat{y}} + z_{18}c \, \mathbf{\hat{z}} & \left(4e\right) & \mbox{O VI} \\ 
\mathbf{B}_{62} & = & -x_{18} \, \mathbf{a}_{1}-y_{18} \, \mathbf{a}_{2} + z_{18} \, \mathbf{a}_{3} & = & -x_{18}a \, \mathbf{\hat{x}}-y_{18}b \, \mathbf{\hat{y}} + z_{18}c \, \mathbf{\hat{z}} & \left(4e\right) & \mbox{O VI} \\ 
\mathbf{B}_{63} & = & x_{18} \, \mathbf{a}_{1}-y_{18} \, \mathbf{a}_{2} + \left(\frac{1}{2} +z_{18}\right) \, \mathbf{a}_{3} & = & x_{18}a \, \mathbf{\hat{x}}-y_{18}b \, \mathbf{\hat{y}} + \left(\frac{1}{2} +z_{18}\right)c \, \mathbf{\hat{z}} & \left(4e\right) & \mbox{O VI} \\ 
\mathbf{B}_{64} & = & -x_{18} \, \mathbf{a}_{1} + y_{18} \, \mathbf{a}_{2} + \left(\frac{1}{2} +z_{18}\right) \, \mathbf{a}_{3} & = & -x_{18}a \, \mathbf{\hat{x}} + y_{18}b \, \mathbf{\hat{y}} + \left(\frac{1}{2} +z_{18}\right)c \, \mathbf{\hat{z}} & \left(4e\right) & \mbox{O VI} \\ 
\mathbf{B}_{65} & = & x_{19} \, \mathbf{a}_{1} + y_{19} \, \mathbf{a}_{2} + z_{19} \, \mathbf{a}_{3} & = & x_{19}a \, \mathbf{\hat{x}} + y_{19}b \, \mathbf{\hat{y}} + z_{19}c \, \mathbf{\hat{z}} & \left(4e\right) & \mbox{O VII} \\ 
\mathbf{B}_{66} & = & -x_{19} \, \mathbf{a}_{1}-y_{19} \, \mathbf{a}_{2} + z_{19} \, \mathbf{a}_{3} & = & -x_{19}a \, \mathbf{\hat{x}}-y_{19}b \, \mathbf{\hat{y}} + z_{19}c \, \mathbf{\hat{z}} & \left(4e\right) & \mbox{O VII} \\ 
\mathbf{B}_{67} & = & x_{19} \, \mathbf{a}_{1}-y_{19} \, \mathbf{a}_{2} + \left(\frac{1}{2} +z_{19}\right) \, \mathbf{a}_{3} & = & x_{19}a \, \mathbf{\hat{x}}-y_{19}b \, \mathbf{\hat{y}} + \left(\frac{1}{2} +z_{19}\right)c \, \mathbf{\hat{z}} & \left(4e\right) & \mbox{O VII} \\ 
\mathbf{B}_{68} & = & -x_{19} \, \mathbf{a}_{1} + y_{19} \, \mathbf{a}_{2} + \left(\frac{1}{2} +z_{19}\right) \, \mathbf{a}_{3} & = & -x_{19}a \, \mathbf{\hat{x}} + y_{19}b \, \mathbf{\hat{y}} + \left(\frac{1}{2} +z_{19}\right)c \, \mathbf{\hat{z}} & \left(4e\right) & \mbox{O VII} \\ 
\mathbf{B}_{69} & = & x_{20} \, \mathbf{a}_{1} + y_{20} \, \mathbf{a}_{2} + z_{20} \, \mathbf{a}_{3} & = & x_{20}a \, \mathbf{\hat{x}} + y_{20}b \, \mathbf{\hat{y}} + z_{20}c \, \mathbf{\hat{z}} & \left(4e\right) & \mbox{O VIII} \\ 
\mathbf{B}_{70} & = & -x_{20} \, \mathbf{a}_{1}-y_{20} \, \mathbf{a}_{2} + z_{20} \, \mathbf{a}_{3} & = & -x_{20}a \, \mathbf{\hat{x}}-y_{20}b \, \mathbf{\hat{y}} + z_{20}c \, \mathbf{\hat{z}} & \left(4e\right) & \mbox{O VIII} \\ 
\mathbf{B}_{71} & = & x_{20} \, \mathbf{a}_{1}-y_{20} \, \mathbf{a}_{2} + \left(\frac{1}{2} +z_{20}\right) \, \mathbf{a}_{3} & = & x_{20}a \, \mathbf{\hat{x}}-y_{20}b \, \mathbf{\hat{y}} + \left(\frac{1}{2} +z_{20}\right)c \, \mathbf{\hat{z}} & \left(4e\right) & \mbox{O VIII} \\ 
\mathbf{B}_{72} & = & -x_{20} \, \mathbf{a}_{1} + y_{20} \, \mathbf{a}_{2} + \left(\frac{1}{2} +z_{20}\right) \, \mathbf{a}_{3} & = & -x_{20}a \, \mathbf{\hat{x}} + y_{20}b \, \mathbf{\hat{y}} + \left(\frac{1}{2} +z_{20}\right)c \, \mathbf{\hat{z}} & \left(4e\right) & \mbox{O VIII} \\ 
\mathbf{B}_{73} & = & x_{21} \, \mathbf{a}_{1} + y_{21} \, \mathbf{a}_{2} + z_{21} \, \mathbf{a}_{3} & = & x_{21}a \, \mathbf{\hat{x}} + y_{21}b \, \mathbf{\hat{y}} + z_{21}c \, \mathbf{\hat{z}} & \left(4e\right) & \mbox{O IX} \\ 
\mathbf{B}_{74} & = & -x_{21} \, \mathbf{a}_{1}-y_{21} \, \mathbf{a}_{2} + z_{21} \, \mathbf{a}_{3} & = & -x_{21}a \, \mathbf{\hat{x}}-y_{21}b \, \mathbf{\hat{y}} + z_{21}c \, \mathbf{\hat{z}} & \left(4e\right) & \mbox{O IX} \\ 
\mathbf{B}_{75} & = & x_{21} \, \mathbf{a}_{1}-y_{21} \, \mathbf{a}_{2} + \left(\frac{1}{2} +z_{21}\right) \, \mathbf{a}_{3} & = & x_{21}a \, \mathbf{\hat{x}}-y_{21}b \, \mathbf{\hat{y}} + \left(\frac{1}{2} +z_{21}\right)c \, \mathbf{\hat{z}} & \left(4e\right) & \mbox{O IX} \\ 
\mathbf{B}_{76} & = & -x_{21} \, \mathbf{a}_{1} + y_{21} \, \mathbf{a}_{2} + \left(\frac{1}{2} +z_{21}\right) \, \mathbf{a}_{3} & = & -x_{21}a \, \mathbf{\hat{x}} + y_{21}b \, \mathbf{\hat{y}} + \left(\frac{1}{2} +z_{21}\right)c \, \mathbf{\hat{z}} & \left(4e\right) & \mbox{O IX} \\ 
\mathbf{B}_{77} & = & x_{22} \, \mathbf{a}_{1} + y_{22} \, \mathbf{a}_{2} + z_{22} \, \mathbf{a}_{3} & = & x_{22}a \, \mathbf{\hat{x}} + y_{22}b \, \mathbf{\hat{y}} + z_{22}c \, \mathbf{\hat{z}} & \left(4e\right) & \mbox{O X} \\ 
\mathbf{B}_{78} & = & -x_{22} \, \mathbf{a}_{1}-y_{22} \, \mathbf{a}_{2} + z_{22} \, \mathbf{a}_{3} & = & -x_{22}a \, \mathbf{\hat{x}}-y_{22}b \, \mathbf{\hat{y}} + z_{22}c \, \mathbf{\hat{z}} & \left(4e\right) & \mbox{O X} \\ 
\mathbf{B}_{79} & = & x_{22} \, \mathbf{a}_{1}-y_{22} \, \mathbf{a}_{2} + \left(\frac{1}{2} +z_{22}\right) \, \mathbf{a}_{3} & = & x_{22}a \, \mathbf{\hat{x}}-y_{22}b \, \mathbf{\hat{y}} + \left(\frac{1}{2} +z_{22}\right)c \, \mathbf{\hat{z}} & \left(4e\right) & \mbox{O X} \\ 
\mathbf{B}_{80} & = & -x_{22} \, \mathbf{a}_{1} + y_{22} \, \mathbf{a}_{2} + \left(\frac{1}{2} +z_{22}\right) \, \mathbf{a}_{3} & = & -x_{22}a \, \mathbf{\hat{x}} + y_{22}b \, \mathbf{\hat{y}} + \left(\frac{1}{2} +z_{22}\right)c \, \mathbf{\hat{z}} & \left(4e\right) & \mbox{O X} \\ 
\mathbf{B}_{81} & = & x_{23} \, \mathbf{a}_{1} + y_{23} \, \mathbf{a}_{2} + z_{23} \, \mathbf{a}_{3} & = & x_{23}a \, \mathbf{\hat{x}} + y_{23}b \, \mathbf{\hat{y}} + z_{23}c \, \mathbf{\hat{z}} & \left(4e\right) & \mbox{O XI} \\ 
\mathbf{B}_{82} & = & -x_{23} \, \mathbf{a}_{1}-y_{23} \, \mathbf{a}_{2} + z_{23} \, \mathbf{a}_{3} & = & -x_{23}a \, \mathbf{\hat{x}}-y_{23}b \, \mathbf{\hat{y}} + z_{23}c \, \mathbf{\hat{z}} & \left(4e\right) & \mbox{O XI} \\ 
\mathbf{B}_{83} & = & x_{23} \, \mathbf{a}_{1}-y_{23} \, \mathbf{a}_{2} + \left(\frac{1}{2} +z_{23}\right) \, \mathbf{a}_{3} & = & x_{23}a \, \mathbf{\hat{x}}-y_{23}b \, \mathbf{\hat{y}} + \left(\frac{1}{2} +z_{23}\right)c \, \mathbf{\hat{z}} & \left(4e\right) & \mbox{O XI} \\ 
\mathbf{B}_{84} & = & -x_{23} \, \mathbf{a}_{1} + y_{23} \, \mathbf{a}_{2} + \left(\frac{1}{2} +z_{23}\right) \, \mathbf{a}_{3} & = & -x_{23}a \, \mathbf{\hat{x}} + y_{23}b \, \mathbf{\hat{y}} + \left(\frac{1}{2} +z_{23}\right)c \, \mathbf{\hat{z}} & \left(4e\right) & \mbox{O XI} \\ 
\mathbf{B}_{85} & = & x_{24} \, \mathbf{a}_{1} + y_{24} \, \mathbf{a}_{2} + z_{24} \, \mathbf{a}_{3} & = & x_{24}a \, \mathbf{\hat{x}} + y_{24}b \, \mathbf{\hat{y}} + z_{24}c \, \mathbf{\hat{z}} & \left(4e\right) & \mbox{O XII} \\ 
\mathbf{B}_{86} & = & -x_{24} \, \mathbf{a}_{1}-y_{24} \, \mathbf{a}_{2} + z_{24} \, \mathbf{a}_{3} & = & -x_{24}a \, \mathbf{\hat{x}}-y_{24}b \, \mathbf{\hat{y}} + z_{24}c \, \mathbf{\hat{z}} & \left(4e\right) & \mbox{O XII} \\ 
\mathbf{B}_{87} & = & x_{24} \, \mathbf{a}_{1}-y_{24} \, \mathbf{a}_{2} + \left(\frac{1}{2} +z_{24}\right) \, \mathbf{a}_{3} & = & x_{24}a \, \mathbf{\hat{x}}-y_{24}b \, \mathbf{\hat{y}} + \left(\frac{1}{2} +z_{24}\right)c \, \mathbf{\hat{z}} & \left(4e\right) & \mbox{O XII} \\ 
\mathbf{B}_{88} & = & -x_{24} \, \mathbf{a}_{1} + y_{24} \, \mathbf{a}_{2} + \left(\frac{1}{2} +z_{24}\right) \, \mathbf{a}_{3} & = & -x_{24}a \, \mathbf{\hat{x}} + y_{24}b \, \mathbf{\hat{y}} + \left(\frac{1}{2} +z_{24}\right)c \, \mathbf{\hat{z}} & \left(4e\right) & \mbox{O XII} \\ 
\mathbf{B}_{89} & = & x_{25} \, \mathbf{a}_{1} + y_{25} \, \mathbf{a}_{2} + z_{25} \, \mathbf{a}_{3} & = & x_{25}a \, \mathbf{\hat{x}} + y_{25}b \, \mathbf{\hat{y}} + z_{25}c \, \mathbf{\hat{z}} & \left(4e\right) & \mbox{O XIII} \\ 
\mathbf{B}_{90} & = & -x_{25} \, \mathbf{a}_{1}-y_{25} \, \mathbf{a}_{2} + z_{25} \, \mathbf{a}_{3} & = & -x_{25}a \, \mathbf{\hat{x}}-y_{25}b \, \mathbf{\hat{y}} + z_{25}c \, \mathbf{\hat{z}} & \left(4e\right) & \mbox{O XIII} \\ 
\mathbf{B}_{91} & = & x_{25} \, \mathbf{a}_{1}-y_{25} \, \mathbf{a}_{2} + \left(\frac{1}{2} +z_{25}\right) \, \mathbf{a}_{3} & = & x_{25}a \, \mathbf{\hat{x}}-y_{25}b \, \mathbf{\hat{y}} + \left(\frac{1}{2} +z_{25}\right)c \, \mathbf{\hat{z}} & \left(4e\right) & \mbox{O XIII} \\ 
\mathbf{B}_{92} & = & -x_{25} \, \mathbf{a}_{1} + y_{25} \, \mathbf{a}_{2} + \left(\frac{1}{2} +z_{25}\right) \, \mathbf{a}_{3} & = & -x_{25}a \, \mathbf{\hat{x}} + y_{25}b \, \mathbf{\hat{y}} + \left(\frac{1}{2} +z_{25}\right)c \, \mathbf{\hat{z}} & \left(4e\right) & \mbox{O XIII} \\ 
\mathbf{B}_{93} & = & x_{26} \, \mathbf{a}_{1} + y_{26} \, \mathbf{a}_{2} + z_{26} \, \mathbf{a}_{3} & = & x_{26}a \, \mathbf{\hat{x}} + y_{26}b \, \mathbf{\hat{y}} + z_{26}c \, \mathbf{\hat{z}} & \left(4e\right) & \mbox{O XIV} \\ 
\mathbf{B}_{94} & = & -x_{26} \, \mathbf{a}_{1}-y_{26} \, \mathbf{a}_{2} + z_{26} \, \mathbf{a}_{3} & = & -x_{26}a \, \mathbf{\hat{x}}-y_{26}b \, \mathbf{\hat{y}} + z_{26}c \, \mathbf{\hat{z}} & \left(4e\right) & \mbox{O XIV} \\ 
\mathbf{B}_{95} & = & x_{26} \, \mathbf{a}_{1}-y_{26} \, \mathbf{a}_{2} + \left(\frac{1}{2} +z_{26}\right) \, \mathbf{a}_{3} & = & x_{26}a \, \mathbf{\hat{x}}-y_{26}b \, \mathbf{\hat{y}} + \left(\frac{1}{2} +z_{26}\right)c \, \mathbf{\hat{z}} & \left(4e\right) & \mbox{O XIV} \\ 
\mathbf{B}_{96} & = & -x_{26} \, \mathbf{a}_{1} + y_{26} \, \mathbf{a}_{2} + \left(\frac{1}{2} +z_{26}\right) \, \mathbf{a}_{3} & = & -x_{26}a \, \mathbf{\hat{x}} + y_{26}b \, \mathbf{\hat{y}} + \left(\frac{1}{2} +z_{26}\right)c \, \mathbf{\hat{z}} & \left(4e\right) & \mbox{O XIV} \\ 
\mathbf{B}_{97} & = & x_{27} \, \mathbf{a}_{1} + y_{27} \, \mathbf{a}_{2} + z_{27} \, \mathbf{a}_{3} & = & x_{27}a \, \mathbf{\hat{x}} + y_{27}b \, \mathbf{\hat{y}} + z_{27}c \, \mathbf{\hat{z}} & \left(4e\right) & \mbox{O XV} \\ 
\mathbf{B}_{98} & = & -x_{27} \, \mathbf{a}_{1}-y_{27} \, \mathbf{a}_{2} + z_{27} \, \mathbf{a}_{3} & = & -x_{27}a \, \mathbf{\hat{x}}-y_{27}b \, \mathbf{\hat{y}} + z_{27}c \, \mathbf{\hat{z}} & \left(4e\right) & \mbox{O XV} \\ 
\mathbf{B}_{99} & = & x_{27} \, \mathbf{a}_{1}-y_{27} \, \mathbf{a}_{2} + \left(\frac{1}{2} +z_{27}\right) \, \mathbf{a}_{3} & = & x_{27}a \, \mathbf{\hat{x}}-y_{27}b \, \mathbf{\hat{y}} + \left(\frac{1}{2} +z_{27}\right)c \, \mathbf{\hat{z}} & \left(4e\right) & \mbox{O XV} \\ 
\mathbf{B}_{100} & = & -x_{27} \, \mathbf{a}_{1} + y_{27} \, \mathbf{a}_{2} + \left(\frac{1}{2} +z_{27}\right) \, \mathbf{a}_{3} & = & -x_{27}a \, \mathbf{\hat{x}} + y_{27}b \, \mathbf{\hat{y}} + \left(\frac{1}{2} +z_{27}\right)c \, \mathbf{\hat{z}} & \left(4e\right) & \mbox{O XV} \\ 
\mathbf{B}_{101} & = & x_{28} \, \mathbf{a}_{1} + y_{28} \, \mathbf{a}_{2} + z_{28} \, \mathbf{a}_{3} & = & x_{28}a \, \mathbf{\hat{x}} + y_{28}b \, \mathbf{\hat{y}} + z_{28}c \, \mathbf{\hat{z}} & \left(4e\right) & \mbox{O XVI} \\ 
\mathbf{B}_{102} & = & -x_{28} \, \mathbf{a}_{1}-y_{28} \, \mathbf{a}_{2} + z_{28} \, \mathbf{a}_{3} & = & -x_{28}a \, \mathbf{\hat{x}}-y_{28}b \, \mathbf{\hat{y}} + z_{28}c \, \mathbf{\hat{z}} & \left(4e\right) & \mbox{O XVI} \\ 
\mathbf{B}_{103} & = & x_{28} \, \mathbf{a}_{1}-y_{28} \, \mathbf{a}_{2} + \left(\frac{1}{2} +z_{28}\right) \, \mathbf{a}_{3} & = & x_{28}a \, \mathbf{\hat{x}}-y_{28}b \, \mathbf{\hat{y}} + \left(\frac{1}{2} +z_{28}\right)c \, \mathbf{\hat{z}} & \left(4e\right) & \mbox{O XVI} \\ 
\mathbf{B}_{104} & = & -x_{28} \, \mathbf{a}_{1} + y_{28} \, \mathbf{a}_{2} + \left(\frac{1}{2} +z_{28}\right) \, \mathbf{a}_{3} & = & -x_{28}a \, \mathbf{\hat{x}} + y_{28}b \, \mathbf{\hat{y}} + \left(\frac{1}{2} +z_{28}\right)c \, \mathbf{\hat{z}} & \left(4e\right) & \mbox{O XVI} \\ 
\mathbf{B}_{105} & = & x_{29} \, \mathbf{a}_{1} + y_{29} \, \mathbf{a}_{2} + z_{29} \, \mathbf{a}_{3} & = & x_{29}a \, \mathbf{\hat{x}} + y_{29}b \, \mathbf{\hat{y}} + z_{29}c \, \mathbf{\hat{z}} & \left(4e\right) & \mbox{S} \\ 
\mathbf{B}_{106} & = & -x_{29} \, \mathbf{a}_{1}-y_{29} \, \mathbf{a}_{2} + z_{29} \, \mathbf{a}_{3} & = & -x_{29}a \, \mathbf{\hat{x}}-y_{29}b \, \mathbf{\hat{y}} + z_{29}c \, \mathbf{\hat{z}} & \left(4e\right) & \mbox{S} \\ 
\mathbf{B}_{107} & = & x_{29} \, \mathbf{a}_{1}-y_{29} \, \mathbf{a}_{2} + \left(\frac{1}{2} +z_{29}\right) \, \mathbf{a}_{3} & = & x_{29}a \, \mathbf{\hat{x}}-y_{29}b \, \mathbf{\hat{y}} + \left(\frac{1}{2} +z_{29}\right)c \, \mathbf{\hat{z}} & \left(4e\right) & \mbox{S} \\ 
\mathbf{B}_{108} & = & -x_{29} \, \mathbf{a}_{1} + y_{29} \, \mathbf{a}_{2} + \left(\frac{1}{2} +z_{29}\right) \, \mathbf{a}_{3} & = & -x_{29}a \, \mathbf{\hat{x}} + y_{29}b \, \mathbf{\hat{y}} + \left(\frac{1}{2} +z_{29}\right)c \, \mathbf{\hat{z}} & \left(4e\right) & \mbox{S} \\ 
\end{longtabu}
\renewcommand{\arraystretch}{1.0}
\noindent \hrulefill
\\
\textbf{References:}
\vspace*{-0.25cm}
\begin{flushleft}
  - \bibentry{Calos_Ca4Al6SO4O12_JSolStateChem_1995}. \\
\end{flushleft}
\textbf{Found in:}
\vspace*{-0.25cm}
\begin{flushleft}
  - \bibentry{Villars_PearsonsCrystalData_2013}. \\
\end{flushleft}
\noindent \hrulefill
\\
\textbf{Geometry files:}
\\
\noindent  - CIF: pp. {\hyperref[A6B4C16D_oP108_27_abcd4e_4e_16e_e_cif]{\pageref{A6B4C16D_oP108_27_abcd4e_4e_16e_e_cif}}} \\
\noindent  - POSCAR: pp. {\hyperref[A6B4C16D_oP108_27_abcd4e_4e_16e_e_poscar]{\pageref{A6B4C16D_oP108_27_abcd4e_4e_16e_e_poscar}}} \\
\onecolumn
{\phantomsection\label{A2B_oP12_29_2a_a}}
\subsection*{\huge \textbf{{\normalfont ZrO$_{2}$ Structure: A2B\_oP12\_29\_2a\_a}}}
\noindent \hrulefill
\vspace*{0.25cm}
\begin{figure}[htp]
  \centering
  \vspace{-1em}
  {\includegraphics[width=1\textwidth]{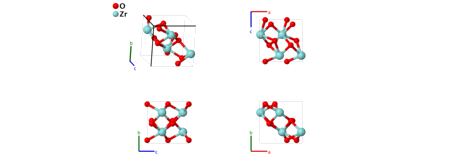}}
\end{figure}
\vspace*{-0.5cm}
\renewcommand{\arraystretch}{1.5}
\begin{equation*}
  \begin{array}{>{$\hspace{-0.15cm}}l<{$}>{$}p{0.5cm}<{$}>{$}p{18.5cm}<{$}}
    \mbox{\large \textbf{Prototype}} &\colon & \ce{ZrO2} \\
    \mbox{\large \textbf{\AFLOW\ prototype label}} &\colon & \mbox{A2B\_oP12\_29\_2a\_a} \\
    \mbox{\large \textbf{\textit{Strukturbericht} designation}} &\colon & \mbox{None} \\
    \mbox{\large \textbf{Pearson symbol}} &\colon & \mbox{oP12} \\
    \mbox{\large \textbf{Space group number}} &\colon & 29 \\
    \mbox{\large \textbf{Space group symbol}} &\colon & Pca2_{1} \\
    \mbox{\large \textbf{\AFLOW\ prototype command}} &\colon &  \texttt{aflow} \,  \, \texttt{-{}-proto=A2B\_oP12\_29\_2a\_a } \, \newline \texttt{-{}-params=}{a,b/a,c/a,x_{1},y_{1},z_{1},x_{2},y_{2},z_{2},x_{3},y_{3},z_{3} }
  \end{array}
\end{equation*}
\renewcommand{\arraystretch}{1.0}

\vspace*{-0.25cm}
\noindent \hrulefill
\begin{itemize}
  \item{ZrO$_{2}$ (pp. {\hyperref[A2B_oP12_29_2a_a]{\pageref{A2B_oP12_29_2a_a}}}) and 
Pyrite (pp. {\hyperref[AB2_oP12_29_a_2a]{\pageref{AB2_oP12_29_a_2a}}}) 
have similar \AFLOW\ prototype labels ({\it{i.e.}}, same symmetry and set of 
Wyckoff positions with different stoichiometry labels due to alphabetic ordering of atomic species). 
They are generated by the same symmetry operations with different sets of parameters 
(\texttt{-{}-params}) specified in their corresponding \CIF\ files.
}
\end{itemize}

\noindent \parbox{1 \linewidth}{
\noindent \hrulefill
\\
\textbf{Simple Orthorhombic primitive vectors:} \\
\vspace*{-0.25cm}
\begin{tabular}{cc}
  \begin{tabular}{c}
    \parbox{0.6 \linewidth}{
      \renewcommand{\arraystretch}{1.5}
      \begin{equation*}
        \centering
        \begin{array}{ccc}
              \mathbf{a}_1 & = & a \, \mathbf{\hat{x}} \\
    \mathbf{a}_2 & = & b \, \mathbf{\hat{y}} \\
    \mathbf{a}_3 & = & c \, \mathbf{\hat{z}} \\

        \end{array}
      \end{equation*}
    }
    \renewcommand{\arraystretch}{1.0}
  \end{tabular}
  \begin{tabular}{c}
    \includegraphics[width=0.3\linewidth]{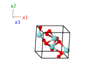} \\
  \end{tabular}
\end{tabular}

}
\vspace*{-0.25cm}

\noindent \hrulefill
\\
\textbf{Basis vectors:}
\vspace*{-0.25cm}
\renewcommand{\arraystretch}{1.5}
\begin{longtabu} to \textwidth{>{\centering $}X[-1,c,c]<{$}>{\centering $}X[-1,c,c]<{$}>{\centering $}X[-1,c,c]<{$}>{\centering $}X[-1,c,c]<{$}>{\centering $}X[-1,c,c]<{$}>{\centering $}X[-1,c,c]<{$}>{\centering $}X[-1,c,c]<{$}}
  & & \mbox{Lattice Coordinates} & & \mbox{Cartesian Coordinates} &\mbox{Wyckoff Position} & \mbox{Atom Type} \\  
  \mathbf{B}_{1} & = & x_{1} \, \mathbf{a}_{1} + y_{1} \, \mathbf{a}_{2} + z_{1} \, \mathbf{a}_{3} & = & x_{1}a \, \mathbf{\hat{x}} + y_{1}b \, \mathbf{\hat{y}} + z_{1}c \, \mathbf{\hat{z}} & \left(4a\right) & \mbox{O I} \\ 
\mathbf{B}_{2} & = & -x_{1} \, \mathbf{a}_{1}-y_{1} \, \mathbf{a}_{2} + \left(\frac{1}{2} +z_{1}\right) \, \mathbf{a}_{3} & = & -x_{1}a \, \mathbf{\hat{x}}-y_{1}b \, \mathbf{\hat{y}} + \left(\frac{1}{2} +z_{1}\right)c \, \mathbf{\hat{z}} & \left(4a\right) & \mbox{O I} \\ 
\mathbf{B}_{3} & = & \left(\frac{1}{2} +x_{1}\right) \, \mathbf{a}_{1}-y_{1} \, \mathbf{a}_{2} + z_{1} \, \mathbf{a}_{3} & = & \left(\frac{1}{2} +x_{1}\right)a \, \mathbf{\hat{x}}-y_{1}b \, \mathbf{\hat{y}} + z_{1}c \, \mathbf{\hat{z}} & \left(4a\right) & \mbox{O I} \\ 
\mathbf{B}_{4} & = & \left(\frac{1}{2} - x_{1}\right) \, \mathbf{a}_{1} + y_{1} \, \mathbf{a}_{2} + \left(\frac{1}{2} +z_{1}\right) \, \mathbf{a}_{3} & = & \left(\frac{1}{2} - x_{1}\right)a \, \mathbf{\hat{x}} + y_{1}b \, \mathbf{\hat{y}} + \left(\frac{1}{2} +z_{1}\right)c \, \mathbf{\hat{z}} & \left(4a\right) & \mbox{O I} \\ 
\mathbf{B}_{5} & = & x_{2} \, \mathbf{a}_{1} + y_{2} \, \mathbf{a}_{2} + z_{2} \, \mathbf{a}_{3} & = & x_{2}a \, \mathbf{\hat{x}} + y_{2}b \, \mathbf{\hat{y}} + z_{2}c \, \mathbf{\hat{z}} & \left(4a\right) & \mbox{O II} \\ 
\mathbf{B}_{6} & = & -x_{2} \, \mathbf{a}_{1}-y_{2} \, \mathbf{a}_{2} + \left(\frac{1}{2} +z_{2}\right) \, \mathbf{a}_{3} & = & -x_{2}a \, \mathbf{\hat{x}}-y_{2}b \, \mathbf{\hat{y}} + \left(\frac{1}{2} +z_{2}\right)c \, \mathbf{\hat{z}} & \left(4a\right) & \mbox{O II} \\ 
\mathbf{B}_{7} & = & \left(\frac{1}{2} +x_{2}\right) \, \mathbf{a}_{1}-y_{2} \, \mathbf{a}_{2} + z_{2} \, \mathbf{a}_{3} & = & \left(\frac{1}{2} +x_{2}\right)a \, \mathbf{\hat{x}}-y_{2}b \, \mathbf{\hat{y}} + z_{2}c \, \mathbf{\hat{z}} & \left(4a\right) & \mbox{O II} \\ 
\mathbf{B}_{8} & = & \left(\frac{1}{2} - x_{2}\right) \, \mathbf{a}_{1} + y_{2} \, \mathbf{a}_{2} + \left(\frac{1}{2} +z_{2}\right) \, \mathbf{a}_{3} & = & \left(\frac{1}{2} - x_{2}\right)a \, \mathbf{\hat{x}} + y_{2}b \, \mathbf{\hat{y}} + \left(\frac{1}{2} +z_{2}\right)c \, \mathbf{\hat{z}} & \left(4a\right) & \mbox{O II} \\ 
\mathbf{B}_{9} & = & x_{3} \, \mathbf{a}_{1} + y_{3} \, \mathbf{a}_{2} + z_{3} \, \mathbf{a}_{3} & = & x_{3}a \, \mathbf{\hat{x}} + y_{3}b \, \mathbf{\hat{y}} + z_{3}c \, \mathbf{\hat{z}} & \left(4a\right) & \mbox{Zr} \\ 
\mathbf{B}_{10} & = & -x_{3} \, \mathbf{a}_{1}-y_{3} \, \mathbf{a}_{2} + \left(\frac{1}{2} +z_{3}\right) \, \mathbf{a}_{3} & = & -x_{3}a \, \mathbf{\hat{x}}-y_{3}b \, \mathbf{\hat{y}} + \left(\frac{1}{2} +z_{3}\right)c \, \mathbf{\hat{z}} & \left(4a\right) & \mbox{Zr} \\ 
\mathbf{B}_{11} & = & \left(\frac{1}{2} +x_{3}\right) \, \mathbf{a}_{1}-y_{3} \, \mathbf{a}_{2} + z_{3} \, \mathbf{a}_{3} & = & \left(\frac{1}{2} +x_{3}\right)a \, \mathbf{\hat{x}}-y_{3}b \, \mathbf{\hat{y}} + z_{3}c \, \mathbf{\hat{z}} & \left(4a\right) & \mbox{Zr} \\ 
\mathbf{B}_{12} & = & \left(\frac{1}{2} - x_{3}\right) \, \mathbf{a}_{1} + y_{3} \, \mathbf{a}_{2} + \left(\frac{1}{2} +z_{3}\right) \, \mathbf{a}_{3} & = & \left(\frac{1}{2} - x_{3}\right)a \, \mathbf{\hat{x}} + y_{3}b \, \mathbf{\hat{y}} + \left(\frac{1}{2} +z_{3}\right)c \, \mathbf{\hat{z}} & \left(4a\right) & \mbox{Zr} \\ 
\end{longtabu}
\renewcommand{\arraystretch}{1.0}
\noindent \hrulefill
\\
\textbf{References:}
\vspace*{-0.25cm}
\begin{flushleft}
  - \bibentry{Grins_O2Zr_JMatChem_1994}. \\
\end{flushleft}
\textbf{Found in:}
\vspace*{-0.25cm}
\begin{flushleft}
  - \bibentry{Villars_PearsonsCrystalData_2013}. \\
\end{flushleft}
\noindent \hrulefill
\\
\textbf{Geometry files:}
\\
\noindent  - CIF: pp. {\hyperref[A2B_oP12_29_2a_a_cif]{\pageref{A2B_oP12_29_2a_a_cif}}} \\
\noindent  - POSCAR: pp. {\hyperref[A2B_oP12_29_2a_a_poscar]{\pageref{A2B_oP12_29_2a_a_poscar}}} \\
\onecolumn
{\phantomsection\label{AB2_oP12_29_a_2a}}
\subsection*{\huge \textbf{{\normalfont \begin{raggedleft}Pyrite (FeS$_{2}$, Low-temperature) Structure: \end{raggedleft} \\ AB2\_oP12\_29\_a\_2a}}}
\noindent \hrulefill
\vspace*{0.25cm}
\begin{figure}[htp]
  \centering
  \vspace{-1em}
  {\includegraphics[width=1\textwidth]{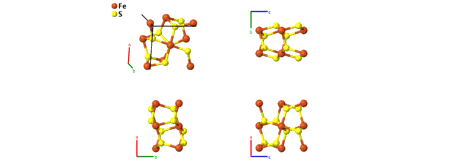}}
\end{figure}
\vspace*{-0.5cm}
\renewcommand{\arraystretch}{1.5}
\begin{equation*}
  \begin{array}{>{$\hspace{-0.15cm}}l<{$}>{$}p{0.5cm}<{$}>{$}p{18.5cm}<{$}}
    \mbox{\large \textbf{Prototype}} &\colon & \ce{FeS2} \\
    \mbox{\large \textbf{\AFLOW\ prototype label}} &\colon & \mbox{AB2\_oP12\_29\_a\_2a} \\
    \mbox{\large \textbf{\textit{Strukturbericht} designation}} &\colon & \mbox{None} \\
    \mbox{\large \textbf{Pearson symbol}} &\colon & \mbox{oP12} \\
    \mbox{\large \textbf{Space group number}} &\colon & 29 \\
    \mbox{\large \textbf{Space group symbol}} &\colon & Pca2_{1} \\
    \mbox{\large \textbf{\AFLOW\ prototype command}} &\colon &  \texttt{aflow} \,  \, \texttt{-{}-proto=AB2\_oP12\_29\_a\_2a } \, \newline \texttt{-{}-params=}{a,b/a,c/a,x_{1},y_{1},z_{1},x_{2},y_{2},z_{2},x_{3},y_{3},z_{3} }
  \end{array}
\end{equation*}
\renewcommand{\arraystretch}{1.0}

\vspace*{-0.25cm}
\noindent \hrulefill
\begin{itemize}
  \item{ZrO$_{2}$ (pp. {\hyperref[A2B_oP12_29_2a_a]{\pageref{A2B_oP12_29_2a_a}}}) and
Pyrite (pp. {\hyperref[AB2_oP12_29_a_2a]{\pageref{AB2_oP12_29_a_2a}}})
have similar \AFLOW\ prototype labels ({\it{i.e.}}, same symmetry and set of
Wyckoff positions with different stoichiometry labels due to alphabetic ordering of atomic species).
They are generated by the same symmetry operations with different sets of parameters
(\texttt{-{}-params}) specified in their corresponding \CIF\ files.
}
\end{itemize}

\noindent \parbox{1 \linewidth}{
\noindent \hrulefill
\\
\textbf{Simple Orthorhombic primitive vectors:} \\
\vspace*{-0.25cm}
\begin{tabular}{cc}
  \begin{tabular}{c}
    \parbox{0.6 \linewidth}{
      \renewcommand{\arraystretch}{1.5}
      \begin{equation*}
        \centering
        \begin{array}{ccc}
              \mathbf{a}_1 & = & a \, \mathbf{\hat{x}} \\
    \mathbf{a}_2 & = & b \, \mathbf{\hat{y}} \\
    \mathbf{a}_3 & = & c \, \mathbf{\hat{z}} \\

        \end{array}
      \end{equation*}
    }
    \renewcommand{\arraystretch}{1.0}
  \end{tabular}
  \begin{tabular}{c}
    \includegraphics[width=0.3\linewidth]{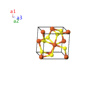} \\
  \end{tabular}
\end{tabular}

}
\vspace*{-0.25cm}

\noindent \hrulefill
\\
\textbf{Basis vectors:}
\vspace*{-0.25cm}
\renewcommand{\arraystretch}{1.5}
\begin{longtabu} to \textwidth{>{\centering $}X[-1,c,c]<{$}>{\centering $}X[-1,c,c]<{$}>{\centering $}X[-1,c,c]<{$}>{\centering $}X[-1,c,c]<{$}>{\centering $}X[-1,c,c]<{$}>{\centering $}X[-1,c,c]<{$}>{\centering $}X[-1,c,c]<{$}}
  & & \mbox{Lattice Coordinates} & & \mbox{Cartesian Coordinates} &\mbox{Wyckoff Position} & \mbox{Atom Type} \\  
  \mathbf{B}_{1} & = & x_{1} \, \mathbf{a}_{1} + y_{1} \, \mathbf{a}_{2} + z_{1} \, \mathbf{a}_{3} & = & x_{1}a \, \mathbf{\hat{x}} + y_{1}b \, \mathbf{\hat{y}} + z_{1}c \, \mathbf{\hat{z}} & \left(4a\right) & \mbox{Fe} \\ 
\mathbf{B}_{2} & = & -x_{1} \, \mathbf{a}_{1}-y_{1} \, \mathbf{a}_{2} + \left(\frac{1}{2} +z_{1}\right) \, \mathbf{a}_{3} & = & -x_{1}a \, \mathbf{\hat{x}}-y_{1}b \, \mathbf{\hat{y}} + \left(\frac{1}{2} +z_{1}\right)c \, \mathbf{\hat{z}} & \left(4a\right) & \mbox{Fe} \\ 
\mathbf{B}_{3} & = & \left(\frac{1}{2} +x_{1}\right) \, \mathbf{a}_{1}-y_{1} \, \mathbf{a}_{2} + z_{1} \, \mathbf{a}_{3} & = & \left(\frac{1}{2} +x_{1}\right)a \, \mathbf{\hat{x}}-y_{1}b \, \mathbf{\hat{y}} + z_{1}c \, \mathbf{\hat{z}} & \left(4a\right) & \mbox{Fe} \\ 
\mathbf{B}_{4} & = & \left(\frac{1}{2} - x_{1}\right) \, \mathbf{a}_{1} + y_{1} \, \mathbf{a}_{2} + \left(\frac{1}{2} +z_{1}\right) \, \mathbf{a}_{3} & = & \left(\frac{1}{2} - x_{1}\right)a \, \mathbf{\hat{x}} + y_{1}b \, \mathbf{\hat{y}} + \left(\frac{1}{2} +z_{1}\right)c \, \mathbf{\hat{z}} & \left(4a\right) & \mbox{Fe} \\ 
\mathbf{B}_{5} & = & x_{2} \, \mathbf{a}_{1} + y_{2} \, \mathbf{a}_{2} + z_{2} \, \mathbf{a}_{3} & = & x_{2}a \, \mathbf{\hat{x}} + y_{2}b \, \mathbf{\hat{y}} + z_{2}c \, \mathbf{\hat{z}} & \left(4a\right) & \mbox{S I} \\ 
\mathbf{B}_{6} & = & -x_{2} \, \mathbf{a}_{1}-y_{2} \, \mathbf{a}_{2} + \left(\frac{1}{2} +z_{2}\right) \, \mathbf{a}_{3} & = & -x_{2}a \, \mathbf{\hat{x}}-y_{2}b \, \mathbf{\hat{y}} + \left(\frac{1}{2} +z_{2}\right)c \, \mathbf{\hat{z}} & \left(4a\right) & \mbox{S I} \\ 
\mathbf{B}_{7} & = & \left(\frac{1}{2} +x_{2}\right) \, \mathbf{a}_{1}-y_{2} \, \mathbf{a}_{2} + z_{2} \, \mathbf{a}_{3} & = & \left(\frac{1}{2} +x_{2}\right)a \, \mathbf{\hat{x}}-y_{2}b \, \mathbf{\hat{y}} + z_{2}c \, \mathbf{\hat{z}} & \left(4a\right) & \mbox{S I} \\ 
\mathbf{B}_{8} & = & \left(\frac{1}{2} - x_{2}\right) \, \mathbf{a}_{1} + y_{2} \, \mathbf{a}_{2} + \left(\frac{1}{2} +z_{2}\right) \, \mathbf{a}_{3} & = & \left(\frac{1}{2} - x_{2}\right)a \, \mathbf{\hat{x}} + y_{2}b \, \mathbf{\hat{y}} + \left(\frac{1}{2} +z_{2}\right)c \, \mathbf{\hat{z}} & \left(4a\right) & \mbox{S I} \\ 
\mathbf{B}_{9} & = & x_{3} \, \mathbf{a}_{1} + y_{3} \, \mathbf{a}_{2} + z_{3} \, \mathbf{a}_{3} & = & x_{3}a \, \mathbf{\hat{x}} + y_{3}b \, \mathbf{\hat{y}} + z_{3}c \, \mathbf{\hat{z}} & \left(4a\right) & \mbox{S II} \\ 
\mathbf{B}_{10} & = & -x_{3} \, \mathbf{a}_{1}-y_{3} \, \mathbf{a}_{2} + \left(\frac{1}{2} +z_{3}\right) \, \mathbf{a}_{3} & = & -x_{3}a \, \mathbf{\hat{x}}-y_{3}b \, \mathbf{\hat{y}} + \left(\frac{1}{2} +z_{3}\right)c \, \mathbf{\hat{z}} & \left(4a\right) & \mbox{S II} \\ 
\mathbf{B}_{11} & = & \left(\frac{1}{2} +x_{3}\right) \, \mathbf{a}_{1}-y_{3} \, \mathbf{a}_{2} + z_{3} \, \mathbf{a}_{3} & = & \left(\frac{1}{2} +x_{3}\right)a \, \mathbf{\hat{x}}-y_{3}b \, \mathbf{\hat{y}} + z_{3}c \, \mathbf{\hat{z}} & \left(4a\right) & \mbox{S II} \\ 
\mathbf{B}_{12} & = & \left(\frac{1}{2} - x_{3}\right) \, \mathbf{a}_{1} + y_{3} \, \mathbf{a}_{2} + \left(\frac{1}{2} +z_{3}\right) \, \mathbf{a}_{3} & = & \left(\frac{1}{2} - x_{3}\right)a \, \mathbf{\hat{x}} + y_{3}b \, \mathbf{\hat{y}} + \left(\frac{1}{2} +z_{3}\right)c \, \mathbf{\hat{z}} & \left(4a\right) & \mbox{S II} \\ 
\end{longtabu}
\renewcommand{\arraystretch}{1.0}
\noindent \hrulefill
\\
\textbf{References:}
\vspace*{-0.25cm}
\begin{flushleft}
  - \bibentry{Bayliss_FeS2_AmerMin_1989}. \\
\end{flushleft}
\textbf{Found in:}
\vspace*{-0.25cm}
\begin{flushleft}
  - \bibentry{Villars_PearsonsCrystalData_2013}. \\
\end{flushleft}
\noindent \hrulefill
\\
\textbf{Geometry files:}
\\
\noindent  - CIF: pp. {\hyperref[AB2_oP12_29_a_2a_cif]{\pageref{AB2_oP12_29_a_2a_cif}}} \\
\noindent  - POSCAR: pp. {\hyperref[AB2_oP12_29_a_2a_poscar]{\pageref{AB2_oP12_29_a_2a_poscar}}} \\
\onecolumn
{\phantomsection\label{ABC_oP12_29_a_a_a}}
\subsection*{\huge \textbf{{\normalfont Cobaltite (CoAsS) Structure: ABC\_oP12\_29\_a\_a\_a}}}
\noindent \hrulefill
\vspace*{0.25cm}
\begin{figure}[htp]
  \centering
  \vspace{-1em}
  {\includegraphics[width=1\textwidth]{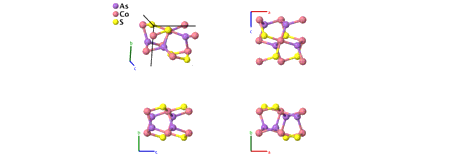}}
\end{figure}
\vspace*{-0.5cm}
\renewcommand{\arraystretch}{1.5}
\begin{equation*}
  \begin{array}{>{$\hspace{-0.15cm}}l<{$}>{$}p{0.5cm}<{$}>{$}p{18.5cm}<{$}}
    \mbox{\large \textbf{Prototype}} &\colon & \ce{CoAsS} \\
    \mbox{\large \textbf{\AFLOW\ prototype label}} &\colon & \mbox{ABC\_oP12\_29\_a\_a\_a} \\
    \mbox{\large \textbf{\textit{Strukturbericht} designation}} &\colon & \mbox{None} \\
    \mbox{\large \textbf{Pearson symbol}} &\colon & \mbox{oP12} \\
    \mbox{\large \textbf{Space group number}} &\colon & 29 \\
    \mbox{\large \textbf{Space group symbol}} &\colon & Pca2_{1} \\
    \mbox{\large \textbf{\AFLOW\ prototype command}} &\colon &  \texttt{aflow} \,  \, \texttt{-{}-proto=ABC\_oP12\_29\_a\_a\_a } \, \newline \texttt{-{}-params=}{a,b/a,c/a,x_{1},y_{1},z_{1},x_{2},y_{2},z_{2},x_{3},y_{3},z_{3} }
  \end{array}
\end{equation*}
\renewcommand{\arraystretch}{1.0}

\noindent \parbox{1 \linewidth}{
\noindent \hrulefill
\\
\textbf{Simple Orthorhombic primitive vectors:} \\
\vspace*{-0.25cm}
\begin{tabular}{cc}
  \begin{tabular}{c}
    \parbox{0.6 \linewidth}{
      \renewcommand{\arraystretch}{1.5}
      \begin{equation*}
        \centering
        \begin{array}{ccc}
              \mathbf{a}_1 & = & a \, \mathbf{\hat{x}} \\
    \mathbf{a}_2 & = & b \, \mathbf{\hat{y}} \\
    \mathbf{a}_3 & = & c \, \mathbf{\hat{z}} \\

        \end{array}
      \end{equation*}
    }
    \renewcommand{\arraystretch}{1.0}
  \end{tabular}
  \begin{tabular}{c}
    \includegraphics[width=0.3\linewidth]{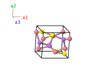} \\
  \end{tabular}
\end{tabular}

}
\vspace*{-0.25cm}

\noindent \hrulefill
\\
\textbf{Basis vectors:}
\vspace*{-0.25cm}
\renewcommand{\arraystretch}{1.5}
\begin{longtabu} to \textwidth{>{\centering $}X[-1,c,c]<{$}>{\centering $}X[-1,c,c]<{$}>{\centering $}X[-1,c,c]<{$}>{\centering $}X[-1,c,c]<{$}>{\centering $}X[-1,c,c]<{$}>{\centering $}X[-1,c,c]<{$}>{\centering $}X[-1,c,c]<{$}}
  & & \mbox{Lattice Coordinates} & & \mbox{Cartesian Coordinates} &\mbox{Wyckoff Position} & \mbox{Atom Type} \\  
  \mathbf{B}_{1} & = & x_{1} \, \mathbf{a}_{1} + y_{1} \, \mathbf{a}_{2} + z_{1} \, \mathbf{a}_{3} & = & x_{1}a \, \mathbf{\hat{x}} + y_{1}b \, \mathbf{\hat{y}} + z_{1}c \, \mathbf{\hat{z}} & \left(4a\right) & \mbox{As} \\ 
\mathbf{B}_{2} & = & -x_{1} \, \mathbf{a}_{1}-y_{1} \, \mathbf{a}_{2} + \left(\frac{1}{2} +z_{1}\right) \, \mathbf{a}_{3} & = & -x_{1}a \, \mathbf{\hat{x}}-y_{1}b \, \mathbf{\hat{y}} + \left(\frac{1}{2} +z_{1}\right)c \, \mathbf{\hat{z}} & \left(4a\right) & \mbox{As} \\ 
\mathbf{B}_{3} & = & \left(\frac{1}{2} +x_{1}\right) \, \mathbf{a}_{1}-y_{1} \, \mathbf{a}_{2} + z_{1} \, \mathbf{a}_{3} & = & \left(\frac{1}{2} +x_{1}\right)a \, \mathbf{\hat{x}}-y_{1}b \, \mathbf{\hat{y}} + z_{1}c \, \mathbf{\hat{z}} & \left(4a\right) & \mbox{As} \\ 
\mathbf{B}_{4} & = & \left(\frac{1}{2} - x_{1}\right) \, \mathbf{a}_{1} + y_{1} \, \mathbf{a}_{2} + \left(\frac{1}{2} +z_{1}\right) \, \mathbf{a}_{3} & = & \left(\frac{1}{2} - x_{1}\right)a \, \mathbf{\hat{x}} + y_{1}b \, \mathbf{\hat{y}} + \left(\frac{1}{2} +z_{1}\right)c \, \mathbf{\hat{z}} & \left(4a\right) & \mbox{As} \\ 
\mathbf{B}_{5} & = & x_{2} \, \mathbf{a}_{1} + y_{2} \, \mathbf{a}_{2} + z_{2} \, \mathbf{a}_{3} & = & x_{2}a \, \mathbf{\hat{x}} + y_{2}b \, \mathbf{\hat{y}} + z_{2}c \, \mathbf{\hat{z}} & \left(4a\right) & \mbox{Co} \\ 
\mathbf{B}_{6} & = & -x_{2} \, \mathbf{a}_{1}-y_{2} \, \mathbf{a}_{2} + \left(\frac{1}{2} +z_{2}\right) \, \mathbf{a}_{3} & = & -x_{2}a \, \mathbf{\hat{x}}-y_{2}b \, \mathbf{\hat{y}} + \left(\frac{1}{2} +z_{2}\right)c \, \mathbf{\hat{z}} & \left(4a\right) & \mbox{Co} \\ 
\mathbf{B}_{7} & = & \left(\frac{1}{2} +x_{2}\right) \, \mathbf{a}_{1}-y_{2} \, \mathbf{a}_{2} + z_{2} \, \mathbf{a}_{3} & = & \left(\frac{1}{2} +x_{2}\right)a \, \mathbf{\hat{x}}-y_{2}b \, \mathbf{\hat{y}} + z_{2}c \, \mathbf{\hat{z}} & \left(4a\right) & \mbox{Co} \\ 
\mathbf{B}_{8} & = & \left(\frac{1}{2} - x_{2}\right) \, \mathbf{a}_{1} + y_{2} \, \mathbf{a}_{2} + \left(\frac{1}{2} +z_{2}\right) \, \mathbf{a}_{3} & = & \left(\frac{1}{2} - x_{2}\right)a \, \mathbf{\hat{x}} + y_{2}b \, \mathbf{\hat{y}} + \left(\frac{1}{2} +z_{2}\right)c \, \mathbf{\hat{z}} & \left(4a\right) & \mbox{Co} \\ 
\mathbf{B}_{9} & = & x_{3} \, \mathbf{a}_{1} + y_{3} \, \mathbf{a}_{2} + z_{3} \, \mathbf{a}_{3} & = & x_{3}a \, \mathbf{\hat{x}} + y_{3}b \, \mathbf{\hat{y}} + z_{3}c \, \mathbf{\hat{z}} & \left(4a\right) & \mbox{S} \\ 
\mathbf{B}_{10} & = & -x_{3} \, \mathbf{a}_{1}-y_{3} \, \mathbf{a}_{2} + \left(\frac{1}{2} +z_{3}\right) \, \mathbf{a}_{3} & = & -x_{3}a \, \mathbf{\hat{x}}-y_{3}b \, \mathbf{\hat{y}} + \left(\frac{1}{2} +z_{3}\right)c \, \mathbf{\hat{z}} & \left(4a\right) & \mbox{S} \\ 
\mathbf{B}_{11} & = & \left(\frac{1}{2} +x_{3}\right) \, \mathbf{a}_{1}-y_{3} \, \mathbf{a}_{2} + z_{3} \, \mathbf{a}_{3} & = & \left(\frac{1}{2} +x_{3}\right)a \, \mathbf{\hat{x}}-y_{3}b \, \mathbf{\hat{y}} + z_{3}c \, \mathbf{\hat{z}} & \left(4a\right) & \mbox{S} \\ 
\mathbf{B}_{12} & = & \left(\frac{1}{2} - x_{3}\right) \, \mathbf{a}_{1} + y_{3} \, \mathbf{a}_{2} + \left(\frac{1}{2} +z_{3}\right) \, \mathbf{a}_{3} & = & \left(\frac{1}{2} - x_{3}\right)a \, \mathbf{\hat{x}} + y_{3}b \, \mathbf{\hat{y}} + \left(\frac{1}{2} +z_{3}\right)c \, \mathbf{\hat{z}} & \left(4a\right) & \mbox{S} \\ 
\end{longtabu}
\renewcommand{\arraystretch}{1.0}
\noindent \hrulefill
\\
\textbf{References:}
\vspace*{-0.25cm}
\begin{flushleft}
  - \bibentry{fleet_CoAsS_CanMin_1990}. \\
\end{flushleft}
\textbf{Found in:}
\vspace*{-0.25cm}
\begin{flushleft}
  - \bibentry{Villars_PearsonsCrystalData_2013}. \\
\end{flushleft}
\noindent \hrulefill
\\
\textbf{Geometry files:}
\\
\noindent  - CIF: pp. {\hyperref[ABC_oP12_29_a_a_a_cif]{\pageref{ABC_oP12_29_a_a_a_cif}}} \\
\noindent  - POSCAR: pp. {\hyperref[ABC_oP12_29_a_a_a_poscar]{\pageref{ABC_oP12_29_a_a_a_poscar}}} \\
\onecolumn
{\phantomsection\label{A5B3C15_oP46_30_a2c_bc_a7c}}
\subsection*{\huge \textbf{{\normalfont Bi$_{5}$Nb$_{3}$O$_{15}$ Structure: A5B3C15\_oP46\_30\_a2c\_bc\_a7c}}}
\noindent \hrulefill
\vspace*{0.25cm}
\begin{figure}[htp]
  \centering
  \vspace{-1em}
  {\includegraphics[width=1\textwidth]{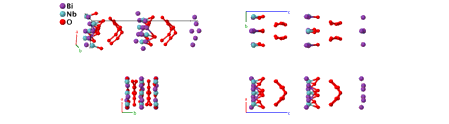}}
\end{figure}
\vspace*{-0.5cm}
\renewcommand{\arraystretch}{1.5}
\begin{equation*}
  \begin{array}{>{$\hspace{-0.15cm}}l<{$}>{$}p{0.5cm}<{$}>{$}p{18.5cm}<{$}}
    \mbox{\large \textbf{Prototype}} &\colon & \ce{Bi5Nb3O15} \\
    \mbox{\large \textbf{\AFLOW\ prototype label}} &\colon & \mbox{A5B3C15\_oP46\_30\_a2c\_bc\_a7c} \\
    \mbox{\large \textbf{\textit{Strukturbericht} designation}} &\colon & \mbox{None} \\
    \mbox{\large \textbf{Pearson symbol}} &\colon & \mbox{oP46} \\
    \mbox{\large \textbf{Space group number}} &\colon & 30 \\
    \mbox{\large \textbf{Space group symbol}} &\colon & Pnc2 \\
    \mbox{\large \textbf{\AFLOW\ prototype command}} &\colon &  \texttt{aflow} \,  \, \texttt{-{}-proto=A5B3C15\_oP46\_30\_a2c\_bc\_a7c } \, \newline \texttt{-{}-params=}{a,b/a,c/a,z_{1},z_{2},z_{3},x_{4},y_{4},z_{4},x_{5},y_{5},z_{5},x_{6},y_{6},z_{6},x_{7},y_{7},z_{7},x_{8},y_{8},} \newline {z_{8},x_{9},y_{9},z_{9},x_{10},y_{10},z_{10},x_{11},y_{11},z_{11},x_{12},y_{12},z_{12},x_{13},y_{13},z_{13} }
  \end{array}
\end{equation*}
\renewcommand{\arraystretch}{1.0}

\noindent \parbox{1 \linewidth}{
\noindent \hrulefill
\\
\textbf{Simple Orthorhombic primitive vectors:} \\
\vspace*{-0.25cm}
\begin{tabular}{cc}
  \begin{tabular}{c}
    \parbox{0.6 \linewidth}{
      \renewcommand{\arraystretch}{1.5}
      \begin{equation*}
        \centering
        \begin{array}{ccc}
              \mathbf{a}_1 & = & a \, \mathbf{\hat{x}} \\
    \mathbf{a}_2 & = & b \, \mathbf{\hat{y}} \\
    \mathbf{a}_3 & = & c \, \mathbf{\hat{z}} \\

        \end{array}
      \end{equation*}
    }
    \renewcommand{\arraystretch}{1.0}
  \end{tabular}
  \begin{tabular}{c}
    \includegraphics[width=0.3\linewidth]{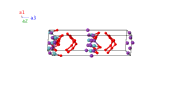} \\
  \end{tabular}
\end{tabular}

}
\vspace*{-0.25cm}

\noindent \hrulefill
\\
\textbf{Basis vectors:}
\vspace*{-0.25cm}
\renewcommand{\arraystretch}{1.5}
\begin{longtabu} to \textwidth{>{\centering $}X[-1,c,c]<{$}>{\centering $}X[-1,c,c]<{$}>{\centering $}X[-1,c,c]<{$}>{\centering $}X[-1,c,c]<{$}>{\centering $}X[-1,c,c]<{$}>{\centering $}X[-1,c,c]<{$}>{\centering $}X[-1,c,c]<{$}}
  & & \mbox{Lattice Coordinates} & & \mbox{Cartesian Coordinates} &\mbox{Wyckoff Position} & \mbox{Atom Type} \\  
  \mathbf{B}_{1} & = & z_{1} \, \mathbf{a}_{3} & = & z_{1}c \, \mathbf{\hat{z}} & \left(2a\right) & \mbox{Bi I} \\ 
\mathbf{B}_{2} & = & \frac{1}{2} \, \mathbf{a}_{2} + \left(\frac{1}{2} +z_{1}\right) \, \mathbf{a}_{3} & = & \frac{1}{2}b \, \mathbf{\hat{y}} + \left(\frac{1}{2} +z_{1}\right)c \, \mathbf{\hat{z}} & \left(2a\right) & \mbox{Bi I} \\ 
\mathbf{B}_{3} & = & z_{2} \, \mathbf{a}_{3} & = & z_{2}c \, \mathbf{\hat{z}} & \left(2a\right) & \mbox{O I} \\ 
\mathbf{B}_{4} & = & \frac{1}{2} \, \mathbf{a}_{2} + \left(\frac{1}{2} +z_{2}\right) \, \mathbf{a}_{3} & = & \frac{1}{2}b \, \mathbf{\hat{y}} + \left(\frac{1}{2} +z_{2}\right)c \, \mathbf{\hat{z}} & \left(2a\right) & \mbox{O I} \\ 
\mathbf{B}_{5} & = & \frac{1}{2} \, \mathbf{a}_{1} + z_{3} \, \mathbf{a}_{3} & = & \frac{1}{2}a \, \mathbf{\hat{x}} + z_{3}c \, \mathbf{\hat{z}} & \left(2b\right) & \mbox{Nb I} \\ 
\mathbf{B}_{6} & = & \frac{1}{2} \, \mathbf{a}_{1} + \frac{1}{2} \, \mathbf{a}_{2} + \left(\frac{1}{2} +z_{3}\right) \, \mathbf{a}_{3} & = & \frac{1}{2}a \, \mathbf{\hat{x}} + \frac{1}{2}b \, \mathbf{\hat{y}} + \left(\frac{1}{2} +z_{3}\right)c \, \mathbf{\hat{z}} & \left(2b\right) & \mbox{Nb I} \\ 
\mathbf{B}_{7} & = & x_{4} \, \mathbf{a}_{1} + y_{4} \, \mathbf{a}_{2} + z_{4} \, \mathbf{a}_{3} & = & x_{4}a \, \mathbf{\hat{x}} + y_{4}b \, \mathbf{\hat{y}} + z_{4}c \, \mathbf{\hat{z}} & \left(4c\right) & \mbox{Bi II} \\ 
\mathbf{B}_{8} & = & -x_{4} \, \mathbf{a}_{1}-y_{4} \, \mathbf{a}_{2} + z_{4} \, \mathbf{a}_{3} & = & -x_{4}a \, \mathbf{\hat{x}}-y_{4}b \, \mathbf{\hat{y}} + z_{4}c \, \mathbf{\hat{z}} & \left(4c\right) & \mbox{Bi II} \\ 
\mathbf{B}_{9} & = & x_{4} \, \mathbf{a}_{1} + \left(\frac{1}{2} - y_{4}\right) \, \mathbf{a}_{2} + \left(\frac{1}{2} +z_{4}\right) \, \mathbf{a}_{3} & = & x_{4}a \, \mathbf{\hat{x}} + \left(\frac{1}{2} - y_{4}\right)b \, \mathbf{\hat{y}} + \left(\frac{1}{2} +z_{4}\right)c \, \mathbf{\hat{z}} & \left(4c\right) & \mbox{Bi II} \\ 
\mathbf{B}_{10} & = & -x_{4} \, \mathbf{a}_{1} + \left(\frac{1}{2} +y_{4}\right) \, \mathbf{a}_{2} + \left(\frac{1}{2} +z_{4}\right) \, \mathbf{a}_{3} & = & -x_{4}a \, \mathbf{\hat{x}} + \left(\frac{1}{2} +y_{4}\right)b \, \mathbf{\hat{y}} + \left(\frac{1}{2} +z_{4}\right)c \, \mathbf{\hat{z}} & \left(4c\right) & \mbox{Bi II} \\ 
\mathbf{B}_{11} & = & x_{5} \, \mathbf{a}_{1} + y_{5} \, \mathbf{a}_{2} + z_{5} \, \mathbf{a}_{3} & = & x_{5}a \, \mathbf{\hat{x}} + y_{5}b \, \mathbf{\hat{y}} + z_{5}c \, \mathbf{\hat{z}} & \left(4c\right) & \mbox{Bi III} \\ 
\mathbf{B}_{12} & = & -x_{5} \, \mathbf{a}_{1}-y_{5} \, \mathbf{a}_{2} + z_{5} \, \mathbf{a}_{3} & = & -x_{5}a \, \mathbf{\hat{x}}-y_{5}b \, \mathbf{\hat{y}} + z_{5}c \, \mathbf{\hat{z}} & \left(4c\right) & \mbox{Bi III} \\ 
\mathbf{B}_{13} & = & x_{5} \, \mathbf{a}_{1} + \left(\frac{1}{2} - y_{5}\right) \, \mathbf{a}_{2} + \left(\frac{1}{2} +z_{5}\right) \, \mathbf{a}_{3} & = & x_{5}a \, \mathbf{\hat{x}} + \left(\frac{1}{2} - y_{5}\right)b \, \mathbf{\hat{y}} + \left(\frac{1}{2} +z_{5}\right)c \, \mathbf{\hat{z}} & \left(4c\right) & \mbox{Bi III} \\ 
\mathbf{B}_{14} & = & -x_{5} \, \mathbf{a}_{1} + \left(\frac{1}{2} +y_{5}\right) \, \mathbf{a}_{2} + \left(\frac{1}{2} +z_{5}\right) \, \mathbf{a}_{3} & = & -x_{5}a \, \mathbf{\hat{x}} + \left(\frac{1}{2} +y_{5}\right)b \, \mathbf{\hat{y}} + \left(\frac{1}{2} +z_{5}\right)c \, \mathbf{\hat{z}} & \left(4c\right) & \mbox{Bi III} \\ 
\mathbf{B}_{15} & = & x_{6} \, \mathbf{a}_{1} + y_{6} \, \mathbf{a}_{2} + z_{6} \, \mathbf{a}_{3} & = & x_{6}a \, \mathbf{\hat{x}} + y_{6}b \, \mathbf{\hat{y}} + z_{6}c \, \mathbf{\hat{z}} & \left(4c\right) & \mbox{Nb II} \\ 
\mathbf{B}_{16} & = & -x_{6} \, \mathbf{a}_{1}-y_{6} \, \mathbf{a}_{2} + z_{6} \, \mathbf{a}_{3} & = & -x_{6}a \, \mathbf{\hat{x}}-y_{6}b \, \mathbf{\hat{y}} + z_{6}c \, \mathbf{\hat{z}} & \left(4c\right) & \mbox{Nb II} \\ 
\mathbf{B}_{17} & = & x_{6} \, \mathbf{a}_{1} + \left(\frac{1}{2} - y_{6}\right) \, \mathbf{a}_{2} + \left(\frac{1}{2} +z_{6}\right) \, \mathbf{a}_{3} & = & x_{6}a \, \mathbf{\hat{x}} + \left(\frac{1}{2} - y_{6}\right)b \, \mathbf{\hat{y}} + \left(\frac{1}{2} +z_{6}\right)c \, \mathbf{\hat{z}} & \left(4c\right) & \mbox{Nb II} \\ 
\mathbf{B}_{18} & = & -x_{6} \, \mathbf{a}_{1} + \left(\frac{1}{2} +y_{6}\right) \, \mathbf{a}_{2} + \left(\frac{1}{2} +z_{6}\right) \, \mathbf{a}_{3} & = & -x_{6}a \, \mathbf{\hat{x}} + \left(\frac{1}{2} +y_{6}\right)b \, \mathbf{\hat{y}} + \left(\frac{1}{2} +z_{6}\right)c \, \mathbf{\hat{z}} & \left(4c\right) & \mbox{Nb II} \\ 
\mathbf{B}_{19} & = & x_{7} \, \mathbf{a}_{1} + y_{7} \, \mathbf{a}_{2} + z_{7} \, \mathbf{a}_{3} & = & x_{7}a \, \mathbf{\hat{x}} + y_{7}b \, \mathbf{\hat{y}} + z_{7}c \, \mathbf{\hat{z}} & \left(4c\right) & \mbox{O II} \\ 
\mathbf{B}_{20} & = & -x_{7} \, \mathbf{a}_{1}-y_{7} \, \mathbf{a}_{2} + z_{7} \, \mathbf{a}_{3} & = & -x_{7}a \, \mathbf{\hat{x}}-y_{7}b \, \mathbf{\hat{y}} + z_{7}c \, \mathbf{\hat{z}} & \left(4c\right) & \mbox{O II} \\ 
\mathbf{B}_{21} & = & x_{7} \, \mathbf{a}_{1} + \left(\frac{1}{2} - y_{7}\right) \, \mathbf{a}_{2} + \left(\frac{1}{2} +z_{7}\right) \, \mathbf{a}_{3} & = & x_{7}a \, \mathbf{\hat{x}} + \left(\frac{1}{2} - y_{7}\right)b \, \mathbf{\hat{y}} + \left(\frac{1}{2} +z_{7}\right)c \, \mathbf{\hat{z}} & \left(4c\right) & \mbox{O II} \\ 
\mathbf{B}_{22} & = & -x_{7} \, \mathbf{a}_{1} + \left(\frac{1}{2} +y_{7}\right) \, \mathbf{a}_{2} + \left(\frac{1}{2} +z_{7}\right) \, \mathbf{a}_{3} & = & -x_{7}a \, \mathbf{\hat{x}} + \left(\frac{1}{2} +y_{7}\right)b \, \mathbf{\hat{y}} + \left(\frac{1}{2} +z_{7}\right)c \, \mathbf{\hat{z}} & \left(4c\right) & \mbox{O II} \\ 
\mathbf{B}_{23} & = & x_{8} \, \mathbf{a}_{1} + y_{8} \, \mathbf{a}_{2} + z_{8} \, \mathbf{a}_{3} & = & x_{8}a \, \mathbf{\hat{x}} + y_{8}b \, \mathbf{\hat{y}} + z_{8}c \, \mathbf{\hat{z}} & \left(4c\right) & \mbox{O III} \\ 
\mathbf{B}_{24} & = & -x_{8} \, \mathbf{a}_{1}-y_{8} \, \mathbf{a}_{2} + z_{8} \, \mathbf{a}_{3} & = & -x_{8}a \, \mathbf{\hat{x}}-y_{8}b \, \mathbf{\hat{y}} + z_{8}c \, \mathbf{\hat{z}} & \left(4c\right) & \mbox{O III} \\ 
\mathbf{B}_{25} & = & x_{8} \, \mathbf{a}_{1} + \left(\frac{1}{2} - y_{8}\right) \, \mathbf{a}_{2} + \left(\frac{1}{2} +z_{8}\right) \, \mathbf{a}_{3} & = & x_{8}a \, \mathbf{\hat{x}} + \left(\frac{1}{2} - y_{8}\right)b \, \mathbf{\hat{y}} + \left(\frac{1}{2} +z_{8}\right)c \, \mathbf{\hat{z}} & \left(4c\right) & \mbox{O III} \\ 
\mathbf{B}_{26} & = & -x_{8} \, \mathbf{a}_{1} + \left(\frac{1}{2} +y_{8}\right) \, \mathbf{a}_{2} + \left(\frac{1}{2} +z_{8}\right) \, \mathbf{a}_{3} & = & -x_{8}a \, \mathbf{\hat{x}} + \left(\frac{1}{2} +y_{8}\right)b \, \mathbf{\hat{y}} + \left(\frac{1}{2} +z_{8}\right)c \, \mathbf{\hat{z}} & \left(4c\right) & \mbox{O III} \\ 
\mathbf{B}_{27} & = & x_{9} \, \mathbf{a}_{1} + y_{9} \, \mathbf{a}_{2} + z_{9} \, \mathbf{a}_{3} & = & x_{9}a \, \mathbf{\hat{x}} + y_{9}b \, \mathbf{\hat{y}} + z_{9}c \, \mathbf{\hat{z}} & \left(4c\right) & \mbox{O IV} \\ 
\mathbf{B}_{28} & = & -x_{9} \, \mathbf{a}_{1}-y_{9} \, \mathbf{a}_{2} + z_{9} \, \mathbf{a}_{3} & = & -x_{9}a \, \mathbf{\hat{x}}-y_{9}b \, \mathbf{\hat{y}} + z_{9}c \, \mathbf{\hat{z}} & \left(4c\right) & \mbox{O IV} \\ 
\mathbf{B}_{29} & = & x_{9} \, \mathbf{a}_{1} + \left(\frac{1}{2} - y_{9}\right) \, \mathbf{a}_{2} + \left(\frac{1}{2} +z_{9}\right) \, \mathbf{a}_{3} & = & x_{9}a \, \mathbf{\hat{x}} + \left(\frac{1}{2} - y_{9}\right)b \, \mathbf{\hat{y}} + \left(\frac{1}{2} +z_{9}\right)c \, \mathbf{\hat{z}} & \left(4c\right) & \mbox{O IV} \\ 
\mathbf{B}_{30} & = & -x_{9} \, \mathbf{a}_{1} + \left(\frac{1}{2} +y_{9}\right) \, \mathbf{a}_{2} + \left(\frac{1}{2} +z_{9}\right) \, \mathbf{a}_{3} & = & -x_{9}a \, \mathbf{\hat{x}} + \left(\frac{1}{2} +y_{9}\right)b \, \mathbf{\hat{y}} + \left(\frac{1}{2} +z_{9}\right)c \, \mathbf{\hat{z}} & \left(4c\right) & \mbox{O IV} \\ 
\mathbf{B}_{31} & = & x_{10} \, \mathbf{a}_{1} + y_{10} \, \mathbf{a}_{2} + z_{10} \, \mathbf{a}_{3} & = & x_{10}a \, \mathbf{\hat{x}} + y_{10}b \, \mathbf{\hat{y}} + z_{10}c \, \mathbf{\hat{z}} & \left(4c\right) & \mbox{O V} \\ 
\mathbf{B}_{32} & = & -x_{10} \, \mathbf{a}_{1}-y_{10} \, \mathbf{a}_{2} + z_{10} \, \mathbf{a}_{3} & = & -x_{10}a \, \mathbf{\hat{x}}-y_{10}b \, \mathbf{\hat{y}} + z_{10}c \, \mathbf{\hat{z}} & \left(4c\right) & \mbox{O V} \\ 
\mathbf{B}_{33} & = & x_{10} \, \mathbf{a}_{1} + \left(\frac{1}{2} - y_{10}\right) \, \mathbf{a}_{2} + \left(\frac{1}{2} +z_{10}\right) \, \mathbf{a}_{3} & = & x_{10}a \, \mathbf{\hat{x}} + \left(\frac{1}{2} - y_{10}\right)b \, \mathbf{\hat{y}} + \left(\frac{1}{2} +z_{10}\right)c \, \mathbf{\hat{z}} & \left(4c\right) & \mbox{O V} \\ 
\mathbf{B}_{34} & = & -x_{10} \, \mathbf{a}_{1} + \left(\frac{1}{2} +y_{10}\right) \, \mathbf{a}_{2} + \left(\frac{1}{2} +z_{10}\right) \, \mathbf{a}_{3} & = & -x_{10}a \, \mathbf{\hat{x}} + \left(\frac{1}{2} +y_{10}\right)b \, \mathbf{\hat{y}} + \left(\frac{1}{2} +z_{10}\right)c \, \mathbf{\hat{z}} & \left(4c\right) & \mbox{O V} \\ 
\mathbf{B}_{35} & = & x_{11} \, \mathbf{a}_{1} + y_{11} \, \mathbf{a}_{2} + z_{11} \, \mathbf{a}_{3} & = & x_{11}a \, \mathbf{\hat{x}} + y_{11}b \, \mathbf{\hat{y}} + z_{11}c \, \mathbf{\hat{z}} & \left(4c\right) & \mbox{O VI} \\ 
\mathbf{B}_{36} & = & -x_{11} \, \mathbf{a}_{1}-y_{11} \, \mathbf{a}_{2} + z_{11} \, \mathbf{a}_{3} & = & -x_{11}a \, \mathbf{\hat{x}}-y_{11}b \, \mathbf{\hat{y}} + z_{11}c \, \mathbf{\hat{z}} & \left(4c\right) & \mbox{O VI} \\ 
\mathbf{B}_{37} & = & x_{11} \, \mathbf{a}_{1} + \left(\frac{1}{2} - y_{11}\right) \, \mathbf{a}_{2} + \left(\frac{1}{2} +z_{11}\right) \, \mathbf{a}_{3} & = & x_{11}a \, \mathbf{\hat{x}} + \left(\frac{1}{2} - y_{11}\right)b \, \mathbf{\hat{y}} + \left(\frac{1}{2} +z_{11}\right)c \, \mathbf{\hat{z}} & \left(4c\right) & \mbox{O VI} \\ 
\mathbf{B}_{38} & = & -x_{11} \, \mathbf{a}_{1} + \left(\frac{1}{2} +y_{11}\right) \, \mathbf{a}_{2} + \left(\frac{1}{2} +z_{11}\right) \, \mathbf{a}_{3} & = & -x_{11}a \, \mathbf{\hat{x}} + \left(\frac{1}{2} +y_{11}\right)b \, \mathbf{\hat{y}} + \left(\frac{1}{2} +z_{11}\right)c \, \mathbf{\hat{z}} & \left(4c\right) & \mbox{O VI} \\ 
\mathbf{B}_{39} & = & x_{12} \, \mathbf{a}_{1} + y_{12} \, \mathbf{a}_{2} + z_{12} \, \mathbf{a}_{3} & = & x_{12}a \, \mathbf{\hat{x}} + y_{12}b \, \mathbf{\hat{y}} + z_{12}c \, \mathbf{\hat{z}} & \left(4c\right) & \mbox{O VII} \\ 
\mathbf{B}_{40} & = & -x_{12} \, \mathbf{a}_{1}-y_{12} \, \mathbf{a}_{2} + z_{12} \, \mathbf{a}_{3} & = & -x_{12}a \, \mathbf{\hat{x}}-y_{12}b \, \mathbf{\hat{y}} + z_{12}c \, \mathbf{\hat{z}} & \left(4c\right) & \mbox{O VII} \\ 
\mathbf{B}_{41} & = & x_{12} \, \mathbf{a}_{1} + \left(\frac{1}{2} - y_{12}\right) \, \mathbf{a}_{2} + \left(\frac{1}{2} +z_{12}\right) \, \mathbf{a}_{3} & = & x_{12}a \, \mathbf{\hat{x}} + \left(\frac{1}{2} - y_{12}\right)b \, \mathbf{\hat{y}} + \left(\frac{1}{2} +z_{12}\right)c \, \mathbf{\hat{z}} & \left(4c\right) & \mbox{O VII} \\ 
\mathbf{B}_{42} & = & -x_{12} \, \mathbf{a}_{1} + \left(\frac{1}{2} +y_{12}\right) \, \mathbf{a}_{2} + \left(\frac{1}{2} +z_{12}\right) \, \mathbf{a}_{3} & = & -x_{12}a \, \mathbf{\hat{x}} + \left(\frac{1}{2} +y_{12}\right)b \, \mathbf{\hat{y}} + \left(\frac{1}{2} +z_{12}\right)c \, \mathbf{\hat{z}} & \left(4c\right) & \mbox{O VII} \\ 
\mathbf{B}_{43} & = & x_{13} \, \mathbf{a}_{1} + y_{13} \, \mathbf{a}_{2} + z_{13} \, \mathbf{a}_{3} & = & x_{13}a \, \mathbf{\hat{x}} + y_{13}b \, \mathbf{\hat{y}} + z_{13}c \, \mathbf{\hat{z}} & \left(4c\right) & \mbox{O VIII} \\ 
\mathbf{B}_{44} & = & -x_{13} \, \mathbf{a}_{1}-y_{13} \, \mathbf{a}_{2} + z_{13} \, \mathbf{a}_{3} & = & -x_{13}a \, \mathbf{\hat{x}}-y_{13}b \, \mathbf{\hat{y}} + z_{13}c \, \mathbf{\hat{z}} & \left(4c\right) & \mbox{O VIII} \\ 
\mathbf{B}_{45} & = & x_{13} \, \mathbf{a}_{1} + \left(\frac{1}{2} - y_{13}\right) \, \mathbf{a}_{2} + \left(\frac{1}{2} +z_{13}\right) \, \mathbf{a}_{3} & = & x_{13}a \, \mathbf{\hat{x}} + \left(\frac{1}{2} - y_{13}\right)b \, \mathbf{\hat{y}} + \left(\frac{1}{2} +z_{13}\right)c \, \mathbf{\hat{z}} & \left(4c\right) & \mbox{O VIII} \\ 
\mathbf{B}_{46} & = & -x_{13} \, \mathbf{a}_{1} + \left(\frac{1}{2} +y_{13}\right) \, \mathbf{a}_{2} + \left(\frac{1}{2} +z_{13}\right) \, \mathbf{a}_{3} & = & -x_{13}a \, \mathbf{\hat{x}} + \left(\frac{1}{2} +y_{13}\right)b \, \mathbf{\hat{y}} + \left(\frac{1}{2} +z_{13}\right)c \, \mathbf{\hat{z}} & \left(4c\right) & \mbox{O VIII} \\ 
\end{longtabu}
\renewcommand{\arraystretch}{1.0}
\noindent \hrulefill
\\
\textbf{References:}
\vspace*{-0.25cm}
\begin{flushleft}
  - \bibentry{Tahara_Bi5Nb3O15_JSolStateChem_2007}. \\
\end{flushleft}
\textbf{Found in:}
\vspace*{-0.25cm}
\begin{flushleft}
  - \bibentry{Villars_PearsonsCrystalData_2013}. \\
\end{flushleft}
\noindent \hrulefill
\\
\textbf{Geometry files:}
\\
\noindent  - CIF: pp. {\hyperref[A5B3C15_oP46_30_a2c_bc_a7c_cif]{\pageref{A5B3C15_oP46_30_a2c_bc_a7c_cif}}} \\
\noindent  - POSCAR: pp. {\hyperref[A5B3C15_oP46_30_a2c_bc_a7c_poscar]{\pageref{A5B3C15_oP46_30_a2c_bc_a7c_poscar}}} \\
\onecolumn
{\phantomsection\label{ABC3_oP20_30_2a_c_3c}}
\subsection*{\huge \textbf{{\normalfont CuBrSe$_{3}$ Structure: ABC3\_oP20\_30\_2a\_c\_3c}}}
\noindent \hrulefill
\vspace*{0.25cm}
\begin{figure}[htp]
  \centering
  \vspace{-1em}
  {\includegraphics[width=1\textwidth]{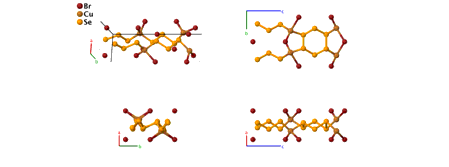}}
\end{figure}
\vspace*{-0.5cm}
\renewcommand{\arraystretch}{1.5}
\begin{equation*}
  \begin{array}{>{$\hspace{-0.15cm}}l<{$}>{$}p{0.5cm}<{$}>{$}p{18.5cm}<{$}}
    \mbox{\large \textbf{Prototype}} &\colon & \ce{CuBrSe3} \\
    \mbox{\large \textbf{\AFLOW\ prototype label}} &\colon & \mbox{ABC3\_oP20\_30\_2a\_c\_3c} \\
    \mbox{\large \textbf{\textit{Strukturbericht} designation}} &\colon & \mbox{None} \\
    \mbox{\large \textbf{Pearson symbol}} &\colon & \mbox{oP20} \\
    \mbox{\large \textbf{Space group number}} &\colon & 30 \\
    \mbox{\large \textbf{Space group symbol}} &\colon & Pnc2 \\
    \mbox{\large \textbf{\AFLOW\ prototype command}} &\colon &  \texttt{aflow} \,  \, \texttt{-{}-proto=ABC3\_oP20\_30\_2a\_c\_3c } \, \newline \texttt{-{}-params=}{a,b/a,c/a,z_{1},z_{2},x_{3},y_{3},z_{3},x_{4},y_{4},z_{4},x_{5},y_{5},z_{5},x_{6},y_{6},z_{6} }
  \end{array}
\end{equation*}
\renewcommand{\arraystretch}{1.0}

\noindent \parbox{1 \linewidth}{
\noindent \hrulefill
\\
\textbf{Simple Orthorhombic primitive vectors:} \\
\vspace*{-0.25cm}
\begin{tabular}{cc}
  \begin{tabular}{c}
    \parbox{0.6 \linewidth}{
      \renewcommand{\arraystretch}{1.5}
      \begin{equation*}
        \centering
        \begin{array}{ccc}
              \mathbf{a}_1 & = & a \, \mathbf{\hat{x}} \\
    \mathbf{a}_2 & = & b \, \mathbf{\hat{y}} \\
    \mathbf{a}_3 & = & c \, \mathbf{\hat{z}} \\

        \end{array}
      \end{equation*}
    }
    \renewcommand{\arraystretch}{1.0}
  \end{tabular}
  \begin{tabular}{c}
    \includegraphics[width=0.3\linewidth]{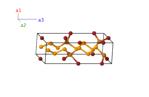} \\
  \end{tabular}
\end{tabular}

}
\vspace*{-0.25cm}

\noindent \hrulefill
\\
\textbf{Basis vectors:}
\vspace*{-0.25cm}
\renewcommand{\arraystretch}{1.5}
\begin{longtabu} to \textwidth{>{\centering $}X[-1,c,c]<{$}>{\centering $}X[-1,c,c]<{$}>{\centering $}X[-1,c,c]<{$}>{\centering $}X[-1,c,c]<{$}>{\centering $}X[-1,c,c]<{$}>{\centering $}X[-1,c,c]<{$}>{\centering $}X[-1,c,c]<{$}}
  & & \mbox{Lattice Coordinates} & & \mbox{Cartesian Coordinates} &\mbox{Wyckoff Position} & \mbox{Atom Type} \\  
  \mathbf{B}_{1} & = & z_{1} \, \mathbf{a}_{3} & = & z_{1}c \, \mathbf{\hat{z}} & \left(2a\right) & \mbox{Br I} \\ 
\mathbf{B}_{2} & = & \frac{1}{2} \, \mathbf{a}_{2} + \left(\frac{1}{2} +z_{1}\right) \, \mathbf{a}_{3} & = & \frac{1}{2}b \, \mathbf{\hat{y}} + \left(\frac{1}{2} +z_{1}\right)c \, \mathbf{\hat{z}} & \left(2a\right) & \mbox{Br I} \\ 
\mathbf{B}_{3} & = & z_{2} \, \mathbf{a}_{3} & = & z_{2}c \, \mathbf{\hat{z}} & \left(2a\right) & \mbox{Br II} \\ 
\mathbf{B}_{4} & = & \frac{1}{2} \, \mathbf{a}_{2} + \left(\frac{1}{2} +z_{2}\right) \, \mathbf{a}_{3} & = & \frac{1}{2}b \, \mathbf{\hat{y}} + \left(\frac{1}{2} +z_{2}\right)c \, \mathbf{\hat{z}} & \left(2a\right) & \mbox{Br II} \\ 
\mathbf{B}_{5} & = & x_{3} \, \mathbf{a}_{1} + y_{3} \, \mathbf{a}_{2} + z_{3} \, \mathbf{a}_{3} & = & x_{3}a \, \mathbf{\hat{x}} + y_{3}b \, \mathbf{\hat{y}} + z_{3}c \, \mathbf{\hat{z}} & \left(4c\right) & \mbox{Cu} \\ 
\mathbf{B}_{6} & = & -x_{3} \, \mathbf{a}_{1}-y_{3} \, \mathbf{a}_{2} + z_{3} \, \mathbf{a}_{3} & = & -x_{3}a \, \mathbf{\hat{x}}-y_{3}b \, \mathbf{\hat{y}} + z_{3}c \, \mathbf{\hat{z}} & \left(4c\right) & \mbox{Cu} \\ 
\mathbf{B}_{7} & = & x_{3} \, \mathbf{a}_{1} + \left(\frac{1}{2} - y_{3}\right) \, \mathbf{a}_{2} + \left(\frac{1}{2} +z_{3}\right) \, \mathbf{a}_{3} & = & x_{3}a \, \mathbf{\hat{x}} + \left(\frac{1}{2} - y_{3}\right)b \, \mathbf{\hat{y}} + \left(\frac{1}{2} +z_{3}\right)c \, \mathbf{\hat{z}} & \left(4c\right) & \mbox{Cu} \\ 
\mathbf{B}_{8} & = & -x_{3} \, \mathbf{a}_{1} + \left(\frac{1}{2} +y_{3}\right) \, \mathbf{a}_{2} + \left(\frac{1}{2} +z_{3}\right) \, \mathbf{a}_{3} & = & -x_{3}a \, \mathbf{\hat{x}} + \left(\frac{1}{2} +y_{3}\right)b \, \mathbf{\hat{y}} + \left(\frac{1}{2} +z_{3}\right)c \, \mathbf{\hat{z}} & \left(4c\right) & \mbox{Cu} \\ 
\mathbf{B}_{9} & = & x_{4} \, \mathbf{a}_{1} + y_{4} \, \mathbf{a}_{2} + z_{4} \, \mathbf{a}_{3} & = & x_{4}a \, \mathbf{\hat{x}} + y_{4}b \, \mathbf{\hat{y}} + z_{4}c \, \mathbf{\hat{z}} & \left(4c\right) & \mbox{Se I} \\ 
\mathbf{B}_{10} & = & -x_{4} \, \mathbf{a}_{1}-y_{4} \, \mathbf{a}_{2} + z_{4} \, \mathbf{a}_{3} & = & -x_{4}a \, \mathbf{\hat{x}}-y_{4}b \, \mathbf{\hat{y}} + z_{4}c \, \mathbf{\hat{z}} & \left(4c\right) & \mbox{Se I} \\ 
\mathbf{B}_{11} & = & x_{4} \, \mathbf{a}_{1} + \left(\frac{1}{2} - y_{4}\right) \, \mathbf{a}_{2} + \left(\frac{1}{2} +z_{4}\right) \, \mathbf{a}_{3} & = & x_{4}a \, \mathbf{\hat{x}} + \left(\frac{1}{2} - y_{4}\right)b \, \mathbf{\hat{y}} + \left(\frac{1}{2} +z_{4}\right)c \, \mathbf{\hat{z}} & \left(4c\right) & \mbox{Se I} \\ 
\mathbf{B}_{12} & = & -x_{4} \, \mathbf{a}_{1} + \left(\frac{1}{2} +y_{4}\right) \, \mathbf{a}_{2} + \left(\frac{1}{2} +z_{4}\right) \, \mathbf{a}_{3} & = & -x_{4}a \, \mathbf{\hat{x}} + \left(\frac{1}{2} +y_{4}\right)b \, \mathbf{\hat{y}} + \left(\frac{1}{2} +z_{4}\right)c \, \mathbf{\hat{z}} & \left(4c\right) & \mbox{Se I} \\ 
\mathbf{B}_{13} & = & x_{5} \, \mathbf{a}_{1} + y_{5} \, \mathbf{a}_{2} + z_{5} \, \mathbf{a}_{3} & = & x_{5}a \, \mathbf{\hat{x}} + y_{5}b \, \mathbf{\hat{y}} + z_{5}c \, \mathbf{\hat{z}} & \left(4c\right) & \mbox{Se II} \\ 
\mathbf{B}_{14} & = & -x_{5} \, \mathbf{a}_{1}-y_{5} \, \mathbf{a}_{2} + z_{5} \, \mathbf{a}_{3} & = & -x_{5}a \, \mathbf{\hat{x}}-y_{5}b \, \mathbf{\hat{y}} + z_{5}c \, \mathbf{\hat{z}} & \left(4c\right) & \mbox{Se II} \\ 
\mathbf{B}_{15} & = & x_{5} \, \mathbf{a}_{1} + \left(\frac{1}{2} - y_{5}\right) \, \mathbf{a}_{2} + \left(\frac{1}{2} +z_{5}\right) \, \mathbf{a}_{3} & = & x_{5}a \, \mathbf{\hat{x}} + \left(\frac{1}{2} - y_{5}\right)b \, \mathbf{\hat{y}} + \left(\frac{1}{2} +z_{5}\right)c \, \mathbf{\hat{z}} & \left(4c\right) & \mbox{Se II} \\ 
\mathbf{B}_{16} & = & -x_{5} \, \mathbf{a}_{1} + \left(\frac{1}{2} +y_{5}\right) \, \mathbf{a}_{2} + \left(\frac{1}{2} +z_{5}\right) \, \mathbf{a}_{3} & = & -x_{5}a \, \mathbf{\hat{x}} + \left(\frac{1}{2} +y_{5}\right)b \, \mathbf{\hat{y}} + \left(\frac{1}{2} +z_{5}\right)c \, \mathbf{\hat{z}} & \left(4c\right) & \mbox{Se II} \\ 
\mathbf{B}_{17} & = & x_{6} \, \mathbf{a}_{1} + y_{6} \, \mathbf{a}_{2} + z_{6} \, \mathbf{a}_{3} & = & x_{6}a \, \mathbf{\hat{x}} + y_{6}b \, \mathbf{\hat{y}} + z_{6}c \, \mathbf{\hat{z}} & \left(4c\right) & \mbox{Se III} \\ 
\mathbf{B}_{18} & = & -x_{6} \, \mathbf{a}_{1}-y_{6} \, \mathbf{a}_{2} + z_{6} \, \mathbf{a}_{3} & = & -x_{6}a \, \mathbf{\hat{x}}-y_{6}b \, \mathbf{\hat{y}} + z_{6}c \, \mathbf{\hat{z}} & \left(4c\right) & \mbox{Se III} \\ 
\mathbf{B}_{19} & = & x_{6} \, \mathbf{a}_{1} + \left(\frac{1}{2} - y_{6}\right) \, \mathbf{a}_{2} + \left(\frac{1}{2} +z_{6}\right) \, \mathbf{a}_{3} & = & x_{6}a \, \mathbf{\hat{x}} + \left(\frac{1}{2} - y_{6}\right)b \, \mathbf{\hat{y}} + \left(\frac{1}{2} +z_{6}\right)c \, \mathbf{\hat{z}} & \left(4c\right) & \mbox{Se III} \\ 
\mathbf{B}_{20} & = & -x_{6} \, \mathbf{a}_{1} + \left(\frac{1}{2} +y_{6}\right) \, \mathbf{a}_{2} + \left(\frac{1}{2} +z_{6}\right) \, \mathbf{a}_{3} & = & -x_{6}a \, \mathbf{\hat{x}} + \left(\frac{1}{2} +y_{6}\right)b \, \mathbf{\hat{y}} + \left(\frac{1}{2} +z_{6}\right)c \, \mathbf{\hat{z}} & \left(4c\right) & \mbox{Se III} \\ 
\end{longtabu}
\renewcommand{\arraystretch}{1.0}
\noindent \hrulefill
\\
\textbf{References:}
\vspace*{-0.25cm}
\begin{flushleft}
  - \bibentry{Sakuma_BrCuSe3_JPhysSocJpn_1991}. \\
\end{flushleft}
\textbf{Found in:}
\vspace*{-0.25cm}
\begin{flushleft}
  - \bibentry{Villars_PearsonsCrystalData_2013}. \\
\end{flushleft}
\noindent \hrulefill
\\
\textbf{Geometry files:}
\\
\noindent  - CIF: pp. {\hyperref[ABC3_oP20_30_2a_c_3c_cif]{\pageref{ABC3_oP20_30_2a_c_3c_cif}}} \\
\noindent  - POSCAR: pp. {\hyperref[ABC3_oP20_30_2a_c_3c_poscar]{\pageref{ABC3_oP20_30_2a_c_3c_poscar}}} \\
\onecolumn
{\phantomsection\label{A13B2C2_oP34_32_a6c_c_c}}
\subsection*{\huge \textbf{{\normalfont Re$_{2}$O$_{5}$[SO$_{4}$]$_{2}$ Structure: A13B2C2\_oP34\_32\_a6c\_c\_c}}}
\noindent \hrulefill
\vspace*{0.25cm}
\begin{figure}[htp]
  \centering
  \vspace{-1em}
  {\includegraphics[width=1\textwidth]{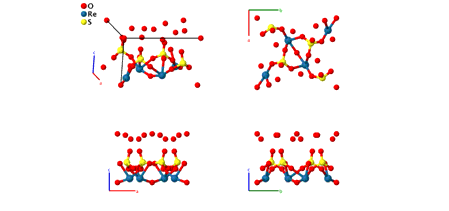}}
\end{figure}
\vspace*{-0.5cm}
\renewcommand{\arraystretch}{1.5}
\begin{equation*}
  \begin{array}{>{$\hspace{-0.15cm}}l<{$}>{$}p{0.5cm}<{$}>{$}p{18.5cm}<{$}}
    \mbox{\large \textbf{Prototype}} &\colon & \ce{Re2O5[SO4]2} \\
    \mbox{\large \textbf{\AFLOW\ prototype label}} &\colon & \mbox{A13B2C2\_oP34\_32\_a6c\_c\_c} \\
    \mbox{\large \textbf{\textit{Strukturbericht} designation}} &\colon & \mbox{None} \\
    \mbox{\large \textbf{Pearson symbol}} &\colon & \mbox{oP34} \\
    \mbox{\large \textbf{Space group number}} &\colon & 32 \\
    \mbox{\large \textbf{Space group symbol}} &\colon & Pba2 \\
    \mbox{\large \textbf{\AFLOW\ prototype command}} &\colon &  \texttt{aflow} \,  \, \texttt{-{}-proto=A13B2C2\_oP34\_32\_a6c\_c\_c } \, \newline \texttt{-{}-params=}{a,b/a,c/a,z_{1},x_{2},y_{2},z_{2},x_{3},y_{3},z_{3},x_{4},y_{4},z_{4},x_{5},y_{5},z_{5},x_{6},y_{6},z_{6},x_{7},} \newline {y_{7},z_{7},x_{8},y_{8},z_{8},x_{9},y_{9},z_{9} }
  \end{array}
\end{equation*}
\renewcommand{\arraystretch}{1.0}

\noindent \parbox{1 \linewidth}{
\noindent \hrulefill
\\
\textbf{Simple Orthorhombic primitive vectors:} \\
\vspace*{-0.25cm}
\begin{tabular}{cc}
  \begin{tabular}{c}
    \parbox{0.6 \linewidth}{
      \renewcommand{\arraystretch}{1.5}
      \begin{equation*}
        \centering
        \begin{array}{ccc}
              \mathbf{a}_1 & = & a \, \mathbf{\hat{x}} \\
    \mathbf{a}_2 & = & b \, \mathbf{\hat{y}} \\
    \mathbf{a}_3 & = & c \, \mathbf{\hat{z}} \\

        \end{array}
      \end{equation*}
    }
    \renewcommand{\arraystretch}{1.0}
  \end{tabular}
  \begin{tabular}{c}
    \includegraphics[width=0.3\linewidth]{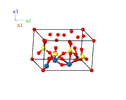} \\
  \end{tabular}
\end{tabular}

}
\vspace*{-0.25cm}

\noindent \hrulefill
\\
\textbf{Basis vectors:}
\vspace*{-0.25cm}
\renewcommand{\arraystretch}{1.5}
\begin{longtabu} to \textwidth{>{\centering $}X[-1,c,c]<{$}>{\centering $}X[-1,c,c]<{$}>{\centering $}X[-1,c,c]<{$}>{\centering $}X[-1,c,c]<{$}>{\centering $}X[-1,c,c]<{$}>{\centering $}X[-1,c,c]<{$}>{\centering $}X[-1,c,c]<{$}}
  & & \mbox{Lattice Coordinates} & & \mbox{Cartesian Coordinates} &\mbox{Wyckoff Position} & \mbox{Atom Type} \\  
  \mathbf{B}_{1} & = & z_{1} \, \mathbf{a}_{3} & = & z_{1}c \, \mathbf{\hat{z}} & \left(2a\right) & \mbox{O I} \\ 
\mathbf{B}_{2} & = & \frac{1}{2} \, \mathbf{a}_{1} + \frac{1}{2} \, \mathbf{a}_{2} + z_{1} \, \mathbf{a}_{3} & = & \frac{1}{2}a \, \mathbf{\hat{x}} + \frac{1}{2}b \, \mathbf{\hat{y}} + z_{1}c \, \mathbf{\hat{z}} & \left(2a\right) & \mbox{O I} \\ 
\mathbf{B}_{3} & = & x_{2} \, \mathbf{a}_{1} + y_{2} \, \mathbf{a}_{2} + z_{2} \, \mathbf{a}_{3} & = & x_{2}a \, \mathbf{\hat{x}} + y_{2}b \, \mathbf{\hat{y}} + z_{2}c \, \mathbf{\hat{z}} & \left(4c\right) & \mbox{O II} \\ 
\mathbf{B}_{4} & = & -x_{2} \, \mathbf{a}_{1}-y_{2} \, \mathbf{a}_{2} + z_{2} \, \mathbf{a}_{3} & = & -x_{2}a \, \mathbf{\hat{x}}-y_{2}b \, \mathbf{\hat{y}} + z_{2}c \, \mathbf{\hat{z}} & \left(4c\right) & \mbox{O II} \\ 
\mathbf{B}_{5} & = & \left(\frac{1}{2} +x_{2}\right) \, \mathbf{a}_{1} + \left(\frac{1}{2} - y_{2}\right) \, \mathbf{a}_{2} + z_{2} \, \mathbf{a}_{3} & = & \left(\frac{1}{2} +x_{2}\right)a \, \mathbf{\hat{x}} + \left(\frac{1}{2} - y_{2}\right)b \, \mathbf{\hat{y}} + z_{2}c \, \mathbf{\hat{z}} & \left(4c\right) & \mbox{O II} \\ 
\mathbf{B}_{6} & = & \left(\frac{1}{2} - x_{2}\right) \, \mathbf{a}_{1} + \left(\frac{1}{2} +y_{2}\right) \, \mathbf{a}_{2} + z_{2} \, \mathbf{a}_{3} & = & \left(\frac{1}{2} - x_{2}\right)a \, \mathbf{\hat{x}} + \left(\frac{1}{2} +y_{2}\right)b \, \mathbf{\hat{y}} + z_{2}c \, \mathbf{\hat{z}} & \left(4c\right) & \mbox{O II} \\ 
\mathbf{B}_{7} & = & x_{3} \, \mathbf{a}_{1} + y_{3} \, \mathbf{a}_{2} + z_{3} \, \mathbf{a}_{3} & = & x_{3}a \, \mathbf{\hat{x}} + y_{3}b \, \mathbf{\hat{y}} + z_{3}c \, \mathbf{\hat{z}} & \left(4c\right) & \mbox{O III} \\ 
\mathbf{B}_{8} & = & -x_{3} \, \mathbf{a}_{1}-y_{3} \, \mathbf{a}_{2} + z_{3} \, \mathbf{a}_{3} & = & -x_{3}a \, \mathbf{\hat{x}}-y_{3}b \, \mathbf{\hat{y}} + z_{3}c \, \mathbf{\hat{z}} & \left(4c\right) & \mbox{O III} \\ 
\mathbf{B}_{9} & = & \left(\frac{1}{2} +x_{3}\right) \, \mathbf{a}_{1} + \left(\frac{1}{2} - y_{3}\right) \, \mathbf{a}_{2} + z_{3} \, \mathbf{a}_{3} & = & \left(\frac{1}{2} +x_{3}\right)a \, \mathbf{\hat{x}} + \left(\frac{1}{2} - y_{3}\right)b \, \mathbf{\hat{y}} + z_{3}c \, \mathbf{\hat{z}} & \left(4c\right) & \mbox{O III} \\ 
\mathbf{B}_{10} & = & \left(\frac{1}{2} - x_{3}\right) \, \mathbf{a}_{1} + \left(\frac{1}{2} +y_{3}\right) \, \mathbf{a}_{2} + z_{3} \, \mathbf{a}_{3} & = & \left(\frac{1}{2} - x_{3}\right)a \, \mathbf{\hat{x}} + \left(\frac{1}{2} +y_{3}\right)b \, \mathbf{\hat{y}} + z_{3}c \, \mathbf{\hat{z}} & \left(4c\right) & \mbox{O III} \\ 
\mathbf{B}_{11} & = & x_{4} \, \mathbf{a}_{1} + y_{4} \, \mathbf{a}_{2} + z_{4} \, \mathbf{a}_{3} & = & x_{4}a \, \mathbf{\hat{x}} + y_{4}b \, \mathbf{\hat{y}} + z_{4}c \, \mathbf{\hat{z}} & \left(4c\right) & \mbox{O IV} \\ 
\mathbf{B}_{12} & = & -x_{4} \, \mathbf{a}_{1}-y_{4} \, \mathbf{a}_{2} + z_{4} \, \mathbf{a}_{3} & = & -x_{4}a \, \mathbf{\hat{x}}-y_{4}b \, \mathbf{\hat{y}} + z_{4}c \, \mathbf{\hat{z}} & \left(4c\right) & \mbox{O IV} \\ 
\mathbf{B}_{13} & = & \left(\frac{1}{2} +x_{4}\right) \, \mathbf{a}_{1} + \left(\frac{1}{2} - y_{4}\right) \, \mathbf{a}_{2} + z_{4} \, \mathbf{a}_{3} & = & \left(\frac{1}{2} +x_{4}\right)a \, \mathbf{\hat{x}} + \left(\frac{1}{2} - y_{4}\right)b \, \mathbf{\hat{y}} + z_{4}c \, \mathbf{\hat{z}} & \left(4c\right) & \mbox{O IV} \\ 
\mathbf{B}_{14} & = & \left(\frac{1}{2} - x_{4}\right) \, \mathbf{a}_{1} + \left(\frac{1}{2} +y_{4}\right) \, \mathbf{a}_{2} + z_{4} \, \mathbf{a}_{3} & = & \left(\frac{1}{2} - x_{4}\right)a \, \mathbf{\hat{x}} + \left(\frac{1}{2} +y_{4}\right)b \, \mathbf{\hat{y}} + z_{4}c \, \mathbf{\hat{z}} & \left(4c\right) & \mbox{O IV} \\ 
\mathbf{B}_{15} & = & x_{5} \, \mathbf{a}_{1} + y_{5} \, \mathbf{a}_{2} + z_{5} \, \mathbf{a}_{3} & = & x_{5}a \, \mathbf{\hat{x}} + y_{5}b \, \mathbf{\hat{y}} + z_{5}c \, \mathbf{\hat{z}} & \left(4c\right) & \mbox{O V} \\ 
\mathbf{B}_{16} & = & -x_{5} \, \mathbf{a}_{1}-y_{5} \, \mathbf{a}_{2} + z_{5} \, \mathbf{a}_{3} & = & -x_{5}a \, \mathbf{\hat{x}}-y_{5}b \, \mathbf{\hat{y}} + z_{5}c \, \mathbf{\hat{z}} & \left(4c\right) & \mbox{O V} \\ 
\mathbf{B}_{17} & = & \left(\frac{1}{2} +x_{5}\right) \, \mathbf{a}_{1} + \left(\frac{1}{2} - y_{5}\right) \, \mathbf{a}_{2} + z_{5} \, \mathbf{a}_{3} & = & \left(\frac{1}{2} +x_{5}\right)a \, \mathbf{\hat{x}} + \left(\frac{1}{2} - y_{5}\right)b \, \mathbf{\hat{y}} + z_{5}c \, \mathbf{\hat{z}} & \left(4c\right) & \mbox{O V} \\ 
\mathbf{B}_{18} & = & \left(\frac{1}{2} - x_{5}\right) \, \mathbf{a}_{1} + \left(\frac{1}{2} +y_{5}\right) \, \mathbf{a}_{2} + z_{5} \, \mathbf{a}_{3} & = & \left(\frac{1}{2} - x_{5}\right)a \, \mathbf{\hat{x}} + \left(\frac{1}{2} +y_{5}\right)b \, \mathbf{\hat{y}} + z_{5}c \, \mathbf{\hat{z}} & \left(4c\right) & \mbox{O V} \\ 
\mathbf{B}_{19} & = & x_{6} \, \mathbf{a}_{1} + y_{6} \, \mathbf{a}_{2} + z_{6} \, \mathbf{a}_{3} & = & x_{6}a \, \mathbf{\hat{x}} + y_{6}b \, \mathbf{\hat{y}} + z_{6}c \, \mathbf{\hat{z}} & \left(4c\right) & \mbox{O VI} \\ 
\mathbf{B}_{20} & = & -x_{6} \, \mathbf{a}_{1}-y_{6} \, \mathbf{a}_{2} + z_{6} \, \mathbf{a}_{3} & = & -x_{6}a \, \mathbf{\hat{x}}-y_{6}b \, \mathbf{\hat{y}} + z_{6}c \, \mathbf{\hat{z}} & \left(4c\right) & \mbox{O VI} \\ 
\mathbf{B}_{21} & = & \left(\frac{1}{2} +x_{6}\right) \, \mathbf{a}_{1} + \left(\frac{1}{2} - y_{6}\right) \, \mathbf{a}_{2} + z_{6} \, \mathbf{a}_{3} & = & \left(\frac{1}{2} +x_{6}\right)a \, \mathbf{\hat{x}} + \left(\frac{1}{2} - y_{6}\right)b \, \mathbf{\hat{y}} + z_{6}c \, \mathbf{\hat{z}} & \left(4c\right) & \mbox{O VI} \\ 
\mathbf{B}_{22} & = & \left(\frac{1}{2} - x_{6}\right) \, \mathbf{a}_{1} + \left(\frac{1}{2} +y_{6}\right) \, \mathbf{a}_{2} + z_{6} \, \mathbf{a}_{3} & = & \left(\frac{1}{2} - x_{6}\right)a \, \mathbf{\hat{x}} + \left(\frac{1}{2} +y_{6}\right)b \, \mathbf{\hat{y}} + z_{6}c \, \mathbf{\hat{z}} & \left(4c\right) & \mbox{O VI} \\ 
\mathbf{B}_{23} & = & x_{7} \, \mathbf{a}_{1} + y_{7} \, \mathbf{a}_{2} + z_{7} \, \mathbf{a}_{3} & = & x_{7}a \, \mathbf{\hat{x}} + y_{7}b \, \mathbf{\hat{y}} + z_{7}c \, \mathbf{\hat{z}} & \left(4c\right) & \mbox{O VII} \\ 
\mathbf{B}_{24} & = & -x_{7} \, \mathbf{a}_{1}-y_{7} \, \mathbf{a}_{2} + z_{7} \, \mathbf{a}_{3} & = & -x_{7}a \, \mathbf{\hat{x}}-y_{7}b \, \mathbf{\hat{y}} + z_{7}c \, \mathbf{\hat{z}} & \left(4c\right) & \mbox{O VII} \\ 
\mathbf{B}_{25} & = & \left(\frac{1}{2} +x_{7}\right) \, \mathbf{a}_{1} + \left(\frac{1}{2} - y_{7}\right) \, \mathbf{a}_{2} + z_{7} \, \mathbf{a}_{3} & = & \left(\frac{1}{2} +x_{7}\right)a \, \mathbf{\hat{x}} + \left(\frac{1}{2} - y_{7}\right)b \, \mathbf{\hat{y}} + z_{7}c \, \mathbf{\hat{z}} & \left(4c\right) & \mbox{O VII} \\ 
\mathbf{B}_{26} & = & \left(\frac{1}{2} - x_{7}\right) \, \mathbf{a}_{1} + \left(\frac{1}{2} +y_{7}\right) \, \mathbf{a}_{2} + z_{7} \, \mathbf{a}_{3} & = & \left(\frac{1}{2} - x_{7}\right)a \, \mathbf{\hat{x}} + \left(\frac{1}{2} +y_{7}\right)b \, \mathbf{\hat{y}} + z_{7}c \, \mathbf{\hat{z}} & \left(4c\right) & \mbox{O VII} \\ 
\mathbf{B}_{27} & = & x_{8} \, \mathbf{a}_{1} + y_{8} \, \mathbf{a}_{2} + z_{8} \, \mathbf{a}_{3} & = & x_{8}a \, \mathbf{\hat{x}} + y_{8}b \, \mathbf{\hat{y}} + z_{8}c \, \mathbf{\hat{z}} & \left(4c\right) & \mbox{Re} \\ 
\mathbf{B}_{28} & = & -x_{8} \, \mathbf{a}_{1}-y_{8} \, \mathbf{a}_{2} + z_{8} \, \mathbf{a}_{3} & = & -x_{8}a \, \mathbf{\hat{x}}-y_{8}b \, \mathbf{\hat{y}} + z_{8}c \, \mathbf{\hat{z}} & \left(4c\right) & \mbox{Re} \\ 
\mathbf{B}_{29} & = & \left(\frac{1}{2} +x_{8}\right) \, \mathbf{a}_{1} + \left(\frac{1}{2} - y_{8}\right) \, \mathbf{a}_{2} + z_{8} \, \mathbf{a}_{3} & = & \left(\frac{1}{2} +x_{8}\right)a \, \mathbf{\hat{x}} + \left(\frac{1}{2} - y_{8}\right)b \, \mathbf{\hat{y}} + z_{8}c \, \mathbf{\hat{z}} & \left(4c\right) & \mbox{Re} \\ 
\mathbf{B}_{30} & = & \left(\frac{1}{2} - x_{8}\right) \, \mathbf{a}_{1} + \left(\frac{1}{2} +y_{8}\right) \, \mathbf{a}_{2} + z_{8} \, \mathbf{a}_{3} & = & \left(\frac{1}{2} - x_{8}\right)a \, \mathbf{\hat{x}} + \left(\frac{1}{2} +y_{8}\right)b \, \mathbf{\hat{y}} + z_{8}c \, \mathbf{\hat{z}} & \left(4c\right) & \mbox{Re} \\ 
\mathbf{B}_{31} & = & x_{9} \, \mathbf{a}_{1} + y_{9} \, \mathbf{a}_{2} + z_{9} \, \mathbf{a}_{3} & = & x_{9}a \, \mathbf{\hat{x}} + y_{9}b \, \mathbf{\hat{y}} + z_{9}c \, \mathbf{\hat{z}} & \left(4c\right) & \mbox{S} \\ 
\mathbf{B}_{32} & = & -x_{9} \, \mathbf{a}_{1}-y_{9} \, \mathbf{a}_{2} + z_{9} \, \mathbf{a}_{3} & = & -x_{9}a \, \mathbf{\hat{x}}-y_{9}b \, \mathbf{\hat{y}} + z_{9}c \, \mathbf{\hat{z}} & \left(4c\right) & \mbox{S} \\ 
\mathbf{B}_{33} & = & \left(\frac{1}{2} +x_{9}\right) \, \mathbf{a}_{1} + \left(\frac{1}{2} - y_{9}\right) \, \mathbf{a}_{2} + z_{9} \, \mathbf{a}_{3} & = & \left(\frac{1}{2} +x_{9}\right)a \, \mathbf{\hat{x}} + \left(\frac{1}{2} - y_{9}\right)b \, \mathbf{\hat{y}} + z_{9}c \, \mathbf{\hat{z}} & \left(4c\right) & \mbox{S} \\ 
\mathbf{B}_{34} & = & \left(\frac{1}{2} - x_{9}\right) \, \mathbf{a}_{1} + \left(\frac{1}{2} +y_{9}\right) \, \mathbf{a}_{2} + z_{9} \, \mathbf{a}_{3} & = & \left(\frac{1}{2} - x_{9}\right)a \, \mathbf{\hat{x}} + \left(\frac{1}{2} +y_{9}\right)b \, \mathbf{\hat{y}} + z_{9}c \, \mathbf{\hat{z}} & \left(4c\right) & \mbox{S} \\ 
\end{longtabu}
\renewcommand{\arraystretch}{1.0}
\noindent \hrulefill
\\
\textbf{References:}
\vspace*{-0.25cm}
\begin{flushleft}
  - \bibentry{Betke_Re2SO42O5_IorgChem_2011}. \\
\end{flushleft}
\textbf{Found in:}
\vspace*{-0.25cm}
\begin{flushleft}
  - \bibentry{Villars_PearsonsCrystalData_2013}. \\
\end{flushleft}
\noindent \hrulefill
\\
\textbf{Geometry files:}
\\
\noindent  - CIF: pp. {\hyperref[A13B2C2_oP34_32_a6c_c_c_cif]{\pageref{A13B2C2_oP34_32_a6c_c_c_cif}}} \\
\noindent  - POSCAR: pp. {\hyperref[A13B2C2_oP34_32_a6c_c_c_poscar]{\pageref{A13B2C2_oP34_32_a6c_c_c_poscar}}} \\
\onecolumn
{\phantomsection\label{A2B3_oP40_33_4a_6a}}
\subsection*{\huge \textbf{{\normalfont $\kappa$-alumina (Al$_{2}$O$_{3}$) Structure: A2B3\_oP40\_33\_4a\_6a}}}
\noindent \hrulefill
\vspace*{0.25cm}
\begin{figure}[htp]
  \centering
  \vspace{-1em}
  {\includegraphics[width=1\textwidth]{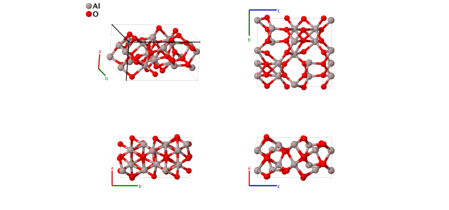}}
\end{figure}
\vspace*{-0.5cm}
\renewcommand{\arraystretch}{1.5}
\begin{equation*}
  \begin{array}{>{$\hspace{-0.15cm}}l<{$}>{$}p{0.5cm}<{$}>{$}p{18.5cm}<{$}}
    \mbox{\large \textbf{Prototype}} &\colon & \ce{$\kappa$-Al2O3} \\
    \mbox{\large \textbf{\AFLOW\ prototype label}} &\colon & \mbox{A2B3\_oP40\_33\_4a\_6a} \\
    \mbox{\large \textbf{\textit{Strukturbericht} designation}} &\colon & \mbox{None} \\
    \mbox{\large \textbf{Pearson symbol}} &\colon & \mbox{oP40} \\
    \mbox{\large \textbf{Space group number}} &\colon & 33 \\
    \mbox{\large \textbf{Space group symbol}} &\colon & Pna2_{1} \\
    \mbox{\large \textbf{\AFLOW\ prototype command}} &\colon &  \texttt{aflow} \,  \, \texttt{-{}-proto=A2B3\_oP40\_33\_4a\_6a } \, \newline \texttt{-{}-params=}{a,b/a,c/a,x_{1},y_{1},z_{1},x_{2},y_{2},z_{2},x_{3},y_{3},z_{3},x_{4},y_{4},z_{4},x_{5},y_{5},z_{5},x_{6},y_{6},} \newline {z_{6},x_{7},y_{7},z_{7},x_{8},y_{8},z_{8},x_{9},y_{9},z_{9},x_{10},y_{10},z_{10} }
  \end{array}
\end{equation*}
\renewcommand{\arraystretch}{1.0}

\vspace*{-0.25cm}
\noindent \hrulefill
\\
\textbf{ \textbf{Other compounds with this structure:}}
\begin{itemize}
   \item{ $(1-x)$Fe$_{2}$O$_{3} \cdot x$Al$_{2}$O$_{3}$,  $(1-x)$Fe$_{2}$O$_{3} \cdot x$Ga$_{2}$O$_{3}$. An approximation to the true crystal structure of $\epsilon$-Ga$_{2}$O$_{3}$ (Yoshioka, 2007).  }
\end{itemize}
\noindent \parbox{1 \linewidth}{
\noindent \hrulefill
\\
\textbf{Simple Orthorhombic primitive vectors:} \\
\vspace*{-0.25cm}
\begin{tabular}{cc}
  \begin{tabular}{c}
    \parbox{0.6 \linewidth}{
      \renewcommand{\arraystretch}{1.5}
      \begin{equation*}
        \centering
        \begin{array}{ccc}
              \mathbf{a}_1 & = & a \, \mathbf{\hat{x}} \\
    \mathbf{a}_2 & = & b \, \mathbf{\hat{y}} \\
    \mathbf{a}_3 & = & c \, \mathbf{\hat{z}} \\

        \end{array}
      \end{equation*}
    }
    \renewcommand{\arraystretch}{1.0}
  \end{tabular}
  \begin{tabular}{c}
    \includegraphics[width=0.3\linewidth]{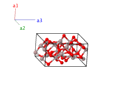} \\
  \end{tabular}
\end{tabular}

}
\vspace*{-0.25cm}

\noindent \hrulefill
\\
\textbf{Basis vectors:}
\vspace*{-0.25cm}
\renewcommand{\arraystretch}{1.5}
\begin{longtabu} to \textwidth{>{\centering $}X[-1,c,c]<{$}>{\centering $}X[-1,c,c]<{$}>{\centering $}X[-1,c,c]<{$}>{\centering $}X[-1,c,c]<{$}>{\centering $}X[-1,c,c]<{$}>{\centering $}X[-1,c,c]<{$}>{\centering $}X[-1,c,c]<{$}}
  & & \mbox{Lattice Coordinates} & & \mbox{Cartesian Coordinates} &\mbox{Wyckoff Position} & \mbox{Atom Type} \\  
  \mathbf{B}_{1} & = & x_{1} \, \mathbf{a}_{1} + y_{1} \, \mathbf{a}_{2} + z_{1} \, \mathbf{a}_{3} & = & x_{1}a \, \mathbf{\hat{x}} + y_{1}b \, \mathbf{\hat{y}} + z_{1}c \, \mathbf{\hat{z}} & \left(4a\right) & \mbox{Al I} \\ 
\mathbf{B}_{2} & = & -x_{1} \, \mathbf{a}_{1}-y_{1} \, \mathbf{a}_{2} + \left(\frac{1}{2} +z_{1}\right) \, \mathbf{a}_{3} & = & -x_{1}a \, \mathbf{\hat{x}}-y_{1}b \, \mathbf{\hat{y}} + \left(\frac{1}{2} +z_{1}\right)c \, \mathbf{\hat{z}} & \left(4a\right) & \mbox{Al I} \\ 
\mathbf{B}_{3} & = & \left(\frac{1}{2} +x_{1}\right) \, \mathbf{a}_{1} + \left(\frac{1}{2} - y_{1}\right) \, \mathbf{a}_{2} + z_{1} \, \mathbf{a}_{3} & = & \left(\frac{1}{2} +x_{1}\right)a \, \mathbf{\hat{x}} + \left(\frac{1}{2} - y_{1}\right)b \, \mathbf{\hat{y}} + z_{1}c \, \mathbf{\hat{z}} & \left(4a\right) & \mbox{Al I} \\ 
\mathbf{B}_{4} & = & \left(\frac{1}{2} - x_{1}\right) \, \mathbf{a}_{1} + \left(\frac{1}{2} +y_{1}\right) \, \mathbf{a}_{2} + \left(\frac{1}{2} +z_{1}\right) \, \mathbf{a}_{3} & = & \left(\frac{1}{2} - x_{1}\right)a \, \mathbf{\hat{x}} + \left(\frac{1}{2} +y_{1}\right)b \, \mathbf{\hat{y}} + \left(\frac{1}{2} +z_{1}\right)c \, \mathbf{\hat{z}} & \left(4a\right) & \mbox{Al I} \\ 
\mathbf{B}_{5} & = & x_{2} \, \mathbf{a}_{1} + y_{2} \, \mathbf{a}_{2} + z_{2} \, \mathbf{a}_{3} & = & x_{2}a \, \mathbf{\hat{x}} + y_{2}b \, \mathbf{\hat{y}} + z_{2}c \, \mathbf{\hat{z}} & \left(4a\right) & \mbox{Al II} \\ 
\mathbf{B}_{6} & = & -x_{2} \, \mathbf{a}_{1}-y_{2} \, \mathbf{a}_{2} + \left(\frac{1}{2} +z_{2}\right) \, \mathbf{a}_{3} & = & -x_{2}a \, \mathbf{\hat{x}}-y_{2}b \, \mathbf{\hat{y}} + \left(\frac{1}{2} +z_{2}\right)c \, \mathbf{\hat{z}} & \left(4a\right) & \mbox{Al II} \\ 
\mathbf{B}_{7} & = & \left(\frac{1}{2} +x_{2}\right) \, \mathbf{a}_{1} + \left(\frac{1}{2} - y_{2}\right) \, \mathbf{a}_{2} + z_{2} \, \mathbf{a}_{3} & = & \left(\frac{1}{2} +x_{2}\right)a \, \mathbf{\hat{x}} + \left(\frac{1}{2} - y_{2}\right)b \, \mathbf{\hat{y}} + z_{2}c \, \mathbf{\hat{z}} & \left(4a\right) & \mbox{Al II} \\ 
\mathbf{B}_{8} & = & \left(\frac{1}{2} - x_{2}\right) \, \mathbf{a}_{1} + \left(\frac{1}{2} +y_{2}\right) \, \mathbf{a}_{2} + \left(\frac{1}{2} +z_{2}\right) \, \mathbf{a}_{3} & = & \left(\frac{1}{2} - x_{2}\right)a \, \mathbf{\hat{x}} + \left(\frac{1}{2} +y_{2}\right)b \, \mathbf{\hat{y}} + \left(\frac{1}{2} +z_{2}\right)c \, \mathbf{\hat{z}} & \left(4a\right) & \mbox{Al II} \\ 
\mathbf{B}_{9} & = & x_{3} \, \mathbf{a}_{1} + y_{3} \, \mathbf{a}_{2} + z_{3} \, \mathbf{a}_{3} & = & x_{3}a \, \mathbf{\hat{x}} + y_{3}b \, \mathbf{\hat{y}} + z_{3}c \, \mathbf{\hat{z}} & \left(4a\right) & \mbox{Al III} \\ 
\mathbf{B}_{10} & = & -x_{3} \, \mathbf{a}_{1}-y_{3} \, \mathbf{a}_{2} + \left(\frac{1}{2} +z_{3}\right) \, \mathbf{a}_{3} & = & -x_{3}a \, \mathbf{\hat{x}}-y_{3}b \, \mathbf{\hat{y}} + \left(\frac{1}{2} +z_{3}\right)c \, \mathbf{\hat{z}} & \left(4a\right) & \mbox{Al III} \\ 
\mathbf{B}_{11} & = & \left(\frac{1}{2} +x_{3}\right) \, \mathbf{a}_{1} + \left(\frac{1}{2} - y_{3}\right) \, \mathbf{a}_{2} + z_{3} \, \mathbf{a}_{3} & = & \left(\frac{1}{2} +x_{3}\right)a \, \mathbf{\hat{x}} + \left(\frac{1}{2} - y_{3}\right)b \, \mathbf{\hat{y}} + z_{3}c \, \mathbf{\hat{z}} & \left(4a\right) & \mbox{Al III} \\ 
\mathbf{B}_{12} & = & \left(\frac{1}{2} - x_{3}\right) \, \mathbf{a}_{1} + \left(\frac{1}{2} +y_{3}\right) \, \mathbf{a}_{2} + \left(\frac{1}{2} +z_{3}\right) \, \mathbf{a}_{3} & = & \left(\frac{1}{2} - x_{3}\right)a \, \mathbf{\hat{x}} + \left(\frac{1}{2} +y_{3}\right)b \, \mathbf{\hat{y}} + \left(\frac{1}{2} +z_{3}\right)c \, \mathbf{\hat{z}} & \left(4a\right) & \mbox{Al III} \\ 
\mathbf{B}_{13} & = & x_{4} \, \mathbf{a}_{1} + y_{4} \, \mathbf{a}_{2} + z_{4} \, \mathbf{a}_{3} & = & x_{4}a \, \mathbf{\hat{x}} + y_{4}b \, \mathbf{\hat{y}} + z_{4}c \, \mathbf{\hat{z}} & \left(4a\right) & \mbox{Al IV} \\ 
\mathbf{B}_{14} & = & -x_{4} \, \mathbf{a}_{1}-y_{4} \, \mathbf{a}_{2} + \left(\frac{1}{2} +z_{4}\right) \, \mathbf{a}_{3} & = & -x_{4}a \, \mathbf{\hat{x}}-y_{4}b \, \mathbf{\hat{y}} + \left(\frac{1}{2} +z_{4}\right)c \, \mathbf{\hat{z}} & \left(4a\right) & \mbox{Al IV} \\ 
\mathbf{B}_{15} & = & \left(\frac{1}{2} +x_{4}\right) \, \mathbf{a}_{1} + \left(\frac{1}{2} - y_{4}\right) \, \mathbf{a}_{2} + z_{4} \, \mathbf{a}_{3} & = & \left(\frac{1}{2} +x_{4}\right)a \, \mathbf{\hat{x}} + \left(\frac{1}{2} - y_{4}\right)b \, \mathbf{\hat{y}} + z_{4}c \, \mathbf{\hat{z}} & \left(4a\right) & \mbox{Al IV} \\ 
\mathbf{B}_{16} & = & \left(\frac{1}{2} - x_{4}\right) \, \mathbf{a}_{1} + \left(\frac{1}{2} +y_{4}\right) \, \mathbf{a}_{2} + \left(\frac{1}{2} +z_{4}\right) \, \mathbf{a}_{3} & = & \left(\frac{1}{2} - x_{4}\right)a \, \mathbf{\hat{x}} + \left(\frac{1}{2} +y_{4}\right)b \, \mathbf{\hat{y}} + \left(\frac{1}{2} +z_{4}\right)c \, \mathbf{\hat{z}} & \left(4a\right) & \mbox{Al IV} \\ 
\mathbf{B}_{17} & = & x_{5} \, \mathbf{a}_{1} + y_{5} \, \mathbf{a}_{2} + z_{5} \, \mathbf{a}_{3} & = & x_{5}a \, \mathbf{\hat{x}} + y_{5}b \, \mathbf{\hat{y}} + z_{5}c \, \mathbf{\hat{z}} & \left(4a\right) & \mbox{O I} \\ 
\mathbf{B}_{18} & = & -x_{5} \, \mathbf{a}_{1}-y_{5} \, \mathbf{a}_{2} + \left(\frac{1}{2} +z_{5}\right) \, \mathbf{a}_{3} & = & -x_{5}a \, \mathbf{\hat{x}}-y_{5}b \, \mathbf{\hat{y}} + \left(\frac{1}{2} +z_{5}\right)c \, \mathbf{\hat{z}} & \left(4a\right) & \mbox{O I} \\ 
\mathbf{B}_{19} & = & \left(\frac{1}{2} +x_{5}\right) \, \mathbf{a}_{1} + \left(\frac{1}{2} - y_{5}\right) \, \mathbf{a}_{2} + z_{5} \, \mathbf{a}_{3} & = & \left(\frac{1}{2} +x_{5}\right)a \, \mathbf{\hat{x}} + \left(\frac{1}{2} - y_{5}\right)b \, \mathbf{\hat{y}} + z_{5}c \, \mathbf{\hat{z}} & \left(4a\right) & \mbox{O I} \\ 
\mathbf{B}_{20} & = & \left(\frac{1}{2} - x_{5}\right) \, \mathbf{a}_{1} + \left(\frac{1}{2} +y_{5}\right) \, \mathbf{a}_{2} + \left(\frac{1}{2} +z_{5}\right) \, \mathbf{a}_{3} & = & \left(\frac{1}{2} - x_{5}\right)a \, \mathbf{\hat{x}} + \left(\frac{1}{2} +y_{5}\right)b \, \mathbf{\hat{y}} + \left(\frac{1}{2} +z_{5}\right)c \, \mathbf{\hat{z}} & \left(4a\right) & \mbox{O I} \\ 
\mathbf{B}_{21} & = & x_{6} \, \mathbf{a}_{1} + y_{6} \, \mathbf{a}_{2} + z_{6} \, \mathbf{a}_{3} & = & x_{6}a \, \mathbf{\hat{x}} + y_{6}b \, \mathbf{\hat{y}} + z_{6}c \, \mathbf{\hat{z}} & \left(4a\right) & \mbox{O II} \\ 
\mathbf{B}_{22} & = & -x_{6} \, \mathbf{a}_{1}-y_{6} \, \mathbf{a}_{2} + \left(\frac{1}{2} +z_{6}\right) \, \mathbf{a}_{3} & = & -x_{6}a \, \mathbf{\hat{x}}-y_{6}b \, \mathbf{\hat{y}} + \left(\frac{1}{2} +z_{6}\right)c \, \mathbf{\hat{z}} & \left(4a\right) & \mbox{O II} \\ 
\mathbf{B}_{23} & = & \left(\frac{1}{2} +x_{6}\right) \, \mathbf{a}_{1} + \left(\frac{1}{2} - y_{6}\right) \, \mathbf{a}_{2} + z_{6} \, \mathbf{a}_{3} & = & \left(\frac{1}{2} +x_{6}\right)a \, \mathbf{\hat{x}} + \left(\frac{1}{2} - y_{6}\right)b \, \mathbf{\hat{y}} + z_{6}c \, \mathbf{\hat{z}} & \left(4a\right) & \mbox{O II} \\ 
\mathbf{B}_{24} & = & \left(\frac{1}{2} - x_{6}\right) \, \mathbf{a}_{1} + \left(\frac{1}{2} +y_{6}\right) \, \mathbf{a}_{2} + \left(\frac{1}{2} +z_{6}\right) \, \mathbf{a}_{3} & = & \left(\frac{1}{2} - x_{6}\right)a \, \mathbf{\hat{x}} + \left(\frac{1}{2} +y_{6}\right)b \, \mathbf{\hat{y}} + \left(\frac{1}{2} +z_{6}\right)c \, \mathbf{\hat{z}} & \left(4a\right) & \mbox{O II} \\ 
\mathbf{B}_{25} & = & x_{7} \, \mathbf{a}_{1} + y_{7} \, \mathbf{a}_{2} + z_{7} \, \mathbf{a}_{3} & = & x_{7}a \, \mathbf{\hat{x}} + y_{7}b \, \mathbf{\hat{y}} + z_{7}c \, \mathbf{\hat{z}} & \left(4a\right) & \mbox{O III} \\ 
\mathbf{B}_{26} & = & -x_{7} \, \mathbf{a}_{1}-y_{7} \, \mathbf{a}_{2} + \left(\frac{1}{2} +z_{7}\right) \, \mathbf{a}_{3} & = & -x_{7}a \, \mathbf{\hat{x}}-y_{7}b \, \mathbf{\hat{y}} + \left(\frac{1}{2} +z_{7}\right)c \, \mathbf{\hat{z}} & \left(4a\right) & \mbox{O III} \\ 
\mathbf{B}_{27} & = & \left(\frac{1}{2} +x_{7}\right) \, \mathbf{a}_{1} + \left(\frac{1}{2} - y_{7}\right) \, \mathbf{a}_{2} + z_{7} \, \mathbf{a}_{3} & = & \left(\frac{1}{2} +x_{7}\right)a \, \mathbf{\hat{x}} + \left(\frac{1}{2} - y_{7}\right)b \, \mathbf{\hat{y}} + z_{7}c \, \mathbf{\hat{z}} & \left(4a\right) & \mbox{O III} \\ 
\mathbf{B}_{28} & = & \left(\frac{1}{2} - x_{7}\right) \, \mathbf{a}_{1} + \left(\frac{1}{2} +y_{7}\right) \, \mathbf{a}_{2} + \left(\frac{1}{2} +z_{7}\right) \, \mathbf{a}_{3} & = & \left(\frac{1}{2} - x_{7}\right)a \, \mathbf{\hat{x}} + \left(\frac{1}{2} +y_{7}\right)b \, \mathbf{\hat{y}} + \left(\frac{1}{2} +z_{7}\right)c \, \mathbf{\hat{z}} & \left(4a\right) & \mbox{O III} \\ 
\mathbf{B}_{29} & = & x_{8} \, \mathbf{a}_{1} + y_{8} \, \mathbf{a}_{2} + z_{8} \, \mathbf{a}_{3} & = & x_{8}a \, \mathbf{\hat{x}} + y_{8}b \, \mathbf{\hat{y}} + z_{8}c \, \mathbf{\hat{z}} & \left(4a\right) & \mbox{O IV} \\ 
\mathbf{B}_{30} & = & -x_{8} \, \mathbf{a}_{1}-y_{8} \, \mathbf{a}_{2} + \left(\frac{1}{2} +z_{8}\right) \, \mathbf{a}_{3} & = & -x_{8}a \, \mathbf{\hat{x}}-y_{8}b \, \mathbf{\hat{y}} + \left(\frac{1}{2} +z_{8}\right)c \, \mathbf{\hat{z}} & \left(4a\right) & \mbox{O IV} \\ 
\mathbf{B}_{31} & = & \left(\frac{1}{2} +x_{8}\right) \, \mathbf{a}_{1} + \left(\frac{1}{2} - y_{8}\right) \, \mathbf{a}_{2} + z_{8} \, \mathbf{a}_{3} & = & \left(\frac{1}{2} +x_{8}\right)a \, \mathbf{\hat{x}} + \left(\frac{1}{2} - y_{8}\right)b \, \mathbf{\hat{y}} + z_{8}c \, \mathbf{\hat{z}} & \left(4a\right) & \mbox{O IV} \\ 
\mathbf{B}_{32} & = & \left(\frac{1}{2} - x_{8}\right) \, \mathbf{a}_{1} + \left(\frac{1}{2} +y_{8}\right) \, \mathbf{a}_{2} + \left(\frac{1}{2} +z_{8}\right) \, \mathbf{a}_{3} & = & \left(\frac{1}{2} - x_{8}\right)a \, \mathbf{\hat{x}} + \left(\frac{1}{2} +y_{8}\right)b \, \mathbf{\hat{y}} + \left(\frac{1}{2} +z_{8}\right)c \, \mathbf{\hat{z}} & \left(4a\right) & \mbox{O IV} \\ 
\mathbf{B}_{33} & = & x_{9} \, \mathbf{a}_{1} + y_{9} \, \mathbf{a}_{2} + z_{9} \, \mathbf{a}_{3} & = & x_{9}a \, \mathbf{\hat{x}} + y_{9}b \, \mathbf{\hat{y}} + z_{9}c \, \mathbf{\hat{z}} & \left(4a\right) & \mbox{O V} \\ 
\mathbf{B}_{34} & = & -x_{9} \, \mathbf{a}_{1}-y_{9} \, \mathbf{a}_{2} + \left(\frac{1}{2} +z_{9}\right) \, \mathbf{a}_{3} & = & -x_{9}a \, \mathbf{\hat{x}}-y_{9}b \, \mathbf{\hat{y}} + \left(\frac{1}{2} +z_{9}\right)c \, \mathbf{\hat{z}} & \left(4a\right) & \mbox{O V} \\ 
\mathbf{B}_{35} & = & \left(\frac{1}{2} +x_{9}\right) \, \mathbf{a}_{1} + \left(\frac{1}{2} - y_{9}\right) \, \mathbf{a}_{2} + z_{9} \, \mathbf{a}_{3} & = & \left(\frac{1}{2} +x_{9}\right)a \, \mathbf{\hat{x}} + \left(\frac{1}{2} - y_{9}\right)b \, \mathbf{\hat{y}} + z_{9}c \, \mathbf{\hat{z}} & \left(4a\right) & \mbox{O V} \\ 
\mathbf{B}_{36} & = & \left(\frac{1}{2} - x_{9}\right) \, \mathbf{a}_{1} + \left(\frac{1}{2} +y_{9}\right) \, \mathbf{a}_{2} + \left(\frac{1}{2} +z_{9}\right) \, \mathbf{a}_{3} & = & \left(\frac{1}{2} - x_{9}\right)a \, \mathbf{\hat{x}} + \left(\frac{1}{2} +y_{9}\right)b \, \mathbf{\hat{y}} + \left(\frac{1}{2} +z_{9}\right)c \, \mathbf{\hat{z}} & \left(4a\right) & \mbox{O V} \\ 
\mathbf{B}_{37} & = & x_{10} \, \mathbf{a}_{1} + y_{10} \, \mathbf{a}_{2} + z_{10} \, \mathbf{a}_{3} & = & x_{10}a \, \mathbf{\hat{x}} + y_{10}b \, \mathbf{\hat{y}} + z_{10}c \, \mathbf{\hat{z}} & \left(4a\right) & \mbox{O VI} \\ 
\mathbf{B}_{38} & = & -x_{10} \, \mathbf{a}_{1}-y_{10} \, \mathbf{a}_{2} + \left(\frac{1}{2} +z_{10}\right) \, \mathbf{a}_{3} & = & -x_{10}a \, \mathbf{\hat{x}}-y_{10}b \, \mathbf{\hat{y}} + \left(\frac{1}{2} +z_{10}\right)c \, \mathbf{\hat{z}} & \left(4a\right) & \mbox{O VI} \\ 
\mathbf{B}_{39} & = & \left(\frac{1}{2} +x_{10}\right) \, \mathbf{a}_{1} + \left(\frac{1}{2} - y_{10}\right) \, \mathbf{a}_{2} + z_{10} \, \mathbf{a}_{3} & = & \left(\frac{1}{2} +x_{10}\right)a \, \mathbf{\hat{x}} + \left(\frac{1}{2} - y_{10}\right)b \, \mathbf{\hat{y}} + z_{10}c \, \mathbf{\hat{z}} & \left(4a\right) & \mbox{O VI} \\ 
\mathbf{B}_{40} & = & \left(\frac{1}{2} - x_{10}\right) \, \mathbf{a}_{1} + \left(\frac{1}{2} +y_{10}\right) \, \mathbf{a}_{2} + \left(\frac{1}{2} +z_{10}\right) \, \mathbf{a}_{3} & = & \left(\frac{1}{2} - x_{10}\right)a \, \mathbf{\hat{x}} + \left(\frac{1}{2} +y_{10}\right)b \, \mathbf{\hat{y}} + \left(\frac{1}{2} +z_{10}\right)c \, \mathbf{\hat{z}} & \left(4a\right) & \mbox{O VI} \\ 
\end{longtabu}
\renewcommand{\arraystretch}{1.0}
\noindent \hrulefill
\\
\textbf{References:}
\vspace*{-0.25cm}
\begin{flushleft}
  - \bibentry{Ollivier_JMC_7_1997}. \\
\end{flushleft}
\textbf{Found in:}
\vspace*{-0.25cm}
\begin{flushleft}
  - \bibentry{Matsuzaki_APL_88_2006}. \\
  - \bibentry{Yoshioka_JPCM_19_2007}. \\
\end{flushleft}
\noindent \hrulefill
\\
\textbf{Geometry files:}
\\
\noindent  - CIF: pp. {\hyperref[A2B3_oP40_33_4a_6a_cif]{\pageref{A2B3_oP40_33_4a_6a_cif}}} \\
\noindent  - POSCAR: pp. {\hyperref[A2B3_oP40_33_4a_6a_poscar]{\pageref{A2B3_oP40_33_4a_6a_poscar}}} \\
\onecolumn
{\phantomsection\label{A2B8C_oP22_34_c_4c_a}}
\subsection*{\huge \textbf{{\normalfont TiAl$_{2}$Br$_{8}$ Structure: A2B8C\_oP22\_34\_c\_4c\_a}}}
\noindent \hrulefill
\vspace*{0.25cm}
\begin{figure}[htp]
  \centering
  \vspace{-1em}
  {\includegraphics[width=1\textwidth]{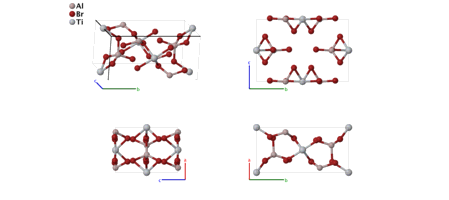}}
\end{figure}
\vspace*{-0.5cm}
\renewcommand{\arraystretch}{1.5}
\begin{equation*}
  \begin{array}{>{$\hspace{-0.15cm}}l<{$}>{$}p{0.5cm}<{$}>{$}p{18.5cm}<{$}}
    \mbox{\large \textbf{Prototype}} &\colon & \ce{TiAl2Br8} \\
    \mbox{\large \textbf{\AFLOW\ prototype label}} &\colon & \mbox{A2B8C\_oP22\_34\_c\_4c\_a} \\
    \mbox{\large \textbf{\textit{Strukturbericht} designation}} &\colon & \mbox{None} \\
    \mbox{\large \textbf{Pearson symbol}} &\colon & \mbox{oP22} \\
    \mbox{\large \textbf{Space group number}} &\colon & 34 \\
    \mbox{\large \textbf{Space group symbol}} &\colon & Pnn2 \\
    \mbox{\large \textbf{\AFLOW\ prototype command}} &\colon &  \texttt{aflow} \,  \, \texttt{-{}-proto=A2B8C\_oP22\_34\_c\_4c\_a } \, \newline \texttt{-{}-params=}{a,b/a,c/a,z_{1},x_{2},y_{2},z_{2},x_{3},y_{3},z_{3},x_{4},y_{4},z_{4},x_{5},y_{5},z_{5},x_{6},y_{6},z_{6} }
  \end{array}
\end{equation*}
\renewcommand{\arraystretch}{1.0}

\noindent \parbox{1 \linewidth}{
\noindent \hrulefill
\\
\textbf{Simple Orthorhombic primitive vectors:} \\
\vspace*{-0.25cm}
\begin{tabular}{cc}
  \begin{tabular}{c}
    \parbox{0.6 \linewidth}{
      \renewcommand{\arraystretch}{1.5}
      \begin{equation*}
        \centering
        \begin{array}{ccc}
              \mathbf{a}_1 & = & a \, \mathbf{\hat{x}} \\
    \mathbf{a}_2 & = & b \, \mathbf{\hat{y}} \\
    \mathbf{a}_3 & = & c \, \mathbf{\hat{z}} \\

        \end{array}
      \end{equation*}
    }
    \renewcommand{\arraystretch}{1.0}
  \end{tabular}
  \begin{tabular}{c}
    \includegraphics[width=0.3\linewidth]{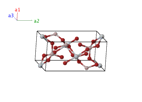} \\
  \end{tabular}
\end{tabular}

}
\vspace*{-0.25cm}

\noindent \hrulefill
\\
\textbf{Basis vectors:}
\vspace*{-0.25cm}
\renewcommand{\arraystretch}{1.5}
\begin{longtabu} to \textwidth{>{\centering $}X[-1,c,c]<{$}>{\centering $}X[-1,c,c]<{$}>{\centering $}X[-1,c,c]<{$}>{\centering $}X[-1,c,c]<{$}>{\centering $}X[-1,c,c]<{$}>{\centering $}X[-1,c,c]<{$}>{\centering $}X[-1,c,c]<{$}}
  & & \mbox{Lattice Coordinates} & & \mbox{Cartesian Coordinates} &\mbox{Wyckoff Position} & \mbox{Atom Type} \\  
  \mathbf{B}_{1} & = & z_{1} \, \mathbf{a}_{3} & = & z_{1}c \, \mathbf{\hat{z}} & \left(2a\right) & \mbox{Ti} \\ 
\mathbf{B}_{2} & = & \frac{1}{2} \, \mathbf{a}_{1} + \frac{1}{2} \, \mathbf{a}_{2} + \left(\frac{1}{2} +z_{1}\right) \, \mathbf{a}_{3} & = & \frac{1}{2}a \, \mathbf{\hat{x}} + \frac{1}{2}b \, \mathbf{\hat{y}} + \left(\frac{1}{2} +z_{1}\right)c \, \mathbf{\hat{z}} & \left(2a\right) & \mbox{Ti} \\ 
\mathbf{B}_{3} & = & x_{2} \, \mathbf{a}_{1} + y_{2} \, \mathbf{a}_{2} + z_{2} \, \mathbf{a}_{3} & = & x_{2}a \, \mathbf{\hat{x}} + y_{2}b \, \mathbf{\hat{y}} + z_{2}c \, \mathbf{\hat{z}} & \left(4c\right) & \mbox{Al} \\ 
\mathbf{B}_{4} & = & -x_{2} \, \mathbf{a}_{1}-y_{2} \, \mathbf{a}_{2} + z_{2} \, \mathbf{a}_{3} & = & -x_{2}a \, \mathbf{\hat{x}}-y_{2}b \, \mathbf{\hat{y}} + z_{2}c \, \mathbf{\hat{z}} & \left(4c\right) & \mbox{Al} \\ 
\mathbf{B}_{5} & = & \left(\frac{1}{2} +x_{2}\right) \, \mathbf{a}_{1} + \left(\frac{1}{2} - y_{2}\right) \, \mathbf{a}_{2} + \left(\frac{1}{2} +z_{2}\right) \, \mathbf{a}_{3} & = & \left(\frac{1}{2} +x_{2}\right)a \, \mathbf{\hat{x}} + \left(\frac{1}{2} - y_{2}\right)b \, \mathbf{\hat{y}} + \left(\frac{1}{2} +z_{2}\right)c \, \mathbf{\hat{z}} & \left(4c\right) & \mbox{Al} \\ 
\mathbf{B}_{6} & = & \left(\frac{1}{2} - x_{2}\right) \, \mathbf{a}_{1} + \left(\frac{1}{2} +y_{2}\right) \, \mathbf{a}_{2} + \left(\frac{1}{2} +z_{2}\right) \, \mathbf{a}_{3} & = & \left(\frac{1}{2} - x_{2}\right)a \, \mathbf{\hat{x}} + \left(\frac{1}{2} +y_{2}\right)b \, \mathbf{\hat{y}} + \left(\frac{1}{2} +z_{2}\right)c \, \mathbf{\hat{z}} & \left(4c\right) & \mbox{Al} \\ 
\mathbf{B}_{7} & = & x_{3} \, \mathbf{a}_{1} + y_{3} \, \mathbf{a}_{2} + z_{3} \, \mathbf{a}_{3} & = & x_{3}a \, \mathbf{\hat{x}} + y_{3}b \, \mathbf{\hat{y}} + z_{3}c \, \mathbf{\hat{z}} & \left(4c\right) & \mbox{Br I} \\ 
\mathbf{B}_{8} & = & -x_{3} \, \mathbf{a}_{1}-y_{3} \, \mathbf{a}_{2} + z_{3} \, \mathbf{a}_{3} & = & -x_{3}a \, \mathbf{\hat{x}}-y_{3}b \, \mathbf{\hat{y}} + z_{3}c \, \mathbf{\hat{z}} & \left(4c\right) & \mbox{Br I} \\ 
\mathbf{B}_{9} & = & \left(\frac{1}{2} +x_{3}\right) \, \mathbf{a}_{1} + \left(\frac{1}{2} - y_{3}\right) \, \mathbf{a}_{2} + \left(\frac{1}{2} +z_{3}\right) \, \mathbf{a}_{3} & = & \left(\frac{1}{2} +x_{3}\right)a \, \mathbf{\hat{x}} + \left(\frac{1}{2} - y_{3}\right)b \, \mathbf{\hat{y}} + \left(\frac{1}{2} +z_{3}\right)c \, \mathbf{\hat{z}} & \left(4c\right) & \mbox{Br I} \\ 
\mathbf{B}_{10} & = & \left(\frac{1}{2} - x_{3}\right) \, \mathbf{a}_{1} + \left(\frac{1}{2} +y_{3}\right) \, \mathbf{a}_{2} + \left(\frac{1}{2} +z_{3}\right) \, \mathbf{a}_{3} & = & \left(\frac{1}{2} - x_{3}\right)a \, \mathbf{\hat{x}} + \left(\frac{1}{2} +y_{3}\right)b \, \mathbf{\hat{y}} + \left(\frac{1}{2} +z_{3}\right)c \, \mathbf{\hat{z}} & \left(4c\right) & \mbox{Br I} \\ 
\mathbf{B}_{11} & = & x_{4} \, \mathbf{a}_{1} + y_{4} \, \mathbf{a}_{2} + z_{4} \, \mathbf{a}_{3} & = & x_{4}a \, \mathbf{\hat{x}} + y_{4}b \, \mathbf{\hat{y}} + z_{4}c \, \mathbf{\hat{z}} & \left(4c\right) & \mbox{Br II} \\ 
\mathbf{B}_{12} & = & -x_{4} \, \mathbf{a}_{1}-y_{4} \, \mathbf{a}_{2} + z_{4} \, \mathbf{a}_{3} & = & -x_{4}a \, \mathbf{\hat{x}}-y_{4}b \, \mathbf{\hat{y}} + z_{4}c \, \mathbf{\hat{z}} & \left(4c\right) & \mbox{Br II} \\ 
\mathbf{B}_{13} & = & \left(\frac{1}{2} +x_{4}\right) \, \mathbf{a}_{1} + \left(\frac{1}{2} - y_{4}\right) \, \mathbf{a}_{2} + \left(\frac{1}{2} +z_{4}\right) \, \mathbf{a}_{3} & = & \left(\frac{1}{2} +x_{4}\right)a \, \mathbf{\hat{x}} + \left(\frac{1}{2} - y_{4}\right)b \, \mathbf{\hat{y}} + \left(\frac{1}{2} +z_{4}\right)c \, \mathbf{\hat{z}} & \left(4c\right) & \mbox{Br II} \\ 
\mathbf{B}_{14} & = & \left(\frac{1}{2} - x_{4}\right) \, \mathbf{a}_{1} + \left(\frac{1}{2} +y_{4}\right) \, \mathbf{a}_{2} + \left(\frac{1}{2} +z_{4}\right) \, \mathbf{a}_{3} & = & \left(\frac{1}{2} - x_{4}\right)a \, \mathbf{\hat{x}} + \left(\frac{1}{2} +y_{4}\right)b \, \mathbf{\hat{y}} + \left(\frac{1}{2} +z_{4}\right)c \, \mathbf{\hat{z}} & \left(4c\right) & \mbox{Br II} \\ 
\mathbf{B}_{15} & = & x_{5} \, \mathbf{a}_{1} + y_{5} \, \mathbf{a}_{2} + z_{5} \, \mathbf{a}_{3} & = & x_{5}a \, \mathbf{\hat{x}} + y_{5}b \, \mathbf{\hat{y}} + z_{5}c \, \mathbf{\hat{z}} & \left(4c\right) & \mbox{Br III} \\ 
\mathbf{B}_{16} & = & -x_{5} \, \mathbf{a}_{1}-y_{5} \, \mathbf{a}_{2} + z_{5} \, \mathbf{a}_{3} & = & -x_{5}a \, \mathbf{\hat{x}}-y_{5}b \, \mathbf{\hat{y}} + z_{5}c \, \mathbf{\hat{z}} & \left(4c\right) & \mbox{Br III} \\ 
\mathbf{B}_{17} & = & \left(\frac{1}{2} +x_{5}\right) \, \mathbf{a}_{1} + \left(\frac{1}{2} - y_{5}\right) \, \mathbf{a}_{2} + \left(\frac{1}{2} +z_{5}\right) \, \mathbf{a}_{3} & = & \left(\frac{1}{2} +x_{5}\right)a \, \mathbf{\hat{x}} + \left(\frac{1}{2} - y_{5}\right)b \, \mathbf{\hat{y}} + \left(\frac{1}{2} +z_{5}\right)c \, \mathbf{\hat{z}} & \left(4c\right) & \mbox{Br III} \\ 
\mathbf{B}_{18} & = & \left(\frac{1}{2} - x_{5}\right) \, \mathbf{a}_{1} + \left(\frac{1}{2} +y_{5}\right) \, \mathbf{a}_{2} + \left(\frac{1}{2} +z_{5}\right) \, \mathbf{a}_{3} & = & \left(\frac{1}{2} - x_{5}\right)a \, \mathbf{\hat{x}} + \left(\frac{1}{2} +y_{5}\right)b \, \mathbf{\hat{y}} + \left(\frac{1}{2} +z_{5}\right)c \, \mathbf{\hat{z}} & \left(4c\right) & \mbox{Br III} \\ 
\mathbf{B}_{19} & = & x_{6} \, \mathbf{a}_{1} + y_{6} \, \mathbf{a}_{2} + z_{6} \, \mathbf{a}_{3} & = & x_{6}a \, \mathbf{\hat{x}} + y_{6}b \, \mathbf{\hat{y}} + z_{6}c \, \mathbf{\hat{z}} & \left(4c\right) & \mbox{Br IV} \\ 
\mathbf{B}_{20} & = & -x_{6} \, \mathbf{a}_{1}-y_{6} \, \mathbf{a}_{2} + z_{6} \, \mathbf{a}_{3} & = & -x_{6}a \, \mathbf{\hat{x}}-y_{6}b \, \mathbf{\hat{y}} + z_{6}c \, \mathbf{\hat{z}} & \left(4c\right) & \mbox{Br IV} \\ 
\mathbf{B}_{21} & = & \left(\frac{1}{2} +x_{6}\right) \, \mathbf{a}_{1} + \left(\frac{1}{2} - y_{6}\right) \, \mathbf{a}_{2} + \left(\frac{1}{2} +z_{6}\right) \, \mathbf{a}_{3} & = & \left(\frac{1}{2} +x_{6}\right)a \, \mathbf{\hat{x}} + \left(\frac{1}{2} - y_{6}\right)b \, \mathbf{\hat{y}} + \left(\frac{1}{2} +z_{6}\right)c \, \mathbf{\hat{z}} & \left(4c\right) & \mbox{Br IV} \\ 
\mathbf{B}_{22} & = & \left(\frac{1}{2} - x_{6}\right) \, \mathbf{a}_{1} + \left(\frac{1}{2} +y_{6}\right) \, \mathbf{a}_{2} + \left(\frac{1}{2} +z_{6}\right) \, \mathbf{a}_{3} & = & \left(\frac{1}{2} - x_{6}\right)a \, \mathbf{\hat{x}} + \left(\frac{1}{2} +y_{6}\right)b \, \mathbf{\hat{y}} + \left(\frac{1}{2} +z_{6}\right)c \, \mathbf{\hat{z}} & \left(4c\right) & \mbox{Br IV} \\ 
\end{longtabu}
\renewcommand{\arraystretch}{1.0}
\noindent \hrulefill
\\
\textbf{References:}
\vspace*{-0.25cm}
\begin{flushleft}
  - \bibentry{Troyanov_Al2Br8Ti_RussJInorgChem_1990}. \\
\end{flushleft}
\textbf{Found in:}
\vspace*{-0.25cm}
\begin{flushleft}
  - \bibentry{Villars_PearsonsCrystalData_2013}. \\
\end{flushleft}
\noindent \hrulefill
\\
\textbf{Geometry files:}
\\
\noindent  - CIF: pp. {\hyperref[A2B8C_oP22_34_c_4c_a_cif]{\pageref{A2B8C_oP22_34_c_4c_a_cif}}} \\
\noindent  - POSCAR: pp. {\hyperref[A2B8C_oP22_34_c_4c_a_poscar]{\pageref{A2B8C_oP22_34_c_4c_a_poscar}}} \\
\onecolumn
{\phantomsection\label{AB2_oP6_34_a_c}}
\subsection*{\huge \textbf{{\normalfont FeSb$_{2}$ Structure: AB2\_oP6\_34\_a\_c}}}
\noindent \hrulefill
\vspace*{0.25cm}
\begin{figure}[htp]
  \centering
  \vspace{-1em}
  {\includegraphics[width=1\textwidth]{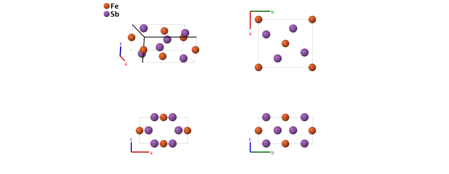}}
\end{figure}
\vspace*{-0.5cm}
\renewcommand{\arraystretch}{1.5}
\begin{equation*}
  \begin{array}{>{$\hspace{-0.15cm}}l<{$}>{$}p{0.5cm}<{$}>{$}p{18.5cm}<{$}}
    \mbox{\large \textbf{Prototype}} &\colon & \ce{FeSb2} \\
    \mbox{\large \textbf{\AFLOW\ prototype label}} &\colon & \mbox{AB2\_oP6\_34\_a\_c} \\
    \mbox{\large \textbf{\textit{Strukturbericht} designation}} &\colon & \mbox{None} \\
    \mbox{\large \textbf{Pearson symbol}} &\colon & \mbox{oP6} \\
    \mbox{\large \textbf{Space group number}} &\colon & 34 \\
    \mbox{\large \textbf{Space group symbol}} &\colon & Pnn2 \\
    \mbox{\large \textbf{\AFLOW\ prototype command}} &\colon &  \texttt{aflow} \,  \, \texttt{-{}-proto=AB2\_oP6\_34\_a\_c } \, \newline \texttt{-{}-params=}{a,b/a,c/a,z_{1},x_{2},y_{2},z_{2} }
  \end{array}
\end{equation*}
\renewcommand{\arraystretch}{1.0}

\noindent \parbox{1 \linewidth}{
\noindent \hrulefill
\\
\textbf{Simple Orthorhombic primitive vectors:} \\
\vspace*{-0.25cm}
\begin{tabular}{cc}
  \begin{tabular}{c}
    \parbox{0.6 \linewidth}{
      \renewcommand{\arraystretch}{1.5}
      \begin{equation*}
        \centering
        \begin{array}{ccc}
              \mathbf{a}_1 & = & a \, \mathbf{\hat{x}} \\
    \mathbf{a}_2 & = & b \, \mathbf{\hat{y}} \\
    \mathbf{a}_3 & = & c \, \mathbf{\hat{z}} \\

        \end{array}
      \end{equation*}
    }
    \renewcommand{\arraystretch}{1.0}
  \end{tabular}
  \begin{tabular}{c}
    \includegraphics[width=0.3\linewidth]{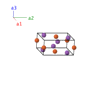} \\
  \end{tabular}
\end{tabular}

}
\vspace*{-0.25cm}

\noindent \hrulefill
\\
\textbf{Basis vectors:}
\vspace*{-0.25cm}
\renewcommand{\arraystretch}{1.5}
\begin{longtabu} to \textwidth{>{\centering $}X[-1,c,c]<{$}>{\centering $}X[-1,c,c]<{$}>{\centering $}X[-1,c,c]<{$}>{\centering $}X[-1,c,c]<{$}>{\centering $}X[-1,c,c]<{$}>{\centering $}X[-1,c,c]<{$}>{\centering $}X[-1,c,c]<{$}}
  & & \mbox{Lattice Coordinates} & & \mbox{Cartesian Coordinates} &\mbox{Wyckoff Position} & \mbox{Atom Type} \\  
  \mathbf{B}_{1} & = & z_{1} \, \mathbf{a}_{3} & = & z_{1}c \, \mathbf{\hat{z}} & \left(2a\right) & \mbox{Fe} \\ 
\mathbf{B}_{2} & = & \frac{1}{2} \, \mathbf{a}_{1} + \frac{1}{2} \, \mathbf{a}_{2} + \left(\frac{1}{2} +z_{1}\right) \, \mathbf{a}_{3} & = & \frac{1}{2}a \, \mathbf{\hat{x}} + \frac{1}{2}b \, \mathbf{\hat{y}} + \left(\frac{1}{2} +z_{1}\right)c \, \mathbf{\hat{z}} & \left(2a\right) & \mbox{Fe} \\ 
\mathbf{B}_{3} & = & x_{2} \, \mathbf{a}_{1} + y_{2} \, \mathbf{a}_{2} + z_{2} \, \mathbf{a}_{3} & = & x_{2}a \, \mathbf{\hat{x}} + y_{2}b \, \mathbf{\hat{y}} + z_{2}c \, \mathbf{\hat{z}} & \left(4c\right) & \mbox{Sb} \\ 
\mathbf{B}_{4} & = & -x_{2} \, \mathbf{a}_{1}-y_{2} \, \mathbf{a}_{2} + z_{2} \, \mathbf{a}_{3} & = & -x_{2}a \, \mathbf{\hat{x}}-y_{2}b \, \mathbf{\hat{y}} + z_{2}c \, \mathbf{\hat{z}} & \left(4c\right) & \mbox{Sb} \\ 
\mathbf{B}_{5} & = & \left(\frac{1}{2} +x_{2}\right) \, \mathbf{a}_{1} + \left(\frac{1}{2} - y_{2}\right) \, \mathbf{a}_{2} + \left(\frac{1}{2} +z_{2}\right) \, \mathbf{a}_{3} & = & \left(\frac{1}{2} +x_{2}\right)a \, \mathbf{\hat{x}} + \left(\frac{1}{2} - y_{2}\right)b \, \mathbf{\hat{y}} + \left(\frac{1}{2} +z_{2}\right)c \, \mathbf{\hat{z}} & \left(4c\right) & \mbox{Sb} \\ 
\mathbf{B}_{6} & = & \left(\frac{1}{2} - x_{2}\right) \, \mathbf{a}_{1} + \left(\frac{1}{2} +y_{2}\right) \, \mathbf{a}_{2} + \left(\frac{1}{2} +z_{2}\right) \, \mathbf{a}_{3} & = & \left(\frac{1}{2} - x_{2}\right)a \, \mathbf{\hat{x}} + \left(\frac{1}{2} +y_{2}\right)b \, \mathbf{\hat{y}} + \left(\frac{1}{2} +z_{2}\right)c \, \mathbf{\hat{z}} & \left(4c\right) & \mbox{Sb} \\ 
\end{longtabu}
\renewcommand{\arraystretch}{1.0}
\noindent \hrulefill
\\
\textbf{References:}
\vspace*{-0.25cm}
\begin{flushleft}
  - \bibentry{Holseth_FeSb2_ActChemScand_1969}. \\
\end{flushleft}
\textbf{Found in:}
\vspace*{-0.25cm}
\begin{flushleft}
  - \bibentry{Villars_PearsonsCrystalData_2013}. \\
\end{flushleft}
\noindent \hrulefill
\\
\textbf{Geometry files:}
\\
\noindent  - CIF: pp. {\hyperref[AB2_oP6_34_a_c_cif]{\pageref{AB2_oP6_34_a_c_cif}}} \\
\noindent  - POSCAR: pp. {\hyperref[AB2_oP6_34_a_c_poscar]{\pageref{AB2_oP6_34_a_c_poscar}}} \\
\onecolumn
{\phantomsection\label{AB8C2_oC22_35_a_ab3e_e}}
\subsection*{\huge \textbf{{\normalfont V$_{2}$MoO$_{8}$ Structure: AB8C2\_oC22\_35\_a\_ab3e\_e}}}
\noindent \hrulefill
\vspace*{0.25cm}
\begin{figure}[htp]
  \centering
  \vspace{-1em}
  {\includegraphics[width=1\textwidth]{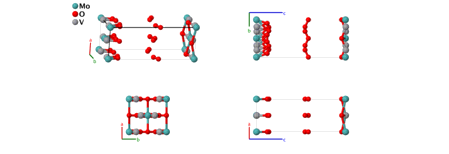}}
\end{figure}
\vspace*{-0.5cm}
\renewcommand{\arraystretch}{1.5}
\begin{equation*}
  \begin{array}{>{$\hspace{-0.15cm}}l<{$}>{$}p{0.5cm}<{$}>{$}p{18.5cm}<{$}}
    \mbox{\large \textbf{Prototype}} &\colon & \ce{V2MoO8} \\
    \mbox{\large \textbf{\AFLOW\ prototype label}} &\colon & \mbox{AB8C2\_oC22\_35\_a\_ab3e\_e} \\
    \mbox{\large \textbf{\textit{Strukturbericht} designation}} &\colon & \mbox{None} \\
    \mbox{\large \textbf{Pearson symbol}} &\colon & \mbox{oC22} \\
    \mbox{\large \textbf{Space group number}} &\colon & 35 \\
    \mbox{\large \textbf{Space group symbol}} &\colon & Cmm2 \\
    \mbox{\large \textbf{\AFLOW\ prototype command}} &\colon &  \texttt{aflow} \,  \, \texttt{-{}-proto=AB8C2\_oC22\_35\_a\_ab3e\_e } \, \newline \texttt{-{}-params=}{a,b/a,c/a,z_{1},z_{2},z_{3},y_{4},z_{4},y_{5},z_{5},y_{6},z_{6},y_{7},z_{7} }
  \end{array}
\end{equation*}
\renewcommand{\arraystretch}{1.0}

\noindent \parbox{1 \linewidth}{
\noindent \hrulefill
\\
\textbf{Base-centered Orthorhombic primitive vectors:} \\
\vspace*{-0.25cm}
\begin{tabular}{cc}
  \begin{tabular}{c}
    \parbox{0.6 \linewidth}{
      \renewcommand{\arraystretch}{1.5}
      \begin{equation*}
        \centering
        \begin{array}{ccc}
              \mathbf{a}_1 & = & \frac12 \, a \, \mathbf{\hat{x}} - \frac12 \, b \, \mathbf{\hat{y}} \\
    \mathbf{a}_2 & = & \frac12 \, a \, \mathbf{\hat{x}} + \frac12 \, b \, \mathbf{\hat{y}} \\
    \mathbf{a}_3 & = & c \, \mathbf{\hat{z}} \\

        \end{array}
      \end{equation*}
    }
    \renewcommand{\arraystretch}{1.0}
  \end{tabular}
  \begin{tabular}{c}
    \includegraphics[width=0.3\linewidth]{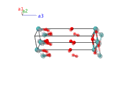} \\
  \end{tabular}
\end{tabular}

}
\vspace*{-0.25cm}

\noindent \hrulefill
\\
\textbf{Basis vectors:}
\vspace*{-0.25cm}
\renewcommand{\arraystretch}{1.5}
\begin{longtabu} to \textwidth{>{\centering $}X[-1,c,c]<{$}>{\centering $}X[-1,c,c]<{$}>{\centering $}X[-1,c,c]<{$}>{\centering $}X[-1,c,c]<{$}>{\centering $}X[-1,c,c]<{$}>{\centering $}X[-1,c,c]<{$}>{\centering $}X[-1,c,c]<{$}}
  & & \mbox{Lattice Coordinates} & & \mbox{Cartesian Coordinates} &\mbox{Wyckoff Position} & \mbox{Atom Type} \\  
  \mathbf{B}_{1} & = & z_{1} \, \mathbf{a}_{3} & = & z_{1}c \, \mathbf{\hat{z}} & \left(2a\right) & \mbox{Mo} \\ 
\mathbf{B}_{2} & = & z_{2} \, \mathbf{a}_{3} & = & z_{2}c \, \mathbf{\hat{z}} & \left(2a\right) & \mbox{O I} \\ 
\mathbf{B}_{3} & = & \frac{1}{2} \, \mathbf{a}_{1} + \frac{1}{2} \, \mathbf{a}_{2} + z_{3} \, \mathbf{a}_{3} & = & \frac{1}{2}a \, \mathbf{\hat{x}} + z_{3}c \, \mathbf{\hat{z}} & \left(2b\right) & \mbox{O II} \\ 
\mathbf{B}_{4} & = & -y_{4} \, \mathbf{a}_{1} + y_{4} \, \mathbf{a}_{2} + z_{4} \, \mathbf{a}_{3} & = & y_{4}b \, \mathbf{\hat{y}} + z_{4}c \, \mathbf{\hat{z}} & \left(4e\right) & \mbox{O III} \\ 
\mathbf{B}_{5} & = & y_{4} \, \mathbf{a}_{1}-y_{4} \, \mathbf{a}_{2} + z_{4} \, \mathbf{a}_{3} & = & -y_{4}b \, \mathbf{\hat{y}} + z_{4}c \, \mathbf{\hat{z}} & \left(4e\right) & \mbox{O III} \\ 
\mathbf{B}_{6} & = & -y_{5} \, \mathbf{a}_{1} + y_{5} \, \mathbf{a}_{2} + z_{5} \, \mathbf{a}_{3} & = & y_{5}b \, \mathbf{\hat{y}} + z_{5}c \, \mathbf{\hat{z}} & \left(4e\right) & \mbox{O IV} \\ 
\mathbf{B}_{7} & = & y_{5} \, \mathbf{a}_{1}-y_{5} \, \mathbf{a}_{2} + z_{5} \, \mathbf{a}_{3} & = & -y_{5}b \, \mathbf{\hat{y}} + z_{5}c \, \mathbf{\hat{z}} & \left(4e\right) & \mbox{O IV} \\ 
\mathbf{B}_{8} & = & -y_{6} \, \mathbf{a}_{1} + y_{6} \, \mathbf{a}_{2} + z_{6} \, \mathbf{a}_{3} & = & y_{6}b \, \mathbf{\hat{y}} + z_{6}c \, \mathbf{\hat{z}} & \left(4e\right) & \mbox{O V} \\ 
\mathbf{B}_{9} & = & y_{6} \, \mathbf{a}_{1}-y_{6} \, \mathbf{a}_{2} + z_{6} \, \mathbf{a}_{3} & = & -y_{6}b \, \mathbf{\hat{y}} + z_{6}c \, \mathbf{\hat{z}} & \left(4e\right) & \mbox{O V} \\ 
\mathbf{B}_{10} & = & -y_{7} \, \mathbf{a}_{1} + y_{7} \, \mathbf{a}_{2} + z_{7} \, \mathbf{a}_{3} & = & y_{7}b \, \mathbf{\hat{y}} + z_{7}c \, \mathbf{\hat{z}} & \left(4e\right) & \mbox{V} \\ 
\mathbf{B}_{11} & = & y_{7} \, \mathbf{a}_{1}-y_{7} \, \mathbf{a}_{2} + z_{7} \, \mathbf{a}_{3} & = & -y_{7}b \, \mathbf{\hat{y}} + z_{7}c \, \mathbf{\hat{z}} & \left(4e\right) & \mbox{V} \\ 
\end{longtabu}
\renewcommand{\arraystretch}{1.0}
\noindent \hrulefill
\\
\textbf{References:}
\vspace*{-0.25cm}
\begin{flushleft}
  - \bibentry{Mahe_V2MoO8_1970}. \\
\end{flushleft}
\textbf{Found in:}
\vspace*{-0.25cm}
\begin{flushleft}
  - \bibentry{Villars_PearsonsCrystalData_2013}. \\
\end{flushleft}
\noindent \hrulefill
\\
\textbf{Geometry files:}
\\
\noindent  - CIF: pp. {\hyperref[AB8C2_oC22_35_a_ab3e_e_cif]{\pageref{AB8C2_oC22_35_a_ab3e_e_cif}}} \\
\noindent  - POSCAR: pp. {\hyperref[AB8C2_oC22_35_a_ab3e_e_poscar]{\pageref{AB8C2_oC22_35_a_ab3e_e_poscar}}} \\
\onecolumn
{\phantomsection\label{AB_oC8_36_a_a}}
\subsection*{\huge \textbf{{\normalfont HCl Structure: AB\_oC8\_36\_a\_a}}}
\noindent \hrulefill
\vspace*{0.25cm}
\begin{figure}[htp]
  \centering
  \vspace{-1em}
  {\includegraphics[width=1\textwidth]{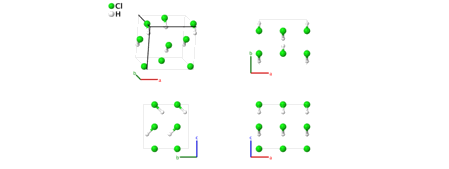}}
\end{figure}
\vspace*{-0.5cm}
\renewcommand{\arraystretch}{1.5}
\begin{equation*}
  \begin{array}{>{$\hspace{-0.15cm}}l<{$}>{$}p{0.5cm}<{$}>{$}p{18.5cm}<{$}}
    \mbox{\large \textbf{Prototype}} &\colon & \ce{HCl} \\
    \mbox{\large \textbf{\AFLOW\ prototype label}} &\colon & \mbox{AB\_oC8\_36\_a\_a} \\
    \mbox{\large \textbf{\textit{Strukturbericht} designation}} &\colon & \mbox{None} \\
    \mbox{\large \textbf{Pearson symbol}} &\colon & \mbox{oC8} \\
    \mbox{\large \textbf{Space group number}} &\colon & 36 \\
    \mbox{\large \textbf{Space group symbol}} &\colon & Cmc2_{1} \\
    \mbox{\large \textbf{\AFLOW\ prototype command}} &\colon &  \texttt{aflow} \,  \, \texttt{-{}-proto=AB\_oC8\_36\_a\_a } \, \newline \texttt{-{}-params=}{a,b/a,c/a,y_{1},z_{1},y_{2},z_{2} }
  \end{array}
\end{equation*}
\renewcommand{\arraystretch}{1.0}

\vspace*{-0.25cm}
\noindent \hrulefill
\begin{itemize}
  \item{The original reference gives the positions of the atoms in the
$Bb2_1m$ setting of space group \#36.  We have transformed this into
the standard $Cmc2_1$ setting.
}
\end{itemize}

\noindent \parbox{1 \linewidth}{
\noindent \hrulefill
\\
\textbf{Base-centered Orthorhombic primitive vectors:} \\
\vspace*{-0.25cm}
\begin{tabular}{cc}
  \begin{tabular}{c}
    \parbox{0.6 \linewidth}{
      \renewcommand{\arraystretch}{1.5}
      \begin{equation*}
        \centering
        \begin{array}{ccc}
              \mathbf{a}_1 & = & \frac12 \, a \, \mathbf{\hat{x}} - \frac12 \, b \, \mathbf{\hat{y}} \\
    \mathbf{a}_2 & = & \frac12 \, a \, \mathbf{\hat{x}} + \frac12 \, b \, \mathbf{\hat{y}} \\
    \mathbf{a}_3 & = & c \, \mathbf{\hat{z}} \\

        \end{array}
      \end{equation*}
    }
    \renewcommand{\arraystretch}{1.0}
  \end{tabular}
  \begin{tabular}{c}
    \includegraphics[width=0.3\linewidth]{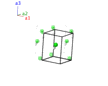} \\
  \end{tabular}
\end{tabular}

}
\vspace*{-0.25cm}

\noindent \hrulefill
\\
\textbf{Basis vectors:}
\vspace*{-0.25cm}
\renewcommand{\arraystretch}{1.5}
\begin{longtabu} to \textwidth{>{\centering $}X[-1,c,c]<{$}>{\centering $}X[-1,c,c]<{$}>{\centering $}X[-1,c,c]<{$}>{\centering $}X[-1,c,c]<{$}>{\centering $}X[-1,c,c]<{$}>{\centering $}X[-1,c,c]<{$}>{\centering $}X[-1,c,c]<{$}}
  & & \mbox{Lattice Coordinates} & & \mbox{Cartesian Coordinates} &\mbox{Wyckoff Position} & \mbox{Atom Type} \\  
  \mathbf{B}_{1} & = & -y_{1} \, \mathbf{a}_{1} + y_{1} \, \mathbf{a}_{2} + z_{1} \, \mathbf{a}_{3} & = & y_{1}b \, \mathbf{\hat{y}} + z_{1}c \, \mathbf{\hat{z}} & \left(4a\right) & \mbox{Cl} \\ 
\mathbf{B}_{2} & = & y_{1} \, \mathbf{a}_{1}-y_{1} \, \mathbf{a}_{2} + \left(\frac{1}{2} +z_{1}\right) \, \mathbf{a}_{3} & = & -y_{1}b \, \mathbf{\hat{y}} + \left(\frac{1}{2} +z_{1}\right)c \, \mathbf{\hat{z}} & \left(4a\right) & \mbox{Cl} \\ 
\mathbf{B}_{3} & = & -y_{2} \, \mathbf{a}_{1} + y_{2} \, \mathbf{a}_{2} + z_{2} \, \mathbf{a}_{3} & = & y_{2}b \, \mathbf{\hat{y}} + z_{2}c \, \mathbf{\hat{z}} & \left(4a\right) & \mbox{H} \\ 
\mathbf{B}_{4} & = & y_{2} \, \mathbf{a}_{1}-y_{2} \, \mathbf{a}_{2} + \left(\frac{1}{2} +z_{2}\right) \, \mathbf{a}_{3} & = & -y_{2}b \, \mathbf{\hat{y}} + \left(\frac{1}{2} +z_{2}\right)c \, \mathbf{\hat{z}} & \left(4a\right) & \mbox{H} \\ 
\end{longtabu}
\renewcommand{\arraystretch}{1.0}
\noindent \hrulefill
\\
\textbf{References:}
\vspace*{-0.25cm}
\begin{flushleft}
  - \bibentry{Sandor_Nature_213_1967}. \\
\end{flushleft}
\noindent \hrulefill
\\
\textbf{Geometry files:}
\\
\noindent  - CIF: pp. {\hyperref[AB_oC8_36_a_a_cif]{\pageref{AB_oC8_36_a_a_cif}}} \\
\noindent  - POSCAR: pp. {\hyperref[AB_oC8_36_a_a_poscar]{\pageref{AB_oC8_36_a_a_poscar}}} \\
\onecolumn
{\phantomsection\label{A2B5C2_oC36_37_d_c2d_d}}
\subsection*{\huge \textbf{{\normalfont Li$_{2}$Si$_{2}$O$_{5}$ Structure: A2B5C2\_oC36\_37\_d\_c2d\_d}}}
\noindent \hrulefill
\vspace*{0.25cm}
\begin{figure}[htp]
  \centering
  \vspace{-1em}
  {\includegraphics[width=1\textwidth]{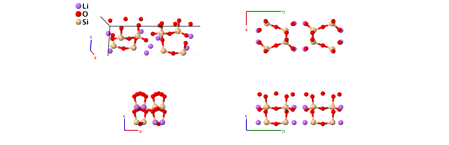}}
\end{figure}
\vspace*{-0.5cm}
\renewcommand{\arraystretch}{1.5}
\begin{equation*}
  \begin{array}{>{$\hspace{-0.15cm}}l<{$}>{$}p{0.5cm}<{$}>{$}p{18.5cm}<{$}}
    \mbox{\large \textbf{Prototype}} &\colon & \ce{Li2Si2O5} \\
    \mbox{\large \textbf{\AFLOW\ prototype label}} &\colon & \mbox{A2B5C2\_oC36\_37\_d\_c2d\_d} \\
    \mbox{\large \textbf{\textit{Strukturbericht} designation}} &\colon & \mbox{None} \\
    \mbox{\large \textbf{Pearson symbol}} &\colon & \mbox{oC36} \\
    \mbox{\large \textbf{Space group number}} &\colon & 37 \\
    \mbox{\large \textbf{Space group symbol}} &\colon & Ccc2 \\
    \mbox{\large \textbf{\AFLOW\ prototype command}} &\colon &  \texttt{aflow} \,  \, \texttt{-{}-proto=A2B5C2\_oC36\_37\_d\_c2d\_d } \, \newline \texttt{-{}-params=}{a,b/a,c/a,z_{1},x_{2},y_{2},z_{2},x_{3},y_{3},z_{3},x_{4},y_{4},z_{4},x_{5},y_{5},z_{5} }
  \end{array}
\end{equation*}
\renewcommand{\arraystretch}{1.0}

\noindent \parbox{1 \linewidth}{
\noindent \hrulefill
\\
\textbf{Base-centered Orthorhombic primitive vectors:} \\
\vspace*{-0.25cm}
\begin{tabular}{cc}
  \begin{tabular}{c}
    \parbox{0.6 \linewidth}{
      \renewcommand{\arraystretch}{1.5}
      \begin{equation*}
        \centering
        \begin{array}{ccc}
              \mathbf{a}_1 & = & \frac12 \, a \, \mathbf{\hat{x}} - \frac12 \, b \, \mathbf{\hat{y}} \\
    \mathbf{a}_2 & = & \frac12 \, a \, \mathbf{\hat{x}} + \frac12 \, b \, \mathbf{\hat{y}} \\
    \mathbf{a}_3 & = & c \, \mathbf{\hat{z}} \\

        \end{array}
      \end{equation*}
    }
    \renewcommand{\arraystretch}{1.0}
  \end{tabular}
  \begin{tabular}{c}
    \includegraphics[width=0.3\linewidth]{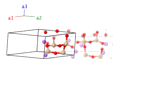} \\
  \end{tabular}
\end{tabular}

}
\vspace*{-0.25cm}

\noindent \hrulefill
\\
\textbf{Basis vectors:}
\vspace*{-0.25cm}
\renewcommand{\arraystretch}{1.5}
\begin{longtabu} to \textwidth{>{\centering $}X[-1,c,c]<{$}>{\centering $}X[-1,c,c]<{$}>{\centering $}X[-1,c,c]<{$}>{\centering $}X[-1,c,c]<{$}>{\centering $}X[-1,c,c]<{$}>{\centering $}X[-1,c,c]<{$}>{\centering $}X[-1,c,c]<{$}}
  & & \mbox{Lattice Coordinates} & & \mbox{Cartesian Coordinates} &\mbox{Wyckoff Position} & \mbox{Atom Type} \\  
  \mathbf{B}_{1} & = & \frac{1}{2} \, \mathbf{a}_{2} + z_{1} \, \mathbf{a}_{3} & = & \frac{1}{4}a \, \mathbf{\hat{x}} + \frac{1}{4}b \, \mathbf{\hat{y}} + z_{1}c \, \mathbf{\hat{z}} & \left(4c\right) & \mbox{O I} \\ 
\mathbf{B}_{2} & = & \frac{1}{2} \, \mathbf{a}_{1} + \left(\frac{1}{2} +z_{1}\right) \, \mathbf{a}_{3} & = & \frac{1}{4}a \, \mathbf{\hat{x}} + \frac{3}{4}b \, \mathbf{\hat{y}} + \left(\frac{1}{2} +z_{1}\right)c \, \mathbf{\hat{z}} & \left(4c\right) & \mbox{O I} \\ 
\mathbf{B}_{3} & = & \left(x_{2}-y_{2}\right) \, \mathbf{a}_{1} + \left(x_{2}+y_{2}\right) \, \mathbf{a}_{2} + z_{2} \, \mathbf{a}_{3} & = & x_{2}a \, \mathbf{\hat{x}} + y_{2}b \, \mathbf{\hat{y}} + z_{2}c \, \mathbf{\hat{z}} & \left(8d\right) & \mbox{Li} \\ 
\mathbf{B}_{4} & = & \left(-x_{2}+y_{2}\right) \, \mathbf{a}_{1} + \left(-x_{2}-y_{2}\right) \, \mathbf{a}_{2} + z_{2} \, \mathbf{a}_{3} & = & -x_{2}a \, \mathbf{\hat{x}}-y_{2}b \, \mathbf{\hat{y}} + z_{2}c \, \mathbf{\hat{z}} & \left(8d\right) & \mbox{Li} \\ 
\mathbf{B}_{5} & = & \left(x_{2}+y_{2}\right) \, \mathbf{a}_{1} + \left(x_{2}-y_{2}\right) \, \mathbf{a}_{2} + \left(\frac{1}{2} +z_{2}\right) \, \mathbf{a}_{3} & = & x_{2}a \, \mathbf{\hat{x}}-y_{2}b \, \mathbf{\hat{y}} + \left(\frac{1}{2} +z_{2}\right)c \, \mathbf{\hat{z}} & \left(8d\right) & \mbox{Li} \\ 
\mathbf{B}_{6} & = & \left(-x_{2}-y_{2}\right) \, \mathbf{a}_{1} + \left(-x_{2}+y_{2}\right) \, \mathbf{a}_{2} + \left(\frac{1}{2} +z_{2}\right) \, \mathbf{a}_{3} & = & -x_{2}a \, \mathbf{\hat{x}} + y_{2}b \, \mathbf{\hat{y}} + \left(\frac{1}{2} +z_{2}\right)c \, \mathbf{\hat{z}} & \left(8d\right) & \mbox{Li} \\ 
\mathbf{B}_{7} & = & \left(x_{3}-y_{3}\right) \, \mathbf{a}_{1} + \left(x_{3}+y_{3}\right) \, \mathbf{a}_{2} + z_{3} \, \mathbf{a}_{3} & = & x_{3}a \, \mathbf{\hat{x}} + y_{3}b \, \mathbf{\hat{y}} + z_{3}c \, \mathbf{\hat{z}} & \left(8d\right) & \mbox{O II} \\ 
\mathbf{B}_{8} & = & \left(-x_{3}+y_{3}\right) \, \mathbf{a}_{1} + \left(-x_{3}-y_{3}\right) \, \mathbf{a}_{2} + z_{3} \, \mathbf{a}_{3} & = & -x_{3}a \, \mathbf{\hat{x}}-y_{3}b \, \mathbf{\hat{y}} + z_{3}c \, \mathbf{\hat{z}} & \left(8d\right) & \mbox{O II} \\ 
\mathbf{B}_{9} & = & \left(x_{3}+y_{3}\right) \, \mathbf{a}_{1} + \left(x_{3}-y_{3}\right) \, \mathbf{a}_{2} + \left(\frac{1}{2} +z_{3}\right) \, \mathbf{a}_{3} & = & x_{3}a \, \mathbf{\hat{x}}-y_{3}b \, \mathbf{\hat{y}} + \left(\frac{1}{2} +z_{3}\right)c \, \mathbf{\hat{z}} & \left(8d\right) & \mbox{O II} \\ 
\mathbf{B}_{10} & = & \left(-x_{3}-y_{3}\right) \, \mathbf{a}_{1} + \left(-x_{3}+y_{3}\right) \, \mathbf{a}_{2} + \left(\frac{1}{2} +z_{3}\right) \, \mathbf{a}_{3} & = & -x_{3}a \, \mathbf{\hat{x}} + y_{3}b \, \mathbf{\hat{y}} + \left(\frac{1}{2} +z_{3}\right)c \, \mathbf{\hat{z}} & \left(8d\right) & \mbox{O II} \\ 
\mathbf{B}_{11} & = & \left(x_{4}-y_{4}\right) \, \mathbf{a}_{1} + \left(x_{4}+y_{4}\right) \, \mathbf{a}_{2} + z_{4} \, \mathbf{a}_{3} & = & x_{4}a \, \mathbf{\hat{x}} + y_{4}b \, \mathbf{\hat{y}} + z_{4}c \, \mathbf{\hat{z}} & \left(8d\right) & \mbox{O III} \\ 
\mathbf{B}_{12} & = & \left(-x_{4}+y_{4}\right) \, \mathbf{a}_{1} + \left(-x_{4}-y_{4}\right) \, \mathbf{a}_{2} + z_{4} \, \mathbf{a}_{3} & = & -x_{4}a \, \mathbf{\hat{x}}-y_{4}b \, \mathbf{\hat{y}} + z_{4}c \, \mathbf{\hat{z}} & \left(8d\right) & \mbox{O III} \\ 
\mathbf{B}_{13} & = & \left(x_{4}+y_{4}\right) \, \mathbf{a}_{1} + \left(x_{4}-y_{4}\right) \, \mathbf{a}_{2} + \left(\frac{1}{2} +z_{4}\right) \, \mathbf{a}_{3} & = & x_{4}a \, \mathbf{\hat{x}}-y_{4}b \, \mathbf{\hat{y}} + \left(\frac{1}{2} +z_{4}\right)c \, \mathbf{\hat{z}} & \left(8d\right) & \mbox{O III} \\ 
\mathbf{B}_{14} & = & \left(-x_{4}-y_{4}\right) \, \mathbf{a}_{1} + \left(-x_{4}+y_{4}\right) \, \mathbf{a}_{2} + \left(\frac{1}{2} +z_{4}\right) \, \mathbf{a}_{3} & = & -x_{4}a \, \mathbf{\hat{x}} + y_{4}b \, \mathbf{\hat{y}} + \left(\frac{1}{2} +z_{4}\right)c \, \mathbf{\hat{z}} & \left(8d\right) & \mbox{O III} \\ 
\mathbf{B}_{15} & = & \left(x_{5}-y_{5}\right) \, \mathbf{a}_{1} + \left(x_{5}+y_{5}\right) \, \mathbf{a}_{2} + z_{5} \, \mathbf{a}_{3} & = & x_{5}a \, \mathbf{\hat{x}} + y_{5}b \, \mathbf{\hat{y}} + z_{5}c \, \mathbf{\hat{z}} & \left(8d\right) & \mbox{Si} \\ 
\mathbf{B}_{16} & = & \left(-x_{5}+y_{5}\right) \, \mathbf{a}_{1} + \left(-x_{5}-y_{5}\right) \, \mathbf{a}_{2} + z_{5} \, \mathbf{a}_{3} & = & -x_{5}a \, \mathbf{\hat{x}}-y_{5}b \, \mathbf{\hat{y}} + z_{5}c \, \mathbf{\hat{z}} & \left(8d\right) & \mbox{Si} \\ 
\mathbf{B}_{17} & = & \left(x_{5}+y_{5}\right) \, \mathbf{a}_{1} + \left(x_{5}-y_{5}\right) \, \mathbf{a}_{2} + \left(\frac{1}{2} +z_{5}\right) \, \mathbf{a}_{3} & = & x_{5}a \, \mathbf{\hat{x}}-y_{5}b \, \mathbf{\hat{y}} + \left(\frac{1}{2} +z_{5}\right)c \, \mathbf{\hat{z}} & \left(8d\right) & \mbox{Si} \\ 
\mathbf{B}_{18} & = & \left(-x_{5}-y_{5}\right) \, \mathbf{a}_{1} + \left(-x_{5}+y_{5}\right) \, \mathbf{a}_{2} + \left(\frac{1}{2} +z_{5}\right) \, \mathbf{a}_{3} & = & -x_{5}a \, \mathbf{\hat{x}} + y_{5}b \, \mathbf{\hat{y}} + \left(\frac{1}{2} +z_{5}\right)c \, \mathbf{\hat{z}} & \left(8d\right) & \mbox{Si} \\ 
\end{longtabu}
\renewcommand{\arraystretch}{1.0}
\noindent \hrulefill
\\
\textbf{References:}
\vspace*{-0.25cm}
\begin{flushleft}
  - \bibentry{DeJong_Li2Si2O5_JNonCrystSol_1994}. \\
\end{flushleft}
\textbf{Found in:}
\vspace*{-0.25cm}
\begin{flushleft}
  - \bibentry{Villars_PearsonsCrystalData_2013}. \\
\end{flushleft}
\noindent \hrulefill
\\
\textbf{Geometry files:}
\\
\noindent  - CIF: pp. {\hyperref[A2B5C2_oC36_37_d_c2d_d_cif]{\pageref{A2B5C2_oC36_37_d_c2d_d_cif}}} \\
\noindent  - POSCAR: pp. {\hyperref[A2B5C2_oC36_37_d_c2d_d_poscar]{\pageref{A2B5C2_oC36_37_d_c2d_d_poscar}}} \\
\onecolumn
{\phantomsection\label{A2B3_oC40_39_2d_2c2d}}
\subsection*{\huge \textbf{{\normalfont Ta$_{3}$S$_{2}$ Structure: A2B3\_oC40\_39\_2d\_2c2d}}}
\noindent \hrulefill
\vspace*{0.25cm}
\begin{figure}[htp]
  \centering
  \vspace{-1em}
  {\includegraphics[width=1\textwidth]{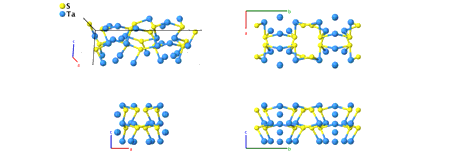}}
\end{figure}
\vspace*{-0.5cm}
\renewcommand{\arraystretch}{1.5}
\begin{equation*}
  \begin{array}{>{$\hspace{-0.15cm}}l<{$}>{$}p{0.5cm}<{$}>{$}p{18.5cm}<{$}}
    \mbox{\large \textbf{Prototype}} &\colon & \ce{Ta3S2} \\
    \mbox{\large \textbf{\AFLOW\ prototype label}} &\colon & \mbox{A2B3\_oC40\_39\_2d\_2c2d} \\
    \mbox{\large \textbf{\textit{Strukturbericht} designation}} &\colon & \mbox{None} \\
    \mbox{\large \textbf{Pearson symbol}} &\colon & \mbox{oC40} \\
    \mbox{\large \textbf{Space group number}} &\colon & 39 \\
    \mbox{\large \textbf{Space group symbol}} &\colon & Abm2 \\
    \mbox{\large \textbf{\AFLOW\ prototype command}} &\colon &  \texttt{aflow} \,  \, \texttt{-{}-proto=A2B3\_oC40\_39\_2d\_2c2d } \, \newline \texttt{-{}-params=}{a,b/a,c/a,x_{1},z_{1},x_{2},z_{2},x_{3},y_{3},z_{3},x_{4},y_{4},z_{4},x_{5},y_{5},z_{5},x_{6},y_{6},z_{6} }
  \end{array}
\end{equation*}
\renewcommand{\arraystretch}{1.0}

\noindent \parbox{1 \linewidth}{
\noindent \hrulefill
\\
\textbf{Base-centered Orthorhombic primitive vectors:} \\
\vspace*{-0.25cm}
\begin{tabular}{cc}
  \begin{tabular}{c}
    \parbox{0.6 \linewidth}{
      \renewcommand{\arraystretch}{1.5}
      \begin{equation*}
        \centering
        \begin{array}{ccc}
              \mathbf{a}_1 & = & a \, \mathbf{\hat{x}} \\
    \mathbf{a}_2 & = & \frac12 \, b \, \mathbf{\hat{y}} - \frac12 \, c \, \mathbf{\hat{z}} \\
    \mathbf{a}_3 & = & \frac12 \, b \, \mathbf{\hat{y}} + \frac12 \, c \, \mathbf{\hat{z}} \\

        \end{array}
      \end{equation*}
    }
    \renewcommand{\arraystretch}{1.0}
  \end{tabular}
  \begin{tabular}{c}
    \includegraphics[width=0.3\linewidth]{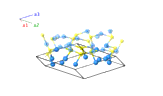} \\
  \end{tabular}
\end{tabular}

}
\vspace*{-0.25cm}

\noindent \hrulefill
\\
\textbf{Basis vectors:}
\vspace*{-0.25cm}
\renewcommand{\arraystretch}{1.5}
\begin{longtabu} to \textwidth{>{\centering $}X[-1,c,c]<{$}>{\centering $}X[-1,c,c]<{$}>{\centering $}X[-1,c,c]<{$}>{\centering $}X[-1,c,c]<{$}>{\centering $}X[-1,c,c]<{$}>{\centering $}X[-1,c,c]<{$}>{\centering $}X[-1,c,c]<{$}}
  & & \mbox{Lattice Coordinates} & & \mbox{Cartesian Coordinates} &\mbox{Wyckoff Position} & \mbox{Atom Type} \\  
  \mathbf{B}_{1} & = & x_{1} \, \mathbf{a}_{1} + \left(\frac{1}{4} - z_{1}\right) \, \mathbf{a}_{2} + \left(\frac{1}{4} +z_{1}\right) \, \mathbf{a}_{3} & = & x_{1}a \, \mathbf{\hat{x}} + \frac{1}{4}b \, \mathbf{\hat{y}} + z_{1}c \, \mathbf{\hat{z}} & \left(4c\right) & \mbox{Ta I} \\ 
\mathbf{B}_{2} & = & -x_{1} \, \mathbf{a}_{1} + \left(\frac{3}{4} - z_{1}\right) \, \mathbf{a}_{2} + \left(\frac{3}{4} +z_{1}\right) \, \mathbf{a}_{3} & = & -x_{1}a \, \mathbf{\hat{x}} + \frac{3}{4}b \, \mathbf{\hat{y}} + z_{1}c \, \mathbf{\hat{z}} & \left(4c\right) & \mbox{Ta I} \\ 
\mathbf{B}_{3} & = & x_{2} \, \mathbf{a}_{1} + \left(\frac{1}{4} - z_{2}\right) \, \mathbf{a}_{2} + \left(\frac{1}{4} +z_{2}\right) \, \mathbf{a}_{3} & = & x_{2}a \, \mathbf{\hat{x}} + \frac{1}{4}b \, \mathbf{\hat{y}} + z_{2}c \, \mathbf{\hat{z}} & \left(4c\right) & \mbox{Ta II} \\ 
\mathbf{B}_{4} & = & -x_{2} \, \mathbf{a}_{1} + \left(\frac{3}{4} - z_{2}\right) \, \mathbf{a}_{2} + \left(\frac{3}{4} +z_{2}\right) \, \mathbf{a}_{3} & = & -x_{2}a \, \mathbf{\hat{x}} + \frac{3}{4}b \, \mathbf{\hat{y}} + z_{2}c \, \mathbf{\hat{z}} & \left(4c\right) & \mbox{Ta II} \\ 
\mathbf{B}_{5} & = & x_{3} \, \mathbf{a}_{1} + \left(y_{3}-z_{3}\right) \, \mathbf{a}_{2} + \left(y_{3}+z_{3}\right) \, \mathbf{a}_{3} & = & x_{3}a \, \mathbf{\hat{x}} + y_{3}b \, \mathbf{\hat{y}} + z_{3}c \, \mathbf{\hat{z}} & \left(8d\right) & \mbox{S I} \\ 
\mathbf{B}_{6} & = & -x_{3} \, \mathbf{a}_{1} + \left(-y_{3}-z_{3}\right) \, \mathbf{a}_{2} + \left(-y_{3}+z_{3}\right) \, \mathbf{a}_{3} & = & -x_{3}a \, \mathbf{\hat{x}}-y_{3}b \, \mathbf{\hat{y}} + z_{3}c \, \mathbf{\hat{z}} & \left(8d\right) & \mbox{S I} \\ 
\mathbf{B}_{7} & = & x_{3} \, \mathbf{a}_{1} + \left(\frac{1}{2} - y_{3} - z_{3}\right) \, \mathbf{a}_{2} + \left(\frac{1}{2} - y_{3} + z_{3}\right) \, \mathbf{a}_{3} & = & x_{3}a \, \mathbf{\hat{x}} + \left(\frac{1}{2} - y_{3}\right)b \, \mathbf{\hat{y}} + z_{3}c \, \mathbf{\hat{z}} & \left(8d\right) & \mbox{S I} \\ 
\mathbf{B}_{8} & = & -x_{3} \, \mathbf{a}_{1} + \left(\frac{1}{2} +y_{3} - z_{3}\right) \, \mathbf{a}_{2} + \left(\frac{1}{2} +y_{3} + z_{3}\right) \, \mathbf{a}_{3} & = & -x_{3}a \, \mathbf{\hat{x}} + \left(\frac{1}{2} +y_{3}\right)b \, \mathbf{\hat{y}} + z_{3}c \, \mathbf{\hat{z}} & \left(8d\right) & \mbox{S I} \\ 
\mathbf{B}_{9} & = & x_{4} \, \mathbf{a}_{1} + \left(y_{4}-z_{4}\right) \, \mathbf{a}_{2} + \left(y_{4}+z_{4}\right) \, \mathbf{a}_{3} & = & x_{4}a \, \mathbf{\hat{x}} + y_{4}b \, \mathbf{\hat{y}} + z_{4}c \, \mathbf{\hat{z}} & \left(8d\right) & \mbox{S II} \\ 
\mathbf{B}_{10} & = & -x_{4} \, \mathbf{a}_{1} + \left(-y_{4}-z_{4}\right) \, \mathbf{a}_{2} + \left(-y_{4}+z_{4}\right) \, \mathbf{a}_{3} & = & -x_{4}a \, \mathbf{\hat{x}}-y_{4}b \, \mathbf{\hat{y}} + z_{4}c \, \mathbf{\hat{z}} & \left(8d\right) & \mbox{S II} \\ 
\mathbf{B}_{11} & = & x_{4} \, \mathbf{a}_{1} + \left(\frac{1}{2} - y_{4} - z_{4}\right) \, \mathbf{a}_{2} + \left(\frac{1}{2} - y_{4} + z_{4}\right) \, \mathbf{a}_{3} & = & x_{4}a \, \mathbf{\hat{x}} + \left(\frac{1}{2} - y_{4}\right)b \, \mathbf{\hat{y}} + z_{4}c \, \mathbf{\hat{z}} & \left(8d\right) & \mbox{S II} \\ 
\mathbf{B}_{12} & = & -x_{4} \, \mathbf{a}_{1} + \left(\frac{1}{2} +y_{4} - z_{4}\right) \, \mathbf{a}_{2} + \left(\frac{1}{2} +y_{4} + z_{4}\right) \, \mathbf{a}_{3} & = & -x_{4}a \, \mathbf{\hat{x}} + \left(\frac{1}{2} +y_{4}\right)b \, \mathbf{\hat{y}} + z_{4}c \, \mathbf{\hat{z}} & \left(8d\right) & \mbox{S II} \\ 
\mathbf{B}_{13} & = & x_{5} \, \mathbf{a}_{1} + \left(y_{5}-z_{5}\right) \, \mathbf{a}_{2} + \left(y_{5}+z_{5}\right) \, \mathbf{a}_{3} & = & x_{5}a \, \mathbf{\hat{x}} + y_{5}b \, \mathbf{\hat{y}} + z_{5}c \, \mathbf{\hat{z}} & \left(8d\right) & \mbox{Ta III} \\ 
\mathbf{B}_{14} & = & -x_{5} \, \mathbf{a}_{1} + \left(-y_{5}-z_{5}\right) \, \mathbf{a}_{2} + \left(-y_{5}+z_{5}\right) \, \mathbf{a}_{3} & = & -x_{5}a \, \mathbf{\hat{x}}-y_{5}b \, \mathbf{\hat{y}} + z_{5}c \, \mathbf{\hat{z}} & \left(8d\right) & \mbox{Ta III} \\ 
\mathbf{B}_{15} & = & x_{5} \, \mathbf{a}_{1} + \left(\frac{1}{2} - y_{5} - z_{5}\right) \, \mathbf{a}_{2} + \left(\frac{1}{2} - y_{5} + z_{5}\right) \, \mathbf{a}_{3} & = & x_{5}a \, \mathbf{\hat{x}} + \left(\frac{1}{2} - y_{5}\right)b \, \mathbf{\hat{y}} + z_{5}c \, \mathbf{\hat{z}} & \left(8d\right) & \mbox{Ta III} \\ 
\mathbf{B}_{16} & = & -x_{5} \, \mathbf{a}_{1} + \left(\frac{1}{2} +y_{5} - z_{5}\right) \, \mathbf{a}_{2} + \left(\frac{1}{2} +y_{5} + z_{5}\right) \, \mathbf{a}_{3} & = & -x_{5}a \, \mathbf{\hat{x}} + \left(\frac{1}{2} +y_{5}\right)b \, \mathbf{\hat{y}} + z_{5}c \, \mathbf{\hat{z}} & \left(8d\right) & \mbox{Ta III} \\ 
\mathbf{B}_{17} & = & x_{6} \, \mathbf{a}_{1} + \left(y_{6}-z_{6}\right) \, \mathbf{a}_{2} + \left(y_{6}+z_{6}\right) \, \mathbf{a}_{3} & = & x_{6}a \, \mathbf{\hat{x}} + y_{6}b \, \mathbf{\hat{y}} + z_{6}c \, \mathbf{\hat{z}} & \left(8d\right) & \mbox{Ta IV} \\ 
\mathbf{B}_{18} & = & -x_{6} \, \mathbf{a}_{1} + \left(-y_{6}-z_{6}\right) \, \mathbf{a}_{2} + \left(-y_{6}+z_{6}\right) \, \mathbf{a}_{3} & = & -x_{6}a \, \mathbf{\hat{x}}-y_{6}b \, \mathbf{\hat{y}} + z_{6}c \, \mathbf{\hat{z}} & \left(8d\right) & \mbox{Ta IV} \\ 
\mathbf{B}_{19} & = & x_{6} \, \mathbf{a}_{1} + \left(\frac{1}{2} - y_{6} - z_{6}\right) \, \mathbf{a}_{2} + \left(\frac{1}{2} - y_{6} + z_{6}\right) \, \mathbf{a}_{3} & = & x_{6}a \, \mathbf{\hat{x}} + \left(\frac{1}{2} - y_{6}\right)b \, \mathbf{\hat{y}} + z_{6}c \, \mathbf{\hat{z}} & \left(8d\right) & \mbox{Ta IV} \\ 
\mathbf{B}_{20} & = & -x_{6} \, \mathbf{a}_{1} + \left(\frac{1}{2} +y_{6} - z_{6}\right) \, \mathbf{a}_{2} + \left(\frac{1}{2} +y_{6} + z_{6}\right) \, \mathbf{a}_{3} & = & -x_{6}a \, \mathbf{\hat{x}} + \left(\frac{1}{2} +y_{6}\right)b \, \mathbf{\hat{y}} + z_{6}c \, \mathbf{\hat{z}} & \left(8d\right) & \mbox{Ta IV} \\ 
\end{longtabu}
\renewcommand{\arraystretch}{1.0}
\noindent \hrulefill
\\
\textbf{References:}
\vspace*{-0.25cm}
\begin{flushleft}
  - \bibentry{Kim_Ta3S2_InorgChem_1991}. \\
\end{flushleft}
\textbf{Found in:}
\vspace*{-0.25cm}
\begin{flushleft}
  - \bibentry{Villars_PearsonsCrystalData_2013}. \\
\end{flushleft}
\noindent \hrulefill
\\
\textbf{Geometry files:}
\\
\noindent  - CIF: pp. {\hyperref[A2B3_oC40_39_2d_2c2d_cif]{\pageref{A2B3_oC40_39_2d_2c2d_cif}}} \\
\noindent  - POSCAR: pp. {\hyperref[A2B3_oC40_39_2d_2c2d_poscar]{\pageref{A2B3_oC40_39_2d_2c2d_poscar}}} \\
\onecolumn
{\phantomsection\label{A9BC_oC44_39_3c3d_a_c}}
\subsection*{\huge \textbf{{\normalfont VPCl$_{9}$ Structure: A9BC\_oC44\_39\_3c3d\_a\_c}}}
\noindent \hrulefill
\vspace*{0.25cm}
\begin{figure}[htp]
  \centering
  \vspace{-1em}
  {\includegraphics[width=1\textwidth]{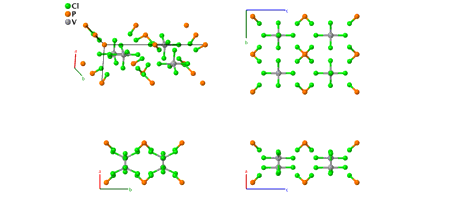}}
\end{figure}
\vspace*{-0.5cm}
\renewcommand{\arraystretch}{1.5}
\begin{equation*}
  \begin{array}{>{$\hspace{-0.15cm}}l<{$}>{$}p{0.5cm}<{$}>{$}p{18.5cm}<{$}}
    \mbox{\large \textbf{Prototype}} &\colon & \ce{VPCl9} \\
    \mbox{\large \textbf{\AFLOW\ prototype label}} &\colon & \mbox{A9BC\_oC44\_39\_3c3d\_a\_c} \\
    \mbox{\large \textbf{\textit{Strukturbericht} designation}} &\colon & \mbox{None} \\
    \mbox{\large \textbf{Pearson symbol}} &\colon & \mbox{oC44} \\
    \mbox{\large \textbf{Space group number}} &\colon & 39 \\
    \mbox{\large \textbf{Space group symbol}} &\colon & Abm2 \\
    \mbox{\large \textbf{\AFLOW\ prototype command}} &\colon &  \texttt{aflow} \,  \, \texttt{-{}-proto=A9BC\_oC44\_39\_3c3d\_a\_c } \, \newline \texttt{-{}-params=}{a,b/a,c/a,z_{1},x_{2},z_{2},x_{3},z_{3},x_{4},z_{4},x_{5},z_{5},x_{6},y_{6},z_{6},x_{7},y_{7},z_{7},x_{8},y_{8},} \newline {z_{8} }
  \end{array}
\end{equation*}
\renewcommand{\arraystretch}{1.0}

\noindent \parbox{1 \linewidth}{
\noindent \hrulefill
\\
\textbf{Base-centered Orthorhombic primitive vectors:} \\
\vspace*{-0.25cm}
\begin{tabular}{cc}
  \begin{tabular}{c}
    \parbox{0.6 \linewidth}{
      \renewcommand{\arraystretch}{1.5}
      \begin{equation*}
        \centering
        \begin{array}{ccc}
              \mathbf{a}_1 & = & a \, \mathbf{\hat{x}} \\
    \mathbf{a}_2 & = & \frac12 \, b \, \mathbf{\hat{y}} - \frac12 \, c \, \mathbf{\hat{z}} \\
    \mathbf{a}_3 & = & \frac12 \, b \, \mathbf{\hat{y}} + \frac12 \, c \, \mathbf{\hat{z}} \\

        \end{array}
      \end{equation*}
    }
    \renewcommand{\arraystretch}{1.0}
  \end{tabular}
  \begin{tabular}{c}
    \includegraphics[width=0.3\linewidth]{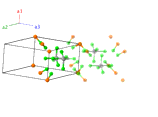} \\
  \end{tabular}
\end{tabular}

}
\vspace*{-0.25cm}

\noindent \hrulefill
\\
\textbf{Basis vectors:}
\vspace*{-0.25cm}
\renewcommand{\arraystretch}{1.5}
\begin{longtabu} to \textwidth{>{\centering $}X[-1,c,c]<{$}>{\centering $}X[-1,c,c]<{$}>{\centering $}X[-1,c,c]<{$}>{\centering $}X[-1,c,c]<{$}>{\centering $}X[-1,c,c]<{$}>{\centering $}X[-1,c,c]<{$}>{\centering $}X[-1,c,c]<{$}}
  & & \mbox{Lattice Coordinates} & & \mbox{Cartesian Coordinates} &\mbox{Wyckoff Position} & \mbox{Atom Type} \\  
  \mathbf{B}_{1} & = & -z_{1} \, \mathbf{a}_{2} + z_{1} \, \mathbf{a}_{3} & = & z_{1}c \, \mathbf{\hat{z}} & \left(4a\right) & \mbox{P} \\ 
\mathbf{B}_{2} & = & \left(\frac{1}{2} - z_{1}\right) \, \mathbf{a}_{2} + \left(\frac{1}{2} +z_{1}\right) \, \mathbf{a}_{3} & = & \frac{1}{2}b \, \mathbf{\hat{y}} + z_{1}c \, \mathbf{\hat{z}} & \left(4a\right) & \mbox{P} \\ 
\mathbf{B}_{3} & = & x_{2} \, \mathbf{a}_{1} + \left(\frac{1}{4} - z_{2}\right) \, \mathbf{a}_{2} + \left(\frac{1}{4} +z_{2}\right) \, \mathbf{a}_{3} & = & x_{2}a \, \mathbf{\hat{x}} + \frac{1}{4}b \, \mathbf{\hat{y}} + z_{2}c \, \mathbf{\hat{z}} & \left(4c\right) & \mbox{Cl I} \\ 
\mathbf{B}_{4} & = & -x_{2} \, \mathbf{a}_{1} + \left(\frac{3}{4} - z_{2}\right) \, \mathbf{a}_{2} + \left(\frac{3}{4} +z_{2}\right) \, \mathbf{a}_{3} & = & -x_{2}a \, \mathbf{\hat{x}} + \frac{3}{4}b \, \mathbf{\hat{y}} + z_{2}c \, \mathbf{\hat{z}} & \left(4c\right) & \mbox{Cl I} \\ 
\mathbf{B}_{5} & = & x_{3} \, \mathbf{a}_{1} + \left(\frac{1}{4} - z_{3}\right) \, \mathbf{a}_{2} + \left(\frac{1}{4} +z_{3}\right) \, \mathbf{a}_{3} & = & x_{3}a \, \mathbf{\hat{x}} + \frac{1}{4}b \, \mathbf{\hat{y}} + z_{3}c \, \mathbf{\hat{z}} & \left(4c\right) & \mbox{Cl II} \\ 
\mathbf{B}_{6} & = & -x_{3} \, \mathbf{a}_{1} + \left(\frac{3}{4} - z_{3}\right) \, \mathbf{a}_{2} + \left(\frac{3}{4} +z_{3}\right) \, \mathbf{a}_{3} & = & -x_{3}a \, \mathbf{\hat{x}} + \frac{3}{4}b \, \mathbf{\hat{y}} + z_{3}c \, \mathbf{\hat{z}} & \left(4c\right) & \mbox{Cl II} \\ 
\mathbf{B}_{7} & = & x_{4} \, \mathbf{a}_{1} + \left(\frac{1}{4} - z_{4}\right) \, \mathbf{a}_{2} + \left(\frac{1}{4} +z_{4}\right) \, \mathbf{a}_{3} & = & x_{4}a \, \mathbf{\hat{x}} + \frac{1}{4}b \, \mathbf{\hat{y}} + z_{4}c \, \mathbf{\hat{z}} & \left(4c\right) & \mbox{Cl III} \\ 
\mathbf{B}_{8} & = & -x_{4} \, \mathbf{a}_{1} + \left(\frac{3}{4} - z_{4}\right) \, \mathbf{a}_{2} + \left(\frac{3}{4} +z_{4}\right) \, \mathbf{a}_{3} & = & -x_{4}a \, \mathbf{\hat{x}} + \frac{3}{4}b \, \mathbf{\hat{y}} + z_{4}c \, \mathbf{\hat{z}} & \left(4c\right) & \mbox{Cl III} \\ 
\mathbf{B}_{9} & = & x_{5} \, \mathbf{a}_{1} + \left(\frac{1}{4} - z_{5}\right) \, \mathbf{a}_{2} + \left(\frac{1}{4} +z_{5}\right) \, \mathbf{a}_{3} & = & x_{5}a \, \mathbf{\hat{x}} + \frac{1}{4}b \, \mathbf{\hat{y}} + z_{5}c \, \mathbf{\hat{z}} & \left(4c\right) & \mbox{V} \\ 
\mathbf{B}_{10} & = & -x_{5} \, \mathbf{a}_{1} + \left(\frac{3}{4} - z_{5}\right) \, \mathbf{a}_{2} + \left(\frac{3}{4} +z_{5}\right) \, \mathbf{a}_{3} & = & -x_{5}a \, \mathbf{\hat{x}} + \frac{3}{4}b \, \mathbf{\hat{y}} + z_{5}c \, \mathbf{\hat{z}} & \left(4c\right) & \mbox{V} \\ 
\mathbf{B}_{11} & = & x_{6} \, \mathbf{a}_{1} + \left(y_{6}-z_{6}\right) \, \mathbf{a}_{2} + \left(y_{6}+z_{6}\right) \, \mathbf{a}_{3} & = & x_{6}a \, \mathbf{\hat{x}} + y_{6}b \, \mathbf{\hat{y}} + z_{6}c \, \mathbf{\hat{z}} & \left(8d\right) & \mbox{Cl IV} \\ 
\mathbf{B}_{12} & = & -x_{6} \, \mathbf{a}_{1} + \left(-y_{6}-z_{6}\right) \, \mathbf{a}_{2} + \left(-y_{6}+z_{6}\right) \, \mathbf{a}_{3} & = & -x_{6}a \, \mathbf{\hat{x}}-y_{6}b \, \mathbf{\hat{y}} + z_{6}c \, \mathbf{\hat{z}} & \left(8d\right) & \mbox{Cl IV} \\ 
\mathbf{B}_{13} & = & x_{6} \, \mathbf{a}_{1} + \left(\frac{1}{2} - y_{6} - z_{6}\right) \, \mathbf{a}_{2} + \left(\frac{1}{2} - y_{6} + z_{6}\right) \, \mathbf{a}_{3} & = & x_{6}a \, \mathbf{\hat{x}} + \left(\frac{1}{2} - y_{6}\right)b \, \mathbf{\hat{y}} + z_{6}c \, \mathbf{\hat{z}} & \left(8d\right) & \mbox{Cl IV} \\ 
\mathbf{B}_{14} & = & -x_{6} \, \mathbf{a}_{1} + \left(\frac{1}{2} +y_{6} - z_{6}\right) \, \mathbf{a}_{2} + \left(\frac{1}{2} +y_{6} + z_{6}\right) \, \mathbf{a}_{3} & = & -x_{6}a \, \mathbf{\hat{x}} + \left(\frac{1}{2} +y_{6}\right)b \, \mathbf{\hat{y}} + z_{6}c \, \mathbf{\hat{z}} & \left(8d\right) & \mbox{Cl IV} \\ 
\mathbf{B}_{15} & = & x_{7} \, \mathbf{a}_{1} + \left(y_{7}-z_{7}\right) \, \mathbf{a}_{2} + \left(y_{7}+z_{7}\right) \, \mathbf{a}_{3} & = & x_{7}a \, \mathbf{\hat{x}} + y_{7}b \, \mathbf{\hat{y}} + z_{7}c \, \mathbf{\hat{z}} & \left(8d\right) & \mbox{Cl V} \\ 
\mathbf{B}_{16} & = & -x_{7} \, \mathbf{a}_{1} + \left(-y_{7}-z_{7}\right) \, \mathbf{a}_{2} + \left(-y_{7}+z_{7}\right) \, \mathbf{a}_{3} & = & -x_{7}a \, \mathbf{\hat{x}}-y_{7}b \, \mathbf{\hat{y}} + z_{7}c \, \mathbf{\hat{z}} & \left(8d\right) & \mbox{Cl V} \\ 
\mathbf{B}_{17} & = & x_{7} \, \mathbf{a}_{1} + \left(\frac{1}{2} - y_{7} - z_{7}\right) \, \mathbf{a}_{2} + \left(\frac{1}{2} - y_{7} + z_{7}\right) \, \mathbf{a}_{3} & = & x_{7}a \, \mathbf{\hat{x}} + \left(\frac{1}{2} - y_{7}\right)b \, \mathbf{\hat{y}} + z_{7}c \, \mathbf{\hat{z}} & \left(8d\right) & \mbox{Cl V} \\ 
\mathbf{B}_{18} & = & -x_{7} \, \mathbf{a}_{1} + \left(\frac{1}{2} +y_{7} - z_{7}\right) \, \mathbf{a}_{2} + \left(\frac{1}{2} +y_{7} + z_{7}\right) \, \mathbf{a}_{3} & = & -x_{7}a \, \mathbf{\hat{x}} + \left(\frac{1}{2} +y_{7}\right)b \, \mathbf{\hat{y}} + z_{7}c \, \mathbf{\hat{z}} & \left(8d\right) & \mbox{Cl V} \\ 
\mathbf{B}_{19} & = & x_{8} \, \mathbf{a}_{1} + \left(y_{8}-z_{8}\right) \, \mathbf{a}_{2} + \left(y_{8}+z_{8}\right) \, \mathbf{a}_{3} & = & x_{8}a \, \mathbf{\hat{x}} + y_{8}b \, \mathbf{\hat{y}} + z_{8}c \, \mathbf{\hat{z}} & \left(8d\right) & \mbox{Cl VI} \\ 
\mathbf{B}_{20} & = & -x_{8} \, \mathbf{a}_{1} + \left(-y_{8}-z_{8}\right) \, \mathbf{a}_{2} + \left(-y_{8}+z_{8}\right) \, \mathbf{a}_{3} & = & -x_{8}a \, \mathbf{\hat{x}}-y_{8}b \, \mathbf{\hat{y}} + z_{8}c \, \mathbf{\hat{z}} & \left(8d\right) & \mbox{Cl VI} \\ 
\mathbf{B}_{21} & = & x_{8} \, \mathbf{a}_{1} + \left(\frac{1}{2} - y_{8} - z_{8}\right) \, \mathbf{a}_{2} + \left(\frac{1}{2} - y_{8} + z_{8}\right) \, \mathbf{a}_{3} & = & x_{8}a \, \mathbf{\hat{x}} + \left(\frac{1}{2} - y_{8}\right)b \, \mathbf{\hat{y}} + z_{8}c \, \mathbf{\hat{z}} & \left(8d\right) & \mbox{Cl VI} \\ 
\mathbf{B}_{22} & = & -x_{8} \, \mathbf{a}_{1} + \left(\frac{1}{2} +y_{8} - z_{8}\right) \, \mathbf{a}_{2} + \left(\frac{1}{2} +y_{8} + z_{8}\right) \, \mathbf{a}_{3} & = & -x_{8}a \, \mathbf{\hat{x}} + \left(\frac{1}{2} +y_{8}\right)b \, \mathbf{\hat{y}} + z_{8}c \, \mathbf{\hat{z}} & \left(8d\right) & \mbox{Cl VI} \\ 
\end{longtabu}
\renewcommand{\arraystretch}{1.0}
\noindent \hrulefill
\\
\textbf{References:}
\vspace*{-0.25cm}
\begin{flushleft}
  - \bibentry{Ziegler_VPCl9_ZNaturB_1977}. \\
\end{flushleft}
\textbf{Found in:}
\vspace*{-0.25cm}
\begin{flushleft}
  - \bibentry{Villars_PearsonsCrystalData_2013}. \\
\end{flushleft}
\noindent \hrulefill
\\
\textbf{Geometry files:}
\\
\noindent  - CIF: pp. {\hyperref[A9BC_oC44_39_3c3d_a_c_cif]{\pageref{A9BC_oC44_39_3c3d_a_c_cif}}} \\
\noindent  - POSCAR: pp. {\hyperref[A9BC_oC44_39_3c3d_a_c_poscar]{\pageref{A9BC_oC44_39_3c3d_a_c_poscar}}} \\
\onecolumn
{\phantomsection\label{AB2C_oC16_40_a_2b_b}}
\subsection*{\huge \textbf{{\normalfont K$_{2}$CdPb Structure: AB2C\_oC16\_40\_a\_2b\_b}}}
\noindent \hrulefill
\vspace*{0.25cm}
\begin{figure}[htp]
  \centering
  \vspace{-1em}
  {\includegraphics[width=1\textwidth]{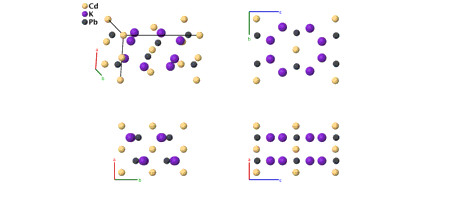}}
\end{figure}
\vspace*{-0.5cm}
\renewcommand{\arraystretch}{1.5}
\begin{equation*}
  \begin{array}{>{$\hspace{-0.15cm}}l<{$}>{$}p{0.5cm}<{$}>{$}p{18.5cm}<{$}}
    \mbox{\large \textbf{Prototype}} &\colon & \ce{K2CdPb} \\
    \mbox{\large \textbf{\AFLOW\ prototype label}} &\colon & \mbox{AB2C\_oC16\_40\_a\_2b\_b} \\
    \mbox{\large \textbf{\textit{Strukturbericht} designation}} &\colon & \mbox{None} \\
    \mbox{\large \textbf{Pearson symbol}} &\colon & \mbox{oC16} \\
    \mbox{\large \textbf{Space group number}} &\colon & 40 \\
    \mbox{\large \textbf{Space group symbol}} &\colon & Ama2 \\
    \mbox{\large \textbf{\AFLOW\ prototype command}} &\colon &  \texttt{aflow} \,  \, \texttt{-{}-proto=AB2C\_oC16\_40\_a\_2b\_b } \, \newline \texttt{-{}-params=}{a,b/a,c/a,z_{1},y_{2},z_{2},y_{3},z_{3},y_{4},z_{4} }
  \end{array}
\end{equation*}
\renewcommand{\arraystretch}{1.0}

\noindent \parbox{1 \linewidth}{
\noindent \hrulefill
\\
\textbf{Base-centered Orthorhombic primitive vectors:} \\
\vspace*{-0.25cm}
\begin{tabular}{cc}
  \begin{tabular}{c}
    \parbox{0.6 \linewidth}{
      \renewcommand{\arraystretch}{1.5}
      \begin{equation*}
        \centering
        \begin{array}{ccc}
              \mathbf{a}_1 & = & a \, \mathbf{\hat{x}} \\
    \mathbf{a}_2 & = & \frac12 \, b \, \mathbf{\hat{y}} - \frac12 \, c \, \mathbf{\hat{z}} \\
    \mathbf{a}_3 & = & \frac12 \, b \, \mathbf{\hat{y}} + \frac12 \, c \, \mathbf{\hat{z}} \\

        \end{array}
      \end{equation*}
    }
    \renewcommand{\arraystretch}{1.0}
  \end{tabular}
  \begin{tabular}{c}
    \includegraphics[width=0.3\linewidth]{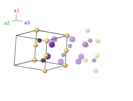} \\
  \end{tabular}
\end{tabular}

}
\vspace*{-0.25cm}

\noindent \hrulefill
\\
\textbf{Basis vectors:}
\vspace*{-0.25cm}
\renewcommand{\arraystretch}{1.5}
\begin{longtabu} to \textwidth{>{\centering $}X[-1,c,c]<{$}>{\centering $}X[-1,c,c]<{$}>{\centering $}X[-1,c,c]<{$}>{\centering $}X[-1,c,c]<{$}>{\centering $}X[-1,c,c]<{$}>{\centering $}X[-1,c,c]<{$}>{\centering $}X[-1,c,c]<{$}}
  & & \mbox{Lattice Coordinates} & & \mbox{Cartesian Coordinates} &\mbox{Wyckoff Position} & \mbox{Atom Type} \\  
  \mathbf{B}_{1} & = & -z_{1} \, \mathbf{a}_{2} + z_{1} \, \mathbf{a}_{3} & = & z_{1}c \, \mathbf{\hat{z}} & \left(4a\right) & \mbox{Cd} \\ 
\mathbf{B}_{2} & = & \frac{1}{2} \, \mathbf{a}_{1}-z_{1} \, \mathbf{a}_{2} + z_{1} \, \mathbf{a}_{3} & = & \frac{1}{2}a \, \mathbf{\hat{x}} + z_{1}c \, \mathbf{\hat{z}} & \left(4a\right) & \mbox{Cd} \\ 
\mathbf{B}_{3} & = & \frac{1}{4} \, \mathbf{a}_{1} + \left(y_{2}-z_{2}\right) \, \mathbf{a}_{2} + \left(y_{2}+z_{2}\right) \, \mathbf{a}_{3} & = & \frac{1}{4}a \, \mathbf{\hat{x}} + y_{2}b \, \mathbf{\hat{y}} + z_{2}c \, \mathbf{\hat{z}} & \left(4b\right) & \mbox{K I} \\ 
\mathbf{B}_{4} & = & \frac{3}{4} \, \mathbf{a}_{1} + \left(-y_{2}-z_{2}\right) \, \mathbf{a}_{2} + \left(-y_{2}+z_{2}\right) \, \mathbf{a}_{3} & = & \frac{3}{4}a \, \mathbf{\hat{x}}-y_{2}b \, \mathbf{\hat{y}} + z_{2}c \, \mathbf{\hat{z}} & \left(4b\right) & \mbox{K I} \\ 
\mathbf{B}_{5} & = & \frac{1}{4} \, \mathbf{a}_{1} + \left(y_{3}-z_{3}\right) \, \mathbf{a}_{2} + \left(y_{3}+z_{3}\right) \, \mathbf{a}_{3} & = & \frac{1}{4}a \, \mathbf{\hat{x}} + y_{3}b \, \mathbf{\hat{y}} + z_{3}c \, \mathbf{\hat{z}} & \left(4b\right) & \mbox{K II} \\ 
\mathbf{B}_{6} & = & \frac{3}{4} \, \mathbf{a}_{1} + \left(-y_{3}-z_{3}\right) \, \mathbf{a}_{2} + \left(-y_{3}+z_{3}\right) \, \mathbf{a}_{3} & = & \frac{3}{4}a \, \mathbf{\hat{x}}-y_{3}b \, \mathbf{\hat{y}} + z_{3}c \, \mathbf{\hat{z}} & \left(4b\right) & \mbox{K II} \\ 
\mathbf{B}_{7} & = & \frac{1}{4} \, \mathbf{a}_{1} + \left(y_{4}-z_{4}\right) \, \mathbf{a}_{2} + \left(y_{4}+z_{4}\right) \, \mathbf{a}_{3} & = & \frac{1}{4}a \, \mathbf{\hat{x}} + y_{4}b \, \mathbf{\hat{y}} + z_{4}c \, \mathbf{\hat{z}} & \left(4b\right) & \mbox{Pb} \\ 
\mathbf{B}_{8} & = & \frac{3}{4} \, \mathbf{a}_{1} + \left(-y_{4}-z_{4}\right) \, \mathbf{a}_{2} + \left(-y_{4}+z_{4}\right) \, \mathbf{a}_{3} & = & \frac{3}{4}a \, \mathbf{\hat{x}}-y_{4}b \, \mathbf{\hat{y}} + z_{4}c \, \mathbf{\hat{z}} & \left(4b\right) & \mbox{Pb} \\ 
\end{longtabu}
\renewcommand{\arraystretch}{1.0}
\noindent \hrulefill
\\
\textbf{References:}
\vspace*{-0.25cm}
\begin{flushleft}
  - \bibentry{Matthes_CdK2Pb_ZNaturB_1979}. \\
\end{flushleft}
\textbf{Found in:}
\vspace*{-0.25cm}
\begin{flushleft}
  - \bibentry{Villars_PearsonsCrystalData_2013}. \\
\end{flushleft}
\noindent \hrulefill
\\
\textbf{Geometry files:}
\\
\noindent  - CIF: pp. {\hyperref[AB2C_oC16_40_a_2b_b_cif]{\pageref{AB2C_oC16_40_a_2b_b_cif}}} \\
\noindent  - POSCAR: pp. {\hyperref[AB2C_oC16_40_a_2b_b_poscar]{\pageref{AB2C_oC16_40_a_2b_b_poscar}}} \\
\onecolumn
{\phantomsection\label{AB3_oC16_40_b_3b}}
\subsection*{\huge \textbf{{\normalfont CeTe$_{3}$ Structure: AB3\_oC16\_40\_b\_3b}}}
\noindent \hrulefill
\vspace*{0.25cm}
\begin{figure}[htp]
  \centering
  \vspace{-1em}
  {\includegraphics[width=1\textwidth]{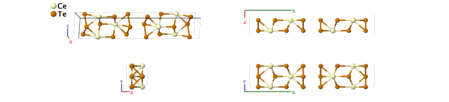}}
\end{figure}
\vspace*{-0.5cm}
\renewcommand{\arraystretch}{1.5}
\begin{equation*}
  \begin{array}{>{$\hspace{-0.15cm}}l<{$}>{$}p{0.5cm}<{$}>{$}p{18.5cm}<{$}}
    \mbox{\large \textbf{Prototype}} &\colon & \ce{CeTe3} \\
    \mbox{\large \textbf{\AFLOW\ prototype label}} &\colon & \mbox{AB3\_oC16\_40\_b\_3b} \\
    \mbox{\large \textbf{\textit{Strukturbericht} designation}} &\colon & \mbox{None} \\
    \mbox{\large \textbf{Pearson symbol}} &\colon & \mbox{oC16} \\
    \mbox{\large \textbf{Space group number}} &\colon & 40 \\
    \mbox{\large \textbf{Space group symbol}} &\colon & Ama2 \\
    \mbox{\large \textbf{\AFLOW\ prototype command}} &\colon &  \texttt{aflow} \,  \, \texttt{-{}-proto=AB3\_oC16\_40\_b\_3b } \, \newline \texttt{-{}-params=}{a,b/a,c/a,y_{1},z_{1},y_{2},z_{2},y_{3},z_{3},y_{4},z_{4} }
  \end{array}
\end{equation*}
\renewcommand{\arraystretch}{1.0}

\noindent \parbox{1 \linewidth}{
\noindent \hrulefill
\\
\textbf{Base-centered Orthorhombic primitive vectors:} \\
\vspace*{-0.25cm}
\begin{tabular}{cc}
  \begin{tabular}{c}
    \parbox{0.6 \linewidth}{
      \renewcommand{\arraystretch}{1.5}
      \begin{equation*}
        \centering
        \begin{array}{ccc}
              \mathbf{a}_1 & = & a \, \mathbf{\hat{x}} \\
    \mathbf{a}_2 & = & \frac12 \, b \, \mathbf{\hat{y}} - \frac12 \, c \, \mathbf{\hat{z}} \\
    \mathbf{a}_3 & = & \frac12 \, b \, \mathbf{\hat{y}} + \frac12 \, c \, \mathbf{\hat{z}} \\

        \end{array}
      \end{equation*}
    }
    \renewcommand{\arraystretch}{1.0}
  \end{tabular}
  \begin{tabular}{c}
    \includegraphics[width=0.3\linewidth]{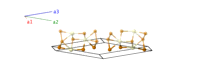} \\
  \end{tabular}
\end{tabular}

}
\vspace*{-0.25cm}

\noindent \hrulefill
\\
\textbf{Basis vectors:}
\vspace*{-0.25cm}
\renewcommand{\arraystretch}{1.5}
\begin{longtabu} to \textwidth{>{\centering $}X[-1,c,c]<{$}>{\centering $}X[-1,c,c]<{$}>{\centering $}X[-1,c,c]<{$}>{\centering $}X[-1,c,c]<{$}>{\centering $}X[-1,c,c]<{$}>{\centering $}X[-1,c,c]<{$}>{\centering $}X[-1,c,c]<{$}}
  & & \mbox{Lattice Coordinates} & & \mbox{Cartesian Coordinates} &\mbox{Wyckoff Position} & \mbox{Atom Type} \\  
  \mathbf{B}_{1} & = & \frac{1}{4} \, \mathbf{a}_{1} + \left(y_{1}-z_{1}\right) \, \mathbf{a}_{2} + \left(y_{1}+z_{1}\right) \, \mathbf{a}_{3} & = & \frac{1}{4}a \, \mathbf{\hat{x}} + y_{1}b \, \mathbf{\hat{y}} + z_{1}c \, \mathbf{\hat{z}} & \left(4b\right) & \mbox{Ce} \\ 
\mathbf{B}_{2} & = & \frac{3}{4} \, \mathbf{a}_{1} + \left(-y_{1}-z_{1}\right) \, \mathbf{a}_{2} + \left(-y_{1}+z_{1}\right) \, \mathbf{a}_{3} & = & \frac{3}{4}a \, \mathbf{\hat{x}}-y_{1}b \, \mathbf{\hat{y}} + z_{1}c \, \mathbf{\hat{z}} & \left(4b\right) & \mbox{Ce} \\ 
\mathbf{B}_{3} & = & \frac{1}{4} \, \mathbf{a}_{1} + \left(y_{2}-z_{2}\right) \, \mathbf{a}_{2} + \left(y_{2}+z_{2}\right) \, \mathbf{a}_{3} & = & \frac{1}{4}a \, \mathbf{\hat{x}} + y_{2}b \, \mathbf{\hat{y}} + z_{2}c \, \mathbf{\hat{z}} & \left(4b\right) & \mbox{Te I} \\ 
\mathbf{B}_{4} & = & \frac{3}{4} \, \mathbf{a}_{1} + \left(-y_{2}-z_{2}\right) \, \mathbf{a}_{2} + \left(-y_{2}+z_{2}\right) \, \mathbf{a}_{3} & = & \frac{3}{4}a \, \mathbf{\hat{x}}-y_{2}b \, \mathbf{\hat{y}} + z_{2}c \, \mathbf{\hat{z}} & \left(4b\right) & \mbox{Te I} \\ 
\mathbf{B}_{5} & = & \frac{1}{4} \, \mathbf{a}_{1} + \left(y_{3}-z_{3}\right) \, \mathbf{a}_{2} + \left(y_{3}+z_{3}\right) \, \mathbf{a}_{3} & = & \frac{1}{4}a \, \mathbf{\hat{x}} + y_{3}b \, \mathbf{\hat{y}} + z_{3}c \, \mathbf{\hat{z}} & \left(4b\right) & \mbox{Te II} \\ 
\mathbf{B}_{6} & = & \frac{3}{4} \, \mathbf{a}_{1} + \left(-y_{3}-z_{3}\right) \, \mathbf{a}_{2} + \left(-y_{3}+z_{3}\right) \, \mathbf{a}_{3} & = & \frac{3}{4}a \, \mathbf{\hat{x}}-y_{3}b \, \mathbf{\hat{y}} + z_{3}c \, \mathbf{\hat{z}} & \left(4b\right) & \mbox{Te II} \\ 
\mathbf{B}_{7} & = & \frac{1}{4} \, \mathbf{a}_{1} + \left(y_{4}-z_{4}\right) \, \mathbf{a}_{2} + \left(y_{4}+z_{4}\right) \, \mathbf{a}_{3} & = & \frac{1}{4}a \, \mathbf{\hat{x}} + y_{4}b \, \mathbf{\hat{y}} + z_{4}c \, \mathbf{\hat{z}} & \left(4b\right) & \mbox{Te III} \\ 
\mathbf{B}_{8} & = & \frac{3}{4} \, \mathbf{a}_{1} + \left(-y_{4}-z_{4}\right) \, \mathbf{a}_{2} + \left(-y_{4}+z_{4}\right) \, \mathbf{a}_{3} & = & \frac{3}{4}a \, \mathbf{\hat{x}}-y_{4}b \, \mathbf{\hat{y}} + z_{4}c \, \mathbf{\hat{z}} & \left(4b\right) & \mbox{Te III} \\ 
\end{longtabu}
\renewcommand{\arraystretch}{1.0}
\noindent \hrulefill
\\
\textbf{References:}
\vspace*{-0.25cm}
\begin{flushleft}
  - \bibentry{Malliakas_CeTe3_JAmerChemSoc_2005}. \\
\end{flushleft}
\textbf{Found in:}
\vspace*{-0.25cm}
\begin{flushleft}
  - \bibentry{Villars_PearsonsCrystalData_2013}. \\
\end{flushleft}
\noindent \hrulefill
\\
\textbf{Geometry files:}
\\
\noindent  - CIF: pp. {\hyperref[AB3_oC16_40_b_3b_cif]{\pageref{AB3_oC16_40_b_3b_cif}}} \\
\noindent  - POSCAR: pp. {\hyperref[AB3_oC16_40_b_3b_poscar]{\pageref{AB3_oC16_40_b_3b_poscar}}} \\
\onecolumn
{\phantomsection\label{A10B3_oF52_42_2abce_ab}}
\subsection*{\huge \textbf{{\normalfont W$_{3}$O$_{10}$ Structure: A10B3\_oF52\_42\_2abce\_ab}}}
\noindent \hrulefill
\vspace*{0.25cm}
\begin{figure}[htp]
  \centering
  \vspace{-1em}
  {\includegraphics[width=1\textwidth]{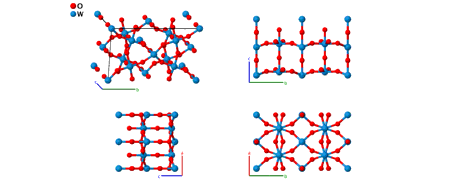}}
\end{figure}
\vspace*{-0.5cm}
\renewcommand{\arraystretch}{1.5}
\begin{equation*}
  \begin{array}{>{$\hspace{-0.15cm}}l<{$}>{$}p{0.5cm}<{$}>{$}p{18.5cm}<{$}}
    \mbox{\large \textbf{Prototype}} &\colon & \ce{W3O10} \\
    \mbox{\large \textbf{\AFLOW\ prototype label}} &\colon & \mbox{A10B3\_oF52\_42\_2abce\_ab} \\
    \mbox{\large \textbf{\textit{Strukturbericht} designation}} &\colon & \mbox{None} \\
    \mbox{\large \textbf{Pearson symbol}} &\colon & \mbox{oF52} \\
    \mbox{\large \textbf{Space group number}} &\colon & 42 \\
    \mbox{\large \textbf{Space group symbol}} &\colon & Fmm2 \\
    \mbox{\large \textbf{\AFLOW\ prototype command}} &\colon &  \texttt{aflow} \,  \, \texttt{-{}-proto=A10B3\_oF52\_42\_2abce\_ab } \, \newline \texttt{-{}-params=}{a,b/a,c/a,z_{1},z_{2},z_{3},z_{4},z_{5},y_{6},z_{6},x_{7},y_{7},z_{7} }
  \end{array}
\end{equation*}
\renewcommand{\arraystretch}{1.0}

\noindent \parbox{1 \linewidth}{
\noindent \hrulefill
\\
\textbf{Face-centered Orthorhombic primitive vectors:} \\
\vspace*{-0.25cm}
\begin{tabular}{cc}
  \begin{tabular}{c}
    \parbox{0.6 \linewidth}{
      \renewcommand{\arraystretch}{1.5}
      \begin{equation*}
        \centering
        \begin{array}{ccc}
              \mathbf{a}_1 & = & \frac12 \, b \, \mathbf{\hat{y}} + \frac12 \, c \, \mathbf{\hat{z}} \\
    \mathbf{a}_2 & = & \frac12 \, a \, \mathbf{\hat{x}} + \frac12 \, c \, \mathbf{\hat{z}} \\
    \mathbf{a}_3 & = & \frac12 \, a \, \mathbf{\hat{x}} + \frac12 \, b \, \mathbf{\hat{y}} \\

        \end{array}
      \end{equation*}
    }
    \renewcommand{\arraystretch}{1.0}
  \end{tabular}
  \begin{tabular}{c}
    \includegraphics[width=0.3\linewidth]{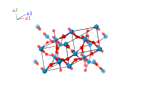} \\
  \end{tabular}
\end{tabular}

}
\vspace*{-0.25cm}

\noindent \hrulefill
\\
\textbf{Basis vectors:}
\vspace*{-0.25cm}
\renewcommand{\arraystretch}{1.5}
\begin{longtabu} to \textwidth{>{\centering $}X[-1,c,c]<{$}>{\centering $}X[-1,c,c]<{$}>{\centering $}X[-1,c,c]<{$}>{\centering $}X[-1,c,c]<{$}>{\centering $}X[-1,c,c]<{$}>{\centering $}X[-1,c,c]<{$}>{\centering $}X[-1,c,c]<{$}}
  & & \mbox{Lattice Coordinates} & & \mbox{Cartesian Coordinates} &\mbox{Wyckoff Position} & \mbox{Atom Type} \\  
  \mathbf{B}_{1} & = & z_{1} \, \mathbf{a}_{1} + z_{1} \, \mathbf{a}_{2}-z_{1} \, \mathbf{a}_{3} & = & z_{1}c \, \mathbf{\hat{z}} & \left(4a\right) & \mbox{O I} \\ 
\mathbf{B}_{2} & = & z_{2} \, \mathbf{a}_{1} + z_{2} \, \mathbf{a}_{2}-z_{2} \, \mathbf{a}_{3} & = & z_{2}c \, \mathbf{\hat{z}} & \left(4a\right) & \mbox{O II} \\ 
\mathbf{B}_{3} & = & z_{3} \, \mathbf{a}_{1} + z_{3} \, \mathbf{a}_{2}-z_{3} \, \mathbf{a}_{3} & = & z_{3}c \, \mathbf{\hat{z}} & \left(4a\right) & \mbox{W I} \\ 
\mathbf{B}_{4} & = & z_{4} \, \mathbf{a}_{1} + z_{4} \, \mathbf{a}_{2} + \left(\frac{1}{2} - z_{4}\right) \, \mathbf{a}_{3} & = & \frac{1}{4}a \, \mathbf{\hat{x}} + \frac{1}{4}b \, \mathbf{\hat{y}} + z_{4}c \, \mathbf{\hat{z}} & \left(8b\right) & \mbox{O III} \\ 
\mathbf{B}_{5} & = & \left(\frac{1}{2} +z_{4}\right) \, \mathbf{a}_{1} + \left(\frac{1}{2} +z_{4}\right) \, \mathbf{a}_{2}-z_{4} \, \mathbf{a}_{3} & = & \frac{1}{4}a \, \mathbf{\hat{x}} + \frac{1}{4}b \, \mathbf{\hat{y}} + \left(\frac{1}{2} +z_{4}\right)c \, \mathbf{\hat{z}} & \left(8b\right) & \mbox{O III} \\ 
\mathbf{B}_{6} & = & z_{5} \, \mathbf{a}_{1} + z_{5} \, \mathbf{a}_{2} + \left(\frac{1}{2} - z_{5}\right) \, \mathbf{a}_{3} & = & \frac{1}{4}a \, \mathbf{\hat{x}} + \frac{1}{4}b \, \mathbf{\hat{y}} + z_{5}c \, \mathbf{\hat{z}} & \left(8b\right) & \mbox{W II} \\ 
\mathbf{B}_{7} & = & \left(\frac{1}{2} +z_{5}\right) \, \mathbf{a}_{1} + \left(\frac{1}{2} +z_{5}\right) \, \mathbf{a}_{2}-z_{5} \, \mathbf{a}_{3} & = & \frac{1}{4}a \, \mathbf{\hat{x}} + \frac{1}{4}b \, \mathbf{\hat{y}} + \left(\frac{1}{2} +z_{5}\right)c \, \mathbf{\hat{z}} & \left(8b\right) & \mbox{W II} \\ 
\mathbf{B}_{8} & = & \left(y_{6}+z_{6}\right) \, \mathbf{a}_{1} + \left(-y_{6}+z_{6}\right) \, \mathbf{a}_{2} + \left(y_{6}-z_{6}\right) \, \mathbf{a}_{3} & = & y_{6}b \, \mathbf{\hat{y}} + z_{6}c \, \mathbf{\hat{z}} & \left(8c\right) & \mbox{O IV} \\ 
\mathbf{B}_{9} & = & \left(-y_{6}+z_{6}\right) \, \mathbf{a}_{1} + \left(y_{6}+z_{6}\right) \, \mathbf{a}_{2} + \left(-y_{6}-z_{6}\right) \, \mathbf{a}_{3} & = & -y_{6}b \, \mathbf{\hat{y}} + z_{6}c \, \mathbf{\hat{z}} & \left(8c\right) & \mbox{O IV} \\ 
\mathbf{B}_{10} & = & \left(-x_{7}+y_{7}+z_{7}\right) \, \mathbf{a}_{1} + \left(x_{7}-y_{7}+z_{7}\right) \, \mathbf{a}_{2} + \left(x_{7}+y_{7}-z_{7}\right) \, \mathbf{a}_{3} & = & x_{7}a \, \mathbf{\hat{x}} + y_{7}b \, \mathbf{\hat{y}} + z_{7}c \, \mathbf{\hat{z}} & \left(16e\right) & \mbox{O V} \\ 
\mathbf{B}_{11} & = & \left(x_{7}-y_{7}+z_{7}\right) \, \mathbf{a}_{1} + \left(-x_{7}+y_{7}+z_{7}\right) \, \mathbf{a}_{2} + \left(-x_{7}-y_{7}-z_{7}\right) \, \mathbf{a}_{3} & = & -x_{7}a \, \mathbf{\hat{x}}-y_{7}b \, \mathbf{\hat{y}} + z_{7}c \, \mathbf{\hat{z}} & \left(16e\right) & \mbox{O V} \\ 
\mathbf{B}_{12} & = & \left(-x_{7}-y_{7}+z_{7}\right) \, \mathbf{a}_{1} + \left(x_{7}+y_{7}+z_{7}\right) \, \mathbf{a}_{2} + \left(x_{7}-y_{7}-z_{7}\right) \, \mathbf{a}_{3} & = & x_{7}a \, \mathbf{\hat{x}}-y_{7}b \, \mathbf{\hat{y}} + z_{7}c \, \mathbf{\hat{z}} & \left(16e\right) & \mbox{O V} \\ 
\mathbf{B}_{13} & = & \left(x_{7}+y_{7}+z_{7}\right) \, \mathbf{a}_{1} + \left(-x_{7}-y_{7}+z_{7}\right) \, \mathbf{a}_{2} + \left(-x_{7}+y_{7}-z_{7}\right) \, \mathbf{a}_{3} & = & -x_{7}a \, \mathbf{\hat{x}} + y_{7}b \, \mathbf{\hat{y}} + z_{7}c \, \mathbf{\hat{z}} & \left(16e\right) & \mbox{O V} \\ 
\end{longtabu}
\renewcommand{\arraystretch}{1.0}
\noindent \hrulefill
\\
\textbf{References:}
\vspace*{-0.25cm}
\begin{flushleft}
  - \bibentry{Gerand_W3O9H2O_JSolStateChem_1981}. \\
\end{flushleft}
\textbf{Found in:}
\vspace*{-0.25cm}
\begin{flushleft}
  - \bibentry{Villars_PearsonsCrystalData_2013}. \\
\end{flushleft}
\noindent \hrulefill
\\
\textbf{Geometry files:}
\\
\noindent  - CIF: pp. {\hyperref[A10B3_oF52_42_2abce_ab_cif]{\pageref{A10B3_oF52_42_2abce_ab_cif}}} \\
\noindent  - POSCAR: pp. {\hyperref[A10B3_oF52_42_2abce_ab_poscar]{\pageref{A10B3_oF52_42_2abce_ab_poscar}}} \\
\onecolumn
{\phantomsection\label{AB_oF8_42_a_a}}
\subsection*{\huge \textbf{{\normalfont \begin{raggedleft}BN (High-pressure, high-temperature) Structure: \end{raggedleft} \\ AB\_oF8\_42\_a\_a}}}
\noindent \hrulefill
\vspace*{0.25cm}
\begin{figure}[htp]
  \centering
  \vspace{-1em}
  {\includegraphics[width=1\textwidth]{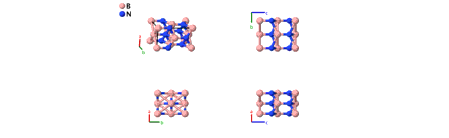}}
\end{figure}
\vspace*{-0.5cm}
\renewcommand{\arraystretch}{1.5}
\begin{equation*}
  \begin{array}{>{$\hspace{-0.15cm}}l<{$}>{$}p{0.5cm}<{$}>{$}p{18.5cm}<{$}}
    \mbox{\large \textbf{Prototype}} &\colon & \ce{BN} \\
    \mbox{\large \textbf{\AFLOW\ prototype label}} &\colon & \mbox{AB\_oF8\_42\_a\_a} \\
    \mbox{\large \textbf{\textit{Strukturbericht} designation}} &\colon & \mbox{None} \\
    \mbox{\large \textbf{Pearson symbol}} &\colon & \mbox{oF8} \\
    \mbox{\large \textbf{Space group number}} &\colon & 42 \\
    \mbox{\large \textbf{Space group symbol}} &\colon & Fmm2 \\
    \mbox{\large \textbf{\AFLOW\ prototype command}} &\colon &  \texttt{aflow} \,  \, \texttt{-{}-proto=AB\_oF8\_42\_a\_a } \, \newline \texttt{-{}-params=}{a,b/a,c/a,z_{1},z_{2} }
  \end{array}
\end{equation*}
\renewcommand{\arraystretch}{1.0}

\noindent \parbox{1 \linewidth}{
\noindent \hrulefill
\\
\textbf{Face-centered Orthorhombic primitive vectors:} \\
\vspace*{-0.25cm}
\begin{tabular}{cc}
  \begin{tabular}{c}
    \parbox{0.6 \linewidth}{
      \renewcommand{\arraystretch}{1.5}
      \begin{equation*}
        \centering
        \begin{array}{ccc}
              \mathbf{a}_1 & = & \frac12 \, b \, \mathbf{\hat{y}} + \frac12 \, c \, \mathbf{\hat{z}} \\
    \mathbf{a}_2 & = & \frac12 \, a \, \mathbf{\hat{x}} + \frac12 \, c \, \mathbf{\hat{z}} \\
    \mathbf{a}_3 & = & \frac12 \, a \, \mathbf{\hat{x}} + \frac12 \, b \, \mathbf{\hat{y}} \\

        \end{array}
      \end{equation*}
    }
    \renewcommand{\arraystretch}{1.0}
  \end{tabular}
  \begin{tabular}{c}
    \includegraphics[width=0.3\linewidth]{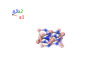} \\
  \end{tabular}
\end{tabular}

}
\vspace*{-0.25cm}

\noindent \hrulefill
\\
\textbf{Basis vectors:}
\vspace*{-0.25cm}
\renewcommand{\arraystretch}{1.5}
\begin{longtabu} to \textwidth{>{\centering $}X[-1,c,c]<{$}>{\centering $}X[-1,c,c]<{$}>{\centering $}X[-1,c,c]<{$}>{\centering $}X[-1,c,c]<{$}>{\centering $}X[-1,c,c]<{$}>{\centering $}X[-1,c,c]<{$}>{\centering $}X[-1,c,c]<{$}}
  & & \mbox{Lattice Coordinates} & & \mbox{Cartesian Coordinates} &\mbox{Wyckoff Position} & \mbox{Atom Type} \\  
  \mathbf{B}_{1} & = & z_{1} \, \mathbf{a}_{1} + z_{1} \, \mathbf{a}_{2}-z_{1} \, \mathbf{a}_{3} & = & z_{1}c \, \mathbf{\hat{z}} & \left(4a\right) & \mbox{B} \\ 
\mathbf{B}_{2} & = & z_{2} \, \mathbf{a}_{1} + z_{2} \, \mathbf{a}_{2}-z_{2} \, \mathbf{a}_{3} & = & z_{2}c \, \mathbf{\hat{z}} & \left(4a\right) & \mbox{N} \\ 
\end{longtabu}
\renewcommand{\arraystretch}{1.0}
\noindent \hrulefill
\\
\textbf{References:}
\vspace*{-0.25cm}
\begin{flushleft}
  - \bibentry{Kurdyumov_BN_Kristallogr_1984}. \\
\end{flushleft}
\textbf{Found in:}
\vspace*{-0.25cm}
\begin{flushleft}
  - \bibentry{Villars_PearsonsCrystalData_2013}. \\
\end{flushleft}
\noindent \hrulefill
\\
\textbf{Geometry files:}
\\
\noindent  - CIF: pp. {\hyperref[AB_oF8_42_a_a_cif]{\pageref{AB_oF8_42_a_a_cif}}} \\
\noindent  - POSCAR: pp. {\hyperref[AB_oF8_42_a_a_poscar]{\pageref{AB_oF8_42_a_a_poscar}}} \\
\onecolumn
{\phantomsection\label{A2BC2_oI20_45_c_b_c}}
\subsection*{\huge \textbf{{\normalfont MnGa$_{2}$Sb$_{2}$ Structure: A2BC2\_oI20\_45\_c\_b\_c}}}
\noindent \hrulefill
\vspace*{0.25cm}
\begin{figure}[htp]
  \centering
  \vspace{-1em}
  {\includegraphics[width=1\textwidth]{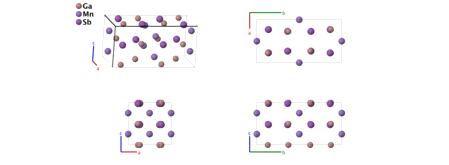}}
\end{figure}
\vspace*{-0.5cm}
\renewcommand{\arraystretch}{1.5}
\begin{equation*}
  \begin{array}{>{$\hspace{-0.15cm}}l<{$}>{$}p{0.5cm}<{$}>{$}p{18.5cm}<{$}}
    \mbox{\large \textbf{Prototype}} &\colon & \ce{MnGa2Sb2} \\
    \mbox{\large \textbf{\AFLOW\ prototype label}} &\colon & \mbox{A2BC2\_oI20\_45\_c\_b\_c} \\
    \mbox{\large \textbf{\textit{Strukturbericht} designation}} &\colon & \mbox{None} \\
    \mbox{\large \textbf{Pearson symbol}} &\colon & \mbox{oI20} \\
    \mbox{\large \textbf{Space group number}} &\colon & 45 \\
    \mbox{\large \textbf{Space group symbol}} &\colon & Iba2 \\
    \mbox{\large \textbf{\AFLOW\ prototype command}} &\colon &  \texttt{aflow} \,  \, \texttt{-{}-proto=A2BC2\_oI20\_45\_c\_b\_c } \, \newline \texttt{-{}-params=}{a,b/a,c/a,z_{1},x_{2},y_{2},z_{2},x_{3},y_{3},z_{3} }
  \end{array}
\end{equation*}
\renewcommand{\arraystretch}{1.0}

\vspace*{-0.25cm}
\noindent \hrulefill
\begin{itemize}
  \item{The Mn site (4b) is reported with an occupation of 0.94.
}
\end{itemize}

\noindent \parbox{1 \linewidth}{
\noindent \hrulefill
\\
\textbf{Body-centered Orthorhombic primitive vectors:} \\
\vspace*{-0.25cm}
\begin{tabular}{cc}
  \begin{tabular}{c}
    \parbox{0.6 \linewidth}{
      \renewcommand{\arraystretch}{1.5}
      \begin{equation*}
        \centering
        \begin{array}{ccc}
              \mathbf{a}_1 & = & - \frac12 \, a \, \mathbf{\hat{x}} + \frac12 \, b \, \mathbf{\hat{y}} + \frac12 \, c \, \mathbf{\hat{z}} \\
    \mathbf{a}_2 & = & ~ \frac12 \, a \, \mathbf{\hat{x}} - \frac12 \, b \, \mathbf{\hat{y}} + \frac12 \, c \, \mathbf{\hat{z}} \\
    \mathbf{a}_3 & = & ~ \frac12 \, a \, \mathbf{\hat{x}} + \frac12 \, b \, \mathbf{\hat{y}} - \frac12 \, c \, \mathbf{\hat{z}} \\

        \end{array}
      \end{equation*}
    }
    \renewcommand{\arraystretch}{1.0}
  \end{tabular}
  \begin{tabular}{c}
    \includegraphics[width=0.3\linewidth]{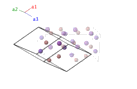} \\
  \end{tabular}
\end{tabular}

}
\vspace*{-0.25cm}

\noindent \hrulefill
\\
\textbf{Basis vectors:}
\vspace*{-0.25cm}
\renewcommand{\arraystretch}{1.5}
\begin{longtabu} to \textwidth{>{\centering $}X[-1,c,c]<{$}>{\centering $}X[-1,c,c]<{$}>{\centering $}X[-1,c,c]<{$}>{\centering $}X[-1,c,c]<{$}>{\centering $}X[-1,c,c]<{$}>{\centering $}X[-1,c,c]<{$}>{\centering $}X[-1,c,c]<{$}}
  & & \mbox{Lattice Coordinates} & & \mbox{Cartesian Coordinates} &\mbox{Wyckoff Position} & \mbox{Atom Type} \\  
  \mathbf{B}_{1} & = & \left(\frac{1}{2} +z_{1}\right) \, \mathbf{a}_{1} + z_{1} \, \mathbf{a}_{2} + \frac{1}{2} \, \mathbf{a}_{3} & = & \frac{1}{2}b \, \mathbf{\hat{y}} + z_{1}c \, \mathbf{\hat{z}} & \left(4b\right) & \mbox{Mn} \\ 
\mathbf{B}_{2} & = & z_{1} \, \mathbf{a}_{1} + \left(\frac{1}{2} +z_{1}\right) \, \mathbf{a}_{2} + \frac{1}{2} \, \mathbf{a}_{3} & = & \frac{1}{2}a \, \mathbf{\hat{x}} + z_{1}c \, \mathbf{\hat{z}} & \left(4b\right) & \mbox{Mn} \\ 
\mathbf{B}_{3} & = & \left(y_{2}+z_{2}\right) \, \mathbf{a}_{1} + \left(x_{2}+z_{2}\right) \, \mathbf{a}_{2} + \left(x_{2}+y_{2}\right) \, \mathbf{a}_{3} & = & x_{2}a \, \mathbf{\hat{x}} + y_{2}b \, \mathbf{\hat{y}} + z_{2}c \, \mathbf{\hat{z}} & \left(8c\right) & \mbox{Ga} \\ 
\mathbf{B}_{4} & = & \left(-y_{2}+z_{2}\right) \, \mathbf{a}_{1} + \left(-x_{2}+z_{2}\right) \, \mathbf{a}_{2} + \left(-x_{2}-y_{2}\right) \, \mathbf{a}_{3} & = & -x_{2}a \, \mathbf{\hat{x}}-y_{2}b \, \mathbf{\hat{y}} + z_{2}c \, \mathbf{\hat{z}} & \left(8c\right) & \mbox{Ga} \\ 
\mathbf{B}_{5} & = & \left(\frac{1}{2} - y_{2} + z_{2}\right) \, \mathbf{a}_{1} + \left(\frac{1}{2} +x_{2} + z_{2}\right) \, \mathbf{a}_{2} + \left(x_{2}-y_{2}\right) \, \mathbf{a}_{3} & = & x_{2}a \, \mathbf{\hat{x}}-y_{2}b \, \mathbf{\hat{y}} + \left(\frac{1}{2} +z_{2}\right)c \, \mathbf{\hat{z}} & \left(8c\right) & \mbox{Ga} \\ 
\mathbf{B}_{6} & = & \left(\frac{1}{2} +y_{2} + z_{2}\right) \, \mathbf{a}_{1} + \left(\frac{1}{2} - x_{2} + z_{2}\right) \, \mathbf{a}_{2} + \left(-x_{2}+y_{2}\right) \, \mathbf{a}_{3} & = & -x_{2}a \, \mathbf{\hat{x}} + y_{2}b \, \mathbf{\hat{y}} + \left(\frac{1}{2} +z_{2}\right)c \, \mathbf{\hat{z}} & \left(8c\right) & \mbox{Ga} \\ 
\mathbf{B}_{7} & = & \left(y_{3}+z_{3}\right) \, \mathbf{a}_{1} + \left(x_{3}+z_{3}\right) \, \mathbf{a}_{2} + \left(x_{3}+y_{3}\right) \, \mathbf{a}_{3} & = & x_{3}a \, \mathbf{\hat{x}} + y_{3}b \, \mathbf{\hat{y}} + z_{3}c \, \mathbf{\hat{z}} & \left(8c\right) & \mbox{Sb} \\ 
\mathbf{B}_{8} & = & \left(-y_{3}+z_{3}\right) \, \mathbf{a}_{1} + \left(-x_{3}+z_{3}\right) \, \mathbf{a}_{2} + \left(-x_{3}-y_{3}\right) \, \mathbf{a}_{3} & = & -x_{3}a \, \mathbf{\hat{x}}-y_{3}b \, \mathbf{\hat{y}} + z_{3}c \, \mathbf{\hat{z}} & \left(8c\right) & \mbox{Sb} \\ 
\mathbf{B}_{9} & = & \left(\frac{1}{2} - y_{3} + z_{3}\right) \, \mathbf{a}_{1} + \left(\frac{1}{2} +x_{3} + z_{3}\right) \, \mathbf{a}_{2} + \left(x_{3}-y_{3}\right) \, \mathbf{a}_{3} & = & x_{3}a \, \mathbf{\hat{x}}-y_{3}b \, \mathbf{\hat{y}} + \left(\frac{1}{2} +z_{3}\right)c \, \mathbf{\hat{z}} & \left(8c\right) & \mbox{Sb} \\ 
\mathbf{B}_{10} & = & \left(\frac{1}{2} +y_{3} + z_{3}\right) \, \mathbf{a}_{1} + \left(\frac{1}{2} - x_{3} + z_{3}\right) \, \mathbf{a}_{2} + \left(-x_{3}+y_{3}\right) \, \mathbf{a}_{3} & = & -x_{3}a \, \mathbf{\hat{x}} + y_{3}b \, \mathbf{\hat{y}} + \left(\frac{1}{2} +z_{3}\right)c \, \mathbf{\hat{z}} & \left(8c\right) & \mbox{Sb} \\ 
\end{longtabu}
\renewcommand{\arraystretch}{1.0}
\noindent \hrulefill
\\
\textbf{References:}
\vspace*{-0.25cm}
\begin{flushleft}
  - \bibentry{Sakakibara_Ga2MnSb2_JCeramScJapn_2009}. \\
\end{flushleft}
\textbf{Found in:}
\vspace*{-0.25cm}
\begin{flushleft}
  - \bibentry{Villars_PearsonsCrystalData_2013}. \\
\end{flushleft}
\noindent \hrulefill
\\
\textbf{Geometry files:}
\\
\noindent  - CIF: pp. {\hyperref[A2BC2_oI20_45_c_b_c_cif]{\pageref{A2BC2_oI20_45_c_b_c_cif}}} \\
\noindent  - POSCAR: pp. {\hyperref[A2BC2_oI20_45_c_b_c_poscar]{\pageref{A2BC2_oI20_45_c_b_c_poscar}}} \\
\onecolumn
{\phantomsection\label{ABC_oI36_46_ac_bc_3b}}
\subsection*{\huge \textbf{{\normalfont TiFeSi Structure: ABC\_oI36\_46\_ac\_bc\_3b}}}
\noindent \hrulefill
\vspace*{0.25cm}
\begin{figure}[htp]
  \centering
  \vspace{-1em}
  {\includegraphics[width=1\textwidth]{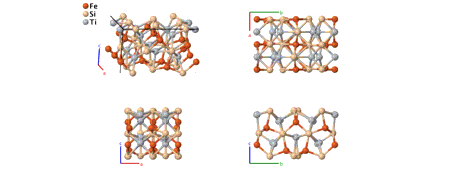}}
\end{figure}
\vspace*{-0.5cm}
\renewcommand{\arraystretch}{1.5}
\begin{equation*}
  \begin{array}{>{$\hspace{-0.15cm}}l<{$}>{$}p{0.5cm}<{$}>{$}p{18.5cm}<{$}}
    \mbox{\large \textbf{Prototype}} &\colon & \ce{TiFeSi} \\
    \mbox{\large \textbf{\AFLOW\ prototype label}} &\colon & \mbox{ABC\_oI36\_46\_ac\_bc\_3b} \\
    \mbox{\large \textbf{\textit{Strukturbericht} designation}} &\colon & \mbox{None} \\
    \mbox{\large \textbf{Pearson symbol}} &\colon & \mbox{oI36} \\
    \mbox{\large \textbf{Space group number}} &\colon & 46 \\
    \mbox{\large \textbf{Space group symbol}} &\colon & Ima2 \\
    \mbox{\large \textbf{\AFLOW\ prototype command}} &\colon &  \texttt{aflow} \,  \, \texttt{-{}-proto=ABC\_oI36\_46\_ac\_bc\_3b } \, \newline \texttt{-{}-params=}{a,b/a,c/a,z_{1},y_{2},z_{2},y_{3},z_{3},y_{4},z_{4},y_{5},z_{5},x_{6},y_{6},z_{6},x_{7},y_{7},z_{7} }
  \end{array}
\end{equation*}
\renewcommand{\arraystretch}{1.0}

\noindent \parbox{1 \linewidth}{
\noindent \hrulefill
\\
\textbf{Body-centered Orthorhombic primitive vectors:} \\
\vspace*{-0.25cm}
\begin{tabular}{cc}
  \begin{tabular}{c}
    \parbox{0.6 \linewidth}{
      \renewcommand{\arraystretch}{1.5}
      \begin{equation*}
        \centering
        \begin{array}{ccc}
              \mathbf{a}_1 & = & - \frac12 \, a \, \mathbf{\hat{x}} + \frac12 \, b \, \mathbf{\hat{y}} + \frac12 \, c \, \mathbf{\hat{z}} \\
    \mathbf{a}_2 & = & ~ \frac12 \, a \, \mathbf{\hat{x}} - \frac12 \, b \, \mathbf{\hat{y}} + \frac12 \, c \, \mathbf{\hat{z}} \\
    \mathbf{a}_3 & = & ~ \frac12 \, a \, \mathbf{\hat{x}} + \frac12 \, b \, \mathbf{\hat{y}} - \frac12 \, c \, \mathbf{\hat{z}} \\

        \end{array}
      \end{equation*}
    }
    \renewcommand{\arraystretch}{1.0}
  \end{tabular}
  \begin{tabular}{c}
    \includegraphics[width=0.3\linewidth]{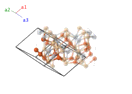} \\
  \end{tabular}
\end{tabular}

}
\vspace*{-0.25cm}

\noindent \hrulefill
\\
\textbf{Basis vectors:}
\vspace*{-0.25cm}
\renewcommand{\arraystretch}{1.5}
\begin{longtabu} to \textwidth{>{\centering $}X[-1,c,c]<{$}>{\centering $}X[-1,c,c]<{$}>{\centering $}X[-1,c,c]<{$}>{\centering $}X[-1,c,c]<{$}>{\centering $}X[-1,c,c]<{$}>{\centering $}X[-1,c,c]<{$}>{\centering $}X[-1,c,c]<{$}}
  & & \mbox{Lattice Coordinates} & & \mbox{Cartesian Coordinates} &\mbox{Wyckoff Position} & \mbox{Atom Type} \\  
  \mathbf{B}_{1} & = & z_{1} \, \mathbf{a}_{1} + z_{1} \, \mathbf{a}_{2} & = & z_{1}c \, \mathbf{\hat{z}} & \left(4a\right) & \mbox{Fe I} \\ 
\mathbf{B}_{2} & = & z_{1} \, \mathbf{a}_{1} + \left(\frac{1}{2} +z_{1}\right) \, \mathbf{a}_{2} + \frac{1}{2} \, \mathbf{a}_{3} & = & \frac{1}{2}a \, \mathbf{\hat{x}} + z_{1}c \, \mathbf{\hat{z}} & \left(4a\right) & \mbox{Fe I} \\ 
\mathbf{B}_{3} & = & \left(y_{2}+z_{2}\right) \, \mathbf{a}_{1} + \left(\frac{1}{4} +z_{2}\right) \, \mathbf{a}_{2} + \left(\frac{1}{4} +y_{2}\right) \, \mathbf{a}_{3} & = & \frac{1}{4}a \, \mathbf{\hat{x}} + y_{2}b \, \mathbf{\hat{y}} + z_{2}c \, \mathbf{\hat{z}} & \left(4b\right) & \mbox{Si I} \\ 
\mathbf{B}_{4} & = & \left(-y_{2}+z_{2}\right) \, \mathbf{a}_{1} + \left(\frac{3}{4} +z_{2}\right) \, \mathbf{a}_{2} + \left(\frac{3}{4} - y_{2}\right) \, \mathbf{a}_{3} & = & \frac{3}{4}a \, \mathbf{\hat{x}}-y_{2}b \, \mathbf{\hat{y}} + z_{2}c \, \mathbf{\hat{z}} & \left(4b\right) & \mbox{Si I} \\ 
\mathbf{B}_{5} & = & \left(y_{3}+z_{3}\right) \, \mathbf{a}_{1} + \left(\frac{1}{4} +z_{3}\right) \, \mathbf{a}_{2} + \left(\frac{1}{4} +y_{3}\right) \, \mathbf{a}_{3} & = & \frac{1}{4}a \, \mathbf{\hat{x}} + y_{3}b \, \mathbf{\hat{y}} + z_{3}c \, \mathbf{\hat{z}} & \left(4b\right) & \mbox{Ti I} \\ 
\mathbf{B}_{6} & = & \left(-y_{3}+z_{3}\right) \, \mathbf{a}_{1} + \left(\frac{3}{4} +z_{3}\right) \, \mathbf{a}_{2} + \left(\frac{3}{4} - y_{3}\right) \, \mathbf{a}_{3} & = & \frac{3}{4}a \, \mathbf{\hat{x}}-y_{3}b \, \mathbf{\hat{y}} + z_{3}c \, \mathbf{\hat{z}} & \left(4b\right) & \mbox{Ti I} \\ 
\mathbf{B}_{7} & = & \left(y_{4}+z_{4}\right) \, \mathbf{a}_{1} + \left(\frac{1}{4} +z_{4}\right) \, \mathbf{a}_{2} + \left(\frac{1}{4} +y_{4}\right) \, \mathbf{a}_{3} & = & \frac{1}{4}a \, \mathbf{\hat{x}} + y_{4}b \, \mathbf{\hat{y}} + z_{4}c \, \mathbf{\hat{z}} & \left(4b\right) & \mbox{Ti II} \\ 
\mathbf{B}_{8} & = & \left(-y_{4}+z_{4}\right) \, \mathbf{a}_{1} + \left(\frac{3}{4} +z_{4}\right) \, \mathbf{a}_{2} + \left(\frac{3}{4} - y_{4}\right) \, \mathbf{a}_{3} & = & \frac{3}{4}a \, \mathbf{\hat{x}}-y_{4}b \, \mathbf{\hat{y}} + z_{4}c \, \mathbf{\hat{z}} & \left(4b\right) & \mbox{Ti II} \\ 
\mathbf{B}_{9} & = & \left(y_{5}+z_{5}\right) \, \mathbf{a}_{1} + \left(\frac{1}{4} +z_{5}\right) \, \mathbf{a}_{2} + \left(\frac{1}{4} +y_{5}\right) \, \mathbf{a}_{3} & = & \frac{1}{4}a \, \mathbf{\hat{x}} + y_{5}b \, \mathbf{\hat{y}} + z_{5}c \, \mathbf{\hat{z}} & \left(4b\right) & \mbox{Ti III} \\ 
\mathbf{B}_{10} & = & \left(-y_{5}+z_{5}\right) \, \mathbf{a}_{1} + \left(\frac{3}{4} +z_{5}\right) \, \mathbf{a}_{2} + \left(\frac{3}{4} - y_{5}\right) \, \mathbf{a}_{3} & = & \frac{3}{4}a \, \mathbf{\hat{x}}-y_{5}b \, \mathbf{\hat{y}} + z_{5}c \, \mathbf{\hat{z}} & \left(4b\right) & \mbox{Ti III} \\ 
\mathbf{B}_{11} & = & \left(y_{6}+z_{6}\right) \, \mathbf{a}_{1} + \left(x_{6}+z_{6}\right) \, \mathbf{a}_{2} + \left(x_{6}+y_{6}\right) \, \mathbf{a}_{3} & = & x_{6}a \, \mathbf{\hat{x}} + y_{6}b \, \mathbf{\hat{y}} + z_{6}c \, \mathbf{\hat{z}} & \left(8c\right) & \mbox{Fe II} \\ 
\mathbf{B}_{12} & = & \left(-y_{6}+z_{6}\right) \, \mathbf{a}_{1} + \left(-x_{6}+z_{6}\right) \, \mathbf{a}_{2} + \left(-x_{6}-y_{6}\right) \, \mathbf{a}_{3} & = & -x_{6}a \, \mathbf{\hat{x}}-y_{6}b \, \mathbf{\hat{y}} + z_{6}c \, \mathbf{\hat{z}} & \left(8c\right) & \mbox{Fe II} \\ 
\mathbf{B}_{13} & = & \left(-y_{6}+z_{6}\right) \, \mathbf{a}_{1} + \left(\frac{1}{2} +x_{6} + z_{6}\right) \, \mathbf{a}_{2} + \left(\frac{1}{2} +x_{6} - y_{6}\right) \, \mathbf{a}_{3} & = & \left(\frac{1}{2} +x_{6}\right)a \, \mathbf{\hat{x}}-y_{6}b \, \mathbf{\hat{y}} + z_{6}c \, \mathbf{\hat{z}} & \left(8c\right) & \mbox{Fe II} \\ 
\mathbf{B}_{14} & = & \left(y_{6}+z_{6}\right) \, \mathbf{a}_{1} + \left(\frac{1}{2} - x_{6} + z_{6}\right) \, \mathbf{a}_{2} + \left(\frac{1}{2} - x_{6} + y_{6}\right) \, \mathbf{a}_{3} & = & \left(\frac{1}{2} - x_{6}\right)a \, \mathbf{\hat{x}} + y_{6}b \, \mathbf{\hat{y}} + z_{6}c \, \mathbf{\hat{z}} & \left(8c\right) & \mbox{Fe II} \\ 
\mathbf{B}_{15} & = & \left(y_{7}+z_{7}\right) \, \mathbf{a}_{1} + \left(x_{7}+z_{7}\right) \, \mathbf{a}_{2} + \left(x_{7}+y_{7}\right) \, \mathbf{a}_{3} & = & x_{7}a \, \mathbf{\hat{x}} + y_{7}b \, \mathbf{\hat{y}} + z_{7}c \, \mathbf{\hat{z}} & \left(8c\right) & \mbox{Si II} \\ 
\mathbf{B}_{16} & = & \left(-y_{7}+z_{7}\right) \, \mathbf{a}_{1} + \left(-x_{7}+z_{7}\right) \, \mathbf{a}_{2} + \left(-x_{7}-y_{7}\right) \, \mathbf{a}_{3} & = & -x_{7}a \, \mathbf{\hat{x}}-y_{7}b \, \mathbf{\hat{y}} + z_{7}c \, \mathbf{\hat{z}} & \left(8c\right) & \mbox{Si II} \\ 
\mathbf{B}_{17} & = & \left(-y_{7}+z_{7}\right) \, \mathbf{a}_{1} + \left(\frac{1}{2} +x_{7} + z_{7}\right) \, \mathbf{a}_{2} + \left(\frac{1}{2} +x_{7} - y_{7}\right) \, \mathbf{a}_{3} & = & \left(\frac{1}{2} +x_{7}\right)a \, \mathbf{\hat{x}}-y_{7}b \, \mathbf{\hat{y}} + z_{7}c \, \mathbf{\hat{z}} & \left(8c\right) & \mbox{Si II} \\ 
\mathbf{B}_{18} & = & \left(y_{7}+z_{7}\right) \, \mathbf{a}_{1} + \left(\frac{1}{2} - x_{7} + z_{7}\right) \, \mathbf{a}_{2} + \left(\frac{1}{2} - x_{7} + y_{7}\right) \, \mathbf{a}_{3} & = & \left(\frac{1}{2} - x_{7}\right)a \, \mathbf{\hat{x}} + y_{7}b \, \mathbf{\hat{y}} + z_{7}c \, \mathbf{\hat{z}} & \left(8c\right) & \mbox{Si II} \\ 
\end{longtabu}
\renewcommand{\arraystretch}{1.0}
\noindent \hrulefill
\\
\textbf{References:}
\vspace*{-0.25cm}
\begin{flushleft}
  - \bibentry{Jeitschko_TiFeSi_ActCrystallogSecB_1970}. \\
\end{flushleft}
\textbf{Found in:}
\vspace*{-0.25cm}
\begin{flushleft}
  - \bibentry{Villars_PearsonsCrystalData_2013}. \\
\end{flushleft}
\noindent \hrulefill
\\
\textbf{Geometry files:}
\\
\noindent  - CIF: pp. {\hyperref[ABC_oI36_46_ac_bc_3b_cif]{\pageref{ABC_oI36_46_ac_bc_3b_cif}}} \\
\noindent  - POSCAR: pp. {\hyperref[ABC_oI36_46_ac_bc_3b_poscar]{\pageref{ABC_oI36_46_ac_bc_3b_poscar}}} \\
\onecolumn
{\phantomsection\label{A2B8CD_oP24_48_k_2m_d_b}}
\subsection*{\huge \textbf{{\normalfont $\alpha$-RbPr[MoO$_{4}$]$_{2}$ Structure: A2B8CD\_oP24\_48\_k\_2m\_d\_b}}}
\noindent \hrulefill
\vspace*{0.25cm}
\begin{figure}[htp]
  \centering
  \vspace{-1em}
  {\includegraphics[width=1\textwidth]{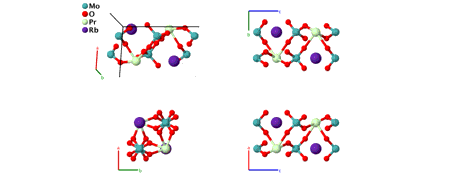}}
\end{figure}
\vspace*{-0.5cm}
\renewcommand{\arraystretch}{1.5}
\begin{equation*}
  \begin{array}{>{$\hspace{-0.15cm}}l<{$}>{$}p{0.5cm}<{$}>{$}p{18.5cm}<{$}}
    \mbox{\large \textbf{Prototype}} &\colon & \ce{$\alpha$-RbPr[MoO4]2} \\
    \mbox{\large \textbf{\AFLOW\ prototype label}} &\colon & \mbox{A2B8CD\_oP24\_48\_k\_2m\_d\_b} \\
    \mbox{\large \textbf{\textit{Strukturbericht} designation}} &\colon & \mbox{None} \\
    \mbox{\large \textbf{Pearson symbol}} &\colon & \mbox{oP24} \\
    \mbox{\large \textbf{Space group number}} &\colon & 48 \\
    \mbox{\large \textbf{Space group symbol}} &\colon & Pnnn \\
    \mbox{\large \textbf{\AFLOW\ prototype command}} &\colon &  \texttt{aflow} \,  \, \texttt{-{}-proto=A2B8CD\_oP24\_48\_k\_2m\_d\_b } \, \newline \texttt{-{}-params=}{a,b/a,c/a,z_{3},x_{4},y_{4},z_{4},x_{5},y_{5},z_{5} }
  \end{array}
\end{equation*}
\renewcommand{\arraystretch}{1.0}

\noindent \parbox{1 \linewidth}{
\noindent \hrulefill
\\
\textbf{Simple Orthorhombic primitive vectors:} \\
\vspace*{-0.25cm}
\begin{tabular}{cc}
  \begin{tabular}{c}
    \parbox{0.6 \linewidth}{
      \renewcommand{\arraystretch}{1.5}
      \begin{equation*}
        \centering
        \begin{array}{ccc}
              \mathbf{a}_1 & = & a \, \mathbf{\hat{x}} \\
    \mathbf{a}_2 & = & b \, \mathbf{\hat{y}} \\
    \mathbf{a}_3 & = & c \, \mathbf{\hat{z}} \\

        \end{array}
      \end{equation*}
    }
    \renewcommand{\arraystretch}{1.0}
  \end{tabular}
  \begin{tabular}{c}
    \includegraphics[width=0.3\linewidth]{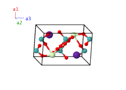} \\
  \end{tabular}
\end{tabular}

}
\vspace*{-0.25cm}

\noindent \hrulefill
\\
\textbf{Basis vectors:}
\vspace*{-0.25cm}
\renewcommand{\arraystretch}{1.5}
\begin{longtabu} to \textwidth{>{\centering $}X[-1,c,c]<{$}>{\centering $}X[-1,c,c]<{$}>{\centering $}X[-1,c,c]<{$}>{\centering $}X[-1,c,c]<{$}>{\centering $}X[-1,c,c]<{$}>{\centering $}X[-1,c,c]<{$}>{\centering $}X[-1,c,c]<{$}}
  & & \mbox{Lattice Coordinates} & & \mbox{Cartesian Coordinates} &\mbox{Wyckoff Position} & \mbox{Atom Type} \\  
  \mathbf{B}_{1} & = & \frac{3}{4} \, \mathbf{a}_{1} + \frac{1}{4} \, \mathbf{a}_{2} + \frac{1}{4} \, \mathbf{a}_{3} & = & \frac{3}{4}a \, \mathbf{\hat{x}} + \frac{1}{4}b \, \mathbf{\hat{y}} + \frac{1}{4}c \, \mathbf{\hat{z}} & \left(2b\right) & \mbox{Rb} \\ 
\mathbf{B}_{2} & = & \frac{1}{4} \, \mathbf{a}_{1} + \frac{3}{4} \, \mathbf{a}_{2} + \frac{3}{4} \, \mathbf{a}_{3} & = & \frac{1}{4}a \, \mathbf{\hat{x}} + \frac{3}{4}b \, \mathbf{\hat{y}} + \frac{3}{4}c \, \mathbf{\hat{z}} & \left(2b\right) & \mbox{Rb} \\ 
\mathbf{B}_{3} & = & \frac{1}{4} \, \mathbf{a}_{1} + \frac{3}{4} \, \mathbf{a}_{2} + \frac{1}{4} \, \mathbf{a}_{3} & = & \frac{1}{4}a \, \mathbf{\hat{x}} + \frac{3}{4}b \, \mathbf{\hat{y}} + \frac{1}{4}c \, \mathbf{\hat{z}} & \left(2d\right) & \mbox{Pr} \\ 
\mathbf{B}_{4} & = & \frac{3}{4} \, \mathbf{a}_{1} + \frac{1}{4} \, \mathbf{a}_{2} + \frac{3}{4} \, \mathbf{a}_{3} & = & \frac{3}{4}a \, \mathbf{\hat{x}} + \frac{1}{4}b \, \mathbf{\hat{y}} + \frac{3}{4}c \, \mathbf{\hat{z}} & \left(2d\right) & \mbox{Pr} \\ 
\mathbf{B}_{5} & = & \frac{1}{4} \, \mathbf{a}_{1} + \frac{1}{4} \, \mathbf{a}_{2} + z_{3} \, \mathbf{a}_{3} & = & \frac{1}{4}a \, \mathbf{\hat{x}} + \frac{1}{4}b \, \mathbf{\hat{y}} + z_{3}c \, \mathbf{\hat{z}} & \left(4k\right) & \mbox{Mo} \\ 
\mathbf{B}_{6} & = & \frac{1}{4} \, \mathbf{a}_{1} + \frac{1}{4} \, \mathbf{a}_{2} + \left(\frac{1}{2} - z_{3}\right) \, \mathbf{a}_{3} & = & \frac{1}{4}a \, \mathbf{\hat{x}} + \frac{1}{4}b \, \mathbf{\hat{y}} + \left(\frac{1}{2} - z_{3}\right)c \, \mathbf{\hat{z}} & \left(4k\right) & \mbox{Mo} \\ 
\mathbf{B}_{7} & = & \frac{3}{4} \, \mathbf{a}_{1} + \frac{3}{4} \, \mathbf{a}_{2}-z_{3} \, \mathbf{a}_{3} & = & \frac{3}{4}a \, \mathbf{\hat{x}} + \frac{3}{4}b \, \mathbf{\hat{y}}-z_{3}c \, \mathbf{\hat{z}} & \left(4k\right) & \mbox{Mo} \\ 
\mathbf{B}_{8} & = & \frac{3}{4} \, \mathbf{a}_{1} + \frac{3}{4} \, \mathbf{a}_{2} + \left(\frac{1}{2} +z_{3}\right) \, \mathbf{a}_{3} & = & \frac{3}{4}a \, \mathbf{\hat{x}} + \frac{3}{4}b \, \mathbf{\hat{y}} + \left(\frac{1}{2} +z_{3}\right)c \, \mathbf{\hat{z}} & \left(4k\right) & \mbox{Mo} \\ 
\mathbf{B}_{9} & = & x_{4} \, \mathbf{a}_{1} + y_{4} \, \mathbf{a}_{2} + z_{4} \, \mathbf{a}_{3} & = & x_{4}a \, \mathbf{\hat{x}} + y_{4}b \, \mathbf{\hat{y}} + z_{4}c \, \mathbf{\hat{z}} & \left(8m\right) & \mbox{O I} \\ 
\mathbf{B}_{10} & = & \left(\frac{1}{2} - x_{4}\right) \, \mathbf{a}_{1} + \left(\frac{1}{2} - y_{4}\right) \, \mathbf{a}_{2} + z_{4} \, \mathbf{a}_{3} & = & \left(\frac{1}{2} - x_{4}\right)a \, \mathbf{\hat{x}} + \left(\frac{1}{2} - y_{4}\right)b \, \mathbf{\hat{y}} + z_{4}c \, \mathbf{\hat{z}} & \left(8m\right) & \mbox{O I} \\ 
\mathbf{B}_{11} & = & \left(\frac{1}{2} - x_{4}\right) \, \mathbf{a}_{1} + y_{4} \, \mathbf{a}_{2} + \left(\frac{1}{2} - z_{4}\right) \, \mathbf{a}_{3} & = & \left(\frac{1}{2} - x_{4}\right)a \, \mathbf{\hat{x}} + y_{4}b \, \mathbf{\hat{y}} + \left(\frac{1}{2} - z_{4}\right)c \, \mathbf{\hat{z}} & \left(8m\right) & \mbox{O I} \\ 
\mathbf{B}_{12} & = & x_{4} \, \mathbf{a}_{1} + \left(\frac{1}{2} - y_{4}\right) \, \mathbf{a}_{2} + \left(\frac{1}{2} - z_{4}\right) \, \mathbf{a}_{3} & = & x_{4}a \, \mathbf{\hat{x}} + \left(\frac{1}{2} - y_{4}\right)b \, \mathbf{\hat{y}} + \left(\frac{1}{2} - z_{4}\right)c \, \mathbf{\hat{z}} & \left(8m\right) & \mbox{O I} \\ 
\mathbf{B}_{13} & = & -x_{4} \, \mathbf{a}_{1}-y_{4} \, \mathbf{a}_{2}-z_{4} \, \mathbf{a}_{3} & = & -x_{4}a \, \mathbf{\hat{x}}-y_{4}b \, \mathbf{\hat{y}}-z_{4}c \, \mathbf{\hat{z}} & \left(8m\right) & \mbox{O I} \\ 
\mathbf{B}_{14} & = & \left(\frac{1}{2} +x_{4}\right) \, \mathbf{a}_{1} + \left(\frac{1}{2} +y_{4}\right) \, \mathbf{a}_{2}-z_{4} \, \mathbf{a}_{3} & = & \left(\frac{1}{2} +x_{4}\right)a \, \mathbf{\hat{x}} + \left(\frac{1}{2} +y_{4}\right)b \, \mathbf{\hat{y}}-z_{4}c \, \mathbf{\hat{z}} & \left(8m\right) & \mbox{O I} \\ 
\mathbf{B}_{15} & = & \left(\frac{1}{2} +x_{4}\right) \, \mathbf{a}_{1}-y_{4} \, \mathbf{a}_{2} + \left(\frac{1}{2} +z_{4}\right) \, \mathbf{a}_{3} & = & \left(\frac{1}{2} +x_{4}\right)a \, \mathbf{\hat{x}}-y_{4}b \, \mathbf{\hat{y}} + \left(\frac{1}{2} +z_{4}\right)c \, \mathbf{\hat{z}} & \left(8m\right) & \mbox{O I} \\ 
\mathbf{B}_{16} & = & -x_{4} \, \mathbf{a}_{1} + \left(\frac{1}{2} +y_{4}\right) \, \mathbf{a}_{2} + \left(\frac{1}{2} +z_{4}\right) \, \mathbf{a}_{3} & = & -x_{4}a \, \mathbf{\hat{x}} + \left(\frac{1}{2} +y_{4}\right)b \, \mathbf{\hat{y}} + \left(\frac{1}{2} +z_{4}\right)c \, \mathbf{\hat{z}} & \left(8m\right) & \mbox{O I} \\ 
\mathbf{B}_{17} & = & x_{5} \, \mathbf{a}_{1} + y_{5} \, \mathbf{a}_{2} + z_{5} \, \mathbf{a}_{3} & = & x_{5}a \, \mathbf{\hat{x}} + y_{5}b \, \mathbf{\hat{y}} + z_{5}c \, \mathbf{\hat{z}} & \left(8m\right) & \mbox{O II} \\ 
\mathbf{B}_{18} & = & \left(\frac{1}{2} - x_{5}\right) \, \mathbf{a}_{1} + \left(\frac{1}{2} - y_{5}\right) \, \mathbf{a}_{2} + z_{5} \, \mathbf{a}_{3} & = & \left(\frac{1}{2} - x_{5}\right)a \, \mathbf{\hat{x}} + \left(\frac{1}{2} - y_{5}\right)b \, \mathbf{\hat{y}} + z_{5}c \, \mathbf{\hat{z}} & \left(8m\right) & \mbox{O II} \\ 
\mathbf{B}_{19} & = & \left(\frac{1}{2} - x_{5}\right) \, \mathbf{a}_{1} + y_{5} \, \mathbf{a}_{2} + \left(\frac{1}{2} - z_{5}\right) \, \mathbf{a}_{3} & = & \left(\frac{1}{2} - x_{5}\right)a \, \mathbf{\hat{x}} + y_{5}b \, \mathbf{\hat{y}} + \left(\frac{1}{2} - z_{5}\right)c \, \mathbf{\hat{z}} & \left(8m\right) & \mbox{O II} \\ 
\mathbf{B}_{20} & = & x_{5} \, \mathbf{a}_{1} + \left(\frac{1}{2} - y_{5}\right) \, \mathbf{a}_{2} + \left(\frac{1}{2} - z_{5}\right) \, \mathbf{a}_{3} & = & x_{5}a \, \mathbf{\hat{x}} + \left(\frac{1}{2} - y_{5}\right)b \, \mathbf{\hat{y}} + \left(\frac{1}{2} - z_{5}\right)c \, \mathbf{\hat{z}} & \left(8m\right) & \mbox{O II} \\ 
\mathbf{B}_{21} & = & -x_{5} \, \mathbf{a}_{1}-y_{5} \, \mathbf{a}_{2}-z_{5} \, \mathbf{a}_{3} & = & -x_{5}a \, \mathbf{\hat{x}}-y_{5}b \, \mathbf{\hat{y}}-z_{5}c \, \mathbf{\hat{z}} & \left(8m\right) & \mbox{O II} \\ 
\mathbf{B}_{22} & = & \left(\frac{1}{2} +x_{5}\right) \, \mathbf{a}_{1} + \left(\frac{1}{2} +y_{5}\right) \, \mathbf{a}_{2}-z_{5} \, \mathbf{a}_{3} & = & \left(\frac{1}{2} +x_{5}\right)a \, \mathbf{\hat{x}} + \left(\frac{1}{2} +y_{5}\right)b \, \mathbf{\hat{y}}-z_{5}c \, \mathbf{\hat{z}} & \left(8m\right) & \mbox{O II} \\ 
\mathbf{B}_{23} & = & \left(\frac{1}{2} +x_{5}\right) \, \mathbf{a}_{1}-y_{5} \, \mathbf{a}_{2} + \left(\frac{1}{2} +z_{5}\right) \, \mathbf{a}_{3} & = & \left(\frac{1}{2} +x_{5}\right)a \, \mathbf{\hat{x}}-y_{5}b \, \mathbf{\hat{y}} + \left(\frac{1}{2} +z_{5}\right)c \, \mathbf{\hat{z}} & \left(8m\right) & \mbox{O II} \\ 
\mathbf{B}_{24} & = & -x_{5} \, \mathbf{a}_{1} + \left(\frac{1}{2} +y_{5}\right) \, \mathbf{a}_{2} + \left(\frac{1}{2} +z_{5}\right) \, \mathbf{a}_{3} & = & -x_{5}a \, \mathbf{\hat{x}} + \left(\frac{1}{2} +y_{5}\right)b \, \mathbf{\hat{y}} + \left(\frac{1}{2} +z_{5}\right)c \, \mathbf{\hat{z}} & \left(8m\right) & \mbox{O II} \\ 
\end{longtabu}
\renewcommand{\arraystretch}{1.0}
\noindent \hrulefill
\\
\textbf{References:}
\vspace*{-0.25cm}
\begin{flushleft}
  - \bibentry{Klevtsov_RbPrMoO42_SovPhysCrystallog_1970}. \\
\end{flushleft}
\textbf{Found in:}
\vspace*{-0.25cm}
\begin{flushleft}
  - \bibentry{Villars_PearsonsCrystalData_2013}. \\
\end{flushleft}
\noindent \hrulefill
\\
\textbf{Geometry files:}
\\
\noindent  - CIF: pp. {\hyperref[A2B8CD_oP24_48_k_2m_d_b_cif]{\pageref{A2B8CD_oP24_48_k_2m_d_b_cif}}} \\
\noindent  - POSCAR: pp. {\hyperref[A2B8CD_oP24_48_k_2m_d_b_poscar]{\pageref{A2B8CD_oP24_48_k_2m_d_b_poscar}}} \\
\onecolumn
{\phantomsection\label{A5B2_oP14_49_dehq_ab}}
\subsection*{\huge \textbf{{\normalfont $\beta$-Ta$_{2}$O$_{5}$ Structure: A5B2\_oP14\_49\_dehq\_ab}}}
\noindent \hrulefill
\vspace*{0.25cm}
\begin{figure}[htp]
  \centering
  \vspace{-1em}
  {\includegraphics[width=1\textwidth]{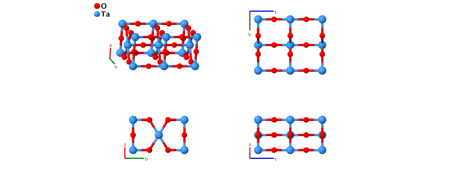}}
\end{figure}
\vspace*{-0.5cm}
\renewcommand{\arraystretch}{1.5}
\begin{equation*}
  \begin{array}{>{$\hspace{-0.15cm}}l<{$}>{$}p{0.5cm}<{$}>{$}p{18.5cm}<{$}}
    \mbox{\large \textbf{Prototype}} &\colon & \ce{$\beta$-Ta2O5} \\
    \mbox{\large \textbf{\AFLOW\ prototype label}} &\colon & \mbox{A5B2\_oP14\_49\_dehq\_ab} \\
    \mbox{\large \textbf{\textit{Strukturbericht} designation}} &\colon & \mbox{None} \\
    \mbox{\large \textbf{Pearson symbol}} &\colon & \mbox{oP14} \\
    \mbox{\large \textbf{Space group number}} &\colon & 49 \\
    \mbox{\large \textbf{Space group symbol}} &\colon & Pccm \\
    \mbox{\large \textbf{\AFLOW\ prototype command}} &\colon &  \texttt{aflow} \,  \, \texttt{-{}-proto=A5B2\_oP14\_49\_dehq\_ab } \, \newline \texttt{-{}-params=}{a,b/a,c/a,x_{6},y_{6} }
  \end{array}
\end{equation*}
\renewcommand{\arraystretch}{1.0}

\vspace*{-0.25cm}
\noindent \hrulefill
\begin{itemize}
  \item{While {\small FINDSYM} identifies space group \#49 for this structure (consistent with the reference), {\small AFLOW-SYM} and Platon identify \#47.
Lowering the tolerance value for {\small AFLOW-SYM} resolves the expected space group \#49.
Space groups \#47 and \#49 are both reasonable classifications since they are commensurate with subgroup relations.
}
\end{itemize}

\noindent \parbox{1 \linewidth}{
\noindent \hrulefill
\\
\textbf{Simple Orthorhombic primitive vectors:} \\
\vspace*{-0.25cm}
\begin{tabular}{cc}
  \begin{tabular}{c}
    \parbox{0.6 \linewidth}{
      \renewcommand{\arraystretch}{1.5}
      \begin{equation*}
        \centering
        \begin{array}{ccc}
              \mathbf{a}_1 & = & a \, \mathbf{\hat{x}} \\
    \mathbf{a}_2 & = & b \, \mathbf{\hat{y}} \\
    \mathbf{a}_3 & = & c \, \mathbf{\hat{z}} \\

        \end{array}
      \end{equation*}
    }
    \renewcommand{\arraystretch}{1.0}
  \end{tabular}
  \begin{tabular}{c}
    \includegraphics[width=0.3\linewidth]{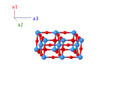} \\
  \end{tabular}
\end{tabular}

}
\vspace*{-0.25cm}

\noindent \hrulefill
\\
\textbf{Basis vectors:}
\vspace*{-0.25cm}
\renewcommand{\arraystretch}{1.5}
\begin{longtabu} to \textwidth{>{\centering $}X[-1,c,c]<{$}>{\centering $}X[-1,c,c]<{$}>{\centering $}X[-1,c,c]<{$}>{\centering $}X[-1,c,c]<{$}>{\centering $}X[-1,c,c]<{$}>{\centering $}X[-1,c,c]<{$}>{\centering $}X[-1,c,c]<{$}}
  & & \mbox{Lattice Coordinates} & & \mbox{Cartesian Coordinates} &\mbox{Wyckoff Position} & \mbox{Atom Type} \\  
  \mathbf{B}_{1} & = & 0 \, \mathbf{a}_{1} + 0 \, \mathbf{a}_{2} + 0 \, \mathbf{a}_{3} & = & 0 \, \mathbf{\hat{x}} + 0 \, \mathbf{\hat{y}} + 0 \, \mathbf{\hat{z}} & \left(2a\right) & \mbox{Ta I} \\ 
\mathbf{B}_{2} & = & \frac{1}{2} \, \mathbf{a}_{3} & = & \frac{1}{2}c \, \mathbf{\hat{z}} & \left(2a\right) & \mbox{Ta I} \\ 
\mathbf{B}_{3} & = & \frac{1}{2} \, \mathbf{a}_{1} + \frac{1}{2} \, \mathbf{a}_{2} & = & \frac{1}{2}a \, \mathbf{\hat{x}} + \frac{1}{2}b \, \mathbf{\hat{y}} & \left(2b\right) & \mbox{Ta II} \\ 
\mathbf{B}_{4} & = & \frac{1}{2} \, \mathbf{a}_{1} + \frac{1}{2} \, \mathbf{a}_{2} + \frac{1}{2} \, \mathbf{a}_{3} & = & \frac{1}{2}a \, \mathbf{\hat{x}} + \frac{1}{2}b \, \mathbf{\hat{y}} + \frac{1}{2}c \, \mathbf{\hat{z}} & \left(2b\right) & \mbox{Ta II} \\ 
\mathbf{B}_{5} & = & \frac{1}{2} \, \mathbf{a}_{1} & = & \frac{1}{2}a \, \mathbf{\hat{x}} & \left(2d\right) & \mbox{O I} \\ 
\mathbf{B}_{6} & = & \frac{1}{2} \, \mathbf{a}_{1} + \frac{1}{2} \, \mathbf{a}_{3} & = & \frac{1}{2}a \, \mathbf{\hat{x}} + \frac{1}{2}c \, \mathbf{\hat{z}} & \left(2d\right) & \mbox{O I} \\ 
\mathbf{B}_{7} & = & \frac{1}{4} \, \mathbf{a}_{3} & = & \frac{1}{4}c \, \mathbf{\hat{z}} & \left(2e\right) & \mbox{O II} \\ 
\mathbf{B}_{8} & = & \frac{3}{4} \, \mathbf{a}_{3} & = & \frac{3}{4}c \, \mathbf{\hat{z}} & \left(2e\right) & \mbox{O II} \\ 
\mathbf{B}_{9} & = & \frac{1}{2} \, \mathbf{a}_{1} + \frac{1}{2} \, \mathbf{a}_{2} + \frac{1}{4} \, \mathbf{a}_{3} & = & \frac{1}{2}a \, \mathbf{\hat{x}} + \frac{1}{2}b \, \mathbf{\hat{y}} + \frac{1}{4}c \, \mathbf{\hat{z}} & \left(2h\right) & \mbox{O III} \\ 
\mathbf{B}_{10} & = & \frac{1}{2} \, \mathbf{a}_{1} + \frac{1}{2} \, \mathbf{a}_{2} + \frac{3}{4} \, \mathbf{a}_{3} & = & \frac{1}{2}a \, \mathbf{\hat{x}} + \frac{1}{2}b \, \mathbf{\hat{y}} + \frac{3}{4}c \, \mathbf{\hat{z}} & \left(2h\right) & \mbox{O III} \\ 
\mathbf{B}_{11} & = & x_{6} \, \mathbf{a}_{1} + y_{6} \, \mathbf{a}_{2} & = & x_{6}a \, \mathbf{\hat{x}} + y_{6}b \, \mathbf{\hat{y}} & \left(4q\right) & \mbox{O IV} \\ 
\mathbf{B}_{12} & = & -x_{6} \, \mathbf{a}_{1}-y_{6} \, \mathbf{a}_{2} & = & -x_{6}a \, \mathbf{\hat{x}}-y_{6}b \, \mathbf{\hat{y}} & \left(4q\right) & \mbox{O IV} \\ 
\mathbf{B}_{13} & = & -x_{6} \, \mathbf{a}_{1} + y_{6} \, \mathbf{a}_{2} + \frac{1}{2} \, \mathbf{a}_{3} & = & -x_{6}a \, \mathbf{\hat{x}} + y_{6}b \, \mathbf{\hat{y}} + \frac{1}{2}c \, \mathbf{\hat{z}} & \left(4q\right) & \mbox{O IV} \\ 
\mathbf{B}_{14} & = & x_{6} \, \mathbf{a}_{1}-y_{6} \, \mathbf{a}_{2} + \frac{1}{2} \, \mathbf{a}_{3} & = & x_{6}a \, \mathbf{\hat{x}}-y_{6}b \, \mathbf{\hat{y}} + \frac{1}{2}c \, \mathbf{\hat{z}} & \left(4q\right) & \mbox{O IV} \\ 
\end{longtabu}
\renewcommand{\arraystretch}{1.0}
\noindent \hrulefill
\\
\textbf{References:}
\vspace*{-0.25cm}
\begin{flushleft}
  - \bibentry{Aleshina_Ta2O5_CrystallogRep_2002}. \\
  - \bibentry{stokes_findsym}. \\
  - \bibentry{aflowsym_2018}. \\
  - \bibentry{platon_2003}. \\
\end{flushleft}
\textbf{Found in:}
\vspace*{-0.25cm}
\begin{flushleft}
  - \bibentry{Villars_PearsonsCrystalData_2013}. \\
\end{flushleft}
\noindent \hrulefill
\\
\textbf{Geometry files:}
\\
\noindent  - CIF: pp. {\hyperref[A5B2_oP14_49_dehq_ab_cif]{\pageref{A5B2_oP14_49_dehq_ab_cif}}} \\
\noindent  - POSCAR: pp. {\hyperref[A5B2_oP14_49_dehq_ab_poscar]{\pageref{A5B2_oP14_49_dehq_ab_poscar}}} \\
\onecolumn
{\phantomsection\label{AB2C8D_oP24_49_g_q_2qr_e}}
\subsection*{\huge \textbf{{\normalfont CsPr[MoO$_{4}$]$_{2}$ Structure: AB2C8D\_oP24\_49\_g\_q\_2qr\_e}}}
\noindent \hrulefill
\vspace*{0.25cm}
\begin{figure}[htp]
  \centering
  \vspace{-1em}
  {\includegraphics[width=1\textwidth]{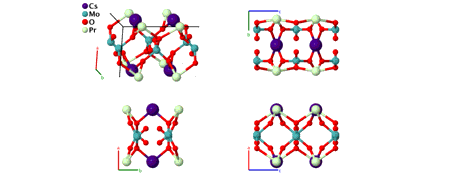}}
\end{figure}
\vspace*{-0.5cm}
\renewcommand{\arraystretch}{1.5}
\begin{equation*}
  \begin{array}{>{$\hspace{-0.15cm}}l<{$}>{$}p{0.5cm}<{$}>{$}p{18.5cm}<{$}}
    \mbox{\large \textbf{Prototype}} &\colon & \ce{CsPr[MoO4]2} \\
    \mbox{\large \textbf{\AFLOW\ prototype label}} &\colon & \mbox{AB2C8D\_oP24\_49\_g\_q\_2qr\_e} \\
    \mbox{\large \textbf{\textit{Strukturbericht} designation}} &\colon & \mbox{None} \\
    \mbox{\large \textbf{Pearson symbol}} &\colon & \mbox{oP24} \\
    \mbox{\large \textbf{Space group number}} &\colon & 49 \\
    \mbox{\large \textbf{Space group symbol}} &\colon & Pccm \\
    \mbox{\large \textbf{\AFLOW\ prototype command}} &\colon &  \texttt{aflow} \,  \, \texttt{-{}-proto=AB2C8D\_oP24\_49\_g\_q\_2qr\_e } \, \newline \texttt{-{}-params=}{a,b/a,c/a,x_{3},y_{3},x_{4},y_{4},x_{5},y_{5},x_{6},y_{6},z_{6} }
  \end{array}
\end{equation*}
\renewcommand{\arraystretch}{1.0}

\noindent \parbox{1 \linewidth}{
\noindent \hrulefill
\\
\textbf{Simple Orthorhombic primitive vectors:} \\
\vspace*{-0.25cm}
\begin{tabular}{cc}
  \begin{tabular}{c}
    \parbox{0.6 \linewidth}{
      \renewcommand{\arraystretch}{1.5}
      \begin{equation*}
        \centering
        \begin{array}{ccc}
              \mathbf{a}_1 & = & a \, \mathbf{\hat{x}} \\
    \mathbf{a}_2 & = & b \, \mathbf{\hat{y}} \\
    \mathbf{a}_3 & = & c \, \mathbf{\hat{z}} \\

        \end{array}
      \end{equation*}
    }
    \renewcommand{\arraystretch}{1.0}
  \end{tabular}
  \begin{tabular}{c}
    \includegraphics[width=0.3\linewidth]{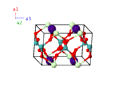} \\
  \end{tabular}
\end{tabular}

}
\vspace*{-0.25cm}

\noindent \hrulefill
\\
\textbf{Basis vectors:}
\vspace*{-0.25cm}
\renewcommand{\arraystretch}{1.5}
\begin{longtabu} to \textwidth{>{\centering $}X[-1,c,c]<{$}>{\centering $}X[-1,c,c]<{$}>{\centering $}X[-1,c,c]<{$}>{\centering $}X[-1,c,c]<{$}>{\centering $}X[-1,c,c]<{$}>{\centering $}X[-1,c,c]<{$}>{\centering $}X[-1,c,c]<{$}}
  & & \mbox{Lattice Coordinates} & & \mbox{Cartesian Coordinates} &\mbox{Wyckoff Position} & \mbox{Atom Type} \\  
  \mathbf{B}_{1} & = & \frac{1}{4} \, \mathbf{a}_{3} & = & \frac{1}{4}c \, \mathbf{\hat{z}} & \left(2e\right) & \mbox{Pr} \\ 
\mathbf{B}_{2} & = & \frac{3}{4} \, \mathbf{a}_{3} & = & \frac{3}{4}c \, \mathbf{\hat{z}} & \left(2e\right) & \mbox{Pr} \\ 
\mathbf{B}_{3} & = & \frac{1}{2} \, \mathbf{a}_{2} + \frac{1}{4} \, \mathbf{a}_{3} & = & \frac{1}{2}b \, \mathbf{\hat{y}} + \frac{1}{4}c \, \mathbf{\hat{z}} & \left(2g\right) & \mbox{Cs} \\ 
\mathbf{B}_{4} & = & \frac{1}{2} \, \mathbf{a}_{2} + \frac{3}{4} \, \mathbf{a}_{3} & = & \frac{1}{2}b \, \mathbf{\hat{y}} + \frac{3}{4}c \, \mathbf{\hat{z}} & \left(2g\right) & \mbox{Cs} \\ 
\mathbf{B}_{5} & = & x_{3} \, \mathbf{a}_{1} + y_{3} \, \mathbf{a}_{2} & = & x_{3}a \, \mathbf{\hat{x}} + y_{3}b \, \mathbf{\hat{y}} & \left(4q\right) & \mbox{Mo} \\ 
\mathbf{B}_{6} & = & -x_{3} \, \mathbf{a}_{1}-y_{3} \, \mathbf{a}_{2} & = & -x_{3}a \, \mathbf{\hat{x}}-y_{3}b \, \mathbf{\hat{y}} & \left(4q\right) & \mbox{Mo} \\ 
\mathbf{B}_{7} & = & -x_{3} \, \mathbf{a}_{1} + y_{3} \, \mathbf{a}_{2} + \frac{1}{2} \, \mathbf{a}_{3} & = & -x_{3}a \, \mathbf{\hat{x}} + y_{3}b \, \mathbf{\hat{y}} + \frac{1}{2}c \, \mathbf{\hat{z}} & \left(4q\right) & \mbox{Mo} \\ 
\mathbf{B}_{8} & = & x_{3} \, \mathbf{a}_{1}-y_{3} \, \mathbf{a}_{2} + \frac{1}{2} \, \mathbf{a}_{3} & = & x_{3}a \, \mathbf{\hat{x}}-y_{3}b \, \mathbf{\hat{y}} + \frac{1}{2}c \, \mathbf{\hat{z}} & \left(4q\right) & \mbox{Mo} \\ 
\mathbf{B}_{9} & = & x_{4} \, \mathbf{a}_{1} + y_{4} \, \mathbf{a}_{2} & = & x_{4}a \, \mathbf{\hat{x}} + y_{4}b \, \mathbf{\hat{y}} & \left(4q\right) & \mbox{O I} \\ 
\mathbf{B}_{10} & = & -x_{4} \, \mathbf{a}_{1}-y_{4} \, \mathbf{a}_{2} & = & -x_{4}a \, \mathbf{\hat{x}}-y_{4}b \, \mathbf{\hat{y}} & \left(4q\right) & \mbox{O I} \\ 
\mathbf{B}_{11} & = & -x_{4} \, \mathbf{a}_{1} + y_{4} \, \mathbf{a}_{2} + \frac{1}{2} \, \mathbf{a}_{3} & = & -x_{4}a \, \mathbf{\hat{x}} + y_{4}b \, \mathbf{\hat{y}} + \frac{1}{2}c \, \mathbf{\hat{z}} & \left(4q\right) & \mbox{O I} \\ 
\mathbf{B}_{12} & = & x_{4} \, \mathbf{a}_{1}-y_{4} \, \mathbf{a}_{2} + \frac{1}{2} \, \mathbf{a}_{3} & = & x_{4}a \, \mathbf{\hat{x}}-y_{4}b \, \mathbf{\hat{y}} + \frac{1}{2}c \, \mathbf{\hat{z}} & \left(4q\right) & \mbox{O I} \\ 
\mathbf{B}_{13} & = & x_{5} \, \mathbf{a}_{1} + y_{5} \, \mathbf{a}_{2} & = & x_{5}a \, \mathbf{\hat{x}} + y_{5}b \, \mathbf{\hat{y}} & \left(4q\right) & \mbox{O II} \\ 
\mathbf{B}_{14} & = & -x_{5} \, \mathbf{a}_{1}-y_{5} \, \mathbf{a}_{2} & = & -x_{5}a \, \mathbf{\hat{x}}-y_{5}b \, \mathbf{\hat{y}} & \left(4q\right) & \mbox{O II} \\ 
\mathbf{B}_{15} & = & -x_{5} \, \mathbf{a}_{1} + y_{5} \, \mathbf{a}_{2} + \frac{1}{2} \, \mathbf{a}_{3} & = & -x_{5}a \, \mathbf{\hat{x}} + y_{5}b \, \mathbf{\hat{y}} + \frac{1}{2}c \, \mathbf{\hat{z}} & \left(4q\right) & \mbox{O II} \\ 
\mathbf{B}_{16} & = & x_{5} \, \mathbf{a}_{1}-y_{5} \, \mathbf{a}_{2} + \frac{1}{2} \, \mathbf{a}_{3} & = & x_{5}a \, \mathbf{\hat{x}}-y_{5}b \, \mathbf{\hat{y}} + \frac{1}{2}c \, \mathbf{\hat{z}} & \left(4q\right) & \mbox{O II} \\ 
\mathbf{B}_{17} & = & x_{6} \, \mathbf{a}_{1} + y_{6} \, \mathbf{a}_{2} + z_{6} \, \mathbf{a}_{3} & = & x_{6}a \, \mathbf{\hat{x}} + y_{6}b \, \mathbf{\hat{y}} + z_{6}c \, \mathbf{\hat{z}} & \left(8r\right) & \mbox{O III} \\ 
\mathbf{B}_{18} & = & -x_{6} \, \mathbf{a}_{1}-y_{6} \, \mathbf{a}_{2} + z_{6} \, \mathbf{a}_{3} & = & -x_{6}a \, \mathbf{\hat{x}}-y_{6}b \, \mathbf{\hat{y}} + z_{6}c \, \mathbf{\hat{z}} & \left(8r\right) & \mbox{O III} \\ 
\mathbf{B}_{19} & = & -x_{6} \, \mathbf{a}_{1} + y_{6} \, \mathbf{a}_{2} + \left(\frac{1}{2} - z_{6}\right) \, \mathbf{a}_{3} & = & -x_{6}a \, \mathbf{\hat{x}} + y_{6}b \, \mathbf{\hat{y}} + \left(\frac{1}{2} - z_{6}\right)c \, \mathbf{\hat{z}} & \left(8r\right) & \mbox{O III} \\ 
\mathbf{B}_{20} & = & x_{6} \, \mathbf{a}_{1}-y_{6} \, \mathbf{a}_{2} + \left(\frac{1}{2} - z_{6}\right) \, \mathbf{a}_{3} & = & x_{6}a \, \mathbf{\hat{x}}-y_{6}b \, \mathbf{\hat{y}} + \left(\frac{1}{2} - z_{6}\right)c \, \mathbf{\hat{z}} & \left(8r\right) & \mbox{O III} \\ 
\mathbf{B}_{21} & = & -x_{6} \, \mathbf{a}_{1}-y_{6} \, \mathbf{a}_{2}-z_{6} \, \mathbf{a}_{3} & = & -x_{6}a \, \mathbf{\hat{x}}-y_{6}b \, \mathbf{\hat{y}}-z_{6}c \, \mathbf{\hat{z}} & \left(8r\right) & \mbox{O III} \\ 
\mathbf{B}_{22} & = & x_{6} \, \mathbf{a}_{1} + y_{6} \, \mathbf{a}_{2}-z_{6} \, \mathbf{a}_{3} & = & x_{6}a \, \mathbf{\hat{x}} + y_{6}b \, \mathbf{\hat{y}}-z_{6}c \, \mathbf{\hat{z}} & \left(8r\right) & \mbox{O III} \\ 
\mathbf{B}_{23} & = & x_{6} \, \mathbf{a}_{1}-y_{6} \, \mathbf{a}_{2} + \left(\frac{1}{2} +z_{6}\right) \, \mathbf{a}_{3} & = & x_{6}a \, \mathbf{\hat{x}}-y_{6}b \, \mathbf{\hat{y}} + \left(\frac{1}{2} +z_{6}\right)c \, \mathbf{\hat{z}} & \left(8r\right) & \mbox{O III} \\ 
\mathbf{B}_{24} & = & -x_{6} \, \mathbf{a}_{1} + y_{6} \, \mathbf{a}_{2} + \left(\frac{1}{2} +z_{6}\right) \, \mathbf{a}_{3} & = & -x_{6}a \, \mathbf{\hat{x}} + y_{6}b \, \mathbf{\hat{y}} + \left(\frac{1}{2} +z_{6}\right)c \, \mathbf{\hat{z}} & \left(8r\right) & \mbox{O III} \\ 
\end{longtabu}
\renewcommand{\arraystretch}{1.0}
\noindent \hrulefill
\\
\textbf{References:}
\vspace*{-0.25cm}
\begin{flushleft}
  - \bibentry{Vinokurov_SovPhysCryst_1972}. \\
\end{flushleft}
\textbf{Found in:}
\vspace*{-0.25cm}
\begin{flushleft}
  - \bibentry{Villars_PearsonsCrystalData_2013}. \\
\end{flushleft}
\noindent \hrulefill
\\
\textbf{Geometry files:}
\\
\noindent  - CIF: pp. {\hyperref[AB2C8D_oP24_49_g_q_2qr_e_cif]{\pageref{AB2C8D_oP24_49_g_q_2qr_e_cif}}} \\
\noindent  - POSCAR: pp. {\hyperref[AB2C8D_oP24_49_g_q_2qr_e_poscar]{\pageref{AB2C8D_oP24_49_g_q_2qr_e_poscar}}} \\
\onecolumn
{\phantomsection\label{A2BC4_oP28_50_ij_ac_ijm}}
\subsection*{\huge \textbf{{\normalfont La$_{2}$NiO$_{4}$ Structure: A2BC4\_oP28\_50\_ij\_ac\_ijm}}}
\noindent \hrulefill
\vspace*{0.25cm}
\begin{figure}[htp]
  \centering
  \vspace{-1em}
  {\includegraphics[width=1\textwidth]{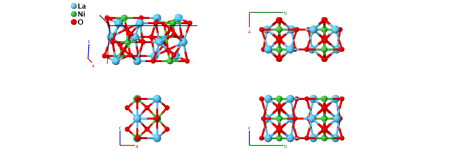}}
\end{figure}
\vspace*{-0.5cm}
\renewcommand{\arraystretch}{1.5}
\begin{equation*}
  \begin{array}{>{$\hspace{-0.15cm}}l<{$}>{$}p{0.5cm}<{$}>{$}p{18.5cm}<{$}}
    \mbox{\large \textbf{Prototype}} &\colon & \ce{La2NiO4} \\
    \mbox{\large \textbf{\AFLOW\ prototype label}} &\colon & \mbox{A2BC4\_oP28\_50\_ij\_ac\_ijm} \\
    \mbox{\large \textbf{\textit{Strukturbericht} designation}} &\colon & \mbox{None} \\
    \mbox{\large \textbf{Pearson symbol}} &\colon & \mbox{oP28} \\
    \mbox{\large \textbf{Space group number}} &\colon & 50 \\
    \mbox{\large \textbf{Space group symbol}} &\colon & Pban \\
    \mbox{\large \textbf{\AFLOW\ prototype command}} &\colon &  \texttt{aflow} \,  \, \texttt{-{}-proto=A2BC4\_oP28\_50\_ij\_ac\_ijm } \, \newline \texttt{-{}-params=}{a,b/a,c/a,y_{3},y_{4},y_{5},y_{6},x_{7},y_{7},z_{7} }
  \end{array}
\end{equation*}
\renewcommand{\arraystretch}{1.0}

\noindent \parbox{1 \linewidth}{
\noindent \hrulefill
\\
\textbf{Simple Orthorhombic primitive vectors:} \\
\vspace*{-0.25cm}
\begin{tabular}{cc}
  \begin{tabular}{c}
    \parbox{0.6 \linewidth}{
      \renewcommand{\arraystretch}{1.5}
      \begin{equation*}
        \centering
        \begin{array}{ccc}
              \mathbf{a}_1 & = & a \, \mathbf{\hat{x}} \\
    \mathbf{a}_2 & = & b \, \mathbf{\hat{y}} \\
    \mathbf{a}_3 & = & c \, \mathbf{\hat{z}} \\

        \end{array}
      \end{equation*}
    }
    \renewcommand{\arraystretch}{1.0}
  \end{tabular}
  \begin{tabular}{c}
    \includegraphics[width=0.3\linewidth]{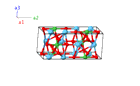} \\
  \end{tabular}
\end{tabular}

}
\vspace*{-0.25cm}

\noindent \hrulefill
\\
\textbf{Basis vectors:}
\vspace*{-0.25cm}
\renewcommand{\arraystretch}{1.5}
\begin{longtabu} to \textwidth{>{\centering $}X[-1,c,c]<{$}>{\centering $}X[-1,c,c]<{$}>{\centering $}X[-1,c,c]<{$}>{\centering $}X[-1,c,c]<{$}>{\centering $}X[-1,c,c]<{$}>{\centering $}X[-1,c,c]<{$}>{\centering $}X[-1,c,c]<{$}}
  & & \mbox{Lattice Coordinates} & & \mbox{Cartesian Coordinates} &\mbox{Wyckoff Position} & \mbox{Atom Type} \\  
  \mathbf{B}_{1} & = & \frac{1}{4} \, \mathbf{a}_{1} + \frac{1}{4} \, \mathbf{a}_{2} & = & \frac{1}{4}a \, \mathbf{\hat{x}} + \frac{1}{4}b \, \mathbf{\hat{y}} & \left(2a\right) & \mbox{Ni I} \\ 
\mathbf{B}_{2} & = & \frac{3}{4} \, \mathbf{a}_{1} + \frac{3}{4} \, \mathbf{a}_{2} & = & \frac{3}{4}a \, \mathbf{\hat{x}} + \frac{3}{4}b \, \mathbf{\hat{y}} & \left(2a\right) & \mbox{Ni I} \\ 
\mathbf{B}_{3} & = & \frac{3}{4} \, \mathbf{a}_{1} + \frac{1}{4} \, \mathbf{a}_{2} + \frac{1}{2} \, \mathbf{a}_{3} & = & \frac{3}{4}a \, \mathbf{\hat{x}} + \frac{1}{4}b \, \mathbf{\hat{y}} + \frac{1}{2}c \, \mathbf{\hat{z}} & \left(2c\right) & \mbox{Ni II} \\ 
\mathbf{B}_{4} & = & \frac{1}{4} \, \mathbf{a}_{1} + \frac{3}{4} \, \mathbf{a}_{2} + \frac{1}{2} \, \mathbf{a}_{3} & = & \frac{1}{4}a \, \mathbf{\hat{x}} + \frac{3}{4}b \, \mathbf{\hat{y}} + \frac{1}{2}c \, \mathbf{\hat{z}} & \left(2c\right) & \mbox{Ni II} \\ 
\mathbf{B}_{5} & = & \frac{1}{4} \, \mathbf{a}_{1} + y_{3} \, \mathbf{a}_{2} & = & \frac{1}{4}a \, \mathbf{\hat{x}} + y_{3}b \, \mathbf{\hat{y}} & \left(4i\right) & \mbox{La I} \\ 
\mathbf{B}_{6} & = & \frac{1}{4} \, \mathbf{a}_{1} + \left(\frac{1}{2} - y_{3}\right) \, \mathbf{a}_{2} & = & \frac{1}{4}a \, \mathbf{\hat{x}} + \left(\frac{1}{2} - y_{3}\right)b \, \mathbf{\hat{y}} & \left(4i\right) & \mbox{La I} \\ 
\mathbf{B}_{7} & = & \frac{3}{4} \, \mathbf{a}_{1}-y_{3} \, \mathbf{a}_{2} & = & \frac{3}{4}a \, \mathbf{\hat{x}}-y_{3}b \, \mathbf{\hat{y}} & \left(4i\right) & \mbox{La I} \\ 
\mathbf{B}_{8} & = & \frac{3}{4} \, \mathbf{a}_{1} + \left(\frac{1}{2} +y_{3}\right) \, \mathbf{a}_{2} & = & \frac{3}{4}a \, \mathbf{\hat{x}} + \left(\frac{1}{2} +y_{3}\right)b \, \mathbf{\hat{y}} & \left(4i\right) & \mbox{La I} \\ 
\mathbf{B}_{9} & = & \frac{1}{4} \, \mathbf{a}_{1} + y_{4} \, \mathbf{a}_{2} & = & \frac{1}{4}a \, \mathbf{\hat{x}} + y_{4}b \, \mathbf{\hat{y}} & \left(4i\right) & \mbox{O I} \\ 
\mathbf{B}_{10} & = & \frac{1}{4} \, \mathbf{a}_{1} + \left(\frac{1}{2} - y_{4}\right) \, \mathbf{a}_{2} & = & \frac{1}{4}a \, \mathbf{\hat{x}} + \left(\frac{1}{2} - y_{4}\right)b \, \mathbf{\hat{y}} & \left(4i\right) & \mbox{O I} \\ 
\mathbf{B}_{11} & = & \frac{3}{4} \, \mathbf{a}_{1}-y_{4} \, \mathbf{a}_{2} & = & \frac{3}{4}a \, \mathbf{\hat{x}}-y_{4}b \, \mathbf{\hat{y}} & \left(4i\right) & \mbox{O I} \\ 
\mathbf{B}_{12} & = & \frac{3}{4} \, \mathbf{a}_{1} + \left(\frac{1}{2} +y_{4}\right) \, \mathbf{a}_{2} & = & \frac{3}{4}a \, \mathbf{\hat{x}} + \left(\frac{1}{2} +y_{4}\right)b \, \mathbf{\hat{y}} & \left(4i\right) & \mbox{O I} \\ 
\mathbf{B}_{13} & = & \frac{1}{4} \, \mathbf{a}_{1} + y_{5} \, \mathbf{a}_{2} + \frac{1}{2} \, \mathbf{a}_{3} & = & \frac{1}{4}a \, \mathbf{\hat{x}} + y_{5}b \, \mathbf{\hat{y}} + \frac{1}{2}c \, \mathbf{\hat{z}} & \left(4j\right) & \mbox{La II} \\ 
\mathbf{B}_{14} & = & \frac{1}{4} \, \mathbf{a}_{1} + \left(\frac{1}{2} - y_{5}\right) \, \mathbf{a}_{2} + \frac{1}{2} \, \mathbf{a}_{3} & = & \frac{1}{4}a \, \mathbf{\hat{x}} + \left(\frac{1}{2} - y_{5}\right)b \, \mathbf{\hat{y}} + \frac{1}{2}c \, \mathbf{\hat{z}} & \left(4j\right) & \mbox{La II} \\ 
\mathbf{B}_{15} & = & \frac{3}{4} \, \mathbf{a}_{1}-y_{5} \, \mathbf{a}_{2} + \frac{1}{2} \, \mathbf{a}_{3} & = & \frac{3}{4}a \, \mathbf{\hat{x}}-y_{5}b \, \mathbf{\hat{y}} + \frac{1}{2}c \, \mathbf{\hat{z}} & \left(4j\right) & \mbox{La II} \\ 
\mathbf{B}_{16} & = & \frac{3}{4} \, \mathbf{a}_{1} + \left(\frac{1}{2} +y_{5}\right) \, \mathbf{a}_{2} + \frac{1}{2} \, \mathbf{a}_{3} & = & \frac{3}{4}a \, \mathbf{\hat{x}} + \left(\frac{1}{2} +y_{5}\right)b \, \mathbf{\hat{y}} + \frac{1}{2}c \, \mathbf{\hat{z}} & \left(4j\right) & \mbox{La II} \\ 
\mathbf{B}_{17} & = & \frac{1}{4} \, \mathbf{a}_{1} + y_{6} \, \mathbf{a}_{2} + \frac{1}{2} \, \mathbf{a}_{3} & = & \frac{1}{4}a \, \mathbf{\hat{x}} + y_{6}b \, \mathbf{\hat{y}} + \frac{1}{2}c \, \mathbf{\hat{z}} & \left(4j\right) & \mbox{O II} \\ 
\mathbf{B}_{18} & = & \frac{1}{4} \, \mathbf{a}_{1} + \left(\frac{1}{2} - y_{6}\right) \, \mathbf{a}_{2} + \frac{1}{2} \, \mathbf{a}_{3} & = & \frac{1}{4}a \, \mathbf{\hat{x}} + \left(\frac{1}{2} - y_{6}\right)b \, \mathbf{\hat{y}} + \frac{1}{2}c \, \mathbf{\hat{z}} & \left(4j\right) & \mbox{O II} \\ 
\mathbf{B}_{19} & = & \frac{3}{4} \, \mathbf{a}_{1}-y_{6} \, \mathbf{a}_{2} + \frac{1}{2} \, \mathbf{a}_{3} & = & \frac{3}{4}a \, \mathbf{\hat{x}}-y_{6}b \, \mathbf{\hat{y}} + \frac{1}{2}c \, \mathbf{\hat{z}} & \left(4j\right) & \mbox{O II} \\ 
\mathbf{B}_{20} & = & \frac{3}{4} \, \mathbf{a}_{1} + \left(\frac{1}{2} +y_{6}\right) \, \mathbf{a}_{2} + \frac{1}{2} \, \mathbf{a}_{3} & = & \frac{3}{4}a \, \mathbf{\hat{x}} + \left(\frac{1}{2} +y_{6}\right)b \, \mathbf{\hat{y}} + \frac{1}{2}c \, \mathbf{\hat{z}} & \left(4j\right) & \mbox{O II} \\ 
\mathbf{B}_{21} & = & x_{7} \, \mathbf{a}_{1} + y_{7} \, \mathbf{a}_{2} + z_{7} \, \mathbf{a}_{3} & = & x_{7}a \, \mathbf{\hat{x}} + y_{7}b \, \mathbf{\hat{y}} + z_{7}c \, \mathbf{\hat{z}} & \left(8m\right) & \mbox{O III} \\ 
\mathbf{B}_{22} & = & \left(\frac{1}{2} - x_{7}\right) \, \mathbf{a}_{1} + \left(\frac{1}{2} - y_{7}\right) \, \mathbf{a}_{2} + z_{7} \, \mathbf{a}_{3} & = & \left(\frac{1}{2} - x_{7}\right)a \, \mathbf{\hat{x}} + \left(\frac{1}{2} - y_{7}\right)b \, \mathbf{\hat{y}} + z_{7}c \, \mathbf{\hat{z}} & \left(8m\right) & \mbox{O III} \\ 
\mathbf{B}_{23} & = & \left(\frac{1}{2} - x_{7}\right) \, \mathbf{a}_{1} + y_{7} \, \mathbf{a}_{2}-z_{7} \, \mathbf{a}_{3} & = & \left(\frac{1}{2} - x_{7}\right)a \, \mathbf{\hat{x}} + y_{7}b \, \mathbf{\hat{y}}-z_{7}c \, \mathbf{\hat{z}} & \left(8m\right) & \mbox{O III} \\ 
\mathbf{B}_{24} & = & x_{7} \, \mathbf{a}_{1} + \left(\frac{1}{2} - y_{7}\right) \, \mathbf{a}_{2}-z_{7} \, \mathbf{a}_{3} & = & x_{7}a \, \mathbf{\hat{x}} + \left(\frac{1}{2} - y_{7}\right)b \, \mathbf{\hat{y}}-z_{7}c \, \mathbf{\hat{z}} & \left(8m\right) & \mbox{O III} \\ 
\mathbf{B}_{25} & = & -x_{7} \, \mathbf{a}_{1}-y_{7} \, \mathbf{a}_{2}-z_{7} \, \mathbf{a}_{3} & = & -x_{7}a \, \mathbf{\hat{x}}-y_{7}b \, \mathbf{\hat{y}}-z_{7}c \, \mathbf{\hat{z}} & \left(8m\right) & \mbox{O III} \\ 
\mathbf{B}_{26} & = & \left(\frac{1}{2} +x_{7}\right) \, \mathbf{a}_{1} + \left(\frac{1}{2} +y_{7}\right) \, \mathbf{a}_{2}-z_{7} \, \mathbf{a}_{3} & = & \left(\frac{1}{2} +x_{7}\right)a \, \mathbf{\hat{x}} + \left(\frac{1}{2} +y_{7}\right)b \, \mathbf{\hat{y}}-z_{7}c \, \mathbf{\hat{z}} & \left(8m\right) & \mbox{O III} \\ 
\mathbf{B}_{27} & = & \left(\frac{1}{2} +x_{7}\right) \, \mathbf{a}_{1}-y_{7} \, \mathbf{a}_{2} + z_{7} \, \mathbf{a}_{3} & = & \left(\frac{1}{2} +x_{7}\right)a \, \mathbf{\hat{x}}-y_{7}b \, \mathbf{\hat{y}} + z_{7}c \, \mathbf{\hat{z}} & \left(8m\right) & \mbox{O III} \\ 
\mathbf{B}_{28} & = & -x_{7} \, \mathbf{a}_{1} + \left(\frac{1}{2} +y_{7}\right) \, \mathbf{a}_{2} + z_{7} \, \mathbf{a}_{3} & = & -x_{7}a \, \mathbf{\hat{x}} + \left(\frac{1}{2} +y_{7}\right)b \, \mathbf{\hat{y}} + z_{7}c \, \mathbf{\hat{z}} & \left(8m\right) & \mbox{O III} \\ 
\end{longtabu}
\renewcommand{\arraystretch}{1.0}
\noindent \hrulefill
\\
\textbf{References:}
\vspace*{-0.25cm}
\begin{flushleft}
  - \bibentry{Odier_La2NiO4_MatResBull_1986}. \\
\end{flushleft}
\textbf{Found in:}
\vspace*{-0.25cm}
\begin{flushleft}
  - \bibentry{Villars_PearsonsCrystalData_2013}. \\
\end{flushleft}
\noindent \hrulefill
\\
\textbf{Geometry files:}
\\
\noindent  - CIF: pp. {\hyperref[A2BC4_oP28_50_ij_ac_ijm_cif]{\pageref{A2BC4_oP28_50_ij_ac_ijm_cif}}} \\
\noindent  - POSCAR: pp. {\hyperref[A2BC4_oP28_50_ij_ac_ijm_poscar]{\pageref{A2BC4_oP28_50_ij_ac_ijm_poscar}}} \\
\onecolumn
{\phantomsection\label{A3BC2_oP48_50_3m_m_2m}}
\subsection*{\huge \textbf{{\normalfont $\alpha$-Tl$_{2}$TeO$_{3}$ Structure: A3BC2\_oP48\_50\_3m\_m\_2m}}}
\noindent \hrulefill
\vspace*{0.25cm}
\begin{figure}[htp]
  \centering
  \vspace{-1em}
  {\includegraphics[width=1\textwidth]{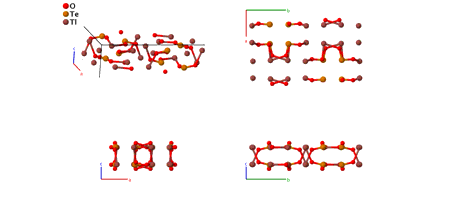}}
\end{figure}
\vspace*{-0.5cm}
\renewcommand{\arraystretch}{1.5}
\begin{equation*}
  \begin{array}{>{$\hspace{-0.15cm}}l<{$}>{$}p{0.5cm}<{$}>{$}p{18.5cm}<{$}}
    \mbox{\large \textbf{Prototype}} &\colon & \ce{$\alpha$-Tl2TeO3} \\
    \mbox{\large \textbf{\AFLOW\ prototype label}} &\colon & \mbox{A3BC2\_oP48\_50\_3m\_m\_2m} \\
    \mbox{\large \textbf{\textit{Strukturbericht} designation}} &\colon & \mbox{None} \\
    \mbox{\large \textbf{Pearson symbol}} &\colon & \mbox{oP48} \\
    \mbox{\large \textbf{Space group number}} &\colon & 50 \\
    \mbox{\large \textbf{Space group symbol}} &\colon & Pban \\
    \mbox{\large \textbf{\AFLOW\ prototype command}} &\colon &  \texttt{aflow} \,  \, \texttt{-{}-proto=A3BC2\_oP48\_50\_3m\_m\_2m } \, \newline \texttt{-{}-params=}{a,b/a,c/a,x_{1},y_{1},z_{1},x_{2},y_{2},z_{2},x_{3},y_{3},z_{3},x_{4},y_{4},z_{4},x_{5},y_{5},z_{5},x_{6},y_{6},} \newline {z_{6} }
  \end{array}
\end{equation*}
\renewcommand{\arraystretch}{1.0}

\noindent \parbox{1 \linewidth}{
\noindent \hrulefill
\\
\textbf{Simple Orthorhombic primitive vectors:} \\
\vspace*{-0.25cm}
\begin{tabular}{cc}
  \begin{tabular}{c}
    \parbox{0.6 \linewidth}{
      \renewcommand{\arraystretch}{1.5}
      \begin{equation*}
        \centering
        \begin{array}{ccc}
              \mathbf{a}_1 & = & a \, \mathbf{\hat{x}} \\
    \mathbf{a}_2 & = & b \, \mathbf{\hat{y}} \\
    \mathbf{a}_3 & = & c \, \mathbf{\hat{z}} \\

        \end{array}
      \end{equation*}
    }
    \renewcommand{\arraystretch}{1.0}
  \end{tabular}
  \begin{tabular}{c}
    \includegraphics[width=0.3\linewidth]{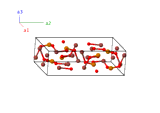} \\
  \end{tabular}
\end{tabular}

}
\vspace*{-0.25cm}

\noindent \hrulefill
\\
\textbf{Basis vectors:}
\vspace*{-0.25cm}
\renewcommand{\arraystretch}{1.5}
\begin{longtabu} to \textwidth{>{\centering $}X[-1,c,c]<{$}>{\centering $}X[-1,c,c]<{$}>{\centering $}X[-1,c,c]<{$}>{\centering $}X[-1,c,c]<{$}>{\centering $}X[-1,c,c]<{$}>{\centering $}X[-1,c,c]<{$}>{\centering $}X[-1,c,c]<{$}}
  & & \mbox{Lattice Coordinates} & & \mbox{Cartesian Coordinates} &\mbox{Wyckoff Position} & \mbox{Atom Type} \\  
  \mathbf{B}_{1} & = & x_{1} \, \mathbf{a}_{1} + y_{1} \, \mathbf{a}_{2} + z_{1} \, \mathbf{a}_{3} & = & x_{1}a \, \mathbf{\hat{x}} + y_{1}b \, \mathbf{\hat{y}} + z_{1}c \, \mathbf{\hat{z}} & \left(8m\right) & \mbox{O I} \\ 
\mathbf{B}_{2} & = & \left(\frac{1}{2} - x_{1}\right) \, \mathbf{a}_{1} + \left(\frac{1}{2} - y_{1}\right) \, \mathbf{a}_{2} + z_{1} \, \mathbf{a}_{3} & = & \left(\frac{1}{2} - x_{1}\right)a \, \mathbf{\hat{x}} + \left(\frac{1}{2} - y_{1}\right)b \, \mathbf{\hat{y}} + z_{1}c \, \mathbf{\hat{z}} & \left(8m\right) & \mbox{O I} \\ 
\mathbf{B}_{3} & = & \left(\frac{1}{2} - x_{1}\right) \, \mathbf{a}_{1} + y_{1} \, \mathbf{a}_{2}-z_{1} \, \mathbf{a}_{3} & = & \left(\frac{1}{2} - x_{1}\right)a \, \mathbf{\hat{x}} + y_{1}b \, \mathbf{\hat{y}}-z_{1}c \, \mathbf{\hat{z}} & \left(8m\right) & \mbox{O I} \\ 
\mathbf{B}_{4} & = & x_{1} \, \mathbf{a}_{1} + \left(\frac{1}{2} - y_{1}\right) \, \mathbf{a}_{2}-z_{1} \, \mathbf{a}_{3} & = & x_{1}a \, \mathbf{\hat{x}} + \left(\frac{1}{2} - y_{1}\right)b \, \mathbf{\hat{y}}-z_{1}c \, \mathbf{\hat{z}} & \left(8m\right) & \mbox{O I} \\ 
\mathbf{B}_{5} & = & -x_{1} \, \mathbf{a}_{1}-y_{1} \, \mathbf{a}_{2}-z_{1} \, \mathbf{a}_{3} & = & -x_{1}a \, \mathbf{\hat{x}}-y_{1}b \, \mathbf{\hat{y}}-z_{1}c \, \mathbf{\hat{z}} & \left(8m\right) & \mbox{O I} \\ 
\mathbf{B}_{6} & = & \left(\frac{1}{2} +x_{1}\right) \, \mathbf{a}_{1} + \left(\frac{1}{2} +y_{1}\right) \, \mathbf{a}_{2}-z_{1} \, \mathbf{a}_{3} & = & \left(\frac{1}{2} +x_{1}\right)a \, \mathbf{\hat{x}} + \left(\frac{1}{2} +y_{1}\right)b \, \mathbf{\hat{y}}-z_{1}c \, \mathbf{\hat{z}} & \left(8m\right) & \mbox{O I} \\ 
\mathbf{B}_{7} & = & \left(\frac{1}{2} +x_{1}\right) \, \mathbf{a}_{1}-y_{1} \, \mathbf{a}_{2} + z_{1} \, \mathbf{a}_{3} & = & \left(\frac{1}{2} +x_{1}\right)a \, \mathbf{\hat{x}}-y_{1}b \, \mathbf{\hat{y}} + z_{1}c \, \mathbf{\hat{z}} & \left(8m\right) & \mbox{O I} \\ 
\mathbf{B}_{8} & = & -x_{1} \, \mathbf{a}_{1} + \left(\frac{1}{2} +y_{1}\right) \, \mathbf{a}_{2} + z_{1} \, \mathbf{a}_{3} & = & -x_{1}a \, \mathbf{\hat{x}} + \left(\frac{1}{2} +y_{1}\right)b \, \mathbf{\hat{y}} + z_{1}c \, \mathbf{\hat{z}} & \left(8m\right) & \mbox{O I} \\ 
\mathbf{B}_{9} & = & x_{2} \, \mathbf{a}_{1} + y_{2} \, \mathbf{a}_{2} + z_{2} \, \mathbf{a}_{3} & = & x_{2}a \, \mathbf{\hat{x}} + y_{2}b \, \mathbf{\hat{y}} + z_{2}c \, \mathbf{\hat{z}} & \left(8m\right) & \mbox{O II} \\ 
\mathbf{B}_{10} & = & \left(\frac{1}{2} - x_{2}\right) \, \mathbf{a}_{1} + \left(\frac{1}{2} - y_{2}\right) \, \mathbf{a}_{2} + z_{2} \, \mathbf{a}_{3} & = & \left(\frac{1}{2} - x_{2}\right)a \, \mathbf{\hat{x}} + \left(\frac{1}{2} - y_{2}\right)b \, \mathbf{\hat{y}} + z_{2}c \, \mathbf{\hat{z}} & \left(8m\right) & \mbox{O II} \\ 
\mathbf{B}_{11} & = & \left(\frac{1}{2} - x_{2}\right) \, \mathbf{a}_{1} + y_{2} \, \mathbf{a}_{2}-z_{2} \, \mathbf{a}_{3} & = & \left(\frac{1}{2} - x_{2}\right)a \, \mathbf{\hat{x}} + y_{2}b \, \mathbf{\hat{y}}-z_{2}c \, \mathbf{\hat{z}} & \left(8m\right) & \mbox{O II} \\ 
\mathbf{B}_{12} & = & x_{2} \, \mathbf{a}_{1} + \left(\frac{1}{2} - y_{2}\right) \, \mathbf{a}_{2}-z_{2} \, \mathbf{a}_{3} & = & x_{2}a \, \mathbf{\hat{x}} + \left(\frac{1}{2} - y_{2}\right)b \, \mathbf{\hat{y}}-z_{2}c \, \mathbf{\hat{z}} & \left(8m\right) & \mbox{O II} \\ 
\mathbf{B}_{13} & = & -x_{2} \, \mathbf{a}_{1}-y_{2} \, \mathbf{a}_{2}-z_{2} \, \mathbf{a}_{3} & = & -x_{2}a \, \mathbf{\hat{x}}-y_{2}b \, \mathbf{\hat{y}}-z_{2}c \, \mathbf{\hat{z}} & \left(8m\right) & \mbox{O II} \\ 
\mathbf{B}_{14} & = & \left(\frac{1}{2} +x_{2}\right) \, \mathbf{a}_{1} + \left(\frac{1}{2} +y_{2}\right) \, \mathbf{a}_{2}-z_{2} \, \mathbf{a}_{3} & = & \left(\frac{1}{2} +x_{2}\right)a \, \mathbf{\hat{x}} + \left(\frac{1}{2} +y_{2}\right)b \, \mathbf{\hat{y}}-z_{2}c \, \mathbf{\hat{z}} & \left(8m\right) & \mbox{O II} \\ 
\mathbf{B}_{15} & = & \left(\frac{1}{2} +x_{2}\right) \, \mathbf{a}_{1}-y_{2} \, \mathbf{a}_{2} + z_{2} \, \mathbf{a}_{3} & = & \left(\frac{1}{2} +x_{2}\right)a \, \mathbf{\hat{x}}-y_{2}b \, \mathbf{\hat{y}} + z_{2}c \, \mathbf{\hat{z}} & \left(8m\right) & \mbox{O II} \\ 
\mathbf{B}_{16} & = & -x_{2} \, \mathbf{a}_{1} + \left(\frac{1}{2} +y_{2}\right) \, \mathbf{a}_{2} + z_{2} \, \mathbf{a}_{3} & = & -x_{2}a \, \mathbf{\hat{x}} + \left(\frac{1}{2} +y_{2}\right)b \, \mathbf{\hat{y}} + z_{2}c \, \mathbf{\hat{z}} & \left(8m\right) & \mbox{O II} \\ 
\mathbf{B}_{17} & = & x_{3} \, \mathbf{a}_{1} + y_{3} \, \mathbf{a}_{2} + z_{3} \, \mathbf{a}_{3} & = & x_{3}a \, \mathbf{\hat{x}} + y_{3}b \, \mathbf{\hat{y}} + z_{3}c \, \mathbf{\hat{z}} & \left(8m\right) & \mbox{O III} \\ 
\mathbf{B}_{18} & = & \left(\frac{1}{2} - x_{3}\right) \, \mathbf{a}_{1} + \left(\frac{1}{2} - y_{3}\right) \, \mathbf{a}_{2} + z_{3} \, \mathbf{a}_{3} & = & \left(\frac{1}{2} - x_{3}\right)a \, \mathbf{\hat{x}} + \left(\frac{1}{2} - y_{3}\right)b \, \mathbf{\hat{y}} + z_{3}c \, \mathbf{\hat{z}} & \left(8m\right) & \mbox{O III} \\ 
\mathbf{B}_{19} & = & \left(\frac{1}{2} - x_{3}\right) \, \mathbf{a}_{1} + y_{3} \, \mathbf{a}_{2}-z_{3} \, \mathbf{a}_{3} & = & \left(\frac{1}{2} - x_{3}\right)a \, \mathbf{\hat{x}} + y_{3}b \, \mathbf{\hat{y}}-z_{3}c \, \mathbf{\hat{z}} & \left(8m\right) & \mbox{O III} \\ 
\mathbf{B}_{20} & = & x_{3} \, \mathbf{a}_{1} + \left(\frac{1}{2} - y_{3}\right) \, \mathbf{a}_{2}-z_{3} \, \mathbf{a}_{3} & = & x_{3}a \, \mathbf{\hat{x}} + \left(\frac{1}{2} - y_{3}\right)b \, \mathbf{\hat{y}}-z_{3}c \, \mathbf{\hat{z}} & \left(8m\right) & \mbox{O III} \\ 
\mathbf{B}_{21} & = & -x_{3} \, \mathbf{a}_{1}-y_{3} \, \mathbf{a}_{2}-z_{3} \, \mathbf{a}_{3} & = & -x_{3}a \, \mathbf{\hat{x}}-y_{3}b \, \mathbf{\hat{y}}-z_{3}c \, \mathbf{\hat{z}} & \left(8m\right) & \mbox{O III} \\ 
\mathbf{B}_{22} & = & \left(\frac{1}{2} +x_{3}\right) \, \mathbf{a}_{1} + \left(\frac{1}{2} +y_{3}\right) \, \mathbf{a}_{2}-z_{3} \, \mathbf{a}_{3} & = & \left(\frac{1}{2} +x_{3}\right)a \, \mathbf{\hat{x}} + \left(\frac{1}{2} +y_{3}\right)b \, \mathbf{\hat{y}}-z_{3}c \, \mathbf{\hat{z}} & \left(8m\right) & \mbox{O III} \\ 
\mathbf{B}_{23} & = & \left(\frac{1}{2} +x_{3}\right) \, \mathbf{a}_{1}-y_{3} \, \mathbf{a}_{2} + z_{3} \, \mathbf{a}_{3} & = & \left(\frac{1}{2} +x_{3}\right)a \, \mathbf{\hat{x}}-y_{3}b \, \mathbf{\hat{y}} + z_{3}c \, \mathbf{\hat{z}} & \left(8m\right) & \mbox{O III} \\ 
\mathbf{B}_{24} & = & -x_{3} \, \mathbf{a}_{1} + \left(\frac{1}{2} +y_{3}\right) \, \mathbf{a}_{2} + z_{3} \, \mathbf{a}_{3} & = & -x_{3}a \, \mathbf{\hat{x}} + \left(\frac{1}{2} +y_{3}\right)b \, \mathbf{\hat{y}} + z_{3}c \, \mathbf{\hat{z}} & \left(8m\right) & \mbox{O III} \\ 
\mathbf{B}_{25} & = & x_{4} \, \mathbf{a}_{1} + y_{4} \, \mathbf{a}_{2} + z_{4} \, \mathbf{a}_{3} & = & x_{4}a \, \mathbf{\hat{x}} + y_{4}b \, \mathbf{\hat{y}} + z_{4}c \, \mathbf{\hat{z}} & \left(8m\right) & \mbox{Te} \\ 
\mathbf{B}_{26} & = & \left(\frac{1}{2} - x_{4}\right) \, \mathbf{a}_{1} + \left(\frac{1}{2} - y_{4}\right) \, \mathbf{a}_{2} + z_{4} \, \mathbf{a}_{3} & = & \left(\frac{1}{2} - x_{4}\right)a \, \mathbf{\hat{x}} + \left(\frac{1}{2} - y_{4}\right)b \, \mathbf{\hat{y}} + z_{4}c \, \mathbf{\hat{z}} & \left(8m\right) & \mbox{Te} \\ 
\mathbf{B}_{27} & = & \left(\frac{1}{2} - x_{4}\right) \, \mathbf{a}_{1} + y_{4} \, \mathbf{a}_{2}-z_{4} \, \mathbf{a}_{3} & = & \left(\frac{1}{2} - x_{4}\right)a \, \mathbf{\hat{x}} + y_{4}b \, \mathbf{\hat{y}}-z_{4}c \, \mathbf{\hat{z}} & \left(8m\right) & \mbox{Te} \\ 
\mathbf{B}_{28} & = & x_{4} \, \mathbf{a}_{1} + \left(\frac{1}{2} - y_{4}\right) \, \mathbf{a}_{2}-z_{4} \, \mathbf{a}_{3} & = & x_{4}a \, \mathbf{\hat{x}} + \left(\frac{1}{2} - y_{4}\right)b \, \mathbf{\hat{y}}-z_{4}c \, \mathbf{\hat{z}} & \left(8m\right) & \mbox{Te} \\ 
\mathbf{B}_{29} & = & -x_{4} \, \mathbf{a}_{1}-y_{4} \, \mathbf{a}_{2}-z_{4} \, \mathbf{a}_{3} & = & -x_{4}a \, \mathbf{\hat{x}}-y_{4}b \, \mathbf{\hat{y}}-z_{4}c \, \mathbf{\hat{z}} & \left(8m\right) & \mbox{Te} \\ 
\mathbf{B}_{30} & = & \left(\frac{1}{2} +x_{4}\right) \, \mathbf{a}_{1} + \left(\frac{1}{2} +y_{4}\right) \, \mathbf{a}_{2}-z_{4} \, \mathbf{a}_{3} & = & \left(\frac{1}{2} +x_{4}\right)a \, \mathbf{\hat{x}} + \left(\frac{1}{2} +y_{4}\right)b \, \mathbf{\hat{y}}-z_{4}c \, \mathbf{\hat{z}} & \left(8m\right) & \mbox{Te} \\ 
\mathbf{B}_{31} & = & \left(\frac{1}{2} +x_{4}\right) \, \mathbf{a}_{1}-y_{4} \, \mathbf{a}_{2} + z_{4} \, \mathbf{a}_{3} & = & \left(\frac{1}{2} +x_{4}\right)a \, \mathbf{\hat{x}}-y_{4}b \, \mathbf{\hat{y}} + z_{4}c \, \mathbf{\hat{z}} & \left(8m\right) & \mbox{Te} \\ 
\mathbf{B}_{32} & = & -x_{4} \, \mathbf{a}_{1} + \left(\frac{1}{2} +y_{4}\right) \, \mathbf{a}_{2} + z_{4} \, \mathbf{a}_{3} & = & -x_{4}a \, \mathbf{\hat{x}} + \left(\frac{1}{2} +y_{4}\right)b \, \mathbf{\hat{y}} + z_{4}c \, \mathbf{\hat{z}} & \left(8m\right) & \mbox{Te} \\ 
\mathbf{B}_{33} & = & x_{5} \, \mathbf{a}_{1} + y_{5} \, \mathbf{a}_{2} + z_{5} \, \mathbf{a}_{3} & = & x_{5}a \, \mathbf{\hat{x}} + y_{5}b \, \mathbf{\hat{y}} + z_{5}c \, \mathbf{\hat{z}} & \left(8m\right) & \mbox{Tl I} \\ 
\mathbf{B}_{34} & = & \left(\frac{1}{2} - x_{5}\right) \, \mathbf{a}_{1} + \left(\frac{1}{2} - y_{5}\right) \, \mathbf{a}_{2} + z_{5} \, \mathbf{a}_{3} & = & \left(\frac{1}{2} - x_{5}\right)a \, \mathbf{\hat{x}} + \left(\frac{1}{2} - y_{5}\right)b \, \mathbf{\hat{y}} + z_{5}c \, \mathbf{\hat{z}} & \left(8m\right) & \mbox{Tl I} \\ 
\mathbf{B}_{35} & = & \left(\frac{1}{2} - x_{5}\right) \, \mathbf{a}_{1} + y_{5} \, \mathbf{a}_{2}-z_{5} \, \mathbf{a}_{3} & = & \left(\frac{1}{2} - x_{5}\right)a \, \mathbf{\hat{x}} + y_{5}b \, \mathbf{\hat{y}}-z_{5}c \, \mathbf{\hat{z}} & \left(8m\right) & \mbox{Tl I} \\ 
\mathbf{B}_{36} & = & x_{5} \, \mathbf{a}_{1} + \left(\frac{1}{2} - y_{5}\right) \, \mathbf{a}_{2}-z_{5} \, \mathbf{a}_{3} & = & x_{5}a \, \mathbf{\hat{x}} + \left(\frac{1}{2} - y_{5}\right)b \, \mathbf{\hat{y}}-z_{5}c \, \mathbf{\hat{z}} & \left(8m\right) & \mbox{Tl I} \\ 
\mathbf{B}_{37} & = & -x_{5} \, \mathbf{a}_{1}-y_{5} \, \mathbf{a}_{2}-z_{5} \, \mathbf{a}_{3} & = & -x_{5}a \, \mathbf{\hat{x}}-y_{5}b \, \mathbf{\hat{y}}-z_{5}c \, \mathbf{\hat{z}} & \left(8m\right) & \mbox{Tl I} \\ 
\mathbf{B}_{38} & = & \left(\frac{1}{2} +x_{5}\right) \, \mathbf{a}_{1} + \left(\frac{1}{2} +y_{5}\right) \, \mathbf{a}_{2}-z_{5} \, \mathbf{a}_{3} & = & \left(\frac{1}{2} +x_{5}\right)a \, \mathbf{\hat{x}} + \left(\frac{1}{2} +y_{5}\right)b \, \mathbf{\hat{y}}-z_{5}c \, \mathbf{\hat{z}} & \left(8m\right) & \mbox{Tl I} \\ 
\mathbf{B}_{39} & = & \left(\frac{1}{2} +x_{5}\right) \, \mathbf{a}_{1}-y_{5} \, \mathbf{a}_{2} + z_{5} \, \mathbf{a}_{3} & = & \left(\frac{1}{2} +x_{5}\right)a \, \mathbf{\hat{x}}-y_{5}b \, \mathbf{\hat{y}} + z_{5}c \, \mathbf{\hat{z}} & \left(8m\right) & \mbox{Tl I} \\ 
\mathbf{B}_{40} & = & -x_{5} \, \mathbf{a}_{1} + \left(\frac{1}{2} +y_{5}\right) \, \mathbf{a}_{2} + z_{5} \, \mathbf{a}_{3} & = & -x_{5}a \, \mathbf{\hat{x}} + \left(\frac{1}{2} +y_{5}\right)b \, \mathbf{\hat{y}} + z_{5}c \, \mathbf{\hat{z}} & \left(8m\right) & \mbox{Tl I} \\ 
\mathbf{B}_{41} & = & x_{6} \, \mathbf{a}_{1} + y_{6} \, \mathbf{a}_{2} + z_{6} \, \mathbf{a}_{3} & = & x_{6}a \, \mathbf{\hat{x}} + y_{6}b \, \mathbf{\hat{y}} + z_{6}c \, \mathbf{\hat{z}} & \left(8m\right) & \mbox{Tl II} \\ 
\mathbf{B}_{42} & = & \left(\frac{1}{2} - x_{6}\right) \, \mathbf{a}_{1} + \left(\frac{1}{2} - y_{6}\right) \, \mathbf{a}_{2} + z_{6} \, \mathbf{a}_{3} & = & \left(\frac{1}{2} - x_{6}\right)a \, \mathbf{\hat{x}} + \left(\frac{1}{2} - y_{6}\right)b \, \mathbf{\hat{y}} + z_{6}c \, \mathbf{\hat{z}} & \left(8m\right) & \mbox{Tl II} \\ 
\mathbf{B}_{43} & = & \left(\frac{1}{2} - x_{6}\right) \, \mathbf{a}_{1} + y_{6} \, \mathbf{a}_{2}-z_{6} \, \mathbf{a}_{3} & = & \left(\frac{1}{2} - x_{6}\right)a \, \mathbf{\hat{x}} + y_{6}b \, \mathbf{\hat{y}}-z_{6}c \, \mathbf{\hat{z}} & \left(8m\right) & \mbox{Tl II} \\ 
\mathbf{B}_{44} & = & x_{6} \, \mathbf{a}_{1} + \left(\frac{1}{2} - y_{6}\right) \, \mathbf{a}_{2}-z_{6} \, \mathbf{a}_{3} & = & x_{6}a \, \mathbf{\hat{x}} + \left(\frac{1}{2} - y_{6}\right)b \, \mathbf{\hat{y}}-z_{6}c \, \mathbf{\hat{z}} & \left(8m\right) & \mbox{Tl II} \\ 
\mathbf{B}_{45} & = & -x_{6} \, \mathbf{a}_{1}-y_{6} \, \mathbf{a}_{2}-z_{6} \, \mathbf{a}_{3} & = & -x_{6}a \, \mathbf{\hat{x}}-y_{6}b \, \mathbf{\hat{y}}-z_{6}c \, \mathbf{\hat{z}} & \left(8m\right) & \mbox{Tl II} \\ 
\mathbf{B}_{46} & = & \left(\frac{1}{2} +x_{6}\right) \, \mathbf{a}_{1} + \left(\frac{1}{2} +y_{6}\right) \, \mathbf{a}_{2}-z_{6} \, \mathbf{a}_{3} & = & \left(\frac{1}{2} +x_{6}\right)a \, \mathbf{\hat{x}} + \left(\frac{1}{2} +y_{6}\right)b \, \mathbf{\hat{y}}-z_{6}c \, \mathbf{\hat{z}} & \left(8m\right) & \mbox{Tl II} \\ 
\mathbf{B}_{47} & = & \left(\frac{1}{2} +x_{6}\right) \, \mathbf{a}_{1}-y_{6} \, \mathbf{a}_{2} + z_{6} \, \mathbf{a}_{3} & = & \left(\frac{1}{2} +x_{6}\right)a \, \mathbf{\hat{x}}-y_{6}b \, \mathbf{\hat{y}} + z_{6}c \, \mathbf{\hat{z}} & \left(8m\right) & \mbox{Tl II} \\ 
\mathbf{B}_{48} & = & -x_{6} \, \mathbf{a}_{1} + \left(\frac{1}{2} +y_{6}\right) \, \mathbf{a}_{2} + z_{6} \, \mathbf{a}_{3} & = & -x_{6}a \, \mathbf{\hat{x}} + \left(\frac{1}{2} +y_{6}\right)b \, \mathbf{\hat{y}} + z_{6}c \, \mathbf{\hat{z}} & \left(8m\right) & \mbox{Tl II} \\ 
\end{longtabu}
\renewcommand{\arraystretch}{1.0}
\noindent \hrulefill
\\
\textbf{References:}
\vspace*{-0.25cm}
\begin{flushleft}
  - \bibentry{Rieger_Tl2TeO3_InorgChem_2007}. \\
\end{flushleft}
\textbf{Found in:}
\vspace*{-0.25cm}
\begin{flushleft}
  - \bibentry{Villars_PearsonsCrystalData_2013}. \\
\end{flushleft}
\noindent \hrulefill
\\
\textbf{Geometry files:}
\\
\noindent  - CIF: pp. {\hyperref[A3BC2_oP48_50_3m_m_2m_cif]{\pageref{A3BC2_oP48_50_3m_m_2m_cif}}} \\
\noindent  - POSCAR: pp. {\hyperref[A3BC2_oP48_50_3m_m_2m_poscar]{\pageref{A3BC2_oP48_50_3m_m_2m_poscar}}} \\
\onecolumn
{\phantomsection\label{A2B_oP24_52_2e_cd}}
\subsection*{\huge \textbf{{\normalfont \begin{raggedleft}GaCl$_{2}$ (High-temperature) Structure: \end{raggedleft} \\ A2B\_oP24\_52\_2e\_cd}}}
\noindent \hrulefill
\vspace*{0.25cm}
\begin{figure}[htp]
  \centering
  \vspace{-1em}
  {\includegraphics[width=1\textwidth]{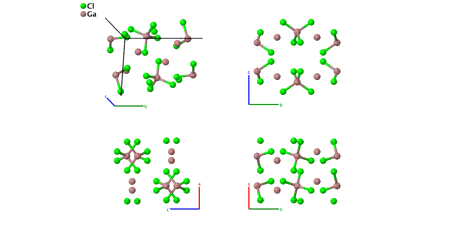}}
\end{figure}
\vspace*{-0.5cm}
\renewcommand{\arraystretch}{1.5}
\begin{equation*}
  \begin{array}{>{$\hspace{-0.15cm}}l<{$}>{$}p{0.5cm}<{$}>{$}p{18.5cm}<{$}}
    \mbox{\large \textbf{Prototype}} &\colon & \ce{GaCl2} \\
    \mbox{\large \textbf{\AFLOW\ prototype label}} &\colon & \mbox{A2B\_oP24\_52\_2e\_cd} \\
    \mbox{\large \textbf{\textit{Strukturbericht} designation}} &\colon & \mbox{None} \\
    \mbox{\large \textbf{Pearson symbol}} &\colon & \mbox{oP24} \\
    \mbox{\large \textbf{Space group number}} &\colon & 52 \\
    \mbox{\large \textbf{Space group symbol}} &\colon & Pnna \\
    \mbox{\large \textbf{\AFLOW\ prototype command}} &\colon &  \texttt{aflow} \,  \, \texttt{-{}-proto=A2B\_oP24\_52\_2e\_cd } \, \newline \texttt{-{}-params=}{a,b/a,c/a,z_{1},x_{2},x_{3},y_{3},z_{3},x_{4},y_{4},z_{4} }
  \end{array}
\end{equation*}
\renewcommand{\arraystretch}{1.0}

\noindent \parbox{1 \linewidth}{
\noindent \hrulefill
\\
\textbf{Simple Orthorhombic primitive vectors:} \\
\vspace*{-0.25cm}
\begin{tabular}{cc}
  \begin{tabular}{c}
    \parbox{0.6 \linewidth}{
      \renewcommand{\arraystretch}{1.5}
      \begin{equation*}
        \centering
        \begin{array}{ccc}
              \mathbf{a}_1 & = & a \, \mathbf{\hat{x}} \\
    \mathbf{a}_2 & = & b \, \mathbf{\hat{y}} \\
    \mathbf{a}_3 & = & c \, \mathbf{\hat{z}} \\

        \end{array}
      \end{equation*}
    }
    \renewcommand{\arraystretch}{1.0}
  \end{tabular}
  \begin{tabular}{c}
    \includegraphics[width=0.3\linewidth]{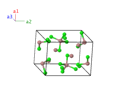} \\
  \end{tabular}
\end{tabular}

}
\vspace*{-0.25cm}

\noindent \hrulefill
\\
\textbf{Basis vectors:}
\vspace*{-0.25cm}
\renewcommand{\arraystretch}{1.5}
\begin{longtabu} to \textwidth{>{\centering $}X[-1,c,c]<{$}>{\centering $}X[-1,c,c]<{$}>{\centering $}X[-1,c,c]<{$}>{\centering $}X[-1,c,c]<{$}>{\centering $}X[-1,c,c]<{$}>{\centering $}X[-1,c,c]<{$}>{\centering $}X[-1,c,c]<{$}}
  & & \mbox{Lattice Coordinates} & & \mbox{Cartesian Coordinates} &\mbox{Wyckoff Position} & \mbox{Atom Type} \\  
  \mathbf{B}_{1} & = & \frac{1}{4} \, \mathbf{a}_{1} + z_{1} \, \mathbf{a}_{3} & = & \frac{1}{4}a \, \mathbf{\hat{x}} + z_{1}c \, \mathbf{\hat{z}} & \left(4c\right) & \mbox{Ga I} \\ 
\mathbf{B}_{2} & = & \frac{1}{4} \, \mathbf{a}_{1} + \frac{1}{2} \, \mathbf{a}_{2} + \left(\frac{1}{2} - z_{1}\right) \, \mathbf{a}_{3} & = & \frac{1}{4}a \, \mathbf{\hat{x}} + \frac{1}{2}b \, \mathbf{\hat{y}} + \left(\frac{1}{2} - z_{1}\right)c \, \mathbf{\hat{z}} & \left(4c\right) & \mbox{Ga I} \\ 
\mathbf{B}_{3} & = & \frac{3}{4} \, \mathbf{a}_{1} + -z_{1} \, \mathbf{a}_{3} & = & \frac{3}{4}a \, \mathbf{\hat{x}} + -z_{1}c \, \mathbf{\hat{z}} & \left(4c\right) & \mbox{Ga I} \\ 
\mathbf{B}_{4} & = & \frac{3}{4} \, \mathbf{a}_{1} + \frac{1}{2} \, \mathbf{a}_{2} + \left(\frac{1}{2} +z_{1}\right) \, \mathbf{a}_{3} & = & \frac{3}{4}a \, \mathbf{\hat{x}} + \frac{1}{2}b \, \mathbf{\hat{y}} + \left(\frac{1}{2} +z_{1}\right)c \, \mathbf{\hat{z}} & \left(4c\right) & \mbox{Ga I} \\ 
\mathbf{B}_{5} & = & x_{2} \, \mathbf{a}_{1} + \frac{1}{4} \, \mathbf{a}_{2} + \frac{1}{4} \, \mathbf{a}_{3} & = & x_{2}a \, \mathbf{\hat{x}} + \frac{1}{4}b \, \mathbf{\hat{y}} + \frac{1}{4}c \, \mathbf{\hat{z}} & \left(4d\right) & \mbox{Ga II} \\ 
\mathbf{B}_{6} & = & \left(\frac{1}{2} - x_{2}\right) \, \mathbf{a}_{1} + \frac{3}{4} \, \mathbf{a}_{2} + \frac{1}{4} \, \mathbf{a}_{3} & = & \left(\frac{1}{2} - x_{2}\right)a \, \mathbf{\hat{x}} + \frac{3}{4}b \, \mathbf{\hat{y}} + \frac{1}{4}c \, \mathbf{\hat{z}} & \left(4d\right) & \mbox{Ga II} \\ 
\mathbf{B}_{7} & = & -x_{2} \, \mathbf{a}_{1} + \frac{3}{4} \, \mathbf{a}_{2} + \frac{3}{4} \, \mathbf{a}_{3} & = & -x_{2}a \, \mathbf{\hat{x}} + \frac{3}{4}b \, \mathbf{\hat{y}} + \frac{3}{4}c \, \mathbf{\hat{z}} & \left(4d\right) & \mbox{Ga II} \\ 
\mathbf{B}_{8} & = & \left(\frac{1}{2} +x_{2}\right) \, \mathbf{a}_{1} + \frac{1}{4} \, \mathbf{a}_{2} + \frac{3}{4} \, \mathbf{a}_{3} & = & \left(\frac{1}{2} +x_{2}\right)a \, \mathbf{\hat{x}} + \frac{1}{4}b \, \mathbf{\hat{y}} + \frac{3}{4}c \, \mathbf{\hat{z}} & \left(4d\right) & \mbox{Ga II} \\ 
\mathbf{B}_{9} & = & x_{3} \, \mathbf{a}_{1} + y_{3} \, \mathbf{a}_{2} + z_{3} \, \mathbf{a}_{3} & = & x_{3}a \, \mathbf{\hat{x}} + y_{3}b \, \mathbf{\hat{y}} + z_{3}c \, \mathbf{\hat{z}} & \left(8e\right) & \mbox{Cl I} \\ 
\mathbf{B}_{10} & = & \left(\frac{1}{2} - x_{3}\right) \, \mathbf{a}_{1}-y_{3} \, \mathbf{a}_{2} + z_{3} \, \mathbf{a}_{3} & = & \left(\frac{1}{2} - x_{3}\right)a \, \mathbf{\hat{x}}-y_{3}b \, \mathbf{\hat{y}} + z_{3}c \, \mathbf{\hat{z}} & \left(8e\right) & \mbox{Cl I} \\ 
\mathbf{B}_{11} & = & \left(\frac{1}{2} - x_{3}\right) \, \mathbf{a}_{1} + \left(\frac{1}{2} +y_{3}\right) \, \mathbf{a}_{2} + \left(\frac{1}{2} - z_{3}\right) \, \mathbf{a}_{3} & = & \left(\frac{1}{2} - x_{3}\right)a \, \mathbf{\hat{x}} + \left(\frac{1}{2} +y_{3}\right)b \, \mathbf{\hat{y}} + \left(\frac{1}{2} - z_{3}\right)c \, \mathbf{\hat{z}} & \left(8e\right) & \mbox{Cl I} \\ 
\mathbf{B}_{12} & = & x_{3} \, \mathbf{a}_{1} + \left(\frac{1}{2} - y_{3}\right) \, \mathbf{a}_{2} + \left(\frac{1}{2} - z_{3}\right) \, \mathbf{a}_{3} & = & x_{3}a \, \mathbf{\hat{x}} + \left(\frac{1}{2} - y_{3}\right)b \, \mathbf{\hat{y}} + \left(\frac{1}{2} - z_{3}\right)c \, \mathbf{\hat{z}} & \left(8e\right) & \mbox{Cl I} \\ 
\mathbf{B}_{13} & = & -x_{3} \, \mathbf{a}_{1}-y_{3} \, \mathbf{a}_{2}-z_{3} \, \mathbf{a}_{3} & = & -x_{3}a \, \mathbf{\hat{x}}-y_{3}b \, \mathbf{\hat{y}}-z_{3}c \, \mathbf{\hat{z}} & \left(8e\right) & \mbox{Cl I} \\ 
\mathbf{B}_{14} & = & \left(\frac{1}{2} +x_{3}\right) \, \mathbf{a}_{1} + y_{3} \, \mathbf{a}_{2}-z_{3} \, \mathbf{a}_{3} & = & \left(\frac{1}{2} +x_{3}\right)a \, \mathbf{\hat{x}} + y_{3}b \, \mathbf{\hat{y}}-z_{3}c \, \mathbf{\hat{z}} & \left(8e\right) & \mbox{Cl I} \\ 
\mathbf{B}_{15} & = & \left(\frac{1}{2} +x_{3}\right) \, \mathbf{a}_{1} + \left(\frac{1}{2} - y_{3}\right) \, \mathbf{a}_{2} + \left(\frac{1}{2} +z_{3}\right) \, \mathbf{a}_{3} & = & \left(\frac{1}{2} +x_{3}\right)a \, \mathbf{\hat{x}} + \left(\frac{1}{2} - y_{3}\right)b \, \mathbf{\hat{y}} + \left(\frac{1}{2} +z_{3}\right)c \, \mathbf{\hat{z}} & \left(8e\right) & \mbox{Cl I} \\ 
\mathbf{B}_{16} & = & -x_{3} \, \mathbf{a}_{1} + \left(\frac{1}{2} +y_{3}\right) \, \mathbf{a}_{2} + \left(\frac{1}{2} +z_{3}\right) \, \mathbf{a}_{3} & = & -x_{3}a \, \mathbf{\hat{x}} + \left(\frac{1}{2} +y_{3}\right)b \, \mathbf{\hat{y}} + \left(\frac{1}{2} +z_{3}\right)c \, \mathbf{\hat{z}} & \left(8e\right) & \mbox{Cl I} \\ 
\mathbf{B}_{17} & = & x_{4} \, \mathbf{a}_{1} + y_{4} \, \mathbf{a}_{2} + z_{4} \, \mathbf{a}_{3} & = & x_{4}a \, \mathbf{\hat{x}} + y_{4}b \, \mathbf{\hat{y}} + z_{4}c \, \mathbf{\hat{z}} & \left(8e\right) & \mbox{Cl II} \\ 
\mathbf{B}_{18} & = & \left(\frac{1}{2} - x_{4}\right) \, \mathbf{a}_{1}-y_{4} \, \mathbf{a}_{2} + z_{4} \, \mathbf{a}_{3} & = & \left(\frac{1}{2} - x_{4}\right)a \, \mathbf{\hat{x}}-y_{4}b \, \mathbf{\hat{y}} + z_{4}c \, \mathbf{\hat{z}} & \left(8e\right) & \mbox{Cl II} \\ 
\mathbf{B}_{19} & = & \left(\frac{1}{2} - x_{4}\right) \, \mathbf{a}_{1} + \left(\frac{1}{2} +y_{4}\right) \, \mathbf{a}_{2} + \left(\frac{1}{2} - z_{4}\right) \, \mathbf{a}_{3} & = & \left(\frac{1}{2} - x_{4}\right)a \, \mathbf{\hat{x}} + \left(\frac{1}{2} +y_{4}\right)b \, \mathbf{\hat{y}} + \left(\frac{1}{2} - z_{4}\right)c \, \mathbf{\hat{z}} & \left(8e\right) & \mbox{Cl II} \\ 
\mathbf{B}_{20} & = & x_{4} \, \mathbf{a}_{1} + \left(\frac{1}{2} - y_{4}\right) \, \mathbf{a}_{2} + \left(\frac{1}{2} - z_{4}\right) \, \mathbf{a}_{3} & = & x_{4}a \, \mathbf{\hat{x}} + \left(\frac{1}{2} - y_{4}\right)b \, \mathbf{\hat{y}} + \left(\frac{1}{2} - z_{4}\right)c \, \mathbf{\hat{z}} & \left(8e\right) & \mbox{Cl II} \\ 
\mathbf{B}_{21} & = & -x_{4} \, \mathbf{a}_{1}-y_{4} \, \mathbf{a}_{2}-z_{4} \, \mathbf{a}_{3} & = & -x_{4}a \, \mathbf{\hat{x}}-y_{4}b \, \mathbf{\hat{y}}-z_{4}c \, \mathbf{\hat{z}} & \left(8e\right) & \mbox{Cl II} \\ 
\mathbf{B}_{22} & = & \left(\frac{1}{2} +x_{4}\right) \, \mathbf{a}_{1} + y_{4} \, \mathbf{a}_{2}-z_{4} \, \mathbf{a}_{3} & = & \left(\frac{1}{2} +x_{4}\right)a \, \mathbf{\hat{x}} + y_{4}b \, \mathbf{\hat{y}}-z_{4}c \, \mathbf{\hat{z}} & \left(8e\right) & \mbox{Cl II} \\ 
\mathbf{B}_{23} & = & \left(\frac{1}{2} +x_{4}\right) \, \mathbf{a}_{1} + \left(\frac{1}{2} - y_{4}\right) \, \mathbf{a}_{2} + \left(\frac{1}{2} +z_{4}\right) \, \mathbf{a}_{3} & = & \left(\frac{1}{2} +x_{4}\right)a \, \mathbf{\hat{x}} + \left(\frac{1}{2} - y_{4}\right)b \, \mathbf{\hat{y}} + \left(\frac{1}{2} +z_{4}\right)c \, \mathbf{\hat{z}} & \left(8e\right) & \mbox{Cl II} \\ 
\mathbf{B}_{24} & = & -x_{4} \, \mathbf{a}_{1} + \left(\frac{1}{2} +y_{4}\right) \, \mathbf{a}_{2} + \left(\frac{1}{2} +z_{4}\right) \, \mathbf{a}_{3} & = & -x_{4}a \, \mathbf{\hat{x}} + \left(\frac{1}{2} +y_{4}\right)b \, \mathbf{\hat{y}} + \left(\frac{1}{2} +z_{4}\right)c \, \mathbf{\hat{z}} & \left(8e\right) & \mbox{Cl II} \\ 
\end{longtabu}
\renewcommand{\arraystretch}{1.0}
\noindent \hrulefill
\\
\textbf{References:}
\vspace*{-0.25cm}
\begin{flushleft}
  - \bibentry{Wilkinson_GaCl2_ActaCrystSecB_1991}. \\
\end{flushleft}
\textbf{Found in:}
\vspace*{-0.25cm}
\begin{flushleft}
  - \bibentry{Villars_PearsonsCrystalData_2013}. \\
\end{flushleft}
\noindent \hrulefill
\\
\textbf{Geometry files:}
\\
\noindent  - CIF: pp. {\hyperref[A2B_oP24_52_2e_cd_cif]{\pageref{A2B_oP24_52_2e_cd_cif}}} \\
\noindent  - POSCAR: pp. {\hyperref[A2B_oP24_52_2e_cd_poscar]{\pageref{A2B_oP24_52_2e_cd_poscar}}} \\
\onecolumn
{\phantomsection\label{A3B2_oP20_52_de_cd}}
\subsection*{\huge \textbf{{\normalfont Sr$_{2}$Bi$_{3}$ Structure: A3B2\_oP20\_52\_de\_cd}}}
\noindent \hrulefill
\vspace*{0.25cm}
\begin{figure}[htp]
  \centering
  \vspace{-1em}
  {\includegraphics[width=1\textwidth]{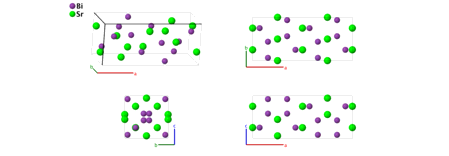}}
\end{figure}
\vspace*{-0.5cm}
\renewcommand{\arraystretch}{1.5}
\begin{equation*}
  \begin{array}{>{$\hspace{-0.15cm}}l<{$}>{$}p{0.5cm}<{$}>{$}p{18.5cm}<{$}}
    \mbox{\large \textbf{Prototype}} &\colon & \ce{Sr2Bi3} \\
    \mbox{\large \textbf{\AFLOW\ prototype label}} &\colon & \mbox{A3B2\_oP20\_52\_de\_cd} \\
    \mbox{\large \textbf{\textit{Strukturbericht} designation}} &\colon & \mbox{None} \\
    \mbox{\large \textbf{Pearson symbol}} &\colon & \mbox{oP20} \\
    \mbox{\large \textbf{Space group number}} &\colon & 52 \\
    \mbox{\large \textbf{Space group symbol}} &\colon & Pnna \\
    \mbox{\large \textbf{\AFLOW\ prototype command}} &\colon &  \texttt{aflow} \,  \, \texttt{-{}-proto=A3B2\_oP20\_52\_de\_cd } \, \newline \texttt{-{}-params=}{a,b/a,c/a,z_{1},x_{2},x_{3},x_{4},y_{4},z_{4} }
  \end{array}
\end{equation*}
\renewcommand{\arraystretch}{1.0}

\noindent \parbox{1 \linewidth}{
\noindent \hrulefill
\\
\textbf{Simple Orthorhombic primitive vectors:} \\
\vspace*{-0.25cm}
\begin{tabular}{cc}
  \begin{tabular}{c}
    \parbox{0.6 \linewidth}{
      \renewcommand{\arraystretch}{1.5}
      \begin{equation*}
        \centering
        \begin{array}{ccc}
              \mathbf{a}_1 & = & a \, \mathbf{\hat{x}} \\
    \mathbf{a}_2 & = & b \, \mathbf{\hat{y}} \\
    \mathbf{a}_3 & = & c \, \mathbf{\hat{z}} \\

        \end{array}
      \end{equation*}
    }
    \renewcommand{\arraystretch}{1.0}
  \end{tabular}
  \begin{tabular}{c}
    \includegraphics[width=0.3\linewidth]{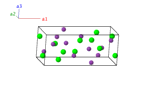} \\
  \end{tabular}
\end{tabular}

}
\vspace*{-0.25cm}

\noindent \hrulefill
\\
\textbf{Basis vectors:}
\vspace*{-0.25cm}
\renewcommand{\arraystretch}{1.5}
\begin{longtabu} to \textwidth{>{\centering $}X[-1,c,c]<{$}>{\centering $}X[-1,c,c]<{$}>{\centering $}X[-1,c,c]<{$}>{\centering $}X[-1,c,c]<{$}>{\centering $}X[-1,c,c]<{$}>{\centering $}X[-1,c,c]<{$}>{\centering $}X[-1,c,c]<{$}}
  & & \mbox{Lattice Coordinates} & & \mbox{Cartesian Coordinates} &\mbox{Wyckoff Position} & \mbox{Atom Type} \\  
  \mathbf{B}_{1} & = & \frac{1}{4} \, \mathbf{a}_{1} + z_{1} \, \mathbf{a}_{3} & = & \frac{1}{4}a \, \mathbf{\hat{x}} + z_{1}c \, \mathbf{\hat{z}} & \left(4c\right) & \mbox{Sr I} \\ 
\mathbf{B}_{2} & = & \frac{1}{4} \, \mathbf{a}_{1} + \frac{1}{2} \, \mathbf{a}_{2} + \left(\frac{1}{2} - z_{1}\right) \, \mathbf{a}_{3} & = & \frac{1}{4}a \, \mathbf{\hat{x}} + \frac{1}{2}b \, \mathbf{\hat{y}} + \left(\frac{1}{2} - z_{1}\right)c \, \mathbf{\hat{z}} & \left(4c\right) & \mbox{Sr I} \\ 
\mathbf{B}_{3} & = & \frac{3}{4} \, \mathbf{a}_{1} + -z_{1} \, \mathbf{a}_{3} & = & \frac{3}{4}a \, \mathbf{\hat{x}} + -z_{1}c \, \mathbf{\hat{z}} & \left(4c\right) & \mbox{Sr I} \\ 
\mathbf{B}_{4} & = & \frac{3}{4} \, \mathbf{a}_{1} + \frac{1}{2} \, \mathbf{a}_{2} + \left(\frac{1}{2} +z_{1}\right) \, \mathbf{a}_{3} & = & \frac{3}{4}a \, \mathbf{\hat{x}} + \frac{1}{2}b \, \mathbf{\hat{y}} + \left(\frac{1}{2} +z_{1}\right)c \, \mathbf{\hat{z}} & \left(4c\right) & \mbox{Sr I} \\ 
\mathbf{B}_{5} & = & x_{2} \, \mathbf{a}_{1} + \frac{1}{4} \, \mathbf{a}_{2} + \frac{1}{4} \, \mathbf{a}_{3} & = & x_{2}a \, \mathbf{\hat{x}} + \frac{1}{4}b \, \mathbf{\hat{y}} + \frac{1}{4}c \, \mathbf{\hat{z}} & \left(4d\right) & \mbox{Bi I} \\ 
\mathbf{B}_{6} & = & \left(\frac{1}{2} - x_{2}\right) \, \mathbf{a}_{1} + \frac{3}{4} \, \mathbf{a}_{2} + \frac{1}{4} \, \mathbf{a}_{3} & = & \left(\frac{1}{2} - x_{2}\right)a \, \mathbf{\hat{x}} + \frac{3}{4}b \, \mathbf{\hat{y}} + \frac{1}{4}c \, \mathbf{\hat{z}} & \left(4d\right) & \mbox{Bi I} \\ 
\mathbf{B}_{7} & = & -x_{2} \, \mathbf{a}_{1} + \frac{3}{4} \, \mathbf{a}_{2} + \frac{3}{4} \, \mathbf{a}_{3} & = & -x_{2}a \, \mathbf{\hat{x}} + \frac{3}{4}b \, \mathbf{\hat{y}} + \frac{3}{4}c \, \mathbf{\hat{z}} & \left(4d\right) & \mbox{Bi I} \\ 
\mathbf{B}_{8} & = & \left(\frac{1}{2} +x_{2}\right) \, \mathbf{a}_{1} + \frac{1}{4} \, \mathbf{a}_{2} + \frac{3}{4} \, \mathbf{a}_{3} & = & \left(\frac{1}{2} +x_{2}\right)a \, \mathbf{\hat{x}} + \frac{1}{4}b \, \mathbf{\hat{y}} + \frac{3}{4}c \, \mathbf{\hat{z}} & \left(4d\right) & \mbox{Bi I} \\ 
\mathbf{B}_{9} & = & x_{3} \, \mathbf{a}_{1} + \frac{1}{4} \, \mathbf{a}_{2} + \frac{1}{4} \, \mathbf{a}_{3} & = & x_{3}a \, \mathbf{\hat{x}} + \frac{1}{4}b \, \mathbf{\hat{y}} + \frac{1}{4}c \, \mathbf{\hat{z}} & \left(4d\right) & \mbox{Sr II} \\ 
\mathbf{B}_{10} & = & \left(\frac{1}{2} - x_{3}\right) \, \mathbf{a}_{1} + \frac{3}{4} \, \mathbf{a}_{2} + \frac{1}{4} \, \mathbf{a}_{3} & = & \left(\frac{1}{2} - x_{3}\right)a \, \mathbf{\hat{x}} + \frac{3}{4}b \, \mathbf{\hat{y}} + \frac{1}{4}c \, \mathbf{\hat{z}} & \left(4d\right) & \mbox{Sr II} \\ 
\mathbf{B}_{11} & = & -x_{3} \, \mathbf{a}_{1} + \frac{3}{4} \, \mathbf{a}_{2} + \frac{3}{4} \, \mathbf{a}_{3} & = & -x_{3}a \, \mathbf{\hat{x}} + \frac{3}{4}b \, \mathbf{\hat{y}} + \frac{3}{4}c \, \mathbf{\hat{z}} & \left(4d\right) & \mbox{Sr II} \\ 
\mathbf{B}_{12} & = & \left(\frac{1}{2} +x_{3}\right) \, \mathbf{a}_{1} + \frac{1}{4} \, \mathbf{a}_{2} + \frac{3}{4} \, \mathbf{a}_{3} & = & \left(\frac{1}{2} +x_{3}\right)a \, \mathbf{\hat{x}} + \frac{1}{4}b \, \mathbf{\hat{y}} + \frac{3}{4}c \, \mathbf{\hat{z}} & \left(4d\right) & \mbox{Sr II} \\ 
\mathbf{B}_{13} & = & x_{4} \, \mathbf{a}_{1} + y_{4} \, \mathbf{a}_{2} + z_{4} \, \mathbf{a}_{3} & = & x_{4}a \, \mathbf{\hat{x}} + y_{4}b \, \mathbf{\hat{y}} + z_{4}c \, \mathbf{\hat{z}} & \left(8e\right) & \mbox{Bi II} \\ 
\mathbf{B}_{14} & = & \left(\frac{1}{2} - x_{4}\right) \, \mathbf{a}_{1}-y_{4} \, \mathbf{a}_{2} + z_{4} \, \mathbf{a}_{3} & = & \left(\frac{1}{2} - x_{4}\right)a \, \mathbf{\hat{x}}-y_{4}b \, \mathbf{\hat{y}} + z_{4}c \, \mathbf{\hat{z}} & \left(8e\right) & \mbox{Bi II} \\ 
\mathbf{B}_{15} & = & \left(\frac{1}{2} - x_{4}\right) \, \mathbf{a}_{1} + \left(\frac{1}{2} +y_{4}\right) \, \mathbf{a}_{2} + \left(\frac{1}{2} - z_{4}\right) \, \mathbf{a}_{3} & = & \left(\frac{1}{2} - x_{4}\right)a \, \mathbf{\hat{x}} + \left(\frac{1}{2} +y_{4}\right)b \, \mathbf{\hat{y}} + \left(\frac{1}{2} - z_{4}\right)c \, \mathbf{\hat{z}} & \left(8e\right) & \mbox{Bi II} \\ 
\mathbf{B}_{16} & = & x_{4} \, \mathbf{a}_{1} + \left(\frac{1}{2} - y_{4}\right) \, \mathbf{a}_{2} + \left(\frac{1}{2} - z_{4}\right) \, \mathbf{a}_{3} & = & x_{4}a \, \mathbf{\hat{x}} + \left(\frac{1}{2} - y_{4}\right)b \, \mathbf{\hat{y}} + \left(\frac{1}{2} - z_{4}\right)c \, \mathbf{\hat{z}} & \left(8e\right) & \mbox{Bi II} \\ 
\mathbf{B}_{17} & = & -x_{4} \, \mathbf{a}_{1}-y_{4} \, \mathbf{a}_{2}-z_{4} \, \mathbf{a}_{3} & = & -x_{4}a \, \mathbf{\hat{x}}-y_{4}b \, \mathbf{\hat{y}}-z_{4}c \, \mathbf{\hat{z}} & \left(8e\right) & \mbox{Bi II} \\ 
\mathbf{B}_{18} & = & \left(\frac{1}{2} +x_{4}\right) \, \mathbf{a}_{1} + y_{4} \, \mathbf{a}_{2}-z_{4} \, \mathbf{a}_{3} & = & \left(\frac{1}{2} +x_{4}\right)a \, \mathbf{\hat{x}} + y_{4}b \, \mathbf{\hat{y}}-z_{4}c \, \mathbf{\hat{z}} & \left(8e\right) & \mbox{Bi II} \\ 
\mathbf{B}_{19} & = & \left(\frac{1}{2} +x_{4}\right) \, \mathbf{a}_{1} + \left(\frac{1}{2} - y_{4}\right) \, \mathbf{a}_{2} + \left(\frac{1}{2} +z_{4}\right) \, \mathbf{a}_{3} & = & \left(\frac{1}{2} +x_{4}\right)a \, \mathbf{\hat{x}} + \left(\frac{1}{2} - y_{4}\right)b \, \mathbf{\hat{y}} + \left(\frac{1}{2} +z_{4}\right)c \, \mathbf{\hat{z}} & \left(8e\right) & \mbox{Bi II} \\ 
\mathbf{B}_{20} & = & -x_{4} \, \mathbf{a}_{1} + \left(\frac{1}{2} +y_{4}\right) \, \mathbf{a}_{2} + \left(\frac{1}{2} +z_{4}\right) \, \mathbf{a}_{3} & = & -x_{4}a \, \mathbf{\hat{x}} + \left(\frac{1}{2} +y_{4}\right)b \, \mathbf{\hat{y}} + \left(\frac{1}{2} +z_{4}\right)c \, \mathbf{\hat{z}} & \left(8e\right) & \mbox{Bi II} \\ 
\end{longtabu}
\renewcommand{\arraystretch}{1.0}
\noindent \hrulefill
\\
\textbf{References:}
\vspace*{-0.25cm}
\begin{flushleft}
  - \bibentry{Merlo_Bi3Sr2_MatResBul_1994}. \\
\end{flushleft}
\textbf{Found in:}
\vspace*{-0.25cm}
\begin{flushleft}
  - \bibentry{Villars_PearsonsCrystalData_2013}. \\
\end{flushleft}
\noindent \hrulefill
\\
\textbf{Geometry files:}
\\
\noindent  - CIF: pp. {\hyperref[A3B2_oP20_52_de_cd_cif]{\pageref{A3B2_oP20_52_de_cd_cif}}} \\
\noindent  - POSCAR: pp. {\hyperref[A3B2_oP20_52_de_cd_poscar]{\pageref{A3B2_oP20_52_de_cd_poscar}}} \\
\onecolumn
{\phantomsection\label{ABC2_oP16_53_h_e_gh}}
\subsection*{\huge \textbf{{\normalfont TaNiTe$_{2}$ Structure: ABC2\_oP16\_53\_h\_e\_gh}}}
\noindent \hrulefill
\vspace*{0.25cm}
\begin{figure}[htp]
  \centering
  \vspace{-1em}
  {\includegraphics[width=1\textwidth]{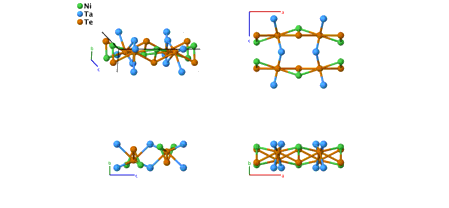}}
\end{figure}
\vspace*{-0.5cm}
\renewcommand{\arraystretch}{1.5}
\begin{equation*}
  \begin{array}{>{$\hspace{-0.15cm}}l<{$}>{$}p{0.5cm}<{$}>{$}p{18.5cm}<{$}}
    \mbox{\large \textbf{Prototype}} &\colon & \ce{TaNiTe2} \\
    \mbox{\large \textbf{\AFLOW\ prototype label}} &\colon & \mbox{ABC2\_oP16\_53\_h\_e\_gh} \\
    \mbox{\large \textbf{\textit{Strukturbericht} designation}} &\colon & \mbox{None} \\
    \mbox{\large \textbf{Pearson symbol}} &\colon & \mbox{oP16} \\
    \mbox{\large \textbf{Space group number}} &\colon & 53 \\
    \mbox{\large \textbf{Space group symbol}} &\colon & Pmna \\
    \mbox{\large \textbf{\AFLOW\ prototype command}} &\colon &  \texttt{aflow} \,  \, \texttt{-{}-proto=ABC2\_oP16\_53\_h\_e\_gh } \, \newline \texttt{-{}-params=}{a,b/a,c/a,x_{1},y_{2},y_{3},z_{3},y_{4},z_{4} }
  \end{array}
\end{equation*}
\renewcommand{\arraystretch}{1.0}

\noindent \parbox{1 \linewidth}{
\noindent \hrulefill
\\
\textbf{Simple Orthorhombic primitive vectors:} \\
\vspace*{-0.25cm}
\begin{tabular}{cc}
  \begin{tabular}{c}
    \parbox{0.6 \linewidth}{
      \renewcommand{\arraystretch}{1.5}
      \begin{equation*}
        \centering
        \begin{array}{ccc}
              \mathbf{a}_1 & = & a \, \mathbf{\hat{x}} \\
    \mathbf{a}_2 & = & b \, \mathbf{\hat{y}} \\
    \mathbf{a}_3 & = & c \, \mathbf{\hat{z}} \\

        \end{array}
      \end{equation*}
    }
    \renewcommand{\arraystretch}{1.0}
  \end{tabular}
  \begin{tabular}{c}
    \includegraphics[width=0.3\linewidth]{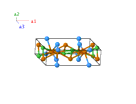} \\
  \end{tabular}
\end{tabular}

}
\vspace*{-0.25cm}

\noindent \hrulefill
\\
\textbf{Basis vectors:}
\vspace*{-0.25cm}
\renewcommand{\arraystretch}{1.5}
\begin{longtabu} to \textwidth{>{\centering $}X[-1,c,c]<{$}>{\centering $}X[-1,c,c]<{$}>{\centering $}X[-1,c,c]<{$}>{\centering $}X[-1,c,c]<{$}>{\centering $}X[-1,c,c]<{$}>{\centering $}X[-1,c,c]<{$}>{\centering $}X[-1,c,c]<{$}}
  & & \mbox{Lattice Coordinates} & & \mbox{Cartesian Coordinates} &\mbox{Wyckoff Position} & \mbox{Atom Type} \\  
  \mathbf{B}_{1} & = & x_{1} \, \mathbf{a}_{1} & = & x_{1}a \, \mathbf{\hat{x}} & \left(4e\right) & \mbox{Ta} \\ 
\mathbf{B}_{2} & = & \left(\frac{1}{2} - x_{1}\right) \, \mathbf{a}_{1} + \frac{1}{2} \, \mathbf{a}_{3} & = & \left(\frac{1}{2} - x_{1}\right)a \, \mathbf{\hat{x}} + \frac{1}{2}c \, \mathbf{\hat{z}} & \left(4e\right) & \mbox{Ta} \\ 
\mathbf{B}_{3} & = & -x_{1} \, \mathbf{a}_{1} & = & -x_{1}a \, \mathbf{\hat{x}} & \left(4e\right) & \mbox{Ta} \\ 
\mathbf{B}_{4} & = & \left(\frac{1}{2} +x_{1}\right) \, \mathbf{a}_{1} + \frac{1}{2} \, \mathbf{a}_{3} & = & \left(\frac{1}{2} +x_{1}\right)a \, \mathbf{\hat{x}} + \frac{1}{2}c \, \mathbf{\hat{z}} & \left(4e\right) & \mbox{Ta} \\ 
\mathbf{B}_{5} & = & \frac{1}{4} \, \mathbf{a}_{1} + y_{2} \, \mathbf{a}_{2} + \frac{1}{4} \, \mathbf{a}_{3} & = & \frac{1}{4}a \, \mathbf{\hat{x}} + y_{2}b \, \mathbf{\hat{y}} + \frac{1}{4}c \, \mathbf{\hat{z}} & \left(4g\right) & \mbox{Te I} \\ 
\mathbf{B}_{6} & = & \frac{1}{4} \, \mathbf{a}_{1}-y_{2} \, \mathbf{a}_{2} + \frac{3}{4} \, \mathbf{a}_{3} & = & \frac{1}{4}a \, \mathbf{\hat{x}}-y_{2}b \, \mathbf{\hat{y}} + \frac{3}{4}c \, \mathbf{\hat{z}} & \left(4g\right) & \mbox{Te I} \\ 
\mathbf{B}_{7} & = & \frac{3}{4} \, \mathbf{a}_{1}-y_{2} \, \mathbf{a}_{2} + \frac{3}{4} \, \mathbf{a}_{3} & = & \frac{3}{4}a \, \mathbf{\hat{x}}-y_{2}b \, \mathbf{\hat{y}} + \frac{3}{4}c \, \mathbf{\hat{z}} & \left(4g\right) & \mbox{Te I} \\ 
\mathbf{B}_{8} & = & \frac{3}{4} \, \mathbf{a}_{1} + y_{2} \, \mathbf{a}_{2} + \frac{1}{4} \, \mathbf{a}_{3} & = & \frac{3}{4}a \, \mathbf{\hat{x}} + y_{2}b \, \mathbf{\hat{y}} + \frac{1}{4}c \, \mathbf{\hat{z}} & \left(4g\right) & \mbox{Te I} \\ 
\mathbf{B}_{9} & = & y_{3} \, \mathbf{a}_{2} + z_{3} \, \mathbf{a}_{3} & = & y_{3}b \, \mathbf{\hat{y}} + z_{3}c \, \mathbf{\hat{z}} & \left(4h\right) & \mbox{Ni} \\ 
\mathbf{B}_{10} & = & \frac{1}{2} \, \mathbf{a}_{1}-y_{3} \, \mathbf{a}_{2} + \left(\frac{1}{2} +z_{3}\right) \, \mathbf{a}_{3} & = & \frac{1}{2}a \, \mathbf{\hat{x}}-y_{3}b \, \mathbf{\hat{y}} + \left(\frac{1}{2} +z_{3}\right)c \, \mathbf{\hat{z}} & \left(4h\right) & \mbox{Ni} \\ 
\mathbf{B}_{11} & = & \frac{1}{2} \, \mathbf{a}_{1} + y_{3} \, \mathbf{a}_{2} + \left(\frac{1}{2} - z_{3}\right) \, \mathbf{a}_{3} & = & \frac{1}{2}a \, \mathbf{\hat{x}} + y_{3}b \, \mathbf{\hat{y}} + \left(\frac{1}{2} - z_{3}\right)c \, \mathbf{\hat{z}} & \left(4h\right) & \mbox{Ni} \\ 
\mathbf{B}_{12} & = & -y_{3} \, \mathbf{a}_{2}-z_{3} \, \mathbf{a}_{3} & = & -y_{3}b \, \mathbf{\hat{y}}-z_{3}c \, \mathbf{\hat{z}} & \left(4h\right) & \mbox{Ni} \\ 
\mathbf{B}_{13} & = & y_{4} \, \mathbf{a}_{2} + z_{4} \, \mathbf{a}_{3} & = & y_{4}b \, \mathbf{\hat{y}} + z_{4}c \, \mathbf{\hat{z}} & \left(4h\right) & \mbox{Te II} \\ 
\mathbf{B}_{14} & = & \frac{1}{2} \, \mathbf{a}_{1}-y_{4} \, \mathbf{a}_{2} + \left(\frac{1}{2} +z_{4}\right) \, \mathbf{a}_{3} & = & \frac{1}{2}a \, \mathbf{\hat{x}}-y_{4}b \, \mathbf{\hat{y}} + \left(\frac{1}{2} +z_{4}\right)c \, \mathbf{\hat{z}} & \left(4h\right) & \mbox{Te II} \\ 
\mathbf{B}_{15} & = & \frac{1}{2} \, \mathbf{a}_{1} + y_{4} \, \mathbf{a}_{2} + \left(\frac{1}{2} - z_{4}\right) \, \mathbf{a}_{3} & = & \frac{1}{2}a \, \mathbf{\hat{x}} + y_{4}b \, \mathbf{\hat{y}} + \left(\frac{1}{2} - z_{4}\right)c \, \mathbf{\hat{z}} & \left(4h\right) & \mbox{Te II} \\ 
\mathbf{B}_{16} & = & -y_{4} \, \mathbf{a}_{2}-z_{4} \, \mathbf{a}_{3} & = & -y_{4}b \, \mathbf{\hat{y}}-z_{4}c \, \mathbf{\hat{z}} & \left(4h\right) & \mbox{Te II} \\ 
\end{longtabu}
\renewcommand{\arraystretch}{1.0}
\noindent \hrulefill
\\
\textbf{References:}
\vspace*{-0.25cm}
\begin{flushleft}
  - \bibentry{Tremel_NiTaTe2_AngewChemIntEd_1991}. \\
\end{flushleft}
\textbf{Found in:}
\vspace*{-0.25cm}
\begin{flushleft}
  - \bibentry{Villars_PearsonsCrystalData_2013}. \\
\end{flushleft}
\noindent \hrulefill
\\
\textbf{Geometry files:}
\\
\noindent  - CIF: pp. {\hyperref[ABC2_oP16_53_h_e_gh_cif]{\pageref{ABC2_oP16_53_h_e_gh_cif}}} \\
\noindent  - POSCAR: pp. {\hyperref[ABC2_oP16_53_h_e_gh_poscar]{\pageref{ABC2_oP16_53_h_e_gh_poscar}}} \\
\onecolumn
{\phantomsection\label{ABC3_oP20_53_e_g_hi}}
\subsection*{\huge \textbf{{\normalfont CuBrSe$_{3}$ Structure: ABC3\_oP20\_53\_e\_g\_hi}}}
\noindent \hrulefill
\vspace*{0.25cm}
\begin{figure}[htp]
  \centering
  \vspace{-1em}
  {\includegraphics[width=1\textwidth]{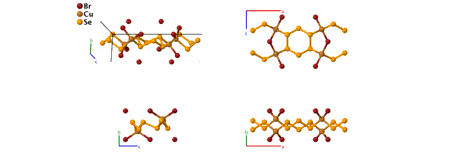}}
\end{figure}
\vspace*{-0.5cm}
\renewcommand{\arraystretch}{1.5}
\begin{equation*}
  \begin{array}{>{$\hspace{-0.15cm}}l<{$}>{$}p{0.5cm}<{$}>{$}p{18.5cm}<{$}}
    \mbox{\large \textbf{Prototype}} &\colon & \ce{CuBrSe3} \\
    \mbox{\large \textbf{\AFLOW\ prototype label}} &\colon & \mbox{ABC3\_oP20\_53\_e\_g\_hi} \\
    \mbox{\large \textbf{\textit{Strukturbericht} designation}} &\colon & \mbox{None} \\
    \mbox{\large \textbf{Pearson symbol}} &\colon & \mbox{oP20} \\
    \mbox{\large \textbf{Space group number}} &\colon & 53 \\
    \mbox{\large \textbf{Space group symbol}} &\colon & Pmna \\
    \mbox{\large \textbf{\AFLOW\ prototype command}} &\colon &  \texttt{aflow} \,  \, \texttt{-{}-proto=ABC3\_oP20\_53\_e\_g\_hi } \, \newline \texttt{-{}-params=}{a,b/a,c/a,x_{1},y_{2},y_{3},z_{3},x_{4},y_{4},z_{4} }
  \end{array}
\end{equation*}
\renewcommand{\arraystretch}{1.0}

\noindent \parbox{1 \linewidth}{
\noindent \hrulefill
\\
\textbf{Simple Orthorhombic primitive vectors:} \\
\vspace*{-0.25cm}
\begin{tabular}{cc}
  \begin{tabular}{c}
    \parbox{0.6 \linewidth}{
      \renewcommand{\arraystretch}{1.5}
      \begin{equation*}
        \centering
        \begin{array}{ccc}
              \mathbf{a}_1 & = & a \, \mathbf{\hat{x}} \\
    \mathbf{a}_2 & = & b \, \mathbf{\hat{y}} \\
    \mathbf{a}_3 & = & c \, \mathbf{\hat{z}} \\

        \end{array}
      \end{equation*}
    }
    \renewcommand{\arraystretch}{1.0}
  \end{tabular}
  \begin{tabular}{c}
    \includegraphics[width=0.3\linewidth]{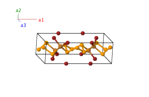} \\
  \end{tabular}
\end{tabular}

}
\vspace*{-0.25cm}

\noindent \hrulefill
\\
\textbf{Basis vectors:}
\vspace*{-0.25cm}
\renewcommand{\arraystretch}{1.5}
\begin{longtabu} to \textwidth{>{\centering $}X[-1,c,c]<{$}>{\centering $}X[-1,c,c]<{$}>{\centering $}X[-1,c,c]<{$}>{\centering $}X[-1,c,c]<{$}>{\centering $}X[-1,c,c]<{$}>{\centering $}X[-1,c,c]<{$}>{\centering $}X[-1,c,c]<{$}}
  & & \mbox{Lattice Coordinates} & & \mbox{Cartesian Coordinates} &\mbox{Wyckoff Position} & \mbox{Atom Type} \\  
  \mathbf{B}_{1} & = & x_{1} \, \mathbf{a}_{1} & = & x_{1}a \, \mathbf{\hat{x}} & \left(4e\right) & \mbox{Br} \\ 
\mathbf{B}_{2} & = & \left(\frac{1}{2} - x_{1}\right) \, \mathbf{a}_{1} + \frac{1}{2} \, \mathbf{a}_{3} & = & \left(\frac{1}{2} - x_{1}\right)a \, \mathbf{\hat{x}} + \frac{1}{2}c \, \mathbf{\hat{z}} & \left(4e\right) & \mbox{Br} \\ 
\mathbf{B}_{3} & = & -x_{1} \, \mathbf{a}_{1} & = & -x_{1}a \, \mathbf{\hat{x}} & \left(4e\right) & \mbox{Br} \\ 
\mathbf{B}_{4} & = & \left(\frac{1}{2} +x_{1}\right) \, \mathbf{a}_{1} + \frac{1}{2} \, \mathbf{a}_{3} & = & \left(\frac{1}{2} +x_{1}\right)a \, \mathbf{\hat{x}} + \frac{1}{2}c \, \mathbf{\hat{z}} & \left(4e\right) & \mbox{Br} \\ 
\mathbf{B}_{5} & = & \frac{1}{4} \, \mathbf{a}_{1} + y_{2} \, \mathbf{a}_{2} + \frac{1}{4} \, \mathbf{a}_{3} & = & \frac{1}{4}a \, \mathbf{\hat{x}} + y_{2}b \, \mathbf{\hat{y}} + \frac{1}{4}c \, \mathbf{\hat{z}} & \left(4g\right) & \mbox{Cu} \\ 
\mathbf{B}_{6} & = & \frac{1}{4} \, \mathbf{a}_{1}-y_{2} \, \mathbf{a}_{2} + \frac{3}{4} \, \mathbf{a}_{3} & = & \frac{1}{4}a \, \mathbf{\hat{x}}-y_{2}b \, \mathbf{\hat{y}} + \frac{3}{4}c \, \mathbf{\hat{z}} & \left(4g\right) & \mbox{Cu} \\ 
\mathbf{B}_{7} & = & \frac{3}{4} \, \mathbf{a}_{1}-y_{2} \, \mathbf{a}_{2} + \frac{3}{4} \, \mathbf{a}_{3} & = & \frac{3}{4}a \, \mathbf{\hat{x}}-y_{2}b \, \mathbf{\hat{y}} + \frac{3}{4}c \, \mathbf{\hat{z}} & \left(4g\right) & \mbox{Cu} \\ 
\mathbf{B}_{8} & = & \frac{3}{4} \, \mathbf{a}_{1} + y_{2} \, \mathbf{a}_{2} + \frac{1}{4} \, \mathbf{a}_{3} & = & \frac{3}{4}a \, \mathbf{\hat{x}} + y_{2}b \, \mathbf{\hat{y}} + \frac{1}{4}c \, \mathbf{\hat{z}} & \left(4g\right) & \mbox{Cu} \\ 
\mathbf{B}_{9} & = & y_{3} \, \mathbf{a}_{2} + z_{3} \, \mathbf{a}_{3} & = & y_{3}b \, \mathbf{\hat{y}} + z_{3}c \, \mathbf{\hat{z}} & \left(4h\right) & \mbox{Se I} \\ 
\mathbf{B}_{10} & = & \frac{1}{2} \, \mathbf{a}_{1}-y_{3} \, \mathbf{a}_{2} + \left(\frac{1}{2} +z_{3}\right) \, \mathbf{a}_{3} & = & \frac{1}{2}a \, \mathbf{\hat{x}}-y_{3}b \, \mathbf{\hat{y}} + \left(\frac{1}{2} +z_{3}\right)c \, \mathbf{\hat{z}} & \left(4h\right) & \mbox{Se I} \\ 
\mathbf{B}_{11} & = & \frac{1}{2} \, \mathbf{a}_{1} + y_{3} \, \mathbf{a}_{2} + \left(\frac{1}{2} - z_{3}\right) \, \mathbf{a}_{3} & = & \frac{1}{2}a \, \mathbf{\hat{x}} + y_{3}b \, \mathbf{\hat{y}} + \left(\frac{1}{2} - z_{3}\right)c \, \mathbf{\hat{z}} & \left(4h\right) & \mbox{Se I} \\ 
\mathbf{B}_{12} & = & -y_{3} \, \mathbf{a}_{2}-z_{3} \, \mathbf{a}_{3} & = & -y_{3}b \, \mathbf{\hat{y}}-z_{3}c \, \mathbf{\hat{z}} & \left(4h\right) & \mbox{Se I} \\ 
\mathbf{B}_{13} & = & x_{4} \, \mathbf{a}_{1} + y_{4} \, \mathbf{a}_{2} + z_{4} \, \mathbf{a}_{3} & = & x_{4}a \, \mathbf{\hat{x}} + y_{4}b \, \mathbf{\hat{y}} + z_{4}c \, \mathbf{\hat{z}} & \left(8i\right) & \mbox{Se II} \\ 
\mathbf{B}_{14} & = & \left(\frac{1}{2} - x_{4}\right) \, \mathbf{a}_{1}-y_{4} \, \mathbf{a}_{2} + \left(\frac{1}{2} +z_{4}\right) \, \mathbf{a}_{3} & = & \left(\frac{1}{2} - x_{4}\right)a \, \mathbf{\hat{x}}-y_{4}b \, \mathbf{\hat{y}} + \left(\frac{1}{2} +z_{4}\right)c \, \mathbf{\hat{z}} & \left(8i\right) & \mbox{Se II} \\ 
\mathbf{B}_{15} & = & \left(\frac{1}{2} - x_{4}\right) \, \mathbf{a}_{1} + y_{4} \, \mathbf{a}_{2} + \left(\frac{1}{2} - z_{4}\right) \, \mathbf{a}_{3} & = & \left(\frac{1}{2} - x_{4}\right)a \, \mathbf{\hat{x}} + y_{4}b \, \mathbf{\hat{y}} + \left(\frac{1}{2} - z_{4}\right)c \, \mathbf{\hat{z}} & \left(8i\right) & \mbox{Se II} \\ 
\mathbf{B}_{16} & = & x_{4} \, \mathbf{a}_{1}-y_{4} \, \mathbf{a}_{2}-z_{4} \, \mathbf{a}_{3} & = & x_{4}a \, \mathbf{\hat{x}}-y_{4}b \, \mathbf{\hat{y}}-z_{4}c \, \mathbf{\hat{z}} & \left(8i\right) & \mbox{Se II} \\ 
\mathbf{B}_{17} & = & -x_{4} \, \mathbf{a}_{1}-y_{4} \, \mathbf{a}_{2}-z_{4} \, \mathbf{a}_{3} & = & -x_{4}a \, \mathbf{\hat{x}}-y_{4}b \, \mathbf{\hat{y}}-z_{4}c \, \mathbf{\hat{z}} & \left(8i\right) & \mbox{Se II} \\ 
\mathbf{B}_{18} & = & \left(\frac{1}{2} +x_{4}\right) \, \mathbf{a}_{1} + y_{4} \, \mathbf{a}_{2} + \left(\frac{1}{2} - z_{4}\right) \, \mathbf{a}_{3} & = & \left(\frac{1}{2} +x_{4}\right)a \, \mathbf{\hat{x}} + y_{4}b \, \mathbf{\hat{y}} + \left(\frac{1}{2} - z_{4}\right)c \, \mathbf{\hat{z}} & \left(8i\right) & \mbox{Se II} \\ 
\mathbf{B}_{19} & = & \left(\frac{1}{2} +x_{4}\right) \, \mathbf{a}_{1}-y_{4} \, \mathbf{a}_{2} + \left(\frac{1}{2} +z_{4}\right) \, \mathbf{a}_{3} & = & \left(\frac{1}{2} +x_{4}\right)a \, \mathbf{\hat{x}}-y_{4}b \, \mathbf{\hat{y}} + \left(\frac{1}{2} +z_{4}\right)c \, \mathbf{\hat{z}} & \left(8i\right) & \mbox{Se II} \\ 
\mathbf{B}_{20} & = & -x_{4} \, \mathbf{a}_{1} + y_{4} \, \mathbf{a}_{2} + z_{4} \, \mathbf{a}_{3} & = & -x_{4}a \, \mathbf{\hat{x}} + y_{4}b \, \mathbf{\hat{y}} + z_{4}c \, \mathbf{\hat{z}} & \left(8i\right) & \mbox{Se II} \\ 
\end{longtabu}
\renewcommand{\arraystretch}{1.0}
\noindent \hrulefill
\\
\textbf{References:}
\vspace*{-0.25cm}
\begin{flushleft}
  - \bibentry{Haendler_BrCuSe3_JSolStateChem_1979}. \\
\end{flushleft}
\textbf{Found in:}
\vspace*{-0.25cm}
\begin{flushleft}
  - \bibentry{Villars_PearsonsCrystalData_2013}. \\
\end{flushleft}
\noindent \hrulefill
\\
\textbf{Geometry files:}
\\
\noindent  - CIF: pp. {\hyperref[ABC3_oP20_53_e_g_hi_cif]{\pageref{ABC3_oP20_53_e_g_hi_cif}}} \\
\noindent  - POSCAR: pp. {\hyperref[ABC3_oP20_53_e_g_hi_poscar]{\pageref{ABC3_oP20_53_e_g_hi_poscar}}} \\
\onecolumn
{\phantomsection\label{ABC3_oP20_54_e_d_cf}}
\subsection*{\huge \textbf{{\normalfont BiGaO$_{3}$ Structure: ABC3\_oP20\_54\_e\_d\_cf}}}
\noindent \hrulefill
\vspace*{0.25cm}
\begin{figure}[htp]
  \centering
  \vspace{-1em}
  {\includegraphics[width=1\textwidth]{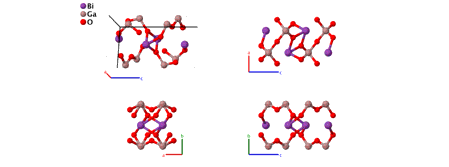}}
\end{figure}
\vspace*{-0.5cm}
\renewcommand{\arraystretch}{1.5}
\begin{equation*}
  \begin{array}{>{$\hspace{-0.15cm}}l<{$}>{$}p{0.5cm}<{$}>{$}p{18.5cm}<{$}}
    \mbox{\large \textbf{Prototype}} &\colon & \ce{BiGaO3} \\
    \mbox{\large \textbf{\AFLOW\ prototype label}} &\colon & \mbox{ABC3\_oP20\_54\_e\_d\_cf} \\
    \mbox{\large \textbf{\textit{Strukturbericht} designation}} &\colon & \mbox{None} \\
    \mbox{\large \textbf{Pearson symbol}} &\colon & \mbox{oP20} \\
    \mbox{\large \textbf{Space group number}} &\colon & 54 \\
    \mbox{\large \textbf{Space group symbol}} &\colon & Pcca \\
    \mbox{\large \textbf{\AFLOW\ prototype command}} &\colon &  \texttt{aflow} \,  \, \texttt{-{}-proto=ABC3\_oP20\_54\_e\_d\_cf } \, \newline \texttt{-{}-params=}{a,b/a,c/a,y_{1},z_{2},z_{3},x_{4},y_{4},z_{4} }
  \end{array}
\end{equation*}
\renewcommand{\arraystretch}{1.0}

\noindent \parbox{1 \linewidth}{
\noindent \hrulefill
\\
\textbf{Simple Orthorhombic primitive vectors:} \\
\vspace*{-0.25cm}
\begin{tabular}{cc}
  \begin{tabular}{c}
    \parbox{0.6 \linewidth}{
      \renewcommand{\arraystretch}{1.5}
      \begin{equation*}
        \centering
        \begin{array}{ccc}
              \mathbf{a}_1 & = & a \, \mathbf{\hat{x}} \\
    \mathbf{a}_2 & = & b \, \mathbf{\hat{y}} \\
    \mathbf{a}_3 & = & c \, \mathbf{\hat{z}} \\

        \end{array}
      \end{equation*}
    }
    \renewcommand{\arraystretch}{1.0}
  \end{tabular}
  \begin{tabular}{c}
    \includegraphics[width=0.3\linewidth]{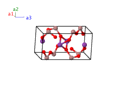} \\
  \end{tabular}
\end{tabular}

}
\vspace*{-0.25cm}

\noindent \hrulefill
\\
\textbf{Basis vectors:}
\vspace*{-0.25cm}
\renewcommand{\arraystretch}{1.5}
\begin{longtabu} to \textwidth{>{\centering $}X[-1,c,c]<{$}>{\centering $}X[-1,c,c]<{$}>{\centering $}X[-1,c,c]<{$}>{\centering $}X[-1,c,c]<{$}>{\centering $}X[-1,c,c]<{$}>{\centering $}X[-1,c,c]<{$}>{\centering $}X[-1,c,c]<{$}}
  & & \mbox{Lattice Coordinates} & & \mbox{Cartesian Coordinates} &\mbox{Wyckoff Position} & \mbox{Atom Type} \\  
  \mathbf{B}_{1} & = & y_{1} \, \mathbf{a}_{2} + \frac{1}{4} \, \mathbf{a}_{3} & = & y_{1}b \, \mathbf{\hat{y}} + \frac{1}{4}c \, \mathbf{\hat{z}} & \left(4c\right) & \mbox{O I} \\ 
\mathbf{B}_{2} & = & \frac{1}{2} \, \mathbf{a}_{1}-y_{1} \, \mathbf{a}_{2} + \frac{1}{4} \, \mathbf{a}_{3} & = & \frac{1}{2}a \, \mathbf{\hat{x}}-y_{1}b \, \mathbf{\hat{y}} + \frac{1}{4}c \, \mathbf{\hat{z}} & \left(4c\right) & \mbox{O I} \\ 
\mathbf{B}_{3} & = & -y_{1} \, \mathbf{a}_{2} + \frac{3}{4} \, \mathbf{a}_{3} & = & -y_{1}b \, \mathbf{\hat{y}} + \frac{3}{4}c \, \mathbf{\hat{z}} & \left(4c\right) & \mbox{O I} \\ 
\mathbf{B}_{4} & = & \frac{1}{2} \, \mathbf{a}_{1} + y_{1} \, \mathbf{a}_{2} + \frac{3}{4} \, \mathbf{a}_{3} & = & \frac{1}{2}a \, \mathbf{\hat{x}} + y_{1}b \, \mathbf{\hat{y}} + \frac{3}{4}c \, \mathbf{\hat{z}} & \left(4c\right) & \mbox{O I} \\ 
\mathbf{B}_{5} & = & \frac{1}{4} \, \mathbf{a}_{1} + z_{2} \, \mathbf{a}_{3} & = & \frac{1}{4}a \, \mathbf{\hat{x}} + z_{2}c \, \mathbf{\hat{z}} & \left(4d\right) & \mbox{Ga} \\ 
\mathbf{B}_{6} & = & \frac{3}{4} \, \mathbf{a}_{1} + \left(\frac{1}{2} - z_{2}\right) \, \mathbf{a}_{3} & = & \frac{3}{4}a \, \mathbf{\hat{x}} + \left(\frac{1}{2} - z_{2}\right)c \, \mathbf{\hat{z}} & \left(4d\right) & \mbox{Ga} \\ 
\mathbf{B}_{7} & = & \frac{3}{4} \, \mathbf{a}_{1} + -z_{2} \, \mathbf{a}_{3} & = & \frac{3}{4}a \, \mathbf{\hat{x}} + -z_{2}c \, \mathbf{\hat{z}} & \left(4d\right) & \mbox{Ga} \\ 
\mathbf{B}_{8} & = & \frac{1}{4} \, \mathbf{a}_{1} + \left(\frac{1}{2} +z_{2}\right) \, \mathbf{a}_{3} & = & \frac{1}{4}a \, \mathbf{\hat{x}} + \left(\frac{1}{2} +z_{2}\right)c \, \mathbf{\hat{z}} & \left(4d\right) & \mbox{Ga} \\ 
\mathbf{B}_{9} & = & \frac{1}{4} \, \mathbf{a}_{1} + \frac{1}{2} \, \mathbf{a}_{2} + z_{3} \, \mathbf{a}_{3} & = & \frac{1}{4}a \, \mathbf{\hat{x}} + \frac{1}{2}b \, \mathbf{\hat{y}} + z_{3}c \, \mathbf{\hat{z}} & \left(4e\right) & \mbox{Bi} \\ 
\mathbf{B}_{10} & = & \frac{3}{4} \, \mathbf{a}_{1} + \frac{1}{2} \, \mathbf{a}_{2} + \left(\frac{1}{2} - z_{3}\right) \, \mathbf{a}_{3} & = & \frac{3}{4}a \, \mathbf{\hat{x}} + \frac{1}{2}b \, \mathbf{\hat{y}} + \left(\frac{1}{2} - z_{3}\right)c \, \mathbf{\hat{z}} & \left(4e\right) & \mbox{Bi} \\ 
\mathbf{B}_{11} & = & \frac{3}{4} \, \mathbf{a}_{1} + \frac{1}{2} \, \mathbf{a}_{2}-z_{3} \, \mathbf{a}_{3} & = & \frac{3}{4}a \, \mathbf{\hat{x}} + \frac{1}{2}b \, \mathbf{\hat{y}}-z_{3}c \, \mathbf{\hat{z}} & \left(4e\right) & \mbox{Bi} \\ 
\mathbf{B}_{12} & = & \frac{1}{4} \, \mathbf{a}_{1} + \frac{1}{2} \, \mathbf{a}_{2} + \left(\frac{1}{2} +z_{3}\right) \, \mathbf{a}_{3} & = & \frac{1}{4}a \, \mathbf{\hat{x}} + \frac{1}{2}b \, \mathbf{\hat{y}} + \left(\frac{1}{2} +z_{3}\right)c \, \mathbf{\hat{z}} & \left(4e\right) & \mbox{Bi} \\ 
\mathbf{B}_{13} & = & x_{4} \, \mathbf{a}_{1} + y_{4} \, \mathbf{a}_{2} + z_{4} \, \mathbf{a}_{3} & = & x_{4}a \, \mathbf{\hat{x}} + y_{4}b \, \mathbf{\hat{y}} + z_{4}c \, \mathbf{\hat{z}} & \left(8f\right) & \mbox{O II} \\ 
\mathbf{B}_{14} & = & \left(\frac{1}{2} - x_{4}\right) \, \mathbf{a}_{1}-y_{4} \, \mathbf{a}_{2} + z_{4} \, \mathbf{a}_{3} & = & \left(\frac{1}{2} - x_{4}\right)a \, \mathbf{\hat{x}}-y_{4}b \, \mathbf{\hat{y}} + z_{4}c \, \mathbf{\hat{z}} & \left(8f\right) & \mbox{O II} \\ 
\mathbf{B}_{15} & = & -x_{4} \, \mathbf{a}_{1} + y_{4} \, \mathbf{a}_{2} + \left(\frac{1}{2} - z_{4}\right) \, \mathbf{a}_{3} & = & -x_{4}a \, \mathbf{\hat{x}} + y_{4}b \, \mathbf{\hat{y}} + \left(\frac{1}{2} - z_{4}\right)c \, \mathbf{\hat{z}} & \left(8f\right) & \mbox{O II} \\ 
\mathbf{B}_{16} & = & \left(\frac{1}{2} +x_{4}\right) \, \mathbf{a}_{1}-y_{4} \, \mathbf{a}_{2} + \left(\frac{1}{2} - z_{4}\right) \, \mathbf{a}_{3} & = & \left(\frac{1}{2} +x_{4}\right)a \, \mathbf{\hat{x}}-y_{4}b \, \mathbf{\hat{y}} + \left(\frac{1}{2} - z_{4}\right)c \, \mathbf{\hat{z}} & \left(8f\right) & \mbox{O II} \\ 
\mathbf{B}_{17} & = & -x_{4} \, \mathbf{a}_{1}-y_{4} \, \mathbf{a}_{2}-z_{4} \, \mathbf{a}_{3} & = & -x_{4}a \, \mathbf{\hat{x}}-y_{4}b \, \mathbf{\hat{y}}-z_{4}c \, \mathbf{\hat{z}} & \left(8f\right) & \mbox{O II} \\ 
\mathbf{B}_{18} & = & \left(\frac{1}{2} +x_{4}\right) \, \mathbf{a}_{1} + y_{4} \, \mathbf{a}_{2}-z_{4} \, \mathbf{a}_{3} & = & \left(\frac{1}{2} +x_{4}\right)a \, \mathbf{\hat{x}} + y_{4}b \, \mathbf{\hat{y}}-z_{4}c \, \mathbf{\hat{z}} & \left(8f\right) & \mbox{O II} \\ 
\mathbf{B}_{19} & = & x_{4} \, \mathbf{a}_{1}-y_{4} \, \mathbf{a}_{2} + \left(\frac{1}{2} +z_{4}\right) \, \mathbf{a}_{3} & = & x_{4}a \, \mathbf{\hat{x}}-y_{4}b \, \mathbf{\hat{y}} + \left(\frac{1}{2} +z_{4}\right)c \, \mathbf{\hat{z}} & \left(8f\right) & \mbox{O II} \\ 
\mathbf{B}_{20} & = & \left(\frac{1}{2} - x_{4}\right) \, \mathbf{a}_{1} + y_{4} \, \mathbf{a}_{2} + \left(\frac{1}{2} +z_{4}\right) \, \mathbf{a}_{3} & = & \left(\frac{1}{2} - x_{4}\right)a \, \mathbf{\hat{x}} + y_{4}b \, \mathbf{\hat{y}} + \left(\frac{1}{2} +z_{4}\right)c \, \mathbf{\hat{z}} & \left(8f\right) & \mbox{O II} \\ 
\end{longtabu}
\renewcommand{\arraystretch}{1.0}
\noindent \hrulefill
\\
\textbf{References:}
\vspace*{-0.25cm}
\begin{flushleft}
  - \bibentry{Yusa_BiGaO3_PhysRevB_2009}. \\
\end{flushleft}
\textbf{Found in:}
\vspace*{-0.25cm}
\begin{flushleft}
  - \bibentry{Villars_PearsonsCrystalData_2013}. \\
\end{flushleft}
\noindent \hrulefill
\\
\textbf{Geometry files:}
\\
\noindent  - CIF: pp. {\hyperref[ABC3_oP20_54_e_d_cf_cif]{\pageref{ABC3_oP20_54_e_d_cf_cif}}} \\
\noindent  - POSCAR: pp. {\hyperref[ABC3_oP20_54_e_d_cf_poscar]{\pageref{ABC3_oP20_54_e_d_cf_poscar}}} \\
\onecolumn
{\phantomsection\label{A2B_oP24_55_2g2h_gh}}
\subsection*{\huge \textbf{{\normalfont GeAs$_{2}$ Structure: A2B\_oP24\_55\_2g2h\_gh}}}
\noindent \hrulefill
\vspace*{0.25cm}
\begin{figure}[htp]
  \centering
  \vspace{-1em}
  {\includegraphics[width=1\textwidth]{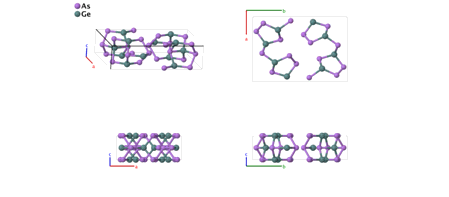}}
\end{figure}
\vspace*{-0.5cm}
\renewcommand{\arraystretch}{1.5}
\begin{equation*}
  \begin{array}{>{$\hspace{-0.15cm}}l<{$}>{$}p{0.5cm}<{$}>{$}p{18.5cm}<{$}}
    \mbox{\large \textbf{Prototype}} &\colon & \ce{GeAs2} \\
    \mbox{\large \textbf{\AFLOW\ prototype label}} &\colon & \mbox{A2B\_oP24\_55\_2g2h\_gh} \\
    \mbox{\large \textbf{\textit{Strukturbericht} designation}} &\colon & \mbox{None} \\
    \mbox{\large \textbf{Pearson symbol}} &\colon & \mbox{oP24} \\
    \mbox{\large \textbf{Space group number}} &\colon & 55 \\
    \mbox{\large \textbf{Space group symbol}} &\colon & Pbam \\
    \mbox{\large \textbf{\AFLOW\ prototype command}} &\colon &  \texttt{aflow} \,  \, \texttt{-{}-proto=A2B\_oP24\_55\_2g2h\_gh } \, \newline \texttt{-{}-params=}{a,b/a,c/a,x_{1},y_{1},x_{2},y_{2},x_{3},y_{3},x_{4},y_{4},x_{5},y_{5},x_{6},y_{6} }
  \end{array}
\end{equation*}
\renewcommand{\arraystretch}{1.0}

\noindent \parbox{1 \linewidth}{
\noindent \hrulefill
\\
\textbf{Simple Orthorhombic primitive vectors:} \\
\vspace*{-0.25cm}
\begin{tabular}{cc}
  \begin{tabular}{c}
    \parbox{0.6 \linewidth}{
      \renewcommand{\arraystretch}{1.5}
      \begin{equation*}
        \centering
        \begin{array}{ccc}
              \mathbf{a}_1 & = & a \, \mathbf{\hat{x}} \\
    \mathbf{a}_2 & = & b \, \mathbf{\hat{y}} \\
    \mathbf{a}_3 & = & c \, \mathbf{\hat{z}} \\

        \end{array}
      \end{equation*}
    }
    \renewcommand{\arraystretch}{1.0}
  \end{tabular}
  \begin{tabular}{c}
    \includegraphics[width=0.3\linewidth]{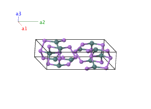} \\
  \end{tabular}
\end{tabular}

}
\vspace*{-0.25cm}

\noindent \hrulefill
\\
\textbf{Basis vectors:}
\vspace*{-0.25cm}
\renewcommand{\arraystretch}{1.5}
\begin{longtabu} to \textwidth{>{\centering $}X[-1,c,c]<{$}>{\centering $}X[-1,c,c]<{$}>{\centering $}X[-1,c,c]<{$}>{\centering $}X[-1,c,c]<{$}>{\centering $}X[-1,c,c]<{$}>{\centering $}X[-1,c,c]<{$}>{\centering $}X[-1,c,c]<{$}}
  & & \mbox{Lattice Coordinates} & & \mbox{Cartesian Coordinates} &\mbox{Wyckoff Position} & \mbox{Atom Type} \\  
  \mathbf{B}_{1} & = & x_{1} \, \mathbf{a}_{1} + y_{1} \, \mathbf{a}_{2} & = & x_{1}a \, \mathbf{\hat{x}} + y_{1}b \, \mathbf{\hat{y}} & \left(4g\right) & \mbox{As I} \\ 
\mathbf{B}_{2} & = & -x_{1} \, \mathbf{a}_{1}-y_{1} \, \mathbf{a}_{2} & = & -x_{1}a \, \mathbf{\hat{x}}-y_{1}b \, \mathbf{\hat{y}} & \left(4g\right) & \mbox{As I} \\ 
\mathbf{B}_{3} & = & \left(\frac{1}{2} - x_{1}\right) \, \mathbf{a}_{1} + \left(\frac{1}{2} +y_{1}\right) \, \mathbf{a}_{2} & = & \left(\frac{1}{2} - x_{1}\right)a \, \mathbf{\hat{x}} + \left(\frac{1}{2} +y_{1}\right)b \, \mathbf{\hat{y}} & \left(4g\right) & \mbox{As I} \\ 
\mathbf{B}_{4} & = & \left(\frac{1}{2} +x_{1}\right) \, \mathbf{a}_{1} + \left(\frac{1}{2} - y_{1}\right) \, \mathbf{a}_{2} & = & \left(\frac{1}{2} +x_{1}\right)a \, \mathbf{\hat{x}} + \left(\frac{1}{2} - y_{1}\right)b \, \mathbf{\hat{y}} & \left(4g\right) & \mbox{As I} \\ 
\mathbf{B}_{5} & = & x_{2} \, \mathbf{a}_{1} + y_{2} \, \mathbf{a}_{2} & = & x_{2}a \, \mathbf{\hat{x}} + y_{2}b \, \mathbf{\hat{y}} & \left(4g\right) & \mbox{As II} \\ 
\mathbf{B}_{6} & = & -x_{2} \, \mathbf{a}_{1}-y_{2} \, \mathbf{a}_{2} & = & -x_{2}a \, \mathbf{\hat{x}}-y_{2}b \, \mathbf{\hat{y}} & \left(4g\right) & \mbox{As II} \\ 
\mathbf{B}_{7} & = & \left(\frac{1}{2} - x_{2}\right) \, \mathbf{a}_{1} + \left(\frac{1}{2} +y_{2}\right) \, \mathbf{a}_{2} & = & \left(\frac{1}{2} - x_{2}\right)a \, \mathbf{\hat{x}} + \left(\frac{1}{2} +y_{2}\right)b \, \mathbf{\hat{y}} & \left(4g\right) & \mbox{As II} \\ 
\mathbf{B}_{8} & = & \left(\frac{1}{2} +x_{2}\right) \, \mathbf{a}_{1} + \left(\frac{1}{2} - y_{2}\right) \, \mathbf{a}_{2} & = & \left(\frac{1}{2} +x_{2}\right)a \, \mathbf{\hat{x}} + \left(\frac{1}{2} - y_{2}\right)b \, \mathbf{\hat{y}} & \left(4g\right) & \mbox{As II} \\ 
\mathbf{B}_{9} & = & x_{3} \, \mathbf{a}_{1} + y_{3} \, \mathbf{a}_{2} & = & x_{3}a \, \mathbf{\hat{x}} + y_{3}b \, \mathbf{\hat{y}} & \left(4g\right) & \mbox{Ge I} \\ 
\mathbf{B}_{10} & = & -x_{3} \, \mathbf{a}_{1}-y_{3} \, \mathbf{a}_{2} & = & -x_{3}a \, \mathbf{\hat{x}}-y_{3}b \, \mathbf{\hat{y}} & \left(4g\right) & \mbox{Ge I} \\ 
\mathbf{B}_{11} & = & \left(\frac{1}{2} - x_{3}\right) \, \mathbf{a}_{1} + \left(\frac{1}{2} +y_{3}\right) \, \mathbf{a}_{2} & = & \left(\frac{1}{2} - x_{3}\right)a \, \mathbf{\hat{x}} + \left(\frac{1}{2} +y_{3}\right)b \, \mathbf{\hat{y}} & \left(4g\right) & \mbox{Ge I} \\ 
\mathbf{B}_{12} & = & \left(\frac{1}{2} +x_{3}\right) \, \mathbf{a}_{1} + \left(\frac{1}{2} - y_{3}\right) \, \mathbf{a}_{2} & = & \left(\frac{1}{2} +x_{3}\right)a \, \mathbf{\hat{x}} + \left(\frac{1}{2} - y_{3}\right)b \, \mathbf{\hat{y}} & \left(4g\right) & \mbox{Ge I} \\ 
\mathbf{B}_{13} & = & x_{4} \, \mathbf{a}_{1} + y_{4} \, \mathbf{a}_{2} + \frac{1}{2} \, \mathbf{a}_{3} & = & x_{4}a \, \mathbf{\hat{x}} + y_{4}b \, \mathbf{\hat{y}} + \frac{1}{2}c \, \mathbf{\hat{z}} & \left(4h\right) & \mbox{As III} \\ 
\mathbf{B}_{14} & = & -x_{4} \, \mathbf{a}_{1}-y_{4} \, \mathbf{a}_{2} + \frac{1}{2} \, \mathbf{a}_{3} & = & -x_{4}a \, \mathbf{\hat{x}}-y_{4}b \, \mathbf{\hat{y}} + \frac{1}{2}c \, \mathbf{\hat{z}} & \left(4h\right) & \mbox{As III} \\ 
\mathbf{B}_{15} & = & \left(\frac{1}{2} - x_{4}\right) \, \mathbf{a}_{1} + \left(\frac{1}{2} +y_{4}\right) \, \mathbf{a}_{2} + \frac{1}{2} \, \mathbf{a}_{3} & = & \left(\frac{1}{2} - x_{4}\right)a \, \mathbf{\hat{x}} + \left(\frac{1}{2} +y_{4}\right)b \, \mathbf{\hat{y}} + \frac{1}{2}c \, \mathbf{\hat{z}} & \left(4h\right) & \mbox{As III} \\ 
\mathbf{B}_{16} & = & \left(\frac{1}{2} +x_{4}\right) \, \mathbf{a}_{1} + \left(\frac{1}{2} - y_{4}\right) \, \mathbf{a}_{2} + \frac{1}{2} \, \mathbf{a}_{3} & = & \left(\frac{1}{2} +x_{4}\right)a \, \mathbf{\hat{x}} + \left(\frac{1}{2} - y_{4}\right)b \, \mathbf{\hat{y}} + \frac{1}{2}c \, \mathbf{\hat{z}} & \left(4h\right) & \mbox{As III} \\ 
\mathbf{B}_{17} & = & x_{5} \, \mathbf{a}_{1} + y_{5} \, \mathbf{a}_{2} + \frac{1}{2} \, \mathbf{a}_{3} & = & x_{5}a \, \mathbf{\hat{x}} + y_{5}b \, \mathbf{\hat{y}} + \frac{1}{2}c \, \mathbf{\hat{z}} & \left(4h\right) & \mbox{As IV} \\ 
\mathbf{B}_{18} & = & -x_{5} \, \mathbf{a}_{1}-y_{5} \, \mathbf{a}_{2} + \frac{1}{2} \, \mathbf{a}_{3} & = & -x_{5}a \, \mathbf{\hat{x}}-y_{5}b \, \mathbf{\hat{y}} + \frac{1}{2}c \, \mathbf{\hat{z}} & \left(4h\right) & \mbox{As IV} \\ 
\mathbf{B}_{19} & = & \left(\frac{1}{2} - x_{5}\right) \, \mathbf{a}_{1} + \left(\frac{1}{2} +y_{5}\right) \, \mathbf{a}_{2} + \frac{1}{2} \, \mathbf{a}_{3} & = & \left(\frac{1}{2} - x_{5}\right)a \, \mathbf{\hat{x}} + \left(\frac{1}{2} +y_{5}\right)b \, \mathbf{\hat{y}} + \frac{1}{2}c \, \mathbf{\hat{z}} & \left(4h\right) & \mbox{As IV} \\ 
\mathbf{B}_{20} & = & \left(\frac{1}{2} +x_{5}\right) \, \mathbf{a}_{1} + \left(\frac{1}{2} - y_{5}\right) \, \mathbf{a}_{2} + \frac{1}{2} \, \mathbf{a}_{3} & = & \left(\frac{1}{2} +x_{5}\right)a \, \mathbf{\hat{x}} + \left(\frac{1}{2} - y_{5}\right)b \, \mathbf{\hat{y}} + \frac{1}{2}c \, \mathbf{\hat{z}} & \left(4h\right) & \mbox{As IV} \\ 
\mathbf{B}_{21} & = & x_{6} \, \mathbf{a}_{1} + y_{6} \, \mathbf{a}_{2} + \frac{1}{2} \, \mathbf{a}_{3} & = & x_{6}a \, \mathbf{\hat{x}} + y_{6}b \, \mathbf{\hat{y}} + \frac{1}{2}c \, \mathbf{\hat{z}} & \left(4h\right) & \mbox{Ge II} \\ 
\mathbf{B}_{22} & = & -x_{6} \, \mathbf{a}_{1}-y_{6} \, \mathbf{a}_{2} + \frac{1}{2} \, \mathbf{a}_{3} & = & -x_{6}a \, \mathbf{\hat{x}}-y_{6}b \, \mathbf{\hat{y}} + \frac{1}{2}c \, \mathbf{\hat{z}} & \left(4h\right) & \mbox{Ge II} \\ 
\mathbf{B}_{23} & = & \left(\frac{1}{2} - x_{6}\right) \, \mathbf{a}_{1} + \left(\frac{1}{2} +y_{6}\right) \, \mathbf{a}_{2} + \frac{1}{2} \, \mathbf{a}_{3} & = & \left(\frac{1}{2} - x_{6}\right)a \, \mathbf{\hat{x}} + \left(\frac{1}{2} +y_{6}\right)b \, \mathbf{\hat{y}} + \frac{1}{2}c \, \mathbf{\hat{z}} & \left(4h\right) & \mbox{Ge II} \\ 
\mathbf{B}_{24} & = & \left(\frac{1}{2} +x_{6}\right) \, \mathbf{a}_{1} + \left(\frac{1}{2} - y_{6}\right) \, \mathbf{a}_{2} + \frac{1}{2} \, \mathbf{a}_{3} & = & \left(\frac{1}{2} +x_{6}\right)a \, \mathbf{\hat{x}} + \left(\frac{1}{2} - y_{6}\right)b \, \mathbf{\hat{y}} + \frac{1}{2}c \, \mathbf{\hat{z}} & \left(4h\right) & \mbox{Ge II} \\ 
\end{longtabu}
\renewcommand{\arraystretch}{1.0}
\noindent \hrulefill
\\
\textbf{References:}
\vspace*{-0.25cm}
\begin{flushleft}
  - \bibentry{Wadsten_GeAs2_ActChemScand_1967}. \\
\end{flushleft}
\textbf{Found in:}
\vspace*{-0.25cm}
\begin{flushleft}
  - \bibentry{Villars_PearsonsCrystalData_2013}. \\
\end{flushleft}
\noindent \hrulefill
\\
\textbf{Geometry files:}
\\
\noindent  - CIF: pp. {\hyperref[A2B_oP24_55_2g2h_gh_cif]{\pageref{A2B_oP24_55_2g2h_gh_cif}}} \\
\noindent  - POSCAR: pp. {\hyperref[A2B_oP24_55_2g2h_gh_poscar]{\pageref{A2B_oP24_55_2g2h_gh_poscar}}} \\
\onecolumn
{\phantomsection\label{A3B5_oP16_55_ch_agh}}
\subsection*{\huge \textbf{{\normalfont Rh$_{5}$Ge$_{3}$ Structure: A3B5\_oP16\_55\_ch\_agh}}}
\noindent \hrulefill
\vspace*{0.25cm}
\begin{figure}[htp]
  \centering
  \vspace{-1em}
  {\includegraphics[width=1\textwidth]{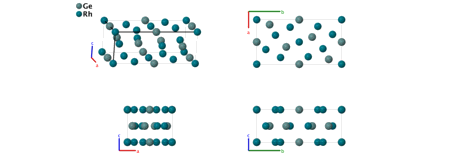}}
\end{figure}
\vspace*{-0.5cm}
\renewcommand{\arraystretch}{1.5}
\begin{equation*}
  \begin{array}{>{$\hspace{-0.15cm}}l<{$}>{$}p{0.5cm}<{$}>{$}p{18.5cm}<{$}}
    \mbox{\large \textbf{Prototype}} &\colon & \ce{Rh5Ge3} \\
    \mbox{\large \textbf{\AFLOW\ prototype label}} &\colon & \mbox{A3B5\_oP16\_55\_ch\_agh} \\
    \mbox{\large \textbf{\textit{Strukturbericht} designation}} &\colon & \mbox{None} \\
    \mbox{\large \textbf{Pearson symbol}} &\colon & \mbox{oP16} \\
    \mbox{\large \textbf{Space group number}} &\colon & 55 \\
    \mbox{\large \textbf{Space group symbol}} &\colon & Pbam \\
    \mbox{\large \textbf{\AFLOW\ prototype command}} &\colon &  \texttt{aflow} \,  \, \texttt{-{}-proto=A3B5\_oP16\_55\_ch\_agh } \, \newline \texttt{-{}-params=}{a,b/a,c/a,x_{3},y_{3},x_{4},y_{4},x_{5},y_{5} }
  \end{array}
\end{equation*}
\renewcommand{\arraystretch}{1.0}

\noindent \parbox{1 \linewidth}{
\noindent \hrulefill
\\
\textbf{Simple Orthorhombic primitive vectors:} \\
\vspace*{-0.25cm}
\begin{tabular}{cc}
  \begin{tabular}{c}
    \parbox{0.6 \linewidth}{
      \renewcommand{\arraystretch}{1.5}
      \begin{equation*}
        \centering
        \begin{array}{ccc}
              \mathbf{a}_1 & = & a \, \mathbf{\hat{x}} \\
    \mathbf{a}_2 & = & b \, \mathbf{\hat{y}} \\
    \mathbf{a}_3 & = & c \, \mathbf{\hat{z}} \\

        \end{array}
      \end{equation*}
    }
    \renewcommand{\arraystretch}{1.0}
  \end{tabular}
  \begin{tabular}{c}
    \includegraphics[width=0.3\linewidth]{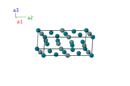} \\
  \end{tabular}
\end{tabular}

}
\vspace*{-0.25cm}

\noindent \hrulefill
\\
\textbf{Basis vectors:}
\vspace*{-0.25cm}
\renewcommand{\arraystretch}{1.5}
\begin{longtabu} to \textwidth{>{\centering $}X[-1,c,c]<{$}>{\centering $}X[-1,c,c]<{$}>{\centering $}X[-1,c,c]<{$}>{\centering $}X[-1,c,c]<{$}>{\centering $}X[-1,c,c]<{$}>{\centering $}X[-1,c,c]<{$}>{\centering $}X[-1,c,c]<{$}}
  & & \mbox{Lattice Coordinates} & & \mbox{Cartesian Coordinates} &\mbox{Wyckoff Position} & \mbox{Atom Type} \\  
  \mathbf{B}_{1} & = & 0 \, \mathbf{a}_{1} + 0 \, \mathbf{a}_{2} + 0 \, \mathbf{a}_{3} & = & 0 \, \mathbf{\hat{x}} + 0 \, \mathbf{\hat{y}} + 0 \, \mathbf{\hat{z}} & \left(2a\right) & \mbox{Rh I} \\ 
\mathbf{B}_{2} & = & \frac{1}{2} \, \mathbf{a}_{1} + \frac{1}{2} \, \mathbf{a}_{2} & = & \frac{1}{2}a \, \mathbf{\hat{x}} + \frac{1}{2}b \, \mathbf{\hat{y}} & \left(2a\right) & \mbox{Rh I} \\ 
\mathbf{B}_{3} & = & \frac{1}{2} \, \mathbf{a}_{2} & = & \frac{1}{2}b \, \mathbf{\hat{y}} & \left(2c\right) & \mbox{Ge I} \\ 
\mathbf{B}_{4} & = & \frac{1}{2} \, \mathbf{a}_{1} & = & \frac{1}{2}a \, \mathbf{\hat{x}} & \left(2c\right) & \mbox{Ge I} \\ 
\mathbf{B}_{5} & = & x_{3} \, \mathbf{a}_{1} + y_{3} \, \mathbf{a}_{2} & = & x_{3}a \, \mathbf{\hat{x}} + y_{3}b \, \mathbf{\hat{y}} & \left(4g\right) & \mbox{Rh II} \\ 
\mathbf{B}_{6} & = & -x_{3} \, \mathbf{a}_{1}-y_{3} \, \mathbf{a}_{2} & = & -x_{3}a \, \mathbf{\hat{x}}-y_{3}b \, \mathbf{\hat{y}} & \left(4g\right) & \mbox{Rh II} \\ 
\mathbf{B}_{7} & = & \left(\frac{1}{2} - x_{3}\right) \, \mathbf{a}_{1} + \left(\frac{1}{2} +y_{3}\right) \, \mathbf{a}_{2} & = & \left(\frac{1}{2} - x_{3}\right)a \, \mathbf{\hat{x}} + \left(\frac{1}{2} +y_{3}\right)b \, \mathbf{\hat{y}} & \left(4g\right) & \mbox{Rh II} \\ 
\mathbf{B}_{8} & = & \left(\frac{1}{2} +x_{3}\right) \, \mathbf{a}_{1} + \left(\frac{1}{2} - y_{3}\right) \, \mathbf{a}_{2} & = & \left(\frac{1}{2} +x_{3}\right)a \, \mathbf{\hat{x}} + \left(\frac{1}{2} - y_{3}\right)b \, \mathbf{\hat{y}} & \left(4g\right) & \mbox{Rh II} \\ 
\mathbf{B}_{9} & = & x_{4} \, \mathbf{a}_{1} + y_{4} \, \mathbf{a}_{2} + \frac{1}{2} \, \mathbf{a}_{3} & = & x_{4}a \, \mathbf{\hat{x}} + y_{4}b \, \mathbf{\hat{y}} + \frac{1}{2}c \, \mathbf{\hat{z}} & \left(4h\right) & \mbox{Ge II} \\ 
\mathbf{B}_{10} & = & -x_{4} \, \mathbf{a}_{1}-y_{4} \, \mathbf{a}_{2} + \frac{1}{2} \, \mathbf{a}_{3} & = & -x_{4}a \, \mathbf{\hat{x}}-y_{4}b \, \mathbf{\hat{y}} + \frac{1}{2}c \, \mathbf{\hat{z}} & \left(4h\right) & \mbox{Ge II} \\ 
\mathbf{B}_{11} & = & \left(\frac{1}{2} - x_{4}\right) \, \mathbf{a}_{1} + \left(\frac{1}{2} +y_{4}\right) \, \mathbf{a}_{2} + \frac{1}{2} \, \mathbf{a}_{3} & = & \left(\frac{1}{2} - x_{4}\right)a \, \mathbf{\hat{x}} + \left(\frac{1}{2} +y_{4}\right)b \, \mathbf{\hat{y}} + \frac{1}{2}c \, \mathbf{\hat{z}} & \left(4h\right) & \mbox{Ge II} \\ 
\mathbf{B}_{12} & = & \left(\frac{1}{2} +x_{4}\right) \, \mathbf{a}_{1} + \left(\frac{1}{2} - y_{4}\right) \, \mathbf{a}_{2} + \frac{1}{2} \, \mathbf{a}_{3} & = & \left(\frac{1}{2} +x_{4}\right)a \, \mathbf{\hat{x}} + \left(\frac{1}{2} - y_{4}\right)b \, \mathbf{\hat{y}} + \frac{1}{2}c \, \mathbf{\hat{z}} & \left(4h\right) & \mbox{Ge II} \\ 
\mathbf{B}_{13} & = & x_{5} \, \mathbf{a}_{1} + y_{5} \, \mathbf{a}_{2} + \frac{1}{2} \, \mathbf{a}_{3} & = & x_{5}a \, \mathbf{\hat{x}} + y_{5}b \, \mathbf{\hat{y}} + \frac{1}{2}c \, \mathbf{\hat{z}} & \left(4h\right) & \mbox{Rh III} \\ 
\mathbf{B}_{14} & = & -x_{5} \, \mathbf{a}_{1}-y_{5} \, \mathbf{a}_{2} + \frac{1}{2} \, \mathbf{a}_{3} & = & -x_{5}a \, \mathbf{\hat{x}}-y_{5}b \, \mathbf{\hat{y}} + \frac{1}{2}c \, \mathbf{\hat{z}} & \left(4h\right) & \mbox{Rh III} \\ 
\mathbf{B}_{15} & = & \left(\frac{1}{2} - x_{5}\right) \, \mathbf{a}_{1} + \left(\frac{1}{2} +y_{5}\right) \, \mathbf{a}_{2} + \frac{1}{2} \, \mathbf{a}_{3} & = & \left(\frac{1}{2} - x_{5}\right)a \, \mathbf{\hat{x}} + \left(\frac{1}{2} +y_{5}\right)b \, \mathbf{\hat{y}} + \frac{1}{2}c \, \mathbf{\hat{z}} & \left(4h\right) & \mbox{Rh III} \\ 
\mathbf{B}_{16} & = & \left(\frac{1}{2} +x_{5}\right) \, \mathbf{a}_{1} + \left(\frac{1}{2} - y_{5}\right) \, \mathbf{a}_{2} + \frac{1}{2} \, \mathbf{a}_{3} & = & \left(\frac{1}{2} +x_{5}\right)a \, \mathbf{\hat{x}} + \left(\frac{1}{2} - y_{5}\right)b \, \mathbf{\hat{y}} + \frac{1}{2}c \, \mathbf{\hat{z}} & \left(4h\right) & \mbox{Rh III} \\ 
\end{longtabu}
\renewcommand{\arraystretch}{1.0}
\noindent \hrulefill
\\
\textbf{References:}
\vspace*{-0.25cm}
\begin{flushleft}
  - \bibentry{Geller_Ga3Rh5_ActCrystallograp_1955}. \\
\end{flushleft}
\textbf{Found in:}
\vspace*{-0.25cm}
\begin{flushleft}
  - \bibentry{Villars_PearsonsCrystalData_2013}. \\
\end{flushleft}
\noindent \hrulefill
\\
\textbf{Geometry files:}
\\
\noindent  - CIF: pp. {\hyperref[A3B5_oP16_55_ch_agh_cif]{\pageref{A3B5_oP16_55_ch_agh_cif}}} \\
\noindent  - POSCAR: pp. {\hyperref[A3B5_oP16_55_ch_agh_poscar]{\pageref{A3B5_oP16_55_ch_agh_poscar}}} \\
\onecolumn
{\phantomsection\label{A_oP16_55_2g2h}}
\subsection*{\huge \textbf{{\normalfont R-carbon Structure: A\_oP16\_55\_2g2h}}}
\noindent \hrulefill
\vspace*{0.25cm}
\begin{figure}[htp]
  \centering
  \vspace{-1em}
  {\includegraphics[width=1\textwidth]{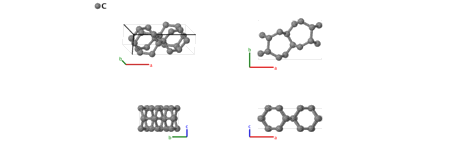}}
\end{figure}
\vspace*{-0.5cm}
\renewcommand{\arraystretch}{1.5}
\begin{equation*}
  \begin{array}{>{$\hspace{-0.15cm}}l<{$}>{$}p{0.5cm}<{$}>{$}p{18.5cm}<{$}}
    \mbox{\large \textbf{Prototype}} &\colon & \ce{C} \\
    \mbox{\large \textbf{\AFLOW\ prototype label}} &\colon & \mbox{A\_oP16\_55\_2g2h} \\
    \mbox{\large \textbf{\textit{Strukturbericht} designation}} &\colon & \mbox{None} \\
    \mbox{\large \textbf{Pearson symbol}} &\colon & \mbox{oP16} \\
    \mbox{\large \textbf{Space group number}} &\colon & 55 \\
    \mbox{\large \textbf{Space group symbol}} &\colon & Pbam \\
    \mbox{\large \textbf{\AFLOW\ prototype command}} &\colon &  \texttt{aflow} \,  \, \texttt{-{}-proto=A\_oP16\_55\_2g2h } \, \newline \texttt{-{}-params=}{a,b/a,c/a,x_{1},y_{1},x_{2},y_{2},x_{3},y_{3},x_{4},y_{4} }
  \end{array}
\end{equation*}
\renewcommand{\arraystretch}{1.0}

\vspace*{-0.25cm}
\noindent \hrulefill
\begin{itemize}
  \item{This is a predicted ``superhard'' allotrope of carbon.  Shortly after
this paper was published, another paper predicted a similar phase,
called ``H-carbon'' (He, 2012). The similarity
between the two structures can be seen by shifting the origin by
$\left(1/2\right) \mathbf{a}_{1}$.  Other sources (Zhao, 2012) refer
to this structure as ``O-carbon.''
}
\end{itemize}

\noindent \parbox{1 \linewidth}{
\noindent \hrulefill
\\
\textbf{Simple Orthorhombic primitive vectors:} \\
\vspace*{-0.25cm}
\begin{tabular}{cc}
  \begin{tabular}{c}
    \parbox{0.6 \linewidth}{
      \renewcommand{\arraystretch}{1.5}
      \begin{equation*}
        \centering
        \begin{array}{ccc}
              \mathbf{a}_1 & = & a \, \mathbf{\hat{x}} \\
    \mathbf{a}_2 & = & b \, \mathbf{\hat{y}} \\
    \mathbf{a}_3 & = & c \, \mathbf{\hat{z}} \\

        \end{array}
      \end{equation*}
    }
    \renewcommand{\arraystretch}{1.0}
  \end{tabular}
  \begin{tabular}{c}
    \includegraphics[width=0.3\linewidth]{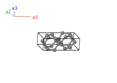} \\
  \end{tabular}
\end{tabular}

}
\vspace*{-0.25cm}

\noindent \hrulefill
\\
\textbf{Basis vectors:}
\vspace*{-0.25cm}
\renewcommand{\arraystretch}{1.5}
\begin{longtabu} to \textwidth{>{\centering $}X[-1,c,c]<{$}>{\centering $}X[-1,c,c]<{$}>{\centering $}X[-1,c,c]<{$}>{\centering $}X[-1,c,c]<{$}>{\centering $}X[-1,c,c]<{$}>{\centering $}X[-1,c,c]<{$}>{\centering $}X[-1,c,c]<{$}}
  & & \mbox{Lattice Coordinates} & & \mbox{Cartesian Coordinates} &\mbox{Wyckoff Position} & \mbox{Atom Type} \\  
  \mathbf{B}_{1} & = & x_{1} \, \mathbf{a}_{1} + y_{1} \, \mathbf{a}_{2} & = & x_{1}a \, \mathbf{\hat{x}} + y_{1}b \, \mathbf{\hat{y}} & \left(4g\right) & \mbox{C I} \\ 
\mathbf{B}_{2} & = & -x_{1} \, \mathbf{a}_{1}-y_{1} \, \mathbf{a}_{2} & = & -x_{1}a \, \mathbf{\hat{x}}-y_{1}b \, \mathbf{\hat{y}} & \left(4g\right) & \mbox{C I} \\ 
\mathbf{B}_{3} & = & \left(\frac{1}{2} - x_{1}\right) \, \mathbf{a}_{1} + \left(\frac{1}{2} +y_{1}\right) \, \mathbf{a}_{2} & = & \left(\frac{1}{2} - x_{1}\right)a \, \mathbf{\hat{x}} + \left(\frac{1}{2} +y_{1}\right)b \, \mathbf{\hat{y}} & \left(4g\right) & \mbox{C I} \\ 
\mathbf{B}_{4} & = & \left(\frac{1}{2} +x_{1}\right) \, \mathbf{a}_{1} + \left(\frac{1}{2} - y_{1}\right) \, \mathbf{a}_{2} & = & \left(\frac{1}{2} +x_{1}\right)a \, \mathbf{\hat{x}} + \left(\frac{1}{2} - y_{1}\right)b \, \mathbf{\hat{y}} & \left(4g\right) & \mbox{C I} \\ 
\mathbf{B}_{5} & = & x_{2} \, \mathbf{a}_{1} + y_{2} \, \mathbf{a}_{2} & = & x_{2}a \, \mathbf{\hat{x}} + y_{2}b \, \mathbf{\hat{y}} & \left(4g\right) & \mbox{C II} \\ 
\mathbf{B}_{6} & = & -x_{2} \, \mathbf{a}_{1}-y_{2} \, \mathbf{a}_{2} & = & -x_{2}a \, \mathbf{\hat{x}}-y_{2}b \, \mathbf{\hat{y}} & \left(4g\right) & \mbox{C II} \\ 
\mathbf{B}_{7} & = & \left(\frac{1}{2} - x_{2}\right) \, \mathbf{a}_{1} + \left(\frac{1}{2} +y_{2}\right) \, \mathbf{a}_{2} & = & \left(\frac{1}{2} - x_{2}\right)a \, \mathbf{\hat{x}} + \left(\frac{1}{2} +y_{2}\right)b \, \mathbf{\hat{y}} & \left(4g\right) & \mbox{C II} \\ 
\mathbf{B}_{8} & = & \left(\frac{1}{2} +x_{2}\right) \, \mathbf{a}_{1} + \left(\frac{1}{2} - y_{2}\right) \, \mathbf{a}_{2} & = & \left(\frac{1}{2} +x_{2}\right)a \, \mathbf{\hat{x}} + \left(\frac{1}{2} - y_{2}\right)b \, \mathbf{\hat{y}} & \left(4g\right) & \mbox{C II} \\ 
\mathbf{B}_{9} & = & x_{3} \, \mathbf{a}_{1} + y_{3} \, \mathbf{a}_{2} + \frac{1}{2} \, \mathbf{a}_{3} & = & x_{3}a \, \mathbf{\hat{x}} + y_{3}b \, \mathbf{\hat{y}} + \frac{1}{2}c \, \mathbf{\hat{z}} & \left(4h\right) & \mbox{C III} \\ 
\mathbf{B}_{10} & = & -x_{3} \, \mathbf{a}_{1}-y_{3} \, \mathbf{a}_{2} + \frac{1}{2} \, \mathbf{a}_{3} & = & -x_{3}a \, \mathbf{\hat{x}}-y_{3}b \, \mathbf{\hat{y}} + \frac{1}{2}c \, \mathbf{\hat{z}} & \left(4h\right) & \mbox{C III} \\ 
\mathbf{B}_{11} & = & \left(\frac{1}{2} - x_{3}\right) \, \mathbf{a}_{1} + \left(\frac{1}{2} +y_{3}\right) \, \mathbf{a}_{2} + \frac{1}{2} \, \mathbf{a}_{3} & = & \left(\frac{1}{2} - x_{3}\right)a \, \mathbf{\hat{x}} + \left(\frac{1}{2} +y_{3}\right)b \, \mathbf{\hat{y}} + \frac{1}{2}c \, \mathbf{\hat{z}} & \left(4h\right) & \mbox{C III} \\ 
\mathbf{B}_{12} & = & \left(\frac{1}{2} +x_{3}\right) \, \mathbf{a}_{1} + \left(\frac{1}{2} - y_{3}\right) \, \mathbf{a}_{2} + \frac{1}{2} \, \mathbf{a}_{3} & = & \left(\frac{1}{2} +x_{3}\right)a \, \mathbf{\hat{x}} + \left(\frac{1}{2} - y_{3}\right)b \, \mathbf{\hat{y}} + \frac{1}{2}c \, \mathbf{\hat{z}} & \left(4h\right) & \mbox{C III} \\ 
\mathbf{B}_{13} & = & x_{4} \, \mathbf{a}_{1} + y_{4} \, \mathbf{a}_{2} + \frac{1}{2} \, \mathbf{a}_{3} & = & x_{4}a \, \mathbf{\hat{x}} + y_{4}b \, \mathbf{\hat{y}} + \frac{1}{2}c \, \mathbf{\hat{z}} & \left(4h\right) & \mbox{C IV} \\ 
\mathbf{B}_{14} & = & -x_{4} \, \mathbf{a}_{1}-y_{4} \, \mathbf{a}_{2} + \frac{1}{2} \, \mathbf{a}_{3} & = & -x_{4}a \, \mathbf{\hat{x}}-y_{4}b \, \mathbf{\hat{y}} + \frac{1}{2}c \, \mathbf{\hat{z}} & \left(4h\right) & \mbox{C IV} \\ 
\mathbf{B}_{15} & = & \left(\frac{1}{2} - x_{4}\right) \, \mathbf{a}_{1} + \left(\frac{1}{2} +y_{4}\right) \, \mathbf{a}_{2} + \frac{1}{2} \, \mathbf{a}_{3} & = & \left(\frac{1}{2} - x_{4}\right)a \, \mathbf{\hat{x}} + \left(\frac{1}{2} +y_{4}\right)b \, \mathbf{\hat{y}} + \frac{1}{2}c \, \mathbf{\hat{z}} & \left(4h\right) & \mbox{C IV} \\ 
\mathbf{B}_{16} & = & \left(\frac{1}{2} +x_{4}\right) \, \mathbf{a}_{1} + \left(\frac{1}{2} - y_{4}\right) \, \mathbf{a}_{2} + \frac{1}{2} \, \mathbf{a}_{3} & = & \left(\frac{1}{2} +x_{4}\right)a \, \mathbf{\hat{x}} + \left(\frac{1}{2} - y_{4}\right)b \, \mathbf{\hat{y}} + \frac{1}{2}c \, \mathbf{\hat{z}} & \left(4h\right) & \mbox{C IV} \\ 
\end{longtabu}
\renewcommand{\arraystretch}{1.0}
\noindent \hrulefill
\\
\textbf{References:}
\vspace*{-0.25cm}
\begin{flushleft}
  - \bibentry{Niu_PRL_108_2012}. \\
  - \bibentry{He_SSC_152_2012}. \\
  - \bibentry{He_JSM_34_2012}. \\
  - \bibentry{Zhao_JACS_134_2012}. \\
\end{flushleft}
\noindent \hrulefill
\\
\textbf{Geometry files:}
\\
\noindent  - CIF: pp. {\hyperref[A_oP16_55_2g2h_cif]{\pageref{A_oP16_55_2g2h_cif}}} \\
\noindent  - POSCAR: pp. {\hyperref[A_oP16_55_2g2h_poscar]{\pageref{A_oP16_55_2g2h_poscar}}} \\
\onecolumn
{\phantomsection\label{A2B_oP6_58_g_a}}
\subsection*{\huge \textbf{{\normalfont $\alpha$-PdCl$_{2}$ ($C50$) Structure: A2B\_oP6\_58\_g\_a}}}
\noindent \hrulefill
\vspace*{0.25cm}
\begin{figure}[htp]
  \centering
  \vspace{-1em}
  {\includegraphics[width=1\textwidth]{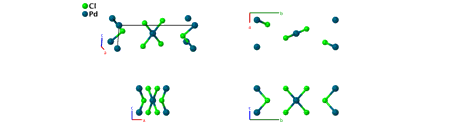}}
\end{figure}
\vspace*{-0.5cm}
\renewcommand{\arraystretch}{1.5}
\begin{equation*}
  \begin{array}{>{$\hspace{-0.15cm}}l<{$}>{$}p{0.5cm}<{$}>{$}p{18.5cm}<{$}}
    \mbox{\large \textbf{Prototype}} &\colon & \ce{$\alpha$-PdCl2} \\
    \mbox{\large \textbf{\AFLOW\ prototype label}} &\colon & \mbox{A2B\_oP6\_58\_g\_a} \\
    \mbox{\large \textbf{\textit{Strukturbericht} designation}} &\colon & \mbox{$C50$} \\
    \mbox{\large \textbf{Pearson symbol}} &\colon & \mbox{oP6} \\
    \mbox{\large \textbf{Space group number}} &\colon & 58 \\
    \mbox{\large \textbf{Space group symbol}} &\colon & Pnnm \\
    \mbox{\large \textbf{\AFLOW\ prototype command}} &\colon &  \texttt{aflow} \,  \, \texttt{-{}-proto=A2B\_oP6\_58\_g\_a } \, \newline \texttt{-{}-params=}{a,b/a,c/a,x_{2},y_{2} }
  \end{array}
\end{equation*}
\renewcommand{\arraystretch}{1.0}

\vspace*{-0.25cm}
\noindent \hrulefill
\begin{itemize}
  \item{(Evers, 2010) implicitly places the Pd atoms at the (2b) Wyckoff
position.  We have shifted the Pd atoms to the (2a) site.
}
\end{itemize}

\noindent \parbox{1 \linewidth}{
\noindent \hrulefill
\\
\textbf{Simple Orthorhombic primitive vectors:} \\
\vspace*{-0.25cm}
\begin{tabular}{cc}
  \begin{tabular}{c}
    \parbox{0.6 \linewidth}{
      \renewcommand{\arraystretch}{1.5}
      \begin{equation*}
        \centering
        \begin{array}{ccc}
              \mathbf{a}_1 & = & a \, \mathbf{\hat{x}} \\
    \mathbf{a}_2 & = & b \, \mathbf{\hat{y}} \\
    \mathbf{a}_3 & = & c \, \mathbf{\hat{z}} \\

        \end{array}
      \end{equation*}
    }
    \renewcommand{\arraystretch}{1.0}
  \end{tabular}
  \begin{tabular}{c}
    \includegraphics[width=0.3\linewidth]{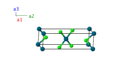} \\
  \end{tabular}
\end{tabular}

}
\vspace*{-0.25cm}

\noindent \hrulefill
\\
\textbf{Basis vectors:}
\vspace*{-0.25cm}
\renewcommand{\arraystretch}{1.5}
\begin{longtabu} to \textwidth{>{\centering $}X[-1,c,c]<{$}>{\centering $}X[-1,c,c]<{$}>{\centering $}X[-1,c,c]<{$}>{\centering $}X[-1,c,c]<{$}>{\centering $}X[-1,c,c]<{$}>{\centering $}X[-1,c,c]<{$}>{\centering $}X[-1,c,c]<{$}}
  & & \mbox{Lattice Coordinates} & & \mbox{Cartesian Coordinates} &\mbox{Wyckoff Position} & \mbox{Atom Type} \\  
  \mathbf{B}_{1} & = & 0 \, \mathbf{a}_{1} + 0 \, \mathbf{a}_{2} + 0 \, \mathbf{a}_{3} & = & 0 \, \mathbf{\hat{x}} + 0 \, \mathbf{\hat{y}} + 0 \, \mathbf{\hat{z}} & \left(2a\right) & \mbox{Pd} \\ 
\mathbf{B}_{2} & = & \frac{1}{2} \, \mathbf{a}_{1} + \frac{1}{2} \, \mathbf{a}_{2} + \frac{1}{2} \, \mathbf{a}_{3} & = & \frac{1}{2}a \, \mathbf{\hat{x}} + \frac{1}{2}b \, \mathbf{\hat{y}} + \frac{1}{2}c \, \mathbf{\hat{z}} & \left(2a\right) & \mbox{Pd} \\ 
\mathbf{B}_{3} & = & x_{2} \, \mathbf{a}_{1} + y_{2} \, \mathbf{a}_{2} & = & x_{2}a \, \mathbf{\hat{x}} + y_{2}b \, \mathbf{\hat{y}} & \left(4g\right) & \mbox{Cl} \\ 
\mathbf{B}_{4} & = & -x_{2} \, \mathbf{a}_{1}-y_{2} \, \mathbf{a}_{2} & = & -x_{2}a \, \mathbf{\hat{x}}-y_{2}b \, \mathbf{\hat{y}} & \left(4g\right) & \mbox{Cl} \\ 
\mathbf{B}_{5} & = & \left(\frac{1}{2} - x_{2}\right) \, \mathbf{a}_{1} + \left(\frac{1}{2} +y_{2}\right) \, \mathbf{a}_{2} + \frac{1}{2} \, \mathbf{a}_{3} & = & \left(\frac{1}{2} - x_{2}\right)a \, \mathbf{\hat{x}} + \left(\frac{1}{2} +y_{2}\right)b \, \mathbf{\hat{y}} + \frac{1}{2}c \, \mathbf{\hat{z}} & \left(4g\right) & \mbox{Cl} \\ 
\mathbf{B}_{6} & = & \left(\frac{1}{2} +x_{2}\right) \, \mathbf{a}_{1} + \left(\frac{1}{2} - y_{2}\right) \, \mathbf{a}_{2} + \frac{1}{2} \, \mathbf{a}_{3} & = & \left(\frac{1}{2} +x_{2}\right)a \, \mathbf{\hat{x}} + \left(\frac{1}{2} - y_{2}\right)b \, \mathbf{\hat{y}} + \frac{1}{2}c \, \mathbf{\hat{z}} & \left(4g\right) & \mbox{Cl} \\ 
\end{longtabu}
\renewcommand{\arraystretch}{1.0}
\noindent \hrulefill
\\
\textbf{References:}
\vspace*{-0.25cm}
\begin{flushleft}
  - \bibentry{Evers_AngewChemIntEd_49_5677_2010}. \\
\end{flushleft}
\noindent \hrulefill
\\
\textbf{Geometry files:}
\\
\noindent  - CIF: pp. {\hyperref[A2B_oP6_58_g_a_cif]{\pageref{A2B_oP6_58_g_a_cif}}} \\
\noindent  - POSCAR: pp. {\hyperref[A2B_oP6_58_g_a_poscar]{\pageref{A2B_oP6_58_g_a_poscar}}} \\
\onecolumn
{\phantomsection\label{ABC_oP6_59_a_b_a}}
\subsection*{\huge \textbf{{\normalfont FeOCl Structure: ABC\_oP6\_59\_a\_b\_a}}}
\noindent \hrulefill
\vspace*{0.25cm}
\begin{figure}[htp]
  \centering
  \vspace{-1em}
  {\includegraphics[width=1\textwidth]{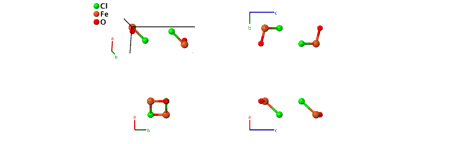}}
\end{figure}
\vspace*{-0.5cm}
\renewcommand{\arraystretch}{1.5}
\begin{equation*}
  \begin{array}{>{$\hspace{-0.15cm}}l<{$}>{$}p{0.5cm}<{$}>{$}p{18.5cm}<{$}}
    \mbox{\large \textbf{Prototype}} &\colon & \ce{FeOCl} \\
    \mbox{\large \textbf{\AFLOW\ prototype label}} &\colon & \mbox{ABC\_oP6\_59\_a\_b\_a} \\
    \mbox{\large \textbf{\textit{Strukturbericht} designation}} &\colon & \mbox{None} \\
    \mbox{\large \textbf{Pearson symbol}} &\colon & \mbox{oP6} \\
    \mbox{\large \textbf{Space group number}} &\colon & 59 \\
    \mbox{\large \textbf{Space group symbol}} &\colon & Pmmn \\
    \mbox{\large \textbf{\AFLOW\ prototype command}} &\colon &  \texttt{aflow} \,  \, \texttt{-{}-proto=ABC\_oP6\_59\_a\_b\_a } \, \newline \texttt{-{}-params=}{a,b/a,c/a,z_{1},z_{2},z_{3} }
  \end{array}
\end{equation*}
\renewcommand{\arraystretch}{1.0}

\vspace*{-0.25cm}
\noindent \hrulefill
\\
\textbf{ Other compounds with this structure:}
\begin{itemize}
   \item{  $M$OCl, ($M$ = Ti, V, Cr, Fe), the superconducting alkali metal intercalcates $\alpha$-$M$N$X$ ($M$ = Ti, Zr, Hf; $X$ = Cl, Br, I).  }
\end{itemize}
\noindent \parbox{1 \linewidth}{
\noindent \hrulefill
\\
\textbf{Simple Orthorhombic primitive vectors:} \\
\vspace*{-0.25cm}
\begin{tabular}{cc}
  \begin{tabular}{c}
    \parbox{0.6 \linewidth}{
      \renewcommand{\arraystretch}{1.5}
      \begin{equation*}
        \centering
        \begin{array}{ccc}
              \mathbf{a}_1 & = & a \, \mathbf{\hat{x}} \\
    \mathbf{a}_2 & = & b \, \mathbf{\hat{y}} \\
    \mathbf{a}_3 & = & c \, \mathbf{\hat{z}} \\

        \end{array}
      \end{equation*}
    }
    \renewcommand{\arraystretch}{1.0}
  \end{tabular}
  \begin{tabular}{c}
    \includegraphics[width=0.3\linewidth]{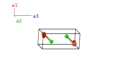} \\
  \end{tabular}
\end{tabular}

}
\vspace*{-0.25cm}

\noindent \hrulefill
\\
\textbf{Basis vectors:}
\vspace*{-0.25cm}
\renewcommand{\arraystretch}{1.5}
\begin{longtabu} to \textwidth{>{\centering $}X[-1,c,c]<{$}>{\centering $}X[-1,c,c]<{$}>{\centering $}X[-1,c,c]<{$}>{\centering $}X[-1,c,c]<{$}>{\centering $}X[-1,c,c]<{$}>{\centering $}X[-1,c,c]<{$}>{\centering $}X[-1,c,c]<{$}}
  & & \mbox{Lattice Coordinates} & & \mbox{Cartesian Coordinates} &\mbox{Wyckoff Position} & \mbox{Atom Type} \\  
  \mathbf{B}_{1} & = & \frac{1}{4} \, \mathbf{a}_{1} + \frac{1}{4} \, \mathbf{a}_{2} + z_{1} \, \mathbf{a}_{3} & = & \frac{1}{4}a \, \mathbf{\hat{x}} + \frac{1}{4}b \, \mathbf{\hat{y}} + z_{1}c \, \mathbf{\hat{z}} & \left(2a\right) & \mbox{Cl} \\ 
\mathbf{B}_{2} & = & \frac{3}{4} \, \mathbf{a}_{1} + \frac{3}{4} \, \mathbf{a}_{2}-z_{1} \, \mathbf{a}_{3} & = & \frac{3}{4}a \, \mathbf{\hat{x}} + \frac{3}{4}b \, \mathbf{\hat{y}}-z_{1}c \, \mathbf{\hat{z}} & \left(2a\right) & \mbox{Cl} \\ 
\mathbf{B}_{3} & = & \frac{1}{4} \, \mathbf{a}_{1} + \frac{1}{4} \, \mathbf{a}_{2} + z_{2} \, \mathbf{a}_{3} & = & \frac{1}{4}a \, \mathbf{\hat{x}} + \frac{1}{4}b \, \mathbf{\hat{y}} + z_{2}c \, \mathbf{\hat{z}} & \left(2a\right) & \mbox{O} \\ 
\mathbf{B}_{4} & = & \frac{3}{4} \, \mathbf{a}_{1} + \frac{3}{4} \, \mathbf{a}_{2}-z_{2} \, \mathbf{a}_{3} & = & \frac{3}{4}a \, \mathbf{\hat{x}} + \frac{3}{4}b \, \mathbf{\hat{y}}-z_{2}c \, \mathbf{\hat{z}} & \left(2a\right) & \mbox{O} \\ 
\mathbf{B}_{5} & = & \frac{1}{4} \, \mathbf{a}_{1} + \frac{3}{4} \, \mathbf{a}_{2} + z_{3} \, \mathbf{a}_{3} & = & \frac{1}{4}a \, \mathbf{\hat{x}} + \frac{3}{4}b \, \mathbf{\hat{y}} + z_{3}c \, \mathbf{\hat{z}} & \left(2b\right) & \mbox{Fe} \\ 
\mathbf{B}_{6} & = & \frac{3}{4} \, \mathbf{a}_{1} + \frac{1}{4} \, \mathbf{a}_{2}-z_{3} \, \mathbf{a}_{3} & = & \frac{3}{4}a \, \mathbf{\hat{x}} + \frac{1}{4}b \, \mathbf{\hat{y}}-z_{3}c \, \mathbf{\hat{z}} & \left(2b\right) & \mbox{Fe} \\ 
\end{longtabu}
\renewcommand{\arraystretch}{1.0}
\noindent \hrulefill
\\
\textbf{References:}
\vspace*{-0.25cm}
\begin{flushleft}
  - \bibentry{Kauzlarich_JACS_108_1986}. \\
\end{flushleft}
\noindent \hrulefill
\\
\textbf{Geometry files:}
\\
\noindent  - CIF: pp. {\hyperref[ABC_oP6_59_a_b_a_cif]{\pageref{ABC_oP6_59_a_b_a_cif}}} \\
\noindent  - POSCAR: pp. {\hyperref[ABC_oP6_59_a_b_a_poscar]{\pageref{ABC_oP6_59_a_b_a_poscar}}} \\
\onecolumn
{\phantomsection\label{A2B3_oP20_60_d_cd}}
\subsection*{\huge \textbf{{\normalfont Rh$_{2}$S$_{3}$ Structure: A2B3\_oP20\_60\_d\_cd}}}
\noindent \hrulefill
\vspace*{0.25cm}
\begin{figure}[htp]
  \centering
  \vspace{-1em}
  {\includegraphics[width=1\textwidth]{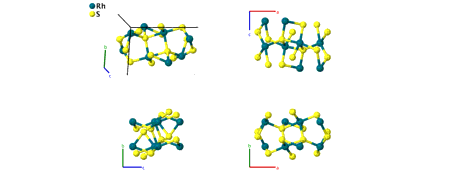}}
\end{figure}
\vspace*{-0.5cm}
\renewcommand{\arraystretch}{1.5}
\begin{equation*}
  \begin{array}{>{$\hspace{-0.15cm}}l<{$}>{$}p{0.5cm}<{$}>{$}p{18.5cm}<{$}}
    \mbox{\large \textbf{Prototype}} &\colon & \ce{Rh2S3} \\
    \mbox{\large \textbf{\AFLOW\ prototype label}} &\colon & \mbox{A2B3\_oP20\_60\_d\_cd} \\
    \mbox{\large \textbf{\textit{Strukturbericht} designation}} &\colon & \mbox{None} \\
    \mbox{\large \textbf{Pearson symbol}} &\colon & \mbox{oP20} \\
    \mbox{\large \textbf{Space group number}} &\colon & 60 \\
    \mbox{\large \textbf{Space group symbol}} &\colon & Pbcn \\
    \mbox{\large \textbf{\AFLOW\ prototype command}} &\colon &  \texttt{aflow} \,  \, \texttt{-{}-proto=A2B3\_oP20\_60\_d\_cd } \, \newline \texttt{-{}-params=}{a,b/a,c/a,y_{1},x_{2},y_{2},z_{2},x_{3},y_{3},z_{3} }
  \end{array}
\end{equation*}
\renewcommand{\arraystretch}{1.0}

\noindent \parbox{1 \linewidth}{
\noindent \hrulefill
\\
\textbf{Simple Orthorhombic primitive vectors:} \\
\vspace*{-0.25cm}
\begin{tabular}{cc}
  \begin{tabular}{c}
    \parbox{0.6 \linewidth}{
      \renewcommand{\arraystretch}{1.5}
      \begin{equation*}
        \centering
        \begin{array}{ccc}
              \mathbf{a}_1 & = & a \, \mathbf{\hat{x}} \\
    \mathbf{a}_2 & = & b \, \mathbf{\hat{y}} \\
    \mathbf{a}_3 & = & c \, \mathbf{\hat{z}} \\

        \end{array}
      \end{equation*}
    }
    \renewcommand{\arraystretch}{1.0}
  \end{tabular}
  \begin{tabular}{c}
    \includegraphics[width=0.3\linewidth]{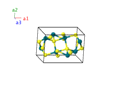} \\
  \end{tabular}
\end{tabular}

}
\vspace*{-0.25cm}

\noindent \hrulefill
\\
\textbf{Basis vectors:}
\vspace*{-0.25cm}
\renewcommand{\arraystretch}{1.5}
\begin{longtabu} to \textwidth{>{\centering $}X[-1,c,c]<{$}>{\centering $}X[-1,c,c]<{$}>{\centering $}X[-1,c,c]<{$}>{\centering $}X[-1,c,c]<{$}>{\centering $}X[-1,c,c]<{$}>{\centering $}X[-1,c,c]<{$}>{\centering $}X[-1,c,c]<{$}}
  & & \mbox{Lattice Coordinates} & & \mbox{Cartesian Coordinates} &\mbox{Wyckoff Position} & \mbox{Atom Type} \\  
  \mathbf{B}_{1} & = & y_{1} \, \mathbf{a}_{2} + \frac{1}{4} \, \mathbf{a}_{3} & = & y_{1}b \, \mathbf{\hat{y}} + \frac{1}{4}c \, \mathbf{\hat{z}} & \left(4c\right) & \mbox{S I} \\ 
\mathbf{B}_{2} & = & \frac{1}{2} \, \mathbf{a}_{1} + \left(\frac{1}{2} - y_{1}\right) \, \mathbf{a}_{2} + \frac{3}{4} \, \mathbf{a}_{3} & = & \frac{1}{2}a \, \mathbf{\hat{x}} + \left(\frac{1}{2} - y_{1}\right)b \, \mathbf{\hat{y}} + \frac{3}{4}c \, \mathbf{\hat{z}} & \left(4c\right) & \mbox{S I} \\ 
\mathbf{B}_{3} & = & -y_{1} \, \mathbf{a}_{2} + \frac{3}{4} \, \mathbf{a}_{3} & = & -y_{1}b \, \mathbf{\hat{y}} + \frac{3}{4}c \, \mathbf{\hat{z}} & \left(4c\right) & \mbox{S I} \\ 
\mathbf{B}_{4} & = & \frac{1}{2} \, \mathbf{a}_{1} + \left(\frac{1}{2} +y_{1}\right) \, \mathbf{a}_{2} + \frac{1}{4} \, \mathbf{a}_{3} & = & \frac{1}{2}a \, \mathbf{\hat{x}} + \left(\frac{1}{2} +y_{1}\right)b \, \mathbf{\hat{y}} + \frac{1}{4}c \, \mathbf{\hat{z}} & \left(4c\right) & \mbox{S I} \\ 
\mathbf{B}_{5} & = & x_{2} \, \mathbf{a}_{1} + y_{2} \, \mathbf{a}_{2} + z_{2} \, \mathbf{a}_{3} & = & x_{2}a \, \mathbf{\hat{x}} + y_{2}b \, \mathbf{\hat{y}} + z_{2}c \, \mathbf{\hat{z}} & \left(8d\right) & \mbox{Rh} \\ 
\mathbf{B}_{6} & = & \left(\frac{1}{2} - x_{2}\right) \, \mathbf{a}_{1} + \left(\frac{1}{2} - y_{2}\right) \, \mathbf{a}_{2} + \left(\frac{1}{2} +z_{2}\right) \, \mathbf{a}_{3} & = & \left(\frac{1}{2} - x_{2}\right)a \, \mathbf{\hat{x}} + \left(\frac{1}{2} - y_{2}\right)b \, \mathbf{\hat{y}} + \left(\frac{1}{2} +z_{2}\right)c \, \mathbf{\hat{z}} & \left(8d\right) & \mbox{Rh} \\ 
\mathbf{B}_{7} & = & -x_{2} \, \mathbf{a}_{1} + y_{2} \, \mathbf{a}_{2} + \left(\frac{1}{2} - z_{2}\right) \, \mathbf{a}_{3} & = & -x_{2}a \, \mathbf{\hat{x}} + y_{2}b \, \mathbf{\hat{y}} + \left(\frac{1}{2} - z_{2}\right)c \, \mathbf{\hat{z}} & \left(8d\right) & \mbox{Rh} \\ 
\mathbf{B}_{8} & = & \left(\frac{1}{2} +x_{2}\right) \, \mathbf{a}_{1} + \left(\frac{1}{2} - y_{2}\right) \, \mathbf{a}_{2}-z_{2} \, \mathbf{a}_{3} & = & \left(\frac{1}{2} +x_{2}\right)a \, \mathbf{\hat{x}} + \left(\frac{1}{2} - y_{2}\right)b \, \mathbf{\hat{y}}-z_{2}c \, \mathbf{\hat{z}} & \left(8d\right) & \mbox{Rh} \\ 
\mathbf{B}_{9} & = & -x_{2} \, \mathbf{a}_{1}-y_{2} \, \mathbf{a}_{2}-z_{2} \, \mathbf{a}_{3} & = & -x_{2}a \, \mathbf{\hat{x}}-y_{2}b \, \mathbf{\hat{y}}-z_{2}c \, \mathbf{\hat{z}} & \left(8d\right) & \mbox{Rh} \\ 
\mathbf{B}_{10} & = & \left(\frac{1}{2} +x_{2}\right) \, \mathbf{a}_{1} + \left(\frac{1}{2} +y_{2}\right) \, \mathbf{a}_{2} + \left(\frac{1}{2} - z_{2}\right) \, \mathbf{a}_{3} & = & \left(\frac{1}{2} +x_{2}\right)a \, \mathbf{\hat{x}} + \left(\frac{1}{2} +y_{2}\right)b \, \mathbf{\hat{y}} + \left(\frac{1}{2} - z_{2}\right)c \, \mathbf{\hat{z}} & \left(8d\right) & \mbox{Rh} \\ 
\mathbf{B}_{11} & = & x_{2} \, \mathbf{a}_{1}-y_{2} \, \mathbf{a}_{2} + \left(\frac{1}{2} +z_{2}\right) \, \mathbf{a}_{3} & = & x_{2}a \, \mathbf{\hat{x}}-y_{2}b \, \mathbf{\hat{y}} + \left(\frac{1}{2} +z_{2}\right)c \, \mathbf{\hat{z}} & \left(8d\right) & \mbox{Rh} \\ 
\mathbf{B}_{12} & = & \left(\frac{1}{2} - x_{2}\right) \, \mathbf{a}_{1} + \left(\frac{1}{2} +y_{2}\right) \, \mathbf{a}_{2} + z_{2} \, \mathbf{a}_{3} & = & \left(\frac{1}{2} - x_{2}\right)a \, \mathbf{\hat{x}} + \left(\frac{1}{2} +y_{2}\right)b \, \mathbf{\hat{y}} + z_{2}c \, \mathbf{\hat{z}} & \left(8d\right) & \mbox{Rh} \\ 
\mathbf{B}_{13} & = & x_{3} \, \mathbf{a}_{1} + y_{3} \, \mathbf{a}_{2} + z_{3} \, \mathbf{a}_{3} & = & x_{3}a \, \mathbf{\hat{x}} + y_{3}b \, \mathbf{\hat{y}} + z_{3}c \, \mathbf{\hat{z}} & \left(8d\right) & \mbox{S II} \\ 
\mathbf{B}_{14} & = & \left(\frac{1}{2} - x_{3}\right) \, \mathbf{a}_{1} + \left(\frac{1}{2} - y_{3}\right) \, \mathbf{a}_{2} + \left(\frac{1}{2} +z_{3}\right) \, \mathbf{a}_{3} & = & \left(\frac{1}{2} - x_{3}\right)a \, \mathbf{\hat{x}} + \left(\frac{1}{2} - y_{3}\right)b \, \mathbf{\hat{y}} + \left(\frac{1}{2} +z_{3}\right)c \, \mathbf{\hat{z}} & \left(8d\right) & \mbox{S II} \\ 
\mathbf{B}_{15} & = & -x_{3} \, \mathbf{a}_{1} + y_{3} \, \mathbf{a}_{2} + \left(\frac{1}{2} - z_{3}\right) \, \mathbf{a}_{3} & = & -x_{3}a \, \mathbf{\hat{x}} + y_{3}b \, \mathbf{\hat{y}} + \left(\frac{1}{2} - z_{3}\right)c \, \mathbf{\hat{z}} & \left(8d\right) & \mbox{S II} \\ 
\mathbf{B}_{16} & = & \left(\frac{1}{2} +x_{3}\right) \, \mathbf{a}_{1} + \left(\frac{1}{2} - y_{3}\right) \, \mathbf{a}_{2}-z_{3} \, \mathbf{a}_{3} & = & \left(\frac{1}{2} +x_{3}\right)a \, \mathbf{\hat{x}} + \left(\frac{1}{2} - y_{3}\right)b \, \mathbf{\hat{y}}-z_{3}c \, \mathbf{\hat{z}} & \left(8d\right) & \mbox{S II} \\ 
\mathbf{B}_{17} & = & -x_{3} \, \mathbf{a}_{1}-y_{3} \, \mathbf{a}_{2}-z_{3} \, \mathbf{a}_{3} & = & -x_{3}a \, \mathbf{\hat{x}}-y_{3}b \, \mathbf{\hat{y}}-z_{3}c \, \mathbf{\hat{z}} & \left(8d\right) & \mbox{S II} \\ 
\mathbf{B}_{18} & = & \left(\frac{1}{2} +x_{3}\right) \, \mathbf{a}_{1} + \left(\frac{1}{2} +y_{3}\right) \, \mathbf{a}_{2} + \left(\frac{1}{2} - z_{3}\right) \, \mathbf{a}_{3} & = & \left(\frac{1}{2} +x_{3}\right)a \, \mathbf{\hat{x}} + \left(\frac{1}{2} +y_{3}\right)b \, \mathbf{\hat{y}} + \left(\frac{1}{2} - z_{3}\right)c \, \mathbf{\hat{z}} & \left(8d\right) & \mbox{S II} \\ 
\mathbf{B}_{19} & = & x_{3} \, \mathbf{a}_{1}-y_{3} \, \mathbf{a}_{2} + \left(\frac{1}{2} +z_{3}\right) \, \mathbf{a}_{3} & = & x_{3}a \, \mathbf{\hat{x}}-y_{3}b \, \mathbf{\hat{y}} + \left(\frac{1}{2} +z_{3}\right)c \, \mathbf{\hat{z}} & \left(8d\right) & \mbox{S II} \\ 
\mathbf{B}_{20} & = & \left(\frac{1}{2} - x_{3}\right) \, \mathbf{a}_{1} + \left(\frac{1}{2} +y_{3}\right) \, \mathbf{a}_{2} + z_{3} \, \mathbf{a}_{3} & = & \left(\frac{1}{2} - x_{3}\right)a \, \mathbf{\hat{x}} + \left(\frac{1}{2} +y_{3}\right)b \, \mathbf{\hat{y}} + z_{3}c \, \mathbf{\hat{z}} & \left(8d\right) & \mbox{S II} \\ 
\end{longtabu}
\renewcommand{\arraystretch}{1.0}
\noindent \hrulefill
\\
\textbf{References:}
\vspace*{-0.25cm}
\begin{flushleft}
  - \bibentry{Parthe_Rh2S3_ActCryst_1966}. \\
\end{flushleft}
\textbf{Found in:}
\vspace*{-0.25cm}
\begin{flushleft}
  - \bibentry{Villars_PearsonsCrystalData_2013}. \\
\end{flushleft}
\noindent \hrulefill
\\
\textbf{Geometry files:}
\\
\noindent  - CIF: pp. {\hyperref[A2B3_oP20_60_d_cd_cif]{\pageref{A2B3_oP20_60_d_cd_cif}}} \\
\noindent  - POSCAR: pp. {\hyperref[A2B3_oP20_60_d_cd_poscar]{\pageref{A2B3_oP20_60_d_cd_poscar}}} \\
\onecolumn
{\phantomsection\label{A3B_oP32_60_3d_d}}
\subsection*{\huge \textbf{{\normalfont WO$_{3}$ Structure: A3B\_oP32\_60\_3d\_d}}}
\noindent \hrulefill
\vspace*{0.25cm}
\begin{figure}[htp]
  \centering
  \vspace{-1em}
  {\includegraphics[width=1\textwidth]{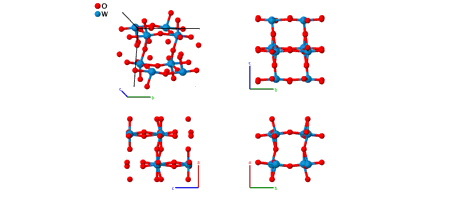}}
\end{figure}
\vspace*{-0.5cm}
\renewcommand{\arraystretch}{1.5}
\begin{equation*}
  \begin{array}{>{$\hspace{-0.15cm}}l<{$}>{$}p{0.5cm}<{$}>{$}p{18.5cm}<{$}}
    \mbox{\large \textbf{Prototype}} &\colon & \ce{WO3} \\
    \mbox{\large \textbf{\AFLOW\ prototype label}} &\colon & \mbox{A3B\_oP32\_60\_3d\_d} \\
    \mbox{\large \textbf{\textit{Strukturbericht} designation}} &\colon & \mbox{None} \\
    \mbox{\large \textbf{Pearson symbol}} &\colon & \mbox{oP32} \\
    \mbox{\large \textbf{Space group number}} &\colon & 60 \\
    \mbox{\large \textbf{Space group symbol}} &\colon & Pbcn \\
    \mbox{\large \textbf{\AFLOW\ prototype command}} &\colon &  \texttt{aflow} \,  \, \texttt{-{}-proto=A3B\_oP32\_60\_3d\_d } \, \newline \texttt{-{}-params=}{a,b/a,c/a,x_{1},y_{1},z_{1},x_{2},y_{2},z_{2},x_{3},y_{3},z_{3},x_{4},y_{4},z_{4} }
  \end{array}
\end{equation*}
\renewcommand{\arraystretch}{1.0}

\noindent \parbox{1 \linewidth}{
\noindent \hrulefill
\\
\textbf{Simple Orthorhombic primitive vectors:} \\
\vspace*{-0.25cm}
\begin{tabular}{cc}
  \begin{tabular}{c}
    \parbox{0.6 \linewidth}{
      \renewcommand{\arraystretch}{1.5}
      \begin{equation*}
        \centering
        \begin{array}{ccc}
              \mathbf{a}_1 & = & a \, \mathbf{\hat{x}} \\
    \mathbf{a}_2 & = & b \, \mathbf{\hat{y}} \\
    \mathbf{a}_3 & = & c \, \mathbf{\hat{z}} \\

        \end{array}
      \end{equation*}
    }
    \renewcommand{\arraystretch}{1.0}
  \end{tabular}
  \begin{tabular}{c}
    \includegraphics[width=0.3\linewidth]{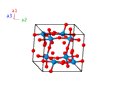} \\
  \end{tabular}
\end{tabular}

}
\vspace*{-0.25cm}

\noindent \hrulefill
\\
\textbf{Basis vectors:}
\vspace*{-0.25cm}
\renewcommand{\arraystretch}{1.5}
\begin{longtabu} to \textwidth{>{\centering $}X[-1,c,c]<{$}>{\centering $}X[-1,c,c]<{$}>{\centering $}X[-1,c,c]<{$}>{\centering $}X[-1,c,c]<{$}>{\centering $}X[-1,c,c]<{$}>{\centering $}X[-1,c,c]<{$}>{\centering $}X[-1,c,c]<{$}}
  & & \mbox{Lattice Coordinates} & & \mbox{Cartesian Coordinates} &\mbox{Wyckoff Position} & \mbox{Atom Type} \\  
  \mathbf{B}_{1} & = & x_{1} \, \mathbf{a}_{1} + y_{1} \, \mathbf{a}_{2} + z_{1} \, \mathbf{a}_{3} & = & x_{1}a \, \mathbf{\hat{x}} + y_{1}b \, \mathbf{\hat{y}} + z_{1}c \, \mathbf{\hat{z}} & \left(8d\right) & \mbox{O I} \\ 
\mathbf{B}_{2} & = & \left(\frac{1}{2} - x_{1}\right) \, \mathbf{a}_{1} + \left(\frac{1}{2} - y_{1}\right) \, \mathbf{a}_{2} + \left(\frac{1}{2} +z_{1}\right) \, \mathbf{a}_{3} & = & \left(\frac{1}{2} - x_{1}\right)a \, \mathbf{\hat{x}} + \left(\frac{1}{2} - y_{1}\right)b \, \mathbf{\hat{y}} + \left(\frac{1}{2} +z_{1}\right)c \, \mathbf{\hat{z}} & \left(8d\right) & \mbox{O I} \\ 
\mathbf{B}_{3} & = & -x_{1} \, \mathbf{a}_{1} + y_{1} \, \mathbf{a}_{2} + \left(\frac{1}{2} - z_{1}\right) \, \mathbf{a}_{3} & = & -x_{1}a \, \mathbf{\hat{x}} + y_{1}b \, \mathbf{\hat{y}} + \left(\frac{1}{2} - z_{1}\right)c \, \mathbf{\hat{z}} & \left(8d\right) & \mbox{O I} \\ 
\mathbf{B}_{4} & = & \left(\frac{1}{2} +x_{1}\right) \, \mathbf{a}_{1} + \left(\frac{1}{2} - y_{1}\right) \, \mathbf{a}_{2}-z_{1} \, \mathbf{a}_{3} & = & \left(\frac{1}{2} +x_{1}\right)a \, \mathbf{\hat{x}} + \left(\frac{1}{2} - y_{1}\right)b \, \mathbf{\hat{y}}-z_{1}c \, \mathbf{\hat{z}} & \left(8d\right) & \mbox{O I} \\ 
\mathbf{B}_{5} & = & -x_{1} \, \mathbf{a}_{1}-y_{1} \, \mathbf{a}_{2}-z_{1} \, \mathbf{a}_{3} & = & -x_{1}a \, \mathbf{\hat{x}}-y_{1}b \, \mathbf{\hat{y}}-z_{1}c \, \mathbf{\hat{z}} & \left(8d\right) & \mbox{O I} \\ 
\mathbf{B}_{6} & = & \left(\frac{1}{2} +x_{1}\right) \, \mathbf{a}_{1} + \left(\frac{1}{2} +y_{1}\right) \, \mathbf{a}_{2} + \left(\frac{1}{2} - z_{1}\right) \, \mathbf{a}_{3} & = & \left(\frac{1}{2} +x_{1}\right)a \, \mathbf{\hat{x}} + \left(\frac{1}{2} +y_{1}\right)b \, \mathbf{\hat{y}} + \left(\frac{1}{2} - z_{1}\right)c \, \mathbf{\hat{z}} & \left(8d\right) & \mbox{O I} \\ 
\mathbf{B}_{7} & = & x_{1} \, \mathbf{a}_{1}-y_{1} \, \mathbf{a}_{2} + \left(\frac{1}{2} +z_{1}\right) \, \mathbf{a}_{3} & = & x_{1}a \, \mathbf{\hat{x}}-y_{1}b \, \mathbf{\hat{y}} + \left(\frac{1}{2} +z_{1}\right)c \, \mathbf{\hat{z}} & \left(8d\right) & \mbox{O I} \\ 
\mathbf{B}_{8} & = & \left(\frac{1}{2} - x_{1}\right) \, \mathbf{a}_{1} + \left(\frac{1}{2} +y_{1}\right) \, \mathbf{a}_{2} + z_{1} \, \mathbf{a}_{3} & = & \left(\frac{1}{2} - x_{1}\right)a \, \mathbf{\hat{x}} + \left(\frac{1}{2} +y_{1}\right)b \, \mathbf{\hat{y}} + z_{1}c \, \mathbf{\hat{z}} & \left(8d\right) & \mbox{O I} \\ 
\mathbf{B}_{9} & = & x_{2} \, \mathbf{a}_{1} + y_{2} \, \mathbf{a}_{2} + z_{2} \, \mathbf{a}_{3} & = & x_{2}a \, \mathbf{\hat{x}} + y_{2}b \, \mathbf{\hat{y}} + z_{2}c \, \mathbf{\hat{z}} & \left(8d\right) & \mbox{O II} \\ 
\mathbf{B}_{10} & = & \left(\frac{1}{2} - x_{2}\right) \, \mathbf{a}_{1} + \left(\frac{1}{2} - y_{2}\right) \, \mathbf{a}_{2} + \left(\frac{1}{2} +z_{2}\right) \, \mathbf{a}_{3} & = & \left(\frac{1}{2} - x_{2}\right)a \, \mathbf{\hat{x}} + \left(\frac{1}{2} - y_{2}\right)b \, \mathbf{\hat{y}} + \left(\frac{1}{2} +z_{2}\right)c \, \mathbf{\hat{z}} & \left(8d\right) & \mbox{O II} \\ 
\mathbf{B}_{11} & = & -x_{2} \, \mathbf{a}_{1} + y_{2} \, \mathbf{a}_{2} + \left(\frac{1}{2} - z_{2}\right) \, \mathbf{a}_{3} & = & -x_{2}a \, \mathbf{\hat{x}} + y_{2}b \, \mathbf{\hat{y}} + \left(\frac{1}{2} - z_{2}\right)c \, \mathbf{\hat{z}} & \left(8d\right) & \mbox{O II} \\ 
\mathbf{B}_{12} & = & \left(\frac{1}{2} +x_{2}\right) \, \mathbf{a}_{1} + \left(\frac{1}{2} - y_{2}\right) \, \mathbf{a}_{2}-z_{2} \, \mathbf{a}_{3} & = & \left(\frac{1}{2} +x_{2}\right)a \, \mathbf{\hat{x}} + \left(\frac{1}{2} - y_{2}\right)b \, \mathbf{\hat{y}}-z_{2}c \, \mathbf{\hat{z}} & \left(8d\right) & \mbox{O II} \\ 
\mathbf{B}_{13} & = & -x_{2} \, \mathbf{a}_{1}-y_{2} \, \mathbf{a}_{2}-z_{2} \, \mathbf{a}_{3} & = & -x_{2}a \, \mathbf{\hat{x}}-y_{2}b \, \mathbf{\hat{y}}-z_{2}c \, \mathbf{\hat{z}} & \left(8d\right) & \mbox{O II} \\ 
\mathbf{B}_{14} & = & \left(\frac{1}{2} +x_{2}\right) \, \mathbf{a}_{1} + \left(\frac{1}{2} +y_{2}\right) \, \mathbf{a}_{2} + \left(\frac{1}{2} - z_{2}\right) \, \mathbf{a}_{3} & = & \left(\frac{1}{2} +x_{2}\right)a \, \mathbf{\hat{x}} + \left(\frac{1}{2} +y_{2}\right)b \, \mathbf{\hat{y}} + \left(\frac{1}{2} - z_{2}\right)c \, \mathbf{\hat{z}} & \left(8d\right) & \mbox{O II} \\ 
\mathbf{B}_{15} & = & x_{2} \, \mathbf{a}_{1}-y_{2} \, \mathbf{a}_{2} + \left(\frac{1}{2} +z_{2}\right) \, \mathbf{a}_{3} & = & x_{2}a \, \mathbf{\hat{x}}-y_{2}b \, \mathbf{\hat{y}} + \left(\frac{1}{2} +z_{2}\right)c \, \mathbf{\hat{z}} & \left(8d\right) & \mbox{O II} \\ 
\mathbf{B}_{16} & = & \left(\frac{1}{2} - x_{2}\right) \, \mathbf{a}_{1} + \left(\frac{1}{2} +y_{2}\right) \, \mathbf{a}_{2} + z_{2} \, \mathbf{a}_{3} & = & \left(\frac{1}{2} - x_{2}\right)a \, \mathbf{\hat{x}} + \left(\frac{1}{2} +y_{2}\right)b \, \mathbf{\hat{y}} + z_{2}c \, \mathbf{\hat{z}} & \left(8d\right) & \mbox{O II} \\ 
\mathbf{B}_{17} & = & x_{3} \, \mathbf{a}_{1} + y_{3} \, \mathbf{a}_{2} + z_{3} \, \mathbf{a}_{3} & = & x_{3}a \, \mathbf{\hat{x}} + y_{3}b \, \mathbf{\hat{y}} + z_{3}c \, \mathbf{\hat{z}} & \left(8d\right) & \mbox{O III} \\ 
\mathbf{B}_{18} & = & \left(\frac{1}{2} - x_{3}\right) \, \mathbf{a}_{1} + \left(\frac{1}{2} - y_{3}\right) \, \mathbf{a}_{2} + \left(\frac{1}{2} +z_{3}\right) \, \mathbf{a}_{3} & = & \left(\frac{1}{2} - x_{3}\right)a \, \mathbf{\hat{x}} + \left(\frac{1}{2} - y_{3}\right)b \, \mathbf{\hat{y}} + \left(\frac{1}{2} +z_{3}\right)c \, \mathbf{\hat{z}} & \left(8d\right) & \mbox{O III} \\ 
\mathbf{B}_{19} & = & -x_{3} \, \mathbf{a}_{1} + y_{3} \, \mathbf{a}_{2} + \left(\frac{1}{2} - z_{3}\right) \, \mathbf{a}_{3} & = & -x_{3}a \, \mathbf{\hat{x}} + y_{3}b \, \mathbf{\hat{y}} + \left(\frac{1}{2} - z_{3}\right)c \, \mathbf{\hat{z}} & \left(8d\right) & \mbox{O III} \\ 
\mathbf{B}_{20} & = & \left(\frac{1}{2} +x_{3}\right) \, \mathbf{a}_{1} + \left(\frac{1}{2} - y_{3}\right) \, \mathbf{a}_{2}-z_{3} \, \mathbf{a}_{3} & = & \left(\frac{1}{2} +x_{3}\right)a \, \mathbf{\hat{x}} + \left(\frac{1}{2} - y_{3}\right)b \, \mathbf{\hat{y}}-z_{3}c \, \mathbf{\hat{z}} & \left(8d\right) & \mbox{O III} \\ 
\mathbf{B}_{21} & = & -x_{3} \, \mathbf{a}_{1}-y_{3} \, \mathbf{a}_{2}-z_{3} \, \mathbf{a}_{3} & = & -x_{3}a \, \mathbf{\hat{x}}-y_{3}b \, \mathbf{\hat{y}}-z_{3}c \, \mathbf{\hat{z}} & \left(8d\right) & \mbox{O III} \\ 
\mathbf{B}_{22} & = & \left(\frac{1}{2} +x_{3}\right) \, \mathbf{a}_{1} + \left(\frac{1}{2} +y_{3}\right) \, \mathbf{a}_{2} + \left(\frac{1}{2} - z_{3}\right) \, \mathbf{a}_{3} & = & \left(\frac{1}{2} +x_{3}\right)a \, \mathbf{\hat{x}} + \left(\frac{1}{2} +y_{3}\right)b \, \mathbf{\hat{y}} + \left(\frac{1}{2} - z_{3}\right)c \, \mathbf{\hat{z}} & \left(8d\right) & \mbox{O III} \\ 
\mathbf{B}_{23} & = & x_{3} \, \mathbf{a}_{1}-y_{3} \, \mathbf{a}_{2} + \left(\frac{1}{2} +z_{3}\right) \, \mathbf{a}_{3} & = & x_{3}a \, \mathbf{\hat{x}}-y_{3}b \, \mathbf{\hat{y}} + \left(\frac{1}{2} +z_{3}\right)c \, \mathbf{\hat{z}} & \left(8d\right) & \mbox{O III} \\ 
\mathbf{B}_{24} & = & \left(\frac{1}{2} - x_{3}\right) \, \mathbf{a}_{1} + \left(\frac{1}{2} +y_{3}\right) \, \mathbf{a}_{2} + z_{3} \, \mathbf{a}_{3} & = & \left(\frac{1}{2} - x_{3}\right)a \, \mathbf{\hat{x}} + \left(\frac{1}{2} +y_{3}\right)b \, \mathbf{\hat{y}} + z_{3}c \, \mathbf{\hat{z}} & \left(8d\right) & \mbox{O III} \\ 
\mathbf{B}_{25} & = & x_{4} \, \mathbf{a}_{1} + y_{4} \, \mathbf{a}_{2} + z_{4} \, \mathbf{a}_{3} & = & x_{4}a \, \mathbf{\hat{x}} + y_{4}b \, \mathbf{\hat{y}} + z_{4}c \, \mathbf{\hat{z}} & \left(8d\right) & \mbox{W} \\ 
\mathbf{B}_{26} & = & \left(\frac{1}{2} - x_{4}\right) \, \mathbf{a}_{1} + \left(\frac{1}{2} - y_{4}\right) \, \mathbf{a}_{2} + \left(\frac{1}{2} +z_{4}\right) \, \mathbf{a}_{3} & = & \left(\frac{1}{2} - x_{4}\right)a \, \mathbf{\hat{x}} + \left(\frac{1}{2} - y_{4}\right)b \, \mathbf{\hat{y}} + \left(\frac{1}{2} +z_{4}\right)c \, \mathbf{\hat{z}} & \left(8d\right) & \mbox{W} \\ 
\mathbf{B}_{27} & = & -x_{4} \, \mathbf{a}_{1} + y_{4} \, \mathbf{a}_{2} + \left(\frac{1}{2} - z_{4}\right) \, \mathbf{a}_{3} & = & -x_{4}a \, \mathbf{\hat{x}} + y_{4}b \, \mathbf{\hat{y}} + \left(\frac{1}{2} - z_{4}\right)c \, \mathbf{\hat{z}} & \left(8d\right) & \mbox{W} \\ 
\mathbf{B}_{28} & = & \left(\frac{1}{2} +x_{4}\right) \, \mathbf{a}_{1} + \left(\frac{1}{2} - y_{4}\right) \, \mathbf{a}_{2}-z_{4} \, \mathbf{a}_{3} & = & \left(\frac{1}{2} +x_{4}\right)a \, \mathbf{\hat{x}} + \left(\frac{1}{2} - y_{4}\right)b \, \mathbf{\hat{y}}-z_{4}c \, \mathbf{\hat{z}} & \left(8d\right) & \mbox{W} \\ 
\mathbf{B}_{29} & = & -x_{4} \, \mathbf{a}_{1}-y_{4} \, \mathbf{a}_{2}-z_{4} \, \mathbf{a}_{3} & = & -x_{4}a \, \mathbf{\hat{x}}-y_{4}b \, \mathbf{\hat{y}}-z_{4}c \, \mathbf{\hat{z}} & \left(8d\right) & \mbox{W} \\ 
\mathbf{B}_{30} & = & \left(\frac{1}{2} +x_{4}\right) \, \mathbf{a}_{1} + \left(\frac{1}{2} +y_{4}\right) \, \mathbf{a}_{2} + \left(\frac{1}{2} - z_{4}\right) \, \mathbf{a}_{3} & = & \left(\frac{1}{2} +x_{4}\right)a \, \mathbf{\hat{x}} + \left(\frac{1}{2} +y_{4}\right)b \, \mathbf{\hat{y}} + \left(\frac{1}{2} - z_{4}\right)c \, \mathbf{\hat{z}} & \left(8d\right) & \mbox{W} \\ 
\mathbf{B}_{31} & = & x_{4} \, \mathbf{a}_{1}-y_{4} \, \mathbf{a}_{2} + \left(\frac{1}{2} +z_{4}\right) \, \mathbf{a}_{3} & = & x_{4}a \, \mathbf{\hat{x}}-y_{4}b \, \mathbf{\hat{y}} + \left(\frac{1}{2} +z_{4}\right)c \, \mathbf{\hat{z}} & \left(8d\right) & \mbox{W} \\ 
\mathbf{B}_{32} & = & \left(\frac{1}{2} - x_{4}\right) \, \mathbf{a}_{1} + \left(\frac{1}{2} +y_{4}\right) \, \mathbf{a}_{2} + z_{4} \, \mathbf{a}_{3} & = & \left(\frac{1}{2} - x_{4}\right)a \, \mathbf{\hat{x}} + \left(\frac{1}{2} +y_{4}\right)b \, \mathbf{\hat{y}} + z_{4}c \, \mathbf{\hat{z}} & \left(8d\right) & \mbox{W} \\ 
\end{longtabu}
\renewcommand{\arraystretch}{1.0}
\noindent \hrulefill
\\
\textbf{References:}
\vspace*{-0.25cm}
\begin{flushleft}
  - \bibentry{Vogt_WO3_JSolStateChem_1999}. \\
\end{flushleft}
\textbf{Found in:}
\vspace*{-0.25cm}
\begin{flushleft}
  - \bibentry{Villars_PearsonsCrystalData_2013}. \\
\end{flushleft}
\noindent \hrulefill
\\
\textbf{Geometry files:}
\\
\noindent  - CIF: pp. {\hyperref[A3B_oP32_60_3d_d_cif]{\pageref{A3B_oP32_60_3d_d_cif}}} \\
\noindent  - POSCAR: pp. {\hyperref[A3B_oP32_60_3d_d_poscar]{\pageref{A3B_oP32_60_3d_d_poscar}}} \\
\onecolumn
{\phantomsection\label{A7B8_oP120_60_7d_8d}}
\subsection*{\huge \textbf{{\normalfont $\beta$-Toluene Structure: A7B8\_oP120\_60\_7d\_8d}}}
\noindent \hrulefill
\vspace*{0.25cm}
\begin{figure}[htp]
  \centering
  \vspace{-1em}
  {\includegraphics[width=1\textwidth]{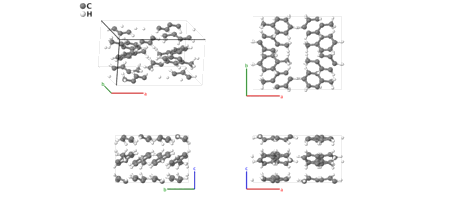}}
\end{figure}
\vspace*{-0.5cm}
\renewcommand{\arraystretch}{1.5}
\begin{equation*}
  \begin{array}{>{$\hspace{-0.15cm}}l<{$}>{$}p{0.5cm}<{$}>{$}p{18.5cm}<{$}}
    \mbox{\large \textbf{Prototype}} &\colon & \ce{$\beta$-C7H8} \\
    \mbox{\large \textbf{\AFLOW\ prototype label}} &\colon & \mbox{A7B8\_oP120\_60\_7d\_8d} \\
    \mbox{\large \textbf{\textit{Strukturbericht} designation}} &\colon & \mbox{None} \\
    \mbox{\large \textbf{Pearson symbol}} &\colon & \mbox{oP120} \\
    \mbox{\large \textbf{Space group number}} &\colon & 60 \\
    \mbox{\large \textbf{Space group symbol}} &\colon & Pbcn \\
    \mbox{\large \textbf{\AFLOW\ prototype command}} &\colon &  \texttt{aflow} \,  \, \texttt{-{}-proto=A7B8\_oP120\_60\_7d\_8d } \, \newline \texttt{-{}-params=}{a,b/a,c/a,x_{1},y_{1},z_{1},x_{2},y_{2},z_{2},x_{3},y_{3},z_{3},x_{4},y_{4},z_{4},x_{5},y_{5},z_{5},x_{6},y_{6},} \newline {z_{6},x_{7},y_{7},z_{7},x_{8},y_{8},z_{8},x_{9},y_{9},z_{9},x_{10},y_{10},z_{10},x_{11},y_{11},z_{11},x_{12},y_{12},z_{12},x_{13},y_{13},z_{13},} \newline {x_{14},y_{14},z_{14},x_{15},y_{15},z_{15} }
  \end{array}
\end{equation*}
\renewcommand{\arraystretch}{1.0}

\vspace*{-0.25cm}
\noindent \hrulefill
\begin{itemize}
  \item{$\beta$-Toluene is a metastable crystalline structure of the toluene
molecule, C$_7$H$_8$, which crystallizes below 178~K.  This data was
constructed from experiments at 105~K.}
  \item{The hydrogen atomic positions were approximated to agree
with the chemistry of the toluene molecule.
}
\end{itemize}

\noindent \parbox{1 \linewidth}{
\noindent \hrulefill
\\
\textbf{Simple Orthorhombic primitive vectors:} \\
\vspace*{-0.25cm}
\begin{tabular}{cc}
  \begin{tabular}{c}
    \parbox{0.6 \linewidth}{
      \renewcommand{\arraystretch}{1.5}
      \begin{equation*}
        \centering
        \begin{array}{ccc}
              \mathbf{a}_1 & = & a \, \mathbf{\hat{x}} \\
    \mathbf{a}_2 & = & b \, \mathbf{\hat{y}} \\
    \mathbf{a}_3 & = & c \, \mathbf{\hat{z}} \\

        \end{array}
      \end{equation*}
    }
    \renewcommand{\arraystretch}{1.0}
  \end{tabular}
  \begin{tabular}{c}
    \includegraphics[width=0.3\linewidth]{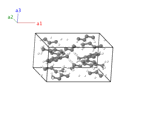} \\
  \end{tabular}
\end{tabular}

}
\vspace*{-0.25cm}

\noindent \hrulefill
\\
\textbf{Basis vectors:}
\vspace*{-0.25cm}
\renewcommand{\arraystretch}{1.5}
\begin{longtabu} to \textwidth{>{\centering $}X[-1,c,c]<{$}>{\centering $}X[-1,c,c]<{$}>{\centering $}X[-1,c,c]<{$}>{\centering $}X[-1,c,c]<{$}>{\centering $}X[-1,c,c]<{$}>{\centering $}X[-1,c,c]<{$}>{\centering $}X[-1,c,c]<{$}}
  & & \mbox{Lattice Coordinates} & & \mbox{Cartesian Coordinates} &\mbox{Wyckoff Position} & \mbox{Atom Type} \\  
  \mathbf{B}_{1} & = & x_{1} \, \mathbf{a}_{1} + y_{1} \, \mathbf{a}_{2} + z_{1} \, \mathbf{a}_{3} & = & x_{1}a \, \mathbf{\hat{x}} + y_{1}b \, \mathbf{\hat{y}} + z_{1}c \, \mathbf{\hat{z}} & \left(8d\right) & \mbox{C I} \\ 
\mathbf{B}_{2} & = & \left(\frac{1}{2} - x_{1}\right) \, \mathbf{a}_{1} + \left(\frac{1}{2} - y_{1}\right) \, \mathbf{a}_{2} + \left(\frac{1}{2} +z_{1}\right) \, \mathbf{a}_{3} & = & \left(\frac{1}{2} - x_{1}\right)a \, \mathbf{\hat{x}} + \left(\frac{1}{2} - y_{1}\right)b \, \mathbf{\hat{y}} + \left(\frac{1}{2} +z_{1}\right)c \, \mathbf{\hat{z}} & \left(8d\right) & \mbox{C I} \\ 
\mathbf{B}_{3} & = & -x_{1} \, \mathbf{a}_{1} + y_{1} \, \mathbf{a}_{2} + \left(\frac{1}{2} - z_{1}\right) \, \mathbf{a}_{3} & = & -x_{1}a \, \mathbf{\hat{x}} + y_{1}b \, \mathbf{\hat{y}} + \left(\frac{1}{2} - z_{1}\right)c \, \mathbf{\hat{z}} & \left(8d\right) & \mbox{C I} \\ 
\mathbf{B}_{4} & = & \left(\frac{1}{2} +x_{1}\right) \, \mathbf{a}_{1} + \left(\frac{1}{2} - y_{1}\right) \, \mathbf{a}_{2}-z_{1} \, \mathbf{a}_{3} & = & \left(\frac{1}{2} +x_{1}\right)a \, \mathbf{\hat{x}} + \left(\frac{1}{2} - y_{1}\right)b \, \mathbf{\hat{y}}-z_{1}c \, \mathbf{\hat{z}} & \left(8d\right) & \mbox{C I} \\ 
\mathbf{B}_{5} & = & -x_{1} \, \mathbf{a}_{1}-y_{1} \, \mathbf{a}_{2}-z_{1} \, \mathbf{a}_{3} & = & -x_{1}a \, \mathbf{\hat{x}}-y_{1}b \, \mathbf{\hat{y}}-z_{1}c \, \mathbf{\hat{z}} & \left(8d\right) & \mbox{C I} \\ 
\mathbf{B}_{6} & = & \left(\frac{1}{2} +x_{1}\right) \, \mathbf{a}_{1} + \left(\frac{1}{2} +y_{1}\right) \, \mathbf{a}_{2} + \left(\frac{1}{2} - z_{1}\right) \, \mathbf{a}_{3} & = & \left(\frac{1}{2} +x_{1}\right)a \, \mathbf{\hat{x}} + \left(\frac{1}{2} +y_{1}\right)b \, \mathbf{\hat{y}} + \left(\frac{1}{2} - z_{1}\right)c \, \mathbf{\hat{z}} & \left(8d\right) & \mbox{C I} \\ 
\mathbf{B}_{7} & = & x_{1} \, \mathbf{a}_{1}-y_{1} \, \mathbf{a}_{2} + \left(\frac{1}{2} +z_{1}\right) \, \mathbf{a}_{3} & = & x_{1}a \, \mathbf{\hat{x}}-y_{1}b \, \mathbf{\hat{y}} + \left(\frac{1}{2} +z_{1}\right)c \, \mathbf{\hat{z}} & \left(8d\right) & \mbox{C I} \\ 
\mathbf{B}_{8} & = & \left(\frac{1}{2} - x_{1}\right) \, \mathbf{a}_{1} + \left(\frac{1}{2} +y_{1}\right) \, \mathbf{a}_{2} + z_{1} \, \mathbf{a}_{3} & = & \left(\frac{1}{2} - x_{1}\right)a \, \mathbf{\hat{x}} + \left(\frac{1}{2} +y_{1}\right)b \, \mathbf{\hat{y}} + z_{1}c \, \mathbf{\hat{z}} & \left(8d\right) & \mbox{C I} \\ 
\mathbf{B}_{9} & = & x_{2} \, \mathbf{a}_{1} + y_{2} \, \mathbf{a}_{2} + z_{2} \, \mathbf{a}_{3} & = & x_{2}a \, \mathbf{\hat{x}} + y_{2}b \, \mathbf{\hat{y}} + z_{2}c \, \mathbf{\hat{z}} & \left(8d\right) & \mbox{C II} \\ 
\mathbf{B}_{10} & = & \left(\frac{1}{2} - x_{2}\right) \, \mathbf{a}_{1} + \left(\frac{1}{2} - y_{2}\right) \, \mathbf{a}_{2} + \left(\frac{1}{2} +z_{2}\right) \, \mathbf{a}_{3} & = & \left(\frac{1}{2} - x_{2}\right)a \, \mathbf{\hat{x}} + \left(\frac{1}{2} - y_{2}\right)b \, \mathbf{\hat{y}} + \left(\frac{1}{2} +z_{2}\right)c \, \mathbf{\hat{z}} & \left(8d\right) & \mbox{C II} \\ 
\mathbf{B}_{11} & = & -x_{2} \, \mathbf{a}_{1} + y_{2} \, \mathbf{a}_{2} + \left(\frac{1}{2} - z_{2}\right) \, \mathbf{a}_{3} & = & -x_{2}a \, \mathbf{\hat{x}} + y_{2}b \, \mathbf{\hat{y}} + \left(\frac{1}{2} - z_{2}\right)c \, \mathbf{\hat{z}} & \left(8d\right) & \mbox{C II} \\ 
\mathbf{B}_{12} & = & \left(\frac{1}{2} +x_{2}\right) \, \mathbf{a}_{1} + \left(\frac{1}{2} - y_{2}\right) \, \mathbf{a}_{2}-z_{2} \, \mathbf{a}_{3} & = & \left(\frac{1}{2} +x_{2}\right)a \, \mathbf{\hat{x}} + \left(\frac{1}{2} - y_{2}\right)b \, \mathbf{\hat{y}}-z_{2}c \, \mathbf{\hat{z}} & \left(8d\right) & \mbox{C II} \\ 
\mathbf{B}_{13} & = & -x_{2} \, \mathbf{a}_{1}-y_{2} \, \mathbf{a}_{2}-z_{2} \, \mathbf{a}_{3} & = & -x_{2}a \, \mathbf{\hat{x}}-y_{2}b \, \mathbf{\hat{y}}-z_{2}c \, \mathbf{\hat{z}} & \left(8d\right) & \mbox{C II} \\ 
\mathbf{B}_{14} & = & \left(\frac{1}{2} +x_{2}\right) \, \mathbf{a}_{1} + \left(\frac{1}{2} +y_{2}\right) \, \mathbf{a}_{2} + \left(\frac{1}{2} - z_{2}\right) \, \mathbf{a}_{3} & = & \left(\frac{1}{2} +x_{2}\right)a \, \mathbf{\hat{x}} + \left(\frac{1}{2} +y_{2}\right)b \, \mathbf{\hat{y}} + \left(\frac{1}{2} - z_{2}\right)c \, \mathbf{\hat{z}} & \left(8d\right) & \mbox{C II} \\ 
\mathbf{B}_{15} & = & x_{2} \, \mathbf{a}_{1}-y_{2} \, \mathbf{a}_{2} + \left(\frac{1}{2} +z_{2}\right) \, \mathbf{a}_{3} & = & x_{2}a \, \mathbf{\hat{x}}-y_{2}b \, \mathbf{\hat{y}} + \left(\frac{1}{2} +z_{2}\right)c \, \mathbf{\hat{z}} & \left(8d\right) & \mbox{C II} \\ 
\mathbf{B}_{16} & = & \left(\frac{1}{2} - x_{2}\right) \, \mathbf{a}_{1} + \left(\frac{1}{2} +y_{2}\right) \, \mathbf{a}_{2} + z_{2} \, \mathbf{a}_{3} & = & \left(\frac{1}{2} - x_{2}\right)a \, \mathbf{\hat{x}} + \left(\frac{1}{2} +y_{2}\right)b \, \mathbf{\hat{y}} + z_{2}c \, \mathbf{\hat{z}} & \left(8d\right) & \mbox{C II} \\ 
\mathbf{B}_{17} & = & x_{3} \, \mathbf{a}_{1} + y_{3} \, \mathbf{a}_{2} + z_{3} \, \mathbf{a}_{3} & = & x_{3}a \, \mathbf{\hat{x}} + y_{3}b \, \mathbf{\hat{y}} + z_{3}c \, \mathbf{\hat{z}} & \left(8d\right) & \mbox{C III} \\ 
\mathbf{B}_{18} & = & \left(\frac{1}{2} - x_{3}\right) \, \mathbf{a}_{1} + \left(\frac{1}{2} - y_{3}\right) \, \mathbf{a}_{2} + \left(\frac{1}{2} +z_{3}\right) \, \mathbf{a}_{3} & = & \left(\frac{1}{2} - x_{3}\right)a \, \mathbf{\hat{x}} + \left(\frac{1}{2} - y_{3}\right)b \, \mathbf{\hat{y}} + \left(\frac{1}{2} +z_{3}\right)c \, \mathbf{\hat{z}} & \left(8d\right) & \mbox{C III} \\ 
\mathbf{B}_{19} & = & -x_{3} \, \mathbf{a}_{1} + y_{3} \, \mathbf{a}_{2} + \left(\frac{1}{2} - z_{3}\right) \, \mathbf{a}_{3} & = & -x_{3}a \, \mathbf{\hat{x}} + y_{3}b \, \mathbf{\hat{y}} + \left(\frac{1}{2} - z_{3}\right)c \, \mathbf{\hat{z}} & \left(8d\right) & \mbox{C III} \\ 
\mathbf{B}_{20} & = & \left(\frac{1}{2} +x_{3}\right) \, \mathbf{a}_{1} + \left(\frac{1}{2} - y_{3}\right) \, \mathbf{a}_{2}-z_{3} \, \mathbf{a}_{3} & = & \left(\frac{1}{2} +x_{3}\right)a \, \mathbf{\hat{x}} + \left(\frac{1}{2} - y_{3}\right)b \, \mathbf{\hat{y}}-z_{3}c \, \mathbf{\hat{z}} & \left(8d\right) & \mbox{C III} \\ 
\mathbf{B}_{21} & = & -x_{3} \, \mathbf{a}_{1}-y_{3} \, \mathbf{a}_{2}-z_{3} \, \mathbf{a}_{3} & = & -x_{3}a \, \mathbf{\hat{x}}-y_{3}b \, \mathbf{\hat{y}}-z_{3}c \, \mathbf{\hat{z}} & \left(8d\right) & \mbox{C III} \\ 
\mathbf{B}_{22} & = & \left(\frac{1}{2} +x_{3}\right) \, \mathbf{a}_{1} + \left(\frac{1}{2} +y_{3}\right) \, \mathbf{a}_{2} + \left(\frac{1}{2} - z_{3}\right) \, \mathbf{a}_{3} & = & \left(\frac{1}{2} +x_{3}\right)a \, \mathbf{\hat{x}} + \left(\frac{1}{2} +y_{3}\right)b \, \mathbf{\hat{y}} + \left(\frac{1}{2} - z_{3}\right)c \, \mathbf{\hat{z}} & \left(8d\right) & \mbox{C III} \\ 
\mathbf{B}_{23} & = & x_{3} \, \mathbf{a}_{1}-y_{3} \, \mathbf{a}_{2} + \left(\frac{1}{2} +z_{3}\right) \, \mathbf{a}_{3} & = & x_{3}a \, \mathbf{\hat{x}}-y_{3}b \, \mathbf{\hat{y}} + \left(\frac{1}{2} +z_{3}\right)c \, \mathbf{\hat{z}} & \left(8d\right) & \mbox{C III} \\ 
\mathbf{B}_{24} & = & \left(\frac{1}{2} - x_{3}\right) \, \mathbf{a}_{1} + \left(\frac{1}{2} +y_{3}\right) \, \mathbf{a}_{2} + z_{3} \, \mathbf{a}_{3} & = & \left(\frac{1}{2} - x_{3}\right)a \, \mathbf{\hat{x}} + \left(\frac{1}{2} +y_{3}\right)b \, \mathbf{\hat{y}} + z_{3}c \, \mathbf{\hat{z}} & \left(8d\right) & \mbox{C III} \\ 
\mathbf{B}_{25} & = & x_{4} \, \mathbf{a}_{1} + y_{4} \, \mathbf{a}_{2} + z_{4} \, \mathbf{a}_{3} & = & x_{4}a \, \mathbf{\hat{x}} + y_{4}b \, \mathbf{\hat{y}} + z_{4}c \, \mathbf{\hat{z}} & \left(8d\right) & \mbox{C IV} \\ 
\mathbf{B}_{26} & = & \left(\frac{1}{2} - x_{4}\right) \, \mathbf{a}_{1} + \left(\frac{1}{2} - y_{4}\right) \, \mathbf{a}_{2} + \left(\frac{1}{2} +z_{4}\right) \, \mathbf{a}_{3} & = & \left(\frac{1}{2} - x_{4}\right)a \, \mathbf{\hat{x}} + \left(\frac{1}{2} - y_{4}\right)b \, \mathbf{\hat{y}} + \left(\frac{1}{2} +z_{4}\right)c \, \mathbf{\hat{z}} & \left(8d\right) & \mbox{C IV} \\ 
\mathbf{B}_{27} & = & -x_{4} \, \mathbf{a}_{1} + y_{4} \, \mathbf{a}_{2} + \left(\frac{1}{2} - z_{4}\right) \, \mathbf{a}_{3} & = & -x_{4}a \, \mathbf{\hat{x}} + y_{4}b \, \mathbf{\hat{y}} + \left(\frac{1}{2} - z_{4}\right)c \, \mathbf{\hat{z}} & \left(8d\right) & \mbox{C IV} \\ 
\mathbf{B}_{28} & = & \left(\frac{1}{2} +x_{4}\right) \, \mathbf{a}_{1} + \left(\frac{1}{2} - y_{4}\right) \, \mathbf{a}_{2}-z_{4} \, \mathbf{a}_{3} & = & \left(\frac{1}{2} +x_{4}\right)a \, \mathbf{\hat{x}} + \left(\frac{1}{2} - y_{4}\right)b \, \mathbf{\hat{y}}-z_{4}c \, \mathbf{\hat{z}} & \left(8d\right) & \mbox{C IV} \\ 
\mathbf{B}_{29} & = & -x_{4} \, \mathbf{a}_{1}-y_{4} \, \mathbf{a}_{2}-z_{4} \, \mathbf{a}_{3} & = & -x_{4}a \, \mathbf{\hat{x}}-y_{4}b \, \mathbf{\hat{y}}-z_{4}c \, \mathbf{\hat{z}} & \left(8d\right) & \mbox{C IV} \\ 
\mathbf{B}_{30} & = & \left(\frac{1}{2} +x_{4}\right) \, \mathbf{a}_{1} + \left(\frac{1}{2} +y_{4}\right) \, \mathbf{a}_{2} + \left(\frac{1}{2} - z_{4}\right) \, \mathbf{a}_{3} & = & \left(\frac{1}{2} +x_{4}\right)a \, \mathbf{\hat{x}} + \left(\frac{1}{2} +y_{4}\right)b \, \mathbf{\hat{y}} + \left(\frac{1}{2} - z_{4}\right)c \, \mathbf{\hat{z}} & \left(8d\right) & \mbox{C IV} \\ 
\mathbf{B}_{31} & = & x_{4} \, \mathbf{a}_{1}-y_{4} \, \mathbf{a}_{2} + \left(\frac{1}{2} +z_{4}\right) \, \mathbf{a}_{3} & = & x_{4}a \, \mathbf{\hat{x}}-y_{4}b \, \mathbf{\hat{y}} + \left(\frac{1}{2} +z_{4}\right)c \, \mathbf{\hat{z}} & \left(8d\right) & \mbox{C IV} \\ 
\mathbf{B}_{32} & = & \left(\frac{1}{2} - x_{4}\right) \, \mathbf{a}_{1} + \left(\frac{1}{2} +y_{4}\right) \, \mathbf{a}_{2} + z_{4} \, \mathbf{a}_{3} & = & \left(\frac{1}{2} - x_{4}\right)a \, \mathbf{\hat{x}} + \left(\frac{1}{2} +y_{4}\right)b \, \mathbf{\hat{y}} + z_{4}c \, \mathbf{\hat{z}} & \left(8d\right) & \mbox{C IV} \\ 
\mathbf{B}_{33} & = & x_{5} \, \mathbf{a}_{1} + y_{5} \, \mathbf{a}_{2} + z_{5} \, \mathbf{a}_{3} & = & x_{5}a \, \mathbf{\hat{x}} + y_{5}b \, \mathbf{\hat{y}} + z_{5}c \, \mathbf{\hat{z}} & \left(8d\right) & \mbox{C V} \\ 
\mathbf{B}_{34} & = & \left(\frac{1}{2} - x_{5}\right) \, \mathbf{a}_{1} + \left(\frac{1}{2} - y_{5}\right) \, \mathbf{a}_{2} + \left(\frac{1}{2} +z_{5}\right) \, \mathbf{a}_{3} & = & \left(\frac{1}{2} - x_{5}\right)a \, \mathbf{\hat{x}} + \left(\frac{1}{2} - y_{5}\right)b \, \mathbf{\hat{y}} + \left(\frac{1}{2} +z_{5}\right)c \, \mathbf{\hat{z}} & \left(8d\right) & \mbox{C V} \\ 
\mathbf{B}_{35} & = & -x_{5} \, \mathbf{a}_{1} + y_{5} \, \mathbf{a}_{2} + \left(\frac{1}{2} - z_{5}\right) \, \mathbf{a}_{3} & = & -x_{5}a \, \mathbf{\hat{x}} + y_{5}b \, \mathbf{\hat{y}} + \left(\frac{1}{2} - z_{5}\right)c \, \mathbf{\hat{z}} & \left(8d\right) & \mbox{C V} \\ 
\mathbf{B}_{36} & = & \left(\frac{1}{2} +x_{5}\right) \, \mathbf{a}_{1} + \left(\frac{1}{2} - y_{5}\right) \, \mathbf{a}_{2}-z_{5} \, \mathbf{a}_{3} & = & \left(\frac{1}{2} +x_{5}\right)a \, \mathbf{\hat{x}} + \left(\frac{1}{2} - y_{5}\right)b \, \mathbf{\hat{y}}-z_{5}c \, \mathbf{\hat{z}} & \left(8d\right) & \mbox{C V} \\ 
\mathbf{B}_{37} & = & -x_{5} \, \mathbf{a}_{1}-y_{5} \, \mathbf{a}_{2}-z_{5} \, \mathbf{a}_{3} & = & -x_{5}a \, \mathbf{\hat{x}}-y_{5}b \, \mathbf{\hat{y}}-z_{5}c \, \mathbf{\hat{z}} & \left(8d\right) & \mbox{C V} \\ 
\mathbf{B}_{38} & = & \left(\frac{1}{2} +x_{5}\right) \, \mathbf{a}_{1} + \left(\frac{1}{2} +y_{5}\right) \, \mathbf{a}_{2} + \left(\frac{1}{2} - z_{5}\right) \, \mathbf{a}_{3} & = & \left(\frac{1}{2} +x_{5}\right)a \, \mathbf{\hat{x}} + \left(\frac{1}{2} +y_{5}\right)b \, \mathbf{\hat{y}} + \left(\frac{1}{2} - z_{5}\right)c \, \mathbf{\hat{z}} & \left(8d\right) & \mbox{C V} \\ 
\mathbf{B}_{39} & = & x_{5} \, \mathbf{a}_{1}-y_{5} \, \mathbf{a}_{2} + \left(\frac{1}{2} +z_{5}\right) \, \mathbf{a}_{3} & = & x_{5}a \, \mathbf{\hat{x}}-y_{5}b \, \mathbf{\hat{y}} + \left(\frac{1}{2} +z_{5}\right)c \, \mathbf{\hat{z}} & \left(8d\right) & \mbox{C V} \\ 
\mathbf{B}_{40} & = & \left(\frac{1}{2} - x_{5}\right) \, \mathbf{a}_{1} + \left(\frac{1}{2} +y_{5}\right) \, \mathbf{a}_{2} + z_{5} \, \mathbf{a}_{3} & = & \left(\frac{1}{2} - x_{5}\right)a \, \mathbf{\hat{x}} + \left(\frac{1}{2} +y_{5}\right)b \, \mathbf{\hat{y}} + z_{5}c \, \mathbf{\hat{z}} & \left(8d\right) & \mbox{C V} \\ 
\mathbf{B}_{41} & = & x_{6} \, \mathbf{a}_{1} + y_{6} \, \mathbf{a}_{2} + z_{6} \, \mathbf{a}_{3} & = & x_{6}a \, \mathbf{\hat{x}} + y_{6}b \, \mathbf{\hat{y}} + z_{6}c \, \mathbf{\hat{z}} & \left(8d\right) & \mbox{C VI} \\ 
\mathbf{B}_{42} & = & \left(\frac{1}{2} - x_{6}\right) \, \mathbf{a}_{1} + \left(\frac{1}{2} - y_{6}\right) \, \mathbf{a}_{2} + \left(\frac{1}{2} +z_{6}\right) \, \mathbf{a}_{3} & = & \left(\frac{1}{2} - x_{6}\right)a \, \mathbf{\hat{x}} + \left(\frac{1}{2} - y_{6}\right)b \, \mathbf{\hat{y}} + \left(\frac{1}{2} +z_{6}\right)c \, \mathbf{\hat{z}} & \left(8d\right) & \mbox{C VI} \\ 
\mathbf{B}_{43} & = & -x_{6} \, \mathbf{a}_{1} + y_{6} \, \mathbf{a}_{2} + \left(\frac{1}{2} - z_{6}\right) \, \mathbf{a}_{3} & = & -x_{6}a \, \mathbf{\hat{x}} + y_{6}b \, \mathbf{\hat{y}} + \left(\frac{1}{2} - z_{6}\right)c \, \mathbf{\hat{z}} & \left(8d\right) & \mbox{C VI} \\ 
\mathbf{B}_{44} & = & \left(\frac{1}{2} +x_{6}\right) \, \mathbf{a}_{1} + \left(\frac{1}{2} - y_{6}\right) \, \mathbf{a}_{2}-z_{6} \, \mathbf{a}_{3} & = & \left(\frac{1}{2} +x_{6}\right)a \, \mathbf{\hat{x}} + \left(\frac{1}{2} - y_{6}\right)b \, \mathbf{\hat{y}}-z_{6}c \, \mathbf{\hat{z}} & \left(8d\right) & \mbox{C VI} \\ 
\mathbf{B}_{45} & = & -x_{6} \, \mathbf{a}_{1}-y_{6} \, \mathbf{a}_{2}-z_{6} \, \mathbf{a}_{3} & = & -x_{6}a \, \mathbf{\hat{x}}-y_{6}b \, \mathbf{\hat{y}}-z_{6}c \, \mathbf{\hat{z}} & \left(8d\right) & \mbox{C VI} \\ 
\mathbf{B}_{46} & = & \left(\frac{1}{2} +x_{6}\right) \, \mathbf{a}_{1} + \left(\frac{1}{2} +y_{6}\right) \, \mathbf{a}_{2} + \left(\frac{1}{2} - z_{6}\right) \, \mathbf{a}_{3} & = & \left(\frac{1}{2} +x_{6}\right)a \, \mathbf{\hat{x}} + \left(\frac{1}{2} +y_{6}\right)b \, \mathbf{\hat{y}} + \left(\frac{1}{2} - z_{6}\right)c \, \mathbf{\hat{z}} & \left(8d\right) & \mbox{C VI} \\ 
\mathbf{B}_{47} & = & x_{6} \, \mathbf{a}_{1}-y_{6} \, \mathbf{a}_{2} + \left(\frac{1}{2} +z_{6}\right) \, \mathbf{a}_{3} & = & x_{6}a \, \mathbf{\hat{x}}-y_{6}b \, \mathbf{\hat{y}} + \left(\frac{1}{2} +z_{6}\right)c \, \mathbf{\hat{z}} & \left(8d\right) & \mbox{C VI} \\ 
\mathbf{B}_{48} & = & \left(\frac{1}{2} - x_{6}\right) \, \mathbf{a}_{1} + \left(\frac{1}{2} +y_{6}\right) \, \mathbf{a}_{2} + z_{6} \, \mathbf{a}_{3} & = & \left(\frac{1}{2} - x_{6}\right)a \, \mathbf{\hat{x}} + \left(\frac{1}{2} +y_{6}\right)b \, \mathbf{\hat{y}} + z_{6}c \, \mathbf{\hat{z}} & \left(8d\right) & \mbox{C VI} \\ 
\mathbf{B}_{49} & = & x_{7} \, \mathbf{a}_{1} + y_{7} \, \mathbf{a}_{2} + z_{7} \, \mathbf{a}_{3} & = & x_{7}a \, \mathbf{\hat{x}} + y_{7}b \, \mathbf{\hat{y}} + z_{7}c \, \mathbf{\hat{z}} & \left(8d\right) & \mbox{C VII} \\ 
\mathbf{B}_{50} & = & \left(\frac{1}{2} - x_{7}\right) \, \mathbf{a}_{1} + \left(\frac{1}{2} - y_{7}\right) \, \mathbf{a}_{2} + \left(\frac{1}{2} +z_{7}\right) \, \mathbf{a}_{3} & = & \left(\frac{1}{2} - x_{7}\right)a \, \mathbf{\hat{x}} + \left(\frac{1}{2} - y_{7}\right)b \, \mathbf{\hat{y}} + \left(\frac{1}{2} +z_{7}\right)c \, \mathbf{\hat{z}} & \left(8d\right) & \mbox{C VII} \\ 
\mathbf{B}_{51} & = & -x_{7} \, \mathbf{a}_{1} + y_{7} \, \mathbf{a}_{2} + \left(\frac{1}{2} - z_{7}\right) \, \mathbf{a}_{3} & = & -x_{7}a \, \mathbf{\hat{x}} + y_{7}b \, \mathbf{\hat{y}} + \left(\frac{1}{2} - z_{7}\right)c \, \mathbf{\hat{z}} & \left(8d\right) & \mbox{C VII} \\ 
\mathbf{B}_{52} & = & \left(\frac{1}{2} +x_{7}\right) \, \mathbf{a}_{1} + \left(\frac{1}{2} - y_{7}\right) \, \mathbf{a}_{2}-z_{7} \, \mathbf{a}_{3} & = & \left(\frac{1}{2} +x_{7}\right)a \, \mathbf{\hat{x}} + \left(\frac{1}{2} - y_{7}\right)b \, \mathbf{\hat{y}}-z_{7}c \, \mathbf{\hat{z}} & \left(8d\right) & \mbox{C VII} \\ 
\mathbf{B}_{53} & = & -x_{7} \, \mathbf{a}_{1}-y_{7} \, \mathbf{a}_{2}-z_{7} \, \mathbf{a}_{3} & = & -x_{7}a \, \mathbf{\hat{x}}-y_{7}b \, \mathbf{\hat{y}}-z_{7}c \, \mathbf{\hat{z}} & \left(8d\right) & \mbox{C VII} \\ 
\mathbf{B}_{54} & = & \left(\frac{1}{2} +x_{7}\right) \, \mathbf{a}_{1} + \left(\frac{1}{2} +y_{7}\right) \, \mathbf{a}_{2} + \left(\frac{1}{2} - z_{7}\right) \, \mathbf{a}_{3} & = & \left(\frac{1}{2} +x_{7}\right)a \, \mathbf{\hat{x}} + \left(\frac{1}{2} +y_{7}\right)b \, \mathbf{\hat{y}} + \left(\frac{1}{2} - z_{7}\right)c \, \mathbf{\hat{z}} & \left(8d\right) & \mbox{C VII} \\ 
\mathbf{B}_{55} & = & x_{7} \, \mathbf{a}_{1}-y_{7} \, \mathbf{a}_{2} + \left(\frac{1}{2} +z_{7}\right) \, \mathbf{a}_{3} & = & x_{7}a \, \mathbf{\hat{x}}-y_{7}b \, \mathbf{\hat{y}} + \left(\frac{1}{2} +z_{7}\right)c \, \mathbf{\hat{z}} & \left(8d\right) & \mbox{C VII} \\ 
\mathbf{B}_{56} & = & \left(\frac{1}{2} - x_{7}\right) \, \mathbf{a}_{1} + \left(\frac{1}{2} +y_{7}\right) \, \mathbf{a}_{2} + z_{7} \, \mathbf{a}_{3} & = & \left(\frac{1}{2} - x_{7}\right)a \, \mathbf{\hat{x}} + \left(\frac{1}{2} +y_{7}\right)b \, \mathbf{\hat{y}} + z_{7}c \, \mathbf{\hat{z}} & \left(8d\right) & \mbox{C VII} \\ 
\mathbf{B}_{57} & = & x_{8} \, \mathbf{a}_{1} + y_{8} \, \mathbf{a}_{2} + z_{8} \, \mathbf{a}_{3} & = & x_{8}a \, \mathbf{\hat{x}} + y_{8}b \, \mathbf{\hat{y}} + z_{8}c \, \mathbf{\hat{z}} & \left(8d\right) & \mbox{H I} \\ 
\mathbf{B}_{58} & = & \left(\frac{1}{2} - x_{8}\right) \, \mathbf{a}_{1} + \left(\frac{1}{2} - y_{8}\right) \, \mathbf{a}_{2} + \left(\frac{1}{2} +z_{8}\right) \, \mathbf{a}_{3} & = & \left(\frac{1}{2} - x_{8}\right)a \, \mathbf{\hat{x}} + \left(\frac{1}{2} - y_{8}\right)b \, \mathbf{\hat{y}} + \left(\frac{1}{2} +z_{8}\right)c \, \mathbf{\hat{z}} & \left(8d\right) & \mbox{H I} \\ 
\mathbf{B}_{59} & = & -x_{8} \, \mathbf{a}_{1} + y_{8} \, \mathbf{a}_{2} + \left(\frac{1}{2} - z_{8}\right) \, \mathbf{a}_{3} & = & -x_{8}a \, \mathbf{\hat{x}} + y_{8}b \, \mathbf{\hat{y}} + \left(\frac{1}{2} - z_{8}\right)c \, \mathbf{\hat{z}} & \left(8d\right) & \mbox{H I} \\ 
\mathbf{B}_{60} & = & \left(\frac{1}{2} +x_{8}\right) \, \mathbf{a}_{1} + \left(\frac{1}{2} - y_{8}\right) \, \mathbf{a}_{2}-z_{8} \, \mathbf{a}_{3} & = & \left(\frac{1}{2} +x_{8}\right)a \, \mathbf{\hat{x}} + \left(\frac{1}{2} - y_{8}\right)b \, \mathbf{\hat{y}}-z_{8}c \, \mathbf{\hat{z}} & \left(8d\right) & \mbox{H I} \\ 
\mathbf{B}_{61} & = & -x_{8} \, \mathbf{a}_{1}-y_{8} \, \mathbf{a}_{2}-z_{8} \, \mathbf{a}_{3} & = & -x_{8}a \, \mathbf{\hat{x}}-y_{8}b \, \mathbf{\hat{y}}-z_{8}c \, \mathbf{\hat{z}} & \left(8d\right) & \mbox{H I} \\ 
\mathbf{B}_{62} & = & \left(\frac{1}{2} +x_{8}\right) \, \mathbf{a}_{1} + \left(\frac{1}{2} +y_{8}\right) \, \mathbf{a}_{2} + \left(\frac{1}{2} - z_{8}\right) \, \mathbf{a}_{3} & = & \left(\frac{1}{2} +x_{8}\right)a \, \mathbf{\hat{x}} + \left(\frac{1}{2} +y_{8}\right)b \, \mathbf{\hat{y}} + \left(\frac{1}{2} - z_{8}\right)c \, \mathbf{\hat{z}} & \left(8d\right) & \mbox{H I} \\ 
\mathbf{B}_{63} & = & x_{8} \, \mathbf{a}_{1}-y_{8} \, \mathbf{a}_{2} + \left(\frac{1}{2} +z_{8}\right) \, \mathbf{a}_{3} & = & x_{8}a \, \mathbf{\hat{x}}-y_{8}b \, \mathbf{\hat{y}} + \left(\frac{1}{2} +z_{8}\right)c \, \mathbf{\hat{z}} & \left(8d\right) & \mbox{H I} \\ 
\mathbf{B}_{64} & = & \left(\frac{1}{2} - x_{8}\right) \, \mathbf{a}_{1} + \left(\frac{1}{2} +y_{8}\right) \, \mathbf{a}_{2} + z_{8} \, \mathbf{a}_{3} & = & \left(\frac{1}{2} - x_{8}\right)a \, \mathbf{\hat{x}} + \left(\frac{1}{2} +y_{8}\right)b \, \mathbf{\hat{y}} + z_{8}c \, \mathbf{\hat{z}} & \left(8d\right) & \mbox{H I} \\ 
\mathbf{B}_{65} & = & x_{9} \, \mathbf{a}_{1} + y_{9} \, \mathbf{a}_{2} + z_{9} \, \mathbf{a}_{3} & = & x_{9}a \, \mathbf{\hat{x}} + y_{9}b \, \mathbf{\hat{y}} + z_{9}c \, \mathbf{\hat{z}} & \left(8d\right) & \mbox{H II} \\ 
\mathbf{B}_{66} & = & \left(\frac{1}{2} - x_{9}\right) \, \mathbf{a}_{1} + \left(\frac{1}{2} - y_{9}\right) \, \mathbf{a}_{2} + \left(\frac{1}{2} +z_{9}\right) \, \mathbf{a}_{3} & = & \left(\frac{1}{2} - x_{9}\right)a \, \mathbf{\hat{x}} + \left(\frac{1}{2} - y_{9}\right)b \, \mathbf{\hat{y}} + \left(\frac{1}{2} +z_{9}\right)c \, \mathbf{\hat{z}} & \left(8d\right) & \mbox{H II} \\ 
\mathbf{B}_{67} & = & -x_{9} \, \mathbf{a}_{1} + y_{9} \, \mathbf{a}_{2} + \left(\frac{1}{2} - z_{9}\right) \, \mathbf{a}_{3} & = & -x_{9}a \, \mathbf{\hat{x}} + y_{9}b \, \mathbf{\hat{y}} + \left(\frac{1}{2} - z_{9}\right)c \, \mathbf{\hat{z}} & \left(8d\right) & \mbox{H II} \\ 
\mathbf{B}_{68} & = & \left(\frac{1}{2} +x_{9}\right) \, \mathbf{a}_{1} + \left(\frac{1}{2} - y_{9}\right) \, \mathbf{a}_{2}-z_{9} \, \mathbf{a}_{3} & = & \left(\frac{1}{2} +x_{9}\right)a \, \mathbf{\hat{x}} + \left(\frac{1}{2} - y_{9}\right)b \, \mathbf{\hat{y}}-z_{9}c \, \mathbf{\hat{z}} & \left(8d\right) & \mbox{H II} \\ 
\mathbf{B}_{69} & = & -x_{9} \, \mathbf{a}_{1}-y_{9} \, \mathbf{a}_{2}-z_{9} \, \mathbf{a}_{3} & = & -x_{9}a \, \mathbf{\hat{x}}-y_{9}b \, \mathbf{\hat{y}}-z_{9}c \, \mathbf{\hat{z}} & \left(8d\right) & \mbox{H II} \\ 
\mathbf{B}_{70} & = & \left(\frac{1}{2} +x_{9}\right) \, \mathbf{a}_{1} + \left(\frac{1}{2} +y_{9}\right) \, \mathbf{a}_{2} + \left(\frac{1}{2} - z_{9}\right) \, \mathbf{a}_{3} & = & \left(\frac{1}{2} +x_{9}\right)a \, \mathbf{\hat{x}} + \left(\frac{1}{2} +y_{9}\right)b \, \mathbf{\hat{y}} + \left(\frac{1}{2} - z_{9}\right)c \, \mathbf{\hat{z}} & \left(8d\right) & \mbox{H II} \\ 
\mathbf{B}_{71} & = & x_{9} \, \mathbf{a}_{1}-y_{9} \, \mathbf{a}_{2} + \left(\frac{1}{2} +z_{9}\right) \, \mathbf{a}_{3} & = & x_{9}a \, \mathbf{\hat{x}}-y_{9}b \, \mathbf{\hat{y}} + \left(\frac{1}{2} +z_{9}\right)c \, \mathbf{\hat{z}} & \left(8d\right) & \mbox{H II} \\ 
\mathbf{B}_{72} & = & \left(\frac{1}{2} - x_{9}\right) \, \mathbf{a}_{1} + \left(\frac{1}{2} +y_{9}\right) \, \mathbf{a}_{2} + z_{9} \, \mathbf{a}_{3} & = & \left(\frac{1}{2} - x_{9}\right)a \, \mathbf{\hat{x}} + \left(\frac{1}{2} +y_{9}\right)b \, \mathbf{\hat{y}} + z_{9}c \, \mathbf{\hat{z}} & \left(8d\right) & \mbox{H II} \\ 
\mathbf{B}_{73} & = & x_{10} \, \mathbf{a}_{1} + y_{10} \, \mathbf{a}_{2} + z_{10} \, \mathbf{a}_{3} & = & x_{10}a \, \mathbf{\hat{x}} + y_{10}b \, \mathbf{\hat{y}} + z_{10}c \, \mathbf{\hat{z}} & \left(8d\right) & \mbox{H III} \\ 
\mathbf{B}_{74} & = & \left(\frac{1}{2} - x_{10}\right) \, \mathbf{a}_{1} + \left(\frac{1}{2} - y_{10}\right) \, \mathbf{a}_{2} + \left(\frac{1}{2} +z_{10}\right) \, \mathbf{a}_{3} & = & \left(\frac{1}{2} - x_{10}\right)a \, \mathbf{\hat{x}} + \left(\frac{1}{2} - y_{10}\right)b \, \mathbf{\hat{y}} + \left(\frac{1}{2} +z_{10}\right)c \, \mathbf{\hat{z}} & \left(8d\right) & \mbox{H III} \\ 
\mathbf{B}_{75} & = & -x_{10} \, \mathbf{a}_{1} + y_{10} \, \mathbf{a}_{2} + \left(\frac{1}{2} - z_{10}\right) \, \mathbf{a}_{3} & = & -x_{10}a \, \mathbf{\hat{x}} + y_{10}b \, \mathbf{\hat{y}} + \left(\frac{1}{2} - z_{10}\right)c \, \mathbf{\hat{z}} & \left(8d\right) & \mbox{H III} \\ 
\mathbf{B}_{76} & = & \left(\frac{1}{2} +x_{10}\right) \, \mathbf{a}_{1} + \left(\frac{1}{2} - y_{10}\right) \, \mathbf{a}_{2}-z_{10} \, \mathbf{a}_{3} & = & \left(\frac{1}{2} +x_{10}\right)a \, \mathbf{\hat{x}} + \left(\frac{1}{2} - y_{10}\right)b \, \mathbf{\hat{y}}-z_{10}c \, \mathbf{\hat{z}} & \left(8d\right) & \mbox{H III} \\ 
\mathbf{B}_{77} & = & -x_{10} \, \mathbf{a}_{1}-y_{10} \, \mathbf{a}_{2}-z_{10} \, \mathbf{a}_{3} & = & -x_{10}a \, \mathbf{\hat{x}}-y_{10}b \, \mathbf{\hat{y}}-z_{10}c \, \mathbf{\hat{z}} & \left(8d\right) & \mbox{H III} \\ 
\mathbf{B}_{78} & = & \left(\frac{1}{2} +x_{10}\right) \, \mathbf{a}_{1} + \left(\frac{1}{2} +y_{10}\right) \, \mathbf{a}_{2} + \left(\frac{1}{2} - z_{10}\right) \, \mathbf{a}_{3} & = & \left(\frac{1}{2} +x_{10}\right)a \, \mathbf{\hat{x}} + \left(\frac{1}{2} +y_{10}\right)b \, \mathbf{\hat{y}} + \left(\frac{1}{2} - z_{10}\right)c \, \mathbf{\hat{z}} & \left(8d\right) & \mbox{H III} \\ 
\mathbf{B}_{79} & = & x_{10} \, \mathbf{a}_{1}-y_{10} \, \mathbf{a}_{2} + \left(\frac{1}{2} +z_{10}\right) \, \mathbf{a}_{3} & = & x_{10}a \, \mathbf{\hat{x}}-y_{10}b \, \mathbf{\hat{y}} + \left(\frac{1}{2} +z_{10}\right)c \, \mathbf{\hat{z}} & \left(8d\right) & \mbox{H III} \\ 
\mathbf{B}_{80} & = & \left(\frac{1}{2} - x_{10}\right) \, \mathbf{a}_{1} + \left(\frac{1}{2} +y_{10}\right) \, \mathbf{a}_{2} + z_{10} \, \mathbf{a}_{3} & = & \left(\frac{1}{2} - x_{10}\right)a \, \mathbf{\hat{x}} + \left(\frac{1}{2} +y_{10}\right)b \, \mathbf{\hat{y}} + z_{10}c \, \mathbf{\hat{z}} & \left(8d\right) & \mbox{H III} \\ 
\mathbf{B}_{81} & = & x_{11} \, \mathbf{a}_{1} + y_{11} \, \mathbf{a}_{2} + z_{11} \, \mathbf{a}_{3} & = & x_{11}a \, \mathbf{\hat{x}} + y_{11}b \, \mathbf{\hat{y}} + z_{11}c \, \mathbf{\hat{z}} & \left(8d\right) & \mbox{H IV} \\ 
\mathbf{B}_{82} & = & \left(\frac{1}{2} - x_{11}\right) \, \mathbf{a}_{1} + \left(\frac{1}{2} - y_{11}\right) \, \mathbf{a}_{2} + \left(\frac{1}{2} +z_{11}\right) \, \mathbf{a}_{3} & = & \left(\frac{1}{2} - x_{11}\right)a \, \mathbf{\hat{x}} + \left(\frac{1}{2} - y_{11}\right)b \, \mathbf{\hat{y}} + \left(\frac{1}{2} +z_{11}\right)c \, \mathbf{\hat{z}} & \left(8d\right) & \mbox{H IV} \\ 
\mathbf{B}_{83} & = & -x_{11} \, \mathbf{a}_{1} + y_{11} \, \mathbf{a}_{2} + \left(\frac{1}{2} - z_{11}\right) \, \mathbf{a}_{3} & = & -x_{11}a \, \mathbf{\hat{x}} + y_{11}b \, \mathbf{\hat{y}} + \left(\frac{1}{2} - z_{11}\right)c \, \mathbf{\hat{z}} & \left(8d\right) & \mbox{H IV} \\ 
\mathbf{B}_{84} & = & \left(\frac{1}{2} +x_{11}\right) \, \mathbf{a}_{1} + \left(\frac{1}{2} - y_{11}\right) \, \mathbf{a}_{2}-z_{11} \, \mathbf{a}_{3} & = & \left(\frac{1}{2} +x_{11}\right)a \, \mathbf{\hat{x}} + \left(\frac{1}{2} - y_{11}\right)b \, \mathbf{\hat{y}}-z_{11}c \, \mathbf{\hat{z}} & \left(8d\right) & \mbox{H IV} \\ 
\mathbf{B}_{85} & = & -x_{11} \, \mathbf{a}_{1}-y_{11} \, \mathbf{a}_{2}-z_{11} \, \mathbf{a}_{3} & = & -x_{11}a \, \mathbf{\hat{x}}-y_{11}b \, \mathbf{\hat{y}}-z_{11}c \, \mathbf{\hat{z}} & \left(8d\right) & \mbox{H IV} \\ 
\mathbf{B}_{86} & = & \left(\frac{1}{2} +x_{11}\right) \, \mathbf{a}_{1} + \left(\frac{1}{2} +y_{11}\right) \, \mathbf{a}_{2} + \left(\frac{1}{2} - z_{11}\right) \, \mathbf{a}_{3} & = & \left(\frac{1}{2} +x_{11}\right)a \, \mathbf{\hat{x}} + \left(\frac{1}{2} +y_{11}\right)b \, \mathbf{\hat{y}} + \left(\frac{1}{2} - z_{11}\right)c \, \mathbf{\hat{z}} & \left(8d\right) & \mbox{H IV} \\ 
\mathbf{B}_{87} & = & x_{11} \, \mathbf{a}_{1}-y_{11} \, \mathbf{a}_{2} + \left(\frac{1}{2} +z_{11}\right) \, \mathbf{a}_{3} & = & x_{11}a \, \mathbf{\hat{x}}-y_{11}b \, \mathbf{\hat{y}} + \left(\frac{1}{2} +z_{11}\right)c \, \mathbf{\hat{z}} & \left(8d\right) & \mbox{H IV} \\ 
\mathbf{B}_{88} & = & \left(\frac{1}{2} - x_{11}\right) \, \mathbf{a}_{1} + \left(\frac{1}{2} +y_{11}\right) \, \mathbf{a}_{2} + z_{11} \, \mathbf{a}_{3} & = & \left(\frac{1}{2} - x_{11}\right)a \, \mathbf{\hat{x}} + \left(\frac{1}{2} +y_{11}\right)b \, \mathbf{\hat{y}} + z_{11}c \, \mathbf{\hat{z}} & \left(8d\right) & \mbox{H IV} \\ 
\mathbf{B}_{89} & = & x_{12} \, \mathbf{a}_{1} + y_{12} \, \mathbf{a}_{2} + z_{12} \, \mathbf{a}_{3} & = & x_{12}a \, \mathbf{\hat{x}} + y_{12}b \, \mathbf{\hat{y}} + z_{12}c \, \mathbf{\hat{z}} & \left(8d\right) & \mbox{H V} \\ 
\mathbf{B}_{90} & = & \left(\frac{1}{2} - x_{12}\right) \, \mathbf{a}_{1} + \left(\frac{1}{2} - y_{12}\right) \, \mathbf{a}_{2} + \left(\frac{1}{2} +z_{12}\right) \, \mathbf{a}_{3} & = & \left(\frac{1}{2} - x_{12}\right)a \, \mathbf{\hat{x}} + \left(\frac{1}{2} - y_{12}\right)b \, \mathbf{\hat{y}} + \left(\frac{1}{2} +z_{12}\right)c \, \mathbf{\hat{z}} & \left(8d\right) & \mbox{H V} \\ 
\mathbf{B}_{91} & = & -x_{12} \, \mathbf{a}_{1} + y_{12} \, \mathbf{a}_{2} + \left(\frac{1}{2} - z_{12}\right) \, \mathbf{a}_{3} & = & -x_{12}a \, \mathbf{\hat{x}} + y_{12}b \, \mathbf{\hat{y}} + \left(\frac{1}{2} - z_{12}\right)c \, \mathbf{\hat{z}} & \left(8d\right) & \mbox{H V} \\ 
\mathbf{B}_{92} & = & \left(\frac{1}{2} +x_{12}\right) \, \mathbf{a}_{1} + \left(\frac{1}{2} - y_{12}\right) \, \mathbf{a}_{2}-z_{12} \, \mathbf{a}_{3} & = & \left(\frac{1}{2} +x_{12}\right)a \, \mathbf{\hat{x}} + \left(\frac{1}{2} - y_{12}\right)b \, \mathbf{\hat{y}}-z_{12}c \, \mathbf{\hat{z}} & \left(8d\right) & \mbox{H V} \\ 
\mathbf{B}_{93} & = & -x_{12} \, \mathbf{a}_{1}-y_{12} \, \mathbf{a}_{2}-z_{12} \, \mathbf{a}_{3} & = & -x_{12}a \, \mathbf{\hat{x}}-y_{12}b \, \mathbf{\hat{y}}-z_{12}c \, \mathbf{\hat{z}} & \left(8d\right) & \mbox{H V} \\ 
\mathbf{B}_{94} & = & \left(\frac{1}{2} +x_{12}\right) \, \mathbf{a}_{1} + \left(\frac{1}{2} +y_{12}\right) \, \mathbf{a}_{2} + \left(\frac{1}{2} - z_{12}\right) \, \mathbf{a}_{3} & = & \left(\frac{1}{2} +x_{12}\right)a \, \mathbf{\hat{x}} + \left(\frac{1}{2} +y_{12}\right)b \, \mathbf{\hat{y}} + \left(\frac{1}{2} - z_{12}\right)c \, \mathbf{\hat{z}} & \left(8d\right) & \mbox{H V} \\ 
\mathbf{B}_{95} & = & x_{12} \, \mathbf{a}_{1}-y_{12} \, \mathbf{a}_{2} + \left(\frac{1}{2} +z_{12}\right) \, \mathbf{a}_{3} & = & x_{12}a \, \mathbf{\hat{x}}-y_{12}b \, \mathbf{\hat{y}} + \left(\frac{1}{2} +z_{12}\right)c \, \mathbf{\hat{z}} & \left(8d\right) & \mbox{H V} \\ 
\mathbf{B}_{96} & = & \left(\frac{1}{2} - x_{12}\right) \, \mathbf{a}_{1} + \left(\frac{1}{2} +y_{12}\right) \, \mathbf{a}_{2} + z_{12} \, \mathbf{a}_{3} & = & \left(\frac{1}{2} - x_{12}\right)a \, \mathbf{\hat{x}} + \left(\frac{1}{2} +y_{12}\right)b \, \mathbf{\hat{y}} + z_{12}c \, \mathbf{\hat{z}} & \left(8d\right) & \mbox{H V} \\ 
\mathbf{B}_{97} & = & x_{13} \, \mathbf{a}_{1} + y_{13} \, \mathbf{a}_{2} + z_{13} \, \mathbf{a}_{3} & = & x_{13}a \, \mathbf{\hat{x}} + y_{13}b \, \mathbf{\hat{y}} + z_{13}c \, \mathbf{\hat{z}} & \left(8d\right) & \mbox{H VI} \\ 
\mathbf{B}_{98} & = & \left(\frac{1}{2} - x_{13}\right) \, \mathbf{a}_{1} + \left(\frac{1}{2} - y_{13}\right) \, \mathbf{a}_{2} + \left(\frac{1}{2} +z_{13}\right) \, \mathbf{a}_{3} & = & \left(\frac{1}{2} - x_{13}\right)a \, \mathbf{\hat{x}} + \left(\frac{1}{2} - y_{13}\right)b \, \mathbf{\hat{y}} + \left(\frac{1}{2} +z_{13}\right)c \, \mathbf{\hat{z}} & \left(8d\right) & \mbox{H VI} \\ 
\mathbf{B}_{99} & = & -x_{13} \, \mathbf{a}_{1} + y_{13} \, \mathbf{a}_{2} + \left(\frac{1}{2} - z_{13}\right) \, \mathbf{a}_{3} & = & -x_{13}a \, \mathbf{\hat{x}} + y_{13}b \, \mathbf{\hat{y}} + \left(\frac{1}{2} - z_{13}\right)c \, \mathbf{\hat{z}} & \left(8d\right) & \mbox{H VI} \\ 
\mathbf{B}_{100} & = & \left(\frac{1}{2} +x_{13}\right) \, \mathbf{a}_{1} + \left(\frac{1}{2} - y_{13}\right) \, \mathbf{a}_{2}-z_{13} \, \mathbf{a}_{3} & = & \left(\frac{1}{2} +x_{13}\right)a \, \mathbf{\hat{x}} + \left(\frac{1}{2} - y_{13}\right)b \, \mathbf{\hat{y}}-z_{13}c \, \mathbf{\hat{z}} & \left(8d\right) & \mbox{H VI} \\ 
\mathbf{B}_{101} & = & -x_{13} \, \mathbf{a}_{1}-y_{13} \, \mathbf{a}_{2}-z_{13} \, \mathbf{a}_{3} & = & -x_{13}a \, \mathbf{\hat{x}}-y_{13}b \, \mathbf{\hat{y}}-z_{13}c \, \mathbf{\hat{z}} & \left(8d\right) & \mbox{H VI} \\ 
\mathbf{B}_{102} & = & \left(\frac{1}{2} +x_{13}\right) \, \mathbf{a}_{1} + \left(\frac{1}{2} +y_{13}\right) \, \mathbf{a}_{2} + \left(\frac{1}{2} - z_{13}\right) \, \mathbf{a}_{3} & = & \left(\frac{1}{2} +x_{13}\right)a \, \mathbf{\hat{x}} + \left(\frac{1}{2} +y_{13}\right)b \, \mathbf{\hat{y}} + \left(\frac{1}{2} - z_{13}\right)c \, \mathbf{\hat{z}} & \left(8d\right) & \mbox{H VI} \\ 
\mathbf{B}_{103} & = & x_{13} \, \mathbf{a}_{1}-y_{13} \, \mathbf{a}_{2} + \left(\frac{1}{2} +z_{13}\right) \, \mathbf{a}_{3} & = & x_{13}a \, \mathbf{\hat{x}}-y_{13}b \, \mathbf{\hat{y}} + \left(\frac{1}{2} +z_{13}\right)c \, \mathbf{\hat{z}} & \left(8d\right) & \mbox{H VI} \\ 
\mathbf{B}_{104} & = & \left(\frac{1}{2} - x_{13}\right) \, \mathbf{a}_{1} + \left(\frac{1}{2} +y_{13}\right) \, \mathbf{a}_{2} + z_{13} \, \mathbf{a}_{3} & = & \left(\frac{1}{2} - x_{13}\right)a \, \mathbf{\hat{x}} + \left(\frac{1}{2} +y_{13}\right)b \, \mathbf{\hat{y}} + z_{13}c \, \mathbf{\hat{z}} & \left(8d\right) & \mbox{H VI} \\ 
\mathbf{B}_{105} & = & x_{14} \, \mathbf{a}_{1} + y_{14} \, \mathbf{a}_{2} + z_{14} \, \mathbf{a}_{3} & = & x_{14}a \, \mathbf{\hat{x}} + y_{14}b \, \mathbf{\hat{y}} + z_{14}c \, \mathbf{\hat{z}} & \left(8d\right) & \mbox{H VII} \\ 
\mathbf{B}_{106} & = & \left(\frac{1}{2} - x_{14}\right) \, \mathbf{a}_{1} + \left(\frac{1}{2} - y_{14}\right) \, \mathbf{a}_{2} + \left(\frac{1}{2} +z_{14}\right) \, \mathbf{a}_{3} & = & \left(\frac{1}{2} - x_{14}\right)a \, \mathbf{\hat{x}} + \left(\frac{1}{2} - y_{14}\right)b \, \mathbf{\hat{y}} + \left(\frac{1}{2} +z_{14}\right)c \, \mathbf{\hat{z}} & \left(8d\right) & \mbox{H VII} \\ 
\mathbf{B}_{107} & = & -x_{14} \, \mathbf{a}_{1} + y_{14} \, \mathbf{a}_{2} + \left(\frac{1}{2} - z_{14}\right) \, \mathbf{a}_{3} & = & -x_{14}a \, \mathbf{\hat{x}} + y_{14}b \, \mathbf{\hat{y}} + \left(\frac{1}{2} - z_{14}\right)c \, \mathbf{\hat{z}} & \left(8d\right) & \mbox{H VII} \\ 
\mathbf{B}_{108} & = & \left(\frac{1}{2} +x_{14}\right) \, \mathbf{a}_{1} + \left(\frac{1}{2} - y_{14}\right) \, \mathbf{a}_{2}-z_{14} \, \mathbf{a}_{3} & = & \left(\frac{1}{2} +x_{14}\right)a \, \mathbf{\hat{x}} + \left(\frac{1}{2} - y_{14}\right)b \, \mathbf{\hat{y}}-z_{14}c \, \mathbf{\hat{z}} & \left(8d\right) & \mbox{H VII} \\ 
\mathbf{B}_{109} & = & -x_{14} \, \mathbf{a}_{1}-y_{14} \, \mathbf{a}_{2}-z_{14} \, \mathbf{a}_{3} & = & -x_{14}a \, \mathbf{\hat{x}}-y_{14}b \, \mathbf{\hat{y}}-z_{14}c \, \mathbf{\hat{z}} & \left(8d\right) & \mbox{H VII} \\ 
\mathbf{B}_{110} & = & \left(\frac{1}{2} +x_{14}\right) \, \mathbf{a}_{1} + \left(\frac{1}{2} +y_{14}\right) \, \mathbf{a}_{2} + \left(\frac{1}{2} - z_{14}\right) \, \mathbf{a}_{3} & = & \left(\frac{1}{2} +x_{14}\right)a \, \mathbf{\hat{x}} + \left(\frac{1}{2} +y_{14}\right)b \, \mathbf{\hat{y}} + \left(\frac{1}{2} - z_{14}\right)c \, \mathbf{\hat{z}} & \left(8d\right) & \mbox{H VII} \\ 
\mathbf{B}_{111} & = & x_{14} \, \mathbf{a}_{1}-y_{14} \, \mathbf{a}_{2} + \left(\frac{1}{2} +z_{14}\right) \, \mathbf{a}_{3} & = & x_{14}a \, \mathbf{\hat{x}}-y_{14}b \, \mathbf{\hat{y}} + \left(\frac{1}{2} +z_{14}\right)c \, \mathbf{\hat{z}} & \left(8d\right) & \mbox{H VII} \\ 
\mathbf{B}_{112} & = & \left(\frac{1}{2} - x_{14}\right) \, \mathbf{a}_{1} + \left(\frac{1}{2} +y_{14}\right) \, \mathbf{a}_{2} + z_{14} \, \mathbf{a}_{3} & = & \left(\frac{1}{2} - x_{14}\right)a \, \mathbf{\hat{x}} + \left(\frac{1}{2} +y_{14}\right)b \, \mathbf{\hat{y}} + z_{14}c \, \mathbf{\hat{z}} & \left(8d\right) & \mbox{H VII} \\ 
\mathbf{B}_{113} & = & x_{15} \, \mathbf{a}_{1} + y_{15} \, \mathbf{a}_{2} + z_{15} \, \mathbf{a}_{3} & = & x_{15}a \, \mathbf{\hat{x}} + y_{15}b \, \mathbf{\hat{y}} + z_{15}c \, \mathbf{\hat{z}} & \left(8d\right) & \mbox{H VIII} \\ 
\mathbf{B}_{114} & = & \left(\frac{1}{2} - x_{15}\right) \, \mathbf{a}_{1} + \left(\frac{1}{2} - y_{15}\right) \, \mathbf{a}_{2} + \left(\frac{1}{2} +z_{15}\right) \, \mathbf{a}_{3} & = & \left(\frac{1}{2} - x_{15}\right)a \, \mathbf{\hat{x}} + \left(\frac{1}{2} - y_{15}\right)b \, \mathbf{\hat{y}} + \left(\frac{1}{2} +z_{15}\right)c \, \mathbf{\hat{z}} & \left(8d\right) & \mbox{H VIII} \\ 
\mathbf{B}_{115} & = & -x_{15} \, \mathbf{a}_{1} + y_{15} \, \mathbf{a}_{2} + \left(\frac{1}{2} - z_{15}\right) \, \mathbf{a}_{3} & = & -x_{15}a \, \mathbf{\hat{x}} + y_{15}b \, \mathbf{\hat{y}} + \left(\frac{1}{2} - z_{15}\right)c \, \mathbf{\hat{z}} & \left(8d\right) & \mbox{H VIII} \\ 
\mathbf{B}_{116} & = & \left(\frac{1}{2} +x_{15}\right) \, \mathbf{a}_{1} + \left(\frac{1}{2} - y_{15}\right) \, \mathbf{a}_{2}-z_{15} \, \mathbf{a}_{3} & = & \left(\frac{1}{2} +x_{15}\right)a \, \mathbf{\hat{x}} + \left(\frac{1}{2} - y_{15}\right)b \, \mathbf{\hat{y}}-z_{15}c \, \mathbf{\hat{z}} & \left(8d\right) & \mbox{H VIII} \\ 
\mathbf{B}_{117} & = & -x_{15} \, \mathbf{a}_{1}-y_{15} \, \mathbf{a}_{2}-z_{15} \, \mathbf{a}_{3} & = & -x_{15}a \, \mathbf{\hat{x}}-y_{15}b \, \mathbf{\hat{y}}-z_{15}c \, \mathbf{\hat{z}} & \left(8d\right) & \mbox{H VIII} \\ 
\mathbf{B}_{118} & = & \left(\frac{1}{2} +x_{15}\right) \, \mathbf{a}_{1} + \left(\frac{1}{2} +y_{15}\right) \, \mathbf{a}_{2} + \left(\frac{1}{2} - z_{15}\right) \, \mathbf{a}_{3} & = & \left(\frac{1}{2} +x_{15}\right)a \, \mathbf{\hat{x}} + \left(\frac{1}{2} +y_{15}\right)b \, \mathbf{\hat{y}} + \left(\frac{1}{2} - z_{15}\right)c \, \mathbf{\hat{z}} & \left(8d\right) & \mbox{H VIII} \\ 
\mathbf{B}_{119} & = & x_{15} \, \mathbf{a}_{1}-y_{15} \, \mathbf{a}_{2} + \left(\frac{1}{2} +z_{15}\right) \, \mathbf{a}_{3} & = & x_{15}a \, \mathbf{\hat{x}}-y_{15}b \, \mathbf{\hat{y}} + \left(\frac{1}{2} +z_{15}\right)c \, \mathbf{\hat{z}} & \left(8d\right) & \mbox{H VIII} \\ 
\mathbf{B}_{120} & = & \left(\frac{1}{2} - x_{15}\right) \, \mathbf{a}_{1} + \left(\frac{1}{2} +y_{15}\right) \, \mathbf{a}_{2} + z_{15} \, \mathbf{a}_{3} & = & \left(\frac{1}{2} - x_{15}\right)a \, \mathbf{\hat{x}} + \left(\frac{1}{2} +y_{15}\right)b \, \mathbf{\hat{y}} + z_{15}c \, \mathbf{\hat{z}} & \left(8d\right) & \mbox{H VIII} \\ 
\end{longtabu}
\renewcommand{\arraystretch}{1.0}
\noindent \hrulefill
\\
\textbf{References:}
\vspace*{-0.25cm}
\begin{flushleft}
  - \bibentry{Andre_1982}. \\
\end{flushleft}
\noindent \hrulefill
\\
\textbf{Geometry files:}
\\
\noindent  - CIF: pp. {\hyperref[A7B8_oP120_60_7d_8d_cif]{\pageref{A7B8_oP120_60_7d_8d_cif}}} \\
\noindent  - POSCAR: pp. {\hyperref[A7B8_oP120_60_7d_8d_poscar]{\pageref{A7B8_oP120_60_7d_8d_poscar}}} \\
\onecolumn
{\phantomsection\label{AB_oP48_61_3c_3c}}
\subsection*{\huge \textbf{{\normalfont Benzene Structure: AB\_oP48\_61\_3c\_3c}}}
\noindent \hrulefill
\vspace*{0.25cm}
\begin{figure}[htp]
  \centering
  \vspace{-1em}
  {\includegraphics[width=1\textwidth]{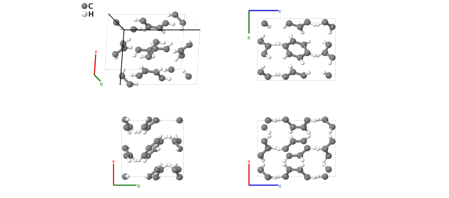}}
\end{figure}
\vspace*{-0.5cm}
\renewcommand{\arraystretch}{1.5}
\begin{equation*}
  \begin{array}{>{$\hspace{-0.15cm}}l<{$}>{$}p{0.5cm}<{$}>{$}p{18.5cm}<{$}}
    \mbox{\large \textbf{Prototype}} &\colon & \ce{Benzene} \\
    \mbox{\large \textbf{\AFLOW\ prototype label}} &\colon & \mbox{AB\_oP48\_61\_3c\_3c} \\
    \mbox{\large \textbf{\textit{Strukturbericht} designation}} &\colon & \mbox{None} \\
    \mbox{\large \textbf{Pearson symbol}} &\colon & \mbox{oP48} \\
    \mbox{\large \textbf{Space group number}} &\colon & 61 \\
    \mbox{\large \textbf{Space group symbol}} &\colon & Pbca \\
    \mbox{\large \textbf{\AFLOW\ prototype command}} &\colon &  \texttt{aflow} \,  \, \texttt{-{}-proto=AB\_oP48\_61\_3c\_3c } \, \newline \texttt{-{}-params=}{a,b/a,c/a,x_{1},y_{1},z_{1},x_{2},y_{2},z_{2},x_{3},y_{3},z_{3},x_{4},y_{4},z_{4},x_{5},y_{5},z_{5},x_{6},y_{6},} \newline {z_{6} }
  \end{array}
\end{equation*}
\renewcommand{\arraystretch}{1.0}

\vspace*{-0.25cm}
\noindent \hrulefill
\begin{itemize}
  \item{Benzene is a liquid at temperatures above 6$^\circ$C (279~K).
This data was constructed from experiments at 150~K.}
  \item{The hydrogen atomic positions were approximated to agree
with the chemistry of the benzene molecule.}
\end{itemize}

\noindent \parbox{1 \linewidth}{
\noindent \hrulefill
\\
\textbf{Simple Orthorhombic primitive vectors:} \\
\vspace*{-0.25cm}
\begin{tabular}{cc}
  \begin{tabular}{c}
    \parbox{0.6 \linewidth}{
      \renewcommand{\arraystretch}{1.5}
      \begin{equation*}
        \centering
        \begin{array}{ccc}
              \mathbf{a}_1 & = & a \, \mathbf{\hat{x}} \\
    \mathbf{a}_2 & = & b \, \mathbf{\hat{y}} \\
    \mathbf{a}_3 & = & c \, \mathbf{\hat{z}} \\

        \end{array}
      \end{equation*}
    }
    \renewcommand{\arraystretch}{1.0}
  \end{tabular}
  \begin{tabular}{c}
    \includegraphics[width=0.3\linewidth]{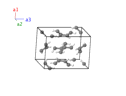} \\
  \end{tabular}
\end{tabular}

}
\vspace*{-0.25cm}

\noindent \hrulefill
\\
\textbf{Basis vectors:}
\vspace*{-0.25cm}
\renewcommand{\arraystretch}{1.5}
\begin{longtabu} to \textwidth{>{\centering $}X[-1,c,c]<{$}>{\centering $}X[-1,c,c]<{$}>{\centering $}X[-1,c,c]<{$}>{\centering $}X[-1,c,c]<{$}>{\centering $}X[-1,c,c]<{$}>{\centering $}X[-1,c,c]<{$}>{\centering $}X[-1,c,c]<{$}}
  & & \mbox{Lattice Coordinates} & & \mbox{Cartesian Coordinates} &\mbox{Wyckoff Position} & \mbox{Atom Type} \\  
  \mathbf{B}_{1} & = & x_{1} \, \mathbf{a}_{1} + y_{1} \, \mathbf{a}_{2} + z_{1} \, \mathbf{a}_{3} & = & x_{1}a \, \mathbf{\hat{x}} + y_{1}b \, \mathbf{\hat{y}} + z_{1}c \, \mathbf{\hat{z}} & \left(8c\right) & \mbox{C I} \\ 
\mathbf{B}_{2} & = & \left(\frac{1}{2} - x_{1}\right) \, \mathbf{a}_{1}-y_{1} \, \mathbf{a}_{2} + \left(\frac{1}{2} +z_{1}\right) \, \mathbf{a}_{3} & = & \left(\frac{1}{2} - x_{1}\right)a \, \mathbf{\hat{x}}-y_{1}b \, \mathbf{\hat{y}} + \left(\frac{1}{2} +z_{1}\right)c \, \mathbf{\hat{z}} & \left(8c\right) & \mbox{C I} \\ 
\mathbf{B}_{3} & = & -x_{1} \, \mathbf{a}_{1} + \left(\frac{1}{2} +y_{1}\right) \, \mathbf{a}_{2} + \left(\frac{1}{2} - z_{1}\right) \, \mathbf{a}_{3} & = & -x_{1}a \, \mathbf{\hat{x}} + \left(\frac{1}{2} +y_{1}\right)b \, \mathbf{\hat{y}} + \left(\frac{1}{2} - z_{1}\right)c \, \mathbf{\hat{z}} & \left(8c\right) & \mbox{C I} \\ 
\mathbf{B}_{4} & = & \left(\frac{1}{2} +x_{1}\right) \, \mathbf{a}_{1} + \left(\frac{1}{2} - y_{1}\right) \, \mathbf{a}_{2}-z_{1} \, \mathbf{a}_{3} & = & \left(\frac{1}{2} +x_{1}\right)a \, \mathbf{\hat{x}} + \left(\frac{1}{2} - y_{1}\right)b \, \mathbf{\hat{y}}-z_{1}c \, \mathbf{\hat{z}} & \left(8c\right) & \mbox{C I} \\ 
\mathbf{B}_{5} & = & -x_{1} \, \mathbf{a}_{1}-y_{1} \, \mathbf{a}_{2}-z_{1} \, \mathbf{a}_{3} & = & -x_{1}a \, \mathbf{\hat{x}}-y_{1}b \, \mathbf{\hat{y}}-z_{1}c \, \mathbf{\hat{z}} & \left(8c\right) & \mbox{C I} \\ 
\mathbf{B}_{6} & = & \left(\frac{1}{2} +x_{1}\right) \, \mathbf{a}_{1} + y_{1} \, \mathbf{a}_{2} + \left(\frac{1}{2} - z_{1}\right) \, \mathbf{a}_{3} & = & \left(\frac{1}{2} +x_{1}\right)a \, \mathbf{\hat{x}} + y_{1}b \, \mathbf{\hat{y}} + \left(\frac{1}{2} - z_{1}\right)c \, \mathbf{\hat{z}} & \left(8c\right) & \mbox{C I} \\ 
\mathbf{B}_{7} & = & x_{1} \, \mathbf{a}_{1} + \left(\frac{1}{2} - y_{1}\right) \, \mathbf{a}_{2} + \left(\frac{1}{2} +z_{1}\right) \, \mathbf{a}_{3} & = & x_{1}a \, \mathbf{\hat{x}} + \left(\frac{1}{2} - y_{1}\right)b \, \mathbf{\hat{y}} + \left(\frac{1}{2} +z_{1}\right)c \, \mathbf{\hat{z}} & \left(8c\right) & \mbox{C I} \\ 
\mathbf{B}_{8} & = & \left(\frac{1}{2} - x_{1}\right) \, \mathbf{a}_{1} + \left(\frac{1}{2} +y_{1}\right) \, \mathbf{a}_{2} + z_{1} \, \mathbf{a}_{3} & = & \left(\frac{1}{2} - x_{1}\right)a \, \mathbf{\hat{x}} + \left(\frac{1}{2} +y_{1}\right)b \, \mathbf{\hat{y}} + z_{1}c \, \mathbf{\hat{z}} & \left(8c\right) & \mbox{C I} \\ 
\mathbf{B}_{9} & = & x_{2} \, \mathbf{a}_{1} + y_{2} \, \mathbf{a}_{2} + z_{2} \, \mathbf{a}_{3} & = & x_{2}a \, \mathbf{\hat{x}} + y_{2}b \, \mathbf{\hat{y}} + z_{2}c \, \mathbf{\hat{z}} & \left(8c\right) & \mbox{C II} \\ 
\mathbf{B}_{10} & = & \left(\frac{1}{2} - x_{2}\right) \, \mathbf{a}_{1}-y_{2} \, \mathbf{a}_{2} + \left(\frac{1}{2} +z_{2}\right) \, \mathbf{a}_{3} & = & \left(\frac{1}{2} - x_{2}\right)a \, \mathbf{\hat{x}}-y_{2}b \, \mathbf{\hat{y}} + \left(\frac{1}{2} +z_{2}\right)c \, \mathbf{\hat{z}} & \left(8c\right) & \mbox{C II} \\ 
\mathbf{B}_{11} & = & -x_{2} \, \mathbf{a}_{1} + \left(\frac{1}{2} +y_{2}\right) \, \mathbf{a}_{2} + \left(\frac{1}{2} - z_{2}\right) \, \mathbf{a}_{3} & = & -x_{2}a \, \mathbf{\hat{x}} + \left(\frac{1}{2} +y_{2}\right)b \, \mathbf{\hat{y}} + \left(\frac{1}{2} - z_{2}\right)c \, \mathbf{\hat{z}} & \left(8c\right) & \mbox{C II} \\ 
\mathbf{B}_{12} & = & \left(\frac{1}{2} +x_{2}\right) \, \mathbf{a}_{1} + \left(\frac{1}{2} - y_{2}\right) \, \mathbf{a}_{2}-z_{2} \, \mathbf{a}_{3} & = & \left(\frac{1}{2} +x_{2}\right)a \, \mathbf{\hat{x}} + \left(\frac{1}{2} - y_{2}\right)b \, \mathbf{\hat{y}}-z_{2}c \, \mathbf{\hat{z}} & \left(8c\right) & \mbox{C II} \\ 
\mathbf{B}_{13} & = & -x_{2} \, \mathbf{a}_{1}-y_{2} \, \mathbf{a}_{2}-z_{2} \, \mathbf{a}_{3} & = & -x_{2}a \, \mathbf{\hat{x}}-y_{2}b \, \mathbf{\hat{y}}-z_{2}c \, \mathbf{\hat{z}} & \left(8c\right) & \mbox{C II} \\ 
\mathbf{B}_{14} & = & \left(\frac{1}{2} +x_{2}\right) \, \mathbf{a}_{1} + y_{2} \, \mathbf{a}_{2} + \left(\frac{1}{2} - z_{2}\right) \, \mathbf{a}_{3} & = & \left(\frac{1}{2} +x_{2}\right)a \, \mathbf{\hat{x}} + y_{2}b \, \mathbf{\hat{y}} + \left(\frac{1}{2} - z_{2}\right)c \, \mathbf{\hat{z}} & \left(8c\right) & \mbox{C II} \\ 
\mathbf{B}_{15} & = & x_{2} \, \mathbf{a}_{1} + \left(\frac{1}{2} - y_{2}\right) \, \mathbf{a}_{2} + \left(\frac{1}{2} +z_{2}\right) \, \mathbf{a}_{3} & = & x_{2}a \, \mathbf{\hat{x}} + \left(\frac{1}{2} - y_{2}\right)b \, \mathbf{\hat{y}} + \left(\frac{1}{2} +z_{2}\right)c \, \mathbf{\hat{z}} & \left(8c\right) & \mbox{C II} \\ 
\mathbf{B}_{16} & = & \left(\frac{1}{2} - x_{2}\right) \, \mathbf{a}_{1} + \left(\frac{1}{2} +y_{2}\right) \, \mathbf{a}_{2} + z_{2} \, \mathbf{a}_{3} & = & \left(\frac{1}{2} - x_{2}\right)a \, \mathbf{\hat{x}} + \left(\frac{1}{2} +y_{2}\right)b \, \mathbf{\hat{y}} + z_{2}c \, \mathbf{\hat{z}} & \left(8c\right) & \mbox{C II} \\ 
\mathbf{B}_{17} & = & x_{3} \, \mathbf{a}_{1} + y_{3} \, \mathbf{a}_{2} + z_{3} \, \mathbf{a}_{3} & = & x_{3}a \, \mathbf{\hat{x}} + y_{3}b \, \mathbf{\hat{y}} + z_{3}c \, \mathbf{\hat{z}} & \left(8c\right) & \mbox{C III} \\ 
\mathbf{B}_{18} & = & \left(\frac{1}{2} - x_{3}\right) \, \mathbf{a}_{1}-y_{3} \, \mathbf{a}_{2} + \left(\frac{1}{2} +z_{3}\right) \, \mathbf{a}_{3} & = & \left(\frac{1}{2} - x_{3}\right)a \, \mathbf{\hat{x}}-y_{3}b \, \mathbf{\hat{y}} + \left(\frac{1}{2} +z_{3}\right)c \, \mathbf{\hat{z}} & \left(8c\right) & \mbox{C III} \\ 
\mathbf{B}_{19} & = & -x_{3} \, \mathbf{a}_{1} + \left(\frac{1}{2} +y_{3}\right) \, \mathbf{a}_{2} + \left(\frac{1}{2} - z_{3}\right) \, \mathbf{a}_{3} & = & -x_{3}a \, \mathbf{\hat{x}} + \left(\frac{1}{2} +y_{3}\right)b \, \mathbf{\hat{y}} + \left(\frac{1}{2} - z_{3}\right)c \, \mathbf{\hat{z}} & \left(8c\right) & \mbox{C III} \\ 
\mathbf{B}_{20} & = & \left(\frac{1}{2} +x_{3}\right) \, \mathbf{a}_{1} + \left(\frac{1}{2} - y_{3}\right) \, \mathbf{a}_{2}-z_{3} \, \mathbf{a}_{3} & = & \left(\frac{1}{2} +x_{3}\right)a \, \mathbf{\hat{x}} + \left(\frac{1}{2} - y_{3}\right)b \, \mathbf{\hat{y}}-z_{3}c \, \mathbf{\hat{z}} & \left(8c\right) & \mbox{C III} \\ 
\mathbf{B}_{21} & = & -x_{3} \, \mathbf{a}_{1}-y_{3} \, \mathbf{a}_{2}-z_{3} \, \mathbf{a}_{3} & = & -x_{3}a \, \mathbf{\hat{x}}-y_{3}b \, \mathbf{\hat{y}}-z_{3}c \, \mathbf{\hat{z}} & \left(8c\right) & \mbox{C III} \\ 
\mathbf{B}_{22} & = & \left(\frac{1}{2} +x_{3}\right) \, \mathbf{a}_{1} + y_{3} \, \mathbf{a}_{2} + \left(\frac{1}{2} - z_{3}\right) \, \mathbf{a}_{3} & = & \left(\frac{1}{2} +x_{3}\right)a \, \mathbf{\hat{x}} + y_{3}b \, \mathbf{\hat{y}} + \left(\frac{1}{2} - z_{3}\right)c \, \mathbf{\hat{z}} & \left(8c\right) & \mbox{C III} \\ 
\mathbf{B}_{23} & = & x_{3} \, \mathbf{a}_{1} + \left(\frac{1}{2} - y_{3}\right) \, \mathbf{a}_{2} + \left(\frac{1}{2} +z_{3}\right) \, \mathbf{a}_{3} & = & x_{3}a \, \mathbf{\hat{x}} + \left(\frac{1}{2} - y_{3}\right)b \, \mathbf{\hat{y}} + \left(\frac{1}{2} +z_{3}\right)c \, \mathbf{\hat{z}} & \left(8c\right) & \mbox{C III} \\ 
\mathbf{B}_{24} & = & \left(\frac{1}{2} - x_{3}\right) \, \mathbf{a}_{1} + \left(\frac{1}{2} +y_{3}\right) \, \mathbf{a}_{2} + z_{3} \, \mathbf{a}_{3} & = & \left(\frac{1}{2} - x_{3}\right)a \, \mathbf{\hat{x}} + \left(\frac{1}{2} +y_{3}\right)b \, \mathbf{\hat{y}} + z_{3}c \, \mathbf{\hat{z}} & \left(8c\right) & \mbox{C III} \\ 
\mathbf{B}_{25} & = & x_{4} \, \mathbf{a}_{1} + y_{4} \, \mathbf{a}_{2} + z_{4} \, \mathbf{a}_{3} & = & x_{4}a \, \mathbf{\hat{x}} + y_{4}b \, \mathbf{\hat{y}} + z_{4}c \, \mathbf{\hat{z}} & \left(8c\right) & \mbox{H I} \\ 
\mathbf{B}_{26} & = & \left(\frac{1}{2} - x_{4}\right) \, \mathbf{a}_{1}-y_{4} \, \mathbf{a}_{2} + \left(\frac{1}{2} +z_{4}\right) \, \mathbf{a}_{3} & = & \left(\frac{1}{2} - x_{4}\right)a \, \mathbf{\hat{x}}-y_{4}b \, \mathbf{\hat{y}} + \left(\frac{1}{2} +z_{4}\right)c \, \mathbf{\hat{z}} & \left(8c\right) & \mbox{H I} \\ 
\mathbf{B}_{27} & = & -x_{4} \, \mathbf{a}_{1} + \left(\frac{1}{2} +y_{4}\right) \, \mathbf{a}_{2} + \left(\frac{1}{2} - z_{4}\right) \, \mathbf{a}_{3} & = & -x_{4}a \, \mathbf{\hat{x}} + \left(\frac{1}{2} +y_{4}\right)b \, \mathbf{\hat{y}} + \left(\frac{1}{2} - z_{4}\right)c \, \mathbf{\hat{z}} & \left(8c\right) & \mbox{H I} \\ 
\mathbf{B}_{28} & = & \left(\frac{1}{2} +x_{4}\right) \, \mathbf{a}_{1} + \left(\frac{1}{2} - y_{4}\right) \, \mathbf{a}_{2}-z_{4} \, \mathbf{a}_{3} & = & \left(\frac{1}{2} +x_{4}\right)a \, \mathbf{\hat{x}} + \left(\frac{1}{2} - y_{4}\right)b \, \mathbf{\hat{y}}-z_{4}c \, \mathbf{\hat{z}} & \left(8c\right) & \mbox{H I} \\ 
\mathbf{B}_{29} & = & -x_{4} \, \mathbf{a}_{1}-y_{4} \, \mathbf{a}_{2}-z_{4} \, \mathbf{a}_{3} & = & -x_{4}a \, \mathbf{\hat{x}}-y_{4}b \, \mathbf{\hat{y}}-z_{4}c \, \mathbf{\hat{z}} & \left(8c\right) & \mbox{H I} \\ 
\mathbf{B}_{30} & = & \left(\frac{1}{2} +x_{4}\right) \, \mathbf{a}_{1} + y_{4} \, \mathbf{a}_{2} + \left(\frac{1}{2} - z_{4}\right) \, \mathbf{a}_{3} & = & \left(\frac{1}{2} +x_{4}\right)a \, \mathbf{\hat{x}} + y_{4}b \, \mathbf{\hat{y}} + \left(\frac{1}{2} - z_{4}\right)c \, \mathbf{\hat{z}} & \left(8c\right) & \mbox{H I} \\ 
\mathbf{B}_{31} & = & x_{4} \, \mathbf{a}_{1} + \left(\frac{1}{2} - y_{4}\right) \, \mathbf{a}_{2} + \left(\frac{1}{2} +z_{4}\right) \, \mathbf{a}_{3} & = & x_{4}a \, \mathbf{\hat{x}} + \left(\frac{1}{2} - y_{4}\right)b \, \mathbf{\hat{y}} + \left(\frac{1}{2} +z_{4}\right)c \, \mathbf{\hat{z}} & \left(8c\right) & \mbox{H I} \\ 
\mathbf{B}_{32} & = & \left(\frac{1}{2} - x_{4}\right) \, \mathbf{a}_{1} + \left(\frac{1}{2} +y_{4}\right) \, \mathbf{a}_{2} + z_{4} \, \mathbf{a}_{3} & = & \left(\frac{1}{2} - x_{4}\right)a \, \mathbf{\hat{x}} + \left(\frac{1}{2} +y_{4}\right)b \, \mathbf{\hat{y}} + z_{4}c \, \mathbf{\hat{z}} & \left(8c\right) & \mbox{H I} \\ 
\mathbf{B}_{33} & = & x_{5} \, \mathbf{a}_{1} + y_{5} \, \mathbf{a}_{2} + z_{5} \, \mathbf{a}_{3} & = & x_{5}a \, \mathbf{\hat{x}} + y_{5}b \, \mathbf{\hat{y}} + z_{5}c \, \mathbf{\hat{z}} & \left(8c\right) & \mbox{H II} \\ 
\mathbf{B}_{34} & = & \left(\frac{1}{2} - x_{5}\right) \, \mathbf{a}_{1}-y_{5} \, \mathbf{a}_{2} + \left(\frac{1}{2} +z_{5}\right) \, \mathbf{a}_{3} & = & \left(\frac{1}{2} - x_{5}\right)a \, \mathbf{\hat{x}}-y_{5}b \, \mathbf{\hat{y}} + \left(\frac{1}{2} +z_{5}\right)c \, \mathbf{\hat{z}} & \left(8c\right) & \mbox{H II} \\ 
\mathbf{B}_{35} & = & -x_{5} \, \mathbf{a}_{1} + \left(\frac{1}{2} +y_{5}\right) \, \mathbf{a}_{2} + \left(\frac{1}{2} - z_{5}\right) \, \mathbf{a}_{3} & = & -x_{5}a \, \mathbf{\hat{x}} + \left(\frac{1}{2} +y_{5}\right)b \, \mathbf{\hat{y}} + \left(\frac{1}{2} - z_{5}\right)c \, \mathbf{\hat{z}} & \left(8c\right) & \mbox{H II} \\ 
\mathbf{B}_{36} & = & \left(\frac{1}{2} +x_{5}\right) \, \mathbf{a}_{1} + \left(\frac{1}{2} - y_{5}\right) \, \mathbf{a}_{2}-z_{5} \, \mathbf{a}_{3} & = & \left(\frac{1}{2} +x_{5}\right)a \, \mathbf{\hat{x}} + \left(\frac{1}{2} - y_{5}\right)b \, \mathbf{\hat{y}}-z_{5}c \, \mathbf{\hat{z}} & \left(8c\right) & \mbox{H II} \\ 
\mathbf{B}_{37} & = & -x_{5} \, \mathbf{a}_{1}-y_{5} \, \mathbf{a}_{2}-z_{5} \, \mathbf{a}_{3} & = & -x_{5}a \, \mathbf{\hat{x}}-y_{5}b \, \mathbf{\hat{y}}-z_{5}c \, \mathbf{\hat{z}} & \left(8c\right) & \mbox{H II} \\ 
\mathbf{B}_{38} & = & \left(\frac{1}{2} +x_{5}\right) \, \mathbf{a}_{1} + y_{5} \, \mathbf{a}_{2} + \left(\frac{1}{2} - z_{5}\right) \, \mathbf{a}_{3} & = & \left(\frac{1}{2} +x_{5}\right)a \, \mathbf{\hat{x}} + y_{5}b \, \mathbf{\hat{y}} + \left(\frac{1}{2} - z_{5}\right)c \, \mathbf{\hat{z}} & \left(8c\right) & \mbox{H II} \\ 
\mathbf{B}_{39} & = & x_{5} \, \mathbf{a}_{1} + \left(\frac{1}{2} - y_{5}\right) \, \mathbf{a}_{2} + \left(\frac{1}{2} +z_{5}\right) \, \mathbf{a}_{3} & = & x_{5}a \, \mathbf{\hat{x}} + \left(\frac{1}{2} - y_{5}\right)b \, \mathbf{\hat{y}} + \left(\frac{1}{2} +z_{5}\right)c \, \mathbf{\hat{z}} & \left(8c\right) & \mbox{H II} \\ 
\mathbf{B}_{40} & = & \left(\frac{1}{2} - x_{5}\right) \, \mathbf{a}_{1} + \left(\frac{1}{2} +y_{5}\right) \, \mathbf{a}_{2} + z_{5} \, \mathbf{a}_{3} & = & \left(\frac{1}{2} - x_{5}\right)a \, \mathbf{\hat{x}} + \left(\frac{1}{2} +y_{5}\right)b \, \mathbf{\hat{y}} + z_{5}c \, \mathbf{\hat{z}} & \left(8c\right) & \mbox{H II} \\ 
\mathbf{B}_{41} & = & x_{6} \, \mathbf{a}_{1} + y_{6} \, \mathbf{a}_{2} + z_{6} \, \mathbf{a}_{3} & = & x_{6}a \, \mathbf{\hat{x}} + y_{6}b \, \mathbf{\hat{y}} + z_{6}c \, \mathbf{\hat{z}} & \left(8c\right) & \mbox{H III} \\ 
\mathbf{B}_{42} & = & \left(\frac{1}{2} - x_{6}\right) \, \mathbf{a}_{1}-y_{6} \, \mathbf{a}_{2} + \left(\frac{1}{2} +z_{6}\right) \, \mathbf{a}_{3} & = & \left(\frac{1}{2} - x_{6}\right)a \, \mathbf{\hat{x}}-y_{6}b \, \mathbf{\hat{y}} + \left(\frac{1}{2} +z_{6}\right)c \, \mathbf{\hat{z}} & \left(8c\right) & \mbox{H III} \\ 
\mathbf{B}_{43} & = & -x_{6} \, \mathbf{a}_{1} + \left(\frac{1}{2} +y_{6}\right) \, \mathbf{a}_{2} + \left(\frac{1}{2} - z_{6}\right) \, \mathbf{a}_{3} & = & -x_{6}a \, \mathbf{\hat{x}} + \left(\frac{1}{2} +y_{6}\right)b \, \mathbf{\hat{y}} + \left(\frac{1}{2} - z_{6}\right)c \, \mathbf{\hat{z}} & \left(8c\right) & \mbox{H III} \\ 
\mathbf{B}_{44} & = & \left(\frac{1}{2} +x_{6}\right) \, \mathbf{a}_{1} + \left(\frac{1}{2} - y_{6}\right) \, \mathbf{a}_{2}-z_{6} \, \mathbf{a}_{3} & = & \left(\frac{1}{2} +x_{6}\right)a \, \mathbf{\hat{x}} + \left(\frac{1}{2} - y_{6}\right)b \, \mathbf{\hat{y}}-z_{6}c \, \mathbf{\hat{z}} & \left(8c\right) & \mbox{H III} \\ 
\mathbf{B}_{45} & = & -x_{6} \, \mathbf{a}_{1}-y_{6} \, \mathbf{a}_{2}-z_{6} \, \mathbf{a}_{3} & = & -x_{6}a \, \mathbf{\hat{x}}-y_{6}b \, \mathbf{\hat{y}}-z_{6}c \, \mathbf{\hat{z}} & \left(8c\right) & \mbox{H III} \\ 
\mathbf{B}_{46} & = & \left(\frac{1}{2} +x_{6}\right) \, \mathbf{a}_{1} + y_{6} \, \mathbf{a}_{2} + \left(\frac{1}{2} - z_{6}\right) \, \mathbf{a}_{3} & = & \left(\frac{1}{2} +x_{6}\right)a \, \mathbf{\hat{x}} + y_{6}b \, \mathbf{\hat{y}} + \left(\frac{1}{2} - z_{6}\right)c \, \mathbf{\hat{z}} & \left(8c\right) & \mbox{H III} \\ 
\mathbf{B}_{47} & = & x_{6} \, \mathbf{a}_{1} + \left(\frac{1}{2} - y_{6}\right) \, \mathbf{a}_{2} + \left(\frac{1}{2} +z_{6}\right) \, \mathbf{a}_{3} & = & x_{6}a \, \mathbf{\hat{x}} + \left(\frac{1}{2} - y_{6}\right)b \, \mathbf{\hat{y}} + \left(\frac{1}{2} +z_{6}\right)c \, \mathbf{\hat{z}} & \left(8c\right) & \mbox{H III} \\ 
\mathbf{B}_{48} & = & \left(\frac{1}{2} - x_{6}\right) \, \mathbf{a}_{1} + \left(\frac{1}{2} +y_{6}\right) \, \mathbf{a}_{2} + z_{6} \, \mathbf{a}_{3} & = & \left(\frac{1}{2} - x_{6}\right)a \, \mathbf{\hat{x}} + \left(\frac{1}{2} +y_{6}\right)b \, \mathbf{\hat{y}} + z_{6}c \, \mathbf{\hat{z}} & \left(8c\right) & \mbox{H III} \\ 
\end{longtabu}
\renewcommand{\arraystretch}{1.0}
\noindent \hrulefill
\\
\textbf{References:}
\vspace*{-0.25cm}
\begin{flushleft}
  - \bibentry{Nayak_2010}. \\
\end{flushleft}
\noindent \hrulefill
\\
\textbf{Geometry files:}
\\
\noindent  - CIF: pp. {\hyperref[AB_oP48_61_3c_3c_cif]{\pageref{AB_oP48_61_3c_3c_cif}}} \\
\noindent  - POSCAR: pp. {\hyperref[AB_oP48_61_3c_3c_poscar]{\pageref{AB_oP48_61_3c_3c_poscar}}} \\
\onecolumn
{\phantomsection\label{A2B3_oP20_62_2c_3c}}
\subsection*{\huge \textbf{{\normalfont \begin{raggedleft}Tongbaite (Cr$_{3}$C$_{2}$, $D5_{10}$) Structure: \end{raggedleft} \\ A2B3\_oP20\_62\_2c\_3c}}}
\noindent \hrulefill
\vspace*{0.25cm}
\begin{figure}[htp]
  \centering
  \vspace{-1em}
  {\includegraphics[width=1\textwidth]{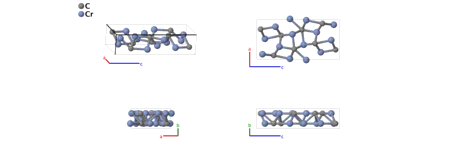}}
\end{figure}
\vspace*{-0.5cm}
\renewcommand{\arraystretch}{1.5}
\begin{equation*}
  \begin{array}{>{$\hspace{-0.15cm}}l<{$}>{$}p{0.5cm}<{$}>{$}p{18.5cm}<{$}}
    \mbox{\large \textbf{Prototype}} &\colon & \ce{Cr3C2} \\
    \mbox{\large \textbf{\AFLOW\ prototype label}} &\colon & \mbox{A2B3\_oP20\_62\_2c\_3c} \\
    \mbox{\large \textbf{\textit{Strukturbericht} designation}} &\colon & \mbox{$D5_{10}$} \\
    \mbox{\large \textbf{Pearson symbol}} &\colon & \mbox{oP20} \\
    \mbox{\large \textbf{Space group number}} &\colon & 62 \\
    \mbox{\large \textbf{Space group symbol}} &\colon & Pnma \\
    \mbox{\large \textbf{\AFLOW\ prototype command}} &\colon &  \texttt{aflow} \,  \, \texttt{-{}-proto=A2B3\_oP20\_62\_2c\_3c } \, \newline \texttt{-{}-params=}{a,b/a,c/a,x_{1},z_{1},x_{2},z_{2},x_{3},z_{3},x_{4},z_{4},x_{5},z_{5} }
  \end{array}
\end{equation*}
\renewcommand{\arraystretch}{1.0}

\vspace*{-0.25cm}
\noindent \hrulefill
\\
\textbf{ Other compounds with this structure:}
\begin{itemize}
   \item{ Hf$_{3}$P$_{2}$  }
\end{itemize}
\vspace*{-0.25cm}
\noindent \hrulefill
\begin{itemize}
  \item{Several authors remark that this is the anti-type of
\href{http://aflow.org/CrystalDatabase/A3B2_oP20_62_3c_2c.html}{Stibnite (Sb$_{2}$S$_{3}$, $D5_{8}$, A3B2\_oP20\_62\_3c\_2c)}.
}
\end{itemize}

\noindent \parbox{1 \linewidth}{
\noindent \hrulefill
\\
\textbf{Simple Orthorhombic primitive vectors:} \\
\vspace*{-0.25cm}
\begin{tabular}{cc}
  \begin{tabular}{c}
    \parbox{0.6 \linewidth}{
      \renewcommand{\arraystretch}{1.5}
      \begin{equation*}
        \centering
        \begin{array}{ccc}
              \mathbf{a}_1 & = & a \, \mathbf{\hat{x}} \\
    \mathbf{a}_2 & = & b \, \mathbf{\hat{y}} \\
    \mathbf{a}_3 & = & c \, \mathbf{\hat{z}} \\

        \end{array}
      \end{equation*}
    }
    \renewcommand{\arraystretch}{1.0}
  \end{tabular}
  \begin{tabular}{c}
    \includegraphics[width=0.3\linewidth]{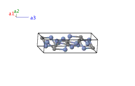} \\
  \end{tabular}
\end{tabular}

}
\vspace*{-0.25cm}

\noindent \hrulefill
\\
\textbf{Basis vectors:}
\vspace*{-0.25cm}
\renewcommand{\arraystretch}{1.5}
\begin{longtabu} to \textwidth{>{\centering $}X[-1,c,c]<{$}>{\centering $}X[-1,c,c]<{$}>{\centering $}X[-1,c,c]<{$}>{\centering $}X[-1,c,c]<{$}>{\centering $}X[-1,c,c]<{$}>{\centering $}X[-1,c,c]<{$}>{\centering $}X[-1,c,c]<{$}}
  & & \mbox{Lattice Coordinates} & & \mbox{Cartesian Coordinates} &\mbox{Wyckoff Position} & \mbox{Atom Type} \\  
  \mathbf{B}_{1} & = & x_{1} \, \mathbf{a}_{1} + \frac{1}{4} \, \mathbf{a}_{2} + z_{1} \, \mathbf{a}_{3} & = & x_{1}a \, \mathbf{\hat{x}} + \frac{1}{4}b \, \mathbf{\hat{y}} + z_{1}c \, \mathbf{\hat{z}} & \left(4c\right) & \mbox{C I} \\ 
\mathbf{B}_{2} & = & \left(\frac{1}{2} - x_{1}\right) \, \mathbf{a}_{1} + \frac{3}{4} \, \mathbf{a}_{2} + \left(\frac{1}{2} +z_{1}\right) \, \mathbf{a}_{3} & = & \left(\frac{1}{2} - x_{1}\right)a \, \mathbf{\hat{x}} + \frac{3}{4}b \, \mathbf{\hat{y}} + \left(\frac{1}{2} +z_{1}\right)c \, \mathbf{\hat{z}} & \left(4c\right) & \mbox{C I} \\ 
\mathbf{B}_{3} & = & -x_{1} \, \mathbf{a}_{1} + \frac{3}{4} \, \mathbf{a}_{2}-z_{1} \, \mathbf{a}_{3} & = & -x_{1}a \, \mathbf{\hat{x}} + \frac{3}{4}b \, \mathbf{\hat{y}}-z_{1}c \, \mathbf{\hat{z}} & \left(4c\right) & \mbox{C I} \\ 
\mathbf{B}_{4} & = & \left(\frac{1}{2} +x_{1}\right) \, \mathbf{a}_{1} + \frac{1}{4} \, \mathbf{a}_{2} + \left(\frac{1}{2} - z_{1}\right) \, \mathbf{a}_{3} & = & \left(\frac{1}{2} +x_{1}\right)a \, \mathbf{\hat{x}} + \frac{1}{4}b \, \mathbf{\hat{y}} + \left(\frac{1}{2} - z_{1}\right)c \, \mathbf{\hat{z}} & \left(4c\right) & \mbox{C I} \\ 
\mathbf{B}_{5} & = & x_{2} \, \mathbf{a}_{1} + \frac{1}{4} \, \mathbf{a}_{2} + z_{2} \, \mathbf{a}_{3} & = & x_{2}a \, \mathbf{\hat{x}} + \frac{1}{4}b \, \mathbf{\hat{y}} + z_{2}c \, \mathbf{\hat{z}} & \left(4c\right) & \mbox{C II} \\ 
\mathbf{B}_{6} & = & \left(\frac{1}{2} - x_{2}\right) \, \mathbf{a}_{1} + \frac{3}{4} \, \mathbf{a}_{2} + \left(\frac{1}{2} +z_{2}\right) \, \mathbf{a}_{3} & = & \left(\frac{1}{2} - x_{2}\right)a \, \mathbf{\hat{x}} + \frac{3}{4}b \, \mathbf{\hat{y}} + \left(\frac{1}{2} +z_{2}\right)c \, \mathbf{\hat{z}} & \left(4c\right) & \mbox{C II} \\ 
\mathbf{B}_{7} & = & -x_{2} \, \mathbf{a}_{1} + \frac{3}{4} \, \mathbf{a}_{2}-z_{2} \, \mathbf{a}_{3} & = & -x_{2}a \, \mathbf{\hat{x}} + \frac{3}{4}b \, \mathbf{\hat{y}}-z_{2}c \, \mathbf{\hat{z}} & \left(4c\right) & \mbox{C II} \\ 
\mathbf{B}_{8} & = & \left(\frac{1}{2} +x_{2}\right) \, \mathbf{a}_{1} + \frac{1}{4} \, \mathbf{a}_{2} + \left(\frac{1}{2} - z_{2}\right) \, \mathbf{a}_{3} & = & \left(\frac{1}{2} +x_{2}\right)a \, \mathbf{\hat{x}} + \frac{1}{4}b \, \mathbf{\hat{y}} + \left(\frac{1}{2} - z_{2}\right)c \, \mathbf{\hat{z}} & \left(4c\right) & \mbox{C II} \\ 
\mathbf{B}_{9} & = & x_{3} \, \mathbf{a}_{1} + \frac{1}{4} \, \mathbf{a}_{2} + z_{3} \, \mathbf{a}_{3} & = & x_{3}a \, \mathbf{\hat{x}} + \frac{1}{4}b \, \mathbf{\hat{y}} + z_{3}c \, \mathbf{\hat{z}} & \left(4c\right) & \mbox{Cr I} \\ 
\mathbf{B}_{10} & = & \left(\frac{1}{2} - x_{3}\right) \, \mathbf{a}_{1} + \frac{3}{4} \, \mathbf{a}_{2} + \left(\frac{1}{2} +z_{3}\right) \, \mathbf{a}_{3} & = & \left(\frac{1}{2} - x_{3}\right)a \, \mathbf{\hat{x}} + \frac{3}{4}b \, \mathbf{\hat{y}} + \left(\frac{1}{2} +z_{3}\right)c \, \mathbf{\hat{z}} & \left(4c\right) & \mbox{Cr I} \\ 
\mathbf{B}_{11} & = & -x_{3} \, \mathbf{a}_{1} + \frac{3}{4} \, \mathbf{a}_{2}-z_{3} \, \mathbf{a}_{3} & = & -x_{3}a \, \mathbf{\hat{x}} + \frac{3}{4}b \, \mathbf{\hat{y}}-z_{3}c \, \mathbf{\hat{z}} & \left(4c\right) & \mbox{Cr I} \\ 
\mathbf{B}_{12} & = & \left(\frac{1}{2} +x_{3}\right) \, \mathbf{a}_{1} + \frac{1}{4} \, \mathbf{a}_{2} + \left(\frac{1}{2} - z_{3}\right) \, \mathbf{a}_{3} & = & \left(\frac{1}{2} +x_{3}\right)a \, \mathbf{\hat{x}} + \frac{1}{4}b \, \mathbf{\hat{y}} + \left(\frac{1}{2} - z_{3}\right)c \, \mathbf{\hat{z}} & \left(4c\right) & \mbox{Cr I} \\ 
\mathbf{B}_{13} & = & x_{4} \, \mathbf{a}_{1} + \frac{1}{4} \, \mathbf{a}_{2} + z_{4} \, \mathbf{a}_{3} & = & x_{4}a \, \mathbf{\hat{x}} + \frac{1}{4}b \, \mathbf{\hat{y}} + z_{4}c \, \mathbf{\hat{z}} & \left(4c\right) & \mbox{Cr II} \\ 
\mathbf{B}_{14} & = & \left(\frac{1}{2} - x_{4}\right) \, \mathbf{a}_{1} + \frac{3}{4} \, \mathbf{a}_{2} + \left(\frac{1}{2} +z_{4}\right) \, \mathbf{a}_{3} & = & \left(\frac{1}{2} - x_{4}\right)a \, \mathbf{\hat{x}} + \frac{3}{4}b \, \mathbf{\hat{y}} + \left(\frac{1}{2} +z_{4}\right)c \, \mathbf{\hat{z}} & \left(4c\right) & \mbox{Cr II} \\ 
\mathbf{B}_{15} & = & -x_{4} \, \mathbf{a}_{1} + \frac{3}{4} \, \mathbf{a}_{2}-z_{4} \, \mathbf{a}_{3} & = & -x_{4}a \, \mathbf{\hat{x}} + \frac{3}{4}b \, \mathbf{\hat{y}}-z_{4}c \, \mathbf{\hat{z}} & \left(4c\right) & \mbox{Cr II} \\ 
\mathbf{B}_{16} & = & \left(\frac{1}{2} +x_{4}\right) \, \mathbf{a}_{1} + \frac{1}{4} \, \mathbf{a}_{2} + \left(\frac{1}{2} - z_{4}\right) \, \mathbf{a}_{3} & = & \left(\frac{1}{2} +x_{4}\right)a \, \mathbf{\hat{x}} + \frac{1}{4}b \, \mathbf{\hat{y}} + \left(\frac{1}{2} - z_{4}\right)c \, \mathbf{\hat{z}} & \left(4c\right) & \mbox{Cr II} \\ 
\mathbf{B}_{17} & = & x_{5} \, \mathbf{a}_{1} + \frac{1}{4} \, \mathbf{a}_{2} + z_{5} \, \mathbf{a}_{3} & = & x_{5}a \, \mathbf{\hat{x}} + \frac{1}{4}b \, \mathbf{\hat{y}} + z_{5}c \, \mathbf{\hat{z}} & \left(4c\right) & \mbox{Cr III} \\ 
\mathbf{B}_{18} & = & \left(\frac{1}{2} - x_{5}\right) \, \mathbf{a}_{1} + \frac{3}{4} \, \mathbf{a}_{2} + \left(\frac{1}{2} +z_{5}\right) \, \mathbf{a}_{3} & = & \left(\frac{1}{2} - x_{5}\right)a \, \mathbf{\hat{x}} + \frac{3}{4}b \, \mathbf{\hat{y}} + \left(\frac{1}{2} +z_{5}\right)c \, \mathbf{\hat{z}} & \left(4c\right) & \mbox{Cr III} \\ 
\mathbf{B}_{19} & = & -x_{5} \, \mathbf{a}_{1} + \frac{3}{4} \, \mathbf{a}_{2}-z_{5} \, \mathbf{a}_{3} & = & -x_{5}a \, \mathbf{\hat{x}} + \frac{3}{4}b \, \mathbf{\hat{y}}-z_{5}c \, \mathbf{\hat{z}} & \left(4c\right) & \mbox{Cr III} \\ 
\mathbf{B}_{20} & = & \left(\frac{1}{2} +x_{5}\right) \, \mathbf{a}_{1} + \frac{1}{4} \, \mathbf{a}_{2} + \left(\frac{1}{2} - z_{5}\right) \, \mathbf{a}_{3} & = & \left(\frac{1}{2} +x_{5}\right)a \, \mathbf{\hat{x}} + \frac{1}{4}b \, \mathbf{\hat{y}} + \left(\frac{1}{2} - z_{5}\right)c \, \mathbf{\hat{z}} & \left(4c\right) & \mbox{Cr III} \\ 
\end{longtabu}
\renewcommand{\arraystretch}{1.0}
\noindent \hrulefill
\\
\textbf{References:}
\vspace*{-0.25cm}
\begin{flushleft}
  - \bibentry{Rundqvist_Acta_Chem_Scand_23_1969}. \\
\end{flushleft}
\noindent \hrulefill
\\
\textbf{Geometry files:}
\\
\noindent  - CIF: pp. {\hyperref[A2B3_oP20_62_2c_3c_cif]{\pageref{A2B3_oP20_62_2c_3c_cif}}} \\
\noindent  - POSCAR: pp. {\hyperref[A2B3_oP20_62_2c_3c_poscar]{\pageref{A2B3_oP20_62_2c_3c_poscar}}} \\
\onecolumn
{\phantomsection\label{A2B4C_oP28_62_ac_2cd_c}}
\subsection*{\huge \textbf{{\normalfont \begin{raggedleft}Forsterite (Mg$_{2}$SiO$_{4}$, $S1_{2}$) Structure: \end{raggedleft} \\ A2B4C\_oP28\_62\_ac\_2cd\_c}}}
\noindent \hrulefill
\vspace*{0.25cm}
\begin{figure}[htp]
  \centering
  \vspace{-1em}
  {\includegraphics[width=1\textwidth]{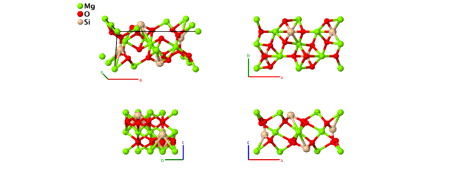}}
\end{figure}
\vspace*{-0.5cm}
\renewcommand{\arraystretch}{1.5}
\begin{equation*}
  \begin{array}{>{$\hspace{-0.15cm}}l<{$}>{$}p{0.5cm}<{$}>{$}p{18.5cm}<{$}}
    \mbox{\large \textbf{Prototype}} &\colon & \ce{Mg2SiO4} \\
    \mbox{\large \textbf{\AFLOW\ prototype label}} &\colon & \mbox{A2B4C\_oP28\_62\_ac\_2cd\_c} \\
    \mbox{\large \textbf{\textit{Strukturbericht} designation}} &\colon & \mbox{$S1_{2}$} \\
    \mbox{\large \textbf{Pearson symbol}} &\colon & \mbox{oP28} \\
    \mbox{\large \textbf{Space group number}} &\colon & 62 \\
    \mbox{\large \textbf{Space group symbol}} &\colon & Pnma \\
    \mbox{\large \textbf{\AFLOW\ prototype command}} &\colon &  \texttt{aflow} \,  \, \texttt{-{}-proto=A2B4C\_oP28\_62\_ac\_2cd\_c } \, \newline \texttt{-{}-params=}{a,b/a,c/a,x_{2},z_{2},x_{3},z_{3},x_{4},z_{4},x_{5},z_{5},x_{6},y_{6},z_{6} }
  \end{array}
\end{equation*}
\renewcommand{\arraystretch}{1.0}

\vspace*{-0.25cm}
\noindent \hrulefill
\\
\textbf{ Other compounds with this structure:}
\begin{itemize}
   \item{ Fe$_{2}$SiO$_{4}$ (fayalite), (Mg,Fe)CaSiO$_{4}$, (Mg,Mn)SiO$_{4}$, Al$_{2}$BeO$_{4}$, Fe$_{2}$SiS$_{4}$, Mg$_{2}$GeS$_{4}$, Mg$_{2}$GeS$_{4}$, Mn$_{2}$GeS$_{4}$, Tm$_{2}$ZnS$_{4}$  }
\end{itemize}
\vspace*{-0.25cm}
\noindent \hrulefill
\begin{itemize}
  \item{This structure is the magnesium end-point of olivine,
(Mg,Fe)$_{2}$SiO$_{4}$.
(Hazen, 1976) reports the structure in the $Pbnm$ setting of space group
\#62.  We have transformed this into the standard $Pnma$ setting.
We use the structural data at 23$^\circ$~C.
}
\end{itemize}

\noindent \parbox{1 \linewidth}{
\noindent \hrulefill
\\
\textbf{Simple Orthorhombic primitive vectors:} \\
\vspace*{-0.25cm}
\begin{tabular}{cc}
  \begin{tabular}{c}
    \parbox{0.6 \linewidth}{
      \renewcommand{\arraystretch}{1.5}
      \begin{equation*}
        \centering
        \begin{array}{ccc}
              \mathbf{a}_1 & = & a \, \mathbf{\hat{x}} \\
    \mathbf{a}_2 & = & b \, \mathbf{\hat{y}} \\
    \mathbf{a}_3 & = & c \, \mathbf{\hat{z}} \\

        \end{array}
      \end{equation*}
    }
    \renewcommand{\arraystretch}{1.0}
  \end{tabular}
  \begin{tabular}{c}
    \includegraphics[width=0.3\linewidth]{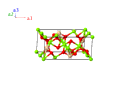} \\
  \end{tabular}
\end{tabular}

}
\vspace*{-0.25cm}

\noindent \hrulefill
\\
\textbf{Basis vectors:}
\vspace*{-0.25cm}
\renewcommand{\arraystretch}{1.5}
\begin{longtabu} to \textwidth{>{\centering $}X[-1,c,c]<{$}>{\centering $}X[-1,c,c]<{$}>{\centering $}X[-1,c,c]<{$}>{\centering $}X[-1,c,c]<{$}>{\centering $}X[-1,c,c]<{$}>{\centering $}X[-1,c,c]<{$}>{\centering $}X[-1,c,c]<{$}}
  & & \mbox{Lattice Coordinates} & & \mbox{Cartesian Coordinates} &\mbox{Wyckoff Position} & \mbox{Atom Type} \\  
  \mathbf{B}_{1} & = & 0 \, \mathbf{a}_{1} + 0 \, \mathbf{a}_{2} + 0 \, \mathbf{a}_{3} & = & 0 \, \mathbf{\hat{x}} + 0 \, \mathbf{\hat{y}} + 0 \, \mathbf{\hat{z}} & \left(4a\right) & \mbox{Mg I} \\ 
\mathbf{B}_{2} & = & \frac{1}{2} \, \mathbf{a}_{1} + \frac{1}{2} \, \mathbf{a}_{3} & = & \frac{1}{2}a \, \mathbf{\hat{x}} + \frac{1}{2}c \, \mathbf{\hat{z}} & \left(4a\right) & \mbox{Mg I} \\ 
\mathbf{B}_{3} & = & \frac{1}{2} \, \mathbf{a}_{2} & = & \frac{1}{2}b \, \mathbf{\hat{y}} & \left(4a\right) & \mbox{Mg I} \\ 
\mathbf{B}_{4} & = & \frac{1}{2} \, \mathbf{a}_{1} + \frac{1}{2} \, \mathbf{a}_{2} + \frac{1}{2} \, \mathbf{a}_{3} & = & \frac{1}{2}a \, \mathbf{\hat{x}} + \frac{1}{2}b \, \mathbf{\hat{y}} + \frac{1}{2}c \, \mathbf{\hat{z}} & \left(4a\right) & \mbox{Mg I} \\ 
\mathbf{B}_{5} & = & x_{2} \, \mathbf{a}_{1} + \frac{1}{4} \, \mathbf{a}_{2} + z_{2} \, \mathbf{a}_{3} & = & x_{2}a \, \mathbf{\hat{x}} + \frac{1}{4}b \, \mathbf{\hat{y}} + z_{2}c \, \mathbf{\hat{z}} & \left(4c\right) & \mbox{Mg II} \\ 
\mathbf{B}_{6} & = & \left(\frac{1}{2} - x_{2}\right) \, \mathbf{a}_{1} + \frac{3}{4} \, \mathbf{a}_{2} + \left(\frac{1}{2} +z_{2}\right) \, \mathbf{a}_{3} & = & \left(\frac{1}{2} - x_{2}\right)a \, \mathbf{\hat{x}} + \frac{3}{4}b \, \mathbf{\hat{y}} + \left(\frac{1}{2} +z_{2}\right)c \, \mathbf{\hat{z}} & \left(4c\right) & \mbox{Mg II} \\ 
\mathbf{B}_{7} & = & -x_{2} \, \mathbf{a}_{1} + \frac{3}{4} \, \mathbf{a}_{2}-z_{2} \, \mathbf{a}_{3} & = & -x_{2}a \, \mathbf{\hat{x}} + \frac{3}{4}b \, \mathbf{\hat{y}}-z_{2}c \, \mathbf{\hat{z}} & \left(4c\right) & \mbox{Mg II} \\ 
\mathbf{B}_{8} & = & \left(\frac{1}{2} +x_{2}\right) \, \mathbf{a}_{1} + \frac{1}{4} \, \mathbf{a}_{2} + \left(\frac{1}{2} - z_{2}\right) \, \mathbf{a}_{3} & = & \left(\frac{1}{2} +x_{2}\right)a \, \mathbf{\hat{x}} + \frac{1}{4}b \, \mathbf{\hat{y}} + \left(\frac{1}{2} - z_{2}\right)c \, \mathbf{\hat{z}} & \left(4c\right) & \mbox{Mg II} \\ 
\mathbf{B}_{9} & = & x_{3} \, \mathbf{a}_{1} + \frac{1}{4} \, \mathbf{a}_{2} + z_{3} \, \mathbf{a}_{3} & = & x_{3}a \, \mathbf{\hat{x}} + \frac{1}{4}b \, \mathbf{\hat{y}} + z_{3}c \, \mathbf{\hat{z}} & \left(4c\right) & \mbox{O I} \\ 
\mathbf{B}_{10} & = & \left(\frac{1}{2} - x_{3}\right) \, \mathbf{a}_{1} + \frac{3}{4} \, \mathbf{a}_{2} + \left(\frac{1}{2} +z_{3}\right) \, \mathbf{a}_{3} & = & \left(\frac{1}{2} - x_{3}\right)a \, \mathbf{\hat{x}} + \frac{3}{4}b \, \mathbf{\hat{y}} + \left(\frac{1}{2} +z_{3}\right)c \, \mathbf{\hat{z}} & \left(4c\right) & \mbox{O I} \\ 
\mathbf{B}_{11} & = & -x_{3} \, \mathbf{a}_{1} + \frac{3}{4} \, \mathbf{a}_{2}-z_{3} \, \mathbf{a}_{3} & = & -x_{3}a \, \mathbf{\hat{x}} + \frac{3}{4}b \, \mathbf{\hat{y}}-z_{3}c \, \mathbf{\hat{z}} & \left(4c\right) & \mbox{O I} \\ 
\mathbf{B}_{12} & = & \left(\frac{1}{2} +x_{3}\right) \, \mathbf{a}_{1} + \frac{1}{4} \, \mathbf{a}_{2} + \left(\frac{1}{2} - z_{3}\right) \, \mathbf{a}_{3} & = & \left(\frac{1}{2} +x_{3}\right)a \, \mathbf{\hat{x}} + \frac{1}{4}b \, \mathbf{\hat{y}} + \left(\frac{1}{2} - z_{3}\right)c \, \mathbf{\hat{z}} & \left(4c\right) & \mbox{O I} \\ 
\mathbf{B}_{13} & = & x_{4} \, \mathbf{a}_{1} + \frac{1}{4} \, \mathbf{a}_{2} + z_{4} \, \mathbf{a}_{3} & = & x_{4}a \, \mathbf{\hat{x}} + \frac{1}{4}b \, \mathbf{\hat{y}} + z_{4}c \, \mathbf{\hat{z}} & \left(4c\right) & \mbox{O II} \\ 
\mathbf{B}_{14} & = & \left(\frac{1}{2} - x_{4}\right) \, \mathbf{a}_{1} + \frac{3}{4} \, \mathbf{a}_{2} + \left(\frac{1}{2} +z_{4}\right) \, \mathbf{a}_{3} & = & \left(\frac{1}{2} - x_{4}\right)a \, \mathbf{\hat{x}} + \frac{3}{4}b \, \mathbf{\hat{y}} + \left(\frac{1}{2} +z_{4}\right)c \, \mathbf{\hat{z}} & \left(4c\right) & \mbox{O II} \\ 
\mathbf{B}_{15} & = & -x_{4} \, \mathbf{a}_{1} + \frac{3}{4} \, \mathbf{a}_{2}-z_{4} \, \mathbf{a}_{3} & = & -x_{4}a \, \mathbf{\hat{x}} + \frac{3}{4}b \, \mathbf{\hat{y}}-z_{4}c \, \mathbf{\hat{z}} & \left(4c\right) & \mbox{O II} \\ 
\mathbf{B}_{16} & = & \left(\frac{1}{2} +x_{4}\right) \, \mathbf{a}_{1} + \frac{1}{4} \, \mathbf{a}_{2} + \left(\frac{1}{2} - z_{4}\right) \, \mathbf{a}_{3} & = & \left(\frac{1}{2} +x_{4}\right)a \, \mathbf{\hat{x}} + \frac{1}{4}b \, \mathbf{\hat{y}} + \left(\frac{1}{2} - z_{4}\right)c \, \mathbf{\hat{z}} & \left(4c\right) & \mbox{O II} \\ 
\mathbf{B}_{17} & = & x_{5} \, \mathbf{a}_{1} + \frac{1}{4} \, \mathbf{a}_{2} + z_{5} \, \mathbf{a}_{3} & = & x_{5}a \, \mathbf{\hat{x}} + \frac{1}{4}b \, \mathbf{\hat{y}} + z_{5}c \, \mathbf{\hat{z}} & \left(4c\right) & \mbox{Si} \\ 
\mathbf{B}_{18} & = & \left(\frac{1}{2} - x_{5}\right) \, \mathbf{a}_{1} + \frac{3}{4} \, \mathbf{a}_{2} + \left(\frac{1}{2} +z_{5}\right) \, \mathbf{a}_{3} & = & \left(\frac{1}{2} - x_{5}\right)a \, \mathbf{\hat{x}} + \frac{3}{4}b \, \mathbf{\hat{y}} + \left(\frac{1}{2} +z_{5}\right)c \, \mathbf{\hat{z}} & \left(4c\right) & \mbox{Si} \\ 
\mathbf{B}_{19} & = & -x_{5} \, \mathbf{a}_{1} + \frac{3}{4} \, \mathbf{a}_{2}-z_{5} \, \mathbf{a}_{3} & = & -x_{5}a \, \mathbf{\hat{x}} + \frac{3}{4}b \, \mathbf{\hat{y}}-z_{5}c \, \mathbf{\hat{z}} & \left(4c\right) & \mbox{Si} \\ 
\mathbf{B}_{20} & = & \left(\frac{1}{2} +x_{5}\right) \, \mathbf{a}_{1} + \frac{1}{4} \, \mathbf{a}_{2} + \left(\frac{1}{2} - z_{5}\right) \, \mathbf{a}_{3} & = & \left(\frac{1}{2} +x_{5}\right)a \, \mathbf{\hat{x}} + \frac{1}{4}b \, \mathbf{\hat{y}} + \left(\frac{1}{2} - z_{5}\right)c \, \mathbf{\hat{z}} & \left(4c\right) & \mbox{Si} \\ 
\mathbf{B}_{21} & = & x_{6} \, \mathbf{a}_{1} + y_{6} \, \mathbf{a}_{2} + z_{6} \, \mathbf{a}_{3} & = & x_{6}a \, \mathbf{\hat{x}} + y_{6}b \, \mathbf{\hat{y}} + z_{6}c \, \mathbf{\hat{z}} & \left(8d\right) & \mbox{O III} \\ 
\mathbf{B}_{22} & = & \left(\frac{1}{2} - x_{6}\right) \, \mathbf{a}_{1}-y_{6} \, \mathbf{a}_{2} + \left(\frac{1}{2} +z_{6}\right) \, \mathbf{a}_{3} & = & \left(\frac{1}{2} - x_{6}\right)a \, \mathbf{\hat{x}}-y_{6}b \, \mathbf{\hat{y}} + \left(\frac{1}{2} +z_{6}\right)c \, \mathbf{\hat{z}} & \left(8d\right) & \mbox{O III} \\ 
\mathbf{B}_{23} & = & -x_{6} \, \mathbf{a}_{1} + \left(\frac{1}{2} +y_{6}\right) \, \mathbf{a}_{2}-z_{6} \, \mathbf{a}_{3} & = & -x_{6}a \, \mathbf{\hat{x}} + \left(\frac{1}{2} +y_{6}\right)b \, \mathbf{\hat{y}}-z_{6}c \, \mathbf{\hat{z}} & \left(8d\right) & \mbox{O III} \\ 
\mathbf{B}_{24} & = & \left(\frac{1}{2} +x_{6}\right) \, \mathbf{a}_{1} + \left(\frac{1}{2} - y_{6}\right) \, \mathbf{a}_{2} + \left(\frac{1}{2} - z_{6}\right) \, \mathbf{a}_{3} & = & \left(\frac{1}{2} +x_{6}\right)a \, \mathbf{\hat{x}} + \left(\frac{1}{2} - y_{6}\right)b \, \mathbf{\hat{y}} + \left(\frac{1}{2} - z_{6}\right)c \, \mathbf{\hat{z}} & \left(8d\right) & \mbox{O III} \\ 
\mathbf{B}_{25} & = & -x_{6} \, \mathbf{a}_{1}-y_{6} \, \mathbf{a}_{2}-z_{6} \, \mathbf{a}_{3} & = & -x_{6}a \, \mathbf{\hat{x}}-y_{6}b \, \mathbf{\hat{y}}-z_{6}c \, \mathbf{\hat{z}} & \left(8d\right) & \mbox{O III} \\ 
\mathbf{B}_{26} & = & \left(\frac{1}{2} +x_{6}\right) \, \mathbf{a}_{1} + y_{6} \, \mathbf{a}_{2} + \left(\frac{1}{2} - z_{6}\right) \, \mathbf{a}_{3} & = & \left(\frac{1}{2} +x_{6}\right)a \, \mathbf{\hat{x}} + y_{6}b \, \mathbf{\hat{y}} + \left(\frac{1}{2} - z_{6}\right)c \, \mathbf{\hat{z}} & \left(8d\right) & \mbox{O III} \\ 
\mathbf{B}_{27} & = & x_{6} \, \mathbf{a}_{1} + \left(\frac{1}{2} - y_{6}\right) \, \mathbf{a}_{2} + z_{6} \, \mathbf{a}_{3} & = & x_{6}a \, \mathbf{\hat{x}} + \left(\frac{1}{2} - y_{6}\right)b \, \mathbf{\hat{y}} + z_{6}c \, \mathbf{\hat{z}} & \left(8d\right) & \mbox{O III} \\ 
\mathbf{B}_{28} & = & \left(\frac{1}{2} - x_{6}\right) \, \mathbf{a}_{1} + \left(\frac{1}{2} +y_{6}\right) \, \mathbf{a}_{2} + \left(\frac{1}{2} +z_{6}\right) \, \mathbf{a}_{3} & = & \left(\frac{1}{2} - x_{6}\right)a \, \mathbf{\hat{x}} + \left(\frac{1}{2} +y_{6}\right)b \, \mathbf{\hat{y}} + \left(\frac{1}{2} +z_{6}\right)c \, \mathbf{\hat{z}} & \left(8d\right) & \mbox{O III} \\ 
\end{longtabu}
\renewcommand{\arraystretch}{1.0}
\noindent \hrulefill
\\
\textbf{References:}
\vspace*{-0.25cm}
\begin{flushleft}
  - \bibentry{Hazen_Am_Min_61_1976}. \\
\end{flushleft}
\noindent \hrulefill
\\
\textbf{Geometry files:}
\\
\noindent  - CIF: pp. {\hyperref[A2B4C_oP28_62_ac_2cd_c_cif]{\pageref{A2B4C_oP28_62_ac_2cd_c_cif}}} \\
\noindent  - POSCAR: pp. {\hyperref[A2B4C_oP28_62_ac_2cd_c_poscar]{\pageref{A2B4C_oP28_62_ac_2cd_c_poscar}}} \\
\onecolumn
{\phantomsection\label{A2B_oP12_62_2c_c}}
\subsection*{\huge \textbf{{\normalfont SrH$_{2}$ ($C29$) Structure: A2B\_oP12\_62\_2c\_c}}}
\noindent \hrulefill
\vspace*{0.25cm}
\begin{figure}[htp]
  \centering
  \vspace{-1em}
  {\includegraphics[width=1\textwidth]{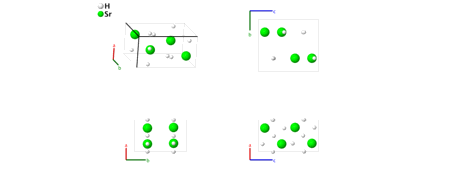}}
\end{figure}
\vspace*{-0.5cm}
\renewcommand{\arraystretch}{1.5}
\begin{equation*}
  \begin{array}{>{$\hspace{-0.15cm}}l<{$}>{$}p{0.5cm}<{$}>{$}p{18.5cm}<{$}}
    \mbox{\large \textbf{Prototype}} &\colon & \ce{SrH2} \\
    \mbox{\large \textbf{\AFLOW\ prototype label}} &\colon & \mbox{A2B\_oP12\_62\_2c\_c} \\
    \mbox{\large \textbf{\textit{Strukturbericht} designation}} &\colon & \mbox{$C29$} \\
    \mbox{\large \textbf{Pearson symbol}} &\colon & \mbox{oP12} \\
    \mbox{\large \textbf{Space group number}} &\colon & 62 \\
    \mbox{\large \textbf{Space group symbol}} &\colon & Pnma \\
    \mbox{\large \textbf{\AFLOW\ prototype command}} &\colon &  \texttt{aflow} \,  \, \texttt{-{}-proto=A2B\_oP12\_62\_2c\_c } \, \newline \texttt{-{}-params=}{a,b/a,c/a,x_{1},z_{1},x_{2},z_{2},x_{3},z_{3} }
  \end{array}
\end{equation*}
\renewcommand{\arraystretch}{1.0}

\vspace*{-0.25cm}
\noindent \hrulefill
\\
\textbf{ Other compounds with this structure:}
\begin{itemize}
   \item{ CaH$_{2}$  }
\end{itemize}
\noindent \parbox{1 \linewidth}{
\noindent \hrulefill
\\
\textbf{Simple Orthorhombic primitive vectors:} \\
\vspace*{-0.25cm}
\begin{tabular}{cc}
  \begin{tabular}{c}
    \parbox{0.6 \linewidth}{
      \renewcommand{\arraystretch}{1.5}
      \begin{equation*}
        \centering
        \begin{array}{ccc}
              \mathbf{a}_1 & = & a \, \mathbf{\hat{x}} \\
    \mathbf{a}_2 & = & b \, \mathbf{\hat{y}} \\
    \mathbf{a}_3 & = & c \, \mathbf{\hat{z}} \\

        \end{array}
      \end{equation*}
    }
    \renewcommand{\arraystretch}{1.0}
  \end{tabular}
  \begin{tabular}{c}
    \includegraphics[width=0.3\linewidth]{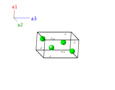} \\
  \end{tabular}
\end{tabular}

}
\vspace*{-0.25cm}

\noindent \hrulefill
\\
\textbf{Basis vectors:}
\vspace*{-0.25cm}
\renewcommand{\arraystretch}{1.5}
\begin{longtabu} to \textwidth{>{\centering $}X[-1,c,c]<{$}>{\centering $}X[-1,c,c]<{$}>{\centering $}X[-1,c,c]<{$}>{\centering $}X[-1,c,c]<{$}>{\centering $}X[-1,c,c]<{$}>{\centering $}X[-1,c,c]<{$}>{\centering $}X[-1,c,c]<{$}}
  & & \mbox{Lattice Coordinates} & & \mbox{Cartesian Coordinates} &\mbox{Wyckoff Position} & \mbox{Atom Type} \\  
  \mathbf{B}_{1} & = & x_{1} \, \mathbf{a}_{1} + \frac{1}{4} \, \mathbf{a}_{2} + z_{1} \, \mathbf{a}_{3} & = & x_{1}a \, \mathbf{\hat{x}} + \frac{1}{4}b \, \mathbf{\hat{y}} + z_{1}c \, \mathbf{\hat{z}} & \left(4c\right) & \mbox{H I} \\ 
\mathbf{B}_{2} & = & \left(\frac{1}{2} - x_{1}\right) \, \mathbf{a}_{1} + \frac{3}{4} \, \mathbf{a}_{2} + \left(\frac{1}{2} +z_{1}\right) \, \mathbf{a}_{3} & = & \left(\frac{1}{2} - x_{1}\right)a \, \mathbf{\hat{x}} + \frac{3}{4}b \, \mathbf{\hat{y}} + \left(\frac{1}{2} +z_{1}\right)c \, \mathbf{\hat{z}} & \left(4c\right) & \mbox{H I} \\ 
\mathbf{B}_{3} & = & -x_{1} \, \mathbf{a}_{1} + \frac{3}{4} \, \mathbf{a}_{2}-z_{1} \, \mathbf{a}_{3} & = & -x_{1}a \, \mathbf{\hat{x}} + \frac{3}{4}b \, \mathbf{\hat{y}}-z_{1}c \, \mathbf{\hat{z}} & \left(4c\right) & \mbox{H I} \\ 
\mathbf{B}_{4} & = & \left(\frac{1}{2} +x_{1}\right) \, \mathbf{a}_{1} + \frac{1}{4} \, \mathbf{a}_{2} + \left(\frac{1}{2} - z_{1}\right) \, \mathbf{a}_{3} & = & \left(\frac{1}{2} +x_{1}\right)a \, \mathbf{\hat{x}} + \frac{1}{4}b \, \mathbf{\hat{y}} + \left(\frac{1}{2} - z_{1}\right)c \, \mathbf{\hat{z}} & \left(4c\right) & \mbox{H I} \\ 
\mathbf{B}_{5} & = & x_{2} \, \mathbf{a}_{1} + \frac{1}{4} \, \mathbf{a}_{2} + z_{2} \, \mathbf{a}_{3} & = & x_{2}a \, \mathbf{\hat{x}} + \frac{1}{4}b \, \mathbf{\hat{y}} + z_{2}c \, \mathbf{\hat{z}} & \left(4c\right) & \mbox{H II} \\ 
\mathbf{B}_{6} & = & \left(\frac{1}{2} - x_{2}\right) \, \mathbf{a}_{1} + \frac{3}{4} \, \mathbf{a}_{2} + \left(\frac{1}{2} +z_{2}\right) \, \mathbf{a}_{3} & = & \left(\frac{1}{2} - x_{2}\right)a \, \mathbf{\hat{x}} + \frac{3}{4}b \, \mathbf{\hat{y}} + \left(\frac{1}{2} +z_{2}\right)c \, \mathbf{\hat{z}} & \left(4c\right) & \mbox{H II} \\ 
\mathbf{B}_{7} & = & -x_{2} \, \mathbf{a}_{1} + \frac{3}{4} \, \mathbf{a}_{2}-z_{2} \, \mathbf{a}_{3} & = & -x_{2}a \, \mathbf{\hat{x}} + \frac{3}{4}b \, \mathbf{\hat{y}}-z_{2}c \, \mathbf{\hat{z}} & \left(4c\right) & \mbox{H II} \\ 
\mathbf{B}_{8} & = & \left(\frac{1}{2} +x_{2}\right) \, \mathbf{a}_{1} + \frac{1}{4} \, \mathbf{a}_{2} + \left(\frac{1}{2} - z_{2}\right) \, \mathbf{a}_{3} & = & \left(\frac{1}{2} +x_{2}\right)a \, \mathbf{\hat{x}} + \frac{1}{4}b \, \mathbf{\hat{y}} + \left(\frac{1}{2} - z_{2}\right)c \, \mathbf{\hat{z}} & \left(4c\right) & \mbox{H II} \\ 
\mathbf{B}_{9} & = & x_{3} \, \mathbf{a}_{1} + \frac{1}{4} \, \mathbf{a}_{2} + z_{3} \, \mathbf{a}_{3} & = & x_{3}a \, \mathbf{\hat{x}} + \frac{1}{4}b \, \mathbf{\hat{y}} + z_{3}c \, \mathbf{\hat{z}} & \left(4c\right) & \mbox{Sr} \\ 
\mathbf{B}_{10} & = & \left(\frac{1}{2} - x_{3}\right) \, \mathbf{a}_{1} + \frac{3}{4} \, \mathbf{a}_{2} + \left(\frac{1}{2} +z_{3}\right) \, \mathbf{a}_{3} & = & \left(\frac{1}{2} - x_{3}\right)a \, \mathbf{\hat{x}} + \frac{3}{4}b \, \mathbf{\hat{y}} + \left(\frac{1}{2} +z_{3}\right)c \, \mathbf{\hat{z}} & \left(4c\right) & \mbox{Sr} \\ 
\mathbf{B}_{11} & = & -x_{3} \, \mathbf{a}_{1} + \frac{3}{4} \, \mathbf{a}_{2}-z_{3} \, \mathbf{a}_{3} & = & -x_{3}a \, \mathbf{\hat{x}} + \frac{3}{4}b \, \mathbf{\hat{y}}-z_{3}c \, \mathbf{\hat{z}} & \left(4c\right) & \mbox{Sr} \\ 
\mathbf{B}_{12} & = & \left(\frac{1}{2} +x_{3}\right) \, \mathbf{a}_{1} + \frac{1}{4} \, \mathbf{a}_{2} + \left(\frac{1}{2} - z_{3}\right) \, \mathbf{a}_{3} & = & \left(\frac{1}{2} +x_{3}\right)a \, \mathbf{\hat{x}} + \frac{1}{4}b \, \mathbf{\hat{y}} + \left(\frac{1}{2} - z_{3}\right)c \, \mathbf{\hat{z}} & \left(4c\right) & \mbox{Sr} \\ 
\end{longtabu}
\renewcommand{\arraystretch}{1.0}
\noindent \hrulefill
\\
\textbf{References:}
\vspace*{-0.25cm}
\begin{flushleft}
  - \bibentry{Roop_Encyc_2013}. \\
\end{flushleft}
\noindent \hrulefill
\\
\textbf{Geometry files:}
\\
\noindent  - CIF: pp. {\hyperref[A2B_oP12_62_2c_c_cif]{\pageref{A2B_oP12_62_2c_c_cif}}} \\
\noindent  - POSCAR: pp. {\hyperref[A2B_oP12_62_2c_c_poscar]{\pageref{A2B_oP12_62_2c_c_poscar}}} \\
\onecolumn
{\phantomsection\label{A3B_oP16_62_cd_c}}
\subsection*{\huge \textbf{{\normalfont $\epsilon$-NiAl$_{3}$ ($D0_{20}$) Structure: A3B\_oP16\_62\_cd\_c}}}
\noindent \hrulefill
\vspace*{0.25cm}
\begin{figure}[htp]
  \centering
  \vspace{-1em}
  {\includegraphics[width=1\textwidth]{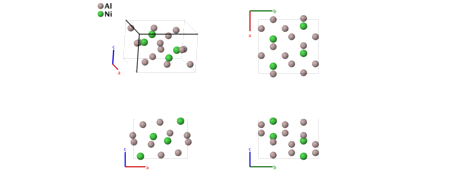}}
\end{figure}
\vspace*{-0.5cm}
\renewcommand{\arraystretch}{1.5}
\begin{equation*}
  \begin{array}{>{$\hspace{-0.15cm}}l<{$}>{$}p{0.5cm}<{$}>{$}p{18.5cm}<{$}}
    \mbox{\large \textbf{Prototype}} &\colon & \ce{$\epsilon$-NiAl3} \\
    \mbox{\large \textbf{\AFLOW\ prototype label}} &\colon & \mbox{A3B\_oP16\_62\_cd\_c} \\
    \mbox{\large \textbf{\textit{Strukturbericht} designation}} &\colon & \mbox{$D0_{20}$} \\
    \mbox{\large \textbf{Pearson symbol}} &\colon & \mbox{oP16} \\
    \mbox{\large \textbf{Space group number}} &\colon & 62 \\
    \mbox{\large \textbf{Space group symbol}} &\colon & Pnma \\
    \mbox{\large \textbf{\AFLOW\ prototype command}} &\colon &  \texttt{aflow} \,  \, \texttt{-{}-proto=A3B\_oP16\_62\_cd\_c } \, \newline \texttt{-{}-params=}{a,b/a,c/a,x_{1},z_{1},x_{2},z_{2},x_{3},y_{3},z_{3} }
  \end{array}
\end{equation*}
\renewcommand{\arraystretch}{1.0}

\noindent \parbox{1 \linewidth}{
\noindent \hrulefill
\\
\textbf{Simple Orthorhombic primitive vectors:} \\
\vspace*{-0.25cm}
\begin{tabular}{cc}
  \begin{tabular}{c}
    \parbox{0.6 \linewidth}{
      \renewcommand{\arraystretch}{1.5}
      \begin{equation*}
        \centering
        \begin{array}{ccc}
              \mathbf{a}_1 & = & a \, \mathbf{\hat{x}} \\
    \mathbf{a}_2 & = & b \, \mathbf{\hat{y}} \\
    \mathbf{a}_3 & = & c \, \mathbf{\hat{z}} \\

        \end{array}
      \end{equation*}
    }
    \renewcommand{\arraystretch}{1.0}
  \end{tabular}
  \begin{tabular}{c}
    \includegraphics[width=0.3\linewidth]{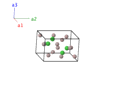} \\
  \end{tabular}
\end{tabular}

}
\vspace*{-0.25cm}

\noindent \hrulefill
\\
\textbf{Basis vectors:}
\vspace*{-0.25cm}
\renewcommand{\arraystretch}{1.5}
\begin{longtabu} to \textwidth{>{\centering $}X[-1,c,c]<{$}>{\centering $}X[-1,c,c]<{$}>{\centering $}X[-1,c,c]<{$}>{\centering $}X[-1,c,c]<{$}>{\centering $}X[-1,c,c]<{$}>{\centering $}X[-1,c,c]<{$}>{\centering $}X[-1,c,c]<{$}}
  & & \mbox{Lattice Coordinates} & & \mbox{Cartesian Coordinates} &\mbox{Wyckoff Position} & \mbox{Atom Type} \\  
  \mathbf{B}_{1} & = & x_{1} \, \mathbf{a}_{1} + \frac{1}{4} \, \mathbf{a}_{2} + z_{1} \, \mathbf{a}_{3} & = & x_{1}a \, \mathbf{\hat{x}} + \frac{1}{4}b \, \mathbf{\hat{y}} + z_{1}c \, \mathbf{\hat{z}} & \left(4c\right) & \mbox{Al I} \\ 
\mathbf{B}_{2} & = & \left(\frac{1}{2} - x_{1}\right) \, \mathbf{a}_{1} + \frac{3}{4} \, \mathbf{a}_{2} + \left(\frac{1}{2} +z_{1}\right) \, \mathbf{a}_{3} & = & \left(\frac{1}{2} - x_{1}\right)a \, \mathbf{\hat{x}} + \frac{3}{4}b \, \mathbf{\hat{y}} + \left(\frac{1}{2} +z_{1}\right)c \, \mathbf{\hat{z}} & \left(4c\right) & \mbox{Al I} \\ 
\mathbf{B}_{3} & = & -x_{1} \, \mathbf{a}_{1} + \frac{3}{4} \, \mathbf{a}_{2}-z_{1} \, \mathbf{a}_{3} & = & -x_{1}a \, \mathbf{\hat{x}} + \frac{3}{4}b \, \mathbf{\hat{y}}-z_{1}c \, \mathbf{\hat{z}} & \left(4c\right) & \mbox{Al I} \\ 
\mathbf{B}_{4} & = & \left(\frac{1}{2} +x_{1}\right) \, \mathbf{a}_{1} + \frac{1}{4} \, \mathbf{a}_{2} + \left(\frac{1}{2} - z_{1}\right) \, \mathbf{a}_{3} & = & \left(\frac{1}{2} +x_{1}\right)a \, \mathbf{\hat{x}} + \frac{1}{4}b \, \mathbf{\hat{y}} + \left(\frac{1}{2} - z_{1}\right)c \, \mathbf{\hat{z}} & \left(4c\right) & \mbox{Al I} \\ 
\mathbf{B}_{5} & = & x_{2} \, \mathbf{a}_{1} + \frac{1}{4} \, \mathbf{a}_{2} + z_{2} \, \mathbf{a}_{3} & = & x_{2}a \, \mathbf{\hat{x}} + \frac{1}{4}b \, \mathbf{\hat{y}} + z_{2}c \, \mathbf{\hat{z}} & \left(4c\right) & \mbox{Ni} \\ 
\mathbf{B}_{6} & = & \left(\frac{1}{2} - x_{2}\right) \, \mathbf{a}_{1} + \frac{3}{4} \, \mathbf{a}_{2} + \left(\frac{1}{2} +z_{2}\right) \, \mathbf{a}_{3} & = & \left(\frac{1}{2} - x_{2}\right)a \, \mathbf{\hat{x}} + \frac{3}{4}b \, \mathbf{\hat{y}} + \left(\frac{1}{2} +z_{2}\right)c \, \mathbf{\hat{z}} & \left(4c\right) & \mbox{Ni} \\ 
\mathbf{B}_{7} & = & -x_{2} \, \mathbf{a}_{1} + \frac{3}{4} \, \mathbf{a}_{2}-z_{2} \, \mathbf{a}_{3} & = & -x_{2}a \, \mathbf{\hat{x}} + \frac{3}{4}b \, \mathbf{\hat{y}}-z_{2}c \, \mathbf{\hat{z}} & \left(4c\right) & \mbox{Ni} \\ 
\mathbf{B}_{8} & = & \left(\frac{1}{2} +x_{2}\right) \, \mathbf{a}_{1} + \frac{1}{4} \, \mathbf{a}_{2} + \left(\frac{1}{2} - z_{2}\right) \, \mathbf{a}_{3} & = & \left(\frac{1}{2} +x_{2}\right)a \, \mathbf{\hat{x}} + \frac{1}{4}b \, \mathbf{\hat{y}} + \left(\frac{1}{2} - z_{2}\right)c \, \mathbf{\hat{z}} & \left(4c\right) & \mbox{Ni} \\ 
\mathbf{B}_{9} & = & x_{3} \, \mathbf{a}_{1} + y_{3} \, \mathbf{a}_{2} + z_{3} \, \mathbf{a}_{3} & = & x_{3}a \, \mathbf{\hat{x}} + y_{3}b \, \mathbf{\hat{y}} + z_{3}c \, \mathbf{\hat{z}} & \left(8d\right) & \mbox{Al II} \\ 
\mathbf{B}_{10} & = & \left(\frac{1}{2} - x_{3}\right) \, \mathbf{a}_{1}-y_{3} \, \mathbf{a}_{2} + \left(\frac{1}{2} +z_{3}\right) \, \mathbf{a}_{3} & = & \left(\frac{1}{2} - x_{3}\right)a \, \mathbf{\hat{x}}-y_{3}b \, \mathbf{\hat{y}} + \left(\frac{1}{2} +z_{3}\right)c \, \mathbf{\hat{z}} & \left(8d\right) & \mbox{Al II} \\ 
\mathbf{B}_{11} & = & -x_{3} \, \mathbf{a}_{1} + \left(\frac{1}{2} +y_{3}\right) \, \mathbf{a}_{2}-z_{3} \, \mathbf{a}_{3} & = & -x_{3}a \, \mathbf{\hat{x}} + \left(\frac{1}{2} +y_{3}\right)b \, \mathbf{\hat{y}}-z_{3}c \, \mathbf{\hat{z}} & \left(8d\right) & \mbox{Al II} \\ 
\mathbf{B}_{12} & = & \left(\frac{1}{2} +x_{3}\right) \, \mathbf{a}_{1} + \left(\frac{1}{2} - y_{3}\right) \, \mathbf{a}_{2} + \left(\frac{1}{2} - z_{3}\right) \, \mathbf{a}_{3} & = & \left(\frac{1}{2} +x_{3}\right)a \, \mathbf{\hat{x}} + \left(\frac{1}{2} - y_{3}\right)b \, \mathbf{\hat{y}} + \left(\frac{1}{2} - z_{3}\right)c \, \mathbf{\hat{z}} & \left(8d\right) & \mbox{Al II} \\ 
\mathbf{B}_{13} & = & -x_{3} \, \mathbf{a}_{1}-y_{3} \, \mathbf{a}_{2}-z_{3} \, \mathbf{a}_{3} & = & -x_{3}a \, \mathbf{\hat{x}}-y_{3}b \, \mathbf{\hat{y}}-z_{3}c \, \mathbf{\hat{z}} & \left(8d\right) & \mbox{Al II} \\ 
\mathbf{B}_{14} & = & \left(\frac{1}{2} +x_{3}\right) \, \mathbf{a}_{1} + y_{3} \, \mathbf{a}_{2} + \left(\frac{1}{2} - z_{3}\right) \, \mathbf{a}_{3} & = & \left(\frac{1}{2} +x_{3}\right)a \, \mathbf{\hat{x}} + y_{3}b \, \mathbf{\hat{y}} + \left(\frac{1}{2} - z_{3}\right)c \, \mathbf{\hat{z}} & \left(8d\right) & \mbox{Al II} \\ 
\mathbf{B}_{15} & = & x_{3} \, \mathbf{a}_{1} + \left(\frac{1}{2} - y_{3}\right) \, \mathbf{a}_{2} + z_{3} \, \mathbf{a}_{3} & = & x_{3}a \, \mathbf{\hat{x}} + \left(\frac{1}{2} - y_{3}\right)b \, \mathbf{\hat{y}} + z_{3}c \, \mathbf{\hat{z}} & \left(8d\right) & \mbox{Al II} \\ 
\mathbf{B}_{16} & = & \left(\frac{1}{2} - x_{3}\right) \, \mathbf{a}_{1} + \left(\frac{1}{2} +y_{3}\right) \, \mathbf{a}_{2} + \left(\frac{1}{2} +z_{3}\right) \, \mathbf{a}_{3} & = & \left(\frac{1}{2} - x_{3}\right)a \, \mathbf{\hat{x}} + \left(\frac{1}{2} +y_{3}\right)b \, \mathbf{\hat{y}} + \left(\frac{1}{2} +z_{3}\right)c \, \mathbf{\hat{z}} & \left(8d\right) & \mbox{Al II} \\ 
\end{longtabu}
\renewcommand{\arraystretch}{1.0}
\noindent \hrulefill
\\
\textbf{References:}
\vspace*{-0.25cm}
\begin{flushleft}
  - \bibentry{bradley37:D0_20}. \\
\end{flushleft}
\textbf{Found in:}
\vspace*{-0.25cm}
\begin{flushleft}
  - \bibentry{Pearson_NRC_1958}. \\
\end{flushleft}
\noindent \hrulefill
\\
\textbf{Geometry files:}
\\
\noindent  - CIF: pp. {\hyperref[A3B_oP16_62_cd_c_cif]{\pageref{A3B_oP16_62_cd_c_cif}}} \\
\noindent  - POSCAR: pp. {\hyperref[A3B_oP16_62_cd_c_poscar]{\pageref{A3B_oP16_62_cd_c_poscar}}} \\
\onecolumn
{\phantomsection\label{AB2C3_oP24_62_c_d_cd}}
\subsection*{\huge \textbf{{\normalfont \begin{raggedleft}Cubanite (CuFe$_{2}$S$_{3}$, $E9_{e}$) Structure: \end{raggedleft} \\ AB2C3\_oP24\_62\_c\_d\_cd}}}
\noindent \hrulefill
\vspace*{0.25cm}
\begin{figure}[htp]
  \centering
  \vspace{-1em}
  {\includegraphics[width=1\textwidth]{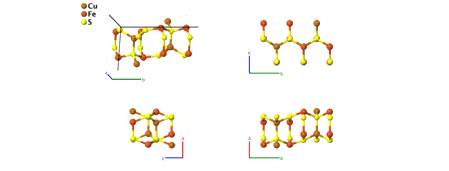}}
\end{figure}
\vspace*{-0.5cm}
\renewcommand{\arraystretch}{1.5}
\begin{equation*}
  \begin{array}{>{$\hspace{-0.15cm}}l<{$}>{$}p{0.5cm}<{$}>{$}p{18.5cm}<{$}}
    \mbox{\large \textbf{Prototype}} &\colon & \ce{CuFe2S3} \\
    \mbox{\large \textbf{\AFLOW\ prototype label}} &\colon & \mbox{AB2C3\_oP24\_62\_c\_d\_cd} \\
    \mbox{\large \textbf{\textit{Strukturbericht} designation}} &\colon & \mbox{$E9_{e}$} \\
    \mbox{\large \textbf{Pearson symbol}} &\colon & \mbox{oP24} \\
    \mbox{\large \textbf{Space group number}} &\colon & 62 \\
    \mbox{\large \textbf{Space group symbol}} &\colon & Pnma \\
    \mbox{\large \textbf{\AFLOW\ prototype command}} &\colon &  \texttt{aflow} \,  \, \texttt{-{}-proto=AB2C3\_oP24\_62\_c\_d\_cd } \, \newline \texttt{-{}-params=}{a,b/a,c/a,x_{1},z_{1},x_{2},z_{2},x_{3},y_{3},z_{3},x_{4},y_{4},z_{4} }
  \end{array}
\end{equation*}
\renewcommand{\arraystretch}{1.0}

\vspace*{-0.25cm}
\noindent \hrulefill
\begin{itemize}
  \item{(Szyma\'{n}ski, 1974) uses the $Pcmn$ orientation of space group \#62
to describe this structure.  We have swapped the $\hat{{\bf x}}$ and
$\hat{\bf{z}}$ axis to transform this into the standard $Pnma$
orientation.
}
\end{itemize}

\noindent \parbox{1 \linewidth}{
\noindent \hrulefill
\\
\textbf{Simple Orthorhombic primitive vectors:} \\
\vspace*{-0.25cm}
\begin{tabular}{cc}
  \begin{tabular}{c}
    \parbox{0.6 \linewidth}{
      \renewcommand{\arraystretch}{1.5}
      \begin{equation*}
        \centering
        \begin{array}{ccc}
              \mathbf{a}_1 & = & a \, \mathbf{\hat{x}} \\
    \mathbf{a}_2 & = & b \, \mathbf{\hat{y}} \\
    \mathbf{a}_3 & = & c \, \mathbf{\hat{z}} \\

        \end{array}
      \end{equation*}
    }
    \renewcommand{\arraystretch}{1.0}
  \end{tabular}
  \begin{tabular}{c}
    \includegraphics[width=0.3\linewidth]{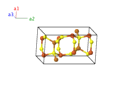} \\
  \end{tabular}
\end{tabular}

}
\vspace*{-0.25cm}

\noindent \hrulefill
\\
\textbf{Basis vectors:}
\vspace*{-0.25cm}
\renewcommand{\arraystretch}{1.5}
\begin{longtabu} to \textwidth{>{\centering $}X[-1,c,c]<{$}>{\centering $}X[-1,c,c]<{$}>{\centering $}X[-1,c,c]<{$}>{\centering $}X[-1,c,c]<{$}>{\centering $}X[-1,c,c]<{$}>{\centering $}X[-1,c,c]<{$}>{\centering $}X[-1,c,c]<{$}}
  & & \mbox{Lattice Coordinates} & & \mbox{Cartesian Coordinates} &\mbox{Wyckoff Position} & \mbox{Atom Type} \\  
  \mathbf{B}_{1} & = & x_{1} \, \mathbf{a}_{1} + \frac{1}{4} \, \mathbf{a}_{2} + z_{1} \, \mathbf{a}_{3} & = & x_{1}a \, \mathbf{\hat{x}} + \frac{1}{4}b \, \mathbf{\hat{y}} + z_{1}c \, \mathbf{\hat{z}} & \left(4c\right) & \mbox{Cu} \\ 
\mathbf{B}_{2} & = & \left(\frac{1}{2} - x_{1}\right) \, \mathbf{a}_{1} + \frac{3}{4} \, \mathbf{a}_{2} + \left(\frac{1}{2} +z_{1}\right) \, \mathbf{a}_{3} & = & \left(\frac{1}{2} - x_{1}\right)a \, \mathbf{\hat{x}} + \frac{3}{4}b \, \mathbf{\hat{y}} + \left(\frac{1}{2} +z_{1}\right)c \, \mathbf{\hat{z}} & \left(4c\right) & \mbox{Cu} \\ 
\mathbf{B}_{3} & = & -x_{1} \, \mathbf{a}_{1} + \frac{3}{4} \, \mathbf{a}_{2}-z_{1} \, \mathbf{a}_{3} & = & -x_{1}a \, \mathbf{\hat{x}} + \frac{3}{4}b \, \mathbf{\hat{y}}-z_{1}c \, \mathbf{\hat{z}} & \left(4c\right) & \mbox{Cu} \\ 
\mathbf{B}_{4} & = & \left(\frac{1}{2} +x_{1}\right) \, \mathbf{a}_{1} + \frac{1}{4} \, \mathbf{a}_{2} + \left(\frac{1}{2} - z_{1}\right) \, \mathbf{a}_{3} & = & \left(\frac{1}{2} +x_{1}\right)a \, \mathbf{\hat{x}} + \frac{1}{4}b \, \mathbf{\hat{y}} + \left(\frac{1}{2} - z_{1}\right)c \, \mathbf{\hat{z}} & \left(4c\right) & \mbox{Cu} \\ 
\mathbf{B}_{5} & = & x_{2} \, \mathbf{a}_{1} + \frac{1}{4} \, \mathbf{a}_{2} + z_{2} \, \mathbf{a}_{3} & = & x_{2}a \, \mathbf{\hat{x}} + \frac{1}{4}b \, \mathbf{\hat{y}} + z_{2}c \, \mathbf{\hat{z}} & \left(4c\right) & \mbox{S I} \\ 
\mathbf{B}_{6} & = & \left(\frac{1}{2} - x_{2}\right) \, \mathbf{a}_{1} + \frac{3}{4} \, \mathbf{a}_{2} + \left(\frac{1}{2} +z_{2}\right) \, \mathbf{a}_{3} & = & \left(\frac{1}{2} - x_{2}\right)a \, \mathbf{\hat{x}} + \frac{3}{4}b \, \mathbf{\hat{y}} + \left(\frac{1}{2} +z_{2}\right)c \, \mathbf{\hat{z}} & \left(4c\right) & \mbox{S I} \\ 
\mathbf{B}_{7} & = & -x_{2} \, \mathbf{a}_{1} + \frac{3}{4} \, \mathbf{a}_{2}-z_{2} \, \mathbf{a}_{3} & = & -x_{2}a \, \mathbf{\hat{x}} + \frac{3}{4}b \, \mathbf{\hat{y}}-z_{2}c \, \mathbf{\hat{z}} & \left(4c\right) & \mbox{S I} \\ 
\mathbf{B}_{8} & = & \left(\frac{1}{2} +x_{2}\right) \, \mathbf{a}_{1} + \frac{1}{4} \, \mathbf{a}_{2} + \left(\frac{1}{2} - z_{2}\right) \, \mathbf{a}_{3} & = & \left(\frac{1}{2} +x_{2}\right)a \, \mathbf{\hat{x}} + \frac{1}{4}b \, \mathbf{\hat{y}} + \left(\frac{1}{2} - z_{2}\right)c \, \mathbf{\hat{z}} & \left(4c\right) & \mbox{S I} \\ 
\mathbf{B}_{9} & = & x_{3} \, \mathbf{a}_{1} + y_{3} \, \mathbf{a}_{2} + z_{3} \, \mathbf{a}_{3} & = & x_{3}a \, \mathbf{\hat{x}} + y_{3}b \, \mathbf{\hat{y}} + z_{3}c \, \mathbf{\hat{z}} & \left(8d\right) & \mbox{Fe} \\ 
\mathbf{B}_{10} & = & \left(\frac{1}{2} - x_{3}\right) \, \mathbf{a}_{1}-y_{3} \, \mathbf{a}_{2} + \left(\frac{1}{2} +z_{3}\right) \, \mathbf{a}_{3} & = & \left(\frac{1}{2} - x_{3}\right)a \, \mathbf{\hat{x}}-y_{3}b \, \mathbf{\hat{y}} + \left(\frac{1}{2} +z_{3}\right)c \, \mathbf{\hat{z}} & \left(8d\right) & \mbox{Fe} \\ 
\mathbf{B}_{11} & = & -x_{3} \, \mathbf{a}_{1} + \left(\frac{1}{2} +y_{3}\right) \, \mathbf{a}_{2}-z_{3} \, \mathbf{a}_{3} & = & -x_{3}a \, \mathbf{\hat{x}} + \left(\frac{1}{2} +y_{3}\right)b \, \mathbf{\hat{y}}-z_{3}c \, \mathbf{\hat{z}} & \left(8d\right) & \mbox{Fe} \\ 
\mathbf{B}_{12} & = & \left(\frac{1}{2} +x_{3}\right) \, \mathbf{a}_{1} + \left(\frac{1}{2} - y_{3}\right) \, \mathbf{a}_{2} + \left(\frac{1}{2} - z_{3}\right) \, \mathbf{a}_{3} & = & \left(\frac{1}{2} +x_{3}\right)a \, \mathbf{\hat{x}} + \left(\frac{1}{2} - y_{3}\right)b \, \mathbf{\hat{y}} + \left(\frac{1}{2} - z_{3}\right)c \, \mathbf{\hat{z}} & \left(8d\right) & \mbox{Fe} \\ 
\mathbf{B}_{13} & = & -x_{3} \, \mathbf{a}_{1}-y_{3} \, \mathbf{a}_{2}-z_{3} \, \mathbf{a}_{3} & = & -x_{3}a \, \mathbf{\hat{x}}-y_{3}b \, \mathbf{\hat{y}}-z_{3}c \, \mathbf{\hat{z}} & \left(8d\right) & \mbox{Fe} \\ 
\mathbf{B}_{14} & = & \left(\frac{1}{2} +x_{3}\right) \, \mathbf{a}_{1} + y_{3} \, \mathbf{a}_{2} + \left(\frac{1}{2} - z_{3}\right) \, \mathbf{a}_{3} & = & \left(\frac{1}{2} +x_{3}\right)a \, \mathbf{\hat{x}} + y_{3}b \, \mathbf{\hat{y}} + \left(\frac{1}{2} - z_{3}\right)c \, \mathbf{\hat{z}} & \left(8d\right) & \mbox{Fe} \\ 
\mathbf{B}_{15} & = & x_{3} \, \mathbf{a}_{1} + \left(\frac{1}{2} - y_{3}\right) \, \mathbf{a}_{2} + z_{3} \, \mathbf{a}_{3} & = & x_{3}a \, \mathbf{\hat{x}} + \left(\frac{1}{2} - y_{3}\right)b \, \mathbf{\hat{y}} + z_{3}c \, \mathbf{\hat{z}} & \left(8d\right) & \mbox{Fe} \\ 
\mathbf{B}_{16} & = & \left(\frac{1}{2} - x_{3}\right) \, \mathbf{a}_{1} + \left(\frac{1}{2} +y_{3}\right) \, \mathbf{a}_{2} + \left(\frac{1}{2} +z_{3}\right) \, \mathbf{a}_{3} & = & \left(\frac{1}{2} - x_{3}\right)a \, \mathbf{\hat{x}} + \left(\frac{1}{2} +y_{3}\right)b \, \mathbf{\hat{y}} + \left(\frac{1}{2} +z_{3}\right)c \, \mathbf{\hat{z}} & \left(8d\right) & \mbox{Fe} \\ 
\mathbf{B}_{17} & = & x_{4} \, \mathbf{a}_{1} + y_{4} \, \mathbf{a}_{2} + z_{4} \, \mathbf{a}_{3} & = & x_{4}a \, \mathbf{\hat{x}} + y_{4}b \, \mathbf{\hat{y}} + z_{4}c \, \mathbf{\hat{z}} & \left(8d\right) & \mbox{S II} \\ 
\mathbf{B}_{18} & = & \left(\frac{1}{2} - x_{4}\right) \, \mathbf{a}_{1}-y_{4} \, \mathbf{a}_{2} + \left(\frac{1}{2} +z_{4}\right) \, \mathbf{a}_{3} & = & \left(\frac{1}{2} - x_{4}\right)a \, \mathbf{\hat{x}}-y_{4}b \, \mathbf{\hat{y}} + \left(\frac{1}{2} +z_{4}\right)c \, \mathbf{\hat{z}} & \left(8d\right) & \mbox{S II} \\ 
\mathbf{B}_{19} & = & -x_{4} \, \mathbf{a}_{1} + \left(\frac{1}{2} +y_{4}\right) \, \mathbf{a}_{2}-z_{4} \, \mathbf{a}_{3} & = & -x_{4}a \, \mathbf{\hat{x}} + \left(\frac{1}{2} +y_{4}\right)b \, \mathbf{\hat{y}}-z_{4}c \, \mathbf{\hat{z}} & \left(8d\right) & \mbox{S II} \\ 
\mathbf{B}_{20} & = & \left(\frac{1}{2} +x_{4}\right) \, \mathbf{a}_{1} + \left(\frac{1}{2} - y_{4}\right) \, \mathbf{a}_{2} + \left(\frac{1}{2} - z_{4}\right) \, \mathbf{a}_{3} & = & \left(\frac{1}{2} +x_{4}\right)a \, \mathbf{\hat{x}} + \left(\frac{1}{2} - y_{4}\right)b \, \mathbf{\hat{y}} + \left(\frac{1}{2} - z_{4}\right)c \, \mathbf{\hat{z}} & \left(8d\right) & \mbox{S II} \\ 
\mathbf{B}_{21} & = & -x_{4} \, \mathbf{a}_{1}-y_{4} \, \mathbf{a}_{2}-z_{4} \, \mathbf{a}_{3} & = & -x_{4}a \, \mathbf{\hat{x}}-y_{4}b \, \mathbf{\hat{y}}-z_{4}c \, \mathbf{\hat{z}} & \left(8d\right) & \mbox{S II} \\ 
\mathbf{B}_{22} & = & \left(\frac{1}{2} +x_{4}\right) \, \mathbf{a}_{1} + y_{4} \, \mathbf{a}_{2} + \left(\frac{1}{2} - z_{4}\right) \, \mathbf{a}_{3} & = & \left(\frac{1}{2} +x_{4}\right)a \, \mathbf{\hat{x}} + y_{4}b \, \mathbf{\hat{y}} + \left(\frac{1}{2} - z_{4}\right)c \, \mathbf{\hat{z}} & \left(8d\right) & \mbox{S II} \\ 
\mathbf{B}_{23} & = & x_{4} \, \mathbf{a}_{1} + \left(\frac{1}{2} - y_{4}\right) \, \mathbf{a}_{2} + z_{4} \, \mathbf{a}_{3} & = & x_{4}a \, \mathbf{\hat{x}} + \left(\frac{1}{2} - y_{4}\right)b \, \mathbf{\hat{y}} + z_{4}c \, \mathbf{\hat{z}} & \left(8d\right) & \mbox{S II} \\ 
\mathbf{B}_{24} & = & \left(\frac{1}{2} - x_{4}\right) \, \mathbf{a}_{1} + \left(\frac{1}{2} +y_{4}\right) \, \mathbf{a}_{2} + \left(\frac{1}{2} +z_{4}\right) \, \mathbf{a}_{3} & = & \left(\frac{1}{2} - x_{4}\right)a \, \mathbf{\hat{x}} + \left(\frac{1}{2} +y_{4}\right)b \, \mathbf{\hat{y}} + \left(\frac{1}{2} +z_{4}\right)c \, \mathbf{\hat{z}} & \left(8d\right) & \mbox{S II} \\ 
\end{longtabu}
\renewcommand{\arraystretch}{1.0}
\noindent \hrulefill
\\
\textbf{References:}
\vspace*{-0.25cm}
\begin{flushleft}
  - \bibentry{Szymanski74_Z_Krist_140_1974}. \\
\end{flushleft}
\noindent \hrulefill
\\
\textbf{Geometry files:}
\\
\noindent  - CIF: pp. {\hyperref[AB2C3_oP24_62_c_d_cd_cif]{\pageref{AB2C3_oP24_62_c_d_cd_cif}}} \\
\noindent  - POSCAR: pp. {\hyperref[AB2C3_oP24_62_c_d_cd_poscar]{\pageref{AB2C3_oP24_62_c_d_cd_poscar}}} \\
\onecolumn
{\phantomsection\label{AB3_oP16_62_c_3c}}
\subsection*{\huge \textbf{{\normalfont \begin{raggedleft}Molybdite (MoO$_{3}$, $D0_{8}$) Structure: \end{raggedleft} \\ AB3\_oP16\_62\_c\_3c}}}
\noindent \hrulefill
\vspace*{0.25cm}
\begin{figure}[htp]
  \centering
  \vspace{-1em}
  {\includegraphics[width=1\textwidth]{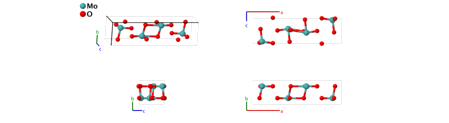}}
\end{figure}
\vspace*{-0.5cm}
\renewcommand{\arraystretch}{1.5}
\begin{equation*}
  \begin{array}{>{$\hspace{-0.15cm}}l<{$}>{$}p{0.5cm}<{$}>{$}p{18.5cm}<{$}}
    \mbox{\large \textbf{Prototype}} &\colon & \ce{MoO3} \\
    \mbox{\large \textbf{\AFLOW\ prototype label}} &\colon & \mbox{AB3\_oP16\_62\_c\_3c} \\
    \mbox{\large \textbf{\textit{Strukturbericht} designation}} &\colon & \mbox{$D0_{8}$} \\
    \mbox{\large \textbf{Pearson symbol}} &\colon & \mbox{oP16} \\
    \mbox{\large \textbf{Space group number}} &\colon & 62 \\
    \mbox{\large \textbf{Space group symbol}} &\colon & Pnma \\
    \mbox{\large \textbf{\AFLOW\ prototype command}} &\colon &  \texttt{aflow} \,  \, \texttt{-{}-proto=AB3\_oP16\_62\_c\_3c } \, \newline \texttt{-{}-params=}{a,b/a,c/a,x_{1},z_{1},x_{2},z_{2},x_{3},z_{3},x_{4},z_{4} }
  \end{array}
\end{equation*}
\renewcommand{\arraystretch}{1.0}

\vspace*{-0.25cm}
\noindent \hrulefill
\begin{itemize}
  \item{The unit cell and atomic positions were originally given in the $Pbnm$
orientation of space group \#62. We have rotated the crystal axis so
that $\mathbf{\hat{y}} \rightarrow \mathbf{\hat{x}} \rightarrow
\mathbf{\hat{z}}$ to put the system in the standard $Pnma$
representation.
}
\end{itemize}

\noindent \parbox{1 \linewidth}{
\noindent \hrulefill
\\
\textbf{Simple Orthorhombic primitive vectors:} \\
\vspace*{-0.25cm}
\begin{tabular}{cc}
  \begin{tabular}{c}
    \parbox{0.6 \linewidth}{
      \renewcommand{\arraystretch}{1.5}
      \begin{equation*}
        \centering
        \begin{array}{ccc}
              \mathbf{a}_1 & = & a \, \mathbf{\hat{x}} \\
    \mathbf{a}_2 & = & b \, \mathbf{\hat{y}} \\
    \mathbf{a}_3 & = & c \, \mathbf{\hat{z}} \\

        \end{array}
      \end{equation*}
    }
    \renewcommand{\arraystretch}{1.0}
  \end{tabular}
  \begin{tabular}{c}
    \includegraphics[width=0.3\linewidth]{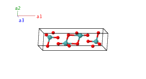} \\
  \end{tabular}
\end{tabular}

}
\vspace*{-0.25cm}

\noindent \hrulefill
\\
\textbf{Basis vectors:}
\vspace*{-0.25cm}
\renewcommand{\arraystretch}{1.5}
\begin{longtabu} to \textwidth{>{\centering $}X[-1,c,c]<{$}>{\centering $}X[-1,c,c]<{$}>{\centering $}X[-1,c,c]<{$}>{\centering $}X[-1,c,c]<{$}>{\centering $}X[-1,c,c]<{$}>{\centering $}X[-1,c,c]<{$}>{\centering $}X[-1,c,c]<{$}}
  & & \mbox{Lattice Coordinates} & & \mbox{Cartesian Coordinates} &\mbox{Wyckoff Position} & \mbox{Atom Type} \\  
  \mathbf{B}_{1} & = & x_{1} \, \mathbf{a}_{1} + \frac{1}{4} \, \mathbf{a}_{2} + z_{1} \, \mathbf{a}_{3} & = & x_{1}a \, \mathbf{\hat{x}} + \frac{1}{4}b \, \mathbf{\hat{y}} + z_{1}c \, \mathbf{\hat{z}} & \left(4c\right) & \mbox{Mo} \\ 
\mathbf{B}_{2} & = & \left(\frac{1}{2} - x_{1}\right) \, \mathbf{a}_{1} + \frac{3}{4} \, \mathbf{a}_{2} + \left(\frac{1}{2} +z_{1}\right) \, \mathbf{a}_{3} & = & \left(\frac{1}{2} - x_{1}\right)a \, \mathbf{\hat{x}} + \frac{3}{4}b \, \mathbf{\hat{y}} + \left(\frac{1}{2} +z_{1}\right)c \, \mathbf{\hat{z}} & \left(4c\right) & \mbox{Mo} \\ 
\mathbf{B}_{3} & = & -x_{1} \, \mathbf{a}_{1} + \frac{3}{4} \, \mathbf{a}_{2}-z_{1} \, \mathbf{a}_{3} & = & -x_{1}a \, \mathbf{\hat{x}} + \frac{3}{4}b \, \mathbf{\hat{y}}-z_{1}c \, \mathbf{\hat{z}} & \left(4c\right) & \mbox{Mo} \\ 
\mathbf{B}_{4} & = & \left(\frac{1}{2} +x_{1}\right) \, \mathbf{a}_{1} + \frac{1}{4} \, \mathbf{a}_{2} + \left(\frac{1}{2} - z_{1}\right) \, \mathbf{a}_{3} & = & \left(\frac{1}{2} +x_{1}\right)a \, \mathbf{\hat{x}} + \frac{1}{4}b \, \mathbf{\hat{y}} + \left(\frac{1}{2} - z_{1}\right)c \, \mathbf{\hat{z}} & \left(4c\right) & \mbox{Mo} \\ 
\mathbf{B}_{5} & = & x_{2} \, \mathbf{a}_{1} + \frac{1}{4} \, \mathbf{a}_{2} + z_{2} \, \mathbf{a}_{3} & = & x_{2}a \, \mathbf{\hat{x}} + \frac{1}{4}b \, \mathbf{\hat{y}} + z_{2}c \, \mathbf{\hat{z}} & \left(4c\right) & \mbox{O I} \\ 
\mathbf{B}_{6} & = & \left(\frac{1}{2} - x_{2}\right) \, \mathbf{a}_{1} + \frac{3}{4} \, \mathbf{a}_{2} + \left(\frac{1}{2} +z_{2}\right) \, \mathbf{a}_{3} & = & \left(\frac{1}{2} - x_{2}\right)a \, \mathbf{\hat{x}} + \frac{3}{4}b \, \mathbf{\hat{y}} + \left(\frac{1}{2} +z_{2}\right)c \, \mathbf{\hat{z}} & \left(4c\right) & \mbox{O I} \\ 
\mathbf{B}_{7} & = & -x_{2} \, \mathbf{a}_{1} + \frac{3}{4} \, \mathbf{a}_{2}-z_{2} \, \mathbf{a}_{3} & = & -x_{2}a \, \mathbf{\hat{x}} + \frac{3}{4}b \, \mathbf{\hat{y}}-z_{2}c \, \mathbf{\hat{z}} & \left(4c\right) & \mbox{O I} \\ 
\mathbf{B}_{8} & = & \left(\frac{1}{2} +x_{2}\right) \, \mathbf{a}_{1} + \frac{1}{4} \, \mathbf{a}_{2} + \left(\frac{1}{2} - z_{2}\right) \, \mathbf{a}_{3} & = & \left(\frac{1}{2} +x_{2}\right)a \, \mathbf{\hat{x}} + \frac{1}{4}b \, \mathbf{\hat{y}} + \left(\frac{1}{2} - z_{2}\right)c \, \mathbf{\hat{z}} & \left(4c\right) & \mbox{O I} \\ 
\mathbf{B}_{9} & = & x_{3} \, \mathbf{a}_{1} + \frac{1}{4} \, \mathbf{a}_{2} + z_{3} \, \mathbf{a}_{3} & = & x_{3}a \, \mathbf{\hat{x}} + \frac{1}{4}b \, \mathbf{\hat{y}} + z_{3}c \, \mathbf{\hat{z}} & \left(4c\right) & \mbox{O II} \\ 
\mathbf{B}_{10} & = & \left(\frac{1}{2} - x_{3}\right) \, \mathbf{a}_{1} + \frac{3}{4} \, \mathbf{a}_{2} + \left(\frac{1}{2} +z_{3}\right) \, \mathbf{a}_{3} & = & \left(\frac{1}{2} - x_{3}\right)a \, \mathbf{\hat{x}} + \frac{3}{4}b \, \mathbf{\hat{y}} + \left(\frac{1}{2} +z_{3}\right)c \, \mathbf{\hat{z}} & \left(4c\right) & \mbox{O II} \\ 
\mathbf{B}_{11} & = & -x_{3} \, \mathbf{a}_{1} + \frac{3}{4} \, \mathbf{a}_{2}-z_{3} \, \mathbf{a}_{3} & = & -x_{3}a \, \mathbf{\hat{x}} + \frac{3}{4}b \, \mathbf{\hat{y}}-z_{3}c \, \mathbf{\hat{z}} & \left(4c\right) & \mbox{O II} \\ 
\mathbf{B}_{12} & = & \left(\frac{1}{2} +x_{3}\right) \, \mathbf{a}_{1} + \frac{1}{4} \, \mathbf{a}_{2} + \left(\frac{1}{2} - z_{3}\right) \, \mathbf{a}_{3} & = & \left(\frac{1}{2} +x_{3}\right)a \, \mathbf{\hat{x}} + \frac{1}{4}b \, \mathbf{\hat{y}} + \left(\frac{1}{2} - z_{3}\right)c \, \mathbf{\hat{z}} & \left(4c\right) & \mbox{O II} \\ 
\mathbf{B}_{13} & = & x_{4} \, \mathbf{a}_{1} + \frac{1}{4} \, \mathbf{a}_{2} + z_{4} \, \mathbf{a}_{3} & = & x_{4}a \, \mathbf{\hat{x}} + \frac{1}{4}b \, \mathbf{\hat{y}} + z_{4}c \, \mathbf{\hat{z}} & \left(4c\right) & \mbox{O III} \\ 
\mathbf{B}_{14} & = & \left(\frac{1}{2} - x_{4}\right) \, \mathbf{a}_{1} + \frac{3}{4} \, \mathbf{a}_{2} + \left(\frac{1}{2} +z_{4}\right) \, \mathbf{a}_{3} & = & \left(\frac{1}{2} - x_{4}\right)a \, \mathbf{\hat{x}} + \frac{3}{4}b \, \mathbf{\hat{y}} + \left(\frac{1}{2} +z_{4}\right)c \, \mathbf{\hat{z}} & \left(4c\right) & \mbox{O III} \\ 
\mathbf{B}_{15} & = & -x_{4} \, \mathbf{a}_{1} + \frac{3}{4} \, \mathbf{a}_{2}-z_{4} \, \mathbf{a}_{3} & = & -x_{4}a \, \mathbf{\hat{x}} + \frac{3}{4}b \, \mathbf{\hat{y}}-z_{4}c \, \mathbf{\hat{z}} & \left(4c\right) & \mbox{O III} \\ 
\mathbf{B}_{16} & = & \left(\frac{1}{2} +x_{4}\right) \, \mathbf{a}_{1} + \frac{1}{4} \, \mathbf{a}_{2} + \left(\frac{1}{2} - z_{4}\right) \, \mathbf{a}_{3} & = & \left(\frac{1}{2} +x_{4}\right)a \, \mathbf{\hat{x}} + \frac{1}{4}b \, \mathbf{\hat{y}} + \left(\frac{1}{2} - z_{4}\right)c \, \mathbf{\hat{z}} & \left(4c\right) & \mbox{O III} \\ 
\end{longtabu}
\renewcommand{\arraystretch}{1.0}
\noindent \hrulefill
\\
\textbf{References:}
\vspace*{-0.25cm}
\begin{flushleft}
  - \bibentry{Sitepu_JAppCryst_38_2005}. \\
\end{flushleft}
\noindent \hrulefill
\\
\textbf{Geometry files:}
\\
\noindent  - CIF: pp. {\hyperref[AB3_oP16_62_c_3c_cif]{\pageref{AB3_oP16_62_c_3c_cif}}} \\
\noindent  - POSCAR: pp. {\hyperref[AB3_oP16_62_c_3c_poscar]{\pageref{AB3_oP16_62_c_3c_poscar}}} \\
\onecolumn
{\phantomsection\label{AB4C_oP24_62_c_2cd_c}}
\subsection*{\huge \textbf{{\normalfont \begin{raggedleft}Barite (BaSO$_{4}$, $H0_{2}$) Structure: \end{raggedleft} \\ AB4C\_oP24\_62\_c\_2cd\_c}}}
\noindent \hrulefill
\vspace*{0.25cm}
\begin{figure}[htp]
  \centering
  \vspace{-1em}
  {\includegraphics[width=1\textwidth]{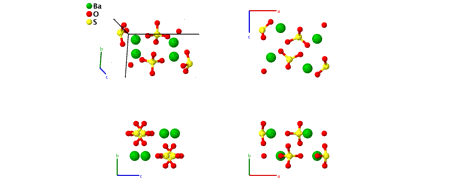}}
\end{figure}
\vspace*{-0.5cm}
\renewcommand{\arraystretch}{1.5}
\begin{equation*}
  \begin{array}{>{$\hspace{-0.15cm}}l<{$}>{$}p{0.5cm}<{$}>{$}p{18.5cm}<{$}}
    \mbox{\large \textbf{Prototype}} &\colon & \ce{BaSO4} \\
    \mbox{\large \textbf{\AFLOW\ prototype label}} &\colon & \mbox{AB4C\_oP24\_62\_c\_2cd\_c} \\
    \mbox{\large \textbf{\textit{Strukturbericht} designation}} &\colon & \mbox{$H0_{2}$} \\
    \mbox{\large \textbf{Pearson symbol}} &\colon & \mbox{oP24} \\
    \mbox{\large \textbf{Space group number}} &\colon & 62 \\
    \mbox{\large \textbf{Space group symbol}} &\colon & Pnma \\
    \mbox{\large \textbf{\AFLOW\ prototype command}} &\colon &  \texttt{aflow} \,  \, \texttt{-{}-proto=AB4C\_oP24\_62\_c\_2cd\_c } \, \newline \texttt{-{}-params=}{a,b/a,c/a,x_{1},z_{1},x_{2},z_{2},x_{3},z_{3},x_{4},z_{4},x_{5},y_{5},z_{5} }
  \end{array}
\end{equation*}
\renewcommand{\arraystretch}{1.0}

\vspace*{-0.25cm}
\noindent \hrulefill
\\
\textbf{ Other compounds with this structure:}
\begin{itemize}
   \item{ SrSO$_{4}$ (celestite), PbSO$_{4}$ (anglesite), KGaH$_{4}$  }
\end{itemize}
\noindent \parbox{1 \linewidth}{
\noindent \hrulefill
\\
\textbf{Simple Orthorhombic primitive vectors:} \\
\vspace*{-0.25cm}
\begin{tabular}{cc}
  \begin{tabular}{c}
    \parbox{0.6 \linewidth}{
      \renewcommand{\arraystretch}{1.5}
      \begin{equation*}
        \centering
        \begin{array}{ccc}
              \mathbf{a}_1 & = & a \, \mathbf{\hat{x}} \\
    \mathbf{a}_2 & = & b \, \mathbf{\hat{y}} \\
    \mathbf{a}_3 & = & c \, \mathbf{\hat{z}} \\

        \end{array}
      \end{equation*}
    }
    \renewcommand{\arraystretch}{1.0}
  \end{tabular}
  \begin{tabular}{c}
    \includegraphics[width=0.3\linewidth]{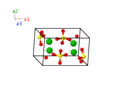} \\
  \end{tabular}
\end{tabular}

}
\vspace*{-0.25cm}

\noindent \hrulefill
\\
\textbf{Basis vectors:}
\vspace*{-0.25cm}
\renewcommand{\arraystretch}{1.5}
\begin{longtabu} to \textwidth{>{\centering $}X[-1,c,c]<{$}>{\centering $}X[-1,c,c]<{$}>{\centering $}X[-1,c,c]<{$}>{\centering $}X[-1,c,c]<{$}>{\centering $}X[-1,c,c]<{$}>{\centering $}X[-1,c,c]<{$}>{\centering $}X[-1,c,c]<{$}}
  & & \mbox{Lattice Coordinates} & & \mbox{Cartesian Coordinates} &\mbox{Wyckoff Position} & \mbox{Atom Type} \\  
  \mathbf{B}_{1} & = & x_{1} \, \mathbf{a}_{1} + \frac{1}{4} \, \mathbf{a}_{2} + z_{1} \, \mathbf{a}_{3} & = & x_{1}a \, \mathbf{\hat{x}} + \frac{1}{4}b \, \mathbf{\hat{y}} + z_{1}c \, \mathbf{\hat{z}} & \left(4c\right) & \mbox{Ba} \\ 
\mathbf{B}_{2} & = & \left(\frac{1}{2} - x_{1}\right) \, \mathbf{a}_{1} + \frac{3}{4} \, \mathbf{a}_{2} + \left(\frac{1}{2} +z_{1}\right) \, \mathbf{a}_{3} & = & \left(\frac{1}{2} - x_{1}\right)a \, \mathbf{\hat{x}} + \frac{3}{4}b \, \mathbf{\hat{y}} + \left(\frac{1}{2} +z_{1}\right)c \, \mathbf{\hat{z}} & \left(4c\right) & \mbox{Ba} \\ 
\mathbf{B}_{3} & = & -x_{1} \, \mathbf{a}_{1} + \frac{3}{4} \, \mathbf{a}_{2}-z_{1} \, \mathbf{a}_{3} & = & -x_{1}a \, \mathbf{\hat{x}} + \frac{3}{4}b \, \mathbf{\hat{y}}-z_{1}c \, \mathbf{\hat{z}} & \left(4c\right) & \mbox{Ba} \\ 
\mathbf{B}_{4} & = & \left(\frac{1}{2} +x_{1}\right) \, \mathbf{a}_{1} + \frac{1}{4} \, \mathbf{a}_{2} + \left(\frac{1}{2} - z_{1}\right) \, \mathbf{a}_{3} & = & \left(\frac{1}{2} +x_{1}\right)a \, \mathbf{\hat{x}} + \frac{1}{4}b \, \mathbf{\hat{y}} + \left(\frac{1}{2} - z_{1}\right)c \, \mathbf{\hat{z}} & \left(4c\right) & \mbox{Ba} \\ 
\mathbf{B}_{5} & = & x_{2} \, \mathbf{a}_{1} + \frac{1}{4} \, \mathbf{a}_{2} + z_{2} \, \mathbf{a}_{3} & = & x_{2}a \, \mathbf{\hat{x}} + \frac{1}{4}b \, \mathbf{\hat{y}} + z_{2}c \, \mathbf{\hat{z}} & \left(4c\right) & \mbox{O I} \\ 
\mathbf{B}_{6} & = & \left(\frac{1}{2} - x_{2}\right) \, \mathbf{a}_{1} + \frac{3}{4} \, \mathbf{a}_{2} + \left(\frac{1}{2} +z_{2}\right) \, \mathbf{a}_{3} & = & \left(\frac{1}{2} - x_{2}\right)a \, \mathbf{\hat{x}} + \frac{3}{4}b \, \mathbf{\hat{y}} + \left(\frac{1}{2} +z_{2}\right)c \, \mathbf{\hat{z}} & \left(4c\right) & \mbox{O I} \\ 
\mathbf{B}_{7} & = & -x_{2} \, \mathbf{a}_{1} + \frac{3}{4} \, \mathbf{a}_{2}-z_{2} \, \mathbf{a}_{3} & = & -x_{2}a \, \mathbf{\hat{x}} + \frac{3}{4}b \, \mathbf{\hat{y}}-z_{2}c \, \mathbf{\hat{z}} & \left(4c\right) & \mbox{O I} \\ 
\mathbf{B}_{8} & = & \left(\frac{1}{2} +x_{2}\right) \, \mathbf{a}_{1} + \frac{1}{4} \, \mathbf{a}_{2} + \left(\frac{1}{2} - z_{2}\right) \, \mathbf{a}_{3} & = & \left(\frac{1}{2} +x_{2}\right)a \, \mathbf{\hat{x}} + \frac{1}{4}b \, \mathbf{\hat{y}} + \left(\frac{1}{2} - z_{2}\right)c \, \mathbf{\hat{z}} & \left(4c\right) & \mbox{O I} \\ 
\mathbf{B}_{9} & = & x_{3} \, \mathbf{a}_{1} + \frac{1}{4} \, \mathbf{a}_{2} + z_{3} \, \mathbf{a}_{3} & = & x_{3}a \, \mathbf{\hat{x}} + \frac{1}{4}b \, \mathbf{\hat{y}} + z_{3}c \, \mathbf{\hat{z}} & \left(4c\right) & \mbox{O II} \\ 
\mathbf{B}_{10} & = & \left(\frac{1}{2} - x_{3}\right) \, \mathbf{a}_{1} + \frac{3}{4} \, \mathbf{a}_{2} + \left(\frac{1}{2} +z_{3}\right) \, \mathbf{a}_{3} & = & \left(\frac{1}{2} - x_{3}\right)a \, \mathbf{\hat{x}} + \frac{3}{4}b \, \mathbf{\hat{y}} + \left(\frac{1}{2} +z_{3}\right)c \, \mathbf{\hat{z}} & \left(4c\right) & \mbox{O II} \\ 
\mathbf{B}_{11} & = & -x_{3} \, \mathbf{a}_{1} + \frac{3}{4} \, \mathbf{a}_{2}-z_{3} \, \mathbf{a}_{3} & = & -x_{3}a \, \mathbf{\hat{x}} + \frac{3}{4}b \, \mathbf{\hat{y}}-z_{3}c \, \mathbf{\hat{z}} & \left(4c\right) & \mbox{O II} \\ 
\mathbf{B}_{12} & = & \left(\frac{1}{2} +x_{3}\right) \, \mathbf{a}_{1} + \frac{1}{4} \, \mathbf{a}_{2} + \left(\frac{1}{2} - z_{3}\right) \, \mathbf{a}_{3} & = & \left(\frac{1}{2} +x_{3}\right)a \, \mathbf{\hat{x}} + \frac{1}{4}b \, \mathbf{\hat{y}} + \left(\frac{1}{2} - z_{3}\right)c \, \mathbf{\hat{z}} & \left(4c\right) & \mbox{O II} \\ 
\mathbf{B}_{13} & = & x_{4} \, \mathbf{a}_{1} + \frac{1}{4} \, \mathbf{a}_{2} + z_{4} \, \mathbf{a}_{3} & = & x_{4}a \, \mathbf{\hat{x}} + \frac{1}{4}b \, \mathbf{\hat{y}} + z_{4}c \, \mathbf{\hat{z}} & \left(4c\right) & \mbox{S} \\ 
\mathbf{B}_{14} & = & \left(\frac{1}{2} - x_{4}\right) \, \mathbf{a}_{1} + \frac{3}{4} \, \mathbf{a}_{2} + \left(\frac{1}{2} +z_{4}\right) \, \mathbf{a}_{3} & = & \left(\frac{1}{2} - x_{4}\right)a \, \mathbf{\hat{x}} + \frac{3}{4}b \, \mathbf{\hat{y}} + \left(\frac{1}{2} +z_{4}\right)c \, \mathbf{\hat{z}} & \left(4c\right) & \mbox{S} \\ 
\mathbf{B}_{15} & = & -x_{4} \, \mathbf{a}_{1} + \frac{3}{4} \, \mathbf{a}_{2}-z_{4} \, \mathbf{a}_{3} & = & -x_{4}a \, \mathbf{\hat{x}} + \frac{3}{4}b \, \mathbf{\hat{y}}-z_{4}c \, \mathbf{\hat{z}} & \left(4c\right) & \mbox{S} \\ 
\mathbf{B}_{16} & = & \left(\frac{1}{2} +x_{4}\right) \, \mathbf{a}_{1} + \frac{1}{4} \, \mathbf{a}_{2} + \left(\frac{1}{2} - z_{4}\right) \, \mathbf{a}_{3} & = & \left(\frac{1}{2} +x_{4}\right)a \, \mathbf{\hat{x}} + \frac{1}{4}b \, \mathbf{\hat{y}} + \left(\frac{1}{2} - z_{4}\right)c \, \mathbf{\hat{z}} & \left(4c\right) & \mbox{S} \\ 
\mathbf{B}_{17} & = & x_{5} \, \mathbf{a}_{1} + y_{5} \, \mathbf{a}_{2} + z_{5} \, \mathbf{a}_{3} & = & x_{5}a \, \mathbf{\hat{x}} + y_{5}b \, \mathbf{\hat{y}} + z_{5}c \, \mathbf{\hat{z}} & \left(8d\right) & \mbox{O III} \\ 
\mathbf{B}_{18} & = & \left(\frac{1}{2} - x_{5}\right) \, \mathbf{a}_{1}-y_{5} \, \mathbf{a}_{2} + \left(\frac{1}{2} +z_{5}\right) \, \mathbf{a}_{3} & = & \left(\frac{1}{2} - x_{5}\right)a \, \mathbf{\hat{x}}-y_{5}b \, \mathbf{\hat{y}} + \left(\frac{1}{2} +z_{5}\right)c \, \mathbf{\hat{z}} & \left(8d\right) & \mbox{O III} \\ 
\mathbf{B}_{19} & = & -x_{5} \, \mathbf{a}_{1} + \left(\frac{1}{2} +y_{5}\right) \, \mathbf{a}_{2}-z_{5} \, \mathbf{a}_{3} & = & -x_{5}a \, \mathbf{\hat{x}} + \left(\frac{1}{2} +y_{5}\right)b \, \mathbf{\hat{y}}-z_{5}c \, \mathbf{\hat{z}} & \left(8d\right) & \mbox{O III} \\ 
\mathbf{B}_{20} & = & \left(\frac{1}{2} +x_{5}\right) \, \mathbf{a}_{1} + \left(\frac{1}{2} - y_{5}\right) \, \mathbf{a}_{2} + \left(\frac{1}{2} - z_{5}\right) \, \mathbf{a}_{3} & = & \left(\frac{1}{2} +x_{5}\right)a \, \mathbf{\hat{x}} + \left(\frac{1}{2} - y_{5}\right)b \, \mathbf{\hat{y}} + \left(\frac{1}{2} - z_{5}\right)c \, \mathbf{\hat{z}} & \left(8d\right) & \mbox{O III} \\ 
\mathbf{B}_{21} & = & -x_{5} \, \mathbf{a}_{1}-y_{5} \, \mathbf{a}_{2}-z_{5} \, \mathbf{a}_{3} & = & -x_{5}a \, \mathbf{\hat{x}}-y_{5}b \, \mathbf{\hat{y}}-z_{5}c \, \mathbf{\hat{z}} & \left(8d\right) & \mbox{O III} \\ 
\mathbf{B}_{22} & = & \left(\frac{1}{2} +x_{5}\right) \, \mathbf{a}_{1} + y_{5} \, \mathbf{a}_{2} + \left(\frac{1}{2} - z_{5}\right) \, \mathbf{a}_{3} & = & \left(\frac{1}{2} +x_{5}\right)a \, \mathbf{\hat{x}} + y_{5}b \, \mathbf{\hat{y}} + \left(\frac{1}{2} - z_{5}\right)c \, \mathbf{\hat{z}} & \left(8d\right) & \mbox{O III} \\ 
\mathbf{B}_{23} & = & x_{5} \, \mathbf{a}_{1} + \left(\frac{1}{2} - y_{5}\right) \, \mathbf{a}_{2} + z_{5} \, \mathbf{a}_{3} & = & x_{5}a \, \mathbf{\hat{x}} + \left(\frac{1}{2} - y_{5}\right)b \, \mathbf{\hat{y}} + z_{5}c \, \mathbf{\hat{z}} & \left(8d\right) & \mbox{O III} \\ 
\mathbf{B}_{24} & = & \left(\frac{1}{2} - x_{5}\right) \, \mathbf{a}_{1} + \left(\frac{1}{2} +y_{5}\right) \, \mathbf{a}_{2} + \left(\frac{1}{2} +z_{5}\right) \, \mathbf{a}_{3} & = & \left(\frac{1}{2} - x_{5}\right)a \, \mathbf{\hat{x}} + \left(\frac{1}{2} +y_{5}\right)b \, \mathbf{\hat{y}} + \left(\frac{1}{2} +z_{5}\right)c \, \mathbf{\hat{z}} & \left(8d\right) & \mbox{O III} \\ 
\end{longtabu}
\renewcommand{\arraystretch}{1.0}
\noindent \hrulefill
\\
\textbf{References:}
\vspace*{-0.25cm}
\begin{flushleft}
  - \bibentry{Colville_Am_Min_52_1967}. \\
\end{flushleft}
\textbf{Found in:}
\vspace*{-0.25cm}
\begin{flushleft}
  - \bibentry{Barthelmy_MinData_2012}. \\
\end{flushleft}
\noindent \hrulefill
\\
\textbf{Geometry files:}
\\
\noindent  - CIF: pp. {\hyperref[AB4C_oP24_62_c_2cd_c_cif]{\pageref{AB4C_oP24_62_c_2cd_c_cif}}} \\
\noindent  - POSCAR: pp. {\hyperref[AB4C_oP24_62_c_2cd_c_poscar]{\pageref{AB4C_oP24_62_c_2cd_c_poscar}}} \\
\onecolumn
{\phantomsection\label{AB_oP8_62_c_c}}
\subsection*{\huge \textbf{{\normalfont \begin{raggedleft}Westerveldite (FeAs, $B14$) Structure: \end{raggedleft} \\ AB\_oP8\_62\_c\_c}}}
\noindent \hrulefill
\vspace*{0.25cm}
\begin{figure}[htp]
  \centering
  \vspace{-1em}
  {\includegraphics[width=1\textwidth]{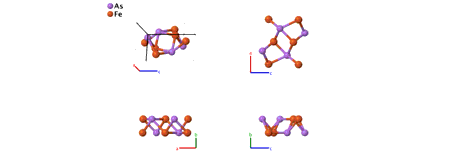}}
\end{figure}
\vspace*{-0.5cm}
\renewcommand{\arraystretch}{1.5}
\begin{equation*}
  \begin{array}{>{$\hspace{-0.15cm}}l<{$}>{$}p{0.5cm}<{$}>{$}p{18.5cm}<{$}}
    \mbox{\large \textbf{Prototype}} &\colon & \ce{FeAs} \\
    \mbox{\large \textbf{\AFLOW\ prototype label}} &\colon & \mbox{AB\_oP8\_62\_c\_c} \\
    \mbox{\large \textbf{\textit{Strukturbericht} designation}} &\colon & \mbox{$B14$} \\
    \mbox{\large \textbf{Pearson symbol}} &\colon & \mbox{oP8} \\
    \mbox{\large \textbf{Space group number}} &\colon & 62 \\
    \mbox{\large \textbf{Space group symbol}} &\colon & Pnma \\
    \mbox{\large \textbf{\AFLOW\ prototype command}} &\colon &  \texttt{aflow} \,  \, \texttt{-{}-proto=AB\_oP8\_62\_c\_c } \, \newline \texttt{-{}-params=}{a,b/a,c/a,x_{1},z_{1},x_{2},z_{2} }
  \end{array}
\end{equation*}
\renewcommand{\arraystretch}{1.0}

\vspace*{-0.25cm}
\noindent \hrulefill
\\
\textbf{ Other compounds with this structure:}
\begin{itemize}
   \item{ CoAs  }
\end{itemize}
\vspace*{-0.25cm}
\noindent \hrulefill
\begin{itemize}
  \item{The $B31$ (\href{http://aflow.org/CrystalDatabase/AB_oP8_62_c_c.MnP.html}{MnP, AB\_oP8\_62\_c\_c}) structure is similar to this one.
(Brandes, 1992) lists $B31$ as the primary structure, but we include
FeAs here for completeness.
We use the data (Selte, 1972) reported at 14~K.
}
\end{itemize}

\noindent \parbox{1 \linewidth}{
\noindent \hrulefill
\\
\textbf{Simple Orthorhombic primitive vectors:} \\
\vspace*{-0.25cm}
\begin{tabular}{cc}
  \begin{tabular}{c}
    \parbox{0.6 \linewidth}{
      \renewcommand{\arraystretch}{1.5}
      \begin{equation*}
        \centering
        \begin{array}{ccc}
              \mathbf{a}_1 & = & a \, \mathbf{\hat{x}} \\
    \mathbf{a}_2 & = & b \, \mathbf{\hat{y}} \\
    \mathbf{a}_3 & = & c \, \mathbf{\hat{z}} \\

        \end{array}
      \end{equation*}
    }
    \renewcommand{\arraystretch}{1.0}
  \end{tabular}
  \begin{tabular}{c}
    \includegraphics[width=0.3\linewidth]{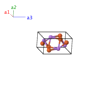} \\
  \end{tabular}
\end{tabular}

}
\vspace*{-0.25cm}

\noindent \hrulefill
\\
\textbf{Basis vectors:}
\vspace*{-0.25cm}
\renewcommand{\arraystretch}{1.5}
\begin{longtabu} to \textwidth{>{\centering $}X[-1,c,c]<{$}>{\centering $}X[-1,c,c]<{$}>{\centering $}X[-1,c,c]<{$}>{\centering $}X[-1,c,c]<{$}>{\centering $}X[-1,c,c]<{$}>{\centering $}X[-1,c,c]<{$}>{\centering $}X[-1,c,c]<{$}}
  & & \mbox{Lattice Coordinates} & & \mbox{Cartesian Coordinates} &\mbox{Wyckoff Position} & \mbox{Atom Type} \\  
  \mathbf{B}_{1} & = & x_{1} \, \mathbf{a}_{1} + \frac{1}{4} \, \mathbf{a}_{2} + z_{1} \, \mathbf{a}_{3} & = & x_{1}a \, \mathbf{\hat{x}} + \frac{1}{4}b \, \mathbf{\hat{y}} + z_{1}c \, \mathbf{\hat{z}} & \left(4c\right) & \mbox{As} \\ 
\mathbf{B}_{2} & = & \left(\frac{1}{2} - x_{1}\right) \, \mathbf{a}_{1} + \frac{3}{4} \, \mathbf{a}_{2} + \left(\frac{1}{2} +z_{1}\right) \, \mathbf{a}_{3} & = & \left(\frac{1}{2} - x_{1}\right)a \, \mathbf{\hat{x}} + \frac{3}{4}b \, \mathbf{\hat{y}} + \left(\frac{1}{2} +z_{1}\right)c \, \mathbf{\hat{z}} & \left(4c\right) & \mbox{As} \\ 
\mathbf{B}_{3} & = & -x_{1} \, \mathbf{a}_{1} + \frac{3}{4} \, \mathbf{a}_{2}-z_{1} \, \mathbf{a}_{3} & = & -x_{1}a \, \mathbf{\hat{x}} + \frac{3}{4}b \, \mathbf{\hat{y}}-z_{1}c \, \mathbf{\hat{z}} & \left(4c\right) & \mbox{As} \\ 
\mathbf{B}_{4} & = & \left(\frac{1}{2} +x_{1}\right) \, \mathbf{a}_{1} + \frac{1}{4} \, \mathbf{a}_{2} + \left(\frac{1}{2} - z_{1}\right) \, \mathbf{a}_{3} & = & \left(\frac{1}{2} +x_{1}\right)a \, \mathbf{\hat{x}} + \frac{1}{4}b \, \mathbf{\hat{y}} + \left(\frac{1}{2} - z_{1}\right)c \, \mathbf{\hat{z}} & \left(4c\right) & \mbox{As} \\ 
\mathbf{B}_{5} & = & x_{2} \, \mathbf{a}_{1} + \frac{1}{4} \, \mathbf{a}_{2} + z_{2} \, \mathbf{a}_{3} & = & x_{2}a \, \mathbf{\hat{x}} + \frac{1}{4}b \, \mathbf{\hat{y}} + z_{2}c \, \mathbf{\hat{z}} & \left(4c\right) & \mbox{Fe} \\ 
\mathbf{B}_{6} & = & \left(\frac{1}{2} - x_{2}\right) \, \mathbf{a}_{1} + \frac{3}{4} \, \mathbf{a}_{2} + \left(\frac{1}{2} +z_{2}\right) \, \mathbf{a}_{3} & = & \left(\frac{1}{2} - x_{2}\right)a \, \mathbf{\hat{x}} + \frac{3}{4}b \, \mathbf{\hat{y}} + \left(\frac{1}{2} +z_{2}\right)c \, \mathbf{\hat{z}} & \left(4c\right) & \mbox{Fe} \\ 
\mathbf{B}_{7} & = & -x_{2} \, \mathbf{a}_{1} + \frac{3}{4} \, \mathbf{a}_{2}-z_{2} \, \mathbf{a}_{3} & = & -x_{2}a \, \mathbf{\hat{x}} + \frac{3}{4}b \, \mathbf{\hat{y}}-z_{2}c \, \mathbf{\hat{z}} & \left(4c\right) & \mbox{Fe} \\ 
\mathbf{B}_{8} & = & \left(\frac{1}{2} +x_{2}\right) \, \mathbf{a}_{1} + \frac{1}{4} \, \mathbf{a}_{2} + \left(\frac{1}{2} - z_{2}\right) \, \mathbf{a}_{3} & = & \left(\frac{1}{2} +x_{2}\right)a \, \mathbf{\hat{x}} + \frac{1}{4}b \, \mathbf{\hat{y}} + \left(\frac{1}{2} - z_{2}\right)c \, \mathbf{\hat{z}} & \left(4c\right) & \mbox{Fe} \\ 
\end{longtabu}
\renewcommand{\arraystretch}{1.0}
\noindent \hrulefill
\\
\textbf{References:}
\vspace*{-0.25cm}
\begin{flushleft}
  - \bibentry{Selte_ActaChemScand_26_3101_1972}. \\
  - \bibentry{Parthe_Gmelin_Handbook_1993}. \\
\end{flushleft}
\textbf{Found in:}
\vspace*{-0.25cm}
\begin{flushleft}
  - \bibentry{Jeffries_PRB_83_134520_2011}. \\
\end{flushleft}
\noindent \hrulefill
\\
\textbf{Geometry files:}
\\
\noindent  - CIF: pp. {\hyperref[AB_oP8_62_c_c_cif]{\pageref{AB_oP8_62_c_c_cif}}} \\
\noindent  - POSCAR: pp. {\hyperref[AB_oP8_62_c_c_poscar]{\pageref{AB_oP8_62_c_c_poscar}}} \\
\onecolumn
{\phantomsection\label{A2BC3_oC24_63_e_c_cg}}
\subsection*{\huge \textbf{{\normalfont Rasvumite (KFe$_{2}$S$_{3}$) Structure: A2BC3\_oC24\_63\_e\_c\_cg}}}
\noindent \hrulefill
\vspace*{0.25cm}
\begin{figure}[htp]
  \centering
  \vspace{-1em}
  {\includegraphics[width=1\textwidth]{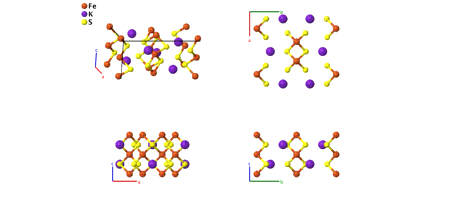}}
\end{figure}
\vspace*{-0.5cm}
\renewcommand{\arraystretch}{1.5}
\begin{equation*}
  \begin{array}{>{$\hspace{-0.15cm}}l<{$}>{$}p{0.5cm}<{$}>{$}p{18.5cm}<{$}}
    \mbox{\large \textbf{Prototype}} &\colon & \ce{KFe2S3} \\
    \mbox{\large \textbf{\AFLOW\ prototype label}} &\colon & \mbox{A2BC3\_oC24\_63\_e\_c\_cg} \\
    \mbox{\large \textbf{\textit{Strukturbericht} designation}} &\colon & \mbox{None} \\
    \mbox{\large \textbf{Pearson symbol}} &\colon & \mbox{oC24} \\
    \mbox{\large \textbf{Space group number}} &\colon & 63 \\
    \mbox{\large \textbf{Space group symbol}} &\colon & Cmcm \\
    \mbox{\large \textbf{\AFLOW\ prototype command}} &\colon &  \texttt{aflow} \,  \, \texttt{-{}-proto=A2BC3\_oC24\_63\_e\_c\_cg } \, \newline \texttt{-{}-params=}{a,b/a,c/a,y_{1},y_{2},x_{3},x_{4},y_{4} }
  \end{array}
\end{equation*}
\renewcommand{\arraystretch}{1.0}

\vspace*{-0.25cm}
\noindent \hrulefill
\\
\textbf{ Other compounds with this structure:}
\begin{itemize}
   \item{ BaFe$_{2}$S$_{3}$, RbFe$_{2}$S$_{3}$, CsFe$_{2}$S$_{3}$  }
\end{itemize}
\noindent \parbox{1 \linewidth}{
\noindent \hrulefill
\\
\textbf{Base-centered Orthorhombic primitive vectors:} \\
\vspace*{-0.25cm}
\begin{tabular}{cc}
  \begin{tabular}{c}
    \parbox{0.6 \linewidth}{
      \renewcommand{\arraystretch}{1.5}
      \begin{equation*}
        \centering
        \begin{array}{ccc}
              \mathbf{a}_1 & = & \frac12 \, a \, \mathbf{\hat{x}} - \frac12 \, b \, \mathbf{\hat{y}} \\
    \mathbf{a}_2 & = & \frac12 \, a \, \mathbf{\hat{x}} + \frac12 \, b \, \mathbf{\hat{y}} \\
    \mathbf{a}_3 & = & c \, \mathbf{\hat{z}} \\

        \end{array}
      \end{equation*}
    }
    \renewcommand{\arraystretch}{1.0}
  \end{tabular}
  \begin{tabular}{c}
    \includegraphics[width=0.3\linewidth]{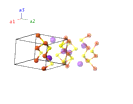} \\
  \end{tabular}
\end{tabular}

}
\vspace*{-0.25cm}

\noindent \hrulefill
\\
\textbf{Basis vectors:}
\vspace*{-0.25cm}
\renewcommand{\arraystretch}{1.5}
\begin{longtabu} to \textwidth{>{\centering $}X[-1,c,c]<{$}>{\centering $}X[-1,c,c]<{$}>{\centering $}X[-1,c,c]<{$}>{\centering $}X[-1,c,c]<{$}>{\centering $}X[-1,c,c]<{$}>{\centering $}X[-1,c,c]<{$}>{\centering $}X[-1,c,c]<{$}}
  & & \mbox{Lattice Coordinates} & & \mbox{Cartesian Coordinates} &\mbox{Wyckoff Position} & \mbox{Atom Type} \\  
  \mathbf{B}_{1} & = & -y_{1} \, \mathbf{a}_{1} + y_{1} \, \mathbf{a}_{2} + \frac{1}{4} \, \mathbf{a}_{3} & = & y_{1}b \, \mathbf{\hat{y}} + \frac{1}{4}c \, \mathbf{\hat{z}} & \left(4c\right) & \mbox{K} \\ 
\mathbf{B}_{2} & = & y_{1} \, \mathbf{a}_{1}-y_{1} \, \mathbf{a}_{2} + \frac{3}{4} \, \mathbf{a}_{3} & = & -y_{1}b \, \mathbf{\hat{y}} + \frac{3}{4}c \, \mathbf{\hat{z}} & \left(4c\right) & \mbox{K} \\ 
\mathbf{B}_{3} & = & -y_{2} \, \mathbf{a}_{1} + y_{2} \, \mathbf{a}_{2} + \frac{1}{4} \, \mathbf{a}_{3} & = & y_{2}b \, \mathbf{\hat{y}} + \frac{1}{4}c \, \mathbf{\hat{z}} & \left(4c\right) & \mbox{S I} \\ 
\mathbf{B}_{4} & = & y_{2} \, \mathbf{a}_{1}-y_{2} \, \mathbf{a}_{2} + \frac{3}{4} \, \mathbf{a}_{3} & = & -y_{2}b \, \mathbf{\hat{y}} + \frac{3}{4}c \, \mathbf{\hat{z}} & \left(4c\right) & \mbox{S I} \\ 
\mathbf{B}_{5} & = & x_{3} \, \mathbf{a}_{1} + x_{3} \, \mathbf{a}_{2} & = & x_{3}a \, \mathbf{\hat{x}} & \left(8e\right) & \mbox{Fe} \\ 
\mathbf{B}_{6} & = & -x_{3} \, \mathbf{a}_{1}-x_{3} \, \mathbf{a}_{2} + \frac{1}{2} \, \mathbf{a}_{3} & = & -x_{3}a \, \mathbf{\hat{x}} + \frac{1}{2}c \, \mathbf{\hat{z}} & \left(8e\right) & \mbox{Fe} \\ 
\mathbf{B}_{7} & = & -x_{3} \, \mathbf{a}_{1}-x_{3} \, \mathbf{a}_{2} & = & -x_{3}a \, \mathbf{\hat{x}} & \left(8e\right) & \mbox{Fe} \\ 
\mathbf{B}_{8} & = & x_{3} \, \mathbf{a}_{1} + x_{3} \, \mathbf{a}_{2} + \frac{1}{2} \, \mathbf{a}_{3} & = & x_{3}a \, \mathbf{\hat{x}} + \frac{1}{2}c \, \mathbf{\hat{z}} & \left(8e\right) & \mbox{Fe} \\ 
\mathbf{B}_{9} & = & \left(x_{4}-y_{4}\right) \, \mathbf{a}_{1} + \left(x_{4}+y_{4}\right) \, \mathbf{a}_{2} + \frac{1}{4} \, \mathbf{a}_{3} & = & x_{4}a \, \mathbf{\hat{x}} + y_{4}b \, \mathbf{\hat{y}} + \frac{1}{4}c \, \mathbf{\hat{z}} & \left(8g\right) & \mbox{S II} \\ 
\mathbf{B}_{10} & = & \left(-x_{4}+y_{4}\right) \, \mathbf{a}_{1} + \left(-x_{4}-y_{4}\right) \, \mathbf{a}_{2} + \frac{3}{4} \, \mathbf{a}_{3} & = & -x_{4}a \, \mathbf{\hat{x}}-y_{4}b \, \mathbf{\hat{y}} + \frac{3}{4}c \, \mathbf{\hat{z}} & \left(8g\right) & \mbox{S II} \\ 
\mathbf{B}_{11} & = & \left(-x_{4}-y_{4}\right) \, \mathbf{a}_{1} + \left(-x_{4}+y_{4}\right) \, \mathbf{a}_{2} + \frac{1}{4} \, \mathbf{a}_{3} & = & -x_{4}a \, \mathbf{\hat{x}} + y_{4}b \, \mathbf{\hat{y}} + \frac{1}{4}c \, \mathbf{\hat{z}} & \left(8g\right) & \mbox{S II} \\ 
\mathbf{B}_{12} & = & \left(x_{4}+y_{4}\right) \, \mathbf{a}_{1} + \left(x_{4}-y_{4}\right) \, \mathbf{a}_{2} + \frac{3}{4} \, \mathbf{a}_{3} & = & x_{4}a \, \mathbf{\hat{x}}-y_{4}b \, \mathbf{\hat{y}} + \frac{3}{4}c \, \mathbf{\hat{z}} & \left(8g\right) & \mbox{S II} \\ 
\end{longtabu}
\renewcommand{\arraystretch}{1.0}
\noindent \hrulefill
\\
\textbf{References:}
\vspace*{-0.25cm}
\begin{flushleft}
  - \bibentry{Clark_Am_Min_65_1980}. \\
\end{flushleft}
\textbf{Found in:}
\vspace*{-0.25cm}
\begin{flushleft}
  - \bibentry{Downs_Am_Min_88_2003}. \\
\end{flushleft}
\noindent \hrulefill
\\
\textbf{Geometry files:}
\\
\noindent  - CIF: pp. {\hyperref[A2BC3_oC24_63_e_c_cg_cif]{\pageref{A2BC3_oC24_63_e_c_cg_cif}}} \\
\noindent  - POSCAR: pp. {\hyperref[A2BC3_oC24_63_e_c_cg_poscar]{\pageref{A2BC3_oC24_63_e_c_cg_poscar}}} \\
\onecolumn
{\phantomsection\label{A43B5C17_oC260_63_c8fg6h_cfg_ce3f2h}}
\subsection*{\huge \textbf{{\normalfont \begin{raggedleft}La$_{43}$Ni$_{17}$Mg$_{5}$ Structure: \end{raggedleft} \\ A43B5C17\_oC260\_63\_c8fg6h\_cfg\_ce3f2h}}}
\noindent \hrulefill
\vspace*{0.25cm}
\begin{figure}[htp]
  \centering
  \vspace{-1em}
  {\includegraphics[width=1\textwidth]{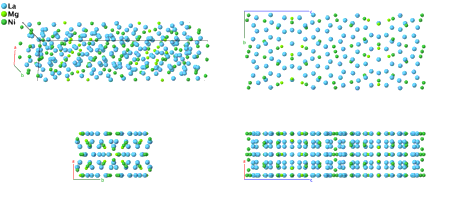}}
\end{figure}
\vspace*{-0.5cm}
\renewcommand{\arraystretch}{1.5}
\begin{equation*}
  \begin{array}{>{$\hspace{-0.15cm}}l<{$}>{$}p{0.5cm}<{$}>{$}p{18.5cm}<{$}}
    \mbox{\large \textbf{Prototype}} &\colon & \ce{La43Ni17Mg5} \\
    \mbox{\large \textbf{\AFLOW\ prototype label}} &\colon & \mbox{A43B5C17\_oC260\_63\_c8fg6h\_cfg\_ce3f2h} \\
    \mbox{\large \textbf{\textit{Strukturbericht} designation}} &\colon & \mbox{None} \\
    \mbox{\large \textbf{Pearson symbol}} &\colon & \mbox{oC260} \\
    \mbox{\large \textbf{Space group number}} &\colon & 63 \\
    \mbox{\large \textbf{Space group symbol}} &\colon & Cmcm \\
    \mbox{\large \textbf{\AFLOW\ prototype command}} &\colon &  \texttt{aflow} \,  \, \texttt{-{}-proto=A43B5C17\_oC260\_63\_c8fg6h\_cfg\_ce3f2h } \, \newline \texttt{-{}-params=}{a,b/a,c/a,y_{1},y_{2},y_{3},x_{4},y_{5},z_{5},y_{6},z_{6},y_{7},z_{7},y_{8},z_{8},y_{9},z_{9},y_{10},z_{10},y_{11},} \newline {z_{11},y_{12},z_{12},y_{13},z_{13},y_{14},z_{14},y_{15},z_{15},y_{16},z_{16},x_{17},y_{17},x_{18},y_{18},x_{19},y_{19},z_{19},x_{20},y_{20},} \newline {z_{20},x_{21},y_{21},z_{21},x_{22},y_{22},z_{22},x_{23},y_{23},z_{23},x_{24},y_{24},z_{24},x_{25},y_{25},z_{25},x_{26},y_{26},z_{26} }
  \end{array}
\end{equation*}
\renewcommand{\arraystretch}{1.0}

\noindent \parbox{1 \linewidth}{
\noindent \hrulefill
\\
\textbf{Base-centered Orthorhombic primitive vectors:} \\
\vspace*{-0.25cm}
\begin{tabular}{cc}
  \begin{tabular}{c}
    \parbox{0.6 \linewidth}{
      \renewcommand{\arraystretch}{1.5}
      \begin{equation*}
        \centering
        \begin{array}{ccc}
              \mathbf{a}_1 & = & \frac12 \, a \, \mathbf{\hat{x}} - \frac12 \, b \, \mathbf{\hat{y}} \\
    \mathbf{a}_2 & = & \frac12 \, a \, \mathbf{\hat{x}} + \frac12 \, b \, \mathbf{\hat{y}} \\
    \mathbf{a}_3 & = & c \, \mathbf{\hat{z}} \\

        \end{array}
      \end{equation*}
    }
    \renewcommand{\arraystretch}{1.0}
  \end{tabular}
  \begin{tabular}{c}
    \includegraphics[width=0.3\linewidth]{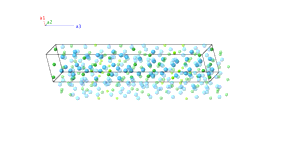} \\
  \end{tabular}
\end{tabular}

}
\vspace*{-0.25cm}

\noindent \hrulefill
\\
\textbf{Basis vectors:}
\vspace*{-0.25cm}
\renewcommand{\arraystretch}{1.5}
\begin{longtabu} to \textwidth{>{\centering $}X[-1,c,c]<{$}>{\centering $}X[-1,c,c]<{$}>{\centering $}X[-1,c,c]<{$}>{\centering $}X[-1,c,c]<{$}>{\centering $}X[-1,c,c]<{$}>{\centering $}X[-1,c,c]<{$}>{\centering $}X[-1,c,c]<{$}}
  & & \mbox{Lattice Coordinates} & & \mbox{Cartesian Coordinates} &\mbox{Wyckoff Position} & \mbox{Atom Type} \\  
  \mathbf{B}_{1} & = & -y_{1} \, \mathbf{a}_{1} + y_{1} \, \mathbf{a}_{2} + \frac{1}{4} \, \mathbf{a}_{3} & = & y_{1}b \, \mathbf{\hat{y}} + \frac{1}{4}c \, \mathbf{\hat{z}} & \left(4c\right) & \mbox{La I} \\ 
\mathbf{B}_{2} & = & y_{1} \, \mathbf{a}_{1}-y_{1} \, \mathbf{a}_{2} + \frac{3}{4} \, \mathbf{a}_{3} & = & -y_{1}b \, \mathbf{\hat{y}} + \frac{3}{4}c \, \mathbf{\hat{z}} & \left(4c\right) & \mbox{La I} \\ 
\mathbf{B}_{3} & = & -y_{2} \, \mathbf{a}_{1} + y_{2} \, \mathbf{a}_{2} + \frac{1}{4} \, \mathbf{a}_{3} & = & y_{2}b \, \mathbf{\hat{y}} + \frac{1}{4}c \, \mathbf{\hat{z}} & \left(4c\right) & \mbox{Mg I} \\ 
\mathbf{B}_{4} & = & y_{2} \, \mathbf{a}_{1}-y_{2} \, \mathbf{a}_{2} + \frac{3}{4} \, \mathbf{a}_{3} & = & -y_{2}b \, \mathbf{\hat{y}} + \frac{3}{4}c \, \mathbf{\hat{z}} & \left(4c\right) & \mbox{Mg I} \\ 
\mathbf{B}_{5} & = & -y_{3} \, \mathbf{a}_{1} + y_{3} \, \mathbf{a}_{2} + \frac{1}{4} \, \mathbf{a}_{3} & = & y_{3}b \, \mathbf{\hat{y}} + \frac{1}{4}c \, \mathbf{\hat{z}} & \left(4c\right) & \mbox{Ni I} \\ 
\mathbf{B}_{6} & = & y_{3} \, \mathbf{a}_{1}-y_{3} \, \mathbf{a}_{2} + \frac{3}{4} \, \mathbf{a}_{3} & = & -y_{3}b \, \mathbf{\hat{y}} + \frac{3}{4}c \, \mathbf{\hat{z}} & \left(4c\right) & \mbox{Ni I} \\ 
\mathbf{B}_{7} & = & x_{4} \, \mathbf{a}_{1} + x_{4} \, \mathbf{a}_{2} & = & x_{4}a \, \mathbf{\hat{x}} & \left(8e\right) & \mbox{Ni II} \\ 
\mathbf{B}_{8} & = & -x_{4} \, \mathbf{a}_{1}-x_{4} \, \mathbf{a}_{2} + \frac{1}{2} \, \mathbf{a}_{3} & = & -x_{4}a \, \mathbf{\hat{x}} + \frac{1}{2}c \, \mathbf{\hat{z}} & \left(8e\right) & \mbox{Ni II} \\ 
\mathbf{B}_{9} & = & -x_{4} \, \mathbf{a}_{1}-x_{4} \, \mathbf{a}_{2} & = & -x_{4}a \, \mathbf{\hat{x}} & \left(8e\right) & \mbox{Ni II} \\ 
\mathbf{B}_{10} & = & x_{4} \, \mathbf{a}_{1} + x_{4} \, \mathbf{a}_{2} + \frac{1}{2} \, \mathbf{a}_{3} & = & x_{4}a \, \mathbf{\hat{x}} + \frac{1}{2}c \, \mathbf{\hat{z}} & \left(8e\right) & \mbox{Ni II} \\ 
\mathbf{B}_{11} & = & -y_{5} \, \mathbf{a}_{1} + y_{5} \, \mathbf{a}_{2} + z_{5} \, \mathbf{a}_{3} & = & y_{5}b \, \mathbf{\hat{y}} + z_{5}c \, \mathbf{\hat{z}} & \left(8f\right) & \mbox{La II} \\ 
\mathbf{B}_{12} & = & y_{5} \, \mathbf{a}_{1}-y_{5} \, \mathbf{a}_{2} + \left(\frac{1}{2} +z_{5}\right) \, \mathbf{a}_{3} & = & -y_{5}b \, \mathbf{\hat{y}} + \left(\frac{1}{2} +z_{5}\right)c \, \mathbf{\hat{z}} & \left(8f\right) & \mbox{La II} \\ 
\mathbf{B}_{13} & = & -y_{5} \, \mathbf{a}_{1} + y_{5} \, \mathbf{a}_{2} + \left(\frac{1}{2} - z_{5}\right) \, \mathbf{a}_{3} & = & y_{5}b \, \mathbf{\hat{y}} + \left(\frac{1}{2} - z_{5}\right)c \, \mathbf{\hat{z}} & \left(8f\right) & \mbox{La II} \\ 
\mathbf{B}_{14} & = & y_{5} \, \mathbf{a}_{1}-y_{5} \, \mathbf{a}_{2}-z_{5} \, \mathbf{a}_{3} & = & -y_{5}b \, \mathbf{\hat{y}}-z_{5}c \, \mathbf{\hat{z}} & \left(8f\right) & \mbox{La II} \\ 
\mathbf{B}_{15} & = & -y_{6} \, \mathbf{a}_{1} + y_{6} \, \mathbf{a}_{2} + z_{6} \, \mathbf{a}_{3} & = & y_{6}b \, \mathbf{\hat{y}} + z_{6}c \, \mathbf{\hat{z}} & \left(8f\right) & \mbox{La III} \\ 
\mathbf{B}_{16} & = & y_{6} \, \mathbf{a}_{1}-y_{6} \, \mathbf{a}_{2} + \left(\frac{1}{2} +z_{6}\right) \, \mathbf{a}_{3} & = & -y_{6}b \, \mathbf{\hat{y}} + \left(\frac{1}{2} +z_{6}\right)c \, \mathbf{\hat{z}} & \left(8f\right) & \mbox{La III} \\ 
\mathbf{B}_{17} & = & -y_{6} \, \mathbf{a}_{1} + y_{6} \, \mathbf{a}_{2} + \left(\frac{1}{2} - z_{6}\right) \, \mathbf{a}_{3} & = & y_{6}b \, \mathbf{\hat{y}} + \left(\frac{1}{2} - z_{6}\right)c \, \mathbf{\hat{z}} & \left(8f\right) & \mbox{La III} \\ 
\mathbf{B}_{18} & = & y_{6} \, \mathbf{a}_{1}-y_{6} \, \mathbf{a}_{2}-z_{6} \, \mathbf{a}_{3} & = & -y_{6}b \, \mathbf{\hat{y}}-z_{6}c \, \mathbf{\hat{z}} & \left(8f\right) & \mbox{La III} \\ 
\mathbf{B}_{19} & = & -y_{7} \, \mathbf{a}_{1} + y_{7} \, \mathbf{a}_{2} + z_{7} \, \mathbf{a}_{3} & = & y_{7}b \, \mathbf{\hat{y}} + z_{7}c \, \mathbf{\hat{z}} & \left(8f\right) & \mbox{La IV} \\ 
\mathbf{B}_{20} & = & y_{7} \, \mathbf{a}_{1}-y_{7} \, \mathbf{a}_{2} + \left(\frac{1}{2} +z_{7}\right) \, \mathbf{a}_{3} & = & -y_{7}b \, \mathbf{\hat{y}} + \left(\frac{1}{2} +z_{7}\right)c \, \mathbf{\hat{z}} & \left(8f\right) & \mbox{La IV} \\ 
\mathbf{B}_{21} & = & -y_{7} \, \mathbf{a}_{1} + y_{7} \, \mathbf{a}_{2} + \left(\frac{1}{2} - z_{7}\right) \, \mathbf{a}_{3} & = & y_{7}b \, \mathbf{\hat{y}} + \left(\frac{1}{2} - z_{7}\right)c \, \mathbf{\hat{z}} & \left(8f\right) & \mbox{La IV} \\ 
\mathbf{B}_{22} & = & y_{7} \, \mathbf{a}_{1}-y_{7} \, \mathbf{a}_{2}-z_{7} \, \mathbf{a}_{3} & = & -y_{7}b \, \mathbf{\hat{y}}-z_{7}c \, \mathbf{\hat{z}} & \left(8f\right) & \mbox{La IV} \\ 
\mathbf{B}_{23} & = & -y_{8} \, \mathbf{a}_{1} + y_{8} \, \mathbf{a}_{2} + z_{8} \, \mathbf{a}_{3} & = & y_{8}b \, \mathbf{\hat{y}} + z_{8}c \, \mathbf{\hat{z}} & \left(8f\right) & \mbox{La V} \\ 
\mathbf{B}_{24} & = & y_{8} \, \mathbf{a}_{1}-y_{8} \, \mathbf{a}_{2} + \left(\frac{1}{2} +z_{8}\right) \, \mathbf{a}_{3} & = & -y_{8}b \, \mathbf{\hat{y}} + \left(\frac{1}{2} +z_{8}\right)c \, \mathbf{\hat{z}} & \left(8f\right) & \mbox{La V} \\ 
\mathbf{B}_{25} & = & -y_{8} \, \mathbf{a}_{1} + y_{8} \, \mathbf{a}_{2} + \left(\frac{1}{2} - z_{8}\right) \, \mathbf{a}_{3} & = & y_{8}b \, \mathbf{\hat{y}} + \left(\frac{1}{2} - z_{8}\right)c \, \mathbf{\hat{z}} & \left(8f\right) & \mbox{La V} \\ 
\mathbf{B}_{26} & = & y_{8} \, \mathbf{a}_{1}-y_{8} \, \mathbf{a}_{2}-z_{8} \, \mathbf{a}_{3} & = & -y_{8}b \, \mathbf{\hat{y}}-z_{8}c \, \mathbf{\hat{z}} & \left(8f\right) & \mbox{La V} \\ 
\mathbf{B}_{27} & = & -y_{9} \, \mathbf{a}_{1} + y_{9} \, \mathbf{a}_{2} + z_{9} \, \mathbf{a}_{3} & = & y_{9}b \, \mathbf{\hat{y}} + z_{9}c \, \mathbf{\hat{z}} & \left(8f\right) & \mbox{La VI} \\ 
\mathbf{B}_{28} & = & y_{9} \, \mathbf{a}_{1}-y_{9} \, \mathbf{a}_{2} + \left(\frac{1}{2} +z_{9}\right) \, \mathbf{a}_{3} & = & -y_{9}b \, \mathbf{\hat{y}} + \left(\frac{1}{2} +z_{9}\right)c \, \mathbf{\hat{z}} & \left(8f\right) & \mbox{La VI} \\ 
\mathbf{B}_{29} & = & -y_{9} \, \mathbf{a}_{1} + y_{9} \, \mathbf{a}_{2} + \left(\frac{1}{2} - z_{9}\right) \, \mathbf{a}_{3} & = & y_{9}b \, \mathbf{\hat{y}} + \left(\frac{1}{2} - z_{9}\right)c \, \mathbf{\hat{z}} & \left(8f\right) & \mbox{La VI} \\ 
\mathbf{B}_{30} & = & y_{9} \, \mathbf{a}_{1}-y_{9} \, \mathbf{a}_{2}-z_{9} \, \mathbf{a}_{3} & = & -y_{9}b \, \mathbf{\hat{y}}-z_{9}c \, \mathbf{\hat{z}} & \left(8f\right) & \mbox{La VI} \\ 
\mathbf{B}_{31} & = & -y_{10} \, \mathbf{a}_{1} + y_{10} \, \mathbf{a}_{2} + z_{10} \, \mathbf{a}_{3} & = & y_{10}b \, \mathbf{\hat{y}} + z_{10}c \, \mathbf{\hat{z}} & \left(8f\right) & \mbox{La VII} \\ 
\mathbf{B}_{32} & = & y_{10} \, \mathbf{a}_{1}-y_{10} \, \mathbf{a}_{2} + \left(\frac{1}{2} +z_{10}\right) \, \mathbf{a}_{3} & = & -y_{10}b \, \mathbf{\hat{y}} + \left(\frac{1}{2} +z_{10}\right)c \, \mathbf{\hat{z}} & \left(8f\right) & \mbox{La VII} \\ 
\mathbf{B}_{33} & = & -y_{10} \, \mathbf{a}_{1} + y_{10} \, \mathbf{a}_{2} + \left(\frac{1}{2} - z_{10}\right) \, \mathbf{a}_{3} & = & y_{10}b \, \mathbf{\hat{y}} + \left(\frac{1}{2} - z_{10}\right)c \, \mathbf{\hat{z}} & \left(8f\right) & \mbox{La VII} \\ 
\mathbf{B}_{34} & = & y_{10} \, \mathbf{a}_{1}-y_{10} \, \mathbf{a}_{2}-z_{10} \, \mathbf{a}_{3} & = & -y_{10}b \, \mathbf{\hat{y}}-z_{10}c \, \mathbf{\hat{z}} & \left(8f\right) & \mbox{La VII} \\ 
\mathbf{B}_{35} & = & -y_{11} \, \mathbf{a}_{1} + y_{11} \, \mathbf{a}_{2} + z_{11} \, \mathbf{a}_{3} & = & y_{11}b \, \mathbf{\hat{y}} + z_{11}c \, \mathbf{\hat{z}} & \left(8f\right) & \mbox{La VIII} \\ 
\mathbf{B}_{36} & = & y_{11} \, \mathbf{a}_{1}-y_{11} \, \mathbf{a}_{2} + \left(\frac{1}{2} +z_{11}\right) \, \mathbf{a}_{3} & = & -y_{11}b \, \mathbf{\hat{y}} + \left(\frac{1}{2} +z_{11}\right)c \, \mathbf{\hat{z}} & \left(8f\right) & \mbox{La VIII} \\ 
\mathbf{B}_{37} & = & -y_{11} \, \mathbf{a}_{1} + y_{11} \, \mathbf{a}_{2} + \left(\frac{1}{2} - z_{11}\right) \, \mathbf{a}_{3} & = & y_{11}b \, \mathbf{\hat{y}} + \left(\frac{1}{2} - z_{11}\right)c \, \mathbf{\hat{z}} & \left(8f\right) & \mbox{La VIII} \\ 
\mathbf{B}_{38} & = & y_{11} \, \mathbf{a}_{1}-y_{11} \, \mathbf{a}_{2}-z_{11} \, \mathbf{a}_{3} & = & -y_{11}b \, \mathbf{\hat{y}}-z_{11}c \, \mathbf{\hat{z}} & \left(8f\right) & \mbox{La VIII} \\ 
\mathbf{B}_{39} & = & -y_{12} \, \mathbf{a}_{1} + y_{12} \, \mathbf{a}_{2} + z_{12} \, \mathbf{a}_{3} & = & y_{12}b \, \mathbf{\hat{y}} + z_{12}c \, \mathbf{\hat{z}} & \left(8f\right) & \mbox{La IX} \\ 
\mathbf{B}_{40} & = & y_{12} \, \mathbf{a}_{1}-y_{12} \, \mathbf{a}_{2} + \left(\frac{1}{2} +z_{12}\right) \, \mathbf{a}_{3} & = & -y_{12}b \, \mathbf{\hat{y}} + \left(\frac{1}{2} +z_{12}\right)c \, \mathbf{\hat{z}} & \left(8f\right) & \mbox{La IX} \\ 
\mathbf{B}_{41} & = & -y_{12} \, \mathbf{a}_{1} + y_{12} \, \mathbf{a}_{2} + \left(\frac{1}{2} - z_{12}\right) \, \mathbf{a}_{3} & = & y_{12}b \, \mathbf{\hat{y}} + \left(\frac{1}{2} - z_{12}\right)c \, \mathbf{\hat{z}} & \left(8f\right) & \mbox{La IX} \\ 
\mathbf{B}_{42} & = & y_{12} \, \mathbf{a}_{1}-y_{12} \, \mathbf{a}_{2}-z_{12} \, \mathbf{a}_{3} & = & -y_{12}b \, \mathbf{\hat{y}}-z_{12}c \, \mathbf{\hat{z}} & \left(8f\right) & \mbox{La IX} \\ 
\mathbf{B}_{43} & = & -y_{13} \, \mathbf{a}_{1} + y_{13} \, \mathbf{a}_{2} + z_{13} \, \mathbf{a}_{3} & = & y_{13}b \, \mathbf{\hat{y}} + z_{13}c \, \mathbf{\hat{z}} & \left(8f\right) & \mbox{Mg II} \\ 
\mathbf{B}_{44} & = & y_{13} \, \mathbf{a}_{1}-y_{13} \, \mathbf{a}_{2} + \left(\frac{1}{2} +z_{13}\right) \, \mathbf{a}_{3} & = & -y_{13}b \, \mathbf{\hat{y}} + \left(\frac{1}{2} +z_{13}\right)c \, \mathbf{\hat{z}} & \left(8f\right) & \mbox{Mg II} \\ 
\mathbf{B}_{45} & = & -y_{13} \, \mathbf{a}_{1} + y_{13} \, \mathbf{a}_{2} + \left(\frac{1}{2} - z_{13}\right) \, \mathbf{a}_{3} & = & y_{13}b \, \mathbf{\hat{y}} + \left(\frac{1}{2} - z_{13}\right)c \, \mathbf{\hat{z}} & \left(8f\right) & \mbox{Mg II} \\ 
\mathbf{B}_{46} & = & y_{13} \, \mathbf{a}_{1}-y_{13} \, \mathbf{a}_{2}-z_{13} \, \mathbf{a}_{3} & = & -y_{13}b \, \mathbf{\hat{y}}-z_{13}c \, \mathbf{\hat{z}} & \left(8f\right) & \mbox{Mg II} \\ 
\mathbf{B}_{47} & = & -y_{14} \, \mathbf{a}_{1} + y_{14} \, \mathbf{a}_{2} + z_{14} \, \mathbf{a}_{3} & = & y_{14}b \, \mathbf{\hat{y}} + z_{14}c \, \mathbf{\hat{z}} & \left(8f\right) & \mbox{Ni III} \\ 
\mathbf{B}_{48} & = & y_{14} \, \mathbf{a}_{1}-y_{14} \, \mathbf{a}_{2} + \left(\frac{1}{2} +z_{14}\right) \, \mathbf{a}_{3} & = & -y_{14}b \, \mathbf{\hat{y}} + \left(\frac{1}{2} +z_{14}\right)c \, \mathbf{\hat{z}} & \left(8f\right) & \mbox{Ni III} \\ 
\mathbf{B}_{49} & = & -y_{14} \, \mathbf{a}_{1} + y_{14} \, \mathbf{a}_{2} + \left(\frac{1}{2} - z_{14}\right) \, \mathbf{a}_{3} & = & y_{14}b \, \mathbf{\hat{y}} + \left(\frac{1}{2} - z_{14}\right)c \, \mathbf{\hat{z}} & \left(8f\right) & \mbox{Ni III} \\ 
\mathbf{B}_{50} & = & y_{14} \, \mathbf{a}_{1}-y_{14} \, \mathbf{a}_{2}-z_{14} \, \mathbf{a}_{3} & = & -y_{14}b \, \mathbf{\hat{y}}-z_{14}c \, \mathbf{\hat{z}} & \left(8f\right) & \mbox{Ni III} \\ 
\mathbf{B}_{51} & = & -y_{15} \, \mathbf{a}_{1} + y_{15} \, \mathbf{a}_{2} + z_{15} \, \mathbf{a}_{3} & = & y_{15}b \, \mathbf{\hat{y}} + z_{15}c \, \mathbf{\hat{z}} & \left(8f\right) & \mbox{Ni IV} \\ 
\mathbf{B}_{52} & = & y_{15} \, \mathbf{a}_{1}-y_{15} \, \mathbf{a}_{2} + \left(\frac{1}{2} +z_{15}\right) \, \mathbf{a}_{3} & = & -y_{15}b \, \mathbf{\hat{y}} + \left(\frac{1}{2} +z_{15}\right)c \, \mathbf{\hat{z}} & \left(8f\right) & \mbox{Ni IV} \\ 
\mathbf{B}_{53} & = & -y_{15} \, \mathbf{a}_{1} + y_{15} \, \mathbf{a}_{2} + \left(\frac{1}{2} - z_{15}\right) \, \mathbf{a}_{3} & = & y_{15}b \, \mathbf{\hat{y}} + \left(\frac{1}{2} - z_{15}\right)c \, \mathbf{\hat{z}} & \left(8f\right) & \mbox{Ni IV} \\ 
\mathbf{B}_{54} & = & y_{15} \, \mathbf{a}_{1}-y_{15} \, \mathbf{a}_{2}-z_{15} \, \mathbf{a}_{3} & = & -y_{15}b \, \mathbf{\hat{y}}-z_{15}c \, \mathbf{\hat{z}} & \left(8f\right) & \mbox{Ni IV} \\ 
\mathbf{B}_{55} & = & -y_{16} \, \mathbf{a}_{1} + y_{16} \, \mathbf{a}_{2} + z_{16} \, \mathbf{a}_{3} & = & y_{16}b \, \mathbf{\hat{y}} + z_{16}c \, \mathbf{\hat{z}} & \left(8f\right) & \mbox{Ni V} \\ 
\mathbf{B}_{56} & = & y_{16} \, \mathbf{a}_{1}-y_{16} \, \mathbf{a}_{2} + \left(\frac{1}{2} +z_{16}\right) \, \mathbf{a}_{3} & = & -y_{16}b \, \mathbf{\hat{y}} + \left(\frac{1}{2} +z_{16}\right)c \, \mathbf{\hat{z}} & \left(8f\right) & \mbox{Ni V} \\ 
\mathbf{B}_{57} & = & -y_{16} \, \mathbf{a}_{1} + y_{16} \, \mathbf{a}_{2} + \left(\frac{1}{2} - z_{16}\right) \, \mathbf{a}_{3} & = & y_{16}b \, \mathbf{\hat{y}} + \left(\frac{1}{2} - z_{16}\right)c \, \mathbf{\hat{z}} & \left(8f\right) & \mbox{Ni V} \\ 
\mathbf{B}_{58} & = & y_{16} \, \mathbf{a}_{1}-y_{16} \, \mathbf{a}_{2}-z_{16} \, \mathbf{a}_{3} & = & -y_{16}b \, \mathbf{\hat{y}}-z_{16}c \, \mathbf{\hat{z}} & \left(8f\right) & \mbox{Ni V} \\ 
\mathbf{B}_{59} & = & \left(x_{17}-y_{17}\right) \, \mathbf{a}_{1} + \left(x_{17}+y_{17}\right) \, \mathbf{a}_{2} + \frac{1}{4} \, \mathbf{a}_{3} & = & x_{17}a \, \mathbf{\hat{x}} + y_{17}b \, \mathbf{\hat{y}} + \frac{1}{4}c \, \mathbf{\hat{z}} & \left(8g\right) & \mbox{La X} \\ 
\mathbf{B}_{60} & = & \left(-x_{17}+y_{17}\right) \, \mathbf{a}_{1} + \left(-x_{17}-y_{17}\right) \, \mathbf{a}_{2} + \frac{3}{4} \, \mathbf{a}_{3} & = & -x_{17}a \, \mathbf{\hat{x}}-y_{17}b \, \mathbf{\hat{y}} + \frac{3}{4}c \, \mathbf{\hat{z}} & \left(8g\right) & \mbox{La X} \\ 
\mathbf{B}_{61} & = & \left(-x_{17}-y_{17}\right) \, \mathbf{a}_{1} + \left(-x_{17}+y_{17}\right) \, \mathbf{a}_{2} + \frac{1}{4} \, \mathbf{a}_{3} & = & -x_{17}a \, \mathbf{\hat{x}} + y_{17}b \, \mathbf{\hat{y}} + \frac{1}{4}c \, \mathbf{\hat{z}} & \left(8g\right) & \mbox{La X} \\ 
\mathbf{B}_{62} & = & \left(x_{17}+y_{17}\right) \, \mathbf{a}_{1} + \left(x_{17}-y_{17}\right) \, \mathbf{a}_{2} + \frac{3}{4} \, \mathbf{a}_{3} & = & x_{17}a \, \mathbf{\hat{x}}-y_{17}b \, \mathbf{\hat{y}} + \frac{3}{4}c \, \mathbf{\hat{z}} & \left(8g\right) & \mbox{La X} \\ 
\mathbf{B}_{63} & = & \left(x_{18}-y_{18}\right) \, \mathbf{a}_{1} + \left(x_{18}+y_{18}\right) \, \mathbf{a}_{2} + \frac{1}{4} \, \mathbf{a}_{3} & = & x_{18}a \, \mathbf{\hat{x}} + y_{18}b \, \mathbf{\hat{y}} + \frac{1}{4}c \, \mathbf{\hat{z}} & \left(8g\right) & \mbox{Mg III} \\ 
\mathbf{B}_{64} & = & \left(-x_{18}+y_{18}\right) \, \mathbf{a}_{1} + \left(-x_{18}-y_{18}\right) \, \mathbf{a}_{2} + \frac{3}{4} \, \mathbf{a}_{3} & = & -x_{18}a \, \mathbf{\hat{x}}-y_{18}b \, \mathbf{\hat{y}} + \frac{3}{4}c \, \mathbf{\hat{z}} & \left(8g\right) & \mbox{Mg III} \\ 
\mathbf{B}_{65} & = & \left(-x_{18}-y_{18}\right) \, \mathbf{a}_{1} + \left(-x_{18}+y_{18}\right) \, \mathbf{a}_{2} + \frac{1}{4} \, \mathbf{a}_{3} & = & -x_{18}a \, \mathbf{\hat{x}} + y_{18}b \, \mathbf{\hat{y}} + \frac{1}{4}c \, \mathbf{\hat{z}} & \left(8g\right) & \mbox{Mg III} \\ 
\mathbf{B}_{66} & = & \left(x_{18}+y_{18}\right) \, \mathbf{a}_{1} + \left(x_{18}-y_{18}\right) \, \mathbf{a}_{2} + \frac{3}{4} \, \mathbf{a}_{3} & = & x_{18}a \, \mathbf{\hat{x}}-y_{18}b \, \mathbf{\hat{y}} + \frac{3}{4}c \, \mathbf{\hat{z}} & \left(8g\right) & \mbox{Mg III} \\ 
\mathbf{B}_{67} & = & \left(x_{19}-y_{19}\right) \, \mathbf{a}_{1} + \left(x_{19}+y_{19}\right) \, \mathbf{a}_{2} + z_{19} \, \mathbf{a}_{3} & = & x_{19}a \, \mathbf{\hat{x}} + y_{19}b \, \mathbf{\hat{y}} + z_{19}c \, \mathbf{\hat{z}} & \left(16h\right) & \mbox{La XI} \\ 
\mathbf{B}_{68} & = & \left(-x_{19}+y_{19}\right) \, \mathbf{a}_{1} + \left(-x_{19}-y_{19}\right) \, \mathbf{a}_{2} + \left(\frac{1}{2} +z_{19}\right) \, \mathbf{a}_{3} & = & -x_{19}a \, \mathbf{\hat{x}}-y_{19}b \, \mathbf{\hat{y}} + \left(\frac{1}{2} +z_{19}\right)c \, \mathbf{\hat{z}} & \left(16h\right) & \mbox{La XI} \\ 
\mathbf{B}_{69} & = & \left(-x_{19}-y_{19}\right) \, \mathbf{a}_{1} + \left(-x_{19}+y_{19}\right) \, \mathbf{a}_{2} + \left(\frac{1}{2} - z_{19}\right) \, \mathbf{a}_{3} & = & -x_{19}a \, \mathbf{\hat{x}} + y_{19}b \, \mathbf{\hat{y}} + \left(\frac{1}{2} - z_{19}\right)c \, \mathbf{\hat{z}} & \left(16h\right) & \mbox{La XI} \\ 
\mathbf{B}_{70} & = & \left(x_{19}+y_{19}\right) \, \mathbf{a}_{1} + \left(x_{19}-y_{19}\right) \, \mathbf{a}_{2}-z_{19} \, \mathbf{a}_{3} & = & x_{19}a \, \mathbf{\hat{x}}-y_{19}b \, \mathbf{\hat{y}}-z_{19}c \, \mathbf{\hat{z}} & \left(16h\right) & \mbox{La XI} \\ 
\mathbf{B}_{71} & = & \left(-x_{19}+y_{19}\right) \, \mathbf{a}_{1} + \left(-x_{19}-y_{19}\right) \, \mathbf{a}_{2}-z_{19} \, \mathbf{a}_{3} & = & -x_{19}a \, \mathbf{\hat{x}}-y_{19}b \, \mathbf{\hat{y}}-z_{19}c \, \mathbf{\hat{z}} & \left(16h\right) & \mbox{La XI} \\ 
\mathbf{B}_{72} & = & \left(x_{19}-y_{19}\right) \, \mathbf{a}_{1} + \left(x_{19}+y_{19}\right) \, \mathbf{a}_{2} + \left(\frac{1}{2} - z_{19}\right) \, \mathbf{a}_{3} & = & x_{19}a \, \mathbf{\hat{x}} + y_{19}b \, \mathbf{\hat{y}} + \left(\frac{1}{2} - z_{19}\right)c \, \mathbf{\hat{z}} & \left(16h\right) & \mbox{La XI} \\ 
\mathbf{B}_{73} & = & \left(x_{19}+y_{19}\right) \, \mathbf{a}_{1} + \left(x_{19}-y_{19}\right) \, \mathbf{a}_{2} + \left(\frac{1}{2} +z_{19}\right) \, \mathbf{a}_{3} & = & x_{19}a \, \mathbf{\hat{x}}-y_{19}b \, \mathbf{\hat{y}} + \left(\frac{1}{2} +z_{19}\right)c \, \mathbf{\hat{z}} & \left(16h\right) & \mbox{La XI} \\ 
\mathbf{B}_{74} & = & \left(-x_{19}-y_{19}\right) \, \mathbf{a}_{1} + \left(-x_{19}+y_{19}\right) \, \mathbf{a}_{2} + z_{19} \, \mathbf{a}_{3} & = & -x_{19}a \, \mathbf{\hat{x}} + y_{19}b \, \mathbf{\hat{y}} + z_{19}c \, \mathbf{\hat{z}} & \left(16h\right) & \mbox{La XI} \\ 
\mathbf{B}_{75} & = & \left(x_{20}-y_{20}\right) \, \mathbf{a}_{1} + \left(x_{20}+y_{20}\right) \, \mathbf{a}_{2} + z_{20} \, \mathbf{a}_{3} & = & x_{20}a \, \mathbf{\hat{x}} + y_{20}b \, \mathbf{\hat{y}} + z_{20}c \, \mathbf{\hat{z}} & \left(16h\right) & \mbox{La XII} \\ 
\mathbf{B}_{76} & = & \left(-x_{20}+y_{20}\right) \, \mathbf{a}_{1} + \left(-x_{20}-y_{20}\right) \, \mathbf{a}_{2} + \left(\frac{1}{2} +z_{20}\right) \, \mathbf{a}_{3} & = & -x_{20}a \, \mathbf{\hat{x}}-y_{20}b \, \mathbf{\hat{y}} + \left(\frac{1}{2} +z_{20}\right)c \, \mathbf{\hat{z}} & \left(16h\right) & \mbox{La XII} \\ 
\mathbf{B}_{77} & = & \left(-x_{20}-y_{20}\right) \, \mathbf{a}_{1} + \left(-x_{20}+y_{20}\right) \, \mathbf{a}_{2} + \left(\frac{1}{2} - z_{20}\right) \, \mathbf{a}_{3} & = & -x_{20}a \, \mathbf{\hat{x}} + y_{20}b \, \mathbf{\hat{y}} + \left(\frac{1}{2} - z_{20}\right)c \, \mathbf{\hat{z}} & \left(16h\right) & \mbox{La XII} \\ 
\mathbf{B}_{78} & = & \left(x_{20}+y_{20}\right) \, \mathbf{a}_{1} + \left(x_{20}-y_{20}\right) \, \mathbf{a}_{2}-z_{20} \, \mathbf{a}_{3} & = & x_{20}a \, \mathbf{\hat{x}}-y_{20}b \, \mathbf{\hat{y}}-z_{20}c \, \mathbf{\hat{z}} & \left(16h\right) & \mbox{La XII} \\ 
\mathbf{B}_{79} & = & \left(-x_{20}+y_{20}\right) \, \mathbf{a}_{1} + \left(-x_{20}-y_{20}\right) \, \mathbf{a}_{2}-z_{20} \, \mathbf{a}_{3} & = & -x_{20}a \, \mathbf{\hat{x}}-y_{20}b \, \mathbf{\hat{y}}-z_{20}c \, \mathbf{\hat{z}} & \left(16h\right) & \mbox{La XII} \\ 
\mathbf{B}_{80} & = & \left(x_{20}-y_{20}\right) \, \mathbf{a}_{1} + \left(x_{20}+y_{20}\right) \, \mathbf{a}_{2} + \left(\frac{1}{2} - z_{20}\right) \, \mathbf{a}_{3} & = & x_{20}a \, \mathbf{\hat{x}} + y_{20}b \, \mathbf{\hat{y}} + \left(\frac{1}{2} - z_{20}\right)c \, \mathbf{\hat{z}} & \left(16h\right) & \mbox{La XII} \\ 
\mathbf{B}_{81} & = & \left(x_{20}+y_{20}\right) \, \mathbf{a}_{1} + \left(x_{20}-y_{20}\right) \, \mathbf{a}_{2} + \left(\frac{1}{2} +z_{20}\right) \, \mathbf{a}_{3} & = & x_{20}a \, \mathbf{\hat{x}}-y_{20}b \, \mathbf{\hat{y}} + \left(\frac{1}{2} +z_{20}\right)c \, \mathbf{\hat{z}} & \left(16h\right) & \mbox{La XII} \\ 
\mathbf{B}_{82} & = & \left(-x_{20}-y_{20}\right) \, \mathbf{a}_{1} + \left(-x_{20}+y_{20}\right) \, \mathbf{a}_{2} + z_{20} \, \mathbf{a}_{3} & = & -x_{20}a \, \mathbf{\hat{x}} + y_{20}b \, \mathbf{\hat{y}} + z_{20}c \, \mathbf{\hat{z}} & \left(16h\right) & \mbox{La XII} \\ 
\mathbf{B}_{83} & = & \left(x_{21}-y_{21}\right) \, \mathbf{a}_{1} + \left(x_{21}+y_{21}\right) \, \mathbf{a}_{2} + z_{21} \, \mathbf{a}_{3} & = & x_{21}a \, \mathbf{\hat{x}} + y_{21}b \, \mathbf{\hat{y}} + z_{21}c \, \mathbf{\hat{z}} & \left(16h\right) & \mbox{La XIII} \\ 
\mathbf{B}_{84} & = & \left(-x_{21}+y_{21}\right) \, \mathbf{a}_{1} + \left(-x_{21}-y_{21}\right) \, \mathbf{a}_{2} + \left(\frac{1}{2} +z_{21}\right) \, \mathbf{a}_{3} & = & -x_{21}a \, \mathbf{\hat{x}}-y_{21}b \, \mathbf{\hat{y}} + \left(\frac{1}{2} +z_{21}\right)c \, \mathbf{\hat{z}} & \left(16h\right) & \mbox{La XIII} \\ 
\mathbf{B}_{85} & = & \left(-x_{21}-y_{21}\right) \, \mathbf{a}_{1} + \left(-x_{21}+y_{21}\right) \, \mathbf{a}_{2} + \left(\frac{1}{2} - z_{21}\right) \, \mathbf{a}_{3} & = & -x_{21}a \, \mathbf{\hat{x}} + y_{21}b \, \mathbf{\hat{y}} + \left(\frac{1}{2} - z_{21}\right)c \, \mathbf{\hat{z}} & \left(16h\right) & \mbox{La XIII} \\ 
\mathbf{B}_{86} & = & \left(x_{21}+y_{21}\right) \, \mathbf{a}_{1} + \left(x_{21}-y_{21}\right) \, \mathbf{a}_{2}-z_{21} \, \mathbf{a}_{3} & = & x_{21}a \, \mathbf{\hat{x}}-y_{21}b \, \mathbf{\hat{y}}-z_{21}c \, \mathbf{\hat{z}} & \left(16h\right) & \mbox{La XIII} \\ 
\mathbf{B}_{87} & = & \left(-x_{21}+y_{21}\right) \, \mathbf{a}_{1} + \left(-x_{21}-y_{21}\right) \, \mathbf{a}_{2}-z_{21} \, \mathbf{a}_{3} & = & -x_{21}a \, \mathbf{\hat{x}}-y_{21}b \, \mathbf{\hat{y}}-z_{21}c \, \mathbf{\hat{z}} & \left(16h\right) & \mbox{La XIII} \\ 
\mathbf{B}_{88} & = & \left(x_{21}-y_{21}\right) \, \mathbf{a}_{1} + \left(x_{21}+y_{21}\right) \, \mathbf{a}_{2} + \left(\frac{1}{2} - z_{21}\right) \, \mathbf{a}_{3} & = & x_{21}a \, \mathbf{\hat{x}} + y_{21}b \, \mathbf{\hat{y}} + \left(\frac{1}{2} - z_{21}\right)c \, \mathbf{\hat{z}} & \left(16h\right) & \mbox{La XIII} \\ 
\mathbf{B}_{89} & = & \left(x_{21}+y_{21}\right) \, \mathbf{a}_{1} + \left(x_{21}-y_{21}\right) \, \mathbf{a}_{2} + \left(\frac{1}{2} +z_{21}\right) \, \mathbf{a}_{3} & = & x_{21}a \, \mathbf{\hat{x}}-y_{21}b \, \mathbf{\hat{y}} + \left(\frac{1}{2} +z_{21}\right)c \, \mathbf{\hat{z}} & \left(16h\right) & \mbox{La XIII} \\ 
\mathbf{B}_{90} & = & \left(-x_{21}-y_{21}\right) \, \mathbf{a}_{1} + \left(-x_{21}+y_{21}\right) \, \mathbf{a}_{2} + z_{21} \, \mathbf{a}_{3} & = & -x_{21}a \, \mathbf{\hat{x}} + y_{21}b \, \mathbf{\hat{y}} + z_{21}c \, \mathbf{\hat{z}} & \left(16h\right) & \mbox{La XIII} \\ 
\mathbf{B}_{91} & = & \left(x_{22}-y_{22}\right) \, \mathbf{a}_{1} + \left(x_{22}+y_{22}\right) \, \mathbf{a}_{2} + z_{22} \, \mathbf{a}_{3} & = & x_{22}a \, \mathbf{\hat{x}} + y_{22}b \, \mathbf{\hat{y}} + z_{22}c \, \mathbf{\hat{z}} & \left(16h\right) & \mbox{La XIV} \\ 
\mathbf{B}_{92} & = & \left(-x_{22}+y_{22}\right) \, \mathbf{a}_{1} + \left(-x_{22}-y_{22}\right) \, \mathbf{a}_{2} + \left(\frac{1}{2} +z_{22}\right) \, \mathbf{a}_{3} & = & -x_{22}a \, \mathbf{\hat{x}}-y_{22}b \, \mathbf{\hat{y}} + \left(\frac{1}{2} +z_{22}\right)c \, \mathbf{\hat{z}} & \left(16h\right) & \mbox{La XIV} \\ 
\mathbf{B}_{93} & = & \left(-x_{22}-y_{22}\right) \, \mathbf{a}_{1} + \left(-x_{22}+y_{22}\right) \, \mathbf{a}_{2} + \left(\frac{1}{2} - z_{22}\right) \, \mathbf{a}_{3} & = & -x_{22}a \, \mathbf{\hat{x}} + y_{22}b \, \mathbf{\hat{y}} + \left(\frac{1}{2} - z_{22}\right)c \, \mathbf{\hat{z}} & \left(16h\right) & \mbox{La XIV} \\ 
\mathbf{B}_{94} & = & \left(x_{22}+y_{22}\right) \, \mathbf{a}_{1} + \left(x_{22}-y_{22}\right) \, \mathbf{a}_{2}-z_{22} \, \mathbf{a}_{3} & = & x_{22}a \, \mathbf{\hat{x}}-y_{22}b \, \mathbf{\hat{y}}-z_{22}c \, \mathbf{\hat{z}} & \left(16h\right) & \mbox{La XIV} \\ 
\mathbf{B}_{95} & = & \left(-x_{22}+y_{22}\right) \, \mathbf{a}_{1} + \left(-x_{22}-y_{22}\right) \, \mathbf{a}_{2}-z_{22} \, \mathbf{a}_{3} & = & -x_{22}a \, \mathbf{\hat{x}}-y_{22}b \, \mathbf{\hat{y}}-z_{22}c \, \mathbf{\hat{z}} & \left(16h\right) & \mbox{La XIV} \\ 
\mathbf{B}_{96} & = & \left(x_{22}-y_{22}\right) \, \mathbf{a}_{1} + \left(x_{22}+y_{22}\right) \, \mathbf{a}_{2} + \left(\frac{1}{2} - z_{22}\right) \, \mathbf{a}_{3} & = & x_{22}a \, \mathbf{\hat{x}} + y_{22}b \, \mathbf{\hat{y}} + \left(\frac{1}{2} - z_{22}\right)c \, \mathbf{\hat{z}} & \left(16h\right) & \mbox{La XIV} \\ 
\mathbf{B}_{97} & = & \left(x_{22}+y_{22}\right) \, \mathbf{a}_{1} + \left(x_{22}-y_{22}\right) \, \mathbf{a}_{2} + \left(\frac{1}{2} +z_{22}\right) \, \mathbf{a}_{3} & = & x_{22}a \, \mathbf{\hat{x}}-y_{22}b \, \mathbf{\hat{y}} + \left(\frac{1}{2} +z_{22}\right)c \, \mathbf{\hat{z}} & \left(16h\right) & \mbox{La XIV} \\ 
\mathbf{B}_{98} & = & \left(-x_{22}-y_{22}\right) \, \mathbf{a}_{1} + \left(-x_{22}+y_{22}\right) \, \mathbf{a}_{2} + z_{22} \, \mathbf{a}_{3} & = & -x_{22}a \, \mathbf{\hat{x}} + y_{22}b \, \mathbf{\hat{y}} + z_{22}c \, \mathbf{\hat{z}} & \left(16h\right) & \mbox{La XIV} \\ 
\mathbf{B}_{99} & = & \left(x_{23}-y_{23}\right) \, \mathbf{a}_{1} + \left(x_{23}+y_{23}\right) \, \mathbf{a}_{2} + z_{23} \, \mathbf{a}_{3} & = & x_{23}a \, \mathbf{\hat{x}} + y_{23}b \, \mathbf{\hat{y}} + z_{23}c \, \mathbf{\hat{z}} & \left(16h\right) & \mbox{La XV} \\ 
\mathbf{B}_{100} & = & \left(-x_{23}+y_{23}\right) \, \mathbf{a}_{1} + \left(-x_{23}-y_{23}\right) \, \mathbf{a}_{2} + \left(\frac{1}{2} +z_{23}\right) \, \mathbf{a}_{3} & = & -x_{23}a \, \mathbf{\hat{x}}-y_{23}b \, \mathbf{\hat{y}} + \left(\frac{1}{2} +z_{23}\right)c \, \mathbf{\hat{z}} & \left(16h\right) & \mbox{La XV} \\ 
\mathbf{B}_{101} & = & \left(-x_{23}-y_{23}\right) \, \mathbf{a}_{1} + \left(-x_{23}+y_{23}\right) \, \mathbf{a}_{2} + \left(\frac{1}{2} - z_{23}\right) \, \mathbf{a}_{3} & = & -x_{23}a \, \mathbf{\hat{x}} + y_{23}b \, \mathbf{\hat{y}} + \left(\frac{1}{2} - z_{23}\right)c \, \mathbf{\hat{z}} & \left(16h\right) & \mbox{La XV} \\ 
\mathbf{B}_{102} & = & \left(x_{23}+y_{23}\right) \, \mathbf{a}_{1} + \left(x_{23}-y_{23}\right) \, \mathbf{a}_{2}-z_{23} \, \mathbf{a}_{3} & = & x_{23}a \, \mathbf{\hat{x}}-y_{23}b \, \mathbf{\hat{y}}-z_{23}c \, \mathbf{\hat{z}} & \left(16h\right) & \mbox{La XV} \\ 
\mathbf{B}_{103} & = & \left(-x_{23}+y_{23}\right) \, \mathbf{a}_{1} + \left(-x_{23}-y_{23}\right) \, \mathbf{a}_{2}-z_{23} \, \mathbf{a}_{3} & = & -x_{23}a \, \mathbf{\hat{x}}-y_{23}b \, \mathbf{\hat{y}}-z_{23}c \, \mathbf{\hat{z}} & \left(16h\right) & \mbox{La XV} \\ 
\mathbf{B}_{104} & = & \left(x_{23}-y_{23}\right) \, \mathbf{a}_{1} + \left(x_{23}+y_{23}\right) \, \mathbf{a}_{2} + \left(\frac{1}{2} - z_{23}\right) \, \mathbf{a}_{3} & = & x_{23}a \, \mathbf{\hat{x}} + y_{23}b \, \mathbf{\hat{y}} + \left(\frac{1}{2} - z_{23}\right)c \, \mathbf{\hat{z}} & \left(16h\right) & \mbox{La XV} \\ 
\mathbf{B}_{105} & = & \left(x_{23}+y_{23}\right) \, \mathbf{a}_{1} + \left(x_{23}-y_{23}\right) \, \mathbf{a}_{2} + \left(\frac{1}{2} +z_{23}\right) \, \mathbf{a}_{3} & = & x_{23}a \, \mathbf{\hat{x}}-y_{23}b \, \mathbf{\hat{y}} + \left(\frac{1}{2} +z_{23}\right)c \, \mathbf{\hat{z}} & \left(16h\right) & \mbox{La XV} \\ 
\mathbf{B}_{106} & = & \left(-x_{23}-y_{23}\right) \, \mathbf{a}_{1} + \left(-x_{23}+y_{23}\right) \, \mathbf{a}_{2} + z_{23} \, \mathbf{a}_{3} & = & -x_{23}a \, \mathbf{\hat{x}} + y_{23}b \, \mathbf{\hat{y}} + z_{23}c \, \mathbf{\hat{z}} & \left(16h\right) & \mbox{La XV} \\ 
\mathbf{B}_{107} & = & \left(x_{24}-y_{24}\right) \, \mathbf{a}_{1} + \left(x_{24}+y_{24}\right) \, \mathbf{a}_{2} + z_{24} \, \mathbf{a}_{3} & = & x_{24}a \, \mathbf{\hat{x}} + y_{24}b \, \mathbf{\hat{y}} + z_{24}c \, \mathbf{\hat{z}} & \left(16h\right) & \mbox{La XVI} \\ 
\mathbf{B}_{108} & = & \left(-x_{24}+y_{24}\right) \, \mathbf{a}_{1} + \left(-x_{24}-y_{24}\right) \, \mathbf{a}_{2} + \left(\frac{1}{2} +z_{24}\right) \, \mathbf{a}_{3} & = & -x_{24}a \, \mathbf{\hat{x}}-y_{24}b \, \mathbf{\hat{y}} + \left(\frac{1}{2} +z_{24}\right)c \, \mathbf{\hat{z}} & \left(16h\right) & \mbox{La XVI} \\ 
\mathbf{B}_{109} & = & \left(-x_{24}-y_{24}\right) \, \mathbf{a}_{1} + \left(-x_{24}+y_{24}\right) \, \mathbf{a}_{2} + \left(\frac{1}{2} - z_{24}\right) \, \mathbf{a}_{3} & = & -x_{24}a \, \mathbf{\hat{x}} + y_{24}b \, \mathbf{\hat{y}} + \left(\frac{1}{2} - z_{24}\right)c \, \mathbf{\hat{z}} & \left(16h\right) & \mbox{La XVI} \\ 
\mathbf{B}_{110} & = & \left(x_{24}+y_{24}\right) \, \mathbf{a}_{1} + \left(x_{24}-y_{24}\right) \, \mathbf{a}_{2}-z_{24} \, \mathbf{a}_{3} & = & x_{24}a \, \mathbf{\hat{x}}-y_{24}b \, \mathbf{\hat{y}}-z_{24}c \, \mathbf{\hat{z}} & \left(16h\right) & \mbox{La XVI} \\ 
\mathbf{B}_{111} & = & \left(-x_{24}+y_{24}\right) \, \mathbf{a}_{1} + \left(-x_{24}-y_{24}\right) \, \mathbf{a}_{2}-z_{24} \, \mathbf{a}_{3} & = & -x_{24}a \, \mathbf{\hat{x}}-y_{24}b \, \mathbf{\hat{y}}-z_{24}c \, \mathbf{\hat{z}} & \left(16h\right) & \mbox{La XVI} \\ 
\mathbf{B}_{112} & = & \left(x_{24}-y_{24}\right) \, \mathbf{a}_{1} + \left(x_{24}+y_{24}\right) \, \mathbf{a}_{2} + \left(\frac{1}{2} - z_{24}\right) \, \mathbf{a}_{3} & = & x_{24}a \, \mathbf{\hat{x}} + y_{24}b \, \mathbf{\hat{y}} + \left(\frac{1}{2} - z_{24}\right)c \, \mathbf{\hat{z}} & \left(16h\right) & \mbox{La XVI} \\ 
\mathbf{B}_{113} & = & \left(x_{24}+y_{24}\right) \, \mathbf{a}_{1} + \left(x_{24}-y_{24}\right) \, \mathbf{a}_{2} + \left(\frac{1}{2} +z_{24}\right) \, \mathbf{a}_{3} & = & x_{24}a \, \mathbf{\hat{x}}-y_{24}b \, \mathbf{\hat{y}} + \left(\frac{1}{2} +z_{24}\right)c \, \mathbf{\hat{z}} & \left(16h\right) & \mbox{La XVI} \\ 
\mathbf{B}_{114} & = & \left(-x_{24}-y_{24}\right) \, \mathbf{a}_{1} + \left(-x_{24}+y_{24}\right) \, \mathbf{a}_{2} + z_{24} \, \mathbf{a}_{3} & = & -x_{24}a \, \mathbf{\hat{x}} + y_{24}b \, \mathbf{\hat{y}} + z_{24}c \, \mathbf{\hat{z}} & \left(16h\right) & \mbox{La XVI} \\ 
\mathbf{B}_{115} & = & \left(x_{25}-y_{25}\right) \, \mathbf{a}_{1} + \left(x_{25}+y_{25}\right) \, \mathbf{a}_{2} + z_{25} \, \mathbf{a}_{3} & = & x_{25}a \, \mathbf{\hat{x}} + y_{25}b \, \mathbf{\hat{y}} + z_{25}c \, \mathbf{\hat{z}} & \left(16h\right) & \mbox{Ni VI} \\ 
\mathbf{B}_{116} & = & \left(-x_{25}+y_{25}\right) \, \mathbf{a}_{1} + \left(-x_{25}-y_{25}\right) \, \mathbf{a}_{2} + \left(\frac{1}{2} +z_{25}\right) \, \mathbf{a}_{3} & = & -x_{25}a \, \mathbf{\hat{x}}-y_{25}b \, \mathbf{\hat{y}} + \left(\frac{1}{2} +z_{25}\right)c \, \mathbf{\hat{z}} & \left(16h\right) & \mbox{Ni VI} \\ 
\mathbf{B}_{117} & = & \left(-x_{25}-y_{25}\right) \, \mathbf{a}_{1} + \left(-x_{25}+y_{25}\right) \, \mathbf{a}_{2} + \left(\frac{1}{2} - z_{25}\right) \, \mathbf{a}_{3} & = & -x_{25}a \, \mathbf{\hat{x}} + y_{25}b \, \mathbf{\hat{y}} + \left(\frac{1}{2} - z_{25}\right)c \, \mathbf{\hat{z}} & \left(16h\right) & \mbox{Ni VI} \\ 
\mathbf{B}_{118} & = & \left(x_{25}+y_{25}\right) \, \mathbf{a}_{1} + \left(x_{25}-y_{25}\right) \, \mathbf{a}_{2}-z_{25} \, \mathbf{a}_{3} & = & x_{25}a \, \mathbf{\hat{x}}-y_{25}b \, \mathbf{\hat{y}}-z_{25}c \, \mathbf{\hat{z}} & \left(16h\right) & \mbox{Ni VI} \\ 
\mathbf{B}_{119} & = & \left(-x_{25}+y_{25}\right) \, \mathbf{a}_{1} + \left(-x_{25}-y_{25}\right) \, \mathbf{a}_{2}-z_{25} \, \mathbf{a}_{3} & = & -x_{25}a \, \mathbf{\hat{x}}-y_{25}b \, \mathbf{\hat{y}}-z_{25}c \, \mathbf{\hat{z}} & \left(16h\right) & \mbox{Ni VI} \\ 
\mathbf{B}_{120} & = & \left(x_{25}-y_{25}\right) \, \mathbf{a}_{1} + \left(x_{25}+y_{25}\right) \, \mathbf{a}_{2} + \left(\frac{1}{2} - z_{25}\right) \, \mathbf{a}_{3} & = & x_{25}a \, \mathbf{\hat{x}} + y_{25}b \, \mathbf{\hat{y}} + \left(\frac{1}{2} - z_{25}\right)c \, \mathbf{\hat{z}} & \left(16h\right) & \mbox{Ni VI} \\ 
\mathbf{B}_{121} & = & \left(x_{25}+y_{25}\right) \, \mathbf{a}_{1} + \left(x_{25}-y_{25}\right) \, \mathbf{a}_{2} + \left(\frac{1}{2} +z_{25}\right) \, \mathbf{a}_{3} & = & x_{25}a \, \mathbf{\hat{x}}-y_{25}b \, \mathbf{\hat{y}} + \left(\frac{1}{2} +z_{25}\right)c \, \mathbf{\hat{z}} & \left(16h\right) & \mbox{Ni VI} \\ 
\mathbf{B}_{122} & = & \left(-x_{25}-y_{25}\right) \, \mathbf{a}_{1} + \left(-x_{25}+y_{25}\right) \, \mathbf{a}_{2} + z_{25} \, \mathbf{a}_{3} & = & -x_{25}a \, \mathbf{\hat{x}} + y_{25}b \, \mathbf{\hat{y}} + z_{25}c \, \mathbf{\hat{z}} & \left(16h\right) & \mbox{Ni VI} \\ 
\mathbf{B}_{123} & = & \left(x_{26}-y_{26}\right) \, \mathbf{a}_{1} + \left(x_{26}+y_{26}\right) \, \mathbf{a}_{2} + z_{26} \, \mathbf{a}_{3} & = & x_{26}a \, \mathbf{\hat{x}} + y_{26}b \, \mathbf{\hat{y}} + z_{26}c \, \mathbf{\hat{z}} & \left(16h\right) & \mbox{Ni VII} \\ 
\mathbf{B}_{124} & = & \left(-x_{26}+y_{26}\right) \, \mathbf{a}_{1} + \left(-x_{26}-y_{26}\right) \, \mathbf{a}_{2} + \left(\frac{1}{2} +z_{26}\right) \, \mathbf{a}_{3} & = & -x_{26}a \, \mathbf{\hat{x}}-y_{26}b \, \mathbf{\hat{y}} + \left(\frac{1}{2} +z_{26}\right)c \, \mathbf{\hat{z}} & \left(16h\right) & \mbox{Ni VII} \\ 
\mathbf{B}_{125} & = & \left(-x_{26}-y_{26}\right) \, \mathbf{a}_{1} + \left(-x_{26}+y_{26}\right) \, \mathbf{a}_{2} + \left(\frac{1}{2} - z_{26}\right) \, \mathbf{a}_{3} & = & -x_{26}a \, \mathbf{\hat{x}} + y_{26}b \, \mathbf{\hat{y}} + \left(\frac{1}{2} - z_{26}\right)c \, \mathbf{\hat{z}} & \left(16h\right) & \mbox{Ni VII} \\ 
\mathbf{B}_{126} & = & \left(x_{26}+y_{26}\right) \, \mathbf{a}_{1} + \left(x_{26}-y_{26}\right) \, \mathbf{a}_{2}-z_{26} \, \mathbf{a}_{3} & = & x_{26}a \, \mathbf{\hat{x}}-y_{26}b \, \mathbf{\hat{y}}-z_{26}c \, \mathbf{\hat{z}} & \left(16h\right) & \mbox{Ni VII} \\ 
\mathbf{B}_{127} & = & \left(-x_{26}+y_{26}\right) \, \mathbf{a}_{1} + \left(-x_{26}-y_{26}\right) \, \mathbf{a}_{2}-z_{26} \, \mathbf{a}_{3} & = & -x_{26}a \, \mathbf{\hat{x}}-y_{26}b \, \mathbf{\hat{y}}-z_{26}c \, \mathbf{\hat{z}} & \left(16h\right) & \mbox{Ni VII} \\ 
\mathbf{B}_{128} & = & \left(x_{26}-y_{26}\right) \, \mathbf{a}_{1} + \left(x_{26}+y_{26}\right) \, \mathbf{a}_{2} + \left(\frac{1}{2} - z_{26}\right) \, \mathbf{a}_{3} & = & x_{26}a \, \mathbf{\hat{x}} + y_{26}b \, \mathbf{\hat{y}} + \left(\frac{1}{2} - z_{26}\right)c \, \mathbf{\hat{z}} & \left(16h\right) & \mbox{Ni VII} \\ 
\mathbf{B}_{129} & = & \left(x_{26}+y_{26}\right) \, \mathbf{a}_{1} + \left(x_{26}-y_{26}\right) \, \mathbf{a}_{2} + \left(\frac{1}{2} +z_{26}\right) \, \mathbf{a}_{3} & = & x_{26}a \, \mathbf{\hat{x}}-y_{26}b \, \mathbf{\hat{y}} + \left(\frac{1}{2} +z_{26}\right)c \, \mathbf{\hat{z}} & \left(16h\right) & \mbox{Ni VII} \\ 
\mathbf{B}_{130} & = & \left(-x_{26}-y_{26}\right) \, \mathbf{a}_{1} + \left(-x_{26}+y_{26}\right) \, \mathbf{a}_{2} + z_{26} \, \mathbf{a}_{3} & = & -x_{26}a \, \mathbf{\hat{x}} + y_{26}b \, \mathbf{\hat{y}} + z_{26}c \, \mathbf{\hat{z}} & \left(16h\right) & \mbox{Ni VII} \\ 
\end{longtabu}
\renewcommand{\arraystretch}{1.0}
\noindent \hrulefill
\\
\textbf{References:}
\vspace*{-0.25cm}
\begin{flushleft}
  - \bibentry{Solokha_Inorg_Chem_48_11586_2009}. \\
\end{flushleft}
\noindent \hrulefill
\\
\textbf{Geometry files:}
\\
\noindent  - CIF: pp. {\hyperref[A43B5C17_oC260_63_c8fg6h_cfg_ce3f2h_cif]{\pageref{A43B5C17_oC260_63_c8fg6h_cfg_ce3f2h_cif}}} \\
\noindent  - POSCAR: pp. {\hyperref[A43B5C17_oC260_63_c8fg6h_cfg_ce3f2h_poscar]{\pageref{A43B5C17_oC260_63_c8fg6h_cfg_ce3f2h_poscar}}} \\
\onecolumn
{\phantomsection\label{A6B_oC28_63_efg_c}}
\subsection*{\huge \textbf{{\normalfont MnAl$_{6}$ ($D2_{h}$) Structure: A6B\_oC28\_63\_efg\_c}}}
\noindent \hrulefill
\vspace*{0.25cm}
\begin{figure}[htp]
  \centering
  \vspace{-1em}
  {\includegraphics[width=1\textwidth]{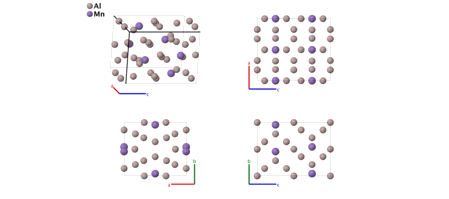}}
\end{figure}
\vspace*{-0.5cm}
\renewcommand{\arraystretch}{1.5}
\begin{equation*}
  \begin{array}{>{$\hspace{-0.15cm}}l<{$}>{$}p{0.5cm}<{$}>{$}p{18.5cm}<{$}}
    \mbox{\large \textbf{Prototype}} &\colon & \ce{MnAl6} \\
    \mbox{\large \textbf{\AFLOW\ prototype label}} &\colon & \mbox{A6B\_oC28\_63\_efg\_c} \\
    \mbox{\large \textbf{\textit{Strukturbericht} designation}} &\colon & \mbox{$D2_{h}$} \\
    \mbox{\large \textbf{Pearson symbol}} &\colon & \mbox{oC28} \\
    \mbox{\large \textbf{Space group number}} &\colon & 63 \\
    \mbox{\large \textbf{Space group symbol}} &\colon & Cmcm \\
    \mbox{\large \textbf{\AFLOW\ prototype command}} &\colon &  \texttt{aflow} \,  \, \texttt{-{}-proto=A6B\_oC28\_63\_efg\_c } \, \newline \texttt{-{}-params=}{a,b/a,c/a,y_{1},x_{2},y_{3},z_{3},x_{4},y_{4} }
  \end{array}
\end{equation*}
\renewcommand{\arraystretch}{1.0}

\noindent \parbox{1 \linewidth}{
\noindent \hrulefill
\\
\textbf{Base-centered Orthorhombic primitive vectors:} \\
\vspace*{-0.25cm}
\begin{tabular}{cc}
  \begin{tabular}{c}
    \parbox{0.6 \linewidth}{
      \renewcommand{\arraystretch}{1.5}
      \begin{equation*}
        \centering
        \begin{array}{ccc}
              \mathbf{a}_1 & = & \frac12 \, a \, \mathbf{\hat{x}} - \frac12 \, b \, \mathbf{\hat{y}} \\
    \mathbf{a}_2 & = & \frac12 \, a \, \mathbf{\hat{x}} + \frac12 \, b \, \mathbf{\hat{y}} \\
    \mathbf{a}_3 & = & c \, \mathbf{\hat{z}} \\

        \end{array}
      \end{equation*}
    }
    \renewcommand{\arraystretch}{1.0}
  \end{tabular}
  \begin{tabular}{c}
    \includegraphics[width=0.3\linewidth]{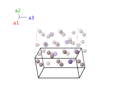} \\
  \end{tabular}
\end{tabular}

}
\vspace*{-0.25cm}

\noindent \hrulefill
\\
\textbf{Basis vectors:}
\vspace*{-0.25cm}
\renewcommand{\arraystretch}{1.5}
\begin{longtabu} to \textwidth{>{\centering $}X[-1,c,c]<{$}>{\centering $}X[-1,c,c]<{$}>{\centering $}X[-1,c,c]<{$}>{\centering $}X[-1,c,c]<{$}>{\centering $}X[-1,c,c]<{$}>{\centering $}X[-1,c,c]<{$}>{\centering $}X[-1,c,c]<{$}}
  & & \mbox{Lattice Coordinates} & & \mbox{Cartesian Coordinates} &\mbox{Wyckoff Position} & \mbox{Atom Type} \\  
  \mathbf{B}_{1} & = & -y_{1} \, \mathbf{a}_{1} + y_{1} \, \mathbf{a}_{2} + \frac{1}{4} \, \mathbf{a}_{3} & = & y_{1}b \, \mathbf{\hat{y}} + \frac{1}{4}c \, \mathbf{\hat{z}} & \left(4c\right) & \mbox{Mn} \\ 
\mathbf{B}_{2} & = & y_{1} \, \mathbf{a}_{1}-y_{1} \, \mathbf{a}_{2} + \frac{3}{4} \, \mathbf{a}_{3} & = & -y_{1}b \, \mathbf{\hat{y}} + \frac{3}{4}c \, \mathbf{\hat{z}} & \left(4c\right) & \mbox{Mn} \\ 
\mathbf{B}_{3} & = & x_{2} \, \mathbf{a}_{1} + x_{2} \, \mathbf{a}_{2} & = & x_{2}a \, \mathbf{\hat{x}} & \left(8e\right) & \mbox{Al I} \\ 
\mathbf{B}_{4} & = & -x_{2} \, \mathbf{a}_{1}-x_{2} \, \mathbf{a}_{2} + \frac{1}{2} \, \mathbf{a}_{3} & = & -x_{2}a \, \mathbf{\hat{x}} + \frac{1}{2}c \, \mathbf{\hat{z}} & \left(8e\right) & \mbox{Al I} \\ 
\mathbf{B}_{5} & = & -x_{2} \, \mathbf{a}_{1}-x_{2} \, \mathbf{a}_{2} & = & -x_{2}a \, \mathbf{\hat{x}} & \left(8e\right) & \mbox{Al I} \\ 
\mathbf{B}_{6} & = & x_{2} \, \mathbf{a}_{1} + x_{2} \, \mathbf{a}_{2} + \frac{1}{2} \, \mathbf{a}_{3} & = & x_{2}a \, \mathbf{\hat{x}} + \frac{1}{2}c \, \mathbf{\hat{z}} & \left(8e\right) & \mbox{Al I} \\ 
\mathbf{B}_{7} & = & -y_{3} \, \mathbf{a}_{1} + y_{3} \, \mathbf{a}_{2} + z_{3} \, \mathbf{a}_{3} & = & y_{3}b \, \mathbf{\hat{y}} + z_{3}c \, \mathbf{\hat{z}} & \left(8f\right) & \mbox{Al II} \\ 
\mathbf{B}_{8} & = & y_{3} \, \mathbf{a}_{1}-y_{3} \, \mathbf{a}_{2} + \left(\frac{1}{2} +z_{3}\right) \, \mathbf{a}_{3} & = & -y_{3}b \, \mathbf{\hat{y}} + \left(\frac{1}{2} +z_{3}\right)c \, \mathbf{\hat{z}} & \left(8f\right) & \mbox{Al II} \\ 
\mathbf{B}_{9} & = & -y_{3} \, \mathbf{a}_{1} + y_{3} \, \mathbf{a}_{2} + \left(\frac{1}{2} - z_{3}\right) \, \mathbf{a}_{3} & = & y_{3}b \, \mathbf{\hat{y}} + \left(\frac{1}{2} - z_{3}\right)c \, \mathbf{\hat{z}} & \left(8f\right) & \mbox{Al II} \\ 
\mathbf{B}_{10} & = & y_{3} \, \mathbf{a}_{1}-y_{3} \, \mathbf{a}_{2}-z_{3} \, \mathbf{a}_{3} & = & -y_{3}b \, \mathbf{\hat{y}}-z_{3}c \, \mathbf{\hat{z}} & \left(8f\right) & \mbox{Al II} \\ 
\mathbf{B}_{11} & = & \left(x_{4}-y_{4}\right) \, \mathbf{a}_{1} + \left(x_{4}+y_{4}\right) \, \mathbf{a}_{2} + \frac{1}{4} \, \mathbf{a}_{3} & = & x_{4}a \, \mathbf{\hat{x}} + y_{4}b \, \mathbf{\hat{y}} + \frac{1}{4}c \, \mathbf{\hat{z}} & \left(8g\right) & \mbox{Al III} \\ 
\mathbf{B}_{12} & = & \left(-x_{4}+y_{4}\right) \, \mathbf{a}_{1} + \left(-x_{4}-y_{4}\right) \, \mathbf{a}_{2} + \frac{3}{4} \, \mathbf{a}_{3} & = & -x_{4}a \, \mathbf{\hat{x}}-y_{4}b \, \mathbf{\hat{y}} + \frac{3}{4}c \, \mathbf{\hat{z}} & \left(8g\right) & \mbox{Al III} \\ 
\mathbf{B}_{13} & = & \left(-x_{4}-y_{4}\right) \, \mathbf{a}_{1} + \left(-x_{4}+y_{4}\right) \, \mathbf{a}_{2} + \frac{1}{4} \, \mathbf{a}_{3} & = & -x_{4}a \, \mathbf{\hat{x}} + y_{4}b \, \mathbf{\hat{y}} + \frac{1}{4}c \, \mathbf{\hat{z}} & \left(8g\right) & \mbox{Al III} \\ 
\mathbf{B}_{14} & = & \left(x_{4}+y_{4}\right) \, \mathbf{a}_{1} + \left(x_{4}-y_{4}\right) \, \mathbf{a}_{2} + \frac{3}{4} \, \mathbf{a}_{3} & = & x_{4}a \, \mathbf{\hat{x}}-y_{4}b \, \mathbf{\hat{y}} + \frac{3}{4}c \, \mathbf{\hat{z}} & \left(8g\right) & \mbox{Al III} \\ 
\end{longtabu}
\renewcommand{\arraystretch}{1.0}
\noindent \hrulefill
\\
\textbf{References:}
\vspace*{-0.25cm}
\begin{flushleft}
  - \bibentry{kontio81:D2_h}. \\
\end{flushleft}
\noindent \hrulefill
\\
\textbf{Geometry files:}
\\
\noindent  - CIF: pp. {\hyperref[A6B_oC28_63_efg_c_cif]{\pageref{A6B_oC28_63_efg_c_cif}}} \\
\noindent  - POSCAR: pp. {\hyperref[A6B_oC28_63_efg_c_poscar]{\pageref{A6B_oC28_63_efg_c_poscar}}} \\
\onecolumn
{\phantomsection\label{AB3C_oC20_63_a_cf_c}}
\subsection*{\huge \textbf{{\normalfont \begin{raggedleft}Post-perovskite (MgSiO$_{3}$) Structure: \end{raggedleft} \\ AB3C\_oC20\_63\_a\_cf\_c}}}
\noindent \hrulefill
\vspace*{0.25cm}
\begin{figure}[htp]
  \centering
  \vspace{-1em}
  {\includegraphics[width=1\textwidth]{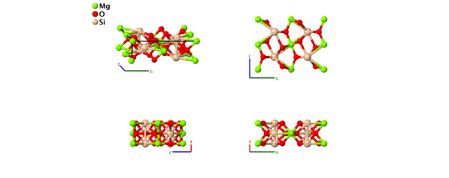}}
\end{figure}
\vspace*{-0.5cm}
\renewcommand{\arraystretch}{1.5}
\begin{equation*}
  \begin{array}{>{$\hspace{-0.15cm}}l<{$}>{$}p{0.5cm}<{$}>{$}p{18.5cm}<{$}}
    \mbox{\large \textbf{Prototype}} &\colon & \ce{MgSiO3} \\
    \mbox{\large \textbf{\AFLOW\ prototype label}} &\colon & \mbox{AB3C\_oC20\_63\_a\_cf\_c} \\
    \mbox{\large \textbf{\textit{Strukturbericht} designation}} &\colon & \mbox{None} \\
    \mbox{\large \textbf{Pearson symbol}} &\colon & \mbox{oC20} \\
    \mbox{\large \textbf{Space group number}} &\colon & 63 \\
    \mbox{\large \textbf{Space group symbol}} &\colon & Cmcm \\
    \mbox{\large \textbf{\AFLOW\ prototype command}} &\colon &  \texttt{aflow} \,  \, \texttt{-{}-proto=AB3C\_oC20\_63\_a\_cf\_c } \, \newline \texttt{-{}-params=}{a,b/a,c/a,y_{2},y_{3},y_{4},z_{4} }
  \end{array}
\end{equation*}
\renewcommand{\arraystretch}{1.0}

\vspace*{-0.25cm}
\noindent \hrulefill
\\
\textbf{ Other compounds with this structure:}
\begin{itemize}
   \item{ CaIrO$_{3}$, MgGeO$_{3}$, NaMgF$_{3}$  }
\end{itemize}
\vspace*{-0.25cm}
\noindent \hrulefill
\begin{itemize}
  \item{This structure was determined by a combination of x-ray diffraction
measurements and atomistic simulations of MgSiO$_{3}$ at a pressure of
121~GPa and a temperature of 300~K.  This approximates the conditions
at the Earth's core-mantle boundary.
}
\end{itemize}

\noindent \parbox{1 \linewidth}{
\noindent \hrulefill
\\
\textbf{Base-centered Orthorhombic primitive vectors:} \\
\vspace*{-0.25cm}
\begin{tabular}{cc}
  \begin{tabular}{c}
    \parbox{0.6 \linewidth}{
      \renewcommand{\arraystretch}{1.5}
      \begin{equation*}
        \centering
        \begin{array}{ccc}
              \mathbf{a}_1 & = & \frac12 \, a \, \mathbf{\hat{x}} - \frac12 \, b \, \mathbf{\hat{y}} \\
    \mathbf{a}_2 & = & \frac12 \, a \, \mathbf{\hat{x}} + \frac12 \, b \, \mathbf{\hat{y}} \\
    \mathbf{a}_3 & = & c \, \mathbf{\hat{z}} \\

        \end{array}
      \end{equation*}
    }
    \renewcommand{\arraystretch}{1.0}
  \end{tabular}
  \begin{tabular}{c}
    \includegraphics[width=0.3\linewidth]{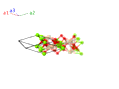} \\
  \end{tabular}
\end{tabular}

}
\vspace*{-0.25cm}

\noindent \hrulefill
\\
\textbf{Basis vectors:}
\vspace*{-0.25cm}
\renewcommand{\arraystretch}{1.5}
\begin{longtabu} to \textwidth{>{\centering $}X[-1,c,c]<{$}>{\centering $}X[-1,c,c]<{$}>{\centering $}X[-1,c,c]<{$}>{\centering $}X[-1,c,c]<{$}>{\centering $}X[-1,c,c]<{$}>{\centering $}X[-1,c,c]<{$}>{\centering $}X[-1,c,c]<{$}}
  & & \mbox{Lattice Coordinates} & & \mbox{Cartesian Coordinates} &\mbox{Wyckoff Position} & \mbox{Atom Type} \\  
  \mathbf{B}_{1} & = & 0 \, \mathbf{a}_{1} + 0 \, \mathbf{a}_{2} + 0 \, \mathbf{a}_{3} & = & 0 \, \mathbf{\hat{x}} + 0 \, \mathbf{\hat{y}} + 0 \, \mathbf{\hat{z}} & \left(4a\right) & \mbox{Mg} \\ 
\mathbf{B}_{2} & = & \frac{1}{2} \, \mathbf{a}_{3} & = & \frac{1}{2}c \, \mathbf{\hat{z}} & \left(4a\right) & \mbox{Mg} \\ 
\mathbf{B}_{3} & = & -y_{2} \, \mathbf{a}_{1} + y_{2} \, \mathbf{a}_{2} + \frac{1}{4} \, \mathbf{a}_{3} & = & y_{2}b \, \mathbf{\hat{y}} + \frac{1}{4}c \, \mathbf{\hat{z}} & \left(4c\right) & \mbox{O I} \\ 
\mathbf{B}_{4} & = & y_{2} \, \mathbf{a}_{1}-y_{2} \, \mathbf{a}_{2} + \frac{3}{4} \, \mathbf{a}_{3} & = & -y_{2}b \, \mathbf{\hat{y}} + \frac{3}{4}c \, \mathbf{\hat{z}} & \left(4c\right) & \mbox{O I} \\ 
\mathbf{B}_{5} & = & -y_{3} \, \mathbf{a}_{1} + y_{3} \, \mathbf{a}_{2} + \frac{1}{4} \, \mathbf{a}_{3} & = & y_{3}b \, \mathbf{\hat{y}} + \frac{1}{4}c \, \mathbf{\hat{z}} & \left(4c\right) & \mbox{Si} \\ 
\mathbf{B}_{6} & = & y_{3} \, \mathbf{a}_{1}-y_{3} \, \mathbf{a}_{2} + \frac{3}{4} \, \mathbf{a}_{3} & = & -y_{3}b \, \mathbf{\hat{y}} + \frac{3}{4}c \, \mathbf{\hat{z}} & \left(4c\right) & \mbox{Si} \\ 
\mathbf{B}_{7} & = & -y_{4} \, \mathbf{a}_{1} + y_{4} \, \mathbf{a}_{2} + z_{4} \, \mathbf{a}_{3} & = & y_{4}b \, \mathbf{\hat{y}} + z_{4}c \, \mathbf{\hat{z}} & \left(8f\right) & \mbox{O II} \\ 
\mathbf{B}_{8} & = & y_{4} \, \mathbf{a}_{1}-y_{4} \, \mathbf{a}_{2} + \left(\frac{1}{2} +z_{4}\right) \, \mathbf{a}_{3} & = & -y_{4}b \, \mathbf{\hat{y}} + \left(\frac{1}{2} +z_{4}\right)c \, \mathbf{\hat{z}} & \left(8f\right) & \mbox{O II} \\ 
\mathbf{B}_{9} & = & -y_{4} \, \mathbf{a}_{1} + y_{4} \, \mathbf{a}_{2} + \left(\frac{1}{2} - z_{4}\right) \, \mathbf{a}_{3} & = & y_{4}b \, \mathbf{\hat{y}} + \left(\frac{1}{2} - z_{4}\right)c \, \mathbf{\hat{z}} & \left(8f\right) & \mbox{O II} \\ 
\mathbf{B}_{10} & = & y_{4} \, \mathbf{a}_{1}-y_{4} \, \mathbf{a}_{2}-z_{4} \, \mathbf{a}_{3} & = & -y_{4}b \, \mathbf{\hat{y}}-z_{4}c \, \mathbf{\hat{z}} & \left(8f\right) & \mbox{O II} \\ 
\end{longtabu}
\renewcommand{\arraystretch}{1.0}
\noindent \hrulefill
\\
\textbf{References:}
\vspace*{-0.25cm}
\begin{flushleft}
  - \bibentry{Murakami_Science_304_2005}. \\
\end{flushleft}
\noindent \hrulefill
\\
\textbf{Geometry files:}
\\
\noindent  - CIF: pp. {\hyperref[AB3C_oC20_63_a_cf_c_cif]{\pageref{AB3C_oC20_63_a_cf_c_cif}}} \\
\noindent  - POSCAR: pp. {\hyperref[AB3C_oC20_63_a_cf_c_poscar]{\pageref{AB3C_oC20_63_a_cf_c_poscar}}} \\
\onecolumn
{\phantomsection\label{AB4C_oC24_63_a_fg_c}}
\subsection*{\huge \textbf{{\normalfont MgSO$_{4}$ Structure: AB4C\_oC24\_63\_a\_fg\_c}}}
\noindent \hrulefill
\vspace*{0.25cm}
\begin{figure}[htp]
  \centering
  \vspace{-1em}
  {\includegraphics[width=1\textwidth]{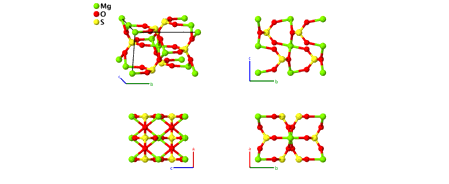}}
\end{figure}
\vspace*{-0.5cm}
\renewcommand{\arraystretch}{1.5}
\begin{equation*}
  \begin{array}{>{$\hspace{-0.15cm}}l<{$}>{$}p{0.5cm}<{$}>{$}p{18.5cm}<{$}}
    \mbox{\large \textbf{Prototype}} &\colon & \ce{MgO$_{4}$S} \\
    \mbox{\large \textbf{\AFLOW\ prototype label}} &\colon & \mbox{AB4C\_oC24\_63\_a\_fg\_c} \\
    \mbox{\large \textbf{\textit{Strukturbericht} designation}} &\colon & \mbox{None} \\
    \mbox{\large \textbf{Pearson symbol}} &\colon & \mbox{oC24} \\
    \mbox{\large \textbf{Space group number}} &\colon & 63 \\
    \mbox{\large \textbf{Space group symbol}} &\colon & Cmcm \\
    \mbox{\large \textbf{\AFLOW\ prototype command}} &\colon &  \texttt{aflow} \,  \, \texttt{-{}-proto=AB4C\_oC24\_63\_a\_fg\_c } \, \newline \texttt{-{}-params=}{a,b/a,c/a,y_{2},y_{3},z_{3},x_{4},y_{4} }
  \end{array}
\end{equation*}
\renewcommand{\arraystretch}{1.0}

\vspace*{-0.25cm}
\noindent \hrulefill
\\
\textbf{ Other compounds with this structure:}
\begin{itemize}
   \item{ CdCrO$_{4}$, CoCrO$_{4}$, MgCrO$_{4}$, NiCrO$_{4}$, NiSO$_{4}$  }
\end{itemize}
\noindent \parbox{1 \linewidth}{
\noindent \hrulefill
\\
\textbf{Base-centered Orthorhombic primitive vectors:} \\
\vspace*{-0.25cm}
\begin{tabular}{cc}
  \begin{tabular}{c}
    \parbox{0.6 \linewidth}{
      \renewcommand{\arraystretch}{1.5}
      \begin{equation*}
        \centering
        \begin{array}{ccc}
              \mathbf{a}_1 & = & \frac12 \, a \, \mathbf{\hat{x}} - \frac12 \, b \, \mathbf{\hat{y}} \\
    \mathbf{a}_2 & = & \frac12 \, a \, \mathbf{\hat{x}} + \frac12 \, b \, \mathbf{\hat{y}} \\
    \mathbf{a}_3 & = & c \, \mathbf{\hat{z}} \\

        \end{array}
      \end{equation*}
    }
    \renewcommand{\arraystretch}{1.0}
  \end{tabular}
  \begin{tabular}{c}
    \includegraphics[width=0.3\linewidth]{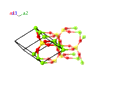} \\
  \end{tabular}
\end{tabular}

}
\vspace*{-0.25cm}

\noindent \hrulefill
\\
\textbf{Basis vectors:}
\vspace*{-0.25cm}
\renewcommand{\arraystretch}{1.5}
\begin{longtabu} to \textwidth{>{\centering $}X[-1,c,c]<{$}>{\centering $}X[-1,c,c]<{$}>{\centering $}X[-1,c,c]<{$}>{\centering $}X[-1,c,c]<{$}>{\centering $}X[-1,c,c]<{$}>{\centering $}X[-1,c,c]<{$}>{\centering $}X[-1,c,c]<{$}}
  & & \mbox{Lattice Coordinates} & & \mbox{Cartesian Coordinates} &\mbox{Wyckoff Position} & \mbox{Atom Type} \\  
  \mathbf{B}_{1} & = & 0 \, \mathbf{a}_{1} + 0 \, \mathbf{a}_{2} + 0 \, \mathbf{a}_{3} & = & 0 \, \mathbf{\hat{x}} + 0 \, \mathbf{\hat{y}} + 0 \, \mathbf{\hat{z}} & \left(4a\right) & \mbox{Mg} \\ 
\mathbf{B}_{2} & = & \frac{1}{2} \, \mathbf{a}_{3} & = & \frac{1}{2}c \, \mathbf{\hat{z}} & \left(4a\right) & \mbox{Mg} \\ 
\mathbf{B}_{3} & = & -y_{2} \, \mathbf{a}_{1} + y_{2} \, \mathbf{a}_{2} + \frac{1}{4} \, \mathbf{a}_{3} & = & y_{2}b \, \mathbf{\hat{y}} + \frac{1}{4}c \, \mathbf{\hat{z}} & \left(4c\right) & \mbox{S} \\ 
\mathbf{B}_{4} & = & y_{2} \, \mathbf{a}_{1}-y_{2} \, \mathbf{a}_{2} + \frac{3}{4} \, \mathbf{a}_{3} & = & -y_{2}b \, \mathbf{\hat{y}} + \frac{3}{4}c \, \mathbf{\hat{z}} & \left(4c\right) & \mbox{S} \\ 
\mathbf{B}_{5} & = & -y_{3} \, \mathbf{a}_{1} + y_{3} \, \mathbf{a}_{2} + z_{3} \, \mathbf{a}_{3} & = & y_{3}b \, \mathbf{\hat{y}} + z_{3}c \, \mathbf{\hat{z}} & \left(8f\right) & \mbox{O I} \\ 
\mathbf{B}_{6} & = & y_{3} \, \mathbf{a}_{1}-y_{3} \, \mathbf{a}_{2} + \left(\frac{1}{2} +z_{3}\right) \, \mathbf{a}_{3} & = & -y_{3}b \, \mathbf{\hat{y}} + \left(\frac{1}{2} +z_{3}\right)c \, \mathbf{\hat{z}} & \left(8f\right) & \mbox{O I} \\ 
\mathbf{B}_{7} & = & -y_{3} \, \mathbf{a}_{1} + y_{3} \, \mathbf{a}_{2} + \left(\frac{1}{2} - z_{3}\right) \, \mathbf{a}_{3} & = & y_{3}b \, \mathbf{\hat{y}} + \left(\frac{1}{2} - z_{3}\right)c \, \mathbf{\hat{z}} & \left(8f\right) & \mbox{O I} \\ 
\mathbf{B}_{8} & = & y_{3} \, \mathbf{a}_{1}-y_{3} \, \mathbf{a}_{2}-z_{3} \, \mathbf{a}_{3} & = & -y_{3}b \, \mathbf{\hat{y}}-z_{3}c \, \mathbf{\hat{z}} & \left(8f\right) & \mbox{O I} \\ 
\mathbf{B}_{9} & = & \left(x_{4}-y_{4}\right) \, \mathbf{a}_{1} + \left(x_{4}+y_{4}\right) \, \mathbf{a}_{2} + \frac{1}{4} \, \mathbf{a}_{3} & = & x_{4}a \, \mathbf{\hat{x}} + y_{4}b \, \mathbf{\hat{y}} + \frac{1}{4}c \, \mathbf{\hat{z}} & \left(8g\right) & \mbox{O II} \\ 
\mathbf{B}_{10} & = & \left(-x_{4}+y_{4}\right) \, \mathbf{a}_{1} + \left(-x_{4}-y_{4}\right) \, \mathbf{a}_{2} + \frac{3}{4} \, \mathbf{a}_{3} & = & -x_{4}a \, \mathbf{\hat{x}}-y_{4}b \, \mathbf{\hat{y}} + \frac{3}{4}c \, \mathbf{\hat{z}} & \left(8g\right) & \mbox{O II} \\ 
\mathbf{B}_{11} & = & \left(-x_{4}-y_{4}\right) \, \mathbf{a}_{1} + \left(-x_{4}+y_{4}\right) \, \mathbf{a}_{2} + \frac{1}{4} \, \mathbf{a}_{3} & = & -x_{4}a \, \mathbf{\hat{x}} + y_{4}b \, \mathbf{\hat{y}} + \frac{1}{4}c \, \mathbf{\hat{z}} & \left(8g\right) & \mbox{O II} \\ 
\mathbf{B}_{12} & = & \left(x_{4}+y_{4}\right) \, \mathbf{a}_{1} + \left(x_{4}-y_{4}\right) \, \mathbf{a}_{2} + \frac{3}{4} \, \mathbf{a}_{3} & = & x_{4}a \, \mathbf{\hat{x}}-y_{4}b \, \mathbf{\hat{y}} + \frac{3}{4}c \, \mathbf{\hat{z}} & \left(8g\right) & \mbox{O II} \\ 
\end{longtabu}
\renewcommand{\arraystretch}{1.0}
\noindent \hrulefill
\\
\textbf{References:}
\vspace*{-0.25cm}
\begin{flushleft}
  - \bibentry{Rentzeperis_Acta_Cryst_11_1958}. \\
\end{flushleft}
\textbf{Found in:}
\vspace*{-0.25cm}
\begin{flushleft}
  - \bibentry{Downs_Am_Min_88_2003}. \\
\end{flushleft}
\noindent \hrulefill
\\
\textbf{Geometry files:}
\\
\noindent  - CIF: pp. {\hyperref[AB4C_oC24_63_a_fg_c_cif]{\pageref{AB4C_oC24_63_a_fg_c_cif}}} \\
\noindent  - POSCAR: pp. {\hyperref[AB4C_oC24_63_a_fg_c_poscar]{\pageref{AB4C_oC24_63_a_fg_c_poscar}}} \\
\onecolumn
{\phantomsection\label{AB4C_oC24_63_c_fg_c}}
\subsection*{\huge \textbf{{\normalfont \begin{raggedleft}Anhydrite (CaSO$_{4}$, $H0_{1}$) Structure: \end{raggedleft} \\ AB4C\_oC24\_63\_c\_fg\_c}}}
\noindent \hrulefill
\vspace*{0.25cm}
\begin{figure}[htp]
  \centering
  \vspace{-1em}
  {\includegraphics[width=1\textwidth]{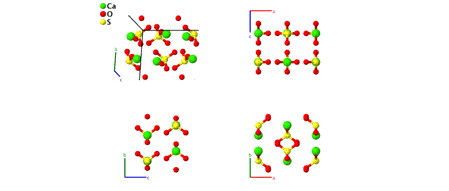}}
\end{figure}
\vspace*{-0.5cm}
\renewcommand{\arraystretch}{1.5}
\begin{equation*}
  \begin{array}{>{$\hspace{-0.15cm}}l<{$}>{$}p{0.5cm}<{$}>{$}p{18.5cm}<{$}}
    \mbox{\large \textbf{Prototype}} &\colon & \ce{CaSO4} \\
    \mbox{\large \textbf{\AFLOW\ prototype label}} &\colon & \mbox{AB4C\_oC24\_63\_c\_fg\_c} \\
    \mbox{\large \textbf{\textit{Strukturbericht} designation}} &\colon & \mbox{$H0_{1}$} \\
    \mbox{\large \textbf{Pearson symbol}} &\colon & \mbox{oC24} \\
    \mbox{\large \textbf{Space group number}} &\colon & 63 \\
    \mbox{\large \textbf{Space group symbol}} &\colon & Cmcm \\
    \mbox{\large \textbf{\AFLOW\ prototype command}} &\colon &  \texttt{aflow} \,  \, \texttt{-{}-proto=AB4C\_oC24\_63\_c\_fg\_c } \, \newline \texttt{-{}-params=}{a,b/a,c/a,y_{1},y_{2},y_{3},z_{3},x_{4},y_{4} }
  \end{array}
\end{equation*}
\renewcommand{\arraystretch}{1.0}

\vspace*{-0.25cm}
\noindent \hrulefill
\\
\textbf{ Other compounds with this structure:}
\begin{itemize}
   \item{ CdCrO$_{4}$, MgCrO$_{4}$, MgSO$_{4}$, NaBF$_{4}$, NaCrO$_{4}$, NiSO$_{4}$  }
\end{itemize}
\vspace*{-0.25cm}
\noindent \hrulefill
\begin{itemize}
  \item{(Hawthorne, 1975) give the structure in the $Amma$ setting of space
group \#63.  We have transformed this to the standard $Cmcm$ setting.
The addition of water into the anhydrite crystal transforms it into
gypsum.
}
\end{itemize}

\noindent \parbox{1 \linewidth}{
\noindent \hrulefill
\\
\textbf{Base-centered Orthorhombic primitive vectors:} \\
\vspace*{-0.25cm}
\begin{tabular}{cc}
  \begin{tabular}{c}
    \parbox{0.6 \linewidth}{
      \renewcommand{\arraystretch}{1.5}
      \begin{equation*}
        \centering
        \begin{array}{ccc}
              \mathbf{a}_1 & = & \frac12 \, a \, \mathbf{\hat{x}} - \frac12 \, b \, \mathbf{\hat{y}} \\
    \mathbf{a}_2 & = & \frac12 \, a \, \mathbf{\hat{x}} + \frac12 \, b \, \mathbf{\hat{y}} \\
    \mathbf{a}_3 & = & c \, \mathbf{\hat{z}} \\

        \end{array}
      \end{equation*}
    }
    \renewcommand{\arraystretch}{1.0}
  \end{tabular}
  \begin{tabular}{c}
    \includegraphics[width=0.3\linewidth]{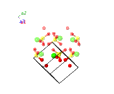} \\
  \end{tabular}
\end{tabular}

}
\vspace*{-0.25cm}

\noindent \hrulefill
\\
\textbf{Basis vectors:}
\vspace*{-0.25cm}
\renewcommand{\arraystretch}{1.5}
\begin{longtabu} to \textwidth{>{\centering $}X[-1,c,c]<{$}>{\centering $}X[-1,c,c]<{$}>{\centering $}X[-1,c,c]<{$}>{\centering $}X[-1,c,c]<{$}>{\centering $}X[-1,c,c]<{$}>{\centering $}X[-1,c,c]<{$}>{\centering $}X[-1,c,c]<{$}}
  & & \mbox{Lattice Coordinates} & & \mbox{Cartesian Coordinates} &\mbox{Wyckoff Position} & \mbox{Atom Type} \\  
  \mathbf{B}_{1} & = & -y_{1} \, \mathbf{a}_{1} + y_{1} \, \mathbf{a}_{2} + \frac{1}{4} \, \mathbf{a}_{3} & = & y_{1}b \, \mathbf{\hat{y}} + \frac{1}{4}c \, \mathbf{\hat{z}} & \left(4c\right) & \mbox{Ca} \\ 
\mathbf{B}_{2} & = & y_{1} \, \mathbf{a}_{1}-y_{1} \, \mathbf{a}_{2} + \frac{3}{4} \, \mathbf{a}_{3} & = & -y_{1}b \, \mathbf{\hat{y}} + \frac{3}{4}c \, \mathbf{\hat{z}} & \left(4c\right) & \mbox{Ca} \\ 
\mathbf{B}_{3} & = & -y_{2} \, \mathbf{a}_{1} + y_{2} \, \mathbf{a}_{2} + \frac{1}{4} \, \mathbf{a}_{3} & = & y_{2}b \, \mathbf{\hat{y}} + \frac{1}{4}c \, \mathbf{\hat{z}} & \left(4c\right) & \mbox{S} \\ 
\mathbf{B}_{4} & = & y_{2} \, \mathbf{a}_{1}-y_{2} \, \mathbf{a}_{2} + \frac{3}{4} \, \mathbf{a}_{3} & = & -y_{2}b \, \mathbf{\hat{y}} + \frac{3}{4}c \, \mathbf{\hat{z}} & \left(4c\right) & \mbox{S} \\ 
\mathbf{B}_{5} & = & -y_{3} \, \mathbf{a}_{1} + y_{3} \, \mathbf{a}_{2} + z_{3} \, \mathbf{a}_{3} & = & y_{3}b \, \mathbf{\hat{y}} + z_{3}c \, \mathbf{\hat{z}} & \left(8f\right) & \mbox{O I} \\ 
\mathbf{B}_{6} & = & y_{3} \, \mathbf{a}_{1}-y_{3} \, \mathbf{a}_{2} + \left(\frac{1}{2} +z_{3}\right) \, \mathbf{a}_{3} & = & -y_{3}b \, \mathbf{\hat{y}} + \left(\frac{1}{2} +z_{3}\right)c \, \mathbf{\hat{z}} & \left(8f\right) & \mbox{O I} \\ 
\mathbf{B}_{7} & = & -y_{3} \, \mathbf{a}_{1} + y_{3} \, \mathbf{a}_{2} + \left(\frac{1}{2} - z_{3}\right) \, \mathbf{a}_{3} & = & y_{3}b \, \mathbf{\hat{y}} + \left(\frac{1}{2} - z_{3}\right)c \, \mathbf{\hat{z}} & \left(8f\right) & \mbox{O I} \\ 
\mathbf{B}_{8} & = & y_{3} \, \mathbf{a}_{1}-y_{3} \, \mathbf{a}_{2}-z_{3} \, \mathbf{a}_{3} & = & -y_{3}b \, \mathbf{\hat{y}}-z_{3}c \, \mathbf{\hat{z}} & \left(8f\right) & \mbox{O I} \\ 
\mathbf{B}_{9} & = & \left(x_{4}-y_{4}\right) \, \mathbf{a}_{1} + \left(x_{4}+y_{4}\right) \, \mathbf{a}_{2} + \frac{1}{4} \, \mathbf{a}_{3} & = & x_{4}a \, \mathbf{\hat{x}} + y_{4}b \, \mathbf{\hat{y}} + \frac{1}{4}c \, \mathbf{\hat{z}} & \left(8g\right) & \mbox{O II} \\ 
\mathbf{B}_{10} & = & \left(-x_{4}+y_{4}\right) \, \mathbf{a}_{1} + \left(-x_{4}-y_{4}\right) \, \mathbf{a}_{2} + \frac{3}{4} \, \mathbf{a}_{3} & = & -x_{4}a \, \mathbf{\hat{x}}-y_{4}b \, \mathbf{\hat{y}} + \frac{3}{4}c \, \mathbf{\hat{z}} & \left(8g\right) & \mbox{O II} \\ 
\mathbf{B}_{11} & = & \left(-x_{4}-y_{4}\right) \, \mathbf{a}_{1} + \left(-x_{4}+y_{4}\right) \, \mathbf{a}_{2} + \frac{1}{4} \, \mathbf{a}_{3} & = & -x_{4}a \, \mathbf{\hat{x}} + y_{4}b \, \mathbf{\hat{y}} + \frac{1}{4}c \, \mathbf{\hat{z}} & \left(8g\right) & \mbox{O II} \\ 
\mathbf{B}_{12} & = & \left(x_{4}+y_{4}\right) \, \mathbf{a}_{1} + \left(x_{4}-y_{4}\right) \, \mathbf{a}_{2} + \frac{3}{4} \, \mathbf{a}_{3} & = & x_{4}a \, \mathbf{\hat{x}}-y_{4}b \, \mathbf{\hat{y}} + \frac{3}{4}c \, \mathbf{\hat{z}} & \left(8g\right) & \mbox{O II} \\ 
\end{longtabu}
\renewcommand{\arraystretch}{1.0}
\noindent \hrulefill
\\
\textbf{References:}
\vspace*{-0.25cm}
\begin{flushleft}
  - \bibentry{Hawthorne_Can_Min_13_1975}. \\
\end{flushleft}
\noindent \hrulefill
\\
\textbf{Geometry files:}
\\
\noindent  - CIF: pp. {\hyperref[AB4C_oC24_63_c_fg_c_cif]{\pageref{AB4C_oC24_63_c_fg_c_cif}}} \\
\noindent  - POSCAR: pp. {\hyperref[AB4C_oC24_63_c_fg_c_poscar]{\pageref{AB4C_oC24_63_c_fg_c_poscar}}} \\
\onecolumn
{\phantomsection\label{A2B_oC24_64_2f_f}}
\subsection*{\huge \textbf{{\normalfont H$_{2}$S (170~GPa) Structure: A2B\_oC24\_64\_2f\_f}}}
\noindent \hrulefill
\vspace*{0.25cm}
\begin{figure}[htp]
  \centering
  \vspace{-1em}
  {\includegraphics[width=1\textwidth]{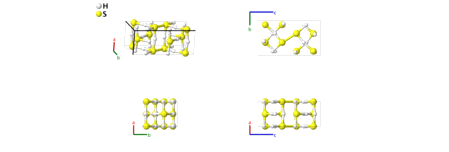}}
\end{figure}
\vspace*{-0.5cm}
\renewcommand{\arraystretch}{1.5}
\begin{equation*}
  \begin{array}{>{$\hspace{-0.15cm}}l<{$}>{$}p{0.5cm}<{$}>{$}p{18.5cm}<{$}}
    \mbox{\large \textbf{Prototype}} &\colon & \ce{H2S} \\
    \mbox{\large \textbf{\AFLOW\ prototype label}} &\colon & \mbox{A2B\_oC24\_64\_2f\_f} \\
    \mbox{\large \textbf{\textit{Strukturbericht} designation}} &\colon & \mbox{None} \\
    \mbox{\large \textbf{Pearson symbol}} &\colon & \mbox{oC24} \\
    \mbox{\large \textbf{Space group number}} &\colon & 64 \\
    \mbox{\large \textbf{Space group symbol}} &\colon & Cmca \\
    \mbox{\large \textbf{\AFLOW\ prototype command}} &\colon &  \texttt{aflow} \,  \, \texttt{-{}-proto=A2B\_oC24\_64\_2f\_f } \, \newline \texttt{-{}-params=}{a,b/a,c/a,y_{1},z_{1},y_{2},z_{2},y_{3},z_{3} }
  \end{array}
\end{equation*}
\renewcommand{\arraystretch}{1.0}

\vspace*{-0.25cm}
\noindent \hrulefill
\begin{itemize}
  \item{This structure was found by first-principles electronic structure
calculations and is predicted to be the stable structure of H$_{2}$S
for pressures $> 140$~GPa.  At 160~GPa it is predicted to be a conventional
superconductor with an approximate transition temperature of 80~K,
however it is unlikely that this is the crystal structure of the 190~K
superconductor, which is likely the
\href{http://aflow.org/CrystalDatabase/A3B_cI8_229_b_a.html}{A3B\_cI8\_229\_b\_a phase of H$_{3}$S} (Bernstein, 2015).
The data presented here was computed at 170~GPa.
}
\end{itemize}

\noindent \parbox{1 \linewidth}{
\noindent \hrulefill
\\
\textbf{Base-centered Orthorhombic primitive vectors:} \\
\vspace*{-0.25cm}
\begin{tabular}{cc}
  \begin{tabular}{c}
    \parbox{0.6 \linewidth}{
      \renewcommand{\arraystretch}{1.5}
      \begin{equation*}
        \centering
        \begin{array}{ccc}
              \mathbf{a}_1 & = & \frac12 \, a \, \mathbf{\hat{x}} - \frac12 \, b \, \mathbf{\hat{y}} \\
    \mathbf{a}_2 & = & \frac12 \, a \, \mathbf{\hat{x}} + \frac12 \, b \, \mathbf{\hat{y}} \\
    \mathbf{a}_3 & = & c \, \mathbf{\hat{z}} \\

        \end{array}
      \end{equation*}
    }
    \renewcommand{\arraystretch}{1.0}
  \end{tabular}
  \begin{tabular}{c}
    \includegraphics[width=0.3\linewidth]{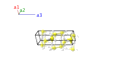} \\
  \end{tabular}
\end{tabular}

}
\vspace*{-0.25cm}

\noindent \hrulefill
\\
\textbf{Basis vectors:}
\vspace*{-0.25cm}
\renewcommand{\arraystretch}{1.5}
\begin{longtabu} to \textwidth{>{\centering $}X[-1,c,c]<{$}>{\centering $}X[-1,c,c]<{$}>{\centering $}X[-1,c,c]<{$}>{\centering $}X[-1,c,c]<{$}>{\centering $}X[-1,c,c]<{$}>{\centering $}X[-1,c,c]<{$}>{\centering $}X[-1,c,c]<{$}}
  & & \mbox{Lattice Coordinates} & & \mbox{Cartesian Coordinates} &\mbox{Wyckoff Position} & \mbox{Atom Type} \\  
  \mathbf{B}_{1} & = & -y_{1} \, \mathbf{a}_{1} + y_{1} \, \mathbf{a}_{2} + z_{1} \, \mathbf{a}_{3} & = & y_{1}b \, \mathbf{\hat{y}} + z_{1}c \, \mathbf{\hat{z}} & \left(8f\right) & \mbox{H I} \\ 
\mathbf{B}_{2} & = & \left(\frac{1}{2} +y_{1}\right) \, \mathbf{a}_{1} + \left(\frac{1}{2} - y_{1}\right) \, \mathbf{a}_{2} + \left(\frac{1}{2} +z_{1}\right) \, \mathbf{a}_{3} & = & \frac{1}{2}a \, \mathbf{\hat{x}}-y_{1}b \, \mathbf{\hat{y}} + \left(\frac{1}{2} +z_{1}\right)c \, \mathbf{\hat{z}} & \left(8f\right) & \mbox{H I} \\ 
\mathbf{B}_{3} & = & \left(\frac{1}{2} - y_{1}\right) \, \mathbf{a}_{1} + \left(\frac{1}{2} +y_{1}\right) \, \mathbf{a}_{2} + \left(\frac{1}{2} - z_{1}\right) \, \mathbf{a}_{3} & = & \frac{1}{2}a \, \mathbf{\hat{x}} + y_{1}b \, \mathbf{\hat{y}} + \left(\frac{1}{2} - z_{1}\right)c \, \mathbf{\hat{z}} & \left(8f\right) & \mbox{H I} \\ 
\mathbf{B}_{4} & = & y_{1} \, \mathbf{a}_{1}-y_{1} \, \mathbf{a}_{2}-z_{1} \, \mathbf{a}_{3} & = & -y_{1}b \, \mathbf{\hat{y}}-z_{1}c \, \mathbf{\hat{z}} & \left(8f\right) & \mbox{H I} \\ 
\mathbf{B}_{5} & = & -y_{2} \, \mathbf{a}_{1} + y_{2} \, \mathbf{a}_{2} + z_{2} \, \mathbf{a}_{3} & = & y_{2}b \, \mathbf{\hat{y}} + z_{2}c \, \mathbf{\hat{z}} & \left(8f\right) & \mbox{H II} \\ 
\mathbf{B}_{6} & = & \left(\frac{1}{2} +y_{2}\right) \, \mathbf{a}_{1} + \left(\frac{1}{2} - y_{2}\right) \, \mathbf{a}_{2} + \left(\frac{1}{2} +z_{2}\right) \, \mathbf{a}_{3} & = & \frac{1}{2}a \, \mathbf{\hat{x}}-y_{2}b \, \mathbf{\hat{y}} + \left(\frac{1}{2} +z_{2}\right)c \, \mathbf{\hat{z}} & \left(8f\right) & \mbox{H II} \\ 
\mathbf{B}_{7} & = & \left(\frac{1}{2} - y_{2}\right) \, \mathbf{a}_{1} + \left(\frac{1}{2} +y_{2}\right) \, \mathbf{a}_{2} + \left(\frac{1}{2} - z_{2}\right) \, \mathbf{a}_{3} & = & \frac{1}{2}a \, \mathbf{\hat{x}} + y_{2}b \, \mathbf{\hat{y}} + \left(\frac{1}{2} - z_{2}\right)c \, \mathbf{\hat{z}} & \left(8f\right) & \mbox{H II} \\ 
\mathbf{B}_{8} & = & y_{2} \, \mathbf{a}_{1}-y_{2} \, \mathbf{a}_{2}-z_{2} \, \mathbf{a}_{3} & = & -y_{2}b \, \mathbf{\hat{y}}-z_{2}c \, \mathbf{\hat{z}} & \left(8f\right) & \mbox{H II} \\ 
\mathbf{B}_{9} & = & -y_{3} \, \mathbf{a}_{1} + y_{3} \, \mathbf{a}_{2} + z_{3} \, \mathbf{a}_{3} & = & y_{3}b \, \mathbf{\hat{y}} + z_{3}c \, \mathbf{\hat{z}} & \left(8f\right) & \mbox{S} \\ 
\mathbf{B}_{10} & = & \left(\frac{1}{2} +y_{3}\right) \, \mathbf{a}_{1} + \left(\frac{1}{2} - y_{3}\right) \, \mathbf{a}_{2} + \left(\frac{1}{2} +z_{3}\right) \, \mathbf{a}_{3} & = & \frac{1}{2}a \, \mathbf{\hat{x}}-y_{3}b \, \mathbf{\hat{y}} + \left(\frac{1}{2} +z_{3}\right)c \, \mathbf{\hat{z}} & \left(8f\right) & \mbox{S} \\ 
\mathbf{B}_{11} & = & \left(\frac{1}{2} - y_{3}\right) \, \mathbf{a}_{1} + \left(\frac{1}{2} +y_{3}\right) \, \mathbf{a}_{2} + \left(\frac{1}{2} - z_{3}\right) \, \mathbf{a}_{3} & = & \frac{1}{2}a \, \mathbf{\hat{x}} + y_{3}b \, \mathbf{\hat{y}} + \left(\frac{1}{2} - z_{3}\right)c \, \mathbf{\hat{z}} & \left(8f\right) & \mbox{S} \\ 
\mathbf{B}_{12} & = & y_{3} \, \mathbf{a}_{1}-y_{3} \, \mathbf{a}_{2}-z_{3} \, \mathbf{a}_{3} & = & -y_{3}b \, \mathbf{\hat{y}}-z_{3}c \, \mathbf{\hat{z}} & \left(8f\right) & \mbox{S} \\ 
\end{longtabu}
\renewcommand{\arraystretch}{1.0}
\noindent \hrulefill
\\
\textbf{References:}
\vspace*{-0.25cm}
\begin{flushleft}
  - \bibentry{Li_JCP_140_2014}. \\
\end{flushleft}
\noindent \hrulefill
\\
\textbf{Geometry files:}
\\
\noindent  - CIF: pp. {\hyperref[A2B_oC24_64_2f_f_cif]{\pageref{A2B_oC24_64_2f_f_cif}}} \\
\noindent  - POSCAR: pp. {\hyperref[A2B_oC24_64_2f_f_poscar]{\pageref{A2B_oC24_64_2f_f_poscar}}} \\
\onecolumn
{\phantomsection\label{A2B4C_oC28_66_l_kl_a}}
\subsection*{\huge \textbf{{\normalfont SrAl$_{2}$Se$_{4}$ Structure: A2B4C\_oC28\_66\_l\_kl\_a}}}
\noindent \hrulefill
\vspace*{0.25cm}
\begin{figure}[htp]
  \centering
  \vspace{-1em}
  {\includegraphics[width=1\textwidth]{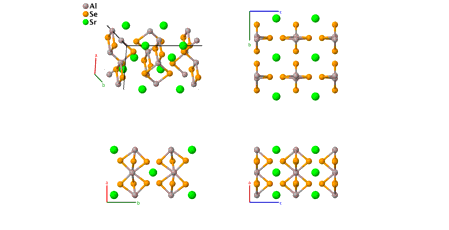}}
\end{figure}
\vspace*{-0.5cm}
\renewcommand{\arraystretch}{1.5}
\begin{equation*}
  \begin{array}{>{$\hspace{-0.15cm}}l<{$}>{$}p{0.5cm}<{$}>{$}p{18.5cm}<{$}}
    \mbox{\large \textbf{Prototype}} &\colon & \ce{SrAl2Se4} \\
    \mbox{\large \textbf{\AFLOW\ prototype label}} &\colon & \mbox{A2B4C\_oC28\_66\_l\_kl\_a} \\
    \mbox{\large \textbf{\textit{Strukturbericht} designation}} &\colon & \mbox{None} \\
    \mbox{\large \textbf{Pearson symbol}} &\colon & \mbox{oC28} \\
    \mbox{\large \textbf{Space group number}} &\colon & 66 \\
    \mbox{\large \textbf{Space group symbol}} &\colon & Cccm \\
    \mbox{\large \textbf{\AFLOW\ prototype command}} &\colon &  \texttt{aflow} \,  \, \texttt{-{}-proto=A2B4C\_oC28\_66\_l\_kl\_a } \, \newline \texttt{-{}-params=}{a,b/a,c/a,z_{2},x_{3},y_{3},x_{4},y_{4} }
  \end{array}
\end{equation*}
\renewcommand{\arraystretch}{1.0}

\noindent \parbox{1 \linewidth}{
\noindent \hrulefill
\\
\textbf{Base-centered Orthorhombic primitive vectors:} \\
\vspace*{-0.25cm}
\begin{tabular}{cc}
  \begin{tabular}{c}
    \parbox{0.6 \linewidth}{
      \renewcommand{\arraystretch}{1.5}
      \begin{equation*}
        \centering
        \begin{array}{ccc}
              \mathbf{a}_1 & = & \frac12 \, a \, \mathbf{\hat{x}} - \frac12 \, b \, \mathbf{\hat{y}} \\
    \mathbf{a}_2 & = & \frac12 \, a \, \mathbf{\hat{x}} + \frac12 \, b \, \mathbf{\hat{y}} \\
    \mathbf{a}_3 & = & c \, \mathbf{\hat{z}} \\

        \end{array}
      \end{equation*}
    }
    \renewcommand{\arraystretch}{1.0}
  \end{tabular}
  \begin{tabular}{c}
    \includegraphics[width=0.3\linewidth]{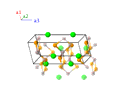} \\
  \end{tabular}
\end{tabular}

}
\vspace*{-0.25cm}

\noindent \hrulefill
\\
\textbf{Basis vectors:}
\vspace*{-0.25cm}
\renewcommand{\arraystretch}{1.5}
\begin{longtabu} to \textwidth{>{\centering $}X[-1,c,c]<{$}>{\centering $}X[-1,c,c]<{$}>{\centering $}X[-1,c,c]<{$}>{\centering $}X[-1,c,c]<{$}>{\centering $}X[-1,c,c]<{$}>{\centering $}X[-1,c,c]<{$}>{\centering $}X[-1,c,c]<{$}}
  & & \mbox{Lattice Coordinates} & & \mbox{Cartesian Coordinates} &\mbox{Wyckoff Position} & \mbox{Atom Type} \\  
  \mathbf{B}_{1} & = & \frac{1}{4} \, \mathbf{a}_{3} & = & \frac{1}{4}c \, \mathbf{\hat{z}} & \left(4a\right) & \mbox{Sr} \\ 
\mathbf{B}_{2} & = & \frac{3}{4} \, \mathbf{a}_{3} & = & \frac{3}{4}c \, \mathbf{\hat{z}} & \left(4a\right) & \mbox{Sr} \\ 
\mathbf{B}_{3} & = & \frac{1}{2} \, \mathbf{a}_{2} + z_{2} \, \mathbf{a}_{3} & = & \frac{1}{4}a \, \mathbf{\hat{x}} + \frac{1}{4}b \, \mathbf{\hat{y}} + z_{2}c \, \mathbf{\hat{z}} & \left(8k\right) & \mbox{Se I} \\ 
\mathbf{B}_{4} & = & \frac{1}{2} \, \mathbf{a}_{1} + \left(\frac{1}{2} - z_{2}\right) \, \mathbf{a}_{3} & = & \frac{1}{4}a \, \mathbf{\hat{x}}- \frac{1}{4}b  \, \mathbf{\hat{y}} + \left(\frac{1}{2} - z_{2}\right)c \, \mathbf{\hat{z}} & \left(8k\right) & \mbox{Se I} \\ 
\mathbf{B}_{5} & = & \frac{1}{2} \, \mathbf{a}_{2}-z_{2} \, \mathbf{a}_{3} & = & \frac{1}{4}a \, \mathbf{\hat{x}} + \frac{1}{4}b \, \mathbf{\hat{y}}-z_{2}c \, \mathbf{\hat{z}} & \left(8k\right) & \mbox{Se I} \\ 
\mathbf{B}_{6} & = & \frac{1}{2} \, \mathbf{a}_{1} + \left(\frac{1}{2} +z_{2}\right) \, \mathbf{a}_{3} & = & \frac{1}{4}a \, \mathbf{\hat{x}} + \frac{3}{4}b \, \mathbf{\hat{y}} + \left(\frac{1}{2} +z_{2}\right)c \, \mathbf{\hat{z}} & \left(8k\right) & \mbox{Se I} \\ 
\mathbf{B}_{7} & = & \left(x_{3}-y_{3}\right) \, \mathbf{a}_{1} + \left(x_{3}+y_{3}\right) \, \mathbf{a}_{2} & = & x_{3}a \, \mathbf{\hat{x}} + y_{3}b \, \mathbf{\hat{y}} & \left(8l\right) & \mbox{Al} \\ 
\mathbf{B}_{8} & = & \left(-x_{3}+y_{3}\right) \, \mathbf{a}_{1} + \left(-x_{3}-y_{3}\right) \, \mathbf{a}_{2} & = & -x_{3}a \, \mathbf{\hat{x}}-y_{3}b \, \mathbf{\hat{y}} & \left(8l\right) & \mbox{Al} \\ 
\mathbf{B}_{9} & = & \left(-x_{3}-y_{3}\right) \, \mathbf{a}_{1} + \left(-x_{3}+y_{3}\right) \, \mathbf{a}_{2} + \frac{1}{2} \, \mathbf{a}_{3} & = & -x_{3}a \, \mathbf{\hat{x}} + y_{3}b \, \mathbf{\hat{y}} + \frac{1}{2}c \, \mathbf{\hat{z}} & \left(8l\right) & \mbox{Al} \\ 
\mathbf{B}_{10} & = & \left(x_{3}+y_{3}\right) \, \mathbf{a}_{1} + \left(x_{3}-y_{3}\right) \, \mathbf{a}_{2} + \frac{1}{2} \, \mathbf{a}_{3} & = & x_{3}a \, \mathbf{\hat{x}}-y_{3}b \, \mathbf{\hat{y}} + \frac{1}{2}c \, \mathbf{\hat{z}} & \left(8l\right) & \mbox{Al} \\ 
\mathbf{B}_{11} & = & \left(x_{4}-y_{4}\right) \, \mathbf{a}_{1} + \left(x_{4}+y_{4}\right) \, \mathbf{a}_{2} & = & x_{4}a \, \mathbf{\hat{x}} + y_{4}b \, \mathbf{\hat{y}} & \left(8l\right) & \mbox{Se II} \\ 
\mathbf{B}_{12} & = & \left(-x_{4}+y_{4}\right) \, \mathbf{a}_{1} + \left(-x_{4}-y_{4}\right) \, \mathbf{a}_{2} & = & -x_{4}a \, \mathbf{\hat{x}}-y_{4}b \, \mathbf{\hat{y}} & \left(8l\right) & \mbox{Se II} \\ 
\mathbf{B}_{13} & = & \left(-x_{4}-y_{4}\right) \, \mathbf{a}_{1} + \left(-x_{4}+y_{4}\right) \, \mathbf{a}_{2} + \frac{1}{2} \, \mathbf{a}_{3} & = & -x_{4}a \, \mathbf{\hat{x}} + y_{4}b \, \mathbf{\hat{y}} + \frac{1}{2}c \, \mathbf{\hat{z}} & \left(8l\right) & \mbox{Se II} \\ 
\mathbf{B}_{14} & = & \left(x_{4}+y_{4}\right) \, \mathbf{a}_{1} + \left(x_{4}-y_{4}\right) \, \mathbf{a}_{2} + \frac{1}{2} \, \mathbf{a}_{3} & = & x_{4}a \, \mathbf{\hat{x}}-y_{4}b \, \mathbf{\hat{y}} + \frac{1}{2}c \, \mathbf{\hat{z}} & \left(8l\right) & \mbox{Se II} \\ 
\end{longtabu}
\renewcommand{\arraystretch}{1.0}
\noindent \hrulefill
\\
\textbf{References:}
\vspace*{-0.25cm}
\begin{flushleft}
  - \bibentry{Klee_Al2Se4Sr_ZNaturforsch_1978}. \\
\end{flushleft}
\textbf{Found in:}
\vspace*{-0.25cm}
\begin{flushleft}
  - \bibentry{Villars_PearsonsCrystalData_2013}. \\
\end{flushleft}
\noindent \hrulefill
\\
\textbf{Geometry files:}
\\
\noindent  - CIF: pp. {\hyperref[A2B4C_oC28_66_l_kl_a_cif]{\pageref{A2B4C_oC28_66_l_kl_a_cif}}} \\
\noindent  - POSCAR: pp. {\hyperref[A2B4C_oC28_66_l_kl_a_poscar]{\pageref{A2B4C_oC28_66_l_kl_a_poscar}}} \\
\onecolumn
{\phantomsection\label{A3B_oC64_66_gi2lm_2l}}
\subsection*{\huge \textbf{{\normalfont H$_{3}$S (60~GPa) Structure: A3B\_oC64\_66\_gi2lm\_2l}}}
\noindent \hrulefill
\vspace*{0.25cm}
\begin{figure}[htp]
  \centering
  \vspace{-1em}
  {\includegraphics[width=1\textwidth]{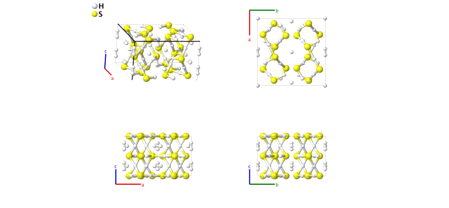}}
\end{figure}
\vspace*{-0.5cm}
\renewcommand{\arraystretch}{1.5}
\begin{equation*}
  \begin{array}{>{$\hspace{-0.15cm}}l<{$}>{$}p{0.5cm}<{$}>{$}p{18.5cm}<{$}}
    \mbox{\large \textbf{Prototype}} &\colon & \ce{H3S} \\
    \mbox{\large \textbf{\AFLOW\ prototype label}} &\colon & \mbox{A3B\_oC64\_66\_gi2lm\_2l} \\
    \mbox{\large \textbf{\textit{Strukturbericht} designation}} &\colon & \mbox{None} \\
    \mbox{\large \textbf{Pearson symbol}} &\colon & \mbox{oC64} \\
    \mbox{\large \textbf{Space group number}} &\colon & 66 \\
    \mbox{\large \textbf{Space group symbol}} &\colon & Cccm \\
    \mbox{\large \textbf{\AFLOW\ prototype command}} &\colon &  \texttt{aflow} \,  \, \texttt{-{}-proto=A3B\_oC64\_66\_gi2lm\_2l } \, \newline \texttt{-{}-params=}{a,b/a,c/a,x_{1},z_{2},x_{3},y_{3},x_{4},y_{4},x_{5},y_{5},x_{6},y_{6},x_{7},y_{7},z_{7} }
  \end{array}
\end{equation*}
\renewcommand{\arraystretch}{1.0}

\vspace*{-0.25cm}
\noindent \hrulefill
\begin{itemize}
  \item{This structure was found by first-principles electronic structure
calculations and is predicted to be the stable structure of H$_{3}$S
for pressures between 40 and 90~GPa.}
  \item{The data presented here was computed at 60~GPa.
}
\end{itemize}

\noindent \parbox{1 \linewidth}{
\noindent \hrulefill
\\
\textbf{Base-centered Orthorhombic primitive vectors:} \\
\vspace*{-0.25cm}
\begin{tabular}{cc}
  \begin{tabular}{c}
    \parbox{0.6 \linewidth}{
      \renewcommand{\arraystretch}{1.5}
      \begin{equation*}
        \centering
        \begin{array}{ccc}
              \mathbf{a}_1 & = & \frac12 \, a \, \mathbf{\hat{x}} - \frac12 \, b \, \mathbf{\hat{y}} \\
    \mathbf{a}_2 & = & \frac12 \, a \, \mathbf{\hat{x}} + \frac12 \, b \, \mathbf{\hat{y}} \\
    \mathbf{a}_3 & = & c \, \mathbf{\hat{z}} \\

        \end{array}
      \end{equation*}
    }
    \renewcommand{\arraystretch}{1.0}
  \end{tabular}
  \begin{tabular}{c}
    \includegraphics[width=0.3\linewidth]{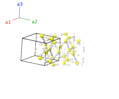} \\
  \end{tabular}
\end{tabular}

}
\vspace*{-0.25cm}

\noindent \hrulefill
\\
\textbf{Basis vectors:}
\vspace*{-0.25cm}
\renewcommand{\arraystretch}{1.5}
\begin{longtabu} to \textwidth{>{\centering $}X[-1,c,c]<{$}>{\centering $}X[-1,c,c]<{$}>{\centering $}X[-1,c,c]<{$}>{\centering $}X[-1,c,c]<{$}>{\centering $}X[-1,c,c]<{$}>{\centering $}X[-1,c,c]<{$}>{\centering $}X[-1,c,c]<{$}}
  & & \mbox{Lattice Coordinates} & & \mbox{Cartesian Coordinates} &\mbox{Wyckoff Position} & \mbox{Atom Type} \\  
  \mathbf{B}_{1} & = & x_{1} \, \mathbf{a}_{1} + x_{1} \, \mathbf{a}_{2} + \frac{1}{4} \, \mathbf{a}_{3} & = & x_{1}a \, \mathbf{\hat{x}} + \frac{1}{4}c \, \mathbf{\hat{z}} & \left(8g\right) & \mbox{H I} \\ 
\mathbf{B}_{2} & = & -x_{1} \, \mathbf{a}_{1}-x_{1} \, \mathbf{a}_{2} + \frac{1}{4} \, \mathbf{a}_{3} & = & -x_{1}a \, \mathbf{\hat{x}} + \frac{1}{4}c \, \mathbf{\hat{z}} & \left(8g\right) & \mbox{H I} \\ 
\mathbf{B}_{3} & = & -x_{1} \, \mathbf{a}_{1}-x_{1} \, \mathbf{a}_{2} + \frac{3}{4} \, \mathbf{a}_{3} & = & -x_{1}a \, \mathbf{\hat{x}} + \frac{3}{4}c \, \mathbf{\hat{z}} & \left(8g\right) & \mbox{H I} \\ 
\mathbf{B}_{4} & = & x_{1} \, \mathbf{a}_{1} + x_{1} \, \mathbf{a}_{2} + \frac{3}{4} \, \mathbf{a}_{3} & = & x_{1}a \, \mathbf{\hat{x}} + \frac{3}{4}c \, \mathbf{\hat{z}} & \left(8g\right) & \mbox{H I} \\ 
\mathbf{B}_{5} & = & z_{2} \, \mathbf{a}_{3} & = & z_{2}c \, \mathbf{\hat{z}} & \left(8i\right) & \mbox{H II} \\ 
\mathbf{B}_{6} & = & \left(\frac{1}{2} - z_{2}\right) \, \mathbf{a}_{3} & = & \left(\frac{1}{2} - z_{2}\right)c \, \mathbf{\hat{z}} & \left(8i\right) & \mbox{H II} \\ 
\mathbf{B}_{7} & = & -z_{2} \, \mathbf{a}_{3} & = & -z_{2}c \, \mathbf{\hat{z}} & \left(8i\right) & \mbox{H II} \\ 
\mathbf{B}_{8} & = & \left(\frac{1}{2} +z_{2}\right) \, \mathbf{a}_{3} & = & \left(\frac{1}{2} +z_{2}\right)c \, \mathbf{\hat{z}} & \left(8i\right) & \mbox{H II} \\ 
\mathbf{B}_{9} & = & \left(x_{3}-y_{3}\right) \, \mathbf{a}_{1} + \left(x_{3}+y_{3}\right) \, \mathbf{a}_{2} & = & x_{3}a \, \mathbf{\hat{x}} + y_{3}b \, \mathbf{\hat{y}} & \left(8l\right) & \mbox{H III} \\ 
\mathbf{B}_{10} & = & \left(-x_{3}+y_{3}\right) \, \mathbf{a}_{1} + \left(-x_{3}-y_{3}\right) \, \mathbf{a}_{2} & = & -x_{3}a \, \mathbf{\hat{x}}-y_{3}b \, \mathbf{\hat{y}} & \left(8l\right) & \mbox{H III} \\ 
\mathbf{B}_{11} & = & \left(-x_{3}-y_{3}\right) \, \mathbf{a}_{1} + \left(-x_{3}+y_{3}\right) \, \mathbf{a}_{2} + \frac{1}{2} \, \mathbf{a}_{3} & = & -x_{3}a \, \mathbf{\hat{x}} + y_{3}b \, \mathbf{\hat{y}} + \frac{1}{2}c \, \mathbf{\hat{z}} & \left(8l\right) & \mbox{H III} \\ 
\mathbf{B}_{12} & = & \left(x_{3}+y_{3}\right) \, \mathbf{a}_{1} + \left(x_{3}-y_{3}\right) \, \mathbf{a}_{2} + \frac{1}{2} \, \mathbf{a}_{3} & = & x_{3}a \, \mathbf{\hat{x}}-y_{3}b \, \mathbf{\hat{y}} + \frac{1}{2}c \, \mathbf{\hat{z}} & \left(8l\right) & \mbox{H III} \\ 
\mathbf{B}_{13} & = & \left(x_{4}-y_{4}\right) \, \mathbf{a}_{1} + \left(x_{4}+y_{4}\right) \, \mathbf{a}_{2} & = & x_{4}a \, \mathbf{\hat{x}} + y_{4}b \, \mathbf{\hat{y}} & \left(8l\right) & \mbox{H IV} \\ 
\mathbf{B}_{14} & = & \left(-x_{4}+y_{4}\right) \, \mathbf{a}_{1} + \left(-x_{4}-y_{4}\right) \, \mathbf{a}_{2} & = & -x_{4}a \, \mathbf{\hat{x}}-y_{4}b \, \mathbf{\hat{y}} & \left(8l\right) & \mbox{H IV} \\ 
\mathbf{B}_{15} & = & \left(-x_{4}-y_{4}\right) \, \mathbf{a}_{1} + \left(-x_{4}+y_{4}\right) \, \mathbf{a}_{2} + \frac{1}{2} \, \mathbf{a}_{3} & = & -x_{4}a \, \mathbf{\hat{x}} + y_{4}b \, \mathbf{\hat{y}} + \frac{1}{2}c \, \mathbf{\hat{z}} & \left(8l\right) & \mbox{H IV} \\ 
\mathbf{B}_{16} & = & \left(x_{4}+y_{4}\right) \, \mathbf{a}_{1} + \left(x_{4}-y_{4}\right) \, \mathbf{a}_{2} + \frac{1}{2} \, \mathbf{a}_{3} & = & x_{4}a \, \mathbf{\hat{x}}-y_{4}b \, \mathbf{\hat{y}} + \frac{1}{2}c \, \mathbf{\hat{z}} & \left(8l\right) & \mbox{H IV} \\ 
\mathbf{B}_{17} & = & \left(x_{5}-y_{5}\right) \, \mathbf{a}_{1} + \left(x_{5}+y_{5}\right) \, \mathbf{a}_{2} & = & x_{5}a \, \mathbf{\hat{x}} + y_{5}b \, \mathbf{\hat{y}} & \left(8l\right) & \mbox{S I} \\ 
\mathbf{B}_{18} & = & \left(-x_{5}+y_{5}\right) \, \mathbf{a}_{1} + \left(-x_{5}-y_{5}\right) \, \mathbf{a}_{2} & = & -x_{5}a \, \mathbf{\hat{x}}-y_{5}b \, \mathbf{\hat{y}} & \left(8l\right) & \mbox{S I} \\ 
\mathbf{B}_{19} & = & \left(-x_{5}-y_{5}\right) \, \mathbf{a}_{1} + \left(-x_{5}+y_{5}\right) \, \mathbf{a}_{2} + \frac{1}{2} \, \mathbf{a}_{3} & = & -x_{5}a \, \mathbf{\hat{x}} + y_{5}b \, \mathbf{\hat{y}} + \frac{1}{2}c \, \mathbf{\hat{z}} & \left(8l\right) & \mbox{S I} \\ 
\mathbf{B}_{20} & = & \left(x_{5}+y_{5}\right) \, \mathbf{a}_{1} + \left(x_{5}-y_{5}\right) \, \mathbf{a}_{2} + \frac{1}{2} \, \mathbf{a}_{3} & = & x_{5}a \, \mathbf{\hat{x}}-y_{5}b \, \mathbf{\hat{y}} + \frac{1}{2}c \, \mathbf{\hat{z}} & \left(8l\right) & \mbox{S I} \\ 
\mathbf{B}_{21} & = & \left(x_{6}-y_{6}\right) \, \mathbf{a}_{1} + \left(x_{6}+y_{6}\right) \, \mathbf{a}_{2} & = & x_{6}a \, \mathbf{\hat{x}} + y_{6}b \, \mathbf{\hat{y}} & \left(8l\right) & \mbox{S II} \\ 
\mathbf{B}_{22} & = & \left(-x_{6}+y_{6}\right) \, \mathbf{a}_{1} + \left(-x_{6}-y_{6}\right) \, \mathbf{a}_{2} & = & -x_{6}a \, \mathbf{\hat{x}}-y_{6}b \, \mathbf{\hat{y}} & \left(8l\right) & \mbox{S II} \\ 
\mathbf{B}_{23} & = & \left(-x_{6}-y_{6}\right) \, \mathbf{a}_{1} + \left(-x_{6}+y_{6}\right) \, \mathbf{a}_{2} + \frac{1}{2} \, \mathbf{a}_{3} & = & -x_{6}a \, \mathbf{\hat{x}} + y_{6}b \, \mathbf{\hat{y}} + \frac{1}{2}c \, \mathbf{\hat{z}} & \left(8l\right) & \mbox{S II} \\ 
\mathbf{B}_{24} & = & \left(x_{6}+y_{6}\right) \, \mathbf{a}_{1} + \left(x_{6}-y_{6}\right) \, \mathbf{a}_{2} + \frac{1}{2} \, \mathbf{a}_{3} & = & x_{6}a \, \mathbf{\hat{x}}-y_{6}b \, \mathbf{\hat{y}} + \frac{1}{2}c \, \mathbf{\hat{z}} & \left(8l\right) & \mbox{S II} \\ 
\mathbf{B}_{25} & = & \left(x_{7}-y_{7}\right) \, \mathbf{a}_{1} + \left(x_{7}+y_{7}\right) \, \mathbf{a}_{2} + z_{7} \, \mathbf{a}_{3} & = & x_{7}a \, \mathbf{\hat{x}} + y_{7}b \, \mathbf{\hat{y}} + z_{7}c \, \mathbf{\hat{z}} & \left(16m\right) & \mbox{H V} \\ 
\mathbf{B}_{26} & = & \left(-x_{7}+y_{7}\right) \, \mathbf{a}_{1} + \left(-x_{7}-y_{7}\right) \, \mathbf{a}_{2} + z_{7} \, \mathbf{a}_{3} & = & -x_{7}a \, \mathbf{\hat{x}}-y_{7}b \, \mathbf{\hat{y}} + z_{7}c \, \mathbf{\hat{z}} & \left(16m\right) & \mbox{H V} \\ 
\mathbf{B}_{27} & = & \left(-x_{7}-y_{7}\right) \, \mathbf{a}_{1} + \left(-x_{7}+y_{7}\right) \, \mathbf{a}_{2} + \left(\frac{1}{2} - z_{7}\right) \, \mathbf{a}_{3} & = & -x_{7}a \, \mathbf{\hat{x}} + y_{7}b \, \mathbf{\hat{y}} + \left(\frac{1}{2} - z_{7}\right)c \, \mathbf{\hat{z}} & \left(16m\right) & \mbox{H V} \\ 
\mathbf{B}_{28} & = & \left(x_{7}+y_{7}\right) \, \mathbf{a}_{1} + \left(x_{7}-y_{7}\right) \, \mathbf{a}_{2} + \left(\frac{1}{2} - z_{7}\right) \, \mathbf{a}_{3} & = & x_{7}a \, \mathbf{\hat{x}}-y_{7}b \, \mathbf{\hat{y}} + \left(\frac{1}{2} - z_{7}\right)c \, \mathbf{\hat{z}} & \left(16m\right) & \mbox{H V} \\ 
\mathbf{B}_{29} & = & \left(-x_{7}+y_{7}\right) \, \mathbf{a}_{1} + \left(-x_{7}-y_{7}\right) \, \mathbf{a}_{2}-z_{7} \, \mathbf{a}_{3} & = & -x_{7}a \, \mathbf{\hat{x}}-y_{7}b \, \mathbf{\hat{y}}-z_{7}c \, \mathbf{\hat{z}} & \left(16m\right) & \mbox{H V} \\ 
\mathbf{B}_{30} & = & \left(x_{7}-y_{7}\right) \, \mathbf{a}_{1} + \left(x_{7}+y_{7}\right) \, \mathbf{a}_{2}-z_{7} \, \mathbf{a}_{3} & = & x_{7}a \, \mathbf{\hat{x}} + y_{7}b \, \mathbf{\hat{y}}-z_{7}c \, \mathbf{\hat{z}} & \left(16m\right) & \mbox{H V} \\ 
\mathbf{B}_{31} & = & \left(x_{7}+y_{7}\right) \, \mathbf{a}_{1} + \left(x_{7}-y_{7}\right) \, \mathbf{a}_{2} + \left(\frac{1}{2} +z_{7}\right) \, \mathbf{a}_{3} & = & x_{7}a \, \mathbf{\hat{x}}-y_{7}b \, \mathbf{\hat{y}} + \left(\frac{1}{2} +z_{7}\right)c \, \mathbf{\hat{z}} & \left(16m\right) & \mbox{H V} \\ 
\mathbf{B}_{32} & = & \left(-x_{7}-y_{7}\right) \, \mathbf{a}_{1} + \left(-x_{7}+y_{7}\right) \, \mathbf{a}_{2} + \left(\frac{1}{2} +z_{7}\right) \, \mathbf{a}_{3} & = & -x_{7}a \, \mathbf{\hat{x}} + y_{7}b \, \mathbf{\hat{y}} + \left(\frac{1}{2} +z_{7}\right)c \, \mathbf{\hat{z}} & \left(16m\right) & \mbox{H V} \\ 
\end{longtabu}
\renewcommand{\arraystretch}{1.0}
\noindent \hrulefill
\\
\textbf{References:}
\vspace*{-0.25cm}
\begin{flushleft}
  - \bibentry{Duan_SR_4_2014}. \\
\end{flushleft}
\noindent \hrulefill
\\
\textbf{Geometry files:}
\\
\noindent  - CIF: pp. {\hyperref[A3B_oC64_66_gi2lm_2l_cif]{\pageref{A3B_oC64_66_gi2lm_2l_cif}}} \\
\noindent  - POSCAR: pp. {\hyperref[A3B_oC64_66_gi2lm_2l_poscar]{\pageref{A3B_oC64_66_gi2lm_2l_poscar}}} \\
\onecolumn
{\phantomsection\label{A3B_oC64_66_kl2m_bdl}}
\subsection*{\huge \textbf{{\normalfont $\beta$-ThI$_{3}$ Structure: A3B\_oC64\_66\_kl2m\_bdl}}}
\noindent \hrulefill
\vspace*{0.25cm}
\begin{figure}[htp]
  \centering
  \vspace{-1em}
  {\includegraphics[width=1\textwidth]{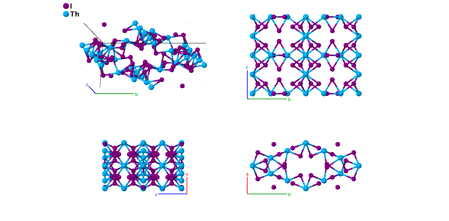}}
\end{figure}
\vspace*{-0.5cm}
\renewcommand{\arraystretch}{1.5}
\begin{equation*}
  \begin{array}{>{$\hspace{-0.15cm}}l<{$}>{$}p{0.5cm}<{$}>{$}p{18.5cm}<{$}}
    \mbox{\large \textbf{Prototype}} &\colon & \ce{$\beta$-ThI3} \\
    \mbox{\large \textbf{\AFLOW\ prototype label}} &\colon & \mbox{A3B\_oC64\_66\_kl2m\_bdl} \\
    \mbox{\large \textbf{\textit{Strukturbericht} designation}} &\colon & \mbox{None} \\
    \mbox{\large \textbf{Pearson symbol}} &\colon & \mbox{oC64} \\
    \mbox{\large \textbf{Space group number}} &\colon & 66 \\
    \mbox{\large \textbf{Space group symbol}} &\colon & Cccm \\
    \mbox{\large \textbf{\AFLOW\ prototype command}} &\colon &  \texttt{aflow} \,  \, \texttt{-{}-proto=A3B\_oC64\_66\_kl2m\_bdl } \, \newline \texttt{-{}-params=}{a,b/a,c/a,z_{3},x_{4},y_{4},x_{5},y_{5},x_{6},y_{6},z_{6},x_{7},y_{7},z_{7} }
  \end{array}
\end{equation*}
\renewcommand{\arraystretch}{1.0}

\noindent \parbox{1 \linewidth}{
\noindent \hrulefill
\\
\textbf{Base-centered Orthorhombic primitive vectors:} \\
\vspace*{-0.25cm}
\begin{tabular}{cc}
  \begin{tabular}{c}
    \parbox{0.6 \linewidth}{
      \renewcommand{\arraystretch}{1.5}
      \begin{equation*}
        \centering
        \begin{array}{ccc}
              \mathbf{a}_1 & = & \frac12 \, a \, \mathbf{\hat{x}} - \frac12 \, b \, \mathbf{\hat{y}} \\
    \mathbf{a}_2 & = & \frac12 \, a \, \mathbf{\hat{x}} + \frac12 \, b \, \mathbf{\hat{y}} \\
    \mathbf{a}_3 & = & c \, \mathbf{\hat{z}} \\

        \end{array}
      \end{equation*}
    }
    \renewcommand{\arraystretch}{1.0}
  \end{tabular}
  \begin{tabular}{c}
    \includegraphics[width=0.3\linewidth]{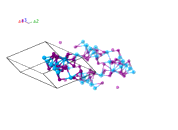} \\
  \end{tabular}
\end{tabular}

}
\vspace*{-0.25cm}

\noindent \hrulefill
\\
\textbf{Basis vectors:}
\vspace*{-0.25cm}
\renewcommand{\arraystretch}{1.5}
\begin{longtabu} to \textwidth{>{\centering $}X[-1,c,c]<{$}>{\centering $}X[-1,c,c]<{$}>{\centering $}X[-1,c,c]<{$}>{\centering $}X[-1,c,c]<{$}>{\centering $}X[-1,c,c]<{$}>{\centering $}X[-1,c,c]<{$}>{\centering $}X[-1,c,c]<{$}}
  & & \mbox{Lattice Coordinates} & & \mbox{Cartesian Coordinates} &\mbox{Wyckoff Position} & \mbox{Atom Type} \\  
  \mathbf{B}_{1} & = & \frac{1}{2} \, \mathbf{a}_{1} + \frac{1}{2} \, \mathbf{a}_{2} + \frac{1}{4} \, \mathbf{a}_{3} & = & \frac{1}{2}a \, \mathbf{\hat{x}} + \frac{1}{4}c \, \mathbf{\hat{z}} & \left(4b\right) & \mbox{Th I} \\ 
\mathbf{B}_{2} & = & \frac{1}{2} \, \mathbf{a}_{1} + \frac{1}{2} \, \mathbf{a}_{2} + \frac{3}{4} \, \mathbf{a}_{3} & = & \frac{1}{2}a \, \mathbf{\hat{x}} + \frac{3}{4}c \, \mathbf{\hat{z}} & \left(4b\right) & \mbox{Th I} \\ 
\mathbf{B}_{3} & = & \frac{1}{2} \, \mathbf{a}_{1} + \frac{1}{2} \, \mathbf{a}_{2} & = & \frac{1}{2}a \, \mathbf{\hat{x}} & \left(4d\right) & \mbox{Th II} \\ 
\mathbf{B}_{4} & = & \frac{1}{2} \, \mathbf{a}_{1} + \frac{1}{2} \, \mathbf{a}_{2} + \frac{1}{2} \, \mathbf{a}_{3} & = & \frac{1}{2}a \, \mathbf{\hat{x}} + \frac{1}{2}c \, \mathbf{\hat{z}} & \left(4d\right) & \mbox{Th II} \\ 
\mathbf{B}_{5} & = & \frac{1}{2} \, \mathbf{a}_{2} + z_{3} \, \mathbf{a}_{3} & = & \frac{1}{4}a \, \mathbf{\hat{x}} + \frac{1}{4}b \, \mathbf{\hat{y}} + z_{3}c \, \mathbf{\hat{z}} & \left(8k\right) & \mbox{I I} \\ 
\mathbf{B}_{6} & = & \frac{1}{2} \, \mathbf{a}_{1} + \left(\frac{1}{2} - z_{3}\right) \, \mathbf{a}_{3} & = & \frac{1}{4}a \, \mathbf{\hat{x}}- \frac{1}{4}b  \, \mathbf{\hat{y}} + \left(\frac{1}{2} - z_{3}\right)c \, \mathbf{\hat{z}} & \left(8k\right) & \mbox{I I} \\ 
\mathbf{B}_{7} & = & \frac{1}{2} \, \mathbf{a}_{2}-z_{3} \, \mathbf{a}_{3} & = & \frac{1}{4}a \, \mathbf{\hat{x}} + \frac{1}{4}b \, \mathbf{\hat{y}}-z_{3}c \, \mathbf{\hat{z}} & \left(8k\right) & \mbox{I I} \\ 
\mathbf{B}_{8} & = & \frac{1}{2} \, \mathbf{a}_{1} + \left(\frac{1}{2} +z_{3}\right) \, \mathbf{a}_{3} & = & \frac{1}{4}a \, \mathbf{\hat{x}} + \frac{3}{4}b \, \mathbf{\hat{y}} + \left(\frac{1}{2} +z_{3}\right)c \, \mathbf{\hat{z}} & \left(8k\right) & \mbox{I I} \\ 
\mathbf{B}_{9} & = & \left(x_{4}-y_{4}\right) \, \mathbf{a}_{1} + \left(x_{4}+y_{4}\right) \, \mathbf{a}_{2} & = & x_{4}a \, \mathbf{\hat{x}} + y_{4}b \, \mathbf{\hat{y}} & \left(8l\right) & \mbox{I II} \\ 
\mathbf{B}_{10} & = & \left(-x_{4}+y_{4}\right) \, \mathbf{a}_{1} + \left(-x_{4}-y_{4}\right) \, \mathbf{a}_{2} & = & -x_{4}a \, \mathbf{\hat{x}}-y_{4}b \, \mathbf{\hat{y}} & \left(8l\right) & \mbox{I II} \\ 
\mathbf{B}_{11} & = & \left(-x_{4}-y_{4}\right) \, \mathbf{a}_{1} + \left(-x_{4}+y_{4}\right) \, \mathbf{a}_{2} + \frac{1}{2} \, \mathbf{a}_{3} & = & -x_{4}a \, \mathbf{\hat{x}} + y_{4}b \, \mathbf{\hat{y}} + \frac{1}{2}c \, \mathbf{\hat{z}} & \left(8l\right) & \mbox{I II} \\ 
\mathbf{B}_{12} & = & \left(x_{4}+y_{4}\right) \, \mathbf{a}_{1} + \left(x_{4}-y_{4}\right) \, \mathbf{a}_{2} + \frac{1}{2} \, \mathbf{a}_{3} & = & x_{4}a \, \mathbf{\hat{x}}-y_{4}b \, \mathbf{\hat{y}} + \frac{1}{2}c \, \mathbf{\hat{z}} & \left(8l\right) & \mbox{I II} \\ 
\mathbf{B}_{13} & = & \left(x_{5}-y_{5}\right) \, \mathbf{a}_{1} + \left(x_{5}+y_{5}\right) \, \mathbf{a}_{2} & = & x_{5}a \, \mathbf{\hat{x}} + y_{5}b \, \mathbf{\hat{y}} & \left(8l\right) & \mbox{Th III} \\ 
\mathbf{B}_{14} & = & \left(-x_{5}+y_{5}\right) \, \mathbf{a}_{1} + \left(-x_{5}-y_{5}\right) \, \mathbf{a}_{2} & = & -x_{5}a \, \mathbf{\hat{x}}-y_{5}b \, \mathbf{\hat{y}} & \left(8l\right) & \mbox{Th III} \\ 
\mathbf{B}_{15} & = & \left(-x_{5}-y_{5}\right) \, \mathbf{a}_{1} + \left(-x_{5}+y_{5}\right) \, \mathbf{a}_{2} + \frac{1}{2} \, \mathbf{a}_{3} & = & -x_{5}a \, \mathbf{\hat{x}} + y_{5}b \, \mathbf{\hat{y}} + \frac{1}{2}c \, \mathbf{\hat{z}} & \left(8l\right) & \mbox{Th III} \\ 
\mathbf{B}_{16} & = & \left(x_{5}+y_{5}\right) \, \mathbf{a}_{1} + \left(x_{5}-y_{5}\right) \, \mathbf{a}_{2} + \frac{1}{2} \, \mathbf{a}_{3} & = & x_{5}a \, \mathbf{\hat{x}}-y_{5}b \, \mathbf{\hat{y}} + \frac{1}{2}c \, \mathbf{\hat{z}} & \left(8l\right) & \mbox{Th III} \\ 
\mathbf{B}_{17} & = & \left(x_{6}-y_{6}\right) \, \mathbf{a}_{1} + \left(x_{6}+y_{6}\right) \, \mathbf{a}_{2} + z_{6} \, \mathbf{a}_{3} & = & x_{6}a \, \mathbf{\hat{x}} + y_{6}b \, \mathbf{\hat{y}} + z_{6}c \, \mathbf{\hat{z}} & \left(16m\right) & \mbox{I III} \\ 
\mathbf{B}_{18} & = & \left(-x_{6}+y_{6}\right) \, \mathbf{a}_{1} + \left(-x_{6}-y_{6}\right) \, \mathbf{a}_{2} + z_{6} \, \mathbf{a}_{3} & = & -x_{6}a \, \mathbf{\hat{x}}-y_{6}b \, \mathbf{\hat{y}} + z_{6}c \, \mathbf{\hat{z}} & \left(16m\right) & \mbox{I III} \\ 
\mathbf{B}_{19} & = & \left(-x_{6}-y_{6}\right) \, \mathbf{a}_{1} + \left(-x_{6}+y_{6}\right) \, \mathbf{a}_{2} + \left(\frac{1}{2} - z_{6}\right) \, \mathbf{a}_{3} & = & -x_{6}a \, \mathbf{\hat{x}} + y_{6}b \, \mathbf{\hat{y}} + \left(\frac{1}{2} - z_{6}\right)c \, \mathbf{\hat{z}} & \left(16m\right) & \mbox{I III} \\ 
\mathbf{B}_{20} & = & \left(x_{6}+y_{6}\right) \, \mathbf{a}_{1} + \left(x_{6}-y_{6}\right) \, \mathbf{a}_{2} + \left(\frac{1}{2} - z_{6}\right) \, \mathbf{a}_{3} & = & x_{6}a \, \mathbf{\hat{x}}-y_{6}b \, \mathbf{\hat{y}} + \left(\frac{1}{2} - z_{6}\right)c \, \mathbf{\hat{z}} & \left(16m\right) & \mbox{I III} \\ 
\mathbf{B}_{21} & = & \left(-x_{6}+y_{6}\right) \, \mathbf{a}_{1} + \left(-x_{6}-y_{6}\right) \, \mathbf{a}_{2}-z_{6} \, \mathbf{a}_{3} & = & -x_{6}a \, \mathbf{\hat{x}}-y_{6}b \, \mathbf{\hat{y}}-z_{6}c \, \mathbf{\hat{z}} & \left(16m\right) & \mbox{I III} \\ 
\mathbf{B}_{22} & = & \left(x_{6}-y_{6}\right) \, \mathbf{a}_{1} + \left(x_{6}+y_{6}\right) \, \mathbf{a}_{2}-z_{6} \, \mathbf{a}_{3} & = & x_{6}a \, \mathbf{\hat{x}} + y_{6}b \, \mathbf{\hat{y}}-z_{6}c \, \mathbf{\hat{z}} & \left(16m\right) & \mbox{I III} \\ 
\mathbf{B}_{23} & = & \left(x_{6}+y_{6}\right) \, \mathbf{a}_{1} + \left(x_{6}-y_{6}\right) \, \mathbf{a}_{2} + \left(\frac{1}{2} +z_{6}\right) \, \mathbf{a}_{3} & = & x_{6}a \, \mathbf{\hat{x}}-y_{6}b \, \mathbf{\hat{y}} + \left(\frac{1}{2} +z_{6}\right)c \, \mathbf{\hat{z}} & \left(16m\right) & \mbox{I III} \\ 
\mathbf{B}_{24} & = & \left(-x_{6}-y_{6}\right) \, \mathbf{a}_{1} + \left(-x_{6}+y_{6}\right) \, \mathbf{a}_{2} + \left(\frac{1}{2} +z_{6}\right) \, \mathbf{a}_{3} & = & -x_{6}a \, \mathbf{\hat{x}} + y_{6}b \, \mathbf{\hat{y}} + \left(\frac{1}{2} +z_{6}\right)c \, \mathbf{\hat{z}} & \left(16m\right) & \mbox{I III} \\ 
\mathbf{B}_{25} & = & \left(x_{7}-y_{7}\right) \, \mathbf{a}_{1} + \left(x_{7}+y_{7}\right) \, \mathbf{a}_{2} + z_{7} \, \mathbf{a}_{3} & = & x_{7}a \, \mathbf{\hat{x}} + y_{7}b \, \mathbf{\hat{y}} + z_{7}c \, \mathbf{\hat{z}} & \left(16m\right) & \mbox{I IV} \\ 
\mathbf{B}_{26} & = & \left(-x_{7}+y_{7}\right) \, \mathbf{a}_{1} + \left(-x_{7}-y_{7}\right) \, \mathbf{a}_{2} + z_{7} \, \mathbf{a}_{3} & = & -x_{7}a \, \mathbf{\hat{x}}-y_{7}b \, \mathbf{\hat{y}} + z_{7}c \, \mathbf{\hat{z}} & \left(16m\right) & \mbox{I IV} \\ 
\mathbf{B}_{27} & = & \left(-x_{7}-y_{7}\right) \, \mathbf{a}_{1} + \left(-x_{7}+y_{7}\right) \, \mathbf{a}_{2} + \left(\frac{1}{2} - z_{7}\right) \, \mathbf{a}_{3} & = & -x_{7}a \, \mathbf{\hat{x}} + y_{7}b \, \mathbf{\hat{y}} + \left(\frac{1}{2} - z_{7}\right)c \, \mathbf{\hat{z}} & \left(16m\right) & \mbox{I IV} \\ 
\mathbf{B}_{28} & = & \left(x_{7}+y_{7}\right) \, \mathbf{a}_{1} + \left(x_{7}-y_{7}\right) \, \mathbf{a}_{2} + \left(\frac{1}{2} - z_{7}\right) \, \mathbf{a}_{3} & = & x_{7}a \, \mathbf{\hat{x}}-y_{7}b \, \mathbf{\hat{y}} + \left(\frac{1}{2} - z_{7}\right)c \, \mathbf{\hat{z}} & \left(16m\right) & \mbox{I IV} \\ 
\mathbf{B}_{29} & = & \left(-x_{7}+y_{7}\right) \, \mathbf{a}_{1} + \left(-x_{7}-y_{7}\right) \, \mathbf{a}_{2}-z_{7} \, \mathbf{a}_{3} & = & -x_{7}a \, \mathbf{\hat{x}}-y_{7}b \, \mathbf{\hat{y}}-z_{7}c \, \mathbf{\hat{z}} & \left(16m\right) & \mbox{I IV} \\ 
\mathbf{B}_{30} & = & \left(x_{7}-y_{7}\right) \, \mathbf{a}_{1} + \left(x_{7}+y_{7}\right) \, \mathbf{a}_{2}-z_{7} \, \mathbf{a}_{3} & = & x_{7}a \, \mathbf{\hat{x}} + y_{7}b \, \mathbf{\hat{y}}-z_{7}c \, \mathbf{\hat{z}} & \left(16m\right) & \mbox{I IV} \\ 
\mathbf{B}_{31} & = & \left(x_{7}+y_{7}\right) \, \mathbf{a}_{1} + \left(x_{7}-y_{7}\right) \, \mathbf{a}_{2} + \left(\frac{1}{2} +z_{7}\right) \, \mathbf{a}_{3} & = & x_{7}a \, \mathbf{\hat{x}}-y_{7}b \, \mathbf{\hat{y}} + \left(\frac{1}{2} +z_{7}\right)c \, \mathbf{\hat{z}} & \left(16m\right) & \mbox{I IV} \\ 
\mathbf{B}_{32} & = & \left(-x_{7}-y_{7}\right) \, \mathbf{a}_{1} + \left(-x_{7}+y_{7}\right) \, \mathbf{a}_{2} + \left(\frac{1}{2} +z_{7}\right) \, \mathbf{a}_{3} & = & -x_{7}a \, \mathbf{\hat{x}} + y_{7}b \, \mathbf{\hat{y}} + \left(\frac{1}{2} +z_{7}\right)c \, \mathbf{\hat{z}} & \left(16m\right) & \mbox{I IV} \\ 
\end{longtabu}
\renewcommand{\arraystretch}{1.0}
\noindent \hrulefill
\\
\textbf{References:}
\vspace*{-0.25cm}
\begin{flushleft}
  - \bibentry{Beck_Ang_Chem_94_1982}. \\
\end{flushleft}
\textbf{Found in:}
\vspace*{-0.25cm}
\begin{flushleft}
  - \bibentry{Downs_AM_88_2003}. \\
\end{flushleft}
\noindent \hrulefill
\\
\textbf{Geometry files:}
\\
\noindent  - CIF: pp. {\hyperref[A3B_oC64_66_kl2m_bdl_cif]{\pageref{A3B_oC64_66_kl2m_bdl_cif}}} \\
\noindent  - POSCAR: pp. {\hyperref[A3B_oC64_66_kl2m_bdl_poscar]{\pageref{A3B_oC64_66_kl2m_bdl_poscar}}} \\
\onecolumn
{\phantomsection\label{A2BC_oC16_67_ag_b_g}}
\subsection*{\huge \textbf{{\normalfont Al$_{2}$CuIr Structure: A2BC\_oC16\_67\_ag\_b\_g}}}
\noindent \hrulefill
\vspace*{0.25cm}
\begin{figure}[htp]
  \centering
  \vspace{-1em}
  {\includegraphics[width=1\textwidth]{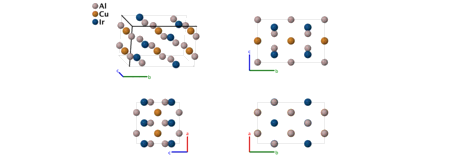}}
\end{figure}
\vspace*{-0.5cm}
\renewcommand{\arraystretch}{1.5}
\begin{equation*}
  \begin{array}{>{$\hspace{-0.15cm}}l<{$}>{$}p{0.5cm}<{$}>{$}p{18.5cm}<{$}}
    \mbox{\large \textbf{Prototype}} &\colon & \ce{Al2CuIr} \\
    \mbox{\large \textbf{\AFLOW\ prototype label}} &\colon & \mbox{A2BC\_oC16\_67\_ag\_b\_g} \\
    \mbox{\large \textbf{\textit{Strukturbericht} designation}} &\colon & \mbox{None} \\
    \mbox{\large \textbf{Pearson symbol}} &\colon & \mbox{oC16} \\
    \mbox{\large \textbf{Space group number}} &\colon & 67 \\
    \mbox{\large \textbf{Space group symbol}} &\colon & Cmma \\
    \mbox{\large \textbf{\AFLOW\ prototype command}} &\colon &  \texttt{aflow} \,  \, \texttt{-{}-proto=A2BC\_oC16\_67\_ag\_b\_g } \, \newline \texttt{-{}-params=}{a,b/a,c/a,z_{3},z_{4} }
  \end{array}
\end{equation*}
\renewcommand{\arraystretch}{1.0}

\vspace*{-0.25cm}
\noindent \hrulefill
\begin{itemize}
  \item{Al$_{2}$CuIr (pp. {\hyperref[A2BC_oC16_67_ag_b_g]{\pageref{A2BC_oC16_67_ag_b_g}}}) and 
CuHoP$_{2}$ (pp. {\hyperref[ABC2_oC16_67_b_g_ag]{\pageref{ABC2_oC16_67_b_g_ag}}})
have similar \AFLOW\ prototype labels ({\it{i.e.}}, same symmetry and set of
Wyckoff positions with different stoichiometry labels due to alphabetic ordering of atomic species).
They are generated by the same symmetry operations with different sets of parameters
(\texttt{-{}-params}) specified in their corresponding \CIF\ files.
}
\end{itemize}

\noindent \parbox{1 \linewidth}{
\noindent \hrulefill
\\
\textbf{Base-centered Orthorhombic primitive vectors:} \\
\vspace*{-0.25cm}
\begin{tabular}{cc}
  \begin{tabular}{c}
    \parbox{0.6 \linewidth}{
      \renewcommand{\arraystretch}{1.5}
      \begin{equation*}
        \centering
        \begin{array}{ccc}
              \mathbf{a}_1 & = & \frac12 \, a \, \mathbf{\hat{x}} - \frac12 \, b \, \mathbf{\hat{y}} \\
    \mathbf{a}_2 & = & \frac12 \, a \, \mathbf{\hat{x}} + \frac12 \, b \, \mathbf{\hat{y}} \\
    \mathbf{a}_3 & = & c \, \mathbf{\hat{z}} \\

        \end{array}
      \end{equation*}
    }
    \renewcommand{\arraystretch}{1.0}
  \end{tabular}
  \begin{tabular}{c}
    \includegraphics[width=0.3\linewidth]{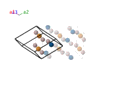} \\
  \end{tabular}
\end{tabular}

}
\vspace*{-0.25cm}

\noindent \hrulefill
\\
\textbf{Basis vectors:}
\vspace*{-0.25cm}
\renewcommand{\arraystretch}{1.5}
\begin{longtabu} to \textwidth{>{\centering $}X[-1,c,c]<{$}>{\centering $}X[-1,c,c]<{$}>{\centering $}X[-1,c,c]<{$}>{\centering $}X[-1,c,c]<{$}>{\centering $}X[-1,c,c]<{$}>{\centering $}X[-1,c,c]<{$}>{\centering $}X[-1,c,c]<{$}}
  & & \mbox{Lattice Coordinates} & & \mbox{Cartesian Coordinates} &\mbox{Wyckoff Position} & \mbox{Atom Type} \\  
  \mathbf{B}_{1} & = & \frac{1}{4} \, \mathbf{a}_{1} + \frac{1}{4} \, \mathbf{a}_{2} & = & \frac{1}{4}a \, \mathbf{\hat{x}} & \left(4a\right) & \mbox{Al I} \\ 
\mathbf{B}_{2} & = & \frac{3}{4} \, \mathbf{a}_{1} + \frac{3}{4} \, \mathbf{a}_{2} & = & \frac{3}{4}a \, \mathbf{\hat{x}} & \left(4a\right) & \mbox{Al I} \\ 
\mathbf{B}_{3} & = & \frac{1}{4} \, \mathbf{a}_{1} + \frac{1}{4} \, \mathbf{a}_{2} + \frac{1}{2} \, \mathbf{a}_{3} & = & \frac{1}{4}a \, \mathbf{\hat{x}} + \frac{1}{2}c \, \mathbf{\hat{z}} & \left(4b\right) & \mbox{Cu} \\ 
\mathbf{B}_{4} & = & \frac{3}{4} \, \mathbf{a}_{1} + \frac{3}{4} \, \mathbf{a}_{2} + \frac{1}{2} \, \mathbf{a}_{3} & = & \frac{3}{4}a \, \mathbf{\hat{x}} + \frac{1}{2}c \, \mathbf{\hat{z}} & \left(4b\right) & \mbox{Cu} \\ 
\mathbf{B}_{5} & = & \frac{3}{4} \, \mathbf{a}_{1} + \frac{1}{4} \, \mathbf{a}_{2} + z_{3} \, \mathbf{a}_{3} & = & \frac{1}{2}a \, \mathbf{\hat{x}}- \frac{1}{4}b  \, \mathbf{\hat{y}} + z_{3}c \, \mathbf{\hat{z}} & \left(4g\right) & \mbox{Al II} \\ 
\mathbf{B}_{6} & = & \frac{1}{4} \, \mathbf{a}_{1} + \frac{3}{4} \, \mathbf{a}_{2}-z_{3} \, \mathbf{a}_{3} & = & \frac{1}{2}a \, \mathbf{\hat{x}} + \frac{1}{4}b \, \mathbf{\hat{y}}-z_{3}c \, \mathbf{\hat{z}} & \left(4g\right) & \mbox{Al II} \\ 
\mathbf{B}_{7} & = & \frac{3}{4} \, \mathbf{a}_{1} + \frac{1}{4} \, \mathbf{a}_{2} + z_{4} \, \mathbf{a}_{3} & = & \frac{1}{2}a \, \mathbf{\hat{x}}- \frac{1}{4}b  \, \mathbf{\hat{y}} + z_{4}c \, \mathbf{\hat{z}} & \left(4g\right) & \mbox{Ir} \\ 
\mathbf{B}_{8} & = & \frac{1}{4} \, \mathbf{a}_{1} + \frac{3}{4} \, \mathbf{a}_{2}-z_{4} \, \mathbf{a}_{3} & = & \frac{1}{2}a \, \mathbf{\hat{x}} + \frac{1}{4}b \, \mathbf{\hat{y}}-z_{4}c \, \mathbf{\hat{z}} & \left(4g\right) & \mbox{Ir} \\ 
\end{longtabu}
\renewcommand{\arraystretch}{1.0}
\noindent \hrulefill
\\
\textbf{References:}
\vspace*{-0.25cm}
\begin{flushleft}
  - \bibentry{Meshi_Al2CuIr_JAlloyComp_2010}. \\
\end{flushleft}
\textbf{Found in:}
\vspace*{-0.25cm}
\begin{flushleft}
  - \bibentry{Villars_PearsonsCrystalData_2013}. \\
\end{flushleft}
\noindent \hrulefill
\\
\textbf{Geometry files:}
\\
\noindent  - CIF: pp. {\hyperref[A2BC_oC16_67_ag_b_g_cif]{\pageref{A2BC_oC16_67_ag_b_g_cif}}} \\
\noindent  - POSCAR: pp. {\hyperref[A2BC_oC16_67_ag_b_g_poscar]{\pageref{A2BC_oC16_67_ag_b_g_poscar}}} \\
\onecolumn
{\phantomsection\label{ABC2_oC16_67_b_g_ag}}
\subsection*{\huge \textbf{{\normalfont HoCuP$_{2}$ Structure: ABC2\_oC16\_67\_b\_g\_ag}}}
\noindent \hrulefill
\vspace*{0.25cm}
\begin{figure}[htp]
  \centering
  \vspace{-1em}
  {\includegraphics[width=1\textwidth]{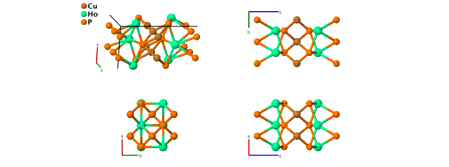}}
\end{figure}
\vspace*{-0.5cm}
\renewcommand{\arraystretch}{1.5}
\begin{equation*}
  \begin{array}{>{$\hspace{-0.15cm}}l<{$}>{$}p{0.5cm}<{$}>{$}p{18.5cm}<{$}}
    \mbox{\large \textbf{Prototype}} &\colon & \ce{HoCuP2} \\
    \mbox{\large \textbf{\AFLOW\ prototype label}} &\colon & \mbox{ABC2\_oC16\_67\_b\_g\_ag} \\
    \mbox{\large \textbf{\textit{Strukturbericht} designation}} &\colon & \mbox{None} \\
    \mbox{\large \textbf{Pearson symbol}} &\colon & \mbox{oC16} \\
    \mbox{\large \textbf{Space group number}} &\colon & 67 \\
    \mbox{\large \textbf{Space group symbol}} &\colon & Cmma \\
    \mbox{\large \textbf{\AFLOW\ prototype command}} &\colon &  \texttt{aflow} \,  \, \texttt{-{}-proto=ABC2\_oC16\_67\_b\_g\_ag } \, \newline \texttt{-{}-params=}{a,b/a,c/a,z_{3},z_{4} }
  \end{array}
\end{equation*}
\renewcommand{\arraystretch}{1.0}

\vspace*{-0.25cm}
\noindent \hrulefill
\begin{itemize}
  \item{Al$_{2}$CuIr (pp. {\hyperref[A2BC_oC16_67_ag_b_g]{\pageref{A2BC_oC16_67_ag_b_g}}}) and
CuHoP$_{2}$ (pp. {\hyperref[ABC2_oC16_67_b_g_ag]{\pageref{ABC2_oC16_67_b_g_ag}}})
have similar \AFLOW\ prototype labels ({\it{i.e.}}, same symmetry and set of
Wyckoff positions with different stoichiometry labels due to alphabetic ordering of atomic species).
They are generated by the same symmetry operations with different sets of parameters
(\texttt{-{}-params}) specified in their corresponding \CIF\ files.
}
\end{itemize}

\noindent \parbox{1 \linewidth}{
\noindent \hrulefill
\\
\textbf{Base-centered Orthorhombic primitive vectors:} \\
\vspace*{-0.25cm}
\begin{tabular}{cc}
  \begin{tabular}{c}
    \parbox{0.6 \linewidth}{
      \renewcommand{\arraystretch}{1.5}
      \begin{equation*}
        \centering
        \begin{array}{ccc}
              \mathbf{a}_1 & = & \frac12 \, a \, \mathbf{\hat{x}} - \frac12 \, b \, \mathbf{\hat{y}} \\
    \mathbf{a}_2 & = & \frac12 \, a \, \mathbf{\hat{x}} + \frac12 \, b \, \mathbf{\hat{y}} \\
    \mathbf{a}_3 & = & c \, \mathbf{\hat{z}} \\

        \end{array}
      \end{equation*}
    }
    \renewcommand{\arraystretch}{1.0}
  \end{tabular}
  \begin{tabular}{c}
    \includegraphics[width=0.3\linewidth]{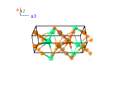} \\
  \end{tabular}
\end{tabular}

}
\vspace*{-0.25cm}

\noindent \hrulefill
\\
\textbf{Basis vectors:}
\vspace*{-0.25cm}
\renewcommand{\arraystretch}{1.5}
\begin{longtabu} to \textwidth{>{\centering $}X[-1,c,c]<{$}>{\centering $}X[-1,c,c]<{$}>{\centering $}X[-1,c,c]<{$}>{\centering $}X[-1,c,c]<{$}>{\centering $}X[-1,c,c]<{$}>{\centering $}X[-1,c,c]<{$}>{\centering $}X[-1,c,c]<{$}}
  & & \mbox{Lattice Coordinates} & & \mbox{Cartesian Coordinates} &\mbox{Wyckoff Position} & \mbox{Atom Type} \\  
  \mathbf{B}_{1} & = & \frac{1}{4} \, \mathbf{a}_{1} + \frac{1}{4} \, \mathbf{a}_{2} & = & \frac{1}{4}a \, \mathbf{\hat{x}} & \left(4a\right) & \mbox{P I} \\ 
\mathbf{B}_{2} & = & \frac{3}{4} \, \mathbf{a}_{1} + \frac{3}{4} \, \mathbf{a}_{2} & = & \frac{3}{4}a \, \mathbf{\hat{x}} & \left(4a\right) & \mbox{P I} \\ 
\mathbf{B}_{3} & = & \frac{1}{4} \, \mathbf{a}_{1} + \frac{1}{4} \, \mathbf{a}_{2} + \frac{1}{2} \, \mathbf{a}_{3} & = & \frac{1}{4}a \, \mathbf{\hat{x}} + \frac{1}{2}c \, \mathbf{\hat{z}} & \left(4b\right) & \mbox{Cu} \\ 
\mathbf{B}_{4} & = & \frac{3}{4} \, \mathbf{a}_{1} + \frac{3}{4} \, \mathbf{a}_{2} + \frac{1}{2} \, \mathbf{a}_{3} & = & \frac{3}{4}a \, \mathbf{\hat{x}} + \frac{1}{2}c \, \mathbf{\hat{z}} & \left(4b\right) & \mbox{Cu} \\ 
\mathbf{B}_{5} & = & \frac{3}{4} \, \mathbf{a}_{1} + \frac{1}{4} \, \mathbf{a}_{2} + z_{3} \, \mathbf{a}_{3} & = & \frac{1}{2}a \, \mathbf{\hat{x}}- \frac{1}{4}b  \, \mathbf{\hat{y}} + z_{3}c \, \mathbf{\hat{z}} & \left(4g\right) & \mbox{Ho} \\ 
\mathbf{B}_{6} & = & \frac{1}{4} \, \mathbf{a}_{1} + \frac{3}{4} \, \mathbf{a}_{2}-z_{3} \, \mathbf{a}_{3} & = & \frac{1}{2}a \, \mathbf{\hat{x}} + \frac{1}{4}b \, \mathbf{\hat{y}}-z_{3}c \, \mathbf{\hat{z}} & \left(4g\right) & \mbox{Ho} \\ 
\mathbf{B}_{7} & = & \frac{3}{4} \, \mathbf{a}_{1} + \frac{1}{4} \, \mathbf{a}_{2} + z_{4} \, \mathbf{a}_{3} & = & \frac{1}{2}a \, \mathbf{\hat{x}}- \frac{1}{4}b  \, \mathbf{\hat{y}} + z_{4}c \, \mathbf{\hat{z}} & \left(4g\right) & \mbox{P II} \\ 
\mathbf{B}_{8} & = & \frac{1}{4} \, \mathbf{a}_{1} + \frac{3}{4} \, \mathbf{a}_{2}-z_{4} \, \mathbf{a}_{3} & = & \frac{1}{2}a \, \mathbf{\hat{x}} + \frac{1}{4}b \, \mathbf{\hat{y}}-z_{4}c \, \mathbf{\hat{z}} & \left(4g\right) & \mbox{P II} \\ 
\end{longtabu}
\renewcommand{\arraystretch}{1.0}
\noindent \hrulefill
\\
\textbf{References:}
\vspace*{-0.25cm}
\begin{flushleft}
  - \bibentry{Mozharivsky_CuHoP2_ZAnorganAllgeChem_2001}. \\
\end{flushleft}
\textbf{Found in:}
\vspace*{-0.25cm}
\begin{flushleft}
  - \bibentry{Villars_PearsonsCrystalData_2013}. \\
\end{flushleft}
\noindent \hrulefill
\\
\textbf{Geometry files:}
\\
\noindent  - CIF: pp. {\hyperref[ABC2_oC16_67_b_g_ag_cif]{\pageref{ABC2_oC16_67_b_g_ag_cif}}} \\
\noindent  - POSCAR: pp. {\hyperref[ABC2_oC16_67_b_g_ag_poscar]{\pageref{ABC2_oC16_67_b_g_ag_poscar}}} \\
\onecolumn
{\phantomsection\label{AB_oC8_67_a_g-FeSe}}
\subsection*{\huge \textbf{{\normalfont $\alpha$-FeSe Structure: AB\_oC8\_67\_a\_g}}}
\noindent \hrulefill
\vspace*{0.25cm}
\begin{figure}[htp]
  \centering
  \vspace{-1em}
  {\includegraphics[width=1\textwidth]{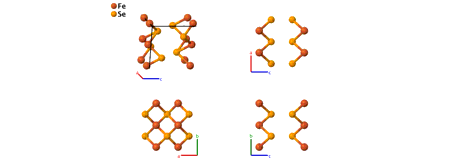}}
\end{figure}
\vspace*{-0.5cm}
\renewcommand{\arraystretch}{1.5}
\begin{equation*}
  \begin{array}{>{$\hspace{-0.15cm}}l<{$}>{$}p{0.5cm}<{$}>{$}p{18.5cm}<{$}}
    \mbox{\large \textbf{Prototype}} &\colon & \ce{$\alpha$-FeSe} \\
    \mbox{\large \textbf{\AFLOW\ prototype label}} &\colon & \mbox{AB\_oC8\_67\_a\_g} \\
    \mbox{\large \textbf{\textit{Strukturbericht} designation}} &\colon & \mbox{None} \\
    \mbox{\large \textbf{Pearson symbol}} &\colon & \mbox{oC8} \\
    \mbox{\large \textbf{Space group number}} &\colon & 67 \\
    \mbox{\large \textbf{Space group symbol}} &\colon & Cmma \\
    \mbox{\large \textbf{\AFLOW\ prototype command}} &\colon &  \texttt{aflow} \,  \, \texttt{-{}-proto=AB\_oC8\_67\_a\_g } \, \newline \texttt{-{}-params=}{a,b/a,c/a,z_{2} }
  \end{array}
\end{equation*}
\renewcommand{\arraystretch}{1.0}

\vspace*{-0.25cm}
\noindent \hrulefill
\begin{itemize}
  \item{We follow the reference in calling this $\alpha$-FeSe.  Some other
authorities refer to the \href{http://aflow.org/CrystalDatabase/AB_tP4_129_a_c.html}{$B10$ (PbO)}-like 
phase of FeSe as $\alpha$-FeSe, calling this phase $\beta$-FeSe.
The authors note that ``the Se ion concentration is close to 1.''
The data is presented for the structure at 7~K.
$\alpha$-FeSe (pp. {\hyperref[AB_oC8_67_a_g-FeSe]{\pageref{AB_oC8_67_a_g-FeSe}}}) 
and $\alpha$-PbO (pp. {\hyperref[AB_oC8_67_a_g-PbO]{\pageref{AB_oC8_67_a_g-PbO}}}))
have the same \AFLOW\ prototype label.
They are generated by the same symmetry operations with different sets of parameters
(\texttt{-{}-params}) specified in their corresponding \CIF\ files.
}
\end{itemize}

\noindent \parbox{1 \linewidth}{
\noindent \hrulefill
\\
\textbf{Base-centered Orthorhombic primitive vectors:} \\
\vspace*{-0.25cm}
\begin{tabular}{cc}
  \begin{tabular}{c}
    \parbox{0.6 \linewidth}{
      \renewcommand{\arraystretch}{1.5}
      \begin{equation*}
        \centering
        \begin{array}{ccc}
              \mathbf{a}_1 & = & \frac12 \, a \, \mathbf{\hat{x}} - \frac12 \, b \, \mathbf{\hat{y}} \\
    \mathbf{a}_2 & = & \frac12 \, a \, \mathbf{\hat{x}} + \frac12 \, b \, \mathbf{\hat{y}} \\
    \mathbf{a}_3 & = & c \, \mathbf{\hat{z}} \\

        \end{array}
      \end{equation*}
    }
    \renewcommand{\arraystretch}{1.0}
  \end{tabular}
  \begin{tabular}{c}
    \includegraphics[width=0.3\linewidth]{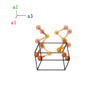} \\
  \end{tabular}
\end{tabular}

}
\vspace*{-0.25cm}

\noindent \hrulefill
\\
\textbf{Basis vectors:}
\vspace*{-0.25cm}
\renewcommand{\arraystretch}{1.5}
\begin{longtabu} to \textwidth{>{\centering $}X[-1,c,c]<{$}>{\centering $}X[-1,c,c]<{$}>{\centering $}X[-1,c,c]<{$}>{\centering $}X[-1,c,c]<{$}>{\centering $}X[-1,c,c]<{$}>{\centering $}X[-1,c,c]<{$}>{\centering $}X[-1,c,c]<{$}}
  & & \mbox{Lattice Coordinates} & & \mbox{Cartesian Coordinates} &\mbox{Wyckoff Position} & \mbox{Atom Type} \\  
  \mathbf{B}_{1} & = & \frac{1}{4} \, \mathbf{a}_{1} + \frac{1}{4} \, \mathbf{a}_{2} & = & \frac{1}{4}a \, \mathbf{\hat{x}} & \left(4a\right) & \mbox{Fe} \\ 
\mathbf{B}_{2} & = & \frac{3}{4} \, \mathbf{a}_{1} + \frac{3}{4} \, \mathbf{a}_{2} & = & \frac{3}{4}a \, \mathbf{\hat{x}} & \left(4a\right) & \mbox{Fe} \\ 
\mathbf{B}_{3} & = & \frac{3}{4} \, \mathbf{a}_{1} + \frac{1}{4} \, \mathbf{a}_{2} + z_{2} \, \mathbf{a}_{3} & = & \frac{1}{2}a \, \mathbf{\hat{x}}- \frac{1}{4}b  \, \mathbf{\hat{y}} + z_{2}c \, \mathbf{\hat{z}} & \left(4g\right) & \mbox{Se} \\ 
\mathbf{B}_{4} & = & \frac{1}{4} \, \mathbf{a}_{1} + \frac{3}{4} \, \mathbf{a}_{2}-z_{2} \, \mathbf{a}_{3} & = & \frac{1}{2}a \, \mathbf{\hat{x}} + \frac{1}{4}b \, \mathbf{\hat{y}}-z_{2}c \, \mathbf{\hat{z}} & \left(4g\right) & \mbox{Se} \\ 
\end{longtabu}
\renewcommand{\arraystretch}{1.0}
\noindent \hrulefill
\\
\textbf{References:}
\vspace*{-0.25cm}
\begin{flushleft}
  - \bibentry{Louca_Phys_Rev_B_81_2010}. \\
\end{flushleft}
\noindent \hrulefill
\\
\textbf{Geometry files:}
\\
\noindent  - CIF: pp. {\hyperref[AB_oC8_67_a_g-FeSe_cif]{\pageref{AB_oC8_67_a_g-FeSe_cif}}} \\
\noindent  - POSCAR: pp. {\hyperref[AB_oC8_67_a_g-FeSe_poscar]{\pageref{AB_oC8_67_a_g-FeSe_poscar}}} \\
\onecolumn
{\phantomsection\label{AB_oC8_67_a_g-PbO}}
\subsection*{\huge \textbf{{\normalfont $\alpha$-PbO Structure: AB\_oC8\_67\_a\_g}}}
\noindent \hrulefill
\vspace*{0.25cm}
\begin{figure}[htp]
  \centering
  \vspace{-1em}
  {\includegraphics[width=1\textwidth]{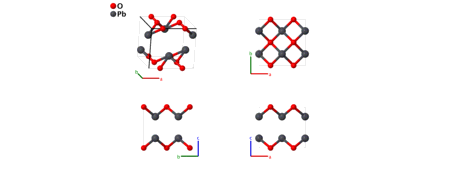}}
\end{figure}
\vspace*{-0.5cm}
\renewcommand{\arraystretch}{1.5}
\begin{equation*}
  \begin{array}{>{$\hspace{-0.15cm}}l<{$}>{$}p{0.5cm}<{$}>{$}p{18.5cm}<{$}}
    \mbox{\large \textbf{Prototype}} &\colon & \ce{$\alpha$-PbO} \\
    \mbox{\large \textbf{\AFLOW\ prototype label}} &\colon & \mbox{AB\_oC8\_67\_a\_g} \\
    \mbox{\large \textbf{\textit{Strukturbericht} designation}} &\colon & \mbox{None} \\
    \mbox{\large \textbf{Pearson symbol}} &\colon & \mbox{oC8} \\
    \mbox{\large \textbf{Space group number}} &\colon & 67 \\
    \mbox{\large \textbf{Space group symbol}} &\colon & Cmma \\
    \mbox{\large \textbf{\AFLOW\ prototype command}} &\colon &  \texttt{aflow} \,  \, \texttt{-{}-proto=AB\_oC8\_67\_a\_g } \, \newline \texttt{-{}-params=}{a,b/a,c/a,z_{2} }
  \end{array}
\end{equation*}
\renewcommand{\arraystretch}{1.0}

\vspace*{-0.25cm}
\noindent \hrulefill
\begin{itemize}
  \item{{\small FINDSYM} identifies space group \#67 for this structure (consistent with the reference); however, since $b/a \approx 1$,
{\small AFLOW-SYM} and Platon identify \#129.
Lowering the tolerance value for {\small AFLOW-SYM} resolves the expected space group \#67.
Space groups \#67 and \#129 are both reasonable classifications since they are commensurate with subgroup relations.
$\alpha$-FeSe (pp. {\hyperref[AB_oC8_67_a_g-FeSe]{\pageref{AB_oC8_67_a_g-FeSe}}})
and $\alpha$-PbO (pp. {\hyperref[AB_oC8_67_a_g-PbO]{\pageref{AB_oC8_67_a_g-PbO}}}))
have the same \AFLOW\ prototype label.
They are generated by the same symmetry operations with different sets of parameters
(\texttt{-{}-params}) specified in their corresponding \CIF\ files.
}
\end{itemize}

\noindent \parbox{1 \linewidth}{
\noindent \hrulefill
\\
\textbf{Base-centered Orthorhombic primitive vectors:} \\
\vspace*{-0.25cm}
\begin{tabular}{cc}
  \begin{tabular}{c}
    \parbox{0.6 \linewidth}{
      \renewcommand{\arraystretch}{1.5}
      \begin{equation*}
        \centering
        \begin{array}{ccc}
              \mathbf{a}_1 & = & \frac12 \, a \, \mathbf{\hat{x}} - \frac12 \, b \, \mathbf{\hat{y}} \\
    \mathbf{a}_2 & = & \frac12 \, a \, \mathbf{\hat{x}} + \frac12 \, b \, \mathbf{\hat{y}} \\
    \mathbf{a}_3 & = & c \, \mathbf{\hat{z}} \\

        \end{array}
      \end{equation*}
    }
    \renewcommand{\arraystretch}{1.0}
  \end{tabular}
  \begin{tabular}{c}
    \includegraphics[width=0.3\linewidth]{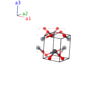} \\
  \end{tabular}
\end{tabular}

}
\vspace*{-0.25cm}

\noindent \hrulefill
\\
\textbf{Basis vectors:}
\vspace*{-0.25cm}
\renewcommand{\arraystretch}{1.5}
\begin{longtabu} to \textwidth{>{\centering $}X[-1,c,c]<{$}>{\centering $}X[-1,c,c]<{$}>{\centering $}X[-1,c,c]<{$}>{\centering $}X[-1,c,c]<{$}>{\centering $}X[-1,c,c]<{$}>{\centering $}X[-1,c,c]<{$}>{\centering $}X[-1,c,c]<{$}}
  & & \mbox{Lattice Coordinates} & & \mbox{Cartesian Coordinates} &\mbox{Wyckoff Position} & \mbox{Atom Type} \\  
  \mathbf{B}_{1} & = & \frac{1}{4} \, \mathbf{a}_{1} + \frac{1}{4} \, \mathbf{a}_{2} & = & \frac{1}{4}a \, \mathbf{\hat{x}} & \left(4a\right) & \mbox{O} \\ 
\mathbf{B}_{2} & = & \frac{3}{4} \, \mathbf{a}_{1} + \frac{3}{4} \, \mathbf{a}_{2} & = & \frac{3}{4}a \, \mathbf{\hat{x}} & \left(4a\right) & \mbox{O} \\ 
\mathbf{B}_{3} & = & \frac{3}{4} \, \mathbf{a}_{1} + \frac{1}{4} \, \mathbf{a}_{2} + z_{2} \, \mathbf{a}_{3} & = & \frac{1}{2}a \, \mathbf{\hat{x}}- \frac{1}{4}b  \, \mathbf{\hat{y}} + z_{2}c \, \mathbf{\hat{z}} & \left(4g\right) & \mbox{Pb} \\ 
\mathbf{B}_{4} & = & \frac{1}{4} \, \mathbf{a}_{1} + \frac{3}{4} \, \mathbf{a}_{2}-z_{2} \, \mathbf{a}_{3} & = & \frac{1}{2}a \, \mathbf{\hat{x}} + \frac{1}{4}b \, \mathbf{\hat{y}}-z_{2}c \, \mathbf{\hat{z}} & \left(4g\right) & \mbox{Pb} \\ 
\end{longtabu}
\renewcommand{\arraystretch}{1.0}
\noindent \hrulefill
\\
\textbf{References:}
\vspace*{-0.25cm}
\begin{flushleft}
  - \bibentry{Boher_PbO_JSolStateChem_1985}. \\
  - \bibentry{stokes_findsym}. \\
  - \bibentry{aflowsym_2018}. \\
  - \bibentry{platon_2003}. \\
\end{flushleft}
\textbf{Found in:}
\vspace*{-0.25cm}
\begin{flushleft}
  - \bibentry{Villars_PearsonsCrystalData_2013}. \\
\end{flushleft}
\noindent \hrulefill
\\
\textbf{Geometry files:}
\\
\noindent  - CIF: pp. {\hyperref[AB_oC8_67_a_g-PbO_cif]{\pageref{AB_oC8_67_a_g-PbO_cif}}} \\
\noindent  - POSCAR: pp. {\hyperref[AB_oC8_67_a_g-PbO_poscar]{\pageref{AB_oC8_67_a_g-PbO_poscar}}} \\
\onecolumn
{\phantomsection\label{AB4_oC20_68_a_i}}
\subsection*{\huge \textbf{{\normalfont PdSn$_{4}$ Structure: AB4\_oC20\_68\_a\_i}}}
\noindent \hrulefill
\vspace*{0.25cm}
\begin{figure}[htp]
  \centering
  \vspace{-1em}
  {\includegraphics[width=1\textwidth]{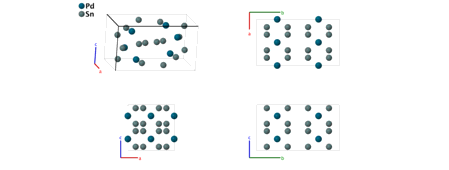}}
\end{figure}
\vspace*{-0.5cm}
\renewcommand{\arraystretch}{1.5}
\begin{equation*}
  \begin{array}{>{$\hspace{-0.15cm}}l<{$}>{$}p{0.5cm}<{$}>{$}p{18.5cm}<{$}}
    \mbox{\large \textbf{Prototype}} &\colon & \ce{PdSn4} \\
    \mbox{\large \textbf{\AFLOW\ prototype label}} &\colon & \mbox{AB4\_oC20\_68\_a\_i} \\
    \mbox{\large \textbf{\textit{Strukturbericht} designation}} &\colon & \mbox{None} \\
    \mbox{\large \textbf{Pearson symbol}} &\colon & \mbox{oC20} \\
    \mbox{\large \textbf{Space group number}} &\colon & 68 \\
    \mbox{\large \textbf{Space group symbol}} &\colon & Ccca \\
    \mbox{\large \textbf{\AFLOW\ prototype command}} &\colon &  \texttt{aflow} \,  \, \texttt{-{}-proto=AB4\_oC20\_68\_a\_i } \, \newline \texttt{-{}-params=}{a,b/a,c/a,x_{2},y_{2},z_{2} }
  \end{array}
\end{equation*}
\renewcommand{\arraystretch}{1.0}

\noindent \parbox{1 \linewidth}{
\noindent \hrulefill
\\
\textbf{Base-centered Orthorhombic primitive vectors:} \\
\vspace*{-0.25cm}
\begin{tabular}{cc}
  \begin{tabular}{c}
    \parbox{0.6 \linewidth}{
      \renewcommand{\arraystretch}{1.5}
      \begin{equation*}
        \centering
        \begin{array}{ccc}
              \mathbf{a}_1 & = & \frac12 \, a \, \mathbf{\hat{x}} - \frac12 \, b \, \mathbf{\hat{y}} \\
    \mathbf{a}_2 & = & \frac12 \, a \, \mathbf{\hat{x}} + \frac12 \, b \, \mathbf{\hat{y}} \\
    \mathbf{a}_3 & = & c \, \mathbf{\hat{z}} \\

        \end{array}
      \end{equation*}
    }
    \renewcommand{\arraystretch}{1.0}
  \end{tabular}
  \begin{tabular}{c}
    \includegraphics[width=0.3\linewidth]{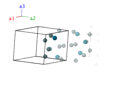} \\
  \end{tabular}
\end{tabular}

}
\vspace*{-0.25cm}

\noindent \hrulefill
\\
\textbf{Basis vectors:}
\vspace*{-0.25cm}
\renewcommand{\arraystretch}{1.5}
\begin{longtabu} to \textwidth{>{\centering $}X[-1,c,c]<{$}>{\centering $}X[-1,c,c]<{$}>{\centering $}X[-1,c,c]<{$}>{\centering $}X[-1,c,c]<{$}>{\centering $}X[-1,c,c]<{$}>{\centering $}X[-1,c,c]<{$}>{\centering $}X[-1,c,c]<{$}}
  & & \mbox{Lattice Coordinates} & & \mbox{Cartesian Coordinates} &\mbox{Wyckoff Position} & \mbox{Atom Type} \\  
  \mathbf{B}_{1} & = & \frac{3}{4} \, \mathbf{a}_{1} + \frac{1}{4} \, \mathbf{a}_{2} + \frac{1}{4} \, \mathbf{a}_{3} & = & \frac{1}{2}a \, \mathbf{\hat{x}}- \frac{1}{4}b  \, \mathbf{\hat{y}} + \frac{1}{4}c \, \mathbf{\hat{z}} & \left(4a\right) & \mbox{Pd} \\ 
\mathbf{B}_{2} & = & \frac{1}{4} \, \mathbf{a}_{1} + \frac{3}{4} \, \mathbf{a}_{2} + \frac{3}{4} \, \mathbf{a}_{3} & = & \frac{1}{2}a \, \mathbf{\hat{x}} + \frac{1}{4}b \, \mathbf{\hat{y}} + \frac{3}{4}c \, \mathbf{\hat{z}} & \left(4a\right) & \mbox{Pd} \\ 
\mathbf{B}_{3} & = & \left(x_{2}-y_{2}\right) \, \mathbf{a}_{1} + \left(x_{2}+y_{2}\right) \, \mathbf{a}_{2} + z_{2} \, \mathbf{a}_{3} & = & x_{2}a \, \mathbf{\hat{x}} + y_{2}b \, \mathbf{\hat{y}} + z_{2}c \, \mathbf{\hat{z}} & \left(16i\right) & \mbox{Sn} \\ 
\mathbf{B}_{4} & = & \left(\frac{1}{2} - x_{2} + y_{2}\right) \, \mathbf{a}_{1} + \left(\frac{1}{2} - x_{2} - y_{2}\right) \, \mathbf{a}_{2} + z_{2} \, \mathbf{a}_{3} & = & \left(\frac{1}{2} - x_{2}\right)a \, \mathbf{\hat{x}}-y_{2}b \, \mathbf{\hat{y}} + z_{2}c \, \mathbf{\hat{z}} & \left(16i\right) & \mbox{Sn} \\ 
\mathbf{B}_{5} & = & \left(-x_{2}-y_{2}\right) \, \mathbf{a}_{1} + \left(-x_{2}+y_{2}\right) \, \mathbf{a}_{2} + \left(\frac{1}{2} - z_{2}\right) \, \mathbf{a}_{3} & = & -x_{2}a \, \mathbf{\hat{x}} + y_{2}b \, \mathbf{\hat{y}} + \left(\frac{1}{2} - z_{2}\right)c \, \mathbf{\hat{z}} & \left(16i\right) & \mbox{Sn} \\ 
\mathbf{B}_{6} & = & \left(\frac{1}{2} +x_{2} + y_{2}\right) \, \mathbf{a}_{1} + \left(\frac{1}{2} +x_{2} - y_{2}\right) \, \mathbf{a}_{2} + \left(\frac{1}{2} - z_{2}\right) \, \mathbf{a}_{3} & = & \left(\frac{1}{2} +x_{2}\right)a \, \mathbf{\hat{x}}-y_{2}b \, \mathbf{\hat{y}} + \left(\frac{1}{2} - z_{2}\right)c \, \mathbf{\hat{z}} & \left(16i\right) & \mbox{Sn} \\ 
\mathbf{B}_{7} & = & \left(-x_{2}+y_{2}\right) \, \mathbf{a}_{1} + \left(-x_{2}-y_{2}\right) \, \mathbf{a}_{2}-z_{2} \, \mathbf{a}_{3} & = & -x_{2}a \, \mathbf{\hat{x}}-y_{2}b \, \mathbf{\hat{y}}-z_{2}c \, \mathbf{\hat{z}} & \left(16i\right) & \mbox{Sn} \\ 
\mathbf{B}_{8} & = & \left(\frac{1}{2} +x_{2} - y_{2}\right) \, \mathbf{a}_{1} + \left(\frac{1}{2} +x_{2} + y_{2}\right) \, \mathbf{a}_{2}-z_{2} \, \mathbf{a}_{3} & = & \left(\frac{1}{2} +x_{2}\right)a \, \mathbf{\hat{x}} + y_{2}b \, \mathbf{\hat{y}}-z_{2}c \, \mathbf{\hat{z}} & \left(16i\right) & \mbox{Sn} \\ 
\mathbf{B}_{9} & = & \left(x_{2}+y_{2}\right) \, \mathbf{a}_{1} + \left(x_{2}-y_{2}\right) \, \mathbf{a}_{2} + \left(\frac{1}{2} +z_{2}\right) \, \mathbf{a}_{3} & = & x_{2}a \, \mathbf{\hat{x}}-y_{2}b \, \mathbf{\hat{y}} + \left(\frac{1}{2} +z_{2}\right)c \, \mathbf{\hat{z}} & \left(16i\right) & \mbox{Sn} \\ 
\mathbf{B}_{10} & = & \left(\frac{1}{2} - x_{2} - y_{2}\right) \, \mathbf{a}_{1} + \left(\frac{1}{2} - x_{2} + y_{2}\right) \, \mathbf{a}_{2} + \left(\frac{1}{2} +z_{2}\right) \, \mathbf{a}_{3} & = & \left(\frac{1}{2} - x_{2}\right)a \, \mathbf{\hat{x}} + y_{2}b \, \mathbf{\hat{y}} + \left(\frac{1}{2} +z_{2}\right)c \, \mathbf{\hat{z}} & \left(16i\right) & \mbox{Sn} \\ 
\end{longtabu}
\renewcommand{\arraystretch}{1.0}
\noindent \hrulefill
\\
\textbf{References:}
\vspace*{-0.25cm}
\begin{flushleft}
  - \bibentry{Yylen_PtSn4_SolStateSci_2004}. \\
\end{flushleft}
\textbf{Found in:}
\vspace*{-0.25cm}
\begin{flushleft}
  - \bibentry{Villars_PearsonsCrystalData_2013}. \\
\end{flushleft}
\noindent \hrulefill
\\
\textbf{Geometry files:}
\\
\noindent  - CIF: pp. {\hyperref[AB4_oC20_68_a_i_cif]{\pageref{AB4_oC20_68_a_i_cif}}} \\
\noindent  - POSCAR: pp. {\hyperref[AB4_oC20_68_a_i_poscar]{\pageref{AB4_oC20_68_a_i_poscar}}} \\
\onecolumn
{\phantomsection\label{AB2_oF48_70_f_fg}}
\subsection*{\huge \textbf{{\normalfont Mn$_{2}$B ($D1_{f}$) Structure: AB2\_oF48\_70\_f\_fg}}}
\noindent \hrulefill
\vspace*{0.25cm}
\begin{figure}[htp]
  \centering
  \vspace{-1em}
  {\includegraphics[width=1\textwidth]{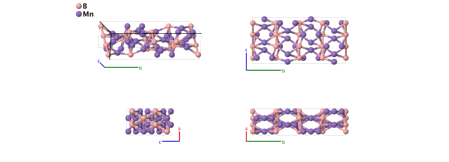}}
\end{figure}
\vspace*{-0.5cm}
\renewcommand{\arraystretch}{1.5}
\begin{equation*}
  \begin{array}{>{$\hspace{-0.15cm}}l<{$}>{$}p{0.5cm}<{$}>{$}p{18.5cm}<{$}}
    \mbox{\large \textbf{Prototype}} &\colon & \ce{Mn2B} \\
    \mbox{\large \textbf{\AFLOW\ prototype label}} &\colon & \mbox{AB2\_oF48\_70\_f\_fg} \\
    \mbox{\large \textbf{\textit{Strukturbericht} designation}} &\colon & \mbox{$D1_{f}$} \\
    \mbox{\large \textbf{Pearson symbol}} &\colon & \mbox{oF48} \\
    \mbox{\large \textbf{Space group number}} &\colon & 70 \\
    \mbox{\large \textbf{Space group symbol}} &\colon & Fddd \\
    \mbox{\large \textbf{\AFLOW\ prototype command}} &\colon &  \texttt{aflow} \,  \, \texttt{-{}-proto=AB2\_oF48\_70\_f\_fg } \, \newline \texttt{-{}-params=}{a,b/a,c/a,y_{1},y_{2},z_{3} }
  \end{array}
\end{equation*}
\renewcommand{\arraystretch}{1.0}

\vspace*{-0.25cm}
\noindent \hrulefill
\\
\textbf{ Other compounds with this structure:}
\begin{itemize}
   \item{ Cr$_{2}$B  }
\end{itemize}
\vspace*{-0.25cm}
\noindent \hrulefill
\begin{itemize}
  \item{Early works, {\em e.g.} (Pearson, 1958) referred to this structure as
Mn$_{4}$B, with the same space group and Wyckoff positions.  The
stoichiometry was fixed by assuming that the (16e) Boron positions
were only half-occupied. Tergenius's 1981 refinement of the structure
showed that the (16e) sites were totally filled, fixing the
stoichiometry to Mn$_{2}$B.  A similar reanalysis showed that the
similar structure known has Cr$_{4}$B also had composition Cr$_{2}$B.
Tergenius gives the atomic positions using the first setting of space
group $Fddd$ \#70. We have translated this into the second setting,
where the origin is on an inversion site.  As a part of this process
the primitive axes were also rotated compared to Tergenius.
}
\end{itemize}

\noindent \parbox{1 \linewidth}{
\noindent \hrulefill
\\
\textbf{Face-centered Orthorhombic primitive vectors:} \\
\vspace*{-0.25cm}
\begin{tabular}{cc}
  \begin{tabular}{c}
    \parbox{0.6 \linewidth}{
      \renewcommand{\arraystretch}{1.5}
      \begin{equation*}
        \centering
        \begin{array}{ccc}
              \mathbf{a}_1 & = & \frac12 \, b \, \mathbf{\hat{y}} + \frac12 \, c \, \mathbf{\hat{z}} \\
    \mathbf{a}_2 & = & \frac12 \, a \, \mathbf{\hat{x}} + \frac12 \, c \, \mathbf{\hat{z}} \\
    \mathbf{a}_3 & = & \frac12 \, a \, \mathbf{\hat{x}} + \frac12 \, b \, \mathbf{\hat{y}} \\

        \end{array}
      \end{equation*}
    }
    \renewcommand{\arraystretch}{1.0}
  \end{tabular}
  \begin{tabular}{c}
    \includegraphics[width=0.3\linewidth]{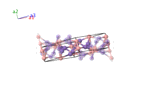} \\
  \end{tabular}
\end{tabular}

}
\vspace*{-0.25cm}

\noindent \hrulefill
\\
\textbf{Basis vectors:}
\vspace*{-0.25cm}
\renewcommand{\arraystretch}{1.5}
\begin{longtabu} to \textwidth{>{\centering $}X[-1,c,c]<{$}>{\centering $}X[-1,c,c]<{$}>{\centering $}X[-1,c,c]<{$}>{\centering $}X[-1,c,c]<{$}>{\centering $}X[-1,c,c]<{$}>{\centering $}X[-1,c,c]<{$}>{\centering $}X[-1,c,c]<{$}}
  & & \mbox{Lattice Coordinates} & & \mbox{Cartesian Coordinates} &\mbox{Wyckoff Position} & \mbox{Atom Type} \\  
  \mathbf{B}_{1} & = & y_{1} \, \mathbf{a}_{1} + \left(\frac{1}{4} - y_{1}\right) \, \mathbf{a}_{2} + y_{1} \, \mathbf{a}_{3} & = & \frac{1}{8}a \, \mathbf{\hat{x}} + y_{1}b \, \mathbf{\hat{y}} + \frac{1}{8}c \, \mathbf{\hat{z}} & \left(16f\right) & \mbox{B} \\ 
\mathbf{B}_{2} & = & \left(\frac{1}{4} - y_{1}\right) \, \mathbf{a}_{1} + y_{1} \, \mathbf{a}_{2} + \left(\frac{1}{4} - y_{1}\right) \, \mathbf{a}_{3} & = & \frac{1}{8}a \, \mathbf{\hat{x}} + \left(\frac{1}{4} - y_{1}\right)b \, \mathbf{\hat{y}} + \frac{1}{8}c \, \mathbf{\hat{z}} & \left(16f\right) & \mbox{B} \\ 
\mathbf{B}_{3} & = & -y_{1} \, \mathbf{a}_{1} + \left(\frac{3}{4} +y_{1}\right) \, \mathbf{a}_{2}-y_{1} \, \mathbf{a}_{3} & = & \frac{3}{8}a \, \mathbf{\hat{x}}-y_{1}b \, \mathbf{\hat{y}} + \frac{3}{8}c \, \mathbf{\hat{z}} & \left(16f\right) & \mbox{B} \\ 
\mathbf{B}_{4} & = & \left(\frac{3}{4} +y_{1}\right) \, \mathbf{a}_{1}-y_{1} \, \mathbf{a}_{2} + \left(\frac{3}{4} +y_{1}\right) \, \mathbf{a}_{3} & = & \frac{3}{8}a \, \mathbf{\hat{x}} + \left(\frac{3}{4} +y_{1}\right)b \, \mathbf{\hat{y}} + \frac{3}{8}c \, \mathbf{\hat{z}} & \left(16f\right) & \mbox{B} \\ 
\mathbf{B}_{5} & = & y_{2} \, \mathbf{a}_{1} + \left(\frac{1}{4} - y_{2}\right) \, \mathbf{a}_{2} + y_{2} \, \mathbf{a}_{3} & = & \frac{1}{8}a \, \mathbf{\hat{x}} + y_{2}b \, \mathbf{\hat{y}} + \frac{1}{8}c \, \mathbf{\hat{z}} & \left(16f\right) & \mbox{Mn I} \\ 
\mathbf{B}_{6} & = & \left(\frac{1}{4} - y_{2}\right) \, \mathbf{a}_{1} + y_{2} \, \mathbf{a}_{2} + \left(\frac{1}{4} - y_{2}\right) \, \mathbf{a}_{3} & = & \frac{1}{8}a \, \mathbf{\hat{x}} + \left(\frac{1}{4} - y_{2}\right)b \, \mathbf{\hat{y}} + \frac{1}{8}c \, \mathbf{\hat{z}} & \left(16f\right) & \mbox{Mn I} \\ 
\mathbf{B}_{7} & = & -y_{2} \, \mathbf{a}_{1} + \left(\frac{3}{4} +y_{2}\right) \, \mathbf{a}_{2}-y_{2} \, \mathbf{a}_{3} & = & \frac{3}{8}a \, \mathbf{\hat{x}}-y_{2}b \, \mathbf{\hat{y}} + \frac{3}{8}c \, \mathbf{\hat{z}} & \left(16f\right) & \mbox{Mn I} \\ 
\mathbf{B}_{8} & = & \left(\frac{3}{4} +y_{2}\right) \, \mathbf{a}_{1}-y_{2} \, \mathbf{a}_{2} + \left(\frac{3}{4} +y_{2}\right) \, \mathbf{a}_{3} & = & \frac{3}{8}a \, \mathbf{\hat{x}} + \left(\frac{3}{4} +y_{2}\right)b \, \mathbf{\hat{y}} + \frac{3}{8}c \, \mathbf{\hat{z}} & \left(16f\right) & \mbox{Mn I} \\ 
\mathbf{B}_{9} & = & z_{3} \, \mathbf{a}_{1} + z_{3} \, \mathbf{a}_{2} + \left(\frac{1}{4} - z_{3}\right) \, \mathbf{a}_{3} & = & \frac{1}{8}a \, \mathbf{\hat{x}} + \frac{1}{8}b \, \mathbf{\hat{y}} + z_{3}c \, \mathbf{\hat{z}} & \left(16g\right) & \mbox{Mn II} \\ 
\mathbf{B}_{10} & = & \left(\frac{1}{4} - z_{3}\right) \, \mathbf{a}_{1} + \left(\frac{1}{4} - z_{3}\right) \, \mathbf{a}_{2} + z_{3} \, \mathbf{a}_{3} & = & \frac{1}{8}a \, \mathbf{\hat{x}} + \frac{1}{8}b \, \mathbf{\hat{y}} + \left(\frac{1}{4} - z_{3}\right)c \, \mathbf{\hat{z}} & \left(16g\right) & \mbox{Mn II} \\ 
\mathbf{B}_{11} & = & -z_{3} \, \mathbf{a}_{1}-z_{3} \, \mathbf{a}_{2} + \left(\frac{3}{4} +z_{3}\right) \, \mathbf{a}_{3} & = & \frac{3}{8}a \, \mathbf{\hat{x}} + \frac{3}{8}b \, \mathbf{\hat{y}}-z_{3}c \, \mathbf{\hat{z}} & \left(16g\right) & \mbox{Mn II} \\ 
\mathbf{B}_{12} & = & \left(\frac{3}{4} +z_{3}\right) \, \mathbf{a}_{1} + \left(\frac{3}{4} +z_{3}\right) \, \mathbf{a}_{2}-z_{3} \, \mathbf{a}_{3} & = & \frac{3}{8}a \, \mathbf{\hat{x}} + \frac{3}{8}b \, \mathbf{\hat{y}} + \left(\frac{3}{4} +z_{3}\right)c \, \mathbf{\hat{z}} & \left(16g\right) & \mbox{Mn II} \\ 
\end{longtabu}
\renewcommand{\arraystretch}{1.0}
\noindent \hrulefill
\\
\textbf{References:}
\vspace*{-0.25cm}
\begin{flushleft}
  - \bibentry{Tergenius_JLCM_82_1981}. \\
  - \bibentry{Pearson_NRC_1958}. \\
\end{flushleft}
\noindent \hrulefill
\\
\textbf{Geometry files:}
\\
\noindent  - CIF: pp. {\hyperref[AB2_oF48_70_f_fg_cif]{\pageref{AB2_oF48_70_f_fg_cif}}} \\
\noindent  - POSCAR: pp. {\hyperref[AB2_oF48_70_f_fg_poscar]{\pageref{AB2_oF48_70_f_fg_poscar}}} \\
\onecolumn
{\phantomsection\label{A4B3_oI14_71_gh_cg}}
\subsection*{\huge \textbf{{\normalfont Ta$_{3}$B$_{4}$ ($D7_{b}$) Structure: A4B3\_oI14\_71\_gh\_cg}}}
\noindent \hrulefill
\vspace*{0.25cm}
\begin{figure}[htp]
  \centering
  \vspace{-1em}
  {\includegraphics[width=1\textwidth]{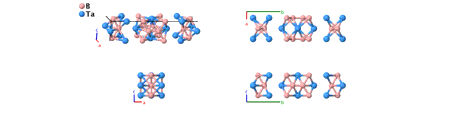}}
\end{figure}
\vspace*{-0.5cm}
\renewcommand{\arraystretch}{1.5}
\begin{equation*}
  \begin{array}{>{$\hspace{-0.15cm}}l<{$}>{$}p{0.5cm}<{$}>{$}p{18.5cm}<{$}}
    \mbox{\large \textbf{Prototype}} &\colon & \ce{Ta3B4} \\
    \mbox{\large \textbf{\AFLOW\ prototype label}} &\colon & \mbox{A4B3\_oI14\_71\_gh\_cg} \\
    \mbox{\large \textbf{\textit{Strukturbericht} designation}} &\colon & \mbox{$D7_{b}$} \\
    \mbox{\large \textbf{Pearson symbol}} &\colon & \mbox{oI14} \\
    \mbox{\large \textbf{Space group number}} &\colon & 71 \\
    \mbox{\large \textbf{Space group symbol}} &\colon & Immm \\
    \mbox{\large \textbf{\AFLOW\ prototype command}} &\colon &  \texttt{aflow} \,  \, \texttt{-{}-proto=A4B3\_oI14\_71\_gh\_cg } \, \newline \texttt{-{}-params=}{a,b/a,c/a,y_{2},y_{3},y_{4} }
  \end{array}
\end{equation*}
\renewcommand{\arraystretch}{1.0}

\vspace*{-0.25cm}
\noindent \hrulefill
\\
\textbf{ Other compounds with this structure:}
\begin{itemize}
   \item{ B$_{4}$CoMo$_{2}$, B$_{4}$Cr$_{3}$, B$_{4}$FeMo$_{2}$, B$_{4}$Mn$_{3}$, B$_{4}$Mo$_{2}$Ni, B$_{4}$Nb$_{3}$, B$_{4}$Ta$_{3}$, B$_{4}$V$_{3}$   }
\end{itemize}
\noindent \parbox{1 \linewidth}{
\noindent \hrulefill
\\
\textbf{Body-centered Orthorhombic primitive vectors:} \\
\vspace*{-0.25cm}
\begin{tabular}{cc}
  \begin{tabular}{c}
    \parbox{0.6 \linewidth}{
      \renewcommand{\arraystretch}{1.5}
      \begin{equation*}
        \centering
        \begin{array}{ccc}
              \mathbf{a}_1 & = & - \frac12 \, a \, \mathbf{\hat{x}} + \frac12 \, b \, \mathbf{\hat{y}} + \frac12 \, c \, \mathbf{\hat{z}} \\
    \mathbf{a}_2 & = & ~ \frac12 \, a \, \mathbf{\hat{x}} - \frac12 \, b \, \mathbf{\hat{y}} + \frac12 \, c \, \mathbf{\hat{z}} \\
    \mathbf{a}_3 & = & ~ \frac12 \, a \, \mathbf{\hat{x}} + \frac12 \, b \, \mathbf{\hat{y}} - \frac12 \, c \, \mathbf{\hat{z}} \\

        \end{array}
      \end{equation*}
    }
    \renewcommand{\arraystretch}{1.0}
  \end{tabular}
  \begin{tabular}{c}
    \includegraphics[width=0.3\linewidth]{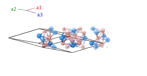} \\
  \end{tabular}
\end{tabular}

}
\vspace*{-0.25cm}

\noindent \hrulefill
\\
\textbf{Basis vectors:}
\vspace*{-0.25cm}
\renewcommand{\arraystretch}{1.5}
\begin{longtabu} to \textwidth{>{\centering $}X[-1,c,c]<{$}>{\centering $}X[-1,c,c]<{$}>{\centering $}X[-1,c,c]<{$}>{\centering $}X[-1,c,c]<{$}>{\centering $}X[-1,c,c]<{$}>{\centering $}X[-1,c,c]<{$}>{\centering $}X[-1,c,c]<{$}}
  & & \mbox{Lattice Coordinates} & & \mbox{Cartesian Coordinates} &\mbox{Wyckoff Position} & \mbox{Atom Type} \\  
  \mathbf{B}_{1} & = & \frac{1}{2} \, \mathbf{a}_{1} + \frac{1}{2} \, \mathbf{a}_{2} & = & \frac{1}{2}c \, \mathbf{\hat{z}} & \left(2c\right) & \mbox{Ta I} \\ 
\mathbf{B}_{2} & = & y_{2} \, \mathbf{a}_{1} + y_{2} \, \mathbf{a}_{3} & = & y_{2}b \, \mathbf{\hat{y}} & \left(4g\right) & \mbox{B I} \\ 
\mathbf{B}_{3} & = & -y_{2} \, \mathbf{a}_{1} + -y_{2} \, \mathbf{a}_{3} & = & -y_{2}b \, \mathbf{\hat{y}} & \left(4g\right) & \mbox{B I} \\ 
\mathbf{B}_{4} & = & y_{3} \, \mathbf{a}_{1} + y_{3} \, \mathbf{a}_{3} & = & y_{3}b \, \mathbf{\hat{y}} & \left(4g\right) & \mbox{Ta II} \\ 
\mathbf{B}_{5} & = & -y_{3} \, \mathbf{a}_{1} + -y_{3} \, \mathbf{a}_{3} & = & -y_{3}b \, \mathbf{\hat{y}} & \left(4g\right) & \mbox{Ta II} \\ 
\mathbf{B}_{6} & = & \left(\frac{1}{2} +y_{4}\right) \, \mathbf{a}_{1} + \frac{1}{2} \, \mathbf{a}_{2} + y_{4} \, \mathbf{a}_{3} & = & y_{4}b \, \mathbf{\hat{y}} + \frac{1}{2}c \, \mathbf{\hat{z}} & \left(4h\right) & \mbox{B II} \\ 
\mathbf{B}_{7} & = & \left(\frac{1}{2} - y_{4}\right) \, \mathbf{a}_{1} + \frac{1}{2} \, \mathbf{a}_{2}-y_{4} \, \mathbf{a}_{3} & = & -y_{4}b \, \mathbf{\hat{y}} + \frac{1}{2}c \, \mathbf{\hat{z}} & \left(4h\right) & \mbox{B II} \\ 
\end{longtabu}
\renewcommand{\arraystretch}{1.0}
\noindent \hrulefill
\\
\textbf{References:}
\vspace*{-0.25cm}
\begin{flushleft}
  - \bibentry{Kiessling_Acta_Chem_Scand_3_603_1949}. \\
\end{flushleft}
\textbf{Found in:}
\vspace*{-0.25cm}
\begin{flushleft}
  - \bibentry{Minyaev_Chem_Mater_3_1991}. \\
\end{flushleft}
\noindent \hrulefill
\\
\textbf{Geometry files:}
\\
\noindent  - CIF: pp. {\hyperref[A4B3_oI14_71_gh_cg_cif]{\pageref{A4B3_oI14_71_gh_cg_cif}}} \\
\noindent  - POSCAR: pp. {\hyperref[A4B3_oI14_71_gh_cg_poscar]{\pageref{A4B3_oI14_71_gh_cg_poscar}}} \\
\onecolumn
{\phantomsection\label{ABC_oI12_71_h_j_g}}
\subsection*{\huge \textbf{{\normalfont NbPS Structure: ABC\_oI12\_71\_h\_j\_g}}}
\noindent \hrulefill
\vspace*{0.25cm}
\begin{figure}[htp]
  \centering
  \vspace{-1em}
  {\includegraphics[width=1\textwidth]{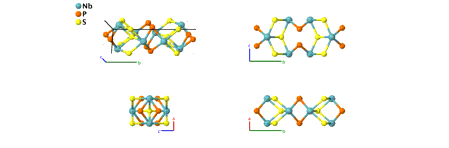}}
\end{figure}
\vspace*{-0.5cm}
\renewcommand{\arraystretch}{1.5}
\begin{equation*}
  \begin{array}{>{$\hspace{-0.15cm}}l<{$}>{$}p{0.5cm}<{$}>{$}p{18.5cm}<{$}}
    \mbox{\large \textbf{Prototype}} &\colon & \ce{NbPS} \\
    \mbox{\large \textbf{\AFLOW\ prototype label}} &\colon & \mbox{ABC\_oI12\_71\_h\_j\_g} \\
    \mbox{\large \textbf{\textit{Strukturbericht} designation}} &\colon & \mbox{None} \\
    \mbox{\large \textbf{Pearson symbol}} &\colon & \mbox{oI12} \\
    \mbox{\large \textbf{Space group number}} &\colon & 71 \\
    \mbox{\large \textbf{Space group symbol}} &\colon & Immm \\
    \mbox{\large \textbf{\AFLOW\ prototype command}} &\colon &  \texttt{aflow} \,  \, \texttt{-{}-proto=ABC\_oI12\_71\_h\_j\_g } \, \newline \texttt{-{}-params=}{a,b/a,c/a,y_{1},y_{2},z_{3} }
  \end{array}
\end{equation*}
\renewcommand{\arraystretch}{1.0}

\vspace*{-0.25cm}
\noindent \hrulefill
\\
\textbf{ Other compounds with this structure:}
\begin{itemize}
   \item{ TaPS, NbPSe, HfSbTe, ZrSbTe  }
\end{itemize}
\noindent \parbox{1 \linewidth}{
\noindent \hrulefill
\\
\textbf{Body-centered Orthorhombic primitive vectors:} \\
\vspace*{-0.25cm}
\begin{tabular}{cc}
  \begin{tabular}{c}
    \parbox{0.6 \linewidth}{
      \renewcommand{\arraystretch}{1.5}
      \begin{equation*}
        \centering
        \begin{array}{ccc}
              \mathbf{a}_1 & = & - \frac12 \, a \, \mathbf{\hat{x}} + \frac12 \, b \, \mathbf{\hat{y}} + \frac12 \, c \, \mathbf{\hat{z}} \\
    \mathbf{a}_2 & = & ~ \frac12 \, a \, \mathbf{\hat{x}} - \frac12 \, b \, \mathbf{\hat{y}} + \frac12 \, c \, \mathbf{\hat{z}} \\
    \mathbf{a}_3 & = & ~ \frac12 \, a \, \mathbf{\hat{x}} + \frac12 \, b \, \mathbf{\hat{y}} - \frac12 \, c \, \mathbf{\hat{z}} \\

        \end{array}
      \end{equation*}
    }
    \renewcommand{\arraystretch}{1.0}
  \end{tabular}
  \begin{tabular}{c}
    \includegraphics[width=0.3\linewidth]{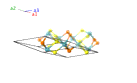} \\
  \end{tabular}
\end{tabular}

}
\vspace*{-0.25cm}

\noindent \hrulefill
\\
\textbf{Basis vectors:}
\vspace*{-0.25cm}
\renewcommand{\arraystretch}{1.5}
\begin{longtabu} to \textwidth{>{\centering $}X[-1,c,c]<{$}>{\centering $}X[-1,c,c]<{$}>{\centering $}X[-1,c,c]<{$}>{\centering $}X[-1,c,c]<{$}>{\centering $}X[-1,c,c]<{$}>{\centering $}X[-1,c,c]<{$}>{\centering $}X[-1,c,c]<{$}}
  & & \mbox{Lattice Coordinates} & & \mbox{Cartesian Coordinates} &\mbox{Wyckoff Position} & \mbox{Atom Type} \\  
  \mathbf{B}_{1} & = & y_{1} \, \mathbf{a}_{1} + y_{1} \, \mathbf{a}_{3} & = & y_{1}b \, \mathbf{\hat{y}} & \left(4g\right) & \mbox{S} \\ 
\mathbf{B}_{2} & = & -y_{1} \, \mathbf{a}_{1} + -y_{1} \, \mathbf{a}_{3} & = & -y_{1}b \, \mathbf{\hat{y}} & \left(4g\right) & \mbox{S} \\ 
\mathbf{B}_{3} & = & \left(\frac{1}{2} +y_{2}\right) \, \mathbf{a}_{1} + \frac{1}{2} \, \mathbf{a}_{2} + y_{2} \, \mathbf{a}_{3} & = & y_{2}b \, \mathbf{\hat{y}} + \frac{1}{2}c \, \mathbf{\hat{z}} & \left(4h\right) & \mbox{Nb} \\ 
\mathbf{B}_{4} & = & \left(\frac{1}{2} - y_{2}\right) \, \mathbf{a}_{1} + \frac{1}{2} \, \mathbf{a}_{2}-y_{2} \, \mathbf{a}_{3} & = & -y_{2}b \, \mathbf{\hat{y}} + \frac{1}{2}c \, \mathbf{\hat{z}} & \left(4h\right) & \mbox{Nb} \\ 
\mathbf{B}_{5} & = & z_{3} \, \mathbf{a}_{1} + \left(\frac{1}{2} +z_{3}\right) \, \mathbf{a}_{2} + \frac{1}{2} \, \mathbf{a}_{3} & = & \frac{1}{2}a \, \mathbf{\hat{x}} + z_{3}c \, \mathbf{\hat{z}} & \left(4j\right) & \mbox{P} \\ 
\mathbf{B}_{6} & = & -z_{3} \, \mathbf{a}_{1} + \left(\frac{1}{2} - z_{3}\right) \, \mathbf{a}_{2} + \frac{1}{2} \, \mathbf{a}_{3} & = & \frac{1}{2}a \, \mathbf{\hat{x}} + -z_{3}c \, \mathbf{\hat{z}} & \left(4j\right) & \mbox{P} \\ 
\end{longtabu}
\renewcommand{\arraystretch}{1.0}
\noindent \hrulefill
\\
\textbf{References:}
\vspace*{-0.25cm}
\begin{flushleft}
  - \bibentry{Donohue_Inorg_Chem_8_1969}. \\
\end{flushleft}
\noindent \hrulefill
\\
\textbf{Geometry files:}
\\
\noindent  - CIF: pp. {\hyperref[ABC_oI12_71_h_j_g_cif]{\pageref{ABC_oI12_71_h_j_g_cif}}} \\
\noindent  - POSCAR: pp. {\hyperref[ABC_oI12_71_h_j_g_poscar]{\pageref{ABC_oI12_71_h_j_g_poscar}}} \\
\onecolumn
{\phantomsection\label{ABCD3_oI48_73_d_e_e_ef}}
\subsection*{\huge \textbf{{\normalfont KAg[CO$_{3}$] Structure: ABCD3\_oI48\_73\_d\_e\_e\_ef}}}
\noindent \hrulefill
\vspace*{0.25cm}
\begin{figure}[htp]
  \centering
  \vspace{-1em}
  {\includegraphics[width=1\textwidth]{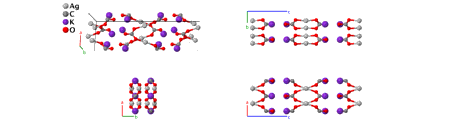}}
\end{figure}
\vspace*{-0.5cm}
\renewcommand{\arraystretch}{1.5}
\begin{equation*}
  \begin{array}{>{$\hspace{-0.15cm}}l<{$}>{$}p{0.5cm}<{$}>{$}p{18.5cm}<{$}}
    \mbox{\large \textbf{Prototype}} &\colon & \ce{KAg[CO3]} \\
    \mbox{\large \textbf{\AFLOW\ prototype label}} &\colon & \mbox{ABCD3\_oI48\_73\_d\_e\_e\_ef} \\
    \mbox{\large \textbf{\textit{Strukturbericht} designation}} &\colon & \mbox{None} \\
    \mbox{\large \textbf{Pearson symbol}} &\colon & \mbox{oI48} \\
    \mbox{\large \textbf{Space group number}} &\colon & 73 \\
    \mbox{\large \textbf{Space group symbol}} &\colon & Ibca \\
    \mbox{\large \textbf{\AFLOW\ prototype command}} &\colon &  \texttt{aflow} \,  \, \texttt{-{}-proto=ABCD3\_oI48\_73\_d\_e\_e\_ef } \, \newline \texttt{-{}-params=}{a,b/a,c/a,y_{1},z_{2},z_{3},z_{4},x_{5},y_{5},z_{5} }
  \end{array}
\end{equation*}
\renewcommand{\arraystretch}{1.0}

\noindent \parbox{1 \linewidth}{
\noindent \hrulefill
\\
\textbf{Body-centered Orthorhombic primitive vectors:} \\
\vspace*{-0.25cm}
\begin{tabular}{cc}
  \begin{tabular}{c}
    \parbox{0.6 \linewidth}{
      \renewcommand{\arraystretch}{1.5}
      \begin{equation*}
        \centering
        \begin{array}{ccc}
              \mathbf{a}_1 & = & - \frac12 \, a \, \mathbf{\hat{x}} + \frac12 \, b \, \mathbf{\hat{y}} + \frac12 \, c \, \mathbf{\hat{z}} \\
    \mathbf{a}_2 & = & ~ \frac12 \, a \, \mathbf{\hat{x}} - \frac12 \, b \, \mathbf{\hat{y}} + \frac12 \, c \, \mathbf{\hat{z}} \\
    \mathbf{a}_3 & = & ~ \frac12 \, a \, \mathbf{\hat{x}} + \frac12 \, b \, \mathbf{\hat{y}} - \frac12 \, c \, \mathbf{\hat{z}} \\

        \end{array}
      \end{equation*}
    }
    \renewcommand{\arraystretch}{1.0}
  \end{tabular}
  \begin{tabular}{c}
    \includegraphics[width=0.3\linewidth]{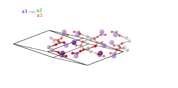} \\
  \end{tabular}
\end{tabular}

}
\vspace*{-0.25cm}

\noindent \hrulefill
\\
\textbf{Basis vectors:}
\vspace*{-0.25cm}
\renewcommand{\arraystretch}{1.5}
\begin{longtabu} to \textwidth{>{\centering $}X[-1,c,c]<{$}>{\centering $}X[-1,c,c]<{$}>{\centering $}X[-1,c,c]<{$}>{\centering $}X[-1,c,c]<{$}>{\centering $}X[-1,c,c]<{$}>{\centering $}X[-1,c,c]<{$}>{\centering $}X[-1,c,c]<{$}}
  & & \mbox{Lattice Coordinates} & & \mbox{Cartesian Coordinates} &\mbox{Wyckoff Position} & \mbox{Atom Type} \\  
  \mathbf{B}_{1} & = & y_{1} \, \mathbf{a}_{1} + \frac{1}{4} \, \mathbf{a}_{2} + \left(\frac{1}{4} +y_{1}\right) \, \mathbf{a}_{3} & = & \frac{1}{4}a \, \mathbf{\hat{x}} + y_{1}b \, \mathbf{\hat{y}} & \left(8d\right) & \mbox{Ag} \\ 
\mathbf{B}_{2} & = & \left(\frac{1}{2} - y_{1}\right) \, \mathbf{a}_{1} + \frac{3}{4} \, \mathbf{a}_{2} + \left(\frac{1}{4} - y_{1}\right) \, \mathbf{a}_{3} & = & \frac{1}{4}a \, \mathbf{\hat{x}}-y_{1}b \, \mathbf{\hat{y}} + \frac{1}{2}c \, \mathbf{\hat{z}} & \left(8d\right) & \mbox{Ag} \\ 
\mathbf{B}_{3} & = & -y_{1} \, \mathbf{a}_{1} + \frac{3}{4} \, \mathbf{a}_{2} + \left(\frac{3}{4} - y_{1}\right) \, \mathbf{a}_{3} & = & \frac{3}{4}a \, \mathbf{\hat{x}}-y_{1}b \, \mathbf{\hat{y}} & \left(8d\right) & \mbox{Ag} \\ 
\mathbf{B}_{4} & = & \left(\frac{1}{2} +y_{1}\right) \, \mathbf{a}_{1} + \frac{1}{4} \, \mathbf{a}_{2} + \left(\frac{3}{4} +y_{1}\right) \, \mathbf{a}_{3} & = & \frac{1}{4}a \, \mathbf{\hat{x}} + \left(\frac{1}{2} +y_{1}\right)b \, \mathbf{\hat{y}} & \left(8d\right) & \mbox{Ag} \\ 
\mathbf{B}_{5} & = & \left(\frac{1}{4} +z_{2}\right) \, \mathbf{a}_{1} + z_{2} \, \mathbf{a}_{2} + \frac{1}{4} \, \mathbf{a}_{3} & = & \frac{1}{4}b \, \mathbf{\hat{y}} + z_{2}c \, \mathbf{\hat{z}} & \left(8e\right) & \mbox{C} \\ 
\mathbf{B}_{6} & = & \left(\frac{1}{4} - z_{2}\right) \, \mathbf{a}_{1} + \left(\frac{1}{2} - z_{2}\right) \, \mathbf{a}_{2} + \frac{3}{4} \, \mathbf{a}_{3} & = & \frac{1}{2}a \, \mathbf{\hat{x}} + \frac{1}{4}b \, \mathbf{\hat{y}}-z_{2}c \, \mathbf{\hat{z}} & \left(8e\right) & \mbox{C} \\ 
\mathbf{B}_{7} & = & \left(\frac{3}{4} - z_{2}\right) \, \mathbf{a}_{1}-z_{2} \, \mathbf{a}_{2} + \frac{3}{4} \, \mathbf{a}_{3} & = & \frac{3}{4}b \, \mathbf{\hat{y}}-z_{2}c \, \mathbf{\hat{z}} & \left(8e\right) & \mbox{C} \\ 
\mathbf{B}_{8} & = & \left(\frac{3}{4} +z_{2}\right) \, \mathbf{a}_{1} + \left(\frac{1}{2} +z_{2}\right) \, \mathbf{a}_{2} + \frac{1}{4} \, \mathbf{a}_{3} & = & \frac{1}{4}b \, \mathbf{\hat{y}} + \left(\frac{1}{2} +z_{2}\right)c \, \mathbf{\hat{z}} & \left(8e\right) & \mbox{C} \\ 
\mathbf{B}_{9} & = & \left(\frac{1}{4} +z_{3}\right) \, \mathbf{a}_{1} + z_{3} \, \mathbf{a}_{2} + \frac{1}{4} \, \mathbf{a}_{3} & = & \frac{1}{4}b \, \mathbf{\hat{y}} + z_{3}c \, \mathbf{\hat{z}} & \left(8e\right) & \mbox{K} \\ 
\mathbf{B}_{10} & = & \left(\frac{1}{4} - z_{3}\right) \, \mathbf{a}_{1} + \left(\frac{1}{2} - z_{3}\right) \, \mathbf{a}_{2} + \frac{3}{4} \, \mathbf{a}_{3} & = & \frac{1}{2}a \, \mathbf{\hat{x}} + \frac{1}{4}b \, \mathbf{\hat{y}}-z_{3}c \, \mathbf{\hat{z}} & \left(8e\right) & \mbox{K} \\ 
\mathbf{B}_{11} & = & \left(\frac{3}{4} - z_{3}\right) \, \mathbf{a}_{1}-z_{3} \, \mathbf{a}_{2} + \frac{3}{4} \, \mathbf{a}_{3} & = & \frac{3}{4}b \, \mathbf{\hat{y}}-z_{3}c \, \mathbf{\hat{z}} & \left(8e\right) & \mbox{K} \\ 
\mathbf{B}_{12} & = & \left(\frac{3}{4} +z_{3}\right) \, \mathbf{a}_{1} + \left(\frac{1}{2} +z_{3}\right) \, \mathbf{a}_{2} + \frac{1}{4} \, \mathbf{a}_{3} & = & \frac{1}{4}b \, \mathbf{\hat{y}} + \left(\frac{1}{2} +z_{3}\right)c \, \mathbf{\hat{z}} & \left(8e\right) & \mbox{K} \\ 
\mathbf{B}_{13} & = & \left(\frac{1}{4} +z_{4}\right) \, \mathbf{a}_{1} + z_{4} \, \mathbf{a}_{2} + \frac{1}{4} \, \mathbf{a}_{3} & = & \frac{1}{4}b \, \mathbf{\hat{y}} + z_{4}c \, \mathbf{\hat{z}} & \left(8e\right) & \mbox{O I} \\ 
\mathbf{B}_{14} & = & \left(\frac{1}{4} - z_{4}\right) \, \mathbf{a}_{1} + \left(\frac{1}{2} - z_{4}\right) \, \mathbf{a}_{2} + \frac{3}{4} \, \mathbf{a}_{3} & = & \frac{1}{2}a \, \mathbf{\hat{x}} + \frac{1}{4}b \, \mathbf{\hat{y}}-z_{4}c \, \mathbf{\hat{z}} & \left(8e\right) & \mbox{O I} \\ 
\mathbf{B}_{15} & = & \left(\frac{3}{4} - z_{4}\right) \, \mathbf{a}_{1}-z_{4} \, \mathbf{a}_{2} + \frac{3}{4} \, \mathbf{a}_{3} & = & \frac{3}{4}b \, \mathbf{\hat{y}}-z_{4}c \, \mathbf{\hat{z}} & \left(8e\right) & \mbox{O I} \\ 
\mathbf{B}_{16} & = & \left(\frac{3}{4} +z_{4}\right) \, \mathbf{a}_{1} + \left(\frac{1}{2} +z_{4}\right) \, \mathbf{a}_{2} + \frac{1}{4} \, \mathbf{a}_{3} & = & \frac{1}{4}b \, \mathbf{\hat{y}} + \left(\frac{1}{2} +z_{4}\right)c \, \mathbf{\hat{z}} & \left(8e\right) & \mbox{O I} \\ 
\mathbf{B}_{17} & = & \left(y_{5}+z_{5}\right) \, \mathbf{a}_{1} + \left(x_{5}+z_{5}\right) \, \mathbf{a}_{2} + \left(x_{5}+y_{5}\right) \, \mathbf{a}_{3} & = & x_{5}a \, \mathbf{\hat{x}} + y_{5}b \, \mathbf{\hat{y}} + z_{5}c \, \mathbf{\hat{z}} & \left(16f\right) & \mbox{O II} \\ 
\mathbf{B}_{18} & = & \left(\frac{1}{2} - y_{5} + z_{5}\right) \, \mathbf{a}_{1} + \left(-x_{5}+z_{5}\right) \, \mathbf{a}_{2} + \left(\frac{1}{2} - x_{5} - y_{5}\right) \, \mathbf{a}_{3} & = & -x_{5}a \, \mathbf{\hat{x}} + \left(\frac{1}{2} - y_{5}\right)b \, \mathbf{\hat{y}} + z_{5}c \, \mathbf{\hat{z}} & \left(16f\right) & \mbox{O II} \\ 
\mathbf{B}_{19} & = & \left(y_{5}-z_{5}\right) \, \mathbf{a}_{1} + \left(\frac{1}{2} - x_{5} - z_{5}\right) \, \mathbf{a}_{2} + \left(\frac{1}{2} - x_{5} + y_{5}\right) \, \mathbf{a}_{3} & = & \left(\frac{1}{2} - x_{5}\right)a \, \mathbf{\hat{x}} + y_{5}b \, \mathbf{\hat{y}}-z_{5}c \, \mathbf{\hat{z}} & \left(16f\right) & \mbox{O II} \\ 
\mathbf{B}_{20} & = & \left(\frac{1}{2} - y_{5} - z_{5}\right) \, \mathbf{a}_{1} + \left(\frac{1}{2} +x_{5} - z_{5}\right) \, \mathbf{a}_{2} + \left(x_{5}-y_{5}\right) \, \mathbf{a}_{3} & = & x_{5}a \, \mathbf{\hat{x}}-y_{5}b \, \mathbf{\hat{y}} + \left(\frac{1}{2} - z_{5}\right)c \, \mathbf{\hat{z}} & \left(16f\right) & \mbox{O II} \\ 
\mathbf{B}_{21} & = & \left(-y_{5}-z_{5}\right) \, \mathbf{a}_{1} + \left(-x_{5}-z_{5}\right) \, \mathbf{a}_{2} + \left(-x_{5}-y_{5}\right) \, \mathbf{a}_{3} & = & -x_{5}a \, \mathbf{\hat{x}}-y_{5}b \, \mathbf{\hat{y}}-z_{5}c \, \mathbf{\hat{z}} & \left(16f\right) & \mbox{O II} \\ 
\mathbf{B}_{22} & = & \left(\frac{1}{2} +y_{5} - z_{5}\right) \, \mathbf{a}_{1} + \left(x_{5}-z_{5}\right) \, \mathbf{a}_{2} + \left(\frac{1}{2} +x_{5} + y_{5}\right) \, \mathbf{a}_{3} & = & x_{5}a \, \mathbf{\hat{x}} + \left(\frac{1}{2} +y_{5}\right)b \, \mathbf{\hat{y}}-z_{5}c \, \mathbf{\hat{z}} & \left(16f\right) & \mbox{O II} \\ 
\mathbf{B}_{23} & = & \left(-y_{5}+z_{5}\right) \, \mathbf{a}_{1} + \left(\frac{1}{2} +x_{5} + z_{5}\right) \, \mathbf{a}_{2} + \left(\frac{1}{2} +x_{5} - y_{5}\right) \, \mathbf{a}_{3} & = & \left(\frac{1}{2} +x_{5}\right)a \, \mathbf{\hat{x}}-y_{5}b \, \mathbf{\hat{y}} + z_{5}c \, \mathbf{\hat{z}} & \left(16f\right) & \mbox{O II} \\ 
\mathbf{B}_{24} & = & \left(\frac{1}{2} +y_{5} + z_{5}\right) \, \mathbf{a}_{1} + \left(\frac{1}{2} - x_{5} + z_{5}\right) \, \mathbf{a}_{2} + \left(-x_{5}+y_{5}\right) \, \mathbf{a}_{3} & = & -x_{5}a \, \mathbf{\hat{x}} + y_{5}b \, \mathbf{\hat{y}} + \left(\frac{1}{2} +z_{5}\right)c \, \mathbf{\hat{z}} & \left(16f\right) & \mbox{O II} \\ 
\end{longtabu}
\renewcommand{\arraystretch}{1.0}
\noindent \hrulefill
\\
\textbf{References:}
\vspace*{-0.25cm}
\begin{flushleft}
  - \bibentry{Zheng_AgCO3K_ZKrist_2000}. \\
\end{flushleft}
\textbf{Found in:}
\vspace*{-0.25cm}
\begin{flushleft}
  - \bibentry{Villars_PearsonsCrystalData_2013}. \\
\end{flushleft}
\noindent \hrulefill
\\
\textbf{Geometry files:}
\\
\noindent  - CIF: pp. {\hyperref[ABCD3_oI48_73_d_e_e_ef_cif]{\pageref{ABCD3_oI48_73_d_e_e_ef_cif}}} \\
\noindent  - POSCAR: pp. {\hyperref[ABCD3_oI48_73_d_e_e_ef_poscar]{\pageref{ABCD3_oI48_73_d_e_e_ef_poscar}}} \\
\onecolumn
{\phantomsection\label{A2B_oI12_74_h_e}}
\subsection*{\huge \textbf{{\normalfont KHg$_{2}$ Structure: A2B\_oI12\_74\_h\_e}}}
\noindent \hrulefill
\vspace*{0.25cm}
\begin{figure}[htp]
  \centering
  \vspace{-1em}
  {\includegraphics[width=1\textwidth]{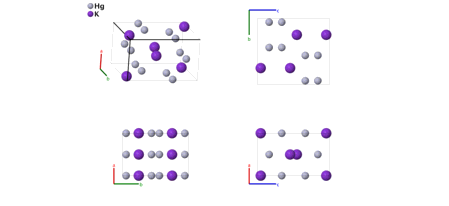}}
\end{figure}
\vspace*{-0.5cm}
\renewcommand{\arraystretch}{1.5}
\begin{equation*}
  \begin{array}{>{$\hspace{-0.15cm}}l<{$}>{$}p{0.5cm}<{$}>{$}p{18.5cm}<{$}}
    \mbox{\large \textbf{Prototype}} &\colon & \ce{KHg2} \\
    \mbox{\large \textbf{\AFLOW\ prototype label}} &\colon & \mbox{A2B\_oI12\_74\_h\_e} \\
    \mbox{\large \textbf{\textit{Strukturbericht} designation}} &\colon & \mbox{None} \\
    \mbox{\large \textbf{Pearson symbol}} &\colon & \mbox{oI12} \\
    \mbox{\large \textbf{Space group number}} &\colon & 74 \\
    \mbox{\large \textbf{Space group symbol}} &\colon & Imma \\
    \mbox{\large \textbf{\AFLOW\ prototype command}} &\colon &  \texttt{aflow} \,  \, \texttt{-{}-proto=A2B\_oI12\_74\_h\_e } \, \newline \texttt{-{}-params=}{a,b/a,c/a,z_{1},y_{2},z_{2} }
  \end{array}
\end{equation*}
\renewcommand{\arraystretch}{1.0}

\noindent \parbox{1 \linewidth}{
\noindent \hrulefill
\\
\textbf{Body-centered Orthorhombic primitive vectors:} \\
\vspace*{-0.25cm}
\begin{tabular}{cc}
  \begin{tabular}{c}
    \parbox{0.6 \linewidth}{
      \renewcommand{\arraystretch}{1.5}
      \begin{equation*}
        \centering
        \begin{array}{ccc}
              \mathbf{a}_1 & = & - \frac12 \, a \, \mathbf{\hat{x}} + \frac12 \, b \, \mathbf{\hat{y}} + \frac12 \, c \, \mathbf{\hat{z}} \\
    \mathbf{a}_2 & = & ~ \frac12 \, a \, \mathbf{\hat{x}} - \frac12 \, b \, \mathbf{\hat{y}} + \frac12 \, c \, \mathbf{\hat{z}} \\
    \mathbf{a}_3 & = & ~ \frac12 \, a \, \mathbf{\hat{x}} + \frac12 \, b \, \mathbf{\hat{y}} - \frac12 \, c \, \mathbf{\hat{z}} \\

        \end{array}
      \end{equation*}
    }
    \renewcommand{\arraystretch}{1.0}
  \end{tabular}
  \begin{tabular}{c}
    \includegraphics[width=0.3\linewidth]{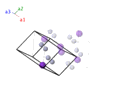} \\
  \end{tabular}
\end{tabular}

}
\vspace*{-0.25cm}

\noindent \hrulefill
\\
\textbf{Basis vectors:}
\vspace*{-0.25cm}
\renewcommand{\arraystretch}{1.5}
\begin{longtabu} to \textwidth{>{\centering $}X[-1,c,c]<{$}>{\centering $}X[-1,c,c]<{$}>{\centering $}X[-1,c,c]<{$}>{\centering $}X[-1,c,c]<{$}>{\centering $}X[-1,c,c]<{$}>{\centering $}X[-1,c,c]<{$}>{\centering $}X[-1,c,c]<{$}}
  & & \mbox{Lattice Coordinates} & & \mbox{Cartesian Coordinates} &\mbox{Wyckoff Position} & \mbox{Atom Type} \\  
  \mathbf{B}_{1} & = & \left(\frac{1}{4} +z_{1}\right) \, \mathbf{a}_{1} + z_{1} \, \mathbf{a}_{2} + \frac{1}{4} \, \mathbf{a}_{3} & = & \frac{1}{4}b \, \mathbf{\hat{y}} + z_{1}c \, \mathbf{\hat{z}} & \left(4e\right) & \mbox{K} \\ 
\mathbf{B}_{2} & = & \left(\frac{3}{4} - z_{1}\right) \, \mathbf{a}_{1}-z_{1} \, \mathbf{a}_{2} + \frac{3}{4} \, \mathbf{a}_{3} & = & \frac{3}{4}b \, \mathbf{\hat{y}}-z_{1}c \, \mathbf{\hat{z}} & \left(4e\right) & \mbox{K} \\ 
\mathbf{B}_{3} & = & \left(y_{2}+z_{2}\right) \, \mathbf{a}_{1} + z_{2} \, \mathbf{a}_{2} + y_{2} \, \mathbf{a}_{3} & = & y_{2}b \, \mathbf{\hat{y}} + z_{2}c \, \mathbf{\hat{z}} & \left(8h\right) & \mbox{Hg} \\ 
\mathbf{B}_{4} & = & \left(\frac{1}{2} - y_{2} + z_{2}\right) \, \mathbf{a}_{1} + z_{2} \, \mathbf{a}_{2} + \left(\frac{1}{2} - y_{2}\right) \, \mathbf{a}_{3} & = & \left(\frac{1}{2} - y_{2}\right)b \, \mathbf{\hat{y}} + z_{2}c \, \mathbf{\hat{z}} & \left(8h\right) & \mbox{Hg} \\ 
\mathbf{B}_{5} & = & \left(\frac{1}{2} +y_{2} - z_{2}\right) \, \mathbf{a}_{1}-z_{2} \, \mathbf{a}_{2} + \left(\frac{1}{2} +y_{2}\right) \, \mathbf{a}_{3} & = & \left(\frac{1}{2} +y_{2}\right)b \, \mathbf{\hat{y}}-z_{2}c \, \mathbf{\hat{z}} & \left(8h\right) & \mbox{Hg} \\ 
\mathbf{B}_{6} & = & \left(-y_{2}-z_{2}\right) \, \mathbf{a}_{1}-z_{2} \, \mathbf{a}_{2}-y_{2} \, \mathbf{a}_{3} & = & -y_{2}b \, \mathbf{\hat{y}}-z_{2}c \, \mathbf{\hat{z}} & \left(8h\right) & \mbox{Hg} \\ 
\end{longtabu}
\renewcommand{\arraystretch}{1.0}
\noindent \hrulefill
\\
\textbf{References:}
\vspace*{-0.25cm}
\begin{flushleft}
  - \bibentry{Duwell_KHg2_ActaCryst_1955}. \\
\end{flushleft}
\textbf{Found in:}
\vspace*{-0.25cm}
\begin{flushleft}
  - \bibentry{Villars_PearsonsCrystalData_2013}. \\
\end{flushleft}
\noindent \hrulefill
\\
\textbf{Geometry files:}
\\
\noindent  - CIF: pp. {\hyperref[A2B_oI12_74_h_e_cif]{\pageref{A2B_oI12_74_h_e_cif}}} \\
\noindent  - POSCAR: pp. {\hyperref[A2B_oI12_74_h_e_poscar]{\pageref{A2B_oI12_74_h_e_poscar}}} \\
\onecolumn
{\phantomsection\label{A4B_oI20_74_beh_e}}
\subsection*{\huge \textbf{{\normalfont Al$_{4}$U ($D1_{b}$) Structure: A4B\_oI20\_74\_beh\_e}}}
\noindent \hrulefill
\vspace*{0.25cm}
\begin{figure}[htp]
  \centering
  \vspace{-1em}
  {\includegraphics[width=1\textwidth]{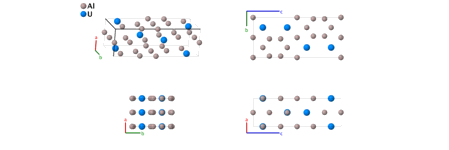}}
\end{figure}
\vspace*{-0.5cm}
\renewcommand{\arraystretch}{1.5}
\begin{equation*}
  \begin{array}{>{$\hspace{-0.15cm}}l<{$}>{$}p{0.5cm}<{$}>{$}p{18.5cm}<{$}}
    \mbox{\large \textbf{Prototype}} &\colon & \ce{Al$_{4}$U} \\
    \mbox{\large \textbf{\AFLOW\ prototype label}} &\colon & \mbox{A4B\_oI20\_74\_beh\_e} \\
    \mbox{\large \textbf{\textit{Strukturbericht} designation}} &\colon & \mbox{$D1_{b}$} \\
    \mbox{\large \textbf{Pearson symbol}} &\colon & \mbox{oI20} \\
    \mbox{\large \textbf{Space group number}} &\colon & 74 \\
    \mbox{\large \textbf{Space group symbol}} &\colon & Imma \\
    \mbox{\large \textbf{\AFLOW\ prototype command}} &\colon &  \texttt{aflow} \,  \, \texttt{-{}-proto=A4B\_oI20\_74\_beh\_e } \, \newline \texttt{-{}-params=}{a,b/a,c/a,z_{2},z_{3},y_{4},z_{4} }
  \end{array}
\end{equation*}
\renewcommand{\arraystretch}{1.0}

\vspace*{-0.25cm}
\noindent \hrulefill
\\
\textbf{ Other compounds with this structure:}
\begin{itemize}
   \item{ Al$_{4}$Gd, Al$_{4}$Np, Al$_{4}$Pu  }
\end{itemize}
\noindent \parbox{1 \linewidth}{
\noindent \hrulefill
\\
\textbf{Body-centered Orthorhombic primitive vectors:} \\
\vspace*{-0.25cm}
\begin{tabular}{cc}
  \begin{tabular}{c}
    \parbox{0.6 \linewidth}{
      \renewcommand{\arraystretch}{1.5}
      \begin{equation*}
        \centering
        \begin{array}{ccc}
              \mathbf{a}_1 & = & - \frac12 \, a \, \mathbf{\hat{x}} + \frac12 \, b \, \mathbf{\hat{y}} + \frac12 \, c \, \mathbf{\hat{z}} \\
    \mathbf{a}_2 & = & ~ \frac12 \, a \, \mathbf{\hat{x}} - \frac12 \, b \, \mathbf{\hat{y}} + \frac12 \, c \, \mathbf{\hat{z}} \\
    \mathbf{a}_3 & = & ~ \frac12 \, a \, \mathbf{\hat{x}} + \frac12 \, b \, \mathbf{\hat{y}} - \frac12 \, c \, \mathbf{\hat{z}} \\

        \end{array}
      \end{equation*}
    }
    \renewcommand{\arraystretch}{1.0}
  \end{tabular}
  \begin{tabular}{c}
    \includegraphics[width=0.3\linewidth]{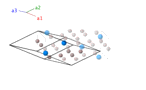} \\
  \end{tabular}
\end{tabular}

}
\vspace*{-0.25cm}

\noindent \hrulefill
\\
\textbf{Basis vectors:}
\vspace*{-0.25cm}
\renewcommand{\arraystretch}{1.5}
\begin{longtabu} to \textwidth{>{\centering $}X[-1,c,c]<{$}>{\centering $}X[-1,c,c]<{$}>{\centering $}X[-1,c,c]<{$}>{\centering $}X[-1,c,c]<{$}>{\centering $}X[-1,c,c]<{$}>{\centering $}X[-1,c,c]<{$}>{\centering $}X[-1,c,c]<{$}}
  & & \mbox{Lattice Coordinates} & & \mbox{Cartesian Coordinates} &\mbox{Wyckoff Position} & \mbox{Atom Type} \\  
  \mathbf{B}_{1} & = & \frac{1}{2} \, \mathbf{a}_{1} + \frac{1}{2} \, \mathbf{a}_{2} & = & \frac{1}{2}c \, \mathbf{\hat{z}} & \left(4b\right) & \mbox{Al I} \\ 
\mathbf{B}_{2} & = & \frac{1}{2} \, \mathbf{a}_{2} + \frac{1}{2} \, \mathbf{a}_{3} & = & \frac{1}{2}a \, \mathbf{\hat{x}} & \left(4b\right) & \mbox{Al I} \\ 
\mathbf{B}_{3} & = & \left(\frac{1}{4} +z_{2}\right) \, \mathbf{a}_{1} + z_{2} \, \mathbf{a}_{2} + \frac{1}{4} \, \mathbf{a}_{3} & = & \frac{1}{4}b \, \mathbf{\hat{y}} + z_{2}c \, \mathbf{\hat{z}} & \left(4e\right) & \mbox{Al II} \\ 
\mathbf{B}_{4} & = & \left(\frac{3}{4} - z_{2}\right) \, \mathbf{a}_{1}-z_{2} \, \mathbf{a}_{2} + \frac{3}{4} \, \mathbf{a}_{3} & = & \frac{3}{4}b \, \mathbf{\hat{y}}-z_{2}c \, \mathbf{\hat{z}} & \left(4e\right) & \mbox{Al II} \\ 
\mathbf{B}_{5} & = & \left(\frac{1}{4} +z_{3}\right) \, \mathbf{a}_{1} + z_{3} \, \mathbf{a}_{2} + \frac{1}{4} \, \mathbf{a}_{3} & = & \frac{1}{4}b \, \mathbf{\hat{y}} + z_{3}c \, \mathbf{\hat{z}} & \left(4e\right) & \mbox{U} \\ 
\mathbf{B}_{6} & = & \left(\frac{3}{4} - z_{3}\right) \, \mathbf{a}_{1}-z_{3} \, \mathbf{a}_{2} + \frac{3}{4} \, \mathbf{a}_{3} & = & \frac{3}{4}b \, \mathbf{\hat{y}}-z_{3}c \, \mathbf{\hat{z}} & \left(4e\right) & \mbox{U} \\ 
\mathbf{B}_{7} & = & \left(y_{4}+z_{4}\right) \, \mathbf{a}_{1} + z_{4} \, \mathbf{a}_{2} + y_{4} \, \mathbf{a}_{3} & = & y_{4}b \, \mathbf{\hat{y}} + z_{4}c \, \mathbf{\hat{z}} & \left(8h\right) & \mbox{Al III} \\ 
\mathbf{B}_{8} & = & \left(\frac{1}{2} - y_{4} + z_{4}\right) \, \mathbf{a}_{1} + z_{4} \, \mathbf{a}_{2} + \left(\frac{1}{2} - y_{4}\right) \, \mathbf{a}_{3} & = & \left(\frac{1}{2} - y_{4}\right)b \, \mathbf{\hat{y}} + z_{4}c \, \mathbf{\hat{z}} & \left(8h\right) & \mbox{Al III} \\ 
\mathbf{B}_{9} & = & \left(\frac{1}{2} +y_{4} - z_{4}\right) \, \mathbf{a}_{1}-z_{4} \, \mathbf{a}_{2} + \left(\frac{1}{2} +y_{4}\right) \, \mathbf{a}_{3} & = & \left(\frac{1}{2} +y_{4}\right)b \, \mathbf{\hat{y}}-z_{4}c \, \mathbf{\hat{z}} & \left(8h\right) & \mbox{Al III} \\ 
\mathbf{B}_{10} & = & \left(-y_{4}-z_{4}\right) \, \mathbf{a}_{1}-z_{4} \, \mathbf{a}_{2}-y_{4} \, \mathbf{a}_{3} & = & -y_{4}b \, \mathbf{\hat{y}}-z_{4}c \, \mathbf{\hat{z}} & \left(8h\right) & \mbox{Al III} \\ 
\end{longtabu}
\renewcommand{\arraystretch}{1.0}
\noindent \hrulefill
\\
\textbf{References:}
\vspace*{-0.25cm}
\begin{flushleft}
  - \bibentry{Borgsted_Gmelin_1989}. \\
\end{flushleft}
\noindent \hrulefill
\\
\textbf{Geometry files:}
\\
\noindent  - CIF: pp. {\hyperref[A4B_oI20_74_beh_e_cif]{\pageref{A4B_oI20_74_beh_e_cif}}} \\
\noindent  - POSCAR: pp. {\hyperref[A4B_oI20_74_beh_e_poscar]{\pageref{A4B_oI20_74_beh_e_poscar}}} \\
\onecolumn
{\phantomsection\label{AB2C12D4_tP76_75_2a2b_2d_12d_4d}}
\subsection*{\huge \textbf{{\normalfont \begin{raggedleft}BaCr$_{2}$Ru$_{4}$O$_{12}$ Structure: \end{raggedleft} \\ AB2C12D4\_tP76\_75\_2a2b\_2d\_12d\_4d}}}
\noindent \hrulefill
\vspace*{0.25cm}
\begin{figure}[htp]
  \centering
  \vspace{-1em}
  {\includegraphics[width=1\textwidth]{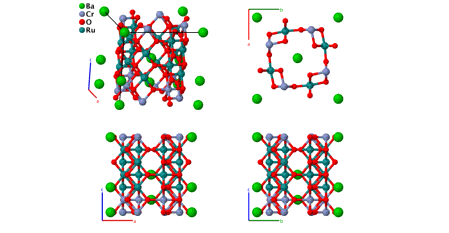}}
\end{figure}
\vspace*{-0.5cm}
\renewcommand{\arraystretch}{1.5}
\begin{equation*}
  \begin{array}{>{$\hspace{-0.15cm}}l<{$}>{$}p{0.5cm}<{$}>{$}p{18.5cm}<{$}}
    \mbox{\large \textbf{Prototype}} &\colon & \ce{BaCr2Ru4O12} \\
    \mbox{\large \textbf{\AFLOW\ prototype label}} &\colon & \mbox{AB2C12D4\_tP76\_75\_2a2b\_2d\_12d\_4d} \\
    \mbox{\large \textbf{\textit{Strukturbericht} designation}} &\colon & \mbox{None} \\
    \mbox{\large \textbf{Pearson symbol}} &\colon & \mbox{tP76} \\
    \mbox{\large \textbf{Space group number}} &\colon & 75 \\
    \mbox{\large \textbf{Space group symbol}} &\colon & P4 \\
    \mbox{\large \textbf{\AFLOW\ prototype command}} &\colon &  \texttt{aflow} \,  \, \texttt{-{}-proto=AB2C12D4\_tP76\_75\_2a2b\_2d\_12d\_4d } \, \newline \texttt{-{}-params=}{a,c/a,z_{1},z_{2},z_{3},z_{4},x_{5},y_{5},z_{5},x_{6},y_{6},z_{6},x_{7},y_{7},z_{7},x_{8},y_{8},z_{8},x_{9},y_{9},z_{9},} \newline {x_{10},y_{10},z_{10},x_{11},y_{11},z_{11},x_{12},y_{12},z_{12},x_{13},y_{13},z_{13},x_{14},y_{14},z_{14},x_{15},y_{15},z_{15},x_{16},y_{16},} \newline {z_{16},x_{17},y_{17},z_{17},x_{18},y_{18},z_{18},x_{19},y_{19},z_{19},x_{20},y_{20},z_{20},x_{21},y_{21},z_{21},x_{22},y_{22},z_{22} }
  \end{array}
\end{equation*}
\renewcommand{\arraystretch}{1.0}

\noindent \parbox{1 \linewidth}{
\noindent \hrulefill
\\
\textbf{Simple Tetragonal primitive vectors:} \\
\vspace*{-0.25cm}
\begin{tabular}{cc}
  \begin{tabular}{c}
    \parbox{0.6 \linewidth}{
      \renewcommand{\arraystretch}{1.5}
      \begin{equation*}
        \centering
        \begin{array}{ccc}
              \mathbf{a}_1 & = & a \, \mathbf{\hat{x}} \\
    \mathbf{a}_2 & = & a \, \mathbf{\hat{y}} \\
    \mathbf{a}_3 & = & c \, \mathbf{\hat{z}} \\

        \end{array}
      \end{equation*}
    }
    \renewcommand{\arraystretch}{1.0}
  \end{tabular}
  \begin{tabular}{c}
    \includegraphics[width=0.3\linewidth]{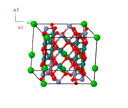} \\
  \end{tabular}
\end{tabular}

}
\vspace*{-0.25cm}

\noindent \hrulefill
\\
\textbf{Basis vectors:}
\vspace*{-0.25cm}
\renewcommand{\arraystretch}{1.5}
\begin{longtabu} to \textwidth{>{\centering $}X[-1,c,c]<{$}>{\centering $}X[-1,c,c]<{$}>{\centering $}X[-1,c,c]<{$}>{\centering $}X[-1,c,c]<{$}>{\centering $}X[-1,c,c]<{$}>{\centering $}X[-1,c,c]<{$}>{\centering $}X[-1,c,c]<{$}}
  & & \mbox{Lattice Coordinates} & & \mbox{Cartesian Coordinates} &\mbox{Wyckoff Position} & \mbox{Atom Type} \\  
  \mathbf{B}_{1} & = & z_{1} \, \mathbf{a}_{3} & = & z_{1}c \, \mathbf{\hat{z}} & \left(1a\right) & \mbox{Ba I} \\ 
\mathbf{B}_{2} & = & z_{2} \, \mathbf{a}_{3} & = & z_{2}c \, \mathbf{\hat{z}} & \left(1a\right) & \mbox{Ba II} \\ 
\mathbf{B}_{3} & = & \frac{1}{2} \, \mathbf{a}_{1} + \frac{1}{2} \, \mathbf{a}_{2} + z_{3} \, \mathbf{a}_{3} & = & \frac{1}{2}a \, \mathbf{\hat{x}} + \frac{1}{2}a \, \mathbf{\hat{y}} + z_{3}c \, \mathbf{\hat{z}} & \left(1b\right) & \mbox{Ba III} \\ 
\mathbf{B}_{4} & = & \frac{1}{2} \, \mathbf{a}_{1} + \frac{1}{2} \, \mathbf{a}_{2} + z_{4} \, \mathbf{a}_{3} & = & \frac{1}{2}a \, \mathbf{\hat{x}} + \frac{1}{2}a \, \mathbf{\hat{y}} + z_{4}c \, \mathbf{\hat{z}} & \left(1b\right) & \mbox{Ba IV} \\ 
\mathbf{B}_{5} & = & x_{5} \, \mathbf{a}_{1} + y_{5} \, \mathbf{a}_{2} + z_{5} \, \mathbf{a}_{3} & = & x_{5}a \, \mathbf{\hat{x}} + y_{5}a \, \mathbf{\hat{y}} + z_{5}c \, \mathbf{\hat{z}} & \left(4d\right) & \mbox{Cr I} \\ 
\mathbf{B}_{6} & = & -x_{5} \, \mathbf{a}_{1}-y_{5} \, \mathbf{a}_{2} + z_{5} \, \mathbf{a}_{3} & = & -x_{5}a \, \mathbf{\hat{x}}-y_{5}a \, \mathbf{\hat{y}} + z_{5}c \, \mathbf{\hat{z}} & \left(4d\right) & \mbox{Cr I} \\ 
\mathbf{B}_{7} & = & -y_{5} \, \mathbf{a}_{1} + x_{5} \, \mathbf{a}_{2} + z_{5} \, \mathbf{a}_{3} & = & -y_{5}a \, \mathbf{\hat{x}} + x_{5}a \, \mathbf{\hat{y}} + z_{5}c \, \mathbf{\hat{z}} & \left(4d\right) & \mbox{Cr I} \\ 
\mathbf{B}_{8} & = & y_{5} \, \mathbf{a}_{1}-x_{5} \, \mathbf{a}_{2} + z_{5} \, \mathbf{a}_{3} & = & y_{5}a \, \mathbf{\hat{x}}-x_{5}a \, \mathbf{\hat{y}} + z_{5}c \, \mathbf{\hat{z}} & \left(4d\right) & \mbox{Cr I} \\ 
\mathbf{B}_{9} & = & x_{6} \, \mathbf{a}_{1} + y_{6} \, \mathbf{a}_{2} + z_{6} \, \mathbf{a}_{3} & = & x_{6}a \, \mathbf{\hat{x}} + y_{6}a \, \mathbf{\hat{y}} + z_{6}c \, \mathbf{\hat{z}} & \left(4d\right) & \mbox{Cr II} \\ 
\mathbf{B}_{10} & = & -x_{6} \, \mathbf{a}_{1}-y_{6} \, \mathbf{a}_{2} + z_{6} \, \mathbf{a}_{3} & = & -x_{6}a \, \mathbf{\hat{x}}-y_{6}a \, \mathbf{\hat{y}} + z_{6}c \, \mathbf{\hat{z}} & \left(4d\right) & \mbox{Cr II} \\ 
\mathbf{B}_{11} & = & -y_{6} \, \mathbf{a}_{1} + x_{6} \, \mathbf{a}_{2} + z_{6} \, \mathbf{a}_{3} & = & -y_{6}a \, \mathbf{\hat{x}} + x_{6}a \, \mathbf{\hat{y}} + z_{6}c \, \mathbf{\hat{z}} & \left(4d\right) & \mbox{Cr II} \\ 
\mathbf{B}_{12} & = & y_{6} \, \mathbf{a}_{1}-x_{6} \, \mathbf{a}_{2} + z_{6} \, \mathbf{a}_{3} & = & y_{6}a \, \mathbf{\hat{x}}-x_{6}a \, \mathbf{\hat{y}} + z_{6}c \, \mathbf{\hat{z}} & \left(4d\right) & \mbox{Cr II} \\ 
\mathbf{B}_{13} & = & x_{7} \, \mathbf{a}_{1} + y_{7} \, \mathbf{a}_{2} + z_{7} \, \mathbf{a}_{3} & = & x_{7}a \, \mathbf{\hat{x}} + y_{7}a \, \mathbf{\hat{y}} + z_{7}c \, \mathbf{\hat{z}} & \left(4d\right) & \mbox{O I} \\ 
\mathbf{B}_{14} & = & -x_{7} \, \mathbf{a}_{1}-y_{7} \, \mathbf{a}_{2} + z_{7} \, \mathbf{a}_{3} & = & -x_{7}a \, \mathbf{\hat{x}}-y_{7}a \, \mathbf{\hat{y}} + z_{7}c \, \mathbf{\hat{z}} & \left(4d\right) & \mbox{O I} \\ 
\mathbf{B}_{15} & = & -y_{7} \, \mathbf{a}_{1} + x_{7} \, \mathbf{a}_{2} + z_{7} \, \mathbf{a}_{3} & = & -y_{7}a \, \mathbf{\hat{x}} + x_{7}a \, \mathbf{\hat{y}} + z_{7}c \, \mathbf{\hat{z}} & \left(4d\right) & \mbox{O I} \\ 
\mathbf{B}_{16} & = & y_{7} \, \mathbf{a}_{1}-x_{7} \, \mathbf{a}_{2} + z_{7} \, \mathbf{a}_{3} & = & y_{7}a \, \mathbf{\hat{x}}-x_{7}a \, \mathbf{\hat{y}} + z_{7}c \, \mathbf{\hat{z}} & \left(4d\right) & \mbox{O I} \\ 
\mathbf{B}_{17} & = & x_{8} \, \mathbf{a}_{1} + y_{8} \, \mathbf{a}_{2} + z_{8} \, \mathbf{a}_{3} & = & x_{8}a \, \mathbf{\hat{x}} + y_{8}a \, \mathbf{\hat{y}} + z_{8}c \, \mathbf{\hat{z}} & \left(4d\right) & \mbox{O II} \\ 
\mathbf{B}_{18} & = & -x_{8} \, \mathbf{a}_{1}-y_{8} \, \mathbf{a}_{2} + z_{8} \, \mathbf{a}_{3} & = & -x_{8}a \, \mathbf{\hat{x}}-y_{8}a \, \mathbf{\hat{y}} + z_{8}c \, \mathbf{\hat{z}} & \left(4d\right) & \mbox{O II} \\ 
\mathbf{B}_{19} & = & -y_{8} \, \mathbf{a}_{1} + x_{8} \, \mathbf{a}_{2} + z_{8} \, \mathbf{a}_{3} & = & -y_{8}a \, \mathbf{\hat{x}} + x_{8}a \, \mathbf{\hat{y}} + z_{8}c \, \mathbf{\hat{z}} & \left(4d\right) & \mbox{O II} \\ 
\mathbf{B}_{20} & = & y_{8} \, \mathbf{a}_{1}-x_{8} \, \mathbf{a}_{2} + z_{8} \, \mathbf{a}_{3} & = & y_{8}a \, \mathbf{\hat{x}}-x_{8}a \, \mathbf{\hat{y}} + z_{8}c \, \mathbf{\hat{z}} & \left(4d\right) & \mbox{O II} \\ 
\mathbf{B}_{21} & = & x_{9} \, \mathbf{a}_{1} + y_{9} \, \mathbf{a}_{2} + z_{9} \, \mathbf{a}_{3} & = & x_{9}a \, \mathbf{\hat{x}} + y_{9}a \, \mathbf{\hat{y}} + z_{9}c \, \mathbf{\hat{z}} & \left(4d\right) & \mbox{O III} \\ 
\mathbf{B}_{22} & = & -x_{9} \, \mathbf{a}_{1}-y_{9} \, \mathbf{a}_{2} + z_{9} \, \mathbf{a}_{3} & = & -x_{9}a \, \mathbf{\hat{x}}-y_{9}a \, \mathbf{\hat{y}} + z_{9}c \, \mathbf{\hat{z}} & \left(4d\right) & \mbox{O III} \\ 
\mathbf{B}_{23} & = & -y_{9} \, \mathbf{a}_{1} + x_{9} \, \mathbf{a}_{2} + z_{9} \, \mathbf{a}_{3} & = & -y_{9}a \, \mathbf{\hat{x}} + x_{9}a \, \mathbf{\hat{y}} + z_{9}c \, \mathbf{\hat{z}} & \left(4d\right) & \mbox{O III} \\ 
\mathbf{B}_{24} & = & y_{9} \, \mathbf{a}_{1}-x_{9} \, \mathbf{a}_{2} + z_{9} \, \mathbf{a}_{3} & = & y_{9}a \, \mathbf{\hat{x}}-x_{9}a \, \mathbf{\hat{y}} + z_{9}c \, \mathbf{\hat{z}} & \left(4d\right) & \mbox{O III} \\ 
\mathbf{B}_{25} & = & x_{10} \, \mathbf{a}_{1} + y_{10} \, \mathbf{a}_{2} + z_{10} \, \mathbf{a}_{3} & = & x_{10}a \, \mathbf{\hat{x}} + y_{10}a \, \mathbf{\hat{y}} + z_{10}c \, \mathbf{\hat{z}} & \left(4d\right) & \mbox{O IV} \\ 
\mathbf{B}_{26} & = & -x_{10} \, \mathbf{a}_{1}-y_{10} \, \mathbf{a}_{2} + z_{10} \, \mathbf{a}_{3} & = & -x_{10}a \, \mathbf{\hat{x}}-y_{10}a \, \mathbf{\hat{y}} + z_{10}c \, \mathbf{\hat{z}} & \left(4d\right) & \mbox{O IV} \\ 
\mathbf{B}_{27} & = & -y_{10} \, \mathbf{a}_{1} + x_{10} \, \mathbf{a}_{2} + z_{10} \, \mathbf{a}_{3} & = & -y_{10}a \, \mathbf{\hat{x}} + x_{10}a \, \mathbf{\hat{y}} + z_{10}c \, \mathbf{\hat{z}} & \left(4d\right) & \mbox{O IV} \\ 
\mathbf{B}_{28} & = & y_{10} \, \mathbf{a}_{1}-x_{10} \, \mathbf{a}_{2} + z_{10} \, \mathbf{a}_{3} & = & y_{10}a \, \mathbf{\hat{x}}-x_{10}a \, \mathbf{\hat{y}} + z_{10}c \, \mathbf{\hat{z}} & \left(4d\right) & \mbox{O IV} \\ 
\mathbf{B}_{29} & = & x_{11} \, \mathbf{a}_{1} + y_{11} \, \mathbf{a}_{2} + z_{11} \, \mathbf{a}_{3} & = & x_{11}a \, \mathbf{\hat{x}} + y_{11}a \, \mathbf{\hat{y}} + z_{11}c \, \mathbf{\hat{z}} & \left(4d\right) & \mbox{O V} \\ 
\mathbf{B}_{30} & = & -x_{11} \, \mathbf{a}_{1}-y_{11} \, \mathbf{a}_{2} + z_{11} \, \mathbf{a}_{3} & = & -x_{11}a \, \mathbf{\hat{x}}-y_{11}a \, \mathbf{\hat{y}} + z_{11}c \, \mathbf{\hat{z}} & \left(4d\right) & \mbox{O V} \\ 
\mathbf{B}_{31} & = & -y_{11} \, \mathbf{a}_{1} + x_{11} \, \mathbf{a}_{2} + z_{11} \, \mathbf{a}_{3} & = & -y_{11}a \, \mathbf{\hat{x}} + x_{11}a \, \mathbf{\hat{y}} + z_{11}c \, \mathbf{\hat{z}} & \left(4d\right) & \mbox{O V} \\ 
\mathbf{B}_{32} & = & y_{11} \, \mathbf{a}_{1}-x_{11} \, \mathbf{a}_{2} + z_{11} \, \mathbf{a}_{3} & = & y_{11}a \, \mathbf{\hat{x}}-x_{11}a \, \mathbf{\hat{y}} + z_{11}c \, \mathbf{\hat{z}} & \left(4d\right) & \mbox{O V} \\ 
\mathbf{B}_{33} & = & x_{12} \, \mathbf{a}_{1} + y_{12} \, \mathbf{a}_{2} + z_{12} \, \mathbf{a}_{3} & = & x_{12}a \, \mathbf{\hat{x}} + y_{12}a \, \mathbf{\hat{y}} + z_{12}c \, \mathbf{\hat{z}} & \left(4d\right) & \mbox{O VI} \\ 
\mathbf{B}_{34} & = & -x_{12} \, \mathbf{a}_{1}-y_{12} \, \mathbf{a}_{2} + z_{12} \, \mathbf{a}_{3} & = & -x_{12}a \, \mathbf{\hat{x}}-y_{12}a \, \mathbf{\hat{y}} + z_{12}c \, \mathbf{\hat{z}} & \left(4d\right) & \mbox{O VI} \\ 
\mathbf{B}_{35} & = & -y_{12} \, \mathbf{a}_{1} + x_{12} \, \mathbf{a}_{2} + z_{12} \, \mathbf{a}_{3} & = & -y_{12}a \, \mathbf{\hat{x}} + x_{12}a \, \mathbf{\hat{y}} + z_{12}c \, \mathbf{\hat{z}} & \left(4d\right) & \mbox{O VI} \\ 
\mathbf{B}_{36} & = & y_{12} \, \mathbf{a}_{1}-x_{12} \, \mathbf{a}_{2} + z_{12} \, \mathbf{a}_{3} & = & y_{12}a \, \mathbf{\hat{x}}-x_{12}a \, \mathbf{\hat{y}} + z_{12}c \, \mathbf{\hat{z}} & \left(4d\right) & \mbox{O VI} \\ 
\mathbf{B}_{37} & = & x_{13} \, \mathbf{a}_{1} + y_{13} \, \mathbf{a}_{2} + z_{13} \, \mathbf{a}_{3} & = & x_{13}a \, \mathbf{\hat{x}} + y_{13}a \, \mathbf{\hat{y}} + z_{13}c \, \mathbf{\hat{z}} & \left(4d\right) & \mbox{O VII} \\ 
\mathbf{B}_{38} & = & -x_{13} \, \mathbf{a}_{1}-y_{13} \, \mathbf{a}_{2} + z_{13} \, \mathbf{a}_{3} & = & -x_{13}a \, \mathbf{\hat{x}}-y_{13}a \, \mathbf{\hat{y}} + z_{13}c \, \mathbf{\hat{z}} & \left(4d\right) & \mbox{O VII} \\ 
\mathbf{B}_{39} & = & -y_{13} \, \mathbf{a}_{1} + x_{13} \, \mathbf{a}_{2} + z_{13} \, \mathbf{a}_{3} & = & -y_{13}a \, \mathbf{\hat{x}} + x_{13}a \, \mathbf{\hat{y}} + z_{13}c \, \mathbf{\hat{z}} & \left(4d\right) & \mbox{O VII} \\ 
\mathbf{B}_{40} & = & y_{13} \, \mathbf{a}_{1}-x_{13} \, \mathbf{a}_{2} + z_{13} \, \mathbf{a}_{3} & = & y_{13}a \, \mathbf{\hat{x}}-x_{13}a \, \mathbf{\hat{y}} + z_{13}c \, \mathbf{\hat{z}} & \left(4d\right) & \mbox{O VII} \\ 
\mathbf{B}_{41} & = & x_{14} \, \mathbf{a}_{1} + y_{14} \, \mathbf{a}_{2} + z_{14} \, \mathbf{a}_{3} & = & x_{14}a \, \mathbf{\hat{x}} + y_{14}a \, \mathbf{\hat{y}} + z_{14}c \, \mathbf{\hat{z}} & \left(4d\right) & \mbox{O VIII} \\ 
\mathbf{B}_{42} & = & -x_{14} \, \mathbf{a}_{1}-y_{14} \, \mathbf{a}_{2} + z_{14} \, \mathbf{a}_{3} & = & -x_{14}a \, \mathbf{\hat{x}}-y_{14}a \, \mathbf{\hat{y}} + z_{14}c \, \mathbf{\hat{z}} & \left(4d\right) & \mbox{O VIII} \\ 
\mathbf{B}_{43} & = & -y_{14} \, \mathbf{a}_{1} + x_{14} \, \mathbf{a}_{2} + z_{14} \, \mathbf{a}_{3} & = & -y_{14}a \, \mathbf{\hat{x}} + x_{14}a \, \mathbf{\hat{y}} + z_{14}c \, \mathbf{\hat{z}} & \left(4d\right) & \mbox{O VIII} \\ 
\mathbf{B}_{44} & = & y_{14} \, \mathbf{a}_{1}-x_{14} \, \mathbf{a}_{2} + z_{14} \, \mathbf{a}_{3} & = & y_{14}a \, \mathbf{\hat{x}}-x_{14}a \, \mathbf{\hat{y}} + z_{14}c \, \mathbf{\hat{z}} & \left(4d\right) & \mbox{O VIII} \\ 
\mathbf{B}_{45} & = & x_{15} \, \mathbf{a}_{1} + y_{15} \, \mathbf{a}_{2} + z_{15} \, \mathbf{a}_{3} & = & x_{15}a \, \mathbf{\hat{x}} + y_{15}a \, \mathbf{\hat{y}} + z_{15}c \, \mathbf{\hat{z}} & \left(4d\right) & \mbox{O IX} \\ 
\mathbf{B}_{46} & = & -x_{15} \, \mathbf{a}_{1}-y_{15} \, \mathbf{a}_{2} + z_{15} \, \mathbf{a}_{3} & = & -x_{15}a \, \mathbf{\hat{x}}-y_{15}a \, \mathbf{\hat{y}} + z_{15}c \, \mathbf{\hat{z}} & \left(4d\right) & \mbox{O IX} \\ 
\mathbf{B}_{47} & = & -y_{15} \, \mathbf{a}_{1} + x_{15} \, \mathbf{a}_{2} + z_{15} \, \mathbf{a}_{3} & = & -y_{15}a \, \mathbf{\hat{x}} + x_{15}a \, \mathbf{\hat{y}} + z_{15}c \, \mathbf{\hat{z}} & \left(4d\right) & \mbox{O IX} \\ 
\mathbf{B}_{48} & = & y_{15} \, \mathbf{a}_{1}-x_{15} \, \mathbf{a}_{2} + z_{15} \, \mathbf{a}_{3} & = & y_{15}a \, \mathbf{\hat{x}}-x_{15}a \, \mathbf{\hat{y}} + z_{15}c \, \mathbf{\hat{z}} & \left(4d\right) & \mbox{O IX} \\ 
\mathbf{B}_{49} & = & x_{16} \, \mathbf{a}_{1} + y_{16} \, \mathbf{a}_{2} + z_{16} \, \mathbf{a}_{3} & = & x_{16}a \, \mathbf{\hat{x}} + y_{16}a \, \mathbf{\hat{y}} + z_{16}c \, \mathbf{\hat{z}} & \left(4d\right) & \mbox{O X} \\ 
\mathbf{B}_{50} & = & -x_{16} \, \mathbf{a}_{1}-y_{16} \, \mathbf{a}_{2} + z_{16} \, \mathbf{a}_{3} & = & -x_{16}a \, \mathbf{\hat{x}}-y_{16}a \, \mathbf{\hat{y}} + z_{16}c \, \mathbf{\hat{z}} & \left(4d\right) & \mbox{O X} \\ 
\mathbf{B}_{51} & = & -y_{16} \, \mathbf{a}_{1} + x_{16} \, \mathbf{a}_{2} + z_{16} \, \mathbf{a}_{3} & = & -y_{16}a \, \mathbf{\hat{x}} + x_{16}a \, \mathbf{\hat{y}} + z_{16}c \, \mathbf{\hat{z}} & \left(4d\right) & \mbox{O X} \\ 
\mathbf{B}_{52} & = & y_{16} \, \mathbf{a}_{1}-x_{16} \, \mathbf{a}_{2} + z_{16} \, \mathbf{a}_{3} & = & y_{16}a \, \mathbf{\hat{x}}-x_{16}a \, \mathbf{\hat{y}} + z_{16}c \, \mathbf{\hat{z}} & \left(4d\right) & \mbox{O X} \\ 
\mathbf{B}_{53} & = & x_{17} \, \mathbf{a}_{1} + y_{17} \, \mathbf{a}_{2} + z_{17} \, \mathbf{a}_{3} & = & x_{17}a \, \mathbf{\hat{x}} + y_{17}a \, \mathbf{\hat{y}} + z_{17}c \, \mathbf{\hat{z}} & \left(4d\right) & \mbox{O XI} \\ 
\mathbf{B}_{54} & = & -x_{17} \, \mathbf{a}_{1}-y_{17} \, \mathbf{a}_{2} + z_{17} \, \mathbf{a}_{3} & = & -x_{17}a \, \mathbf{\hat{x}}-y_{17}a \, \mathbf{\hat{y}} + z_{17}c \, \mathbf{\hat{z}} & \left(4d\right) & \mbox{O XI} \\ 
\mathbf{B}_{55} & = & -y_{17} \, \mathbf{a}_{1} + x_{17} \, \mathbf{a}_{2} + z_{17} \, \mathbf{a}_{3} & = & -y_{17}a \, \mathbf{\hat{x}} + x_{17}a \, \mathbf{\hat{y}} + z_{17}c \, \mathbf{\hat{z}} & \left(4d\right) & \mbox{O XI} \\ 
\mathbf{B}_{56} & = & y_{17} \, \mathbf{a}_{1}-x_{17} \, \mathbf{a}_{2} + z_{17} \, \mathbf{a}_{3} & = & y_{17}a \, \mathbf{\hat{x}}-x_{17}a \, \mathbf{\hat{y}} + z_{17}c \, \mathbf{\hat{z}} & \left(4d\right) & \mbox{O XI} \\ 
\mathbf{B}_{57} & = & x_{18} \, \mathbf{a}_{1} + y_{18} \, \mathbf{a}_{2} + z_{18} \, \mathbf{a}_{3} & = & x_{18}a \, \mathbf{\hat{x}} + y_{18}a \, \mathbf{\hat{y}} + z_{18}c \, \mathbf{\hat{z}} & \left(4d\right) & \mbox{O XII} \\ 
\mathbf{B}_{58} & = & -x_{18} \, \mathbf{a}_{1}-y_{18} \, \mathbf{a}_{2} + z_{18} \, \mathbf{a}_{3} & = & -x_{18}a \, \mathbf{\hat{x}}-y_{18}a \, \mathbf{\hat{y}} + z_{18}c \, \mathbf{\hat{z}} & \left(4d\right) & \mbox{O XII} \\ 
\mathbf{B}_{59} & = & -y_{18} \, \mathbf{a}_{1} + x_{18} \, \mathbf{a}_{2} + z_{18} \, \mathbf{a}_{3} & = & -y_{18}a \, \mathbf{\hat{x}} + x_{18}a \, \mathbf{\hat{y}} + z_{18}c \, \mathbf{\hat{z}} & \left(4d\right) & \mbox{O XII} \\ 
\mathbf{B}_{60} & = & y_{18} \, \mathbf{a}_{1}-x_{18} \, \mathbf{a}_{2} + z_{18} \, \mathbf{a}_{3} & = & y_{18}a \, \mathbf{\hat{x}}-x_{18}a \, \mathbf{\hat{y}} + z_{18}c \, \mathbf{\hat{z}} & \left(4d\right) & \mbox{O XII} \\ 
\mathbf{B}_{61} & = & x_{19} \, \mathbf{a}_{1} + y_{19} \, \mathbf{a}_{2} + z_{19} \, \mathbf{a}_{3} & = & x_{19}a \, \mathbf{\hat{x}} + y_{19}a \, \mathbf{\hat{y}} + z_{19}c \, \mathbf{\hat{z}} & \left(4d\right) & \mbox{Ru I} \\ 
\mathbf{B}_{62} & = & -x_{19} \, \mathbf{a}_{1}-y_{19} \, \mathbf{a}_{2} + z_{19} \, \mathbf{a}_{3} & = & -x_{19}a \, \mathbf{\hat{x}}-y_{19}a \, \mathbf{\hat{y}} + z_{19}c \, \mathbf{\hat{z}} & \left(4d\right) & \mbox{Ru I} \\ 
\mathbf{B}_{63} & = & -y_{19} \, \mathbf{a}_{1} + x_{19} \, \mathbf{a}_{2} + z_{19} \, \mathbf{a}_{3} & = & -y_{19}a \, \mathbf{\hat{x}} + x_{19}a \, \mathbf{\hat{y}} + z_{19}c \, \mathbf{\hat{z}} & \left(4d\right) & \mbox{Ru I} \\ 
\mathbf{B}_{64} & = & y_{19} \, \mathbf{a}_{1}-x_{19} \, \mathbf{a}_{2} + z_{19} \, \mathbf{a}_{3} & = & y_{19}a \, \mathbf{\hat{x}}-x_{19}a \, \mathbf{\hat{y}} + z_{19}c \, \mathbf{\hat{z}} & \left(4d\right) & \mbox{Ru I} \\ 
\mathbf{B}_{65} & = & x_{20} \, \mathbf{a}_{1} + y_{20} \, \mathbf{a}_{2} + z_{20} \, \mathbf{a}_{3} & = & x_{20}a \, \mathbf{\hat{x}} + y_{20}a \, \mathbf{\hat{y}} + z_{20}c \, \mathbf{\hat{z}} & \left(4d\right) & \mbox{Ru II} \\ 
\mathbf{B}_{66} & = & -x_{20} \, \mathbf{a}_{1}-y_{20} \, \mathbf{a}_{2} + z_{20} \, \mathbf{a}_{3} & = & -x_{20}a \, \mathbf{\hat{x}}-y_{20}a \, \mathbf{\hat{y}} + z_{20}c \, \mathbf{\hat{z}} & \left(4d\right) & \mbox{Ru II} \\ 
\mathbf{B}_{67} & = & -y_{20} \, \mathbf{a}_{1} + x_{20} \, \mathbf{a}_{2} + z_{20} \, \mathbf{a}_{3} & = & -y_{20}a \, \mathbf{\hat{x}} + x_{20}a \, \mathbf{\hat{y}} + z_{20}c \, \mathbf{\hat{z}} & \left(4d\right) & \mbox{Ru II} \\ 
\mathbf{B}_{68} & = & y_{20} \, \mathbf{a}_{1}-x_{20} \, \mathbf{a}_{2} + z_{20} \, \mathbf{a}_{3} & = & y_{20}a \, \mathbf{\hat{x}}-x_{20}a \, \mathbf{\hat{y}} + z_{20}c \, \mathbf{\hat{z}} & \left(4d\right) & \mbox{Ru II} \\ 
\mathbf{B}_{69} & = & x_{21} \, \mathbf{a}_{1} + y_{21} \, \mathbf{a}_{2} + z_{21} \, \mathbf{a}_{3} & = & x_{21}a \, \mathbf{\hat{x}} + y_{21}a \, \mathbf{\hat{y}} + z_{21}c \, \mathbf{\hat{z}} & \left(4d\right) & \mbox{Ru III} \\ 
\mathbf{B}_{70} & = & -x_{21} \, \mathbf{a}_{1}-y_{21} \, \mathbf{a}_{2} + z_{21} \, \mathbf{a}_{3} & = & -x_{21}a \, \mathbf{\hat{x}}-y_{21}a \, \mathbf{\hat{y}} + z_{21}c \, \mathbf{\hat{z}} & \left(4d\right) & \mbox{Ru III} \\ 
\mathbf{B}_{71} & = & -y_{21} \, \mathbf{a}_{1} + x_{21} \, \mathbf{a}_{2} + z_{21} \, \mathbf{a}_{3} & = & -y_{21}a \, \mathbf{\hat{x}} + x_{21}a \, \mathbf{\hat{y}} + z_{21}c \, \mathbf{\hat{z}} & \left(4d\right) & \mbox{Ru III} \\ 
\mathbf{B}_{72} & = & y_{21} \, \mathbf{a}_{1}-x_{21} \, \mathbf{a}_{2} + z_{21} \, \mathbf{a}_{3} & = & y_{21}a \, \mathbf{\hat{x}}-x_{21}a \, \mathbf{\hat{y}} + z_{21}c \, \mathbf{\hat{z}} & \left(4d\right) & \mbox{Ru III} \\ 
\mathbf{B}_{73} & = & x_{22} \, \mathbf{a}_{1} + y_{22} \, \mathbf{a}_{2} + z_{22} \, \mathbf{a}_{3} & = & x_{22}a \, \mathbf{\hat{x}} + y_{22}a \, \mathbf{\hat{y}} + z_{22}c \, \mathbf{\hat{z}} & \left(4d\right) & \mbox{Ru IV} \\ 
\mathbf{B}_{74} & = & -x_{22} \, \mathbf{a}_{1}-y_{22} \, \mathbf{a}_{2} + z_{22} \, \mathbf{a}_{3} & = & -x_{22}a \, \mathbf{\hat{x}}-y_{22}a \, \mathbf{\hat{y}} + z_{22}c \, \mathbf{\hat{z}} & \left(4d\right) & \mbox{Ru IV} \\ 
\mathbf{B}_{75} & = & -y_{22} \, \mathbf{a}_{1} + x_{22} \, \mathbf{a}_{2} + z_{22} \, \mathbf{a}_{3} & = & -y_{22}a \, \mathbf{\hat{x}} + x_{22}a \, \mathbf{\hat{y}} + z_{22}c \, \mathbf{\hat{z}} & \left(4d\right) & \mbox{Ru IV} \\ 
\mathbf{B}_{76} & = & y_{22} \, \mathbf{a}_{1}-x_{22} \, \mathbf{a}_{2} + z_{22} \, \mathbf{a}_{3} & = & y_{22}a \, \mathbf{\hat{x}}-x_{22}a \, \mathbf{\hat{y}} + z_{22}c \, \mathbf{\hat{z}} & \left(4d\right) & \mbox{Ru IV} \\ 
\end{longtabu}
\renewcommand{\arraystretch}{1.0}
\noindent \hrulefill
\\
\textbf{References:}
\vspace*{-0.25cm}
\begin{flushleft}
  - \bibentry{Cadee_BaCr2Ru4O12_MatResBull_1979}. \\
\end{flushleft}
\textbf{Found in:}
\vspace*{-0.25cm}
\begin{flushleft}
  - \bibentry{Villars_PearsonsCrystalData_2013}. \\
\end{flushleft}
\noindent \hrulefill
\\
\textbf{Geometry files:}
\\
\noindent  - CIF: pp. {\hyperref[AB2C12D4_tP76_75_2a2b_2d_12d_4d_cif]{\pageref{AB2C12D4_tP76_75_2a2b_2d_12d_4d_cif}}} \\
\noindent  - POSCAR: pp. {\hyperref[AB2C12D4_tP76_75_2a2b_2d_12d_4d_poscar]{\pageref{AB2C12D4_tP76_75_2a2b_2d_12d_4d_poscar}}} \\
\onecolumn
{\phantomsection\label{A2BC_tP16_76_2a_a_a}}
\subsection*{\huge \textbf{{\normalfont LaRhC$_{2}$ Structure: A2BC\_tP16\_76\_2a\_a\_a}}}
\noindent \hrulefill
\vspace*{0.25cm}
\begin{figure}[htp]
  \centering
  \vspace{-1em}
  {\includegraphics[width=1\textwidth]{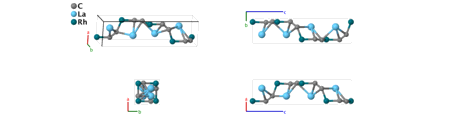}}
\end{figure}
\vspace*{-0.5cm}
\renewcommand{\arraystretch}{1.5}
\begin{equation*}
  \begin{array}{>{$\hspace{-0.15cm}}l<{$}>{$}p{0.5cm}<{$}>{$}p{18.5cm}<{$}}
    \mbox{\large \textbf{Prototype}} &\colon & \ce{LaRhC2} \\
    \mbox{\large \textbf{\AFLOW\ prototype label}} &\colon & \mbox{A2BC\_tP16\_76\_2a\_a\_a} \\
    \mbox{\large \textbf{\textit{Strukturbericht} designation}} &\colon & \mbox{None} \\
    \mbox{\large \textbf{Pearson symbol}} &\colon & \mbox{tP16} \\
    \mbox{\large \textbf{Space group number}} &\colon & 76 \\
    \mbox{\large \textbf{Space group symbol}} &\colon & P4_{1} \\
    \mbox{\large \textbf{\AFLOW\ prototype command}} &\colon &  \texttt{aflow} \,  \, \texttt{-{}-proto=A2BC\_tP16\_76\_2a\_a\_a } \, \newline \texttt{-{}-params=}{a,c/a,x_{1},y_{1},z_{1},x_{2},y_{2},z_{2},x_{3},y_{3},z_{3},x_{4},y_{4},z_{4} }
  \end{array}
\end{equation*}
\renewcommand{\arraystretch}{1.0}

\noindent \parbox{1 \linewidth}{
\noindent \hrulefill
\\
\textbf{Simple Tetragonal primitive vectors:} \\
\vspace*{-0.25cm}
\begin{tabular}{cc}
  \begin{tabular}{c}
    \parbox{0.6 \linewidth}{
      \renewcommand{\arraystretch}{1.5}
      \begin{equation*}
        \centering
        \begin{array}{ccc}
              \mathbf{a}_1 & = & a \, \mathbf{\hat{x}} \\
    \mathbf{a}_2 & = & a \, \mathbf{\hat{y}} \\
    \mathbf{a}_3 & = & c \, \mathbf{\hat{z}} \\

        \end{array}
      \end{equation*}
    }
    \renewcommand{\arraystretch}{1.0}
  \end{tabular}
  \begin{tabular}{c}
    \includegraphics[width=0.3\linewidth]{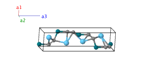} \\
  \end{tabular}
\end{tabular}

}
\vspace*{-0.25cm}

\noindent \hrulefill
\\
\textbf{Basis vectors:}
\vspace*{-0.25cm}
\renewcommand{\arraystretch}{1.5}
\begin{longtabu} to \textwidth{>{\centering $}X[-1,c,c]<{$}>{\centering $}X[-1,c,c]<{$}>{\centering $}X[-1,c,c]<{$}>{\centering $}X[-1,c,c]<{$}>{\centering $}X[-1,c,c]<{$}>{\centering $}X[-1,c,c]<{$}>{\centering $}X[-1,c,c]<{$}}
  & & \mbox{Lattice Coordinates} & & \mbox{Cartesian Coordinates} &\mbox{Wyckoff Position} & \mbox{Atom Type} \\  
  \mathbf{B}_{1} & = & x_{1} \, \mathbf{a}_{1} + y_{1} \, \mathbf{a}_{2} + z_{1} \, \mathbf{a}_{3} & = & x_{1}a \, \mathbf{\hat{x}} + y_{1}a \, \mathbf{\hat{y}} + z_{1}c \, \mathbf{\hat{z}} & \left(4a\right) & \mbox{C I} \\ 
\mathbf{B}_{2} & = & -x_{1} \, \mathbf{a}_{1}-y_{1} \, \mathbf{a}_{2} + \left(\frac{1}{2} +z_{1}\right) \, \mathbf{a}_{3} & = & -x_{1}a \, \mathbf{\hat{x}}-y_{1}a \, \mathbf{\hat{y}} + \left(\frac{1}{2} +z_{1}\right)c \, \mathbf{\hat{z}} & \left(4a\right) & \mbox{C I} \\ 
\mathbf{B}_{3} & = & -y_{1} \, \mathbf{a}_{1} + x_{1} \, \mathbf{a}_{2} + \left(\frac{1}{4} +z_{1}\right) \, \mathbf{a}_{3} & = & -y_{1}a \, \mathbf{\hat{x}} + x_{1}a \, \mathbf{\hat{y}} + \left(\frac{1}{4} +z_{1}\right)c \, \mathbf{\hat{z}} & \left(4a\right) & \mbox{C I} \\ 
\mathbf{B}_{4} & = & y_{1} \, \mathbf{a}_{1}-x_{1} \, \mathbf{a}_{2} + \left(\frac{3}{4} +z_{1}\right) \, \mathbf{a}_{3} & = & y_{1}a \, \mathbf{\hat{x}}-x_{1}a \, \mathbf{\hat{y}} + \left(\frac{3}{4} +z_{1}\right)c \, \mathbf{\hat{z}} & \left(4a\right) & \mbox{C I} \\ 
\mathbf{B}_{5} & = & x_{2} \, \mathbf{a}_{1} + y_{2} \, \mathbf{a}_{2} + z_{2} \, \mathbf{a}_{3} & = & x_{2}a \, \mathbf{\hat{x}} + y_{2}a \, \mathbf{\hat{y}} + z_{2}c \, \mathbf{\hat{z}} & \left(4a\right) & \mbox{C II} \\ 
\mathbf{B}_{6} & = & -x_{2} \, \mathbf{a}_{1}-y_{2} \, \mathbf{a}_{2} + \left(\frac{1}{2} +z_{2}\right) \, \mathbf{a}_{3} & = & -x_{2}a \, \mathbf{\hat{x}}-y_{2}a \, \mathbf{\hat{y}} + \left(\frac{1}{2} +z_{2}\right)c \, \mathbf{\hat{z}} & \left(4a\right) & \mbox{C II} \\ 
\mathbf{B}_{7} & = & -y_{2} \, \mathbf{a}_{1} + x_{2} \, \mathbf{a}_{2} + \left(\frac{1}{4} +z_{2}\right) \, \mathbf{a}_{3} & = & -y_{2}a \, \mathbf{\hat{x}} + x_{2}a \, \mathbf{\hat{y}} + \left(\frac{1}{4} +z_{2}\right)c \, \mathbf{\hat{z}} & \left(4a\right) & \mbox{C II} \\ 
\mathbf{B}_{8} & = & y_{2} \, \mathbf{a}_{1}-x_{2} \, \mathbf{a}_{2} + \left(\frac{3}{4} +z_{2}\right) \, \mathbf{a}_{3} & = & y_{2}a \, \mathbf{\hat{x}}-x_{2}a \, \mathbf{\hat{y}} + \left(\frac{3}{4} +z_{2}\right)c \, \mathbf{\hat{z}} & \left(4a\right) & \mbox{C II} \\ 
\mathbf{B}_{9} & = & x_{3} \, \mathbf{a}_{1} + y_{3} \, \mathbf{a}_{2} + z_{3} \, \mathbf{a}_{3} & = & x_{3}a \, \mathbf{\hat{x}} + y_{3}a \, \mathbf{\hat{y}} + z_{3}c \, \mathbf{\hat{z}} & \left(4a\right) & \mbox{La} \\ 
\mathbf{B}_{10} & = & -x_{3} \, \mathbf{a}_{1}-y_{3} \, \mathbf{a}_{2} + \left(\frac{1}{2} +z_{3}\right) \, \mathbf{a}_{3} & = & -x_{3}a \, \mathbf{\hat{x}}-y_{3}a \, \mathbf{\hat{y}} + \left(\frac{1}{2} +z_{3}\right)c \, \mathbf{\hat{z}} & \left(4a\right) & \mbox{La} \\ 
\mathbf{B}_{11} & = & -y_{3} \, \mathbf{a}_{1} + x_{3} \, \mathbf{a}_{2} + \left(\frac{1}{4} +z_{3}\right) \, \mathbf{a}_{3} & = & -y_{3}a \, \mathbf{\hat{x}} + x_{3}a \, \mathbf{\hat{y}} + \left(\frac{1}{4} +z_{3}\right)c \, \mathbf{\hat{z}} & \left(4a\right) & \mbox{La} \\ 
\mathbf{B}_{12} & = & y_{3} \, \mathbf{a}_{1}-x_{3} \, \mathbf{a}_{2} + \left(\frac{3}{4} +z_{3}\right) \, \mathbf{a}_{3} & = & y_{3}a \, \mathbf{\hat{x}}-x_{3}a \, \mathbf{\hat{y}} + \left(\frac{3}{4} +z_{3}\right)c \, \mathbf{\hat{z}} & \left(4a\right) & \mbox{La} \\ 
\mathbf{B}_{13} & = & x_{4} \, \mathbf{a}_{1} + y_{4} \, \mathbf{a}_{2} + z_{4} \, \mathbf{a}_{3} & = & x_{4}a \, \mathbf{\hat{x}} + y_{4}a \, \mathbf{\hat{y}} + z_{4}c \, \mathbf{\hat{z}} & \left(4a\right) & \mbox{Rh} \\ 
\mathbf{B}_{14} & = & -x_{4} \, \mathbf{a}_{1}-y_{4} \, \mathbf{a}_{2} + \left(\frac{1}{2} +z_{4}\right) \, \mathbf{a}_{3} & = & -x_{4}a \, \mathbf{\hat{x}}-y_{4}a \, \mathbf{\hat{y}} + \left(\frac{1}{2} +z_{4}\right)c \, \mathbf{\hat{z}} & \left(4a\right) & \mbox{Rh} \\ 
\mathbf{B}_{15} & = & -y_{4} \, \mathbf{a}_{1} + x_{4} \, \mathbf{a}_{2} + \left(\frac{1}{4} +z_{4}\right) \, \mathbf{a}_{3} & = & -y_{4}a \, \mathbf{\hat{x}} + x_{4}a \, \mathbf{\hat{y}} + \left(\frac{1}{4} +z_{4}\right)c \, \mathbf{\hat{z}} & \left(4a\right) & \mbox{Rh} \\ 
\mathbf{B}_{16} & = & y_{4} \, \mathbf{a}_{1}-x_{4} \, \mathbf{a}_{2} + \left(\frac{3}{4} +z_{4}\right) \, \mathbf{a}_{3} & = & y_{4}a \, \mathbf{\hat{x}}-x_{4}a \, \mathbf{\hat{y}} + \left(\frac{3}{4} +z_{4}\right)c \, \mathbf{\hat{z}} & \left(4a\right) & \mbox{Rh} \\ 
\end{longtabu}
\renewcommand{\arraystretch}{1.0}
\noindent \hrulefill
\\
\textbf{References:}
\vspace*{-0.25cm}
\begin{flushleft}
  - \bibentry{Tsokol_LaRhC2_SovPhysCrystallog_1988}. \\
\end{flushleft}
\textbf{Found in:}
\vspace*{-0.25cm}
\begin{flushleft}
  - \bibentry{Villars_PearsonsCrystalData_2013}. \\
\end{flushleft}
\noindent \hrulefill
\\
\textbf{Geometry files:}
\\
\noindent  - CIF: pp. {\hyperref[A2BC_tP16_76_2a_a_a_cif]{\pageref{A2BC_tP16_76_2a_a_a_cif}}} \\
\noindent  - POSCAR: pp. {\hyperref[A2BC_tP16_76_2a_a_a_poscar]{\pageref{A2BC_tP16_76_2a_a_a_poscar}}} \\
\onecolumn
{\phantomsection\label{A3B7_tP40_76_3a_7a}}
\subsection*{\huge \textbf{{\normalfont Cs$_{3}$P$_{7}$ Structure: A3B7\_tP40\_76\_3a\_7a}}}
\noindent \hrulefill
\vspace*{0.25cm}
\begin{figure}[htp]
  \centering
  \vspace{-1em}
  {\includegraphics[width=1\textwidth]{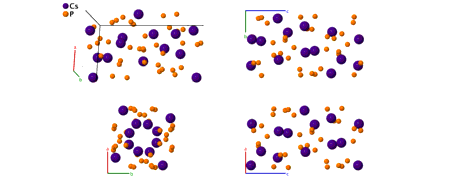}}
\end{figure}
\vspace*{-0.5cm}
\renewcommand{\arraystretch}{1.5}
\begin{equation*}
  \begin{array}{>{$\hspace{-0.15cm}}l<{$}>{$}p{0.5cm}<{$}>{$}p{18.5cm}<{$}}
    \mbox{\large \textbf{Prototype}} &\colon & \ce{Cs$_{3}$P$_{7}$} \\
    \mbox{\large \textbf{\AFLOW\ prototype label}} &\colon & \mbox{A3B7\_tP40\_76\_3a\_7a} \\
    \mbox{\large \textbf{\textit{Strukturbericht} designation}} &\colon & \mbox{None} \\
    \mbox{\large \textbf{Pearson symbol}} &\colon & \mbox{tP40} \\
    \mbox{\large \textbf{Space group number}} &\colon & 76 \\
    \mbox{\large \textbf{Space group symbol}} &\colon & P4_{1} \\
    \mbox{\large \textbf{\AFLOW\ prototype command}} &\colon &  \texttt{aflow} \,  \, \texttt{-{}-proto=A3B7\_tP40\_76\_3a\_7a } \, \newline \texttt{-{}-params=}{a,c/a,x_{1},y_{1},z_{1},x_{2},y_{2},z_{2},x_{3},y_{3},z_{3},x_{4},y_{4},z_{4},x_{5},y_{5},z_{5},x_{6},y_{6},z_{6},x_{7},} \newline {y_{7},z_{7},x_{8},y_{8},z_{8},x_{9},y_{9},z_{9},x_{10},y_{10},z_{10} }
  \end{array}
\end{equation*}
\renewcommand{\arraystretch}{1.0}

\noindent \parbox{1 \linewidth}{
\noindent \hrulefill
\\
\textbf{Simple Tetragonal primitive vectors:} \\
\vspace*{-0.25cm}
\begin{tabular}{cc}
  \begin{tabular}{c}
    \parbox{0.6 \linewidth}{
      \renewcommand{\arraystretch}{1.5}
      \begin{equation*}
        \centering
        \begin{array}{ccc}
              \mathbf{a}_1 & = & a \, \mathbf{\hat{x}} \\
    \mathbf{a}_2 & = & a \, \mathbf{\hat{y}} \\
    \mathbf{a}_3 & = & c \, \mathbf{\hat{z}} \\

        \end{array}
      \end{equation*}
    }
    \renewcommand{\arraystretch}{1.0}
  \end{tabular}
  \begin{tabular}{c}
    \includegraphics[width=0.3\linewidth]{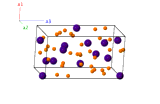} \\
  \end{tabular}
\end{tabular}

}
\vspace*{-0.25cm}

\noindent \hrulefill
\\
\textbf{Basis vectors:}
\vspace*{-0.25cm}
\renewcommand{\arraystretch}{1.5}
\begin{longtabu} to \textwidth{>{\centering $}X[-1,c,c]<{$}>{\centering $}X[-1,c,c]<{$}>{\centering $}X[-1,c,c]<{$}>{\centering $}X[-1,c,c]<{$}>{\centering $}X[-1,c,c]<{$}>{\centering $}X[-1,c,c]<{$}>{\centering $}X[-1,c,c]<{$}}
  & & \mbox{Lattice Coordinates} & & \mbox{Cartesian Coordinates} &\mbox{Wyckoff Position} & \mbox{Atom Type} \\  
  \mathbf{B}_{1} & = & x_{1} \, \mathbf{a}_{1} + y_{1} \, \mathbf{a}_{2} + z_{1} \, \mathbf{a}_{3} & = & x_{1}a \, \mathbf{\hat{x}} + y_{1}a \, \mathbf{\hat{y}} + z_{1}c \, \mathbf{\hat{z}} & \left(4a\right) & \mbox{Cs I} \\ 
\mathbf{B}_{2} & = & -x_{1} \, \mathbf{a}_{1}-y_{1} \, \mathbf{a}_{2} + \left(\frac{1}{2} +z_{1}\right) \, \mathbf{a}_{3} & = & -x_{1}a \, \mathbf{\hat{x}}-y_{1}a \, \mathbf{\hat{y}} + \left(\frac{1}{2} +z_{1}\right)c \, \mathbf{\hat{z}} & \left(4a\right) & \mbox{Cs I} \\ 
\mathbf{B}_{3} & = & -y_{1} \, \mathbf{a}_{1} + x_{1} \, \mathbf{a}_{2} + \left(\frac{1}{4} +z_{1}\right) \, \mathbf{a}_{3} & = & -y_{1}a \, \mathbf{\hat{x}} + x_{1}a \, \mathbf{\hat{y}} + \left(\frac{1}{4} +z_{1}\right)c \, \mathbf{\hat{z}} & \left(4a\right) & \mbox{Cs I} \\ 
\mathbf{B}_{4} & = & y_{1} \, \mathbf{a}_{1}-x_{1} \, \mathbf{a}_{2} + \left(\frac{3}{4} +z_{1}\right) \, \mathbf{a}_{3} & = & y_{1}a \, \mathbf{\hat{x}}-x_{1}a \, \mathbf{\hat{y}} + \left(\frac{3}{4} +z_{1}\right)c \, \mathbf{\hat{z}} & \left(4a\right) & \mbox{Cs I} \\ 
\mathbf{B}_{5} & = & x_{2} \, \mathbf{a}_{1} + y_{2} \, \mathbf{a}_{2} + z_{2} \, \mathbf{a}_{3} & = & x_{2}a \, \mathbf{\hat{x}} + y_{2}a \, \mathbf{\hat{y}} + z_{2}c \, \mathbf{\hat{z}} & \left(4a\right) & \mbox{Cs II} \\ 
\mathbf{B}_{6} & = & -x_{2} \, \mathbf{a}_{1}-y_{2} \, \mathbf{a}_{2} + \left(\frac{1}{2} +z_{2}\right) \, \mathbf{a}_{3} & = & -x_{2}a \, \mathbf{\hat{x}}-y_{2}a \, \mathbf{\hat{y}} + \left(\frac{1}{2} +z_{2}\right)c \, \mathbf{\hat{z}} & \left(4a\right) & \mbox{Cs II} \\ 
\mathbf{B}_{7} & = & -y_{2} \, \mathbf{a}_{1} + x_{2} \, \mathbf{a}_{2} + \left(\frac{1}{4} +z_{2}\right) \, \mathbf{a}_{3} & = & -y_{2}a \, \mathbf{\hat{x}} + x_{2}a \, \mathbf{\hat{y}} + \left(\frac{1}{4} +z_{2}\right)c \, \mathbf{\hat{z}} & \left(4a\right) & \mbox{Cs II} \\ 
\mathbf{B}_{8} & = & y_{2} \, \mathbf{a}_{1}-x_{2} \, \mathbf{a}_{2} + \left(\frac{3}{4} +z_{2}\right) \, \mathbf{a}_{3} & = & y_{2}a \, \mathbf{\hat{x}}-x_{2}a \, \mathbf{\hat{y}} + \left(\frac{3}{4} +z_{2}\right)c \, \mathbf{\hat{z}} & \left(4a\right) & \mbox{Cs II} \\ 
\mathbf{B}_{9} & = & x_{3} \, \mathbf{a}_{1} + y_{3} \, \mathbf{a}_{2} + z_{3} \, \mathbf{a}_{3} & = & x_{3}a \, \mathbf{\hat{x}} + y_{3}a \, \mathbf{\hat{y}} + z_{3}c \, \mathbf{\hat{z}} & \left(4a\right) & \mbox{Cs III} \\ 
\mathbf{B}_{10} & = & -x_{3} \, \mathbf{a}_{1}-y_{3} \, \mathbf{a}_{2} + \left(\frac{1}{2} +z_{3}\right) \, \mathbf{a}_{3} & = & -x_{3}a \, \mathbf{\hat{x}}-y_{3}a \, \mathbf{\hat{y}} + \left(\frac{1}{2} +z_{3}\right)c \, \mathbf{\hat{z}} & \left(4a\right) & \mbox{Cs III} \\ 
\mathbf{B}_{11} & = & -y_{3} \, \mathbf{a}_{1} + x_{3} \, \mathbf{a}_{2} + \left(\frac{1}{4} +z_{3}\right) \, \mathbf{a}_{3} & = & -y_{3}a \, \mathbf{\hat{x}} + x_{3}a \, \mathbf{\hat{y}} + \left(\frac{1}{4} +z_{3}\right)c \, \mathbf{\hat{z}} & \left(4a\right) & \mbox{Cs III} \\ 
\mathbf{B}_{12} & = & y_{3} \, \mathbf{a}_{1}-x_{3} \, \mathbf{a}_{2} + \left(\frac{3}{4} +z_{3}\right) \, \mathbf{a}_{3} & = & y_{3}a \, \mathbf{\hat{x}}-x_{3}a \, \mathbf{\hat{y}} + \left(\frac{3}{4} +z_{3}\right)c \, \mathbf{\hat{z}} & \left(4a\right) & \mbox{Cs III} \\ 
\mathbf{B}_{13} & = & x_{4} \, \mathbf{a}_{1} + y_{4} \, \mathbf{a}_{2} + z_{4} \, \mathbf{a}_{3} & = & x_{4}a \, \mathbf{\hat{x}} + y_{4}a \, \mathbf{\hat{y}} + z_{4}c \, \mathbf{\hat{z}} & \left(4a\right) & \mbox{P I} \\ 
\mathbf{B}_{14} & = & -x_{4} \, \mathbf{a}_{1}-y_{4} \, \mathbf{a}_{2} + \left(\frac{1}{2} +z_{4}\right) \, \mathbf{a}_{3} & = & -x_{4}a \, \mathbf{\hat{x}}-y_{4}a \, \mathbf{\hat{y}} + \left(\frac{1}{2} +z_{4}\right)c \, \mathbf{\hat{z}} & \left(4a\right) & \mbox{P I} \\ 
\mathbf{B}_{15} & = & -y_{4} \, \mathbf{a}_{1} + x_{4} \, \mathbf{a}_{2} + \left(\frac{1}{4} +z_{4}\right) \, \mathbf{a}_{3} & = & -y_{4}a \, \mathbf{\hat{x}} + x_{4}a \, \mathbf{\hat{y}} + \left(\frac{1}{4} +z_{4}\right)c \, \mathbf{\hat{z}} & \left(4a\right) & \mbox{P I} \\ 
\mathbf{B}_{16} & = & y_{4} \, \mathbf{a}_{1}-x_{4} \, \mathbf{a}_{2} + \left(\frac{3}{4} +z_{4}\right) \, \mathbf{a}_{3} & = & y_{4}a \, \mathbf{\hat{x}}-x_{4}a \, \mathbf{\hat{y}} + \left(\frac{3}{4} +z_{4}\right)c \, \mathbf{\hat{z}} & \left(4a\right) & \mbox{P I} \\ 
\mathbf{B}_{17} & = & x_{5} \, \mathbf{a}_{1} + y_{5} \, \mathbf{a}_{2} + z_{5} \, \mathbf{a}_{3} & = & x_{5}a \, \mathbf{\hat{x}} + y_{5}a \, \mathbf{\hat{y}} + z_{5}c \, \mathbf{\hat{z}} & \left(4a\right) & \mbox{P II} \\ 
\mathbf{B}_{18} & = & -x_{5} \, \mathbf{a}_{1}-y_{5} \, \mathbf{a}_{2} + \left(\frac{1}{2} +z_{5}\right) \, \mathbf{a}_{3} & = & -x_{5}a \, \mathbf{\hat{x}}-y_{5}a \, \mathbf{\hat{y}} + \left(\frac{1}{2} +z_{5}\right)c \, \mathbf{\hat{z}} & \left(4a\right) & \mbox{P II} \\ 
\mathbf{B}_{19} & = & -y_{5} \, \mathbf{a}_{1} + x_{5} \, \mathbf{a}_{2} + \left(\frac{1}{4} +z_{5}\right) \, \mathbf{a}_{3} & = & -y_{5}a \, \mathbf{\hat{x}} + x_{5}a \, \mathbf{\hat{y}} + \left(\frac{1}{4} +z_{5}\right)c \, \mathbf{\hat{z}} & \left(4a\right) & \mbox{P II} \\ 
\mathbf{B}_{20} & = & y_{5} \, \mathbf{a}_{1}-x_{5} \, \mathbf{a}_{2} + \left(\frac{3}{4} +z_{5}\right) \, \mathbf{a}_{3} & = & y_{5}a \, \mathbf{\hat{x}}-x_{5}a \, \mathbf{\hat{y}} + \left(\frac{3}{4} +z_{5}\right)c \, \mathbf{\hat{z}} & \left(4a\right) & \mbox{P II} \\ 
\mathbf{B}_{21} & = & x_{6} \, \mathbf{a}_{1} + y_{6} \, \mathbf{a}_{2} + z_{6} \, \mathbf{a}_{3} & = & x_{6}a \, \mathbf{\hat{x}} + y_{6}a \, \mathbf{\hat{y}} + z_{6}c \, \mathbf{\hat{z}} & \left(4a\right) & \mbox{P III} \\ 
\mathbf{B}_{22} & = & -x_{6} \, \mathbf{a}_{1}-y_{6} \, \mathbf{a}_{2} + \left(\frac{1}{2} +z_{6}\right) \, \mathbf{a}_{3} & = & -x_{6}a \, \mathbf{\hat{x}}-y_{6}a \, \mathbf{\hat{y}} + \left(\frac{1}{2} +z_{6}\right)c \, \mathbf{\hat{z}} & \left(4a\right) & \mbox{P III} \\ 
\mathbf{B}_{23} & = & -y_{6} \, \mathbf{a}_{1} + x_{6} \, \mathbf{a}_{2} + \left(\frac{1}{4} +z_{6}\right) \, \mathbf{a}_{3} & = & -y_{6}a \, \mathbf{\hat{x}} + x_{6}a \, \mathbf{\hat{y}} + \left(\frac{1}{4} +z_{6}\right)c \, \mathbf{\hat{z}} & \left(4a\right) & \mbox{P III} \\ 
\mathbf{B}_{24} & = & y_{6} \, \mathbf{a}_{1}-x_{6} \, \mathbf{a}_{2} + \left(\frac{3}{4} +z_{6}\right) \, \mathbf{a}_{3} & = & y_{6}a \, \mathbf{\hat{x}}-x_{6}a \, \mathbf{\hat{y}} + \left(\frac{3}{4} +z_{6}\right)c \, \mathbf{\hat{z}} & \left(4a\right) & \mbox{P III} \\ 
\mathbf{B}_{25} & = & x_{7} \, \mathbf{a}_{1} + y_{7} \, \mathbf{a}_{2} + z_{7} \, \mathbf{a}_{3} & = & x_{7}a \, \mathbf{\hat{x}} + y_{7}a \, \mathbf{\hat{y}} + z_{7}c \, \mathbf{\hat{z}} & \left(4a\right) & \mbox{P IV} \\ 
\mathbf{B}_{26} & = & -x_{7} \, \mathbf{a}_{1}-y_{7} \, \mathbf{a}_{2} + \left(\frac{1}{2} +z_{7}\right) \, \mathbf{a}_{3} & = & -x_{7}a \, \mathbf{\hat{x}}-y_{7}a \, \mathbf{\hat{y}} + \left(\frac{1}{2} +z_{7}\right)c \, \mathbf{\hat{z}} & \left(4a\right) & \mbox{P IV} \\ 
\mathbf{B}_{27} & = & -y_{7} \, \mathbf{a}_{1} + x_{7} \, \mathbf{a}_{2} + \left(\frac{1}{4} +z_{7}\right) \, \mathbf{a}_{3} & = & -y_{7}a \, \mathbf{\hat{x}} + x_{7}a \, \mathbf{\hat{y}} + \left(\frac{1}{4} +z_{7}\right)c \, \mathbf{\hat{z}} & \left(4a\right) & \mbox{P IV} \\ 
\mathbf{B}_{28} & = & y_{7} \, \mathbf{a}_{1}-x_{7} \, \mathbf{a}_{2} + \left(\frac{3}{4} +z_{7}\right) \, \mathbf{a}_{3} & = & y_{7}a \, \mathbf{\hat{x}}-x_{7}a \, \mathbf{\hat{y}} + \left(\frac{3}{4} +z_{7}\right)c \, \mathbf{\hat{z}} & \left(4a\right) & \mbox{P IV} \\ 
\mathbf{B}_{29} & = & x_{8} \, \mathbf{a}_{1} + y_{8} \, \mathbf{a}_{2} + z_{8} \, \mathbf{a}_{3} & = & x_{8}a \, \mathbf{\hat{x}} + y_{8}a \, \mathbf{\hat{y}} + z_{8}c \, \mathbf{\hat{z}} & \left(4a\right) & \mbox{P V} \\ 
\mathbf{B}_{30} & = & -x_{8} \, \mathbf{a}_{1}-y_{8} \, \mathbf{a}_{2} + \left(\frac{1}{2} +z_{8}\right) \, \mathbf{a}_{3} & = & -x_{8}a \, \mathbf{\hat{x}}-y_{8}a \, \mathbf{\hat{y}} + \left(\frac{1}{2} +z_{8}\right)c \, \mathbf{\hat{z}} & \left(4a\right) & \mbox{P V} \\ 
\mathbf{B}_{31} & = & -y_{8} \, \mathbf{a}_{1} + x_{8} \, \mathbf{a}_{2} + \left(\frac{1}{4} +z_{8}\right) \, \mathbf{a}_{3} & = & -y_{8}a \, \mathbf{\hat{x}} + x_{8}a \, \mathbf{\hat{y}} + \left(\frac{1}{4} +z_{8}\right)c \, \mathbf{\hat{z}} & \left(4a\right) & \mbox{P V} \\ 
\mathbf{B}_{32} & = & y_{8} \, \mathbf{a}_{1}-x_{8} \, \mathbf{a}_{2} + \left(\frac{3}{4} +z_{8}\right) \, \mathbf{a}_{3} & = & y_{8}a \, \mathbf{\hat{x}}-x_{8}a \, \mathbf{\hat{y}} + \left(\frac{3}{4} +z_{8}\right)c \, \mathbf{\hat{z}} & \left(4a\right) & \mbox{P V} \\ 
\mathbf{B}_{33} & = & x_{9} \, \mathbf{a}_{1} + y_{9} \, \mathbf{a}_{2} + z_{9} \, \mathbf{a}_{3} & = & x_{9}a \, \mathbf{\hat{x}} + y_{9}a \, \mathbf{\hat{y}} + z_{9}c \, \mathbf{\hat{z}} & \left(4a\right) & \mbox{P VI} \\ 
\mathbf{B}_{34} & = & -x_{9} \, \mathbf{a}_{1}-y_{9} \, \mathbf{a}_{2} + \left(\frac{1}{2} +z_{9}\right) \, \mathbf{a}_{3} & = & -x_{9}a \, \mathbf{\hat{x}}-y_{9}a \, \mathbf{\hat{y}} + \left(\frac{1}{2} +z_{9}\right)c \, \mathbf{\hat{z}} & \left(4a\right) & \mbox{P VI} \\ 
\mathbf{B}_{35} & = & -y_{9} \, \mathbf{a}_{1} + x_{9} \, \mathbf{a}_{2} + \left(\frac{1}{4} +z_{9}\right) \, \mathbf{a}_{3} & = & -y_{9}a \, \mathbf{\hat{x}} + x_{9}a \, \mathbf{\hat{y}} + \left(\frac{1}{4} +z_{9}\right)c \, \mathbf{\hat{z}} & \left(4a\right) & \mbox{P VI} \\ 
\mathbf{B}_{36} & = & y_{9} \, \mathbf{a}_{1}-x_{9} \, \mathbf{a}_{2} + \left(\frac{3}{4} +z_{9}\right) \, \mathbf{a}_{3} & = & y_{9}a \, \mathbf{\hat{x}}-x_{9}a \, \mathbf{\hat{y}} + \left(\frac{3}{4} +z_{9}\right)c \, \mathbf{\hat{z}} & \left(4a\right) & \mbox{P VI} \\ 
\mathbf{B}_{37} & = & x_{10} \, \mathbf{a}_{1} + y_{10} \, \mathbf{a}_{2} + z_{10} \, \mathbf{a}_{3} & = & x_{10}a \, \mathbf{\hat{x}} + y_{10}a \, \mathbf{\hat{y}} + z_{10}c \, \mathbf{\hat{z}} & \left(4a\right) & \mbox{P VII} \\ 
\mathbf{B}_{38} & = & -x_{10} \, \mathbf{a}_{1}-y_{10} \, \mathbf{a}_{2} + \left(\frac{1}{2} +z_{10}\right) \, \mathbf{a}_{3} & = & -x_{10}a \, \mathbf{\hat{x}}-y_{10}a \, \mathbf{\hat{y}} + \left(\frac{1}{2} +z_{10}\right)c \, \mathbf{\hat{z}} & \left(4a\right) & \mbox{P VII} \\ 
\mathbf{B}_{39} & = & -y_{10} \, \mathbf{a}_{1} + x_{10} \, \mathbf{a}_{2} + \left(\frac{1}{4} +z_{10}\right) \, \mathbf{a}_{3} & = & -y_{10}a \, \mathbf{\hat{x}} + x_{10}a \, \mathbf{\hat{y}} + \left(\frac{1}{4} +z_{10}\right)c \, \mathbf{\hat{z}} & \left(4a\right) & \mbox{P VII} \\ 
\mathbf{B}_{40} & = & y_{10} \, \mathbf{a}_{1}-x_{10} \, \mathbf{a}_{2} + \left(\frac{3}{4} +z_{10}\right) \, \mathbf{a}_{3} & = & y_{10}a \, \mathbf{\hat{x}}-x_{10}a \, \mathbf{\hat{y}} + \left(\frac{3}{4} +z_{10}\right)c \, \mathbf{\hat{z}} & \left(4a\right) & \mbox{P VII} \\ 
\end{longtabu}
\renewcommand{\arraystretch}{1.0}
\noindent \hrulefill
\\
\textbf{References:}
\vspace*{-0.25cm}
\begin{flushleft}
  - \bibentry{Meyer_ZAAC_552_1987}. \\
\end{flushleft}
\textbf{Found in:}
\vspace*{-0.25cm}
\begin{flushleft}
  - \bibentry{Tilley_Crystals_and_Crystal_Structures_2006}. \\
\end{flushleft}
\noindent \hrulefill
\\
\textbf{Geometry files:}
\\
\noindent  - CIF: pp. {\hyperref[A3B7_tP40_76_3a_7a_cif]{\pageref{A3B7_tP40_76_3a_7a_cif}}} \\
\noindent  - POSCAR: pp. {\hyperref[A3B7_tP40_76_3a_7a_poscar]{\pageref{A3B7_tP40_76_3a_7a_poscar}}} \\
\onecolumn
{\phantomsection\label{A2B6CD7_tP64_77_2d_6d_d_ab6d}}
\subsection*{\huge \textbf{{\normalfont \begin{raggedleft}Pinnoite (MgB$_{2}$O(OH)$_{6}$) Structure: \end{raggedleft} \\ A2B6CD7\_tP64\_77\_2d\_6d\_d\_ab6d}}}
\noindent \hrulefill
\vspace*{0.25cm}
\begin{figure}[htp]
  \centering
  \vspace{-1em}
  {\includegraphics[width=1\textwidth]{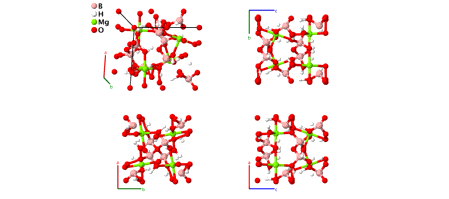}}
\end{figure}
\vspace*{-0.5cm}
\renewcommand{\arraystretch}{1.5}
\begin{equation*}
  \begin{array}{>{$\hspace{-0.15cm}}l<{$}>{$}p{0.5cm}<{$}>{$}p{18.5cm}<{$}}
    \mbox{\large \textbf{Prototype}} &\colon & \ce{MgB2O(OH)6} \\
    \mbox{\large \textbf{\AFLOW\ prototype label}} &\colon & \mbox{A2B6CD7\_tP64\_77\_2d\_6d\_d\_ab6d} \\
    \mbox{\large \textbf{\textit{Strukturbericht} designation}} &\colon & \mbox{None} \\
    \mbox{\large \textbf{Pearson symbol}} &\colon & \mbox{tP64} \\
    \mbox{\large \textbf{Space group number}} &\colon & 77 \\
    \mbox{\large \textbf{Space group symbol}} &\colon & P4_{2} \\
    \mbox{\large \textbf{\AFLOW\ prototype command}} &\colon &  \texttt{aflow} \,  \, \texttt{-{}-proto=A2B6CD7\_tP64\_77\_2d\_6d\_d\_ab6d } \, \newline \texttt{-{}-params=}{a,c/a,z_{1},z_{2},x_{3},y_{3},z_{3},x_{4},y_{4},z_{4},x_{5},y_{5},z_{5},x_{6},y_{6},z_{6},x_{7},y_{7},z_{7},x_{8},y_{8},} \newline {z_{8},x_{9},y_{9},z_{9},x_{10},y_{10},z_{10},x_{11},y_{11},z_{11},x_{12},y_{12},z_{12},x_{13},y_{13},z_{13},x_{14},y_{14},z_{14},x_{15},y_{15},z_{15},} \newline {x_{16},y_{16},z_{16},x_{17},y_{17},z_{17} }
  \end{array}
\end{equation*}
\renewcommand{\arraystretch}{1.0}

\noindent \parbox{1 \linewidth}{
\noindent \hrulefill
\\
\textbf{Simple Tetragonal primitive vectors:} \\
\vspace*{-0.25cm}
\begin{tabular}{cc}
  \begin{tabular}{c}
    \parbox{0.6 \linewidth}{
      \renewcommand{\arraystretch}{1.5}
      \begin{equation*}
        \centering
        \begin{array}{ccc}
              \mathbf{a}_1 & = & a \, \mathbf{\hat{x}} \\
    \mathbf{a}_2 & = & a \, \mathbf{\hat{y}} \\
    \mathbf{a}_3 & = & c \, \mathbf{\hat{z}} \\

        \end{array}
      \end{equation*}
    }
    \renewcommand{\arraystretch}{1.0}
  \end{tabular}
  \begin{tabular}{c}
    \includegraphics[width=0.3\linewidth]{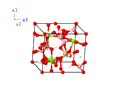} \\
  \end{tabular}
\end{tabular}

}
\vspace*{-0.25cm}

\noindent \hrulefill
\\
\textbf{Basis vectors:}
\vspace*{-0.25cm}
\renewcommand{\arraystretch}{1.5}
\begin{longtabu} to \textwidth{>{\centering $}X[-1,c,c]<{$}>{\centering $}X[-1,c,c]<{$}>{\centering $}X[-1,c,c]<{$}>{\centering $}X[-1,c,c]<{$}>{\centering $}X[-1,c,c]<{$}>{\centering $}X[-1,c,c]<{$}>{\centering $}X[-1,c,c]<{$}}
  & & \mbox{Lattice Coordinates} & & \mbox{Cartesian Coordinates} &\mbox{Wyckoff Position} & \mbox{Atom Type} \\  
  \mathbf{B}_{1} & = & z_{1} \, \mathbf{a}_{3} & = & z_{1}c \, \mathbf{\hat{z}} & \left(2a\right) & \mbox{O I} \\ 
\mathbf{B}_{2} & = & \left(\frac{1}{2} +z_{1}\right) \, \mathbf{a}_{3} & = & \left(\frac{1}{2} +z_{1}\right)c \, \mathbf{\hat{z}} & \left(2a\right) & \mbox{O I} \\ 
\mathbf{B}_{3} & = & \frac{1}{2} \, \mathbf{a}_{1} + \frac{1}{2} \, \mathbf{a}_{2} + z_{2} \, \mathbf{a}_{3} & = & \frac{1}{2}a \, \mathbf{\hat{x}} + \frac{1}{2}a \, \mathbf{\hat{y}} + z_{2}c \, \mathbf{\hat{z}} & \left(2b\right) & \mbox{O II} \\ 
\mathbf{B}_{4} & = & \frac{1}{2} \, \mathbf{a}_{1} + \frac{1}{2} \, \mathbf{a}_{2} + \left(\frac{1}{2} +z_{2}\right) \, \mathbf{a}_{3} & = & \frac{1}{2}a \, \mathbf{\hat{x}} + \frac{1}{2}a \, \mathbf{\hat{y}} + \left(\frac{1}{2} +z_{2}\right)c \, \mathbf{\hat{z}} & \left(2b\right) & \mbox{O II} \\ 
\mathbf{B}_{5} & = & x_{3} \, \mathbf{a}_{1} + y_{3} \, \mathbf{a}_{2} + z_{3} \, \mathbf{a}_{3} & = & x_{3}a \, \mathbf{\hat{x}} + y_{3}a \, \mathbf{\hat{y}} + z_{3}c \, \mathbf{\hat{z}} & \left(4d\right) & \mbox{B I} \\ 
\mathbf{B}_{6} & = & -x_{3} \, \mathbf{a}_{1}-y_{3} \, \mathbf{a}_{2} + z_{3} \, \mathbf{a}_{3} & = & -x_{3}a \, \mathbf{\hat{x}}-y_{3}a \, \mathbf{\hat{y}} + z_{3}c \, \mathbf{\hat{z}} & \left(4d\right) & \mbox{B I} \\ 
\mathbf{B}_{7} & = & -y_{3} \, \mathbf{a}_{1} + x_{3} \, \mathbf{a}_{2} + \left(\frac{1}{2} +z_{3}\right) \, \mathbf{a}_{3} & = & -y_{3}a \, \mathbf{\hat{x}} + x_{3}a \, \mathbf{\hat{y}} + \left(\frac{1}{2} +z_{3}\right)c \, \mathbf{\hat{z}} & \left(4d\right) & \mbox{B I} \\ 
\mathbf{B}_{8} & = & y_{3} \, \mathbf{a}_{1}-x_{3} \, \mathbf{a}_{2} + \left(\frac{1}{2} +z_{3}\right) \, \mathbf{a}_{3} & = & y_{3}a \, \mathbf{\hat{x}}-x_{3}a \, \mathbf{\hat{y}} + \left(\frac{1}{2} +z_{3}\right)c \, \mathbf{\hat{z}} & \left(4d\right) & \mbox{B I} \\ 
\mathbf{B}_{9} & = & x_{4} \, \mathbf{a}_{1} + y_{4} \, \mathbf{a}_{2} + z_{4} \, \mathbf{a}_{3} & = & x_{4}a \, \mathbf{\hat{x}} + y_{4}a \, \mathbf{\hat{y}} + z_{4}c \, \mathbf{\hat{z}} & \left(4d\right) & \mbox{B II} \\ 
\mathbf{B}_{10} & = & -x_{4} \, \mathbf{a}_{1}-y_{4} \, \mathbf{a}_{2} + z_{4} \, \mathbf{a}_{3} & = & -x_{4}a \, \mathbf{\hat{x}}-y_{4}a \, \mathbf{\hat{y}} + z_{4}c \, \mathbf{\hat{z}} & \left(4d\right) & \mbox{B II} \\ 
\mathbf{B}_{11} & = & -y_{4} \, \mathbf{a}_{1} + x_{4} \, \mathbf{a}_{2} + \left(\frac{1}{2} +z_{4}\right) \, \mathbf{a}_{3} & = & -y_{4}a \, \mathbf{\hat{x}} + x_{4}a \, \mathbf{\hat{y}} + \left(\frac{1}{2} +z_{4}\right)c \, \mathbf{\hat{z}} & \left(4d\right) & \mbox{B II} \\ 
\mathbf{B}_{12} & = & y_{4} \, \mathbf{a}_{1}-x_{4} \, \mathbf{a}_{2} + \left(\frac{1}{2} +z_{4}\right) \, \mathbf{a}_{3} & = & y_{4}a \, \mathbf{\hat{x}}-x_{4}a \, \mathbf{\hat{y}} + \left(\frac{1}{2} +z_{4}\right)c \, \mathbf{\hat{z}} & \left(4d\right) & \mbox{B II} \\ 
\mathbf{B}_{13} & = & x_{5} \, \mathbf{a}_{1} + y_{5} \, \mathbf{a}_{2} + z_{5} \, \mathbf{a}_{3} & = & x_{5}a \, \mathbf{\hat{x}} + y_{5}a \, \mathbf{\hat{y}} + z_{5}c \, \mathbf{\hat{z}} & \left(4d\right) & \mbox{H I} \\ 
\mathbf{B}_{14} & = & -x_{5} \, \mathbf{a}_{1}-y_{5} \, \mathbf{a}_{2} + z_{5} \, \mathbf{a}_{3} & = & -x_{5}a \, \mathbf{\hat{x}}-y_{5}a \, \mathbf{\hat{y}} + z_{5}c \, \mathbf{\hat{z}} & \left(4d\right) & \mbox{H I} \\ 
\mathbf{B}_{15} & = & -y_{5} \, \mathbf{a}_{1} + x_{5} \, \mathbf{a}_{2} + \left(\frac{1}{2} +z_{5}\right) \, \mathbf{a}_{3} & = & -y_{5}a \, \mathbf{\hat{x}} + x_{5}a \, \mathbf{\hat{y}} + \left(\frac{1}{2} +z_{5}\right)c \, \mathbf{\hat{z}} & \left(4d\right) & \mbox{H I} \\ 
\mathbf{B}_{16} & = & y_{5} \, \mathbf{a}_{1}-x_{5} \, \mathbf{a}_{2} + \left(\frac{1}{2} +z_{5}\right) \, \mathbf{a}_{3} & = & y_{5}a \, \mathbf{\hat{x}}-x_{5}a \, \mathbf{\hat{y}} + \left(\frac{1}{2} +z_{5}\right)c \, \mathbf{\hat{z}} & \left(4d\right) & \mbox{H I} \\ 
\mathbf{B}_{17} & = & x_{6} \, \mathbf{a}_{1} + y_{6} \, \mathbf{a}_{2} + z_{6} \, \mathbf{a}_{3} & = & x_{6}a \, \mathbf{\hat{x}} + y_{6}a \, \mathbf{\hat{y}} + z_{6}c \, \mathbf{\hat{z}} & \left(4d\right) & \mbox{H II} \\ 
\mathbf{B}_{18} & = & -x_{6} \, \mathbf{a}_{1}-y_{6} \, \mathbf{a}_{2} + z_{6} \, \mathbf{a}_{3} & = & -x_{6}a \, \mathbf{\hat{x}}-y_{6}a \, \mathbf{\hat{y}} + z_{6}c \, \mathbf{\hat{z}} & \left(4d\right) & \mbox{H II} \\ 
\mathbf{B}_{19} & = & -y_{6} \, \mathbf{a}_{1} + x_{6} \, \mathbf{a}_{2} + \left(\frac{1}{2} +z_{6}\right) \, \mathbf{a}_{3} & = & -y_{6}a \, \mathbf{\hat{x}} + x_{6}a \, \mathbf{\hat{y}} + \left(\frac{1}{2} +z_{6}\right)c \, \mathbf{\hat{z}} & \left(4d\right) & \mbox{H II} \\ 
\mathbf{B}_{20} & = & y_{6} \, \mathbf{a}_{1}-x_{6} \, \mathbf{a}_{2} + \left(\frac{1}{2} +z_{6}\right) \, \mathbf{a}_{3} & = & y_{6}a \, \mathbf{\hat{x}}-x_{6}a \, \mathbf{\hat{y}} + \left(\frac{1}{2} +z_{6}\right)c \, \mathbf{\hat{z}} & \left(4d\right) & \mbox{H II} \\ 
\mathbf{B}_{21} & = & x_{7} \, \mathbf{a}_{1} + y_{7} \, \mathbf{a}_{2} + z_{7} \, \mathbf{a}_{3} & = & x_{7}a \, \mathbf{\hat{x}} + y_{7}a \, \mathbf{\hat{y}} + z_{7}c \, \mathbf{\hat{z}} & \left(4d\right) & \mbox{H III} \\ 
\mathbf{B}_{22} & = & -x_{7} \, \mathbf{a}_{1}-y_{7} \, \mathbf{a}_{2} + z_{7} \, \mathbf{a}_{3} & = & -x_{7}a \, \mathbf{\hat{x}}-y_{7}a \, \mathbf{\hat{y}} + z_{7}c \, \mathbf{\hat{z}} & \left(4d\right) & \mbox{H III} \\ 
\mathbf{B}_{23} & = & -y_{7} \, \mathbf{a}_{1} + x_{7} \, \mathbf{a}_{2} + \left(\frac{1}{2} +z_{7}\right) \, \mathbf{a}_{3} & = & -y_{7}a \, \mathbf{\hat{x}} + x_{7}a \, \mathbf{\hat{y}} + \left(\frac{1}{2} +z_{7}\right)c \, \mathbf{\hat{z}} & \left(4d\right) & \mbox{H III} \\ 
\mathbf{B}_{24} & = & y_{7} \, \mathbf{a}_{1}-x_{7} \, \mathbf{a}_{2} + \left(\frac{1}{2} +z_{7}\right) \, \mathbf{a}_{3} & = & y_{7}a \, \mathbf{\hat{x}}-x_{7}a \, \mathbf{\hat{y}} + \left(\frac{1}{2} +z_{7}\right)c \, \mathbf{\hat{z}} & \left(4d\right) & \mbox{H III} \\ 
\mathbf{B}_{25} & = & x_{8} \, \mathbf{a}_{1} + y_{8} \, \mathbf{a}_{2} + z_{8} \, \mathbf{a}_{3} & = & x_{8}a \, \mathbf{\hat{x}} + y_{8}a \, \mathbf{\hat{y}} + z_{8}c \, \mathbf{\hat{z}} & \left(4d\right) & \mbox{H IV} \\ 
\mathbf{B}_{26} & = & -x_{8} \, \mathbf{a}_{1}-y_{8} \, \mathbf{a}_{2} + z_{8} \, \mathbf{a}_{3} & = & -x_{8}a \, \mathbf{\hat{x}}-y_{8}a \, \mathbf{\hat{y}} + z_{8}c \, \mathbf{\hat{z}} & \left(4d\right) & \mbox{H IV} \\ 
\mathbf{B}_{27} & = & -y_{8} \, \mathbf{a}_{1} + x_{8} \, \mathbf{a}_{2} + \left(\frac{1}{2} +z_{8}\right) \, \mathbf{a}_{3} & = & -y_{8}a \, \mathbf{\hat{x}} + x_{8}a \, \mathbf{\hat{y}} + \left(\frac{1}{2} +z_{8}\right)c \, \mathbf{\hat{z}} & \left(4d\right) & \mbox{H IV} \\ 
\mathbf{B}_{28} & = & y_{8} \, \mathbf{a}_{1}-x_{8} \, \mathbf{a}_{2} + \left(\frac{1}{2} +z_{8}\right) \, \mathbf{a}_{3} & = & y_{8}a \, \mathbf{\hat{x}}-x_{8}a \, \mathbf{\hat{y}} + \left(\frac{1}{2} +z_{8}\right)c \, \mathbf{\hat{z}} & \left(4d\right) & \mbox{H IV} \\ 
\mathbf{B}_{29} & = & x_{9} \, \mathbf{a}_{1} + y_{9} \, \mathbf{a}_{2} + z_{9} \, \mathbf{a}_{3} & = & x_{9}a \, \mathbf{\hat{x}} + y_{9}a \, \mathbf{\hat{y}} + z_{9}c \, \mathbf{\hat{z}} & \left(4d\right) & \mbox{H V} \\ 
\mathbf{B}_{30} & = & -x_{9} \, \mathbf{a}_{1}-y_{9} \, \mathbf{a}_{2} + z_{9} \, \mathbf{a}_{3} & = & -x_{9}a \, \mathbf{\hat{x}}-y_{9}a \, \mathbf{\hat{y}} + z_{9}c \, \mathbf{\hat{z}} & \left(4d\right) & \mbox{H V} \\ 
\mathbf{B}_{31} & = & -y_{9} \, \mathbf{a}_{1} + x_{9} \, \mathbf{a}_{2} + \left(\frac{1}{2} +z_{9}\right) \, \mathbf{a}_{3} & = & -y_{9}a \, \mathbf{\hat{x}} + x_{9}a \, \mathbf{\hat{y}} + \left(\frac{1}{2} +z_{9}\right)c \, \mathbf{\hat{z}} & \left(4d\right) & \mbox{H V} \\ 
\mathbf{B}_{32} & = & y_{9} \, \mathbf{a}_{1}-x_{9} \, \mathbf{a}_{2} + \left(\frac{1}{2} +z_{9}\right) \, \mathbf{a}_{3} & = & y_{9}a \, \mathbf{\hat{x}}-x_{9}a \, \mathbf{\hat{y}} + \left(\frac{1}{2} +z_{9}\right)c \, \mathbf{\hat{z}} & \left(4d\right) & \mbox{H V} \\ 
\mathbf{B}_{33} & = & x_{10} \, \mathbf{a}_{1} + y_{10} \, \mathbf{a}_{2} + z_{10} \, \mathbf{a}_{3} & = & x_{10}a \, \mathbf{\hat{x}} + y_{10}a \, \mathbf{\hat{y}} + z_{10}c \, \mathbf{\hat{z}} & \left(4d\right) & \mbox{H VI} \\ 
\mathbf{B}_{34} & = & -x_{10} \, \mathbf{a}_{1}-y_{10} \, \mathbf{a}_{2} + z_{10} \, \mathbf{a}_{3} & = & -x_{10}a \, \mathbf{\hat{x}}-y_{10}a \, \mathbf{\hat{y}} + z_{10}c \, \mathbf{\hat{z}} & \left(4d\right) & \mbox{H VI} \\ 
\mathbf{B}_{35} & = & -y_{10} \, \mathbf{a}_{1} + x_{10} \, \mathbf{a}_{2} + \left(\frac{1}{2} +z_{10}\right) \, \mathbf{a}_{3} & = & -y_{10}a \, \mathbf{\hat{x}} + x_{10}a \, \mathbf{\hat{y}} + \left(\frac{1}{2} +z_{10}\right)c \, \mathbf{\hat{z}} & \left(4d\right) & \mbox{H VI} \\ 
\mathbf{B}_{36} & = & y_{10} \, \mathbf{a}_{1}-x_{10} \, \mathbf{a}_{2} + \left(\frac{1}{2} +z_{10}\right) \, \mathbf{a}_{3} & = & y_{10}a \, \mathbf{\hat{x}}-x_{10}a \, \mathbf{\hat{y}} + \left(\frac{1}{2} +z_{10}\right)c \, \mathbf{\hat{z}} & \left(4d\right) & \mbox{H VI} \\ 
\mathbf{B}_{37} & = & x_{11} \, \mathbf{a}_{1} + y_{11} \, \mathbf{a}_{2} + z_{11} \, \mathbf{a}_{3} & = & x_{11}a \, \mathbf{\hat{x}} + y_{11}a \, \mathbf{\hat{y}} + z_{11}c \, \mathbf{\hat{z}} & \left(4d\right) & \mbox{Mg} \\ 
\mathbf{B}_{38} & = & -x_{11} \, \mathbf{a}_{1}-y_{11} \, \mathbf{a}_{2} + z_{11} \, \mathbf{a}_{3} & = & -x_{11}a \, \mathbf{\hat{x}}-y_{11}a \, \mathbf{\hat{y}} + z_{11}c \, \mathbf{\hat{z}} & \left(4d\right) & \mbox{Mg} \\ 
\mathbf{B}_{39} & = & -y_{11} \, \mathbf{a}_{1} + x_{11} \, \mathbf{a}_{2} + \left(\frac{1}{2} +z_{11}\right) \, \mathbf{a}_{3} & = & -y_{11}a \, \mathbf{\hat{x}} + x_{11}a \, \mathbf{\hat{y}} + \left(\frac{1}{2} +z_{11}\right)c \, \mathbf{\hat{z}} & \left(4d\right) & \mbox{Mg} \\ 
\mathbf{B}_{40} & = & y_{11} \, \mathbf{a}_{1}-x_{11} \, \mathbf{a}_{2} + \left(\frac{1}{2} +z_{11}\right) \, \mathbf{a}_{3} & = & y_{11}a \, \mathbf{\hat{x}}-x_{11}a \, \mathbf{\hat{y}} + \left(\frac{1}{2} +z_{11}\right)c \, \mathbf{\hat{z}} & \left(4d\right) & \mbox{Mg} \\ 
\mathbf{B}_{41} & = & x_{12} \, \mathbf{a}_{1} + y_{12} \, \mathbf{a}_{2} + z_{12} \, \mathbf{a}_{3} & = & x_{12}a \, \mathbf{\hat{x}} + y_{12}a \, \mathbf{\hat{y}} + z_{12}c \, \mathbf{\hat{z}} & \left(4d\right) & \mbox{O III} \\ 
\mathbf{B}_{42} & = & -x_{12} \, \mathbf{a}_{1}-y_{12} \, \mathbf{a}_{2} + z_{12} \, \mathbf{a}_{3} & = & -x_{12}a \, \mathbf{\hat{x}}-y_{12}a \, \mathbf{\hat{y}} + z_{12}c \, \mathbf{\hat{z}} & \left(4d\right) & \mbox{O III} \\ 
\mathbf{B}_{43} & = & -y_{12} \, \mathbf{a}_{1} + x_{12} \, \mathbf{a}_{2} + \left(\frac{1}{2} +z_{12}\right) \, \mathbf{a}_{3} & = & -y_{12}a \, \mathbf{\hat{x}} + x_{12}a \, \mathbf{\hat{y}} + \left(\frac{1}{2} +z_{12}\right)c \, \mathbf{\hat{z}} & \left(4d\right) & \mbox{O III} \\ 
\mathbf{B}_{44} & = & y_{12} \, \mathbf{a}_{1}-x_{12} \, \mathbf{a}_{2} + \left(\frac{1}{2} +z_{12}\right) \, \mathbf{a}_{3} & = & y_{12}a \, \mathbf{\hat{x}}-x_{12}a \, \mathbf{\hat{y}} + \left(\frac{1}{2} +z_{12}\right)c \, \mathbf{\hat{z}} & \left(4d\right) & \mbox{O III} \\ 
\mathbf{B}_{45} & = & x_{13} \, \mathbf{a}_{1} + y_{13} \, \mathbf{a}_{2} + z_{13} \, \mathbf{a}_{3} & = & x_{13}a \, \mathbf{\hat{x}} + y_{13}a \, \mathbf{\hat{y}} + z_{13}c \, \mathbf{\hat{z}} & \left(4d\right) & \mbox{O IV} \\ 
\mathbf{B}_{46} & = & -x_{13} \, \mathbf{a}_{1}-y_{13} \, \mathbf{a}_{2} + z_{13} \, \mathbf{a}_{3} & = & -x_{13}a \, \mathbf{\hat{x}}-y_{13}a \, \mathbf{\hat{y}} + z_{13}c \, \mathbf{\hat{z}} & \left(4d\right) & \mbox{O IV} \\ 
\mathbf{B}_{47} & = & -y_{13} \, \mathbf{a}_{1} + x_{13} \, \mathbf{a}_{2} + \left(\frac{1}{2} +z_{13}\right) \, \mathbf{a}_{3} & = & -y_{13}a \, \mathbf{\hat{x}} + x_{13}a \, \mathbf{\hat{y}} + \left(\frac{1}{2} +z_{13}\right)c \, \mathbf{\hat{z}} & \left(4d\right) & \mbox{O IV} \\ 
\mathbf{B}_{48} & = & y_{13} \, \mathbf{a}_{1}-x_{13} \, \mathbf{a}_{2} + \left(\frac{1}{2} +z_{13}\right) \, \mathbf{a}_{3} & = & y_{13}a \, \mathbf{\hat{x}}-x_{13}a \, \mathbf{\hat{y}} + \left(\frac{1}{2} +z_{13}\right)c \, \mathbf{\hat{z}} & \left(4d\right) & \mbox{O IV} \\ 
\mathbf{B}_{49} & = & x_{14} \, \mathbf{a}_{1} + y_{14} \, \mathbf{a}_{2} + z_{14} \, \mathbf{a}_{3} & = & x_{14}a \, \mathbf{\hat{x}} + y_{14}a \, \mathbf{\hat{y}} + z_{14}c \, \mathbf{\hat{z}} & \left(4d\right) & \mbox{O V} \\ 
\mathbf{B}_{50} & = & -x_{14} \, \mathbf{a}_{1}-y_{14} \, \mathbf{a}_{2} + z_{14} \, \mathbf{a}_{3} & = & -x_{14}a \, \mathbf{\hat{x}}-y_{14}a \, \mathbf{\hat{y}} + z_{14}c \, \mathbf{\hat{z}} & \left(4d\right) & \mbox{O V} \\ 
\mathbf{B}_{51} & = & -y_{14} \, \mathbf{a}_{1} + x_{14} \, \mathbf{a}_{2} + \left(\frac{1}{2} +z_{14}\right) \, \mathbf{a}_{3} & = & -y_{14}a \, \mathbf{\hat{x}} + x_{14}a \, \mathbf{\hat{y}} + \left(\frac{1}{2} +z_{14}\right)c \, \mathbf{\hat{z}} & \left(4d\right) & \mbox{O V} \\ 
\mathbf{B}_{52} & = & y_{14} \, \mathbf{a}_{1}-x_{14} \, \mathbf{a}_{2} + \left(\frac{1}{2} +z_{14}\right) \, \mathbf{a}_{3} & = & y_{14}a \, \mathbf{\hat{x}}-x_{14}a \, \mathbf{\hat{y}} + \left(\frac{1}{2} +z_{14}\right)c \, \mathbf{\hat{z}} & \left(4d\right) & \mbox{O V} \\ 
\mathbf{B}_{53} & = & x_{15} \, \mathbf{a}_{1} + y_{15} \, \mathbf{a}_{2} + z_{15} \, \mathbf{a}_{3} & = & x_{15}a \, \mathbf{\hat{x}} + y_{15}a \, \mathbf{\hat{y}} + z_{15}c \, \mathbf{\hat{z}} & \left(4d\right) & \mbox{O VI} \\ 
\mathbf{B}_{54} & = & -x_{15} \, \mathbf{a}_{1}-y_{15} \, \mathbf{a}_{2} + z_{15} \, \mathbf{a}_{3} & = & -x_{15}a \, \mathbf{\hat{x}}-y_{15}a \, \mathbf{\hat{y}} + z_{15}c \, \mathbf{\hat{z}} & \left(4d\right) & \mbox{O VI} \\ 
\mathbf{B}_{55} & = & -y_{15} \, \mathbf{a}_{1} + x_{15} \, \mathbf{a}_{2} + \left(\frac{1}{2} +z_{15}\right) \, \mathbf{a}_{3} & = & -y_{15}a \, \mathbf{\hat{x}} + x_{15}a \, \mathbf{\hat{y}} + \left(\frac{1}{2} +z_{15}\right)c \, \mathbf{\hat{z}} & \left(4d\right) & \mbox{O VI} \\ 
\mathbf{B}_{56} & = & y_{15} \, \mathbf{a}_{1}-x_{15} \, \mathbf{a}_{2} + \left(\frac{1}{2} +z_{15}\right) \, \mathbf{a}_{3} & = & y_{15}a \, \mathbf{\hat{x}}-x_{15}a \, \mathbf{\hat{y}} + \left(\frac{1}{2} +z_{15}\right)c \, \mathbf{\hat{z}} & \left(4d\right) & \mbox{O VI} \\ 
\mathbf{B}_{57} & = & x_{16} \, \mathbf{a}_{1} + y_{16} \, \mathbf{a}_{2} + z_{16} \, \mathbf{a}_{3} & = & x_{16}a \, \mathbf{\hat{x}} + y_{16}a \, \mathbf{\hat{y}} + z_{16}c \, \mathbf{\hat{z}} & \left(4d\right) & \mbox{O VII} \\ 
\mathbf{B}_{58} & = & -x_{16} \, \mathbf{a}_{1}-y_{16} \, \mathbf{a}_{2} + z_{16} \, \mathbf{a}_{3} & = & -x_{16}a \, \mathbf{\hat{x}}-y_{16}a \, \mathbf{\hat{y}} + z_{16}c \, \mathbf{\hat{z}} & \left(4d\right) & \mbox{O VII} \\ 
\mathbf{B}_{59} & = & -y_{16} \, \mathbf{a}_{1} + x_{16} \, \mathbf{a}_{2} + \left(\frac{1}{2} +z_{16}\right) \, \mathbf{a}_{3} & = & -y_{16}a \, \mathbf{\hat{x}} + x_{16}a \, \mathbf{\hat{y}} + \left(\frac{1}{2} +z_{16}\right)c \, \mathbf{\hat{z}} & \left(4d\right) & \mbox{O VII} \\ 
\mathbf{B}_{60} & = & y_{16} \, \mathbf{a}_{1}-x_{16} \, \mathbf{a}_{2} + \left(\frac{1}{2} +z_{16}\right) \, \mathbf{a}_{3} & = & y_{16}a \, \mathbf{\hat{x}}-x_{16}a \, \mathbf{\hat{y}} + \left(\frac{1}{2} +z_{16}\right)c \, \mathbf{\hat{z}} & \left(4d\right) & \mbox{O VII} \\ 
\mathbf{B}_{61} & = & x_{17} \, \mathbf{a}_{1} + y_{17} \, \mathbf{a}_{2} + z_{17} \, \mathbf{a}_{3} & = & x_{17}a \, \mathbf{\hat{x}} + y_{17}a \, \mathbf{\hat{y}} + z_{17}c \, \mathbf{\hat{z}} & \left(4d\right) & \mbox{O VIII} \\ 
\mathbf{B}_{62} & = & -x_{17} \, \mathbf{a}_{1}-y_{17} \, \mathbf{a}_{2} + z_{17} \, \mathbf{a}_{3} & = & -x_{17}a \, \mathbf{\hat{x}}-y_{17}a \, \mathbf{\hat{y}} + z_{17}c \, \mathbf{\hat{z}} & \left(4d\right) & \mbox{O VIII} \\ 
\mathbf{B}_{63} & = & -y_{17} \, \mathbf{a}_{1} + x_{17} \, \mathbf{a}_{2} + \left(\frac{1}{2} +z_{17}\right) \, \mathbf{a}_{3} & = & -y_{17}a \, \mathbf{\hat{x}} + x_{17}a \, \mathbf{\hat{y}} + \left(\frac{1}{2} +z_{17}\right)c \, \mathbf{\hat{z}} & \left(4d\right) & \mbox{O VIII} \\ 
\mathbf{B}_{64} & = & y_{17} \, \mathbf{a}_{1}-x_{17} \, \mathbf{a}_{2} + \left(\frac{1}{2} +z_{17}\right) \, \mathbf{a}_{3} & = & y_{17}a \, \mathbf{\hat{x}}-x_{17}a \, \mathbf{\hat{y}} + \left(\frac{1}{2} +z_{17}\right)c \, \mathbf{\hat{z}} & \left(4d\right) & \mbox{O VIII} \\ 
\end{longtabu}
\renewcommand{\arraystretch}{1.0}
\noindent \hrulefill
\\
\textbf{References:}
\vspace*{-0.25cm}
\begin{flushleft}
  - \bibentry{Genkina_MgB2OOH6_SovPhysCryst_1983}. \\
\end{flushleft}
\textbf{Found in:}
\vspace*{-0.25cm}
\begin{flushleft}
  - \bibentry{Villars_PearsonsCrystalData_2013}. \\
\end{flushleft}
\noindent \hrulefill
\\
\textbf{Geometry files:}
\\
\noindent  - CIF: pp. {\hyperref[A2B6CD7_tP64_77_2d_6d_d_ab6d_cif]{\pageref{A2B6CD7_tP64_77_2d_6d_d_ab6d_cif}}} \\
\noindent  - POSCAR: pp. {\hyperref[A2B6CD7_tP64_77_2d_6d_d_ab6d_poscar]{\pageref{A2B6CD7_tP64_77_2d_6d_d_ab6d_poscar}}} \\
\onecolumn
{\phantomsection\label{A2B_tP48_77_8d_4d}}
\subsection*{\huge \textbf{{\normalfont H$_{2}$S III Structure: A2B\_tP48\_77\_8d\_4d}}}
\noindent \hrulefill
\vspace*{0.25cm}
\begin{figure}[htp]
  \centering
  \vspace{-1em}
  {\includegraphics[width=1\textwidth]{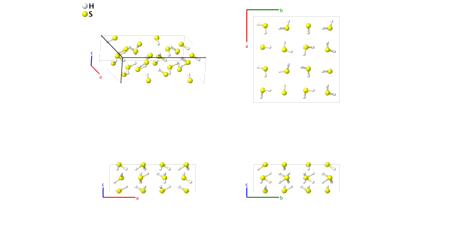}}
\end{figure}
\vspace*{-0.5cm}
\renewcommand{\arraystretch}{1.5}
\begin{equation*}
  \begin{array}{>{$\hspace{-0.15cm}}l<{$}>{$}p{0.5cm}<{$}>{$}p{18.5cm}<{$}}
    \mbox{\large \textbf{Prototype}} &\colon & \ce{H2S} \\
    \mbox{\large \textbf{\AFLOW\ prototype label}} &\colon & \mbox{A2B\_tP48\_77\_8d\_4d} \\
    \mbox{\large \textbf{\textit{Strukturbericht} designation}} &\colon & \mbox{None} \\
    \mbox{\large \textbf{Pearson symbol}} &\colon & \mbox{tP48} \\
    \mbox{\large \textbf{Space group number}} &\colon & 77 \\
    \mbox{\large \textbf{Space group symbol}} &\colon & P4_{2} \\
    \mbox{\large \textbf{\AFLOW\ prototype command}} &\colon &  \texttt{aflow} \,  \, \texttt{-{}-proto=A2B\_tP48\_77\_8d\_4d } \, \newline \texttt{-{}-params=}{a,c/a,x_{1},y_{1},z_{1},x_{2},y_{2},z_{2},x_{3},y_{3},z_{3},x_{4},y_{4},z_{4},x_{5},y_{5},z_{5},x_{6},y_{6},z_{6},x_{7},} \newline {y_{7},z_{7},x_{8},y_{8},z_{8},x_{9},y_{9},z_{9},x_{10},y_{10},z_{10},x_{11},y_{11},z_{11},x_{12},y_{12},z_{12} }
  \end{array}
\end{equation*}
\renewcommand{\arraystretch}{1.0}

\vspace*{-0.25cm}
\noindent \hrulefill
\begin{itemize}
  \item{This is one candidate structure for the H$_{2}$S~III structure, which
is stable at pressures under 4~GPa and temperatures less than $\approx
100~K$ (Shimizu, 1995). The data presented here was for
D$_{2}$S at 102~K and ambient pressure.
}
\end{itemize}

\noindent \parbox{1 \linewidth}{
\noindent \hrulefill
\\
\textbf{Simple Tetragonal primitive vectors:} \\
\vspace*{-0.25cm}
\begin{tabular}{cc}
  \begin{tabular}{c}
    \parbox{0.6 \linewidth}{
      \renewcommand{\arraystretch}{1.5}
      \begin{equation*}
        \centering
        \begin{array}{ccc}
              \mathbf{a}_1 & = & a \, \mathbf{\hat{x}} \\
    \mathbf{a}_2 & = & a \, \mathbf{\hat{y}} \\
    \mathbf{a}_3 & = & c \, \mathbf{\hat{z}} \\

        \end{array}
      \end{equation*}
    }
    \renewcommand{\arraystretch}{1.0}
  \end{tabular}
  \begin{tabular}{c}
    \includegraphics[width=0.3\linewidth]{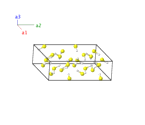} \\
  \end{tabular}
\end{tabular}

}
\vspace*{-0.25cm}

\noindent \hrulefill
\\
\textbf{Basis vectors:}
\vspace*{-0.25cm}
\renewcommand{\arraystretch}{1.5}
\begin{longtabu} to \textwidth{>{\centering $}X[-1,c,c]<{$}>{\centering $}X[-1,c,c]<{$}>{\centering $}X[-1,c,c]<{$}>{\centering $}X[-1,c,c]<{$}>{\centering $}X[-1,c,c]<{$}>{\centering $}X[-1,c,c]<{$}>{\centering $}X[-1,c,c]<{$}}
  & & \mbox{Lattice Coordinates} & & \mbox{Cartesian Coordinates} &\mbox{Wyckoff Position} & \mbox{Atom Type} \\  
  \mathbf{B}_{1} & = & x_{1} \, \mathbf{a}_{1} + y_{1} \, \mathbf{a}_{2} + z_{1} \, \mathbf{a}_{3} & = & x_{1}a \, \mathbf{\hat{x}} + y_{1}a \, \mathbf{\hat{y}} + z_{1}c \, \mathbf{\hat{z}} & \left(4d\right) & \mbox{H I} \\ 
\mathbf{B}_{2} & = & -x_{1} \, \mathbf{a}_{1}-y_{1} \, \mathbf{a}_{2} + z_{1} \, \mathbf{a}_{3} & = & -x_{1}a \, \mathbf{\hat{x}}-y_{1}a \, \mathbf{\hat{y}} + z_{1}c \, \mathbf{\hat{z}} & \left(4d\right) & \mbox{H I} \\ 
\mathbf{B}_{3} & = & -y_{1} \, \mathbf{a}_{1} + x_{1} \, \mathbf{a}_{2} + \left(\frac{1}{2} +z_{1}\right) \, \mathbf{a}_{3} & = & -y_{1}a \, \mathbf{\hat{x}} + x_{1}a \, \mathbf{\hat{y}} + \left(\frac{1}{2} +z_{1}\right)c \, \mathbf{\hat{z}} & \left(4d\right) & \mbox{H I} \\ 
\mathbf{B}_{4} & = & y_{1} \, \mathbf{a}_{1}-x_{1} \, \mathbf{a}_{2} + \left(\frac{1}{2} +z_{1}\right) \, \mathbf{a}_{3} & = & y_{1}a \, \mathbf{\hat{x}}-x_{1}a \, \mathbf{\hat{y}} + \left(\frac{1}{2} +z_{1}\right)c \, \mathbf{\hat{z}} & \left(4d\right) & \mbox{H I} \\ 
\mathbf{B}_{5} & = & x_{2} \, \mathbf{a}_{1} + y_{2} \, \mathbf{a}_{2} + z_{2} \, \mathbf{a}_{3} & = & x_{2}a \, \mathbf{\hat{x}} + y_{2}a \, \mathbf{\hat{y}} + z_{2}c \, \mathbf{\hat{z}} & \left(4d\right) & \mbox{H II} \\ 
\mathbf{B}_{6} & = & -x_{2} \, \mathbf{a}_{1}-y_{2} \, \mathbf{a}_{2} + z_{2} \, \mathbf{a}_{3} & = & -x_{2}a \, \mathbf{\hat{x}}-y_{2}a \, \mathbf{\hat{y}} + z_{2}c \, \mathbf{\hat{z}} & \left(4d\right) & \mbox{H II} \\ 
\mathbf{B}_{7} & = & -y_{2} \, \mathbf{a}_{1} + x_{2} \, \mathbf{a}_{2} + \left(\frac{1}{2} +z_{2}\right) \, \mathbf{a}_{3} & = & -y_{2}a \, \mathbf{\hat{x}} + x_{2}a \, \mathbf{\hat{y}} + \left(\frac{1}{2} +z_{2}\right)c \, \mathbf{\hat{z}} & \left(4d\right) & \mbox{H II} \\ 
\mathbf{B}_{8} & = & y_{2} \, \mathbf{a}_{1}-x_{2} \, \mathbf{a}_{2} + \left(\frac{1}{2} +z_{2}\right) \, \mathbf{a}_{3} & = & y_{2}a \, \mathbf{\hat{x}}-x_{2}a \, \mathbf{\hat{y}} + \left(\frac{1}{2} +z_{2}\right)c \, \mathbf{\hat{z}} & \left(4d\right) & \mbox{H II} \\ 
\mathbf{B}_{9} & = & x_{3} \, \mathbf{a}_{1} + y_{3} \, \mathbf{a}_{2} + z_{3} \, \mathbf{a}_{3} & = & x_{3}a \, \mathbf{\hat{x}} + y_{3}a \, \mathbf{\hat{y}} + z_{3}c \, \mathbf{\hat{z}} & \left(4d\right) & \mbox{H III} \\ 
\mathbf{B}_{10} & = & -x_{3} \, \mathbf{a}_{1}-y_{3} \, \mathbf{a}_{2} + z_{3} \, \mathbf{a}_{3} & = & -x_{3}a \, \mathbf{\hat{x}}-y_{3}a \, \mathbf{\hat{y}} + z_{3}c \, \mathbf{\hat{z}} & \left(4d\right) & \mbox{H III} \\ 
\mathbf{B}_{11} & = & -y_{3} \, \mathbf{a}_{1} + x_{3} \, \mathbf{a}_{2} + \left(\frac{1}{2} +z_{3}\right) \, \mathbf{a}_{3} & = & -y_{3}a \, \mathbf{\hat{x}} + x_{3}a \, \mathbf{\hat{y}} + \left(\frac{1}{2} +z_{3}\right)c \, \mathbf{\hat{z}} & \left(4d\right) & \mbox{H III} \\ 
\mathbf{B}_{12} & = & y_{3} \, \mathbf{a}_{1}-x_{3} \, \mathbf{a}_{2} + \left(\frac{1}{2} +z_{3}\right) \, \mathbf{a}_{3} & = & y_{3}a \, \mathbf{\hat{x}}-x_{3}a \, \mathbf{\hat{y}} + \left(\frac{1}{2} +z_{3}\right)c \, \mathbf{\hat{z}} & \left(4d\right) & \mbox{H III} \\ 
\mathbf{B}_{13} & = & x_{4} \, \mathbf{a}_{1} + y_{4} \, \mathbf{a}_{2} + z_{4} \, \mathbf{a}_{3} & = & x_{4}a \, \mathbf{\hat{x}} + y_{4}a \, \mathbf{\hat{y}} + z_{4}c \, \mathbf{\hat{z}} & \left(4d\right) & \mbox{H IV} \\ 
\mathbf{B}_{14} & = & -x_{4} \, \mathbf{a}_{1}-y_{4} \, \mathbf{a}_{2} + z_{4} \, \mathbf{a}_{3} & = & -x_{4}a \, \mathbf{\hat{x}}-y_{4}a \, \mathbf{\hat{y}} + z_{4}c \, \mathbf{\hat{z}} & \left(4d\right) & \mbox{H IV} \\ 
\mathbf{B}_{15} & = & -y_{4} \, \mathbf{a}_{1} + x_{4} \, \mathbf{a}_{2} + \left(\frac{1}{2} +z_{4}\right) \, \mathbf{a}_{3} & = & -y_{4}a \, \mathbf{\hat{x}} + x_{4}a \, \mathbf{\hat{y}} + \left(\frac{1}{2} +z_{4}\right)c \, \mathbf{\hat{z}} & \left(4d\right) & \mbox{H IV} \\ 
\mathbf{B}_{16} & = & y_{4} \, \mathbf{a}_{1}-x_{4} \, \mathbf{a}_{2} + \left(\frac{1}{2} +z_{4}\right) \, \mathbf{a}_{3} & = & y_{4}a \, \mathbf{\hat{x}}-x_{4}a \, \mathbf{\hat{y}} + \left(\frac{1}{2} +z_{4}\right)c \, \mathbf{\hat{z}} & \left(4d\right) & \mbox{H IV} \\ 
\mathbf{B}_{17} & = & x_{5} \, \mathbf{a}_{1} + y_{5} \, \mathbf{a}_{2} + z_{5} \, \mathbf{a}_{3} & = & x_{5}a \, \mathbf{\hat{x}} + y_{5}a \, \mathbf{\hat{y}} + z_{5}c \, \mathbf{\hat{z}} & \left(4d\right) & \mbox{H V} \\ 
\mathbf{B}_{18} & = & -x_{5} \, \mathbf{a}_{1}-y_{5} \, \mathbf{a}_{2} + z_{5} \, \mathbf{a}_{3} & = & -x_{5}a \, \mathbf{\hat{x}}-y_{5}a \, \mathbf{\hat{y}} + z_{5}c \, \mathbf{\hat{z}} & \left(4d\right) & \mbox{H V} \\ 
\mathbf{B}_{19} & = & -y_{5} \, \mathbf{a}_{1} + x_{5} \, \mathbf{a}_{2} + \left(\frac{1}{2} +z_{5}\right) \, \mathbf{a}_{3} & = & -y_{5}a \, \mathbf{\hat{x}} + x_{5}a \, \mathbf{\hat{y}} + \left(\frac{1}{2} +z_{5}\right)c \, \mathbf{\hat{z}} & \left(4d\right) & \mbox{H V} \\ 
\mathbf{B}_{20} & = & y_{5} \, \mathbf{a}_{1}-x_{5} \, \mathbf{a}_{2} + \left(\frac{1}{2} +z_{5}\right) \, \mathbf{a}_{3} & = & y_{5}a \, \mathbf{\hat{x}}-x_{5}a \, \mathbf{\hat{y}} + \left(\frac{1}{2} +z_{5}\right)c \, \mathbf{\hat{z}} & \left(4d\right) & \mbox{H V} \\ 
\mathbf{B}_{21} & = & x_{6} \, \mathbf{a}_{1} + y_{6} \, \mathbf{a}_{2} + z_{6} \, \mathbf{a}_{3} & = & x_{6}a \, \mathbf{\hat{x}} + y_{6}a \, \mathbf{\hat{y}} + z_{6}c \, \mathbf{\hat{z}} & \left(4d\right) & \mbox{H VI} \\ 
\mathbf{B}_{22} & = & -x_{6} \, \mathbf{a}_{1}-y_{6} \, \mathbf{a}_{2} + z_{6} \, \mathbf{a}_{3} & = & -x_{6}a \, \mathbf{\hat{x}}-y_{6}a \, \mathbf{\hat{y}} + z_{6}c \, \mathbf{\hat{z}} & \left(4d\right) & \mbox{H VI} \\ 
\mathbf{B}_{23} & = & -y_{6} \, \mathbf{a}_{1} + x_{6} \, \mathbf{a}_{2} + \left(\frac{1}{2} +z_{6}\right) \, \mathbf{a}_{3} & = & -y_{6}a \, \mathbf{\hat{x}} + x_{6}a \, \mathbf{\hat{y}} + \left(\frac{1}{2} +z_{6}\right)c \, \mathbf{\hat{z}} & \left(4d\right) & \mbox{H VI} \\ 
\mathbf{B}_{24} & = & y_{6} \, \mathbf{a}_{1}-x_{6} \, \mathbf{a}_{2} + \left(\frac{1}{2} +z_{6}\right) \, \mathbf{a}_{3} & = & y_{6}a \, \mathbf{\hat{x}}-x_{6}a \, \mathbf{\hat{y}} + \left(\frac{1}{2} +z_{6}\right)c \, \mathbf{\hat{z}} & \left(4d\right) & \mbox{H VI} \\ 
\mathbf{B}_{25} & = & x_{7} \, \mathbf{a}_{1} + y_{7} \, \mathbf{a}_{2} + z_{7} \, \mathbf{a}_{3} & = & x_{7}a \, \mathbf{\hat{x}} + y_{7}a \, \mathbf{\hat{y}} + z_{7}c \, \mathbf{\hat{z}} & \left(4d\right) & \mbox{H VII} \\ 
\mathbf{B}_{26} & = & -x_{7} \, \mathbf{a}_{1}-y_{7} \, \mathbf{a}_{2} + z_{7} \, \mathbf{a}_{3} & = & -x_{7}a \, \mathbf{\hat{x}}-y_{7}a \, \mathbf{\hat{y}} + z_{7}c \, \mathbf{\hat{z}} & \left(4d\right) & \mbox{H VII} \\ 
\mathbf{B}_{27} & = & -y_{7} \, \mathbf{a}_{1} + x_{7} \, \mathbf{a}_{2} + \left(\frac{1}{2} +z_{7}\right) \, \mathbf{a}_{3} & = & -y_{7}a \, \mathbf{\hat{x}} + x_{7}a \, \mathbf{\hat{y}} + \left(\frac{1}{2} +z_{7}\right)c \, \mathbf{\hat{z}} & \left(4d\right) & \mbox{H VII} \\ 
\mathbf{B}_{28} & = & y_{7} \, \mathbf{a}_{1}-x_{7} \, \mathbf{a}_{2} + \left(\frac{1}{2} +z_{7}\right) \, \mathbf{a}_{3} & = & y_{7}a \, \mathbf{\hat{x}}-x_{7}a \, \mathbf{\hat{y}} + \left(\frac{1}{2} +z_{7}\right)c \, \mathbf{\hat{z}} & \left(4d\right) & \mbox{H VII} \\ 
\mathbf{B}_{29} & = & x_{8} \, \mathbf{a}_{1} + y_{8} \, \mathbf{a}_{2} + z_{8} \, \mathbf{a}_{3} & = & x_{8}a \, \mathbf{\hat{x}} + y_{8}a \, \mathbf{\hat{y}} + z_{8}c \, \mathbf{\hat{z}} & \left(4d\right) & \mbox{H VIII} \\ 
\mathbf{B}_{30} & = & -x_{8} \, \mathbf{a}_{1}-y_{8} \, \mathbf{a}_{2} + z_{8} \, \mathbf{a}_{3} & = & -x_{8}a \, \mathbf{\hat{x}}-y_{8}a \, \mathbf{\hat{y}} + z_{8}c \, \mathbf{\hat{z}} & \left(4d\right) & \mbox{H VIII} \\ 
\mathbf{B}_{31} & = & -y_{8} \, \mathbf{a}_{1} + x_{8} \, \mathbf{a}_{2} + \left(\frac{1}{2} +z_{8}\right) \, \mathbf{a}_{3} & = & -y_{8}a \, \mathbf{\hat{x}} + x_{8}a \, \mathbf{\hat{y}} + \left(\frac{1}{2} +z_{8}\right)c \, \mathbf{\hat{z}} & \left(4d\right) & \mbox{H VIII} \\ 
\mathbf{B}_{32} & = & y_{8} \, \mathbf{a}_{1}-x_{8} \, \mathbf{a}_{2} + \left(\frac{1}{2} +z_{8}\right) \, \mathbf{a}_{3} & = & y_{8}a \, \mathbf{\hat{x}}-x_{8}a \, \mathbf{\hat{y}} + \left(\frac{1}{2} +z_{8}\right)c \, \mathbf{\hat{z}} & \left(4d\right) & \mbox{H VIII} \\ 
\mathbf{B}_{33} & = & x_{9} \, \mathbf{a}_{1} + y_{9} \, \mathbf{a}_{2} + z_{9} \, \mathbf{a}_{3} & = & x_{9}a \, \mathbf{\hat{x}} + y_{9}a \, \mathbf{\hat{y}} + z_{9}c \, \mathbf{\hat{z}} & \left(4d\right) & \mbox{S I} \\ 
\mathbf{B}_{34} & = & -x_{9} \, \mathbf{a}_{1}-y_{9} \, \mathbf{a}_{2} + z_{9} \, \mathbf{a}_{3} & = & -x_{9}a \, \mathbf{\hat{x}}-y_{9}a \, \mathbf{\hat{y}} + z_{9}c \, \mathbf{\hat{z}} & \left(4d\right) & \mbox{S I} \\ 
\mathbf{B}_{35} & = & -y_{9} \, \mathbf{a}_{1} + x_{9} \, \mathbf{a}_{2} + \left(\frac{1}{2} +z_{9}\right) \, \mathbf{a}_{3} & = & -y_{9}a \, \mathbf{\hat{x}} + x_{9}a \, \mathbf{\hat{y}} + \left(\frac{1}{2} +z_{9}\right)c \, \mathbf{\hat{z}} & \left(4d\right) & \mbox{S I} \\ 
\mathbf{B}_{36} & = & y_{9} \, \mathbf{a}_{1}-x_{9} \, \mathbf{a}_{2} + \left(\frac{1}{2} +z_{9}\right) \, \mathbf{a}_{3} & = & y_{9}a \, \mathbf{\hat{x}}-x_{9}a \, \mathbf{\hat{y}} + \left(\frac{1}{2} +z_{9}\right)c \, \mathbf{\hat{z}} & \left(4d\right) & \mbox{S I} \\ 
\mathbf{B}_{37} & = & x_{10} \, \mathbf{a}_{1} + y_{10} \, \mathbf{a}_{2} + z_{10} \, \mathbf{a}_{3} & = & x_{10}a \, \mathbf{\hat{x}} + y_{10}a \, \mathbf{\hat{y}} + z_{10}c \, \mathbf{\hat{z}} & \left(4d\right) & \mbox{S II} \\ 
\mathbf{B}_{38} & = & -x_{10} \, \mathbf{a}_{1}-y_{10} \, \mathbf{a}_{2} + z_{10} \, \mathbf{a}_{3} & = & -x_{10}a \, \mathbf{\hat{x}}-y_{10}a \, \mathbf{\hat{y}} + z_{10}c \, \mathbf{\hat{z}} & \left(4d\right) & \mbox{S II} \\ 
\mathbf{B}_{39} & = & -y_{10} \, \mathbf{a}_{1} + x_{10} \, \mathbf{a}_{2} + \left(\frac{1}{2} +z_{10}\right) \, \mathbf{a}_{3} & = & -y_{10}a \, \mathbf{\hat{x}} + x_{10}a \, \mathbf{\hat{y}} + \left(\frac{1}{2} +z_{10}\right)c \, \mathbf{\hat{z}} & \left(4d\right) & \mbox{S II} \\ 
\mathbf{B}_{40} & = & y_{10} \, \mathbf{a}_{1}-x_{10} \, \mathbf{a}_{2} + \left(\frac{1}{2} +z_{10}\right) \, \mathbf{a}_{3} & = & y_{10}a \, \mathbf{\hat{x}}-x_{10}a \, \mathbf{\hat{y}} + \left(\frac{1}{2} +z_{10}\right)c \, \mathbf{\hat{z}} & \left(4d\right) & \mbox{S II} \\ 
\mathbf{B}_{41} & = & x_{11} \, \mathbf{a}_{1} + y_{11} \, \mathbf{a}_{2} + z_{11} \, \mathbf{a}_{3} & = & x_{11}a \, \mathbf{\hat{x}} + y_{11}a \, \mathbf{\hat{y}} + z_{11}c \, \mathbf{\hat{z}} & \left(4d\right) & \mbox{S III} \\ 
\mathbf{B}_{42} & = & -x_{11} \, \mathbf{a}_{1}-y_{11} \, \mathbf{a}_{2} + z_{11} \, \mathbf{a}_{3} & = & -x_{11}a \, \mathbf{\hat{x}}-y_{11}a \, \mathbf{\hat{y}} + z_{11}c \, \mathbf{\hat{z}} & \left(4d\right) & \mbox{S III} \\ 
\mathbf{B}_{43} & = & -y_{11} \, \mathbf{a}_{1} + x_{11} \, \mathbf{a}_{2} + \left(\frac{1}{2} +z_{11}\right) \, \mathbf{a}_{3} & = & -y_{11}a \, \mathbf{\hat{x}} + x_{11}a \, \mathbf{\hat{y}} + \left(\frac{1}{2} +z_{11}\right)c \, \mathbf{\hat{z}} & \left(4d\right) & \mbox{S III} \\ 
\mathbf{B}_{44} & = & y_{11} \, \mathbf{a}_{1}-x_{11} \, \mathbf{a}_{2} + \left(\frac{1}{2} +z_{11}\right) \, \mathbf{a}_{3} & = & y_{11}a \, \mathbf{\hat{x}}-x_{11}a \, \mathbf{\hat{y}} + \left(\frac{1}{2} +z_{11}\right)c \, \mathbf{\hat{z}} & \left(4d\right) & \mbox{S III} \\ 
\mathbf{B}_{45} & = & x_{12} \, \mathbf{a}_{1} + y_{12} \, \mathbf{a}_{2} + z_{12} \, \mathbf{a}_{3} & = & x_{12}a \, \mathbf{\hat{x}} + y_{12}a \, \mathbf{\hat{y}} + z_{12}c \, \mathbf{\hat{z}} & \left(4d\right) & \mbox{S IV} \\ 
\mathbf{B}_{46} & = & -x_{12} \, \mathbf{a}_{1}-y_{12} \, \mathbf{a}_{2} + z_{12} \, \mathbf{a}_{3} & = & -x_{12}a \, \mathbf{\hat{x}}-y_{12}a \, \mathbf{\hat{y}} + z_{12}c \, \mathbf{\hat{z}} & \left(4d\right) & \mbox{S IV} \\ 
\mathbf{B}_{47} & = & -y_{12} \, \mathbf{a}_{1} + x_{12} \, \mathbf{a}_{2} + \left(\frac{1}{2} +z_{12}\right) \, \mathbf{a}_{3} & = & -y_{12}a \, \mathbf{\hat{x}} + x_{12}a \, \mathbf{\hat{y}} + \left(\frac{1}{2} +z_{12}\right)c \, \mathbf{\hat{z}} & \left(4d\right) & \mbox{S IV} \\ 
\mathbf{B}_{48} & = & y_{12} \, \mathbf{a}_{1}-x_{12} \, \mathbf{a}_{2} + \left(\frac{1}{2} +z_{12}\right) \, \mathbf{a}_{3} & = & y_{12}a \, \mathbf{\hat{x}}-x_{12}a \, \mathbf{\hat{y}} + \left(\frac{1}{2} +z_{12}\right)c \, \mathbf{\hat{z}} & \left(4d\right) & \mbox{S IV} \\ 
\end{longtabu}
\renewcommand{\arraystretch}{1.0}
\noindent \hrulefill
\\
\textbf{References:}
\vspace*{-0.25cm}
\begin{flushleft}
  - \bibentry{Sandor_Nature_224_1969}. \\
\end{flushleft}
\textbf{Found in:}
\vspace*{-0.25cm}
\begin{flushleft}
  - \bibentry{Shimizu_PRB_51_1995}. \\
\end{flushleft}
\noindent \hrulefill
\\
\textbf{Geometry files:}
\\
\noindent  - CIF: pp. {\hyperref[A2B_tP48_77_8d_4d_cif]{\pageref{A2B_tP48_77_8d_4d_cif}}} \\
\noindent  - POSCAR: pp. {\hyperref[A2B_tP48_77_8d_4d_poscar]{\pageref{A2B_tP48_77_8d_4d_poscar}}} \\
\onecolumn
{\phantomsection\label{A2B7C2_tP88_78_4a_14a_4a}}
\subsection*{\huge \textbf{{\normalfont Sr$_{2}$As$_{2}$O$_{7}$ Structure: A2B7C2\_tP88\_78\_4a\_14a\_4a}}}
\noindent \hrulefill
\vspace*{0.25cm}
\begin{figure}[htp]
  \centering
  \vspace{-1em}
  {\includegraphics[width=1\textwidth]{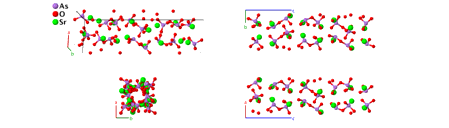}}
\end{figure}
\vspace*{-0.5cm}
\renewcommand{\arraystretch}{1.5}
\begin{equation*}
  \begin{array}{>{$\hspace{-0.15cm}}l<{$}>{$}p{0.5cm}<{$}>{$}p{18.5cm}<{$}}
    \mbox{\large \textbf{Prototype}} &\colon & \ce{Sr2As2O7} \\
    \mbox{\large \textbf{\AFLOW\ prototype label}} &\colon & \mbox{A2B7C2\_tP88\_78\_4a\_14a\_4a} \\
    \mbox{\large \textbf{\textit{Strukturbericht} designation}} &\colon & \mbox{None} \\
    \mbox{\large \textbf{Pearson symbol}} &\colon & \mbox{tP88} \\
    \mbox{\large \textbf{Space group number}} &\colon & 78 \\
    \mbox{\large \textbf{Space group symbol}} &\colon & P4_{3} \\
    \mbox{\large \textbf{\AFLOW\ prototype command}} &\colon &  \texttt{aflow} \,  \, \texttt{-{}-proto=A2B7C2\_tP88\_78\_4a\_14a\_4a } \, \newline \texttt{-{}-params=}{a,c/a,x_{1},y_{1},z_{1},x_{2},y_{2},z_{2},x_{3},y_{3},z_{3},x_{4},y_{4},z_{4},x_{5},y_{5},z_{5},x_{6},y_{6},z_{6},x_{7},} \newline {y_{7},z_{7},x_{8},y_{8},z_{8},x_{9},y_{9},z_{9},x_{10},y_{10},z_{10},x_{11},y_{11},z_{11},x_{12},y_{12},z_{12},x_{13},y_{13},z_{13},x_{14},y_{14},} \newline {z_{14},x_{15},y_{15},z_{15},x_{16},y_{16},z_{16},x_{17},y_{17},z_{17},x_{18},y_{18},z_{18},x_{19},y_{19},z_{19},x_{20},y_{20},z_{20},x_{21},} \newline {y_{21},z_{21},x_{22},y_{22},z_{22} }
  \end{array}
\end{equation*}
\renewcommand{\arraystretch}{1.0}

\noindent \parbox{1 \linewidth}{
\noindent \hrulefill
\\
\textbf{Simple Tetragonal primitive vectors:} \\
\vspace*{-0.25cm}
\begin{tabular}{cc}
  \begin{tabular}{c}
    \parbox{0.6 \linewidth}{
      \renewcommand{\arraystretch}{1.5}
      \begin{equation*}
        \centering
        \begin{array}{ccc}
              \mathbf{a}_1 & = & a \, \mathbf{\hat{x}} \\
    \mathbf{a}_2 & = & a \, \mathbf{\hat{y}} \\
    \mathbf{a}_3 & = & c \, \mathbf{\hat{z}} \\

        \end{array}
      \end{equation*}
    }
    \renewcommand{\arraystretch}{1.0}
  \end{tabular}
  \begin{tabular}{c}
    \includegraphics[width=0.3\linewidth]{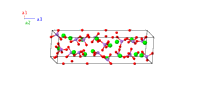} \\
  \end{tabular}
\end{tabular}

}
\vspace*{-0.25cm}

\noindent \hrulefill
\\
\textbf{Basis vectors:}
\vspace*{-0.25cm}
\renewcommand{\arraystretch}{1.5}
\begin{longtabu} to \textwidth{>{\centering $}X[-1,c,c]<{$}>{\centering $}X[-1,c,c]<{$}>{\centering $}X[-1,c,c]<{$}>{\centering $}X[-1,c,c]<{$}>{\centering $}X[-1,c,c]<{$}>{\centering $}X[-1,c,c]<{$}>{\centering $}X[-1,c,c]<{$}}
  & & \mbox{Lattice Coordinates} & & \mbox{Cartesian Coordinates} &\mbox{Wyckoff Position} & \mbox{Atom Type} \\  
  \mathbf{B}_{1} & = & x_{1} \, \mathbf{a}_{1} + y_{1} \, \mathbf{a}_{2} + z_{1} \, \mathbf{a}_{3} & = & x_{1}a \, \mathbf{\hat{x}} + y_{1}a \, \mathbf{\hat{y}} + z_{1}c \, \mathbf{\hat{z}} & \left(4a\right) & \mbox{As I} \\ 
\mathbf{B}_{2} & = & -x_{1} \, \mathbf{a}_{1}-y_{1} \, \mathbf{a}_{2} + \left(\frac{1}{2} +z_{1}\right) \, \mathbf{a}_{3} & = & -x_{1}a \, \mathbf{\hat{x}}-y_{1}a \, \mathbf{\hat{y}} + \left(\frac{1}{2} +z_{1}\right)c \, \mathbf{\hat{z}} & \left(4a\right) & \mbox{As I} \\ 
\mathbf{B}_{3} & = & -y_{1} \, \mathbf{a}_{1} + x_{1} \, \mathbf{a}_{2} + \left(\frac{3}{4} +z_{1}\right) \, \mathbf{a}_{3} & = & -y_{1}a \, \mathbf{\hat{x}} + x_{1}a \, \mathbf{\hat{y}} + \left(\frac{3}{4} +z_{1}\right)c \, \mathbf{\hat{z}} & \left(4a\right) & \mbox{As I} \\ 
\mathbf{B}_{4} & = & y_{1} \, \mathbf{a}_{1}-x_{1} \, \mathbf{a}_{2} + \left(\frac{1}{4} +z_{1}\right) \, \mathbf{a}_{3} & = & y_{1}a \, \mathbf{\hat{x}}-x_{1}a \, \mathbf{\hat{y}} + \left(\frac{1}{4} +z_{1}\right)c \, \mathbf{\hat{z}} & \left(4a\right) & \mbox{As I} \\ 
\mathbf{B}_{5} & = & x_{2} \, \mathbf{a}_{1} + y_{2} \, \mathbf{a}_{2} + z_{2} \, \mathbf{a}_{3} & = & x_{2}a \, \mathbf{\hat{x}} + y_{2}a \, \mathbf{\hat{y}} + z_{2}c \, \mathbf{\hat{z}} & \left(4a\right) & \mbox{As II} \\ 
\mathbf{B}_{6} & = & -x_{2} \, \mathbf{a}_{1}-y_{2} \, \mathbf{a}_{2} + \left(\frac{1}{2} +z_{2}\right) \, \mathbf{a}_{3} & = & -x_{2}a \, \mathbf{\hat{x}}-y_{2}a \, \mathbf{\hat{y}} + \left(\frac{1}{2} +z_{2}\right)c \, \mathbf{\hat{z}} & \left(4a\right) & \mbox{As II} \\ 
\mathbf{B}_{7} & = & -y_{2} \, \mathbf{a}_{1} + x_{2} \, \mathbf{a}_{2} + \left(\frac{3}{4} +z_{2}\right) \, \mathbf{a}_{3} & = & -y_{2}a \, \mathbf{\hat{x}} + x_{2}a \, \mathbf{\hat{y}} + \left(\frac{3}{4} +z_{2}\right)c \, \mathbf{\hat{z}} & \left(4a\right) & \mbox{As II} \\ 
\mathbf{B}_{8} & = & y_{2} \, \mathbf{a}_{1}-x_{2} \, \mathbf{a}_{2} + \left(\frac{1}{4} +z_{2}\right) \, \mathbf{a}_{3} & = & y_{2}a \, \mathbf{\hat{x}}-x_{2}a \, \mathbf{\hat{y}} + \left(\frac{1}{4} +z_{2}\right)c \, \mathbf{\hat{z}} & \left(4a\right) & \mbox{As II} \\ 
\mathbf{B}_{9} & = & x_{3} \, \mathbf{a}_{1} + y_{3} \, \mathbf{a}_{2} + z_{3} \, \mathbf{a}_{3} & = & x_{3}a \, \mathbf{\hat{x}} + y_{3}a \, \mathbf{\hat{y}} + z_{3}c \, \mathbf{\hat{z}} & \left(4a\right) & \mbox{As III} \\ 
\mathbf{B}_{10} & = & -x_{3} \, \mathbf{a}_{1}-y_{3} \, \mathbf{a}_{2} + \left(\frac{1}{2} +z_{3}\right) \, \mathbf{a}_{3} & = & -x_{3}a \, \mathbf{\hat{x}}-y_{3}a \, \mathbf{\hat{y}} + \left(\frac{1}{2} +z_{3}\right)c \, \mathbf{\hat{z}} & \left(4a\right) & \mbox{As III} \\ 
\mathbf{B}_{11} & = & -y_{3} \, \mathbf{a}_{1} + x_{3} \, \mathbf{a}_{2} + \left(\frac{3}{4} +z_{3}\right) \, \mathbf{a}_{3} & = & -y_{3}a \, \mathbf{\hat{x}} + x_{3}a \, \mathbf{\hat{y}} + \left(\frac{3}{4} +z_{3}\right)c \, \mathbf{\hat{z}} & \left(4a\right) & \mbox{As III} \\ 
\mathbf{B}_{12} & = & y_{3} \, \mathbf{a}_{1}-x_{3} \, \mathbf{a}_{2} + \left(\frac{1}{4} +z_{3}\right) \, \mathbf{a}_{3} & = & y_{3}a \, \mathbf{\hat{x}}-x_{3}a \, \mathbf{\hat{y}} + \left(\frac{1}{4} +z_{3}\right)c \, \mathbf{\hat{z}} & \left(4a\right) & \mbox{As III} \\ 
\mathbf{B}_{13} & = & x_{4} \, \mathbf{a}_{1} + y_{4} \, \mathbf{a}_{2} + z_{4} \, \mathbf{a}_{3} & = & x_{4}a \, \mathbf{\hat{x}} + y_{4}a \, \mathbf{\hat{y}} + z_{4}c \, \mathbf{\hat{z}} & \left(4a\right) & \mbox{As IV} \\ 
\mathbf{B}_{14} & = & -x_{4} \, \mathbf{a}_{1}-y_{4} \, \mathbf{a}_{2} + \left(\frac{1}{2} +z_{4}\right) \, \mathbf{a}_{3} & = & -x_{4}a \, \mathbf{\hat{x}}-y_{4}a \, \mathbf{\hat{y}} + \left(\frac{1}{2} +z_{4}\right)c \, \mathbf{\hat{z}} & \left(4a\right) & \mbox{As IV} \\ 
\mathbf{B}_{15} & = & -y_{4} \, \mathbf{a}_{1} + x_{4} \, \mathbf{a}_{2} + \left(\frac{3}{4} +z_{4}\right) \, \mathbf{a}_{3} & = & -y_{4}a \, \mathbf{\hat{x}} + x_{4}a \, \mathbf{\hat{y}} + \left(\frac{3}{4} +z_{4}\right)c \, \mathbf{\hat{z}} & \left(4a\right) & \mbox{As IV} \\ 
\mathbf{B}_{16} & = & y_{4} \, \mathbf{a}_{1}-x_{4} \, \mathbf{a}_{2} + \left(\frac{1}{4} +z_{4}\right) \, \mathbf{a}_{3} & = & y_{4}a \, \mathbf{\hat{x}}-x_{4}a \, \mathbf{\hat{y}} + \left(\frac{1}{4} +z_{4}\right)c \, \mathbf{\hat{z}} & \left(4a\right) & \mbox{As IV} \\ 
\mathbf{B}_{17} & = & x_{5} \, \mathbf{a}_{1} + y_{5} \, \mathbf{a}_{2} + z_{5} \, \mathbf{a}_{3} & = & x_{5}a \, \mathbf{\hat{x}} + y_{5}a \, \mathbf{\hat{y}} + z_{5}c \, \mathbf{\hat{z}} & \left(4a\right) & \mbox{O I} \\ 
\mathbf{B}_{18} & = & -x_{5} \, \mathbf{a}_{1}-y_{5} \, \mathbf{a}_{2} + \left(\frac{1}{2} +z_{5}\right) \, \mathbf{a}_{3} & = & -x_{5}a \, \mathbf{\hat{x}}-y_{5}a \, \mathbf{\hat{y}} + \left(\frac{1}{2} +z_{5}\right)c \, \mathbf{\hat{z}} & \left(4a\right) & \mbox{O I} \\ 
\mathbf{B}_{19} & = & -y_{5} \, \mathbf{a}_{1} + x_{5} \, \mathbf{a}_{2} + \left(\frac{3}{4} +z_{5}\right) \, \mathbf{a}_{3} & = & -y_{5}a \, \mathbf{\hat{x}} + x_{5}a \, \mathbf{\hat{y}} + \left(\frac{3}{4} +z_{5}\right)c \, \mathbf{\hat{z}} & \left(4a\right) & \mbox{O I} \\ 
\mathbf{B}_{20} & = & y_{5} \, \mathbf{a}_{1}-x_{5} \, \mathbf{a}_{2} + \left(\frac{1}{4} +z_{5}\right) \, \mathbf{a}_{3} & = & y_{5}a \, \mathbf{\hat{x}}-x_{5}a \, \mathbf{\hat{y}} + \left(\frac{1}{4} +z_{5}\right)c \, \mathbf{\hat{z}} & \left(4a\right) & \mbox{O I} \\ 
\mathbf{B}_{21} & = & x_{6} \, \mathbf{a}_{1} + y_{6} \, \mathbf{a}_{2} + z_{6} \, \mathbf{a}_{3} & = & x_{6}a \, \mathbf{\hat{x}} + y_{6}a \, \mathbf{\hat{y}} + z_{6}c \, \mathbf{\hat{z}} & \left(4a\right) & \mbox{O II} \\ 
\mathbf{B}_{22} & = & -x_{6} \, \mathbf{a}_{1}-y_{6} \, \mathbf{a}_{2} + \left(\frac{1}{2} +z_{6}\right) \, \mathbf{a}_{3} & = & -x_{6}a \, \mathbf{\hat{x}}-y_{6}a \, \mathbf{\hat{y}} + \left(\frac{1}{2} +z_{6}\right)c \, \mathbf{\hat{z}} & \left(4a\right) & \mbox{O II} \\ 
\mathbf{B}_{23} & = & -y_{6} \, \mathbf{a}_{1} + x_{6} \, \mathbf{a}_{2} + \left(\frac{3}{4} +z_{6}\right) \, \mathbf{a}_{3} & = & -y_{6}a \, \mathbf{\hat{x}} + x_{6}a \, \mathbf{\hat{y}} + \left(\frac{3}{4} +z_{6}\right)c \, \mathbf{\hat{z}} & \left(4a\right) & \mbox{O II} \\ 
\mathbf{B}_{24} & = & y_{6} \, \mathbf{a}_{1}-x_{6} \, \mathbf{a}_{2} + \left(\frac{1}{4} +z_{6}\right) \, \mathbf{a}_{3} & = & y_{6}a \, \mathbf{\hat{x}}-x_{6}a \, \mathbf{\hat{y}} + \left(\frac{1}{4} +z_{6}\right)c \, \mathbf{\hat{z}} & \left(4a\right) & \mbox{O II} \\ 
\mathbf{B}_{25} & = & x_{7} \, \mathbf{a}_{1} + y_{7} \, \mathbf{a}_{2} + z_{7} \, \mathbf{a}_{3} & = & x_{7}a \, \mathbf{\hat{x}} + y_{7}a \, \mathbf{\hat{y}} + z_{7}c \, \mathbf{\hat{z}} & \left(4a\right) & \mbox{O III} \\ 
\mathbf{B}_{26} & = & -x_{7} \, \mathbf{a}_{1}-y_{7} \, \mathbf{a}_{2} + \left(\frac{1}{2} +z_{7}\right) \, \mathbf{a}_{3} & = & -x_{7}a \, \mathbf{\hat{x}}-y_{7}a \, \mathbf{\hat{y}} + \left(\frac{1}{2} +z_{7}\right)c \, \mathbf{\hat{z}} & \left(4a\right) & \mbox{O III} \\ 
\mathbf{B}_{27} & = & -y_{7} \, \mathbf{a}_{1} + x_{7} \, \mathbf{a}_{2} + \left(\frac{3}{4} +z_{7}\right) \, \mathbf{a}_{3} & = & -y_{7}a \, \mathbf{\hat{x}} + x_{7}a \, \mathbf{\hat{y}} + \left(\frac{3}{4} +z_{7}\right)c \, \mathbf{\hat{z}} & \left(4a\right) & \mbox{O III} \\ 
\mathbf{B}_{28} & = & y_{7} \, \mathbf{a}_{1}-x_{7} \, \mathbf{a}_{2} + \left(\frac{1}{4} +z_{7}\right) \, \mathbf{a}_{3} & = & y_{7}a \, \mathbf{\hat{x}}-x_{7}a \, \mathbf{\hat{y}} + \left(\frac{1}{4} +z_{7}\right)c \, \mathbf{\hat{z}} & \left(4a\right) & \mbox{O III} \\ 
\mathbf{B}_{29} & = & x_{8} \, \mathbf{a}_{1} + y_{8} \, \mathbf{a}_{2} + z_{8} \, \mathbf{a}_{3} & = & x_{8}a \, \mathbf{\hat{x}} + y_{8}a \, \mathbf{\hat{y}} + z_{8}c \, \mathbf{\hat{z}} & \left(4a\right) & \mbox{O IV} \\ 
\mathbf{B}_{30} & = & -x_{8} \, \mathbf{a}_{1}-y_{8} \, \mathbf{a}_{2} + \left(\frac{1}{2} +z_{8}\right) \, \mathbf{a}_{3} & = & -x_{8}a \, \mathbf{\hat{x}}-y_{8}a \, \mathbf{\hat{y}} + \left(\frac{1}{2} +z_{8}\right)c \, \mathbf{\hat{z}} & \left(4a\right) & \mbox{O IV} \\ 
\mathbf{B}_{31} & = & -y_{8} \, \mathbf{a}_{1} + x_{8} \, \mathbf{a}_{2} + \left(\frac{3}{4} +z_{8}\right) \, \mathbf{a}_{3} & = & -y_{8}a \, \mathbf{\hat{x}} + x_{8}a \, \mathbf{\hat{y}} + \left(\frac{3}{4} +z_{8}\right)c \, \mathbf{\hat{z}} & \left(4a\right) & \mbox{O IV} \\ 
\mathbf{B}_{32} & = & y_{8} \, \mathbf{a}_{1}-x_{8} \, \mathbf{a}_{2} + \left(\frac{1}{4} +z_{8}\right) \, \mathbf{a}_{3} & = & y_{8}a \, \mathbf{\hat{x}}-x_{8}a \, \mathbf{\hat{y}} + \left(\frac{1}{4} +z_{8}\right)c \, \mathbf{\hat{z}} & \left(4a\right) & \mbox{O IV} \\ 
\mathbf{B}_{33} & = & x_{9} \, \mathbf{a}_{1} + y_{9} \, \mathbf{a}_{2} + z_{9} \, \mathbf{a}_{3} & = & x_{9}a \, \mathbf{\hat{x}} + y_{9}a \, \mathbf{\hat{y}} + z_{9}c \, \mathbf{\hat{z}} & \left(4a\right) & \mbox{O V} \\ 
\mathbf{B}_{34} & = & -x_{9} \, \mathbf{a}_{1}-y_{9} \, \mathbf{a}_{2} + \left(\frac{1}{2} +z_{9}\right) \, \mathbf{a}_{3} & = & -x_{9}a \, \mathbf{\hat{x}}-y_{9}a \, \mathbf{\hat{y}} + \left(\frac{1}{2} +z_{9}\right)c \, \mathbf{\hat{z}} & \left(4a\right) & \mbox{O V} \\ 
\mathbf{B}_{35} & = & -y_{9} \, \mathbf{a}_{1} + x_{9} \, \mathbf{a}_{2} + \left(\frac{3}{4} +z_{9}\right) \, \mathbf{a}_{3} & = & -y_{9}a \, \mathbf{\hat{x}} + x_{9}a \, \mathbf{\hat{y}} + \left(\frac{3}{4} +z_{9}\right)c \, \mathbf{\hat{z}} & \left(4a\right) & \mbox{O V} \\ 
\mathbf{B}_{36} & = & y_{9} \, \mathbf{a}_{1}-x_{9} \, \mathbf{a}_{2} + \left(\frac{1}{4} +z_{9}\right) \, \mathbf{a}_{3} & = & y_{9}a \, \mathbf{\hat{x}}-x_{9}a \, \mathbf{\hat{y}} + \left(\frac{1}{4} +z_{9}\right)c \, \mathbf{\hat{z}} & \left(4a\right) & \mbox{O V} \\ 
\mathbf{B}_{37} & = & x_{10} \, \mathbf{a}_{1} + y_{10} \, \mathbf{a}_{2} + z_{10} \, \mathbf{a}_{3} & = & x_{10}a \, \mathbf{\hat{x}} + y_{10}a \, \mathbf{\hat{y}} + z_{10}c \, \mathbf{\hat{z}} & \left(4a\right) & \mbox{O VI} \\ 
\mathbf{B}_{38} & = & -x_{10} \, \mathbf{a}_{1}-y_{10} \, \mathbf{a}_{2} + \left(\frac{1}{2} +z_{10}\right) \, \mathbf{a}_{3} & = & -x_{10}a \, \mathbf{\hat{x}}-y_{10}a \, \mathbf{\hat{y}} + \left(\frac{1}{2} +z_{10}\right)c \, \mathbf{\hat{z}} & \left(4a\right) & \mbox{O VI} \\ 
\mathbf{B}_{39} & = & -y_{10} \, \mathbf{a}_{1} + x_{10} \, \mathbf{a}_{2} + \left(\frac{3}{4} +z_{10}\right) \, \mathbf{a}_{3} & = & -y_{10}a \, \mathbf{\hat{x}} + x_{10}a \, \mathbf{\hat{y}} + \left(\frac{3}{4} +z_{10}\right)c \, \mathbf{\hat{z}} & \left(4a\right) & \mbox{O VI} \\ 
\mathbf{B}_{40} & = & y_{10} \, \mathbf{a}_{1}-x_{10} \, \mathbf{a}_{2} + \left(\frac{1}{4} +z_{10}\right) \, \mathbf{a}_{3} & = & y_{10}a \, \mathbf{\hat{x}}-x_{10}a \, \mathbf{\hat{y}} + \left(\frac{1}{4} +z_{10}\right)c \, \mathbf{\hat{z}} & \left(4a\right) & \mbox{O VI} \\ 
\mathbf{B}_{41} & = & x_{11} \, \mathbf{a}_{1} + y_{11} \, \mathbf{a}_{2} + z_{11} \, \mathbf{a}_{3} & = & x_{11}a \, \mathbf{\hat{x}} + y_{11}a \, \mathbf{\hat{y}} + z_{11}c \, \mathbf{\hat{z}} & \left(4a\right) & \mbox{O VII} \\ 
\mathbf{B}_{42} & = & -x_{11} \, \mathbf{a}_{1}-y_{11} \, \mathbf{a}_{2} + \left(\frac{1}{2} +z_{11}\right) \, \mathbf{a}_{3} & = & -x_{11}a \, \mathbf{\hat{x}}-y_{11}a \, \mathbf{\hat{y}} + \left(\frac{1}{2} +z_{11}\right)c \, \mathbf{\hat{z}} & \left(4a\right) & \mbox{O VII} \\ 
\mathbf{B}_{43} & = & -y_{11} \, \mathbf{a}_{1} + x_{11} \, \mathbf{a}_{2} + \left(\frac{3}{4} +z_{11}\right) \, \mathbf{a}_{3} & = & -y_{11}a \, \mathbf{\hat{x}} + x_{11}a \, \mathbf{\hat{y}} + \left(\frac{3}{4} +z_{11}\right)c \, \mathbf{\hat{z}} & \left(4a\right) & \mbox{O VII} \\ 
\mathbf{B}_{44} & = & y_{11} \, \mathbf{a}_{1}-x_{11} \, \mathbf{a}_{2} + \left(\frac{1}{4} +z_{11}\right) \, \mathbf{a}_{3} & = & y_{11}a \, \mathbf{\hat{x}}-x_{11}a \, \mathbf{\hat{y}} + \left(\frac{1}{4} +z_{11}\right)c \, \mathbf{\hat{z}} & \left(4a\right) & \mbox{O VII} \\ 
\mathbf{B}_{45} & = & x_{12} \, \mathbf{a}_{1} + y_{12} \, \mathbf{a}_{2} + z_{12} \, \mathbf{a}_{3} & = & x_{12}a \, \mathbf{\hat{x}} + y_{12}a \, \mathbf{\hat{y}} + z_{12}c \, \mathbf{\hat{z}} & \left(4a\right) & \mbox{O VIII} \\ 
\mathbf{B}_{46} & = & -x_{12} \, \mathbf{a}_{1}-y_{12} \, \mathbf{a}_{2} + \left(\frac{1}{2} +z_{12}\right) \, \mathbf{a}_{3} & = & -x_{12}a \, \mathbf{\hat{x}}-y_{12}a \, \mathbf{\hat{y}} + \left(\frac{1}{2} +z_{12}\right)c \, \mathbf{\hat{z}} & \left(4a\right) & \mbox{O VIII} \\ 
\mathbf{B}_{47} & = & -y_{12} \, \mathbf{a}_{1} + x_{12} \, \mathbf{a}_{2} + \left(\frac{3}{4} +z_{12}\right) \, \mathbf{a}_{3} & = & -y_{12}a \, \mathbf{\hat{x}} + x_{12}a \, \mathbf{\hat{y}} + \left(\frac{3}{4} +z_{12}\right)c \, \mathbf{\hat{z}} & \left(4a\right) & \mbox{O VIII} \\ 
\mathbf{B}_{48} & = & y_{12} \, \mathbf{a}_{1}-x_{12} \, \mathbf{a}_{2} + \left(\frac{1}{4} +z_{12}\right) \, \mathbf{a}_{3} & = & y_{12}a \, \mathbf{\hat{x}}-x_{12}a \, \mathbf{\hat{y}} + \left(\frac{1}{4} +z_{12}\right)c \, \mathbf{\hat{z}} & \left(4a\right) & \mbox{O VIII} \\ 
\mathbf{B}_{49} & = & x_{13} \, \mathbf{a}_{1} + y_{13} \, \mathbf{a}_{2} + z_{13} \, \mathbf{a}_{3} & = & x_{13}a \, \mathbf{\hat{x}} + y_{13}a \, \mathbf{\hat{y}} + z_{13}c \, \mathbf{\hat{z}} & \left(4a\right) & \mbox{O IX} \\ 
\mathbf{B}_{50} & = & -x_{13} \, \mathbf{a}_{1}-y_{13} \, \mathbf{a}_{2} + \left(\frac{1}{2} +z_{13}\right) \, \mathbf{a}_{3} & = & -x_{13}a \, \mathbf{\hat{x}}-y_{13}a \, \mathbf{\hat{y}} + \left(\frac{1}{2} +z_{13}\right)c \, \mathbf{\hat{z}} & \left(4a\right) & \mbox{O IX} \\ 
\mathbf{B}_{51} & = & -y_{13} \, \mathbf{a}_{1} + x_{13} \, \mathbf{a}_{2} + \left(\frac{3}{4} +z_{13}\right) \, \mathbf{a}_{3} & = & -y_{13}a \, \mathbf{\hat{x}} + x_{13}a \, \mathbf{\hat{y}} + \left(\frac{3}{4} +z_{13}\right)c \, \mathbf{\hat{z}} & \left(4a\right) & \mbox{O IX} \\ 
\mathbf{B}_{52} & = & y_{13} \, \mathbf{a}_{1}-x_{13} \, \mathbf{a}_{2} + \left(\frac{1}{4} +z_{13}\right) \, \mathbf{a}_{3} & = & y_{13}a \, \mathbf{\hat{x}}-x_{13}a \, \mathbf{\hat{y}} + \left(\frac{1}{4} +z_{13}\right)c \, \mathbf{\hat{z}} & \left(4a\right) & \mbox{O IX} \\ 
\mathbf{B}_{53} & = & x_{14} \, \mathbf{a}_{1} + y_{14} \, \mathbf{a}_{2} + z_{14} \, \mathbf{a}_{3} & = & x_{14}a \, \mathbf{\hat{x}} + y_{14}a \, \mathbf{\hat{y}} + z_{14}c \, \mathbf{\hat{z}} & \left(4a\right) & \mbox{O X} \\ 
\mathbf{B}_{54} & = & -x_{14} \, \mathbf{a}_{1}-y_{14} \, \mathbf{a}_{2} + \left(\frac{1}{2} +z_{14}\right) \, \mathbf{a}_{3} & = & -x_{14}a \, \mathbf{\hat{x}}-y_{14}a \, \mathbf{\hat{y}} + \left(\frac{1}{2} +z_{14}\right)c \, \mathbf{\hat{z}} & \left(4a\right) & \mbox{O X} \\ 
\mathbf{B}_{55} & = & -y_{14} \, \mathbf{a}_{1} + x_{14} \, \mathbf{a}_{2} + \left(\frac{3}{4} +z_{14}\right) \, \mathbf{a}_{3} & = & -y_{14}a \, \mathbf{\hat{x}} + x_{14}a \, \mathbf{\hat{y}} + \left(\frac{3}{4} +z_{14}\right)c \, \mathbf{\hat{z}} & \left(4a\right) & \mbox{O X} \\ 
\mathbf{B}_{56} & = & y_{14} \, \mathbf{a}_{1}-x_{14} \, \mathbf{a}_{2} + \left(\frac{1}{4} +z_{14}\right) \, \mathbf{a}_{3} & = & y_{14}a \, \mathbf{\hat{x}}-x_{14}a \, \mathbf{\hat{y}} + \left(\frac{1}{4} +z_{14}\right)c \, \mathbf{\hat{z}} & \left(4a\right) & \mbox{O X} \\ 
\mathbf{B}_{57} & = & x_{15} \, \mathbf{a}_{1} + y_{15} \, \mathbf{a}_{2} + z_{15} \, \mathbf{a}_{3} & = & x_{15}a \, \mathbf{\hat{x}} + y_{15}a \, \mathbf{\hat{y}} + z_{15}c \, \mathbf{\hat{z}} & \left(4a\right) & \mbox{O XI} \\ 
\mathbf{B}_{58} & = & -x_{15} \, \mathbf{a}_{1}-y_{15} \, \mathbf{a}_{2} + \left(\frac{1}{2} +z_{15}\right) \, \mathbf{a}_{3} & = & -x_{15}a \, \mathbf{\hat{x}}-y_{15}a \, \mathbf{\hat{y}} + \left(\frac{1}{2} +z_{15}\right)c \, \mathbf{\hat{z}} & \left(4a\right) & \mbox{O XI} \\ 
\mathbf{B}_{59} & = & -y_{15} \, \mathbf{a}_{1} + x_{15} \, \mathbf{a}_{2} + \left(\frac{3}{4} +z_{15}\right) \, \mathbf{a}_{3} & = & -y_{15}a \, \mathbf{\hat{x}} + x_{15}a \, \mathbf{\hat{y}} + \left(\frac{3}{4} +z_{15}\right)c \, \mathbf{\hat{z}} & \left(4a\right) & \mbox{O XI} \\ 
\mathbf{B}_{60} & = & y_{15} \, \mathbf{a}_{1}-x_{15} \, \mathbf{a}_{2} + \left(\frac{1}{4} +z_{15}\right) \, \mathbf{a}_{3} & = & y_{15}a \, \mathbf{\hat{x}}-x_{15}a \, \mathbf{\hat{y}} + \left(\frac{1}{4} +z_{15}\right)c \, \mathbf{\hat{z}} & \left(4a\right) & \mbox{O XI} \\ 
\mathbf{B}_{61} & = & x_{16} \, \mathbf{a}_{1} + y_{16} \, \mathbf{a}_{2} + z_{16} \, \mathbf{a}_{3} & = & x_{16}a \, \mathbf{\hat{x}} + y_{16}a \, \mathbf{\hat{y}} + z_{16}c \, \mathbf{\hat{z}} & \left(4a\right) & \mbox{O XII} \\ 
\mathbf{B}_{62} & = & -x_{16} \, \mathbf{a}_{1}-y_{16} \, \mathbf{a}_{2} + \left(\frac{1}{2} +z_{16}\right) \, \mathbf{a}_{3} & = & -x_{16}a \, \mathbf{\hat{x}}-y_{16}a \, \mathbf{\hat{y}} + \left(\frac{1}{2} +z_{16}\right)c \, \mathbf{\hat{z}} & \left(4a\right) & \mbox{O XII} \\ 
\mathbf{B}_{63} & = & -y_{16} \, \mathbf{a}_{1} + x_{16} \, \mathbf{a}_{2} + \left(\frac{3}{4} +z_{16}\right) \, \mathbf{a}_{3} & = & -y_{16}a \, \mathbf{\hat{x}} + x_{16}a \, \mathbf{\hat{y}} + \left(\frac{3}{4} +z_{16}\right)c \, \mathbf{\hat{z}} & \left(4a\right) & \mbox{O XII} \\ 
\mathbf{B}_{64} & = & y_{16} \, \mathbf{a}_{1}-x_{16} \, \mathbf{a}_{2} + \left(\frac{1}{4} +z_{16}\right) \, \mathbf{a}_{3} & = & y_{16}a \, \mathbf{\hat{x}}-x_{16}a \, \mathbf{\hat{y}} + \left(\frac{1}{4} +z_{16}\right)c \, \mathbf{\hat{z}} & \left(4a\right) & \mbox{O XII} \\ 
\mathbf{B}_{65} & = & x_{17} \, \mathbf{a}_{1} + y_{17} \, \mathbf{a}_{2} + z_{17} \, \mathbf{a}_{3} & = & x_{17}a \, \mathbf{\hat{x}} + y_{17}a \, \mathbf{\hat{y}} + z_{17}c \, \mathbf{\hat{z}} & \left(4a\right) & \mbox{O XIII} \\ 
\mathbf{B}_{66} & = & -x_{17} \, \mathbf{a}_{1}-y_{17} \, \mathbf{a}_{2} + \left(\frac{1}{2} +z_{17}\right) \, \mathbf{a}_{3} & = & -x_{17}a \, \mathbf{\hat{x}}-y_{17}a \, \mathbf{\hat{y}} + \left(\frac{1}{2} +z_{17}\right)c \, \mathbf{\hat{z}} & \left(4a\right) & \mbox{O XIII} \\ 
\mathbf{B}_{67} & = & -y_{17} \, \mathbf{a}_{1} + x_{17} \, \mathbf{a}_{2} + \left(\frac{3}{4} +z_{17}\right) \, \mathbf{a}_{3} & = & -y_{17}a \, \mathbf{\hat{x}} + x_{17}a \, \mathbf{\hat{y}} + \left(\frac{3}{4} +z_{17}\right)c \, \mathbf{\hat{z}} & \left(4a\right) & \mbox{O XIII} \\ 
\mathbf{B}_{68} & = & y_{17} \, \mathbf{a}_{1}-x_{17} \, \mathbf{a}_{2} + \left(\frac{1}{4} +z_{17}\right) \, \mathbf{a}_{3} & = & y_{17}a \, \mathbf{\hat{x}}-x_{17}a \, \mathbf{\hat{y}} + \left(\frac{1}{4} +z_{17}\right)c \, \mathbf{\hat{z}} & \left(4a\right) & \mbox{O XIII} \\ 
\mathbf{B}_{69} & = & x_{18} \, \mathbf{a}_{1} + y_{18} \, \mathbf{a}_{2} + z_{18} \, \mathbf{a}_{3} & = & x_{18}a \, \mathbf{\hat{x}} + y_{18}a \, \mathbf{\hat{y}} + z_{18}c \, \mathbf{\hat{z}} & \left(4a\right) & \mbox{O XIV} \\ 
\mathbf{B}_{70} & = & -x_{18} \, \mathbf{a}_{1}-y_{18} \, \mathbf{a}_{2} + \left(\frac{1}{2} +z_{18}\right) \, \mathbf{a}_{3} & = & -x_{18}a \, \mathbf{\hat{x}}-y_{18}a \, \mathbf{\hat{y}} + \left(\frac{1}{2} +z_{18}\right)c \, \mathbf{\hat{z}} & \left(4a\right) & \mbox{O XIV} \\ 
\mathbf{B}_{71} & = & -y_{18} \, \mathbf{a}_{1} + x_{18} \, \mathbf{a}_{2} + \left(\frac{3}{4} +z_{18}\right) \, \mathbf{a}_{3} & = & -y_{18}a \, \mathbf{\hat{x}} + x_{18}a \, \mathbf{\hat{y}} + \left(\frac{3}{4} +z_{18}\right)c \, \mathbf{\hat{z}} & \left(4a\right) & \mbox{O XIV} \\ 
\mathbf{B}_{72} & = & y_{18} \, \mathbf{a}_{1}-x_{18} \, \mathbf{a}_{2} + \left(\frac{1}{4} +z_{18}\right) \, \mathbf{a}_{3} & = & y_{18}a \, \mathbf{\hat{x}}-x_{18}a \, \mathbf{\hat{y}} + \left(\frac{1}{4} +z_{18}\right)c \, \mathbf{\hat{z}} & \left(4a\right) & \mbox{O XIV} \\ 
\mathbf{B}_{73} & = & x_{19} \, \mathbf{a}_{1} + y_{19} \, \mathbf{a}_{2} + z_{19} \, \mathbf{a}_{3} & = & x_{19}a \, \mathbf{\hat{x}} + y_{19}a \, \mathbf{\hat{y}} + z_{19}c \, \mathbf{\hat{z}} & \left(4a\right) & \mbox{Sr I} \\ 
\mathbf{B}_{74} & = & -x_{19} \, \mathbf{a}_{1}-y_{19} \, \mathbf{a}_{2} + \left(\frac{1}{2} +z_{19}\right) \, \mathbf{a}_{3} & = & -x_{19}a \, \mathbf{\hat{x}}-y_{19}a \, \mathbf{\hat{y}} + \left(\frac{1}{2} +z_{19}\right)c \, \mathbf{\hat{z}} & \left(4a\right) & \mbox{Sr I} \\ 
\mathbf{B}_{75} & = & -y_{19} \, \mathbf{a}_{1} + x_{19} \, \mathbf{a}_{2} + \left(\frac{3}{4} +z_{19}\right) \, \mathbf{a}_{3} & = & -y_{19}a \, \mathbf{\hat{x}} + x_{19}a \, \mathbf{\hat{y}} + \left(\frac{3}{4} +z_{19}\right)c \, \mathbf{\hat{z}} & \left(4a\right) & \mbox{Sr I} \\ 
\mathbf{B}_{76} & = & y_{19} \, \mathbf{a}_{1}-x_{19} \, \mathbf{a}_{2} + \left(\frac{1}{4} +z_{19}\right) \, \mathbf{a}_{3} & = & y_{19}a \, \mathbf{\hat{x}}-x_{19}a \, \mathbf{\hat{y}} + \left(\frac{1}{4} +z_{19}\right)c \, \mathbf{\hat{z}} & \left(4a\right) & \mbox{Sr I} \\ 
\mathbf{B}_{77} & = & x_{20} \, \mathbf{a}_{1} + y_{20} \, \mathbf{a}_{2} + z_{20} \, \mathbf{a}_{3} & = & x_{20}a \, \mathbf{\hat{x}} + y_{20}a \, \mathbf{\hat{y}} + z_{20}c \, \mathbf{\hat{z}} & \left(4a\right) & \mbox{Sr II} \\ 
\mathbf{B}_{78} & = & -x_{20} \, \mathbf{a}_{1}-y_{20} \, \mathbf{a}_{2} + \left(\frac{1}{2} +z_{20}\right) \, \mathbf{a}_{3} & = & -x_{20}a \, \mathbf{\hat{x}}-y_{20}a \, \mathbf{\hat{y}} + \left(\frac{1}{2} +z_{20}\right)c \, \mathbf{\hat{z}} & \left(4a\right) & \mbox{Sr II} \\ 
\mathbf{B}_{79} & = & -y_{20} \, \mathbf{a}_{1} + x_{20} \, \mathbf{a}_{2} + \left(\frac{3}{4} +z_{20}\right) \, \mathbf{a}_{3} & = & -y_{20}a \, \mathbf{\hat{x}} + x_{20}a \, \mathbf{\hat{y}} + \left(\frac{3}{4} +z_{20}\right)c \, \mathbf{\hat{z}} & \left(4a\right) & \mbox{Sr II} \\ 
\mathbf{B}_{80} & = & y_{20} \, \mathbf{a}_{1}-x_{20} \, \mathbf{a}_{2} + \left(\frac{1}{4} +z_{20}\right) \, \mathbf{a}_{3} & = & y_{20}a \, \mathbf{\hat{x}}-x_{20}a \, \mathbf{\hat{y}} + \left(\frac{1}{4} +z_{20}\right)c \, \mathbf{\hat{z}} & \left(4a\right) & \mbox{Sr II} \\ 
\mathbf{B}_{81} & = & x_{21} \, \mathbf{a}_{1} + y_{21} \, \mathbf{a}_{2} + z_{21} \, \mathbf{a}_{3} & = & x_{21}a \, \mathbf{\hat{x}} + y_{21}a \, \mathbf{\hat{y}} + z_{21}c \, \mathbf{\hat{z}} & \left(4a\right) & \mbox{Sr III} \\ 
\mathbf{B}_{82} & = & -x_{21} \, \mathbf{a}_{1}-y_{21} \, \mathbf{a}_{2} + \left(\frac{1}{2} +z_{21}\right) \, \mathbf{a}_{3} & = & -x_{21}a \, \mathbf{\hat{x}}-y_{21}a \, \mathbf{\hat{y}} + \left(\frac{1}{2} +z_{21}\right)c \, \mathbf{\hat{z}} & \left(4a\right) & \mbox{Sr III} \\ 
\mathbf{B}_{83} & = & -y_{21} \, \mathbf{a}_{1} + x_{21} \, \mathbf{a}_{2} + \left(\frac{3}{4} +z_{21}\right) \, \mathbf{a}_{3} & = & -y_{21}a \, \mathbf{\hat{x}} + x_{21}a \, \mathbf{\hat{y}} + \left(\frac{3}{4} +z_{21}\right)c \, \mathbf{\hat{z}} & \left(4a\right) & \mbox{Sr III} \\ 
\mathbf{B}_{84} & = & y_{21} \, \mathbf{a}_{1}-x_{21} \, \mathbf{a}_{2} + \left(\frac{1}{4} +z_{21}\right) \, \mathbf{a}_{3} & = & y_{21}a \, \mathbf{\hat{x}}-x_{21}a \, \mathbf{\hat{y}} + \left(\frac{1}{4} +z_{21}\right)c \, \mathbf{\hat{z}} & \left(4a\right) & \mbox{Sr III} \\ 
\mathbf{B}_{85} & = & x_{22} \, \mathbf{a}_{1} + y_{22} \, \mathbf{a}_{2} + z_{22} \, \mathbf{a}_{3} & = & x_{22}a \, \mathbf{\hat{x}} + y_{22}a \, \mathbf{\hat{y}} + z_{22}c \, \mathbf{\hat{z}} & \left(4a\right) & \mbox{Sr IV} \\ 
\mathbf{B}_{86} & = & -x_{22} \, \mathbf{a}_{1}-y_{22} \, \mathbf{a}_{2} + \left(\frac{1}{2} +z_{22}\right) \, \mathbf{a}_{3} & = & -x_{22}a \, \mathbf{\hat{x}}-y_{22}a \, \mathbf{\hat{y}} + \left(\frac{1}{2} +z_{22}\right)c \, \mathbf{\hat{z}} & \left(4a\right) & \mbox{Sr IV} \\ 
\mathbf{B}_{87} & = & -y_{22} \, \mathbf{a}_{1} + x_{22} \, \mathbf{a}_{2} + \left(\frac{3}{4} +z_{22}\right) \, \mathbf{a}_{3} & = & -y_{22}a \, \mathbf{\hat{x}} + x_{22}a \, \mathbf{\hat{y}} + \left(\frac{3}{4} +z_{22}\right)c \, \mathbf{\hat{z}} & \left(4a\right) & \mbox{Sr IV} \\ 
\mathbf{B}_{88} & = & y_{22} \, \mathbf{a}_{1}-x_{22} \, \mathbf{a}_{2} + \left(\frac{1}{4} +z_{22}\right) \, \mathbf{a}_{3} & = & y_{22}a \, \mathbf{\hat{x}}-x_{22}a \, \mathbf{\hat{y}} + \left(\frac{1}{4} +z_{22}\right)c \, \mathbf{\hat{z}} & \left(4a\right) & \mbox{Sr IV} \\ 
\end{longtabu}
\renewcommand{\arraystretch}{1.0}
\noindent \hrulefill
\\
\textbf{References:}
\vspace*{-0.25cm}
\begin{flushleft}
  - \bibentry{Mbarek_Sr2As2O7_ActCrystallogSecE_2013}. \\
\end{flushleft}
\textbf{Found in:}
\vspace*{-0.25cm}
\begin{flushleft}
  - \bibentry{Villars_PearsonsCrystalData_2013}. \\
\end{flushleft}
\noindent \hrulefill
\\
\textbf{Geometry files:}
\\
\noindent  - CIF: pp. {\hyperref[A2B7C2_tP88_78_4a_14a_4a_cif]{\pageref{A2B7C2_tP88_78_4a_14a_4a_cif}}} \\
\noindent  - POSCAR: pp. {\hyperref[A2B7C2_tP88_78_4a_14a_4a_poscar]{\pageref{A2B7C2_tP88_78_4a_14a_4a_poscar}}} \\
\onecolumn
{\phantomsection\label{A2BC2_tI20_79_c_2a_c}}
\subsection*{\huge \textbf{{\normalfont TlZn$_{2}$Sb$_{2}$ Structure: A2BC2\_tI20\_79\_c\_2a\_c}}}
\noindent \hrulefill
\vspace*{0.25cm}
\begin{figure}[htp]
  \centering
  \vspace{-1em}
  {\includegraphics[width=1\textwidth]{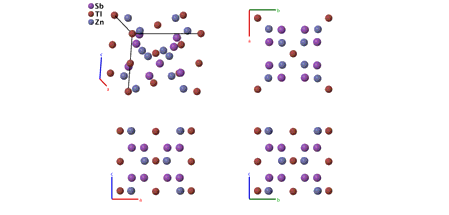}}
\end{figure}
\vspace*{-0.5cm}
\renewcommand{\arraystretch}{1.5}
\begin{equation*}
  \begin{array}{>{$\hspace{-0.15cm}}l<{$}>{$}p{0.5cm}<{$}>{$}p{18.5cm}<{$}}
    \mbox{\large \textbf{Prototype}} &\colon & \ce{TlZn2Sb2} \\
    \mbox{\large \textbf{\AFLOW\ prototype label}} &\colon & \mbox{A2BC2\_tI20\_79\_c\_2a\_c} \\
    \mbox{\large \textbf{\textit{Strukturbericht} designation}} &\colon & \mbox{None} \\
    \mbox{\large \textbf{Pearson symbol}} &\colon & \mbox{tI20} \\
    \mbox{\large \textbf{Space group number}} &\colon & 79 \\
    \mbox{\large \textbf{Space group symbol}} &\colon & I4 \\
    \mbox{\large \textbf{\AFLOW\ prototype command}} &\colon &  \texttt{aflow} \,  \, \texttt{-{}-proto=A2BC2\_tI20\_79\_c\_2a\_c } \, \newline \texttt{-{}-params=}{a,c/a,z_{1},z_{2},x_{3},y_{3},z_{3},x_{4},y_{4},z_{4} }
  \end{array}
\end{equation*}
\renewcommand{\arraystretch}{1.0}

\noindent \parbox{1 \linewidth}{
\noindent \hrulefill
\\
\textbf{Body-centered Tetragonal primitive vectors:} \\
\vspace*{-0.25cm}
\begin{tabular}{cc}
  \begin{tabular}{c}
    \parbox{0.6 \linewidth}{
      \renewcommand{\arraystretch}{1.5}
      \begin{equation*}
        \centering
        \begin{array}{ccc}
              \mathbf{a}_1 & = & - \frac12 \, a \, \mathbf{\hat{x}} + \frac12 \, a \, \mathbf{\hat{y}} + \frac12 \, c \, \mathbf{\hat{z}} \\
    \mathbf{a}_2 & = & ~ \frac12 \, a \, \mathbf{\hat{x}} - \frac12 \, a \, \mathbf{\hat{y}} + \frac12 \, c \, \mathbf{\hat{z}} \\
    \mathbf{a}_3 & = & ~ \frac12 \, a \, \mathbf{\hat{x}} + \frac12 \, a \, \mathbf{\hat{y}} - \frac12 \, c \, \mathbf{\hat{z}} \\

        \end{array}
      \end{equation*}
    }
    \renewcommand{\arraystretch}{1.0}
  \end{tabular}
  \begin{tabular}{c}
    \includegraphics[width=0.3\linewidth]{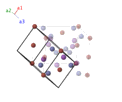} \\
  \end{tabular}
\end{tabular}

}
\vspace*{-0.25cm}

\noindent \hrulefill
\\
\textbf{Basis vectors:}
\vspace*{-0.25cm}
\renewcommand{\arraystretch}{1.5}
\begin{longtabu} to \textwidth{>{\centering $}X[-1,c,c]<{$}>{\centering $}X[-1,c,c]<{$}>{\centering $}X[-1,c,c]<{$}>{\centering $}X[-1,c,c]<{$}>{\centering $}X[-1,c,c]<{$}>{\centering $}X[-1,c,c]<{$}>{\centering $}X[-1,c,c]<{$}}
  & & \mbox{Lattice Coordinates} & & \mbox{Cartesian Coordinates} &\mbox{Wyckoff Position} & \mbox{Atom Type} \\  
  \mathbf{B}_{1} & = & z_{1} \, \mathbf{a}_{1} + z_{1} \, \mathbf{a}_{2} & = & z_{1}c \, \mathbf{\hat{z}} & \left(2a\right) & \mbox{Tl I} \\ 
\mathbf{B}_{2} & = & z_{2} \, \mathbf{a}_{1} + z_{2} \, \mathbf{a}_{2} & = & z_{2}c \, \mathbf{\hat{z}} & \left(2a\right) & \mbox{Tl II} \\ 
\mathbf{B}_{3} & = & \left(y_{3}+z_{3}\right) \, \mathbf{a}_{1} + \left(x_{3}+z_{3}\right) \, \mathbf{a}_{2} + \left(x_{3}+y_{3}\right) \, \mathbf{a}_{3} & = & x_{3}a \, \mathbf{\hat{x}} + y_{3}a \, \mathbf{\hat{y}} + z_{3}c \, \mathbf{\hat{z}} & \left(8c\right) & \mbox{Sb} \\ 
\mathbf{B}_{4} & = & \left(-y_{3}+z_{3}\right) \, \mathbf{a}_{1} + \left(-x_{3}+z_{3}\right) \, \mathbf{a}_{2} + \left(-x_{3}-y_{3}\right) \, \mathbf{a}_{3} & = & -x_{3}a \, \mathbf{\hat{x}}-y_{3}a \, \mathbf{\hat{y}} + z_{3}c \, \mathbf{\hat{z}} & \left(8c\right) & \mbox{Sb} \\ 
\mathbf{B}_{5} & = & \left(x_{3}+z_{3}\right) \, \mathbf{a}_{1} + \left(-y_{3}+z_{3}\right) \, \mathbf{a}_{2} + \left(x_{3}-y_{3}\right) \, \mathbf{a}_{3} & = & -y_{3}a \, \mathbf{\hat{x}} + x_{3}a \, \mathbf{\hat{y}} + z_{3}c \, \mathbf{\hat{z}} & \left(8c\right) & \mbox{Sb} \\ 
\mathbf{B}_{6} & = & \left(-x_{3}+z_{3}\right) \, \mathbf{a}_{1} + \left(y_{3}+z_{3}\right) \, \mathbf{a}_{2} + \left(-x_{3}+y_{3}\right) \, \mathbf{a}_{3} & = & y_{3}a \, \mathbf{\hat{x}}-x_{3}a \, \mathbf{\hat{y}} + z_{3}c \, \mathbf{\hat{z}} & \left(8c\right) & \mbox{Sb} \\ 
\mathbf{B}_{7} & = & \left(y_{4}+z_{4}\right) \, \mathbf{a}_{1} + \left(x_{4}+z_{4}\right) \, \mathbf{a}_{2} + \left(x_{4}+y_{4}\right) \, \mathbf{a}_{3} & = & x_{4}a \, \mathbf{\hat{x}} + y_{4}a \, \mathbf{\hat{y}} + z_{4}c \, \mathbf{\hat{z}} & \left(8c\right) & \mbox{Zn} \\ 
\mathbf{B}_{8} & = & \left(-y_{4}+z_{4}\right) \, \mathbf{a}_{1} + \left(-x_{4}+z_{4}\right) \, \mathbf{a}_{2} + \left(-x_{4}-y_{4}\right) \, \mathbf{a}_{3} & = & -x_{4}a \, \mathbf{\hat{x}}-y_{4}a \, \mathbf{\hat{y}} + z_{4}c \, \mathbf{\hat{z}} & \left(8c\right) & \mbox{Zn} \\ 
\mathbf{B}_{9} & = & \left(x_{4}+z_{4}\right) \, \mathbf{a}_{1} + \left(-y_{4}+z_{4}\right) \, \mathbf{a}_{2} + \left(x_{4}-y_{4}\right) \, \mathbf{a}_{3} & = & -y_{4}a \, \mathbf{\hat{x}} + x_{4}a \, \mathbf{\hat{y}} + z_{4}c \, \mathbf{\hat{z}} & \left(8c\right) & \mbox{Zn} \\ 
\mathbf{B}_{10} & = & \left(-x_{4}+z_{4}\right) \, \mathbf{a}_{1} + \left(y_{4}+z_{4}\right) \, \mathbf{a}_{2} + \left(-x_{4}+y_{4}\right) \, \mathbf{a}_{3} & = & y_{4}a \, \mathbf{\hat{x}}-x_{4}a \, \mathbf{\hat{y}} + z_{4}c \, \mathbf{\hat{z}} & \left(8c\right) & \mbox{Zn} \\ 
\end{longtabu}
\renewcommand{\arraystretch}{1.0}
\noindent \hrulefill
\\
\textbf{References:}
\vspace*{-0.25cm}
\begin{flushleft}
  - \bibentry{Czybulka_TlZn2Sb2_JLessCommMet_1988}. \\
\end{flushleft}
\textbf{Found in:}
\vspace*{-0.25cm}
\begin{flushleft}
  - \bibentry{Villars_PearsonsCrystalData_2013}. \\
\end{flushleft}
\noindent \hrulefill
\\
\textbf{Geometry files:}
\\
\noindent  - CIF: pp. {\hyperref[A2BC2_tI20_79_c_2a_c_cif]{\pageref{A2BC2_tI20_79_c_2a_c_cif}}} \\
\noindent  - POSCAR: pp. {\hyperref[A2BC2_tI20_79_c_2a_c_poscar]{\pageref{A2BC2_tI20_79_c_2a_c_poscar}}} \\
\onecolumn
{\phantomsection\label{AB2_tI48_80_2b_4b}}
\subsection*{\huge \textbf{{\normalfont $\beta$-NbO$_{2}$ Structure: AB2\_tI48\_80\_2b\_4b}}}
\noindent \hrulefill
\vspace*{0.25cm}
\begin{figure}[htp]
  \centering
  \vspace{-1em}
  {\includegraphics[width=1\textwidth]{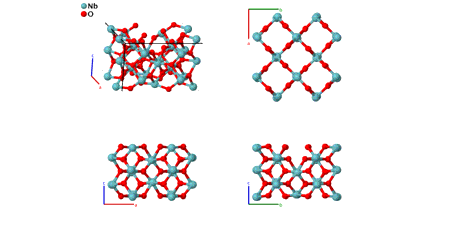}}
\end{figure}
\vspace*{-0.5cm}
\renewcommand{\arraystretch}{1.5}
\begin{equation*}
  \begin{array}{>{$\hspace{-0.15cm}}l<{$}>{$}p{0.5cm}<{$}>{$}p{18.5cm}<{$}}
    \mbox{\large \textbf{Prototype}} &\colon & \ce{$\beta$-NbO2} \\
    \mbox{\large \textbf{\AFLOW\ prototype label}} &\colon & \mbox{AB2\_tI48\_80\_2b\_4b} \\
    \mbox{\large \textbf{\textit{Strukturbericht} designation}} &\colon & \mbox{None} \\
    \mbox{\large \textbf{Pearson symbol}} &\colon & \mbox{tI48} \\
    \mbox{\large \textbf{Space group number}} &\colon & 80 \\
    \mbox{\large \textbf{Space group symbol}} &\colon & I4_{1} \\
    \mbox{\large \textbf{\AFLOW\ prototype command}} &\colon &  \texttt{aflow} \,  \, \texttt{-{}-proto=AB2\_tI48\_80\_2b\_4b } \, \newline \texttt{-{}-params=}{a,c/a,x_{1},y_{1},z_{1},x_{2},y_{2},z_{2},x_{3},y_{3},z_{3},x_{4},y_{4},z_{4},x_{5},y_{5},z_{5},x_{6},y_{6},z_{6} }
  \end{array}
\end{equation*}
\renewcommand{\arraystretch}{1.0}

\vspace*{-0.25cm}
\noindent \hrulefill
\begin{itemize}
  \item{This crystal is not quite stoichiometric. The actual composition was
found to be NbO$_{2-x}$, where $0.002 \le x \le 0.01$.
}
\end{itemize}

\noindent \parbox{1 \linewidth}{
\noindent \hrulefill
\\
\textbf{Body-centered Tetragonal primitive vectors:} \\
\vspace*{-0.25cm}
\begin{tabular}{cc}
  \begin{tabular}{c}
    \parbox{0.6 \linewidth}{
      \renewcommand{\arraystretch}{1.5}
      \begin{equation*}
        \centering
        \begin{array}{ccc}
              \mathbf{a}_1 & = & - \frac12 \, a \, \mathbf{\hat{x}} + \frac12 \, a \, \mathbf{\hat{y}} + \frac12 \, c \, \mathbf{\hat{z}} \\
    \mathbf{a}_2 & = & ~ \frac12 \, a \, \mathbf{\hat{x}} - \frac12 \, a \, \mathbf{\hat{y}} + \frac12 \, c \, \mathbf{\hat{z}} \\
    \mathbf{a}_3 & = & ~ \frac12 \, a \, \mathbf{\hat{x}} + \frac12 \, a \, \mathbf{\hat{y}} - \frac12 \, c \, \mathbf{\hat{z}} \\

        \end{array}
      \end{equation*}
    }
    \renewcommand{\arraystretch}{1.0}
  \end{tabular}
  \begin{tabular}{c}
    \includegraphics[width=0.3\linewidth]{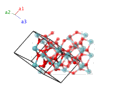} \\
  \end{tabular}
\end{tabular}

}
\vspace*{-0.25cm}

\noindent \hrulefill
\\
\textbf{Basis vectors:}
\vspace*{-0.25cm}
\renewcommand{\arraystretch}{1.5}
\begin{longtabu} to \textwidth{>{\centering $}X[-1,c,c]<{$}>{\centering $}X[-1,c,c]<{$}>{\centering $}X[-1,c,c]<{$}>{\centering $}X[-1,c,c]<{$}>{\centering $}X[-1,c,c]<{$}>{\centering $}X[-1,c,c]<{$}>{\centering $}X[-1,c,c]<{$}}
  & & \mbox{Lattice Coordinates} & & \mbox{Cartesian Coordinates} &\mbox{Wyckoff Position} & \mbox{Atom Type} \\  
  \mathbf{B}_{1} & = & \left(y_{1}+z_{1}\right) \, \mathbf{a}_{1} + \left(x_{1}+z_{1}\right) \, \mathbf{a}_{2} + \left(x_{1}+y_{1}\right) \, \mathbf{a}_{3} & = & x_{1}a \, \mathbf{\hat{x}} + y_{1}a \, \mathbf{\hat{y}} + z_{1}c \, \mathbf{\hat{z}} & \left(8b\right) & \mbox{Nb I} \\ 
\mathbf{B}_{2} & = & \left(-y_{1}+z_{1}\right) \, \mathbf{a}_{1} + \left(-x_{1}+z_{1}\right) \, \mathbf{a}_{2} + \left(-x_{1}-y_{1}\right) \, \mathbf{a}_{3} & = & -x_{1}a \, \mathbf{\hat{x}}-y_{1}a \, \mathbf{\hat{y}} + z_{1}c \, \mathbf{\hat{z}} & \left(8b\right) & \mbox{Nb I} \\ 
\mathbf{B}_{3} & = & \left(\frac{3}{4} +x_{1} + z_{1}\right) \, \mathbf{a}_{1} + \left(\frac{1}{4} - y_{1} + z_{1}\right) \, \mathbf{a}_{2} + \left(\frac{1}{2} +x_{1} - y_{1}\right) \, \mathbf{a}_{3} & = & -y_{1}a \, \mathbf{\hat{x}} + \left(\frac{1}{2} +x_{1}\right)a \, \mathbf{\hat{y}} + \left(\frac{1}{4} +z_{1}\right)c \, \mathbf{\hat{z}} & \left(8b\right) & \mbox{Nb I} \\ 
\mathbf{B}_{4} & = & \left(\frac{3}{4} - x_{1} + z_{1}\right) \, \mathbf{a}_{1} + \left(\frac{1}{4} +y_{1} + z_{1}\right) \, \mathbf{a}_{2} + \left(\frac{1}{2} - x_{1} + y_{1}\right) \, \mathbf{a}_{3} & = & y_{1}a \, \mathbf{\hat{x}} + \left(\frac{1}{2} - x_{1}\right)a \, \mathbf{\hat{y}} + \left(\frac{1}{4} +z_{1}\right)c \, \mathbf{\hat{z}} & \left(8b\right) & \mbox{Nb I} \\ 
\mathbf{B}_{5} & = & \left(y_{2}+z_{2}\right) \, \mathbf{a}_{1} + \left(x_{2}+z_{2}\right) \, \mathbf{a}_{2} + \left(x_{2}+y_{2}\right) \, \mathbf{a}_{3} & = & x_{2}a \, \mathbf{\hat{x}} + y_{2}a \, \mathbf{\hat{y}} + z_{2}c \, \mathbf{\hat{z}} & \left(8b\right) & \mbox{Nb II} \\ 
\mathbf{B}_{6} & = & \left(-y_{2}+z_{2}\right) \, \mathbf{a}_{1} + \left(-x_{2}+z_{2}\right) \, \mathbf{a}_{2} + \left(-x_{2}-y_{2}\right) \, \mathbf{a}_{3} & = & -x_{2}a \, \mathbf{\hat{x}}-y_{2}a \, \mathbf{\hat{y}} + z_{2}c \, \mathbf{\hat{z}} & \left(8b\right) & \mbox{Nb II} \\ 
\mathbf{B}_{7} & = & \left(\frac{3}{4} +x_{2} + z_{2}\right) \, \mathbf{a}_{1} + \left(\frac{1}{4} - y_{2} + z_{2}\right) \, \mathbf{a}_{2} + \left(\frac{1}{2} +x_{2} - y_{2}\right) \, \mathbf{a}_{3} & = & -y_{2}a \, \mathbf{\hat{x}} + \left(\frac{1}{2} +x_{2}\right)a \, \mathbf{\hat{y}} + \left(\frac{1}{4} +z_{2}\right)c \, \mathbf{\hat{z}} & \left(8b\right) & \mbox{Nb II} \\ 
\mathbf{B}_{8} & = & \left(\frac{3}{4} - x_{2} + z_{2}\right) \, \mathbf{a}_{1} + \left(\frac{1}{4} +y_{2} + z_{2}\right) \, \mathbf{a}_{2} + \left(\frac{1}{2} - x_{2} + y_{2}\right) \, \mathbf{a}_{3} & = & y_{2}a \, \mathbf{\hat{x}} + \left(\frac{1}{2} - x_{2}\right)a \, \mathbf{\hat{y}} + \left(\frac{1}{4} +z_{2}\right)c \, \mathbf{\hat{z}} & \left(8b\right) & \mbox{Nb II} \\ 
\mathbf{B}_{9} & = & \left(y_{3}+z_{3}\right) \, \mathbf{a}_{1} + \left(x_{3}+z_{3}\right) \, \mathbf{a}_{2} + \left(x_{3}+y_{3}\right) \, \mathbf{a}_{3} & = & x_{3}a \, \mathbf{\hat{x}} + y_{3}a \, \mathbf{\hat{y}} + z_{3}c \, \mathbf{\hat{z}} & \left(8b\right) & \mbox{O I} \\ 
\mathbf{B}_{10} & = & \left(-y_{3}+z_{3}\right) \, \mathbf{a}_{1} + \left(-x_{3}+z_{3}\right) \, \mathbf{a}_{2} + \left(-x_{3}-y_{3}\right) \, \mathbf{a}_{3} & = & -x_{3}a \, \mathbf{\hat{x}}-y_{3}a \, \mathbf{\hat{y}} + z_{3}c \, \mathbf{\hat{z}} & \left(8b\right) & \mbox{O I} \\ 
\mathbf{B}_{11} & = & \left(\frac{3}{4} +x_{3} + z_{3}\right) \, \mathbf{a}_{1} + \left(\frac{1}{4} - y_{3} + z_{3}\right) \, \mathbf{a}_{2} + \left(\frac{1}{2} +x_{3} - y_{3}\right) \, \mathbf{a}_{3} & = & -y_{3}a \, \mathbf{\hat{x}} + \left(\frac{1}{2} +x_{3}\right)a \, \mathbf{\hat{y}} + \left(\frac{1}{4} +z_{3}\right)c \, \mathbf{\hat{z}} & \left(8b\right) & \mbox{O I} \\ 
\mathbf{B}_{12} & = & \left(\frac{3}{4} - x_{3} + z_{3}\right) \, \mathbf{a}_{1} + \left(\frac{1}{4} +y_{3} + z_{3}\right) \, \mathbf{a}_{2} + \left(\frac{1}{2} - x_{3} + y_{3}\right) \, \mathbf{a}_{3} & = & y_{3}a \, \mathbf{\hat{x}} + \left(\frac{1}{2} - x_{3}\right)a \, \mathbf{\hat{y}} + \left(\frac{1}{4} +z_{3}\right)c \, \mathbf{\hat{z}} & \left(8b\right) & \mbox{O I} \\ 
\mathbf{B}_{13} & = & \left(y_{4}+z_{4}\right) \, \mathbf{a}_{1} + \left(x_{4}+z_{4}\right) \, \mathbf{a}_{2} + \left(x_{4}+y_{4}\right) \, \mathbf{a}_{3} & = & x_{4}a \, \mathbf{\hat{x}} + y_{4}a \, \mathbf{\hat{y}} + z_{4}c \, \mathbf{\hat{z}} & \left(8b\right) & \mbox{O II} \\ 
\mathbf{B}_{14} & = & \left(-y_{4}+z_{4}\right) \, \mathbf{a}_{1} + \left(-x_{4}+z_{4}\right) \, \mathbf{a}_{2} + \left(-x_{4}-y_{4}\right) \, \mathbf{a}_{3} & = & -x_{4}a \, \mathbf{\hat{x}}-y_{4}a \, \mathbf{\hat{y}} + z_{4}c \, \mathbf{\hat{z}} & \left(8b\right) & \mbox{O II} \\ 
\mathbf{B}_{15} & = & \left(\frac{3}{4} +x_{4} + z_{4}\right) \, \mathbf{a}_{1} + \left(\frac{1}{4} - y_{4} + z_{4}\right) \, \mathbf{a}_{2} + \left(\frac{1}{2} +x_{4} - y_{4}\right) \, \mathbf{a}_{3} & = & -y_{4}a \, \mathbf{\hat{x}} + \left(\frac{1}{2} +x_{4}\right)a \, \mathbf{\hat{y}} + \left(\frac{1}{4} +z_{4}\right)c \, \mathbf{\hat{z}} & \left(8b\right) & \mbox{O II} \\ 
\mathbf{B}_{16} & = & \left(\frac{3}{4} - x_{4} + z_{4}\right) \, \mathbf{a}_{1} + \left(\frac{1}{4} +y_{4} + z_{4}\right) \, \mathbf{a}_{2} + \left(\frac{1}{2} - x_{4} + y_{4}\right) \, \mathbf{a}_{3} & = & y_{4}a \, \mathbf{\hat{x}} + \left(\frac{1}{2} - x_{4}\right)a \, \mathbf{\hat{y}} + \left(\frac{1}{4} +z_{4}\right)c \, \mathbf{\hat{z}} & \left(8b\right) & \mbox{O II} \\ 
\mathbf{B}_{17} & = & \left(y_{5}+z_{5}\right) \, \mathbf{a}_{1} + \left(x_{5}+z_{5}\right) \, \mathbf{a}_{2} + \left(x_{5}+y_{5}\right) \, \mathbf{a}_{3} & = & x_{5}a \, \mathbf{\hat{x}} + y_{5}a \, \mathbf{\hat{y}} + z_{5}c \, \mathbf{\hat{z}} & \left(8b\right) & \mbox{O III} \\ 
\mathbf{B}_{18} & = & \left(-y_{5}+z_{5}\right) \, \mathbf{a}_{1} + \left(-x_{5}+z_{5}\right) \, \mathbf{a}_{2} + \left(-x_{5}-y_{5}\right) \, \mathbf{a}_{3} & = & -x_{5}a \, \mathbf{\hat{x}}-y_{5}a \, \mathbf{\hat{y}} + z_{5}c \, \mathbf{\hat{z}} & \left(8b\right) & \mbox{O III} \\ 
\mathbf{B}_{19} & = & \left(\frac{3}{4} +x_{5} + z_{5}\right) \, \mathbf{a}_{1} + \left(\frac{1}{4} - y_{5} + z_{5}\right) \, \mathbf{a}_{2} + \left(\frac{1}{2} +x_{5} - y_{5}\right) \, \mathbf{a}_{3} & = & -y_{5}a \, \mathbf{\hat{x}} + \left(\frac{1}{2} +x_{5}\right)a \, \mathbf{\hat{y}} + \left(\frac{1}{4} +z_{5}\right)c \, \mathbf{\hat{z}} & \left(8b\right) & \mbox{O III} \\ 
\mathbf{B}_{20} & = & \left(\frac{3}{4} - x_{5} + z_{5}\right) \, \mathbf{a}_{1} + \left(\frac{1}{4} +y_{5} + z_{5}\right) \, \mathbf{a}_{2} + \left(\frac{1}{2} - x_{5} + y_{5}\right) \, \mathbf{a}_{3} & = & y_{5}a \, \mathbf{\hat{x}} + \left(\frac{1}{2} - x_{5}\right)a \, \mathbf{\hat{y}} + \left(\frac{1}{4} +z_{5}\right)c \, \mathbf{\hat{z}} & \left(8b\right) & \mbox{O III} \\ 
\mathbf{B}_{21} & = & \left(y_{6}+z_{6}\right) \, \mathbf{a}_{1} + \left(x_{6}+z_{6}\right) \, \mathbf{a}_{2} + \left(x_{6}+y_{6}\right) \, \mathbf{a}_{3} & = & x_{6}a \, \mathbf{\hat{x}} + y_{6}a \, \mathbf{\hat{y}} + z_{6}c \, \mathbf{\hat{z}} & \left(8b\right) & \mbox{O IV} \\ 
\mathbf{B}_{22} & = & \left(-y_{6}+z_{6}\right) \, \mathbf{a}_{1} + \left(-x_{6}+z_{6}\right) \, \mathbf{a}_{2} + \left(-x_{6}-y_{6}\right) \, \mathbf{a}_{3} & = & -x_{6}a \, \mathbf{\hat{x}}-y_{6}a \, \mathbf{\hat{y}} + z_{6}c \, \mathbf{\hat{z}} & \left(8b\right) & \mbox{O IV} \\ 
\mathbf{B}_{23} & = & \left(\frac{3}{4} +x_{6} + z_{6}\right) \, \mathbf{a}_{1} + \left(\frac{1}{4} - y_{6} + z_{6}\right) \, \mathbf{a}_{2} + \left(\frac{1}{2} +x_{6} - y_{6}\right) \, \mathbf{a}_{3} & = & -y_{6}a \, \mathbf{\hat{x}} + \left(\frac{1}{2} +x_{6}\right)a \, \mathbf{\hat{y}} + \left(\frac{1}{4} +z_{6}\right)c \, \mathbf{\hat{z}} & \left(8b\right) & \mbox{O IV} \\ 
\mathbf{B}_{24} & = & \left(\frac{3}{4} - x_{6} + z_{6}\right) \, \mathbf{a}_{1} + \left(\frac{1}{4} +y_{6} + z_{6}\right) \, \mathbf{a}_{2} + \left(\frac{1}{2} - x_{6} + y_{6}\right) \, \mathbf{a}_{3} & = & y_{6}a \, \mathbf{\hat{x}} + \left(\frac{1}{2} - x_{6}\right)a \, \mathbf{\hat{y}} + \left(\frac{1}{4} +z_{6}\right)c \, \mathbf{\hat{z}} & \left(8b\right) & \mbox{O IV} \\ 
\end{longtabu}
\renewcommand{\arraystretch}{1.0}
\noindent \hrulefill
\\
\textbf{References:}
\vspace*{-0.25cm}
\begin{flushleft}
  - \bibentry{Schweizer_Z_f_Natb_37_1982}. \\
\end{flushleft}
\textbf{Found in:}
\vspace*{-0.25cm}
\begin{flushleft}
  - \bibentry{Villars_Pearsons_Handbook_of_Crystallographic_Data_IV_1991}. \\
\end{flushleft}
\noindent \hrulefill
\\
\textbf{Geometry files:}
\\
\noindent  - CIF: pp. {\hyperref[AB2_tI48_80_2b_4b_cif]{\pageref{AB2_tI48_80_2b_4b_cif}}} \\
\noindent  - POSCAR: pp. {\hyperref[AB2_tI48_80_2b_4b_poscar]{\pageref{AB2_tI48_80_2b_4b_poscar}}} \\
\onecolumn
{\phantomsection\label{AB2_tP12_81_adg_2h}}
\subsection*{\huge \textbf{{\normalfont GeSe$_{2}$ (High-pressure) Structure: AB2\_tP12\_81\_adg\_2h}}}
\noindent \hrulefill
\vspace*{0.25cm}
\begin{figure}[htp]
  \centering
  \vspace{-1em}
  {\includegraphics[width=1\textwidth]{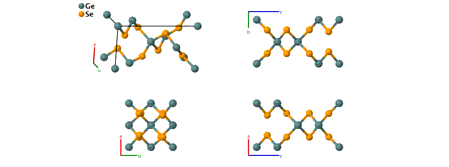}}
\end{figure}
\vspace*{-0.5cm}
\renewcommand{\arraystretch}{1.5}
\begin{equation*}
  \begin{array}{>{$\hspace{-0.15cm}}l<{$}>{$}p{0.5cm}<{$}>{$}p{18.5cm}<{$}}
    \mbox{\large \textbf{Prototype}} &\colon & \ce{GeSe2} \\
    \mbox{\large \textbf{\AFLOW\ prototype label}} &\colon & \mbox{AB2\_tP12\_81\_adg\_2h} \\
    \mbox{\large \textbf{\textit{Strukturbericht} designation}} &\colon & \mbox{None} \\
    \mbox{\large \textbf{Pearson symbol}} &\colon & \mbox{tP12} \\
    \mbox{\large \textbf{Space group number}} &\colon & 81 \\
    \mbox{\large \textbf{Space group symbol}} &\colon & P\bar{4} \\
    \mbox{\large \textbf{\AFLOW\ prototype command}} &\colon &  \texttt{aflow} \,  \, \texttt{-{}-proto=AB2\_tP12\_81\_adg\_2h } \, \newline \texttt{-{}-params=}{a,c/a,z_{3},x_{4},y_{4},z_{4},x_{5},y_{5},z_{5} }
  \end{array}
\end{equation*}
\renewcommand{\arraystretch}{1.0}

\noindent \parbox{1 \linewidth}{
\noindent \hrulefill
\\
\textbf{Simple Tetragonal primitive vectors:} \\
\vspace*{-0.25cm}
\begin{tabular}{cc}
  \begin{tabular}{c}
    \parbox{0.6 \linewidth}{
      \renewcommand{\arraystretch}{1.5}
      \begin{equation*}
        \centering
        \begin{array}{ccc}
              \mathbf{a}_1 & = & a \, \mathbf{\hat{x}} \\
    \mathbf{a}_2 & = & a \, \mathbf{\hat{y}} \\
    \mathbf{a}_3 & = & c \, \mathbf{\hat{z}} \\

        \end{array}
      \end{equation*}
    }
    \renewcommand{\arraystretch}{1.0}
  \end{tabular}
  \begin{tabular}{c}
    \includegraphics[width=0.3\linewidth]{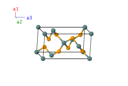} \\
  \end{tabular}
\end{tabular}

}
\vspace*{-0.25cm}

\noindent \hrulefill
\\
\textbf{Basis vectors:}
\vspace*{-0.25cm}
\renewcommand{\arraystretch}{1.5}
\begin{longtabu} to \textwidth{>{\centering $}X[-1,c,c]<{$}>{\centering $}X[-1,c,c]<{$}>{\centering $}X[-1,c,c]<{$}>{\centering $}X[-1,c,c]<{$}>{\centering $}X[-1,c,c]<{$}>{\centering $}X[-1,c,c]<{$}>{\centering $}X[-1,c,c]<{$}}
  & & \mbox{Lattice Coordinates} & & \mbox{Cartesian Coordinates} &\mbox{Wyckoff Position} & \mbox{Atom Type} \\  
  \mathbf{B}_{1} & = & 0 \, \mathbf{a}_{1} + 0 \, \mathbf{a}_{2} + 0 \, \mathbf{a}_{3} & = & 0 \, \mathbf{\hat{x}} + 0 \, \mathbf{\hat{y}} + 0 \, \mathbf{\hat{z}} & \left(1a\right) & \mbox{Ge I} \\ 
\mathbf{B}_{2} & = & \frac{1}{2} \, \mathbf{a}_{1} + \frac{1}{2} \, \mathbf{a}_{2} + \frac{1}{2} \, \mathbf{a}_{3} & = & \frac{1}{2}a \, \mathbf{\hat{x}} + \frac{1}{2}a \, \mathbf{\hat{y}} + \frac{1}{2}c \, \mathbf{\hat{z}} & \left(1d\right) & \mbox{Ge II} \\ 
\mathbf{B}_{3} & = & \frac{1}{2} \, \mathbf{a}_{2} + z_{3} \, \mathbf{a}_{3} & = & \frac{1}{2}a \, \mathbf{\hat{y}} + z_{3}c \, \mathbf{\hat{z}} & \left(2g\right) & \mbox{Ge III} \\ 
\mathbf{B}_{4} & = & \frac{1}{2} \, \mathbf{a}_{1} + -z_{3} \, \mathbf{a}_{3} & = & \frac{1}{2}a \, \mathbf{\hat{x}} + -z_{3}c \, \mathbf{\hat{z}} & \left(2g\right) & \mbox{Ge III} \\ 
\mathbf{B}_{5} & = & x_{4} \, \mathbf{a}_{1} + y_{4} \, \mathbf{a}_{2} + z_{4} \, \mathbf{a}_{3} & = & x_{4}a \, \mathbf{\hat{x}} + y_{4}a \, \mathbf{\hat{y}} + z_{4}c \, \mathbf{\hat{z}} & \left(4h\right) & \mbox{Se I} \\ 
\mathbf{B}_{6} & = & -x_{4} \, \mathbf{a}_{1}-y_{4} \, \mathbf{a}_{2} + z_{4} \, \mathbf{a}_{3} & = & -x_{4}a \, \mathbf{\hat{x}}-y_{4}a \, \mathbf{\hat{y}} + z_{4}c \, \mathbf{\hat{z}} & \left(4h\right) & \mbox{Se I} \\ 
\mathbf{B}_{7} & = & y_{4} \, \mathbf{a}_{1}-x_{4} \, \mathbf{a}_{2}-z_{4} \, \mathbf{a}_{3} & = & y_{4}a \, \mathbf{\hat{x}}-x_{4}a \, \mathbf{\hat{y}}-z_{4}c \, \mathbf{\hat{z}} & \left(4h\right) & \mbox{Se I} \\ 
\mathbf{B}_{8} & = & -y_{4} \, \mathbf{a}_{1} + x_{4} \, \mathbf{a}_{2}-z_{4} \, \mathbf{a}_{3} & = & -y_{4}a \, \mathbf{\hat{x}} + x_{4}a \, \mathbf{\hat{y}}-z_{4}c \, \mathbf{\hat{z}} & \left(4h\right) & \mbox{Se I} \\ 
\mathbf{B}_{9} & = & x_{5} \, \mathbf{a}_{1} + y_{5} \, \mathbf{a}_{2} + z_{5} \, \mathbf{a}_{3} & = & x_{5}a \, \mathbf{\hat{x}} + y_{5}a \, \mathbf{\hat{y}} + z_{5}c \, \mathbf{\hat{z}} & \left(4h\right) & \mbox{Se II} \\ 
\mathbf{B}_{10} & = & -x_{5} \, \mathbf{a}_{1}-y_{5} \, \mathbf{a}_{2} + z_{5} \, \mathbf{a}_{3} & = & -x_{5}a \, \mathbf{\hat{x}}-y_{5}a \, \mathbf{\hat{y}} + z_{5}c \, \mathbf{\hat{z}} & \left(4h\right) & \mbox{Se II} \\ 
\mathbf{B}_{11} & = & y_{5} \, \mathbf{a}_{1}-x_{5} \, \mathbf{a}_{2}-z_{5} \, \mathbf{a}_{3} & = & y_{5}a \, \mathbf{\hat{x}}-x_{5}a \, \mathbf{\hat{y}}-z_{5}c \, \mathbf{\hat{z}} & \left(4h\right) & \mbox{Se II} \\ 
\mathbf{B}_{12} & = & -y_{5} \, \mathbf{a}_{1} + x_{5} \, \mathbf{a}_{2}-z_{5} \, \mathbf{a}_{3} & = & -y_{5}a \, \mathbf{\hat{x}} + x_{5}a \, \mathbf{\hat{y}}-z_{5}c \, \mathbf{\hat{z}} & \left(4h\right) & \mbox{Se II} \\ 
\end{longtabu}
\renewcommand{\arraystretch}{1.0}
\noindent \hrulefill
\\
\textbf{References:}
\vspace*{-0.25cm}
\begin{flushleft}
  - \bibentry{Grzechnik_GeSe2_JSolStateChem_2000}. \\
\end{flushleft}
\textbf{Found in:}
\vspace*{-0.25cm}
\begin{flushleft}
  - \bibentry{Villars_PearsonsCrystalData_2013}. \\
\end{flushleft}
\noindent \hrulefill
\\
\textbf{Geometry files:}
\\
\noindent  - CIF: pp. {\hyperref[AB2_tP12_81_adg_2h_cif]{\pageref{AB2_tP12_81_adg_2h_cif}}} \\
\noindent  - POSCAR: pp. {\hyperref[AB2_tP12_81_adg_2h_poscar]{\pageref{AB2_tP12_81_adg_2h_poscar}}} \\
\onecolumn
{\phantomsection\label{A3B_tI32_82_3g_g}}
\subsection*{\huge \textbf{{\normalfont Ni$_{3}$P ($D0_{e}$) Structure: A3B\_tI32\_82\_3g\_g}}}
\noindent \hrulefill
\vspace*{0.25cm}
\begin{figure}[htp]
  \centering
  \vspace{-1em}
  {\includegraphics[width=1\textwidth]{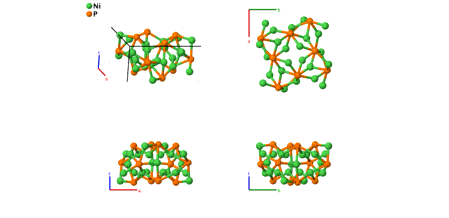}}
\end{figure}
\vspace*{-0.5cm}
\renewcommand{\arraystretch}{1.5}
\begin{equation*}
  \begin{array}{>{$\hspace{-0.15cm}}l<{$}>{$}p{0.5cm}<{$}>{$}p{18.5cm}<{$}}
    \mbox{\large \textbf{Prototype}} &\colon & \ce{Ni$_{3}$P} \\
    \mbox{\large \textbf{\AFLOW\ prototype label}} &\colon & \mbox{A3B\_tI32\_82\_3g\_g} \\
    \mbox{\large \textbf{\textit{Strukturbericht} designation}} &\colon & \mbox{$D0_{e}$} \\
    \mbox{\large \textbf{Pearson symbol}} &\colon & \mbox{tI32} \\
    \mbox{\large \textbf{Space group number}} &\colon & 82 \\
    \mbox{\large \textbf{Space group symbol}} &\colon & I\bar{4} \\
    \mbox{\large \textbf{\AFLOW\ prototype command}} &\colon &  \texttt{aflow} \,  \, \texttt{-{}-proto=A3B\_tI32\_82\_3g\_g } \, \newline \texttt{-{}-params=}{a,c/a,x_{1},y_{1},z_{1},x_{2},y_{2},z_{2},x_{3},y_{3},z_{3},x_{4},y_{4},z_{4} }
  \end{array}
\end{equation*}
\renewcommand{\arraystretch}{1.0}

\vspace*{-0.25cm}
\noindent \hrulefill
\\
\textbf{ Other compounds with this structure:}
\begin{itemize}
   \item{ Cr$_{3}$P, Fe$_{3}$P, Mn$_{3}$P, Mo$_{3}$P, Ti$_{3}$P, V$_{3}$P, BFe$_{3}$  }
\end{itemize}
\noindent \parbox{1 \linewidth}{
\noindent \hrulefill
\\
\textbf{Body-centered Tetragonal primitive vectors:} \\
\vspace*{-0.25cm}
\begin{tabular}{cc}
  \begin{tabular}{c}
    \parbox{0.6 \linewidth}{
      \renewcommand{\arraystretch}{1.5}
      \begin{equation*}
        \centering
        \begin{array}{ccc}
              \mathbf{a}_1 & = & - \frac12 \, a \, \mathbf{\hat{x}} + \frac12 \, a \, \mathbf{\hat{y}} + \frac12 \, c \, \mathbf{\hat{z}} \\
    \mathbf{a}_2 & = & ~ \frac12 \, a \, \mathbf{\hat{x}} - \frac12 \, a \, \mathbf{\hat{y}} + \frac12 \, c \, \mathbf{\hat{z}} \\
    \mathbf{a}_3 & = & ~ \frac12 \, a \, \mathbf{\hat{x}} + \frac12 \, a \, \mathbf{\hat{y}} - \frac12 \, c \, \mathbf{\hat{z}} \\

        \end{array}
      \end{equation*}
    }
    \renewcommand{\arraystretch}{1.0}
  \end{tabular}
  \begin{tabular}{c}
    \includegraphics[width=0.3\linewidth]{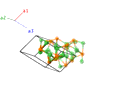} \\
  \end{tabular}
\end{tabular}

}
\vspace*{-0.25cm}

\noindent \hrulefill
\\
\textbf{Basis vectors:}
\vspace*{-0.25cm}
\renewcommand{\arraystretch}{1.5}
\begin{longtabu} to \textwidth{>{\centering $}X[-1,c,c]<{$}>{\centering $}X[-1,c,c]<{$}>{\centering $}X[-1,c,c]<{$}>{\centering $}X[-1,c,c]<{$}>{\centering $}X[-1,c,c]<{$}>{\centering $}X[-1,c,c]<{$}>{\centering $}X[-1,c,c]<{$}}
  & & \mbox{Lattice Coordinates} & & \mbox{Cartesian Coordinates} &\mbox{Wyckoff Position} & \mbox{Atom Type} \\  
  \mathbf{B}_{1} & = & \left(y_{1}+z_{1}\right) \, \mathbf{a}_{1} + \left(x_{1}+z_{1}\right) \, \mathbf{a}_{2} + \left(x_{1}+y_{1}\right) \, \mathbf{a}_{3} & = & x_{1}a \, \mathbf{\hat{x}} + y_{1}a \, \mathbf{\hat{y}} + z_{1}c \, \mathbf{\hat{z}} & \left(8g\right) & \mbox{Ni I} \\ 
\mathbf{B}_{2} & = & \left(-y_{1}+z_{1}\right) \, \mathbf{a}_{1} + \left(-x_{1}+z_{1}\right) \, \mathbf{a}_{2} + \left(-x_{1}-y_{1}\right) \, \mathbf{a}_{3} & = & -x_{1}a \, \mathbf{\hat{x}}-y_{1}a \, \mathbf{\hat{y}} + z_{1}c \, \mathbf{\hat{z}} & \left(8g\right) & \mbox{Ni I} \\ 
\mathbf{B}_{3} & = & \left(-x_{1}-z_{1}\right) \, \mathbf{a}_{1} + \left(y_{1}-z_{1}\right) \, \mathbf{a}_{2} + \left(-x_{1}+y_{1}\right) \, \mathbf{a}_{3} & = & y_{1}a \, \mathbf{\hat{x}}-x_{1}a \, \mathbf{\hat{y}}-z_{1}c \, \mathbf{\hat{z}} & \left(8g\right) & \mbox{Ni I} \\ 
\mathbf{B}_{4} & = & \left(x_{1}-z_{1}\right) \, \mathbf{a}_{1} + \left(-y_{1}-z_{1}\right) \, \mathbf{a}_{2} + \left(x_{1}-y_{1}\right) \, \mathbf{a}_{3} & = & -y_{1}a \, \mathbf{\hat{x}} + x_{1}a \, \mathbf{\hat{y}}-z_{1}c \, \mathbf{\hat{z}} & \left(8g\right) & \mbox{Ni I} \\ 
\mathbf{B}_{5} & = & \left(y_{2}+z_{2}\right) \, \mathbf{a}_{1} + \left(x_{2}+z_{2}\right) \, \mathbf{a}_{2} + \left(x_{2}+y_{2}\right) \, \mathbf{a}_{3} & = & x_{2}a \, \mathbf{\hat{x}} + y_{2}a \, \mathbf{\hat{y}} + z_{2}c \, \mathbf{\hat{z}} & \left(8g\right) & \mbox{Ni II} \\ 
\mathbf{B}_{6} & = & \left(-y_{2}+z_{2}\right) \, \mathbf{a}_{1} + \left(-x_{2}+z_{2}\right) \, \mathbf{a}_{2} + \left(-x_{2}-y_{2}\right) \, \mathbf{a}_{3} & = & -x_{2}a \, \mathbf{\hat{x}}-y_{2}a \, \mathbf{\hat{y}} + z_{2}c \, \mathbf{\hat{z}} & \left(8g\right) & \mbox{Ni II} \\ 
\mathbf{B}_{7} & = & \left(-x_{2}-z_{2}\right) \, \mathbf{a}_{1} + \left(y_{2}-z_{2}\right) \, \mathbf{a}_{2} + \left(-x_{2}+y_{2}\right) \, \mathbf{a}_{3} & = & y_{2}a \, \mathbf{\hat{x}}-x_{2}a \, \mathbf{\hat{y}}-z_{2}c \, \mathbf{\hat{z}} & \left(8g\right) & \mbox{Ni II} \\ 
\mathbf{B}_{8} & = & \left(x_{2}-z_{2}\right) \, \mathbf{a}_{1} + \left(-y_{2}-z_{2}\right) \, \mathbf{a}_{2} + \left(x_{2}-y_{2}\right) \, \mathbf{a}_{3} & = & -y_{2}a \, \mathbf{\hat{x}} + x_{2}a \, \mathbf{\hat{y}}-z_{2}c \, \mathbf{\hat{z}} & \left(8g\right) & \mbox{Ni II} \\ 
\mathbf{B}_{9} & = & \left(y_{3}+z_{3}\right) \, \mathbf{a}_{1} + \left(x_{3}+z_{3}\right) \, \mathbf{a}_{2} + \left(x_{3}+y_{3}\right) \, \mathbf{a}_{3} & = & x_{3}a \, \mathbf{\hat{x}} + y_{3}a \, \mathbf{\hat{y}} + z_{3}c \, \mathbf{\hat{z}} & \left(8g\right) & \mbox{Ni III} \\ 
\mathbf{B}_{10} & = & \left(-y_{3}+z_{3}\right) \, \mathbf{a}_{1} + \left(-x_{3}+z_{3}\right) \, \mathbf{a}_{2} + \left(-x_{3}-y_{3}\right) \, \mathbf{a}_{3} & = & -x_{3}a \, \mathbf{\hat{x}}-y_{3}a \, \mathbf{\hat{y}} + z_{3}c \, \mathbf{\hat{z}} & \left(8g\right) & \mbox{Ni III} \\ 
\mathbf{B}_{11} & = & \left(-x_{3}-z_{3}\right) \, \mathbf{a}_{1} + \left(y_{3}-z_{3}\right) \, \mathbf{a}_{2} + \left(-x_{3}+y_{3}\right) \, \mathbf{a}_{3} & = & y_{3}a \, \mathbf{\hat{x}}-x_{3}a \, \mathbf{\hat{y}}-z_{3}c \, \mathbf{\hat{z}} & \left(8g\right) & \mbox{Ni III} \\ 
\mathbf{B}_{12} & = & \left(x_{3}-z_{3}\right) \, \mathbf{a}_{1} + \left(-y_{3}-z_{3}\right) \, \mathbf{a}_{2} + \left(x_{3}-y_{3}\right) \, \mathbf{a}_{3} & = & -y_{3}a \, \mathbf{\hat{x}} + x_{3}a \, \mathbf{\hat{y}}-z_{3}c \, \mathbf{\hat{z}} & \left(8g\right) & \mbox{Ni III} \\ 
\mathbf{B}_{13} & = & \left(y_{4}+z_{4}\right) \, \mathbf{a}_{1} + \left(x_{4}+z_{4}\right) \, \mathbf{a}_{2} + \left(x_{4}+y_{4}\right) \, \mathbf{a}_{3} & = & x_{4}a \, \mathbf{\hat{x}} + y_{4}a \, \mathbf{\hat{y}} + z_{4}c \, \mathbf{\hat{z}} & \left(8g\right) & \mbox{P} \\ 
\mathbf{B}_{14} & = & \left(-y_{4}+z_{4}\right) \, \mathbf{a}_{1} + \left(-x_{4}+z_{4}\right) \, \mathbf{a}_{2} + \left(-x_{4}-y_{4}\right) \, \mathbf{a}_{3} & = & -x_{4}a \, \mathbf{\hat{x}}-y_{4}a \, \mathbf{\hat{y}} + z_{4}c \, \mathbf{\hat{z}} & \left(8g\right) & \mbox{P} \\ 
\mathbf{B}_{15} & = & \left(-x_{4}-z_{4}\right) \, \mathbf{a}_{1} + \left(y_{4}-z_{4}\right) \, \mathbf{a}_{2} + \left(-x_{4}+y_{4}\right) \, \mathbf{a}_{3} & = & y_{4}a \, \mathbf{\hat{x}}-x_{4}a \, \mathbf{\hat{y}}-z_{4}c \, \mathbf{\hat{z}} & \left(8g\right) & \mbox{P} \\ 
\mathbf{B}_{16} & = & \left(x_{4}-z_{4}\right) \, \mathbf{a}_{1} + \left(-y_{4}-z_{4}\right) \, \mathbf{a}_{2} + \left(x_{4}-y_{4}\right) \, \mathbf{a}_{3} & = & -y_{4}a \, \mathbf{\hat{x}} + x_{4}a \, \mathbf{\hat{y}}-z_{4}c \, \mathbf{\hat{z}} & \left(8g\right) & \mbox{P} \\ 
\end{longtabu}
\renewcommand{\arraystretch}{1.0}
\noindent \hrulefill
\\
\textbf{References:}
\vspace*{-0.25cm}
\begin{flushleft}
  - \bibentry{Rundqvist_Acta_Chem_Scand_16_1962}. \\
\end{flushleft}
\noindent \hrulefill
\\
\textbf{Geometry files:}
\\
\noindent  - CIF: pp. {\hyperref[A3B_tI32_82_3g_g_cif]{\pageref{A3B_tI32_82_3g_g_cif}}} \\
\noindent  - POSCAR: pp. {\hyperref[A3B_tI32_82_3g_g_poscar]{\pageref{A3B_tI32_82_3g_g_poscar}}} \\
\onecolumn
{\phantomsection\label{A3B2_tP10_83_adk_j}}
\subsection*{\huge \textbf{{\normalfont Ti$_{2}$Ge$_{3}$ Structure: A3B2\_tP10\_83\_adk\_j}}}
\noindent \hrulefill
\vspace*{0.25cm}
\begin{figure}[htp]
  \centering
  \vspace{-1em}
  {\includegraphics[width=1\textwidth]{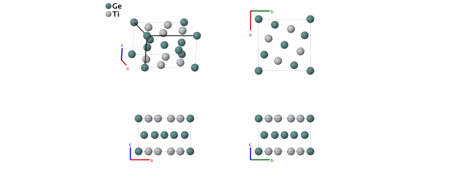}}
\end{figure}
\vspace*{-0.5cm}
\renewcommand{\arraystretch}{1.5}
\begin{equation*}
  \begin{array}{>{$\hspace{-0.15cm}}l<{$}>{$}p{0.5cm}<{$}>{$}p{18.5cm}<{$}}
    \mbox{\large \textbf{Prototype}} &\colon & \ce{Ti2Ge3} \\
    \mbox{\large \textbf{\AFLOW\ prototype label}} &\colon & \mbox{A3B2\_tP10\_83\_adk\_j} \\
    \mbox{\large \textbf{\textit{Strukturbericht} designation}} &\colon & \mbox{None} \\
    \mbox{\large \textbf{Pearson symbol}} &\colon & \mbox{tP10} \\
    \mbox{\large \textbf{Space group number}} &\colon & 83 \\
    \mbox{\large \textbf{Space group symbol}} &\colon & P4/m \\
    \mbox{\large \textbf{\AFLOW\ prototype command}} &\colon &  \texttt{aflow} \,  \, \texttt{-{}-proto=A3B2\_tP10\_83\_adk\_j } \, \newline \texttt{-{}-params=}{a,c/a,x_{3},y_{3},x_{4},y_{4} }
  \end{array}
\end{equation*}
\renewcommand{\arraystretch}{1.0}

\noindent \parbox{1 \linewidth}{
\noindent \hrulefill
\\
\textbf{Simple Tetragonal primitive vectors:} \\
\vspace*{-0.25cm}
\begin{tabular}{cc}
  \begin{tabular}{c}
    \parbox{0.6 \linewidth}{
      \renewcommand{\arraystretch}{1.5}
      \begin{equation*}
        \centering
        \begin{array}{ccc}
              \mathbf{a}_1 & = & a \, \mathbf{\hat{x}} \\
    \mathbf{a}_2 & = & a \, \mathbf{\hat{y}} \\
    \mathbf{a}_3 & = & c \, \mathbf{\hat{z}} \\

        \end{array}
      \end{equation*}
    }
    \renewcommand{\arraystretch}{1.0}
  \end{tabular}
  \begin{tabular}{c}
    \includegraphics[width=0.3\linewidth]{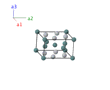} \\
  \end{tabular}
\end{tabular}

}
\vspace*{-0.25cm}

\noindent \hrulefill
\\
\textbf{Basis vectors:}
\vspace*{-0.25cm}
\renewcommand{\arraystretch}{1.5}
\begin{longtabu} to \textwidth{>{\centering $}X[-1,c,c]<{$}>{\centering $}X[-1,c,c]<{$}>{\centering $}X[-1,c,c]<{$}>{\centering $}X[-1,c,c]<{$}>{\centering $}X[-1,c,c]<{$}>{\centering $}X[-1,c,c]<{$}>{\centering $}X[-1,c,c]<{$}}
  & & \mbox{Lattice Coordinates} & & \mbox{Cartesian Coordinates} &\mbox{Wyckoff Position} & \mbox{Atom Type} \\  
  \mathbf{B}_{1} & = & 0 \, \mathbf{a}_{1} + 0 \, \mathbf{a}_{2} + 0 \, \mathbf{a}_{3} & = & 0 \, \mathbf{\hat{x}} + 0 \, \mathbf{\hat{y}} + 0 \, \mathbf{\hat{z}} & \left(1a\right) & \mbox{Ge I} \\ 
\mathbf{B}_{2} & = & \frac{1}{2} \, \mathbf{a}_{1} + \frac{1}{2} \, \mathbf{a}_{2} + \frac{1}{2} \, \mathbf{a}_{3} & = & \frac{1}{2}a \, \mathbf{\hat{x}} + \frac{1}{2}a \, \mathbf{\hat{y}} + \frac{1}{2}c \, \mathbf{\hat{z}} & \left(1d\right) & \mbox{Ge II} \\ 
\mathbf{B}_{3} & = & x_{3} \, \mathbf{a}_{1} + y_{3} \, \mathbf{a}_{2} & = & x_{3}a \, \mathbf{\hat{x}} + y_{3}a \, \mathbf{\hat{y}} & \left(4j\right) & \mbox{Ti} \\ 
\mathbf{B}_{4} & = & -x_{3} \, \mathbf{a}_{1}-y_{3} \, \mathbf{a}_{2} & = & -x_{3}a \, \mathbf{\hat{x}}-y_{3}a \, \mathbf{\hat{y}} & \left(4j\right) & \mbox{Ti} \\ 
\mathbf{B}_{5} & = & -y_{3} \, \mathbf{a}_{1} + x_{3} \, \mathbf{a}_{2} & = & -y_{3}a \, \mathbf{\hat{x}} + x_{3}a \, \mathbf{\hat{y}} & \left(4j\right) & \mbox{Ti} \\ 
\mathbf{B}_{6} & = & y_{3} \, \mathbf{a}_{1}-x_{3} \, \mathbf{a}_{2} & = & y_{3}a \, \mathbf{\hat{x}}-x_{3}a \, \mathbf{\hat{y}} & \left(4j\right) & \mbox{Ti} \\ 
\mathbf{B}_{7} & = & x_{4} \, \mathbf{a}_{1} + y_{4} \, \mathbf{a}_{2} + \frac{1}{2} \, \mathbf{a}_{3} & = & x_{4}a \, \mathbf{\hat{x}} + y_{4}a \, \mathbf{\hat{y}} + \frac{1}{2}c \, \mathbf{\hat{z}} & \left(4k\right) & \mbox{Ge III} \\ 
\mathbf{B}_{8} & = & -x_{4} \, \mathbf{a}_{1}-y_{4} \, \mathbf{a}_{2} + \frac{1}{2} \, \mathbf{a}_{3} & = & -x_{4}a \, \mathbf{\hat{x}}-y_{4}a \, \mathbf{\hat{y}} + \frac{1}{2}c \, \mathbf{\hat{z}} & \left(4k\right) & \mbox{Ge III} \\ 
\mathbf{B}_{9} & = & -y_{4} \, \mathbf{a}_{1} + x_{4} \, \mathbf{a}_{2} + \frac{1}{2} \, \mathbf{a}_{3} & = & -y_{4}a \, \mathbf{\hat{x}} + x_{4}a \, \mathbf{\hat{y}} + \frac{1}{2}c \, \mathbf{\hat{z}} & \left(4k\right) & \mbox{Ge III} \\ 
\mathbf{B}_{10} & = & y_{4} \, \mathbf{a}_{1}-x_{4} \, \mathbf{a}_{2} + \frac{1}{2} \, \mathbf{a}_{3} & = & y_{4}a \, \mathbf{\hat{x}}-x_{4}a \, \mathbf{\hat{y}} + \frac{1}{2}c \, \mathbf{\hat{z}} & \left(4k\right) & \mbox{Ge III} \\ 
\end{longtabu}
\renewcommand{\arraystretch}{1.0}
\noindent \hrulefill
\\
\textbf{References:}
\vspace*{-0.25cm}
\begin{flushleft}
  - \bibentry{Schubert_Ti2Ge3_Naturwissen_1962}. \\
\end{flushleft}
\textbf{Found in:}
\vspace*{-0.25cm}
\begin{flushleft}
  - \bibentry{Villars_PearsonsCrystalData_2013}. \\
\end{flushleft}
\noindent \hrulefill
\\
\textbf{Geometry files:}
\\
\noindent  - CIF: pp. {\hyperref[A3B2_tP10_83_adk_j_cif]{\pageref{A3B2_tP10_83_adk_j_cif}}} \\
\noindent  - POSCAR: pp. {\hyperref[A3B2_tP10_83_adk_j_poscar]{\pageref{A3B2_tP10_83_adk_j_poscar}}} \\
\onecolumn
{\phantomsection\label{A2B_tP30_85_ab2g_cg}}
\subsection*{\huge \textbf{{\normalfont SrBr$_{2}$ Structure: A2B\_tP30\_85\_ab2g\_cg}}}
\noindent \hrulefill
\vspace*{0.25cm}
\begin{figure}[htp]
  \centering
  \vspace{-1em}
  {\includegraphics[width=1\textwidth]{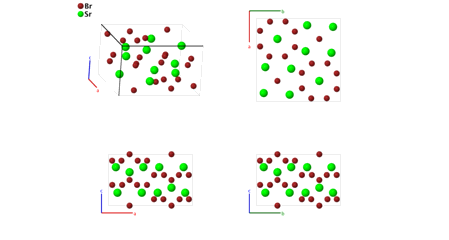}}
\end{figure}
\vspace*{-0.5cm}
\renewcommand{\arraystretch}{1.5}
\begin{equation*}
  \begin{array}{>{$\hspace{-0.15cm}}l<{$}>{$}p{0.5cm}<{$}>{$}p{18.5cm}<{$}}
    \mbox{\large \textbf{Prototype}} &\colon & \ce{SrBr2} \\
    \mbox{\large \textbf{\AFLOW\ prototype label}} &\colon & \mbox{A2B\_tP30\_85\_ab2g\_cg} \\
    \mbox{\large \textbf{\textit{Strukturbericht} designation}} &\colon & \mbox{None} \\
    \mbox{\large \textbf{Pearson symbol}} &\colon & \mbox{tP30} \\
    \mbox{\large \textbf{Space group number}} &\colon & 85 \\
    \mbox{\large \textbf{Space group symbol}} &\colon & P4/n \\
    \mbox{\large \textbf{\AFLOW\ prototype command}} &\colon &  \texttt{aflow} \,  \, \texttt{-{}-proto=A2B\_tP30\_85\_ab2g\_cg } \, \newline \texttt{-{}-params=}{a,c/a,z_{3},x_{4},y_{4},z_{4},x_{5},y_{5},z_{5},x_{6},y_{6},z_{6} }
  \end{array}
\end{equation*}
\renewcommand{\arraystretch}{1.0}

\vspace*{-0.25cm}
\noindent \hrulefill
\begin{itemize}
  \item{Using the work of (Kamermans, 1939), (Herrmann, 1943) designated the structure SrBr$_{2}$ as {\it{Strukturbericht}} $C53$, 
and placed it in space group $Pnma$ \#62.  
(Sass, 1963) pointed out that the structure proposed by Kamermans did not agree with powder diffraction data, 
and proposed this structure, also found by (Frit, 1969).  
(Parth\'{e}, 1993) gives the current structure the $C53$ designation.  
We will follow the original {\it{Strukturbericht}}, and give the $C53$ designation to the $Pnma$ structure.
}
\end{itemize}

\noindent \parbox{1 \linewidth}{
\noindent \hrulefill
\\
\textbf{Simple Tetragonal primitive vectors:} \\
\vspace*{-0.25cm}
\begin{tabular}{cc}
  \begin{tabular}{c}
    \parbox{0.6 \linewidth}{
      \renewcommand{\arraystretch}{1.5}
      \begin{equation*}
        \centering
        \begin{array}{ccc}
              \mathbf{a}_1 & = & a \, \mathbf{\hat{x}} \\
    \mathbf{a}_2 & = & a \, \mathbf{\hat{y}} \\
    \mathbf{a}_3 & = & c \, \mathbf{\hat{z}} \\

        \end{array}
      \end{equation*}
    }
    \renewcommand{\arraystretch}{1.0}
  \end{tabular}
  \begin{tabular}{c}
    \includegraphics[width=0.3\linewidth]{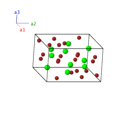} \\
  \end{tabular}
\end{tabular}

}
\vspace*{-0.25cm}

\noindent \hrulefill
\\
\textbf{Basis vectors:}
\vspace*{-0.25cm}
\renewcommand{\arraystretch}{1.5}
\begin{longtabu} to \textwidth{>{\centering $}X[-1,c,c]<{$}>{\centering $}X[-1,c,c]<{$}>{\centering $}X[-1,c,c]<{$}>{\centering $}X[-1,c,c]<{$}>{\centering $}X[-1,c,c]<{$}>{\centering $}X[-1,c,c]<{$}>{\centering $}X[-1,c,c]<{$}}
  & & \mbox{Lattice Coordinates} & & \mbox{Cartesian Coordinates} &\mbox{Wyckoff Position} & \mbox{Atom Type} \\  
  \mathbf{B}_{1} & = & \frac{1}{4} \, \mathbf{a}_{1} + \frac{3}{4} \, \mathbf{a}_{2} & = & \frac{1}{4}a \, \mathbf{\hat{x}} + \frac{3}{4}a \, \mathbf{\hat{y}} & \left(2a\right) & \mbox{Br I} \\ 
\mathbf{B}_{2} & = & \frac{3}{4} \, \mathbf{a}_{1} + \frac{1}{4} \, \mathbf{a}_{2} & = & \frac{3}{4}a \, \mathbf{\hat{x}} + \frac{1}{4}a \, \mathbf{\hat{y}} & \left(2a\right) & \mbox{Br I} \\ 
\mathbf{B}_{3} & = & \frac{1}{4} \, \mathbf{a}_{1} + \frac{3}{4} \, \mathbf{a}_{2} + \frac{1}{2} \, \mathbf{a}_{3} & = & \frac{1}{4}a \, \mathbf{\hat{x}} + \frac{3}{4}a \, \mathbf{\hat{y}} + \frac{1}{2}c \, \mathbf{\hat{z}} & \left(2b\right) & \mbox{Br II} \\ 
\mathbf{B}_{4} & = & \frac{3}{4} \, \mathbf{a}_{1} + \frac{1}{4} \, \mathbf{a}_{2} + \frac{1}{2} \, \mathbf{a}_{3} & = & \frac{3}{4}a \, \mathbf{\hat{x}} + \frac{1}{4}a \, \mathbf{\hat{y}} + \frac{1}{2}c \, \mathbf{\hat{z}} & \left(2b\right) & \mbox{Br II} \\ 
\mathbf{B}_{5} & = & \frac{1}{4} \, \mathbf{a}_{1} + \frac{1}{4} \, \mathbf{a}_{2} + z_{3} \, \mathbf{a}_{3} & = & \frac{1}{4}a \, \mathbf{\hat{x}} + \frac{1}{4}a \, \mathbf{\hat{y}} + z_{3}c \, \mathbf{\hat{z}} & \left(2c\right) & \mbox{Sr I} \\ 
\mathbf{B}_{6} & = & \frac{3}{4} \, \mathbf{a}_{1} + \frac{3}{4} \, \mathbf{a}_{2}-z_{3} \, \mathbf{a}_{3} & = & \frac{3}{4}a \, \mathbf{\hat{x}} + \frac{3}{4}a \, \mathbf{\hat{y}}-z_{3}c \, \mathbf{\hat{z}} & \left(2c\right) & \mbox{Sr I} \\ 
\mathbf{B}_{7} & = & x_{4} \, \mathbf{a}_{1} + y_{4} \, \mathbf{a}_{2} + z_{4} \, \mathbf{a}_{3} & = & x_{4}a \, \mathbf{\hat{x}} + y_{4}a \, \mathbf{\hat{y}} + z_{4}c \, \mathbf{\hat{z}} & \left(8g\right) & \mbox{Br III} \\ 
\mathbf{B}_{8} & = & \left(\frac{1}{2} - x_{4}\right) \, \mathbf{a}_{1} + \left(\frac{1}{2} - y_{4}\right) \, \mathbf{a}_{2} + z_{4} \, \mathbf{a}_{3} & = & \left(\frac{1}{2} - x_{4}\right)a \, \mathbf{\hat{x}} + \left(\frac{1}{2} - y_{4}\right)a \, \mathbf{\hat{y}} + z_{4}c \, \mathbf{\hat{z}} & \left(8g\right) & \mbox{Br III} \\ 
\mathbf{B}_{9} & = & \left(\frac{1}{2} - y_{4}\right) \, \mathbf{a}_{1} + x_{4} \, \mathbf{a}_{2} + z_{4} \, \mathbf{a}_{3} & = & \left(\frac{1}{2} - y_{4}\right)a \, \mathbf{\hat{x}} + x_{4}a \, \mathbf{\hat{y}} + z_{4}c \, \mathbf{\hat{z}} & \left(8g\right) & \mbox{Br III} \\ 
\mathbf{B}_{10} & = & y_{4} \, \mathbf{a}_{1} + \left(\frac{1}{2} - x_{4}\right) \, \mathbf{a}_{2} + z_{4} \, \mathbf{a}_{3} & = & y_{4}a \, \mathbf{\hat{x}} + \left(\frac{1}{2} - x_{4}\right)a \, \mathbf{\hat{y}} + z_{4}c \, \mathbf{\hat{z}} & \left(8g\right) & \mbox{Br III} \\ 
\mathbf{B}_{11} & = & -x_{4} \, \mathbf{a}_{1}-y_{4} \, \mathbf{a}_{2}-z_{4} \, \mathbf{a}_{3} & = & -x_{4}a \, \mathbf{\hat{x}}-y_{4}a \, \mathbf{\hat{y}}-z_{4}c \, \mathbf{\hat{z}} & \left(8g\right) & \mbox{Br III} \\ 
\mathbf{B}_{12} & = & \left(\frac{1}{2} +x_{4}\right) \, \mathbf{a}_{1} + \left(\frac{1}{2} +y_{4}\right) \, \mathbf{a}_{2}-z_{4} \, \mathbf{a}_{3} & = & \left(\frac{1}{2} +x_{4}\right)a \, \mathbf{\hat{x}} + \left(\frac{1}{2} +y_{4}\right)a \, \mathbf{\hat{y}}-z_{4}c \, \mathbf{\hat{z}} & \left(8g\right) & \mbox{Br III} \\ 
\mathbf{B}_{13} & = & \left(\frac{1}{2} +y_{4}\right) \, \mathbf{a}_{1}-x_{4} \, \mathbf{a}_{2}-z_{4} \, \mathbf{a}_{3} & = & \left(\frac{1}{2} +y_{4}\right)a \, \mathbf{\hat{x}}-x_{4}a \, \mathbf{\hat{y}}-z_{4}c \, \mathbf{\hat{z}} & \left(8g\right) & \mbox{Br III} \\ 
\mathbf{B}_{14} & = & -y_{4} \, \mathbf{a}_{1} + \left(\frac{1}{2} +x_{4}\right) \, \mathbf{a}_{2}-z_{4} \, \mathbf{a}_{3} & = & -y_{4}a \, \mathbf{\hat{x}} + \left(\frac{1}{2} +x_{4}\right)a \, \mathbf{\hat{y}}-z_{4}c \, \mathbf{\hat{z}} & \left(8g\right) & \mbox{Br III} \\ 
\mathbf{B}_{15} & = & x_{5} \, \mathbf{a}_{1} + y_{5} \, \mathbf{a}_{2} + z_{5} \, \mathbf{a}_{3} & = & x_{5}a \, \mathbf{\hat{x}} + y_{5}a \, \mathbf{\hat{y}} + z_{5}c \, \mathbf{\hat{z}} & \left(8g\right) & \mbox{Br IV} \\ 
\mathbf{B}_{16} & = & \left(\frac{1}{2} - x_{5}\right) \, \mathbf{a}_{1} + \left(\frac{1}{2} - y_{5}\right) \, \mathbf{a}_{2} + z_{5} \, \mathbf{a}_{3} & = & \left(\frac{1}{2} - x_{5}\right)a \, \mathbf{\hat{x}} + \left(\frac{1}{2} - y_{5}\right)a \, \mathbf{\hat{y}} + z_{5}c \, \mathbf{\hat{z}} & \left(8g\right) & \mbox{Br IV} \\ 
\mathbf{B}_{17} & = & \left(\frac{1}{2} - y_{5}\right) \, \mathbf{a}_{1} + x_{5} \, \mathbf{a}_{2} + z_{5} \, \mathbf{a}_{3} & = & \left(\frac{1}{2} - y_{5}\right)a \, \mathbf{\hat{x}} + x_{5}a \, \mathbf{\hat{y}} + z_{5}c \, \mathbf{\hat{z}} & \left(8g\right) & \mbox{Br IV} \\ 
\mathbf{B}_{18} & = & y_{5} \, \mathbf{a}_{1} + \left(\frac{1}{2} - x_{5}\right) \, \mathbf{a}_{2} + z_{5} \, \mathbf{a}_{3} & = & y_{5}a \, \mathbf{\hat{x}} + \left(\frac{1}{2} - x_{5}\right)a \, \mathbf{\hat{y}} + z_{5}c \, \mathbf{\hat{z}} & \left(8g\right) & \mbox{Br IV} \\ 
\mathbf{B}_{19} & = & -x_{5} \, \mathbf{a}_{1}-y_{5} \, \mathbf{a}_{2}-z_{5} \, \mathbf{a}_{3} & = & -x_{5}a \, \mathbf{\hat{x}}-y_{5}a \, \mathbf{\hat{y}}-z_{5}c \, \mathbf{\hat{z}} & \left(8g\right) & \mbox{Br IV} \\ 
\mathbf{B}_{20} & = & \left(\frac{1}{2} +x_{5}\right) \, \mathbf{a}_{1} + \left(\frac{1}{2} +y_{5}\right) \, \mathbf{a}_{2}-z_{5} \, \mathbf{a}_{3} & = & \left(\frac{1}{2} +x_{5}\right)a \, \mathbf{\hat{x}} + \left(\frac{1}{2} +y_{5}\right)a \, \mathbf{\hat{y}}-z_{5}c \, \mathbf{\hat{z}} & \left(8g\right) & \mbox{Br IV} \\ 
\mathbf{B}_{21} & = & \left(\frac{1}{2} +y_{5}\right) \, \mathbf{a}_{1}-x_{5} \, \mathbf{a}_{2}-z_{5} \, \mathbf{a}_{3} & = & \left(\frac{1}{2} +y_{5}\right)a \, \mathbf{\hat{x}}-x_{5}a \, \mathbf{\hat{y}}-z_{5}c \, \mathbf{\hat{z}} & \left(8g\right) & \mbox{Br IV} \\ 
\mathbf{B}_{22} & = & -y_{5} \, \mathbf{a}_{1} + \left(\frac{1}{2} +x_{5}\right) \, \mathbf{a}_{2}-z_{5} \, \mathbf{a}_{3} & = & -y_{5}a \, \mathbf{\hat{x}} + \left(\frac{1}{2} +x_{5}\right)a \, \mathbf{\hat{y}}-z_{5}c \, \mathbf{\hat{z}} & \left(8g\right) & \mbox{Br IV} \\ 
\mathbf{B}_{23} & = & x_{6} \, \mathbf{a}_{1} + y_{6} \, \mathbf{a}_{2} + z_{6} \, \mathbf{a}_{3} & = & x_{6}a \, \mathbf{\hat{x}} + y_{6}a \, \mathbf{\hat{y}} + z_{6}c \, \mathbf{\hat{z}} & \left(8g\right) & \mbox{Sr II} \\ 
\mathbf{B}_{24} & = & \left(\frac{1}{2} - x_{6}\right) \, \mathbf{a}_{1} + \left(\frac{1}{2} - y_{6}\right) \, \mathbf{a}_{2} + z_{6} \, \mathbf{a}_{3} & = & \left(\frac{1}{2} - x_{6}\right)a \, \mathbf{\hat{x}} + \left(\frac{1}{2} - y_{6}\right)a \, \mathbf{\hat{y}} + z_{6}c \, \mathbf{\hat{z}} & \left(8g\right) & \mbox{Sr II} \\ 
\mathbf{B}_{25} & = & \left(\frac{1}{2} - y_{6}\right) \, \mathbf{a}_{1} + x_{6} \, \mathbf{a}_{2} + z_{6} \, \mathbf{a}_{3} & = & \left(\frac{1}{2} - y_{6}\right)a \, \mathbf{\hat{x}} + x_{6}a \, \mathbf{\hat{y}} + z_{6}c \, \mathbf{\hat{z}} & \left(8g\right) & \mbox{Sr II} \\ 
\mathbf{B}_{26} & = & y_{6} \, \mathbf{a}_{1} + \left(\frac{1}{2} - x_{6}\right) \, \mathbf{a}_{2} + z_{6} \, \mathbf{a}_{3} & = & y_{6}a \, \mathbf{\hat{x}} + \left(\frac{1}{2} - x_{6}\right)a \, \mathbf{\hat{y}} + z_{6}c \, \mathbf{\hat{z}} & \left(8g\right) & \mbox{Sr II} \\ 
\mathbf{B}_{27} & = & -x_{6} \, \mathbf{a}_{1}-y_{6} \, \mathbf{a}_{2}-z_{6} \, \mathbf{a}_{3} & = & -x_{6}a \, \mathbf{\hat{x}}-y_{6}a \, \mathbf{\hat{y}}-z_{6}c \, \mathbf{\hat{z}} & \left(8g\right) & \mbox{Sr II} \\ 
\mathbf{B}_{28} & = & \left(\frac{1}{2} +x_{6}\right) \, \mathbf{a}_{1} + \left(\frac{1}{2} +y_{6}\right) \, \mathbf{a}_{2}-z_{6} \, \mathbf{a}_{3} & = & \left(\frac{1}{2} +x_{6}\right)a \, \mathbf{\hat{x}} + \left(\frac{1}{2} +y_{6}\right)a \, \mathbf{\hat{y}}-z_{6}c \, \mathbf{\hat{z}} & \left(8g\right) & \mbox{Sr II} \\ 
\mathbf{B}_{29} & = & \left(\frac{1}{2} +y_{6}\right) \, \mathbf{a}_{1}-x_{6} \, \mathbf{a}_{2}-z_{6} \, \mathbf{a}_{3} & = & \left(\frac{1}{2} +y_{6}\right)a \, \mathbf{\hat{x}}-x_{6}a \, \mathbf{\hat{y}}-z_{6}c \, \mathbf{\hat{z}} & \left(8g\right) & \mbox{Sr II} \\ 
\mathbf{B}_{30} & = & -y_{6} \, \mathbf{a}_{1} + \left(\frac{1}{2} +x_{6}\right) \, \mathbf{a}_{2}-z_{6} \, \mathbf{a}_{3} & = & -y_{6}a \, \mathbf{\hat{x}} + \left(\frac{1}{2} +x_{6}\right)a \, \mathbf{\hat{y}}-z_{6}c \, \mathbf{\hat{z}} & \left(8g\right) & \mbox{Sr II} \\ 
\end{longtabu}
\renewcommand{\arraystretch}{1.0}
\noindent \hrulefill
\\
\textbf{References:}
\vspace*{-0.25cm}
\begin{flushleft}
  - \bibentry{Frit_SrBr2_JInorgNucChem_1969}. \\
  - \bibentry{Herrmann_Struk_VII_1939}. \\
  - \bibentry{Kamermans_Z_Krist_101_406_1939}. \\
  - \bibentry{Sass_J_Phys_Chem_67_2826_1963}. \\
  - \bibentry{Parthe_Gmlen_1993}. \\
\end{flushleft}
\textbf{Found in:}
\vspace*{-0.25cm}
\begin{flushleft}
  - \bibentry{Villars_PearsonsCrystalData_2013}. \\
\end{flushleft}
\noindent \hrulefill
\\
\textbf{Geometry files:}
\\
\noindent  - CIF: pp. {\hyperref[A2B_tP30_85_ab2g_cg_cif]{\pageref{A2B_tP30_85_ab2g_cg_cif}}} \\
\noindent  - POSCAR: pp. {\hyperref[A2B_tP30_85_ab2g_cg_poscar]{\pageref{A2B_tP30_85_ab2g_cg_poscar}}} \\
\onecolumn
{\phantomsection\label{AB3_tP32_86_g_3g}}
\subsection*{\huge \textbf{{\normalfont Ti$_{3}$P Structure: AB3\_tP32\_86\_g\_3g}}}
\noindent \hrulefill
\vspace*{0.25cm}
\begin{figure}[htp]
  \centering
  \vspace{-1em}
  {\includegraphics[width=1\textwidth]{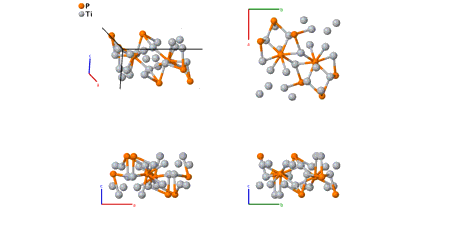}}
\end{figure}
\vspace*{-0.5cm}
\renewcommand{\arraystretch}{1.5}
\begin{equation*}
  \begin{array}{>{$\hspace{-0.15cm}}l<{$}>{$}p{0.5cm}<{$}>{$}p{18.5cm}<{$}}
    \mbox{\large \textbf{Prototype}} &\colon & \ce{Ti3P} \\
    \mbox{\large \textbf{\AFLOW\ prototype label}} &\colon & \mbox{AB3\_tP32\_86\_g\_3g} \\
    \mbox{\large \textbf{\textit{Strukturbericht} designation}} &\colon & \mbox{None} \\
    \mbox{\large \textbf{Pearson symbol}} &\colon & \mbox{tP32} \\
    \mbox{\large \textbf{Space group number}} &\colon & 86 \\
    \mbox{\large \textbf{Space group symbol}} &\colon & P4_{2}/n \\
    \mbox{\large \textbf{\AFLOW\ prototype command}} &\colon &  \texttt{aflow} \,  \, \texttt{-{}-proto=AB3\_tP32\_86\_g\_3g } \, \newline \texttt{-{}-params=}{a,c/a,x_{1},y_{1},z_{1},x_{2},y_{2},z_{2},x_{3},y_{3},z_{3},x_{4},y_{4},z_{4} }
  \end{array}
\end{equation*}
\renewcommand{\arraystretch}{1.0}

\noindent \parbox{1 \linewidth}{
\noindent \hrulefill
\\
\textbf{Simple Tetragonal primitive vectors:} \\
\vspace*{-0.25cm}
\begin{tabular}{cc}
  \begin{tabular}{c}
    \parbox{0.6 \linewidth}{
      \renewcommand{\arraystretch}{1.5}
      \begin{equation*}
        \centering
        \begin{array}{ccc}
              \mathbf{a}_1 & = & a \, \mathbf{\hat{x}} \\
    \mathbf{a}_2 & = & a \, \mathbf{\hat{y}} \\
    \mathbf{a}_3 & = & c \, \mathbf{\hat{z}} \\

        \end{array}
      \end{equation*}
    }
    \renewcommand{\arraystretch}{1.0}
  \end{tabular}
  \begin{tabular}{c}
    \includegraphics[width=0.3\linewidth]{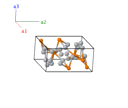} \\
  \end{tabular}
\end{tabular}

}
\vspace*{-0.25cm}

\noindent \hrulefill
\\
\textbf{Basis vectors:}
\vspace*{-0.25cm}
\renewcommand{\arraystretch}{1.5}
\begin{longtabu} to \textwidth{>{\centering $}X[-1,c,c]<{$}>{\centering $}X[-1,c,c]<{$}>{\centering $}X[-1,c,c]<{$}>{\centering $}X[-1,c,c]<{$}>{\centering $}X[-1,c,c]<{$}>{\centering $}X[-1,c,c]<{$}>{\centering $}X[-1,c,c]<{$}}
  & & \mbox{Lattice Coordinates} & & \mbox{Cartesian Coordinates} &\mbox{Wyckoff Position} & \mbox{Atom Type} \\  
  \mathbf{B}_{1} & = & x_{1} \, \mathbf{a}_{1} + y_{1} \, \mathbf{a}_{2} + z_{1} \, \mathbf{a}_{3} & = & x_{1}a \, \mathbf{\hat{x}} + y_{1}a \, \mathbf{\hat{y}} + z_{1}c \, \mathbf{\hat{z}} & \left(8g\right) & \mbox{P} \\ 
\mathbf{B}_{2} & = & \left(\frac{1}{2} - x_{1}\right) \, \mathbf{a}_{1} + \left(\frac{1}{2} - y_{1}\right) \, \mathbf{a}_{2} + z_{1} \, \mathbf{a}_{3} & = & \left(\frac{1}{2} - x_{1}\right)a \, \mathbf{\hat{x}} + \left(\frac{1}{2} - y_{1}\right)a \, \mathbf{\hat{y}} + z_{1}c \, \mathbf{\hat{z}} & \left(8g\right) & \mbox{P} \\ 
\mathbf{B}_{3} & = & -y_{1} \, \mathbf{a}_{1} + \left(\frac{1}{2} +x_{1}\right) \, \mathbf{a}_{2} + \left(\frac{1}{2} +z_{1}\right) \, \mathbf{a}_{3} & = & -y_{1}a \, \mathbf{\hat{x}} + \left(\frac{1}{2} +x_{1}\right)a \, \mathbf{\hat{y}} + \left(\frac{1}{2} +z_{1}\right)c \, \mathbf{\hat{z}} & \left(8g\right) & \mbox{P} \\ 
\mathbf{B}_{4} & = & \left(\frac{1}{2} +y_{1}\right) \, \mathbf{a}_{1}-x_{1} \, \mathbf{a}_{2} + \left(\frac{1}{2} +z_{1}\right) \, \mathbf{a}_{3} & = & \left(\frac{1}{2} +y_{1}\right)a \, \mathbf{\hat{x}}-x_{1}a \, \mathbf{\hat{y}} + \left(\frac{1}{2} +z_{1}\right)c \, \mathbf{\hat{z}} & \left(8g\right) & \mbox{P} \\ 
\mathbf{B}_{5} & = & -x_{1} \, \mathbf{a}_{1}-y_{1} \, \mathbf{a}_{2}-z_{1} \, \mathbf{a}_{3} & = & -x_{1}a \, \mathbf{\hat{x}}-y_{1}a \, \mathbf{\hat{y}}-z_{1}c \, \mathbf{\hat{z}} & \left(8g\right) & \mbox{P} \\ 
\mathbf{B}_{6} & = & \left(\frac{1}{2} +x_{1}\right) \, \mathbf{a}_{1} + \left(\frac{1}{2} +y_{1}\right) \, \mathbf{a}_{2}-z_{1} \, \mathbf{a}_{3} & = & \left(\frac{1}{2} +x_{1}\right)a \, \mathbf{\hat{x}} + \left(\frac{1}{2} +y_{1}\right)a \, \mathbf{\hat{y}}-z_{1}c \, \mathbf{\hat{z}} & \left(8g\right) & \mbox{P} \\ 
\mathbf{B}_{7} & = & y_{1} \, \mathbf{a}_{1} + \left(\frac{1}{2} - x_{1}\right) \, \mathbf{a}_{2} + \left(\frac{1}{2} - z_{1}\right) \, \mathbf{a}_{3} & = & y_{1}a \, \mathbf{\hat{x}} + \left(\frac{1}{2} - x_{1}\right)a \, \mathbf{\hat{y}} + \left(\frac{1}{2} - z_{1}\right)c \, \mathbf{\hat{z}} & \left(8g\right) & \mbox{P} \\ 
\mathbf{B}_{8} & = & \left(\frac{1}{2} - y_{1}\right) \, \mathbf{a}_{1} + x_{1} \, \mathbf{a}_{2} + \left(\frac{1}{2} - z_{1}\right) \, \mathbf{a}_{3} & = & \left(\frac{1}{2} - y_{1}\right)a \, \mathbf{\hat{x}} + x_{1}a \, \mathbf{\hat{y}} + \left(\frac{1}{2} - z_{1}\right)c \, \mathbf{\hat{z}} & \left(8g\right) & \mbox{P} \\ 
\mathbf{B}_{9} & = & x_{2} \, \mathbf{a}_{1} + y_{2} \, \mathbf{a}_{2} + z_{2} \, \mathbf{a}_{3} & = & x_{2}a \, \mathbf{\hat{x}} + y_{2}a \, \mathbf{\hat{y}} + z_{2}c \, \mathbf{\hat{z}} & \left(8g\right) & \mbox{Ti I} \\ 
\mathbf{B}_{10} & = & \left(\frac{1}{2} - x_{2}\right) \, \mathbf{a}_{1} + \left(\frac{1}{2} - y_{2}\right) \, \mathbf{a}_{2} + z_{2} \, \mathbf{a}_{3} & = & \left(\frac{1}{2} - x_{2}\right)a \, \mathbf{\hat{x}} + \left(\frac{1}{2} - y_{2}\right)a \, \mathbf{\hat{y}} + z_{2}c \, \mathbf{\hat{z}} & \left(8g\right) & \mbox{Ti I} \\ 
\mathbf{B}_{11} & = & -y_{2} \, \mathbf{a}_{1} + \left(\frac{1}{2} +x_{2}\right) \, \mathbf{a}_{2} + \left(\frac{1}{2} +z_{2}\right) \, \mathbf{a}_{3} & = & -y_{2}a \, \mathbf{\hat{x}} + \left(\frac{1}{2} +x_{2}\right)a \, \mathbf{\hat{y}} + \left(\frac{1}{2} +z_{2}\right)c \, \mathbf{\hat{z}} & \left(8g\right) & \mbox{Ti I} \\ 
\mathbf{B}_{12} & = & \left(\frac{1}{2} +y_{2}\right) \, \mathbf{a}_{1}-x_{2} \, \mathbf{a}_{2} + \left(\frac{1}{2} +z_{2}\right) \, \mathbf{a}_{3} & = & \left(\frac{1}{2} +y_{2}\right)a \, \mathbf{\hat{x}}-x_{2}a \, \mathbf{\hat{y}} + \left(\frac{1}{2} +z_{2}\right)c \, \mathbf{\hat{z}} & \left(8g\right) & \mbox{Ti I} \\ 
\mathbf{B}_{13} & = & -x_{2} \, \mathbf{a}_{1}-y_{2} \, \mathbf{a}_{2}-z_{2} \, \mathbf{a}_{3} & = & -x_{2}a \, \mathbf{\hat{x}}-y_{2}a \, \mathbf{\hat{y}}-z_{2}c \, \mathbf{\hat{z}} & \left(8g\right) & \mbox{Ti I} \\ 
\mathbf{B}_{14} & = & \left(\frac{1}{2} +x_{2}\right) \, \mathbf{a}_{1} + \left(\frac{1}{2} +y_{2}\right) \, \mathbf{a}_{2}-z_{2} \, \mathbf{a}_{3} & = & \left(\frac{1}{2} +x_{2}\right)a \, \mathbf{\hat{x}} + \left(\frac{1}{2} +y_{2}\right)a \, \mathbf{\hat{y}}-z_{2}c \, \mathbf{\hat{z}} & \left(8g\right) & \mbox{Ti I} \\ 
\mathbf{B}_{15} & = & y_{2} \, \mathbf{a}_{1} + \left(\frac{1}{2} - x_{2}\right) \, \mathbf{a}_{2} + \left(\frac{1}{2} - z_{2}\right) \, \mathbf{a}_{3} & = & y_{2}a \, \mathbf{\hat{x}} + \left(\frac{1}{2} - x_{2}\right)a \, \mathbf{\hat{y}} + \left(\frac{1}{2} - z_{2}\right)c \, \mathbf{\hat{z}} & \left(8g\right) & \mbox{Ti I} \\ 
\mathbf{B}_{16} & = & \left(\frac{1}{2} - y_{2}\right) \, \mathbf{a}_{1} + x_{2} \, \mathbf{a}_{2} + \left(\frac{1}{2} - z_{2}\right) \, \mathbf{a}_{3} & = & \left(\frac{1}{2} - y_{2}\right)a \, \mathbf{\hat{x}} + x_{2}a \, \mathbf{\hat{y}} + \left(\frac{1}{2} - z_{2}\right)c \, \mathbf{\hat{z}} & \left(8g\right) & \mbox{Ti I} \\ 
\mathbf{B}_{17} & = & x_{3} \, \mathbf{a}_{1} + y_{3} \, \mathbf{a}_{2} + z_{3} \, \mathbf{a}_{3} & = & x_{3}a \, \mathbf{\hat{x}} + y_{3}a \, \mathbf{\hat{y}} + z_{3}c \, \mathbf{\hat{z}} & \left(8g\right) & \mbox{Ti II} \\ 
\mathbf{B}_{18} & = & \left(\frac{1}{2} - x_{3}\right) \, \mathbf{a}_{1} + \left(\frac{1}{2} - y_{3}\right) \, \mathbf{a}_{2} + z_{3} \, \mathbf{a}_{3} & = & \left(\frac{1}{2} - x_{3}\right)a \, \mathbf{\hat{x}} + \left(\frac{1}{2} - y_{3}\right)a \, \mathbf{\hat{y}} + z_{3}c \, \mathbf{\hat{z}} & \left(8g\right) & \mbox{Ti II} \\ 
\mathbf{B}_{19} & = & -y_{3} \, \mathbf{a}_{1} + \left(\frac{1}{2} +x_{3}\right) \, \mathbf{a}_{2} + \left(\frac{1}{2} +z_{3}\right) \, \mathbf{a}_{3} & = & -y_{3}a \, \mathbf{\hat{x}} + \left(\frac{1}{2} +x_{3}\right)a \, \mathbf{\hat{y}} + \left(\frac{1}{2} +z_{3}\right)c \, \mathbf{\hat{z}} & \left(8g\right) & \mbox{Ti II} \\ 
\mathbf{B}_{20} & = & \left(\frac{1}{2} +y_{3}\right) \, \mathbf{a}_{1}-x_{3} \, \mathbf{a}_{2} + \left(\frac{1}{2} +z_{3}\right) \, \mathbf{a}_{3} & = & \left(\frac{1}{2} +y_{3}\right)a \, \mathbf{\hat{x}}-x_{3}a \, \mathbf{\hat{y}} + \left(\frac{1}{2} +z_{3}\right)c \, \mathbf{\hat{z}} & \left(8g\right) & \mbox{Ti II} \\ 
\mathbf{B}_{21} & = & -x_{3} \, \mathbf{a}_{1}-y_{3} \, \mathbf{a}_{2}-z_{3} \, \mathbf{a}_{3} & = & -x_{3}a \, \mathbf{\hat{x}}-y_{3}a \, \mathbf{\hat{y}}-z_{3}c \, \mathbf{\hat{z}} & \left(8g\right) & \mbox{Ti II} \\ 
\mathbf{B}_{22} & = & \left(\frac{1}{2} +x_{3}\right) \, \mathbf{a}_{1} + \left(\frac{1}{2} +y_{3}\right) \, \mathbf{a}_{2}-z_{3} \, \mathbf{a}_{3} & = & \left(\frac{1}{2} +x_{3}\right)a \, \mathbf{\hat{x}} + \left(\frac{1}{2} +y_{3}\right)a \, \mathbf{\hat{y}}-z_{3}c \, \mathbf{\hat{z}} & \left(8g\right) & \mbox{Ti II} \\ 
\mathbf{B}_{23} & = & y_{3} \, \mathbf{a}_{1} + \left(\frac{1}{2} - x_{3}\right) \, \mathbf{a}_{2} + \left(\frac{1}{2} - z_{3}\right) \, \mathbf{a}_{3} & = & y_{3}a \, \mathbf{\hat{x}} + \left(\frac{1}{2} - x_{3}\right)a \, \mathbf{\hat{y}} + \left(\frac{1}{2} - z_{3}\right)c \, \mathbf{\hat{z}} & \left(8g\right) & \mbox{Ti II} \\ 
\mathbf{B}_{24} & = & \left(\frac{1}{2} - y_{3}\right) \, \mathbf{a}_{1} + x_{3} \, \mathbf{a}_{2} + \left(\frac{1}{2} - z_{3}\right) \, \mathbf{a}_{3} & = & \left(\frac{1}{2} - y_{3}\right)a \, \mathbf{\hat{x}} + x_{3}a \, \mathbf{\hat{y}} + \left(\frac{1}{2} - z_{3}\right)c \, \mathbf{\hat{z}} & \left(8g\right) & \mbox{Ti II} \\ 
\mathbf{B}_{25} & = & x_{4} \, \mathbf{a}_{1} + y_{4} \, \mathbf{a}_{2} + z_{4} \, \mathbf{a}_{3} & = & x_{4}a \, \mathbf{\hat{x}} + y_{4}a \, \mathbf{\hat{y}} + z_{4}c \, \mathbf{\hat{z}} & \left(8g\right) & \mbox{Ti III} \\ 
\mathbf{B}_{26} & = & \left(\frac{1}{2} - x_{4}\right) \, \mathbf{a}_{1} + \left(\frac{1}{2} - y_{4}\right) \, \mathbf{a}_{2} + z_{4} \, \mathbf{a}_{3} & = & \left(\frac{1}{2} - x_{4}\right)a \, \mathbf{\hat{x}} + \left(\frac{1}{2} - y_{4}\right)a \, \mathbf{\hat{y}} + z_{4}c \, \mathbf{\hat{z}} & \left(8g\right) & \mbox{Ti III} \\ 
\mathbf{B}_{27} & = & -y_{4} \, \mathbf{a}_{1} + \left(\frac{1}{2} +x_{4}\right) \, \mathbf{a}_{2} + \left(\frac{1}{2} +z_{4}\right) \, \mathbf{a}_{3} & = & -y_{4}a \, \mathbf{\hat{x}} + \left(\frac{1}{2} +x_{4}\right)a \, \mathbf{\hat{y}} + \left(\frac{1}{2} +z_{4}\right)c \, \mathbf{\hat{z}} & \left(8g\right) & \mbox{Ti III} \\ 
\mathbf{B}_{28} & = & \left(\frac{1}{2} +y_{4}\right) \, \mathbf{a}_{1}-x_{4} \, \mathbf{a}_{2} + \left(\frac{1}{2} +z_{4}\right) \, \mathbf{a}_{3} & = & \left(\frac{1}{2} +y_{4}\right)a \, \mathbf{\hat{x}}-x_{4}a \, \mathbf{\hat{y}} + \left(\frac{1}{2} +z_{4}\right)c \, \mathbf{\hat{z}} & \left(8g\right) & \mbox{Ti III} \\ 
\mathbf{B}_{29} & = & -x_{4} \, \mathbf{a}_{1}-y_{4} \, \mathbf{a}_{2}-z_{4} \, \mathbf{a}_{3} & = & -x_{4}a \, \mathbf{\hat{x}}-y_{4}a \, \mathbf{\hat{y}}-z_{4}c \, \mathbf{\hat{z}} & \left(8g\right) & \mbox{Ti III} \\ 
\mathbf{B}_{30} & = & \left(\frac{1}{2} +x_{4}\right) \, \mathbf{a}_{1} + \left(\frac{1}{2} +y_{4}\right) \, \mathbf{a}_{2}-z_{4} \, \mathbf{a}_{3} & = & \left(\frac{1}{2} +x_{4}\right)a \, \mathbf{\hat{x}} + \left(\frac{1}{2} +y_{4}\right)a \, \mathbf{\hat{y}}-z_{4}c \, \mathbf{\hat{z}} & \left(8g\right) & \mbox{Ti III} \\ 
\mathbf{B}_{31} & = & y_{4} \, \mathbf{a}_{1} + \left(\frac{1}{2} - x_{4}\right) \, \mathbf{a}_{2} + \left(\frac{1}{2} - z_{4}\right) \, \mathbf{a}_{3} & = & y_{4}a \, \mathbf{\hat{x}} + \left(\frac{1}{2} - x_{4}\right)a \, \mathbf{\hat{y}} + \left(\frac{1}{2} - z_{4}\right)c \, \mathbf{\hat{z}} & \left(8g\right) & \mbox{Ti III} \\ 
\mathbf{B}_{32} & = & \left(\frac{1}{2} - y_{4}\right) \, \mathbf{a}_{1} + x_{4} \, \mathbf{a}_{2} + \left(\frac{1}{2} - z_{4}\right) \, \mathbf{a}_{3} & = & \left(\frac{1}{2} - y_{4}\right)a \, \mathbf{\hat{x}} + x_{4}a \, \mathbf{\hat{y}} + \left(\frac{1}{2} - z_{4}\right)c \, \mathbf{\hat{z}} & \left(8g\right) & \mbox{Ti III} \\ 
\end{longtabu}
\renewcommand{\arraystretch}{1.0}
\noindent \hrulefill
\\
\textbf{References:}
\vspace*{-0.25cm}
\begin{flushleft}
  - \bibentry{Eremenko_PTi3_DopovAkadNaukUkrRSR_1965}. \\
\end{flushleft}
\textbf{Found in:}
\vspace*{-0.25cm}
\begin{flushleft}
  - \bibentry{Villars_PearsonsCrystalData_2013}. \\
\end{flushleft}
\noindent \hrulefill
\\
\textbf{Geometry files:}
\\
\noindent  - CIF: pp. {\hyperref[AB3_tP32_86_g_3g_cif]{\pageref{AB3_tP32_86_g_3g_cif}}} \\
\noindent  - POSCAR: pp. {\hyperref[AB3_tP32_86_g_3g_poscar]{\pageref{AB3_tP32_86_g_3g_poscar}}} \\
\onecolumn
{\phantomsection\label{A4B_tI20_88_f_a}}
\subsection*{\huge \textbf{{\normalfont ThCl$_{4}$ Structure: A4B\_tI20\_88\_f\_a}}}
\noindent \hrulefill
\vspace*{0.25cm}
\begin{figure}[htp]
  \centering
  \vspace{-1em}
  {\includegraphics[width=1\textwidth]{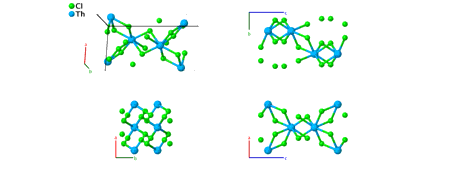}}
\end{figure}
\vspace*{-0.5cm}
\renewcommand{\arraystretch}{1.5}
\begin{equation*}
  \begin{array}{>{$\hspace{-0.15cm}}l<{$}>{$}p{0.5cm}<{$}>{$}p{18.5cm}<{$}}
    \mbox{\large \textbf{Prototype}} &\colon & \ce{ThCl4} \\
    \mbox{\large \textbf{\AFLOW\ prototype label}} &\colon & \mbox{A4B\_tI20\_88\_f\_a} \\
    \mbox{\large \textbf{\textit{Strukturbericht} designation}} &\colon & \mbox{None} \\
    \mbox{\large \textbf{Pearson symbol}} &\colon & \mbox{tI20} \\
    \mbox{\large \textbf{Space group number}} &\colon & 88 \\
    \mbox{\large \textbf{Space group symbol}} &\colon & I4_{1}/a \\
    \mbox{\large \textbf{\AFLOW\ prototype command}} &\colon &  \texttt{aflow} \,  \, \texttt{-{}-proto=A4B\_tI20\_88\_f\_a } \, \newline \texttt{-{}-params=}{a,c/a,x_{2},y_{2},z_{2} }
  \end{array}
\end{equation*}
\renewcommand{\arraystretch}{1.0}

\noindent \parbox{1 \linewidth}{
\noindent \hrulefill
\\
\textbf{Body-centered Tetragonal primitive vectors:} \\
\vspace*{-0.25cm}
\begin{tabular}{cc}
  \begin{tabular}{c}
    \parbox{0.6 \linewidth}{
      \renewcommand{\arraystretch}{1.5}
      \begin{equation*}
        \centering
        \begin{array}{ccc}
              \mathbf{a}_1 & = & - \frac12 \, a \, \mathbf{\hat{x}} + \frac12 \, a \, \mathbf{\hat{y}} + \frac12 \, c \, \mathbf{\hat{z}} \\
    \mathbf{a}_2 & = & ~ \frac12 \, a \, \mathbf{\hat{x}} - \frac12 \, a \, \mathbf{\hat{y}} + \frac12 \, c \, \mathbf{\hat{z}} \\
    \mathbf{a}_3 & = & ~ \frac12 \, a \, \mathbf{\hat{x}} + \frac12 \, a \, \mathbf{\hat{y}} - \frac12 \, c \, \mathbf{\hat{z}} \\

        \end{array}
      \end{equation*}
    }
    \renewcommand{\arraystretch}{1.0}
  \end{tabular}
  \begin{tabular}{c}
    \includegraphics[width=0.3\linewidth]{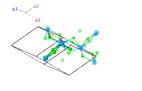} \\
  \end{tabular}
\end{tabular}

}
\vspace*{-0.25cm}

\noindent \hrulefill
\\
\textbf{Basis vectors:}
\vspace*{-0.25cm}
\renewcommand{\arraystretch}{1.5}
\begin{longtabu} to \textwidth{>{\centering $}X[-1,c,c]<{$}>{\centering $}X[-1,c,c]<{$}>{\centering $}X[-1,c,c]<{$}>{\centering $}X[-1,c,c]<{$}>{\centering $}X[-1,c,c]<{$}>{\centering $}X[-1,c,c]<{$}>{\centering $}X[-1,c,c]<{$}}
  & & \mbox{Lattice Coordinates} & & \mbox{Cartesian Coordinates} &\mbox{Wyckoff Position} & \mbox{Atom Type} \\  
  \mathbf{B}_{1} & = & \frac{3}{8} \, \mathbf{a}_{1} + \frac{1}{8} \, \mathbf{a}_{2} + \frac{1}{4} \, \mathbf{a}_{3} & = & \frac{1}{4}a \, \mathbf{\hat{y}} + \frac{1}{8}c \, \mathbf{\hat{z}} & \left(4a\right) & \mbox{Th} \\ 
\mathbf{B}_{2} & = & \frac{5}{8} \, \mathbf{a}_{1} + \frac{7}{8} \, \mathbf{a}_{2} + \frac{3}{4} \, \mathbf{a}_{3} & = & \frac{1}{2}a \, \mathbf{\hat{x}} + \frac{1}{4}a \, \mathbf{\hat{y}} + \frac{3}{8}c \, \mathbf{\hat{z}} & \left(4a\right) & \mbox{Th} \\ 
\mathbf{B}_{3} & = & \left(y_{2}+z_{2}\right) \, \mathbf{a}_{1} + \left(x_{2}+z_{2}\right) \, \mathbf{a}_{2} + \left(x_{2}+y_{2}\right) \, \mathbf{a}_{3} & = & x_{2}a \, \mathbf{\hat{x}} + y_{2}a \, \mathbf{\hat{y}} + z_{2}c \, \mathbf{\hat{z}} & \left(16f\right) & \mbox{Cl} \\ 
\mathbf{B}_{4} & = & \left(\frac{1}{2} - y_{2} + z_{2}\right) \, \mathbf{a}_{1} + \left(-x_{2}+z_{2}\right) \, \mathbf{a}_{2} + \left(\frac{1}{2} - x_{2} - y_{2}\right) \, \mathbf{a}_{3} & = & -x_{2}a \, \mathbf{\hat{x}} + \left(\frac{1}{2} - y_{2}\right)a \, \mathbf{\hat{y}} + z_{2}c \, \mathbf{\hat{z}} & \left(16f\right) & \mbox{Cl} \\ 
\mathbf{B}_{5} & = & \left(\frac{1}{2} +x_{2} + z_{2}\right) \, \mathbf{a}_{1} + \left(-y_{2}+z_{2}\right) \, \mathbf{a}_{2} + \left(x_{2}-y_{2}\right) \, \mathbf{a}_{3} & = & \left(\frac{3}{4} - y_{2}\right)a \, \mathbf{\hat{x}} + \left(\frac{1}{4} +x_{2}\right)a \, \mathbf{\hat{y}} + \left(\frac{1}{4} +z_{2}\right)c \, \mathbf{\hat{z}} & \left(16f\right) & \mbox{Cl} \\ 
\mathbf{B}_{6} & = & \left(\frac{1}{2} - x_{2} + z_{2}\right) \, \mathbf{a}_{1} + \left(\frac{1}{2} +y_{2} + z_{2}\right) \, \mathbf{a}_{2} + \left(\frac{1}{2} - x_{2} + y_{2}\right) \, \mathbf{a}_{3} & = & \left(\frac{1}{4} +y_{2}\right)a \, \mathbf{\hat{x}} + \left(\frac{1}{4} - x_{2}\right)a \, \mathbf{\hat{y}} + \left(\frac{1}{4} +z_{2}\right)c \, \mathbf{\hat{z}} & \left(16f\right) & \mbox{Cl} \\ 
\mathbf{B}_{7} & = & \left(-y_{2}-z_{2}\right) \, \mathbf{a}_{1} + \left(-x_{2}-z_{2}\right) \, \mathbf{a}_{2} + \left(-x_{2}-y_{2}\right) \, \mathbf{a}_{3} & = & -x_{2}a \, \mathbf{\hat{x}}-y_{2}a \, \mathbf{\hat{y}}-z_{2}c \, \mathbf{\hat{z}} & \left(16f\right) & \mbox{Cl} \\ 
\mathbf{B}_{8} & = & \left(\frac{1}{2} +y_{2} - z_{2}\right) \, \mathbf{a}_{1} + \left(x_{2}-z_{2}\right) \, \mathbf{a}_{2} + \left(\frac{1}{2} +x_{2} + y_{2}\right) \, \mathbf{a}_{3} & = & x_{2}a \, \mathbf{\hat{x}} + \left(\frac{1}{2} +y_{2}\right)a \, \mathbf{\hat{y}}-z_{2}c \, \mathbf{\hat{z}} & \left(16f\right) & \mbox{Cl} \\ 
\mathbf{B}_{9} & = & \left(\frac{1}{2} - x_{2} - z_{2}\right) \, \mathbf{a}_{1} + \left(y_{2}-z_{2}\right) \, \mathbf{a}_{2} + \left(-x_{2}+y_{2}\right) \, \mathbf{a}_{3} & = & \left(- \frac{1}{4} +y_{2}\right)a \, \mathbf{\hat{x}} + \left(\frac{1}{4} - x_{2}\right)a \, \mathbf{\hat{y}} + \left(\frac{1}{4} - z_{2}\right)c \, \mathbf{\hat{z}} & \left(16f\right) & \mbox{Cl} \\ 
\mathbf{B}_{10} & = & \left(\frac{1}{2} +x_{2} - z_{2}\right) \, \mathbf{a}_{1} + \left(\frac{1}{2} - y_{2} - z_{2}\right) \, \mathbf{a}_{2} + \left(\frac{1}{2} +x_{2} - y_{2}\right) \, \mathbf{a}_{3} & = & \left(\frac{1}{4} - y_{2}\right)a \, \mathbf{\hat{x}} + \left(\frac{1}{4} +x_{2}\right)a \, \mathbf{\hat{y}} + \left(\frac{1}{4} - z_{2}\right)c \, \mathbf{\hat{z}} & \left(16f\right) & \mbox{Cl} \\ 
\end{longtabu}
\renewcommand{\arraystretch}{1.0}
\noindent \hrulefill
\\
\textbf{References:}
\vspace*{-0.25cm}
\begin{flushleft}
  - \bibentry{Mason_ThCl4_JLessCommMat_1974}. \\
\end{flushleft}
\textbf{Found in:}
\vspace*{-0.25cm}
\begin{flushleft}
  - \bibentry{Villars_PearsonsCrystalData_2013}. \\
\end{flushleft}
\noindent \hrulefill
\\
\textbf{Geometry files:}
\\
\noindent  - CIF: pp. {\hyperref[A4B_tI20_88_f_a_cif]{\pageref{A4B_tI20_88_f_a_cif}}} \\
\noindent  - POSCAR: pp. {\hyperref[A4B_tI20_88_f_a_poscar]{\pageref{A4B_tI20_88_f_a_poscar}}} \\
\onecolumn
{\phantomsection\label{AB2_tI96_88_2f_4f}}
\subsection*{\huge \textbf{{\normalfont $\alpha$-NbO$_{2}$ Structure: AB2\_tI96\_88\_2f\_4f}}}
\noindent \hrulefill
\vspace*{0.25cm}
\begin{figure}[htp]
  \centering
  \vspace{-1em}
  {\includegraphics[width=1\textwidth]{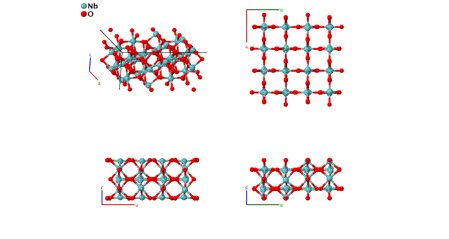}}
\end{figure}
\vspace*{-0.5cm}
\renewcommand{\arraystretch}{1.5}
\begin{equation*}
  \begin{array}{>{$\hspace{-0.15cm}}l<{$}>{$}p{0.5cm}<{$}>{$}p{18.5cm}<{$}}
    \mbox{\large \textbf{Prototype}} &\colon & \ce{$\alpha$-NbO2} \\
    \mbox{\large \textbf{\AFLOW\ prototype label}} &\colon & \mbox{AB2\_tI96\_88\_2f\_4f} \\
    \mbox{\large \textbf{\textit{Strukturbericht} designation}} &\colon & \mbox{None} \\
    \mbox{\large \textbf{Pearson symbol}} &\colon & \mbox{tI96} \\
    \mbox{\large \textbf{Space group number}} &\colon & 88 \\
    \mbox{\large \textbf{Space group symbol}} &\colon & I4_{1}/a \\
    \mbox{\large \textbf{\AFLOW\ prototype command}} &\colon &  \texttt{aflow} \,  \, \texttt{-{}-proto=AB2\_tI96\_88\_2f\_4f } \, \newline \texttt{-{}-params=}{a,c/a,x_{1},y_{1},z_{1},x_{2},y_{2},z_{2},x_{3},y_{3},z_{3},x_{4},y_{4},z_{4},x_{5},y_{5},z_{5},x_{6},y_{6},z_{6} }
  \end{array}
\end{equation*}
\renewcommand{\arraystretch}{1.0}

\vspace*{-0.25cm}
\noindent \hrulefill
\begin{itemize}
  \item{Although Bolzan {\em et al.} (Bolzan, 1994) also gives
structural information $\alpha$-NbO$_{2}$, Pynn {\em et
  al.} (Pynn, 1996) is the only reference we found which
unambiguously states that this structure is reported in setting 2 of
space group \#88.
}
\end{itemize}

\noindent \parbox{1 \linewidth}{
\noindent \hrulefill
\\
\textbf{Body-centered Tetragonal primitive vectors:} \\
\vspace*{-0.25cm}
\begin{tabular}{cc}
  \begin{tabular}{c}
    \parbox{0.6 \linewidth}{
      \renewcommand{\arraystretch}{1.5}
      \begin{equation*}
        \centering
        \begin{array}{ccc}
              \mathbf{a}_1 & = & - \frac12 \, a \, \mathbf{\hat{x}} + \frac12 \, a \, \mathbf{\hat{y}} + \frac12 \, c \, \mathbf{\hat{z}} \\
    \mathbf{a}_2 & = & ~ \frac12 \, a \, \mathbf{\hat{x}} - \frac12 \, a \, \mathbf{\hat{y}} + \frac12 \, c \, \mathbf{\hat{z}} \\
    \mathbf{a}_3 & = & ~ \frac12 \, a \, \mathbf{\hat{x}} + \frac12 \, a \, \mathbf{\hat{y}} - \frac12 \, c \, \mathbf{\hat{z}} \\

        \end{array}
      \end{equation*}
    }
    \renewcommand{\arraystretch}{1.0}
  \end{tabular}
  \begin{tabular}{c}
    \includegraphics[width=0.3\linewidth]{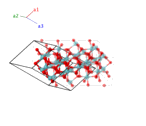} \\
  \end{tabular}
\end{tabular}

}
\vspace*{-0.25cm}

\noindent \hrulefill
\\
\textbf{Basis vectors:}
\vspace*{-0.25cm}
\renewcommand{\arraystretch}{1.5}
\begin{longtabu} to \textwidth{>{\centering $}X[-1,c,c]<{$}>{\centering $}X[-1,c,c]<{$}>{\centering $}X[-1,c,c]<{$}>{\centering $}X[-1,c,c]<{$}>{\centering $}X[-1,c,c]<{$}>{\centering $}X[-1,c,c]<{$}>{\centering $}X[-1,c,c]<{$}}
  & & \mbox{Lattice Coordinates} & & \mbox{Cartesian Coordinates} &\mbox{Wyckoff Position} & \mbox{Atom Type} \\  
  \mathbf{B}_{1} & = & \left(y_{1}+z_{1}\right) \, \mathbf{a}_{1} + \left(x_{1}+z_{1}\right) \, \mathbf{a}_{2} + \left(x_{1}+y_{1}\right) \, \mathbf{a}_{3} & = & x_{1}a \, \mathbf{\hat{x}} + y_{1}a \, \mathbf{\hat{y}} + z_{1}c \, \mathbf{\hat{z}} & \left(16f\right) & \mbox{Nb I} \\ 
\mathbf{B}_{2} & = & \left(\frac{1}{2} - y_{1} + z_{1}\right) \, \mathbf{a}_{1} + \left(-x_{1}+z_{1}\right) \, \mathbf{a}_{2} + \left(\frac{1}{2} - x_{1} - y_{1}\right) \, \mathbf{a}_{3} & = & -x_{1}a \, \mathbf{\hat{x}} + \left(\frac{1}{2} - y_{1}\right)a \, \mathbf{\hat{y}} + z_{1}c \, \mathbf{\hat{z}} & \left(16f\right) & \mbox{Nb I} \\ 
\mathbf{B}_{3} & = & \left(\frac{1}{2} +x_{1} + z_{1}\right) \, \mathbf{a}_{1} + \left(-y_{1}+z_{1}\right) \, \mathbf{a}_{2} + \left(x_{1}-y_{1}\right) \, \mathbf{a}_{3} & = & \left(\frac{3}{4} - y_{1}\right)a \, \mathbf{\hat{x}} + \left(\frac{1}{4} +x_{1}\right)a \, \mathbf{\hat{y}} + \left(\frac{1}{4} +z_{1}\right)c \, \mathbf{\hat{z}} & \left(16f\right) & \mbox{Nb I} \\ 
\mathbf{B}_{4} & = & \left(\frac{1}{2} - x_{1} + z_{1}\right) \, \mathbf{a}_{1} + \left(\frac{1}{2} +y_{1} + z_{1}\right) \, \mathbf{a}_{2} + \left(\frac{1}{2} - x_{1} + y_{1}\right) \, \mathbf{a}_{3} & = & \left(\frac{1}{4} +y_{1}\right)a \, \mathbf{\hat{x}} + \left(\frac{1}{4} - x_{1}\right)a \, \mathbf{\hat{y}} + \left(\frac{1}{4} +z_{1}\right)c \, \mathbf{\hat{z}} & \left(16f\right) & \mbox{Nb I} \\ 
\mathbf{B}_{5} & = & \left(-y_{1}-z_{1}\right) \, \mathbf{a}_{1} + \left(-x_{1}-z_{1}\right) \, \mathbf{a}_{2} + \left(-x_{1}-y_{1}\right) \, \mathbf{a}_{3} & = & -x_{1}a \, \mathbf{\hat{x}}-y_{1}a \, \mathbf{\hat{y}}-z_{1}c \, \mathbf{\hat{z}} & \left(16f\right) & \mbox{Nb I} \\ 
\mathbf{B}_{6} & = & \left(\frac{1}{2} +y_{1} - z_{1}\right) \, \mathbf{a}_{1} + \left(x_{1}-z_{1}\right) \, \mathbf{a}_{2} + \left(\frac{1}{2} +x_{1} + y_{1}\right) \, \mathbf{a}_{3} & = & x_{1}a \, \mathbf{\hat{x}} + \left(\frac{1}{2} +y_{1}\right)a \, \mathbf{\hat{y}}-z_{1}c \, \mathbf{\hat{z}} & \left(16f\right) & \mbox{Nb I} \\ 
\mathbf{B}_{7} & = & \left(\frac{1}{2} - x_{1} - z_{1}\right) \, \mathbf{a}_{1} + \left(y_{1}-z_{1}\right) \, \mathbf{a}_{2} + \left(-x_{1}+y_{1}\right) \, \mathbf{a}_{3} & = & \left(- \frac{1}{4} +y_{1}\right)a \, \mathbf{\hat{x}} + \left(\frac{1}{4} - x_{1}\right)a \, \mathbf{\hat{y}} + \left(\frac{1}{4} - z_{1}\right)c \, \mathbf{\hat{z}} & \left(16f\right) & \mbox{Nb I} \\ 
\mathbf{B}_{8} & = & \left(\frac{1}{2} +x_{1} - z_{1}\right) \, \mathbf{a}_{1} + \left(\frac{1}{2} - y_{1} - z_{1}\right) \, \mathbf{a}_{2} + \left(\frac{1}{2} +x_{1} - y_{1}\right) \, \mathbf{a}_{3} & = & \left(\frac{1}{4} - y_{1}\right)a \, \mathbf{\hat{x}} + \left(\frac{1}{4} +x_{1}\right)a \, \mathbf{\hat{y}} + \left(\frac{1}{4} - z_{1}\right)c \, \mathbf{\hat{z}} & \left(16f\right) & \mbox{Nb I} \\ 
\mathbf{B}_{9} & = & \left(y_{2}+z_{2}\right) \, \mathbf{a}_{1} + \left(x_{2}+z_{2}\right) \, \mathbf{a}_{2} + \left(x_{2}+y_{2}\right) \, \mathbf{a}_{3} & = & x_{2}a \, \mathbf{\hat{x}} + y_{2}a \, \mathbf{\hat{y}} + z_{2}c \, \mathbf{\hat{z}} & \left(16f\right) & \mbox{Nb II} \\ 
\mathbf{B}_{10} & = & \left(\frac{1}{2} - y_{2} + z_{2}\right) \, \mathbf{a}_{1} + \left(-x_{2}+z_{2}\right) \, \mathbf{a}_{2} + \left(\frac{1}{2} - x_{2} - y_{2}\right) \, \mathbf{a}_{3} & = & -x_{2}a \, \mathbf{\hat{x}} + \left(\frac{1}{2} - y_{2}\right)a \, \mathbf{\hat{y}} + z_{2}c \, \mathbf{\hat{z}} & \left(16f\right) & \mbox{Nb II} \\ 
\mathbf{B}_{11} & = & \left(\frac{1}{2} +x_{2} + z_{2}\right) \, \mathbf{a}_{1} + \left(-y_{2}+z_{2}\right) \, \mathbf{a}_{2} + \left(x_{2}-y_{2}\right) \, \mathbf{a}_{3} & = & \left(\frac{3}{4} - y_{2}\right)a \, \mathbf{\hat{x}} + \left(\frac{1}{4} +x_{2}\right)a \, \mathbf{\hat{y}} + \left(\frac{1}{4} +z_{2}\right)c \, \mathbf{\hat{z}} & \left(16f\right) & \mbox{Nb II} \\ 
\mathbf{B}_{12} & = & \left(\frac{1}{2} - x_{2} + z_{2}\right) \, \mathbf{a}_{1} + \left(\frac{1}{2} +y_{2} + z_{2}\right) \, \mathbf{a}_{2} + \left(\frac{1}{2} - x_{2} + y_{2}\right) \, \mathbf{a}_{3} & = & \left(\frac{1}{4} +y_{2}\right)a \, \mathbf{\hat{x}} + \left(\frac{1}{4} - x_{2}\right)a \, \mathbf{\hat{y}} + \left(\frac{1}{4} +z_{2}\right)c \, \mathbf{\hat{z}} & \left(16f\right) & \mbox{Nb II} \\ 
\mathbf{B}_{13} & = & \left(-y_{2}-z_{2}\right) \, \mathbf{a}_{1} + \left(-x_{2}-z_{2}\right) \, \mathbf{a}_{2} + \left(-x_{2}-y_{2}\right) \, \mathbf{a}_{3} & = & -x_{2}a \, \mathbf{\hat{x}}-y_{2}a \, \mathbf{\hat{y}}-z_{2}c \, \mathbf{\hat{z}} & \left(16f\right) & \mbox{Nb II} \\ 
\mathbf{B}_{14} & = & \left(\frac{1}{2} +y_{2} - z_{2}\right) \, \mathbf{a}_{1} + \left(x_{2}-z_{2}\right) \, \mathbf{a}_{2} + \left(\frac{1}{2} +x_{2} + y_{2}\right) \, \mathbf{a}_{3} & = & x_{2}a \, \mathbf{\hat{x}} + \left(\frac{1}{2} +y_{2}\right)a \, \mathbf{\hat{y}}-z_{2}c \, \mathbf{\hat{z}} & \left(16f\right) & \mbox{Nb II} \\ 
\mathbf{B}_{15} & = & \left(\frac{1}{2} - x_{2} - z_{2}\right) \, \mathbf{a}_{1} + \left(y_{2}-z_{2}\right) \, \mathbf{a}_{2} + \left(-x_{2}+y_{2}\right) \, \mathbf{a}_{3} & = & \left(- \frac{1}{4} +y_{2}\right)a \, \mathbf{\hat{x}} + \left(\frac{1}{4} - x_{2}\right)a \, \mathbf{\hat{y}} + \left(\frac{1}{4} - z_{2}\right)c \, \mathbf{\hat{z}} & \left(16f\right) & \mbox{Nb II} \\ 
\mathbf{B}_{16} & = & \left(\frac{1}{2} +x_{2} - z_{2}\right) \, \mathbf{a}_{1} + \left(\frac{1}{2} - y_{2} - z_{2}\right) \, \mathbf{a}_{2} + \left(\frac{1}{2} +x_{2} - y_{2}\right) \, \mathbf{a}_{3} & = & \left(\frac{1}{4} - y_{2}\right)a \, \mathbf{\hat{x}} + \left(\frac{1}{4} +x_{2}\right)a \, \mathbf{\hat{y}} + \left(\frac{1}{4} - z_{2}\right)c \, \mathbf{\hat{z}} & \left(16f\right) & \mbox{Nb II} \\ 
\mathbf{B}_{17} & = & \left(y_{3}+z_{3}\right) \, \mathbf{a}_{1} + \left(x_{3}+z_{3}\right) \, \mathbf{a}_{2} + \left(x_{3}+y_{3}\right) \, \mathbf{a}_{3} & = & x_{3}a \, \mathbf{\hat{x}} + y_{3}a \, \mathbf{\hat{y}} + z_{3}c \, \mathbf{\hat{z}} & \left(16f\right) & \mbox{O I} \\ 
\mathbf{B}_{18} & = & \left(\frac{1}{2} - y_{3} + z_{3}\right) \, \mathbf{a}_{1} + \left(-x_{3}+z_{3}\right) \, \mathbf{a}_{2} + \left(\frac{1}{2} - x_{3} - y_{3}\right) \, \mathbf{a}_{3} & = & -x_{3}a \, \mathbf{\hat{x}} + \left(\frac{1}{2} - y_{3}\right)a \, \mathbf{\hat{y}} + z_{3}c \, \mathbf{\hat{z}} & \left(16f\right) & \mbox{O I} \\ 
\mathbf{B}_{19} & = & \left(\frac{1}{2} +x_{3} + z_{3}\right) \, \mathbf{a}_{1} + \left(-y_{3}+z_{3}\right) \, \mathbf{a}_{2} + \left(x_{3}-y_{3}\right) \, \mathbf{a}_{3} & = & \left(\frac{3}{4} - y_{3}\right)a \, \mathbf{\hat{x}} + \left(\frac{1}{4} +x_{3}\right)a \, \mathbf{\hat{y}} + \left(\frac{1}{4} +z_{3}\right)c \, \mathbf{\hat{z}} & \left(16f\right) & \mbox{O I} \\ 
\mathbf{B}_{20} & = & \left(\frac{1}{2} - x_{3} + z_{3}\right) \, \mathbf{a}_{1} + \left(\frac{1}{2} +y_{3} + z_{3}\right) \, \mathbf{a}_{2} + \left(\frac{1}{2} - x_{3} + y_{3}\right) \, \mathbf{a}_{3} & = & \left(\frac{1}{4} +y_{3}\right)a \, \mathbf{\hat{x}} + \left(\frac{1}{4} - x_{3}\right)a \, \mathbf{\hat{y}} + \left(\frac{1}{4} +z_{3}\right)c \, \mathbf{\hat{z}} & \left(16f\right) & \mbox{O I} \\ 
\mathbf{B}_{21} & = & \left(-y_{3}-z_{3}\right) \, \mathbf{a}_{1} + \left(-x_{3}-z_{3}\right) \, \mathbf{a}_{2} + \left(-x_{3}-y_{3}\right) \, \mathbf{a}_{3} & = & -x_{3}a \, \mathbf{\hat{x}}-y_{3}a \, \mathbf{\hat{y}}-z_{3}c \, \mathbf{\hat{z}} & \left(16f\right) & \mbox{O I} \\ 
\mathbf{B}_{22} & = & \left(\frac{1}{2} +y_{3} - z_{3}\right) \, \mathbf{a}_{1} + \left(x_{3}-z_{3}\right) \, \mathbf{a}_{2} + \left(\frac{1}{2} +x_{3} + y_{3}\right) \, \mathbf{a}_{3} & = & x_{3}a \, \mathbf{\hat{x}} + \left(\frac{1}{2} +y_{3}\right)a \, \mathbf{\hat{y}}-z_{3}c \, \mathbf{\hat{z}} & \left(16f\right) & \mbox{O I} \\ 
\mathbf{B}_{23} & = & \left(\frac{1}{2} - x_{3} - z_{3}\right) \, \mathbf{a}_{1} + \left(y_{3}-z_{3}\right) \, \mathbf{a}_{2} + \left(-x_{3}+y_{3}\right) \, \mathbf{a}_{3} & = & \left(- \frac{1}{4} +y_{3}\right)a \, \mathbf{\hat{x}} + \left(\frac{1}{4} - x_{3}\right)a \, \mathbf{\hat{y}} + \left(\frac{1}{4} - z_{3}\right)c \, \mathbf{\hat{z}} & \left(16f\right) & \mbox{O I} \\ 
\mathbf{B}_{24} & = & \left(\frac{1}{2} +x_{3} - z_{3}\right) \, \mathbf{a}_{1} + \left(\frac{1}{2} - y_{3} - z_{3}\right) \, \mathbf{a}_{2} + \left(\frac{1}{2} +x_{3} - y_{3}\right) \, \mathbf{a}_{3} & = & \left(\frac{1}{4} - y_{3}\right)a \, \mathbf{\hat{x}} + \left(\frac{1}{4} +x_{3}\right)a \, \mathbf{\hat{y}} + \left(\frac{1}{4} - z_{3}\right)c \, \mathbf{\hat{z}} & \left(16f\right) & \mbox{O I} \\ 
\mathbf{B}_{25} & = & \left(y_{4}+z_{4}\right) \, \mathbf{a}_{1} + \left(x_{4}+z_{4}\right) \, \mathbf{a}_{2} + \left(x_{4}+y_{4}\right) \, \mathbf{a}_{3} & = & x_{4}a \, \mathbf{\hat{x}} + y_{4}a \, \mathbf{\hat{y}} + z_{4}c \, \mathbf{\hat{z}} & \left(16f\right) & \mbox{O II} \\ 
\mathbf{B}_{26} & = & \left(\frac{1}{2} - y_{4} + z_{4}\right) \, \mathbf{a}_{1} + \left(-x_{4}+z_{4}\right) \, \mathbf{a}_{2} + \left(\frac{1}{2} - x_{4} - y_{4}\right) \, \mathbf{a}_{3} & = & -x_{4}a \, \mathbf{\hat{x}} + \left(\frac{1}{2} - y_{4}\right)a \, \mathbf{\hat{y}} + z_{4}c \, \mathbf{\hat{z}} & \left(16f\right) & \mbox{O II} \\ 
\mathbf{B}_{27} & = & \left(\frac{1}{2} +x_{4} + z_{4}\right) \, \mathbf{a}_{1} + \left(-y_{4}+z_{4}\right) \, \mathbf{a}_{2} + \left(x_{4}-y_{4}\right) \, \mathbf{a}_{3} & = & \left(\frac{3}{4} - y_{4}\right)a \, \mathbf{\hat{x}} + \left(\frac{1}{4} +x_{4}\right)a \, \mathbf{\hat{y}} + \left(\frac{1}{4} +z_{4}\right)c \, \mathbf{\hat{z}} & \left(16f\right) & \mbox{O II} \\ 
\mathbf{B}_{28} & = & \left(\frac{1}{2} - x_{4} + z_{4}\right) \, \mathbf{a}_{1} + \left(\frac{1}{2} +y_{4} + z_{4}\right) \, \mathbf{a}_{2} + \left(\frac{1}{2} - x_{4} + y_{4}\right) \, \mathbf{a}_{3} & = & \left(\frac{1}{4} +y_{4}\right)a \, \mathbf{\hat{x}} + \left(\frac{1}{4} - x_{4}\right)a \, \mathbf{\hat{y}} + \left(\frac{1}{4} +z_{4}\right)c \, \mathbf{\hat{z}} & \left(16f\right) & \mbox{O II} \\ 
\mathbf{B}_{29} & = & \left(-y_{4}-z_{4}\right) \, \mathbf{a}_{1} + \left(-x_{4}-z_{4}\right) \, \mathbf{a}_{2} + \left(-x_{4}-y_{4}\right) \, \mathbf{a}_{3} & = & -x_{4}a \, \mathbf{\hat{x}}-y_{4}a \, \mathbf{\hat{y}}-z_{4}c \, \mathbf{\hat{z}} & \left(16f\right) & \mbox{O II} \\ 
\mathbf{B}_{30} & = & \left(\frac{1}{2} +y_{4} - z_{4}\right) \, \mathbf{a}_{1} + \left(x_{4}-z_{4}\right) \, \mathbf{a}_{2} + \left(\frac{1}{2} +x_{4} + y_{4}\right) \, \mathbf{a}_{3} & = & x_{4}a \, \mathbf{\hat{x}} + \left(\frac{1}{2} +y_{4}\right)a \, \mathbf{\hat{y}}-z_{4}c \, \mathbf{\hat{z}} & \left(16f\right) & \mbox{O II} \\ 
\mathbf{B}_{31} & = & \left(\frac{1}{2} - x_{4} - z_{4}\right) \, \mathbf{a}_{1} + \left(y_{4}-z_{4}\right) \, \mathbf{a}_{2} + \left(-x_{4}+y_{4}\right) \, \mathbf{a}_{3} & = & \left(- \frac{1}{4} +y_{4}\right)a \, \mathbf{\hat{x}} + \left(\frac{1}{4} - x_{4}\right)a \, \mathbf{\hat{y}} + \left(\frac{1}{4} - z_{4}\right)c \, \mathbf{\hat{z}} & \left(16f\right) & \mbox{O II} \\ 
\mathbf{B}_{32} & = & \left(\frac{1}{2} +x_{4} - z_{4}\right) \, \mathbf{a}_{1} + \left(\frac{1}{2} - y_{4} - z_{4}\right) \, \mathbf{a}_{2} + \left(\frac{1}{2} +x_{4} - y_{4}\right) \, \mathbf{a}_{3} & = & \left(\frac{1}{4} - y_{4}\right)a \, \mathbf{\hat{x}} + \left(\frac{1}{4} +x_{4}\right)a \, \mathbf{\hat{y}} + \left(\frac{1}{4} - z_{4}\right)c \, \mathbf{\hat{z}} & \left(16f\right) & \mbox{O II} \\ 
\mathbf{B}_{33} & = & \left(y_{5}+z_{5}\right) \, \mathbf{a}_{1} + \left(x_{5}+z_{5}\right) \, \mathbf{a}_{2} + \left(x_{5}+y_{5}\right) \, \mathbf{a}_{3} & = & x_{5}a \, \mathbf{\hat{x}} + y_{5}a \, \mathbf{\hat{y}} + z_{5}c \, \mathbf{\hat{z}} & \left(16f\right) & \mbox{O III} \\ 
\mathbf{B}_{34} & = & \left(\frac{1}{2} - y_{5} + z_{5}\right) \, \mathbf{a}_{1} + \left(-x_{5}+z_{5}\right) \, \mathbf{a}_{2} + \left(\frac{1}{2} - x_{5} - y_{5}\right) \, \mathbf{a}_{3} & = & -x_{5}a \, \mathbf{\hat{x}} + \left(\frac{1}{2} - y_{5}\right)a \, \mathbf{\hat{y}} + z_{5}c \, \mathbf{\hat{z}} & \left(16f\right) & \mbox{O III} \\ 
\mathbf{B}_{35} & = & \left(\frac{1}{2} +x_{5} + z_{5}\right) \, \mathbf{a}_{1} + \left(-y_{5}+z_{5}\right) \, \mathbf{a}_{2} + \left(x_{5}-y_{5}\right) \, \mathbf{a}_{3} & = & \left(\frac{3}{4} - y_{5}\right)a \, \mathbf{\hat{x}} + \left(\frac{1}{4} +x_{5}\right)a \, \mathbf{\hat{y}} + \left(\frac{1}{4} +z_{5}\right)c \, \mathbf{\hat{z}} & \left(16f\right) & \mbox{O III} \\ 
\mathbf{B}_{36} & = & \left(\frac{1}{2} - x_{5} + z_{5}\right) \, \mathbf{a}_{1} + \left(\frac{1}{2} +y_{5} + z_{5}\right) \, \mathbf{a}_{2} + \left(\frac{1}{2} - x_{5} + y_{5}\right) \, \mathbf{a}_{3} & = & \left(\frac{1}{4} +y_{5}\right)a \, \mathbf{\hat{x}} + \left(\frac{1}{4} - x_{5}\right)a \, \mathbf{\hat{y}} + \left(\frac{1}{4} +z_{5}\right)c \, \mathbf{\hat{z}} & \left(16f\right) & \mbox{O III} \\ 
\mathbf{B}_{37} & = & \left(-y_{5}-z_{5}\right) \, \mathbf{a}_{1} + \left(-x_{5}-z_{5}\right) \, \mathbf{a}_{2} + \left(-x_{5}-y_{5}\right) \, \mathbf{a}_{3} & = & -x_{5}a \, \mathbf{\hat{x}}-y_{5}a \, \mathbf{\hat{y}}-z_{5}c \, \mathbf{\hat{z}} & \left(16f\right) & \mbox{O III} \\ 
\mathbf{B}_{38} & = & \left(\frac{1}{2} +y_{5} - z_{5}\right) \, \mathbf{a}_{1} + \left(x_{5}-z_{5}\right) \, \mathbf{a}_{2} + \left(\frac{1}{2} +x_{5} + y_{5}\right) \, \mathbf{a}_{3} & = & x_{5}a \, \mathbf{\hat{x}} + \left(\frac{1}{2} +y_{5}\right)a \, \mathbf{\hat{y}}-z_{5}c \, \mathbf{\hat{z}} & \left(16f\right) & \mbox{O III} \\ 
\mathbf{B}_{39} & = & \left(\frac{1}{2} - x_{5} - z_{5}\right) \, \mathbf{a}_{1} + \left(y_{5}-z_{5}\right) \, \mathbf{a}_{2} + \left(-x_{5}+y_{5}\right) \, \mathbf{a}_{3} & = & \left(- \frac{1}{4} +y_{5}\right)a \, \mathbf{\hat{x}} + \left(\frac{1}{4} - x_{5}\right)a \, \mathbf{\hat{y}} + \left(\frac{1}{4} - z_{5}\right)c \, \mathbf{\hat{z}} & \left(16f\right) & \mbox{O III} \\ 
\mathbf{B}_{40} & = & \left(\frac{1}{2} +x_{5} - z_{5}\right) \, \mathbf{a}_{1} + \left(\frac{1}{2} - y_{5} - z_{5}\right) \, \mathbf{a}_{2} + \left(\frac{1}{2} +x_{5} - y_{5}\right) \, \mathbf{a}_{3} & = & \left(\frac{1}{4} - y_{5}\right)a \, \mathbf{\hat{x}} + \left(\frac{1}{4} +x_{5}\right)a \, \mathbf{\hat{y}} + \left(\frac{1}{4} - z_{5}\right)c \, \mathbf{\hat{z}} & \left(16f\right) & \mbox{O III} \\ 
\mathbf{B}_{41} & = & \left(y_{6}+z_{6}\right) \, \mathbf{a}_{1} + \left(x_{6}+z_{6}\right) \, \mathbf{a}_{2} + \left(x_{6}+y_{6}\right) \, \mathbf{a}_{3} & = & x_{6}a \, \mathbf{\hat{x}} + y_{6}a \, \mathbf{\hat{y}} + z_{6}c \, \mathbf{\hat{z}} & \left(16f\right) & \mbox{O IV} \\ 
\mathbf{B}_{42} & = & \left(\frac{1}{2} - y_{6} + z_{6}\right) \, \mathbf{a}_{1} + \left(-x_{6}+z_{6}\right) \, \mathbf{a}_{2} + \left(\frac{1}{2} - x_{6} - y_{6}\right) \, \mathbf{a}_{3} & = & -x_{6}a \, \mathbf{\hat{x}} + \left(\frac{1}{2} - y_{6}\right)a \, \mathbf{\hat{y}} + z_{6}c \, \mathbf{\hat{z}} & \left(16f\right) & \mbox{O IV} \\ 
\mathbf{B}_{43} & = & \left(\frac{1}{2} +x_{6} + z_{6}\right) \, \mathbf{a}_{1} + \left(-y_{6}+z_{6}\right) \, \mathbf{a}_{2} + \left(x_{6}-y_{6}\right) \, \mathbf{a}_{3} & = & \left(\frac{3}{4} - y_{6}\right)a \, \mathbf{\hat{x}} + \left(\frac{1}{4} +x_{6}\right)a \, \mathbf{\hat{y}} + \left(\frac{1}{4} +z_{6}\right)c \, \mathbf{\hat{z}} & \left(16f\right) & \mbox{O IV} \\ 
\mathbf{B}_{44} & = & \left(\frac{1}{2} - x_{6} + z_{6}\right) \, \mathbf{a}_{1} + \left(\frac{1}{2} +y_{6} + z_{6}\right) \, \mathbf{a}_{2} + \left(\frac{1}{2} - x_{6} + y_{6}\right) \, \mathbf{a}_{3} & = & \left(\frac{1}{4} +y_{6}\right)a \, \mathbf{\hat{x}} + \left(\frac{1}{4} - x_{6}\right)a \, \mathbf{\hat{y}} + \left(\frac{1}{4} +z_{6}\right)c \, \mathbf{\hat{z}} & \left(16f\right) & \mbox{O IV} \\ 
\mathbf{B}_{45} & = & \left(-y_{6}-z_{6}\right) \, \mathbf{a}_{1} + \left(-x_{6}-z_{6}\right) \, \mathbf{a}_{2} + \left(-x_{6}-y_{6}\right) \, \mathbf{a}_{3} & = & -x_{6}a \, \mathbf{\hat{x}}-y_{6}a \, \mathbf{\hat{y}}-z_{6}c \, \mathbf{\hat{z}} & \left(16f\right) & \mbox{O IV} \\ 
\mathbf{B}_{46} & = & \left(\frac{1}{2} +y_{6} - z_{6}\right) \, \mathbf{a}_{1} + \left(x_{6}-z_{6}\right) \, \mathbf{a}_{2} + \left(\frac{1}{2} +x_{6} + y_{6}\right) \, \mathbf{a}_{3} & = & x_{6}a \, \mathbf{\hat{x}} + \left(\frac{1}{2} +y_{6}\right)a \, \mathbf{\hat{y}}-z_{6}c \, \mathbf{\hat{z}} & \left(16f\right) & \mbox{O IV} \\ 
\mathbf{B}_{47} & = & \left(\frac{1}{2} - x_{6} - z_{6}\right) \, \mathbf{a}_{1} + \left(y_{6}-z_{6}\right) \, \mathbf{a}_{2} + \left(-x_{6}+y_{6}\right) \, \mathbf{a}_{3} & = & \left(- \frac{1}{4} +y_{6}\right)a \, \mathbf{\hat{x}} + \left(\frac{1}{4} - x_{6}\right)a \, \mathbf{\hat{y}} + \left(\frac{1}{4} - z_{6}\right)c \, \mathbf{\hat{z}} & \left(16f\right) & \mbox{O IV} \\ 
\mathbf{B}_{48} & = & \left(\frac{1}{2} +x_{6} - z_{6}\right) \, \mathbf{a}_{1} + \left(\frac{1}{2} - y_{6} - z_{6}\right) \, \mathbf{a}_{2} + \left(\frac{1}{2} +x_{6} - y_{6}\right) \, \mathbf{a}_{3} & = & \left(\frac{1}{4} - y_{6}\right)a \, \mathbf{\hat{x}} + \left(\frac{1}{4} +x_{6}\right)a \, \mathbf{\hat{y}} + \left(\frac{1}{4} - z_{6}\right)c \, \mathbf{\hat{z}} & \left(16f\right) & \mbox{O IV} \\ 
\end{longtabu}
\renewcommand{\arraystretch}{1.0}
\noindent \hrulefill
\\
\textbf{References:}
\vspace*{-0.25cm}
\begin{flushleft}
  - \bibentry{Pynn_PRB_13_1996}. \\
\end{flushleft}
\textbf{Found in:}
\vspace*{-0.25cm}
\begin{flushleft}
  - \bibentry{Bolzan_JSSC_113_1994}. \\
\end{flushleft}
\noindent \hrulefill
\\
\textbf{Geometry files:}
\\
\noindent  - CIF: pp. {\hyperref[AB2_tI96_88_2f_4f_cif]{\pageref{AB2_tI96_88_2f_4f_cif}}} \\
\noindent  - POSCAR: pp. {\hyperref[AB2_tI96_88_2f_4f_poscar]{\pageref{AB2_tI96_88_2f_4f_poscar}}} \\
\onecolumn
{\phantomsection\label{A17BC4D_tP184_89_17p_p_4p_io}}
\subsection*{\huge \textbf{{\normalfont C$_{17}$FeO$_{4}$Pt Structure: A17BC4D\_tP184\_89\_17p\_p\_4p\_io}}}
\noindent \hrulefill
\vspace*{0.25cm}
\begin{figure}[htp]
  \centering
  \vspace{-1em}
  {\includegraphics[width=1\textwidth]{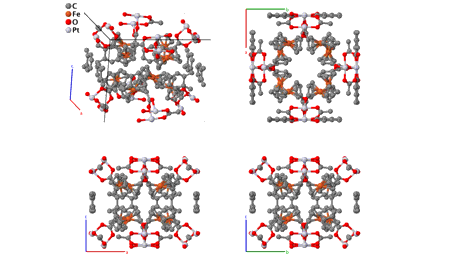}}
\end{figure}
\vspace*{-0.5cm}
\renewcommand{\arraystretch}{1.5}
\begin{equation*}
  \begin{array}{>{$\hspace{-0.15cm}}l<{$}>{$}p{0.5cm}<{$}>{$}p{18.5cm}<{$}}
    \mbox{\large \textbf{Prototype}} &\colon & \ce{C17FeO4Pt} \\
    \mbox{\large \textbf{\AFLOW\ prototype label}} &\colon & \mbox{A17BC4D\_tP184\_89\_17p\_p\_4p\_io} \\
    \mbox{\large \textbf{\textit{Strukturbericht} designation}} &\colon & \mbox{None} \\
    \mbox{\large \textbf{Pearson symbol}} &\colon & \mbox{tP184} \\
    \mbox{\large \textbf{Space group number}} &\colon & 89 \\
    \mbox{\large \textbf{Space group symbol}} &\colon & P422 \\
    \mbox{\large \textbf{\AFLOW\ prototype command}} &\colon &  \texttt{aflow} \,  \, \texttt{-{}-proto=A17BC4D\_tP184\_89\_17p\_p\_4p\_io } \, \newline \texttt{-{}-params=}{a,c/a,z_{1},x_{2},x_{3},y_{3},z_{3},x_{4},y_{4},z_{4},x_{5},y_{5},z_{5},x_{6},y_{6},z_{6},x_{7},y_{7},z_{7},x_{8},y_{8},} \newline {z_{8},x_{9},y_{9},z_{9},x_{10},y_{10},z_{10},x_{11},y_{11},z_{11},x_{12},y_{12},z_{12},x_{13},y_{13},z_{13},x_{14},y_{14},z_{14},x_{15},y_{15},z_{15},} \newline {x_{16},y_{16},z_{16},x_{17},y_{17},z_{17},x_{18},y_{18},z_{18},x_{19},y_{19},z_{19},x_{20},y_{20},z_{20},x_{21},y_{21},z_{21},x_{22},y_{22},} \newline {z_{22},x_{23},y_{23},z_{23},x_{24},y_{24},z_{24} }
  \end{array}
\end{equation*}
\renewcommand{\arraystretch}{1.0}

\vspace*{-0.25cm}
\noindent \hrulefill
\begin{itemize}
  \item{Structures exhibiting space group \#89 are quite rare.  
According to (Hoffmann, 2014), there are only two entries in the Inorganic Crystal Structure Database with space group \#89; however, they are incorrectly classified. 
This structure is listed in the Cambridge Structure Database (ID=863010).  Only the non-hydrogen atoms are listed.
}
\end{itemize}

\noindent \parbox{1 \linewidth}{
\noindent \hrulefill
\\
\textbf{Simple Tetragonal primitive vectors:} \\
\vspace*{-0.25cm}
\begin{tabular}{cc}
  \begin{tabular}{c}
    \parbox{0.6 \linewidth}{
      \renewcommand{\arraystretch}{1.5}
      \begin{equation*}
        \centering
        \begin{array}{ccc}
              \mathbf{a}_1 & = & a \, \mathbf{\hat{x}} \\
    \mathbf{a}_2 & = & a \, \mathbf{\hat{y}} \\
    \mathbf{a}_3 & = & c \, \mathbf{\hat{z}} \\

        \end{array}
      \end{equation*}
    }
    \renewcommand{\arraystretch}{1.0}
  \end{tabular}
  \begin{tabular}{c}
    \includegraphics[width=0.3\linewidth]{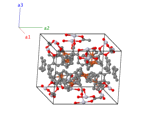} \\
  \end{tabular}
\end{tabular}

}
\vspace*{-0.25cm}

\noindent \hrulefill
\\
\textbf{Basis vectors:}
\vspace*{-0.25cm}
\renewcommand{\arraystretch}{1.5}
\begin{longtabu} to \textwidth{>{\centering $}X[-1,c,c]<{$}>{\centering $}X[-1,c,c]<{$}>{\centering $}X[-1,c,c]<{$}>{\centering $}X[-1,c,c]<{$}>{\centering $}X[-1,c,c]<{$}>{\centering $}X[-1,c,c]<{$}>{\centering $}X[-1,c,c]<{$}}
  & & \mbox{Lattice Coordinates} & & \mbox{Cartesian Coordinates} &\mbox{Wyckoff Position} & \mbox{Atom Type} \\  
  \mathbf{B}_{1} & = & \frac{1}{2} \, \mathbf{a}_{2} + z_{1} \, \mathbf{a}_{3} & = & \frac{1}{2}a \, \mathbf{\hat{y}} + z_{1}c \, \mathbf{\hat{z}} & \left(4i\right) & \mbox{Pt I} \\ 
\mathbf{B}_{2} & = & \frac{1}{2} \, \mathbf{a}_{1} + z_{1} \, \mathbf{a}_{3} & = & \frac{1}{2}a \, \mathbf{\hat{x}} + z_{1}c \, \mathbf{\hat{z}} & \left(4i\right) & \mbox{Pt I} \\ 
\mathbf{B}_{3} & = & \frac{1}{2} \, \mathbf{a}_{2}-z_{1} \, \mathbf{a}_{3} & = & \frac{1}{2}a \, \mathbf{\hat{y}}-z_{1}c \, \mathbf{\hat{z}} & \left(4i\right) & \mbox{Pt I} \\ 
\mathbf{B}_{4} & = & \frac{1}{2} \, \mathbf{a}_{1} + -z_{1} \, \mathbf{a}_{3} & = & \frac{1}{2}a \, \mathbf{\hat{x}} + -z_{1}c \, \mathbf{\hat{z}} & \left(4i\right) & \mbox{Pt I} \\ 
\mathbf{B}_{5} & = & x_{2} \, \mathbf{a}_{1} + \frac{1}{2} \, \mathbf{a}_{2} & = & x_{2}a \, \mathbf{\hat{x}} + \frac{1}{2}a \, \mathbf{\hat{y}} & \left(4o\right) & \mbox{Pt II} \\ 
\mathbf{B}_{6} & = & -x_{2} \, \mathbf{a}_{1} + \frac{1}{2} \, \mathbf{a}_{2} & = & -x_{2}a \, \mathbf{\hat{x}} + \frac{1}{2}a \, \mathbf{\hat{y}} & \left(4o\right) & \mbox{Pt II} \\ 
\mathbf{B}_{7} & = & \frac{1}{2} \, \mathbf{a}_{1} + x_{2} \, \mathbf{a}_{2} & = & \frac{1}{2}a \, \mathbf{\hat{x}} + x_{2}a \, \mathbf{\hat{y}} & \left(4o\right) & \mbox{Pt II} \\ 
\mathbf{B}_{8} & = & \frac{1}{2} \, \mathbf{a}_{1}-x_{2} \, \mathbf{a}_{2} & = & \frac{1}{2}a \, \mathbf{\hat{x}}-x_{2}a \, \mathbf{\hat{y}} & \left(4o\right) & \mbox{Pt II} \\ 
\mathbf{B}_{9} & = & x_{3} \, \mathbf{a}_{1} + y_{3} \, \mathbf{a}_{2} + z_{3} \, \mathbf{a}_{3} & = & x_{3}a \, \mathbf{\hat{x}} + y_{3}a \, \mathbf{\hat{y}} + z_{3}c \, \mathbf{\hat{z}} & \left(8p\right) & \mbox{C I} \\ 
\mathbf{B}_{10} & = & -x_{3} \, \mathbf{a}_{1}-y_{3} \, \mathbf{a}_{2} + z_{3} \, \mathbf{a}_{3} & = & -x_{3}a \, \mathbf{\hat{x}}-y_{3}a \, \mathbf{\hat{y}} + z_{3}c \, \mathbf{\hat{z}} & \left(8p\right) & \mbox{C I} \\ 
\mathbf{B}_{11} & = & -y_{3} \, \mathbf{a}_{1} + x_{3} \, \mathbf{a}_{2} + z_{3} \, \mathbf{a}_{3} & = & -y_{3}a \, \mathbf{\hat{x}} + x_{3}a \, \mathbf{\hat{y}} + z_{3}c \, \mathbf{\hat{z}} & \left(8p\right) & \mbox{C I} \\ 
\mathbf{B}_{12} & = & y_{3} \, \mathbf{a}_{1}-x_{3} \, \mathbf{a}_{2} + z_{3} \, \mathbf{a}_{3} & = & y_{3}a \, \mathbf{\hat{x}}-x_{3}a \, \mathbf{\hat{y}} + z_{3}c \, \mathbf{\hat{z}} & \left(8p\right) & \mbox{C I} \\ 
\mathbf{B}_{13} & = & -x_{3} \, \mathbf{a}_{1} + y_{3} \, \mathbf{a}_{2}-z_{3} \, \mathbf{a}_{3} & = & -x_{3}a \, \mathbf{\hat{x}} + y_{3}a \, \mathbf{\hat{y}}-z_{3}c \, \mathbf{\hat{z}} & \left(8p\right) & \mbox{C I} \\ 
\mathbf{B}_{14} & = & x_{3} \, \mathbf{a}_{1}-y_{3} \, \mathbf{a}_{2}-z_{3} \, \mathbf{a}_{3} & = & x_{3}a \, \mathbf{\hat{x}}-y_{3}a \, \mathbf{\hat{y}}-z_{3}c \, \mathbf{\hat{z}} & \left(8p\right) & \mbox{C I} \\ 
\mathbf{B}_{15} & = & y_{3} \, \mathbf{a}_{1} + x_{3} \, \mathbf{a}_{2}-z_{3} \, \mathbf{a}_{3} & = & y_{3}a \, \mathbf{\hat{x}} + x_{3}a \, \mathbf{\hat{y}}-z_{3}c \, \mathbf{\hat{z}} & \left(8p\right) & \mbox{C I} \\ 
\mathbf{B}_{16} & = & -y_{3} \, \mathbf{a}_{1}-x_{3} \, \mathbf{a}_{2}-z_{3} \, \mathbf{a}_{3} & = & -y_{3}a \, \mathbf{\hat{x}}-x_{3}a \, \mathbf{\hat{y}}-z_{3}c \, \mathbf{\hat{z}} & \left(8p\right) & \mbox{C I} \\ 
\mathbf{B}_{17} & = & x_{4} \, \mathbf{a}_{1} + y_{4} \, \mathbf{a}_{2} + z_{4} \, \mathbf{a}_{3} & = & x_{4}a \, \mathbf{\hat{x}} + y_{4}a \, \mathbf{\hat{y}} + z_{4}c \, \mathbf{\hat{z}} & \left(8p\right) & \mbox{C II} \\ 
\mathbf{B}_{18} & = & -x_{4} \, \mathbf{a}_{1}-y_{4} \, \mathbf{a}_{2} + z_{4} \, \mathbf{a}_{3} & = & -x_{4}a \, \mathbf{\hat{x}}-y_{4}a \, \mathbf{\hat{y}} + z_{4}c \, \mathbf{\hat{z}} & \left(8p\right) & \mbox{C II} \\ 
\mathbf{B}_{19} & = & -y_{4} \, \mathbf{a}_{1} + x_{4} \, \mathbf{a}_{2} + z_{4} \, \mathbf{a}_{3} & = & -y_{4}a \, \mathbf{\hat{x}} + x_{4}a \, \mathbf{\hat{y}} + z_{4}c \, \mathbf{\hat{z}} & \left(8p\right) & \mbox{C II} \\ 
\mathbf{B}_{20} & = & y_{4} \, \mathbf{a}_{1}-x_{4} \, \mathbf{a}_{2} + z_{4} \, \mathbf{a}_{3} & = & y_{4}a \, \mathbf{\hat{x}}-x_{4}a \, \mathbf{\hat{y}} + z_{4}c \, \mathbf{\hat{z}} & \left(8p\right) & \mbox{C II} \\ 
\mathbf{B}_{21} & = & -x_{4} \, \mathbf{a}_{1} + y_{4} \, \mathbf{a}_{2}-z_{4} \, \mathbf{a}_{3} & = & -x_{4}a \, \mathbf{\hat{x}} + y_{4}a \, \mathbf{\hat{y}}-z_{4}c \, \mathbf{\hat{z}} & \left(8p\right) & \mbox{C II} \\ 
\mathbf{B}_{22} & = & x_{4} \, \mathbf{a}_{1}-y_{4} \, \mathbf{a}_{2}-z_{4} \, \mathbf{a}_{3} & = & x_{4}a \, \mathbf{\hat{x}}-y_{4}a \, \mathbf{\hat{y}}-z_{4}c \, \mathbf{\hat{z}} & \left(8p\right) & \mbox{C II} \\ 
\mathbf{B}_{23} & = & y_{4} \, \mathbf{a}_{1} + x_{4} \, \mathbf{a}_{2}-z_{4} \, \mathbf{a}_{3} & = & y_{4}a \, \mathbf{\hat{x}} + x_{4}a \, \mathbf{\hat{y}}-z_{4}c \, \mathbf{\hat{z}} & \left(8p\right) & \mbox{C II} \\ 
\mathbf{B}_{24} & = & -y_{4} \, \mathbf{a}_{1}-x_{4} \, \mathbf{a}_{2}-z_{4} \, \mathbf{a}_{3} & = & -y_{4}a \, \mathbf{\hat{x}}-x_{4}a \, \mathbf{\hat{y}}-z_{4}c \, \mathbf{\hat{z}} & \left(8p\right) & \mbox{C II} \\ 
\mathbf{B}_{25} & = & x_{5} \, \mathbf{a}_{1} + y_{5} \, \mathbf{a}_{2} + z_{5} \, \mathbf{a}_{3} & = & x_{5}a \, \mathbf{\hat{x}} + y_{5}a \, \mathbf{\hat{y}} + z_{5}c \, \mathbf{\hat{z}} & \left(8p\right) & \mbox{C III} \\ 
\mathbf{B}_{26} & = & -x_{5} \, \mathbf{a}_{1}-y_{5} \, \mathbf{a}_{2} + z_{5} \, \mathbf{a}_{3} & = & -x_{5}a \, \mathbf{\hat{x}}-y_{5}a \, \mathbf{\hat{y}} + z_{5}c \, \mathbf{\hat{z}} & \left(8p\right) & \mbox{C III} \\ 
\mathbf{B}_{27} & = & -y_{5} \, \mathbf{a}_{1} + x_{5} \, \mathbf{a}_{2} + z_{5} \, \mathbf{a}_{3} & = & -y_{5}a \, \mathbf{\hat{x}} + x_{5}a \, \mathbf{\hat{y}} + z_{5}c \, \mathbf{\hat{z}} & \left(8p\right) & \mbox{C III} \\ 
\mathbf{B}_{28} & = & y_{5} \, \mathbf{a}_{1}-x_{5} \, \mathbf{a}_{2} + z_{5} \, \mathbf{a}_{3} & = & y_{5}a \, \mathbf{\hat{x}}-x_{5}a \, \mathbf{\hat{y}} + z_{5}c \, \mathbf{\hat{z}} & \left(8p\right) & \mbox{C III} \\ 
\mathbf{B}_{29} & = & -x_{5} \, \mathbf{a}_{1} + y_{5} \, \mathbf{a}_{2}-z_{5} \, \mathbf{a}_{3} & = & -x_{5}a \, \mathbf{\hat{x}} + y_{5}a \, \mathbf{\hat{y}}-z_{5}c \, \mathbf{\hat{z}} & \left(8p\right) & \mbox{C III} \\ 
\mathbf{B}_{30} & = & x_{5} \, \mathbf{a}_{1}-y_{5} \, \mathbf{a}_{2}-z_{5} \, \mathbf{a}_{3} & = & x_{5}a \, \mathbf{\hat{x}}-y_{5}a \, \mathbf{\hat{y}}-z_{5}c \, \mathbf{\hat{z}} & \left(8p\right) & \mbox{C III} \\ 
\mathbf{B}_{31} & = & y_{5} \, \mathbf{a}_{1} + x_{5} \, \mathbf{a}_{2}-z_{5} \, \mathbf{a}_{3} & = & y_{5}a \, \mathbf{\hat{x}} + x_{5}a \, \mathbf{\hat{y}}-z_{5}c \, \mathbf{\hat{z}} & \left(8p\right) & \mbox{C III} \\ 
\mathbf{B}_{32} & = & -y_{5} \, \mathbf{a}_{1}-x_{5} \, \mathbf{a}_{2}-z_{5} \, \mathbf{a}_{3} & = & -y_{5}a \, \mathbf{\hat{x}}-x_{5}a \, \mathbf{\hat{y}}-z_{5}c \, \mathbf{\hat{z}} & \left(8p\right) & \mbox{C III} \\ 
\mathbf{B}_{33} & = & x_{6} \, \mathbf{a}_{1} + y_{6} \, \mathbf{a}_{2} + z_{6} \, \mathbf{a}_{3} & = & x_{6}a \, \mathbf{\hat{x}} + y_{6}a \, \mathbf{\hat{y}} + z_{6}c \, \mathbf{\hat{z}} & \left(8p\right) & \mbox{C IV} \\ 
\mathbf{B}_{34} & = & -x_{6} \, \mathbf{a}_{1}-y_{6} \, \mathbf{a}_{2} + z_{6} \, \mathbf{a}_{3} & = & -x_{6}a \, \mathbf{\hat{x}}-y_{6}a \, \mathbf{\hat{y}} + z_{6}c \, \mathbf{\hat{z}} & \left(8p\right) & \mbox{C IV} \\ 
\mathbf{B}_{35} & = & -y_{6} \, \mathbf{a}_{1} + x_{6} \, \mathbf{a}_{2} + z_{6} \, \mathbf{a}_{3} & = & -y_{6}a \, \mathbf{\hat{x}} + x_{6}a \, \mathbf{\hat{y}} + z_{6}c \, \mathbf{\hat{z}} & \left(8p\right) & \mbox{C IV} \\ 
\mathbf{B}_{36} & = & y_{6} \, \mathbf{a}_{1}-x_{6} \, \mathbf{a}_{2} + z_{6} \, \mathbf{a}_{3} & = & y_{6}a \, \mathbf{\hat{x}}-x_{6}a \, \mathbf{\hat{y}} + z_{6}c \, \mathbf{\hat{z}} & \left(8p\right) & \mbox{C IV} \\ 
\mathbf{B}_{37} & = & -x_{6} \, \mathbf{a}_{1} + y_{6} \, \mathbf{a}_{2}-z_{6} \, \mathbf{a}_{3} & = & -x_{6}a \, \mathbf{\hat{x}} + y_{6}a \, \mathbf{\hat{y}}-z_{6}c \, \mathbf{\hat{z}} & \left(8p\right) & \mbox{C IV} \\ 
\mathbf{B}_{38} & = & x_{6} \, \mathbf{a}_{1}-y_{6} \, \mathbf{a}_{2}-z_{6} \, \mathbf{a}_{3} & = & x_{6}a \, \mathbf{\hat{x}}-y_{6}a \, \mathbf{\hat{y}}-z_{6}c \, \mathbf{\hat{z}} & \left(8p\right) & \mbox{C IV} \\ 
\mathbf{B}_{39} & = & y_{6} \, \mathbf{a}_{1} + x_{6} \, \mathbf{a}_{2}-z_{6} \, \mathbf{a}_{3} & = & y_{6}a \, \mathbf{\hat{x}} + x_{6}a \, \mathbf{\hat{y}}-z_{6}c \, \mathbf{\hat{z}} & \left(8p\right) & \mbox{C IV} \\ 
\mathbf{B}_{40} & = & -y_{6} \, \mathbf{a}_{1}-x_{6} \, \mathbf{a}_{2}-z_{6} \, \mathbf{a}_{3} & = & -y_{6}a \, \mathbf{\hat{x}}-x_{6}a \, \mathbf{\hat{y}}-z_{6}c \, \mathbf{\hat{z}} & \left(8p\right) & \mbox{C IV} \\ 
\mathbf{B}_{41} & = & x_{7} \, \mathbf{a}_{1} + y_{7} \, \mathbf{a}_{2} + z_{7} \, \mathbf{a}_{3} & = & x_{7}a \, \mathbf{\hat{x}} + y_{7}a \, \mathbf{\hat{y}} + z_{7}c \, \mathbf{\hat{z}} & \left(8p\right) & \mbox{C V} \\ 
\mathbf{B}_{42} & = & -x_{7} \, \mathbf{a}_{1}-y_{7} \, \mathbf{a}_{2} + z_{7} \, \mathbf{a}_{3} & = & -x_{7}a \, \mathbf{\hat{x}}-y_{7}a \, \mathbf{\hat{y}} + z_{7}c \, \mathbf{\hat{z}} & \left(8p\right) & \mbox{C V} \\ 
\mathbf{B}_{43} & = & -y_{7} \, \mathbf{a}_{1} + x_{7} \, \mathbf{a}_{2} + z_{7} \, \mathbf{a}_{3} & = & -y_{7}a \, \mathbf{\hat{x}} + x_{7}a \, \mathbf{\hat{y}} + z_{7}c \, \mathbf{\hat{z}} & \left(8p\right) & \mbox{C V} \\ 
\mathbf{B}_{44} & = & y_{7} \, \mathbf{a}_{1}-x_{7} \, \mathbf{a}_{2} + z_{7} \, \mathbf{a}_{3} & = & y_{7}a \, \mathbf{\hat{x}}-x_{7}a \, \mathbf{\hat{y}} + z_{7}c \, \mathbf{\hat{z}} & \left(8p\right) & \mbox{C V} \\ 
\mathbf{B}_{45} & = & -x_{7} \, \mathbf{a}_{1} + y_{7} \, \mathbf{a}_{2}-z_{7} \, \mathbf{a}_{3} & = & -x_{7}a \, \mathbf{\hat{x}} + y_{7}a \, \mathbf{\hat{y}}-z_{7}c \, \mathbf{\hat{z}} & \left(8p\right) & \mbox{C V} \\ 
\mathbf{B}_{46} & = & x_{7} \, \mathbf{a}_{1}-y_{7} \, \mathbf{a}_{2}-z_{7} \, \mathbf{a}_{3} & = & x_{7}a \, \mathbf{\hat{x}}-y_{7}a \, \mathbf{\hat{y}}-z_{7}c \, \mathbf{\hat{z}} & \left(8p\right) & \mbox{C V} \\ 
\mathbf{B}_{47} & = & y_{7} \, \mathbf{a}_{1} + x_{7} \, \mathbf{a}_{2}-z_{7} \, \mathbf{a}_{3} & = & y_{7}a \, \mathbf{\hat{x}} + x_{7}a \, \mathbf{\hat{y}}-z_{7}c \, \mathbf{\hat{z}} & \left(8p\right) & \mbox{C V} \\ 
\mathbf{B}_{48} & = & -y_{7} \, \mathbf{a}_{1}-x_{7} \, \mathbf{a}_{2}-z_{7} \, \mathbf{a}_{3} & = & -y_{7}a \, \mathbf{\hat{x}}-x_{7}a \, \mathbf{\hat{y}}-z_{7}c \, \mathbf{\hat{z}} & \left(8p\right) & \mbox{C V} \\ 
\mathbf{B}_{49} & = & x_{8} \, \mathbf{a}_{1} + y_{8} \, \mathbf{a}_{2} + z_{8} \, \mathbf{a}_{3} & = & x_{8}a \, \mathbf{\hat{x}} + y_{8}a \, \mathbf{\hat{y}} + z_{8}c \, \mathbf{\hat{z}} & \left(8p\right) & \mbox{C VI} \\ 
\mathbf{B}_{50} & = & -x_{8} \, \mathbf{a}_{1}-y_{8} \, \mathbf{a}_{2} + z_{8} \, \mathbf{a}_{3} & = & -x_{8}a \, \mathbf{\hat{x}}-y_{8}a \, \mathbf{\hat{y}} + z_{8}c \, \mathbf{\hat{z}} & \left(8p\right) & \mbox{C VI} \\ 
\mathbf{B}_{51} & = & -y_{8} \, \mathbf{a}_{1} + x_{8} \, \mathbf{a}_{2} + z_{8} \, \mathbf{a}_{3} & = & -y_{8}a \, \mathbf{\hat{x}} + x_{8}a \, \mathbf{\hat{y}} + z_{8}c \, \mathbf{\hat{z}} & \left(8p\right) & \mbox{C VI} \\ 
\mathbf{B}_{52} & = & y_{8} \, \mathbf{a}_{1}-x_{8} \, \mathbf{a}_{2} + z_{8} \, \mathbf{a}_{3} & = & y_{8}a \, \mathbf{\hat{x}}-x_{8}a \, \mathbf{\hat{y}} + z_{8}c \, \mathbf{\hat{z}} & \left(8p\right) & \mbox{C VI} \\ 
\mathbf{B}_{53} & = & -x_{8} \, \mathbf{a}_{1} + y_{8} \, \mathbf{a}_{2}-z_{8} \, \mathbf{a}_{3} & = & -x_{8}a \, \mathbf{\hat{x}} + y_{8}a \, \mathbf{\hat{y}}-z_{8}c \, \mathbf{\hat{z}} & \left(8p\right) & \mbox{C VI} \\ 
\mathbf{B}_{54} & = & x_{8} \, \mathbf{a}_{1}-y_{8} \, \mathbf{a}_{2}-z_{8} \, \mathbf{a}_{3} & = & x_{8}a \, \mathbf{\hat{x}}-y_{8}a \, \mathbf{\hat{y}}-z_{8}c \, \mathbf{\hat{z}} & \left(8p\right) & \mbox{C VI} \\ 
\mathbf{B}_{55} & = & y_{8} \, \mathbf{a}_{1} + x_{8} \, \mathbf{a}_{2}-z_{8} \, \mathbf{a}_{3} & = & y_{8}a \, \mathbf{\hat{x}} + x_{8}a \, \mathbf{\hat{y}}-z_{8}c \, \mathbf{\hat{z}} & \left(8p\right) & \mbox{C VI} \\ 
\mathbf{B}_{56} & = & -y_{8} \, \mathbf{a}_{1}-x_{8} \, \mathbf{a}_{2}-z_{8} \, \mathbf{a}_{3} & = & -y_{8}a \, \mathbf{\hat{x}}-x_{8}a \, \mathbf{\hat{y}}-z_{8}c \, \mathbf{\hat{z}} & \left(8p\right) & \mbox{C VI} \\ 
\mathbf{B}_{57} & = & x_{9} \, \mathbf{a}_{1} + y_{9} \, \mathbf{a}_{2} + z_{9} \, \mathbf{a}_{3} & = & x_{9}a \, \mathbf{\hat{x}} + y_{9}a \, \mathbf{\hat{y}} + z_{9}c \, \mathbf{\hat{z}} & \left(8p\right) & \mbox{C VII} \\ 
\mathbf{B}_{58} & = & -x_{9} \, \mathbf{a}_{1}-y_{9} \, \mathbf{a}_{2} + z_{9} \, \mathbf{a}_{3} & = & -x_{9}a \, \mathbf{\hat{x}}-y_{9}a \, \mathbf{\hat{y}} + z_{9}c \, \mathbf{\hat{z}} & \left(8p\right) & \mbox{C VII} \\ 
\mathbf{B}_{59} & = & -y_{9} \, \mathbf{a}_{1} + x_{9} \, \mathbf{a}_{2} + z_{9} \, \mathbf{a}_{3} & = & -y_{9}a \, \mathbf{\hat{x}} + x_{9}a \, \mathbf{\hat{y}} + z_{9}c \, \mathbf{\hat{z}} & \left(8p\right) & \mbox{C VII} \\ 
\mathbf{B}_{60} & = & y_{9} \, \mathbf{a}_{1}-x_{9} \, \mathbf{a}_{2} + z_{9} \, \mathbf{a}_{3} & = & y_{9}a \, \mathbf{\hat{x}}-x_{9}a \, \mathbf{\hat{y}} + z_{9}c \, \mathbf{\hat{z}} & \left(8p\right) & \mbox{C VII} \\ 
\mathbf{B}_{61} & = & -x_{9} \, \mathbf{a}_{1} + y_{9} \, \mathbf{a}_{2}-z_{9} \, \mathbf{a}_{3} & = & -x_{9}a \, \mathbf{\hat{x}} + y_{9}a \, \mathbf{\hat{y}}-z_{9}c \, \mathbf{\hat{z}} & \left(8p\right) & \mbox{C VII} \\ 
\mathbf{B}_{62} & = & x_{9} \, \mathbf{a}_{1}-y_{9} \, \mathbf{a}_{2}-z_{9} \, \mathbf{a}_{3} & = & x_{9}a \, \mathbf{\hat{x}}-y_{9}a \, \mathbf{\hat{y}}-z_{9}c \, \mathbf{\hat{z}} & \left(8p\right) & \mbox{C VII} \\ 
\mathbf{B}_{63} & = & y_{9} \, \mathbf{a}_{1} + x_{9} \, \mathbf{a}_{2}-z_{9} \, \mathbf{a}_{3} & = & y_{9}a \, \mathbf{\hat{x}} + x_{9}a \, \mathbf{\hat{y}}-z_{9}c \, \mathbf{\hat{z}} & \left(8p\right) & \mbox{C VII} \\ 
\mathbf{B}_{64} & = & -y_{9} \, \mathbf{a}_{1}-x_{9} \, \mathbf{a}_{2}-z_{9} \, \mathbf{a}_{3} & = & -y_{9}a \, \mathbf{\hat{x}}-x_{9}a \, \mathbf{\hat{y}}-z_{9}c \, \mathbf{\hat{z}} & \left(8p\right) & \mbox{C VII} \\ 
\mathbf{B}_{65} & = & x_{10} \, \mathbf{a}_{1} + y_{10} \, \mathbf{a}_{2} + z_{10} \, \mathbf{a}_{3} & = & x_{10}a \, \mathbf{\hat{x}} + y_{10}a \, \mathbf{\hat{y}} + z_{10}c \, \mathbf{\hat{z}} & \left(8p\right) & \mbox{C VIII} \\ 
\mathbf{B}_{66} & = & -x_{10} \, \mathbf{a}_{1}-y_{10} \, \mathbf{a}_{2} + z_{10} \, \mathbf{a}_{3} & = & -x_{10}a \, \mathbf{\hat{x}}-y_{10}a \, \mathbf{\hat{y}} + z_{10}c \, \mathbf{\hat{z}} & \left(8p\right) & \mbox{C VIII} \\ 
\mathbf{B}_{67} & = & -y_{10} \, \mathbf{a}_{1} + x_{10} \, \mathbf{a}_{2} + z_{10} \, \mathbf{a}_{3} & = & -y_{10}a \, \mathbf{\hat{x}} + x_{10}a \, \mathbf{\hat{y}} + z_{10}c \, \mathbf{\hat{z}} & \left(8p\right) & \mbox{C VIII} \\ 
\mathbf{B}_{68} & = & y_{10} \, \mathbf{a}_{1}-x_{10} \, \mathbf{a}_{2} + z_{10} \, \mathbf{a}_{3} & = & y_{10}a \, \mathbf{\hat{x}}-x_{10}a \, \mathbf{\hat{y}} + z_{10}c \, \mathbf{\hat{z}} & \left(8p\right) & \mbox{C VIII} \\ 
\mathbf{B}_{69} & = & -x_{10} \, \mathbf{a}_{1} + y_{10} \, \mathbf{a}_{2}-z_{10} \, \mathbf{a}_{3} & = & -x_{10}a \, \mathbf{\hat{x}} + y_{10}a \, \mathbf{\hat{y}}-z_{10}c \, \mathbf{\hat{z}} & \left(8p\right) & \mbox{C VIII} \\ 
\mathbf{B}_{70} & = & x_{10} \, \mathbf{a}_{1}-y_{10} \, \mathbf{a}_{2}-z_{10} \, \mathbf{a}_{3} & = & x_{10}a \, \mathbf{\hat{x}}-y_{10}a \, \mathbf{\hat{y}}-z_{10}c \, \mathbf{\hat{z}} & \left(8p\right) & \mbox{C VIII} \\ 
\mathbf{B}_{71} & = & y_{10} \, \mathbf{a}_{1} + x_{10} \, \mathbf{a}_{2}-z_{10} \, \mathbf{a}_{3} & = & y_{10}a \, \mathbf{\hat{x}} + x_{10}a \, \mathbf{\hat{y}}-z_{10}c \, \mathbf{\hat{z}} & \left(8p\right) & \mbox{C VIII} \\ 
\mathbf{B}_{72} & = & -y_{10} \, \mathbf{a}_{1}-x_{10} \, \mathbf{a}_{2}-z_{10} \, \mathbf{a}_{3} & = & -y_{10}a \, \mathbf{\hat{x}}-x_{10}a \, \mathbf{\hat{y}}-z_{10}c \, \mathbf{\hat{z}} & \left(8p\right) & \mbox{C VIII} \\ 
\mathbf{B}_{73} & = & x_{11} \, \mathbf{a}_{1} + y_{11} \, \mathbf{a}_{2} + z_{11} \, \mathbf{a}_{3} & = & x_{11}a \, \mathbf{\hat{x}} + y_{11}a \, \mathbf{\hat{y}} + z_{11}c \, \mathbf{\hat{z}} & \left(8p\right) & \mbox{C IX} \\ 
\mathbf{B}_{74} & = & -x_{11} \, \mathbf{a}_{1}-y_{11} \, \mathbf{a}_{2} + z_{11} \, \mathbf{a}_{3} & = & -x_{11}a \, \mathbf{\hat{x}}-y_{11}a \, \mathbf{\hat{y}} + z_{11}c \, \mathbf{\hat{z}} & \left(8p\right) & \mbox{C IX} \\ 
\mathbf{B}_{75} & = & -y_{11} \, \mathbf{a}_{1} + x_{11} \, \mathbf{a}_{2} + z_{11} \, \mathbf{a}_{3} & = & -y_{11}a \, \mathbf{\hat{x}} + x_{11}a \, \mathbf{\hat{y}} + z_{11}c \, \mathbf{\hat{z}} & \left(8p\right) & \mbox{C IX} \\ 
\mathbf{B}_{76} & = & y_{11} \, \mathbf{a}_{1}-x_{11} \, \mathbf{a}_{2} + z_{11} \, \mathbf{a}_{3} & = & y_{11}a \, \mathbf{\hat{x}}-x_{11}a \, \mathbf{\hat{y}} + z_{11}c \, \mathbf{\hat{z}} & \left(8p\right) & \mbox{C IX} \\ 
\mathbf{B}_{77} & = & -x_{11} \, \mathbf{a}_{1} + y_{11} \, \mathbf{a}_{2}-z_{11} \, \mathbf{a}_{3} & = & -x_{11}a \, \mathbf{\hat{x}} + y_{11}a \, \mathbf{\hat{y}}-z_{11}c \, \mathbf{\hat{z}} & \left(8p\right) & \mbox{C IX} \\ 
\mathbf{B}_{78} & = & x_{11} \, \mathbf{a}_{1}-y_{11} \, \mathbf{a}_{2}-z_{11} \, \mathbf{a}_{3} & = & x_{11}a \, \mathbf{\hat{x}}-y_{11}a \, \mathbf{\hat{y}}-z_{11}c \, \mathbf{\hat{z}} & \left(8p\right) & \mbox{C IX} \\ 
\mathbf{B}_{79} & = & y_{11} \, \mathbf{a}_{1} + x_{11} \, \mathbf{a}_{2}-z_{11} \, \mathbf{a}_{3} & = & y_{11}a \, \mathbf{\hat{x}} + x_{11}a \, \mathbf{\hat{y}}-z_{11}c \, \mathbf{\hat{z}} & \left(8p\right) & \mbox{C IX} \\ 
\mathbf{B}_{80} & = & -y_{11} \, \mathbf{a}_{1}-x_{11} \, \mathbf{a}_{2}-z_{11} \, \mathbf{a}_{3} & = & -y_{11}a \, \mathbf{\hat{x}}-x_{11}a \, \mathbf{\hat{y}}-z_{11}c \, \mathbf{\hat{z}} & \left(8p\right) & \mbox{C IX} \\ 
\mathbf{B}_{81} & = & x_{12} \, \mathbf{a}_{1} + y_{12} \, \mathbf{a}_{2} + z_{12} \, \mathbf{a}_{3} & = & x_{12}a \, \mathbf{\hat{x}} + y_{12}a \, \mathbf{\hat{y}} + z_{12}c \, \mathbf{\hat{z}} & \left(8p\right) & \mbox{C X} \\ 
\mathbf{B}_{82} & = & -x_{12} \, \mathbf{a}_{1}-y_{12} \, \mathbf{a}_{2} + z_{12} \, \mathbf{a}_{3} & = & -x_{12}a \, \mathbf{\hat{x}}-y_{12}a \, \mathbf{\hat{y}} + z_{12}c \, \mathbf{\hat{z}} & \left(8p\right) & \mbox{C X} \\ 
\mathbf{B}_{83} & = & -y_{12} \, \mathbf{a}_{1} + x_{12} \, \mathbf{a}_{2} + z_{12} \, \mathbf{a}_{3} & = & -y_{12}a \, \mathbf{\hat{x}} + x_{12}a \, \mathbf{\hat{y}} + z_{12}c \, \mathbf{\hat{z}} & \left(8p\right) & \mbox{C X} \\ 
\mathbf{B}_{84} & = & y_{12} \, \mathbf{a}_{1}-x_{12} \, \mathbf{a}_{2} + z_{12} \, \mathbf{a}_{3} & = & y_{12}a \, \mathbf{\hat{x}}-x_{12}a \, \mathbf{\hat{y}} + z_{12}c \, \mathbf{\hat{z}} & \left(8p\right) & \mbox{C X} \\ 
\mathbf{B}_{85} & = & -x_{12} \, \mathbf{a}_{1} + y_{12} \, \mathbf{a}_{2}-z_{12} \, \mathbf{a}_{3} & = & -x_{12}a \, \mathbf{\hat{x}} + y_{12}a \, \mathbf{\hat{y}}-z_{12}c \, \mathbf{\hat{z}} & \left(8p\right) & \mbox{C X} \\ 
\mathbf{B}_{86} & = & x_{12} \, \mathbf{a}_{1}-y_{12} \, \mathbf{a}_{2}-z_{12} \, \mathbf{a}_{3} & = & x_{12}a \, \mathbf{\hat{x}}-y_{12}a \, \mathbf{\hat{y}}-z_{12}c \, \mathbf{\hat{z}} & \left(8p\right) & \mbox{C X} \\ 
\mathbf{B}_{87} & = & y_{12} \, \mathbf{a}_{1} + x_{12} \, \mathbf{a}_{2}-z_{12} \, \mathbf{a}_{3} & = & y_{12}a \, \mathbf{\hat{x}} + x_{12}a \, \mathbf{\hat{y}}-z_{12}c \, \mathbf{\hat{z}} & \left(8p\right) & \mbox{C X} \\ 
\mathbf{B}_{88} & = & -y_{12} \, \mathbf{a}_{1}-x_{12} \, \mathbf{a}_{2}-z_{12} \, \mathbf{a}_{3} & = & -y_{12}a \, \mathbf{\hat{x}}-x_{12}a \, \mathbf{\hat{y}}-z_{12}c \, \mathbf{\hat{z}} & \left(8p\right) & \mbox{C X} \\ 
\mathbf{B}_{89} & = & x_{13} \, \mathbf{a}_{1} + y_{13} \, \mathbf{a}_{2} + z_{13} \, \mathbf{a}_{3} & = & x_{13}a \, \mathbf{\hat{x}} + y_{13}a \, \mathbf{\hat{y}} + z_{13}c \, \mathbf{\hat{z}} & \left(8p\right) & \mbox{C XI} \\ 
\mathbf{B}_{90} & = & -x_{13} \, \mathbf{a}_{1}-y_{13} \, \mathbf{a}_{2} + z_{13} \, \mathbf{a}_{3} & = & -x_{13}a \, \mathbf{\hat{x}}-y_{13}a \, \mathbf{\hat{y}} + z_{13}c \, \mathbf{\hat{z}} & \left(8p\right) & \mbox{C XI} \\ 
\mathbf{B}_{91} & = & -y_{13} \, \mathbf{a}_{1} + x_{13} \, \mathbf{a}_{2} + z_{13} \, \mathbf{a}_{3} & = & -y_{13}a \, \mathbf{\hat{x}} + x_{13}a \, \mathbf{\hat{y}} + z_{13}c \, \mathbf{\hat{z}} & \left(8p\right) & \mbox{C XI} \\ 
\mathbf{B}_{92} & = & y_{13} \, \mathbf{a}_{1}-x_{13} \, \mathbf{a}_{2} + z_{13} \, \mathbf{a}_{3} & = & y_{13}a \, \mathbf{\hat{x}}-x_{13}a \, \mathbf{\hat{y}} + z_{13}c \, \mathbf{\hat{z}} & \left(8p\right) & \mbox{C XI} \\ 
\mathbf{B}_{93} & = & -x_{13} \, \mathbf{a}_{1} + y_{13} \, \mathbf{a}_{2}-z_{13} \, \mathbf{a}_{3} & = & -x_{13}a \, \mathbf{\hat{x}} + y_{13}a \, \mathbf{\hat{y}}-z_{13}c \, \mathbf{\hat{z}} & \left(8p\right) & \mbox{C XI} \\ 
\mathbf{B}_{94} & = & x_{13} \, \mathbf{a}_{1}-y_{13} \, \mathbf{a}_{2}-z_{13} \, \mathbf{a}_{3} & = & x_{13}a \, \mathbf{\hat{x}}-y_{13}a \, \mathbf{\hat{y}}-z_{13}c \, \mathbf{\hat{z}} & \left(8p\right) & \mbox{C XI} \\ 
\mathbf{B}_{95} & = & y_{13} \, \mathbf{a}_{1} + x_{13} \, \mathbf{a}_{2}-z_{13} \, \mathbf{a}_{3} & = & y_{13}a \, \mathbf{\hat{x}} + x_{13}a \, \mathbf{\hat{y}}-z_{13}c \, \mathbf{\hat{z}} & \left(8p\right) & \mbox{C XI} \\ 
\mathbf{B}_{96} & = & -y_{13} \, \mathbf{a}_{1}-x_{13} \, \mathbf{a}_{2}-z_{13} \, \mathbf{a}_{3} & = & -y_{13}a \, \mathbf{\hat{x}}-x_{13}a \, \mathbf{\hat{y}}-z_{13}c \, \mathbf{\hat{z}} & \left(8p\right) & \mbox{C XI} \\ 
\mathbf{B}_{97} & = & x_{14} \, \mathbf{a}_{1} + y_{14} \, \mathbf{a}_{2} + z_{14} \, \mathbf{a}_{3} & = & x_{14}a \, \mathbf{\hat{x}} + y_{14}a \, \mathbf{\hat{y}} + z_{14}c \, \mathbf{\hat{z}} & \left(8p\right) & \mbox{C XII} \\ 
\mathbf{B}_{98} & = & -x_{14} \, \mathbf{a}_{1}-y_{14} \, \mathbf{a}_{2} + z_{14} \, \mathbf{a}_{3} & = & -x_{14}a \, \mathbf{\hat{x}}-y_{14}a \, \mathbf{\hat{y}} + z_{14}c \, \mathbf{\hat{z}} & \left(8p\right) & \mbox{C XII} \\ 
\mathbf{B}_{99} & = & -y_{14} \, \mathbf{a}_{1} + x_{14} \, \mathbf{a}_{2} + z_{14} \, \mathbf{a}_{3} & = & -y_{14}a \, \mathbf{\hat{x}} + x_{14}a \, \mathbf{\hat{y}} + z_{14}c \, \mathbf{\hat{z}} & \left(8p\right) & \mbox{C XII} \\ 
\mathbf{B}_{100} & = & y_{14} \, \mathbf{a}_{1}-x_{14} \, \mathbf{a}_{2} + z_{14} \, \mathbf{a}_{3} & = & y_{14}a \, \mathbf{\hat{x}}-x_{14}a \, \mathbf{\hat{y}} + z_{14}c \, \mathbf{\hat{z}} & \left(8p\right) & \mbox{C XII} \\ 
\mathbf{B}_{101} & = & -x_{14} \, \mathbf{a}_{1} + y_{14} \, \mathbf{a}_{2}-z_{14} \, \mathbf{a}_{3} & = & -x_{14}a \, \mathbf{\hat{x}} + y_{14}a \, \mathbf{\hat{y}}-z_{14}c \, \mathbf{\hat{z}} & \left(8p\right) & \mbox{C XII} \\ 
\mathbf{B}_{102} & = & x_{14} \, \mathbf{a}_{1}-y_{14} \, \mathbf{a}_{2}-z_{14} \, \mathbf{a}_{3} & = & x_{14}a \, \mathbf{\hat{x}}-y_{14}a \, \mathbf{\hat{y}}-z_{14}c \, \mathbf{\hat{z}} & \left(8p\right) & \mbox{C XII} \\ 
\mathbf{B}_{103} & = & y_{14} \, \mathbf{a}_{1} + x_{14} \, \mathbf{a}_{2}-z_{14} \, \mathbf{a}_{3} & = & y_{14}a \, \mathbf{\hat{x}} + x_{14}a \, \mathbf{\hat{y}}-z_{14}c \, \mathbf{\hat{z}} & \left(8p\right) & \mbox{C XII} \\ 
\mathbf{B}_{104} & = & -y_{14} \, \mathbf{a}_{1}-x_{14} \, \mathbf{a}_{2}-z_{14} \, \mathbf{a}_{3} & = & -y_{14}a \, \mathbf{\hat{x}}-x_{14}a \, \mathbf{\hat{y}}-z_{14}c \, \mathbf{\hat{z}} & \left(8p\right) & \mbox{C XII} \\ 
\mathbf{B}_{105} & = & x_{15} \, \mathbf{a}_{1} + y_{15} \, \mathbf{a}_{2} + z_{15} \, \mathbf{a}_{3} & = & x_{15}a \, \mathbf{\hat{x}} + y_{15}a \, \mathbf{\hat{y}} + z_{15}c \, \mathbf{\hat{z}} & \left(8p\right) & \mbox{C XIII} \\ 
\mathbf{B}_{106} & = & -x_{15} \, \mathbf{a}_{1}-y_{15} \, \mathbf{a}_{2} + z_{15} \, \mathbf{a}_{3} & = & -x_{15}a \, \mathbf{\hat{x}}-y_{15}a \, \mathbf{\hat{y}} + z_{15}c \, \mathbf{\hat{z}} & \left(8p\right) & \mbox{C XIII} \\ 
\mathbf{B}_{107} & = & -y_{15} \, \mathbf{a}_{1} + x_{15} \, \mathbf{a}_{2} + z_{15} \, \mathbf{a}_{3} & = & -y_{15}a \, \mathbf{\hat{x}} + x_{15}a \, \mathbf{\hat{y}} + z_{15}c \, \mathbf{\hat{z}} & \left(8p\right) & \mbox{C XIII} \\ 
\mathbf{B}_{108} & = & y_{15} \, \mathbf{a}_{1}-x_{15} \, \mathbf{a}_{2} + z_{15} \, \mathbf{a}_{3} & = & y_{15}a \, \mathbf{\hat{x}}-x_{15}a \, \mathbf{\hat{y}} + z_{15}c \, \mathbf{\hat{z}} & \left(8p\right) & \mbox{C XIII} \\ 
\mathbf{B}_{109} & = & -x_{15} \, \mathbf{a}_{1} + y_{15} \, \mathbf{a}_{2}-z_{15} \, \mathbf{a}_{3} & = & -x_{15}a \, \mathbf{\hat{x}} + y_{15}a \, \mathbf{\hat{y}}-z_{15}c \, \mathbf{\hat{z}} & \left(8p\right) & \mbox{C XIII} \\ 
\mathbf{B}_{110} & = & x_{15} \, \mathbf{a}_{1}-y_{15} \, \mathbf{a}_{2}-z_{15} \, \mathbf{a}_{3} & = & x_{15}a \, \mathbf{\hat{x}}-y_{15}a \, \mathbf{\hat{y}}-z_{15}c \, \mathbf{\hat{z}} & \left(8p\right) & \mbox{C XIII} \\ 
\mathbf{B}_{111} & = & y_{15} \, \mathbf{a}_{1} + x_{15} \, \mathbf{a}_{2}-z_{15} \, \mathbf{a}_{3} & = & y_{15}a \, \mathbf{\hat{x}} + x_{15}a \, \mathbf{\hat{y}}-z_{15}c \, \mathbf{\hat{z}} & \left(8p\right) & \mbox{C XIII} \\ 
\mathbf{B}_{112} & = & -y_{15} \, \mathbf{a}_{1}-x_{15} \, \mathbf{a}_{2}-z_{15} \, \mathbf{a}_{3} & = & -y_{15}a \, \mathbf{\hat{x}}-x_{15}a \, \mathbf{\hat{y}}-z_{15}c \, \mathbf{\hat{z}} & \left(8p\right) & \mbox{C XIII} \\ 
\mathbf{B}_{113} & = & x_{16} \, \mathbf{a}_{1} + y_{16} \, \mathbf{a}_{2} + z_{16} \, \mathbf{a}_{3} & = & x_{16}a \, \mathbf{\hat{x}} + y_{16}a \, \mathbf{\hat{y}} + z_{16}c \, \mathbf{\hat{z}} & \left(8p\right) & \mbox{C XIV} \\ 
\mathbf{B}_{114} & = & -x_{16} \, \mathbf{a}_{1}-y_{16} \, \mathbf{a}_{2} + z_{16} \, \mathbf{a}_{3} & = & -x_{16}a \, \mathbf{\hat{x}}-y_{16}a \, \mathbf{\hat{y}} + z_{16}c \, \mathbf{\hat{z}} & \left(8p\right) & \mbox{C XIV} \\ 
\mathbf{B}_{115} & = & -y_{16} \, \mathbf{a}_{1} + x_{16} \, \mathbf{a}_{2} + z_{16} \, \mathbf{a}_{3} & = & -y_{16}a \, \mathbf{\hat{x}} + x_{16}a \, \mathbf{\hat{y}} + z_{16}c \, \mathbf{\hat{z}} & \left(8p\right) & \mbox{C XIV} \\ 
\mathbf{B}_{116} & = & y_{16} \, \mathbf{a}_{1}-x_{16} \, \mathbf{a}_{2} + z_{16} \, \mathbf{a}_{3} & = & y_{16}a \, \mathbf{\hat{x}}-x_{16}a \, \mathbf{\hat{y}} + z_{16}c \, \mathbf{\hat{z}} & \left(8p\right) & \mbox{C XIV} \\ 
\mathbf{B}_{117} & = & -x_{16} \, \mathbf{a}_{1} + y_{16} \, \mathbf{a}_{2}-z_{16} \, \mathbf{a}_{3} & = & -x_{16}a \, \mathbf{\hat{x}} + y_{16}a \, \mathbf{\hat{y}}-z_{16}c \, \mathbf{\hat{z}} & \left(8p\right) & \mbox{C XIV} \\ 
\mathbf{B}_{118} & = & x_{16} \, \mathbf{a}_{1}-y_{16} \, \mathbf{a}_{2}-z_{16} \, \mathbf{a}_{3} & = & x_{16}a \, \mathbf{\hat{x}}-y_{16}a \, \mathbf{\hat{y}}-z_{16}c \, \mathbf{\hat{z}} & \left(8p\right) & \mbox{C XIV} \\ 
\mathbf{B}_{119} & = & y_{16} \, \mathbf{a}_{1} + x_{16} \, \mathbf{a}_{2}-z_{16} \, \mathbf{a}_{3} & = & y_{16}a \, \mathbf{\hat{x}} + x_{16}a \, \mathbf{\hat{y}}-z_{16}c \, \mathbf{\hat{z}} & \left(8p\right) & \mbox{C XIV} \\ 
\mathbf{B}_{120} & = & -y_{16} \, \mathbf{a}_{1}-x_{16} \, \mathbf{a}_{2}-z_{16} \, \mathbf{a}_{3} & = & -y_{16}a \, \mathbf{\hat{x}}-x_{16}a \, \mathbf{\hat{y}}-z_{16}c \, \mathbf{\hat{z}} & \left(8p\right) & \mbox{C XIV} \\ 
\mathbf{B}_{121} & = & x_{17} \, \mathbf{a}_{1} + y_{17} \, \mathbf{a}_{2} + z_{17} \, \mathbf{a}_{3} & = & x_{17}a \, \mathbf{\hat{x}} + y_{17}a \, \mathbf{\hat{y}} + z_{17}c \, \mathbf{\hat{z}} & \left(8p\right) & \mbox{C XV} \\ 
\mathbf{B}_{122} & = & -x_{17} \, \mathbf{a}_{1}-y_{17} \, \mathbf{a}_{2} + z_{17} \, \mathbf{a}_{3} & = & -x_{17}a \, \mathbf{\hat{x}}-y_{17}a \, \mathbf{\hat{y}} + z_{17}c \, \mathbf{\hat{z}} & \left(8p\right) & \mbox{C XV} \\ 
\mathbf{B}_{123} & = & -y_{17} \, \mathbf{a}_{1} + x_{17} \, \mathbf{a}_{2} + z_{17} \, \mathbf{a}_{3} & = & -y_{17}a \, \mathbf{\hat{x}} + x_{17}a \, \mathbf{\hat{y}} + z_{17}c \, \mathbf{\hat{z}} & \left(8p\right) & \mbox{C XV} \\ 
\mathbf{B}_{124} & = & y_{17} \, \mathbf{a}_{1}-x_{17} \, \mathbf{a}_{2} + z_{17} \, \mathbf{a}_{3} & = & y_{17}a \, \mathbf{\hat{x}}-x_{17}a \, \mathbf{\hat{y}} + z_{17}c \, \mathbf{\hat{z}} & \left(8p\right) & \mbox{C XV} \\ 
\mathbf{B}_{125} & = & -x_{17} \, \mathbf{a}_{1} + y_{17} \, \mathbf{a}_{2}-z_{17} \, \mathbf{a}_{3} & = & -x_{17}a \, \mathbf{\hat{x}} + y_{17}a \, \mathbf{\hat{y}}-z_{17}c \, \mathbf{\hat{z}} & \left(8p\right) & \mbox{C XV} \\ 
\mathbf{B}_{126} & = & x_{17} \, \mathbf{a}_{1}-y_{17} \, \mathbf{a}_{2}-z_{17} \, \mathbf{a}_{3} & = & x_{17}a \, \mathbf{\hat{x}}-y_{17}a \, \mathbf{\hat{y}}-z_{17}c \, \mathbf{\hat{z}} & \left(8p\right) & \mbox{C XV} \\ 
\mathbf{B}_{127} & = & y_{17} \, \mathbf{a}_{1} + x_{17} \, \mathbf{a}_{2}-z_{17} \, \mathbf{a}_{3} & = & y_{17}a \, \mathbf{\hat{x}} + x_{17}a \, \mathbf{\hat{y}}-z_{17}c \, \mathbf{\hat{z}} & \left(8p\right) & \mbox{C XV} \\ 
\mathbf{B}_{128} & = & -y_{17} \, \mathbf{a}_{1}-x_{17} \, \mathbf{a}_{2}-z_{17} \, \mathbf{a}_{3} & = & -y_{17}a \, \mathbf{\hat{x}}-x_{17}a \, \mathbf{\hat{y}}-z_{17}c \, \mathbf{\hat{z}} & \left(8p\right) & \mbox{C XV} \\ 
\mathbf{B}_{129} & = & x_{18} \, \mathbf{a}_{1} + y_{18} \, \mathbf{a}_{2} + z_{18} \, \mathbf{a}_{3} & = & x_{18}a \, \mathbf{\hat{x}} + y_{18}a \, \mathbf{\hat{y}} + z_{18}c \, \mathbf{\hat{z}} & \left(8p\right) & \mbox{C XVI} \\ 
\mathbf{B}_{130} & = & -x_{18} \, \mathbf{a}_{1}-y_{18} \, \mathbf{a}_{2} + z_{18} \, \mathbf{a}_{3} & = & -x_{18}a \, \mathbf{\hat{x}}-y_{18}a \, \mathbf{\hat{y}} + z_{18}c \, \mathbf{\hat{z}} & \left(8p\right) & \mbox{C XVI} \\ 
\mathbf{B}_{131} & = & -y_{18} \, \mathbf{a}_{1} + x_{18} \, \mathbf{a}_{2} + z_{18} \, \mathbf{a}_{3} & = & -y_{18}a \, \mathbf{\hat{x}} + x_{18}a \, \mathbf{\hat{y}} + z_{18}c \, \mathbf{\hat{z}} & \left(8p\right) & \mbox{C XVI} \\ 
\mathbf{B}_{132} & = & y_{18} \, \mathbf{a}_{1}-x_{18} \, \mathbf{a}_{2} + z_{18} \, \mathbf{a}_{3} & = & y_{18}a \, \mathbf{\hat{x}}-x_{18}a \, \mathbf{\hat{y}} + z_{18}c \, \mathbf{\hat{z}} & \left(8p\right) & \mbox{C XVI} \\ 
\mathbf{B}_{133} & = & -x_{18} \, \mathbf{a}_{1} + y_{18} \, \mathbf{a}_{2}-z_{18} \, \mathbf{a}_{3} & = & -x_{18}a \, \mathbf{\hat{x}} + y_{18}a \, \mathbf{\hat{y}}-z_{18}c \, \mathbf{\hat{z}} & \left(8p\right) & \mbox{C XVI} \\ 
\mathbf{B}_{134} & = & x_{18} \, \mathbf{a}_{1}-y_{18} \, \mathbf{a}_{2}-z_{18} \, \mathbf{a}_{3} & = & x_{18}a \, \mathbf{\hat{x}}-y_{18}a \, \mathbf{\hat{y}}-z_{18}c \, \mathbf{\hat{z}} & \left(8p\right) & \mbox{C XVI} \\ 
\mathbf{B}_{135} & = & y_{18} \, \mathbf{a}_{1} + x_{18} \, \mathbf{a}_{2}-z_{18} \, \mathbf{a}_{3} & = & y_{18}a \, \mathbf{\hat{x}} + x_{18}a \, \mathbf{\hat{y}}-z_{18}c \, \mathbf{\hat{z}} & \left(8p\right) & \mbox{C XVI} \\ 
\mathbf{B}_{136} & = & -y_{18} \, \mathbf{a}_{1}-x_{18} \, \mathbf{a}_{2}-z_{18} \, \mathbf{a}_{3} & = & -y_{18}a \, \mathbf{\hat{x}}-x_{18}a \, \mathbf{\hat{y}}-z_{18}c \, \mathbf{\hat{z}} & \left(8p\right) & \mbox{C XVI} \\ 
\mathbf{B}_{137} & = & x_{19} \, \mathbf{a}_{1} + y_{19} \, \mathbf{a}_{2} + z_{19} \, \mathbf{a}_{3} & = & x_{19}a \, \mathbf{\hat{x}} + y_{19}a \, \mathbf{\hat{y}} + z_{19}c \, \mathbf{\hat{z}} & \left(8p\right) & \mbox{C XVII} \\ 
\mathbf{B}_{138} & = & -x_{19} \, \mathbf{a}_{1}-y_{19} \, \mathbf{a}_{2} + z_{19} \, \mathbf{a}_{3} & = & -x_{19}a \, \mathbf{\hat{x}}-y_{19}a \, \mathbf{\hat{y}} + z_{19}c \, \mathbf{\hat{z}} & \left(8p\right) & \mbox{C XVII} \\ 
\mathbf{B}_{139} & = & -y_{19} \, \mathbf{a}_{1} + x_{19} \, \mathbf{a}_{2} + z_{19} \, \mathbf{a}_{3} & = & -y_{19}a \, \mathbf{\hat{x}} + x_{19}a \, \mathbf{\hat{y}} + z_{19}c \, \mathbf{\hat{z}} & \left(8p\right) & \mbox{C XVII} \\ 
\mathbf{B}_{140} & = & y_{19} \, \mathbf{a}_{1}-x_{19} \, \mathbf{a}_{2} + z_{19} \, \mathbf{a}_{3} & = & y_{19}a \, \mathbf{\hat{x}}-x_{19}a \, \mathbf{\hat{y}} + z_{19}c \, \mathbf{\hat{z}} & \left(8p\right) & \mbox{C XVII} \\ 
\mathbf{B}_{141} & = & -x_{19} \, \mathbf{a}_{1} + y_{19} \, \mathbf{a}_{2}-z_{19} \, \mathbf{a}_{3} & = & -x_{19}a \, \mathbf{\hat{x}} + y_{19}a \, \mathbf{\hat{y}}-z_{19}c \, \mathbf{\hat{z}} & \left(8p\right) & \mbox{C XVII} \\ 
\mathbf{B}_{142} & = & x_{19} \, \mathbf{a}_{1}-y_{19} \, \mathbf{a}_{2}-z_{19} \, \mathbf{a}_{3} & = & x_{19}a \, \mathbf{\hat{x}}-y_{19}a \, \mathbf{\hat{y}}-z_{19}c \, \mathbf{\hat{z}} & \left(8p\right) & \mbox{C XVII} \\ 
\mathbf{B}_{143} & = & y_{19} \, \mathbf{a}_{1} + x_{19} \, \mathbf{a}_{2}-z_{19} \, \mathbf{a}_{3} & = & y_{19}a \, \mathbf{\hat{x}} + x_{19}a \, \mathbf{\hat{y}}-z_{19}c \, \mathbf{\hat{z}} & \left(8p\right) & \mbox{C XVII} \\ 
\mathbf{B}_{144} & = & -y_{19} \, \mathbf{a}_{1}-x_{19} \, \mathbf{a}_{2}-z_{19} \, \mathbf{a}_{3} & = & -y_{19}a \, \mathbf{\hat{x}}-x_{19}a \, \mathbf{\hat{y}}-z_{19}c \, \mathbf{\hat{z}} & \left(8p\right) & \mbox{C XVII} \\ 
\mathbf{B}_{145} & = & x_{20} \, \mathbf{a}_{1} + y_{20} \, \mathbf{a}_{2} + z_{20} \, \mathbf{a}_{3} & = & x_{20}a \, \mathbf{\hat{x}} + y_{20}a \, \mathbf{\hat{y}} + z_{20}c \, \mathbf{\hat{z}} & \left(8p\right) & \mbox{Fe} \\ 
\mathbf{B}_{146} & = & -x_{20} \, \mathbf{a}_{1}-y_{20} \, \mathbf{a}_{2} + z_{20} \, \mathbf{a}_{3} & = & -x_{20}a \, \mathbf{\hat{x}}-y_{20}a \, \mathbf{\hat{y}} + z_{20}c \, \mathbf{\hat{z}} & \left(8p\right) & \mbox{Fe} \\ 
\mathbf{B}_{147} & = & -y_{20} \, \mathbf{a}_{1} + x_{20} \, \mathbf{a}_{2} + z_{20} \, \mathbf{a}_{3} & = & -y_{20}a \, \mathbf{\hat{x}} + x_{20}a \, \mathbf{\hat{y}} + z_{20}c \, \mathbf{\hat{z}} & \left(8p\right) & \mbox{Fe} \\ 
\mathbf{B}_{148} & = & y_{20} \, \mathbf{a}_{1}-x_{20} \, \mathbf{a}_{2} + z_{20} \, \mathbf{a}_{3} & = & y_{20}a \, \mathbf{\hat{x}}-x_{20}a \, \mathbf{\hat{y}} + z_{20}c \, \mathbf{\hat{z}} & \left(8p\right) & \mbox{Fe} \\ 
\mathbf{B}_{149} & = & -x_{20} \, \mathbf{a}_{1} + y_{20} \, \mathbf{a}_{2}-z_{20} \, \mathbf{a}_{3} & = & -x_{20}a \, \mathbf{\hat{x}} + y_{20}a \, \mathbf{\hat{y}}-z_{20}c \, \mathbf{\hat{z}} & \left(8p\right) & \mbox{Fe} \\ 
\mathbf{B}_{150} & = & x_{20} \, \mathbf{a}_{1}-y_{20} \, \mathbf{a}_{2}-z_{20} \, \mathbf{a}_{3} & = & x_{20}a \, \mathbf{\hat{x}}-y_{20}a \, \mathbf{\hat{y}}-z_{20}c \, \mathbf{\hat{z}} & \left(8p\right) & \mbox{Fe} \\ 
\mathbf{B}_{151} & = & y_{20} \, \mathbf{a}_{1} + x_{20} \, \mathbf{a}_{2}-z_{20} \, \mathbf{a}_{3} & = & y_{20}a \, \mathbf{\hat{x}} + x_{20}a \, \mathbf{\hat{y}}-z_{20}c \, \mathbf{\hat{z}} & \left(8p\right) & \mbox{Fe} \\ 
\mathbf{B}_{152} & = & -y_{20} \, \mathbf{a}_{1}-x_{20} \, \mathbf{a}_{2}-z_{20} \, \mathbf{a}_{3} & = & -y_{20}a \, \mathbf{\hat{x}}-x_{20}a \, \mathbf{\hat{y}}-z_{20}c \, \mathbf{\hat{z}} & \left(8p\right) & \mbox{Fe} \\ 
\mathbf{B}_{153} & = & x_{21} \, \mathbf{a}_{1} + y_{21} \, \mathbf{a}_{2} + z_{21} \, \mathbf{a}_{3} & = & x_{21}a \, \mathbf{\hat{x}} + y_{21}a \, \mathbf{\hat{y}} + z_{21}c \, \mathbf{\hat{z}} & \left(8p\right) & \mbox{O I} \\ 
\mathbf{B}_{154} & = & -x_{21} \, \mathbf{a}_{1}-y_{21} \, \mathbf{a}_{2} + z_{21} \, \mathbf{a}_{3} & = & -x_{21}a \, \mathbf{\hat{x}}-y_{21}a \, \mathbf{\hat{y}} + z_{21}c \, \mathbf{\hat{z}} & \left(8p\right) & \mbox{O I} \\ 
\mathbf{B}_{155} & = & -y_{21} \, \mathbf{a}_{1} + x_{21} \, \mathbf{a}_{2} + z_{21} \, \mathbf{a}_{3} & = & -y_{21}a \, \mathbf{\hat{x}} + x_{21}a \, \mathbf{\hat{y}} + z_{21}c \, \mathbf{\hat{z}} & \left(8p\right) & \mbox{O I} \\ 
\mathbf{B}_{156} & = & y_{21} \, \mathbf{a}_{1}-x_{21} \, \mathbf{a}_{2} + z_{21} \, \mathbf{a}_{3} & = & y_{21}a \, \mathbf{\hat{x}}-x_{21}a \, \mathbf{\hat{y}} + z_{21}c \, \mathbf{\hat{z}} & \left(8p\right) & \mbox{O I} \\ 
\mathbf{B}_{157} & = & -x_{21} \, \mathbf{a}_{1} + y_{21} \, \mathbf{a}_{2}-z_{21} \, \mathbf{a}_{3} & = & -x_{21}a \, \mathbf{\hat{x}} + y_{21}a \, \mathbf{\hat{y}}-z_{21}c \, \mathbf{\hat{z}} & \left(8p\right) & \mbox{O I} \\ 
\mathbf{B}_{158} & = & x_{21} \, \mathbf{a}_{1}-y_{21} \, \mathbf{a}_{2}-z_{21} \, \mathbf{a}_{3} & = & x_{21}a \, \mathbf{\hat{x}}-y_{21}a \, \mathbf{\hat{y}}-z_{21}c \, \mathbf{\hat{z}} & \left(8p\right) & \mbox{O I} \\ 
\mathbf{B}_{159} & = & y_{21} \, \mathbf{a}_{1} + x_{21} \, \mathbf{a}_{2}-z_{21} \, \mathbf{a}_{3} & = & y_{21}a \, \mathbf{\hat{x}} + x_{21}a \, \mathbf{\hat{y}}-z_{21}c \, \mathbf{\hat{z}} & \left(8p\right) & \mbox{O I} \\ 
\mathbf{B}_{160} & = & -y_{21} \, \mathbf{a}_{1}-x_{21} \, \mathbf{a}_{2}-z_{21} \, \mathbf{a}_{3} & = & -y_{21}a \, \mathbf{\hat{x}}-x_{21}a \, \mathbf{\hat{y}}-z_{21}c \, \mathbf{\hat{z}} & \left(8p\right) & \mbox{O I} \\ 
\mathbf{B}_{161} & = & x_{22} \, \mathbf{a}_{1} + y_{22} \, \mathbf{a}_{2} + z_{22} \, \mathbf{a}_{3} & = & x_{22}a \, \mathbf{\hat{x}} + y_{22}a \, \mathbf{\hat{y}} + z_{22}c \, \mathbf{\hat{z}} & \left(8p\right) & \mbox{O II} \\ 
\mathbf{B}_{162} & = & -x_{22} \, \mathbf{a}_{1}-y_{22} \, \mathbf{a}_{2} + z_{22} \, \mathbf{a}_{3} & = & -x_{22}a \, \mathbf{\hat{x}}-y_{22}a \, \mathbf{\hat{y}} + z_{22}c \, \mathbf{\hat{z}} & \left(8p\right) & \mbox{O II} \\ 
\mathbf{B}_{163} & = & -y_{22} \, \mathbf{a}_{1} + x_{22} \, \mathbf{a}_{2} + z_{22} \, \mathbf{a}_{3} & = & -y_{22}a \, \mathbf{\hat{x}} + x_{22}a \, \mathbf{\hat{y}} + z_{22}c \, \mathbf{\hat{z}} & \left(8p\right) & \mbox{O II} \\ 
\mathbf{B}_{164} & = & y_{22} \, \mathbf{a}_{1}-x_{22} \, \mathbf{a}_{2} + z_{22} \, \mathbf{a}_{3} & = & y_{22}a \, \mathbf{\hat{x}}-x_{22}a \, \mathbf{\hat{y}} + z_{22}c \, \mathbf{\hat{z}} & \left(8p\right) & \mbox{O II} \\ 
\mathbf{B}_{165} & = & -x_{22} \, \mathbf{a}_{1} + y_{22} \, \mathbf{a}_{2}-z_{22} \, \mathbf{a}_{3} & = & -x_{22}a \, \mathbf{\hat{x}} + y_{22}a \, \mathbf{\hat{y}}-z_{22}c \, \mathbf{\hat{z}} & \left(8p\right) & \mbox{O II} \\ 
\mathbf{B}_{166} & = & x_{22} \, \mathbf{a}_{1}-y_{22} \, \mathbf{a}_{2}-z_{22} \, \mathbf{a}_{3} & = & x_{22}a \, \mathbf{\hat{x}}-y_{22}a \, \mathbf{\hat{y}}-z_{22}c \, \mathbf{\hat{z}} & \left(8p\right) & \mbox{O II} \\ 
\mathbf{B}_{167} & = & y_{22} \, \mathbf{a}_{1} + x_{22} \, \mathbf{a}_{2}-z_{22} \, \mathbf{a}_{3} & = & y_{22}a \, \mathbf{\hat{x}} + x_{22}a \, \mathbf{\hat{y}}-z_{22}c \, \mathbf{\hat{z}} & \left(8p\right) & \mbox{O II} \\ 
\mathbf{B}_{168} & = & -y_{22} \, \mathbf{a}_{1}-x_{22} \, \mathbf{a}_{2}-z_{22} \, \mathbf{a}_{3} & = & -y_{22}a \, \mathbf{\hat{x}}-x_{22}a \, \mathbf{\hat{y}}-z_{22}c \, \mathbf{\hat{z}} & \left(8p\right) & \mbox{O II} \\ 
\mathbf{B}_{169} & = & x_{23} \, \mathbf{a}_{1} + y_{23} \, \mathbf{a}_{2} + z_{23} \, \mathbf{a}_{3} & = & x_{23}a \, \mathbf{\hat{x}} + y_{23}a \, \mathbf{\hat{y}} + z_{23}c \, \mathbf{\hat{z}} & \left(8p\right) & \mbox{O III} \\ 
\mathbf{B}_{170} & = & -x_{23} \, \mathbf{a}_{1}-y_{23} \, \mathbf{a}_{2} + z_{23} \, \mathbf{a}_{3} & = & -x_{23}a \, \mathbf{\hat{x}}-y_{23}a \, \mathbf{\hat{y}} + z_{23}c \, \mathbf{\hat{z}} & \left(8p\right) & \mbox{O III} \\ 
\mathbf{B}_{171} & = & -y_{23} \, \mathbf{a}_{1} + x_{23} \, \mathbf{a}_{2} + z_{23} \, \mathbf{a}_{3} & = & -y_{23}a \, \mathbf{\hat{x}} + x_{23}a \, \mathbf{\hat{y}} + z_{23}c \, \mathbf{\hat{z}} & \left(8p\right) & \mbox{O III} \\ 
\mathbf{B}_{172} & = & y_{23} \, \mathbf{a}_{1}-x_{23} \, \mathbf{a}_{2} + z_{23} \, \mathbf{a}_{3} & = & y_{23}a \, \mathbf{\hat{x}}-x_{23}a \, \mathbf{\hat{y}} + z_{23}c \, \mathbf{\hat{z}} & \left(8p\right) & \mbox{O III} \\ 
\mathbf{B}_{173} & = & -x_{23} \, \mathbf{a}_{1} + y_{23} \, \mathbf{a}_{2}-z_{23} \, \mathbf{a}_{3} & = & -x_{23}a \, \mathbf{\hat{x}} + y_{23}a \, \mathbf{\hat{y}}-z_{23}c \, \mathbf{\hat{z}} & \left(8p\right) & \mbox{O III} \\ 
\mathbf{B}_{174} & = & x_{23} \, \mathbf{a}_{1}-y_{23} \, \mathbf{a}_{2}-z_{23} \, \mathbf{a}_{3} & = & x_{23}a \, \mathbf{\hat{x}}-y_{23}a \, \mathbf{\hat{y}}-z_{23}c \, \mathbf{\hat{z}} & \left(8p\right) & \mbox{O III} \\ 
\mathbf{B}_{175} & = & y_{23} \, \mathbf{a}_{1} + x_{23} \, \mathbf{a}_{2}-z_{23} \, \mathbf{a}_{3} & = & y_{23}a \, \mathbf{\hat{x}} + x_{23}a \, \mathbf{\hat{y}}-z_{23}c \, \mathbf{\hat{z}} & \left(8p\right) & \mbox{O III} \\ 
\mathbf{B}_{176} & = & -y_{23} \, \mathbf{a}_{1}-x_{23} \, \mathbf{a}_{2}-z_{23} \, \mathbf{a}_{3} & = & -y_{23}a \, \mathbf{\hat{x}}-x_{23}a \, \mathbf{\hat{y}}-z_{23}c \, \mathbf{\hat{z}} & \left(8p\right) & \mbox{O III} \\ 
\mathbf{B}_{177} & = & x_{24} \, \mathbf{a}_{1} + y_{24} \, \mathbf{a}_{2} + z_{24} \, \mathbf{a}_{3} & = & x_{24}a \, \mathbf{\hat{x}} + y_{24}a \, \mathbf{\hat{y}} + z_{24}c \, \mathbf{\hat{z}} & \left(8p\right) & \mbox{O IV} \\ 
\mathbf{B}_{178} & = & -x_{24} \, \mathbf{a}_{1}-y_{24} \, \mathbf{a}_{2} + z_{24} \, \mathbf{a}_{3} & = & -x_{24}a \, \mathbf{\hat{x}}-y_{24}a \, \mathbf{\hat{y}} + z_{24}c \, \mathbf{\hat{z}} & \left(8p\right) & \mbox{O IV} \\ 
\mathbf{B}_{179} & = & -y_{24} \, \mathbf{a}_{1} + x_{24} \, \mathbf{a}_{2} + z_{24} \, \mathbf{a}_{3} & = & -y_{24}a \, \mathbf{\hat{x}} + x_{24}a \, \mathbf{\hat{y}} + z_{24}c \, \mathbf{\hat{z}} & \left(8p\right) & \mbox{O IV} \\ 
\mathbf{B}_{180} & = & y_{24} \, \mathbf{a}_{1}-x_{24} \, \mathbf{a}_{2} + z_{24} \, \mathbf{a}_{3} & = & y_{24}a \, \mathbf{\hat{x}}-x_{24}a \, \mathbf{\hat{y}} + z_{24}c \, \mathbf{\hat{z}} & \left(8p\right) & \mbox{O IV} \\ 
\mathbf{B}_{181} & = & -x_{24} \, \mathbf{a}_{1} + y_{24} \, \mathbf{a}_{2}-z_{24} \, \mathbf{a}_{3} & = & -x_{24}a \, \mathbf{\hat{x}} + y_{24}a \, \mathbf{\hat{y}}-z_{24}c \, \mathbf{\hat{z}} & \left(8p\right) & \mbox{O IV} \\ 
\mathbf{B}_{182} & = & x_{24} \, \mathbf{a}_{1}-y_{24} \, \mathbf{a}_{2}-z_{24} \, \mathbf{a}_{3} & = & x_{24}a \, \mathbf{\hat{x}}-y_{24}a \, \mathbf{\hat{y}}-z_{24}c \, \mathbf{\hat{z}} & \left(8p\right) & \mbox{O IV} \\ 
\mathbf{B}_{183} & = & y_{24} \, \mathbf{a}_{1} + x_{24} \, \mathbf{a}_{2}-z_{24} \, \mathbf{a}_{3} & = & y_{24}a \, \mathbf{\hat{x}} + x_{24}a \, \mathbf{\hat{y}}-z_{24}c \, \mathbf{\hat{z}} & \left(8p\right) & \mbox{O IV} \\ 
\mathbf{B}_{184} & = & -y_{24} \, \mathbf{a}_{1}-x_{24} \, \mathbf{a}_{2}-z_{24} \, \mathbf{a}_{3} & = & -y_{24}a \, \mathbf{\hat{x}}-x_{24}a \, \mathbf{\hat{y}}-z_{24}c \, \mathbf{\hat{z}} & \left(8p\right) & \mbox{O IV} \\ 
\end{longtabu}
\renewcommand{\arraystretch}{1.0}
\noindent \hrulefill
\\
\textbf{References:}
\vspace*{-0.25cm}
\begin{flushleft}
  - \bibentry{Tanaka_Pt4FenComplex_ic_22_2011}. \\
\end{flushleft}
\textbf{Found in:}
\vspace*{-0.25cm}
\begin{flushleft}
  - \bibentry{Groom_CSD_72_2016}. \\
  - \bibentry{Hoffmann_SpaceGroupProject_2014}. \\
\end{flushleft}
\noindent \hrulefill
\\
\textbf{Geometry files:}
\\
\noindent  - CIF: pp. {\hyperref[A17BC4D_tP184_89_17p_p_4p_io_cif]{\pageref{A17BC4D_tP184_89_17p_p_4p_io_cif}}} \\
\noindent  - POSCAR: pp. {\hyperref[A17BC4D_tP184_89_17p_p_4p_io_poscar]{\pageref{A17BC4D_tP184_89_17p_p_4p_io_poscar}}} \\
\onecolumn
{\phantomsection\label{A4B2C13D_tP40_90_g_d_cef2g_c}}
\subsection*{\huge \textbf{{\normalfont \begin{raggedleft}Na$_{4}$Ti$_{2}$Si$_{8}$O$_{22}$[H$_{2}$O]$_{4}$ Structure: \end{raggedleft} \\ A4B2C13D\_tP40\_90\_g\_d\_cef2g\_c}}}
\noindent \hrulefill
\vspace*{0.25cm}
\begin{figure}[htp]
  \centering
  \vspace{-1em}
  {\includegraphics[width=1\textwidth]{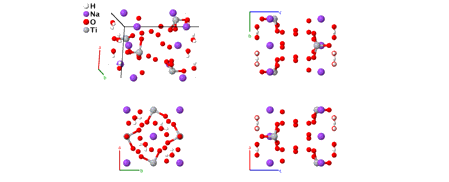}}
\end{figure}
\vspace*{-0.5cm}
\renewcommand{\arraystretch}{1.5}
\begin{equation*}
  \begin{array}{>{$\hspace{-0.15cm}}l<{$}>{$}p{0.5cm}<{$}>{$}p{18.5cm}<{$}}
    \mbox{\large \textbf{Prototype}} &\colon & \ce{Na4Ti2Si8O22[H2O]4} \\
    \mbox{\large \textbf{\AFLOW\ prototype label}} &\colon & \mbox{A4B2C13D\_tP40\_90\_g\_d\_cef2g\_c} \\
    \mbox{\large \textbf{\textit{Strukturbericht} designation}} &\colon & \mbox{None} \\
    \mbox{\large \textbf{Pearson symbol}} &\colon & \mbox{tP40} \\
    \mbox{\large \textbf{Space group number}} &\colon & 90 \\
    \mbox{\large \textbf{Space group symbol}} &\colon & P42_{1}2 \\
    \mbox{\large \textbf{\AFLOW\ prototype command}} &\colon &  \texttt{aflow} \,  \, \texttt{-{}-proto=A4B2C13D\_tP40\_90\_g\_d\_cef2g\_c } \, \newline \texttt{-{}-params=}{a,c/a,z_{1},z_{2},z_{3},x_{4},x_{5},x_{6},y_{6},z_{6},x_{7},y_{7},z_{7},x_{8},y_{8},z_{8} }
  \end{array}
\end{equation*}
\renewcommand{\arraystretch}{1.0}

\noindent \parbox{1 \linewidth}{
\noindent \hrulefill
\\
\textbf{Simple Tetragonal primitive vectors:} \\
\vspace*{-0.25cm}
\begin{tabular}{cc}
  \begin{tabular}{c}
    \parbox{0.6 \linewidth}{
      \renewcommand{\arraystretch}{1.5}
      \begin{equation*}
        \centering
        \begin{array}{ccc}
              \mathbf{a}_1 & = & a \, \mathbf{\hat{x}} \\
    \mathbf{a}_2 & = & a \, \mathbf{\hat{y}} \\
    \mathbf{a}_3 & = & c \, \mathbf{\hat{z}} \\

        \end{array}
      \end{equation*}
    }
    \renewcommand{\arraystretch}{1.0}
  \end{tabular}
  \begin{tabular}{c}
    \includegraphics[width=0.3\linewidth]{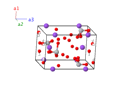} \\
  \end{tabular}
\end{tabular}

}
\vspace*{-0.25cm}

\noindent \hrulefill
\\
\textbf{Basis vectors:}
\vspace*{-0.25cm}
\renewcommand{\arraystretch}{1.5}
\begin{longtabu} to \textwidth{>{\centering $}X[-1,c,c]<{$}>{\centering $}X[-1,c,c]<{$}>{\centering $}X[-1,c,c]<{$}>{\centering $}X[-1,c,c]<{$}>{\centering $}X[-1,c,c]<{$}>{\centering $}X[-1,c,c]<{$}>{\centering $}X[-1,c,c]<{$}}
  & & \mbox{Lattice Coordinates} & & \mbox{Cartesian Coordinates} &\mbox{Wyckoff Position} & \mbox{Atom Type} \\  
  \mathbf{B}_{1} & = & \frac{1}{2} \, \mathbf{a}_{2} + z_{1} \, \mathbf{a}_{3} & = & \frac{1}{2}a \, \mathbf{\hat{y}} + z_{1}c \, \mathbf{\hat{z}} & \left(2c\right) & \mbox{O I} \\ 
\mathbf{B}_{2} & = & \frac{1}{2} \, \mathbf{a}_{1} + -z_{1} \, \mathbf{a}_{3} & = & \frac{1}{2}a \, \mathbf{\hat{x}} + -z_{1}c \, \mathbf{\hat{z}} & \left(2c\right) & \mbox{O I} \\ 
\mathbf{B}_{3} & = & \frac{1}{2} \, \mathbf{a}_{2} + z_{2} \, \mathbf{a}_{3} & = & \frac{1}{2}a \, \mathbf{\hat{y}} + z_{2}c \, \mathbf{\hat{z}} & \left(2c\right) & \mbox{Ti} \\ 
\mathbf{B}_{4} & = & \frac{1}{2} \, \mathbf{a}_{1} + -z_{2} \, \mathbf{a}_{3} & = & \frac{1}{2}a \, \mathbf{\hat{x}} + -z_{2}c \, \mathbf{\hat{z}} & \left(2c\right) & \mbox{Ti} \\ 
\mathbf{B}_{5} & = & z_{3} \, \mathbf{a}_{3} & = & z_{3}c \, \mathbf{\hat{z}} & \left(4d\right) & \mbox{Na} \\ 
\mathbf{B}_{6} & = & \frac{1}{2} \, \mathbf{a}_{1} + \frac{1}{2} \, \mathbf{a}_{2} + z_{3} \, \mathbf{a}_{3} & = & \frac{1}{2}a \, \mathbf{\hat{x}} + \frac{1}{2}a \, \mathbf{\hat{y}} + z_{3}c \, \mathbf{\hat{z}} & \left(4d\right) & \mbox{Na} \\ 
\mathbf{B}_{7} & = & \frac{1}{2} \, \mathbf{a}_{1} + \frac{1}{2} \, \mathbf{a}_{2}-z_{3} \, \mathbf{a}_{3} & = & \frac{1}{2}a \, \mathbf{\hat{x}} + \frac{1}{2}a \, \mathbf{\hat{y}}-z_{3}c \, \mathbf{\hat{z}} & \left(4d\right) & \mbox{Na} \\ 
\mathbf{B}_{8} & = & -z_{3} \, \mathbf{a}_{3} & = & -z_{3}c \, \mathbf{\hat{z}} & \left(4d\right) & \mbox{Na} \\ 
\mathbf{B}_{9} & = & x_{4} \, \mathbf{a}_{1} + x_{4} \, \mathbf{a}_{2} & = & x_{4}a \, \mathbf{\hat{x}} + x_{4}a \, \mathbf{\hat{y}} & \left(4e\right) & \mbox{O II} \\ 
\mathbf{B}_{10} & = & -x_{4} \, \mathbf{a}_{1}-x_{4} \, \mathbf{a}_{2} & = & -x_{4}a \, \mathbf{\hat{x}}-x_{4}a \, \mathbf{\hat{y}} & \left(4e\right) & \mbox{O II} \\ 
\mathbf{B}_{11} & = & \left(\frac{1}{2} - x_{4}\right) \, \mathbf{a}_{1} + \left(\frac{1}{2} +x_{4}\right) \, \mathbf{a}_{2} & = & \left(\frac{1}{2} - x_{4}\right)a \, \mathbf{\hat{x}} + \left(\frac{1}{2} +x_{4}\right)a \, \mathbf{\hat{y}} & \left(4e\right) & \mbox{O II} \\ 
\mathbf{B}_{12} & = & \left(\frac{1}{2} +x_{4}\right) \, \mathbf{a}_{1} + \left(\frac{1}{2} - x_{4}\right) \, \mathbf{a}_{2} & = & \left(\frac{1}{2} +x_{4}\right)a \, \mathbf{\hat{x}} + \left(\frac{1}{2} - x_{4}\right)a \, \mathbf{\hat{y}} & \left(4e\right) & \mbox{O II} \\ 
\mathbf{B}_{13} & = & x_{5} \, \mathbf{a}_{1} + x_{5} \, \mathbf{a}_{2} + \frac{1}{2} \, \mathbf{a}_{3} & = & x_{5}a \, \mathbf{\hat{x}} + x_{5}a \, \mathbf{\hat{y}} + \frac{1}{2}c \, \mathbf{\hat{z}} & \left(4f\right) & \mbox{O III} \\ 
\mathbf{B}_{14} & = & -x_{5} \, \mathbf{a}_{1}-x_{5} \, \mathbf{a}_{2} + \frac{1}{2} \, \mathbf{a}_{3} & = & -x_{5}a \, \mathbf{\hat{x}}-x_{5}a \, \mathbf{\hat{y}} + \frac{1}{2}c \, \mathbf{\hat{z}} & \left(4f\right) & \mbox{O III} \\ 
\mathbf{B}_{15} & = & \left(\frac{1}{2} - x_{5}\right) \, \mathbf{a}_{1} + \left(\frac{1}{2} +x_{5}\right) \, \mathbf{a}_{2} + \frac{1}{2} \, \mathbf{a}_{3} & = & \left(\frac{1}{2} - x_{5}\right)a \, \mathbf{\hat{x}} + \left(\frac{1}{2} +x_{5}\right)a \, \mathbf{\hat{y}} + \frac{1}{2}c \, \mathbf{\hat{z}} & \left(4f\right) & \mbox{O III} \\ 
\mathbf{B}_{16} & = & \left(\frac{1}{2} +x_{5}\right) \, \mathbf{a}_{1} + \left(\frac{1}{2} - x_{5}\right) \, \mathbf{a}_{2} + \frac{1}{2} \, \mathbf{a}_{3} & = & \left(\frac{1}{2} +x_{5}\right)a \, \mathbf{\hat{x}} + \left(\frac{1}{2} - x_{5}\right)a \, \mathbf{\hat{y}} + \frac{1}{2}c \, \mathbf{\hat{z}} & \left(4f\right) & \mbox{O III} \\ 
\mathbf{B}_{17} & = & x_{6} \, \mathbf{a}_{1} + y_{6} \, \mathbf{a}_{2} + z_{6} \, \mathbf{a}_{3} & = & x_{6}a \, \mathbf{\hat{x}} + y_{6}a \, \mathbf{\hat{y}} + z_{6}c \, \mathbf{\hat{z}} & \left(8g\right) & \mbox{H} \\ 
\mathbf{B}_{18} & = & -x_{6} \, \mathbf{a}_{1}-y_{6} \, \mathbf{a}_{2} + z_{6} \, \mathbf{a}_{3} & = & -x_{6}a \, \mathbf{\hat{x}}-y_{6}a \, \mathbf{\hat{y}} + z_{6}c \, \mathbf{\hat{z}} & \left(8g\right) & \mbox{H} \\ 
\mathbf{B}_{19} & = & \left(\frac{1}{2} - y_{6}\right) \, \mathbf{a}_{1} + \left(\frac{1}{2} +x_{6}\right) \, \mathbf{a}_{2} + z_{6} \, \mathbf{a}_{3} & = & \left(\frac{1}{2} - y_{6}\right)a \, \mathbf{\hat{x}} + \left(\frac{1}{2} +x_{6}\right)a \, \mathbf{\hat{y}} + z_{6}c \, \mathbf{\hat{z}} & \left(8g\right) & \mbox{H} \\ 
\mathbf{B}_{20} & = & \left(\frac{1}{2} +y_{6}\right) \, \mathbf{a}_{1} + \left(\frac{1}{2} - x_{6}\right) \, \mathbf{a}_{2} + z_{6} \, \mathbf{a}_{3} & = & \left(\frac{1}{2} +y_{6}\right)a \, \mathbf{\hat{x}} + \left(\frac{1}{2} - x_{6}\right)a \, \mathbf{\hat{y}} + z_{6}c \, \mathbf{\hat{z}} & \left(8g\right) & \mbox{H} \\ 
\mathbf{B}_{21} & = & \left(\frac{1}{2} - x_{6}\right) \, \mathbf{a}_{1} + \left(\frac{1}{2} +y_{6}\right) \, \mathbf{a}_{2}-z_{6} \, \mathbf{a}_{3} & = & \left(\frac{1}{2} - x_{6}\right)a \, \mathbf{\hat{x}} + \left(\frac{1}{2} +y_{6}\right)a \, \mathbf{\hat{y}}-z_{6}c \, \mathbf{\hat{z}} & \left(8g\right) & \mbox{H} \\ 
\mathbf{B}_{22} & = & \left(\frac{1}{2} +x_{6}\right) \, \mathbf{a}_{1} + \left(\frac{1}{2} - y_{6}\right) \, \mathbf{a}_{2}-z_{6} \, \mathbf{a}_{3} & = & \left(\frac{1}{2} +x_{6}\right)a \, \mathbf{\hat{x}} + \left(\frac{1}{2} - y_{6}\right)a \, \mathbf{\hat{y}}-z_{6}c \, \mathbf{\hat{z}} & \left(8g\right) & \mbox{H} \\ 
\mathbf{B}_{23} & = & y_{6} \, \mathbf{a}_{1} + x_{6} \, \mathbf{a}_{2}-z_{6} \, \mathbf{a}_{3} & = & y_{6}a \, \mathbf{\hat{x}} + x_{6}a \, \mathbf{\hat{y}}-z_{6}c \, \mathbf{\hat{z}} & \left(8g\right) & \mbox{H} \\ 
\mathbf{B}_{24} & = & -y_{6} \, \mathbf{a}_{1}-x_{6} \, \mathbf{a}_{2}-z_{6} \, \mathbf{a}_{3} & = & -y_{6}a \, \mathbf{\hat{x}}-x_{6}a \, \mathbf{\hat{y}}-z_{6}c \, \mathbf{\hat{z}} & \left(8g\right) & \mbox{H} \\ 
\mathbf{B}_{25} & = & x_{7} \, \mathbf{a}_{1} + y_{7} \, \mathbf{a}_{2} + z_{7} \, \mathbf{a}_{3} & = & x_{7}a \, \mathbf{\hat{x}} + y_{7}a \, \mathbf{\hat{y}} + z_{7}c \, \mathbf{\hat{z}} & \left(8g\right) & \mbox{O IV} \\ 
\mathbf{B}_{26} & = & -x_{7} \, \mathbf{a}_{1}-y_{7} \, \mathbf{a}_{2} + z_{7} \, \mathbf{a}_{3} & = & -x_{7}a \, \mathbf{\hat{x}}-y_{7}a \, \mathbf{\hat{y}} + z_{7}c \, \mathbf{\hat{z}} & \left(8g\right) & \mbox{O IV} \\ 
\mathbf{B}_{27} & = & \left(\frac{1}{2} - y_{7}\right) \, \mathbf{a}_{1} + \left(\frac{1}{2} +x_{7}\right) \, \mathbf{a}_{2} + z_{7} \, \mathbf{a}_{3} & = & \left(\frac{1}{2} - y_{7}\right)a \, \mathbf{\hat{x}} + \left(\frac{1}{2} +x_{7}\right)a \, \mathbf{\hat{y}} + z_{7}c \, \mathbf{\hat{z}} & \left(8g\right) & \mbox{O IV} \\ 
\mathbf{B}_{28} & = & \left(\frac{1}{2} +y_{7}\right) \, \mathbf{a}_{1} + \left(\frac{1}{2} - x_{7}\right) \, \mathbf{a}_{2} + z_{7} \, \mathbf{a}_{3} & = & \left(\frac{1}{2} +y_{7}\right)a \, \mathbf{\hat{x}} + \left(\frac{1}{2} - x_{7}\right)a \, \mathbf{\hat{y}} + z_{7}c \, \mathbf{\hat{z}} & \left(8g\right) & \mbox{O IV} \\ 
\mathbf{B}_{29} & = & \left(\frac{1}{2} - x_{7}\right) \, \mathbf{a}_{1} + \left(\frac{1}{2} +y_{7}\right) \, \mathbf{a}_{2}-z_{7} \, \mathbf{a}_{3} & = & \left(\frac{1}{2} - x_{7}\right)a \, \mathbf{\hat{x}} + \left(\frac{1}{2} +y_{7}\right)a \, \mathbf{\hat{y}}-z_{7}c \, \mathbf{\hat{z}} & \left(8g\right) & \mbox{O IV} \\ 
\mathbf{B}_{30} & = & \left(\frac{1}{2} +x_{7}\right) \, \mathbf{a}_{1} + \left(\frac{1}{2} - y_{7}\right) \, \mathbf{a}_{2}-z_{7} \, \mathbf{a}_{3} & = & \left(\frac{1}{2} +x_{7}\right)a \, \mathbf{\hat{x}} + \left(\frac{1}{2} - y_{7}\right)a \, \mathbf{\hat{y}}-z_{7}c \, \mathbf{\hat{z}} & \left(8g\right) & \mbox{O IV} \\ 
\mathbf{B}_{31} & = & y_{7} \, \mathbf{a}_{1} + x_{7} \, \mathbf{a}_{2}-z_{7} \, \mathbf{a}_{3} & = & y_{7}a \, \mathbf{\hat{x}} + x_{7}a \, \mathbf{\hat{y}}-z_{7}c \, \mathbf{\hat{z}} & \left(8g\right) & \mbox{O IV} \\ 
\mathbf{B}_{32} & = & -y_{7} \, \mathbf{a}_{1}-x_{7} \, \mathbf{a}_{2}-z_{7} \, \mathbf{a}_{3} & = & -y_{7}a \, \mathbf{\hat{x}}-x_{7}a \, \mathbf{\hat{y}}-z_{7}c \, \mathbf{\hat{z}} & \left(8g\right) & \mbox{O IV} \\ 
\mathbf{B}_{33} & = & x_{8} \, \mathbf{a}_{1} + y_{8} \, \mathbf{a}_{2} + z_{8} \, \mathbf{a}_{3} & = & x_{8}a \, \mathbf{\hat{x}} + y_{8}a \, \mathbf{\hat{y}} + z_{8}c \, \mathbf{\hat{z}} & \left(8g\right) & \mbox{O V} \\ 
\mathbf{B}_{34} & = & -x_{8} \, \mathbf{a}_{1}-y_{8} \, \mathbf{a}_{2} + z_{8} \, \mathbf{a}_{3} & = & -x_{8}a \, \mathbf{\hat{x}}-y_{8}a \, \mathbf{\hat{y}} + z_{8}c \, \mathbf{\hat{z}} & \left(8g\right) & \mbox{O V} \\ 
\mathbf{B}_{35} & = & \left(\frac{1}{2} - y_{8}\right) \, \mathbf{a}_{1} + \left(\frac{1}{2} +x_{8}\right) \, \mathbf{a}_{2} + z_{8} \, \mathbf{a}_{3} & = & \left(\frac{1}{2} - y_{8}\right)a \, \mathbf{\hat{x}} + \left(\frac{1}{2} +x_{8}\right)a \, \mathbf{\hat{y}} + z_{8}c \, \mathbf{\hat{z}} & \left(8g\right) & \mbox{O V} \\ 
\mathbf{B}_{36} & = & \left(\frac{1}{2} +y_{8}\right) \, \mathbf{a}_{1} + \left(\frac{1}{2} - x_{8}\right) \, \mathbf{a}_{2} + z_{8} \, \mathbf{a}_{3} & = & \left(\frac{1}{2} +y_{8}\right)a \, \mathbf{\hat{x}} + \left(\frac{1}{2} - x_{8}\right)a \, \mathbf{\hat{y}} + z_{8}c \, \mathbf{\hat{z}} & \left(8g\right) & \mbox{O V} \\ 
\mathbf{B}_{37} & = & \left(\frac{1}{2} - x_{8}\right) \, \mathbf{a}_{1} + \left(\frac{1}{2} +y_{8}\right) \, \mathbf{a}_{2}-z_{8} \, \mathbf{a}_{3} & = & \left(\frac{1}{2} - x_{8}\right)a \, \mathbf{\hat{x}} + \left(\frac{1}{2} +y_{8}\right)a \, \mathbf{\hat{y}}-z_{8}c \, \mathbf{\hat{z}} & \left(8g\right) & \mbox{O V} \\ 
\mathbf{B}_{38} & = & \left(\frac{1}{2} +x_{8}\right) \, \mathbf{a}_{1} + \left(\frac{1}{2} - y_{8}\right) \, \mathbf{a}_{2}-z_{8} \, \mathbf{a}_{3} & = & \left(\frac{1}{2} +x_{8}\right)a \, \mathbf{\hat{x}} + \left(\frac{1}{2} - y_{8}\right)a \, \mathbf{\hat{y}}-z_{8}c \, \mathbf{\hat{z}} & \left(8g\right) & \mbox{O V} \\ 
\mathbf{B}_{39} & = & y_{8} \, \mathbf{a}_{1} + x_{8} \, \mathbf{a}_{2}-z_{8} \, \mathbf{a}_{3} & = & y_{8}a \, \mathbf{\hat{x}} + x_{8}a \, \mathbf{\hat{y}}-z_{8}c \, \mathbf{\hat{z}} & \left(8g\right) & \mbox{O V} \\ 
\mathbf{B}_{40} & = & -y_{8} \, \mathbf{a}_{1}-x_{8} \, \mathbf{a}_{2}-z_{8} \, \mathbf{a}_{3} & = & -y_{8}a \, \mathbf{\hat{x}}-x_{8}a \, \mathbf{\hat{y}}-z_{8}c \, \mathbf{\hat{z}} & \left(8g\right) & \mbox{O V} \\ 
\end{longtabu}
\renewcommand{\arraystretch}{1.0}
\noindent \hrulefill
\\
\textbf{References:}
\vspace*{-0.25cm}
\begin{flushleft}
  - \bibentry{Ferdov_Na4Ti2Si8O22H2O4_ActCrystallogrSecE_2007}. \\
\end{flushleft}
\textbf{Found in:}
\vspace*{-0.25cm}
\begin{flushleft}
  - \bibentry{Villars_PearsonsCrystalData_2013}. \\
\end{flushleft}
\noindent \hrulefill
\\
\textbf{Geometry files:}
\\
\noindent  - CIF: pp. {\hyperref[A4B2C13D_tP40_90_g_d_cef2g_c_cif]{\pageref{A4B2C13D_tP40_90_g_d_cef2g_c_cif}}} \\
\noindent  - POSCAR: pp. {\hyperref[A4B2C13D_tP40_90_g_d_cef2g_c_poscar]{\pageref{A4B2C13D_tP40_90_g_d_cef2g_c_poscar}}} \\
\onecolumn
{\phantomsection\label{AB4C17D4E_tP54_90_a_g_c4g_g_c}}
\subsection*{\huge \textbf{{\normalfont \begin{raggedleft}BaCu$_{4}$[VO][PO$_{4}$]$_{4}$ Structure: \end{raggedleft} \\ AB4C17D4E\_tP54\_90\_a\_g\_c4g\_g\_c}}}
\noindent \hrulefill
\vspace*{0.25cm}
\begin{figure}[htp]
  \centering
  \vspace{-1em}
  {\includegraphics[width=1\textwidth]{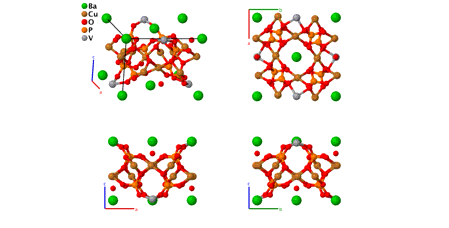}}
\end{figure}
\vspace*{-0.5cm}
\renewcommand{\arraystretch}{1.5}
\begin{equation*}
  \begin{array}{>{$\hspace{-0.15cm}}l<{$}>{$}p{0.5cm}<{$}>{$}p{18.5cm}<{$}}
    \mbox{\large \textbf{Prototype}} &\colon & \ce{BaCu4[VO][PO4]4} \\
    \mbox{\large \textbf{\AFLOW\ prototype label}} &\colon & \mbox{AB4C17D4E\_tP54\_90\_a\_g\_c4g\_g\_c} \\
    \mbox{\large \textbf{\textit{Strukturbericht} designation}} &\colon & \mbox{None} \\
    \mbox{\large \textbf{Pearson symbol}} &\colon & \mbox{tP54} \\
    \mbox{\large \textbf{Space group number}} &\colon & 90 \\
    \mbox{\large \textbf{Space group symbol}} &\colon & P42_{1}2 \\
    \mbox{\large \textbf{\AFLOW\ prototype command}} &\colon &  \texttt{aflow} \,  \, \texttt{-{}-proto=AB4C17D4E\_tP54\_90\_a\_g\_c4g\_g\_c } \, \newline \texttt{-{}-params=}{a,c/a,z_{2},z_{3},x_{4},y_{4},z_{4},x_{5},y_{5},z_{5},x_{6},y_{6},z_{6},x_{7},y_{7},z_{7},x_{8},y_{8},z_{8},x_{9},y_{9},} \newline {z_{9} }
  \end{array}
\end{equation*}
\renewcommand{\arraystretch}{1.0}

\noindent \parbox{1 \linewidth}{
\noindent \hrulefill
\\
\textbf{Simple Tetragonal primitive vectors:} \\
\vspace*{-0.25cm}
\begin{tabular}{cc}
  \begin{tabular}{c}
    \parbox{0.6 \linewidth}{
      \renewcommand{\arraystretch}{1.5}
      \begin{equation*}
        \centering
        \begin{array}{ccc}
              \mathbf{a}_1 & = & a \, \mathbf{\hat{x}} \\
    \mathbf{a}_2 & = & a \, \mathbf{\hat{y}} \\
    \mathbf{a}_3 & = & c \, \mathbf{\hat{z}} \\

        \end{array}
      \end{equation*}
    }
    \renewcommand{\arraystretch}{1.0}
  \end{tabular}
  \begin{tabular}{c}
    \includegraphics[width=0.3\linewidth]{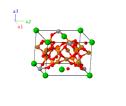} \\
  \end{tabular}
\end{tabular}

}
\vspace*{-0.25cm}

\noindent \hrulefill
\\
\textbf{Basis vectors:}
\vspace*{-0.25cm}
\renewcommand{\arraystretch}{1.5}
\begin{longtabu} to \textwidth{>{\centering $}X[-1,c,c]<{$}>{\centering $}X[-1,c,c]<{$}>{\centering $}X[-1,c,c]<{$}>{\centering $}X[-1,c,c]<{$}>{\centering $}X[-1,c,c]<{$}>{\centering $}X[-1,c,c]<{$}>{\centering $}X[-1,c,c]<{$}}
  & & \mbox{Lattice Coordinates} & & \mbox{Cartesian Coordinates} &\mbox{Wyckoff Position} & \mbox{Atom Type} \\  
  \mathbf{B}_{1} & = & 0 \, \mathbf{a}_{1} + 0 \, \mathbf{a}_{2} + 0 \, \mathbf{a}_{3} & = & 0 \, \mathbf{\hat{x}} + 0 \, \mathbf{\hat{y}} + 0 \, \mathbf{\hat{z}} & \left(2a\right) & \mbox{Ba} \\ 
\mathbf{B}_{2} & = & \frac{1}{2} \, \mathbf{a}_{1} + \frac{1}{2} \, \mathbf{a}_{2} & = & \frac{1}{2}a \, \mathbf{\hat{x}} + \frac{1}{2}a \, \mathbf{\hat{y}} & \left(2a\right) & \mbox{Ba} \\ 
\mathbf{B}_{3} & = & \frac{1}{2} \, \mathbf{a}_{2} + z_{2} \, \mathbf{a}_{3} & = & \frac{1}{2}a \, \mathbf{\hat{y}} + z_{2}c \, \mathbf{\hat{z}} & \left(2c\right) & \mbox{O I} \\ 
\mathbf{B}_{4} & = & \frac{1}{2} \, \mathbf{a}_{1} + -z_{2} \, \mathbf{a}_{3} & = & \frac{1}{2}a \, \mathbf{\hat{x}} + -z_{2}c \, \mathbf{\hat{z}} & \left(2c\right) & \mbox{O I} \\ 
\mathbf{B}_{5} & = & \frac{1}{2} \, \mathbf{a}_{2} + z_{3} \, \mathbf{a}_{3} & = & \frac{1}{2}a \, \mathbf{\hat{y}} + z_{3}c \, \mathbf{\hat{z}} & \left(2c\right) & \mbox{V} \\ 
\mathbf{B}_{6} & = & \frac{1}{2} \, \mathbf{a}_{1} + -z_{3} \, \mathbf{a}_{3} & = & \frac{1}{2}a \, \mathbf{\hat{x}} + -z_{3}c \, \mathbf{\hat{z}} & \left(2c\right) & \mbox{V} \\ 
\mathbf{B}_{7} & = & x_{4} \, \mathbf{a}_{1} + y_{4} \, \mathbf{a}_{2} + z_{4} \, \mathbf{a}_{3} & = & x_{4}a \, \mathbf{\hat{x}} + y_{4}a \, \mathbf{\hat{y}} + z_{4}c \, \mathbf{\hat{z}} & \left(8g\right) & \mbox{Cu} \\ 
\mathbf{B}_{8} & = & -x_{4} \, \mathbf{a}_{1}-y_{4} \, \mathbf{a}_{2} + z_{4} \, \mathbf{a}_{3} & = & -x_{4}a \, \mathbf{\hat{x}}-y_{4}a \, \mathbf{\hat{y}} + z_{4}c \, \mathbf{\hat{z}} & \left(8g\right) & \mbox{Cu} \\ 
\mathbf{B}_{9} & = & \left(\frac{1}{2} - y_{4}\right) \, \mathbf{a}_{1} + \left(\frac{1}{2} +x_{4}\right) \, \mathbf{a}_{2} + z_{4} \, \mathbf{a}_{3} & = & \left(\frac{1}{2} - y_{4}\right)a \, \mathbf{\hat{x}} + \left(\frac{1}{2} +x_{4}\right)a \, \mathbf{\hat{y}} + z_{4}c \, \mathbf{\hat{z}} & \left(8g\right) & \mbox{Cu} \\ 
\mathbf{B}_{10} & = & \left(\frac{1}{2} +y_{4}\right) \, \mathbf{a}_{1} + \left(\frac{1}{2} - x_{4}\right) \, \mathbf{a}_{2} + z_{4} \, \mathbf{a}_{3} & = & \left(\frac{1}{2} +y_{4}\right)a \, \mathbf{\hat{x}} + \left(\frac{1}{2} - x_{4}\right)a \, \mathbf{\hat{y}} + z_{4}c \, \mathbf{\hat{z}} & \left(8g\right) & \mbox{Cu} \\ 
\mathbf{B}_{11} & = & \left(\frac{1}{2} - x_{4}\right) \, \mathbf{a}_{1} + \left(\frac{1}{2} +y_{4}\right) \, \mathbf{a}_{2}-z_{4} \, \mathbf{a}_{3} & = & \left(\frac{1}{2} - x_{4}\right)a \, \mathbf{\hat{x}} + \left(\frac{1}{2} +y_{4}\right)a \, \mathbf{\hat{y}}-z_{4}c \, \mathbf{\hat{z}} & \left(8g\right) & \mbox{Cu} \\ 
\mathbf{B}_{12} & = & \left(\frac{1}{2} +x_{4}\right) \, \mathbf{a}_{1} + \left(\frac{1}{2} - y_{4}\right) \, \mathbf{a}_{2}-z_{4} \, \mathbf{a}_{3} & = & \left(\frac{1}{2} +x_{4}\right)a \, \mathbf{\hat{x}} + \left(\frac{1}{2} - y_{4}\right)a \, \mathbf{\hat{y}}-z_{4}c \, \mathbf{\hat{z}} & \left(8g\right) & \mbox{Cu} \\ 
\mathbf{B}_{13} & = & y_{4} \, \mathbf{a}_{1} + x_{4} \, \mathbf{a}_{2}-z_{4} \, \mathbf{a}_{3} & = & y_{4}a \, \mathbf{\hat{x}} + x_{4}a \, \mathbf{\hat{y}}-z_{4}c \, \mathbf{\hat{z}} & \left(8g\right) & \mbox{Cu} \\ 
\mathbf{B}_{14} & = & -y_{4} \, \mathbf{a}_{1}-x_{4} \, \mathbf{a}_{2}-z_{4} \, \mathbf{a}_{3} & = & -y_{4}a \, \mathbf{\hat{x}}-x_{4}a \, \mathbf{\hat{y}}-z_{4}c \, \mathbf{\hat{z}} & \left(8g\right) & \mbox{Cu} \\ 
\mathbf{B}_{15} & = & x_{5} \, \mathbf{a}_{1} + y_{5} \, \mathbf{a}_{2} + z_{5} \, \mathbf{a}_{3} & = & x_{5}a \, \mathbf{\hat{x}} + y_{5}a \, \mathbf{\hat{y}} + z_{5}c \, \mathbf{\hat{z}} & \left(8g\right) & \mbox{O II} \\ 
\mathbf{B}_{16} & = & -x_{5} \, \mathbf{a}_{1}-y_{5} \, \mathbf{a}_{2} + z_{5} \, \mathbf{a}_{3} & = & -x_{5}a \, \mathbf{\hat{x}}-y_{5}a \, \mathbf{\hat{y}} + z_{5}c \, \mathbf{\hat{z}} & \left(8g\right) & \mbox{O II} \\ 
\mathbf{B}_{17} & = & \left(\frac{1}{2} - y_{5}\right) \, \mathbf{a}_{1} + \left(\frac{1}{2} +x_{5}\right) \, \mathbf{a}_{2} + z_{5} \, \mathbf{a}_{3} & = & \left(\frac{1}{2} - y_{5}\right)a \, \mathbf{\hat{x}} + \left(\frac{1}{2} +x_{5}\right)a \, \mathbf{\hat{y}} + z_{5}c \, \mathbf{\hat{z}} & \left(8g\right) & \mbox{O II} \\ 
\mathbf{B}_{18} & = & \left(\frac{1}{2} +y_{5}\right) \, \mathbf{a}_{1} + \left(\frac{1}{2} - x_{5}\right) \, \mathbf{a}_{2} + z_{5} \, \mathbf{a}_{3} & = & \left(\frac{1}{2} +y_{5}\right)a \, \mathbf{\hat{x}} + \left(\frac{1}{2} - x_{5}\right)a \, \mathbf{\hat{y}} + z_{5}c \, \mathbf{\hat{z}} & \left(8g\right) & \mbox{O II} \\ 
\mathbf{B}_{19} & = & \left(\frac{1}{2} - x_{5}\right) \, \mathbf{a}_{1} + \left(\frac{1}{2} +y_{5}\right) \, \mathbf{a}_{2}-z_{5} \, \mathbf{a}_{3} & = & \left(\frac{1}{2} - x_{5}\right)a \, \mathbf{\hat{x}} + \left(\frac{1}{2} +y_{5}\right)a \, \mathbf{\hat{y}}-z_{5}c \, \mathbf{\hat{z}} & \left(8g\right) & \mbox{O II} \\ 
\mathbf{B}_{20} & = & \left(\frac{1}{2} +x_{5}\right) \, \mathbf{a}_{1} + \left(\frac{1}{2} - y_{5}\right) \, \mathbf{a}_{2}-z_{5} \, \mathbf{a}_{3} & = & \left(\frac{1}{2} +x_{5}\right)a \, \mathbf{\hat{x}} + \left(\frac{1}{2} - y_{5}\right)a \, \mathbf{\hat{y}}-z_{5}c \, \mathbf{\hat{z}} & \left(8g\right) & \mbox{O II} \\ 
\mathbf{B}_{21} & = & y_{5} \, \mathbf{a}_{1} + x_{5} \, \mathbf{a}_{2}-z_{5} \, \mathbf{a}_{3} & = & y_{5}a \, \mathbf{\hat{x}} + x_{5}a \, \mathbf{\hat{y}}-z_{5}c \, \mathbf{\hat{z}} & \left(8g\right) & \mbox{O II} \\ 
\mathbf{B}_{22} & = & -y_{5} \, \mathbf{a}_{1}-x_{5} \, \mathbf{a}_{2}-z_{5} \, \mathbf{a}_{3} & = & -y_{5}a \, \mathbf{\hat{x}}-x_{5}a \, \mathbf{\hat{y}}-z_{5}c \, \mathbf{\hat{z}} & \left(8g\right) & \mbox{O II} \\ 
\mathbf{B}_{23} & = & x_{6} \, \mathbf{a}_{1} + y_{6} \, \mathbf{a}_{2} + z_{6} \, \mathbf{a}_{3} & = & x_{6}a \, \mathbf{\hat{x}} + y_{6}a \, \mathbf{\hat{y}} + z_{6}c \, \mathbf{\hat{z}} & \left(8g\right) & \mbox{O III} \\ 
\mathbf{B}_{24} & = & -x_{6} \, \mathbf{a}_{1}-y_{6} \, \mathbf{a}_{2} + z_{6} \, \mathbf{a}_{3} & = & -x_{6}a \, \mathbf{\hat{x}}-y_{6}a \, \mathbf{\hat{y}} + z_{6}c \, \mathbf{\hat{z}} & \left(8g\right) & \mbox{O III} \\ 
\mathbf{B}_{25} & = & \left(\frac{1}{2} - y_{6}\right) \, \mathbf{a}_{1} + \left(\frac{1}{2} +x_{6}\right) \, \mathbf{a}_{2} + z_{6} \, \mathbf{a}_{3} & = & \left(\frac{1}{2} - y_{6}\right)a \, \mathbf{\hat{x}} + \left(\frac{1}{2} +x_{6}\right)a \, \mathbf{\hat{y}} + z_{6}c \, \mathbf{\hat{z}} & \left(8g\right) & \mbox{O III} \\ 
\mathbf{B}_{26} & = & \left(\frac{1}{2} +y_{6}\right) \, \mathbf{a}_{1} + \left(\frac{1}{2} - x_{6}\right) \, \mathbf{a}_{2} + z_{6} \, \mathbf{a}_{3} & = & \left(\frac{1}{2} +y_{6}\right)a \, \mathbf{\hat{x}} + \left(\frac{1}{2} - x_{6}\right)a \, \mathbf{\hat{y}} + z_{6}c \, \mathbf{\hat{z}} & \left(8g\right) & \mbox{O III} \\ 
\mathbf{B}_{27} & = & \left(\frac{1}{2} - x_{6}\right) \, \mathbf{a}_{1} + \left(\frac{1}{2} +y_{6}\right) \, \mathbf{a}_{2}-z_{6} \, \mathbf{a}_{3} & = & \left(\frac{1}{2} - x_{6}\right)a \, \mathbf{\hat{x}} + \left(\frac{1}{2} +y_{6}\right)a \, \mathbf{\hat{y}}-z_{6}c \, \mathbf{\hat{z}} & \left(8g\right) & \mbox{O III} \\ 
\mathbf{B}_{28} & = & \left(\frac{1}{2} +x_{6}\right) \, \mathbf{a}_{1} + \left(\frac{1}{2} - y_{6}\right) \, \mathbf{a}_{2}-z_{6} \, \mathbf{a}_{3} & = & \left(\frac{1}{2} +x_{6}\right)a \, \mathbf{\hat{x}} + \left(\frac{1}{2} - y_{6}\right)a \, \mathbf{\hat{y}}-z_{6}c \, \mathbf{\hat{z}} & \left(8g\right) & \mbox{O III} \\ 
\mathbf{B}_{29} & = & y_{6} \, \mathbf{a}_{1} + x_{6} \, \mathbf{a}_{2}-z_{6} \, \mathbf{a}_{3} & = & y_{6}a \, \mathbf{\hat{x}} + x_{6}a \, \mathbf{\hat{y}}-z_{6}c \, \mathbf{\hat{z}} & \left(8g\right) & \mbox{O III} \\ 
\mathbf{B}_{30} & = & -y_{6} \, \mathbf{a}_{1}-x_{6} \, \mathbf{a}_{2}-z_{6} \, \mathbf{a}_{3} & = & -y_{6}a \, \mathbf{\hat{x}}-x_{6}a \, \mathbf{\hat{y}}-z_{6}c \, \mathbf{\hat{z}} & \left(8g\right) & \mbox{O III} \\ 
\mathbf{B}_{31} & = & x_{7} \, \mathbf{a}_{1} + y_{7} \, \mathbf{a}_{2} + z_{7} \, \mathbf{a}_{3} & = & x_{7}a \, \mathbf{\hat{x}} + y_{7}a \, \mathbf{\hat{y}} + z_{7}c \, \mathbf{\hat{z}} & \left(8g\right) & \mbox{O IV} \\ 
\mathbf{B}_{32} & = & -x_{7} \, \mathbf{a}_{1}-y_{7} \, \mathbf{a}_{2} + z_{7} \, \mathbf{a}_{3} & = & -x_{7}a \, \mathbf{\hat{x}}-y_{7}a \, \mathbf{\hat{y}} + z_{7}c \, \mathbf{\hat{z}} & \left(8g\right) & \mbox{O IV} \\ 
\mathbf{B}_{33} & = & \left(\frac{1}{2} - y_{7}\right) \, \mathbf{a}_{1} + \left(\frac{1}{2} +x_{7}\right) \, \mathbf{a}_{2} + z_{7} \, \mathbf{a}_{3} & = & \left(\frac{1}{2} - y_{7}\right)a \, \mathbf{\hat{x}} + \left(\frac{1}{2} +x_{7}\right)a \, \mathbf{\hat{y}} + z_{7}c \, \mathbf{\hat{z}} & \left(8g\right) & \mbox{O IV} \\ 
\mathbf{B}_{34} & = & \left(\frac{1}{2} +y_{7}\right) \, \mathbf{a}_{1} + \left(\frac{1}{2} - x_{7}\right) \, \mathbf{a}_{2} + z_{7} \, \mathbf{a}_{3} & = & \left(\frac{1}{2} +y_{7}\right)a \, \mathbf{\hat{x}} + \left(\frac{1}{2} - x_{7}\right)a \, \mathbf{\hat{y}} + z_{7}c \, \mathbf{\hat{z}} & \left(8g\right) & \mbox{O IV} \\ 
\mathbf{B}_{35} & = & \left(\frac{1}{2} - x_{7}\right) \, \mathbf{a}_{1} + \left(\frac{1}{2} +y_{7}\right) \, \mathbf{a}_{2}-z_{7} \, \mathbf{a}_{3} & = & \left(\frac{1}{2} - x_{7}\right)a \, \mathbf{\hat{x}} + \left(\frac{1}{2} +y_{7}\right)a \, \mathbf{\hat{y}}-z_{7}c \, \mathbf{\hat{z}} & \left(8g\right) & \mbox{O IV} \\ 
\mathbf{B}_{36} & = & \left(\frac{1}{2} +x_{7}\right) \, \mathbf{a}_{1} + \left(\frac{1}{2} - y_{7}\right) \, \mathbf{a}_{2}-z_{7} \, \mathbf{a}_{3} & = & \left(\frac{1}{2} +x_{7}\right)a \, \mathbf{\hat{x}} + \left(\frac{1}{2} - y_{7}\right)a \, \mathbf{\hat{y}}-z_{7}c \, \mathbf{\hat{z}} & \left(8g\right) & \mbox{O IV} \\ 
\mathbf{B}_{37} & = & y_{7} \, \mathbf{a}_{1} + x_{7} \, \mathbf{a}_{2}-z_{7} \, \mathbf{a}_{3} & = & y_{7}a \, \mathbf{\hat{x}} + x_{7}a \, \mathbf{\hat{y}}-z_{7}c \, \mathbf{\hat{z}} & \left(8g\right) & \mbox{O IV} \\ 
\mathbf{B}_{38} & = & -y_{7} \, \mathbf{a}_{1}-x_{7} \, \mathbf{a}_{2}-z_{7} \, \mathbf{a}_{3} & = & -y_{7}a \, \mathbf{\hat{x}}-x_{7}a \, \mathbf{\hat{y}}-z_{7}c \, \mathbf{\hat{z}} & \left(8g\right) & \mbox{O IV} \\ 
\mathbf{B}_{39} & = & x_{8} \, \mathbf{a}_{1} + y_{8} \, \mathbf{a}_{2} + z_{8} \, \mathbf{a}_{3} & = & x_{8}a \, \mathbf{\hat{x}} + y_{8}a \, \mathbf{\hat{y}} + z_{8}c \, \mathbf{\hat{z}} & \left(8g\right) & \mbox{O V} \\ 
\mathbf{B}_{40} & = & -x_{8} \, \mathbf{a}_{1}-y_{8} \, \mathbf{a}_{2} + z_{8} \, \mathbf{a}_{3} & = & -x_{8}a \, \mathbf{\hat{x}}-y_{8}a \, \mathbf{\hat{y}} + z_{8}c \, \mathbf{\hat{z}} & \left(8g\right) & \mbox{O V} \\ 
\mathbf{B}_{41} & = & \left(\frac{1}{2} - y_{8}\right) \, \mathbf{a}_{1} + \left(\frac{1}{2} +x_{8}\right) \, \mathbf{a}_{2} + z_{8} \, \mathbf{a}_{3} & = & \left(\frac{1}{2} - y_{8}\right)a \, \mathbf{\hat{x}} + \left(\frac{1}{2} +x_{8}\right)a \, \mathbf{\hat{y}} + z_{8}c \, \mathbf{\hat{z}} & \left(8g\right) & \mbox{O V} \\ 
\mathbf{B}_{42} & = & \left(\frac{1}{2} +y_{8}\right) \, \mathbf{a}_{1} + \left(\frac{1}{2} - x_{8}\right) \, \mathbf{a}_{2} + z_{8} \, \mathbf{a}_{3} & = & \left(\frac{1}{2} +y_{8}\right)a \, \mathbf{\hat{x}} + \left(\frac{1}{2} - x_{8}\right)a \, \mathbf{\hat{y}} + z_{8}c \, \mathbf{\hat{z}} & \left(8g\right) & \mbox{O V} \\ 
\mathbf{B}_{43} & = & \left(\frac{1}{2} - x_{8}\right) \, \mathbf{a}_{1} + \left(\frac{1}{2} +y_{8}\right) \, \mathbf{a}_{2}-z_{8} \, \mathbf{a}_{3} & = & \left(\frac{1}{2} - x_{8}\right)a \, \mathbf{\hat{x}} + \left(\frac{1}{2} +y_{8}\right)a \, \mathbf{\hat{y}}-z_{8}c \, \mathbf{\hat{z}} & \left(8g\right) & \mbox{O V} \\ 
\mathbf{B}_{44} & = & \left(\frac{1}{2} +x_{8}\right) \, \mathbf{a}_{1} + \left(\frac{1}{2} - y_{8}\right) \, \mathbf{a}_{2}-z_{8} \, \mathbf{a}_{3} & = & \left(\frac{1}{2} +x_{8}\right)a \, \mathbf{\hat{x}} + \left(\frac{1}{2} - y_{8}\right)a \, \mathbf{\hat{y}}-z_{8}c \, \mathbf{\hat{z}} & \left(8g\right) & \mbox{O V} \\ 
\mathbf{B}_{45} & = & y_{8} \, \mathbf{a}_{1} + x_{8} \, \mathbf{a}_{2}-z_{8} \, \mathbf{a}_{3} & = & y_{8}a \, \mathbf{\hat{x}} + x_{8}a \, \mathbf{\hat{y}}-z_{8}c \, \mathbf{\hat{z}} & \left(8g\right) & \mbox{O V} \\ 
\mathbf{B}_{46} & = & -y_{8} \, \mathbf{a}_{1}-x_{8} \, \mathbf{a}_{2}-z_{8} \, \mathbf{a}_{3} & = & -y_{8}a \, \mathbf{\hat{x}}-x_{8}a \, \mathbf{\hat{y}}-z_{8}c \, \mathbf{\hat{z}} & \left(8g\right) & \mbox{O V} \\ 
\mathbf{B}_{47} & = & x_{9} \, \mathbf{a}_{1} + y_{9} \, \mathbf{a}_{2} + z_{9} \, \mathbf{a}_{3} & = & x_{9}a \, \mathbf{\hat{x}} + y_{9}a \, \mathbf{\hat{y}} + z_{9}c \, \mathbf{\hat{z}} & \left(8g\right) & \mbox{P} \\ 
\mathbf{B}_{48} & = & -x_{9} \, \mathbf{a}_{1}-y_{9} \, \mathbf{a}_{2} + z_{9} \, \mathbf{a}_{3} & = & -x_{9}a \, \mathbf{\hat{x}}-y_{9}a \, \mathbf{\hat{y}} + z_{9}c \, \mathbf{\hat{z}} & \left(8g\right) & \mbox{P} \\ 
\mathbf{B}_{49} & = & \left(\frac{1}{2} - y_{9}\right) \, \mathbf{a}_{1} + \left(\frac{1}{2} +x_{9}\right) \, \mathbf{a}_{2} + z_{9} \, \mathbf{a}_{3} & = & \left(\frac{1}{2} - y_{9}\right)a \, \mathbf{\hat{x}} + \left(\frac{1}{2} +x_{9}\right)a \, \mathbf{\hat{y}} + z_{9}c \, \mathbf{\hat{z}} & \left(8g\right) & \mbox{P} \\ 
\mathbf{B}_{50} & = & \left(\frac{1}{2} +y_{9}\right) \, \mathbf{a}_{1} + \left(\frac{1}{2} - x_{9}\right) \, \mathbf{a}_{2} + z_{9} \, \mathbf{a}_{3} & = & \left(\frac{1}{2} +y_{9}\right)a \, \mathbf{\hat{x}} + \left(\frac{1}{2} - x_{9}\right)a \, \mathbf{\hat{y}} + z_{9}c \, \mathbf{\hat{z}} & \left(8g\right) & \mbox{P} \\ 
\mathbf{B}_{51} & = & \left(\frac{1}{2} - x_{9}\right) \, \mathbf{a}_{1} + \left(\frac{1}{2} +y_{9}\right) \, \mathbf{a}_{2}-z_{9} \, \mathbf{a}_{3} & = & \left(\frac{1}{2} - x_{9}\right)a \, \mathbf{\hat{x}} + \left(\frac{1}{2} +y_{9}\right)a \, \mathbf{\hat{y}}-z_{9}c \, \mathbf{\hat{z}} & \left(8g\right) & \mbox{P} \\ 
\mathbf{B}_{52} & = & \left(\frac{1}{2} +x_{9}\right) \, \mathbf{a}_{1} + \left(\frac{1}{2} - y_{9}\right) \, \mathbf{a}_{2}-z_{9} \, \mathbf{a}_{3} & = & \left(\frac{1}{2} +x_{9}\right)a \, \mathbf{\hat{x}} + \left(\frac{1}{2} - y_{9}\right)a \, \mathbf{\hat{y}}-z_{9}c \, \mathbf{\hat{z}} & \left(8g\right) & \mbox{P} \\ 
\mathbf{B}_{53} & = & y_{9} \, \mathbf{a}_{1} + x_{9} \, \mathbf{a}_{2}-z_{9} \, \mathbf{a}_{3} & = & y_{9}a \, \mathbf{\hat{x}} + x_{9}a \, \mathbf{\hat{y}}-z_{9}c \, \mathbf{\hat{z}} & \left(8g\right) & \mbox{P} \\ 
\mathbf{B}_{54} & = & -y_{9} \, \mathbf{a}_{1}-x_{9} \, \mathbf{a}_{2}-z_{9} \, \mathbf{a}_{3} & = & -y_{9}a \, \mathbf{\hat{x}}-x_{9}a \, \mathbf{\hat{y}}-z_{9}c \, \mathbf{\hat{z}} & \left(8g\right) & \mbox{P} \\ 
\end{longtabu}
\renewcommand{\arraystretch}{1.0}
\noindent \hrulefill
\\
\textbf{References:}
\vspace*{-0.25cm}
\begin{flushleft}
  - \bibentry{Meyer_BaCuVOPO44_ZAnorgAllgChemie_1997}. \\
\end{flushleft}
\textbf{Found in:}
\vspace*{-0.25cm}
\begin{flushleft}
  - \bibentry{Villars_PearsonsCrystalData_2013}. \\
\end{flushleft}
\noindent \hrulefill
\\
\textbf{Geometry files:}
\\
\noindent  - CIF: pp. {\hyperref[AB4C17D4E_tP54_90_a_g_c4g_g_c_cif]{\pageref{AB4C17D4E_tP54_90_a_g_c4g_g_c_cif}}} \\
\noindent  - POSCAR: pp. {\hyperref[AB4C17D4E_tP54_90_a_g_c4g_g_c_poscar]{\pageref{AB4C17D4E_tP54_90_a_g_c4g_g_c_poscar}}} \\
\onecolumn
{\phantomsection\label{ABC_tP24_91_d_d_d}}
\subsection*{\huge \textbf{{\normalfont ThBC Structure: ABC\_tP24\_91\_d\_d\_d}}}
\noindent \hrulefill
\vspace*{0.25cm}
\begin{figure}[htp]
  \centering
  \vspace{-1em}
  {\includegraphics[width=1\textwidth]{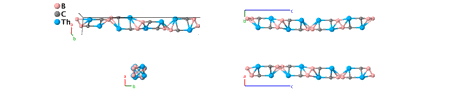}}
\end{figure}
\vspace*{-0.5cm}
\renewcommand{\arraystretch}{1.5}
\begin{equation*}
  \begin{array}{>{$\hspace{-0.15cm}}l<{$}>{$}p{0.5cm}<{$}>{$}p{18.5cm}<{$}}
    \mbox{\large \textbf{Prototype}} &\colon & \ce{ThBC} \\
    \mbox{\large \textbf{\AFLOW\ prototype label}} &\colon & \mbox{ABC\_tP24\_91\_d\_d\_d} \\
    \mbox{\large \textbf{\textit{Strukturbericht} designation}} &\colon & \mbox{None} \\
    \mbox{\large \textbf{Pearson symbol}} &\colon & \mbox{tP24} \\
    \mbox{\large \textbf{Space group number}} &\colon & 91 \\
    \mbox{\large \textbf{Space group symbol}} &\colon & P4_{1}22 \\
    \mbox{\large \textbf{\AFLOW\ prototype command}} &\colon &  \texttt{aflow} \,  \, \texttt{-{}-proto=ABC\_tP24\_91\_d\_d\_d } \, \newline \texttt{-{}-params=}{a,c/a,x_{1},y_{1},z_{1},x_{2},y_{2},z_{2},x_{3},y_{3},z_{3} }
  \end{array}
\end{equation*}
\renewcommand{\arraystretch}{1.0}

\noindent \parbox{1 \linewidth}{
\noindent \hrulefill
\\
\textbf{Simple Tetragonal primitive vectors:} \\
\vspace*{-0.25cm}
\begin{tabular}{cc}
  \begin{tabular}{c}
    \parbox{0.6 \linewidth}{
      \renewcommand{\arraystretch}{1.5}
      \begin{equation*}
        \centering
        \begin{array}{ccc}
              \mathbf{a}_1 & = & a \, \mathbf{\hat{x}} \\
    \mathbf{a}_2 & = & a \, \mathbf{\hat{y}} \\
    \mathbf{a}_3 & = & c \, \mathbf{\hat{z}} \\

        \end{array}
      \end{equation*}
    }
    \renewcommand{\arraystretch}{1.0}
  \end{tabular}
  \begin{tabular}{c}
    \includegraphics[width=0.3\linewidth]{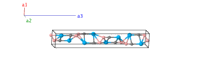} \\
  \end{tabular}
\end{tabular}

}
\vspace*{-0.25cm}

\noindent \hrulefill
\\
\textbf{Basis vectors:}
\vspace*{-0.25cm}
\renewcommand{\arraystretch}{1.5}
\begin{longtabu} to \textwidth{>{\centering $}X[-1,c,c]<{$}>{\centering $}X[-1,c,c]<{$}>{\centering $}X[-1,c,c]<{$}>{\centering $}X[-1,c,c]<{$}>{\centering $}X[-1,c,c]<{$}>{\centering $}X[-1,c,c]<{$}>{\centering $}X[-1,c,c]<{$}}
  & & \mbox{Lattice Coordinates} & & \mbox{Cartesian Coordinates} &\mbox{Wyckoff Position} & \mbox{Atom Type} \\  
  \mathbf{B}_{1} & = & x_{1} \, \mathbf{a}_{1} + y_{1} \, \mathbf{a}_{2} + z_{1} \, \mathbf{a}_{3} & = & x_{1}a \, \mathbf{\hat{x}} + y_{1}a \, \mathbf{\hat{y}} + z_{1}c \, \mathbf{\hat{z}} & \left(8d\right) & \mbox{B} \\ 
\mathbf{B}_{2} & = & -x_{1} \, \mathbf{a}_{1}-y_{1} \, \mathbf{a}_{2} + \left(\frac{1}{2} +z_{1}\right) \, \mathbf{a}_{3} & = & -x_{1}a \, \mathbf{\hat{x}}-y_{1}a \, \mathbf{\hat{y}} + \left(\frac{1}{2} +z_{1}\right)c \, \mathbf{\hat{z}} & \left(8d\right) & \mbox{B} \\ 
\mathbf{B}_{3} & = & -y_{1} \, \mathbf{a}_{1} + x_{1} \, \mathbf{a}_{2} + \left(\frac{1}{4} +z_{1}\right) \, \mathbf{a}_{3} & = & -y_{1}a \, \mathbf{\hat{x}} + x_{1}a \, \mathbf{\hat{y}} + \left(\frac{1}{4} +z_{1}\right)c \, \mathbf{\hat{z}} & \left(8d\right) & \mbox{B} \\ 
\mathbf{B}_{4} & = & y_{1} \, \mathbf{a}_{1}-x_{1} \, \mathbf{a}_{2} + \left(\frac{3}{4} +z_{1}\right) \, \mathbf{a}_{3} & = & y_{1}a \, \mathbf{\hat{x}}-x_{1}a \, \mathbf{\hat{y}} + \left(\frac{3}{4} +z_{1}\right)c \, \mathbf{\hat{z}} & \left(8d\right) & \mbox{B} \\ 
\mathbf{B}_{5} & = & -x_{1} \, \mathbf{a}_{1} + y_{1} \, \mathbf{a}_{2}-z_{1} \, \mathbf{a}_{3} & = & -x_{1}a \, \mathbf{\hat{x}} + y_{1}a \, \mathbf{\hat{y}}-z_{1}c \, \mathbf{\hat{z}} & \left(8d\right) & \mbox{B} \\ 
\mathbf{B}_{6} & = & x_{1} \, \mathbf{a}_{1}-y_{1} \, \mathbf{a}_{2} + \left(\frac{1}{2} - z_{1}\right) \, \mathbf{a}_{3} & = & x_{1}a \, \mathbf{\hat{x}}-y_{1}a \, \mathbf{\hat{y}} + \left(\frac{1}{2} - z_{1}\right)c \, \mathbf{\hat{z}} & \left(8d\right) & \mbox{B} \\ 
\mathbf{B}_{7} & = & y_{1} \, \mathbf{a}_{1} + x_{1} \, \mathbf{a}_{2} + \left(\frac{3}{4} - z_{1}\right) \, \mathbf{a}_{3} & = & y_{1}a \, \mathbf{\hat{x}} + x_{1}a \, \mathbf{\hat{y}} + \left(\frac{3}{4} - z_{1}\right)c \, \mathbf{\hat{z}} & \left(8d\right) & \mbox{B} \\ 
\mathbf{B}_{8} & = & -y_{1} \, \mathbf{a}_{1}-x_{1} \, \mathbf{a}_{2} + \left(\frac{1}{4} - z_{1}\right) \, \mathbf{a}_{3} & = & -y_{1}a \, \mathbf{\hat{x}}-x_{1}a \, \mathbf{\hat{y}} + \left(\frac{1}{4} - z_{1}\right)c \, \mathbf{\hat{z}} & \left(8d\right) & \mbox{B} \\ 
\mathbf{B}_{9} & = & x_{2} \, \mathbf{a}_{1} + y_{2} \, \mathbf{a}_{2} + z_{2} \, \mathbf{a}_{3} & = & x_{2}a \, \mathbf{\hat{x}} + y_{2}a \, \mathbf{\hat{y}} + z_{2}c \, \mathbf{\hat{z}} & \left(8d\right) & \mbox{C} \\ 
\mathbf{B}_{10} & = & -x_{2} \, \mathbf{a}_{1}-y_{2} \, \mathbf{a}_{2} + \left(\frac{1}{2} +z_{2}\right) \, \mathbf{a}_{3} & = & -x_{2}a \, \mathbf{\hat{x}}-y_{2}a \, \mathbf{\hat{y}} + \left(\frac{1}{2} +z_{2}\right)c \, \mathbf{\hat{z}} & \left(8d\right) & \mbox{C} \\ 
\mathbf{B}_{11} & = & -y_{2} \, \mathbf{a}_{1} + x_{2} \, \mathbf{a}_{2} + \left(\frac{1}{4} +z_{2}\right) \, \mathbf{a}_{3} & = & -y_{2}a \, \mathbf{\hat{x}} + x_{2}a \, \mathbf{\hat{y}} + \left(\frac{1}{4} +z_{2}\right)c \, \mathbf{\hat{z}} & \left(8d\right) & \mbox{C} \\ 
\mathbf{B}_{12} & = & y_{2} \, \mathbf{a}_{1}-x_{2} \, \mathbf{a}_{2} + \left(\frac{3}{4} +z_{2}\right) \, \mathbf{a}_{3} & = & y_{2}a \, \mathbf{\hat{x}}-x_{2}a \, \mathbf{\hat{y}} + \left(\frac{3}{4} +z_{2}\right)c \, \mathbf{\hat{z}} & \left(8d\right) & \mbox{C} \\ 
\mathbf{B}_{13} & = & -x_{2} \, \mathbf{a}_{1} + y_{2} \, \mathbf{a}_{2}-z_{2} \, \mathbf{a}_{3} & = & -x_{2}a \, \mathbf{\hat{x}} + y_{2}a \, \mathbf{\hat{y}}-z_{2}c \, \mathbf{\hat{z}} & \left(8d\right) & \mbox{C} \\ 
\mathbf{B}_{14} & = & x_{2} \, \mathbf{a}_{1}-y_{2} \, \mathbf{a}_{2} + \left(\frac{1}{2} - z_{2}\right) \, \mathbf{a}_{3} & = & x_{2}a \, \mathbf{\hat{x}}-y_{2}a \, \mathbf{\hat{y}} + \left(\frac{1}{2} - z_{2}\right)c \, \mathbf{\hat{z}} & \left(8d\right) & \mbox{C} \\ 
\mathbf{B}_{15} & = & y_{2} \, \mathbf{a}_{1} + x_{2} \, \mathbf{a}_{2} + \left(\frac{3}{4} - z_{2}\right) \, \mathbf{a}_{3} & = & y_{2}a \, \mathbf{\hat{x}} + x_{2}a \, \mathbf{\hat{y}} + \left(\frac{3}{4} - z_{2}\right)c \, \mathbf{\hat{z}} & \left(8d\right) & \mbox{C} \\ 
\mathbf{B}_{16} & = & -y_{2} \, \mathbf{a}_{1}-x_{2} \, \mathbf{a}_{2} + \left(\frac{1}{4} - z_{2}\right) \, \mathbf{a}_{3} & = & -y_{2}a \, \mathbf{\hat{x}}-x_{2}a \, \mathbf{\hat{y}} + \left(\frac{1}{4} - z_{2}\right)c \, \mathbf{\hat{z}} & \left(8d\right) & \mbox{C} \\ 
\mathbf{B}_{17} & = & x_{3} \, \mathbf{a}_{1} + y_{3} \, \mathbf{a}_{2} + z_{3} \, \mathbf{a}_{3} & = & x_{3}a \, \mathbf{\hat{x}} + y_{3}a \, \mathbf{\hat{y}} + z_{3}c \, \mathbf{\hat{z}} & \left(8d\right) & \mbox{Th} \\ 
\mathbf{B}_{18} & = & -x_{3} \, \mathbf{a}_{1}-y_{3} \, \mathbf{a}_{2} + \left(\frac{1}{2} +z_{3}\right) \, \mathbf{a}_{3} & = & -x_{3}a \, \mathbf{\hat{x}}-y_{3}a \, \mathbf{\hat{y}} + \left(\frac{1}{2} +z_{3}\right)c \, \mathbf{\hat{z}} & \left(8d\right) & \mbox{Th} \\ 
\mathbf{B}_{19} & = & -y_{3} \, \mathbf{a}_{1} + x_{3} \, \mathbf{a}_{2} + \left(\frac{1}{4} +z_{3}\right) \, \mathbf{a}_{3} & = & -y_{3}a \, \mathbf{\hat{x}} + x_{3}a \, \mathbf{\hat{y}} + \left(\frac{1}{4} +z_{3}\right)c \, \mathbf{\hat{z}} & \left(8d\right) & \mbox{Th} \\ 
\mathbf{B}_{20} & = & y_{3} \, \mathbf{a}_{1}-x_{3} \, \mathbf{a}_{2} + \left(\frac{3}{4} +z_{3}\right) \, \mathbf{a}_{3} & = & y_{3}a \, \mathbf{\hat{x}}-x_{3}a \, \mathbf{\hat{y}} + \left(\frac{3}{4} +z_{3}\right)c \, \mathbf{\hat{z}} & \left(8d\right) & \mbox{Th} \\ 
\mathbf{B}_{21} & = & -x_{3} \, \mathbf{a}_{1} + y_{3} \, \mathbf{a}_{2}-z_{3} \, \mathbf{a}_{3} & = & -x_{3}a \, \mathbf{\hat{x}} + y_{3}a \, \mathbf{\hat{y}}-z_{3}c \, \mathbf{\hat{z}} & \left(8d\right) & \mbox{Th} \\ 
\mathbf{B}_{22} & = & x_{3} \, \mathbf{a}_{1}-y_{3} \, \mathbf{a}_{2} + \left(\frac{1}{2} - z_{3}\right) \, \mathbf{a}_{3} & = & x_{3}a \, \mathbf{\hat{x}}-y_{3}a \, \mathbf{\hat{y}} + \left(\frac{1}{2} - z_{3}\right)c \, \mathbf{\hat{z}} & \left(8d\right) & \mbox{Th} \\ 
\mathbf{B}_{23} & = & y_{3} \, \mathbf{a}_{1} + x_{3} \, \mathbf{a}_{2} + \left(\frac{3}{4} - z_{3}\right) \, \mathbf{a}_{3} & = & y_{3}a \, \mathbf{\hat{x}} + x_{3}a \, \mathbf{\hat{y}} + \left(\frac{3}{4} - z_{3}\right)c \, \mathbf{\hat{z}} & \left(8d\right) & \mbox{Th} \\ 
\mathbf{B}_{24} & = & -y_{3} \, \mathbf{a}_{1}-x_{3} \, \mathbf{a}_{2} + \left(\frac{1}{4} - z_{3}\right) \, \mathbf{a}_{3} & = & -y_{3}a \, \mathbf{\hat{x}}-x_{3}a \, \mathbf{\hat{y}} + \left(\frac{1}{4} - z_{3}\right)c \, \mathbf{\hat{z}} & \left(8d\right) & \mbox{Th} \\ 
\end{longtabu}
\renewcommand{\arraystretch}{1.0}
\noindent \hrulefill
\\
\textbf{References:}
\vspace*{-0.25cm}
\begin{flushleft}
  - \bibentry{Rogl_ThBC_JNucMat_1978}. \\
\end{flushleft}
\textbf{Found in:}
\vspace*{-0.25cm}
\begin{flushleft}
  - \bibentry{Villars_PearsonsCrystalData_2013}. \\
\end{flushleft}
\noindent \hrulefill
\\
\textbf{Geometry files:}
\\
\noindent  - CIF: pp. {\hyperref[ABC_tP24_91_d_d_d_cif]{\pageref{ABC_tP24_91_d_d_d_cif}}} \\
\noindent  - POSCAR: pp. {\hyperref[ABC_tP24_91_d_d_d_poscar]{\pageref{ABC_tP24_91_d_d_d_poscar}}} \\
\onecolumn
{\phantomsection\label{AB32CD4E8_tP184_93_i_16p_af_2p_4p}}
\subsection*{\huge \textbf{{\normalfont \begin{raggedleft}AsPh$_{4}$CeS$_{8}$P$_{4}$Me$_{8}$ Structure: \end{raggedleft} \\ AB32CD4E8\_tP184\_93\_i\_16p\_af\_2p\_4p}}}
\noindent \hrulefill
\vspace*{0.25cm}
\begin{figure}[htp]
  \centering
  \vspace{-1em}
  {\includegraphics[width=1\textwidth]{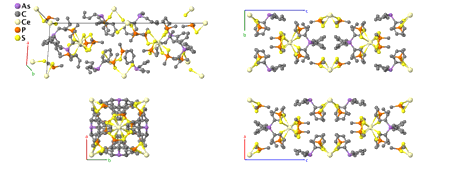}}
\end{figure}
\vspace*{-0.5cm}
\renewcommand{\arraystretch}{1.5}
\begin{equation*}
  \begin{array}{>{$\hspace{-0.15cm}}l<{$}>{$}p{0.5cm}<{$}>{$}p{18.5cm}<{$}}
    \mbox{\large \textbf{Prototype}} &\colon & \ce{AsPh4CeS8P4Me8} \\
    \mbox{\large \textbf{\AFLOW\ prototype label}} &\colon & \mbox{AB32CD4E8\_tP184\_93\_i\_16p\_af\_2p\_4p} \\
    \mbox{\large \textbf{\textit{Strukturbericht} designation}} &\colon & \mbox{None} \\
    \mbox{\large \textbf{Pearson symbol}} &\colon & \mbox{tP184} \\
    \mbox{\large \textbf{Space group number}} &\colon & 93 \\
    \mbox{\large \textbf{Space group symbol}} &\colon & P4_{2}22 \\
    \mbox{\large \textbf{\AFLOW\ prototype command}} &\colon &  \texttt{aflow} \,  \, \texttt{-{}-proto=AB32CD4E8\_tP184\_93\_i\_16p\_af\_2p\_4p } \, \newline \texttt{-{}-params=}{a,c/a,z_{3},x_{4},y_{4},z_{4},x_{5},y_{5},z_{5},x_{6},y_{6},z_{6},x_{7},y_{7},z_{7},x_{8},y_{8},z_{8},x_{9},y_{9},z_{9},} \newline {x_{10},y_{10},z_{10},x_{11},y_{11},z_{11},x_{12},y_{12},z_{12},x_{13},y_{13},z_{13},x_{14},y_{14},z_{14},x_{15},y_{15},z_{15},x_{16},y_{16},} \newline {z_{16},x_{17},y_{17},z_{17},x_{18},y_{18},z_{18},x_{19},y_{19},z_{19},x_{20},y_{20},z_{20},x_{21},y_{21},z_{21},x_{22},y_{22},z_{22},x_{23},} \newline {y_{23},z_{23},x_{24},y_{24},z_{24},x_{25},y_{25},z_{25} }
  \end{array}
\end{equation*}
\renewcommand{\arraystretch}{1.0}

\vspace*{-0.25cm}
\noindent \hrulefill
\begin{itemize}
  \item{Structures exhibiting space group \#93 are quite rare.  
According to (Hoffmann, 2014), there are no entries in the Inorganic Crystal Structure Database with space group \#93.
The hydrogen atoms are not included in this prototype. (Ph = Phenyl and Me = Methyl)
}
\end{itemize}

\noindent \parbox{1 \linewidth}{
\noindent \hrulefill
\\
\textbf{Simple Tetragonal primitive vectors:} \\
\vspace*{-0.25cm}
\begin{tabular}{cc}
  \begin{tabular}{c}
    \parbox{0.6 \linewidth}{
      \renewcommand{\arraystretch}{1.5}
      \begin{equation*}
        \centering
        \begin{array}{ccc}
              \mathbf{a}_1 & = & a \, \mathbf{\hat{x}} \\
    \mathbf{a}_2 & = & a \, \mathbf{\hat{y}} \\
    \mathbf{a}_3 & = & c \, \mathbf{\hat{z}} \\

        \end{array}
      \end{equation*}
    }
    \renewcommand{\arraystretch}{1.0}
  \end{tabular}
  \begin{tabular}{c}
    \includegraphics[width=0.3\linewidth]{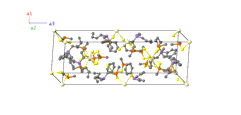} \\
  \end{tabular}
\end{tabular}

}
\vspace*{-0.25cm}

\noindent \hrulefill
\\
\textbf{Basis vectors:}
\vspace*{-0.25cm}
\renewcommand{\arraystretch}{1.5}
\begin{longtabu} to \textwidth{>{\centering $}X[-1,c,c]<{$}>{\centering $}X[-1,c,c]<{$}>{\centering $}X[-1,c,c]<{$}>{\centering $}X[-1,c,c]<{$}>{\centering $}X[-1,c,c]<{$}>{\centering $}X[-1,c,c]<{$}>{\centering $}X[-1,c,c]<{$}}
  & & \mbox{Lattice Coordinates} & & \mbox{Cartesian Coordinates} &\mbox{Wyckoff Position} & \mbox{Atom Type} \\  
  \mathbf{B}_{1} & = & 0 \, \mathbf{a}_{1} + 0 \, \mathbf{a}_{2} + 0 \, \mathbf{a}_{3} & = & 0 \, \mathbf{\hat{x}} + 0 \, \mathbf{\hat{y}} + 0 \, \mathbf{\hat{z}} & \left(2a\right) & \mbox{Ce I} \\ 
\mathbf{B}_{2} & = & \frac{1}{2} \, \mathbf{a}_{3} & = & \frac{1}{2}c \, \mathbf{\hat{z}} & \left(2a\right) & \mbox{Ce I} \\ 
\mathbf{B}_{3} & = & \frac{1}{2} \, \mathbf{a}_{1} + \frac{1}{2} \, \mathbf{a}_{2} + \frac{1}{4} \, \mathbf{a}_{3} & = & \frac{1}{2}a \, \mathbf{\hat{x}} + \frac{1}{2}a \, \mathbf{\hat{y}} + \frac{1}{4}c \, \mathbf{\hat{z}} & \left(2f\right) & \mbox{Ce II} \\ 
\mathbf{B}_{4} & = & \frac{1}{2} \, \mathbf{a}_{1} + \frac{1}{2} \, \mathbf{a}_{2} + \frac{3}{4} \, \mathbf{a}_{3} & = & \frac{1}{2}a \, \mathbf{\hat{x}} + \frac{1}{2}a \, \mathbf{\hat{y}} + \frac{3}{4}c \, \mathbf{\hat{z}} & \left(2f\right) & \mbox{Ce II} \\ 
\mathbf{B}_{5} & = & \frac{1}{2} \, \mathbf{a}_{2} + z_{3} \, \mathbf{a}_{3} & = & \frac{1}{2}a \, \mathbf{\hat{y}} + z_{3}c \, \mathbf{\hat{z}} & \left(4i\right) & \mbox{As} \\ 
\mathbf{B}_{6} & = & \frac{1}{2} \, \mathbf{a}_{1} + \left(\frac{1}{2} +z_{3}\right) \, \mathbf{a}_{3} & = & \frac{1}{2}a \, \mathbf{\hat{x}} + \left(\frac{1}{2} +z_{3}\right)c \, \mathbf{\hat{z}} & \left(4i\right) & \mbox{As} \\ 
\mathbf{B}_{7} & = & \frac{1}{2} \, \mathbf{a}_{2}-z_{3} \, \mathbf{a}_{3} & = & \frac{1}{2}a \, \mathbf{\hat{y}}-z_{3}c \, \mathbf{\hat{z}} & \left(4i\right) & \mbox{As} \\ 
\mathbf{B}_{8} & = & \frac{1}{2} \, \mathbf{a}_{1} + \left(\frac{1}{2} - z_{3}\right) \, \mathbf{a}_{3} & = & \frac{1}{2}a \, \mathbf{\hat{x}} + \left(\frac{1}{2} - z_{3}\right)c \, \mathbf{\hat{z}} & \left(4i\right) & \mbox{As} \\ 
\mathbf{B}_{9} & = & x_{4} \, \mathbf{a}_{1} + y_{4} \, \mathbf{a}_{2} + z_{4} \, \mathbf{a}_{3} & = & x_{4}a \, \mathbf{\hat{x}} + y_{4}a \, \mathbf{\hat{y}} + z_{4}c \, \mathbf{\hat{z}} & \left(8p\right) & \mbox{C I} \\ 
\mathbf{B}_{10} & = & -x_{4} \, \mathbf{a}_{1}-y_{4} \, \mathbf{a}_{2} + z_{4} \, \mathbf{a}_{3} & = & -x_{4}a \, \mathbf{\hat{x}}-y_{4}a \, \mathbf{\hat{y}} + z_{4}c \, \mathbf{\hat{z}} & \left(8p\right) & \mbox{C I} \\ 
\mathbf{B}_{11} & = & -y_{4} \, \mathbf{a}_{1} + x_{4} \, \mathbf{a}_{2} + \left(\frac{1}{2} +z_{4}\right) \, \mathbf{a}_{3} & = & -y_{4}a \, \mathbf{\hat{x}} + x_{4}a \, \mathbf{\hat{y}} + \left(\frac{1}{2} +z_{4}\right)c \, \mathbf{\hat{z}} & \left(8p\right) & \mbox{C I} \\ 
\mathbf{B}_{12} & = & y_{4} \, \mathbf{a}_{1}-x_{4} \, \mathbf{a}_{2} + \left(\frac{1}{2} +z_{4}\right) \, \mathbf{a}_{3} & = & y_{4}a \, \mathbf{\hat{x}}-x_{4}a \, \mathbf{\hat{y}} + \left(\frac{1}{2} +z_{4}\right)c \, \mathbf{\hat{z}} & \left(8p\right) & \mbox{C I} \\ 
\mathbf{B}_{13} & = & -x_{4} \, \mathbf{a}_{1} + y_{4} \, \mathbf{a}_{2}-z_{4} \, \mathbf{a}_{3} & = & -x_{4}a \, \mathbf{\hat{x}} + y_{4}a \, \mathbf{\hat{y}}-z_{4}c \, \mathbf{\hat{z}} & \left(8p\right) & \mbox{C I} \\ 
\mathbf{B}_{14} & = & x_{4} \, \mathbf{a}_{1}-y_{4} \, \mathbf{a}_{2}-z_{4} \, \mathbf{a}_{3} & = & x_{4}a \, \mathbf{\hat{x}}-y_{4}a \, \mathbf{\hat{y}}-z_{4}c \, \mathbf{\hat{z}} & \left(8p\right) & \mbox{C I} \\ 
\mathbf{B}_{15} & = & y_{4} \, \mathbf{a}_{1} + x_{4} \, \mathbf{a}_{2} + \left(\frac{1}{2} - z_{4}\right) \, \mathbf{a}_{3} & = & y_{4}a \, \mathbf{\hat{x}} + x_{4}a \, \mathbf{\hat{y}} + \left(\frac{1}{2} - z_{4}\right)c \, \mathbf{\hat{z}} & \left(8p\right) & \mbox{C I} \\ 
\mathbf{B}_{16} & = & -y_{4} \, \mathbf{a}_{1}-x_{4} \, \mathbf{a}_{2} + \left(\frac{1}{2} - z_{4}\right) \, \mathbf{a}_{3} & = & -y_{4}a \, \mathbf{\hat{x}}-x_{4}a \, \mathbf{\hat{y}} + \left(\frac{1}{2} - z_{4}\right)c \, \mathbf{\hat{z}} & \left(8p\right) & \mbox{C I} \\ 
\mathbf{B}_{17} & = & x_{5} \, \mathbf{a}_{1} + y_{5} \, \mathbf{a}_{2} + z_{5} \, \mathbf{a}_{3} & = & x_{5}a \, \mathbf{\hat{x}} + y_{5}a \, \mathbf{\hat{y}} + z_{5}c \, \mathbf{\hat{z}} & \left(8p\right) & \mbox{C II} \\ 
\mathbf{B}_{18} & = & -x_{5} \, \mathbf{a}_{1}-y_{5} \, \mathbf{a}_{2} + z_{5} \, \mathbf{a}_{3} & = & -x_{5}a \, \mathbf{\hat{x}}-y_{5}a \, \mathbf{\hat{y}} + z_{5}c \, \mathbf{\hat{z}} & \left(8p\right) & \mbox{C II} \\ 
\mathbf{B}_{19} & = & -y_{5} \, \mathbf{a}_{1} + x_{5} \, \mathbf{a}_{2} + \left(\frac{1}{2} +z_{5}\right) \, \mathbf{a}_{3} & = & -y_{5}a \, \mathbf{\hat{x}} + x_{5}a \, \mathbf{\hat{y}} + \left(\frac{1}{2} +z_{5}\right)c \, \mathbf{\hat{z}} & \left(8p\right) & \mbox{C II} \\ 
\mathbf{B}_{20} & = & y_{5} \, \mathbf{a}_{1}-x_{5} \, \mathbf{a}_{2} + \left(\frac{1}{2} +z_{5}\right) \, \mathbf{a}_{3} & = & y_{5}a \, \mathbf{\hat{x}}-x_{5}a \, \mathbf{\hat{y}} + \left(\frac{1}{2} +z_{5}\right)c \, \mathbf{\hat{z}} & \left(8p\right) & \mbox{C II} \\ 
\mathbf{B}_{21} & = & -x_{5} \, \mathbf{a}_{1} + y_{5} \, \mathbf{a}_{2}-z_{5} \, \mathbf{a}_{3} & = & -x_{5}a \, \mathbf{\hat{x}} + y_{5}a \, \mathbf{\hat{y}}-z_{5}c \, \mathbf{\hat{z}} & \left(8p\right) & \mbox{C II} \\ 
\mathbf{B}_{22} & = & x_{5} \, \mathbf{a}_{1}-y_{5} \, \mathbf{a}_{2}-z_{5} \, \mathbf{a}_{3} & = & x_{5}a \, \mathbf{\hat{x}}-y_{5}a \, \mathbf{\hat{y}}-z_{5}c \, \mathbf{\hat{z}} & \left(8p\right) & \mbox{C II} \\ 
\mathbf{B}_{23} & = & y_{5} \, \mathbf{a}_{1} + x_{5} \, \mathbf{a}_{2} + \left(\frac{1}{2} - z_{5}\right) \, \mathbf{a}_{3} & = & y_{5}a \, \mathbf{\hat{x}} + x_{5}a \, \mathbf{\hat{y}} + \left(\frac{1}{2} - z_{5}\right)c \, \mathbf{\hat{z}} & \left(8p\right) & \mbox{C II} \\ 
\mathbf{B}_{24} & = & -y_{5} \, \mathbf{a}_{1}-x_{5} \, \mathbf{a}_{2} + \left(\frac{1}{2} - z_{5}\right) \, \mathbf{a}_{3} & = & -y_{5}a \, \mathbf{\hat{x}}-x_{5}a \, \mathbf{\hat{y}} + \left(\frac{1}{2} - z_{5}\right)c \, \mathbf{\hat{z}} & \left(8p\right) & \mbox{C II} \\ 
\mathbf{B}_{25} & = & x_{6} \, \mathbf{a}_{1} + y_{6} \, \mathbf{a}_{2} + z_{6} \, \mathbf{a}_{3} & = & x_{6}a \, \mathbf{\hat{x}} + y_{6}a \, \mathbf{\hat{y}} + z_{6}c \, \mathbf{\hat{z}} & \left(8p\right) & \mbox{C III} \\ 
\mathbf{B}_{26} & = & -x_{6} \, \mathbf{a}_{1}-y_{6} \, \mathbf{a}_{2} + z_{6} \, \mathbf{a}_{3} & = & -x_{6}a \, \mathbf{\hat{x}}-y_{6}a \, \mathbf{\hat{y}} + z_{6}c \, \mathbf{\hat{z}} & \left(8p\right) & \mbox{C III} \\ 
\mathbf{B}_{27} & = & -y_{6} \, \mathbf{a}_{1} + x_{6} \, \mathbf{a}_{2} + \left(\frac{1}{2} +z_{6}\right) \, \mathbf{a}_{3} & = & -y_{6}a \, \mathbf{\hat{x}} + x_{6}a \, \mathbf{\hat{y}} + \left(\frac{1}{2} +z_{6}\right)c \, \mathbf{\hat{z}} & \left(8p\right) & \mbox{C III} \\ 
\mathbf{B}_{28} & = & y_{6} \, \mathbf{a}_{1}-x_{6} \, \mathbf{a}_{2} + \left(\frac{1}{2} +z_{6}\right) \, \mathbf{a}_{3} & = & y_{6}a \, \mathbf{\hat{x}}-x_{6}a \, \mathbf{\hat{y}} + \left(\frac{1}{2} +z_{6}\right)c \, \mathbf{\hat{z}} & \left(8p\right) & \mbox{C III} \\ 
\mathbf{B}_{29} & = & -x_{6} \, \mathbf{a}_{1} + y_{6} \, \mathbf{a}_{2}-z_{6} \, \mathbf{a}_{3} & = & -x_{6}a \, \mathbf{\hat{x}} + y_{6}a \, \mathbf{\hat{y}}-z_{6}c \, \mathbf{\hat{z}} & \left(8p\right) & \mbox{C III} \\ 
\mathbf{B}_{30} & = & x_{6} \, \mathbf{a}_{1}-y_{6} \, \mathbf{a}_{2}-z_{6} \, \mathbf{a}_{3} & = & x_{6}a \, \mathbf{\hat{x}}-y_{6}a \, \mathbf{\hat{y}}-z_{6}c \, \mathbf{\hat{z}} & \left(8p\right) & \mbox{C III} \\ 
\mathbf{B}_{31} & = & y_{6} \, \mathbf{a}_{1} + x_{6} \, \mathbf{a}_{2} + \left(\frac{1}{2} - z_{6}\right) \, \mathbf{a}_{3} & = & y_{6}a \, \mathbf{\hat{x}} + x_{6}a \, \mathbf{\hat{y}} + \left(\frac{1}{2} - z_{6}\right)c \, \mathbf{\hat{z}} & \left(8p\right) & \mbox{C III} \\ 
\mathbf{B}_{32} & = & -y_{6} \, \mathbf{a}_{1}-x_{6} \, \mathbf{a}_{2} + \left(\frac{1}{2} - z_{6}\right) \, \mathbf{a}_{3} & = & -y_{6}a \, \mathbf{\hat{x}}-x_{6}a \, \mathbf{\hat{y}} + \left(\frac{1}{2} - z_{6}\right)c \, \mathbf{\hat{z}} & \left(8p\right) & \mbox{C III} \\ 
\mathbf{B}_{33} & = & x_{7} \, \mathbf{a}_{1} + y_{7} \, \mathbf{a}_{2} + z_{7} \, \mathbf{a}_{3} & = & x_{7}a \, \mathbf{\hat{x}} + y_{7}a \, \mathbf{\hat{y}} + z_{7}c \, \mathbf{\hat{z}} & \left(8p\right) & \mbox{C IV} \\ 
\mathbf{B}_{34} & = & -x_{7} \, \mathbf{a}_{1}-y_{7} \, \mathbf{a}_{2} + z_{7} \, \mathbf{a}_{3} & = & -x_{7}a \, \mathbf{\hat{x}}-y_{7}a \, \mathbf{\hat{y}} + z_{7}c \, \mathbf{\hat{z}} & \left(8p\right) & \mbox{C IV} \\ 
\mathbf{B}_{35} & = & -y_{7} \, \mathbf{a}_{1} + x_{7} \, \mathbf{a}_{2} + \left(\frac{1}{2} +z_{7}\right) \, \mathbf{a}_{3} & = & -y_{7}a \, \mathbf{\hat{x}} + x_{7}a \, \mathbf{\hat{y}} + \left(\frac{1}{2} +z_{7}\right)c \, \mathbf{\hat{z}} & \left(8p\right) & \mbox{C IV} \\ 
\mathbf{B}_{36} & = & y_{7} \, \mathbf{a}_{1}-x_{7} \, \mathbf{a}_{2} + \left(\frac{1}{2} +z_{7}\right) \, \mathbf{a}_{3} & = & y_{7}a \, \mathbf{\hat{x}}-x_{7}a \, \mathbf{\hat{y}} + \left(\frac{1}{2} +z_{7}\right)c \, \mathbf{\hat{z}} & \left(8p\right) & \mbox{C IV} \\ 
\mathbf{B}_{37} & = & -x_{7} \, \mathbf{a}_{1} + y_{7} \, \mathbf{a}_{2}-z_{7} \, \mathbf{a}_{3} & = & -x_{7}a \, \mathbf{\hat{x}} + y_{7}a \, \mathbf{\hat{y}}-z_{7}c \, \mathbf{\hat{z}} & \left(8p\right) & \mbox{C IV} \\ 
\mathbf{B}_{38} & = & x_{7} \, \mathbf{a}_{1}-y_{7} \, \mathbf{a}_{2}-z_{7} \, \mathbf{a}_{3} & = & x_{7}a \, \mathbf{\hat{x}}-y_{7}a \, \mathbf{\hat{y}}-z_{7}c \, \mathbf{\hat{z}} & \left(8p\right) & \mbox{C IV} \\ 
\mathbf{B}_{39} & = & y_{7} \, \mathbf{a}_{1} + x_{7} \, \mathbf{a}_{2} + \left(\frac{1}{2} - z_{7}\right) \, \mathbf{a}_{3} & = & y_{7}a \, \mathbf{\hat{x}} + x_{7}a \, \mathbf{\hat{y}} + \left(\frac{1}{2} - z_{7}\right)c \, \mathbf{\hat{z}} & \left(8p\right) & \mbox{C IV} \\ 
\mathbf{B}_{40} & = & -y_{7} \, \mathbf{a}_{1}-x_{7} \, \mathbf{a}_{2} + \left(\frac{1}{2} - z_{7}\right) \, \mathbf{a}_{3} & = & -y_{7}a \, \mathbf{\hat{x}}-x_{7}a \, \mathbf{\hat{y}} + \left(\frac{1}{2} - z_{7}\right)c \, \mathbf{\hat{z}} & \left(8p\right) & \mbox{C IV} \\ 
\mathbf{B}_{41} & = & x_{8} \, \mathbf{a}_{1} + y_{8} \, \mathbf{a}_{2} + z_{8} \, \mathbf{a}_{3} & = & x_{8}a \, \mathbf{\hat{x}} + y_{8}a \, \mathbf{\hat{y}} + z_{8}c \, \mathbf{\hat{z}} & \left(8p\right) & \mbox{C V} \\ 
\mathbf{B}_{42} & = & -x_{8} \, \mathbf{a}_{1}-y_{8} \, \mathbf{a}_{2} + z_{8} \, \mathbf{a}_{3} & = & -x_{8}a \, \mathbf{\hat{x}}-y_{8}a \, \mathbf{\hat{y}} + z_{8}c \, \mathbf{\hat{z}} & \left(8p\right) & \mbox{C V} \\ 
\mathbf{B}_{43} & = & -y_{8} \, \mathbf{a}_{1} + x_{8} \, \mathbf{a}_{2} + \left(\frac{1}{2} +z_{8}\right) \, \mathbf{a}_{3} & = & -y_{8}a \, \mathbf{\hat{x}} + x_{8}a \, \mathbf{\hat{y}} + \left(\frac{1}{2} +z_{8}\right)c \, \mathbf{\hat{z}} & \left(8p\right) & \mbox{C V} \\ 
\mathbf{B}_{44} & = & y_{8} \, \mathbf{a}_{1}-x_{8} \, \mathbf{a}_{2} + \left(\frac{1}{2} +z_{8}\right) \, \mathbf{a}_{3} & = & y_{8}a \, \mathbf{\hat{x}}-x_{8}a \, \mathbf{\hat{y}} + \left(\frac{1}{2} +z_{8}\right)c \, \mathbf{\hat{z}} & \left(8p\right) & \mbox{C V} \\ 
\mathbf{B}_{45} & = & -x_{8} \, \mathbf{a}_{1} + y_{8} \, \mathbf{a}_{2}-z_{8} \, \mathbf{a}_{3} & = & -x_{8}a \, \mathbf{\hat{x}} + y_{8}a \, \mathbf{\hat{y}}-z_{8}c \, \mathbf{\hat{z}} & \left(8p\right) & \mbox{C V} \\ 
\mathbf{B}_{46} & = & x_{8} \, \mathbf{a}_{1}-y_{8} \, \mathbf{a}_{2}-z_{8} \, \mathbf{a}_{3} & = & x_{8}a \, \mathbf{\hat{x}}-y_{8}a \, \mathbf{\hat{y}}-z_{8}c \, \mathbf{\hat{z}} & \left(8p\right) & \mbox{C V} \\ 
\mathbf{B}_{47} & = & y_{8} \, \mathbf{a}_{1} + x_{8} \, \mathbf{a}_{2} + \left(\frac{1}{2} - z_{8}\right) \, \mathbf{a}_{3} & = & y_{8}a \, \mathbf{\hat{x}} + x_{8}a \, \mathbf{\hat{y}} + \left(\frac{1}{2} - z_{8}\right)c \, \mathbf{\hat{z}} & \left(8p\right) & \mbox{C V} \\ 
\mathbf{B}_{48} & = & -y_{8} \, \mathbf{a}_{1}-x_{8} \, \mathbf{a}_{2} + \left(\frac{1}{2} - z_{8}\right) \, \mathbf{a}_{3} & = & -y_{8}a \, \mathbf{\hat{x}}-x_{8}a \, \mathbf{\hat{y}} + \left(\frac{1}{2} - z_{8}\right)c \, \mathbf{\hat{z}} & \left(8p\right) & \mbox{C V} \\ 
\mathbf{B}_{49} & = & x_{9} \, \mathbf{a}_{1} + y_{9} \, \mathbf{a}_{2} + z_{9} \, \mathbf{a}_{3} & = & x_{9}a \, \mathbf{\hat{x}} + y_{9}a \, \mathbf{\hat{y}} + z_{9}c \, \mathbf{\hat{z}} & \left(8p\right) & \mbox{C VI} \\ 
\mathbf{B}_{50} & = & -x_{9} \, \mathbf{a}_{1}-y_{9} \, \mathbf{a}_{2} + z_{9} \, \mathbf{a}_{3} & = & -x_{9}a \, \mathbf{\hat{x}}-y_{9}a \, \mathbf{\hat{y}} + z_{9}c \, \mathbf{\hat{z}} & \left(8p\right) & \mbox{C VI} \\ 
\mathbf{B}_{51} & = & -y_{9} \, \mathbf{a}_{1} + x_{9} \, \mathbf{a}_{2} + \left(\frac{1}{2} +z_{9}\right) \, \mathbf{a}_{3} & = & -y_{9}a \, \mathbf{\hat{x}} + x_{9}a \, \mathbf{\hat{y}} + \left(\frac{1}{2} +z_{9}\right)c \, \mathbf{\hat{z}} & \left(8p\right) & \mbox{C VI} \\ 
\mathbf{B}_{52} & = & y_{9} \, \mathbf{a}_{1}-x_{9} \, \mathbf{a}_{2} + \left(\frac{1}{2} +z_{9}\right) \, \mathbf{a}_{3} & = & y_{9}a \, \mathbf{\hat{x}}-x_{9}a \, \mathbf{\hat{y}} + \left(\frac{1}{2} +z_{9}\right)c \, \mathbf{\hat{z}} & \left(8p\right) & \mbox{C VI} \\ 
\mathbf{B}_{53} & = & -x_{9} \, \mathbf{a}_{1} + y_{9} \, \mathbf{a}_{2}-z_{9} \, \mathbf{a}_{3} & = & -x_{9}a \, \mathbf{\hat{x}} + y_{9}a \, \mathbf{\hat{y}}-z_{9}c \, \mathbf{\hat{z}} & \left(8p\right) & \mbox{C VI} \\ 
\mathbf{B}_{54} & = & x_{9} \, \mathbf{a}_{1}-y_{9} \, \mathbf{a}_{2}-z_{9} \, \mathbf{a}_{3} & = & x_{9}a \, \mathbf{\hat{x}}-y_{9}a \, \mathbf{\hat{y}}-z_{9}c \, \mathbf{\hat{z}} & \left(8p\right) & \mbox{C VI} \\ 
\mathbf{B}_{55} & = & y_{9} \, \mathbf{a}_{1} + x_{9} \, \mathbf{a}_{2} + \left(\frac{1}{2} - z_{9}\right) \, \mathbf{a}_{3} & = & y_{9}a \, \mathbf{\hat{x}} + x_{9}a \, \mathbf{\hat{y}} + \left(\frac{1}{2} - z_{9}\right)c \, \mathbf{\hat{z}} & \left(8p\right) & \mbox{C VI} \\ 
\mathbf{B}_{56} & = & -y_{9} \, \mathbf{a}_{1}-x_{9} \, \mathbf{a}_{2} + \left(\frac{1}{2} - z_{9}\right) \, \mathbf{a}_{3} & = & -y_{9}a \, \mathbf{\hat{x}}-x_{9}a \, \mathbf{\hat{y}} + \left(\frac{1}{2} - z_{9}\right)c \, \mathbf{\hat{z}} & \left(8p\right) & \mbox{C VI} \\ 
\mathbf{B}_{57} & = & x_{10} \, \mathbf{a}_{1} + y_{10} \, \mathbf{a}_{2} + z_{10} \, \mathbf{a}_{3} & = & x_{10}a \, \mathbf{\hat{x}} + y_{10}a \, \mathbf{\hat{y}} + z_{10}c \, \mathbf{\hat{z}} & \left(8p\right) & \mbox{C VII} \\ 
\mathbf{B}_{58} & = & -x_{10} \, \mathbf{a}_{1}-y_{10} \, \mathbf{a}_{2} + z_{10} \, \mathbf{a}_{3} & = & -x_{10}a \, \mathbf{\hat{x}}-y_{10}a \, \mathbf{\hat{y}} + z_{10}c \, \mathbf{\hat{z}} & \left(8p\right) & \mbox{C VII} \\ 
\mathbf{B}_{59} & = & -y_{10} \, \mathbf{a}_{1} + x_{10} \, \mathbf{a}_{2} + \left(\frac{1}{2} +z_{10}\right) \, \mathbf{a}_{3} & = & -y_{10}a \, \mathbf{\hat{x}} + x_{10}a \, \mathbf{\hat{y}} + \left(\frac{1}{2} +z_{10}\right)c \, \mathbf{\hat{z}} & \left(8p\right) & \mbox{C VII} \\ 
\mathbf{B}_{60} & = & y_{10} \, \mathbf{a}_{1}-x_{10} \, \mathbf{a}_{2} + \left(\frac{1}{2} +z_{10}\right) \, \mathbf{a}_{3} & = & y_{10}a \, \mathbf{\hat{x}}-x_{10}a \, \mathbf{\hat{y}} + \left(\frac{1}{2} +z_{10}\right)c \, \mathbf{\hat{z}} & \left(8p\right) & \mbox{C VII} \\ 
\mathbf{B}_{61} & = & -x_{10} \, \mathbf{a}_{1} + y_{10} \, \mathbf{a}_{2}-z_{10} \, \mathbf{a}_{3} & = & -x_{10}a \, \mathbf{\hat{x}} + y_{10}a \, \mathbf{\hat{y}}-z_{10}c \, \mathbf{\hat{z}} & \left(8p\right) & \mbox{C VII} \\ 
\mathbf{B}_{62} & = & x_{10} \, \mathbf{a}_{1}-y_{10} \, \mathbf{a}_{2}-z_{10} \, \mathbf{a}_{3} & = & x_{10}a \, \mathbf{\hat{x}}-y_{10}a \, \mathbf{\hat{y}}-z_{10}c \, \mathbf{\hat{z}} & \left(8p\right) & \mbox{C VII} \\ 
\mathbf{B}_{63} & = & y_{10} \, \mathbf{a}_{1} + x_{10} \, \mathbf{a}_{2} + \left(\frac{1}{2} - z_{10}\right) \, \mathbf{a}_{3} & = & y_{10}a \, \mathbf{\hat{x}} + x_{10}a \, \mathbf{\hat{y}} + \left(\frac{1}{2} - z_{10}\right)c \, \mathbf{\hat{z}} & \left(8p\right) & \mbox{C VII} \\ 
\mathbf{B}_{64} & = & -y_{10} \, \mathbf{a}_{1}-x_{10} \, \mathbf{a}_{2} + \left(\frac{1}{2} - z_{10}\right) \, \mathbf{a}_{3} & = & -y_{10}a \, \mathbf{\hat{x}}-x_{10}a \, \mathbf{\hat{y}} + \left(\frac{1}{2} - z_{10}\right)c \, \mathbf{\hat{z}} & \left(8p\right) & \mbox{C VII} \\ 
\mathbf{B}_{65} & = & x_{11} \, \mathbf{a}_{1} + y_{11} \, \mathbf{a}_{2} + z_{11} \, \mathbf{a}_{3} & = & x_{11}a \, \mathbf{\hat{x}} + y_{11}a \, \mathbf{\hat{y}} + z_{11}c \, \mathbf{\hat{z}} & \left(8p\right) & \mbox{C VIII} \\ 
\mathbf{B}_{66} & = & -x_{11} \, \mathbf{a}_{1}-y_{11} \, \mathbf{a}_{2} + z_{11} \, \mathbf{a}_{3} & = & -x_{11}a \, \mathbf{\hat{x}}-y_{11}a \, \mathbf{\hat{y}} + z_{11}c \, \mathbf{\hat{z}} & \left(8p\right) & \mbox{C VIII} \\ 
\mathbf{B}_{67} & = & -y_{11} \, \mathbf{a}_{1} + x_{11} \, \mathbf{a}_{2} + \left(\frac{1}{2} +z_{11}\right) \, \mathbf{a}_{3} & = & -y_{11}a \, \mathbf{\hat{x}} + x_{11}a \, \mathbf{\hat{y}} + \left(\frac{1}{2} +z_{11}\right)c \, \mathbf{\hat{z}} & \left(8p\right) & \mbox{C VIII} \\ 
\mathbf{B}_{68} & = & y_{11} \, \mathbf{a}_{1}-x_{11} \, \mathbf{a}_{2} + \left(\frac{1}{2} +z_{11}\right) \, \mathbf{a}_{3} & = & y_{11}a \, \mathbf{\hat{x}}-x_{11}a \, \mathbf{\hat{y}} + \left(\frac{1}{2} +z_{11}\right)c \, \mathbf{\hat{z}} & \left(8p\right) & \mbox{C VIII} \\ 
\mathbf{B}_{69} & = & -x_{11} \, \mathbf{a}_{1} + y_{11} \, \mathbf{a}_{2}-z_{11} \, \mathbf{a}_{3} & = & -x_{11}a \, \mathbf{\hat{x}} + y_{11}a \, \mathbf{\hat{y}}-z_{11}c \, \mathbf{\hat{z}} & \left(8p\right) & \mbox{C VIII} \\ 
\mathbf{B}_{70} & = & x_{11} \, \mathbf{a}_{1}-y_{11} \, \mathbf{a}_{2}-z_{11} \, \mathbf{a}_{3} & = & x_{11}a \, \mathbf{\hat{x}}-y_{11}a \, \mathbf{\hat{y}}-z_{11}c \, \mathbf{\hat{z}} & \left(8p\right) & \mbox{C VIII} \\ 
\mathbf{B}_{71} & = & y_{11} \, \mathbf{a}_{1} + x_{11} \, \mathbf{a}_{2} + \left(\frac{1}{2} - z_{11}\right) \, \mathbf{a}_{3} & = & y_{11}a \, \mathbf{\hat{x}} + x_{11}a \, \mathbf{\hat{y}} + \left(\frac{1}{2} - z_{11}\right)c \, \mathbf{\hat{z}} & \left(8p\right) & \mbox{C VIII} \\ 
\mathbf{B}_{72} & = & -y_{11} \, \mathbf{a}_{1}-x_{11} \, \mathbf{a}_{2} + \left(\frac{1}{2} - z_{11}\right) \, \mathbf{a}_{3} & = & -y_{11}a \, \mathbf{\hat{x}}-x_{11}a \, \mathbf{\hat{y}} + \left(\frac{1}{2} - z_{11}\right)c \, \mathbf{\hat{z}} & \left(8p\right) & \mbox{C VIII} \\ 
\mathbf{B}_{73} & = & x_{12} \, \mathbf{a}_{1} + y_{12} \, \mathbf{a}_{2} + z_{12} \, \mathbf{a}_{3} & = & x_{12}a \, \mathbf{\hat{x}} + y_{12}a \, \mathbf{\hat{y}} + z_{12}c \, \mathbf{\hat{z}} & \left(8p\right) & \mbox{C IX} \\ 
\mathbf{B}_{74} & = & -x_{12} \, \mathbf{a}_{1}-y_{12} \, \mathbf{a}_{2} + z_{12} \, \mathbf{a}_{3} & = & -x_{12}a \, \mathbf{\hat{x}}-y_{12}a \, \mathbf{\hat{y}} + z_{12}c \, \mathbf{\hat{z}} & \left(8p\right) & \mbox{C IX} \\ 
\mathbf{B}_{75} & = & -y_{12} \, \mathbf{a}_{1} + x_{12} \, \mathbf{a}_{2} + \left(\frac{1}{2} +z_{12}\right) \, \mathbf{a}_{3} & = & -y_{12}a \, \mathbf{\hat{x}} + x_{12}a \, \mathbf{\hat{y}} + \left(\frac{1}{2} +z_{12}\right)c \, \mathbf{\hat{z}} & \left(8p\right) & \mbox{C IX} \\ 
\mathbf{B}_{76} & = & y_{12} \, \mathbf{a}_{1}-x_{12} \, \mathbf{a}_{2} + \left(\frac{1}{2} +z_{12}\right) \, \mathbf{a}_{3} & = & y_{12}a \, \mathbf{\hat{x}}-x_{12}a \, \mathbf{\hat{y}} + \left(\frac{1}{2} +z_{12}\right)c \, \mathbf{\hat{z}} & \left(8p\right) & \mbox{C IX} \\ 
\mathbf{B}_{77} & = & -x_{12} \, \mathbf{a}_{1} + y_{12} \, \mathbf{a}_{2}-z_{12} \, \mathbf{a}_{3} & = & -x_{12}a \, \mathbf{\hat{x}} + y_{12}a \, \mathbf{\hat{y}}-z_{12}c \, \mathbf{\hat{z}} & \left(8p\right) & \mbox{C IX} \\ 
\mathbf{B}_{78} & = & x_{12} \, \mathbf{a}_{1}-y_{12} \, \mathbf{a}_{2}-z_{12} \, \mathbf{a}_{3} & = & x_{12}a \, \mathbf{\hat{x}}-y_{12}a \, \mathbf{\hat{y}}-z_{12}c \, \mathbf{\hat{z}} & \left(8p\right) & \mbox{C IX} \\ 
\mathbf{B}_{79} & = & y_{12} \, \mathbf{a}_{1} + x_{12} \, \mathbf{a}_{2} + \left(\frac{1}{2} - z_{12}\right) \, \mathbf{a}_{3} & = & y_{12}a \, \mathbf{\hat{x}} + x_{12}a \, \mathbf{\hat{y}} + \left(\frac{1}{2} - z_{12}\right)c \, \mathbf{\hat{z}} & \left(8p\right) & \mbox{C IX} \\ 
\mathbf{B}_{80} & = & -y_{12} \, \mathbf{a}_{1}-x_{12} \, \mathbf{a}_{2} + \left(\frac{1}{2} - z_{12}\right) \, \mathbf{a}_{3} & = & -y_{12}a \, \mathbf{\hat{x}}-x_{12}a \, \mathbf{\hat{y}} + \left(\frac{1}{2} - z_{12}\right)c \, \mathbf{\hat{z}} & \left(8p\right) & \mbox{C IX} \\ 
\mathbf{B}_{81} & = & x_{13} \, \mathbf{a}_{1} + y_{13} \, \mathbf{a}_{2} + z_{13} \, \mathbf{a}_{3} & = & x_{13}a \, \mathbf{\hat{x}} + y_{13}a \, \mathbf{\hat{y}} + z_{13}c \, \mathbf{\hat{z}} & \left(8p\right) & \mbox{C X} \\ 
\mathbf{B}_{82} & = & -x_{13} \, \mathbf{a}_{1}-y_{13} \, \mathbf{a}_{2} + z_{13} \, \mathbf{a}_{3} & = & -x_{13}a \, \mathbf{\hat{x}}-y_{13}a \, \mathbf{\hat{y}} + z_{13}c \, \mathbf{\hat{z}} & \left(8p\right) & \mbox{C X} \\ 
\mathbf{B}_{83} & = & -y_{13} \, \mathbf{a}_{1} + x_{13} \, \mathbf{a}_{2} + \left(\frac{1}{2} +z_{13}\right) \, \mathbf{a}_{3} & = & -y_{13}a \, \mathbf{\hat{x}} + x_{13}a \, \mathbf{\hat{y}} + \left(\frac{1}{2} +z_{13}\right)c \, \mathbf{\hat{z}} & \left(8p\right) & \mbox{C X} \\ 
\mathbf{B}_{84} & = & y_{13} \, \mathbf{a}_{1}-x_{13} \, \mathbf{a}_{2} + \left(\frac{1}{2} +z_{13}\right) \, \mathbf{a}_{3} & = & y_{13}a \, \mathbf{\hat{x}}-x_{13}a \, \mathbf{\hat{y}} + \left(\frac{1}{2} +z_{13}\right)c \, \mathbf{\hat{z}} & \left(8p\right) & \mbox{C X} \\ 
\mathbf{B}_{85} & = & -x_{13} \, \mathbf{a}_{1} + y_{13} \, \mathbf{a}_{2}-z_{13} \, \mathbf{a}_{3} & = & -x_{13}a \, \mathbf{\hat{x}} + y_{13}a \, \mathbf{\hat{y}}-z_{13}c \, \mathbf{\hat{z}} & \left(8p\right) & \mbox{C X} \\ 
\mathbf{B}_{86} & = & x_{13} \, \mathbf{a}_{1}-y_{13} \, \mathbf{a}_{2}-z_{13} \, \mathbf{a}_{3} & = & x_{13}a \, \mathbf{\hat{x}}-y_{13}a \, \mathbf{\hat{y}}-z_{13}c \, \mathbf{\hat{z}} & \left(8p\right) & \mbox{C X} \\ 
\mathbf{B}_{87} & = & y_{13} \, \mathbf{a}_{1} + x_{13} \, \mathbf{a}_{2} + \left(\frac{1}{2} - z_{13}\right) \, \mathbf{a}_{3} & = & y_{13}a \, \mathbf{\hat{x}} + x_{13}a \, \mathbf{\hat{y}} + \left(\frac{1}{2} - z_{13}\right)c \, \mathbf{\hat{z}} & \left(8p\right) & \mbox{C X} \\ 
\mathbf{B}_{88} & = & -y_{13} \, \mathbf{a}_{1}-x_{13} \, \mathbf{a}_{2} + \left(\frac{1}{2} - z_{13}\right) \, \mathbf{a}_{3} & = & -y_{13}a \, \mathbf{\hat{x}}-x_{13}a \, \mathbf{\hat{y}} + \left(\frac{1}{2} - z_{13}\right)c \, \mathbf{\hat{z}} & \left(8p\right) & \mbox{C X} \\ 
\mathbf{B}_{89} & = & x_{14} \, \mathbf{a}_{1} + y_{14} \, \mathbf{a}_{2} + z_{14} \, \mathbf{a}_{3} & = & x_{14}a \, \mathbf{\hat{x}} + y_{14}a \, \mathbf{\hat{y}} + z_{14}c \, \mathbf{\hat{z}} & \left(8p\right) & \mbox{C XI} \\ 
\mathbf{B}_{90} & = & -x_{14} \, \mathbf{a}_{1}-y_{14} \, \mathbf{a}_{2} + z_{14} \, \mathbf{a}_{3} & = & -x_{14}a \, \mathbf{\hat{x}}-y_{14}a \, \mathbf{\hat{y}} + z_{14}c \, \mathbf{\hat{z}} & \left(8p\right) & \mbox{C XI} \\ 
\mathbf{B}_{91} & = & -y_{14} \, \mathbf{a}_{1} + x_{14} \, \mathbf{a}_{2} + \left(\frac{1}{2} +z_{14}\right) \, \mathbf{a}_{3} & = & -y_{14}a \, \mathbf{\hat{x}} + x_{14}a \, \mathbf{\hat{y}} + \left(\frac{1}{2} +z_{14}\right)c \, \mathbf{\hat{z}} & \left(8p\right) & \mbox{C XI} \\ 
\mathbf{B}_{92} & = & y_{14} \, \mathbf{a}_{1}-x_{14} \, \mathbf{a}_{2} + \left(\frac{1}{2} +z_{14}\right) \, \mathbf{a}_{3} & = & y_{14}a \, \mathbf{\hat{x}}-x_{14}a \, \mathbf{\hat{y}} + \left(\frac{1}{2} +z_{14}\right)c \, \mathbf{\hat{z}} & \left(8p\right) & \mbox{C XI} \\ 
\mathbf{B}_{93} & = & -x_{14} \, \mathbf{a}_{1} + y_{14} \, \mathbf{a}_{2}-z_{14} \, \mathbf{a}_{3} & = & -x_{14}a \, \mathbf{\hat{x}} + y_{14}a \, \mathbf{\hat{y}}-z_{14}c \, \mathbf{\hat{z}} & \left(8p\right) & \mbox{C XI} \\ 
\mathbf{B}_{94} & = & x_{14} \, \mathbf{a}_{1}-y_{14} \, \mathbf{a}_{2}-z_{14} \, \mathbf{a}_{3} & = & x_{14}a \, \mathbf{\hat{x}}-y_{14}a \, \mathbf{\hat{y}}-z_{14}c \, \mathbf{\hat{z}} & \left(8p\right) & \mbox{C XI} \\ 
\mathbf{B}_{95} & = & y_{14} \, \mathbf{a}_{1} + x_{14} \, \mathbf{a}_{2} + \left(\frac{1}{2} - z_{14}\right) \, \mathbf{a}_{3} & = & y_{14}a \, \mathbf{\hat{x}} + x_{14}a \, \mathbf{\hat{y}} + \left(\frac{1}{2} - z_{14}\right)c \, \mathbf{\hat{z}} & \left(8p\right) & \mbox{C XI} \\ 
\mathbf{B}_{96} & = & -y_{14} \, \mathbf{a}_{1}-x_{14} \, \mathbf{a}_{2} + \left(\frac{1}{2} - z_{14}\right) \, \mathbf{a}_{3} & = & -y_{14}a \, \mathbf{\hat{x}}-x_{14}a \, \mathbf{\hat{y}} + \left(\frac{1}{2} - z_{14}\right)c \, \mathbf{\hat{z}} & \left(8p\right) & \mbox{C XI} \\ 
\mathbf{B}_{97} & = & x_{15} \, \mathbf{a}_{1} + y_{15} \, \mathbf{a}_{2} + z_{15} \, \mathbf{a}_{3} & = & x_{15}a \, \mathbf{\hat{x}} + y_{15}a \, \mathbf{\hat{y}} + z_{15}c \, \mathbf{\hat{z}} & \left(8p\right) & \mbox{C XII} \\ 
\mathbf{B}_{98} & = & -x_{15} \, \mathbf{a}_{1}-y_{15} \, \mathbf{a}_{2} + z_{15} \, \mathbf{a}_{3} & = & -x_{15}a \, \mathbf{\hat{x}}-y_{15}a \, \mathbf{\hat{y}} + z_{15}c \, \mathbf{\hat{z}} & \left(8p\right) & \mbox{C XII} \\ 
\mathbf{B}_{99} & = & -y_{15} \, \mathbf{a}_{1} + x_{15} \, \mathbf{a}_{2} + \left(\frac{1}{2} +z_{15}\right) \, \mathbf{a}_{3} & = & -y_{15}a \, \mathbf{\hat{x}} + x_{15}a \, \mathbf{\hat{y}} + \left(\frac{1}{2} +z_{15}\right)c \, \mathbf{\hat{z}} & \left(8p\right) & \mbox{C XII} \\ 
\mathbf{B}_{100} & = & y_{15} \, \mathbf{a}_{1}-x_{15} \, \mathbf{a}_{2} + \left(\frac{1}{2} +z_{15}\right) \, \mathbf{a}_{3} & = & y_{15}a \, \mathbf{\hat{x}}-x_{15}a \, \mathbf{\hat{y}} + \left(\frac{1}{2} +z_{15}\right)c \, \mathbf{\hat{z}} & \left(8p\right) & \mbox{C XII} \\ 
\mathbf{B}_{101} & = & -x_{15} \, \mathbf{a}_{1} + y_{15} \, \mathbf{a}_{2}-z_{15} \, \mathbf{a}_{3} & = & -x_{15}a \, \mathbf{\hat{x}} + y_{15}a \, \mathbf{\hat{y}}-z_{15}c \, \mathbf{\hat{z}} & \left(8p\right) & \mbox{C XII} \\ 
\mathbf{B}_{102} & = & x_{15} \, \mathbf{a}_{1}-y_{15} \, \mathbf{a}_{2}-z_{15} \, \mathbf{a}_{3} & = & x_{15}a \, \mathbf{\hat{x}}-y_{15}a \, \mathbf{\hat{y}}-z_{15}c \, \mathbf{\hat{z}} & \left(8p\right) & \mbox{C XII} \\ 
\mathbf{B}_{103} & = & y_{15} \, \mathbf{a}_{1} + x_{15} \, \mathbf{a}_{2} + \left(\frac{1}{2} - z_{15}\right) \, \mathbf{a}_{3} & = & y_{15}a \, \mathbf{\hat{x}} + x_{15}a \, \mathbf{\hat{y}} + \left(\frac{1}{2} - z_{15}\right)c \, \mathbf{\hat{z}} & \left(8p\right) & \mbox{C XII} \\ 
\mathbf{B}_{104} & = & -y_{15} \, \mathbf{a}_{1}-x_{15} \, \mathbf{a}_{2} + \left(\frac{1}{2} - z_{15}\right) \, \mathbf{a}_{3} & = & -y_{15}a \, \mathbf{\hat{x}}-x_{15}a \, \mathbf{\hat{y}} + \left(\frac{1}{2} - z_{15}\right)c \, \mathbf{\hat{z}} & \left(8p\right) & \mbox{C XII} \\ 
\mathbf{B}_{105} & = & x_{16} \, \mathbf{a}_{1} + y_{16} \, \mathbf{a}_{2} + z_{16} \, \mathbf{a}_{3} & = & x_{16}a \, \mathbf{\hat{x}} + y_{16}a \, \mathbf{\hat{y}} + z_{16}c \, \mathbf{\hat{z}} & \left(8p\right) & \mbox{C XIII} \\ 
\mathbf{B}_{106} & = & -x_{16} \, \mathbf{a}_{1}-y_{16} \, \mathbf{a}_{2} + z_{16} \, \mathbf{a}_{3} & = & -x_{16}a \, \mathbf{\hat{x}}-y_{16}a \, \mathbf{\hat{y}} + z_{16}c \, \mathbf{\hat{z}} & \left(8p\right) & \mbox{C XIII} \\ 
\mathbf{B}_{107} & = & -y_{16} \, \mathbf{a}_{1} + x_{16} \, \mathbf{a}_{2} + \left(\frac{1}{2} +z_{16}\right) \, \mathbf{a}_{3} & = & -y_{16}a \, \mathbf{\hat{x}} + x_{16}a \, \mathbf{\hat{y}} + \left(\frac{1}{2} +z_{16}\right)c \, \mathbf{\hat{z}} & \left(8p\right) & \mbox{C XIII} \\ 
\mathbf{B}_{108} & = & y_{16} \, \mathbf{a}_{1}-x_{16} \, \mathbf{a}_{2} + \left(\frac{1}{2} +z_{16}\right) \, \mathbf{a}_{3} & = & y_{16}a \, \mathbf{\hat{x}}-x_{16}a \, \mathbf{\hat{y}} + \left(\frac{1}{2} +z_{16}\right)c \, \mathbf{\hat{z}} & \left(8p\right) & \mbox{C XIII} \\ 
\mathbf{B}_{109} & = & -x_{16} \, \mathbf{a}_{1} + y_{16} \, \mathbf{a}_{2}-z_{16} \, \mathbf{a}_{3} & = & -x_{16}a \, \mathbf{\hat{x}} + y_{16}a \, \mathbf{\hat{y}}-z_{16}c \, \mathbf{\hat{z}} & \left(8p\right) & \mbox{C XIII} \\ 
\mathbf{B}_{110} & = & x_{16} \, \mathbf{a}_{1}-y_{16} \, \mathbf{a}_{2}-z_{16} \, \mathbf{a}_{3} & = & x_{16}a \, \mathbf{\hat{x}}-y_{16}a \, \mathbf{\hat{y}}-z_{16}c \, \mathbf{\hat{z}} & \left(8p\right) & \mbox{C XIII} \\ 
\mathbf{B}_{111} & = & y_{16} \, \mathbf{a}_{1} + x_{16} \, \mathbf{a}_{2} + \left(\frac{1}{2} - z_{16}\right) \, \mathbf{a}_{3} & = & y_{16}a \, \mathbf{\hat{x}} + x_{16}a \, \mathbf{\hat{y}} + \left(\frac{1}{2} - z_{16}\right)c \, \mathbf{\hat{z}} & \left(8p\right) & \mbox{C XIII} \\ 
\mathbf{B}_{112} & = & -y_{16} \, \mathbf{a}_{1}-x_{16} \, \mathbf{a}_{2} + \left(\frac{1}{2} - z_{16}\right) \, \mathbf{a}_{3} & = & -y_{16}a \, \mathbf{\hat{x}}-x_{16}a \, \mathbf{\hat{y}} + \left(\frac{1}{2} - z_{16}\right)c \, \mathbf{\hat{z}} & \left(8p\right) & \mbox{C XIII} \\ 
\mathbf{B}_{113} & = & x_{17} \, \mathbf{a}_{1} + y_{17} \, \mathbf{a}_{2} + z_{17} \, \mathbf{a}_{3} & = & x_{17}a \, \mathbf{\hat{x}} + y_{17}a \, \mathbf{\hat{y}} + z_{17}c \, \mathbf{\hat{z}} & \left(8p\right) & \mbox{C XIV} \\ 
\mathbf{B}_{114} & = & -x_{17} \, \mathbf{a}_{1}-y_{17} \, \mathbf{a}_{2} + z_{17} \, \mathbf{a}_{3} & = & -x_{17}a \, \mathbf{\hat{x}}-y_{17}a \, \mathbf{\hat{y}} + z_{17}c \, \mathbf{\hat{z}} & \left(8p\right) & \mbox{C XIV} \\ 
\mathbf{B}_{115} & = & -y_{17} \, \mathbf{a}_{1} + x_{17} \, \mathbf{a}_{2} + \left(\frac{1}{2} +z_{17}\right) \, \mathbf{a}_{3} & = & -y_{17}a \, \mathbf{\hat{x}} + x_{17}a \, \mathbf{\hat{y}} + \left(\frac{1}{2} +z_{17}\right)c \, \mathbf{\hat{z}} & \left(8p\right) & \mbox{C XIV} \\ 
\mathbf{B}_{116} & = & y_{17} \, \mathbf{a}_{1}-x_{17} \, \mathbf{a}_{2} + \left(\frac{1}{2} +z_{17}\right) \, \mathbf{a}_{3} & = & y_{17}a \, \mathbf{\hat{x}}-x_{17}a \, \mathbf{\hat{y}} + \left(\frac{1}{2} +z_{17}\right)c \, \mathbf{\hat{z}} & \left(8p\right) & \mbox{C XIV} \\ 
\mathbf{B}_{117} & = & -x_{17} \, \mathbf{a}_{1} + y_{17} \, \mathbf{a}_{2}-z_{17} \, \mathbf{a}_{3} & = & -x_{17}a \, \mathbf{\hat{x}} + y_{17}a \, \mathbf{\hat{y}}-z_{17}c \, \mathbf{\hat{z}} & \left(8p\right) & \mbox{C XIV} \\ 
\mathbf{B}_{118} & = & x_{17} \, \mathbf{a}_{1}-y_{17} \, \mathbf{a}_{2}-z_{17} \, \mathbf{a}_{3} & = & x_{17}a \, \mathbf{\hat{x}}-y_{17}a \, \mathbf{\hat{y}}-z_{17}c \, \mathbf{\hat{z}} & \left(8p\right) & \mbox{C XIV} \\ 
\mathbf{B}_{119} & = & y_{17} \, \mathbf{a}_{1} + x_{17} \, \mathbf{a}_{2} + \left(\frac{1}{2} - z_{17}\right) \, \mathbf{a}_{3} & = & y_{17}a \, \mathbf{\hat{x}} + x_{17}a \, \mathbf{\hat{y}} + \left(\frac{1}{2} - z_{17}\right)c \, \mathbf{\hat{z}} & \left(8p\right) & \mbox{C XIV} \\ 
\mathbf{B}_{120} & = & -y_{17} \, \mathbf{a}_{1}-x_{17} \, \mathbf{a}_{2} + \left(\frac{1}{2} - z_{17}\right) \, \mathbf{a}_{3} & = & -y_{17}a \, \mathbf{\hat{x}}-x_{17}a \, \mathbf{\hat{y}} + \left(\frac{1}{2} - z_{17}\right)c \, \mathbf{\hat{z}} & \left(8p\right) & \mbox{C XIV} \\ 
\mathbf{B}_{121} & = & x_{18} \, \mathbf{a}_{1} + y_{18} \, \mathbf{a}_{2} + z_{18} \, \mathbf{a}_{3} & = & x_{18}a \, \mathbf{\hat{x}} + y_{18}a \, \mathbf{\hat{y}} + z_{18}c \, \mathbf{\hat{z}} & \left(8p\right) & \mbox{C XV} \\ 
\mathbf{B}_{122} & = & -x_{18} \, \mathbf{a}_{1}-y_{18} \, \mathbf{a}_{2} + z_{18} \, \mathbf{a}_{3} & = & -x_{18}a \, \mathbf{\hat{x}}-y_{18}a \, \mathbf{\hat{y}} + z_{18}c \, \mathbf{\hat{z}} & \left(8p\right) & \mbox{C XV} \\ 
\mathbf{B}_{123} & = & -y_{18} \, \mathbf{a}_{1} + x_{18} \, \mathbf{a}_{2} + \left(\frac{1}{2} +z_{18}\right) \, \mathbf{a}_{3} & = & -y_{18}a \, \mathbf{\hat{x}} + x_{18}a \, \mathbf{\hat{y}} + \left(\frac{1}{2} +z_{18}\right)c \, \mathbf{\hat{z}} & \left(8p\right) & \mbox{C XV} \\ 
\mathbf{B}_{124} & = & y_{18} \, \mathbf{a}_{1}-x_{18} \, \mathbf{a}_{2} + \left(\frac{1}{2} +z_{18}\right) \, \mathbf{a}_{3} & = & y_{18}a \, \mathbf{\hat{x}}-x_{18}a \, \mathbf{\hat{y}} + \left(\frac{1}{2} +z_{18}\right)c \, \mathbf{\hat{z}} & \left(8p\right) & \mbox{C XV} \\ 
\mathbf{B}_{125} & = & -x_{18} \, \mathbf{a}_{1} + y_{18} \, \mathbf{a}_{2}-z_{18} \, \mathbf{a}_{3} & = & -x_{18}a \, \mathbf{\hat{x}} + y_{18}a \, \mathbf{\hat{y}}-z_{18}c \, \mathbf{\hat{z}} & \left(8p\right) & \mbox{C XV} \\ 
\mathbf{B}_{126} & = & x_{18} \, \mathbf{a}_{1}-y_{18} \, \mathbf{a}_{2}-z_{18} \, \mathbf{a}_{3} & = & x_{18}a \, \mathbf{\hat{x}}-y_{18}a \, \mathbf{\hat{y}}-z_{18}c \, \mathbf{\hat{z}} & \left(8p\right) & \mbox{C XV} \\ 
\mathbf{B}_{127} & = & y_{18} \, \mathbf{a}_{1} + x_{18} \, \mathbf{a}_{2} + \left(\frac{1}{2} - z_{18}\right) \, \mathbf{a}_{3} & = & y_{18}a \, \mathbf{\hat{x}} + x_{18}a \, \mathbf{\hat{y}} + \left(\frac{1}{2} - z_{18}\right)c \, \mathbf{\hat{z}} & \left(8p\right) & \mbox{C XV} \\ 
\mathbf{B}_{128} & = & -y_{18} \, \mathbf{a}_{1}-x_{18} \, \mathbf{a}_{2} + \left(\frac{1}{2} - z_{18}\right) \, \mathbf{a}_{3} & = & -y_{18}a \, \mathbf{\hat{x}}-x_{18}a \, \mathbf{\hat{y}} + \left(\frac{1}{2} - z_{18}\right)c \, \mathbf{\hat{z}} & \left(8p\right) & \mbox{C XV} \\ 
\mathbf{B}_{129} & = & x_{19} \, \mathbf{a}_{1} + y_{19} \, \mathbf{a}_{2} + z_{19} \, \mathbf{a}_{3} & = & x_{19}a \, \mathbf{\hat{x}} + y_{19}a \, \mathbf{\hat{y}} + z_{19}c \, \mathbf{\hat{z}} & \left(8p\right) & \mbox{C XVI} \\ 
\mathbf{B}_{130} & = & -x_{19} \, \mathbf{a}_{1}-y_{19} \, \mathbf{a}_{2} + z_{19} \, \mathbf{a}_{3} & = & -x_{19}a \, \mathbf{\hat{x}}-y_{19}a \, \mathbf{\hat{y}} + z_{19}c \, \mathbf{\hat{z}} & \left(8p\right) & \mbox{C XVI} \\ 
\mathbf{B}_{131} & = & -y_{19} \, \mathbf{a}_{1} + x_{19} \, \mathbf{a}_{2} + \left(\frac{1}{2} +z_{19}\right) \, \mathbf{a}_{3} & = & -y_{19}a \, \mathbf{\hat{x}} + x_{19}a \, \mathbf{\hat{y}} + \left(\frac{1}{2} +z_{19}\right)c \, \mathbf{\hat{z}} & \left(8p\right) & \mbox{C XVI} \\ 
\mathbf{B}_{132} & = & y_{19} \, \mathbf{a}_{1}-x_{19} \, \mathbf{a}_{2} + \left(\frac{1}{2} +z_{19}\right) \, \mathbf{a}_{3} & = & y_{19}a \, \mathbf{\hat{x}}-x_{19}a \, \mathbf{\hat{y}} + \left(\frac{1}{2} +z_{19}\right)c \, \mathbf{\hat{z}} & \left(8p\right) & \mbox{C XVI} \\ 
\mathbf{B}_{133} & = & -x_{19} \, \mathbf{a}_{1} + y_{19} \, \mathbf{a}_{2}-z_{19} \, \mathbf{a}_{3} & = & -x_{19}a \, \mathbf{\hat{x}} + y_{19}a \, \mathbf{\hat{y}}-z_{19}c \, \mathbf{\hat{z}} & \left(8p\right) & \mbox{C XVI} \\ 
\mathbf{B}_{134} & = & x_{19} \, \mathbf{a}_{1}-y_{19} \, \mathbf{a}_{2}-z_{19} \, \mathbf{a}_{3} & = & x_{19}a \, \mathbf{\hat{x}}-y_{19}a \, \mathbf{\hat{y}}-z_{19}c \, \mathbf{\hat{z}} & \left(8p\right) & \mbox{C XVI} \\ 
\mathbf{B}_{135} & = & y_{19} \, \mathbf{a}_{1} + x_{19} \, \mathbf{a}_{2} + \left(\frac{1}{2} - z_{19}\right) \, \mathbf{a}_{3} & = & y_{19}a \, \mathbf{\hat{x}} + x_{19}a \, \mathbf{\hat{y}} + \left(\frac{1}{2} - z_{19}\right)c \, \mathbf{\hat{z}} & \left(8p\right) & \mbox{C XVI} \\ 
\mathbf{B}_{136} & = & -y_{19} \, \mathbf{a}_{1}-x_{19} \, \mathbf{a}_{2} + \left(\frac{1}{2} - z_{19}\right) \, \mathbf{a}_{3} & = & -y_{19}a \, \mathbf{\hat{x}}-x_{19}a \, \mathbf{\hat{y}} + \left(\frac{1}{2} - z_{19}\right)c \, \mathbf{\hat{z}} & \left(8p\right) & \mbox{C XVI} \\ 
\mathbf{B}_{137} & = & x_{20} \, \mathbf{a}_{1} + y_{20} \, \mathbf{a}_{2} + z_{20} \, \mathbf{a}_{3} & = & x_{20}a \, \mathbf{\hat{x}} + y_{20}a \, \mathbf{\hat{y}} + z_{20}c \, \mathbf{\hat{z}} & \left(8p\right) & \mbox{P I} \\ 
\mathbf{B}_{138} & = & -x_{20} \, \mathbf{a}_{1}-y_{20} \, \mathbf{a}_{2} + z_{20} \, \mathbf{a}_{3} & = & -x_{20}a \, \mathbf{\hat{x}}-y_{20}a \, \mathbf{\hat{y}} + z_{20}c \, \mathbf{\hat{z}} & \left(8p\right) & \mbox{P I} \\ 
\mathbf{B}_{139} & = & -y_{20} \, \mathbf{a}_{1} + x_{20} \, \mathbf{a}_{2} + \left(\frac{1}{2} +z_{20}\right) \, \mathbf{a}_{3} & = & -y_{20}a \, \mathbf{\hat{x}} + x_{20}a \, \mathbf{\hat{y}} + \left(\frac{1}{2} +z_{20}\right)c \, \mathbf{\hat{z}} & \left(8p\right) & \mbox{P I} \\ 
\mathbf{B}_{140} & = & y_{20} \, \mathbf{a}_{1}-x_{20} \, \mathbf{a}_{2} + \left(\frac{1}{2} +z_{20}\right) \, \mathbf{a}_{3} & = & y_{20}a \, \mathbf{\hat{x}}-x_{20}a \, \mathbf{\hat{y}} + \left(\frac{1}{2} +z_{20}\right)c \, \mathbf{\hat{z}} & \left(8p\right) & \mbox{P I} \\ 
\mathbf{B}_{141} & = & -x_{20} \, \mathbf{a}_{1} + y_{20} \, \mathbf{a}_{2}-z_{20} \, \mathbf{a}_{3} & = & -x_{20}a \, \mathbf{\hat{x}} + y_{20}a \, \mathbf{\hat{y}}-z_{20}c \, \mathbf{\hat{z}} & \left(8p\right) & \mbox{P I} \\ 
\mathbf{B}_{142} & = & x_{20} \, \mathbf{a}_{1}-y_{20} \, \mathbf{a}_{2}-z_{20} \, \mathbf{a}_{3} & = & x_{20}a \, \mathbf{\hat{x}}-y_{20}a \, \mathbf{\hat{y}}-z_{20}c \, \mathbf{\hat{z}} & \left(8p\right) & \mbox{P I} \\ 
\mathbf{B}_{143} & = & y_{20} \, \mathbf{a}_{1} + x_{20} \, \mathbf{a}_{2} + \left(\frac{1}{2} - z_{20}\right) \, \mathbf{a}_{3} & = & y_{20}a \, \mathbf{\hat{x}} + x_{20}a \, \mathbf{\hat{y}} + \left(\frac{1}{2} - z_{20}\right)c \, \mathbf{\hat{z}} & \left(8p\right) & \mbox{P I} \\ 
\mathbf{B}_{144} & = & -y_{20} \, \mathbf{a}_{1}-x_{20} \, \mathbf{a}_{2} + \left(\frac{1}{2} - z_{20}\right) \, \mathbf{a}_{3} & = & -y_{20}a \, \mathbf{\hat{x}}-x_{20}a \, \mathbf{\hat{y}} + \left(\frac{1}{2} - z_{20}\right)c \, \mathbf{\hat{z}} & \left(8p\right) & \mbox{P I} \\ 
\mathbf{B}_{145} & = & x_{21} \, \mathbf{a}_{1} + y_{21} \, \mathbf{a}_{2} + z_{21} \, \mathbf{a}_{3} & = & x_{21}a \, \mathbf{\hat{x}} + y_{21}a \, \mathbf{\hat{y}} + z_{21}c \, \mathbf{\hat{z}} & \left(8p\right) & \mbox{P II} \\ 
\mathbf{B}_{146} & = & -x_{21} \, \mathbf{a}_{1}-y_{21} \, \mathbf{a}_{2} + z_{21} \, \mathbf{a}_{3} & = & -x_{21}a \, \mathbf{\hat{x}}-y_{21}a \, \mathbf{\hat{y}} + z_{21}c \, \mathbf{\hat{z}} & \left(8p\right) & \mbox{P II} \\ 
\mathbf{B}_{147} & = & -y_{21} \, \mathbf{a}_{1} + x_{21} \, \mathbf{a}_{2} + \left(\frac{1}{2} +z_{21}\right) \, \mathbf{a}_{3} & = & -y_{21}a \, \mathbf{\hat{x}} + x_{21}a \, \mathbf{\hat{y}} + \left(\frac{1}{2} +z_{21}\right)c \, \mathbf{\hat{z}} & \left(8p\right) & \mbox{P II} \\ 
\mathbf{B}_{148} & = & y_{21} \, \mathbf{a}_{1}-x_{21} \, \mathbf{a}_{2} + \left(\frac{1}{2} +z_{21}\right) \, \mathbf{a}_{3} & = & y_{21}a \, \mathbf{\hat{x}}-x_{21}a \, \mathbf{\hat{y}} + \left(\frac{1}{2} +z_{21}\right)c \, \mathbf{\hat{z}} & \left(8p\right) & \mbox{P II} \\ 
\mathbf{B}_{149} & = & -x_{21} \, \mathbf{a}_{1} + y_{21} \, \mathbf{a}_{2}-z_{21} \, \mathbf{a}_{3} & = & -x_{21}a \, \mathbf{\hat{x}} + y_{21}a \, \mathbf{\hat{y}}-z_{21}c \, \mathbf{\hat{z}} & \left(8p\right) & \mbox{P II} \\ 
\mathbf{B}_{150} & = & x_{21} \, \mathbf{a}_{1}-y_{21} \, \mathbf{a}_{2}-z_{21} \, \mathbf{a}_{3} & = & x_{21}a \, \mathbf{\hat{x}}-y_{21}a \, \mathbf{\hat{y}}-z_{21}c \, \mathbf{\hat{z}} & \left(8p\right) & \mbox{P II} \\ 
\mathbf{B}_{151} & = & y_{21} \, \mathbf{a}_{1} + x_{21} \, \mathbf{a}_{2} + \left(\frac{1}{2} - z_{21}\right) \, \mathbf{a}_{3} & = & y_{21}a \, \mathbf{\hat{x}} + x_{21}a \, \mathbf{\hat{y}} + \left(\frac{1}{2} - z_{21}\right)c \, \mathbf{\hat{z}} & \left(8p\right) & \mbox{P II} \\ 
\mathbf{B}_{152} & = & -y_{21} \, \mathbf{a}_{1}-x_{21} \, \mathbf{a}_{2} + \left(\frac{1}{2} - z_{21}\right) \, \mathbf{a}_{3} & = & -y_{21}a \, \mathbf{\hat{x}}-x_{21}a \, \mathbf{\hat{y}} + \left(\frac{1}{2} - z_{21}\right)c \, \mathbf{\hat{z}} & \left(8p\right) & \mbox{P II} \\ 
\mathbf{B}_{153} & = & x_{22} \, \mathbf{a}_{1} + y_{22} \, \mathbf{a}_{2} + z_{22} \, \mathbf{a}_{3} & = & x_{22}a \, \mathbf{\hat{x}} + y_{22}a \, \mathbf{\hat{y}} + z_{22}c \, \mathbf{\hat{z}} & \left(8p\right) & \mbox{S I} \\ 
\mathbf{B}_{154} & = & -x_{22} \, \mathbf{a}_{1}-y_{22} \, \mathbf{a}_{2} + z_{22} \, \mathbf{a}_{3} & = & -x_{22}a \, \mathbf{\hat{x}}-y_{22}a \, \mathbf{\hat{y}} + z_{22}c \, \mathbf{\hat{z}} & \left(8p\right) & \mbox{S I} \\ 
\mathbf{B}_{155} & = & -y_{22} \, \mathbf{a}_{1} + x_{22} \, \mathbf{a}_{2} + \left(\frac{1}{2} +z_{22}\right) \, \mathbf{a}_{3} & = & -y_{22}a \, \mathbf{\hat{x}} + x_{22}a \, \mathbf{\hat{y}} + \left(\frac{1}{2} +z_{22}\right)c \, \mathbf{\hat{z}} & \left(8p\right) & \mbox{S I} \\ 
\mathbf{B}_{156} & = & y_{22} \, \mathbf{a}_{1}-x_{22} \, \mathbf{a}_{2} + \left(\frac{1}{2} +z_{22}\right) \, \mathbf{a}_{3} & = & y_{22}a \, \mathbf{\hat{x}}-x_{22}a \, \mathbf{\hat{y}} + \left(\frac{1}{2} +z_{22}\right)c \, \mathbf{\hat{z}} & \left(8p\right) & \mbox{S I} \\ 
\mathbf{B}_{157} & = & -x_{22} \, \mathbf{a}_{1} + y_{22} \, \mathbf{a}_{2}-z_{22} \, \mathbf{a}_{3} & = & -x_{22}a \, \mathbf{\hat{x}} + y_{22}a \, \mathbf{\hat{y}}-z_{22}c \, \mathbf{\hat{z}} & \left(8p\right) & \mbox{S I} \\ 
\mathbf{B}_{158} & = & x_{22} \, \mathbf{a}_{1}-y_{22} \, \mathbf{a}_{2}-z_{22} \, \mathbf{a}_{3} & = & x_{22}a \, \mathbf{\hat{x}}-y_{22}a \, \mathbf{\hat{y}}-z_{22}c \, \mathbf{\hat{z}} & \left(8p\right) & \mbox{S I} \\ 
\mathbf{B}_{159} & = & y_{22} \, \mathbf{a}_{1} + x_{22} \, \mathbf{a}_{2} + \left(\frac{1}{2} - z_{22}\right) \, \mathbf{a}_{3} & = & y_{22}a \, \mathbf{\hat{x}} + x_{22}a \, \mathbf{\hat{y}} + \left(\frac{1}{2} - z_{22}\right)c \, \mathbf{\hat{z}} & \left(8p\right) & \mbox{S I} \\ 
\mathbf{B}_{160} & = & -y_{22} \, \mathbf{a}_{1}-x_{22} \, \mathbf{a}_{2} + \left(\frac{1}{2} - z_{22}\right) \, \mathbf{a}_{3} & = & -y_{22}a \, \mathbf{\hat{x}}-x_{22}a \, \mathbf{\hat{y}} + \left(\frac{1}{2} - z_{22}\right)c \, \mathbf{\hat{z}} & \left(8p\right) & \mbox{S I} \\ 
\mathbf{B}_{161} & = & x_{23} \, \mathbf{a}_{1} + y_{23} \, \mathbf{a}_{2} + z_{23} \, \mathbf{a}_{3} & = & x_{23}a \, \mathbf{\hat{x}} + y_{23}a \, \mathbf{\hat{y}} + z_{23}c \, \mathbf{\hat{z}} & \left(8p\right) & \mbox{S II} \\ 
\mathbf{B}_{162} & = & -x_{23} \, \mathbf{a}_{1}-y_{23} \, \mathbf{a}_{2} + z_{23} \, \mathbf{a}_{3} & = & -x_{23}a \, \mathbf{\hat{x}}-y_{23}a \, \mathbf{\hat{y}} + z_{23}c \, \mathbf{\hat{z}} & \left(8p\right) & \mbox{S II} \\ 
\mathbf{B}_{163} & = & -y_{23} \, \mathbf{a}_{1} + x_{23} \, \mathbf{a}_{2} + \left(\frac{1}{2} +z_{23}\right) \, \mathbf{a}_{3} & = & -y_{23}a \, \mathbf{\hat{x}} + x_{23}a \, \mathbf{\hat{y}} + \left(\frac{1}{2} +z_{23}\right)c \, \mathbf{\hat{z}} & \left(8p\right) & \mbox{S II} \\ 
\mathbf{B}_{164} & = & y_{23} \, \mathbf{a}_{1}-x_{23} \, \mathbf{a}_{2} + \left(\frac{1}{2} +z_{23}\right) \, \mathbf{a}_{3} & = & y_{23}a \, \mathbf{\hat{x}}-x_{23}a \, \mathbf{\hat{y}} + \left(\frac{1}{2} +z_{23}\right)c \, \mathbf{\hat{z}} & \left(8p\right) & \mbox{S II} \\ 
\mathbf{B}_{165} & = & -x_{23} \, \mathbf{a}_{1} + y_{23} \, \mathbf{a}_{2}-z_{23} \, \mathbf{a}_{3} & = & -x_{23}a \, \mathbf{\hat{x}} + y_{23}a \, \mathbf{\hat{y}}-z_{23}c \, \mathbf{\hat{z}} & \left(8p\right) & \mbox{S II} \\ 
\mathbf{B}_{166} & = & x_{23} \, \mathbf{a}_{1}-y_{23} \, \mathbf{a}_{2}-z_{23} \, \mathbf{a}_{3} & = & x_{23}a \, \mathbf{\hat{x}}-y_{23}a \, \mathbf{\hat{y}}-z_{23}c \, \mathbf{\hat{z}} & \left(8p\right) & \mbox{S II} \\ 
\mathbf{B}_{167} & = & y_{23} \, \mathbf{a}_{1} + x_{23} \, \mathbf{a}_{2} + \left(\frac{1}{2} - z_{23}\right) \, \mathbf{a}_{3} & = & y_{23}a \, \mathbf{\hat{x}} + x_{23}a \, \mathbf{\hat{y}} + \left(\frac{1}{2} - z_{23}\right)c \, \mathbf{\hat{z}} & \left(8p\right) & \mbox{S II} \\ 
\mathbf{B}_{168} & = & -y_{23} \, \mathbf{a}_{1}-x_{23} \, \mathbf{a}_{2} + \left(\frac{1}{2} - z_{23}\right) \, \mathbf{a}_{3} & = & -y_{23}a \, \mathbf{\hat{x}}-x_{23}a \, \mathbf{\hat{y}} + \left(\frac{1}{2} - z_{23}\right)c \, \mathbf{\hat{z}} & \left(8p\right) & \mbox{S II} \\ 
\mathbf{B}_{169} & = & x_{24} \, \mathbf{a}_{1} + y_{24} \, \mathbf{a}_{2} + z_{24} \, \mathbf{a}_{3} & = & x_{24}a \, \mathbf{\hat{x}} + y_{24}a \, \mathbf{\hat{y}} + z_{24}c \, \mathbf{\hat{z}} & \left(8p\right) & \mbox{S III} \\ 
\mathbf{B}_{170} & = & -x_{24} \, \mathbf{a}_{1}-y_{24} \, \mathbf{a}_{2} + z_{24} \, \mathbf{a}_{3} & = & -x_{24}a \, \mathbf{\hat{x}}-y_{24}a \, \mathbf{\hat{y}} + z_{24}c \, \mathbf{\hat{z}} & \left(8p\right) & \mbox{S III} \\ 
\mathbf{B}_{171} & = & -y_{24} \, \mathbf{a}_{1} + x_{24} \, \mathbf{a}_{2} + \left(\frac{1}{2} +z_{24}\right) \, \mathbf{a}_{3} & = & -y_{24}a \, \mathbf{\hat{x}} + x_{24}a \, \mathbf{\hat{y}} + \left(\frac{1}{2} +z_{24}\right)c \, \mathbf{\hat{z}} & \left(8p\right) & \mbox{S III} \\ 
\mathbf{B}_{172} & = & y_{24} \, \mathbf{a}_{1}-x_{24} \, \mathbf{a}_{2} + \left(\frac{1}{2} +z_{24}\right) \, \mathbf{a}_{3} & = & y_{24}a \, \mathbf{\hat{x}}-x_{24}a \, \mathbf{\hat{y}} + \left(\frac{1}{2} +z_{24}\right)c \, \mathbf{\hat{z}} & \left(8p\right) & \mbox{S III} \\ 
\mathbf{B}_{173} & = & -x_{24} \, \mathbf{a}_{1} + y_{24} \, \mathbf{a}_{2}-z_{24} \, \mathbf{a}_{3} & = & -x_{24}a \, \mathbf{\hat{x}} + y_{24}a \, \mathbf{\hat{y}}-z_{24}c \, \mathbf{\hat{z}} & \left(8p\right) & \mbox{S III} \\ 
\mathbf{B}_{174} & = & x_{24} \, \mathbf{a}_{1}-y_{24} \, \mathbf{a}_{2}-z_{24} \, \mathbf{a}_{3} & = & x_{24}a \, \mathbf{\hat{x}}-y_{24}a \, \mathbf{\hat{y}}-z_{24}c \, \mathbf{\hat{z}} & \left(8p\right) & \mbox{S III} \\ 
\mathbf{B}_{175} & = & y_{24} \, \mathbf{a}_{1} + x_{24} \, \mathbf{a}_{2} + \left(\frac{1}{2} - z_{24}\right) \, \mathbf{a}_{3} & = & y_{24}a \, \mathbf{\hat{x}} + x_{24}a \, \mathbf{\hat{y}} + \left(\frac{1}{2} - z_{24}\right)c \, \mathbf{\hat{z}} & \left(8p\right) & \mbox{S III} \\ 
\mathbf{B}_{176} & = & -y_{24} \, \mathbf{a}_{1}-x_{24} \, \mathbf{a}_{2} + \left(\frac{1}{2} - z_{24}\right) \, \mathbf{a}_{3} & = & -y_{24}a \, \mathbf{\hat{x}}-x_{24}a \, \mathbf{\hat{y}} + \left(\frac{1}{2} - z_{24}\right)c \, \mathbf{\hat{z}} & \left(8p\right) & \mbox{S III} \\ 
\mathbf{B}_{177} & = & x_{25} \, \mathbf{a}_{1} + y_{25} \, \mathbf{a}_{2} + z_{25} \, \mathbf{a}_{3} & = & x_{25}a \, \mathbf{\hat{x}} + y_{25}a \, \mathbf{\hat{y}} + z_{25}c \, \mathbf{\hat{z}} & \left(8p\right) & \mbox{S IV} \\ 
\mathbf{B}_{178} & = & -x_{25} \, \mathbf{a}_{1}-y_{25} \, \mathbf{a}_{2} + z_{25} \, \mathbf{a}_{3} & = & -x_{25}a \, \mathbf{\hat{x}}-y_{25}a \, \mathbf{\hat{y}} + z_{25}c \, \mathbf{\hat{z}} & \left(8p\right) & \mbox{S IV} \\ 
\mathbf{B}_{179} & = & -y_{25} \, \mathbf{a}_{1} + x_{25} \, \mathbf{a}_{2} + \left(\frac{1}{2} +z_{25}\right) \, \mathbf{a}_{3} & = & -y_{25}a \, \mathbf{\hat{x}} + x_{25}a \, \mathbf{\hat{y}} + \left(\frac{1}{2} +z_{25}\right)c \, \mathbf{\hat{z}} & \left(8p\right) & \mbox{S IV} \\ 
\mathbf{B}_{180} & = & y_{25} \, \mathbf{a}_{1}-x_{25} \, \mathbf{a}_{2} + \left(\frac{1}{2} +z_{25}\right) \, \mathbf{a}_{3} & = & y_{25}a \, \mathbf{\hat{x}}-x_{25}a \, \mathbf{\hat{y}} + \left(\frac{1}{2} +z_{25}\right)c \, \mathbf{\hat{z}} & \left(8p\right) & \mbox{S IV} \\ 
\mathbf{B}_{181} & = & -x_{25} \, \mathbf{a}_{1} + y_{25} \, \mathbf{a}_{2}-z_{25} \, \mathbf{a}_{3} & = & -x_{25}a \, \mathbf{\hat{x}} + y_{25}a \, \mathbf{\hat{y}}-z_{25}c \, \mathbf{\hat{z}} & \left(8p\right) & \mbox{S IV} \\ 
\mathbf{B}_{182} & = & x_{25} \, \mathbf{a}_{1}-y_{25} \, \mathbf{a}_{2}-z_{25} \, \mathbf{a}_{3} & = & x_{25}a \, \mathbf{\hat{x}}-y_{25}a \, \mathbf{\hat{y}}-z_{25}c \, \mathbf{\hat{z}} & \left(8p\right) & \mbox{S IV} \\ 
\mathbf{B}_{183} & = & y_{25} \, \mathbf{a}_{1} + x_{25} \, \mathbf{a}_{2} + \left(\frac{1}{2} - z_{25}\right) \, \mathbf{a}_{3} & = & y_{25}a \, \mathbf{\hat{x}} + x_{25}a \, \mathbf{\hat{y}} + \left(\frac{1}{2} - z_{25}\right)c \, \mathbf{\hat{z}} & \left(8p\right) & \mbox{S IV} \\ 
\mathbf{B}_{184} & = & -y_{25} \, \mathbf{a}_{1}-x_{25} \, \mathbf{a}_{2} + \left(\frac{1}{2} - z_{25}\right) \, \mathbf{a}_{3} & = & -y_{25}a \, \mathbf{\hat{x}}-x_{25}a \, \mathbf{\hat{y}} + \left(\frac{1}{2} - z_{25}\right)c \, \mathbf{\hat{z}} & \left(8p\right) & \mbox{S IV} \\ 
\end{longtabu}
\renewcommand{\arraystretch}{1.0}
\noindent \hrulefill
\\
\textbf{References:}
\vspace*{-0.25cm}
\begin{flushleft}
  - \bibentry{Spiliadis_SG93_1982}. \\
\end{flushleft}
\textbf{Found in:}
\vspace*{-0.25cm}
\begin{flushleft}
  - \bibentry{Hoffmann_SpaceGroupProject_2014}. \\
\end{flushleft}
\noindent \hrulefill
\\
\textbf{Geometry files:}
\\
\noindent  - CIF: pp. {\hyperref[AB32CD4E8_tP184_93_i_16p_af_2p_4p_cif]{\pageref{AB32CD4E8_tP184_93_i_16p_af_2p_4p_cif}}} \\
\noindent  - POSCAR: pp. {\hyperref[AB32CD4E8_tP184_93_i_16p_af_2p_4p_poscar]{\pageref{AB32CD4E8_tP184_93_i_16p_af_2p_4p_poscar}}} \\
\onecolumn
{\phantomsection\label{A14B3C5_tP44_94_c3g_ad_bg}}
\subsection*{\huge \textbf{{\normalfont \begin{raggedleft}Na$_{5}$Fe$_{3}$F$_{14}$ (High-temperature) Structure: \end{raggedleft} \\ A14B3C5\_tP44\_94\_c3g\_ad\_bg}}}
\noindent \hrulefill
\vspace*{0.25cm}
\begin{figure}[htp]
  \centering
  \vspace{-1em}
  {\includegraphics[width=1\textwidth]{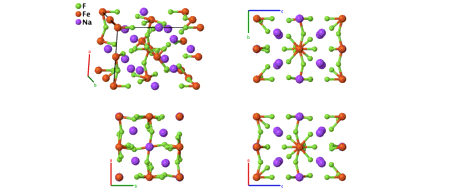}}
\end{figure}
\vspace*{-0.5cm}
\renewcommand{\arraystretch}{1.5}
\begin{equation*}
  \begin{array}{>{$\hspace{-0.15cm}}l<{$}>{$}p{0.5cm}<{$}>{$}p{18.5cm}<{$}}
    \mbox{\large \textbf{Prototype}} &\colon & \ce{Na5Fe3F14} \\
    \mbox{\large \textbf{\AFLOW\ prototype label}} &\colon & \mbox{A14B3C5\_tP44\_94\_c3g\_ad\_bg} \\
    \mbox{\large \textbf{\textit{Strukturbericht} designation}} &\colon & \mbox{None} \\
    \mbox{\large \textbf{Pearson symbol}} &\colon & \mbox{tP44} \\
    \mbox{\large \textbf{Space group number}} &\colon & 94 \\
    \mbox{\large \textbf{Space group symbol}} &\colon & P4_{2}2_{1}2 \\
    \mbox{\large \textbf{\AFLOW\ prototype command}} &\colon &  \texttt{aflow} \,  \, \texttt{-{}-proto=A14B3C5\_tP44\_94\_c3g\_ad\_bg } \, \newline \texttt{-{}-params=}{a,c/a,z_{3},z_{4},x_{5},y_{5},z_{5},x_{6},y_{6},z_{6},x_{7},y_{7},z_{7},x_{8},y_{8},z_{8} }
  \end{array}
\end{equation*}
\renewcommand{\arraystretch}{1.0}

\noindent \parbox{1 \linewidth}{
\noindent \hrulefill
\\
\textbf{Simple Tetragonal primitive vectors:} \\
\vspace*{-0.25cm}
\begin{tabular}{cc}
  \begin{tabular}{c}
    \parbox{0.6 \linewidth}{
      \renewcommand{\arraystretch}{1.5}
      \begin{equation*}
        \centering
        \begin{array}{ccc}
              \mathbf{a}_1 & = & a \, \mathbf{\hat{x}} \\
    \mathbf{a}_2 & = & a \, \mathbf{\hat{y}} \\
    \mathbf{a}_3 & = & c \, \mathbf{\hat{z}} \\

        \end{array}
      \end{equation*}
    }
    \renewcommand{\arraystretch}{1.0}
  \end{tabular}
  \begin{tabular}{c}
    \includegraphics[width=0.3\linewidth]{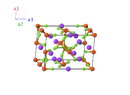} \\
  \end{tabular}
\end{tabular}

}
\vspace*{-0.25cm}

\noindent \hrulefill
\\
\textbf{Basis vectors:}
\vspace*{-0.25cm}
\renewcommand{\arraystretch}{1.5}
\begin{longtabu} to \textwidth{>{\centering $}X[-1,c,c]<{$}>{\centering $}X[-1,c,c]<{$}>{\centering $}X[-1,c,c]<{$}>{\centering $}X[-1,c,c]<{$}>{\centering $}X[-1,c,c]<{$}>{\centering $}X[-1,c,c]<{$}>{\centering $}X[-1,c,c]<{$}}
  & & \mbox{Lattice Coordinates} & & \mbox{Cartesian Coordinates} &\mbox{Wyckoff Position} & \mbox{Atom Type} \\  
  \mathbf{B}_{1} & = & 0 \, \mathbf{a}_{1} + 0 \, \mathbf{a}_{2} + 0 \, \mathbf{a}_{3} & = & 0 \, \mathbf{\hat{x}} + 0 \, \mathbf{\hat{y}} + 0 \, \mathbf{\hat{z}} & \left(2a\right) & \mbox{Fe I} \\ 
\mathbf{B}_{2} & = & \frac{1}{2} \, \mathbf{a}_{1} + \frac{1}{2} \, \mathbf{a}_{2} + \frac{1}{2} \, \mathbf{a}_{3} & = & \frac{1}{2}a \, \mathbf{\hat{x}} + \frac{1}{2}a \, \mathbf{\hat{y}} + \frac{1}{2}c \, \mathbf{\hat{z}} & \left(2a\right) & \mbox{Fe I} \\ 
\mathbf{B}_{3} & = & \frac{1}{2} \, \mathbf{a}_{3} & = & \frac{1}{2}c \, \mathbf{\hat{z}} & \left(2b\right) & \mbox{Na I} \\ 
\mathbf{B}_{4} & = & \frac{1}{2} \, \mathbf{a}_{1} + \frac{1}{2} \, \mathbf{a}_{2} & = & \frac{1}{2}a \, \mathbf{\hat{x}} + \frac{1}{2}a \, \mathbf{\hat{y}} & \left(2b\right) & \mbox{Na I} \\ 
\mathbf{B}_{5} & = & z_{3} \, \mathbf{a}_{3} & = & z_{3}c \, \mathbf{\hat{z}} & \left(4c\right) & \mbox{F I} \\ 
\mathbf{B}_{6} & = & \frac{1}{2} \, \mathbf{a}_{1} + \frac{1}{2} \, \mathbf{a}_{2} + \left(\frac{1}{2} +z_{3}\right) \, \mathbf{a}_{3} & = & \frac{1}{2}a \, \mathbf{\hat{x}} + \frac{1}{2}a \, \mathbf{\hat{y}} + \left(\frac{1}{2} +z_{3}\right)c \, \mathbf{\hat{z}} & \left(4c\right) & \mbox{F I} \\ 
\mathbf{B}_{7} & = & \frac{1}{2} \, \mathbf{a}_{1} + \frac{1}{2} \, \mathbf{a}_{2} + \left(\frac{1}{2} - z_{3}\right) \, \mathbf{a}_{3} & = & \frac{1}{2}a \, \mathbf{\hat{x}} + \frac{1}{2}a \, \mathbf{\hat{y}} + \left(\frac{1}{2} - z_{3}\right)c \, \mathbf{\hat{z}} & \left(4c\right) & \mbox{F I} \\ 
\mathbf{B}_{8} & = & -z_{3} \, \mathbf{a}_{3} & = & -z_{3}c \, \mathbf{\hat{z}} & \left(4c\right) & \mbox{F I} \\ 
\mathbf{B}_{9} & = & \frac{1}{2} \, \mathbf{a}_{2} + z_{4} \, \mathbf{a}_{3} & = & \frac{1}{2}a \, \mathbf{\hat{y}} + z_{4}c \, \mathbf{\hat{z}} & \left(4d\right) & \mbox{Fe II} \\ 
\mathbf{B}_{10} & = & \frac{1}{2} \, \mathbf{a}_{2} + \left(\frac{1}{2} +z_{4}\right) \, \mathbf{a}_{3} & = & \frac{1}{2}a \, \mathbf{\hat{y}} + \left(\frac{1}{2} +z_{4}\right)c \, \mathbf{\hat{z}} & \left(4d\right) & \mbox{Fe II} \\ 
\mathbf{B}_{11} & = & \frac{1}{2} \, \mathbf{a}_{1} + \left(\frac{1}{2} - z_{4}\right) \, \mathbf{a}_{3} & = & \frac{1}{2}a \, \mathbf{\hat{x}} + \left(\frac{1}{2} - z_{4}\right)c \, \mathbf{\hat{z}} & \left(4d\right) & \mbox{Fe II} \\ 
\mathbf{B}_{12} & = & \frac{1}{2} \, \mathbf{a}_{1} + -z_{4} \, \mathbf{a}_{3} & = & \frac{1}{2}a \, \mathbf{\hat{x}} + -z_{4}c \, \mathbf{\hat{z}} & \left(4d\right) & \mbox{Fe II} \\ 
\mathbf{B}_{13} & = & x_{5} \, \mathbf{a}_{1} + y_{5} \, \mathbf{a}_{2} + z_{5} \, \mathbf{a}_{3} & = & x_{5}a \, \mathbf{\hat{x}} + y_{5}a \, \mathbf{\hat{y}} + z_{5}c \, \mathbf{\hat{z}} & \left(8g\right) & \mbox{F II} \\ 
\mathbf{B}_{14} & = & -x_{5} \, \mathbf{a}_{1}-y_{5} \, \mathbf{a}_{2} + z_{5} \, \mathbf{a}_{3} & = & -x_{5}a \, \mathbf{\hat{x}}-y_{5}a \, \mathbf{\hat{y}} + z_{5}c \, \mathbf{\hat{z}} & \left(8g\right) & \mbox{F II} \\ 
\mathbf{B}_{15} & = & \left(\frac{1}{2} - y_{5}\right) \, \mathbf{a}_{1} + \left(\frac{1}{2} +x_{5}\right) \, \mathbf{a}_{2} + \left(\frac{1}{2} +z_{5}\right) \, \mathbf{a}_{3} & = & \left(\frac{1}{2} - y_{5}\right)a \, \mathbf{\hat{x}} + \left(\frac{1}{2} +x_{5}\right)a \, \mathbf{\hat{y}} + \left(\frac{1}{2} +z_{5}\right)c \, \mathbf{\hat{z}} & \left(8g\right) & \mbox{F II} \\ 
\mathbf{B}_{16} & = & \left(\frac{1}{2} +y_{5}\right) \, \mathbf{a}_{1} + \left(\frac{1}{2} - x_{5}\right) \, \mathbf{a}_{2} + \left(\frac{1}{2} +z_{5}\right) \, \mathbf{a}_{3} & = & \left(\frac{1}{2} +y_{5}\right)a \, \mathbf{\hat{x}} + \left(\frac{1}{2} - x_{5}\right)a \, \mathbf{\hat{y}} + \left(\frac{1}{2} +z_{5}\right)c \, \mathbf{\hat{z}} & \left(8g\right) & \mbox{F II} \\ 
\mathbf{B}_{17} & = & \left(\frac{1}{2} - x_{5}\right) \, \mathbf{a}_{1} + \left(\frac{1}{2} +y_{5}\right) \, \mathbf{a}_{2} + \left(\frac{1}{2} - z_{5}\right) \, \mathbf{a}_{3} & = & \left(\frac{1}{2} - x_{5}\right)a \, \mathbf{\hat{x}} + \left(\frac{1}{2} +y_{5}\right)a \, \mathbf{\hat{y}} + \left(\frac{1}{2} - z_{5}\right)c \, \mathbf{\hat{z}} & \left(8g\right) & \mbox{F II} \\ 
\mathbf{B}_{18} & = & \left(\frac{1}{2} +x_{5}\right) \, \mathbf{a}_{1} + \left(\frac{1}{2} - y_{5}\right) \, \mathbf{a}_{2} + \left(\frac{1}{2} - z_{5}\right) \, \mathbf{a}_{3} & = & \left(\frac{1}{2} +x_{5}\right)a \, \mathbf{\hat{x}} + \left(\frac{1}{2} - y_{5}\right)a \, \mathbf{\hat{y}} + \left(\frac{1}{2} - z_{5}\right)c \, \mathbf{\hat{z}} & \left(8g\right) & \mbox{F II} \\ 
\mathbf{B}_{19} & = & y_{5} \, \mathbf{a}_{1} + x_{5} \, \mathbf{a}_{2}-z_{5} \, \mathbf{a}_{3} & = & y_{5}a \, \mathbf{\hat{x}} + x_{5}a \, \mathbf{\hat{y}}-z_{5}c \, \mathbf{\hat{z}} & \left(8g\right) & \mbox{F II} \\ 
\mathbf{B}_{20} & = & -y_{5} \, \mathbf{a}_{1}-x_{5} \, \mathbf{a}_{2}-z_{5} \, \mathbf{a}_{3} & = & -y_{5}a \, \mathbf{\hat{x}}-x_{5}a \, \mathbf{\hat{y}}-z_{5}c \, \mathbf{\hat{z}} & \left(8g\right) & \mbox{F II} \\ 
\mathbf{B}_{21} & = & x_{6} \, \mathbf{a}_{1} + y_{6} \, \mathbf{a}_{2} + z_{6} \, \mathbf{a}_{3} & = & x_{6}a \, \mathbf{\hat{x}} + y_{6}a \, \mathbf{\hat{y}} + z_{6}c \, \mathbf{\hat{z}} & \left(8g\right) & \mbox{F III} \\ 
\mathbf{B}_{22} & = & -x_{6} \, \mathbf{a}_{1}-y_{6} \, \mathbf{a}_{2} + z_{6} \, \mathbf{a}_{3} & = & -x_{6}a \, \mathbf{\hat{x}}-y_{6}a \, \mathbf{\hat{y}} + z_{6}c \, \mathbf{\hat{z}} & \left(8g\right) & \mbox{F III} \\ 
\mathbf{B}_{23} & = & \left(\frac{1}{2} - y_{6}\right) \, \mathbf{a}_{1} + \left(\frac{1}{2} +x_{6}\right) \, \mathbf{a}_{2} + \left(\frac{1}{2} +z_{6}\right) \, \mathbf{a}_{3} & = & \left(\frac{1}{2} - y_{6}\right)a \, \mathbf{\hat{x}} + \left(\frac{1}{2} +x_{6}\right)a \, \mathbf{\hat{y}} + \left(\frac{1}{2} +z_{6}\right)c \, \mathbf{\hat{z}} & \left(8g\right) & \mbox{F III} \\ 
\mathbf{B}_{24} & = & \left(\frac{1}{2} +y_{6}\right) \, \mathbf{a}_{1} + \left(\frac{1}{2} - x_{6}\right) \, \mathbf{a}_{2} + \left(\frac{1}{2} +z_{6}\right) \, \mathbf{a}_{3} & = & \left(\frac{1}{2} +y_{6}\right)a \, \mathbf{\hat{x}} + \left(\frac{1}{2} - x_{6}\right)a \, \mathbf{\hat{y}} + \left(\frac{1}{2} +z_{6}\right)c \, \mathbf{\hat{z}} & \left(8g\right) & \mbox{F III} \\ 
\mathbf{B}_{25} & = & \left(\frac{1}{2} - x_{6}\right) \, \mathbf{a}_{1} + \left(\frac{1}{2} +y_{6}\right) \, \mathbf{a}_{2} + \left(\frac{1}{2} - z_{6}\right) \, \mathbf{a}_{3} & = & \left(\frac{1}{2} - x_{6}\right)a \, \mathbf{\hat{x}} + \left(\frac{1}{2} +y_{6}\right)a \, \mathbf{\hat{y}} + \left(\frac{1}{2} - z_{6}\right)c \, \mathbf{\hat{z}} & \left(8g\right) & \mbox{F III} \\ 
\mathbf{B}_{26} & = & \left(\frac{1}{2} +x_{6}\right) \, \mathbf{a}_{1} + \left(\frac{1}{2} - y_{6}\right) \, \mathbf{a}_{2} + \left(\frac{1}{2} - z_{6}\right) \, \mathbf{a}_{3} & = & \left(\frac{1}{2} +x_{6}\right)a \, \mathbf{\hat{x}} + \left(\frac{1}{2} - y_{6}\right)a \, \mathbf{\hat{y}} + \left(\frac{1}{2} - z_{6}\right)c \, \mathbf{\hat{z}} & \left(8g\right) & \mbox{F III} \\ 
\mathbf{B}_{27} & = & y_{6} \, \mathbf{a}_{1} + x_{6} \, \mathbf{a}_{2}-z_{6} \, \mathbf{a}_{3} & = & y_{6}a \, \mathbf{\hat{x}} + x_{6}a \, \mathbf{\hat{y}}-z_{6}c \, \mathbf{\hat{z}} & \left(8g\right) & \mbox{F III} \\ 
\mathbf{B}_{28} & = & -y_{6} \, \mathbf{a}_{1}-x_{6} \, \mathbf{a}_{2}-z_{6} \, \mathbf{a}_{3} & = & -y_{6}a \, \mathbf{\hat{x}}-x_{6}a \, \mathbf{\hat{y}}-z_{6}c \, \mathbf{\hat{z}} & \left(8g\right) & \mbox{F III} \\ 
\mathbf{B}_{29} & = & x_{7} \, \mathbf{a}_{1} + y_{7} \, \mathbf{a}_{2} + z_{7} \, \mathbf{a}_{3} & = & x_{7}a \, \mathbf{\hat{x}} + y_{7}a \, \mathbf{\hat{y}} + z_{7}c \, \mathbf{\hat{z}} & \left(8g\right) & \mbox{F IV} \\ 
\mathbf{B}_{30} & = & -x_{7} \, \mathbf{a}_{1}-y_{7} \, \mathbf{a}_{2} + z_{7} \, \mathbf{a}_{3} & = & -x_{7}a \, \mathbf{\hat{x}}-y_{7}a \, \mathbf{\hat{y}} + z_{7}c \, \mathbf{\hat{z}} & \left(8g\right) & \mbox{F IV} \\ 
\mathbf{B}_{31} & = & \left(\frac{1}{2} - y_{7}\right) \, \mathbf{a}_{1} + \left(\frac{1}{2} +x_{7}\right) \, \mathbf{a}_{2} + \left(\frac{1}{2} +z_{7}\right) \, \mathbf{a}_{3} & = & \left(\frac{1}{2} - y_{7}\right)a \, \mathbf{\hat{x}} + \left(\frac{1}{2} +x_{7}\right)a \, \mathbf{\hat{y}} + \left(\frac{1}{2} +z_{7}\right)c \, \mathbf{\hat{z}} & \left(8g\right) & \mbox{F IV} \\ 
\mathbf{B}_{32} & = & \left(\frac{1}{2} +y_{7}\right) \, \mathbf{a}_{1} + \left(\frac{1}{2} - x_{7}\right) \, \mathbf{a}_{2} + \left(\frac{1}{2} +z_{7}\right) \, \mathbf{a}_{3} & = & \left(\frac{1}{2} +y_{7}\right)a \, \mathbf{\hat{x}} + \left(\frac{1}{2} - x_{7}\right)a \, \mathbf{\hat{y}} + \left(\frac{1}{2} +z_{7}\right)c \, \mathbf{\hat{z}} & \left(8g\right) & \mbox{F IV} \\ 
\mathbf{B}_{33} & = & \left(\frac{1}{2} - x_{7}\right) \, \mathbf{a}_{1} + \left(\frac{1}{2} +y_{7}\right) \, \mathbf{a}_{2} + \left(\frac{1}{2} - z_{7}\right) \, \mathbf{a}_{3} & = & \left(\frac{1}{2} - x_{7}\right)a \, \mathbf{\hat{x}} + \left(\frac{1}{2} +y_{7}\right)a \, \mathbf{\hat{y}} + \left(\frac{1}{2} - z_{7}\right)c \, \mathbf{\hat{z}} & \left(8g\right) & \mbox{F IV} \\ 
\mathbf{B}_{34} & = & \left(\frac{1}{2} +x_{7}\right) \, \mathbf{a}_{1} + \left(\frac{1}{2} - y_{7}\right) \, \mathbf{a}_{2} + \left(\frac{1}{2} - z_{7}\right) \, \mathbf{a}_{3} & = & \left(\frac{1}{2} +x_{7}\right)a \, \mathbf{\hat{x}} + \left(\frac{1}{2} - y_{7}\right)a \, \mathbf{\hat{y}} + \left(\frac{1}{2} - z_{7}\right)c \, \mathbf{\hat{z}} & \left(8g\right) & \mbox{F IV} \\ 
\mathbf{B}_{35} & = & y_{7} \, \mathbf{a}_{1} + x_{7} \, \mathbf{a}_{2}-z_{7} \, \mathbf{a}_{3} & = & y_{7}a \, \mathbf{\hat{x}} + x_{7}a \, \mathbf{\hat{y}}-z_{7}c \, \mathbf{\hat{z}} & \left(8g\right) & \mbox{F IV} \\ 
\mathbf{B}_{36} & = & -y_{7} \, \mathbf{a}_{1}-x_{7} \, \mathbf{a}_{2}-z_{7} \, \mathbf{a}_{3} & = & -y_{7}a \, \mathbf{\hat{x}}-x_{7}a \, \mathbf{\hat{y}}-z_{7}c \, \mathbf{\hat{z}} & \left(8g\right) & \mbox{F IV} \\ 
\mathbf{B}_{37} & = & x_{8} \, \mathbf{a}_{1} + y_{8} \, \mathbf{a}_{2} + z_{8} \, \mathbf{a}_{3} & = & x_{8}a \, \mathbf{\hat{x}} + y_{8}a \, \mathbf{\hat{y}} + z_{8}c \, \mathbf{\hat{z}} & \left(8g\right) & \mbox{Na II} \\ 
\mathbf{B}_{38} & = & -x_{8} \, \mathbf{a}_{1}-y_{8} \, \mathbf{a}_{2} + z_{8} \, \mathbf{a}_{3} & = & -x_{8}a \, \mathbf{\hat{x}}-y_{8}a \, \mathbf{\hat{y}} + z_{8}c \, \mathbf{\hat{z}} & \left(8g\right) & \mbox{Na II} \\ 
\mathbf{B}_{39} & = & \left(\frac{1}{2} - y_{8}\right) \, \mathbf{a}_{1} + \left(\frac{1}{2} +x_{8}\right) \, \mathbf{a}_{2} + \left(\frac{1}{2} +z_{8}\right) \, \mathbf{a}_{3} & = & \left(\frac{1}{2} - y_{8}\right)a \, \mathbf{\hat{x}} + \left(\frac{1}{2} +x_{8}\right)a \, \mathbf{\hat{y}} + \left(\frac{1}{2} +z_{8}\right)c \, \mathbf{\hat{z}} & \left(8g\right) & \mbox{Na II} \\ 
\mathbf{B}_{40} & = & \left(\frac{1}{2} +y_{8}\right) \, \mathbf{a}_{1} + \left(\frac{1}{2} - x_{8}\right) \, \mathbf{a}_{2} + \left(\frac{1}{2} +z_{8}\right) \, \mathbf{a}_{3} & = & \left(\frac{1}{2} +y_{8}\right)a \, \mathbf{\hat{x}} + \left(\frac{1}{2} - x_{8}\right)a \, \mathbf{\hat{y}} + \left(\frac{1}{2} +z_{8}\right)c \, \mathbf{\hat{z}} & \left(8g\right) & \mbox{Na II} \\ 
\mathbf{B}_{41} & = & \left(\frac{1}{2} - x_{8}\right) \, \mathbf{a}_{1} + \left(\frac{1}{2} +y_{8}\right) \, \mathbf{a}_{2} + \left(\frac{1}{2} - z_{8}\right) \, \mathbf{a}_{3} & = & \left(\frac{1}{2} - x_{8}\right)a \, \mathbf{\hat{x}} + \left(\frac{1}{2} +y_{8}\right)a \, \mathbf{\hat{y}} + \left(\frac{1}{2} - z_{8}\right)c \, \mathbf{\hat{z}} & \left(8g\right) & \mbox{Na II} \\ 
\mathbf{B}_{42} & = & \left(\frac{1}{2} +x_{8}\right) \, \mathbf{a}_{1} + \left(\frac{1}{2} - y_{8}\right) \, \mathbf{a}_{2} + \left(\frac{1}{2} - z_{8}\right) \, \mathbf{a}_{3} & = & \left(\frac{1}{2} +x_{8}\right)a \, \mathbf{\hat{x}} + \left(\frac{1}{2} - y_{8}\right)a \, \mathbf{\hat{y}} + \left(\frac{1}{2} - z_{8}\right)c \, \mathbf{\hat{z}} & \left(8g\right) & \mbox{Na II} \\ 
\mathbf{B}_{43} & = & y_{8} \, \mathbf{a}_{1} + x_{8} \, \mathbf{a}_{2}-z_{8} \, \mathbf{a}_{3} & = & y_{8}a \, \mathbf{\hat{x}} + x_{8}a \, \mathbf{\hat{y}}-z_{8}c \, \mathbf{\hat{z}} & \left(8g\right) & \mbox{Na II} \\ 
\mathbf{B}_{44} & = & -y_{8} \, \mathbf{a}_{1}-x_{8} \, \mathbf{a}_{2}-z_{8} \, \mathbf{a}_{3} & = & -y_{8}a \, \mathbf{\hat{x}}-x_{8}a \, \mathbf{\hat{y}}-z_{8}c \, \mathbf{\hat{z}} & \left(8g\right) & \mbox{Na II} \\ 
\end{longtabu}
\renewcommand{\arraystretch}{1.0}
\noindent \hrulefill
\\
\textbf{References:}
\vspace*{-0.25cm}
\begin{flushleft}
  - \bibentry{Vlasse_Na5Fe3F14_JSolStateChem_1976}. \\
\end{flushleft}
\textbf{Found in:}
\vspace*{-0.25cm}
\begin{flushleft}
  - \bibentry{Villars_PearsonsCrystalData_2013}. \\
\end{flushleft}
\noindent \hrulefill
\\
\textbf{Geometry files:}
\\
\noindent  - CIF: pp. {\hyperref[A14B3C5_tP44_94_c3g_ad_bg_cif]{\pageref{A14B3C5_tP44_94_c3g_ad_bg_cif}}} \\
\noindent  - POSCAR: pp. {\hyperref[A14B3C5_tP44_94_c3g_ad_bg_poscar]{\pageref{A14B3C5_tP44_94_c3g_ad_bg_poscar}}} \\
\onecolumn
{\phantomsection\label{A6B2C_tP18_94_eg_c_a}}
\subsection*{\huge \textbf{{\normalfont Li$_{2}$MoF$_{6}$ Structure: A6B2C\_tP18\_94\_eg\_c\_a}}}
\noindent \hrulefill
\vspace*{0.25cm}
\begin{figure}[htp]
  \centering
  \vspace{-1em}
  {\includegraphics[width=1\textwidth]{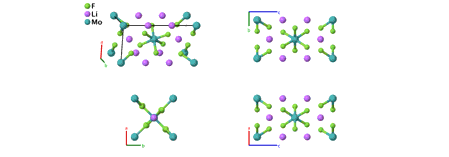}}
\end{figure}
\vspace*{-0.5cm}
\renewcommand{\arraystretch}{1.5}
\begin{equation*}
  \begin{array}{>{$\hspace{-0.15cm}}l<{$}>{$}p{0.5cm}<{$}>{$}p{18.5cm}<{$}}
    \mbox{\large \textbf{Prototype}} &\colon & \ce{Li2MoF6} \\
    \mbox{\large \textbf{\AFLOW\ prototype label}} &\colon & \mbox{A6B2C\_tP18\_94\_eg\_c\_a} \\
    \mbox{\large \textbf{\textit{Strukturbericht} designation}} &\colon & \mbox{None} \\
    \mbox{\large \textbf{Pearson symbol}} &\colon & \mbox{tP18} \\
    \mbox{\large \textbf{Space group number}} &\colon & 94 \\
    \mbox{\large \textbf{Space group symbol}} &\colon & P4_{2}2_{1}2 \\
    \mbox{\large \textbf{\AFLOW\ prototype command}} &\colon &  \texttt{aflow} \,  \, \texttt{-{}-proto=A6B2C\_tP18\_94\_eg\_c\_a } \, \newline \texttt{-{}-params=}{a,c/a,z_{2},x_{3},x_{4},y_{4},z_{4} }
  \end{array}
\end{equation*}
\renewcommand{\arraystretch}{1.0}

\noindent \parbox{1 \linewidth}{
\noindent \hrulefill
\\
\textbf{Simple Tetragonal primitive vectors:} \\
\vspace*{-0.25cm}
\begin{tabular}{cc}
  \begin{tabular}{c}
    \parbox{0.6 \linewidth}{
      \renewcommand{\arraystretch}{1.5}
      \begin{equation*}
        \centering
        \begin{array}{ccc}
              \mathbf{a}_1 & = & a \, \mathbf{\hat{x}} \\
    \mathbf{a}_2 & = & a \, \mathbf{\hat{y}} \\
    \mathbf{a}_3 & = & c \, \mathbf{\hat{z}} \\

        \end{array}
      \end{equation*}
    }
    \renewcommand{\arraystretch}{1.0}
  \end{tabular}
  \begin{tabular}{c}
    \includegraphics[width=0.3\linewidth]{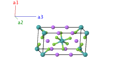} \\
  \end{tabular}
\end{tabular}

}
\vspace*{-0.25cm}

\noindent \hrulefill
\\
\textbf{Basis vectors:}
\vspace*{-0.25cm}
\renewcommand{\arraystretch}{1.5}
\begin{longtabu} to \textwidth{>{\centering $}X[-1,c,c]<{$}>{\centering $}X[-1,c,c]<{$}>{\centering $}X[-1,c,c]<{$}>{\centering $}X[-1,c,c]<{$}>{\centering $}X[-1,c,c]<{$}>{\centering $}X[-1,c,c]<{$}>{\centering $}X[-1,c,c]<{$}}
  & & \mbox{Lattice Coordinates} & & \mbox{Cartesian Coordinates} &\mbox{Wyckoff Position} & \mbox{Atom Type} \\  
  \mathbf{B}_{1} & = & 0 \, \mathbf{a}_{1} + 0 \, \mathbf{a}_{2} + 0 \, \mathbf{a}_{3} & = & 0 \, \mathbf{\hat{x}} + 0 \, \mathbf{\hat{y}} + 0 \, \mathbf{\hat{z}} & \left(2a\right) & \mbox{Mo} \\ 
\mathbf{B}_{2} & = & \frac{1}{2} \, \mathbf{a}_{1} + \frac{1}{2} \, \mathbf{a}_{2} + \frac{1}{2} \, \mathbf{a}_{3} & = & \frac{1}{2}a \, \mathbf{\hat{x}} + \frac{1}{2}a \, \mathbf{\hat{y}} + \frac{1}{2}c \, \mathbf{\hat{z}} & \left(2a\right) & \mbox{Mo} \\ 
\mathbf{B}_{3} & = & z_{2} \, \mathbf{a}_{3} & = & z_{2}c \, \mathbf{\hat{z}} & \left(4c\right) & \mbox{Li} \\ 
\mathbf{B}_{4} & = & \frac{1}{2} \, \mathbf{a}_{1} + \frac{1}{2} \, \mathbf{a}_{2} + \left(\frac{1}{2} +z_{2}\right) \, \mathbf{a}_{3} & = & \frac{1}{2}a \, \mathbf{\hat{x}} + \frac{1}{2}a \, \mathbf{\hat{y}} + \left(\frac{1}{2} +z_{2}\right)c \, \mathbf{\hat{z}} & \left(4c\right) & \mbox{Li} \\ 
\mathbf{B}_{5} & = & \frac{1}{2} \, \mathbf{a}_{1} + \frac{1}{2} \, \mathbf{a}_{2} + \left(\frac{1}{2} - z_{2}\right) \, \mathbf{a}_{3} & = & \frac{1}{2}a \, \mathbf{\hat{x}} + \frac{1}{2}a \, \mathbf{\hat{y}} + \left(\frac{1}{2} - z_{2}\right)c \, \mathbf{\hat{z}} & \left(4c\right) & \mbox{Li} \\ 
\mathbf{B}_{6} & = & -z_{2} \, \mathbf{a}_{3} & = & -z_{2}c \, \mathbf{\hat{z}} & \left(4c\right) & \mbox{Li} \\ 
\mathbf{B}_{7} & = & x_{3} \, \mathbf{a}_{1} + x_{3} \, \mathbf{a}_{2} & = & x_{3}a \, \mathbf{\hat{x}} + x_{3}a \, \mathbf{\hat{y}} & \left(4e\right) & \mbox{F I} \\ 
\mathbf{B}_{8} & = & -x_{3} \, \mathbf{a}_{1}-x_{3} \, \mathbf{a}_{2} & = & -x_{3}a \, \mathbf{\hat{x}}-x_{3}a \, \mathbf{\hat{y}} & \left(4e\right) & \mbox{F I} \\ 
\mathbf{B}_{9} & = & \left(\frac{1}{2} - x_{3}\right) \, \mathbf{a}_{1} + \left(\frac{1}{2} +x_{3}\right) \, \mathbf{a}_{2} + \frac{1}{2} \, \mathbf{a}_{3} & = & \left(\frac{1}{2} - x_{3}\right)a \, \mathbf{\hat{x}} + \left(\frac{1}{2} +x_{3}\right)a \, \mathbf{\hat{y}} + \frac{1}{2}c \, \mathbf{\hat{z}} & \left(4e\right) & \mbox{F I} \\ 
\mathbf{B}_{10} & = & \left(\frac{1}{2} +x_{3}\right) \, \mathbf{a}_{1} + \left(\frac{1}{2} - x_{3}\right) \, \mathbf{a}_{2} + \frac{1}{2} \, \mathbf{a}_{3} & = & \left(\frac{1}{2} +x_{3}\right)a \, \mathbf{\hat{x}} + \left(\frac{1}{2} - x_{3}\right)a \, \mathbf{\hat{y}} + \frac{1}{2}c \, \mathbf{\hat{z}} & \left(4e\right) & \mbox{F I} \\ 
\mathbf{B}_{11} & = & x_{4} \, \mathbf{a}_{1} + y_{4} \, \mathbf{a}_{2} + z_{4} \, \mathbf{a}_{3} & = & x_{4}a \, \mathbf{\hat{x}} + y_{4}a \, \mathbf{\hat{y}} + z_{4}c \, \mathbf{\hat{z}} & \left(8g\right) & \mbox{F II} \\ 
\mathbf{B}_{12} & = & -x_{4} \, \mathbf{a}_{1}-y_{4} \, \mathbf{a}_{2} + z_{4} \, \mathbf{a}_{3} & = & -x_{4}a \, \mathbf{\hat{x}}-y_{4}a \, \mathbf{\hat{y}} + z_{4}c \, \mathbf{\hat{z}} & \left(8g\right) & \mbox{F II} \\ 
\mathbf{B}_{13} & = & \left(\frac{1}{2} - y_{4}\right) \, \mathbf{a}_{1} + \left(\frac{1}{2} +x_{4}\right) \, \mathbf{a}_{2} + \left(\frac{1}{2} +z_{4}\right) \, \mathbf{a}_{3} & = & \left(\frac{1}{2} - y_{4}\right)a \, \mathbf{\hat{x}} + \left(\frac{1}{2} +x_{4}\right)a \, \mathbf{\hat{y}} + \left(\frac{1}{2} +z_{4}\right)c \, \mathbf{\hat{z}} & \left(8g\right) & \mbox{F II} \\ 
\mathbf{B}_{14} & = & \left(\frac{1}{2} +y_{4}\right) \, \mathbf{a}_{1} + \left(\frac{1}{2} - x_{4}\right) \, \mathbf{a}_{2} + \left(\frac{1}{2} +z_{4}\right) \, \mathbf{a}_{3} & = & \left(\frac{1}{2} +y_{4}\right)a \, \mathbf{\hat{x}} + \left(\frac{1}{2} - x_{4}\right)a \, \mathbf{\hat{y}} + \left(\frac{1}{2} +z_{4}\right)c \, \mathbf{\hat{z}} & \left(8g\right) & \mbox{F II} \\ 
\mathbf{B}_{15} & = & \left(\frac{1}{2} - x_{4}\right) \, \mathbf{a}_{1} + \left(\frac{1}{2} +y_{4}\right) \, \mathbf{a}_{2} + \left(\frac{1}{2} - z_{4}\right) \, \mathbf{a}_{3} & = & \left(\frac{1}{2} - x_{4}\right)a \, \mathbf{\hat{x}} + \left(\frac{1}{2} +y_{4}\right)a \, \mathbf{\hat{y}} + \left(\frac{1}{2} - z_{4}\right)c \, \mathbf{\hat{z}} & \left(8g\right) & \mbox{F II} \\ 
\mathbf{B}_{16} & = & \left(\frac{1}{2} +x_{4}\right) \, \mathbf{a}_{1} + \left(\frac{1}{2} - y_{4}\right) \, \mathbf{a}_{2} + \left(\frac{1}{2} - z_{4}\right) \, \mathbf{a}_{3} & = & \left(\frac{1}{2} +x_{4}\right)a \, \mathbf{\hat{x}} + \left(\frac{1}{2} - y_{4}\right)a \, \mathbf{\hat{y}} + \left(\frac{1}{2} - z_{4}\right)c \, \mathbf{\hat{z}} & \left(8g\right) & \mbox{F II} \\ 
\mathbf{B}_{17} & = & y_{4} \, \mathbf{a}_{1} + x_{4} \, \mathbf{a}_{2}-z_{4} \, \mathbf{a}_{3} & = & y_{4}a \, \mathbf{\hat{x}} + x_{4}a \, \mathbf{\hat{y}}-z_{4}c \, \mathbf{\hat{z}} & \left(8g\right) & \mbox{F II} \\ 
\mathbf{B}_{18} & = & -y_{4} \, \mathbf{a}_{1}-x_{4} \, \mathbf{a}_{2}-z_{4} \, \mathbf{a}_{3} & = & -y_{4}a \, \mathbf{\hat{x}}-x_{4}a \, \mathbf{\hat{y}}-z_{4}c \, \mathbf{\hat{z}} & \left(8g\right) & \mbox{F II} \\ 
\end{longtabu}
\renewcommand{\arraystretch}{1.0}
\noindent \hrulefill
\\
\textbf{References:}
\vspace*{-0.25cm}
\begin{flushleft}
  - \bibentry{Brunton_Li2MoF6_MatResBull_1971}. \\
\end{flushleft}
\textbf{Found in:}
\vspace*{-0.25cm}
\begin{flushleft}
  - \bibentry{Villars_PearsonsCrystalData_2013}. \\
\end{flushleft}
\noindent \hrulefill
\\
\textbf{Geometry files:}
\\
\noindent  - CIF: pp. {\hyperref[A6B2C_tP18_94_eg_c_a_cif]{\pageref{A6B2C_tP18_94_eg_c_a_cif}}} \\
\noindent  - POSCAR: pp. {\hyperref[A6B2C_tP18_94_eg_c_a_poscar]{\pageref{A6B2C_tP18_94_eg_c_a_poscar}}} \\
\onecolumn
{\phantomsection\label{ABC_tP24_95_d_d_d}}
\subsection*{\huge \textbf{{\normalfont ThBC Structure: ABC\_tP24\_95\_d\_d\_d}}}
\noindent \hrulefill
\vspace*{0.25cm}
\begin{figure}[htp]
  \centering
  \vspace{-1em}
  {\includegraphics[width=1\textwidth]{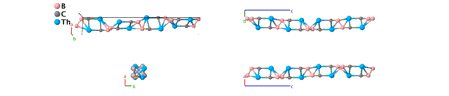}}
\end{figure}
\vspace*{-0.5cm}
\renewcommand{\arraystretch}{1.5}
\begin{equation*}
  \begin{array}{>{$\hspace{-0.15cm}}l<{$}>{$}p{0.5cm}<{$}>{$}p{18.5cm}<{$}}
    \mbox{\large \textbf{Prototype}} &\colon & \ce{ThBC} \\
    \mbox{\large \textbf{\AFLOW\ prototype label}} &\colon & \mbox{ABC\_tP24\_95\_d\_d\_d} \\
    \mbox{\large \textbf{\textit{Strukturbericht} designation}} &\colon & \mbox{None} \\
    \mbox{\large \textbf{Pearson symbol}} &\colon & \mbox{tP24} \\
    \mbox{\large \textbf{Space group number}} &\colon & 95 \\
    \mbox{\large \textbf{Space group symbol}} &\colon & P4_{3}22 \\
    \mbox{\large \textbf{\AFLOW\ prototype command}} &\colon &  \texttt{aflow} \,  \, \texttt{-{}-proto=ABC\_tP24\_95\_d\_d\_d } \, \newline \texttt{-{}-params=}{a,c/a,x_{1},y_{1},z_{1},x_{2},y_{2},z_{2},x_{3},y_{3},z_{3} }
  \end{array}
\end{equation*}
\renewcommand{\arraystretch}{1.0}

\vspace*{-0.25cm}
\noindent \hrulefill
\begin{itemize}
  \item{This structure is the enantiomorph of the \hyperref[ABC_tP24_91_d_d_d]{ThBC (ABC\_tP24\_91\_d\_d\_d) structure},
and was generated by reflecting the coordinates of the space group \#91 structure through the $z=0$ plane.
}
\end{itemize}

\noindent \parbox{1 \linewidth}{
\noindent \hrulefill
\\
\textbf{Simple Tetragonal primitive vectors:} \\
\vspace*{-0.25cm}
\begin{tabular}{cc}
  \begin{tabular}{c}
    \parbox{0.6 \linewidth}{
      \renewcommand{\arraystretch}{1.5}
      \begin{equation*}
        \centering
        \begin{array}{ccc}
              \mathbf{a}_1 & = & a \, \mathbf{\hat{x}} \\
    \mathbf{a}_2 & = & a \, \mathbf{\hat{y}} \\
    \mathbf{a}_3 & = & c \, \mathbf{\hat{z}} \\

        \end{array}
      \end{equation*}
    }
    \renewcommand{\arraystretch}{1.0}
  \end{tabular}
  \begin{tabular}{c}
    \includegraphics[width=0.3\linewidth]{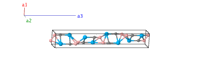} \\
  \end{tabular}
\end{tabular}

}
\vspace*{-0.25cm}

\noindent \hrulefill
\\
\textbf{Basis vectors:}
\vspace*{-0.25cm}
\renewcommand{\arraystretch}{1.5}
\begin{longtabu} to \textwidth{>{\centering $}X[-1,c,c]<{$}>{\centering $}X[-1,c,c]<{$}>{\centering $}X[-1,c,c]<{$}>{\centering $}X[-1,c,c]<{$}>{\centering $}X[-1,c,c]<{$}>{\centering $}X[-1,c,c]<{$}>{\centering $}X[-1,c,c]<{$}}
  & & \mbox{Lattice Coordinates} & & \mbox{Cartesian Coordinates} &\mbox{Wyckoff Position} & \mbox{Atom Type} \\  
  \mathbf{B}_{1} & = & x_{1} \, \mathbf{a}_{1} + y_{1} \, \mathbf{a}_{2} + z_{1} \, \mathbf{a}_{3} & = & x_{1}a \, \mathbf{\hat{x}} + y_{1}a \, \mathbf{\hat{y}} + z_{1}c \, \mathbf{\hat{z}} & \left(8d\right) & \mbox{B} \\ 
\mathbf{B}_{2} & = & -x_{1} \, \mathbf{a}_{1}-y_{1} \, \mathbf{a}_{2} + \left(\frac{1}{2} +z_{1}\right) \, \mathbf{a}_{3} & = & -x_{1}a \, \mathbf{\hat{x}}-y_{1}a \, \mathbf{\hat{y}} + \left(\frac{1}{2} +z_{1}\right)c \, \mathbf{\hat{z}} & \left(8d\right) & \mbox{B} \\ 
\mathbf{B}_{3} & = & -y_{1} \, \mathbf{a}_{1} + x_{1} \, \mathbf{a}_{2} + \left(\frac{3}{4} +z_{1}\right) \, \mathbf{a}_{3} & = & -y_{1}a \, \mathbf{\hat{x}} + x_{1}a \, \mathbf{\hat{y}} + \left(\frac{3}{4} +z_{1}\right)c \, \mathbf{\hat{z}} & \left(8d\right) & \mbox{B} \\ 
\mathbf{B}_{4} & = & y_{1} \, \mathbf{a}_{1}-x_{1} \, \mathbf{a}_{2} + \left(\frac{1}{4} +z_{1}\right) \, \mathbf{a}_{3} & = & y_{1}a \, \mathbf{\hat{x}}-x_{1}a \, \mathbf{\hat{y}} + \left(\frac{1}{4} +z_{1}\right)c \, \mathbf{\hat{z}} & \left(8d\right) & \mbox{B} \\ 
\mathbf{B}_{5} & = & -x_{1} \, \mathbf{a}_{1} + y_{1} \, \mathbf{a}_{2}-z_{1} \, \mathbf{a}_{3} & = & -x_{1}a \, \mathbf{\hat{x}} + y_{1}a \, \mathbf{\hat{y}}-z_{1}c \, \mathbf{\hat{z}} & \left(8d\right) & \mbox{B} \\ 
\mathbf{B}_{6} & = & x_{1} \, \mathbf{a}_{1}-y_{1} \, \mathbf{a}_{2} + \left(\frac{1}{2} - z_{1}\right) \, \mathbf{a}_{3} & = & x_{1}a \, \mathbf{\hat{x}}-y_{1}a \, \mathbf{\hat{y}} + \left(\frac{1}{2} - z_{1}\right)c \, \mathbf{\hat{z}} & \left(8d\right) & \mbox{B} \\ 
\mathbf{B}_{7} & = & y_{1} \, \mathbf{a}_{1} + x_{1} \, \mathbf{a}_{2} + \left(\frac{1}{4} - z_{1}\right) \, \mathbf{a}_{3} & = & y_{1}a \, \mathbf{\hat{x}} + x_{1}a \, \mathbf{\hat{y}} + \left(\frac{1}{4} - z_{1}\right)c \, \mathbf{\hat{z}} & \left(8d\right) & \mbox{B} \\ 
\mathbf{B}_{8} & = & -y_{1} \, \mathbf{a}_{1}-x_{1} \, \mathbf{a}_{2} + \left(\frac{3}{4} - z_{1}\right) \, \mathbf{a}_{3} & = & -y_{1}a \, \mathbf{\hat{x}}-x_{1}a \, \mathbf{\hat{y}} + \left(\frac{3}{4} - z_{1}\right)c \, \mathbf{\hat{z}} & \left(8d\right) & \mbox{B} \\ 
\mathbf{B}_{9} & = & x_{2} \, \mathbf{a}_{1} + y_{2} \, \mathbf{a}_{2} + z_{2} \, \mathbf{a}_{3} & = & x_{2}a \, \mathbf{\hat{x}} + y_{2}a \, \mathbf{\hat{y}} + z_{2}c \, \mathbf{\hat{z}} & \left(8d\right) & \mbox{C} \\ 
\mathbf{B}_{10} & = & -x_{2} \, \mathbf{a}_{1}-y_{2} \, \mathbf{a}_{2} + \left(\frac{1}{2} +z_{2}\right) \, \mathbf{a}_{3} & = & -x_{2}a \, \mathbf{\hat{x}}-y_{2}a \, \mathbf{\hat{y}} + \left(\frac{1}{2} +z_{2}\right)c \, \mathbf{\hat{z}} & \left(8d\right) & \mbox{C} \\ 
\mathbf{B}_{11} & = & -y_{2} \, \mathbf{a}_{1} + x_{2} \, \mathbf{a}_{2} + \left(\frac{3}{4} +z_{2}\right) \, \mathbf{a}_{3} & = & -y_{2}a \, \mathbf{\hat{x}} + x_{2}a \, \mathbf{\hat{y}} + \left(\frac{3}{4} +z_{2}\right)c \, \mathbf{\hat{z}} & \left(8d\right) & \mbox{C} \\ 
\mathbf{B}_{12} & = & y_{2} \, \mathbf{a}_{1}-x_{2} \, \mathbf{a}_{2} + \left(\frac{1}{4} +z_{2}\right) \, \mathbf{a}_{3} & = & y_{2}a \, \mathbf{\hat{x}}-x_{2}a \, \mathbf{\hat{y}} + \left(\frac{1}{4} +z_{2}\right)c \, \mathbf{\hat{z}} & \left(8d\right) & \mbox{C} \\ 
\mathbf{B}_{13} & = & -x_{2} \, \mathbf{a}_{1} + y_{2} \, \mathbf{a}_{2}-z_{2} \, \mathbf{a}_{3} & = & -x_{2}a \, \mathbf{\hat{x}} + y_{2}a \, \mathbf{\hat{y}}-z_{2}c \, \mathbf{\hat{z}} & \left(8d\right) & \mbox{C} \\ 
\mathbf{B}_{14} & = & x_{2} \, \mathbf{a}_{1}-y_{2} \, \mathbf{a}_{2} + \left(\frac{1}{2} - z_{2}\right) \, \mathbf{a}_{3} & = & x_{2}a \, \mathbf{\hat{x}}-y_{2}a \, \mathbf{\hat{y}} + \left(\frac{1}{2} - z_{2}\right)c \, \mathbf{\hat{z}} & \left(8d\right) & \mbox{C} \\ 
\mathbf{B}_{15} & = & y_{2} \, \mathbf{a}_{1} + x_{2} \, \mathbf{a}_{2} + \left(\frac{1}{4} - z_{2}\right) \, \mathbf{a}_{3} & = & y_{2}a \, \mathbf{\hat{x}} + x_{2}a \, \mathbf{\hat{y}} + \left(\frac{1}{4} - z_{2}\right)c \, \mathbf{\hat{z}} & \left(8d\right) & \mbox{C} \\ 
\mathbf{B}_{16} & = & -y_{2} \, \mathbf{a}_{1}-x_{2} \, \mathbf{a}_{2} + \left(\frac{3}{4} - z_{2}\right) \, \mathbf{a}_{3} & = & -y_{2}a \, \mathbf{\hat{x}}-x_{2}a \, \mathbf{\hat{y}} + \left(\frac{3}{4} - z_{2}\right)c \, \mathbf{\hat{z}} & \left(8d\right) & \mbox{C} \\ 
\mathbf{B}_{17} & = & x_{3} \, \mathbf{a}_{1} + y_{3} \, \mathbf{a}_{2} + z_{3} \, \mathbf{a}_{3} & = & x_{3}a \, \mathbf{\hat{x}} + y_{3}a \, \mathbf{\hat{y}} + z_{3}c \, \mathbf{\hat{z}} & \left(8d\right) & \mbox{Th} \\ 
\mathbf{B}_{18} & = & -x_{3} \, \mathbf{a}_{1}-y_{3} \, \mathbf{a}_{2} + \left(\frac{1}{2} +z_{3}\right) \, \mathbf{a}_{3} & = & -x_{3}a \, \mathbf{\hat{x}}-y_{3}a \, \mathbf{\hat{y}} + \left(\frac{1}{2} +z_{3}\right)c \, \mathbf{\hat{z}} & \left(8d\right) & \mbox{Th} \\ 
\mathbf{B}_{19} & = & -y_{3} \, \mathbf{a}_{1} + x_{3} \, \mathbf{a}_{2} + \left(\frac{3}{4} +z_{3}\right) \, \mathbf{a}_{3} & = & -y_{3}a \, \mathbf{\hat{x}} + x_{3}a \, \mathbf{\hat{y}} + \left(\frac{3}{4} +z_{3}\right)c \, \mathbf{\hat{z}} & \left(8d\right) & \mbox{Th} \\ 
\mathbf{B}_{20} & = & y_{3} \, \mathbf{a}_{1}-x_{3} \, \mathbf{a}_{2} + \left(\frac{1}{4} +z_{3}\right) \, \mathbf{a}_{3} & = & y_{3}a \, \mathbf{\hat{x}}-x_{3}a \, \mathbf{\hat{y}} + \left(\frac{1}{4} +z_{3}\right)c \, \mathbf{\hat{z}} & \left(8d\right) & \mbox{Th} \\ 
\mathbf{B}_{21} & = & -x_{3} \, \mathbf{a}_{1} + y_{3} \, \mathbf{a}_{2}-z_{3} \, \mathbf{a}_{3} & = & -x_{3}a \, \mathbf{\hat{x}} + y_{3}a \, \mathbf{\hat{y}}-z_{3}c \, \mathbf{\hat{z}} & \left(8d\right) & \mbox{Th} \\ 
\mathbf{B}_{22} & = & x_{3} \, \mathbf{a}_{1}-y_{3} \, \mathbf{a}_{2} + \left(\frac{1}{2} - z_{3}\right) \, \mathbf{a}_{3} & = & x_{3}a \, \mathbf{\hat{x}}-y_{3}a \, \mathbf{\hat{y}} + \left(\frac{1}{2} - z_{3}\right)c \, \mathbf{\hat{z}} & \left(8d\right) & \mbox{Th} \\ 
\mathbf{B}_{23} & = & y_{3} \, \mathbf{a}_{1} + x_{3} \, \mathbf{a}_{2} + \left(\frac{1}{4} - z_{3}\right) \, \mathbf{a}_{3} & = & y_{3}a \, \mathbf{\hat{x}} + x_{3}a \, \mathbf{\hat{y}} + \left(\frac{1}{4} - z_{3}\right)c \, \mathbf{\hat{z}} & \left(8d\right) & \mbox{Th} \\ 
\mathbf{B}_{24} & = & -y_{3} \, \mathbf{a}_{1}-x_{3} \, \mathbf{a}_{2} + \left(\frac{3}{4} - z_{3}\right) \, \mathbf{a}_{3} & = & -y_{3}a \, \mathbf{\hat{x}}-x_{3}a \, \mathbf{\hat{y}} + \left(\frac{3}{4} - z_{3}\right)c \, \mathbf{\hat{z}} & \left(8d\right) & \mbox{Th} \\ 
\end{longtabu}
\renewcommand{\arraystretch}{1.0}
\noindent \hrulefill
\\
\textbf{References:}
\vspace*{-0.25cm}
\begin{flushleft}
  - \bibentry{Rogl_ThBC_JNucMat_1978}. \\
\end{flushleft}
\noindent \hrulefill
\\
\textbf{Geometry files:}
\\
\noindent  - CIF: pp. {\hyperref[ABC_tP24_95_d_d_d_cif]{\pageref{ABC_tP24_95_d_d_d_cif}}} \\
\noindent  - POSCAR: pp. {\hyperref[ABC_tP24_95_d_d_d_poscar]{\pageref{ABC_tP24_95_d_d_d_poscar}}} \\
\onecolumn
{\phantomsection\label{A2B8CD_tI24_97_d_k_a_b}}
\subsection*{\huge \textbf{{\normalfont NaGdCu$_{2}$F$_{8}$ Structure: A2B8CD\_tI24\_97\_d\_k\_a\_b}}}
\noindent \hrulefill
\vspace*{0.25cm}
\begin{figure}[htp]
  \centering
  \vspace{-1em}
  {\includegraphics[width=1\textwidth]{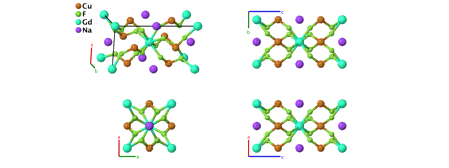}}
\end{figure}
\vspace*{-0.5cm}
\renewcommand{\arraystretch}{1.5}
\begin{equation*}
  \begin{array}{>{$\hspace{-0.15cm}}l<{$}>{$}p{0.5cm}<{$}>{$}p{18.5cm}<{$}}
    \mbox{\large \textbf{Prototype}} &\colon & \ce{NaGdCu2F8} \\
    \mbox{\large \textbf{\AFLOW\ prototype label}} &\colon & \mbox{A2B8CD\_tI24\_97\_d\_k\_a\_b} \\
    \mbox{\large \textbf{\textit{Strukturbericht} designation}} &\colon & \mbox{None} \\
    \mbox{\large \textbf{Pearson symbol}} &\colon & \mbox{tI24} \\
    \mbox{\large \textbf{Space group number}} &\colon & 97 \\
    \mbox{\large \textbf{Space group symbol}} &\colon & I422 \\
    \mbox{\large \textbf{\AFLOW\ prototype command}} &\colon &  \texttt{aflow} \,  \, \texttt{-{}-proto=A2B8CD\_tI24\_97\_d\_k\_a\_b } \, \newline \texttt{-{}-params=}{a,c/a,x_{4},y_{4},z_{4} }
  \end{array}
\end{equation*}
\renewcommand{\arraystretch}{1.0}

\noindent \parbox{1 \linewidth}{
\noindent \hrulefill
\\
\textbf{Body-centered Tetragonal primitive vectors:} \\
\vspace*{-0.25cm}
\begin{tabular}{cc}
  \begin{tabular}{c}
    \parbox{0.6 \linewidth}{
      \renewcommand{\arraystretch}{1.5}
      \begin{equation*}
        \centering
        \begin{array}{ccc}
              \mathbf{a}_1 & = & - \frac12 \, a \, \mathbf{\hat{x}} + \frac12 \, a \, \mathbf{\hat{y}} + \frac12 \, c \, \mathbf{\hat{z}} \\
    \mathbf{a}_2 & = & ~ \frac12 \, a \, \mathbf{\hat{x}} - \frac12 \, a \, \mathbf{\hat{y}} + \frac12 \, c \, \mathbf{\hat{z}} \\
    \mathbf{a}_3 & = & ~ \frac12 \, a \, \mathbf{\hat{x}} + \frac12 \, a \, \mathbf{\hat{y}} - \frac12 \, c \, \mathbf{\hat{z}} \\

        \end{array}
      \end{equation*}
    }
    \renewcommand{\arraystretch}{1.0}
  \end{tabular}
  \begin{tabular}{c}
    \includegraphics[width=0.3\linewidth]{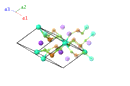} \\
  \end{tabular}
\end{tabular}

}
\vspace*{-0.25cm}

\noindent \hrulefill
\\
\textbf{Basis vectors:}
\vspace*{-0.25cm}
\renewcommand{\arraystretch}{1.5}
\begin{longtabu} to \textwidth{>{\centering $}X[-1,c,c]<{$}>{\centering $}X[-1,c,c]<{$}>{\centering $}X[-1,c,c]<{$}>{\centering $}X[-1,c,c]<{$}>{\centering $}X[-1,c,c]<{$}>{\centering $}X[-1,c,c]<{$}>{\centering $}X[-1,c,c]<{$}}
  & & \mbox{Lattice Coordinates} & & \mbox{Cartesian Coordinates} &\mbox{Wyckoff Position} & \mbox{Atom Type} \\  
  \mathbf{B}_{1} & = & 0 \, \mathbf{a}_{1} + 0 \, \mathbf{a}_{2} + 0 \, \mathbf{a}_{3} & = & 0 \, \mathbf{\hat{x}} + 0 \, \mathbf{\hat{y}} + 0 \, \mathbf{\hat{z}} & \left(2a\right) & \mbox{Gd} \\ 
\mathbf{B}_{2} & = & \frac{1}{2} \, \mathbf{a}_{1} + \frac{1}{2} \, \mathbf{a}_{2} & = & \frac{1}{2}c \, \mathbf{\hat{z}} & \left(2b\right) & \mbox{Na} \\ 
\mathbf{B}_{3} & = & \frac{3}{4} \, \mathbf{a}_{1} + \frac{1}{4} \, \mathbf{a}_{2} + \frac{1}{2} \, \mathbf{a}_{3} & = & \frac{1}{2}a \, \mathbf{\hat{y}} + \frac{1}{4}c \, \mathbf{\hat{z}} & \left(4d\right) & \mbox{Cu} \\ 
\mathbf{B}_{4} & = & \frac{1}{4} \, \mathbf{a}_{1} + \frac{3}{4} \, \mathbf{a}_{2} + \frac{1}{2} \, \mathbf{a}_{3} & = & \frac{1}{2}a \, \mathbf{\hat{x}} + \frac{1}{4}c \, \mathbf{\hat{z}} & \left(4d\right) & \mbox{Cu} \\ 
\mathbf{B}_{5} & = & \left(y_{4}+z_{4}\right) \, \mathbf{a}_{1} + \left(x_{4}+z_{4}\right) \, \mathbf{a}_{2} + \left(x_{4}+y_{4}\right) \, \mathbf{a}_{3} & = & x_{4}a \, \mathbf{\hat{x}} + y_{4}a \, \mathbf{\hat{y}} + z_{4}c \, \mathbf{\hat{z}} & \left(16k\right) & \mbox{F} \\ 
\mathbf{B}_{6} & = & \left(-y_{4}+z_{4}\right) \, \mathbf{a}_{1} + \left(-x_{4}+z_{4}\right) \, \mathbf{a}_{2} + \left(-x_{4}-y_{4}\right) \, \mathbf{a}_{3} & = & -x_{4}a \, \mathbf{\hat{x}}-y_{4}a \, \mathbf{\hat{y}} + z_{4}c \, \mathbf{\hat{z}} & \left(16k\right) & \mbox{F} \\ 
\mathbf{B}_{7} & = & \left(x_{4}+z_{4}\right) \, \mathbf{a}_{1} + \left(-y_{4}+z_{4}\right) \, \mathbf{a}_{2} + \left(x_{4}-y_{4}\right) \, \mathbf{a}_{3} & = & -y_{4}a \, \mathbf{\hat{x}} + x_{4}a \, \mathbf{\hat{y}} + z_{4}c \, \mathbf{\hat{z}} & \left(16k\right) & \mbox{F} \\ 
\mathbf{B}_{8} & = & \left(-x_{4}+z_{4}\right) \, \mathbf{a}_{1} + \left(y_{4}+z_{4}\right) \, \mathbf{a}_{2} + \left(-x_{4}+y_{4}\right) \, \mathbf{a}_{3} & = & y_{4}a \, \mathbf{\hat{x}}-x_{4}a \, \mathbf{\hat{y}} + z_{4}c \, \mathbf{\hat{z}} & \left(16k\right) & \mbox{F} \\ 
\mathbf{B}_{9} & = & \left(y_{4}-z_{4}\right) \, \mathbf{a}_{1} + \left(-x_{4}-z_{4}\right) \, \mathbf{a}_{2} + \left(-x_{4}+y_{4}\right) \, \mathbf{a}_{3} & = & -x_{4}a \, \mathbf{\hat{x}} + y_{4}a \, \mathbf{\hat{y}}-z_{4}c \, \mathbf{\hat{z}} & \left(16k\right) & \mbox{F} \\ 
\mathbf{B}_{10} & = & \left(-y_{4}-z_{4}\right) \, \mathbf{a}_{1} + \left(x_{4}-z_{4}\right) \, \mathbf{a}_{2} + \left(x_{4}-y_{4}\right) \, \mathbf{a}_{3} & = & x_{4}a \, \mathbf{\hat{x}}-y_{4}a \, \mathbf{\hat{y}}-z_{4}c \, \mathbf{\hat{z}} & \left(16k\right) & \mbox{F} \\ 
\mathbf{B}_{11} & = & \left(x_{4}-z_{4}\right) \, \mathbf{a}_{1} + \left(y_{4}-z_{4}\right) \, \mathbf{a}_{2} + \left(x_{4}+y_{4}\right) \, \mathbf{a}_{3} & = & y_{4}a \, \mathbf{\hat{x}} + x_{4}a \, \mathbf{\hat{y}}-z_{4}c \, \mathbf{\hat{z}} & \left(16k\right) & \mbox{F} \\ 
\mathbf{B}_{12} & = & \left(-x_{4}-z_{4}\right) \, \mathbf{a}_{1} + \left(-y_{4}-z_{4}\right) \, \mathbf{a}_{2} + \left(-x_{4}-y_{4}\right) \, \mathbf{a}_{3} & = & -y_{4}a \, \mathbf{\hat{x}}-x_{4}a \, \mathbf{\hat{y}}-z_{4}c \, \mathbf{\hat{z}} & \left(16k\right) & \mbox{F} \\ 
\end{longtabu}
\renewcommand{\arraystretch}{1.0}
\noindent \hrulefill
\\
\textbf{References:}
\vspace*{-0.25cm}
\begin{flushleft}
  - \bibentry{DeNadai_NaCu2GdF8_JMatChem_1998}. \\
\end{flushleft}
\textbf{Found in:}
\vspace*{-0.25cm}
\begin{flushleft}
  - \bibentry{Villars_PearsonsCrystalData_2013}. \\
\end{flushleft}
\noindent \hrulefill
\\
\textbf{Geometry files:}
\\
\noindent  - CIF: pp. {\hyperref[A2B8CD_tI24_97_d_k_a_b_cif]{\pageref{A2B8CD_tI24_97_d_k_a_b_cif}}} \\
\noindent  - POSCAR: pp. {\hyperref[A2B8CD_tI24_97_d_k_a_b_poscar]{\pageref{A2B8CD_tI24_97_d_k_a_b_poscar}}} \\
\onecolumn
{\phantomsection\label{AB8C2_tI44_97_e_2k_cd}}
\subsection*{\huge \textbf{{\normalfont Ta$_{2}$Se$_{8}$I Structure: AB8C2\_tI44\_97\_e\_2k\_cd}}}
\noindent \hrulefill
\vspace*{0.25cm}
\begin{figure}[htp]
  \centering
  \vspace{-1em}
  {\includegraphics[width=1\textwidth]{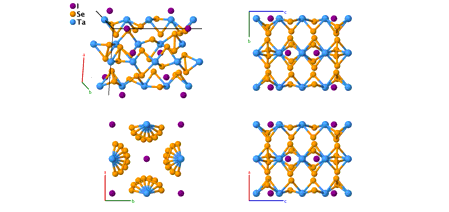}}
\end{figure}
\vspace*{-0.5cm}
\renewcommand{\arraystretch}{1.5}
\begin{equation*}
  \begin{array}{>{$\hspace{-0.15cm}}l<{$}>{$}p{0.5cm}<{$}>{$}p{18.5cm}<{$}}
    \mbox{\large \textbf{Prototype}} &\colon & \ce{Ta2Se8I} \\
    \mbox{\large \textbf{\AFLOW\ prototype label}} &\colon & \mbox{AB8C2\_tI44\_97\_e\_2k\_cd} \\
    \mbox{\large \textbf{\textit{Strukturbericht} designation}} &\colon & \mbox{None} \\
    \mbox{\large \textbf{Pearson symbol}} &\colon & \mbox{tI44} \\
    \mbox{\large \textbf{Space group number}} &\colon & 97 \\
    \mbox{\large \textbf{Space group symbol}} &\colon & I422 \\
    \mbox{\large \textbf{\AFLOW\ prototype command}} &\colon &  \texttt{aflow} \,  \, \texttt{-{}-proto=AB8C2\_tI44\_97\_e\_2k\_cd } \, \newline \texttt{-{}-params=}{a,c/a,z_{3},x_{4},y_{4},z_{4},x_{5},y_{5},z_{5} }
  \end{array}
\end{equation*}
\renewcommand{\arraystretch}{1.0}

\noindent \parbox{1 \linewidth}{
\noindent \hrulefill
\\
\textbf{Body-centered Tetragonal primitive vectors:} \\
\vspace*{-0.25cm}
\begin{tabular}{cc}
  \begin{tabular}{c}
    \parbox{0.6 \linewidth}{
      \renewcommand{\arraystretch}{1.5}
      \begin{equation*}
        \centering
        \begin{array}{ccc}
              \mathbf{a}_1 & = & - \frac12 \, a \, \mathbf{\hat{x}} + \frac12 \, a \, \mathbf{\hat{y}} + \frac12 \, c \, \mathbf{\hat{z}} \\
    \mathbf{a}_2 & = & ~ \frac12 \, a \, \mathbf{\hat{x}} - \frac12 \, a \, \mathbf{\hat{y}} + \frac12 \, c \, \mathbf{\hat{z}} \\
    \mathbf{a}_3 & = & ~ \frac12 \, a \, \mathbf{\hat{x}} + \frac12 \, a \, \mathbf{\hat{y}} - \frac12 \, c \, \mathbf{\hat{z}} \\

        \end{array}
      \end{equation*}
    }
    \renewcommand{\arraystretch}{1.0}
  \end{tabular}
  \begin{tabular}{c}
    \includegraphics[width=0.3\linewidth]{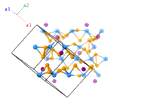} \\
  \end{tabular}
\end{tabular}

}
\vspace*{-0.25cm}

\noindent \hrulefill
\\
\textbf{Basis vectors:}
\vspace*{-0.25cm}
\renewcommand{\arraystretch}{1.5}
\begin{longtabu} to \textwidth{>{\centering $}X[-1,c,c]<{$}>{\centering $}X[-1,c,c]<{$}>{\centering $}X[-1,c,c]<{$}>{\centering $}X[-1,c,c]<{$}>{\centering $}X[-1,c,c]<{$}>{\centering $}X[-1,c,c]<{$}>{\centering $}X[-1,c,c]<{$}}
  & & \mbox{Lattice Coordinates} & & \mbox{Cartesian Coordinates} &\mbox{Wyckoff Position} & \mbox{Atom Type} \\  
  \mathbf{B}_{1} & = & \frac{1}{2} \, \mathbf{a}_{1} + \frac{1}{2} \, \mathbf{a}_{3} & = & \frac{1}{2}a \, \mathbf{\hat{y}} & \left(4c\right) & \mbox{Ta I} \\ 
\mathbf{B}_{2} & = & \frac{1}{2} \, \mathbf{a}_{2} + \frac{1}{2} \, \mathbf{a}_{3} & = & \frac{1}{2}a \, \mathbf{\hat{x}} & \left(4c\right) & \mbox{Ta I} \\ 
\mathbf{B}_{3} & = & \frac{3}{4} \, \mathbf{a}_{1} + \frac{1}{4} \, \mathbf{a}_{2} + \frac{1}{2} \, \mathbf{a}_{3} & = & \frac{1}{2}a \, \mathbf{\hat{y}} + \frac{1}{4}c \, \mathbf{\hat{z}} & \left(4d\right) & \mbox{Ta II} \\ 
\mathbf{B}_{4} & = & \frac{1}{4} \, \mathbf{a}_{1} + \frac{3}{4} \, \mathbf{a}_{2} + \frac{1}{2} \, \mathbf{a}_{3} & = & \frac{1}{2}a \, \mathbf{\hat{x}} + \frac{1}{4}c \, \mathbf{\hat{z}} & \left(4d\right) & \mbox{Ta II} \\ 
\mathbf{B}_{5} & = & z_{3} \, \mathbf{a}_{1} + z_{3} \, \mathbf{a}_{2} & = & z_{3}c \, \mathbf{\hat{z}} & \left(4e\right) & \mbox{I} \\ 
\mathbf{B}_{6} & = & -z_{3} \, \mathbf{a}_{1}-z_{3} \, \mathbf{a}_{2} & = & -z_{3}c \, \mathbf{\hat{z}} & \left(4e\right) & \mbox{I} \\ 
\mathbf{B}_{7} & = & \left(y_{4}+z_{4}\right) \, \mathbf{a}_{1} + \left(x_{4}+z_{4}\right) \, \mathbf{a}_{2} + \left(x_{4}+y_{4}\right) \, \mathbf{a}_{3} & = & x_{4}a \, \mathbf{\hat{x}} + y_{4}a \, \mathbf{\hat{y}} + z_{4}c \, \mathbf{\hat{z}} & \left(16k\right) & \mbox{Se I} \\ 
\mathbf{B}_{8} & = & \left(-y_{4}+z_{4}\right) \, \mathbf{a}_{1} + \left(-x_{4}+z_{4}\right) \, \mathbf{a}_{2} + \left(-x_{4}-y_{4}\right) \, \mathbf{a}_{3} & = & -x_{4}a \, \mathbf{\hat{x}}-y_{4}a \, \mathbf{\hat{y}} + z_{4}c \, \mathbf{\hat{z}} & \left(16k\right) & \mbox{Se I} \\ 
\mathbf{B}_{9} & = & \left(x_{4}+z_{4}\right) \, \mathbf{a}_{1} + \left(-y_{4}+z_{4}\right) \, \mathbf{a}_{2} + \left(x_{4}-y_{4}\right) \, \mathbf{a}_{3} & = & -y_{4}a \, \mathbf{\hat{x}} + x_{4}a \, \mathbf{\hat{y}} + z_{4}c \, \mathbf{\hat{z}} & \left(16k\right) & \mbox{Se I} \\ 
\mathbf{B}_{10} & = & \left(-x_{4}+z_{4}\right) \, \mathbf{a}_{1} + \left(y_{4}+z_{4}\right) \, \mathbf{a}_{2} + \left(-x_{4}+y_{4}\right) \, \mathbf{a}_{3} & = & y_{4}a \, \mathbf{\hat{x}}-x_{4}a \, \mathbf{\hat{y}} + z_{4}c \, \mathbf{\hat{z}} & \left(16k\right) & \mbox{Se I} \\ 
\mathbf{B}_{11} & = & \left(y_{4}-z_{4}\right) \, \mathbf{a}_{1} + \left(-x_{4}-z_{4}\right) \, \mathbf{a}_{2} + \left(-x_{4}+y_{4}\right) \, \mathbf{a}_{3} & = & -x_{4}a \, \mathbf{\hat{x}} + y_{4}a \, \mathbf{\hat{y}}-z_{4}c \, \mathbf{\hat{z}} & \left(16k\right) & \mbox{Se I} \\ 
\mathbf{B}_{12} & = & \left(-y_{4}-z_{4}\right) \, \mathbf{a}_{1} + \left(x_{4}-z_{4}\right) \, \mathbf{a}_{2} + \left(x_{4}-y_{4}\right) \, \mathbf{a}_{3} & = & x_{4}a \, \mathbf{\hat{x}}-y_{4}a \, \mathbf{\hat{y}}-z_{4}c \, \mathbf{\hat{z}} & \left(16k\right) & \mbox{Se I} \\ 
\mathbf{B}_{13} & = & \left(x_{4}-z_{4}\right) \, \mathbf{a}_{1} + \left(y_{4}-z_{4}\right) \, \mathbf{a}_{2} + \left(x_{4}+y_{4}\right) \, \mathbf{a}_{3} & = & y_{4}a \, \mathbf{\hat{x}} + x_{4}a \, \mathbf{\hat{y}}-z_{4}c \, \mathbf{\hat{z}} & \left(16k\right) & \mbox{Se I} \\ 
\mathbf{B}_{14} & = & \left(-x_{4}-z_{4}\right) \, \mathbf{a}_{1} + \left(-y_{4}-z_{4}\right) \, \mathbf{a}_{2} + \left(-x_{4}-y_{4}\right) \, \mathbf{a}_{3} & = & -y_{4}a \, \mathbf{\hat{x}}-x_{4}a \, \mathbf{\hat{y}}-z_{4}c \, \mathbf{\hat{z}} & \left(16k\right) & \mbox{Se I} \\ 
\mathbf{B}_{15} & = & \left(y_{5}+z_{5}\right) \, \mathbf{a}_{1} + \left(x_{5}+z_{5}\right) \, \mathbf{a}_{2} + \left(x_{5}+y_{5}\right) \, \mathbf{a}_{3} & = & x_{5}a \, \mathbf{\hat{x}} + y_{5}a \, \mathbf{\hat{y}} + z_{5}c \, \mathbf{\hat{z}} & \left(16k\right) & \mbox{Se II} \\ 
\mathbf{B}_{16} & = & \left(-y_{5}+z_{5}\right) \, \mathbf{a}_{1} + \left(-x_{5}+z_{5}\right) \, \mathbf{a}_{2} + \left(-x_{5}-y_{5}\right) \, \mathbf{a}_{3} & = & -x_{5}a \, \mathbf{\hat{x}}-y_{5}a \, \mathbf{\hat{y}} + z_{5}c \, \mathbf{\hat{z}} & \left(16k\right) & \mbox{Se II} \\ 
\mathbf{B}_{17} & = & \left(x_{5}+z_{5}\right) \, \mathbf{a}_{1} + \left(-y_{5}+z_{5}\right) \, \mathbf{a}_{2} + \left(x_{5}-y_{5}\right) \, \mathbf{a}_{3} & = & -y_{5}a \, \mathbf{\hat{x}} + x_{5}a \, \mathbf{\hat{y}} + z_{5}c \, \mathbf{\hat{z}} & \left(16k\right) & \mbox{Se II} \\ 
\mathbf{B}_{18} & = & \left(-x_{5}+z_{5}\right) \, \mathbf{a}_{1} + \left(y_{5}+z_{5}\right) \, \mathbf{a}_{2} + \left(-x_{5}+y_{5}\right) \, \mathbf{a}_{3} & = & y_{5}a \, \mathbf{\hat{x}}-x_{5}a \, \mathbf{\hat{y}} + z_{5}c \, \mathbf{\hat{z}} & \left(16k\right) & \mbox{Se II} \\ 
\mathbf{B}_{19} & = & \left(y_{5}-z_{5}\right) \, \mathbf{a}_{1} + \left(-x_{5}-z_{5}\right) \, \mathbf{a}_{2} + \left(-x_{5}+y_{5}\right) \, \mathbf{a}_{3} & = & -x_{5}a \, \mathbf{\hat{x}} + y_{5}a \, \mathbf{\hat{y}}-z_{5}c \, \mathbf{\hat{z}} & \left(16k\right) & \mbox{Se II} \\ 
\mathbf{B}_{20} & = & \left(-y_{5}-z_{5}\right) \, \mathbf{a}_{1} + \left(x_{5}-z_{5}\right) \, \mathbf{a}_{2} + \left(x_{5}-y_{5}\right) \, \mathbf{a}_{3} & = & x_{5}a \, \mathbf{\hat{x}}-y_{5}a \, \mathbf{\hat{y}}-z_{5}c \, \mathbf{\hat{z}} & \left(16k\right) & \mbox{Se II} \\ 
\mathbf{B}_{21} & = & \left(x_{5}-z_{5}\right) \, \mathbf{a}_{1} + \left(y_{5}-z_{5}\right) \, \mathbf{a}_{2} + \left(x_{5}+y_{5}\right) \, \mathbf{a}_{3} & = & y_{5}a \, \mathbf{\hat{x}} + x_{5}a \, \mathbf{\hat{y}}-z_{5}c \, \mathbf{\hat{z}} & \left(16k\right) & \mbox{Se II} \\ 
\mathbf{B}_{22} & = & \left(-x_{5}-z_{5}\right) \, \mathbf{a}_{1} + \left(-y_{5}-z_{5}\right) \, \mathbf{a}_{2} + \left(-x_{5}-y_{5}\right) \, \mathbf{a}_{3} & = & -y_{5}a \, \mathbf{\hat{x}}-x_{5}a \, \mathbf{\hat{y}}-z_{5}c \, \mathbf{\hat{z}} & \left(16k\right) & \mbox{Se II} \\ 
\end{longtabu}
\renewcommand{\arraystretch}{1.0}
\noindent \hrulefill
\\
\textbf{References:}
\vspace*{-0.25cm}
\begin{flushleft}
  - \bibentry{Gressier_ISe8Ta2_JSolStateChem_1984}. \\
\end{flushleft}
\textbf{Found in:}
\vspace*{-0.25cm}
\begin{flushleft}
  - \bibentry{Villars_PearsonsCrystalData_2013}. \\
\end{flushleft}
\noindent \hrulefill
\\
\textbf{Geometry files:}
\\
\noindent  - CIF: pp. {\hyperref[AB8C2_tI44_97_e_2k_cd_cif]{\pageref{AB8C2_tI44_97_e_2k_cd_cif}}} \\
\noindent  - POSCAR: pp. {\hyperref[AB8C2_tI44_97_e_2k_cd_poscar]{\pageref{AB8C2_tI44_97_e_2k_cd_poscar}}} \\
\onecolumn
{\phantomsection\label{A2B_tI12_98_f_a}}
\subsection*{\huge \textbf{{\normalfont CdAs$_{2}$ Structure: A2B\_tI12\_98\_f\_a}}}
\noindent \hrulefill
\vspace*{0.25cm}
\begin{figure}[htp]
  \centering
  \vspace{-1em}
  {\includegraphics[width=1\textwidth]{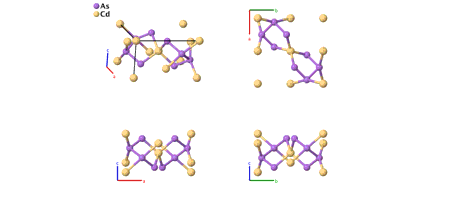}}
\end{figure}
\vspace*{-0.5cm}
\renewcommand{\arraystretch}{1.5}
\begin{equation*}
  \begin{array}{>{$\hspace{-0.15cm}}l<{$}>{$}p{0.5cm}<{$}>{$}p{18.5cm}<{$}}
    \mbox{\large \textbf{Prototype}} &\colon & \ce{CdAs2} \\
    \mbox{\large \textbf{\AFLOW\ prototype label}} &\colon & \mbox{A2B\_tI12\_98\_f\_a} \\
    \mbox{\large \textbf{\textit{Strukturbericht} designation}} &\colon & \mbox{None} \\
    \mbox{\large \textbf{Pearson symbol}} &\colon & \mbox{tI12} \\
    \mbox{\large \textbf{Space group number}} &\colon & 98 \\
    \mbox{\large \textbf{Space group symbol}} &\colon & I4_{1}22 \\
    \mbox{\large \textbf{\AFLOW\ prototype command}} &\colon &  \texttt{aflow} \,  \, \texttt{-{}-proto=A2B\_tI12\_98\_f\_a } \, \newline \texttt{-{}-params=}{a,c/a,x_{2} }
  \end{array}
\end{equation*}
\renewcommand{\arraystretch}{1.0}

\noindent \parbox{1 \linewidth}{
\noindent \hrulefill
\\
\textbf{Body-centered Tetragonal primitive vectors:} \\
\vspace*{-0.25cm}
\begin{tabular}{cc}
  \begin{tabular}{c}
    \parbox{0.6 \linewidth}{
      \renewcommand{\arraystretch}{1.5}
      \begin{equation*}
        \centering
        \begin{array}{ccc}
              \mathbf{a}_1 & = & - \frac12 \, a \, \mathbf{\hat{x}} + \frac12 \, a \, \mathbf{\hat{y}} + \frac12 \, c \, \mathbf{\hat{z}} \\
    \mathbf{a}_2 & = & ~ \frac12 \, a \, \mathbf{\hat{x}} - \frac12 \, a \, \mathbf{\hat{y}} + \frac12 \, c \, \mathbf{\hat{z}} \\
    \mathbf{a}_3 & = & ~ \frac12 \, a \, \mathbf{\hat{x}} + \frac12 \, a \, \mathbf{\hat{y}} - \frac12 \, c \, \mathbf{\hat{z}} \\

        \end{array}
      \end{equation*}
    }
    \renewcommand{\arraystretch}{1.0}
  \end{tabular}
  \begin{tabular}{c}
    \includegraphics[width=0.3\linewidth]{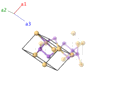} \\
  \end{tabular}
\end{tabular}

}
\vspace*{-0.25cm}

\noindent \hrulefill
\\
\textbf{Basis vectors:}
\vspace*{-0.25cm}
\renewcommand{\arraystretch}{1.5}
\begin{longtabu} to \textwidth{>{\centering $}X[-1,c,c]<{$}>{\centering $}X[-1,c,c]<{$}>{\centering $}X[-1,c,c]<{$}>{\centering $}X[-1,c,c]<{$}>{\centering $}X[-1,c,c]<{$}>{\centering $}X[-1,c,c]<{$}>{\centering $}X[-1,c,c]<{$}}
  & & \mbox{Lattice Coordinates} & & \mbox{Cartesian Coordinates} &\mbox{Wyckoff Position} & \mbox{Atom Type} \\  
  \mathbf{B}_{1} & = & 0 \, \mathbf{a}_{1} + 0 \, \mathbf{a}_{2} + 0 \, \mathbf{a}_{3} & = & 0 \, \mathbf{\hat{x}} + 0 \, \mathbf{\hat{y}} + 0 \, \mathbf{\hat{z}} & \left(4a\right) & \mbox{Cd} \\ 
\mathbf{B}_{2} & = & \frac{3}{4} \, \mathbf{a}_{1} + \frac{1}{4} \, \mathbf{a}_{2} + \frac{1}{2} \, \mathbf{a}_{3} & = & \frac{1}{2}a \, \mathbf{\hat{y}} + \frac{1}{4}c \, \mathbf{\hat{z}} & \left(4a\right) & \mbox{Cd} \\ 
\mathbf{B}_{3} & = & \frac{3}{8} \, \mathbf{a}_{1} + \left(\frac{1}{8} +x_{2}\right) \, \mathbf{a}_{2} + \left(\frac{1}{4} +x_{2}\right) \, \mathbf{a}_{3} & = & x_{2}a \, \mathbf{\hat{x}} + \frac{1}{4}a \, \mathbf{\hat{y}} + \frac{1}{8}c \, \mathbf{\hat{z}} & \left(8f\right) & \mbox{As} \\ 
\mathbf{B}_{4} & = & \frac{7}{8} \, \mathbf{a}_{1} + \left(\frac{1}{8} - x_{2}\right) \, \mathbf{a}_{2} + \left(\frac{3}{4} - x_{2}\right) \, \mathbf{a}_{3} & = & -x_{2}a \, \mathbf{\hat{x}} + \frac{3}{4}a \, \mathbf{\hat{y}} + \frac{1}{8}c \, \mathbf{\hat{z}} & \left(8f\right) & \mbox{As} \\ 
\mathbf{B}_{5} & = & \left(\frac{7}{8} +x_{2}\right) \, \mathbf{a}_{1} + \frac{1}{8} \, \mathbf{a}_{2} + \left(\frac{1}{4} +x_{2}\right) \, \mathbf{a}_{3} & = & \frac{3}{4}a \, \mathbf{\hat{x}} + \left(\frac{1}{2} +x_{2}\right)a \, \mathbf{\hat{y}} + \frac{3}{8}c \, \mathbf{\hat{z}} & \left(8f\right) & \mbox{As} \\ 
\mathbf{B}_{6} & = & \left(\frac{7}{8} - x_{2}\right) \, \mathbf{a}_{1} + \frac{5}{8} \, \mathbf{a}_{2} + \left(\frac{3}{4} - x_{2}\right) \, \mathbf{a}_{3} & = & \frac{1}{4}a \, \mathbf{\hat{x}} + \left(\frac{1}{2} - x_{2}\right)a \, \mathbf{\hat{y}} + \frac{3}{8}c \, \mathbf{\hat{z}} & \left(8f\right) & \mbox{As} \\ 
\end{longtabu}
\renewcommand{\arraystretch}{1.0}
\noindent \hrulefill
\\
\textbf{References:}
\vspace*{-0.25cm}
\begin{flushleft}
  - \bibentry{Yakimovich_CdAs2_InorgMat_1996}. \\
\end{flushleft}
\textbf{Found in:}
\vspace*{-0.25cm}
\begin{flushleft}
  - \bibentry{Villars_PearsonsCrystalData_2013}. \\
\end{flushleft}
\noindent \hrulefill
\\
\textbf{Geometry files:}
\\
\noindent  - CIF: pp. {\hyperref[A2B_tI12_98_f_a_cif]{\pageref{A2B_tI12_98_f_a_cif}}} \\
\noindent  - POSCAR: pp. {\hyperref[A2B_tI12_98_f_a_poscar]{\pageref{A2B_tI12_98_f_a_poscar}}} \\
\onecolumn
{\phantomsection\label{A2B8C2D_tP26_100_c_abcd_c_a}}
\subsection*{\huge \textbf{{\normalfont \begin{raggedleft}Fresnoite (Ba$_{2}$TiSi$_{2}$O$_{8}$) Structure: \end{raggedleft} \\ A2B8C2D\_tP26\_100\_c\_abcd\_c\_a}}}
\noindent \hrulefill
\vspace*{0.25cm}
\begin{figure}[htp]
  \centering
  \vspace{-1em}
  {\includegraphics[width=1\textwidth]{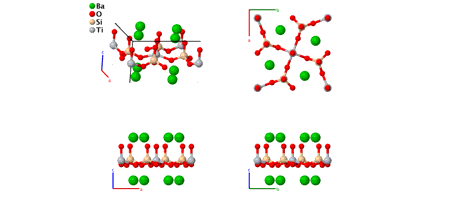}}
\end{figure}
\vspace*{-0.5cm}
\renewcommand{\arraystretch}{1.5}
\begin{equation*}
  \begin{array}{>{$\hspace{-0.15cm}}l<{$}>{$}p{0.5cm}<{$}>{$}p{18.5cm}<{$}}
    \mbox{\large \textbf{Prototype}} &\colon & \ce{Ba2TiSi2O8} \\
    \mbox{\large \textbf{\AFLOW\ prototype label}} &\colon & \mbox{A2B8C2D\_tP26\_100\_c\_abcd\_c\_a} \\
    \mbox{\large \textbf{\textit{Strukturbericht} designation}} &\colon & \mbox{None} \\
    \mbox{\large \textbf{Pearson symbol}} &\colon & \mbox{tP26} \\
    \mbox{\large \textbf{Space group number}} &\colon & 100 \\
    \mbox{\large \textbf{Space group symbol}} &\colon & P4bm \\
    \mbox{\large \textbf{\AFLOW\ prototype command}} &\colon &  \texttt{aflow} \,  \, \texttt{-{}-proto=A2B8C2D\_tP26\_100\_c\_abcd\_c\_a } \, \newline \texttt{-{}-params=}{a,c/a,z_{1},z_{2},z_{3},x_{4},z_{4},x_{5},z_{5},x_{6},z_{6},x_{7},y_{7},z_{7} }
  \end{array}
\end{equation*}
\renewcommand{\arraystretch}{1.0}

\vspace*{-0.25cm}
\noindent \hrulefill
\begin{itemize}
  \item{Found in the
\href{http://webmineral.com/data/Fresnoite.shtml#.WV1oNnUrKUE}{Big
  Creek-Rush Creek sanbornite deposit}, 5 miles NE of Trimmer, Fresno
  Co. California.
}
\end{itemize}

\noindent \parbox{1 \linewidth}{
\noindent \hrulefill
\\
\textbf{Simple Tetragonal primitive vectors:} \\
\vspace*{-0.25cm}
\begin{tabular}{cc}
  \begin{tabular}{c}
    \parbox{0.6 \linewidth}{
      \renewcommand{\arraystretch}{1.5}
      \begin{equation*}
        \centering
        \begin{array}{ccc}
              \mathbf{a}_1 & = & a \, \mathbf{\hat{x}} \\
    \mathbf{a}_2 & = & a \, \mathbf{\hat{y}} \\
    \mathbf{a}_3 & = & c \, \mathbf{\hat{z}} \\

        \end{array}
      \end{equation*}
    }
    \renewcommand{\arraystretch}{1.0}
  \end{tabular}
  \begin{tabular}{c}
    \includegraphics[width=0.3\linewidth]{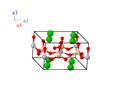} \\
  \end{tabular}
\end{tabular}

}
\vspace*{-0.25cm}

\noindent \hrulefill
\\
\textbf{Basis vectors:}
\vspace*{-0.25cm}
\renewcommand{\arraystretch}{1.5}
\begin{longtabu} to \textwidth{>{\centering $}X[-1,c,c]<{$}>{\centering $}X[-1,c,c]<{$}>{\centering $}X[-1,c,c]<{$}>{\centering $}X[-1,c,c]<{$}>{\centering $}X[-1,c,c]<{$}>{\centering $}X[-1,c,c]<{$}>{\centering $}X[-1,c,c]<{$}}
  & & \mbox{Lattice Coordinates} & & \mbox{Cartesian Coordinates} &\mbox{Wyckoff Position} & \mbox{Atom Type} \\  
  \mathbf{B}_{1} & = & z_{1} \, \mathbf{a}_{3} & = & z_{1}c \, \mathbf{\hat{z}} & \left(2a\right) & \mbox{O I} \\ 
\mathbf{B}_{2} & = & \frac{1}{2} \, \mathbf{a}_{1} + \frac{1}{2} \, \mathbf{a}_{2} + z_{1} \, \mathbf{a}_{3} & = & \frac{1}{2}a \, \mathbf{\hat{x}} + \frac{1}{2}a \, \mathbf{\hat{y}} + z_{1}c \, \mathbf{\hat{z}} & \left(2a\right) & \mbox{O I} \\ 
\mathbf{B}_{3} & = & z_{2} \, \mathbf{a}_{3} & = & z_{2}c \, \mathbf{\hat{z}} & \left(2a\right) & \mbox{Ti} \\ 
\mathbf{B}_{4} & = & \frac{1}{2} \, \mathbf{a}_{1} + \frac{1}{2} \, \mathbf{a}_{2} + z_{2} \, \mathbf{a}_{3} & = & \frac{1}{2}a \, \mathbf{\hat{x}} + \frac{1}{2}a \, \mathbf{\hat{y}} + z_{2}c \, \mathbf{\hat{z}} & \left(2a\right) & \mbox{Ti} \\ 
\mathbf{B}_{5} & = & \frac{1}{2} \, \mathbf{a}_{1} + z_{3} \, \mathbf{a}_{3} & = & \frac{1}{2}a \, \mathbf{\hat{x}} + z_{3}c \, \mathbf{\hat{z}} & \left(2b\right) & \mbox{O II} \\ 
\mathbf{B}_{6} & = & \frac{1}{2} \, \mathbf{a}_{2} + z_{3} \, \mathbf{a}_{3} & = & \frac{1}{2}a \, \mathbf{\hat{y}} + z_{3}c \, \mathbf{\hat{z}} & \left(2b\right) & \mbox{O II} \\ 
\mathbf{B}_{7} & = & x_{4} \, \mathbf{a}_{1} + \left(\frac{1}{2} +x_{4}\right) \, \mathbf{a}_{2} + z_{4} \, \mathbf{a}_{3} & = & x_{4}a \, \mathbf{\hat{x}} + \left(\frac{1}{2} +x_{4}\right)a \, \mathbf{\hat{y}} + z_{4}c \, \mathbf{\hat{z}} & \left(4c\right) & \mbox{Ba} \\ 
\mathbf{B}_{8} & = & -x_{4} \, \mathbf{a}_{1} + \left(\frac{1}{2} - x_{4}\right) \, \mathbf{a}_{2} + z_{4} \, \mathbf{a}_{3} & = & -x_{4}a \, \mathbf{\hat{x}} + \left(\frac{1}{2} - x_{4}\right)a \, \mathbf{\hat{y}} + z_{4}c \, \mathbf{\hat{z}} & \left(4c\right) & \mbox{Ba} \\ 
\mathbf{B}_{9} & = & \left(\frac{1}{2} - x_{4}\right) \, \mathbf{a}_{1} + x_{4} \, \mathbf{a}_{2} + z_{4} \, \mathbf{a}_{3} & = & \left(\frac{1}{2} - x_{4}\right)a \, \mathbf{\hat{x}} + x_{4}a \, \mathbf{\hat{y}} + z_{4}c \, \mathbf{\hat{z}} & \left(4c\right) & \mbox{Ba} \\ 
\mathbf{B}_{10} & = & \left(\frac{1}{2} +x_{4}\right) \, \mathbf{a}_{1}-x_{4} \, \mathbf{a}_{2} + z_{4} \, \mathbf{a}_{3} & = & \left(\frac{1}{2} +x_{4}\right)a \, \mathbf{\hat{x}}-x_{4}a \, \mathbf{\hat{y}} + z_{4}c \, \mathbf{\hat{z}} & \left(4c\right) & \mbox{Ba} \\ 
\mathbf{B}_{11} & = & x_{5} \, \mathbf{a}_{1} + \left(\frac{1}{2} +x_{5}\right) \, \mathbf{a}_{2} + z_{5} \, \mathbf{a}_{3} & = & x_{5}a \, \mathbf{\hat{x}} + \left(\frac{1}{2} +x_{5}\right)a \, \mathbf{\hat{y}} + z_{5}c \, \mathbf{\hat{z}} & \left(4c\right) & \mbox{O III} \\ 
\mathbf{B}_{12} & = & -x_{5} \, \mathbf{a}_{1} + \left(\frac{1}{2} - x_{5}\right) \, \mathbf{a}_{2} + z_{5} \, \mathbf{a}_{3} & = & -x_{5}a \, \mathbf{\hat{x}} + \left(\frac{1}{2} - x_{5}\right)a \, \mathbf{\hat{y}} + z_{5}c \, \mathbf{\hat{z}} & \left(4c\right) & \mbox{O III} \\ 
\mathbf{B}_{13} & = & \left(\frac{1}{2} - x_{5}\right) \, \mathbf{a}_{1} + x_{5} \, \mathbf{a}_{2} + z_{5} \, \mathbf{a}_{3} & = & \left(\frac{1}{2} - x_{5}\right)a \, \mathbf{\hat{x}} + x_{5}a \, \mathbf{\hat{y}} + z_{5}c \, \mathbf{\hat{z}} & \left(4c\right) & \mbox{O III} \\ 
\mathbf{B}_{14} & = & \left(\frac{1}{2} +x_{5}\right) \, \mathbf{a}_{1}-x_{5} \, \mathbf{a}_{2} + z_{5} \, \mathbf{a}_{3} & = & \left(\frac{1}{2} +x_{5}\right)a \, \mathbf{\hat{x}}-x_{5}a \, \mathbf{\hat{y}} + z_{5}c \, \mathbf{\hat{z}} & \left(4c\right) & \mbox{O III} \\ 
\mathbf{B}_{15} & = & x_{6} \, \mathbf{a}_{1} + \left(\frac{1}{2} +x_{6}\right) \, \mathbf{a}_{2} + z_{6} \, \mathbf{a}_{3} & = & x_{6}a \, \mathbf{\hat{x}} + \left(\frac{1}{2} +x_{6}\right)a \, \mathbf{\hat{y}} + z_{6}c \, \mathbf{\hat{z}} & \left(4c\right) & \mbox{Si} \\ 
\mathbf{B}_{16} & = & -x_{6} \, \mathbf{a}_{1} + \left(\frac{1}{2} - x_{6}\right) \, \mathbf{a}_{2} + z_{6} \, \mathbf{a}_{3} & = & -x_{6}a \, \mathbf{\hat{x}} + \left(\frac{1}{2} - x_{6}\right)a \, \mathbf{\hat{y}} + z_{6}c \, \mathbf{\hat{z}} & \left(4c\right) & \mbox{Si} \\ 
\mathbf{B}_{17} & = & \left(\frac{1}{2} - x_{6}\right) \, \mathbf{a}_{1} + x_{6} \, \mathbf{a}_{2} + z_{6} \, \mathbf{a}_{3} & = & \left(\frac{1}{2} - x_{6}\right)a \, \mathbf{\hat{x}} + x_{6}a \, \mathbf{\hat{y}} + z_{6}c \, \mathbf{\hat{z}} & \left(4c\right) & \mbox{Si} \\ 
\mathbf{B}_{18} & = & \left(\frac{1}{2} +x_{6}\right) \, \mathbf{a}_{1}-x_{6} \, \mathbf{a}_{2} + z_{6} \, \mathbf{a}_{3} & = & \left(\frac{1}{2} +x_{6}\right)a \, \mathbf{\hat{x}}-x_{6}a \, \mathbf{\hat{y}} + z_{6}c \, \mathbf{\hat{z}} & \left(4c\right) & \mbox{Si} \\ 
\mathbf{B}_{19} & = & x_{7} \, \mathbf{a}_{1} + y_{7} \, \mathbf{a}_{2} + z_{7} \, \mathbf{a}_{3} & = & x_{7}a \, \mathbf{\hat{x}} + y_{7}a \, \mathbf{\hat{y}} + z_{7}c \, \mathbf{\hat{z}} & \left(8d\right) & \mbox{O IV} \\ 
\mathbf{B}_{20} & = & -x_{7} \, \mathbf{a}_{1}-y_{7} \, \mathbf{a}_{2} + z_{7} \, \mathbf{a}_{3} & = & -x_{7}a \, \mathbf{\hat{x}}-y_{7}a \, \mathbf{\hat{y}} + z_{7}c \, \mathbf{\hat{z}} & \left(8d\right) & \mbox{O IV} \\ 
\mathbf{B}_{21} & = & -y_{7} \, \mathbf{a}_{1} + x_{7} \, \mathbf{a}_{2} + z_{7} \, \mathbf{a}_{3} & = & -y_{7}a \, \mathbf{\hat{x}} + x_{7}a \, \mathbf{\hat{y}} + z_{7}c \, \mathbf{\hat{z}} & \left(8d\right) & \mbox{O IV} \\ 
\mathbf{B}_{22} & = & y_{7} \, \mathbf{a}_{1}-x_{7} \, \mathbf{a}_{2} + z_{7} \, \mathbf{a}_{3} & = & y_{7}a \, \mathbf{\hat{x}}-x_{7}a \, \mathbf{\hat{y}} + z_{7}c \, \mathbf{\hat{z}} & \left(8d\right) & \mbox{O IV} \\ 
\mathbf{B}_{23} & = & \left(\frac{1}{2} +x_{7}\right) \, \mathbf{a}_{1} + \left(\frac{1}{2} - y_{7}\right) \, \mathbf{a}_{2} + z_{7} \, \mathbf{a}_{3} & = & \left(\frac{1}{2} +x_{7}\right)a \, \mathbf{\hat{x}} + \left(\frac{1}{2} - y_{7}\right)a \, \mathbf{\hat{y}} + z_{7}c \, \mathbf{\hat{z}} & \left(8d\right) & \mbox{O IV} \\ 
\mathbf{B}_{24} & = & \left(\frac{1}{2} - x_{7}\right) \, \mathbf{a}_{1} + \left(\frac{1}{2} +y_{7}\right) \, \mathbf{a}_{2} + z_{7} \, \mathbf{a}_{3} & = & \left(\frac{1}{2} - x_{7}\right)a \, \mathbf{\hat{x}} + \left(\frac{1}{2} +y_{7}\right)a \, \mathbf{\hat{y}} + z_{7}c \, \mathbf{\hat{z}} & \left(8d\right) & \mbox{O IV} \\ 
\mathbf{B}_{25} & = & \left(\frac{1}{2} - y_{7}\right) \, \mathbf{a}_{1} + \left(\frac{1}{2} - x_{7}\right) \, \mathbf{a}_{2} + z_{7} \, \mathbf{a}_{3} & = & \left(\frac{1}{2} - y_{7}\right)a \, \mathbf{\hat{x}} + \left(\frac{1}{2} - x_{7}\right)a \, \mathbf{\hat{y}} + z_{7}c \, \mathbf{\hat{z}} & \left(8d\right) & \mbox{O IV} \\ 
\mathbf{B}_{26} & = & \left(\frac{1}{2} +y_{7}\right) \, \mathbf{a}_{1} + \left(\frac{1}{2} +x_{7}\right) \, \mathbf{a}_{2} + z_{7} \, \mathbf{a}_{3} & = & \left(\frac{1}{2} +y_{7}\right)a \, \mathbf{\hat{x}} + \left(\frac{1}{2} +x_{7}\right)a \, \mathbf{\hat{y}} + z_{7}c \, \mathbf{\hat{z}} & \left(8d\right) & \mbox{O IV} \\ 
\end{longtabu}
\renewcommand{\arraystretch}{1.0}
\noindent \hrulefill
\\
\textbf{References:}
\vspace*{-0.25cm}
\begin{flushleft}
  - \bibentry{Markgraf_1985}. \\
\end{flushleft}
\noindent \hrulefill
\\
\textbf{Geometry files:}
\\
\noindent  - CIF: pp. {\hyperref[A2B8C2D_tP26_100_c_abcd_c_a_cif]{\pageref{A2B8C2D_tP26_100_c_abcd_c_a_cif}}} \\
\noindent  - POSCAR: pp. {\hyperref[A2B8C2D_tP26_100_c_abcd_c_a_poscar]{\pageref{A2B8C2D_tP26_100_c_abcd_c_a_poscar}}} \\
\onecolumn
{\phantomsection\label{A3B11C6_tP40_100_ac_bc2d_cd}}
\subsection*{\huge \textbf{{\normalfont Ce$_{3}$Si$_{6}$N$_{11}$ Structure: A3B11C6\_tP40\_100\_ac\_bc2d\_cd}}}
\noindent \hrulefill
\vspace*{0.25cm}
\begin{figure}[htp]
  \centering
  \vspace{-1em}
  {\includegraphics[width=1\textwidth]{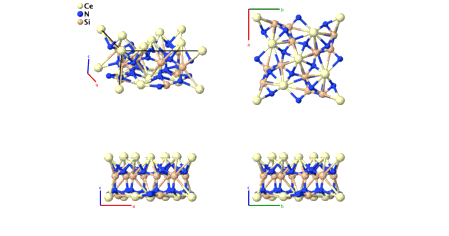}}
\end{figure}
\vspace*{-0.5cm}
\renewcommand{\arraystretch}{1.5}
\begin{equation*}
  \begin{array}{>{$\hspace{-0.15cm}}l<{$}>{$}p{0.5cm}<{$}>{$}p{18.5cm}<{$}}
    \mbox{\large \textbf{Prototype}} &\colon & \ce{Ce3Si6N11} \\
    \mbox{\large \textbf{\AFLOW\ prototype label}} &\colon & \mbox{A3B11C6\_tP40\_100\_ac\_bc2d\_cd} \\
    \mbox{\large \textbf{\textit{Strukturbericht} designation}} &\colon & \mbox{None} \\
    \mbox{\large \textbf{Pearson symbol}} &\colon & \mbox{tP40} \\
    \mbox{\large \textbf{Space group number}} &\colon & 100 \\
    \mbox{\large \textbf{Space group symbol}} &\colon & P4bm \\
    \mbox{\large \textbf{\AFLOW\ prototype command}} &\colon &  \texttt{aflow} \,  \, \texttt{-{}-proto=A3B11C6\_tP40\_100\_ac\_bc2d\_cd } \, \newline \texttt{-{}-params=}{a,c/a,z_{1},z_{2},x_{3},z_{3},x_{4},z_{4},x_{5},z_{5},x_{6},y_{6},z_{6},x_{7},y_{7},z_{7},x_{8},y_{8},z_{8} }
  \end{array}
\end{equation*}
\renewcommand{\arraystretch}{1.0}

\noindent \parbox{1 \linewidth}{
\noindent \hrulefill
\\
\textbf{Simple Tetragonal primitive vectors:} \\
\vspace*{-0.25cm}
\begin{tabular}{cc}
  \begin{tabular}{c}
    \parbox{0.6 \linewidth}{
      \renewcommand{\arraystretch}{1.5}
      \begin{equation*}
        \centering
        \begin{array}{ccc}
              \mathbf{a}_1 & = & a \, \mathbf{\hat{x}} \\
    \mathbf{a}_2 & = & a \, \mathbf{\hat{y}} \\
    \mathbf{a}_3 & = & c \, \mathbf{\hat{z}} \\

        \end{array}
      \end{equation*}
    }
    \renewcommand{\arraystretch}{1.0}
  \end{tabular}
  \begin{tabular}{c}
    \includegraphics[width=0.3\linewidth]{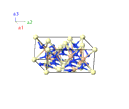} \\
  \end{tabular}
\end{tabular}

}
\vspace*{-0.25cm}

\noindent \hrulefill
\\
\textbf{Basis vectors:}
\vspace*{-0.25cm}
\renewcommand{\arraystretch}{1.5}
\begin{longtabu} to \textwidth{>{\centering $}X[-1,c,c]<{$}>{\centering $}X[-1,c,c]<{$}>{\centering $}X[-1,c,c]<{$}>{\centering $}X[-1,c,c]<{$}>{\centering $}X[-1,c,c]<{$}>{\centering $}X[-1,c,c]<{$}>{\centering $}X[-1,c,c]<{$}}
  & & \mbox{Lattice Coordinates} & & \mbox{Cartesian Coordinates} &\mbox{Wyckoff Position} & \mbox{Atom Type} \\  
  \mathbf{B}_{1} & = & z_{1} \, \mathbf{a}_{3} & = & z_{1}c \, \mathbf{\hat{z}} & \left(2a\right) & \mbox{Ce I} \\ 
\mathbf{B}_{2} & = & \frac{1}{2} \, \mathbf{a}_{1} + \frac{1}{2} \, \mathbf{a}_{2} + z_{1} \, \mathbf{a}_{3} & = & \frac{1}{2}a \, \mathbf{\hat{x}} + \frac{1}{2}a \, \mathbf{\hat{y}} + z_{1}c \, \mathbf{\hat{z}} & \left(2a\right) & \mbox{Ce I} \\ 
\mathbf{B}_{3} & = & \frac{1}{2} \, \mathbf{a}_{1} + z_{2} \, \mathbf{a}_{3} & = & \frac{1}{2}a \, \mathbf{\hat{x}} + z_{2}c \, \mathbf{\hat{z}} & \left(2b\right) & \mbox{N I} \\ 
\mathbf{B}_{4} & = & \frac{1}{2} \, \mathbf{a}_{2} + z_{2} \, \mathbf{a}_{3} & = & \frac{1}{2}a \, \mathbf{\hat{y}} + z_{2}c \, \mathbf{\hat{z}} & \left(2b\right) & \mbox{N I} \\ 
\mathbf{B}_{5} & = & x_{3} \, \mathbf{a}_{1} + \left(\frac{1}{2} +x_{3}\right) \, \mathbf{a}_{2} + z_{3} \, \mathbf{a}_{3} & = & x_{3}a \, \mathbf{\hat{x}} + \left(\frac{1}{2} +x_{3}\right)a \, \mathbf{\hat{y}} + z_{3}c \, \mathbf{\hat{z}} & \left(4c\right) & \mbox{Ce II} \\ 
\mathbf{B}_{6} & = & -x_{3} \, \mathbf{a}_{1} + \left(\frac{1}{2} - x_{3}\right) \, \mathbf{a}_{2} + z_{3} \, \mathbf{a}_{3} & = & -x_{3}a \, \mathbf{\hat{x}} + \left(\frac{1}{2} - x_{3}\right)a \, \mathbf{\hat{y}} + z_{3}c \, \mathbf{\hat{z}} & \left(4c\right) & \mbox{Ce II} \\ 
\mathbf{B}_{7} & = & \left(\frac{1}{2} - x_{3}\right) \, \mathbf{a}_{1} + x_{3} \, \mathbf{a}_{2} + z_{3} \, \mathbf{a}_{3} & = & \left(\frac{1}{2} - x_{3}\right)a \, \mathbf{\hat{x}} + x_{3}a \, \mathbf{\hat{y}} + z_{3}c \, \mathbf{\hat{z}} & \left(4c\right) & \mbox{Ce II} \\ 
\mathbf{B}_{8} & = & \left(\frac{1}{2} +x_{3}\right) \, \mathbf{a}_{1}-x_{3} \, \mathbf{a}_{2} + z_{3} \, \mathbf{a}_{3} & = & \left(\frac{1}{2} +x_{3}\right)a \, \mathbf{\hat{x}}-x_{3}a \, \mathbf{\hat{y}} + z_{3}c \, \mathbf{\hat{z}} & \left(4c\right) & \mbox{Ce II} \\ 
\mathbf{B}_{9} & = & x_{4} \, \mathbf{a}_{1} + \left(\frac{1}{2} +x_{4}\right) \, \mathbf{a}_{2} + z_{4} \, \mathbf{a}_{3} & = & x_{4}a \, \mathbf{\hat{x}} + \left(\frac{1}{2} +x_{4}\right)a \, \mathbf{\hat{y}} + z_{4}c \, \mathbf{\hat{z}} & \left(4c\right) & \mbox{N II} \\ 
\mathbf{B}_{10} & = & -x_{4} \, \mathbf{a}_{1} + \left(\frac{1}{2} - x_{4}\right) \, \mathbf{a}_{2} + z_{4} \, \mathbf{a}_{3} & = & -x_{4}a \, \mathbf{\hat{x}} + \left(\frac{1}{2} - x_{4}\right)a \, \mathbf{\hat{y}} + z_{4}c \, \mathbf{\hat{z}} & \left(4c\right) & \mbox{N II} \\ 
\mathbf{B}_{11} & = & \left(\frac{1}{2} - x_{4}\right) \, \mathbf{a}_{1} + x_{4} \, \mathbf{a}_{2} + z_{4} \, \mathbf{a}_{3} & = & \left(\frac{1}{2} - x_{4}\right)a \, \mathbf{\hat{x}} + x_{4}a \, \mathbf{\hat{y}} + z_{4}c \, \mathbf{\hat{z}} & \left(4c\right) & \mbox{N II} \\ 
\mathbf{B}_{12} & = & \left(\frac{1}{2} +x_{4}\right) \, \mathbf{a}_{1}-x_{4} \, \mathbf{a}_{2} + z_{4} \, \mathbf{a}_{3} & = & \left(\frac{1}{2} +x_{4}\right)a \, \mathbf{\hat{x}}-x_{4}a \, \mathbf{\hat{y}} + z_{4}c \, \mathbf{\hat{z}} & \left(4c\right) & \mbox{N II} \\ 
\mathbf{B}_{13} & = & x_{5} \, \mathbf{a}_{1} + \left(\frac{1}{2} +x_{5}\right) \, \mathbf{a}_{2} + z_{5} \, \mathbf{a}_{3} & = & x_{5}a \, \mathbf{\hat{x}} + \left(\frac{1}{2} +x_{5}\right)a \, \mathbf{\hat{y}} + z_{5}c \, \mathbf{\hat{z}} & \left(4c\right) & \mbox{Si I} \\ 
\mathbf{B}_{14} & = & -x_{5} \, \mathbf{a}_{1} + \left(\frac{1}{2} - x_{5}\right) \, \mathbf{a}_{2} + z_{5} \, \mathbf{a}_{3} & = & -x_{5}a \, \mathbf{\hat{x}} + \left(\frac{1}{2} - x_{5}\right)a \, \mathbf{\hat{y}} + z_{5}c \, \mathbf{\hat{z}} & \left(4c\right) & \mbox{Si I} \\ 
\mathbf{B}_{15} & = & \left(\frac{1}{2} - x_{5}\right) \, \mathbf{a}_{1} + x_{5} \, \mathbf{a}_{2} + z_{5} \, \mathbf{a}_{3} & = & \left(\frac{1}{2} - x_{5}\right)a \, \mathbf{\hat{x}} + x_{5}a \, \mathbf{\hat{y}} + z_{5}c \, \mathbf{\hat{z}} & \left(4c\right) & \mbox{Si I} \\ 
\mathbf{B}_{16} & = & \left(\frac{1}{2} +x_{5}\right) \, \mathbf{a}_{1}-x_{5} \, \mathbf{a}_{2} + z_{5} \, \mathbf{a}_{3} & = & \left(\frac{1}{2} +x_{5}\right)a \, \mathbf{\hat{x}}-x_{5}a \, \mathbf{\hat{y}} + z_{5}c \, \mathbf{\hat{z}} & \left(4c\right) & \mbox{Si I} \\ 
\mathbf{B}_{17} & = & x_{6} \, \mathbf{a}_{1} + y_{6} \, \mathbf{a}_{2} + z_{6} \, \mathbf{a}_{3} & = & x_{6}a \, \mathbf{\hat{x}} + y_{6}a \, \mathbf{\hat{y}} + z_{6}c \, \mathbf{\hat{z}} & \left(8d\right) & \mbox{N III} \\ 
\mathbf{B}_{18} & = & -x_{6} \, \mathbf{a}_{1}-y_{6} \, \mathbf{a}_{2} + z_{6} \, \mathbf{a}_{3} & = & -x_{6}a \, \mathbf{\hat{x}}-y_{6}a \, \mathbf{\hat{y}} + z_{6}c \, \mathbf{\hat{z}} & \left(8d\right) & \mbox{N III} \\ 
\mathbf{B}_{19} & = & -y_{6} \, \mathbf{a}_{1} + x_{6} \, \mathbf{a}_{2} + z_{6} \, \mathbf{a}_{3} & = & -y_{6}a \, \mathbf{\hat{x}} + x_{6}a \, \mathbf{\hat{y}} + z_{6}c \, \mathbf{\hat{z}} & \left(8d\right) & \mbox{N III} \\ 
\mathbf{B}_{20} & = & y_{6} \, \mathbf{a}_{1}-x_{6} \, \mathbf{a}_{2} + z_{6} \, \mathbf{a}_{3} & = & y_{6}a \, \mathbf{\hat{x}}-x_{6}a \, \mathbf{\hat{y}} + z_{6}c \, \mathbf{\hat{z}} & \left(8d\right) & \mbox{N III} \\ 
\mathbf{B}_{21} & = & \left(\frac{1}{2} +x_{6}\right) \, \mathbf{a}_{1} + \left(\frac{1}{2} - y_{6}\right) \, \mathbf{a}_{2} + z_{6} \, \mathbf{a}_{3} & = & \left(\frac{1}{2} +x_{6}\right)a \, \mathbf{\hat{x}} + \left(\frac{1}{2} - y_{6}\right)a \, \mathbf{\hat{y}} + z_{6}c \, \mathbf{\hat{z}} & \left(8d\right) & \mbox{N III} \\ 
\mathbf{B}_{22} & = & \left(\frac{1}{2} - x_{6}\right) \, \mathbf{a}_{1} + \left(\frac{1}{2} +y_{6}\right) \, \mathbf{a}_{2} + z_{6} \, \mathbf{a}_{3} & = & \left(\frac{1}{2} - x_{6}\right)a \, \mathbf{\hat{x}} + \left(\frac{1}{2} +y_{6}\right)a \, \mathbf{\hat{y}} + z_{6}c \, \mathbf{\hat{z}} & \left(8d\right) & \mbox{N III} \\ 
\mathbf{B}_{23} & = & \left(\frac{1}{2} - y_{6}\right) \, \mathbf{a}_{1} + \left(\frac{1}{2} - x_{6}\right) \, \mathbf{a}_{2} + z_{6} \, \mathbf{a}_{3} & = & \left(\frac{1}{2} - y_{6}\right)a \, \mathbf{\hat{x}} + \left(\frac{1}{2} - x_{6}\right)a \, \mathbf{\hat{y}} + z_{6}c \, \mathbf{\hat{z}} & \left(8d\right) & \mbox{N III} \\ 
\mathbf{B}_{24} & = & \left(\frac{1}{2} +y_{6}\right) \, \mathbf{a}_{1} + \left(\frac{1}{2} +x_{6}\right) \, \mathbf{a}_{2} + z_{6} \, \mathbf{a}_{3} & = & \left(\frac{1}{2} +y_{6}\right)a \, \mathbf{\hat{x}} + \left(\frac{1}{2} +x_{6}\right)a \, \mathbf{\hat{y}} + z_{6}c \, \mathbf{\hat{z}} & \left(8d\right) & \mbox{N III} \\ 
\mathbf{B}_{25} & = & x_{7} \, \mathbf{a}_{1} + y_{7} \, \mathbf{a}_{2} + z_{7} \, \mathbf{a}_{3} & = & x_{7}a \, \mathbf{\hat{x}} + y_{7}a \, \mathbf{\hat{y}} + z_{7}c \, \mathbf{\hat{z}} & \left(8d\right) & \mbox{N IV} \\ 
\mathbf{B}_{26} & = & -x_{7} \, \mathbf{a}_{1}-y_{7} \, \mathbf{a}_{2} + z_{7} \, \mathbf{a}_{3} & = & -x_{7}a \, \mathbf{\hat{x}}-y_{7}a \, \mathbf{\hat{y}} + z_{7}c \, \mathbf{\hat{z}} & \left(8d\right) & \mbox{N IV} \\ 
\mathbf{B}_{27} & = & -y_{7} \, \mathbf{a}_{1} + x_{7} \, \mathbf{a}_{2} + z_{7} \, \mathbf{a}_{3} & = & -y_{7}a \, \mathbf{\hat{x}} + x_{7}a \, \mathbf{\hat{y}} + z_{7}c \, \mathbf{\hat{z}} & \left(8d\right) & \mbox{N IV} \\ 
\mathbf{B}_{28} & = & y_{7} \, \mathbf{a}_{1}-x_{7} \, \mathbf{a}_{2} + z_{7} \, \mathbf{a}_{3} & = & y_{7}a \, \mathbf{\hat{x}}-x_{7}a \, \mathbf{\hat{y}} + z_{7}c \, \mathbf{\hat{z}} & \left(8d\right) & \mbox{N IV} \\ 
\mathbf{B}_{29} & = & \left(\frac{1}{2} +x_{7}\right) \, \mathbf{a}_{1} + \left(\frac{1}{2} - y_{7}\right) \, \mathbf{a}_{2} + z_{7} \, \mathbf{a}_{3} & = & \left(\frac{1}{2} +x_{7}\right)a \, \mathbf{\hat{x}} + \left(\frac{1}{2} - y_{7}\right)a \, \mathbf{\hat{y}} + z_{7}c \, \mathbf{\hat{z}} & \left(8d\right) & \mbox{N IV} \\ 
\mathbf{B}_{30} & = & \left(\frac{1}{2} - x_{7}\right) \, \mathbf{a}_{1} + \left(\frac{1}{2} +y_{7}\right) \, \mathbf{a}_{2} + z_{7} \, \mathbf{a}_{3} & = & \left(\frac{1}{2} - x_{7}\right)a \, \mathbf{\hat{x}} + \left(\frac{1}{2} +y_{7}\right)a \, \mathbf{\hat{y}} + z_{7}c \, \mathbf{\hat{z}} & \left(8d\right) & \mbox{N IV} \\ 
\mathbf{B}_{31} & = & \left(\frac{1}{2} - y_{7}\right) \, \mathbf{a}_{1} + \left(\frac{1}{2} - x_{7}\right) \, \mathbf{a}_{2} + z_{7} \, \mathbf{a}_{3} & = & \left(\frac{1}{2} - y_{7}\right)a \, \mathbf{\hat{x}} + \left(\frac{1}{2} - x_{7}\right)a \, \mathbf{\hat{y}} + z_{7}c \, \mathbf{\hat{z}} & \left(8d\right) & \mbox{N IV} \\ 
\mathbf{B}_{32} & = & \left(\frac{1}{2} +y_{7}\right) \, \mathbf{a}_{1} + \left(\frac{1}{2} +x_{7}\right) \, \mathbf{a}_{2} + z_{7} \, \mathbf{a}_{3} & = & \left(\frac{1}{2} +y_{7}\right)a \, \mathbf{\hat{x}} + \left(\frac{1}{2} +x_{7}\right)a \, \mathbf{\hat{y}} + z_{7}c \, \mathbf{\hat{z}} & \left(8d\right) & \mbox{N IV} \\ 
\mathbf{B}_{33} & = & x_{8} \, \mathbf{a}_{1} + y_{8} \, \mathbf{a}_{2} + z_{8} \, \mathbf{a}_{3} & = & x_{8}a \, \mathbf{\hat{x}} + y_{8}a \, \mathbf{\hat{y}} + z_{8}c \, \mathbf{\hat{z}} & \left(8d\right) & \mbox{Si II} \\ 
\mathbf{B}_{34} & = & -x_{8} \, \mathbf{a}_{1}-y_{8} \, \mathbf{a}_{2} + z_{8} \, \mathbf{a}_{3} & = & -x_{8}a \, \mathbf{\hat{x}}-y_{8}a \, \mathbf{\hat{y}} + z_{8}c \, \mathbf{\hat{z}} & \left(8d\right) & \mbox{Si II} \\ 
\mathbf{B}_{35} & = & -y_{8} \, \mathbf{a}_{1} + x_{8} \, \mathbf{a}_{2} + z_{8} \, \mathbf{a}_{3} & = & -y_{8}a \, \mathbf{\hat{x}} + x_{8}a \, \mathbf{\hat{y}} + z_{8}c \, \mathbf{\hat{z}} & \left(8d\right) & \mbox{Si II} \\ 
\mathbf{B}_{36} & = & y_{8} \, \mathbf{a}_{1}-x_{8} \, \mathbf{a}_{2} + z_{8} \, \mathbf{a}_{3} & = & y_{8}a \, \mathbf{\hat{x}}-x_{8}a \, \mathbf{\hat{y}} + z_{8}c \, \mathbf{\hat{z}} & \left(8d\right) & \mbox{Si II} \\ 
\mathbf{B}_{37} & = & \left(\frac{1}{2} +x_{8}\right) \, \mathbf{a}_{1} + \left(\frac{1}{2} - y_{8}\right) \, \mathbf{a}_{2} + z_{8} \, \mathbf{a}_{3} & = & \left(\frac{1}{2} +x_{8}\right)a \, \mathbf{\hat{x}} + \left(\frac{1}{2} - y_{8}\right)a \, \mathbf{\hat{y}} + z_{8}c \, \mathbf{\hat{z}} & \left(8d\right) & \mbox{Si II} \\ 
\mathbf{B}_{38} & = & \left(\frac{1}{2} - x_{8}\right) \, \mathbf{a}_{1} + \left(\frac{1}{2} +y_{8}\right) \, \mathbf{a}_{2} + z_{8} \, \mathbf{a}_{3} & = & \left(\frac{1}{2} - x_{8}\right)a \, \mathbf{\hat{x}} + \left(\frac{1}{2} +y_{8}\right)a \, \mathbf{\hat{y}} + z_{8}c \, \mathbf{\hat{z}} & \left(8d\right) & \mbox{Si II} \\ 
\mathbf{B}_{39} & = & \left(\frac{1}{2} - y_{8}\right) \, \mathbf{a}_{1} + \left(\frac{1}{2} - x_{8}\right) \, \mathbf{a}_{2} + z_{8} \, \mathbf{a}_{3} & = & \left(\frac{1}{2} - y_{8}\right)a \, \mathbf{\hat{x}} + \left(\frac{1}{2} - x_{8}\right)a \, \mathbf{\hat{y}} + z_{8}c \, \mathbf{\hat{z}} & \left(8d\right) & \mbox{Si II} \\ 
\mathbf{B}_{40} & = & \left(\frac{1}{2} +y_{8}\right) \, \mathbf{a}_{1} + \left(\frac{1}{2} +x_{8}\right) \, \mathbf{a}_{2} + z_{8} \, \mathbf{a}_{3} & = & \left(\frac{1}{2} +y_{8}\right)a \, \mathbf{\hat{x}} + \left(\frac{1}{2} +x_{8}\right)a \, \mathbf{\hat{y}} + z_{8}c \, \mathbf{\hat{z}} & \left(8d\right) & \mbox{Si II} \\ 
\end{longtabu}
\renewcommand{\arraystretch}{1.0}
\noindent \hrulefill
\\
\textbf{References:}
\vspace*{-0.25cm}
\begin{flushleft}
  - \bibentry{Woike_Ce3Si6N11_InorgChem_1995}. \\
\end{flushleft}
\textbf{Found in:}
\vspace*{-0.25cm}
\begin{flushleft}
  - \bibentry{Villars_PearsonsCrystalData_2013}. \\
\end{flushleft}
\noindent \hrulefill
\\
\textbf{Geometry files:}
\\
\noindent  - CIF: pp. {\hyperref[A3B11C6_tP40_100_ac_bc2d_cd_cif]{\pageref{A3B11C6_tP40_100_ac_bc2d_cd_cif}}} \\
\noindent  - POSCAR: pp. {\hyperref[A3B11C6_tP40_100_ac_bc2d_cd_poscar]{\pageref{A3B11C6_tP40_100_ac_bc2d_cd_poscar}}} \\
\onecolumn
{\phantomsection\label{A7B7C2_tP32_101_bde_ade_d}}
\subsection*{\huge \textbf{{\normalfont $\gamma$-MgNiSn Structure: A7B7C2\_tP32\_101\_bde\_ade\_d}}}
\noindent \hrulefill
\vspace*{0.25cm}
\begin{figure}[htp]
  \centering
  \vspace{-1em}
  {\includegraphics[width=1\textwidth]{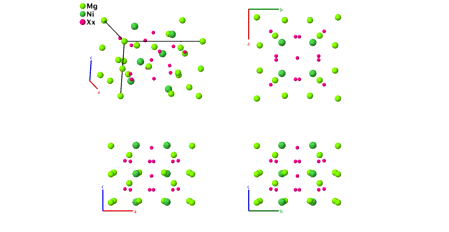}}
\end{figure}
\vspace*{-0.5cm}
\renewcommand{\arraystretch}{1.5}
\begin{equation*}
  \begin{array}{>{$\hspace{-0.15cm}}l<{$}>{$}p{0.5cm}<{$}>{$}p{18.5cm}<{$}}
    \mbox{\large \textbf{Prototype}} &\colon & \ce{$\gamma$-MgNiSn} \\
    \mbox{\large \textbf{\AFLOW\ prototype label}} &\colon & \mbox{A7B7C2\_tP32\_101\_bde\_ade\_d} \\
    \mbox{\large \textbf{\textit{Strukturbericht} designation}} &\colon & \mbox{None} \\
    \mbox{\large \textbf{Pearson symbol}} &\colon & \mbox{tP32} \\
    \mbox{\large \textbf{Space group number}} &\colon & 101 \\
    \mbox{\large \textbf{Space group symbol}} &\colon & P4_{2}cm \\
    \mbox{\large \textbf{\AFLOW\ prototype command}} &\colon &  \texttt{aflow} \,  \, \texttt{-{}-proto=A7B7C2\_tP32\_101\_bde\_ade\_d } \, \newline \texttt{-{}-params=}{a,c/a,z_{1},z_{2},x_{3},z_{3},x_{4},z_{4},x_{5},z_{5},x_{6},y_{6},z_{6},x_{7},y_{7},z_{7} }
  \end{array}
\end{equation*}
\renewcommand{\arraystretch}{1.0}

\vspace*{-0.25cm}
\noindent \hrulefill
\begin{itemize}
  \item{This is the $\gamma$ phase of the Mg-Ni-Sn ternary system.  
The (2b), (8e) and (4d) Wyckoff positions are partially occupied and are 
represented by the labels M I, M II, and M III, respectively.
Here, M I is 0.88Mg+0.12Ni, M II is 0.96Mg+0.05Ni, and M III is 0.88Sn+0.12Mg.
The Jmol image does not distinguish between the different M labels and is represented 
by the "Xx" atoms.
}
\end{itemize}

\noindent \parbox{1 \linewidth}{
\noindent \hrulefill
\\
\textbf{Simple Tetragonal primitive vectors:} \\
\vspace*{-0.25cm}
\begin{tabular}{cc}
  \begin{tabular}{c}
    \parbox{0.6 \linewidth}{
      \renewcommand{\arraystretch}{1.5}
      \begin{equation*}
        \centering
        \begin{array}{ccc}
              \mathbf{a}_1 & = & a \, \mathbf{\hat{x}} \\
    \mathbf{a}_2 & = & a \, \mathbf{\hat{y}} \\
    \mathbf{a}_3 & = & c \, \mathbf{\hat{z}} \\

        \end{array}
      \end{equation*}
    }
    \renewcommand{\arraystretch}{1.0}
  \end{tabular}
  \begin{tabular}{c}
    \includegraphics[width=0.3\linewidth]{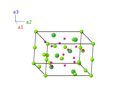} \\
  \end{tabular}
\end{tabular}

}
\vspace*{-0.25cm}

\noindent \hrulefill
\\
\textbf{Basis vectors:}
\vspace*{-0.25cm}
\renewcommand{\arraystretch}{1.5}
\begin{longtabu} to \textwidth{>{\centering $}X[-1,c,c]<{$}>{\centering $}X[-1,c,c]<{$}>{\centering $}X[-1,c,c]<{$}>{\centering $}X[-1,c,c]<{$}>{\centering $}X[-1,c,c]<{$}>{\centering $}X[-1,c,c]<{$}>{\centering $}X[-1,c,c]<{$}}
  & & \mbox{Lattice Coordinates} & & \mbox{Cartesian Coordinates} &\mbox{Wyckoff Position} & \mbox{Atom Type} \\  
  \mathbf{B}_{1} & = & z_{1} \, \mathbf{a}_{3} & = & z_{1}c \, \mathbf{\hat{z}} & \left(2a\right) & \mbox{Mg I} \\ 
\mathbf{B}_{2} & = & \left(\frac{1}{2} +z_{1}\right) \, \mathbf{a}_{3} & = & \left(\frac{1}{2} +z_{1}\right)c \, \mathbf{\hat{z}} & \left(2a\right) & \mbox{Mg I} \\ 
\mathbf{B}_{3} & = & \frac{1}{2} \, \mathbf{a}_{1} + \frac{1}{2} \, \mathbf{a}_{2} + z_{2} \, \mathbf{a}_{3} & = & \frac{1}{2}a \, \mathbf{\hat{x}} + \frac{1}{2}a \, \mathbf{\hat{y}} + z_{2}c \, \mathbf{\hat{z}} & \left(2b\right) & \mbox{M I} \\ 
\mathbf{B}_{4} & = & \frac{1}{2} \, \mathbf{a}_{1} + \frac{1}{2} \, \mathbf{a}_{2} + \left(\frac{1}{2} +z_{2}\right) \, \mathbf{a}_{3} & = & \frac{1}{2}a \, \mathbf{\hat{x}} + \frac{1}{2}a \, \mathbf{\hat{y}} + \left(\frac{1}{2} +z_{2}\right)c \, \mathbf{\hat{z}} & \left(2b\right) & \mbox{M I} \\ 
\mathbf{B}_{5} & = & x_{3} \, \mathbf{a}_{1} + x_{3} \, \mathbf{a}_{2} + z_{3} \, \mathbf{a}_{3} & = & x_{3}a \, \mathbf{\hat{x}} + x_{3}a \, \mathbf{\hat{y}} + z_{3}c \, \mathbf{\hat{z}} & \left(4d\right) & \mbox{M II} \\ 
\mathbf{B}_{6} & = & -x_{3} \, \mathbf{a}_{1}-x_{3} \, \mathbf{a}_{2} + z_{3} \, \mathbf{a}_{3} & = & -x_{3}a \, \mathbf{\hat{x}}-x_{3}a \, \mathbf{\hat{y}} + z_{3}c \, \mathbf{\hat{z}} & \left(4d\right) & \mbox{M II} \\ 
\mathbf{B}_{7} & = & -x_{3} \, \mathbf{a}_{1} + x_{3} \, \mathbf{a}_{2} + \left(\frac{1}{2} +z_{3}\right) \, \mathbf{a}_{3} & = & -x_{3}a \, \mathbf{\hat{x}} + x_{3}a \, \mathbf{\hat{y}} + \left(\frac{1}{2} +z_{3}\right)c \, \mathbf{\hat{z}} & \left(4d\right) & \mbox{M II} \\ 
\mathbf{B}_{8} & = & x_{3} \, \mathbf{a}_{1}-x_{3} \, \mathbf{a}_{2} + \left(\frac{1}{2} +z_{3}\right) \, \mathbf{a}_{3} & = & x_{3}a \, \mathbf{\hat{x}}-x_{3}a \, \mathbf{\hat{y}} + \left(\frac{1}{2} +z_{3}\right)c \, \mathbf{\hat{z}} & \left(4d\right) & \mbox{M II} \\ 
\mathbf{B}_{9} & = & x_{4} \, \mathbf{a}_{1} + x_{4} \, \mathbf{a}_{2} + z_{4} \, \mathbf{a}_{3} & = & x_{4}a \, \mathbf{\hat{x}} + x_{4}a \, \mathbf{\hat{y}} + z_{4}c \, \mathbf{\hat{z}} & \left(4d\right) & \mbox{Mg II} \\ 
\mathbf{B}_{10} & = & -x_{4} \, \mathbf{a}_{1}-x_{4} \, \mathbf{a}_{2} + z_{4} \, \mathbf{a}_{3} & = & -x_{4}a \, \mathbf{\hat{x}}-x_{4}a \, \mathbf{\hat{y}} + z_{4}c \, \mathbf{\hat{z}} & \left(4d\right) & \mbox{Mg II} \\ 
\mathbf{B}_{11} & = & -x_{4} \, \mathbf{a}_{1} + x_{4} \, \mathbf{a}_{2} + \left(\frac{1}{2} +z_{4}\right) \, \mathbf{a}_{3} & = & -x_{4}a \, \mathbf{\hat{x}} + x_{4}a \, \mathbf{\hat{y}} + \left(\frac{1}{2} +z_{4}\right)c \, \mathbf{\hat{z}} & \left(4d\right) & \mbox{Mg II} \\ 
\mathbf{B}_{12} & = & x_{4} \, \mathbf{a}_{1}-x_{4} \, \mathbf{a}_{2} + \left(\frac{1}{2} +z_{4}\right) \, \mathbf{a}_{3} & = & x_{4}a \, \mathbf{\hat{x}}-x_{4}a \, \mathbf{\hat{y}} + \left(\frac{1}{2} +z_{4}\right)c \, \mathbf{\hat{z}} & \left(4d\right) & \mbox{Mg II} \\ 
\mathbf{B}_{13} & = & x_{5} \, \mathbf{a}_{1} + x_{5} \, \mathbf{a}_{2} + z_{5} \, \mathbf{a}_{3} & = & x_{5}a \, \mathbf{\hat{x}} + x_{5}a \, \mathbf{\hat{y}} + z_{5}c \, \mathbf{\hat{z}} & \left(4d\right) & \mbox{Ni} \\ 
\mathbf{B}_{14} & = & -x_{5} \, \mathbf{a}_{1}-x_{5} \, \mathbf{a}_{2} + z_{5} \, \mathbf{a}_{3} & = & -x_{5}a \, \mathbf{\hat{x}}-x_{5}a \, \mathbf{\hat{y}} + z_{5}c \, \mathbf{\hat{z}} & \left(4d\right) & \mbox{Ni} \\ 
\mathbf{B}_{15} & = & -x_{5} \, \mathbf{a}_{1} + x_{5} \, \mathbf{a}_{2} + \left(\frac{1}{2} +z_{5}\right) \, \mathbf{a}_{3} & = & -x_{5}a \, \mathbf{\hat{x}} + x_{5}a \, \mathbf{\hat{y}} + \left(\frac{1}{2} +z_{5}\right)c \, \mathbf{\hat{z}} & \left(4d\right) & \mbox{Ni} \\ 
\mathbf{B}_{16} & = & x_{5} \, \mathbf{a}_{1}-x_{5} \, \mathbf{a}_{2} + \left(\frac{1}{2} +z_{5}\right) \, \mathbf{a}_{3} & = & x_{5}a \, \mathbf{\hat{x}}-x_{5}a \, \mathbf{\hat{y}} + \left(\frac{1}{2} +z_{5}\right)c \, \mathbf{\hat{z}} & \left(4d\right) & \mbox{Ni} \\ 
\mathbf{B}_{17} & = & x_{6} \, \mathbf{a}_{1} + y_{6} \, \mathbf{a}_{2} + z_{6} \, \mathbf{a}_{3} & = & x_{6}a \, \mathbf{\hat{x}} + y_{6}a \, \mathbf{\hat{y}} + z_{6}c \, \mathbf{\hat{z}} & \left(8e\right) & \mbox{M III} \\ 
\mathbf{B}_{18} & = & -x_{6} \, \mathbf{a}_{1}-y_{6} \, \mathbf{a}_{2} + z_{6} \, \mathbf{a}_{3} & = & -x_{6}a \, \mathbf{\hat{x}}-y_{6}a \, \mathbf{\hat{y}} + z_{6}c \, \mathbf{\hat{z}} & \left(8e\right) & \mbox{M III} \\ 
\mathbf{B}_{19} & = & -y_{6} \, \mathbf{a}_{1} + x_{6} \, \mathbf{a}_{2} + \left(\frac{1}{2} +z_{6}\right) \, \mathbf{a}_{3} & = & -y_{6}a \, \mathbf{\hat{x}} + x_{6}a \, \mathbf{\hat{y}} + \left(\frac{1}{2} +z_{6}\right)c \, \mathbf{\hat{z}} & \left(8e\right) & \mbox{M III} \\ 
\mathbf{B}_{20} & = & y_{6} \, \mathbf{a}_{1}-x_{6} \, \mathbf{a}_{2} + \left(\frac{1}{2} +z_{6}\right) \, \mathbf{a}_{3} & = & y_{6}a \, \mathbf{\hat{x}}-x_{6}a \, \mathbf{\hat{y}} + \left(\frac{1}{2} +z_{6}\right)c \, \mathbf{\hat{z}} & \left(8e\right) & \mbox{M III} \\ 
\mathbf{B}_{21} & = & x_{6} \, \mathbf{a}_{1}-y_{6} \, \mathbf{a}_{2} + \left(\frac{1}{2} +z_{6}\right) \, \mathbf{a}_{3} & = & x_{6}a \, \mathbf{\hat{x}}-y_{6}a \, \mathbf{\hat{y}} + \left(\frac{1}{2} +z_{6}\right)c \, \mathbf{\hat{z}} & \left(8e\right) & \mbox{M III} \\ 
\mathbf{B}_{22} & = & -x_{6} \, \mathbf{a}_{1} + y_{6} \, \mathbf{a}_{2} + \left(\frac{1}{2} +z_{6}\right) \, \mathbf{a}_{3} & = & -x_{6}a \, \mathbf{\hat{x}} + y_{6}a \, \mathbf{\hat{y}} + \left(\frac{1}{2} +z_{6}\right)c \, \mathbf{\hat{z}} & \left(8e\right) & \mbox{M III} \\ 
\mathbf{B}_{23} & = & -y_{6} \, \mathbf{a}_{1}-x_{6} \, \mathbf{a}_{2} + z_{6} \, \mathbf{a}_{3} & = & -y_{6}a \, \mathbf{\hat{x}}-x_{6}a \, \mathbf{\hat{y}} + z_{6}c \, \mathbf{\hat{z}} & \left(8e\right) & \mbox{M III} \\ 
\mathbf{B}_{24} & = & y_{6} \, \mathbf{a}_{1} + x_{6} \, \mathbf{a}_{2} + z_{6} \, \mathbf{a}_{3} & = & y_{6}a \, \mathbf{\hat{x}} + x_{6}a \, \mathbf{\hat{y}} + z_{6}c \, \mathbf{\hat{z}} & \left(8e\right) & \mbox{M III} \\ 
\mathbf{B}_{25} & = & x_{7} \, \mathbf{a}_{1} + y_{7} \, \mathbf{a}_{2} + z_{7} \, \mathbf{a}_{3} & = & x_{7}a \, \mathbf{\hat{x}} + y_{7}a \, \mathbf{\hat{y}} + z_{7}c \, \mathbf{\hat{z}} & \left(8e\right) & \mbox{Mg III} \\ 
\mathbf{B}_{26} & = & -x_{7} \, \mathbf{a}_{1}-y_{7} \, \mathbf{a}_{2} + z_{7} \, \mathbf{a}_{3} & = & -x_{7}a \, \mathbf{\hat{x}}-y_{7}a \, \mathbf{\hat{y}} + z_{7}c \, \mathbf{\hat{z}} & \left(8e\right) & \mbox{Mg III} \\ 
\mathbf{B}_{27} & = & -y_{7} \, \mathbf{a}_{1} + x_{7} \, \mathbf{a}_{2} + \left(\frac{1}{2} +z_{7}\right) \, \mathbf{a}_{3} & = & -y_{7}a \, \mathbf{\hat{x}} + x_{7}a \, \mathbf{\hat{y}} + \left(\frac{1}{2} +z_{7}\right)c \, \mathbf{\hat{z}} & \left(8e\right) & \mbox{Mg III} \\ 
\mathbf{B}_{28} & = & y_{7} \, \mathbf{a}_{1}-x_{7} \, \mathbf{a}_{2} + \left(\frac{1}{2} +z_{7}\right) \, \mathbf{a}_{3} & = & y_{7}a \, \mathbf{\hat{x}}-x_{7}a \, \mathbf{\hat{y}} + \left(\frac{1}{2} +z_{7}\right)c \, \mathbf{\hat{z}} & \left(8e\right) & \mbox{Mg III} \\ 
\mathbf{B}_{29} & = & x_{7} \, \mathbf{a}_{1}-y_{7} \, \mathbf{a}_{2} + \left(\frac{1}{2} +z_{7}\right) \, \mathbf{a}_{3} & = & x_{7}a \, \mathbf{\hat{x}}-y_{7}a \, \mathbf{\hat{y}} + \left(\frac{1}{2} +z_{7}\right)c \, \mathbf{\hat{z}} & \left(8e\right) & \mbox{Mg III} \\ 
\mathbf{B}_{30} & = & -x_{7} \, \mathbf{a}_{1} + y_{7} \, \mathbf{a}_{2} + \left(\frac{1}{2} +z_{7}\right) \, \mathbf{a}_{3} & = & -x_{7}a \, \mathbf{\hat{x}} + y_{7}a \, \mathbf{\hat{y}} + \left(\frac{1}{2} +z_{7}\right)c \, \mathbf{\hat{z}} & \left(8e\right) & \mbox{Mg III} \\ 
\mathbf{B}_{31} & = & -y_{7} \, \mathbf{a}_{1}-x_{7} \, \mathbf{a}_{2} + z_{7} \, \mathbf{a}_{3} & = & -y_{7}a \, \mathbf{\hat{x}}-x_{7}a \, \mathbf{\hat{y}} + z_{7}c \, \mathbf{\hat{z}} & \left(8e\right) & \mbox{Mg III} \\ 
\mathbf{B}_{32} & = & y_{7} \, \mathbf{a}_{1} + x_{7} \, \mathbf{a}_{2} + z_{7} \, \mathbf{a}_{3} & = & y_{7}a \, \mathbf{\hat{x}} + x_{7}a \, \mathbf{\hat{y}} + z_{7}c \, \mathbf{\hat{z}} & \left(8e\right) & \mbox{Mg III} \\ 
\end{longtabu}
\renewcommand{\arraystretch}{1.0}
\noindent \hrulefill
\\
\textbf{References:}
\vspace*{-0.25cm}
\begin{flushleft}
  - \bibentry{Boudard_MgNiSn_JAllComp_2004}. \\
\end{flushleft}
\textbf{Found in:}
\vspace*{-0.25cm}
\begin{flushleft}
  - \bibentry{Villars_PearsonsCrystalData_2013}. \\
\end{flushleft}
\noindent \hrulefill
\\
\textbf{Geometry files:}
\\
\noindent  - CIF: pp. {\hyperref[A7B7C2_tP32_101_bde_ade_d_cif]{\pageref{A7B7C2_tP32_101_bde_ade_d_cif}}} \\
\noindent  - POSCAR: pp. {\hyperref[A7B7C2_tP32_101_bde_ade_d_poscar]{\pageref{A7B7C2_tP32_101_bde_ade_d_poscar}}} \\
\onecolumn
{\phantomsection\label{A2B3_tP20_102_2c_b2c}}
\subsection*{\huge \textbf{{\normalfont Gd$_{3}$Al$_{2}$ Structure: A2B3\_tP20\_102\_2c\_b2c}}}
\noindent \hrulefill
\vspace*{0.25cm}
\begin{figure}[htp]
  \centering
  \vspace{-1em}
  {\includegraphics[width=1\textwidth]{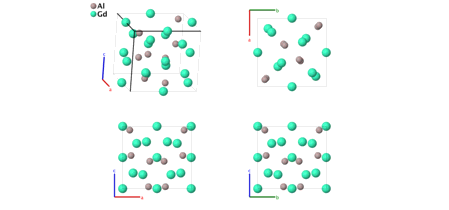}}
\end{figure}
\vspace*{-0.5cm}
\renewcommand{\arraystretch}{1.5}
\begin{equation*}
  \begin{array}{>{$\hspace{-0.15cm}}l<{$}>{$}p{0.5cm}<{$}>{$}p{18.5cm}<{$}}
    \mbox{\large \textbf{Prototype}} &\colon & \ce{Gd3Al2} \\
    \mbox{\large \textbf{\AFLOW\ prototype label}} &\colon & \mbox{A2B3\_tP20\_102\_2c\_b2c} \\
    \mbox{\large \textbf{\textit{Strukturbericht} designation}} &\colon & \mbox{None} \\
    \mbox{\large \textbf{Pearson symbol}} &\colon & \mbox{tP20} \\
    \mbox{\large \textbf{Space group number}} &\colon & 102 \\
    \mbox{\large \textbf{Space group symbol}} &\colon & P4_{2}nm \\
    \mbox{\large \textbf{\AFLOW\ prototype command}} &\colon &  \texttt{aflow} \,  \, \texttt{-{}-proto=A2B3\_tP20\_102\_2c\_b2c } \, \newline \texttt{-{}-params=}{a,c/a,z_{1},x_{2},z_{2},x_{3},z_{3},x_{4},z_{4},x_{5},z_{5} }
  \end{array}
\end{equation*}
\renewcommand{\arraystretch}{1.0}

\noindent \parbox{1 \linewidth}{
\noindent \hrulefill
\\
\textbf{Simple Tetragonal primitive vectors:} \\
\vspace*{-0.25cm}
\begin{tabular}{cc}
  \begin{tabular}{c}
    \parbox{0.6 \linewidth}{
      \renewcommand{\arraystretch}{1.5}
      \begin{equation*}
        \centering
        \begin{array}{ccc}
              \mathbf{a}_1 & = & a \, \mathbf{\hat{x}} \\
    \mathbf{a}_2 & = & a \, \mathbf{\hat{y}} \\
    \mathbf{a}_3 & = & c \, \mathbf{\hat{z}} \\

        \end{array}
      \end{equation*}
    }
    \renewcommand{\arraystretch}{1.0}
  \end{tabular}
  \begin{tabular}{c}
    \includegraphics[width=0.3\linewidth]{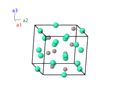} \\
  \end{tabular}
\end{tabular}

}
\vspace*{-0.25cm}

\noindent \hrulefill
\\
\textbf{Basis vectors:}
\vspace*{-0.25cm}
\renewcommand{\arraystretch}{1.5}
\begin{longtabu} to \textwidth{>{\centering $}X[-1,c,c]<{$}>{\centering $}X[-1,c,c]<{$}>{\centering $}X[-1,c,c]<{$}>{\centering $}X[-1,c,c]<{$}>{\centering $}X[-1,c,c]<{$}>{\centering $}X[-1,c,c]<{$}>{\centering $}X[-1,c,c]<{$}}
  & & \mbox{Lattice Coordinates} & & \mbox{Cartesian Coordinates} &\mbox{Wyckoff Position} & \mbox{Atom Type} \\  
  \mathbf{B}_{1} & = & \frac{1}{2} \, \mathbf{a}_{2} + z_{1} \, \mathbf{a}_{3} & = & \frac{1}{2}a \, \mathbf{\hat{y}} + z_{1}c \, \mathbf{\hat{z}} & \left(4b\right) & \mbox{Gd I} \\ 
\mathbf{B}_{2} & = & \frac{1}{2} \, \mathbf{a}_{2} + \left(\frac{1}{2} +z_{1}\right) \, \mathbf{a}_{3} & = & \frac{1}{2}a \, \mathbf{\hat{y}} + \left(\frac{1}{2} +z_{1}\right)c \, \mathbf{\hat{z}} & \left(4b\right) & \mbox{Gd I} \\ 
\mathbf{B}_{3} & = & \frac{1}{2} \, \mathbf{a}_{1} + \left(\frac{1}{2} +z_{1}\right) \, \mathbf{a}_{3} & = & \frac{1}{2}a \, \mathbf{\hat{x}} + \left(\frac{1}{2} +z_{1}\right)c \, \mathbf{\hat{z}} & \left(4b\right) & \mbox{Gd I} \\ 
\mathbf{B}_{4} & = & \frac{1}{2} \, \mathbf{a}_{1} + z_{1} \, \mathbf{a}_{3} & = & \frac{1}{2}a \, \mathbf{\hat{x}} + z_{1}c \, \mathbf{\hat{z}} & \left(4b\right) & \mbox{Gd I} \\ 
\mathbf{B}_{5} & = & x_{2} \, \mathbf{a}_{1} + x_{2} \, \mathbf{a}_{2} + z_{2} \, \mathbf{a}_{3} & = & x_{2}a \, \mathbf{\hat{x}} + x_{2}a \, \mathbf{\hat{y}} + z_{2}c \, \mathbf{\hat{z}} & \left(4c\right) & \mbox{Al I} \\ 
\mathbf{B}_{6} & = & -x_{2} \, \mathbf{a}_{1}-x_{2} \, \mathbf{a}_{2} + z_{2} \, \mathbf{a}_{3} & = & -x_{2}a \, \mathbf{\hat{x}}-x_{2}a \, \mathbf{\hat{y}} + z_{2}c \, \mathbf{\hat{z}} & \left(4c\right) & \mbox{Al I} \\ 
\mathbf{B}_{7} & = & \left(\frac{1}{2} - x_{2}\right) \, \mathbf{a}_{1} + \left(\frac{1}{2} +x_{2}\right) \, \mathbf{a}_{2} + \left(\frac{1}{2} +z_{2}\right) \, \mathbf{a}_{3} & = & \left(\frac{1}{2} - x_{2}\right)a \, \mathbf{\hat{x}} + \left(\frac{1}{2} +x_{2}\right)a \, \mathbf{\hat{y}} + \left(\frac{1}{2} +z_{2}\right)c \, \mathbf{\hat{z}} & \left(4c\right) & \mbox{Al I} \\ 
\mathbf{B}_{8} & = & \left(\frac{1}{2} +x_{2}\right) \, \mathbf{a}_{1} + \left(\frac{1}{2} - x_{2}\right) \, \mathbf{a}_{2} + \left(\frac{1}{2} +z_{2}\right) \, \mathbf{a}_{3} & = & \left(\frac{1}{2} +x_{2}\right)a \, \mathbf{\hat{x}} + \left(\frac{1}{2} - x_{2}\right)a \, \mathbf{\hat{y}} + \left(\frac{1}{2} +z_{2}\right)c \, \mathbf{\hat{z}} & \left(4c\right) & \mbox{Al I} \\ 
\mathbf{B}_{9} & = & x_{3} \, \mathbf{a}_{1} + x_{3} \, \mathbf{a}_{2} + z_{3} \, \mathbf{a}_{3} & = & x_{3}a \, \mathbf{\hat{x}} + x_{3}a \, \mathbf{\hat{y}} + z_{3}c \, \mathbf{\hat{z}} & \left(4c\right) & \mbox{Al II} \\ 
\mathbf{B}_{10} & = & -x_{3} \, \mathbf{a}_{1}-x_{3} \, \mathbf{a}_{2} + z_{3} \, \mathbf{a}_{3} & = & -x_{3}a \, \mathbf{\hat{x}}-x_{3}a \, \mathbf{\hat{y}} + z_{3}c \, \mathbf{\hat{z}} & \left(4c\right) & \mbox{Al II} \\ 
\mathbf{B}_{11} & = & \left(\frac{1}{2} - x_{3}\right) \, \mathbf{a}_{1} + \left(\frac{1}{2} +x_{3}\right) \, \mathbf{a}_{2} + \left(\frac{1}{2} +z_{3}\right) \, \mathbf{a}_{3} & = & \left(\frac{1}{2} - x_{3}\right)a \, \mathbf{\hat{x}} + \left(\frac{1}{2} +x_{3}\right)a \, \mathbf{\hat{y}} + \left(\frac{1}{2} +z_{3}\right)c \, \mathbf{\hat{z}} & \left(4c\right) & \mbox{Al II} \\ 
\mathbf{B}_{12} & = & \left(\frac{1}{2} +x_{3}\right) \, \mathbf{a}_{1} + \left(\frac{1}{2} - x_{3}\right) \, \mathbf{a}_{2} + \left(\frac{1}{2} +z_{3}\right) \, \mathbf{a}_{3} & = & \left(\frac{1}{2} +x_{3}\right)a \, \mathbf{\hat{x}} + \left(\frac{1}{2} - x_{3}\right)a \, \mathbf{\hat{y}} + \left(\frac{1}{2} +z_{3}\right)c \, \mathbf{\hat{z}} & \left(4c\right) & \mbox{Al II} \\ 
\mathbf{B}_{13} & = & x_{4} \, \mathbf{a}_{1} + x_{4} \, \mathbf{a}_{2} + z_{4} \, \mathbf{a}_{3} & = & x_{4}a \, \mathbf{\hat{x}} + x_{4}a \, \mathbf{\hat{y}} + z_{4}c \, \mathbf{\hat{z}} & \left(4c\right) & \mbox{Gd II} \\ 
\mathbf{B}_{14} & = & -x_{4} \, \mathbf{a}_{1}-x_{4} \, \mathbf{a}_{2} + z_{4} \, \mathbf{a}_{3} & = & -x_{4}a \, \mathbf{\hat{x}}-x_{4}a \, \mathbf{\hat{y}} + z_{4}c \, \mathbf{\hat{z}} & \left(4c\right) & \mbox{Gd II} \\ 
\mathbf{B}_{15} & = & \left(\frac{1}{2} - x_{4}\right) \, \mathbf{a}_{1} + \left(\frac{1}{2} +x_{4}\right) \, \mathbf{a}_{2} + \left(\frac{1}{2} +z_{4}\right) \, \mathbf{a}_{3} & = & \left(\frac{1}{2} - x_{4}\right)a \, \mathbf{\hat{x}} + \left(\frac{1}{2} +x_{4}\right)a \, \mathbf{\hat{y}} + \left(\frac{1}{2} +z_{4}\right)c \, \mathbf{\hat{z}} & \left(4c\right) & \mbox{Gd II} \\ 
\mathbf{B}_{16} & = & \left(\frac{1}{2} +x_{4}\right) \, \mathbf{a}_{1} + \left(\frac{1}{2} - x_{4}\right) \, \mathbf{a}_{2} + \left(\frac{1}{2} +z_{4}\right) \, \mathbf{a}_{3} & = & \left(\frac{1}{2} +x_{4}\right)a \, \mathbf{\hat{x}} + \left(\frac{1}{2} - x_{4}\right)a \, \mathbf{\hat{y}} + \left(\frac{1}{2} +z_{4}\right)c \, \mathbf{\hat{z}} & \left(4c\right) & \mbox{Gd II} \\ 
\mathbf{B}_{17} & = & x_{5} \, \mathbf{a}_{1} + x_{5} \, \mathbf{a}_{2} + z_{5} \, \mathbf{a}_{3} & = & x_{5}a \, \mathbf{\hat{x}} + x_{5}a \, \mathbf{\hat{y}} + z_{5}c \, \mathbf{\hat{z}} & \left(4c\right) & \mbox{Gd III} \\ 
\mathbf{B}_{18} & = & -x_{5} \, \mathbf{a}_{1}-x_{5} \, \mathbf{a}_{2} + z_{5} \, \mathbf{a}_{3} & = & -x_{5}a \, \mathbf{\hat{x}}-x_{5}a \, \mathbf{\hat{y}} + z_{5}c \, \mathbf{\hat{z}} & \left(4c\right) & \mbox{Gd III} \\ 
\mathbf{B}_{19} & = & \left(\frac{1}{2} - x_{5}\right) \, \mathbf{a}_{1} + \left(\frac{1}{2} +x_{5}\right) \, \mathbf{a}_{2} + \left(\frac{1}{2} +z_{5}\right) \, \mathbf{a}_{3} & = & \left(\frac{1}{2} - x_{5}\right)a \, \mathbf{\hat{x}} + \left(\frac{1}{2} +x_{5}\right)a \, \mathbf{\hat{y}} + \left(\frac{1}{2} +z_{5}\right)c \, \mathbf{\hat{z}} & \left(4c\right) & \mbox{Gd III} \\ 
\mathbf{B}_{20} & = & \left(\frac{1}{2} +x_{5}\right) \, \mathbf{a}_{1} + \left(\frac{1}{2} - x_{5}\right) \, \mathbf{a}_{2} + \left(\frac{1}{2} +z_{5}\right) \, \mathbf{a}_{3} & = & \left(\frac{1}{2} +x_{5}\right)a \, \mathbf{\hat{x}} + \left(\frac{1}{2} - x_{5}\right)a \, \mathbf{\hat{y}} + \left(\frac{1}{2} +z_{5}\right)c \, \mathbf{\hat{z}} & \left(4c\right) & \mbox{Gd III} \\ 
\end{longtabu}
\renewcommand{\arraystretch}{1.0}
\noindent \hrulefill
\\
\textbf{References:}
\vspace*{-0.25cm}
\begin{flushleft}
  - \bibentry{Buschow_Al2Gd3_JLessCommMetals_1965}. \\
\end{flushleft}
\textbf{Found in:}
\vspace*{-0.25cm}
\begin{flushleft}
  - \bibentry{Villars_PearsonsCrystalData_2013}. \\
\end{flushleft}
\noindent \hrulefill
\\
\textbf{Geometry files:}
\\
\noindent  - CIF: pp. {\hyperref[A2B3_tP20_102_2c_b2c_cif]{\pageref{A2B3_tP20_102_2c_b2c_cif}}} \\
\noindent  - POSCAR: pp. {\hyperref[A2B3_tP20_102_2c_b2c_poscar]{\pageref{A2B3_tP20_102_2c_b2c_poscar}}} \\
\onecolumn
{\phantomsection\label{AB4_tP10_103_a_d}}
\subsection*{\huge \textbf{{\normalfont NbTe$_{4}$ Structure: AB4\_tP10\_103\_a\_d}}}
\noindent \hrulefill
\vspace*{0.25cm}
\begin{figure}[htp]
  \centering
  \vspace{-1em}
  {\includegraphics[width=1\textwidth]{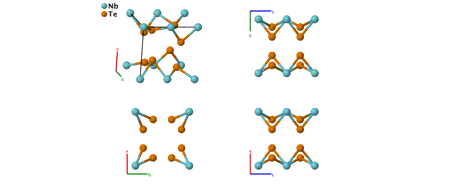}}
\end{figure}
\vspace*{-0.5cm}
\renewcommand{\arraystretch}{1.5}
\begin{equation*}
  \begin{array}{>{$\hspace{-0.15cm}}l<{$}>{$}p{0.5cm}<{$}>{$}p{18.5cm}<{$}}
    \mbox{\large \textbf{Prototype}} &\colon & \ce{NbTe4} \\
    \mbox{\large \textbf{\AFLOW\ prototype label}} &\colon & \mbox{AB4\_tP10\_103\_a\_d} \\
    \mbox{\large \textbf{\textit{Strukturbericht} designation}} &\colon & \mbox{None} \\
    \mbox{\large \textbf{Pearson symbol}} &\colon & \mbox{tP10} \\
    \mbox{\large \textbf{Space group number}} &\colon & 103 \\
    \mbox{\large \textbf{Space group symbol}} &\colon & P4cc \\
    \mbox{\large \textbf{\AFLOW\ prototype command}} &\colon &  \texttt{aflow} \,  \, \texttt{-{}-proto=AB4\_tP10\_103\_a\_d } \, \newline \texttt{-{}-params=}{a,c/a,z_{1},x_{2},y_{2},z_{2} }
  \end{array}
\end{equation*}
\renewcommand{\arraystretch}{1.0}

\noindent \parbox{1 \linewidth}{
\noindent \hrulefill
\\
\textbf{Simple Tetragonal primitive vectors:} \\
\vspace*{-0.25cm}
\begin{tabular}{cc}
  \begin{tabular}{c}
    \parbox{0.6 \linewidth}{
      \renewcommand{\arraystretch}{1.5}
      \begin{equation*}
        \centering
        \begin{array}{ccc}
              \mathbf{a}_1 & = & a \, \mathbf{\hat{x}} \\
    \mathbf{a}_2 & = & a \, \mathbf{\hat{y}} \\
    \mathbf{a}_3 & = & c \, \mathbf{\hat{z}} \\

        \end{array}
      \end{equation*}
    }
    \renewcommand{\arraystretch}{1.0}
  \end{tabular}
  \begin{tabular}{c}
    \includegraphics[width=0.3\linewidth]{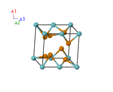} \\
  \end{tabular}
\end{tabular}

}
\vspace*{-0.25cm}

\noindent \hrulefill
\\
\textbf{Basis vectors:}
\vspace*{-0.25cm}
\renewcommand{\arraystretch}{1.5}
\begin{longtabu} to \textwidth{>{\centering $}X[-1,c,c]<{$}>{\centering $}X[-1,c,c]<{$}>{\centering $}X[-1,c,c]<{$}>{\centering $}X[-1,c,c]<{$}>{\centering $}X[-1,c,c]<{$}>{\centering $}X[-1,c,c]<{$}>{\centering $}X[-1,c,c]<{$}}
  & & \mbox{Lattice Coordinates} & & \mbox{Cartesian Coordinates} &\mbox{Wyckoff Position} & \mbox{Atom Type} \\  
  \mathbf{B}_{1} & = & z_{1} \, \mathbf{a}_{3} & = & z_{1}c \, \mathbf{\hat{z}} & \left(2a\right) & \mbox{Nb} \\ 
\mathbf{B}_{2} & = & \left(\frac{1}{2} +z_{1}\right) \, \mathbf{a}_{3} & = & \left(\frac{1}{2} +z_{1}\right)c \, \mathbf{\hat{z}} & \left(2a\right) & \mbox{Nb} \\ 
\mathbf{B}_{3} & = & x_{2} \, \mathbf{a}_{1} + y_{2} \, \mathbf{a}_{2} + z_{2} \, \mathbf{a}_{3} & = & x_{2}a \, \mathbf{\hat{x}} + y_{2}a \, \mathbf{\hat{y}} + z_{2}c \, \mathbf{\hat{z}} & \left(8d\right) & \mbox{Te} \\ 
\mathbf{B}_{4} & = & -x_{2} \, \mathbf{a}_{1}-y_{2} \, \mathbf{a}_{2} + z_{2} \, \mathbf{a}_{3} & = & -x_{2}a \, \mathbf{\hat{x}}-y_{2}a \, \mathbf{\hat{y}} + z_{2}c \, \mathbf{\hat{z}} & \left(8d\right) & \mbox{Te} \\ 
\mathbf{B}_{5} & = & -y_{2} \, \mathbf{a}_{1} + x_{2} \, \mathbf{a}_{2} + z_{2} \, \mathbf{a}_{3} & = & -y_{2}a \, \mathbf{\hat{x}} + x_{2}a \, \mathbf{\hat{y}} + z_{2}c \, \mathbf{\hat{z}} & \left(8d\right) & \mbox{Te} \\ 
\mathbf{B}_{6} & = & y_{2} \, \mathbf{a}_{1}-x_{2} \, \mathbf{a}_{2} + z_{2} \, \mathbf{a}_{3} & = & y_{2}a \, \mathbf{\hat{x}}-x_{2}a \, \mathbf{\hat{y}} + z_{2}c \, \mathbf{\hat{z}} & \left(8d\right) & \mbox{Te} \\ 
\mathbf{B}_{7} & = & x_{2} \, \mathbf{a}_{1}-y_{2} \, \mathbf{a}_{2} + \left(\frac{1}{2} +z_{2}\right) \, \mathbf{a}_{3} & = & x_{2}a \, \mathbf{\hat{x}}-y_{2}a \, \mathbf{\hat{y}} + \left(\frac{1}{2} +z_{2}\right)c \, \mathbf{\hat{z}} & \left(8d\right) & \mbox{Te} \\ 
\mathbf{B}_{8} & = & -x_{2} \, \mathbf{a}_{1} + y_{2} \, \mathbf{a}_{2} + \left(\frac{1}{2} +z_{2}\right) \, \mathbf{a}_{3} & = & -x_{2}a \, \mathbf{\hat{x}} + y_{2}a \, \mathbf{\hat{y}} + \left(\frac{1}{2} +z_{2}\right)c \, \mathbf{\hat{z}} & \left(8d\right) & \mbox{Te} \\ 
\mathbf{B}_{9} & = & -y_{2} \, \mathbf{a}_{1}-x_{2} \, \mathbf{a}_{2} + \left(\frac{1}{2} +z_{2}\right) \, \mathbf{a}_{3} & = & -y_{2}a \, \mathbf{\hat{x}}-x_{2}a \, \mathbf{\hat{y}} + \left(\frac{1}{2} +z_{2}\right)c \, \mathbf{\hat{z}} & \left(8d\right) & \mbox{Te} \\ 
\mathbf{B}_{10} & = & y_{2} \, \mathbf{a}_{1} + x_{2} \, \mathbf{a}_{2} + \left(\frac{1}{2} +z_{2}\right) \, \mathbf{a}_{3} & = & y_{2}a \, \mathbf{\hat{x}} + x_{2}a \, \mathbf{\hat{y}} + \left(\frac{1}{2} +z_{2}\right)c \, \mathbf{\hat{z}} & \left(8d\right) & \mbox{Te} \\ 
\end{longtabu}
\renewcommand{\arraystretch}{1.0}
\noindent \hrulefill
\\
\textbf{References:}
\vspace*{-0.25cm}
\begin{flushleft}
  - \bibentry{Bohm_NbTe4_ZKristallogrCrysMat_1987}. \\
\end{flushleft}
\textbf{Found in:}
\vspace*{-0.25cm}
\begin{flushleft}
  - \bibentry{Villars_PearsonsCrystalData_2013}. \\
\end{flushleft}
\noindent \hrulefill
\\
\textbf{Geometry files:}
\\
\noindent  - CIF: pp. {\hyperref[AB4_tP10_103_a_d_cif]{\pageref{AB4_tP10_103_a_d_cif}}} \\
\noindent  - POSCAR: pp. {\hyperref[AB4_tP10_103_a_d_poscar]{\pageref{AB4_tP10_103_a_d_poscar}}} \\
\onecolumn
{\phantomsection\label{A5B5C4_tP28_104_ac_ac_c}}
\subsection*{\huge \textbf{{\normalfont Ba$_{5}$In$_{4}$Bi$_{5}$ Structure: A5B5C4\_tP28\_104\_ac\_ac\_c}}}
\noindent \hrulefill
\vspace*{0.25cm}
\begin{figure}[htp]
  \centering
  \vspace{-1em}
  {\includegraphics[width=1\textwidth]{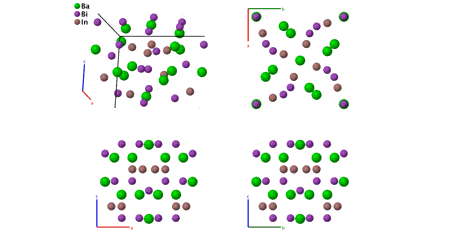}}
\end{figure}
\vspace*{-0.5cm}
\renewcommand{\arraystretch}{1.5}
\begin{equation*}
  \begin{array}{>{$\hspace{-0.15cm}}l<{$}>{$}p{0.5cm}<{$}>{$}p{18.5cm}<{$}}
    \mbox{\large \textbf{Prototype}} &\colon & \ce{Ba5In4Bi5} \\
    \mbox{\large \textbf{\AFLOW\ prototype label}} &\colon & \mbox{A5B5C4\_tP28\_104\_ac\_ac\_c} \\
    \mbox{\large \textbf{\textit{Strukturbericht} designation}} &\colon & \mbox{None} \\
    \mbox{\large \textbf{Pearson symbol}} &\colon & \mbox{tP28} \\
    \mbox{\large \textbf{Space group number}} &\colon & 104 \\
    \mbox{\large \textbf{Space group symbol}} &\colon & P4nc \\
    \mbox{\large \textbf{\AFLOW\ prototype command}} &\colon &  \texttt{aflow} \,  \, \texttt{-{}-proto=A5B5C4\_tP28\_104\_ac\_ac\_c } \, \newline \texttt{-{}-params=}{a,c/a,z_{1},z_{2},x_{3},y_{3},z_{3},x_{4},y_{4},z_{4},x_{5},y_{5},z_{5} }
  \end{array}
\end{equation*}
\renewcommand{\arraystretch}{1.0}

\noindent \parbox{1 \linewidth}{
\noindent \hrulefill
\\
\textbf{Simple Tetragonal primitive vectors:} \\
\vspace*{-0.25cm}
\begin{tabular}{cc}
  \begin{tabular}{c}
    \parbox{0.6 \linewidth}{
      \renewcommand{\arraystretch}{1.5}
      \begin{equation*}
        \centering
        \begin{array}{ccc}
              \mathbf{a}_1 & = & a \, \mathbf{\hat{x}} \\
    \mathbf{a}_2 & = & a \, \mathbf{\hat{y}} \\
    \mathbf{a}_3 & = & c \, \mathbf{\hat{z}} \\

        \end{array}
      \end{equation*}
    }
    \renewcommand{\arraystretch}{1.0}
  \end{tabular}
  \begin{tabular}{c}
    \includegraphics[width=0.3\linewidth]{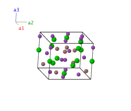} \\
  \end{tabular}
\end{tabular}

}
\vspace*{-0.25cm}

\noindent \hrulefill
\\
\textbf{Basis vectors:}
\vspace*{-0.25cm}
\renewcommand{\arraystretch}{1.5}
\begin{longtabu} to \textwidth{>{\centering $}X[-1,c,c]<{$}>{\centering $}X[-1,c,c]<{$}>{\centering $}X[-1,c,c]<{$}>{\centering $}X[-1,c,c]<{$}>{\centering $}X[-1,c,c]<{$}>{\centering $}X[-1,c,c]<{$}>{\centering $}X[-1,c,c]<{$}}
  & & \mbox{Lattice Coordinates} & & \mbox{Cartesian Coordinates} &\mbox{Wyckoff Position} & \mbox{Atom Type} \\  
  \mathbf{B}_{1} & = & z_{1} \, \mathbf{a}_{3} & = & z_{1}c \, \mathbf{\hat{z}} & \left(2a\right) & \mbox{Ba I} \\ 
\mathbf{B}_{2} & = & \frac{1}{2} \, \mathbf{a}_{1} + \frac{1}{2} \, \mathbf{a}_{2} + \left(\frac{1}{2} +z_{1}\right) \, \mathbf{a}_{3} & = & \frac{1}{2}a \, \mathbf{\hat{x}} + \frac{1}{2}a \, \mathbf{\hat{y}} + \left(\frac{1}{2} +z_{1}\right)c \, \mathbf{\hat{z}} & \left(2a\right) & \mbox{Ba I} \\ 
\mathbf{B}_{3} & = & z_{2} \, \mathbf{a}_{3} & = & z_{2}c \, \mathbf{\hat{z}} & \left(2a\right) & \mbox{Bi I} \\ 
\mathbf{B}_{4} & = & \frac{1}{2} \, \mathbf{a}_{1} + \frac{1}{2} \, \mathbf{a}_{2} + \left(\frac{1}{2} +z_{2}\right) \, \mathbf{a}_{3} & = & \frac{1}{2}a \, \mathbf{\hat{x}} + \frac{1}{2}a \, \mathbf{\hat{y}} + \left(\frac{1}{2} +z_{2}\right)c \, \mathbf{\hat{z}} & \left(2a\right) & \mbox{Bi I} \\ 
\mathbf{B}_{5} & = & x_{3} \, \mathbf{a}_{1} + y_{3} \, \mathbf{a}_{2} + z_{3} \, \mathbf{a}_{3} & = & x_{3}a \, \mathbf{\hat{x}} + y_{3}a \, \mathbf{\hat{y}} + z_{3}c \, \mathbf{\hat{z}} & \left(8c\right) & \mbox{Ba II} \\ 
\mathbf{B}_{6} & = & -x_{3} \, \mathbf{a}_{1}-y_{3} \, \mathbf{a}_{2} + z_{3} \, \mathbf{a}_{3} & = & -x_{3}a \, \mathbf{\hat{x}}-y_{3}a \, \mathbf{\hat{y}} + z_{3}c \, \mathbf{\hat{z}} & \left(8c\right) & \mbox{Ba II} \\ 
\mathbf{B}_{7} & = & -y_{3} \, \mathbf{a}_{1} + x_{3} \, \mathbf{a}_{2} + z_{3} \, \mathbf{a}_{3} & = & -y_{3}a \, \mathbf{\hat{x}} + x_{3}a \, \mathbf{\hat{y}} + z_{3}c \, \mathbf{\hat{z}} & \left(8c\right) & \mbox{Ba II} \\ 
\mathbf{B}_{8} & = & y_{3} \, \mathbf{a}_{1}-x_{3} \, \mathbf{a}_{2} + z_{3} \, \mathbf{a}_{3} & = & y_{3}a \, \mathbf{\hat{x}}-x_{3}a \, \mathbf{\hat{y}} + z_{3}c \, \mathbf{\hat{z}} & \left(8c\right) & \mbox{Ba II} \\ 
\mathbf{B}_{9} & = & \left(\frac{1}{2} +x_{3}\right) \, \mathbf{a}_{1} + \left(\frac{1}{2} - y_{3}\right) \, \mathbf{a}_{2} + \left(\frac{1}{2} +z_{3}\right) \, \mathbf{a}_{3} & = & \left(\frac{1}{2} +x_{3}\right)a \, \mathbf{\hat{x}} + \left(\frac{1}{2} - y_{3}\right)a \, \mathbf{\hat{y}} + \left(\frac{1}{2} +z_{3}\right)c \, \mathbf{\hat{z}} & \left(8c\right) & \mbox{Ba II} \\ 
\mathbf{B}_{10} & = & \left(\frac{1}{2} - x_{3}\right) \, \mathbf{a}_{1} + \left(\frac{1}{2} +y_{3}\right) \, \mathbf{a}_{2} + \left(\frac{1}{2} +z_{3}\right) \, \mathbf{a}_{3} & = & \left(\frac{1}{2} - x_{3}\right)a \, \mathbf{\hat{x}} + \left(\frac{1}{2} +y_{3}\right)a \, \mathbf{\hat{y}} + \left(\frac{1}{2} +z_{3}\right)c \, \mathbf{\hat{z}} & \left(8c\right) & \mbox{Ba II} \\ 
\mathbf{B}_{11} & = & \left(\frac{1}{2} - y_{3}\right) \, \mathbf{a}_{1} + \left(\frac{1}{2} - x_{3}\right) \, \mathbf{a}_{2} + \left(\frac{1}{2} +z_{3}\right) \, \mathbf{a}_{3} & = & \left(\frac{1}{2} - y_{3}\right)a \, \mathbf{\hat{x}} + \left(\frac{1}{2} - x_{3}\right)a \, \mathbf{\hat{y}} + \left(\frac{1}{2} +z_{3}\right)c \, \mathbf{\hat{z}} & \left(8c\right) & \mbox{Ba II} \\ 
\mathbf{B}_{12} & = & \left(\frac{1}{2} +y_{3}\right) \, \mathbf{a}_{1} + \left(\frac{1}{2} +x_{3}\right) \, \mathbf{a}_{2} + \left(\frac{1}{2} +z_{3}\right) \, \mathbf{a}_{3} & = & \left(\frac{1}{2} +y_{3}\right)a \, \mathbf{\hat{x}} + \left(\frac{1}{2} +x_{3}\right)a \, \mathbf{\hat{y}} + \left(\frac{1}{2} +z_{3}\right)c \, \mathbf{\hat{z}} & \left(8c\right) & \mbox{Ba II} \\ 
\mathbf{B}_{13} & = & x_{4} \, \mathbf{a}_{1} + y_{4} \, \mathbf{a}_{2} + z_{4} \, \mathbf{a}_{3} & = & x_{4}a \, \mathbf{\hat{x}} + y_{4}a \, \mathbf{\hat{y}} + z_{4}c \, \mathbf{\hat{z}} & \left(8c\right) & \mbox{Bi II} \\ 
\mathbf{B}_{14} & = & -x_{4} \, \mathbf{a}_{1}-y_{4} \, \mathbf{a}_{2} + z_{4} \, \mathbf{a}_{3} & = & -x_{4}a \, \mathbf{\hat{x}}-y_{4}a \, \mathbf{\hat{y}} + z_{4}c \, \mathbf{\hat{z}} & \left(8c\right) & \mbox{Bi II} \\ 
\mathbf{B}_{15} & = & -y_{4} \, \mathbf{a}_{1} + x_{4} \, \mathbf{a}_{2} + z_{4} \, \mathbf{a}_{3} & = & -y_{4}a \, \mathbf{\hat{x}} + x_{4}a \, \mathbf{\hat{y}} + z_{4}c \, \mathbf{\hat{z}} & \left(8c\right) & \mbox{Bi II} \\ 
\mathbf{B}_{16} & = & y_{4} \, \mathbf{a}_{1}-x_{4} \, \mathbf{a}_{2} + z_{4} \, \mathbf{a}_{3} & = & y_{4}a \, \mathbf{\hat{x}}-x_{4}a \, \mathbf{\hat{y}} + z_{4}c \, \mathbf{\hat{z}} & \left(8c\right) & \mbox{Bi II} \\ 
\mathbf{B}_{17} & = & \left(\frac{1}{2} +x_{4}\right) \, \mathbf{a}_{1} + \left(\frac{1}{2} - y_{4}\right) \, \mathbf{a}_{2} + \left(\frac{1}{2} +z_{4}\right) \, \mathbf{a}_{3} & = & \left(\frac{1}{2} +x_{4}\right)a \, \mathbf{\hat{x}} + \left(\frac{1}{2} - y_{4}\right)a \, \mathbf{\hat{y}} + \left(\frac{1}{2} +z_{4}\right)c \, \mathbf{\hat{z}} & \left(8c\right) & \mbox{Bi II} \\ 
\mathbf{B}_{18} & = & \left(\frac{1}{2} - x_{4}\right) \, \mathbf{a}_{1} + \left(\frac{1}{2} +y_{4}\right) \, \mathbf{a}_{2} + \left(\frac{1}{2} +z_{4}\right) \, \mathbf{a}_{3} & = & \left(\frac{1}{2} - x_{4}\right)a \, \mathbf{\hat{x}} + \left(\frac{1}{2} +y_{4}\right)a \, \mathbf{\hat{y}} + \left(\frac{1}{2} +z_{4}\right)c \, \mathbf{\hat{z}} & \left(8c\right) & \mbox{Bi II} \\ 
\mathbf{B}_{19} & = & \left(\frac{1}{2} - y_{4}\right) \, \mathbf{a}_{1} + \left(\frac{1}{2} - x_{4}\right) \, \mathbf{a}_{2} + \left(\frac{1}{2} +z_{4}\right) \, \mathbf{a}_{3} & = & \left(\frac{1}{2} - y_{4}\right)a \, \mathbf{\hat{x}} + \left(\frac{1}{2} - x_{4}\right)a \, \mathbf{\hat{y}} + \left(\frac{1}{2} +z_{4}\right)c \, \mathbf{\hat{z}} & \left(8c\right) & \mbox{Bi II} \\ 
\mathbf{B}_{20} & = & \left(\frac{1}{2} +y_{4}\right) \, \mathbf{a}_{1} + \left(\frac{1}{2} +x_{4}\right) \, \mathbf{a}_{2} + \left(\frac{1}{2} +z_{4}\right) \, \mathbf{a}_{3} & = & \left(\frac{1}{2} +y_{4}\right)a \, \mathbf{\hat{x}} + \left(\frac{1}{2} +x_{4}\right)a \, \mathbf{\hat{y}} + \left(\frac{1}{2} +z_{4}\right)c \, \mathbf{\hat{z}} & \left(8c\right) & \mbox{Bi II} \\ 
\mathbf{B}_{21} & = & x_{5} \, \mathbf{a}_{1} + y_{5} \, \mathbf{a}_{2} + z_{5} \, \mathbf{a}_{3} & = & x_{5}a \, \mathbf{\hat{x}} + y_{5}a \, \mathbf{\hat{y}} + z_{5}c \, \mathbf{\hat{z}} & \left(8c\right) & \mbox{In} \\ 
\mathbf{B}_{22} & = & -x_{5} \, \mathbf{a}_{1}-y_{5} \, \mathbf{a}_{2} + z_{5} \, \mathbf{a}_{3} & = & -x_{5}a \, \mathbf{\hat{x}}-y_{5}a \, \mathbf{\hat{y}} + z_{5}c \, \mathbf{\hat{z}} & \left(8c\right) & \mbox{In} \\ 
\mathbf{B}_{23} & = & -y_{5} \, \mathbf{a}_{1} + x_{5} \, \mathbf{a}_{2} + z_{5} \, \mathbf{a}_{3} & = & -y_{5}a \, \mathbf{\hat{x}} + x_{5}a \, \mathbf{\hat{y}} + z_{5}c \, \mathbf{\hat{z}} & \left(8c\right) & \mbox{In} \\ 
\mathbf{B}_{24} & = & y_{5} \, \mathbf{a}_{1}-x_{5} \, \mathbf{a}_{2} + z_{5} \, \mathbf{a}_{3} & = & y_{5}a \, \mathbf{\hat{x}}-x_{5}a \, \mathbf{\hat{y}} + z_{5}c \, \mathbf{\hat{z}} & \left(8c\right) & \mbox{In} \\ 
\mathbf{B}_{25} & = & \left(\frac{1}{2} +x_{5}\right) \, \mathbf{a}_{1} + \left(\frac{1}{2} - y_{5}\right) \, \mathbf{a}_{2} + \left(\frac{1}{2} +z_{5}\right) \, \mathbf{a}_{3} & = & \left(\frac{1}{2} +x_{5}\right)a \, \mathbf{\hat{x}} + \left(\frac{1}{2} - y_{5}\right)a \, \mathbf{\hat{y}} + \left(\frac{1}{2} +z_{5}\right)c \, \mathbf{\hat{z}} & \left(8c\right) & \mbox{In} \\ 
\mathbf{B}_{26} & = & \left(\frac{1}{2} - x_{5}\right) \, \mathbf{a}_{1} + \left(\frac{1}{2} +y_{5}\right) \, \mathbf{a}_{2} + \left(\frac{1}{2} +z_{5}\right) \, \mathbf{a}_{3} & = & \left(\frac{1}{2} - x_{5}\right)a \, \mathbf{\hat{x}} + \left(\frac{1}{2} +y_{5}\right)a \, \mathbf{\hat{y}} + \left(\frac{1}{2} +z_{5}\right)c \, \mathbf{\hat{z}} & \left(8c\right) & \mbox{In} \\ 
\mathbf{B}_{27} & = & \left(\frac{1}{2} - y_{5}\right) \, \mathbf{a}_{1} + \left(\frac{1}{2} - x_{5}\right) \, \mathbf{a}_{2} + \left(\frac{1}{2} +z_{5}\right) \, \mathbf{a}_{3} & = & \left(\frac{1}{2} - y_{5}\right)a \, \mathbf{\hat{x}} + \left(\frac{1}{2} - x_{5}\right)a \, \mathbf{\hat{y}} + \left(\frac{1}{2} +z_{5}\right)c \, \mathbf{\hat{z}} & \left(8c\right) & \mbox{In} \\ 
\mathbf{B}_{28} & = & \left(\frac{1}{2} +y_{5}\right) \, \mathbf{a}_{1} + \left(\frac{1}{2} +x_{5}\right) \, \mathbf{a}_{2} + \left(\frac{1}{2} +z_{5}\right) \, \mathbf{a}_{3} & = & \left(\frac{1}{2} +y_{5}\right)a \, \mathbf{\hat{x}} + \left(\frac{1}{2} +x_{5}\right)a \, \mathbf{\hat{y}} + \left(\frac{1}{2} +z_{5}\right)c \, \mathbf{\hat{z}} & \left(8c\right) & \mbox{In} \\ 
\end{longtabu}
\renewcommand{\arraystretch}{1.0}
\noindent \hrulefill
\\
\textbf{References:}
\vspace*{-0.25cm}
\begin{flushleft}
  - \bibentry{Ponou_Ba5Bi5In4_ChemAEuroJ_2004}. \\
\end{flushleft}
\textbf{Found in:}
\vspace*{-0.25cm}
\begin{flushleft}
  - \bibentry{Villars_PearsonsCrystalData_2013}. \\
\end{flushleft}
\noindent \hrulefill
\\
\textbf{Geometry files:}
\\
\noindent  - CIF: pp. {\hyperref[A5B5C4_tP28_104_ac_ac_c_cif]{\pageref{A5B5C4_tP28_104_ac_ac_c_cif}}} \\
\noindent  - POSCAR: pp. {\hyperref[A5B5C4_tP28_104_ac_ac_c_poscar]{\pageref{A5B5C4_tP28_104_ac_ac_c_poscar}}} \\
\onecolumn
{\phantomsection\label{AB6C4_tP22_104_a_2ac_c}}
\subsection*{\huge \textbf{{\normalfont Tl$_{4}$HgI$_{6}$ Structure: AB6C4\_tP22\_104\_a\_2ac\_c}}}
\noindent \hrulefill
\vspace*{0.25cm}
\begin{figure}[htp]
  \centering
  \vspace{-1em}
  {\includegraphics[width=1\textwidth]{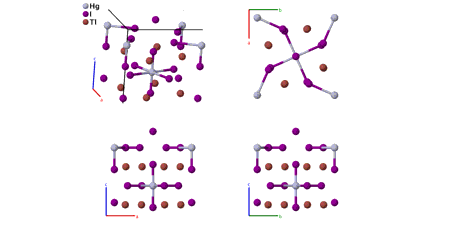}}
\end{figure}
\vspace*{-0.5cm}
\renewcommand{\arraystretch}{1.5}
\begin{equation*}
  \begin{array}{>{$\hspace{-0.15cm}}l<{$}>{$}p{0.5cm}<{$}>{$}p{18.5cm}<{$}}
    \mbox{\large \textbf{Prototype}} &\colon & \ce{Tl4HgI6} \\
    \mbox{\large \textbf{\AFLOW\ prototype label}} &\colon & \mbox{AB6C4\_tP22\_104\_a\_2ac\_c} \\
    \mbox{\large \textbf{\textit{Strukturbericht} designation}} &\colon & \mbox{None} \\
    \mbox{\large \textbf{Pearson symbol}} &\colon & \mbox{tP22} \\
    \mbox{\large \textbf{Space group number}} &\colon & 104 \\
    \mbox{\large \textbf{Space group symbol}} &\colon & P4nc \\
    \mbox{\large \textbf{\AFLOW\ prototype command}} &\colon &  \texttt{aflow} \,  \, \texttt{-{}-proto=AB6C4\_tP22\_104\_a\_2ac\_c } \, \newline \texttt{-{}-params=}{a,c/a,z_{1},z_{2},z_{3},x_{4},y_{4},z_{4},x_{5},y_{5},z_{5} }
  \end{array}
\end{equation*}
\renewcommand{\arraystretch}{1.0}

\vspace*{-0.25cm}
\noindent \hrulefill
\begin{itemize}
  \item{The second I site is reported with an occupancy 0.92.
}
\end{itemize}

\noindent \parbox{1 \linewidth}{
\noindent \hrulefill
\\
\textbf{Simple Tetragonal primitive vectors:} \\
\vspace*{-0.25cm}
\begin{tabular}{cc}
  \begin{tabular}{c}
    \parbox{0.6 \linewidth}{
      \renewcommand{\arraystretch}{1.5}
      \begin{equation*}
        \centering
        \begin{array}{ccc}
              \mathbf{a}_1 & = & a \, \mathbf{\hat{x}} \\
    \mathbf{a}_2 & = & a \, \mathbf{\hat{y}} \\
    \mathbf{a}_3 & = & c \, \mathbf{\hat{z}} \\

        \end{array}
      \end{equation*}
    }
    \renewcommand{\arraystretch}{1.0}
  \end{tabular}
  \begin{tabular}{c}
    \includegraphics[width=0.3\linewidth]{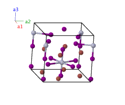} \\
  \end{tabular}
\end{tabular}

}
\vspace*{-0.25cm}

\noindent \hrulefill
\\
\textbf{Basis vectors:}
\vspace*{-0.25cm}
\renewcommand{\arraystretch}{1.5}
\begin{longtabu} to \textwidth{>{\centering $}X[-1,c,c]<{$}>{\centering $}X[-1,c,c]<{$}>{\centering $}X[-1,c,c]<{$}>{\centering $}X[-1,c,c]<{$}>{\centering $}X[-1,c,c]<{$}>{\centering $}X[-1,c,c]<{$}>{\centering $}X[-1,c,c]<{$}}
  & & \mbox{Lattice Coordinates} & & \mbox{Cartesian Coordinates} &\mbox{Wyckoff Position} & \mbox{Atom Type} \\  
  \mathbf{B}_{1} & = & z_{1} \, \mathbf{a}_{3} & = & z_{1}c \, \mathbf{\hat{z}} & \left(2a\right) & \mbox{Hg} \\ 
\mathbf{B}_{2} & = & \frac{1}{2} \, \mathbf{a}_{1} + \frac{1}{2} \, \mathbf{a}_{2} + \left(\frac{1}{2} +z_{1}\right) \, \mathbf{a}_{3} & = & \frac{1}{2}a \, \mathbf{\hat{x}} + \frac{1}{2}a \, \mathbf{\hat{y}} + \left(\frac{1}{2} +z_{1}\right)c \, \mathbf{\hat{z}} & \left(2a\right) & \mbox{Hg} \\ 
\mathbf{B}_{3} & = & z_{2} \, \mathbf{a}_{3} & = & z_{2}c \, \mathbf{\hat{z}} & \left(2a\right) & \mbox{I I} \\ 
\mathbf{B}_{4} & = & \frac{1}{2} \, \mathbf{a}_{1} + \frac{1}{2} \, \mathbf{a}_{2} + \left(\frac{1}{2} +z_{2}\right) \, \mathbf{a}_{3} & = & \frac{1}{2}a \, \mathbf{\hat{x}} + \frac{1}{2}a \, \mathbf{\hat{y}} + \left(\frac{1}{2} +z_{2}\right)c \, \mathbf{\hat{z}} & \left(2a\right) & \mbox{I I} \\ 
\mathbf{B}_{5} & = & z_{3} \, \mathbf{a}_{3} & = & z_{3}c \, \mathbf{\hat{z}} & \left(2a\right) & \mbox{I II} \\ 
\mathbf{B}_{6} & = & \frac{1}{2} \, \mathbf{a}_{1} + \frac{1}{2} \, \mathbf{a}_{2} + \left(\frac{1}{2} +z_{3}\right) \, \mathbf{a}_{3} & = & \frac{1}{2}a \, \mathbf{\hat{x}} + \frac{1}{2}a \, \mathbf{\hat{y}} + \left(\frac{1}{2} +z_{3}\right)c \, \mathbf{\hat{z}} & \left(2a\right) & \mbox{I II} \\ 
\mathbf{B}_{7} & = & x_{4} \, \mathbf{a}_{1} + y_{4} \, \mathbf{a}_{2} + z_{4} \, \mathbf{a}_{3} & = & x_{4}a \, \mathbf{\hat{x}} + y_{4}a \, \mathbf{\hat{y}} + z_{4}c \, \mathbf{\hat{z}} & \left(8c\right) & \mbox{I III} \\ 
\mathbf{B}_{8} & = & -x_{4} \, \mathbf{a}_{1}-y_{4} \, \mathbf{a}_{2} + z_{4} \, \mathbf{a}_{3} & = & -x_{4}a \, \mathbf{\hat{x}}-y_{4}a \, \mathbf{\hat{y}} + z_{4}c \, \mathbf{\hat{z}} & \left(8c\right) & \mbox{I III} \\ 
\mathbf{B}_{9} & = & -y_{4} \, \mathbf{a}_{1} + x_{4} \, \mathbf{a}_{2} + z_{4} \, \mathbf{a}_{3} & = & -y_{4}a \, \mathbf{\hat{x}} + x_{4}a \, \mathbf{\hat{y}} + z_{4}c \, \mathbf{\hat{z}} & \left(8c\right) & \mbox{I III} \\ 
\mathbf{B}_{10} & = & y_{4} \, \mathbf{a}_{1}-x_{4} \, \mathbf{a}_{2} + z_{4} \, \mathbf{a}_{3} & = & y_{4}a \, \mathbf{\hat{x}}-x_{4}a \, \mathbf{\hat{y}} + z_{4}c \, \mathbf{\hat{z}} & \left(8c\right) & \mbox{I III} \\ 
\mathbf{B}_{11} & = & \left(\frac{1}{2} +x_{4}\right) \, \mathbf{a}_{1} + \left(\frac{1}{2} - y_{4}\right) \, \mathbf{a}_{2} + \left(\frac{1}{2} +z_{4}\right) \, \mathbf{a}_{3} & = & \left(\frac{1}{2} +x_{4}\right)a \, \mathbf{\hat{x}} + \left(\frac{1}{2} - y_{4}\right)a \, \mathbf{\hat{y}} + \left(\frac{1}{2} +z_{4}\right)c \, \mathbf{\hat{z}} & \left(8c\right) & \mbox{I III} \\ 
\mathbf{B}_{12} & = & \left(\frac{1}{2} - x_{4}\right) \, \mathbf{a}_{1} + \left(\frac{1}{2} +y_{4}\right) \, \mathbf{a}_{2} + \left(\frac{1}{2} +z_{4}\right) \, \mathbf{a}_{3} & = & \left(\frac{1}{2} - x_{4}\right)a \, \mathbf{\hat{x}} + \left(\frac{1}{2} +y_{4}\right)a \, \mathbf{\hat{y}} + \left(\frac{1}{2} +z_{4}\right)c \, \mathbf{\hat{z}} & \left(8c\right) & \mbox{I III} \\ 
\mathbf{B}_{13} & = & \left(\frac{1}{2} - y_{4}\right) \, \mathbf{a}_{1} + \left(\frac{1}{2} - x_{4}\right) \, \mathbf{a}_{2} + \left(\frac{1}{2} +z_{4}\right) \, \mathbf{a}_{3} & = & \left(\frac{1}{2} - y_{4}\right)a \, \mathbf{\hat{x}} + \left(\frac{1}{2} - x_{4}\right)a \, \mathbf{\hat{y}} + \left(\frac{1}{2} +z_{4}\right)c \, \mathbf{\hat{z}} & \left(8c\right) & \mbox{I III} \\ 
\mathbf{B}_{14} & = & \left(\frac{1}{2} +y_{4}\right) \, \mathbf{a}_{1} + \left(\frac{1}{2} +x_{4}\right) \, \mathbf{a}_{2} + \left(\frac{1}{2} +z_{4}\right) \, \mathbf{a}_{3} & = & \left(\frac{1}{2} +y_{4}\right)a \, \mathbf{\hat{x}} + \left(\frac{1}{2} +x_{4}\right)a \, \mathbf{\hat{y}} + \left(\frac{1}{2} +z_{4}\right)c \, \mathbf{\hat{z}} & \left(8c\right) & \mbox{I III} \\ 
\mathbf{B}_{15} & = & x_{5} \, \mathbf{a}_{1} + y_{5} \, \mathbf{a}_{2} + z_{5} \, \mathbf{a}_{3} & = & x_{5}a \, \mathbf{\hat{x}} + y_{5}a \, \mathbf{\hat{y}} + z_{5}c \, \mathbf{\hat{z}} & \left(8c\right) & \mbox{Tl} \\ 
\mathbf{B}_{16} & = & -x_{5} \, \mathbf{a}_{1}-y_{5} \, \mathbf{a}_{2} + z_{5} \, \mathbf{a}_{3} & = & -x_{5}a \, \mathbf{\hat{x}}-y_{5}a \, \mathbf{\hat{y}} + z_{5}c \, \mathbf{\hat{z}} & \left(8c\right) & \mbox{Tl} \\ 
\mathbf{B}_{17} & = & -y_{5} \, \mathbf{a}_{1} + x_{5} \, \mathbf{a}_{2} + z_{5} \, \mathbf{a}_{3} & = & -y_{5}a \, \mathbf{\hat{x}} + x_{5}a \, \mathbf{\hat{y}} + z_{5}c \, \mathbf{\hat{z}} & \left(8c\right) & \mbox{Tl} \\ 
\mathbf{B}_{18} & = & y_{5} \, \mathbf{a}_{1}-x_{5} \, \mathbf{a}_{2} + z_{5} \, \mathbf{a}_{3} & = & y_{5}a \, \mathbf{\hat{x}}-x_{5}a \, \mathbf{\hat{y}} + z_{5}c \, \mathbf{\hat{z}} & \left(8c\right) & \mbox{Tl} \\ 
\mathbf{B}_{19} & = & \left(\frac{1}{2} +x_{5}\right) \, \mathbf{a}_{1} + \left(\frac{1}{2} - y_{5}\right) \, \mathbf{a}_{2} + \left(\frac{1}{2} +z_{5}\right) \, \mathbf{a}_{3} & = & \left(\frac{1}{2} +x_{5}\right)a \, \mathbf{\hat{x}} + \left(\frac{1}{2} - y_{5}\right)a \, \mathbf{\hat{y}} + \left(\frac{1}{2} +z_{5}\right)c \, \mathbf{\hat{z}} & \left(8c\right) & \mbox{Tl} \\ 
\mathbf{B}_{20} & = & \left(\frac{1}{2} - x_{5}\right) \, \mathbf{a}_{1} + \left(\frac{1}{2} +y_{5}\right) \, \mathbf{a}_{2} + \left(\frac{1}{2} +z_{5}\right) \, \mathbf{a}_{3} & = & \left(\frac{1}{2} - x_{5}\right)a \, \mathbf{\hat{x}} + \left(\frac{1}{2} +y_{5}\right)a \, \mathbf{\hat{y}} + \left(\frac{1}{2} +z_{5}\right)c \, \mathbf{\hat{z}} & \left(8c\right) & \mbox{Tl} \\ 
\mathbf{B}_{21} & = & \left(\frac{1}{2} - y_{5}\right) \, \mathbf{a}_{1} + \left(\frac{1}{2} - x_{5}\right) \, \mathbf{a}_{2} + \left(\frac{1}{2} +z_{5}\right) \, \mathbf{a}_{3} & = & \left(\frac{1}{2} - y_{5}\right)a \, \mathbf{\hat{x}} + \left(\frac{1}{2} - x_{5}\right)a \, \mathbf{\hat{y}} + \left(\frac{1}{2} +z_{5}\right)c \, \mathbf{\hat{z}} & \left(8c\right) & \mbox{Tl} \\ 
\mathbf{B}_{22} & = & \left(\frac{1}{2} +y_{5}\right) \, \mathbf{a}_{1} + \left(\frac{1}{2} +x_{5}\right) \, \mathbf{a}_{2} + \left(\frac{1}{2} +z_{5}\right) \, \mathbf{a}_{3} & = & \left(\frac{1}{2} +y_{5}\right)a \, \mathbf{\hat{x}} + \left(\frac{1}{2} +x_{5}\right)a \, \mathbf{\hat{y}} + \left(\frac{1}{2} +z_{5}\right)c \, \mathbf{\hat{z}} & \left(8c\right) & \mbox{Tl} \\ 
\end{longtabu}
\renewcommand{\arraystretch}{1.0}
\noindent \hrulefill
\\
\textbf{References:}
\vspace*{-0.25cm}
\begin{flushleft}
  - \bibentry{Badikov_Tl4HgI6_InorgMat_2004}. \\
\end{flushleft}
\textbf{Found in:}
\vspace*{-0.25cm}
\begin{flushleft}
  - \bibentry{Villars_PearsonsCrystalData_2013}. \\
\end{flushleft}
\noindent \hrulefill
\\
\textbf{Geometry files:}
\\
\noindent  - CIF: pp. {\hyperref[AB6C4_tP22_104_a_2ac_c_cif]{\pageref{AB6C4_tP22_104_a_2ac_c_cif}}} \\
\noindent  - POSCAR: pp. {\hyperref[AB6C4_tP22_104_a_2ac_c_poscar]{\pageref{AB6C4_tP22_104_a_2ac_c_poscar}}} \\
\onecolumn
{\phantomsection\label{A2BC2_tP20_105_f_ac_2e}}
\subsection*{\huge \textbf{{\normalfont BaGe$_{2}$As$_{2}$ Structure: A2BC2\_tP20\_105\_f\_ac\_2e}}}
\noindent \hrulefill
\vspace*{0.25cm}
\begin{figure}[htp]
  \centering
  \vspace{-1em}
  {\includegraphics[width=1\textwidth]{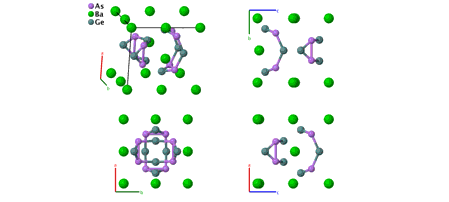}}
\end{figure}
\vspace*{-0.5cm}
\renewcommand{\arraystretch}{1.5}
\begin{equation*}
  \begin{array}{>{$\hspace{-0.15cm}}l<{$}>{$}p{0.5cm}<{$}>{$}p{18.5cm}<{$}}
    \mbox{\large \textbf{Prototype}} &\colon & \ce{BaGe2As2} \\
    \mbox{\large \textbf{\AFLOW\ prototype label}} &\colon & \mbox{A2BC2\_tP20\_105\_f\_ac\_2e} \\
    \mbox{\large \textbf{\textit{Strukturbericht} designation}} &\colon & \mbox{None} \\
    \mbox{\large \textbf{Pearson symbol}} &\colon & \mbox{tP20} \\
    \mbox{\large \textbf{Space group number}} &\colon & 105 \\
    \mbox{\large \textbf{Space group symbol}} &\colon & P4_{2}mc \\
    \mbox{\large \textbf{\AFLOW\ prototype command}} &\colon &  \texttt{aflow} \,  \, \texttt{-{}-proto=A2BC2\_tP20\_105\_f\_ac\_2e } \, \newline \texttt{-{}-params=}{a,c/a,z_{1},z_{2},x_{3},z_{3},x_{4},z_{4},x_{5},y_{5},z_{5} }
  \end{array}
\end{equation*}
\renewcommand{\arraystretch}{1.0}

\noindent \parbox{1 \linewidth}{
\noindent \hrulefill
\\
\textbf{Simple Tetragonal primitive vectors:} \\
\vspace*{-0.25cm}
\begin{tabular}{cc}
  \begin{tabular}{c}
    \parbox{0.6 \linewidth}{
      \renewcommand{\arraystretch}{1.5}
      \begin{equation*}
        \centering
        \begin{array}{ccc}
              \mathbf{a}_1 & = & a \, \mathbf{\hat{x}} \\
    \mathbf{a}_2 & = & a \, \mathbf{\hat{y}} \\
    \mathbf{a}_3 & = & c \, \mathbf{\hat{z}} \\

        \end{array}
      \end{equation*}
    }
    \renewcommand{\arraystretch}{1.0}
  \end{tabular}
  \begin{tabular}{c}
    \includegraphics[width=0.3\linewidth]{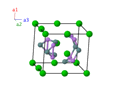} \\
  \end{tabular}
\end{tabular}

}
\vspace*{-0.25cm}

\noindent \hrulefill
\\
\textbf{Basis vectors:}
\vspace*{-0.25cm}
\renewcommand{\arraystretch}{1.5}
\begin{longtabu} to \textwidth{>{\centering $}X[-1,c,c]<{$}>{\centering $}X[-1,c,c]<{$}>{\centering $}X[-1,c,c]<{$}>{\centering $}X[-1,c,c]<{$}>{\centering $}X[-1,c,c]<{$}>{\centering $}X[-1,c,c]<{$}>{\centering $}X[-1,c,c]<{$}}
  & & \mbox{Lattice Coordinates} & & \mbox{Cartesian Coordinates} &\mbox{Wyckoff Position} & \mbox{Atom Type} \\  
  \mathbf{B}_{1} & = & z_{1} \, \mathbf{a}_{3} & = & z_{1}c \, \mathbf{\hat{z}} & \left(2a\right) & \mbox{Ba I} \\ 
\mathbf{B}_{2} & = & \left(\frac{1}{2} +z_{1}\right) \, \mathbf{a}_{3} & = & \left(\frac{1}{2} +z_{1}\right)c \, \mathbf{\hat{z}} & \left(2a\right) & \mbox{Ba I} \\ 
\mathbf{B}_{3} & = & \frac{1}{2} \, \mathbf{a}_{2} + z_{2} \, \mathbf{a}_{3} & = & \frac{1}{2}a \, \mathbf{\hat{y}} + z_{2}c \, \mathbf{\hat{z}} & \left(2c\right) & \mbox{Ba II} \\ 
\mathbf{B}_{4} & = & \frac{1}{2} \, \mathbf{a}_{1} + \left(\frac{1}{2} +z_{2}\right) \, \mathbf{a}_{3} & = & \frac{1}{2}a \, \mathbf{\hat{x}} + \left(\frac{1}{2} +z_{2}\right)c \, \mathbf{\hat{z}} & \left(2c\right) & \mbox{Ba II} \\ 
\mathbf{B}_{5} & = & x_{3} \, \mathbf{a}_{1} + \frac{1}{2} \, \mathbf{a}_{2} + z_{3} \, \mathbf{a}_{3} & = & x_{3}a \, \mathbf{\hat{x}} + \frac{1}{2}a \, \mathbf{\hat{y}} + z_{3}c \, \mathbf{\hat{z}} & \left(4e\right) & \mbox{Ge I} \\ 
\mathbf{B}_{6} & = & -x_{3} \, \mathbf{a}_{1} + \frac{1}{2} \, \mathbf{a}_{2} + z_{3} \, \mathbf{a}_{3} & = & -x_{3}a \, \mathbf{\hat{x}} + \frac{1}{2}a \, \mathbf{\hat{y}} + z_{3}c \, \mathbf{\hat{z}} & \left(4e\right) & \mbox{Ge I} \\ 
\mathbf{B}_{7} & = & \frac{1}{2} \, \mathbf{a}_{1} + x_{3} \, \mathbf{a}_{2} + \left(\frac{1}{2} +z_{3}\right) \, \mathbf{a}_{3} & = & \frac{1}{2}a \, \mathbf{\hat{x}} + x_{3}a \, \mathbf{\hat{y}} + \left(\frac{1}{2} +z_{3}\right)c \, \mathbf{\hat{z}} & \left(4e\right) & \mbox{Ge I} \\ 
\mathbf{B}_{8} & = & \frac{1}{2} \, \mathbf{a}_{1}-x_{3} \, \mathbf{a}_{2} + \left(\frac{1}{2} +z_{3}\right) \, \mathbf{a}_{3} & = & \frac{1}{2}a \, \mathbf{\hat{x}}-x_{3}a \, \mathbf{\hat{y}} + \left(\frac{1}{2} +z_{3}\right)c \, \mathbf{\hat{z}} & \left(4e\right) & \mbox{Ge I} \\ 
\mathbf{B}_{9} & = & x_{4} \, \mathbf{a}_{1} + \frac{1}{2} \, \mathbf{a}_{2} + z_{4} \, \mathbf{a}_{3} & = & x_{4}a \, \mathbf{\hat{x}} + \frac{1}{2}a \, \mathbf{\hat{y}} + z_{4}c \, \mathbf{\hat{z}} & \left(4e\right) & \mbox{Ge II} \\ 
\mathbf{B}_{10} & = & -x_{4} \, \mathbf{a}_{1} + \frac{1}{2} \, \mathbf{a}_{2} + z_{4} \, \mathbf{a}_{3} & = & -x_{4}a \, \mathbf{\hat{x}} + \frac{1}{2}a \, \mathbf{\hat{y}} + z_{4}c \, \mathbf{\hat{z}} & \left(4e\right) & \mbox{Ge II} \\ 
\mathbf{B}_{11} & = & \frac{1}{2} \, \mathbf{a}_{1} + x_{4} \, \mathbf{a}_{2} + \left(\frac{1}{2} +z_{4}\right) \, \mathbf{a}_{3} & = & \frac{1}{2}a \, \mathbf{\hat{x}} + x_{4}a \, \mathbf{\hat{y}} + \left(\frac{1}{2} +z_{4}\right)c \, \mathbf{\hat{z}} & \left(4e\right) & \mbox{Ge II} \\ 
\mathbf{B}_{12} & = & \frac{1}{2} \, \mathbf{a}_{1}-x_{4} \, \mathbf{a}_{2} + \left(\frac{1}{2} +z_{4}\right) \, \mathbf{a}_{3} & = & \frac{1}{2}a \, \mathbf{\hat{x}}-x_{4}a \, \mathbf{\hat{y}} + \left(\frac{1}{2} +z_{4}\right)c \, \mathbf{\hat{z}} & \left(4e\right) & \mbox{Ge II} \\ 
\mathbf{B}_{13} & = & x_{5} \, \mathbf{a}_{1} + y_{5} \, \mathbf{a}_{2} + z_{5} \, \mathbf{a}_{3} & = & x_{5}a \, \mathbf{\hat{x}} + y_{5}a \, \mathbf{\hat{y}} + z_{5}c \, \mathbf{\hat{z}} & \left(8f\right) & \mbox{As} \\ 
\mathbf{B}_{14} & = & -x_{5} \, \mathbf{a}_{1}-y_{5} \, \mathbf{a}_{2} + z_{5} \, \mathbf{a}_{3} & = & -x_{5}a \, \mathbf{\hat{x}}-y_{5}a \, \mathbf{\hat{y}} + z_{5}c \, \mathbf{\hat{z}} & \left(8f\right) & \mbox{As} \\ 
\mathbf{B}_{15} & = & -y_{5} \, \mathbf{a}_{1} + x_{5} \, \mathbf{a}_{2} + \left(\frac{1}{2} +z_{5}\right) \, \mathbf{a}_{3} & = & -y_{5}a \, \mathbf{\hat{x}} + x_{5}a \, \mathbf{\hat{y}} + \left(\frac{1}{2} +z_{5}\right)c \, \mathbf{\hat{z}} & \left(8f\right) & \mbox{As} \\ 
\mathbf{B}_{16} & = & y_{5} \, \mathbf{a}_{1}-x_{5} \, \mathbf{a}_{2} + \left(\frac{1}{2} +z_{5}\right) \, \mathbf{a}_{3} & = & y_{5}a \, \mathbf{\hat{x}}-x_{5}a \, \mathbf{\hat{y}} + \left(\frac{1}{2} +z_{5}\right)c \, \mathbf{\hat{z}} & \left(8f\right) & \mbox{As} \\ 
\mathbf{B}_{17} & = & x_{5} \, \mathbf{a}_{1}-y_{5} \, \mathbf{a}_{2} + z_{5} \, \mathbf{a}_{3} & = & x_{5}a \, \mathbf{\hat{x}}-y_{5}a \, \mathbf{\hat{y}} + z_{5}c \, \mathbf{\hat{z}} & \left(8f\right) & \mbox{As} \\ 
\mathbf{B}_{18} & = & -x_{5} \, \mathbf{a}_{1} + y_{5} \, \mathbf{a}_{2} + z_{5} \, \mathbf{a}_{3} & = & -x_{5}a \, \mathbf{\hat{x}} + y_{5}a \, \mathbf{\hat{y}} + z_{5}c \, \mathbf{\hat{z}} & \left(8f\right) & \mbox{As} \\ 
\mathbf{B}_{19} & = & -y_{5} \, \mathbf{a}_{1}-x_{5} \, \mathbf{a}_{2} + \left(\frac{1}{2} +z_{5}\right) \, \mathbf{a}_{3} & = & -y_{5}a \, \mathbf{\hat{x}}-x_{5}a \, \mathbf{\hat{y}} + \left(\frac{1}{2} +z_{5}\right)c \, \mathbf{\hat{z}} & \left(8f\right) & \mbox{As} \\ 
\mathbf{B}_{20} & = & y_{5} \, \mathbf{a}_{1} + x_{5} \, \mathbf{a}_{2} + \left(\frac{1}{2} +z_{5}\right) \, \mathbf{a}_{3} & = & y_{5}a \, \mathbf{\hat{x}} + x_{5}a \, \mathbf{\hat{y}} + \left(\frac{1}{2} +z_{5}\right)c \, \mathbf{\hat{z}} & \left(8f\right) & \mbox{As} \\ 
\end{longtabu}
\renewcommand{\arraystretch}{1.0}
\noindent \hrulefill
\\
\textbf{References:}
\vspace*{-0.25cm}
\begin{flushleft}
  - \bibentry{Eisenmann_BaGe2As2_ZNaturforschung_1981}. \\
\end{flushleft}
\textbf{Found in:}
\vspace*{-0.25cm}
\begin{flushleft}
  - \bibentry{Villars_PearsonsCrystalData_2013}. \\
\end{flushleft}
\noindent \hrulefill
\\
\textbf{Geometry files:}
\\
\noindent  - CIF: pp. {\hyperref[A2BC2_tP20_105_f_ac_2e_cif]{\pageref{A2BC2_tP20_105_f_ac_2e_cif}}} \\
\noindent  - POSCAR: pp. {\hyperref[A2BC2_tP20_105_f_ac_2e_poscar]{\pageref{A2BC2_tP20_105_f_ac_2e_poscar}}} \\
\onecolumn
{\phantomsection\label{A3BC3D_tP64_106_3c_c_3c_c}}
\subsection*{\huge \textbf{{\normalfont NaZn[OH]$_{3}$ Structure: A3BC3D\_tP64\_106\_3c\_c\_3c\_c}}}
\noindent \hrulefill
\vspace*{0.25cm}
\begin{figure}[htp]
  \centering
  \vspace{-1em}
  {\includegraphics[width=1\textwidth]{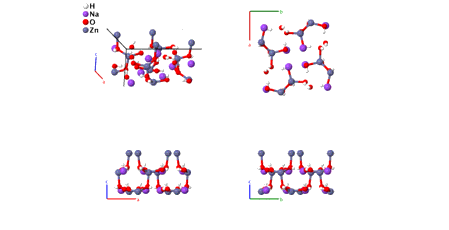}}
\end{figure}
\vspace*{-0.5cm}
\renewcommand{\arraystretch}{1.5}
\begin{equation*}
  \begin{array}{>{$\hspace{-0.15cm}}l<{$}>{$}p{0.5cm}<{$}>{$}p{18.5cm}<{$}}
    \mbox{\large \textbf{Prototype}} &\colon & \ce{NaZn[OH]3} \\
    \mbox{\large \textbf{\AFLOW\ prototype label}} &\colon & \mbox{A3BC3D\_tP64\_106\_3c\_c\_3c\_c} \\
    \mbox{\large \textbf{\textit{Strukturbericht} designation}} &\colon & \mbox{None} \\
    \mbox{\large \textbf{Pearson symbol}} &\colon & \mbox{tP64} \\
    \mbox{\large \textbf{Space group number}} &\colon & 106 \\
    \mbox{\large \textbf{Space group symbol}} &\colon & P4_{2}bc \\
    \mbox{\large \textbf{\AFLOW\ prototype command}} &\colon &  \texttt{aflow} \,  \, \texttt{-{}-proto=A3BC3D\_tP64\_106\_3c\_c\_3c\_c } \, \newline \texttt{-{}-params=}{a,c/a,x_{1},y_{1},z_{1},x_{2},y_{2},z_{2},x_{3},y_{3},z_{3},x_{4},y_{4},z_{4},x_{5},y_{5},z_{5},x_{6},y_{6},z_{6},x_{7},} \newline {y_{7},z_{7},x_{8},y_{8},z_{8} }
  \end{array}
\end{equation*}
\renewcommand{\arraystretch}{1.0}

\noindent \parbox{1 \linewidth}{
\noindent \hrulefill
\\
\textbf{Simple Tetragonal primitive vectors:} \\
\vspace*{-0.25cm}
\begin{tabular}{cc}
  \begin{tabular}{c}
    \parbox{0.6 \linewidth}{
      \renewcommand{\arraystretch}{1.5}
      \begin{equation*}
        \centering
        \begin{array}{ccc}
              \mathbf{a}_1 & = & a \, \mathbf{\hat{x}} \\
    \mathbf{a}_2 & = & a \, \mathbf{\hat{y}} \\
    \mathbf{a}_3 & = & c \, \mathbf{\hat{z}} \\

        \end{array}
      \end{equation*}
    }
    \renewcommand{\arraystretch}{1.0}
  \end{tabular}
  \begin{tabular}{c}
    \includegraphics[width=0.3\linewidth]{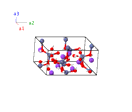} \\
  \end{tabular}
\end{tabular}

}
\vspace*{-0.25cm}

\noindent \hrulefill
\\
\textbf{Basis vectors:}
\vspace*{-0.25cm}
\renewcommand{\arraystretch}{1.5}
\begin{longtabu} to \textwidth{>{\centering $}X[-1,c,c]<{$}>{\centering $}X[-1,c,c]<{$}>{\centering $}X[-1,c,c]<{$}>{\centering $}X[-1,c,c]<{$}>{\centering $}X[-1,c,c]<{$}>{\centering $}X[-1,c,c]<{$}>{\centering $}X[-1,c,c]<{$}}
  & & \mbox{Lattice Coordinates} & & \mbox{Cartesian Coordinates} &\mbox{Wyckoff Position} & \mbox{Atom Type} \\  
  \mathbf{B}_{1} & = & x_{1} \, \mathbf{a}_{1} + y_{1} \, \mathbf{a}_{2} + z_{1} \, \mathbf{a}_{3} & = & x_{1}a \, \mathbf{\hat{x}} + y_{1}a \, \mathbf{\hat{y}} + z_{1}c \, \mathbf{\hat{z}} & \left(8c\right) & \mbox{H I} \\ 
\mathbf{B}_{2} & = & -x_{1} \, \mathbf{a}_{1}-y_{1} \, \mathbf{a}_{2} + z_{1} \, \mathbf{a}_{3} & = & -x_{1}a \, \mathbf{\hat{x}}-y_{1}a \, \mathbf{\hat{y}} + z_{1}c \, \mathbf{\hat{z}} & \left(8c\right) & \mbox{H I} \\ 
\mathbf{B}_{3} & = & -y_{1} \, \mathbf{a}_{1} + x_{1} \, \mathbf{a}_{2} + \left(\frac{1}{2} +z_{1}\right) \, \mathbf{a}_{3} & = & -y_{1}a \, \mathbf{\hat{x}} + x_{1}a \, \mathbf{\hat{y}} + \left(\frac{1}{2} +z_{1}\right)c \, \mathbf{\hat{z}} & \left(8c\right) & \mbox{H I} \\ 
\mathbf{B}_{4} & = & y_{1} \, \mathbf{a}_{1}-x_{1} \, \mathbf{a}_{2} + \left(\frac{1}{2} +z_{1}\right) \, \mathbf{a}_{3} & = & y_{1}a \, \mathbf{\hat{x}}-x_{1}a \, \mathbf{\hat{y}} + \left(\frac{1}{2} +z_{1}\right)c \, \mathbf{\hat{z}} & \left(8c\right) & \mbox{H I} \\ 
\mathbf{B}_{5} & = & \left(\frac{1}{2} +x_{1}\right) \, \mathbf{a}_{1} + \left(\frac{1}{2} - y_{1}\right) \, \mathbf{a}_{2} + z_{1} \, \mathbf{a}_{3} & = & \left(\frac{1}{2} +x_{1}\right)a \, \mathbf{\hat{x}} + \left(\frac{1}{2} - y_{1}\right)a \, \mathbf{\hat{y}} + z_{1}c \, \mathbf{\hat{z}} & \left(8c\right) & \mbox{H I} \\ 
\mathbf{B}_{6} & = & \left(\frac{1}{2} - x_{1}\right) \, \mathbf{a}_{1} + \left(\frac{1}{2} +y_{1}\right) \, \mathbf{a}_{2} + z_{1} \, \mathbf{a}_{3} & = & \left(\frac{1}{2} - x_{1}\right)a \, \mathbf{\hat{x}} + \left(\frac{1}{2} +y_{1}\right)a \, \mathbf{\hat{y}} + z_{1}c \, \mathbf{\hat{z}} & \left(8c\right) & \mbox{H I} \\ 
\mathbf{B}_{7} & = & \left(\frac{1}{2} - y_{1}\right) \, \mathbf{a}_{1} + \left(\frac{1}{2} - x_{1}\right) \, \mathbf{a}_{2} + \left(\frac{1}{2} +z_{1}\right) \, \mathbf{a}_{3} & = & \left(\frac{1}{2} - y_{1}\right)a \, \mathbf{\hat{x}} + \left(\frac{1}{2} - x_{1}\right)a \, \mathbf{\hat{y}} + \left(\frac{1}{2} +z_{1}\right)c \, \mathbf{\hat{z}} & \left(8c\right) & \mbox{H I} \\ 
\mathbf{B}_{8} & = & \left(\frac{1}{2} +y_{1}\right) \, \mathbf{a}_{1} + \left(\frac{1}{2} +x_{1}\right) \, \mathbf{a}_{2} + \left(\frac{1}{2} +z_{1}\right) \, \mathbf{a}_{3} & = & \left(\frac{1}{2} +y_{1}\right)a \, \mathbf{\hat{x}} + \left(\frac{1}{2} +x_{1}\right)a \, \mathbf{\hat{y}} + \left(\frac{1}{2} +z_{1}\right)c \, \mathbf{\hat{z}} & \left(8c\right) & \mbox{H I} \\ 
\mathbf{B}_{9} & = & x_{2} \, \mathbf{a}_{1} + y_{2} \, \mathbf{a}_{2} + z_{2} \, \mathbf{a}_{3} & = & x_{2}a \, \mathbf{\hat{x}} + y_{2}a \, \mathbf{\hat{y}} + z_{2}c \, \mathbf{\hat{z}} & \left(8c\right) & \mbox{H II} \\ 
\mathbf{B}_{10} & = & -x_{2} \, \mathbf{a}_{1}-y_{2} \, \mathbf{a}_{2} + z_{2} \, \mathbf{a}_{3} & = & -x_{2}a \, \mathbf{\hat{x}}-y_{2}a \, \mathbf{\hat{y}} + z_{2}c \, \mathbf{\hat{z}} & \left(8c\right) & \mbox{H II} \\ 
\mathbf{B}_{11} & = & -y_{2} \, \mathbf{a}_{1} + x_{2} \, \mathbf{a}_{2} + \left(\frac{1}{2} +z_{2}\right) \, \mathbf{a}_{3} & = & -y_{2}a \, \mathbf{\hat{x}} + x_{2}a \, \mathbf{\hat{y}} + \left(\frac{1}{2} +z_{2}\right)c \, \mathbf{\hat{z}} & \left(8c\right) & \mbox{H II} \\ 
\mathbf{B}_{12} & = & y_{2} \, \mathbf{a}_{1}-x_{2} \, \mathbf{a}_{2} + \left(\frac{1}{2} +z_{2}\right) \, \mathbf{a}_{3} & = & y_{2}a \, \mathbf{\hat{x}}-x_{2}a \, \mathbf{\hat{y}} + \left(\frac{1}{2} +z_{2}\right)c \, \mathbf{\hat{z}} & \left(8c\right) & \mbox{H II} \\ 
\mathbf{B}_{13} & = & \left(\frac{1}{2} +x_{2}\right) \, \mathbf{a}_{1} + \left(\frac{1}{2} - y_{2}\right) \, \mathbf{a}_{2} + z_{2} \, \mathbf{a}_{3} & = & \left(\frac{1}{2} +x_{2}\right)a \, \mathbf{\hat{x}} + \left(\frac{1}{2} - y_{2}\right)a \, \mathbf{\hat{y}} + z_{2}c \, \mathbf{\hat{z}} & \left(8c\right) & \mbox{H II} \\ 
\mathbf{B}_{14} & = & \left(\frac{1}{2} - x_{2}\right) \, \mathbf{a}_{1} + \left(\frac{1}{2} +y_{2}\right) \, \mathbf{a}_{2} + z_{2} \, \mathbf{a}_{3} & = & \left(\frac{1}{2} - x_{2}\right)a \, \mathbf{\hat{x}} + \left(\frac{1}{2} +y_{2}\right)a \, \mathbf{\hat{y}} + z_{2}c \, \mathbf{\hat{z}} & \left(8c\right) & \mbox{H II} \\ 
\mathbf{B}_{15} & = & \left(\frac{1}{2} - y_{2}\right) \, \mathbf{a}_{1} + \left(\frac{1}{2} - x_{2}\right) \, \mathbf{a}_{2} + \left(\frac{1}{2} +z_{2}\right) \, \mathbf{a}_{3} & = & \left(\frac{1}{2} - y_{2}\right)a \, \mathbf{\hat{x}} + \left(\frac{1}{2} - x_{2}\right)a \, \mathbf{\hat{y}} + \left(\frac{1}{2} +z_{2}\right)c \, \mathbf{\hat{z}} & \left(8c\right) & \mbox{H II} \\ 
\mathbf{B}_{16} & = & \left(\frac{1}{2} +y_{2}\right) \, \mathbf{a}_{1} + \left(\frac{1}{2} +x_{2}\right) \, \mathbf{a}_{2} + \left(\frac{1}{2} +z_{2}\right) \, \mathbf{a}_{3} & = & \left(\frac{1}{2} +y_{2}\right)a \, \mathbf{\hat{x}} + \left(\frac{1}{2} +x_{2}\right)a \, \mathbf{\hat{y}} + \left(\frac{1}{2} +z_{2}\right)c \, \mathbf{\hat{z}} & \left(8c\right) & \mbox{H II} \\ 
\mathbf{B}_{17} & = & x_{3} \, \mathbf{a}_{1} + y_{3} \, \mathbf{a}_{2} + z_{3} \, \mathbf{a}_{3} & = & x_{3}a \, \mathbf{\hat{x}} + y_{3}a \, \mathbf{\hat{y}} + z_{3}c \, \mathbf{\hat{z}} & \left(8c\right) & \mbox{H III} \\ 
\mathbf{B}_{18} & = & -x_{3} \, \mathbf{a}_{1}-y_{3} \, \mathbf{a}_{2} + z_{3} \, \mathbf{a}_{3} & = & -x_{3}a \, \mathbf{\hat{x}}-y_{3}a \, \mathbf{\hat{y}} + z_{3}c \, \mathbf{\hat{z}} & \left(8c\right) & \mbox{H III} \\ 
\mathbf{B}_{19} & = & -y_{3} \, \mathbf{a}_{1} + x_{3} \, \mathbf{a}_{2} + \left(\frac{1}{2} +z_{3}\right) \, \mathbf{a}_{3} & = & -y_{3}a \, \mathbf{\hat{x}} + x_{3}a \, \mathbf{\hat{y}} + \left(\frac{1}{2} +z_{3}\right)c \, \mathbf{\hat{z}} & \left(8c\right) & \mbox{H III} \\ 
\mathbf{B}_{20} & = & y_{3} \, \mathbf{a}_{1}-x_{3} \, \mathbf{a}_{2} + \left(\frac{1}{2} +z_{3}\right) \, \mathbf{a}_{3} & = & y_{3}a \, \mathbf{\hat{x}}-x_{3}a \, \mathbf{\hat{y}} + \left(\frac{1}{2} +z_{3}\right)c \, \mathbf{\hat{z}} & \left(8c\right) & \mbox{H III} \\ 
\mathbf{B}_{21} & = & \left(\frac{1}{2} +x_{3}\right) \, \mathbf{a}_{1} + \left(\frac{1}{2} - y_{3}\right) \, \mathbf{a}_{2} + z_{3} \, \mathbf{a}_{3} & = & \left(\frac{1}{2} +x_{3}\right)a \, \mathbf{\hat{x}} + \left(\frac{1}{2} - y_{3}\right)a \, \mathbf{\hat{y}} + z_{3}c \, \mathbf{\hat{z}} & \left(8c\right) & \mbox{H III} \\ 
\mathbf{B}_{22} & = & \left(\frac{1}{2} - x_{3}\right) \, \mathbf{a}_{1} + \left(\frac{1}{2} +y_{3}\right) \, \mathbf{a}_{2} + z_{3} \, \mathbf{a}_{3} & = & \left(\frac{1}{2} - x_{3}\right)a \, \mathbf{\hat{x}} + \left(\frac{1}{2} +y_{3}\right)a \, \mathbf{\hat{y}} + z_{3}c \, \mathbf{\hat{z}} & \left(8c\right) & \mbox{H III} \\ 
\mathbf{B}_{23} & = & \left(\frac{1}{2} - y_{3}\right) \, \mathbf{a}_{1} + \left(\frac{1}{2} - x_{3}\right) \, \mathbf{a}_{2} + \left(\frac{1}{2} +z_{3}\right) \, \mathbf{a}_{3} & = & \left(\frac{1}{2} - y_{3}\right)a \, \mathbf{\hat{x}} + \left(\frac{1}{2} - x_{3}\right)a \, \mathbf{\hat{y}} + \left(\frac{1}{2} +z_{3}\right)c \, \mathbf{\hat{z}} & \left(8c\right) & \mbox{H III} \\ 
\mathbf{B}_{24} & = & \left(\frac{1}{2} +y_{3}\right) \, \mathbf{a}_{1} + \left(\frac{1}{2} +x_{3}\right) \, \mathbf{a}_{2} + \left(\frac{1}{2} +z_{3}\right) \, \mathbf{a}_{3} & = & \left(\frac{1}{2} +y_{3}\right)a \, \mathbf{\hat{x}} + \left(\frac{1}{2} +x_{3}\right)a \, \mathbf{\hat{y}} + \left(\frac{1}{2} +z_{3}\right)c \, \mathbf{\hat{z}} & \left(8c\right) & \mbox{H III} \\ 
\mathbf{B}_{25} & = & x_{4} \, \mathbf{a}_{1} + y_{4} \, \mathbf{a}_{2} + z_{4} \, \mathbf{a}_{3} & = & x_{4}a \, \mathbf{\hat{x}} + y_{4}a \, \mathbf{\hat{y}} + z_{4}c \, \mathbf{\hat{z}} & \left(8c\right) & \mbox{Na} \\ 
\mathbf{B}_{26} & = & -x_{4} \, \mathbf{a}_{1}-y_{4} \, \mathbf{a}_{2} + z_{4} \, \mathbf{a}_{3} & = & -x_{4}a \, \mathbf{\hat{x}}-y_{4}a \, \mathbf{\hat{y}} + z_{4}c \, \mathbf{\hat{z}} & \left(8c\right) & \mbox{Na} \\ 
\mathbf{B}_{27} & = & -y_{4} \, \mathbf{a}_{1} + x_{4} \, \mathbf{a}_{2} + \left(\frac{1}{2} +z_{4}\right) \, \mathbf{a}_{3} & = & -y_{4}a \, \mathbf{\hat{x}} + x_{4}a \, \mathbf{\hat{y}} + \left(\frac{1}{2} +z_{4}\right)c \, \mathbf{\hat{z}} & \left(8c\right) & \mbox{Na} \\ 
\mathbf{B}_{28} & = & y_{4} \, \mathbf{a}_{1}-x_{4} \, \mathbf{a}_{2} + \left(\frac{1}{2} +z_{4}\right) \, \mathbf{a}_{3} & = & y_{4}a \, \mathbf{\hat{x}}-x_{4}a \, \mathbf{\hat{y}} + \left(\frac{1}{2} +z_{4}\right)c \, \mathbf{\hat{z}} & \left(8c\right) & \mbox{Na} \\ 
\mathbf{B}_{29} & = & \left(\frac{1}{2} +x_{4}\right) \, \mathbf{a}_{1} + \left(\frac{1}{2} - y_{4}\right) \, \mathbf{a}_{2} + z_{4} \, \mathbf{a}_{3} & = & \left(\frac{1}{2} +x_{4}\right)a \, \mathbf{\hat{x}} + \left(\frac{1}{2} - y_{4}\right)a \, \mathbf{\hat{y}} + z_{4}c \, \mathbf{\hat{z}} & \left(8c\right) & \mbox{Na} \\ 
\mathbf{B}_{30} & = & \left(\frac{1}{2} - x_{4}\right) \, \mathbf{a}_{1} + \left(\frac{1}{2} +y_{4}\right) \, \mathbf{a}_{2} + z_{4} \, \mathbf{a}_{3} & = & \left(\frac{1}{2} - x_{4}\right)a \, \mathbf{\hat{x}} + \left(\frac{1}{2} +y_{4}\right)a \, \mathbf{\hat{y}} + z_{4}c \, \mathbf{\hat{z}} & \left(8c\right) & \mbox{Na} \\ 
\mathbf{B}_{31} & = & \left(\frac{1}{2} - y_{4}\right) \, \mathbf{a}_{1} + \left(\frac{1}{2} - x_{4}\right) \, \mathbf{a}_{2} + \left(\frac{1}{2} +z_{4}\right) \, \mathbf{a}_{3} & = & \left(\frac{1}{2} - y_{4}\right)a \, \mathbf{\hat{x}} + \left(\frac{1}{2} - x_{4}\right)a \, \mathbf{\hat{y}} + \left(\frac{1}{2} +z_{4}\right)c \, \mathbf{\hat{z}} & \left(8c\right) & \mbox{Na} \\ 
\mathbf{B}_{32} & = & \left(\frac{1}{2} +y_{4}\right) \, \mathbf{a}_{1} + \left(\frac{1}{2} +x_{4}\right) \, \mathbf{a}_{2} + \left(\frac{1}{2} +z_{4}\right) \, \mathbf{a}_{3} & = & \left(\frac{1}{2} +y_{4}\right)a \, \mathbf{\hat{x}} + \left(\frac{1}{2} +x_{4}\right)a \, \mathbf{\hat{y}} + \left(\frac{1}{2} +z_{4}\right)c \, \mathbf{\hat{z}} & \left(8c\right) & \mbox{Na} \\ 
\mathbf{B}_{33} & = & x_{5} \, \mathbf{a}_{1} + y_{5} \, \mathbf{a}_{2} + z_{5} \, \mathbf{a}_{3} & = & x_{5}a \, \mathbf{\hat{x}} + y_{5}a \, \mathbf{\hat{y}} + z_{5}c \, \mathbf{\hat{z}} & \left(8c\right) & \mbox{O I} \\ 
\mathbf{B}_{34} & = & -x_{5} \, \mathbf{a}_{1}-y_{5} \, \mathbf{a}_{2} + z_{5} \, \mathbf{a}_{3} & = & -x_{5}a \, \mathbf{\hat{x}}-y_{5}a \, \mathbf{\hat{y}} + z_{5}c \, \mathbf{\hat{z}} & \left(8c\right) & \mbox{O I} \\ 
\mathbf{B}_{35} & = & -y_{5} \, \mathbf{a}_{1} + x_{5} \, \mathbf{a}_{2} + \left(\frac{1}{2} +z_{5}\right) \, \mathbf{a}_{3} & = & -y_{5}a \, \mathbf{\hat{x}} + x_{5}a \, \mathbf{\hat{y}} + \left(\frac{1}{2} +z_{5}\right)c \, \mathbf{\hat{z}} & \left(8c\right) & \mbox{O I} \\ 
\mathbf{B}_{36} & = & y_{5} \, \mathbf{a}_{1}-x_{5} \, \mathbf{a}_{2} + \left(\frac{1}{2} +z_{5}\right) \, \mathbf{a}_{3} & = & y_{5}a \, \mathbf{\hat{x}}-x_{5}a \, \mathbf{\hat{y}} + \left(\frac{1}{2} +z_{5}\right)c \, \mathbf{\hat{z}} & \left(8c\right) & \mbox{O I} \\ 
\mathbf{B}_{37} & = & \left(\frac{1}{2} +x_{5}\right) \, \mathbf{a}_{1} + \left(\frac{1}{2} - y_{5}\right) \, \mathbf{a}_{2} + z_{5} \, \mathbf{a}_{3} & = & \left(\frac{1}{2} +x_{5}\right)a \, \mathbf{\hat{x}} + \left(\frac{1}{2} - y_{5}\right)a \, \mathbf{\hat{y}} + z_{5}c \, \mathbf{\hat{z}} & \left(8c\right) & \mbox{O I} \\ 
\mathbf{B}_{38} & = & \left(\frac{1}{2} - x_{5}\right) \, \mathbf{a}_{1} + \left(\frac{1}{2} +y_{5}\right) \, \mathbf{a}_{2} + z_{5} \, \mathbf{a}_{3} & = & \left(\frac{1}{2} - x_{5}\right)a \, \mathbf{\hat{x}} + \left(\frac{1}{2} +y_{5}\right)a \, \mathbf{\hat{y}} + z_{5}c \, \mathbf{\hat{z}} & \left(8c\right) & \mbox{O I} \\ 
\mathbf{B}_{39} & = & \left(\frac{1}{2} - y_{5}\right) \, \mathbf{a}_{1} + \left(\frac{1}{2} - x_{5}\right) \, \mathbf{a}_{2} + \left(\frac{1}{2} +z_{5}\right) \, \mathbf{a}_{3} & = & \left(\frac{1}{2} - y_{5}\right)a \, \mathbf{\hat{x}} + \left(\frac{1}{2} - x_{5}\right)a \, \mathbf{\hat{y}} + \left(\frac{1}{2} +z_{5}\right)c \, \mathbf{\hat{z}} & \left(8c\right) & \mbox{O I} \\ 
\mathbf{B}_{40} & = & \left(\frac{1}{2} +y_{5}\right) \, \mathbf{a}_{1} + \left(\frac{1}{2} +x_{5}\right) \, \mathbf{a}_{2} + \left(\frac{1}{2} +z_{5}\right) \, \mathbf{a}_{3} & = & \left(\frac{1}{2} +y_{5}\right)a \, \mathbf{\hat{x}} + \left(\frac{1}{2} +x_{5}\right)a \, \mathbf{\hat{y}} + \left(\frac{1}{2} +z_{5}\right)c \, \mathbf{\hat{z}} & \left(8c\right) & \mbox{O I} \\ 
\mathbf{B}_{41} & = & x_{6} \, \mathbf{a}_{1} + y_{6} \, \mathbf{a}_{2} + z_{6} \, \mathbf{a}_{3} & = & x_{6}a \, \mathbf{\hat{x}} + y_{6}a \, \mathbf{\hat{y}} + z_{6}c \, \mathbf{\hat{z}} & \left(8c\right) & \mbox{O II} \\ 
\mathbf{B}_{42} & = & -x_{6} \, \mathbf{a}_{1}-y_{6} \, \mathbf{a}_{2} + z_{6} \, \mathbf{a}_{3} & = & -x_{6}a \, \mathbf{\hat{x}}-y_{6}a \, \mathbf{\hat{y}} + z_{6}c \, \mathbf{\hat{z}} & \left(8c\right) & \mbox{O II} \\ 
\mathbf{B}_{43} & = & -y_{6} \, \mathbf{a}_{1} + x_{6} \, \mathbf{a}_{2} + \left(\frac{1}{2} +z_{6}\right) \, \mathbf{a}_{3} & = & -y_{6}a \, \mathbf{\hat{x}} + x_{6}a \, \mathbf{\hat{y}} + \left(\frac{1}{2} +z_{6}\right)c \, \mathbf{\hat{z}} & \left(8c\right) & \mbox{O II} \\ 
\mathbf{B}_{44} & = & y_{6} \, \mathbf{a}_{1}-x_{6} \, \mathbf{a}_{2} + \left(\frac{1}{2} +z_{6}\right) \, \mathbf{a}_{3} & = & y_{6}a \, \mathbf{\hat{x}}-x_{6}a \, \mathbf{\hat{y}} + \left(\frac{1}{2} +z_{6}\right)c \, \mathbf{\hat{z}} & \left(8c\right) & \mbox{O II} \\ 
\mathbf{B}_{45} & = & \left(\frac{1}{2} +x_{6}\right) \, \mathbf{a}_{1} + \left(\frac{1}{2} - y_{6}\right) \, \mathbf{a}_{2} + z_{6} \, \mathbf{a}_{3} & = & \left(\frac{1}{2} +x_{6}\right)a \, \mathbf{\hat{x}} + \left(\frac{1}{2} - y_{6}\right)a \, \mathbf{\hat{y}} + z_{6}c \, \mathbf{\hat{z}} & \left(8c\right) & \mbox{O II} \\ 
\mathbf{B}_{46} & = & \left(\frac{1}{2} - x_{6}\right) \, \mathbf{a}_{1} + \left(\frac{1}{2} +y_{6}\right) \, \mathbf{a}_{2} + z_{6} \, \mathbf{a}_{3} & = & \left(\frac{1}{2} - x_{6}\right)a \, \mathbf{\hat{x}} + \left(\frac{1}{2} +y_{6}\right)a \, \mathbf{\hat{y}} + z_{6}c \, \mathbf{\hat{z}} & \left(8c\right) & \mbox{O II} \\ 
\mathbf{B}_{47} & = & \left(\frac{1}{2} - y_{6}\right) \, \mathbf{a}_{1} + \left(\frac{1}{2} - x_{6}\right) \, \mathbf{a}_{2} + \left(\frac{1}{2} +z_{6}\right) \, \mathbf{a}_{3} & = & \left(\frac{1}{2} - y_{6}\right)a \, \mathbf{\hat{x}} + \left(\frac{1}{2} - x_{6}\right)a \, \mathbf{\hat{y}} + \left(\frac{1}{2} +z_{6}\right)c \, \mathbf{\hat{z}} & \left(8c\right) & \mbox{O II} \\ 
\mathbf{B}_{48} & = & \left(\frac{1}{2} +y_{6}\right) \, \mathbf{a}_{1} + \left(\frac{1}{2} +x_{6}\right) \, \mathbf{a}_{2} + \left(\frac{1}{2} +z_{6}\right) \, \mathbf{a}_{3} & = & \left(\frac{1}{2} +y_{6}\right)a \, \mathbf{\hat{x}} + \left(\frac{1}{2} +x_{6}\right)a \, \mathbf{\hat{y}} + \left(\frac{1}{2} +z_{6}\right)c \, \mathbf{\hat{z}} & \left(8c\right) & \mbox{O II} \\ 
\mathbf{B}_{49} & = & x_{7} \, \mathbf{a}_{1} + y_{7} \, \mathbf{a}_{2} + z_{7} \, \mathbf{a}_{3} & = & x_{7}a \, \mathbf{\hat{x}} + y_{7}a \, \mathbf{\hat{y}} + z_{7}c \, \mathbf{\hat{z}} & \left(8c\right) & \mbox{O III} \\ 
\mathbf{B}_{50} & = & -x_{7} \, \mathbf{a}_{1}-y_{7} \, \mathbf{a}_{2} + z_{7} \, \mathbf{a}_{3} & = & -x_{7}a \, \mathbf{\hat{x}}-y_{7}a \, \mathbf{\hat{y}} + z_{7}c \, \mathbf{\hat{z}} & \left(8c\right) & \mbox{O III} \\ 
\mathbf{B}_{51} & = & -y_{7} \, \mathbf{a}_{1} + x_{7} \, \mathbf{a}_{2} + \left(\frac{1}{2} +z_{7}\right) \, \mathbf{a}_{3} & = & -y_{7}a \, \mathbf{\hat{x}} + x_{7}a \, \mathbf{\hat{y}} + \left(\frac{1}{2} +z_{7}\right)c \, \mathbf{\hat{z}} & \left(8c\right) & \mbox{O III} \\ 
\mathbf{B}_{52} & = & y_{7} \, \mathbf{a}_{1}-x_{7} \, \mathbf{a}_{2} + \left(\frac{1}{2} +z_{7}\right) \, \mathbf{a}_{3} & = & y_{7}a \, \mathbf{\hat{x}}-x_{7}a \, \mathbf{\hat{y}} + \left(\frac{1}{2} +z_{7}\right)c \, \mathbf{\hat{z}} & \left(8c\right) & \mbox{O III} \\ 
\mathbf{B}_{53} & = & \left(\frac{1}{2} +x_{7}\right) \, \mathbf{a}_{1} + \left(\frac{1}{2} - y_{7}\right) \, \mathbf{a}_{2} + z_{7} \, \mathbf{a}_{3} & = & \left(\frac{1}{2} +x_{7}\right)a \, \mathbf{\hat{x}} + \left(\frac{1}{2} - y_{7}\right)a \, \mathbf{\hat{y}} + z_{7}c \, \mathbf{\hat{z}} & \left(8c\right) & \mbox{O III} \\ 
\mathbf{B}_{54} & = & \left(\frac{1}{2} - x_{7}\right) \, \mathbf{a}_{1} + \left(\frac{1}{2} +y_{7}\right) \, \mathbf{a}_{2} + z_{7} \, \mathbf{a}_{3} & = & \left(\frac{1}{2} - x_{7}\right)a \, \mathbf{\hat{x}} + \left(\frac{1}{2} +y_{7}\right)a \, \mathbf{\hat{y}} + z_{7}c \, \mathbf{\hat{z}} & \left(8c\right) & \mbox{O III} \\ 
\mathbf{B}_{55} & = & \left(\frac{1}{2} - y_{7}\right) \, \mathbf{a}_{1} + \left(\frac{1}{2} - x_{7}\right) \, \mathbf{a}_{2} + \left(\frac{1}{2} +z_{7}\right) \, \mathbf{a}_{3} & = & \left(\frac{1}{2} - y_{7}\right)a \, \mathbf{\hat{x}} + \left(\frac{1}{2} - x_{7}\right)a \, \mathbf{\hat{y}} + \left(\frac{1}{2} +z_{7}\right)c \, \mathbf{\hat{z}} & \left(8c\right) & \mbox{O III} \\ 
\mathbf{B}_{56} & = & \left(\frac{1}{2} +y_{7}\right) \, \mathbf{a}_{1} + \left(\frac{1}{2} +x_{7}\right) \, \mathbf{a}_{2} + \left(\frac{1}{2} +z_{7}\right) \, \mathbf{a}_{3} & = & \left(\frac{1}{2} +y_{7}\right)a \, \mathbf{\hat{x}} + \left(\frac{1}{2} +x_{7}\right)a \, \mathbf{\hat{y}} + \left(\frac{1}{2} +z_{7}\right)c \, \mathbf{\hat{z}} & \left(8c\right) & \mbox{O III} \\ 
\mathbf{B}_{57} & = & x_{8} \, \mathbf{a}_{1} + y_{8} \, \mathbf{a}_{2} + z_{8} \, \mathbf{a}_{3} & = & x_{8}a \, \mathbf{\hat{x}} + y_{8}a \, \mathbf{\hat{y}} + z_{8}c \, \mathbf{\hat{z}} & \left(8c\right) & \mbox{Zn} \\ 
\mathbf{B}_{58} & = & -x_{8} \, \mathbf{a}_{1}-y_{8} \, \mathbf{a}_{2} + z_{8} \, \mathbf{a}_{3} & = & -x_{8}a \, \mathbf{\hat{x}}-y_{8}a \, \mathbf{\hat{y}} + z_{8}c \, \mathbf{\hat{z}} & \left(8c\right) & \mbox{Zn} \\ 
\mathbf{B}_{59} & = & -y_{8} \, \mathbf{a}_{1} + x_{8} \, \mathbf{a}_{2} + \left(\frac{1}{2} +z_{8}\right) \, \mathbf{a}_{3} & = & -y_{8}a \, \mathbf{\hat{x}} + x_{8}a \, \mathbf{\hat{y}} + \left(\frac{1}{2} +z_{8}\right)c \, \mathbf{\hat{z}} & \left(8c\right) & \mbox{Zn} \\ 
\mathbf{B}_{60} & = & y_{8} \, \mathbf{a}_{1}-x_{8} \, \mathbf{a}_{2} + \left(\frac{1}{2} +z_{8}\right) \, \mathbf{a}_{3} & = & y_{8}a \, \mathbf{\hat{x}}-x_{8}a \, \mathbf{\hat{y}} + \left(\frac{1}{2} +z_{8}\right)c \, \mathbf{\hat{z}} & \left(8c\right) & \mbox{Zn} \\ 
\mathbf{B}_{61} & = & \left(\frac{1}{2} +x_{8}\right) \, \mathbf{a}_{1} + \left(\frac{1}{2} - y_{8}\right) \, \mathbf{a}_{2} + z_{8} \, \mathbf{a}_{3} & = & \left(\frac{1}{2} +x_{8}\right)a \, \mathbf{\hat{x}} + \left(\frac{1}{2} - y_{8}\right)a \, \mathbf{\hat{y}} + z_{8}c \, \mathbf{\hat{z}} & \left(8c\right) & \mbox{Zn} \\ 
\mathbf{B}_{62} & = & \left(\frac{1}{2} - x_{8}\right) \, \mathbf{a}_{1} + \left(\frac{1}{2} +y_{8}\right) \, \mathbf{a}_{2} + z_{8} \, \mathbf{a}_{3} & = & \left(\frac{1}{2} - x_{8}\right)a \, \mathbf{\hat{x}} + \left(\frac{1}{2} +y_{8}\right)a \, \mathbf{\hat{y}} + z_{8}c \, \mathbf{\hat{z}} & \left(8c\right) & \mbox{Zn} \\ 
\mathbf{B}_{63} & = & \left(\frac{1}{2} - y_{8}\right) \, \mathbf{a}_{1} + \left(\frac{1}{2} - x_{8}\right) \, \mathbf{a}_{2} + \left(\frac{1}{2} +z_{8}\right) \, \mathbf{a}_{3} & = & \left(\frac{1}{2} - y_{8}\right)a \, \mathbf{\hat{x}} + \left(\frac{1}{2} - x_{8}\right)a \, \mathbf{\hat{y}} + \left(\frac{1}{2} +z_{8}\right)c \, \mathbf{\hat{z}} & \left(8c\right) & \mbox{Zn} \\ 
\mathbf{B}_{64} & = & \left(\frac{1}{2} +y_{8}\right) \, \mathbf{a}_{1} + \left(\frac{1}{2} +x_{8}\right) \, \mathbf{a}_{2} + \left(\frac{1}{2} +z_{8}\right) \, \mathbf{a}_{3} & = & \left(\frac{1}{2} +y_{8}\right)a \, \mathbf{\hat{x}} + \left(\frac{1}{2} +x_{8}\right)a \, \mathbf{\hat{y}} + \left(\frac{1}{2} +z_{8}\right)c \, \mathbf{\hat{z}} & \left(8c\right) & \mbox{Zn} \\ 
\end{longtabu}
\renewcommand{\arraystretch}{1.0}
\noindent \hrulefill
\\
\textbf{References:}
\vspace*{-0.25cm}
\begin{flushleft}
  - \bibentry{Stahl_NaZnOH3_ZAnorganAllgChem_1998}. \\
\end{flushleft}
\textbf{Found in:}
\vspace*{-0.25cm}
\begin{flushleft}
  - \bibentry{Villars_PearsonsCrystalData_2013}. \\
\end{flushleft}
\noindent \hrulefill
\\
\textbf{Geometry files:}
\\
\noindent  - CIF: pp. {\hyperref[A3BC3D_tP64_106_3c_c_3c_c_cif]{\pageref{A3BC3D_tP64_106_3c_c_3c_c_cif}}} \\
\noindent  - POSCAR: pp. {\hyperref[A3BC3D_tP64_106_3c_c_3c_c_poscar]{\pageref{A3BC3D_tP64_106_3c_c_3c_c_poscar}}} \\
\onecolumn
{\phantomsection\label{A5B7_tI24_107_ac_abd}}
\subsection*{\huge \textbf{{\normalfont Co$_{5}$Ge$_{7}$ Structure: A5B7\_tI24\_107\_ac\_abd}}}
\noindent \hrulefill
\vspace*{0.25cm}
\begin{figure}[htp]
  \centering
  \vspace{-1em}
  {\includegraphics[width=1\textwidth]{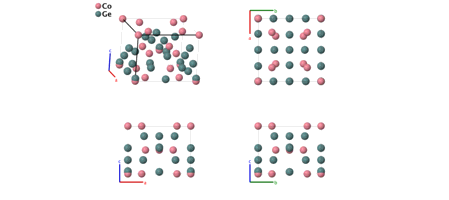}}
\end{figure}
\vspace*{-0.5cm}
\renewcommand{\arraystretch}{1.5}
\begin{equation*}
  \begin{array}{>{$\hspace{-0.15cm}}l<{$}>{$}p{0.5cm}<{$}>{$}p{18.5cm}<{$}}
    \mbox{\large \textbf{Prototype}} &\colon & \ce{Co5Ge7} \\
    \mbox{\large \textbf{\AFLOW\ prototype label}} &\colon & \mbox{A5B7\_tI24\_107\_ac\_abd} \\
    \mbox{\large \textbf{\textit{Strukturbericht} designation}} &\colon & \mbox{None} \\
    \mbox{\large \textbf{Pearson symbol}} &\colon & \mbox{tI24} \\
    \mbox{\large \textbf{Space group number}} &\colon & 107 \\
    \mbox{\large \textbf{Space group symbol}} &\colon & I4mm \\
    \mbox{\large \textbf{\AFLOW\ prototype command}} &\colon &  \texttt{aflow} \,  \, \texttt{-{}-proto=A5B7\_tI24\_107\_ac\_abd } \, \newline \texttt{-{}-params=}{a,c/a,z_{1},z_{2},z_{3},x_{4},z_{4},x_{5},z_{5} }
  \end{array}
\end{equation*}
\renewcommand{\arraystretch}{1.0}

\noindent \parbox{1 \linewidth}{
\noindent \hrulefill
\\
\textbf{Body-centered Tetragonal primitive vectors:} \\
\vspace*{-0.25cm}
\begin{tabular}{cc}
  \begin{tabular}{c}
    \parbox{0.6 \linewidth}{
      \renewcommand{\arraystretch}{1.5}
      \begin{equation*}
        \centering
        \begin{array}{ccc}
              \mathbf{a}_1 & = & - \frac12 \, a \, \mathbf{\hat{x}} + \frac12 \, a \, \mathbf{\hat{y}} + \frac12 \, c \, \mathbf{\hat{z}} \\
    \mathbf{a}_2 & = & ~ \frac12 \, a \, \mathbf{\hat{x}} - \frac12 \, a \, \mathbf{\hat{y}} + \frac12 \, c \, \mathbf{\hat{z}} \\
    \mathbf{a}_3 & = & ~ \frac12 \, a \, \mathbf{\hat{x}} + \frac12 \, a \, \mathbf{\hat{y}} - \frac12 \, c \, \mathbf{\hat{z}} \\

        \end{array}
      \end{equation*}
    }
    \renewcommand{\arraystretch}{1.0}
  \end{tabular}
  \begin{tabular}{c}
    \includegraphics[width=0.3\linewidth]{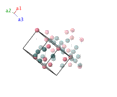} \\
  \end{tabular}
\end{tabular}

}
\vspace*{-0.25cm}

\noindent \hrulefill
\\
\textbf{Basis vectors:}
\vspace*{-0.25cm}
\renewcommand{\arraystretch}{1.5}
\begin{longtabu} to \textwidth{>{\centering $}X[-1,c,c]<{$}>{\centering $}X[-1,c,c]<{$}>{\centering $}X[-1,c,c]<{$}>{\centering $}X[-1,c,c]<{$}>{\centering $}X[-1,c,c]<{$}>{\centering $}X[-1,c,c]<{$}>{\centering $}X[-1,c,c]<{$}}
  & & \mbox{Lattice Coordinates} & & \mbox{Cartesian Coordinates} &\mbox{Wyckoff Position} & \mbox{Atom Type} \\  
  \mathbf{B}_{1} & = & z_{1} \, \mathbf{a}_{1} + z_{1} \, \mathbf{a}_{2} & = & z_{1}c \, \mathbf{\hat{z}} & \left(2a\right) & \mbox{Co I} \\ 
\mathbf{B}_{2} & = & z_{2} \, \mathbf{a}_{1} + z_{2} \, \mathbf{a}_{2} & = & z_{2}c \, \mathbf{\hat{z}} & \left(2a\right) & \mbox{Ge I} \\ 
\mathbf{B}_{3} & = & \left(\frac{1}{2} +z_{3}\right) \, \mathbf{a}_{1} + z_{3} \, \mathbf{a}_{2} + \frac{1}{2} \, \mathbf{a}_{3} & = & \frac{1}{2}a \, \mathbf{\hat{y}} + z_{3}c \, \mathbf{\hat{z}} & \left(4b\right) & \mbox{Ge II} \\ 
\mathbf{B}_{4} & = & z_{3} \, \mathbf{a}_{1} + \left(\frac{1}{2} +z_{3}\right) \, \mathbf{a}_{2} + \frac{1}{2} \, \mathbf{a}_{3} & = & \frac{1}{2}a \, \mathbf{\hat{x}} + z_{3}c \, \mathbf{\hat{z}} & \left(4b\right) & \mbox{Ge II} \\ 
\mathbf{B}_{5} & = & \left(x_{4}+z_{4}\right) \, \mathbf{a}_{1} + \left(x_{4}+z_{4}\right) \, \mathbf{a}_{2} + 2x_{4} \, \mathbf{a}_{3} & = & x_{4}a \, \mathbf{\hat{x}} + x_{4}a \, \mathbf{\hat{y}} + z_{4}c \, \mathbf{\hat{z}} & \left(8c\right) & \mbox{Co II} \\ 
\mathbf{B}_{6} & = & \left(-x_{4}+z_{4}\right) \, \mathbf{a}_{1} + \left(-x_{4}+z_{4}\right) \, \mathbf{a}_{2}-2x_{4} \, \mathbf{a}_{3} & = & -x_{4}a \, \mathbf{\hat{x}}-x_{4}a \, \mathbf{\hat{y}} + z_{4}c \, \mathbf{\hat{z}} & \left(8c\right) & \mbox{Co II} \\ 
\mathbf{B}_{7} & = & \left(x_{4}+z_{4}\right) \, \mathbf{a}_{1} + \left(-x_{4}+z_{4}\right) \, \mathbf{a}_{2} & = & -x_{4}a \, \mathbf{\hat{x}} + x_{4}a \, \mathbf{\hat{y}} + z_{4}c \, \mathbf{\hat{z}} & \left(8c\right) & \mbox{Co II} \\ 
\mathbf{B}_{8} & = & \left(-x_{4}+z_{4}\right) \, \mathbf{a}_{1} + \left(x_{4}+z_{4}\right) \, \mathbf{a}_{2} & = & x_{4}a \, \mathbf{\hat{x}}-x_{4}a \, \mathbf{\hat{y}} + z_{4}c \, \mathbf{\hat{z}} & \left(8c\right) & \mbox{Co II} \\ 
\mathbf{B}_{9} & = & z_{5} \, \mathbf{a}_{1} + \left(x_{5}+z_{5}\right) \, \mathbf{a}_{2} + x_{5} \, \mathbf{a}_{3} & = & x_{5}a \, \mathbf{\hat{x}} + z_{5}c \, \mathbf{\hat{z}} & \left(8d\right) & \mbox{Ge III} \\ 
\mathbf{B}_{10} & = & z_{5} \, \mathbf{a}_{1} + \left(-x_{5}+z_{5}\right) \, \mathbf{a}_{2}-x_{5} \, \mathbf{a}_{3} & = & -x_{5}a \, \mathbf{\hat{x}} + z_{5}c \, \mathbf{\hat{z}} & \left(8d\right) & \mbox{Ge III} \\ 
\mathbf{B}_{11} & = & \left(x_{5}+z_{5}\right) \, \mathbf{a}_{1} + z_{5} \, \mathbf{a}_{2} + x_{5} \, \mathbf{a}_{3} & = & x_{5}a \, \mathbf{\hat{y}} + z_{5}c \, \mathbf{\hat{z}} & \left(8d\right) & \mbox{Ge III} \\ 
\mathbf{B}_{12} & = & \left(-x_{5}+z_{5}\right) \, \mathbf{a}_{1} + z_{5} \, \mathbf{a}_{2}-x_{5} \, \mathbf{a}_{3} & = & -x_{5}a \, \mathbf{\hat{y}} + z_{5}c \, \mathbf{\hat{z}} & \left(8d\right) & \mbox{Ge III} \\ 
\end{longtabu}
\renewcommand{\arraystretch}{1.0}
\noindent \hrulefill
\\
\textbf{References:}
\vspace*{-0.25cm}
\begin{flushleft}
  - \bibentry{Schubert_Co5Ge7_Naturwissen_1960}. \\
\end{flushleft}
\textbf{Found in:}
\vspace*{-0.25cm}
\begin{flushleft}
  - \bibentry{Villars_PearsonsCrystalData_2013}. \\
\end{flushleft}
\noindent \hrulefill
\\
\textbf{Geometry files:}
\\
\noindent  - CIF: pp. {\hyperref[A5B7_tI24_107_ac_abd_cif]{\pageref{A5B7_tI24_107_ac_abd_cif}}} \\
\noindent  - POSCAR: pp. {\hyperref[A5B7_tI24_107_ac_abd_poscar]{\pageref{A5B7_tI24_107_ac_abd_poscar}}} \\
\onecolumn
{\phantomsection\label{AB_tI4_107_a_a}}
\subsection*{\huge \textbf{{\normalfont \begin{raggedleft}GeP (High-pressure, superconducting) Structure: \end{raggedleft} \\ AB\_tI4\_107\_a\_a}}}
\noindent \hrulefill
\vspace*{0.25cm}
\begin{figure}[htp]
  \centering
  \vspace{-1em}
  {\includegraphics[width=1\textwidth]{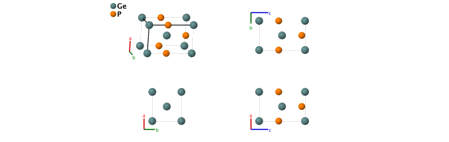}}
\end{figure}
\vspace*{-0.5cm}
\renewcommand{\arraystretch}{1.5}
\begin{equation*}
  \begin{array}{>{$\hspace{-0.15cm}}l<{$}>{$}p{0.5cm}<{$}>{$}p{18.5cm}<{$}}
    \mbox{\large \textbf{Prototype}} &\colon & \ce{GeP} \\
    \mbox{\large \textbf{\AFLOW\ prototype label}} &\colon & \mbox{AB\_tI4\_107\_a\_a} \\
    \mbox{\large \textbf{\textit{Strukturbericht} designation}} &\colon & \mbox{None} \\
    \mbox{\large \textbf{Pearson symbol}} &\colon & \mbox{tI4} \\
    \mbox{\large \textbf{Space group number}} &\colon & 107 \\
    \mbox{\large \textbf{Space group symbol}} &\colon & I4mm \\
    \mbox{\large \textbf{\AFLOW\ prototype command}} &\colon &  \texttt{aflow} \,  \, \texttt{-{}-proto=AB\_tI4\_107\_a\_a } \, \newline \texttt{-{}-params=}{a,c/a,z_{1},z_{2} }
  \end{array}
\end{equation*}
\renewcommand{\arraystretch}{1.0}

\noindent \parbox{1 \linewidth}{
\noindent \hrulefill
\\
\textbf{Body-centered Tetragonal primitive vectors:} \\
\vspace*{-0.25cm}
\begin{tabular}{cc}
  \begin{tabular}{c}
    \parbox{0.6 \linewidth}{
      \renewcommand{\arraystretch}{1.5}
      \begin{equation*}
        \centering
        \begin{array}{ccc}
              \mathbf{a}_1 & = & - \frac12 \, a \, \mathbf{\hat{x}} + \frac12 \, a \, \mathbf{\hat{y}} + \frac12 \, c \, \mathbf{\hat{z}} \\
    \mathbf{a}_2 & = & ~ \frac12 \, a \, \mathbf{\hat{x}} - \frac12 \, a \, \mathbf{\hat{y}} + \frac12 \, c \, \mathbf{\hat{z}} \\
    \mathbf{a}_3 & = & ~ \frac12 \, a \, \mathbf{\hat{x}} + \frac12 \, a \, \mathbf{\hat{y}} - \frac12 \, c \, \mathbf{\hat{z}} \\

        \end{array}
      \end{equation*}
    }
    \renewcommand{\arraystretch}{1.0}
  \end{tabular}
  \begin{tabular}{c}
    \includegraphics[width=0.3\linewidth]{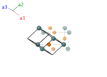} \\
  \end{tabular}
\end{tabular}

}
\vspace*{-0.25cm}

\noindent \hrulefill
\\
\textbf{Basis vectors:}
\vspace*{-0.25cm}
\renewcommand{\arraystretch}{1.5}
\begin{longtabu} to \textwidth{>{\centering $}X[-1,c,c]<{$}>{\centering $}X[-1,c,c]<{$}>{\centering $}X[-1,c,c]<{$}>{\centering $}X[-1,c,c]<{$}>{\centering $}X[-1,c,c]<{$}>{\centering $}X[-1,c,c]<{$}>{\centering $}X[-1,c,c]<{$}}
  & & \mbox{Lattice Coordinates} & & \mbox{Cartesian Coordinates} &\mbox{Wyckoff Position} & \mbox{Atom Type} \\  
  \mathbf{B}_{1} & = & z_{1} \, \mathbf{a}_{1} + z_{1} \, \mathbf{a}_{2} & = & z_{1}c \, \mathbf{\hat{z}} & \left(2a\right) & \mbox{Ge} \\ 
\mathbf{B}_{2} & = & z_{2} \, \mathbf{a}_{1} + z_{2} \, \mathbf{a}_{2} & = & z_{2}c \, \mathbf{\hat{z}} & \left(2a\right) & \mbox{P} \\ 
\end{longtabu}
\renewcommand{\arraystretch}{1.0}
\noindent \hrulefill
\\
\textbf{References:}
\vspace*{-0.25cm}
\begin{flushleft}
  - \bibentry{Donohue_GeP_JSolStateChem_1970}. \\
\end{flushleft}
\textbf{Found in:}
\vspace*{-0.25cm}
\begin{flushleft}
  - \bibentry{Villars_PearsonsCrystalData_2013}. \\
\end{flushleft}
\noindent \hrulefill
\\
\textbf{Geometry files:}
\\
\noindent  - CIF: pp. {\hyperref[AB_tI4_107_a_a_cif]{\pageref{AB_tI4_107_a_a_cif}}} \\
\noindent  - POSCAR: pp. {\hyperref[AB_tI4_107_a_a_poscar]{\pageref{AB_tI4_107_a_a_poscar}}} \\
\onecolumn
{\phantomsection\label{A3B5_tI32_108_ac_a2c}}
\subsection*{\huge \textbf{{\normalfont Sr$_{5}$Si$_{3}$ Structure: A3B5\_tI32\_108\_ac\_a2c}}}
\noindent \hrulefill
\vspace*{0.25cm}
\begin{figure}[htp]
  \centering
  \vspace{-1em}
  {\includegraphics[width=1\textwidth]{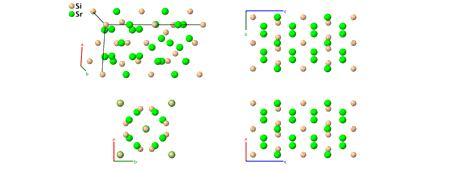}}
\end{figure}
\vspace*{-0.5cm}
\renewcommand{\arraystretch}{1.5}
\begin{equation*}
  \begin{array}{>{$\hspace{-0.15cm}}l<{$}>{$}p{0.5cm}<{$}>{$}p{18.5cm}<{$}}
    \mbox{\large \textbf{Prototype}} &\colon & \ce{Sr5Si3} \\
    \mbox{\large \textbf{\AFLOW\ prototype label}} &\colon & \mbox{A3B5\_tI32\_108\_ac\_a2c} \\
    \mbox{\large \textbf{\textit{Strukturbericht} designation}} &\colon & \mbox{None} \\
    \mbox{\large \textbf{Pearson symbol}} &\colon & \mbox{tI32} \\
    \mbox{\large \textbf{Space group number}} &\colon & 108 \\
    \mbox{\large \textbf{Space group symbol}} &\colon & I4cm \\
    \mbox{\large \textbf{\AFLOW\ prototype command}} &\colon &  \texttt{aflow} \,  \, \texttt{-{}-proto=A3B5\_tI32\_108\_ac\_a2c } \, \newline \texttt{-{}-params=}{a,c/a,z_{1},z_{2},x_{3},z_{3},x_{4},z_{4},x_{5},z_{5} }
  \end{array}
\end{equation*}
\renewcommand{\arraystretch}{1.0}

\noindent \parbox{1 \linewidth}{
\noindent \hrulefill
\\
\textbf{Body-centered Tetragonal primitive vectors:} \\
\vspace*{-0.25cm}
\begin{tabular}{cc}
  \begin{tabular}{c}
    \parbox{0.6 \linewidth}{
      \renewcommand{\arraystretch}{1.5}
      \begin{equation*}
        \centering
        \begin{array}{ccc}
              \mathbf{a}_1 & = & - \frac12 \, a \, \mathbf{\hat{x}} + \frac12 \, a \, \mathbf{\hat{y}} + \frac12 \, c \, \mathbf{\hat{z}} \\
    \mathbf{a}_2 & = & ~ \frac12 \, a \, \mathbf{\hat{x}} - \frac12 \, a \, \mathbf{\hat{y}} + \frac12 \, c \, \mathbf{\hat{z}} \\
    \mathbf{a}_3 & = & ~ \frac12 \, a \, \mathbf{\hat{x}} + \frac12 \, a \, \mathbf{\hat{y}} - \frac12 \, c \, \mathbf{\hat{z}} \\

        \end{array}
      \end{equation*}
    }
    \renewcommand{\arraystretch}{1.0}
  \end{tabular}
  \begin{tabular}{c}
    \includegraphics[width=0.3\linewidth]{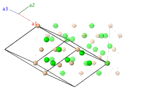} \\
  \end{tabular}
\end{tabular}

}
\vspace*{-0.25cm}

\noindent \hrulefill
\\
\textbf{Basis vectors:}
\vspace*{-0.25cm}
\renewcommand{\arraystretch}{1.5}
\begin{longtabu} to \textwidth{>{\centering $}X[-1,c,c]<{$}>{\centering $}X[-1,c,c]<{$}>{\centering $}X[-1,c,c]<{$}>{\centering $}X[-1,c,c]<{$}>{\centering $}X[-1,c,c]<{$}>{\centering $}X[-1,c,c]<{$}>{\centering $}X[-1,c,c]<{$}}
  & & \mbox{Lattice Coordinates} & & \mbox{Cartesian Coordinates} &\mbox{Wyckoff Position} & \mbox{Atom Type} \\  
  \mathbf{B}_{1} & = & z_{1} \, \mathbf{a}_{1} + z_{1} \, \mathbf{a}_{2} & = & z_{1}c \, \mathbf{\hat{z}} & \left(4a\right) & \mbox{Si I} \\ 
\mathbf{B}_{2} & = & \left(\frac{1}{2} +z_{1}\right) \, \mathbf{a}_{1} + \left(\frac{1}{2} +z_{1}\right) \, \mathbf{a}_{2} & = & \left(\frac{1}{2} +z_{1}\right)c \, \mathbf{\hat{z}} & \left(4a\right) & \mbox{Si I} \\ 
\mathbf{B}_{3} & = & z_{2} \, \mathbf{a}_{1} + z_{2} \, \mathbf{a}_{2} & = & z_{2}c \, \mathbf{\hat{z}} & \left(4a\right) & \mbox{Sr I} \\ 
\mathbf{B}_{4} & = & \left(\frac{1}{2} +z_{2}\right) \, \mathbf{a}_{1} + \left(\frac{1}{2} +z_{2}\right) \, \mathbf{a}_{2} & = & \left(\frac{1}{2} +z_{2}\right)c \, \mathbf{\hat{z}} & \left(4a\right) & \mbox{Sr I} \\ 
\mathbf{B}_{5} & = & \left(\frac{1}{2} +x_{3} + z_{3}\right) \, \mathbf{a}_{1} + \left(x_{3}+z_{3}\right) \, \mathbf{a}_{2} + \left(\frac{1}{2} +2x_{3}\right) \, \mathbf{a}_{3} & = & x_{3}a \, \mathbf{\hat{x}} + \left(\frac{1}{2} +x_{3}\right)a \, \mathbf{\hat{y}} + z_{3}c \, \mathbf{\hat{z}} & \left(8c\right) & \mbox{Si II} \\ 
\mathbf{B}_{6} & = & \left(\frac{1}{2} - x_{3} + z_{3}\right) \, \mathbf{a}_{1} + \left(-x_{3}+z_{3}\right) \, \mathbf{a}_{2} + \left(\frac{1}{2} - 2x_{3}\right) \, \mathbf{a}_{3} & = & -x_{3}a \, \mathbf{\hat{x}} + \left(\frac{1}{2} - x_{3}\right)a \, \mathbf{\hat{y}} + z_{3}c \, \mathbf{\hat{z}} & \left(8c\right) & \mbox{Si II} \\ 
\mathbf{B}_{7} & = & \left(x_{3}+z_{3}\right) \, \mathbf{a}_{1} + \left(\frac{1}{2} - x_{3} + z_{3}\right) \, \mathbf{a}_{2} + \frac{1}{2} \, \mathbf{a}_{3} & = & \left(\frac{1}{2} - x_{3}\right)a \, \mathbf{\hat{x}} + x_{3}a \, \mathbf{\hat{y}} + z_{3}c \, \mathbf{\hat{z}} & \left(8c\right) & \mbox{Si II} \\ 
\mathbf{B}_{8} & = & \left(-x_{3}+z_{3}\right) \, \mathbf{a}_{1} + \left(\frac{1}{2} +x_{3} + z_{3}\right) \, \mathbf{a}_{2} + \frac{1}{2} \, \mathbf{a}_{3} & = & \left(\frac{1}{2} +x_{3}\right)a \, \mathbf{\hat{x}}-x_{3}a \, \mathbf{\hat{y}} + z_{3}c \, \mathbf{\hat{z}} & \left(8c\right) & \mbox{Si II} \\ 
\mathbf{B}_{9} & = & \left(\frac{1}{2} +x_{4} + z_{4}\right) \, \mathbf{a}_{1} + \left(x_{4}+z_{4}\right) \, \mathbf{a}_{2} + \left(\frac{1}{2} +2x_{4}\right) \, \mathbf{a}_{3} & = & x_{4}a \, \mathbf{\hat{x}} + \left(\frac{1}{2} +x_{4}\right)a \, \mathbf{\hat{y}} + z_{4}c \, \mathbf{\hat{z}} & \left(8c\right) & \mbox{Sr II} \\ 
\mathbf{B}_{10} & = & \left(\frac{1}{2} - x_{4} + z_{4}\right) \, \mathbf{a}_{1} + \left(-x_{4}+z_{4}\right) \, \mathbf{a}_{2} + \left(\frac{1}{2} - 2x_{4}\right) \, \mathbf{a}_{3} & = & -x_{4}a \, \mathbf{\hat{x}} + \left(\frac{1}{2} - x_{4}\right)a \, \mathbf{\hat{y}} + z_{4}c \, \mathbf{\hat{z}} & \left(8c\right) & \mbox{Sr II} \\ 
\mathbf{B}_{11} & = & \left(x_{4}+z_{4}\right) \, \mathbf{a}_{1} + \left(\frac{1}{2} - x_{4} + z_{4}\right) \, \mathbf{a}_{2} + \frac{1}{2} \, \mathbf{a}_{3} & = & \left(\frac{1}{2} - x_{4}\right)a \, \mathbf{\hat{x}} + x_{4}a \, \mathbf{\hat{y}} + z_{4}c \, \mathbf{\hat{z}} & \left(8c\right) & \mbox{Sr II} \\ 
\mathbf{B}_{12} & = & \left(-x_{4}+z_{4}\right) \, \mathbf{a}_{1} + \left(\frac{1}{2} +x_{4} + z_{4}\right) \, \mathbf{a}_{2} + \frac{1}{2} \, \mathbf{a}_{3} & = & \left(\frac{1}{2} +x_{4}\right)a \, \mathbf{\hat{x}}-x_{4}a \, \mathbf{\hat{y}} + z_{4}c \, \mathbf{\hat{z}} & \left(8c\right) & \mbox{Sr II} \\ 
\mathbf{B}_{13} & = & \left(\frac{1}{2} +x_{5} + z_{5}\right) \, \mathbf{a}_{1} + \left(x_{5}+z_{5}\right) \, \mathbf{a}_{2} + \left(\frac{1}{2} +2x_{5}\right) \, \mathbf{a}_{3} & = & x_{5}a \, \mathbf{\hat{x}} + \left(\frac{1}{2} +x_{5}\right)a \, \mathbf{\hat{y}} + z_{5}c \, \mathbf{\hat{z}} & \left(8c\right) & \mbox{Sr III} \\ 
\mathbf{B}_{14} & = & \left(\frac{1}{2} - x_{5} + z_{5}\right) \, \mathbf{a}_{1} + \left(-x_{5}+z_{5}\right) \, \mathbf{a}_{2} + \left(\frac{1}{2} - 2x_{5}\right) \, \mathbf{a}_{3} & = & -x_{5}a \, \mathbf{\hat{x}} + \left(\frac{1}{2} - x_{5}\right)a \, \mathbf{\hat{y}} + z_{5}c \, \mathbf{\hat{z}} & \left(8c\right) & \mbox{Sr III} \\ 
\mathbf{B}_{15} & = & \left(x_{5}+z_{5}\right) \, \mathbf{a}_{1} + \left(\frac{1}{2} - x_{5} + z_{5}\right) \, \mathbf{a}_{2} + \frac{1}{2} \, \mathbf{a}_{3} & = & \left(\frac{1}{2} - x_{5}\right)a \, \mathbf{\hat{x}} + x_{5}a \, \mathbf{\hat{y}} + z_{5}c \, \mathbf{\hat{z}} & \left(8c\right) & \mbox{Sr III} \\ 
\mathbf{B}_{16} & = & \left(-x_{5}+z_{5}\right) \, \mathbf{a}_{1} + \left(\frac{1}{2} +x_{5} + z_{5}\right) \, \mathbf{a}_{2} + \frac{1}{2} \, \mathbf{a}_{3} & = & \left(\frac{1}{2} +x_{5}\right)a \, \mathbf{\hat{x}}-x_{5}a \, \mathbf{\hat{y}} + z_{5}c \, \mathbf{\hat{z}} & \left(8c\right) & \mbox{Sr III} \\ 
\end{longtabu}
\renewcommand{\arraystretch}{1.0}
\noindent \hrulefill
\\
\textbf{References:}
\vspace*{-0.25cm}
\begin{flushleft}
  - \bibentry{Nagorsen_Sr5Si3_ZNaturfB_1967}. \\
\end{flushleft}
\textbf{Found in:}
\vspace*{-0.25cm}
\begin{flushleft}
  - \bibentry{Villars_PearsonsCrystalData_2013}. \\
\end{flushleft}
\noindent \hrulefill
\\
\textbf{Geometry files:}
\\
\noindent  - CIF: pp. {\hyperref[A3B5_tI32_108_ac_a2c_cif]{\pageref{A3B5_tI32_108_ac_a2c_cif}}} \\
\noindent  - POSCAR: pp. {\hyperref[A3B5_tI32_108_ac_a2c_poscar]{\pageref{A3B5_tI32_108_ac_a2c_poscar}}} \\
\onecolumn
{\phantomsection\label{ABC_tI12_109_a_a_a}}
\subsection*{\huge \textbf{{\normalfont LaPtSi Structure: ABC\_tI12\_109\_a\_a\_a}}}
\noindent \hrulefill
\vspace*{0.25cm}
\begin{figure}[htp]
  \centering
  \vspace{-1em}
  {\includegraphics[width=1\textwidth]{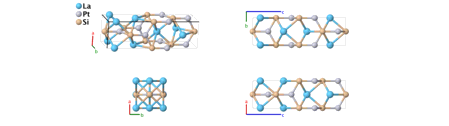}}
\end{figure}
\vspace*{-0.5cm}
\renewcommand{\arraystretch}{1.5}
\begin{equation*}
  \begin{array}{>{$\hspace{-0.15cm}}l<{$}>{$}p{0.5cm}<{$}>{$}p{18.5cm}<{$}}
    \mbox{\large \textbf{Prototype}} &\colon & \ce{LaPtSi} \\
    \mbox{\large \textbf{\AFLOW\ prototype label}} &\colon & \mbox{ABC\_tI12\_109\_a\_a\_a} \\
    \mbox{\large \textbf{\textit{Strukturbericht} designation}} &\colon & \mbox{None} \\
    \mbox{\large \textbf{Pearson symbol}} &\colon & \mbox{tI12} \\
    \mbox{\large \textbf{Space group number}} &\colon & 109 \\
    \mbox{\large \textbf{Space group symbol}} &\colon & I4_{1}md \\
    \mbox{\large \textbf{\AFLOW\ prototype command}} &\colon &  \texttt{aflow} \,  \, \texttt{-{}-proto=ABC\_tI12\_109\_a\_a\_a } \, \newline \texttt{-{}-params=}{a,c/a,z_{1},z_{2},z_{3} }
  \end{array}
\end{equation*}
\renewcommand{\arraystretch}{1.0}

\noindent \parbox{1 \linewidth}{
\noindent \hrulefill
\\
\textbf{Body-centered Tetragonal primitive vectors:} \\
\vspace*{-0.25cm}
\begin{tabular}{cc}
  \begin{tabular}{c}
    \parbox{0.6 \linewidth}{
      \renewcommand{\arraystretch}{1.5}
      \begin{equation*}
        \centering
        \begin{array}{ccc}
              \mathbf{a}_1 & = & - \frac12 \, a \, \mathbf{\hat{x}} + \frac12 \, a \, \mathbf{\hat{y}} + \frac12 \, c \, \mathbf{\hat{z}} \\
    \mathbf{a}_2 & = & ~ \frac12 \, a \, \mathbf{\hat{x}} - \frac12 \, a \, \mathbf{\hat{y}} + \frac12 \, c \, \mathbf{\hat{z}} \\
    \mathbf{a}_3 & = & ~ \frac12 \, a \, \mathbf{\hat{x}} + \frac12 \, a \, \mathbf{\hat{y}} - \frac12 \, c \, \mathbf{\hat{z}} \\

        \end{array}
      \end{equation*}
    }
    \renewcommand{\arraystretch}{1.0}
  \end{tabular}
  \begin{tabular}{c}
    \includegraphics[width=0.3\linewidth]{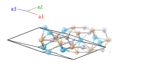} \\
  \end{tabular}
\end{tabular}

}
\vspace*{-0.25cm}

\noindent \hrulefill
\\
\textbf{Basis vectors:}
\vspace*{-0.25cm}
\renewcommand{\arraystretch}{1.5}
\begin{longtabu} to \textwidth{>{\centering $}X[-1,c,c]<{$}>{\centering $}X[-1,c,c]<{$}>{\centering $}X[-1,c,c]<{$}>{\centering $}X[-1,c,c]<{$}>{\centering $}X[-1,c,c]<{$}>{\centering $}X[-1,c,c]<{$}>{\centering $}X[-1,c,c]<{$}}
  & & \mbox{Lattice Coordinates} & & \mbox{Cartesian Coordinates} &\mbox{Wyckoff Position} & \mbox{Atom Type} \\  
  \mathbf{B}_{1} & = & z_{1} \, \mathbf{a}_{1} + z_{1} \, \mathbf{a}_{2} & = & z_{1}c \, \mathbf{\hat{z}} & \left(4a\right) & \mbox{La} \\ 
\mathbf{B}_{2} & = & \left(\frac{3}{4} +z_{1}\right) \, \mathbf{a}_{1} + \left(\frac{1}{4} +z_{1}\right) \, \mathbf{a}_{2} + \frac{1}{2} \, \mathbf{a}_{3} & = & \frac{1}{2}a \, \mathbf{\hat{y}} + \left(\frac{1}{4} +z_{1}\right)c \, \mathbf{\hat{z}} & \left(4a\right) & \mbox{La} \\ 
\mathbf{B}_{3} & = & z_{2} \, \mathbf{a}_{1} + z_{2} \, \mathbf{a}_{2} & = & z_{2}c \, \mathbf{\hat{z}} & \left(4a\right) & \mbox{Pt} \\ 
\mathbf{B}_{4} & = & \left(\frac{3}{4} +z_{2}\right) \, \mathbf{a}_{1} + \left(\frac{1}{4} +z_{2}\right) \, \mathbf{a}_{2} + \frac{1}{2} \, \mathbf{a}_{3} & = & \frac{1}{2}a \, \mathbf{\hat{y}} + \left(\frac{1}{4} +z_{2}\right)c \, \mathbf{\hat{z}} & \left(4a\right) & \mbox{Pt} \\ 
\mathbf{B}_{5} & = & z_{3} \, \mathbf{a}_{1} + z_{3} \, \mathbf{a}_{2} & = & z_{3}c \, \mathbf{\hat{z}} & \left(4a\right) & \mbox{Si} \\ 
\mathbf{B}_{6} & = & \left(\frac{3}{4} +z_{3}\right) \, \mathbf{a}_{1} + \left(\frac{1}{4} +z_{3}\right) \, \mathbf{a}_{2} + \frac{1}{2} \, \mathbf{a}_{3} & = & \frac{1}{2}a \, \mathbf{\hat{y}} + \left(\frac{1}{4} +z_{3}\right)c \, \mathbf{\hat{z}} & \left(4a\right) & \mbox{Si} \\ 
\end{longtabu}
\renewcommand{\arraystretch}{1.0}
\noindent \hrulefill
\\
\textbf{References:}
\vspace*{-0.25cm}
\begin{flushleft}
  - \bibentry{Klepp_LaPtSi_ActaCrystallogSecB_1982}. \\
\end{flushleft}
\textbf{Found in:}
\vspace*{-0.25cm}
\begin{flushleft}
  - \bibentry{Villars_PearsonsCrystalData_2013}. \\
\end{flushleft}
\noindent \hrulefill
\\
\textbf{Geometry files:}
\\
\noindent  - CIF: pp. {\hyperref[ABC_tI12_109_a_a_a_cif]{\pageref{ABC_tI12_109_a_a_a_cif}}} \\
\noindent  - POSCAR: pp. {\hyperref[ABC_tI12_109_a_a_a_poscar]{\pageref{ABC_tI12_109_a_a_a_poscar}}} \\
\onecolumn
{\phantomsection\label{AB_tI8_109_a_a}}
\subsection*{\huge \textbf{{\normalfont NbAs Structure: AB\_tI8\_109\_a\_a}}}
\noindent \hrulefill
\vspace*{0.25cm}
\begin{figure}[htp]
  \centering
  \vspace{-1em}
  {\includegraphics[width=1\textwidth]{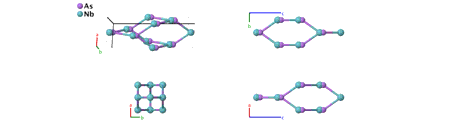}}
\end{figure}
\vspace*{-0.5cm}
\renewcommand{\arraystretch}{1.5}
\begin{equation*}
  \begin{array}{>{$\hspace{-0.15cm}}l<{$}>{$}p{0.5cm}<{$}>{$}p{18.5cm}<{$}}
    \mbox{\large \textbf{Prototype}} &\colon & \ce{NbAs} \\
    \mbox{\large \textbf{\AFLOW\ prototype label}} &\colon & \mbox{AB\_tI8\_109\_a\_a} \\
    \mbox{\large \textbf{\textit{Strukturbericht} designation}} &\colon & \mbox{None} \\
    \mbox{\large \textbf{Pearson symbol}} &\colon & \mbox{tI8} \\
    \mbox{\large \textbf{Space group number}} &\colon & 109 \\
    \mbox{\large \textbf{Space group symbol}} &\colon & I4_{1}md \\
    \mbox{\large \textbf{\AFLOW\ prototype command}} &\colon &  \texttt{aflow} \,  \, \texttt{-{}-proto=AB\_tI8\_109\_a\_a } \, \newline \texttt{-{}-params=}{a,c/a,z_{1},z_{2} }
  \end{array}
\end{equation*}
\renewcommand{\arraystretch}{1.0}

\noindent \parbox{1 \linewidth}{
\noindent \hrulefill
\\
\textbf{Body-centered Tetragonal primitive vectors:} \\
\vspace*{-0.25cm}
\begin{tabular}{cc}
  \begin{tabular}{c}
    \parbox{0.6 \linewidth}{
      \renewcommand{\arraystretch}{1.5}
      \begin{equation*}
        \centering
        \begin{array}{ccc}
              \mathbf{a}_1 & = & - \frac12 \, a \, \mathbf{\hat{x}} + \frac12 \, a \, \mathbf{\hat{y}} + \frac12 \, c \, \mathbf{\hat{z}} \\
    \mathbf{a}_2 & = & ~ \frac12 \, a \, \mathbf{\hat{x}} - \frac12 \, a \, \mathbf{\hat{y}} + \frac12 \, c \, \mathbf{\hat{z}} \\
    \mathbf{a}_3 & = & ~ \frac12 \, a \, \mathbf{\hat{x}} + \frac12 \, a \, \mathbf{\hat{y}} - \frac12 \, c \, \mathbf{\hat{z}} \\

        \end{array}
      \end{equation*}
    }
    \renewcommand{\arraystretch}{1.0}
  \end{tabular}
  \begin{tabular}{c}
    \includegraphics[width=0.3\linewidth]{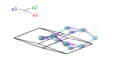} \\
  \end{tabular}
\end{tabular}

}
\vspace*{-0.25cm}

\noindent \hrulefill
\\
\textbf{Basis vectors:}
\vspace*{-0.25cm}
\renewcommand{\arraystretch}{1.5}
\begin{longtabu} to \textwidth{>{\centering $}X[-1,c,c]<{$}>{\centering $}X[-1,c,c]<{$}>{\centering $}X[-1,c,c]<{$}>{\centering $}X[-1,c,c]<{$}>{\centering $}X[-1,c,c]<{$}>{\centering $}X[-1,c,c]<{$}>{\centering $}X[-1,c,c]<{$}}
  & & \mbox{Lattice Coordinates} & & \mbox{Cartesian Coordinates} &\mbox{Wyckoff Position} & \mbox{Atom Type} \\  
  \mathbf{B}_{1} & = & z_{1} \, \mathbf{a}_{1} + z_{1} \, \mathbf{a}_{2} & = & z_{1}c \, \mathbf{\hat{z}} & \left(4a\right) & \mbox{As} \\ 
\mathbf{B}_{2} & = & \left(\frac{3}{4} +z_{1}\right) \, \mathbf{a}_{1} + \left(\frac{1}{4} +z_{1}\right) \, \mathbf{a}_{2} + \frac{1}{2} \, \mathbf{a}_{3} & = & \frac{1}{2}a \, \mathbf{\hat{y}} + \left(\frac{1}{4} +z_{1}\right)c \, \mathbf{\hat{z}} & \left(4a\right) & \mbox{As} \\ 
\mathbf{B}_{3} & = & z_{2} \, \mathbf{a}_{1} + z_{2} \, \mathbf{a}_{2} & = & z_{2}c \, \mathbf{\hat{z}} & \left(4a\right) & \mbox{Nb} \\ 
\mathbf{B}_{4} & = & \left(\frac{3}{4} +z_{2}\right) \, \mathbf{a}_{1} + \left(\frac{1}{4} +z_{2}\right) \, \mathbf{a}_{2} + \frac{1}{2} \, \mathbf{a}_{3} & = & \frac{1}{2}a \, \mathbf{\hat{y}} + \left(\frac{1}{4} +z_{2}\right)c \, \mathbf{\hat{z}} & \left(4a\right) & \mbox{Nb} \\ 
\end{longtabu}
\renewcommand{\arraystretch}{1.0}
\noindent \hrulefill
\\
\textbf{References:}
\vspace*{-0.25cm}
\begin{flushleft}
  - \bibentry{Furuseth_NbAs_ActChemScand_1964}. \\
\end{flushleft}
\textbf{Found in:}
\vspace*{-0.25cm}
\begin{flushleft}
  - \bibentry{Villars_PearsonsCrystalData_2013}. \\
\end{flushleft}
\noindent \hrulefill
\\
\textbf{Geometry files:}
\\
\noindent  - CIF: pp. {\hyperref[AB_tI8_109_a_a_cif]{\pageref{AB_tI8_109_a_a_cif}}} \\
\noindent  - POSCAR: pp. {\hyperref[AB_tI8_109_a_a_poscar]{\pageref{AB_tI8_109_a_a_poscar}}} \\
\onecolumn
{\phantomsection\label{A2BC8_tI176_110_2b_b_8b}}
\subsection*{\huge \textbf{{\normalfont Be[BH$_{4}$]$_{2}$ Structure: A2BC8\_tI176\_110\_2b\_b\_8b}}}
\noindent \hrulefill
\vspace*{0.25cm}
\begin{figure}[htp]
  \centering
  \vspace{-1em}
  {\includegraphics[width=1\textwidth]{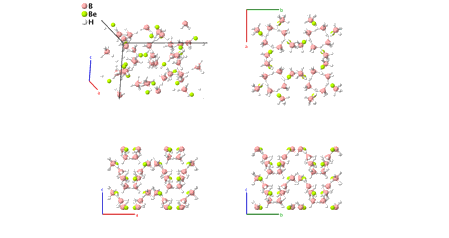}}
\end{figure}
\vspace*{-0.5cm}
\renewcommand{\arraystretch}{1.5}
\begin{equation*}
  \begin{array}{>{$\hspace{-0.15cm}}l<{$}>{$}p{0.5cm}<{$}>{$}p{18.5cm}<{$}}
    \mbox{\large \textbf{Prototype}} &\colon & \ce{Be[BH4]2} \\
    \mbox{\large \textbf{\AFLOW\ prototype label}} &\colon & \mbox{A2BC8\_tI176\_110\_2b\_b\_8b} \\
    \mbox{\large \textbf{\textit{Strukturbericht} designation}} &\colon & \mbox{None} \\
    \mbox{\large \textbf{Pearson symbol}} &\colon & \mbox{tI176} \\
    \mbox{\large \textbf{Space group number}} &\colon & 110 \\
    \mbox{\large \textbf{Space group symbol}} &\colon & I4_{1}cd \\
    \mbox{\large \textbf{\AFLOW\ prototype command}} &\colon &  \texttt{aflow} \,  \, \texttt{-{}-proto=A2BC8\_tI176\_110\_2b\_b\_8b } \, \newline \texttt{-{}-params=}{a,c/a,x_{1},y_{1},z_{1},x_{2},y_{2},z_{2},x_{3},y_{3},z_{3},x_{4},y_{4},z_{4},x_{5},y_{5},z_{5},x_{6},y_{6},z_{6},x_{7},} \newline {y_{7},z_{7},x_{8},y_{8},z_{8},x_{9},y_{9},z_{9},x_{10},y_{10},z_{10},x_{11},y_{11},z_{11} }
  \end{array}
\end{equation*}
\renewcommand{\arraystretch}{1.0}

\noindent \parbox{1 \linewidth}{
\noindent \hrulefill
\\
\textbf{Body-centered Tetragonal primitive vectors:} \\
\vspace*{-0.25cm}
\begin{tabular}{cc}
  \begin{tabular}{c}
    \parbox{0.6 \linewidth}{
      \renewcommand{\arraystretch}{1.5}
      \begin{equation*}
        \centering
        \begin{array}{ccc}
              \mathbf{a}_1 & = & - \frac12 \, a \, \mathbf{\hat{x}} + \frac12 \, a \, \mathbf{\hat{y}} + \frac12 \, c \, \mathbf{\hat{z}} \\
    \mathbf{a}_2 & = & ~ \frac12 \, a \, \mathbf{\hat{x}} - \frac12 \, a \, \mathbf{\hat{y}} + \frac12 \, c \, \mathbf{\hat{z}} \\
    \mathbf{a}_3 & = & ~ \frac12 \, a \, \mathbf{\hat{x}} + \frac12 \, a \, \mathbf{\hat{y}} - \frac12 \, c \, \mathbf{\hat{z}} \\

        \end{array}
      \end{equation*}
    }
    \renewcommand{\arraystretch}{1.0}
  \end{tabular}
  \begin{tabular}{c}
    \includegraphics[width=0.3\linewidth]{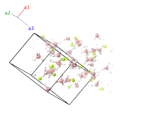} \\
  \end{tabular}
\end{tabular}

}
\vspace*{-0.25cm}

\noindent \hrulefill
\\
\textbf{Basis vectors:}
\vspace*{-0.25cm}
\renewcommand{\arraystretch}{1.5}
\begin{longtabu} to \textwidth{>{\centering $}X[-1,c,c]<{$}>{\centering $}X[-1,c,c]<{$}>{\centering $}X[-1,c,c]<{$}>{\centering $}X[-1,c,c]<{$}>{\centering $}X[-1,c,c]<{$}>{\centering $}X[-1,c,c]<{$}>{\centering $}X[-1,c,c]<{$}}
  & & \mbox{Lattice Coordinates} & & \mbox{Cartesian Coordinates} &\mbox{Wyckoff Position} & \mbox{Atom Type} \\  
  \mathbf{B}_{1} & = & \left(y_{1}+z_{1}\right) \, \mathbf{a}_{1} + \left(x_{1}+z_{1}\right) \, \mathbf{a}_{2} + \left(x_{1}+y_{1}\right) \, \mathbf{a}_{3} & = & x_{1}a \, \mathbf{\hat{x}} + y_{1}a \, \mathbf{\hat{y}} + z_{1}c \, \mathbf{\hat{z}} & \left(16b\right) & \mbox{B I} \\ 
\mathbf{B}_{2} & = & \left(-y_{1}+z_{1}\right) \, \mathbf{a}_{1} + \left(-x_{1}+z_{1}\right) \, \mathbf{a}_{2} + \left(-x_{1}-y_{1}\right) \, \mathbf{a}_{3} & = & -x_{1}a \, \mathbf{\hat{x}}-y_{1}a \, \mathbf{\hat{y}} + z_{1}c \, \mathbf{\hat{z}} & \left(16b\right) & \mbox{B I} \\ 
\mathbf{B}_{3} & = & \left(\frac{3}{4} +x_{1} + z_{1}\right) \, \mathbf{a}_{1} + \left(\frac{1}{4} - y_{1} + z_{1}\right) \, \mathbf{a}_{2} + \left(\frac{1}{2} +x_{1} - y_{1}\right) \, \mathbf{a}_{3} & = & -y_{1}a \, \mathbf{\hat{x}} + \left(\frac{1}{2} +x_{1}\right)a \, \mathbf{\hat{y}} + \left(\frac{1}{4} +z_{1}\right)c \, \mathbf{\hat{z}} & \left(16b\right) & \mbox{B I} \\ 
\mathbf{B}_{4} & = & \left(\frac{3}{4} - x_{1} + z_{1}\right) \, \mathbf{a}_{1} + \left(\frac{1}{4} +y_{1} + z_{1}\right) \, \mathbf{a}_{2} + \left(\frac{1}{2} - x_{1} + y_{1}\right) \, \mathbf{a}_{3} & = & y_{1}a \, \mathbf{\hat{x}} + \left(\frac{1}{2} - x_{1}\right)a \, \mathbf{\hat{y}} + \left(\frac{1}{4} +z_{1}\right)c \, \mathbf{\hat{z}} & \left(16b\right) & \mbox{B I} \\ 
\mathbf{B}_{5} & = & \left(\frac{1}{2} - y_{1} + z_{1}\right) \, \mathbf{a}_{1} + \left(\frac{1}{2} +x_{1} + z_{1}\right) \, \mathbf{a}_{2} + \left(x_{1}-y_{1}\right) \, \mathbf{a}_{3} & = & x_{1}a \, \mathbf{\hat{x}}-y_{1}a \, \mathbf{\hat{y}} + \left(\frac{1}{2} +z_{1}\right)c \, \mathbf{\hat{z}} & \left(16b\right) & \mbox{B I} \\ 
\mathbf{B}_{6} & = & \left(\frac{1}{2} +y_{1} + z_{1}\right) \, \mathbf{a}_{1} + \left(\frac{1}{2} - x_{1} + z_{1}\right) \, \mathbf{a}_{2} + \left(-x_{1}+y_{1}\right) \, \mathbf{a}_{3} & = & -x_{1}a \, \mathbf{\hat{x}} + y_{1}a \, \mathbf{\hat{y}} + \left(\frac{1}{2} +z_{1}\right)c \, \mathbf{\hat{z}} & \left(16b\right) & \mbox{B I} \\ 
\mathbf{B}_{7} & = & \left(\frac{1}{4} - x_{1} + z_{1}\right) \, \mathbf{a}_{1} + \left(\frac{3}{4} - y_{1} + z_{1}\right) \, \mathbf{a}_{2} + \left(\frac{1}{2} - x_{1} - y_{1}\right) \, \mathbf{a}_{3} & = & \left(\frac{1}{2} - y_{1}\right)a \, \mathbf{\hat{x}}-x_{1}a \, \mathbf{\hat{y}} + \left(\frac{1}{4} +z_{1}\right)c \, \mathbf{\hat{z}} & \left(16b\right) & \mbox{B I} \\ 
\mathbf{B}_{8} & = & \left(\frac{1}{4} +x_{1} + z_{1}\right) \, \mathbf{a}_{1} + \left(\frac{3}{4} +y_{1} + z_{1}\right) \, \mathbf{a}_{2} + \left(\frac{1}{2} +x_{1} + y_{1}\right) \, \mathbf{a}_{3} & = & \left(\frac{1}{2} +y_{1}\right)a \, \mathbf{\hat{x}} + x_{1}a \, \mathbf{\hat{y}} + \left(\frac{1}{4} +z_{1}\right)c \, \mathbf{\hat{z}} & \left(16b\right) & \mbox{B I} \\ 
\mathbf{B}_{9} & = & \left(y_{2}+z_{2}\right) \, \mathbf{a}_{1} + \left(x_{2}+z_{2}\right) \, \mathbf{a}_{2} + \left(x_{2}+y_{2}\right) \, \mathbf{a}_{3} & = & x_{2}a \, \mathbf{\hat{x}} + y_{2}a \, \mathbf{\hat{y}} + z_{2}c \, \mathbf{\hat{z}} & \left(16b\right) & \mbox{B II} \\ 
\mathbf{B}_{10} & = & \left(-y_{2}+z_{2}\right) \, \mathbf{a}_{1} + \left(-x_{2}+z_{2}\right) \, \mathbf{a}_{2} + \left(-x_{2}-y_{2}\right) \, \mathbf{a}_{3} & = & -x_{2}a \, \mathbf{\hat{x}}-y_{2}a \, \mathbf{\hat{y}} + z_{2}c \, \mathbf{\hat{z}} & \left(16b\right) & \mbox{B II} \\ 
\mathbf{B}_{11} & = & \left(\frac{3}{4} +x_{2} + z_{2}\right) \, \mathbf{a}_{1} + \left(\frac{1}{4} - y_{2} + z_{2}\right) \, \mathbf{a}_{2} + \left(\frac{1}{2} +x_{2} - y_{2}\right) \, \mathbf{a}_{3} & = & -y_{2}a \, \mathbf{\hat{x}} + \left(\frac{1}{2} +x_{2}\right)a \, \mathbf{\hat{y}} + \left(\frac{1}{4} +z_{2}\right)c \, \mathbf{\hat{z}} & \left(16b\right) & \mbox{B II} \\ 
\mathbf{B}_{12} & = & \left(\frac{3}{4} - x_{2} + z_{2}\right) \, \mathbf{a}_{1} + \left(\frac{1}{4} +y_{2} + z_{2}\right) \, \mathbf{a}_{2} + \left(\frac{1}{2} - x_{2} + y_{2}\right) \, \mathbf{a}_{3} & = & y_{2}a \, \mathbf{\hat{x}} + \left(\frac{1}{2} - x_{2}\right)a \, \mathbf{\hat{y}} + \left(\frac{1}{4} +z_{2}\right)c \, \mathbf{\hat{z}} & \left(16b\right) & \mbox{B II} \\ 
\mathbf{B}_{13} & = & \left(\frac{1}{2} - y_{2} + z_{2}\right) \, \mathbf{a}_{1} + \left(\frac{1}{2} +x_{2} + z_{2}\right) \, \mathbf{a}_{2} + \left(x_{2}-y_{2}\right) \, \mathbf{a}_{3} & = & x_{2}a \, \mathbf{\hat{x}}-y_{2}a \, \mathbf{\hat{y}} + \left(\frac{1}{2} +z_{2}\right)c \, \mathbf{\hat{z}} & \left(16b\right) & \mbox{B II} \\ 
\mathbf{B}_{14} & = & \left(\frac{1}{2} +y_{2} + z_{2}\right) \, \mathbf{a}_{1} + \left(\frac{1}{2} - x_{2} + z_{2}\right) \, \mathbf{a}_{2} + \left(-x_{2}+y_{2}\right) \, \mathbf{a}_{3} & = & -x_{2}a \, \mathbf{\hat{x}} + y_{2}a \, \mathbf{\hat{y}} + \left(\frac{1}{2} +z_{2}\right)c \, \mathbf{\hat{z}} & \left(16b\right) & \mbox{B II} \\ 
\mathbf{B}_{15} & = & \left(\frac{1}{4} - x_{2} + z_{2}\right) \, \mathbf{a}_{1} + \left(\frac{3}{4} - y_{2} + z_{2}\right) \, \mathbf{a}_{2} + \left(\frac{1}{2} - x_{2} - y_{2}\right) \, \mathbf{a}_{3} & = & \left(\frac{1}{2} - y_{2}\right)a \, \mathbf{\hat{x}}-x_{2}a \, \mathbf{\hat{y}} + \left(\frac{1}{4} +z_{2}\right)c \, \mathbf{\hat{z}} & \left(16b\right) & \mbox{B II} \\ 
\mathbf{B}_{16} & = & \left(\frac{1}{4} +x_{2} + z_{2}\right) \, \mathbf{a}_{1} + \left(\frac{3}{4} +y_{2} + z_{2}\right) \, \mathbf{a}_{2} + \left(\frac{1}{2} +x_{2} + y_{2}\right) \, \mathbf{a}_{3} & = & \left(\frac{1}{2} +y_{2}\right)a \, \mathbf{\hat{x}} + x_{2}a \, \mathbf{\hat{y}} + \left(\frac{1}{4} +z_{2}\right)c \, \mathbf{\hat{z}} & \left(16b\right) & \mbox{B II} \\ 
\mathbf{B}_{17} & = & \left(y_{3}+z_{3}\right) \, \mathbf{a}_{1} + \left(x_{3}+z_{3}\right) \, \mathbf{a}_{2} + \left(x_{3}+y_{3}\right) \, \mathbf{a}_{3} & = & x_{3}a \, \mathbf{\hat{x}} + y_{3}a \, \mathbf{\hat{y}} + z_{3}c \, \mathbf{\hat{z}} & \left(16b\right) & \mbox{Be} \\ 
\mathbf{B}_{18} & = & \left(-y_{3}+z_{3}\right) \, \mathbf{a}_{1} + \left(-x_{3}+z_{3}\right) \, \mathbf{a}_{2} + \left(-x_{3}-y_{3}\right) \, \mathbf{a}_{3} & = & -x_{3}a \, \mathbf{\hat{x}}-y_{3}a \, \mathbf{\hat{y}} + z_{3}c \, \mathbf{\hat{z}} & \left(16b\right) & \mbox{Be} \\ 
\mathbf{B}_{19} & = & \left(\frac{3}{4} +x_{3} + z_{3}\right) \, \mathbf{a}_{1} + \left(\frac{1}{4} - y_{3} + z_{3}\right) \, \mathbf{a}_{2} + \left(\frac{1}{2} +x_{3} - y_{3}\right) \, \mathbf{a}_{3} & = & -y_{3}a \, \mathbf{\hat{x}} + \left(\frac{1}{2} +x_{3}\right)a \, \mathbf{\hat{y}} + \left(\frac{1}{4} +z_{3}\right)c \, \mathbf{\hat{z}} & \left(16b\right) & \mbox{Be} \\ 
\mathbf{B}_{20} & = & \left(\frac{3}{4} - x_{3} + z_{3}\right) \, \mathbf{a}_{1} + \left(\frac{1}{4} +y_{3} + z_{3}\right) \, \mathbf{a}_{2} + \left(\frac{1}{2} - x_{3} + y_{3}\right) \, \mathbf{a}_{3} & = & y_{3}a \, \mathbf{\hat{x}} + \left(\frac{1}{2} - x_{3}\right)a \, \mathbf{\hat{y}} + \left(\frac{1}{4} +z_{3}\right)c \, \mathbf{\hat{z}} & \left(16b\right) & \mbox{Be} \\ 
\mathbf{B}_{21} & = & \left(\frac{1}{2} - y_{3} + z_{3}\right) \, \mathbf{a}_{1} + \left(\frac{1}{2} +x_{3} + z_{3}\right) \, \mathbf{a}_{2} + \left(x_{3}-y_{3}\right) \, \mathbf{a}_{3} & = & x_{3}a \, \mathbf{\hat{x}}-y_{3}a \, \mathbf{\hat{y}} + \left(\frac{1}{2} +z_{3}\right)c \, \mathbf{\hat{z}} & \left(16b\right) & \mbox{Be} \\ 
\mathbf{B}_{22} & = & \left(\frac{1}{2} +y_{3} + z_{3}\right) \, \mathbf{a}_{1} + \left(\frac{1}{2} - x_{3} + z_{3}\right) \, \mathbf{a}_{2} + \left(-x_{3}+y_{3}\right) \, \mathbf{a}_{3} & = & -x_{3}a \, \mathbf{\hat{x}} + y_{3}a \, \mathbf{\hat{y}} + \left(\frac{1}{2} +z_{3}\right)c \, \mathbf{\hat{z}} & \left(16b\right) & \mbox{Be} \\ 
\mathbf{B}_{23} & = & \left(\frac{1}{4} - x_{3} + z_{3}\right) \, \mathbf{a}_{1} + \left(\frac{3}{4} - y_{3} + z_{3}\right) \, \mathbf{a}_{2} + \left(\frac{1}{2} - x_{3} - y_{3}\right) \, \mathbf{a}_{3} & = & \left(\frac{1}{2} - y_{3}\right)a \, \mathbf{\hat{x}}-x_{3}a \, \mathbf{\hat{y}} + \left(\frac{1}{4} +z_{3}\right)c \, \mathbf{\hat{z}} & \left(16b\right) & \mbox{Be} \\ 
\mathbf{B}_{24} & = & \left(\frac{1}{4} +x_{3} + z_{3}\right) \, \mathbf{a}_{1} + \left(\frac{3}{4} +y_{3} + z_{3}\right) \, \mathbf{a}_{2} + \left(\frac{1}{2} +x_{3} + y_{3}\right) \, \mathbf{a}_{3} & = & \left(\frac{1}{2} +y_{3}\right)a \, \mathbf{\hat{x}} + x_{3}a \, \mathbf{\hat{y}} + \left(\frac{1}{4} +z_{3}\right)c \, \mathbf{\hat{z}} & \left(16b\right) & \mbox{Be} \\ 
\mathbf{B}_{25} & = & \left(y_{4}+z_{4}\right) \, \mathbf{a}_{1} + \left(x_{4}+z_{4}\right) \, \mathbf{a}_{2} + \left(x_{4}+y_{4}\right) \, \mathbf{a}_{3} & = & x_{4}a \, \mathbf{\hat{x}} + y_{4}a \, \mathbf{\hat{y}} + z_{4}c \, \mathbf{\hat{z}} & \left(16b\right) & \mbox{H I} \\ 
\mathbf{B}_{26} & = & \left(-y_{4}+z_{4}\right) \, \mathbf{a}_{1} + \left(-x_{4}+z_{4}\right) \, \mathbf{a}_{2} + \left(-x_{4}-y_{4}\right) \, \mathbf{a}_{3} & = & -x_{4}a \, \mathbf{\hat{x}}-y_{4}a \, \mathbf{\hat{y}} + z_{4}c \, \mathbf{\hat{z}} & \left(16b\right) & \mbox{H I} \\ 
\mathbf{B}_{27} & = & \left(\frac{3}{4} +x_{4} + z_{4}\right) \, \mathbf{a}_{1} + \left(\frac{1}{4} - y_{4} + z_{4}\right) \, \mathbf{a}_{2} + \left(\frac{1}{2} +x_{4} - y_{4}\right) \, \mathbf{a}_{3} & = & -y_{4}a \, \mathbf{\hat{x}} + \left(\frac{1}{2} +x_{4}\right)a \, \mathbf{\hat{y}} + \left(\frac{1}{4} +z_{4}\right)c \, \mathbf{\hat{z}} & \left(16b\right) & \mbox{H I} \\ 
\mathbf{B}_{28} & = & \left(\frac{3}{4} - x_{4} + z_{4}\right) \, \mathbf{a}_{1} + \left(\frac{1}{4} +y_{4} + z_{4}\right) \, \mathbf{a}_{2} + \left(\frac{1}{2} - x_{4} + y_{4}\right) \, \mathbf{a}_{3} & = & y_{4}a \, \mathbf{\hat{x}} + \left(\frac{1}{2} - x_{4}\right)a \, \mathbf{\hat{y}} + \left(\frac{1}{4} +z_{4}\right)c \, \mathbf{\hat{z}} & \left(16b\right) & \mbox{H I} \\ 
\mathbf{B}_{29} & = & \left(\frac{1}{2} - y_{4} + z_{4}\right) \, \mathbf{a}_{1} + \left(\frac{1}{2} +x_{4} + z_{4}\right) \, \mathbf{a}_{2} + \left(x_{4}-y_{4}\right) \, \mathbf{a}_{3} & = & x_{4}a \, \mathbf{\hat{x}}-y_{4}a \, \mathbf{\hat{y}} + \left(\frac{1}{2} +z_{4}\right)c \, \mathbf{\hat{z}} & \left(16b\right) & \mbox{H I} \\ 
\mathbf{B}_{30} & = & \left(\frac{1}{2} +y_{4} + z_{4}\right) \, \mathbf{a}_{1} + \left(\frac{1}{2} - x_{4} + z_{4}\right) \, \mathbf{a}_{2} + \left(-x_{4}+y_{4}\right) \, \mathbf{a}_{3} & = & -x_{4}a \, \mathbf{\hat{x}} + y_{4}a \, \mathbf{\hat{y}} + \left(\frac{1}{2} +z_{4}\right)c \, \mathbf{\hat{z}} & \left(16b\right) & \mbox{H I} \\ 
\mathbf{B}_{31} & = & \left(\frac{1}{4} - x_{4} + z_{4}\right) \, \mathbf{a}_{1} + \left(\frac{3}{4} - y_{4} + z_{4}\right) \, \mathbf{a}_{2} + \left(\frac{1}{2} - x_{4} - y_{4}\right) \, \mathbf{a}_{3} & = & \left(\frac{1}{2} - y_{4}\right)a \, \mathbf{\hat{x}}-x_{4}a \, \mathbf{\hat{y}} + \left(\frac{1}{4} +z_{4}\right)c \, \mathbf{\hat{z}} & \left(16b\right) & \mbox{H I} \\ 
\mathbf{B}_{32} & = & \left(\frac{1}{4} +x_{4} + z_{4}\right) \, \mathbf{a}_{1} + \left(\frac{3}{4} +y_{4} + z_{4}\right) \, \mathbf{a}_{2} + \left(\frac{1}{2} +x_{4} + y_{4}\right) \, \mathbf{a}_{3} & = & \left(\frac{1}{2} +y_{4}\right)a \, \mathbf{\hat{x}} + x_{4}a \, \mathbf{\hat{y}} + \left(\frac{1}{4} +z_{4}\right)c \, \mathbf{\hat{z}} & \left(16b\right) & \mbox{H I} \\ 
\mathbf{B}_{33} & = & \left(y_{5}+z_{5}\right) \, \mathbf{a}_{1} + \left(x_{5}+z_{5}\right) \, \mathbf{a}_{2} + \left(x_{5}+y_{5}\right) \, \mathbf{a}_{3} & = & x_{5}a \, \mathbf{\hat{x}} + y_{5}a \, \mathbf{\hat{y}} + z_{5}c \, \mathbf{\hat{z}} & \left(16b\right) & \mbox{H II} \\ 
\mathbf{B}_{34} & = & \left(-y_{5}+z_{5}\right) \, \mathbf{a}_{1} + \left(-x_{5}+z_{5}\right) \, \mathbf{a}_{2} + \left(-x_{5}-y_{5}\right) \, \mathbf{a}_{3} & = & -x_{5}a \, \mathbf{\hat{x}}-y_{5}a \, \mathbf{\hat{y}} + z_{5}c \, \mathbf{\hat{z}} & \left(16b\right) & \mbox{H II} \\ 
\mathbf{B}_{35} & = & \left(\frac{3}{4} +x_{5} + z_{5}\right) \, \mathbf{a}_{1} + \left(\frac{1}{4} - y_{5} + z_{5}\right) \, \mathbf{a}_{2} + \left(\frac{1}{2} +x_{5} - y_{5}\right) \, \mathbf{a}_{3} & = & -y_{5}a \, \mathbf{\hat{x}} + \left(\frac{1}{2} +x_{5}\right)a \, \mathbf{\hat{y}} + \left(\frac{1}{4} +z_{5}\right)c \, \mathbf{\hat{z}} & \left(16b\right) & \mbox{H II} \\ 
\mathbf{B}_{36} & = & \left(\frac{3}{4} - x_{5} + z_{5}\right) \, \mathbf{a}_{1} + \left(\frac{1}{4} +y_{5} + z_{5}\right) \, \mathbf{a}_{2} + \left(\frac{1}{2} - x_{5} + y_{5}\right) \, \mathbf{a}_{3} & = & y_{5}a \, \mathbf{\hat{x}} + \left(\frac{1}{2} - x_{5}\right)a \, \mathbf{\hat{y}} + \left(\frac{1}{4} +z_{5}\right)c \, \mathbf{\hat{z}} & \left(16b\right) & \mbox{H II} \\ 
\mathbf{B}_{37} & = & \left(\frac{1}{2} - y_{5} + z_{5}\right) \, \mathbf{a}_{1} + \left(\frac{1}{2} +x_{5} + z_{5}\right) \, \mathbf{a}_{2} + \left(x_{5}-y_{5}\right) \, \mathbf{a}_{3} & = & x_{5}a \, \mathbf{\hat{x}}-y_{5}a \, \mathbf{\hat{y}} + \left(\frac{1}{2} +z_{5}\right)c \, \mathbf{\hat{z}} & \left(16b\right) & \mbox{H II} \\ 
\mathbf{B}_{38} & = & \left(\frac{1}{2} +y_{5} + z_{5}\right) \, \mathbf{a}_{1} + \left(\frac{1}{2} - x_{5} + z_{5}\right) \, \mathbf{a}_{2} + \left(-x_{5}+y_{5}\right) \, \mathbf{a}_{3} & = & -x_{5}a \, \mathbf{\hat{x}} + y_{5}a \, \mathbf{\hat{y}} + \left(\frac{1}{2} +z_{5}\right)c \, \mathbf{\hat{z}} & \left(16b\right) & \mbox{H II} \\ 
\mathbf{B}_{39} & = & \left(\frac{1}{4} - x_{5} + z_{5}\right) \, \mathbf{a}_{1} + \left(\frac{3}{4} - y_{5} + z_{5}\right) \, \mathbf{a}_{2} + \left(\frac{1}{2} - x_{5} - y_{5}\right) \, \mathbf{a}_{3} & = & \left(\frac{1}{2} - y_{5}\right)a \, \mathbf{\hat{x}}-x_{5}a \, \mathbf{\hat{y}} + \left(\frac{1}{4} +z_{5}\right)c \, \mathbf{\hat{z}} & \left(16b\right) & \mbox{H II} \\ 
\mathbf{B}_{40} & = & \left(\frac{1}{4} +x_{5} + z_{5}\right) \, \mathbf{a}_{1} + \left(\frac{3}{4} +y_{5} + z_{5}\right) \, \mathbf{a}_{2} + \left(\frac{1}{2} +x_{5} + y_{5}\right) \, \mathbf{a}_{3} & = & \left(\frac{1}{2} +y_{5}\right)a \, \mathbf{\hat{x}} + x_{5}a \, \mathbf{\hat{y}} + \left(\frac{1}{4} +z_{5}\right)c \, \mathbf{\hat{z}} & \left(16b\right) & \mbox{H II} \\ 
\mathbf{B}_{41} & = & \left(y_{6}+z_{6}\right) \, \mathbf{a}_{1} + \left(x_{6}+z_{6}\right) \, \mathbf{a}_{2} + \left(x_{6}+y_{6}\right) \, \mathbf{a}_{3} & = & x_{6}a \, \mathbf{\hat{x}} + y_{6}a \, \mathbf{\hat{y}} + z_{6}c \, \mathbf{\hat{z}} & \left(16b\right) & \mbox{H III} \\ 
\mathbf{B}_{42} & = & \left(-y_{6}+z_{6}\right) \, \mathbf{a}_{1} + \left(-x_{6}+z_{6}\right) \, \mathbf{a}_{2} + \left(-x_{6}-y_{6}\right) \, \mathbf{a}_{3} & = & -x_{6}a \, \mathbf{\hat{x}}-y_{6}a \, \mathbf{\hat{y}} + z_{6}c \, \mathbf{\hat{z}} & \left(16b\right) & \mbox{H III} \\ 
\mathbf{B}_{43} & = & \left(\frac{3}{4} +x_{6} + z_{6}\right) \, \mathbf{a}_{1} + \left(\frac{1}{4} - y_{6} + z_{6}\right) \, \mathbf{a}_{2} + \left(\frac{1}{2} +x_{6} - y_{6}\right) \, \mathbf{a}_{3} & = & -y_{6}a \, \mathbf{\hat{x}} + \left(\frac{1}{2} +x_{6}\right)a \, \mathbf{\hat{y}} + \left(\frac{1}{4} +z_{6}\right)c \, \mathbf{\hat{z}} & \left(16b\right) & \mbox{H III} \\ 
\mathbf{B}_{44} & = & \left(\frac{3}{4} - x_{6} + z_{6}\right) \, \mathbf{a}_{1} + \left(\frac{1}{4} +y_{6} + z_{6}\right) \, \mathbf{a}_{2} + \left(\frac{1}{2} - x_{6} + y_{6}\right) \, \mathbf{a}_{3} & = & y_{6}a \, \mathbf{\hat{x}} + \left(\frac{1}{2} - x_{6}\right)a \, \mathbf{\hat{y}} + \left(\frac{1}{4} +z_{6}\right)c \, \mathbf{\hat{z}} & \left(16b\right) & \mbox{H III} \\ 
\mathbf{B}_{45} & = & \left(\frac{1}{2} - y_{6} + z_{6}\right) \, \mathbf{a}_{1} + \left(\frac{1}{2} +x_{6} + z_{6}\right) \, \mathbf{a}_{2} + \left(x_{6}-y_{6}\right) \, \mathbf{a}_{3} & = & x_{6}a \, \mathbf{\hat{x}}-y_{6}a \, \mathbf{\hat{y}} + \left(\frac{1}{2} +z_{6}\right)c \, \mathbf{\hat{z}} & \left(16b\right) & \mbox{H III} \\ 
\mathbf{B}_{46} & = & \left(\frac{1}{2} +y_{6} + z_{6}\right) \, \mathbf{a}_{1} + \left(\frac{1}{2} - x_{6} + z_{6}\right) \, \mathbf{a}_{2} + \left(-x_{6}+y_{6}\right) \, \mathbf{a}_{3} & = & -x_{6}a \, \mathbf{\hat{x}} + y_{6}a \, \mathbf{\hat{y}} + \left(\frac{1}{2} +z_{6}\right)c \, \mathbf{\hat{z}} & \left(16b\right) & \mbox{H III} \\ 
\mathbf{B}_{47} & = & \left(\frac{1}{4} - x_{6} + z_{6}\right) \, \mathbf{a}_{1} + \left(\frac{3}{4} - y_{6} + z_{6}\right) \, \mathbf{a}_{2} + \left(\frac{1}{2} - x_{6} - y_{6}\right) \, \mathbf{a}_{3} & = & \left(\frac{1}{2} - y_{6}\right)a \, \mathbf{\hat{x}}-x_{6}a \, \mathbf{\hat{y}} + \left(\frac{1}{4} +z_{6}\right)c \, \mathbf{\hat{z}} & \left(16b\right) & \mbox{H III} \\ 
\mathbf{B}_{48} & = & \left(\frac{1}{4} +x_{6} + z_{6}\right) \, \mathbf{a}_{1} + \left(\frac{3}{4} +y_{6} + z_{6}\right) \, \mathbf{a}_{2} + \left(\frac{1}{2} +x_{6} + y_{6}\right) \, \mathbf{a}_{3} & = & \left(\frac{1}{2} +y_{6}\right)a \, \mathbf{\hat{x}} + x_{6}a \, \mathbf{\hat{y}} + \left(\frac{1}{4} +z_{6}\right)c \, \mathbf{\hat{z}} & \left(16b\right) & \mbox{H III} \\ 
\mathbf{B}_{49} & = & \left(y_{7}+z_{7}\right) \, \mathbf{a}_{1} + \left(x_{7}+z_{7}\right) \, \mathbf{a}_{2} + \left(x_{7}+y_{7}\right) \, \mathbf{a}_{3} & = & x_{7}a \, \mathbf{\hat{x}} + y_{7}a \, \mathbf{\hat{y}} + z_{7}c \, \mathbf{\hat{z}} & \left(16b\right) & \mbox{H IV} \\ 
\mathbf{B}_{50} & = & \left(-y_{7}+z_{7}\right) \, \mathbf{a}_{1} + \left(-x_{7}+z_{7}\right) \, \mathbf{a}_{2} + \left(-x_{7}-y_{7}\right) \, \mathbf{a}_{3} & = & -x_{7}a \, \mathbf{\hat{x}}-y_{7}a \, \mathbf{\hat{y}} + z_{7}c \, \mathbf{\hat{z}} & \left(16b\right) & \mbox{H IV} \\ 
\mathbf{B}_{51} & = & \left(\frac{3}{4} +x_{7} + z_{7}\right) \, \mathbf{a}_{1} + \left(\frac{1}{4} - y_{7} + z_{7}\right) \, \mathbf{a}_{2} + \left(\frac{1}{2} +x_{7} - y_{7}\right) \, \mathbf{a}_{3} & = & -y_{7}a \, \mathbf{\hat{x}} + \left(\frac{1}{2} +x_{7}\right)a \, \mathbf{\hat{y}} + \left(\frac{1}{4} +z_{7}\right)c \, \mathbf{\hat{z}} & \left(16b\right) & \mbox{H IV} \\ 
\mathbf{B}_{52} & = & \left(\frac{3}{4} - x_{7} + z_{7}\right) \, \mathbf{a}_{1} + \left(\frac{1}{4} +y_{7} + z_{7}\right) \, \mathbf{a}_{2} + \left(\frac{1}{2} - x_{7} + y_{7}\right) \, \mathbf{a}_{3} & = & y_{7}a \, \mathbf{\hat{x}} + \left(\frac{1}{2} - x_{7}\right)a \, \mathbf{\hat{y}} + \left(\frac{1}{4} +z_{7}\right)c \, \mathbf{\hat{z}} & \left(16b\right) & \mbox{H IV} \\ 
\mathbf{B}_{53} & = & \left(\frac{1}{2} - y_{7} + z_{7}\right) \, \mathbf{a}_{1} + \left(\frac{1}{2} +x_{7} + z_{7}\right) \, \mathbf{a}_{2} + \left(x_{7}-y_{7}\right) \, \mathbf{a}_{3} & = & x_{7}a \, \mathbf{\hat{x}}-y_{7}a \, \mathbf{\hat{y}} + \left(\frac{1}{2} +z_{7}\right)c \, \mathbf{\hat{z}} & \left(16b\right) & \mbox{H IV} \\ 
\mathbf{B}_{54} & = & \left(\frac{1}{2} +y_{7} + z_{7}\right) \, \mathbf{a}_{1} + \left(\frac{1}{2} - x_{7} + z_{7}\right) \, \mathbf{a}_{2} + \left(-x_{7}+y_{7}\right) \, \mathbf{a}_{3} & = & -x_{7}a \, \mathbf{\hat{x}} + y_{7}a \, \mathbf{\hat{y}} + \left(\frac{1}{2} +z_{7}\right)c \, \mathbf{\hat{z}} & \left(16b\right) & \mbox{H IV} \\ 
\mathbf{B}_{55} & = & \left(\frac{1}{4} - x_{7} + z_{7}\right) \, \mathbf{a}_{1} + \left(\frac{3}{4} - y_{7} + z_{7}\right) \, \mathbf{a}_{2} + \left(\frac{1}{2} - x_{7} - y_{7}\right) \, \mathbf{a}_{3} & = & \left(\frac{1}{2} - y_{7}\right)a \, \mathbf{\hat{x}}-x_{7}a \, \mathbf{\hat{y}} + \left(\frac{1}{4} +z_{7}\right)c \, \mathbf{\hat{z}} & \left(16b\right) & \mbox{H IV} \\ 
\mathbf{B}_{56} & = & \left(\frac{1}{4} +x_{7} + z_{7}\right) \, \mathbf{a}_{1} + \left(\frac{3}{4} +y_{7} + z_{7}\right) \, \mathbf{a}_{2} + \left(\frac{1}{2} +x_{7} + y_{7}\right) \, \mathbf{a}_{3} & = & \left(\frac{1}{2} +y_{7}\right)a \, \mathbf{\hat{x}} + x_{7}a \, \mathbf{\hat{y}} + \left(\frac{1}{4} +z_{7}\right)c \, \mathbf{\hat{z}} & \left(16b\right) & \mbox{H IV} \\ 
\mathbf{B}_{57} & = & \left(y_{8}+z_{8}\right) \, \mathbf{a}_{1} + \left(x_{8}+z_{8}\right) \, \mathbf{a}_{2} + \left(x_{8}+y_{8}\right) \, \mathbf{a}_{3} & = & x_{8}a \, \mathbf{\hat{x}} + y_{8}a \, \mathbf{\hat{y}} + z_{8}c \, \mathbf{\hat{z}} & \left(16b\right) & \mbox{H V} \\ 
\mathbf{B}_{58} & = & \left(-y_{8}+z_{8}\right) \, \mathbf{a}_{1} + \left(-x_{8}+z_{8}\right) \, \mathbf{a}_{2} + \left(-x_{8}-y_{8}\right) \, \mathbf{a}_{3} & = & -x_{8}a \, \mathbf{\hat{x}}-y_{8}a \, \mathbf{\hat{y}} + z_{8}c \, \mathbf{\hat{z}} & \left(16b\right) & \mbox{H V} \\ 
\mathbf{B}_{59} & = & \left(\frac{3}{4} +x_{8} + z_{8}\right) \, \mathbf{a}_{1} + \left(\frac{1}{4} - y_{8} + z_{8}\right) \, \mathbf{a}_{2} + \left(\frac{1}{2} +x_{8} - y_{8}\right) \, \mathbf{a}_{3} & = & -y_{8}a \, \mathbf{\hat{x}} + \left(\frac{1}{2} +x_{8}\right)a \, \mathbf{\hat{y}} + \left(\frac{1}{4} +z_{8}\right)c \, \mathbf{\hat{z}} & \left(16b\right) & \mbox{H V} \\ 
\mathbf{B}_{60} & = & \left(\frac{3}{4} - x_{8} + z_{8}\right) \, \mathbf{a}_{1} + \left(\frac{1}{4} +y_{8} + z_{8}\right) \, \mathbf{a}_{2} + \left(\frac{1}{2} - x_{8} + y_{8}\right) \, \mathbf{a}_{3} & = & y_{8}a \, \mathbf{\hat{x}} + \left(\frac{1}{2} - x_{8}\right)a \, \mathbf{\hat{y}} + \left(\frac{1}{4} +z_{8}\right)c \, \mathbf{\hat{z}} & \left(16b\right) & \mbox{H V} \\ 
\mathbf{B}_{61} & = & \left(\frac{1}{2} - y_{8} + z_{8}\right) \, \mathbf{a}_{1} + \left(\frac{1}{2} +x_{8} + z_{8}\right) \, \mathbf{a}_{2} + \left(x_{8}-y_{8}\right) \, \mathbf{a}_{3} & = & x_{8}a \, \mathbf{\hat{x}}-y_{8}a \, \mathbf{\hat{y}} + \left(\frac{1}{2} +z_{8}\right)c \, \mathbf{\hat{z}} & \left(16b\right) & \mbox{H V} \\ 
\mathbf{B}_{62} & = & \left(\frac{1}{2} +y_{8} + z_{8}\right) \, \mathbf{a}_{1} + \left(\frac{1}{2} - x_{8} + z_{8}\right) \, \mathbf{a}_{2} + \left(-x_{8}+y_{8}\right) \, \mathbf{a}_{3} & = & -x_{8}a \, \mathbf{\hat{x}} + y_{8}a \, \mathbf{\hat{y}} + \left(\frac{1}{2} +z_{8}\right)c \, \mathbf{\hat{z}} & \left(16b\right) & \mbox{H V} \\ 
\mathbf{B}_{63} & = & \left(\frac{1}{4} - x_{8} + z_{8}\right) \, \mathbf{a}_{1} + \left(\frac{3}{4} - y_{8} + z_{8}\right) \, \mathbf{a}_{2} + \left(\frac{1}{2} - x_{8} - y_{8}\right) \, \mathbf{a}_{3} & = & \left(\frac{1}{2} - y_{8}\right)a \, \mathbf{\hat{x}}-x_{8}a \, \mathbf{\hat{y}} + \left(\frac{1}{4} +z_{8}\right)c \, \mathbf{\hat{z}} & \left(16b\right) & \mbox{H V} \\ 
\mathbf{B}_{64} & = & \left(\frac{1}{4} +x_{8} + z_{8}\right) \, \mathbf{a}_{1} + \left(\frac{3}{4} +y_{8} + z_{8}\right) \, \mathbf{a}_{2} + \left(\frac{1}{2} +x_{8} + y_{8}\right) \, \mathbf{a}_{3} & = & \left(\frac{1}{2} +y_{8}\right)a \, \mathbf{\hat{x}} + x_{8}a \, \mathbf{\hat{y}} + \left(\frac{1}{4} +z_{8}\right)c \, \mathbf{\hat{z}} & \left(16b\right) & \mbox{H V} \\ 
\mathbf{B}_{65} & = & \left(y_{9}+z_{9}\right) \, \mathbf{a}_{1} + \left(x_{9}+z_{9}\right) \, \mathbf{a}_{2} + \left(x_{9}+y_{9}\right) \, \mathbf{a}_{3} & = & x_{9}a \, \mathbf{\hat{x}} + y_{9}a \, \mathbf{\hat{y}} + z_{9}c \, \mathbf{\hat{z}} & \left(16b\right) & \mbox{H VI} \\ 
\mathbf{B}_{66} & = & \left(-y_{9}+z_{9}\right) \, \mathbf{a}_{1} + \left(-x_{9}+z_{9}\right) \, \mathbf{a}_{2} + \left(-x_{9}-y_{9}\right) \, \mathbf{a}_{3} & = & -x_{9}a \, \mathbf{\hat{x}}-y_{9}a \, \mathbf{\hat{y}} + z_{9}c \, \mathbf{\hat{z}} & \left(16b\right) & \mbox{H VI} \\ 
\mathbf{B}_{67} & = & \left(\frac{3}{4} +x_{9} + z_{9}\right) \, \mathbf{a}_{1} + \left(\frac{1}{4} - y_{9} + z_{9}\right) \, \mathbf{a}_{2} + \left(\frac{1}{2} +x_{9} - y_{9}\right) \, \mathbf{a}_{3} & = & -y_{9}a \, \mathbf{\hat{x}} + \left(\frac{1}{2} +x_{9}\right)a \, \mathbf{\hat{y}} + \left(\frac{1}{4} +z_{9}\right)c \, \mathbf{\hat{z}} & \left(16b\right) & \mbox{H VI} \\ 
\mathbf{B}_{68} & = & \left(\frac{3}{4} - x_{9} + z_{9}\right) \, \mathbf{a}_{1} + \left(\frac{1}{4} +y_{9} + z_{9}\right) \, \mathbf{a}_{2} + \left(\frac{1}{2} - x_{9} + y_{9}\right) \, \mathbf{a}_{3} & = & y_{9}a \, \mathbf{\hat{x}} + \left(\frac{1}{2} - x_{9}\right)a \, \mathbf{\hat{y}} + \left(\frac{1}{4} +z_{9}\right)c \, \mathbf{\hat{z}} & \left(16b\right) & \mbox{H VI} \\ 
\mathbf{B}_{69} & = & \left(\frac{1}{2} - y_{9} + z_{9}\right) \, \mathbf{a}_{1} + \left(\frac{1}{2} +x_{9} + z_{9}\right) \, \mathbf{a}_{2} + \left(x_{9}-y_{9}\right) \, \mathbf{a}_{3} & = & x_{9}a \, \mathbf{\hat{x}}-y_{9}a \, \mathbf{\hat{y}} + \left(\frac{1}{2} +z_{9}\right)c \, \mathbf{\hat{z}} & \left(16b\right) & \mbox{H VI} \\ 
\mathbf{B}_{70} & = & \left(\frac{1}{2} +y_{9} + z_{9}\right) \, \mathbf{a}_{1} + \left(\frac{1}{2} - x_{9} + z_{9}\right) \, \mathbf{a}_{2} + \left(-x_{9}+y_{9}\right) \, \mathbf{a}_{3} & = & -x_{9}a \, \mathbf{\hat{x}} + y_{9}a \, \mathbf{\hat{y}} + \left(\frac{1}{2} +z_{9}\right)c \, \mathbf{\hat{z}} & \left(16b\right) & \mbox{H VI} \\ 
\mathbf{B}_{71} & = & \left(\frac{1}{4} - x_{9} + z_{9}\right) \, \mathbf{a}_{1} + \left(\frac{3}{4} - y_{9} + z_{9}\right) \, \mathbf{a}_{2} + \left(\frac{1}{2} - x_{9} - y_{9}\right) \, \mathbf{a}_{3} & = & \left(\frac{1}{2} - y_{9}\right)a \, \mathbf{\hat{x}}-x_{9}a \, \mathbf{\hat{y}} + \left(\frac{1}{4} +z_{9}\right)c \, \mathbf{\hat{z}} & \left(16b\right) & \mbox{H VI} \\ 
\mathbf{B}_{72} & = & \left(\frac{1}{4} +x_{9} + z_{9}\right) \, \mathbf{a}_{1} + \left(\frac{3}{4} +y_{9} + z_{9}\right) \, \mathbf{a}_{2} + \left(\frac{1}{2} +x_{9} + y_{9}\right) \, \mathbf{a}_{3} & = & \left(\frac{1}{2} +y_{9}\right)a \, \mathbf{\hat{x}} + x_{9}a \, \mathbf{\hat{y}} + \left(\frac{1}{4} +z_{9}\right)c \, \mathbf{\hat{z}} & \left(16b\right) & \mbox{H VI} \\ 
\mathbf{B}_{73} & = & \left(y_{10}+z_{10}\right) \, \mathbf{a}_{1} + \left(x_{10}+z_{10}\right) \, \mathbf{a}_{2} + \left(x_{10}+y_{10}\right) \, \mathbf{a}_{3} & = & x_{10}a \, \mathbf{\hat{x}} + y_{10}a \, \mathbf{\hat{y}} + z_{10}c \, \mathbf{\hat{z}} & \left(16b\right) & \mbox{H VII} \\ 
\mathbf{B}_{74} & = & \left(-y_{10}+z_{10}\right) \, \mathbf{a}_{1} + \left(-x_{10}+z_{10}\right) \, \mathbf{a}_{2} + \left(-x_{10}-y_{10}\right) \, \mathbf{a}_{3} & = & -x_{10}a \, \mathbf{\hat{x}}-y_{10}a \, \mathbf{\hat{y}} + z_{10}c \, \mathbf{\hat{z}} & \left(16b\right) & \mbox{H VII} \\ 
\mathbf{B}_{75} & = & \left(\frac{3}{4} +x_{10} + z_{10}\right) \, \mathbf{a}_{1} + \left(\frac{1}{4} - y_{10} + z_{10}\right) \, \mathbf{a}_{2} + \left(\frac{1}{2} +x_{10} - y_{10}\right) \, \mathbf{a}_{3} & = & -y_{10}a \, \mathbf{\hat{x}} + \left(\frac{1}{2} +x_{10}\right)a \, \mathbf{\hat{y}} + \left(\frac{1}{4} +z_{10}\right)c \, \mathbf{\hat{z}} & \left(16b\right) & \mbox{H VII} \\ 
\mathbf{B}_{76} & = & \left(\frac{3}{4} - x_{10} + z_{10}\right) \, \mathbf{a}_{1} + \left(\frac{1}{4} +y_{10} + z_{10}\right) \, \mathbf{a}_{2} + \left(\frac{1}{2} - x_{10} + y_{10}\right) \, \mathbf{a}_{3} & = & y_{10}a \, \mathbf{\hat{x}} + \left(\frac{1}{2} - x_{10}\right)a \, \mathbf{\hat{y}} + \left(\frac{1}{4} +z_{10}\right)c \, \mathbf{\hat{z}} & \left(16b\right) & \mbox{H VII} \\ 
\mathbf{B}_{77} & = & \left(\frac{1}{2} - y_{10} + z_{10}\right) \, \mathbf{a}_{1} + \left(\frac{1}{2} +x_{10} + z_{10}\right) \, \mathbf{a}_{2} + \left(x_{10}-y_{10}\right) \, \mathbf{a}_{3} & = & x_{10}a \, \mathbf{\hat{x}}-y_{10}a \, \mathbf{\hat{y}} + \left(\frac{1}{2} +z_{10}\right)c \, \mathbf{\hat{z}} & \left(16b\right) & \mbox{H VII} \\ 
\mathbf{B}_{78} & = & \left(\frac{1}{2} +y_{10} + z_{10}\right) \, \mathbf{a}_{1} + \left(\frac{1}{2} - x_{10} + z_{10}\right) \, \mathbf{a}_{2} + \left(-x_{10}+y_{10}\right) \, \mathbf{a}_{3} & = & -x_{10}a \, \mathbf{\hat{x}} + y_{10}a \, \mathbf{\hat{y}} + \left(\frac{1}{2} +z_{10}\right)c \, \mathbf{\hat{z}} & \left(16b\right) & \mbox{H VII} \\ 
\mathbf{B}_{79} & = & \left(\frac{1}{4} - x_{10} + z_{10}\right) \, \mathbf{a}_{1} + \left(\frac{3}{4} - y_{10} + z_{10}\right) \, \mathbf{a}_{2} + \left(\frac{1}{2} - x_{10} - y_{10}\right) \, \mathbf{a}_{3} & = & \left(\frac{1}{2} - y_{10}\right)a \, \mathbf{\hat{x}}-x_{10}a \, \mathbf{\hat{y}} + \left(\frac{1}{4} +z_{10}\right)c \, \mathbf{\hat{z}} & \left(16b\right) & \mbox{H VII} \\ 
\mathbf{B}_{80} & = & \left(\frac{1}{4} +x_{10} + z_{10}\right) \, \mathbf{a}_{1} + \left(\frac{3}{4} +y_{10} + z_{10}\right) \, \mathbf{a}_{2} + \left(\frac{1}{2} +x_{10} + y_{10}\right) \, \mathbf{a}_{3} & = & \left(\frac{1}{2} +y_{10}\right)a \, \mathbf{\hat{x}} + x_{10}a \, \mathbf{\hat{y}} + \left(\frac{1}{4} +z_{10}\right)c \, \mathbf{\hat{z}} & \left(16b\right) & \mbox{H VII} \\ 
\mathbf{B}_{81} & = & \left(y_{11}+z_{11}\right) \, \mathbf{a}_{1} + \left(x_{11}+z_{11}\right) \, \mathbf{a}_{2} + \left(x_{11}+y_{11}\right) \, \mathbf{a}_{3} & = & x_{11}a \, \mathbf{\hat{x}} + y_{11}a \, \mathbf{\hat{y}} + z_{11}c \, \mathbf{\hat{z}} & \left(16b\right) & \mbox{H VIII} \\ 
\mathbf{B}_{82} & = & \left(-y_{11}+z_{11}\right) \, \mathbf{a}_{1} + \left(-x_{11}+z_{11}\right) \, \mathbf{a}_{2} + \left(-x_{11}-y_{11}\right) \, \mathbf{a}_{3} & = & -x_{11}a \, \mathbf{\hat{x}}-y_{11}a \, \mathbf{\hat{y}} + z_{11}c \, \mathbf{\hat{z}} & \left(16b\right) & \mbox{H VIII} \\ 
\mathbf{B}_{83} & = & \left(\frac{3}{4} +x_{11} + z_{11}\right) \, \mathbf{a}_{1} + \left(\frac{1}{4} - y_{11} + z_{11}\right) \, \mathbf{a}_{2} + \left(\frac{1}{2} +x_{11} - y_{11}\right) \, \mathbf{a}_{3} & = & -y_{11}a \, \mathbf{\hat{x}} + \left(\frac{1}{2} +x_{11}\right)a \, \mathbf{\hat{y}} + \left(\frac{1}{4} +z_{11}\right)c \, \mathbf{\hat{z}} & \left(16b\right) & \mbox{H VIII} \\ 
\mathbf{B}_{84} & = & \left(\frac{3}{4} - x_{11} + z_{11}\right) \, \mathbf{a}_{1} + \left(\frac{1}{4} +y_{11} + z_{11}\right) \, \mathbf{a}_{2} + \left(\frac{1}{2} - x_{11} + y_{11}\right) \, \mathbf{a}_{3} & = & y_{11}a \, \mathbf{\hat{x}} + \left(\frac{1}{2} - x_{11}\right)a \, \mathbf{\hat{y}} + \left(\frac{1}{4} +z_{11}\right)c \, \mathbf{\hat{z}} & \left(16b\right) & \mbox{H VIII} \\ 
\mathbf{B}_{85} & = & \left(\frac{1}{2} - y_{11} + z_{11}\right) \, \mathbf{a}_{1} + \left(\frac{1}{2} +x_{11} + z_{11}\right) \, \mathbf{a}_{2} + \left(x_{11}-y_{11}\right) \, \mathbf{a}_{3} & = & x_{11}a \, \mathbf{\hat{x}}-y_{11}a \, \mathbf{\hat{y}} + \left(\frac{1}{2} +z_{11}\right)c \, \mathbf{\hat{z}} & \left(16b\right) & \mbox{H VIII} \\ 
\mathbf{B}_{86} & = & \left(\frac{1}{2} +y_{11} + z_{11}\right) \, \mathbf{a}_{1} + \left(\frac{1}{2} - x_{11} + z_{11}\right) \, \mathbf{a}_{2} + \left(-x_{11}+y_{11}\right) \, \mathbf{a}_{3} & = & -x_{11}a \, \mathbf{\hat{x}} + y_{11}a \, \mathbf{\hat{y}} + \left(\frac{1}{2} +z_{11}\right)c \, \mathbf{\hat{z}} & \left(16b\right) & \mbox{H VIII} \\ 
\mathbf{B}_{87} & = & \left(\frac{1}{4} - x_{11} + z_{11}\right) \, \mathbf{a}_{1} + \left(\frac{3}{4} - y_{11} + z_{11}\right) \, \mathbf{a}_{2} + \left(\frac{1}{2} - x_{11} - y_{11}\right) \, \mathbf{a}_{3} & = & \left(\frac{1}{2} - y_{11}\right)a \, \mathbf{\hat{x}}-x_{11}a \, \mathbf{\hat{y}} + \left(\frac{1}{4} +z_{11}\right)c \, \mathbf{\hat{z}} & \left(16b\right) & \mbox{H VIII} \\ 
\mathbf{B}_{88} & = & \left(\frac{1}{4} +x_{11} + z_{11}\right) \, \mathbf{a}_{1} + \left(\frac{3}{4} +y_{11} + z_{11}\right) \, \mathbf{a}_{2} + \left(\frac{1}{2} +x_{11} + y_{11}\right) \, \mathbf{a}_{3} & = & \left(\frac{1}{2} +y_{11}\right)a \, \mathbf{\hat{x}} + x_{11}a \, \mathbf{\hat{y}} + \left(\frac{1}{4} +z_{11}\right)c \, \mathbf{\hat{z}} & \left(16b\right) & \mbox{H VIII} \\ 
\end{longtabu}
\renewcommand{\arraystretch}{1.0}
\noindent \hrulefill
\\
\textbf{References:}
\vspace*{-0.25cm}
\begin{flushleft}
  - \bibentry{Marynick_BeBH42_InorgChem_1972}. \\
\end{flushleft}
\textbf{Found in:}
\vspace*{-0.25cm}
\begin{flushleft}
  - \bibentry{Villars_PearsonsCrystalData_2013}. \\
\end{flushleft}
\noindent \hrulefill
\\
\textbf{Geometry files:}
\\
\noindent  - CIF: pp. {\hyperref[A2BC8_tI176_110_2b_b_8b_cif]{\pageref{A2BC8_tI176_110_2b_b_8b_cif}}} \\
\noindent  - POSCAR: pp. {\hyperref[A2BC8_tI176_110_2b_b_8b_poscar]{\pageref{A2BC8_tI176_110_2b_b_8b_poscar}}} \\
\onecolumn
{\phantomsection\label{A2B_tP12_111_2n_adf}}
\subsection*{\huge \textbf{{\normalfont MnF$_{2}$ Structure: A2B\_tP12\_111\_2n\_adf}}}
\noindent \hrulefill
\vspace*{0.25cm}
\begin{figure}[htp]
  \centering
  \vspace{-1em}
  {\includegraphics[width=1\textwidth]{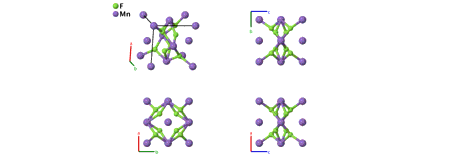}}
\end{figure}
\vspace*{-0.5cm}
\renewcommand{\arraystretch}{1.5}
\begin{equation*}
  \begin{array}{>{$\hspace{-0.15cm}}l<{$}>{$}p{0.5cm}<{$}>{$}p{18.5cm}<{$}}
    \mbox{\large \textbf{Prototype}} &\colon & \ce{MnF2} \\
    \mbox{\large \textbf{\AFLOW\ prototype label}} &\colon & \mbox{A2B\_tP12\_111\_2n\_adf} \\
    \mbox{\large \textbf{\textit{Strukturbericht} designation}} &\colon & \mbox{None} \\
    \mbox{\large \textbf{Pearson symbol}} &\colon & \mbox{tP12} \\
    \mbox{\large \textbf{Space group number}} &\colon & 111 \\
    \mbox{\large \textbf{Space group symbol}} &\colon & P\bar{4}2m \\
    \mbox{\large \textbf{\AFLOW\ prototype command}} &\colon &  \texttt{aflow} \,  \, \texttt{-{}-proto=A2B\_tP12\_111\_2n\_adf } \, \newline \texttt{-{}-params=}{a,c/a,x_{4},z_{4},x_{5},z_{5} }
  \end{array}
\end{equation*}
\renewcommand{\arraystretch}{1.0}

\noindent \parbox{1 \linewidth}{
\noindent \hrulefill
\\
\textbf{Simple Tetragonal primitive vectors:} \\
\vspace*{-0.25cm}
\begin{tabular}{cc}
  \begin{tabular}{c}
    \parbox{0.6 \linewidth}{
      \renewcommand{\arraystretch}{1.5}
      \begin{equation*}
        \centering
        \begin{array}{ccc}
              \mathbf{a}_1 & = & a \, \mathbf{\hat{x}} \\
    \mathbf{a}_2 & = & a \, \mathbf{\hat{y}} \\
    \mathbf{a}_3 & = & c \, \mathbf{\hat{z}} \\

        \end{array}
      \end{equation*}
    }
    \renewcommand{\arraystretch}{1.0}
  \end{tabular}
  \begin{tabular}{c}
    \includegraphics[width=0.3\linewidth]{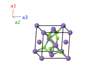} \\
  \end{tabular}
\end{tabular}

}
\vspace*{-0.25cm}

\noindent \hrulefill
\\
\textbf{Basis vectors:}
\vspace*{-0.25cm}
\renewcommand{\arraystretch}{1.5}
\begin{longtabu} to \textwidth{>{\centering $}X[-1,c,c]<{$}>{\centering $}X[-1,c,c]<{$}>{\centering $}X[-1,c,c]<{$}>{\centering $}X[-1,c,c]<{$}>{\centering $}X[-1,c,c]<{$}>{\centering $}X[-1,c,c]<{$}>{\centering $}X[-1,c,c]<{$}}
  & & \mbox{Lattice Coordinates} & & \mbox{Cartesian Coordinates} &\mbox{Wyckoff Position} & \mbox{Atom Type} \\  
  \mathbf{B}_{1} & = & 0 \, \mathbf{a}_{1} + 0 \, \mathbf{a}_{2} + 0 \, \mathbf{a}_{3} & = & 0 \, \mathbf{\hat{x}} + 0 \, \mathbf{\hat{y}} + 0 \, \mathbf{\hat{z}} & \left(1a\right) & \mbox{Mn I} \\ 
\mathbf{B}_{2} & = & \frac{1}{2} \, \mathbf{a}_{1} + \frac{1}{2} \, \mathbf{a}_{2} & = & \frac{1}{2}a \, \mathbf{\hat{x}} + \frac{1}{2}a \, \mathbf{\hat{y}} & \left(1d\right) & \mbox{Mn II} \\ 
\mathbf{B}_{3} & = & \frac{1}{2} \, \mathbf{a}_{1} + \frac{1}{2} \, \mathbf{a}_{3} & = & \frac{1}{2}a \, \mathbf{\hat{x}} + \frac{1}{2}c \, \mathbf{\hat{z}} & \left(2f\right) & \mbox{Mn III} \\ 
\mathbf{B}_{4} & = & \frac{1}{2} \, \mathbf{a}_{2} + \frac{1}{2} \, \mathbf{a}_{3} & = & \frac{1}{2}a \, \mathbf{\hat{y}} + \frac{1}{2}c \, \mathbf{\hat{z}} & \left(2f\right) & \mbox{Mn III} \\ 
\mathbf{B}_{5} & = & x_{4} \, \mathbf{a}_{1} + x_{4} \, \mathbf{a}_{2} + z_{4} \, \mathbf{a}_{3} & = & x_{4}a \, \mathbf{\hat{x}} + x_{4}a \, \mathbf{\hat{y}} + z_{4}c \, \mathbf{\hat{z}} & \left(4n\right) & \mbox{F I} \\ 
\mathbf{B}_{6} & = & -x_{4} \, \mathbf{a}_{1}-x_{4} \, \mathbf{a}_{2} + z_{4} \, \mathbf{a}_{3} & = & -x_{4}a \, \mathbf{\hat{x}}-x_{4}a \, \mathbf{\hat{y}} + z_{4}c \, \mathbf{\hat{z}} & \left(4n\right) & \mbox{F I} \\ 
\mathbf{B}_{7} & = & x_{4} \, \mathbf{a}_{1}-x_{4} \, \mathbf{a}_{2}-z_{4} \, \mathbf{a}_{3} & = & x_{4}a \, \mathbf{\hat{x}}-x_{4}a \, \mathbf{\hat{y}}-z_{4}c \, \mathbf{\hat{z}} & \left(4n\right) & \mbox{F I} \\ 
\mathbf{B}_{8} & = & -x_{4} \, \mathbf{a}_{1} + x_{4} \, \mathbf{a}_{2}-z_{4} \, \mathbf{a}_{3} & = & -x_{4}a \, \mathbf{\hat{x}} + x_{4}a \, \mathbf{\hat{y}}-z_{4}c \, \mathbf{\hat{z}} & \left(4n\right) & \mbox{F I} \\ 
\mathbf{B}_{9} & = & x_{5} \, \mathbf{a}_{1} + x_{5} \, \mathbf{a}_{2} + z_{5} \, \mathbf{a}_{3} & = & x_{5}a \, \mathbf{\hat{x}} + x_{5}a \, \mathbf{\hat{y}} + z_{5}c \, \mathbf{\hat{z}} & \left(4n\right) & \mbox{F II} \\ 
\mathbf{B}_{10} & = & -x_{5} \, \mathbf{a}_{1}-x_{5} \, \mathbf{a}_{2} + z_{5} \, \mathbf{a}_{3} & = & -x_{5}a \, \mathbf{\hat{x}}-x_{5}a \, \mathbf{\hat{y}} + z_{5}c \, \mathbf{\hat{z}} & \left(4n\right) & \mbox{F II} \\ 
\mathbf{B}_{11} & = & x_{5} \, \mathbf{a}_{1}-x_{5} \, \mathbf{a}_{2}-z_{5} \, \mathbf{a}_{3} & = & x_{5}a \, \mathbf{\hat{x}}-x_{5}a \, \mathbf{\hat{y}}-z_{5}c \, \mathbf{\hat{z}} & \left(4n\right) & \mbox{F II} \\ 
\mathbf{B}_{12} & = & -x_{5} \, \mathbf{a}_{1} + x_{5} \, \mathbf{a}_{2}-z_{5} \, \mathbf{a}_{3} & = & -x_{5}a \, \mathbf{\hat{x}} + x_{5}a \, \mathbf{\hat{y}}-z_{5}c \, \mathbf{\hat{z}} & \left(4n\right) & \mbox{F II} \\ 
\end{longtabu}
\renewcommand{\arraystretch}{1.0}
\noindent \hrulefill
\\
\textbf{References:}
\vspace*{-0.25cm}
\begin{flushleft}
  - \bibentry{Yagi_JGeoPhysResSolidEarth_MnF2_1979}. \\
\end{flushleft}
\textbf{Found in:}
\vspace*{-0.25cm}
\begin{flushleft}
  - \bibentry{Villars_PearsonsCrystalData_2013}. \\
\end{flushleft}
\noindent \hrulefill
\\
\textbf{Geometry files:}
\\
\noindent  - CIF: pp. {\hyperref[A2B_tP12_111_2n_adf_cif]{\pageref{A2B_tP12_111_2n_adf_cif}}} \\
\noindent  - POSCAR: pp. {\hyperref[A2B_tP12_111_2n_adf_poscar]{\pageref{A2B_tP12_111_2n_adf_poscar}}} \\
\onecolumn
{\phantomsection\label{AB_tP8_111_n_n}}
\subsection*{\huge \textbf{{\normalfont NV (Low-temperature) Structure: AB\_tP8\_111\_n\_n}}}
\noindent \hrulefill
\vspace*{0.25cm}
\begin{figure}[htp]
  \centering
  \vspace{-1em}
  {\includegraphics[width=1\textwidth]{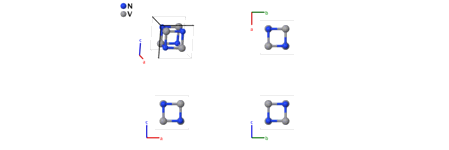}}
\end{figure}
\vspace*{-0.5cm}
\renewcommand{\arraystretch}{1.5}
\begin{equation*}
  \begin{array}{>{$\hspace{-0.15cm}}l<{$}>{$}p{0.5cm}<{$}>{$}p{18.5cm}<{$}}
    \mbox{\large \textbf{Prototype}} &\colon & \ce{VN} \\
    \mbox{\large \textbf{\AFLOW\ prototype label}} &\colon & \mbox{AB\_tP8\_111\_n\_n} \\
    \mbox{\large \textbf{\textit{Strukturbericht} designation}} &\colon & \mbox{None} \\
    \mbox{\large \textbf{Pearson symbol}} &\colon & \mbox{tP8} \\
    \mbox{\large \textbf{Space group number}} &\colon & 111 \\
    \mbox{\large \textbf{Space group symbol}} &\colon & P\bar{4}2m \\
    \mbox{\large \textbf{\AFLOW\ prototype command}} &\colon &  \texttt{aflow} \,  \, \texttt{-{}-proto=AB\_tP8\_111\_n\_n } \, \newline \texttt{-{}-params=}{a,c/a,x_{1},z_{1},x_{2},z_{2} }
  \end{array}
\end{equation*}
\renewcommand{\arraystretch}{1.0}

\vspace*{-0.25cm}
\noindent \hrulefill
\begin{itemize}
  \item{{\small FINDSYM} identifies space group \#111 for this structure (consistent with the reference); however, since $c/a \approx 1$, 
{\small AFLOW-SYM} and Platon identify \#215.
Lowering the tolerance value for {\small AFLOW-SYM} resolves the expected space group \#111.
Space groups \#111 and \#215 are both reasonable classifications since they are commensurate with subgroup relations.
}
\end{itemize}

\noindent \parbox{1 \linewidth}{
\noindent \hrulefill
\\
\textbf{Simple Tetragonal primitive vectors:} \\
\vspace*{-0.25cm}
\begin{tabular}{cc}
  \begin{tabular}{c}
    \parbox{0.6 \linewidth}{
      \renewcommand{\arraystretch}{1.5}
      \begin{equation*}
        \centering
        \begin{array}{ccc}
              \mathbf{a}_1 & = & a \, \mathbf{\hat{x}} \\
    \mathbf{a}_2 & = & a \, \mathbf{\hat{y}} \\
    \mathbf{a}_3 & = & c \, \mathbf{\hat{z}} \\

        \end{array}
      \end{equation*}
    }
    \renewcommand{\arraystretch}{1.0}
  \end{tabular}
  \begin{tabular}{c}
    \includegraphics[width=0.3\linewidth]{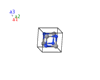} \\
  \end{tabular}
\end{tabular}

}
\vspace*{-0.25cm}

\noindent \hrulefill
\\
\textbf{Basis vectors:}
\vspace*{-0.25cm}
\renewcommand{\arraystretch}{1.5}
\begin{longtabu} to \textwidth{>{\centering $}X[-1,c,c]<{$}>{\centering $}X[-1,c,c]<{$}>{\centering $}X[-1,c,c]<{$}>{\centering $}X[-1,c,c]<{$}>{\centering $}X[-1,c,c]<{$}>{\centering $}X[-1,c,c]<{$}>{\centering $}X[-1,c,c]<{$}}
  & & \mbox{Lattice Coordinates} & & \mbox{Cartesian Coordinates} &\mbox{Wyckoff Position} & \mbox{Atom Type} \\  
  \mathbf{B}_{1} & = & x_{1} \, \mathbf{a}_{1} + x_{1} \, \mathbf{a}_{2} + z_{1} \, \mathbf{a}_{3} & = & x_{1}a \, \mathbf{\hat{x}} + x_{1}a \, \mathbf{\hat{y}} + z_{1}c \, \mathbf{\hat{z}} & \left(4n\right) & \mbox{N} \\ 
\mathbf{B}_{2} & = & -x_{1} \, \mathbf{a}_{1}-x_{1} \, \mathbf{a}_{2} + z_{1} \, \mathbf{a}_{3} & = & -x_{1}a \, \mathbf{\hat{x}}-x_{1}a \, \mathbf{\hat{y}} + z_{1}c \, \mathbf{\hat{z}} & \left(4n\right) & \mbox{N} \\ 
\mathbf{B}_{3} & = & x_{1} \, \mathbf{a}_{1}-x_{1} \, \mathbf{a}_{2}-z_{1} \, \mathbf{a}_{3} & = & x_{1}a \, \mathbf{\hat{x}}-x_{1}a \, \mathbf{\hat{y}}-z_{1}c \, \mathbf{\hat{z}} & \left(4n\right) & \mbox{N} \\ 
\mathbf{B}_{4} & = & -x_{1} \, \mathbf{a}_{1} + x_{1} \, \mathbf{a}_{2}-z_{1} \, \mathbf{a}_{3} & = & -x_{1}a \, \mathbf{\hat{x}} + x_{1}a \, \mathbf{\hat{y}}-z_{1}c \, \mathbf{\hat{z}} & \left(4n\right) & \mbox{N} \\ 
\mathbf{B}_{5} & = & x_{2} \, \mathbf{a}_{1} + x_{2} \, \mathbf{a}_{2} + z_{2} \, \mathbf{a}_{3} & = & x_{2}a \, \mathbf{\hat{x}} + x_{2}a \, \mathbf{\hat{y}} + z_{2}c \, \mathbf{\hat{z}} & \left(4n\right) & \mbox{V} \\ 
\mathbf{B}_{6} & = & -x_{2} \, \mathbf{a}_{1}-x_{2} \, \mathbf{a}_{2} + z_{2} \, \mathbf{a}_{3} & = & -x_{2}a \, \mathbf{\hat{x}}-x_{2}a \, \mathbf{\hat{y}} + z_{2}c \, \mathbf{\hat{z}} & \left(4n\right) & \mbox{V} \\ 
\mathbf{B}_{7} & = & x_{2} \, \mathbf{a}_{1}-x_{2} \, \mathbf{a}_{2}-z_{2} \, \mathbf{a}_{3} & = & x_{2}a \, \mathbf{\hat{x}}-x_{2}a \, \mathbf{\hat{y}}-z_{2}c \, \mathbf{\hat{z}} & \left(4n\right) & \mbox{V} \\ 
\mathbf{B}_{8} & = & -x_{2} \, \mathbf{a}_{1} + x_{2} \, \mathbf{a}_{2}-z_{2} \, \mathbf{a}_{3} & = & -x_{2}a \, \mathbf{\hat{x}} + x_{2}a \, \mathbf{\hat{y}}-z_{2}c \, \mathbf{\hat{z}} & \left(4n\right) & \mbox{V} \\ 
\end{longtabu}
\renewcommand{\arraystretch}{1.0}
\noindent \hrulefill
\\
\textbf{References:}
\vspace*{-0.25cm}
\begin{flushleft}
  - \bibentry{Kubel_NV_PhysRevB_1988}. \\
  - \bibentry{stokes_findsym}. \\
  - \bibentry{aflowsym_2018}. \\
  - \bibentry{platon_2003}. \\
\end{flushleft}
\textbf{Found in:}
\vspace*{-0.25cm}
\begin{flushleft}
  - \bibentry{Villars_PearsonsCrystalData_2013}. \\
\end{flushleft}
\noindent \hrulefill
\\
\textbf{Geometry files:}
\\
\noindent  - CIF: pp. {\hyperref[AB_tP8_111_n_n_cif]{\pageref{AB_tP8_111_n_n_cif}}} \\
\noindent  - POSCAR: pp. {\hyperref[AB_tP8_111_n_n_poscar]{\pageref{AB_tP8_111_n_n_poscar}}} \\
\onecolumn
{\phantomsection\label{AB4C_tP12_112_b_n_e}}
\subsection*{\huge \textbf{{\normalfont $\alpha$-CuAlCl$_{4}$ Structure: AB4C\_tP12\_112\_b\_n\_e}}}
\noindent \hrulefill
\vspace*{0.25cm}
\begin{figure}[htp]
  \centering
  \vspace{-1em}
  {\includegraphics[width=1\textwidth]{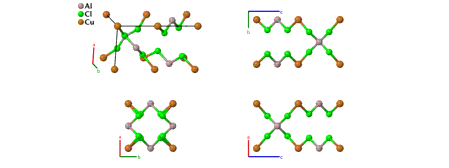}}
\end{figure}
\vspace*{-0.5cm}
\renewcommand{\arraystretch}{1.5}
\begin{equation*}
  \begin{array}{>{$\hspace{-0.15cm}}l<{$}>{$}p{0.5cm}<{$}>{$}p{18.5cm}<{$}}
    \mbox{\large \textbf{Prototype}} &\colon & \ce{$\alpha$-CuAlCl4} \\
    \mbox{\large \textbf{\AFLOW\ prototype label}} &\colon & \mbox{AB4C\_tP12\_112\_b\_n\_e} \\
    \mbox{\large \textbf{\textit{Strukturbericht} designation}} &\colon & \mbox{None} \\
    \mbox{\large \textbf{Pearson symbol}} &\colon & \mbox{tP12} \\
    \mbox{\large \textbf{Space group number}} &\colon & 112 \\
    \mbox{\large \textbf{Space group symbol}} &\colon & P\bar{4}2c \\
    \mbox{\large \textbf{\AFLOW\ prototype command}} &\colon &  \texttt{aflow} \,  \, \texttt{-{}-proto=AB4C\_tP12\_112\_b\_n\_e } \, \newline \texttt{-{}-params=}{a,c/a,x_{3},y_{3},z_{3} }
  \end{array}
\end{equation*}
\renewcommand{\arraystretch}{1.0}

\noindent \parbox{1 \linewidth}{
\noindent \hrulefill
\\
\textbf{Simple Tetragonal primitive vectors:} \\
\vspace*{-0.25cm}
\begin{tabular}{cc}
  \begin{tabular}{c}
    \parbox{0.6 \linewidth}{
      \renewcommand{\arraystretch}{1.5}
      \begin{equation*}
        \centering
        \begin{array}{ccc}
              \mathbf{a}_1 & = & a \, \mathbf{\hat{x}} \\
    \mathbf{a}_2 & = & a \, \mathbf{\hat{y}} \\
    \mathbf{a}_3 & = & c \, \mathbf{\hat{z}} \\

        \end{array}
      \end{equation*}
    }
    \renewcommand{\arraystretch}{1.0}
  \end{tabular}
  \begin{tabular}{c}
    \includegraphics[width=0.3\linewidth]{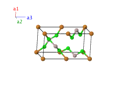} \\
  \end{tabular}
\end{tabular}

}
\vspace*{-0.25cm}

\noindent \hrulefill
\\
\textbf{Basis vectors:}
\vspace*{-0.25cm}
\renewcommand{\arraystretch}{1.5}
\begin{longtabu} to \textwidth{>{\centering $}X[-1,c,c]<{$}>{\centering $}X[-1,c,c]<{$}>{\centering $}X[-1,c,c]<{$}>{\centering $}X[-1,c,c]<{$}>{\centering $}X[-1,c,c]<{$}>{\centering $}X[-1,c,c]<{$}>{\centering $}X[-1,c,c]<{$}}
  & & \mbox{Lattice Coordinates} & & \mbox{Cartesian Coordinates} &\mbox{Wyckoff Position} & \mbox{Atom Type} \\  
  \mathbf{B}_{1} & = & \frac{1}{2} \, \mathbf{a}_{1} + \frac{1}{4} \, \mathbf{a}_{3} & = & \frac{1}{2}a \, \mathbf{\hat{x}} + \frac{1}{4}c \, \mathbf{\hat{z}} & \left(2b\right) & \mbox{Al} \\ 
\mathbf{B}_{2} & = & \frac{1}{2} \, \mathbf{a}_{2} + \frac{3}{4} \, \mathbf{a}_{3} & = & \frac{1}{2}a \, \mathbf{\hat{y}} + \frac{3}{4}c \, \mathbf{\hat{z}} & \left(2b\right) & \mbox{Al} \\ 
\mathbf{B}_{3} & = & 0 \, \mathbf{a}_{1} + 0 \, \mathbf{a}_{2} + 0 \, \mathbf{a}_{3} & = & 0 \, \mathbf{\hat{x}} + 0 \, \mathbf{\hat{y}} + 0 \, \mathbf{\hat{z}} & \left(2e\right) & \mbox{Cu} \\ 
\mathbf{B}_{4} & = & \frac{1}{2} \, \mathbf{a}_{3} & = & \frac{1}{2}c \, \mathbf{\hat{z}} & \left(2e\right) & \mbox{Cu} \\ 
\mathbf{B}_{5} & = & x_{3} \, \mathbf{a}_{1} + y_{3} \, \mathbf{a}_{2} + z_{3} \, \mathbf{a}_{3} & = & x_{3}a \, \mathbf{\hat{x}} + y_{3}a \, \mathbf{\hat{y}} + z_{3}c \, \mathbf{\hat{z}} & \left(8n\right) & \mbox{Cl} \\ 
\mathbf{B}_{6} & = & -x_{3} \, \mathbf{a}_{1}-y_{3} \, \mathbf{a}_{2} + z_{3} \, \mathbf{a}_{3} & = & -x_{3}a \, \mathbf{\hat{x}}-y_{3}a \, \mathbf{\hat{y}} + z_{3}c \, \mathbf{\hat{z}} & \left(8n\right) & \mbox{Cl} \\ 
\mathbf{B}_{7} & = & y_{3} \, \mathbf{a}_{1}-x_{3} \, \mathbf{a}_{2}-z_{3} \, \mathbf{a}_{3} & = & y_{3}a \, \mathbf{\hat{x}}-x_{3}a \, \mathbf{\hat{y}}-z_{3}c \, \mathbf{\hat{z}} & \left(8n\right) & \mbox{Cl} \\ 
\mathbf{B}_{8} & = & -y_{3} \, \mathbf{a}_{1} + x_{3} \, \mathbf{a}_{2}-z_{3} \, \mathbf{a}_{3} & = & -y_{3}a \, \mathbf{\hat{x}} + x_{3}a \, \mathbf{\hat{y}}-z_{3}c \, \mathbf{\hat{z}} & \left(8n\right) & \mbox{Cl} \\ 
\mathbf{B}_{9} & = & -x_{3} \, \mathbf{a}_{1} + y_{3} \, \mathbf{a}_{2} + \left(\frac{1}{2} - z_{3}\right) \, \mathbf{a}_{3} & = & -x_{3}a \, \mathbf{\hat{x}} + y_{3}a \, \mathbf{\hat{y}} + \left(\frac{1}{2} - z_{3}\right)c \, \mathbf{\hat{z}} & \left(8n\right) & \mbox{Cl} \\ 
\mathbf{B}_{10} & = & x_{3} \, \mathbf{a}_{1}-y_{3} \, \mathbf{a}_{2} + \left(\frac{1}{2} - z_{3}\right) \, \mathbf{a}_{3} & = & x_{3}a \, \mathbf{\hat{x}}-y_{3}a \, \mathbf{\hat{y}} + \left(\frac{1}{2} - z_{3}\right)c \, \mathbf{\hat{z}} & \left(8n\right) & \mbox{Cl} \\ 
\mathbf{B}_{11} & = & -y_{3} \, \mathbf{a}_{1}-x_{3} \, \mathbf{a}_{2} + \left(\frac{1}{2} +z_{3}\right) \, \mathbf{a}_{3} & = & -y_{3}a \, \mathbf{\hat{x}}-x_{3}a \, \mathbf{\hat{y}} + \left(\frac{1}{2} +z_{3}\right)c \, \mathbf{\hat{z}} & \left(8n\right) & \mbox{Cl} \\ 
\mathbf{B}_{12} & = & y_{3} \, \mathbf{a}_{1} + x_{3} \, \mathbf{a}_{2} + \left(\frac{1}{2} +z_{3}\right) \, \mathbf{a}_{3} & = & y_{3}a \, \mathbf{\hat{x}} + x_{3}a \, \mathbf{\hat{y}} + \left(\frac{1}{2} +z_{3}\right)c \, \mathbf{\hat{z}} & \left(8n\right) & \mbox{Cl} \\ 
\end{longtabu}
\renewcommand{\arraystretch}{1.0}
\noindent \hrulefill
\\
\textbf{References:}
\vspace*{-0.25cm}
\begin{flushleft}
  - \bibentry{Martin_AlCl4Cu_InorgChem_1998}. \\
\end{flushleft}
\textbf{Found in:}
\vspace*{-0.25cm}
\begin{flushleft}
  - \bibentry{Villars_PearsonsCrystalData_2013}. \\
\end{flushleft}
\noindent \hrulefill
\\
\textbf{Geometry files:}
\\
\noindent  - CIF: pp. {\hyperref[AB4C_tP12_112_b_n_e_cif]{\pageref{AB4C_tP12_112_b_n_e_cif}}} \\
\noindent  - POSCAR: pp. {\hyperref[AB4C_tP12_112_b_n_e_poscar]{\pageref{AB4C_tP12_112_b_n_e_poscar}}} \\
\onecolumn
{\phantomsection\label{A2BC7D2_tP24_113_e_a_cef_e}}
\subsection*{\huge \textbf{{\normalfont \begin{raggedleft}Akermanite (Ca$_2$MgSi$_2$O$_7$, $S5_{3}$) Structure: \end{raggedleft} \\ A2BC7D2\_tP24\_113\_e\_a\_cef\_e}}}
\noindent \hrulefill
\vspace*{0.25cm}
\begin{figure}[htp]
  \centering
  \vspace{-1em}
  {\includegraphics[width=1\textwidth]{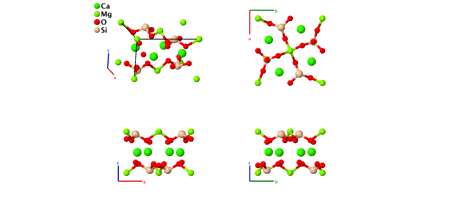}}
\end{figure}
\vspace*{-0.5cm}
\renewcommand{\arraystretch}{1.5}
\begin{equation*}
  \begin{array}{>{$\hspace{-0.15cm}}l<{$}>{$}p{0.5cm}<{$}>{$}p{18.5cm}<{$}}
    \mbox{\large \textbf{Prototype}} &\colon & \ce{Ca2MgSi2O7} \\
    \mbox{\large \textbf{\AFLOW\ prototype label}} &\colon & \mbox{A2BC7D2\_tP24\_113\_e\_a\_cef\_e} \\
    \mbox{\large \textbf{\textit{Strukturbericht} designation}} &\colon & \mbox{$S5_{3}$} \\
    \mbox{\large \textbf{Pearson symbol}} &\colon & \mbox{tP24} \\
    \mbox{\large \textbf{Space group number}} &\colon & 113 \\
    \mbox{\large \textbf{Space group symbol}} &\colon & P\bar{4}2_{1}m \\
    \mbox{\large \textbf{\AFLOW\ prototype command}} &\colon &  \texttt{aflow} \,  \, \texttt{-{}-proto=A2BC7D2\_tP24\_113\_e\_a\_cef\_e } \, \newline \texttt{-{}-params=}{a,c/a,z_{2},x_{3},z_{3},x_{4},z_{4},x_{5},z_{5},x_{6},y_{6},z_{6} }
  \end{array}
\end{equation*}
\renewcommand{\arraystretch}{1.0}

\vspace*{-0.25cm}
\noindent \hrulefill
\\
\textbf{ Other compounds with this structure:}
\begin{itemize}
   \item{ (Ca,Na)$_{2}$(Al,Mg,Fe)(Si,Al)$_{2}$O$_{7}$ (melilite), (Ca,Na)$_{2}$(Al,Mg,Fe)(Si$_{2}$O$_{7}$) (alumoakermanite), Ca$_{2}$Al$_{2}$SiO$_{7}$ (gehlenite), Ca$_{2}$BeSi$_{2}$O$_{7}$ (gugiaite), Ca$_{2}$ZnSi$_{2}$O$_{7}$ (hardystonite), Ca$_{2}$B$_{2}$SiO$_{7}$ (okayamalite), (Ca,Na)$_{2}$(Mg,Al,Si)$_{3}$O$_{7}$, Sr$_{2}$ZrSi$_{2}$O$_{7}$, Sr$_{2}$MnGe$_{2}$O$_{7}$, Sr$_{2}$MnGe$_{2}$S$_{6}$O     }
\end{itemize}
\vspace*{-0.25cm}
\noindent \hrulefill
\begin{itemize}
  \item{Akermanite is an end point of the mineral melilite, which, like
\hyperref[A7B2C2_mC22_12_aij_h_i]{thortveitite ($S2_{1}$)}, is a sorosilicate, a mineral containing
isolated Si$_{2}$O$_{7}$ or related groups.  We have followed
(Parth\'{e}, 1997) and use akermanite to represent the entire class of
materials.
We use the ambient pressure data from (Yang, 1997) to describe the
structure.
}
\end{itemize}

\noindent \parbox{1 \linewidth}{
\noindent \hrulefill
\\
\textbf{Simple Tetragonal primitive vectors:} \\
\vspace*{-0.25cm}
\begin{tabular}{cc}
  \begin{tabular}{c}
    \parbox{0.6 \linewidth}{
      \renewcommand{\arraystretch}{1.5}
      \begin{equation*}
        \centering
        \begin{array}{ccc}
              \mathbf{a}_1 & = & a \, \mathbf{\hat{x}} \\
    \mathbf{a}_2 & = & a \, \mathbf{\hat{y}} \\
    \mathbf{a}_3 & = & c \, \mathbf{\hat{z}} \\

        \end{array}
      \end{equation*}
    }
    \renewcommand{\arraystretch}{1.0}
  \end{tabular}
  \begin{tabular}{c}
    \includegraphics[width=0.3\linewidth]{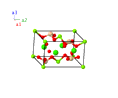} \\
  \end{tabular}
\end{tabular}

}
\vspace*{-0.25cm}

\noindent \hrulefill
\\
\textbf{Basis vectors:}
\vspace*{-0.25cm}
\renewcommand{\arraystretch}{1.5}
\begin{longtabu} to \textwidth{>{\centering $}X[-1,c,c]<{$}>{\centering $}X[-1,c,c]<{$}>{\centering $}X[-1,c,c]<{$}>{\centering $}X[-1,c,c]<{$}>{\centering $}X[-1,c,c]<{$}>{\centering $}X[-1,c,c]<{$}>{\centering $}X[-1,c,c]<{$}}
  & & \mbox{Lattice Coordinates} & & \mbox{Cartesian Coordinates} &\mbox{Wyckoff Position} & \mbox{Atom Type} \\  
  \mathbf{B}_{1} & = & 0 \, \mathbf{a}_{1} + 0 \, \mathbf{a}_{2} + 0 \, \mathbf{a}_{3} & = & 0 \, \mathbf{\hat{x}} + 0 \, \mathbf{\hat{y}} + 0 \, \mathbf{\hat{z}} & \left(2a\right) & \mbox{Mg} \\ 
\mathbf{B}_{2} & = & \frac{1}{2} \, \mathbf{a}_{1} + \frac{1}{2} \, \mathbf{a}_{2} & = & \frac{1}{2}a \, \mathbf{\hat{x}} + \frac{1}{2}a \, \mathbf{\hat{y}} & \left(2a\right) & \mbox{Mg} \\ 
\mathbf{B}_{3} & = & \frac{1}{2} \, \mathbf{a}_{2} + z_{2} \, \mathbf{a}_{3} & = & \frac{1}{2}a \, \mathbf{\hat{y}} + z_{2}c \, \mathbf{\hat{z}} & \left(2c\right) & \mbox{O I} \\ 
\mathbf{B}_{4} & = & \frac{1}{2} \, \mathbf{a}_{1} + -z_{2} \, \mathbf{a}_{3} & = & \frac{1}{2}a \, \mathbf{\hat{x}} + -z_{2}c \, \mathbf{\hat{z}} & \left(2c\right) & \mbox{O I} \\ 
\mathbf{B}_{5} & = & x_{3} \, \mathbf{a}_{1} + \left(\frac{1}{2} +x_{3}\right) \, \mathbf{a}_{2} + z_{3} \, \mathbf{a}_{3} & = & x_{3}a \, \mathbf{\hat{x}} + \left(\frac{1}{2} +x_{3}\right)a \, \mathbf{\hat{y}} + z_{3}c \, \mathbf{\hat{z}} & \left(4e\right) & \mbox{Ca} \\ 
\mathbf{B}_{6} & = & -x_{3} \, \mathbf{a}_{1} + \left(\frac{1}{2} - x_{3}\right) \, \mathbf{a}_{2} + z_{3} \, \mathbf{a}_{3} & = & -x_{3}a \, \mathbf{\hat{x}} + \left(\frac{1}{2} - x_{3}\right)a \, \mathbf{\hat{y}} + z_{3}c \, \mathbf{\hat{z}} & \left(4e\right) & \mbox{Ca} \\ 
\mathbf{B}_{7} & = & \left(\frac{1}{2} +x_{3}\right) \, \mathbf{a}_{1}-x_{3} \, \mathbf{a}_{2}-z_{3} \, \mathbf{a}_{3} & = & \left(\frac{1}{2} +x_{3}\right)a \, \mathbf{\hat{x}}-x_{3}a \, \mathbf{\hat{y}}-z_{3}c \, \mathbf{\hat{z}} & \left(4e\right) & \mbox{Ca} \\ 
\mathbf{B}_{8} & = & \left(\frac{1}{2} - x_{3}\right) \, \mathbf{a}_{1} + x_{3} \, \mathbf{a}_{2}-z_{3} \, \mathbf{a}_{3} & = & \left(\frac{1}{2} - x_{3}\right)a \, \mathbf{\hat{x}} + x_{3}a \, \mathbf{\hat{y}}-z_{3}c \, \mathbf{\hat{z}} & \left(4e\right) & \mbox{Ca} \\ 
\mathbf{B}_{9} & = & x_{4} \, \mathbf{a}_{1} + \left(\frac{1}{2} +x_{4}\right) \, \mathbf{a}_{2} + z_{4} \, \mathbf{a}_{3} & = & x_{4}a \, \mathbf{\hat{x}} + \left(\frac{1}{2} +x_{4}\right)a \, \mathbf{\hat{y}} + z_{4}c \, \mathbf{\hat{z}} & \left(4e\right) & \mbox{O II} \\ 
\mathbf{B}_{10} & = & -x_{4} \, \mathbf{a}_{1} + \left(\frac{1}{2} - x_{4}\right) \, \mathbf{a}_{2} + z_{4} \, \mathbf{a}_{3} & = & -x_{4}a \, \mathbf{\hat{x}} + \left(\frac{1}{2} - x_{4}\right)a \, \mathbf{\hat{y}} + z_{4}c \, \mathbf{\hat{z}} & \left(4e\right) & \mbox{O II} \\ 
\mathbf{B}_{11} & = & \left(\frac{1}{2} +x_{4}\right) \, \mathbf{a}_{1}-x_{4} \, \mathbf{a}_{2}-z_{4} \, \mathbf{a}_{3} & = & \left(\frac{1}{2} +x_{4}\right)a \, \mathbf{\hat{x}}-x_{4}a \, \mathbf{\hat{y}}-z_{4}c \, \mathbf{\hat{z}} & \left(4e\right) & \mbox{O II} \\ 
\mathbf{B}_{12} & = & \left(\frac{1}{2} - x_{4}\right) \, \mathbf{a}_{1} + x_{4} \, \mathbf{a}_{2}-z_{4} \, \mathbf{a}_{3} & = & \left(\frac{1}{2} - x_{4}\right)a \, \mathbf{\hat{x}} + x_{4}a \, \mathbf{\hat{y}}-z_{4}c \, \mathbf{\hat{z}} & \left(4e\right) & \mbox{O II} \\ 
\mathbf{B}_{13} & = & x_{5} \, \mathbf{a}_{1} + \left(\frac{1}{2} +x_{5}\right) \, \mathbf{a}_{2} + z_{5} \, \mathbf{a}_{3} & = & x_{5}a \, \mathbf{\hat{x}} + \left(\frac{1}{2} +x_{5}\right)a \, \mathbf{\hat{y}} + z_{5}c \, \mathbf{\hat{z}} & \left(4e\right) & \mbox{Si} \\ 
\mathbf{B}_{14} & = & -x_{5} \, \mathbf{a}_{1} + \left(\frac{1}{2} - x_{5}\right) \, \mathbf{a}_{2} + z_{5} \, \mathbf{a}_{3} & = & -x_{5}a \, \mathbf{\hat{x}} + \left(\frac{1}{2} - x_{5}\right)a \, \mathbf{\hat{y}} + z_{5}c \, \mathbf{\hat{z}} & \left(4e\right) & \mbox{Si} \\ 
\mathbf{B}_{15} & = & \left(\frac{1}{2} +x_{5}\right) \, \mathbf{a}_{1}-x_{5} \, \mathbf{a}_{2}-z_{5} \, \mathbf{a}_{3} & = & \left(\frac{1}{2} +x_{5}\right)a \, \mathbf{\hat{x}}-x_{5}a \, \mathbf{\hat{y}}-z_{5}c \, \mathbf{\hat{z}} & \left(4e\right) & \mbox{Si} \\ 
\mathbf{B}_{16} & = & \left(\frac{1}{2} - x_{5}\right) \, \mathbf{a}_{1} + x_{5} \, \mathbf{a}_{2}-z_{5} \, \mathbf{a}_{3} & = & \left(\frac{1}{2} - x_{5}\right)a \, \mathbf{\hat{x}} + x_{5}a \, \mathbf{\hat{y}}-z_{5}c \, \mathbf{\hat{z}} & \left(4e\right) & \mbox{Si} \\ 
\mathbf{B}_{17} & = & x_{6} \, \mathbf{a}_{1} + y_{6} \, \mathbf{a}_{2} + z_{6} \, \mathbf{a}_{3} & = & x_{6}a \, \mathbf{\hat{x}} + y_{6}a \, \mathbf{\hat{y}} + z_{6}c \, \mathbf{\hat{z}} & \left(8f\right) & \mbox{O III} \\ 
\mathbf{B}_{18} & = & -x_{6} \, \mathbf{a}_{1}-y_{6} \, \mathbf{a}_{2} + z_{6} \, \mathbf{a}_{3} & = & -x_{6}a \, \mathbf{\hat{x}}-y_{6}a \, \mathbf{\hat{y}} + z_{6}c \, \mathbf{\hat{z}} & \left(8f\right) & \mbox{O III} \\ 
\mathbf{B}_{19} & = & y_{6} \, \mathbf{a}_{1}-x_{6} \, \mathbf{a}_{2}-z_{6} \, \mathbf{a}_{3} & = & y_{6}a \, \mathbf{\hat{x}}-x_{6}a \, \mathbf{\hat{y}}-z_{6}c \, \mathbf{\hat{z}} & \left(8f\right) & \mbox{O III} \\ 
\mathbf{B}_{20} & = & -y_{6} \, \mathbf{a}_{1} + x_{6} \, \mathbf{a}_{2}-z_{6} \, \mathbf{a}_{3} & = & -y_{6}a \, \mathbf{\hat{x}} + x_{6}a \, \mathbf{\hat{y}}-z_{6}c \, \mathbf{\hat{z}} & \left(8f\right) & \mbox{O III} \\ 
\mathbf{B}_{21} & = & \left(\frac{1}{2} - x_{6}\right) \, \mathbf{a}_{1} + \left(\frac{1}{2} +y_{6}\right) \, \mathbf{a}_{2}-z_{6} \, \mathbf{a}_{3} & = & \left(\frac{1}{2} - x_{6}\right)a \, \mathbf{\hat{x}} + \left(\frac{1}{2} +y_{6}\right)a \, \mathbf{\hat{y}}-z_{6}c \, \mathbf{\hat{z}} & \left(8f\right) & \mbox{O III} \\ 
\mathbf{B}_{22} & = & \left(\frac{1}{2} +x_{6}\right) \, \mathbf{a}_{1} + \left(\frac{1}{2} - y_{6}\right) \, \mathbf{a}_{2}-z_{6} \, \mathbf{a}_{3} & = & \left(\frac{1}{2} +x_{6}\right)a \, \mathbf{\hat{x}} + \left(\frac{1}{2} - y_{6}\right)a \, \mathbf{\hat{y}}-z_{6}c \, \mathbf{\hat{z}} & \left(8f\right) & \mbox{O III} \\ 
\mathbf{B}_{23} & = & \left(\frac{1}{2} - y_{6}\right) \, \mathbf{a}_{1} + \left(\frac{1}{2} - x_{6}\right) \, \mathbf{a}_{2} + z_{6} \, \mathbf{a}_{3} & = & \left(\frac{1}{2} - y_{6}\right)a \, \mathbf{\hat{x}} + \left(\frac{1}{2} - x_{6}\right)a \, \mathbf{\hat{y}} + z_{6}c \, \mathbf{\hat{z}} & \left(8f\right) & \mbox{O III} \\ 
\mathbf{B}_{24} & = & \left(\frac{1}{2} +y_{6}\right) \, \mathbf{a}_{1} + \left(\frac{1}{2} +x_{6}\right) \, \mathbf{a}_{2} + z_{6} \, \mathbf{a}_{3} & = & \left(\frac{1}{2} +y_{6}\right)a \, \mathbf{\hat{x}} + \left(\frac{1}{2} +x_{6}\right)a \, \mathbf{\hat{y}} + z_{6}c \, \mathbf{\hat{z}} & \left(8f\right) & \mbox{O III} \\ 
\end{longtabu}
\renewcommand{\arraystretch}{1.0}
\noindent \hrulefill
\\
\textbf{References:}
\vspace*{-0.25cm}
\begin{flushleft}
  - \bibentry{Yang_PCM_24_510_1997}. \\
  - \bibentry{parthe93:TYPIX}. \\
\end{flushleft}
\noindent \hrulefill
\\
\textbf{Geometry files:}
\\
\noindent  - CIF: pp. {\hyperref[A2BC7D2_tP24_113_e_a_cef_e_cif]{\pageref{A2BC7D2_tP24_113_e_a_cef_e_cif}}} \\
\noindent  - POSCAR: pp. {\hyperref[A2BC7D2_tP24_113_e_a_cef_e_poscar]{\pageref{A2BC7D2_tP24_113_e_a_cef_e_poscar}}} \\
\onecolumn
{\phantomsection\label{A3B_tP32_114_3e_e}}
\subsection*{\huge \textbf{{\normalfont SeO$_{3}$ Structure: A3B\_tP32\_114\_3e\_e}}}
\noindent \hrulefill
\vspace*{0.25cm}
\begin{figure}[htp]
  \centering
  \vspace{-1em}
  {\includegraphics[width=1\textwidth]{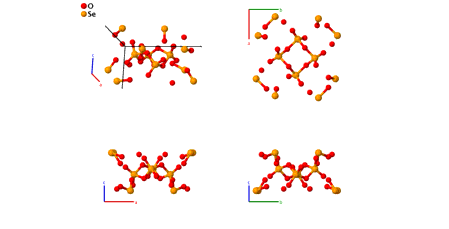}}
\end{figure}
\vspace*{-0.5cm}
\renewcommand{\arraystretch}{1.5}
\begin{equation*}
  \begin{array}{>{$\hspace{-0.15cm}}l<{$}>{$}p{0.5cm}<{$}>{$}p{18.5cm}<{$}}
    \mbox{\large \textbf{Prototype}} &\colon & \ce{SeO3} \\
    \mbox{\large \textbf{\AFLOW\ prototype label}} &\colon & \mbox{A3B\_tP32\_114\_3e\_e} \\
    \mbox{\large \textbf{\textit{Strukturbericht} designation}} &\colon & \mbox{None} \\
    \mbox{\large \textbf{Pearson symbol}} &\colon & \mbox{tP32} \\
    \mbox{\large \textbf{Space group number}} &\colon & 114 \\
    \mbox{\large \textbf{Space group symbol}} &\colon & P\bar{4}2_{1}c \\
    \mbox{\large \textbf{\AFLOW\ prototype command}} &\colon &  \texttt{aflow} \,  \, \texttt{-{}-proto=A3B\_tP32\_114\_3e\_e } \, \newline \texttt{-{}-params=}{a,c/a,x_{1},y_{1},z_{1},x_{2},y_{2},z_{2},x_{3},y_{3},z_{3},x_{4},y_{4},z_{4} }
  \end{array}
\end{equation*}
\renewcommand{\arraystretch}{1.0}

\noindent \parbox{1 \linewidth}{
\noindent \hrulefill
\\
\textbf{Simple Tetragonal primitive vectors:} \\
\vspace*{-0.25cm}
\begin{tabular}{cc}
  \begin{tabular}{c}
    \parbox{0.6 \linewidth}{
      \renewcommand{\arraystretch}{1.5}
      \begin{equation*}
        \centering
        \begin{array}{ccc}
              \mathbf{a}_1 & = & a \, \mathbf{\hat{x}} \\
    \mathbf{a}_2 & = & a \, \mathbf{\hat{y}} \\
    \mathbf{a}_3 & = & c \, \mathbf{\hat{z}} \\

        \end{array}
      \end{equation*}
    }
    \renewcommand{\arraystretch}{1.0}
  \end{tabular}
  \begin{tabular}{c}
    \includegraphics[width=0.3\linewidth]{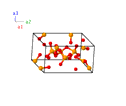} \\
  \end{tabular}
\end{tabular}

}
\vspace*{-0.25cm}

\noindent \hrulefill
\\
\textbf{Basis vectors:}
\vspace*{-0.25cm}
\renewcommand{\arraystretch}{1.5}
\begin{longtabu} to \textwidth{>{\centering $}X[-1,c,c]<{$}>{\centering $}X[-1,c,c]<{$}>{\centering $}X[-1,c,c]<{$}>{\centering $}X[-1,c,c]<{$}>{\centering $}X[-1,c,c]<{$}>{\centering $}X[-1,c,c]<{$}>{\centering $}X[-1,c,c]<{$}}
  & & \mbox{Lattice Coordinates} & & \mbox{Cartesian Coordinates} &\mbox{Wyckoff Position} & \mbox{Atom Type} \\  
  \mathbf{B}_{1} & = & x_{1} \, \mathbf{a}_{1} + y_{1} \, \mathbf{a}_{2} + z_{1} \, \mathbf{a}_{3} & = & x_{1}a \, \mathbf{\hat{x}} + y_{1}a \, \mathbf{\hat{y}} + z_{1}c \, \mathbf{\hat{z}} & \left(8e\right) & \mbox{O I} \\ 
\mathbf{B}_{2} & = & -x_{1} \, \mathbf{a}_{1}-y_{1} \, \mathbf{a}_{2} + z_{1} \, \mathbf{a}_{3} & = & -x_{1}a \, \mathbf{\hat{x}}-y_{1}a \, \mathbf{\hat{y}} + z_{1}c \, \mathbf{\hat{z}} & \left(8e\right) & \mbox{O I} \\ 
\mathbf{B}_{3} & = & y_{1} \, \mathbf{a}_{1}-x_{1} \, \mathbf{a}_{2}-z_{1} \, \mathbf{a}_{3} & = & y_{1}a \, \mathbf{\hat{x}}-x_{1}a \, \mathbf{\hat{y}}-z_{1}c \, \mathbf{\hat{z}} & \left(8e\right) & \mbox{O I} \\ 
\mathbf{B}_{4} & = & -y_{1} \, \mathbf{a}_{1} + x_{1} \, \mathbf{a}_{2}-z_{1} \, \mathbf{a}_{3} & = & -y_{1}a \, \mathbf{\hat{x}} + x_{1}a \, \mathbf{\hat{y}}-z_{1}c \, \mathbf{\hat{z}} & \left(8e\right) & \mbox{O I} \\ 
\mathbf{B}_{5} & = & \left(\frac{1}{2} - x_{1}\right) \, \mathbf{a}_{1} + \left(\frac{1}{2} +y_{1}\right) \, \mathbf{a}_{2} + \left(\frac{1}{2} - z_{1}\right) \, \mathbf{a}_{3} & = & \left(\frac{1}{2} - x_{1}\right)a \, \mathbf{\hat{x}} + \left(\frac{1}{2} +y_{1}\right)a \, \mathbf{\hat{y}} + \left(\frac{1}{2} - z_{1}\right)c \, \mathbf{\hat{z}} & \left(8e\right) & \mbox{O I} \\ 
\mathbf{B}_{6} & = & \left(\frac{1}{2} +x_{1}\right) \, \mathbf{a}_{1} + \left(\frac{1}{2} - y_{1}\right) \, \mathbf{a}_{2} + \left(\frac{1}{2} - z_{1}\right) \, \mathbf{a}_{3} & = & \left(\frac{1}{2} +x_{1}\right)a \, \mathbf{\hat{x}} + \left(\frac{1}{2} - y_{1}\right)a \, \mathbf{\hat{y}} + \left(\frac{1}{2} - z_{1}\right)c \, \mathbf{\hat{z}} & \left(8e\right) & \mbox{O I} \\ 
\mathbf{B}_{7} & = & \left(\frac{1}{2} - y_{1}\right) \, \mathbf{a}_{1} + \left(\frac{1}{2} - x_{1}\right) \, \mathbf{a}_{2} + \left(\frac{1}{2} +z_{1}\right) \, \mathbf{a}_{3} & = & \left(\frac{1}{2} - y_{1}\right)a \, \mathbf{\hat{x}} + \left(\frac{1}{2} - x_{1}\right)a \, \mathbf{\hat{y}} + \left(\frac{1}{2} +z_{1}\right)c \, \mathbf{\hat{z}} & \left(8e\right) & \mbox{O I} \\ 
\mathbf{B}_{8} & = & \left(\frac{1}{2} +y_{1}\right) \, \mathbf{a}_{1} + \left(\frac{1}{2} +x_{1}\right) \, \mathbf{a}_{2} + \left(\frac{1}{2} +z_{1}\right) \, \mathbf{a}_{3} & = & \left(\frac{1}{2} +y_{1}\right)a \, \mathbf{\hat{x}} + \left(\frac{1}{2} +x_{1}\right)a \, \mathbf{\hat{y}} + \left(\frac{1}{2} +z_{1}\right)c \, \mathbf{\hat{z}} & \left(8e\right) & \mbox{O I} \\ 
\mathbf{B}_{9} & = & x_{2} \, \mathbf{a}_{1} + y_{2} \, \mathbf{a}_{2} + z_{2} \, \mathbf{a}_{3} & = & x_{2}a \, \mathbf{\hat{x}} + y_{2}a \, \mathbf{\hat{y}} + z_{2}c \, \mathbf{\hat{z}} & \left(8e\right) & \mbox{O II} \\ 
\mathbf{B}_{10} & = & -x_{2} \, \mathbf{a}_{1}-y_{2} \, \mathbf{a}_{2} + z_{2} \, \mathbf{a}_{3} & = & -x_{2}a \, \mathbf{\hat{x}}-y_{2}a \, \mathbf{\hat{y}} + z_{2}c \, \mathbf{\hat{z}} & \left(8e\right) & \mbox{O II} \\ 
\mathbf{B}_{11} & = & y_{2} \, \mathbf{a}_{1}-x_{2} \, \mathbf{a}_{2}-z_{2} \, \mathbf{a}_{3} & = & y_{2}a \, \mathbf{\hat{x}}-x_{2}a \, \mathbf{\hat{y}}-z_{2}c \, \mathbf{\hat{z}} & \left(8e\right) & \mbox{O II} \\ 
\mathbf{B}_{12} & = & -y_{2} \, \mathbf{a}_{1} + x_{2} \, \mathbf{a}_{2}-z_{2} \, \mathbf{a}_{3} & = & -y_{2}a \, \mathbf{\hat{x}} + x_{2}a \, \mathbf{\hat{y}}-z_{2}c \, \mathbf{\hat{z}} & \left(8e\right) & \mbox{O II} \\ 
\mathbf{B}_{13} & = & \left(\frac{1}{2} - x_{2}\right) \, \mathbf{a}_{1} + \left(\frac{1}{2} +y_{2}\right) \, \mathbf{a}_{2} + \left(\frac{1}{2} - z_{2}\right) \, \mathbf{a}_{3} & = & \left(\frac{1}{2} - x_{2}\right)a \, \mathbf{\hat{x}} + \left(\frac{1}{2} +y_{2}\right)a \, \mathbf{\hat{y}} + \left(\frac{1}{2} - z_{2}\right)c \, \mathbf{\hat{z}} & \left(8e\right) & \mbox{O II} \\ 
\mathbf{B}_{14} & = & \left(\frac{1}{2} +x_{2}\right) \, \mathbf{a}_{1} + \left(\frac{1}{2} - y_{2}\right) \, \mathbf{a}_{2} + \left(\frac{1}{2} - z_{2}\right) \, \mathbf{a}_{3} & = & \left(\frac{1}{2} +x_{2}\right)a \, \mathbf{\hat{x}} + \left(\frac{1}{2} - y_{2}\right)a \, \mathbf{\hat{y}} + \left(\frac{1}{2} - z_{2}\right)c \, \mathbf{\hat{z}} & \left(8e\right) & \mbox{O II} \\ 
\mathbf{B}_{15} & = & \left(\frac{1}{2} - y_{2}\right) \, \mathbf{a}_{1} + \left(\frac{1}{2} - x_{2}\right) \, \mathbf{a}_{2} + \left(\frac{1}{2} +z_{2}\right) \, \mathbf{a}_{3} & = & \left(\frac{1}{2} - y_{2}\right)a \, \mathbf{\hat{x}} + \left(\frac{1}{2} - x_{2}\right)a \, \mathbf{\hat{y}} + \left(\frac{1}{2} +z_{2}\right)c \, \mathbf{\hat{z}} & \left(8e\right) & \mbox{O II} \\ 
\mathbf{B}_{16} & = & \left(\frac{1}{2} +y_{2}\right) \, \mathbf{a}_{1} + \left(\frac{1}{2} +x_{2}\right) \, \mathbf{a}_{2} + \left(\frac{1}{2} +z_{2}\right) \, \mathbf{a}_{3} & = & \left(\frac{1}{2} +y_{2}\right)a \, \mathbf{\hat{x}} + \left(\frac{1}{2} +x_{2}\right)a \, \mathbf{\hat{y}} + \left(\frac{1}{2} +z_{2}\right)c \, \mathbf{\hat{z}} & \left(8e\right) & \mbox{O II} \\ 
\mathbf{B}_{17} & = & x_{3} \, \mathbf{a}_{1} + y_{3} \, \mathbf{a}_{2} + z_{3} \, \mathbf{a}_{3} & = & x_{3}a \, \mathbf{\hat{x}} + y_{3}a \, \mathbf{\hat{y}} + z_{3}c \, \mathbf{\hat{z}} & \left(8e\right) & \mbox{O III} \\ 
\mathbf{B}_{18} & = & -x_{3} \, \mathbf{a}_{1}-y_{3} \, \mathbf{a}_{2} + z_{3} \, \mathbf{a}_{3} & = & -x_{3}a \, \mathbf{\hat{x}}-y_{3}a \, \mathbf{\hat{y}} + z_{3}c \, \mathbf{\hat{z}} & \left(8e\right) & \mbox{O III} \\ 
\mathbf{B}_{19} & = & y_{3} \, \mathbf{a}_{1}-x_{3} \, \mathbf{a}_{2}-z_{3} \, \mathbf{a}_{3} & = & y_{3}a \, \mathbf{\hat{x}}-x_{3}a \, \mathbf{\hat{y}}-z_{3}c \, \mathbf{\hat{z}} & \left(8e\right) & \mbox{O III} \\ 
\mathbf{B}_{20} & = & -y_{3} \, \mathbf{a}_{1} + x_{3} \, \mathbf{a}_{2}-z_{3} \, \mathbf{a}_{3} & = & -y_{3}a \, \mathbf{\hat{x}} + x_{3}a \, \mathbf{\hat{y}}-z_{3}c \, \mathbf{\hat{z}} & \left(8e\right) & \mbox{O III} \\ 
\mathbf{B}_{21} & = & \left(\frac{1}{2} - x_{3}\right) \, \mathbf{a}_{1} + \left(\frac{1}{2} +y_{3}\right) \, \mathbf{a}_{2} + \left(\frac{1}{2} - z_{3}\right) \, \mathbf{a}_{3} & = & \left(\frac{1}{2} - x_{3}\right)a \, \mathbf{\hat{x}} + \left(\frac{1}{2} +y_{3}\right)a \, \mathbf{\hat{y}} + \left(\frac{1}{2} - z_{3}\right)c \, \mathbf{\hat{z}} & \left(8e\right) & \mbox{O III} \\ 
\mathbf{B}_{22} & = & \left(\frac{1}{2} +x_{3}\right) \, \mathbf{a}_{1} + \left(\frac{1}{2} - y_{3}\right) \, \mathbf{a}_{2} + \left(\frac{1}{2} - z_{3}\right) \, \mathbf{a}_{3} & = & \left(\frac{1}{2} +x_{3}\right)a \, \mathbf{\hat{x}} + \left(\frac{1}{2} - y_{3}\right)a \, \mathbf{\hat{y}} + \left(\frac{1}{2} - z_{3}\right)c \, \mathbf{\hat{z}} & \left(8e\right) & \mbox{O III} \\ 
\mathbf{B}_{23} & = & \left(\frac{1}{2} - y_{3}\right) \, \mathbf{a}_{1} + \left(\frac{1}{2} - x_{3}\right) \, \mathbf{a}_{2} + \left(\frac{1}{2} +z_{3}\right) \, \mathbf{a}_{3} & = & \left(\frac{1}{2} - y_{3}\right)a \, \mathbf{\hat{x}} + \left(\frac{1}{2} - x_{3}\right)a \, \mathbf{\hat{y}} + \left(\frac{1}{2} +z_{3}\right)c \, \mathbf{\hat{z}} & \left(8e\right) & \mbox{O III} \\ 
\mathbf{B}_{24} & = & \left(\frac{1}{2} +y_{3}\right) \, \mathbf{a}_{1} + \left(\frac{1}{2} +x_{3}\right) \, \mathbf{a}_{2} + \left(\frac{1}{2} +z_{3}\right) \, \mathbf{a}_{3} & = & \left(\frac{1}{2} +y_{3}\right)a \, \mathbf{\hat{x}} + \left(\frac{1}{2} +x_{3}\right)a \, \mathbf{\hat{y}} + \left(\frac{1}{2} +z_{3}\right)c \, \mathbf{\hat{z}} & \left(8e\right) & \mbox{O III} \\ 
\mathbf{B}_{25} & = & x_{4} \, \mathbf{a}_{1} + y_{4} \, \mathbf{a}_{2} + z_{4} \, \mathbf{a}_{3} & = & x_{4}a \, \mathbf{\hat{x}} + y_{4}a \, \mathbf{\hat{y}} + z_{4}c \, \mathbf{\hat{z}} & \left(8e\right) & \mbox{Se} \\ 
\mathbf{B}_{26} & = & -x_{4} \, \mathbf{a}_{1}-y_{4} \, \mathbf{a}_{2} + z_{4} \, \mathbf{a}_{3} & = & -x_{4}a \, \mathbf{\hat{x}}-y_{4}a \, \mathbf{\hat{y}} + z_{4}c \, \mathbf{\hat{z}} & \left(8e\right) & \mbox{Se} \\ 
\mathbf{B}_{27} & = & y_{4} \, \mathbf{a}_{1}-x_{4} \, \mathbf{a}_{2}-z_{4} \, \mathbf{a}_{3} & = & y_{4}a \, \mathbf{\hat{x}}-x_{4}a \, \mathbf{\hat{y}}-z_{4}c \, \mathbf{\hat{z}} & \left(8e\right) & \mbox{Se} \\ 
\mathbf{B}_{28} & = & -y_{4} \, \mathbf{a}_{1} + x_{4} \, \mathbf{a}_{2}-z_{4} \, \mathbf{a}_{3} & = & -y_{4}a \, \mathbf{\hat{x}} + x_{4}a \, \mathbf{\hat{y}}-z_{4}c \, \mathbf{\hat{z}} & \left(8e\right) & \mbox{Se} \\ 
\mathbf{B}_{29} & = & \left(\frac{1}{2} - x_{4}\right) \, \mathbf{a}_{1} + \left(\frac{1}{2} +y_{4}\right) \, \mathbf{a}_{2} + \left(\frac{1}{2} - z_{4}\right) \, \mathbf{a}_{3} & = & \left(\frac{1}{2} - x_{4}\right)a \, \mathbf{\hat{x}} + \left(\frac{1}{2} +y_{4}\right)a \, \mathbf{\hat{y}} + \left(\frac{1}{2} - z_{4}\right)c \, \mathbf{\hat{z}} & \left(8e\right) & \mbox{Se} \\ 
\mathbf{B}_{30} & = & \left(\frac{1}{2} +x_{4}\right) \, \mathbf{a}_{1} + \left(\frac{1}{2} - y_{4}\right) \, \mathbf{a}_{2} + \left(\frac{1}{2} - z_{4}\right) \, \mathbf{a}_{3} & = & \left(\frac{1}{2} +x_{4}\right)a \, \mathbf{\hat{x}} + \left(\frac{1}{2} - y_{4}\right)a \, \mathbf{\hat{y}} + \left(\frac{1}{2} - z_{4}\right)c \, \mathbf{\hat{z}} & \left(8e\right) & \mbox{Se} \\ 
\mathbf{B}_{31} & = & \left(\frac{1}{2} - y_{4}\right) \, \mathbf{a}_{1} + \left(\frac{1}{2} - x_{4}\right) \, \mathbf{a}_{2} + \left(\frac{1}{2} +z_{4}\right) \, \mathbf{a}_{3} & = & \left(\frac{1}{2} - y_{4}\right)a \, \mathbf{\hat{x}} + \left(\frac{1}{2} - x_{4}\right)a \, \mathbf{\hat{y}} + \left(\frac{1}{2} +z_{4}\right)c \, \mathbf{\hat{z}} & \left(8e\right) & \mbox{Se} \\ 
\mathbf{B}_{32} & = & \left(\frac{1}{2} +y_{4}\right) \, \mathbf{a}_{1} + \left(\frac{1}{2} +x_{4}\right) \, \mathbf{a}_{2} + \left(\frac{1}{2} +z_{4}\right) \, \mathbf{a}_{3} & = & \left(\frac{1}{2} +y_{4}\right)a \, \mathbf{\hat{x}} + \left(\frac{1}{2} +x_{4}\right)a \, \mathbf{\hat{y}} + \left(\frac{1}{2} +z_{4}\right)c \, \mathbf{\hat{z}} & \left(8e\right) & \mbox{Se} \\ 
\end{longtabu}
\renewcommand{\arraystretch}{1.0}
\noindent \hrulefill
\\
\textbf{References:}
\vspace*{-0.25cm}
\begin{flushleft}
  - \bibentry{Mijlhoff_SeO3_ActCrystallogr_1962}. \\
\end{flushleft}
\textbf{Found in:}
\vspace*{-0.25cm}
\begin{flushleft}
  - \bibentry{Villars_PearsonsCrystalData_2013}. \\
\end{flushleft}
\noindent \hrulefill
\\
\textbf{Geometry files:}
\\
\noindent  - CIF: pp. {\hyperref[A3B_tP32_114_3e_e_cif]{\pageref{A3B_tP32_114_3e_e_cif}}} \\
\noindent  - POSCAR: pp. {\hyperref[A3B_tP32_114_3e_e_poscar]{\pageref{A3B_tP32_114_3e_e_poscar}}} \\
\onecolumn
{\phantomsection\label{A4B_tP10_114_e_a}}
\subsection*{\huge \textbf{{\normalfont Pd$_{4}$Se Structure: A4B\_tP10\_114\_e\_a}}}
\noindent \hrulefill
\vspace*{0.25cm}
\begin{figure}[htp]
  \centering
  \vspace{-1em}
  {\includegraphics[width=1\textwidth]{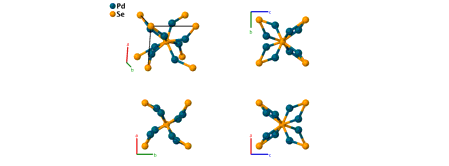}}
\end{figure}
\vspace*{-0.5cm}
\renewcommand{\arraystretch}{1.5}
\begin{equation*}
  \begin{array}{>{$\hspace{-0.15cm}}l<{$}>{$}p{0.5cm}<{$}>{$}p{18.5cm}<{$}}
    \mbox{\large \textbf{Prototype}} &\colon & \ce{Pd4Se} \\
    \mbox{\large \textbf{\AFLOW\ prototype label}} &\colon & \mbox{A4B\_tP10\_114\_e\_a} \\
    \mbox{\large \textbf{\textit{Strukturbericht} designation}} &\colon & \mbox{None} \\
    \mbox{\large \textbf{Pearson symbol}} &\colon & \mbox{tP10} \\
    \mbox{\large \textbf{Space group number}} &\colon & 114 \\
    \mbox{\large \textbf{Space group symbol}} &\colon & P\bar{4}2_{1}c \\
    \mbox{\large \textbf{\AFLOW\ prototype command}} &\colon &  \texttt{aflow} \,  \, \texttt{-{}-proto=A4B\_tP10\_114\_e\_a } \, \newline \texttt{-{}-params=}{a,c/a,x_{2},y_{2},z_{2} }
  \end{array}
\end{equation*}
\renewcommand{\arraystretch}{1.0}

\noindent \parbox{1 \linewidth}{
\noindent \hrulefill
\\
\textbf{Simple Tetragonal primitive vectors:} \\
\vspace*{-0.25cm}
\begin{tabular}{cc}
  \begin{tabular}{c}
    \parbox{0.6 \linewidth}{
      \renewcommand{\arraystretch}{1.5}
      \begin{equation*}
        \centering
        \begin{array}{ccc}
              \mathbf{a}_1 & = & a \, \mathbf{\hat{x}} \\
    \mathbf{a}_2 & = & a \, \mathbf{\hat{y}} \\
    \mathbf{a}_3 & = & c \, \mathbf{\hat{z}} \\

        \end{array}
      \end{equation*}
    }
    \renewcommand{\arraystretch}{1.0}
  \end{tabular}
  \begin{tabular}{c}
    \includegraphics[width=0.3\linewidth]{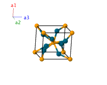} \\
  \end{tabular}
\end{tabular}

}
\vspace*{-0.25cm}

\noindent \hrulefill
\\
\textbf{Basis vectors:}
\vspace*{-0.25cm}
\renewcommand{\arraystretch}{1.5}
\begin{longtabu} to \textwidth{>{\centering $}X[-1,c,c]<{$}>{\centering $}X[-1,c,c]<{$}>{\centering $}X[-1,c,c]<{$}>{\centering $}X[-1,c,c]<{$}>{\centering $}X[-1,c,c]<{$}>{\centering $}X[-1,c,c]<{$}>{\centering $}X[-1,c,c]<{$}}
  & & \mbox{Lattice Coordinates} & & \mbox{Cartesian Coordinates} &\mbox{Wyckoff Position} & \mbox{Atom Type} \\  
  \mathbf{B}_{1} & = & 0 \, \mathbf{a}_{1} + 0 \, \mathbf{a}_{2} + 0 \, \mathbf{a}_{3} & = & 0 \, \mathbf{\hat{x}} + 0 \, \mathbf{\hat{y}} + 0 \, \mathbf{\hat{z}} & \left(2a\right) & \mbox{Se} \\ 
\mathbf{B}_{2} & = & \frac{1}{2} \, \mathbf{a}_{1} + \frac{1}{2} \, \mathbf{a}_{2} + \frac{1}{2} \, \mathbf{a}_{3} & = & \frac{1}{2}a \, \mathbf{\hat{x}} + \frac{1}{2}a \, \mathbf{\hat{y}} + \frac{1}{2}c \, \mathbf{\hat{z}} & \left(2a\right) & \mbox{Se} \\ 
\mathbf{B}_{3} & = & x_{2} \, \mathbf{a}_{1} + y_{2} \, \mathbf{a}_{2} + z_{2} \, \mathbf{a}_{3} & = & x_{2}a \, \mathbf{\hat{x}} + y_{2}a \, \mathbf{\hat{y}} + z_{2}c \, \mathbf{\hat{z}} & \left(8e\right) & \mbox{Pd} \\ 
\mathbf{B}_{4} & = & -x_{2} \, \mathbf{a}_{1}-y_{2} \, \mathbf{a}_{2} + z_{2} \, \mathbf{a}_{3} & = & -x_{2}a \, \mathbf{\hat{x}}-y_{2}a \, \mathbf{\hat{y}} + z_{2}c \, \mathbf{\hat{z}} & \left(8e\right) & \mbox{Pd} \\ 
\mathbf{B}_{5} & = & y_{2} \, \mathbf{a}_{1}-x_{2} \, \mathbf{a}_{2}-z_{2} \, \mathbf{a}_{3} & = & y_{2}a \, \mathbf{\hat{x}}-x_{2}a \, \mathbf{\hat{y}}-z_{2}c \, \mathbf{\hat{z}} & \left(8e\right) & \mbox{Pd} \\ 
\mathbf{B}_{6} & = & -y_{2} \, \mathbf{a}_{1} + x_{2} \, \mathbf{a}_{2}-z_{2} \, \mathbf{a}_{3} & = & -y_{2}a \, \mathbf{\hat{x}} + x_{2}a \, \mathbf{\hat{y}}-z_{2}c \, \mathbf{\hat{z}} & \left(8e\right) & \mbox{Pd} \\ 
\mathbf{B}_{7} & = & \left(\frac{1}{2} - x_{2}\right) \, \mathbf{a}_{1} + \left(\frac{1}{2} +y_{2}\right) \, \mathbf{a}_{2} + \left(\frac{1}{2} - z_{2}\right) \, \mathbf{a}_{3} & = & \left(\frac{1}{2} - x_{2}\right)a \, \mathbf{\hat{x}} + \left(\frac{1}{2} +y_{2}\right)a \, \mathbf{\hat{y}} + \left(\frac{1}{2} - z_{2}\right)c \, \mathbf{\hat{z}} & \left(8e\right) & \mbox{Pd} \\ 
\mathbf{B}_{8} & = & \left(\frac{1}{2} +x_{2}\right) \, \mathbf{a}_{1} + \left(\frac{1}{2} - y_{2}\right) \, \mathbf{a}_{2} + \left(\frac{1}{2} - z_{2}\right) \, \mathbf{a}_{3} & = & \left(\frac{1}{2} +x_{2}\right)a \, \mathbf{\hat{x}} + \left(\frac{1}{2} - y_{2}\right)a \, \mathbf{\hat{y}} + \left(\frac{1}{2} - z_{2}\right)c \, \mathbf{\hat{z}} & \left(8e\right) & \mbox{Pd} \\ 
\mathbf{B}_{9} & = & \left(\frac{1}{2} - y_{2}\right) \, \mathbf{a}_{1} + \left(\frac{1}{2} - x_{2}\right) \, \mathbf{a}_{2} + \left(\frac{1}{2} +z_{2}\right) \, \mathbf{a}_{3} & = & \left(\frac{1}{2} - y_{2}\right)a \, \mathbf{\hat{x}} + \left(\frac{1}{2} - x_{2}\right)a \, \mathbf{\hat{y}} + \left(\frac{1}{2} +z_{2}\right)c \, \mathbf{\hat{z}} & \left(8e\right) & \mbox{Pd} \\ 
\mathbf{B}_{10} & = & \left(\frac{1}{2} +y_{2}\right) \, \mathbf{a}_{1} + \left(\frac{1}{2} +x_{2}\right) \, \mathbf{a}_{2} + \left(\frac{1}{2} +z_{2}\right) \, \mathbf{a}_{3} & = & \left(\frac{1}{2} +y_{2}\right)a \, \mathbf{\hat{x}} + \left(\frac{1}{2} +x_{2}\right)a \, \mathbf{\hat{y}} + \left(\frac{1}{2} +z_{2}\right)c \, \mathbf{\hat{z}} & \left(8e\right) & \mbox{Pd} \\ 
\end{longtabu}
\renewcommand{\arraystretch}{1.0}
\noindent \hrulefill
\\
\textbf{References:}
\vspace*{-0.25cm}
\begin{flushleft}
  - \bibentry{Gronvold_Pd4Se_ActChemScand_1956}. \\
\end{flushleft}
\textbf{Found in:}
\vspace*{-0.25cm}
\begin{flushleft}
  - \bibentry{Villars_PearsonsCrystalData_2013}. \\
\end{flushleft}
\noindent \hrulefill
\\
\textbf{Geometry files:}
\\
\noindent  - CIF: pp. {\hyperref[A4B_tP10_114_e_a_cif]{\pageref{A4B_tP10_114_e_a_cif}}} \\
\noindent  - POSCAR: pp. {\hyperref[A4B_tP10_114_e_a_poscar]{\pageref{A4B_tP10_114_e_a_poscar}}} \\
\onecolumn
{\phantomsection\label{A2B3_tP5_115_g_ag}}
\subsection*{\huge \textbf{{\normalfont Rh$_{3}$P$_{2}$ Structure: A2B3\_tP5\_115\_g\_ag}}}
\noindent \hrulefill
\vspace*{0.25cm}
\begin{figure}[htp]
  \centering
  \vspace{-1em}
  {\includegraphics[width=1\textwidth]{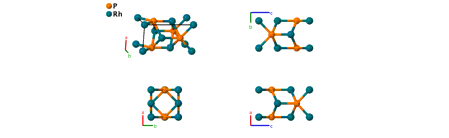}}
\end{figure}
\vspace*{-0.5cm}
\renewcommand{\arraystretch}{1.5}
\begin{equation*}
  \begin{array}{>{$\hspace{-0.15cm}}l<{$}>{$}p{0.5cm}<{$}>{$}p{18.5cm}<{$}}
    \mbox{\large \textbf{Prototype}} &\colon & \ce{Rh3P2} \\
    \mbox{\large \textbf{\AFLOW\ prototype label}} &\colon & \mbox{A2B3\_tP5\_115\_g\_ag} \\
    \mbox{\large \textbf{\textit{Strukturbericht} designation}} &\colon & \mbox{None} \\
    \mbox{\large \textbf{Pearson symbol}} &\colon & \mbox{tP5} \\
    \mbox{\large \textbf{Space group number}} &\colon & 115 \\
    \mbox{\large \textbf{Space group symbol}} &\colon & P\bar{4}m2 \\
    \mbox{\large \textbf{\AFLOW\ prototype command}} &\colon &  \texttt{aflow} \,  \, \texttt{-{}-proto=A2B3\_tP5\_115\_g\_ag } \, \newline \texttt{-{}-params=}{a,c/a,z_{2},z_{3} }
  \end{array}
\end{equation*}
\renewcommand{\arraystretch}{1.0}

\noindent \parbox{1 \linewidth}{
\noindent \hrulefill
\\
\textbf{Simple Tetragonal primitive vectors:} \\
\vspace*{-0.25cm}
\begin{tabular}{cc}
  \begin{tabular}{c}
    \parbox{0.6 \linewidth}{
      \renewcommand{\arraystretch}{1.5}
      \begin{equation*}
        \centering
        \begin{array}{ccc}
              \mathbf{a}_1 & = & a \, \mathbf{\hat{x}} \\
    \mathbf{a}_2 & = & a \, \mathbf{\hat{y}} \\
    \mathbf{a}_3 & = & c \, \mathbf{\hat{z}} \\

        \end{array}
      \end{equation*}
    }
    \renewcommand{\arraystretch}{1.0}
  \end{tabular}
  \begin{tabular}{c}
    \includegraphics[width=0.3\linewidth]{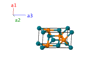} \\
  \end{tabular}
\end{tabular}

}
\vspace*{-0.25cm}

\noindent \hrulefill
\\
\textbf{Basis vectors:}
\vspace*{-0.25cm}
\renewcommand{\arraystretch}{1.5}
\begin{longtabu} to \textwidth{>{\centering $}X[-1,c,c]<{$}>{\centering $}X[-1,c,c]<{$}>{\centering $}X[-1,c,c]<{$}>{\centering $}X[-1,c,c]<{$}>{\centering $}X[-1,c,c]<{$}>{\centering $}X[-1,c,c]<{$}>{\centering $}X[-1,c,c]<{$}}
  & & \mbox{Lattice Coordinates} & & \mbox{Cartesian Coordinates} &\mbox{Wyckoff Position} & \mbox{Atom Type} \\  
  \mathbf{B}_{1} & = & 0 \, \mathbf{a}_{1} + 0 \, \mathbf{a}_{2} + 0 \, \mathbf{a}_{3} & = & 0 \, \mathbf{\hat{x}} + 0 \, \mathbf{\hat{y}} + 0 \, \mathbf{\hat{z}} & \left(1a\right) & \mbox{Rh I} \\ 
\mathbf{B}_{2} & = & \frac{1}{2} \, \mathbf{a}_{2} + z_{2} \, \mathbf{a}_{3} & = & \frac{1}{2}a \, \mathbf{\hat{y}} + z_{2}c \, \mathbf{\hat{z}} & \left(2g\right) & \mbox{P} \\ 
\mathbf{B}_{3} & = & \frac{1}{2} \, \mathbf{a}_{1} + -z_{2} \, \mathbf{a}_{3} & = & \frac{1}{2}a \, \mathbf{\hat{x}} + -z_{2}c \, \mathbf{\hat{z}} & \left(2g\right) & \mbox{P} \\ 
\mathbf{B}_{4} & = & \frac{1}{2} \, \mathbf{a}_{2} + z_{3} \, \mathbf{a}_{3} & = & \frac{1}{2}a \, \mathbf{\hat{y}} + z_{3}c \, \mathbf{\hat{z}} & \left(2g\right) & \mbox{Rh II} \\ 
\mathbf{B}_{5} & = & \frac{1}{2} \, \mathbf{a}_{1} + -z_{3} \, \mathbf{a}_{3} & = & \frac{1}{2}a \, \mathbf{\hat{x}} + -z_{3}c \, \mathbf{\hat{z}} & \left(2g\right) & \mbox{Rh II} \\ 
\end{longtabu}
\renewcommand{\arraystretch}{1.0}
\noindent \hrulefill
\\
\textbf{References:}
\vspace*{-0.25cm}
\begin{flushleft}
  - \bibentry{Ghadraoui_Rh3P2_ActaCrystallogrSecC_1983}. \\
\end{flushleft}
\textbf{Found in:}
\vspace*{-0.25cm}
\begin{flushleft}
  - \bibentry{Villars_PearsonsCrystalData_2013}. \\
\end{flushleft}
\noindent \hrulefill
\\
\textbf{Geometry files:}
\\
\noindent  - CIF: pp. {\hyperref[A2B3_tP5_115_g_ag_cif]{\pageref{A2B3_tP5_115_g_ag_cif}}} \\
\noindent  - POSCAR: pp. {\hyperref[A2B3_tP5_115_g_ag_poscar]{\pageref{A2B3_tP5_115_g_ag_poscar}}} \\
\onecolumn
{\phantomsection\label{AB2_tP12_115_j_egi}}
\subsection*{\huge \textbf{{\normalfont HgI$_{2}$ Structure: AB2\_tP12\_115\_j\_egi}}}
\noindent \hrulefill
\vspace*{0.25cm}
\begin{figure}[htp]
  \centering
  \vspace{-1em}
  {\includegraphics[width=1\textwidth]{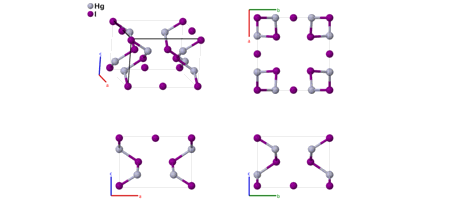}}
\end{figure}
\vspace*{-0.5cm}
\renewcommand{\arraystretch}{1.5}
\begin{equation*}
  \begin{array}{>{$\hspace{-0.15cm}}l<{$}>{$}p{0.5cm}<{$}>{$}p{18.5cm}<{$}}
    \mbox{\large \textbf{Prototype}} &\colon & \ce{HgI2} \\
    \mbox{\large \textbf{\AFLOW\ prototype label}} &\colon & \mbox{AB2\_tP12\_115\_j\_egi} \\
    \mbox{\large \textbf{\textit{Strukturbericht} designation}} &\colon & \mbox{None} \\
    \mbox{\large \textbf{Pearson symbol}} &\colon & \mbox{tP12} \\
    \mbox{\large \textbf{Space group number}} &\colon & 115 \\
    \mbox{\large \textbf{Space group symbol}} &\colon & P\bar{4}m2 \\
    \mbox{\large \textbf{\AFLOW\ prototype command}} &\colon &  \texttt{aflow} \,  \, \texttt{-{}-proto=AB2\_tP12\_115\_j\_egi } \, \newline \texttt{-{}-params=}{a,c/a,z_{1},z_{2},x_{3},x_{4},z_{4} }
  \end{array}
\end{equation*}
\renewcommand{\arraystretch}{1.0}

\noindent \parbox{1 \linewidth}{
\noindent \hrulefill
\\
\textbf{Simple Tetragonal primitive vectors:} \\
\vspace*{-0.25cm}
\begin{tabular}{cc}
  \begin{tabular}{c}
    \parbox{0.6 \linewidth}{
      \renewcommand{\arraystretch}{1.5}
      \begin{equation*}
        \centering
        \begin{array}{ccc}
              \mathbf{a}_1 & = & a \, \mathbf{\hat{x}} \\
    \mathbf{a}_2 & = & a \, \mathbf{\hat{y}} \\
    \mathbf{a}_3 & = & c \, \mathbf{\hat{z}} \\

        \end{array}
      \end{equation*}
    }
    \renewcommand{\arraystretch}{1.0}
  \end{tabular}
  \begin{tabular}{c}
    \includegraphics[width=0.3\linewidth]{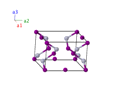} \\
  \end{tabular}
\end{tabular}

}
\vspace*{-0.25cm}

\noindent \hrulefill
\\
\textbf{Basis vectors:}
\vspace*{-0.25cm}
\renewcommand{\arraystretch}{1.5}
\begin{longtabu} to \textwidth{>{\centering $}X[-1,c,c]<{$}>{\centering $}X[-1,c,c]<{$}>{\centering $}X[-1,c,c]<{$}>{\centering $}X[-1,c,c]<{$}>{\centering $}X[-1,c,c]<{$}>{\centering $}X[-1,c,c]<{$}>{\centering $}X[-1,c,c]<{$}}
  & & \mbox{Lattice Coordinates} & & \mbox{Cartesian Coordinates} &\mbox{Wyckoff Position} & \mbox{Atom Type} \\  
  \mathbf{B}_{1} & = & z_{1} \, \mathbf{a}_{3} & = & z_{1}c \, \mathbf{\hat{z}} & \left(2e\right) & \mbox{I I} \\ 
\mathbf{B}_{2} & = & -z_{1} \, \mathbf{a}_{3} & = & -z_{1}c \, \mathbf{\hat{z}} & \left(2e\right) & \mbox{I I} \\ 
\mathbf{B}_{3} & = & \frac{1}{2} \, \mathbf{a}_{2} + z_{2} \, \mathbf{a}_{3} & = & \frac{1}{2}a \, \mathbf{\hat{y}} + z_{2}c \, \mathbf{\hat{z}} & \left(2g\right) & \mbox{I II} \\ 
\mathbf{B}_{4} & = & \frac{1}{2} \, \mathbf{a}_{1} + -z_{2} \, \mathbf{a}_{3} & = & \frac{1}{2}a \, \mathbf{\hat{x}} + -z_{2}c \, \mathbf{\hat{z}} & \left(2g\right) & \mbox{I II} \\ 
\mathbf{B}_{5} & = & x_{3} \, \mathbf{a}_{1} + x_{3} \, \mathbf{a}_{2} + \frac{1}{2} \, \mathbf{a}_{3} & = & x_{3}a \, \mathbf{\hat{x}} + x_{3}a \, \mathbf{\hat{y}} + \frac{1}{2}c \, \mathbf{\hat{z}} & \left(4i\right) & \mbox{I III} \\ 
\mathbf{B}_{6} & = & -x_{3} \, \mathbf{a}_{1}-x_{3} \, \mathbf{a}_{2} + \frac{1}{2} \, \mathbf{a}_{3} & = & -x_{3}a \, \mathbf{\hat{x}}-x_{3}a \, \mathbf{\hat{y}} + \frac{1}{2}c \, \mathbf{\hat{z}} & \left(4i\right) & \mbox{I III} \\ 
\mathbf{B}_{7} & = & x_{3} \, \mathbf{a}_{1}-x_{3} \, \mathbf{a}_{2} + \frac{1}{2} \, \mathbf{a}_{3} & = & x_{3}a \, \mathbf{\hat{x}}-x_{3}a \, \mathbf{\hat{y}} + \frac{1}{2}c \, \mathbf{\hat{z}} & \left(4i\right) & \mbox{I III} \\ 
\mathbf{B}_{8} & = & -x_{3} \, \mathbf{a}_{1} + x_{3} \, \mathbf{a}_{2} + \frac{1}{2} \, \mathbf{a}_{3} & = & -x_{3}a \, \mathbf{\hat{x}} + x_{3}a \, \mathbf{\hat{y}} + \frac{1}{2}c \, \mathbf{\hat{z}} & \left(4i\right) & \mbox{I III} \\ 
\mathbf{B}_{9} & = & x_{4} \, \mathbf{a}_{1} + z_{4} \, \mathbf{a}_{3} & = & x_{4}a \, \mathbf{\hat{x}} + z_{4}c \, \mathbf{\hat{z}} & \left(4j\right) & \mbox{Hg} \\ 
\mathbf{B}_{10} & = & -x_{4} \, \mathbf{a}_{1} + z_{4} \, \mathbf{a}_{3} & = & -x_{4}a \, \mathbf{\hat{x}} + z_{4}c \, \mathbf{\hat{z}} & \left(4j\right) & \mbox{Hg} \\ 
\mathbf{B}_{11} & = & -x_{4} \, \mathbf{a}_{2}-z_{4} \, \mathbf{a}_{3} & = & -x_{4}a \, \mathbf{\hat{y}}-z_{4}c \, \mathbf{\hat{z}} & \left(4j\right) & \mbox{Hg} \\ 
\mathbf{B}_{12} & = & x_{4} \, \mathbf{a}_{2}-z_{4} \, \mathbf{a}_{3} & = & x_{4}a \, \mathbf{\hat{y}}-z_{4}c \, \mathbf{\hat{z}} & \left(4j\right) & \mbox{Hg} \\ 
\end{longtabu}
\renewcommand{\arraystretch}{1.0}
\noindent \hrulefill
\\
\textbf{References:}
\vspace*{-0.25cm}
\begin{flushleft}
  - \bibentry{Hostettler_HgI2_ActCrystallogrSecB_2002}. \\
\end{flushleft}
\textbf{Found in:}
\vspace*{-0.25cm}
\begin{flushleft}
  - \bibentry{Villars_PearsonsCrystalData_2013}. \\
\end{flushleft}
\noindent \hrulefill
\\
\textbf{Geometry files:}
\\
\noindent  - CIF: pp. {\hyperref[AB2_tP12_115_j_egi_cif]{\pageref{AB2_tP12_115_j_egi_cif}}} \\
\noindent  - POSCAR: pp. {\hyperref[AB2_tP12_115_j_egi_poscar]{\pageref{AB2_tP12_115_j_egi_poscar}}} \\
\onecolumn
{\phantomsection\label{A2B3_tP20_116_bci_fj}}
\subsection*{\huge \textbf{{\normalfont Ru$_{2}$Sn$_{3}$ Structure: A2B3\_tP20\_116\_bci\_fj}}}
\noindent \hrulefill
\vspace*{0.25cm}
\begin{figure}[htp]
  \centering
  \vspace{-1em}
  {\includegraphics[width=1\textwidth]{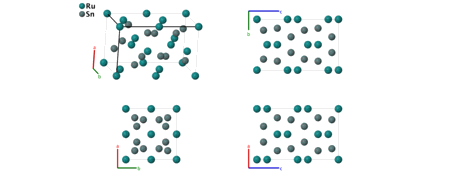}}
\end{figure}
\vspace*{-0.5cm}
\renewcommand{\arraystretch}{1.5}
\begin{equation*}
  \begin{array}{>{$\hspace{-0.15cm}}l<{$}>{$}p{0.5cm}<{$}>{$}p{18.5cm}<{$}}
    \mbox{\large \textbf{Prototype}} &\colon & \ce{Ru2Sn3} \\
    \mbox{\large \textbf{\AFLOW\ prototype label}} &\colon & \mbox{A2B3\_tP20\_116\_bci\_fj} \\
    \mbox{\large \textbf{\textit{Strukturbericht} designation}} &\colon & \mbox{None} \\
    \mbox{\large \textbf{Pearson symbol}} &\colon & \mbox{tP20} \\
    \mbox{\large \textbf{Space group number}} &\colon & 116 \\
    \mbox{\large \textbf{Space group symbol}} &\colon & P\bar{4}c2 \\
    \mbox{\large \textbf{\AFLOW\ prototype command}} &\colon &  \texttt{aflow} \,  \, \texttt{-{}-proto=A2B3\_tP20\_116\_bci\_fj } \, \newline \texttt{-{}-params=}{a,c/a,x_{3},z_{4},x_{5},y_{5},z_{5} }
  \end{array}
\end{equation*}
\renewcommand{\arraystretch}{1.0}

\noindent \parbox{1 \linewidth}{
\noindent \hrulefill
\\
\textbf{Simple Tetragonal primitive vectors:} \\
\vspace*{-0.25cm}
\begin{tabular}{cc}
  \begin{tabular}{c}
    \parbox{0.6 \linewidth}{
      \renewcommand{\arraystretch}{1.5}
      \begin{equation*}
        \centering
        \begin{array}{ccc}
              \mathbf{a}_1 & = & a \, \mathbf{\hat{x}} \\
    \mathbf{a}_2 & = & a \, \mathbf{\hat{y}} \\
    \mathbf{a}_3 & = & c \, \mathbf{\hat{z}} \\

        \end{array}
      \end{equation*}
    }
    \renewcommand{\arraystretch}{1.0}
  \end{tabular}
  \begin{tabular}{c}
    \includegraphics[width=0.3\linewidth]{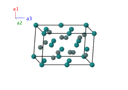} \\
  \end{tabular}
\end{tabular}

}
\vspace*{-0.25cm}

\noindent \hrulefill
\\
\textbf{Basis vectors:}
\vspace*{-0.25cm}
\renewcommand{\arraystretch}{1.5}
\begin{longtabu} to \textwidth{>{\centering $}X[-1,c,c]<{$}>{\centering $}X[-1,c,c]<{$}>{\centering $}X[-1,c,c]<{$}>{\centering $}X[-1,c,c]<{$}>{\centering $}X[-1,c,c]<{$}>{\centering $}X[-1,c,c]<{$}>{\centering $}X[-1,c,c]<{$}}
  & & \mbox{Lattice Coordinates} & & \mbox{Cartesian Coordinates} &\mbox{Wyckoff Position} & \mbox{Atom Type} \\  
  \mathbf{B}_{1} & = & \frac{1}{2} \, \mathbf{a}_{1} + \frac{1}{2} \, \mathbf{a}_{2} + \frac{1}{4} \, \mathbf{a}_{3} & = & \frac{1}{2}a \, \mathbf{\hat{x}} + \frac{1}{2}a \, \mathbf{\hat{y}} + \frac{1}{4}c \, \mathbf{\hat{z}} & \left(2b\right) & \mbox{Ru I} \\ 
\mathbf{B}_{2} & = & \frac{1}{2} \, \mathbf{a}_{1} + \frac{1}{2} \, \mathbf{a}_{2} + \frac{3}{4} \, \mathbf{a}_{3} & = & \frac{1}{2}a \, \mathbf{\hat{x}} + \frac{1}{2}a \, \mathbf{\hat{y}} + \frac{3}{4}c \, \mathbf{\hat{z}} & \left(2b\right) & \mbox{Ru I} \\ 
\mathbf{B}_{3} & = & 0 \, \mathbf{a}_{1} + 0 \, \mathbf{a}_{2} + 0 \, \mathbf{a}_{3} & = & 0 \, \mathbf{\hat{x}} + 0 \, \mathbf{\hat{y}} + 0 \, \mathbf{\hat{z}} & \left(2c\right) & \mbox{Ru II} \\ 
\mathbf{B}_{4} & = & \frac{1}{2} \, \mathbf{a}_{3} & = & \frac{1}{2}c \, \mathbf{\hat{z}} & \left(2c\right) & \mbox{Ru II} \\ 
\mathbf{B}_{5} & = & x_{3} \, \mathbf{a}_{1} + x_{3} \, \mathbf{a}_{2} + \frac{3}{4} \, \mathbf{a}_{3} & = & x_{3}a \, \mathbf{\hat{x}} + x_{3}a \, \mathbf{\hat{y}} + \frac{3}{4}c \, \mathbf{\hat{z}} & \left(4f\right) & \mbox{Sn I} \\ 
\mathbf{B}_{6} & = & -x_{3} \, \mathbf{a}_{1}-x_{3} \, \mathbf{a}_{2} + \frac{3}{4} \, \mathbf{a}_{3} & = & -x_{3}a \, \mathbf{\hat{x}}-x_{3}a \, \mathbf{\hat{y}} + \frac{3}{4}c \, \mathbf{\hat{z}} & \left(4f\right) & \mbox{Sn I} \\ 
\mathbf{B}_{7} & = & x_{3} \, \mathbf{a}_{1}-x_{3} \, \mathbf{a}_{2} + \frac{1}{4} \, \mathbf{a}_{3} & = & x_{3}a \, \mathbf{\hat{x}}-x_{3}a \, \mathbf{\hat{y}} + \frac{1}{4}c \, \mathbf{\hat{z}} & \left(4f\right) & \mbox{Sn I} \\ 
\mathbf{B}_{8} & = & -x_{3} \, \mathbf{a}_{1} + x_{3} \, \mathbf{a}_{2} + \frac{1}{4} \, \mathbf{a}_{3} & = & -x_{3}a \, \mathbf{\hat{x}} + x_{3}a \, \mathbf{\hat{y}} + \frac{1}{4}c \, \mathbf{\hat{z}} & \left(4f\right) & \mbox{Sn I} \\ 
\mathbf{B}_{9} & = & \frac{1}{2} \, \mathbf{a}_{2} + z_{4} \, \mathbf{a}_{3} & = & \frac{1}{2}a \, \mathbf{\hat{y}} + z_{4}c \, \mathbf{\hat{z}} & \left(4i\right) & \mbox{Ru III} \\ 
\mathbf{B}_{10} & = & \frac{1}{2} \, \mathbf{a}_{1} + -z_{4} \, \mathbf{a}_{3} & = & \frac{1}{2}a \, \mathbf{\hat{x}} + -z_{4}c \, \mathbf{\hat{z}} & \left(4i\right) & \mbox{Ru III} \\ 
\mathbf{B}_{11} & = & \frac{1}{2} \, \mathbf{a}_{2} + \left(\frac{1}{2} +z_{4}\right) \, \mathbf{a}_{3} & = & \frac{1}{2}a \, \mathbf{\hat{y}} + \left(\frac{1}{2} +z_{4}\right)c \, \mathbf{\hat{z}} & \left(4i\right) & \mbox{Ru III} \\ 
\mathbf{B}_{12} & = & \frac{1}{2} \, \mathbf{a}_{1} + \left(\frac{1}{2} - z_{4}\right) \, \mathbf{a}_{3} & = & \frac{1}{2}a \, \mathbf{\hat{x}} + \left(\frac{1}{2} - z_{4}\right)c \, \mathbf{\hat{z}} & \left(4i\right) & \mbox{Ru III} \\ 
\mathbf{B}_{13} & = & x_{5} \, \mathbf{a}_{1} + y_{5} \, \mathbf{a}_{2} + z_{5} \, \mathbf{a}_{3} & = & x_{5}a \, \mathbf{\hat{x}} + y_{5}a \, \mathbf{\hat{y}} + z_{5}c \, \mathbf{\hat{z}} & \left(8j\right) & \mbox{Sn II} \\ 
\mathbf{B}_{14} & = & -x_{5} \, \mathbf{a}_{1}-y_{5} \, \mathbf{a}_{2} + z_{5} \, \mathbf{a}_{3} & = & -x_{5}a \, \mathbf{\hat{x}}-y_{5}a \, \mathbf{\hat{y}} + z_{5}c \, \mathbf{\hat{z}} & \left(8j\right) & \mbox{Sn II} \\ 
\mathbf{B}_{15} & = & y_{5} \, \mathbf{a}_{1}-x_{5} \, \mathbf{a}_{2}-z_{5} \, \mathbf{a}_{3} & = & y_{5}a \, \mathbf{\hat{x}}-x_{5}a \, \mathbf{\hat{y}}-z_{5}c \, \mathbf{\hat{z}} & \left(8j\right) & \mbox{Sn II} \\ 
\mathbf{B}_{16} & = & -y_{5} \, \mathbf{a}_{1} + x_{5} \, \mathbf{a}_{2}-z_{5} \, \mathbf{a}_{3} & = & -y_{5}a \, \mathbf{\hat{x}} + x_{5}a \, \mathbf{\hat{y}}-z_{5}c \, \mathbf{\hat{z}} & \left(8j\right) & \mbox{Sn II} \\ 
\mathbf{B}_{17} & = & x_{5} \, \mathbf{a}_{1}-y_{5} \, \mathbf{a}_{2} + \left(\frac{1}{2} +z_{5}\right) \, \mathbf{a}_{3} & = & x_{5}a \, \mathbf{\hat{x}}-y_{5}a \, \mathbf{\hat{y}} + \left(\frac{1}{2} +z_{5}\right)c \, \mathbf{\hat{z}} & \left(8j\right) & \mbox{Sn II} \\ 
\mathbf{B}_{18} & = & -x_{5} \, \mathbf{a}_{1} + y_{5} \, \mathbf{a}_{2} + \left(\frac{1}{2} +z_{5}\right) \, \mathbf{a}_{3} & = & -x_{5}a \, \mathbf{\hat{x}} + y_{5}a \, \mathbf{\hat{y}} + \left(\frac{1}{2} +z_{5}\right)c \, \mathbf{\hat{z}} & \left(8j\right) & \mbox{Sn II} \\ 
\mathbf{B}_{19} & = & y_{5} \, \mathbf{a}_{1} + x_{5} \, \mathbf{a}_{2} + \left(\frac{1}{2} - z_{5}\right) \, \mathbf{a}_{3} & = & y_{5}a \, \mathbf{\hat{x}} + x_{5}a \, \mathbf{\hat{y}} + \left(\frac{1}{2} - z_{5}\right)c \, \mathbf{\hat{z}} & \left(8j\right) & \mbox{Sn II} \\ 
\mathbf{B}_{20} & = & -y_{5} \, \mathbf{a}_{1}-x_{5} \, \mathbf{a}_{2} + \left(\frac{1}{2} - z_{5}\right) \, \mathbf{a}_{3} & = & -y_{5}a \, \mathbf{\hat{x}}-x_{5}a \, \mathbf{\hat{y}} + \left(\frac{1}{2} - z_{5}\right)c \, \mathbf{\hat{z}} & \left(8j\right) & \mbox{Sn II} \\ 
\end{longtabu}
\renewcommand{\arraystretch}{1.0}
\noindent \hrulefill
\\
\textbf{References:}
\vspace*{-0.25cm}
\begin{flushleft}
  - \bibentry{Schwomma_Ru2Sn3_MonatshChem_1964}. \\
\end{flushleft}
\textbf{Found in:}
\vspace*{-0.25cm}
\begin{flushleft}
  - \bibentry{Villars_PearsonsCrystalData_2013}. \\
\end{flushleft}
\noindent \hrulefill
\\
\textbf{Geometry files:}
\\
\noindent  - CIF: pp. {\hyperref[A2B3_tP20_116_bci_fj_cif]{\pageref{A2B3_tP20_116_bci_fj_cif}}} \\
\noindent  - POSCAR: pp. {\hyperref[A2B3_tP20_116_bci_fj_poscar]{\pageref{A2B3_tP20_116_bci_fj_poscar}}} \\
\onecolumn
{\phantomsection\label{A2B3_tP20_117_i_adgh}}
\subsection*{\huge \textbf{{\normalfont \begin{raggedleft}$\beta$-Bi$_{2}$O$_{3}$ (High-temperature) Structure: \end{raggedleft} \\ A2B3\_tP20\_117\_i\_adgh}}}
\noindent \hrulefill
\vspace*{0.25cm}
\begin{figure}[htp]
  \centering
  \vspace{-1em}
  {\includegraphics[width=1\textwidth]{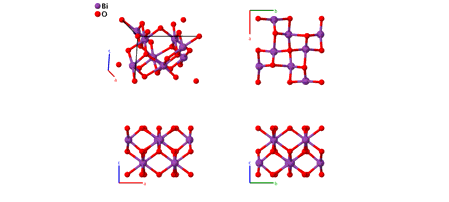}}
\end{figure}
\vspace*{-0.5cm}
\renewcommand{\arraystretch}{1.5}
\begin{equation*}
  \begin{array}{>{$\hspace{-0.15cm}}l<{$}>{$}p{0.5cm}<{$}>{$}p{18.5cm}<{$}}
    \mbox{\large \textbf{Prototype}} &\colon & \ce{$\beta$-Bi2O3} \\
    \mbox{\large \textbf{\AFLOW\ prototype label}} &\colon & \mbox{A2B3\_tP20\_117\_i\_adgh} \\
    \mbox{\large \textbf{\textit{Strukturbericht} designation}} &\colon & \mbox{None} \\
    \mbox{\large \textbf{Pearson symbol}} &\colon & \mbox{tP20} \\
    \mbox{\large \textbf{Space group number}} &\colon & 117 \\
    \mbox{\large \textbf{Space group symbol}} &\colon & P\bar{4}b2 \\
    \mbox{\large \textbf{\AFLOW\ prototype command}} &\colon &  \texttt{aflow} \,  \, \texttt{-{}-proto=A2B3\_tP20\_117\_i\_adgh } \, \newline \texttt{-{}-params=}{a,c/a,x_{3},x_{4},x_{5},y_{5},z_{5} }
  \end{array}
\end{equation*}
\renewcommand{\arraystretch}{1.0}

\noindent \parbox{1 \linewidth}{
\noindent \hrulefill
\\
\textbf{Simple Tetragonal primitive vectors:} \\
\vspace*{-0.25cm}
\begin{tabular}{cc}
  \begin{tabular}{c}
    \parbox{0.6 \linewidth}{
      \renewcommand{\arraystretch}{1.5}
      \begin{equation*}
        \centering
        \begin{array}{ccc}
              \mathbf{a}_1 & = & a \, \mathbf{\hat{x}} \\
    \mathbf{a}_2 & = & a \, \mathbf{\hat{y}} \\
    \mathbf{a}_3 & = & c \, \mathbf{\hat{z}} \\

        \end{array}
      \end{equation*}
    }
    \renewcommand{\arraystretch}{1.0}
  \end{tabular}
  \begin{tabular}{c}
    \includegraphics[width=0.3\linewidth]{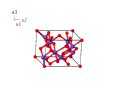} \\
  \end{tabular}
\end{tabular}

}
\vspace*{-0.25cm}

\noindent \hrulefill
\\
\textbf{Basis vectors:}
\vspace*{-0.25cm}
\renewcommand{\arraystretch}{1.5}
\begin{longtabu} to \textwidth{>{\centering $}X[-1,c,c]<{$}>{\centering $}X[-1,c,c]<{$}>{\centering $}X[-1,c,c]<{$}>{\centering $}X[-1,c,c]<{$}>{\centering $}X[-1,c,c]<{$}>{\centering $}X[-1,c,c]<{$}>{\centering $}X[-1,c,c]<{$}}
  & & \mbox{Lattice Coordinates} & & \mbox{Cartesian Coordinates} &\mbox{Wyckoff Position} & \mbox{Atom Type} \\  
  \mathbf{B}_{1} & = & 0 \, \mathbf{a}_{1} + 0 \, \mathbf{a}_{2} + 0 \, \mathbf{a}_{3} & = & 0 \, \mathbf{\hat{x}} + 0 \, \mathbf{\hat{y}} + 0 \, \mathbf{\hat{z}} & \left(2a\right) & \mbox{O I} \\ 
\mathbf{B}_{2} & = & \frac{1}{2} \, \mathbf{a}_{1} + \frac{1}{2} \, \mathbf{a}_{2} & = & \frac{1}{2}a \, \mathbf{\hat{x}} + \frac{1}{2}a \, \mathbf{\hat{y}} & \left(2a\right) & \mbox{O I} \\ 
\mathbf{B}_{3} & = & \frac{1}{2} \, \mathbf{a}_{2} + \frac{1}{2} \, \mathbf{a}_{3} & = & \frac{1}{2}a \, \mathbf{\hat{y}} + \frac{1}{2}c \, \mathbf{\hat{z}} & \left(2d\right) & \mbox{O II} \\ 
\mathbf{B}_{4} & = & \frac{1}{2} \, \mathbf{a}_{1} + \frac{1}{2} \, \mathbf{a}_{3} & = & \frac{1}{2}a \, \mathbf{\hat{x}} + \frac{1}{2}c \, \mathbf{\hat{z}} & \left(2d\right) & \mbox{O II} \\ 
\mathbf{B}_{5} & = & x_{3} \, \mathbf{a}_{1} + \left(\frac{1}{2} +x_{3}\right) \, \mathbf{a}_{2} & = & x_{3}a \, \mathbf{\hat{x}} + \left(\frac{1}{2} +x_{3}\right)a \, \mathbf{\hat{y}} & \left(4g\right) & \mbox{O III} \\ 
\mathbf{B}_{6} & = & -x_{3} \, \mathbf{a}_{1} + \left(\frac{1}{2} - x_{3}\right) \, \mathbf{a}_{2} & = & -x_{3}a \, \mathbf{\hat{x}} + \left(\frac{1}{2} - x_{3}\right)a \, \mathbf{\hat{y}} & \left(4g\right) & \mbox{O III} \\ 
\mathbf{B}_{7} & = & \left(\frac{1}{2} +x_{3}\right) \, \mathbf{a}_{1}-x_{3} \, \mathbf{a}_{2} & = & \left(\frac{1}{2} +x_{3}\right)a \, \mathbf{\hat{x}}-x_{3}a \, \mathbf{\hat{y}} & \left(4g\right) & \mbox{O III} \\ 
\mathbf{B}_{8} & = & \left(\frac{1}{2} - x_{3}\right) \, \mathbf{a}_{1} + x_{3} \, \mathbf{a}_{2} & = & \left(\frac{1}{2} - x_{3}\right)a \, \mathbf{\hat{x}} + x_{3}a \, \mathbf{\hat{y}} & \left(4g\right) & \mbox{O III} \\ 
\mathbf{B}_{9} & = & x_{4} \, \mathbf{a}_{1} + \left(\frac{1}{2} +x_{4}\right) \, \mathbf{a}_{2} + \frac{1}{2} \, \mathbf{a}_{3} & = & x_{4}a \, \mathbf{\hat{x}} + \left(\frac{1}{2} +x_{4}\right)a \, \mathbf{\hat{y}} + \frac{1}{2}c \, \mathbf{\hat{z}} & \left(4h\right) & \mbox{O IV} \\ 
\mathbf{B}_{10} & = & -x_{4} \, \mathbf{a}_{1} + \left(\frac{1}{2} - x_{4}\right) \, \mathbf{a}_{2} + \frac{1}{2} \, \mathbf{a}_{3} & = & -x_{4}a \, \mathbf{\hat{x}} + \left(\frac{1}{2} - x_{4}\right)a \, \mathbf{\hat{y}} + \frac{1}{2}c \, \mathbf{\hat{z}} & \left(4h\right) & \mbox{O IV} \\ 
\mathbf{B}_{11} & = & \left(\frac{1}{2} +x_{4}\right) \, \mathbf{a}_{1}-x_{4} \, \mathbf{a}_{2} + \frac{1}{2} \, \mathbf{a}_{3} & = & \left(\frac{1}{2} +x_{4}\right)a \, \mathbf{\hat{x}}-x_{4}a \, \mathbf{\hat{y}} + \frac{1}{2}c \, \mathbf{\hat{z}} & \left(4h\right) & \mbox{O IV} \\ 
\mathbf{B}_{12} & = & \left(\frac{1}{2} - x_{4}\right) \, \mathbf{a}_{1} + x_{4} \, \mathbf{a}_{2} + \frac{1}{2} \, \mathbf{a}_{3} & = & \left(\frac{1}{2} - x_{4}\right)a \, \mathbf{\hat{x}} + x_{4}a \, \mathbf{\hat{y}} + \frac{1}{2}c \, \mathbf{\hat{z}} & \left(4h\right) & \mbox{O IV} \\ 
\mathbf{B}_{13} & = & x_{5} \, \mathbf{a}_{1} + y_{5} \, \mathbf{a}_{2} + z_{5} \, \mathbf{a}_{3} & = & x_{5}a \, \mathbf{\hat{x}} + y_{5}a \, \mathbf{\hat{y}} + z_{5}c \, \mathbf{\hat{z}} & \left(8i\right) & \mbox{Bi} \\ 
\mathbf{B}_{14} & = & -x_{5} \, \mathbf{a}_{1}-y_{5} \, \mathbf{a}_{2} + z_{5} \, \mathbf{a}_{3} & = & -x_{5}a \, \mathbf{\hat{x}}-y_{5}a \, \mathbf{\hat{y}} + z_{5}c \, \mathbf{\hat{z}} & \left(8i\right) & \mbox{Bi} \\ 
\mathbf{B}_{15} & = & y_{5} \, \mathbf{a}_{1}-x_{5} \, \mathbf{a}_{2}-z_{5} \, \mathbf{a}_{3} & = & y_{5}a \, \mathbf{\hat{x}}-x_{5}a \, \mathbf{\hat{y}}-z_{5}c \, \mathbf{\hat{z}} & \left(8i\right) & \mbox{Bi} \\ 
\mathbf{B}_{16} & = & -y_{5} \, \mathbf{a}_{1} + x_{5} \, \mathbf{a}_{2}-z_{5} \, \mathbf{a}_{3} & = & -y_{5}a \, \mathbf{\hat{x}} + x_{5}a \, \mathbf{\hat{y}}-z_{5}c \, \mathbf{\hat{z}} & \left(8i\right) & \mbox{Bi} \\ 
\mathbf{B}_{17} & = & \left(\frac{1}{2} +x_{5}\right) \, \mathbf{a}_{1} + \left(\frac{1}{2} - y_{5}\right) \, \mathbf{a}_{2} + z_{5} \, \mathbf{a}_{3} & = & \left(\frac{1}{2} +x_{5}\right)a \, \mathbf{\hat{x}} + \left(\frac{1}{2} - y_{5}\right)a \, \mathbf{\hat{y}} + z_{5}c \, \mathbf{\hat{z}} & \left(8i\right) & \mbox{Bi} \\ 
\mathbf{B}_{18} & = & \left(\frac{1}{2} - x_{5}\right) \, \mathbf{a}_{1} + \left(\frac{1}{2} +y_{5}\right) \, \mathbf{a}_{2} + z_{5} \, \mathbf{a}_{3} & = & \left(\frac{1}{2} - x_{5}\right)a \, \mathbf{\hat{x}} + \left(\frac{1}{2} +y_{5}\right)a \, \mathbf{\hat{y}} + z_{5}c \, \mathbf{\hat{z}} & \left(8i\right) & \mbox{Bi} \\ 
\mathbf{B}_{19} & = & \left(\frac{1}{2} +y_{5}\right) \, \mathbf{a}_{1} + \left(\frac{1}{2} +x_{5}\right) \, \mathbf{a}_{2}-z_{5} \, \mathbf{a}_{3} & = & \left(\frac{1}{2} +y_{5}\right)a \, \mathbf{\hat{x}} + \left(\frac{1}{2} +x_{5}\right)a \, \mathbf{\hat{y}}-z_{5}c \, \mathbf{\hat{z}} & \left(8i\right) & \mbox{Bi} \\ 
\mathbf{B}_{20} & = & \left(\frac{1}{2} - y_{5}\right) \, \mathbf{a}_{1} + \left(\frac{1}{2} - x_{5}\right) \, \mathbf{a}_{2}-z_{5} \, \mathbf{a}_{3} & = & \left(\frac{1}{2} - y_{5}\right)a \, \mathbf{\hat{x}} + \left(\frac{1}{2} - x_{5}\right)a \, \mathbf{\hat{y}}-z_{5}c \, \mathbf{\hat{z}} & \left(8i\right) & \mbox{Bi} \\ 
\end{longtabu}
\renewcommand{\arraystretch}{1.0}
\noindent \hrulefill
\\
\textbf{References:}
\vspace*{-0.25cm}
\begin{flushleft}
  - \bibentry{Sillen_Bi2O3_ArkKemiMineralGeol_1937}. \\
\end{flushleft}
\textbf{Found in:}
\vspace*{-0.25cm}
\begin{flushleft}
  - \bibentry{Villars_PearsonsCrystalData_2013}. \\
\end{flushleft}
\noindent \hrulefill
\\
\textbf{Geometry files:}
\\
\noindent  - CIF: pp. {\hyperref[A2B3_tP20_117_i_adgh_cif]{\pageref{A2B3_tP20_117_i_adgh_cif}}} \\
\noindent  - POSCAR: pp. {\hyperref[A2B3_tP20_117_i_adgh_poscar]{\pageref{A2B3_tP20_117_i_adgh_poscar}}} \\
\onecolumn
{\phantomsection\label{A3B_tP16_118_ei_f}}
\subsection*{\huge \textbf{{\normalfont RuIn$_{3}$ Structure: A3B\_tP16\_118\_ei\_f}}}
\noindent \hrulefill
\vspace*{0.25cm}
\begin{figure}[htp]
  \centering
  \vspace{-1em}
  {\includegraphics[width=1\textwidth]{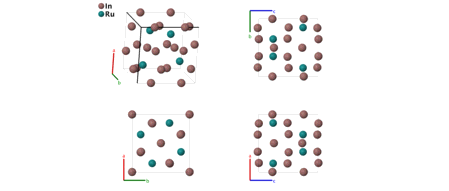}}
\end{figure}
\vspace*{-0.5cm}
\renewcommand{\arraystretch}{1.5}
\begin{equation*}
  \begin{array}{>{$\hspace{-0.15cm}}l<{$}>{$}p{0.5cm}<{$}>{$}p{18.5cm}<{$}}
    \mbox{\large \textbf{Prototype}} &\colon & \ce{RuIn3} \\
    \mbox{\large \textbf{\AFLOW\ prototype label}} &\colon & \mbox{A3B\_tP16\_118\_ei\_f} \\
    \mbox{\large \textbf{\textit{Strukturbericht} designation}} &\colon & \mbox{None} \\
    \mbox{\large \textbf{Pearson symbol}} &\colon & \mbox{tP16} \\
    \mbox{\large \textbf{Space group number}} &\colon & 118 \\
    \mbox{\large \textbf{Space group symbol}} &\colon & P\bar{4}n2 \\
    \mbox{\large \textbf{\AFLOW\ prototype command}} &\colon &  \texttt{aflow} \,  \, \texttt{-{}-proto=A3B\_tP16\_118\_ei\_f } \, \newline \texttt{-{}-params=}{a,c/a,z_{1},x_{2},x_{3},y_{3},z_{3} }
  \end{array}
\end{equation*}
\renewcommand{\arraystretch}{1.0}

\noindent \parbox{1 \linewidth}{
\noindent \hrulefill
\\
\textbf{Simple Tetragonal primitive vectors:} \\
\vspace*{-0.25cm}
\begin{tabular}{cc}
  \begin{tabular}{c}
    \parbox{0.6 \linewidth}{
      \renewcommand{\arraystretch}{1.5}
      \begin{equation*}
        \centering
        \begin{array}{ccc}
              \mathbf{a}_1 & = & a \, \mathbf{\hat{x}} \\
    \mathbf{a}_2 & = & a \, \mathbf{\hat{y}} \\
    \mathbf{a}_3 & = & c \, \mathbf{\hat{z}} \\

        \end{array}
      \end{equation*}
    }
    \renewcommand{\arraystretch}{1.0}
  \end{tabular}
  \begin{tabular}{c}
    \includegraphics[width=0.3\linewidth]{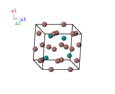} \\
  \end{tabular}
\end{tabular}

}
\vspace*{-0.25cm}

\noindent \hrulefill
\\
\textbf{Basis vectors:}
\vspace*{-0.25cm}
\renewcommand{\arraystretch}{1.5}
\begin{longtabu} to \textwidth{>{\centering $}X[-1,c,c]<{$}>{\centering $}X[-1,c,c]<{$}>{\centering $}X[-1,c,c]<{$}>{\centering $}X[-1,c,c]<{$}>{\centering $}X[-1,c,c]<{$}>{\centering $}X[-1,c,c]<{$}>{\centering $}X[-1,c,c]<{$}}
  & & \mbox{Lattice Coordinates} & & \mbox{Cartesian Coordinates} &\mbox{Wyckoff Position} & \mbox{Atom Type} \\  
  \mathbf{B}_{1} & = & z_{1} \, \mathbf{a}_{3} & = & z_{1}c \, \mathbf{\hat{z}} & \left(4e\right) & \mbox{In I} \\ 
\mathbf{B}_{2} & = & -z_{1} \, \mathbf{a}_{3} & = & -z_{1}c \, \mathbf{\hat{z}} & \left(4e\right) & \mbox{In I} \\ 
\mathbf{B}_{3} & = & \frac{1}{2} \, \mathbf{a}_{1} + \frac{1}{2} \, \mathbf{a}_{2} + \left(\frac{1}{2} +z_{1}\right) \, \mathbf{a}_{3} & = & \frac{1}{2}a \, \mathbf{\hat{x}} + \frac{1}{2}a \, \mathbf{\hat{y}} + \left(\frac{1}{2} +z_{1}\right)c \, \mathbf{\hat{z}} & \left(4e\right) & \mbox{In I} \\ 
\mathbf{B}_{4} & = & \frac{1}{2} \, \mathbf{a}_{1} + \frac{1}{2} \, \mathbf{a}_{2} + \left(\frac{1}{2} - z_{1}\right) \, \mathbf{a}_{3} & = & \frac{1}{2}a \, \mathbf{\hat{x}} + \frac{1}{2}a \, \mathbf{\hat{y}} + \left(\frac{1}{2} - z_{1}\right)c \, \mathbf{\hat{z}} & \left(4e\right) & \mbox{In I} \\ 
\mathbf{B}_{5} & = & x_{2} \, \mathbf{a}_{1} + \left(\frac{1}{2} - x_{2}\right) \, \mathbf{a}_{2} + \frac{1}{4} \, \mathbf{a}_{3} & = & x_{2}a \, \mathbf{\hat{x}} + \left(\frac{1}{2} - x_{2}\right)a \, \mathbf{\hat{y}} + \frac{1}{4}c \, \mathbf{\hat{z}} & \left(4f\right) & \mbox{Ru} \\ 
\mathbf{B}_{6} & = & -x_{2} \, \mathbf{a}_{1} + \left(\frac{1}{2} +x_{2}\right) \, \mathbf{a}_{2} + \frac{1}{4} \, \mathbf{a}_{3} & = & -x_{2}a \, \mathbf{\hat{x}} + \left(\frac{1}{2} +x_{2}\right)a \, \mathbf{\hat{y}} + \frac{1}{4}c \, \mathbf{\hat{z}} & \left(4f\right) & \mbox{Ru} \\ 
\mathbf{B}_{7} & = & \left(\frac{1}{2} - x_{2}\right) \, \mathbf{a}_{1}-x_{2} \, \mathbf{a}_{2} + \frac{3}{4} \, \mathbf{a}_{3} & = & \left(\frac{1}{2} - x_{2}\right)a \, \mathbf{\hat{x}}-x_{2}a \, \mathbf{\hat{y}} + \frac{3}{4}c \, \mathbf{\hat{z}} & \left(4f\right) & \mbox{Ru} \\ 
\mathbf{B}_{8} & = & \left(\frac{1}{2} +x_{2}\right) \, \mathbf{a}_{1} + x_{2} \, \mathbf{a}_{2} + \frac{3}{4} \, \mathbf{a}_{3} & = & \left(\frac{1}{2} +x_{2}\right)a \, \mathbf{\hat{x}} + x_{2}a \, \mathbf{\hat{y}} + \frac{3}{4}c \, \mathbf{\hat{z}} & \left(4f\right) & \mbox{Ru} \\ 
\mathbf{B}_{9} & = & x_{3} \, \mathbf{a}_{1} + y_{3} \, \mathbf{a}_{2} + z_{3} \, \mathbf{a}_{3} & = & x_{3}a \, \mathbf{\hat{x}} + y_{3}a \, \mathbf{\hat{y}} + z_{3}c \, \mathbf{\hat{z}} & \left(8i\right) & \mbox{In II} \\ 
\mathbf{B}_{10} & = & -x_{3} \, \mathbf{a}_{1}-y_{3} \, \mathbf{a}_{2} + z_{3} \, \mathbf{a}_{3} & = & -x_{3}a \, \mathbf{\hat{x}}-y_{3}a \, \mathbf{\hat{y}} + z_{3}c \, \mathbf{\hat{z}} & \left(8i\right) & \mbox{In II} \\ 
\mathbf{B}_{11} & = & y_{3} \, \mathbf{a}_{1}-x_{3} \, \mathbf{a}_{2}-z_{3} \, \mathbf{a}_{3} & = & y_{3}a \, \mathbf{\hat{x}}-x_{3}a \, \mathbf{\hat{y}}-z_{3}c \, \mathbf{\hat{z}} & \left(8i\right) & \mbox{In II} \\ 
\mathbf{B}_{12} & = & -y_{3} \, \mathbf{a}_{1} + x_{3} \, \mathbf{a}_{2}-z_{3} \, \mathbf{a}_{3} & = & -y_{3}a \, \mathbf{\hat{x}} + x_{3}a \, \mathbf{\hat{y}}-z_{3}c \, \mathbf{\hat{z}} & \left(8i\right) & \mbox{In II} \\ 
\mathbf{B}_{13} & = & \left(\frac{1}{2} +x_{3}\right) \, \mathbf{a}_{1} + \left(\frac{1}{2} - y_{3}\right) \, \mathbf{a}_{2} + \left(\frac{1}{2} +z_{3}\right) \, \mathbf{a}_{3} & = & \left(\frac{1}{2} +x_{3}\right)a \, \mathbf{\hat{x}} + \left(\frac{1}{2} - y_{3}\right)a \, \mathbf{\hat{y}} + \left(\frac{1}{2} +z_{3}\right)c \, \mathbf{\hat{z}} & \left(8i\right) & \mbox{In II} \\ 
\mathbf{B}_{14} & = & \left(\frac{1}{2} - x_{3}\right) \, \mathbf{a}_{1} + \left(\frac{1}{2} +y_{3}\right) \, \mathbf{a}_{2} + \left(\frac{1}{2} +z_{3}\right) \, \mathbf{a}_{3} & = & \left(\frac{1}{2} - x_{3}\right)a \, \mathbf{\hat{x}} + \left(\frac{1}{2} +y_{3}\right)a \, \mathbf{\hat{y}} + \left(\frac{1}{2} +z_{3}\right)c \, \mathbf{\hat{z}} & \left(8i\right) & \mbox{In II} \\ 
\mathbf{B}_{15} & = & \left(\frac{1}{2} +y_{3}\right) \, \mathbf{a}_{1} + \left(\frac{1}{2} +x_{3}\right) \, \mathbf{a}_{2} + \left(\frac{1}{2} - z_{3}\right) \, \mathbf{a}_{3} & = & \left(\frac{1}{2} +y_{3}\right)a \, \mathbf{\hat{x}} + \left(\frac{1}{2} +x_{3}\right)a \, \mathbf{\hat{y}} + \left(\frac{1}{2} - z_{3}\right)c \, \mathbf{\hat{z}} & \left(8i\right) & \mbox{In II} \\ 
\mathbf{B}_{16} & = & \left(\frac{1}{2} - y_{3}\right) \, \mathbf{a}_{1} + \left(\frac{1}{2} - x_{3}\right) \, \mathbf{a}_{2} + \left(\frac{1}{2} - z_{3}\right) \, \mathbf{a}_{3} & = & \left(\frac{1}{2} - y_{3}\right)a \, \mathbf{\hat{x}} + \left(\frac{1}{2} - x_{3}\right)a \, \mathbf{\hat{y}} + \left(\frac{1}{2} - z_{3}\right)c \, \mathbf{\hat{z}} & \left(8i\right) & \mbox{In II} \\ 
\end{longtabu}
\renewcommand{\arraystretch}{1.0}
\noindent \hrulefill
\\
\textbf{References:}
\vspace*{-0.25cm}
\begin{flushleft}
  - \bibentry{Roof_RuIn3_PowderDiff_1986}. \\
\end{flushleft}
\textbf{Found in:}
\vspace*{-0.25cm}
\begin{flushleft}
  - \bibentry{Villars_PearsonsCrystalData_2013}. \\
\end{flushleft}
\noindent \hrulefill
\\
\textbf{Geometry files:}
\\
\noindent  - CIF: pp. {\hyperref[A3B_tP16_118_ei_f_cif]{\pageref{A3B_tP16_118_ei_f_cif}}} \\
\noindent  - POSCAR: pp. {\hyperref[A3B_tP16_118_ei_f_poscar]{\pageref{A3B_tP16_118_ei_f_poscar}}} \\
\onecolumn
{\phantomsection\label{A5B3_tP32_118_g2i_aceh}}
\subsection*{\huge \textbf{{\normalfont Ir$_{3}$Ga$_{5}$ Structure: A5B3\_tP32\_118\_g2i\_aceh}}}
\noindent \hrulefill
\vspace*{0.25cm}
\begin{figure}[htp]
  \centering
  \vspace{-1em}
  {\includegraphics[width=1\textwidth]{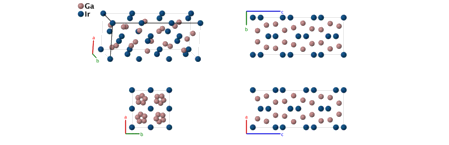}}
\end{figure}
\vspace*{-0.5cm}
\renewcommand{\arraystretch}{1.5}
\begin{equation*}
  \begin{array}{>{$\hspace{-0.15cm}}l<{$}>{$}p{0.5cm}<{$}>{$}p{18.5cm}<{$}}
    \mbox{\large \textbf{Prototype}} &\colon & \ce{Ir3Ga5} \\
    \mbox{\large \textbf{\AFLOW\ prototype label}} &\colon & \mbox{A5B3\_tP32\_118\_g2i\_aceh} \\
    \mbox{\large \textbf{\textit{Strukturbericht} designation}} &\colon & \mbox{None} \\
    \mbox{\large \textbf{Pearson symbol}} &\colon & \mbox{tP32} \\
    \mbox{\large \textbf{Space group number}} &\colon & 118 \\
    \mbox{\large \textbf{Space group symbol}} &\colon & P\bar{4}n2 \\
    \mbox{\large \textbf{\AFLOW\ prototype command}} &\colon &  \texttt{aflow} \,  \, \texttt{-{}-proto=A5B3\_tP32\_118\_g2i\_aceh } \, \newline \texttt{-{}-params=}{a,c/a,z_{3},x_{4},z_{5},x_{6},y_{6},z_{6},x_{7},y_{7},z_{7} }
  \end{array}
\end{equation*}
\renewcommand{\arraystretch}{1.0}

\noindent \parbox{1 \linewidth}{
\noindent \hrulefill
\\
\textbf{Simple Tetragonal primitive vectors:} \\
\vspace*{-0.25cm}
\begin{tabular}{cc}
  \begin{tabular}{c}
    \parbox{0.6 \linewidth}{
      \renewcommand{\arraystretch}{1.5}
      \begin{equation*}
        \centering
        \begin{array}{ccc}
              \mathbf{a}_1 & = & a \, \mathbf{\hat{x}} \\
    \mathbf{a}_2 & = & a \, \mathbf{\hat{y}} \\
    \mathbf{a}_3 & = & c \, \mathbf{\hat{z}} \\

        \end{array}
      \end{equation*}
    }
    \renewcommand{\arraystretch}{1.0}
  \end{tabular}
  \begin{tabular}{c}
    \includegraphics[width=0.3\linewidth]{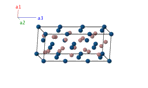} \\
  \end{tabular}
\end{tabular}

}
\vspace*{-0.25cm}

\noindent \hrulefill
\\
\textbf{Basis vectors:}
\vspace*{-0.25cm}
\renewcommand{\arraystretch}{1.5}
\begin{longtabu} to \textwidth{>{\centering $}X[-1,c,c]<{$}>{\centering $}X[-1,c,c]<{$}>{\centering $}X[-1,c,c]<{$}>{\centering $}X[-1,c,c]<{$}>{\centering $}X[-1,c,c]<{$}>{\centering $}X[-1,c,c]<{$}>{\centering $}X[-1,c,c]<{$}}
  & & \mbox{Lattice Coordinates} & & \mbox{Cartesian Coordinates} &\mbox{Wyckoff Position} & \mbox{Atom Type} \\  
  \mathbf{B}_{1} & = & 0 \, \mathbf{a}_{1} + 0 \, \mathbf{a}_{2} + 0 \, \mathbf{a}_{3} & = & 0 \, \mathbf{\hat{x}} + 0 \, \mathbf{\hat{y}} + 0 \, \mathbf{\hat{z}} & \left(2a\right) & \mbox{Ir I} \\ 
\mathbf{B}_{2} & = & \frac{1}{2} \, \mathbf{a}_{1} + \frac{1}{2} \, \mathbf{a}_{2} + \frac{1}{2} \, \mathbf{a}_{3} & = & \frac{1}{2}a \, \mathbf{\hat{x}} + \frac{1}{2}a \, \mathbf{\hat{y}} + \frac{1}{2}c \, \mathbf{\hat{z}} & \left(2a\right) & \mbox{Ir I} \\ 
\mathbf{B}_{3} & = & \frac{1}{2} \, \mathbf{a}_{2} + \frac{1}{4} \, \mathbf{a}_{3} & = & \frac{1}{2}a \, \mathbf{\hat{y}} + \frac{1}{4}c \, \mathbf{\hat{z}} & \left(2c\right) & \mbox{Ir II} \\ 
\mathbf{B}_{4} & = & \frac{1}{2} \, \mathbf{a}_{1} + \frac{3}{4} \, \mathbf{a}_{3} & = & \frac{1}{2}a \, \mathbf{\hat{x}} + \frac{3}{4}c \, \mathbf{\hat{z}} & \left(2c\right) & \mbox{Ir II} \\ 
\mathbf{B}_{5} & = & z_{3} \, \mathbf{a}_{3} & = & z_{3}c \, \mathbf{\hat{z}} & \left(4e\right) & \mbox{Ir III} \\ 
\mathbf{B}_{6} & = & -z_{3} \, \mathbf{a}_{3} & = & -z_{3}c \, \mathbf{\hat{z}} & \left(4e\right) & \mbox{Ir III} \\ 
\mathbf{B}_{7} & = & \frac{1}{2} \, \mathbf{a}_{1} + \frac{1}{2} \, \mathbf{a}_{2} + \left(\frac{1}{2} +z_{3}\right) \, \mathbf{a}_{3} & = & \frac{1}{2}a \, \mathbf{\hat{x}} + \frac{1}{2}a \, \mathbf{\hat{y}} + \left(\frac{1}{2} +z_{3}\right)c \, \mathbf{\hat{z}} & \left(4e\right) & \mbox{Ir III} \\ 
\mathbf{B}_{8} & = & \frac{1}{2} \, \mathbf{a}_{1} + \frac{1}{2} \, \mathbf{a}_{2} + \left(\frac{1}{2} - z_{3}\right) \, \mathbf{a}_{3} & = & \frac{1}{2}a \, \mathbf{\hat{x}} + \frac{1}{2}a \, \mathbf{\hat{y}} + \left(\frac{1}{2} - z_{3}\right)c \, \mathbf{\hat{z}} & \left(4e\right) & \mbox{Ir III} \\ 
\mathbf{B}_{9} & = & x_{4} \, \mathbf{a}_{1} + \left(\frac{1}{2} +x_{4}\right) \, \mathbf{a}_{2} + \frac{1}{4} \, \mathbf{a}_{3} & = & x_{4}a \, \mathbf{\hat{x}} + \left(\frac{1}{2} +x_{4}\right)a \, \mathbf{\hat{y}} + \frac{1}{4}c \, \mathbf{\hat{z}} & \left(4g\right) & \mbox{Ga I} \\ 
\mathbf{B}_{10} & = & -x_{4} \, \mathbf{a}_{1} + \left(\frac{1}{2} - x_{4}\right) \, \mathbf{a}_{2} + \frac{1}{4} \, \mathbf{a}_{3} & = & -x_{4}a \, \mathbf{\hat{x}} + \left(\frac{1}{2} - x_{4}\right)a \, \mathbf{\hat{y}} + \frac{1}{4}c \, \mathbf{\hat{z}} & \left(4g\right) & \mbox{Ga I} \\ 
\mathbf{B}_{11} & = & \left(\frac{1}{2} +x_{4}\right) \, \mathbf{a}_{1}-x_{4} \, \mathbf{a}_{2} + \frac{3}{4} \, \mathbf{a}_{3} & = & \left(\frac{1}{2} +x_{4}\right)a \, \mathbf{\hat{x}}-x_{4}a \, \mathbf{\hat{y}} + \frac{3}{4}c \, \mathbf{\hat{z}} & \left(4g\right) & \mbox{Ga I} \\ 
\mathbf{B}_{12} & = & \left(\frac{1}{2} - x_{4}\right) \, \mathbf{a}_{1} + x_{4} \, \mathbf{a}_{2} + \frac{3}{4} \, \mathbf{a}_{3} & = & \left(\frac{1}{2} - x_{4}\right)a \, \mathbf{\hat{x}} + x_{4}a \, \mathbf{\hat{y}} + \frac{3}{4}c \, \mathbf{\hat{z}} & \left(4g\right) & \mbox{Ga I} \\ 
\mathbf{B}_{13} & = & \frac{1}{2} \, \mathbf{a}_{2} + z_{5} \, \mathbf{a}_{3} & = & \frac{1}{2}a \, \mathbf{\hat{y}} + z_{5}c \, \mathbf{\hat{z}} & \left(4h\right) & \mbox{Ir IV} \\ 
\mathbf{B}_{14} & = & \frac{1}{2} \, \mathbf{a}_{1} + -z_{5} \, \mathbf{a}_{3} & = & \frac{1}{2}a \, \mathbf{\hat{x}} + -z_{5}c \, \mathbf{\hat{z}} & \left(4h\right) & \mbox{Ir IV} \\ 
\mathbf{B}_{15} & = & \frac{1}{2} \, \mathbf{a}_{1} + \left(\frac{1}{2} +z_{5}\right) \, \mathbf{a}_{3} & = & \frac{1}{2}a \, \mathbf{\hat{x}} + \left(\frac{1}{2} +z_{5}\right)c \, \mathbf{\hat{z}} & \left(4h\right) & \mbox{Ir IV} \\ 
\mathbf{B}_{16} & = & \frac{1}{2} \, \mathbf{a}_{2} + \left(\frac{1}{2} - z_{5}\right) \, \mathbf{a}_{3} & = & \frac{1}{2}a \, \mathbf{\hat{y}} + \left(\frac{1}{2} - z_{5}\right)c \, \mathbf{\hat{z}} & \left(4h\right) & \mbox{Ir IV} \\ 
\mathbf{B}_{17} & = & x_{6} \, \mathbf{a}_{1} + y_{6} \, \mathbf{a}_{2} + z_{6} \, \mathbf{a}_{3} & = & x_{6}a \, \mathbf{\hat{x}} + y_{6}a \, \mathbf{\hat{y}} + z_{6}c \, \mathbf{\hat{z}} & \left(8i\right) & \mbox{Ga II} \\ 
\mathbf{B}_{18} & = & -x_{6} \, \mathbf{a}_{1}-y_{6} \, \mathbf{a}_{2} + z_{6} \, \mathbf{a}_{3} & = & -x_{6}a \, \mathbf{\hat{x}}-y_{6}a \, \mathbf{\hat{y}} + z_{6}c \, \mathbf{\hat{z}} & \left(8i\right) & \mbox{Ga II} \\ 
\mathbf{B}_{19} & = & y_{6} \, \mathbf{a}_{1}-x_{6} \, \mathbf{a}_{2}-z_{6} \, \mathbf{a}_{3} & = & y_{6}a \, \mathbf{\hat{x}}-x_{6}a \, \mathbf{\hat{y}}-z_{6}c \, \mathbf{\hat{z}} & \left(8i\right) & \mbox{Ga II} \\ 
\mathbf{B}_{20} & = & -y_{6} \, \mathbf{a}_{1} + x_{6} \, \mathbf{a}_{2}-z_{6} \, \mathbf{a}_{3} & = & -y_{6}a \, \mathbf{\hat{x}} + x_{6}a \, \mathbf{\hat{y}}-z_{6}c \, \mathbf{\hat{z}} & \left(8i\right) & \mbox{Ga II} \\ 
\mathbf{B}_{21} & = & \left(\frac{1}{2} +x_{6}\right) \, \mathbf{a}_{1} + \left(\frac{1}{2} - y_{6}\right) \, \mathbf{a}_{2} + \left(\frac{1}{2} +z_{6}\right) \, \mathbf{a}_{3} & = & \left(\frac{1}{2} +x_{6}\right)a \, \mathbf{\hat{x}} + \left(\frac{1}{2} - y_{6}\right)a \, \mathbf{\hat{y}} + \left(\frac{1}{2} +z_{6}\right)c \, \mathbf{\hat{z}} & \left(8i\right) & \mbox{Ga II} \\ 
\mathbf{B}_{22} & = & \left(\frac{1}{2} - x_{6}\right) \, \mathbf{a}_{1} + \left(\frac{1}{2} +y_{6}\right) \, \mathbf{a}_{2} + \left(\frac{1}{2} +z_{6}\right) \, \mathbf{a}_{3} & = & \left(\frac{1}{2} - x_{6}\right)a \, \mathbf{\hat{x}} + \left(\frac{1}{2} +y_{6}\right)a \, \mathbf{\hat{y}} + \left(\frac{1}{2} +z_{6}\right)c \, \mathbf{\hat{z}} & \left(8i\right) & \mbox{Ga II} \\ 
\mathbf{B}_{23} & = & \left(\frac{1}{2} +y_{6}\right) \, \mathbf{a}_{1} + \left(\frac{1}{2} +x_{6}\right) \, \mathbf{a}_{2} + \left(\frac{1}{2} - z_{6}\right) \, \mathbf{a}_{3} & = & \left(\frac{1}{2} +y_{6}\right)a \, \mathbf{\hat{x}} + \left(\frac{1}{2} +x_{6}\right)a \, \mathbf{\hat{y}} + \left(\frac{1}{2} - z_{6}\right)c \, \mathbf{\hat{z}} & \left(8i\right) & \mbox{Ga II} \\ 
\mathbf{B}_{24} & = & \left(\frac{1}{2} - y_{6}\right) \, \mathbf{a}_{1} + \left(\frac{1}{2} - x_{6}\right) \, \mathbf{a}_{2} + \left(\frac{1}{2} - z_{6}\right) \, \mathbf{a}_{3} & = & \left(\frac{1}{2} - y_{6}\right)a \, \mathbf{\hat{x}} + \left(\frac{1}{2} - x_{6}\right)a \, \mathbf{\hat{y}} + \left(\frac{1}{2} - z_{6}\right)c \, \mathbf{\hat{z}} & \left(8i\right) & \mbox{Ga II} \\ 
\mathbf{B}_{25} & = & x_{7} \, \mathbf{a}_{1} + y_{7} \, \mathbf{a}_{2} + z_{7} \, \mathbf{a}_{3} & = & x_{7}a \, \mathbf{\hat{x}} + y_{7}a \, \mathbf{\hat{y}} + z_{7}c \, \mathbf{\hat{z}} & \left(8i\right) & \mbox{Ga III} \\ 
\mathbf{B}_{26} & = & -x_{7} \, \mathbf{a}_{1}-y_{7} \, \mathbf{a}_{2} + z_{7} \, \mathbf{a}_{3} & = & -x_{7}a \, \mathbf{\hat{x}}-y_{7}a \, \mathbf{\hat{y}} + z_{7}c \, \mathbf{\hat{z}} & \left(8i\right) & \mbox{Ga III} \\ 
\mathbf{B}_{27} & = & y_{7} \, \mathbf{a}_{1}-x_{7} \, \mathbf{a}_{2}-z_{7} \, \mathbf{a}_{3} & = & y_{7}a \, \mathbf{\hat{x}}-x_{7}a \, \mathbf{\hat{y}}-z_{7}c \, \mathbf{\hat{z}} & \left(8i\right) & \mbox{Ga III} \\ 
\mathbf{B}_{28} & = & -y_{7} \, \mathbf{a}_{1} + x_{7} \, \mathbf{a}_{2}-z_{7} \, \mathbf{a}_{3} & = & -y_{7}a \, \mathbf{\hat{x}} + x_{7}a \, \mathbf{\hat{y}}-z_{7}c \, \mathbf{\hat{z}} & \left(8i\right) & \mbox{Ga III} \\ 
\mathbf{B}_{29} & = & \left(\frac{1}{2} +x_{7}\right) \, \mathbf{a}_{1} + \left(\frac{1}{2} - y_{7}\right) \, \mathbf{a}_{2} + \left(\frac{1}{2} +z_{7}\right) \, \mathbf{a}_{3} & = & \left(\frac{1}{2} +x_{7}\right)a \, \mathbf{\hat{x}} + \left(\frac{1}{2} - y_{7}\right)a \, \mathbf{\hat{y}} + \left(\frac{1}{2} +z_{7}\right)c \, \mathbf{\hat{z}} & \left(8i\right) & \mbox{Ga III} \\ 
\mathbf{B}_{30} & = & \left(\frac{1}{2} - x_{7}\right) \, \mathbf{a}_{1} + \left(\frac{1}{2} +y_{7}\right) \, \mathbf{a}_{2} + \left(\frac{1}{2} +z_{7}\right) \, \mathbf{a}_{3} & = & \left(\frac{1}{2} - x_{7}\right)a \, \mathbf{\hat{x}} + \left(\frac{1}{2} +y_{7}\right)a \, \mathbf{\hat{y}} + \left(\frac{1}{2} +z_{7}\right)c \, \mathbf{\hat{z}} & \left(8i\right) & \mbox{Ga III} \\ 
\mathbf{B}_{31} & = & \left(\frac{1}{2} +y_{7}\right) \, \mathbf{a}_{1} + \left(\frac{1}{2} +x_{7}\right) \, \mathbf{a}_{2} + \left(\frac{1}{2} - z_{7}\right) \, \mathbf{a}_{3} & = & \left(\frac{1}{2} +y_{7}\right)a \, \mathbf{\hat{x}} + \left(\frac{1}{2} +x_{7}\right)a \, \mathbf{\hat{y}} + \left(\frac{1}{2} - z_{7}\right)c \, \mathbf{\hat{z}} & \left(8i\right) & \mbox{Ga III} \\ 
\mathbf{B}_{32} & = & \left(\frac{1}{2} - y_{7}\right) \, \mathbf{a}_{1} + \left(\frac{1}{2} - x_{7}\right) \, \mathbf{a}_{2} + \left(\frac{1}{2} - z_{7}\right) \, \mathbf{a}_{3} & = & \left(\frac{1}{2} - y_{7}\right)a \, \mathbf{\hat{x}} + \left(\frac{1}{2} - x_{7}\right)a \, \mathbf{\hat{y}} + \left(\frac{1}{2} - z_{7}\right)c \, \mathbf{\hat{z}} & \left(8i\right) & \mbox{Ga III} \\ 
\end{longtabu}
\renewcommand{\arraystretch}{1.0}
\noindent \hrulefill
\\
\textbf{References:}
\vspace*{-0.25cm}
\begin{flushleft}
  - \bibentry{Vollenkle_Ir3Ga5_MonatshChem_1967}. \\
\end{flushleft}
\textbf{Found in:}
\vspace*{-0.25cm}
\begin{flushleft}
  - \bibentry{Villars_PearsonsCrystalData_2013}. \\
\end{flushleft}
\noindent \hrulefill
\\
\textbf{Geometry files:}
\\
\noindent  - CIF: pp. {\hyperref[A5B3_tP32_118_g2i_aceh_cif]{\pageref{A5B3_tP32_118_g2i_aceh_cif}}} \\
\noindent  - POSCAR: pp. {\hyperref[A5B3_tP32_118_g2i_aceh_poscar]{\pageref{A5B3_tP32_118_g2i_aceh_poscar}}} \\
\onecolumn
{\phantomsection\label{A3B_tI24_119_b2i_af}}
\subsection*{\huge \textbf{{\normalfont RbGa$_{3}$ Structure: A3B\_tI24\_119\_b2i\_af}}}
\noindent \hrulefill
\vspace*{0.25cm}
\begin{figure}[htp]
  \centering
  \vspace{-1em}
  {\includegraphics[width=1\textwidth]{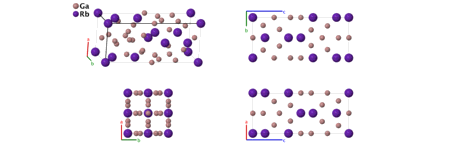}}
\end{figure}
\vspace*{-0.5cm}
\renewcommand{\arraystretch}{1.5}
\begin{equation*}
  \begin{array}{>{$\hspace{-0.15cm}}l<{$}>{$}p{0.5cm}<{$}>{$}p{18.5cm}<{$}}
    \mbox{\large \textbf{Prototype}} &\colon & \ce{RbGa3} \\
    \mbox{\large \textbf{\AFLOW\ prototype label}} &\colon & \mbox{A3B\_tI24\_119\_b2i\_af} \\
    \mbox{\large \textbf{\textit{Strukturbericht} designation}} &\colon & \mbox{None} \\
    \mbox{\large \textbf{Pearson symbol}} &\colon & \mbox{tI24} \\
    \mbox{\large \textbf{Space group number}} &\colon & 119 \\
    \mbox{\large \textbf{Space group symbol}} &\colon & I\bar{4}m2 \\
    \mbox{\large \textbf{\AFLOW\ prototype command}} &\colon &  \texttt{aflow} \,  \, \texttt{-{}-proto=A3B\_tI24\_119\_b2i\_af } \, \newline \texttt{-{}-params=}{a,c/a,z_{3},x_{4},z_{4},x_{5},z_{5} }
  \end{array}
\end{equation*}
\renewcommand{\arraystretch}{1.0}

\vspace*{-0.25cm}
\noindent \hrulefill
\\
\textbf{ Other compounds with this structure:}
\begin{itemize}
   \item{ CsGa$_{3}$, KGa$_{3}$  }
\end{itemize}
\noindent \parbox{1 \linewidth}{
\noindent \hrulefill
\\
\textbf{Body-centered Tetragonal primitive vectors:} \\
\vspace*{-0.25cm}
\begin{tabular}{cc}
  \begin{tabular}{c}
    \parbox{0.6 \linewidth}{
      \renewcommand{\arraystretch}{1.5}
      \begin{equation*}
        \centering
        \begin{array}{ccc}
              \mathbf{a}_1 & = & - \frac12 \, a \, \mathbf{\hat{x}} + \frac12 \, a \, \mathbf{\hat{y}} + \frac12 \, c \, \mathbf{\hat{z}} \\
    \mathbf{a}_2 & = & ~ \frac12 \, a \, \mathbf{\hat{x}} - \frac12 \, a \, \mathbf{\hat{y}} + \frac12 \, c \, \mathbf{\hat{z}} \\
    \mathbf{a}_3 & = & ~ \frac12 \, a \, \mathbf{\hat{x}} + \frac12 \, a \, \mathbf{\hat{y}} - \frac12 \, c \, \mathbf{\hat{z}} \\

        \end{array}
      \end{equation*}
    }
    \renewcommand{\arraystretch}{1.0}
  \end{tabular}
  \begin{tabular}{c}
    \includegraphics[width=0.3\linewidth]{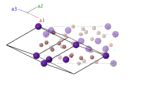} \\
  \end{tabular}
\end{tabular}

}
\vspace*{-0.25cm}

\noindent \hrulefill
\\
\textbf{Basis vectors:}
\vspace*{-0.25cm}
\renewcommand{\arraystretch}{1.5}
\begin{longtabu} to \textwidth{>{\centering $}X[-1,c,c]<{$}>{\centering $}X[-1,c,c]<{$}>{\centering $}X[-1,c,c]<{$}>{\centering $}X[-1,c,c]<{$}>{\centering $}X[-1,c,c]<{$}>{\centering $}X[-1,c,c]<{$}>{\centering $}X[-1,c,c]<{$}}
  & & \mbox{Lattice Coordinates} & & \mbox{Cartesian Coordinates} &\mbox{Wyckoff Position} & \mbox{Atom Type} \\  
  \mathbf{B}_{1} & = & 0 \, \mathbf{a}_{1} + 0 \, \mathbf{a}_{2} + 0 \, \mathbf{a}_{3} & = & 0 \, \mathbf{\hat{x}} + 0 \, \mathbf{\hat{y}} + 0 \, \mathbf{\hat{z}} & \left(2a\right) & \mbox{Rb I} \\ 
\mathbf{B}_{2} & = & \frac{1}{2} \, \mathbf{a}_{1} + \frac{1}{2} \, \mathbf{a}_{2} & = & \frac{1}{2}c \, \mathbf{\hat{z}} & \left(2b\right) & \mbox{Ga I} \\ 
\mathbf{B}_{3} & = & \left(\frac{1}{2} +z_{3}\right) \, \mathbf{a}_{1} + z_{3} \, \mathbf{a}_{2} + \frac{1}{2} \, \mathbf{a}_{3} & = & \frac{1}{2}a \, \mathbf{\hat{y}} + z_{3}c \, \mathbf{\hat{z}} & \left(4f\right) & \mbox{Rb II} \\ 
\mathbf{B}_{4} & = & -z_{3} \, \mathbf{a}_{1} + \left(\frac{1}{2} - z_{3}\right) \, \mathbf{a}_{2} + \frac{1}{2} \, \mathbf{a}_{3} & = & \frac{1}{2}a \, \mathbf{\hat{x}} + -z_{3}c \, \mathbf{\hat{z}} & \left(4f\right) & \mbox{Rb II} \\ 
\mathbf{B}_{5} & = & z_{4} \, \mathbf{a}_{1} + \left(x_{4}+z_{4}\right) \, \mathbf{a}_{2} + x_{4} \, \mathbf{a}_{3} & = & x_{4}a \, \mathbf{\hat{x}} + z_{4}c \, \mathbf{\hat{z}} & \left(8i\right) & \mbox{Ga II} \\ 
\mathbf{B}_{6} & = & z_{4} \, \mathbf{a}_{1} + \left(-x_{4}+z_{4}\right) \, \mathbf{a}_{2}-x_{4} \, \mathbf{a}_{3} & = & -x_{4}a \, \mathbf{\hat{x}} + z_{4}c \, \mathbf{\hat{z}} & \left(8i\right) & \mbox{Ga II} \\ 
\mathbf{B}_{7} & = & \left(-x_{4}-z_{4}\right) \, \mathbf{a}_{1}-z_{4} \, \mathbf{a}_{2}-x_{4} \, \mathbf{a}_{3} & = & -x_{4}a \, \mathbf{\hat{y}}-z_{4}c \, \mathbf{\hat{z}} & \left(8i\right) & \mbox{Ga II} \\ 
\mathbf{B}_{8} & = & \left(x_{4}-z_{4}\right) \, \mathbf{a}_{1}-z_{4} \, \mathbf{a}_{2} + x_{4} \, \mathbf{a}_{3} & = & x_{4}a \, \mathbf{\hat{y}}-z_{4}c \, \mathbf{\hat{z}} & \left(8i\right) & \mbox{Ga II} \\ 
\mathbf{B}_{9} & = & z_{5} \, \mathbf{a}_{1} + \left(x_{5}+z_{5}\right) \, \mathbf{a}_{2} + x_{5} \, \mathbf{a}_{3} & = & x_{5}a \, \mathbf{\hat{x}} + z_{5}c \, \mathbf{\hat{z}} & \left(8i\right) & \mbox{Ga III} \\ 
\mathbf{B}_{10} & = & z_{5} \, \mathbf{a}_{1} + \left(-x_{5}+z_{5}\right) \, \mathbf{a}_{2}-x_{5} \, \mathbf{a}_{3} & = & -x_{5}a \, \mathbf{\hat{x}} + z_{5}c \, \mathbf{\hat{z}} & \left(8i\right) & \mbox{Ga III} \\ 
\mathbf{B}_{11} & = & \left(-x_{5}-z_{5}\right) \, \mathbf{a}_{1}-z_{5} \, \mathbf{a}_{2}-x_{5} \, \mathbf{a}_{3} & = & -x_{5}a \, \mathbf{\hat{y}}-z_{5}c \, \mathbf{\hat{z}} & \left(8i\right) & \mbox{Ga III} \\ 
\mathbf{B}_{12} & = & \left(x_{5}-z_{5}\right) \, \mathbf{a}_{1}-z_{5} \, \mathbf{a}_{2} + x_{5} \, \mathbf{a}_{3} & = & x_{5}a \, \mathbf{\hat{y}}-z_{5}c \, \mathbf{\hat{z}} & \left(8i\right) & \mbox{Ga III} \\ 
\end{longtabu}
\renewcommand{\arraystretch}{1.0}
\noindent \hrulefill
\\
\textbf{References:}
\vspace*{-0.25cm}
\begin{flushleft}
  - \bibentry{Ling_ZAAC_480_1981}. \\
\end{flushleft}
\textbf{Found in:}
\vspace*{-0.25cm}
\begin{flushleft}
  - \bibentry{Villars_Pearson_Handbook_III_1991}. \\
\end{flushleft}
\noindent \hrulefill
\\
\textbf{Geometry files:}
\\
\noindent  - CIF: pp. {\hyperref[A3B_tI24_119_b2i_af_cif]{\pageref{A3B_tI24_119_b2i_af_cif}}} \\
\noindent  - POSCAR: pp. {\hyperref[A3B_tI24_119_b2i_af_poscar]{\pageref{A3B_tI24_119_b2i_af_poscar}}} \\
\onecolumn
{\phantomsection\label{AB_tI4_119_c_a}}
\subsection*{\huge \textbf{{\normalfont GaSb Structure: AB\_tI4\_119\_c\_a}}}
\noindent \hrulefill
\vspace*{0.25cm}
\begin{figure}[htp]
  \centering
  \vspace{-1em}
  {\includegraphics[width=1\textwidth]{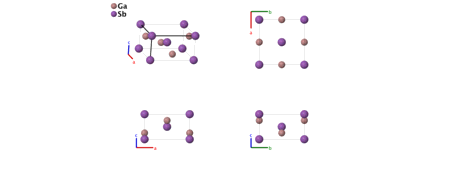}}
\end{figure}
\vspace*{-0.5cm}
\renewcommand{\arraystretch}{1.5}
\begin{equation*}
  \begin{array}{>{$\hspace{-0.15cm}}l<{$}>{$}p{0.5cm}<{$}>{$}p{18.5cm}<{$}}
    \mbox{\large \textbf{Prototype}} &\colon & \ce{GaSb} \\
    \mbox{\large \textbf{\AFLOW\ prototype label}} &\colon & \mbox{AB\_tI4\_119\_c\_a} \\
    \mbox{\large \textbf{\textit{Strukturbericht} designation}} &\colon & \mbox{None} \\
    \mbox{\large \textbf{Pearson symbol}} &\colon & \mbox{tI4} \\
    \mbox{\large \textbf{Space group number}} &\colon & 119 \\
    \mbox{\large \textbf{Space group symbol}} &\colon & I\bar{4}m2 \\
    \mbox{\large \textbf{\AFLOW\ prototype command}} &\colon &  \texttt{aflow} \,  \, \texttt{-{}-proto=AB\_tI4\_119\_c\_a } \, \newline \texttt{-{}-params=}{a,c/a }
  \end{array}
\end{equation*}
\renewcommand{\arraystretch}{1.0}

\noindent \parbox{1 \linewidth}{
\noindent \hrulefill
\\
\textbf{Body-centered Tetragonal primitive vectors:} \\
\vspace*{-0.25cm}
\begin{tabular}{cc}
  \begin{tabular}{c}
    \parbox{0.6 \linewidth}{
      \renewcommand{\arraystretch}{1.5}
      \begin{equation*}
        \centering
        \begin{array}{ccc}
              \mathbf{a}_1 & = & - \frac12 \, a \, \mathbf{\hat{x}} + \frac12 \, a \, \mathbf{\hat{y}} + \frac12 \, c \, \mathbf{\hat{z}} \\
    \mathbf{a}_2 & = & ~ \frac12 \, a \, \mathbf{\hat{x}} - \frac12 \, a \, \mathbf{\hat{y}} + \frac12 \, c \, \mathbf{\hat{z}} \\
    \mathbf{a}_3 & = & ~ \frac12 \, a \, \mathbf{\hat{x}} + \frac12 \, a \, \mathbf{\hat{y}} - \frac12 \, c \, \mathbf{\hat{z}} \\

        \end{array}
      \end{equation*}
    }
    \renewcommand{\arraystretch}{1.0}
  \end{tabular}
  \begin{tabular}{c}
    \includegraphics[width=0.3\linewidth]{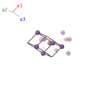} \\
  \end{tabular}
\end{tabular}

}
\vspace*{-0.25cm}

\noindent \hrulefill
\\
\textbf{Basis vectors:}
\vspace*{-0.25cm}
\renewcommand{\arraystretch}{1.5}
\begin{longtabu} to \textwidth{>{\centering $}X[-1,c,c]<{$}>{\centering $}X[-1,c,c]<{$}>{\centering $}X[-1,c,c]<{$}>{\centering $}X[-1,c,c]<{$}>{\centering $}X[-1,c,c]<{$}>{\centering $}X[-1,c,c]<{$}>{\centering $}X[-1,c,c]<{$}}
  & & \mbox{Lattice Coordinates} & & \mbox{Cartesian Coordinates} &\mbox{Wyckoff Position} & \mbox{Atom Type} \\  
  \mathbf{B}_{1} & = & 0 \, \mathbf{a}_{1} + 0 \, \mathbf{a}_{2} + 0 \, \mathbf{a}_{3} & = & 0 \, \mathbf{\hat{x}} + 0 \, \mathbf{\hat{y}} + 0 \, \mathbf{\hat{z}} & \left(2a\right) & \mbox{Sb} \\ 
\mathbf{B}_{2} & = & \frac{3}{4} \, \mathbf{a}_{1} + \frac{1}{4} \, \mathbf{a}_{2} + \frac{1}{2} \, \mathbf{a}_{3} & = & \frac{1}{2}a \, \mathbf{\hat{y}} + \frac{1}{4}c \, \mathbf{\hat{z}} & \left(2c\right) & \mbox{Ga} \\ 
\end{longtabu}
\renewcommand{\arraystretch}{1.0}
\noindent \hrulefill
\\
\textbf{References:}
\vspace*{-0.25cm}
\begin{flushleft}
  - \bibentry{Mcdonald_GaSb_JApplPhys_1965}. \\
\end{flushleft}
\textbf{Found in:}
\vspace*{-0.25cm}
\begin{flushleft}
  - \bibentry{Villars_PearsonsCrystalData_2013}. \\
\end{flushleft}
\noindent \hrulefill
\\
\textbf{Geometry files:}
\\
\noindent  - CIF: pp. {\hyperref[AB_tI4_119_c_a_cif]{\pageref{AB_tI4_119_c_a_cif}}} \\
\noindent  - POSCAR: pp. {\hyperref[AB_tI4_119_c_a_poscar]{\pageref{AB_tI4_119_c_a_poscar}}} \\
\onecolumn
{\phantomsection\label{A4BC2_tI28_120_i_d_e}}
\subsection*{\huge \textbf{{\normalfont KAu$_{4}$Sn$_{2}$ Structure: A4BC2\_tI28\_120\_i\_d\_e}}}
\noindent \hrulefill
\vspace*{0.25cm}
\begin{figure}[htp]
  \centering
  \vspace{-1em}
  {\includegraphics[width=1\textwidth]{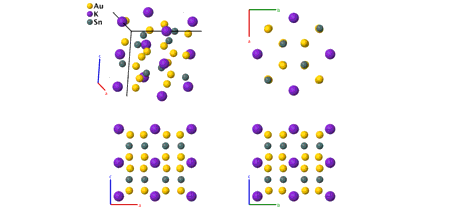}}
\end{figure}
\vspace*{-0.5cm}
\renewcommand{\arraystretch}{1.5}
\begin{equation*}
  \begin{array}{>{$\hspace{-0.15cm}}l<{$}>{$}p{0.5cm}<{$}>{$}p{18.5cm}<{$}}
    \mbox{\large \textbf{Prototype}} &\colon & \ce{KAu4Sn2} \\
    \mbox{\large \textbf{\AFLOW\ prototype label}} &\colon & \mbox{A4BC2\_tI28\_120\_i\_d\_e} \\
    \mbox{\large \textbf{\textit{Strukturbericht} designation}} &\colon & \mbox{None} \\
    \mbox{\large \textbf{Pearson symbol}} &\colon & \mbox{tI28} \\
    \mbox{\large \textbf{Space group number}} &\colon & 120 \\
    \mbox{\large \textbf{Space group symbol}} &\colon & I\bar{4}c2 \\
    \mbox{\large \textbf{\AFLOW\ prototype command}} &\colon &  \texttt{aflow} \,  \, \texttt{-{}-proto=A4BC2\_tI28\_120\_i\_d\_e } \, \newline \texttt{-{}-params=}{a,c/a,x_{2},x_{3},y_{3},z_{3} }
  \end{array}
\end{equation*}
\renewcommand{\arraystretch}{1.0}

\noindent \parbox{1 \linewidth}{
\noindent \hrulefill
\\
\textbf{Body-centered Tetragonal primitive vectors:} \\
\vspace*{-0.25cm}
\begin{tabular}{cc}
  \begin{tabular}{c}
    \parbox{0.6 \linewidth}{
      \renewcommand{\arraystretch}{1.5}
      \begin{equation*}
        \centering
        \begin{array}{ccc}
              \mathbf{a}_1 & = & - \frac12 \, a \, \mathbf{\hat{x}} + \frac12 \, a \, \mathbf{\hat{y}} + \frac12 \, c \, \mathbf{\hat{z}} \\
    \mathbf{a}_2 & = & ~ \frac12 \, a \, \mathbf{\hat{x}} - \frac12 \, a \, \mathbf{\hat{y}} + \frac12 \, c \, \mathbf{\hat{z}} \\
    \mathbf{a}_3 & = & ~ \frac12 \, a \, \mathbf{\hat{x}} + \frac12 \, a \, \mathbf{\hat{y}} - \frac12 \, c \, \mathbf{\hat{z}} \\

        \end{array}
      \end{equation*}
    }
    \renewcommand{\arraystretch}{1.0}
  \end{tabular}
  \begin{tabular}{c}
    \includegraphics[width=0.3\linewidth]{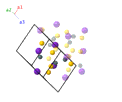} \\
  \end{tabular}
\end{tabular}

}
\vspace*{-0.25cm}

\noindent \hrulefill
\\
\textbf{Basis vectors:}
\vspace*{-0.25cm}
\renewcommand{\arraystretch}{1.5}
\begin{longtabu} to \textwidth{>{\centering $}X[-1,c,c]<{$}>{\centering $}X[-1,c,c]<{$}>{\centering $}X[-1,c,c]<{$}>{\centering $}X[-1,c,c]<{$}>{\centering $}X[-1,c,c]<{$}>{\centering $}X[-1,c,c]<{$}>{\centering $}X[-1,c,c]<{$}}
  & & \mbox{Lattice Coordinates} & & \mbox{Cartesian Coordinates} &\mbox{Wyckoff Position} & \mbox{Atom Type} \\  
  \mathbf{B}_{1} & = & \frac{1}{2} \, \mathbf{a}_{1} + \frac{1}{2} \, \mathbf{a}_{3} & = & \frac{1}{2}a \, \mathbf{\hat{y}} & \left(4d\right) & \mbox{K} \\ 
\mathbf{B}_{2} & = & \frac{1}{2} \, \mathbf{a}_{2} + \frac{1}{2} \, \mathbf{a}_{3} & = & \frac{1}{2}a \, \mathbf{\hat{x}} & \left(4d\right) & \mbox{K} \\ 
\mathbf{B}_{3} & = & \left(\frac{1}{4} +x_{2}\right) \, \mathbf{a}_{1} + \left(\frac{1}{4} +x_{2}\right) \, \mathbf{a}_{2} + 2x_{2} \, \mathbf{a}_{3} & = & x_{2}a \, \mathbf{\hat{x}} + x_{2}a \, \mathbf{\hat{y}} + \frac{1}{4}c \, \mathbf{\hat{z}} & \left(8e\right) & \mbox{Sn} \\ 
\mathbf{B}_{4} & = & \left(\frac{1}{4} - x_{2}\right) \, \mathbf{a}_{1} + \left(\frac{1}{4} - x_{2}\right) \, \mathbf{a}_{2}-2x_{2} \, \mathbf{a}_{3} & = & -x_{2}a \, \mathbf{\hat{x}}-x_{2}a \, \mathbf{\hat{y}} + \frac{1}{4}c \, \mathbf{\hat{z}} & \left(8e\right) & \mbox{Sn} \\ 
\mathbf{B}_{5} & = & \left(\frac{3}{4} - x_{2}\right) \, \mathbf{a}_{1} + \left(\frac{3}{4} +x_{2}\right) \, \mathbf{a}_{2} & = & x_{2}a \, \mathbf{\hat{x}}-x_{2}a \, \mathbf{\hat{y}} + \frac{3}{4}c \, \mathbf{\hat{z}} & \left(8e\right) & \mbox{Sn} \\ 
\mathbf{B}_{6} & = & \left(\frac{3}{4} +x_{2}\right) \, \mathbf{a}_{1} + \left(\frac{3}{4} - x_{2}\right) \, \mathbf{a}_{2} & = & -x_{2}a \, \mathbf{\hat{x}} + x_{2}a \, \mathbf{\hat{y}} + \frac{3}{4}c \, \mathbf{\hat{z}} & \left(8e\right) & \mbox{Sn} \\ 
\mathbf{B}_{7} & = & \left(y_{3}+z_{3}\right) \, \mathbf{a}_{1} + \left(x_{3}+z_{3}\right) \, \mathbf{a}_{2} + \left(x_{3}+y_{3}\right) \, \mathbf{a}_{3} & = & x_{3}a \, \mathbf{\hat{x}} + y_{3}a \, \mathbf{\hat{y}} + z_{3}c \, \mathbf{\hat{z}} & \left(16i\right) & \mbox{Au} \\ 
\mathbf{B}_{8} & = & \left(-y_{3}+z_{3}\right) \, \mathbf{a}_{1} + \left(-x_{3}+z_{3}\right) \, \mathbf{a}_{2} + \left(-x_{3}-y_{3}\right) \, \mathbf{a}_{3} & = & -x_{3}a \, \mathbf{\hat{x}}-y_{3}a \, \mathbf{\hat{y}} + z_{3}c \, \mathbf{\hat{z}} & \left(16i\right) & \mbox{Au} \\ 
\mathbf{B}_{9} & = & \left(-x_{3}-z_{3}\right) \, \mathbf{a}_{1} + \left(y_{3}-z_{3}\right) \, \mathbf{a}_{2} + \left(-x_{3}+y_{3}\right) \, \mathbf{a}_{3} & = & y_{3}a \, \mathbf{\hat{x}}-x_{3}a \, \mathbf{\hat{y}}-z_{3}c \, \mathbf{\hat{z}} & \left(16i\right) & \mbox{Au} \\ 
\mathbf{B}_{10} & = & \left(x_{3}-z_{3}\right) \, \mathbf{a}_{1} + \left(-y_{3}-z_{3}\right) \, \mathbf{a}_{2} + \left(x_{3}-y_{3}\right) \, \mathbf{a}_{3} & = & -y_{3}a \, \mathbf{\hat{x}} + x_{3}a \, \mathbf{\hat{y}}-z_{3}c \, \mathbf{\hat{z}} & \left(16i\right) & \mbox{Au} \\ 
\mathbf{B}_{11} & = & \left(\frac{1}{2} - y_{3} + z_{3}\right) \, \mathbf{a}_{1} + \left(\frac{1}{2} +x_{3} + z_{3}\right) \, \mathbf{a}_{2} + \left(x_{3}-y_{3}\right) \, \mathbf{a}_{3} & = & x_{3}a \, \mathbf{\hat{x}}-y_{3}a \, \mathbf{\hat{y}} + \left(\frac{1}{2} +z_{3}\right)c \, \mathbf{\hat{z}} & \left(16i\right) & \mbox{Au} \\ 
\mathbf{B}_{12} & = & \left(\frac{1}{2} +y_{3} + z_{3}\right) \, \mathbf{a}_{1} + \left(\frac{1}{2} - x_{3} + z_{3}\right) \, \mathbf{a}_{2} + \left(-x_{3}+y_{3}\right) \, \mathbf{a}_{3} & = & -x_{3}a \, \mathbf{\hat{x}} + y_{3}a \, \mathbf{\hat{y}} + \left(\frac{1}{2} +z_{3}\right)c \, \mathbf{\hat{z}} & \left(16i\right) & \mbox{Au} \\ 
\mathbf{B}_{13} & = & \left(\frac{1}{2} +x_{3} - z_{3}\right) \, \mathbf{a}_{1} + \left(\frac{1}{2} +y_{3} - z_{3}\right) \, \mathbf{a}_{2} + \left(x_{3}+y_{3}\right) \, \mathbf{a}_{3} & = & y_{3}a \, \mathbf{\hat{x}} + x_{3}a \, \mathbf{\hat{y}} + \left(\frac{1}{2} - z_{3}\right)c \, \mathbf{\hat{z}} & \left(16i\right) & \mbox{Au} \\ 
\mathbf{B}_{14} & = & \left(\frac{1}{2} - x_{3} - z_{3}\right) \, \mathbf{a}_{1} + \left(\frac{1}{2} - y_{3} - z_{3}\right) \, \mathbf{a}_{2} + \left(-x_{3}-y_{3}\right) \, \mathbf{a}_{3} & = & -y_{3}a \, \mathbf{\hat{x}}-x_{3}a \, \mathbf{\hat{y}} + \left(\frac{1}{2} - z_{3}\right)c \, \mathbf{\hat{z}} & \left(16i\right) & \mbox{Au} \\ 
\end{longtabu}
\renewcommand{\arraystretch}{1.0}
\noindent \hrulefill
\\
\textbf{References:}
\vspace*{-0.25cm}
\begin{flushleft}
  - \bibentry{Sinnen_KAu4Sn2_ZNaturB_1978}. \\
\end{flushleft}
\textbf{Found in:}
\vspace*{-0.25cm}
\begin{flushleft}
  - \bibentry{Villars_PearsonsCrystalData_2013}. \\
\end{flushleft}
\noindent \hrulefill
\\
\textbf{Geometry files:}
\\
\noindent  - CIF: pp. {\hyperref[A4BC2_tI28_120_i_d_e_cif]{\pageref{A4BC2_tI28_120_i_d_e_cif}}} \\
\noindent  - POSCAR: pp. {\hyperref[A4BC2_tI28_120_i_d_e_poscar]{\pageref{A4BC2_tI28_120_i_d_e_poscar}}} \\
\onecolumn
{\phantomsection\label{A4BC4D_tP10_123_gh_a_i_d}}
\subsection*{\huge \textbf{{\normalfont \begin{raggedleft}CaRbFe$_{4}$As$_{4}$ (Superconducting) Structure: \end{raggedleft} \\ A4BC4D\_tP10\_123\_gh\_a\_i\_d}}}
\noindent \hrulefill
\vspace*{0.25cm}
\begin{figure}[htp]
  \centering
  \vspace{-1em}
  {\includegraphics[width=1\textwidth]{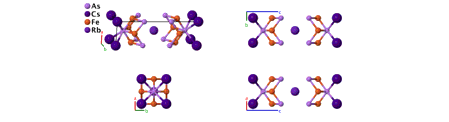}}
\end{figure}
\vspace*{-0.5cm}
\renewcommand{\arraystretch}{1.5}
\begin{equation*}
  \begin{array}{>{$\hspace{-0.15cm}}l<{$}>{$}p{0.5cm}<{$}>{$}p{18.5cm}<{$}}
    \mbox{\large \textbf{Prototype}} &\colon & \ce{CsRbFe4As4} \\
    \mbox{\large \textbf{\AFLOW\ prototype label}} &\colon & \mbox{A4BC4D\_tP10\_123\_gh\_a\_i\_d} \\
    \mbox{\large \textbf{\textit{Strukturbericht} designation}} &\colon & \mbox{None} \\
    \mbox{\large \textbf{Pearson symbol}} &\colon & \mbox{tP10} \\
    \mbox{\large \textbf{Space group number}} &\colon & 123 \\
    \mbox{\large \textbf{Space group symbol}} &\colon & P4/mmm \\
    \mbox{\large \textbf{\AFLOW\ prototype command}} &\colon &  \texttt{aflow} \,  \, \texttt{-{}-proto=A4BC4D\_tP10\_123\_gh\_a\_i\_d } \, \newline \texttt{-{}-params=}{a,c/a,z_{3},z_{4},z_{5} }
  \end{array}
\end{equation*}
\renewcommand{\arraystretch}{1.0}

\vspace*{-0.25cm}
\noindent \hrulefill
\\
\textbf{ Other compounds with this structure:}
\begin{itemize}
   \item{ CaKFe$_{4}$As$_{4}$, CaCsFe$_{4}$As$_{4}$, SrRbFe$_{4}$As$_{4}$, SrCsFe$_{4}$As$_{4}$, BaCsFe$_{4}$As$_{4}$  }
\end{itemize}
\vspace*{-0.25cm}
\noindent \hrulefill
\begin{itemize}
  \item{These compounds form a family of stoichiometric superconductors with
transition temperatures $T_{\mathrm{c}}$ ranging from 26--37~K.
}
\end{itemize}

\noindent \parbox{1 \linewidth}{
\noindent \hrulefill
\\
\textbf{Simple Tetragonal primitive vectors:} \\
\vspace*{-0.25cm}
\begin{tabular}{cc}
  \begin{tabular}{c}
    \parbox{0.6 \linewidth}{
      \renewcommand{\arraystretch}{1.5}
      \begin{equation*}
        \centering
        \begin{array}{ccc}
              \mathbf{a}_1 & = & a \, \mathbf{\hat{x}} \\
    \mathbf{a}_2 & = & a \, \mathbf{\hat{y}} \\
    \mathbf{a}_3 & = & c \, \mathbf{\hat{z}} \\

        \end{array}
      \end{equation*}
    }
    \renewcommand{\arraystretch}{1.0}
  \end{tabular}
  \begin{tabular}{c}
    \includegraphics[width=0.3\linewidth]{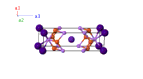} \\
  \end{tabular}
\end{tabular}

}
\vspace*{-0.25cm}

\noindent \hrulefill
\\
\textbf{Basis vectors:}
\vspace*{-0.25cm}
\renewcommand{\arraystretch}{1.5}
\begin{longtabu} to \textwidth{>{\centering $}X[-1,c,c]<{$}>{\centering $}X[-1,c,c]<{$}>{\centering $}X[-1,c,c]<{$}>{\centering $}X[-1,c,c]<{$}>{\centering $}X[-1,c,c]<{$}>{\centering $}X[-1,c,c]<{$}>{\centering $}X[-1,c,c]<{$}}
  & & \mbox{Lattice Coordinates} & & \mbox{Cartesian Coordinates} &\mbox{Wyckoff Position} & \mbox{Atom Type} \\  
  \mathbf{B}_{1} & = & 0 \, \mathbf{a}_{1} + 0 \, \mathbf{a}_{2} + 0 \, \mathbf{a}_{3} & = & 0 \, \mathbf{\hat{x}} + 0 \, \mathbf{\hat{y}} + 0 \, \mathbf{\hat{z}} & \left(1a\right) & \mbox{Cs} \\ 
\mathbf{B}_{2} & = & \frac{1}{2} \, \mathbf{a}_{1} + \frac{1}{2} \, \mathbf{a}_{2} + \frac{1}{2} \, \mathbf{a}_{3} & = & \frac{1}{2}a \, \mathbf{\hat{x}} + \frac{1}{2}a \, \mathbf{\hat{y}} + \frac{1}{2}c \, \mathbf{\hat{z}} & \left(1d\right) & \mbox{Rb} \\ 
\mathbf{B}_{3} & = & z_{3} \, \mathbf{a}_{3} & = & z_{3}c \, \mathbf{\hat{z}} & \left(2g\right) & \mbox{As I} \\ 
\mathbf{B}_{4} & = & -z_{3} \, \mathbf{a}_{3} & = & -z_{3}c \, \mathbf{\hat{z}} & \left(2g\right) & \mbox{As I} \\ 
\mathbf{B}_{5} & = & \frac{1}{2} \, \mathbf{a}_{1} + \frac{1}{2} \, \mathbf{a}_{2} + z_{4} \, \mathbf{a}_{3} & = & \frac{1}{2}a \, \mathbf{\hat{x}} + \frac{1}{2}a \, \mathbf{\hat{y}} + z_{4}c \, \mathbf{\hat{z}} & \left(2h\right) & \mbox{As II} \\ 
\mathbf{B}_{6} & = & \frac{1}{2} \, \mathbf{a}_{1} + \frac{1}{2} \, \mathbf{a}_{2}-z_{4} \, \mathbf{a}_{3} & = & \frac{1}{2}a \, \mathbf{\hat{x}} + \frac{1}{2}a \, \mathbf{\hat{y}}-z_{4}c \, \mathbf{\hat{z}} & \left(2h\right) & \mbox{As II} \\ 
\mathbf{B}_{7} & = & \frac{1}{2} \, \mathbf{a}_{2} + z_{5} \, \mathbf{a}_{3} & = & \frac{1}{2}a \, \mathbf{\hat{y}} + z_{5}c \, \mathbf{\hat{z}} & \left(4i\right) & \mbox{Fe} \\ 
\mathbf{B}_{8} & = & \frac{1}{2} \, \mathbf{a}_{1} + z_{5} \, \mathbf{a}_{3} & = & \frac{1}{2}a \, \mathbf{\hat{x}} + z_{5}c \, \mathbf{\hat{z}} & \left(4i\right) & \mbox{Fe} \\ 
\mathbf{B}_{9} & = & \frac{1}{2} \, \mathbf{a}_{2}-z_{5} \, \mathbf{a}_{3} & = & \frac{1}{2}a \, \mathbf{\hat{y}}-z_{5}c \, \mathbf{\hat{z}} & \left(4i\right) & \mbox{Fe} \\ 
\mathbf{B}_{10} & = & \frac{1}{2} \, \mathbf{a}_{1} + -z_{5} \, \mathbf{a}_{3} & = & \frac{1}{2}a \, \mathbf{\hat{x}} + -z_{5}c \, \mathbf{\hat{z}} & \left(4i\right) & \mbox{Fe} \\ 
\end{longtabu}
\renewcommand{\arraystretch}{1.0}
\noindent \hrulefill
\\
\textbf{References:}
\vspace*{-0.25cm}
\begin{flushleft}
  - \bibentry{Iyo_JACS_138_2016}. \\
\end{flushleft}
\noindent \hrulefill
\\
\textbf{Geometry files:}
\\
\noindent  - CIF: pp. {\hyperref[A4BC4D_tP10_123_gh_a_i_d_cif]{\pageref{A4BC4D_tP10_123_gh_a_i_d_cif}}} \\
\noindent  - POSCAR: pp. {\hyperref[A4BC4D_tP10_123_gh_a_i_d_poscar]{\pageref{A4BC4D_tP10_123_gh_a_i_d_poscar}}} \\
\onecolumn
{\phantomsection\label{AB4C_tP12_124_a_m_c}}
\subsection*{\huge \textbf{{\normalfont Nb$_{4}$CoSi Structure: AB4C\_tP12\_124\_a\_m\_c}}}
\noindent \hrulefill
\vspace*{0.25cm}
\begin{figure}[htp]
  \centering
  \vspace{-1em}
  {\includegraphics[width=1\textwidth]{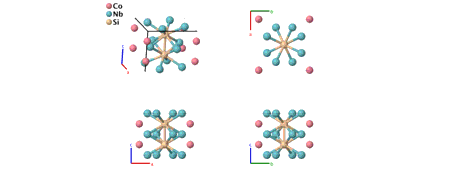}}
\end{figure}
\vspace*{-0.5cm}
\renewcommand{\arraystretch}{1.5}
\begin{equation*}
  \begin{array}{>{$\hspace{-0.15cm}}l<{$}>{$}p{0.5cm}<{$}>{$}p{18.5cm}<{$}}
    \mbox{\large \textbf{Prototype}} &\colon & \ce{Nb4CoSi} \\
    \mbox{\large \textbf{\AFLOW\ prototype label}} &\colon & \mbox{AB4C\_tP12\_124\_a\_m\_c} \\
    \mbox{\large \textbf{\textit{Strukturbericht} designation}} &\colon & \mbox{None} \\
    \mbox{\large \textbf{Pearson symbol}} &\colon & \mbox{tP12} \\
    \mbox{\large \textbf{Space group number}} &\colon & 124 \\
    \mbox{\large \textbf{Space group symbol}} &\colon & P4/mcc \\
    \mbox{\large \textbf{\AFLOW\ prototype command}} &\colon &  \texttt{aflow} \,  \, \texttt{-{}-proto=AB4C\_tP12\_124\_a\_m\_c } \, \newline \texttt{-{}-params=}{a,c/a,x_{3},y_{3} }
  \end{array}
\end{equation*}
\renewcommand{\arraystretch}{1.0}

\noindent \parbox{1 \linewidth}{
\noindent \hrulefill
\\
\textbf{Simple Tetragonal primitive vectors:} \\
\vspace*{-0.25cm}
\begin{tabular}{cc}
  \begin{tabular}{c}
    \parbox{0.6 \linewidth}{
      \renewcommand{\arraystretch}{1.5}
      \begin{equation*}
        \centering
        \begin{array}{ccc}
              \mathbf{a}_1 & = & a \, \mathbf{\hat{x}} \\
    \mathbf{a}_2 & = & a \, \mathbf{\hat{y}} \\
    \mathbf{a}_3 & = & c \, \mathbf{\hat{z}} \\

        \end{array}
      \end{equation*}
    }
    \renewcommand{\arraystretch}{1.0}
  \end{tabular}
  \begin{tabular}{c}
    \includegraphics[width=0.3\linewidth]{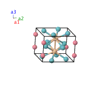} \\
  \end{tabular}
\end{tabular}

}
\vspace*{-0.25cm}

\noindent \hrulefill
\\
\textbf{Basis vectors:}
\vspace*{-0.25cm}
\renewcommand{\arraystretch}{1.5}
\begin{longtabu} to \textwidth{>{\centering $}X[-1,c,c]<{$}>{\centering $}X[-1,c,c]<{$}>{\centering $}X[-1,c,c]<{$}>{\centering $}X[-1,c,c]<{$}>{\centering $}X[-1,c,c]<{$}>{\centering $}X[-1,c,c]<{$}>{\centering $}X[-1,c,c]<{$}}
  & & \mbox{Lattice Coordinates} & & \mbox{Cartesian Coordinates} &\mbox{Wyckoff Position} & \mbox{Atom Type} \\  
  \mathbf{B}_{1} & = & \frac{1}{4} \, \mathbf{a}_{3} & = & \frac{1}{4}c \, \mathbf{\hat{z}} & \left(2a\right) & \mbox{Co} \\ 
\mathbf{B}_{2} & = & \frac{3}{4} \, \mathbf{a}_{3} & = & \frac{3}{4}c \, \mathbf{\hat{z}} & \left(2a\right) & \mbox{Co} \\ 
\mathbf{B}_{3} & = & \frac{1}{2} \, \mathbf{a}_{1} + \frac{1}{2} \, \mathbf{a}_{2} + \frac{1}{4} \, \mathbf{a}_{3} & = & \frac{1}{2}a \, \mathbf{\hat{x}} + \frac{1}{2}a \, \mathbf{\hat{y}} + \frac{1}{4}c \, \mathbf{\hat{z}} & \left(2c\right) & \mbox{Si} \\ 
\mathbf{B}_{4} & = & \frac{1}{2} \, \mathbf{a}_{1} + \frac{1}{2} \, \mathbf{a}_{2} + \frac{3}{4} \, \mathbf{a}_{3} & = & \frac{1}{2}a \, \mathbf{\hat{x}} + \frac{1}{2}a \, \mathbf{\hat{y}} + \frac{3}{4}c \, \mathbf{\hat{z}} & \left(2c\right) & \mbox{Si} \\ 
\mathbf{B}_{5} & = & x_{3} \, \mathbf{a}_{1} + y_{3} \, \mathbf{a}_{2} & = & x_{3}a \, \mathbf{\hat{x}} + y_{3}a \, \mathbf{\hat{y}} & \left(8m\right) & \mbox{Nb} \\ 
\mathbf{B}_{6} & = & -x_{3} \, \mathbf{a}_{1}-y_{3} \, \mathbf{a}_{2} & = & -x_{3}a \, \mathbf{\hat{x}}-y_{3}a \, \mathbf{\hat{y}} & \left(8m\right) & \mbox{Nb} \\ 
\mathbf{B}_{7} & = & -y_{3} \, \mathbf{a}_{1} + x_{3} \, \mathbf{a}_{2} & = & -y_{3}a \, \mathbf{\hat{x}} + x_{3}a \, \mathbf{\hat{y}} & \left(8m\right) & \mbox{Nb} \\ 
\mathbf{B}_{8} & = & y_{3} \, \mathbf{a}_{1}-x_{3} \, \mathbf{a}_{2} & = & y_{3}a \, \mathbf{\hat{x}}-x_{3}a \, \mathbf{\hat{y}} & \left(8m\right) & \mbox{Nb} \\ 
\mathbf{B}_{9} & = & -x_{3} \, \mathbf{a}_{1} + y_{3} \, \mathbf{a}_{2} + \frac{1}{2} \, \mathbf{a}_{3} & = & -x_{3}a \, \mathbf{\hat{x}} + y_{3}a \, \mathbf{\hat{y}} + \frac{1}{2}c \, \mathbf{\hat{z}} & \left(8m\right) & \mbox{Nb} \\ 
\mathbf{B}_{10} & = & x_{3} \, \mathbf{a}_{1}-y_{3} \, \mathbf{a}_{2} + \frac{1}{2} \, \mathbf{a}_{3} & = & x_{3}a \, \mathbf{\hat{x}}-y_{3}a \, \mathbf{\hat{y}} + \frac{1}{2}c \, \mathbf{\hat{z}} & \left(8m\right) & \mbox{Nb} \\ 
\mathbf{B}_{11} & = & y_{3} \, \mathbf{a}_{1} + x_{3} \, \mathbf{a}_{2} + \frac{1}{2} \, \mathbf{a}_{3} & = & y_{3}a \, \mathbf{\hat{x}} + x_{3}a \, \mathbf{\hat{y}} + \frac{1}{2}c \, \mathbf{\hat{z}} & \left(8m\right) & \mbox{Nb} \\ 
\mathbf{B}_{12} & = & -y_{3} \, \mathbf{a}_{1}-x_{3} \, \mathbf{a}_{2} + \frac{1}{2} \, \mathbf{a}_{3} & = & -y_{3}a \, \mathbf{\hat{x}}-x_{3}a \, \mathbf{\hat{y}} + \frac{1}{2}c \, \mathbf{\hat{z}} & \left(8m\right) & \mbox{Nb} \\ 
\end{longtabu}
\renewcommand{\arraystretch}{1.0}
\noindent \hrulefill
\\
\textbf{References:}
\vspace*{-0.25cm}
\begin{flushleft}
  - \bibentry{Gladyshevskii_Nb4CoSi_JStructChem_1965}. \\
\end{flushleft}
\textbf{Found in:}
\vspace*{-0.25cm}
\begin{flushleft}
  - \bibentry{Villars_PearsonsCrystalData_2013}. \\
\end{flushleft}
\noindent \hrulefill
\\
\textbf{Geometry files:}
\\
\noindent  - CIF: pp. {\hyperref[AB4C_tP12_124_a_m_c_cif]{\pageref{AB4C_tP12_124_a_m_c_cif}}} \\
\noindent  - POSCAR: pp. {\hyperref[AB4C_tP12_124_a_m_c_poscar]{\pageref{AB4C_tP12_124_a_m_c_poscar}}} \\
\onecolumn
{\phantomsection\label{AB4_tP10_124_a_m}}
\subsection*{\huge \textbf{{\normalfont NbTe$_{4}$ Structure: AB4\_tP10\_124\_a\_m}}}
\noindent \hrulefill
\vspace*{0.25cm}
\begin{figure}[htp]
  \centering
  \vspace{-1em}
  {\includegraphics[width=1\textwidth]{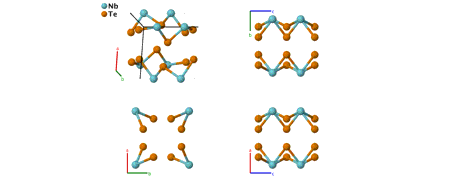}}
\end{figure}
\vspace*{-0.5cm}
\renewcommand{\arraystretch}{1.5}
\begin{equation*}
  \begin{array}{>{$\hspace{-0.15cm}}l<{$}>{$}p{0.5cm}<{$}>{$}p{18.5cm}<{$}}
    \mbox{\large \textbf{Prototype}} &\colon & \ce{NbTe4} \\
    \mbox{\large \textbf{\AFLOW\ prototype label}} &\colon & \mbox{AB4\_tP10\_124\_a\_m} \\
    \mbox{\large \textbf{\textit{Strukturbericht} designation}} &\colon & \mbox{None} \\
    \mbox{\large \textbf{Pearson symbol}} &\colon & \mbox{tP10} \\
    \mbox{\large \textbf{Space group number}} &\colon & 124 \\
    \mbox{\large \textbf{Space group symbol}} &\colon & P4/mcc \\
    \mbox{\large \textbf{\AFLOW\ prototype command}} &\colon &  \texttt{aflow} \,  \, \texttt{-{}-proto=AB4\_tP10\_124\_a\_m } \, \newline \texttt{-{}-params=}{a,c/a,x_{2},y_{2} }
  \end{array}
\end{equation*}
\renewcommand{\arraystretch}{1.0}

\noindent \parbox{1 \linewidth}{
\noindent \hrulefill
\\
\textbf{Simple Tetragonal primitive vectors:} \\
\vspace*{-0.25cm}
\begin{tabular}{cc}
  \begin{tabular}{c}
    \parbox{0.6 \linewidth}{
      \renewcommand{\arraystretch}{1.5}
      \begin{equation*}
        \centering
        \begin{array}{ccc}
              \mathbf{a}_1 & = & a \, \mathbf{\hat{x}} \\
    \mathbf{a}_2 & = & a \, \mathbf{\hat{y}} \\
    \mathbf{a}_3 & = & c \, \mathbf{\hat{z}} \\

        \end{array}
      \end{equation*}
    }
    \renewcommand{\arraystretch}{1.0}
  \end{tabular}
  \begin{tabular}{c}
    \includegraphics[width=0.3\linewidth]{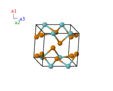} \\
  \end{tabular}
\end{tabular}

}
\vspace*{-0.25cm}

\noindent \hrulefill
\\
\textbf{Basis vectors:}
\vspace*{-0.25cm}
\renewcommand{\arraystretch}{1.5}
\begin{longtabu} to \textwidth{>{\centering $}X[-1,c,c]<{$}>{\centering $}X[-1,c,c]<{$}>{\centering $}X[-1,c,c]<{$}>{\centering $}X[-1,c,c]<{$}>{\centering $}X[-1,c,c]<{$}>{\centering $}X[-1,c,c]<{$}>{\centering $}X[-1,c,c]<{$}}
  & & \mbox{Lattice Coordinates} & & \mbox{Cartesian Coordinates} &\mbox{Wyckoff Position} & \mbox{Atom Type} \\  
  \mathbf{B}_{1} & = & \frac{1}{4} \, \mathbf{a}_{3} & = & \frac{1}{4}c \, \mathbf{\hat{z}} & \left(2a\right) & \mbox{Nb} \\ 
\mathbf{B}_{2} & = & \frac{3}{4} \, \mathbf{a}_{3} & = & \frac{3}{4}c \, \mathbf{\hat{z}} & \left(2a\right) & \mbox{Nb} \\ 
\mathbf{B}_{3} & = & x_{2} \, \mathbf{a}_{1} + y_{2} \, \mathbf{a}_{2} & = & x_{2}a \, \mathbf{\hat{x}} + y_{2}a \, \mathbf{\hat{y}} & \left(8m\right) & \mbox{Te} \\ 
\mathbf{B}_{4} & = & -x_{2} \, \mathbf{a}_{1}-y_{2} \, \mathbf{a}_{2} & = & -x_{2}a \, \mathbf{\hat{x}}-y_{2}a \, \mathbf{\hat{y}} & \left(8m\right) & \mbox{Te} \\ 
\mathbf{B}_{5} & = & -y_{2} \, \mathbf{a}_{1} + x_{2} \, \mathbf{a}_{2} & = & -y_{2}a \, \mathbf{\hat{x}} + x_{2}a \, \mathbf{\hat{y}} & \left(8m\right) & \mbox{Te} \\ 
\mathbf{B}_{6} & = & y_{2} \, \mathbf{a}_{1}-x_{2} \, \mathbf{a}_{2} & = & y_{2}a \, \mathbf{\hat{x}}-x_{2}a \, \mathbf{\hat{y}} & \left(8m\right) & \mbox{Te} \\ 
\mathbf{B}_{7} & = & -x_{2} \, \mathbf{a}_{1} + y_{2} \, \mathbf{a}_{2} + \frac{1}{2} \, \mathbf{a}_{3} & = & -x_{2}a \, \mathbf{\hat{x}} + y_{2}a \, \mathbf{\hat{y}} + \frac{1}{2}c \, \mathbf{\hat{z}} & \left(8m\right) & \mbox{Te} \\ 
\mathbf{B}_{8} & = & x_{2} \, \mathbf{a}_{1}-y_{2} \, \mathbf{a}_{2} + \frac{1}{2} \, \mathbf{a}_{3} & = & x_{2}a \, \mathbf{\hat{x}}-y_{2}a \, \mathbf{\hat{y}} + \frac{1}{2}c \, \mathbf{\hat{z}} & \left(8m\right) & \mbox{Te} \\ 
\mathbf{B}_{9} & = & y_{2} \, \mathbf{a}_{1} + x_{2} \, \mathbf{a}_{2} + \frac{1}{2} \, \mathbf{a}_{3} & = & y_{2}a \, \mathbf{\hat{x}} + x_{2}a \, \mathbf{\hat{y}} + \frac{1}{2}c \, \mathbf{\hat{z}} & \left(8m\right) & \mbox{Te} \\ 
\mathbf{B}_{10} & = & -y_{2} \, \mathbf{a}_{1}-x_{2} \, \mathbf{a}_{2} + \frac{1}{2} \, \mathbf{a}_{3} & = & -y_{2}a \, \mathbf{\hat{x}}-x_{2}a \, \mathbf{\hat{y}} + \frac{1}{2}c \, \mathbf{\hat{z}} & \left(8m\right) & \mbox{Te} \\ 
\end{longtabu}
\renewcommand{\arraystretch}{1.0}
\noindent \hrulefill
\\
\textbf{References:}
\vspace*{-0.25cm}
\begin{flushleft}
  - \bibentry{Selte_NbTe4_ActaChemScand_1964}. \\
\end{flushleft}
\textbf{Found in:}
\vspace*{-0.25cm}
\begin{flushleft}
  - \bibentry{Villars_PearsonsCrystalData_2013}. \\
\end{flushleft}
\noindent \hrulefill
\\
\textbf{Geometry files:}
\\
\noindent  - CIF: pp. {\hyperref[AB4_tP10_124_a_m_cif]{\pageref{AB4_tP10_124_a_m_cif}}} \\
\noindent  - POSCAR: pp. {\hyperref[AB4_tP10_124_a_m_poscar]{\pageref{AB4_tP10_124_a_m_poscar}}} \\
\onecolumn
{\phantomsection\label{A4B_tP10_125_m_a}}
\subsection*{\huge \textbf{{\normalfont PtPb$_{4}$ Structure: A4B\_tP10\_125\_m\_a}}}
\noindent \hrulefill
\vspace*{0.25cm}
\begin{figure}[htp]
  \centering
  \vspace{-1em}
  {\includegraphics[width=1\textwidth]{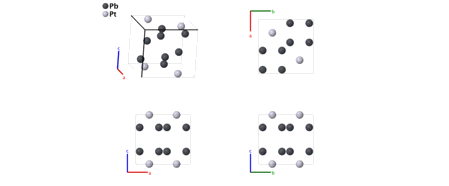}}
\end{figure}
\vspace*{-0.5cm}
\renewcommand{\arraystretch}{1.5}
\begin{equation*}
  \begin{array}{>{$\hspace{-0.15cm}}l<{$}>{$}p{0.5cm}<{$}>{$}p{18.5cm}<{$}}
    \mbox{\large \textbf{Prototype}} &\colon & \ce{PtPb4} \\
    \mbox{\large \textbf{\AFLOW\ prototype label}} &\colon & \mbox{A4B\_tP10\_125\_m\_a} \\
    \mbox{\large \textbf{\textit{Strukturbericht} designation}} &\colon & \mbox{None} \\
    \mbox{\large \textbf{Pearson symbol}} &\colon & \mbox{tP10} \\
    \mbox{\large \textbf{Space group number}} &\colon & 125 \\
    \mbox{\large \textbf{Space group symbol}} &\colon & P4/nbm \\
    \mbox{\large \textbf{\AFLOW\ prototype command}} &\colon &  \texttt{aflow} \,  \, \texttt{-{}-proto=A4B\_tP10\_125\_m\_a } \, \newline \texttt{-{}-params=}{a,c/a,x_{2},z_{2} }
  \end{array}
\end{equation*}
\renewcommand{\arraystretch}{1.0}

\noindent \parbox{1 \linewidth}{
\noindent \hrulefill
\\
\textbf{Simple Tetragonal primitive vectors:} \\
\vspace*{-0.25cm}
\begin{tabular}{cc}
  \begin{tabular}{c}
    \parbox{0.6 \linewidth}{
      \renewcommand{\arraystretch}{1.5}
      \begin{equation*}
        \centering
        \begin{array}{ccc}
              \mathbf{a}_1 & = & a \, \mathbf{\hat{x}} \\
    \mathbf{a}_2 & = & a \, \mathbf{\hat{y}} \\
    \mathbf{a}_3 & = & c \, \mathbf{\hat{z}} \\

        \end{array}
      \end{equation*}
    }
    \renewcommand{\arraystretch}{1.0}
  \end{tabular}
  \begin{tabular}{c}
    \includegraphics[width=0.3\linewidth]{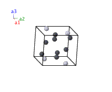} \\
  \end{tabular}
\end{tabular}

}
\vspace*{-0.25cm}

\noindent \hrulefill
\\
\textbf{Basis vectors:}
\vspace*{-0.25cm}
\renewcommand{\arraystretch}{1.5}
\begin{longtabu} to \textwidth{>{\centering $}X[-1,c,c]<{$}>{\centering $}X[-1,c,c]<{$}>{\centering $}X[-1,c,c]<{$}>{\centering $}X[-1,c,c]<{$}>{\centering $}X[-1,c,c]<{$}>{\centering $}X[-1,c,c]<{$}>{\centering $}X[-1,c,c]<{$}}
  & & \mbox{Lattice Coordinates} & & \mbox{Cartesian Coordinates} &\mbox{Wyckoff Position} & \mbox{Atom Type} \\  
  \mathbf{B}_{1} & = & \frac{1}{4} \, \mathbf{a}_{1} + \frac{1}{4} \, \mathbf{a}_{2} & = & \frac{1}{4}a \, \mathbf{\hat{x}} + \frac{1}{4}a \, \mathbf{\hat{y}} & \left(2a\right) & \mbox{Pt} \\ 
\mathbf{B}_{2} & = & \frac{3}{4} \, \mathbf{a}_{1} + \frac{3}{4} \, \mathbf{a}_{2} & = & \frac{3}{4}a \, \mathbf{\hat{x}} + \frac{3}{4}a \, \mathbf{\hat{y}} & \left(2a\right) & \mbox{Pt} \\ 
\mathbf{B}_{3} & = & x_{2} \, \mathbf{a}_{1}-x_{2} \, \mathbf{a}_{2} + z_{2} \, \mathbf{a}_{3} & = & x_{2}a \, \mathbf{\hat{x}}-x_{2}a \, \mathbf{\hat{y}} + z_{2}c \, \mathbf{\hat{z}} & \left(8m\right) & \mbox{Pb} \\ 
\mathbf{B}_{4} & = & \left(\frac{1}{2} - x_{2}\right) \, \mathbf{a}_{1} + \left(\frac{1}{2} +x_{2}\right) \, \mathbf{a}_{2} + z_{2} \, \mathbf{a}_{3} & = & \left(\frac{1}{2} - x_{2}\right)a \, \mathbf{\hat{x}} + \left(\frac{1}{2} +x_{2}\right)a \, \mathbf{\hat{y}} + z_{2}c \, \mathbf{\hat{z}} & \left(8m\right) & \mbox{Pb} \\ 
\mathbf{B}_{5} & = & \left(\frac{1}{2} +x_{2}\right) \, \mathbf{a}_{1} + x_{2} \, \mathbf{a}_{2} + z_{2} \, \mathbf{a}_{3} & = & \left(\frac{1}{2} +x_{2}\right)a \, \mathbf{\hat{x}} + x_{2}a \, \mathbf{\hat{y}} + z_{2}c \, \mathbf{\hat{z}} & \left(8m\right) & \mbox{Pb} \\ 
\mathbf{B}_{6} & = & -x_{2} \, \mathbf{a}_{1} + \left(\frac{1}{2} - x_{2}\right) \, \mathbf{a}_{2} + z_{2} \, \mathbf{a}_{3} & = & -x_{2}a \, \mathbf{\hat{x}} + \left(\frac{1}{2} - x_{2}\right)a \, \mathbf{\hat{y}} + z_{2}c \, \mathbf{\hat{z}} & \left(8m\right) & \mbox{Pb} \\ 
\mathbf{B}_{7} & = & \left(\frac{1}{2} - x_{2}\right) \, \mathbf{a}_{1}-x_{2} \, \mathbf{a}_{2}-z_{2} \, \mathbf{a}_{3} & = & \left(\frac{1}{2} - x_{2}\right)a \, \mathbf{\hat{x}}-x_{2}a \, \mathbf{\hat{y}}-z_{2}c \, \mathbf{\hat{z}} & \left(8m\right) & \mbox{Pb} \\ 
\mathbf{B}_{8} & = & x_{2} \, \mathbf{a}_{1} + \left(\frac{1}{2} +x_{2}\right) \, \mathbf{a}_{2}-z_{2} \, \mathbf{a}_{3} & = & x_{2}a \, \mathbf{\hat{x}} + \left(\frac{1}{2} +x_{2}\right)a \, \mathbf{\hat{y}}-z_{2}c \, \mathbf{\hat{z}} & \left(8m\right) & \mbox{Pb} \\ 
\mathbf{B}_{9} & = & -x_{2} \, \mathbf{a}_{1} + x_{2} \, \mathbf{a}_{2}-z_{2} \, \mathbf{a}_{3} & = & -x_{2}a \, \mathbf{\hat{x}} + x_{2}a \, \mathbf{\hat{y}}-z_{2}c \, \mathbf{\hat{z}} & \left(8m\right) & \mbox{Pb} \\ 
\mathbf{B}_{10} & = & \left(\frac{1}{2} +x_{2}\right) \, \mathbf{a}_{1} + \left(\frac{1}{2} - x_{2}\right) \, \mathbf{a}_{2}-z_{2} \, \mathbf{a}_{3} & = & \left(\frac{1}{2} +x_{2}\right)a \, \mathbf{\hat{x}} + \left(\frac{1}{2} - x_{2}\right)a \, \mathbf{\hat{y}}-z_{2}c \, \mathbf{\hat{z}} & \left(8m\right) & \mbox{Pb} \\ 
\end{longtabu}
\renewcommand{\arraystretch}{1.0}
\noindent \hrulefill
\\
\textbf{References:}
\vspace*{-0.25cm}
\begin{flushleft}
  - \bibentry{Graham_PtPb4_ActCrystallogr_1954}. \\
\end{flushleft}
\textbf{Found in:}
\vspace*{-0.25cm}
\begin{flushleft}
  - \bibentry{Villars_PearsonsCrystalData_2013}. \\
\end{flushleft}
\noindent \hrulefill
\\
\textbf{Geometry files:}
\\
\noindent  - CIF: pp. {\hyperref[A4B_tP10_125_m_a_cif]{\pageref{A4B_tP10_125_m_a_cif}}} \\
\noindent  - POSCAR: pp. {\hyperref[A4B_tP10_125_m_a_poscar]{\pageref{A4B_tP10_125_m_a_poscar}}} \\
\onecolumn
{\phantomsection\label{ABC4_tP12_125_a_b_m}}
\subsection*{\huge \textbf{{\normalfont KCeSe$_{4}$ Structure: ABC4\_tP12\_125\_a\_b\_m}}}
\noindent \hrulefill
\vspace*{0.25cm}
\begin{figure}[htp]
  \centering
  \vspace{-1em}
  {\includegraphics[width=1\textwidth]{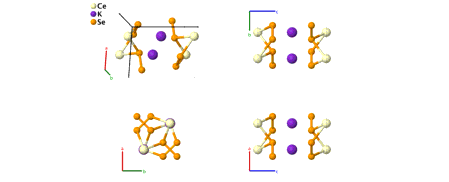}}
\end{figure}
\vspace*{-0.5cm}
\renewcommand{\arraystretch}{1.5}
\begin{equation*}
  \begin{array}{>{$\hspace{-0.15cm}}l<{$}>{$}p{0.5cm}<{$}>{$}p{18.5cm}<{$}}
    \mbox{\large \textbf{Prototype}} &\colon & \ce{KCeSe4} \\
    \mbox{\large \textbf{\AFLOW\ prototype label}} &\colon & \mbox{ABC4\_tP12\_125\_a\_b\_m} \\
    \mbox{\large \textbf{\textit{Strukturbericht} designation}} &\colon & \mbox{None} \\
    \mbox{\large \textbf{Pearson symbol}} &\colon & \mbox{tP12} \\
    \mbox{\large \textbf{Space group number}} &\colon & 125 \\
    \mbox{\large \textbf{Space group symbol}} &\colon & P4/nbm \\
    \mbox{\large \textbf{\AFLOW\ prototype command}} &\colon &  \texttt{aflow} \,  \, \texttt{-{}-proto=ABC4\_tP12\_125\_a\_b\_m } \, \newline \texttt{-{}-params=}{a,c/a,x_{3},z_{3} }
  \end{array}
\end{equation*}
\renewcommand{\arraystretch}{1.0}

\noindent \parbox{1 \linewidth}{
\noindent \hrulefill
\\
\textbf{Simple Tetragonal primitive vectors:} \\
\vspace*{-0.25cm}
\begin{tabular}{cc}
  \begin{tabular}{c}
    \parbox{0.6 \linewidth}{
      \renewcommand{\arraystretch}{1.5}
      \begin{equation*}
        \centering
        \begin{array}{ccc}
              \mathbf{a}_1 & = & a \, \mathbf{\hat{x}} \\
    \mathbf{a}_2 & = & a \, \mathbf{\hat{y}} \\
    \mathbf{a}_3 & = & c \, \mathbf{\hat{z}} \\

        \end{array}
      \end{equation*}
    }
    \renewcommand{\arraystretch}{1.0}
  \end{tabular}
  \begin{tabular}{c}
    \includegraphics[width=0.3\linewidth]{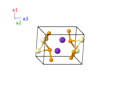} \\
  \end{tabular}
\end{tabular}

}
\vspace*{-0.25cm}

\noindent \hrulefill
\\
\textbf{Basis vectors:}
\vspace*{-0.25cm}
\renewcommand{\arraystretch}{1.5}
\begin{longtabu} to \textwidth{>{\centering $}X[-1,c,c]<{$}>{\centering $}X[-1,c,c]<{$}>{\centering $}X[-1,c,c]<{$}>{\centering $}X[-1,c,c]<{$}>{\centering $}X[-1,c,c]<{$}>{\centering $}X[-1,c,c]<{$}>{\centering $}X[-1,c,c]<{$}}
  & & \mbox{Lattice Coordinates} & & \mbox{Cartesian Coordinates} &\mbox{Wyckoff Position} & \mbox{Atom Type} \\  
  \mathbf{B}_{1} & = & \frac{1}{4} \, \mathbf{a}_{1} + \frac{1}{4} \, \mathbf{a}_{2} & = & \frac{1}{4}a \, \mathbf{\hat{x}} + \frac{1}{4}a \, \mathbf{\hat{y}} & \left(2a\right) & \mbox{Ce} \\ 
\mathbf{B}_{2} & = & \frac{3}{4} \, \mathbf{a}_{1} + \frac{3}{4} \, \mathbf{a}_{2} & = & \frac{3}{4}a \, \mathbf{\hat{x}} + \frac{3}{4}a \, \mathbf{\hat{y}} & \left(2a\right) & \mbox{Ce} \\ 
\mathbf{B}_{3} & = & \frac{1}{4} \, \mathbf{a}_{1} + \frac{1}{4} \, \mathbf{a}_{2} + \frac{1}{2} \, \mathbf{a}_{3} & = & \frac{1}{4}a \, \mathbf{\hat{x}} + \frac{1}{4}a \, \mathbf{\hat{y}} + \frac{1}{2}c \, \mathbf{\hat{z}} & \left(2b\right) & \mbox{K} \\ 
\mathbf{B}_{4} & = & \frac{3}{4} \, \mathbf{a}_{1} + \frac{3}{4} \, \mathbf{a}_{2} + \frac{1}{2} \, \mathbf{a}_{3} & = & \frac{3}{4}a \, \mathbf{\hat{x}} + \frac{3}{4}a \, \mathbf{\hat{y}} + \frac{1}{2}c \, \mathbf{\hat{z}} & \left(2b\right) & \mbox{K} \\ 
\mathbf{B}_{5} & = & x_{3} \, \mathbf{a}_{1}-x_{3} \, \mathbf{a}_{2} + z_{3} \, \mathbf{a}_{3} & = & x_{3}a \, \mathbf{\hat{x}}-x_{3}a \, \mathbf{\hat{y}} + z_{3}c \, \mathbf{\hat{z}} & \left(8m\right) & \mbox{Se} \\ 
\mathbf{B}_{6} & = & \left(\frac{1}{2} - x_{3}\right) \, \mathbf{a}_{1} + \left(\frac{1}{2} +x_{3}\right) \, \mathbf{a}_{2} + z_{3} \, \mathbf{a}_{3} & = & \left(\frac{1}{2} - x_{3}\right)a \, \mathbf{\hat{x}} + \left(\frac{1}{2} +x_{3}\right)a \, \mathbf{\hat{y}} + z_{3}c \, \mathbf{\hat{z}} & \left(8m\right) & \mbox{Se} \\ 
\mathbf{B}_{7} & = & \left(\frac{1}{2} +x_{3}\right) \, \mathbf{a}_{1} + x_{3} \, \mathbf{a}_{2} + z_{3} \, \mathbf{a}_{3} & = & \left(\frac{1}{2} +x_{3}\right)a \, \mathbf{\hat{x}} + x_{3}a \, \mathbf{\hat{y}} + z_{3}c \, \mathbf{\hat{z}} & \left(8m\right) & \mbox{Se} \\ 
\mathbf{B}_{8} & = & -x_{3} \, \mathbf{a}_{1} + \left(\frac{1}{2} - x_{3}\right) \, \mathbf{a}_{2} + z_{3} \, \mathbf{a}_{3} & = & -x_{3}a \, \mathbf{\hat{x}} + \left(\frac{1}{2} - x_{3}\right)a \, \mathbf{\hat{y}} + z_{3}c \, \mathbf{\hat{z}} & \left(8m\right) & \mbox{Se} \\ 
\mathbf{B}_{9} & = & \left(\frac{1}{2} - x_{3}\right) \, \mathbf{a}_{1}-x_{3} \, \mathbf{a}_{2}-z_{3} \, \mathbf{a}_{3} & = & \left(\frac{1}{2} - x_{3}\right)a \, \mathbf{\hat{x}}-x_{3}a \, \mathbf{\hat{y}}-z_{3}c \, \mathbf{\hat{z}} & \left(8m\right) & \mbox{Se} \\ 
\mathbf{B}_{10} & = & x_{3} \, \mathbf{a}_{1} + \left(\frac{1}{2} +x_{3}\right) \, \mathbf{a}_{2}-z_{3} \, \mathbf{a}_{3} & = & x_{3}a \, \mathbf{\hat{x}} + \left(\frac{1}{2} +x_{3}\right)a \, \mathbf{\hat{y}}-z_{3}c \, \mathbf{\hat{z}} & \left(8m\right) & \mbox{Se} \\ 
\mathbf{B}_{11} & = & -x_{3} \, \mathbf{a}_{1} + x_{3} \, \mathbf{a}_{2}-z_{3} \, \mathbf{a}_{3} & = & -x_{3}a \, \mathbf{\hat{x}} + x_{3}a \, \mathbf{\hat{y}}-z_{3}c \, \mathbf{\hat{z}} & \left(8m\right) & \mbox{Se} \\ 
\mathbf{B}_{12} & = & \left(\frac{1}{2} +x_{3}\right) \, \mathbf{a}_{1} + \left(\frac{1}{2} - x_{3}\right) \, \mathbf{a}_{2}-z_{3} \, \mathbf{a}_{3} & = & \left(\frac{1}{2} +x_{3}\right)a \, \mathbf{\hat{x}} + \left(\frac{1}{2} - x_{3}\right)a \, \mathbf{\hat{y}}-z_{3}c \, \mathbf{\hat{z}} & \left(8m\right) & \mbox{Se} \\ 
\end{longtabu}
\renewcommand{\arraystretch}{1.0}
\noindent \hrulefill
\\
\textbf{References:}
\vspace*{-0.25cm}
\begin{flushleft}
  - \bibentry{Sutorik_KCeSe4_AngeChemIntEd_1992}. \\
\end{flushleft}
\textbf{Found in:}
\vspace*{-0.25cm}
\begin{flushleft}
  - \bibentry{Villars_PearsonsCrystalData_2013}. \\
\end{flushleft}
\noindent \hrulefill
\\
\textbf{Geometry files:}
\\
\noindent  - CIF: pp. {\hyperref[ABC4_tP12_125_a_b_m_cif]{\pageref{ABC4_tP12_125_a_b_m_cif}}} \\
\noindent  - POSCAR: pp. {\hyperref[ABC4_tP12_125_a_b_m_poscar]{\pageref{ABC4_tP12_125_a_b_m_poscar}}} \\
\onecolumn
{\phantomsection\label{A2BC4_tP28_126_cd_e_k}}
\subsection*{\huge \textbf{{\normalfont BiAl$_{2}$S$_{4}$ Structure: A2BC4\_tP28\_126\_cd\_e\_k}}}
\noindent \hrulefill
\vspace*{0.25cm}
\begin{figure}[htp]
  \centering
  \vspace{-1em}
  {\includegraphics[width=1\textwidth]{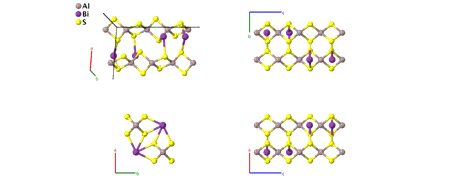}}
\end{figure}
\vspace*{-0.5cm}
\renewcommand{\arraystretch}{1.5}
\begin{equation*}
  \begin{array}{>{$\hspace{-0.15cm}}l<{$}>{$}p{0.5cm}<{$}>{$}p{18.5cm}<{$}}
    \mbox{\large \textbf{Prototype}} &\colon & \ce{BiAl2S4} \\
    \mbox{\large \textbf{\AFLOW\ prototype label}} &\colon & \mbox{A2BC4\_tP28\_126\_cd\_e\_k} \\
    \mbox{\large \textbf{\textit{Strukturbericht} designation}} &\colon & \mbox{None} \\
    \mbox{\large \textbf{Pearson symbol}} &\colon & \mbox{tP28} \\
    \mbox{\large \textbf{Space group number}} &\colon & 126 \\
    \mbox{\large \textbf{Space group symbol}} &\colon & P4/nnc \\
    \mbox{\large \textbf{\AFLOW\ prototype command}} &\colon &  \texttt{aflow} \,  \, \texttt{-{}-proto=A2BC4\_tP28\_126\_cd\_e\_k } \, \newline \texttt{-{}-params=}{a,c/a,z_{3},x_{4},y_{4},z_{4} }
  \end{array}
\end{equation*}
\renewcommand{\arraystretch}{1.0}

\noindent \parbox{1 \linewidth}{
\noindent \hrulefill
\\
\textbf{Simple Tetragonal primitive vectors:} \\
\vspace*{-0.25cm}
\begin{tabular}{cc}
  \begin{tabular}{c}
    \parbox{0.6 \linewidth}{
      \renewcommand{\arraystretch}{1.5}
      \begin{equation*}
        \centering
        \begin{array}{ccc}
              \mathbf{a}_1 & = & a \, \mathbf{\hat{x}} \\
    \mathbf{a}_2 & = & a \, \mathbf{\hat{y}} \\
    \mathbf{a}_3 & = & c \, \mathbf{\hat{z}} \\

        \end{array}
      \end{equation*}
    }
    \renewcommand{\arraystretch}{1.0}
  \end{tabular}
  \begin{tabular}{c}
    \includegraphics[width=0.3\linewidth]{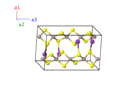} \\
  \end{tabular}
\end{tabular}

}
\vspace*{-0.25cm}

\noindent \hrulefill
\\
\textbf{Basis vectors:}
\vspace*{-0.25cm}
\renewcommand{\arraystretch}{1.5}
\begin{longtabu} to \textwidth{>{\centering $}X[-1,c,c]<{$}>{\centering $}X[-1,c,c]<{$}>{\centering $}X[-1,c,c]<{$}>{\centering $}X[-1,c,c]<{$}>{\centering $}X[-1,c,c]<{$}>{\centering $}X[-1,c,c]<{$}>{\centering $}X[-1,c,c]<{$}}
  & & \mbox{Lattice Coordinates} & & \mbox{Cartesian Coordinates} &\mbox{Wyckoff Position} & \mbox{Atom Type} \\  
  \mathbf{B}_{1} & = & \frac{1}{4} \, \mathbf{a}_{1} + \frac{3}{4} \, \mathbf{a}_{2} + \frac{3}{4} \, \mathbf{a}_{3} & = & \frac{1}{4}a \, \mathbf{\hat{x}} + \frac{3}{4}a \, \mathbf{\hat{y}} + \frac{3}{4}c \, \mathbf{\hat{z}} & \left(4c\right) & \mbox{Al I} \\ 
\mathbf{B}_{2} & = & \frac{3}{4} \, \mathbf{a}_{1} + \frac{1}{4} \, \mathbf{a}_{2} + \frac{3}{4} \, \mathbf{a}_{3} & = & \frac{3}{4}a \, \mathbf{\hat{x}} + \frac{1}{4}a \, \mathbf{\hat{y}} + \frac{3}{4}c \, \mathbf{\hat{z}} & \left(4c\right) & \mbox{Al I} \\ 
\mathbf{B}_{3} & = & \frac{3}{4} \, \mathbf{a}_{1} + \frac{1}{4} \, \mathbf{a}_{2} + \frac{1}{4} \, \mathbf{a}_{3} & = & \frac{3}{4}a \, \mathbf{\hat{x}} + \frac{1}{4}a \, \mathbf{\hat{y}} + \frac{1}{4}c \, \mathbf{\hat{z}} & \left(4c\right) & \mbox{Al I} \\ 
\mathbf{B}_{4} & = & \frac{1}{4} \, \mathbf{a}_{1} + \frac{3}{4} \, \mathbf{a}_{2} + \frac{1}{4} \, \mathbf{a}_{3} & = & \frac{1}{4}a \, \mathbf{\hat{x}} + \frac{3}{4}a \, \mathbf{\hat{y}} + \frac{1}{4}c \, \mathbf{\hat{z}} & \left(4c\right) & \mbox{Al I} \\ 
\mathbf{B}_{5} & = & \frac{1}{4} \, \mathbf{a}_{1} + \frac{3}{4} \, \mathbf{a}_{2} & = & \frac{1}{4}a \, \mathbf{\hat{x}} + \frac{3}{4}a \, \mathbf{\hat{y}} & \left(4d\right) & \mbox{Al II} \\ 
\mathbf{B}_{6} & = & \frac{3}{4} \, \mathbf{a}_{1} + \frac{1}{4} \, \mathbf{a}_{2} & = & \frac{3}{4}a \, \mathbf{\hat{x}} + \frac{1}{4}a \, \mathbf{\hat{y}} & \left(4d\right) & \mbox{Al II} \\ 
\mathbf{B}_{7} & = & \frac{1}{4} \, \mathbf{a}_{1} + \frac{3}{4} \, \mathbf{a}_{2} + \frac{1}{2} \, \mathbf{a}_{3} & = & \frac{1}{4}a \, \mathbf{\hat{x}} + \frac{3}{4}a \, \mathbf{\hat{y}} + \frac{1}{2}c \, \mathbf{\hat{z}} & \left(4d\right) & \mbox{Al II} \\ 
\mathbf{B}_{8} & = & \frac{3}{4} \, \mathbf{a}_{1} + \frac{1}{4} \, \mathbf{a}_{2} + \frac{1}{2} \, \mathbf{a}_{3} & = & \frac{3}{4}a \, \mathbf{\hat{x}} + \frac{1}{4}a \, \mathbf{\hat{y}} + \frac{1}{2}c \, \mathbf{\hat{z}} & \left(4d\right) & \mbox{Al II} \\ 
\mathbf{B}_{9} & = & \frac{1}{4} \, \mathbf{a}_{1} + \frac{1}{4} \, \mathbf{a}_{2} + z_{3} \, \mathbf{a}_{3} & = & \frac{1}{4}a \, \mathbf{\hat{x}} + \frac{1}{4}a \, \mathbf{\hat{y}} + z_{3}c \, \mathbf{\hat{z}} & \left(4e\right) & \mbox{Bi} \\ 
\mathbf{B}_{10} & = & \frac{1}{4} \, \mathbf{a}_{1} + \frac{1}{4} \, \mathbf{a}_{2} + \left(\frac{1}{2} - z_{3}\right) \, \mathbf{a}_{3} & = & \frac{1}{4}a \, \mathbf{\hat{x}} + \frac{1}{4}a \, \mathbf{\hat{y}} + \left(\frac{1}{2} - z_{3}\right)c \, \mathbf{\hat{z}} & \left(4e\right) & \mbox{Bi} \\ 
\mathbf{B}_{11} & = & \frac{3}{4} \, \mathbf{a}_{1} + \frac{3}{4} \, \mathbf{a}_{2}-z_{3} \, \mathbf{a}_{3} & = & \frac{3}{4}a \, \mathbf{\hat{x}} + \frac{3}{4}a \, \mathbf{\hat{y}}-z_{3}c \, \mathbf{\hat{z}} & \left(4e\right) & \mbox{Bi} \\ 
\mathbf{B}_{12} & = & \frac{3}{4} \, \mathbf{a}_{1} + \frac{3}{4} \, \mathbf{a}_{2} + \left(\frac{1}{2} +z_{3}\right) \, \mathbf{a}_{3} & = & \frac{3}{4}a \, \mathbf{\hat{x}} + \frac{3}{4}a \, \mathbf{\hat{y}} + \left(\frac{1}{2} +z_{3}\right)c \, \mathbf{\hat{z}} & \left(4e\right) & \mbox{Bi} \\ 
\mathbf{B}_{13} & = & x_{4} \, \mathbf{a}_{1} + y_{4} \, \mathbf{a}_{2} + z_{4} \, \mathbf{a}_{3} & = & x_{4}a \, \mathbf{\hat{x}} + y_{4}a \, \mathbf{\hat{y}} + z_{4}c \, \mathbf{\hat{z}} & \left(16k\right) & \mbox{S} \\ 
\mathbf{B}_{14} & = & \left(\frac{1}{2} - x_{4}\right) \, \mathbf{a}_{1} + \left(\frac{1}{2} - y_{4}\right) \, \mathbf{a}_{2} + z_{4} \, \mathbf{a}_{3} & = & \left(\frac{1}{2} - x_{4}\right)a \, \mathbf{\hat{x}} + \left(\frac{1}{2} - y_{4}\right)a \, \mathbf{\hat{y}} + z_{4}c \, \mathbf{\hat{z}} & \left(16k\right) & \mbox{S} \\ 
\mathbf{B}_{15} & = & \left(\frac{1}{2} - y_{4}\right) \, \mathbf{a}_{1} + x_{4} \, \mathbf{a}_{2} + z_{4} \, \mathbf{a}_{3} & = & \left(\frac{1}{2} - y_{4}\right)a \, \mathbf{\hat{x}} + x_{4}a \, \mathbf{\hat{y}} + z_{4}c \, \mathbf{\hat{z}} & \left(16k\right) & \mbox{S} \\ 
\mathbf{B}_{16} & = & y_{4} \, \mathbf{a}_{1} + \left(\frac{1}{2} - x_{4}\right) \, \mathbf{a}_{2} + z_{4} \, \mathbf{a}_{3} & = & y_{4}a \, \mathbf{\hat{x}} + \left(\frac{1}{2} - x_{4}\right)a \, \mathbf{\hat{y}} + z_{4}c \, \mathbf{\hat{z}} & \left(16k\right) & \mbox{S} \\ 
\mathbf{B}_{17} & = & \left(\frac{1}{2} - x_{4}\right) \, \mathbf{a}_{1} + y_{4} \, \mathbf{a}_{2} + \left(\frac{1}{2} - z_{4}\right) \, \mathbf{a}_{3} & = & \left(\frac{1}{2} - x_{4}\right)a \, \mathbf{\hat{x}} + y_{4}a \, \mathbf{\hat{y}} + \left(\frac{1}{2} - z_{4}\right)c \, \mathbf{\hat{z}} & \left(16k\right) & \mbox{S} \\ 
\mathbf{B}_{18} & = & x_{4} \, \mathbf{a}_{1} + \left(\frac{1}{2} - y_{4}\right) \, \mathbf{a}_{2} + \left(\frac{1}{2} - z_{4}\right) \, \mathbf{a}_{3} & = & x_{4}a \, \mathbf{\hat{x}} + \left(\frac{1}{2} - y_{4}\right)a \, \mathbf{\hat{y}} + \left(\frac{1}{2} - z_{4}\right)c \, \mathbf{\hat{z}} & \left(16k\right) & \mbox{S} \\ 
\mathbf{B}_{19} & = & y_{4} \, \mathbf{a}_{1} + x_{4} \, \mathbf{a}_{2} + \left(\frac{1}{2} - z_{4}\right) \, \mathbf{a}_{3} & = & y_{4}a \, \mathbf{\hat{x}} + x_{4}a \, \mathbf{\hat{y}} + \left(\frac{1}{2} - z_{4}\right)c \, \mathbf{\hat{z}} & \left(16k\right) & \mbox{S} \\ 
\mathbf{B}_{20} & = & \left(\frac{1}{2} - y_{4}\right) \, \mathbf{a}_{1} + \left(\frac{1}{2} - x_{4}\right) \, \mathbf{a}_{2} + \left(\frac{1}{2} - z_{4}\right) \, \mathbf{a}_{3} & = & \left(\frac{1}{2} - y_{4}\right)a \, \mathbf{\hat{x}} + \left(\frac{1}{2} - x_{4}\right)a \, \mathbf{\hat{y}} + \left(\frac{1}{2} - z_{4}\right)c \, \mathbf{\hat{z}} & \left(16k\right) & \mbox{S} \\ 
\mathbf{B}_{21} & = & -x_{4} \, \mathbf{a}_{1}-y_{4} \, \mathbf{a}_{2}-z_{4} \, \mathbf{a}_{3} & = & -x_{4}a \, \mathbf{\hat{x}}-y_{4}a \, \mathbf{\hat{y}}-z_{4}c \, \mathbf{\hat{z}} & \left(16k\right) & \mbox{S} \\ 
\mathbf{B}_{22} & = & \left(\frac{1}{2} +x_{4}\right) \, \mathbf{a}_{1} + \left(\frac{1}{2} +y_{4}\right) \, \mathbf{a}_{2}-z_{4} \, \mathbf{a}_{3} & = & \left(\frac{1}{2} +x_{4}\right)a \, \mathbf{\hat{x}} + \left(\frac{1}{2} +y_{4}\right)a \, \mathbf{\hat{y}}-z_{4}c \, \mathbf{\hat{z}} & \left(16k\right) & \mbox{S} \\ 
\mathbf{B}_{23} & = & \left(\frac{1}{2} +y_{4}\right) \, \mathbf{a}_{1}-x_{4} \, \mathbf{a}_{2}-z_{4} \, \mathbf{a}_{3} & = & \left(\frac{1}{2} +y_{4}\right)a \, \mathbf{\hat{x}}-x_{4}a \, \mathbf{\hat{y}}-z_{4}c \, \mathbf{\hat{z}} & \left(16k\right) & \mbox{S} \\ 
\mathbf{B}_{24} & = & -y_{4} \, \mathbf{a}_{1} + \left(\frac{1}{2} +x_{4}\right) \, \mathbf{a}_{2}-z_{4} \, \mathbf{a}_{3} & = & -y_{4}a \, \mathbf{\hat{x}} + \left(\frac{1}{2} +x_{4}\right)a \, \mathbf{\hat{y}}-z_{4}c \, \mathbf{\hat{z}} & \left(16k\right) & \mbox{S} \\ 
\mathbf{B}_{25} & = & \left(\frac{1}{2} +x_{4}\right) \, \mathbf{a}_{1}-y_{4} \, \mathbf{a}_{2} + \left(\frac{1}{2} +z_{4}\right) \, \mathbf{a}_{3} & = & \left(\frac{1}{2} +x_{4}\right)a \, \mathbf{\hat{x}}-y_{4}a \, \mathbf{\hat{y}} + \left(\frac{1}{2} +z_{4}\right)c \, \mathbf{\hat{z}} & \left(16k\right) & \mbox{S} \\ 
\mathbf{B}_{26} & = & -x_{4} \, \mathbf{a}_{1} + \left(\frac{1}{2} +y_{4}\right) \, \mathbf{a}_{2} + \left(\frac{1}{2} +z_{4}\right) \, \mathbf{a}_{3} & = & -x_{4}a \, \mathbf{\hat{x}} + \left(\frac{1}{2} +y_{4}\right)a \, \mathbf{\hat{y}} + \left(\frac{1}{2} +z_{4}\right)c \, \mathbf{\hat{z}} & \left(16k\right) & \mbox{S} \\ 
\mathbf{B}_{27} & = & -y_{4} \, \mathbf{a}_{1}-x_{4} \, \mathbf{a}_{2} + \left(\frac{1}{2} +z_{4}\right) \, \mathbf{a}_{3} & = & -y_{4}a \, \mathbf{\hat{x}}-x_{4}a \, \mathbf{\hat{y}} + \left(\frac{1}{2} +z_{4}\right)c \, \mathbf{\hat{z}} & \left(16k\right) & \mbox{S} \\ 
\mathbf{B}_{28} & = & \left(\frac{1}{2} +y_{4}\right) \, \mathbf{a}_{1} + \left(\frac{1}{2} +x_{4}\right) \, \mathbf{a}_{2} + \left(\frac{1}{2} +z_{4}\right) \, \mathbf{a}_{3} & = & \left(\frac{1}{2} +y_{4}\right)a \, \mathbf{\hat{x}} + \left(\frac{1}{2} +x_{4}\right)a \, \mathbf{\hat{y}} + \left(\frac{1}{2} +z_{4}\right)c \, \mathbf{\hat{z}} & \left(16k\right) & \mbox{S} \\ 
\end{longtabu}
\renewcommand{\arraystretch}{1.0}
\noindent \hrulefill
\\
\textbf{References:}
\vspace*{-0.25cm}
\begin{flushleft}
  - \bibentry{Kalpen_Al2BiS4_ZAnorgAllgChem_1998}. \\
\end{flushleft}
\textbf{Found in:}
\vspace*{-0.25cm}
\begin{flushleft}
  - \bibentry{Villars_PearsonsCrystalData_2013}. \\
\end{flushleft}
\noindent \hrulefill
\\
\textbf{Geometry files:}
\\
\noindent  - CIF: pp. {\hyperref[A2BC4_tP28_126_cd_e_k_cif]{\pageref{A2BC4_tP28_126_cd_e_k_cif}}} \\
\noindent  - POSCAR: pp. {\hyperref[A2BC4_tP28_126_cd_e_k_poscar]{\pageref{A2BC4_tP28_126_cd_e_k_poscar}}} \\
\onecolumn
{\phantomsection\label{A4B_tP20_127_ehj_g}}
\subsection*{\huge \textbf{{\normalfont ThB$_{4}$ ($D1_{e}$) Structure: A4B\_tP20\_127\_ehj\_g}}}
\noindent \hrulefill
\vspace*{0.25cm}
\begin{figure}[htp]
  \centering
  \vspace{-1em}
  {\includegraphics[width=1\textwidth]{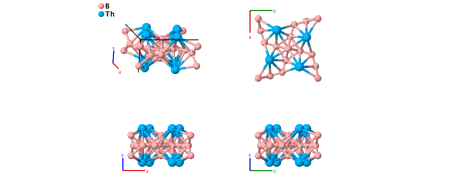}}
\end{figure}
\vspace*{-0.5cm}
\renewcommand{\arraystretch}{1.5}
\begin{equation*}
  \begin{array}{>{$\hspace{-0.15cm}}l<{$}>{$}p{0.5cm}<{$}>{$}p{18.5cm}<{$}}
    \mbox{\large \textbf{Prototype}} &\colon & \ce{ThB4} \\
    \mbox{\large \textbf{\AFLOW\ prototype label}} &\colon & \mbox{A4B\_tP20\_127\_ehj\_g} \\
    \mbox{\large \textbf{\textit{Strukturbericht} designation}} &\colon & \mbox{$D1_{e}$} \\
    \mbox{\large \textbf{Pearson symbol}} &\colon & \mbox{tP20} \\
    \mbox{\large \textbf{Space group number}} &\colon & 127 \\
    \mbox{\large \textbf{Space group symbol}} &\colon & P4/mbm \\
    \mbox{\large \textbf{\AFLOW\ prototype command}} &\colon &  \texttt{aflow} \,  \, \texttt{-{}-proto=A4B\_tP20\_127\_ehj\_g } \, \newline \texttt{-{}-params=}{a,c/a,z_{1},x_{2},x_{3},x_{4},y_{4} }
  \end{array}
\end{equation*}
\renewcommand{\arraystretch}{1.0}

\vspace*{-0.25cm}
\noindent \hrulefill
\\
\textbf{ Other compounds with this structure:}
\begin{itemize}
   \item{ B$_{4}$Ce, B$_{4}$U, B$_{4}$Y  }
\end{itemize}
\noindent \parbox{1 \linewidth}{
\noindent \hrulefill
\\
\textbf{Simple Tetragonal primitive vectors:} \\
\vspace*{-0.25cm}
\begin{tabular}{cc}
  \begin{tabular}{c}
    \parbox{0.6 \linewidth}{
      \renewcommand{\arraystretch}{1.5}
      \begin{equation*}
        \centering
        \begin{array}{ccc}
              \mathbf{a}_1 & = & a \, \mathbf{\hat{x}} \\
    \mathbf{a}_2 & = & a \, \mathbf{\hat{y}} \\
    \mathbf{a}_3 & = & c \, \mathbf{\hat{z}} \\

        \end{array}
      \end{equation*}
    }
    \renewcommand{\arraystretch}{1.0}
  \end{tabular}
  \begin{tabular}{c}
    \includegraphics[width=0.3\linewidth]{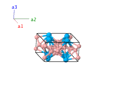} \\
  \end{tabular}
\end{tabular}

}
\vspace*{-0.25cm}

\noindent \hrulefill
\\
\textbf{Basis vectors:}
\vspace*{-0.25cm}
\renewcommand{\arraystretch}{1.5}
\begin{longtabu} to \textwidth{>{\centering $}X[-1,c,c]<{$}>{\centering $}X[-1,c,c]<{$}>{\centering $}X[-1,c,c]<{$}>{\centering $}X[-1,c,c]<{$}>{\centering $}X[-1,c,c]<{$}>{\centering $}X[-1,c,c]<{$}>{\centering $}X[-1,c,c]<{$}}
  & & \mbox{Lattice Coordinates} & & \mbox{Cartesian Coordinates} &\mbox{Wyckoff Position} & \mbox{Atom Type} \\  
  \mathbf{B}_{1} & = & z_{1} \, \mathbf{a}_{3} & = & z_{1}c \, \mathbf{\hat{z}} & \left(4e\right) & \mbox{B I} \\ 
\mathbf{B}_{2} & = & \frac{1}{2} \, \mathbf{a}_{1} + \frac{1}{2} \, \mathbf{a}_{2}-z_{1} \, \mathbf{a}_{3} & = & \frac{1}{2}a \, \mathbf{\hat{x}} + \frac{1}{2}a \, \mathbf{\hat{y}}-z_{1}c \, \mathbf{\hat{z}} & \left(4e\right) & \mbox{B I} \\ 
\mathbf{B}_{3} & = & -z_{1} \, \mathbf{a}_{3} & = & -z_{1}c \, \mathbf{\hat{z}} & \left(4e\right) & \mbox{B I} \\ 
\mathbf{B}_{4} & = & \frac{1}{2} \, \mathbf{a}_{1} + \frac{1}{2} \, \mathbf{a}_{2} + z_{1} \, \mathbf{a}_{3} & = & \frac{1}{2}a \, \mathbf{\hat{x}} + \frac{1}{2}a \, \mathbf{\hat{y}} + z_{1}c \, \mathbf{\hat{z}} & \left(4e\right) & \mbox{B I} \\ 
\mathbf{B}_{5} & = & x_{2} \, \mathbf{a}_{1} + \left(\frac{1}{2} +x_{2}\right) \, \mathbf{a}_{2} & = & x_{2}a \, \mathbf{\hat{x}} + \left(\frac{1}{2} +x_{2}\right)a \, \mathbf{\hat{y}} & \left(4g\right) & \mbox{Th} \\ 
\mathbf{B}_{6} & = & -x_{2} \, \mathbf{a}_{1} + \left(\frac{1}{2} - x_{2}\right) \, \mathbf{a}_{2} & = & -x_{2}a \, \mathbf{\hat{x}} + \left(\frac{1}{2} - x_{2}\right)a \, \mathbf{\hat{y}} & \left(4g\right) & \mbox{Th} \\ 
\mathbf{B}_{7} & = & \left(\frac{1}{2} - x_{2}\right) \, \mathbf{a}_{1} + x_{2} \, \mathbf{a}_{2} & = & \left(\frac{1}{2} - x_{2}\right)a \, \mathbf{\hat{x}} + x_{2}a \, \mathbf{\hat{y}} & \left(4g\right) & \mbox{Th} \\ 
\mathbf{B}_{8} & = & \left(\frac{1}{2} +x_{2}\right) \, \mathbf{a}_{1}-x_{2} \, \mathbf{a}_{2} & = & \left(\frac{1}{2} +x_{2}\right)a \, \mathbf{\hat{x}}-x_{2}a \, \mathbf{\hat{y}} & \left(4g\right) & \mbox{Th} \\ 
\mathbf{B}_{9} & = & x_{3} \, \mathbf{a}_{1} + \left(\frac{1}{2} +x_{3}\right) \, \mathbf{a}_{2} + \frac{1}{2} \, \mathbf{a}_{3} & = & x_{3}a \, \mathbf{\hat{x}} + \left(\frac{1}{2} +x_{3}\right)a \, \mathbf{\hat{y}} + \frac{1}{2}c \, \mathbf{\hat{z}} & \left(4h\right) & \mbox{B II} \\ 
\mathbf{B}_{10} & = & -x_{3} \, \mathbf{a}_{1} + \left(\frac{1}{2} - x_{3}\right) \, \mathbf{a}_{2} + \frac{1}{2} \, \mathbf{a}_{3} & = & -x_{3}a \, \mathbf{\hat{x}} + \left(\frac{1}{2} - x_{3}\right)a \, \mathbf{\hat{y}} + \frac{1}{2}c \, \mathbf{\hat{z}} & \left(4h\right) & \mbox{B II} \\ 
\mathbf{B}_{11} & = & \left(\frac{1}{2} - x_{3}\right) \, \mathbf{a}_{1} + x_{3} \, \mathbf{a}_{2} + \frac{1}{2} \, \mathbf{a}_{3} & = & \left(\frac{1}{2} - x_{3}\right)a \, \mathbf{\hat{x}} + x_{3}a \, \mathbf{\hat{y}} + \frac{1}{2}c \, \mathbf{\hat{z}} & \left(4h\right) & \mbox{B II} \\ 
\mathbf{B}_{12} & = & \left(\frac{1}{2} +x_{3}\right) \, \mathbf{a}_{1}-x_{3} \, \mathbf{a}_{2} + \frac{1}{2} \, \mathbf{a}_{3} & = & \left(\frac{1}{2} +x_{3}\right)a \, \mathbf{\hat{x}}-x_{3}a \, \mathbf{\hat{y}} + \frac{1}{2}c \, \mathbf{\hat{z}} & \left(4h\right) & \mbox{B II} \\ 
\mathbf{B}_{13} & = & x_{4} \, \mathbf{a}_{1} + y_{4} \, \mathbf{a}_{2} + \frac{1}{2} \, \mathbf{a}_{3} & = & x_{4}a \, \mathbf{\hat{x}} + y_{4}a \, \mathbf{\hat{y}} + \frac{1}{2}c \, \mathbf{\hat{z}} & \left(8j\right) & \mbox{B III} \\ 
\mathbf{B}_{14} & = & -x_{4} \, \mathbf{a}_{1}-y_{4} \, \mathbf{a}_{2} + \frac{1}{2} \, \mathbf{a}_{3} & = & -x_{4}a \, \mathbf{\hat{x}}-y_{4}a \, \mathbf{\hat{y}} + \frac{1}{2}c \, \mathbf{\hat{z}} & \left(8j\right) & \mbox{B III} \\ 
\mathbf{B}_{15} & = & -y_{4} \, \mathbf{a}_{1} + x_{4} \, \mathbf{a}_{2} + \frac{1}{2} \, \mathbf{a}_{3} & = & -y_{4}a \, \mathbf{\hat{x}} + x_{4}a \, \mathbf{\hat{y}} + \frac{1}{2}c \, \mathbf{\hat{z}} & \left(8j\right) & \mbox{B III} \\ 
\mathbf{B}_{16} & = & y_{4} \, \mathbf{a}_{1}-x_{4} \, \mathbf{a}_{2} + \frac{1}{2} \, \mathbf{a}_{3} & = & y_{4}a \, \mathbf{\hat{x}}-x_{4}a \, \mathbf{\hat{y}} + \frac{1}{2}c \, \mathbf{\hat{z}} & \left(8j\right) & \mbox{B III} \\ 
\mathbf{B}_{17} & = & \left(\frac{1}{2} - x_{4}\right) \, \mathbf{a}_{1} + \left(\frac{1}{2} +y_{4}\right) \, \mathbf{a}_{2} + \frac{1}{2} \, \mathbf{a}_{3} & = & \left(\frac{1}{2} - x_{4}\right)a \, \mathbf{\hat{x}} + \left(\frac{1}{2} +y_{4}\right)a \, \mathbf{\hat{y}} + \frac{1}{2}c \, \mathbf{\hat{z}} & \left(8j\right) & \mbox{B III} \\ 
\mathbf{B}_{18} & = & \left(\frac{1}{2} +x_{4}\right) \, \mathbf{a}_{1} + \left(\frac{1}{2} - y_{4}\right) \, \mathbf{a}_{2} + \frac{1}{2} \, \mathbf{a}_{3} & = & \left(\frac{1}{2} +x_{4}\right)a \, \mathbf{\hat{x}} + \left(\frac{1}{2} - y_{4}\right)a \, \mathbf{\hat{y}} + \frac{1}{2}c \, \mathbf{\hat{z}} & \left(8j\right) & \mbox{B III} \\ 
\mathbf{B}_{19} & = & \left(\frac{1}{2} +y_{4}\right) \, \mathbf{a}_{1} + \left(\frac{1}{2} +x_{4}\right) \, \mathbf{a}_{2} + \frac{1}{2} \, \mathbf{a}_{3} & = & \left(\frac{1}{2} +y_{4}\right)a \, \mathbf{\hat{x}} + \left(\frac{1}{2} +x_{4}\right)a \, \mathbf{\hat{y}} + \frac{1}{2}c \, \mathbf{\hat{z}} & \left(8j\right) & \mbox{B III} \\ 
\mathbf{B}_{20} & = & \left(\frac{1}{2} - y_{4}\right) \, \mathbf{a}_{1} + \left(\frac{1}{2} - x_{4}\right) \, \mathbf{a}_{2} + \frac{1}{2} \, \mathbf{a}_{3} & = & \left(\frac{1}{2} - y_{4}\right)a \, \mathbf{\hat{x}} + \left(\frac{1}{2} - x_{4}\right)a \, \mathbf{\hat{y}} + \frac{1}{2}c \, \mathbf{\hat{z}} & \left(8j\right) & \mbox{B III} \\ 
\end{longtabu}
\renewcommand{\arraystretch}{1.0}
\noindent \hrulefill
\\
\textbf{References:}
\vspace*{-0.25cm}
\begin{flushleft}
  - \bibentry{Zalkin_JCP_18_1950}. \\
\end{flushleft}
\noindent \hrulefill
\\
\textbf{Geometry files:}
\\
\noindent  - CIF: pp. {\hyperref[A4B_tP20_127_ehj_g_cif]{\pageref{A4B_tP20_127_ehj_g_cif}}} \\
\noindent  - POSCAR: pp. {\hyperref[A4B_tP20_127_ehj_g_poscar]{\pageref{A4B_tP20_127_ehj_g_poscar}}} \\
\onecolumn
{\phantomsection\label{A6B2C_tP18_128_eh_d_b}}
\subsection*{\huge \textbf{{\normalfont \begin{raggedleft}K$_{2}$SnCl$_{6}$ (Low-temperature) Structure: \end{raggedleft} \\ A6B2C\_tP18\_128\_eh\_d\_b}}}
\noindent \hrulefill
\vspace*{0.25cm}
\begin{figure}[htp]
  \centering
  \vspace{-1em}
  {\includegraphics[width=1\textwidth]{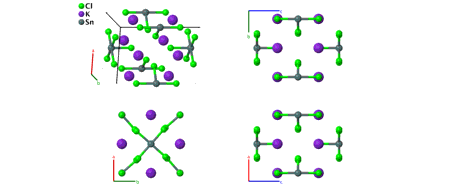}}
\end{figure}
\vspace*{-0.5cm}
\renewcommand{\arraystretch}{1.5}
\begin{equation*}
  \begin{array}{>{$\hspace{-0.15cm}}l<{$}>{$}p{0.5cm}<{$}>{$}p{18.5cm}<{$}}
    \mbox{\large \textbf{Prototype}} &\colon & \ce{K2SnCl6} \\
    \mbox{\large \textbf{\AFLOW\ prototype label}} &\colon & \mbox{A6B2C\_tP18\_128\_eh\_d\_b} \\
    \mbox{\large \textbf{\textit{Strukturbericht} designation}} &\colon & \mbox{None} \\
    \mbox{\large \textbf{Pearson symbol}} &\colon & \mbox{tP18} \\
    \mbox{\large \textbf{Space group number}} &\colon & 128 \\
    \mbox{\large \textbf{Space group symbol}} &\colon & P4/mnc \\
    \mbox{\large \textbf{\AFLOW\ prototype command}} &\colon &  \texttt{aflow} \,  \, \texttt{-{}-proto=A6B2C\_tP18\_128\_eh\_d\_b } \, \newline \texttt{-{}-params=}{a,c/a,z_{3},x_{4},y_{4} }
  \end{array}
\end{equation*}
\renewcommand{\arraystretch}{1.0}

\noindent \parbox{1 \linewidth}{
\noindent \hrulefill
\\
\textbf{Simple Tetragonal primitive vectors:} \\
\vspace*{-0.25cm}
\begin{tabular}{cc}
  \begin{tabular}{c}
    \parbox{0.6 \linewidth}{
      \renewcommand{\arraystretch}{1.5}
      \begin{equation*}
        \centering
        \begin{array}{ccc}
              \mathbf{a}_1 & = & a \, \mathbf{\hat{x}} \\
    \mathbf{a}_2 & = & a \, \mathbf{\hat{y}} \\
    \mathbf{a}_3 & = & c \, \mathbf{\hat{z}} \\

        \end{array}
      \end{equation*}
    }
    \renewcommand{\arraystretch}{1.0}
  \end{tabular}
  \begin{tabular}{c}
    \includegraphics[width=0.3\linewidth]{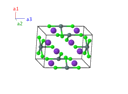} \\
  \end{tabular}
\end{tabular}

}
\vspace*{-0.25cm}

\noindent \hrulefill
\\
\textbf{Basis vectors:}
\vspace*{-0.25cm}
\renewcommand{\arraystretch}{1.5}
\begin{longtabu} to \textwidth{>{\centering $}X[-1,c,c]<{$}>{\centering $}X[-1,c,c]<{$}>{\centering $}X[-1,c,c]<{$}>{\centering $}X[-1,c,c]<{$}>{\centering $}X[-1,c,c]<{$}>{\centering $}X[-1,c,c]<{$}>{\centering $}X[-1,c,c]<{$}}
  & & \mbox{Lattice Coordinates} & & \mbox{Cartesian Coordinates} &\mbox{Wyckoff Position} & \mbox{Atom Type} \\  
  \mathbf{B}_{1} & = & \frac{1}{2} \, \mathbf{a}_{3} & = & \frac{1}{2}c \, \mathbf{\hat{z}} & \left(2b\right) & \mbox{Sn} \\ 
\mathbf{B}_{2} & = & \frac{1}{2} \, \mathbf{a}_{1} + \frac{1}{2} \, \mathbf{a}_{2} & = & \frac{1}{2}a \, \mathbf{\hat{x}} + \frac{1}{2}a \, \mathbf{\hat{y}} & \left(2b\right) & \mbox{Sn} \\ 
\mathbf{B}_{3} & = & \frac{1}{2} \, \mathbf{a}_{2} + \frac{1}{4} \, \mathbf{a}_{3} & = & \frac{1}{2}a \, \mathbf{\hat{y}} + \frac{1}{4}c \, \mathbf{\hat{z}} & \left(4d\right) & \mbox{K} \\ 
\mathbf{B}_{4} & = & \frac{1}{2} \, \mathbf{a}_{1} + \frac{1}{4} \, \mathbf{a}_{3} & = & \frac{1}{2}a \, \mathbf{\hat{x}} + \frac{1}{4}c \, \mathbf{\hat{z}} & \left(4d\right) & \mbox{K} \\ 
\mathbf{B}_{5} & = & \frac{1}{2} \, \mathbf{a}_{2} + \frac{3}{4} \, \mathbf{a}_{3} & = & \frac{1}{2}a \, \mathbf{\hat{y}} + \frac{3}{4}c \, \mathbf{\hat{z}} & \left(4d\right) & \mbox{K} \\ 
\mathbf{B}_{6} & = & \frac{1}{2} \, \mathbf{a}_{1} + \frac{3}{4} \, \mathbf{a}_{3} & = & \frac{1}{2}a \, \mathbf{\hat{x}} + \frac{3}{4}c \, \mathbf{\hat{z}} & \left(4d\right) & \mbox{K} \\ 
\mathbf{B}_{7} & = & z_{3} \, \mathbf{a}_{3} & = & z_{3}c \, \mathbf{\hat{z}} & \left(4e\right) & \mbox{Cl I} \\ 
\mathbf{B}_{8} & = & \frac{1}{2} \, \mathbf{a}_{1} + \frac{1}{2} \, \mathbf{a}_{2} + \left(\frac{1}{2} - z_{3}\right) \, \mathbf{a}_{3} & = & \frac{1}{2}a \, \mathbf{\hat{x}} + \frac{1}{2}a \, \mathbf{\hat{y}} + \left(\frac{1}{2} - z_{3}\right)c \, \mathbf{\hat{z}} & \left(4e\right) & \mbox{Cl I} \\ 
\mathbf{B}_{9} & = & -z_{3} \, \mathbf{a}_{3} & = & -z_{3}c \, \mathbf{\hat{z}} & \left(4e\right) & \mbox{Cl I} \\ 
\mathbf{B}_{10} & = & \frac{1}{2} \, \mathbf{a}_{1} + \frac{1}{2} \, \mathbf{a}_{2} + \left(\frac{1}{2} +z_{3}\right) \, \mathbf{a}_{3} & = & \frac{1}{2}a \, \mathbf{\hat{x}} + \frac{1}{2}a \, \mathbf{\hat{y}} + \left(\frac{1}{2} +z_{3}\right)c \, \mathbf{\hat{z}} & \left(4e\right) & \mbox{Cl I} \\ 
\mathbf{B}_{11} & = & x_{4} \, \mathbf{a}_{1} + y_{4} \, \mathbf{a}_{2} & = & x_{4}a \, \mathbf{\hat{x}} + y_{4}a \, \mathbf{\hat{y}} & \left(8h\right) & \mbox{Cl II} \\ 
\mathbf{B}_{12} & = & -x_{4} \, \mathbf{a}_{1}-y_{4} \, \mathbf{a}_{2} & = & -x_{4}a \, \mathbf{\hat{x}}-y_{4}a \, \mathbf{\hat{y}} & \left(8h\right) & \mbox{Cl II} \\ 
\mathbf{B}_{13} & = & -y_{4} \, \mathbf{a}_{1} + x_{4} \, \mathbf{a}_{2} & = & -y_{4}a \, \mathbf{\hat{x}} + x_{4}a \, \mathbf{\hat{y}} & \left(8h\right) & \mbox{Cl II} \\ 
\mathbf{B}_{14} & = & y_{4} \, \mathbf{a}_{1}-x_{4} \, \mathbf{a}_{2} & = & y_{4}a \, \mathbf{\hat{x}}-x_{4}a \, \mathbf{\hat{y}} & \left(8h\right) & \mbox{Cl II} \\ 
\mathbf{B}_{15} & = & \left(\frac{1}{2} - x_{4}\right) \, \mathbf{a}_{1} + \left(\frac{1}{2} +y_{4}\right) \, \mathbf{a}_{2} + \frac{1}{2} \, \mathbf{a}_{3} & = & \left(\frac{1}{2} - x_{4}\right)a \, \mathbf{\hat{x}} + \left(\frac{1}{2} +y_{4}\right)a \, \mathbf{\hat{y}} + \frac{1}{2}c \, \mathbf{\hat{z}} & \left(8h\right) & \mbox{Cl II} \\ 
\mathbf{B}_{16} & = & \left(\frac{1}{2} +x_{4}\right) \, \mathbf{a}_{1} + \left(\frac{1}{2} - y_{4}\right) \, \mathbf{a}_{2} + \frac{1}{2} \, \mathbf{a}_{3} & = & \left(\frac{1}{2} +x_{4}\right)a \, \mathbf{\hat{x}} + \left(\frac{1}{2} - y_{4}\right)a \, \mathbf{\hat{y}} + \frac{1}{2}c \, \mathbf{\hat{z}} & \left(8h\right) & \mbox{Cl II} \\ 
\mathbf{B}_{17} & = & \left(\frac{1}{2} +y_{4}\right) \, \mathbf{a}_{1} + \left(\frac{1}{2} +x_{4}\right) \, \mathbf{a}_{2} + \frac{1}{2} \, \mathbf{a}_{3} & = & \left(\frac{1}{2} +y_{4}\right)a \, \mathbf{\hat{x}} + \left(\frac{1}{2} +x_{4}\right)a \, \mathbf{\hat{y}} + \frac{1}{2}c \, \mathbf{\hat{z}} & \left(8h\right) & \mbox{Cl II} \\ 
\mathbf{B}_{18} & = & \left(\frac{1}{2} - y_{4}\right) \, \mathbf{a}_{1} + \left(\frac{1}{2} - x_{4}\right) \, \mathbf{a}_{2} + \frac{1}{2} \, \mathbf{a}_{3} & = & \left(\frac{1}{2} - y_{4}\right)a \, \mathbf{\hat{x}} + \left(\frac{1}{2} - x_{4}\right)a \, \mathbf{\hat{y}} + \frac{1}{2}c \, \mathbf{\hat{z}} & \left(8h\right) & \mbox{Cl II} \\ 
\end{longtabu}
\renewcommand{\arraystretch}{1.0}
\noindent \hrulefill
\\
\textbf{References:}
\vspace*{-0.25cm}
\begin{flushleft}
  - \bibentry{Boysen_K2SnCl6_ActCrystallogrSecB_1978}. \\
\end{flushleft}
\textbf{Found in:}
\vspace*{-0.25cm}
\begin{flushleft}
  - \bibentry{Villars_PearsonsCrystalData_2013}. \\
\end{flushleft}
\noindent \hrulefill
\\
\textbf{Geometry files:}
\\
\noindent  - CIF: pp. {\hyperref[A6B2C_tP18_128_eh_d_b_cif]{\pageref{A6B2C_tP18_128_eh_d_b_cif}}} \\
\noindent  - POSCAR: pp. {\hyperref[A6B2C_tP18_128_eh_d_b_poscar]{\pageref{A6B2C_tP18_128_eh_d_b_poscar}}} \\
\onecolumn
{\phantomsection\label{A7B2C_tP40_128_egi_h_e}}
\subsection*{\huge \textbf{{\normalfont FeCu$_{2}$Al$_{7}$ ($E9_{a}$) Structure: A7B2C\_tP40\_128\_egi\_h\_e}}}
\noindent \hrulefill
\vspace*{0.25cm}
\begin{figure}[htp]
  \centering
  \vspace{-1em}
  {\includegraphics[width=1\textwidth]{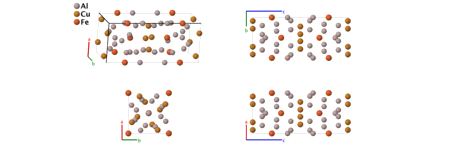}}
\end{figure}
\vspace*{-0.5cm}
\renewcommand{\arraystretch}{1.5}
\begin{equation*}
  \begin{array}{>{$\hspace{-0.15cm}}l<{$}>{$}p{0.5cm}<{$}>{$}p{18.5cm}<{$}}
    \mbox{\large \textbf{Prototype}} &\colon & \ce{FeCu2Al7} \\
    \mbox{\large \textbf{\AFLOW\ prototype label}} &\colon & \mbox{A7B2C\_tP40\_128\_egi\_h\_e} \\
    \mbox{\large \textbf{\textit{Strukturbericht} designation}} &\colon & \mbox{$E9_{a}$} \\
    \mbox{\large \textbf{Pearson symbol}} &\colon & \mbox{tP40} \\
    \mbox{\large \textbf{Space group number}} &\colon & 128 \\
    \mbox{\large \textbf{Space group symbol}} &\colon & P4/mnc \\
    \mbox{\large \textbf{\AFLOW\ prototype command}} &\colon &  \texttt{aflow} \,  \, \texttt{-{}-proto=A7B2C\_tP40\_128\_egi\_h\_e } \, \newline \texttt{-{}-params=}{a,c/a,z_{1},z_{2},x_{3},x_{4},y_{4},x_{5},y_{5},z_{5} }
  \end{array}
\end{equation*}
\renewcommand{\arraystretch}{1.0}

\vspace*{-0.25cm}
\noindent \hrulefill
\\
\textbf{ Other compounds with this structure:}
\begin{itemize}
   \item{ T(CoCuAl), an alloy with an approximate composition of Co$_{2}$Cu$_{4.9}$Al$_{17.7}$, NiCu$_{3}$Al$_{6}$  }
\end{itemize}
\noindent \parbox{1 \linewidth}{
\noindent \hrulefill
\\
\textbf{Simple Tetragonal primitive vectors:} \\
\vspace*{-0.25cm}
\begin{tabular}{cc}
  \begin{tabular}{c}
    \parbox{0.6 \linewidth}{
      \renewcommand{\arraystretch}{1.5}
      \begin{equation*}
        \centering
        \begin{array}{ccc}
              \mathbf{a}_1 & = & a \, \mathbf{\hat{x}} \\
    \mathbf{a}_2 & = & a \, \mathbf{\hat{y}} \\
    \mathbf{a}_3 & = & c \, \mathbf{\hat{z}} \\

        \end{array}
      \end{equation*}
    }
    \renewcommand{\arraystretch}{1.0}
  \end{tabular}
  \begin{tabular}{c}
    \includegraphics[width=0.3\linewidth]{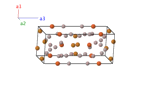} \\
  \end{tabular}
\end{tabular}

}
\vspace*{-0.25cm}

\noindent \hrulefill
\\
\textbf{Basis vectors:}
\vspace*{-0.25cm}
\renewcommand{\arraystretch}{1.5}
\begin{longtabu} to \textwidth{>{\centering $}X[-1,c,c]<{$}>{\centering $}X[-1,c,c]<{$}>{\centering $}X[-1,c,c]<{$}>{\centering $}X[-1,c,c]<{$}>{\centering $}X[-1,c,c]<{$}>{\centering $}X[-1,c,c]<{$}>{\centering $}X[-1,c,c]<{$}}
  & & \mbox{Lattice Coordinates} & & \mbox{Cartesian Coordinates} &\mbox{Wyckoff Position} & \mbox{Atom Type} \\  
  \mathbf{B}_{1} & = & z_{1} \, \mathbf{a}_{3} & = & z_{1}c \, \mathbf{\hat{z}} & \left(4e\right) & \mbox{Al I} \\ 
\mathbf{B}_{2} & = & \frac{1}{2} \, \mathbf{a}_{1} + \frac{1}{2} \, \mathbf{a}_{2} + \left(\frac{1}{2} - z_{1}\right) \, \mathbf{a}_{3} & = & \frac{1}{2}a \, \mathbf{\hat{x}} + \frac{1}{2}a \, \mathbf{\hat{y}} + \left(\frac{1}{2} - z_{1}\right)c \, \mathbf{\hat{z}} & \left(4e\right) & \mbox{Al I} \\ 
\mathbf{B}_{3} & = & -z_{1} \, \mathbf{a}_{3} & = & -z_{1}c \, \mathbf{\hat{z}} & \left(4e\right) & \mbox{Al I} \\ 
\mathbf{B}_{4} & = & \frac{1}{2} \, \mathbf{a}_{1} + \frac{1}{2} \, \mathbf{a}_{2} + \left(\frac{1}{2} +z_{1}\right) \, \mathbf{a}_{3} & = & \frac{1}{2}a \, \mathbf{\hat{x}} + \frac{1}{2}a \, \mathbf{\hat{y}} + \left(\frac{1}{2} +z_{1}\right)c \, \mathbf{\hat{z}} & \left(4e\right) & \mbox{Al I} \\ 
\mathbf{B}_{5} & = & z_{2} \, \mathbf{a}_{3} & = & z_{2}c \, \mathbf{\hat{z}} & \left(4e\right) & \mbox{Fe} \\ 
\mathbf{B}_{6} & = & \frac{1}{2} \, \mathbf{a}_{1} + \frac{1}{2} \, \mathbf{a}_{2} + \left(\frac{1}{2} - z_{2}\right) \, \mathbf{a}_{3} & = & \frac{1}{2}a \, \mathbf{\hat{x}} + \frac{1}{2}a \, \mathbf{\hat{y}} + \left(\frac{1}{2} - z_{2}\right)c \, \mathbf{\hat{z}} & \left(4e\right) & \mbox{Fe} \\ 
\mathbf{B}_{7} & = & -z_{2} \, \mathbf{a}_{3} & = & -z_{2}c \, \mathbf{\hat{z}} & \left(4e\right) & \mbox{Fe} \\ 
\mathbf{B}_{8} & = & \frac{1}{2} \, \mathbf{a}_{1} + \frac{1}{2} \, \mathbf{a}_{2} + \left(\frac{1}{2} +z_{2}\right) \, \mathbf{a}_{3} & = & \frac{1}{2}a \, \mathbf{\hat{x}} + \frac{1}{2}a \, \mathbf{\hat{y}} + \left(\frac{1}{2} +z_{2}\right)c \, \mathbf{\hat{z}} & \left(4e\right) & \mbox{Fe} \\ 
\mathbf{B}_{9} & = & x_{3} \, \mathbf{a}_{1} + \left(\frac{1}{2} +x_{3}\right) \, \mathbf{a}_{2} + \frac{1}{4} \, \mathbf{a}_{3} & = & x_{3}a \, \mathbf{\hat{x}} + \left(\frac{1}{2} +x_{3}\right)a \, \mathbf{\hat{y}} + \frac{1}{4}c \, \mathbf{\hat{z}} & \left(8g\right) & \mbox{Al II} \\ 
\mathbf{B}_{10} & = & -x_{3} \, \mathbf{a}_{1} + \left(\frac{1}{2} - x_{3}\right) \, \mathbf{a}_{2} + \frac{1}{4} \, \mathbf{a}_{3} & = & -x_{3}a \, \mathbf{\hat{x}} + \left(\frac{1}{2} - x_{3}\right)a \, \mathbf{\hat{y}} + \frac{1}{4}c \, \mathbf{\hat{z}} & \left(8g\right) & \mbox{Al II} \\ 
\mathbf{B}_{11} & = & \left(\frac{1}{2} - x_{3}\right) \, \mathbf{a}_{1} + x_{3} \, \mathbf{a}_{2} + \frac{1}{4} \, \mathbf{a}_{3} & = & \left(\frac{1}{2} - x_{3}\right)a \, \mathbf{\hat{x}} + x_{3}a \, \mathbf{\hat{y}} + \frac{1}{4}c \, \mathbf{\hat{z}} & \left(8g\right) & \mbox{Al II} \\ 
\mathbf{B}_{12} & = & \left(\frac{1}{2} +x_{3}\right) \, \mathbf{a}_{1}-x_{3} \, \mathbf{a}_{2} + \frac{1}{4} \, \mathbf{a}_{3} & = & \left(\frac{1}{2} +x_{3}\right)a \, \mathbf{\hat{x}}-x_{3}a \, \mathbf{\hat{y}} + \frac{1}{4}c \, \mathbf{\hat{z}} & \left(8g\right) & \mbox{Al II} \\ 
\mathbf{B}_{13} & = & -x_{3} \, \mathbf{a}_{1} + \left(\frac{1}{2} - x_{3}\right) \, \mathbf{a}_{2} + \frac{3}{4} \, \mathbf{a}_{3} & = & -x_{3}a \, \mathbf{\hat{x}} + \left(\frac{1}{2} - x_{3}\right)a \, \mathbf{\hat{y}} + \frac{3}{4}c \, \mathbf{\hat{z}} & \left(8g\right) & \mbox{Al II} \\ 
\mathbf{B}_{14} & = & x_{3} \, \mathbf{a}_{1} + \left(\frac{1}{2} +x_{3}\right) \, \mathbf{a}_{2} + \frac{3}{4} \, \mathbf{a}_{3} & = & x_{3}a \, \mathbf{\hat{x}} + \left(\frac{1}{2} +x_{3}\right)a \, \mathbf{\hat{y}} + \frac{3}{4}c \, \mathbf{\hat{z}} & \left(8g\right) & \mbox{Al II} \\ 
\mathbf{B}_{15} & = & \left(\frac{1}{2} +x_{3}\right) \, \mathbf{a}_{1}-x_{3} \, \mathbf{a}_{2} + \frac{3}{4} \, \mathbf{a}_{3} & = & \left(\frac{1}{2} +x_{3}\right)a \, \mathbf{\hat{x}}-x_{3}a \, \mathbf{\hat{y}} + \frac{3}{4}c \, \mathbf{\hat{z}} & \left(8g\right) & \mbox{Al II} \\ 
\mathbf{B}_{16} & = & \left(\frac{1}{2} - x_{3}\right) \, \mathbf{a}_{1} + x_{3} \, \mathbf{a}_{2} + \frac{3}{4} \, \mathbf{a}_{3} & = & \left(\frac{1}{2} - x_{3}\right)a \, \mathbf{\hat{x}} + x_{3}a \, \mathbf{\hat{y}} + \frac{3}{4}c \, \mathbf{\hat{z}} & \left(8g\right) & \mbox{Al II} \\ 
\mathbf{B}_{17} & = & x_{4} \, \mathbf{a}_{1} + y_{4} \, \mathbf{a}_{2} & = & x_{4}a \, \mathbf{\hat{x}} + y_{4}a \, \mathbf{\hat{y}} & \left(8h\right) & \mbox{Cu} \\ 
\mathbf{B}_{18} & = & -x_{4} \, \mathbf{a}_{1}-y_{4} \, \mathbf{a}_{2} & = & -x_{4}a \, \mathbf{\hat{x}}-y_{4}a \, \mathbf{\hat{y}} & \left(8h\right) & \mbox{Cu} \\ 
\mathbf{B}_{19} & = & -y_{4} \, \mathbf{a}_{1} + x_{4} \, \mathbf{a}_{2} & = & -y_{4}a \, \mathbf{\hat{x}} + x_{4}a \, \mathbf{\hat{y}} & \left(8h\right) & \mbox{Cu} \\ 
\mathbf{B}_{20} & = & y_{4} \, \mathbf{a}_{1}-x_{4} \, \mathbf{a}_{2} & = & y_{4}a \, \mathbf{\hat{x}}-x_{4}a \, \mathbf{\hat{y}} & \left(8h\right) & \mbox{Cu} \\ 
\mathbf{B}_{21} & = & \left(\frac{1}{2} - x_{4}\right) \, \mathbf{a}_{1} + \left(\frac{1}{2} +y_{4}\right) \, \mathbf{a}_{2} + \frac{1}{2} \, \mathbf{a}_{3} & = & \left(\frac{1}{2} - x_{4}\right)a \, \mathbf{\hat{x}} + \left(\frac{1}{2} +y_{4}\right)a \, \mathbf{\hat{y}} + \frac{1}{2}c \, \mathbf{\hat{z}} & \left(8h\right) & \mbox{Cu} \\ 
\mathbf{B}_{22} & = & \left(\frac{1}{2} +x_{4}\right) \, \mathbf{a}_{1} + \left(\frac{1}{2} - y_{4}\right) \, \mathbf{a}_{2} + \frac{1}{2} \, \mathbf{a}_{3} & = & \left(\frac{1}{2} +x_{4}\right)a \, \mathbf{\hat{x}} + \left(\frac{1}{2} - y_{4}\right)a \, \mathbf{\hat{y}} + \frac{1}{2}c \, \mathbf{\hat{z}} & \left(8h\right) & \mbox{Cu} \\ 
\mathbf{B}_{23} & = & \left(\frac{1}{2} +y_{4}\right) \, \mathbf{a}_{1} + \left(\frac{1}{2} +x_{4}\right) \, \mathbf{a}_{2} + \frac{1}{2} \, \mathbf{a}_{3} & = & \left(\frac{1}{2} +y_{4}\right)a \, \mathbf{\hat{x}} + \left(\frac{1}{2} +x_{4}\right)a \, \mathbf{\hat{y}} + \frac{1}{2}c \, \mathbf{\hat{z}} & \left(8h\right) & \mbox{Cu} \\ 
\mathbf{B}_{24} & = & \left(\frac{1}{2} - y_{4}\right) \, \mathbf{a}_{1} + \left(\frac{1}{2} - x_{4}\right) \, \mathbf{a}_{2} + \frac{1}{2} \, \mathbf{a}_{3} & = & \left(\frac{1}{2} - y_{4}\right)a \, \mathbf{\hat{x}} + \left(\frac{1}{2} - x_{4}\right)a \, \mathbf{\hat{y}} + \frac{1}{2}c \, \mathbf{\hat{z}} & \left(8h\right) & \mbox{Cu} \\ 
\mathbf{B}_{25} & = & x_{5} \, \mathbf{a}_{1} + y_{5} \, \mathbf{a}_{2} + z_{5} \, \mathbf{a}_{3} & = & x_{5}a \, \mathbf{\hat{x}} + y_{5}a \, \mathbf{\hat{y}} + z_{5}c \, \mathbf{\hat{z}} & \left(16i\right) & \mbox{Al III} \\ 
\mathbf{B}_{26} & = & -x_{5} \, \mathbf{a}_{1}-y_{5} \, \mathbf{a}_{2} + z_{5} \, \mathbf{a}_{3} & = & -x_{5}a \, \mathbf{\hat{x}}-y_{5}a \, \mathbf{\hat{y}} + z_{5}c \, \mathbf{\hat{z}} & \left(16i\right) & \mbox{Al III} \\ 
\mathbf{B}_{27} & = & -y_{5} \, \mathbf{a}_{1} + x_{5} \, \mathbf{a}_{2} + z_{5} \, \mathbf{a}_{3} & = & -y_{5}a \, \mathbf{\hat{x}} + x_{5}a \, \mathbf{\hat{y}} + z_{5}c \, \mathbf{\hat{z}} & \left(16i\right) & \mbox{Al III} \\ 
\mathbf{B}_{28} & = & y_{5} \, \mathbf{a}_{1}-x_{5} \, \mathbf{a}_{2} + z_{5} \, \mathbf{a}_{3} & = & y_{5}a \, \mathbf{\hat{x}}-x_{5}a \, \mathbf{\hat{y}} + z_{5}c \, \mathbf{\hat{z}} & \left(16i\right) & \mbox{Al III} \\ 
\mathbf{B}_{29} & = & \left(\frac{1}{2} - x_{5}\right) \, \mathbf{a}_{1} + \left(\frac{1}{2} +y_{5}\right) \, \mathbf{a}_{2} + \left(\frac{1}{2} - z_{5}\right) \, \mathbf{a}_{3} & = & \left(\frac{1}{2} - x_{5}\right)a \, \mathbf{\hat{x}} + \left(\frac{1}{2} +y_{5}\right)a \, \mathbf{\hat{y}} + \left(\frac{1}{2} - z_{5}\right)c \, \mathbf{\hat{z}} & \left(16i\right) & \mbox{Al III} \\ 
\mathbf{B}_{30} & = & \left(\frac{1}{2} +x_{5}\right) \, \mathbf{a}_{1} + \left(\frac{1}{2} - y_{5}\right) \, \mathbf{a}_{2} + \left(\frac{1}{2} - z_{5}\right) \, \mathbf{a}_{3} & = & \left(\frac{1}{2} +x_{5}\right)a \, \mathbf{\hat{x}} + \left(\frac{1}{2} - y_{5}\right)a \, \mathbf{\hat{y}} + \left(\frac{1}{2} - z_{5}\right)c \, \mathbf{\hat{z}} & \left(16i\right) & \mbox{Al III} \\ 
\mathbf{B}_{31} & = & \left(\frac{1}{2} +y_{5}\right) \, \mathbf{a}_{1} + \left(\frac{1}{2} +x_{5}\right) \, \mathbf{a}_{2} + \left(\frac{1}{2} - z_{5}\right) \, \mathbf{a}_{3} & = & \left(\frac{1}{2} +y_{5}\right)a \, \mathbf{\hat{x}} + \left(\frac{1}{2} +x_{5}\right)a \, \mathbf{\hat{y}} + \left(\frac{1}{2} - z_{5}\right)c \, \mathbf{\hat{z}} & \left(16i\right) & \mbox{Al III} \\ 
\mathbf{B}_{32} & = & \left(\frac{1}{2} - y_{5}\right) \, \mathbf{a}_{1} + \left(\frac{1}{2} - x_{5}\right) \, \mathbf{a}_{2} + \left(\frac{1}{2} - z_{5}\right) \, \mathbf{a}_{3} & = & \left(\frac{1}{2} - y_{5}\right)a \, \mathbf{\hat{x}} + \left(\frac{1}{2} - x_{5}\right)a \, \mathbf{\hat{y}} + \left(\frac{1}{2} - z_{5}\right)c \, \mathbf{\hat{z}} & \left(16i\right) & \mbox{Al III} \\ 
\mathbf{B}_{33} & = & -x_{5} \, \mathbf{a}_{1}-y_{5} \, \mathbf{a}_{2}-z_{5} \, \mathbf{a}_{3} & = & -x_{5}a \, \mathbf{\hat{x}}-y_{5}a \, \mathbf{\hat{y}}-z_{5}c \, \mathbf{\hat{z}} & \left(16i\right) & \mbox{Al III} \\ 
\mathbf{B}_{34} & = & x_{5} \, \mathbf{a}_{1} + y_{5} \, \mathbf{a}_{2}-z_{5} \, \mathbf{a}_{3} & = & x_{5}a \, \mathbf{\hat{x}} + y_{5}a \, \mathbf{\hat{y}}-z_{5}c \, \mathbf{\hat{z}} & \left(16i\right) & \mbox{Al III} \\ 
\mathbf{B}_{35} & = & y_{5} \, \mathbf{a}_{1}-x_{5} \, \mathbf{a}_{2}-z_{5} \, \mathbf{a}_{3} & = & y_{5}a \, \mathbf{\hat{x}}-x_{5}a \, \mathbf{\hat{y}}-z_{5}c \, \mathbf{\hat{z}} & \left(16i\right) & \mbox{Al III} \\ 
\mathbf{B}_{36} & = & -y_{5} \, \mathbf{a}_{1} + x_{5} \, \mathbf{a}_{2}-z_{5} \, \mathbf{a}_{3} & = & -y_{5}a \, \mathbf{\hat{x}} + x_{5}a \, \mathbf{\hat{y}}-z_{5}c \, \mathbf{\hat{z}} & \left(16i\right) & \mbox{Al III} \\ 
\mathbf{B}_{37} & = & \left(\frac{1}{2} +x_{5}\right) \, \mathbf{a}_{1} + \left(\frac{1}{2} - y_{5}\right) \, \mathbf{a}_{2} + \left(\frac{1}{2} +z_{5}\right) \, \mathbf{a}_{3} & = & \left(\frac{1}{2} +x_{5}\right)a \, \mathbf{\hat{x}} + \left(\frac{1}{2} - y_{5}\right)a \, \mathbf{\hat{y}} + \left(\frac{1}{2} +z_{5}\right)c \, \mathbf{\hat{z}} & \left(16i\right) & \mbox{Al III} \\ 
\mathbf{B}_{38} & = & \left(\frac{1}{2} - x_{5}\right) \, \mathbf{a}_{1} + \left(\frac{1}{2} +y_{5}\right) \, \mathbf{a}_{2} + \left(\frac{1}{2} +z_{5}\right) \, \mathbf{a}_{3} & = & \left(\frac{1}{2} - x_{5}\right)a \, \mathbf{\hat{x}} + \left(\frac{1}{2} +y_{5}\right)a \, \mathbf{\hat{y}} + \left(\frac{1}{2} +z_{5}\right)c \, \mathbf{\hat{z}} & \left(16i\right) & \mbox{Al III} \\ 
\mathbf{B}_{39} & = & \left(\frac{1}{2} - y_{5}\right) \, \mathbf{a}_{1} + \left(\frac{1}{2} - x_{5}\right) \, \mathbf{a}_{2} + \left(\frac{1}{2} +z_{5}\right) \, \mathbf{a}_{3} & = & \left(\frac{1}{2} - y_{5}\right)a \, \mathbf{\hat{x}} + \left(\frac{1}{2} - x_{5}\right)a \, \mathbf{\hat{y}} + \left(\frac{1}{2} +z_{5}\right)c \, \mathbf{\hat{z}} & \left(16i\right) & \mbox{Al III} \\ 
\mathbf{B}_{40} & = & \left(\frac{1}{2} +y_{5}\right) \, \mathbf{a}_{1} + \left(\frac{1}{2} +x_{5}\right) \, \mathbf{a}_{2} + \left(\frac{1}{2} +z_{5}\right) \, \mathbf{a}_{3} & = & \left(\frac{1}{2} +y_{5}\right)a \, \mathbf{\hat{x}} + \left(\frac{1}{2} +x_{5}\right)a \, \mathbf{\hat{y}} + \left(\frac{1}{2} +z_{5}\right)c \, \mathbf{\hat{z}} & \left(16i\right) & \mbox{Al III} \\ 
\end{longtabu}
\renewcommand{\arraystretch}{1.0}
\noindent \hrulefill
\\
\textbf{References:}
\vspace*{-0.25cm}
\begin{flushleft}
  - \bibentry{Bown_Acta_Cryst_9_911_1956}. \\
\end{flushleft}
\noindent \hrulefill
\\
\textbf{Geometry files:}
\\
\noindent  - CIF: pp. {\hyperref[A7B2C_tP40_128_egi_h_e_cif]{\pageref{A7B2C_tP40_128_egi_h_e_cif}}} \\
\noindent  - POSCAR: pp. {\hyperref[A7B2C_tP40_128_egi_h_e_poscar]{\pageref{A7B2C_tP40_128_egi_h_e_poscar}}} \\
\onecolumn
{\phantomsection\label{A2BC4_tP28_130_f_c_g}}
\subsection*{\huge \textbf{{\normalfont CuBi$_{2}$O$_{4}$ Structure: A2BC4\_tP28\_130\_f\_c\_g}}}
\noindent \hrulefill
\vspace*{0.25cm}
\begin{figure}[htp]
  \centering
  \vspace{-1em}
  {\includegraphics[width=1\textwidth]{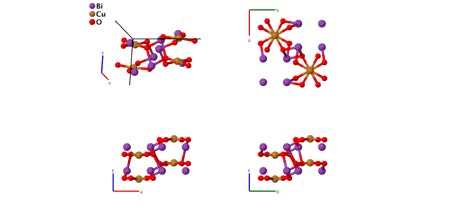}}
\end{figure}
\vspace*{-0.5cm}
\renewcommand{\arraystretch}{1.5}
\begin{equation*}
  \begin{array}{>{$\hspace{-0.15cm}}l<{$}>{$}p{0.5cm}<{$}>{$}p{18.5cm}<{$}}
    \mbox{\large \textbf{Prototype}} &\colon & \ce{CuBi2O4} \\
    \mbox{\large \textbf{\AFLOW\ prototype label}} &\colon & \mbox{A2BC4\_tP28\_130\_f\_c\_g} \\
    \mbox{\large \textbf{\textit{Strukturbericht} designation}} &\colon & \mbox{None} \\
    \mbox{\large \textbf{Pearson symbol}} &\colon & \mbox{tP28} \\
    \mbox{\large \textbf{Space group number}} &\colon & 130 \\
    \mbox{\large \textbf{Space group symbol}} &\colon & P4/ncc \\
    \mbox{\large \textbf{\AFLOW\ prototype command}} &\colon &  \texttt{aflow} \,  \, \texttt{-{}-proto=A2BC4\_tP28\_130\_f\_c\_g } \, \newline \texttt{-{}-params=}{a,c/a,z_{1},x_{2},x_{3},y_{3},z_{3} }
  \end{array}
\end{equation*}
\renewcommand{\arraystretch}{1.0}

\noindent \parbox{1 \linewidth}{
\noindent \hrulefill
\\
\textbf{Simple Tetragonal primitive vectors:} \\
\vspace*{-0.25cm}
\begin{tabular}{cc}
  \begin{tabular}{c}
    \parbox{0.6 \linewidth}{
      \renewcommand{\arraystretch}{1.5}
      \begin{equation*}
        \centering
        \begin{array}{ccc}
              \mathbf{a}_1 & = & a \, \mathbf{\hat{x}} \\
    \mathbf{a}_2 & = & a \, \mathbf{\hat{y}} \\
    \mathbf{a}_3 & = & c \, \mathbf{\hat{z}} \\

        \end{array}
      \end{equation*}
    }
    \renewcommand{\arraystretch}{1.0}
  \end{tabular}
  \begin{tabular}{c}
    \includegraphics[width=0.3\linewidth]{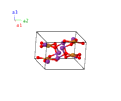} \\
  \end{tabular}
\end{tabular}

}
\vspace*{-0.25cm}

\noindent \hrulefill
\\
\textbf{Basis vectors:}
\vspace*{-0.25cm}
\renewcommand{\arraystretch}{1.5}
\begin{longtabu} to \textwidth{>{\centering $}X[-1,c,c]<{$}>{\centering $}X[-1,c,c]<{$}>{\centering $}X[-1,c,c]<{$}>{\centering $}X[-1,c,c]<{$}>{\centering $}X[-1,c,c]<{$}>{\centering $}X[-1,c,c]<{$}>{\centering $}X[-1,c,c]<{$}}
  & & \mbox{Lattice Coordinates} & & \mbox{Cartesian Coordinates} &\mbox{Wyckoff Position} & \mbox{Atom Type} \\  
  \mathbf{B}_{1} & = & \frac{1}{4} \, \mathbf{a}_{1} + \frac{1}{4} \, \mathbf{a}_{2} + z_{1} \, \mathbf{a}_{3} & = & \frac{1}{4}a \, \mathbf{\hat{x}} + \frac{1}{4}a \, \mathbf{\hat{y}} + z_{1}c \, \mathbf{\hat{z}} & \left(4c\right) & \mbox{Cu} \\ 
\mathbf{B}_{2} & = & \frac{3}{4} \, \mathbf{a}_{1} + \frac{3}{4} \, \mathbf{a}_{2} + \left(\frac{1}{2} - z_{1}\right) \, \mathbf{a}_{3} & = & \frac{3}{4}a \, \mathbf{\hat{x}} + \frac{3}{4}a \, \mathbf{\hat{y}} + \left(\frac{1}{2} - z_{1}\right)c \, \mathbf{\hat{z}} & \left(4c\right) & \mbox{Cu} \\ 
\mathbf{B}_{3} & = & \frac{3}{4} \, \mathbf{a}_{1} + \frac{3}{4} \, \mathbf{a}_{2}-z_{1} \, \mathbf{a}_{3} & = & \frac{3}{4}a \, \mathbf{\hat{x}} + \frac{3}{4}a \, \mathbf{\hat{y}}-z_{1}c \, \mathbf{\hat{z}} & \left(4c\right) & \mbox{Cu} \\ 
\mathbf{B}_{4} & = & \frac{1}{4} \, \mathbf{a}_{1} + \frac{1}{4} \, \mathbf{a}_{2} + \left(\frac{1}{2} +z_{1}\right) \, \mathbf{a}_{3} & = & \frac{1}{4}a \, \mathbf{\hat{x}} + \frac{1}{4}a \, \mathbf{\hat{y}} + \left(\frac{1}{2} +z_{1}\right)c \, \mathbf{\hat{z}} & \left(4c\right) & \mbox{Cu} \\ 
\mathbf{B}_{5} & = & x_{2} \, \mathbf{a}_{1}-x_{2} \, \mathbf{a}_{2} + \frac{1}{4} \, \mathbf{a}_{3} & = & x_{2}a \, \mathbf{\hat{x}}-x_{2}a \, \mathbf{\hat{y}} + \frac{1}{4}c \, \mathbf{\hat{z}} & \left(8f\right) & \mbox{Bi} \\ 
\mathbf{B}_{6} & = & \left(\frac{1}{2} - x_{2}\right) \, \mathbf{a}_{1} + \left(\frac{1}{2} - x_{2}\right) \, \mathbf{a}_{2} + \frac{1}{4} \, \mathbf{a}_{3} & = & \left(\frac{1}{2} - x_{2}\right)a \, \mathbf{\hat{x}} + \left(\frac{1}{2} - x_{2}\right)a \, \mathbf{\hat{y}} + \frac{1}{4}c \, \mathbf{\hat{z}} & \left(8f\right) & \mbox{Bi} \\ 
\mathbf{B}_{7} & = & \left(\frac{1}{2} +x_{2}\right) \, \mathbf{a}_{1} + x_{2} \, \mathbf{a}_{2} + \frac{1}{4} \, \mathbf{a}_{3} & = & \left(\frac{1}{2} +x_{2}\right)a \, \mathbf{\hat{x}} + x_{2}a \, \mathbf{\hat{y}} + \frac{1}{4}c \, \mathbf{\hat{z}} & \left(8f\right) & \mbox{Bi} \\ 
\mathbf{B}_{8} & = & -x_{2} \, \mathbf{a}_{1} + \left(\frac{1}{2} - x_{2}\right) \, \mathbf{a}_{2} + \frac{1}{4} \, \mathbf{a}_{3} & = & -x_{2}a \, \mathbf{\hat{x}} + \left(\frac{1}{2} - x_{2}\right)a \, \mathbf{\hat{y}} + \frac{1}{4}c \, \mathbf{\hat{z}} & \left(8f\right) & \mbox{Bi} \\ 
\mathbf{B}_{9} & = & -x_{2} \, \mathbf{a}_{1} + x_{2} \, \mathbf{a}_{2} + \frac{3}{4} \, \mathbf{a}_{3} & = & -x_{2}a \, \mathbf{\hat{x}} + x_{2}a \, \mathbf{\hat{y}} + \frac{3}{4}c \, \mathbf{\hat{z}} & \left(8f\right) & \mbox{Bi} \\ 
\mathbf{B}_{10} & = & \left(\frac{1}{2} +x_{2}\right) \, \mathbf{a}_{1} + \left(\frac{1}{2} - x_{2}\right) \, \mathbf{a}_{2} + \frac{3}{4} \, \mathbf{a}_{3} & = & \left(\frac{1}{2} +x_{2}\right)a \, \mathbf{\hat{x}} + \left(\frac{1}{2} - x_{2}\right)a \, \mathbf{\hat{y}} + \frac{3}{4}c \, \mathbf{\hat{z}} & \left(8f\right) & \mbox{Bi} \\ 
\mathbf{B}_{11} & = & \left(\frac{1}{2} - x_{2}\right) \, \mathbf{a}_{1}-x_{2} \, \mathbf{a}_{2} + \frac{3}{4} \, \mathbf{a}_{3} & = & \left(\frac{1}{2} - x_{2}\right)a \, \mathbf{\hat{x}}-x_{2}a \, \mathbf{\hat{y}} + \frac{3}{4}c \, \mathbf{\hat{z}} & \left(8f\right) & \mbox{Bi} \\ 
\mathbf{B}_{12} & = & x_{2} \, \mathbf{a}_{1} + \left(\frac{1}{2} +x_{2}\right) \, \mathbf{a}_{2} + \frac{3}{4} \, \mathbf{a}_{3} & = & x_{2}a \, \mathbf{\hat{x}} + \left(\frac{1}{2} +x_{2}\right)a \, \mathbf{\hat{y}} + \frac{3}{4}c \, \mathbf{\hat{z}} & \left(8f\right) & \mbox{Bi} \\ 
\mathbf{B}_{13} & = & x_{3} \, \mathbf{a}_{1} + y_{3} \, \mathbf{a}_{2} + z_{3} \, \mathbf{a}_{3} & = & x_{3}a \, \mathbf{\hat{x}} + y_{3}a \, \mathbf{\hat{y}} + z_{3}c \, \mathbf{\hat{z}} & \left(16g\right) & \mbox{O} \\ 
\mathbf{B}_{14} & = & \left(\frac{1}{2} - x_{3}\right) \, \mathbf{a}_{1} + \left(\frac{1}{2} - y_{3}\right) \, \mathbf{a}_{2} + z_{3} \, \mathbf{a}_{3} & = & \left(\frac{1}{2} - x_{3}\right)a \, \mathbf{\hat{x}} + \left(\frac{1}{2} - y_{3}\right)a \, \mathbf{\hat{y}} + z_{3}c \, \mathbf{\hat{z}} & \left(16g\right) & \mbox{O} \\ 
\mathbf{B}_{15} & = & \left(\frac{1}{2} - y_{3}\right) \, \mathbf{a}_{1} + x_{3} \, \mathbf{a}_{2} + z_{3} \, \mathbf{a}_{3} & = & \left(\frac{1}{2} - y_{3}\right)a \, \mathbf{\hat{x}} + x_{3}a \, \mathbf{\hat{y}} + z_{3}c \, \mathbf{\hat{z}} & \left(16g\right) & \mbox{O} \\ 
\mathbf{B}_{16} & = & y_{3} \, \mathbf{a}_{1} + \left(\frac{1}{2} - x_{3}\right) \, \mathbf{a}_{2} + z_{3} \, \mathbf{a}_{3} & = & y_{3}a \, \mathbf{\hat{x}} + \left(\frac{1}{2} - x_{3}\right)a \, \mathbf{\hat{y}} + z_{3}c \, \mathbf{\hat{z}} & \left(16g\right) & \mbox{O} \\ 
\mathbf{B}_{17} & = & -x_{3} \, \mathbf{a}_{1} + \left(\frac{1}{2} +y_{3}\right) \, \mathbf{a}_{2} + \left(\frac{1}{2} - z_{3}\right) \, \mathbf{a}_{3} & = & -x_{3}a \, \mathbf{\hat{x}} + \left(\frac{1}{2} +y_{3}\right)a \, \mathbf{\hat{y}} + \left(\frac{1}{2} - z_{3}\right)c \, \mathbf{\hat{z}} & \left(16g\right) & \mbox{O} \\ 
\mathbf{B}_{18} & = & \left(\frac{1}{2} +x_{3}\right) \, \mathbf{a}_{1}-y_{3} \, \mathbf{a}_{2} + \left(\frac{1}{2} - z_{3}\right) \, \mathbf{a}_{3} & = & \left(\frac{1}{2} +x_{3}\right)a \, \mathbf{\hat{x}}-y_{3}a \, \mathbf{\hat{y}} + \left(\frac{1}{2} - z_{3}\right)c \, \mathbf{\hat{z}} & \left(16g\right) & \mbox{O} \\ 
\mathbf{B}_{19} & = & \left(\frac{1}{2} +y_{3}\right) \, \mathbf{a}_{1} + \left(\frac{1}{2} +x_{3}\right) \, \mathbf{a}_{2} + \left(\frac{1}{2} - z_{3}\right) \, \mathbf{a}_{3} & = & \left(\frac{1}{2} +y_{3}\right)a \, \mathbf{\hat{x}} + \left(\frac{1}{2} +x_{3}\right)a \, \mathbf{\hat{y}} + \left(\frac{1}{2} - z_{3}\right)c \, \mathbf{\hat{z}} & \left(16g\right) & \mbox{O} \\ 
\mathbf{B}_{20} & = & -y_{3} \, \mathbf{a}_{1}-x_{3} \, \mathbf{a}_{2} + \left(\frac{1}{2} - z_{3}\right) \, \mathbf{a}_{3} & = & -y_{3}a \, \mathbf{\hat{x}}-x_{3}a \, \mathbf{\hat{y}} + \left(\frac{1}{2} - z_{3}\right)c \, \mathbf{\hat{z}} & \left(16g\right) & \mbox{O} \\ 
\mathbf{B}_{21} & = & -x_{3} \, \mathbf{a}_{1}-y_{3} \, \mathbf{a}_{2}-z_{3} \, \mathbf{a}_{3} & = & -x_{3}a \, \mathbf{\hat{x}}-y_{3}a \, \mathbf{\hat{y}}-z_{3}c \, \mathbf{\hat{z}} & \left(16g\right) & \mbox{O} \\ 
\mathbf{B}_{22} & = & \left(\frac{1}{2} +x_{3}\right) \, \mathbf{a}_{1} + \left(\frac{1}{2} +y_{3}\right) \, \mathbf{a}_{2}-z_{3} \, \mathbf{a}_{3} & = & \left(\frac{1}{2} +x_{3}\right)a \, \mathbf{\hat{x}} + \left(\frac{1}{2} +y_{3}\right)a \, \mathbf{\hat{y}}-z_{3}c \, \mathbf{\hat{z}} & \left(16g\right) & \mbox{O} \\ 
\mathbf{B}_{23} & = & \left(\frac{1}{2} +y_{3}\right) \, \mathbf{a}_{1}-x_{3} \, \mathbf{a}_{2}-z_{3} \, \mathbf{a}_{3} & = & \left(\frac{1}{2} +y_{3}\right)a \, \mathbf{\hat{x}}-x_{3}a \, \mathbf{\hat{y}}-z_{3}c \, \mathbf{\hat{z}} & \left(16g\right) & \mbox{O} \\ 
\mathbf{B}_{24} & = & -y_{3} \, \mathbf{a}_{1} + \left(\frac{1}{2} +x_{3}\right) \, \mathbf{a}_{2}-z_{3} \, \mathbf{a}_{3} & = & -y_{3}a \, \mathbf{\hat{x}} + \left(\frac{1}{2} +x_{3}\right)a \, \mathbf{\hat{y}}-z_{3}c \, \mathbf{\hat{z}} & \left(16g\right) & \mbox{O} \\ 
\mathbf{B}_{25} & = & x_{3} \, \mathbf{a}_{1} + \left(\frac{1}{2} - y_{3}\right) \, \mathbf{a}_{2} + \left(\frac{1}{2} +z_{3}\right) \, \mathbf{a}_{3} & = & x_{3}a \, \mathbf{\hat{x}} + \left(\frac{1}{2} - y_{3}\right)a \, \mathbf{\hat{y}} + \left(\frac{1}{2} +z_{3}\right)c \, \mathbf{\hat{z}} & \left(16g\right) & \mbox{O} \\ 
\mathbf{B}_{26} & = & \left(\frac{1}{2} - x_{3}\right) \, \mathbf{a}_{1} + y_{3} \, \mathbf{a}_{2} + \left(\frac{1}{2} +z_{3}\right) \, \mathbf{a}_{3} & = & \left(\frac{1}{2} - x_{3}\right)a \, \mathbf{\hat{x}} + y_{3}a \, \mathbf{\hat{y}} + \left(\frac{1}{2} +z_{3}\right)c \, \mathbf{\hat{z}} & \left(16g\right) & \mbox{O} \\ 
\mathbf{B}_{27} & = & \left(\frac{1}{2} - y_{3}\right) \, \mathbf{a}_{1} + \left(\frac{1}{2} - x_{3}\right) \, \mathbf{a}_{2} + \left(\frac{1}{2} +z_{3}\right) \, \mathbf{a}_{3} & = & \left(\frac{1}{2} - y_{3}\right)a \, \mathbf{\hat{x}} + \left(\frac{1}{2} - x_{3}\right)a \, \mathbf{\hat{y}} + \left(\frac{1}{2} +z_{3}\right)c \, \mathbf{\hat{z}} & \left(16g\right) & \mbox{O} \\ 
\mathbf{B}_{28} & = & y_{3} \, \mathbf{a}_{1} + x_{3} \, \mathbf{a}_{2} + \left(\frac{1}{2} +z_{3}\right) \, \mathbf{a}_{3} & = & y_{3}a \, \mathbf{\hat{x}} + x_{3}a \, \mathbf{\hat{y}} + \left(\frac{1}{2} +z_{3}\right)c \, \mathbf{\hat{z}} & \left(16g\right) & \mbox{O} \\ 
\end{longtabu}
\renewcommand{\arraystretch}{1.0}
\noindent \hrulefill
\\
\textbf{References:}
\vspace*{-0.25cm}
\begin{flushleft}
  - \bibentry{Boivin_Bi2CuO4_BullSocFrMineralCristallogr_1976}. \\
\end{flushleft}
\textbf{Found in:}
\vspace*{-0.25cm}
\begin{flushleft}
  - \bibentry{Villars_PearsonsCrystalData_2013}. \\
\end{flushleft}
\noindent \hrulefill
\\
\textbf{Geometry files:}
\\
\noindent  - CIF: pp. {\hyperref[A2BC4_tP28_130_f_c_g_cif]{\pageref{A2BC4_tP28_130_f_c_g_cif}}} \\
\noindent  - POSCAR: pp. {\hyperref[A2BC4_tP28_130_f_c_g_poscar]{\pageref{A2BC4_tP28_130_f_c_g_poscar}}} \\
\onecolumn
{\phantomsection\label{A5B3_tP32_130_cg_cf}}
\subsection*{\huge \textbf{{\normalfont Ba$_{5}$Si$_{3}$ Structure: A5B3\_tP32\_130\_cg\_cf}}}
\noindent \hrulefill
\vspace*{0.25cm}
\begin{figure}[htp]
  \centering
  \vspace{-1em}
  {\includegraphics[width=1\textwidth]{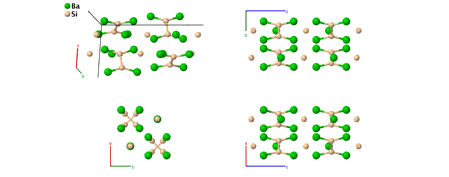}}
\end{figure}
\vspace*{-0.5cm}
\renewcommand{\arraystretch}{1.5}
\begin{equation*}
  \begin{array}{>{$\hspace{-0.15cm}}l<{$}>{$}p{0.5cm}<{$}>{$}p{18.5cm}<{$}}
    \mbox{\large \textbf{Prototype}} &\colon & \ce{Ba$_{5}$Si$_{3}$} \\
    \mbox{\large \textbf{\AFLOW\ prototype label}} &\colon & \mbox{A5B3\_tP32\_130\_cg\_cf} \\
    \mbox{\large \textbf{\textit{Strukturbericht} designation}} &\colon & \mbox{None} \\
    \mbox{\large \textbf{Pearson symbol}} &\colon & \mbox{tP32} \\
    \mbox{\large \textbf{Space group number}} &\colon & 130 \\
    \mbox{\large \textbf{Space group symbol}} &\colon & P4/ncc \\
    \mbox{\large \textbf{\AFLOW\ prototype command}} &\colon &  \texttt{aflow} \,  \, \texttt{-{}-proto=A5B3\_tP32\_130\_cg\_cf } \, \newline \texttt{-{}-params=}{a,c/a,z_{1},z_{2},x_{3},x_{4},y_{4},z_{4} }
  \end{array}
\end{equation*}
\renewcommand{\arraystretch}{1.0}

\noindent \parbox{1 \linewidth}{
\noindent \hrulefill
\\
\textbf{Simple Tetragonal primitive vectors:} \\
\vspace*{-0.25cm}
\begin{tabular}{cc}
  \begin{tabular}{c}
    \parbox{0.6 \linewidth}{
      \renewcommand{\arraystretch}{1.5}
      \begin{equation*}
        \centering
        \begin{array}{ccc}
              \mathbf{a}_1 & = & a \, \mathbf{\hat{x}} \\
    \mathbf{a}_2 & = & a \, \mathbf{\hat{y}} \\
    \mathbf{a}_3 & = & c \, \mathbf{\hat{z}} \\

        \end{array}
      \end{equation*}
    }
    \renewcommand{\arraystretch}{1.0}
  \end{tabular}
  \begin{tabular}{c}
    \includegraphics[width=0.3\linewidth]{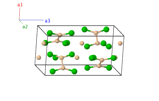} \\
  \end{tabular}
\end{tabular}

}
\vspace*{-0.25cm}

\noindent \hrulefill
\\
\textbf{Basis vectors:}
\vspace*{-0.25cm}
\renewcommand{\arraystretch}{1.5}
\begin{longtabu} to \textwidth{>{\centering $}X[-1,c,c]<{$}>{\centering $}X[-1,c,c]<{$}>{\centering $}X[-1,c,c]<{$}>{\centering $}X[-1,c,c]<{$}>{\centering $}X[-1,c,c]<{$}>{\centering $}X[-1,c,c]<{$}>{\centering $}X[-1,c,c]<{$}}
  & & \mbox{Lattice Coordinates} & & \mbox{Cartesian Coordinates} &\mbox{Wyckoff Position} & \mbox{Atom Type} \\  
  \mathbf{B}_{1} & = & \frac{1}{4} \, \mathbf{a}_{1} + \frac{1}{4} \, \mathbf{a}_{2} + z_{1} \, \mathbf{a}_{3} & = & \frac{1}{4}a \, \mathbf{\hat{x}} + \frac{1}{4}a \, \mathbf{\hat{y}} + z_{1}c \, \mathbf{\hat{z}} & \left(4c\right) & \mbox{Ba I} \\ 
\mathbf{B}_{2} & = & \frac{3}{4} \, \mathbf{a}_{1} + \frac{3}{4} \, \mathbf{a}_{2} + \left(\frac{1}{2} - z_{1}\right) \, \mathbf{a}_{3} & = & \frac{3}{4}a \, \mathbf{\hat{x}} + \frac{3}{4}a \, \mathbf{\hat{y}} + \left(\frac{1}{2} - z_{1}\right)c \, \mathbf{\hat{z}} & \left(4c\right) & \mbox{Ba I} \\ 
\mathbf{B}_{3} & = & \frac{3}{4} \, \mathbf{a}_{1} + \frac{3}{4} \, \mathbf{a}_{2}-z_{1} \, \mathbf{a}_{3} & = & \frac{3}{4}a \, \mathbf{\hat{x}} + \frac{3}{4}a \, \mathbf{\hat{y}}-z_{1}c \, \mathbf{\hat{z}} & \left(4c\right) & \mbox{Ba I} \\ 
\mathbf{B}_{4} & = & \frac{1}{4} \, \mathbf{a}_{1} + \frac{1}{4} \, \mathbf{a}_{2} + \left(\frac{1}{2} +z_{1}\right) \, \mathbf{a}_{3} & = & \frac{1}{4}a \, \mathbf{\hat{x}} + \frac{1}{4}a \, \mathbf{\hat{y}} + \left(\frac{1}{2} +z_{1}\right)c \, \mathbf{\hat{z}} & \left(4c\right) & \mbox{Ba I} \\ 
\mathbf{B}_{5} & = & \frac{1}{4} \, \mathbf{a}_{1} + \frac{1}{4} \, \mathbf{a}_{2} + z_{2} \, \mathbf{a}_{3} & = & \frac{1}{4}a \, \mathbf{\hat{x}} + \frac{1}{4}a \, \mathbf{\hat{y}} + z_{2}c \, \mathbf{\hat{z}} & \left(4c\right) & \mbox{Si I} \\ 
\mathbf{B}_{6} & = & \frac{3}{4} \, \mathbf{a}_{1} + \frac{3}{4} \, \mathbf{a}_{2} + \left(\frac{1}{2} - z_{2}\right) \, \mathbf{a}_{3} & = & \frac{3}{4}a \, \mathbf{\hat{x}} + \frac{3}{4}a \, \mathbf{\hat{y}} + \left(\frac{1}{2} - z_{2}\right)c \, \mathbf{\hat{z}} & \left(4c\right) & \mbox{Si I} \\ 
\mathbf{B}_{7} & = & \frac{3}{4} \, \mathbf{a}_{1} + \frac{3}{4} \, \mathbf{a}_{2}-z_{2} \, \mathbf{a}_{3} & = & \frac{3}{4}a \, \mathbf{\hat{x}} + \frac{3}{4}a \, \mathbf{\hat{y}}-z_{2}c \, \mathbf{\hat{z}} & \left(4c\right) & \mbox{Si I} \\ 
\mathbf{B}_{8} & = & \frac{1}{4} \, \mathbf{a}_{1} + \frac{1}{4} \, \mathbf{a}_{2} + \left(\frac{1}{2} +z_{2}\right) \, \mathbf{a}_{3} & = & \frac{1}{4}a \, \mathbf{\hat{x}} + \frac{1}{4}a \, \mathbf{\hat{y}} + \left(\frac{1}{2} +z_{2}\right)c \, \mathbf{\hat{z}} & \left(4c\right) & \mbox{Si I} \\ 
\mathbf{B}_{9} & = & x_{3} \, \mathbf{a}_{1}-x_{3} \, \mathbf{a}_{2} + \frac{1}{4} \, \mathbf{a}_{3} & = & x_{3}a \, \mathbf{\hat{x}}-x_{3}a \, \mathbf{\hat{y}} + \frac{1}{4}c \, \mathbf{\hat{z}} & \left(8f\right) & \mbox{Si II} \\ 
\mathbf{B}_{10} & = & \left(\frac{1}{2} - x_{3}\right) \, \mathbf{a}_{1} + \left(\frac{1}{2} - x_{3}\right) \, \mathbf{a}_{2} + \frac{1}{4} \, \mathbf{a}_{3} & = & \left(\frac{1}{2} - x_{3}\right)a \, \mathbf{\hat{x}} + \left(\frac{1}{2} - x_{3}\right)a \, \mathbf{\hat{y}} + \frac{1}{4}c \, \mathbf{\hat{z}} & \left(8f\right) & \mbox{Si II} \\ 
\mathbf{B}_{11} & = & \left(\frac{1}{2} +x_{3}\right) \, \mathbf{a}_{1} + x_{3} \, \mathbf{a}_{2} + \frac{1}{4} \, \mathbf{a}_{3} & = & \left(\frac{1}{2} +x_{3}\right)a \, \mathbf{\hat{x}} + x_{3}a \, \mathbf{\hat{y}} + \frac{1}{4}c \, \mathbf{\hat{z}} & \left(8f\right) & \mbox{Si II} \\ 
\mathbf{B}_{12} & = & -x_{3} \, \mathbf{a}_{1} + \left(\frac{1}{2} - x_{3}\right) \, \mathbf{a}_{2} + \frac{1}{4} \, \mathbf{a}_{3} & = & -x_{3}a \, \mathbf{\hat{x}} + \left(\frac{1}{2} - x_{3}\right)a \, \mathbf{\hat{y}} + \frac{1}{4}c \, \mathbf{\hat{z}} & \left(8f\right) & \mbox{Si II} \\ 
\mathbf{B}_{13} & = & -x_{3} \, \mathbf{a}_{1} + x_{3} \, \mathbf{a}_{2} + \frac{3}{4} \, \mathbf{a}_{3} & = & -x_{3}a \, \mathbf{\hat{x}} + x_{3}a \, \mathbf{\hat{y}} + \frac{3}{4}c \, \mathbf{\hat{z}} & \left(8f\right) & \mbox{Si II} \\ 
\mathbf{B}_{14} & = & \left(\frac{1}{2} +x_{3}\right) \, \mathbf{a}_{1} + \left(\frac{1}{2} - x_{3}\right) \, \mathbf{a}_{2} + \frac{3}{4} \, \mathbf{a}_{3} & = & \left(\frac{1}{2} +x_{3}\right)a \, \mathbf{\hat{x}} + \left(\frac{1}{2} - x_{3}\right)a \, \mathbf{\hat{y}} + \frac{3}{4}c \, \mathbf{\hat{z}} & \left(8f\right) & \mbox{Si II} \\ 
\mathbf{B}_{15} & = & \left(\frac{1}{2} - x_{3}\right) \, \mathbf{a}_{1}-x_{3} \, \mathbf{a}_{2} + \frac{3}{4} \, \mathbf{a}_{3} & = & \left(\frac{1}{2} - x_{3}\right)a \, \mathbf{\hat{x}}-x_{3}a \, \mathbf{\hat{y}} + \frac{3}{4}c \, \mathbf{\hat{z}} & \left(8f\right) & \mbox{Si II} \\ 
\mathbf{B}_{16} & = & x_{3} \, \mathbf{a}_{1} + \left(\frac{1}{2} +x_{3}\right) \, \mathbf{a}_{2} + \frac{3}{4} \, \mathbf{a}_{3} & = & x_{3}a \, \mathbf{\hat{x}} + \left(\frac{1}{2} +x_{3}\right)a \, \mathbf{\hat{y}} + \frac{3}{4}c \, \mathbf{\hat{z}} & \left(8f\right) & \mbox{Si II} \\ 
\mathbf{B}_{17} & = & x_{4} \, \mathbf{a}_{1} + y_{4} \, \mathbf{a}_{2} + z_{4} \, \mathbf{a}_{3} & = & x_{4}a \, \mathbf{\hat{x}} + y_{4}a \, \mathbf{\hat{y}} + z_{4}c \, \mathbf{\hat{z}} & \left(16g\right) & \mbox{Ba II} \\ 
\mathbf{B}_{18} & = & \left(\frac{1}{2} - x_{4}\right) \, \mathbf{a}_{1} + \left(\frac{1}{2} - y_{4}\right) \, \mathbf{a}_{2} + z_{4} \, \mathbf{a}_{3} & = & \left(\frac{1}{2} - x_{4}\right)a \, \mathbf{\hat{x}} + \left(\frac{1}{2} - y_{4}\right)a \, \mathbf{\hat{y}} + z_{4}c \, \mathbf{\hat{z}} & \left(16g\right) & \mbox{Ba II} \\ 
\mathbf{B}_{19} & = & \left(\frac{1}{2} - y_{4}\right) \, \mathbf{a}_{1} + x_{4} \, \mathbf{a}_{2} + z_{4} \, \mathbf{a}_{3} & = & \left(\frac{1}{2} - y_{4}\right)a \, \mathbf{\hat{x}} + x_{4}a \, \mathbf{\hat{y}} + z_{4}c \, \mathbf{\hat{z}} & \left(16g\right) & \mbox{Ba II} \\ 
\mathbf{B}_{20} & = & y_{4} \, \mathbf{a}_{1} + \left(\frac{1}{2} - x_{4}\right) \, \mathbf{a}_{2} + z_{4} \, \mathbf{a}_{3} & = & y_{4}a \, \mathbf{\hat{x}} + \left(\frac{1}{2} - x_{4}\right)a \, \mathbf{\hat{y}} + z_{4}c \, \mathbf{\hat{z}} & \left(16g\right) & \mbox{Ba II} \\ 
\mathbf{B}_{21} & = & -x_{4} \, \mathbf{a}_{1} + \left(\frac{1}{2} +y_{4}\right) \, \mathbf{a}_{2} + \left(\frac{1}{2} - z_{4}\right) \, \mathbf{a}_{3} & = & -x_{4}a \, \mathbf{\hat{x}} + \left(\frac{1}{2} +y_{4}\right)a \, \mathbf{\hat{y}} + \left(\frac{1}{2} - z_{4}\right)c \, \mathbf{\hat{z}} & \left(16g\right) & \mbox{Ba II} \\ 
\mathbf{B}_{22} & = & \left(\frac{1}{2} +x_{4}\right) \, \mathbf{a}_{1}-y_{4} \, \mathbf{a}_{2} + \left(\frac{1}{2} - z_{4}\right) \, \mathbf{a}_{3} & = & \left(\frac{1}{2} +x_{4}\right)a \, \mathbf{\hat{x}}-y_{4}a \, \mathbf{\hat{y}} + \left(\frac{1}{2} - z_{4}\right)c \, \mathbf{\hat{z}} & \left(16g\right) & \mbox{Ba II} \\ 
\mathbf{B}_{23} & = & \left(\frac{1}{2} +y_{4}\right) \, \mathbf{a}_{1} + \left(\frac{1}{2} +x_{4}\right) \, \mathbf{a}_{2} + \left(\frac{1}{2} - z_{4}\right) \, \mathbf{a}_{3} & = & \left(\frac{1}{2} +y_{4}\right)a \, \mathbf{\hat{x}} + \left(\frac{1}{2} +x_{4}\right)a \, \mathbf{\hat{y}} + \left(\frac{1}{2} - z_{4}\right)c \, \mathbf{\hat{z}} & \left(16g\right) & \mbox{Ba II} \\ 
\mathbf{B}_{24} & = & -y_{4} \, \mathbf{a}_{1}-x_{4} \, \mathbf{a}_{2} + \left(\frac{1}{2} - z_{4}\right) \, \mathbf{a}_{3} & = & -y_{4}a \, \mathbf{\hat{x}}-x_{4}a \, \mathbf{\hat{y}} + \left(\frac{1}{2} - z_{4}\right)c \, \mathbf{\hat{z}} & \left(16g\right) & \mbox{Ba II} \\ 
\mathbf{B}_{25} & = & -x_{4} \, \mathbf{a}_{1}-y_{4} \, \mathbf{a}_{2}-z_{4} \, \mathbf{a}_{3} & = & -x_{4}a \, \mathbf{\hat{x}}-y_{4}a \, \mathbf{\hat{y}}-z_{4}c \, \mathbf{\hat{z}} & \left(16g\right) & \mbox{Ba II} \\ 
\mathbf{B}_{26} & = & \left(\frac{1}{2} +x_{4}\right) \, \mathbf{a}_{1} + \left(\frac{1}{2} +y_{4}\right) \, \mathbf{a}_{2}-z_{4} \, \mathbf{a}_{3} & = & \left(\frac{1}{2} +x_{4}\right)a \, \mathbf{\hat{x}} + \left(\frac{1}{2} +y_{4}\right)a \, \mathbf{\hat{y}}-z_{4}c \, \mathbf{\hat{z}} & \left(16g\right) & \mbox{Ba II} \\ 
\mathbf{B}_{27} & = & \left(\frac{1}{2} +y_{4}\right) \, \mathbf{a}_{1}-x_{4} \, \mathbf{a}_{2}-z_{4} \, \mathbf{a}_{3} & = & \left(\frac{1}{2} +y_{4}\right)a \, \mathbf{\hat{x}}-x_{4}a \, \mathbf{\hat{y}}-z_{4}c \, \mathbf{\hat{z}} & \left(16g\right) & \mbox{Ba II} \\ 
\mathbf{B}_{28} & = & -y_{4} \, \mathbf{a}_{1} + \left(\frac{1}{2} +x_{4}\right) \, \mathbf{a}_{2}-z_{4} \, \mathbf{a}_{3} & = & -y_{4}a \, \mathbf{\hat{x}} + \left(\frac{1}{2} +x_{4}\right)a \, \mathbf{\hat{y}}-z_{4}c \, \mathbf{\hat{z}} & \left(16g\right) & \mbox{Ba II} \\ 
\mathbf{B}_{29} & = & x_{4} \, \mathbf{a}_{1} + \left(\frac{1}{2} - y_{4}\right) \, \mathbf{a}_{2} + \left(\frac{1}{2} +z_{4}\right) \, \mathbf{a}_{3} & = & x_{4}a \, \mathbf{\hat{x}} + \left(\frac{1}{2} - y_{4}\right)a \, \mathbf{\hat{y}} + \left(\frac{1}{2} +z_{4}\right)c \, \mathbf{\hat{z}} & \left(16g\right) & \mbox{Ba II} \\ 
\mathbf{B}_{30} & = & \left(\frac{1}{2} - x_{4}\right) \, \mathbf{a}_{1} + y_{4} \, \mathbf{a}_{2} + \left(\frac{1}{2} +z_{4}\right) \, \mathbf{a}_{3} & = & \left(\frac{1}{2} - x_{4}\right)a \, \mathbf{\hat{x}} + y_{4}a \, \mathbf{\hat{y}} + \left(\frac{1}{2} +z_{4}\right)c \, \mathbf{\hat{z}} & \left(16g\right) & \mbox{Ba II} \\ 
\mathbf{B}_{31} & = & \left(\frac{1}{2} - y_{4}\right) \, \mathbf{a}_{1} + \left(\frac{1}{2} - x_{4}\right) \, \mathbf{a}_{2} + \left(\frac{1}{2} +z_{4}\right) \, \mathbf{a}_{3} & = & \left(\frac{1}{2} - y_{4}\right)a \, \mathbf{\hat{x}} + \left(\frac{1}{2} - x_{4}\right)a \, \mathbf{\hat{y}} + \left(\frac{1}{2} +z_{4}\right)c \, \mathbf{\hat{z}} & \left(16g\right) & \mbox{Ba II} \\ 
\mathbf{B}_{32} & = & y_{4} \, \mathbf{a}_{1} + x_{4} \, \mathbf{a}_{2} + \left(\frac{1}{2} +z_{4}\right) \, \mathbf{a}_{3} & = & y_{4}a \, \mathbf{\hat{x}} + x_{4}a \, \mathbf{\hat{y}} + \left(\frac{1}{2} +z_{4}\right)c \, \mathbf{\hat{z}} & \left(16g\right) & \mbox{Ba II} \\ 
\end{longtabu}
\renewcommand{\arraystretch}{1.0}
\noindent \hrulefill
\\
\textbf{References:}
\vspace*{-0.25cm}
\begin{flushleft}
  - \bibentry{Nesper_1966}. \\
\end{flushleft}
\noindent \hrulefill
\\
\textbf{Geometry files:}
\\
\noindent  - CIF: pp. {\hyperref[A5B3_tP32_130_cg_cf_cif]{\pageref{A5B3_tP32_130_cg_cf_cif}}} \\
\noindent  - POSCAR: pp. {\hyperref[A5B3_tP32_130_cg_cf_poscar]{\pageref{A5B3_tP32_130_cg_cf_poscar}}} \\
\onecolumn
{\phantomsection\label{A2B2C4D_tP18_132_e_i_o_d}}
\subsection*{\huge \textbf{{\normalfont Rb$_{2}$TiCu$_{2}$S$_{4}$ Structure: A2B2C4D\_tP18\_132\_e\_i\_o\_d}}}
\noindent \hrulefill
\vspace*{0.25cm}
\begin{figure}[htp]
  \centering
  \vspace{-1em}
  {\includegraphics[width=1\textwidth]{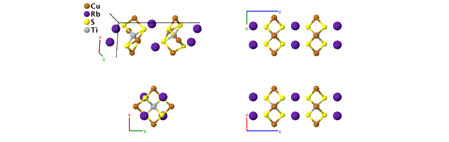}}
\end{figure}
\vspace*{-0.5cm}
\renewcommand{\arraystretch}{1.5}
\begin{equation*}
  \begin{array}{>{$\hspace{-0.15cm}}l<{$}>{$}p{0.5cm}<{$}>{$}p{18.5cm}<{$}}
    \mbox{\large \textbf{Prototype}} &\colon & \ce{Rb2TiCu2Se4} \\
    \mbox{\large \textbf{\AFLOW\ prototype label}} &\colon & \mbox{A2B2C4D\_tP18\_132\_e\_i\_o\_d} \\
    \mbox{\large \textbf{\textit{Strukturbericht} designation}} &\colon & \mbox{None} \\
    \mbox{\large \textbf{Pearson symbol}} &\colon & \mbox{tP18} \\
    \mbox{\large \textbf{Space group number}} &\colon & 132 \\
    \mbox{\large \textbf{Space group symbol}} &\colon & P4_{2}/mcm \\
    \mbox{\large \textbf{\AFLOW\ prototype command}} &\colon &  \texttt{aflow} \,  \, \texttt{-{}-proto=A2B2C4D\_tP18\_132\_e\_i\_o\_d } \, \newline \texttt{-{}-params=}{a,c/a,x_{3},x_{4},z_{4} }
  \end{array}
\end{equation*}
\renewcommand{\arraystretch}{1.0}

\vspace*{-0.25cm}
\noindent \hrulefill
\\
\textbf{ Other compounds with this structure:}
\begin{itemize}
   \item{ Cs$_{2}$TiAg$_{2}$S$_{4}$, Cs$_{2}$TiCu$_{2}$Se$_{4}$   }
\end{itemize}
\vspace*{-0.25cm}
\noindent \hrulefill
\begin{itemize}
  \item{The atomic positions for this structure are not given in the main
text, but are provided as a Crystallographic Information File (\CIF),
\hyperref[http://pubs.acs.org/doi/suppl/10.1021/ic001346d/suppl_file/ic001346d.cif]{ic001346d.cif}.
}
\end{itemize}

\noindent \parbox{1 \linewidth}{
\noindent \hrulefill
\\
\textbf{Simple Tetragonal primitive vectors:} \\
\vspace*{-0.25cm}
\begin{tabular}{cc}
  \begin{tabular}{c}
    \parbox{0.6 \linewidth}{
      \renewcommand{\arraystretch}{1.5}
      \begin{equation*}
        \centering
        \begin{array}{ccc}
              \mathbf{a}_1 & = & a \, \mathbf{\hat{x}} \\
    \mathbf{a}_2 & = & a \, \mathbf{\hat{y}} \\
    \mathbf{a}_3 & = & c \, \mathbf{\hat{z}} \\

        \end{array}
      \end{equation*}
    }
    \renewcommand{\arraystretch}{1.0}
  \end{tabular}
  \begin{tabular}{c}
    \includegraphics[width=0.3\linewidth]{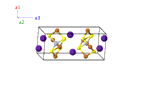} \\
  \end{tabular}
\end{tabular}

}
\vspace*{-0.25cm}

\noindent \hrulefill
\\
\textbf{Basis vectors:}
\vspace*{-0.25cm}
\renewcommand{\arraystretch}{1.5}
\begin{longtabu} to \textwidth{>{\centering $}X[-1,c,c]<{$}>{\centering $}X[-1,c,c]<{$}>{\centering $}X[-1,c,c]<{$}>{\centering $}X[-1,c,c]<{$}>{\centering $}X[-1,c,c]<{$}>{\centering $}X[-1,c,c]<{$}>{\centering $}X[-1,c,c]<{$}}
  & & \mbox{Lattice Coordinates} & & \mbox{Cartesian Coordinates} &\mbox{Wyckoff Position} & \mbox{Atom Type} \\  
  \mathbf{B}_{1} & = & \frac{1}{2} \, \mathbf{a}_{1} + \frac{1}{2} \, \mathbf{a}_{2} + \frac{1}{4} \, \mathbf{a}_{3} & = & \frac{1}{2}a \, \mathbf{\hat{x}} + \frac{1}{2}a \, \mathbf{\hat{y}} + \frac{1}{4}c \, \mathbf{\hat{z}} & \left(2d\right) & \mbox{Ti} \\ 
\mathbf{B}_{2} & = & \frac{1}{2} \, \mathbf{a}_{1} + \frac{1}{2} \, \mathbf{a}_{2} + \frac{3}{4} \, \mathbf{a}_{3} & = & \frac{1}{2}a \, \mathbf{\hat{x}} + \frac{1}{2}a \, \mathbf{\hat{y}} + \frac{3}{4}c \, \mathbf{\hat{z}} & \left(2d\right) & \mbox{Ti} \\ 
\mathbf{B}_{3} & = & \frac{1}{2} \, \mathbf{a}_{2} + \frac{1}{4} \, \mathbf{a}_{3} & = & \frac{1}{2}a \, \mathbf{\hat{y}} + \frac{1}{4}c \, \mathbf{\hat{z}} & \left(4e\right) & \mbox{Cu} \\ 
\mathbf{B}_{4} & = & \frac{1}{2} \, \mathbf{a}_{1} + \frac{3}{4} \, \mathbf{a}_{3} & = & \frac{1}{2}a \, \mathbf{\hat{x}} + \frac{3}{4}c \, \mathbf{\hat{z}} & \left(4e\right) & \mbox{Cu} \\ 
\mathbf{B}_{5} & = & \frac{1}{2} \, \mathbf{a}_{2} + \frac{3}{4} \, \mathbf{a}_{3} & = & \frac{1}{2}a \, \mathbf{\hat{y}} + \frac{3}{4}c \, \mathbf{\hat{z}} & \left(4e\right) & \mbox{Cu} \\ 
\mathbf{B}_{6} & = & \frac{1}{2} \, \mathbf{a}_{1} + \frac{1}{4} \, \mathbf{a}_{3} & = & \frac{1}{2}a \, \mathbf{\hat{x}} + \frac{1}{4}c \, \mathbf{\hat{z}} & \left(4e\right) & \mbox{Cu} \\ 
\mathbf{B}_{7} & = & x_{3} \, \mathbf{a}_{1} + x_{3} \, \mathbf{a}_{2} & = & x_{3}a \, \mathbf{\hat{x}} + x_{3}a \, \mathbf{\hat{y}} & \left(4i\right) & \mbox{Rb} \\ 
\mathbf{B}_{8} & = & -x_{3} \, \mathbf{a}_{1}-x_{3} \, \mathbf{a}_{2} & = & -x_{3}a \, \mathbf{\hat{x}}-x_{3}a \, \mathbf{\hat{y}} & \left(4i\right) & \mbox{Rb} \\ 
\mathbf{B}_{9} & = & -x_{3} \, \mathbf{a}_{1} + x_{3} \, \mathbf{a}_{2} + \frac{1}{2} \, \mathbf{a}_{3} & = & -x_{3}a \, \mathbf{\hat{x}} + x_{3}a \, \mathbf{\hat{y}} + \frac{1}{2}c \, \mathbf{\hat{z}} & \left(4i\right) & \mbox{Rb} \\ 
\mathbf{B}_{10} & = & x_{3} \, \mathbf{a}_{1}-x_{3} \, \mathbf{a}_{2} + \frac{1}{2} \, \mathbf{a}_{3} & = & x_{3}a \, \mathbf{\hat{x}}-x_{3}a \, \mathbf{\hat{y}} + \frac{1}{2}c \, \mathbf{\hat{z}} & \left(4i\right) & \mbox{Rb} \\ 
\mathbf{B}_{11} & = & x_{4} \, \mathbf{a}_{1} + x_{4} \, \mathbf{a}_{2} + z_{4} \, \mathbf{a}_{3} & = & x_{4}a \, \mathbf{\hat{x}} + x_{4}a \, \mathbf{\hat{y}} + z_{4}c \, \mathbf{\hat{z}} & \left(8o\right) & \mbox{S} \\ 
\mathbf{B}_{12} & = & -x_{4} \, \mathbf{a}_{1}-x_{4} \, \mathbf{a}_{2} + z_{4} \, \mathbf{a}_{3} & = & -x_{4}a \, \mathbf{\hat{x}}-x_{4}a \, \mathbf{\hat{y}} + z_{4}c \, \mathbf{\hat{z}} & \left(8o\right) & \mbox{S} \\ 
\mathbf{B}_{13} & = & -x_{4} \, \mathbf{a}_{1} + x_{4} \, \mathbf{a}_{2} + \left(\frac{1}{2} +z_{4}\right) \, \mathbf{a}_{3} & = & -x_{4}a \, \mathbf{\hat{x}} + x_{4}a \, \mathbf{\hat{y}} + \left(\frac{1}{2} +z_{4}\right)c \, \mathbf{\hat{z}} & \left(8o\right) & \mbox{S} \\ 
\mathbf{B}_{14} & = & x_{4} \, \mathbf{a}_{1}-x_{4} \, \mathbf{a}_{2} + \left(\frac{1}{2} +z_{4}\right) \, \mathbf{a}_{3} & = & x_{4}a \, \mathbf{\hat{x}}-x_{4}a \, \mathbf{\hat{y}} + \left(\frac{1}{2} +z_{4}\right)c \, \mathbf{\hat{z}} & \left(8o\right) & \mbox{S} \\ 
\mathbf{B}_{15} & = & -x_{4} \, \mathbf{a}_{1} + x_{4} \, \mathbf{a}_{2} + \left(\frac{1}{2} - z_{4}\right) \, \mathbf{a}_{3} & = & -x_{4}a \, \mathbf{\hat{x}} + x_{4}a \, \mathbf{\hat{y}} + \left(\frac{1}{2} - z_{4}\right)c \, \mathbf{\hat{z}} & \left(8o\right) & \mbox{S} \\ 
\mathbf{B}_{16} & = & x_{4} \, \mathbf{a}_{1}-x_{4} \, \mathbf{a}_{2} + \left(\frac{1}{2} - z_{4}\right) \, \mathbf{a}_{3} & = & x_{4}a \, \mathbf{\hat{x}}-x_{4}a \, \mathbf{\hat{y}} + \left(\frac{1}{2} - z_{4}\right)c \, \mathbf{\hat{z}} & \left(8o\right) & \mbox{S} \\ 
\mathbf{B}_{17} & = & x_{4} \, \mathbf{a}_{1} + x_{4} \, \mathbf{a}_{2}-z_{4} \, \mathbf{a}_{3} & = & x_{4}a \, \mathbf{\hat{x}} + x_{4}a \, \mathbf{\hat{y}}-z_{4}c \, \mathbf{\hat{z}} & \left(8o\right) & \mbox{S} \\ 
\mathbf{B}_{18} & = & -x_{4} \, \mathbf{a}_{1}-x_{4} \, \mathbf{a}_{2}-z_{4} \, \mathbf{a}_{3} & = & -x_{4}a \, \mathbf{\hat{x}}-x_{4}a \, \mathbf{\hat{y}}-z_{4}c \, \mathbf{\hat{z}} & \left(8o\right) & \mbox{S} \\ 
\end{longtabu}
\renewcommand{\arraystretch}{1.0}
\noindent \hrulefill
\\
\textbf{References:}
\vspace*{-0.25cm}
\begin{flushleft}
  - \bibentry{Huang_Inorg_Chem_40_2001}. \\
\end{flushleft}
\noindent \hrulefill
\\
\textbf{Geometry files:}
\\
\noindent  - CIF: pp. {\hyperref[A2B2C4D_tP18_132_e_i_o_d_cif]{\pageref{A2B2C4D_tP18_132_e_i_o_d_cif}}} \\
\noindent  - POSCAR: pp. {\hyperref[A2B2C4D_tP18_132_e_i_o_d_poscar]{\pageref{A2B2C4D_tP18_132_e_i_o_d_poscar}}} \\
\onecolumn
{\phantomsection\label{AB6C_tP16_132_d_io_a}}
\subsection*{\huge \textbf{{\normalfont AgUF$_{6}$ Structure: AB6C\_tP16\_132\_d\_io\_a}}}
\noindent \hrulefill
\vspace*{0.25cm}
\begin{figure}[htp]
  \centering
  \vspace{-1em}
  {\includegraphics[width=1\textwidth]{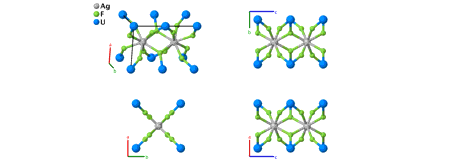}}
\end{figure}
\vspace*{-0.5cm}
\renewcommand{\arraystretch}{1.5}
\begin{equation*}
  \begin{array}{>{$\hspace{-0.15cm}}l<{$}>{$}p{0.5cm}<{$}>{$}p{18.5cm}<{$}}
    \mbox{\large \textbf{Prototype}} &\colon & \ce{AgUF6} \\
    \mbox{\large \textbf{\AFLOW\ prototype label}} &\colon & \mbox{AB6C\_tP16\_132\_d\_io\_a} \\
    \mbox{\large \textbf{\textit{Strukturbericht} designation}} &\colon & \mbox{None} \\
    \mbox{\large \textbf{Pearson symbol}} &\colon & \mbox{tP16} \\
    \mbox{\large \textbf{Space group number}} &\colon & 132 \\
    \mbox{\large \textbf{Space group symbol}} &\colon & P4_{2}/mcm \\
    \mbox{\large \textbf{\AFLOW\ prototype command}} &\colon &  \texttt{aflow} \,  \, \texttt{-{}-proto=AB6C\_tP16\_132\_d\_io\_a } \, \newline \texttt{-{}-params=}{a,c/a,x_{3},x_{4},z_{4} }
  \end{array}
\end{equation*}
\renewcommand{\arraystretch}{1.0}

\noindent \parbox{1 \linewidth}{
\noindent \hrulefill
\\
\textbf{Simple Tetragonal primitive vectors:} \\
\vspace*{-0.25cm}
\begin{tabular}{cc}
  \begin{tabular}{c}
    \parbox{0.6 \linewidth}{
      \renewcommand{\arraystretch}{1.5}
      \begin{equation*}
        \centering
        \begin{array}{ccc}
              \mathbf{a}_1 & = & a \, \mathbf{\hat{x}} \\
    \mathbf{a}_2 & = & a \, \mathbf{\hat{y}} \\
    \mathbf{a}_3 & = & c \, \mathbf{\hat{z}} \\

        \end{array}
      \end{equation*}
    }
    \renewcommand{\arraystretch}{1.0}
  \end{tabular}
  \begin{tabular}{c}
    \includegraphics[width=0.3\linewidth]{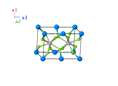} \\
  \end{tabular}
\end{tabular}

}
\vspace*{-0.25cm}

\noindent \hrulefill
\\
\textbf{Basis vectors:}
\vspace*{-0.25cm}
\renewcommand{\arraystretch}{1.5}
\begin{longtabu} to \textwidth{>{\centering $}X[-1,c,c]<{$}>{\centering $}X[-1,c,c]<{$}>{\centering $}X[-1,c,c]<{$}>{\centering $}X[-1,c,c]<{$}>{\centering $}X[-1,c,c]<{$}>{\centering $}X[-1,c,c]<{$}>{\centering $}X[-1,c,c]<{$}}
  & & \mbox{Lattice Coordinates} & & \mbox{Cartesian Coordinates} &\mbox{Wyckoff Position} & \mbox{Atom Type} \\  
  \mathbf{B}_{1} & = & 0 \, \mathbf{a}_{1} + 0 \, \mathbf{a}_{2} + 0 \, \mathbf{a}_{3} & = & 0 \, \mathbf{\hat{x}} + 0 \, \mathbf{\hat{y}} + 0 \, \mathbf{\hat{z}} & \left(2a\right) & \mbox{U} \\ 
\mathbf{B}_{2} & = & \frac{1}{2} \, \mathbf{a}_{3} & = & \frac{1}{2}c \, \mathbf{\hat{z}} & \left(2a\right) & \mbox{U} \\ 
\mathbf{B}_{3} & = & \frac{1}{2} \, \mathbf{a}_{1} + \frac{1}{2} \, \mathbf{a}_{2} + \frac{1}{4} \, \mathbf{a}_{3} & = & \frac{1}{2}a \, \mathbf{\hat{x}} + \frac{1}{2}a \, \mathbf{\hat{y}} + \frac{1}{4}c \, \mathbf{\hat{z}} & \left(2d\right) & \mbox{Ag} \\ 
\mathbf{B}_{4} & = & \frac{1}{2} \, \mathbf{a}_{1} + \frac{1}{2} \, \mathbf{a}_{2} + \frac{3}{4} \, \mathbf{a}_{3} & = & \frac{1}{2}a \, \mathbf{\hat{x}} + \frac{1}{2}a \, \mathbf{\hat{y}} + \frac{3}{4}c \, \mathbf{\hat{z}} & \left(2d\right) & \mbox{Ag} \\ 
\mathbf{B}_{5} & = & x_{3} \, \mathbf{a}_{1} + x_{3} \, \mathbf{a}_{2} & = & x_{3}a \, \mathbf{\hat{x}} + x_{3}a \, \mathbf{\hat{y}} & \left(4i\right) & \mbox{F I} \\ 
\mathbf{B}_{6} & = & -x_{3} \, \mathbf{a}_{1}-x_{3} \, \mathbf{a}_{2} & = & -x_{3}a \, \mathbf{\hat{x}}-x_{3}a \, \mathbf{\hat{y}} & \left(4i\right) & \mbox{F I} \\ 
\mathbf{B}_{7} & = & -x_{3} \, \mathbf{a}_{1} + x_{3} \, \mathbf{a}_{2} + \frac{1}{2} \, \mathbf{a}_{3} & = & -x_{3}a \, \mathbf{\hat{x}} + x_{3}a \, \mathbf{\hat{y}} + \frac{1}{2}c \, \mathbf{\hat{z}} & \left(4i\right) & \mbox{F I} \\ 
\mathbf{B}_{8} & = & x_{3} \, \mathbf{a}_{1}-x_{3} \, \mathbf{a}_{2} + \frac{1}{2} \, \mathbf{a}_{3} & = & x_{3}a \, \mathbf{\hat{x}}-x_{3}a \, \mathbf{\hat{y}} + \frac{1}{2}c \, \mathbf{\hat{z}} & \left(4i\right) & \mbox{F I} \\ 
\mathbf{B}_{9} & = & x_{4} \, \mathbf{a}_{1} + x_{4} \, \mathbf{a}_{2} + z_{4} \, \mathbf{a}_{3} & = & x_{4}a \, \mathbf{\hat{x}} + x_{4}a \, \mathbf{\hat{y}} + z_{4}c \, \mathbf{\hat{z}} & \left(8o\right) & \mbox{F II} \\ 
\mathbf{B}_{10} & = & -x_{4} \, \mathbf{a}_{1}-x_{4} \, \mathbf{a}_{2} + z_{4} \, \mathbf{a}_{3} & = & -x_{4}a \, \mathbf{\hat{x}}-x_{4}a \, \mathbf{\hat{y}} + z_{4}c \, \mathbf{\hat{z}} & \left(8o\right) & \mbox{F II} \\ 
\mathbf{B}_{11} & = & -x_{4} \, \mathbf{a}_{1} + x_{4} \, \mathbf{a}_{2} + \left(\frac{1}{2} +z_{4}\right) \, \mathbf{a}_{3} & = & -x_{4}a \, \mathbf{\hat{x}} + x_{4}a \, \mathbf{\hat{y}} + \left(\frac{1}{2} +z_{4}\right)c \, \mathbf{\hat{z}} & \left(8o\right) & \mbox{F II} \\ 
\mathbf{B}_{12} & = & x_{4} \, \mathbf{a}_{1}-x_{4} \, \mathbf{a}_{2} + \left(\frac{1}{2} +z_{4}\right) \, \mathbf{a}_{3} & = & x_{4}a \, \mathbf{\hat{x}}-x_{4}a \, \mathbf{\hat{y}} + \left(\frac{1}{2} +z_{4}\right)c \, \mathbf{\hat{z}} & \left(8o\right) & \mbox{F II} \\ 
\mathbf{B}_{13} & = & -x_{4} \, \mathbf{a}_{1} + x_{4} \, \mathbf{a}_{2} + \left(\frac{1}{2} - z_{4}\right) \, \mathbf{a}_{3} & = & -x_{4}a \, \mathbf{\hat{x}} + x_{4}a \, \mathbf{\hat{y}} + \left(\frac{1}{2} - z_{4}\right)c \, \mathbf{\hat{z}} & \left(8o\right) & \mbox{F II} \\ 
\mathbf{B}_{14} & = & x_{4} \, \mathbf{a}_{1}-x_{4} \, \mathbf{a}_{2} + \left(\frac{1}{2} - z_{4}\right) \, \mathbf{a}_{3} & = & x_{4}a \, \mathbf{\hat{x}}-x_{4}a \, \mathbf{\hat{y}} + \left(\frac{1}{2} - z_{4}\right)c \, \mathbf{\hat{z}} & \left(8o\right) & \mbox{F II} \\ 
\mathbf{B}_{15} & = & x_{4} \, \mathbf{a}_{1} + x_{4} \, \mathbf{a}_{2}-z_{4} \, \mathbf{a}_{3} & = & x_{4}a \, \mathbf{\hat{x}} + x_{4}a \, \mathbf{\hat{y}}-z_{4}c \, \mathbf{\hat{z}} & \left(8o\right) & \mbox{F II} \\ 
\mathbf{B}_{16} & = & -x_{4} \, \mathbf{a}_{1}-x_{4} \, \mathbf{a}_{2}-z_{4} \, \mathbf{a}_{3} & = & -x_{4}a \, \mathbf{\hat{x}}-x_{4}a \, \mathbf{\hat{y}}-z_{4}c \, \mathbf{\hat{z}} & \left(8o\right) & \mbox{F II} \\ 
\end{longtabu}
\renewcommand{\arraystretch}{1.0}
\noindent \hrulefill
\\
\textbf{References:}
\vspace*{-0.25cm}
\begin{flushleft}
  - \bibentry{charpin_AgUF6_CRHebdSeancesAcadSci_1965}. \\
\end{flushleft}
\textbf{Found in:}
\vspace*{-0.25cm}
\begin{flushleft}
  - \bibentry{Villars_PearsonsCrystalData_2013}. \\
\end{flushleft}
\noindent \hrulefill
\\
\textbf{Geometry files:}
\\
\noindent  - CIF: pp. {\hyperref[AB6C_tP16_132_d_io_a_cif]{\pageref{AB6C_tP16_132_d_io_a_cif}}} \\
\noindent  - POSCAR: pp. {\hyperref[AB6C_tP16_132_d_io_a_poscar]{\pageref{AB6C_tP16_132_d_io_a_poscar}}} \\
\onecolumn
{\phantomsection\label{AB3_tP32_133_h_i2j}}
\subsection*{\huge \textbf{{\normalfont $\beta$-V$_{3}$S Structure: AB3\_tP32\_133\_h\_i2j}}}
\noindent \hrulefill
\vspace*{0.25cm}
\begin{figure}[htp]
  \centering
  \vspace{-1em}
  {\includegraphics[width=1\textwidth]{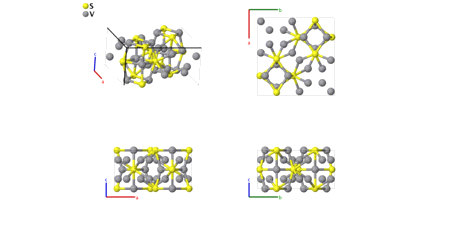}}
\end{figure}
\vspace*{-0.5cm}
\renewcommand{\arraystretch}{1.5}
\begin{equation*}
  \begin{array}{>{$\hspace{-0.15cm}}l<{$}>{$}p{0.5cm}<{$}>{$}p{18.5cm}<{$}}
    \mbox{\large \textbf{Prototype}} &\colon & \ce{$\beta$-V3S} \\
    \mbox{\large \textbf{\AFLOW\ prototype label}} &\colon & \mbox{AB3\_tP32\_133\_h\_i2j} \\
    \mbox{\large \textbf{\textit{Strukturbericht} designation}} &\colon & \mbox{None} \\
    \mbox{\large \textbf{Pearson symbol}} &\colon & \mbox{tP32} \\
    \mbox{\large \textbf{Space group number}} &\colon & 133 \\
    \mbox{\large \textbf{Space group symbol}} &\colon & P4_{2}/nbc \\
    \mbox{\large \textbf{\AFLOW\ prototype command}} &\colon &  \texttt{aflow} \,  \, \texttt{-{}-proto=AB3\_tP32\_133\_h\_i2j } \, \newline \texttt{-{}-params=}{a,c/a,x_{1},x_{2},x_{3},x_{4} }
  \end{array}
\end{equation*}
\renewcommand{\arraystretch}{1.0}

\noindent \parbox{1 \linewidth}{
\noindent \hrulefill
\\
\textbf{Simple Tetragonal primitive vectors:} \\
\vspace*{-0.25cm}
\begin{tabular}{cc}
  \begin{tabular}{c}
    \parbox{0.6 \linewidth}{
      \renewcommand{\arraystretch}{1.5}
      \begin{equation*}
        \centering
        \begin{array}{ccc}
              \mathbf{a}_1 & = & a \, \mathbf{\hat{x}} \\
    \mathbf{a}_2 & = & a \, \mathbf{\hat{y}} \\
    \mathbf{a}_3 & = & c \, \mathbf{\hat{z}} \\

        \end{array}
      \end{equation*}
    }
    \renewcommand{\arraystretch}{1.0}
  \end{tabular}
  \begin{tabular}{c}
    \includegraphics[width=0.3\linewidth]{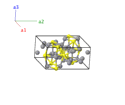} \\
  \end{tabular}
\end{tabular}

}
\vspace*{-0.25cm}

\noindent \hrulefill
\\
\textbf{Basis vectors:}
\vspace*{-0.25cm}
\renewcommand{\arraystretch}{1.5}
\begin{longtabu} to \textwidth{>{\centering $}X[-1,c,c]<{$}>{\centering $}X[-1,c,c]<{$}>{\centering $}X[-1,c,c]<{$}>{\centering $}X[-1,c,c]<{$}>{\centering $}X[-1,c,c]<{$}>{\centering $}X[-1,c,c]<{$}>{\centering $}X[-1,c,c]<{$}}
  & & \mbox{Lattice Coordinates} & & \mbox{Cartesian Coordinates} &\mbox{Wyckoff Position} & \mbox{Atom Type} \\  
  \mathbf{B}_{1} & = & x_{1} \, \mathbf{a}_{1} + \frac{1}{4} \, \mathbf{a}_{2} & = & x_{1}a \, \mathbf{\hat{x}} + \frac{1}{4}a \, \mathbf{\hat{y}} & \left(8h\right) & \mbox{S} \\ 
\mathbf{B}_{2} & = & \left(\frac{1}{2} - x_{1}\right) \, \mathbf{a}_{1} + \frac{1}{4} \, \mathbf{a}_{2} & = & \left(\frac{1}{2} - x_{1}\right)a \, \mathbf{\hat{x}} + \frac{1}{4}a \, \mathbf{\hat{y}} & \left(8h\right) & \mbox{S} \\ 
\mathbf{B}_{3} & = & \frac{1}{4} \, \mathbf{a}_{1} + x_{1} \, \mathbf{a}_{2} + \frac{1}{2} \, \mathbf{a}_{3} & = & \frac{1}{4}a \, \mathbf{\hat{x}} + x_{1}a \, \mathbf{\hat{y}} + \frac{1}{2}c \, \mathbf{\hat{z}} & \left(8h\right) & \mbox{S} \\ 
\mathbf{B}_{4} & = & \frac{1}{4} \, \mathbf{a}_{1} + \left(\frac{1}{2} - x_{1}\right) \, \mathbf{a}_{2} + \frac{1}{2} \, \mathbf{a}_{3} & = & \frac{1}{4}a \, \mathbf{\hat{x}} + \left(\frac{1}{2} - x_{1}\right)a \, \mathbf{\hat{y}} + \frac{1}{2}c \, \mathbf{\hat{z}} & \left(8h\right) & \mbox{S} \\ 
\mathbf{B}_{5} & = & -x_{1} \, \mathbf{a}_{1} + \frac{3}{4} \, \mathbf{a}_{2} & = & -x_{1}a \, \mathbf{\hat{x}} + \frac{3}{4}a \, \mathbf{\hat{y}} & \left(8h\right) & \mbox{S} \\ 
\mathbf{B}_{6} & = & \left(\frac{1}{2} +x_{1}\right) \, \mathbf{a}_{1} + \frac{3}{4} \, \mathbf{a}_{2} & = & \left(\frac{1}{2} +x_{1}\right)a \, \mathbf{\hat{x}} + \frac{3}{4}a \, \mathbf{\hat{y}} & \left(8h\right) & \mbox{S} \\ 
\mathbf{B}_{7} & = & \frac{3}{4} \, \mathbf{a}_{1}-x_{1} \, \mathbf{a}_{2} + \frac{1}{2} \, \mathbf{a}_{3} & = & \frac{3}{4}a \, \mathbf{\hat{x}}-x_{1}a \, \mathbf{\hat{y}} + \frac{1}{2}c \, \mathbf{\hat{z}} & \left(8h\right) & \mbox{S} \\ 
\mathbf{B}_{8} & = & \frac{3}{4} \, \mathbf{a}_{1} + \left(\frac{1}{2} +x_{1}\right) \, \mathbf{a}_{2} + \frac{1}{2} \, \mathbf{a}_{3} & = & \frac{3}{4}a \, \mathbf{\hat{x}} + \left(\frac{1}{2} +x_{1}\right)a \, \mathbf{\hat{y}} + \frac{1}{2}c \, \mathbf{\hat{z}} & \left(8h\right) & \mbox{S} \\ 
\mathbf{B}_{9} & = & x_{2} \, \mathbf{a}_{1} + \frac{1}{4} \, \mathbf{a}_{2} + \frac{1}{2} \, \mathbf{a}_{3} & = & x_{2}a \, \mathbf{\hat{x}} + \frac{1}{4}a \, \mathbf{\hat{y}} + \frac{1}{2}c \, \mathbf{\hat{z}} & \left(8i\right) & \mbox{V I} \\ 
\mathbf{B}_{10} & = & \left(\frac{1}{2} - x_{2}\right) \, \mathbf{a}_{1} + \frac{1}{4} \, \mathbf{a}_{2} + \frac{1}{2} \, \mathbf{a}_{3} & = & \left(\frac{1}{2} - x_{2}\right)a \, \mathbf{\hat{x}} + \frac{1}{4}a \, \mathbf{\hat{y}} + \frac{1}{2}c \, \mathbf{\hat{z}} & \left(8i\right) & \mbox{V I} \\ 
\mathbf{B}_{11} & = & \frac{1}{4} \, \mathbf{a}_{1} + x_{2} \, \mathbf{a}_{2} & = & \frac{1}{4}a \, \mathbf{\hat{x}} + x_{2}a \, \mathbf{\hat{y}} & \left(8i\right) & \mbox{V I} \\ 
\mathbf{B}_{12} & = & \frac{1}{4} \, \mathbf{a}_{1} + \left(\frac{1}{2} - x_{2}\right) \, \mathbf{a}_{2} & = & \frac{1}{4}a \, \mathbf{\hat{x}} + \left(\frac{1}{2} - x_{2}\right)a \, \mathbf{\hat{y}} & \left(8i\right) & \mbox{V I} \\ 
\mathbf{B}_{13} & = & -x_{2} \, \mathbf{a}_{1} + \frac{3}{4} \, \mathbf{a}_{2} + \frac{1}{2} \, \mathbf{a}_{3} & = & -x_{2}a \, \mathbf{\hat{x}} + \frac{3}{4}a \, \mathbf{\hat{y}} + \frac{1}{2}c \, \mathbf{\hat{z}} & \left(8i\right) & \mbox{V I} \\ 
\mathbf{B}_{14} & = & \left(\frac{1}{2} +x_{2}\right) \, \mathbf{a}_{1} + \frac{3}{4} \, \mathbf{a}_{2} + \frac{1}{2} \, \mathbf{a}_{3} & = & \left(\frac{1}{2} +x_{2}\right)a \, \mathbf{\hat{x}} + \frac{3}{4}a \, \mathbf{\hat{y}} + \frac{1}{2}c \, \mathbf{\hat{z}} & \left(8i\right) & \mbox{V I} \\ 
\mathbf{B}_{15} & = & \frac{3}{4} \, \mathbf{a}_{1}-x_{2} \, \mathbf{a}_{2} & = & \frac{3}{4}a \, \mathbf{\hat{x}}-x_{2}a \, \mathbf{\hat{y}} & \left(8i\right) & \mbox{V I} \\ 
\mathbf{B}_{16} & = & \frac{3}{4} \, \mathbf{a}_{1} + \left(\frac{1}{2} +x_{2}\right) \, \mathbf{a}_{2} & = & \frac{3}{4}a \, \mathbf{\hat{x}} + \left(\frac{1}{2} +x_{2}\right)a \, \mathbf{\hat{y}} & \left(8i\right) & \mbox{V I} \\ 
\mathbf{B}_{17} & = & x_{3} \, \mathbf{a}_{1} + x_{3} \, \mathbf{a}_{2} + \frac{1}{4} \, \mathbf{a}_{3} & = & x_{3}a \, \mathbf{\hat{x}} + x_{3}a \, \mathbf{\hat{y}} + \frac{1}{4}c \, \mathbf{\hat{z}} & \left(8j\right) & \mbox{V II} \\ 
\mathbf{B}_{18} & = & \left(\frac{1}{2} - x_{3}\right) \, \mathbf{a}_{1} + \left(\frac{1}{2} - x_{3}\right) \, \mathbf{a}_{2} + \frac{1}{4} \, \mathbf{a}_{3} & = & \left(\frac{1}{2} - x_{3}\right)a \, \mathbf{\hat{x}} + \left(\frac{1}{2} - x_{3}\right)a \, \mathbf{\hat{y}} + \frac{1}{4}c \, \mathbf{\hat{z}} & \left(8j\right) & \mbox{V II} \\ 
\mathbf{B}_{19} & = & \left(\frac{1}{2} - x_{3}\right) \, \mathbf{a}_{1} + x_{3} \, \mathbf{a}_{2} + \frac{3}{4} \, \mathbf{a}_{3} & = & \left(\frac{1}{2} - x_{3}\right)a \, \mathbf{\hat{x}} + x_{3}a \, \mathbf{\hat{y}} + \frac{3}{4}c \, \mathbf{\hat{z}} & \left(8j\right) & \mbox{V II} \\ 
\mathbf{B}_{20} & = & x_{3} \, \mathbf{a}_{1} + \left(\frac{1}{2} - x_{3}\right) \, \mathbf{a}_{2} + \frac{3}{4} \, \mathbf{a}_{3} & = & x_{3}a \, \mathbf{\hat{x}} + \left(\frac{1}{2} - x_{3}\right)a \, \mathbf{\hat{y}} + \frac{3}{4}c \, \mathbf{\hat{z}} & \left(8j\right) & \mbox{V II} \\ 
\mathbf{B}_{21} & = & -x_{3} \, \mathbf{a}_{1}-x_{3} \, \mathbf{a}_{2} + \frac{3}{4} \, \mathbf{a}_{3} & = & -x_{3}a \, \mathbf{\hat{x}}-x_{3}a \, \mathbf{\hat{y}} + \frac{3}{4}c \, \mathbf{\hat{z}} & \left(8j\right) & \mbox{V II} \\ 
\mathbf{B}_{22} & = & \left(\frac{1}{2} +x_{3}\right) \, \mathbf{a}_{1} + \left(\frac{1}{2} +x_{3}\right) \, \mathbf{a}_{2} + \frac{3}{4} \, \mathbf{a}_{3} & = & \left(\frac{1}{2} +x_{3}\right)a \, \mathbf{\hat{x}} + \left(\frac{1}{2} +x_{3}\right)a \, \mathbf{\hat{y}} + \frac{3}{4}c \, \mathbf{\hat{z}} & \left(8j\right) & \mbox{V II} \\ 
\mathbf{B}_{23} & = & \left(\frac{1}{2} +x_{3}\right) \, \mathbf{a}_{1}-x_{3} \, \mathbf{a}_{2} + \frac{1}{4} \, \mathbf{a}_{3} & = & \left(\frac{1}{2} +x_{3}\right)a \, \mathbf{\hat{x}}-x_{3}a \, \mathbf{\hat{y}} + \frac{1}{4}c \, \mathbf{\hat{z}} & \left(8j\right) & \mbox{V II} \\ 
\mathbf{B}_{24} & = & -x_{3} \, \mathbf{a}_{1} + \left(\frac{1}{2} +x_{3}\right) \, \mathbf{a}_{2} + \frac{1}{4} \, \mathbf{a}_{3} & = & -x_{3}a \, \mathbf{\hat{x}} + \left(\frac{1}{2} +x_{3}\right)a \, \mathbf{\hat{y}} + \frac{1}{4}c \, \mathbf{\hat{z}} & \left(8j\right) & \mbox{V II} \\ 
\mathbf{B}_{25} & = & x_{4} \, \mathbf{a}_{1} + x_{4} \, \mathbf{a}_{2} + \frac{1}{4} \, \mathbf{a}_{3} & = & x_{4}a \, \mathbf{\hat{x}} + x_{4}a \, \mathbf{\hat{y}} + \frac{1}{4}c \, \mathbf{\hat{z}} & \left(8j\right) & \mbox{V III} \\ 
\mathbf{B}_{26} & = & \left(\frac{1}{2} - x_{4}\right) \, \mathbf{a}_{1} + \left(\frac{1}{2} - x_{4}\right) \, \mathbf{a}_{2} + \frac{1}{4} \, \mathbf{a}_{3} & = & \left(\frac{1}{2} - x_{4}\right)a \, \mathbf{\hat{x}} + \left(\frac{1}{2} - x_{4}\right)a \, \mathbf{\hat{y}} + \frac{1}{4}c \, \mathbf{\hat{z}} & \left(8j\right) & \mbox{V III} \\ 
\mathbf{B}_{27} & = & \left(\frac{1}{2} - x_{4}\right) \, \mathbf{a}_{1} + x_{4} \, \mathbf{a}_{2} + \frac{3}{4} \, \mathbf{a}_{3} & = & \left(\frac{1}{2} - x_{4}\right)a \, \mathbf{\hat{x}} + x_{4}a \, \mathbf{\hat{y}} + \frac{3}{4}c \, \mathbf{\hat{z}} & \left(8j\right) & \mbox{V III} \\ 
\mathbf{B}_{28} & = & x_{4} \, \mathbf{a}_{1} + \left(\frac{1}{2} - x_{4}\right) \, \mathbf{a}_{2} + \frac{3}{4} \, \mathbf{a}_{3} & = & x_{4}a \, \mathbf{\hat{x}} + \left(\frac{1}{2} - x_{4}\right)a \, \mathbf{\hat{y}} + \frac{3}{4}c \, \mathbf{\hat{z}} & \left(8j\right) & \mbox{V III} \\ 
\mathbf{B}_{29} & = & -x_{4} \, \mathbf{a}_{1}-x_{4} \, \mathbf{a}_{2} + \frac{3}{4} \, \mathbf{a}_{3} & = & -x_{4}a \, \mathbf{\hat{x}}-x_{4}a \, \mathbf{\hat{y}} + \frac{3}{4}c \, \mathbf{\hat{z}} & \left(8j\right) & \mbox{V III} \\ 
\mathbf{B}_{30} & = & \left(\frac{1}{2} +x_{4}\right) \, \mathbf{a}_{1} + \left(\frac{1}{2} +x_{4}\right) \, \mathbf{a}_{2} + \frac{3}{4} \, \mathbf{a}_{3} & = & \left(\frac{1}{2} +x_{4}\right)a \, \mathbf{\hat{x}} + \left(\frac{1}{2} +x_{4}\right)a \, \mathbf{\hat{y}} + \frac{3}{4}c \, \mathbf{\hat{z}} & \left(8j\right) & \mbox{V III} \\ 
\mathbf{B}_{31} & = & \left(\frac{1}{2} +x_{4}\right) \, \mathbf{a}_{1}-x_{4} \, \mathbf{a}_{2} + \frac{1}{4} \, \mathbf{a}_{3} & = & \left(\frac{1}{2} +x_{4}\right)a \, \mathbf{\hat{x}}-x_{4}a \, \mathbf{\hat{y}} + \frac{1}{4}c \, \mathbf{\hat{z}} & \left(8j\right) & \mbox{V III} \\ 
\mathbf{B}_{32} & = & -x_{4} \, \mathbf{a}_{1} + \left(\frac{1}{2} +x_{4}\right) \, \mathbf{a}_{2} + \frac{1}{4} \, \mathbf{a}_{3} & = & -x_{4}a \, \mathbf{\hat{x}} + \left(\frac{1}{2} +x_{4}\right)a \, \mathbf{\hat{y}} + \frac{1}{4}c \, \mathbf{\hat{z}} & \left(8j\right) & \mbox{V III} \\ 
\end{longtabu}
\renewcommand{\arraystretch}{1.0}
\noindent \hrulefill
\\
\textbf{References:}
\vspace*{-0.25cm}
\begin{flushleft}
  - \bibentry{Pedersen_V3S_ActaCrystallogr_1959}. \\
\end{flushleft}
\textbf{Found in:}
\vspace*{-0.25cm}
\begin{flushleft}
  - \bibentry{Villars_PearsonsCrystalData_2013}. \\
\end{flushleft}
\noindent \hrulefill
\\
\textbf{Geometry files:}
\\
\noindent  - CIF: pp. {\hyperref[AB3_tP32_133_h_i2j_cif]{\pageref{AB3_tP32_133_h_i2j_cif}}} \\
\noindent  - POSCAR: pp. {\hyperref[AB3_tP32_133_h_i2j_poscar]{\pageref{AB3_tP32_133_h_i2j_poscar}}} \\
\onecolumn
{\phantomsection\label{A2B_tP24_135_gh_h}}
\subsection*{\huge \textbf{{\normalfont \begin{raggedleft}Downeyite (SeO$_{2}$, $C47$) Structure: \end{raggedleft} \\ A2B\_tP24\_135\_gh\_h}}}
\noindent \hrulefill
\vspace*{0.25cm}
\begin{figure}[htp]
  \centering
  \vspace{-1em}
  {\includegraphics[width=1\textwidth]{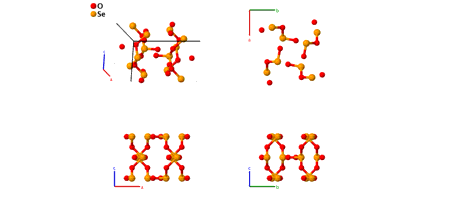}}
\end{figure}
\vspace*{-0.5cm}
\renewcommand{\arraystretch}{1.5}
\begin{equation*}
  \begin{array}{>{$\hspace{-0.15cm}}l<{$}>{$}p{0.5cm}<{$}>{$}p{18.5cm}<{$}}
    \mbox{\large \textbf{Prototype}} &\colon & \ce{SeO2} \\
    \mbox{\large \textbf{\AFLOW\ prototype label}} &\colon & \mbox{A2B\_tP24\_135\_gh\_h} \\
    \mbox{\large \textbf{\textit{Strukturbericht} designation}} &\colon & \mbox{$C47$} \\
    \mbox{\large \textbf{Pearson symbol}} &\colon & \mbox{tP24} \\
    \mbox{\large \textbf{Space group number}} &\colon & 135 \\
    \mbox{\large \textbf{Space group symbol}} &\colon & P4_{2}/mbc \\
    \mbox{\large \textbf{\AFLOW\ prototype command}} &\colon &  \texttt{aflow} \,  \, \texttt{-{}-proto=A2B\_tP24\_135\_gh\_h } \, \newline \texttt{-{}-params=}{a,c/a,x_{1},x_{2},y_{2},x_{3},y_{3} }
  \end{array}
\end{equation*}
\renewcommand{\arraystretch}{1.0}

\vspace*{-0.25cm}
\noindent \hrulefill
\begin{itemize}
  \item{Data for this structure was taken at 139~K.
}
\end{itemize}

\noindent \parbox{1 \linewidth}{
\noindent \hrulefill
\\
\textbf{Simple Tetragonal primitive vectors:} \\
\vspace*{-0.25cm}
\begin{tabular}{cc}
  \begin{tabular}{c}
    \parbox{0.6 \linewidth}{
      \renewcommand{\arraystretch}{1.5}
      \begin{equation*}
        \centering
        \begin{array}{ccc}
              \mathbf{a}_1 & = & a \, \mathbf{\hat{x}} \\
    \mathbf{a}_2 & = & a \, \mathbf{\hat{y}} \\
    \mathbf{a}_3 & = & c \, \mathbf{\hat{z}} \\

        \end{array}
      \end{equation*}
    }
    \renewcommand{\arraystretch}{1.0}
  \end{tabular}
  \begin{tabular}{c}
    \includegraphics[width=0.3\linewidth]{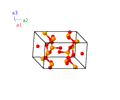} \\
  \end{tabular}
\end{tabular}

}
\vspace*{-0.25cm}

\noindent \hrulefill
\\
\textbf{Basis vectors:}
\vspace*{-0.25cm}
\renewcommand{\arraystretch}{1.5}
\begin{longtabu} to \textwidth{>{\centering $}X[-1,c,c]<{$}>{\centering $}X[-1,c,c]<{$}>{\centering $}X[-1,c,c]<{$}>{\centering $}X[-1,c,c]<{$}>{\centering $}X[-1,c,c]<{$}>{\centering $}X[-1,c,c]<{$}>{\centering $}X[-1,c,c]<{$}}
  & & \mbox{Lattice Coordinates} & & \mbox{Cartesian Coordinates} &\mbox{Wyckoff Position} & \mbox{Atom Type} \\  
  \mathbf{B}_{1} & = & x_{1} \, \mathbf{a}_{1} + \left(\frac{1}{2} +x_{1}\right) \, \mathbf{a}_{2} + \frac{1}{4} \, \mathbf{a}_{3} & = & x_{1}a \, \mathbf{\hat{x}} + \left(\frac{1}{2} +x_{1}\right)a \, \mathbf{\hat{y}} + \frac{1}{4}c \, \mathbf{\hat{z}} & \left(8g\right) & \mbox{O I} \\ 
\mathbf{B}_{2} & = & -x_{1} \, \mathbf{a}_{1} + \left(\frac{1}{2} - x_{1}\right) \, \mathbf{a}_{2} + \frac{1}{4} \, \mathbf{a}_{3} & = & -x_{1}a \, \mathbf{\hat{x}} + \left(\frac{1}{2} - x_{1}\right)a \, \mathbf{\hat{y}} + \frac{1}{4}c \, \mathbf{\hat{z}} & \left(8g\right) & \mbox{O I} \\ 
\mathbf{B}_{3} & = & \left(\frac{1}{2} - x_{1}\right) \, \mathbf{a}_{1} + x_{1} \, \mathbf{a}_{2} + \frac{3}{4} \, \mathbf{a}_{3} & = & \left(\frac{1}{2} - x_{1}\right)a \, \mathbf{\hat{x}} + x_{1}a \, \mathbf{\hat{y}} + \frac{3}{4}c \, \mathbf{\hat{z}} & \left(8g\right) & \mbox{O I} \\ 
\mathbf{B}_{4} & = & \left(\frac{1}{2} +x_{1}\right) \, \mathbf{a}_{1}-x_{1} \, \mathbf{a}_{2} + \frac{3}{4} \, \mathbf{a}_{3} & = & \left(\frac{1}{2} +x_{1}\right)a \, \mathbf{\hat{x}}-x_{1}a \, \mathbf{\hat{y}} + \frac{3}{4}c \, \mathbf{\hat{z}} & \left(8g\right) & \mbox{O I} \\ 
\mathbf{B}_{5} & = & -x_{1} \, \mathbf{a}_{1} + \left(\frac{1}{2} - x_{1}\right) \, \mathbf{a}_{2} + \frac{3}{4} \, \mathbf{a}_{3} & = & -x_{1}a \, \mathbf{\hat{x}} + \left(\frac{1}{2} - x_{1}\right)a \, \mathbf{\hat{y}} + \frac{3}{4}c \, \mathbf{\hat{z}} & \left(8g\right) & \mbox{O I} \\ 
\mathbf{B}_{6} & = & x_{1} \, \mathbf{a}_{1} + \left(\frac{1}{2} +x_{1}\right) \, \mathbf{a}_{2} + \frac{3}{4} \, \mathbf{a}_{3} & = & x_{1}a \, \mathbf{\hat{x}} + \left(\frac{1}{2} +x_{1}\right)a \, \mathbf{\hat{y}} + \frac{3}{4}c \, \mathbf{\hat{z}} & \left(8g\right) & \mbox{O I} \\ 
\mathbf{B}_{7} & = & \left(\frac{1}{2} +x_{1}\right) \, \mathbf{a}_{1}-x_{1} \, \mathbf{a}_{2} + \frac{1}{4} \, \mathbf{a}_{3} & = & \left(\frac{1}{2} +x_{1}\right)a \, \mathbf{\hat{x}}-x_{1}a \, \mathbf{\hat{y}} + \frac{1}{4}c \, \mathbf{\hat{z}} & \left(8g\right) & \mbox{O I} \\ 
\mathbf{B}_{8} & = & \left(\frac{1}{2} - x_{1}\right) \, \mathbf{a}_{1} + x_{1} \, \mathbf{a}_{2} + \frac{1}{4} \, \mathbf{a}_{3} & = & \left(\frac{1}{2} - x_{1}\right)a \, \mathbf{\hat{x}} + x_{1}a \, \mathbf{\hat{y}} + \frac{1}{4}c \, \mathbf{\hat{z}} & \left(8g\right) & \mbox{O I} \\ 
\mathbf{B}_{9} & = & x_{2} \, \mathbf{a}_{1} + y_{2} \, \mathbf{a}_{2} & = & x_{2}a \, \mathbf{\hat{x}} + y_{2}a \, \mathbf{\hat{y}} & \left(8h\right) & \mbox{O II} \\ 
\mathbf{B}_{10} & = & -x_{2} \, \mathbf{a}_{1}-y_{2} \, \mathbf{a}_{2} & = & -x_{2}a \, \mathbf{\hat{x}}-y_{2}a \, \mathbf{\hat{y}} & \left(8h\right) & \mbox{O II} \\ 
\mathbf{B}_{11} & = & -y_{2} \, \mathbf{a}_{1} + x_{2} \, \mathbf{a}_{2} + \frac{1}{2} \, \mathbf{a}_{3} & = & -y_{2}a \, \mathbf{\hat{x}} + x_{2}a \, \mathbf{\hat{y}} + \frac{1}{2}c \, \mathbf{\hat{z}} & \left(8h\right) & \mbox{O II} \\ 
\mathbf{B}_{12} & = & y_{2} \, \mathbf{a}_{1}-x_{2} \, \mathbf{a}_{2} + \frac{1}{2} \, \mathbf{a}_{3} & = & y_{2}a \, \mathbf{\hat{x}}-x_{2}a \, \mathbf{\hat{y}} + \frac{1}{2}c \, \mathbf{\hat{z}} & \left(8h\right) & \mbox{O II} \\ 
\mathbf{B}_{13} & = & \left(\frac{1}{2} - x_{2}\right) \, \mathbf{a}_{1} + \left(\frac{1}{2} +y_{2}\right) \, \mathbf{a}_{2} & = & \left(\frac{1}{2} - x_{2}\right)a \, \mathbf{\hat{x}} + \left(\frac{1}{2} +y_{2}\right)a \, \mathbf{\hat{y}} & \left(8h\right) & \mbox{O II} \\ 
\mathbf{B}_{14} & = & \left(\frac{1}{2} +x_{2}\right) \, \mathbf{a}_{1} + \left(\frac{1}{2} - y_{2}\right) \, \mathbf{a}_{2} & = & \left(\frac{1}{2} +x_{2}\right)a \, \mathbf{\hat{x}} + \left(\frac{1}{2} - y_{2}\right)a \, \mathbf{\hat{y}} & \left(8h\right) & \mbox{O II} \\ 
\mathbf{B}_{15} & = & \left(\frac{1}{2} +y_{2}\right) \, \mathbf{a}_{1} + \left(\frac{1}{2} +x_{2}\right) \, \mathbf{a}_{2} + \frac{1}{2} \, \mathbf{a}_{3} & = & \left(\frac{1}{2} +y_{2}\right)a \, \mathbf{\hat{x}} + \left(\frac{1}{2} +x_{2}\right)a \, \mathbf{\hat{y}} + \frac{1}{2}c \, \mathbf{\hat{z}} & \left(8h\right) & \mbox{O II} \\ 
\mathbf{B}_{16} & = & \left(\frac{1}{2} - y_{2}\right) \, \mathbf{a}_{1} + \left(\frac{1}{2} - x_{2}\right) \, \mathbf{a}_{2} + \frac{1}{2} \, \mathbf{a}_{3} & = & \left(\frac{1}{2} - y_{2}\right)a \, \mathbf{\hat{x}} + \left(\frac{1}{2} - x_{2}\right)a \, \mathbf{\hat{y}} + \frac{1}{2}c \, \mathbf{\hat{z}} & \left(8h\right) & \mbox{O II} \\ 
\mathbf{B}_{17} & = & x_{3} \, \mathbf{a}_{1} + y_{3} \, \mathbf{a}_{2} & = & x_{3}a \, \mathbf{\hat{x}} + y_{3}a \, \mathbf{\hat{y}} & \left(8h\right) & \mbox{Se} \\ 
\mathbf{B}_{18} & = & -x_{3} \, \mathbf{a}_{1}-y_{3} \, \mathbf{a}_{2} & = & -x_{3}a \, \mathbf{\hat{x}}-y_{3}a \, \mathbf{\hat{y}} & \left(8h\right) & \mbox{Se} \\ 
\mathbf{B}_{19} & = & -y_{3} \, \mathbf{a}_{1} + x_{3} \, \mathbf{a}_{2} + \frac{1}{2} \, \mathbf{a}_{3} & = & -y_{3}a \, \mathbf{\hat{x}} + x_{3}a \, \mathbf{\hat{y}} + \frac{1}{2}c \, \mathbf{\hat{z}} & \left(8h\right) & \mbox{Se} \\ 
\mathbf{B}_{20} & = & y_{3} \, \mathbf{a}_{1}-x_{3} \, \mathbf{a}_{2} + \frac{1}{2} \, \mathbf{a}_{3} & = & y_{3}a \, \mathbf{\hat{x}}-x_{3}a \, \mathbf{\hat{y}} + \frac{1}{2}c \, \mathbf{\hat{z}} & \left(8h\right) & \mbox{Se} \\ 
\mathbf{B}_{21} & = & \left(\frac{1}{2} - x_{3}\right) \, \mathbf{a}_{1} + \left(\frac{1}{2} +y_{3}\right) \, \mathbf{a}_{2} & = & \left(\frac{1}{2} - x_{3}\right)a \, \mathbf{\hat{x}} + \left(\frac{1}{2} +y_{3}\right)a \, \mathbf{\hat{y}} & \left(8h\right) & \mbox{Se} \\ 
\mathbf{B}_{22} & = & \left(\frac{1}{2} +x_{3}\right) \, \mathbf{a}_{1} + \left(\frac{1}{2} - y_{3}\right) \, \mathbf{a}_{2} & = & \left(\frac{1}{2} +x_{3}\right)a \, \mathbf{\hat{x}} + \left(\frac{1}{2} - y_{3}\right)a \, \mathbf{\hat{y}} & \left(8h\right) & \mbox{Se} \\ 
\mathbf{B}_{23} & = & \left(\frac{1}{2} +y_{3}\right) \, \mathbf{a}_{1} + \left(\frac{1}{2} +x_{3}\right) \, \mathbf{a}_{2} + \frac{1}{2} \, \mathbf{a}_{3} & = & \left(\frac{1}{2} +y_{3}\right)a \, \mathbf{\hat{x}} + \left(\frac{1}{2} +x_{3}\right)a \, \mathbf{\hat{y}} + \frac{1}{2}c \, \mathbf{\hat{z}} & \left(8h\right) & \mbox{Se} \\ 
\mathbf{B}_{24} & = & \left(\frac{1}{2} - y_{3}\right) \, \mathbf{a}_{1} + \left(\frac{1}{2} - x_{3}\right) \, \mathbf{a}_{2} + \frac{1}{2} \, \mathbf{a}_{3} & = & \left(\frac{1}{2} - y_{3}\right)a \, \mathbf{\hat{x}} + \left(\frac{1}{2} - x_{3}\right)a \, \mathbf{\hat{y}} + \frac{1}{2}c \, \mathbf{\hat{z}} & \left(8h\right) & \mbox{Se} \\ 
\end{longtabu}
\renewcommand{\arraystretch}{1.0}
\noindent \hrulefill
\\
\textbf{References:}
\vspace*{-0.25cm}
\begin{flushleft}
  - \bibentry{Stahl_Z_Kristall_202_1992}. \\
\end{flushleft}
\textbf{Found in:}
\vspace*{-0.25cm}
\begin{flushleft}
  - \bibentry{Downs_Am_Min_88_2003}. \\
\end{flushleft}
\noindent \hrulefill
\\
\textbf{Geometry files:}
\\
\noindent  - CIF: pp. {\hyperref[A2B_tP24_135_gh_h_cif]{\pageref{A2B_tP24_135_gh_h_cif}}} \\
\noindent  - POSCAR: pp. {\hyperref[A2B_tP24_135_gh_h_poscar]{\pageref{A2B_tP24_135_gh_h_poscar}}} \\
\onecolumn
{\phantomsection\label{A4B2C_tP28_135_gh_h_d}}
\subsection*{\huge \textbf{{\normalfont ZnSb$_{2}$O$_{4}$ Structure: A4B2C\_tP28\_135\_gh\_h\_d}}}
\noindent \hrulefill
\vspace*{0.25cm}
\begin{figure}[htp]
  \centering
  \vspace{-1em}
  {\includegraphics[width=1\textwidth]{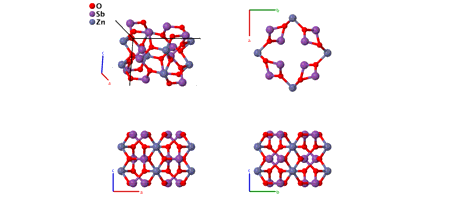}}
\end{figure}
\vspace*{-0.5cm}
\renewcommand{\arraystretch}{1.5}
\begin{equation*}
  \begin{array}{>{$\hspace{-0.15cm}}l<{$}>{$}p{0.5cm}<{$}>{$}p{18.5cm}<{$}}
    \mbox{\large \textbf{Prototype}} &\colon & \ce{ZnSb2O4} \\
    \mbox{\large \textbf{\AFLOW\ prototype label}} &\colon & \mbox{A4B2C\_tP28\_135\_gh\_h\_d} \\
    \mbox{\large \textbf{\textit{Strukturbericht} designation}} &\colon & \mbox{None} \\
    \mbox{\large \textbf{Pearson symbol}} &\colon & \mbox{tP28} \\
    \mbox{\large \textbf{Space group number}} &\colon & 135 \\
    \mbox{\large \textbf{Space group symbol}} &\colon & P4_{2}/mbc \\
    \mbox{\large \textbf{\AFLOW\ prototype command}} &\colon &  \texttt{aflow} \,  \, \texttt{-{}-proto=A4B2C\_tP28\_135\_gh\_h\_d } \, \newline \texttt{-{}-params=}{a,c/a,x_{2},x_{3},y_{3},x_{4},y_{4} }
  \end{array}
\end{equation*}
\renewcommand{\arraystretch}{1.0}

\noindent \parbox{1 \linewidth}{
\noindent \hrulefill
\\
\textbf{Simple Tetragonal primitive vectors:} \\
\vspace*{-0.25cm}
\begin{tabular}{cc}
  \begin{tabular}{c}
    \parbox{0.6 \linewidth}{
      \renewcommand{\arraystretch}{1.5}
      \begin{equation*}
        \centering
        \begin{array}{ccc}
              \mathbf{a}_1 & = & a \, \mathbf{\hat{x}} \\
    \mathbf{a}_2 & = & a \, \mathbf{\hat{y}} \\
    \mathbf{a}_3 & = & c \, \mathbf{\hat{z}} \\

        \end{array}
      \end{equation*}
    }
    \renewcommand{\arraystretch}{1.0}
  \end{tabular}
  \begin{tabular}{c}
    \includegraphics[width=0.3\linewidth]{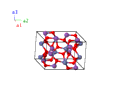} \\
  \end{tabular}
\end{tabular}

}
\vspace*{-0.25cm}

\noindent \hrulefill
\\
\textbf{Basis vectors:}
\vspace*{-0.25cm}
\renewcommand{\arraystretch}{1.5}
\begin{longtabu} to \textwidth{>{\centering $}X[-1,c,c]<{$}>{\centering $}X[-1,c,c]<{$}>{\centering $}X[-1,c,c]<{$}>{\centering $}X[-1,c,c]<{$}>{\centering $}X[-1,c,c]<{$}>{\centering $}X[-1,c,c]<{$}>{\centering $}X[-1,c,c]<{$}}
  & & \mbox{Lattice Coordinates} & & \mbox{Cartesian Coordinates} &\mbox{Wyckoff Position} & \mbox{Atom Type} \\  
  \mathbf{B}_{1} & = & \frac{1}{2} \, \mathbf{a}_{2} + \frac{1}{4} \, \mathbf{a}_{3} & = & \frac{1}{2}a \, \mathbf{\hat{y}} + \frac{1}{4}c \, \mathbf{\hat{z}} & \left(4d\right) & \mbox{Zn} \\ 
\mathbf{B}_{2} & = & \frac{1}{2} \, \mathbf{a}_{1} + \frac{3}{4} \, \mathbf{a}_{3} & = & \frac{1}{2}a \, \mathbf{\hat{x}} + \frac{3}{4}c \, \mathbf{\hat{z}} & \left(4d\right) & \mbox{Zn} \\ 
\mathbf{B}_{3} & = & \frac{1}{2} \, \mathbf{a}_{2} + \frac{3}{4} \, \mathbf{a}_{3} & = & \frac{1}{2}a \, \mathbf{\hat{y}} + \frac{3}{4}c \, \mathbf{\hat{z}} & \left(4d\right) & \mbox{Zn} \\ 
\mathbf{B}_{4} & = & \frac{1}{2} \, \mathbf{a}_{1} + \frac{1}{4} \, \mathbf{a}_{3} & = & \frac{1}{2}a \, \mathbf{\hat{x}} + \frac{1}{4}c \, \mathbf{\hat{z}} & \left(4d\right) & \mbox{Zn} \\ 
\mathbf{B}_{5} & = & x_{2} \, \mathbf{a}_{1} + \left(\frac{1}{2} +x_{2}\right) \, \mathbf{a}_{2} + \frac{1}{4} \, \mathbf{a}_{3} & = & x_{2}a \, \mathbf{\hat{x}} + \left(\frac{1}{2} +x_{2}\right)a \, \mathbf{\hat{y}} + \frac{1}{4}c \, \mathbf{\hat{z}} & \left(8g\right) & \mbox{O I} \\ 
\mathbf{B}_{6} & = & -x_{2} \, \mathbf{a}_{1} + \left(\frac{1}{2} - x_{2}\right) \, \mathbf{a}_{2} + \frac{1}{4} \, \mathbf{a}_{3} & = & -x_{2}a \, \mathbf{\hat{x}} + \left(\frac{1}{2} - x_{2}\right)a \, \mathbf{\hat{y}} + \frac{1}{4}c \, \mathbf{\hat{z}} & \left(8g\right) & \mbox{O I} \\ 
\mathbf{B}_{7} & = & \left(\frac{1}{2} - x_{2}\right) \, \mathbf{a}_{1} + x_{2} \, \mathbf{a}_{2} + \frac{3}{4} \, \mathbf{a}_{3} & = & \left(\frac{1}{2} - x_{2}\right)a \, \mathbf{\hat{x}} + x_{2}a \, \mathbf{\hat{y}} + \frac{3}{4}c \, \mathbf{\hat{z}} & \left(8g\right) & \mbox{O I} \\ 
\mathbf{B}_{8} & = & \left(\frac{1}{2} +x_{2}\right) \, \mathbf{a}_{1}-x_{2} \, \mathbf{a}_{2} + \frac{3}{4} \, \mathbf{a}_{3} & = & \left(\frac{1}{2} +x_{2}\right)a \, \mathbf{\hat{x}}-x_{2}a \, \mathbf{\hat{y}} + \frac{3}{4}c \, \mathbf{\hat{z}} & \left(8g\right) & \mbox{O I} \\ 
\mathbf{B}_{9} & = & -x_{2} \, \mathbf{a}_{1} + \left(\frac{1}{2} - x_{2}\right) \, \mathbf{a}_{2} + \frac{3}{4} \, \mathbf{a}_{3} & = & -x_{2}a \, \mathbf{\hat{x}} + \left(\frac{1}{2} - x_{2}\right)a \, \mathbf{\hat{y}} + \frac{3}{4}c \, \mathbf{\hat{z}} & \left(8g\right) & \mbox{O I} \\ 
\mathbf{B}_{10} & = & x_{2} \, \mathbf{a}_{1} + \left(\frac{1}{2} +x_{2}\right) \, \mathbf{a}_{2} + \frac{3}{4} \, \mathbf{a}_{3} & = & x_{2}a \, \mathbf{\hat{x}} + \left(\frac{1}{2} +x_{2}\right)a \, \mathbf{\hat{y}} + \frac{3}{4}c \, \mathbf{\hat{z}} & \left(8g\right) & \mbox{O I} \\ 
\mathbf{B}_{11} & = & \left(\frac{1}{2} +x_{2}\right) \, \mathbf{a}_{1}-x_{2} \, \mathbf{a}_{2} + \frac{1}{4} \, \mathbf{a}_{3} & = & \left(\frac{1}{2} +x_{2}\right)a \, \mathbf{\hat{x}}-x_{2}a \, \mathbf{\hat{y}} + \frac{1}{4}c \, \mathbf{\hat{z}} & \left(8g\right) & \mbox{O I} \\ 
\mathbf{B}_{12} & = & \left(\frac{1}{2} - x_{2}\right) \, \mathbf{a}_{1} + x_{2} \, \mathbf{a}_{2} + \frac{1}{4} \, \mathbf{a}_{3} & = & \left(\frac{1}{2} - x_{2}\right)a \, \mathbf{\hat{x}} + x_{2}a \, \mathbf{\hat{y}} + \frac{1}{4}c \, \mathbf{\hat{z}} & \left(8g\right) & \mbox{O I} \\ 
\mathbf{B}_{13} & = & x_{3} \, \mathbf{a}_{1} + y_{3} \, \mathbf{a}_{2} & = & x_{3}a \, \mathbf{\hat{x}} + y_{3}a \, \mathbf{\hat{y}} & \left(8h\right) & \mbox{O II} \\ 
\mathbf{B}_{14} & = & -x_{3} \, \mathbf{a}_{1}-y_{3} \, \mathbf{a}_{2} & = & -x_{3}a \, \mathbf{\hat{x}}-y_{3}a \, \mathbf{\hat{y}} & \left(8h\right) & \mbox{O II} \\ 
\mathbf{B}_{15} & = & -y_{3} \, \mathbf{a}_{1} + x_{3} \, \mathbf{a}_{2} + \frac{1}{2} \, \mathbf{a}_{3} & = & -y_{3}a \, \mathbf{\hat{x}} + x_{3}a \, \mathbf{\hat{y}} + \frac{1}{2}c \, \mathbf{\hat{z}} & \left(8h\right) & \mbox{O II} \\ 
\mathbf{B}_{16} & = & y_{3} \, \mathbf{a}_{1}-x_{3} \, \mathbf{a}_{2} + \frac{1}{2} \, \mathbf{a}_{3} & = & y_{3}a \, \mathbf{\hat{x}}-x_{3}a \, \mathbf{\hat{y}} + \frac{1}{2}c \, \mathbf{\hat{z}} & \left(8h\right) & \mbox{O II} \\ 
\mathbf{B}_{17} & = & \left(\frac{1}{2} - x_{3}\right) \, \mathbf{a}_{1} + \left(\frac{1}{2} +y_{3}\right) \, \mathbf{a}_{2} & = & \left(\frac{1}{2} - x_{3}\right)a \, \mathbf{\hat{x}} + \left(\frac{1}{2} +y_{3}\right)a \, \mathbf{\hat{y}} & \left(8h\right) & \mbox{O II} \\ 
\mathbf{B}_{18} & = & \left(\frac{1}{2} +x_{3}\right) \, \mathbf{a}_{1} + \left(\frac{1}{2} - y_{3}\right) \, \mathbf{a}_{2} & = & \left(\frac{1}{2} +x_{3}\right)a \, \mathbf{\hat{x}} + \left(\frac{1}{2} - y_{3}\right)a \, \mathbf{\hat{y}} & \left(8h\right) & \mbox{O II} \\ 
\mathbf{B}_{19} & = & \left(\frac{1}{2} +y_{3}\right) \, \mathbf{a}_{1} + \left(\frac{1}{2} +x_{3}\right) \, \mathbf{a}_{2} + \frac{1}{2} \, \mathbf{a}_{3} & = & \left(\frac{1}{2} +y_{3}\right)a \, \mathbf{\hat{x}} + \left(\frac{1}{2} +x_{3}\right)a \, \mathbf{\hat{y}} + \frac{1}{2}c \, \mathbf{\hat{z}} & \left(8h\right) & \mbox{O II} \\ 
\mathbf{B}_{20} & = & \left(\frac{1}{2} - y_{3}\right) \, \mathbf{a}_{1} + \left(\frac{1}{2} - x_{3}\right) \, \mathbf{a}_{2} + \frac{1}{2} \, \mathbf{a}_{3} & = & \left(\frac{1}{2} - y_{3}\right)a \, \mathbf{\hat{x}} + \left(\frac{1}{2} - x_{3}\right)a \, \mathbf{\hat{y}} + \frac{1}{2}c \, \mathbf{\hat{z}} & \left(8h\right) & \mbox{O II} \\ 
\mathbf{B}_{21} & = & x_{4} \, \mathbf{a}_{1} + y_{4} \, \mathbf{a}_{2} & = & x_{4}a \, \mathbf{\hat{x}} + y_{4}a \, \mathbf{\hat{y}} & \left(8h\right) & \mbox{Sb} \\ 
\mathbf{B}_{22} & = & -x_{4} \, \mathbf{a}_{1}-y_{4} \, \mathbf{a}_{2} & = & -x_{4}a \, \mathbf{\hat{x}}-y_{4}a \, \mathbf{\hat{y}} & \left(8h\right) & \mbox{Sb} \\ 
\mathbf{B}_{23} & = & -y_{4} \, \mathbf{a}_{1} + x_{4} \, \mathbf{a}_{2} + \frac{1}{2} \, \mathbf{a}_{3} & = & -y_{4}a \, \mathbf{\hat{x}} + x_{4}a \, \mathbf{\hat{y}} + \frac{1}{2}c \, \mathbf{\hat{z}} & \left(8h\right) & \mbox{Sb} \\ 
\mathbf{B}_{24} & = & y_{4} \, \mathbf{a}_{1}-x_{4} \, \mathbf{a}_{2} + \frac{1}{2} \, \mathbf{a}_{3} & = & y_{4}a \, \mathbf{\hat{x}}-x_{4}a \, \mathbf{\hat{y}} + \frac{1}{2}c \, \mathbf{\hat{z}} & \left(8h\right) & \mbox{Sb} \\ 
\mathbf{B}_{25} & = & \left(\frac{1}{2} - x_{4}\right) \, \mathbf{a}_{1} + \left(\frac{1}{2} +y_{4}\right) \, \mathbf{a}_{2} & = & \left(\frac{1}{2} - x_{4}\right)a \, \mathbf{\hat{x}} + \left(\frac{1}{2} +y_{4}\right)a \, \mathbf{\hat{y}} & \left(8h\right) & \mbox{Sb} \\ 
\mathbf{B}_{26} & = & \left(\frac{1}{2} +x_{4}\right) \, \mathbf{a}_{1} + \left(\frac{1}{2} - y_{4}\right) \, \mathbf{a}_{2} & = & \left(\frac{1}{2} +x_{4}\right)a \, \mathbf{\hat{x}} + \left(\frac{1}{2} - y_{4}\right)a \, \mathbf{\hat{y}} & \left(8h\right) & \mbox{Sb} \\ 
\mathbf{B}_{27} & = & \left(\frac{1}{2} +y_{4}\right) \, \mathbf{a}_{1} + \left(\frac{1}{2} +x_{4}\right) \, \mathbf{a}_{2} + \frac{1}{2} \, \mathbf{a}_{3} & = & \left(\frac{1}{2} +y_{4}\right)a \, \mathbf{\hat{x}} + \left(\frac{1}{2} +x_{4}\right)a \, \mathbf{\hat{y}} + \frac{1}{2}c \, \mathbf{\hat{z}} & \left(8h\right) & \mbox{Sb} \\ 
\mathbf{B}_{28} & = & \left(\frac{1}{2} - y_{4}\right) \, \mathbf{a}_{1} + \left(\frac{1}{2} - x_{4}\right) \, \mathbf{a}_{2} + \frac{1}{2} \, \mathbf{a}_{3} & = & \left(\frac{1}{2} - y_{4}\right)a \, \mathbf{\hat{x}} + \left(\frac{1}{2} - x_{4}\right)a \, \mathbf{\hat{y}} + \frac{1}{2}c \, \mathbf{\hat{z}} & \left(8h\right) & \mbox{Sb} \\ 
\end{longtabu}
\renewcommand{\arraystretch}{1.0}
\noindent \hrulefill
\\
\textbf{References:}
\vspace*{-0.25cm}
\begin{flushleft}
  - \bibentry{Staahl_ZnSb2O4_1943}. \\
\end{flushleft}
\textbf{Found in:}
\vspace*{-0.25cm}
\begin{flushleft}
  - \bibentry{Villars_PearsonsCrystalData_2013}. \\
\end{flushleft}
\noindent \hrulefill
\\
\textbf{Geometry files:}
\\
\noindent  - CIF: pp. {\hyperref[A4B2C_tP28_135_gh_h_d_cif]{\pageref{A4B2C_tP28_135_gh_h_d_cif}}} \\
\noindent  - POSCAR: pp. {\hyperref[A4B2C_tP28_135_gh_h_d_poscar]{\pageref{A4B2C_tP28_135_gh_h_d_poscar}}} \\
\onecolumn
{\phantomsection\label{A2B3_tP40_137_cdf_3g}}
\subsection*{\huge \textbf{{\normalfont Zn$_{3}$P$_{2}$ ($D5_{9}$) Structure: A2B3\_tP40\_137\_cdf\_3g}}}
\noindent \hrulefill
\vspace*{0.25cm}
\begin{figure}[htp]
  \centering
  \vspace{-1em}
  {\includegraphics[width=1\textwidth]{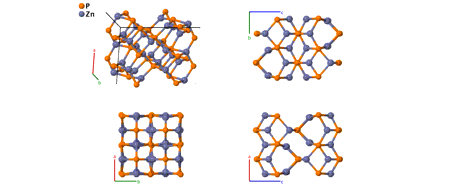}}
\end{figure}
\vspace*{-0.5cm}
\renewcommand{\arraystretch}{1.5}
\begin{equation*}
  \begin{array}{>{$\hspace{-0.15cm}}l<{$}>{$}p{0.5cm}<{$}>{$}p{18.5cm}<{$}}
    \mbox{\large \textbf{Prototype}} &\colon & \ce{Zn3P2} \\
    \mbox{\large \textbf{\AFLOW\ prototype label}} &\colon & \mbox{A2B3\_tP40\_137\_cdf\_3g} \\
    \mbox{\large \textbf{\textit{Strukturbericht} designation}} &\colon & \mbox{$D5_{9}$} \\
    \mbox{\large \textbf{Pearson symbol}} &\colon & \mbox{tP40} \\
    \mbox{\large \textbf{Space group number}} &\colon & 137 \\
    \mbox{\large \textbf{Space group symbol}} &\colon & P4_{2}/nmc \\
    \mbox{\large \textbf{\AFLOW\ prototype command}} &\colon &  \texttt{aflow} \,  \, \texttt{-{}-proto=A2B3\_tP40\_137\_cdf\_3g } \, \newline \texttt{-{}-params=}{a,c/a,z_{1},z_{2},x_{3},y_{4},z_{4},y_{5},z_{5},y_{6},z_{6} }
  \end{array}
\end{equation*}
\renewcommand{\arraystretch}{1.0}

\vspace*{-0.25cm}
\noindent \hrulefill
\\
\textbf{ Other compounds with this structure:}
\begin{itemize}
   \item{ $\alpha$-As$_{2}$Cd$_{3}$, $\alpha$-As$_{2}$Zn$_{3}$,  Cd$_{3}$P$_{2}$  }
\end{itemize}
\vspace*{-0.25cm}
\noindent \hrulefill
\begin{itemize}
  \item{(Stackelberg, 1935) gives the atomic positions in the first setting of
space group $P4_2/nmc$ \#137.  
We have changed this to the second setting, placing the origin of the 
system at the inversion site.
On page 803, (Pearson, 1958) gives the space group as $P4_2/mmc$ 
\#131, but it is correctly given as $P4_2/nmc$ on page 111.
}
\end{itemize}

\noindent \parbox{1 \linewidth}{
\noindent \hrulefill
\\
\textbf{Simple Tetragonal primitive vectors:} \\
\vspace*{-0.25cm}
\begin{tabular}{cc}
  \begin{tabular}{c}
    \parbox{0.6 \linewidth}{
      \renewcommand{\arraystretch}{1.5}
      \begin{equation*}
        \centering
        \begin{array}{ccc}
              \mathbf{a}_1 & = & a \, \mathbf{\hat{x}} \\
    \mathbf{a}_2 & = & a \, \mathbf{\hat{y}} \\
    \mathbf{a}_3 & = & c \, \mathbf{\hat{z}} \\

        \end{array}
      \end{equation*}
    }
    \renewcommand{\arraystretch}{1.0}
  \end{tabular}
  \begin{tabular}{c}
    \includegraphics[width=0.3\linewidth]{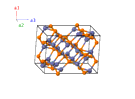} \\
  \end{tabular}
\end{tabular}

}
\vspace*{-0.25cm}

\noindent \hrulefill
\\
\textbf{Basis vectors:}
\vspace*{-0.25cm}
\renewcommand{\arraystretch}{1.5}
\begin{longtabu} to \textwidth{>{\centering $}X[-1,c,c]<{$}>{\centering $}X[-1,c,c]<{$}>{\centering $}X[-1,c,c]<{$}>{\centering $}X[-1,c,c]<{$}>{\centering $}X[-1,c,c]<{$}>{\centering $}X[-1,c,c]<{$}>{\centering $}X[-1,c,c]<{$}}
  & & \mbox{Lattice Coordinates} & & \mbox{Cartesian Coordinates} &\mbox{Wyckoff Position} & \mbox{Atom Type} \\  
  \mathbf{B}_{1} & = & \frac{3}{4} \, \mathbf{a}_{1} + \frac{1}{4} \, \mathbf{a}_{2} + z_{1} \, \mathbf{a}_{3} & = & \frac{3}{4}a \, \mathbf{\hat{x}} + \frac{1}{4}a \, \mathbf{\hat{y}} + z_{1}c \, \mathbf{\hat{z}} & \left(4c\right) & \mbox{P I} \\ 
\mathbf{B}_{2} & = & \frac{1}{4} \, \mathbf{a}_{1} + \frac{3}{4} \, \mathbf{a}_{2} + \left(\frac{1}{2} +z_{1}\right) \, \mathbf{a}_{3} & = & \frac{1}{4}a \, \mathbf{\hat{x}} + \frac{3}{4}a \, \mathbf{\hat{y}} + \left(\frac{1}{2} +z_{1}\right)c \, \mathbf{\hat{z}} & \left(4c\right) & \mbox{P I} \\ 
\mathbf{B}_{3} & = & \frac{1}{4} \, \mathbf{a}_{1} + \frac{3}{4} \, \mathbf{a}_{2}-z_{1} \, \mathbf{a}_{3} & = & \frac{1}{4}a \, \mathbf{\hat{x}} + \frac{3}{4}a \, \mathbf{\hat{y}}-z_{1}c \, \mathbf{\hat{z}} & \left(4c\right) & \mbox{P I} \\ 
\mathbf{B}_{4} & = & \frac{3}{4} \, \mathbf{a}_{1} + \frac{1}{4} \, \mathbf{a}_{2} + \left(\frac{1}{2} - z_{1}\right) \, \mathbf{a}_{3} & = & \frac{3}{4}a \, \mathbf{\hat{x}} + \frac{1}{4}a \, \mathbf{\hat{y}} + \left(\frac{1}{2} - z_{1}\right)c \, \mathbf{\hat{z}} & \left(4c\right) & \mbox{P I} \\ 
\mathbf{B}_{5} & = & \frac{1}{4} \, \mathbf{a}_{1} + \frac{1}{4} \, \mathbf{a}_{2} + z_{2} \, \mathbf{a}_{3} & = & \frac{1}{4}a \, \mathbf{\hat{x}} + \frac{1}{4}a \, \mathbf{\hat{y}} + z_{2}c \, \mathbf{\hat{z}} & \left(4d\right) & \mbox{P II} \\ 
\mathbf{B}_{6} & = & \frac{1}{4} \, \mathbf{a}_{1} + \frac{1}{4} \, \mathbf{a}_{2} + \left(\frac{1}{2} +z_{2}\right) \, \mathbf{a}_{3} & = & \frac{1}{4}a \, \mathbf{\hat{x}} + \frac{1}{4}a \, \mathbf{\hat{y}} + \left(\frac{1}{2} +z_{2}\right)c \, \mathbf{\hat{z}} & \left(4d\right) & \mbox{P II} \\ 
\mathbf{B}_{7} & = & \frac{3}{4} \, \mathbf{a}_{1} + \frac{3}{4} \, \mathbf{a}_{2}-z_{2} \, \mathbf{a}_{3} & = & \frac{3}{4}a \, \mathbf{\hat{x}} + \frac{3}{4}a \, \mathbf{\hat{y}}-z_{2}c \, \mathbf{\hat{z}} & \left(4d\right) & \mbox{P II} \\ 
\mathbf{B}_{8} & = & \frac{3}{4} \, \mathbf{a}_{1} + \frac{3}{4} \, \mathbf{a}_{2} + \left(\frac{1}{2} - z_{2}\right) \, \mathbf{a}_{3} & = & \frac{3}{4}a \, \mathbf{\hat{x}} + \frac{3}{4}a \, \mathbf{\hat{y}} + \left(\frac{1}{2} - z_{2}\right)c \, \mathbf{\hat{z}} & \left(4d\right) & \mbox{P II} \\ 
\mathbf{B}_{9} & = & x_{3} \, \mathbf{a}_{1}-x_{3} \, \mathbf{a}_{2} + \frac{1}{4} \, \mathbf{a}_{3} & = & x_{3}a \, \mathbf{\hat{x}}-x_{3}a \, \mathbf{\hat{y}} + \frac{1}{4}c \, \mathbf{\hat{z}} & \left(8f\right) & \mbox{P III} \\ 
\mathbf{B}_{10} & = & \left(\frac{1}{2} - x_{3}\right) \, \mathbf{a}_{1} + \left(\frac{1}{2} +x_{3}\right) \, \mathbf{a}_{2} + \frac{1}{4} \, \mathbf{a}_{3} & = & \left(\frac{1}{2} - x_{3}\right)a \, \mathbf{\hat{x}} + \left(\frac{1}{2} +x_{3}\right)a \, \mathbf{\hat{y}} + \frac{1}{4}c \, \mathbf{\hat{z}} & \left(8f\right) & \mbox{P III} \\ 
\mathbf{B}_{11} & = & \left(\frac{1}{2} +x_{3}\right) \, \mathbf{a}_{1} + x_{3} \, \mathbf{a}_{2} + \frac{3}{4} \, \mathbf{a}_{3} & = & \left(\frac{1}{2} +x_{3}\right)a \, \mathbf{\hat{x}} + x_{3}a \, \mathbf{\hat{y}} + \frac{3}{4}c \, \mathbf{\hat{z}} & \left(8f\right) & \mbox{P III} \\ 
\mathbf{B}_{12} & = & -x_{3} \, \mathbf{a}_{1} + \left(\frac{1}{2} - x_{3}\right) \, \mathbf{a}_{2} + \frac{3}{4} \, \mathbf{a}_{3} & = & -x_{3}a \, \mathbf{\hat{x}} + \left(\frac{1}{2} - x_{3}\right)a \, \mathbf{\hat{y}} + \frac{3}{4}c \, \mathbf{\hat{z}} & \left(8f\right) & \mbox{P III} \\ 
\mathbf{B}_{13} & = & -x_{3} \, \mathbf{a}_{1} + x_{3} \, \mathbf{a}_{2} + \frac{3}{4} \, \mathbf{a}_{3} & = & -x_{3}a \, \mathbf{\hat{x}} + x_{3}a \, \mathbf{\hat{y}} + \frac{3}{4}c \, \mathbf{\hat{z}} & \left(8f\right) & \mbox{P III} \\ 
\mathbf{B}_{14} & = & \left(\frac{1}{2} +x_{3}\right) \, \mathbf{a}_{1} + \left(\frac{1}{2} - x_{3}\right) \, \mathbf{a}_{2} + \frac{3}{4} \, \mathbf{a}_{3} & = & \left(\frac{1}{2} +x_{3}\right)a \, \mathbf{\hat{x}} + \left(\frac{1}{2} - x_{3}\right)a \, \mathbf{\hat{y}} + \frac{3}{4}c \, \mathbf{\hat{z}} & \left(8f\right) & \mbox{P III} \\ 
\mathbf{B}_{15} & = & \left(\frac{1}{2} - x_{3}\right) \, \mathbf{a}_{1}-x_{3} \, \mathbf{a}_{2} + \frac{1}{4} \, \mathbf{a}_{3} & = & \left(\frac{1}{2} - x_{3}\right)a \, \mathbf{\hat{x}}-x_{3}a \, \mathbf{\hat{y}} + \frac{1}{4}c \, \mathbf{\hat{z}} & \left(8f\right) & \mbox{P III} \\ 
\mathbf{B}_{16} & = & x_{3} \, \mathbf{a}_{1} + \left(\frac{1}{2} +x_{3}\right) \, \mathbf{a}_{2} + \frac{1}{4} \, \mathbf{a}_{3} & = & x_{3}a \, \mathbf{\hat{x}} + \left(\frac{1}{2} +x_{3}\right)a \, \mathbf{\hat{y}} + \frac{1}{4}c \, \mathbf{\hat{z}} & \left(8f\right) & \mbox{P III} \\ 
\mathbf{B}_{17} & = & \frac{1}{4} \, \mathbf{a}_{1} + y_{4} \, \mathbf{a}_{2} + z_{4} \, \mathbf{a}_{3} & = & \frac{1}{4}a \, \mathbf{\hat{x}} + y_{4}a \, \mathbf{\hat{y}} + z_{4}c \, \mathbf{\hat{z}} & \left(8g\right) & \mbox{Zn I} \\ 
\mathbf{B}_{18} & = & \frac{1}{4} \, \mathbf{a}_{1} + \left(\frac{1}{2} - y_{4}\right) \, \mathbf{a}_{2} + z_{4} \, \mathbf{a}_{3} & = & \frac{1}{4}a \, \mathbf{\hat{x}} + \left(\frac{1}{2} - y_{4}\right)a \, \mathbf{\hat{y}} + z_{4}c \, \mathbf{\hat{z}} & \left(8g\right) & \mbox{Zn I} \\ 
\mathbf{B}_{19} & = & \left(\frac{1}{2} - y_{4}\right) \, \mathbf{a}_{1} + \frac{1}{4} \, \mathbf{a}_{2} + \left(\frac{1}{2} +z_{4}\right) \, \mathbf{a}_{3} & = & \left(\frac{1}{2} - y_{4}\right)a \, \mathbf{\hat{x}} + \frac{1}{4}a \, \mathbf{\hat{y}} + \left(\frac{1}{2} +z_{4}\right)c \, \mathbf{\hat{z}} & \left(8g\right) & \mbox{Zn I} \\ 
\mathbf{B}_{20} & = & y_{4} \, \mathbf{a}_{1} + \frac{1}{4} \, \mathbf{a}_{2} + \left(\frac{1}{2} +z_{4}\right) \, \mathbf{a}_{3} & = & y_{4}a \, \mathbf{\hat{x}} + \frac{1}{4}a \, \mathbf{\hat{y}} + \left(\frac{1}{2} +z_{4}\right)c \, \mathbf{\hat{z}} & \left(8g\right) & \mbox{Zn I} \\ 
\mathbf{B}_{21} & = & \frac{3}{4} \, \mathbf{a}_{1} + \left(\frac{1}{2} +y_{4}\right) \, \mathbf{a}_{2}-z_{4} \, \mathbf{a}_{3} & = & \frac{3}{4}a \, \mathbf{\hat{x}} + \left(\frac{1}{2} +y_{4}\right)a \, \mathbf{\hat{y}}-z_{4}c \, \mathbf{\hat{z}} & \left(8g\right) & \mbox{Zn I} \\ 
\mathbf{B}_{22} & = & \frac{3}{4} \, \mathbf{a}_{1}-y_{4} \, \mathbf{a}_{2}-z_{4} \, \mathbf{a}_{3} & = & \frac{3}{4}a \, \mathbf{\hat{x}}-y_{4}a \, \mathbf{\hat{y}}-z_{4}c \, \mathbf{\hat{z}} & \left(8g\right) & \mbox{Zn I} \\ 
\mathbf{B}_{23} & = & \left(\frac{1}{2} +y_{4}\right) \, \mathbf{a}_{1} + \frac{3}{4} \, \mathbf{a}_{2} + \left(\frac{1}{2} - z_{4}\right) \, \mathbf{a}_{3} & = & \left(\frac{1}{2} +y_{4}\right)a \, \mathbf{\hat{x}} + \frac{3}{4}a \, \mathbf{\hat{y}} + \left(\frac{1}{2} - z_{4}\right)c \, \mathbf{\hat{z}} & \left(8g\right) & \mbox{Zn I} \\ 
\mathbf{B}_{24} & = & -y_{4} \, \mathbf{a}_{1} + \frac{3}{4} \, \mathbf{a}_{2} + \left(\frac{1}{2} - z_{4}\right) \, \mathbf{a}_{3} & = & -y_{4}a \, \mathbf{\hat{x}} + \frac{3}{4}a \, \mathbf{\hat{y}} + \left(\frac{1}{2} - z_{4}\right)c \, \mathbf{\hat{z}} & \left(8g\right) & \mbox{Zn I} \\ 
\mathbf{B}_{25} & = & \frac{1}{4} \, \mathbf{a}_{1} + y_{5} \, \mathbf{a}_{2} + z_{5} \, \mathbf{a}_{3} & = & \frac{1}{4}a \, \mathbf{\hat{x}} + y_{5}a \, \mathbf{\hat{y}} + z_{5}c \, \mathbf{\hat{z}} & \left(8g\right) & \mbox{Zn II} \\ 
\mathbf{B}_{26} & = & \frac{1}{4} \, \mathbf{a}_{1} + \left(\frac{1}{2} - y_{5}\right) \, \mathbf{a}_{2} + z_{5} \, \mathbf{a}_{3} & = & \frac{1}{4}a \, \mathbf{\hat{x}} + \left(\frac{1}{2} - y_{5}\right)a \, \mathbf{\hat{y}} + z_{5}c \, \mathbf{\hat{z}} & \left(8g\right) & \mbox{Zn II} \\ 
\mathbf{B}_{27} & = & \left(\frac{1}{2} - y_{5}\right) \, \mathbf{a}_{1} + \frac{1}{4} \, \mathbf{a}_{2} + \left(\frac{1}{2} +z_{5}\right) \, \mathbf{a}_{3} & = & \left(\frac{1}{2} - y_{5}\right)a \, \mathbf{\hat{x}} + \frac{1}{4}a \, \mathbf{\hat{y}} + \left(\frac{1}{2} +z_{5}\right)c \, \mathbf{\hat{z}} & \left(8g\right) & \mbox{Zn II} \\ 
\mathbf{B}_{28} & = & y_{5} \, \mathbf{a}_{1} + \frac{1}{4} \, \mathbf{a}_{2} + \left(\frac{1}{2} +z_{5}\right) \, \mathbf{a}_{3} & = & y_{5}a \, \mathbf{\hat{x}} + \frac{1}{4}a \, \mathbf{\hat{y}} + \left(\frac{1}{2} +z_{5}\right)c \, \mathbf{\hat{z}} & \left(8g\right) & \mbox{Zn II} \\ 
\mathbf{B}_{29} & = & \frac{3}{4} \, \mathbf{a}_{1} + \left(\frac{1}{2} +y_{5}\right) \, \mathbf{a}_{2}-z_{5} \, \mathbf{a}_{3} & = & \frac{3}{4}a \, \mathbf{\hat{x}} + \left(\frac{1}{2} +y_{5}\right)a \, \mathbf{\hat{y}}-z_{5}c \, \mathbf{\hat{z}} & \left(8g\right) & \mbox{Zn II} \\ 
\mathbf{B}_{30} & = & \frac{3}{4} \, \mathbf{a}_{1}-y_{5} \, \mathbf{a}_{2}-z_{5} \, \mathbf{a}_{3} & = & \frac{3}{4}a \, \mathbf{\hat{x}}-y_{5}a \, \mathbf{\hat{y}}-z_{5}c \, \mathbf{\hat{z}} & \left(8g\right) & \mbox{Zn II} \\ 
\mathbf{B}_{31} & = & \left(\frac{1}{2} +y_{5}\right) \, \mathbf{a}_{1} + \frac{3}{4} \, \mathbf{a}_{2} + \left(\frac{1}{2} - z_{5}\right) \, \mathbf{a}_{3} & = & \left(\frac{1}{2} +y_{5}\right)a \, \mathbf{\hat{x}} + \frac{3}{4}a \, \mathbf{\hat{y}} + \left(\frac{1}{2} - z_{5}\right)c \, \mathbf{\hat{z}} & \left(8g\right) & \mbox{Zn II} \\ 
\mathbf{B}_{32} & = & -y_{5} \, \mathbf{a}_{1} + \frac{3}{4} \, \mathbf{a}_{2} + \left(\frac{1}{2} - z_{5}\right) \, \mathbf{a}_{3} & = & -y_{5}a \, \mathbf{\hat{x}} + \frac{3}{4}a \, \mathbf{\hat{y}} + \left(\frac{1}{2} - z_{5}\right)c \, \mathbf{\hat{z}} & \left(8g\right) & \mbox{Zn II} \\ 
\mathbf{B}_{33} & = & \frac{1}{4} \, \mathbf{a}_{1} + y_{6} \, \mathbf{a}_{2} + z_{6} \, \mathbf{a}_{3} & = & \frac{1}{4}a \, \mathbf{\hat{x}} + y_{6}a \, \mathbf{\hat{y}} + z_{6}c \, \mathbf{\hat{z}} & \left(8g\right) & \mbox{Zn III} \\ 
\mathbf{B}_{34} & = & \frac{1}{4} \, \mathbf{a}_{1} + \left(\frac{1}{2} - y_{6}\right) \, \mathbf{a}_{2} + z_{6} \, \mathbf{a}_{3} & = & \frac{1}{4}a \, \mathbf{\hat{x}} + \left(\frac{1}{2} - y_{6}\right)a \, \mathbf{\hat{y}} + z_{6}c \, \mathbf{\hat{z}} & \left(8g\right) & \mbox{Zn III} \\ 
\mathbf{B}_{35} & = & \left(\frac{1}{2} - y_{6}\right) \, \mathbf{a}_{1} + \frac{1}{4} \, \mathbf{a}_{2} + \left(\frac{1}{2} +z_{6}\right) \, \mathbf{a}_{3} & = & \left(\frac{1}{2} - y_{6}\right)a \, \mathbf{\hat{x}} + \frac{1}{4}a \, \mathbf{\hat{y}} + \left(\frac{1}{2} +z_{6}\right)c \, \mathbf{\hat{z}} & \left(8g\right) & \mbox{Zn III} \\ 
\mathbf{B}_{36} & = & y_{6} \, \mathbf{a}_{1} + \frac{1}{4} \, \mathbf{a}_{2} + \left(\frac{1}{2} +z_{6}\right) \, \mathbf{a}_{3} & = & y_{6}a \, \mathbf{\hat{x}} + \frac{1}{4}a \, \mathbf{\hat{y}} + \left(\frac{1}{2} +z_{6}\right)c \, \mathbf{\hat{z}} & \left(8g\right) & \mbox{Zn III} \\ 
\mathbf{B}_{37} & = & \frac{3}{4} \, \mathbf{a}_{1} + \left(\frac{1}{2} +y_{6}\right) \, \mathbf{a}_{2}-z_{6} \, \mathbf{a}_{3} & = & \frac{3}{4}a \, \mathbf{\hat{x}} + \left(\frac{1}{2} +y_{6}\right)a \, \mathbf{\hat{y}}-z_{6}c \, \mathbf{\hat{z}} & \left(8g\right) & \mbox{Zn III} \\ 
\mathbf{B}_{38} & = & \frac{3}{4} \, \mathbf{a}_{1}-y_{6} \, \mathbf{a}_{2}-z_{6} \, \mathbf{a}_{3} & = & \frac{3}{4}a \, \mathbf{\hat{x}}-y_{6}a \, \mathbf{\hat{y}}-z_{6}c \, \mathbf{\hat{z}} & \left(8g\right) & \mbox{Zn III} \\ 
\mathbf{B}_{39} & = & \left(\frac{1}{2} +y_{6}\right) \, \mathbf{a}_{1} + \frac{3}{4} \, \mathbf{a}_{2} + \left(\frac{1}{2} - z_{6}\right) \, \mathbf{a}_{3} & = & \left(\frac{1}{2} +y_{6}\right)a \, \mathbf{\hat{x}} + \frac{3}{4}a \, \mathbf{\hat{y}} + \left(\frac{1}{2} - z_{6}\right)c \, \mathbf{\hat{z}} & \left(8g\right) & \mbox{Zn III} \\ 
\mathbf{B}_{40} & = & -y_{6} \, \mathbf{a}_{1} + \frac{3}{4} \, \mathbf{a}_{2} + \left(\frac{1}{2} - z_{6}\right) \, \mathbf{a}_{3} & = & -y_{6}a \, \mathbf{\hat{x}} + \frac{3}{4}a \, \mathbf{\hat{y}} + \left(\frac{1}{2} - z_{6}\right)c \, \mathbf{\hat{z}} & \left(8g\right) & \mbox{Zn III} \\ 
\end{longtabu}
\renewcommand{\arraystretch}{1.0}
\noindent \hrulefill
\\
\textbf{References:}
\vspace*{-0.25cm}
\begin{flushleft}
  - \bibentry{Stackelberg_ZPhysChem_28_1935}. \\
\end{flushleft}
\textbf{Found in:}
\vspace*{-0.25cm}
\begin{flushleft}
  - \bibentry{Downs_AmMin_2003}. \\
  - \bibentry{Pearson_NRC_1958}. \\
\end{flushleft}
\noindent \hrulefill
\\
\textbf{Geometry files:}
\\
\noindent  - CIF: pp. {\hyperref[A2B3_tP40_137_cdf_3g_cif]{\pageref{A2B3_tP40_137_cdf_3g_cif}}} \\
\noindent  - POSCAR: pp. {\hyperref[A2B3_tP40_137_cdf_3g_poscar]{\pageref{A2B3_tP40_137_cdf_3g_poscar}}} \\
\onecolumn
{\phantomsection\label{A2B_tP6_137_d_a}}
\subsection*{\huge \textbf{{\normalfont ZrO$_{2}$ (High-temperature) Structure: A2B\_tP6\_137\_d\_a}}}
\noindent \hrulefill
\vspace*{0.25cm}
\begin{figure}[htp]
  \centering
  \vspace{-1em}
  {\includegraphics[width=1\textwidth]{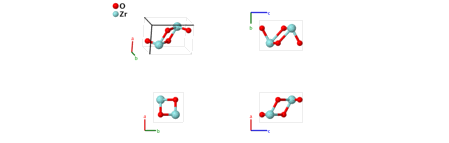}}
\end{figure}
\vspace*{-0.5cm}
\renewcommand{\arraystretch}{1.5}
\begin{equation*}
  \begin{array}{>{$\hspace{-0.15cm}}l<{$}>{$}p{0.5cm}<{$}>{$}p{18.5cm}<{$}}
    \mbox{\large \textbf{Prototype}} &\colon & \ce{ZrO2} \\
    \mbox{\large \textbf{\AFLOW\ prototype label}} &\colon & \mbox{A2B\_tP6\_137\_d\_a} \\
    \mbox{\large \textbf{\textit{Strukturbericht} designation}} &\colon & \mbox{None} \\
    \mbox{\large \textbf{Pearson symbol}} &\colon & \mbox{tP6} \\
    \mbox{\large \textbf{Space group number}} &\colon & 137 \\
    \mbox{\large \textbf{Space group symbol}} &\colon & P4_{2}/nmc \\
    \mbox{\large \textbf{\AFLOW\ prototype command}} &\colon &  \texttt{aflow} \,  \, \texttt{-{}-proto=A2B\_tP6\_137\_d\_a } \, \newline \texttt{-{}-params=}{a,c/a,z_{2} }
  \end{array}
\end{equation*}
\renewcommand{\arraystretch}{1.0}

\vspace*{-0.25cm}
\noindent \hrulefill
\begin{itemize}
  \item{ZrO$_{2}$ (pp. {\hyperref[A2B_tP6_137_d_a]{\pageref{A2B_tP6_137_d_a}}}) and 
HgI$_{2}$ (pp. {\hyperref[AB2_tP6_137_a_d]{\pageref{AB2_tP6_137_a_d}}})
have similar \AFLOW\ prototype labels ({\it{i.e.}}, same symmetry and set of
Wyckoff positions with different stoichiometry labels due to alphabetic ordering of atomic species).
They are generated by the same symmetry operations with different sets of parameters
(\texttt{-{}-params}) specified in their corresponding \CIF\ files.
}
\end{itemize}

\noindent \parbox{1 \linewidth}{
\noindent \hrulefill
\\
\textbf{Simple Tetragonal primitive vectors:} \\
\vspace*{-0.25cm}
\begin{tabular}{cc}
  \begin{tabular}{c}
    \parbox{0.6 \linewidth}{
      \renewcommand{\arraystretch}{1.5}
      \begin{equation*}
        \centering
        \begin{array}{ccc}
              \mathbf{a}_1 & = & a \, \mathbf{\hat{x}} \\
    \mathbf{a}_2 & = & a \, \mathbf{\hat{y}} \\
    \mathbf{a}_3 & = & c \, \mathbf{\hat{z}} \\

        \end{array}
      \end{equation*}
    }
    \renewcommand{\arraystretch}{1.0}
  \end{tabular}
  \begin{tabular}{c}
    \includegraphics[width=0.3\linewidth]{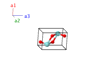} \\
  \end{tabular}
\end{tabular}

}
\vspace*{-0.25cm}

\noindent \hrulefill
\\
\textbf{Basis vectors:}
\vspace*{-0.25cm}
\renewcommand{\arraystretch}{1.5}
\begin{longtabu} to \textwidth{>{\centering $}X[-1,c,c]<{$}>{\centering $}X[-1,c,c]<{$}>{\centering $}X[-1,c,c]<{$}>{\centering $}X[-1,c,c]<{$}>{\centering $}X[-1,c,c]<{$}>{\centering $}X[-1,c,c]<{$}>{\centering $}X[-1,c,c]<{$}}
  & & \mbox{Lattice Coordinates} & & \mbox{Cartesian Coordinates} &\mbox{Wyckoff Position} & \mbox{Atom Type} \\  
  \mathbf{B}_{1} & = & \frac{3}{4} \, \mathbf{a}_{1} + \frac{1}{4} \, \mathbf{a}_{2} + \frac{3}{4} \, \mathbf{a}_{3} & = & \frac{3}{4}a \, \mathbf{\hat{x}} + \frac{1}{4}a \, \mathbf{\hat{y}} + \frac{3}{4}c \, \mathbf{\hat{z}} & \left(2a\right) & \mbox{Zr} \\ 
\mathbf{B}_{2} & = & \frac{1}{4} \, \mathbf{a}_{1} + \frac{3}{4} \, \mathbf{a}_{2} + \frac{1}{4} \, \mathbf{a}_{3} & = & \frac{1}{4}a \, \mathbf{\hat{x}} + \frac{3}{4}a \, \mathbf{\hat{y}} + \frac{1}{4}c \, \mathbf{\hat{z}} & \left(2a\right) & \mbox{Zr} \\ 
\mathbf{B}_{3} & = & \frac{1}{4} \, \mathbf{a}_{1} + \frac{1}{4} \, \mathbf{a}_{2} + z_{2} \, \mathbf{a}_{3} & = & \frac{1}{4}a \, \mathbf{\hat{x}} + \frac{1}{4}a \, \mathbf{\hat{y}} + z_{2}c \, \mathbf{\hat{z}} & \left(4d\right) & \mbox{O} \\ 
\mathbf{B}_{4} & = & \frac{1}{4} \, \mathbf{a}_{1} + \frac{1}{4} \, \mathbf{a}_{2} + \left(\frac{1}{2} +z_{2}\right) \, \mathbf{a}_{3} & = & \frac{1}{4}a \, \mathbf{\hat{x}} + \frac{1}{4}a \, \mathbf{\hat{y}} + \left(\frac{1}{2} +z_{2}\right)c \, \mathbf{\hat{z}} & \left(4d\right) & \mbox{O} \\ 
\mathbf{B}_{5} & = & \frac{3}{4} \, \mathbf{a}_{1} + \frac{3}{4} \, \mathbf{a}_{2}-z_{2} \, \mathbf{a}_{3} & = & \frac{3}{4}a \, \mathbf{\hat{x}} + \frac{3}{4}a \, \mathbf{\hat{y}}-z_{2}c \, \mathbf{\hat{z}} & \left(4d\right) & \mbox{O} \\ 
\mathbf{B}_{6} & = & \frac{3}{4} \, \mathbf{a}_{1} + \frac{3}{4} \, \mathbf{a}_{2} + \left(\frac{1}{2} - z_{2}\right) \, \mathbf{a}_{3} & = & \frac{3}{4}a \, \mathbf{\hat{x}} + \frac{3}{4}a \, \mathbf{\hat{y}} + \left(\frac{1}{2} - z_{2}\right)c \, \mathbf{\hat{z}} & \left(4d\right) & \mbox{O} \\ 
\end{longtabu}
\renewcommand{\arraystretch}{1.0}
\noindent \hrulefill
\\
\textbf{References:}
\vspace*{-0.25cm}
\begin{flushleft}
  - \bibentry{Teufer_ZrO2_Act_Crystallogr_1962}. \\
\end{flushleft}
\textbf{Found in:}
\vspace*{-0.25cm}
\begin{flushleft}
  - \bibentry{Villars_PearsonsCrystalData_2013}. \\
\end{flushleft}
\noindent \hrulefill
\\
\textbf{Geometry files:}
\\
\noindent  - CIF: pp. {\hyperref[A2B_tP6_137_d_a_cif]{\pageref{A2B_tP6_137_d_a_cif}}} \\
\noindent  - POSCAR: pp. {\hyperref[A2B_tP6_137_d_a_poscar]{\pageref{A2B_tP6_137_d_a_poscar}}} \\
\onecolumn
{\phantomsection\label{A4BC4_tP18_137_g_b_g}}
\subsection*{\huge \textbf{{\normalfont CeCo$_{4}$B$_{4}$ Structure: A4BC4\_tP18\_137\_g\_b\_g}}}
\noindent \hrulefill
\vspace*{0.25cm}
\begin{figure}[htp]
  \centering
  \vspace{-1em}
  {\includegraphics[width=1\textwidth]{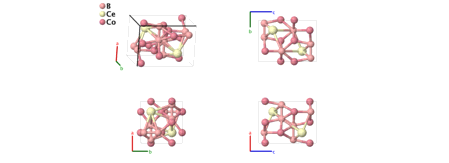}}
\end{figure}
\vspace*{-0.5cm}
\renewcommand{\arraystretch}{1.5}
\begin{equation*}
  \begin{array}{>{$\hspace{-0.15cm}}l<{$}>{$}p{0.5cm}<{$}>{$}p{18.5cm}<{$}}
    \mbox{\large \textbf{Prototype}} &\colon & \ce{CeCo4B4} \\
    \mbox{\large \textbf{\AFLOW\ prototype label}} &\colon & \mbox{A4BC4\_tP18\_137\_g\_b\_g} \\
    \mbox{\large \textbf{\textit{Strukturbericht} designation}} &\colon & \mbox{None} \\
    \mbox{\large \textbf{Pearson symbol}} &\colon & \mbox{tP18} \\
    \mbox{\large \textbf{Space group number}} &\colon & 137 \\
    \mbox{\large \textbf{Space group symbol}} &\colon & P4_{2}/nmc \\
    \mbox{\large \textbf{\AFLOW\ prototype command}} &\colon &  \texttt{aflow} \,  \, \texttt{-{}-proto=A4BC4\_tP18\_137\_g\_b\_g } \, \newline \texttt{-{}-params=}{a,c/a,y_{2},z_{2},y_{3},z_{3} }
  \end{array}
\end{equation*}
\renewcommand{\arraystretch}{1.0}

\noindent \parbox{1 \linewidth}{
\noindent \hrulefill
\\
\textbf{Simple Tetragonal primitive vectors:} \\
\vspace*{-0.25cm}
\begin{tabular}{cc}
  \begin{tabular}{c}
    \parbox{0.6 \linewidth}{
      \renewcommand{\arraystretch}{1.5}
      \begin{equation*}
        \centering
        \begin{array}{ccc}
              \mathbf{a}_1 & = & a \, \mathbf{\hat{x}} \\
    \mathbf{a}_2 & = & a \, \mathbf{\hat{y}} \\
    \mathbf{a}_3 & = & c \, \mathbf{\hat{z}} \\

        \end{array}
      \end{equation*}
    }
    \renewcommand{\arraystretch}{1.0}
  \end{tabular}
  \begin{tabular}{c}
    \includegraphics[width=0.3\linewidth]{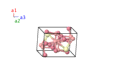} \\
  \end{tabular}
\end{tabular}

}
\vspace*{-0.25cm}

\noindent \hrulefill
\\
\textbf{Basis vectors:}
\vspace*{-0.25cm}
\renewcommand{\arraystretch}{1.5}
\begin{longtabu} to \textwidth{>{\centering $}X[-1,c,c]<{$}>{\centering $}X[-1,c,c]<{$}>{\centering $}X[-1,c,c]<{$}>{\centering $}X[-1,c,c]<{$}>{\centering $}X[-1,c,c]<{$}>{\centering $}X[-1,c,c]<{$}>{\centering $}X[-1,c,c]<{$}}
  & & \mbox{Lattice Coordinates} & & \mbox{Cartesian Coordinates} &\mbox{Wyckoff Position} & \mbox{Atom Type} \\  
  \mathbf{B}_{1} & = & \frac{3}{4} \, \mathbf{a}_{1} + \frac{1}{4} \, \mathbf{a}_{2} + \frac{1}{4} \, \mathbf{a}_{3} & = & \frac{3}{4}a \, \mathbf{\hat{x}} + \frac{1}{4}a \, \mathbf{\hat{y}} + \frac{1}{4}c \, \mathbf{\hat{z}} & \left(2b\right) & \mbox{Ce} \\ 
\mathbf{B}_{2} & = & \frac{1}{4} \, \mathbf{a}_{1} + \frac{3}{4} \, \mathbf{a}_{2} + \frac{3}{4} \, \mathbf{a}_{3} & = & \frac{1}{4}a \, \mathbf{\hat{x}} + \frac{3}{4}a \, \mathbf{\hat{y}} + \frac{3}{4}c \, \mathbf{\hat{z}} & \left(2b\right) & \mbox{Ce} \\ 
\mathbf{B}_{3} & = & \frac{1}{4} \, \mathbf{a}_{1} + y_{2} \, \mathbf{a}_{2} + z_{2} \, \mathbf{a}_{3} & = & \frac{1}{4}a \, \mathbf{\hat{x}} + y_{2}a \, \mathbf{\hat{y}} + z_{2}c \, \mathbf{\hat{z}} & \left(8g\right) & \mbox{B} \\ 
\mathbf{B}_{4} & = & \frac{1}{4} \, \mathbf{a}_{1} + \left(\frac{1}{2} - y_{2}\right) \, \mathbf{a}_{2} + z_{2} \, \mathbf{a}_{3} & = & \frac{1}{4}a \, \mathbf{\hat{x}} + \left(\frac{1}{2} - y_{2}\right)a \, \mathbf{\hat{y}} + z_{2}c \, \mathbf{\hat{z}} & \left(8g\right) & \mbox{B} \\ 
\mathbf{B}_{5} & = & \left(\frac{1}{2} - y_{2}\right) \, \mathbf{a}_{1} + \frac{1}{4} \, \mathbf{a}_{2} + \left(\frac{1}{2} +z_{2}\right) \, \mathbf{a}_{3} & = & \left(\frac{1}{2} - y_{2}\right)a \, \mathbf{\hat{x}} + \frac{1}{4}a \, \mathbf{\hat{y}} + \left(\frac{1}{2} +z_{2}\right)c \, \mathbf{\hat{z}} & \left(8g\right) & \mbox{B} \\ 
\mathbf{B}_{6} & = & y_{2} \, \mathbf{a}_{1} + \frac{1}{4} \, \mathbf{a}_{2} + \left(\frac{1}{2} +z_{2}\right) \, \mathbf{a}_{3} & = & y_{2}a \, \mathbf{\hat{x}} + \frac{1}{4}a \, \mathbf{\hat{y}} + \left(\frac{1}{2} +z_{2}\right)c \, \mathbf{\hat{z}} & \left(8g\right) & \mbox{B} \\ 
\mathbf{B}_{7} & = & \frac{3}{4} \, \mathbf{a}_{1} + \left(\frac{1}{2} +y_{2}\right) \, \mathbf{a}_{2}-z_{2} \, \mathbf{a}_{3} & = & \frac{3}{4}a \, \mathbf{\hat{x}} + \left(\frac{1}{2} +y_{2}\right)a \, \mathbf{\hat{y}}-z_{2}c \, \mathbf{\hat{z}} & \left(8g\right) & \mbox{B} \\ 
\mathbf{B}_{8} & = & \frac{3}{4} \, \mathbf{a}_{1}-y_{2} \, \mathbf{a}_{2}-z_{2} \, \mathbf{a}_{3} & = & \frac{3}{4}a \, \mathbf{\hat{x}}-y_{2}a \, \mathbf{\hat{y}}-z_{2}c \, \mathbf{\hat{z}} & \left(8g\right) & \mbox{B} \\ 
\mathbf{B}_{9} & = & \left(\frac{1}{2} +y_{2}\right) \, \mathbf{a}_{1} + \frac{3}{4} \, \mathbf{a}_{2} + \left(\frac{1}{2} - z_{2}\right) \, \mathbf{a}_{3} & = & \left(\frac{1}{2} +y_{2}\right)a \, \mathbf{\hat{x}} + \frac{3}{4}a \, \mathbf{\hat{y}} + \left(\frac{1}{2} - z_{2}\right)c \, \mathbf{\hat{z}} & \left(8g\right) & \mbox{B} \\ 
\mathbf{B}_{10} & = & -y_{2} \, \mathbf{a}_{1} + \frac{3}{4} \, \mathbf{a}_{2} + \left(\frac{1}{2} - z_{2}\right) \, \mathbf{a}_{3} & = & -y_{2}a \, \mathbf{\hat{x}} + \frac{3}{4}a \, \mathbf{\hat{y}} + \left(\frac{1}{2} - z_{2}\right)c \, \mathbf{\hat{z}} & \left(8g\right) & \mbox{B} \\ 
\mathbf{B}_{11} & = & \frac{1}{4} \, \mathbf{a}_{1} + y_{3} \, \mathbf{a}_{2} + z_{3} \, \mathbf{a}_{3} & = & \frac{1}{4}a \, \mathbf{\hat{x}} + y_{3}a \, \mathbf{\hat{y}} + z_{3}c \, \mathbf{\hat{z}} & \left(8g\right) & \mbox{Co} \\ 
\mathbf{B}_{12} & = & \frac{1}{4} \, \mathbf{a}_{1} + \left(\frac{1}{2} - y_{3}\right) \, \mathbf{a}_{2} + z_{3} \, \mathbf{a}_{3} & = & \frac{1}{4}a \, \mathbf{\hat{x}} + \left(\frac{1}{2} - y_{3}\right)a \, \mathbf{\hat{y}} + z_{3}c \, \mathbf{\hat{z}} & \left(8g\right) & \mbox{Co} \\ 
\mathbf{B}_{13} & = & \left(\frac{1}{2} - y_{3}\right) \, \mathbf{a}_{1} + \frac{1}{4} \, \mathbf{a}_{2} + \left(\frac{1}{2} +z_{3}\right) \, \mathbf{a}_{3} & = & \left(\frac{1}{2} - y_{3}\right)a \, \mathbf{\hat{x}} + \frac{1}{4}a \, \mathbf{\hat{y}} + \left(\frac{1}{2} +z_{3}\right)c \, \mathbf{\hat{z}} & \left(8g\right) & \mbox{Co} \\ 
\mathbf{B}_{14} & = & y_{3} \, \mathbf{a}_{1} + \frac{1}{4} \, \mathbf{a}_{2} + \left(\frac{1}{2} +z_{3}\right) \, \mathbf{a}_{3} & = & y_{3}a \, \mathbf{\hat{x}} + \frac{1}{4}a \, \mathbf{\hat{y}} + \left(\frac{1}{2} +z_{3}\right)c \, \mathbf{\hat{z}} & \left(8g\right) & \mbox{Co} \\ 
\mathbf{B}_{15} & = & \frac{3}{4} \, \mathbf{a}_{1} + \left(\frac{1}{2} +y_{3}\right) \, \mathbf{a}_{2}-z_{3} \, \mathbf{a}_{3} & = & \frac{3}{4}a \, \mathbf{\hat{x}} + \left(\frac{1}{2} +y_{3}\right)a \, \mathbf{\hat{y}}-z_{3}c \, \mathbf{\hat{z}} & \left(8g\right) & \mbox{Co} \\ 
\mathbf{B}_{16} & = & \frac{3}{4} \, \mathbf{a}_{1}-y_{3} \, \mathbf{a}_{2}-z_{3} \, \mathbf{a}_{3} & = & \frac{3}{4}a \, \mathbf{\hat{x}}-y_{3}a \, \mathbf{\hat{y}}-z_{3}c \, \mathbf{\hat{z}} & \left(8g\right) & \mbox{Co} \\ 
\mathbf{B}_{17} & = & \left(\frac{1}{2} +y_{3}\right) \, \mathbf{a}_{1} + \frac{3}{4} \, \mathbf{a}_{2} + \left(\frac{1}{2} - z_{3}\right) \, \mathbf{a}_{3} & = & \left(\frac{1}{2} +y_{3}\right)a \, \mathbf{\hat{x}} + \frac{3}{4}a \, \mathbf{\hat{y}} + \left(\frac{1}{2} - z_{3}\right)c \, \mathbf{\hat{z}} & \left(8g\right) & \mbox{Co} \\ 
\mathbf{B}_{18} & = & -y_{3} \, \mathbf{a}_{1} + \frac{3}{4} \, \mathbf{a}_{2} + \left(\frac{1}{2} - z_{3}\right) \, \mathbf{a}_{3} & = & -y_{3}a \, \mathbf{\hat{x}} + \frac{3}{4}a \, \mathbf{\hat{y}} + \left(\frac{1}{2} - z_{3}\right)c \, \mathbf{\hat{z}} & \left(8g\right) & \mbox{Co} \\ 
\end{longtabu}
\renewcommand{\arraystretch}{1.0}
\noindent \hrulefill
\\
\textbf{References:}
\vspace*{-0.25cm}
\begin{flushleft}
  - \bibentry{Kuzma_CeCo4B4_SovPhysCryst_1972}. \\
\end{flushleft}
\textbf{Found in:}
\vspace*{-0.25cm}
\begin{flushleft}
  - \bibentry{Villars_PearsonsCrystalData_2013}. \\
\end{flushleft}
\noindent \hrulefill
\\
\textbf{Geometry files:}
\\
\noindent  - CIF: pp. {\hyperref[A4BC4_tP18_137_g_b_g_cif]{\pageref{A4BC4_tP18_137_g_b_g_cif}}} \\
\noindent  - POSCAR: pp. {\hyperref[A4BC4_tP18_137_g_b_g_poscar]{\pageref{A4BC4_tP18_137_g_b_g_poscar}}} \\
\onecolumn
{\phantomsection\label{AB2_tP6_137_a_d}}
\subsection*{\huge \textbf{{\normalfont HgI$_{2}$ ($C13$) Structure: AB2\_tP6\_137\_a\_d}}}
\noindent \hrulefill
\vspace*{0.25cm}
\begin{figure}[htp]
  \centering
  \vspace{-1em}
  {\includegraphics[width=1\textwidth]{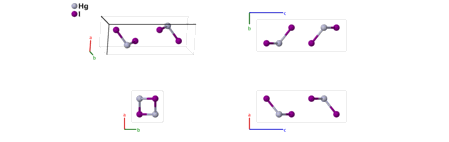}}
\end{figure}
\vspace*{-0.5cm}
\renewcommand{\arraystretch}{1.5}
\begin{equation*}
  \begin{array}{>{$\hspace{-0.15cm}}l<{$}>{$}p{0.5cm}<{$}>{$}p{18.5cm}<{$}}
    \mbox{\large \textbf{Prototype}} &\colon & \ce{HgI$_{2}$} \\
    \mbox{\large \textbf{\AFLOW\ prototype label}} &\colon & \mbox{AB2\_tP6\_137\_a\_d} \\
    \mbox{\large \textbf{\textit{Strukturbericht} designation}} &\colon & \mbox{$C13$} \\
    \mbox{\large \textbf{Pearson symbol}} &\colon & \mbox{tP6} \\
    \mbox{\large \textbf{Space group number}} &\colon & 137 \\
    \mbox{\large \textbf{Space group symbol}} &\colon & P4_{2}/nmc \\
    \mbox{\large \textbf{\AFLOW\ prototype command}} &\colon &  \texttt{aflow} \,  \, \texttt{-{}-proto=AB2\_tP6\_137\_a\_d } \, \newline \texttt{-{}-params=}{a,c/a,z_{2} }
  \end{array}
\end{equation*}
\renewcommand{\arraystretch}{1.0}

\vspace*{-0.25cm}
\noindent \hrulefill
\begin{itemize}
  \item{The \CIF\ and \POSCAR\ files contain the data at room temperature, 293~K.
ZrO$_{2}$ (pp. {\hyperref[A2B_tP6_137_d_a]{\pageref{A2B_tP6_137_d_a}}}) and
HgI$_{2}$ (pp. {\hyperref[AB2_tP6_137_a_d]{\pageref{AB2_tP6_137_a_d}}})
have similar \AFLOW\ prototype labels ({\it{i.e.}}, same symmetry and set of
Wyckoff positions with different stoichiometry labels due to alphabetic ordering of atomic species).
They are generated by the same symmetry operations with different sets of parameters
(\texttt{-{}-params}) specified in their corresponding \CIF\ files.
}
\end{itemize}

\noindent \parbox{1 \linewidth}{
\noindent \hrulefill
\\
\textbf{Simple Tetragonal primitive vectors:} \\
\vspace*{-0.25cm}
\begin{tabular}{cc}
  \begin{tabular}{c}
    \parbox{0.6 \linewidth}{
      \renewcommand{\arraystretch}{1.5}
      \begin{equation*}
        \centering
        \begin{array}{ccc}
              \mathbf{a}_1 & = & a \, \mathbf{\hat{x}} \\
    \mathbf{a}_2 & = & a \, \mathbf{\hat{y}} \\
    \mathbf{a}_3 & = & c \, \mathbf{\hat{z}} \\

        \end{array}
      \end{equation*}
    }
    \renewcommand{\arraystretch}{1.0}
  \end{tabular}
  \begin{tabular}{c}
    \includegraphics[width=0.3\linewidth]{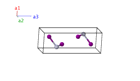} \\
  \end{tabular}
\end{tabular}

}
\vspace*{-0.25cm}

\noindent \hrulefill
\\
\textbf{Basis vectors:}
\vspace*{-0.25cm}
\renewcommand{\arraystretch}{1.5}
\begin{longtabu} to \textwidth{>{\centering $}X[-1,c,c]<{$}>{\centering $}X[-1,c,c]<{$}>{\centering $}X[-1,c,c]<{$}>{\centering $}X[-1,c,c]<{$}>{\centering $}X[-1,c,c]<{$}>{\centering $}X[-1,c,c]<{$}>{\centering $}X[-1,c,c]<{$}}
  & & \mbox{Lattice Coordinates} & & \mbox{Cartesian Coordinates} &\mbox{Wyckoff Position} & \mbox{Atom Type} \\  
  \mathbf{B}_{1} & = & \frac{3}{4} \, \mathbf{a}_{1} + \frac{1}{4} \, \mathbf{a}_{2} + \frac{3}{4} \, \mathbf{a}_{3} & = & \frac{3}{4}a \, \mathbf{\hat{x}} + \frac{1}{4}a \, \mathbf{\hat{y}} + \frac{3}{4}c \, \mathbf{\hat{z}} & \left(2a\right) & \mbox{Hg} \\ 
\mathbf{B}_{2} & = & \frac{1}{4} \, \mathbf{a}_{1} + \frac{3}{4} \, \mathbf{a}_{2} + \frac{1}{4} \, \mathbf{a}_{3} & = & \frac{1}{4}a \, \mathbf{\hat{x}} + \frac{3}{4}a \, \mathbf{\hat{y}} + \frac{1}{4}c \, \mathbf{\hat{z}} & \left(2a\right) & \mbox{Hg} \\ 
\mathbf{B}_{3} & = & \frac{1}{4} \, \mathbf{a}_{1} + \frac{1}{4} \, \mathbf{a}_{2} + z_{2} \, \mathbf{a}_{3} & = & \frac{1}{4}a \, \mathbf{\hat{x}} + \frac{1}{4}a \, \mathbf{\hat{y}} + z_{2}c \, \mathbf{\hat{z}} & \left(4d\right) & \mbox{I} \\ 
\mathbf{B}_{4} & = & \frac{1}{4} \, \mathbf{a}_{1} + \frac{1}{4} \, \mathbf{a}_{2} + \left(\frac{1}{2} +z_{2}\right) \, \mathbf{a}_{3} & = & \frac{1}{4}a \, \mathbf{\hat{x}} + \frac{1}{4}a \, \mathbf{\hat{y}} + \left(\frac{1}{2} +z_{2}\right)c \, \mathbf{\hat{z}} & \left(4d\right) & \mbox{I} \\ 
\mathbf{B}_{5} & = & \frac{3}{4} \, \mathbf{a}_{1} + \frac{3}{4} \, \mathbf{a}_{2}-z_{2} \, \mathbf{a}_{3} & = & \frac{3}{4}a \, \mathbf{\hat{x}} + \frac{3}{4}a \, \mathbf{\hat{y}}-z_{2}c \, \mathbf{\hat{z}} & \left(4d\right) & \mbox{I} \\ 
\mathbf{B}_{6} & = & \frac{3}{4} \, \mathbf{a}_{1} + \frac{3}{4} \, \mathbf{a}_{2} + \left(\frac{1}{2} - z_{2}\right) \, \mathbf{a}_{3} & = & \frac{3}{4}a \, \mathbf{\hat{x}} + \frac{3}{4}a \, \mathbf{\hat{y}} + \left(\frac{1}{2} - z_{2}\right)c \, \mathbf{\hat{z}} & \left(4d\right) & \mbox{I} \\ 
\end{longtabu}
\renewcommand{\arraystretch}{1.0}
\noindent \hrulefill
\\
\textbf{References:}
\vspace*{-0.25cm}
\begin{flushleft}
  - \bibentry{Schwarzenbach_Acta_CristB_63_2007}. \\
\end{flushleft}
\textbf{Found in:}
\vspace*{-0.25cm}
\begin{flushleft}
  - \bibentry{Downs_AM_88_2003}. \\
\end{flushleft}
\noindent \hrulefill
\\
\textbf{Geometry files:}
\\
\noindent  - CIF: pp. {\hyperref[AB2_tP6_137_a_d_cif]{\pageref{AB2_tP6_137_a_d_cif}}} \\
\noindent  - POSCAR: pp. {\hyperref[AB2_tP6_137_a_d_poscar]{\pageref{AB2_tP6_137_a_d_poscar}}} \\
\onecolumn
{\phantomsection\label{A_tP12_138_bi}}
\subsection*{\huge \textbf{{\normalfont C (T12 Group IV) Structure: A\_tP12\_138\_bi}}}
\noindent \hrulefill
\vspace*{0.25cm}
\begin{figure}[htp]
  \centering
  \vspace{-1em}
  {\includegraphics[width=1\textwidth]{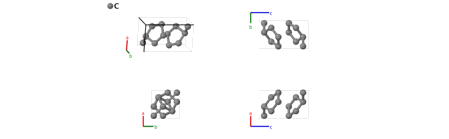}}
\end{figure}
\vspace*{-0.5cm}
\renewcommand{\arraystretch}{1.5}
\begin{equation*}
  \begin{array}{>{$\hspace{-0.15cm}}l<{$}>{$}p{0.5cm}<{$}>{$}p{18.5cm}<{$}}
    \mbox{\large \textbf{Prototype}} &\colon & \ce{C} \\
    \mbox{\large \textbf{\AFLOW\ prototype label}} &\colon & \mbox{A\_tP12\_138\_bi} \\
    \mbox{\large \textbf{\textit{Strukturbericht} designation}} &\colon & \mbox{None} \\
    \mbox{\large \textbf{Pearson symbol}} &\colon & \mbox{tP12} \\
    \mbox{\large \textbf{Space group number}} &\colon & 138 \\
    \mbox{\large \textbf{Space group symbol}} &\colon & P4_{2}/ncm \\
    \mbox{\large \textbf{\AFLOW\ prototype command}} &\colon &  \texttt{aflow} \,  \, \texttt{-{}-proto=A\_tP12\_138\_bi } \, \newline \texttt{-{}-params=}{a,c/a,x_{2},z_{2} }
  \end{array}
\end{equation*}
\renewcommand{\arraystretch}{1.0}

\vspace*{-0.25cm}
\noindent \hrulefill
\\
\textbf{ Other elements with this structure:}
\begin{itemize}
   \item{ Si, Ge  }
\end{itemize}
\vspace*{-0.25cm}
\noindent \hrulefill
\begin{itemize}
  \item{This is a tetragonal allotrope of the diamond structure found
computationally in C, Si, and Ge.  The authors state ``The T12
polymorph naturally accounts for the experimental $d$ spacings and Raman
spectra of synthesized metastable Ge and Si-XIII phases with
long-puzzling unknown structures, respectively.''
}
\end{itemize}

\noindent \parbox{1 \linewidth}{
\noindent \hrulefill
\\
\textbf{Simple Tetragonal primitive vectors:} \\
\vspace*{-0.25cm}
\begin{tabular}{cc}
  \begin{tabular}{c}
    \parbox{0.6 \linewidth}{
      \renewcommand{\arraystretch}{1.5}
      \begin{equation*}
        \centering
        \begin{array}{ccc}
              \mathbf{a}_1 & = & a \, \mathbf{\hat{x}} \\
    \mathbf{a}_2 & = & a \, \mathbf{\hat{y}} \\
    \mathbf{a}_3 & = & c \, \mathbf{\hat{z}} \\

        \end{array}
      \end{equation*}
    }
    \renewcommand{\arraystretch}{1.0}
  \end{tabular}
  \begin{tabular}{c}
    \includegraphics[width=0.3\linewidth]{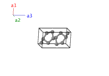} \\
  \end{tabular}
\end{tabular}

}
\vspace*{-0.25cm}

\noindent \hrulefill
\\
\textbf{Basis vectors:}
\vspace*{-0.25cm}
\renewcommand{\arraystretch}{1.5}
\begin{longtabu} to \textwidth{>{\centering $}X[-1,c,c]<{$}>{\centering $}X[-1,c,c]<{$}>{\centering $}X[-1,c,c]<{$}>{\centering $}X[-1,c,c]<{$}>{\centering $}X[-1,c,c]<{$}>{\centering $}X[-1,c,c]<{$}>{\centering $}X[-1,c,c]<{$}}
  & & \mbox{Lattice Coordinates} & & \mbox{Cartesian Coordinates} &\mbox{Wyckoff Position} & \mbox{Atom Type} \\  
  \mathbf{B}_{1} & = & \frac{3}{4} \, \mathbf{a}_{1} + \frac{1}{4} \, \mathbf{a}_{2} + \frac{3}{4} \, \mathbf{a}_{3} & = & \frac{3}{4}a \, \mathbf{\hat{x}} + \frac{1}{4}a \, \mathbf{\hat{y}} + \frac{3}{4}c \, \mathbf{\hat{z}} & \left(4b\right) & \mbox{C I} \\ 
\mathbf{B}_{2} & = & \frac{1}{4} \, \mathbf{a}_{1} + \frac{3}{4} \, \mathbf{a}_{2} + \frac{1}{4} \, \mathbf{a}_{3} & = & \frac{1}{4}a \, \mathbf{\hat{x}} + \frac{3}{4}a \, \mathbf{\hat{y}} + \frac{1}{4}c \, \mathbf{\hat{z}} & \left(4b\right) & \mbox{C I} \\ 
\mathbf{B}_{3} & = & \frac{1}{4} \, \mathbf{a}_{1} + \frac{3}{4} \, \mathbf{a}_{2} + \frac{3}{4} \, \mathbf{a}_{3} & = & \frac{1}{4}a \, \mathbf{\hat{x}} + \frac{3}{4}a \, \mathbf{\hat{y}} + \frac{3}{4}c \, \mathbf{\hat{z}} & \left(4b\right) & \mbox{C I} \\ 
\mathbf{B}_{4} & = & \frac{3}{4} \, \mathbf{a}_{1} + \frac{1}{4} \, \mathbf{a}_{2} + \frac{1}{4} \, \mathbf{a}_{3} & = & \frac{3}{4}a \, \mathbf{\hat{x}} + \frac{1}{4}a \, \mathbf{\hat{y}} + \frac{1}{4}c \, \mathbf{\hat{z}} & \left(4b\right) & \mbox{C I} \\ 
\mathbf{B}_{5} & = & x_{2} \, \mathbf{a}_{1} + x_{2} \, \mathbf{a}_{2} + z_{2} \, \mathbf{a}_{3} & = & x_{2}a \, \mathbf{\hat{x}} + x_{2}a \, \mathbf{\hat{y}} + z_{2}c \, \mathbf{\hat{z}} & \left(8i\right) & \mbox{C II} \\ 
\mathbf{B}_{6} & = & \left(\frac{1}{2} - x_{2}\right) \, \mathbf{a}_{1} + \left(\frac{1}{2} - x_{2}\right) \, \mathbf{a}_{2} + z_{2} \, \mathbf{a}_{3} & = & \left(\frac{1}{2} - x_{2}\right)a \, \mathbf{\hat{x}} + \left(\frac{1}{2} - x_{2}\right)a \, \mathbf{\hat{y}} + z_{2}c \, \mathbf{\hat{z}} & \left(8i\right) & \mbox{C II} \\ 
\mathbf{B}_{7} & = & \left(\frac{1}{2} - x_{2}\right) \, \mathbf{a}_{1} + x_{2} \, \mathbf{a}_{2} + \left(\frac{1}{2} +z_{2}\right) \, \mathbf{a}_{3} & = & \left(\frac{1}{2} - x_{2}\right)a \, \mathbf{\hat{x}} + x_{2}a \, \mathbf{\hat{y}} + \left(\frac{1}{2} +z_{2}\right)c \, \mathbf{\hat{z}} & \left(8i\right) & \mbox{C II} \\ 
\mathbf{B}_{8} & = & x_{2} \, \mathbf{a}_{1} + \left(\frac{1}{2} - x_{2}\right) \, \mathbf{a}_{2} + \left(\frac{1}{2} +z_{2}\right) \, \mathbf{a}_{3} & = & x_{2}a \, \mathbf{\hat{x}} + \left(\frac{1}{2} - x_{2}\right)a \, \mathbf{\hat{y}} + \left(\frac{1}{2} +z_{2}\right)c \, \mathbf{\hat{z}} & \left(8i\right) & \mbox{C II} \\ 
\mathbf{B}_{9} & = & -x_{2} \, \mathbf{a}_{1} + \left(\frac{1}{2} +x_{2}\right) \, \mathbf{a}_{2} + \left(\frac{1}{2} - z_{2}\right) \, \mathbf{a}_{3} & = & -x_{2}a \, \mathbf{\hat{x}} + \left(\frac{1}{2} +x_{2}\right)a \, \mathbf{\hat{y}} + \left(\frac{1}{2} - z_{2}\right)c \, \mathbf{\hat{z}} & \left(8i\right) & \mbox{C II} \\ 
\mathbf{B}_{10} & = & \left(\frac{1}{2} +x_{2}\right) \, \mathbf{a}_{1}-x_{2} \, \mathbf{a}_{2} + \left(\frac{1}{2} - z_{2}\right) \, \mathbf{a}_{3} & = & \left(\frac{1}{2} +x_{2}\right)a \, \mathbf{\hat{x}}-x_{2}a \, \mathbf{\hat{y}} + \left(\frac{1}{2} - z_{2}\right)c \, \mathbf{\hat{z}} & \left(8i\right) & \mbox{C II} \\ 
\mathbf{B}_{11} & = & \left(\frac{1}{2} +x_{2}\right) \, \mathbf{a}_{1} + \left(\frac{1}{2} +x_{2}\right) \, \mathbf{a}_{2}-z_{2} \, \mathbf{a}_{3} & = & \left(\frac{1}{2} +x_{2}\right)a \, \mathbf{\hat{x}} + \left(\frac{1}{2} +x_{2}\right)a \, \mathbf{\hat{y}}-z_{2}c \, \mathbf{\hat{z}} & \left(8i\right) & \mbox{C II} \\ 
\mathbf{B}_{12} & = & -x_{2} \, \mathbf{a}_{1}-x_{2} \, \mathbf{a}_{2}-z_{2} \, \mathbf{a}_{3} & = & -x_{2}a \, \mathbf{\hat{x}}-x_{2}a \, \mathbf{\hat{y}}-z_{2}c \, \mathbf{\hat{z}} & \left(8i\right) & \mbox{C II} \\ 
\end{longtabu}
\renewcommand{\arraystretch}{1.0}
\noindent \hrulefill
\\
\textbf{References:}
\vspace*{-0.25cm}
\begin{flushleft}
  - \bibentry{Zhao_JACS_134_2012}. \\
\end{flushleft}
\noindent \hrulefill
\\
\textbf{Geometry files:}
\\
\noindent  - CIF: pp. {\hyperref[A_tP12_138_bi_cif]{\pageref{A_tP12_138_bi_cif}}} \\
\noindent  - POSCAR: pp. {\hyperref[A_tP12_138_bi_poscar]{\pageref{A_tP12_138_bi_poscar}}} \\
\onecolumn
{\phantomsection\label{AB_tI8_139_e_e}}
\subsection*{\huge \textbf{{\normalfont Calomel (Hg$_{2}$Cl$_{2}$, $D3_{1}$) Structure: AB\_tI8\_139\_e\_e}}}
\noindent \hrulefill
\vspace*{0.25cm}
\begin{figure}[htp]
  \centering
  \vspace{-1em}
  {\includegraphics[width=1\textwidth]{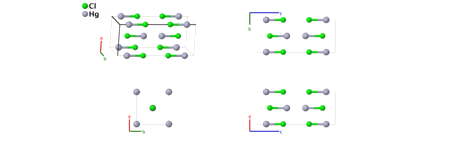}}
\end{figure}
\vspace*{-0.5cm}
\renewcommand{\arraystretch}{1.5}
\begin{equation*}
  \begin{array}{>{$\hspace{-0.15cm}}l<{$}>{$}p{0.5cm}<{$}>{$}p{18.5cm}<{$}}
    \mbox{\large \textbf{Prototype}} &\colon & \ce{Hg2Cl2} \\
    \mbox{\large \textbf{\AFLOW\ prototype label}} &\colon & \mbox{AB\_tI8\_139\_e\_e} \\
    \mbox{\large \textbf{\textit{Strukturbericht} designation}} &\colon & \mbox{$D3_{1}$} \\
    \mbox{\large \textbf{Pearson symbol}} &\colon & \mbox{tI8} \\
    \mbox{\large \textbf{Space group number}} &\colon & 139 \\
    \mbox{\large \textbf{Space group symbol}} &\colon & I4/mmm \\
    \mbox{\large \textbf{\AFLOW\ prototype command}} &\colon &  \texttt{aflow} \,  \, \texttt{-{}-proto=AB\_tI8\_139\_e\_e } \, \newline \texttt{-{}-params=}{a,c/a,z_{1},z_{2} }
  \end{array}
\end{equation*}
\renewcommand{\arraystretch}{1.0}

\noindent \parbox{1 \linewidth}{
\noindent \hrulefill
\\
\textbf{Body-centered Tetragonal primitive vectors:} \\
\vspace*{-0.25cm}
\begin{tabular}{cc}
  \begin{tabular}{c}
    \parbox{0.6 \linewidth}{
      \renewcommand{\arraystretch}{1.5}
      \begin{equation*}
        \centering
        \begin{array}{ccc}
              \mathbf{a}_1 & = & - \frac12 \, a \, \mathbf{\hat{x}} + \frac12 \, a \, \mathbf{\hat{y}} + \frac12 \, c \, \mathbf{\hat{z}} \\
    \mathbf{a}_2 & = & ~ \frac12 \, a \, \mathbf{\hat{x}} - \frac12 \, a \, \mathbf{\hat{y}} + \frac12 \, c \, \mathbf{\hat{z}} \\
    \mathbf{a}_3 & = & ~ \frac12 \, a \, \mathbf{\hat{x}} + \frac12 \, a \, \mathbf{\hat{y}} - \frac12 \, c \, \mathbf{\hat{z}} \\

        \end{array}
      \end{equation*}
    }
    \renewcommand{\arraystretch}{1.0}
  \end{tabular}
  \begin{tabular}{c}
    \includegraphics[width=0.3\linewidth]{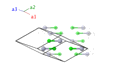} \\
  \end{tabular}
\end{tabular}

}
\vspace*{-0.25cm}

\noindent \hrulefill
\\
\textbf{Basis vectors:}
\vspace*{-0.25cm}
\renewcommand{\arraystretch}{1.5}
\begin{longtabu} to \textwidth{>{\centering $}X[-1,c,c]<{$}>{\centering $}X[-1,c,c]<{$}>{\centering $}X[-1,c,c]<{$}>{\centering $}X[-1,c,c]<{$}>{\centering $}X[-1,c,c]<{$}>{\centering $}X[-1,c,c]<{$}>{\centering $}X[-1,c,c]<{$}}
  & & \mbox{Lattice Coordinates} & & \mbox{Cartesian Coordinates} &\mbox{Wyckoff Position} & \mbox{Atom Type} \\  
  \mathbf{B}_{1} & = & z_{1} \, \mathbf{a}_{1} + z_{1} \, \mathbf{a}_{2} & = & z_{1}c \, \mathbf{\hat{z}} & \left(4e\right) & \mbox{Cl} \\ 
\mathbf{B}_{2} & = & -z_{1} \, \mathbf{a}_{1}-z_{1} \, \mathbf{a}_{2} & = & -z_{1}c \, \mathbf{\hat{z}} & \left(4e\right) & \mbox{Cl} \\ 
\mathbf{B}_{3} & = & z_{2} \, \mathbf{a}_{1} + z_{2} \, \mathbf{a}_{2} & = & z_{2}c \, \mathbf{\hat{z}} & \left(4e\right) & \mbox{Hg} \\ 
\mathbf{B}_{4} & = & -z_{2} \, \mathbf{a}_{1}-z_{2} \, \mathbf{a}_{2} & = & -z_{2}c \, \mathbf{\hat{z}} & \left(4e\right) & \mbox{Hg} \\ 
\end{longtabu}
\renewcommand{\arraystretch}{1.0}
\noindent \hrulefill
\\
\textbf{References:}
\vspace*{-0.25cm}
\begin{flushleft}
  - \bibentry{Calos_Z_Kristall_187_1989}. \\
\end{flushleft}
\textbf{Found in:}
\vspace*{-0.25cm}
\begin{flushleft}
  - \bibentry{Downs_Am_Min_88_2003}. \\
\end{flushleft}
\noindent \hrulefill
\\
\textbf{Geometry files:}
\\
\noindent  - CIF: pp. {\hyperref[AB_tI8_139_e_e_cif]{\pageref{AB_tI8_139_e_e_cif}}} \\
\noindent  - POSCAR: pp. {\hyperref[AB_tI8_139_e_e_poscar]{\pageref{AB_tI8_139_e_e_poscar}}} \\
\onecolumn
{\phantomsection\label{A3B5_tI32_140_ah_bk}}
\subsection*{\huge \textbf{{\normalfont W$_{5}$Si$_{3}$ ($D8_{m}$) Structure: A3B5\_tI32\_140\_ah\_bk}}}
\noindent \hrulefill
\vspace*{0.25cm}
\begin{figure}[htp]
  \centering
  \vspace{-1em}
  {\includegraphics[width=1\textwidth]{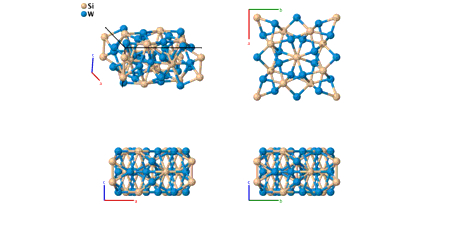}}
\end{figure}
\vspace*{-0.5cm}
\renewcommand{\arraystretch}{1.5}
\begin{equation*}
  \begin{array}{>{$\hspace{-0.15cm}}l<{$}>{$}p{0.5cm}<{$}>{$}p{18.5cm}<{$}}
    \mbox{\large \textbf{Prototype}} &\colon & \ce{W5Si3} \\
    \mbox{\large \textbf{\AFLOW\ prototype label}} &\colon & \mbox{A3B5\_tI32\_140\_ah\_bk} \\
    \mbox{\large \textbf{\textit{Strukturbericht} designation}} &\colon & \mbox{$D8_{m}$} \\
    \mbox{\large \textbf{Pearson symbol}} &\colon & \mbox{tI32} \\
    \mbox{\large \textbf{Space group number}} &\colon & 140 \\
    \mbox{\large \textbf{Space group symbol}} &\colon & I4/mcm \\
    \mbox{\large \textbf{\AFLOW\ prototype command}} &\colon &  \texttt{aflow} \,  \, \texttt{-{}-proto=A3B5\_tI32\_140\_ah\_bk } \, \newline \texttt{-{}-params=}{a,c/a,x_{3},x_{4},y_{4} }
  \end{array}
\end{equation*}
\renewcommand{\arraystretch}{1.0}

\vspace*{-0.25cm}
\noindent \hrulefill
\\
\textbf{ Other compounds with this structure:}
\begin{itemize}
   \item{ Cr$_{5}$Ge$_{3}$, Cr$_{5}$Si$_{3}$, Mo$_{5}$Si$_{3}$, Nb$_{5}$Si$_{3}$, Ta$_{5}$Si$_{3}$, V$_{5}$Si$_{3}$  }
\end{itemize}
\vspace*{-0.25cm}
\noindent \hrulefill
\begin{itemize}
  \item{(Pearson, 1958) refers to this as the ``T1 phase.''
}
\end{itemize}

\noindent \parbox{1 \linewidth}{
\noindent \hrulefill
\\
\textbf{Body-centered Tetragonal primitive vectors:} \\
\vspace*{-0.25cm}
\begin{tabular}{cc}
  \begin{tabular}{c}
    \parbox{0.6 \linewidth}{
      \renewcommand{\arraystretch}{1.5}
      \begin{equation*}
        \centering
        \begin{array}{ccc}
              \mathbf{a}_1 & = & - \frac12 \, a \, \mathbf{\hat{x}} + \frac12 \, a \, \mathbf{\hat{y}} + \frac12 \, c \, \mathbf{\hat{z}} \\
    \mathbf{a}_2 & = & ~ \frac12 \, a \, \mathbf{\hat{x}} - \frac12 \, a \, \mathbf{\hat{y}} + \frac12 \, c \, \mathbf{\hat{z}} \\
    \mathbf{a}_3 & = & ~ \frac12 \, a \, \mathbf{\hat{x}} + \frac12 \, a \, \mathbf{\hat{y}} - \frac12 \, c \, \mathbf{\hat{z}} \\

        \end{array}
      \end{equation*}
    }
    \renewcommand{\arraystretch}{1.0}
  \end{tabular}
  \begin{tabular}{c}
    \includegraphics[width=0.3\linewidth]{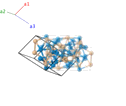} \\
  \end{tabular}
\end{tabular}

}
\vspace*{-0.25cm}

\noindent \hrulefill
\\
\textbf{Basis vectors:}
\vspace*{-0.25cm}
\renewcommand{\arraystretch}{1.5}
\begin{longtabu} to \textwidth{>{\centering $}X[-1,c,c]<{$}>{\centering $}X[-1,c,c]<{$}>{\centering $}X[-1,c,c]<{$}>{\centering $}X[-1,c,c]<{$}>{\centering $}X[-1,c,c]<{$}>{\centering $}X[-1,c,c]<{$}>{\centering $}X[-1,c,c]<{$}}
  & & \mbox{Lattice Coordinates} & & \mbox{Cartesian Coordinates} &\mbox{Wyckoff Position} & \mbox{Atom Type} \\  
  \mathbf{B}_{1} & = & \frac{1}{4} \, \mathbf{a}_{1} + \frac{1}{4} \, \mathbf{a}_{2} & = & \frac{1}{4}c \, \mathbf{\hat{z}} & \left(4a\right) & \mbox{Si I} \\ 
\mathbf{B}_{2} & = & \frac{3}{4} \, \mathbf{a}_{1} + \frac{3}{4} \, \mathbf{a}_{2} & = & \frac{3}{4}c \, \mathbf{\hat{z}} & \left(4a\right) & \mbox{Si I} \\ 
\mathbf{B}_{3} & = & \frac{3}{4} \, \mathbf{a}_{1} + \frac{1}{4} \, \mathbf{a}_{2} + \frac{1}{2} \, \mathbf{a}_{3} & = & \frac{1}{2}a \, \mathbf{\hat{y}} + \frac{1}{4}c \, \mathbf{\hat{z}} & \left(4b\right) & \mbox{W I} \\ 
\mathbf{B}_{4} & = & \frac{1}{4} \, \mathbf{a}_{1} + \frac{3}{4} \, \mathbf{a}_{2} + \frac{1}{2} \, \mathbf{a}_{3} & = & \frac{1}{2}a \, \mathbf{\hat{x}} + \frac{1}{4}c \, \mathbf{\hat{z}} & \left(4b\right) & \mbox{W I} \\ 
\mathbf{B}_{5} & = & \left(\frac{1}{2} +x_{3}\right) \, \mathbf{a}_{1} + x_{3} \, \mathbf{a}_{2} + \left(\frac{1}{2} +2x_{3}\right) \, \mathbf{a}_{3} & = & x_{3}a \, \mathbf{\hat{x}} + \left(\frac{1}{2} +x_{3}\right)a \, \mathbf{\hat{y}} & \left(8h\right) & \mbox{Si II} \\ 
\mathbf{B}_{6} & = & \left(\frac{1}{2} - x_{3}\right) \, \mathbf{a}_{1}-x_{3} \, \mathbf{a}_{2} + \left(\frac{1}{2} - 2x_{3}\right) \, \mathbf{a}_{3} & = & -x_{3}a \, \mathbf{\hat{x}} + \left(\frac{1}{2} - x_{3}\right)a \, \mathbf{\hat{y}} & \left(8h\right) & \mbox{Si II} \\ 
\mathbf{B}_{7} & = & x_{3} \, \mathbf{a}_{1} + \left(\frac{1}{2} - x_{3}\right) \, \mathbf{a}_{2} + \frac{1}{2} \, \mathbf{a}_{3} & = & \left(\frac{1}{2} - x_{3}\right)a \, \mathbf{\hat{x}} + x_{3}a \, \mathbf{\hat{y}} & \left(8h\right) & \mbox{Si II} \\ 
\mathbf{B}_{8} & = & -x_{3} \, \mathbf{a}_{1} + \left(\frac{1}{2} +x_{3}\right) \, \mathbf{a}_{2} + \frac{1}{2} \, \mathbf{a}_{3} & = & \left(\frac{1}{2} +x_{3}\right)a \, \mathbf{\hat{x}}-x_{3}a \, \mathbf{\hat{y}} & \left(8h\right) & \mbox{Si II} \\ 
\mathbf{B}_{9} & = & y_{4} \, \mathbf{a}_{1} + x_{4} \, \mathbf{a}_{2} + \left(x_{4}+y_{4}\right) \, \mathbf{a}_{3} & = & x_{4}a \, \mathbf{\hat{x}} + y_{4}a \, \mathbf{\hat{y}} & \left(16k\right) & \mbox{W II} \\ 
\mathbf{B}_{10} & = & -y_{4} \, \mathbf{a}_{1}-x_{4} \, \mathbf{a}_{2} + \left(-x_{4}-y_{4}\right) \, \mathbf{a}_{3} & = & -x_{4}a \, \mathbf{\hat{x}}-y_{4}a \, \mathbf{\hat{y}} & \left(16k\right) & \mbox{W II} \\ 
\mathbf{B}_{11} & = & x_{4} \, \mathbf{a}_{1}-y_{4} \, \mathbf{a}_{2} + \left(x_{4}-y_{4}\right) \, \mathbf{a}_{3} & = & -y_{4}a \, \mathbf{\hat{x}} + x_{4}a \, \mathbf{\hat{y}} & \left(16k\right) & \mbox{W II} \\ 
\mathbf{B}_{12} & = & -x_{4} \, \mathbf{a}_{1} + y_{4} \, \mathbf{a}_{2} + \left(-x_{4}+y_{4}\right) \, \mathbf{a}_{3} & = & y_{4}a \, \mathbf{\hat{x}}-x_{4}a \, \mathbf{\hat{y}} & \left(16k\right) & \mbox{W II} \\ 
\mathbf{B}_{13} & = & \left(\frac{1}{2} +y_{4}\right) \, \mathbf{a}_{1} + \left(\frac{1}{2} - x_{4}\right) \, \mathbf{a}_{2} + \left(-x_{4}+y_{4}\right) \, \mathbf{a}_{3} & = & -x_{4}a \, \mathbf{\hat{x}} + y_{4}a \, \mathbf{\hat{y}} + \frac{1}{2}c \, \mathbf{\hat{z}} & \left(16k\right) & \mbox{W II} \\ 
\mathbf{B}_{14} & = & \left(\frac{1}{2} - y_{4}\right) \, \mathbf{a}_{1} + \left(\frac{1}{2} +x_{4}\right) \, \mathbf{a}_{2} + \left(x_{4}-y_{4}\right) \, \mathbf{a}_{3} & = & x_{4}a \, \mathbf{\hat{x}}-y_{4}a \, \mathbf{\hat{y}} + \frac{1}{2}c \, \mathbf{\hat{z}} & \left(16k\right) & \mbox{W II} \\ 
\mathbf{B}_{15} & = & \left(\frac{1}{2} +x_{4}\right) \, \mathbf{a}_{1} + \left(\frac{1}{2} +y_{4}\right) \, \mathbf{a}_{2} + \left(x_{4}+y_{4}\right) \, \mathbf{a}_{3} & = & y_{4}a \, \mathbf{\hat{x}} + x_{4}a \, \mathbf{\hat{y}} + \frac{1}{2}c \, \mathbf{\hat{z}} & \left(16k\right) & \mbox{W II} \\ 
\mathbf{B}_{16} & = & \left(\frac{1}{2} - x_{4}\right) \, \mathbf{a}_{1} + \left(\frac{1}{2} - y_{4}\right) \, \mathbf{a}_{2} + \left(-x_{4}-y_{4}\right) \, \mathbf{a}_{3} & = & -y_{4}a \, \mathbf{\hat{x}}-x_{4}a \, \mathbf{\hat{y}} + \frac{1}{2}c \, \mathbf{\hat{z}} & \left(16k\right) & \mbox{W II} \\ 
\end{longtabu}
\renewcommand{\arraystretch}{1.0}
\noindent \hrulefill
\\
\textbf{References:}
\vspace*{-0.25cm}
\begin{flushleft}
  - \bibentry{Aronsson_Acta_Chem_Scand_9_1955}. \\
\end{flushleft}
\noindent \hrulefill
\\
\textbf{Geometry files:}
\\
\noindent  - CIF: pp. {\hyperref[A3B5_tI32_140_ah_bk_cif]{\pageref{A3B5_tI32_140_ah_bk_cif}}} \\
\noindent  - POSCAR: pp. {\hyperref[A3B5_tI32_140_ah_bk_poscar]{\pageref{A3B5_tI32_140_ah_bk_poscar}}} \\
\onecolumn
{\phantomsection\label{A3B5_tI32_140_ah_cl}}
\subsection*{\huge \textbf{{\normalfont Cr$_{5}$B$_{3}$ ($D8_{l}$) Structure: A3B5\_tI32\_140\_ah\_cl}}}
\noindent \hrulefill
\vspace*{0.25cm}
\begin{figure}[htp]
  \centering
  \vspace{-1em}
  {\includegraphics[width=1\textwidth]{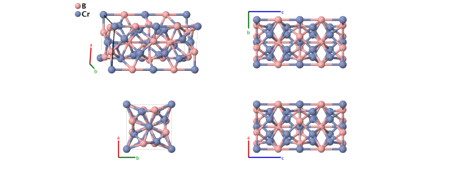}}
\end{figure}
\vspace*{-0.5cm}
\renewcommand{\arraystretch}{1.5}
\begin{equation*}
  \begin{array}{>{$\hspace{-0.15cm}}l<{$}>{$}p{0.5cm}<{$}>{$}p{18.5cm}<{$}}
    \mbox{\large \textbf{Prototype}} &\colon & \ce{Cr5B3} \\
    \mbox{\large \textbf{\AFLOW\ prototype label}} &\colon & \mbox{A3B5\_tI32\_140\_ah\_cl} \\
    \mbox{\large \textbf{\textit{Strukturbericht} designation}} &\colon & \mbox{$D8_{l}$} \\
    \mbox{\large \textbf{Pearson symbol}} &\colon & \mbox{tI32} \\
    \mbox{\large \textbf{Space group number}} &\colon & 140 \\
    \mbox{\large \textbf{Space group symbol}} &\colon & I4/mcm \\
    \mbox{\large \textbf{\AFLOW\ prototype command}} &\colon &  \texttt{aflow} \,  \, \texttt{-{}-proto=A3B5\_tI32\_140\_ah\_cl } \, \newline \texttt{-{}-params=}{a,c/a,x_{3},x_{4},z_{4} }
  \end{array}
\end{equation*}
\renewcommand{\arraystretch}{1.0}

\vspace*{-0.25cm}
\noindent \hrulefill
\\
\textbf{ Other compounds with this structure:}
\begin{itemize}
   \item{ Eu$_{5}$Ge$_{3}$, Pr$_{5}$Si$_{3}$, Ta$_{5}$Ge$_{3}$, Nb$_{5}$Si$_{3}$, Ta$_{5}$Si$_{3}$, EuIrGe$_{2}$, Fe$_{5}$PB$_{2}$, Fe$_{5}$SiB$_{2}$, Fe$_{4}$CoPB$_{2}$, Fe$_{4}$MnPB$_{2}$, Fe$_{4}$CoSiB$_{2}$, Fe$_{4}$MnSiB$_{2}$, Mo$_{5}$SiB$_{2}$, Sr$_{5}$In$_{3}$, Gd$_{5}$CoSi$_{2}$, Ca$_{5}$Si$_{3}$,  }
\end{itemize}
\vspace*{-0.25cm}
\noindent \hrulefill
\begin{itemize}
  \item{We have been unable to obtain a copy of (Bertaut, 1953), so we use the
data found online from (Downs, 2003).
Although Cr$_{5}$B$_{3}$ is universally regarded as the prototype for
$D8_{l}$, it appears that no determination of the internal parameters
has ever been made.  
The values found in (Bertaut, 1953) seem to be
resonable choices.
}
\end{itemize}

\noindent \parbox{1 \linewidth}{
\noindent \hrulefill
\\
\textbf{Body-centered Tetragonal primitive vectors:} \\
\vspace*{-0.25cm}
\begin{tabular}{cc}
  \begin{tabular}{c}
    \parbox{0.6 \linewidth}{
      \renewcommand{\arraystretch}{1.5}
      \begin{equation*}
        \centering
        \begin{array}{ccc}
              \mathbf{a}_1 & = & - \frac12 \, a \, \mathbf{\hat{x}} + \frac12 \, a \, \mathbf{\hat{y}} + \frac12 \, c \, \mathbf{\hat{z}} \\
    \mathbf{a}_2 & = & ~ \frac12 \, a \, \mathbf{\hat{x}} - \frac12 \, a \, \mathbf{\hat{y}} + \frac12 \, c \, \mathbf{\hat{z}} \\
    \mathbf{a}_3 & = & ~ \frac12 \, a \, \mathbf{\hat{x}} + \frac12 \, a \, \mathbf{\hat{y}} - \frac12 \, c \, \mathbf{\hat{z}} \\

        \end{array}
      \end{equation*}
    }
    \renewcommand{\arraystretch}{1.0}
  \end{tabular}
  \begin{tabular}{c}
    \includegraphics[width=0.3\linewidth]{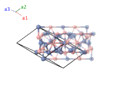} \\
  \end{tabular}
\end{tabular}

}
\vspace*{-0.25cm}

\noindent \hrulefill
\\
\textbf{Basis vectors:}
\vspace*{-0.25cm}
\renewcommand{\arraystretch}{1.5}
\begin{longtabu} to \textwidth{>{\centering $}X[-1,c,c]<{$}>{\centering $}X[-1,c,c]<{$}>{\centering $}X[-1,c,c]<{$}>{\centering $}X[-1,c,c]<{$}>{\centering $}X[-1,c,c]<{$}>{\centering $}X[-1,c,c]<{$}>{\centering $}X[-1,c,c]<{$}}
  & & \mbox{Lattice Coordinates} & & \mbox{Cartesian Coordinates} &\mbox{Wyckoff Position} & \mbox{Atom Type} \\  
  \mathbf{B}_{1} & = & \frac{1}{4} \, \mathbf{a}_{1} + \frac{1}{4} \, \mathbf{a}_{2} & = & \frac{1}{4}c \, \mathbf{\hat{z}} & \left(4a\right) & \mbox{B I} \\ 
\mathbf{B}_{2} & = & \frac{3}{4} \, \mathbf{a}_{1} + \frac{3}{4} \, \mathbf{a}_{2} & = & \frac{3}{4}c \, \mathbf{\hat{z}} & \left(4a\right) & \mbox{B I} \\ 
\mathbf{B}_{3} & = & 0 \, \mathbf{a}_{1} + 0 \, \mathbf{a}_{2} + 0 \, \mathbf{a}_{3} & = & 0 \, \mathbf{\hat{x}} + 0 \, \mathbf{\hat{y}} + 0 \, \mathbf{\hat{z}} & \left(4c\right) & \mbox{Cr I} \\ 
\mathbf{B}_{4} & = & \frac{1}{2} \, \mathbf{a}_{1} + \frac{1}{2} \, \mathbf{a}_{2} & = & \frac{1}{2}c \, \mathbf{\hat{z}} & \left(4c\right) & \mbox{Cr I} \\ 
\mathbf{B}_{5} & = & \left(\frac{1}{2} +x_{3}\right) \, \mathbf{a}_{1} + x_{3} \, \mathbf{a}_{2} + \left(\frac{1}{2} +2x_{3}\right) \, \mathbf{a}_{3} & = & x_{3}a \, \mathbf{\hat{x}} + \left(\frac{1}{2} +x_{3}\right)a \, \mathbf{\hat{y}} & \left(8h\right) & \mbox{B II} \\ 
\mathbf{B}_{6} & = & \left(\frac{1}{2} - x_{3}\right) \, \mathbf{a}_{1}-x_{3} \, \mathbf{a}_{2} + \left(\frac{1}{2} - 2x_{3}\right) \, \mathbf{a}_{3} & = & -x_{3}a \, \mathbf{\hat{x}} + \left(\frac{1}{2} - x_{3}\right)a \, \mathbf{\hat{y}} & \left(8h\right) & \mbox{B II} \\ 
\mathbf{B}_{7} & = & x_{3} \, \mathbf{a}_{1} + \left(\frac{1}{2} - x_{3}\right) \, \mathbf{a}_{2} + \frac{1}{2} \, \mathbf{a}_{3} & = & \left(\frac{1}{2} - x_{3}\right)a \, \mathbf{\hat{x}} + x_{3}a \, \mathbf{\hat{y}} & \left(8h\right) & \mbox{B II} \\ 
\mathbf{B}_{8} & = & -x_{3} \, \mathbf{a}_{1} + \left(\frac{1}{2} +x_{3}\right) \, \mathbf{a}_{2} + \frac{1}{2} \, \mathbf{a}_{3} & = & \left(\frac{1}{2} +x_{3}\right)a \, \mathbf{\hat{x}}-x_{3}a \, \mathbf{\hat{y}} & \left(8h\right) & \mbox{B II} \\ 
\mathbf{B}_{9} & = & \left(\frac{1}{2} +x_{4} + z_{4}\right) \, \mathbf{a}_{1} + \left(x_{4}+z_{4}\right) \, \mathbf{a}_{2} + \left(\frac{1}{2} +2x_{4}\right) \, \mathbf{a}_{3} & = & x_{4}a \, \mathbf{\hat{x}} + \left(\frac{1}{2} +x_{4}\right)a \, \mathbf{\hat{y}} + z_{4}c \, \mathbf{\hat{z}} & \left(16l\right) & \mbox{Cr II} \\ 
\mathbf{B}_{10} & = & \left(\frac{1}{2} - x_{4} + z_{4}\right) \, \mathbf{a}_{1} + \left(-x_{4}+z_{4}\right) \, \mathbf{a}_{2} + \left(\frac{1}{2} - 2x_{4}\right) \, \mathbf{a}_{3} & = & -x_{4}a \, \mathbf{\hat{x}} + \left(\frac{1}{2} - x_{4}\right)a \, \mathbf{\hat{y}} + z_{4}c \, \mathbf{\hat{z}} & \left(16l\right) & \mbox{Cr II} \\ 
\mathbf{B}_{11} & = & \left(x_{4}+z_{4}\right) \, \mathbf{a}_{1} + \left(\frac{1}{2} - x_{4} + z_{4}\right) \, \mathbf{a}_{2} + \frac{1}{2} \, \mathbf{a}_{3} & = & \left(\frac{1}{2} - x_{4}\right)a \, \mathbf{\hat{x}} + x_{4}a \, \mathbf{\hat{y}} + z_{4}c \, \mathbf{\hat{z}} & \left(16l\right) & \mbox{Cr II} \\ 
\mathbf{B}_{12} & = & \left(-x_{4}+z_{4}\right) \, \mathbf{a}_{1} + \left(\frac{1}{2} +x_{4} + z_{4}\right) \, \mathbf{a}_{2} + \frac{1}{2} \, \mathbf{a}_{3} & = & \left(\frac{1}{2} +x_{4}\right)a \, \mathbf{\hat{x}}-x_{4}a \, \mathbf{\hat{y}} + z_{4}c \, \mathbf{\hat{z}} & \left(16l\right) & \mbox{Cr II} \\ 
\mathbf{B}_{13} & = & \left(x_{4}-z_{4}\right) \, \mathbf{a}_{1} + \left(\frac{1}{2} - x_{4} - z_{4}\right) \, \mathbf{a}_{2} + \frac{1}{2} \, \mathbf{a}_{3} & = & \left(\frac{1}{2} - x_{4}\right)a \, \mathbf{\hat{x}} + x_{4}a \, \mathbf{\hat{y}}-z_{4}c \, \mathbf{\hat{z}} & \left(16l\right) & \mbox{Cr II} \\ 
\mathbf{B}_{14} & = & \left(-x_{4}-z_{4}\right) \, \mathbf{a}_{1} + \left(\frac{1}{2} +x_{4} - z_{4}\right) \, \mathbf{a}_{2} + \frac{1}{2} \, \mathbf{a}_{3} & = & \left(\frac{1}{2} +x_{4}\right)a \, \mathbf{\hat{x}}-x_{4}a \, \mathbf{\hat{y}}-z_{4}c \, \mathbf{\hat{z}} & \left(16l\right) & \mbox{Cr II} \\ 
\mathbf{B}_{15} & = & \left(\frac{1}{2} +x_{4} - z_{4}\right) \, \mathbf{a}_{1} + \left(x_{4}-z_{4}\right) \, \mathbf{a}_{2} + \left(\frac{1}{2} +2x_{4}\right) \, \mathbf{a}_{3} & = & x_{4}a \, \mathbf{\hat{x}} + \left(\frac{1}{2} +x_{4}\right)a \, \mathbf{\hat{y}}-z_{4}c \, \mathbf{\hat{z}} & \left(16l\right) & \mbox{Cr II} \\ 
\mathbf{B}_{16} & = & \left(\frac{1}{2} - x_{4} - z_{4}\right) \, \mathbf{a}_{1} + \left(-x_{4}-z_{4}\right) \, \mathbf{a}_{2} + \left(\frac{1}{2} - 2x_{4}\right) \, \mathbf{a}_{3} & = & -x_{4}a \, \mathbf{\hat{x}} + \left(\frac{1}{2} - x_{4}\right)a \, \mathbf{\hat{y}}-z_{4}c \, \mathbf{\hat{z}} & \left(16l\right) & \mbox{Cr II} \\ 
\end{longtabu}
\renewcommand{\arraystretch}{1.0}
\noindent \hrulefill
\\
\textbf{References:}
\vspace*{-0.25cm}
\begin{flushleft}
  - \bibentry{Bertaut_CRAS_236_1953}. \\
\end{flushleft}
\textbf{Found in:}
\vspace*{-0.25cm}
\begin{flushleft}
  - \bibentry{Downs_Am_Min_88_2003}. \\
\end{flushleft}
\noindent \hrulefill
\\
\textbf{Geometry files:}
\\
\noindent  - CIF: pp. {\hyperref[A3B5_tI32_140_ah_cl_cif]{\pageref{A3B5_tI32_140_ah_cl_cif}}} \\
\noindent  - POSCAR: pp. {\hyperref[A3B5_tI32_140_ah_cl_poscar]{\pageref{A3B5_tI32_140_ah_cl_poscar}}} \\
\onecolumn
{\phantomsection\label{A2B_tI12_141_e_a}}
\subsection*{\huge \textbf{{\normalfont $\alpha$-ThSi$_{2}$ ($C_{c}$) Structure: A2B\_tI12\_141\_e\_a}}}
\noindent \hrulefill
\vspace*{0.25cm}
\begin{figure}[htp]
  \centering
  \vspace{-1em}
  {\includegraphics[width=1\textwidth]{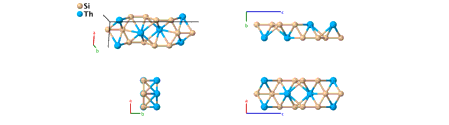}}
\end{figure}
\vspace*{-0.5cm}
\renewcommand{\arraystretch}{1.5}
\begin{equation*}
  \begin{array}{>{$\hspace{-0.15cm}}l<{$}>{$}p{0.5cm}<{$}>{$}p{18.5cm}<{$}}
    \mbox{\large \textbf{Prototype}} &\colon & \ce{$\alpha$-ThSi2} \\
    \mbox{\large \textbf{\AFLOW\ prototype label}} &\colon & \mbox{A2B\_tI12\_141\_e\_a} \\
    \mbox{\large \textbf{\textit{Strukturbericht} designation}} &\colon & \mbox{$C_{c}$} \\
    \mbox{\large \textbf{Pearson symbol}} &\colon & \mbox{tI12} \\
    \mbox{\large \textbf{Space group number}} &\colon & 141 \\
    \mbox{\large \textbf{Space group symbol}} &\colon & I4_{1}/amd \\
    \mbox{\large \textbf{\AFLOW\ prototype command}} &\colon &  \texttt{aflow} \,  \, \texttt{-{}-proto=A2B\_tI12\_141\_e\_a } \, \newline \texttt{-{}-params=}{a,c/a,z_{2} }
  \end{array}
\end{equation*}
\renewcommand{\arraystretch}{1.0}

\noindent \parbox{1 \linewidth}{
\noindent \hrulefill
\\
\textbf{Body-centered Tetragonal primitive vectors:} \\
\vspace*{-0.25cm}
\begin{tabular}{cc}
  \begin{tabular}{c}
    \parbox{0.6 \linewidth}{
      \renewcommand{\arraystretch}{1.5}
      \begin{equation*}
        \centering
        \begin{array}{ccc}
              \mathbf{a}_1 & = & - \frac12 \, a \, \mathbf{\hat{x}} + \frac12 \, a \, \mathbf{\hat{y}} + \frac12 \, c \, \mathbf{\hat{z}} \\
    \mathbf{a}_2 & = & ~ \frac12 \, a \, \mathbf{\hat{x}} - \frac12 \, a \, \mathbf{\hat{y}} + \frac12 \, c \, \mathbf{\hat{z}} \\
    \mathbf{a}_3 & = & ~ \frac12 \, a \, \mathbf{\hat{x}} + \frac12 \, a \, \mathbf{\hat{y}} - \frac12 \, c \, \mathbf{\hat{z}} \\

        \end{array}
      \end{equation*}
    }
    \renewcommand{\arraystretch}{1.0}
  \end{tabular}
  \begin{tabular}{c}
    \includegraphics[width=0.3\linewidth]{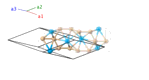} \\
  \end{tabular}
\end{tabular}

}
\vspace*{-0.25cm}

\noindent \hrulefill
\\
\textbf{Basis vectors:}
\vspace*{-0.25cm}
\renewcommand{\arraystretch}{1.5}
\begin{longtabu} to \textwidth{>{\centering $}X[-1,c,c]<{$}>{\centering $}X[-1,c,c]<{$}>{\centering $}X[-1,c,c]<{$}>{\centering $}X[-1,c,c]<{$}>{\centering $}X[-1,c,c]<{$}>{\centering $}X[-1,c,c]<{$}>{\centering $}X[-1,c,c]<{$}}
  & & \mbox{Lattice Coordinates} & & \mbox{Cartesian Coordinates} &\mbox{Wyckoff Position} & \mbox{Atom Type} \\  
  \mathbf{B}_{1} & = & \frac{7}{8} \, \mathbf{a}_{1} + \frac{1}{8} \, \mathbf{a}_{2} + \frac{3}{4} \, \mathbf{a}_{3} & = & \frac{3}{4}a \, \mathbf{\hat{y}} + \frac{1}{8}c \, \mathbf{\hat{z}} & \left(4a\right) & \mbox{Th} \\ 
\mathbf{B}_{2} & = & \frac{1}{8} \, \mathbf{a}_{1} + \frac{7}{8} \, \mathbf{a}_{2} + \frac{1}{4} \, \mathbf{a}_{3} & = & \frac{1}{2}a \, \mathbf{\hat{x}} + \frac{3}{4}a \, \mathbf{\hat{y}} + \frac{3}{8}c \, \mathbf{\hat{z}} & \left(4a\right) & \mbox{Th} \\ 
\mathbf{B}_{3} & = & \left(\frac{1}{4} +z_{2}\right) \, \mathbf{a}_{1} + z_{2} \, \mathbf{a}_{2} + \frac{1}{4} \, \mathbf{a}_{3} & = & \frac{1}{4}a \, \mathbf{\hat{y}} + z_{2}c \, \mathbf{\hat{z}} & \left(8e\right) & \mbox{Si} \\ 
\mathbf{B}_{4} & = & z_{2} \, \mathbf{a}_{1} + \left(\frac{1}{4} +z_{2}\right) \, \mathbf{a}_{2} + \frac{3}{4} \, \mathbf{a}_{3} & = & \frac{1}{2}a \, \mathbf{\hat{x}} + \frac{1}{4}a \, \mathbf{\hat{y}} + \left(- \frac{1}{4} +z_{2}\right)c \, \mathbf{\hat{z}} & \left(8e\right) & \mbox{Si} \\ 
\mathbf{B}_{5} & = & \left(\frac{3}{4} - z_{2}\right) \, \mathbf{a}_{1}-z_{2} \, \mathbf{a}_{2} + \frac{3}{4} \, \mathbf{a}_{3} & = & \frac{3}{4}a \, \mathbf{\hat{y}}-z_{2}c \, \mathbf{\hat{z}} & \left(8e\right) & \mbox{Si} \\ 
\mathbf{B}_{6} & = & -z_{2} \, \mathbf{a}_{1} + \left(\frac{3}{4} - z_{2}\right) \, \mathbf{a}_{2} + \frac{1}{4} \, \mathbf{a}_{3} & = & \frac{1}{2}a \, \mathbf{\hat{x}} + \frac{3}{4}a \, \mathbf{\hat{y}} + \left(\frac{1}{4} - z_{2}\right)c \, \mathbf{\hat{z}} & \left(8e\right) & \mbox{Si} \\ 
\end{longtabu}
\renewcommand{\arraystretch}{1.0}
\noindent \hrulefill
\\
\textbf{References:}
\vspace*{-0.25cm}
\begin{flushleft}
  - \bibentry{Brauer_ZAAC_249_1942}. \\
\end{flushleft}
\textbf{Found in:}
\vspace*{-0.25cm}
\begin{flushleft}
  - \bibentry{Pearson_Alloys_1972}. \\
\end{flushleft}
\noindent \hrulefill
\\
\textbf{Geometry files:}
\\
\noindent  - CIF: pp. {\hyperref[A2B_tI12_141_e_a_cif]{\pageref{A2B_tI12_141_e_a_cif}}} \\
\noindent  - POSCAR: pp. {\hyperref[A2B_tI12_141_e_a_poscar]{\pageref{A2B_tI12_141_e_a_poscar}}} \\
\onecolumn
{\phantomsection\label{A_tI16_142_f}}
\subsection*{\huge \textbf{{\normalfont S-III Structure: A\_tI16\_142\_f}}}
\noindent \hrulefill
\vspace*{0.25cm}
\begin{figure}[htp]
  \centering
  \vspace{-1em}
  {\includegraphics[width=1\textwidth]{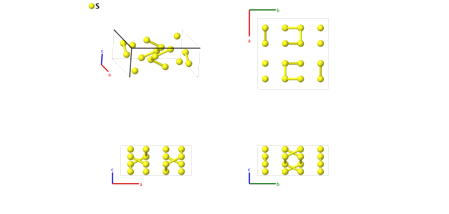}}
\end{figure}
\vspace*{-0.5cm}
\renewcommand{\arraystretch}{1.5}
\begin{equation*}
  \begin{array}{>{$\hspace{-0.15cm}}l<{$}>{$}p{0.5cm}<{$}>{$}p{18.5cm}<{$}}
    \mbox{\large \textbf{Prototype}} &\colon & \ce{S} \\
    \mbox{\large \textbf{\AFLOW\ prototype label}} &\colon & \mbox{A\_tI16\_142\_f} \\
    \mbox{\large \textbf{\textit{Strukturbericht} designation}} &\colon & \mbox{None} \\
    \mbox{\large \textbf{Pearson symbol}} &\colon & \mbox{tI16} \\
    \mbox{\large \textbf{Space group number}} &\colon & 142 \\
    \mbox{\large \textbf{Space group symbol}} &\colon & I4_{1}/acd \\
    \mbox{\large \textbf{\AFLOW\ prototype command}} &\colon &  \texttt{aflow} \,  \, \texttt{-{}-proto=A\_tI16\_142\_f } \, \newline \texttt{-{}-params=}{a,c/a,x_{1} }
  \end{array}
\end{equation*}
\renewcommand{\arraystretch}{1.0}

\vspace*{-0.25cm}
\noindent \hrulefill
\\
\textbf{ Other elements with this structure:}
\begin{itemize}
   \item{  Se (Se-VII, prepared at 450 K and 20 GPa)  }
\end{itemize}
\vspace*{-0.25cm}
\noindent \hrulefill
\begin{itemize}
  \item{The S-III phase is found when sulfur is pressurized above 36~GPa at
300~K.  At 300~K it is stable up to 83~GPa.}
  \item{This data was taken at 12~GPa and 300~K.
}
\end{itemize}

\noindent \parbox{1 \linewidth}{
\noindent \hrulefill
\\
\textbf{Body-centered Tetragonal primitive vectors:} \\
\vspace*{-0.25cm}
\begin{tabular}{cc}
  \begin{tabular}{c}
    \parbox{0.6 \linewidth}{
      \renewcommand{\arraystretch}{1.5}
      \begin{equation*}
        \centering
        \begin{array}{ccc}
              \mathbf{a}_1 & = & - \frac12 \, a \, \mathbf{\hat{x}} + \frac12 \, a \, \mathbf{\hat{y}} + \frac12 \, c \, \mathbf{\hat{z}} \\
    \mathbf{a}_2 & = & ~ \frac12 \, a \, \mathbf{\hat{x}} - \frac12 \, a \, \mathbf{\hat{y}} + \frac12 \, c \, \mathbf{\hat{z}} \\
    \mathbf{a}_3 & = & ~ \frac12 \, a \, \mathbf{\hat{x}} + \frac12 \, a \, \mathbf{\hat{y}} - \frac12 \, c \, \mathbf{\hat{z}} \\

        \end{array}
      \end{equation*}
    }
    \renewcommand{\arraystretch}{1.0}
  \end{tabular}
  \begin{tabular}{c}
    \includegraphics[width=0.3\linewidth]{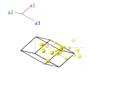} \\
  \end{tabular}
\end{tabular}

}
\vspace*{-0.25cm}

\noindent \hrulefill
\\
\textbf{Basis vectors:}
\vspace*{-0.25cm}
\renewcommand{\arraystretch}{1.5}
\begin{longtabu} to \textwidth{>{\centering $}X[-1,c,c]<{$}>{\centering $}X[-1,c,c]<{$}>{\centering $}X[-1,c,c]<{$}>{\centering $}X[-1,c,c]<{$}>{\centering $}X[-1,c,c]<{$}>{\centering $}X[-1,c,c]<{$}>{\centering $}X[-1,c,c]<{$}}
  & & \mbox{Lattice Coordinates} & & \mbox{Cartesian Coordinates} &\mbox{Wyckoff Position} & \mbox{Atom Type} \\  
  \mathbf{B}_{1} & = & \left(\frac{3}{8} +x_{1}\right) \, \mathbf{a}_{1} + \left(\frac{1}{8} +x_{1}\right) \, \mathbf{a}_{2} + \left(\frac{1}{4} +2x_{1}\right) \, \mathbf{a}_{3} & = & x_{1}a \, \mathbf{\hat{x}} + \left(\frac{1}{4} +x_{1}\right)a \, \mathbf{\hat{y}} + \frac{1}{8}c \, \mathbf{\hat{z}} & \left(16f\right) & \mbox{S} \\ 
\mathbf{B}_{2} & = & \left(\frac{3}{8} - x_{1}\right) \, \mathbf{a}_{1} + \left(\frac{1}{8} - x_{1}\right) \, \mathbf{a}_{2} + \left(\frac{1}{4} - 2x_{1}\right) \, \mathbf{a}_{3} & = & -x_{1}a \, \mathbf{\hat{x}} + \left(\frac{1}{4} - x_{1}\right)a \, \mathbf{\hat{y}} + \frac{1}{8}c \, \mathbf{\hat{z}} & \left(16f\right) & \mbox{S} \\ 
\mathbf{B}_{3} & = & \left(\frac{1}{8} +x_{1}\right) \, \mathbf{a}_{1} + \left(\frac{3}{8} - x_{1}\right) \, \mathbf{a}_{2} + \frac{3}{4} \, \mathbf{a}_{3} & = & \left(\frac{1}{2} - x_{1}\right)a \, \mathbf{\hat{x}} + \left(\frac{1}{4} +x_{1}\right)a \, \mathbf{\hat{y}}- \frac{1}{8}c  \, \mathbf{\hat{z}} & \left(16f\right) & \mbox{S} \\ 
\mathbf{B}_{4} & = & \left(\frac{1}{8} - x_{1}\right) \, \mathbf{a}_{1} + \left(\frac{3}{8} +x_{1}\right) \, \mathbf{a}_{2} + \frac{3}{4} \, \mathbf{a}_{3} & = & \left(\frac{1}{2} +x_{1}\right)a \, \mathbf{\hat{x}} + \left(\frac{1}{4} - x_{1}\right)a \, \mathbf{\hat{y}} + \frac{7}{8}c \, \mathbf{\hat{z}} & \left(16f\right) & \mbox{S} \\ 
\mathbf{B}_{5} & = & \left(\frac{5}{8} - x_{1}\right) \, \mathbf{a}_{1} + \left(\frac{7}{8} - x_{1}\right) \, \mathbf{a}_{2} + \left(\frac{3}{4} - 2x_{1}\right) \, \mathbf{a}_{3} & = & \left(\frac{1}{2} - x_{1}\right)a \, \mathbf{\hat{x}} + \left(\frac{1}{4} - x_{1}\right)a \, \mathbf{\hat{y}} + \frac{3}{8}c \, \mathbf{\hat{z}} & \left(16f\right) & \mbox{S} \\ 
\mathbf{B}_{6} & = & \left(\frac{5}{8} +x_{1}\right) \, \mathbf{a}_{1} + \left(\frac{7}{8} +x_{1}\right) \, \mathbf{a}_{2} + \left(\frac{3}{4} +2x_{1}\right) \, \mathbf{a}_{3} & = & \left(\frac{1}{2} +x_{1}\right)a \, \mathbf{\hat{x}} + \left(\frac{1}{4} +x_{1}\right)a \, \mathbf{\hat{y}} + \frac{3}{8}c \, \mathbf{\hat{z}} & \left(16f\right) & \mbox{S} \\ 
\mathbf{B}_{7} & = & \left(\frac{7}{8} - x_{1}\right) \, \mathbf{a}_{1} + \left(\frac{5}{8} +x_{1}\right) \, \mathbf{a}_{2} + \frac{1}{4} \, \mathbf{a}_{3} & = & x_{1}a \, \mathbf{\hat{x}} + \left(\frac{1}{4} - x_{1}\right)a \, \mathbf{\hat{y}} + \frac{5}{8}c \, \mathbf{\hat{z}} & \left(16f\right) & \mbox{S} \\ 
\mathbf{B}_{8} & = & \left(\frac{7}{8} +x_{1}\right) \, \mathbf{a}_{1} + \left(\frac{5}{8} - x_{1}\right) \, \mathbf{a}_{2} + \frac{1}{4} \, \mathbf{a}_{3} & = & -x_{1}a \, \mathbf{\hat{x}} + \left(\frac{1}{4} +x_{1}\right)a \, \mathbf{\hat{y}} + \frac{5}{8}c \, \mathbf{\hat{z}} & \left(16f\right) & \mbox{S} \\ 
\end{longtabu}
\renewcommand{\arraystretch}{1.0}
\noindent \hrulefill
\\
\textbf{References:}
\vspace*{-0.25cm}
\begin{flushleft}
  - \bibentry{degtyareva05:SIII}. \\
\end{flushleft}
\noindent \hrulefill
\\
\textbf{Geometry files:}
\\
\noindent  - CIF: pp. {\hyperref[A_tI16_142_f_cif]{\pageref{A_tI16_142_f_cif}}} \\
\noindent  - POSCAR: pp. {\hyperref[A_tI16_142_f_poscar]{\pageref{A_tI16_142_f_poscar}}} \\
\onecolumn
{\phantomsection\label{A4B14C3_hP21_143_bd_ac4d_d}}
\subsection*{\huge \textbf{{\normalfont \begin{raggedleft}Simpsonite (Ta$_{3}$Al$_{4}$O$_{13}$[OH]) Structure: \end{raggedleft} \\ A4B14C3\_hP21\_143\_bd\_ac4d\_d}}}
\noindent \hrulefill
\vspace*{0.25cm}
\begin{figure}[htp]
  \centering
  \vspace{-1em}
  {\includegraphics[width=1\textwidth]{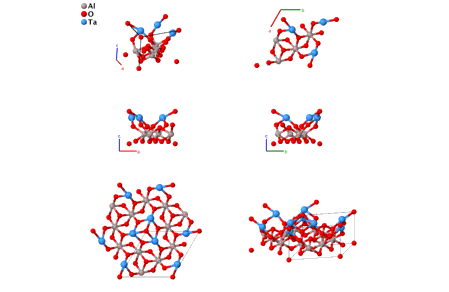}}
\end{figure}
\vspace*{-0.5cm}
\renewcommand{\arraystretch}{1.5}
\begin{equation*}
  \begin{array}{>{$\hspace{-0.15cm}}l<{$}>{$}p{0.5cm}<{$}>{$}p{18.5cm}<{$}}
    \mbox{\large \textbf{Prototype}} &\colon & \ce{Ta3Al4O13[OH]} \\
    \mbox{\large \textbf{\AFLOW\ prototype label}} &\colon & \mbox{A4B14C3\_hP21\_143\_bd\_ac4d\_d} \\
    \mbox{\large \textbf{\textit{Strukturbericht} designation}} &\colon & \mbox{None} \\
    \mbox{\large \textbf{Pearson symbol}} &\colon & \mbox{hP21} \\
    \mbox{\large \textbf{Space group number}} &\colon & 143 \\
    \mbox{\large \textbf{Space group symbol}} &\colon & P3 \\
    \mbox{\large \textbf{\AFLOW\ prototype command}} &\colon &  \texttt{aflow} \,  \, \texttt{-{}-proto=A4B14C3\_hP21\_143\_bd\_ac4d\_d } \, \newline \texttt{-{}-params=}{a,c/a,z_{1},z_{2},z_{3},x_{4},y_{4},z_{4},x_{5},y_{5},z_{5},x_{6},y_{6},z_{6},x_{7},y_{7},z_{7},x_{8},y_{8},z_{8},x_{9},} \newline {y_{9},z_{9} }
  \end{array}
\end{equation*}
\renewcommand{\arraystretch}{1.0}

\vspace*{-0.25cm}
\noindent \hrulefill
\begin{itemize}
  \item{The OH molecule is centered on the (1c) site; however, it is only listed as O in this prototype.
}
\end{itemize}

\noindent \parbox{1 \linewidth}{
\noindent \hrulefill
\\
\textbf{Trigonal Hexagonal primitive vectors:} \\
\vspace*{-0.25cm}
\begin{tabular}{cc}
  \begin{tabular}{c}
    \parbox{0.6 \linewidth}{
      \renewcommand{\arraystretch}{1.5}
      \begin{equation*}
        \centering
        \begin{array}{ccc}
              \mathbf{a}_1 & = & \frac12 \, a \, \mathbf{\hat{x}} - \frac{\sqrt3}2 \, a \, \mathbf{\hat{y}} \\
    \mathbf{a}_2 & = & \frac12 \, a \, \mathbf{\hat{x}} + \frac{\sqrt3}2 \, a \, \mathbf{\hat{y}} \\
    \mathbf{a}_3 & = & c \, \mathbf{\hat{z}} \\

        \end{array}
      \end{equation*}
    }
    \renewcommand{\arraystretch}{1.0}
  \end{tabular}
  \begin{tabular}{c}
    \includegraphics[width=0.3\linewidth]{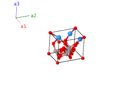} \\
  \end{tabular}
\end{tabular}

}
\vspace*{-0.25cm}

\noindent \hrulefill
\\
\textbf{Basis vectors:}
\vspace*{-0.25cm}
\renewcommand{\arraystretch}{1.5}
\begin{longtabu} to \textwidth{>{\centering $}X[-1,c,c]<{$}>{\centering $}X[-1,c,c]<{$}>{\centering $}X[-1,c,c]<{$}>{\centering $}X[-1,c,c]<{$}>{\centering $}X[-1,c,c]<{$}>{\centering $}X[-1,c,c]<{$}>{\centering $}X[-1,c,c]<{$}}
  & & \mbox{Lattice Coordinates} & & \mbox{Cartesian Coordinates} &\mbox{Wyckoff Position} & \mbox{Atom Type} \\  
  \mathbf{B}_{1} & = & z_{1} \, \mathbf{a}_{3} & = & z_{1}c \, \mathbf{\hat{z}} & \left(1a\right) & \mbox{O I} \\ 
\mathbf{B}_{2} & = & \frac{1}{3} \, \mathbf{a}_{1} + \frac{2}{3} \, \mathbf{a}_{2} + z_{2} \, \mathbf{a}_{3} & = & \frac{1}{2}a \, \mathbf{\hat{x}} + \frac{1}{2\sqrt{3}}a \, \mathbf{\hat{y}} + z_{2}c \, \mathbf{\hat{z}} & \left(1b\right) & \mbox{Al I} \\ 
\mathbf{B}_{3} & = & \frac{2}{3} \, \mathbf{a}_{1} + \frac{1}{3} \, \mathbf{a}_{2} + z_{3} \, \mathbf{a}_{3} & = & \frac{1}{2}a \, \mathbf{\hat{x}}- \frac{1}{2\sqrt{3}}a  \, \mathbf{\hat{y}} + z_{3}c \, \mathbf{\hat{z}} & \left(1c\right) & \mbox{O II} \\ 
\mathbf{B}_{4} & = & x_{4} \, \mathbf{a}_{1} + y_{4} \, \mathbf{a}_{2} + z_{4} \, \mathbf{a}_{3} & = & \frac{1}{2}\left(x_{4}+y_{4}\right)a \, \mathbf{\hat{x}} + \frac{\sqrt{3}}{2}\left(-x_{4}+y_{4}\right)a \, \mathbf{\hat{y}} + z_{4}c \, \mathbf{\hat{z}} & \left(3d\right) & \mbox{Al II} \\ 
\mathbf{B}_{5} & = & -y_{4} \, \mathbf{a}_{1} + \left(x_{4}-y_{4}\right) \, \mathbf{a}_{2} + z_{4} \, \mathbf{a}_{3} & = & \left(\frac{1}{2}x_{4}-y_{4}\right)a \, \mathbf{\hat{x}} + \frac{\sqrt{3}}{2}x_{4}a \, \mathbf{\hat{y}} + z_{4}c \, \mathbf{\hat{z}} & \left(3d\right) & \mbox{Al II} \\ 
\mathbf{B}_{6} & = & \left(-x_{4}+y_{4}\right) \, \mathbf{a}_{1}-x_{4} \, \mathbf{a}_{2} + z_{4} \, \mathbf{a}_{3} & = & \left(-x_{4}+\frac{1}{2}y_{4}\right)a \, \mathbf{\hat{x}}-\frac{\sqrt{3}}{2}y_{4}a \, \mathbf{\hat{y}} + z_{4}c \, \mathbf{\hat{z}} & \left(3d\right) & \mbox{Al II} \\ 
\mathbf{B}_{7} & = & x_{5} \, \mathbf{a}_{1} + y_{5} \, \mathbf{a}_{2} + z_{5} \, \mathbf{a}_{3} & = & \frac{1}{2}\left(x_{5}+y_{5}\right)a \, \mathbf{\hat{x}} + \frac{\sqrt{3}}{2}\left(-x_{5}+y_{5}\right)a \, \mathbf{\hat{y}} + z_{5}c \, \mathbf{\hat{z}} & \left(3d\right) & \mbox{O III} \\ 
\mathbf{B}_{8} & = & -y_{5} \, \mathbf{a}_{1} + \left(x_{5}-y_{5}\right) \, \mathbf{a}_{2} + z_{5} \, \mathbf{a}_{3} & = & \left(\frac{1}{2}x_{5}-y_{5}\right)a \, \mathbf{\hat{x}} + \frac{\sqrt{3}}{2}x_{5}a \, \mathbf{\hat{y}} + z_{5}c \, \mathbf{\hat{z}} & \left(3d\right) & \mbox{O III} \\ 
\mathbf{B}_{9} & = & \left(-x_{5}+y_{5}\right) \, \mathbf{a}_{1}-x_{5} \, \mathbf{a}_{2} + z_{5} \, \mathbf{a}_{3} & = & \left(-x_{5}+\frac{1}{2}y_{5}\right)a \, \mathbf{\hat{x}}-\frac{\sqrt{3}}{2}y_{5}a \, \mathbf{\hat{y}} + z_{5}c \, \mathbf{\hat{z}} & \left(3d\right) & \mbox{O III} \\ 
\mathbf{B}_{10} & = & x_{6} \, \mathbf{a}_{1} + y_{6} \, \mathbf{a}_{2} + z_{6} \, \mathbf{a}_{3} & = & \frac{1}{2}\left(x_{6}+y_{6}\right)a \, \mathbf{\hat{x}} + \frac{\sqrt{3}}{2}\left(-x_{6}+y_{6}\right)a \, \mathbf{\hat{y}} + z_{6}c \, \mathbf{\hat{z}} & \left(3d\right) & \mbox{O IV} \\ 
\mathbf{B}_{11} & = & -y_{6} \, \mathbf{a}_{1} + \left(x_{6}-y_{6}\right) \, \mathbf{a}_{2} + z_{6} \, \mathbf{a}_{3} & = & \left(\frac{1}{2}x_{6}-y_{6}\right)a \, \mathbf{\hat{x}} + \frac{\sqrt{3}}{2}x_{6}a \, \mathbf{\hat{y}} + z_{6}c \, \mathbf{\hat{z}} & \left(3d\right) & \mbox{O IV} \\ 
\mathbf{B}_{12} & = & \left(-x_{6}+y_{6}\right) \, \mathbf{a}_{1}-x_{6} \, \mathbf{a}_{2} + z_{6} \, \mathbf{a}_{3} & = & \left(-x_{6}+\frac{1}{2}y_{6}\right)a \, \mathbf{\hat{x}}-\frac{\sqrt{3}}{2}y_{6}a \, \mathbf{\hat{y}} + z_{6}c \, \mathbf{\hat{z}} & \left(3d\right) & \mbox{O IV} \\ 
\mathbf{B}_{13} & = & x_{7} \, \mathbf{a}_{1} + y_{7} \, \mathbf{a}_{2} + z_{7} \, \mathbf{a}_{3} & = & \frac{1}{2}\left(x_{7}+y_{7}\right)a \, \mathbf{\hat{x}} + \frac{\sqrt{3}}{2}\left(-x_{7}+y_{7}\right)a \, \mathbf{\hat{y}} + z_{7}c \, \mathbf{\hat{z}} & \left(3d\right) & \mbox{O V} \\ 
\mathbf{B}_{14} & = & -y_{7} \, \mathbf{a}_{1} + \left(x_{7}-y_{7}\right) \, \mathbf{a}_{2} + z_{7} \, \mathbf{a}_{3} & = & \left(\frac{1}{2}x_{7}-y_{7}\right)a \, \mathbf{\hat{x}} + \frac{\sqrt{3}}{2}x_{7}a \, \mathbf{\hat{y}} + z_{7}c \, \mathbf{\hat{z}} & \left(3d\right) & \mbox{O V} \\ 
\mathbf{B}_{15} & = & \left(-x_{7}+y_{7}\right) \, \mathbf{a}_{1}-x_{7} \, \mathbf{a}_{2} + z_{7} \, \mathbf{a}_{3} & = & \left(-x_{7}+\frac{1}{2}y_{7}\right)a \, \mathbf{\hat{x}}-\frac{\sqrt{3}}{2}y_{7}a \, \mathbf{\hat{y}} + z_{7}c \, \mathbf{\hat{z}} & \left(3d\right) & \mbox{O V} \\ 
\mathbf{B}_{16} & = & x_{8} \, \mathbf{a}_{1} + y_{8} \, \mathbf{a}_{2} + z_{8} \, \mathbf{a}_{3} & = & \frac{1}{2}\left(x_{8}+y_{8}\right)a \, \mathbf{\hat{x}} + \frac{\sqrt{3}}{2}\left(-x_{8}+y_{8}\right)a \, \mathbf{\hat{y}} + z_{8}c \, \mathbf{\hat{z}} & \left(3d\right) & \mbox{O VI} \\ 
\mathbf{B}_{17} & = & -y_{8} \, \mathbf{a}_{1} + \left(x_{8}-y_{8}\right) \, \mathbf{a}_{2} + z_{8} \, \mathbf{a}_{3} & = & \left(\frac{1}{2}x_{8}-y_{8}\right)a \, \mathbf{\hat{x}} + \frac{\sqrt{3}}{2}x_{8}a \, \mathbf{\hat{y}} + z_{8}c \, \mathbf{\hat{z}} & \left(3d\right) & \mbox{O VI} \\ 
\mathbf{B}_{18} & = & \left(-x_{8}+y_{8}\right) \, \mathbf{a}_{1}-x_{8} \, \mathbf{a}_{2} + z_{8} \, \mathbf{a}_{3} & = & \left(-x_{8}+\frac{1}{2}y_{8}\right)a \, \mathbf{\hat{x}}-\frac{\sqrt{3}}{2}y_{8}a \, \mathbf{\hat{y}} + z_{8}c \, \mathbf{\hat{z}} & \left(3d\right) & \mbox{O VI} \\ 
\mathbf{B}_{19} & = & x_{9} \, \mathbf{a}_{1} + y_{9} \, \mathbf{a}_{2} + z_{9} \, \mathbf{a}_{3} & = & \frac{1}{2}\left(x_{9}+y_{9}\right)a \, \mathbf{\hat{x}} + \frac{\sqrt{3}}{2}\left(-x_{9}+y_{9}\right)a \, \mathbf{\hat{y}} + z_{9}c \, \mathbf{\hat{z}} & \left(3d\right) & \mbox{Ta} \\ 
\mathbf{B}_{20} & = & -y_{9} \, \mathbf{a}_{1} + \left(x_{9}-y_{9}\right) \, \mathbf{a}_{2} + z_{9} \, \mathbf{a}_{3} & = & \left(\frac{1}{2}x_{9}-y_{9}\right)a \, \mathbf{\hat{x}} + \frac{\sqrt{3}}{2}x_{9}a \, \mathbf{\hat{y}} + z_{9}c \, \mathbf{\hat{z}} & \left(3d\right) & \mbox{Ta} \\ 
\mathbf{B}_{21} & = & \left(-x_{9}+y_{9}\right) \, \mathbf{a}_{1}-x_{9} \, \mathbf{a}_{2} + z_{9} \, \mathbf{a}_{3} & = & \left(-x_{9}+\frac{1}{2}y_{9}\right)a \, \mathbf{\hat{x}}-\frac{\sqrt{3}}{2}y_{9}a \, \mathbf{\hat{y}} + z_{9}c \, \mathbf{\hat{z}} & \left(3d\right) & \mbox{Ta} \\ 
\end{longtabu}
\renewcommand{\arraystretch}{1.0}
\noindent \hrulefill
\\
\textbf{References:}
\vspace*{-0.25cm}
\begin{flushleft}
  - \bibentry{Ercit_Ta3Al4O13OH_CanMin_1992}. \\
\end{flushleft}
\textbf{Found in:}
\vspace*{-0.25cm}
\begin{flushleft}
  - \bibentry{Villars_PearsonsCrystalData_2013}. \\
\end{flushleft}
\noindent \hrulefill
\\
\textbf{Geometry files:}
\\
\noindent  - CIF: pp. {\hyperref[A4B14C3_hP21_143_bd_ac4d_d_cif]{\pageref{A4B14C3_hP21_143_bd_ac4d_d_cif}}} \\
\noindent  - POSCAR: pp. {\hyperref[A4B14C3_hP21_143_bd_ac4d_d_poscar]{\pageref{A4B14C3_hP21_143_bd_ac4d_d_poscar}}} \\
\onecolumn
{\phantomsection\label{A4B6C_hP11_143_bd_2d_a}}
\subsection*{\huge \textbf{{\normalfont ScRh$_{6}$P$_{4}$ Structure: A4B6C\_hP11\_143\_bd\_2d\_a}}}
\noindent \hrulefill
\vspace*{0.25cm}
\begin{figure}[htp]
  \centering
  \vspace{-1em}
  {\includegraphics[width=1\textwidth]{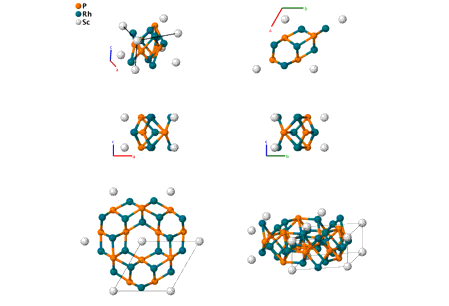}}
\end{figure}
\vspace*{-0.5cm}
\renewcommand{\arraystretch}{1.5}
\begin{equation*}
  \begin{array}{>{$\hspace{-0.15cm}}l<{$}>{$}p{0.5cm}<{$}>{$}p{18.5cm}<{$}}
    \mbox{\large \textbf{Prototype}} &\colon & \ce{ScRh6P4} \\
    \mbox{\large \textbf{\AFLOW\ prototype label}} &\colon & \mbox{A4B6C\_hP11\_143\_bd\_2d\_a} \\
    \mbox{\large \textbf{\textit{Strukturbericht} designation}} &\colon & \mbox{None} \\
    \mbox{\large \textbf{Pearson symbol}} &\colon & \mbox{hP11} \\
    \mbox{\large \textbf{Space group number}} &\colon & 143 \\
    \mbox{\large \textbf{Space group symbol}} &\colon & P3 \\
    \mbox{\large \textbf{\AFLOW\ prototype command}} &\colon &  \texttt{aflow} \,  \, \texttt{-{}-proto=A4B6C\_hP11\_143\_bd\_2d\_a } \, \newline \texttt{-{}-params=}{a,c/a,z_{1},z_{2},x_{3},y_{3},z_{3},x_{4},y_{4},z_{4},x_{5},y_{5},z_{5} }
  \end{array}
\end{equation*}
\renewcommand{\arraystretch}{1.0}

\vspace*{-0.25cm}
\noindent \hrulefill
\begin{itemize}
  \item{While {\small FINDSYM} identifies space group \#143 for this structure (consistent with the reference), {\small AFLOW-SYM} and Platon identify
\#174 and \#187, respectively.
Lowering the tolerance value for {\small AFLOW-SYM} resolves the expected space group \#143.
Space groups \#143, \#174, and \#187 are reasonable classifications since they are commensurate with subgroup relations.
}
\end{itemize}

\noindent \parbox{1 \linewidth}{
\noindent \hrulefill
\\
\textbf{Trigonal Hexagonal primitive vectors:} \\
\vspace*{-0.25cm}
\begin{tabular}{cc}
  \begin{tabular}{c}
    \parbox{0.6 \linewidth}{
      \renewcommand{\arraystretch}{1.5}
      \begin{equation*}
        \centering
        \begin{array}{ccc}
              \mathbf{a}_1 & = & \frac12 \, a \, \mathbf{\hat{x}} - \frac{\sqrt3}2 \, a \, \mathbf{\hat{y}} \\
    \mathbf{a}_2 & = & \frac12 \, a \, \mathbf{\hat{x}} + \frac{\sqrt3}2 \, a \, \mathbf{\hat{y}} \\
    \mathbf{a}_3 & = & c \, \mathbf{\hat{z}} \\

        \end{array}
      \end{equation*}
    }
    \renewcommand{\arraystretch}{1.0}
  \end{tabular}
  \begin{tabular}{c}
    \includegraphics[width=0.3\linewidth]{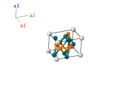} \\
  \end{tabular}
\end{tabular}

}
\vspace*{-0.25cm}

\noindent \hrulefill
\\
\textbf{Basis vectors:}
\vspace*{-0.25cm}
\renewcommand{\arraystretch}{1.5}
\begin{longtabu} to \textwidth{>{\centering $}X[-1,c,c]<{$}>{\centering $}X[-1,c,c]<{$}>{\centering $}X[-1,c,c]<{$}>{\centering $}X[-1,c,c]<{$}>{\centering $}X[-1,c,c]<{$}>{\centering $}X[-1,c,c]<{$}>{\centering $}X[-1,c,c]<{$}}
  & & \mbox{Lattice Coordinates} & & \mbox{Cartesian Coordinates} &\mbox{Wyckoff Position} & \mbox{Atom Type} \\  
  \mathbf{B}_{1} & = & z_{1} \, \mathbf{a}_{3} & = & z_{1}c \, \mathbf{\hat{z}} & \left(1a\right) & \mbox{Sc} \\ 
\mathbf{B}_{2} & = & \frac{1}{3} \, \mathbf{a}_{1} + \frac{2}{3} \, \mathbf{a}_{2} + z_{2} \, \mathbf{a}_{3} & = & \frac{1}{2}a \, \mathbf{\hat{x}} + \frac{1}{2\sqrt{3}}a \, \mathbf{\hat{y}} + z_{2}c \, \mathbf{\hat{z}} & \left(1b\right) & \mbox{P I} \\ 
\mathbf{B}_{3} & = & x_{3} \, \mathbf{a}_{1} + y_{3} \, \mathbf{a}_{2} + z_{3} \, \mathbf{a}_{3} & = & \frac{1}{2}\left(x_{3}+y_{3}\right)a \, \mathbf{\hat{x}} + \frac{\sqrt{3}}{2}\left(-x_{3}+y_{3}\right)a \, \mathbf{\hat{y}} + z_{3}c \, \mathbf{\hat{z}} & \left(3d\right) & \mbox{P II} \\ 
\mathbf{B}_{4} & = & -y_{3} \, \mathbf{a}_{1} + \left(x_{3}-y_{3}\right) \, \mathbf{a}_{2} + z_{3} \, \mathbf{a}_{3} & = & \left(\frac{1}{2}x_{3}-y_{3}\right)a \, \mathbf{\hat{x}} + \frac{\sqrt{3}}{2}x_{3}a \, \mathbf{\hat{y}} + z_{3}c \, \mathbf{\hat{z}} & \left(3d\right) & \mbox{P II} \\ 
\mathbf{B}_{5} & = & \left(-x_{3}+y_{3}\right) \, \mathbf{a}_{1}-x_{3} \, \mathbf{a}_{2} + z_{3} \, \mathbf{a}_{3} & = & \left(-x_{3}+\frac{1}{2}y_{3}\right)a \, \mathbf{\hat{x}}-\frac{\sqrt{3}}{2}y_{3}a \, \mathbf{\hat{y}} + z_{3}c \, \mathbf{\hat{z}} & \left(3d\right) & \mbox{P II} \\ 
\mathbf{B}_{6} & = & x_{4} \, \mathbf{a}_{1} + y_{4} \, \mathbf{a}_{2} + z_{4} \, \mathbf{a}_{3} & = & \frac{1}{2}\left(x_{4}+y_{4}\right)a \, \mathbf{\hat{x}} + \frac{\sqrt{3}}{2}\left(-x_{4}+y_{4}\right)a \, \mathbf{\hat{y}} + z_{4}c \, \mathbf{\hat{z}} & \left(3d\right) & \mbox{Rh I} \\ 
\mathbf{B}_{7} & = & -y_{4} \, \mathbf{a}_{1} + \left(x_{4}-y_{4}\right) \, \mathbf{a}_{2} + z_{4} \, \mathbf{a}_{3} & = & \left(\frac{1}{2}x_{4}-y_{4}\right)a \, \mathbf{\hat{x}} + \frac{\sqrt{3}}{2}x_{4}a \, \mathbf{\hat{y}} + z_{4}c \, \mathbf{\hat{z}} & \left(3d\right) & \mbox{Rh I} \\ 
\mathbf{B}_{8} & = & \left(-x_{4}+y_{4}\right) \, \mathbf{a}_{1}-x_{4} \, \mathbf{a}_{2} + z_{4} \, \mathbf{a}_{3} & = & \left(-x_{4}+\frac{1}{2}y_{4}\right)a \, \mathbf{\hat{x}}-\frac{\sqrt{3}}{2}y_{4}a \, \mathbf{\hat{y}} + z_{4}c \, \mathbf{\hat{z}} & \left(3d\right) & \mbox{Rh I} \\ 
\mathbf{B}_{9} & = & x_{5} \, \mathbf{a}_{1} + y_{5} \, \mathbf{a}_{2} + z_{5} \, \mathbf{a}_{3} & = & \frac{1}{2}\left(x_{5}+y_{5}\right)a \, \mathbf{\hat{x}} + \frac{\sqrt{3}}{2}\left(-x_{5}+y_{5}\right)a \, \mathbf{\hat{y}} + z_{5}c \, \mathbf{\hat{z}} & \left(3d\right) & \mbox{Rh II} \\ 
\mathbf{B}_{10} & = & -y_{5} \, \mathbf{a}_{1} + \left(x_{5}-y_{5}\right) \, \mathbf{a}_{2} + z_{5} \, \mathbf{a}_{3} & = & \left(\frac{1}{2}x_{5}-y_{5}\right)a \, \mathbf{\hat{x}} + \frac{\sqrt{3}}{2}x_{5}a \, \mathbf{\hat{y}} + z_{5}c \, \mathbf{\hat{z}} & \left(3d\right) & \mbox{Rh II} \\ 
\mathbf{B}_{11} & = & \left(-x_{5}+y_{5}\right) \, \mathbf{a}_{1}-x_{5} \, \mathbf{a}_{2} + z_{5} \, \mathbf{a}_{3} & = & \left(-x_{5}+\frac{1}{2}y_{5}\right)a \, \mathbf{\hat{x}}-\frac{\sqrt{3}}{2}y_{5}a \, \mathbf{\hat{y}} + z_{5}c \, \mathbf{\hat{z}} & \left(3d\right) & \mbox{Rh II} \\ 
\end{longtabu}
\renewcommand{\arraystretch}{1.0}
\noindent \hrulefill
\\
\textbf{References:}
\vspace*{-0.25cm}
\begin{flushleft}
  - \bibentry{Pfannenschmidt_ScRh6P4_MonatChemMon_2011}. \\
  - \bibentry{stokes_findsym}. \\
  - \bibentry{aflowsym_2018}. \\
  - \bibentry{platon_2003}. \\
\end{flushleft}
\textbf{Found in:}
\vspace*{-0.25cm}
\begin{flushleft}
  - \bibentry{Villars_PearsonsCrystalData_2013}. \\
\end{flushleft}
\noindent \hrulefill
\\
\textbf{Geometry files:}
\\
\noindent  - CIF: pp. {\hyperref[A4B6C_hP11_143_bd_2d_a_cif]{\pageref{A4B6C_hP11_143_bd_2d_a_cif}}} \\
\noindent  - POSCAR: pp. {\hyperref[A4B6C_hP11_143_bd_2d_a_poscar]{\pageref{A4B6C_hP11_143_bd_2d_a_poscar}}} \\
\onecolumn
{\phantomsection\label{AB2_hP12_143_cd_ab2d}}
\subsection*{\huge \textbf{{\normalfont MoS$_{2}$ Structure: AB2\_hP12\_143\_cd\_ab2d}}}
\noindent \hrulefill
\vspace*{0.25cm}
\begin{figure}[htp]
  \centering
  \vspace{-1em}
  {\includegraphics[width=1\textwidth]{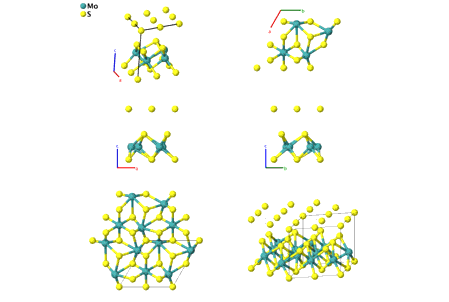}}
\end{figure}
\vspace*{-0.5cm}
\renewcommand{\arraystretch}{1.5}
\begin{equation*}
  \begin{array}{>{$\hspace{-0.15cm}}l<{$}>{$}p{0.5cm}<{$}>{$}p{18.5cm}<{$}}
    \mbox{\large \textbf{Prototype}} &\colon & \ce{MoS2} \\
    \mbox{\large \textbf{\AFLOW\ prototype label}} &\colon & \mbox{AB2\_hP12\_143\_cd\_ab2d} \\
    \mbox{\large \textbf{\textit{Strukturbericht} designation}} &\colon & \mbox{None} \\
    \mbox{\large \textbf{Pearson symbol}} &\colon & \mbox{hP12} \\
    \mbox{\large \textbf{Space group number}} &\colon & 143 \\
    \mbox{\large \textbf{Space group symbol}} &\colon & P3 \\
    \mbox{\large \textbf{\AFLOW\ prototype command}} &\colon &  \texttt{aflow} \,  \, \texttt{-{}-proto=AB2\_hP12\_143\_cd\_ab2d } \, \newline \texttt{-{}-params=}{a,c/a,z_{1},z_{2},z_{3},x_{4},y_{4},z_{4},x_{5},y_{5},z_{5},x_{6},y_{6},z_{6} }
  \end{array}
\end{equation*}
\renewcommand{\arraystretch}{1.0}

\noindent \parbox{1 \linewidth}{
\noindent \hrulefill
\\
\textbf{Trigonal Hexagonal primitive vectors:} \\
\vspace*{-0.25cm}
\begin{tabular}{cc}
  \begin{tabular}{c}
    \parbox{0.6 \linewidth}{
      \renewcommand{\arraystretch}{1.5}
      \begin{equation*}
        \centering
        \begin{array}{ccc}
              \mathbf{a}_1 & = & \frac12 \, a \, \mathbf{\hat{x}} - \frac{\sqrt3}2 \, a \, \mathbf{\hat{y}} \\
    \mathbf{a}_2 & = & \frac12 \, a \, \mathbf{\hat{x}} + \frac{\sqrt3}2 \, a \, \mathbf{\hat{y}} \\
    \mathbf{a}_3 & = & c \, \mathbf{\hat{z}} \\

        \end{array}
      \end{equation*}
    }
    \renewcommand{\arraystretch}{1.0}
  \end{tabular}
  \begin{tabular}{c}
    \includegraphics[width=0.3\linewidth]{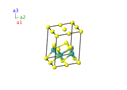} \\
  \end{tabular}
\end{tabular}

}
\vspace*{-0.25cm}

\noindent \hrulefill
\\
\textbf{Basis vectors:}
\vspace*{-0.25cm}
\renewcommand{\arraystretch}{1.5}
\begin{longtabu} to \textwidth{>{\centering $}X[-1,c,c]<{$}>{\centering $}X[-1,c,c]<{$}>{\centering $}X[-1,c,c]<{$}>{\centering $}X[-1,c,c]<{$}>{\centering $}X[-1,c,c]<{$}>{\centering $}X[-1,c,c]<{$}>{\centering $}X[-1,c,c]<{$}}
  & & \mbox{Lattice Coordinates} & & \mbox{Cartesian Coordinates} &\mbox{Wyckoff Position} & \mbox{Atom Type} \\  
  \mathbf{B}_{1} & = & z_{1} \, \mathbf{a}_{3} & = & z_{1}c \, \mathbf{\hat{z}} & \left(1a\right) & \mbox{S I} \\ 
\mathbf{B}_{2} & = & \frac{1}{3} \, \mathbf{a}_{1} + \frac{2}{3} \, \mathbf{a}_{2} + z_{2} \, \mathbf{a}_{3} & = & \frac{1}{2}a \, \mathbf{\hat{x}} + \frac{1}{2\sqrt{3}}a \, \mathbf{\hat{y}} + z_{2}c \, \mathbf{\hat{z}} & \left(1b\right) & \mbox{S II} \\ 
\mathbf{B}_{3} & = & \frac{2}{3} \, \mathbf{a}_{1} + \frac{1}{3} \, \mathbf{a}_{2} + z_{3} \, \mathbf{a}_{3} & = & \frac{1}{2}a \, \mathbf{\hat{x}}- \frac{1}{2\sqrt{3}}a  \, \mathbf{\hat{y}} + z_{3}c \, \mathbf{\hat{z}} & \left(1c\right) & \mbox{Mo I} \\ 
\mathbf{B}_{4} & = & x_{4} \, \mathbf{a}_{1} + y_{4} \, \mathbf{a}_{2} + z_{4} \, \mathbf{a}_{3} & = & \frac{1}{2}\left(x_{4}+y_{4}\right)a \, \mathbf{\hat{x}} + \frac{\sqrt{3}}{2}\left(-x_{4}+y_{4}\right)a \, \mathbf{\hat{y}} + z_{4}c \, \mathbf{\hat{z}} & \left(3d\right) & \mbox{Mo II} \\ 
\mathbf{B}_{5} & = & -y_{4} \, \mathbf{a}_{1} + \left(x_{4}-y_{4}\right) \, \mathbf{a}_{2} + z_{4} \, \mathbf{a}_{3} & = & \left(\frac{1}{2}x_{4}-y_{4}\right)a \, \mathbf{\hat{x}} + \frac{\sqrt{3}}{2}x_{4}a \, \mathbf{\hat{y}} + z_{4}c \, \mathbf{\hat{z}} & \left(3d\right) & \mbox{Mo II} \\ 
\mathbf{B}_{6} & = & \left(-x_{4}+y_{4}\right) \, \mathbf{a}_{1}-x_{4} \, \mathbf{a}_{2} + z_{4} \, \mathbf{a}_{3} & = & \left(-x_{4}+\frac{1}{2}y_{4}\right)a \, \mathbf{\hat{x}}-\frac{\sqrt{3}}{2}y_{4}a \, \mathbf{\hat{y}} + z_{4}c \, \mathbf{\hat{z}} & \left(3d\right) & \mbox{Mo II} \\ 
\mathbf{B}_{7} & = & x_{5} \, \mathbf{a}_{1} + y_{5} \, \mathbf{a}_{2} + z_{5} \, \mathbf{a}_{3} & = & \frac{1}{2}\left(x_{5}+y_{5}\right)a \, \mathbf{\hat{x}} + \frac{\sqrt{3}}{2}\left(-x_{5}+y_{5}\right)a \, \mathbf{\hat{y}} + z_{5}c \, \mathbf{\hat{z}} & \left(3d\right) & \mbox{S III} \\ 
\mathbf{B}_{8} & = & -y_{5} \, \mathbf{a}_{1} + \left(x_{5}-y_{5}\right) \, \mathbf{a}_{2} + z_{5} \, \mathbf{a}_{3} & = & \left(\frac{1}{2}x_{5}-y_{5}\right)a \, \mathbf{\hat{x}} + \frac{\sqrt{3}}{2}x_{5}a \, \mathbf{\hat{y}} + z_{5}c \, \mathbf{\hat{z}} & \left(3d\right) & \mbox{S III} \\ 
\mathbf{B}_{9} & = & \left(-x_{5}+y_{5}\right) \, \mathbf{a}_{1}-x_{5} \, \mathbf{a}_{2} + z_{5} \, \mathbf{a}_{3} & = & \left(-x_{5}+\frac{1}{2}y_{5}\right)a \, \mathbf{\hat{x}}-\frac{\sqrt{3}}{2}y_{5}a \, \mathbf{\hat{y}} + z_{5}c \, \mathbf{\hat{z}} & \left(3d\right) & \mbox{S III} \\ 
\mathbf{B}_{10} & = & x_{6} \, \mathbf{a}_{1} + y_{6} \, \mathbf{a}_{2} + z_{6} \, \mathbf{a}_{3} & = & \frac{1}{2}\left(x_{6}+y_{6}\right)a \, \mathbf{\hat{x}} + \frac{\sqrt{3}}{2}\left(-x_{6}+y_{6}\right)a \, \mathbf{\hat{y}} + z_{6}c \, \mathbf{\hat{z}} & \left(3d\right) & \mbox{S IV} \\ 
\mathbf{B}_{11} & = & -y_{6} \, \mathbf{a}_{1} + \left(x_{6}-y_{6}\right) \, \mathbf{a}_{2} + z_{6} \, \mathbf{a}_{3} & = & \left(\frac{1}{2}x_{6}-y_{6}\right)a \, \mathbf{\hat{x}} + \frac{\sqrt{3}}{2}x_{6}a \, \mathbf{\hat{y}} + z_{6}c \, \mathbf{\hat{z}} & \left(3d\right) & \mbox{S IV} \\ 
\mathbf{B}_{12} & = & \left(-x_{6}+y_{6}\right) \, \mathbf{a}_{1}-x_{6} \, \mathbf{a}_{2} + z_{6} \, \mathbf{a}_{3} & = & \left(-x_{6}+\frac{1}{2}y_{6}\right)a \, \mathbf{\hat{x}}-\frac{\sqrt{3}}{2}y_{6}a \, \mathbf{\hat{y}} + z_{6}c \, \mathbf{\hat{z}} & \left(3d\right) & \mbox{S IV} \\ 
\end{longtabu}
\renewcommand{\arraystretch}{1.0}
\noindent \hrulefill
\\
\textbf{References:}
\vspace*{-0.25cm}
\begin{flushleft}
  - \bibentry{Dungey_MoS2_ChemMat_1998}. \\
\end{flushleft}
\textbf{Found in:}
\vspace*{-0.25cm}
\begin{flushleft}
  - \bibentry{Villars_PearsonsCrystalData_2013}. \\
\end{flushleft}
\noindent \hrulefill
\\
\textbf{Geometry files:}
\\
\noindent  - CIF: pp. {\hyperref[AB2_hP12_143_cd_ab2d_cif]{\pageref{AB2_hP12_143_cd_ab2d_cif}}} \\
\noindent  - POSCAR: pp. {\hyperref[AB2_hP12_143_cd_ab2d_poscar]{\pageref{AB2_hP12_143_cd_ab2d_poscar}}} \\
\onecolumn
{\phantomsection\label{A4B_hP15_144_4a_a}}
\subsection*{\huge \textbf{{\normalfont IrGe$_{4}$ Structure: A4B\_hP15\_144\_4a\_a}}}
\noindent \hrulefill
\vspace*{0.25cm}
\begin{figure}[htp]
  \centering
  \vspace{-1em}
  {\includegraphics[width=1\textwidth]{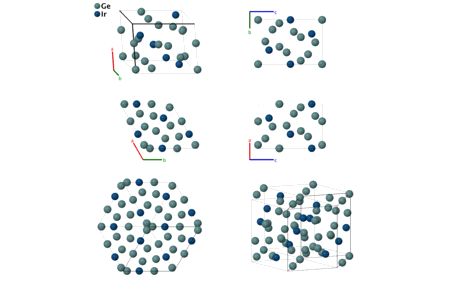}}
\end{figure}
\vspace*{-0.5cm}
\renewcommand{\arraystretch}{1.5}
\begin{equation*}
  \begin{array}{>{$\hspace{-0.15cm}}l<{$}>{$}p{0.5cm}<{$}>{$}p{18.5cm}<{$}}
    \mbox{\large \textbf{Prototype}} &\colon & \ce{IrGe4} \\
    \mbox{\large \textbf{\AFLOW\ prototype label}} &\colon & \mbox{A4B\_hP15\_144\_4a\_a} \\
    \mbox{\large \textbf{\textit{Strukturbericht} designation}} &\colon & \mbox{None} \\
    \mbox{\large \textbf{Pearson symbol}} &\colon & \mbox{hP15} \\
    \mbox{\large \textbf{Space group number}} &\colon & 144 \\
    \mbox{\large \textbf{Space group symbol}} &\colon & P3_{1} \\
    \mbox{\large \textbf{\AFLOW\ prototype command}} &\colon &  \texttt{aflow} \,  \, \texttt{-{}-proto=A4B\_hP15\_144\_4a\_a } \, \newline \texttt{-{}-params=}{a,c/a,x_{1},y_{1},z_{1},x_{2},y_{2},z_{2},x_{3},y_{3},z_{3},x_{4},y_{4},z_{4},x_{5},y_{5},z_{5} }
  \end{array}
\end{equation*}
\renewcommand{\arraystretch}{1.0}

\noindent \parbox{1 \linewidth}{
\noindent \hrulefill
\\
\textbf{Trigonal Hexagonal primitive vectors:} \\
\vspace*{-0.25cm}
\begin{tabular}{cc}
  \begin{tabular}{c}
    \parbox{0.6 \linewidth}{
      \renewcommand{\arraystretch}{1.5}
      \begin{equation*}
        \centering
        \begin{array}{ccc}
              \mathbf{a}_1 & = & \frac12 \, a \, \mathbf{\hat{x}} - \frac{\sqrt3}2 \, a \, \mathbf{\hat{y}} \\
    \mathbf{a}_2 & = & \frac12 \, a \, \mathbf{\hat{x}} + \frac{\sqrt3}2 \, a \, \mathbf{\hat{y}} \\
    \mathbf{a}_3 & = & c \, \mathbf{\hat{z}} \\

        \end{array}
      \end{equation*}
    }
    \renewcommand{\arraystretch}{1.0}
  \end{tabular}
  \begin{tabular}{c}
    \includegraphics[width=0.3\linewidth]{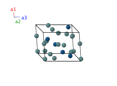} \\
  \end{tabular}
\end{tabular}

}
\vspace*{-0.25cm}

\noindent \hrulefill
\\
\textbf{Basis vectors:}
\vspace*{-0.25cm}
\renewcommand{\arraystretch}{1.5}
\begin{longtabu} to \textwidth{>{\centering $}X[-1,c,c]<{$}>{\centering $}X[-1,c,c]<{$}>{\centering $}X[-1,c,c]<{$}>{\centering $}X[-1,c,c]<{$}>{\centering $}X[-1,c,c]<{$}>{\centering $}X[-1,c,c]<{$}>{\centering $}X[-1,c,c]<{$}}
  & & \mbox{Lattice Coordinates} & & \mbox{Cartesian Coordinates} &\mbox{Wyckoff Position} & \mbox{Atom Type} \\  
  \mathbf{B}_{1} & = & x_{1} \, \mathbf{a}_{1} + y_{1} \, \mathbf{a}_{2} + z_{1} \, \mathbf{a}_{3} & = & \frac{1}{2}\left(x_{1}+y_{1}\right)a \, \mathbf{\hat{x}} + \frac{\sqrt{3}}{2}\left(-x_{1}+y_{1}\right)a \, \mathbf{\hat{y}} + z_{1}c \, \mathbf{\hat{z}} & \left(3a\right) & \mbox{Ge I} \\ 
\mathbf{B}_{2} & = & -y_{1} \, \mathbf{a}_{1} + \left(x_{1}-y_{1}\right) \, \mathbf{a}_{2} + \left(\frac{1}{3} +z_{1}\right) \, \mathbf{a}_{3} & = & \left(\frac{1}{2}x_{1}-y_{1}\right)a \, \mathbf{\hat{x}} + \frac{\sqrt{3}}{2}x_{1}a \, \mathbf{\hat{y}} + \left(\frac{1}{3} +z_{1}\right)c \, \mathbf{\hat{z}} & \left(3a\right) & \mbox{Ge I} \\ 
\mathbf{B}_{3} & = & \left(-x_{1}+y_{1}\right) \, \mathbf{a}_{1}-x_{1} \, \mathbf{a}_{2} + \left(\frac{2}{3} +z_{1}\right) \, \mathbf{a}_{3} & = & \left(-x_{1}+\frac{1}{2}y_{1}\right)a \, \mathbf{\hat{x}}-\frac{\sqrt{3}}{2}y_{1}a \, \mathbf{\hat{y}} + \left(\frac{2}{3} +z_{1}\right)c \, \mathbf{\hat{z}} & \left(3a\right) & \mbox{Ge I} \\ 
\mathbf{B}_{4} & = & x_{2} \, \mathbf{a}_{1} + y_{2} \, \mathbf{a}_{2} + z_{2} \, \mathbf{a}_{3} & = & \frac{1}{2}\left(x_{2}+y_{2}\right)a \, \mathbf{\hat{x}} + \frac{\sqrt{3}}{2}\left(-x_{2}+y_{2}\right)a \, \mathbf{\hat{y}} + z_{2}c \, \mathbf{\hat{z}} & \left(3a\right) & \mbox{Ge II} \\ 
\mathbf{B}_{5} & = & -y_{2} \, \mathbf{a}_{1} + \left(x_{2}-y_{2}\right) \, \mathbf{a}_{2} + \left(\frac{1}{3} +z_{2}\right) \, \mathbf{a}_{3} & = & \left(\frac{1}{2}x_{2}-y_{2}\right)a \, \mathbf{\hat{x}} + \frac{\sqrt{3}}{2}x_{2}a \, \mathbf{\hat{y}} + \left(\frac{1}{3} +z_{2}\right)c \, \mathbf{\hat{z}} & \left(3a\right) & \mbox{Ge II} \\ 
\mathbf{B}_{6} & = & \left(-x_{2}+y_{2}\right) \, \mathbf{a}_{1}-x_{2} \, \mathbf{a}_{2} + \left(\frac{2}{3} +z_{2}\right) \, \mathbf{a}_{3} & = & \left(-x_{2}+\frac{1}{2}y_{2}\right)a \, \mathbf{\hat{x}}-\frac{\sqrt{3}}{2}y_{2}a \, \mathbf{\hat{y}} + \left(\frac{2}{3} +z_{2}\right)c \, \mathbf{\hat{z}} & \left(3a\right) & \mbox{Ge II} \\ 
\mathbf{B}_{7} & = & x_{3} \, \mathbf{a}_{1} + y_{3} \, \mathbf{a}_{2} + z_{3} \, \mathbf{a}_{3} & = & \frac{1}{2}\left(x_{3}+y_{3}\right)a \, \mathbf{\hat{x}} + \frac{\sqrt{3}}{2}\left(-x_{3}+y_{3}\right)a \, \mathbf{\hat{y}} + z_{3}c \, \mathbf{\hat{z}} & \left(3a\right) & \mbox{Ge III} \\ 
\mathbf{B}_{8} & = & -y_{3} \, \mathbf{a}_{1} + \left(x_{3}-y_{3}\right) \, \mathbf{a}_{2} + \left(\frac{1}{3} +z_{3}\right) \, \mathbf{a}_{3} & = & \left(\frac{1}{2}x_{3}-y_{3}\right)a \, \mathbf{\hat{x}} + \frac{\sqrt{3}}{2}x_{3}a \, \mathbf{\hat{y}} + \left(\frac{1}{3} +z_{3}\right)c \, \mathbf{\hat{z}} & \left(3a\right) & \mbox{Ge III} \\ 
\mathbf{B}_{9} & = & \left(-x_{3}+y_{3}\right) \, \mathbf{a}_{1}-x_{3} \, \mathbf{a}_{2} + \left(\frac{2}{3} +z_{3}\right) \, \mathbf{a}_{3} & = & \left(-x_{3}+\frac{1}{2}y_{3}\right)a \, \mathbf{\hat{x}}-\frac{\sqrt{3}}{2}y_{3}a \, \mathbf{\hat{y}} + \left(\frac{2}{3} +z_{3}\right)c \, \mathbf{\hat{z}} & \left(3a\right) & \mbox{Ge III} \\ 
\mathbf{B}_{10} & = & x_{4} \, \mathbf{a}_{1} + y_{4} \, \mathbf{a}_{2} + z_{4} \, \mathbf{a}_{3} & = & \frac{1}{2}\left(x_{4}+y_{4}\right)a \, \mathbf{\hat{x}} + \frac{\sqrt{3}}{2}\left(-x_{4}+y_{4}\right)a \, \mathbf{\hat{y}} + z_{4}c \, \mathbf{\hat{z}} & \left(3a\right) & \mbox{Ge IV} \\ 
\mathbf{B}_{11} & = & -y_{4} \, \mathbf{a}_{1} + \left(x_{4}-y_{4}\right) \, \mathbf{a}_{2} + \left(\frac{1}{3} +z_{4}\right) \, \mathbf{a}_{3} & = & \left(\frac{1}{2}x_{4}-y_{4}\right)a \, \mathbf{\hat{x}} + \frac{\sqrt{3}}{2}x_{4}a \, \mathbf{\hat{y}} + \left(\frac{1}{3} +z_{4}\right)c \, \mathbf{\hat{z}} & \left(3a\right) & \mbox{Ge IV} \\ 
\mathbf{B}_{12} & = & \left(-x_{4}+y_{4}\right) \, \mathbf{a}_{1}-x_{4} \, \mathbf{a}_{2} + \left(\frac{2}{3} +z_{4}\right) \, \mathbf{a}_{3} & = & \left(-x_{4}+\frac{1}{2}y_{4}\right)a \, \mathbf{\hat{x}}-\frac{\sqrt{3}}{2}y_{4}a \, \mathbf{\hat{y}} + \left(\frac{2}{3} +z_{4}\right)c \, \mathbf{\hat{z}} & \left(3a\right) & \mbox{Ge IV} \\ 
\mathbf{B}_{13} & = & x_{5} \, \mathbf{a}_{1} + y_{5} \, \mathbf{a}_{2} + z_{5} \, \mathbf{a}_{3} & = & \frac{1}{2}\left(x_{5}+y_{5}\right)a \, \mathbf{\hat{x}} + \frac{\sqrt{3}}{2}\left(-x_{5}+y_{5}\right)a \, \mathbf{\hat{y}} + z_{5}c \, \mathbf{\hat{z}} & \left(3a\right) & \mbox{Ir} \\ 
\mathbf{B}_{14} & = & -y_{5} \, \mathbf{a}_{1} + \left(x_{5}-y_{5}\right) \, \mathbf{a}_{2} + \left(\frac{1}{3} +z_{5}\right) \, \mathbf{a}_{3} & = & \left(\frac{1}{2}x_{5}-y_{5}\right)a \, \mathbf{\hat{x}} + \frac{\sqrt{3}}{2}x_{5}a \, \mathbf{\hat{y}} + \left(\frac{1}{3} +z_{5}\right)c \, \mathbf{\hat{z}} & \left(3a\right) & \mbox{Ir} \\ 
\mathbf{B}_{15} & = & \left(-x_{5}+y_{5}\right) \, \mathbf{a}_{1}-x_{5} \, \mathbf{a}_{2} + \left(\frac{2}{3} +z_{5}\right) \, \mathbf{a}_{3} & = & \left(-x_{5}+\frac{1}{2}y_{5}\right)a \, \mathbf{\hat{x}}-\frac{\sqrt{3}}{2}y_{5}a \, \mathbf{\hat{y}} + \left(\frac{2}{3} +z_{5}\right)c \, \mathbf{\hat{z}} & \left(3a\right) & \mbox{Ir} \\ 
\end{longtabu}
\renewcommand{\arraystretch}{1.0}
\noindent \hrulefill
\\
\textbf{References:}
\vspace*{-0.25cm}
\begin{flushleft}
  - \bibentry{Schubert_IrGe4_Naturwissen_1968}. \\
\end{flushleft}
\textbf{Found in:}
\vspace*{-0.25cm}
\begin{flushleft}
  - \bibentry{Villars_PearsonsCrystalData_2013}. \\
\end{flushleft}
\noindent \hrulefill
\\
\textbf{Geometry files:}
\\
\noindent  - CIF: pp. {\hyperref[A4B_hP15_144_4a_a_cif]{\pageref{A4B_hP15_144_4a_a_cif}}} \\
\noindent  - POSCAR: pp. {\hyperref[A4B_hP15_144_4a_a_poscar]{\pageref{A4B_hP15_144_4a_a_poscar}}} \\
\onecolumn
{\phantomsection\label{AB_hP6_144_a_a}}
\subsection*{\huge \textbf{{\normalfont TeZn (High-pressure) Structure: AB\_hP6\_144\_a\_a}}}
\noindent \hrulefill
\vspace*{0.25cm}
\begin{figure}[htp]
  \centering
  \vspace{-1em}
  {\includegraphics[width=1\textwidth]{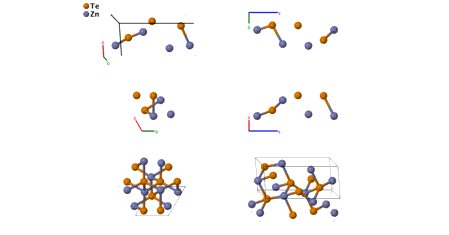}}
\end{figure}
\vspace*{-0.5cm}
\renewcommand{\arraystretch}{1.5}
\begin{equation*}
  \begin{array}{>{$\hspace{-0.15cm}}l<{$}>{$}p{0.5cm}<{$}>{$}p{18.5cm}<{$}}
    \mbox{\large \textbf{Prototype}} &\colon & \ce{ZnTe} \\
    \mbox{\large \textbf{\AFLOW\ prototype label}} &\colon & \mbox{AB\_hP6\_144\_a\_a} \\
    \mbox{\large \textbf{\textit{Strukturbericht} designation}} &\colon & \mbox{None} \\
    \mbox{\large \textbf{Pearson symbol}} &\colon & \mbox{hP6} \\
    \mbox{\large \textbf{Space group number}} &\colon & 144 \\
    \mbox{\large \textbf{Space group symbol}} &\colon & P3_{1} \\
    \mbox{\large \textbf{\AFLOW\ prototype command}} &\colon &  \texttt{aflow} \,  \, \texttt{-{}-proto=AB\_hP6\_144\_a\_a } \, \newline \texttt{-{}-params=}{a,c/a,x_{1},y_{1},z_{1},x_{2},y_{2},z_{2} }
  \end{array}
\end{equation*}
\renewcommand{\arraystretch}{1.0}

\noindent \parbox{1 \linewidth}{
\noindent \hrulefill
\\
\textbf{Trigonal Hexagonal primitive vectors:} \\
\vspace*{-0.25cm}
\begin{tabular}{cc}
  \begin{tabular}{c}
    \parbox{0.6 \linewidth}{
      \renewcommand{\arraystretch}{1.5}
      \begin{equation*}
        \centering
        \begin{array}{ccc}
              \mathbf{a}_1 & = & \frac12 \, a \, \mathbf{\hat{x}} - \frac{\sqrt3}2 \, a \, \mathbf{\hat{y}} \\
    \mathbf{a}_2 & = & \frac12 \, a \, \mathbf{\hat{x}} + \frac{\sqrt3}2 \, a \, \mathbf{\hat{y}} \\
    \mathbf{a}_3 & = & c \, \mathbf{\hat{z}} \\

        \end{array}
      \end{equation*}
    }
    \renewcommand{\arraystretch}{1.0}
  \end{tabular}
  \begin{tabular}{c}
    \includegraphics[width=0.3\linewidth]{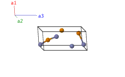} \\
  \end{tabular}
\end{tabular}

}
\vspace*{-0.25cm}

\noindent \hrulefill
\\
\textbf{Basis vectors:}
\vspace*{-0.25cm}
\renewcommand{\arraystretch}{1.5}
\begin{longtabu} to \textwidth{>{\centering $}X[-1,c,c]<{$}>{\centering $}X[-1,c,c]<{$}>{\centering $}X[-1,c,c]<{$}>{\centering $}X[-1,c,c]<{$}>{\centering $}X[-1,c,c]<{$}>{\centering $}X[-1,c,c]<{$}>{\centering $}X[-1,c,c]<{$}}
  & & \mbox{Lattice Coordinates} & & \mbox{Cartesian Coordinates} &\mbox{Wyckoff Position} & \mbox{Atom Type} \\  
  \mathbf{B}_{1} & = & x_{1} \, \mathbf{a}_{1} + y_{1} \, \mathbf{a}_{2} + z_{1} \, \mathbf{a}_{3} & = & \frac{1}{2}\left(x_{1}+y_{1}\right)a \, \mathbf{\hat{x}} + \frac{\sqrt{3}}{2}\left(-x_{1}+y_{1}\right)a \, \mathbf{\hat{y}} + z_{1}c \, \mathbf{\hat{z}} & \left(3a\right) & \mbox{Te} \\ 
\mathbf{B}_{2} & = & -y_{1} \, \mathbf{a}_{1} + \left(x_{1}-y_{1}\right) \, \mathbf{a}_{2} + \left(\frac{1}{3} +z_{1}\right) \, \mathbf{a}_{3} & = & \left(\frac{1}{2}x_{1}-y_{1}\right)a \, \mathbf{\hat{x}} + \frac{\sqrt{3}}{2}x_{1}a \, \mathbf{\hat{y}} + \left(\frac{1}{3} +z_{1}\right)c \, \mathbf{\hat{z}} & \left(3a\right) & \mbox{Te} \\ 
\mathbf{B}_{3} & = & \left(-x_{1}+y_{1}\right) \, \mathbf{a}_{1}-x_{1} \, \mathbf{a}_{2} + \left(\frac{2}{3} +z_{1}\right) \, \mathbf{a}_{3} & = & \left(-x_{1}+\frac{1}{2}y_{1}\right)a \, \mathbf{\hat{x}}-\frac{\sqrt{3}}{2}y_{1}a \, \mathbf{\hat{y}} + \left(\frac{2}{3} +z_{1}\right)c \, \mathbf{\hat{z}} & \left(3a\right) & \mbox{Te} \\ 
\mathbf{B}_{4} & = & x_{2} \, \mathbf{a}_{1} + y_{2} \, \mathbf{a}_{2} + z_{2} \, \mathbf{a}_{3} & = & \frac{1}{2}\left(x_{2}+y_{2}\right)a \, \mathbf{\hat{x}} + \frac{\sqrt{3}}{2}\left(-x_{2}+y_{2}\right)a \, \mathbf{\hat{y}} + z_{2}c \, \mathbf{\hat{z}} & \left(3a\right) & \mbox{Zn} \\ 
\mathbf{B}_{5} & = & -y_{2} \, \mathbf{a}_{1} + \left(x_{2}-y_{2}\right) \, \mathbf{a}_{2} + \left(\frac{1}{3} +z_{2}\right) \, \mathbf{a}_{3} & = & \left(\frac{1}{2}x_{2}-y_{2}\right)a \, \mathbf{\hat{x}} + \frac{\sqrt{3}}{2}x_{2}a \, \mathbf{\hat{y}} + \left(\frac{1}{3} +z_{2}\right)c \, \mathbf{\hat{z}} & \left(3a\right) & \mbox{Zn} \\ 
\mathbf{B}_{6} & = & \left(-x_{2}+y_{2}\right) \, \mathbf{a}_{1}-x_{2} \, \mathbf{a}_{2} + \left(\frac{2}{3} +z_{2}\right) \, \mathbf{a}_{3} & = & \left(-x_{2}+\frac{1}{2}y_{2}\right)a \, \mathbf{\hat{x}}-\frac{\sqrt{3}}{2}y_{2}a \, \mathbf{\hat{y}} + \left(\frac{2}{3} +z_{2}\right)c \, \mathbf{\hat{z}} & \left(3a\right) & \mbox{Zn} \\ 
\end{longtabu}
\renewcommand{\arraystretch}{1.0}
\noindent \hrulefill
\\
\textbf{References:}
\vspace*{-0.25cm}
\begin{flushleft}
  - \bibentry{Kusaba_TeZn_AIPConfPrc_1994}. \\
\end{flushleft}
\textbf{Found in:}
\vspace*{-0.25cm}
\begin{flushleft}
  - \bibentry{Villars_PearsonsCrystalData_2013}. \\
\end{flushleft}
\noindent \hrulefill
\\
\textbf{Geometry files:}
\\
\noindent  - CIF: pp. {\hyperref[AB_hP6_144_a_a_cif]{\pageref{AB_hP6_144_a_a_cif}}} \\
\noindent  - POSCAR: pp. {\hyperref[AB_hP6_144_a_a_poscar]{\pageref{AB_hP6_144_a_a_poscar}}} \\
\onecolumn
{\phantomsection\label{A2B3C3DE7_hP48_145_2a_3a_3a_a_7a}}
\subsection*{\huge \textbf{{\normalfont \begin{raggedleft}Sheldrickite (NaCa$_{3}$[CO$_{3}$]$_{2}$F$_{3}$[H$_{2}$O]) Structure: \end{raggedleft} \\ A2B3C3DE7\_hP48\_145\_2a\_3a\_3a\_a\_7a}}}
\noindent \hrulefill
\vspace*{0.25cm}
\begin{figure}[htp]
  \centering
  \vspace{-1em}
  {\includegraphics[width=1\textwidth]{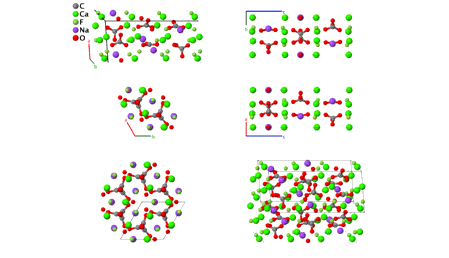}}
\end{figure}
\vspace*{-0.5cm}
\renewcommand{\arraystretch}{1.5}
\begin{equation*}
  \begin{array}{>{$\hspace{-0.15cm}}l<{$}>{$}p{0.5cm}<{$}>{$}p{18.5cm}<{$}}
    \mbox{\large \textbf{Prototype}} &\colon & \ce{NaCa3[CO3]2F3[H2O]} \\
    \mbox{\large \textbf{\AFLOW\ prototype label}} &\colon & \mbox{A2B3C3DE7\_hP48\_145\_2a\_3a\_3a\_a\_7a} \\
    \mbox{\large \textbf{\textit{Strukturbericht} designation}} &\colon & \mbox{None} \\
    \mbox{\large \textbf{Pearson symbol}} &\colon & \mbox{hP48} \\
    \mbox{\large \textbf{Space group number}} &\colon & 145 \\
    \mbox{\large \textbf{Space group symbol}} &\colon & P3_{2} \\
    \mbox{\large \textbf{\AFLOW\ prototype command}} &\colon &  \texttt{aflow} \,  \, \texttt{-{}-proto=A2B3C3DE7\_hP48\_145\_2a\_3a\_3a\_a\_7a } \, \newline \texttt{-{}-params=}{a,c/a,x_{1},y_{1},z_{1},x_{2},y_{2},z_{2},x_{3},y_{3},z_{3},x_{4},y_{4},z_{4},x_{5},y_{5},z_{5},x_{6},y_{6},z_{6},x_{7},} \newline {y_{7},z_{7},x_{8},y_{8},z_{8},x_{9},y_{9},z_{9},x_{10},y_{10},z_{10},x_{11},y_{11},z_{11},x_{12},y_{12},z_{12},x_{13},y_{13},z_{13},x_{14},y_{14},} \newline {z_{14},x_{15},y_{15},z_{15},x_{16},y_{16},z_{16} }
  \end{array}
\end{equation*}
\renewcommand{\arraystretch}{1.0}

\vspace*{-0.25cm}
\noindent \hrulefill
\begin{itemize}
  \item{The H$_{2}$O molecule is centered on one of the (3a) sites; however, it is only listed as O in this prototype.  
}
\end{itemize}

\noindent \parbox{1 \linewidth}{
\noindent \hrulefill
\\
\textbf{Trigonal Hexagonal primitive vectors:} \\
\vspace*{-0.25cm}
\begin{tabular}{cc}
  \begin{tabular}{c}
    \parbox{0.6 \linewidth}{
      \renewcommand{\arraystretch}{1.5}
      \begin{equation*}
        \centering
        \begin{array}{ccc}
              \mathbf{a}_1 & = & \frac12 \, a \, \mathbf{\hat{x}} - \frac{\sqrt3}2 \, a \, \mathbf{\hat{y}} \\
    \mathbf{a}_2 & = & \frac12 \, a \, \mathbf{\hat{x}} + \frac{\sqrt3}2 \, a \, \mathbf{\hat{y}} \\
    \mathbf{a}_3 & = & c \, \mathbf{\hat{z}} \\

        \end{array}
      \end{equation*}
    }
    \renewcommand{\arraystretch}{1.0}
  \end{tabular}
  \begin{tabular}{c}
    \includegraphics[width=0.3\linewidth]{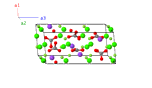} \\
  \end{tabular}
\end{tabular}

}
\vspace*{-0.25cm}

\noindent \hrulefill
\\
\textbf{Basis vectors:}
\vspace*{-0.25cm}
\renewcommand{\arraystretch}{1.5}
\begin{longtabu} to \textwidth{>{\centering $}X[-1,c,c]<{$}>{\centering $}X[-1,c,c]<{$}>{\centering $}X[-1,c,c]<{$}>{\centering $}X[-1,c,c]<{$}>{\centering $}X[-1,c,c]<{$}>{\centering $}X[-1,c,c]<{$}>{\centering $}X[-1,c,c]<{$}}
  & & \mbox{Lattice Coordinates} & & \mbox{Cartesian Coordinates} &\mbox{Wyckoff Position} & \mbox{Atom Type} \\  
  \mathbf{B}_{1} & = & x_{1} \, \mathbf{a}_{1} + y_{1} \, \mathbf{a}_{2} + z_{1} \, \mathbf{a}_{3} & = & \frac{1}{2}\left(x_{1}+y_{1}\right)a \, \mathbf{\hat{x}} + \frac{\sqrt{3}}{2}\left(-x_{1}+y_{1}\right)a \, \mathbf{\hat{y}} + z_{1}c \, \mathbf{\hat{z}} & \left(3a\right) & \mbox{C I} \\ 
\mathbf{B}_{2} & = & -y_{1} \, \mathbf{a}_{1} + \left(x_{1}-y_{1}\right) \, \mathbf{a}_{2} + \left(\frac{2}{3} +z_{1}\right) \, \mathbf{a}_{3} & = & \left(\frac{1}{2}x_{1}-y_{1}\right)a \, \mathbf{\hat{x}} + \frac{\sqrt{3}}{2}x_{1}a \, \mathbf{\hat{y}} + \left(\frac{2}{3} +z_{1}\right)c \, \mathbf{\hat{z}} & \left(3a\right) & \mbox{C I} \\ 
\mathbf{B}_{3} & = & \left(-x_{1}+y_{1}\right) \, \mathbf{a}_{1}-x_{1} \, \mathbf{a}_{2} + \left(\frac{1}{3} +z_{1}\right) \, \mathbf{a}_{3} & = & \left(-x_{1}+\frac{1}{2}y_{1}\right)a \, \mathbf{\hat{x}}-\frac{\sqrt{3}}{2}y_{1}a \, \mathbf{\hat{y}} + \left(\frac{1}{3} +z_{1}\right)c \, \mathbf{\hat{z}} & \left(3a\right) & \mbox{C I} \\ 
\mathbf{B}_{4} & = & x_{2} \, \mathbf{a}_{1} + y_{2} \, \mathbf{a}_{2} + z_{2} \, \mathbf{a}_{3} & = & \frac{1}{2}\left(x_{2}+y_{2}\right)a \, \mathbf{\hat{x}} + \frac{\sqrt{3}}{2}\left(-x_{2}+y_{2}\right)a \, \mathbf{\hat{y}} + z_{2}c \, \mathbf{\hat{z}} & \left(3a\right) & \mbox{C II} \\ 
\mathbf{B}_{5} & = & -y_{2} \, \mathbf{a}_{1} + \left(x_{2}-y_{2}\right) \, \mathbf{a}_{2} + \left(\frac{2}{3} +z_{2}\right) \, \mathbf{a}_{3} & = & \left(\frac{1}{2}x_{2}-y_{2}\right)a \, \mathbf{\hat{x}} + \frac{\sqrt{3}}{2}x_{2}a \, \mathbf{\hat{y}} + \left(\frac{2}{3} +z_{2}\right)c \, \mathbf{\hat{z}} & \left(3a\right) & \mbox{C II} \\ 
\mathbf{B}_{6} & = & \left(-x_{2}+y_{2}\right) \, \mathbf{a}_{1}-x_{2} \, \mathbf{a}_{2} + \left(\frac{1}{3} +z_{2}\right) \, \mathbf{a}_{3} & = & \left(-x_{2}+\frac{1}{2}y_{2}\right)a \, \mathbf{\hat{x}}-\frac{\sqrt{3}}{2}y_{2}a \, \mathbf{\hat{y}} + \left(\frac{1}{3} +z_{2}\right)c \, \mathbf{\hat{z}} & \left(3a\right) & \mbox{C II} \\ 
\mathbf{B}_{7} & = & x_{3} \, \mathbf{a}_{1} + y_{3} \, \mathbf{a}_{2} + z_{3} \, \mathbf{a}_{3} & = & \frac{1}{2}\left(x_{3}+y_{3}\right)a \, \mathbf{\hat{x}} + \frac{\sqrt{3}}{2}\left(-x_{3}+y_{3}\right)a \, \mathbf{\hat{y}} + z_{3}c \, \mathbf{\hat{z}} & \left(3a\right) & \mbox{Ca I} \\ 
\mathbf{B}_{8} & = & -y_{3} \, \mathbf{a}_{1} + \left(x_{3}-y_{3}\right) \, \mathbf{a}_{2} + \left(\frac{2}{3} +z_{3}\right) \, \mathbf{a}_{3} & = & \left(\frac{1}{2}x_{3}-y_{3}\right)a \, \mathbf{\hat{x}} + \frac{\sqrt{3}}{2}x_{3}a \, \mathbf{\hat{y}} + \left(\frac{2}{3} +z_{3}\right)c \, \mathbf{\hat{z}} & \left(3a\right) & \mbox{Ca I} \\ 
\mathbf{B}_{9} & = & \left(-x_{3}+y_{3}\right) \, \mathbf{a}_{1}-x_{3} \, \mathbf{a}_{2} + \left(\frac{1}{3} +z_{3}\right) \, \mathbf{a}_{3} & = & \left(-x_{3}+\frac{1}{2}y_{3}\right)a \, \mathbf{\hat{x}}-\frac{\sqrt{3}}{2}y_{3}a \, \mathbf{\hat{y}} + \left(\frac{1}{3} +z_{3}\right)c \, \mathbf{\hat{z}} & \left(3a\right) & \mbox{Ca I} \\ 
\mathbf{B}_{10} & = & x_{4} \, \mathbf{a}_{1} + y_{4} \, \mathbf{a}_{2} + z_{4} \, \mathbf{a}_{3} & = & \frac{1}{2}\left(x_{4}+y_{4}\right)a \, \mathbf{\hat{x}} + \frac{\sqrt{3}}{2}\left(-x_{4}+y_{4}\right)a \, \mathbf{\hat{y}} + z_{4}c \, \mathbf{\hat{z}} & \left(3a\right) & \mbox{Ca II} \\ 
\mathbf{B}_{11} & = & -y_{4} \, \mathbf{a}_{1} + \left(x_{4}-y_{4}\right) \, \mathbf{a}_{2} + \left(\frac{2}{3} +z_{4}\right) \, \mathbf{a}_{3} & = & \left(\frac{1}{2}x_{4}-y_{4}\right)a \, \mathbf{\hat{x}} + \frac{\sqrt{3}}{2}x_{4}a \, \mathbf{\hat{y}} + \left(\frac{2}{3} +z_{4}\right)c \, \mathbf{\hat{z}} & \left(3a\right) & \mbox{Ca II} \\ 
\mathbf{B}_{12} & = & \left(-x_{4}+y_{4}\right) \, \mathbf{a}_{1}-x_{4} \, \mathbf{a}_{2} + \left(\frac{1}{3} +z_{4}\right) \, \mathbf{a}_{3} & = & \left(-x_{4}+\frac{1}{2}y_{4}\right)a \, \mathbf{\hat{x}}-\frac{\sqrt{3}}{2}y_{4}a \, \mathbf{\hat{y}} + \left(\frac{1}{3} +z_{4}\right)c \, \mathbf{\hat{z}} & \left(3a\right) & \mbox{Ca II} \\ 
\mathbf{B}_{13} & = & x_{5} \, \mathbf{a}_{1} + y_{5} \, \mathbf{a}_{2} + z_{5} \, \mathbf{a}_{3} & = & \frac{1}{2}\left(x_{5}+y_{5}\right)a \, \mathbf{\hat{x}} + \frac{\sqrt{3}}{2}\left(-x_{5}+y_{5}\right)a \, \mathbf{\hat{y}} + z_{5}c \, \mathbf{\hat{z}} & \left(3a\right) & \mbox{Ca III} \\ 
\mathbf{B}_{14} & = & -y_{5} \, \mathbf{a}_{1} + \left(x_{5}-y_{5}\right) \, \mathbf{a}_{2} + \left(\frac{2}{3} +z_{5}\right) \, \mathbf{a}_{3} & = & \left(\frac{1}{2}x_{5}-y_{5}\right)a \, \mathbf{\hat{x}} + \frac{\sqrt{3}}{2}x_{5}a \, \mathbf{\hat{y}} + \left(\frac{2}{3} +z_{5}\right)c \, \mathbf{\hat{z}} & \left(3a\right) & \mbox{Ca III} \\ 
\mathbf{B}_{15} & = & \left(-x_{5}+y_{5}\right) \, \mathbf{a}_{1}-x_{5} \, \mathbf{a}_{2} + \left(\frac{1}{3} +z_{5}\right) \, \mathbf{a}_{3} & = & \left(-x_{5}+\frac{1}{2}y_{5}\right)a \, \mathbf{\hat{x}}-\frac{\sqrt{3}}{2}y_{5}a \, \mathbf{\hat{y}} + \left(\frac{1}{3} +z_{5}\right)c \, \mathbf{\hat{z}} & \left(3a\right) & \mbox{Ca III} \\ 
\mathbf{B}_{16} & = & x_{6} \, \mathbf{a}_{1} + y_{6} \, \mathbf{a}_{2} + z_{6} \, \mathbf{a}_{3} & = & \frac{1}{2}\left(x_{6}+y_{6}\right)a \, \mathbf{\hat{x}} + \frac{\sqrt{3}}{2}\left(-x_{6}+y_{6}\right)a \, \mathbf{\hat{y}} + z_{6}c \, \mathbf{\hat{z}} & \left(3a\right) & \mbox{F I} \\ 
\mathbf{B}_{17} & = & -y_{6} \, \mathbf{a}_{1} + \left(x_{6}-y_{6}\right) \, \mathbf{a}_{2} + \left(\frac{2}{3} +z_{6}\right) \, \mathbf{a}_{3} & = & \left(\frac{1}{2}x_{6}-y_{6}\right)a \, \mathbf{\hat{x}} + \frac{\sqrt{3}}{2}x_{6}a \, \mathbf{\hat{y}} + \left(\frac{2}{3} +z_{6}\right)c \, \mathbf{\hat{z}} & \left(3a\right) & \mbox{F I} \\ 
\mathbf{B}_{18} & = & \left(-x_{6}+y_{6}\right) \, \mathbf{a}_{1}-x_{6} \, \mathbf{a}_{2} + \left(\frac{1}{3} +z_{6}\right) \, \mathbf{a}_{3} & = & \left(-x_{6}+\frac{1}{2}y_{6}\right)a \, \mathbf{\hat{x}}-\frac{\sqrt{3}}{2}y_{6}a \, \mathbf{\hat{y}} + \left(\frac{1}{3} +z_{6}\right)c \, \mathbf{\hat{z}} & \left(3a\right) & \mbox{F I} \\ 
\mathbf{B}_{19} & = & x_{7} \, \mathbf{a}_{1} + y_{7} \, \mathbf{a}_{2} + z_{7} \, \mathbf{a}_{3} & = & \frac{1}{2}\left(x_{7}+y_{7}\right)a \, \mathbf{\hat{x}} + \frac{\sqrt{3}}{2}\left(-x_{7}+y_{7}\right)a \, \mathbf{\hat{y}} + z_{7}c \, \mathbf{\hat{z}} & \left(3a\right) & \mbox{F II} \\ 
\mathbf{B}_{20} & = & -y_{7} \, \mathbf{a}_{1} + \left(x_{7}-y_{7}\right) \, \mathbf{a}_{2} + \left(\frac{2}{3} +z_{7}\right) \, \mathbf{a}_{3} & = & \left(\frac{1}{2}x_{7}-y_{7}\right)a \, \mathbf{\hat{x}} + \frac{\sqrt{3}}{2}x_{7}a \, \mathbf{\hat{y}} + \left(\frac{2}{3} +z_{7}\right)c \, \mathbf{\hat{z}} & \left(3a\right) & \mbox{F II} \\ 
\mathbf{B}_{21} & = & \left(-x_{7}+y_{7}\right) \, \mathbf{a}_{1}-x_{7} \, \mathbf{a}_{2} + \left(\frac{1}{3} +z_{7}\right) \, \mathbf{a}_{3} & = & \left(-x_{7}+\frac{1}{2}y_{7}\right)a \, \mathbf{\hat{x}}-\frac{\sqrt{3}}{2}y_{7}a \, \mathbf{\hat{y}} + \left(\frac{1}{3} +z_{7}\right)c \, \mathbf{\hat{z}} & \left(3a\right) & \mbox{F II} \\ 
\mathbf{B}_{22} & = & x_{8} \, \mathbf{a}_{1} + y_{8} \, \mathbf{a}_{2} + z_{8} \, \mathbf{a}_{3} & = & \frac{1}{2}\left(x_{8}+y_{8}\right)a \, \mathbf{\hat{x}} + \frac{\sqrt{3}}{2}\left(-x_{8}+y_{8}\right)a \, \mathbf{\hat{y}} + z_{8}c \, \mathbf{\hat{z}} & \left(3a\right) & \mbox{F III} \\ 
\mathbf{B}_{23} & = & -y_{8} \, \mathbf{a}_{1} + \left(x_{8}-y_{8}\right) \, \mathbf{a}_{2} + \left(\frac{2}{3} +z_{8}\right) \, \mathbf{a}_{3} & = & \left(\frac{1}{2}x_{8}-y_{8}\right)a \, \mathbf{\hat{x}} + \frac{\sqrt{3}}{2}x_{8}a \, \mathbf{\hat{y}} + \left(\frac{2}{3} +z_{8}\right)c \, \mathbf{\hat{z}} & \left(3a\right) & \mbox{F III} \\ 
\mathbf{B}_{24} & = & \left(-x_{8}+y_{8}\right) \, \mathbf{a}_{1}-x_{8} \, \mathbf{a}_{2} + \left(\frac{1}{3} +z_{8}\right) \, \mathbf{a}_{3} & = & \left(-x_{8}+\frac{1}{2}y_{8}\right)a \, \mathbf{\hat{x}}-\frac{\sqrt{3}}{2}y_{8}a \, \mathbf{\hat{y}} + \left(\frac{1}{3} +z_{8}\right)c \, \mathbf{\hat{z}} & \left(3a\right) & \mbox{F III} \\ 
\mathbf{B}_{25} & = & x_{9} \, \mathbf{a}_{1} + y_{9} \, \mathbf{a}_{2} + z_{9} \, \mathbf{a}_{3} & = & \frac{1}{2}\left(x_{9}+y_{9}\right)a \, \mathbf{\hat{x}} + \frac{\sqrt{3}}{2}\left(-x_{9}+y_{9}\right)a \, \mathbf{\hat{y}} + z_{9}c \, \mathbf{\hat{z}} & \left(3a\right) & \mbox{Na} \\ 
\mathbf{B}_{26} & = & -y_{9} \, \mathbf{a}_{1} + \left(x_{9}-y_{9}\right) \, \mathbf{a}_{2} + \left(\frac{2}{3} +z_{9}\right) \, \mathbf{a}_{3} & = & \left(\frac{1}{2}x_{9}-y_{9}\right)a \, \mathbf{\hat{x}} + \frac{\sqrt{3}}{2}x_{9}a \, \mathbf{\hat{y}} + \left(\frac{2}{3} +z_{9}\right)c \, \mathbf{\hat{z}} & \left(3a\right) & \mbox{Na} \\ 
\mathbf{B}_{27} & = & \left(-x_{9}+y_{9}\right) \, \mathbf{a}_{1}-x_{9} \, \mathbf{a}_{2} + \left(\frac{1}{3} +z_{9}\right) \, \mathbf{a}_{3} & = & \left(-x_{9}+\frac{1}{2}y_{9}\right)a \, \mathbf{\hat{x}}-\frac{\sqrt{3}}{2}y_{9}a \, \mathbf{\hat{y}} + \left(\frac{1}{3} +z_{9}\right)c \, \mathbf{\hat{z}} & \left(3a\right) & \mbox{Na} \\ 
\mathbf{B}_{28} & = & x_{10} \, \mathbf{a}_{1} + y_{10} \, \mathbf{a}_{2} + z_{10} \, \mathbf{a}_{3} & = & \frac{1}{2}\left(x_{10}+y_{10}\right)a \, \mathbf{\hat{x}} + \frac{\sqrt{3}}{2}\left(-x_{10}+y_{10}\right)a \, \mathbf{\hat{y}} + z_{10}c \, \mathbf{\hat{z}} & \left(3a\right) & \mbox{O I} \\ 
\mathbf{B}_{29} & = & -y_{10} \, \mathbf{a}_{1} + \left(x_{10}-y_{10}\right) \, \mathbf{a}_{2} + \left(\frac{2}{3} +z_{10}\right) \, \mathbf{a}_{3} & = & \left(\frac{1}{2}x_{10}-y_{10}\right)a \, \mathbf{\hat{x}} + \frac{\sqrt{3}}{2}x_{10}a \, \mathbf{\hat{y}} + \left(\frac{2}{3} +z_{10}\right)c \, \mathbf{\hat{z}} & \left(3a\right) & \mbox{O I} \\ 
\mathbf{B}_{30} & = & \left(-x_{10}+y_{10}\right) \, \mathbf{a}_{1}-x_{10} \, \mathbf{a}_{2} + \left(\frac{1}{3} +z_{10}\right) \, \mathbf{a}_{3} & = & \left(-x_{10}+\frac{1}{2}y_{10}\right)a \, \mathbf{\hat{x}}-\frac{\sqrt{3}}{2}y_{10}a \, \mathbf{\hat{y}} + \left(\frac{1}{3} +z_{10}\right)c \, \mathbf{\hat{z}} & \left(3a\right) & \mbox{O I} \\ 
\mathbf{B}_{31} & = & x_{11} \, \mathbf{a}_{1} + y_{11} \, \mathbf{a}_{2} + z_{11} \, \mathbf{a}_{3} & = & \frac{1}{2}\left(x_{11}+y_{11}\right)a \, \mathbf{\hat{x}} + \frac{\sqrt{3}}{2}\left(-x_{11}+y_{11}\right)a \, \mathbf{\hat{y}} + z_{11}c \, \mathbf{\hat{z}} & \left(3a\right) & \mbox{O II} \\ 
\mathbf{B}_{32} & = & -y_{11} \, \mathbf{a}_{1} + \left(x_{11}-y_{11}\right) \, \mathbf{a}_{2} + \left(\frac{2}{3} +z_{11}\right) \, \mathbf{a}_{3} & = & \left(\frac{1}{2}x_{11}-y_{11}\right)a \, \mathbf{\hat{x}} + \frac{\sqrt{3}}{2}x_{11}a \, \mathbf{\hat{y}} + \left(\frac{2}{3} +z_{11}\right)c \, \mathbf{\hat{z}} & \left(3a\right) & \mbox{O II} \\ 
\mathbf{B}_{33} & = & \left(-x_{11}+y_{11}\right) \, \mathbf{a}_{1}-x_{11} \, \mathbf{a}_{2} + \left(\frac{1}{3} +z_{11}\right) \, \mathbf{a}_{3} & = & \left(-x_{11}+\frac{1}{2}y_{11}\right)a \, \mathbf{\hat{x}}-\frac{\sqrt{3}}{2}y_{11}a \, \mathbf{\hat{y}} + \left(\frac{1}{3} +z_{11}\right)c \, \mathbf{\hat{z}} & \left(3a\right) & \mbox{O II} \\ 
\mathbf{B}_{34} & = & x_{12} \, \mathbf{a}_{1} + y_{12} \, \mathbf{a}_{2} + z_{12} \, \mathbf{a}_{3} & = & \frac{1}{2}\left(x_{12}+y_{12}\right)a \, \mathbf{\hat{x}} + \frac{\sqrt{3}}{2}\left(-x_{12}+y_{12}\right)a \, \mathbf{\hat{y}} + z_{12}c \, \mathbf{\hat{z}} & \left(3a\right) & \mbox{O III} \\ 
\mathbf{B}_{35} & = & -y_{12} \, \mathbf{a}_{1} + \left(x_{12}-y_{12}\right) \, \mathbf{a}_{2} + \left(\frac{2}{3} +z_{12}\right) \, \mathbf{a}_{3} & = & \left(\frac{1}{2}x_{12}-y_{12}\right)a \, \mathbf{\hat{x}} + \frac{\sqrt{3}}{2}x_{12}a \, \mathbf{\hat{y}} + \left(\frac{2}{3} +z_{12}\right)c \, \mathbf{\hat{z}} & \left(3a\right) & \mbox{O III} \\ 
\mathbf{B}_{36} & = & \left(-x_{12}+y_{12}\right) \, \mathbf{a}_{1}-x_{12} \, \mathbf{a}_{2} + \left(\frac{1}{3} +z_{12}\right) \, \mathbf{a}_{3} & = & \left(-x_{12}+\frac{1}{2}y_{12}\right)a \, \mathbf{\hat{x}}-\frac{\sqrt{3}}{2}y_{12}a \, \mathbf{\hat{y}} + \left(\frac{1}{3} +z_{12}\right)c \, \mathbf{\hat{z}} & \left(3a\right) & \mbox{O III} \\ 
\mathbf{B}_{37} & = & x_{13} \, \mathbf{a}_{1} + y_{13} \, \mathbf{a}_{2} + z_{13} \, \mathbf{a}_{3} & = & \frac{1}{2}\left(x_{13}+y_{13}\right)a \, \mathbf{\hat{x}} + \frac{\sqrt{3}}{2}\left(-x_{13}+y_{13}\right)a \, \mathbf{\hat{y}} + z_{13}c \, \mathbf{\hat{z}} & \left(3a\right) & \mbox{O IV} \\ 
\mathbf{B}_{38} & = & -y_{13} \, \mathbf{a}_{1} + \left(x_{13}-y_{13}\right) \, \mathbf{a}_{2} + \left(\frac{2}{3} +z_{13}\right) \, \mathbf{a}_{3} & = & \left(\frac{1}{2}x_{13}-y_{13}\right)a \, \mathbf{\hat{x}} + \frac{\sqrt{3}}{2}x_{13}a \, \mathbf{\hat{y}} + \left(\frac{2}{3} +z_{13}\right)c \, \mathbf{\hat{z}} & \left(3a\right) & \mbox{O IV} \\ 
\mathbf{B}_{39} & = & \left(-x_{13}+y_{13}\right) \, \mathbf{a}_{1}-x_{13} \, \mathbf{a}_{2} + \left(\frac{1}{3} +z_{13}\right) \, \mathbf{a}_{3} & = & \left(-x_{13}+\frac{1}{2}y_{13}\right)a \, \mathbf{\hat{x}}-\frac{\sqrt{3}}{2}y_{13}a \, \mathbf{\hat{y}} + \left(\frac{1}{3} +z_{13}\right)c \, \mathbf{\hat{z}} & \left(3a\right) & \mbox{O IV} \\ 
\mathbf{B}_{40} & = & x_{14} \, \mathbf{a}_{1} + y_{14} \, \mathbf{a}_{2} + z_{14} \, \mathbf{a}_{3} & = & \frac{1}{2}\left(x_{14}+y_{14}\right)a \, \mathbf{\hat{x}} + \frac{\sqrt{3}}{2}\left(-x_{14}+y_{14}\right)a \, \mathbf{\hat{y}} + z_{14}c \, \mathbf{\hat{z}} & \left(3a\right) & \mbox{O V} \\ 
\mathbf{B}_{41} & = & -y_{14} \, \mathbf{a}_{1} + \left(x_{14}-y_{14}\right) \, \mathbf{a}_{2} + \left(\frac{2}{3} +z_{14}\right) \, \mathbf{a}_{3} & = & \left(\frac{1}{2}x_{14}-y_{14}\right)a \, \mathbf{\hat{x}} + \frac{\sqrt{3}}{2}x_{14}a \, \mathbf{\hat{y}} + \left(\frac{2}{3} +z_{14}\right)c \, \mathbf{\hat{z}} & \left(3a\right) & \mbox{O V} \\ 
\mathbf{B}_{42} & = & \left(-x_{14}+y_{14}\right) \, \mathbf{a}_{1}-x_{14} \, \mathbf{a}_{2} + \left(\frac{1}{3} +z_{14}\right) \, \mathbf{a}_{3} & = & \left(-x_{14}+\frac{1}{2}y_{14}\right)a \, \mathbf{\hat{x}}-\frac{\sqrt{3}}{2}y_{14}a \, \mathbf{\hat{y}} + \left(\frac{1}{3} +z_{14}\right)c \, \mathbf{\hat{z}} & \left(3a\right) & \mbox{O V} \\ 
\mathbf{B}_{43} & = & x_{15} \, \mathbf{a}_{1} + y_{15} \, \mathbf{a}_{2} + z_{15} \, \mathbf{a}_{3} & = & \frac{1}{2}\left(x_{15}+y_{15}\right)a \, \mathbf{\hat{x}} + \frac{\sqrt{3}}{2}\left(-x_{15}+y_{15}\right)a \, \mathbf{\hat{y}} + z_{15}c \, \mathbf{\hat{z}} & \left(3a\right) & \mbox{O VI} \\ 
\mathbf{B}_{44} & = & -y_{15} \, \mathbf{a}_{1} + \left(x_{15}-y_{15}\right) \, \mathbf{a}_{2} + \left(\frac{2}{3} +z_{15}\right) \, \mathbf{a}_{3} & = & \left(\frac{1}{2}x_{15}-y_{15}\right)a \, \mathbf{\hat{x}} + \frac{\sqrt{3}}{2}x_{15}a \, \mathbf{\hat{y}} + \left(\frac{2}{3} +z_{15}\right)c \, \mathbf{\hat{z}} & \left(3a\right) & \mbox{O VI} \\ 
\mathbf{B}_{45} & = & \left(-x_{15}+y_{15}\right) \, \mathbf{a}_{1}-x_{15} \, \mathbf{a}_{2} + \left(\frac{1}{3} +z_{15}\right) \, \mathbf{a}_{3} & = & \left(-x_{15}+\frac{1}{2}y_{15}\right)a \, \mathbf{\hat{x}}-\frac{\sqrt{3}}{2}y_{15}a \, \mathbf{\hat{y}} + \left(\frac{1}{3} +z_{15}\right)c \, \mathbf{\hat{z}} & \left(3a\right) & \mbox{O VI} \\ 
\mathbf{B}_{46} & = & x_{16} \, \mathbf{a}_{1} + y_{16} \, \mathbf{a}_{2} + z_{16} \, \mathbf{a}_{3} & = & \frac{1}{2}\left(x_{16}+y_{16}\right)a \, \mathbf{\hat{x}} + \frac{\sqrt{3}}{2}\left(-x_{16}+y_{16}\right)a \, \mathbf{\hat{y}} + z_{16}c \, \mathbf{\hat{z}} & \left(3a\right) & \mbox{O VII} \\ 
\mathbf{B}_{47} & = & -y_{16} \, \mathbf{a}_{1} + \left(x_{16}-y_{16}\right) \, \mathbf{a}_{2} + \left(\frac{2}{3} +z_{16}\right) \, \mathbf{a}_{3} & = & \left(\frac{1}{2}x_{16}-y_{16}\right)a \, \mathbf{\hat{x}} + \frac{\sqrt{3}}{2}x_{16}a \, \mathbf{\hat{y}} + \left(\frac{2}{3} +z_{16}\right)c \, \mathbf{\hat{z}} & \left(3a\right) & \mbox{O VII} \\ 
\mathbf{B}_{48} & = & \left(-x_{16}+y_{16}\right) \, \mathbf{a}_{1}-x_{16} \, \mathbf{a}_{2} + \left(\frac{1}{3} +z_{16}\right) \, \mathbf{a}_{3} & = & \left(-x_{16}+\frac{1}{2}y_{16}\right)a \, \mathbf{\hat{x}}-\frac{\sqrt{3}}{2}y_{16}a \, \mathbf{\hat{y}} + \left(\frac{1}{3} +z_{16}\right)c \, \mathbf{\hat{z}} & \left(3a\right) & \mbox{O VII} \\ 
\end{longtabu}
\renewcommand{\arraystretch}{1.0}
\noindent \hrulefill
\\
\textbf{References:}
\vspace*{-0.25cm}
\begin{flushleft}
  - \bibentry{Sheldrickite_NaCa3CO32F3H2O_CanMineral_1997}. \\
\end{flushleft}
\textbf{Found in:}
\vspace*{-0.25cm}
\begin{flushleft}
  - \bibentry{Villars_PearsonsCrystalData_2013}. \\
\end{flushleft}
\noindent \hrulefill
\\
\textbf{Geometry files:}
\\
\noindent  - CIF: pp. {\hyperref[A2B3C3DE7_hP48_145_2a_3a_3a_a_7a_cif]{\pageref{A2B3C3DE7_hP48_145_2a_3a_3a_a_7a_cif}}} \\
\noindent  - POSCAR: pp. {\hyperref[A2B3C3DE7_hP48_145_2a_3a_3a_a_7a_poscar]{\pageref{A2B3C3DE7_hP48_145_2a_3a_3a_a_7a_poscar}}} \\
\onecolumn
{\phantomsection\label{A3BC_hR5_146_b_a_a}}
\subsection*{\huge \textbf{{\normalfont \begin{raggedleft}$\gamma$-Ag$_{3}$SI (Low-temperature) Structure: \end{raggedleft} \\ A3BC\_hR5\_146\_b\_a\_a}}}
\noindent \hrulefill
\vspace*{0.25cm}
\begin{figure}[htp]
  \centering
  \vspace{-1em}
  {\includegraphics[width=1\textwidth]{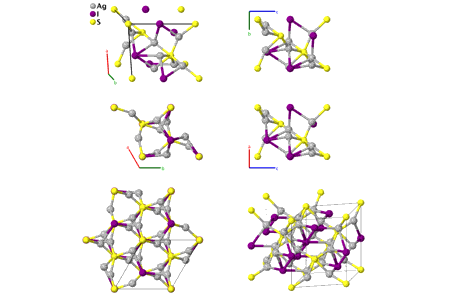}}
\end{figure}
\vspace*{-0.5cm}
\renewcommand{\arraystretch}{1.5}
\begin{equation*}
  \begin{array}{>{$\hspace{-0.15cm}}l<{$}>{$}p{0.5cm}<{$}>{$}p{18.5cm}<{$}}
    \mbox{\large \textbf{Prototype}} &\colon & \ce{$\gamma$-Ag3SI} \\
    \mbox{\large \textbf{\AFLOW\ prototype label}} &\colon & \mbox{A3BC\_hR5\_146\_b\_a\_a} \\
    \mbox{\large \textbf{\textit{Strukturbericht} designation}} &\colon & \mbox{None} \\
    \mbox{\large \textbf{Pearson symbol}} &\colon & \mbox{hR5} \\
    \mbox{\large \textbf{Space group number}} &\colon & 146 \\
    \mbox{\large \textbf{Space group symbol}} &\colon & R3 \\
    \mbox{\large \textbf{\AFLOW\ prototype command}} &\colon &  \texttt{aflow} \,  \, \texttt{-{}-proto=A3BC\_hR5\_146\_b\_a\_a [-{}-hex]} \, \newline \texttt{-{}-params=}{a,c/a,x_{1},x_{2},x_{3},y_{3},z_{3} }
  \end{array}
\end{equation*}
\renewcommand{\arraystretch}{1.0}

\noindent \parbox{1 \linewidth}{
\noindent \hrulefill
\\
\textbf{Rhombohedral primitive vectors:} \\
\vspace*{-0.25cm}
\begin{tabular}{cc}
  \begin{tabular}{c}
    \parbox{0.6 \linewidth}{
      \renewcommand{\arraystretch}{1.5}
      \begin{equation*}
        \centering
        \begin{array}{ccc}
              \mathbf{a}_1 & = & ~ \frac12 \, a \, \mathbf{\hat{x}} - \frac{1}{2\sqrt{3}} \, a \, \mathbf{\hat{y}} + \frac13 \, c \, \mathbf{\hat{z}} \\
    \mathbf{a}_2 & = & \frac{1}{\sqrt{3}} \, a \, \mathbf{\hat{y}} + \frac13 \, c \, \mathbf{\hat{z}} \\
    \mathbf{a}_3 & = & - \frac12 \, a \, \mathbf{\hat{x}} - \frac{1}{2\sqrt{3}} \, a \, \mathbf{\hat{y}} + \frac13 \, c \, \mathbf{\hat{z}} \\

        \end{array}
      \end{equation*}
    }
    \renewcommand{\arraystretch}{1.0}
  \end{tabular}
  \begin{tabular}{c}
    \includegraphics[width=0.3\linewidth]{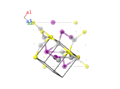} \\
  \end{tabular}
\end{tabular}

}
\vspace*{-0.25cm}

\noindent \hrulefill
\\
\textbf{Basis vectors:}
\vspace*{-0.25cm}
\renewcommand{\arraystretch}{1.5}
\begin{longtabu} to \textwidth{>{\centering $}X[-1,c,c]<{$}>{\centering $}X[-1,c,c]<{$}>{\centering $}X[-1,c,c]<{$}>{\centering $}X[-1,c,c]<{$}>{\centering $}X[-1,c,c]<{$}>{\centering $}X[-1,c,c]<{$}>{\centering $}X[-1,c,c]<{$}}
  & & \mbox{Lattice Coordinates} & & \mbox{Cartesian Coordinates} &\mbox{Wyckoff Position} & \mbox{Atom Type} \\  
  \mathbf{B}_{1} & = & x_{1} \, \mathbf{a}_{1} + x_{1} \, \mathbf{a}_{2} + x_{1} \, \mathbf{a}_{3} & = & x_{1}c \, \mathbf{\hat{z}} & \left(1a\right) & \mbox{I} \\ 
\mathbf{B}_{2} & = & x_{2} \, \mathbf{a}_{1} + x_{2} \, \mathbf{a}_{2} + x_{2} \, \mathbf{a}_{3} & = & x_{2}c \, \mathbf{\hat{z}} & \left(1a\right) & \mbox{S} \\ 
\mathbf{B}_{3} & = & x_{3} \, \mathbf{a}_{1} + y_{3} \, \mathbf{a}_{2} + z_{3} \, \mathbf{a}_{3} & = & \frac{1}{2}\left(x_{3}-z_{3}\right)a \, \mathbf{\hat{x}} + \left(-\frac{1}{2\sqrt{3}}x_{3}+\frac{1}{\sqrt{3}}y_{3}-\frac{1}{2\sqrt{3}}z_{3}\right)a \, \mathbf{\hat{y}} + \frac{1}{3}\left(x_{3}+y_{3}+z_{3}\right)c \, \mathbf{\hat{z}} & \left(3b\right) & \mbox{Ag} \\ 
\mathbf{B}_{4} & = & z_{3} \, \mathbf{a}_{1} + x_{3} \, \mathbf{a}_{2} + y_{3} \, \mathbf{a}_{3} & = & \frac{1}{2}\left(-y_{3}+z_{3}\right)a \, \mathbf{\hat{x}} + \left(\frac{1}{\sqrt{3}}x_{3}-\frac{1}{2\sqrt{3}}y_{3}-\frac{1}{2\sqrt{3}}z_{3}\right)a \, \mathbf{\hat{y}} + \frac{1}{3}\left(x_{3}+y_{3}+z_{3}\right)c \, \mathbf{\hat{z}} & \left(3b\right) & \mbox{Ag} \\ 
\mathbf{B}_{5} & = & y_{3} \, \mathbf{a}_{1} + z_{3} \, \mathbf{a}_{2} + x_{3} \, \mathbf{a}_{3} & = & \frac{1}{2}\left(-x_{3}+y_{3}\right)a \, \mathbf{\hat{x}} + \left(-\frac{1}{2\sqrt{3}}x_{3}-\frac{1}{2\sqrt{3}}y_{3}+\frac{1}{\sqrt{3}}z_{3}\right)a \, \mathbf{\hat{y}} + \frac{1}{3}\left(x_{3}+y_{3}+z_{3}\right)c \, \mathbf{\hat{z}} & \left(3b\right) & \mbox{Ag} \\ 
\end{longtabu}
\renewcommand{\arraystretch}{1.0}
\noindent \hrulefill
\\
\textbf{References:}
\vspace*{-0.25cm}
\begin{flushleft}
  - \bibentry{Hoshino_Ag3SI_JPhysSocJpn_1979}. \\
\end{flushleft}
\textbf{Found in:}
\vspace*{-0.25cm}
\begin{flushleft}
  - \bibentry{Villars_PearsonsCrystalData_2013}. \\
\end{flushleft}
\noindent \hrulefill
\\
\textbf{Geometry files:}
\\
\noindent  - CIF: pp. {\hyperref[A3BC_hR5_146_b_a_a_cif]{\pageref{A3BC_hR5_146_b_a_a_cif}}} \\
\noindent  - POSCAR: pp. {\hyperref[A3BC_hR5_146_b_a_a_poscar]{\pageref{A3BC_hR5_146_b_a_a_poscar}}} \\
\onecolumn
{\phantomsection\label{ABC3_hR10_146_2a_2a_2b}}
\subsection*{\huge \textbf{{\normalfont FePSe$_{3}$ Structure: ABC3\_hR10\_146\_2a\_2a\_2b}}}
\noindent \hrulefill
\vspace*{0.25cm}
\begin{figure}[htp]
  \centering
  \vspace{-1em}
  {\includegraphics[width=1\textwidth]{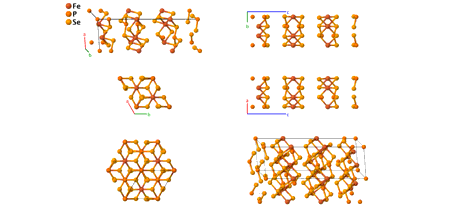}}
\end{figure}
\vspace*{-0.5cm}
\renewcommand{\arraystretch}{1.5}
\begin{equation*}
  \begin{array}{>{$\hspace{-0.15cm}}l<{$}>{$}p{0.5cm}<{$}>{$}p{18.5cm}<{$}}
    \mbox{\large \textbf{Prototype}} &\colon & \ce{FePSe3} \\
    \mbox{\large \textbf{\AFLOW\ prototype label}} &\colon & \mbox{ABC3\_hR10\_146\_2a\_2a\_2b} \\
    \mbox{\large \textbf{\textit{Strukturbericht} designation}} &\colon & \mbox{None} \\
    \mbox{\large \textbf{Pearson symbol}} &\colon & \mbox{hR10} \\
    \mbox{\large \textbf{Space group number}} &\colon & 146 \\
    \mbox{\large \textbf{Space group symbol}} &\colon & R3 \\
    \mbox{\large \textbf{\AFLOW\ prototype command}} &\colon &  \texttt{aflow} \,  \, \texttt{-{}-proto=ABC3\_hR10\_146\_2a\_2a\_2b [-{}-hex]} \, \newline \texttt{-{}-params=}{a,c/a,x_{1},x_{2},x_{3},x_{4},x_{5},y_{5},z_{5},x_{6},y_{6},z_{6} }
  \end{array}
\end{equation*}
\renewcommand{\arraystretch}{1.0}

\noindent \parbox{1 \linewidth}{
\noindent \hrulefill
\\
\textbf{Rhombohedral primitive vectors:} \\
\vspace*{-0.25cm}
\begin{tabular}{cc}
  \begin{tabular}{c}
    \parbox{0.6 \linewidth}{
      \renewcommand{\arraystretch}{1.5}
      \begin{equation*}
        \centering
        \begin{array}{ccc}
              \mathbf{a}_1 & = & ~ \frac12 \, a \, \mathbf{\hat{x}} - \frac{1}{2\sqrt{3}} \, a \, \mathbf{\hat{y}} + \frac13 \, c \, \mathbf{\hat{z}} \\
    \mathbf{a}_2 & = & \frac{1}{\sqrt{3}} \, a \, \mathbf{\hat{y}} + \frac13 \, c \, \mathbf{\hat{z}} \\
    \mathbf{a}_3 & = & - \frac12 \, a \, \mathbf{\hat{x}} - \frac{1}{2\sqrt{3}} \, a \, \mathbf{\hat{y}} + \frac13 \, c \, \mathbf{\hat{z}} \\

        \end{array}
      \end{equation*}
    }
    \renewcommand{\arraystretch}{1.0}
  \end{tabular}
  \begin{tabular}{c}
    \includegraphics[width=0.3\linewidth]{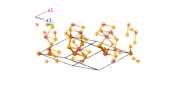} \\
  \end{tabular}
\end{tabular}

}
\vspace*{-0.25cm}

\noindent \hrulefill
\\
\textbf{Basis vectors:}
\vspace*{-0.25cm}
\renewcommand{\arraystretch}{1.5}
\begin{longtabu} to \textwidth{>{\centering $}X[-1,c,c]<{$}>{\centering $}X[-1,c,c]<{$}>{\centering $}X[-1,c,c]<{$}>{\centering $}X[-1,c,c]<{$}>{\centering $}X[-1,c,c]<{$}>{\centering $}X[-1,c,c]<{$}>{\centering $}X[-1,c,c]<{$}}
  & & \mbox{Lattice Coordinates} & & \mbox{Cartesian Coordinates} &\mbox{Wyckoff Position} & \mbox{Atom Type} \\  
  \mathbf{B}_{1} & = & x_{1} \, \mathbf{a}_{1} + x_{1} \, \mathbf{a}_{2} + x_{1} \, \mathbf{a}_{3} & = & x_{1}c \, \mathbf{\hat{z}} & \left(1a\right) & \mbox{Fe I} \\ 
\mathbf{B}_{2} & = & x_{2} \, \mathbf{a}_{1} + x_{2} \, \mathbf{a}_{2} + x_{2} \, \mathbf{a}_{3} & = & x_{2}c \, \mathbf{\hat{z}} & \left(1a\right) & \mbox{Fe II} \\ 
\mathbf{B}_{3} & = & x_{3} \, \mathbf{a}_{1} + x_{3} \, \mathbf{a}_{2} + x_{3} \, \mathbf{a}_{3} & = & x_{3}c \, \mathbf{\hat{z}} & \left(1a\right) & \mbox{P I} \\ 
\mathbf{B}_{4} & = & x_{4} \, \mathbf{a}_{1} + x_{4} \, \mathbf{a}_{2} + x_{4} \, \mathbf{a}_{3} & = & x_{4}c \, \mathbf{\hat{z}} & \left(1a\right) & \mbox{P II} \\ 
\mathbf{B}_{5} & = & x_{5} \, \mathbf{a}_{1} + y_{5} \, \mathbf{a}_{2} + z_{5} \, \mathbf{a}_{3} & = & \frac{1}{2}\left(x_{5}-z_{5}\right)a \, \mathbf{\hat{x}} + \left(-\frac{1}{2\sqrt{3}}x_{5}+\frac{1}{\sqrt{3}}y_{5}-\frac{1}{2\sqrt{3}}z_{5}\right)a \, \mathbf{\hat{y}} + \frac{1}{3}\left(x_{5}+y_{5}+z_{5}\right)c \, \mathbf{\hat{z}} & \left(3b\right) & \mbox{Se I} \\ 
\mathbf{B}_{6} & = & z_{5} \, \mathbf{a}_{1} + x_{5} \, \mathbf{a}_{2} + y_{5} \, \mathbf{a}_{3} & = & \frac{1}{2}\left(-y_{5}+z_{5}\right)a \, \mathbf{\hat{x}} + \left(\frac{1}{\sqrt{3}}x_{5}-\frac{1}{2\sqrt{3}}y_{5}-\frac{1}{2\sqrt{3}}z_{5}\right)a \, \mathbf{\hat{y}} + \frac{1}{3}\left(x_{5}+y_{5}+z_{5}\right)c \, \mathbf{\hat{z}} & \left(3b\right) & \mbox{Se I} \\ 
\mathbf{B}_{7} & = & y_{5} \, \mathbf{a}_{1} + z_{5} \, \mathbf{a}_{2} + x_{5} \, \mathbf{a}_{3} & = & \frac{1}{2}\left(-x_{5}+y_{5}\right)a \, \mathbf{\hat{x}} + \left(-\frac{1}{2\sqrt{3}}x_{5}-\frac{1}{2\sqrt{3}}y_{5}+\frac{1}{\sqrt{3}}z_{5}\right)a \, \mathbf{\hat{y}} + \frac{1}{3}\left(x_{5}+y_{5}+z_{5}\right)c \, \mathbf{\hat{z}} & \left(3b\right) & \mbox{Se I} \\ 
\mathbf{B}_{8} & = & x_{6} \, \mathbf{a}_{1} + y_{6} \, \mathbf{a}_{2} + z_{6} \, \mathbf{a}_{3} & = & \frac{1}{2}\left(x_{6}-z_{6}\right)a \, \mathbf{\hat{x}} + \left(-\frac{1}{2\sqrt{3}}x_{6}+\frac{1}{\sqrt{3}}y_{6}-\frac{1}{2\sqrt{3}}z_{6}\right)a \, \mathbf{\hat{y}} + \frac{1}{3}\left(x_{6}+y_{6}+z_{6}\right)c \, \mathbf{\hat{z}} & \left(3b\right) & \mbox{Se II} \\ 
\mathbf{B}_{9} & = & z_{6} \, \mathbf{a}_{1} + x_{6} \, \mathbf{a}_{2} + y_{6} \, \mathbf{a}_{3} & = & \frac{1}{2}\left(-y_{6}+z_{6}\right)a \, \mathbf{\hat{x}} + \left(\frac{1}{\sqrt{3}}x_{6}-\frac{1}{2\sqrt{3}}y_{6}-\frac{1}{2\sqrt{3}}z_{6}\right)a \, \mathbf{\hat{y}} + \frac{1}{3}\left(x_{6}+y_{6}+z_{6}\right)c \, \mathbf{\hat{z}} & \left(3b\right) & \mbox{Se II} \\ 
\mathbf{B}_{10} & = & y_{6} \, \mathbf{a}_{1} + z_{6} \, \mathbf{a}_{2} + x_{6} \, \mathbf{a}_{3} & = & \frac{1}{2}\left(-x_{6}+y_{6}\right)a \, \mathbf{\hat{x}} + \left(-\frac{1}{2\sqrt{3}}x_{6}-\frac{1}{2\sqrt{3}}y_{6}+\frac{1}{\sqrt{3}}z_{6}\right)a \, \mathbf{\hat{y}} + \frac{1}{3}\left(x_{6}+y_{6}+z_{6}\right)c \, \mathbf{\hat{z}} & \left(3b\right) & \mbox{Se II} \\ 
\end{longtabu}
\renewcommand{\arraystretch}{1.0}
\noindent \hrulefill
\\
\textbf{References:}
\vspace*{-0.25cm}
\begin{flushleft}
  - \bibentry{Klingen_FePSe3_ZAnorgAllgChem_1973}. \\
\end{flushleft}
\textbf{Found in:}
\vspace*{-0.25cm}
\begin{flushleft}
  - \bibentry{Villars_PearsonsCrystalData_2013}. \\
\end{flushleft}
\noindent \hrulefill
\\
\textbf{Geometry files:}
\\
\noindent  - CIF: pp. {\hyperref[ABC3_hR10_146_2a_2a_2b_cif]{\pageref{ABC3_hR10_146_2a_2a_2b_cif}}} \\
\noindent  - POSCAR: pp. {\hyperref[ABC3_hR10_146_2a_2a_2b_poscar]{\pageref{ABC3_hR10_146_2a_2a_2b_poscar}}} \\
\onecolumn
{\phantomsection\label{A2B4C_hR42_148_2f_4f_f}}
\subsection*{\huge \textbf{{\normalfont \begin{raggedleft}Phenakite (Be$_2$SiO$_4$, $S1_{3}$) Structure: \end{raggedleft} \\ A2B4C\_hR42\_148\_2f\_4f\_f}}}
\noindent \hrulefill
\vspace*{0.25cm}
\begin{figure}[htp]
  \centering
  \vspace{-1em}
  {\includegraphics[width=1\textwidth]{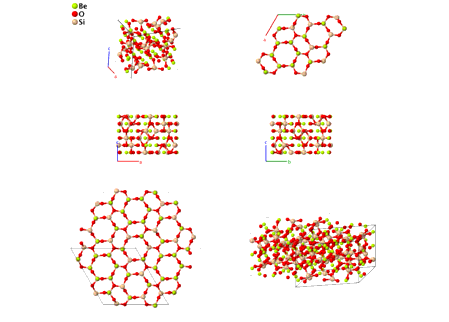}}
\end{figure}
\vspace*{-0.5cm}
\renewcommand{\arraystretch}{1.5}
\begin{equation*}
  \begin{array}{>{$\hspace{-0.15cm}}l<{$}>{$}p{0.5cm}<{$}>{$}p{18.5cm}<{$}}
    \mbox{\large \textbf{Prototype}} &\colon & \ce{Be2SiO4} \\
    \mbox{\large \textbf{\AFLOW\ prototype label}} &\colon & \mbox{A2B4C\_hR42\_148\_2f\_4f\_f} \\
    \mbox{\large \textbf{\textit{Strukturbericht} designation}} &\colon & \mbox{$S1_{3}$} \\
    \mbox{\large \textbf{Pearson symbol}} &\colon & \mbox{hR42} \\
    \mbox{\large \textbf{Space group number}} &\colon & 148 \\
    \mbox{\large \textbf{Space group symbol}} &\colon & R\bar{3} \\
    \mbox{\large \textbf{\AFLOW\ prototype command}} &\colon &  \texttt{aflow} \,  \, \texttt{-{}-proto=A2B4C\_hR42\_148\_2f\_4f\_f [-{}-hex]} \, \newline \texttt{-{}-params=}{a,c/a,x_{1},y_{1},z_{1},x_{2},y_{2},z_{2},x_{3},y_{3},z_{3},x_{4},y_{4},z_{4},x_{5},y_{5},z_{5},x_{6},y_{6},z_{6},x_{7},} \newline {y_{7},z_{7} }
  \end{array}
\end{equation*}
\renewcommand{\arraystretch}{1.0}

\vspace*{-0.25cm}
\noindent \hrulefill
\\
\textbf{ Other compounds with this structure:}
\begin{itemize}
   \item{ Zn$_2$SiO$_4$ (willemite), (Zn,Mn)$_2$SiO$_4$ (troostite), LiZnPO$_4$   }
\end{itemize}
\noindent \parbox{1 \linewidth}{
\noindent \hrulefill
\\
\textbf{Rhombohedral primitive vectors:} \\
\vspace*{-0.25cm}
\begin{tabular}{cc}
  \begin{tabular}{c}
    \parbox{0.6 \linewidth}{
      \renewcommand{\arraystretch}{1.5}
      \begin{equation*}
        \centering
        \begin{array}{ccc}
              \mathbf{a}_1 & = & ~ \frac12 \, a \, \mathbf{\hat{x}} - \frac{1}{2\sqrt{3}} \, a \, \mathbf{\hat{y}} + \frac13 \, c \, \mathbf{\hat{z}} \\
    \mathbf{a}_2 & = & \frac{1}{\sqrt{3}} \, a \, \mathbf{\hat{y}} + \frac13 \, c \, \mathbf{\hat{z}} \\
    \mathbf{a}_3 & = & - \frac12 \, a \, \mathbf{\hat{x}} - \frac{1}{2\sqrt{3}} \, a \, \mathbf{\hat{y}} + \frac13 \, c \, \mathbf{\hat{z}} \\

        \end{array}
      \end{equation*}
    }
    \renewcommand{\arraystretch}{1.0}
  \end{tabular}
  \begin{tabular}{c}
    \includegraphics[width=0.3\linewidth]{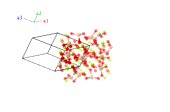} \\
  \end{tabular}
\end{tabular}

}
\vspace*{-0.25cm}

\noindent \hrulefill
\\
\textbf{Basis vectors:}
\vspace*{-0.25cm}
\renewcommand{\arraystretch}{1.5}
\begin{longtabu} to \textwidth{>{\centering $}X[-1,c,c]<{$}>{\centering $}X[-1,c,c]<{$}>{\centering $}X[-1,c,c]<{$}>{\centering $}X[-1,c,c]<{$}>{\centering $}X[-1,c,c]<{$}>{\centering $}X[-1,c,c]<{$}>{\centering $}X[-1,c,c]<{$}}
  & & \mbox{Lattice Coordinates} & & \mbox{Cartesian Coordinates} &\mbox{Wyckoff Position} & \mbox{Atom Type} \\  
  \mathbf{B}_{1} & = & x_{1} \, \mathbf{a}_{1} + y_{1} \, \mathbf{a}_{2} + z_{1} \, \mathbf{a}_{3} & = & \frac{1}{2}\left(x_{1}-z_{1}\right)a \, \mathbf{\hat{x}} + \left(-\frac{1}{2\sqrt{3}}x_{1}+\frac{1}{\sqrt{3}}y_{1}-\frac{1}{2\sqrt{3}}z_{1}\right)a \, \mathbf{\hat{y}} + \frac{1}{3}\left(x_{1}+y_{1}+z_{1}\right)c \, \mathbf{\hat{z}} & \left(6f\right) & \mbox{Be I} \\ 
\mathbf{B}_{2} & = & z_{1} \, \mathbf{a}_{1} + x_{1} \, \mathbf{a}_{2} + y_{1} \, \mathbf{a}_{3} & = & \frac{1}{2}\left(-y_{1}+z_{1}\right)a \, \mathbf{\hat{x}} + \left(\frac{1}{\sqrt{3}}x_{1}-\frac{1}{2\sqrt{3}}y_{1}-\frac{1}{2\sqrt{3}}z_{1}\right)a \, \mathbf{\hat{y}} + \frac{1}{3}\left(x_{1}+y_{1}+z_{1}\right)c \, \mathbf{\hat{z}} & \left(6f\right) & \mbox{Be I} \\ 
\mathbf{B}_{3} & = & y_{1} \, \mathbf{a}_{1} + z_{1} \, \mathbf{a}_{2} + x_{1} \, \mathbf{a}_{3} & = & \frac{1}{2}\left(-x_{1}+y_{1}\right)a \, \mathbf{\hat{x}} + \left(-\frac{1}{2\sqrt{3}}x_{1}-\frac{1}{2\sqrt{3}}y_{1}+\frac{1}{\sqrt{3}}z_{1}\right)a \, \mathbf{\hat{y}} + \frac{1}{3}\left(x_{1}+y_{1}+z_{1}\right)c \, \mathbf{\hat{z}} & \left(6f\right) & \mbox{Be I} \\ 
\mathbf{B}_{4} & = & -x_{1} \, \mathbf{a}_{1}-y_{1} \, \mathbf{a}_{2}-z_{1} \, \mathbf{a}_{3} & = & \frac{1}{2}\left(-x_{1}+z_{1}\right)a \, \mathbf{\hat{x}} + \left(\frac{1}{2\sqrt{3}}x_{1}-\frac{1}{\sqrt{3}}y_{1}+\frac{1}{2\sqrt{3}}z_{1}\right)a \, \mathbf{\hat{y}}-\frac{1}{3}\left(x_{1}+y_{1}+z_{1}\right)c \, \mathbf{\hat{z}} & \left(6f\right) & \mbox{Be I} \\ 
\mathbf{B}_{5} & = & -z_{1} \, \mathbf{a}_{1}-x_{1} \, \mathbf{a}_{2}-y_{1} \, \mathbf{a}_{3} & = & \frac{1}{2}\left(y_{1}-z_{1}\right)a \, \mathbf{\hat{x}} + \left(-\frac{1}{\sqrt{3}}x_{1}+\frac{1}{2\sqrt{3}}y_{1}+\frac{1}{2\sqrt{3}}z_{1}\right)a \, \mathbf{\hat{y}}-\frac{1}{3}\left(x_{1}+y_{1}+z_{1}\right)c \, \mathbf{\hat{z}} & \left(6f\right) & \mbox{Be I} \\ 
\mathbf{B}_{6} & = & -y_{1} \, \mathbf{a}_{1}-z_{1} \, \mathbf{a}_{2}-x_{1} \, \mathbf{a}_{3} & = & \frac{1}{2}\left(x_{1}-y_{1}\right)a \, \mathbf{\hat{x}} + \left(\frac{1}{2\sqrt{3}}x_{1}+\frac{1}{2\sqrt{3}}y_{1}-\frac{1}{\sqrt{3}}z_{1}\right)a \, \mathbf{\hat{y}}-\frac{1}{3}\left(x_{1}+y_{1}+z_{1}\right)c \, \mathbf{\hat{z}} & \left(6f\right) & \mbox{Be I} \\ 
\mathbf{B}_{7} & = & x_{2} \, \mathbf{a}_{1} + y_{2} \, \mathbf{a}_{2} + z_{2} \, \mathbf{a}_{3} & = & \frac{1}{2}\left(x_{2}-z_{2}\right)a \, \mathbf{\hat{x}} + \left(-\frac{1}{2\sqrt{3}}x_{2}+\frac{1}{\sqrt{3}}y_{2}-\frac{1}{2\sqrt{3}}z_{2}\right)a \, \mathbf{\hat{y}} + \frac{1}{3}\left(x_{2}+y_{2}+z_{2}\right)c \, \mathbf{\hat{z}} & \left(6f\right) & \mbox{Be II} \\ 
\mathbf{B}_{8} & = & z_{2} \, \mathbf{a}_{1} + x_{2} \, \mathbf{a}_{2} + y_{2} \, \mathbf{a}_{3} & = & \frac{1}{2}\left(-y_{2}+z_{2}\right)a \, \mathbf{\hat{x}} + \left(\frac{1}{\sqrt{3}}x_{2}-\frac{1}{2\sqrt{3}}y_{2}-\frac{1}{2\sqrt{3}}z_{2}\right)a \, \mathbf{\hat{y}} + \frac{1}{3}\left(x_{2}+y_{2}+z_{2}\right)c \, \mathbf{\hat{z}} & \left(6f\right) & \mbox{Be II} \\ 
\mathbf{B}_{9} & = & y_{2} \, \mathbf{a}_{1} + z_{2} \, \mathbf{a}_{2} + x_{2} \, \mathbf{a}_{3} & = & \frac{1}{2}\left(-x_{2}+y_{2}\right)a \, \mathbf{\hat{x}} + \left(-\frac{1}{2\sqrt{3}}x_{2}-\frac{1}{2\sqrt{3}}y_{2}+\frac{1}{\sqrt{3}}z_{2}\right)a \, \mathbf{\hat{y}} + \frac{1}{3}\left(x_{2}+y_{2}+z_{2}\right)c \, \mathbf{\hat{z}} & \left(6f\right) & \mbox{Be II} \\ 
\mathbf{B}_{10} & = & -x_{2} \, \mathbf{a}_{1}-y_{2} \, \mathbf{a}_{2}-z_{2} \, \mathbf{a}_{3} & = & \frac{1}{2}\left(-x_{2}+z_{2}\right)a \, \mathbf{\hat{x}} + \left(\frac{1}{2\sqrt{3}}x_{2}-\frac{1}{\sqrt{3}}y_{2}+\frac{1}{2\sqrt{3}}z_{2}\right)a \, \mathbf{\hat{y}}-\frac{1}{3}\left(x_{2}+y_{2}+z_{2}\right)c \, \mathbf{\hat{z}} & \left(6f\right) & \mbox{Be II} \\ 
\mathbf{B}_{11} & = & -z_{2} \, \mathbf{a}_{1}-x_{2} \, \mathbf{a}_{2}-y_{2} \, \mathbf{a}_{3} & = & \frac{1}{2}\left(y_{2}-z_{2}\right)a \, \mathbf{\hat{x}} + \left(-\frac{1}{\sqrt{3}}x_{2}+\frac{1}{2\sqrt{3}}y_{2}+\frac{1}{2\sqrt{3}}z_{2}\right)a \, \mathbf{\hat{y}}-\frac{1}{3}\left(x_{2}+y_{2}+z_{2}\right)c \, \mathbf{\hat{z}} & \left(6f\right) & \mbox{Be II} \\ 
\mathbf{B}_{12} & = & -y_{2} \, \mathbf{a}_{1}-z_{2} \, \mathbf{a}_{2}-x_{2} \, \mathbf{a}_{3} & = & \frac{1}{2}\left(x_{2}-y_{2}\right)a \, \mathbf{\hat{x}} + \left(\frac{1}{2\sqrt{3}}x_{2}+\frac{1}{2\sqrt{3}}y_{2}-\frac{1}{\sqrt{3}}z_{2}\right)a \, \mathbf{\hat{y}}-\frac{1}{3}\left(x_{2}+y_{2}+z_{2}\right)c \, \mathbf{\hat{z}} & \left(6f\right) & \mbox{Be II} \\ 
\mathbf{B}_{13} & = & x_{3} \, \mathbf{a}_{1} + y_{3} \, \mathbf{a}_{2} + z_{3} \, \mathbf{a}_{3} & = & \frac{1}{2}\left(x_{3}-z_{3}\right)a \, \mathbf{\hat{x}} + \left(-\frac{1}{2\sqrt{3}}x_{3}+\frac{1}{\sqrt{3}}y_{3}-\frac{1}{2\sqrt{3}}z_{3}\right)a \, \mathbf{\hat{y}} + \frac{1}{3}\left(x_{3}+y_{3}+z_{3}\right)c \, \mathbf{\hat{z}} & \left(6f\right) & \mbox{O I} \\ 
\mathbf{B}_{14} & = & z_{3} \, \mathbf{a}_{1} + x_{3} \, \mathbf{a}_{2} + y_{3} \, \mathbf{a}_{3} & = & \frac{1}{2}\left(-y_{3}+z_{3}\right)a \, \mathbf{\hat{x}} + \left(\frac{1}{\sqrt{3}}x_{3}-\frac{1}{2\sqrt{3}}y_{3}-\frac{1}{2\sqrt{3}}z_{3}\right)a \, \mathbf{\hat{y}} + \frac{1}{3}\left(x_{3}+y_{3}+z_{3}\right)c \, \mathbf{\hat{z}} & \left(6f\right) & \mbox{O I} \\ 
\mathbf{B}_{15} & = & y_{3} \, \mathbf{a}_{1} + z_{3} \, \mathbf{a}_{2} + x_{3} \, \mathbf{a}_{3} & = & \frac{1}{2}\left(-x_{3}+y_{3}\right)a \, \mathbf{\hat{x}} + \left(-\frac{1}{2\sqrt{3}}x_{3}-\frac{1}{2\sqrt{3}}y_{3}+\frac{1}{\sqrt{3}}z_{3}\right)a \, \mathbf{\hat{y}} + \frac{1}{3}\left(x_{3}+y_{3}+z_{3}\right)c \, \mathbf{\hat{z}} & \left(6f\right) & \mbox{O I} \\ 
\mathbf{B}_{16} & = & -x_{3} \, \mathbf{a}_{1}-y_{3} \, \mathbf{a}_{2}-z_{3} \, \mathbf{a}_{3} & = & \frac{1}{2}\left(-x_{3}+z_{3}\right)a \, \mathbf{\hat{x}} + \left(\frac{1}{2\sqrt{3}}x_{3}-\frac{1}{\sqrt{3}}y_{3}+\frac{1}{2\sqrt{3}}z_{3}\right)a \, \mathbf{\hat{y}}-\frac{1}{3}\left(x_{3}+y_{3}+z_{3}\right)c \, \mathbf{\hat{z}} & \left(6f\right) & \mbox{O I} \\ 
\mathbf{B}_{17} & = & -z_{3} \, \mathbf{a}_{1}-x_{3} \, \mathbf{a}_{2}-y_{3} \, \mathbf{a}_{3} & = & \frac{1}{2}\left(y_{3}-z_{3}\right)a \, \mathbf{\hat{x}} + \left(-\frac{1}{\sqrt{3}}x_{3}+\frac{1}{2\sqrt{3}}y_{3}+\frac{1}{2\sqrt{3}}z_{3}\right)a \, \mathbf{\hat{y}}-\frac{1}{3}\left(x_{3}+y_{3}+z_{3}\right)c \, \mathbf{\hat{z}} & \left(6f\right) & \mbox{O I} \\ 
\mathbf{B}_{18} & = & -y_{3} \, \mathbf{a}_{1}-z_{3} \, \mathbf{a}_{2}-x_{3} \, \mathbf{a}_{3} & = & \frac{1}{2}\left(x_{3}-y_{3}\right)a \, \mathbf{\hat{x}} + \left(\frac{1}{2\sqrt{3}}x_{3}+\frac{1}{2\sqrt{3}}y_{3}-\frac{1}{\sqrt{3}}z_{3}\right)a \, \mathbf{\hat{y}}-\frac{1}{3}\left(x_{3}+y_{3}+z_{3}\right)c \, \mathbf{\hat{z}} & \left(6f\right) & \mbox{O I} \\ 
\mathbf{B}_{19} & = & x_{4} \, \mathbf{a}_{1} + y_{4} \, \mathbf{a}_{2} + z_{4} \, \mathbf{a}_{3} & = & \frac{1}{2}\left(x_{4}-z_{4}\right)a \, \mathbf{\hat{x}} + \left(-\frac{1}{2\sqrt{3}}x_{4}+\frac{1}{\sqrt{3}}y_{4}-\frac{1}{2\sqrt{3}}z_{4}\right)a \, \mathbf{\hat{y}} + \frac{1}{3}\left(x_{4}+y_{4}+z_{4}\right)c \, \mathbf{\hat{z}} & \left(6f\right) & \mbox{O II} \\ 
\mathbf{B}_{20} & = & z_{4} \, \mathbf{a}_{1} + x_{4} \, \mathbf{a}_{2} + y_{4} \, \mathbf{a}_{3} & = & \frac{1}{2}\left(-y_{4}+z_{4}\right)a \, \mathbf{\hat{x}} + \left(\frac{1}{\sqrt{3}}x_{4}-\frac{1}{2\sqrt{3}}y_{4}-\frac{1}{2\sqrt{3}}z_{4}\right)a \, \mathbf{\hat{y}} + \frac{1}{3}\left(x_{4}+y_{4}+z_{4}\right)c \, \mathbf{\hat{z}} & \left(6f\right) & \mbox{O II} \\ 
\mathbf{B}_{21} & = & y_{4} \, \mathbf{a}_{1} + z_{4} \, \mathbf{a}_{2} + x_{4} \, \mathbf{a}_{3} & = & \frac{1}{2}\left(-x_{4}+y_{4}\right)a \, \mathbf{\hat{x}} + \left(-\frac{1}{2\sqrt{3}}x_{4}-\frac{1}{2\sqrt{3}}y_{4}+\frac{1}{\sqrt{3}}z_{4}\right)a \, \mathbf{\hat{y}} + \frac{1}{3}\left(x_{4}+y_{4}+z_{4}\right)c \, \mathbf{\hat{z}} & \left(6f\right) & \mbox{O II} \\ 
\mathbf{B}_{22} & = & -x_{4} \, \mathbf{a}_{1}-y_{4} \, \mathbf{a}_{2}-z_{4} \, \mathbf{a}_{3} & = & \frac{1}{2}\left(-x_{4}+z_{4}\right)a \, \mathbf{\hat{x}} + \left(\frac{1}{2\sqrt{3}}x_{4}-\frac{1}{\sqrt{3}}y_{4}+\frac{1}{2\sqrt{3}}z_{4}\right)a \, \mathbf{\hat{y}}-\frac{1}{3}\left(x_{4}+y_{4}+z_{4}\right)c \, \mathbf{\hat{z}} & \left(6f\right) & \mbox{O II} \\ 
\mathbf{B}_{23} & = & -z_{4} \, \mathbf{a}_{1}-x_{4} \, \mathbf{a}_{2}-y_{4} \, \mathbf{a}_{3} & = & \frac{1}{2}\left(y_{4}-z_{4}\right)a \, \mathbf{\hat{x}} + \left(-\frac{1}{\sqrt{3}}x_{4}+\frac{1}{2\sqrt{3}}y_{4}+\frac{1}{2\sqrt{3}}z_{4}\right)a \, \mathbf{\hat{y}}-\frac{1}{3}\left(x_{4}+y_{4}+z_{4}\right)c \, \mathbf{\hat{z}} & \left(6f\right) & \mbox{O II} \\ 
\mathbf{B}_{24} & = & -y_{4} \, \mathbf{a}_{1}-z_{4} \, \mathbf{a}_{2}-x_{4} \, \mathbf{a}_{3} & = & \frac{1}{2}\left(x_{4}-y_{4}\right)a \, \mathbf{\hat{x}} + \left(\frac{1}{2\sqrt{3}}x_{4}+\frac{1}{2\sqrt{3}}y_{4}-\frac{1}{\sqrt{3}}z_{4}\right)a \, \mathbf{\hat{y}}-\frac{1}{3}\left(x_{4}+y_{4}+z_{4}\right)c \, \mathbf{\hat{z}} & \left(6f\right) & \mbox{O II} \\ 
\mathbf{B}_{25} & = & x_{5} \, \mathbf{a}_{1} + y_{5} \, \mathbf{a}_{2} + z_{5} \, \mathbf{a}_{3} & = & \frac{1}{2}\left(x_{5}-z_{5}\right)a \, \mathbf{\hat{x}} + \left(-\frac{1}{2\sqrt{3}}x_{5}+\frac{1}{\sqrt{3}}y_{5}-\frac{1}{2\sqrt{3}}z_{5}\right)a \, \mathbf{\hat{y}} + \frac{1}{3}\left(x_{5}+y_{5}+z_{5}\right)c \, \mathbf{\hat{z}} & \left(6f\right) & \mbox{O III} \\ 
\mathbf{B}_{26} & = & z_{5} \, \mathbf{a}_{1} + x_{5} \, \mathbf{a}_{2} + y_{5} \, \mathbf{a}_{3} & = & \frac{1}{2}\left(-y_{5}+z_{5}\right)a \, \mathbf{\hat{x}} + \left(\frac{1}{\sqrt{3}}x_{5}-\frac{1}{2\sqrt{3}}y_{5}-\frac{1}{2\sqrt{3}}z_{5}\right)a \, \mathbf{\hat{y}} + \frac{1}{3}\left(x_{5}+y_{5}+z_{5}\right)c \, \mathbf{\hat{z}} & \left(6f\right) & \mbox{O III} \\ 
\mathbf{B}_{27} & = & y_{5} \, \mathbf{a}_{1} + z_{5} \, \mathbf{a}_{2} + x_{5} \, \mathbf{a}_{3} & = & \frac{1}{2}\left(-x_{5}+y_{5}\right)a \, \mathbf{\hat{x}} + \left(-\frac{1}{2\sqrt{3}}x_{5}-\frac{1}{2\sqrt{3}}y_{5}+\frac{1}{\sqrt{3}}z_{5}\right)a \, \mathbf{\hat{y}} + \frac{1}{3}\left(x_{5}+y_{5}+z_{5}\right)c \, \mathbf{\hat{z}} & \left(6f\right) & \mbox{O III} \\ 
\mathbf{B}_{28} & = & -x_{5} \, \mathbf{a}_{1}-y_{5} \, \mathbf{a}_{2}-z_{5} \, \mathbf{a}_{3} & = & \frac{1}{2}\left(-x_{5}+z_{5}\right)a \, \mathbf{\hat{x}} + \left(\frac{1}{2\sqrt{3}}x_{5}-\frac{1}{\sqrt{3}}y_{5}+\frac{1}{2\sqrt{3}}z_{5}\right)a \, \mathbf{\hat{y}}-\frac{1}{3}\left(x_{5}+y_{5}+z_{5}\right)c \, \mathbf{\hat{z}} & \left(6f\right) & \mbox{O III} \\ 
\mathbf{B}_{29} & = & -z_{5} \, \mathbf{a}_{1}-x_{5} \, \mathbf{a}_{2}-y_{5} \, \mathbf{a}_{3} & = & \frac{1}{2}\left(y_{5}-z_{5}\right)a \, \mathbf{\hat{x}} + \left(-\frac{1}{\sqrt{3}}x_{5}+\frac{1}{2\sqrt{3}}y_{5}+\frac{1}{2\sqrt{3}}z_{5}\right)a \, \mathbf{\hat{y}}-\frac{1}{3}\left(x_{5}+y_{5}+z_{5}\right)c \, \mathbf{\hat{z}} & \left(6f\right) & \mbox{O III} \\ 
\mathbf{B}_{30} & = & -y_{5} \, \mathbf{a}_{1}-z_{5} \, \mathbf{a}_{2}-x_{5} \, \mathbf{a}_{3} & = & \frac{1}{2}\left(x_{5}-y_{5}\right)a \, \mathbf{\hat{x}} + \left(\frac{1}{2\sqrt{3}}x_{5}+\frac{1}{2\sqrt{3}}y_{5}-\frac{1}{\sqrt{3}}z_{5}\right)a \, \mathbf{\hat{y}}-\frac{1}{3}\left(x_{5}+y_{5}+z_{5}\right)c \, \mathbf{\hat{z}} & \left(6f\right) & \mbox{O III} \\ 
\mathbf{B}_{31} & = & x_{6} \, \mathbf{a}_{1} + y_{6} \, \mathbf{a}_{2} + z_{6} \, \mathbf{a}_{3} & = & \frac{1}{2}\left(x_{6}-z_{6}\right)a \, \mathbf{\hat{x}} + \left(-\frac{1}{2\sqrt{3}}x_{6}+\frac{1}{\sqrt{3}}y_{6}-\frac{1}{2\sqrt{3}}z_{6}\right)a \, \mathbf{\hat{y}} + \frac{1}{3}\left(x_{6}+y_{6}+z_{6}\right)c \, \mathbf{\hat{z}} & \left(6f\right) & \mbox{O IV} \\ 
\mathbf{B}_{32} & = & z_{6} \, \mathbf{a}_{1} + x_{6} \, \mathbf{a}_{2} + y_{6} \, \mathbf{a}_{3} & = & \frac{1}{2}\left(-y_{6}+z_{6}\right)a \, \mathbf{\hat{x}} + \left(\frac{1}{\sqrt{3}}x_{6}-\frac{1}{2\sqrt{3}}y_{6}-\frac{1}{2\sqrt{3}}z_{6}\right)a \, \mathbf{\hat{y}} + \frac{1}{3}\left(x_{6}+y_{6}+z_{6}\right)c \, \mathbf{\hat{z}} & \left(6f\right) & \mbox{O IV} \\ 
\mathbf{B}_{33} & = & y_{6} \, \mathbf{a}_{1} + z_{6} \, \mathbf{a}_{2} + x_{6} \, \mathbf{a}_{3} & = & \frac{1}{2}\left(-x_{6}+y_{6}\right)a \, \mathbf{\hat{x}} + \left(-\frac{1}{2\sqrt{3}}x_{6}-\frac{1}{2\sqrt{3}}y_{6}+\frac{1}{\sqrt{3}}z_{6}\right)a \, \mathbf{\hat{y}} + \frac{1}{3}\left(x_{6}+y_{6}+z_{6}\right)c \, \mathbf{\hat{z}} & \left(6f\right) & \mbox{O IV} \\ 
\mathbf{B}_{34} & = & -x_{6} \, \mathbf{a}_{1}-y_{6} \, \mathbf{a}_{2}-z_{6} \, \mathbf{a}_{3} & = & \frac{1}{2}\left(-x_{6}+z_{6}\right)a \, \mathbf{\hat{x}} + \left(\frac{1}{2\sqrt{3}}x_{6}-\frac{1}{\sqrt{3}}y_{6}+\frac{1}{2\sqrt{3}}z_{6}\right)a \, \mathbf{\hat{y}}-\frac{1}{3}\left(x_{6}+y_{6}+z_{6}\right)c \, \mathbf{\hat{z}} & \left(6f\right) & \mbox{O IV} \\ 
\mathbf{B}_{35} & = & -z_{6} \, \mathbf{a}_{1}-x_{6} \, \mathbf{a}_{2}-y_{6} \, \mathbf{a}_{3} & = & \frac{1}{2}\left(y_{6}-z_{6}\right)a \, \mathbf{\hat{x}} + \left(-\frac{1}{\sqrt{3}}x_{6}+\frac{1}{2\sqrt{3}}y_{6}+\frac{1}{2\sqrt{3}}z_{6}\right)a \, \mathbf{\hat{y}}-\frac{1}{3}\left(x_{6}+y_{6}+z_{6}\right)c \, \mathbf{\hat{z}} & \left(6f\right) & \mbox{O IV} \\ 
\mathbf{B}_{36} & = & -y_{6} \, \mathbf{a}_{1}-z_{6} \, \mathbf{a}_{2}-x_{6} \, \mathbf{a}_{3} & = & \frac{1}{2}\left(x_{6}-y_{6}\right)a \, \mathbf{\hat{x}} + \left(\frac{1}{2\sqrt{3}}x_{6}+\frac{1}{2\sqrt{3}}y_{6}-\frac{1}{\sqrt{3}}z_{6}\right)a \, \mathbf{\hat{y}}-\frac{1}{3}\left(x_{6}+y_{6}+z_{6}\right)c \, \mathbf{\hat{z}} & \left(6f\right) & \mbox{O IV} \\ 
\mathbf{B}_{37} & = & x_{7} \, \mathbf{a}_{1} + y_{7} \, \mathbf{a}_{2} + z_{7} \, \mathbf{a}_{3} & = & \frac{1}{2}\left(x_{7}-z_{7}\right)a \, \mathbf{\hat{x}} + \left(-\frac{1}{2\sqrt{3}}x_{7}+\frac{1}{\sqrt{3}}y_{7}-\frac{1}{2\sqrt{3}}z_{7}\right)a \, \mathbf{\hat{y}} + \frac{1}{3}\left(x_{7}+y_{7}+z_{7}\right)c \, \mathbf{\hat{z}} & \left(6f\right) & \mbox{Si} \\ 
\mathbf{B}_{38} & = & z_{7} \, \mathbf{a}_{1} + x_{7} \, \mathbf{a}_{2} + y_{7} \, \mathbf{a}_{3} & = & \frac{1}{2}\left(-y_{7}+z_{7}\right)a \, \mathbf{\hat{x}} + \left(\frac{1}{\sqrt{3}}x_{7}-\frac{1}{2\sqrt{3}}y_{7}-\frac{1}{2\sqrt{3}}z_{7}\right)a \, \mathbf{\hat{y}} + \frac{1}{3}\left(x_{7}+y_{7}+z_{7}\right)c \, \mathbf{\hat{z}} & \left(6f\right) & \mbox{Si} \\ 
\mathbf{B}_{39} & = & y_{7} \, \mathbf{a}_{1} + z_{7} \, \mathbf{a}_{2} + x_{7} \, \mathbf{a}_{3} & = & \frac{1}{2}\left(-x_{7}+y_{7}\right)a \, \mathbf{\hat{x}} + \left(-\frac{1}{2\sqrt{3}}x_{7}-\frac{1}{2\sqrt{3}}y_{7}+\frac{1}{\sqrt{3}}z_{7}\right)a \, \mathbf{\hat{y}} + \frac{1}{3}\left(x_{7}+y_{7}+z_{7}\right)c \, \mathbf{\hat{z}} & \left(6f\right) & \mbox{Si} \\ 
\mathbf{B}_{40} & = & -x_{7} \, \mathbf{a}_{1}-y_{7} \, \mathbf{a}_{2}-z_{7} \, \mathbf{a}_{3} & = & \frac{1}{2}\left(-x_{7}+z_{7}\right)a \, \mathbf{\hat{x}} + \left(\frac{1}{2\sqrt{3}}x_{7}-\frac{1}{\sqrt{3}}y_{7}+\frac{1}{2\sqrt{3}}z_{7}\right)a \, \mathbf{\hat{y}}-\frac{1}{3}\left(x_{7}+y_{7}+z_{7}\right)c \, \mathbf{\hat{z}} & \left(6f\right) & \mbox{Si} \\ 
\mathbf{B}_{41} & = & -z_{7} \, \mathbf{a}_{1}-x_{7} \, \mathbf{a}_{2}-y_{7} \, \mathbf{a}_{3} & = & \frac{1}{2}\left(y_{7}-z_{7}\right)a \, \mathbf{\hat{x}} + \left(-\frac{1}{\sqrt{3}}x_{7}+\frac{1}{2\sqrt{3}}y_{7}+\frac{1}{2\sqrt{3}}z_{7}\right)a \, \mathbf{\hat{y}}-\frac{1}{3}\left(x_{7}+y_{7}+z_{7}\right)c \, \mathbf{\hat{z}} & \left(6f\right) & \mbox{Si} \\ 
\mathbf{B}_{42} & = & -y_{7} \, \mathbf{a}_{1}-z_{7} \, \mathbf{a}_{2}-x_{7} \, \mathbf{a}_{3} & = & \frac{1}{2}\left(x_{7}-y_{7}\right)a \, \mathbf{\hat{x}} + \left(\frac{1}{2\sqrt{3}}x_{7}+\frac{1}{2\sqrt{3}}y_{7}-\frac{1}{\sqrt{3}}z_{7}\right)a \, \mathbf{\hat{y}}-\frac{1}{3}\left(x_{7}+y_{7}+z_{7}\right)c \, \mathbf{\hat{z}} & \left(6f\right) & \mbox{Si} \\ 
\end{longtabu}
\renewcommand{\arraystretch}{1.0}
\noindent \hrulefill
\\
\textbf{References:}
\vspace*{-0.25cm}
\begin{flushleft}
  - \bibentry{Hazen_PCM_14_426_1987}. \\
\end{flushleft}
\noindent \hrulefill
\\
\textbf{Geometry files:}
\\
\noindent  - CIF: pp. {\hyperref[A2B4C_hR42_148_2f_4f_f_cif]{\pageref{A2B4C_hR42_148_2f_4f_f_cif}}} \\
\noindent  - POSCAR: pp. {\hyperref[A2B4C_hR42_148_2f_4f_f_poscar]{\pageref{A2B4C_hR42_148_2f_4f_f_poscar}}} \\
\onecolumn
{\phantomsection\label{A2B_hR18_148_2f_f}}
\subsection*{\huge \textbf{{\normalfont $\beta$-PdCl$_2$ Structure: A2B\_hR18\_148\_2f\_f}}}
\noindent \hrulefill
\vspace*{0.25cm}
\begin{figure}[htp]
  \centering
  \vspace{-1em}
  {\includegraphics[width=1\textwidth]{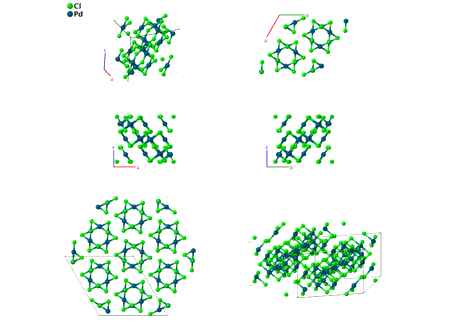}}
\end{figure}
\vspace*{-0.5cm}
\renewcommand{\arraystretch}{1.5}
\begin{equation*}
  \begin{array}{>{$\hspace{-0.15cm}}l<{$}>{$}p{0.5cm}<{$}>{$}p{18.5cm}<{$}}
    \mbox{\large \textbf{Prototype}} &\colon & \ce{$\beta$-PdCl2} \\
    \mbox{\large \textbf{\AFLOW\ prototype label}} &\colon & \mbox{A2B\_hR18\_148\_2f\_f} \\
    \mbox{\large \textbf{\textit{Strukturbericht} designation}} &\colon & \mbox{None} \\
    \mbox{\large \textbf{Pearson symbol}} &\colon & \mbox{hR18} \\
    \mbox{\large \textbf{Space group number}} &\colon & 148 \\
    \mbox{\large \textbf{Space group symbol}} &\colon & R\bar{3} \\
    \mbox{\large \textbf{\AFLOW\ prototype command}} &\colon &  \texttt{aflow} \,  \, \texttt{-{}-proto=A2B\_hR18\_148\_2f\_f [-{}-hex]} \, \newline \texttt{-{}-params=}{a,c/a,x_{1},y_{1},z_{1},x_{2},y_{2},z_{2},x_{3},y_{3},z_{3} }
  \end{array}
\end{equation*}
\renewcommand{\arraystretch}{1.0}

\vspace*{-0.25cm}
\noindent \hrulefill
\begin{itemize}
  \item{(Dell'Amico, 1996) did the original assessment of the crystal
structure of $\beta$-PdCl$_{2}$, but it is difficult to determine the
Wyckoff positions from this paper.  We relied on (Villars, 2010) for
the Wyckoff positions.
}
\end{itemize}

\noindent \parbox{1 \linewidth}{
\noindent \hrulefill
\\
\textbf{Rhombohedral primitive vectors:} \\
\vspace*{-0.25cm}
\begin{tabular}{cc}
  \begin{tabular}{c}
    \parbox{0.6 \linewidth}{
      \renewcommand{\arraystretch}{1.5}
      \begin{equation*}
        \centering
        \begin{array}{ccc}
              \mathbf{a}_1 & = & ~ \frac12 \, a \, \mathbf{\hat{x}} - \frac{1}{2\sqrt{3}} \, a \, \mathbf{\hat{y}} + \frac13 \, c \, \mathbf{\hat{z}} \\
    \mathbf{a}_2 & = & \frac{1}{\sqrt{3}} \, a \, \mathbf{\hat{y}} + \frac13 \, c \, \mathbf{\hat{z}} \\
    \mathbf{a}_3 & = & - \frac12 \, a \, \mathbf{\hat{x}} - \frac{1}{2\sqrt{3}} \, a \, \mathbf{\hat{y}} + \frac13 \, c \, \mathbf{\hat{z}} \\

        \end{array}
      \end{equation*}
    }
    \renewcommand{\arraystretch}{1.0}
  \end{tabular}
  \begin{tabular}{c}
    \includegraphics[width=0.3\linewidth]{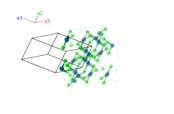} \\
  \end{tabular}
\end{tabular}

}
\vspace*{-0.25cm}

\noindent \hrulefill
\\
\textbf{Basis vectors:}
\vspace*{-0.25cm}
\renewcommand{\arraystretch}{1.5}
\begin{longtabu} to \textwidth{>{\centering $}X[-1,c,c]<{$}>{\centering $}X[-1,c,c]<{$}>{\centering $}X[-1,c,c]<{$}>{\centering $}X[-1,c,c]<{$}>{\centering $}X[-1,c,c]<{$}>{\centering $}X[-1,c,c]<{$}>{\centering $}X[-1,c,c]<{$}}
  & & \mbox{Lattice Coordinates} & & \mbox{Cartesian Coordinates} &\mbox{Wyckoff Position} & \mbox{Atom Type} \\  
  \mathbf{B}_{1} & = & x_{1} \, \mathbf{a}_{1} + y_{1} \, \mathbf{a}_{2} + z_{1} \, \mathbf{a}_{3} & = & \frac{1}{2}\left(x_{1}-z_{1}\right)a \, \mathbf{\hat{x}} + \left(-\frac{1}{2\sqrt{3}}x_{1}+\frac{1}{\sqrt{3}}y_{1}-\frac{1}{2\sqrt{3}}z_{1}\right)a \, \mathbf{\hat{y}} + \frac{1}{3}\left(x_{1}+y_{1}+z_{1}\right)c \, \mathbf{\hat{z}} & \left(6f\right) & \mbox{Cl I} \\ 
\mathbf{B}_{2} & = & z_{1} \, \mathbf{a}_{1} + x_{1} \, \mathbf{a}_{2} + y_{1} \, \mathbf{a}_{3} & = & \frac{1}{2}\left(-y_{1}+z_{1}\right)a \, \mathbf{\hat{x}} + \left(\frac{1}{\sqrt{3}}x_{1}-\frac{1}{2\sqrt{3}}y_{1}-\frac{1}{2\sqrt{3}}z_{1}\right)a \, \mathbf{\hat{y}} + \frac{1}{3}\left(x_{1}+y_{1}+z_{1}\right)c \, \mathbf{\hat{z}} & \left(6f\right) & \mbox{Cl I} \\ 
\mathbf{B}_{3} & = & y_{1} \, \mathbf{a}_{1} + z_{1} \, \mathbf{a}_{2} + x_{1} \, \mathbf{a}_{3} & = & \frac{1}{2}\left(-x_{1}+y_{1}\right)a \, \mathbf{\hat{x}} + \left(-\frac{1}{2\sqrt{3}}x_{1}-\frac{1}{2\sqrt{3}}y_{1}+\frac{1}{\sqrt{3}}z_{1}\right)a \, \mathbf{\hat{y}} + \frac{1}{3}\left(x_{1}+y_{1}+z_{1}\right)c \, \mathbf{\hat{z}} & \left(6f\right) & \mbox{Cl I} \\ 
\mathbf{B}_{4} & = & -x_{1} \, \mathbf{a}_{1}-y_{1} \, \mathbf{a}_{2}-z_{1} \, \mathbf{a}_{3} & = & \frac{1}{2}\left(-x_{1}+z_{1}\right)a \, \mathbf{\hat{x}} + \left(\frac{1}{2\sqrt{3}}x_{1}-\frac{1}{\sqrt{3}}y_{1}+\frac{1}{2\sqrt{3}}z_{1}\right)a \, \mathbf{\hat{y}}-\frac{1}{3}\left(x_{1}+y_{1}+z_{1}\right)c \, \mathbf{\hat{z}} & \left(6f\right) & \mbox{Cl I} \\ 
\mathbf{B}_{5} & = & -z_{1} \, \mathbf{a}_{1}-x_{1} \, \mathbf{a}_{2}-y_{1} \, \mathbf{a}_{3} & = & \frac{1}{2}\left(y_{1}-z_{1}\right)a \, \mathbf{\hat{x}} + \left(-\frac{1}{\sqrt{3}}x_{1}+\frac{1}{2\sqrt{3}}y_{1}+\frac{1}{2\sqrt{3}}z_{1}\right)a \, \mathbf{\hat{y}}-\frac{1}{3}\left(x_{1}+y_{1}+z_{1}\right)c \, \mathbf{\hat{z}} & \left(6f\right) & \mbox{Cl I} \\ 
\mathbf{B}_{6} & = & -y_{1} \, \mathbf{a}_{1}-z_{1} \, \mathbf{a}_{2}-x_{1} \, \mathbf{a}_{3} & = & \frac{1}{2}\left(x_{1}-y_{1}\right)a \, \mathbf{\hat{x}} + \left(\frac{1}{2\sqrt{3}}x_{1}+\frac{1}{2\sqrt{3}}y_{1}-\frac{1}{\sqrt{3}}z_{1}\right)a \, \mathbf{\hat{y}}-\frac{1}{3}\left(x_{1}+y_{1}+z_{1}\right)c \, \mathbf{\hat{z}} & \left(6f\right) & \mbox{Cl I} \\ 
\mathbf{B}_{7} & = & x_{2} \, \mathbf{a}_{1} + y_{2} \, \mathbf{a}_{2} + z_{2} \, \mathbf{a}_{3} & = & \frac{1}{2}\left(x_{2}-z_{2}\right)a \, \mathbf{\hat{x}} + \left(-\frac{1}{2\sqrt{3}}x_{2}+\frac{1}{\sqrt{3}}y_{2}-\frac{1}{2\sqrt{3}}z_{2}\right)a \, \mathbf{\hat{y}} + \frac{1}{3}\left(x_{2}+y_{2}+z_{2}\right)c \, \mathbf{\hat{z}} & \left(6f\right) & \mbox{Cl II} \\ 
\mathbf{B}_{8} & = & z_{2} \, \mathbf{a}_{1} + x_{2} \, \mathbf{a}_{2} + y_{2} \, \mathbf{a}_{3} & = & \frac{1}{2}\left(-y_{2}+z_{2}\right)a \, \mathbf{\hat{x}} + \left(\frac{1}{\sqrt{3}}x_{2}-\frac{1}{2\sqrt{3}}y_{2}-\frac{1}{2\sqrt{3}}z_{2}\right)a \, \mathbf{\hat{y}} + \frac{1}{3}\left(x_{2}+y_{2}+z_{2}\right)c \, \mathbf{\hat{z}} & \left(6f\right) & \mbox{Cl II} \\ 
\mathbf{B}_{9} & = & y_{2} \, \mathbf{a}_{1} + z_{2} \, \mathbf{a}_{2} + x_{2} \, \mathbf{a}_{3} & = & \frac{1}{2}\left(-x_{2}+y_{2}\right)a \, \mathbf{\hat{x}} + \left(-\frac{1}{2\sqrt{3}}x_{2}-\frac{1}{2\sqrt{3}}y_{2}+\frac{1}{\sqrt{3}}z_{2}\right)a \, \mathbf{\hat{y}} + \frac{1}{3}\left(x_{2}+y_{2}+z_{2}\right)c \, \mathbf{\hat{z}} & \left(6f\right) & \mbox{Cl II} \\ 
\mathbf{B}_{10} & = & -x_{2} \, \mathbf{a}_{1}-y_{2} \, \mathbf{a}_{2}-z_{2} \, \mathbf{a}_{3} & = & \frac{1}{2}\left(-x_{2}+z_{2}\right)a \, \mathbf{\hat{x}} + \left(\frac{1}{2\sqrt{3}}x_{2}-\frac{1}{\sqrt{3}}y_{2}+\frac{1}{2\sqrt{3}}z_{2}\right)a \, \mathbf{\hat{y}}-\frac{1}{3}\left(x_{2}+y_{2}+z_{2}\right)c \, \mathbf{\hat{z}} & \left(6f\right) & \mbox{Cl II} \\ 
\mathbf{B}_{11} & = & -z_{2} \, \mathbf{a}_{1}-x_{2} \, \mathbf{a}_{2}-y_{2} \, \mathbf{a}_{3} & = & \frac{1}{2}\left(y_{2}-z_{2}\right)a \, \mathbf{\hat{x}} + \left(-\frac{1}{\sqrt{3}}x_{2}+\frac{1}{2\sqrt{3}}y_{2}+\frac{1}{2\sqrt{3}}z_{2}\right)a \, \mathbf{\hat{y}}-\frac{1}{3}\left(x_{2}+y_{2}+z_{2}\right)c \, \mathbf{\hat{z}} & \left(6f\right) & \mbox{Cl II} \\ 
\mathbf{B}_{12} & = & -y_{2} \, \mathbf{a}_{1}-z_{2} \, \mathbf{a}_{2}-x_{2} \, \mathbf{a}_{3} & = & \frac{1}{2}\left(x_{2}-y_{2}\right)a \, \mathbf{\hat{x}} + \left(\frac{1}{2\sqrt{3}}x_{2}+\frac{1}{2\sqrt{3}}y_{2}-\frac{1}{\sqrt{3}}z_{2}\right)a \, \mathbf{\hat{y}}-\frac{1}{3}\left(x_{2}+y_{2}+z_{2}\right)c \, \mathbf{\hat{z}} & \left(6f\right) & \mbox{Cl II} \\ 
\mathbf{B}_{13} & = & x_{3} \, \mathbf{a}_{1} + y_{3} \, \mathbf{a}_{2} + z_{3} \, \mathbf{a}_{3} & = & \frac{1}{2}\left(x_{3}-z_{3}\right)a \, \mathbf{\hat{x}} + \left(-\frac{1}{2\sqrt{3}}x_{3}+\frac{1}{\sqrt{3}}y_{3}-\frac{1}{2\sqrt{3}}z_{3}\right)a \, \mathbf{\hat{y}} + \frac{1}{3}\left(x_{3}+y_{3}+z_{3}\right)c \, \mathbf{\hat{z}} & \left(6f\right) & \mbox{Pd} \\ 
\mathbf{B}_{14} & = & z_{3} \, \mathbf{a}_{1} + x_{3} \, \mathbf{a}_{2} + y_{3} \, \mathbf{a}_{3} & = & \frac{1}{2}\left(-y_{3}+z_{3}\right)a \, \mathbf{\hat{x}} + \left(\frac{1}{\sqrt{3}}x_{3}-\frac{1}{2\sqrt{3}}y_{3}-\frac{1}{2\sqrt{3}}z_{3}\right)a \, \mathbf{\hat{y}} + \frac{1}{3}\left(x_{3}+y_{3}+z_{3}\right)c \, \mathbf{\hat{z}} & \left(6f\right) & \mbox{Pd} \\ 
\mathbf{B}_{15} & = & y_{3} \, \mathbf{a}_{1} + z_{3} \, \mathbf{a}_{2} + x_{3} \, \mathbf{a}_{3} & = & \frac{1}{2}\left(-x_{3}+y_{3}\right)a \, \mathbf{\hat{x}} + \left(-\frac{1}{2\sqrt{3}}x_{3}-\frac{1}{2\sqrt{3}}y_{3}+\frac{1}{\sqrt{3}}z_{3}\right)a \, \mathbf{\hat{y}} + \frac{1}{3}\left(x_{3}+y_{3}+z_{3}\right)c \, \mathbf{\hat{z}} & \left(6f\right) & \mbox{Pd} \\ 
\mathbf{B}_{16} & = & -x_{3} \, \mathbf{a}_{1}-y_{3} \, \mathbf{a}_{2}-z_{3} \, \mathbf{a}_{3} & = & \frac{1}{2}\left(-x_{3}+z_{3}\right)a \, \mathbf{\hat{x}} + \left(\frac{1}{2\sqrt{3}}x_{3}-\frac{1}{\sqrt{3}}y_{3}+\frac{1}{2\sqrt{3}}z_{3}\right)a \, \mathbf{\hat{y}}-\frac{1}{3}\left(x_{3}+y_{3}+z_{3}\right)c \, \mathbf{\hat{z}} & \left(6f\right) & \mbox{Pd} \\ 
\mathbf{B}_{17} & = & -z_{3} \, \mathbf{a}_{1}-x_{3} \, \mathbf{a}_{2}-y_{3} \, \mathbf{a}_{3} & = & \frac{1}{2}\left(y_{3}-z_{3}\right)a \, \mathbf{\hat{x}} + \left(-\frac{1}{\sqrt{3}}x_{3}+\frac{1}{2\sqrt{3}}y_{3}+\frac{1}{2\sqrt{3}}z_{3}\right)a \, \mathbf{\hat{y}}-\frac{1}{3}\left(x_{3}+y_{3}+z_{3}\right)c \, \mathbf{\hat{z}} & \left(6f\right) & \mbox{Pd} \\ 
\mathbf{B}_{18} & = & -y_{3} \, \mathbf{a}_{1}-z_{3} \, \mathbf{a}_{2}-x_{3} \, \mathbf{a}_{3} & = & \frac{1}{2}\left(x_{3}-y_{3}\right)a \, \mathbf{\hat{x}} + \left(\frac{1}{2\sqrt{3}}x_{3}+\frac{1}{2\sqrt{3}}y_{3}-\frac{1}{\sqrt{3}}z_{3}\right)a \, \mathbf{\hat{y}}-\frac{1}{3}\left(x_{3}+y_{3}+z_{3}\right)c \, \mathbf{\hat{z}} & \left(6f\right) & \mbox{Pd} \\ 
\end{longtabu}
\renewcommand{\arraystretch}{1.0}
\noindent \hrulefill
\\
\textbf{References:}
\vspace*{-0.25cm}
\begin{flushleft}
  - \bibentry{DellAmico_AngewChemIntEd_35_1131_1996}. \\
  - \bibentry{Villars_Landolt_Bornstein_43A8_2010}. \\
\end{flushleft}
\noindent \hrulefill
\\
\textbf{Geometry files:}
\\
\noindent  - CIF: pp. {\hyperref[A2B_hR18_148_2f_f_cif]{\pageref{A2B_hR18_148_2f_f_cif}}} \\
\noindent  - POSCAR: pp. {\hyperref[A2B_hR18_148_2f_f_poscar]{\pageref{A2B_hR18_148_2f_f_poscar}}} \\
\onecolumn
{\phantomsection\label{AB3_hP24_149_acgi_3l}}
\subsection*{\huge \textbf{{\normalfont \begin{raggedleft}Ti$_{3}$O (Room-temperature) Structure: \end{raggedleft} \\ AB3\_hP24\_149\_acgi\_3l}}}
\noindent \hrulefill
\vspace*{0.25cm}
\begin{figure}[htp]
  \centering
  \vspace{-1em}
  {\includegraphics[width=1\textwidth]{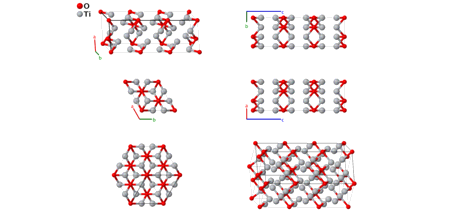}}
\end{figure}
\vspace*{-0.5cm}
\renewcommand{\arraystretch}{1.5}
\begin{equation*}
  \begin{array}{>{$\hspace{-0.15cm}}l<{$}>{$}p{0.5cm}<{$}>{$}p{18.5cm}<{$}}
    \mbox{\large \textbf{Prototype}} &\colon & \ce{Ti3O} \\
    \mbox{\large \textbf{\AFLOW\ prototype label}} &\colon & \mbox{AB3\_hP24\_149\_acgi\_3l} \\
    \mbox{\large \textbf{\textit{Strukturbericht} designation}} &\colon & \mbox{None} \\
    \mbox{\large \textbf{Pearson symbol}} &\colon & \mbox{hP24} \\
    \mbox{\large \textbf{Space group number}} &\colon & 149 \\
    \mbox{\large \textbf{Space group symbol}} &\colon & P312 \\
    \mbox{\large \textbf{\AFLOW\ prototype command}} &\colon &  \texttt{aflow} \,  \, \texttt{-{}-proto=AB3\_hP24\_149\_acgi\_3l } \, \newline \texttt{-{}-params=}{a,c/a,z_{3},z_{4},x_{5},y_{5},z_{5},x_{6},y_{6},z_{6},x_{7},y_{7},z_{7} }
  \end{array}
\end{equation*}
\renewcommand{\arraystretch}{1.0}

\noindent \parbox{1 \linewidth}{
\noindent \hrulefill
\\
\textbf{Trigonal Hexagonal primitive vectors:} \\
\vspace*{-0.25cm}
\begin{tabular}{cc}
  \begin{tabular}{c}
    \parbox{0.6 \linewidth}{
      \renewcommand{\arraystretch}{1.5}
      \begin{equation*}
        \centering
        \begin{array}{ccc}
              \mathbf{a}_1 & = & \frac12 \, a \, \mathbf{\hat{x}} - \frac{\sqrt3}2 \, a \, \mathbf{\hat{y}} \\
    \mathbf{a}_2 & = & \frac12 \, a \, \mathbf{\hat{x}} + \frac{\sqrt3}2 \, a \, \mathbf{\hat{y}} \\
    \mathbf{a}_3 & = & c \, \mathbf{\hat{z}} \\

        \end{array}
      \end{equation*}
    }
    \renewcommand{\arraystretch}{1.0}
  \end{tabular}
  \begin{tabular}{c}
    \includegraphics[width=0.3\linewidth]{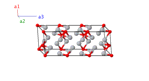} \\
  \end{tabular}
\end{tabular}

}
\vspace*{-0.25cm}

\noindent \hrulefill
\\
\textbf{Basis vectors:}
\vspace*{-0.25cm}
\renewcommand{\arraystretch}{1.5}
\begin{longtabu} to \textwidth{>{\centering $}X[-1,c,c]<{$}>{\centering $}X[-1,c,c]<{$}>{\centering $}X[-1,c,c]<{$}>{\centering $}X[-1,c,c]<{$}>{\centering $}X[-1,c,c]<{$}>{\centering $}X[-1,c,c]<{$}>{\centering $}X[-1,c,c]<{$}}
  & & \mbox{Lattice Coordinates} & & \mbox{Cartesian Coordinates} &\mbox{Wyckoff Position} & \mbox{Atom Type} \\  
  \mathbf{B}_{1} & = & 0 \, \mathbf{a}_{1} + 0 \, \mathbf{a}_{2} + 0 \, \mathbf{a}_{3} & = & 0 \, \mathbf{\hat{x}} + 0 \, \mathbf{\hat{y}} + 0 \, \mathbf{\hat{z}} & \left(1a\right) & \mbox{O I} \\ 
\mathbf{B}_{2} & = & \frac{1}{3} \, \mathbf{a}_{1} + \frac{2}{3} \, \mathbf{a}_{2} & = & \frac{1}{2}a \, \mathbf{\hat{x}} + \frac{1}{2\sqrt{3}}a \, \mathbf{\hat{y}} & \left(1c\right) & \mbox{O II} \\ 
\mathbf{B}_{3} & = & z_{3} \, \mathbf{a}_{3} & = & z_{3}c \, \mathbf{\hat{z}} & \left(2g\right) & \mbox{O III} \\ 
\mathbf{B}_{4} & = & -z_{3} \, \mathbf{a}_{3} & = & -z_{3}c \, \mathbf{\hat{z}} & \left(2g\right) & \mbox{O III} \\ 
\mathbf{B}_{5} & = & \frac{2}{3} \, \mathbf{a}_{1} + \frac{1}{3} \, \mathbf{a}_{2} + z_{4} \, \mathbf{a}_{3} & = & \frac{1}{2}a \, \mathbf{\hat{x}}- \frac{1}{2\sqrt{3}}a  \, \mathbf{\hat{y}} + z_{4}c \, \mathbf{\hat{z}} & \left(2i\right) & \mbox{O IV} \\ 
\mathbf{B}_{6} & = & \frac{2}{3} \, \mathbf{a}_{1} + \frac{1}{3} \, \mathbf{a}_{2}-z_{4} \, \mathbf{a}_{3} & = & \frac{1}{2}a \, \mathbf{\hat{x}}- \frac{1}{2\sqrt{3}}a  \, \mathbf{\hat{y}}-z_{4}c \, \mathbf{\hat{z}} & \left(2i\right) & \mbox{O IV} \\ 
\mathbf{B}_{7} & = & x_{5} \, \mathbf{a}_{1} + y_{5} \, \mathbf{a}_{2} + z_{5} \, \mathbf{a}_{3} & = & \frac{1}{2}\left(x_{5}+y_{5}\right)a \, \mathbf{\hat{x}} + \frac{\sqrt{3}}{2}\left(-x_{5}+y_{5}\right)a \, \mathbf{\hat{y}} + z_{5}c \, \mathbf{\hat{z}} & \left(6l\right) & \mbox{Ti I} \\ 
\mathbf{B}_{8} & = & -y_{5} \, \mathbf{a}_{1} + \left(x_{5}-y_{5}\right) \, \mathbf{a}_{2} + z_{5} \, \mathbf{a}_{3} & = & \left(\frac{1}{2}x_{5}-y_{5}\right)a \, \mathbf{\hat{x}} + \frac{\sqrt{3}}{2}x_{5}a \, \mathbf{\hat{y}} + z_{5}c \, \mathbf{\hat{z}} & \left(6l\right) & \mbox{Ti I} \\ 
\mathbf{B}_{9} & = & \left(-x_{5}+y_{5}\right) \, \mathbf{a}_{1}-x_{5} \, \mathbf{a}_{2} + z_{5} \, \mathbf{a}_{3} & = & \left(-x_{5}+\frac{1}{2}y_{5}\right)a \, \mathbf{\hat{x}}-\frac{\sqrt{3}}{2}y_{5}a \, \mathbf{\hat{y}} + z_{5}c \, \mathbf{\hat{z}} & \left(6l\right) & \mbox{Ti I} \\ 
\mathbf{B}_{10} & = & -y_{5} \, \mathbf{a}_{1}-x_{5} \, \mathbf{a}_{2}-z_{5} \, \mathbf{a}_{3} & = & -\frac{1}{2}\left(x_{5}+y_{5}\right)a \, \mathbf{\hat{x}} + \frac{\sqrt{3}}{2}\left(-x_{5}+y_{5}\right)a \, \mathbf{\hat{y}}-z_{5}c \, \mathbf{\hat{z}} & \left(6l\right) & \mbox{Ti I} \\ 
\mathbf{B}_{11} & = & \left(-x_{5}+y_{5}\right) \, \mathbf{a}_{1} + y_{5} \, \mathbf{a}_{2}-z_{5} \, \mathbf{a}_{3} & = & \left(-\frac{1}{2}x_{5}+y_{5}\right)a \, \mathbf{\hat{x}} + \frac{\sqrt{3}}{2}x_{5}a \, \mathbf{\hat{y}}-z_{5}c \, \mathbf{\hat{z}} & \left(6l\right) & \mbox{Ti I} \\ 
\mathbf{B}_{12} & = & x_{5} \, \mathbf{a}_{1} + \left(x_{5}-y_{5}\right) \, \mathbf{a}_{2}-z_{5} \, \mathbf{a}_{3} & = & \left(x_{5}-\frac{1}{2}y_{5}\right)a \, \mathbf{\hat{x}}-\frac{\sqrt{3}}{2}y_{5}a \, \mathbf{\hat{y}}-z_{5}c \, \mathbf{\hat{z}} & \left(6l\right) & \mbox{Ti I} \\ 
\mathbf{B}_{13} & = & x_{6} \, \mathbf{a}_{1} + y_{6} \, \mathbf{a}_{2} + z_{6} \, \mathbf{a}_{3} & = & \frac{1}{2}\left(x_{6}+y_{6}\right)a \, \mathbf{\hat{x}} + \frac{\sqrt{3}}{2}\left(-x_{6}+y_{6}\right)a \, \mathbf{\hat{y}} + z_{6}c \, \mathbf{\hat{z}} & \left(6l\right) & \mbox{Ti II} \\ 
\mathbf{B}_{14} & = & -y_{6} \, \mathbf{a}_{1} + \left(x_{6}-y_{6}\right) \, \mathbf{a}_{2} + z_{6} \, \mathbf{a}_{3} & = & \left(\frac{1}{2}x_{6}-y_{6}\right)a \, \mathbf{\hat{x}} + \frac{\sqrt{3}}{2}x_{6}a \, \mathbf{\hat{y}} + z_{6}c \, \mathbf{\hat{z}} & \left(6l\right) & \mbox{Ti II} \\ 
\mathbf{B}_{15} & = & \left(-x_{6}+y_{6}\right) \, \mathbf{a}_{1}-x_{6} \, \mathbf{a}_{2} + z_{6} \, \mathbf{a}_{3} & = & \left(-x_{6}+\frac{1}{2}y_{6}\right)a \, \mathbf{\hat{x}}-\frac{\sqrt{3}}{2}y_{6}a \, \mathbf{\hat{y}} + z_{6}c \, \mathbf{\hat{z}} & \left(6l\right) & \mbox{Ti II} \\ 
\mathbf{B}_{16} & = & -y_{6} \, \mathbf{a}_{1}-x_{6} \, \mathbf{a}_{2}-z_{6} \, \mathbf{a}_{3} & = & -\frac{1}{2}\left(x_{6}+y_{6}\right)a \, \mathbf{\hat{x}} + \frac{\sqrt{3}}{2}\left(-x_{6}+y_{6}\right)a \, \mathbf{\hat{y}}-z_{6}c \, \mathbf{\hat{z}} & \left(6l\right) & \mbox{Ti II} \\ 
\mathbf{B}_{17} & = & \left(-x_{6}+y_{6}\right) \, \mathbf{a}_{1} + y_{6} \, \mathbf{a}_{2}-z_{6} \, \mathbf{a}_{3} & = & \left(-\frac{1}{2}x_{6}+y_{6}\right)a \, \mathbf{\hat{x}} + \frac{\sqrt{3}}{2}x_{6}a \, \mathbf{\hat{y}}-z_{6}c \, \mathbf{\hat{z}} & \left(6l\right) & \mbox{Ti II} \\ 
\mathbf{B}_{18} & = & x_{6} \, \mathbf{a}_{1} + \left(x_{6}-y_{6}\right) \, \mathbf{a}_{2}-z_{6} \, \mathbf{a}_{3} & = & \left(x_{6}-\frac{1}{2}y_{6}\right)a \, \mathbf{\hat{x}}-\frac{\sqrt{3}}{2}y_{6}a \, \mathbf{\hat{y}}-z_{6}c \, \mathbf{\hat{z}} & \left(6l\right) & \mbox{Ti II} \\ 
\mathbf{B}_{19} & = & x_{7} \, \mathbf{a}_{1} + y_{7} \, \mathbf{a}_{2} + z_{7} \, \mathbf{a}_{3} & = & \frac{1}{2}\left(x_{7}+y_{7}\right)a \, \mathbf{\hat{x}} + \frac{\sqrt{3}}{2}\left(-x_{7}+y_{7}\right)a \, \mathbf{\hat{y}} + z_{7}c \, \mathbf{\hat{z}} & \left(6l\right) & \mbox{Ti III} \\ 
\mathbf{B}_{20} & = & -y_{7} \, \mathbf{a}_{1} + \left(x_{7}-y_{7}\right) \, \mathbf{a}_{2} + z_{7} \, \mathbf{a}_{3} & = & \left(\frac{1}{2}x_{7}-y_{7}\right)a \, \mathbf{\hat{x}} + \frac{\sqrt{3}}{2}x_{7}a \, \mathbf{\hat{y}} + z_{7}c \, \mathbf{\hat{z}} & \left(6l\right) & \mbox{Ti III} \\ 
\mathbf{B}_{21} & = & \left(-x_{7}+y_{7}\right) \, \mathbf{a}_{1}-x_{7} \, \mathbf{a}_{2} + z_{7} \, \mathbf{a}_{3} & = & \left(-x_{7}+\frac{1}{2}y_{7}\right)a \, \mathbf{\hat{x}}-\frac{\sqrt{3}}{2}y_{7}a \, \mathbf{\hat{y}} + z_{7}c \, \mathbf{\hat{z}} & \left(6l\right) & \mbox{Ti III} \\ 
\mathbf{B}_{22} & = & -y_{7} \, \mathbf{a}_{1}-x_{7} \, \mathbf{a}_{2}-z_{7} \, \mathbf{a}_{3} & = & -\frac{1}{2}\left(x_{7}+y_{7}\right)a \, \mathbf{\hat{x}} + \frac{\sqrt{3}}{2}\left(-x_{7}+y_{7}\right)a \, \mathbf{\hat{y}}-z_{7}c \, \mathbf{\hat{z}} & \left(6l\right) & \mbox{Ti III} \\ 
\mathbf{B}_{23} & = & \left(-x_{7}+y_{7}\right) \, \mathbf{a}_{1} + y_{7} \, \mathbf{a}_{2}-z_{7} \, \mathbf{a}_{3} & = & \left(-\frac{1}{2}x_{7}+y_{7}\right)a \, \mathbf{\hat{x}} + \frac{\sqrt{3}}{2}x_{7}a \, \mathbf{\hat{y}}-z_{7}c \, \mathbf{\hat{z}} & \left(6l\right) & \mbox{Ti III} \\ 
\mathbf{B}_{24} & = & x_{7} \, \mathbf{a}_{1} + \left(x_{7}-y_{7}\right) \, \mathbf{a}_{2}-z_{7} \, \mathbf{a}_{3} & = & \left(x_{7}-\frac{1}{2}y_{7}\right)a \, \mathbf{\hat{x}}-\frac{\sqrt{3}}{2}y_{7}a \, \mathbf{\hat{y}}-z_{7}c \, \mathbf{\hat{z}} & \left(6l\right) & \mbox{Ti III} \\ 
\end{longtabu}
\renewcommand{\arraystretch}{1.0}
\noindent \hrulefill
\\
\textbf{References:}
\vspace*{-0.25cm}
\begin{flushleft}
  - \bibentry{Jostsons_Ti3O_ActCrystallogrSecB_1968}. \\
\end{flushleft}
\textbf{Found in:}
\vspace*{-0.25cm}
\begin{flushleft}
  - \bibentry{Villars_PearsonsCrystalData_2013}. \\
\end{flushleft}
\noindent \hrulefill
\\
\textbf{Geometry files:}
\\
\noindent  - CIF: pp. {\hyperref[AB3_hP24_149_acgi_3l_cif]{\pageref{AB3_hP24_149_acgi_3l_cif}}} \\
\noindent  - POSCAR: pp. {\hyperref[AB3_hP24_149_acgi_3l_poscar]{\pageref{AB3_hP24_149_acgi_3l_poscar}}} \\
\onecolumn
{\phantomsection\label{A3B_hP24_153_3c_2b}}
\subsection*{\huge \textbf{{\normalfont CrCl$_{3}$ Structure: A3B\_hP24\_153\_3c\_2b}}}
\noindent \hrulefill
\vspace*{0.25cm}
\begin{figure}[htp]
  \centering
  \vspace{-1em}
  {\includegraphics[width=1\textwidth]{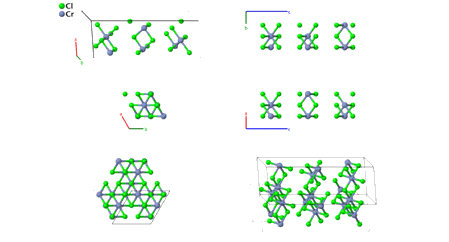}}
\end{figure}
\vspace*{-0.5cm}
\renewcommand{\arraystretch}{1.5}
\begin{equation*}
  \begin{array}{>{$\hspace{-0.15cm}}l<{$}>{$}p{0.5cm}<{$}>{$}p{18.5cm}<{$}}
    \mbox{\large \textbf{Prototype}} &\colon & \ce{CrCl3} \\
    \mbox{\large \textbf{\AFLOW\ prototype label}} &\colon & \mbox{A3B\_hP24\_153\_3c\_2b} \\
    \mbox{\large \textbf{\textit{Strukturbericht} designation}} &\colon & \mbox{None} \\
    \mbox{\large \textbf{Pearson symbol}} &\colon & \mbox{hP24} \\
    \mbox{\large \textbf{Space group number}} &\colon & 153 \\
    \mbox{\large \textbf{Space group symbol}} &\colon & P3_{2}12 \\
    \mbox{\large \textbf{\AFLOW\ prototype command}} &\colon &  \texttt{aflow} \,  \, \texttt{-{}-proto=A3B\_hP24\_153\_3c\_2b } \, \newline \texttt{-{}-params=}{a,c/a,x_{1},x_{2},x_{3},y_{3},z_{3},x_{4},y_{4},z_{4},x_{5},y_{5},z_{5} }
  \end{array}
\end{equation*}
\renewcommand{\arraystretch}{1.0}

\vspace*{-0.25cm}
\noindent \hrulefill
\begin{itemize}
  \item{This structure is the enantiomorph of the \href{http://aflow.org/CrystalDatabase/A3B_hP24_151_3c_2a.html}{CrCl$_{3}$ 
(D0$_{4}$, A3B\_hP24\_151\_3c\_2a) structure}, and was generated by reflecting the coordinates of the space group 
\#151 structure through the $z=0$ plane.
}
\end{itemize}

\noindent \parbox{1 \linewidth}{
\noindent \hrulefill
\\
\textbf{Trigonal Hexagonal primitive vectors:} \\
\vspace*{-0.25cm}
\begin{tabular}{cc}
  \begin{tabular}{c}
    \parbox{0.6 \linewidth}{
      \renewcommand{\arraystretch}{1.5}
      \begin{equation*}
        \centering
        \begin{array}{ccc}
              \mathbf{a}_1 & = & \frac12 \, a \, \mathbf{\hat{x}} - \frac{\sqrt3}2 \, a \, \mathbf{\hat{y}} \\
    \mathbf{a}_2 & = & \frac12 \, a \, \mathbf{\hat{x}} + \frac{\sqrt3}2 \, a \, \mathbf{\hat{y}} \\
    \mathbf{a}_3 & = & c \, \mathbf{\hat{z}} \\

        \end{array}
      \end{equation*}
    }
    \renewcommand{\arraystretch}{1.0}
  \end{tabular}
  \begin{tabular}{c}
    \includegraphics[width=0.3\linewidth]{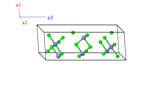} \\
  \end{tabular}
\end{tabular}

}
\vspace*{-0.25cm}

\noindent \hrulefill
\\
\textbf{Basis vectors:}
\vspace*{-0.25cm}
\renewcommand{\arraystretch}{1.5}
\begin{longtabu} to \textwidth{>{\centering $}X[-1,c,c]<{$}>{\centering $}X[-1,c,c]<{$}>{\centering $}X[-1,c,c]<{$}>{\centering $}X[-1,c,c]<{$}>{\centering $}X[-1,c,c]<{$}>{\centering $}X[-1,c,c]<{$}>{\centering $}X[-1,c,c]<{$}}
  & & \mbox{Lattice Coordinates} & & \mbox{Cartesian Coordinates} &\mbox{Wyckoff Position} & \mbox{Atom Type} \\  
  \mathbf{B}_{1} & = & x_{1} \, \mathbf{a}_{1}-x_{1} \, \mathbf{a}_{2} + \frac{1}{6} \, \mathbf{a}_{3} & = & -\sqrt{3}x_{1}a \, \mathbf{\hat{y}} + \frac{1}{6}c \, \mathbf{\hat{z}} & \left(3b\right) & \mbox{Cr I} \\ 
\mathbf{B}_{2} & = & x_{1} \, \mathbf{a}_{1} + 2x_{1} \, \mathbf{a}_{2} + \frac{5}{6} \, \mathbf{a}_{3} & = & \frac{3}{2}x_{1}a \, \mathbf{\hat{x}} + \frac{\sqrt{3}}{2}x_{1}a \, \mathbf{\hat{y}} + \frac{5}{6}c \, \mathbf{\hat{z}} & \left(3b\right) & \mbox{Cr I} \\ 
\mathbf{B}_{3} & = & -2x_{1} \, \mathbf{a}_{1}-x_{1} \, \mathbf{a}_{2} + \frac{1}{2} \, \mathbf{a}_{3} & = & -\frac{3}{2}x_{1}a \, \mathbf{\hat{x}} + \frac{\sqrt{3}}{2}x_{1}a \, \mathbf{\hat{y}} + \frac{1}{2}c \, \mathbf{\hat{z}} & \left(3b\right) & \mbox{Cr I} \\ 
\mathbf{B}_{4} & = & x_{2} \, \mathbf{a}_{1}-x_{2} \, \mathbf{a}_{2} + \frac{1}{6} \, \mathbf{a}_{3} & = & -\sqrt{3}x_{2}a \, \mathbf{\hat{y}} + \frac{1}{6}c \, \mathbf{\hat{z}} & \left(3b\right) & \mbox{Cr II} \\ 
\mathbf{B}_{5} & = & x_{2} \, \mathbf{a}_{1} + 2x_{2} \, \mathbf{a}_{2} + \frac{5}{6} \, \mathbf{a}_{3} & = & \frac{3}{2}x_{2}a \, \mathbf{\hat{x}} + \frac{\sqrt{3}}{2}x_{2}a \, \mathbf{\hat{y}} + \frac{5}{6}c \, \mathbf{\hat{z}} & \left(3b\right) & \mbox{Cr II} \\ 
\mathbf{B}_{6} & = & -2x_{2} \, \mathbf{a}_{1}-x_{2} \, \mathbf{a}_{2} + \frac{1}{2} \, \mathbf{a}_{3} & = & -\frac{3}{2}x_{2}a \, \mathbf{\hat{x}} + \frac{\sqrt{3}}{2}x_{2}a \, \mathbf{\hat{y}} + \frac{1}{2}c \, \mathbf{\hat{z}} & \left(3b\right) & \mbox{Cr II} \\ 
\mathbf{B}_{7} & = & x_{3} \, \mathbf{a}_{1} + y_{3} \, \mathbf{a}_{2} + z_{3} \, \mathbf{a}_{3} & = & \frac{1}{2}\left(x_{3}+y_{3}\right)a \, \mathbf{\hat{x}} + \frac{\sqrt{3}}{2}\left(-x_{3}+y_{3}\right)a \, \mathbf{\hat{y}} + z_{3}c \, \mathbf{\hat{z}} & \left(6c\right) & \mbox{Cl I} \\ 
\mathbf{B}_{8} & = & -y_{3} \, \mathbf{a}_{1} + \left(x_{3}-y_{3}\right) \, \mathbf{a}_{2} + \left(\frac{2}{3} +z_{3}\right) \, \mathbf{a}_{3} & = & \left(\frac{1}{2}x_{3}-y_{3}\right)a \, \mathbf{\hat{x}} + \frac{\sqrt{3}}{2}x_{3}a \, \mathbf{\hat{y}} + \left(\frac{2}{3} +z_{3}\right)c \, \mathbf{\hat{z}} & \left(6c\right) & \mbox{Cl I} \\ 
\mathbf{B}_{9} & = & \left(-x_{3}+y_{3}\right) \, \mathbf{a}_{1}-x_{3} \, \mathbf{a}_{2} + \left(\frac{1}{3} +z_{3}\right) \, \mathbf{a}_{3} & = & \left(-x_{3}+\frac{1}{2}y_{3}\right)a \, \mathbf{\hat{x}}-\frac{\sqrt{3}}{2}y_{3}a \, \mathbf{\hat{y}} + \left(\frac{1}{3} +z_{3}\right)c \, \mathbf{\hat{z}} & \left(6c\right) & \mbox{Cl I} \\ 
\mathbf{B}_{10} & = & -y_{3} \, \mathbf{a}_{1}-x_{3} \, \mathbf{a}_{2} + \left(\frac{1}{3} - z_{3}\right) \, \mathbf{a}_{3} & = & -\frac{1}{2}\left(x_{3}+y_{3}\right)a \, \mathbf{\hat{x}} + \frac{\sqrt{3}}{2}\left(-x_{3}+y_{3}\right)a \, \mathbf{\hat{y}} + \left(\frac{1}{3} - z_{3}\right)c \, \mathbf{\hat{z}} & \left(6c\right) & \mbox{Cl I} \\ 
\mathbf{B}_{11} & = & \left(-x_{3}+y_{3}\right) \, \mathbf{a}_{1} + y_{3} \, \mathbf{a}_{2} + \left(\frac{2}{3} - z_{3}\right) \, \mathbf{a}_{3} & = & \left(-\frac{1}{2}x_{3}+y_{3}\right)a \, \mathbf{\hat{x}} + \frac{\sqrt{3}}{2}x_{3}a \, \mathbf{\hat{y}} + \left(\frac{2}{3} - z_{3}\right)c \, \mathbf{\hat{z}} & \left(6c\right) & \mbox{Cl I} \\ 
\mathbf{B}_{12} & = & x_{3} \, \mathbf{a}_{1} + \left(x_{3}-y_{3}\right) \, \mathbf{a}_{2}-z_{3} \, \mathbf{a}_{3} & = & \left(x_{3}-\frac{1}{2}y_{3}\right)a \, \mathbf{\hat{x}}-\frac{\sqrt{3}}{2}y_{3}a \, \mathbf{\hat{y}}-z_{3}c \, \mathbf{\hat{z}} & \left(6c\right) & \mbox{Cl I} \\ 
\mathbf{B}_{13} & = & x_{4} \, \mathbf{a}_{1} + y_{4} \, \mathbf{a}_{2} + z_{4} \, \mathbf{a}_{3} & = & \frac{1}{2}\left(x_{4}+y_{4}\right)a \, \mathbf{\hat{x}} + \frac{\sqrt{3}}{2}\left(-x_{4}+y_{4}\right)a \, \mathbf{\hat{y}} + z_{4}c \, \mathbf{\hat{z}} & \left(6c\right) & \mbox{Cl II} \\ 
\mathbf{B}_{14} & = & -y_{4} \, \mathbf{a}_{1} + \left(x_{4}-y_{4}\right) \, \mathbf{a}_{2} + \left(\frac{2}{3} +z_{4}\right) \, \mathbf{a}_{3} & = & \left(\frac{1}{2}x_{4}-y_{4}\right)a \, \mathbf{\hat{x}} + \frac{\sqrt{3}}{2}x_{4}a \, \mathbf{\hat{y}} + \left(\frac{2}{3} +z_{4}\right)c \, \mathbf{\hat{z}} & \left(6c\right) & \mbox{Cl II} \\ 
\mathbf{B}_{15} & = & \left(-x_{4}+y_{4}\right) \, \mathbf{a}_{1}-x_{4} \, \mathbf{a}_{2} + \left(\frac{1}{3} +z_{4}\right) \, \mathbf{a}_{3} & = & \left(-x_{4}+\frac{1}{2}y_{4}\right)a \, \mathbf{\hat{x}}-\frac{\sqrt{3}}{2}y_{4}a \, \mathbf{\hat{y}} + \left(\frac{1}{3} +z_{4}\right)c \, \mathbf{\hat{z}} & \left(6c\right) & \mbox{Cl II} \\ 
\mathbf{B}_{16} & = & -y_{4} \, \mathbf{a}_{1}-x_{4} \, \mathbf{a}_{2} + \left(\frac{1}{3} - z_{4}\right) \, \mathbf{a}_{3} & = & -\frac{1}{2}\left(x_{4}+y_{4}\right)a \, \mathbf{\hat{x}} + \frac{\sqrt{3}}{2}\left(-x_{4}+y_{4}\right)a \, \mathbf{\hat{y}} + \left(\frac{1}{3} - z_{4}\right)c \, \mathbf{\hat{z}} & \left(6c\right) & \mbox{Cl II} \\ 
\mathbf{B}_{17} & = & \left(-x_{4}+y_{4}\right) \, \mathbf{a}_{1} + y_{4} \, \mathbf{a}_{2} + \left(\frac{2}{3} - z_{4}\right) \, \mathbf{a}_{3} & = & \left(-\frac{1}{2}x_{4}+y_{4}\right)a \, \mathbf{\hat{x}} + \frac{\sqrt{3}}{2}x_{4}a \, \mathbf{\hat{y}} + \left(\frac{2}{3} - z_{4}\right)c \, \mathbf{\hat{z}} & \left(6c\right) & \mbox{Cl II} \\ 
\mathbf{B}_{18} & = & x_{4} \, \mathbf{a}_{1} + \left(x_{4}-y_{4}\right) \, \mathbf{a}_{2}-z_{4} \, \mathbf{a}_{3} & = & \left(x_{4}-\frac{1}{2}y_{4}\right)a \, \mathbf{\hat{x}}-\frac{\sqrt{3}}{2}y_{4}a \, \mathbf{\hat{y}}-z_{4}c \, \mathbf{\hat{z}} & \left(6c\right) & \mbox{Cl II} \\ 
\mathbf{B}_{19} & = & x_{5} \, \mathbf{a}_{1} + y_{5} \, \mathbf{a}_{2} + z_{5} \, \mathbf{a}_{3} & = & \frac{1}{2}\left(x_{5}+y_{5}\right)a \, \mathbf{\hat{x}} + \frac{\sqrt{3}}{2}\left(-x_{5}+y_{5}\right)a \, \mathbf{\hat{y}} + z_{5}c \, \mathbf{\hat{z}} & \left(6c\right) & \mbox{Cl III} \\ 
\mathbf{B}_{20} & = & -y_{5} \, \mathbf{a}_{1} + \left(x_{5}-y_{5}\right) \, \mathbf{a}_{2} + \left(\frac{2}{3} +z_{5}\right) \, \mathbf{a}_{3} & = & \left(\frac{1}{2}x_{5}-y_{5}\right)a \, \mathbf{\hat{x}} + \frac{\sqrt{3}}{2}x_{5}a \, \mathbf{\hat{y}} + \left(\frac{2}{3} +z_{5}\right)c \, \mathbf{\hat{z}} & \left(6c\right) & \mbox{Cl III} \\ 
\mathbf{B}_{21} & = & \left(-x_{5}+y_{5}\right) \, \mathbf{a}_{1}-x_{5} \, \mathbf{a}_{2} + \left(\frac{1}{3} +z_{5}\right) \, \mathbf{a}_{3} & = & \left(-x_{5}+\frac{1}{2}y_{5}\right)a \, \mathbf{\hat{x}}-\frac{\sqrt{3}}{2}y_{5}a \, \mathbf{\hat{y}} + \left(\frac{1}{3} +z_{5}\right)c \, \mathbf{\hat{z}} & \left(6c\right) & \mbox{Cl III} \\ 
\mathbf{B}_{22} & = & -y_{5} \, \mathbf{a}_{1}-x_{5} \, \mathbf{a}_{2} + \left(\frac{1}{3} - z_{5}\right) \, \mathbf{a}_{3} & = & -\frac{1}{2}\left(x_{5}+y_{5}\right)a \, \mathbf{\hat{x}} + \frac{\sqrt{3}}{2}\left(-x_{5}+y_{5}\right)a \, \mathbf{\hat{y}} + \left(\frac{1}{3} - z_{5}\right)c \, \mathbf{\hat{z}} & \left(6c\right) & \mbox{Cl III} \\ 
\mathbf{B}_{23} & = & \left(-x_{5}+y_{5}\right) \, \mathbf{a}_{1} + y_{5} \, \mathbf{a}_{2} + \left(\frac{2}{3} - z_{5}\right) \, \mathbf{a}_{3} & = & \left(-\frac{1}{2}x_{5}+y_{5}\right)a \, \mathbf{\hat{x}} + \frac{\sqrt{3}}{2}x_{5}a \, \mathbf{\hat{y}} + \left(\frac{2}{3} - z_{5}\right)c \, \mathbf{\hat{z}} & \left(6c\right) & \mbox{Cl III} \\ 
\mathbf{B}_{24} & = & x_{5} \, \mathbf{a}_{1} + \left(x_{5}-y_{5}\right) \, \mathbf{a}_{2}-z_{5} \, \mathbf{a}_{3} & = & \left(x_{5}-\frac{1}{2}y_{5}\right)a \, \mathbf{\hat{x}}-\frac{\sqrt{3}}{2}y_{5}a \, \mathbf{\hat{y}}-z_{5}c \, \mathbf{\hat{z}} & \left(6c\right) & \mbox{Cl III} \\ 
\end{longtabu}
\renewcommand{\arraystretch}{1.0}
\noindent \hrulefill
\\
\textbf{References:}
\vspace*{-0.25cm}
\begin{flushleft}
  - \bibentry{Wooster_CrCl3Structure_Zeitschrift_1930}. \\
\end{flushleft}
\textbf{Found in:}
\vspace*{-0.25cm}
\begin{flushleft}
  - \bibentry{downs03}. \\
\end{flushleft}
\noindent \hrulefill
\\
\textbf{Geometry files:}
\\
\noindent  - CIF: pp. {\hyperref[A3B_hP24_153_3c_2b_cif]{\pageref{A3B_hP24_153_3c_2b_cif}}} \\
\noindent  - POSCAR: pp. {\hyperref[A3B_hP24_153_3c_2b_poscar]{\pageref{A3B_hP24_153_3c_2b_poscar}}} \\
\onecolumn
{\phantomsection\label{A_hP9_154_bc}}
\subsection*{\huge \textbf{{\normalfont S-II Structure: A\_hP9\_154\_bc}}}
\noindent \hrulefill
\vspace*{0.25cm}
\begin{figure}[htp]
  \centering
  \vspace{-1em}
  {\includegraphics[width=1\textwidth]{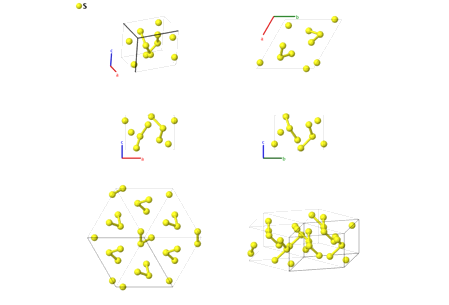}}
\end{figure}
\vspace*{-0.5cm}
\renewcommand{\arraystretch}{1.5}
\begin{equation*}
  \begin{array}{>{$\hspace{-0.15cm}}l<{$}>{$}p{0.5cm}<{$}>{$}p{18.5cm}<{$}}
    \mbox{\large \textbf{Prototype}} &\colon & \ce{S} \\
    \mbox{\large \textbf{\AFLOW\ prototype label}} &\colon & \mbox{A\_hP9\_154\_bc} \\
    \mbox{\large \textbf{\textit{Strukturbericht} designation}} &\colon & \mbox{None} \\
    \mbox{\large \textbf{Pearson symbol}} &\colon & \mbox{hP9} \\
    \mbox{\large \textbf{Space group number}} &\colon & 154 \\
    \mbox{\large \textbf{Space group symbol}} &\colon & P3_{2}21 \\
    \mbox{\large \textbf{\AFLOW\ prototype command}} &\colon &  \texttt{aflow} \,  \, \texttt{-{}-proto=A\_hP9\_154\_bc } \, \newline \texttt{-{}-params=}{a,c/a,x_{1},x_{2},y_{2},z_{2} }
  \end{array}
\end{equation*}
\renewcommand{\arraystretch}{1.0}

\vspace*{-0.25cm}
\noindent \hrulefill
\begin{itemize}
  \item{The S-II phase is found when sulfur heated and pressurized above 3~GPa.
This data was taken at 5.8~GPa and 800~K.
}
\end{itemize}

\noindent \parbox{1 \linewidth}{
\noindent \hrulefill
\\
\textbf{Trigonal Hexagonal primitive vectors:} \\
\vspace*{-0.25cm}
\begin{tabular}{cc}
  \begin{tabular}{c}
    \parbox{0.6 \linewidth}{
      \renewcommand{\arraystretch}{1.5}
      \begin{equation*}
        \centering
        \begin{array}{ccc}
              \mathbf{a}_1 & = & \frac12 \, a \, \mathbf{\hat{x}} - \frac{\sqrt3}2 \, a \, \mathbf{\hat{y}} \\
    \mathbf{a}_2 & = & \frac12 \, a \, \mathbf{\hat{x}} + \frac{\sqrt3}2 \, a \, \mathbf{\hat{y}} \\
    \mathbf{a}_3 & = & c \, \mathbf{\hat{z}} \\

        \end{array}
      \end{equation*}
    }
    \renewcommand{\arraystretch}{1.0}
  \end{tabular}
  \begin{tabular}{c}
    \includegraphics[width=0.3\linewidth]{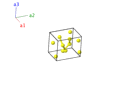} \\
  \end{tabular}
\end{tabular}

}
\vspace*{-0.25cm}

\noindent \hrulefill
\\
\textbf{Basis vectors:}
\vspace*{-0.25cm}
\renewcommand{\arraystretch}{1.5}
\begin{longtabu} to \textwidth{>{\centering $}X[-1,c,c]<{$}>{\centering $}X[-1,c,c]<{$}>{\centering $}X[-1,c,c]<{$}>{\centering $}X[-1,c,c]<{$}>{\centering $}X[-1,c,c]<{$}>{\centering $}X[-1,c,c]<{$}>{\centering $}X[-1,c,c]<{$}}
  & & \mbox{Lattice Coordinates} & & \mbox{Cartesian Coordinates} &\mbox{Wyckoff Position} & \mbox{Atom Type} \\  
  \mathbf{B}_{1} & = & x_{1} \, \mathbf{a}_{1} + \frac{1}{6} \, \mathbf{a}_{3} & = & \frac{1}{2}x_{1}a \, \mathbf{\hat{x}}-\frac{\sqrt{3}}{2}x_{1}a \, \mathbf{\hat{y}} + \frac{1}{6}c \, \mathbf{\hat{z}} & \left(3b\right) & \mbox{S I} \\ 
\mathbf{B}_{2} & = & x_{1} \, \mathbf{a}_{2} + \frac{5}{6} \, \mathbf{a}_{3} & = & \frac{1}{2}x_{1}a \, \mathbf{\hat{x}} + \frac{\sqrt{3}}{2}x_{1}a \, \mathbf{\hat{y}} + \frac{5}{6}c \, \mathbf{\hat{z}} & \left(3b\right) & \mbox{S I} \\ 
\mathbf{B}_{3} & = & -x_{1} \, \mathbf{a}_{1}-x_{1} \, \mathbf{a}_{2} + \frac{1}{2} \, \mathbf{a}_{3} & = & -x_{1}a \, \mathbf{\hat{x}} + \frac{1}{2}c \, \mathbf{\hat{z}} & \left(3b\right) & \mbox{S I} \\ 
\mathbf{B}_{4} & = & x_{2} \, \mathbf{a}_{1} + y_{2} \, \mathbf{a}_{2} + z_{2} \, \mathbf{a}_{3} & = & \frac{1}{2}\left(x_{2}+y_{2}\right)a \, \mathbf{\hat{x}} + \frac{\sqrt{3}}{2}\left(-x_{2}+y_{2}\right)a \, \mathbf{\hat{y}} + z_{2}c \, \mathbf{\hat{z}} & \left(6c\right) & \mbox{S II} \\ 
\mathbf{B}_{5} & = & -y_{2} \, \mathbf{a}_{1} + \left(x_{2}-y_{2}\right) \, \mathbf{a}_{2} + \left(\frac{2}{3} +z_{2}\right) \, \mathbf{a}_{3} & = & \left(\frac{1}{2}x_{2}-y_{2}\right)a \, \mathbf{\hat{x}} + \frac{\sqrt{3}}{2}x_{2}a \, \mathbf{\hat{y}} + \left(\frac{2}{3} +z_{2}\right)c \, \mathbf{\hat{z}} & \left(6c\right) & \mbox{S II} \\ 
\mathbf{B}_{6} & = & \left(-x_{2}+y_{2}\right) \, \mathbf{a}_{1}-x_{2} \, \mathbf{a}_{2} + \left(\frac{1}{3} +z_{2}\right) \, \mathbf{a}_{3} & = & \left(-x_{2}+\frac{1}{2}y_{2}\right)a \, \mathbf{\hat{x}}-\frac{\sqrt{3}}{2}y_{2}a \, \mathbf{\hat{y}} + \left(\frac{1}{3} +z_{2}\right)c \, \mathbf{\hat{z}} & \left(6c\right) & \mbox{S II} \\ 
\mathbf{B}_{7} & = & y_{2} \, \mathbf{a}_{1} + x_{2} \, \mathbf{a}_{2}-z_{2} \, \mathbf{a}_{3} & = & \frac{1}{2}\left(x_{2}+y_{2}\right)a \, \mathbf{\hat{x}} + \frac{\sqrt{3}}{2}\left(x_{2}-y_{2}\right)a \, \mathbf{\hat{y}}-z_{2}c \, \mathbf{\hat{z}} & \left(6c\right) & \mbox{S II} \\ 
\mathbf{B}_{8} & = & \left(x_{2}-y_{2}\right) \, \mathbf{a}_{1}-y_{2} \, \mathbf{a}_{2} + \left(\frac{1}{3} - z_{2}\right) \, \mathbf{a}_{3} & = & \left(\frac{1}{2}x_{2}-y_{2}\right)a \, \mathbf{\hat{x}}-\frac{\sqrt{3}}{2}x_{2}a \, \mathbf{\hat{y}} + \left(\frac{1}{3} - z_{2}\right)c \, \mathbf{\hat{z}} & \left(6c\right) & \mbox{S II} \\ 
\mathbf{B}_{9} & = & -x_{2} \, \mathbf{a}_{1} + \left(-x_{2}+y_{2}\right) \, \mathbf{a}_{2} + \left(\frac{2}{3} - z_{2}\right) \, \mathbf{a}_{3} & = & \left(-x_{2}+\frac{1}{2}y_{2}\right)a \, \mathbf{\hat{x}} + \frac{\sqrt{3}}{2}y_{2}a \, \mathbf{\hat{y}} + \left(\frac{2}{3} - z_{2}\right)c \, \mathbf{\hat{z}} & \left(6c\right) & \mbox{S II} \\ 
\end{longtabu}
\renewcommand{\arraystretch}{1.0}
\noindent \hrulefill
\\
\textbf{References:}
\vspace*{-0.25cm}
\begin{flushleft}
  - \bibentry{degtyareva05:SIII}. \\
\end{flushleft}
\noindent \hrulefill
\\
\textbf{Geometry files:}
\\
\noindent  - CIF: pp. {\hyperref[A_hP9_154_bc_cif]{\pageref{A_hP9_154_bc_cif}}} \\
\noindent  - POSCAR: pp. {\hyperref[A_hP9_154_bc_poscar]{\pageref{A_hP9_154_bc_poscar}}} \\
\onecolumn
{\phantomsection\label{AB2_hP9_156_b2c_3a2bc}}
\subsection*{\huge \textbf{{\normalfont \begin{raggedleft}CdI$_{2}$ (Polytype 6H$_{1}$) Structure: \end{raggedleft} \\ AB2\_hP9\_156\_b2c\_3a2bc}}}
\noindent \hrulefill
\vspace*{0.25cm}
\begin{figure}[htp]
  \centering
  \vspace{-1em}
  {\includegraphics[width=1\textwidth]{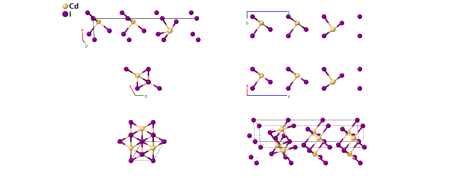}}
\end{figure}
\vspace*{-0.5cm}
\renewcommand{\arraystretch}{1.5}
\begin{equation*}
  \begin{array}{>{$\hspace{-0.15cm}}l<{$}>{$}p{0.5cm}<{$}>{$}p{18.5cm}<{$}}
    \mbox{\large \textbf{Prototype}} &\colon & \ce{CdI2} \\
    \mbox{\large \textbf{\AFLOW\ prototype label}} &\colon & \mbox{AB2\_hP9\_156\_b2c\_3a2bc} \\
    \mbox{\large \textbf{\textit{Strukturbericht} designation}} &\colon & \mbox{None} \\
    \mbox{\large \textbf{Pearson symbol}} &\colon & \mbox{hP9} \\
    \mbox{\large \textbf{Space group number}} &\colon & 156 \\
    \mbox{\large \textbf{Space group symbol}} &\colon & P3m1 \\
    \mbox{\large \textbf{\AFLOW\ prototype command}} &\colon &  \texttt{aflow} \,  \, \texttt{-{}-proto=AB2\_hP9\_156\_b2c\_3a2bc } \, \newline \texttt{-{}-params=}{a,c/a,z_{1},z_{2},z_{3},z_{4},z_{5},z_{6},z_{7},z_{8},z_{9} }
  \end{array}
\end{equation*}
\renewcommand{\arraystretch}{1.0}

\noindent \parbox{1 \linewidth}{
\noindent \hrulefill
\\
\textbf{Trigonal Hexagonal primitive vectors:} \\
\vspace*{-0.25cm}
\begin{tabular}{cc}
  \begin{tabular}{c}
    \parbox{0.6 \linewidth}{
      \renewcommand{\arraystretch}{1.5}
      \begin{equation*}
        \centering
        \begin{array}{ccc}
              \mathbf{a}_1 & = & \frac12 \, a \, \mathbf{\hat{x}} - \frac{\sqrt3}2 \, a \, \mathbf{\hat{y}} \\
    \mathbf{a}_2 & = & \frac12 \, a \, \mathbf{\hat{x}} + \frac{\sqrt3}2 \, a \, \mathbf{\hat{y}} \\
    \mathbf{a}_3 & = & c \, \mathbf{\hat{z}} \\

        \end{array}
      \end{equation*}
    }
    \renewcommand{\arraystretch}{1.0}
  \end{tabular}
  \begin{tabular}{c}
    \includegraphics[width=0.3\linewidth]{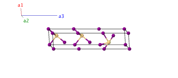} \\
  \end{tabular}
\end{tabular}

}
\vspace*{-0.25cm}

\noindent \hrulefill
\\
\textbf{Basis vectors:}
\vspace*{-0.25cm}
\renewcommand{\arraystretch}{1.5}
\begin{longtabu} to \textwidth{>{\centering $}X[-1,c,c]<{$}>{\centering $}X[-1,c,c]<{$}>{\centering $}X[-1,c,c]<{$}>{\centering $}X[-1,c,c]<{$}>{\centering $}X[-1,c,c]<{$}>{\centering $}X[-1,c,c]<{$}>{\centering $}X[-1,c,c]<{$}}
  & & \mbox{Lattice Coordinates} & & \mbox{Cartesian Coordinates} &\mbox{Wyckoff Position} & \mbox{Atom Type} \\  
  \mathbf{B}_{1} & = & z_{1} \, \mathbf{a}_{3} & = & z_{1}c \, \mathbf{\hat{z}} & \left(1a\right) & \mbox{I I} \\ 
\mathbf{B}_{2} & = & z_{2} \, \mathbf{a}_{3} & = & z_{2}c \, \mathbf{\hat{z}} & \left(1a\right) & \mbox{I II} \\ 
\mathbf{B}_{3} & = & z_{3} \, \mathbf{a}_{3} & = & z_{3}c \, \mathbf{\hat{z}} & \left(1a\right) & \mbox{I III} \\ 
\mathbf{B}_{4} & = & \frac{1}{3} \, \mathbf{a}_{1} + \frac{2}{3} \, \mathbf{a}_{2} + z_{4} \, \mathbf{a}_{3} & = & \frac{1}{2}a \, \mathbf{\hat{x}} + \frac{1}{2\sqrt{3}}a \, \mathbf{\hat{y}} + z_{4}c \, \mathbf{\hat{z}} & \left(1b\right) & \mbox{Cd I} \\ 
\mathbf{B}_{5} & = & \frac{1}{3} \, \mathbf{a}_{1} + \frac{2}{3} \, \mathbf{a}_{2} + z_{5} \, \mathbf{a}_{3} & = & \frac{1}{2}a \, \mathbf{\hat{x}} + \frac{1}{2\sqrt{3}}a \, \mathbf{\hat{y}} + z_{5}c \, \mathbf{\hat{z}} & \left(1b\right) & \mbox{I IV} \\ 
\mathbf{B}_{6} & = & \frac{1}{3} \, \mathbf{a}_{1} + \frac{2}{3} \, \mathbf{a}_{2} + z_{6} \, \mathbf{a}_{3} & = & \frac{1}{2}a \, \mathbf{\hat{x}} + \frac{1}{2\sqrt{3}}a \, \mathbf{\hat{y}} + z_{6}c \, \mathbf{\hat{z}} & \left(1b\right) & \mbox{I V} \\ 
\mathbf{B}_{7} & = & \frac{2}{3} \, \mathbf{a}_{1} + \frac{1}{3} \, \mathbf{a}_{2} + z_{7} \, \mathbf{a}_{3} & = & \frac{1}{2}a \, \mathbf{\hat{x}}- \frac{1}{2\sqrt{3}}a  \, \mathbf{\hat{y}} + z_{7}c \, \mathbf{\hat{z}} & \left(1c\right) & \mbox{Cd II} \\ 
\mathbf{B}_{8} & = & \frac{2}{3} \, \mathbf{a}_{1} + \frac{1}{3} \, \mathbf{a}_{2} + z_{8} \, \mathbf{a}_{3} & = & \frac{1}{2}a \, \mathbf{\hat{x}}- \frac{1}{2\sqrt{3}}a  \, \mathbf{\hat{y}} + z_{8}c \, \mathbf{\hat{z}} & \left(1c\right) & \mbox{Cd III} \\ 
\mathbf{B}_{9} & = & \frac{2}{3} \, \mathbf{a}_{1} + \frac{1}{3} \, \mathbf{a}_{2} + z_{9} \, \mathbf{a}_{3} & = & \frac{1}{2}a \, \mathbf{\hat{x}}- \frac{1}{2\sqrt{3}}a  \, \mathbf{\hat{y}} + z_{9}c \, \mathbf{\hat{z}} & \left(1c\right) & \mbox{I VI} \\ 
\end{longtabu}
\renewcommand{\arraystretch}{1.0}
\noindent \hrulefill
\\
\textbf{References:}
\vspace*{-0.25cm}
\begin{flushleft}
  - \bibentry{Mitchell_CdI2_ZKristallogr_1956}. \\
\end{flushleft}
\textbf{Found in:}
\vspace*{-0.25cm}
\begin{flushleft}
  - \bibentry{Villars_PearsonsCrystalData_2013}. \\
\end{flushleft}
\noindent \hrulefill
\\
\textbf{Geometry files:}
\\
\noindent  - CIF: pp. {\hyperref[AB2_hP9_156_b2c_3a2bc_cif]{\pageref{AB2_hP9_156_b2c_3a2bc_cif}}} \\
\noindent  - POSCAR: pp. {\hyperref[AB2_hP9_156_b2c_3a2bc_poscar]{\pageref{AB2_hP9_156_b2c_3a2bc_poscar}}} \\
\onecolumn
{\phantomsection\label{AB_hP12_156_2ab3c_2ab3c}}
\subsection*{\huge \textbf{{\normalfont CuI Structure: AB\_hP12\_156\_2ab3c\_2ab3c}}}
\noindent \hrulefill
\vspace*{0.25cm}
\begin{figure}[htp]
  \centering
  \vspace{-1em}
  {\includegraphics[width=1\textwidth]{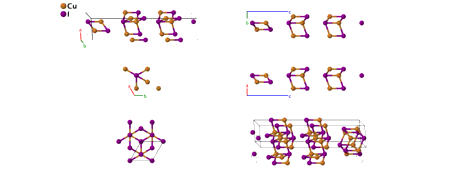}}
\end{figure}
\vspace*{-0.5cm}
\renewcommand{\arraystretch}{1.5}
\begin{equation*}
  \begin{array}{>{$\hspace{-0.15cm}}l<{$}>{$}p{0.5cm}<{$}>{$}p{18.5cm}<{$}}
    \mbox{\large \textbf{Prototype}} &\colon & \ce{CuI} \\
    \mbox{\large \textbf{\AFLOW\ prototype label}} &\colon & \mbox{AB\_hP12\_156\_2ab3c\_2ab3c} \\
    \mbox{\large \textbf{\textit{Strukturbericht} designation}} &\colon & \mbox{None} \\
    \mbox{\large \textbf{Pearson symbol}} &\colon & \mbox{hP12} \\
    \mbox{\large \textbf{Space group number}} &\colon & 156 \\
    \mbox{\large \textbf{Space group symbol}} &\colon & P3m1 \\
    \mbox{\large \textbf{\AFLOW\ prototype command}} &\colon &  \texttt{aflow} \,  \, \texttt{-{}-proto=AB\_hP12\_156\_2ab3c\_2ab3c } \, \newline \texttt{-{}-params=}{a,c/a,z_{1},z_{2},z_{3},z_{4},z_{5},z_{6},z_{7},z_{8},z_{9},z_{10},z_{11},z_{12} }
  \end{array}
\end{equation*}
\renewcommand{\arraystretch}{1.0}

\noindent \parbox{1 \linewidth}{
\noindent \hrulefill
\\
\textbf{Trigonal Hexagonal primitive vectors:} \\
\vspace*{-0.25cm}
\begin{tabular}{cc}
  \begin{tabular}{c}
    \parbox{0.6 \linewidth}{
      \renewcommand{\arraystretch}{1.5}
      \begin{equation*}
        \centering
        \begin{array}{ccc}
              \mathbf{a}_1 & = & \frac12 \, a \, \mathbf{\hat{x}} - \frac{\sqrt3}2 \, a \, \mathbf{\hat{y}} \\
    \mathbf{a}_2 & = & \frac12 \, a \, \mathbf{\hat{x}} + \frac{\sqrt3}2 \, a \, \mathbf{\hat{y}} \\
    \mathbf{a}_3 & = & c \, \mathbf{\hat{z}} \\

        \end{array}
      \end{equation*}
    }
    \renewcommand{\arraystretch}{1.0}
  \end{tabular}
  \begin{tabular}{c}
    \includegraphics[width=0.3\linewidth]{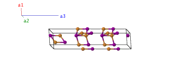} \\
  \end{tabular}
\end{tabular}

}
\vspace*{-0.25cm}

\noindent \hrulefill
\\
\textbf{Basis vectors:}
\vspace*{-0.25cm}
\renewcommand{\arraystretch}{1.5}
\begin{longtabu} to \textwidth{>{\centering $}X[-1,c,c]<{$}>{\centering $}X[-1,c,c]<{$}>{\centering $}X[-1,c,c]<{$}>{\centering $}X[-1,c,c]<{$}>{\centering $}X[-1,c,c]<{$}>{\centering $}X[-1,c,c]<{$}>{\centering $}X[-1,c,c]<{$}}
  & & \mbox{Lattice Coordinates} & & \mbox{Cartesian Coordinates} &\mbox{Wyckoff Position} & \mbox{Atom Type} \\  
  \mathbf{B}_{1} & = & z_{1} \, \mathbf{a}_{3} & = & z_{1}c \, \mathbf{\hat{z}} & \left(1a\right) & \mbox{Cu I} \\ 
\mathbf{B}_{2} & = & z_{2} \, \mathbf{a}_{3} & = & z_{2}c \, \mathbf{\hat{z}} & \left(1a\right) & \mbox{Cu II} \\ 
\mathbf{B}_{3} & = & z_{3} \, \mathbf{a}_{3} & = & z_{3}c \, \mathbf{\hat{z}} & \left(1a\right) & \mbox{I I} \\ 
\mathbf{B}_{4} & = & z_{4} \, \mathbf{a}_{3} & = & z_{4}c \, \mathbf{\hat{z}} & \left(1a\right) & \mbox{I II} \\ 
\mathbf{B}_{5} & = & \frac{1}{3} \, \mathbf{a}_{1} + \frac{2}{3} \, \mathbf{a}_{2} + z_{5} \, \mathbf{a}_{3} & = & \frac{1}{2}a \, \mathbf{\hat{x}} + \frac{1}{2\sqrt{3}}a \, \mathbf{\hat{y}} + z_{5}c \, \mathbf{\hat{z}} & \left(1b\right) & \mbox{Cu III} \\ 
\mathbf{B}_{6} & = & \frac{1}{3} \, \mathbf{a}_{1} + \frac{2}{3} \, \mathbf{a}_{2} + z_{6} \, \mathbf{a}_{3} & = & \frac{1}{2}a \, \mathbf{\hat{x}} + \frac{1}{2\sqrt{3}}a \, \mathbf{\hat{y}} + z_{6}c \, \mathbf{\hat{z}} & \left(1b\right) & \mbox{I III} \\ 
\mathbf{B}_{7} & = & \frac{2}{3} \, \mathbf{a}_{1} + \frac{1}{3} \, \mathbf{a}_{2} + z_{7} \, \mathbf{a}_{3} & = & \frac{1}{2}a \, \mathbf{\hat{x}}- \frac{1}{2\sqrt{3}}a  \, \mathbf{\hat{y}} + z_{7}c \, \mathbf{\hat{z}} & \left(1c\right) & \mbox{Cu IV} \\ 
\mathbf{B}_{8} & = & \frac{2}{3} \, \mathbf{a}_{1} + \frac{1}{3} \, \mathbf{a}_{2} + z_{8} \, \mathbf{a}_{3} & = & \frac{1}{2}a \, \mathbf{\hat{x}}- \frac{1}{2\sqrt{3}}a  \, \mathbf{\hat{y}} + z_{8}c \, \mathbf{\hat{z}} & \left(1c\right) & \mbox{Cu V} \\ 
\mathbf{B}_{9} & = & \frac{2}{3} \, \mathbf{a}_{1} + \frac{1}{3} \, \mathbf{a}_{2} + z_{9} \, \mathbf{a}_{3} & = & \frac{1}{2}a \, \mathbf{\hat{x}}- \frac{1}{2\sqrt{3}}a  \, \mathbf{\hat{y}} + z_{9}c \, \mathbf{\hat{z}} & \left(1c\right) & \mbox{Cu VI} \\ 
\mathbf{B}_{10} & = & \frac{2}{3} \, \mathbf{a}_{1} + \frac{1}{3} \, \mathbf{a}_{2} + z_{10} \, \mathbf{a}_{3} & = & \frac{1}{2}a \, \mathbf{\hat{x}}- \frac{1}{2\sqrt{3}}a  \, \mathbf{\hat{y}} + z_{10}c \, \mathbf{\hat{z}} & \left(1c\right) & \mbox{I IV} \\ 
\mathbf{B}_{11} & = & \frac{2}{3} \, \mathbf{a}_{1} + \frac{1}{3} \, \mathbf{a}_{2} + z_{11} \, \mathbf{a}_{3} & = & \frac{1}{2}a \, \mathbf{\hat{x}}- \frac{1}{2\sqrt{3}}a  \, \mathbf{\hat{y}} + z_{11}c \, \mathbf{\hat{z}} & \left(1c\right) & \mbox{I V} \\ 
\mathbf{B}_{12} & = & \frac{2}{3} \, \mathbf{a}_{1} + \frac{1}{3} \, \mathbf{a}_{2} + z_{12} \, \mathbf{a}_{3} & = & \frac{1}{2}a \, \mathbf{\hat{x}}- \frac{1}{2\sqrt{3}}a  \, \mathbf{\hat{y}} + z_{12}c \, \mathbf{\hat{z}} & \left(1c\right) & \mbox{I VI} \\ 
\end{longtabu}
\renewcommand{\arraystretch}{1.0}
\noindent \hrulefill
\\
\textbf{References:}
\vspace*{-0.25cm}
\begin{flushleft}
  - \bibentry{Kurdyumova_CuI_SovPhysCrys_1970}. \\
\end{flushleft}
\textbf{Found in:}
\vspace*{-0.25cm}
\begin{flushleft}
  - \bibentry{Villars_PearsonsCrystalData_2013}. \\
\end{flushleft}
\noindent \hrulefill
\\
\textbf{Geometry files:}
\\
\noindent  - CIF: pp. {\hyperref[AB_hP12_156_2ab3c_2ab3c_cif]{\pageref{AB_hP12_156_2ab3c_2ab3c_cif}}} \\
\noindent  - POSCAR: pp. {\hyperref[AB_hP12_156_2ab3c_2ab3c_poscar]{\pageref{AB_hP12_156_2ab3c_2ab3c_poscar}}} \\
\onecolumn
{\phantomsection\label{AB_hP4_156_ac_ac}}
\subsection*{\huge \textbf{{\normalfont $\beta$-CuI Structure: AB\_hP4\_156\_ac\_ac}}}
\noindent \hrulefill
\vspace*{0.25cm}
\begin{figure}[htp]
  \centering
  \vspace{-1em}
  {\includegraphics[width=1\textwidth]{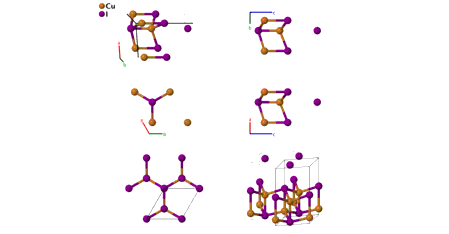}}
\end{figure}
\vspace*{-0.5cm}
\renewcommand{\arraystretch}{1.5}
\begin{equation*}
  \begin{array}{>{$\hspace{-0.15cm}}l<{$}>{$}p{0.5cm}<{$}>{$}p{18.5cm}<{$}}
    \mbox{\large \textbf{Prototype}} &\colon & \ce{$\beta$-CuI} \\
    \mbox{\large \textbf{\AFLOW\ prototype label}} &\colon & \mbox{AB\_hP4\_156\_ac\_ac} \\
    \mbox{\large \textbf{\textit{Strukturbericht} designation}} &\colon & \mbox{None} \\
    \mbox{\large \textbf{Pearson symbol}} &\colon & \mbox{hP4} \\
    \mbox{\large \textbf{Space group number}} &\colon & 156 \\
    \mbox{\large \textbf{Space group symbol}} &\colon & P3m1 \\
    \mbox{\large \textbf{\AFLOW\ prototype command}} &\colon &  \texttt{aflow} \,  \, \texttt{-{}-proto=AB\_hP4\_156\_ac\_ac } \, \newline \texttt{-{}-params=}{a,c/a,z_{1},z_{2},z_{3},z_{4} }
  \end{array}
\end{equation*}
\renewcommand{\arraystretch}{1.0}

\noindent \parbox{1 \linewidth}{
\noindent \hrulefill
\\
\textbf{Trigonal Hexagonal primitive vectors:} \\
\vspace*{-0.25cm}
\begin{tabular}{cc}
  \begin{tabular}{c}
    \parbox{0.6 \linewidth}{
      \renewcommand{\arraystretch}{1.5}
      \begin{equation*}
        \centering
        \begin{array}{ccc}
              \mathbf{a}_1 & = & \frac12 \, a \, \mathbf{\hat{x}} - \frac{\sqrt3}2 \, a \, \mathbf{\hat{y}} \\
    \mathbf{a}_2 & = & \frac12 \, a \, \mathbf{\hat{x}} + \frac{\sqrt3}2 \, a \, \mathbf{\hat{y}} \\
    \mathbf{a}_3 & = & c \, \mathbf{\hat{z}} \\

        \end{array}
      \end{equation*}
    }
    \renewcommand{\arraystretch}{1.0}
  \end{tabular}
  \begin{tabular}{c}
    \includegraphics[width=0.3\linewidth]{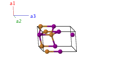} \\
  \end{tabular}
\end{tabular}

}
\vspace*{-0.25cm}

\noindent \hrulefill
\\
\textbf{Basis vectors:}
\vspace*{-0.25cm}
\renewcommand{\arraystretch}{1.5}
\begin{longtabu} to \textwidth{>{\centering $}X[-1,c,c]<{$}>{\centering $}X[-1,c,c]<{$}>{\centering $}X[-1,c,c]<{$}>{\centering $}X[-1,c,c]<{$}>{\centering $}X[-1,c,c]<{$}>{\centering $}X[-1,c,c]<{$}>{\centering $}X[-1,c,c]<{$}}
  & & \mbox{Lattice Coordinates} & & \mbox{Cartesian Coordinates} &\mbox{Wyckoff Position} & \mbox{Atom Type} \\  
  \mathbf{B}_{1} & = & z_{1} \, \mathbf{a}_{3} & = & z_{1}c \, \mathbf{\hat{z}} & \left(1a\right) & \mbox{Cu I} \\ 
\mathbf{B}_{2} & = & z_{2} \, \mathbf{a}_{3} & = & z_{2}c \, \mathbf{\hat{z}} & \left(1a\right) & \mbox{I I} \\ 
\mathbf{B}_{3} & = & \frac{2}{3} \, \mathbf{a}_{1} + \frac{1}{3} \, \mathbf{a}_{2} + z_{3} \, \mathbf{a}_{3} & = & \frac{1}{2}a \, \mathbf{\hat{x}}- \frac{1}{2\sqrt{3}}a  \, \mathbf{\hat{y}} + z_{3}c \, \mathbf{\hat{z}} & \left(1c\right) & \mbox{Cu II} \\ 
\mathbf{B}_{4} & = & \frac{2}{3} \, \mathbf{a}_{1} + \frac{1}{3} \, \mathbf{a}_{2} + z_{4} \, \mathbf{a}_{3} & = & \frac{1}{2}a \, \mathbf{\hat{x}}- \frac{1}{2\sqrt{3}}a  \, \mathbf{\hat{y}} + z_{4}c \, \mathbf{\hat{z}} & \left(1c\right) & \mbox{I II} \\ 
\end{longtabu}
\renewcommand{\arraystretch}{1.0}
\noindent \hrulefill
\\
\textbf{References:}
\vspace*{-0.25cm}
\begin{flushleft}
  - \bibentry{Sakuma_CuI_JPhysSocJpn_1988}. \\
\end{flushleft}
\textbf{Found in:}
\vspace*{-0.25cm}
\begin{flushleft}
  - \bibentry{Villars_PearsonsCrystalData_2013}. \\
\end{flushleft}
\noindent \hrulefill
\\
\textbf{Geometry files:}
\\
\noindent  - CIF: pp. {\hyperref[AB_hP4_156_ac_ac_cif]{\pageref{AB_hP4_156_ac_ac_cif}}} \\
\noindent  - POSCAR: pp. {\hyperref[AB_hP4_156_ac_ac_poscar]{\pageref{AB_hP4_156_ac_ac_poscar}}} \\
\onecolumn
{\phantomsection\label{A5B6C2_hP13_157_2ac_2c_b}}
\subsection*{\huge \textbf{{\normalfont Ag$_{5}$Pb$_{2}$O$_{6}$ Structure: A5B6C2\_hP13\_157\_2ac\_2c\_b}}}
\noindent \hrulefill
\vspace*{0.25cm}
\begin{figure}[htp]
  \centering
  \vspace{-1em}
  {\includegraphics[width=1\textwidth]{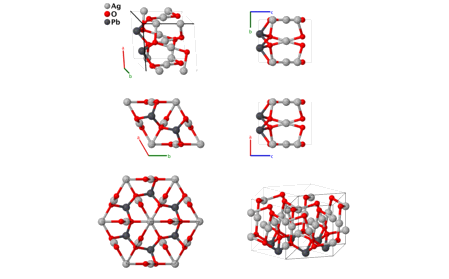}}
\end{figure}
\vspace*{-0.5cm}
\renewcommand{\arraystretch}{1.5}
\begin{equation*}
  \begin{array}{>{$\hspace{-0.15cm}}l<{$}>{$}p{0.5cm}<{$}>{$}p{18.5cm}<{$}}
    \mbox{\large \textbf{Prototype}} &\colon & \ce{Ag5Pb2O6} \\
    \mbox{\large \textbf{\AFLOW\ prototype label}} &\colon & \mbox{A5B6C2\_hP13\_157\_2ac\_2c\_b} \\
    \mbox{\large \textbf{\textit{Strukturbericht} designation}} &\colon & \mbox{None} \\
    \mbox{\large \textbf{Pearson symbol}} &\colon & \mbox{hP13} \\
    \mbox{\large \textbf{Space group number}} &\colon & 157 \\
    \mbox{\large \textbf{Space group symbol}} &\colon & P31m \\
    \mbox{\large \textbf{\AFLOW\ prototype command}} &\colon &  \texttt{aflow} \,  \, \texttt{-{}-proto=A5B6C2\_hP13\_157\_2ac\_2c\_b } \, \newline \texttt{-{}-params=}{a,c/a,z_{1},z_{2},z_{3},x_{4},z_{4},x_{5},z_{5},x_{6},z_{6} }
  \end{array}
\end{equation*}
\renewcommand{\arraystretch}{1.0}

\vspace*{-0.25cm}
\noindent \hrulefill
\begin{itemize}
  \item{The original reference (Bystr{\"o}m, 1950) lists this structure as Ag$_{5}$Pb$_{2}$O$_{6}$, while (Villars, 1985) lists it as Ag$_{2}$PbO$_{3}$.
(Bystr{\"o}m, 1950) provides three Wyckoff positions for Ag (1a, 1a, and 3c), while (Villars, 195) only provides two (1a and 3c), 
giving rise to the stoichiometry discrepancy.
While both descriptions yield space group \#157, the authors use the structure and coordinates provided by the original reference (Bystr{\"o}m, 1950).
}
\end{itemize}

\noindent \parbox{1 \linewidth}{
\noindent \hrulefill
\\
\textbf{Trigonal Hexagonal primitive vectors:} \\
\vspace*{-0.25cm}
\begin{tabular}{cc}
  \begin{tabular}{c}
    \parbox{0.6 \linewidth}{
      \renewcommand{\arraystretch}{1.5}
      \begin{equation*}
        \centering
        \begin{array}{ccc}
              \mathbf{a}_1 & = & \frac12 \, a \, \mathbf{\hat{x}} - \frac{\sqrt3}2 \, a \, \mathbf{\hat{y}} \\
    \mathbf{a}_2 & = & \frac12 \, a \, \mathbf{\hat{x}} + \frac{\sqrt3}2 \, a \, \mathbf{\hat{y}} \\
    \mathbf{a}_3 & = & c \, \mathbf{\hat{z}} \\

        \end{array}
      \end{equation*}
    }
    \renewcommand{\arraystretch}{1.0}
  \end{tabular}
  \begin{tabular}{c}
    \includegraphics[width=0.3\linewidth]{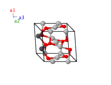} \\
  \end{tabular}
\end{tabular}

}
\vspace*{-0.25cm}

\noindent \hrulefill
\\
\textbf{Basis vectors:}
\vspace*{-0.25cm}
\renewcommand{\arraystretch}{1.5}
\begin{longtabu} to \textwidth{>{\centering $}X[-1,c,c]<{$}>{\centering $}X[-1,c,c]<{$}>{\centering $}X[-1,c,c]<{$}>{\centering $}X[-1,c,c]<{$}>{\centering $}X[-1,c,c]<{$}>{\centering $}X[-1,c,c]<{$}>{\centering $}X[-1,c,c]<{$}}
  & & \mbox{Lattice Coordinates} & & \mbox{Cartesian Coordinates} &\mbox{Wyckoff Position} & \mbox{Atom Type} \\  
  \mathbf{B}_{1} & = & z_{1} \, \mathbf{a}_{3} & = & z_{1}c \, \mathbf{\hat{z}} & \left(1a\right) & \mbox{Ag I} \\ 
\mathbf{B}_{2} & = & z_{2} \, \mathbf{a}_{3} & = & z_{2}c \, \mathbf{\hat{z}} & \left(1a\right) & \mbox{Ag II} \\ 
\mathbf{B}_{3} & = & \frac{1}{3} \, \mathbf{a}_{1} + \frac{2}{3} \, \mathbf{a}_{2} + z_{3} \, \mathbf{a}_{3} & = & \frac{1}{2}a \, \mathbf{\hat{x}} + \frac{1}{2\sqrt{3}}a \, \mathbf{\hat{y}} + z_{3}c \, \mathbf{\hat{z}} & \left(2b\right) & \mbox{Pb} \\ 
\mathbf{B}_{4} & = & \frac{2}{3} \, \mathbf{a}_{1} + \frac{1}{3} \, \mathbf{a}_{2} + z_{3} \, \mathbf{a}_{3} & = & \frac{1}{2}a \, \mathbf{\hat{x}}- \frac{1}{2\sqrt{3}}a  \, \mathbf{\hat{y}} + z_{3}c \, \mathbf{\hat{z}} & \left(2b\right) & \mbox{Pb} \\ 
\mathbf{B}_{5} & = & x_{4} \, \mathbf{a}_{1} + z_{4} \, \mathbf{a}_{3} & = & \frac{1}{2}x_{4}a \, \mathbf{\hat{x}}-\frac{\sqrt{3}}{2}x_{4}a \, \mathbf{\hat{y}} + z_{4}c \, \mathbf{\hat{z}} & \left(3c\right) & \mbox{Ag III} \\ 
\mathbf{B}_{6} & = & x_{4} \, \mathbf{a}_{2} + z_{4} \, \mathbf{a}_{3} & = & \frac{1}{2}x_{4}a \, \mathbf{\hat{x}} + \frac{\sqrt{3}}{2}x_{4}a \, \mathbf{\hat{y}} + z_{4}c \, \mathbf{\hat{z}} & \left(3c\right) & \mbox{Ag III} \\ 
\mathbf{B}_{7} & = & -x_{4} \, \mathbf{a}_{1}-x_{4} \, \mathbf{a}_{2} + z_{4} \, \mathbf{a}_{3} & = & -x_{4}a \, \mathbf{\hat{x}} + z_{4}c \, \mathbf{\hat{z}} & \left(3c\right) & \mbox{Ag III} \\ 
\mathbf{B}_{8} & = & x_{5} \, \mathbf{a}_{1} + z_{5} \, \mathbf{a}_{3} & = & \frac{1}{2}x_{5}a \, \mathbf{\hat{x}}-\frac{\sqrt{3}}{2}x_{5}a \, \mathbf{\hat{y}} + z_{5}c \, \mathbf{\hat{z}} & \left(3c\right) & \mbox{O I} \\ 
\mathbf{B}_{9} & = & x_{5} \, \mathbf{a}_{2} + z_{5} \, \mathbf{a}_{3} & = & \frac{1}{2}x_{5}a \, \mathbf{\hat{x}} + \frac{\sqrt{3}}{2}x_{5}a \, \mathbf{\hat{y}} + z_{5}c \, \mathbf{\hat{z}} & \left(3c\right) & \mbox{O I} \\ 
\mathbf{B}_{10} & = & -x_{5} \, \mathbf{a}_{1}-x_{5} \, \mathbf{a}_{2} + z_{5} \, \mathbf{a}_{3} & = & -x_{5}a \, \mathbf{\hat{x}} + z_{5}c \, \mathbf{\hat{z}} & \left(3c\right) & \mbox{O I} \\ 
\mathbf{B}_{11} & = & x_{6} \, \mathbf{a}_{1} + z_{6} \, \mathbf{a}_{3} & = & \frac{1}{2}x_{6}a \, \mathbf{\hat{x}}-\frac{\sqrt{3}}{2}x_{6}a \, \mathbf{\hat{y}} + z_{6}c \, \mathbf{\hat{z}} & \left(3c\right) & \mbox{O II} \\ 
\mathbf{B}_{12} & = & x_{6} \, \mathbf{a}_{2} + z_{6} \, \mathbf{a}_{3} & = & \frac{1}{2}x_{6}a \, \mathbf{\hat{x}} + \frac{\sqrt{3}}{2}x_{6}a \, \mathbf{\hat{y}} + z_{6}c \, \mathbf{\hat{z}} & \left(3c\right) & \mbox{O II} \\ 
\mathbf{B}_{13} & = & -x_{6} \, \mathbf{a}_{1}-x_{6} \, \mathbf{a}_{2} + z_{6} \, \mathbf{a}_{3} & = & -x_{6}a \, \mathbf{\hat{x}} + z_{6}c \, \mathbf{\hat{z}} & \left(3c\right) & \mbox{O II} \\ 
\end{longtabu}
\renewcommand{\arraystretch}{1.0}
\noindent \hrulefill
\\
\textbf{References:}
\vspace*{-0.25cm}
\begin{flushleft}
  - \bibentry{Bystrom_Ag5Pb2O6_actachems_4_1950}. \\
\end{flushleft}
\textbf{Found in:}
\vspace*{-0.25cm}
\begin{flushleft}
  - \bibentry{villars91:pearson_Ag5Pb2O6}. \\
\end{flushleft}
\noindent \hrulefill
\\
\textbf{Geometry files:}
\\
\noindent  - CIF: pp. {\hyperref[A5B6C2_hP13_157_2ac_2c_b_cif]{\pageref{A5B6C2_hP13_157_2ac_2c_b_cif}}} \\
\noindent  - POSCAR: pp. {\hyperref[A5B6C2_hP13_157_2ac_2c_b_poscar]{\pageref{A5B6C2_hP13_157_2ac_2c_b_poscar}}} \\
\onecolumn
{\phantomsection\label{A3B_hP8_158_d_a}}
\subsection*{\huge \textbf{{\normalfont $\beta$-RuCl$_{3}$ Structure: A3B\_hP8\_158\_d\_a}}}
\noindent \hrulefill
\vspace*{0.25cm}
\begin{figure}[htp]
  \centering
  \vspace{-1em}
  {\includegraphics[width=1\textwidth]{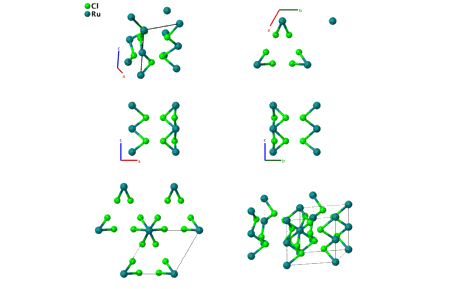}}
\end{figure}
\vspace*{-0.5cm}
\renewcommand{\arraystretch}{1.5}
\begin{equation*}
  \begin{array}{>{$\hspace{-0.15cm}}l<{$}>{$}p{0.5cm}<{$}>{$}p{18.5cm}<{$}}
    \mbox{\large \textbf{Prototype}} &\colon & \ce{$\beta$-RuCl3} \\
    \mbox{\large \textbf{\AFLOW\ prototype label}} &\colon & \mbox{A3B\_hP8\_158\_d\_a} \\
    \mbox{\large \textbf{\textit{Strukturbericht} designation}} &\colon & \mbox{None} \\
    \mbox{\large \textbf{Pearson symbol}} &\colon & \mbox{hP8} \\
    \mbox{\large \textbf{Space group number}} &\colon & 158 \\
    \mbox{\large \textbf{Space group symbol}} &\colon & P3c1 \\
    \mbox{\large \textbf{\AFLOW\ prototype command}} &\colon &  \texttt{aflow} \,  \, \texttt{-{}-proto=A3B\_hP8\_158\_d\_a } \, \newline \texttt{-{}-params=}{a,c/a,z_{1},x_{2},y_{2},z_{2} }
  \end{array}
\end{equation*}
\renewcommand{\arraystretch}{1.0}

\vspace*{-0.25cm}
\noindent \hrulefill
\begin{itemize}
  \item{Pearson comments that space groups \#185, \#188, \#193, could not be rejected, but this structure is consistent with space group \#158.
We also provide the structure with space group \#185: \hyperref[A3B_hP8_185_c_a]{$\beta$-RuCl$_{3}$ (A3B\_hP8\_185\_c\_a) structure}.
}
\end{itemize}

\noindent \parbox{1 \linewidth}{
\noindent \hrulefill
\\
\textbf{Trigonal Hexagonal primitive vectors:} \\
\vspace*{-0.25cm}
\begin{tabular}{cc}
  \begin{tabular}{c}
    \parbox{0.6 \linewidth}{
      \renewcommand{\arraystretch}{1.5}
      \begin{equation*}
        \centering
        \begin{array}{ccc}
              \mathbf{a}_1 & = & \frac12 \, a \, \mathbf{\hat{x}} - \frac{\sqrt3}2 \, a \, \mathbf{\hat{y}} \\
    \mathbf{a}_2 & = & \frac12 \, a \, \mathbf{\hat{x}} + \frac{\sqrt3}2 \, a \, \mathbf{\hat{y}} \\
    \mathbf{a}_3 & = & c \, \mathbf{\hat{z}} \\

        \end{array}
      \end{equation*}
    }
    \renewcommand{\arraystretch}{1.0}
  \end{tabular}
  \begin{tabular}{c}
    \includegraphics[width=0.3\linewidth]{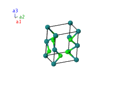} \\
  \end{tabular}
\end{tabular}

}
\vspace*{-0.25cm}

\noindent \hrulefill
\\
\textbf{Basis vectors:}
\vspace*{-0.25cm}
\renewcommand{\arraystretch}{1.5}
\begin{longtabu} to \textwidth{>{\centering $}X[-1,c,c]<{$}>{\centering $}X[-1,c,c]<{$}>{\centering $}X[-1,c,c]<{$}>{\centering $}X[-1,c,c]<{$}>{\centering $}X[-1,c,c]<{$}>{\centering $}X[-1,c,c]<{$}>{\centering $}X[-1,c,c]<{$}}
  & & \mbox{Lattice Coordinates} & & \mbox{Cartesian Coordinates} &\mbox{Wyckoff Position} & \mbox{Atom Type} \\  
  \mathbf{B}_{1} & = & z_{1} \, \mathbf{a}_{3} & = & z_{1}c \, \mathbf{\hat{z}} & \left(2a\right) & \mbox{Ru} \\ 
\mathbf{B}_{2} & = & \left(\frac{1}{2} +z_{1}\right) \, \mathbf{a}_{3} & = & \left(\frac{1}{2} +z_{1}\right)c \, \mathbf{\hat{z}} & \left(2a\right) & \mbox{Ru} \\ 
\mathbf{B}_{3} & = & x_{2} \, \mathbf{a}_{1} + y_{2} \, \mathbf{a}_{2} + z_{2} \, \mathbf{a}_{3} & = & \frac{1}{2}\left(x_{2}+y_{2}\right)a \, \mathbf{\hat{x}} + \frac{\sqrt{3}}{2}\left(-x_{2}+y_{2}\right)a \, \mathbf{\hat{y}} + z_{2}c \, \mathbf{\hat{z}} & \left(6d\right) & \mbox{Cl} \\ 
\mathbf{B}_{4} & = & -y_{2} \, \mathbf{a}_{1} + \left(x_{2}-y_{2}\right) \, \mathbf{a}_{2} + z_{2} \, \mathbf{a}_{3} & = & \left(\frac{1}{2}x_{2}-y_{2}\right)a \, \mathbf{\hat{x}} + \frac{\sqrt{3}}{2}x_{2}a \, \mathbf{\hat{y}} + z_{2}c \, \mathbf{\hat{z}} & \left(6d\right) & \mbox{Cl} \\ 
\mathbf{B}_{5} & = & \left(-x_{2}+y_{2}\right) \, \mathbf{a}_{1}-x_{2} \, \mathbf{a}_{2} + z_{2} \, \mathbf{a}_{3} & = & \left(-x_{2}+\frac{1}{2}y_{2}\right)a \, \mathbf{\hat{x}}-\frac{\sqrt{3}}{2}y_{2}a \, \mathbf{\hat{y}} + z_{2}c \, \mathbf{\hat{z}} & \left(6d\right) & \mbox{Cl} \\ 
\mathbf{B}_{6} & = & -y_{2} \, \mathbf{a}_{1}-x_{2} \, \mathbf{a}_{2} + \left(\frac{1}{2} +z_{2}\right) \, \mathbf{a}_{3} & = & -\frac{1}{2}\left(x_{2}+y_{2}\right)a \, \mathbf{\hat{x}} + \frac{\sqrt{3}}{2}\left(-x_{2}+y_{2}\right)a \, \mathbf{\hat{y}} + \left(\frac{1}{2} +z_{2}\right)c \, \mathbf{\hat{z}} & \left(6d\right) & \mbox{Cl} \\ 
\mathbf{B}_{7} & = & \left(-x_{2}+y_{2}\right) \, \mathbf{a}_{1} + y_{2} \, \mathbf{a}_{2} + \left(\frac{1}{2} +z_{2}\right) \, \mathbf{a}_{3} & = & \left(-\frac{1}{2}x_{2}+y_{2}\right)a \, \mathbf{\hat{x}} + \frac{\sqrt{3}}{2}x_{2}a \, \mathbf{\hat{y}} + \left(\frac{1}{2} +z_{2}\right)c \, \mathbf{\hat{z}} & \left(6d\right) & \mbox{Cl} \\ 
\mathbf{B}_{8} & = & x_{2} \, \mathbf{a}_{1} + \left(x_{2}-y_{2}\right) \, \mathbf{a}_{2} + \left(\frac{1}{2} +z_{2}\right) \, \mathbf{a}_{3} & = & \left(x_{2}-\frac{1}{2}y_{2}\right)a \, \mathbf{\hat{x}}-\frac{\sqrt{3}}{2}y_{2}a \, \mathbf{\hat{y}} + \left(\frac{1}{2} +z_{2}\right)c \, \mathbf{\hat{z}} & \left(6d\right) & \mbox{Cl} \\ 
\end{longtabu}
\renewcommand{\arraystretch}{1.0}
\noindent \hrulefill
\\
\textbf{References:}
\vspace*{-0.25cm}
\begin{flushleft}
  - \bibentry{Fletcher_RuCl3_1967}. \\
\end{flushleft}
\noindent \hrulefill
\\
\textbf{Geometry files:}
\\
\noindent  - CIF: pp. {\hyperref[A3B_hP8_158_d_a_cif]{\pageref{A3B_hP8_158_d_a_cif}}} \\
\noindent  - POSCAR: pp. {\hyperref[A3B_hP8_158_d_a_poscar]{\pageref{A3B_hP8_158_d_a_poscar}}} \\
\onecolumn
{\phantomsection\label{A2B3_hP20_159_bc_2c}}
\subsection*{\huge \textbf{{\normalfont \begin{raggedleft}Bi$_{2}$O$_{3}$ (High-pressure) Structure: \end{raggedleft} \\ A2B3\_hP20\_159\_bc\_2c}}}
\noindent \hrulefill
\vspace*{0.25cm}
\begin{figure}[htp]
  \centering
  \vspace{-1em}
  {\includegraphics[width=1\textwidth]{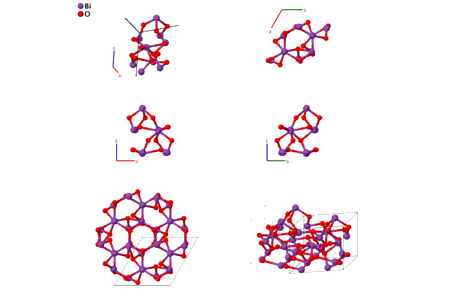}}
\end{figure}
\vspace*{-0.5cm}
\renewcommand{\arraystretch}{1.5}
\begin{equation*}
  \begin{array}{>{$\hspace{-0.15cm}}l<{$}>{$}p{0.5cm}<{$}>{$}p{18.5cm}<{$}}
    \mbox{\large \textbf{Prototype}} &\colon & \ce{Bi2O3} \\
    \mbox{\large \textbf{\AFLOW\ prototype label}} &\colon & \mbox{A2B3\_hP20\_159\_bc\_2c} \\
    \mbox{\large \textbf{\textit{Strukturbericht} designation}} &\colon & \mbox{None} \\
    \mbox{\large \textbf{Pearson symbol}} &\colon & \mbox{hP20} \\
    \mbox{\large \textbf{Space group number}} &\colon & 159 \\
    \mbox{\large \textbf{Space group symbol}} &\colon & P31c \\
    \mbox{\large \textbf{\AFLOW\ prototype command}} &\colon &  \texttt{aflow} \,  \, \texttt{-{}-proto=A2B3\_hP20\_159\_bc\_2c } \, \newline \texttt{-{}-params=}{a,c/a,z_{1},x_{2},y_{2},z_{2},x_{3},y_{3},z_{3},x_{4},y_{4},z_{4} }
  \end{array}
\end{equation*}
\renewcommand{\arraystretch}{1.0}

\noindent \parbox{1 \linewidth}{
\noindent \hrulefill
\\
\textbf{Trigonal Hexagonal primitive vectors:} \\
\vspace*{-0.25cm}
\begin{tabular}{cc}
  \begin{tabular}{c}
    \parbox{0.6 \linewidth}{
      \renewcommand{\arraystretch}{1.5}
      \begin{equation*}
        \centering
        \begin{array}{ccc}
              \mathbf{a}_1 & = & \frac12 \, a \, \mathbf{\hat{x}} - \frac{\sqrt3}2 \, a \, \mathbf{\hat{y}} \\
    \mathbf{a}_2 & = & \frac12 \, a \, \mathbf{\hat{x}} + \frac{\sqrt3}2 \, a \, \mathbf{\hat{y}} \\
    \mathbf{a}_3 & = & c \, \mathbf{\hat{z}} \\

        \end{array}
      \end{equation*}
    }
    \renewcommand{\arraystretch}{1.0}
  \end{tabular}
  \begin{tabular}{c}
    \includegraphics[width=0.3\linewidth]{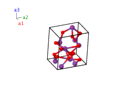} \\
  \end{tabular}
\end{tabular}

}
\vspace*{-0.25cm}

\noindent \hrulefill
\\
\textbf{Basis vectors:}
\vspace*{-0.25cm}
\renewcommand{\arraystretch}{1.5}
\begin{longtabu} to \textwidth{>{\centering $}X[-1,c,c]<{$}>{\centering $}X[-1,c,c]<{$}>{\centering $}X[-1,c,c]<{$}>{\centering $}X[-1,c,c]<{$}>{\centering $}X[-1,c,c]<{$}>{\centering $}X[-1,c,c]<{$}>{\centering $}X[-1,c,c]<{$}}
  & & \mbox{Lattice Coordinates} & & \mbox{Cartesian Coordinates} &\mbox{Wyckoff Position} & \mbox{Atom Type} \\  
  \mathbf{B}_{1} & = & \frac{1}{3} \, \mathbf{a}_{1} + \frac{2}{3} \, \mathbf{a}_{2} + z_{1} \, \mathbf{a}_{3} & = & \frac{1}{2}a \, \mathbf{\hat{x}} + \frac{1}{2\sqrt{3}}a \, \mathbf{\hat{y}} + z_{1}c \, \mathbf{\hat{z}} & \left(2b\right) & \mbox{Bi I} \\ 
\mathbf{B}_{2} & = & \frac{2}{3} \, \mathbf{a}_{1} + \frac{1}{3} \, \mathbf{a}_{2} + \left(\frac{1}{2} +z_{1}\right) \, \mathbf{a}_{3} & = & \frac{1}{2}a \, \mathbf{\hat{x}}- \frac{1}{2\sqrt{3}}a  \, \mathbf{\hat{y}} + \left(\frac{1}{2} +z_{1}\right)c \, \mathbf{\hat{z}} & \left(2b\right) & \mbox{Bi I} \\ 
\mathbf{B}_{3} & = & x_{2} \, \mathbf{a}_{1} + y_{2} \, \mathbf{a}_{2} + z_{2} \, \mathbf{a}_{3} & = & \frac{1}{2}\left(x_{2}+y_{2}\right)a \, \mathbf{\hat{x}} + \frac{\sqrt{3}}{2}\left(-x_{2}+y_{2}\right)a \, \mathbf{\hat{y}} + z_{2}c \, \mathbf{\hat{z}} & \left(6c\right) & \mbox{Bi II} \\ 
\mathbf{B}_{4} & = & -y_{2} \, \mathbf{a}_{1} + \left(x_{2}-y_{2}\right) \, \mathbf{a}_{2} + z_{2} \, \mathbf{a}_{3} & = & \left(\frac{1}{2}x_{2}-y_{2}\right)a \, \mathbf{\hat{x}} + \frac{\sqrt{3}}{2}x_{2}a \, \mathbf{\hat{y}} + z_{2}c \, \mathbf{\hat{z}} & \left(6c\right) & \mbox{Bi II} \\ 
\mathbf{B}_{5} & = & \left(-x_{2}+y_{2}\right) \, \mathbf{a}_{1}-x_{2} \, \mathbf{a}_{2} + z_{2} \, \mathbf{a}_{3} & = & \left(-x_{2}+\frac{1}{2}y_{2}\right)a \, \mathbf{\hat{x}}-\frac{\sqrt{3}}{2}y_{2}a \, \mathbf{\hat{y}} + z_{2}c \, \mathbf{\hat{z}} & \left(6c\right) & \mbox{Bi II} \\ 
\mathbf{B}_{6} & = & y_{2} \, \mathbf{a}_{1} + x_{2} \, \mathbf{a}_{2} + \left(\frac{1}{2} +z_{2}\right) \, \mathbf{a}_{3} & = & \frac{1}{2}\left(x_{2}+y_{2}\right)a \, \mathbf{\hat{x}} + \frac{\sqrt{3}}{2}\left(x_{2}-y_{2}\right)a \, \mathbf{\hat{y}} + \left(\frac{1}{2} +z_{2}\right)c \, \mathbf{\hat{z}} & \left(6c\right) & \mbox{Bi II} \\ 
\mathbf{B}_{7} & = & \left(x_{2}-y_{2}\right) \, \mathbf{a}_{1}-y_{2} \, \mathbf{a}_{2} + \left(\frac{1}{2} +z_{2}\right) \, \mathbf{a}_{3} & = & \left(\frac{1}{2}x_{2}-y_{2}\right)a \, \mathbf{\hat{x}}-\frac{\sqrt{3}}{2}x_{2}a \, \mathbf{\hat{y}} + \left(\frac{1}{2} +z_{2}\right)c \, \mathbf{\hat{z}} & \left(6c\right) & \mbox{Bi II} \\ 
\mathbf{B}_{8} & = & -x_{2} \, \mathbf{a}_{1} + \left(-x_{2}+y_{2}\right) \, \mathbf{a}_{2} + \left(\frac{1}{2} +z_{2}\right) \, \mathbf{a}_{3} & = & \left(-x_{2}+\frac{1}{2}y_{2}\right)a \, \mathbf{\hat{x}} + \frac{\sqrt{3}}{2}y_{2}a \, \mathbf{\hat{y}} + \left(\frac{1}{2} +z_{2}\right)c \, \mathbf{\hat{z}} & \left(6c\right) & \mbox{Bi II} \\ 
\mathbf{B}_{9} & = & x_{3} \, \mathbf{a}_{1} + y_{3} \, \mathbf{a}_{2} + z_{3} \, \mathbf{a}_{3} & = & \frac{1}{2}\left(x_{3}+y_{3}\right)a \, \mathbf{\hat{x}} + \frac{\sqrt{3}}{2}\left(-x_{3}+y_{3}\right)a \, \mathbf{\hat{y}} + z_{3}c \, \mathbf{\hat{z}} & \left(6c\right) & \mbox{O I} \\ 
\mathbf{B}_{10} & = & -y_{3} \, \mathbf{a}_{1} + \left(x_{3}-y_{3}\right) \, \mathbf{a}_{2} + z_{3} \, \mathbf{a}_{3} & = & \left(\frac{1}{2}x_{3}-y_{3}\right)a \, \mathbf{\hat{x}} + \frac{\sqrt{3}}{2}x_{3}a \, \mathbf{\hat{y}} + z_{3}c \, \mathbf{\hat{z}} & \left(6c\right) & \mbox{O I} \\ 
\mathbf{B}_{11} & = & \left(-x_{3}+y_{3}\right) \, \mathbf{a}_{1}-x_{3} \, \mathbf{a}_{2} + z_{3} \, \mathbf{a}_{3} & = & \left(-x_{3}+\frac{1}{2}y_{3}\right)a \, \mathbf{\hat{x}}-\frac{\sqrt{3}}{2}y_{3}a \, \mathbf{\hat{y}} + z_{3}c \, \mathbf{\hat{z}} & \left(6c\right) & \mbox{O I} \\ 
\mathbf{B}_{12} & = & y_{3} \, \mathbf{a}_{1} + x_{3} \, \mathbf{a}_{2} + \left(\frac{1}{2} +z_{3}\right) \, \mathbf{a}_{3} & = & \frac{1}{2}\left(x_{3}+y_{3}\right)a \, \mathbf{\hat{x}} + \frac{\sqrt{3}}{2}\left(x_{3}-y_{3}\right)a \, \mathbf{\hat{y}} + \left(\frac{1}{2} +z_{3}\right)c \, \mathbf{\hat{z}} & \left(6c\right) & \mbox{O I} \\ 
\mathbf{B}_{13} & = & \left(x_{3}-y_{3}\right) \, \mathbf{a}_{1}-y_{3} \, \mathbf{a}_{2} + \left(\frac{1}{2} +z_{3}\right) \, \mathbf{a}_{3} & = & \left(\frac{1}{2}x_{3}-y_{3}\right)a \, \mathbf{\hat{x}}-\frac{\sqrt{3}}{2}x_{3}a \, \mathbf{\hat{y}} + \left(\frac{1}{2} +z_{3}\right)c \, \mathbf{\hat{z}} & \left(6c\right) & \mbox{O I} \\ 
\mathbf{B}_{14} & = & -x_{3} \, \mathbf{a}_{1} + \left(-x_{3}+y_{3}\right) \, \mathbf{a}_{2} + \left(\frac{1}{2} +z_{3}\right) \, \mathbf{a}_{3} & = & \left(-x_{3}+\frac{1}{2}y_{3}\right)a \, \mathbf{\hat{x}} + \frac{\sqrt{3}}{2}y_{3}a \, \mathbf{\hat{y}} + \left(\frac{1}{2} +z_{3}\right)c \, \mathbf{\hat{z}} & \left(6c\right) & \mbox{O I} \\ 
\mathbf{B}_{15} & = & x_{4} \, \mathbf{a}_{1} + y_{4} \, \mathbf{a}_{2} + z_{4} \, \mathbf{a}_{3} & = & \frac{1}{2}\left(x_{4}+y_{4}\right)a \, \mathbf{\hat{x}} + \frac{\sqrt{3}}{2}\left(-x_{4}+y_{4}\right)a \, \mathbf{\hat{y}} + z_{4}c \, \mathbf{\hat{z}} & \left(6c\right) & \mbox{O II} \\ 
\mathbf{B}_{16} & = & -y_{4} \, \mathbf{a}_{1} + \left(x_{4}-y_{4}\right) \, \mathbf{a}_{2} + z_{4} \, \mathbf{a}_{3} & = & \left(\frac{1}{2}x_{4}-y_{4}\right)a \, \mathbf{\hat{x}} + \frac{\sqrt{3}}{2}x_{4}a \, \mathbf{\hat{y}} + z_{4}c \, \mathbf{\hat{z}} & \left(6c\right) & \mbox{O II} \\ 
\mathbf{B}_{17} & = & \left(-x_{4}+y_{4}\right) \, \mathbf{a}_{1}-x_{4} \, \mathbf{a}_{2} + z_{4} \, \mathbf{a}_{3} & = & \left(-x_{4}+\frac{1}{2}y_{4}\right)a \, \mathbf{\hat{x}}-\frac{\sqrt{3}}{2}y_{4}a \, \mathbf{\hat{y}} + z_{4}c \, \mathbf{\hat{z}} & \left(6c\right) & \mbox{O II} \\ 
\mathbf{B}_{18} & = & y_{4} \, \mathbf{a}_{1} + x_{4} \, \mathbf{a}_{2} + \left(\frac{1}{2} +z_{4}\right) \, \mathbf{a}_{3} & = & \frac{1}{2}\left(x_{4}+y_{4}\right)a \, \mathbf{\hat{x}} + \frac{\sqrt{3}}{2}\left(x_{4}-y_{4}\right)a \, \mathbf{\hat{y}} + \left(\frac{1}{2} +z_{4}\right)c \, \mathbf{\hat{z}} & \left(6c\right) & \mbox{O II} \\ 
\mathbf{B}_{19} & = & \left(x_{4}-y_{4}\right) \, \mathbf{a}_{1}-y_{4} \, \mathbf{a}_{2} + \left(\frac{1}{2} +z_{4}\right) \, \mathbf{a}_{3} & = & \left(\frac{1}{2}x_{4}-y_{4}\right)a \, \mathbf{\hat{x}}-\frac{\sqrt{3}}{2}x_{4}a \, \mathbf{\hat{y}} + \left(\frac{1}{2} +z_{4}\right)c \, \mathbf{\hat{z}} & \left(6c\right) & \mbox{O II} \\ 
\mathbf{B}_{20} & = & -x_{4} \, \mathbf{a}_{1} + \left(-x_{4}+y_{4}\right) \, \mathbf{a}_{2} + \left(\frac{1}{2} +z_{4}\right) \, \mathbf{a}_{3} & = & \left(-x_{4}+\frac{1}{2}y_{4}\right)a \, \mathbf{\hat{x}} + \frac{\sqrt{3}}{2}y_{4}a \, \mathbf{\hat{y}} + \left(\frac{1}{2} +z_{4}\right)c \, \mathbf{\hat{z}} & \left(6c\right) & \mbox{O II} \\ 
\end{longtabu}
\renewcommand{\arraystretch}{1.0}
\noindent \hrulefill
\\
\textbf{References:}
\vspace*{-0.25cm}
\begin{flushleft}
  - \bibentry{Locherer_Bi2O3_PhysRevB_2011}. \\
\end{flushleft}
\textbf{Found in:}
\vspace*{-0.25cm}
\begin{flushleft}
  - \bibentry{Villars_PearsonsCrystalData_2013}. \\
\end{flushleft}
\noindent \hrulefill
\\
\textbf{Geometry files:}
\\
\noindent  - CIF: pp. {\hyperref[A2B3_hP20_159_bc_2c_cif]{\pageref{A2B3_hP20_159_bc_2c_cif}}} \\
\noindent  - POSCAR: pp. {\hyperref[A2B3_hP20_159_bc_2c_poscar]{\pageref{A2B3_hP20_159_bc_2c_poscar}}} \\
\onecolumn
{\phantomsection\label{A4B3_hP28_159_ab2c_2c}}
\subsection*{\huge \textbf{{\normalfont Nierite ($\alpha$-Si$_{3}$N$_{4}$) Structure: A4B3\_hP28\_159\_ab2c\_2c}}}
\noindent \hrulefill
\vspace*{0.25cm}
\begin{figure}[htp]
  \centering
  \vspace{-1em}
  {\includegraphics[width=1\textwidth]{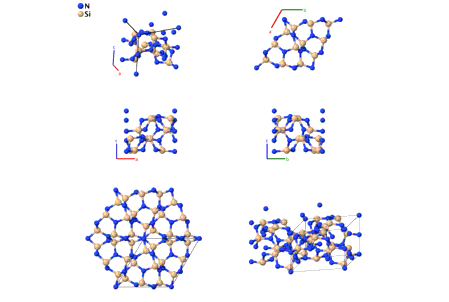}}
\end{figure}
\vspace*{-0.5cm}
\renewcommand{\arraystretch}{1.5}
\begin{equation*}
  \begin{array}{>{$\hspace{-0.15cm}}l<{$}>{$}p{0.5cm}<{$}>{$}p{18.5cm}<{$}}
    \mbox{\large \textbf{Prototype}} &\colon & \ce{$\alpha$-Si3N4} \\
    \mbox{\large \textbf{\AFLOW\ prototype label}} &\colon & \mbox{A4B3\_hP28\_159\_ab2c\_2c} \\
    \mbox{\large \textbf{\textit{Strukturbericht} designation}} &\colon & \mbox{None} \\
    \mbox{\large \textbf{Pearson symbol}} &\colon & \mbox{hP28} \\
    \mbox{\large \textbf{Space group number}} &\colon & 159 \\
    \mbox{\large \textbf{Space group symbol}} &\colon & P31c \\
    \mbox{\large \textbf{\AFLOW\ prototype command}} &\colon &  \texttt{aflow} \,  \, \texttt{-{}-proto=A4B3\_hP28\_159\_ab2c\_2c } \, \newline \texttt{-{}-params=}{a,c/a,z_{1},z_{2},x_{3},y_{3},z_{3},x_{4},y_{4},z_{4},x_{5},y_{5},z_{5},x_{6},y_{6},z_{6} }
  \end{array}
\end{equation*}
\renewcommand{\arraystretch}{1.0}

\noindent \parbox{1 \linewidth}{
\noindent \hrulefill
\\
\textbf{Trigonal Hexagonal primitive vectors:} \\
\vspace*{-0.25cm}
\begin{tabular}{cc}
  \begin{tabular}{c}
    \parbox{0.6 \linewidth}{
      \renewcommand{\arraystretch}{1.5}
      \begin{equation*}
        \centering
        \begin{array}{ccc}
              \mathbf{a}_1 & = & \frac12 \, a \, \mathbf{\hat{x}} - \frac{\sqrt3}2 \, a \, \mathbf{\hat{y}} \\
    \mathbf{a}_2 & = & \frac12 \, a \, \mathbf{\hat{x}} + \frac{\sqrt3}2 \, a \, \mathbf{\hat{y}} \\
    \mathbf{a}_3 & = & c \, \mathbf{\hat{z}} \\

        \end{array}
      \end{equation*}
    }
    \renewcommand{\arraystretch}{1.0}
  \end{tabular}
  \begin{tabular}{c}
    \includegraphics[width=0.3\linewidth]{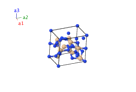} \\
  \end{tabular}
\end{tabular}

}
\vspace*{-0.25cm}

\noindent \hrulefill
\\
\textbf{Basis vectors:}
\vspace*{-0.25cm}
\renewcommand{\arraystretch}{1.5}
\begin{longtabu} to \textwidth{>{\centering $}X[-1,c,c]<{$}>{\centering $}X[-1,c,c]<{$}>{\centering $}X[-1,c,c]<{$}>{\centering $}X[-1,c,c]<{$}>{\centering $}X[-1,c,c]<{$}>{\centering $}X[-1,c,c]<{$}>{\centering $}X[-1,c,c]<{$}}
  & & \mbox{Lattice Coordinates} & & \mbox{Cartesian Coordinates} &\mbox{Wyckoff Position} & \mbox{Atom Type} \\  
  \mathbf{B}_{1} & = & z_{1} \, \mathbf{a}_{3} & = & z_{1}c \, \mathbf{\hat{z}} & \left(2a\right) & \mbox{N I} \\ 
\mathbf{B}_{2} & = & \left(\frac{1}{2} +z_{1}\right) \, \mathbf{a}_{3} & = & \left(\frac{1}{2} +z_{1}\right)c \, \mathbf{\hat{z}} & \left(2a\right) & \mbox{N I} \\ 
\mathbf{B}_{3} & = & \frac{1}{3} \, \mathbf{a}_{1} + \frac{2}{3} \, \mathbf{a}_{2} + z_{2} \, \mathbf{a}_{3} & = & \frac{1}{2}a \, \mathbf{\hat{x}} + \frac{1}{2\sqrt{3}}a \, \mathbf{\hat{y}} + z_{2}c \, \mathbf{\hat{z}} & \left(2b\right) & \mbox{N II} \\ 
\mathbf{B}_{4} & = & \frac{2}{3} \, \mathbf{a}_{1} + \frac{1}{3} \, \mathbf{a}_{2} + \left(\frac{1}{2} +z_{2}\right) \, \mathbf{a}_{3} & = & \frac{1}{2}a \, \mathbf{\hat{x}}- \frac{1}{2\sqrt{3}}a  \, \mathbf{\hat{y}} + \left(\frac{1}{2} +z_{2}\right)c \, \mathbf{\hat{z}} & \left(2b\right) & \mbox{N II} \\ 
\mathbf{B}_{5} & = & x_{3} \, \mathbf{a}_{1} + y_{3} \, \mathbf{a}_{2} + z_{3} \, \mathbf{a}_{3} & = & \frac{1}{2}\left(x_{3}+y_{3}\right)a \, \mathbf{\hat{x}} + \frac{\sqrt{3}}{2}\left(-x_{3}+y_{3}\right)a \, \mathbf{\hat{y}} + z_{3}c \, \mathbf{\hat{z}} & \left(6c\right) & \mbox{N III} \\ 
\mathbf{B}_{6} & = & -y_{3} \, \mathbf{a}_{1} + \left(x_{3}-y_{3}\right) \, \mathbf{a}_{2} + z_{3} \, \mathbf{a}_{3} & = & \left(\frac{1}{2}x_{3}-y_{3}\right)a \, \mathbf{\hat{x}} + \frac{\sqrt{3}}{2}x_{3}a \, \mathbf{\hat{y}} + z_{3}c \, \mathbf{\hat{z}} & \left(6c\right) & \mbox{N III} \\ 
\mathbf{B}_{7} & = & \left(-x_{3}+y_{3}\right) \, \mathbf{a}_{1}-x_{3} \, \mathbf{a}_{2} + z_{3} \, \mathbf{a}_{3} & = & \left(-x_{3}+\frac{1}{2}y_{3}\right)a \, \mathbf{\hat{x}}-\frac{\sqrt{3}}{2}y_{3}a \, \mathbf{\hat{y}} + z_{3}c \, \mathbf{\hat{z}} & \left(6c\right) & \mbox{N III} \\ 
\mathbf{B}_{8} & = & y_{3} \, \mathbf{a}_{1} + x_{3} \, \mathbf{a}_{2} + \left(\frac{1}{2} +z_{3}\right) \, \mathbf{a}_{3} & = & \frac{1}{2}\left(x_{3}+y_{3}\right)a \, \mathbf{\hat{x}} + \frac{\sqrt{3}}{2}\left(x_{3}-y_{3}\right)a \, \mathbf{\hat{y}} + \left(\frac{1}{2} +z_{3}\right)c \, \mathbf{\hat{z}} & \left(6c\right) & \mbox{N III} \\ 
\mathbf{B}_{9} & = & \left(x_{3}-y_{3}\right) \, \mathbf{a}_{1}-y_{3} \, \mathbf{a}_{2} + \left(\frac{1}{2} +z_{3}\right) \, \mathbf{a}_{3} & = & \left(\frac{1}{2}x_{3}-y_{3}\right)a \, \mathbf{\hat{x}}-\frac{\sqrt{3}}{2}x_{3}a \, \mathbf{\hat{y}} + \left(\frac{1}{2} +z_{3}\right)c \, \mathbf{\hat{z}} & \left(6c\right) & \mbox{N III} \\ 
\mathbf{B}_{10} & = & -x_{3} \, \mathbf{a}_{1} + \left(-x_{3}+y_{3}\right) \, \mathbf{a}_{2} + \left(\frac{1}{2} +z_{3}\right) \, \mathbf{a}_{3} & = & \left(-x_{3}+\frac{1}{2}y_{3}\right)a \, \mathbf{\hat{x}} + \frac{\sqrt{3}}{2}y_{3}a \, \mathbf{\hat{y}} + \left(\frac{1}{2} +z_{3}\right)c \, \mathbf{\hat{z}} & \left(6c\right) & \mbox{N III} \\ 
\mathbf{B}_{11} & = & x_{4} \, \mathbf{a}_{1} + y_{4} \, \mathbf{a}_{2} + z_{4} \, \mathbf{a}_{3} & = & \frac{1}{2}\left(x_{4}+y_{4}\right)a \, \mathbf{\hat{x}} + \frac{\sqrt{3}}{2}\left(-x_{4}+y_{4}\right)a \, \mathbf{\hat{y}} + z_{4}c \, \mathbf{\hat{z}} & \left(6c\right) & \mbox{N IV} \\ 
\mathbf{B}_{12} & = & -y_{4} \, \mathbf{a}_{1} + \left(x_{4}-y_{4}\right) \, \mathbf{a}_{2} + z_{4} \, \mathbf{a}_{3} & = & \left(\frac{1}{2}x_{4}-y_{4}\right)a \, \mathbf{\hat{x}} + \frac{\sqrt{3}}{2}x_{4}a \, \mathbf{\hat{y}} + z_{4}c \, \mathbf{\hat{z}} & \left(6c\right) & \mbox{N IV} \\ 
\mathbf{B}_{13} & = & \left(-x_{4}+y_{4}\right) \, \mathbf{a}_{1}-x_{4} \, \mathbf{a}_{2} + z_{4} \, \mathbf{a}_{3} & = & \left(-x_{4}+\frac{1}{2}y_{4}\right)a \, \mathbf{\hat{x}}-\frac{\sqrt{3}}{2}y_{4}a \, \mathbf{\hat{y}} + z_{4}c \, \mathbf{\hat{z}} & \left(6c\right) & \mbox{N IV} \\ 
\mathbf{B}_{14} & = & y_{4} \, \mathbf{a}_{1} + x_{4} \, \mathbf{a}_{2} + \left(\frac{1}{2} +z_{4}\right) \, \mathbf{a}_{3} & = & \frac{1}{2}\left(x_{4}+y_{4}\right)a \, \mathbf{\hat{x}} + \frac{\sqrt{3}}{2}\left(x_{4}-y_{4}\right)a \, \mathbf{\hat{y}} + \left(\frac{1}{2} +z_{4}\right)c \, \mathbf{\hat{z}} & \left(6c\right) & \mbox{N IV} \\ 
\mathbf{B}_{15} & = & \left(x_{4}-y_{4}\right) \, \mathbf{a}_{1}-y_{4} \, \mathbf{a}_{2} + \left(\frac{1}{2} +z_{4}\right) \, \mathbf{a}_{3} & = & \left(\frac{1}{2}x_{4}-y_{4}\right)a \, \mathbf{\hat{x}}-\frac{\sqrt{3}}{2}x_{4}a \, \mathbf{\hat{y}} + \left(\frac{1}{2} +z_{4}\right)c \, \mathbf{\hat{z}} & \left(6c\right) & \mbox{N IV} \\ 
\mathbf{B}_{16} & = & -x_{4} \, \mathbf{a}_{1} + \left(-x_{4}+y_{4}\right) \, \mathbf{a}_{2} + \left(\frac{1}{2} +z_{4}\right) \, \mathbf{a}_{3} & = & \left(-x_{4}+\frac{1}{2}y_{4}\right)a \, \mathbf{\hat{x}} + \frac{\sqrt{3}}{2}y_{4}a \, \mathbf{\hat{y}} + \left(\frac{1}{2} +z_{4}\right)c \, \mathbf{\hat{z}} & \left(6c\right) & \mbox{N IV} \\ 
\mathbf{B}_{17} & = & x_{5} \, \mathbf{a}_{1} + y_{5} \, \mathbf{a}_{2} + z_{5} \, \mathbf{a}_{3} & = & \frac{1}{2}\left(x_{5}+y_{5}\right)a \, \mathbf{\hat{x}} + \frac{\sqrt{3}}{2}\left(-x_{5}+y_{5}\right)a \, \mathbf{\hat{y}} + z_{5}c \, \mathbf{\hat{z}} & \left(6c\right) & \mbox{Si I} \\ 
\mathbf{B}_{18} & = & -y_{5} \, \mathbf{a}_{1} + \left(x_{5}-y_{5}\right) \, \mathbf{a}_{2} + z_{5} \, \mathbf{a}_{3} & = & \left(\frac{1}{2}x_{5}-y_{5}\right)a \, \mathbf{\hat{x}} + \frac{\sqrt{3}}{2}x_{5}a \, \mathbf{\hat{y}} + z_{5}c \, \mathbf{\hat{z}} & \left(6c\right) & \mbox{Si I} \\ 
\mathbf{B}_{19} & = & \left(-x_{5}+y_{5}\right) \, \mathbf{a}_{1}-x_{5} \, \mathbf{a}_{2} + z_{5} \, \mathbf{a}_{3} & = & \left(-x_{5}+\frac{1}{2}y_{5}\right)a \, \mathbf{\hat{x}}-\frac{\sqrt{3}}{2}y_{5}a \, \mathbf{\hat{y}} + z_{5}c \, \mathbf{\hat{z}} & \left(6c\right) & \mbox{Si I} \\ 
\mathbf{B}_{20} & = & y_{5} \, \mathbf{a}_{1} + x_{5} \, \mathbf{a}_{2} + \left(\frac{1}{2} +z_{5}\right) \, \mathbf{a}_{3} & = & \frac{1}{2}\left(x_{5}+y_{5}\right)a \, \mathbf{\hat{x}} + \frac{\sqrt{3}}{2}\left(x_{5}-y_{5}\right)a \, \mathbf{\hat{y}} + \left(\frac{1}{2} +z_{5}\right)c \, \mathbf{\hat{z}} & \left(6c\right) & \mbox{Si I} \\ 
\mathbf{B}_{21} & = & \left(x_{5}-y_{5}\right) \, \mathbf{a}_{1}-y_{5} \, \mathbf{a}_{2} + \left(\frac{1}{2} +z_{5}\right) \, \mathbf{a}_{3} & = & \left(\frac{1}{2}x_{5}-y_{5}\right)a \, \mathbf{\hat{x}}-\frac{\sqrt{3}}{2}x_{5}a \, \mathbf{\hat{y}} + \left(\frac{1}{2} +z_{5}\right)c \, \mathbf{\hat{z}} & \left(6c\right) & \mbox{Si I} \\ 
\mathbf{B}_{22} & = & -x_{5} \, \mathbf{a}_{1} + \left(-x_{5}+y_{5}\right) \, \mathbf{a}_{2} + \left(\frac{1}{2} +z_{5}\right) \, \mathbf{a}_{3} & = & \left(-x_{5}+\frac{1}{2}y_{5}\right)a \, \mathbf{\hat{x}} + \frac{\sqrt{3}}{2}y_{5}a \, \mathbf{\hat{y}} + \left(\frac{1}{2} +z_{5}\right)c \, \mathbf{\hat{z}} & \left(6c\right) & \mbox{Si I} \\ 
\mathbf{B}_{23} & = & x_{6} \, \mathbf{a}_{1} + y_{6} \, \mathbf{a}_{2} + z_{6} \, \mathbf{a}_{3} & = & \frac{1}{2}\left(x_{6}+y_{6}\right)a \, \mathbf{\hat{x}} + \frac{\sqrt{3}}{2}\left(-x_{6}+y_{6}\right)a \, \mathbf{\hat{y}} + z_{6}c \, \mathbf{\hat{z}} & \left(6c\right) & \mbox{Si II} \\ 
\mathbf{B}_{24} & = & -y_{6} \, \mathbf{a}_{1} + \left(x_{6}-y_{6}\right) \, \mathbf{a}_{2} + z_{6} \, \mathbf{a}_{3} & = & \left(\frac{1}{2}x_{6}-y_{6}\right)a \, \mathbf{\hat{x}} + \frac{\sqrt{3}}{2}x_{6}a \, \mathbf{\hat{y}} + z_{6}c \, \mathbf{\hat{z}} & \left(6c\right) & \mbox{Si II} \\ 
\mathbf{B}_{25} & = & \left(-x_{6}+y_{6}\right) \, \mathbf{a}_{1}-x_{6} \, \mathbf{a}_{2} + z_{6} \, \mathbf{a}_{3} & = & \left(-x_{6}+\frac{1}{2}y_{6}\right)a \, \mathbf{\hat{x}}-\frac{\sqrt{3}}{2}y_{6}a \, \mathbf{\hat{y}} + z_{6}c \, \mathbf{\hat{z}} & \left(6c\right) & \mbox{Si II} \\ 
\mathbf{B}_{26} & = & y_{6} \, \mathbf{a}_{1} + x_{6} \, \mathbf{a}_{2} + \left(\frac{1}{2} +z_{6}\right) \, \mathbf{a}_{3} & = & \frac{1}{2}\left(x_{6}+y_{6}\right)a \, \mathbf{\hat{x}} + \frac{\sqrt{3}}{2}\left(x_{6}-y_{6}\right)a \, \mathbf{\hat{y}} + \left(\frac{1}{2} +z_{6}\right)c \, \mathbf{\hat{z}} & \left(6c\right) & \mbox{Si II} \\ 
\mathbf{B}_{27} & = & \left(x_{6}-y_{6}\right) \, \mathbf{a}_{1}-y_{6} \, \mathbf{a}_{2} + \left(\frac{1}{2} +z_{6}\right) \, \mathbf{a}_{3} & = & \left(\frac{1}{2}x_{6}-y_{6}\right)a \, \mathbf{\hat{x}}-\frac{\sqrt{3}}{2}x_{6}a \, \mathbf{\hat{y}} + \left(\frac{1}{2} +z_{6}\right)c \, \mathbf{\hat{z}} & \left(6c\right) & \mbox{Si II} \\ 
\mathbf{B}_{28} & = & -x_{6} \, \mathbf{a}_{1} + \left(-x_{6}+y_{6}\right) \, \mathbf{a}_{2} + \left(\frac{1}{2} +z_{6}\right) \, \mathbf{a}_{3} & = & \left(-x_{6}+\frac{1}{2}y_{6}\right)a \, \mathbf{\hat{x}} + \frac{\sqrt{3}}{2}y_{6}a \, \mathbf{\hat{y}} + \left(\frac{1}{2} +z_{6}\right)c \, \mathbf{\hat{z}} & \left(6c\right) & \mbox{Si II} \\ 
\end{longtabu}
\renewcommand{\arraystretch}{1.0}
\noindent \hrulefill
\\
\textbf{References:}
\vspace*{-0.25cm}
\begin{flushleft}
  - \bibentry{Hardie_Si3N4_Nature_1957}. \\
\end{flushleft}
\textbf{Found in:}
\vspace*{-0.25cm}
\begin{flushleft}
  - \bibentry{Villars_PearsonsCrystalData_2013}. \\
\end{flushleft}
\noindent \hrulefill
\\
\textbf{Geometry files:}
\\
\noindent  - CIF: pp. {\hyperref[A4B3_hP28_159_ab2c_2c_cif]{\pageref{A4B3_hP28_159_ab2c_2c_cif}}} \\
\noindent  - POSCAR: pp. {\hyperref[A4B3_hP28_159_ab2c_2c_poscar]{\pageref{A4B3_hP28_159_ab2c_2c_poscar}}} \\
\onecolumn
{\phantomsection\label{AB4C7D_hP26_159_b_ac_a2c_b}}
\subsection*{\huge \textbf{{\normalfont YbBaCo$_{4}$O$_{7}$ Structure: AB4C7D\_hP26\_159\_b\_ac\_a2c\_b}}}
\noindent \hrulefill
\vspace*{0.25cm}
\begin{figure}[htp]
  \centering
  \vspace{-1em}
  {\includegraphics[width=1\textwidth]{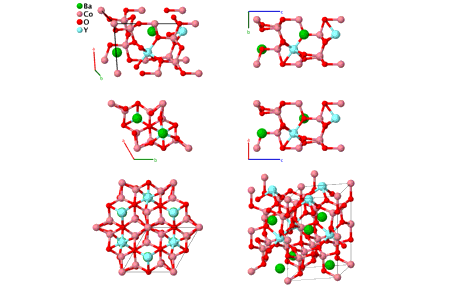}}
\end{figure}
\vspace*{-0.5cm}
\renewcommand{\arraystretch}{1.5}
\begin{equation*}
  \begin{array}{>{$\hspace{-0.15cm}}l<{$}>{$}p{0.5cm}<{$}>{$}p{18.5cm}<{$}}
    \mbox{\large \textbf{Prototype}} &\colon & \ce{YbBaCo4O7} \\
    \mbox{\large \textbf{\AFLOW\ prototype label}} &\colon & \mbox{AB4C7D\_hP26\_159\_b\_ac\_a2c\_b} \\
    \mbox{\large \textbf{\textit{Strukturbericht} designation}} &\colon & \mbox{None} \\
    \mbox{\large \textbf{Pearson symbol}} &\colon & \mbox{hP26} \\
    \mbox{\large \textbf{Space group number}} &\colon & 159 \\
    \mbox{\large \textbf{Space group symbol}} &\colon & P31c \\
    \mbox{\large \textbf{\AFLOW\ prototype command}} &\colon &  \texttt{aflow} \,  \, \texttt{-{}-proto=AB4C7D\_hP26\_159\_b\_ac\_a2c\_b } \, \newline \texttt{-{}-params=}{a,c/a,z_{1},z_{2},z_{3},z_{4},x_{5},y_{5},z_{5},x_{6},y_{6},z_{6},x_{7},y_{7},z_{7} }
  \end{array}
\end{equation*}
\renewcommand{\arraystretch}{1.0}

\noindent \parbox{1 \linewidth}{
\noindent \hrulefill
\\
\textbf{Trigonal Hexagonal primitive vectors:} \\
\vspace*{-0.25cm}
\begin{tabular}{cc}
  \begin{tabular}{c}
    \parbox{0.6 \linewidth}{
      \renewcommand{\arraystretch}{1.5}
      \begin{equation*}
        \centering
        \begin{array}{ccc}
              \mathbf{a}_1 & = & \frac12 \, a \, \mathbf{\hat{x}} - \frac{\sqrt3}2 \, a \, \mathbf{\hat{y}} \\
    \mathbf{a}_2 & = & \frac12 \, a \, \mathbf{\hat{x}} + \frac{\sqrt3}2 \, a \, \mathbf{\hat{y}} \\
    \mathbf{a}_3 & = & c \, \mathbf{\hat{z}} \\

        \end{array}
      \end{equation*}
    }
    \renewcommand{\arraystretch}{1.0}
  \end{tabular}
  \begin{tabular}{c}
    \includegraphics[width=0.3\linewidth]{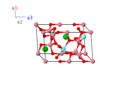} \\
  \end{tabular}
\end{tabular}

}
\vspace*{-0.25cm}

\noindent \hrulefill
\\
\textbf{Basis vectors:}
\vspace*{-0.25cm}
\renewcommand{\arraystretch}{1.5}
\begin{longtabu} to \textwidth{>{\centering $}X[-1,c,c]<{$}>{\centering $}X[-1,c,c]<{$}>{\centering $}X[-1,c,c]<{$}>{\centering $}X[-1,c,c]<{$}>{\centering $}X[-1,c,c]<{$}>{\centering $}X[-1,c,c]<{$}>{\centering $}X[-1,c,c]<{$}}
  & & \mbox{Lattice Coordinates} & & \mbox{Cartesian Coordinates} &\mbox{Wyckoff Position} & \mbox{Atom Type} \\  
  \mathbf{B}_{1} & = & z_{1} \, \mathbf{a}_{3} & = & z_{1}c \, \mathbf{\hat{z}} & \left(2a\right) & \mbox{Co I} \\ 
\mathbf{B}_{2} & = & \left(\frac{1}{2} +z_{1}\right) \, \mathbf{a}_{3} & = & \left(\frac{1}{2} +z_{1}\right)c \, \mathbf{\hat{z}} & \left(2a\right) & \mbox{Co I} \\ 
\mathbf{B}_{3} & = & z_{2} \, \mathbf{a}_{3} & = & z_{2}c \, \mathbf{\hat{z}} & \left(2a\right) & \mbox{O I} \\ 
\mathbf{B}_{4} & = & \left(\frac{1}{2} +z_{2}\right) \, \mathbf{a}_{3} & = & \left(\frac{1}{2} +z_{2}\right)c \, \mathbf{\hat{z}} & \left(2a\right) & \mbox{O I} \\ 
\mathbf{B}_{5} & = & \frac{1}{3} \, \mathbf{a}_{1} + \frac{2}{3} \, \mathbf{a}_{2} + z_{3} \, \mathbf{a}_{3} & = & \frac{1}{2}a \, \mathbf{\hat{x}} + \frac{1}{2\sqrt{3}}a \, \mathbf{\hat{y}} + z_{3}c \, \mathbf{\hat{z}} & \left(2b\right) & \mbox{Ba} \\ 
\mathbf{B}_{6} & = & \frac{2}{3} \, \mathbf{a}_{1} + \frac{1}{3} \, \mathbf{a}_{2} + \left(\frac{1}{2} +z_{3}\right) \, \mathbf{a}_{3} & = & \frac{1}{2}a \, \mathbf{\hat{x}}- \frac{1}{2\sqrt{3}}a  \, \mathbf{\hat{y}} + \left(\frac{1}{2} +z_{3}\right)c \, \mathbf{\hat{z}} & \left(2b\right) & \mbox{Ba} \\ 
\mathbf{B}_{7} & = & \frac{1}{3} \, \mathbf{a}_{1} + \frac{2}{3} \, \mathbf{a}_{2} + z_{4} \, \mathbf{a}_{3} & = & \frac{1}{2}a \, \mathbf{\hat{x}} + \frac{1}{2\sqrt{3}}a \, \mathbf{\hat{y}} + z_{4}c \, \mathbf{\hat{z}} & \left(2b\right) & \mbox{Y} \\ 
\mathbf{B}_{8} & = & \frac{2}{3} \, \mathbf{a}_{1} + \frac{1}{3} \, \mathbf{a}_{2} + \left(\frac{1}{2} +z_{4}\right) \, \mathbf{a}_{3} & = & \frac{1}{2}a \, \mathbf{\hat{x}}- \frac{1}{2\sqrt{3}}a  \, \mathbf{\hat{y}} + \left(\frac{1}{2} +z_{4}\right)c \, \mathbf{\hat{z}} & \left(2b\right) & \mbox{Y} \\ 
\mathbf{B}_{9} & = & x_{5} \, \mathbf{a}_{1} + y_{5} \, \mathbf{a}_{2} + z_{5} \, \mathbf{a}_{3} & = & \frac{1}{2}\left(x_{5}+y_{5}\right)a \, \mathbf{\hat{x}} + \frac{\sqrt{3}}{2}\left(-x_{5}+y_{5}\right)a \, \mathbf{\hat{y}} + z_{5}c \, \mathbf{\hat{z}} & \left(6c\right) & \mbox{Co II} \\ 
\mathbf{B}_{10} & = & -y_{5} \, \mathbf{a}_{1} + \left(x_{5}-y_{5}\right) \, \mathbf{a}_{2} + z_{5} \, \mathbf{a}_{3} & = & \left(\frac{1}{2}x_{5}-y_{5}\right)a \, \mathbf{\hat{x}} + \frac{\sqrt{3}}{2}x_{5}a \, \mathbf{\hat{y}} + z_{5}c \, \mathbf{\hat{z}} & \left(6c\right) & \mbox{Co II} \\ 
\mathbf{B}_{11} & = & \left(-x_{5}+y_{5}\right) \, \mathbf{a}_{1}-x_{5} \, \mathbf{a}_{2} + z_{5} \, \mathbf{a}_{3} & = & \left(-x_{5}+\frac{1}{2}y_{5}\right)a \, \mathbf{\hat{x}}-\frac{\sqrt{3}}{2}y_{5}a \, \mathbf{\hat{y}} + z_{5}c \, \mathbf{\hat{z}} & \left(6c\right) & \mbox{Co II} \\ 
\mathbf{B}_{12} & = & y_{5} \, \mathbf{a}_{1} + x_{5} \, \mathbf{a}_{2} + \left(\frac{1}{2} +z_{5}\right) \, \mathbf{a}_{3} & = & \frac{1}{2}\left(x_{5}+y_{5}\right)a \, \mathbf{\hat{x}} + \frac{\sqrt{3}}{2}\left(x_{5}-y_{5}\right)a \, \mathbf{\hat{y}} + \left(\frac{1}{2} +z_{5}\right)c \, \mathbf{\hat{z}} & \left(6c\right) & \mbox{Co II} \\ 
\mathbf{B}_{13} & = & \left(x_{5}-y_{5}\right) \, \mathbf{a}_{1}-y_{5} \, \mathbf{a}_{2} + \left(\frac{1}{2} +z_{5}\right) \, \mathbf{a}_{3} & = & \left(\frac{1}{2}x_{5}-y_{5}\right)a \, \mathbf{\hat{x}}-\frac{\sqrt{3}}{2}x_{5}a \, \mathbf{\hat{y}} + \left(\frac{1}{2} +z_{5}\right)c \, \mathbf{\hat{z}} & \left(6c\right) & \mbox{Co II} \\ 
\mathbf{B}_{14} & = & -x_{5} \, \mathbf{a}_{1} + \left(-x_{5}+y_{5}\right) \, \mathbf{a}_{2} + \left(\frac{1}{2} +z_{5}\right) \, \mathbf{a}_{3} & = & \left(-x_{5}+\frac{1}{2}y_{5}\right)a \, \mathbf{\hat{x}} + \frac{\sqrt{3}}{2}y_{5}a \, \mathbf{\hat{y}} + \left(\frac{1}{2} +z_{5}\right)c \, \mathbf{\hat{z}} & \left(6c\right) & \mbox{Co II} \\ 
\mathbf{B}_{15} & = & x_{6} \, \mathbf{a}_{1} + y_{6} \, \mathbf{a}_{2} + z_{6} \, \mathbf{a}_{3} & = & \frac{1}{2}\left(x_{6}+y_{6}\right)a \, \mathbf{\hat{x}} + \frac{\sqrt{3}}{2}\left(-x_{6}+y_{6}\right)a \, \mathbf{\hat{y}} + z_{6}c \, \mathbf{\hat{z}} & \left(6c\right) & \mbox{O II} \\ 
\mathbf{B}_{16} & = & -y_{6} \, \mathbf{a}_{1} + \left(x_{6}-y_{6}\right) \, \mathbf{a}_{2} + z_{6} \, \mathbf{a}_{3} & = & \left(\frac{1}{2}x_{6}-y_{6}\right)a \, \mathbf{\hat{x}} + \frac{\sqrt{3}}{2}x_{6}a \, \mathbf{\hat{y}} + z_{6}c \, \mathbf{\hat{z}} & \left(6c\right) & \mbox{O II} \\ 
\mathbf{B}_{17} & = & \left(-x_{6}+y_{6}\right) \, \mathbf{a}_{1}-x_{6} \, \mathbf{a}_{2} + z_{6} \, \mathbf{a}_{3} & = & \left(-x_{6}+\frac{1}{2}y_{6}\right)a \, \mathbf{\hat{x}}-\frac{\sqrt{3}}{2}y_{6}a \, \mathbf{\hat{y}} + z_{6}c \, \mathbf{\hat{z}} & \left(6c\right) & \mbox{O II} \\ 
\mathbf{B}_{18} & = & y_{6} \, \mathbf{a}_{1} + x_{6} \, \mathbf{a}_{2} + \left(\frac{1}{2} +z_{6}\right) \, \mathbf{a}_{3} & = & \frac{1}{2}\left(x_{6}+y_{6}\right)a \, \mathbf{\hat{x}} + \frac{\sqrt{3}}{2}\left(x_{6}-y_{6}\right)a \, \mathbf{\hat{y}} + \left(\frac{1}{2} +z_{6}\right)c \, \mathbf{\hat{z}} & \left(6c\right) & \mbox{O II} \\ 
\mathbf{B}_{19} & = & \left(x_{6}-y_{6}\right) \, \mathbf{a}_{1}-y_{6} \, \mathbf{a}_{2} + \left(\frac{1}{2} +z_{6}\right) \, \mathbf{a}_{3} & = & \left(\frac{1}{2}x_{6}-y_{6}\right)a \, \mathbf{\hat{x}}-\frac{\sqrt{3}}{2}x_{6}a \, \mathbf{\hat{y}} + \left(\frac{1}{2} +z_{6}\right)c \, \mathbf{\hat{z}} & \left(6c\right) & \mbox{O II} \\ 
\mathbf{B}_{20} & = & -x_{6} \, \mathbf{a}_{1} + \left(-x_{6}+y_{6}\right) \, \mathbf{a}_{2} + \left(\frac{1}{2} +z_{6}\right) \, \mathbf{a}_{3} & = & \left(-x_{6}+\frac{1}{2}y_{6}\right)a \, \mathbf{\hat{x}} + \frac{\sqrt{3}}{2}y_{6}a \, \mathbf{\hat{y}} + \left(\frac{1}{2} +z_{6}\right)c \, \mathbf{\hat{z}} & \left(6c\right) & \mbox{O II} \\ 
\mathbf{B}_{21} & = & x_{7} \, \mathbf{a}_{1} + y_{7} \, \mathbf{a}_{2} + z_{7} \, \mathbf{a}_{3} & = & \frac{1}{2}\left(x_{7}+y_{7}\right)a \, \mathbf{\hat{x}} + \frac{\sqrt{3}}{2}\left(-x_{7}+y_{7}\right)a \, \mathbf{\hat{y}} + z_{7}c \, \mathbf{\hat{z}} & \left(6c\right) & \mbox{O III} \\ 
\mathbf{B}_{22} & = & -y_{7} \, \mathbf{a}_{1} + \left(x_{7}-y_{7}\right) \, \mathbf{a}_{2} + z_{7} \, \mathbf{a}_{3} & = & \left(\frac{1}{2}x_{7}-y_{7}\right)a \, \mathbf{\hat{x}} + \frac{\sqrt{3}}{2}x_{7}a \, \mathbf{\hat{y}} + z_{7}c \, \mathbf{\hat{z}} & \left(6c\right) & \mbox{O III} \\ 
\mathbf{B}_{23} & = & \left(-x_{7}+y_{7}\right) \, \mathbf{a}_{1}-x_{7} \, \mathbf{a}_{2} + z_{7} \, \mathbf{a}_{3} & = & \left(-x_{7}+\frac{1}{2}y_{7}\right)a \, \mathbf{\hat{x}}-\frac{\sqrt{3}}{2}y_{7}a \, \mathbf{\hat{y}} + z_{7}c \, \mathbf{\hat{z}} & \left(6c\right) & \mbox{O III} \\ 
\mathbf{B}_{24} & = & y_{7} \, \mathbf{a}_{1} + x_{7} \, \mathbf{a}_{2} + \left(\frac{1}{2} +z_{7}\right) \, \mathbf{a}_{3} & = & \frac{1}{2}\left(x_{7}+y_{7}\right)a \, \mathbf{\hat{x}} + \frac{\sqrt{3}}{2}\left(x_{7}-y_{7}\right)a \, \mathbf{\hat{y}} + \left(\frac{1}{2} +z_{7}\right)c \, \mathbf{\hat{z}} & \left(6c\right) & \mbox{O III} \\ 
\mathbf{B}_{25} & = & \left(x_{7}-y_{7}\right) \, \mathbf{a}_{1}-y_{7} \, \mathbf{a}_{2} + \left(\frac{1}{2} +z_{7}\right) \, \mathbf{a}_{3} & = & \left(\frac{1}{2}x_{7}-y_{7}\right)a \, \mathbf{\hat{x}}-\frac{\sqrt{3}}{2}x_{7}a \, \mathbf{\hat{y}} + \left(\frac{1}{2} +z_{7}\right)c \, \mathbf{\hat{z}} & \left(6c\right) & \mbox{O III} \\ 
\mathbf{B}_{26} & = & -x_{7} \, \mathbf{a}_{1} + \left(-x_{7}+y_{7}\right) \, \mathbf{a}_{2} + \left(\frac{1}{2} +z_{7}\right) \, \mathbf{a}_{3} & = & \left(-x_{7}+\frac{1}{2}y_{7}\right)a \, \mathbf{\hat{x}} + \frac{\sqrt{3}}{2}y_{7}a \, \mathbf{\hat{y}} + \left(\frac{1}{2} +z_{7}\right)c \, \mathbf{\hat{z}} & \left(6c\right) & \mbox{O III} \\ 
\end{longtabu}
\renewcommand{\arraystretch}{1.0}
\noindent \hrulefill
\\
\textbf{References:}
\vspace*{-0.25cm}
\begin{flushleft}
  - \bibentry{Huq_BaYbCo4O7_JSolStateChem_2006}. \\
\end{flushleft}
\textbf{Found in:}
\vspace*{-0.25cm}
\begin{flushleft}
  - \bibentry{Villars_PearsonsCrystalData_2013}. \\
\end{flushleft}
\noindent \hrulefill
\\
\textbf{Geometry files:}
\\
\noindent  - CIF: pp. {\hyperref[AB4C7D_hP26_159_b_ac_a2c_b_cif]{\pageref{AB4C7D_hP26_159_b_ac_a2c_b_cif}}} \\
\noindent  - POSCAR: pp. {\hyperref[AB4C7D_hP26_159_b_ac_a2c_b_poscar]{\pageref{AB4C7D_hP26_159_b_ac_a2c_b_poscar}}} \\
\onecolumn
{\phantomsection\label{A3B_hR4_160_b_a}}
\subsection*{\huge \textbf{{\normalfont H$_{3}$S (130~GPa) Structure: A3B\_hR4\_160\_b\_a}}}
\noindent \hrulefill
\vspace*{0.25cm}
\begin{figure}[htp]
  \centering
  \vspace{-1em}
  {\includegraphics[width=1\textwidth]{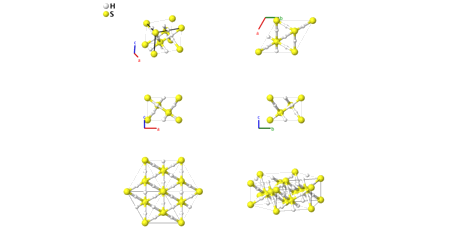}}
\end{figure}
\vspace*{-0.5cm}
\renewcommand{\arraystretch}{1.5}
\begin{equation*}
  \begin{array}{>{$\hspace{-0.15cm}}l<{$}>{$}p{0.5cm}<{$}>{$}p{18.5cm}<{$}}
    \mbox{\large \textbf{Prototype}} &\colon & \ce{H3S} \\
    \mbox{\large \textbf{\AFLOW\ prototype label}} &\colon & \mbox{A3B\_hR4\_160\_b\_a} \\
    \mbox{\large \textbf{\textit{Strukturbericht} designation}} &\colon & \mbox{None} \\
    \mbox{\large \textbf{Pearson symbol}} &\colon & \mbox{hR4} \\
    \mbox{\large \textbf{Space group number}} &\colon & 160 \\
    \mbox{\large \textbf{Space group symbol}} &\colon & R3m \\
    \mbox{\large \textbf{\AFLOW\ prototype command}} &\colon &  \texttt{aflow} \,  \, \texttt{-{}-proto=A3B\_hR4\_160\_b\_a [-{}-hex]} \, \newline \texttt{-{}-params=}{a,c/a,x_{1},x_{2},z_{2} }
  \end{array}
\end{equation*}
\renewcommand{\arraystretch}{1.0}

\vspace*{-0.25cm}
\noindent \hrulefill
\begin{itemize}
  \item{This structure was found by first-principles electronic structure
calculations and is predicted to be the stable structure of H$_{3}$S
for pressures between 90 and 150 ~GPa.
When $c/a \rightarrow \sqrt{8}$, $x_{2} \rightarrow 1/2$ and
$z_{2} \rightarrow 0$ this structure continuously evolves into
the cubic \href{http://aflow.org/CrystalDatabase/A3B_cI8_229_b_a.html}{200~GPa H$_{3}$S Structure}.
The data presented here was computed at 130~GPa.
}
\end{itemize}

\noindent \parbox{1 \linewidth}{
\noindent \hrulefill
\\
\textbf{Rhombohedral primitive vectors:} \\
\vspace*{-0.25cm}
\begin{tabular}{cc}
  \begin{tabular}{c}
    \parbox{0.6 \linewidth}{
      \renewcommand{\arraystretch}{1.5}
      \begin{equation*}
        \centering
        \begin{array}{ccc}
              \mathbf{a}_1 & = & ~ \frac12 \, a \, \mathbf{\hat{x}} - \frac{1}{2\sqrt{3}} \, a \, \mathbf{\hat{y}} + \frac13 \, c \, \mathbf{\hat{z}} \\
    \mathbf{a}_2 & = & \frac{1}{\sqrt{3}} \, a \, \mathbf{\hat{y}} + \frac13 \, c \, \mathbf{\hat{z}} \\
    \mathbf{a}_3 & = & - \frac12 \, a \, \mathbf{\hat{x}} - \frac{1}{2\sqrt{3}} \, a \, \mathbf{\hat{y}} + \frac13 \, c \, \mathbf{\hat{z}} \\

        \end{array}
      \end{equation*}
    }
    \renewcommand{\arraystretch}{1.0}
  \end{tabular}
  \begin{tabular}{c}
    \includegraphics[width=0.3\linewidth]{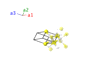} \\
  \end{tabular}
\end{tabular}

}
\vspace*{-0.25cm}

\noindent \hrulefill
\\
\textbf{Basis vectors:}
\vspace*{-0.25cm}
\renewcommand{\arraystretch}{1.5}
\begin{longtabu} to \textwidth{>{\centering $}X[-1,c,c]<{$}>{\centering $}X[-1,c,c]<{$}>{\centering $}X[-1,c,c]<{$}>{\centering $}X[-1,c,c]<{$}>{\centering $}X[-1,c,c]<{$}>{\centering $}X[-1,c,c]<{$}>{\centering $}X[-1,c,c]<{$}}
  & & \mbox{Lattice Coordinates} & & \mbox{Cartesian Coordinates} &\mbox{Wyckoff Position} & \mbox{Atom Type} \\  
  \mathbf{B}_{1} & = & x_{1} \, \mathbf{a}_{1} + x_{1} \, \mathbf{a}_{2} + x_{1} \, \mathbf{a}_{3} & = & x_{1}c \, \mathbf{\hat{z}} & \left(1a\right) & \mbox{S} \\ 
\mathbf{B}_{2} & = & x_{2} \, \mathbf{a}_{1} + x_{2} \, \mathbf{a}_{2} + z_{2} \, \mathbf{a}_{3} & = & \frac{1}{2}\left(x_{2}-z_{2}\right)a \, \mathbf{\hat{x}} + \frac{1}{2\sqrt{3}}\left(x_{2}-z_{2}\right)a \, \mathbf{\hat{y}} + \left(\frac{2}{3}x_{2}+\frac{1}{3}z_{2}\right)c \, \mathbf{\hat{z}} & \left(3b\right) & \mbox{H} \\ 
\mathbf{B}_{3} & = & z_{2} \, \mathbf{a}_{1} + x_{2} \, \mathbf{a}_{2} + x_{2} \, \mathbf{a}_{3} & = & \frac{1}{2}\left(-x_{2}+z_{2}\right)a \, \mathbf{\hat{x}} + \frac{1}{2\sqrt{3}}\left(x_{2}-z_{2}\right)a \, \mathbf{\hat{y}} + \left(\frac{2}{3}x_{2}+\frac{1}{3}z_{2}\right)c \, \mathbf{\hat{z}} & \left(3b\right) & \mbox{H} \\ 
\mathbf{B}_{4} & = & x_{2} \, \mathbf{a}_{1} + z_{2} \, \mathbf{a}_{2} + x_{2} \, \mathbf{a}_{3} & = & \frac{1}{\sqrt{3}}\left(-x_{2}+z_{2}\right)a \, \mathbf{\hat{y}} + \left(\frac{2}{3}x_{2}+\frac{1}{3}z_{2}\right)c \, \mathbf{\hat{z}} & \left(3b\right) & \mbox{H} \\ 
\end{longtabu}
\renewcommand{\arraystretch}{1.0}
\noindent \hrulefill
\\
\textbf{References:}
\vspace*{-0.25cm}
\begin{flushleft}
  - \bibentry{Duan_SR_4_2014}. \\
\end{flushleft}
\noindent \hrulefill
\\
\textbf{Geometry files:}
\\
\noindent  - CIF: pp. {\hyperref[A3B_hR4_160_b_a_cif]{\pageref{A3B_hR4_160_b_a_cif}}} \\
\noindent  - POSCAR: pp. {\hyperref[A3B_hR4_160_b_a_poscar]{\pageref{A3B_hR4_160_b_a_poscar}}} \\
\onecolumn
{\phantomsection\label{A8B5_hR26_160_a3bc_a3b}}
\subsection*{\huge \textbf{{\normalfont Al$_{8}$Cr$_{5}$ ($D8_{10}$) Structure: A8B5\_hR26\_160\_a3bc\_a3b}}}
\noindent \hrulefill
\vspace*{0.25cm}
\begin{figure}[htp]
  \centering
  \vspace{-1em}
  {\includegraphics[width=1\textwidth]{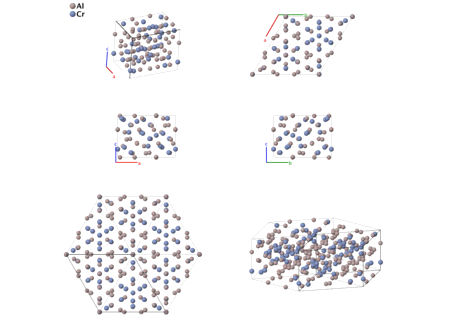}}
\end{figure}
\vspace*{-0.5cm}
\renewcommand{\arraystretch}{1.5}
\begin{equation*}
  \begin{array}{>{$\hspace{-0.15cm}}l<{$}>{$}p{0.5cm}<{$}>{$}p{18.5cm}<{$}}
    \mbox{\large \textbf{Prototype}} &\colon & \ce{Al$_{8}$Cr$_{5}$} \\
    \mbox{\large \textbf{\AFLOW\ prototype label}} &\colon & \mbox{A8B5\_hR26\_160\_a3bc\_a3b} \\
    \mbox{\large \textbf{\textit{Strukturbericht} designation}} &\colon & \mbox{$D8_{10}$} \\
    \mbox{\large \textbf{Pearson symbol}} &\colon & \mbox{hR26} \\
    \mbox{\large \textbf{Space group number}} &\colon & 160 \\
    \mbox{\large \textbf{Space group symbol}} &\colon & R3m \\
    \mbox{\large \textbf{\AFLOW\ prototype command}} &\colon &  \texttt{aflow} \,  \, \texttt{-{}-proto=A8B5\_hR26\_160\_a3bc\_a3b [-{}-hex]} \, \newline \texttt{-{}-params=}{a,c/a,x_{1},x_{2},x_{3},z_{3},x_{4},z_{4},x_{5},z_{5},x_{6},z_{6},x_{7},z_{7},x_{8},z_{8},x_{9},y_{9},z_{9} }
  \end{array}
\end{equation*}
\renewcommand{\arraystretch}{1.0}

\vspace*{-0.25cm}
\noindent \hrulefill
\begin{itemize}
  \item{(Bradley, 1937) notes that the positions here are very close to the
positions of the atoms in the
\href{http://aflow.org/CrystalDatabase/A5B8_cI52_217_ce_cg.html}{$\gamma$-brass (Cu$_{5}$Zn$_{8}$, $D8_{2}$)} structure.
}
\end{itemize}

\noindent \parbox{1 \linewidth}{
\noindent \hrulefill
\\
\textbf{Rhombohedral primitive vectors:} \\
\vspace*{-0.25cm}
\begin{tabular}{cc}
  \begin{tabular}{c}
    \parbox{0.6 \linewidth}{
      \renewcommand{\arraystretch}{1.5}
      \begin{equation*}
        \centering
        \begin{array}{ccc}
              \mathbf{a}_1 & = & ~ \frac12 \, a \, \mathbf{\hat{x}} - \frac{1}{2\sqrt{3}} \, a \, \mathbf{\hat{y}} + \frac13 \, c \, \mathbf{\hat{z}} \\
    \mathbf{a}_2 & = & \frac{1}{\sqrt{3}} \, a \, \mathbf{\hat{y}} + \frac13 \, c \, \mathbf{\hat{z}} \\
    \mathbf{a}_3 & = & - \frac12 \, a \, \mathbf{\hat{x}} - \frac{1}{2\sqrt{3}} \, a \, \mathbf{\hat{y}} + \frac13 \, c \, \mathbf{\hat{z}} \\

        \end{array}
      \end{equation*}
    }
    \renewcommand{\arraystretch}{1.0}
  \end{tabular}
  \begin{tabular}{c}
    \includegraphics[width=0.3\linewidth]{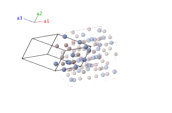} \\
  \end{tabular}
\end{tabular}

}
\vspace*{-0.25cm}

\noindent \hrulefill
\\
\textbf{Basis vectors:}
\vspace*{-0.25cm}
\renewcommand{\arraystretch}{1.5}
\begin{longtabu} to \textwidth{>{\centering $}X[-1,c,c]<{$}>{\centering $}X[-1,c,c]<{$}>{\centering $}X[-1,c,c]<{$}>{\centering $}X[-1,c,c]<{$}>{\centering $}X[-1,c,c]<{$}>{\centering $}X[-1,c,c]<{$}>{\centering $}X[-1,c,c]<{$}}
  & & \mbox{Lattice Coordinates} & & \mbox{Cartesian Coordinates} &\mbox{Wyckoff Position} & \mbox{Atom Type} \\  
  \mathbf{B}_{1} & = & x_{1} \, \mathbf{a}_{1} + x_{1} \, \mathbf{a}_{2} + x_{1} \, \mathbf{a}_{3} & = & x_{1}c \, \mathbf{\hat{z}} & \left(1a\right) & \mbox{Al I} \\ 
\mathbf{B}_{2} & = & x_{2} \, \mathbf{a}_{1} + x_{2} \, \mathbf{a}_{2} + x_{2} \, \mathbf{a}_{3} & = & x_{2}c \, \mathbf{\hat{z}} & \left(1a\right) & \mbox{Cr I} \\ 
\mathbf{B}_{3} & = & x_{3} \, \mathbf{a}_{1} + x_{3} \, \mathbf{a}_{2} + z_{3} \, \mathbf{a}_{3} & = & \frac{1}{2}\left(x_{3}-z_{3}\right)a \, \mathbf{\hat{x}} + \frac{1}{2\sqrt{3}}\left(x_{3}-z_{3}\right)a \, \mathbf{\hat{y}} + \left(\frac{2}{3}x_{3}+\frac{1}{3}z_{3}\right)c \, \mathbf{\hat{z}} & \left(3b\right) & \mbox{Al II} \\ 
\mathbf{B}_{4} & = & z_{3} \, \mathbf{a}_{1} + x_{3} \, \mathbf{a}_{2} + x_{3} \, \mathbf{a}_{3} & = & \frac{1}{2}\left(-x_{3}+z_{3}\right)a \, \mathbf{\hat{x}} + \frac{1}{2\sqrt{3}}\left(x_{3}-z_{3}\right)a \, \mathbf{\hat{y}} + \left(\frac{2}{3}x_{3}+\frac{1}{3}z_{3}\right)c \, \mathbf{\hat{z}} & \left(3b\right) & \mbox{Al II} \\ 
\mathbf{B}_{5} & = & x_{3} \, \mathbf{a}_{1} + z_{3} \, \mathbf{a}_{2} + x_{3} \, \mathbf{a}_{3} & = & \frac{1}{\sqrt{3}}\left(-x_{3}+z_{3}\right)a \, \mathbf{\hat{y}} + \left(\frac{2}{3}x_{3}+\frac{1}{3}z_{3}\right)c \, \mathbf{\hat{z}} & \left(3b\right) & \mbox{Al II} \\ 
\mathbf{B}_{6} & = & x_{4} \, \mathbf{a}_{1} + x_{4} \, \mathbf{a}_{2} + z_{4} \, \mathbf{a}_{3} & = & \frac{1}{2}\left(x_{4}-z_{4}\right)a \, \mathbf{\hat{x}} + \frac{1}{2\sqrt{3}}\left(x_{4}-z_{4}\right)a \, \mathbf{\hat{y}} + \left(\frac{2}{3}x_{4}+\frac{1}{3}z_{4}\right)c \, \mathbf{\hat{z}} & \left(3b\right) & \mbox{Al III} \\ 
\mathbf{B}_{7} & = & z_{4} \, \mathbf{a}_{1} + x_{4} \, \mathbf{a}_{2} + x_{4} \, \mathbf{a}_{3} & = & \frac{1}{2}\left(-x_{4}+z_{4}\right)a \, \mathbf{\hat{x}} + \frac{1}{2\sqrt{3}}\left(x_{4}-z_{4}\right)a \, \mathbf{\hat{y}} + \left(\frac{2}{3}x_{4}+\frac{1}{3}z_{4}\right)c \, \mathbf{\hat{z}} & \left(3b\right) & \mbox{Al III} \\ 
\mathbf{B}_{8} & = & x_{4} \, \mathbf{a}_{1} + z_{4} \, \mathbf{a}_{2} + x_{4} \, \mathbf{a}_{3} & = & \frac{1}{\sqrt{3}}\left(-x_{4}+z_{4}\right)a \, \mathbf{\hat{y}} + \left(\frac{2}{3}x_{4}+\frac{1}{3}z_{4}\right)c \, \mathbf{\hat{z}} & \left(3b\right) & \mbox{Al III} \\ 
\mathbf{B}_{9} & = & x_{5} \, \mathbf{a}_{1} + x_{5} \, \mathbf{a}_{2} + z_{5} \, \mathbf{a}_{3} & = & \frac{1}{2}\left(x_{5}-z_{5}\right)a \, \mathbf{\hat{x}} + \frac{1}{2\sqrt{3}}\left(x_{5}-z_{5}\right)a \, \mathbf{\hat{y}} + \left(\frac{2}{3}x_{5}+\frac{1}{3}z_{5}\right)c \, \mathbf{\hat{z}} & \left(3b\right) & \mbox{Al IV} \\ 
\mathbf{B}_{10} & = & z_{5} \, \mathbf{a}_{1} + x_{5} \, \mathbf{a}_{2} + x_{5} \, \mathbf{a}_{3} & = & \frac{1}{2}\left(-x_{5}+z_{5}\right)a \, \mathbf{\hat{x}} + \frac{1}{2\sqrt{3}}\left(x_{5}-z_{5}\right)a \, \mathbf{\hat{y}} + \left(\frac{2}{3}x_{5}+\frac{1}{3}z_{5}\right)c \, \mathbf{\hat{z}} & \left(3b\right) & \mbox{Al IV} \\ 
\mathbf{B}_{11} & = & x_{5} \, \mathbf{a}_{1} + z_{5} \, \mathbf{a}_{2} + x_{5} \, \mathbf{a}_{3} & = & \frac{1}{\sqrt{3}}\left(-x_{5}+z_{5}\right)a \, \mathbf{\hat{y}} + \left(\frac{2}{3}x_{5}+\frac{1}{3}z_{5}\right)c \, \mathbf{\hat{z}} & \left(3b\right) & \mbox{Al IV} \\ 
\mathbf{B}_{12} & = & x_{6} \, \mathbf{a}_{1} + x_{6} \, \mathbf{a}_{2} + z_{6} \, \mathbf{a}_{3} & = & \frac{1}{2}\left(x_{6}-z_{6}\right)a \, \mathbf{\hat{x}} + \frac{1}{2\sqrt{3}}\left(x_{6}-z_{6}\right)a \, \mathbf{\hat{y}} + \left(\frac{2}{3}x_{6}+\frac{1}{3}z_{6}\right)c \, \mathbf{\hat{z}} & \left(3b\right) & \mbox{Cr II} \\ 
\mathbf{B}_{13} & = & z_{6} \, \mathbf{a}_{1} + x_{6} \, \mathbf{a}_{2} + x_{6} \, \mathbf{a}_{3} & = & \frac{1}{2}\left(-x_{6}+z_{6}\right)a \, \mathbf{\hat{x}} + \frac{1}{2\sqrt{3}}\left(x_{6}-z_{6}\right)a \, \mathbf{\hat{y}} + \left(\frac{2}{3}x_{6}+\frac{1}{3}z_{6}\right)c \, \mathbf{\hat{z}} & \left(3b\right) & \mbox{Cr II} \\ 
\mathbf{B}_{14} & = & x_{6} \, \mathbf{a}_{1} + z_{6} \, \mathbf{a}_{2} + x_{6} \, \mathbf{a}_{3} & = & \frac{1}{\sqrt{3}}\left(-x_{6}+z_{6}\right)a \, \mathbf{\hat{y}} + \left(\frac{2}{3}x_{6}+\frac{1}{3}z_{6}\right)c \, \mathbf{\hat{z}} & \left(3b\right) & \mbox{Cr II} \\ 
\mathbf{B}_{15} & = & x_{7} \, \mathbf{a}_{1} + x_{7} \, \mathbf{a}_{2} + z_{7} \, \mathbf{a}_{3} & = & \frac{1}{2}\left(x_{7}-z_{7}\right)a \, \mathbf{\hat{x}} + \frac{1}{2\sqrt{3}}\left(x_{7}-z_{7}\right)a \, \mathbf{\hat{y}} + \left(\frac{2}{3}x_{7}+\frac{1}{3}z_{7}\right)c \, \mathbf{\hat{z}} & \left(3b\right) & \mbox{Cr III} \\ 
\mathbf{B}_{16} & = & z_{7} \, \mathbf{a}_{1} + x_{7} \, \mathbf{a}_{2} + x_{7} \, \mathbf{a}_{3} & = & \frac{1}{2}\left(-x_{7}+z_{7}\right)a \, \mathbf{\hat{x}} + \frac{1}{2\sqrt{3}}\left(x_{7}-z_{7}\right)a \, \mathbf{\hat{y}} + \left(\frac{2}{3}x_{7}+\frac{1}{3}z_{7}\right)c \, \mathbf{\hat{z}} & \left(3b\right) & \mbox{Cr III} \\ 
\mathbf{B}_{17} & = & x_{7} \, \mathbf{a}_{1} + z_{7} \, \mathbf{a}_{2} + x_{7} \, \mathbf{a}_{3} & = & \frac{1}{\sqrt{3}}\left(-x_{7}+z_{7}\right)a \, \mathbf{\hat{y}} + \left(\frac{2}{3}x_{7}+\frac{1}{3}z_{7}\right)c \, \mathbf{\hat{z}} & \left(3b\right) & \mbox{Cr III} \\ 
\mathbf{B}_{18} & = & x_{8} \, \mathbf{a}_{1} + x_{8} \, \mathbf{a}_{2} + z_{8} \, \mathbf{a}_{3} & = & \frac{1}{2}\left(x_{8}-z_{8}\right)a \, \mathbf{\hat{x}} + \frac{1}{2\sqrt{3}}\left(x_{8}-z_{8}\right)a \, \mathbf{\hat{y}} + \left(\frac{2}{3}x_{8}+\frac{1}{3}z_{8}\right)c \, \mathbf{\hat{z}} & \left(3b\right) & \mbox{Cr IV} \\ 
\mathbf{B}_{19} & = & z_{8} \, \mathbf{a}_{1} + x_{8} \, \mathbf{a}_{2} + x_{8} \, \mathbf{a}_{3} & = & \frac{1}{2}\left(-x_{8}+z_{8}\right)a \, \mathbf{\hat{x}} + \frac{1}{2\sqrt{3}}\left(x_{8}-z_{8}\right)a \, \mathbf{\hat{y}} + \left(\frac{2}{3}x_{8}+\frac{1}{3}z_{8}\right)c \, \mathbf{\hat{z}} & \left(3b\right) & \mbox{Cr IV} \\ 
\mathbf{B}_{20} & = & x_{8} \, \mathbf{a}_{1} + z_{8} \, \mathbf{a}_{2} + x_{8} \, \mathbf{a}_{3} & = & \frac{1}{\sqrt{3}}\left(-x_{8}+z_{8}\right)a \, \mathbf{\hat{y}} + \left(\frac{2}{3}x_{8}+\frac{1}{3}z_{8}\right)c \, \mathbf{\hat{z}} & \left(3b\right) & \mbox{Cr IV} \\ 
\mathbf{B}_{21} & = & x_{9} \, \mathbf{a}_{1} + y_{9} \, \mathbf{a}_{2} + z_{9} \, \mathbf{a}_{3} & = & \frac{1}{2}\left(x_{9}-z_{9}\right)a \, \mathbf{\hat{x}} + \left(-\frac{1}{2\sqrt{3}}x_{9}+\frac{1}{\sqrt{3}}y_{9}-\frac{1}{2\sqrt{3}}z_{9}\right)a \, \mathbf{\hat{y}} + \frac{1}{3}\left(x_{9}+y_{9}+z_{9}\right)c \, \mathbf{\hat{z}} & \left(6c\right) & \mbox{Al V} \\ 
\mathbf{B}_{22} & = & z_{9} \, \mathbf{a}_{1} + x_{9} \, \mathbf{a}_{2} + y_{9} \, \mathbf{a}_{3} & = & \frac{1}{2}\left(-y_{9}+z_{9}\right)a \, \mathbf{\hat{x}} + \left(\frac{1}{\sqrt{3}}x_{9}-\frac{1}{2\sqrt{3}}y_{9}-\frac{1}{2\sqrt{3}}z_{9}\right)a \, \mathbf{\hat{y}} + \frac{1}{3}\left(x_{9}+y_{9}+z_{9}\right)c \, \mathbf{\hat{z}} & \left(6c\right) & \mbox{Al V} \\ 
\mathbf{B}_{23} & = & y_{9} \, \mathbf{a}_{1} + z_{9} \, \mathbf{a}_{2} + x_{9} \, \mathbf{a}_{3} & = & \frac{1}{2}\left(-x_{9}+y_{9}\right)a \, \mathbf{\hat{x}} + \left(-\frac{1}{2\sqrt{3}}x_{9}-\frac{1}{2\sqrt{3}}y_{9}+\frac{1}{\sqrt{3}}z_{9}\right)a \, \mathbf{\hat{y}} + \frac{1}{3}\left(x_{9}+y_{9}+z_{9}\right)c \, \mathbf{\hat{z}} & \left(6c\right) & \mbox{Al V} \\ 
\mathbf{B}_{24} & = & z_{9} \, \mathbf{a}_{1} + y_{9} \, \mathbf{a}_{2} + x_{9} \, \mathbf{a}_{3} & = & \frac{1}{2}\left(-x_{9}+z_{9}\right)a \, \mathbf{\hat{x}} + \left(-\frac{1}{2\sqrt{3}}x_{9}+\frac{1}{\sqrt{3}}y_{9}-\frac{1}{2\sqrt{3}}z_{9}\right)a \, \mathbf{\hat{y}} + \frac{1}{3}\left(x_{9}+y_{9}+z_{9}\right)c \, \mathbf{\hat{z}} & \left(6c\right) & \mbox{Al V} \\ 
\mathbf{B}_{25} & = & y_{9} \, \mathbf{a}_{1} + x_{9} \, \mathbf{a}_{2} + z_{9} \, \mathbf{a}_{3} & = & \frac{1}{2}\left(y_{9}-z_{9}\right)a \, \mathbf{\hat{x}} + \left(\frac{1}{\sqrt{3}}x_{9}-\frac{1}{2\sqrt{3}}y_{9}-\frac{1}{2\sqrt{3}}z_{9}\right)a \, \mathbf{\hat{y}} + \frac{1}{3}\left(x_{9}+y_{9}+z_{9}\right)c \, \mathbf{\hat{z}} & \left(6c\right) & \mbox{Al V} \\ 
\mathbf{B}_{26} & = & x_{9} \, \mathbf{a}_{1} + z_{9} \, \mathbf{a}_{2} + y_{9} \, \mathbf{a}_{3} & = & \frac{1}{2}\left(x_{9}-y_{9}\right)a \, \mathbf{\hat{x}} + \left(-\frac{1}{2\sqrt{3}}x_{9}-\frac{1}{2\sqrt{3}}y_{9}+\frac{1}{\sqrt{3}}z_{9}\right)a \, \mathbf{\hat{y}} + \frac{1}{3}\left(x_{9}+y_{9}+z_{9}\right)c \, \mathbf{\hat{z}} & \left(6c\right) & \mbox{Al V} \\ 
\end{longtabu}
\renewcommand{\arraystretch}{1.0}
\noindent \hrulefill
\\
\textbf{References:}
\vspace*{-0.25cm}
\begin{flushleft}
  - \bibentry{Bradley_Z_Kristallgr_96_1937}. \\
\end{flushleft}
\textbf{Found in:}
\vspace*{-0.25cm}
\begin{flushleft}
  - \bibentry{Grazulis_Cryst_Online_2014}. \\
\end{flushleft}
\noindent \hrulefill
\\
\textbf{Geometry files:}
\\
\noindent  - CIF: pp. {\hyperref[A8B5_hR26_160_a3bc_a3b_cif]{\pageref{A8B5_hR26_160_a3bc_a3b_cif}}} \\
\noindent  - POSCAR: pp. {\hyperref[A8B5_hR26_160_a3bc_a3b_poscar]{\pageref{A8B5_hR26_160_a3bc_a3b_poscar}}} \\
\onecolumn
{\phantomsection\label{ABC_hR3_160_a_a_a}}
\subsection*{\huge \textbf{{\normalfont \begin{raggedleft}Carbonyl Sulphide (COS, $F0_{2}$) Structure: \end{raggedleft} \\ ABC\_hR3\_160\_a\_a\_a}}}
\noindent \hrulefill
\vspace*{0.25cm}
\begin{figure}[htp]
  \centering
  \vspace{-1em}
  {\includegraphics[width=1\textwidth]{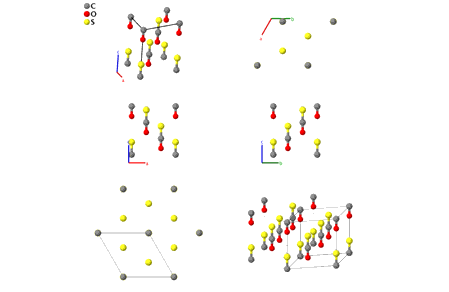}}
\end{figure}
\vspace*{-0.5cm}
\renewcommand{\arraystretch}{1.5}
\begin{equation*}
  \begin{array}{>{$\hspace{-0.15cm}}l<{$}>{$}p{0.5cm}<{$}>{$}p{18.5cm}<{$}}
    \mbox{\large \textbf{Prototype}} &\colon & \ce{COS} \\
    \mbox{\large \textbf{\AFLOW\ prototype label}} &\colon & \mbox{ABC\_hR3\_160\_a\_a\_a} \\
    \mbox{\large \textbf{\textit{Strukturbericht} designation}} &\colon & \mbox{$F0_{2}$} \\
    \mbox{\large \textbf{Pearson symbol}} &\colon & \mbox{hR3} \\
    \mbox{\large \textbf{Space group number}} &\colon & 160 \\
    \mbox{\large \textbf{Space group symbol}} &\colon & R3m \\
    \mbox{\large \textbf{\AFLOW\ prototype command}} &\colon &  \texttt{aflow} \,  \, \texttt{-{}-proto=ABC\_hR3\_160\_a\_a\_a [-{}-hex]} \, \newline \texttt{-{}-params=}{a,c/a,x_{1},x_{2},x_{3} }
  \end{array}
\end{equation*}
\renewcommand{\arraystretch}{1.0}

\vspace*{-0.25cm}
\noindent \hrulefill
\begin{itemize}
  \item{(Overell, 1982) does not give the Wyckoff positions directly.  They
are inferred from the C-O and C-S bond lengths.
The experimental data was obtained at 90~K.
}
\end{itemize}

\noindent \parbox{1 \linewidth}{
\noindent \hrulefill
\\
\textbf{Rhombohedral primitive vectors:} \\
\vspace*{-0.25cm}
\begin{tabular}{cc}
  \begin{tabular}{c}
    \parbox{0.6 \linewidth}{
      \renewcommand{\arraystretch}{1.5}
      \begin{equation*}
        \centering
        \begin{array}{ccc}
              \mathbf{a}_1 & = & ~ \frac12 \, a \, \mathbf{\hat{x}} - \frac{1}{2\sqrt{3}} \, a \, \mathbf{\hat{y}} + \frac13 \, c \, \mathbf{\hat{z}} \\
    \mathbf{a}_2 & = & \frac{1}{\sqrt{3}} \, a \, \mathbf{\hat{y}} + \frac13 \, c \, \mathbf{\hat{z}} \\
    \mathbf{a}_3 & = & - \frac12 \, a \, \mathbf{\hat{x}} - \frac{1}{2\sqrt{3}} \, a \, \mathbf{\hat{y}} + \frac13 \, c \, \mathbf{\hat{z}} \\

        \end{array}
      \end{equation*}
    }
    \renewcommand{\arraystretch}{1.0}
  \end{tabular}
  \begin{tabular}{c}
    \includegraphics[width=0.3\linewidth]{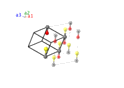} \\
  \end{tabular}
\end{tabular}

}
\vspace*{-0.25cm}

\noindent \hrulefill
\\
\textbf{Basis vectors:}
\vspace*{-0.25cm}
\renewcommand{\arraystretch}{1.5}
\begin{longtabu} to \textwidth{>{\centering $}X[-1,c,c]<{$}>{\centering $}X[-1,c,c]<{$}>{\centering $}X[-1,c,c]<{$}>{\centering $}X[-1,c,c]<{$}>{\centering $}X[-1,c,c]<{$}>{\centering $}X[-1,c,c]<{$}>{\centering $}X[-1,c,c]<{$}}
  & & \mbox{Lattice Coordinates} & & \mbox{Cartesian Coordinates} &\mbox{Wyckoff Position} & \mbox{Atom Type} \\  
  \mathbf{B}_{1} & = & x_{1} \, \mathbf{a}_{1} + x_{1} \, \mathbf{a}_{2} + x_{1} \, \mathbf{a}_{3} & = & x_{1}c \, \mathbf{\hat{z}} & \left(1a\right) & \mbox{C} \\ 
\mathbf{B}_{2} & = & x_{2} \, \mathbf{a}_{1} + x_{2} \, \mathbf{a}_{2} + x_{2} \, \mathbf{a}_{3} & = & x_{2}c \, \mathbf{\hat{z}} & \left(1a\right) & \mbox{O} \\ 
\mathbf{B}_{3} & = & x_{3} \, \mathbf{a}_{1} + x_{3} \, \mathbf{a}_{2} + x_{3} \, \mathbf{a}_{3} & = & x_{3}c \, \mathbf{\hat{z}} & \left(1a\right) & \mbox{S} \\ 
\end{longtabu}
\renewcommand{\arraystretch}{1.0}
\noindent \hrulefill
\\
\textbf{References:}
\vspace*{-0.25cm}
\begin{flushleft}
  - \bibentry{Overell_Acta_Cryst_B_38_1982}. \\
\end{flushleft}
\noindent \hrulefill
\\
\textbf{Geometry files:}
\\
\noindent  - CIF: pp. {\hyperref[ABC_hR3_160_a_a_a_cif]{\pageref{ABC_hR3_160_a_a_a_cif}}} \\
\noindent  - POSCAR: pp. {\hyperref[ABC_hR3_160_a_a_a_poscar]{\pageref{ABC_hR3_160_a_a_a_poscar}}} \\
\onecolumn
{\phantomsection\label{AB_hR10_160_5a_5a}}
\subsection*{\huge \textbf{{\normalfont \begin{raggedleft}Moissanite-15R (SiC, $B7$) Structure: \end{raggedleft} \\ AB\_hR10\_160\_5a\_5a}}}
\noindent \hrulefill
\vspace*{0.25cm}
\begin{figure}[htp]
  \centering
  \vspace{-1em}
  {\includegraphics[width=1\textwidth]{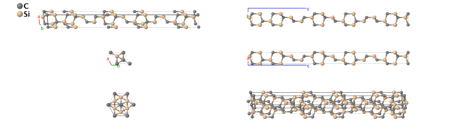}}
\end{figure}
\vspace*{-0.5cm}
\renewcommand{\arraystretch}{1.5}
\begin{equation*}
  \begin{array}{>{$\hspace{-0.15cm}}l<{$}>{$}p{0.5cm}<{$}>{$}p{18.5cm}<{$}}
    \mbox{\large \textbf{Prototype}} &\colon & \ce{SiC} \\
    \mbox{\large \textbf{\AFLOW\ prototype label}} &\colon & \mbox{AB\_hR10\_160\_5a\_5a} \\
    \mbox{\large \textbf{\textit{Strukturbericht} designation}} &\colon & \mbox{$B7$} \\
    \mbox{\large \textbf{Pearson symbol}} &\colon & \mbox{hR10} \\
    \mbox{\large \textbf{Space group number}} &\colon & 160 \\
    \mbox{\large \textbf{Space group symbol}} &\colon & R3m \\
    \mbox{\large \textbf{\AFLOW\ prototype command}} &\colon &  \texttt{aflow} \,  \, \texttt{-{}-proto=AB\_hR10\_160\_5a\_5a [-{}-hex]} \, \newline \texttt{-{}-params=}{a,c/a,x_{1},x_{2},x_{3},x_{4},x_{5},x_{6},x_{7},x_{8},x_{9},x_{10} }
  \end{array}
\end{equation*}
\renewcommand{\arraystretch}{1.0}

\vspace*{-0.25cm}
\noindent \hrulefill
\begin{itemize}
  \item{(Ewald, 1931) and (Thibault, 1944) both call this structure ``Type I
$\alpha$-silicon carbide.''}
  \item{The atomic positions are not well determined.  We follow (Thibolt,
1944) and assume that the (0001) planes of carbon atoms are equally
spaced, and that each carbon atom has a silicon atom at a distance of
$c/20$ along the $\mathbf{\hat{z}}$ axis.
}
\end{itemize}

\noindent \parbox{1 \linewidth}{
\noindent \hrulefill
\\
\textbf{Rhombohedral primitive vectors:} \\
\vspace*{-0.25cm}
\begin{tabular}{cc}
  \begin{tabular}{c}
    \parbox{0.6 \linewidth}{
      \renewcommand{\arraystretch}{1.5}
      \begin{equation*}
        \centering
        \begin{array}{ccc}
              \mathbf{a}_1 & = & ~ \frac12 \, a \, \mathbf{\hat{x}} - \frac{1}{2\sqrt{3}} \, a \, \mathbf{\hat{y}} + \frac13 \, c \, \mathbf{\hat{z}} \\
    \mathbf{a}_2 & = & \frac{1}{\sqrt{3}} \, a \, \mathbf{\hat{y}} + \frac13 \, c \, \mathbf{\hat{z}} \\
    \mathbf{a}_3 & = & - \frac12 \, a \, \mathbf{\hat{x}} - \frac{1}{2\sqrt{3}} \, a \, \mathbf{\hat{y}} + \frac13 \, c \, \mathbf{\hat{z}} \\

        \end{array}
      \end{equation*}
    }
    \renewcommand{\arraystretch}{1.0}
  \end{tabular}
  \begin{tabular}{c}
    \includegraphics[width=0.3\linewidth]{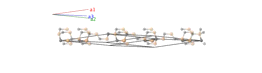} \\
  \end{tabular}
\end{tabular}

}
\vspace*{-0.25cm}

\noindent \hrulefill
\\
\textbf{Basis vectors:}
\vspace*{-0.25cm}
\renewcommand{\arraystretch}{1.5}
\begin{longtabu} to \textwidth{>{\centering $}X[-1,c,c]<{$}>{\centering $}X[-1,c,c]<{$}>{\centering $}X[-1,c,c]<{$}>{\centering $}X[-1,c,c]<{$}>{\centering $}X[-1,c,c]<{$}>{\centering $}X[-1,c,c]<{$}>{\centering $}X[-1,c,c]<{$}}
  & & \mbox{Lattice Coordinates} & & \mbox{Cartesian Coordinates} &\mbox{Wyckoff Position} & \mbox{Atom Type} \\  
  \mathbf{B}_{1} & = & x_{1} \, \mathbf{a}_{1} + x_{1} \, \mathbf{a}_{2} + x_{1} \, \mathbf{a}_{3} & = & x_{1}c \, \mathbf{\hat{z}} & \left(1a\right) & \mbox{C I} \\ 
\mathbf{B}_{2} & = & x_{2} \, \mathbf{a}_{1} + x_{2} \, \mathbf{a}_{2} + x_{2} \, \mathbf{a}_{3} & = & x_{2}c \, \mathbf{\hat{z}} & \left(1a\right) & \mbox{C II} \\ 
\mathbf{B}_{3} & = & x_{3} \, \mathbf{a}_{1} + x_{3} \, \mathbf{a}_{2} + x_{3} \, \mathbf{a}_{3} & = & x_{3}c \, \mathbf{\hat{z}} & \left(1a\right) & \mbox{C III} \\ 
\mathbf{B}_{4} & = & x_{4} \, \mathbf{a}_{1} + x_{4} \, \mathbf{a}_{2} + x_{4} \, \mathbf{a}_{3} & = & x_{4}c \, \mathbf{\hat{z}} & \left(1a\right) & \mbox{C IV} \\ 
\mathbf{B}_{5} & = & x_{5} \, \mathbf{a}_{1} + x_{5} \, \mathbf{a}_{2} + x_{5} \, \mathbf{a}_{3} & = & x_{5}c \, \mathbf{\hat{z}} & \left(1a\right) & \mbox{C V} \\ 
\mathbf{B}_{6} & = & x_{6} \, \mathbf{a}_{1} + x_{6} \, \mathbf{a}_{2} + x_{6} \, \mathbf{a}_{3} & = & x_{6}c \, \mathbf{\hat{z}} & \left(1a\right) & \mbox{Si I} \\ 
\mathbf{B}_{7} & = & x_{7} \, \mathbf{a}_{1} + x_{7} \, \mathbf{a}_{2} + x_{7} \, \mathbf{a}_{3} & = & x_{7}c \, \mathbf{\hat{z}} & \left(1a\right) & \mbox{Si II} \\ 
\mathbf{B}_{8} & = & x_{8} \, \mathbf{a}_{1} + x_{8} \, \mathbf{a}_{2} + x_{8} \, \mathbf{a}_{3} & = & x_{8}c \, \mathbf{\hat{z}} & \left(1a\right) & \mbox{Si III} \\ 
\mathbf{B}_{9} & = & x_{9} \, \mathbf{a}_{1} + x_{9} \, \mathbf{a}_{2} + x_{9} \, \mathbf{a}_{3} & = & x_{9}c \, \mathbf{\hat{z}} & \left(1a\right) & \mbox{Si IV} \\ 
\mathbf{B}_{10} & = & x_{10} \, \mathbf{a}_{1} + x_{10} \, \mathbf{a}_{2} + x_{10} \, \mathbf{a}_{3} & = & x_{10}c \, \mathbf{\hat{z}} & \left(1a\right) & \mbox{Si V} \\ 
\end{longtabu}
\renewcommand{\arraystretch}{1.0}
\noindent \hrulefill
\\
\textbf{References:}
\vspace*{-0.25cm}
\begin{flushleft}
  - \bibentry{Thibault_AmMin_29_327_1944}. \\
  - \bibentry{Ewald_Struk_I_1931}. \\
\end{flushleft}
\textbf{Found in:}
\vspace*{-0.25cm}
\begin{flushleft}
  - \bibentry{Harris_SiC_1995}. \\
\end{flushleft}
\noindent \hrulefill
\\
\textbf{Geometry files:}
\\
\noindent  - CIF: pp. {\hyperref[AB_hR10_160_5a_5a_cif]{\pageref{AB_hR10_160_5a_5a_cif}}} \\
\noindent  - POSCAR: pp. {\hyperref[AB_hR10_160_5a_5a_poscar]{\pageref{AB_hR10_160_5a_5a_poscar}}} \\
\onecolumn
{\phantomsection\label{A2B3_hP5_164_d_ad}}
\subsection*{\huge \textbf{{\normalfont La$_{2}$O$_{3}$ ($D5_{2}$) Structure: A2B3\_hP5\_164\_d\_ad}}}
\noindent \hrulefill
\vspace*{0.25cm}
\begin{figure}[htp]
  \centering
  \vspace{-1em}
  {\includegraphics[width=1\textwidth]{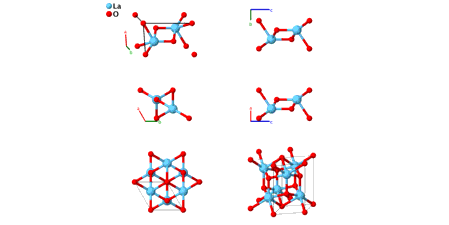}}
\end{figure}
\vspace*{-0.5cm}
\renewcommand{\arraystretch}{1.5}
\begin{equation*}
  \begin{array}{>{$\hspace{-0.15cm}}l<{$}>{$}p{0.5cm}<{$}>{$}p{18.5cm}<{$}}
    \mbox{\large \textbf{Prototype}} &\colon & \ce{La$_{2}$O$_{3}$} \\
    \mbox{\large \textbf{\AFLOW\ prototype label}} &\colon & \mbox{A2B3\_hP5\_164\_d\_ad} \\
    \mbox{\large \textbf{\textit{Strukturbericht} designation}} &\colon & \mbox{$D5_{2}$} \\
    \mbox{\large \textbf{Pearson symbol}} &\colon & \mbox{hP5} \\
    \mbox{\large \textbf{Space group number}} &\colon & 164 \\
    \mbox{\large \textbf{Space group symbol}} &\colon & P\bar{3}m1 \\
    \mbox{\large \textbf{\AFLOW\ prototype command}} &\colon &  \texttt{aflow} \,  \, \texttt{-{}-proto=A2B3\_hP5\_164\_d\_ad } \, \newline \texttt{-{}-params=}{a,c/a,z_{2},z_{3} }
  \end{array}
\end{equation*}
\renewcommand{\arraystretch}{1.0}

\vspace*{-0.25cm}
\noindent \hrulefill
\\
\textbf{ Other compounds with this structure:}
\begin{itemize}
   \item{ Ac$_{2}$O$_{3}$, Ce$_{2}$O$_{3}$, Nd$_{2}$O$_{3}$, Pr$_{2}$O$_{3}$, Th$_{2}$N$_{3}$, U$_{2}$N$_{3}$, $\alpha$-Bi$_{2}$Mg$_{3}$, $\alpha$-Sb$_{2}$Mg$_{3}$  }
\end{itemize}
\vspace*{-0.25cm}
\noindent \hrulefill
\begin{itemize}
  \item{This structure has proved rather controversial.  (Zachariasen 1926,
1929) originally proposed that the crystal structure of
La$_{2}$O$_{3}$ and other lanthanide series oxides belong to trigonal
space group $P321$ \#150.  This was immediately criticized by (Pauling,
1928), who suggested the trigonal space group ($P\bar{3}m1$ \#164).
This was confirmed by (Koehler, 1953).}
  \item{Much later, (Aldebert, 1979) stated that the structure was hexagonal,
space group $P6_3/mmc$ $\#194$.  This would represent a doubling of the
$P\bar{3}m1$ unit cell in the $z$-direction.  However, they give
structural parameters, used by us and by (Villars, 1991, 2016), which
have a density consistent with the $P\bar{3}m1$ structure, and are in
reasonable agreement with lattice parameters given in (Pearson, 1958)
and (Koehler, 1953).  We agree with (Villars, 1991, 2016) that these
are close to the correct values for the structure.}
  \item{This structure is very similar to
\href{http://aflow.org/CrystalDatabase/A3B2_hP5_164_ad_d.html}{Al$_{3}$Ni$_{2}$ ($D5_{13}$, A3B2\_hP5\_164\_ad\_d)}.
We follow (Pearson, 1958) and assign the intermetallics as $D5_{13}$,
keeping $D5_{2}$ for oxides and related compounds.
}
\end{itemize}

\noindent \parbox{1 \linewidth}{
\noindent \hrulefill
\\
\textbf{Trigonal Hexagonal primitive vectors:} \\
\vspace*{-0.25cm}
\begin{tabular}{cc}
  \begin{tabular}{c}
    \parbox{0.6 \linewidth}{
      \renewcommand{\arraystretch}{1.5}
      \begin{equation*}
        \centering
        \begin{array}{ccc}
              \mathbf{a}_1 & = & \frac12 \, a \, \mathbf{\hat{x}} - \frac{\sqrt3}2 \, a \, \mathbf{\hat{y}} \\
    \mathbf{a}_2 & = & \frac12 \, a \, \mathbf{\hat{x}} + \frac{\sqrt3}2 \, a \, \mathbf{\hat{y}} \\
    \mathbf{a}_3 & = & c \, \mathbf{\hat{z}} \\

        \end{array}
      \end{equation*}
    }
    \renewcommand{\arraystretch}{1.0}
  \end{tabular}
  \begin{tabular}{c}
    \includegraphics[width=0.3\linewidth]{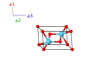} \\
  \end{tabular}
\end{tabular}

}
\vspace*{-0.25cm}

\noindent \hrulefill
\\
\textbf{Basis vectors:}
\vspace*{-0.25cm}
\renewcommand{\arraystretch}{1.5}
\begin{longtabu} to \textwidth{>{\centering $}X[-1,c,c]<{$}>{\centering $}X[-1,c,c]<{$}>{\centering $}X[-1,c,c]<{$}>{\centering $}X[-1,c,c]<{$}>{\centering $}X[-1,c,c]<{$}>{\centering $}X[-1,c,c]<{$}>{\centering $}X[-1,c,c]<{$}}
  & & \mbox{Lattice Coordinates} & & \mbox{Cartesian Coordinates} &\mbox{Wyckoff Position} & \mbox{Atom Type} \\  
  \mathbf{B}_{1} & = & 0 \, \mathbf{a}_{1} + 0 \, \mathbf{a}_{2} + 0 \, \mathbf{a}_{3} & = & 0 \, \mathbf{\hat{x}} + 0 \, \mathbf{\hat{y}} + 0 \, \mathbf{\hat{z}} & \left(1a\right) & \mbox{O I} \\ 
\mathbf{B}_{2} & = & \frac{1}{3} \, \mathbf{a}_{1} + \frac{2}{3} \, \mathbf{a}_{2} + z_{2} \, \mathbf{a}_{3} & = & \frac{1}{2}a \, \mathbf{\hat{x}} + \frac{1}{2\sqrt{3}}a \, \mathbf{\hat{y}} + z_{2}c \, \mathbf{\hat{z}} & \left(2d\right) & \mbox{La} \\ 
\mathbf{B}_{3} & = & \frac{2}{3} \, \mathbf{a}_{1} + \frac{1}{3} \, \mathbf{a}_{2}-z_{2} \, \mathbf{a}_{3} & = & \frac{1}{2}a \, \mathbf{\hat{x}}- \frac{1}{2\sqrt{3}}a  \, \mathbf{\hat{y}}-z_{2}c \, \mathbf{\hat{z}} & \left(2d\right) & \mbox{La} \\ 
\mathbf{B}_{4} & = & \frac{1}{3} \, \mathbf{a}_{1} + \frac{2}{3} \, \mathbf{a}_{2} + z_{3} \, \mathbf{a}_{3} & = & \frac{1}{2}a \, \mathbf{\hat{x}} + \frac{1}{2\sqrt{3}}a \, \mathbf{\hat{y}} + z_{3}c \, \mathbf{\hat{z}} & \left(2d\right) & \mbox{O II} \\ 
\mathbf{B}_{5} & = & \frac{2}{3} \, \mathbf{a}_{1} + \frac{1}{3} \, \mathbf{a}_{2}-z_{3} \, \mathbf{a}_{3} & = & \frac{1}{2}a \, \mathbf{\hat{x}}- \frac{1}{2\sqrt{3}}a  \, \mathbf{\hat{y}}-z_{3}c \, \mathbf{\hat{z}} & \left(2d\right) & \mbox{O II} \\ 
\end{longtabu}
\renewcommand{\arraystretch}{1.0}
\noindent \hrulefill
\\
\textbf{References:}
\vspace*{-0.25cm}
\begin{flushleft}
  - \bibentry{Aldebert_MatResBull_14_1979}. \\
  - \bibentry{Zachariasen_ZphysChem_123_1926}. \\
  - \bibentry{Zachariasen_ZKrist_70_1929}. \\
  - \bibentry{Pauling_ZKrist_69_1928}. \\
  - \bibentry{Koehler_Acta_Cryst_6_741_1953}. \\
  - \bibentry{Villars_Pauling_La2O3_2016}. \\
  - \bibentry{Pearson_NRC_1958}. \\
\end{flushleft}
\noindent \hrulefill
\\
\textbf{Geometry files:}
\\
\noindent  - CIF: pp. {\hyperref[A2B3_hP5_164_d_ad_cif]{\pageref{A2B3_hP5_164_d_ad_cif}}} \\
\noindent  - POSCAR: pp. {\hyperref[A2B3_hP5_164_d_ad_poscar]{\pageref{A2B3_hP5_164_d_ad_poscar}}} \\
\onecolumn
{\phantomsection\label{AB2_hP9_164_bd_c2d}}
\subsection*{\huge \textbf{{\normalfont \begin{raggedleft}$\delta_{H}^{II}$-NW$_2$ Structure: \end{raggedleft} \\ AB2\_hP9\_164\_bd\_c2d}}}
\noindent \hrulefill
\vspace*{0.25cm}
\begin{figure}[htp]
  \centering
  \vspace{-1em}
  {\includegraphics[width=1\textwidth]{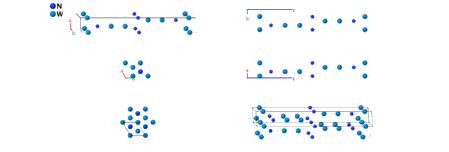}}
\end{figure}
\vspace*{-0.5cm}
\renewcommand{\arraystretch}{1.5}
\begin{equation*}
  \begin{array}{>{$\hspace{-0.15cm}}l<{$}>{$}p{0.5cm}<{$}>{$}p{18.5cm}<{$}}
    \mbox{\large \textbf{Prototype}} &\colon & \ce{$\delta_{H}^{II}$-NW2} \\
    \mbox{\large \textbf{\AFLOW\ prototype label}} &\colon & \mbox{AB2\_hP9\_164\_bd\_c2d} \\
    \mbox{\large \textbf{\textit{Strukturbericht} designation}} &\colon & \mbox{None} \\
    \mbox{\large \textbf{Pearson symbol}} &\colon & \mbox{hP9} \\
    \mbox{\large \textbf{Space group number}} &\colon & 164 \\
    \mbox{\large \textbf{Space group symbol}} &\colon & P\bar{3}m1 \\
    \mbox{\large \textbf{\AFLOW\ prototype command}} &\colon &  \texttt{aflow} \,  \, \texttt{-{}-proto=AB2\_hP9\_164\_bd\_c2d } \, \newline \texttt{-{}-params=}{a,c/a,z_{2},z_{3},z_{4},z_{5} }
  \end{array}
\end{equation*}
\renewcommand{\arraystretch}{1.0}

\vspace*{-0.25cm}
\noindent \hrulefill
\begin{itemize}
  \item{Khitrova and Pinkser put this structure in space group P$\overline{3}$
(\#147), but the Wyckoff positions used are identical with space
group P$\overline{3}$m1 (\#164), so we assign this to the higher
symmetry space group.
}
\end{itemize}

\noindent \parbox{1 \linewidth}{
\noindent \hrulefill
\\
\textbf{Trigonal Hexagonal primitive vectors:} \\
\vspace*{-0.25cm}
\begin{tabular}{cc}
  \begin{tabular}{c}
    \parbox{0.6 \linewidth}{
      \renewcommand{\arraystretch}{1.5}
      \begin{equation*}
        \centering
        \begin{array}{ccc}
              \mathbf{a}_1 & = & \frac12 \, a \, \mathbf{\hat{x}} - \frac{\sqrt3}2 \, a \, \mathbf{\hat{y}} \\
    \mathbf{a}_2 & = & \frac12 \, a \, \mathbf{\hat{x}} + \frac{\sqrt3}2 \, a \, \mathbf{\hat{y}} \\
    \mathbf{a}_3 & = & c \, \mathbf{\hat{z}} \\

        \end{array}
      \end{equation*}
    }
    \renewcommand{\arraystretch}{1.0}
  \end{tabular}
  \begin{tabular}{c}
    \includegraphics[width=0.3\linewidth]{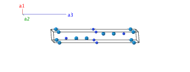} \\
  \end{tabular}
\end{tabular}

}
\vspace*{-0.25cm}

\noindent \hrulefill
\\
\textbf{Basis vectors:}
\vspace*{-0.25cm}
\renewcommand{\arraystretch}{1.5}
\begin{longtabu} to \textwidth{>{\centering $}X[-1,c,c]<{$}>{\centering $}X[-1,c,c]<{$}>{\centering $}X[-1,c,c]<{$}>{\centering $}X[-1,c,c]<{$}>{\centering $}X[-1,c,c]<{$}>{\centering $}X[-1,c,c]<{$}>{\centering $}X[-1,c,c]<{$}}
  & & \mbox{Lattice Coordinates} & & \mbox{Cartesian Coordinates} &\mbox{Wyckoff Position} & \mbox{Atom Type} \\  
  \mathbf{B}_{1} & = & \frac{1}{2} \, \mathbf{a}_{3} & = & \frac{1}{2}c \, \mathbf{\hat{z}} & \left(1b\right) & \mbox{N I} \\ 
\mathbf{B}_{2} & = & z_{2} \, \mathbf{a}_{3} & = & z_{2}c \, \mathbf{\hat{z}} & \left(2c\right) & \mbox{W I} \\ 
\mathbf{B}_{3} & = & -z_{2} \, \mathbf{a}_{3} & = & -z_{2}c \, \mathbf{\hat{z}} & \left(2c\right) & \mbox{W I} \\ 
\mathbf{B}_{4} & = & \frac{1}{3} \, \mathbf{a}_{1} + \frac{2}{3} \, \mathbf{a}_{2} + z_{3} \, \mathbf{a}_{3} & = & \frac{1}{2}a \, \mathbf{\hat{x}} + \frac{1}{2\sqrt{3}}a \, \mathbf{\hat{y}} + z_{3}c \, \mathbf{\hat{z}} & \left(2d\right) & \mbox{N II} \\ 
\mathbf{B}_{5} & = & \frac{2}{3} \, \mathbf{a}_{1} + \frac{1}{3} \, \mathbf{a}_{2}-z_{3} \, \mathbf{a}_{3} & = & \frac{1}{2}a \, \mathbf{\hat{x}}- \frac{1}{2\sqrt{3}}a  \, \mathbf{\hat{y}}-z_{3}c \, \mathbf{\hat{z}} & \left(2d\right) & \mbox{N II} \\ 
\mathbf{B}_{6} & = & \frac{1}{3} \, \mathbf{a}_{1} + \frac{2}{3} \, \mathbf{a}_{2} + z_{4} \, \mathbf{a}_{3} & = & \frac{1}{2}a \, \mathbf{\hat{x}} + \frac{1}{2\sqrt{3}}a \, \mathbf{\hat{y}} + z_{4}c \, \mathbf{\hat{z}} & \left(2d\right) & \mbox{W II} \\ 
\mathbf{B}_{7} & = & \frac{2}{3} \, \mathbf{a}_{1} + \frac{1}{3} \, \mathbf{a}_{2}-z_{4} \, \mathbf{a}_{3} & = & \frac{1}{2}a \, \mathbf{\hat{x}}- \frac{1}{2\sqrt{3}}a  \, \mathbf{\hat{y}}-z_{4}c \, \mathbf{\hat{z}} & \left(2d\right) & \mbox{W II} \\ 
\mathbf{B}_{8} & = & \frac{1}{3} \, \mathbf{a}_{1} + \frac{2}{3} \, \mathbf{a}_{2} + z_{5} \, \mathbf{a}_{3} & = & \frac{1}{2}a \, \mathbf{\hat{x}} + \frac{1}{2\sqrt{3}}a \, \mathbf{\hat{y}} + z_{5}c \, \mathbf{\hat{z}} & \left(2d\right) & \mbox{W III} \\ 
\mathbf{B}_{9} & = & \frac{2}{3} \, \mathbf{a}_{1} + \frac{1}{3} \, \mathbf{a}_{2}-z_{5} \, \mathbf{a}_{3} & = & \frac{1}{2}a \, \mathbf{\hat{x}}- \frac{1}{2\sqrt{3}}a  \, \mathbf{\hat{y}}-z_{5}c \, \mathbf{\hat{z}} & \left(2d\right) & \mbox{W III} \\ 
\end{longtabu}
\renewcommand{\arraystretch}{1.0}
\noindent \hrulefill
\\
\textbf{References:}
\vspace*{-0.25cm}
\begin{flushleft}
  - \bibentry{Khitrova_1962}. \\
\end{flushleft}
\noindent \hrulefill
\\
\textbf{Geometry files:}
\\
\noindent  - CIF: pp. {\hyperref[AB2_hP9_164_bd_c2d_cif]{\pageref{AB2_hP9_164_bd_c2d_cif}}} \\
\noindent  - POSCAR: pp. {\hyperref[AB2_hP9_164_bd_c2d_poscar]{\pageref{AB2_hP9_164_bd_c2d_poscar}}} \\
\onecolumn
{\phantomsection\label{ABC2_hP4_164_a_b_d}}
\subsection*{\huge \textbf{{\normalfont CuNiSb$_{2}$ Structure: ABC2\_hP4\_164\_a\_b\_d}}}
\noindent \hrulefill
\vspace*{0.25cm}
\begin{figure}[htp]
  \centering
  \vspace{-1em}
  {\includegraphics[width=1\textwidth]{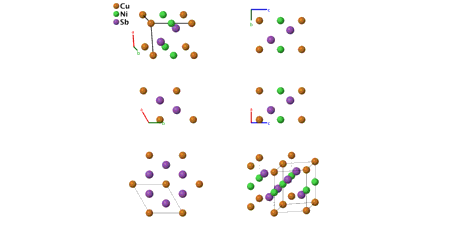}}
\end{figure}
\vspace*{-0.5cm}
\renewcommand{\arraystretch}{1.5}
\begin{equation*}
  \begin{array}{>{$\hspace{-0.15cm}}l<{$}>{$}p{0.5cm}<{$}>{$}p{18.5cm}<{$}}
    \mbox{\large \textbf{Prototype}} &\colon & \ce{CuNiSb$_{2}$} \\
    \mbox{\large \textbf{\AFLOW\ prototype label}} &\colon & \mbox{ABC2\_hP4\_164\_a\_b\_d} \\
    \mbox{\large \textbf{\textit{Strukturbericht} designation}} &\colon & \mbox{None} \\
    \mbox{\large \textbf{Pearson symbol}} &\colon & \mbox{hP4} \\
    \mbox{\large \textbf{Space group number}} &\colon & 164 \\
    \mbox{\large \textbf{Space group symbol}} &\colon & P\bar{3}m1 \\
    \mbox{\large \textbf{\AFLOW\ prototype command}} &\colon &  \texttt{aflow} \,  \, \texttt{-{}-proto=ABC2\_hP4\_164\_a\_b\_d } \, \newline \texttt{-{}-params=}{a,c/a,z_{3} }
  \end{array}
\end{equation*}
\renewcommand{\arraystretch}{1.0}

\noindent \parbox{1 \linewidth}{
\noindent \hrulefill
\\
\textbf{Trigonal Hexagonal primitive vectors:} \\
\vspace*{-0.25cm}
\begin{tabular}{cc}
  \begin{tabular}{c}
    \parbox{0.6 \linewidth}{
      \renewcommand{\arraystretch}{1.5}
      \begin{equation*}
        \centering
        \begin{array}{ccc}
              \mathbf{a}_1 & = & \frac12 \, a \, \mathbf{\hat{x}} - \frac{\sqrt3}2 \, a \, \mathbf{\hat{y}} \\
    \mathbf{a}_2 & = & \frac12 \, a \, \mathbf{\hat{x}} + \frac{\sqrt3}2 \, a \, \mathbf{\hat{y}} \\
    \mathbf{a}_3 & = & c \, \mathbf{\hat{z}} \\

        \end{array}
      \end{equation*}
    }
    \renewcommand{\arraystretch}{1.0}
  \end{tabular}
  \begin{tabular}{c}
    \includegraphics[width=0.3\linewidth]{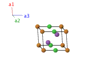} \\
  \end{tabular}
\end{tabular}

}
\vspace*{-0.25cm}

\noindent \hrulefill
\\
\textbf{Basis vectors:}
\vspace*{-0.25cm}
\renewcommand{\arraystretch}{1.5}
\begin{longtabu} to \textwidth{>{\centering $}X[-1,c,c]<{$}>{\centering $}X[-1,c,c]<{$}>{\centering $}X[-1,c,c]<{$}>{\centering $}X[-1,c,c]<{$}>{\centering $}X[-1,c,c]<{$}>{\centering $}X[-1,c,c]<{$}>{\centering $}X[-1,c,c]<{$}}
  & & \mbox{Lattice Coordinates} & & \mbox{Cartesian Coordinates} &\mbox{Wyckoff Position} & \mbox{Atom Type} \\  
  \mathbf{B}_{1} & = & 0 \, \mathbf{a}_{1} + 0 \, \mathbf{a}_{2} + 0 \, \mathbf{a}_{3} & = & 0 \, \mathbf{\hat{x}} + 0 \, \mathbf{\hat{y}} + 0 \, \mathbf{\hat{z}} & \left(1a\right) & \mbox{Cu} \\ 
\mathbf{B}_{2} & = & \frac{1}{2} \, \mathbf{a}_{3} & = & \frac{1}{2}c \, \mathbf{\hat{z}} & \left(1b\right) & \mbox{Ni} \\ 
\mathbf{B}_{3} & = & \frac{1}{3} \, \mathbf{a}_{1} + \frac{2}{3} \, \mathbf{a}_{2} + z_{3} \, \mathbf{a}_{3} & = & \frac{1}{2}a \, \mathbf{\hat{x}} + \frac{1}{2\sqrt{3}}a \, \mathbf{\hat{y}} + z_{3}c \, \mathbf{\hat{z}} & \left(2d\right) & \mbox{Sb} \\ 
\mathbf{B}_{4} & = & \frac{2}{3} \, \mathbf{a}_{1} + \frac{1}{3} \, \mathbf{a}_{2}-z_{3} \, \mathbf{a}_{3} & = & \frac{1}{2}a \, \mathbf{\hat{x}}- \frac{1}{2\sqrt{3}}a  \, \mathbf{\hat{y}}-z_{3}c \, \mathbf{\hat{z}} & \left(2d\right) & \mbox{Sb} \\ 
\end{longtabu}
\renewcommand{\arraystretch}{1.0}
\noindent \hrulefill
\\
\textbf{References:}
\vspace*{-0.25cm}
\begin{flushleft}
  - \bibentry{Kift_Hull_2010}. \\
\end{flushleft}
\noindent \hrulefill
\\
\textbf{Geometry files:}
\\
\noindent  - CIF: pp. {\hyperref[ABC2_hP4_164_a_b_d_cif]{\pageref{ABC2_hP4_164_a_b_d_cif}}} \\
\noindent  - POSCAR: pp. {\hyperref[ABC2_hP4_164_a_b_d_poscar]{\pageref{ABC2_hP4_164_a_b_d_poscar}}} \\
\onecolumn
{\phantomsection\label{A3B_hP24_165_bdg_f}}
\subsection*{\huge \textbf{{\normalfont Cu$_{3}$P ($D0_{21}$) Structure: A3B\_hP24\_165\_bdg\_f}}}
\noindent \hrulefill
\vspace*{0.25cm}
\begin{figure}[htp]
  \centering
  \vspace{-1em}
  {\includegraphics[width=1\textwidth]{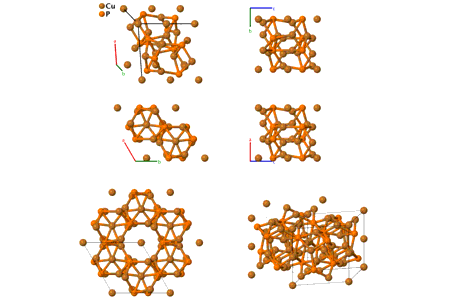}}
\end{figure}
\vspace*{-0.5cm}
\renewcommand{\arraystretch}{1.5}
\begin{equation*}
  \begin{array}{>{$\hspace{-0.15cm}}l<{$}>{$}p{0.5cm}<{$}>{$}p{18.5cm}<{$}}
    \mbox{\large \textbf{Prototype}} &\colon & \ce{Cu$_{3}$P} \\
    \mbox{\large \textbf{\AFLOW\ prototype label}} &\colon & \mbox{A3B\_hP24\_165\_bdg\_f} \\
    \mbox{\large \textbf{\textit{Strukturbericht} designation}} &\colon & \mbox{$D0_{21}$} \\
    \mbox{\large \textbf{Pearson symbol}} &\colon & \mbox{hP24} \\
    \mbox{\large \textbf{Space group number}} &\colon & 165 \\
    \mbox{\large \textbf{Space group symbol}} &\colon & P\bar{3}c1 \\
    \mbox{\large \textbf{\AFLOW\ prototype command}} &\colon &  \texttt{aflow} \,  \, \texttt{-{}-proto=A3B\_hP24\_165\_bdg\_f } \, \newline \texttt{-{}-params=}{a,c/a,z_{2},x_{3},x_{4},y_{4},z_{4} }
  \end{array}
\end{equation*}
\renewcommand{\arraystretch}{1.0}

\vspace*{-0.25cm}
\noindent \hrulefill
\\
\textbf{ Other compounds with this structure:}
\begin{itemize}
   \item{ Cu$_{3}$As, CeF$_{3}$, LaF$_{3}$   }
\end{itemize}
\vspace*{-0.25cm}
\noindent \hrulefill
\begin{itemize}
  \item{Range and Hafner (Range, 1993) and Olofsson (Olofson, 1972) argue that
several compounds which were previously identified as having this
structure actually take on the Cu$_{3}$P structure, space group
$P6_3cm$, \href{http://aflow.org/CrystalDatabase/A3B_hP24_185_ab2c_c.html}{A3B\_hP24\_185\_ab2c\_c}. These
structures include Cu$_{3}$P, Na$_{3}$As, AuMg$_{3}$, Ir$_{3}$ and
Mg$_{3}$Pt.}
  \item{The $D1_{21}$ structure is crystallographically equivalent to the
H$_{3}$Ho structure,
\href{http://aflow.org/CrystalDatabase/A3B_hP24_165_adg_f.html}{A3B\_hP24\_165\_adg\_f}.  We have not found
any evidence in the literature showing that either the hydride
structure or the fluoride structures mentioned above should be in any
other space group.}
  \item{We were unable to obtain the original reference, so we use the
crystallographic information presented in the American Mineralogist
Crystal Structure Database (Downs, 2003).  The original version of
{\em Pearson's Handbook} (Pearson, 1958) states that the lattice
constants for this structure should be slightly smaller.}
  \item{Cu$_{3}$P (pp. {\hyperref[A3B_hP24_185_ab2c_c]{\pageref{A3B_hP24_185_ab2c_c}}}) and
Na$_{3}$As (pp. {\hyperref[AB3_hP24_185_c_ab2c]{\pageref{AB3_hP24_185_c_ab2c}}})
have similar \AFLOW\ prototype labels ({\it{i.e.}}, same symmetry and set of
Wyckoff positions with different stoichiometry labels due to alphabetic ordering of atomic species).
They are generated by the same symmetry operations with different sets of parameters
(\texttt{-{}-params}) specified in their corresponding \CIF\ files.
}
\end{itemize}

\noindent \parbox{1 \linewidth}{
\noindent \hrulefill
\\
\textbf{Trigonal Hexagonal primitive vectors:} \\
\vspace*{-0.25cm}
\begin{tabular}{cc}
  \begin{tabular}{c}
    \parbox{0.6 \linewidth}{
      \renewcommand{\arraystretch}{1.5}
      \begin{equation*}
        \centering
        \begin{array}{ccc}
              \mathbf{a}_1 & = & \frac12 \, a \, \mathbf{\hat{x}} - \frac{\sqrt3}2 \, a \, \mathbf{\hat{y}} \\
    \mathbf{a}_2 & = & \frac12 \, a \, \mathbf{\hat{x}} + \frac{\sqrt3}2 \, a \, \mathbf{\hat{y}} \\
    \mathbf{a}_3 & = & c \, \mathbf{\hat{z}} \\

        \end{array}
      \end{equation*}
    }
    \renewcommand{\arraystretch}{1.0}
  \end{tabular}
  \begin{tabular}{c}
    \includegraphics[width=0.3\linewidth]{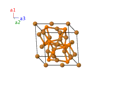} \\
  \end{tabular}
\end{tabular}

}
\vspace*{-0.25cm}

\noindent \hrulefill
\\
\textbf{Basis vectors:}
\vspace*{-0.25cm}
\renewcommand{\arraystretch}{1.5}
\begin{longtabu} to \textwidth{>{\centering $}X[-1,c,c]<{$}>{\centering $}X[-1,c,c]<{$}>{\centering $}X[-1,c,c]<{$}>{\centering $}X[-1,c,c]<{$}>{\centering $}X[-1,c,c]<{$}>{\centering $}X[-1,c,c]<{$}>{\centering $}X[-1,c,c]<{$}}
  & & \mbox{Lattice Coordinates} & & \mbox{Cartesian Coordinates} &\mbox{Wyckoff Position} & \mbox{Atom Type} \\  
  \mathbf{B}_{1} & = & 0 \, \mathbf{a}_{1} + 0 \, \mathbf{a}_{2} + 0 \, \mathbf{a}_{3} & = & 0 \, \mathbf{\hat{x}} + 0 \, \mathbf{\hat{y}} + 0 \, \mathbf{\hat{z}} & \left(2b\right) & \mbox{Cu I} \\ 
\mathbf{B}_{2} & = & \frac{1}{2} \, \mathbf{a}_{3} & = & \frac{1}{2}c \, \mathbf{\hat{z}} & \left(2b\right) & \mbox{Cu I} \\ 
\mathbf{B}_{3} & = & \frac{1}{3} \, \mathbf{a}_{1} + \frac{2}{3} \, \mathbf{a}_{2} + z_{2} \, \mathbf{a}_{3} & = & \frac{1}{2}a \, \mathbf{\hat{x}} + \frac{1}{2\sqrt{3}}a \, \mathbf{\hat{y}} + z_{2}c \, \mathbf{\hat{z}} & \left(4d\right) & \mbox{Cu II} \\ 
\mathbf{B}_{4} & = & \frac{2}{3} \, \mathbf{a}_{1} + \frac{1}{3} \, \mathbf{a}_{2} + \left(\frac{1}{2} - z_{2}\right) \, \mathbf{a}_{3} & = & \frac{1}{2}a \, \mathbf{\hat{x}}- \frac{1}{2\sqrt{3}}a  \, \mathbf{\hat{y}} + \left(\frac{1}{2} - z_{2}\right)c \, \mathbf{\hat{z}} & \left(4d\right) & \mbox{Cu II} \\ 
\mathbf{B}_{5} & = & \frac{2}{3} \, \mathbf{a}_{1} + \frac{1}{3} \, \mathbf{a}_{2}-z_{2} \, \mathbf{a}_{3} & = & \frac{1}{2}a \, \mathbf{\hat{x}}- \frac{1}{2\sqrt{3}}a  \, \mathbf{\hat{y}}-z_{2}c \, \mathbf{\hat{z}} & \left(4d\right) & \mbox{Cu II} \\ 
\mathbf{B}_{6} & = & \frac{1}{3} \, \mathbf{a}_{1} + \frac{2}{3} \, \mathbf{a}_{2} + \left(\frac{1}{2} +z_{2}\right) \, \mathbf{a}_{3} & = & \frac{1}{2}a \, \mathbf{\hat{x}} + \frac{1}{2\sqrt{3}}a \, \mathbf{\hat{y}} + \left(\frac{1}{2} +z_{2}\right)c \, \mathbf{\hat{z}} & \left(4d\right) & \mbox{Cu II} \\ 
\mathbf{B}_{7} & = & x_{3} \, \mathbf{a}_{1} + \frac{1}{4} \, \mathbf{a}_{3} & = & \frac{1}{2}x_{3}a \, \mathbf{\hat{x}}-\frac{\sqrt{3}}{2}x_{3}a \, \mathbf{\hat{y}} + \frac{1}{4}c \, \mathbf{\hat{z}} & \left(6f\right) & \mbox{P} \\ 
\mathbf{B}_{8} & = & x_{3} \, \mathbf{a}_{2} + \frac{1}{4} \, \mathbf{a}_{3} & = & \frac{1}{2}x_{3}a \, \mathbf{\hat{x}} + \frac{\sqrt{3}}{2}x_{3}a \, \mathbf{\hat{y}} + \frac{1}{4}c \, \mathbf{\hat{z}} & \left(6f\right) & \mbox{P} \\ 
\mathbf{B}_{9} & = & -x_{3} \, \mathbf{a}_{1}-x_{3} \, \mathbf{a}_{2} + \frac{1}{4} \, \mathbf{a}_{3} & = & -x_{3}a \, \mathbf{\hat{x}} + \frac{1}{4}c \, \mathbf{\hat{z}} & \left(6f\right) & \mbox{P} \\ 
\mathbf{B}_{10} & = & -x_{3} \, \mathbf{a}_{1} + \frac{3}{4} \, \mathbf{a}_{3} & = & -\frac{1}{2}x_{3}a \, \mathbf{\hat{x}} + \frac{\sqrt{3}}{2}x_{3}a \, \mathbf{\hat{y}} + \frac{3}{4}c \, \mathbf{\hat{z}} & \left(6f\right) & \mbox{P} \\ 
\mathbf{B}_{11} & = & -x_{3} \, \mathbf{a}_{2} + \frac{3}{4} \, \mathbf{a}_{3} & = & -\frac{1}{2}x_{3}a \, \mathbf{\hat{x}}-\frac{\sqrt{3}}{2}x_{3}a \, \mathbf{\hat{y}} + \frac{3}{4}c \, \mathbf{\hat{z}} & \left(6f\right) & \mbox{P} \\ 
\mathbf{B}_{12} & = & x_{3} \, \mathbf{a}_{1} + x_{3} \, \mathbf{a}_{2} + \frac{3}{4} \, \mathbf{a}_{3} & = & x_{3}a \, \mathbf{\hat{x}} + \frac{3}{4}c \, \mathbf{\hat{z}} & \left(6f\right) & \mbox{P} \\ 
\mathbf{B}_{13} & = & x_{4} \, \mathbf{a}_{1} + y_{4} \, \mathbf{a}_{2} + z_{4} \, \mathbf{a}_{3} & = & \frac{1}{2}\left(x_{4}+y_{4}\right)a \, \mathbf{\hat{x}} + \frac{\sqrt{3}}{2}\left(-x_{4}+y_{4}\right)a \, \mathbf{\hat{y}} + z_{4}c \, \mathbf{\hat{z}} & \left(12g\right) & \mbox{Cu III} \\ 
\mathbf{B}_{14} & = & -y_{4} \, \mathbf{a}_{1} + \left(x_{4}-y_{4}\right) \, \mathbf{a}_{2} + z_{4} \, \mathbf{a}_{3} & = & \left(\frac{1}{2}x_{4}-y_{4}\right)a \, \mathbf{\hat{x}} + \frac{\sqrt{3}}{2}x_{4}a \, \mathbf{\hat{y}} + z_{4}c \, \mathbf{\hat{z}} & \left(12g\right) & \mbox{Cu III} \\ 
\mathbf{B}_{15} & = & \left(-x_{4}+y_{4}\right) \, \mathbf{a}_{1}-x_{4} \, \mathbf{a}_{2} + z_{4} \, \mathbf{a}_{3} & = & \left(-x_{4}+\frac{1}{2}y_{4}\right)a \, \mathbf{\hat{x}}-\frac{\sqrt{3}}{2}y_{4}a \, \mathbf{\hat{y}} + z_{4}c \, \mathbf{\hat{z}} & \left(12g\right) & \mbox{Cu III} \\ 
\mathbf{B}_{16} & = & y_{4} \, \mathbf{a}_{1} + x_{4} \, \mathbf{a}_{2} + \left(\frac{1}{2} - z_{4}\right) \, \mathbf{a}_{3} & = & \frac{1}{2}\left(x_{4}+y_{4}\right)a \, \mathbf{\hat{x}} + \frac{\sqrt{3}}{2}\left(x_{4}-y_{4}\right)a \, \mathbf{\hat{y}} + \left(\frac{1}{2} - z_{4}\right)c \, \mathbf{\hat{z}} & \left(12g\right) & \mbox{Cu III} \\ 
\mathbf{B}_{17} & = & \left(x_{4}-y_{4}\right) \, \mathbf{a}_{1}-y_{4} \, \mathbf{a}_{2} + \left(\frac{1}{2} - z_{4}\right) \, \mathbf{a}_{3} & = & \left(\frac{1}{2}x_{4}-y_{4}\right)a \, \mathbf{\hat{x}}-\frac{\sqrt{3}}{2}x_{4}a \, \mathbf{\hat{y}} + \left(\frac{1}{2} - z_{4}\right)c \, \mathbf{\hat{z}} & \left(12g\right) & \mbox{Cu III} \\ 
\mathbf{B}_{18} & = & -x_{4} \, \mathbf{a}_{1} + \left(-x_{4}+y_{4}\right) \, \mathbf{a}_{2} + \left(\frac{1}{2} - z_{4}\right) \, \mathbf{a}_{3} & = & \left(-x_{4}+\frac{1}{2}y_{4}\right)a \, \mathbf{\hat{x}} + \frac{\sqrt{3}}{2}y_{4}a \, \mathbf{\hat{y}} + \left(\frac{1}{2} - z_{4}\right)c \, \mathbf{\hat{z}} & \left(12g\right) & \mbox{Cu III} \\ 
\mathbf{B}_{19} & = & -x_{4} \, \mathbf{a}_{1}-y_{4} \, \mathbf{a}_{2}-z_{4} \, \mathbf{a}_{3} & = & -\frac{1}{2}\left(x_{4}+y_{4}\right)a \, \mathbf{\hat{x}} + \frac{\sqrt{3}}{2}\left(x_{4}-y_{4}\right)a \, \mathbf{\hat{y}}-z_{4}c \, \mathbf{\hat{z}} & \left(12g\right) & \mbox{Cu III} \\ 
\mathbf{B}_{20} & = & y_{4} \, \mathbf{a}_{1} + \left(-x_{4}+y_{4}\right) \, \mathbf{a}_{2}-z_{4} \, \mathbf{a}_{3} & = & \left(-\frac{1}{2}x_{4}+y_{4}\right)a \, \mathbf{\hat{x}}-\frac{\sqrt{3}}{2}x_{4}a \, \mathbf{\hat{y}}-z_{4}c \, \mathbf{\hat{z}} & \left(12g\right) & \mbox{Cu III} \\ 
\mathbf{B}_{21} & = & \left(x_{4}-y_{4}\right) \, \mathbf{a}_{1} + x_{4} \, \mathbf{a}_{2}-z_{4} \, \mathbf{a}_{3} & = & \left(x_{4}-\frac{1}{2}y_{4}\right)a \, \mathbf{\hat{x}} + \frac{\sqrt{3}}{2}y_{4}a \, \mathbf{\hat{y}}-z_{4}c \, \mathbf{\hat{z}} & \left(12g\right) & \mbox{Cu III} \\ 
\mathbf{B}_{22} & = & -y_{4} \, \mathbf{a}_{1}-x_{4} \, \mathbf{a}_{2} + \left(\frac{1}{2} +z_{4}\right) \, \mathbf{a}_{3} & = & -\frac{1}{2}\left(x_{4}+y_{4}\right)a \, \mathbf{\hat{x}} + \frac{\sqrt{3}}{2}\left(-x_{4}+y_{4}\right)a \, \mathbf{\hat{y}} + \left(\frac{1}{2} +z_{4}\right)c \, \mathbf{\hat{z}} & \left(12g\right) & \mbox{Cu III} \\ 
\mathbf{B}_{23} & = & \left(-x_{4}+y_{4}\right) \, \mathbf{a}_{1} + y_{4} \, \mathbf{a}_{2} + \left(\frac{1}{2} +z_{4}\right) \, \mathbf{a}_{3} & = & \left(-\frac{1}{2}x_{4}+y_{4}\right)a \, \mathbf{\hat{x}} + \frac{\sqrt{3}}{2}x_{4}a \, \mathbf{\hat{y}} + \left(\frac{1}{2} +z_{4}\right)c \, \mathbf{\hat{z}} & \left(12g\right) & \mbox{Cu III} \\ 
\mathbf{B}_{24} & = & x_{4} \, \mathbf{a}_{1} + \left(x_{4}-y_{4}\right) \, \mathbf{a}_{2} + \left(\frac{1}{2} +z_{4}\right) \, \mathbf{a}_{3} & = & \left(x_{4}-\frac{1}{2}y_{4}\right)a \, \mathbf{\hat{x}}-\frac{\sqrt{3}}{2}y_{4}a \, \mathbf{\hat{y}} + \left(\frac{1}{2} +z_{4}\right)c \, \mathbf{\hat{z}} & \left(12g\right) & \mbox{Cu III} \\ 
\end{longtabu}
\renewcommand{\arraystretch}{1.0}
\noindent \hrulefill
\\
\textbf{References:}
\vspace*{-0.25cm}
\begin{flushleft}
  - \bibentry{Steenberg_AKMG_A12_1938}. \\
\end{flushleft}
\textbf{Found in:}
\vspace*{-0.25cm}
\begin{flushleft}
  - \bibentry{Pearson_NRC_1958}. \\
  - \bibentry{Downs_2003}. \\
\end{flushleft}
\noindent \hrulefill
\\
\textbf{Geometry files:}
\\
\noindent  - CIF: pp. {\hyperref[A3B_hP24_165_bdg_f_cif]{\pageref{A3B_hP24_165_bdg_f_cif}}} \\
\noindent  - POSCAR: pp. {\hyperref[A3B_hP24_165_bdg_f_poscar]{\pageref{A3B_hP24_165_bdg_f_poscar}}} \\
\onecolumn
{\phantomsection\label{A4B3_hR7_166_2c_ac}}
\subsection*{\huge \textbf{{\normalfont Al$_{4}$C$_{3}$ ($D7_{1}$) Structure: A4B3\_hR7\_166\_2c\_ac}}}
\noindent \hrulefill
\vspace*{0.25cm}
\begin{figure}[htp]
  \centering
  \vspace{-1em}
  {\includegraphics[width=1\textwidth]{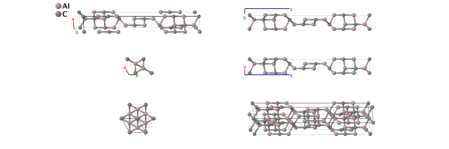}}
\end{figure}
\vspace*{-0.5cm}
\renewcommand{\arraystretch}{1.5}
\begin{equation*}
  \begin{array}{>{$\hspace{-0.15cm}}l<{$}>{$}p{0.5cm}<{$}>{$}p{18.5cm}<{$}}
    \mbox{\large \textbf{Prototype}} &\colon & \ce{Al$_{4}$C$_{3}$} \\
    \mbox{\large \textbf{\AFLOW\ prototype label}} &\colon & \mbox{A4B3\_hR7\_166\_2c\_ac} \\
    \mbox{\large \textbf{\textit{Strukturbericht} designation}} &\colon & \mbox{$D7_{1}$} \\
    \mbox{\large \textbf{Pearson symbol}} &\colon & \mbox{hR7} \\
    \mbox{\large \textbf{Space group number}} &\colon & 166 \\
    \mbox{\large \textbf{Space group symbol}} &\colon & R\bar{3}m \\
    \mbox{\large \textbf{\AFLOW\ prototype command}} &\colon &  \texttt{aflow} \,  \, \texttt{-{}-proto=A4B3\_hR7\_166\_2c\_ac [-{}-hex]} \, \newline \texttt{-{}-params=}{a,c/a,x_{2},x_{3},x_{4} }
  \end{array}
\end{equation*}
\renewcommand{\arraystretch}{1.0}

\noindent \parbox{1 \linewidth}{
\noindent \hrulefill
\\
\textbf{Rhombohedral primitive vectors:} \\
\vspace*{-0.25cm}
\begin{tabular}{cc}
  \begin{tabular}{c}
    \parbox{0.6 \linewidth}{
      \renewcommand{\arraystretch}{1.5}
      \begin{equation*}
        \centering
        \begin{array}{ccc}
              \mathbf{a}_1 & = & ~ \frac12 \, a \, \mathbf{\hat{x}} - \frac{1}{2\sqrt{3}} \, a \, \mathbf{\hat{y}} + \frac13 \, c \, \mathbf{\hat{z}} \\
    \mathbf{a}_2 & = & \frac{1}{\sqrt{3}} \, a \, \mathbf{\hat{y}} + \frac13 \, c \, \mathbf{\hat{z}} \\
    \mathbf{a}_3 & = & - \frac12 \, a \, \mathbf{\hat{x}} - \frac{1}{2\sqrt{3}} \, a \, \mathbf{\hat{y}} + \frac13 \, c \, \mathbf{\hat{z}} \\

        \end{array}
      \end{equation*}
    }
    \renewcommand{\arraystretch}{1.0}
  \end{tabular}
  \begin{tabular}{c}
    \includegraphics[width=0.3\linewidth]{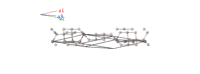} \\
  \end{tabular}
\end{tabular}

}
\vspace*{-0.25cm}

\noindent \hrulefill
\\
\textbf{Basis vectors:}
\vspace*{-0.25cm}
\renewcommand{\arraystretch}{1.5}
\begin{longtabu} to \textwidth{>{\centering $}X[-1,c,c]<{$}>{\centering $}X[-1,c,c]<{$}>{\centering $}X[-1,c,c]<{$}>{\centering $}X[-1,c,c]<{$}>{\centering $}X[-1,c,c]<{$}>{\centering $}X[-1,c,c]<{$}>{\centering $}X[-1,c,c]<{$}}
  & & \mbox{Lattice Coordinates} & & \mbox{Cartesian Coordinates} &\mbox{Wyckoff Position} & \mbox{Atom Type} \\  
  \mathbf{B}_{1} & = & 0 \, \mathbf{a}_{1} + 0 \, \mathbf{a}_{2} + 0 \, \mathbf{a}_{3} & = & 0 \, \mathbf{\hat{x}} + 0 \, \mathbf{\hat{y}} + 0 \, \mathbf{\hat{z}} & \left(1a\right) & \mbox{C I} \\ 
\mathbf{B}_{2} & = & x_{2} \, \mathbf{a}_{1} + x_{2} \, \mathbf{a}_{2} + x_{2} \, \mathbf{a}_{3} & = & x_{2}c \, \mathbf{\hat{z}} & \left(2c\right) & \mbox{Al I} \\ 
\mathbf{B}_{3} & = & -x_{2} \, \mathbf{a}_{1}-x_{2} \, \mathbf{a}_{2}-x_{2} \, \mathbf{a}_{3} & = & -x_{2}c \, \mathbf{\hat{z}} & \left(2c\right) & \mbox{Al I} \\ 
\mathbf{B}_{4} & = & x_{3} \, \mathbf{a}_{1} + x_{3} \, \mathbf{a}_{2} + x_{3} \, \mathbf{a}_{3} & = & x_{3}c \, \mathbf{\hat{z}} & \left(2c\right) & \mbox{Al II} \\ 
\mathbf{B}_{5} & = & -x_{3} \, \mathbf{a}_{1}-x_{3} \, \mathbf{a}_{2}-x_{3} \, \mathbf{a}_{3} & = & -x_{3}c \, \mathbf{\hat{z}} & \left(2c\right) & \mbox{Al II} \\ 
\mathbf{B}_{6} & = & x_{4} \, \mathbf{a}_{1} + x_{4} \, \mathbf{a}_{2} + x_{4} \, \mathbf{a}_{3} & = & x_{4}c \, \mathbf{\hat{z}} & \left(2c\right) & \mbox{C II} \\ 
\mathbf{B}_{7} & = & -x_{4} \, \mathbf{a}_{1}-x_{4} \, \mathbf{a}_{2}-x_{4} \, \mathbf{a}_{3} & = & -x_{4}c \, \mathbf{\hat{z}} & \left(2c\right) & \mbox{C II} \\ 
\end{longtabu}
\renewcommand{\arraystretch}{1.0}
\noindent \hrulefill
\\
\textbf{References:}
\vspace*{-0.25cm}
\begin{flushleft}
  - \bibentry{Gesing_Z_Naturforsh_50b_1995}. \\
\end{flushleft}
\noindent \hrulefill
\\
\textbf{Geometry files:}
\\
\noindent  - CIF: pp. {\hyperref[A4B3_hR7_166_2c_ac_cif]{\pageref{A4B3_hR7_166_2c_ac_cif}}} \\
\noindent  - POSCAR: pp. {\hyperref[A4B3_hR7_166_2c_ac_poscar]{\pageref{A4B3_hR7_166_2c_ac_poscar}}} \\
\onecolumn
{\phantomsection\label{ABC_hR6_166_c_c_c}}
\subsection*{\huge \textbf{{\normalfont SmSI Structure: ABC\_hR6\_166\_c\_c\_c}}}
\noindent \hrulefill
\vspace*{0.25cm}
\begin{figure}[htp]
  \centering
  \vspace{-1em}
  {\includegraphics[width=1\textwidth]{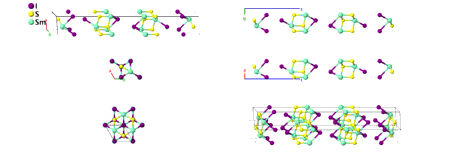}}
\end{figure}
\vspace*{-0.5cm}
\renewcommand{\arraystretch}{1.5}
\begin{equation*}
  \begin{array}{>{$\hspace{-0.15cm}}l<{$}>{$}p{0.5cm}<{$}>{$}p{18.5cm}<{$}}
    \mbox{\large \textbf{Prototype}} &\colon & \ce{SmSI} \\
    \mbox{\large \textbf{\AFLOW\ prototype label}} &\colon & \mbox{ABC\_hR6\_166\_c\_c\_c} \\
    \mbox{\large \textbf{\textit{Strukturbericht} designation}} &\colon & \mbox{None} \\
    \mbox{\large \textbf{Pearson symbol}} &\colon & \mbox{hR6} \\
    \mbox{\large \textbf{Space group number}} &\colon & 166 \\
    \mbox{\large \textbf{Space group symbol}} &\colon & R\bar{3}m \\
    \mbox{\large \textbf{\AFLOW\ prototype command}} &\colon &  \texttt{aflow} \,  \, \texttt{-{}-proto=ABC\_hR6\_166\_c\_c\_c [-{}-hex]} \, \newline \texttt{-{}-params=}{a,c/a,x_{1},x_{2},x_{3} }
  \end{array}
\end{equation*}
\renewcommand{\arraystretch}{1.0}

\vspace*{-0.25cm}
\noindent \hrulefill
\\
\textbf{ Other compounds with this structure:}
\begin{itemize}
   \item{  $\beta$-CeSI, PrSI, NdSI, and the superconducting alkali metal intercalcates $\beta$-$M$N$X$ ($M$ = Zr, Hf; $X$ = Cl, Br, I).  }
\end{itemize}
\vspace*{-0.25cm}
\noindent \hrulefill
\begin{itemize}
  \item{Although this has the same crystallographic structure as
\hyperref[AB2_hR16_166_c_2c.html]{the C12 structure}, the layering is
substantially different.
}
\end{itemize}

\noindent \parbox{1 \linewidth}{
\noindent \hrulefill
\\
\textbf{Rhombohedral primitive vectors:} \\
\vspace*{-0.25cm}
\begin{tabular}{cc}
  \begin{tabular}{c}
    \parbox{0.6 \linewidth}{
      \renewcommand{\arraystretch}{1.5}
      \begin{equation*}
        \centering
        \begin{array}{ccc}
              \mathbf{a}_1 & = & ~ \frac12 \, a \, \mathbf{\hat{x}} - \frac{1}{2\sqrt{3}} \, a \, \mathbf{\hat{y}} + \frac13 \, c \, \mathbf{\hat{z}} \\
    \mathbf{a}_2 & = & \frac{1}{\sqrt{3}} \, a \, \mathbf{\hat{y}} + \frac13 \, c \, \mathbf{\hat{z}} \\
    \mathbf{a}_3 & = & - \frac12 \, a \, \mathbf{\hat{x}} - \frac{1}{2\sqrt{3}} \, a \, \mathbf{\hat{y}} + \frac13 \, c \, \mathbf{\hat{z}} \\

        \end{array}
      \end{equation*}
    }
    \renewcommand{\arraystretch}{1.0}
  \end{tabular}
  \begin{tabular}{c}
    \includegraphics[width=0.3\linewidth]{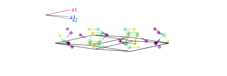} \\
  \end{tabular}
\end{tabular}

}
\vspace*{-0.25cm}

\noindent \hrulefill
\\
\textbf{Basis vectors:}
\vspace*{-0.25cm}
\renewcommand{\arraystretch}{1.5}
\begin{longtabu} to \textwidth{>{\centering $}X[-1,c,c]<{$}>{\centering $}X[-1,c,c]<{$}>{\centering $}X[-1,c,c]<{$}>{\centering $}X[-1,c,c]<{$}>{\centering $}X[-1,c,c]<{$}>{\centering $}X[-1,c,c]<{$}>{\centering $}X[-1,c,c]<{$}}
  & & \mbox{Lattice Coordinates} & & \mbox{Cartesian Coordinates} &\mbox{Wyckoff Position} & \mbox{Atom Type} \\  
  \mathbf{B}_{1} & = & x_{1} \, \mathbf{a}_{1} + x_{1} \, \mathbf{a}_{2} + x_{1} \, \mathbf{a}_{3} & = & x_{1}c \, \mathbf{\hat{z}} & \left(2c\right) & \mbox{I} \\ 
\mathbf{B}_{2} & = & -x_{1} \, \mathbf{a}_{1}-x_{1} \, \mathbf{a}_{2}-x_{1} \, \mathbf{a}_{3} & = & -x_{1}c \, \mathbf{\hat{z}} & \left(2c\right) & \mbox{I} \\ 
\mathbf{B}_{3} & = & x_{2} \, \mathbf{a}_{1} + x_{2} \, \mathbf{a}_{2} + x_{2} \, \mathbf{a}_{3} & = & x_{2}c \, \mathbf{\hat{z}} & \left(2c\right) & \mbox{S} \\ 
\mathbf{B}_{4} & = & -x_{2} \, \mathbf{a}_{1}-x_{2} \, \mathbf{a}_{2}-x_{2} \, \mathbf{a}_{3} & = & -x_{2}c \, \mathbf{\hat{z}} & \left(2c\right) & \mbox{S} \\ 
\mathbf{B}_{5} & = & x_{3} \, \mathbf{a}_{1} + x_{3} \, \mathbf{a}_{2} + x_{3} \, \mathbf{a}_{3} & = & x_{3}c \, \mathbf{\hat{z}} & \left(2c\right) & \mbox{Sm} \\ 
\mathbf{B}_{6} & = & -x_{3} \, \mathbf{a}_{1}-x_{3} \, \mathbf{a}_{2}-x_{3} \, \mathbf{a}_{3} & = & -x_{3}c \, \mathbf{\hat{z}} & \left(2c\right) & \mbox{Sm} \\ 
\end{longtabu}
\renewcommand{\arraystretch}{1.0}
\noindent \hrulefill
\\
\textbf{References:}
\vspace*{-0.25cm}
\begin{flushleft}
  - \bibentry{Beck_ZAAC_535_1986}. \\
\end{flushleft}
\textbf{Found in:}
\vspace*{-0.25cm}
\begin{flushleft}
  - \bibentry{Fogg_ChemComm_0_1998}. \\
\end{flushleft}
\noindent \hrulefill
\\
\textbf{Geometry files:}
\\
\noindent  - CIF: pp. {\hyperref[ABC_hR6_166_c_c_c_cif]{\pageref{ABC_hR6_166_c_c_c_cif}}} \\
\noindent  - POSCAR: pp. {\hyperref[ABC_hR6_166_c_c_c_poscar]{\pageref{ABC_hR6_166_c_c_c_poscar}}} \\
\onecolumn
{\phantomsection\label{AB3C_hR10_167_b_e_a}}
\subsection*{\huge \textbf{{\normalfont PrNiO$_{3}$ Structure: AB3C\_hR10\_167\_b\_e\_a}}}
\noindent \hrulefill
\vspace*{0.25cm}
\begin{figure}[htp]
  \centering
  \vspace{-1em}
  {\includegraphics[width=1\textwidth]{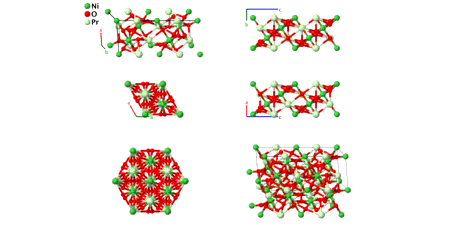}}
\end{figure}
\vspace*{-0.5cm}
\renewcommand{\arraystretch}{1.5}
\begin{equation*}
  \begin{array}{>{$\hspace{-0.15cm}}l<{$}>{$}p{0.5cm}<{$}>{$}p{18.5cm}<{$}}
    \mbox{\large \textbf{Prototype}} &\colon & \ce{PrNiO3} \\
    \mbox{\large \textbf{\AFLOW\ prototype label}} &\colon & \mbox{AB3C\_hR10\_167\_b\_e\_a} \\
    \mbox{\large \textbf{\textit{Strukturbericht} designation}} &\colon & \mbox{None} \\
    \mbox{\large \textbf{Pearson symbol}} &\colon & \mbox{hR10} \\
    \mbox{\large \textbf{Space group number}} &\colon & 167 \\
    \mbox{\large \textbf{Space group symbol}} &\colon & R\bar{3}c \\
    \mbox{\large \textbf{\AFLOW\ prototype command}} &\colon &  \texttt{aflow} \,  \, \texttt{-{}-proto=AB3C\_hR10\_167\_b\_e\_a [-{}-hex]} \, \newline \texttt{-{}-params=}{a,c/a,x_{3} }
  \end{array}
\end{equation*}
\renewcommand{\arraystretch}{1.0}

\vspace*{-0.25cm}
\noindent \hrulefill
\\
\textbf{ Other compounds with this structure:}
\begin{itemize}
   \item{ LaNiO$_{3}$  }
\end{itemize}
\vspace*{-0.25cm}
\noindent \hrulefill
\begin{itemize}
  \item{This structure has the same crystal structure and occupies the same
Wyckoff positions as \href{http://aflow.org/CrystalDatabase/ABC3_hR10_167_a_b_e.CaCO3.html}{Calcite (CaC$O_{3}$ {\em Strukturbericht} $G0_{1}$, ABC3\_hR10\_167\_a\_b\_e)},
but $c/a$ and $x_{3}$ are different enough to warrant calling this a
new structure.
This is the high-temperature form of PrNiO$_{3}$, and we present the data collected at 500$^{\circ}$~C. 
Below 500$^{\circ}$~C this compound transforms to the 
\href{http://aflow.org/CrystalDatabase/AB3C_oP20_62_c_cd_a.html}{orthorhombic perovskite structure (AB3C\_oP20\_62\_c\_cd\_a)}.
}
\end{itemize}

\noindent \parbox{1 \linewidth}{
\noindent \hrulefill
\\
\textbf{Rhombohedral primitive vectors:} \\
\vspace*{-0.25cm}
\begin{tabular}{cc}
  \begin{tabular}{c}
    \parbox{0.6 \linewidth}{
      \renewcommand{\arraystretch}{1.5}
      \begin{equation*}
        \centering
        \begin{array}{ccc}
              \mathbf{a}_1 & = & ~ \frac12 \, a \, \mathbf{\hat{x}} - \frac{1}{2\sqrt{3}} \, a \, \mathbf{\hat{y}} + \frac13 \, c \, \mathbf{\hat{z}} \\
    \mathbf{a}_2 & = & \frac{1}{\sqrt{3}} \, a \, \mathbf{\hat{y}} + \frac13 \, c \, \mathbf{\hat{z}} \\
    \mathbf{a}_3 & = & - \frac12 \, a \, \mathbf{\hat{x}} - \frac{1}{2\sqrt{3}} \, a \, \mathbf{\hat{y}} + \frac13 \, c \, \mathbf{\hat{z}} \\

        \end{array}
      \end{equation*}
    }
    \renewcommand{\arraystretch}{1.0}
  \end{tabular}
  \begin{tabular}{c}
    \includegraphics[width=0.3\linewidth]{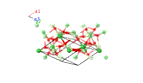} \\
  \end{tabular}
\end{tabular}

}
\vspace*{-0.25cm}

\noindent \hrulefill
\\
\textbf{Basis vectors:}
\vspace*{-0.25cm}
\renewcommand{\arraystretch}{1.5}
\begin{longtabu} to \textwidth{>{\centering $}X[-1,c,c]<{$}>{\centering $}X[-1,c,c]<{$}>{\centering $}X[-1,c,c]<{$}>{\centering $}X[-1,c,c]<{$}>{\centering $}X[-1,c,c]<{$}>{\centering $}X[-1,c,c]<{$}>{\centering $}X[-1,c,c]<{$}}
  & & \mbox{Lattice Coordinates} & & \mbox{Cartesian Coordinates} &\mbox{Wyckoff Position} & \mbox{Atom Type} \\  
  \mathbf{B}_{1} & = & \frac{1}{4} \, \mathbf{a}_{1} + \frac{1}{4} \, \mathbf{a}_{2} + \frac{1}{4} \, \mathbf{a}_{3} & = & \frac{1}{4}c \, \mathbf{\hat{z}} & \left(2a\right) & \mbox{Pr} \\ 
\mathbf{B}_{2} & = & \frac{3}{4} \, \mathbf{a}_{1} + \frac{3}{4} \, \mathbf{a}_{2} + \frac{3}{4} \, \mathbf{a}_{3} & = & \frac{3}{4}c \, \mathbf{\hat{z}} & \left(2a\right) & \mbox{Pr} \\ 
\mathbf{B}_{3} & = & 0 \, \mathbf{a}_{1} + 0 \, \mathbf{a}_{2} + 0 \, \mathbf{a}_{3} & = & 0 \, \mathbf{\hat{x}} + 0 \, \mathbf{\hat{y}} + 0 \, \mathbf{\hat{z}} & \left(2b\right) & \mbox{Ni} \\ 
\mathbf{B}_{4} & = & \frac{1}{2} \, \mathbf{a}_{1} + \frac{1}{2} \, \mathbf{a}_{2} + \frac{1}{2} \, \mathbf{a}_{3} & = & \frac{1}{2}c \, \mathbf{\hat{z}} & \left(2b\right) & \mbox{Ni} \\ 
\mathbf{B}_{5} & = & x_{3} \, \mathbf{a}_{1} + \left(\frac{1}{2} - x_{3}\right) \, \mathbf{a}_{2} + \frac{1}{4} \, \mathbf{a}_{3} & = & \left(- \frac{1}{8} +\frac{1}{2}x_{3}\right)a \, \mathbf{\hat{x}} + \left(\frac{\sqrt{3}}{8} - \frac{\sqrt{3}}{2}x_{3}\right)a \, \mathbf{\hat{y}} + \frac{1}{4}c \, \mathbf{\hat{z}} & \left(6e\right) & \mbox{O} \\ 
\mathbf{B}_{6} & = & \frac{1}{4} \, \mathbf{a}_{1} + x_{3} \, \mathbf{a}_{2} + \left(\frac{1}{2} - x_{3}\right) \, \mathbf{a}_{3} & = & \left(- \frac{1}{8} +\frac{1}{2}x_{3}\right)a \, \mathbf{\hat{x}} + \left(- \frac{\sqrt{3}}{8} +\frac{\sqrt{3}}{2}x_{3}\right)a \, \mathbf{\hat{y}} + \frac{1}{4}c \, \mathbf{\hat{z}} & \left(6e\right) & \mbox{O} \\ 
\mathbf{B}_{7} & = & \left(\frac{1}{2} - x_{3}\right) \, \mathbf{a}_{1} + \frac{1}{4} \, \mathbf{a}_{2} + x_{3} \, \mathbf{a}_{3} & = & \left(\frac{1}{4} - x_{3}\right)a \, \mathbf{\hat{x}} + \frac{1}{4}c \, \mathbf{\hat{z}} & \left(6e\right) & \mbox{O} \\ 
\mathbf{B}_{8} & = & -x_{3} \, \mathbf{a}_{1} + \left(\frac{1}{2} +x_{3}\right) \, \mathbf{a}_{2} + \frac{3}{4} \, \mathbf{a}_{3} & = & -a\left(\frac{1}{2}x_{3}+\frac{3}{8}\right) \, \mathbf{\hat{x}} + \left(\frac{1}{8\sqrt{3}} +\frac{\sqrt{3}}{2}x_{3}\right)a \, \mathbf{\hat{y}} + \frac{5}{12}c \, \mathbf{\hat{z}} & \left(6e\right) & \mbox{O} \\ 
\mathbf{B}_{9} & = & \frac{3}{4} \, \mathbf{a}_{1}-x_{3} \, \mathbf{a}_{2} + \left(\frac{1}{2} +x_{3}\right) \, \mathbf{a}_{3} & = & \left(\frac{1}{8} - \frac{1}{2}x_{3}\right)a \, \mathbf{\hat{x}}-a\left(\frac{\sqrt{3}}{2}x_{3}+\frac{5}{8\sqrt{3}}\right) \, \mathbf{\hat{y}} + \frac{5}{12}c \, \mathbf{\hat{z}} & \left(6e\right) & \mbox{O} \\ 
\mathbf{B}_{10} & = & \left(\frac{1}{2} +x_{3}\right) \, \mathbf{a}_{1} + \frac{3}{4} \, \mathbf{a}_{2}-x_{3} \, \mathbf{a}_{3} & = & \left(\frac{1}{4} +x_{3}\right)a \, \mathbf{\hat{x}} + \frac{1}{2\sqrt{3}}a \, \mathbf{\hat{y}} + \frac{5}{12}c \, \mathbf{\hat{z}} & \left(6e\right) & \mbox{O} \\ 
\end{longtabu}
\renewcommand{\arraystretch}{1.0}
\noindent \hrulefill
\\
\textbf{References:}
\vspace*{-0.25cm}
\begin{flushleft}
  - \bibentry{Huang_Mat_Res_Bull_25_1019_1990}. \\
\end{flushleft}
\noindent \hrulefill
\\
\textbf{Geometry files:}
\\
\noindent  - CIF: pp. {\hyperref[AB3C_hR10_167_b_e_a_cif]{\pageref{AB3C_hR10_167_b_e_a_cif}}} \\
\noindent  - POSCAR: pp. {\hyperref[AB3C_hR10_167_b_e_a_poscar]{\pageref{AB3C_hR10_167_b_e_a_poscar}}} \\
\onecolumn
{\phantomsection\label{ABC2_hR24_167_e_e_2e}}
\subsection*{\huge \textbf{{\normalfont KBO$_{2}$ ($F5_{13}$) Structure: ABC2\_hR24\_167\_e\_e\_2e}}}
\noindent \hrulefill
\vspace*{0.25cm}
\begin{figure}[htp]
  \centering
  \vspace{-1em}
  {\includegraphics[width=1\textwidth]{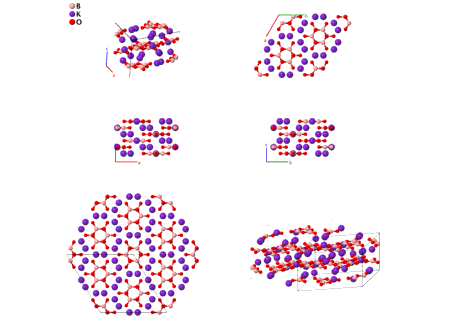}}
\end{figure}
\vspace*{-0.5cm}
\renewcommand{\arraystretch}{1.5}
\begin{equation*}
  \begin{array}{>{$\hspace{-0.15cm}}l<{$}>{$}p{0.5cm}<{$}>{$}p{18.5cm}<{$}}
    \mbox{\large \textbf{Prototype}} &\colon & \ce{KBO2} \\
    \mbox{\large \textbf{\AFLOW\ prototype label}} &\colon & \mbox{ABC2\_hR24\_167\_e\_e\_2e} \\
    \mbox{\large \textbf{\textit{Strukturbericht} designation}} &\colon & \mbox{$F5_{13}$} \\
    \mbox{\large \textbf{Pearson symbol}} &\colon & \mbox{hR24} \\
    \mbox{\large \textbf{Space group number}} &\colon & 167 \\
    \mbox{\large \textbf{Space group symbol}} &\colon & R\bar{3}c \\
    \mbox{\large \textbf{\AFLOW\ prototype command}} &\colon &  \texttt{aflow} \,  \, \texttt{-{}-proto=ABC2\_hR24\_167\_e\_e\_2e [-{}-hex]} \, \newline \texttt{-{}-params=}{a,c/a,x_{1},x_{2},x_{3},x_{4} }
  \end{array}
\end{equation*}
\renewcommand{\arraystretch}{1.0}

\vspace*{-0.25cm}
\noindent \hrulefill
\\
\textbf{ Other compounds with this structure:}
\begin{itemize}
   \item{ NaBO$_{2}$, NaBS$_{2}$   }
\end{itemize}
\noindent \parbox{1 \linewidth}{
\noindent \hrulefill
\\
\textbf{Rhombohedral primitive vectors:} \\
\vspace*{-0.25cm}
\begin{tabular}{cc}
  \begin{tabular}{c}
    \parbox{0.6 \linewidth}{
      \renewcommand{\arraystretch}{1.5}
      \begin{equation*}
        \centering
        \begin{array}{ccc}
              \mathbf{a}_1 & = & ~ \frac12 \, a \, \mathbf{\hat{x}} - \frac{1}{2\sqrt{3}} \, a \, \mathbf{\hat{y}} + \frac13 \, c \, \mathbf{\hat{z}} \\
    \mathbf{a}_2 & = & \frac{1}{\sqrt{3}} \, a \, \mathbf{\hat{y}} + \frac13 \, c \, \mathbf{\hat{z}} \\
    \mathbf{a}_3 & = & - \frac12 \, a \, \mathbf{\hat{x}} - \frac{1}{2\sqrt{3}} \, a \, \mathbf{\hat{y}} + \frac13 \, c \, \mathbf{\hat{z}} \\

        \end{array}
      \end{equation*}
    }
    \renewcommand{\arraystretch}{1.0}
  \end{tabular}
  \begin{tabular}{c}
    \includegraphics[width=0.3\linewidth]{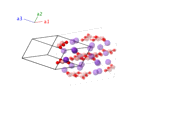} \\
  \end{tabular}
\end{tabular}

}
\vspace*{-0.25cm}

\noindent \hrulefill
\\
\textbf{Basis vectors:}
\vspace*{-0.25cm}
\renewcommand{\arraystretch}{1.5}
\begin{longtabu} to \textwidth{>{\centering $}X[-1,c,c]<{$}>{\centering $}X[-1,c,c]<{$}>{\centering $}X[-1,c,c]<{$}>{\centering $}X[-1,c,c]<{$}>{\centering $}X[-1,c,c]<{$}>{\centering $}X[-1,c,c]<{$}>{\centering $}X[-1,c,c]<{$}}
  & & \mbox{Lattice Coordinates} & & \mbox{Cartesian Coordinates} &\mbox{Wyckoff Position} & \mbox{Atom Type} \\  
  \mathbf{B}_{1} & = & x_{1} \, \mathbf{a}_{1} + \left(\frac{1}{2} - x_{1}\right) \, \mathbf{a}_{2} + \frac{1}{4} \, \mathbf{a}_{3} & = & \left(- \frac{1}{8} +\frac{1}{2}x_{1}\right)a \, \mathbf{\hat{x}} + \left(\frac{\sqrt{3}}{8} - \frac{\sqrt{3}}{2}x_{1}\right)a \, \mathbf{\hat{y}} + \frac{1}{4}c \, \mathbf{\hat{z}} & \left(6e\right) & \mbox{B} \\ 
\mathbf{B}_{2} & = & \frac{1}{4} \, \mathbf{a}_{1} + x_{1} \, \mathbf{a}_{2} + \left(\frac{1}{2} - x_{1}\right) \, \mathbf{a}_{3} & = & \left(- \frac{1}{8} +\frac{1}{2}x_{1}\right)a \, \mathbf{\hat{x}} + \left(- \frac{\sqrt{3}}{8} +\frac{\sqrt{3}}{2}x_{1}\right)a \, \mathbf{\hat{y}} + \frac{1}{4}c \, \mathbf{\hat{z}} & \left(6e\right) & \mbox{B} \\ 
\mathbf{B}_{3} & = & \left(\frac{1}{2} - x_{1}\right) \, \mathbf{a}_{1} + \frac{1}{4} \, \mathbf{a}_{2} + x_{1} \, \mathbf{a}_{3} & = & \left(\frac{1}{4} - x_{1}\right)a \, \mathbf{\hat{x}} + \frac{1}{4}c \, \mathbf{\hat{z}} & \left(6e\right) & \mbox{B} \\ 
\mathbf{B}_{4} & = & -x_{1} \, \mathbf{a}_{1} + \left(\frac{1}{2} +x_{1}\right) \, \mathbf{a}_{2} + \frac{3}{4} \, \mathbf{a}_{3} & = & -a\left(\frac{1}{2}x_{1}+\frac{3}{8}\right) \, \mathbf{\hat{x}} + \left(\frac{1}{8\sqrt{3}} +\frac{\sqrt{3}}{2}x_{1}\right)a \, \mathbf{\hat{y}} + \frac{5}{12}c \, \mathbf{\hat{z}} & \left(6e\right) & \mbox{B} \\ 
\mathbf{B}_{5} & = & \frac{3}{4} \, \mathbf{a}_{1}-x_{1} \, \mathbf{a}_{2} + \left(\frac{1}{2} +x_{1}\right) \, \mathbf{a}_{3} & = & \left(\frac{1}{8} - \frac{1}{2}x_{1}\right)a \, \mathbf{\hat{x}}-a\left(\frac{\sqrt{3}}{2}x_{1}+\frac{5}{8\sqrt{3}}\right) \, \mathbf{\hat{y}} + \frac{5}{12}c \, \mathbf{\hat{z}} & \left(6e\right) & \mbox{B} \\ 
\mathbf{B}_{6} & = & \left(\frac{1}{2} +x_{1}\right) \, \mathbf{a}_{1} + \frac{3}{4} \, \mathbf{a}_{2}-x_{1} \, \mathbf{a}_{3} & = & \left(\frac{1}{4} +x_{1}\right)a \, \mathbf{\hat{x}} + \frac{1}{2\sqrt{3}}a \, \mathbf{\hat{y}} + \frac{5}{12}c \, \mathbf{\hat{z}} & \left(6e\right) & \mbox{B} \\ 
\mathbf{B}_{7} & = & x_{2} \, \mathbf{a}_{1} + \left(\frac{1}{2} - x_{2}\right) \, \mathbf{a}_{2} + \frac{1}{4} \, \mathbf{a}_{3} & = & \left(- \frac{1}{8} +\frac{1}{2}x_{2}\right)a \, \mathbf{\hat{x}} + \left(\frac{\sqrt{3}}{8} - \frac{\sqrt{3}}{2}x_{2}\right)a \, \mathbf{\hat{y}} + \frac{1}{4}c \, \mathbf{\hat{z}} & \left(6e\right) & \mbox{K} \\ 
\mathbf{B}_{8} & = & \frac{1}{4} \, \mathbf{a}_{1} + x_{2} \, \mathbf{a}_{2} + \left(\frac{1}{2} - x_{2}\right) \, \mathbf{a}_{3} & = & \left(- \frac{1}{8} +\frac{1}{2}x_{2}\right)a \, \mathbf{\hat{x}} + \left(- \frac{\sqrt{3}}{8} +\frac{\sqrt{3}}{2}x_{2}\right)a \, \mathbf{\hat{y}} + \frac{1}{4}c \, \mathbf{\hat{z}} & \left(6e\right) & \mbox{K} \\ 
\mathbf{B}_{9} & = & \left(\frac{1}{2} - x_{2}\right) \, \mathbf{a}_{1} + \frac{1}{4} \, \mathbf{a}_{2} + x_{2} \, \mathbf{a}_{3} & = & \left(\frac{1}{4} - x_{2}\right)a \, \mathbf{\hat{x}} + \frac{1}{4}c \, \mathbf{\hat{z}} & \left(6e\right) & \mbox{K} \\ 
\mathbf{B}_{10} & = & -x_{2} \, \mathbf{a}_{1} + \left(\frac{1}{2} +x_{2}\right) \, \mathbf{a}_{2} + \frac{3}{4} \, \mathbf{a}_{3} & = & -a\left(\frac{1}{2}x_{2}+\frac{3}{8}\right) \, \mathbf{\hat{x}} + \left(\frac{1}{8\sqrt{3}} +\frac{\sqrt{3}}{2}x_{2}\right)a \, \mathbf{\hat{y}} + \frac{5}{12}c \, \mathbf{\hat{z}} & \left(6e\right) & \mbox{K} \\ 
\mathbf{B}_{11} & = & \frac{3}{4} \, \mathbf{a}_{1}-x_{2} \, \mathbf{a}_{2} + \left(\frac{1}{2} +x_{2}\right) \, \mathbf{a}_{3} & = & \left(\frac{1}{8} - \frac{1}{2}x_{2}\right)a \, \mathbf{\hat{x}}-a\left(\frac{\sqrt{3}}{2}x_{2}+\frac{5}{8\sqrt{3}}\right) \, \mathbf{\hat{y}} + \frac{5}{12}c \, \mathbf{\hat{z}} & \left(6e\right) & \mbox{K} \\ 
\mathbf{B}_{12} & = & \left(\frac{1}{2} +x_{2}\right) \, \mathbf{a}_{1} + \frac{3}{4} \, \mathbf{a}_{2}-x_{2} \, \mathbf{a}_{3} & = & \left(\frac{1}{4} +x_{2}\right)a \, \mathbf{\hat{x}} + \frac{1}{2\sqrt{3}}a \, \mathbf{\hat{y}} + \frac{5}{12}c \, \mathbf{\hat{z}} & \left(6e\right) & \mbox{K} \\ 
\mathbf{B}_{13} & = & x_{3} \, \mathbf{a}_{1} + \left(\frac{1}{2} - x_{3}\right) \, \mathbf{a}_{2} + \frac{1}{4} \, \mathbf{a}_{3} & = & \left(- \frac{1}{8} +\frac{1}{2}x_{3}\right)a \, \mathbf{\hat{x}} + \left(\frac{\sqrt{3}}{8} - \frac{\sqrt{3}}{2}x_{3}\right)a \, \mathbf{\hat{y}} + \frac{1}{4}c \, \mathbf{\hat{z}} & \left(6e\right) & \mbox{O I} \\ 
\mathbf{B}_{14} & = & \frac{1}{4} \, \mathbf{a}_{1} + x_{3} \, \mathbf{a}_{2} + \left(\frac{1}{2} - x_{3}\right) \, \mathbf{a}_{3} & = & \left(- \frac{1}{8} +\frac{1}{2}x_{3}\right)a \, \mathbf{\hat{x}} + \left(- \frac{\sqrt{3}}{8} +\frac{\sqrt{3}}{2}x_{3}\right)a \, \mathbf{\hat{y}} + \frac{1}{4}c \, \mathbf{\hat{z}} & \left(6e\right) & \mbox{O I} \\ 
\mathbf{B}_{15} & = & \left(\frac{1}{2} - x_{3}\right) \, \mathbf{a}_{1} + \frac{1}{4} \, \mathbf{a}_{2} + x_{3} \, \mathbf{a}_{3} & = & \left(\frac{1}{4} - x_{3}\right)a \, \mathbf{\hat{x}} + \frac{1}{4}c \, \mathbf{\hat{z}} & \left(6e\right) & \mbox{O I} \\ 
\mathbf{B}_{16} & = & -x_{3} \, \mathbf{a}_{1} + \left(\frac{1}{2} +x_{3}\right) \, \mathbf{a}_{2} + \frac{3}{4} \, \mathbf{a}_{3} & = & -a\left(\frac{1}{2}x_{3}+\frac{3}{8}\right) \, \mathbf{\hat{x}} + \left(\frac{1}{8\sqrt{3}} +\frac{\sqrt{3}}{2}x_{3}\right)a \, \mathbf{\hat{y}} + \frac{5}{12}c \, \mathbf{\hat{z}} & \left(6e\right) & \mbox{O I} \\ 
\mathbf{B}_{17} & = & \frac{3}{4} \, \mathbf{a}_{1}-x_{3} \, \mathbf{a}_{2} + \left(\frac{1}{2} +x_{3}\right) \, \mathbf{a}_{3} & = & \left(\frac{1}{8} - \frac{1}{2}x_{3}\right)a \, \mathbf{\hat{x}}-a\left(\frac{\sqrt{3}}{2}x_{3}+\frac{5}{8\sqrt{3}}\right) \, \mathbf{\hat{y}} + \frac{5}{12}c \, \mathbf{\hat{z}} & \left(6e\right) & \mbox{O I} \\ 
\mathbf{B}_{18} & = & \left(\frac{1}{2} +x_{3}\right) \, \mathbf{a}_{1} + \frac{3}{4} \, \mathbf{a}_{2}-x_{3} \, \mathbf{a}_{3} & = & \left(\frac{1}{4} +x_{3}\right)a \, \mathbf{\hat{x}} + \frac{1}{2\sqrt{3}}a \, \mathbf{\hat{y}} + \frac{5}{12}c \, \mathbf{\hat{z}} & \left(6e\right) & \mbox{O I} \\ 
\mathbf{B}_{19} & = & x_{4} \, \mathbf{a}_{1} + \left(\frac{1}{2} - x_{4}\right) \, \mathbf{a}_{2} + \frac{1}{4} \, \mathbf{a}_{3} & = & \left(- \frac{1}{8} +\frac{1}{2}x_{4}\right)a \, \mathbf{\hat{x}} + \left(\frac{\sqrt{3}}{8} - \frac{\sqrt{3}}{2}x_{4}\right)a \, \mathbf{\hat{y}} + \frac{1}{4}c \, \mathbf{\hat{z}} & \left(6e\right) & \mbox{O II} \\ 
\mathbf{B}_{20} & = & \frac{1}{4} \, \mathbf{a}_{1} + x_{4} \, \mathbf{a}_{2} + \left(\frac{1}{2} - x_{4}\right) \, \mathbf{a}_{3} & = & \left(- \frac{1}{8} +\frac{1}{2}x_{4}\right)a \, \mathbf{\hat{x}} + \left(- \frac{\sqrt{3}}{8} +\frac{\sqrt{3}}{2}x_{4}\right)a \, \mathbf{\hat{y}} + \frac{1}{4}c \, \mathbf{\hat{z}} & \left(6e\right) & \mbox{O II} \\ 
\mathbf{B}_{21} & = & \left(\frac{1}{2} - x_{4}\right) \, \mathbf{a}_{1} + \frac{1}{4} \, \mathbf{a}_{2} + x_{4} \, \mathbf{a}_{3} & = & \left(\frac{1}{4} - x_{4}\right)a \, \mathbf{\hat{x}} + \frac{1}{4}c \, \mathbf{\hat{z}} & \left(6e\right) & \mbox{O II} \\ 
\mathbf{B}_{22} & = & -x_{4} \, \mathbf{a}_{1} + \left(\frac{1}{2} +x_{4}\right) \, \mathbf{a}_{2} + \frac{3}{4} \, \mathbf{a}_{3} & = & -a\left(\frac{1}{2}x_{4}+\frac{3}{8}\right) \, \mathbf{\hat{x}} + \left(\frac{1}{8\sqrt{3}} +\frac{\sqrt{3}}{2}x_{4}\right)a \, \mathbf{\hat{y}} + \frac{5}{12}c \, \mathbf{\hat{z}} & \left(6e\right) & \mbox{O II} \\ 
\mathbf{B}_{23} & = & \frac{3}{4} \, \mathbf{a}_{1}-x_{4} \, \mathbf{a}_{2} + \left(\frac{1}{2} +x_{4}\right) \, \mathbf{a}_{3} & = & \left(\frac{1}{8} - \frac{1}{2}x_{4}\right)a \, \mathbf{\hat{x}}-a\left(\frac{\sqrt{3}}{2}x_{4}+\frac{5}{8\sqrt{3}}\right) \, \mathbf{\hat{y}} + \frac{5}{12}c \, \mathbf{\hat{z}} & \left(6e\right) & \mbox{O II} \\ 
\mathbf{B}_{24} & = & \left(\frac{1}{2} +x_{4}\right) \, \mathbf{a}_{1} + \frac{3}{4} \, \mathbf{a}_{2}-x_{4} \, \mathbf{a}_{3} & = & \left(\frac{1}{4} +x_{4}\right)a \, \mathbf{\hat{x}} + \frac{1}{2\sqrt{3}}a \, \mathbf{\hat{y}} + \frac{5}{12}c \, \mathbf{\hat{z}} & \left(6e\right) & \mbox{O II} \\ 
\end{longtabu}
\renewcommand{\arraystretch}{1.0}
\noindent \hrulefill
\\
\textbf{References:}
\vspace*{-0.25cm}
\begin{flushleft}
  - \bibentry{Schneider_Acta_Cryst_B_26_1970}. \\
\end{flushleft}
\textbf{Found in:}
\vspace*{-0.25cm}
\begin{flushleft}
  - \bibentry{Villars_Landolt_2007}. \\
\end{flushleft}
\noindent \hrulefill
\\
\textbf{Geometry files:}
\\
\noindent  - CIF: pp. {\hyperref[ABC2_hR24_167_e_e_2e_cif]{\pageref{ABC2_hR24_167_e_e_2e_cif}}} \\
\noindent  - POSCAR: pp. {\hyperref[ABC2_hR24_167_e_e_2e_poscar]{\pageref{ABC2_hR24_167_e_e_2e_poscar}}} \\
\onecolumn
{\phantomsection\label{A2B13C4_hP57_168_d_c6d_2d}}
\subsection*{\huge \textbf{{\normalfont K$_{2}$Ta$_{4}$O$_{9}$F$_{4}$ Structure: A2B13C4\_hP57\_168\_d\_c6d\_2d}}}
\noindent \hrulefill
\vspace*{0.25cm}
\begin{figure}[htp]
  \centering
  \vspace{-1em}
  {\includegraphics[width=1\textwidth]{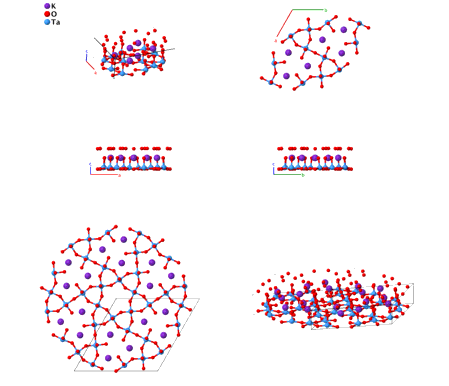}}
\end{figure}
\vspace*{-0.5cm}
\renewcommand{\arraystretch}{1.5}
\begin{equation*}
  \begin{array}{>{$\hspace{-0.15cm}}l<{$}>{$}p{0.5cm}<{$}>{$}p{18.5cm}<{$}}
    \mbox{\large \textbf{Prototype}} &\colon & \ce{K2Ta4O9F4} \\
    \mbox{\large \textbf{\AFLOW\ prototype label}} &\colon & \mbox{A2B13C4\_hP57\_168\_d\_c6d\_2d} \\
    \mbox{\large \textbf{\textit{Strukturbericht} designation}} &\colon & \mbox{None} \\
    \mbox{\large \textbf{Pearson symbol}} &\colon & \mbox{hP57} \\
    \mbox{\large \textbf{Space group number}} &\colon & 168 \\
    \mbox{\large \textbf{Space group symbol}} &\colon & P6 \\
    \mbox{\large \textbf{\AFLOW\ prototype command}} &\colon &  \texttt{aflow} \,  \, \texttt{-{}-proto=A2B13C4\_hP57\_168\_d\_c6d\_2d } \, \newline \texttt{-{}-params=}{a,c/a,z_{1},x_{2},y_{2},z_{2},x_{3},y_{3},z_{3},x_{4},y_{4},z_{4},x_{5},y_{5},z_{5},x_{6},y_{6},z_{6},x_{7},y_{7},z_{7},} \newline {x_{8},y_{8},z_{8},x_{9},y_{9},z_{9},x_{10},y_{10},z_{10} }
  \end{array}
\end{equation*}
\renewcommand{\arraystretch}{1.0}

\vspace*{-0.25cm}
\noindent \hrulefill
\begin{itemize}
  \item{The O sites are partially occupied with the following concentration 0.692O + 0.308F.
}
\end{itemize}

\noindent \parbox{1 \linewidth}{
\noindent \hrulefill
\\
\textbf{Hexagonal primitive vectors:} \\
\vspace*{-0.25cm}
\begin{tabular}{cc}
  \begin{tabular}{c}
    \parbox{0.6 \linewidth}{
      \renewcommand{\arraystretch}{1.5}
      \begin{equation*}
        \centering
        \begin{array}{ccc}
              \mathbf{a}_1 & = & \frac12 \, a \, \mathbf{\hat{x}} - \frac{\sqrt3}2 \, a \, \mathbf{\hat{y}} \\
    \mathbf{a}_2 & = & \frac12 \, a \, \mathbf{\hat{x}} + \frac{\sqrt3}2 \, a \, \mathbf{\hat{y}} \\
    \mathbf{a}_3 & = & c \, \mathbf{\hat{z}} \\

        \end{array}
      \end{equation*}
    }
    \renewcommand{\arraystretch}{1.0}
  \end{tabular}
  \begin{tabular}{c}
    \includegraphics[width=0.3\linewidth]{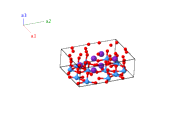} \\
  \end{tabular}
\end{tabular}

}
\vspace*{-0.25cm}

\noindent \hrulefill
\\
\textbf{Basis vectors:}
\vspace*{-0.25cm}
\renewcommand{\arraystretch}{1.5}
\begin{longtabu} to \textwidth{>{\centering $}X[-1,c,c]<{$}>{\centering $}X[-1,c,c]<{$}>{\centering $}X[-1,c,c]<{$}>{\centering $}X[-1,c,c]<{$}>{\centering $}X[-1,c,c]<{$}>{\centering $}X[-1,c,c]<{$}>{\centering $}X[-1,c,c]<{$}}
  & & \mbox{Lattice Coordinates} & & \mbox{Cartesian Coordinates} &\mbox{Wyckoff Position} & \mbox{Atom Type} \\  
  \mathbf{B}_{1} & = & \frac{1}{2} \, \mathbf{a}_{1} + z_{1} \, \mathbf{a}_{3} & = & \frac{1}{4}a \, \mathbf{\hat{x}}- \frac{\sqrt{3}}{4}a  \, \mathbf{\hat{y}} + z_{1}c \, \mathbf{\hat{z}} & \left(3c\right) & \mbox{O I} \\ 
\mathbf{B}_{2} & = & \frac{1}{2} \, \mathbf{a}_{2} + z_{1} \, \mathbf{a}_{3} & = & \frac{1}{4}a \, \mathbf{\hat{x}} + \frac{\sqrt{3}}{4}a \, \mathbf{\hat{y}} + z_{1}c \, \mathbf{\hat{z}} & \left(3c\right) & \mbox{O I} \\ 
\mathbf{B}_{3} & = & \frac{1}{2} \, \mathbf{a}_{1} + \frac{1}{2} \, \mathbf{a}_{2} + z_{1} \, \mathbf{a}_{3} & = & \frac{1}{2}a \, \mathbf{\hat{x}} + z_{1}c \, \mathbf{\hat{z}} & \left(3c\right) & \mbox{O I} \\ 
\mathbf{B}_{4} & = & x_{2} \, \mathbf{a}_{1} + y_{2} \, \mathbf{a}_{2} + z_{2} \, \mathbf{a}_{3} & = & \frac{1}{2}\left(x_{2}+y_{2}\right)a \, \mathbf{\hat{x}} + \frac{\sqrt{3}}{2}\left(-x_{2}+y_{2}\right)a \, \mathbf{\hat{y}} + z_{2}c \, \mathbf{\hat{z}} & \left(6d\right) & \mbox{K} \\ 
\mathbf{B}_{5} & = & -y_{2} \, \mathbf{a}_{1} + \left(x_{2}-y_{2}\right) \, \mathbf{a}_{2} + z_{2} \, \mathbf{a}_{3} & = & \left(\frac{1}{2}x_{2}-y_{2}\right)a \, \mathbf{\hat{x}} + \frac{\sqrt{3}}{2}x_{2}a \, \mathbf{\hat{y}} + z_{2}c \, \mathbf{\hat{z}} & \left(6d\right) & \mbox{K} \\ 
\mathbf{B}_{6} & = & \left(-x_{2}+y_{2}\right) \, \mathbf{a}_{1}-x_{2} \, \mathbf{a}_{2} + z_{2} \, \mathbf{a}_{3} & = & \left(-x_{2}+\frac{1}{2}y_{2}\right)a \, \mathbf{\hat{x}}-\frac{\sqrt{3}}{2}y_{2}a \, \mathbf{\hat{y}} + z_{2}c \, \mathbf{\hat{z}} & \left(6d\right) & \mbox{K} \\ 
\mathbf{B}_{7} & = & -x_{2} \, \mathbf{a}_{1}-y_{2} \, \mathbf{a}_{2} + z_{2} \, \mathbf{a}_{3} & = & -\frac{1}{2}\left(x_{2}+y_{2}\right)a \, \mathbf{\hat{x}} + \frac{\sqrt{3}}{2}\left(x_{2}-y_{2}\right)a \, \mathbf{\hat{y}} + z_{2}c \, \mathbf{\hat{z}} & \left(6d\right) & \mbox{K} \\ 
\mathbf{B}_{8} & = & y_{2} \, \mathbf{a}_{1} + \left(-x_{2}+y_{2}\right) \, \mathbf{a}_{2} + z_{2} \, \mathbf{a}_{3} & = & \left(-\frac{1}{2}x_{2}+y_{2}\right)a \, \mathbf{\hat{x}}-\frac{\sqrt{3}}{2}x_{2}a \, \mathbf{\hat{y}} + z_{2}c \, \mathbf{\hat{z}} & \left(6d\right) & \mbox{K} \\ 
\mathbf{B}_{9} & = & \left(x_{2}-y_{2}\right) \, \mathbf{a}_{1} + x_{2} \, \mathbf{a}_{2} + z_{2} \, \mathbf{a}_{3} & = & \left(x_{2}-\frac{1}{2}y_{2}\right)a \, \mathbf{\hat{x}} + \frac{\sqrt{3}}{2}y_{2}a \, \mathbf{\hat{y}} + z_{2}c \, \mathbf{\hat{z}} & \left(6d\right) & \mbox{K} \\ 
\mathbf{B}_{10} & = & x_{3} \, \mathbf{a}_{1} + y_{3} \, \mathbf{a}_{2} + z_{3} \, \mathbf{a}_{3} & = & \frac{1}{2}\left(x_{3}+y_{3}\right)a \, \mathbf{\hat{x}} + \frac{\sqrt{3}}{2}\left(-x_{3}+y_{3}\right)a \, \mathbf{\hat{y}} + z_{3}c \, \mathbf{\hat{z}} & \left(6d\right) & \mbox{O II} \\ 
\mathbf{B}_{11} & = & -y_{3} \, \mathbf{a}_{1} + \left(x_{3}-y_{3}\right) \, \mathbf{a}_{2} + z_{3} \, \mathbf{a}_{3} & = & \left(\frac{1}{2}x_{3}-y_{3}\right)a \, \mathbf{\hat{x}} + \frac{\sqrt{3}}{2}x_{3}a \, \mathbf{\hat{y}} + z_{3}c \, \mathbf{\hat{z}} & \left(6d\right) & \mbox{O II} \\ 
\mathbf{B}_{12} & = & \left(-x_{3}+y_{3}\right) \, \mathbf{a}_{1}-x_{3} \, \mathbf{a}_{2} + z_{3} \, \mathbf{a}_{3} & = & \left(-x_{3}+\frac{1}{2}y_{3}\right)a \, \mathbf{\hat{x}}-\frac{\sqrt{3}}{2}y_{3}a \, \mathbf{\hat{y}} + z_{3}c \, \mathbf{\hat{z}} & \left(6d\right) & \mbox{O II} \\ 
\mathbf{B}_{13} & = & -x_{3} \, \mathbf{a}_{1}-y_{3} \, \mathbf{a}_{2} + z_{3} \, \mathbf{a}_{3} & = & -\frac{1}{2}\left(x_{3}+y_{3}\right)a \, \mathbf{\hat{x}} + \frac{\sqrt{3}}{2}\left(x_{3}-y_{3}\right)a \, \mathbf{\hat{y}} + z_{3}c \, \mathbf{\hat{z}} & \left(6d\right) & \mbox{O II} \\ 
\mathbf{B}_{14} & = & y_{3} \, \mathbf{a}_{1} + \left(-x_{3}+y_{3}\right) \, \mathbf{a}_{2} + z_{3} \, \mathbf{a}_{3} & = & \left(-\frac{1}{2}x_{3}+y_{3}\right)a \, \mathbf{\hat{x}}-\frac{\sqrt{3}}{2}x_{3}a \, \mathbf{\hat{y}} + z_{3}c \, \mathbf{\hat{z}} & \left(6d\right) & \mbox{O II} \\ 
\mathbf{B}_{15} & = & \left(x_{3}-y_{3}\right) \, \mathbf{a}_{1} + x_{3} \, \mathbf{a}_{2} + z_{3} \, \mathbf{a}_{3} & = & \left(x_{3}-\frac{1}{2}y_{3}\right)a \, \mathbf{\hat{x}} + \frac{\sqrt{3}}{2}y_{3}a \, \mathbf{\hat{y}} + z_{3}c \, \mathbf{\hat{z}} & \left(6d\right) & \mbox{O II} \\ 
\mathbf{B}_{16} & = & x_{4} \, \mathbf{a}_{1} + y_{4} \, \mathbf{a}_{2} + z_{4} \, \mathbf{a}_{3} & = & \frac{1}{2}\left(x_{4}+y_{4}\right)a \, \mathbf{\hat{x}} + \frac{\sqrt{3}}{2}\left(-x_{4}+y_{4}\right)a \, \mathbf{\hat{y}} + z_{4}c \, \mathbf{\hat{z}} & \left(6d\right) & \mbox{O III} \\ 
\mathbf{B}_{17} & = & -y_{4} \, \mathbf{a}_{1} + \left(x_{4}-y_{4}\right) \, \mathbf{a}_{2} + z_{4} \, \mathbf{a}_{3} & = & \left(\frac{1}{2}x_{4}-y_{4}\right)a \, \mathbf{\hat{x}} + \frac{\sqrt{3}}{2}x_{4}a \, \mathbf{\hat{y}} + z_{4}c \, \mathbf{\hat{z}} & \left(6d\right) & \mbox{O III} \\ 
\mathbf{B}_{18} & = & \left(-x_{4}+y_{4}\right) \, \mathbf{a}_{1}-x_{4} \, \mathbf{a}_{2} + z_{4} \, \mathbf{a}_{3} & = & \left(-x_{4}+\frac{1}{2}y_{4}\right)a \, \mathbf{\hat{x}}-\frac{\sqrt{3}}{2}y_{4}a \, \mathbf{\hat{y}} + z_{4}c \, \mathbf{\hat{z}} & \left(6d\right) & \mbox{O III} \\ 
\mathbf{B}_{19} & = & -x_{4} \, \mathbf{a}_{1}-y_{4} \, \mathbf{a}_{2} + z_{4} \, \mathbf{a}_{3} & = & -\frac{1}{2}\left(x_{4}+y_{4}\right)a \, \mathbf{\hat{x}} + \frac{\sqrt{3}}{2}\left(x_{4}-y_{4}\right)a \, \mathbf{\hat{y}} + z_{4}c \, \mathbf{\hat{z}} & \left(6d\right) & \mbox{O III} \\ 
\mathbf{B}_{20} & = & y_{4} \, \mathbf{a}_{1} + \left(-x_{4}+y_{4}\right) \, \mathbf{a}_{2} + z_{4} \, \mathbf{a}_{3} & = & \left(-\frac{1}{2}x_{4}+y_{4}\right)a \, \mathbf{\hat{x}}-\frac{\sqrt{3}}{2}x_{4}a \, \mathbf{\hat{y}} + z_{4}c \, \mathbf{\hat{z}} & \left(6d\right) & \mbox{O III} \\ 
\mathbf{B}_{21} & = & \left(x_{4}-y_{4}\right) \, \mathbf{a}_{1} + x_{4} \, \mathbf{a}_{2} + z_{4} \, \mathbf{a}_{3} & = & \left(x_{4}-\frac{1}{2}y_{4}\right)a \, \mathbf{\hat{x}} + \frac{\sqrt{3}}{2}y_{4}a \, \mathbf{\hat{y}} + z_{4}c \, \mathbf{\hat{z}} & \left(6d\right) & \mbox{O III} \\ 
\mathbf{B}_{22} & = & x_{5} \, \mathbf{a}_{1} + y_{5} \, \mathbf{a}_{2} + z_{5} \, \mathbf{a}_{3} & = & \frac{1}{2}\left(x_{5}+y_{5}\right)a \, \mathbf{\hat{x}} + \frac{\sqrt{3}}{2}\left(-x_{5}+y_{5}\right)a \, \mathbf{\hat{y}} + z_{5}c \, \mathbf{\hat{z}} & \left(6d\right) & \mbox{O IV} \\ 
\mathbf{B}_{23} & = & -y_{5} \, \mathbf{a}_{1} + \left(x_{5}-y_{5}\right) \, \mathbf{a}_{2} + z_{5} \, \mathbf{a}_{3} & = & \left(\frac{1}{2}x_{5}-y_{5}\right)a \, \mathbf{\hat{x}} + \frac{\sqrt{3}}{2}x_{5}a \, \mathbf{\hat{y}} + z_{5}c \, \mathbf{\hat{z}} & \left(6d\right) & \mbox{O IV} \\ 
\mathbf{B}_{24} & = & \left(-x_{5}+y_{5}\right) \, \mathbf{a}_{1}-x_{5} \, \mathbf{a}_{2} + z_{5} \, \mathbf{a}_{3} & = & \left(-x_{5}+\frac{1}{2}y_{5}\right)a \, \mathbf{\hat{x}}-\frac{\sqrt{3}}{2}y_{5}a \, \mathbf{\hat{y}} + z_{5}c \, \mathbf{\hat{z}} & \left(6d\right) & \mbox{O IV} \\ 
\mathbf{B}_{25} & = & -x_{5} \, \mathbf{a}_{1}-y_{5} \, \mathbf{a}_{2} + z_{5} \, \mathbf{a}_{3} & = & -\frac{1}{2}\left(x_{5}+y_{5}\right)a \, \mathbf{\hat{x}} + \frac{\sqrt{3}}{2}\left(x_{5}-y_{5}\right)a \, \mathbf{\hat{y}} + z_{5}c \, \mathbf{\hat{z}} & \left(6d\right) & \mbox{O IV} \\ 
\mathbf{B}_{26} & = & y_{5} \, \mathbf{a}_{1} + \left(-x_{5}+y_{5}\right) \, \mathbf{a}_{2} + z_{5} \, \mathbf{a}_{3} & = & \left(-\frac{1}{2}x_{5}+y_{5}\right)a \, \mathbf{\hat{x}}-\frac{\sqrt{3}}{2}x_{5}a \, \mathbf{\hat{y}} + z_{5}c \, \mathbf{\hat{z}} & \left(6d\right) & \mbox{O IV} \\ 
\mathbf{B}_{27} & = & \left(x_{5}-y_{5}\right) \, \mathbf{a}_{1} + x_{5} \, \mathbf{a}_{2} + z_{5} \, \mathbf{a}_{3} & = & \left(x_{5}-\frac{1}{2}y_{5}\right)a \, \mathbf{\hat{x}} + \frac{\sqrt{3}}{2}y_{5}a \, \mathbf{\hat{y}} + z_{5}c \, \mathbf{\hat{z}} & \left(6d\right) & \mbox{O IV} \\ 
\mathbf{B}_{28} & = & x_{6} \, \mathbf{a}_{1} + y_{6} \, \mathbf{a}_{2} + z_{6} \, \mathbf{a}_{3} & = & \frac{1}{2}\left(x_{6}+y_{6}\right)a \, \mathbf{\hat{x}} + \frac{\sqrt{3}}{2}\left(-x_{6}+y_{6}\right)a \, \mathbf{\hat{y}} + z_{6}c \, \mathbf{\hat{z}} & \left(6d\right) & \mbox{O V} \\ 
\mathbf{B}_{29} & = & -y_{6} \, \mathbf{a}_{1} + \left(x_{6}-y_{6}\right) \, \mathbf{a}_{2} + z_{6} \, \mathbf{a}_{3} & = & \left(\frac{1}{2}x_{6}-y_{6}\right)a \, \mathbf{\hat{x}} + \frac{\sqrt{3}}{2}x_{6}a \, \mathbf{\hat{y}} + z_{6}c \, \mathbf{\hat{z}} & \left(6d\right) & \mbox{O V} \\ 
\mathbf{B}_{30} & = & \left(-x_{6}+y_{6}\right) \, \mathbf{a}_{1}-x_{6} \, \mathbf{a}_{2} + z_{6} \, \mathbf{a}_{3} & = & \left(-x_{6}+\frac{1}{2}y_{6}\right)a \, \mathbf{\hat{x}}-\frac{\sqrt{3}}{2}y_{6}a \, \mathbf{\hat{y}} + z_{6}c \, \mathbf{\hat{z}} & \left(6d\right) & \mbox{O V} \\ 
\mathbf{B}_{31} & = & -x_{6} \, \mathbf{a}_{1}-y_{6} \, \mathbf{a}_{2} + z_{6} \, \mathbf{a}_{3} & = & -\frac{1}{2}\left(x_{6}+y_{6}\right)a \, \mathbf{\hat{x}} + \frac{\sqrt{3}}{2}\left(x_{6}-y_{6}\right)a \, \mathbf{\hat{y}} + z_{6}c \, \mathbf{\hat{z}} & \left(6d\right) & \mbox{O V} \\ 
\mathbf{B}_{32} & = & y_{6} \, \mathbf{a}_{1} + \left(-x_{6}+y_{6}\right) \, \mathbf{a}_{2} + z_{6} \, \mathbf{a}_{3} & = & \left(-\frac{1}{2}x_{6}+y_{6}\right)a \, \mathbf{\hat{x}}-\frac{\sqrt{3}}{2}x_{6}a \, \mathbf{\hat{y}} + z_{6}c \, \mathbf{\hat{z}} & \left(6d\right) & \mbox{O V} \\ 
\mathbf{B}_{33} & = & \left(x_{6}-y_{6}\right) \, \mathbf{a}_{1} + x_{6} \, \mathbf{a}_{2} + z_{6} \, \mathbf{a}_{3} & = & \left(x_{6}-\frac{1}{2}y_{6}\right)a \, \mathbf{\hat{x}} + \frac{\sqrt{3}}{2}y_{6}a \, \mathbf{\hat{y}} + z_{6}c \, \mathbf{\hat{z}} & \left(6d\right) & \mbox{O V} \\ 
\mathbf{B}_{34} & = & x_{7} \, \mathbf{a}_{1} + y_{7} \, \mathbf{a}_{2} + z_{7} \, \mathbf{a}_{3} & = & \frac{1}{2}\left(x_{7}+y_{7}\right)a \, \mathbf{\hat{x}} + \frac{\sqrt{3}}{2}\left(-x_{7}+y_{7}\right)a \, \mathbf{\hat{y}} + z_{7}c \, \mathbf{\hat{z}} & \left(6d\right) & \mbox{O VI} \\ 
\mathbf{B}_{35} & = & -y_{7} \, \mathbf{a}_{1} + \left(x_{7}-y_{7}\right) \, \mathbf{a}_{2} + z_{7} \, \mathbf{a}_{3} & = & \left(\frac{1}{2}x_{7}-y_{7}\right)a \, \mathbf{\hat{x}} + \frac{\sqrt{3}}{2}x_{7}a \, \mathbf{\hat{y}} + z_{7}c \, \mathbf{\hat{z}} & \left(6d\right) & \mbox{O VI} \\ 
\mathbf{B}_{36} & = & \left(-x_{7}+y_{7}\right) \, \mathbf{a}_{1}-x_{7} \, \mathbf{a}_{2} + z_{7} \, \mathbf{a}_{3} & = & \left(-x_{7}+\frac{1}{2}y_{7}\right)a \, \mathbf{\hat{x}}-\frac{\sqrt{3}}{2}y_{7}a \, \mathbf{\hat{y}} + z_{7}c \, \mathbf{\hat{z}} & \left(6d\right) & \mbox{O VI} \\ 
\mathbf{B}_{37} & = & -x_{7} \, \mathbf{a}_{1}-y_{7} \, \mathbf{a}_{2} + z_{7} \, \mathbf{a}_{3} & = & -\frac{1}{2}\left(x_{7}+y_{7}\right)a \, \mathbf{\hat{x}} + \frac{\sqrt{3}}{2}\left(x_{7}-y_{7}\right)a \, \mathbf{\hat{y}} + z_{7}c \, \mathbf{\hat{z}} & \left(6d\right) & \mbox{O VI} \\ 
\mathbf{B}_{38} & = & y_{7} \, \mathbf{a}_{1} + \left(-x_{7}+y_{7}\right) \, \mathbf{a}_{2} + z_{7} \, \mathbf{a}_{3} & = & \left(-\frac{1}{2}x_{7}+y_{7}\right)a \, \mathbf{\hat{x}}-\frac{\sqrt{3}}{2}x_{7}a \, \mathbf{\hat{y}} + z_{7}c \, \mathbf{\hat{z}} & \left(6d\right) & \mbox{O VI} \\ 
\mathbf{B}_{39} & = & \left(x_{7}-y_{7}\right) \, \mathbf{a}_{1} + x_{7} \, \mathbf{a}_{2} + z_{7} \, \mathbf{a}_{3} & = & \left(x_{7}-\frac{1}{2}y_{7}\right)a \, \mathbf{\hat{x}} + \frac{\sqrt{3}}{2}y_{7}a \, \mathbf{\hat{y}} + z_{7}c \, \mathbf{\hat{z}} & \left(6d\right) & \mbox{O VI} \\ 
\mathbf{B}_{40} & = & x_{8} \, \mathbf{a}_{1} + y_{8} \, \mathbf{a}_{2} + z_{8} \, \mathbf{a}_{3} & = & \frac{1}{2}\left(x_{8}+y_{8}\right)a \, \mathbf{\hat{x}} + \frac{\sqrt{3}}{2}\left(-x_{8}+y_{8}\right)a \, \mathbf{\hat{y}} + z_{8}c \, \mathbf{\hat{z}} & \left(6d\right) & \mbox{O VII} \\ 
\mathbf{B}_{41} & = & -y_{8} \, \mathbf{a}_{1} + \left(x_{8}-y_{8}\right) \, \mathbf{a}_{2} + z_{8} \, \mathbf{a}_{3} & = & \left(\frac{1}{2}x_{8}-y_{8}\right)a \, \mathbf{\hat{x}} + \frac{\sqrt{3}}{2}x_{8}a \, \mathbf{\hat{y}} + z_{8}c \, \mathbf{\hat{z}} & \left(6d\right) & \mbox{O VII} \\ 
\mathbf{B}_{42} & = & \left(-x_{8}+y_{8}\right) \, \mathbf{a}_{1}-x_{8} \, \mathbf{a}_{2} + z_{8} \, \mathbf{a}_{3} & = & \left(-x_{8}+\frac{1}{2}y_{8}\right)a \, \mathbf{\hat{x}}-\frac{\sqrt{3}}{2}y_{8}a \, \mathbf{\hat{y}} + z_{8}c \, \mathbf{\hat{z}} & \left(6d\right) & \mbox{O VII} \\ 
\mathbf{B}_{43} & = & -x_{8} \, \mathbf{a}_{1}-y_{8} \, \mathbf{a}_{2} + z_{8} \, \mathbf{a}_{3} & = & -\frac{1}{2}\left(x_{8}+y_{8}\right)a \, \mathbf{\hat{x}} + \frac{\sqrt{3}}{2}\left(x_{8}-y_{8}\right)a \, \mathbf{\hat{y}} + z_{8}c \, \mathbf{\hat{z}} & \left(6d\right) & \mbox{O VII} \\ 
\mathbf{B}_{44} & = & y_{8} \, \mathbf{a}_{1} + \left(-x_{8}+y_{8}\right) \, \mathbf{a}_{2} + z_{8} \, \mathbf{a}_{3} & = & \left(-\frac{1}{2}x_{8}+y_{8}\right)a \, \mathbf{\hat{x}}-\frac{\sqrt{3}}{2}x_{8}a \, \mathbf{\hat{y}} + z_{8}c \, \mathbf{\hat{z}} & \left(6d\right) & \mbox{O VII} \\ 
\mathbf{B}_{45} & = & \left(x_{8}-y_{8}\right) \, \mathbf{a}_{1} + x_{8} \, \mathbf{a}_{2} + z_{8} \, \mathbf{a}_{3} & = & \left(x_{8}-\frac{1}{2}y_{8}\right)a \, \mathbf{\hat{x}} + \frac{\sqrt{3}}{2}y_{8}a \, \mathbf{\hat{y}} + z_{8}c \, \mathbf{\hat{z}} & \left(6d\right) & \mbox{O VII} \\ 
\mathbf{B}_{46} & = & x_{9} \, \mathbf{a}_{1} + y_{9} \, \mathbf{a}_{2} + z_{9} \, \mathbf{a}_{3} & = & \frac{1}{2}\left(x_{9}+y_{9}\right)a \, \mathbf{\hat{x}} + \frac{\sqrt{3}}{2}\left(-x_{9}+y_{9}\right)a \, \mathbf{\hat{y}} + z_{9}c \, \mathbf{\hat{z}} & \left(6d\right) & \mbox{Ta I} \\ 
\mathbf{B}_{47} & = & -y_{9} \, \mathbf{a}_{1} + \left(x_{9}-y_{9}\right) \, \mathbf{a}_{2} + z_{9} \, \mathbf{a}_{3} & = & \left(\frac{1}{2}x_{9}-y_{9}\right)a \, \mathbf{\hat{x}} + \frac{\sqrt{3}}{2}x_{9}a \, \mathbf{\hat{y}} + z_{9}c \, \mathbf{\hat{z}} & \left(6d\right) & \mbox{Ta I} \\ 
\mathbf{B}_{48} & = & \left(-x_{9}+y_{9}\right) \, \mathbf{a}_{1}-x_{9} \, \mathbf{a}_{2} + z_{9} \, \mathbf{a}_{3} & = & \left(-x_{9}+\frac{1}{2}y_{9}\right)a \, \mathbf{\hat{x}}-\frac{\sqrt{3}}{2}y_{9}a \, \mathbf{\hat{y}} + z_{9}c \, \mathbf{\hat{z}} & \left(6d\right) & \mbox{Ta I} \\ 
\mathbf{B}_{49} & = & -x_{9} \, \mathbf{a}_{1}-y_{9} \, \mathbf{a}_{2} + z_{9} \, \mathbf{a}_{3} & = & -\frac{1}{2}\left(x_{9}+y_{9}\right)a \, \mathbf{\hat{x}} + \frac{\sqrt{3}}{2}\left(x_{9}-y_{9}\right)a \, \mathbf{\hat{y}} + z_{9}c \, \mathbf{\hat{z}} & \left(6d\right) & \mbox{Ta I} \\ 
\mathbf{B}_{50} & = & y_{9} \, \mathbf{a}_{1} + \left(-x_{9}+y_{9}\right) \, \mathbf{a}_{2} + z_{9} \, \mathbf{a}_{3} & = & \left(-\frac{1}{2}x_{9}+y_{9}\right)a \, \mathbf{\hat{x}}-\frac{\sqrt{3}}{2}x_{9}a \, \mathbf{\hat{y}} + z_{9}c \, \mathbf{\hat{z}} & \left(6d\right) & \mbox{Ta I} \\ 
\mathbf{B}_{51} & = & \left(x_{9}-y_{9}\right) \, \mathbf{a}_{1} + x_{9} \, \mathbf{a}_{2} + z_{9} \, \mathbf{a}_{3} & = & \left(x_{9}-\frac{1}{2}y_{9}\right)a \, \mathbf{\hat{x}} + \frac{\sqrt{3}}{2}y_{9}a \, \mathbf{\hat{y}} + z_{9}c \, \mathbf{\hat{z}} & \left(6d\right) & \mbox{Ta I} \\ 
\mathbf{B}_{52} & = & x_{10} \, \mathbf{a}_{1} + y_{10} \, \mathbf{a}_{2} + z_{10} \, \mathbf{a}_{3} & = & \frac{1}{2}\left(x_{10}+y_{10}\right)a \, \mathbf{\hat{x}} + \frac{\sqrt{3}}{2}\left(-x_{10}+y_{10}\right)a \, \mathbf{\hat{y}} + z_{10}c \, \mathbf{\hat{z}} & \left(6d\right) & \mbox{Ta II} \\ 
\mathbf{B}_{53} & = & -y_{10} \, \mathbf{a}_{1} + \left(x_{10}-y_{10}\right) \, \mathbf{a}_{2} + z_{10} \, \mathbf{a}_{3} & = & \left(\frac{1}{2}x_{10}-y_{10}\right)a \, \mathbf{\hat{x}} + \frac{\sqrt{3}}{2}x_{10}a \, \mathbf{\hat{y}} + z_{10}c \, \mathbf{\hat{z}} & \left(6d\right) & \mbox{Ta II} \\ 
\mathbf{B}_{54} & = & \left(-x_{10}+y_{10}\right) \, \mathbf{a}_{1}-x_{10} \, \mathbf{a}_{2} + z_{10} \, \mathbf{a}_{3} & = & \left(-x_{10}+\frac{1}{2}y_{10}\right)a \, \mathbf{\hat{x}}-\frac{\sqrt{3}}{2}y_{10}a \, \mathbf{\hat{y}} + z_{10}c \, \mathbf{\hat{z}} & \left(6d\right) & \mbox{Ta II} \\ 
\mathbf{B}_{55} & = & -x_{10} \, \mathbf{a}_{1}-y_{10} \, \mathbf{a}_{2} + z_{10} \, \mathbf{a}_{3} & = & -\frac{1}{2}\left(x_{10}+y_{10}\right)a \, \mathbf{\hat{x}} + \frac{\sqrt{3}}{2}\left(x_{10}-y_{10}\right)a \, \mathbf{\hat{y}} + z_{10}c \, \mathbf{\hat{z}} & \left(6d\right) & \mbox{Ta II} \\ 
\mathbf{B}_{56} & = & y_{10} \, \mathbf{a}_{1} + \left(-x_{10}+y_{10}\right) \, \mathbf{a}_{2} + z_{10} \, \mathbf{a}_{3} & = & \left(-\frac{1}{2}x_{10}+y_{10}\right)a \, \mathbf{\hat{x}}-\frac{\sqrt{3}}{2}x_{10}a \, \mathbf{\hat{y}} + z_{10}c \, \mathbf{\hat{z}} & \left(6d\right) & \mbox{Ta II} \\ 
\mathbf{B}_{57} & = & \left(x_{10}-y_{10}\right) \, \mathbf{a}_{1} + x_{10} \, \mathbf{a}_{2} + z_{10} \, \mathbf{a}_{3} & = & \left(x_{10}-\frac{1}{2}y_{10}\right)a \, \mathbf{\hat{x}} + \frac{\sqrt{3}}{2}y_{10}a \, \mathbf{\hat{y}} + z_{10}c \, \mathbf{\hat{z}} & \left(6d\right) & \mbox{Ta II} \\ 
\end{longtabu}
\renewcommand{\arraystretch}{1.0}
\noindent \hrulefill
\\
\textbf{References:}
\vspace*{-0.25cm}
\begin{flushleft}
  - \bibentry{Boukhari_K2Ta4O9F4_ActCrystallogSecB_1979}. \\
\end{flushleft}
\textbf{Found in:}
\vspace*{-0.25cm}
\begin{flushleft}
  - \bibentry{Villars_PearsonsCrystalData_2013}. \\
\end{flushleft}
\noindent \hrulefill
\\
\textbf{Geometry files:}
\\
\noindent  - CIF: pp. {\hyperref[A2B13C4_hP57_168_d_c6d_2d_cif]{\pageref{A2B13C4_hP57_168_d_c6d_2d_cif}}} \\
\noindent  - POSCAR: pp. {\hyperref[A2B13C4_hP57_168_d_c6d_2d_poscar]{\pageref{A2B13C4_hP57_168_d_c6d_2d_poscar}}} \\
\onecolumn
{\phantomsection\label{AB4C_hP72_168_2d_8d_2d}}
\subsection*{\huge \textbf{{\normalfont Al[PO$_{4}$] Structure: AB4C\_hP72\_168\_2d\_8d\_2d}}}
\noindent \hrulefill
\vspace*{0.25cm}
\begin{figure}[htp]
  \centering
  \vspace{-1em}
  {\includegraphics[width=1\textwidth]{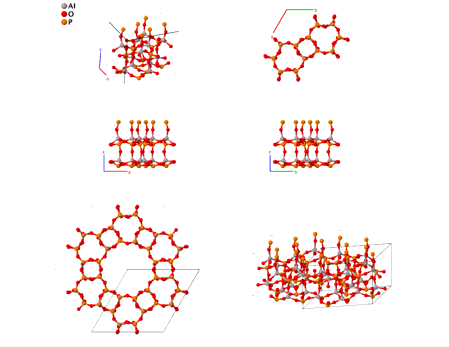}}
\end{figure}
\vspace*{-0.5cm}
\renewcommand{\arraystretch}{1.5}
\begin{equation*}
  \begin{array}{>{$\hspace{-0.15cm}}l<{$}>{$}p{0.5cm}<{$}>{$}p{18.5cm}<{$}}
    \mbox{\large \textbf{Prototype}} &\colon & \ce{Al[PO4]} \\
    \mbox{\large \textbf{\AFLOW\ prototype label}} &\colon & \mbox{AB4C\_hP72\_168\_2d\_8d\_2d} \\
    \mbox{\large \textbf{\textit{Strukturbericht} designation}} &\colon & \mbox{None} \\
    \mbox{\large \textbf{Pearson symbol}} &\colon & \mbox{hP72} \\
    \mbox{\large \textbf{Space group number}} &\colon & 168 \\
    \mbox{\large \textbf{Space group symbol}} &\colon & P6 \\
    \mbox{\large \textbf{\AFLOW\ prototype command}} &\colon &  \texttt{aflow} \,  \, \texttt{-{}-proto=AB4C\_hP72\_168\_2d\_8d\_2d } \, \newline \texttt{-{}-params=}{a,c/a,x_{1},y_{1},z_{1},x_{2},y_{2},z_{2},x_{3},y_{3},z_{3},x_{4},y_{4},z_{4},x_{5},y_{5},z_{5},x_{6},y_{6},z_{6},x_{7},} \newline {y_{7},z_{7},x_{8},y_{8},z_{8},x_{9},y_{9},z_{9},x_{10},y_{10},z_{10},x_{11},y_{11},z_{11},x_{12},y_{12},z_{12} }
  \end{array}
\end{equation*}
\renewcommand{\arraystretch}{1.0}

\vspace*{-0.25cm}
\noindent \hrulefill
\begin{itemize}
  \item{The same compound appears also in a \#184 polytype.
}
\end{itemize}

\noindent \parbox{1 \linewidth}{
\noindent \hrulefill
\\
\textbf{Hexagonal primitive vectors:} \\
\vspace*{-0.25cm}
\begin{tabular}{cc}
  \begin{tabular}{c}
    \parbox{0.6 \linewidth}{
      \renewcommand{\arraystretch}{1.5}
      \begin{equation*}
        \centering
        \begin{array}{ccc}
              \mathbf{a}_1 & = & \frac12 \, a \, \mathbf{\hat{x}} - \frac{\sqrt3}2 \, a \, \mathbf{\hat{y}} \\
    \mathbf{a}_2 & = & \frac12 \, a \, \mathbf{\hat{x}} + \frac{\sqrt3}2 \, a \, \mathbf{\hat{y}} \\
    \mathbf{a}_3 & = & c \, \mathbf{\hat{z}} \\

        \end{array}
      \end{equation*}
    }
    \renewcommand{\arraystretch}{1.0}
  \end{tabular}
  \begin{tabular}{c}
    \includegraphics[width=0.3\linewidth]{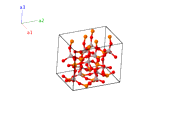} \\
  \end{tabular}
\end{tabular}

}
\vspace*{-0.25cm}

\noindent \hrulefill
\\
\textbf{Basis vectors:}
\vspace*{-0.25cm}
\renewcommand{\arraystretch}{1.5}
\begin{longtabu} to \textwidth{>{\centering $}X[-1,c,c]<{$}>{\centering $}X[-1,c,c]<{$}>{\centering $}X[-1,c,c]<{$}>{\centering $}X[-1,c,c]<{$}>{\centering $}X[-1,c,c]<{$}>{\centering $}X[-1,c,c]<{$}>{\centering $}X[-1,c,c]<{$}}
  & & \mbox{Lattice Coordinates} & & \mbox{Cartesian Coordinates} &\mbox{Wyckoff Position} & \mbox{Atom Type} \\  
  \mathbf{B}_{1} & = & x_{1} \, \mathbf{a}_{1} + y_{1} \, \mathbf{a}_{2} + z_{1} \, \mathbf{a}_{3} & = & \frac{1}{2}\left(x_{1}+y_{1}\right)a \, \mathbf{\hat{x}} + \frac{\sqrt{3}}{2}\left(-x_{1}+y_{1}\right)a \, \mathbf{\hat{y}} + z_{1}c \, \mathbf{\hat{z}} & \left(6d\right) & \mbox{Al I} \\ 
\mathbf{B}_{2} & = & -y_{1} \, \mathbf{a}_{1} + \left(x_{1}-y_{1}\right) \, \mathbf{a}_{2} + z_{1} \, \mathbf{a}_{3} & = & \left(\frac{1}{2}x_{1}-y_{1}\right)a \, \mathbf{\hat{x}} + \frac{\sqrt{3}}{2}x_{1}a \, \mathbf{\hat{y}} + z_{1}c \, \mathbf{\hat{z}} & \left(6d\right) & \mbox{Al I} \\ 
\mathbf{B}_{3} & = & \left(-x_{1}+y_{1}\right) \, \mathbf{a}_{1}-x_{1} \, \mathbf{a}_{2} + z_{1} \, \mathbf{a}_{3} & = & \left(-x_{1}+\frac{1}{2}y_{1}\right)a \, \mathbf{\hat{x}}-\frac{\sqrt{3}}{2}y_{1}a \, \mathbf{\hat{y}} + z_{1}c \, \mathbf{\hat{z}} & \left(6d\right) & \mbox{Al I} \\ 
\mathbf{B}_{4} & = & -x_{1} \, \mathbf{a}_{1}-y_{1} \, \mathbf{a}_{2} + z_{1} \, \mathbf{a}_{3} & = & -\frac{1}{2}\left(x_{1}+y_{1}\right)a \, \mathbf{\hat{x}} + \frac{\sqrt{3}}{2}\left(x_{1}-y_{1}\right)a \, \mathbf{\hat{y}} + z_{1}c \, \mathbf{\hat{z}} & \left(6d\right) & \mbox{Al I} \\ 
\mathbf{B}_{5} & = & y_{1} \, \mathbf{a}_{1} + \left(-x_{1}+y_{1}\right) \, \mathbf{a}_{2} + z_{1} \, \mathbf{a}_{3} & = & \left(-\frac{1}{2}x_{1}+y_{1}\right)a \, \mathbf{\hat{x}}-\frac{\sqrt{3}}{2}x_{1}a \, \mathbf{\hat{y}} + z_{1}c \, \mathbf{\hat{z}} & \left(6d\right) & \mbox{Al I} \\ 
\mathbf{B}_{6} & = & \left(x_{1}-y_{1}\right) \, \mathbf{a}_{1} + x_{1} \, \mathbf{a}_{2} + z_{1} \, \mathbf{a}_{3} & = & \left(x_{1}-\frac{1}{2}y_{1}\right)a \, \mathbf{\hat{x}} + \frac{\sqrt{3}}{2}y_{1}a \, \mathbf{\hat{y}} + z_{1}c \, \mathbf{\hat{z}} & \left(6d\right) & \mbox{Al I} \\ 
\mathbf{B}_{7} & = & x_{2} \, \mathbf{a}_{1} + y_{2} \, \mathbf{a}_{2} + z_{2} \, \mathbf{a}_{3} & = & \frac{1}{2}\left(x_{2}+y_{2}\right)a \, \mathbf{\hat{x}} + \frac{\sqrt{3}}{2}\left(-x_{2}+y_{2}\right)a \, \mathbf{\hat{y}} + z_{2}c \, \mathbf{\hat{z}} & \left(6d\right) & \mbox{Al II} \\ 
\mathbf{B}_{8} & = & -y_{2} \, \mathbf{a}_{1} + \left(x_{2}-y_{2}\right) \, \mathbf{a}_{2} + z_{2} \, \mathbf{a}_{3} & = & \left(\frac{1}{2}x_{2}-y_{2}\right)a \, \mathbf{\hat{x}} + \frac{\sqrt{3}}{2}x_{2}a \, \mathbf{\hat{y}} + z_{2}c \, \mathbf{\hat{z}} & \left(6d\right) & \mbox{Al II} \\ 
\mathbf{B}_{9} & = & \left(-x_{2}+y_{2}\right) \, \mathbf{a}_{1}-x_{2} \, \mathbf{a}_{2} + z_{2} \, \mathbf{a}_{3} & = & \left(-x_{2}+\frac{1}{2}y_{2}\right)a \, \mathbf{\hat{x}}-\frac{\sqrt{3}}{2}y_{2}a \, \mathbf{\hat{y}} + z_{2}c \, \mathbf{\hat{z}} & \left(6d\right) & \mbox{Al II} \\ 
\mathbf{B}_{10} & = & -x_{2} \, \mathbf{a}_{1}-y_{2} \, \mathbf{a}_{2} + z_{2} \, \mathbf{a}_{3} & = & -\frac{1}{2}\left(x_{2}+y_{2}\right)a \, \mathbf{\hat{x}} + \frac{\sqrt{3}}{2}\left(x_{2}-y_{2}\right)a \, \mathbf{\hat{y}} + z_{2}c \, \mathbf{\hat{z}} & \left(6d\right) & \mbox{Al II} \\ 
\mathbf{B}_{11} & = & y_{2} \, \mathbf{a}_{1} + \left(-x_{2}+y_{2}\right) \, \mathbf{a}_{2} + z_{2} \, \mathbf{a}_{3} & = & \left(-\frac{1}{2}x_{2}+y_{2}\right)a \, \mathbf{\hat{x}}-\frac{\sqrt{3}}{2}x_{2}a \, \mathbf{\hat{y}} + z_{2}c \, \mathbf{\hat{z}} & \left(6d\right) & \mbox{Al II} \\ 
\mathbf{B}_{12} & = & \left(x_{2}-y_{2}\right) \, \mathbf{a}_{1} + x_{2} \, \mathbf{a}_{2} + z_{2} \, \mathbf{a}_{3} & = & \left(x_{2}-\frac{1}{2}y_{2}\right)a \, \mathbf{\hat{x}} + \frac{\sqrt{3}}{2}y_{2}a \, \mathbf{\hat{y}} + z_{2}c \, \mathbf{\hat{z}} & \left(6d\right) & \mbox{Al II} \\ 
\mathbf{B}_{13} & = & x_{3} \, \mathbf{a}_{1} + y_{3} \, \mathbf{a}_{2} + z_{3} \, \mathbf{a}_{3} & = & \frac{1}{2}\left(x_{3}+y_{3}\right)a \, \mathbf{\hat{x}} + \frac{\sqrt{3}}{2}\left(-x_{3}+y_{3}\right)a \, \mathbf{\hat{y}} + z_{3}c \, \mathbf{\hat{z}} & \left(6d\right) & \mbox{O I} \\ 
\mathbf{B}_{14} & = & -y_{3} \, \mathbf{a}_{1} + \left(x_{3}-y_{3}\right) \, \mathbf{a}_{2} + z_{3} \, \mathbf{a}_{3} & = & \left(\frac{1}{2}x_{3}-y_{3}\right)a \, \mathbf{\hat{x}} + \frac{\sqrt{3}}{2}x_{3}a \, \mathbf{\hat{y}} + z_{3}c \, \mathbf{\hat{z}} & \left(6d\right) & \mbox{O I} \\ 
\mathbf{B}_{15} & = & \left(-x_{3}+y_{3}\right) \, \mathbf{a}_{1}-x_{3} \, \mathbf{a}_{2} + z_{3} \, \mathbf{a}_{3} & = & \left(-x_{3}+\frac{1}{2}y_{3}\right)a \, \mathbf{\hat{x}}-\frac{\sqrt{3}}{2}y_{3}a \, \mathbf{\hat{y}} + z_{3}c \, \mathbf{\hat{z}} & \left(6d\right) & \mbox{O I} \\ 
\mathbf{B}_{16} & = & -x_{3} \, \mathbf{a}_{1}-y_{3} \, \mathbf{a}_{2} + z_{3} \, \mathbf{a}_{3} & = & -\frac{1}{2}\left(x_{3}+y_{3}\right)a \, \mathbf{\hat{x}} + \frac{\sqrt{3}}{2}\left(x_{3}-y_{3}\right)a \, \mathbf{\hat{y}} + z_{3}c \, \mathbf{\hat{z}} & \left(6d\right) & \mbox{O I} \\ 
\mathbf{B}_{17} & = & y_{3} \, \mathbf{a}_{1} + \left(-x_{3}+y_{3}\right) \, \mathbf{a}_{2} + z_{3} \, \mathbf{a}_{3} & = & \left(-\frac{1}{2}x_{3}+y_{3}\right)a \, \mathbf{\hat{x}}-\frac{\sqrt{3}}{2}x_{3}a \, \mathbf{\hat{y}} + z_{3}c \, \mathbf{\hat{z}} & \left(6d\right) & \mbox{O I} \\ 
\mathbf{B}_{18} & = & \left(x_{3}-y_{3}\right) \, \mathbf{a}_{1} + x_{3} \, \mathbf{a}_{2} + z_{3} \, \mathbf{a}_{3} & = & \left(x_{3}-\frac{1}{2}y_{3}\right)a \, \mathbf{\hat{x}} + \frac{\sqrt{3}}{2}y_{3}a \, \mathbf{\hat{y}} + z_{3}c \, \mathbf{\hat{z}} & \left(6d\right) & \mbox{O I} \\ 
\mathbf{B}_{19} & = & x_{4} \, \mathbf{a}_{1} + y_{4} \, \mathbf{a}_{2} + z_{4} \, \mathbf{a}_{3} & = & \frac{1}{2}\left(x_{4}+y_{4}\right)a \, \mathbf{\hat{x}} + \frac{\sqrt{3}}{2}\left(-x_{4}+y_{4}\right)a \, \mathbf{\hat{y}} + z_{4}c \, \mathbf{\hat{z}} & \left(6d\right) & \mbox{O II} \\ 
\mathbf{B}_{20} & = & -y_{4} \, \mathbf{a}_{1} + \left(x_{4}-y_{4}\right) \, \mathbf{a}_{2} + z_{4} \, \mathbf{a}_{3} & = & \left(\frac{1}{2}x_{4}-y_{4}\right)a \, \mathbf{\hat{x}} + \frac{\sqrt{3}}{2}x_{4}a \, \mathbf{\hat{y}} + z_{4}c \, \mathbf{\hat{z}} & \left(6d\right) & \mbox{O II} \\ 
\mathbf{B}_{21} & = & \left(-x_{4}+y_{4}\right) \, \mathbf{a}_{1}-x_{4} \, \mathbf{a}_{2} + z_{4} \, \mathbf{a}_{3} & = & \left(-x_{4}+\frac{1}{2}y_{4}\right)a \, \mathbf{\hat{x}}-\frac{\sqrt{3}}{2}y_{4}a \, \mathbf{\hat{y}} + z_{4}c \, \mathbf{\hat{z}} & \left(6d\right) & \mbox{O II} \\ 
\mathbf{B}_{22} & = & -x_{4} \, \mathbf{a}_{1}-y_{4} \, \mathbf{a}_{2} + z_{4} \, \mathbf{a}_{3} & = & -\frac{1}{2}\left(x_{4}+y_{4}\right)a \, \mathbf{\hat{x}} + \frac{\sqrt{3}}{2}\left(x_{4}-y_{4}\right)a \, \mathbf{\hat{y}} + z_{4}c \, \mathbf{\hat{z}} & \left(6d\right) & \mbox{O II} \\ 
\mathbf{B}_{23} & = & y_{4} \, \mathbf{a}_{1} + \left(-x_{4}+y_{4}\right) \, \mathbf{a}_{2} + z_{4} \, \mathbf{a}_{3} & = & \left(-\frac{1}{2}x_{4}+y_{4}\right)a \, \mathbf{\hat{x}}-\frac{\sqrt{3}}{2}x_{4}a \, \mathbf{\hat{y}} + z_{4}c \, \mathbf{\hat{z}} & \left(6d\right) & \mbox{O II} \\ 
\mathbf{B}_{24} & = & \left(x_{4}-y_{4}\right) \, \mathbf{a}_{1} + x_{4} \, \mathbf{a}_{2} + z_{4} \, \mathbf{a}_{3} & = & \left(x_{4}-\frac{1}{2}y_{4}\right)a \, \mathbf{\hat{x}} + \frac{\sqrt{3}}{2}y_{4}a \, \mathbf{\hat{y}} + z_{4}c \, \mathbf{\hat{z}} & \left(6d\right) & \mbox{O II} \\ 
\mathbf{B}_{25} & = & x_{5} \, \mathbf{a}_{1} + y_{5} \, \mathbf{a}_{2} + z_{5} \, \mathbf{a}_{3} & = & \frac{1}{2}\left(x_{5}+y_{5}\right)a \, \mathbf{\hat{x}} + \frac{\sqrt{3}}{2}\left(-x_{5}+y_{5}\right)a \, \mathbf{\hat{y}} + z_{5}c \, \mathbf{\hat{z}} & \left(6d\right) & \mbox{O III} \\ 
\mathbf{B}_{26} & = & -y_{5} \, \mathbf{a}_{1} + \left(x_{5}-y_{5}\right) \, \mathbf{a}_{2} + z_{5} \, \mathbf{a}_{3} & = & \left(\frac{1}{2}x_{5}-y_{5}\right)a \, \mathbf{\hat{x}} + \frac{\sqrt{3}}{2}x_{5}a \, \mathbf{\hat{y}} + z_{5}c \, \mathbf{\hat{z}} & \left(6d\right) & \mbox{O III} \\ 
\mathbf{B}_{27} & = & \left(-x_{5}+y_{5}\right) \, \mathbf{a}_{1}-x_{5} \, \mathbf{a}_{2} + z_{5} \, \mathbf{a}_{3} & = & \left(-x_{5}+\frac{1}{2}y_{5}\right)a \, \mathbf{\hat{x}}-\frac{\sqrt{3}}{2}y_{5}a \, \mathbf{\hat{y}} + z_{5}c \, \mathbf{\hat{z}} & \left(6d\right) & \mbox{O III} \\ 
\mathbf{B}_{28} & = & -x_{5} \, \mathbf{a}_{1}-y_{5} \, \mathbf{a}_{2} + z_{5} \, \mathbf{a}_{3} & = & -\frac{1}{2}\left(x_{5}+y_{5}\right)a \, \mathbf{\hat{x}} + \frac{\sqrt{3}}{2}\left(x_{5}-y_{5}\right)a \, \mathbf{\hat{y}} + z_{5}c \, \mathbf{\hat{z}} & \left(6d\right) & \mbox{O III} \\ 
\mathbf{B}_{29} & = & y_{5} \, \mathbf{a}_{1} + \left(-x_{5}+y_{5}\right) \, \mathbf{a}_{2} + z_{5} \, \mathbf{a}_{3} & = & \left(-\frac{1}{2}x_{5}+y_{5}\right)a \, \mathbf{\hat{x}}-\frac{\sqrt{3}}{2}x_{5}a \, \mathbf{\hat{y}} + z_{5}c \, \mathbf{\hat{z}} & \left(6d\right) & \mbox{O III} \\ 
\mathbf{B}_{30} & = & \left(x_{5}-y_{5}\right) \, \mathbf{a}_{1} + x_{5} \, \mathbf{a}_{2} + z_{5} \, \mathbf{a}_{3} & = & \left(x_{5}-\frac{1}{2}y_{5}\right)a \, \mathbf{\hat{x}} + \frac{\sqrt{3}}{2}y_{5}a \, \mathbf{\hat{y}} + z_{5}c \, \mathbf{\hat{z}} & \left(6d\right) & \mbox{O III} \\ 
\mathbf{B}_{31} & = & x_{6} \, \mathbf{a}_{1} + y_{6} \, \mathbf{a}_{2} + z_{6} \, \mathbf{a}_{3} & = & \frac{1}{2}\left(x_{6}+y_{6}\right)a \, \mathbf{\hat{x}} + \frac{\sqrt{3}}{2}\left(-x_{6}+y_{6}\right)a \, \mathbf{\hat{y}} + z_{6}c \, \mathbf{\hat{z}} & \left(6d\right) & \mbox{O IV} \\ 
\mathbf{B}_{32} & = & -y_{6} \, \mathbf{a}_{1} + \left(x_{6}-y_{6}\right) \, \mathbf{a}_{2} + z_{6} \, \mathbf{a}_{3} & = & \left(\frac{1}{2}x_{6}-y_{6}\right)a \, \mathbf{\hat{x}} + \frac{\sqrt{3}}{2}x_{6}a \, \mathbf{\hat{y}} + z_{6}c \, \mathbf{\hat{z}} & \left(6d\right) & \mbox{O IV} \\ 
\mathbf{B}_{33} & = & \left(-x_{6}+y_{6}\right) \, \mathbf{a}_{1}-x_{6} \, \mathbf{a}_{2} + z_{6} \, \mathbf{a}_{3} & = & \left(-x_{6}+\frac{1}{2}y_{6}\right)a \, \mathbf{\hat{x}}-\frac{\sqrt{3}}{2}y_{6}a \, \mathbf{\hat{y}} + z_{6}c \, \mathbf{\hat{z}} & \left(6d\right) & \mbox{O IV} \\ 
\mathbf{B}_{34} & = & -x_{6} \, \mathbf{a}_{1}-y_{6} \, \mathbf{a}_{2} + z_{6} \, \mathbf{a}_{3} & = & -\frac{1}{2}\left(x_{6}+y_{6}\right)a \, \mathbf{\hat{x}} + \frac{\sqrt{3}}{2}\left(x_{6}-y_{6}\right)a \, \mathbf{\hat{y}} + z_{6}c \, \mathbf{\hat{z}} & \left(6d\right) & \mbox{O IV} \\ 
\mathbf{B}_{35} & = & y_{6} \, \mathbf{a}_{1} + \left(-x_{6}+y_{6}\right) \, \mathbf{a}_{2} + z_{6} \, \mathbf{a}_{3} & = & \left(-\frac{1}{2}x_{6}+y_{6}\right)a \, \mathbf{\hat{x}}-\frac{\sqrt{3}}{2}x_{6}a \, \mathbf{\hat{y}} + z_{6}c \, \mathbf{\hat{z}} & \left(6d\right) & \mbox{O IV} \\ 
\mathbf{B}_{36} & = & \left(x_{6}-y_{6}\right) \, \mathbf{a}_{1} + x_{6} \, \mathbf{a}_{2} + z_{6} \, \mathbf{a}_{3} & = & \left(x_{6}-\frac{1}{2}y_{6}\right)a \, \mathbf{\hat{x}} + \frac{\sqrt{3}}{2}y_{6}a \, \mathbf{\hat{y}} + z_{6}c \, \mathbf{\hat{z}} & \left(6d\right) & \mbox{O IV} \\ 
\mathbf{B}_{37} & = & x_{7} \, \mathbf{a}_{1} + y_{7} \, \mathbf{a}_{2} + z_{7} \, \mathbf{a}_{3} & = & \frac{1}{2}\left(x_{7}+y_{7}\right)a \, \mathbf{\hat{x}} + \frac{\sqrt{3}}{2}\left(-x_{7}+y_{7}\right)a \, \mathbf{\hat{y}} + z_{7}c \, \mathbf{\hat{z}} & \left(6d\right) & \mbox{O V} \\ 
\mathbf{B}_{38} & = & -y_{7} \, \mathbf{a}_{1} + \left(x_{7}-y_{7}\right) \, \mathbf{a}_{2} + z_{7} \, \mathbf{a}_{3} & = & \left(\frac{1}{2}x_{7}-y_{7}\right)a \, \mathbf{\hat{x}} + \frac{\sqrt{3}}{2}x_{7}a \, \mathbf{\hat{y}} + z_{7}c \, \mathbf{\hat{z}} & \left(6d\right) & \mbox{O V} \\ 
\mathbf{B}_{39} & = & \left(-x_{7}+y_{7}\right) \, \mathbf{a}_{1}-x_{7} \, \mathbf{a}_{2} + z_{7} \, \mathbf{a}_{3} & = & \left(-x_{7}+\frac{1}{2}y_{7}\right)a \, \mathbf{\hat{x}}-\frac{\sqrt{3}}{2}y_{7}a \, \mathbf{\hat{y}} + z_{7}c \, \mathbf{\hat{z}} & \left(6d\right) & \mbox{O V} \\ 
\mathbf{B}_{40} & = & -x_{7} \, \mathbf{a}_{1}-y_{7} \, \mathbf{a}_{2} + z_{7} \, \mathbf{a}_{3} & = & -\frac{1}{2}\left(x_{7}+y_{7}\right)a \, \mathbf{\hat{x}} + \frac{\sqrt{3}}{2}\left(x_{7}-y_{7}\right)a \, \mathbf{\hat{y}} + z_{7}c \, \mathbf{\hat{z}} & \left(6d\right) & \mbox{O V} \\ 
\mathbf{B}_{41} & = & y_{7} \, \mathbf{a}_{1} + \left(-x_{7}+y_{7}\right) \, \mathbf{a}_{2} + z_{7} \, \mathbf{a}_{3} & = & \left(-\frac{1}{2}x_{7}+y_{7}\right)a \, \mathbf{\hat{x}}-\frac{\sqrt{3}}{2}x_{7}a \, \mathbf{\hat{y}} + z_{7}c \, \mathbf{\hat{z}} & \left(6d\right) & \mbox{O V} \\ 
\mathbf{B}_{42} & = & \left(x_{7}-y_{7}\right) \, \mathbf{a}_{1} + x_{7} \, \mathbf{a}_{2} + z_{7} \, \mathbf{a}_{3} & = & \left(x_{7}-\frac{1}{2}y_{7}\right)a \, \mathbf{\hat{x}} + \frac{\sqrt{3}}{2}y_{7}a \, \mathbf{\hat{y}} + z_{7}c \, \mathbf{\hat{z}} & \left(6d\right) & \mbox{O V} \\ 
\mathbf{B}_{43} & = & x_{8} \, \mathbf{a}_{1} + y_{8} \, \mathbf{a}_{2} + z_{8} \, \mathbf{a}_{3} & = & \frac{1}{2}\left(x_{8}+y_{8}\right)a \, \mathbf{\hat{x}} + \frac{\sqrt{3}}{2}\left(-x_{8}+y_{8}\right)a \, \mathbf{\hat{y}} + z_{8}c \, \mathbf{\hat{z}} & \left(6d\right) & \mbox{O VI} \\ 
\mathbf{B}_{44} & = & -y_{8} \, \mathbf{a}_{1} + \left(x_{8}-y_{8}\right) \, \mathbf{a}_{2} + z_{8} \, \mathbf{a}_{3} & = & \left(\frac{1}{2}x_{8}-y_{8}\right)a \, \mathbf{\hat{x}} + \frac{\sqrt{3}}{2}x_{8}a \, \mathbf{\hat{y}} + z_{8}c \, \mathbf{\hat{z}} & \left(6d\right) & \mbox{O VI} \\ 
\mathbf{B}_{45} & = & \left(-x_{8}+y_{8}\right) \, \mathbf{a}_{1}-x_{8} \, \mathbf{a}_{2} + z_{8} \, \mathbf{a}_{3} & = & \left(-x_{8}+\frac{1}{2}y_{8}\right)a \, \mathbf{\hat{x}}-\frac{\sqrt{3}}{2}y_{8}a \, \mathbf{\hat{y}} + z_{8}c \, \mathbf{\hat{z}} & \left(6d\right) & \mbox{O VI} \\ 
\mathbf{B}_{46} & = & -x_{8} \, \mathbf{a}_{1}-y_{8} \, \mathbf{a}_{2} + z_{8} \, \mathbf{a}_{3} & = & -\frac{1}{2}\left(x_{8}+y_{8}\right)a \, \mathbf{\hat{x}} + \frac{\sqrt{3}}{2}\left(x_{8}-y_{8}\right)a \, \mathbf{\hat{y}} + z_{8}c \, \mathbf{\hat{z}} & \left(6d\right) & \mbox{O VI} \\ 
\mathbf{B}_{47} & = & y_{8} \, \mathbf{a}_{1} + \left(-x_{8}+y_{8}\right) \, \mathbf{a}_{2} + z_{8} \, \mathbf{a}_{3} & = & \left(-\frac{1}{2}x_{8}+y_{8}\right)a \, \mathbf{\hat{x}}-\frac{\sqrt{3}}{2}x_{8}a \, \mathbf{\hat{y}} + z_{8}c \, \mathbf{\hat{z}} & \left(6d\right) & \mbox{O VI} \\ 
\mathbf{B}_{48} & = & \left(x_{8}-y_{8}\right) \, \mathbf{a}_{1} + x_{8} \, \mathbf{a}_{2} + z_{8} \, \mathbf{a}_{3} & = & \left(x_{8}-\frac{1}{2}y_{8}\right)a \, \mathbf{\hat{x}} + \frac{\sqrt{3}}{2}y_{8}a \, \mathbf{\hat{y}} + z_{8}c \, \mathbf{\hat{z}} & \left(6d\right) & \mbox{O VI} \\ 
\mathbf{B}_{49} & = & x_{9} \, \mathbf{a}_{1} + y_{9} \, \mathbf{a}_{2} + z_{9} \, \mathbf{a}_{3} & = & \frac{1}{2}\left(x_{9}+y_{9}\right)a \, \mathbf{\hat{x}} + \frac{\sqrt{3}}{2}\left(-x_{9}+y_{9}\right)a \, \mathbf{\hat{y}} + z_{9}c \, \mathbf{\hat{z}} & \left(6d\right) & \mbox{O VII} \\ 
\mathbf{B}_{50} & = & -y_{9} \, \mathbf{a}_{1} + \left(x_{9}-y_{9}\right) \, \mathbf{a}_{2} + z_{9} \, \mathbf{a}_{3} & = & \left(\frac{1}{2}x_{9}-y_{9}\right)a \, \mathbf{\hat{x}} + \frac{\sqrt{3}}{2}x_{9}a \, \mathbf{\hat{y}} + z_{9}c \, \mathbf{\hat{z}} & \left(6d\right) & \mbox{O VII} \\ 
\mathbf{B}_{51} & = & \left(-x_{9}+y_{9}\right) \, \mathbf{a}_{1}-x_{9} \, \mathbf{a}_{2} + z_{9} \, \mathbf{a}_{3} & = & \left(-x_{9}+\frac{1}{2}y_{9}\right)a \, \mathbf{\hat{x}}-\frac{\sqrt{3}}{2}y_{9}a \, \mathbf{\hat{y}} + z_{9}c \, \mathbf{\hat{z}} & \left(6d\right) & \mbox{O VII} \\ 
\mathbf{B}_{52} & = & -x_{9} \, \mathbf{a}_{1}-y_{9} \, \mathbf{a}_{2} + z_{9} \, \mathbf{a}_{3} & = & -\frac{1}{2}\left(x_{9}+y_{9}\right)a \, \mathbf{\hat{x}} + \frac{\sqrt{3}}{2}\left(x_{9}-y_{9}\right)a \, \mathbf{\hat{y}} + z_{9}c \, \mathbf{\hat{z}} & \left(6d\right) & \mbox{O VII} \\ 
\mathbf{B}_{53} & = & y_{9} \, \mathbf{a}_{1} + \left(-x_{9}+y_{9}\right) \, \mathbf{a}_{2} + z_{9} \, \mathbf{a}_{3} & = & \left(-\frac{1}{2}x_{9}+y_{9}\right)a \, \mathbf{\hat{x}}-\frac{\sqrt{3}}{2}x_{9}a \, \mathbf{\hat{y}} + z_{9}c \, \mathbf{\hat{z}} & \left(6d\right) & \mbox{O VII} \\ 
\mathbf{B}_{54} & = & \left(x_{9}-y_{9}\right) \, \mathbf{a}_{1} + x_{9} \, \mathbf{a}_{2} + z_{9} \, \mathbf{a}_{3} & = & \left(x_{9}-\frac{1}{2}y_{9}\right)a \, \mathbf{\hat{x}} + \frac{\sqrt{3}}{2}y_{9}a \, \mathbf{\hat{y}} + z_{9}c \, \mathbf{\hat{z}} & \left(6d\right) & \mbox{O VII} \\ 
\mathbf{B}_{55} & = & x_{10} \, \mathbf{a}_{1} + y_{10} \, \mathbf{a}_{2} + z_{10} \, \mathbf{a}_{3} & = & \frac{1}{2}\left(x_{10}+y_{10}\right)a \, \mathbf{\hat{x}} + \frac{\sqrt{3}}{2}\left(-x_{10}+y_{10}\right)a \, \mathbf{\hat{y}} + z_{10}c \, \mathbf{\hat{z}} & \left(6d\right) & \mbox{O VIII} \\ 
\mathbf{B}_{56} & = & -y_{10} \, \mathbf{a}_{1} + \left(x_{10}-y_{10}\right) \, \mathbf{a}_{2} + z_{10} \, \mathbf{a}_{3} & = & \left(\frac{1}{2}x_{10}-y_{10}\right)a \, \mathbf{\hat{x}} + \frac{\sqrt{3}}{2}x_{10}a \, \mathbf{\hat{y}} + z_{10}c \, \mathbf{\hat{z}} & \left(6d\right) & \mbox{O VIII} \\ 
\mathbf{B}_{57} & = & \left(-x_{10}+y_{10}\right) \, \mathbf{a}_{1}-x_{10} \, \mathbf{a}_{2} + z_{10} \, \mathbf{a}_{3} & = & \left(-x_{10}+\frac{1}{2}y_{10}\right)a \, \mathbf{\hat{x}}-\frac{\sqrt{3}}{2}y_{10}a \, \mathbf{\hat{y}} + z_{10}c \, \mathbf{\hat{z}} & \left(6d\right) & \mbox{O VIII} \\ 
\mathbf{B}_{58} & = & -x_{10} \, \mathbf{a}_{1}-y_{10} \, \mathbf{a}_{2} + z_{10} \, \mathbf{a}_{3} & = & -\frac{1}{2}\left(x_{10}+y_{10}\right)a \, \mathbf{\hat{x}} + \frac{\sqrt{3}}{2}\left(x_{10}-y_{10}\right)a \, \mathbf{\hat{y}} + z_{10}c \, \mathbf{\hat{z}} & \left(6d\right) & \mbox{O VIII} \\ 
\mathbf{B}_{59} & = & y_{10} \, \mathbf{a}_{1} + \left(-x_{10}+y_{10}\right) \, \mathbf{a}_{2} + z_{10} \, \mathbf{a}_{3} & = & \left(-\frac{1}{2}x_{10}+y_{10}\right)a \, \mathbf{\hat{x}}-\frac{\sqrt{3}}{2}x_{10}a \, \mathbf{\hat{y}} + z_{10}c \, \mathbf{\hat{z}} & \left(6d\right) & \mbox{O VIII} \\ 
\mathbf{B}_{60} & = & \left(x_{10}-y_{10}\right) \, \mathbf{a}_{1} + x_{10} \, \mathbf{a}_{2} + z_{10} \, \mathbf{a}_{3} & = & \left(x_{10}-\frac{1}{2}y_{10}\right)a \, \mathbf{\hat{x}} + \frac{\sqrt{3}}{2}y_{10}a \, \mathbf{\hat{y}} + z_{10}c \, \mathbf{\hat{z}} & \left(6d\right) & \mbox{O VIII} \\ 
\mathbf{B}_{61} & = & x_{11} \, \mathbf{a}_{1} + y_{11} \, \mathbf{a}_{2} + z_{11} \, \mathbf{a}_{3} & = & \frac{1}{2}\left(x_{11}+y_{11}\right)a \, \mathbf{\hat{x}} + \frac{\sqrt{3}}{2}\left(-x_{11}+y_{11}\right)a \, \mathbf{\hat{y}} + z_{11}c \, \mathbf{\hat{z}} & \left(6d\right) & \mbox{P I} \\ 
\mathbf{B}_{62} & = & -y_{11} \, \mathbf{a}_{1} + \left(x_{11}-y_{11}\right) \, \mathbf{a}_{2} + z_{11} \, \mathbf{a}_{3} & = & \left(\frac{1}{2}x_{11}-y_{11}\right)a \, \mathbf{\hat{x}} + \frac{\sqrt{3}}{2}x_{11}a \, \mathbf{\hat{y}} + z_{11}c \, \mathbf{\hat{z}} & \left(6d\right) & \mbox{P I} \\ 
\mathbf{B}_{63} & = & \left(-x_{11}+y_{11}\right) \, \mathbf{a}_{1}-x_{11} \, \mathbf{a}_{2} + z_{11} \, \mathbf{a}_{3} & = & \left(-x_{11}+\frac{1}{2}y_{11}\right)a \, \mathbf{\hat{x}}-\frac{\sqrt{3}}{2}y_{11}a \, \mathbf{\hat{y}} + z_{11}c \, \mathbf{\hat{z}} & \left(6d\right) & \mbox{P I} \\ 
\mathbf{B}_{64} & = & -x_{11} \, \mathbf{a}_{1}-y_{11} \, \mathbf{a}_{2} + z_{11} \, \mathbf{a}_{3} & = & -\frac{1}{2}\left(x_{11}+y_{11}\right)a \, \mathbf{\hat{x}} + \frac{\sqrt{3}}{2}\left(x_{11}-y_{11}\right)a \, \mathbf{\hat{y}} + z_{11}c \, \mathbf{\hat{z}} & \left(6d\right) & \mbox{P I} \\ 
\mathbf{B}_{65} & = & y_{11} \, \mathbf{a}_{1} + \left(-x_{11}+y_{11}\right) \, \mathbf{a}_{2} + z_{11} \, \mathbf{a}_{3} & = & \left(-\frac{1}{2}x_{11}+y_{11}\right)a \, \mathbf{\hat{x}}-\frac{\sqrt{3}}{2}x_{11}a \, \mathbf{\hat{y}} + z_{11}c \, \mathbf{\hat{z}} & \left(6d\right) & \mbox{P I} \\ 
\mathbf{B}_{66} & = & \left(x_{11}-y_{11}\right) \, \mathbf{a}_{1} + x_{11} \, \mathbf{a}_{2} + z_{11} \, \mathbf{a}_{3} & = & \left(x_{11}-\frac{1}{2}y_{11}\right)a \, \mathbf{\hat{x}} + \frac{\sqrt{3}}{2}y_{11}a \, \mathbf{\hat{y}} + z_{11}c \, \mathbf{\hat{z}} & \left(6d\right) & \mbox{P I} \\ 
\mathbf{B}_{67} & = & x_{12} \, \mathbf{a}_{1} + y_{12} \, \mathbf{a}_{2} + z_{12} \, \mathbf{a}_{3} & = & \frac{1}{2}\left(x_{12}+y_{12}\right)a \, \mathbf{\hat{x}} + \frac{\sqrt{3}}{2}\left(-x_{12}+y_{12}\right)a \, \mathbf{\hat{y}} + z_{12}c \, \mathbf{\hat{z}} & \left(6d\right) & \mbox{P II} \\ 
\mathbf{B}_{68} & = & -y_{12} \, \mathbf{a}_{1} + \left(x_{12}-y_{12}\right) \, \mathbf{a}_{2} + z_{12} \, \mathbf{a}_{3} & = & \left(\frac{1}{2}x_{12}-y_{12}\right)a \, \mathbf{\hat{x}} + \frac{\sqrt{3}}{2}x_{12}a \, \mathbf{\hat{y}} + z_{12}c \, \mathbf{\hat{z}} & \left(6d\right) & \mbox{P II} \\ 
\mathbf{B}_{69} & = & \left(-x_{12}+y_{12}\right) \, \mathbf{a}_{1}-x_{12} \, \mathbf{a}_{2} + z_{12} \, \mathbf{a}_{3} & = & \left(-x_{12}+\frac{1}{2}y_{12}\right)a \, \mathbf{\hat{x}}-\frac{\sqrt{3}}{2}y_{12}a \, \mathbf{\hat{y}} + z_{12}c \, \mathbf{\hat{z}} & \left(6d\right) & \mbox{P II} \\ 
\mathbf{B}_{70} & = & -x_{12} \, \mathbf{a}_{1}-y_{12} \, \mathbf{a}_{2} + z_{12} \, \mathbf{a}_{3} & = & -\frac{1}{2}\left(x_{12}+y_{12}\right)a \, \mathbf{\hat{x}} + \frac{\sqrt{3}}{2}\left(x_{12}-y_{12}\right)a \, \mathbf{\hat{y}} + z_{12}c \, \mathbf{\hat{z}} & \left(6d\right) & \mbox{P II} \\ 
\mathbf{B}_{71} & = & y_{12} \, \mathbf{a}_{1} + \left(-x_{12}+y_{12}\right) \, \mathbf{a}_{2} + z_{12} \, \mathbf{a}_{3} & = & \left(-\frac{1}{2}x_{12}+y_{12}\right)a \, \mathbf{\hat{x}}-\frac{\sqrt{3}}{2}x_{12}a \, \mathbf{\hat{y}} + z_{12}c \, \mathbf{\hat{z}} & \left(6d\right) & \mbox{P II} \\ 
\mathbf{B}_{72} & = & \left(x_{12}-y_{12}\right) \, \mathbf{a}_{1} + x_{12} \, \mathbf{a}_{2} + z_{12} \, \mathbf{a}_{3} & = & \left(x_{12}-\frac{1}{2}y_{12}\right)a \, \mathbf{\hat{x}} + \frac{\sqrt{3}}{2}y_{12}a \, \mathbf{\hat{y}} + z_{12}c \, \mathbf{\hat{z}} & \left(6d\right) & \mbox{P II} \\ 
\end{longtabu}
\renewcommand{\arraystretch}{1.0}
\noindent \hrulefill
\\
\textbf{References:}
\vspace*{-0.25cm}
\begin{flushleft}
  - \bibentry{Richardson_AlPO4_ActaCrystallogrSecC_1987}. \\
\end{flushleft}
\textbf{Found in:}
\vspace*{-0.25cm}
\begin{flushleft}
  - \bibentry{Villars_PearsonsCrystalData_2013}. \\
\end{flushleft}
\noindent \hrulefill
\\
\textbf{Geometry files:}
\\
\noindent  - CIF: pp. {\hyperref[AB4C_hP72_168_2d_8d_2d_cif]{\pageref{AB4C_hP72_168_2d_8d_2d_cif}}} \\
\noindent  - POSCAR: pp. {\hyperref[AB4C_hP72_168_2d_8d_2d_poscar]{\pageref{AB4C_hP72_168_2d_8d_2d_poscar}}} \\
\onecolumn
{\phantomsection\label{A2B3_hP30_169_2a_3a}}
\subsection*{\huge \textbf{{\normalfont $\alpha$-Al$_{2}$S$_{3}$ Structure: A2B3\_hP30\_169\_2a\_3a}}}
\noindent \hrulefill
\vspace*{0.25cm}
\begin{figure}[htp]
  \centering
  \vspace{-1em}
  {\includegraphics[width=1\textwidth]{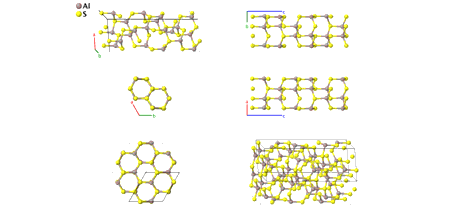}}
\end{figure}
\vspace*{-0.5cm}
\renewcommand{\arraystretch}{1.5}
\begin{equation*}
  \begin{array}{>{$\hspace{-0.15cm}}l<{$}>{$}p{0.5cm}<{$}>{$}p{18.5cm}<{$}}
    \mbox{\large \textbf{Prototype}} &\colon & \ce{$\alpha$-Al2S3} \\
    \mbox{\large \textbf{\AFLOW\ prototype label}} &\colon & \mbox{A2B3\_hP30\_169\_2a\_3a} \\
    \mbox{\large \textbf{\textit{Strukturbericht} designation}} &\colon & \mbox{None} \\
    \mbox{\large \textbf{Pearson symbol}} &\colon & \mbox{hP30} \\
    \mbox{\large \textbf{Space group number}} &\colon & 169 \\
    \mbox{\large \textbf{Space group symbol}} &\colon & P6_{1} \\
    \mbox{\large \textbf{\AFLOW\ prototype command}} &\colon &  \texttt{aflow} \,  \, \texttt{-{}-proto=A2B3\_hP30\_169\_2a\_3a } \, \newline \texttt{-{}-params=}{a,c/a,x_{1},y_{1},z_{1},x_{2},y_{2},z_{2},x_{3},y_{3},z_{3},x_{4},y_{4},z_{4},x_{5},y_{5},z_{5} }
  \end{array}
\end{equation*}
\renewcommand{\arraystretch}{1.0}

\noindent \parbox{1 \linewidth}{
\noindent \hrulefill
\\
\textbf{Hexagonal primitive vectors:} \\
\vspace*{-0.25cm}
\begin{tabular}{cc}
  \begin{tabular}{c}
    \parbox{0.6 \linewidth}{
      \renewcommand{\arraystretch}{1.5}
      \begin{equation*}
        \centering
        \begin{array}{ccc}
              \mathbf{a}_1 & = & \frac12 \, a \, \mathbf{\hat{x}} - \frac{\sqrt3}2 \, a \, \mathbf{\hat{y}} \\
    \mathbf{a}_2 & = & \frac12 \, a \, \mathbf{\hat{x}} + \frac{\sqrt3}2 \, a \, \mathbf{\hat{y}} \\
    \mathbf{a}_3 & = & c \, \mathbf{\hat{z}} \\

        \end{array}
      \end{equation*}
    }
    \renewcommand{\arraystretch}{1.0}
  \end{tabular}
  \begin{tabular}{c}
    \includegraphics[width=0.3\linewidth]{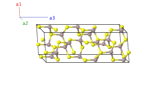} \\
  \end{tabular}
\end{tabular}

}
\vspace*{-0.25cm}

\noindent \hrulefill
\\
\textbf{Basis vectors:}
\vspace*{-0.25cm}
\renewcommand{\arraystretch}{1.5}
\begin{longtabu} to \textwidth{>{\centering $}X[-1,c,c]<{$}>{\centering $}X[-1,c,c]<{$}>{\centering $}X[-1,c,c]<{$}>{\centering $}X[-1,c,c]<{$}>{\centering $}X[-1,c,c]<{$}>{\centering $}X[-1,c,c]<{$}>{\centering $}X[-1,c,c]<{$}}
  & & \mbox{Lattice Coordinates} & & \mbox{Cartesian Coordinates} &\mbox{Wyckoff Position} & \mbox{Atom Type} \\  
  \mathbf{B}_{1} & = & x_{1} \, \mathbf{a}_{1} + y_{1} \, \mathbf{a}_{2} + z_{1} \, \mathbf{a}_{3} & = & \frac{1}{2}\left(x_{1}+y_{1}\right)a \, \mathbf{\hat{x}} + \frac{\sqrt{3}}{2}\left(-x_{1}+y_{1}\right)a \, \mathbf{\hat{y}} + z_{1}c \, \mathbf{\hat{z}} & \left(6a\right) & \mbox{Al I} \\ 
\mathbf{B}_{2} & = & -y_{1} \, \mathbf{a}_{1} + \left(x_{1}-y_{1}\right) \, \mathbf{a}_{2} + \left(\frac{1}{3} +z_{1}\right) \, \mathbf{a}_{3} & = & \left(\frac{1}{2}x_{1}-y_{1}\right)a \, \mathbf{\hat{x}} + \frac{\sqrt{3}}{2}x_{1}a \, \mathbf{\hat{y}} + \left(\frac{1}{3} +z_{1}\right)c \, \mathbf{\hat{z}} & \left(6a\right) & \mbox{Al I} \\ 
\mathbf{B}_{3} & = & \left(-x_{1}+y_{1}\right) \, \mathbf{a}_{1}-x_{1} \, \mathbf{a}_{2} + \left(\frac{2}{3} +z_{1}\right) \, \mathbf{a}_{3} & = & \left(-x_{1}+\frac{1}{2}y_{1}\right)a \, \mathbf{\hat{x}}-\frac{\sqrt{3}}{2}y_{1}a \, \mathbf{\hat{y}} + \left(\frac{2}{3} +z_{1}\right)c \, \mathbf{\hat{z}} & \left(6a\right) & \mbox{Al I} \\ 
\mathbf{B}_{4} & = & -x_{1} \, \mathbf{a}_{1}-y_{1} \, \mathbf{a}_{2} + \left(\frac{1}{2} +z_{1}\right) \, \mathbf{a}_{3} & = & -\frac{1}{2}\left(x_{1}+y_{1}\right)a \, \mathbf{\hat{x}} + \frac{\sqrt{3}}{2}\left(x_{1}-y_{1}\right)a \, \mathbf{\hat{y}} + \left(\frac{1}{2} +z_{1}\right)c \, \mathbf{\hat{z}} & \left(6a\right) & \mbox{Al I} \\ 
\mathbf{B}_{5} & = & y_{1} \, \mathbf{a}_{1} + \left(-x_{1}+y_{1}\right) \, \mathbf{a}_{2} + \left(\frac{5}{6} +z_{1}\right) \, \mathbf{a}_{3} & = & \left(-\frac{1}{2}x_{1}+y_{1}\right)a \, \mathbf{\hat{x}}-\frac{\sqrt{3}}{2}x_{1}a \, \mathbf{\hat{y}} + \left(\frac{5}{6} +z_{1}\right)c \, \mathbf{\hat{z}} & \left(6a\right) & \mbox{Al I} \\ 
\mathbf{B}_{6} & = & \left(x_{1}-y_{1}\right) \, \mathbf{a}_{1} + x_{1} \, \mathbf{a}_{2} + \left(\frac{1}{6} +z_{1}\right) \, \mathbf{a}_{3} & = & \left(x_{1}-\frac{1}{2}y_{1}\right)a \, \mathbf{\hat{x}} + \frac{\sqrt{3}}{2}y_{1}a \, \mathbf{\hat{y}} + \left(\frac{1}{6} +z_{1}\right)c \, \mathbf{\hat{z}} & \left(6a\right) & \mbox{Al I} \\ 
\mathbf{B}_{7} & = & x_{2} \, \mathbf{a}_{1} + y_{2} \, \mathbf{a}_{2} + z_{2} \, \mathbf{a}_{3} & = & \frac{1}{2}\left(x_{2}+y_{2}\right)a \, \mathbf{\hat{x}} + \frac{\sqrt{3}}{2}\left(-x_{2}+y_{2}\right)a \, \mathbf{\hat{y}} + z_{2}c \, \mathbf{\hat{z}} & \left(6a\right) & \mbox{Al II} \\ 
\mathbf{B}_{8} & = & -y_{2} \, \mathbf{a}_{1} + \left(x_{2}-y_{2}\right) \, \mathbf{a}_{2} + \left(\frac{1}{3} +z_{2}\right) \, \mathbf{a}_{3} & = & \left(\frac{1}{2}x_{2}-y_{2}\right)a \, \mathbf{\hat{x}} + \frac{\sqrt{3}}{2}x_{2}a \, \mathbf{\hat{y}} + \left(\frac{1}{3} +z_{2}\right)c \, \mathbf{\hat{z}} & \left(6a\right) & \mbox{Al II} \\ 
\mathbf{B}_{9} & = & \left(-x_{2}+y_{2}\right) \, \mathbf{a}_{1}-x_{2} \, \mathbf{a}_{2} + \left(\frac{2}{3} +z_{2}\right) \, \mathbf{a}_{3} & = & \left(-x_{2}+\frac{1}{2}y_{2}\right)a \, \mathbf{\hat{x}}-\frac{\sqrt{3}}{2}y_{2}a \, \mathbf{\hat{y}} + \left(\frac{2}{3} +z_{2}\right)c \, \mathbf{\hat{z}} & \left(6a\right) & \mbox{Al II} \\ 
\mathbf{B}_{10} & = & -x_{2} \, \mathbf{a}_{1}-y_{2} \, \mathbf{a}_{2} + \left(\frac{1}{2} +z_{2}\right) \, \mathbf{a}_{3} & = & -\frac{1}{2}\left(x_{2}+y_{2}\right)a \, \mathbf{\hat{x}} + \frac{\sqrt{3}}{2}\left(x_{2}-y_{2}\right)a \, \mathbf{\hat{y}} + \left(\frac{1}{2} +z_{2}\right)c \, \mathbf{\hat{z}} & \left(6a\right) & \mbox{Al II} \\ 
\mathbf{B}_{11} & = & y_{2} \, \mathbf{a}_{1} + \left(-x_{2}+y_{2}\right) \, \mathbf{a}_{2} + \left(\frac{5}{6} +z_{2}\right) \, \mathbf{a}_{3} & = & \left(-\frac{1}{2}x_{2}+y_{2}\right)a \, \mathbf{\hat{x}}-\frac{\sqrt{3}}{2}x_{2}a \, \mathbf{\hat{y}} + \left(\frac{5}{6} +z_{2}\right)c \, \mathbf{\hat{z}} & \left(6a\right) & \mbox{Al II} \\ 
\mathbf{B}_{12} & = & \left(x_{2}-y_{2}\right) \, \mathbf{a}_{1} + x_{2} \, \mathbf{a}_{2} + \left(\frac{1}{6} +z_{2}\right) \, \mathbf{a}_{3} & = & \left(x_{2}-\frac{1}{2}y_{2}\right)a \, \mathbf{\hat{x}} + \frac{\sqrt{3}}{2}y_{2}a \, \mathbf{\hat{y}} + \left(\frac{1}{6} +z_{2}\right)c \, \mathbf{\hat{z}} & \left(6a\right) & \mbox{Al II} \\ 
\mathbf{B}_{13} & = & x_{3} \, \mathbf{a}_{1} + y_{3} \, \mathbf{a}_{2} + z_{3} \, \mathbf{a}_{3} & = & \frac{1}{2}\left(x_{3}+y_{3}\right)a \, \mathbf{\hat{x}} + \frac{\sqrt{3}}{2}\left(-x_{3}+y_{3}\right)a \, \mathbf{\hat{y}} + z_{3}c \, \mathbf{\hat{z}} & \left(6a\right) & \mbox{S I} \\ 
\mathbf{B}_{14} & = & -y_{3} \, \mathbf{a}_{1} + \left(x_{3}-y_{3}\right) \, \mathbf{a}_{2} + \left(\frac{1}{3} +z_{3}\right) \, \mathbf{a}_{3} & = & \left(\frac{1}{2}x_{3}-y_{3}\right)a \, \mathbf{\hat{x}} + \frac{\sqrt{3}}{2}x_{3}a \, \mathbf{\hat{y}} + \left(\frac{1}{3} +z_{3}\right)c \, \mathbf{\hat{z}} & \left(6a\right) & \mbox{S I} \\ 
\mathbf{B}_{15} & = & \left(-x_{3}+y_{3}\right) \, \mathbf{a}_{1}-x_{3} \, \mathbf{a}_{2} + \left(\frac{2}{3} +z_{3}\right) \, \mathbf{a}_{3} & = & \left(-x_{3}+\frac{1}{2}y_{3}\right)a \, \mathbf{\hat{x}}-\frac{\sqrt{3}}{2}y_{3}a \, \mathbf{\hat{y}} + \left(\frac{2}{3} +z_{3}\right)c \, \mathbf{\hat{z}} & \left(6a\right) & \mbox{S I} \\ 
\mathbf{B}_{16} & = & -x_{3} \, \mathbf{a}_{1}-y_{3} \, \mathbf{a}_{2} + \left(\frac{1}{2} +z_{3}\right) \, \mathbf{a}_{3} & = & -\frac{1}{2}\left(x_{3}+y_{3}\right)a \, \mathbf{\hat{x}} + \frac{\sqrt{3}}{2}\left(x_{3}-y_{3}\right)a \, \mathbf{\hat{y}} + \left(\frac{1}{2} +z_{3}\right)c \, \mathbf{\hat{z}} & \left(6a\right) & \mbox{S I} \\ 
\mathbf{B}_{17} & = & y_{3} \, \mathbf{a}_{1} + \left(-x_{3}+y_{3}\right) \, \mathbf{a}_{2} + \left(\frac{5}{6} +z_{3}\right) \, \mathbf{a}_{3} & = & \left(-\frac{1}{2}x_{3}+y_{3}\right)a \, \mathbf{\hat{x}}-\frac{\sqrt{3}}{2}x_{3}a \, \mathbf{\hat{y}} + \left(\frac{5}{6} +z_{3}\right)c \, \mathbf{\hat{z}} & \left(6a\right) & \mbox{S I} \\ 
\mathbf{B}_{18} & = & \left(x_{3}-y_{3}\right) \, \mathbf{a}_{1} + x_{3} \, \mathbf{a}_{2} + \left(\frac{1}{6} +z_{3}\right) \, \mathbf{a}_{3} & = & \left(x_{3}-\frac{1}{2}y_{3}\right)a \, \mathbf{\hat{x}} + \frac{\sqrt{3}}{2}y_{3}a \, \mathbf{\hat{y}} + \left(\frac{1}{6} +z_{3}\right)c \, \mathbf{\hat{z}} & \left(6a\right) & \mbox{S I} \\ 
\mathbf{B}_{19} & = & x_{4} \, \mathbf{a}_{1} + y_{4} \, \mathbf{a}_{2} + z_{4} \, \mathbf{a}_{3} & = & \frac{1}{2}\left(x_{4}+y_{4}\right)a \, \mathbf{\hat{x}} + \frac{\sqrt{3}}{2}\left(-x_{4}+y_{4}\right)a \, \mathbf{\hat{y}} + z_{4}c \, \mathbf{\hat{z}} & \left(6a\right) & \mbox{S II} \\ 
\mathbf{B}_{20} & = & -y_{4} \, \mathbf{a}_{1} + \left(x_{4}-y_{4}\right) \, \mathbf{a}_{2} + \left(\frac{1}{3} +z_{4}\right) \, \mathbf{a}_{3} & = & \left(\frac{1}{2}x_{4}-y_{4}\right)a \, \mathbf{\hat{x}} + \frac{\sqrt{3}}{2}x_{4}a \, \mathbf{\hat{y}} + \left(\frac{1}{3} +z_{4}\right)c \, \mathbf{\hat{z}} & \left(6a\right) & \mbox{S II} \\ 
\mathbf{B}_{21} & = & \left(-x_{4}+y_{4}\right) \, \mathbf{a}_{1}-x_{4} \, \mathbf{a}_{2} + \left(\frac{2}{3} +z_{4}\right) \, \mathbf{a}_{3} & = & \left(-x_{4}+\frac{1}{2}y_{4}\right)a \, \mathbf{\hat{x}}-\frac{\sqrt{3}}{2}y_{4}a \, \mathbf{\hat{y}} + \left(\frac{2}{3} +z_{4}\right)c \, \mathbf{\hat{z}} & \left(6a\right) & \mbox{S II} \\ 
\mathbf{B}_{22} & = & -x_{4} \, \mathbf{a}_{1}-y_{4} \, \mathbf{a}_{2} + \left(\frac{1}{2} +z_{4}\right) \, \mathbf{a}_{3} & = & -\frac{1}{2}\left(x_{4}+y_{4}\right)a \, \mathbf{\hat{x}} + \frac{\sqrt{3}}{2}\left(x_{4}-y_{4}\right)a \, \mathbf{\hat{y}} + \left(\frac{1}{2} +z_{4}\right)c \, \mathbf{\hat{z}} & \left(6a\right) & \mbox{S II} \\ 
\mathbf{B}_{23} & = & y_{4} \, \mathbf{a}_{1} + \left(-x_{4}+y_{4}\right) \, \mathbf{a}_{2} + \left(\frac{5}{6} +z_{4}\right) \, \mathbf{a}_{3} & = & \left(-\frac{1}{2}x_{4}+y_{4}\right)a \, \mathbf{\hat{x}}-\frac{\sqrt{3}}{2}x_{4}a \, \mathbf{\hat{y}} + \left(\frac{5}{6} +z_{4}\right)c \, \mathbf{\hat{z}} & \left(6a\right) & \mbox{S II} \\ 
\mathbf{B}_{24} & = & \left(x_{4}-y_{4}\right) \, \mathbf{a}_{1} + x_{4} \, \mathbf{a}_{2} + \left(\frac{1}{6} +z_{4}\right) \, \mathbf{a}_{3} & = & \left(x_{4}-\frac{1}{2}y_{4}\right)a \, \mathbf{\hat{x}} + \frac{\sqrt{3}}{2}y_{4}a \, \mathbf{\hat{y}} + \left(\frac{1}{6} +z_{4}\right)c \, \mathbf{\hat{z}} & \left(6a\right) & \mbox{S II} \\ 
\mathbf{B}_{25} & = & x_{5} \, \mathbf{a}_{1} + y_{5} \, \mathbf{a}_{2} + z_{5} \, \mathbf{a}_{3} & = & \frac{1}{2}\left(x_{5}+y_{5}\right)a \, \mathbf{\hat{x}} + \frac{\sqrt{3}}{2}\left(-x_{5}+y_{5}\right)a \, \mathbf{\hat{y}} + z_{5}c \, \mathbf{\hat{z}} & \left(6a\right) & \mbox{S III} \\ 
\mathbf{B}_{26} & = & -y_{5} \, \mathbf{a}_{1} + \left(x_{5}-y_{5}\right) \, \mathbf{a}_{2} + \left(\frac{1}{3} +z_{5}\right) \, \mathbf{a}_{3} & = & \left(\frac{1}{2}x_{5}-y_{5}\right)a \, \mathbf{\hat{x}} + \frac{\sqrt{3}}{2}x_{5}a \, \mathbf{\hat{y}} + \left(\frac{1}{3} +z_{5}\right)c \, \mathbf{\hat{z}} & \left(6a\right) & \mbox{S III} \\ 
\mathbf{B}_{27} & = & \left(-x_{5}+y_{5}\right) \, \mathbf{a}_{1}-x_{5} \, \mathbf{a}_{2} + \left(\frac{2}{3} +z_{5}\right) \, \mathbf{a}_{3} & = & \left(-x_{5}+\frac{1}{2}y_{5}\right)a \, \mathbf{\hat{x}}-\frac{\sqrt{3}}{2}y_{5}a \, \mathbf{\hat{y}} + \left(\frac{2}{3} +z_{5}\right)c \, \mathbf{\hat{z}} & \left(6a\right) & \mbox{S III} \\ 
\mathbf{B}_{28} & = & -x_{5} \, \mathbf{a}_{1}-y_{5} \, \mathbf{a}_{2} + \left(\frac{1}{2} +z_{5}\right) \, \mathbf{a}_{3} & = & -\frac{1}{2}\left(x_{5}+y_{5}\right)a \, \mathbf{\hat{x}} + \frac{\sqrt{3}}{2}\left(x_{5}-y_{5}\right)a \, \mathbf{\hat{y}} + \left(\frac{1}{2} +z_{5}\right)c \, \mathbf{\hat{z}} & \left(6a\right) & \mbox{S III} \\ 
\mathbf{B}_{29} & = & y_{5} \, \mathbf{a}_{1} + \left(-x_{5}+y_{5}\right) \, \mathbf{a}_{2} + \left(\frac{5}{6} +z_{5}\right) \, \mathbf{a}_{3} & = & \left(-\frac{1}{2}x_{5}+y_{5}\right)a \, \mathbf{\hat{x}}-\frac{\sqrt{3}}{2}x_{5}a \, \mathbf{\hat{y}} + \left(\frac{5}{6} +z_{5}\right)c \, \mathbf{\hat{z}} & \left(6a\right) & \mbox{S III} \\ 
\mathbf{B}_{30} & = & \left(x_{5}-y_{5}\right) \, \mathbf{a}_{1} + x_{5} \, \mathbf{a}_{2} + \left(\frac{1}{6} +z_{5}\right) \, \mathbf{a}_{3} & = & \left(x_{5}-\frac{1}{2}y_{5}\right)a \, \mathbf{\hat{x}} + \frac{\sqrt{3}}{2}y_{5}a \, \mathbf{\hat{y}} + \left(\frac{1}{6} +z_{5}\right)c \, \mathbf{\hat{z}} & \left(6a\right) & \mbox{S III} \\ 
\end{longtabu}
\renewcommand{\arraystretch}{1.0}
\noindent \hrulefill
\\
\textbf{References:}
\vspace*{-0.25cm}
\begin{flushleft}
  - \bibentry{Eisenmann_Al2S3_ZKrist_1992}. \\
\end{flushleft}
\textbf{Found in:}
\vspace*{-0.25cm}
\begin{flushleft}
  - \bibentry{Villars_PearsonsCrystalData_2013}. \\
\end{flushleft}
\noindent \hrulefill
\\
\textbf{Geometry files:}
\\
\noindent  - CIF: pp. {\hyperref[A2B3_hP30_169_2a_3a_cif]{\pageref{A2B3_hP30_169_2a_3a_cif}}} \\
\noindent  - POSCAR: pp. {\hyperref[A2B3_hP30_169_2a_3a_poscar]{\pageref{A2B3_hP30_169_2a_3a_poscar}}} \\
\onecolumn
{\phantomsection\label{A2B3_hP30_170_2a_3a}}
\subsection*{\huge \textbf{{\normalfont Al$_{2}$S$_{3}$ Structure: A2B3\_hP30\_170\_2a\_3a}}}
\noindent \hrulefill
\vspace*{0.25cm}
\begin{figure}[htp]
  \centering
  \vspace{-1em}
  {\includegraphics[width=1\textwidth]{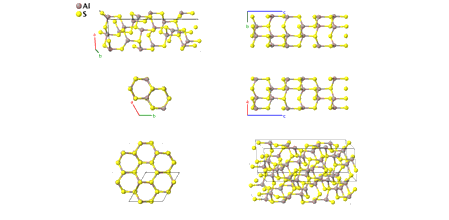}}
\end{figure}
\vspace*{-0.5cm}
\renewcommand{\arraystretch}{1.5}
\begin{equation*}
  \begin{array}{>{$\hspace{-0.15cm}}l<{$}>{$}p{0.5cm}<{$}>{$}p{18.5cm}<{$}}
    \mbox{\large \textbf{Prototype}} &\colon & \ce{Al2S3} \\
    \mbox{\large \textbf{\AFLOW\ prototype label}} &\colon & \mbox{A2B3\_hP30\_170\_2a\_3a} \\
    \mbox{\large \textbf{\textit{Strukturbericht} designation}} &\colon & \mbox{None} \\
    \mbox{\large \textbf{Pearson symbol}} &\colon & \mbox{hP30} \\
    \mbox{\large \textbf{Space group number}} &\colon & 170 \\
    \mbox{\large \textbf{Space group symbol}} &\colon & P6_{5} \\
    \mbox{\large \textbf{\AFLOW\ prototype command}} &\colon &  \texttt{aflow} \,  \, \texttt{-{}-proto=A2B3\_hP30\_170\_2a\_3a } \, \newline \texttt{-{}-params=}{a,c/a,x_{1},y_{1},z_{1},x_{2},y_{2},z_{2},x_{3},y_{3},z_{3},x_{4},y_{4},z_{4},x_{5},y_{5},z_{5} }
  \end{array}
\end{equation*}
\renewcommand{\arraystretch}{1.0}

\vspace*{-0.25cm}
\noindent \hrulefill
\begin{itemize}
  \item{This structure is the enantiomorph of the \hyperref[A2B3_hP30_169_2a_3a]{Al$_{2}$S$_{3}$ (A2B3\_hP30\_169\_2a\_3a) structure},
and was generated by reflecting the coordinates of the space group \#169 structure through the $z=0$ plane.
}
\end{itemize}

\noindent \parbox{1 \linewidth}{
\noindent \hrulefill
\\
\textbf{Hexagonal primitive vectors:} \\
\vspace*{-0.25cm}
\begin{tabular}{cc}
  \begin{tabular}{c}
    \parbox{0.6 \linewidth}{
      \renewcommand{\arraystretch}{1.5}
      \begin{equation*}
        \centering
        \begin{array}{ccc}
              \mathbf{a}_1 & = & \frac12 \, a \, \mathbf{\hat{x}} - \frac{\sqrt3}2 \, a \, \mathbf{\hat{y}} \\
    \mathbf{a}_2 & = & \frac12 \, a \, \mathbf{\hat{x}} + \frac{\sqrt3}2 \, a \, \mathbf{\hat{y}} \\
    \mathbf{a}_3 & = & c \, \mathbf{\hat{z}} \\

        \end{array}
      \end{equation*}
    }
    \renewcommand{\arraystretch}{1.0}
  \end{tabular}
  \begin{tabular}{c}
    \includegraphics[width=0.3\linewidth]{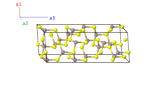} \\
  \end{tabular}
\end{tabular}

}
\vspace*{-0.25cm}

\noindent \hrulefill
\\
\textbf{Basis vectors:}
\vspace*{-0.25cm}
\renewcommand{\arraystretch}{1.5}
\begin{longtabu} to \textwidth{>{\centering $}X[-1,c,c]<{$}>{\centering $}X[-1,c,c]<{$}>{\centering $}X[-1,c,c]<{$}>{\centering $}X[-1,c,c]<{$}>{\centering $}X[-1,c,c]<{$}>{\centering $}X[-1,c,c]<{$}>{\centering $}X[-1,c,c]<{$}}
  & & \mbox{Lattice Coordinates} & & \mbox{Cartesian Coordinates} &\mbox{Wyckoff Position} & \mbox{Atom Type} \\  
  \mathbf{B}_{1} & = & x_{1} \, \mathbf{a}_{1} + y_{1} \, \mathbf{a}_{2} + z_{1} \, \mathbf{a}_{3} & = & \frac{1}{2}\left(x_{1}+y_{1}\right)a \, \mathbf{\hat{x}} + \frac{\sqrt{3}}{2}\left(-x_{1}+y_{1}\right)a \, \mathbf{\hat{y}} + z_{1}c \, \mathbf{\hat{z}} & \left(6a\right) & \mbox{Al I} \\ 
\mathbf{B}_{2} & = & -y_{1} \, \mathbf{a}_{1} + \left(x_{1}-y_{1}\right) \, \mathbf{a}_{2} + \left(\frac{2}{3} +z_{1}\right) \, \mathbf{a}_{3} & = & \left(\frac{1}{2}x_{1}-y_{1}\right)a \, \mathbf{\hat{x}} + \frac{\sqrt{3}}{2}x_{1}a \, \mathbf{\hat{y}} + \left(\frac{2}{3} +z_{1}\right)c \, \mathbf{\hat{z}} & \left(6a\right) & \mbox{Al I} \\ 
\mathbf{B}_{3} & = & \left(-x_{1}+y_{1}\right) \, \mathbf{a}_{1}-x_{1} \, \mathbf{a}_{2} + \left(\frac{1}{3} +z_{1}\right) \, \mathbf{a}_{3} & = & \left(-x_{1}+\frac{1}{2}y_{1}\right)a \, \mathbf{\hat{x}}-\frac{\sqrt{3}}{2}y_{1}a \, \mathbf{\hat{y}} + \left(\frac{1}{3} +z_{1}\right)c \, \mathbf{\hat{z}} & \left(6a\right) & \mbox{Al I} \\ 
\mathbf{B}_{4} & = & -x_{1} \, \mathbf{a}_{1}-y_{1} \, \mathbf{a}_{2} + \left(\frac{1}{2} +z_{1}\right) \, \mathbf{a}_{3} & = & -\frac{1}{2}\left(x_{1}+y_{1}\right)a \, \mathbf{\hat{x}} + \frac{\sqrt{3}}{2}\left(x_{1}-y_{1}\right)a \, \mathbf{\hat{y}} + \left(\frac{1}{2} +z_{1}\right)c \, \mathbf{\hat{z}} & \left(6a\right) & \mbox{Al I} \\ 
\mathbf{B}_{5} & = & y_{1} \, \mathbf{a}_{1} + \left(-x_{1}+y_{1}\right) \, \mathbf{a}_{2} + \left(\frac{1}{6} +z_{1}\right) \, \mathbf{a}_{3} & = & \left(-\frac{1}{2}x_{1}+y_{1}\right)a \, \mathbf{\hat{x}}-\frac{\sqrt{3}}{2}x_{1}a \, \mathbf{\hat{y}} + \left(\frac{1}{6} +z_{1}\right)c \, \mathbf{\hat{z}} & \left(6a\right) & \mbox{Al I} \\ 
\mathbf{B}_{6} & = & \left(x_{1}-y_{1}\right) \, \mathbf{a}_{1} + x_{1} \, \mathbf{a}_{2} + \left(\frac{5}{6} +z_{1}\right) \, \mathbf{a}_{3} & = & \left(x_{1}-\frac{1}{2}y_{1}\right)a \, \mathbf{\hat{x}} + \frac{\sqrt{3}}{2}y_{1}a \, \mathbf{\hat{y}} + \left(\frac{5}{6} +z_{1}\right)c \, \mathbf{\hat{z}} & \left(6a\right) & \mbox{Al I} \\ 
\mathbf{B}_{7} & = & x_{2} \, \mathbf{a}_{1} + y_{2} \, \mathbf{a}_{2} + z_{2} \, \mathbf{a}_{3} & = & \frac{1}{2}\left(x_{2}+y_{2}\right)a \, \mathbf{\hat{x}} + \frac{\sqrt{3}}{2}\left(-x_{2}+y_{2}\right)a \, \mathbf{\hat{y}} + z_{2}c \, \mathbf{\hat{z}} & \left(6a\right) & \mbox{Al II} \\ 
\mathbf{B}_{8} & = & -y_{2} \, \mathbf{a}_{1} + \left(x_{2}-y_{2}\right) \, \mathbf{a}_{2} + \left(\frac{2}{3} +z_{2}\right) \, \mathbf{a}_{3} & = & \left(\frac{1}{2}x_{2}-y_{2}\right)a \, \mathbf{\hat{x}} + \frac{\sqrt{3}}{2}x_{2}a \, \mathbf{\hat{y}} + \left(\frac{2}{3} +z_{2}\right)c \, \mathbf{\hat{z}} & \left(6a\right) & \mbox{Al II} \\ 
\mathbf{B}_{9} & = & \left(-x_{2}+y_{2}\right) \, \mathbf{a}_{1}-x_{2} \, \mathbf{a}_{2} + \left(\frac{1}{3} +z_{2}\right) \, \mathbf{a}_{3} & = & \left(-x_{2}+\frac{1}{2}y_{2}\right)a \, \mathbf{\hat{x}}-\frac{\sqrt{3}}{2}y_{2}a \, \mathbf{\hat{y}} + \left(\frac{1}{3} +z_{2}\right)c \, \mathbf{\hat{z}} & \left(6a\right) & \mbox{Al II} \\ 
\mathbf{B}_{10} & = & -x_{2} \, \mathbf{a}_{1}-y_{2} \, \mathbf{a}_{2} + \left(\frac{1}{2} +z_{2}\right) \, \mathbf{a}_{3} & = & -\frac{1}{2}\left(x_{2}+y_{2}\right)a \, \mathbf{\hat{x}} + \frac{\sqrt{3}}{2}\left(x_{2}-y_{2}\right)a \, \mathbf{\hat{y}} + \left(\frac{1}{2} +z_{2}\right)c \, \mathbf{\hat{z}} & \left(6a\right) & \mbox{Al II} \\ 
\mathbf{B}_{11} & = & y_{2} \, \mathbf{a}_{1} + \left(-x_{2}+y_{2}\right) \, \mathbf{a}_{2} + \left(\frac{1}{6} +z_{2}\right) \, \mathbf{a}_{3} & = & \left(-\frac{1}{2}x_{2}+y_{2}\right)a \, \mathbf{\hat{x}}-\frac{\sqrt{3}}{2}x_{2}a \, \mathbf{\hat{y}} + \left(\frac{1}{6} +z_{2}\right)c \, \mathbf{\hat{z}} & \left(6a\right) & \mbox{Al II} \\ 
\mathbf{B}_{12} & = & \left(x_{2}-y_{2}\right) \, \mathbf{a}_{1} + x_{2} \, \mathbf{a}_{2} + \left(\frac{5}{6} +z_{2}\right) \, \mathbf{a}_{3} & = & \left(x_{2}-\frac{1}{2}y_{2}\right)a \, \mathbf{\hat{x}} + \frac{\sqrt{3}}{2}y_{2}a \, \mathbf{\hat{y}} + \left(\frac{5}{6} +z_{2}\right)c \, \mathbf{\hat{z}} & \left(6a\right) & \mbox{Al II} \\ 
\mathbf{B}_{13} & = & x_{3} \, \mathbf{a}_{1} + y_{3} \, \mathbf{a}_{2} + z_{3} \, \mathbf{a}_{3} & = & \frac{1}{2}\left(x_{3}+y_{3}\right)a \, \mathbf{\hat{x}} + \frac{\sqrt{3}}{2}\left(-x_{3}+y_{3}\right)a \, \mathbf{\hat{y}} + z_{3}c \, \mathbf{\hat{z}} & \left(6a\right) & \mbox{S I} \\ 
\mathbf{B}_{14} & = & -y_{3} \, \mathbf{a}_{1} + \left(x_{3}-y_{3}\right) \, \mathbf{a}_{2} + \left(\frac{2}{3} +z_{3}\right) \, \mathbf{a}_{3} & = & \left(\frac{1}{2}x_{3}-y_{3}\right)a \, \mathbf{\hat{x}} + \frac{\sqrt{3}}{2}x_{3}a \, \mathbf{\hat{y}} + \left(\frac{2}{3} +z_{3}\right)c \, \mathbf{\hat{z}} & \left(6a\right) & \mbox{S I} \\ 
\mathbf{B}_{15} & = & \left(-x_{3}+y_{3}\right) \, \mathbf{a}_{1}-x_{3} \, \mathbf{a}_{2} + \left(\frac{1}{3} +z_{3}\right) \, \mathbf{a}_{3} & = & \left(-x_{3}+\frac{1}{2}y_{3}\right)a \, \mathbf{\hat{x}}-\frac{\sqrt{3}}{2}y_{3}a \, \mathbf{\hat{y}} + \left(\frac{1}{3} +z_{3}\right)c \, \mathbf{\hat{z}} & \left(6a\right) & \mbox{S I} \\ 
\mathbf{B}_{16} & = & -x_{3} \, \mathbf{a}_{1}-y_{3} \, \mathbf{a}_{2} + \left(\frac{1}{2} +z_{3}\right) \, \mathbf{a}_{3} & = & -\frac{1}{2}\left(x_{3}+y_{3}\right)a \, \mathbf{\hat{x}} + \frac{\sqrt{3}}{2}\left(x_{3}-y_{3}\right)a \, \mathbf{\hat{y}} + \left(\frac{1}{2} +z_{3}\right)c \, \mathbf{\hat{z}} & \left(6a\right) & \mbox{S I} \\ 
\mathbf{B}_{17} & = & y_{3} \, \mathbf{a}_{1} + \left(-x_{3}+y_{3}\right) \, \mathbf{a}_{2} + \left(\frac{1}{6} +z_{3}\right) \, \mathbf{a}_{3} & = & \left(-\frac{1}{2}x_{3}+y_{3}\right)a \, \mathbf{\hat{x}}-\frac{\sqrt{3}}{2}x_{3}a \, \mathbf{\hat{y}} + \left(\frac{1}{6} +z_{3}\right)c \, \mathbf{\hat{z}} & \left(6a\right) & \mbox{S I} \\ 
\mathbf{B}_{18} & = & \left(x_{3}-y_{3}\right) \, \mathbf{a}_{1} + x_{3} \, \mathbf{a}_{2} + \left(\frac{5}{6} +z_{3}\right) \, \mathbf{a}_{3} & = & \left(x_{3}-\frac{1}{2}y_{3}\right)a \, \mathbf{\hat{x}} + \frac{\sqrt{3}}{2}y_{3}a \, \mathbf{\hat{y}} + \left(\frac{5}{6} +z_{3}\right)c \, \mathbf{\hat{z}} & \left(6a\right) & \mbox{S I} \\ 
\mathbf{B}_{19} & = & x_{4} \, \mathbf{a}_{1} + y_{4} \, \mathbf{a}_{2} + z_{4} \, \mathbf{a}_{3} & = & \frac{1}{2}\left(x_{4}+y_{4}\right)a \, \mathbf{\hat{x}} + \frac{\sqrt{3}}{2}\left(-x_{4}+y_{4}\right)a \, \mathbf{\hat{y}} + z_{4}c \, \mathbf{\hat{z}} & \left(6a\right) & \mbox{S II} \\ 
\mathbf{B}_{20} & = & -y_{4} \, \mathbf{a}_{1} + \left(x_{4}-y_{4}\right) \, \mathbf{a}_{2} + \left(\frac{2}{3} +z_{4}\right) \, \mathbf{a}_{3} & = & \left(\frac{1}{2}x_{4}-y_{4}\right)a \, \mathbf{\hat{x}} + \frac{\sqrt{3}}{2}x_{4}a \, \mathbf{\hat{y}} + \left(\frac{2}{3} +z_{4}\right)c \, \mathbf{\hat{z}} & \left(6a\right) & \mbox{S II} \\ 
\mathbf{B}_{21} & = & \left(-x_{4}+y_{4}\right) \, \mathbf{a}_{1}-x_{4} \, \mathbf{a}_{2} + \left(\frac{1}{3} +z_{4}\right) \, \mathbf{a}_{3} & = & \left(-x_{4}+\frac{1}{2}y_{4}\right)a \, \mathbf{\hat{x}}-\frac{\sqrt{3}}{2}y_{4}a \, \mathbf{\hat{y}} + \left(\frac{1}{3} +z_{4}\right)c \, \mathbf{\hat{z}} & \left(6a\right) & \mbox{S II} \\ 
\mathbf{B}_{22} & = & -x_{4} \, \mathbf{a}_{1}-y_{4} \, \mathbf{a}_{2} + \left(\frac{1}{2} +z_{4}\right) \, \mathbf{a}_{3} & = & -\frac{1}{2}\left(x_{4}+y_{4}\right)a \, \mathbf{\hat{x}} + \frac{\sqrt{3}}{2}\left(x_{4}-y_{4}\right)a \, \mathbf{\hat{y}} + \left(\frac{1}{2} +z_{4}\right)c \, \mathbf{\hat{z}} & \left(6a\right) & \mbox{S II} \\ 
\mathbf{B}_{23} & = & y_{4} \, \mathbf{a}_{1} + \left(-x_{4}+y_{4}\right) \, \mathbf{a}_{2} + \left(\frac{1}{6} +z_{4}\right) \, \mathbf{a}_{3} & = & \left(-\frac{1}{2}x_{4}+y_{4}\right)a \, \mathbf{\hat{x}}-\frac{\sqrt{3}}{2}x_{4}a \, \mathbf{\hat{y}} + \left(\frac{1}{6} +z_{4}\right)c \, \mathbf{\hat{z}} & \left(6a\right) & \mbox{S II} \\ 
\mathbf{B}_{24} & = & \left(x_{4}-y_{4}\right) \, \mathbf{a}_{1} + x_{4} \, \mathbf{a}_{2} + \left(\frac{5}{6} +z_{4}\right) \, \mathbf{a}_{3} & = & \left(x_{4}-\frac{1}{2}y_{4}\right)a \, \mathbf{\hat{x}} + \frac{\sqrt{3}}{2}y_{4}a \, \mathbf{\hat{y}} + \left(\frac{5}{6} +z_{4}\right)c \, \mathbf{\hat{z}} & \left(6a\right) & \mbox{S II} \\ 
\mathbf{B}_{25} & = & x_{5} \, \mathbf{a}_{1} + y_{5} \, \mathbf{a}_{2} + z_{5} \, \mathbf{a}_{3} & = & \frac{1}{2}\left(x_{5}+y_{5}\right)a \, \mathbf{\hat{x}} + \frac{\sqrt{3}}{2}\left(-x_{5}+y_{5}\right)a \, \mathbf{\hat{y}} + z_{5}c \, \mathbf{\hat{z}} & \left(6a\right) & \mbox{S III} \\ 
\mathbf{B}_{26} & = & -y_{5} \, \mathbf{a}_{1} + \left(x_{5}-y_{5}\right) \, \mathbf{a}_{2} + \left(\frac{2}{3} +z_{5}\right) \, \mathbf{a}_{3} & = & \left(\frac{1}{2}x_{5}-y_{5}\right)a \, \mathbf{\hat{x}} + \frac{\sqrt{3}}{2}x_{5}a \, \mathbf{\hat{y}} + \left(\frac{2}{3} +z_{5}\right)c \, \mathbf{\hat{z}} & \left(6a\right) & \mbox{S III} \\ 
\mathbf{B}_{27} & = & \left(-x_{5}+y_{5}\right) \, \mathbf{a}_{1}-x_{5} \, \mathbf{a}_{2} + \left(\frac{1}{3} +z_{5}\right) \, \mathbf{a}_{3} & = & \left(-x_{5}+\frac{1}{2}y_{5}\right)a \, \mathbf{\hat{x}}-\frac{\sqrt{3}}{2}y_{5}a \, \mathbf{\hat{y}} + \left(\frac{1}{3} +z_{5}\right)c \, \mathbf{\hat{z}} & \left(6a\right) & \mbox{S III} \\ 
\mathbf{B}_{28} & = & -x_{5} \, \mathbf{a}_{1}-y_{5} \, \mathbf{a}_{2} + \left(\frac{1}{2} +z_{5}\right) \, \mathbf{a}_{3} & = & -\frac{1}{2}\left(x_{5}+y_{5}\right)a \, \mathbf{\hat{x}} + \frac{\sqrt{3}}{2}\left(x_{5}-y_{5}\right)a \, \mathbf{\hat{y}} + \left(\frac{1}{2} +z_{5}\right)c \, \mathbf{\hat{z}} & \left(6a\right) & \mbox{S III} \\ 
\mathbf{B}_{29} & = & y_{5} \, \mathbf{a}_{1} + \left(-x_{5}+y_{5}\right) \, \mathbf{a}_{2} + \left(\frac{1}{6} +z_{5}\right) \, \mathbf{a}_{3} & = & \left(-\frac{1}{2}x_{5}+y_{5}\right)a \, \mathbf{\hat{x}}-\frac{\sqrt{3}}{2}x_{5}a \, \mathbf{\hat{y}} + \left(\frac{1}{6} +z_{5}\right)c \, \mathbf{\hat{z}} & \left(6a\right) & \mbox{S III} \\ 
\mathbf{B}_{30} & = & \left(x_{5}-y_{5}\right) \, \mathbf{a}_{1} + x_{5} \, \mathbf{a}_{2} + \left(\frac{5}{6} +z_{5}\right) \, \mathbf{a}_{3} & = & \left(x_{5}-\frac{1}{2}y_{5}\right)a \, \mathbf{\hat{x}} + \frac{\sqrt{3}}{2}y_{5}a \, \mathbf{\hat{y}} + \left(\frac{5}{6} +z_{5}\right)c \, \mathbf{\hat{z}} & \left(6a\right) & \mbox{S III} \\ 
\end{longtabu}
\renewcommand{\arraystretch}{1.0}
\noindent \hrulefill
\\
\textbf{References:}
\vspace*{-0.25cm}
\begin{flushleft}
  - \bibentry{Eisenmann_Al2S3_ZKrist_1992}. \\
\end{flushleft}
\noindent \hrulefill
\\
\textbf{Geometry files:}
\\
\noindent  - CIF: pp. {\hyperref[A2B3_hP30_170_2a_3a_cif]{\pageref{A2B3_hP30_170_2a_3a_cif}}} \\
\noindent  - POSCAR: pp. {\hyperref[A2B3_hP30_170_2a_3a_poscar]{\pageref{A2B3_hP30_170_2a_3a_poscar}}} \\
\onecolumn
{\phantomsection\label{A10B2C_hP39_171_5c_c_a}}
\subsection*{\huge \textbf{{\normalfont Sr[S$_{2}$O$_{6}$][H$_{2}$O]$_{4}$ Structure: A10B2C\_hP39\_171\_5c\_c\_a}}}
\noindent \hrulefill
\vspace*{0.25cm}
\begin{figure}[htp]
  \centering
  \vspace{-1em}
  {\includegraphics[width=1\textwidth]{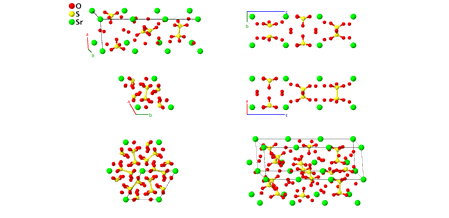}}
\end{figure}
\vspace*{-0.5cm}
\renewcommand{\arraystretch}{1.5}
\begin{equation*}
  \begin{array}{>{$\hspace{-0.15cm}}l<{$}>{$}p{0.5cm}<{$}>{$}p{18.5cm}<{$}}
    \mbox{\large \textbf{Prototype}} &\colon & \ce{Sr[S2O6][H2O]4} \\
    \mbox{\large \textbf{\AFLOW\ prototype label}} &\colon & \mbox{A10B2C\_hP39\_171\_5c\_c\_a} \\
    \mbox{\large \textbf{\textit{Strukturbericht} designation}} &\colon & \mbox{None} \\
    \mbox{\large \textbf{Pearson symbol}} &\colon & \mbox{hP39} \\
    \mbox{\large \textbf{Space group number}} &\colon & 171 \\
    \mbox{\large \textbf{Space group symbol}} &\colon & P6_{2} \\
    \mbox{\large \textbf{\AFLOW\ prototype command}} &\colon &  \texttt{aflow} \,  \, \texttt{-{}-proto=A10B2C\_hP39\_171\_5c\_c\_a } \, \newline \texttt{-{}-params=}{a,c/a,z_{1},x_{2},y_{2},z_{2},x_{3},y_{3},z_{3},x_{4},y_{4},z_{4},x_{5},y_{5},z_{5},x_{6},y_{6},z_{6},x_{7},y_{7},z_{7} }
  \end{array}
\end{equation*}
\renewcommand{\arraystretch}{1.0}

\vspace*{-0.25cm}
\noindent \hrulefill
\begin{itemize}
  \item{This structure is the enantiomorph of the \hyperref[A10B2C_hP39_172_5c_c_a]{Sr[S$_{2}$O$_{6}$][H$_{2}$O]$_{4}$ (A10B2C\_hP39\_172\_5c\_c\_a) structure}.
Only the non-hydrogen atoms are included in the prototype.
}
\end{itemize}

\noindent \parbox{1 \linewidth}{
\noindent \hrulefill
\\
\textbf{Hexagonal primitive vectors:} \\
\vspace*{-0.25cm}
\begin{tabular}{cc}
  \begin{tabular}{c}
    \parbox{0.6 \linewidth}{
      \renewcommand{\arraystretch}{1.5}
      \begin{equation*}
        \centering
        \begin{array}{ccc}
              \mathbf{a}_1 & = & \frac12 \, a \, \mathbf{\hat{x}} - \frac{\sqrt3}2 \, a \, \mathbf{\hat{y}} \\
    \mathbf{a}_2 & = & \frac12 \, a \, \mathbf{\hat{x}} + \frac{\sqrt3}2 \, a \, \mathbf{\hat{y}} \\
    \mathbf{a}_3 & = & c \, \mathbf{\hat{z}} \\

        \end{array}
      \end{equation*}
    }
    \renewcommand{\arraystretch}{1.0}
  \end{tabular}
  \begin{tabular}{c}
    \includegraphics[width=0.3\linewidth]{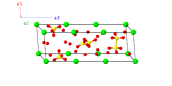} \\
  \end{tabular}
\end{tabular}

}
\vspace*{-0.25cm}

\noindent \hrulefill
\\
\textbf{Basis vectors:}
\vspace*{-0.25cm}
\renewcommand{\arraystretch}{1.5}
\begin{longtabu} to \textwidth{>{\centering $}X[-1,c,c]<{$}>{\centering $}X[-1,c,c]<{$}>{\centering $}X[-1,c,c]<{$}>{\centering $}X[-1,c,c]<{$}>{\centering $}X[-1,c,c]<{$}>{\centering $}X[-1,c,c]<{$}>{\centering $}X[-1,c,c]<{$}}
  & & \mbox{Lattice Coordinates} & & \mbox{Cartesian Coordinates} &\mbox{Wyckoff Position} & \mbox{Atom Type} \\  
  \mathbf{B}_{1} & = & z_{1} \, \mathbf{a}_{3} & = & z_{1}c \, \mathbf{\hat{z}} & \left(3a\right) & \mbox{Sr} \\ 
\mathbf{B}_{2} & = & \left(\frac{2}{3} +z_{1}\right) \, \mathbf{a}_{3} & = & \left(\frac{2}{3} +z_{1}\right)c \, \mathbf{\hat{z}} & \left(3a\right) & \mbox{Sr} \\ 
\mathbf{B}_{3} & = & \left(\frac{1}{3} +z_{1}\right) \, \mathbf{a}_{3} & = & \left(\frac{1}{3} +z_{1}\right)c \, \mathbf{\hat{z}} & \left(3a\right) & \mbox{Sr} \\ 
\mathbf{B}_{4} & = & x_{2} \, \mathbf{a}_{1} + y_{2} \, \mathbf{a}_{2} + z_{2} \, \mathbf{a}_{3} & = & \frac{1}{2}\left(x_{2}+y_{2}\right)a \, \mathbf{\hat{x}} + \frac{\sqrt{3}}{2}\left(-x_{2}+y_{2}\right)a \, \mathbf{\hat{y}} + z_{2}c \, \mathbf{\hat{z}} & \left(6c\right) & \mbox{O I} \\ 
\mathbf{B}_{5} & = & -y_{2} \, \mathbf{a}_{1} + \left(x_{2}-y_{2}\right) \, \mathbf{a}_{2} + \left(\frac{2}{3} +z_{2}\right) \, \mathbf{a}_{3} & = & \left(\frac{1}{2}x_{2}-y_{2}\right)a \, \mathbf{\hat{x}} + \frac{\sqrt{3}}{2}x_{2}a \, \mathbf{\hat{y}} + \left(\frac{2}{3} +z_{2}\right)c \, \mathbf{\hat{z}} & \left(6c\right) & \mbox{O I} \\ 
\mathbf{B}_{6} & = & \left(-x_{2}+y_{2}\right) \, \mathbf{a}_{1}-x_{2} \, \mathbf{a}_{2} + \left(\frac{1}{3} +z_{2}\right) \, \mathbf{a}_{3} & = & \left(-x_{2}+\frac{1}{2}y_{2}\right)a \, \mathbf{\hat{x}}-\frac{\sqrt{3}}{2}y_{2}a \, \mathbf{\hat{y}} + \left(\frac{1}{3} +z_{2}\right)c \, \mathbf{\hat{z}} & \left(6c\right) & \mbox{O I} \\ 
\mathbf{B}_{7} & = & -x_{2} \, \mathbf{a}_{1}-y_{2} \, \mathbf{a}_{2} + z_{2} \, \mathbf{a}_{3} & = & -\frac{1}{2}\left(x_{2}+y_{2}\right)a \, \mathbf{\hat{x}} + \frac{\sqrt{3}}{2}\left(x_{2}-y_{2}\right)a \, \mathbf{\hat{y}} + z_{2}c \, \mathbf{\hat{z}} & \left(6c\right) & \mbox{O I} \\ 
\mathbf{B}_{8} & = & y_{2} \, \mathbf{a}_{1} + \left(-x_{2}+y_{2}\right) \, \mathbf{a}_{2} + \left(\frac{2}{3} +z_{2}\right) \, \mathbf{a}_{3} & = & \left(-\frac{1}{2}x_{2}+y_{2}\right)a \, \mathbf{\hat{x}}-\frac{\sqrt{3}}{2}x_{2}a \, \mathbf{\hat{y}} + \left(\frac{2}{3} +z_{2}\right)c \, \mathbf{\hat{z}} & \left(6c\right) & \mbox{O I} \\ 
\mathbf{B}_{9} & = & \left(x_{2}-y_{2}\right) \, \mathbf{a}_{1} + x_{2} \, \mathbf{a}_{2} + \left(\frac{1}{3} +z_{2}\right) \, \mathbf{a}_{3} & = & \left(x_{2}-\frac{1}{2}y_{2}\right)a \, \mathbf{\hat{x}} + \frac{\sqrt{3}}{2}y_{2}a \, \mathbf{\hat{y}} + \left(\frac{1}{3} +z_{2}\right)c \, \mathbf{\hat{z}} & \left(6c\right) & \mbox{O I} \\ 
\mathbf{B}_{10} & = & x_{3} \, \mathbf{a}_{1} + y_{3} \, \mathbf{a}_{2} + z_{3} \, \mathbf{a}_{3} & = & \frac{1}{2}\left(x_{3}+y_{3}\right)a \, \mathbf{\hat{x}} + \frac{\sqrt{3}}{2}\left(-x_{3}+y_{3}\right)a \, \mathbf{\hat{y}} + z_{3}c \, \mathbf{\hat{z}} & \left(6c\right) & \mbox{O II} \\ 
\mathbf{B}_{11} & = & -y_{3} \, \mathbf{a}_{1} + \left(x_{3}-y_{3}\right) \, \mathbf{a}_{2} + \left(\frac{2}{3} +z_{3}\right) \, \mathbf{a}_{3} & = & \left(\frac{1}{2}x_{3}-y_{3}\right)a \, \mathbf{\hat{x}} + \frac{\sqrt{3}}{2}x_{3}a \, \mathbf{\hat{y}} + \left(\frac{2}{3} +z_{3}\right)c \, \mathbf{\hat{z}} & \left(6c\right) & \mbox{O II} \\ 
\mathbf{B}_{12} & = & \left(-x_{3}+y_{3}\right) \, \mathbf{a}_{1}-x_{3} \, \mathbf{a}_{2} + \left(\frac{1}{3} +z_{3}\right) \, \mathbf{a}_{3} & = & \left(-x_{3}+\frac{1}{2}y_{3}\right)a \, \mathbf{\hat{x}}-\frac{\sqrt{3}}{2}y_{3}a \, \mathbf{\hat{y}} + \left(\frac{1}{3} +z_{3}\right)c \, \mathbf{\hat{z}} & \left(6c\right) & \mbox{O II} \\ 
\mathbf{B}_{13} & = & -x_{3} \, \mathbf{a}_{1}-y_{3} \, \mathbf{a}_{2} + z_{3} \, \mathbf{a}_{3} & = & -\frac{1}{2}\left(x_{3}+y_{3}\right)a \, \mathbf{\hat{x}} + \frac{\sqrt{3}}{2}\left(x_{3}-y_{3}\right)a \, \mathbf{\hat{y}} + z_{3}c \, \mathbf{\hat{z}} & \left(6c\right) & \mbox{O II} \\ 
\mathbf{B}_{14} & = & y_{3} \, \mathbf{a}_{1} + \left(-x_{3}+y_{3}\right) \, \mathbf{a}_{2} + \left(\frac{2}{3} +z_{3}\right) \, \mathbf{a}_{3} & = & \left(-\frac{1}{2}x_{3}+y_{3}\right)a \, \mathbf{\hat{x}}-\frac{\sqrt{3}}{2}x_{3}a \, \mathbf{\hat{y}} + \left(\frac{2}{3} +z_{3}\right)c \, \mathbf{\hat{z}} & \left(6c\right) & \mbox{O II} \\ 
\mathbf{B}_{15} & = & \left(x_{3}-y_{3}\right) \, \mathbf{a}_{1} + x_{3} \, \mathbf{a}_{2} + \left(\frac{1}{3} +z_{3}\right) \, \mathbf{a}_{3} & = & \left(x_{3}-\frac{1}{2}y_{3}\right)a \, \mathbf{\hat{x}} + \frac{\sqrt{3}}{2}y_{3}a \, \mathbf{\hat{y}} + \left(\frac{1}{3} +z_{3}\right)c \, \mathbf{\hat{z}} & \left(6c\right) & \mbox{O II} \\ 
\mathbf{B}_{16} & = & x_{4} \, \mathbf{a}_{1} + y_{4} \, \mathbf{a}_{2} + z_{4} \, \mathbf{a}_{3} & = & \frac{1}{2}\left(x_{4}+y_{4}\right)a \, \mathbf{\hat{x}} + \frac{\sqrt{3}}{2}\left(-x_{4}+y_{4}\right)a \, \mathbf{\hat{y}} + z_{4}c \, \mathbf{\hat{z}} & \left(6c\right) & \mbox{O III} \\ 
\mathbf{B}_{17} & = & -y_{4} \, \mathbf{a}_{1} + \left(x_{4}-y_{4}\right) \, \mathbf{a}_{2} + \left(\frac{2}{3} +z_{4}\right) \, \mathbf{a}_{3} & = & \left(\frac{1}{2}x_{4}-y_{4}\right)a \, \mathbf{\hat{x}} + \frac{\sqrt{3}}{2}x_{4}a \, \mathbf{\hat{y}} + \left(\frac{2}{3} +z_{4}\right)c \, \mathbf{\hat{z}} & \left(6c\right) & \mbox{O III} \\ 
\mathbf{B}_{18} & = & \left(-x_{4}+y_{4}\right) \, \mathbf{a}_{1}-x_{4} \, \mathbf{a}_{2} + \left(\frac{1}{3} +z_{4}\right) \, \mathbf{a}_{3} & = & \left(-x_{4}+\frac{1}{2}y_{4}\right)a \, \mathbf{\hat{x}}-\frac{\sqrt{3}}{2}y_{4}a \, \mathbf{\hat{y}} + \left(\frac{1}{3} +z_{4}\right)c \, \mathbf{\hat{z}} & \left(6c\right) & \mbox{O III} \\ 
\mathbf{B}_{19} & = & -x_{4} \, \mathbf{a}_{1}-y_{4} \, \mathbf{a}_{2} + z_{4} \, \mathbf{a}_{3} & = & -\frac{1}{2}\left(x_{4}+y_{4}\right)a \, \mathbf{\hat{x}} + \frac{\sqrt{3}}{2}\left(x_{4}-y_{4}\right)a \, \mathbf{\hat{y}} + z_{4}c \, \mathbf{\hat{z}} & \left(6c\right) & \mbox{O III} \\ 
\mathbf{B}_{20} & = & y_{4} \, \mathbf{a}_{1} + \left(-x_{4}+y_{4}\right) \, \mathbf{a}_{2} + \left(\frac{2}{3} +z_{4}\right) \, \mathbf{a}_{3} & = & \left(-\frac{1}{2}x_{4}+y_{4}\right)a \, \mathbf{\hat{x}}-\frac{\sqrt{3}}{2}x_{4}a \, \mathbf{\hat{y}} + \left(\frac{2}{3} +z_{4}\right)c \, \mathbf{\hat{z}} & \left(6c\right) & \mbox{O III} \\ 
\mathbf{B}_{21} & = & \left(x_{4}-y_{4}\right) \, \mathbf{a}_{1} + x_{4} \, \mathbf{a}_{2} + \left(\frac{1}{3} +z_{4}\right) \, \mathbf{a}_{3} & = & \left(x_{4}-\frac{1}{2}y_{4}\right)a \, \mathbf{\hat{x}} + \frac{\sqrt{3}}{2}y_{4}a \, \mathbf{\hat{y}} + \left(\frac{1}{3} +z_{4}\right)c \, \mathbf{\hat{z}} & \left(6c\right) & \mbox{O III} \\ 
\mathbf{B}_{22} & = & x_{5} \, \mathbf{a}_{1} + y_{5} \, \mathbf{a}_{2} + z_{5} \, \mathbf{a}_{3} & = & \frac{1}{2}\left(x_{5}+y_{5}\right)a \, \mathbf{\hat{x}} + \frac{\sqrt{3}}{2}\left(-x_{5}+y_{5}\right)a \, \mathbf{\hat{y}} + z_{5}c \, \mathbf{\hat{z}} & \left(6c\right) & \mbox{O IV} \\ 
\mathbf{B}_{23} & = & -y_{5} \, \mathbf{a}_{1} + \left(x_{5}-y_{5}\right) \, \mathbf{a}_{2} + \left(\frac{2}{3} +z_{5}\right) \, \mathbf{a}_{3} & = & \left(\frac{1}{2}x_{5}-y_{5}\right)a \, \mathbf{\hat{x}} + \frac{\sqrt{3}}{2}x_{5}a \, \mathbf{\hat{y}} + \left(\frac{2}{3} +z_{5}\right)c \, \mathbf{\hat{z}} & \left(6c\right) & \mbox{O IV} \\ 
\mathbf{B}_{24} & = & \left(-x_{5}+y_{5}\right) \, \mathbf{a}_{1}-x_{5} \, \mathbf{a}_{2} + \left(\frac{1}{3} +z_{5}\right) \, \mathbf{a}_{3} & = & \left(-x_{5}+\frac{1}{2}y_{5}\right)a \, \mathbf{\hat{x}}-\frac{\sqrt{3}}{2}y_{5}a \, \mathbf{\hat{y}} + \left(\frac{1}{3} +z_{5}\right)c \, \mathbf{\hat{z}} & \left(6c\right) & \mbox{O IV} \\ 
\mathbf{B}_{25} & = & -x_{5} \, \mathbf{a}_{1}-y_{5} \, \mathbf{a}_{2} + z_{5} \, \mathbf{a}_{3} & = & -\frac{1}{2}\left(x_{5}+y_{5}\right)a \, \mathbf{\hat{x}} + \frac{\sqrt{3}}{2}\left(x_{5}-y_{5}\right)a \, \mathbf{\hat{y}} + z_{5}c \, \mathbf{\hat{z}} & \left(6c\right) & \mbox{O IV} \\ 
\mathbf{B}_{26} & = & y_{5} \, \mathbf{a}_{1} + \left(-x_{5}+y_{5}\right) \, \mathbf{a}_{2} + \left(\frac{2}{3} +z_{5}\right) \, \mathbf{a}_{3} & = & \left(-\frac{1}{2}x_{5}+y_{5}\right)a \, \mathbf{\hat{x}}-\frac{\sqrt{3}}{2}x_{5}a \, \mathbf{\hat{y}} + \left(\frac{2}{3} +z_{5}\right)c \, \mathbf{\hat{z}} & \left(6c\right) & \mbox{O IV} \\ 
\mathbf{B}_{27} & = & \left(x_{5}-y_{5}\right) \, \mathbf{a}_{1} + x_{5} \, \mathbf{a}_{2} + \left(\frac{1}{3} +z_{5}\right) \, \mathbf{a}_{3} & = & \left(x_{5}-\frac{1}{2}y_{5}\right)a \, \mathbf{\hat{x}} + \frac{\sqrt{3}}{2}y_{5}a \, \mathbf{\hat{y}} + \left(\frac{1}{3} +z_{5}\right)c \, \mathbf{\hat{z}} & \left(6c\right) & \mbox{O IV} \\ 
\mathbf{B}_{28} & = & x_{6} \, \mathbf{a}_{1} + y_{6} \, \mathbf{a}_{2} + z_{6} \, \mathbf{a}_{3} & = & \frac{1}{2}\left(x_{6}+y_{6}\right)a \, \mathbf{\hat{x}} + \frac{\sqrt{3}}{2}\left(-x_{6}+y_{6}\right)a \, \mathbf{\hat{y}} + z_{6}c \, \mathbf{\hat{z}} & \left(6c\right) & \mbox{O V} \\ 
\mathbf{B}_{29} & = & -y_{6} \, \mathbf{a}_{1} + \left(x_{6}-y_{6}\right) \, \mathbf{a}_{2} + \left(\frac{2}{3} +z_{6}\right) \, \mathbf{a}_{3} & = & \left(\frac{1}{2}x_{6}-y_{6}\right)a \, \mathbf{\hat{x}} + \frac{\sqrt{3}}{2}x_{6}a \, \mathbf{\hat{y}} + \left(\frac{2}{3} +z_{6}\right)c \, \mathbf{\hat{z}} & \left(6c\right) & \mbox{O V} \\ 
\mathbf{B}_{30} & = & \left(-x_{6}+y_{6}\right) \, \mathbf{a}_{1}-x_{6} \, \mathbf{a}_{2} + \left(\frac{1}{3} +z_{6}\right) \, \mathbf{a}_{3} & = & \left(-x_{6}+\frac{1}{2}y_{6}\right)a \, \mathbf{\hat{x}}-\frac{\sqrt{3}}{2}y_{6}a \, \mathbf{\hat{y}} + \left(\frac{1}{3} +z_{6}\right)c \, \mathbf{\hat{z}} & \left(6c\right) & \mbox{O V} \\ 
\mathbf{B}_{31} & = & -x_{6} \, \mathbf{a}_{1}-y_{6} \, \mathbf{a}_{2} + z_{6} \, \mathbf{a}_{3} & = & -\frac{1}{2}\left(x_{6}+y_{6}\right)a \, \mathbf{\hat{x}} + \frac{\sqrt{3}}{2}\left(x_{6}-y_{6}\right)a \, \mathbf{\hat{y}} + z_{6}c \, \mathbf{\hat{z}} & \left(6c\right) & \mbox{O V} \\ 
\mathbf{B}_{32} & = & y_{6} \, \mathbf{a}_{1} + \left(-x_{6}+y_{6}\right) \, \mathbf{a}_{2} + \left(\frac{2}{3} +z_{6}\right) \, \mathbf{a}_{3} & = & \left(-\frac{1}{2}x_{6}+y_{6}\right)a \, \mathbf{\hat{x}}-\frac{\sqrt{3}}{2}x_{6}a \, \mathbf{\hat{y}} + \left(\frac{2}{3} +z_{6}\right)c \, \mathbf{\hat{z}} & \left(6c\right) & \mbox{O V} \\ 
\mathbf{B}_{33} & = & \left(x_{6}-y_{6}\right) \, \mathbf{a}_{1} + x_{6} \, \mathbf{a}_{2} + \left(\frac{1}{3} +z_{6}\right) \, \mathbf{a}_{3} & = & \left(x_{6}-\frac{1}{2}y_{6}\right)a \, \mathbf{\hat{x}} + \frac{\sqrt{3}}{2}y_{6}a \, \mathbf{\hat{y}} + \left(\frac{1}{3} +z_{6}\right)c \, \mathbf{\hat{z}} & \left(6c\right) & \mbox{O V} \\ 
\mathbf{B}_{34} & = & x_{7} \, \mathbf{a}_{1} + y_{7} \, \mathbf{a}_{2} + z_{7} \, \mathbf{a}_{3} & = & \frac{1}{2}\left(x_{7}+y_{7}\right)a \, \mathbf{\hat{x}} + \frac{\sqrt{3}}{2}\left(-x_{7}+y_{7}\right)a \, \mathbf{\hat{y}} + z_{7}c \, \mathbf{\hat{z}} & \left(6c\right) & \mbox{S} \\ 
\mathbf{B}_{35} & = & -y_{7} \, \mathbf{a}_{1} + \left(x_{7}-y_{7}\right) \, \mathbf{a}_{2} + \left(\frac{2}{3} +z_{7}\right) \, \mathbf{a}_{3} & = & \left(\frac{1}{2}x_{7}-y_{7}\right)a \, \mathbf{\hat{x}} + \frac{\sqrt{3}}{2}x_{7}a \, \mathbf{\hat{y}} + \left(\frac{2}{3} +z_{7}\right)c \, \mathbf{\hat{z}} & \left(6c\right) & \mbox{S} \\ 
\mathbf{B}_{36} & = & \left(-x_{7}+y_{7}\right) \, \mathbf{a}_{1}-x_{7} \, \mathbf{a}_{2} + \left(\frac{1}{3} +z_{7}\right) \, \mathbf{a}_{3} & = & \left(-x_{7}+\frac{1}{2}y_{7}\right)a \, \mathbf{\hat{x}}-\frac{\sqrt{3}}{2}y_{7}a \, \mathbf{\hat{y}} + \left(\frac{1}{3} +z_{7}\right)c \, \mathbf{\hat{z}} & \left(6c\right) & \mbox{S} \\ 
\mathbf{B}_{37} & = & -x_{7} \, \mathbf{a}_{1}-y_{7} \, \mathbf{a}_{2} + z_{7} \, \mathbf{a}_{3} & = & -\frac{1}{2}\left(x_{7}+y_{7}\right)a \, \mathbf{\hat{x}} + \frac{\sqrt{3}}{2}\left(x_{7}-y_{7}\right)a \, \mathbf{\hat{y}} + z_{7}c \, \mathbf{\hat{z}} & \left(6c\right) & \mbox{S} \\ 
\mathbf{B}_{38} & = & y_{7} \, \mathbf{a}_{1} + \left(-x_{7}+y_{7}\right) \, \mathbf{a}_{2} + \left(\frac{2}{3} +z_{7}\right) \, \mathbf{a}_{3} & = & \left(-\frac{1}{2}x_{7}+y_{7}\right)a \, \mathbf{\hat{x}}-\frac{\sqrt{3}}{2}x_{7}a \, \mathbf{\hat{y}} + \left(\frac{2}{3} +z_{7}\right)c \, \mathbf{\hat{z}} & \left(6c\right) & \mbox{S} \\ 
\mathbf{B}_{39} & = & \left(x_{7}-y_{7}\right) \, \mathbf{a}_{1} + x_{7} \, \mathbf{a}_{2} + \left(\frac{1}{3} +z_{7}\right) \, \mathbf{a}_{3} & = & \left(x_{7}-\frac{1}{2}y_{7}\right)a \, \mathbf{\hat{x}} + \frac{\sqrt{3}}{2}y_{7}a \, \mathbf{\hat{y}} + \left(\frac{1}{3} +z_{7}\right)c \, \mathbf{\hat{z}} & \left(6c\right) & \mbox{S} \\ 
\end{longtabu}
\renewcommand{\arraystretch}{1.0}
\noindent \hrulefill
\\
\textbf{References:}
\vspace*{-0.25cm}
\begin{flushleft}
  - \bibentry{Hargreaves_SrS2O6H2O4_ZKristallogrCrystalMat_1972}. \\
\end{flushleft}
\textbf{Found in:}
\vspace*{-0.25cm}
\begin{flushleft}
  - \bibentry{Villars_PearsonsCrystalData_2013}. \\
\end{flushleft}
\noindent \hrulefill
\\
\textbf{Geometry files:}
\\
\noindent  - CIF: pp. {\hyperref[A10B2C_hP39_171_5c_c_a_cif]{\pageref{A10B2C_hP39_171_5c_c_a_cif}}} \\
\noindent  - POSCAR: pp. {\hyperref[A10B2C_hP39_171_5c_c_a_poscar]{\pageref{A10B2C_hP39_171_5c_c_a_poscar}}} \\
\onecolumn
{\phantomsection\label{A10B2C_hP39_172_5c_c_a}}
\subsection*{\huge \textbf{{\normalfont Sr[S$_{2}$O$_{6}$][H$_{2}$O]$_{4}$ Structure: A10B2C\_hP39\_172\_5c\_c\_a}}}
\noindent \hrulefill
\vspace*{0.25cm}
\begin{figure}[htp]
  \centering
  \vspace{-1em}
  {\includegraphics[width=1\textwidth]{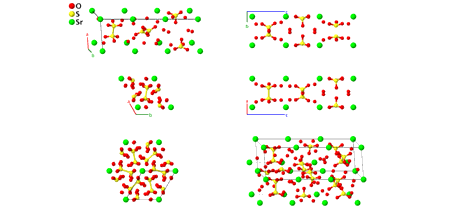}}
\end{figure}
\vspace*{-0.5cm}
\renewcommand{\arraystretch}{1.5}
\begin{equation*}
  \begin{array}{>{$\hspace{-0.15cm}}l<{$}>{$}p{0.5cm}<{$}>{$}p{18.5cm}<{$}}
    \mbox{\large \textbf{Prototype}} &\colon & \ce{Sr[S2O6][H2O]4} \\
    \mbox{\large \textbf{\AFLOW\ prototype label}} &\colon & \mbox{A10B2C\_hP39\_172\_5c\_c\_a} \\
    \mbox{\large \textbf{\textit{Strukturbericht} designation}} &\colon & \mbox{None} \\
    \mbox{\large \textbf{Pearson symbol}} &\colon & \mbox{hP39} \\
    \mbox{\large \textbf{Space group number}} &\colon & 172 \\
    \mbox{\large \textbf{Space group symbol}} &\colon & P6_{4} \\
    \mbox{\large \textbf{\AFLOW\ prototype command}} &\colon &  \texttt{aflow} \,  \, \texttt{-{}-proto=A10B2C\_hP39\_172\_5c\_c\_a } \, \newline \texttt{-{}-params=}{a,c/a,z_{1},x_{2},y_{2},z_{2},x_{3},y_{3},z_{3},x_{4},y_{4},z_{4},x_{5},y_{5},z_{5},x_{6},y_{6},z_{6},x_{7},y_{7},z_{7} }
  \end{array}
\end{equation*}
\renewcommand{\arraystretch}{1.0}

\vspace*{-0.25cm}
\noindent \hrulefill
\begin{itemize}
  \item{This structure is the enantiomorph of the \hyperref[A10B2C_hP39_171_5c_c_a]{Sr[S$_{2}$O$_{6}$][H$_{2}$O]$_{4}$ (A10B2C\_hP39\_171\_5c\_c\_a) structure}, 
and was generated by reflecting the coordinates of the space group \#171 structure through the $z=0$ plane.  
Only the non-hydrogen atoms are included in the prototype.
}
\end{itemize}

\noindent \parbox{1 \linewidth}{
\noindent \hrulefill
\\
\textbf{Hexagonal primitive vectors:} \\
\vspace*{-0.25cm}
\begin{tabular}{cc}
  \begin{tabular}{c}
    \parbox{0.6 \linewidth}{
      \renewcommand{\arraystretch}{1.5}
      \begin{equation*}
        \centering
        \begin{array}{ccc}
              \mathbf{a}_1 & = & \frac12 \, a \, \mathbf{\hat{x}} - \frac{\sqrt3}2 \, a \, \mathbf{\hat{y}} \\
    \mathbf{a}_2 & = & \frac12 \, a \, \mathbf{\hat{x}} + \frac{\sqrt3}2 \, a \, \mathbf{\hat{y}} \\
    \mathbf{a}_3 & = & c \, \mathbf{\hat{z}} \\

        \end{array}
      \end{equation*}
    }
    \renewcommand{\arraystretch}{1.0}
  \end{tabular}
  \begin{tabular}{c}
    \includegraphics[width=0.3\linewidth]{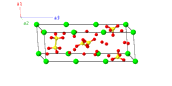} \\
  \end{tabular}
\end{tabular}

}
\vspace*{-0.25cm}

\noindent \hrulefill
\\
\textbf{Basis vectors:}
\vspace*{-0.25cm}
\renewcommand{\arraystretch}{1.5}
\begin{longtabu} to \textwidth{>{\centering $}X[-1,c,c]<{$}>{\centering $}X[-1,c,c]<{$}>{\centering $}X[-1,c,c]<{$}>{\centering $}X[-1,c,c]<{$}>{\centering $}X[-1,c,c]<{$}>{\centering $}X[-1,c,c]<{$}>{\centering $}X[-1,c,c]<{$}}
  & & \mbox{Lattice Coordinates} & & \mbox{Cartesian Coordinates} &\mbox{Wyckoff Position} & \mbox{Atom Type} \\  
  \mathbf{B}_{1} & = & z_{1} \, \mathbf{a}_{3} & = & z_{1}c \, \mathbf{\hat{z}} & \left(3a\right) & \mbox{Sr} \\ 
\mathbf{B}_{2} & = & \left(\frac{1}{3} +z_{1}\right) \, \mathbf{a}_{3} & = & \left(\frac{1}{3} +z_{1}\right)c \, \mathbf{\hat{z}} & \left(3a\right) & \mbox{Sr} \\ 
\mathbf{B}_{3} & = & \left(\frac{2}{3} +z_{1}\right) \, \mathbf{a}_{3} & = & \left(\frac{2}{3} +z_{1}\right)c \, \mathbf{\hat{z}} & \left(3a\right) & \mbox{Sr} \\ 
\mathbf{B}_{4} & = & x_{2} \, \mathbf{a}_{1} + y_{2} \, \mathbf{a}_{2} + z_{2} \, \mathbf{a}_{3} & = & \frac{1}{2}\left(x_{2}+y_{2}\right)a \, \mathbf{\hat{x}} + \frac{\sqrt{3}}{2}\left(-x_{2}+y_{2}\right)a \, \mathbf{\hat{y}} + z_{2}c \, \mathbf{\hat{z}} & \left(6c\right) & \mbox{O I} \\ 
\mathbf{B}_{5} & = & -y_{2} \, \mathbf{a}_{1} + \left(x_{2}-y_{2}\right) \, \mathbf{a}_{2} + \left(\frac{1}{3} +z_{2}\right) \, \mathbf{a}_{3} & = & \left(\frac{1}{2}x_{2}-y_{2}\right)a \, \mathbf{\hat{x}} + \frac{\sqrt{3}}{2}x_{2}a \, \mathbf{\hat{y}} + \left(\frac{1}{3} +z_{2}\right)c \, \mathbf{\hat{z}} & \left(6c\right) & \mbox{O I} \\ 
\mathbf{B}_{6} & = & \left(-x_{2}+y_{2}\right) \, \mathbf{a}_{1}-x_{2} \, \mathbf{a}_{2} + \left(\frac{2}{3} +z_{2}\right) \, \mathbf{a}_{3} & = & \left(-x_{2}+\frac{1}{2}y_{2}\right)a \, \mathbf{\hat{x}}-\frac{\sqrt{3}}{2}y_{2}a \, \mathbf{\hat{y}} + \left(\frac{2}{3} +z_{2}\right)c \, \mathbf{\hat{z}} & \left(6c\right) & \mbox{O I} \\ 
\mathbf{B}_{7} & = & -x_{2} \, \mathbf{a}_{1}-y_{2} \, \mathbf{a}_{2} + z_{2} \, \mathbf{a}_{3} & = & -\frac{1}{2}\left(x_{2}+y_{2}\right)a \, \mathbf{\hat{x}} + \frac{\sqrt{3}}{2}\left(x_{2}-y_{2}\right)a \, \mathbf{\hat{y}} + z_{2}c \, \mathbf{\hat{z}} & \left(6c\right) & \mbox{O I} \\ 
\mathbf{B}_{8} & = & y_{2} \, \mathbf{a}_{1} + \left(-x_{2}+y_{2}\right) \, \mathbf{a}_{2} + \left(\frac{1}{3} +z_{2}\right) \, \mathbf{a}_{3} & = & \left(-\frac{1}{2}x_{2}+y_{2}\right)a \, \mathbf{\hat{x}}-\frac{\sqrt{3}}{2}x_{2}a \, \mathbf{\hat{y}} + \left(\frac{1}{3} +z_{2}\right)c \, \mathbf{\hat{z}} & \left(6c\right) & \mbox{O I} \\ 
\mathbf{B}_{9} & = & \left(x_{2}-y_{2}\right) \, \mathbf{a}_{1} + x_{2} \, \mathbf{a}_{2} + \left(\frac{2}{3} +z_{2}\right) \, \mathbf{a}_{3} & = & \left(x_{2}-\frac{1}{2}y_{2}\right)a \, \mathbf{\hat{x}} + \frac{\sqrt{3}}{2}y_{2}a \, \mathbf{\hat{y}} + \left(\frac{2}{3} +z_{2}\right)c \, \mathbf{\hat{z}} & \left(6c\right) & \mbox{O I} \\ 
\mathbf{B}_{10} & = & x_{3} \, \mathbf{a}_{1} + y_{3} \, \mathbf{a}_{2} + z_{3} \, \mathbf{a}_{3} & = & \frac{1}{2}\left(x_{3}+y_{3}\right)a \, \mathbf{\hat{x}} + \frac{\sqrt{3}}{2}\left(-x_{3}+y_{3}\right)a \, \mathbf{\hat{y}} + z_{3}c \, \mathbf{\hat{z}} & \left(6c\right) & \mbox{O II} \\ 
\mathbf{B}_{11} & = & -y_{3} \, \mathbf{a}_{1} + \left(x_{3}-y_{3}\right) \, \mathbf{a}_{2} + \left(\frac{1}{3} +z_{3}\right) \, \mathbf{a}_{3} & = & \left(\frac{1}{2}x_{3}-y_{3}\right)a \, \mathbf{\hat{x}} + \frac{\sqrt{3}}{2}x_{3}a \, \mathbf{\hat{y}} + \left(\frac{1}{3} +z_{3}\right)c \, \mathbf{\hat{z}} & \left(6c\right) & \mbox{O II} \\ 
\mathbf{B}_{12} & = & \left(-x_{3}+y_{3}\right) \, \mathbf{a}_{1}-x_{3} \, \mathbf{a}_{2} + \left(\frac{2}{3} +z_{3}\right) \, \mathbf{a}_{3} & = & \left(-x_{3}+\frac{1}{2}y_{3}\right)a \, \mathbf{\hat{x}}-\frac{\sqrt{3}}{2}y_{3}a \, \mathbf{\hat{y}} + \left(\frac{2}{3} +z_{3}\right)c \, \mathbf{\hat{z}} & \left(6c\right) & \mbox{O II} \\ 
\mathbf{B}_{13} & = & -x_{3} \, \mathbf{a}_{1}-y_{3} \, \mathbf{a}_{2} + z_{3} \, \mathbf{a}_{3} & = & -\frac{1}{2}\left(x_{3}+y_{3}\right)a \, \mathbf{\hat{x}} + \frac{\sqrt{3}}{2}\left(x_{3}-y_{3}\right)a \, \mathbf{\hat{y}} + z_{3}c \, \mathbf{\hat{z}} & \left(6c\right) & \mbox{O II} \\ 
\mathbf{B}_{14} & = & y_{3} \, \mathbf{a}_{1} + \left(-x_{3}+y_{3}\right) \, \mathbf{a}_{2} + \left(\frac{1}{3} +z_{3}\right) \, \mathbf{a}_{3} & = & \left(-\frac{1}{2}x_{3}+y_{3}\right)a \, \mathbf{\hat{x}}-\frac{\sqrt{3}}{2}x_{3}a \, \mathbf{\hat{y}} + \left(\frac{1}{3} +z_{3}\right)c \, \mathbf{\hat{z}} & \left(6c\right) & \mbox{O II} \\ 
\mathbf{B}_{15} & = & \left(x_{3}-y_{3}\right) \, \mathbf{a}_{1} + x_{3} \, \mathbf{a}_{2} + \left(\frac{2}{3} +z_{3}\right) \, \mathbf{a}_{3} & = & \left(x_{3}-\frac{1}{2}y_{3}\right)a \, \mathbf{\hat{x}} + \frac{\sqrt{3}}{2}y_{3}a \, \mathbf{\hat{y}} + \left(\frac{2}{3} +z_{3}\right)c \, \mathbf{\hat{z}} & \left(6c\right) & \mbox{O II} \\ 
\mathbf{B}_{16} & = & x_{4} \, \mathbf{a}_{1} + y_{4} \, \mathbf{a}_{2} + z_{4} \, \mathbf{a}_{3} & = & \frac{1}{2}\left(x_{4}+y_{4}\right)a \, \mathbf{\hat{x}} + \frac{\sqrt{3}}{2}\left(-x_{4}+y_{4}\right)a \, \mathbf{\hat{y}} + z_{4}c \, \mathbf{\hat{z}} & \left(6c\right) & \mbox{O III} \\ 
\mathbf{B}_{17} & = & -y_{4} \, \mathbf{a}_{1} + \left(x_{4}-y_{4}\right) \, \mathbf{a}_{2} + \left(\frac{1}{3} +z_{4}\right) \, \mathbf{a}_{3} & = & \left(\frac{1}{2}x_{4}-y_{4}\right)a \, \mathbf{\hat{x}} + \frac{\sqrt{3}}{2}x_{4}a \, \mathbf{\hat{y}} + \left(\frac{1}{3} +z_{4}\right)c \, \mathbf{\hat{z}} & \left(6c\right) & \mbox{O III} \\ 
\mathbf{B}_{18} & = & \left(-x_{4}+y_{4}\right) \, \mathbf{a}_{1}-x_{4} \, \mathbf{a}_{2} + \left(\frac{2}{3} +z_{4}\right) \, \mathbf{a}_{3} & = & \left(-x_{4}+\frac{1}{2}y_{4}\right)a \, \mathbf{\hat{x}}-\frac{\sqrt{3}}{2}y_{4}a \, \mathbf{\hat{y}} + \left(\frac{2}{3} +z_{4}\right)c \, \mathbf{\hat{z}} & \left(6c\right) & \mbox{O III} \\ 
\mathbf{B}_{19} & = & -x_{4} \, \mathbf{a}_{1}-y_{4} \, \mathbf{a}_{2} + z_{4} \, \mathbf{a}_{3} & = & -\frac{1}{2}\left(x_{4}+y_{4}\right)a \, \mathbf{\hat{x}} + \frac{\sqrt{3}}{2}\left(x_{4}-y_{4}\right)a \, \mathbf{\hat{y}} + z_{4}c \, \mathbf{\hat{z}} & \left(6c\right) & \mbox{O III} \\ 
\mathbf{B}_{20} & = & y_{4} \, \mathbf{a}_{1} + \left(-x_{4}+y_{4}\right) \, \mathbf{a}_{2} + \left(\frac{1}{3} +z_{4}\right) \, \mathbf{a}_{3} & = & \left(-\frac{1}{2}x_{4}+y_{4}\right)a \, \mathbf{\hat{x}}-\frac{\sqrt{3}}{2}x_{4}a \, \mathbf{\hat{y}} + \left(\frac{1}{3} +z_{4}\right)c \, \mathbf{\hat{z}} & \left(6c\right) & \mbox{O III} \\ 
\mathbf{B}_{21} & = & \left(x_{4}-y_{4}\right) \, \mathbf{a}_{1} + x_{4} \, \mathbf{a}_{2} + \left(\frac{2}{3} +z_{4}\right) \, \mathbf{a}_{3} & = & \left(x_{4}-\frac{1}{2}y_{4}\right)a \, \mathbf{\hat{x}} + \frac{\sqrt{3}}{2}y_{4}a \, \mathbf{\hat{y}} + \left(\frac{2}{3} +z_{4}\right)c \, \mathbf{\hat{z}} & \left(6c\right) & \mbox{O III} \\ 
\mathbf{B}_{22} & = & x_{5} \, \mathbf{a}_{1} + y_{5} \, \mathbf{a}_{2} + z_{5} \, \mathbf{a}_{3} & = & \frac{1}{2}\left(x_{5}+y_{5}\right)a \, \mathbf{\hat{x}} + \frac{\sqrt{3}}{2}\left(-x_{5}+y_{5}\right)a \, \mathbf{\hat{y}} + z_{5}c \, \mathbf{\hat{z}} & \left(6c\right) & \mbox{O IV} \\ 
\mathbf{B}_{23} & = & -y_{5} \, \mathbf{a}_{1} + \left(x_{5}-y_{5}\right) \, \mathbf{a}_{2} + \left(\frac{1}{3} +z_{5}\right) \, \mathbf{a}_{3} & = & \left(\frac{1}{2}x_{5}-y_{5}\right)a \, \mathbf{\hat{x}} + \frac{\sqrt{3}}{2}x_{5}a \, \mathbf{\hat{y}} + \left(\frac{1}{3} +z_{5}\right)c \, \mathbf{\hat{z}} & \left(6c\right) & \mbox{O IV} \\ 
\mathbf{B}_{24} & = & \left(-x_{5}+y_{5}\right) \, \mathbf{a}_{1}-x_{5} \, \mathbf{a}_{2} + \left(\frac{2}{3} +z_{5}\right) \, \mathbf{a}_{3} & = & \left(-x_{5}+\frac{1}{2}y_{5}\right)a \, \mathbf{\hat{x}}-\frac{\sqrt{3}}{2}y_{5}a \, \mathbf{\hat{y}} + \left(\frac{2}{3} +z_{5}\right)c \, \mathbf{\hat{z}} & \left(6c\right) & \mbox{O IV} \\ 
\mathbf{B}_{25} & = & -x_{5} \, \mathbf{a}_{1}-y_{5} \, \mathbf{a}_{2} + z_{5} \, \mathbf{a}_{3} & = & -\frac{1}{2}\left(x_{5}+y_{5}\right)a \, \mathbf{\hat{x}} + \frac{\sqrt{3}}{2}\left(x_{5}-y_{5}\right)a \, \mathbf{\hat{y}} + z_{5}c \, \mathbf{\hat{z}} & \left(6c\right) & \mbox{O IV} \\ 
\mathbf{B}_{26} & = & y_{5} \, \mathbf{a}_{1} + \left(-x_{5}+y_{5}\right) \, \mathbf{a}_{2} + \left(\frac{1}{3} +z_{5}\right) \, \mathbf{a}_{3} & = & \left(-\frac{1}{2}x_{5}+y_{5}\right)a \, \mathbf{\hat{x}}-\frac{\sqrt{3}}{2}x_{5}a \, \mathbf{\hat{y}} + \left(\frac{1}{3} +z_{5}\right)c \, \mathbf{\hat{z}} & \left(6c\right) & \mbox{O IV} \\ 
\mathbf{B}_{27} & = & \left(x_{5}-y_{5}\right) \, \mathbf{a}_{1} + x_{5} \, \mathbf{a}_{2} + \left(\frac{2}{3} +z_{5}\right) \, \mathbf{a}_{3} & = & \left(x_{5}-\frac{1}{2}y_{5}\right)a \, \mathbf{\hat{x}} + \frac{\sqrt{3}}{2}y_{5}a \, \mathbf{\hat{y}} + \left(\frac{2}{3} +z_{5}\right)c \, \mathbf{\hat{z}} & \left(6c\right) & \mbox{O IV} \\ 
\mathbf{B}_{28} & = & x_{6} \, \mathbf{a}_{1} + y_{6} \, \mathbf{a}_{2} + z_{6} \, \mathbf{a}_{3} & = & \frac{1}{2}\left(x_{6}+y_{6}\right)a \, \mathbf{\hat{x}} + \frac{\sqrt{3}}{2}\left(-x_{6}+y_{6}\right)a \, \mathbf{\hat{y}} + z_{6}c \, \mathbf{\hat{z}} & \left(6c\right) & \mbox{O V} \\ 
\mathbf{B}_{29} & = & -y_{6} \, \mathbf{a}_{1} + \left(x_{6}-y_{6}\right) \, \mathbf{a}_{2} + \left(\frac{1}{3} +z_{6}\right) \, \mathbf{a}_{3} & = & \left(\frac{1}{2}x_{6}-y_{6}\right)a \, \mathbf{\hat{x}} + \frac{\sqrt{3}}{2}x_{6}a \, \mathbf{\hat{y}} + \left(\frac{1}{3} +z_{6}\right)c \, \mathbf{\hat{z}} & \left(6c\right) & \mbox{O V} \\ 
\mathbf{B}_{30} & = & \left(-x_{6}+y_{6}\right) \, \mathbf{a}_{1}-x_{6} \, \mathbf{a}_{2} + \left(\frac{2}{3} +z_{6}\right) \, \mathbf{a}_{3} & = & \left(-x_{6}+\frac{1}{2}y_{6}\right)a \, \mathbf{\hat{x}}-\frac{\sqrt{3}}{2}y_{6}a \, \mathbf{\hat{y}} + \left(\frac{2}{3} +z_{6}\right)c \, \mathbf{\hat{z}} & \left(6c\right) & \mbox{O V} \\ 
\mathbf{B}_{31} & = & -x_{6} \, \mathbf{a}_{1}-y_{6} \, \mathbf{a}_{2} + z_{6} \, \mathbf{a}_{3} & = & -\frac{1}{2}\left(x_{6}+y_{6}\right)a \, \mathbf{\hat{x}} + \frac{\sqrt{3}}{2}\left(x_{6}-y_{6}\right)a \, \mathbf{\hat{y}} + z_{6}c \, \mathbf{\hat{z}} & \left(6c\right) & \mbox{O V} \\ 
\mathbf{B}_{32} & = & y_{6} \, \mathbf{a}_{1} + \left(-x_{6}+y_{6}\right) \, \mathbf{a}_{2} + \left(\frac{1}{3} +z_{6}\right) \, \mathbf{a}_{3} & = & \left(-\frac{1}{2}x_{6}+y_{6}\right)a \, \mathbf{\hat{x}}-\frac{\sqrt{3}}{2}x_{6}a \, \mathbf{\hat{y}} + \left(\frac{1}{3} +z_{6}\right)c \, \mathbf{\hat{z}} & \left(6c\right) & \mbox{O V} \\ 
\mathbf{B}_{33} & = & \left(x_{6}-y_{6}\right) \, \mathbf{a}_{1} + x_{6} \, \mathbf{a}_{2} + \left(\frac{2}{3} +z_{6}\right) \, \mathbf{a}_{3} & = & \left(x_{6}-\frac{1}{2}y_{6}\right)a \, \mathbf{\hat{x}} + \frac{\sqrt{3}}{2}y_{6}a \, \mathbf{\hat{y}} + \left(\frac{2}{3} +z_{6}\right)c \, \mathbf{\hat{z}} & \left(6c\right) & \mbox{O V} \\ 
\mathbf{B}_{34} & = & x_{7} \, \mathbf{a}_{1} + y_{7} \, \mathbf{a}_{2} + z_{7} \, \mathbf{a}_{3} & = & \frac{1}{2}\left(x_{7}+y_{7}\right)a \, \mathbf{\hat{x}} + \frac{\sqrt{3}}{2}\left(-x_{7}+y_{7}\right)a \, \mathbf{\hat{y}} + z_{7}c \, \mathbf{\hat{z}} & \left(6c\right) & \mbox{S} \\ 
\mathbf{B}_{35} & = & -y_{7} \, \mathbf{a}_{1} + \left(x_{7}-y_{7}\right) \, \mathbf{a}_{2} + \left(\frac{1}{3} +z_{7}\right) \, \mathbf{a}_{3} & = & \left(\frac{1}{2}x_{7}-y_{7}\right)a \, \mathbf{\hat{x}} + \frac{\sqrt{3}}{2}x_{7}a \, \mathbf{\hat{y}} + \left(\frac{1}{3} +z_{7}\right)c \, \mathbf{\hat{z}} & \left(6c\right) & \mbox{S} \\ 
\mathbf{B}_{36} & = & \left(-x_{7}+y_{7}\right) \, \mathbf{a}_{1}-x_{7} \, \mathbf{a}_{2} + \left(\frac{2}{3} +z_{7}\right) \, \mathbf{a}_{3} & = & \left(-x_{7}+\frac{1}{2}y_{7}\right)a \, \mathbf{\hat{x}}-\frac{\sqrt{3}}{2}y_{7}a \, \mathbf{\hat{y}} + \left(\frac{2}{3} +z_{7}\right)c \, \mathbf{\hat{z}} & \left(6c\right) & \mbox{S} \\ 
\mathbf{B}_{37} & = & -x_{7} \, \mathbf{a}_{1}-y_{7} \, \mathbf{a}_{2} + z_{7} \, \mathbf{a}_{3} & = & -\frac{1}{2}\left(x_{7}+y_{7}\right)a \, \mathbf{\hat{x}} + \frac{\sqrt{3}}{2}\left(x_{7}-y_{7}\right)a \, \mathbf{\hat{y}} + z_{7}c \, \mathbf{\hat{z}} & \left(6c\right) & \mbox{S} \\ 
\mathbf{B}_{38} & = & y_{7} \, \mathbf{a}_{1} + \left(-x_{7}+y_{7}\right) \, \mathbf{a}_{2} + \left(\frac{1}{3} +z_{7}\right) \, \mathbf{a}_{3} & = & \left(-\frac{1}{2}x_{7}+y_{7}\right)a \, \mathbf{\hat{x}}-\frac{\sqrt{3}}{2}x_{7}a \, \mathbf{\hat{y}} + \left(\frac{1}{3} +z_{7}\right)c \, \mathbf{\hat{z}} & \left(6c\right) & \mbox{S} \\ 
\mathbf{B}_{39} & = & \left(x_{7}-y_{7}\right) \, \mathbf{a}_{1} + x_{7} \, \mathbf{a}_{2} + \left(\frac{2}{3} +z_{7}\right) \, \mathbf{a}_{3} & = & \left(x_{7}-\frac{1}{2}y_{7}\right)a \, \mathbf{\hat{x}} + \frac{\sqrt{3}}{2}y_{7}a \, \mathbf{\hat{y}} + \left(\frac{2}{3} +z_{7}\right)c \, \mathbf{\hat{z}} & \left(6c\right) & \mbox{S} \\ 
\end{longtabu}
\renewcommand{\arraystretch}{1.0}
\noindent \hrulefill
\\
\textbf{References:}
\vspace*{-0.25cm}
\begin{flushleft}
  - \bibentry{Hargreaves_SrS2O6H2O4_ZKristallogrCrystalMat_1972}. \\
\end{flushleft}
\noindent \hrulefill
\\
\textbf{Geometry files:}
\\
\noindent  - CIF: pp. {\hyperref[A10B2C_hP39_172_5c_c_a_cif]{\pageref{A10B2C_hP39_172_5c_c_a_cif}}} \\
\noindent  - POSCAR: pp. {\hyperref[A10B2C_hP39_172_5c_c_a_poscar]{\pageref{A10B2C_hP39_172_5c_c_a_poscar}}} \\
\onecolumn
{\phantomsection\label{A3B_hP8_173_c_b}}
\subsection*{\huge \textbf{{\normalfont PI$_{3}$ Structure: A3B\_hP8\_173\_c\_b}}}
\noindent \hrulefill
\vspace*{0.25cm}
\begin{figure}[htp]
  \centering
  \vspace{-1em}
  {\includegraphics[width=1\textwidth]{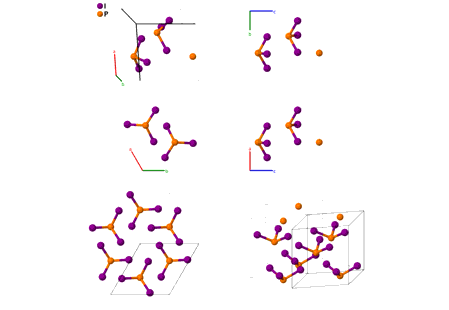}}
\end{figure}
\vspace*{-0.5cm}
\renewcommand{\arraystretch}{1.5}
\begin{equation*}
  \begin{array}{>{$\hspace{-0.15cm}}l<{$}>{$}p{0.5cm}<{$}>{$}p{18.5cm}<{$}}
    \mbox{\large \textbf{Prototype}} &\colon & \ce{PI3} \\
    \mbox{\large \textbf{\AFLOW\ prototype label}} &\colon & \mbox{A3B\_hP8\_173\_c\_b} \\
    \mbox{\large \textbf{\textit{Strukturbericht} designation}} &\colon & \mbox{None} \\
    \mbox{\large \textbf{Pearson symbol}} &\colon & \mbox{hP8} \\
    \mbox{\large \textbf{Space group number}} &\colon & 173 \\
    \mbox{\large \textbf{Space group symbol}} &\colon & P6_{3} \\
    \mbox{\large \textbf{\AFLOW\ prototype command}} &\colon &  \texttt{aflow} \,  \, \texttt{-{}-proto=A3B\_hP8\_173\_c\_b } \, \newline \texttt{-{}-params=}{a,c/a,z_{1},x_{2},y_{2},z_{2} }
  \end{array}
\end{equation*}
\renewcommand{\arraystretch}{1.0}

\noindent \parbox{1 \linewidth}{
\noindent \hrulefill
\\
\textbf{Hexagonal primitive vectors:} \\
\vspace*{-0.25cm}
\begin{tabular}{cc}
  \begin{tabular}{c}
    \parbox{0.6 \linewidth}{
      \renewcommand{\arraystretch}{1.5}
      \begin{equation*}
        \centering
        \begin{array}{ccc}
              \mathbf{a}_1 & = & \frac12 \, a \, \mathbf{\hat{x}} - \frac{\sqrt3}2 \, a \, \mathbf{\hat{y}} \\
    \mathbf{a}_2 & = & \frac12 \, a \, \mathbf{\hat{x}} + \frac{\sqrt3}2 \, a \, \mathbf{\hat{y}} \\
    \mathbf{a}_3 & = & c \, \mathbf{\hat{z}} \\

        \end{array}
      \end{equation*}
    }
    \renewcommand{\arraystretch}{1.0}
  \end{tabular}
  \begin{tabular}{c}
    \includegraphics[width=0.3\linewidth]{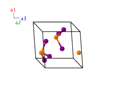} \\
  \end{tabular}
\end{tabular}

}
\vspace*{-0.25cm}

\noindent \hrulefill
\\
\textbf{Basis vectors:}
\vspace*{-0.25cm}
\renewcommand{\arraystretch}{1.5}
\begin{longtabu} to \textwidth{>{\centering $}X[-1,c,c]<{$}>{\centering $}X[-1,c,c]<{$}>{\centering $}X[-1,c,c]<{$}>{\centering $}X[-1,c,c]<{$}>{\centering $}X[-1,c,c]<{$}>{\centering $}X[-1,c,c]<{$}>{\centering $}X[-1,c,c]<{$}}
  & & \mbox{Lattice Coordinates} & & \mbox{Cartesian Coordinates} &\mbox{Wyckoff Position} & \mbox{Atom Type} \\  
  \mathbf{B}_{1} & = & \frac{1}{3} \, \mathbf{a}_{1} + \frac{2}{3} \, \mathbf{a}_{2} + z_{1} \, \mathbf{a}_{3} & = & \frac{1}{2}a \, \mathbf{\hat{x}} + \frac{1}{2\sqrt{3}}a \, \mathbf{\hat{y}} + z_{1}c \, \mathbf{\hat{z}} & \left(2b\right) & \mbox{P} \\ 
\mathbf{B}_{2} & = & \frac{2}{3} \, \mathbf{a}_{1} + \frac{1}{3} \, \mathbf{a}_{2} + \left(\frac{1}{2} +z_{1}\right) \, \mathbf{a}_{3} & = & \frac{1}{2}a \, \mathbf{\hat{x}}- \frac{1}{2\sqrt{3}}a  \, \mathbf{\hat{y}} + \left(\frac{1}{2} +z_{1}\right)c \, \mathbf{\hat{z}} & \left(2b\right) & \mbox{P} \\ 
\mathbf{B}_{3} & = & x_{2} \, \mathbf{a}_{1} + y_{2} \, \mathbf{a}_{2} + z_{2} \, \mathbf{a}_{3} & = & \frac{1}{2}\left(x_{2}+y_{2}\right)a \, \mathbf{\hat{x}} + \frac{\sqrt{3}}{2}\left(-x_{2}+y_{2}\right)a \, \mathbf{\hat{y}} + z_{2}c \, \mathbf{\hat{z}} & \left(6c\right) & \mbox{I} \\ 
\mathbf{B}_{4} & = & -y_{2} \, \mathbf{a}_{1} + \left(x_{2}-y_{2}\right) \, \mathbf{a}_{2} + z_{2} \, \mathbf{a}_{3} & = & \left(\frac{1}{2}x_{2}-y_{2}\right)a \, \mathbf{\hat{x}} + \frac{\sqrt{3}}{2}x_{2}a \, \mathbf{\hat{y}} + z_{2}c \, \mathbf{\hat{z}} & \left(6c\right) & \mbox{I} \\ 
\mathbf{B}_{5} & = & \left(-x_{2}+y_{2}\right) \, \mathbf{a}_{1}-x_{2} \, \mathbf{a}_{2} + z_{2} \, \mathbf{a}_{3} & = & \left(-x_{2}+\frac{1}{2}y_{2}\right)a \, \mathbf{\hat{x}}-\frac{\sqrt{3}}{2}y_{2}a \, \mathbf{\hat{y}} + z_{2}c \, \mathbf{\hat{z}} & \left(6c\right) & \mbox{I} \\ 
\mathbf{B}_{6} & = & -x_{2} \, \mathbf{a}_{1}-y_{2} \, \mathbf{a}_{2} + \left(\frac{1}{2} +z_{2}\right) \, \mathbf{a}_{3} & = & -\frac{1}{2}\left(x_{2}+y_{2}\right)a \, \mathbf{\hat{x}} + \frac{\sqrt{3}}{2}\left(x_{2}-y_{2}\right)a \, \mathbf{\hat{y}} + \left(\frac{1}{2} +z_{2}\right)c \, \mathbf{\hat{z}} & \left(6c\right) & \mbox{I} \\ 
\mathbf{B}_{7} & = & y_{2} \, \mathbf{a}_{1} + \left(-x_{2}+y_{2}\right) \, \mathbf{a}_{2} + \left(\frac{1}{2} +z_{2}\right) \, \mathbf{a}_{3} & = & \left(-\frac{1}{2}x_{2}+y_{2}\right)a \, \mathbf{\hat{x}}-\frac{\sqrt{3}}{2}x_{2}a \, \mathbf{\hat{y}} + \left(\frac{1}{2} +z_{2}\right)c \, \mathbf{\hat{z}} & \left(6c\right) & \mbox{I} \\ 
\mathbf{B}_{8} & = & \left(x_{2}-y_{2}\right) \, \mathbf{a}_{1} + x_{2} \, \mathbf{a}_{2} + \left(\frac{1}{2} +z_{2}\right) \, \mathbf{a}_{3} & = & \left(x_{2}-\frac{1}{2}y_{2}\right)a \, \mathbf{\hat{x}} + \frac{\sqrt{3}}{2}y_{2}a \, \mathbf{\hat{y}} + \left(\frac{1}{2} +z_{2}\right)c \, \mathbf{\hat{z}} & \left(6c\right) & \mbox{I} \\ 
\end{longtabu}
\renewcommand{\arraystretch}{1.0}
\noindent \hrulefill
\\
\textbf{References:}
\vspace*{-0.25cm}
\begin{flushleft}
  - \bibentry{Lance_I3P_InorgChem_1976}. \\
\end{flushleft}
\textbf{Found in:}
\vspace*{-0.25cm}
\begin{flushleft}
  - \bibentry{Villars_PearsonsCrystalData_2013}. \\
\end{flushleft}
\noindent \hrulefill
\\
\textbf{Geometry files:}
\\
\noindent  - CIF: pp. {\hyperref[A3B_hP8_173_c_b_cif]{\pageref{A3B_hP8_173_c_b_cif}}} \\
\noindent  - POSCAR: pp. {\hyperref[A3B_hP8_173_c_b_poscar]{\pageref{A3B_hP8_173_c_b_poscar}}} \\
\onecolumn
{\phantomsection\label{A4B3_hP14_173_bc_c}}
\subsection*{\huge \textbf{{\normalfont $\beta$-Si$_{3}$N$_{4}$ Structure: A4B3\_hP14\_173\_bc\_c}}}
\noindent \hrulefill
\vspace*{0.25cm}
\begin{figure}[htp]
  \centering
  \vspace{-1em}
  {\includegraphics[width=1\textwidth]{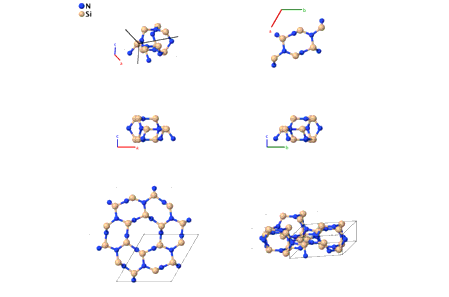}}
\end{figure}
\vspace*{-0.5cm}
\renewcommand{\arraystretch}{1.5}
\begin{equation*}
  \begin{array}{>{$\hspace{-0.15cm}}l<{$}>{$}p{0.5cm}<{$}>{$}p{18.5cm}<{$}}
    \mbox{\large \textbf{Prototype}} &\colon & \ce{$\beta$-Si3N4} \\
    \mbox{\large \textbf{\AFLOW\ prototype label}} &\colon & \mbox{A4B3\_hP14\_173\_bc\_c} \\
    \mbox{\large \textbf{\textit{Strukturbericht} designation}} &\colon & \mbox{None} \\
    \mbox{\large \textbf{Pearson symbol}} &\colon & \mbox{hP14} \\
    \mbox{\large \textbf{Space group number}} &\colon & 173 \\
    \mbox{\large \textbf{Space group symbol}} &\colon & P6_{3} \\
    \mbox{\large \textbf{\AFLOW\ prototype command}} &\colon &  \texttt{aflow} \,  \, \texttt{-{}-proto=A4B3\_hP14\_173\_bc\_c } \, \newline \texttt{-{}-params=}{a,c/a,z_{1},x_{2},y_{2},z_{2},x_{3},y_{3},z_{3} }
  \end{array}
\end{equation*}
\renewcommand{\arraystretch}{1.0}

\noindent \parbox{1 \linewidth}{
\noindent \hrulefill
\\
\textbf{Hexagonal primitive vectors:} \\
\vspace*{-0.25cm}
\begin{tabular}{cc}
  \begin{tabular}{c}
    \parbox{0.6 \linewidth}{
      \renewcommand{\arraystretch}{1.5}
      \begin{equation*}
        \centering
        \begin{array}{ccc}
              \mathbf{a}_1 & = & \frac12 \, a \, \mathbf{\hat{x}} - \frac{\sqrt3}2 \, a \, \mathbf{\hat{y}} \\
    \mathbf{a}_2 & = & \frac12 \, a \, \mathbf{\hat{x}} + \frac{\sqrt3}2 \, a \, \mathbf{\hat{y}} \\
    \mathbf{a}_3 & = & c \, \mathbf{\hat{z}} \\

        \end{array}
      \end{equation*}
    }
    \renewcommand{\arraystretch}{1.0}
  \end{tabular}
  \begin{tabular}{c}
    \includegraphics[width=0.3\linewidth]{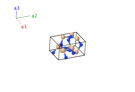} \\
  \end{tabular}
\end{tabular}

}
\vspace*{-0.25cm}

\noindent \hrulefill
\\
\textbf{Basis vectors:}
\vspace*{-0.25cm}
\renewcommand{\arraystretch}{1.5}
\begin{longtabu} to \textwidth{>{\centering $}X[-1,c,c]<{$}>{\centering $}X[-1,c,c]<{$}>{\centering $}X[-1,c,c]<{$}>{\centering $}X[-1,c,c]<{$}>{\centering $}X[-1,c,c]<{$}>{\centering $}X[-1,c,c]<{$}>{\centering $}X[-1,c,c]<{$}}
  & & \mbox{Lattice Coordinates} & & \mbox{Cartesian Coordinates} &\mbox{Wyckoff Position} & \mbox{Atom Type} \\  
  \mathbf{B}_{1} & = & \frac{1}{3} \, \mathbf{a}_{1} + \frac{2}{3} \, \mathbf{a}_{2} + z_{1} \, \mathbf{a}_{3} & = & \frac{1}{2}a \, \mathbf{\hat{x}} + \frac{1}{2\sqrt{3}}a \, \mathbf{\hat{y}} + z_{1}c \, \mathbf{\hat{z}} & \left(2b\right) & \mbox{N I} \\ 
\mathbf{B}_{2} & = & \frac{2}{3} \, \mathbf{a}_{1} + \frac{1}{3} \, \mathbf{a}_{2} + \left(\frac{1}{2} +z_{1}\right) \, \mathbf{a}_{3} & = & \frac{1}{2}a \, \mathbf{\hat{x}}- \frac{1}{2\sqrt{3}}a  \, \mathbf{\hat{y}} + \left(\frac{1}{2} +z_{1}\right)c \, \mathbf{\hat{z}} & \left(2b\right) & \mbox{N I} \\ 
\mathbf{B}_{3} & = & x_{2} \, \mathbf{a}_{1} + y_{2} \, \mathbf{a}_{2} + z_{2} \, \mathbf{a}_{3} & = & \frac{1}{2}\left(x_{2}+y_{2}\right)a \, \mathbf{\hat{x}} + \frac{\sqrt{3}}{2}\left(-x_{2}+y_{2}\right)a \, \mathbf{\hat{y}} + z_{2}c \, \mathbf{\hat{z}} & \left(6c\right) & \mbox{N II} \\ 
\mathbf{B}_{4} & = & -y_{2} \, \mathbf{a}_{1} + \left(x_{2}-y_{2}\right) \, \mathbf{a}_{2} + z_{2} \, \mathbf{a}_{3} & = & \left(\frac{1}{2}x_{2}-y_{2}\right)a \, \mathbf{\hat{x}} + \frac{\sqrt{3}}{2}x_{2}a \, \mathbf{\hat{y}} + z_{2}c \, \mathbf{\hat{z}} & \left(6c\right) & \mbox{N II} \\ 
\mathbf{B}_{5} & = & \left(-x_{2}+y_{2}\right) \, \mathbf{a}_{1}-x_{2} \, \mathbf{a}_{2} + z_{2} \, \mathbf{a}_{3} & = & \left(-x_{2}+\frac{1}{2}y_{2}\right)a \, \mathbf{\hat{x}}-\frac{\sqrt{3}}{2}y_{2}a \, \mathbf{\hat{y}} + z_{2}c \, \mathbf{\hat{z}} & \left(6c\right) & \mbox{N II} \\ 
\mathbf{B}_{6} & = & -x_{2} \, \mathbf{a}_{1}-y_{2} \, \mathbf{a}_{2} + \left(\frac{1}{2} +z_{2}\right) \, \mathbf{a}_{3} & = & -\frac{1}{2}\left(x_{2}+y_{2}\right)a \, \mathbf{\hat{x}} + \frac{\sqrt{3}}{2}\left(x_{2}-y_{2}\right)a \, \mathbf{\hat{y}} + \left(\frac{1}{2} +z_{2}\right)c \, \mathbf{\hat{z}} & \left(6c\right) & \mbox{N II} \\ 
\mathbf{B}_{7} & = & y_{2} \, \mathbf{a}_{1} + \left(-x_{2}+y_{2}\right) \, \mathbf{a}_{2} + \left(\frac{1}{2} +z_{2}\right) \, \mathbf{a}_{3} & = & \left(-\frac{1}{2}x_{2}+y_{2}\right)a \, \mathbf{\hat{x}}-\frac{\sqrt{3}}{2}x_{2}a \, \mathbf{\hat{y}} + \left(\frac{1}{2} +z_{2}\right)c \, \mathbf{\hat{z}} & \left(6c\right) & \mbox{N II} \\ 
\mathbf{B}_{8} & = & \left(x_{2}-y_{2}\right) \, \mathbf{a}_{1} + x_{2} \, \mathbf{a}_{2} + \left(\frac{1}{2} +z_{2}\right) \, \mathbf{a}_{3} & = & \left(x_{2}-\frac{1}{2}y_{2}\right)a \, \mathbf{\hat{x}} + \frac{\sqrt{3}}{2}y_{2}a \, \mathbf{\hat{y}} + \left(\frac{1}{2} +z_{2}\right)c \, \mathbf{\hat{z}} & \left(6c\right) & \mbox{N II} \\ 
\mathbf{B}_{9} & = & x_{3} \, \mathbf{a}_{1} + y_{3} \, \mathbf{a}_{2} + z_{3} \, \mathbf{a}_{3} & = & \frac{1}{2}\left(x_{3}+y_{3}\right)a \, \mathbf{\hat{x}} + \frac{\sqrt{3}}{2}\left(-x_{3}+y_{3}\right)a \, \mathbf{\hat{y}} + z_{3}c \, \mathbf{\hat{z}} & \left(6c\right) & \mbox{Si} \\ 
\mathbf{B}_{10} & = & -y_{3} \, \mathbf{a}_{1} + \left(x_{3}-y_{3}\right) \, \mathbf{a}_{2} + z_{3} \, \mathbf{a}_{3} & = & \left(\frac{1}{2}x_{3}-y_{3}\right)a \, \mathbf{\hat{x}} + \frac{\sqrt{3}}{2}x_{3}a \, \mathbf{\hat{y}} + z_{3}c \, \mathbf{\hat{z}} & \left(6c\right) & \mbox{Si} \\ 
\mathbf{B}_{11} & = & \left(-x_{3}+y_{3}\right) \, \mathbf{a}_{1}-x_{3} \, \mathbf{a}_{2} + z_{3} \, \mathbf{a}_{3} & = & \left(-x_{3}+\frac{1}{2}y_{3}\right)a \, \mathbf{\hat{x}}-\frac{\sqrt{3}}{2}y_{3}a \, \mathbf{\hat{y}} + z_{3}c \, \mathbf{\hat{z}} & \left(6c\right) & \mbox{Si} \\ 
\mathbf{B}_{12} & = & -x_{3} \, \mathbf{a}_{1}-y_{3} \, \mathbf{a}_{2} + \left(\frac{1}{2} +z_{3}\right) \, \mathbf{a}_{3} & = & -\frac{1}{2}\left(x_{3}+y_{3}\right)a \, \mathbf{\hat{x}} + \frac{\sqrt{3}}{2}\left(x_{3}-y_{3}\right)a \, \mathbf{\hat{y}} + \left(\frac{1}{2} +z_{3}\right)c \, \mathbf{\hat{z}} & \left(6c\right) & \mbox{Si} \\ 
\mathbf{B}_{13} & = & y_{3} \, \mathbf{a}_{1} + \left(-x_{3}+y_{3}\right) \, \mathbf{a}_{2} + \left(\frac{1}{2} +z_{3}\right) \, \mathbf{a}_{3} & = & \left(-\frac{1}{2}x_{3}+y_{3}\right)a \, \mathbf{\hat{x}}-\frac{\sqrt{3}}{2}x_{3}a \, \mathbf{\hat{y}} + \left(\frac{1}{2} +z_{3}\right)c \, \mathbf{\hat{z}} & \left(6c\right) & \mbox{Si} \\ 
\mathbf{B}_{14} & = & \left(x_{3}-y_{3}\right) \, \mathbf{a}_{1} + x_{3} \, \mathbf{a}_{2} + \left(\frac{1}{2} +z_{3}\right) \, \mathbf{a}_{3} & = & \left(x_{3}-\frac{1}{2}y_{3}\right)a \, \mathbf{\hat{x}} + \frac{\sqrt{3}}{2}y_{3}a \, \mathbf{\hat{y}} + \left(\frac{1}{2} +z_{3}\right)c \, \mathbf{\hat{z}} & \left(6c\right) & \mbox{Si} \\ 
\end{longtabu}
\renewcommand{\arraystretch}{1.0}
\noindent \hrulefill
\\
\textbf{References:}
\vspace*{-0.25cm}
\begin{flushleft}
  - \bibentry{Forgeng_Si3N4_TransMetallSoc_1958}. \\
\end{flushleft}
\textbf{Found in:}
\vspace*{-0.25cm}
\begin{flushleft}
  - \bibentry{Villars_PearsonsCrystalData_2013}. \\
\end{flushleft}
\noindent \hrulefill
\\
\textbf{Geometry files:}
\\
\noindent  - CIF: pp. {\hyperref[A4B3_hP14_173_bc_c_cif]{\pageref{A4B3_hP14_173_bc_c_cif}}} \\
\noindent  - POSCAR: pp. {\hyperref[A4B3_hP14_173_bc_c_poscar]{\pageref{A4B3_hP14_173_bc_c_poscar}}} \\
\onecolumn
{\phantomsection\label{A12B7C2_hP21_174_2j2k_ajk_cf}}
\subsection*{\huge \textbf{{\normalfont Fe$_{12}$Zr$_{2}$P$_{7}$ Structure: A12B7C2\_hP21\_174\_2j2k\_ajk\_cf}}}
\noindent \hrulefill
\vspace*{0.25cm}
\begin{figure}[htp]
  \centering
  \vspace{-1em}
  {\includegraphics[width=1\textwidth]{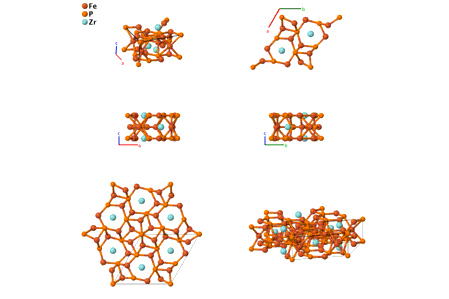}}
\end{figure}
\vspace*{-0.5cm}
\renewcommand{\arraystretch}{1.5}
\begin{equation*}
  \begin{array}{>{$\hspace{-0.15cm}}l<{$}>{$}p{0.5cm}<{$}>{$}p{18.5cm}<{$}}
    \mbox{\large \textbf{Prototype}} &\colon & \ce{Fe12Zr2P7} \\
    \mbox{\large \textbf{\AFLOW\ prototype label}} &\colon & \mbox{A12B7C2\_hP21\_174\_2j2k\_ajk\_cf} \\
    \mbox{\large \textbf{\textit{Strukturbericht} designation}} &\colon & \mbox{None} \\
    \mbox{\large \textbf{Pearson symbol}} &\colon & \mbox{hP21} \\
    \mbox{\large \textbf{Space group number}} &\colon & 174 \\
    \mbox{\large \textbf{Space group symbol}} &\colon & P\bar{6} \\
    \mbox{\large \textbf{\AFLOW\ prototype command}} &\colon &  \texttt{aflow} \,  \, \texttt{-{}-proto=A12B7C2\_hP21\_174\_2j2k\_ajk\_cf } \, \newline \texttt{-{}-params=}{a,c/a,x_{4},y_{4},x_{5},y_{5},x_{6},y_{6},x_{7},y_{7},x_{8},y_{8},x_{9},y_{9} }
  \end{array}
\end{equation*}
\renewcommand{\arraystretch}{1.0}

\noindent \parbox{1 \linewidth}{
\noindent \hrulefill
\\
\textbf{Hexagonal primitive vectors:} \\
\vspace*{-0.25cm}
\begin{tabular}{cc}
  \begin{tabular}{c}
    \parbox{0.6 \linewidth}{
      \renewcommand{\arraystretch}{1.5}
      \begin{equation*}
        \centering
        \begin{array}{ccc}
              \mathbf{a}_1 & = & \frac12 \, a \, \mathbf{\hat{x}} - \frac{\sqrt3}2 \, a \, \mathbf{\hat{y}} \\
    \mathbf{a}_2 & = & \frac12 \, a \, \mathbf{\hat{x}} + \frac{\sqrt3}2 \, a \, \mathbf{\hat{y}} \\
    \mathbf{a}_3 & = & c \, \mathbf{\hat{z}} \\

        \end{array}
      \end{equation*}
    }
    \renewcommand{\arraystretch}{1.0}
  \end{tabular}
  \begin{tabular}{c}
    \includegraphics[width=0.3\linewidth]{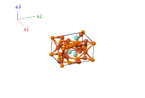} \\
  \end{tabular}
\end{tabular}

}
\vspace*{-0.25cm}

\noindent \hrulefill
\\
\textbf{Basis vectors:}
\vspace*{-0.25cm}
\renewcommand{\arraystretch}{1.5}
\begin{longtabu} to \textwidth{>{\centering $}X[-1,c,c]<{$}>{\centering $}X[-1,c,c]<{$}>{\centering $}X[-1,c,c]<{$}>{\centering $}X[-1,c,c]<{$}>{\centering $}X[-1,c,c]<{$}>{\centering $}X[-1,c,c]<{$}>{\centering $}X[-1,c,c]<{$}}
  & & \mbox{Lattice Coordinates} & & \mbox{Cartesian Coordinates} &\mbox{Wyckoff Position} & \mbox{Atom Type} \\  
  \mathbf{B}_{1} & = & 0 \, \mathbf{a}_{1} + 0 \, \mathbf{a}_{2} + 0 \, \mathbf{a}_{3} & = & 0 \, \mathbf{\hat{x}} + 0 \, \mathbf{\hat{y}} + 0 \, \mathbf{\hat{z}} & \left(1a\right) & \mbox{P I} \\ 
\mathbf{B}_{2} & = & \frac{1}{3} \, \mathbf{a}_{1} + \frac{2}{3} \, \mathbf{a}_{2} & = & \frac{1}{2}a \, \mathbf{\hat{x}} + \frac{1}{2\sqrt{3}}a \, \mathbf{\hat{y}} & \left(1c\right) & \mbox{Zr I} \\ 
\mathbf{B}_{3} & = & \frac{2}{3} \, \mathbf{a}_{1} + \frac{1}{3} \, \mathbf{a}_{2} + \frac{1}{2} \, \mathbf{a}_{3} & = & \frac{1}{2}a \, \mathbf{\hat{x}}- \frac{1}{2\sqrt{3}}a  \, \mathbf{\hat{y}} + \frac{1}{2}c \, \mathbf{\hat{z}} & \left(1f\right) & \mbox{Zr II} \\ 
\mathbf{B}_{4} & = & x_{4} \, \mathbf{a}_{1} + y_{4} \, \mathbf{a}_{2} & = & \frac{1}{2}\left(x_{4}+y_{4}\right)a \, \mathbf{\hat{x}} + \frac{\sqrt{3}}{2}\left(-x_{4}+y_{4}\right)a \, \mathbf{\hat{y}} & \left(3j\right) & \mbox{Fe I} \\ 
\mathbf{B}_{5} & = & -y_{4} \, \mathbf{a}_{1} + \left(x_{4}-y_{4}\right) \, \mathbf{a}_{2} & = & \left(\frac{1}{2}x_{4}-y_{4}\right)a \, \mathbf{\hat{x}} + \frac{\sqrt{3}}{2}x_{4}a \, \mathbf{\hat{y}} & \left(3j\right) & \mbox{Fe I} \\ 
\mathbf{B}_{6} & = & \left(-x_{4}+y_{4}\right) \, \mathbf{a}_{1}-x_{4} \, \mathbf{a}_{2} & = & \left(-x_{4}+\frac{1}{2}y_{4}\right)a \, \mathbf{\hat{x}}-\frac{\sqrt{3}}{2}y_{4}a \, \mathbf{\hat{y}} & \left(3j\right) & \mbox{Fe I} \\ 
\mathbf{B}_{7} & = & x_{5} \, \mathbf{a}_{1} + y_{5} \, \mathbf{a}_{2} & = & \frac{1}{2}\left(x_{5}+y_{5}\right)a \, \mathbf{\hat{x}} + \frac{\sqrt{3}}{2}\left(-x_{5}+y_{5}\right)a \, \mathbf{\hat{y}} & \left(3j\right) & \mbox{Fe II} \\ 
\mathbf{B}_{8} & = & -y_{5} \, \mathbf{a}_{1} + \left(x_{5}-y_{5}\right) \, \mathbf{a}_{2} & = & \left(\frac{1}{2}x_{5}-y_{5}\right)a \, \mathbf{\hat{x}} + \frac{\sqrt{3}}{2}x_{5}a \, \mathbf{\hat{y}} & \left(3j\right) & \mbox{Fe II} \\ 
\mathbf{B}_{9} & = & \left(-x_{5}+y_{5}\right) \, \mathbf{a}_{1}-x_{5} \, \mathbf{a}_{2} & = & \left(-x_{5}+\frac{1}{2}y_{5}\right)a \, \mathbf{\hat{x}}-\frac{\sqrt{3}}{2}y_{5}a \, \mathbf{\hat{y}} & \left(3j\right) & \mbox{Fe II} \\ 
\mathbf{B}_{10} & = & x_{6} \, \mathbf{a}_{1} + y_{6} \, \mathbf{a}_{2} & = & \frac{1}{2}\left(x_{6}+y_{6}\right)a \, \mathbf{\hat{x}} + \frac{\sqrt{3}}{2}\left(-x_{6}+y_{6}\right)a \, \mathbf{\hat{y}} & \left(3j\right) & \mbox{P II} \\ 
\mathbf{B}_{11} & = & -y_{6} \, \mathbf{a}_{1} + \left(x_{6}-y_{6}\right) \, \mathbf{a}_{2} & = & \left(\frac{1}{2}x_{6}-y_{6}\right)a \, \mathbf{\hat{x}} + \frac{\sqrt{3}}{2}x_{6}a \, \mathbf{\hat{y}} & \left(3j\right) & \mbox{P II} \\ 
\mathbf{B}_{12} & = & \left(-x_{6}+y_{6}\right) \, \mathbf{a}_{1}-x_{6} \, \mathbf{a}_{2} & = & \left(-x_{6}+\frac{1}{2}y_{6}\right)a \, \mathbf{\hat{x}}-\frac{\sqrt{3}}{2}y_{6}a \, \mathbf{\hat{y}} & \left(3j\right) & \mbox{P II} \\ 
\mathbf{B}_{13} & = & x_{7} \, \mathbf{a}_{1} + y_{7} \, \mathbf{a}_{2} + \frac{1}{2} \, \mathbf{a}_{3} & = & \frac{1}{2}\left(x_{7}+y_{7}\right)a \, \mathbf{\hat{x}} + \frac{\sqrt{3}}{2}\left(-x_{7}+y_{7}\right)a \, \mathbf{\hat{y}} + \frac{1}{2}c \, \mathbf{\hat{z}} & \left(3k\right) & \mbox{Fe III} \\ 
\mathbf{B}_{14} & = & -y_{7} \, \mathbf{a}_{1} + \left(x_{7}-y_{7}\right) \, \mathbf{a}_{2} + \frac{1}{2} \, \mathbf{a}_{3} & = & \left(\frac{1}{2}x_{7}-y_{7}\right)a \, \mathbf{\hat{x}} + \frac{\sqrt{3}}{2}x_{7}a \, \mathbf{\hat{y}} + \frac{1}{2}c \, \mathbf{\hat{z}} & \left(3k\right) & \mbox{Fe III} \\ 
\mathbf{B}_{15} & = & \left(-x_{7}+y_{7}\right) \, \mathbf{a}_{1}-x_{7} \, \mathbf{a}_{2} + \frac{1}{2} \, \mathbf{a}_{3} & = & \left(-x_{7}+\frac{1}{2}y_{7}\right)a \, \mathbf{\hat{x}}-\frac{\sqrt{3}}{2}y_{7}a \, \mathbf{\hat{y}} + \frac{1}{2}c \, \mathbf{\hat{z}} & \left(3k\right) & \mbox{Fe III} \\ 
\mathbf{B}_{16} & = & x_{8} \, \mathbf{a}_{1} + y_{8} \, \mathbf{a}_{2} + \frac{1}{2} \, \mathbf{a}_{3} & = & \frac{1}{2}\left(x_{8}+y_{8}\right)a \, \mathbf{\hat{x}} + \frac{\sqrt{3}}{2}\left(-x_{8}+y_{8}\right)a \, \mathbf{\hat{y}} + \frac{1}{2}c \, \mathbf{\hat{z}} & \left(3k\right) & \mbox{Fe IV} \\ 
\mathbf{B}_{17} & = & -y_{8} \, \mathbf{a}_{1} + \left(x_{8}-y_{8}\right) \, \mathbf{a}_{2} + \frac{1}{2} \, \mathbf{a}_{3} & = & \left(\frac{1}{2}x_{8}-y_{8}\right)a \, \mathbf{\hat{x}} + \frac{\sqrt{3}}{2}x_{8}a \, \mathbf{\hat{y}} + \frac{1}{2}c \, \mathbf{\hat{z}} & \left(3k\right) & \mbox{Fe IV} \\ 
\mathbf{B}_{18} & = & \left(-x_{8}+y_{8}\right) \, \mathbf{a}_{1}-x_{8} \, \mathbf{a}_{2} + \frac{1}{2} \, \mathbf{a}_{3} & = & \left(-x_{8}+\frac{1}{2}y_{8}\right)a \, \mathbf{\hat{x}}-\frac{\sqrt{3}}{2}y_{8}a \, \mathbf{\hat{y}} + \frac{1}{2}c \, \mathbf{\hat{z}} & \left(3k\right) & \mbox{Fe IV} \\ 
\mathbf{B}_{19} & = & x_{9} \, \mathbf{a}_{1} + y_{9} \, \mathbf{a}_{2} + \frac{1}{2} \, \mathbf{a}_{3} & = & \frac{1}{2}\left(x_{9}+y_{9}\right)a \, \mathbf{\hat{x}} + \frac{\sqrt{3}}{2}\left(-x_{9}+y_{9}\right)a \, \mathbf{\hat{y}} + \frac{1}{2}c \, \mathbf{\hat{z}} & \left(3k\right) & \mbox{P III} \\ 
\mathbf{B}_{20} & = & -y_{9} \, \mathbf{a}_{1} + \left(x_{9}-y_{9}\right) \, \mathbf{a}_{2} + \frac{1}{2} \, \mathbf{a}_{3} & = & \left(\frac{1}{2}x_{9}-y_{9}\right)a \, \mathbf{\hat{x}} + \frac{\sqrt{3}}{2}x_{9}a \, \mathbf{\hat{y}} + \frac{1}{2}c \, \mathbf{\hat{z}} & \left(3k\right) & \mbox{P III} \\ 
\mathbf{B}_{21} & = & \left(-x_{9}+y_{9}\right) \, \mathbf{a}_{1}-x_{9} \, \mathbf{a}_{2} + \frac{1}{2} \, \mathbf{a}_{3} & = & \left(-x_{9}+\frac{1}{2}y_{9}\right)a \, \mathbf{\hat{x}}-\frac{\sqrt{3}}{2}y_{9}a \, \mathbf{\hat{y}} + \frac{1}{2}c \, \mathbf{\hat{z}} & \left(3k\right) & \mbox{P III} \\ 
\end{longtabu}
\renewcommand{\arraystretch}{1.0}
\noindent \hrulefill
\\
\textbf{References:}
\vspace*{-0.25cm}
\begin{flushleft}
  - \bibentry{Ganglberger_Zr2Fe12P7_MonatChemMo_1968}. \\
\end{flushleft}
\textbf{Found in:}
\vspace*{-0.25cm}
\begin{flushleft}
  - \bibentry{Villars_PearsonsCrystalData_2013}. \\
\end{flushleft}
\noindent \hrulefill
\\
\textbf{Geometry files:}
\\
\noindent  - CIF: pp. {\hyperref[A12B7C2_hP21_174_2j2k_ajk_cf_cif]{\pageref{A12B7C2_hP21_174_2j2k_ajk_cf_cif}}} \\
\noindent  - POSCAR: pp. {\hyperref[A12B7C2_hP21_174_2j2k_ajk_cf_poscar]{\pageref{A12B7C2_hP21_174_2j2k_ajk_cf_poscar}}} \\
\onecolumn
{\phantomsection\label{ABC_hP12_174_cj_fk_aj}}
\subsection*{\huge \textbf{{\normalfont GdSI Structure: ABC\_hP12\_174\_cj\_fk\_aj}}}
\noindent \hrulefill
\vspace*{0.25cm}
\begin{figure}[htp]
  \centering
  \vspace{-1em}
  {\includegraphics[width=1\textwidth]{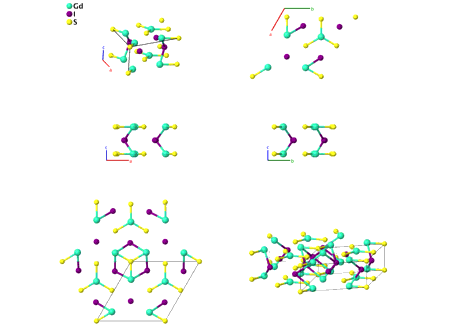}}
\end{figure}
\vspace*{-0.5cm}
\renewcommand{\arraystretch}{1.5}
\begin{equation*}
  \begin{array}{>{$\hspace{-0.15cm}}l<{$}>{$}p{0.5cm}<{$}>{$}p{18.5cm}<{$}}
    \mbox{\large \textbf{Prototype}} &\colon & \ce{GdSI} \\
    \mbox{\large \textbf{\AFLOW\ prototype label}} &\colon & \mbox{ABC\_hP12\_174\_cj\_fk\_aj} \\
    \mbox{\large \textbf{\textit{Strukturbericht} designation}} &\colon & \mbox{None} \\
    \mbox{\large \textbf{Pearson symbol}} &\colon & \mbox{hP12} \\
    \mbox{\large \textbf{Space group number}} &\colon & 174 \\
    \mbox{\large \textbf{Space group symbol}} &\colon & P\bar{6} \\
    \mbox{\large \textbf{\AFLOW\ prototype command}} &\colon &  \texttt{aflow} \,  \, \texttt{-{}-proto=ABC\_hP12\_174\_cj\_fk\_aj } \, \newline \texttt{-{}-params=}{a,c/a,x_{4},y_{4},x_{5},y_{5},x_{6},y_{6} }
  \end{array}
\end{equation*}
\renewcommand{\arraystretch}{1.0}

\noindent \parbox{1 \linewidth}{
\noindent \hrulefill
\\
\textbf{Hexagonal primitive vectors:} \\
\vspace*{-0.25cm}
\begin{tabular}{cc}
  \begin{tabular}{c}
    \parbox{0.6 \linewidth}{
      \renewcommand{\arraystretch}{1.5}
      \begin{equation*}
        \centering
        \begin{array}{ccc}
              \mathbf{a}_1 & = & \frac12 \, a \, \mathbf{\hat{x}} - \frac{\sqrt3}2 \, a \, \mathbf{\hat{y}} \\
    \mathbf{a}_2 & = & \frac12 \, a \, \mathbf{\hat{x}} + \frac{\sqrt3}2 \, a \, \mathbf{\hat{y}} \\
    \mathbf{a}_3 & = & c \, \mathbf{\hat{z}} \\

        \end{array}
      \end{equation*}
    }
    \renewcommand{\arraystretch}{1.0}
  \end{tabular}
  \begin{tabular}{c}
    \includegraphics[width=0.3\linewidth]{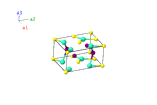} \\
  \end{tabular}
\end{tabular}

}
\vspace*{-0.25cm}

\noindent \hrulefill
\\
\textbf{Basis vectors:}
\vspace*{-0.25cm}
\renewcommand{\arraystretch}{1.5}
\begin{longtabu} to \textwidth{>{\centering $}X[-1,c,c]<{$}>{\centering $}X[-1,c,c]<{$}>{\centering $}X[-1,c,c]<{$}>{\centering $}X[-1,c,c]<{$}>{\centering $}X[-1,c,c]<{$}>{\centering $}X[-1,c,c]<{$}>{\centering $}X[-1,c,c]<{$}}
  & & \mbox{Lattice Coordinates} & & \mbox{Cartesian Coordinates} &\mbox{Wyckoff Position} & \mbox{Atom Type} \\  
  \mathbf{B}_{1} & = & 0 \, \mathbf{a}_{1} + 0 \, \mathbf{a}_{2} + 0 \, \mathbf{a}_{3} & = & 0 \, \mathbf{\hat{x}} + 0 \, \mathbf{\hat{y}} + 0 \, \mathbf{\hat{z}} & \left(1a\right) & \mbox{S I} \\ 
\mathbf{B}_{2} & = & \frac{1}{3} \, \mathbf{a}_{1} + \frac{2}{3} \, \mathbf{a}_{2} & = & \frac{1}{2}a \, \mathbf{\hat{x}} + \frac{1}{2\sqrt{3}}a \, \mathbf{\hat{y}} & \left(1c\right) & \mbox{Gd I} \\ 
\mathbf{B}_{3} & = & \frac{2}{3} \, \mathbf{a}_{1} + \frac{1}{3} \, \mathbf{a}_{2} + \frac{1}{2} \, \mathbf{a}_{3} & = & \frac{1}{2}a \, \mathbf{\hat{x}}- \frac{1}{2\sqrt{3}}a  \, \mathbf{\hat{y}} + \frac{1}{2}c \, \mathbf{\hat{z}} & \left(1f\right) & \mbox{I I} \\ 
\mathbf{B}_{4} & = & x_{4} \, \mathbf{a}_{1} + y_{4} \, \mathbf{a}_{2} & = & \frac{1}{2}\left(x_{4}+y_{4}\right)a \, \mathbf{\hat{x}} + \frac{\sqrt{3}}{2}\left(-x_{4}+y_{4}\right)a \, \mathbf{\hat{y}} & \left(3j\right) & \mbox{Gd II} \\ 
\mathbf{B}_{5} & = & -y_{4} \, \mathbf{a}_{1} + \left(x_{4}-y_{4}\right) \, \mathbf{a}_{2} & = & \left(\frac{1}{2}x_{4}-y_{4}\right)a \, \mathbf{\hat{x}} + \frac{\sqrt{3}}{2}x_{4}a \, \mathbf{\hat{y}} & \left(3j\right) & \mbox{Gd II} \\ 
\mathbf{B}_{6} & = & \left(-x_{4}+y_{4}\right) \, \mathbf{a}_{1}-x_{4} \, \mathbf{a}_{2} & = & \left(-x_{4}+\frac{1}{2}y_{4}\right)a \, \mathbf{\hat{x}}-\frac{\sqrt{3}}{2}y_{4}a \, \mathbf{\hat{y}} & \left(3j\right) & \mbox{Gd II} \\ 
\mathbf{B}_{7} & = & x_{5} \, \mathbf{a}_{1} + y_{5} \, \mathbf{a}_{2} & = & \frac{1}{2}\left(x_{5}+y_{5}\right)a \, \mathbf{\hat{x}} + \frac{\sqrt{3}}{2}\left(-x_{5}+y_{5}\right)a \, \mathbf{\hat{y}} & \left(3j\right) & \mbox{S II} \\ 
\mathbf{B}_{8} & = & -y_{5} \, \mathbf{a}_{1} + \left(x_{5}-y_{5}\right) \, \mathbf{a}_{2} & = & \left(\frac{1}{2}x_{5}-y_{5}\right)a \, \mathbf{\hat{x}} + \frac{\sqrt{3}}{2}x_{5}a \, \mathbf{\hat{y}} & \left(3j\right) & \mbox{S II} \\ 
\mathbf{B}_{9} & = & \left(-x_{5}+y_{5}\right) \, \mathbf{a}_{1}-x_{5} \, \mathbf{a}_{2} & = & \left(-x_{5}+\frac{1}{2}y_{5}\right)a \, \mathbf{\hat{x}}-\frac{\sqrt{3}}{2}y_{5}a \, \mathbf{\hat{y}} & \left(3j\right) & \mbox{S II} \\ 
\mathbf{B}_{10} & = & x_{6} \, \mathbf{a}_{1} + y_{6} \, \mathbf{a}_{2} + \frac{1}{2} \, \mathbf{a}_{3} & = & \frac{1}{2}\left(x_{6}+y_{6}\right)a \, \mathbf{\hat{x}} + \frac{\sqrt{3}}{2}\left(-x_{6}+y_{6}\right)a \, \mathbf{\hat{y}} + \frac{1}{2}c \, \mathbf{\hat{z}} & \left(3k\right) & \mbox{I II} \\ 
\mathbf{B}_{11} & = & -y_{6} \, \mathbf{a}_{1} + \left(x_{6}-y_{6}\right) \, \mathbf{a}_{2} + \frac{1}{2} \, \mathbf{a}_{3} & = & \left(\frac{1}{2}x_{6}-y_{6}\right)a \, \mathbf{\hat{x}} + \frac{\sqrt{3}}{2}x_{6}a \, \mathbf{\hat{y}} + \frac{1}{2}c \, \mathbf{\hat{z}} & \left(3k\right) & \mbox{I II} \\ 
\mathbf{B}_{12} & = & \left(-x_{6}+y_{6}\right) \, \mathbf{a}_{1}-x_{6} \, \mathbf{a}_{2} + \frac{1}{2} \, \mathbf{a}_{3} & = & \left(-x_{6}+\frac{1}{2}y_{6}\right)a \, \mathbf{\hat{x}}-\frac{\sqrt{3}}{2}y_{6}a \, \mathbf{\hat{y}} + \frac{1}{2}c \, \mathbf{\hat{z}} & \left(3k\right) & \mbox{I II} \\ 
\end{longtabu}
\renewcommand{\arraystretch}{1.0}
\noindent \hrulefill
\\
\textbf{References:}
\vspace*{-0.25cm}
\begin{flushleft}
  - \bibentry{Dagron_GdSI_CRAcadSc_1969}. \\
\end{flushleft}
\textbf{Found in:}
\vspace*{-0.25cm}
\begin{flushleft}
  - \bibentry{Villars_PearsonsCrystalData_2013}. \\
\end{flushleft}
\noindent \hrulefill
\\
\textbf{Geometry files:}
\\
\noindent  - CIF: pp. {\hyperref[ABC_hP12_174_cj_fk_aj_cif]{\pageref{ABC_hP12_174_cj_fk_aj_cif}}} \\
\noindent  - POSCAR: pp. {\hyperref[ABC_hP12_174_cj_fk_aj_poscar]{\pageref{ABC_hP12_174_cj_fk_aj_poscar}}} \\
\onecolumn
{\phantomsection\label{A8B7C6_hP21_175_ck_aj_k}}
\subsection*{\huge \textbf{{\normalfont Nb$_{7}$Ru$_{6}$B$_{8}$ Structure: A8B7C6\_hP21\_175\_ck\_aj\_k}}}
\noindent \hrulefill
\vspace*{0.25cm}
\begin{figure}[htp]
  \centering
  \vspace{-1em}
  {\includegraphics[width=1\textwidth]{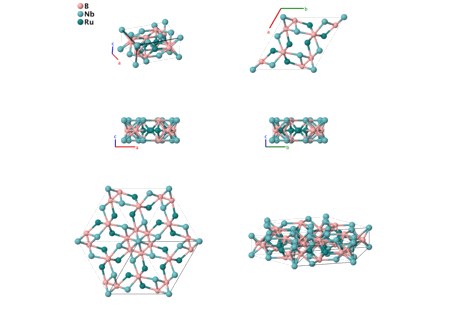}}
\end{figure}
\vspace*{-0.5cm}
\renewcommand{\arraystretch}{1.5}
\begin{equation*}
  \begin{array}{>{$\hspace{-0.15cm}}l<{$}>{$}p{0.5cm}<{$}>{$}p{18.5cm}<{$}}
    \mbox{\large \textbf{Prototype}} &\colon & \ce{Nb7Ru6B8} \\
    \mbox{\large \textbf{\AFLOW\ prototype label}} &\colon & \mbox{A8B7C6\_hP21\_175\_ck\_aj\_k} \\
    \mbox{\large \textbf{\textit{Strukturbericht} designation}} &\colon & \mbox{None} \\
    \mbox{\large \textbf{Pearson symbol}} &\colon & \mbox{hP21} \\
    \mbox{\large \textbf{Space group number}} &\colon & 175 \\
    \mbox{\large \textbf{Space group symbol}} &\colon & P6/m \\
    \mbox{\large \textbf{\AFLOW\ prototype command}} &\colon &  \texttt{aflow} \,  \, \texttt{-{}-proto=A8B7C6\_hP21\_175\_ck\_aj\_k } \, \newline \texttt{-{}-params=}{a,c/a,x_{3},y_{3},x_{4},y_{4},x_{5},y_{5} }
  \end{array}
\end{equation*}
\renewcommand{\arraystretch}{1.0}

\noindent \parbox{1 \linewidth}{
\noindent \hrulefill
\\
\textbf{Hexagonal primitive vectors:} \\
\vspace*{-0.25cm}
\begin{tabular}{cc}
  \begin{tabular}{c}
    \parbox{0.6 \linewidth}{
      \renewcommand{\arraystretch}{1.5}
      \begin{equation*}
        \centering
        \begin{array}{ccc}
              \mathbf{a}_1 & = & \frac12 \, a \, \mathbf{\hat{x}} - \frac{\sqrt3}2 \, a \, \mathbf{\hat{y}} \\
    \mathbf{a}_2 & = & \frac12 \, a \, \mathbf{\hat{x}} + \frac{\sqrt3}2 \, a \, \mathbf{\hat{y}} \\
    \mathbf{a}_3 & = & c \, \mathbf{\hat{z}} \\

        \end{array}
      \end{equation*}
    }
    \renewcommand{\arraystretch}{1.0}
  \end{tabular}
  \begin{tabular}{c}
    \includegraphics[width=0.3\linewidth]{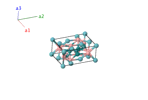} \\
  \end{tabular}
\end{tabular}

}
\vspace*{-0.25cm}

\noindent \hrulefill
\\
\textbf{Basis vectors:}
\vspace*{-0.25cm}
\renewcommand{\arraystretch}{1.5}
\begin{longtabu} to \textwidth{>{\centering $}X[-1,c,c]<{$}>{\centering $}X[-1,c,c]<{$}>{\centering $}X[-1,c,c]<{$}>{\centering $}X[-1,c,c]<{$}>{\centering $}X[-1,c,c]<{$}>{\centering $}X[-1,c,c]<{$}>{\centering $}X[-1,c,c]<{$}}
  & & \mbox{Lattice Coordinates} & & \mbox{Cartesian Coordinates} &\mbox{Wyckoff Position} & \mbox{Atom Type} \\  
  \mathbf{B}_{1} & = & 0 \, \mathbf{a}_{1} + 0 \, \mathbf{a}_{2} + 0 \, \mathbf{a}_{3} & = & 0 \, \mathbf{\hat{x}} + 0 \, \mathbf{\hat{y}} + 0 \, \mathbf{\hat{z}} & \left(1a\right) & \mbox{Nb I} \\ 
\mathbf{B}_{2} & = & \frac{1}{3} \, \mathbf{a}_{1} + \frac{2}{3} \, \mathbf{a}_{2} & = & \frac{1}{2}a \, \mathbf{\hat{x}} + \frac{1}{2\sqrt{3}}a \, \mathbf{\hat{y}} & \left(2c\right) & \mbox{B I} \\ 
\mathbf{B}_{3} & = & \frac{2}{3} \, \mathbf{a}_{1} + \frac{1}{3} \, \mathbf{a}_{2} & = & \frac{1}{2}a \, \mathbf{\hat{x}}- \frac{1}{2\sqrt{3}}a  \, \mathbf{\hat{y}} & \left(2c\right) & \mbox{B I} \\ 
\mathbf{B}_{4} & = & x_{3} \, \mathbf{a}_{1} + y_{3} \, \mathbf{a}_{2} & = & \frac{1}{2}\left(x_{3}+y_{3}\right)a \, \mathbf{\hat{x}} + \frac{\sqrt{3}}{2}\left(-x_{3}+y_{3}\right)a \, \mathbf{\hat{y}} & \left(6j\right) & \mbox{Nb II} \\ 
\mathbf{B}_{5} & = & -y_{3} \, \mathbf{a}_{1} + \left(x_{3}-y_{3}\right) \, \mathbf{a}_{2} & = & \left(\frac{1}{2}x_{3}-y_{3}\right)a \, \mathbf{\hat{x}} + \frac{\sqrt{3}}{2}x_{3}a \, \mathbf{\hat{y}} & \left(6j\right) & \mbox{Nb II} \\ 
\mathbf{B}_{6} & = & \left(-x_{3}+y_{3}\right) \, \mathbf{a}_{1}-x_{3} \, \mathbf{a}_{2} & = & \left(-x_{3}+\frac{1}{2}y_{3}\right)a \, \mathbf{\hat{x}}-\frac{\sqrt{3}}{2}y_{3}a \, \mathbf{\hat{y}} & \left(6j\right) & \mbox{Nb II} \\ 
\mathbf{B}_{7} & = & -x_{3} \, \mathbf{a}_{1}-y_{3} \, \mathbf{a}_{2} & = & -\frac{1}{2}\left(x_{3}+y_{3}\right)a \, \mathbf{\hat{x}} + \frac{\sqrt{3}}{2}\left(x_{3}-y_{3}\right)a \, \mathbf{\hat{y}} & \left(6j\right) & \mbox{Nb II} \\ 
\mathbf{B}_{8} & = & y_{3} \, \mathbf{a}_{1} + \left(-x_{3}+y_{3}\right) \, \mathbf{a}_{2} & = & \left(-\frac{1}{2}x_{3}+y_{3}\right)a \, \mathbf{\hat{x}}-\frac{\sqrt{3}}{2}x_{3}a \, \mathbf{\hat{y}} & \left(6j\right) & \mbox{Nb II} \\ 
\mathbf{B}_{9} & = & \left(x_{3}-y_{3}\right) \, \mathbf{a}_{1} + x_{3} \, \mathbf{a}_{2} & = & \left(x_{3}-\frac{1}{2}y_{3}\right)a \, \mathbf{\hat{x}} + \frac{\sqrt{3}}{2}y_{3}a \, \mathbf{\hat{y}} & \left(6j\right) & \mbox{Nb II} \\ 
\mathbf{B}_{10} & = & x_{4} \, \mathbf{a}_{1} + y_{4} \, \mathbf{a}_{2} + \frac{1}{2} \, \mathbf{a}_{3} & = & \frac{1}{2}\left(x_{4}+y_{4}\right)a \, \mathbf{\hat{x}} + \frac{\sqrt{3}}{2}\left(-x_{4}+y_{4}\right)a \, \mathbf{\hat{y}} + \frac{1}{2}c \, \mathbf{\hat{z}} & \left(6k\right) & \mbox{B II} \\ 
\mathbf{B}_{11} & = & -y_{4} \, \mathbf{a}_{1} + \left(x_{4}-y_{4}\right) \, \mathbf{a}_{2} + \frac{1}{2} \, \mathbf{a}_{3} & = & \left(\frac{1}{2}x_{4}-y_{4}\right)a \, \mathbf{\hat{x}} + \frac{\sqrt{3}}{2}x_{4}a \, \mathbf{\hat{y}} + \frac{1}{2}c \, \mathbf{\hat{z}} & \left(6k\right) & \mbox{B II} \\ 
\mathbf{B}_{12} & = & \left(-x_{4}+y_{4}\right) \, \mathbf{a}_{1}-x_{4} \, \mathbf{a}_{2} + \frac{1}{2} \, \mathbf{a}_{3} & = & \left(-x_{4}+\frac{1}{2}y_{4}\right)a \, \mathbf{\hat{x}}-\frac{\sqrt{3}}{2}y_{4}a \, \mathbf{\hat{y}} + \frac{1}{2}c \, \mathbf{\hat{z}} & \left(6k\right) & \mbox{B II} \\ 
\mathbf{B}_{13} & = & -x_{4} \, \mathbf{a}_{1}-y_{4} \, \mathbf{a}_{2} + \frac{1}{2} \, \mathbf{a}_{3} & = & -\frac{1}{2}\left(x_{4}+y_{4}\right)a \, \mathbf{\hat{x}} + \frac{\sqrt{3}}{2}\left(x_{4}-y_{4}\right)a \, \mathbf{\hat{y}} + \frac{1}{2}c \, \mathbf{\hat{z}} & \left(6k\right) & \mbox{B II} \\ 
\mathbf{B}_{14} & = & y_{4} \, \mathbf{a}_{1} + \left(-x_{4}+y_{4}\right) \, \mathbf{a}_{2} + \frac{1}{2} \, \mathbf{a}_{3} & = & \left(-\frac{1}{2}x_{4}+y_{4}\right)a \, \mathbf{\hat{x}}-\frac{\sqrt{3}}{2}x_{4}a \, \mathbf{\hat{y}} + \frac{1}{2}c \, \mathbf{\hat{z}} & \left(6k\right) & \mbox{B II} \\ 
\mathbf{B}_{15} & = & \left(x_{4}-y_{4}\right) \, \mathbf{a}_{1} + x_{4} \, \mathbf{a}_{2} + \frac{1}{2} \, \mathbf{a}_{3} & = & \left(x_{4}-\frac{1}{2}y_{4}\right)a \, \mathbf{\hat{x}} + \frac{\sqrt{3}}{2}y_{4}a \, \mathbf{\hat{y}} + \frac{1}{2}c \, \mathbf{\hat{z}} & \left(6k\right) & \mbox{B II} \\ 
\mathbf{B}_{16} & = & x_{5} \, \mathbf{a}_{1} + y_{5} \, \mathbf{a}_{2} + \frac{1}{2} \, \mathbf{a}_{3} & = & \frac{1}{2}\left(x_{5}+y_{5}\right)a \, \mathbf{\hat{x}} + \frac{\sqrt{3}}{2}\left(-x_{5}+y_{5}\right)a \, \mathbf{\hat{y}} + \frac{1}{2}c \, \mathbf{\hat{z}} & \left(6k\right) & \mbox{Ru} \\ 
\mathbf{B}_{17} & = & -y_{5} \, \mathbf{a}_{1} + \left(x_{5}-y_{5}\right) \, \mathbf{a}_{2} + \frac{1}{2} \, \mathbf{a}_{3} & = & \left(\frac{1}{2}x_{5}-y_{5}\right)a \, \mathbf{\hat{x}} + \frac{\sqrt{3}}{2}x_{5}a \, \mathbf{\hat{y}} + \frac{1}{2}c \, \mathbf{\hat{z}} & \left(6k\right) & \mbox{Ru} \\ 
\mathbf{B}_{18} & = & \left(-x_{5}+y_{5}\right) \, \mathbf{a}_{1}-x_{5} \, \mathbf{a}_{2} + \frac{1}{2} \, \mathbf{a}_{3} & = & \left(-x_{5}+\frac{1}{2}y_{5}\right)a \, \mathbf{\hat{x}}-\frac{\sqrt{3}}{2}y_{5}a \, \mathbf{\hat{y}} + \frac{1}{2}c \, \mathbf{\hat{z}} & \left(6k\right) & \mbox{Ru} \\ 
\mathbf{B}_{19} & = & -x_{5} \, \mathbf{a}_{1}-y_{5} \, \mathbf{a}_{2} + \frac{1}{2} \, \mathbf{a}_{3} & = & -\frac{1}{2}\left(x_{5}+y_{5}\right)a \, \mathbf{\hat{x}} + \frac{\sqrt{3}}{2}\left(x_{5}-y_{5}\right)a \, \mathbf{\hat{y}} + \frac{1}{2}c \, \mathbf{\hat{z}} & \left(6k\right) & \mbox{Ru} \\ 
\mathbf{B}_{20} & = & y_{5} \, \mathbf{a}_{1} + \left(-x_{5}+y_{5}\right) \, \mathbf{a}_{2} + \frac{1}{2} \, \mathbf{a}_{3} & = & \left(-\frac{1}{2}x_{5}+y_{5}\right)a \, \mathbf{\hat{x}}-\frac{\sqrt{3}}{2}x_{5}a \, \mathbf{\hat{y}} + \frac{1}{2}c \, \mathbf{\hat{z}} & \left(6k\right) & \mbox{Ru} \\ 
\mathbf{B}_{21} & = & \left(x_{5}-y_{5}\right) \, \mathbf{a}_{1} + x_{5} \, \mathbf{a}_{2} + \frac{1}{2} \, \mathbf{a}_{3} & = & \left(x_{5}-\frac{1}{2}y_{5}\right)a \, \mathbf{\hat{x}} + \frac{\sqrt{3}}{2}y_{5}a \, \mathbf{\hat{y}} + \frac{1}{2}c \, \mathbf{\hat{z}} & \left(6k\right) & \mbox{Ru} \\ 
\end{longtabu}
\renewcommand{\arraystretch}{1.0}
\noindent \hrulefill
\\
\textbf{References:}
\vspace*{-0.25cm}
\begin{flushleft}
  - \bibentry{Zheng_Nb7Ru6B8_InorgChem_2012}. \\
\end{flushleft}
\textbf{Found in:}
\vspace*{-0.25cm}
\begin{flushleft}
  - \bibentry{Villars_PearsonsCrystalData_2013}. \\
\end{flushleft}
\noindent \hrulefill
\\
\textbf{Geometry files:}
\\
\noindent  - CIF: pp. {\hyperref[A8B7C6_hP21_175_ck_aj_k_cif]{\pageref{A8B7C6_hP21_175_ck_aj_k_cif}}} \\
\noindent  - POSCAR: pp. {\hyperref[A8B7C6_hP21_175_ck_aj_k_poscar]{\pageref{A8B7C6_hP21_175_ck_aj_k_poscar}}} \\
\onecolumn
{\phantomsection\label{ABC_hP36_175_jk_jk_jk}}
\subsection*{\huge \textbf{{\normalfont Mg[NH] Structure: ABC\_hP36\_175\_jk\_jk\_jk}}}
\noindent \hrulefill
\vspace*{0.25cm}
\begin{figure}[htp]
  \centering
  \vspace{-1em}
  {\includegraphics[width=1\textwidth]{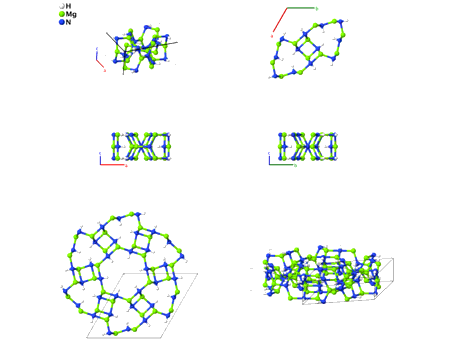}}
\end{figure}
\vspace*{-0.5cm}
\renewcommand{\arraystretch}{1.5}
\begin{equation*}
  \begin{array}{>{$\hspace{-0.15cm}}l<{$}>{$}p{0.5cm}<{$}>{$}p{18.5cm}<{$}}
    \mbox{\large \textbf{Prototype}} &\colon & \ce{Mg[NH]} \\
    \mbox{\large \textbf{\AFLOW\ prototype label}} &\colon & \mbox{ABC\_hP36\_175\_jk\_jk\_jk} \\
    \mbox{\large \textbf{\textit{Strukturbericht} designation}} &\colon & \mbox{None} \\
    \mbox{\large \textbf{Pearson symbol}} &\colon & \mbox{hP36} \\
    \mbox{\large \textbf{Space group number}} &\colon & 175 \\
    \mbox{\large \textbf{Space group symbol}} &\colon & P6/m \\
    \mbox{\large \textbf{\AFLOW\ prototype command}} &\colon &  \texttt{aflow} \,  \, \texttt{-{}-proto=ABC\_hP36\_175\_jk\_jk\_jk } \, \newline \texttt{-{}-params=}{a,c/a,x_{1},y_{1},x_{2},y_{2},x_{3},y_{3},x_{4},y_{4},x_{5},y_{5},x_{6},y_{6} }
  \end{array}
\end{equation*}
\renewcommand{\arraystretch}{1.0}

\noindent \parbox{1 \linewidth}{
\noindent \hrulefill
\\
\textbf{Hexagonal primitive vectors:} \\
\vspace*{-0.25cm}
\begin{tabular}{cc}
  \begin{tabular}{c}
    \parbox{0.6 \linewidth}{
      \renewcommand{\arraystretch}{1.5}
      \begin{equation*}
        \centering
        \begin{array}{ccc}
              \mathbf{a}_1 & = & \frac12 \, a \, \mathbf{\hat{x}} - \frac{\sqrt3}2 \, a \, \mathbf{\hat{y}} \\
    \mathbf{a}_2 & = & \frac12 \, a \, \mathbf{\hat{x}} + \frac{\sqrt3}2 \, a \, \mathbf{\hat{y}} \\
    \mathbf{a}_3 & = & c \, \mathbf{\hat{z}} \\

        \end{array}
      \end{equation*}
    }
    \renewcommand{\arraystretch}{1.0}
  \end{tabular}
  \begin{tabular}{c}
    \includegraphics[width=0.3\linewidth]{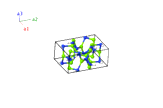} \\
  \end{tabular}
\end{tabular}

}
\vspace*{-0.25cm}

\noindent \hrulefill
\\
\textbf{Basis vectors:}
\vspace*{-0.25cm}
\renewcommand{\arraystretch}{1.5}
\begin{longtabu} to \textwidth{>{\centering $}X[-1,c,c]<{$}>{\centering $}X[-1,c,c]<{$}>{\centering $}X[-1,c,c]<{$}>{\centering $}X[-1,c,c]<{$}>{\centering $}X[-1,c,c]<{$}>{\centering $}X[-1,c,c]<{$}>{\centering $}X[-1,c,c]<{$}}
  & & \mbox{Lattice Coordinates} & & \mbox{Cartesian Coordinates} &\mbox{Wyckoff Position} & \mbox{Atom Type} \\  
  \mathbf{B}_{1} & = & x_{1} \, \mathbf{a}_{1} + y_{1} \, \mathbf{a}_{2} & = & \frac{1}{2}\left(x_{1}+y_{1}\right)a \, \mathbf{\hat{x}} + \frac{\sqrt{3}}{2}\left(-x_{1}+y_{1}\right)a \, \mathbf{\hat{y}} & \left(6j\right) & \mbox{H I} \\ 
\mathbf{B}_{2} & = & -y_{1} \, \mathbf{a}_{1} + \left(x_{1}-y_{1}\right) \, \mathbf{a}_{2} & = & \left(\frac{1}{2}x_{1}-y_{1}\right)a \, \mathbf{\hat{x}} + \frac{\sqrt{3}}{2}x_{1}a \, \mathbf{\hat{y}} & \left(6j\right) & \mbox{H I} \\ 
\mathbf{B}_{3} & = & \left(-x_{1}+y_{1}\right) \, \mathbf{a}_{1}-x_{1} \, \mathbf{a}_{2} & = & \left(-x_{1}+\frac{1}{2}y_{1}\right)a \, \mathbf{\hat{x}}-\frac{\sqrt{3}}{2}y_{1}a \, \mathbf{\hat{y}} & \left(6j\right) & \mbox{H I} \\ 
\mathbf{B}_{4} & = & -x_{1} \, \mathbf{a}_{1}-y_{1} \, \mathbf{a}_{2} & = & -\frac{1}{2}\left(x_{1}+y_{1}\right)a \, \mathbf{\hat{x}} + \frac{\sqrt{3}}{2}\left(x_{1}-y_{1}\right)a \, \mathbf{\hat{y}} & \left(6j\right) & \mbox{H I} \\ 
\mathbf{B}_{5} & = & y_{1} \, \mathbf{a}_{1} + \left(-x_{1}+y_{1}\right) \, \mathbf{a}_{2} & = & \left(-\frac{1}{2}x_{1}+y_{1}\right)a \, \mathbf{\hat{x}}-\frac{\sqrt{3}}{2}x_{1}a \, \mathbf{\hat{y}} & \left(6j\right) & \mbox{H I} \\ 
\mathbf{B}_{6} & = & \left(x_{1}-y_{1}\right) \, \mathbf{a}_{1} + x_{1} \, \mathbf{a}_{2} & = & \left(x_{1}-\frac{1}{2}y_{1}\right)a \, \mathbf{\hat{x}} + \frac{\sqrt{3}}{2}y_{1}a \, \mathbf{\hat{y}} & \left(6j\right) & \mbox{H I} \\ 
\mathbf{B}_{7} & = & x_{2} \, \mathbf{a}_{1} + y_{2} \, \mathbf{a}_{2} & = & \frac{1}{2}\left(x_{2}+y_{2}\right)a \, \mathbf{\hat{x}} + \frac{\sqrt{3}}{2}\left(-x_{2}+y_{2}\right)a \, \mathbf{\hat{y}} & \left(6j\right) & \mbox{Mg I} \\ 
\mathbf{B}_{8} & = & -y_{2} \, \mathbf{a}_{1} + \left(x_{2}-y_{2}\right) \, \mathbf{a}_{2} & = & \left(\frac{1}{2}x_{2}-y_{2}\right)a \, \mathbf{\hat{x}} + \frac{\sqrt{3}}{2}x_{2}a \, \mathbf{\hat{y}} & \left(6j\right) & \mbox{Mg I} \\ 
\mathbf{B}_{9} & = & \left(-x_{2}+y_{2}\right) \, \mathbf{a}_{1}-x_{2} \, \mathbf{a}_{2} & = & \left(-x_{2}+\frac{1}{2}y_{2}\right)a \, \mathbf{\hat{x}}-\frac{\sqrt{3}}{2}y_{2}a \, \mathbf{\hat{y}} & \left(6j\right) & \mbox{Mg I} \\ 
\mathbf{B}_{10} & = & -x_{2} \, \mathbf{a}_{1}-y_{2} \, \mathbf{a}_{2} & = & -\frac{1}{2}\left(x_{2}+y_{2}\right)a \, \mathbf{\hat{x}} + \frac{\sqrt{3}}{2}\left(x_{2}-y_{2}\right)a \, \mathbf{\hat{y}} & \left(6j\right) & \mbox{Mg I} \\ 
\mathbf{B}_{11} & = & y_{2} \, \mathbf{a}_{1} + \left(-x_{2}+y_{2}\right) \, \mathbf{a}_{2} & = & \left(-\frac{1}{2}x_{2}+y_{2}\right)a \, \mathbf{\hat{x}}-\frac{\sqrt{3}}{2}x_{2}a \, \mathbf{\hat{y}} & \left(6j\right) & \mbox{Mg I} \\ 
\mathbf{B}_{12} & = & \left(x_{2}-y_{2}\right) \, \mathbf{a}_{1} + x_{2} \, \mathbf{a}_{2} & = & \left(x_{2}-\frac{1}{2}y_{2}\right)a \, \mathbf{\hat{x}} + \frac{\sqrt{3}}{2}y_{2}a \, \mathbf{\hat{y}} & \left(6j\right) & \mbox{Mg I} \\ 
\mathbf{B}_{13} & = & x_{3} \, \mathbf{a}_{1} + y_{3} \, \mathbf{a}_{2} & = & \frac{1}{2}\left(x_{3}+y_{3}\right)a \, \mathbf{\hat{x}} + \frac{\sqrt{3}}{2}\left(-x_{3}+y_{3}\right)a \, \mathbf{\hat{y}} & \left(6j\right) & \mbox{N I} \\ 
\mathbf{B}_{14} & = & -y_{3} \, \mathbf{a}_{1} + \left(x_{3}-y_{3}\right) \, \mathbf{a}_{2} & = & \left(\frac{1}{2}x_{3}-y_{3}\right)a \, \mathbf{\hat{x}} + \frac{\sqrt{3}}{2}x_{3}a \, \mathbf{\hat{y}} & \left(6j\right) & \mbox{N I} \\ 
\mathbf{B}_{15} & = & \left(-x_{3}+y_{3}\right) \, \mathbf{a}_{1}-x_{3} \, \mathbf{a}_{2} & = & \left(-x_{3}+\frac{1}{2}y_{3}\right)a \, \mathbf{\hat{x}}-\frac{\sqrt{3}}{2}y_{3}a \, \mathbf{\hat{y}} & \left(6j\right) & \mbox{N I} \\ 
\mathbf{B}_{16} & = & -x_{3} \, \mathbf{a}_{1}-y_{3} \, \mathbf{a}_{2} & = & -\frac{1}{2}\left(x_{3}+y_{3}\right)a \, \mathbf{\hat{x}} + \frac{\sqrt{3}}{2}\left(x_{3}-y_{3}\right)a \, \mathbf{\hat{y}} & \left(6j\right) & \mbox{N I} \\ 
\mathbf{B}_{17} & = & y_{3} \, \mathbf{a}_{1} + \left(-x_{3}+y_{3}\right) \, \mathbf{a}_{2} & = & \left(-\frac{1}{2}x_{3}+y_{3}\right)a \, \mathbf{\hat{x}}-\frac{\sqrt{3}}{2}x_{3}a \, \mathbf{\hat{y}} & \left(6j\right) & \mbox{N I} \\ 
\mathbf{B}_{18} & = & \left(x_{3}-y_{3}\right) \, \mathbf{a}_{1} + x_{3} \, \mathbf{a}_{2} & = & \left(x_{3}-\frac{1}{2}y_{3}\right)a \, \mathbf{\hat{x}} + \frac{\sqrt{3}}{2}y_{3}a \, \mathbf{\hat{y}} & \left(6j\right) & \mbox{N I} \\ 
\mathbf{B}_{19} & = & x_{4} \, \mathbf{a}_{1} + y_{4} \, \mathbf{a}_{2} + \frac{1}{2} \, \mathbf{a}_{3} & = & \frac{1}{2}\left(x_{4}+y_{4}\right)a \, \mathbf{\hat{x}} + \frac{\sqrt{3}}{2}\left(-x_{4}+y_{4}\right)a \, \mathbf{\hat{y}} + \frac{1}{2}c \, \mathbf{\hat{z}} & \left(6k\right) & \mbox{H II} \\ 
\mathbf{B}_{20} & = & -y_{4} \, \mathbf{a}_{1} + \left(x_{4}-y_{4}\right) \, \mathbf{a}_{2} + \frac{1}{2} \, \mathbf{a}_{3} & = & \left(\frac{1}{2}x_{4}-y_{4}\right)a \, \mathbf{\hat{x}} + \frac{\sqrt{3}}{2}x_{4}a \, \mathbf{\hat{y}} + \frac{1}{2}c \, \mathbf{\hat{z}} & \left(6k\right) & \mbox{H II} \\ 
\mathbf{B}_{21} & = & \left(-x_{4}+y_{4}\right) \, \mathbf{a}_{1}-x_{4} \, \mathbf{a}_{2} + \frac{1}{2} \, \mathbf{a}_{3} & = & \left(-x_{4}+\frac{1}{2}y_{4}\right)a \, \mathbf{\hat{x}}-\frac{\sqrt{3}}{2}y_{4}a \, \mathbf{\hat{y}} + \frac{1}{2}c \, \mathbf{\hat{z}} & \left(6k\right) & \mbox{H II} \\ 
\mathbf{B}_{22} & = & -x_{4} \, \mathbf{a}_{1}-y_{4} \, \mathbf{a}_{2} + \frac{1}{2} \, \mathbf{a}_{3} & = & -\frac{1}{2}\left(x_{4}+y_{4}\right)a \, \mathbf{\hat{x}} + \frac{\sqrt{3}}{2}\left(x_{4}-y_{4}\right)a \, \mathbf{\hat{y}} + \frac{1}{2}c \, \mathbf{\hat{z}} & \left(6k\right) & \mbox{H II} \\ 
\mathbf{B}_{23} & = & y_{4} \, \mathbf{a}_{1} + \left(-x_{4}+y_{4}\right) \, \mathbf{a}_{2} + \frac{1}{2} \, \mathbf{a}_{3} & = & \left(-\frac{1}{2}x_{4}+y_{4}\right)a \, \mathbf{\hat{x}}-\frac{\sqrt{3}}{2}x_{4}a \, \mathbf{\hat{y}} + \frac{1}{2}c \, \mathbf{\hat{z}} & \left(6k\right) & \mbox{H II} \\ 
\mathbf{B}_{24} & = & \left(x_{4}-y_{4}\right) \, \mathbf{a}_{1} + x_{4} \, \mathbf{a}_{2} + \frac{1}{2} \, \mathbf{a}_{3} & = & \left(x_{4}-\frac{1}{2}y_{4}\right)a \, \mathbf{\hat{x}} + \frac{\sqrt{3}}{2}y_{4}a \, \mathbf{\hat{y}} + \frac{1}{2}c \, \mathbf{\hat{z}} & \left(6k\right) & \mbox{H II} \\ 
\mathbf{B}_{25} & = & x_{5} \, \mathbf{a}_{1} + y_{5} \, \mathbf{a}_{2} + \frac{1}{2} \, \mathbf{a}_{3} & = & \frac{1}{2}\left(x_{5}+y_{5}\right)a \, \mathbf{\hat{x}} + \frac{\sqrt{3}}{2}\left(-x_{5}+y_{5}\right)a \, \mathbf{\hat{y}} + \frac{1}{2}c \, \mathbf{\hat{z}} & \left(6k\right) & \mbox{Mg II} \\ 
\mathbf{B}_{26} & = & -y_{5} \, \mathbf{a}_{1} + \left(x_{5}-y_{5}\right) \, \mathbf{a}_{2} + \frac{1}{2} \, \mathbf{a}_{3} & = & \left(\frac{1}{2}x_{5}-y_{5}\right)a \, \mathbf{\hat{x}} + \frac{\sqrt{3}}{2}x_{5}a \, \mathbf{\hat{y}} + \frac{1}{2}c \, \mathbf{\hat{z}} & \left(6k\right) & \mbox{Mg II} \\ 
\mathbf{B}_{27} & = & \left(-x_{5}+y_{5}\right) \, \mathbf{a}_{1}-x_{5} \, \mathbf{a}_{2} + \frac{1}{2} \, \mathbf{a}_{3} & = & \left(-x_{5}+\frac{1}{2}y_{5}\right)a \, \mathbf{\hat{x}}-\frac{\sqrt{3}}{2}y_{5}a \, \mathbf{\hat{y}} + \frac{1}{2}c \, \mathbf{\hat{z}} & \left(6k\right) & \mbox{Mg II} \\ 
\mathbf{B}_{28} & = & -x_{5} \, \mathbf{a}_{1}-y_{5} \, \mathbf{a}_{2} + \frac{1}{2} \, \mathbf{a}_{3} & = & -\frac{1}{2}\left(x_{5}+y_{5}\right)a \, \mathbf{\hat{x}} + \frac{\sqrt{3}}{2}\left(x_{5}-y_{5}\right)a \, \mathbf{\hat{y}} + \frac{1}{2}c \, \mathbf{\hat{z}} & \left(6k\right) & \mbox{Mg II} \\ 
\mathbf{B}_{29} & = & y_{5} \, \mathbf{a}_{1} + \left(-x_{5}+y_{5}\right) \, \mathbf{a}_{2} + \frac{1}{2} \, \mathbf{a}_{3} & = & \left(-\frac{1}{2}x_{5}+y_{5}\right)a \, \mathbf{\hat{x}}-\frac{\sqrt{3}}{2}x_{5}a \, \mathbf{\hat{y}} + \frac{1}{2}c \, \mathbf{\hat{z}} & \left(6k\right) & \mbox{Mg II} \\ 
\mathbf{B}_{30} & = & \left(x_{5}-y_{5}\right) \, \mathbf{a}_{1} + x_{5} \, \mathbf{a}_{2} + \frac{1}{2} \, \mathbf{a}_{3} & = & \left(x_{5}-\frac{1}{2}y_{5}\right)a \, \mathbf{\hat{x}} + \frac{\sqrt{3}}{2}y_{5}a \, \mathbf{\hat{y}} + \frac{1}{2}c \, \mathbf{\hat{z}} & \left(6k\right) & \mbox{Mg II} \\ 
\mathbf{B}_{31} & = & x_{6} \, \mathbf{a}_{1} + y_{6} \, \mathbf{a}_{2} + \frac{1}{2} \, \mathbf{a}_{3} & = & \frac{1}{2}\left(x_{6}+y_{6}\right)a \, \mathbf{\hat{x}} + \frac{\sqrt{3}}{2}\left(-x_{6}+y_{6}\right)a \, \mathbf{\hat{y}} + \frac{1}{2}c \, \mathbf{\hat{z}} & \left(6k\right) & \mbox{N II} \\ 
\mathbf{B}_{32} & = & -y_{6} \, \mathbf{a}_{1} + \left(x_{6}-y_{6}\right) \, \mathbf{a}_{2} + \frac{1}{2} \, \mathbf{a}_{3} & = & \left(\frac{1}{2}x_{6}-y_{6}\right)a \, \mathbf{\hat{x}} + \frac{\sqrt{3}}{2}x_{6}a \, \mathbf{\hat{y}} + \frac{1}{2}c \, \mathbf{\hat{z}} & \left(6k\right) & \mbox{N II} \\ 
\mathbf{B}_{33} & = & \left(-x_{6}+y_{6}\right) \, \mathbf{a}_{1}-x_{6} \, \mathbf{a}_{2} + \frac{1}{2} \, \mathbf{a}_{3} & = & \left(-x_{6}+\frac{1}{2}y_{6}\right)a \, \mathbf{\hat{x}}-\frac{\sqrt{3}}{2}y_{6}a \, \mathbf{\hat{y}} + \frac{1}{2}c \, \mathbf{\hat{z}} & \left(6k\right) & \mbox{N II} \\ 
\mathbf{B}_{34} & = & -x_{6} \, \mathbf{a}_{1}-y_{6} \, \mathbf{a}_{2} + \frac{1}{2} \, \mathbf{a}_{3} & = & -\frac{1}{2}\left(x_{6}+y_{6}\right)a \, \mathbf{\hat{x}} + \frac{\sqrt{3}}{2}\left(x_{6}-y_{6}\right)a \, \mathbf{\hat{y}} + \frac{1}{2}c \, \mathbf{\hat{z}} & \left(6k\right) & \mbox{N II} \\ 
\mathbf{B}_{35} & = & y_{6} \, \mathbf{a}_{1} + \left(-x_{6}+y_{6}\right) \, \mathbf{a}_{2} + \frac{1}{2} \, \mathbf{a}_{3} & = & \left(-\frac{1}{2}x_{6}+y_{6}\right)a \, \mathbf{\hat{x}}-\frac{\sqrt{3}}{2}x_{6}a \, \mathbf{\hat{y}} + \frac{1}{2}c \, \mathbf{\hat{z}} & \left(6k\right) & \mbox{N II} \\ 
\mathbf{B}_{36} & = & \left(x_{6}-y_{6}\right) \, \mathbf{a}_{1} + x_{6} \, \mathbf{a}_{2} + \frac{1}{2} \, \mathbf{a}_{3} & = & \left(x_{6}-\frac{1}{2}y_{6}\right)a \, \mathbf{\hat{x}} + \frac{\sqrt{3}}{2}y_{6}a \, \mathbf{\hat{y}} + \frac{1}{2}c \, \mathbf{\hat{z}} & \left(6k\right) & \mbox{N II} \\ 
\end{longtabu}
\renewcommand{\arraystretch}{1.0}
\noindent \hrulefill
\\
\textbf{References:}
\vspace*{-0.25cm}
\begin{flushleft}
  - \bibentry{Dolci_MgNH_InorgChem_2010}. \\
\end{flushleft}
\textbf{Found in:}
\vspace*{-0.25cm}
\begin{flushleft}
  - \bibentry{Villars_PearsonsCrystalData_2013}. \\
\end{flushleft}
\noindent \hrulefill
\\
\textbf{Geometry files:}
\\
\noindent  - CIF: pp. {\hyperref[ABC_hP36_175_jk_jk_jk_cif]{\pageref{ABC_hP36_175_jk_jk_jk_cif}}} \\
\noindent  - POSCAR: pp. {\hyperref[ABC_hP36_175_jk_jk_jk_poscar]{\pageref{ABC_hP36_175_jk_jk_jk_poscar}}} \\
\onecolumn
{\phantomsection\label{A3B2_hP10_176_h_bd}}
\subsection*{\huge \textbf{{\normalfont Er$_{3}$Ru$_{2}$ Structure: A3B2\_hP10\_176\_h\_bd}}}
\noindent \hrulefill
\vspace*{0.25cm}
\begin{figure}[htp]
  \centering
  \vspace{-1em}
  {\includegraphics[width=1\textwidth]{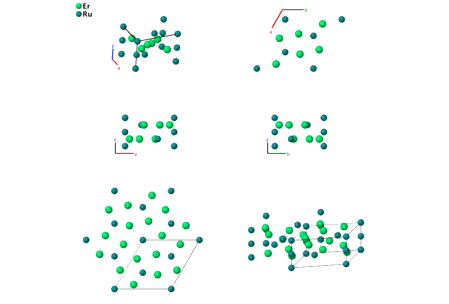}}
\end{figure}
\vspace*{-0.5cm}
\renewcommand{\arraystretch}{1.5}
\begin{equation*}
  \begin{array}{>{$\hspace{-0.15cm}}l<{$}>{$}p{0.5cm}<{$}>{$}p{18.5cm}<{$}}
    \mbox{\large \textbf{Prototype}} &\colon & \ce{Er3Ru2} \\
    \mbox{\large \textbf{\AFLOW\ prototype label}} &\colon & \mbox{A3B2\_hP10\_176\_h\_bd} \\
    \mbox{\large \textbf{\textit{Strukturbericht} designation}} &\colon & \mbox{None} \\
    \mbox{\large \textbf{Pearson symbol}} &\colon & \mbox{hP10} \\
    \mbox{\large \textbf{Space group number}} &\colon & 176 \\
    \mbox{\large \textbf{Space group symbol}} &\colon & P6_{3}/m \\
    \mbox{\large \textbf{\AFLOW\ prototype command}} &\colon &  \texttt{aflow} \,  \, \texttt{-{}-proto=A3B2\_hP10\_176\_h\_bd } \, \newline \texttt{-{}-params=}{a,c/a,x_{3},y_{3} }
  \end{array}
\end{equation*}
\renewcommand{\arraystretch}{1.0}

\noindent \parbox{1 \linewidth}{
\noindent \hrulefill
\\
\textbf{Hexagonal primitive vectors:} \\
\vspace*{-0.25cm}
\begin{tabular}{cc}
  \begin{tabular}{c}
    \parbox{0.6 \linewidth}{
      \renewcommand{\arraystretch}{1.5}
      \begin{equation*}
        \centering
        \begin{array}{ccc}
              \mathbf{a}_1 & = & \frac12 \, a \, \mathbf{\hat{x}} - \frac{\sqrt3}2 \, a \, \mathbf{\hat{y}} \\
    \mathbf{a}_2 & = & \frac12 \, a \, \mathbf{\hat{x}} + \frac{\sqrt3}2 \, a \, \mathbf{\hat{y}} \\
    \mathbf{a}_3 & = & c \, \mathbf{\hat{z}} \\

        \end{array}
      \end{equation*}
    }
    \renewcommand{\arraystretch}{1.0}
  \end{tabular}
  \begin{tabular}{c}
    \includegraphics[width=0.3\linewidth]{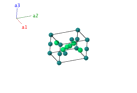} \\
  \end{tabular}
\end{tabular}

}
\vspace*{-0.25cm}

\noindent \hrulefill
\\
\textbf{Basis vectors:}
\vspace*{-0.25cm}
\renewcommand{\arraystretch}{1.5}
\begin{longtabu} to \textwidth{>{\centering $}X[-1,c,c]<{$}>{\centering $}X[-1,c,c]<{$}>{\centering $}X[-1,c,c]<{$}>{\centering $}X[-1,c,c]<{$}>{\centering $}X[-1,c,c]<{$}>{\centering $}X[-1,c,c]<{$}>{\centering $}X[-1,c,c]<{$}}
  & & \mbox{Lattice Coordinates} & & \mbox{Cartesian Coordinates} &\mbox{Wyckoff Position} & \mbox{Atom Type} \\  
  \mathbf{B}_{1} & = & 0 \, \mathbf{a}_{1} + 0 \, \mathbf{a}_{2} + 0 \, \mathbf{a}_{3} & = & 0 \, \mathbf{\hat{x}} + 0 \, \mathbf{\hat{y}} + 0 \, \mathbf{\hat{z}} & \left(2b\right) & \mbox{Ru I} \\ 
\mathbf{B}_{2} & = & \frac{1}{2} \, \mathbf{a}_{3} & = & \frac{1}{2}c \, \mathbf{\hat{z}} & \left(2b\right) & \mbox{Ru I} \\ 
\mathbf{B}_{3} & = & \frac{2}{3} \, \mathbf{a}_{1} + \frac{1}{3} \, \mathbf{a}_{2} + \frac{1}{4} \, \mathbf{a}_{3} & = & \frac{1}{2}a \, \mathbf{\hat{x}}- \frac{1}{2\sqrt{3}}a  \, \mathbf{\hat{y}} + \frac{1}{4}c \, \mathbf{\hat{z}} & \left(2d\right) & \mbox{Ru II} \\ 
\mathbf{B}_{4} & = & \frac{1}{3} \, \mathbf{a}_{1} + \frac{2}{3} \, \mathbf{a}_{2} + \frac{3}{4} \, \mathbf{a}_{3} & = & \frac{1}{2}a \, \mathbf{\hat{x}} + \frac{1}{2\sqrt{3}}a \, \mathbf{\hat{y}} + \frac{3}{4}c \, \mathbf{\hat{z}} & \left(2d\right) & \mbox{Ru II} \\ 
\mathbf{B}_{5} & = & x_{3} \, \mathbf{a}_{1} + y_{3} \, \mathbf{a}_{2} + \frac{1}{4} \, \mathbf{a}_{3} & = & \frac{1}{2}\left(x_{3}+y_{3}\right)a \, \mathbf{\hat{x}} + \frac{\sqrt{3}}{2}\left(-x_{3}+y_{3}\right)a \, \mathbf{\hat{y}} + \frac{1}{4}c \, \mathbf{\hat{z}} & \left(6h\right) & \mbox{Er} \\ 
\mathbf{B}_{6} & = & -y_{3} \, \mathbf{a}_{1} + \left(x_{3}-y_{3}\right) \, \mathbf{a}_{2} + \frac{1}{4} \, \mathbf{a}_{3} & = & \left(\frac{1}{2}x_{3}-y_{3}\right)a \, \mathbf{\hat{x}} + \frac{\sqrt{3}}{2}x_{3}a \, \mathbf{\hat{y}} + \frac{1}{4}c \, \mathbf{\hat{z}} & \left(6h\right) & \mbox{Er} \\ 
\mathbf{B}_{7} & = & \left(-x_{3}+y_{3}\right) \, \mathbf{a}_{1}-x_{3} \, \mathbf{a}_{2} + \frac{1}{4} \, \mathbf{a}_{3} & = & \left(-x_{3}+\frac{1}{2}y_{3}\right)a \, \mathbf{\hat{x}}-\frac{\sqrt{3}}{2}y_{3}a \, \mathbf{\hat{y}} + \frac{1}{4}c \, \mathbf{\hat{z}} & \left(6h\right) & \mbox{Er} \\ 
\mathbf{B}_{8} & = & -x_{3} \, \mathbf{a}_{1}-y_{3} \, \mathbf{a}_{2} + \frac{3}{4} \, \mathbf{a}_{3} & = & -\frac{1}{2}\left(x_{3}+y_{3}\right)a \, \mathbf{\hat{x}} + \frac{\sqrt{3}}{2}\left(x_{3}-y_{3}\right)a \, \mathbf{\hat{y}} + \frac{3}{4}c \, \mathbf{\hat{z}} & \left(6h\right) & \mbox{Er} \\ 
\mathbf{B}_{9} & = & y_{3} \, \mathbf{a}_{1} + \left(-x_{3}+y_{3}\right) \, \mathbf{a}_{2} + \frac{3}{4} \, \mathbf{a}_{3} & = & \left(-\frac{1}{2}x_{3}+y_{3}\right)a \, \mathbf{\hat{x}}-\frac{\sqrt{3}}{2}x_{3}a \, \mathbf{\hat{y}} + \frac{3}{4}c \, \mathbf{\hat{z}} & \left(6h\right) & \mbox{Er} \\ 
\mathbf{B}_{10} & = & \left(x_{3}-y_{3}\right) \, \mathbf{a}_{1} + x_{3} \, \mathbf{a}_{2} + \frac{3}{4} \, \mathbf{a}_{3} & = & \left(x_{3}-\frac{1}{2}y_{3}\right)a \, \mathbf{\hat{x}} + \frac{\sqrt{3}}{2}y_{3}a \, \mathbf{\hat{y}} + \frac{3}{4}c \, \mathbf{\hat{z}} & \left(6h\right) & \mbox{Er} \\ 
\end{longtabu}
\renewcommand{\arraystretch}{1.0}
\noindent \hrulefill
\\
\textbf{References:}
\vspace*{-0.25cm}
\begin{flushleft}
  - \bibentry{Palenzona_Er3Ru2_JLessCommMetals_1990phase}. \\
\end{flushleft}
\textbf{Found in:}
\vspace*{-0.25cm}
\begin{flushleft}
  - \bibentry{Villars_PearsonsCrystalData_2013}. \\
\end{flushleft}
\noindent \hrulefill
\\
\textbf{Geometry files:}
\\
\noindent  - CIF: pp. {\hyperref[A3B2_hP10_176_h_bd_cif]{\pageref{A3B2_hP10_176_h_bd_cif}}} \\
\noindent  - POSCAR: pp. {\hyperref[A3B2_hP10_176_h_bd_poscar]{\pageref{A3B2_hP10_176_h_bd_poscar}}} \\
\onecolumn
{\phantomsection\label{A3B3C_hP14_176_h_h_d}}
\subsection*{\huge \textbf{{\normalfont Fe$_{3}$Te$_{3}$Tl Structure: A3B3C\_hP14\_176\_h\_h\_d}}}
\noindent \hrulefill
\vspace*{0.25cm}
\begin{figure}[htp]
  \centering
  \vspace{-1em}
  {\includegraphics[width=1\textwidth]{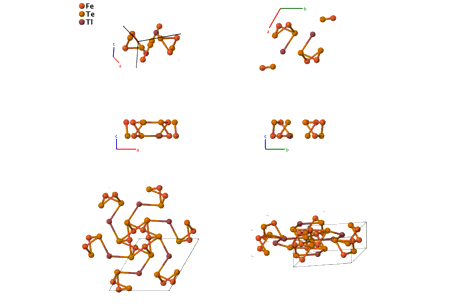}}
\end{figure}
\vspace*{-0.5cm}
\renewcommand{\arraystretch}{1.5}
\begin{equation*}
  \begin{array}{>{$\hspace{-0.15cm}}l<{$}>{$}p{0.5cm}<{$}>{$}p{18.5cm}<{$}}
    \mbox{\large \textbf{Prototype}} &\colon & \ce{Fe3Te3Tl} \\
    \mbox{\large \textbf{\AFLOW\ prototype label}} &\colon & \mbox{A3B3C\_hP14\_176\_h\_h\_d} \\
    \mbox{\large \textbf{\textit{Strukturbericht} designation}} &\colon & \mbox{None} \\
    \mbox{\large \textbf{Pearson symbol}} &\colon & \mbox{hP14} \\
    \mbox{\large \textbf{Space group number}} &\colon & 176 \\
    \mbox{\large \textbf{Space group symbol}} &\colon & P6_{3}/m \\
    \mbox{\large \textbf{\AFLOW\ prototype command}} &\colon &  \texttt{aflow} \,  \, \texttt{-{}-proto=A3B3C\_hP14\_176\_h\_h\_d } \, \newline \texttt{-{}-params=}{a,c/a,x_{2},y_{2},x_{3},y_{3} }
  \end{array}
\end{equation*}
\renewcommand{\arraystretch}{1.0}

\noindent \parbox{1 \linewidth}{
\noindent \hrulefill
\\
\textbf{Hexagonal primitive vectors:} \\
\vspace*{-0.25cm}
\begin{tabular}{cc}
  \begin{tabular}{c}
    \parbox{0.6 \linewidth}{
      \renewcommand{\arraystretch}{1.5}
      \begin{equation*}
        \centering
        \begin{array}{ccc}
              \mathbf{a}_1 & = & \frac12 \, a \, \mathbf{\hat{x}} - \frac{\sqrt3}2 \, a \, \mathbf{\hat{y}} \\
    \mathbf{a}_2 & = & \frac12 \, a \, \mathbf{\hat{x}} + \frac{\sqrt3}2 \, a \, \mathbf{\hat{y}} \\
    \mathbf{a}_3 & = & c \, \mathbf{\hat{z}} \\

        \end{array}
      \end{equation*}
    }
    \renewcommand{\arraystretch}{1.0}
  \end{tabular}
  \begin{tabular}{c}
    \includegraphics[width=0.3\linewidth]{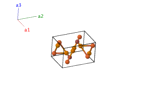} \\
  \end{tabular}
\end{tabular}

}
\vspace*{-0.25cm}

\noindent \hrulefill
\\
\textbf{Basis vectors:}
\vspace*{-0.25cm}
\renewcommand{\arraystretch}{1.5}
\begin{longtabu} to \textwidth{>{\centering $}X[-1,c,c]<{$}>{\centering $}X[-1,c,c]<{$}>{\centering $}X[-1,c,c]<{$}>{\centering $}X[-1,c,c]<{$}>{\centering $}X[-1,c,c]<{$}>{\centering $}X[-1,c,c]<{$}>{\centering $}X[-1,c,c]<{$}}
  & & \mbox{Lattice Coordinates} & & \mbox{Cartesian Coordinates} &\mbox{Wyckoff Position} & \mbox{Atom Type} \\  
  \mathbf{B}_{1} & = & \frac{2}{3} \, \mathbf{a}_{1} + \frac{1}{3} \, \mathbf{a}_{2} + \frac{1}{4} \, \mathbf{a}_{3} & = & \frac{1}{2}a \, \mathbf{\hat{x}}- \frac{1}{2\sqrt{3}}a  \, \mathbf{\hat{y}} + \frac{1}{4}c \, \mathbf{\hat{z}} & \left(2d\right) & \mbox{Tl} \\ 
\mathbf{B}_{2} & = & \frac{1}{3} \, \mathbf{a}_{1} + \frac{2}{3} \, \mathbf{a}_{2} + \frac{3}{4} \, \mathbf{a}_{3} & = & \frac{1}{2}a \, \mathbf{\hat{x}} + \frac{1}{2\sqrt{3}}a \, \mathbf{\hat{y}} + \frac{3}{4}c \, \mathbf{\hat{z}} & \left(2d\right) & \mbox{Tl} \\ 
\mathbf{B}_{3} & = & x_{2} \, \mathbf{a}_{1} + y_{2} \, \mathbf{a}_{2} + \frac{1}{4} \, \mathbf{a}_{3} & = & \frac{1}{2}\left(x_{2}+y_{2}\right)a \, \mathbf{\hat{x}} + \frac{\sqrt{3}}{2}\left(-x_{2}+y_{2}\right)a \, \mathbf{\hat{y}} + \frac{1}{4}c \, \mathbf{\hat{z}} & \left(6h\right) & \mbox{Fe} \\ 
\mathbf{B}_{4} & = & -y_{2} \, \mathbf{a}_{1} + \left(x_{2}-y_{2}\right) \, \mathbf{a}_{2} + \frac{1}{4} \, \mathbf{a}_{3} & = & \left(\frac{1}{2}x_{2}-y_{2}\right)a \, \mathbf{\hat{x}} + \frac{\sqrt{3}}{2}x_{2}a \, \mathbf{\hat{y}} + \frac{1}{4}c \, \mathbf{\hat{z}} & \left(6h\right) & \mbox{Fe} \\ 
\mathbf{B}_{5} & = & \left(-x_{2}+y_{2}\right) \, \mathbf{a}_{1}-x_{2} \, \mathbf{a}_{2} + \frac{1}{4} \, \mathbf{a}_{3} & = & \left(-x_{2}+\frac{1}{2}y_{2}\right)a \, \mathbf{\hat{x}}-\frac{\sqrt{3}}{2}y_{2}a \, \mathbf{\hat{y}} + \frac{1}{4}c \, \mathbf{\hat{z}} & \left(6h\right) & \mbox{Fe} \\ 
\mathbf{B}_{6} & = & -x_{2} \, \mathbf{a}_{1}-y_{2} \, \mathbf{a}_{2} + \frac{3}{4} \, \mathbf{a}_{3} & = & -\frac{1}{2}\left(x_{2}+y_{2}\right)a \, \mathbf{\hat{x}} + \frac{\sqrt{3}}{2}\left(x_{2}-y_{2}\right)a \, \mathbf{\hat{y}} + \frac{3}{4}c \, \mathbf{\hat{z}} & \left(6h\right) & \mbox{Fe} \\ 
\mathbf{B}_{7} & = & y_{2} \, \mathbf{a}_{1} + \left(-x_{2}+y_{2}\right) \, \mathbf{a}_{2} + \frac{3}{4} \, \mathbf{a}_{3} & = & \left(-\frac{1}{2}x_{2}+y_{2}\right)a \, \mathbf{\hat{x}}-\frac{\sqrt{3}}{2}x_{2}a \, \mathbf{\hat{y}} + \frac{3}{4}c \, \mathbf{\hat{z}} & \left(6h\right) & \mbox{Fe} \\ 
\mathbf{B}_{8} & = & \left(x_{2}-y_{2}\right) \, \mathbf{a}_{1} + x_{2} \, \mathbf{a}_{2} + \frac{3}{4} \, \mathbf{a}_{3} & = & \left(x_{2}-\frac{1}{2}y_{2}\right)a \, \mathbf{\hat{x}} + \frac{\sqrt{3}}{2}y_{2}a \, \mathbf{\hat{y}} + \frac{3}{4}c \, \mathbf{\hat{z}} & \left(6h\right) & \mbox{Fe} \\ 
\mathbf{B}_{9} & = & x_{3} \, \mathbf{a}_{1} + y_{3} \, \mathbf{a}_{2} + \frac{1}{4} \, \mathbf{a}_{3} & = & \frac{1}{2}\left(x_{3}+y_{3}\right)a \, \mathbf{\hat{x}} + \frac{\sqrt{3}}{2}\left(-x_{3}+y_{3}\right)a \, \mathbf{\hat{y}} + \frac{1}{4}c \, \mathbf{\hat{z}} & \left(6h\right) & \mbox{Te} \\ 
\mathbf{B}_{10} & = & -y_{3} \, \mathbf{a}_{1} + \left(x_{3}-y_{3}\right) \, \mathbf{a}_{2} + \frac{1}{4} \, \mathbf{a}_{3} & = & \left(\frac{1}{2}x_{3}-y_{3}\right)a \, \mathbf{\hat{x}} + \frac{\sqrt{3}}{2}x_{3}a \, \mathbf{\hat{y}} + \frac{1}{4}c \, \mathbf{\hat{z}} & \left(6h\right) & \mbox{Te} \\ 
\mathbf{B}_{11} & = & \left(-x_{3}+y_{3}\right) \, \mathbf{a}_{1}-x_{3} \, \mathbf{a}_{2} + \frac{1}{4} \, \mathbf{a}_{3} & = & \left(-x_{3}+\frac{1}{2}y_{3}\right)a \, \mathbf{\hat{x}}-\frac{\sqrt{3}}{2}y_{3}a \, \mathbf{\hat{y}} + \frac{1}{4}c \, \mathbf{\hat{z}} & \left(6h\right) & \mbox{Te} \\ 
\mathbf{B}_{12} & = & -x_{3} \, \mathbf{a}_{1}-y_{3} \, \mathbf{a}_{2} + \frac{3}{4} \, \mathbf{a}_{3} & = & -\frac{1}{2}\left(x_{3}+y_{3}\right)a \, \mathbf{\hat{x}} + \frac{\sqrt{3}}{2}\left(x_{3}-y_{3}\right)a \, \mathbf{\hat{y}} + \frac{3}{4}c \, \mathbf{\hat{z}} & \left(6h\right) & \mbox{Te} \\ 
\mathbf{B}_{13} & = & y_{3} \, \mathbf{a}_{1} + \left(-x_{3}+y_{3}\right) \, \mathbf{a}_{2} + \frac{3}{4} \, \mathbf{a}_{3} & = & \left(-\frac{1}{2}x_{3}+y_{3}\right)a \, \mathbf{\hat{x}}-\frac{\sqrt{3}}{2}x_{3}a \, \mathbf{\hat{y}} + \frac{3}{4}c \, \mathbf{\hat{z}} & \left(6h\right) & \mbox{Te} \\ 
\mathbf{B}_{14} & = & \left(x_{3}-y_{3}\right) \, \mathbf{a}_{1} + x_{3} \, \mathbf{a}_{2} + \frac{3}{4} \, \mathbf{a}_{3} & = & \left(x_{3}-\frac{1}{2}y_{3}\right)a \, \mathbf{\hat{x}} + \frac{\sqrt{3}}{2}y_{3}a \, \mathbf{\hat{y}} + \frac{3}{4}c \, \mathbf{\hat{z}} & \left(6h\right) & \mbox{Te} \\ 
\end{longtabu}
\renewcommand{\arraystretch}{1.0}
\noindent \hrulefill
\\
\textbf{References:}
\vspace*{-0.25cm}
\begin{flushleft}
  - \bibentry{Klepp_Fe3Te3Tl_ActCrystallogr_1978}. \\
\end{flushleft}
\textbf{Found in:}
\vspace*{-0.25cm}
\begin{flushleft}
  - \bibentry{Villars_PearsonsCrystalData_2013}. \\
\end{flushleft}
\noindent \hrulefill
\\
\textbf{Geometry files:}
\\
\noindent  - CIF: pp. {\hyperref[A3B3C_hP14_176_h_h_d_cif]{\pageref{A3B3C_hP14_176_h_h_d_cif}}} \\
\noindent  - POSCAR: pp. {\hyperref[A3B3C_hP14_176_h_h_d_poscar]{\pageref{A3B3C_hP14_176_h_h_d_poscar}}} \\
\onecolumn
{\phantomsection\label{A3B_hP8_176_h_d}}
\subsection*{\huge \textbf{{\normalfont UCl$_{3}$ Structure: A3B\_hP8\_176\_h\_d}}}
\noindent \hrulefill
\vspace*{0.25cm}
\begin{figure}[htp]
  \centering
  \vspace{-1em}
  {\includegraphics[width=1\textwidth]{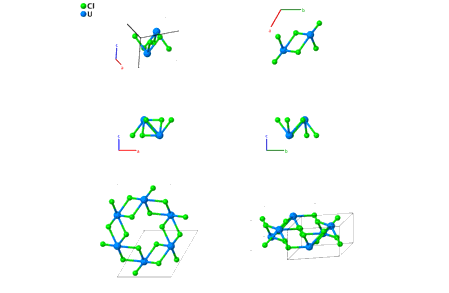}}
\end{figure}
\vspace*{-0.5cm}
\renewcommand{\arraystretch}{1.5}
\begin{equation*}
  \begin{array}{>{$\hspace{-0.15cm}}l<{$}>{$}p{0.5cm}<{$}>{$}p{18.5cm}<{$}}
    \mbox{\large \textbf{Prototype}} &\colon & \ce{UCl3} \\
    \mbox{\large \textbf{\AFLOW\ prototype label}} &\colon & \mbox{A3B\_hP8\_176\_h\_d} \\
    \mbox{\large \textbf{\textit{Strukturbericht} designation}} &\colon & \mbox{None} \\
    \mbox{\large \textbf{Pearson symbol}} &\colon & \mbox{hP8} \\
    \mbox{\large \textbf{Space group number}} &\colon & 176 \\
    \mbox{\large \textbf{Space group symbol}} &\colon & P6_{3}/m \\
    \mbox{\large \textbf{\AFLOW\ prototype command}} &\colon &  \texttt{aflow} \,  \, \texttt{-{}-proto=A3B\_hP8\_176\_h\_d } \, \newline \texttt{-{}-params=}{a,c/a,x_{2},y_{2} }
  \end{array}
\end{equation*}
\renewcommand{\arraystretch}{1.0}

\noindent \parbox{1 \linewidth}{
\noindent \hrulefill
\\
\textbf{Hexagonal primitive vectors:} \\
\vspace*{-0.25cm}
\begin{tabular}{cc}
  \begin{tabular}{c}
    \parbox{0.6 \linewidth}{
      \renewcommand{\arraystretch}{1.5}
      \begin{equation*}
        \centering
        \begin{array}{ccc}
              \mathbf{a}_1 & = & \frac12 \, a \, \mathbf{\hat{x}} - \frac{\sqrt3}2 \, a \, \mathbf{\hat{y}} \\
    \mathbf{a}_2 & = & \frac12 \, a \, \mathbf{\hat{x}} + \frac{\sqrt3}2 \, a \, \mathbf{\hat{y}} \\
    \mathbf{a}_3 & = & c \, \mathbf{\hat{z}} \\

        \end{array}
      \end{equation*}
    }
    \renewcommand{\arraystretch}{1.0}
  \end{tabular}
  \begin{tabular}{c}
    \includegraphics[width=0.3\linewidth]{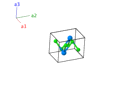} \\
  \end{tabular}
\end{tabular}

}
\vspace*{-0.25cm}

\noindent \hrulefill
\\
\textbf{Basis vectors:}
\vspace*{-0.25cm}
\renewcommand{\arraystretch}{1.5}
\begin{longtabu} to \textwidth{>{\centering $}X[-1,c,c]<{$}>{\centering $}X[-1,c,c]<{$}>{\centering $}X[-1,c,c]<{$}>{\centering $}X[-1,c,c]<{$}>{\centering $}X[-1,c,c]<{$}>{\centering $}X[-1,c,c]<{$}>{\centering $}X[-1,c,c]<{$}}
  & & \mbox{Lattice Coordinates} & & \mbox{Cartesian Coordinates} &\mbox{Wyckoff Position} & \mbox{Atom Type} \\  
  \mathbf{B}_{1} & = & \frac{2}{3} \, \mathbf{a}_{1} + \frac{1}{3} \, \mathbf{a}_{2} + \frac{1}{4} \, \mathbf{a}_{3} & = & \frac{1}{2}a \, \mathbf{\hat{x}}- \frac{1}{2\sqrt{3}}a  \, \mathbf{\hat{y}} + \frac{1}{4}c \, \mathbf{\hat{z}} & \left(2d\right) & \mbox{U} \\ 
\mathbf{B}_{2} & = & \frac{1}{3} \, \mathbf{a}_{1} + \frac{2}{3} \, \mathbf{a}_{2} + \frac{3}{4} \, \mathbf{a}_{3} & = & \frac{1}{2}a \, \mathbf{\hat{x}} + \frac{1}{2\sqrt{3}}a \, \mathbf{\hat{y}} + \frac{3}{4}c \, \mathbf{\hat{z}} & \left(2d\right) & \mbox{U} \\ 
\mathbf{B}_{3} & = & x_{2} \, \mathbf{a}_{1} + y_{2} \, \mathbf{a}_{2} + \frac{1}{4} \, \mathbf{a}_{3} & = & \frac{1}{2}\left(x_{2}+y_{2}\right)a \, \mathbf{\hat{x}} + \frac{\sqrt{3}}{2}\left(-x_{2}+y_{2}\right)a \, \mathbf{\hat{y}} + \frac{1}{4}c \, \mathbf{\hat{z}} & \left(6h\right) & \mbox{Cl} \\ 
\mathbf{B}_{4} & = & -y_{2} \, \mathbf{a}_{1} + \left(x_{2}-y_{2}\right) \, \mathbf{a}_{2} + \frac{1}{4} \, \mathbf{a}_{3} & = & \left(\frac{1}{2}x_{2}-y_{2}\right)a \, \mathbf{\hat{x}} + \frac{\sqrt{3}}{2}x_{2}a \, \mathbf{\hat{y}} + \frac{1}{4}c \, \mathbf{\hat{z}} & \left(6h\right) & \mbox{Cl} \\ 
\mathbf{B}_{5} & = & \left(-x_{2}+y_{2}\right) \, \mathbf{a}_{1}-x_{2} \, \mathbf{a}_{2} + \frac{1}{4} \, \mathbf{a}_{3} & = & \left(-x_{2}+\frac{1}{2}y_{2}\right)a \, \mathbf{\hat{x}}-\frac{\sqrt{3}}{2}y_{2}a \, \mathbf{\hat{y}} + \frac{1}{4}c \, \mathbf{\hat{z}} & \left(6h\right) & \mbox{Cl} \\ 
\mathbf{B}_{6} & = & -x_{2} \, \mathbf{a}_{1}-y_{2} \, \mathbf{a}_{2} + \frac{3}{4} \, \mathbf{a}_{3} & = & -\frac{1}{2}\left(x_{2}+y_{2}\right)a \, \mathbf{\hat{x}} + \frac{\sqrt{3}}{2}\left(x_{2}-y_{2}\right)a \, \mathbf{\hat{y}} + \frac{3}{4}c \, \mathbf{\hat{z}} & \left(6h\right) & \mbox{Cl} \\ 
\mathbf{B}_{7} & = & y_{2} \, \mathbf{a}_{1} + \left(-x_{2}+y_{2}\right) \, \mathbf{a}_{2} + \frac{3}{4} \, \mathbf{a}_{3} & = & \left(-\frac{1}{2}x_{2}+y_{2}\right)a \, \mathbf{\hat{x}}-\frac{\sqrt{3}}{2}x_{2}a \, \mathbf{\hat{y}} + \frac{3}{4}c \, \mathbf{\hat{z}} & \left(6h\right) & \mbox{Cl} \\ 
\mathbf{B}_{8} & = & \left(x_{2}-y_{2}\right) \, \mathbf{a}_{1} + x_{2} \, \mathbf{a}_{2} + \frac{3}{4} \, \mathbf{a}_{3} & = & \left(x_{2}-\frac{1}{2}y_{2}\right)a \, \mathbf{\hat{x}} + \frac{\sqrt{3}}{2}y_{2}a \, \mathbf{\hat{y}} + \frac{3}{4}c \, \mathbf{\hat{z}} & \left(6h\right) & \mbox{Cl} \\ 
\end{longtabu}
\renewcommand{\arraystretch}{1.0}
\noindent \hrulefill
\\
\textbf{References:}
\vspace*{-0.25cm}
\begin{flushleft}
  - \bibentry{Zachariasen_UCl3_ActCrystallog_1948}. \\
\end{flushleft}
\textbf{Found in:}
\vspace*{-0.25cm}
\begin{flushleft}
  - \bibentry{Villars_PearsonsCrystalData_2013}. \\
\end{flushleft}
\noindent \hrulefill
\\
\textbf{Geometry files:}
\\
\noindent  - CIF: pp. {\hyperref[A3B_hP8_176_h_d_cif]{\pageref{A3B_hP8_176_h_d_cif}}} \\
\noindent  - POSCAR: pp. {\hyperref[A3B_hP8_176_h_d_poscar]{\pageref{A3B_hP8_176_h_d_poscar}}} \\
\onecolumn
{\phantomsection\label{A2B_hP36_177_j2lm_n}}
\subsection*{\huge \textbf{{\normalfont SiO$_{2}$ Structure: A2B\_hP36\_177\_j2lm\_n}}}
\noindent \hrulefill
\vspace*{0.25cm}
\begin{figure}[htp]
  \centering
  \vspace{-1em}
  {\includegraphics[width=1\textwidth]{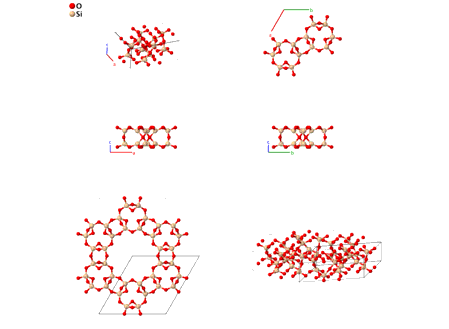}}
\end{figure}
\vspace*{-0.5cm}
\renewcommand{\arraystretch}{1.5}
\begin{equation*}
  \begin{array}{>{$\hspace{-0.15cm}}l<{$}>{$}p{0.5cm}<{$}>{$}p{18.5cm}<{$}}
    \mbox{\large \textbf{Prototype}} &\colon & \ce{SiO2} \\
    \mbox{\large \textbf{\AFLOW\ prototype label}} &\colon & \mbox{A2B\_hP36\_177\_j2lm\_n} \\
    \mbox{\large \textbf{\textit{Strukturbericht} designation}} &\colon & \mbox{None} \\
    \mbox{\large \textbf{Pearson symbol}} &\colon & \mbox{hP36} \\
    \mbox{\large \textbf{Space group number}} &\colon & 177 \\
    \mbox{\large \textbf{Space group symbol}} &\colon & P622 \\
    \mbox{\large \textbf{\AFLOW\ prototype command}} &\colon &  \texttt{aflow} \,  \, \texttt{-{}-proto=A2B\_hP36\_177\_j2lm\_n } \, \newline \texttt{-{}-params=}{a,c/a,x_{1},x_{2},x_{3},x_{4},x_{5},y_{5},z_{5} }
  \end{array}
\end{equation*}
\renewcommand{\arraystretch}{1.0}

\noindent \parbox{1 \linewidth}{
\noindent \hrulefill
\\
\textbf{Hexagonal primitive vectors:} \\
\vspace*{-0.25cm}
\begin{tabular}{cc}
  \begin{tabular}{c}
    \parbox{0.6 \linewidth}{
      \renewcommand{\arraystretch}{1.5}
      \begin{equation*}
        \centering
        \begin{array}{ccc}
              \mathbf{a}_1 & = & \frac12 \, a \, \mathbf{\hat{x}} - \frac{\sqrt3}2 \, a \, \mathbf{\hat{y}} \\
    \mathbf{a}_2 & = & \frac12 \, a \, \mathbf{\hat{x}} + \frac{\sqrt3}2 \, a \, \mathbf{\hat{y}} \\
    \mathbf{a}_3 & = & c \, \mathbf{\hat{z}} \\

        \end{array}
      \end{equation*}
    }
    \renewcommand{\arraystretch}{1.0}
  \end{tabular}
  \begin{tabular}{c}
    \includegraphics[width=0.3\linewidth]{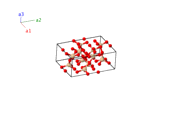} \\
  \end{tabular}
\end{tabular}

}
\vspace*{-0.25cm}

\noindent \hrulefill
\\
\textbf{Basis vectors:}
\vspace*{-0.25cm}
\renewcommand{\arraystretch}{1.5}
\begin{longtabu} to \textwidth{>{\centering $}X[-1,c,c]<{$}>{\centering $}X[-1,c,c]<{$}>{\centering $}X[-1,c,c]<{$}>{\centering $}X[-1,c,c]<{$}>{\centering $}X[-1,c,c]<{$}>{\centering $}X[-1,c,c]<{$}>{\centering $}X[-1,c,c]<{$}}
  & & \mbox{Lattice Coordinates} & & \mbox{Cartesian Coordinates} &\mbox{Wyckoff Position} & \mbox{Atom Type} \\  
  \mathbf{B}_{1} & = & x_{1} \, \mathbf{a}_{1} & = & \frac{1}{2}x_{1}a \, \mathbf{\hat{x}}-\frac{\sqrt{3}}{2}x_{1}a \, \mathbf{\hat{y}} & \left(6j\right) & \mbox{O I} \\ 
\mathbf{B}_{2} & = & x_{1} \, \mathbf{a}_{2} & = & \frac{1}{2}x_{1}a \, \mathbf{\hat{x}} + \frac{\sqrt{3}}{2}x_{1}a \, \mathbf{\hat{y}} & \left(6j\right) & \mbox{O I} \\ 
\mathbf{B}_{3} & = & -x_{1} \, \mathbf{a}_{1}-x_{1} \, \mathbf{a}_{2} & = & -x_{1}a \, \mathbf{\hat{x}} & \left(6j\right) & \mbox{O I} \\ 
\mathbf{B}_{4} & = & -x_{1} \, \mathbf{a}_{1} & = & -\frac{1}{2}x_{1}a \, \mathbf{\hat{x}} + \frac{\sqrt{3}}{2}x_{1}a \, \mathbf{\hat{y}} & \left(6j\right) & \mbox{O I} \\ 
\mathbf{B}_{5} & = & -x_{1} \, \mathbf{a}_{2} & = & -\frac{1}{2}x_{1}a \, \mathbf{\hat{x}}-\frac{\sqrt{3}}{2}x_{1}a \, \mathbf{\hat{y}} & \left(6j\right) & \mbox{O I} \\ 
\mathbf{B}_{6} & = & x_{1} \, \mathbf{a}_{1} + x_{1} \, \mathbf{a}_{2} & = & x_{1}a \, \mathbf{\hat{x}} & \left(6j\right) & \mbox{O I} \\ 
\mathbf{B}_{7} & = & x_{2} \, \mathbf{a}_{1}-x_{2} \, \mathbf{a}_{2} & = & -\sqrt{3}x_{2}a \, \mathbf{\hat{y}} & \left(6l\right) & \mbox{O II} \\ 
\mathbf{B}_{8} & = & x_{2} \, \mathbf{a}_{1} + 2x_{2} \, \mathbf{a}_{2} & = & \frac{3}{2}x_{2}a \, \mathbf{\hat{x}} + \frac{\sqrt{3}}{2}x_{2}a \, \mathbf{\hat{y}} & \left(6l\right) & \mbox{O II} \\ 
\mathbf{B}_{9} & = & -2x_{2} \, \mathbf{a}_{1}-x_{2} \, \mathbf{a}_{2} & = & -\frac{3}{2}x_{2}a \, \mathbf{\hat{x}} + \frac{\sqrt{3}}{2}x_{2}a \, \mathbf{\hat{y}} & \left(6l\right) & \mbox{O II} \\ 
\mathbf{B}_{10} & = & -x_{2} \, \mathbf{a}_{1} + x_{2} \, \mathbf{a}_{2} & = & \sqrt{3}x_{2}a \, \mathbf{\hat{y}} & \left(6l\right) & \mbox{O II} \\ 
\mathbf{B}_{11} & = & -x_{2} \, \mathbf{a}_{1}-2x_{2} \, \mathbf{a}_{2} & = & -\frac{3}{2}x_{2}a \, \mathbf{\hat{x}}-\frac{\sqrt{3}}{2}x_{2}a \, \mathbf{\hat{y}} & \left(6l\right) & \mbox{O II} \\ 
\mathbf{B}_{12} & = & 2x_{2} \, \mathbf{a}_{1} + x_{2} \, \mathbf{a}_{2} & = & \frac{3}{2}x_{2}a \, \mathbf{\hat{x}}-\frac{\sqrt{3}}{2}x_{2}a \, \mathbf{\hat{y}} & \left(6l\right) & \mbox{O II} \\ 
\mathbf{B}_{13} & = & x_{3} \, \mathbf{a}_{1}-x_{3} \, \mathbf{a}_{2} & = & -\sqrt{3}x_{3}a \, \mathbf{\hat{y}} & \left(6l\right) & \mbox{O III} \\ 
\mathbf{B}_{14} & = & x_{3} \, \mathbf{a}_{1} + 2x_{3} \, \mathbf{a}_{2} & = & \frac{3}{2}x_{3}a \, \mathbf{\hat{x}} + \frac{\sqrt{3}}{2}x_{3}a \, \mathbf{\hat{y}} & \left(6l\right) & \mbox{O III} \\ 
\mathbf{B}_{15} & = & -2x_{3} \, \mathbf{a}_{1}-x_{3} \, \mathbf{a}_{2} & = & -\frac{3}{2}x_{3}a \, \mathbf{\hat{x}} + \frac{\sqrt{3}}{2}x_{3}a \, \mathbf{\hat{y}} & \left(6l\right) & \mbox{O III} \\ 
\mathbf{B}_{16} & = & -x_{3} \, \mathbf{a}_{1} + x_{3} \, \mathbf{a}_{2} & = & \sqrt{3}x_{3}a \, \mathbf{\hat{y}} & \left(6l\right) & \mbox{O III} \\ 
\mathbf{B}_{17} & = & -x_{3} \, \mathbf{a}_{1}-2x_{3} \, \mathbf{a}_{2} & = & -\frac{3}{2}x_{3}a \, \mathbf{\hat{x}}-\frac{\sqrt{3}}{2}x_{3}a \, \mathbf{\hat{y}} & \left(6l\right) & \mbox{O III} \\ 
\mathbf{B}_{18} & = & 2x_{3} \, \mathbf{a}_{1} + x_{3} \, \mathbf{a}_{2} & = & \frac{3}{2}x_{3}a \, \mathbf{\hat{x}}-\frac{\sqrt{3}}{2}x_{3}a \, \mathbf{\hat{y}} & \left(6l\right) & \mbox{O III} \\ 
\mathbf{B}_{19} & = & x_{4} \, \mathbf{a}_{1}-x_{4} \, \mathbf{a}_{2} + \frac{1}{2} \, \mathbf{a}_{3} & = & -\sqrt{3}x_{4}a \, \mathbf{\hat{y}} + \frac{1}{2}c \, \mathbf{\hat{z}} & \left(6m\right) & \mbox{O IV} \\ 
\mathbf{B}_{20} & = & x_{4} \, \mathbf{a}_{1} + 2x_{4} \, \mathbf{a}_{2} + \frac{1}{2} \, \mathbf{a}_{3} & = & \frac{3}{2}x_{4}a \, \mathbf{\hat{x}} + \frac{\sqrt{3}}{2}x_{4}a \, \mathbf{\hat{y}} + \frac{1}{2}c \, \mathbf{\hat{z}} & \left(6m\right) & \mbox{O IV} \\ 
\mathbf{B}_{21} & = & -2x_{4} \, \mathbf{a}_{1}-x_{4} \, \mathbf{a}_{2} + \frac{1}{2} \, \mathbf{a}_{3} & = & -\frac{3}{2}x_{4}a \, \mathbf{\hat{x}} + \frac{\sqrt{3}}{2}x_{4}a \, \mathbf{\hat{y}} + \frac{1}{2}c \, \mathbf{\hat{z}} & \left(6m\right) & \mbox{O IV} \\ 
\mathbf{B}_{22} & = & -x_{4} \, \mathbf{a}_{1} + x_{4} \, \mathbf{a}_{2} + \frac{1}{2} \, \mathbf{a}_{3} & = & \sqrt{3}x_{4}a \, \mathbf{\hat{y}} + \frac{1}{2}c \, \mathbf{\hat{z}} & \left(6m\right) & \mbox{O IV} \\ 
\mathbf{B}_{23} & = & -x_{4} \, \mathbf{a}_{1}-2x_{4} \, \mathbf{a}_{2} + \frac{1}{2} \, \mathbf{a}_{3} & = & -\frac{3}{2}x_{4}a \, \mathbf{\hat{x}}-\frac{\sqrt{3}}{2}x_{4}a \, \mathbf{\hat{y}} + \frac{1}{2}c \, \mathbf{\hat{z}} & \left(6m\right) & \mbox{O IV} \\ 
\mathbf{B}_{24} & = & 2x_{4} \, \mathbf{a}_{1} + x_{4} \, \mathbf{a}_{2} + \frac{1}{2} \, \mathbf{a}_{3} & = & \frac{3}{2}x_{4}a \, \mathbf{\hat{x}}-\frac{\sqrt{3}}{2}x_{4}a \, \mathbf{\hat{y}} + \frac{1}{2}c \, \mathbf{\hat{z}} & \left(6m\right) & \mbox{O IV} \\ 
\mathbf{B}_{25} & = & x_{5} \, \mathbf{a}_{1} + y_{5} \, \mathbf{a}_{2} + z_{5} \, \mathbf{a}_{3} & = & \frac{1}{2}\left(x_{5}+y_{5}\right)a \, \mathbf{\hat{x}} + \frac{\sqrt{3}}{2}\left(-x_{5}+y_{5}\right)a \, \mathbf{\hat{y}} + z_{5}c \, \mathbf{\hat{z}} & \left(12n\right) & \mbox{Si} \\ 
\mathbf{B}_{26} & = & -y_{5} \, \mathbf{a}_{1} + \left(x_{5}-y_{5}\right) \, \mathbf{a}_{2} + z_{5} \, \mathbf{a}_{3} & = & \left(\frac{1}{2}x_{5}-y_{5}\right)a \, \mathbf{\hat{x}} + \frac{\sqrt{3}}{2}x_{5}a \, \mathbf{\hat{y}} + z_{5}c \, \mathbf{\hat{z}} & \left(12n\right) & \mbox{Si} \\ 
\mathbf{B}_{27} & = & \left(-x_{5}+y_{5}\right) \, \mathbf{a}_{1}-x_{5} \, \mathbf{a}_{2} + z_{5} \, \mathbf{a}_{3} & = & \left(-x_{5}+\frac{1}{2}y_{5}\right)a \, \mathbf{\hat{x}}-\frac{\sqrt{3}}{2}y_{5}a \, \mathbf{\hat{y}} + z_{5}c \, \mathbf{\hat{z}} & \left(12n\right) & \mbox{Si} \\ 
\mathbf{B}_{28} & = & -x_{5} \, \mathbf{a}_{1}-y_{5} \, \mathbf{a}_{2} + z_{5} \, \mathbf{a}_{3} & = & -\frac{1}{2}\left(x_{5}+y_{5}\right)a \, \mathbf{\hat{x}} + \frac{\sqrt{3}}{2}\left(x_{5}-y_{5}\right)a \, \mathbf{\hat{y}} + z_{5}c \, \mathbf{\hat{z}} & \left(12n\right) & \mbox{Si} \\ 
\mathbf{B}_{29} & = & y_{5} \, \mathbf{a}_{1} + \left(-x_{5}+y_{5}\right) \, \mathbf{a}_{2} + z_{5} \, \mathbf{a}_{3} & = & \left(-\frac{1}{2}x_{5}+y_{5}\right)a \, \mathbf{\hat{x}}-\frac{\sqrt{3}}{2}x_{5}a \, \mathbf{\hat{y}} + z_{5}c \, \mathbf{\hat{z}} & \left(12n\right) & \mbox{Si} \\ 
\mathbf{B}_{30} & = & \left(x_{5}-y_{5}\right) \, \mathbf{a}_{1} + x_{5} \, \mathbf{a}_{2} + z_{5} \, \mathbf{a}_{3} & = & \left(x_{5}-\frac{1}{2}y_{5}\right)a \, \mathbf{\hat{x}} + \frac{\sqrt{3}}{2}y_{5}a \, \mathbf{\hat{y}} + z_{5}c \, \mathbf{\hat{z}} & \left(12n\right) & \mbox{Si} \\ 
\mathbf{B}_{31} & = & y_{5} \, \mathbf{a}_{1} + x_{5} \, \mathbf{a}_{2}-z_{5} \, \mathbf{a}_{3} & = & \frac{1}{2}\left(x_{5}+y_{5}\right)a \, \mathbf{\hat{x}} + \frac{\sqrt{3}}{2}\left(x_{5}-y_{5}\right)a \, \mathbf{\hat{y}}-z_{5}c \, \mathbf{\hat{z}} & \left(12n\right) & \mbox{Si} \\ 
\mathbf{B}_{32} & = & \left(x_{5}-y_{5}\right) \, \mathbf{a}_{1}-y_{5} \, \mathbf{a}_{2}-z_{5} \, \mathbf{a}_{3} & = & \left(\frac{1}{2}x_{5}-y_{5}\right)a \, \mathbf{\hat{x}}-\frac{\sqrt{3}}{2}x_{5}a \, \mathbf{\hat{y}}-z_{5}c \, \mathbf{\hat{z}} & \left(12n\right) & \mbox{Si} \\ 
\mathbf{B}_{33} & = & -x_{5} \, \mathbf{a}_{1} + \left(-x_{5}+y_{5}\right) \, \mathbf{a}_{2}-z_{5} \, \mathbf{a}_{3} & = & \left(-x_{5}+\frac{1}{2}y_{5}\right)a \, \mathbf{\hat{x}} + \frac{\sqrt{3}}{2}y_{5}a \, \mathbf{\hat{y}}-z_{5}c \, \mathbf{\hat{z}} & \left(12n\right) & \mbox{Si} \\ 
\mathbf{B}_{34} & = & -y_{5} \, \mathbf{a}_{1}-x_{5} \, \mathbf{a}_{2}-z_{5} \, \mathbf{a}_{3} & = & -\frac{1}{2}\left(x_{5}+y_{5}\right)a \, \mathbf{\hat{x}} + \frac{\sqrt{3}}{2}\left(-x_{5}+y_{5}\right)a \, \mathbf{\hat{y}}-z_{5}c \, \mathbf{\hat{z}} & \left(12n\right) & \mbox{Si} \\ 
\mathbf{B}_{35} & = & \left(-x_{5}+y_{5}\right) \, \mathbf{a}_{1} + y_{5} \, \mathbf{a}_{2}-z_{5} \, \mathbf{a}_{3} & = & \left(-\frac{1}{2}x_{5}+y_{5}\right)a \, \mathbf{\hat{x}} + \frac{\sqrt{3}}{2}x_{5}a \, \mathbf{\hat{y}}-z_{5}c \, \mathbf{\hat{z}} & \left(12n\right) & \mbox{Si} \\ 
\mathbf{B}_{36} & = & x_{5} \, \mathbf{a}_{1} + \left(x_{5}-y_{5}\right) \, \mathbf{a}_{2}-z_{5} \, \mathbf{a}_{3} & = & \left(x_{5}-\frac{1}{2}y_{5}\right)a \, \mathbf{\hat{x}}-\frac{\sqrt{3}}{2}y_{5}a \, \mathbf{\hat{y}}-z_{5}c \, \mathbf{\hat{z}} & \left(12n\right) & \mbox{Si} \\ 
\end{longtabu}
\renewcommand{\arraystretch}{1.0}
\noindent \hrulefill
\\
\textbf{References:}
\vspace*{-0.25cm}
\begin{flushleft}
  - \bibentry{Foster_SiO2_jacs_126_2004}. \\
\end{flushleft}
\textbf{Found in:}
\vspace*{-0.25cm}
\begin{flushleft}
  - \bibentry{icsd:ICSD_170519}. \\
\end{flushleft}
\noindent \hrulefill
\\
\textbf{Geometry files:}
\\
\noindent  - CIF: pp. {\hyperref[A2B_hP36_177_j2lm_n_cif]{\pageref{A2B_hP36_177_j2lm_n_cif}}} \\
\noindent  - POSCAR: pp. {\hyperref[A2B_hP36_177_j2lm_n_poscar]{\pageref{A2B_hP36_177_j2lm_n_poscar}}} \\
\onecolumn
{\phantomsection\label{AB3_hP24_178_b_ac}}
\subsection*{\huge \textbf{{\normalfont AuF$_{3}$ Structure: AB3\_hP24\_178\_b\_ac}}}
\noindent \hrulefill
\vspace*{0.25cm}
\begin{figure}[htp]
  \centering
  \vspace{-1em}
  {\includegraphics[width=1\textwidth]{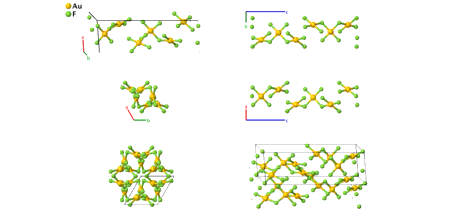}}
\end{figure}
\vspace*{-0.5cm}
\renewcommand{\arraystretch}{1.5}
\begin{equation*}
  \begin{array}{>{$\hspace{-0.15cm}}l<{$}>{$}p{0.5cm}<{$}>{$}p{18.5cm}<{$}}
    \mbox{\large \textbf{Prototype}} &\colon & \ce{AuF3} \\
    \mbox{\large \textbf{\AFLOW\ prototype label}} &\colon & \mbox{AB3\_hP24\_178\_b\_ac} \\
    \mbox{\large \textbf{\textit{Strukturbericht} designation}} &\colon & \mbox{None} \\
    \mbox{\large \textbf{Pearson symbol}} &\colon & \mbox{hP24} \\
    \mbox{\large \textbf{Space group number}} &\colon & 178 \\
    \mbox{\large \textbf{Space group symbol}} &\colon & P6_{1}22 \\
    \mbox{\large \textbf{\AFLOW\ prototype command}} &\colon &  \texttt{aflow} \,  \, \texttt{-{}-proto=AB3\_hP24\_178\_b\_ac } \, \newline \texttt{-{}-params=}{a,c/a,x_{1},x_{2},x_{3},y_{3},z_{3} }
  \end{array}
\end{equation*}
\renewcommand{\arraystretch}{1.0}

\noindent \parbox{1 \linewidth}{
\noindent \hrulefill
\\
\textbf{Hexagonal primitive vectors:} \\
\vspace*{-0.25cm}
\begin{tabular}{cc}
  \begin{tabular}{c}
    \parbox{0.6 \linewidth}{
      \renewcommand{\arraystretch}{1.5}
      \begin{equation*}
        \centering
        \begin{array}{ccc}
              \mathbf{a}_1 & = & \frac12 \, a \, \mathbf{\hat{x}} - \frac{\sqrt3}2 \, a \, \mathbf{\hat{y}} \\
    \mathbf{a}_2 & = & \frac12 \, a \, \mathbf{\hat{x}} + \frac{\sqrt3}2 \, a \, \mathbf{\hat{y}} \\
    \mathbf{a}_3 & = & c \, \mathbf{\hat{z}} \\

        \end{array}
      \end{equation*}
    }
    \renewcommand{\arraystretch}{1.0}
  \end{tabular}
  \begin{tabular}{c}
    \includegraphics[width=0.3\linewidth]{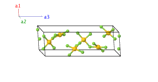} \\
  \end{tabular}
\end{tabular}

}
\vspace*{-0.25cm}

\noindent \hrulefill
\\
\textbf{Basis vectors:}
\vspace*{-0.25cm}
\renewcommand{\arraystretch}{1.5}
\begin{longtabu} to \textwidth{>{\centering $}X[-1,c,c]<{$}>{\centering $}X[-1,c,c]<{$}>{\centering $}X[-1,c,c]<{$}>{\centering $}X[-1,c,c]<{$}>{\centering $}X[-1,c,c]<{$}>{\centering $}X[-1,c,c]<{$}>{\centering $}X[-1,c,c]<{$}}
  & & \mbox{Lattice Coordinates} & & \mbox{Cartesian Coordinates} &\mbox{Wyckoff Position} & \mbox{Atom Type} \\  
  \mathbf{B}_{1} & = & x_{1} \, \mathbf{a}_{1} & = & \frac{1}{2}x_{1}a \, \mathbf{\hat{x}}-\frac{\sqrt{3}}{2}x_{1}a \, \mathbf{\hat{y}} & \left(6a\right) & \mbox{F I} \\ 
\mathbf{B}_{2} & = & x_{1} \, \mathbf{a}_{2} + \frac{1}{3} \, \mathbf{a}_{3} & = & \frac{1}{2}x_{1}a \, \mathbf{\hat{x}} + \frac{\sqrt{3}}{2}x_{1}a \, \mathbf{\hat{y}} + \frac{1}{3}c \, \mathbf{\hat{z}} & \left(6a\right) & \mbox{F I} \\ 
\mathbf{B}_{3} & = & -x_{1} \, \mathbf{a}_{1}-x_{1} \, \mathbf{a}_{2} + \frac{2}{3} \, \mathbf{a}_{3} & = & -x_{1}a \, \mathbf{\hat{x}} + \frac{2}{3}c \, \mathbf{\hat{z}} & \left(6a\right) & \mbox{F I} \\ 
\mathbf{B}_{4} & = & -x_{1} \, \mathbf{a}_{1} + \frac{1}{2} \, \mathbf{a}_{3} & = & -\frac{1}{2}x_{1}a \, \mathbf{\hat{x}} + \frac{\sqrt{3}}{2}x_{1}a \, \mathbf{\hat{y}} + \frac{1}{2}c \, \mathbf{\hat{z}} & \left(6a\right) & \mbox{F I} \\ 
\mathbf{B}_{5} & = & -x_{1} \, \mathbf{a}_{2} + \frac{5}{6} \, \mathbf{a}_{3} & = & -\frac{1}{2}x_{1}a \, \mathbf{\hat{x}}-\frac{\sqrt{3}}{2}x_{1}a \, \mathbf{\hat{y}} + \frac{5}{6}c \, \mathbf{\hat{z}} & \left(6a\right) & \mbox{F I} \\ 
\mathbf{B}_{6} & = & x_{1} \, \mathbf{a}_{1} + x_{1} \, \mathbf{a}_{2} + \frac{1}{6} \, \mathbf{a}_{3} & = & x_{1}a \, \mathbf{\hat{x}} + \frac{1}{6}c \, \mathbf{\hat{z}} & \left(6a\right) & \mbox{F I} \\ 
\mathbf{B}_{7} & = & x_{2} \, \mathbf{a}_{1} + 2x_{2} \, \mathbf{a}_{2} + \frac{1}{4} \, \mathbf{a}_{3} & = & \frac{3}{2}x_{2}a \, \mathbf{\hat{x}} + \frac{\sqrt{3}}{2}x_{2}a \, \mathbf{\hat{y}} + \frac{1}{4}c \, \mathbf{\hat{z}} & \left(6b\right) & \mbox{Au} \\ 
\mathbf{B}_{8} & = & -2x_{2} \, \mathbf{a}_{1}-x_{2} \, \mathbf{a}_{2} + \frac{7}{12} \, \mathbf{a}_{3} & = & -\frac{3}{2}x_{2}a \, \mathbf{\hat{x}} + \frac{\sqrt{3}}{2}x_{2}a \, \mathbf{\hat{y}} + \frac{7}{12}c \, \mathbf{\hat{z}} & \left(6b\right) & \mbox{Au} \\ 
\mathbf{B}_{9} & = & x_{2} \, \mathbf{a}_{1}-x_{2} \, \mathbf{a}_{2} + \frac{11}{12} \, \mathbf{a}_{3} & = & -\sqrt{3}x_{2}a \, \mathbf{\hat{y}} + \frac{11}{12}c \, \mathbf{\hat{z}} & \left(6b\right) & \mbox{Au} \\ 
\mathbf{B}_{10} & = & -x_{2} \, \mathbf{a}_{1}-2x_{2} \, \mathbf{a}_{2} + \frac{3}{4} \, \mathbf{a}_{3} & = & -\frac{3}{2}x_{2}a \, \mathbf{\hat{x}}-\frac{\sqrt{3}}{2}x_{2}a \, \mathbf{\hat{y}} + \frac{3}{4}c \, \mathbf{\hat{z}} & \left(6b\right) & \mbox{Au} \\ 
\mathbf{B}_{11} & = & 2x_{2} \, \mathbf{a}_{1} + x_{2} \, \mathbf{a}_{2} + \frac{1}{12} \, \mathbf{a}_{3} & = & \frac{3}{2}x_{2}a \, \mathbf{\hat{x}}-\frac{\sqrt{3}}{2}x_{2}a \, \mathbf{\hat{y}} + \frac{1}{12}c \, \mathbf{\hat{z}} & \left(6b\right) & \mbox{Au} \\ 
\mathbf{B}_{12} & = & -x_{2} \, \mathbf{a}_{1} + x_{2} \, \mathbf{a}_{2} + \frac{5}{12} \, \mathbf{a}_{3} & = & \sqrt{3}x_{2}a \, \mathbf{\hat{y}} + \frac{5}{12}c \, \mathbf{\hat{z}} & \left(6b\right) & \mbox{Au} \\ 
\mathbf{B}_{13} & = & x_{3} \, \mathbf{a}_{1} + y_{3} \, \mathbf{a}_{2} + z_{3} \, \mathbf{a}_{3} & = & \frac{1}{2}\left(x_{3}+y_{3}\right)a \, \mathbf{\hat{x}} + \frac{\sqrt{3}}{2}\left(-x_{3}+y_{3}\right)a \, \mathbf{\hat{y}} + z_{3}c \, \mathbf{\hat{z}} & \left(12c\right) & \mbox{F II} \\ 
\mathbf{B}_{14} & = & -y_{3} \, \mathbf{a}_{1} + \left(x_{3}-y_{3}\right) \, \mathbf{a}_{2} + \left(\frac{1}{3} +z_{3}\right) \, \mathbf{a}_{3} & = & \left(\frac{1}{2}x_{3}-y_{3}\right)a \, \mathbf{\hat{x}} + \frac{\sqrt{3}}{2}x_{3}a \, \mathbf{\hat{y}} + \left(\frac{1}{3} +z_{3}\right)c \, \mathbf{\hat{z}} & \left(12c\right) & \mbox{F II} \\ 
\mathbf{B}_{15} & = & \left(-x_{3}+y_{3}\right) \, \mathbf{a}_{1}-x_{3} \, \mathbf{a}_{2} + \left(\frac{2}{3} +z_{3}\right) \, \mathbf{a}_{3} & = & \left(-x_{3}+\frac{1}{2}y_{3}\right)a \, \mathbf{\hat{x}}-\frac{\sqrt{3}}{2}y_{3}a \, \mathbf{\hat{y}} + \left(\frac{2}{3} +z_{3}\right)c \, \mathbf{\hat{z}} & \left(12c\right) & \mbox{F II} \\ 
\mathbf{B}_{16} & = & -x_{3} \, \mathbf{a}_{1}-y_{3} \, \mathbf{a}_{2} + \left(\frac{1}{2} +z_{3}\right) \, \mathbf{a}_{3} & = & -\frac{1}{2}\left(x_{3}+y_{3}\right)a \, \mathbf{\hat{x}} + \frac{\sqrt{3}}{2}\left(x_{3}-y_{3}\right)a \, \mathbf{\hat{y}} + \left(\frac{1}{2} +z_{3}\right)c \, \mathbf{\hat{z}} & \left(12c\right) & \mbox{F II} \\ 
\mathbf{B}_{17} & = & y_{3} \, \mathbf{a}_{1} + \left(-x_{3}+y_{3}\right) \, \mathbf{a}_{2} + \left(\frac{5}{6} +z_{3}\right) \, \mathbf{a}_{3} & = & \left(-\frac{1}{2}x_{3}+y_{3}\right)a \, \mathbf{\hat{x}}-\frac{\sqrt{3}}{2}x_{3}a \, \mathbf{\hat{y}} + \left(\frac{5}{6} +z_{3}\right)c \, \mathbf{\hat{z}} & \left(12c\right) & \mbox{F II} \\ 
\mathbf{B}_{18} & = & \left(x_{3}-y_{3}\right) \, \mathbf{a}_{1} + x_{3} \, \mathbf{a}_{2} + \left(\frac{1}{6} +z_{3}\right) \, \mathbf{a}_{3} & = & \left(x_{3}-\frac{1}{2}y_{3}\right)a \, \mathbf{\hat{x}} + \frac{\sqrt{3}}{2}y_{3}a \, \mathbf{\hat{y}} + \left(\frac{1}{6} +z_{3}\right)c \, \mathbf{\hat{z}} & \left(12c\right) & \mbox{F II} \\ 
\mathbf{B}_{19} & = & y_{3} \, \mathbf{a}_{1} + x_{3} \, \mathbf{a}_{2} + \left(\frac{1}{3} - z_{3}\right) \, \mathbf{a}_{3} & = & \frac{1}{2}\left(x_{3}+y_{3}\right)a \, \mathbf{\hat{x}} + \frac{\sqrt{3}}{2}\left(x_{3}-y_{3}\right)a \, \mathbf{\hat{y}} + \left(\frac{1}{3} - z_{3}\right)c \, \mathbf{\hat{z}} & \left(12c\right) & \mbox{F II} \\ 
\mathbf{B}_{20} & = & \left(x_{3}-y_{3}\right) \, \mathbf{a}_{1}-y_{3} \, \mathbf{a}_{2}-z_{3} \, \mathbf{a}_{3} & = & \left(\frac{1}{2}x_{3}-y_{3}\right)a \, \mathbf{\hat{x}}-\frac{\sqrt{3}}{2}x_{3}a \, \mathbf{\hat{y}}-z_{3}c \, \mathbf{\hat{z}} & \left(12c\right) & \mbox{F II} \\ 
\mathbf{B}_{21} & = & -x_{3} \, \mathbf{a}_{1} + \left(-x_{3}+y_{3}\right) \, \mathbf{a}_{2} + \left(\frac{2}{3} - z_{3}\right) \, \mathbf{a}_{3} & = & \left(-x_{3}+\frac{1}{2}y_{3}\right)a \, \mathbf{\hat{x}} + \frac{\sqrt{3}}{2}y_{3}a \, \mathbf{\hat{y}} + \left(\frac{2}{3} - z_{3}\right)c \, \mathbf{\hat{z}} & \left(12c\right) & \mbox{F II} \\ 
\mathbf{B}_{22} & = & -y_{3} \, \mathbf{a}_{1}-x_{3} \, \mathbf{a}_{2} + \left(\frac{5}{6} - z_{3}\right) \, \mathbf{a}_{3} & = & -\frac{1}{2}\left(x_{3}+y_{3}\right)a \, \mathbf{\hat{x}} + \frac{\sqrt{3}}{2}\left(-x_{3}+y_{3}\right)a \, \mathbf{\hat{y}} + \left(\frac{5}{6} - z_{3}\right)c \, \mathbf{\hat{z}} & \left(12c\right) & \mbox{F II} \\ 
\mathbf{B}_{23} & = & \left(-x_{3}+y_{3}\right) \, \mathbf{a}_{1} + y_{3} \, \mathbf{a}_{2} + \left(\frac{1}{2} - z_{3}\right) \, \mathbf{a}_{3} & = & \left(-\frac{1}{2}x_{3}+y_{3}\right)a \, \mathbf{\hat{x}} + \frac{\sqrt{3}}{2}x_{3}a \, \mathbf{\hat{y}} + \left(\frac{1}{2} - z_{3}\right)c \, \mathbf{\hat{z}} & \left(12c\right) & \mbox{F II} \\ 
\mathbf{B}_{24} & = & x_{3} \, \mathbf{a}_{1} + \left(x_{3}-y_{3}\right) \, \mathbf{a}_{2} + \left(\frac{1}{6} - z_{3}\right) \, \mathbf{a}_{3} & = & \left(x_{3}-\frac{1}{2}y_{3}\right)a \, \mathbf{\hat{x}}-\frac{\sqrt{3}}{2}y_{3}a \, \mathbf{\hat{y}} + \left(\frac{1}{6} - z_{3}\right)c \, \mathbf{\hat{z}} & \left(12c\right) & \mbox{F II} \\ 
\end{longtabu}
\renewcommand{\arraystretch}{1.0}
\noindent \hrulefill
\\
\textbf{References:}
\vspace*{-0.25cm}
\begin{flushleft}
  - \bibentry{Asprey_AuF3_InorgChem_1964}. \\
\end{flushleft}
\textbf{Found in:}
\vspace*{-0.25cm}
\begin{flushleft}
  - \bibentry{Villars_PearsonsCrystalData_2013}. \\
\end{flushleft}
\noindent \hrulefill
\\
\textbf{Geometry files:}
\\
\noindent  - CIF: pp. {\hyperref[AB3_hP24_178_b_ac_cif]{\pageref{AB3_hP24_178_b_ac_cif}}} \\
\noindent  - POSCAR: pp. {\hyperref[AB3_hP24_178_b_ac_poscar]{\pageref{AB3_hP24_178_b_ac_poscar}}} \\
\onecolumn
{\phantomsection\label{A_hP6_178_a}}
\subsection*{\huge \textbf{{\normalfont Sc-V (High-pressure) Structure: A\_hP6\_178\_a}}}
\noindent \hrulefill
\vspace*{0.25cm}
\begin{figure}[htp]
  \centering
  \vspace{-1em}
  {\includegraphics[width=1\textwidth]{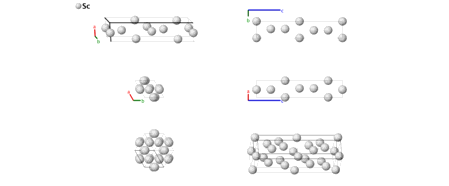}}
\end{figure}
\vspace*{-0.5cm}
\renewcommand{\arraystretch}{1.5}
\begin{equation*}
  \begin{array}{>{$\hspace{-0.15cm}}l<{$}>{$}p{0.5cm}<{$}>{$}p{18.5cm}<{$}}
    \mbox{\large \textbf{Prototype}} &\colon & \ce{Sc} \\
    \mbox{\large \textbf{\AFLOW\ prototype label}} &\colon & \mbox{A\_hP6\_178\_a} \\
    \mbox{\large \textbf{\textit{Strukturbericht} designation}} &\colon & \mbox{None} \\
    \mbox{\large \textbf{Pearson symbol}} &\colon & \mbox{hP6} \\
    \mbox{\large \textbf{Space group number}} &\colon & 178 \\
    \mbox{\large \textbf{Space group symbol}} &\colon & P6_{1}22 \\
    \mbox{\large \textbf{\AFLOW\ prototype command}} &\colon &  \texttt{aflow} \,  \, \texttt{-{}-proto=A\_hP6\_178\_a } \, \newline \texttt{-{}-params=}{a,c/a,x_{1} }
  \end{array}
\end{equation*}
\renewcommand{\arraystretch}{1.0}

\vspace*{-0.25cm}
\noindent \hrulefill
\begin{itemize}
  \item{This high pressure phase of scandium becomes stable at 240~GPa.  We use the
experimental data at 242~GPa and 297~K.
This chiral structure could also have been presented in the enantiomorphic
space group $P6_{5}22$ (\#179).
}
\end{itemize}

\noindent \parbox{1 \linewidth}{
\noindent \hrulefill
\\
\textbf{Hexagonal primitive vectors:} \\
\vspace*{-0.25cm}
\begin{tabular}{cc}
  \begin{tabular}{c}
    \parbox{0.6 \linewidth}{
      \renewcommand{\arraystretch}{1.5}
      \begin{equation*}
        \centering
        \begin{array}{ccc}
              \mathbf{a}_1 & = & \frac12 \, a \, \mathbf{\hat{x}} - \frac{\sqrt3}2 \, a \, \mathbf{\hat{y}} \\
    \mathbf{a}_2 & = & \frac12 \, a \, \mathbf{\hat{x}} + \frac{\sqrt3}2 \, a \, \mathbf{\hat{y}} \\
    \mathbf{a}_3 & = & c \, \mathbf{\hat{z}} \\

        \end{array}
      \end{equation*}
    }
    \renewcommand{\arraystretch}{1.0}
  \end{tabular}
  \begin{tabular}{c}
    \includegraphics[width=0.3\linewidth]{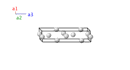} \\
  \end{tabular}
\end{tabular}

}
\vspace*{-0.25cm}

\noindent \hrulefill
\\
\textbf{Basis vectors:}
\vspace*{-0.25cm}
\renewcommand{\arraystretch}{1.5}
\begin{longtabu} to \textwidth{>{\centering $}X[-1,c,c]<{$}>{\centering $}X[-1,c,c]<{$}>{\centering $}X[-1,c,c]<{$}>{\centering $}X[-1,c,c]<{$}>{\centering $}X[-1,c,c]<{$}>{\centering $}X[-1,c,c]<{$}>{\centering $}X[-1,c,c]<{$}}
  & & \mbox{Lattice Coordinates} & & \mbox{Cartesian Coordinates} &\mbox{Wyckoff Position} & \mbox{Atom Type} \\  
  \mathbf{B}_{1} & = & x_{1} \, \mathbf{a}_{1} & = & \frac{1}{2}x_{1}a \, \mathbf{\hat{x}}-\frac{\sqrt{3}}{2}x_{1}a \, \mathbf{\hat{y}} & \left(6a\right) & \mbox{Sc} \\ 
\mathbf{B}_{2} & = & x_{1} \, \mathbf{a}_{2} + \frac{1}{3} \, \mathbf{a}_{3} & = & \frac{1}{2}x_{1}a \, \mathbf{\hat{x}} + \frac{\sqrt{3}}{2}x_{1}a \, \mathbf{\hat{y}} + \frac{1}{3}c \, \mathbf{\hat{z}} & \left(6a\right) & \mbox{Sc} \\ 
\mathbf{B}_{3} & = & -x_{1} \, \mathbf{a}_{1}-x_{1} \, \mathbf{a}_{2} + \frac{2}{3} \, \mathbf{a}_{3} & = & -x_{1}a \, \mathbf{\hat{x}} + \frac{2}{3}c \, \mathbf{\hat{z}} & \left(6a\right) & \mbox{Sc} \\ 
\mathbf{B}_{4} & = & -x_{1} \, \mathbf{a}_{1} + \frac{1}{2} \, \mathbf{a}_{3} & = & -\frac{1}{2}x_{1}a \, \mathbf{\hat{x}} + \frac{\sqrt{3}}{2}x_{1}a \, \mathbf{\hat{y}} + \frac{1}{2}c \, \mathbf{\hat{z}} & \left(6a\right) & \mbox{Sc} \\ 
\mathbf{B}_{5} & = & -x_{1} \, \mathbf{a}_{2} + \frac{5}{6} \, \mathbf{a}_{3} & = & -\frac{1}{2}x_{1}a \, \mathbf{\hat{x}}-\frac{\sqrt{3}}{2}x_{1}a \, \mathbf{\hat{y}} + \frac{5}{6}c \, \mathbf{\hat{z}} & \left(6a\right) & \mbox{Sc} \\ 
\mathbf{B}_{6} & = & x_{1} \, \mathbf{a}_{1} + x_{1} \, \mathbf{a}_{2} + \frac{1}{6} \, \mathbf{a}_{3} & = & x_{1}a \, \mathbf{\hat{x}} + \frac{1}{6}c \, \mathbf{\hat{z}} & \left(6a\right) & \mbox{Sc} \\ 
\end{longtabu}
\renewcommand{\arraystretch}{1.0}
\noindent \hrulefill
\\
\textbf{References:}
\vspace*{-0.25cm}
\begin{flushleft}
  - \bibentry{Akahama_PRL_94_2005}. \\
\end{flushleft}
\noindent \hrulefill
\\
\textbf{Geometry files:}
\\
\noindent  - CIF: pp. {\hyperref[A_hP6_178_a_cif]{\pageref{A_hP6_178_a_cif}}} \\
\noindent  - POSCAR: pp. {\hyperref[A_hP6_178_a_poscar]{\pageref{A_hP6_178_a_poscar}}} \\
\onecolumn
{\phantomsection\label{AB3_hP24_179_b_ac}}
\subsection*{\huge \textbf{{\normalfont AuF$_{3}$ Structure: AB3\_hP24\_179\_b\_ac}}}
\noindent \hrulefill
\vspace*{0.25cm}
\begin{figure}[htp]
  \centering
  \vspace{-1em}
  {\includegraphics[width=1\textwidth]{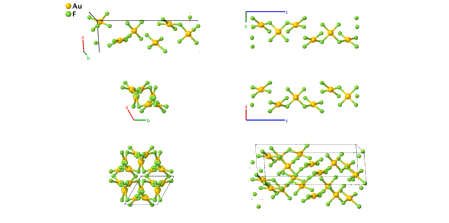}}
\end{figure}
\vspace*{-0.5cm}
\renewcommand{\arraystretch}{1.5}
\begin{equation*}
  \begin{array}{>{$\hspace{-0.15cm}}l<{$}>{$}p{0.5cm}<{$}>{$}p{18.5cm}<{$}}
    \mbox{\large \textbf{Prototype}} &\colon & \ce{AuF3} \\
    \mbox{\large \textbf{\AFLOW\ prototype label}} &\colon & \mbox{AB3\_hP24\_179\_b\_ac} \\
    \mbox{\large \textbf{\textit{Strukturbericht} designation}} &\colon & \mbox{None} \\
    \mbox{\large \textbf{Pearson symbol}} &\colon & \mbox{hP24} \\
    \mbox{\large \textbf{Space group number}} &\colon & 179 \\
    \mbox{\large \textbf{Space group symbol}} &\colon & P6_{5}22 \\
    \mbox{\large \textbf{\AFLOW\ prototype command}} &\colon &  \texttt{aflow} \,  \, \texttt{-{}-proto=AB3\_hP24\_179\_b\_ac } \, \newline \texttt{-{}-params=}{a,c/a,x_{1},x_{2},x_{3},y_{3},z_{3} }
  \end{array}
\end{equation*}
\renewcommand{\arraystretch}{1.0}

\vspace*{-0.25cm}
\noindent \hrulefill
\begin{itemize}
  \item{This structure is the enantiomorph of the \hyperref[AB3_hP24_178_b_ac]{AuF$_{3}$ (AB3\_hP24\_178\_b\_ac) structure}, 
and was generated by reflecting the coordinates of the space group \#178 structure through the $z=0$ plane.
}
\end{itemize}

\noindent \parbox{1 \linewidth}{
\noindent \hrulefill
\\
\textbf{Hexagonal primitive vectors:} \\
\vspace*{-0.25cm}
\begin{tabular}{cc}
  \begin{tabular}{c}
    \parbox{0.6 \linewidth}{
      \renewcommand{\arraystretch}{1.5}
      \begin{equation*}
        \centering
        \begin{array}{ccc}
              \mathbf{a}_1 & = & \frac12 \, a \, \mathbf{\hat{x}} - \frac{\sqrt3}2 \, a \, \mathbf{\hat{y}} \\
    \mathbf{a}_2 & = & \frac12 \, a \, \mathbf{\hat{x}} + \frac{\sqrt3}2 \, a \, \mathbf{\hat{y}} \\
    \mathbf{a}_3 & = & c \, \mathbf{\hat{z}} \\

        \end{array}
      \end{equation*}
    }
    \renewcommand{\arraystretch}{1.0}
  \end{tabular}
  \begin{tabular}{c}
    \includegraphics[width=0.3\linewidth]{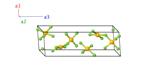} \\
  \end{tabular}
\end{tabular}

}
\vspace*{-0.25cm}

\noindent \hrulefill
\\
\textbf{Basis vectors:}
\vspace*{-0.25cm}
\renewcommand{\arraystretch}{1.5}
\begin{longtabu} to \textwidth{>{\centering $}X[-1,c,c]<{$}>{\centering $}X[-1,c,c]<{$}>{\centering $}X[-1,c,c]<{$}>{\centering $}X[-1,c,c]<{$}>{\centering $}X[-1,c,c]<{$}>{\centering $}X[-1,c,c]<{$}>{\centering $}X[-1,c,c]<{$}}
  & & \mbox{Lattice Coordinates} & & \mbox{Cartesian Coordinates} &\mbox{Wyckoff Position} & \mbox{Atom Type} \\  
  \mathbf{B}_{1} & = & x_{1} \, \mathbf{a}_{1} & = & \frac{1}{2}x_{1}a \, \mathbf{\hat{x}}-\frac{\sqrt{3}}{2}x_{1}a \, \mathbf{\hat{y}} & \left(6a\right) & \mbox{F I} \\ 
\mathbf{B}_{2} & = & x_{1} \, \mathbf{a}_{2} + \frac{2}{3} \, \mathbf{a}_{3} & = & \frac{1}{2}x_{1}a \, \mathbf{\hat{x}} + \frac{\sqrt{3}}{2}x_{1}a \, \mathbf{\hat{y}} + \frac{2}{3}c \, \mathbf{\hat{z}} & \left(6a\right) & \mbox{F I} \\ 
\mathbf{B}_{3} & = & -x_{1} \, \mathbf{a}_{1}-x_{1} \, \mathbf{a}_{2} + \frac{1}{3} \, \mathbf{a}_{3} & = & -x_{1}a \, \mathbf{\hat{x}} + \frac{1}{3}c \, \mathbf{\hat{z}} & \left(6a\right) & \mbox{F I} \\ 
\mathbf{B}_{4} & = & -x_{1} \, \mathbf{a}_{1} + \frac{1}{2} \, \mathbf{a}_{3} & = & -\frac{1}{2}x_{1}a \, \mathbf{\hat{x}} + \frac{\sqrt{3}}{2}x_{1}a \, \mathbf{\hat{y}} + \frac{1}{2}c \, \mathbf{\hat{z}} & \left(6a\right) & \mbox{F I} \\ 
\mathbf{B}_{5} & = & -x_{1} \, \mathbf{a}_{2} + \frac{1}{6} \, \mathbf{a}_{3} & = & -\frac{1}{2}x_{1}a \, \mathbf{\hat{x}}-\frac{\sqrt{3}}{2}x_{1}a \, \mathbf{\hat{y}} + \frac{1}{6}c \, \mathbf{\hat{z}} & \left(6a\right) & \mbox{F I} \\ 
\mathbf{B}_{6} & = & x_{1} \, \mathbf{a}_{1} + x_{1} \, \mathbf{a}_{2} + \frac{5}{6} \, \mathbf{a}_{3} & = & x_{1}a \, \mathbf{\hat{x}} + \frac{5}{6}c \, \mathbf{\hat{z}} & \left(6a\right) & \mbox{F I} \\ 
\mathbf{B}_{7} & = & x_{2} \, \mathbf{a}_{1} + 2x_{2} \, \mathbf{a}_{2} + \frac{3}{4} \, \mathbf{a}_{3} & = & \frac{3}{2}x_{2}a \, \mathbf{\hat{x}} + \frac{\sqrt{3}}{2}x_{2}a \, \mathbf{\hat{y}} + \frac{3}{4}c \, \mathbf{\hat{z}} & \left(6b\right) & \mbox{Au} \\ 
\mathbf{B}_{8} & = & -2x_{2} \, \mathbf{a}_{1}-x_{2} \, \mathbf{a}_{2} + \frac{5}{12} \, \mathbf{a}_{3} & = & -\frac{3}{2}x_{2}a \, \mathbf{\hat{x}} + \frac{\sqrt{3}}{2}x_{2}a \, \mathbf{\hat{y}} + \frac{5}{12}c \, \mathbf{\hat{z}} & \left(6b\right) & \mbox{Au} \\ 
\mathbf{B}_{9} & = & x_{2} \, \mathbf{a}_{1}-x_{2} \, \mathbf{a}_{2} + \frac{1}{12} \, \mathbf{a}_{3} & = & -\sqrt{3}x_{2}a \, \mathbf{\hat{y}} + \frac{1}{12}c \, \mathbf{\hat{z}} & \left(6b\right) & \mbox{Au} \\ 
\mathbf{B}_{10} & = & -x_{2} \, \mathbf{a}_{1}-2x_{2} \, \mathbf{a}_{2} + \frac{1}{4} \, \mathbf{a}_{3} & = & -\frac{3}{2}x_{2}a \, \mathbf{\hat{x}}-\frac{\sqrt{3}}{2}x_{2}a \, \mathbf{\hat{y}} + \frac{1}{4}c \, \mathbf{\hat{z}} & \left(6b\right) & \mbox{Au} \\ 
\mathbf{B}_{11} & = & 2x_{2} \, \mathbf{a}_{1} + x_{2} \, \mathbf{a}_{2} + \frac{11}{12} \, \mathbf{a}_{3} & = & \frac{3}{2}x_{2}a \, \mathbf{\hat{x}}-\frac{\sqrt{3}}{2}x_{2}a \, \mathbf{\hat{y}} + \frac{11}{12}c \, \mathbf{\hat{z}} & \left(6b\right) & \mbox{Au} \\ 
\mathbf{B}_{12} & = & -x_{2} \, \mathbf{a}_{1} + x_{2} \, \mathbf{a}_{2} + \frac{7}{12} \, \mathbf{a}_{3} & = & \sqrt{3}x_{2}a \, \mathbf{\hat{y}} + \frac{7}{12}c \, \mathbf{\hat{z}} & \left(6b\right) & \mbox{Au} \\ 
\mathbf{B}_{13} & = & x_{3} \, \mathbf{a}_{1} + y_{3} \, \mathbf{a}_{2} + z_{3} \, \mathbf{a}_{3} & = & \frac{1}{2}\left(x_{3}+y_{3}\right)a \, \mathbf{\hat{x}} + \frac{\sqrt{3}}{2}\left(-x_{3}+y_{3}\right)a \, \mathbf{\hat{y}} + z_{3}c \, \mathbf{\hat{z}} & \left(12c\right) & \mbox{F II} \\ 
\mathbf{B}_{14} & = & -y_{3} \, \mathbf{a}_{1} + \left(x_{3}-y_{3}\right) \, \mathbf{a}_{2} + \left(\frac{2}{3} +z_{3}\right) \, \mathbf{a}_{3} & = & \left(\frac{1}{2}x_{3}-y_{3}\right)a \, \mathbf{\hat{x}} + \frac{\sqrt{3}}{2}x_{3}a \, \mathbf{\hat{y}} + \left(\frac{2}{3} +z_{3}\right)c \, \mathbf{\hat{z}} & \left(12c\right) & \mbox{F II} \\ 
\mathbf{B}_{15} & = & \left(-x_{3}+y_{3}\right) \, \mathbf{a}_{1}-x_{3} \, \mathbf{a}_{2} + \left(\frac{1}{3} +z_{3}\right) \, \mathbf{a}_{3} & = & \left(-x_{3}+\frac{1}{2}y_{3}\right)a \, \mathbf{\hat{x}}-\frac{\sqrt{3}}{2}y_{3}a \, \mathbf{\hat{y}} + \left(\frac{1}{3} +z_{3}\right)c \, \mathbf{\hat{z}} & \left(12c\right) & \mbox{F II} \\ 
\mathbf{B}_{16} & = & -x_{3} \, \mathbf{a}_{1}-y_{3} \, \mathbf{a}_{2} + \left(\frac{1}{2} +z_{3}\right) \, \mathbf{a}_{3} & = & -\frac{1}{2}\left(x_{3}+y_{3}\right)a \, \mathbf{\hat{x}} + \frac{\sqrt{3}}{2}\left(x_{3}-y_{3}\right)a \, \mathbf{\hat{y}} + \left(\frac{1}{2} +z_{3}\right)c \, \mathbf{\hat{z}} & \left(12c\right) & \mbox{F II} \\ 
\mathbf{B}_{17} & = & y_{3} \, \mathbf{a}_{1} + \left(-x_{3}+y_{3}\right) \, \mathbf{a}_{2} + \left(\frac{1}{6} +z_{3}\right) \, \mathbf{a}_{3} & = & \left(-\frac{1}{2}x_{3}+y_{3}\right)a \, \mathbf{\hat{x}}-\frac{\sqrt{3}}{2}x_{3}a \, \mathbf{\hat{y}} + \left(\frac{1}{6} +z_{3}\right)c \, \mathbf{\hat{z}} & \left(12c\right) & \mbox{F II} \\ 
\mathbf{B}_{18} & = & \left(x_{3}-y_{3}\right) \, \mathbf{a}_{1} + x_{3} \, \mathbf{a}_{2} + \left(\frac{5}{6} +z_{3}\right) \, \mathbf{a}_{3} & = & \left(x_{3}-\frac{1}{2}y_{3}\right)a \, \mathbf{\hat{x}} + \frac{\sqrt{3}}{2}y_{3}a \, \mathbf{\hat{y}} + \left(\frac{5}{6} +z_{3}\right)c \, \mathbf{\hat{z}} & \left(12c\right) & \mbox{F II} \\ 
\mathbf{B}_{19} & = & y_{3} \, \mathbf{a}_{1} + x_{3} \, \mathbf{a}_{2} + \left(\frac{2}{3} - z_{3}\right) \, \mathbf{a}_{3} & = & \frac{1}{2}\left(x_{3}+y_{3}\right)a \, \mathbf{\hat{x}} + \frac{\sqrt{3}}{2}\left(x_{3}-y_{3}\right)a \, \mathbf{\hat{y}} + \left(\frac{2}{3} - z_{3}\right)c \, \mathbf{\hat{z}} & \left(12c\right) & \mbox{F II} \\ 
\mathbf{B}_{20} & = & \left(x_{3}-y_{3}\right) \, \mathbf{a}_{1}-y_{3} \, \mathbf{a}_{2}-z_{3} \, \mathbf{a}_{3} & = & \left(\frac{1}{2}x_{3}-y_{3}\right)a \, \mathbf{\hat{x}}-\frac{\sqrt{3}}{2}x_{3}a \, \mathbf{\hat{y}}-z_{3}c \, \mathbf{\hat{z}} & \left(12c\right) & \mbox{F II} \\ 
\mathbf{B}_{21} & = & -x_{3} \, \mathbf{a}_{1} + \left(-x_{3}+y_{3}\right) \, \mathbf{a}_{2} + \left(\frac{1}{3} - z_{3}\right) \, \mathbf{a}_{3} & = & \left(-x_{3}+\frac{1}{2}y_{3}\right)a \, \mathbf{\hat{x}} + \frac{\sqrt{3}}{2}y_{3}a \, \mathbf{\hat{y}} + \left(\frac{1}{3} - z_{3}\right)c \, \mathbf{\hat{z}} & \left(12c\right) & \mbox{F II} \\ 
\mathbf{B}_{22} & = & -y_{3} \, \mathbf{a}_{1}-x_{3} \, \mathbf{a}_{2} + \left(\frac{1}{6} - z_{3}\right) \, \mathbf{a}_{3} & = & -\frac{1}{2}\left(x_{3}+y_{3}\right)a \, \mathbf{\hat{x}} + \frac{\sqrt{3}}{2}\left(-x_{3}+y_{3}\right)a \, \mathbf{\hat{y}} + \left(\frac{1}{6} - z_{3}\right)c \, \mathbf{\hat{z}} & \left(12c\right) & \mbox{F II} \\ 
\mathbf{B}_{23} & = & \left(-x_{3}+y_{3}\right) \, \mathbf{a}_{1} + y_{3} \, \mathbf{a}_{2} + \left(\frac{1}{2} - z_{3}\right) \, \mathbf{a}_{3} & = & \left(-\frac{1}{2}x_{3}+y_{3}\right)a \, \mathbf{\hat{x}} + \frac{\sqrt{3}}{2}x_{3}a \, \mathbf{\hat{y}} + \left(\frac{1}{2} - z_{3}\right)c \, \mathbf{\hat{z}} & \left(12c\right) & \mbox{F II} \\ 
\mathbf{B}_{24} & = & x_{3} \, \mathbf{a}_{1} + \left(x_{3}-y_{3}\right) \, \mathbf{a}_{2} + \left(\frac{5}{6} - z_{3}\right) \, \mathbf{a}_{3} & = & \left(x_{3}-\frac{1}{2}y_{3}\right)a \, \mathbf{\hat{x}}-\frac{\sqrt{3}}{2}y_{3}a \, \mathbf{\hat{y}} + \left(\frac{5}{6} - z_{3}\right)c \, \mathbf{\hat{z}} & \left(12c\right) & \mbox{F II} \\ 
\end{longtabu}
\renewcommand{\arraystretch}{1.0}
\noindent \hrulefill
\\
\textbf{References:}
\vspace*{-0.25cm}
\begin{flushleft}
  - \bibentry{Asprey_AuF3_InorgChem_1964}. \\
\end{flushleft}
\noindent \hrulefill
\\
\textbf{Geometry files:}
\\
\noindent  - CIF: pp. {\hyperref[AB3_hP24_179_b_ac_cif]{\pageref{AB3_hP24_179_b_ac_cif}}} \\
\noindent  - POSCAR: pp. {\hyperref[AB3_hP24_179_b_ac_poscar]{\pageref{AB3_hP24_179_b_ac_poscar}}} \\
\onecolumn
{\phantomsection\label{A2B_hP9_181_j_c}}
\subsection*{\huge \textbf{{\normalfont $\beta$-SiO$_{2}$ Structure: A2B\_hP9\_181\_j\_c}}}
\noindent \hrulefill
\vspace*{0.25cm}
\begin{figure}[htp]
  \centering
  \vspace{-1em}
  {\includegraphics[width=1\textwidth]{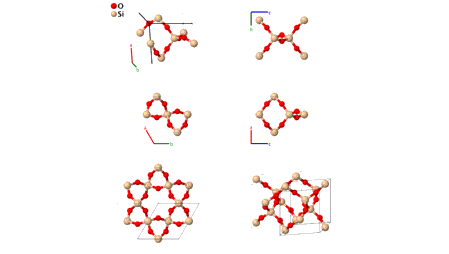}}
\end{figure}
\vspace*{-0.5cm}
\renewcommand{\arraystretch}{1.5}
\begin{equation*}
  \begin{array}{>{$\hspace{-0.15cm}}l<{$}>{$}p{0.5cm}<{$}>{$}p{18.5cm}<{$}}
    \mbox{\large \textbf{Prototype}} &\colon & \ce{$\beta$-SiO2} \\
    \mbox{\large \textbf{\AFLOW\ prototype label}} &\colon & \mbox{A2B\_hP9\_181\_j\_c} \\
    \mbox{\large \textbf{\textit{Strukturbericht} designation}} &\colon & \mbox{None} \\
    \mbox{\large \textbf{Pearson symbol}} &\colon & \mbox{hP9} \\
    \mbox{\large \textbf{Space group number}} &\colon & 181 \\
    \mbox{\large \textbf{Space group symbol}} &\colon & P6_{4}22 \\
    \mbox{\large \textbf{\AFLOW\ prototype command}} &\colon &  \texttt{aflow} \,  \, \texttt{-{}-proto=A2B\_hP9\_181\_j\_c } \, \newline \texttt{-{}-params=}{a,c/a,x_{2} }
  \end{array}
\end{equation*}
\renewcommand{\arraystretch}{1.0}

\vspace*{-0.25cm}
\noindent \hrulefill
\begin{itemize}
  \item{This structure is the enantiomorph of the \href{http://aflow.org/CrystalDatabase/A2B_hP9_180_j_c.html}{$\beta$-Quartz (A2B\_hP9\_180\_j\_c) structure},
and was generated by reflecting the coordinates of the space group \#180 structure through the $z=0$ plane.
}
\end{itemize}

\noindent \parbox{1 \linewidth}{
\noindent \hrulefill
\\
\textbf{Hexagonal primitive vectors:} \\
\vspace*{-0.25cm}
\begin{tabular}{cc}
  \begin{tabular}{c}
    \parbox{0.6 \linewidth}{
      \renewcommand{\arraystretch}{1.5}
      \begin{equation*}
        \centering
        \begin{array}{ccc}
              \mathbf{a}_1 & = & \frac12 \, a \, \mathbf{\hat{x}} - \frac{\sqrt3}2 \, a \, \mathbf{\hat{y}} \\
    \mathbf{a}_2 & = & \frac12 \, a \, \mathbf{\hat{x}} + \frac{\sqrt3}2 \, a \, \mathbf{\hat{y}} \\
    \mathbf{a}_3 & = & c \, \mathbf{\hat{z}} \\

        \end{array}
      \end{equation*}
    }
    \renewcommand{\arraystretch}{1.0}
  \end{tabular}
  \begin{tabular}{c}
    \includegraphics[width=0.3\linewidth]{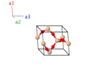} \\
  \end{tabular}
\end{tabular}

}
\vspace*{-0.25cm}

\noindent \hrulefill
\\
\textbf{Basis vectors:}
\vspace*{-0.25cm}
\renewcommand{\arraystretch}{1.5}
\begin{longtabu} to \textwidth{>{\centering $}X[-1,c,c]<{$}>{\centering $}X[-1,c,c]<{$}>{\centering $}X[-1,c,c]<{$}>{\centering $}X[-1,c,c]<{$}>{\centering $}X[-1,c,c]<{$}>{\centering $}X[-1,c,c]<{$}>{\centering $}X[-1,c,c]<{$}}
  & & \mbox{Lattice Coordinates} & & \mbox{Cartesian Coordinates} &\mbox{Wyckoff Position} & \mbox{Atom Type} \\  
  \mathbf{B}_{1} & = & \frac{1}{2} \, \mathbf{a}_{1} & = & \frac{1}{4}a \, \mathbf{\hat{x}}- \frac{\sqrt{3}}{4}a  \, \mathbf{\hat{y}} & \left(3c\right) & \mbox{Si} \\ 
\mathbf{B}_{2} & = & \frac{1}{2} \, \mathbf{a}_{2} + \frac{1}{3} \, \mathbf{a}_{3} & = & \frac{1}{4}a \, \mathbf{\hat{x}} + \frac{\sqrt{3}}{4}a \, \mathbf{\hat{y}} + \frac{1}{3}c \, \mathbf{\hat{z}} & \left(3c\right) & \mbox{Si} \\ 
\mathbf{B}_{3} & = & \frac{1}{2} \, \mathbf{a}_{1} + \frac{1}{2} \, \mathbf{a}_{2} + \frac{2}{3} \, \mathbf{a}_{3} & = & \frac{1}{2}a \, \mathbf{\hat{x}} + \frac{2}{3}c \, \mathbf{\hat{z}} & \left(3c\right) & \mbox{Si} \\ 
\mathbf{B}_{4} & = & x_{2} \, \mathbf{a}_{1} + 2x_{2} \, \mathbf{a}_{2} + \frac{1}{2} \, \mathbf{a}_{3} & = & \frac{3}{2}x_{2}a \, \mathbf{\hat{x}} + \frac{\sqrt{3}}{2}x_{2}a \, \mathbf{\hat{y}} + \frac{1}{2}c \, \mathbf{\hat{z}} & \left(6j\right) & \mbox{O} \\ 
\mathbf{B}_{5} & = & -2x_{2} \, \mathbf{a}_{1}-x_{2} \, \mathbf{a}_{2} + \frac{5}{6} \, \mathbf{a}_{3} & = & -\frac{3}{2}x_{2}a \, \mathbf{\hat{x}} + \frac{\sqrt{3}}{2}x_{2}a \, \mathbf{\hat{y}} + \frac{5}{6}c \, \mathbf{\hat{z}} & \left(6j\right) & \mbox{O} \\ 
\mathbf{B}_{6} & = & x_{2} \, \mathbf{a}_{1}-x_{2} \, \mathbf{a}_{2} + \frac{1}{6} \, \mathbf{a}_{3} & = & -\sqrt{3}x_{2}a \, \mathbf{\hat{y}} + \frac{1}{6}c \, \mathbf{\hat{z}} & \left(6j\right) & \mbox{O} \\ 
\mathbf{B}_{7} & = & -x_{2} \, \mathbf{a}_{1}-2x_{2} \, \mathbf{a}_{2} + \frac{1}{2} \, \mathbf{a}_{3} & = & -\frac{3}{2}x_{2}a \, \mathbf{\hat{x}}-\frac{\sqrt{3}}{2}x_{2}a \, \mathbf{\hat{y}} + \frac{1}{2}c \, \mathbf{\hat{z}} & \left(6j\right) & \mbox{O} \\ 
\mathbf{B}_{8} & = & 2x_{2} \, \mathbf{a}_{1} + x_{2} \, \mathbf{a}_{2} + \frac{5}{6} \, \mathbf{a}_{3} & = & \frac{3}{2}x_{2}a \, \mathbf{\hat{x}}-\frac{\sqrt{3}}{2}x_{2}a \, \mathbf{\hat{y}} + \frac{5}{6}c \, \mathbf{\hat{z}} & \left(6j\right) & \mbox{O} \\ 
\mathbf{B}_{9} & = & -x_{2} \, \mathbf{a}_{1} + x_{2} \, \mathbf{a}_{2} + \frac{1}{6} \, \mathbf{a}_{3} & = & \sqrt{3}x_{2}a \, \mathbf{\hat{y}} + \frac{1}{6}c \, \mathbf{\hat{z}} & \left(6j\right) & \mbox{O} \\ 
\end{longtabu}
\renewcommand{\arraystretch}{1.0}
\noindent \hrulefill
\\
\textbf{References:}
\vspace*{-0.25cm}
\begin{flushleft}
  - \bibentry{Wright_beta_jssc_1981}. \\
\end{flushleft}
\noindent \hrulefill
\\
\textbf{Geometry files:}
\\
\noindent  - CIF: pp. {\hyperref[A2B_hP9_181_j_c_cif]{\pageref{A2B_hP9_181_j_c_cif}}} \\
\noindent  - POSCAR: pp. {\hyperref[A2B_hP9_181_j_c_poscar]{\pageref{A2B_hP9_181_j_c_poscar}}} \\
\onecolumn
{\phantomsection\label{ABC_hP3_183_a_a_a}}
\subsection*{\huge \textbf{{\normalfont AuCN Structure: ABC\_hP3\_183\_a\_a\_a}}}
\noindent \hrulefill
\vspace*{0.25cm}
\begin{figure}[htp]
  \centering
  \vspace{-1em}
  {\includegraphics[width=1\textwidth]{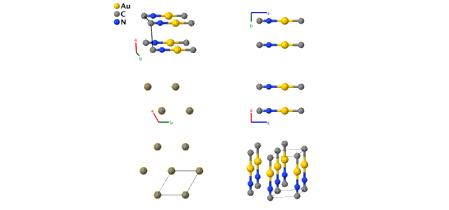}}
\end{figure}
\vspace*{-0.5cm}
\renewcommand{\arraystretch}{1.5}
\begin{equation*}
  \begin{array}{>{$\hspace{-0.15cm}}l<{$}>{$}p{0.5cm}<{$}>{$}p{18.5cm}<{$}}
    \mbox{\large \textbf{Prototype}} &\colon & \ce{AuCN} \\
    \mbox{\large \textbf{\AFLOW\ prototype label}} &\colon & \mbox{ABC\_hP3\_183\_a\_a\_a} \\
    \mbox{\large \textbf{\textit{Strukturbericht} designation}} &\colon & \mbox{None} \\
    \mbox{\large \textbf{Pearson symbol}} &\colon & \mbox{hP3} \\
    \mbox{\large \textbf{Space group number}} &\colon & 183 \\
    \mbox{\large \textbf{Space group symbol}} &\colon & P6mm \\
    \mbox{\large \textbf{\AFLOW\ prototype command}} &\colon &  \texttt{aflow} \,  \, \texttt{-{}-proto=ABC\_hP3\_183\_a\_a\_a } \, \newline \texttt{-{}-params=}{a,c/a,z_{1},z_{2},z_{3} }
  \end{array}
\end{equation*}
\renewcommand{\arraystretch}{1.0}

\noindent \parbox{1 \linewidth}{
\noindent \hrulefill
\\
\textbf{Hexagonal primitive vectors:} \\
\vspace*{-0.25cm}
\begin{tabular}{cc}
  \begin{tabular}{c}
    \parbox{0.6 \linewidth}{
      \renewcommand{\arraystretch}{1.5}
      \begin{equation*}
        \centering
        \begin{array}{ccc}
              \mathbf{a}_1 & = & \frac12 \, a \, \mathbf{\hat{x}} - \frac{\sqrt3}2 \, a \, \mathbf{\hat{y}} \\
    \mathbf{a}_2 & = & \frac12 \, a \, \mathbf{\hat{x}} + \frac{\sqrt3}2 \, a \, \mathbf{\hat{y}} \\
    \mathbf{a}_3 & = & c \, \mathbf{\hat{z}} \\

        \end{array}
      \end{equation*}
    }
    \renewcommand{\arraystretch}{1.0}
  \end{tabular}
  \begin{tabular}{c}
    \includegraphics[width=0.3\linewidth]{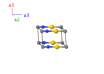} \\
  \end{tabular}
\end{tabular}

}
\vspace*{-0.25cm}

\noindent \hrulefill
\\
\textbf{Basis vectors:}
\vspace*{-0.25cm}
\renewcommand{\arraystretch}{1.5}
\begin{longtabu} to \textwidth{>{\centering $}X[-1,c,c]<{$}>{\centering $}X[-1,c,c]<{$}>{\centering $}X[-1,c,c]<{$}>{\centering $}X[-1,c,c]<{$}>{\centering $}X[-1,c,c]<{$}>{\centering $}X[-1,c,c]<{$}>{\centering $}X[-1,c,c]<{$}}
  & & \mbox{Lattice Coordinates} & & \mbox{Cartesian Coordinates} &\mbox{Wyckoff Position} & \mbox{Atom Type} \\  
  \mathbf{B}_{1} & = & z_{1} \, \mathbf{a}_{3} & = & z_{1}c \, \mathbf{\hat{z}} & \left(1a\right) & \mbox{Au} \\ 
\mathbf{B}_{2} & = & z_{2} \, \mathbf{a}_{3} & = & z_{2}c \, \mathbf{\hat{z}} & \left(1a\right) & \mbox{C} \\ 
\mathbf{B}_{3} & = & z_{3} \, \mathbf{a}_{3} & = & z_{3}c \, \mathbf{\hat{z}} & \left(1a\right) & \mbox{N} \\ 
\end{longtabu}
\renewcommand{\arraystretch}{1.0}
\noindent \hrulefill
\\
\textbf{References:}
\vspace*{-0.25cm}
\begin{flushleft}
  - \bibentry{Hibble_AuCN_InorgChem_2003}. \\
\end{flushleft}
\textbf{Found in:}
\vspace*{-0.25cm}
\begin{flushleft}
  - \bibentry{Villars_PearsonsCrystalData_2013}. \\
\end{flushleft}
\noindent \hrulefill
\\
\textbf{Geometry files:}
\\
\noindent  - CIF: pp. {\hyperref[ABC_hP3_183_a_a_a_cif]{\pageref{ABC_hP3_183_a_a_a_cif}}} \\
\noindent  - POSCAR: pp. {\hyperref[ABC_hP3_183_a_a_a_poscar]{\pageref{ABC_hP3_183_a_a_a_poscar}}} \\
\onecolumn
{\phantomsection\label{AB_hP6_183_c_ab}}
\subsection*{\huge \textbf{{\normalfont CrFe$_{3}$NiSn$_{5}$ Structure: AB\_hP6\_183\_c\_ab}}}
\noindent \hrulefill
\vspace*{0.25cm}
\begin{figure}[htp]
  \centering
  \vspace{-1em}
  {\includegraphics[width=1\textwidth]{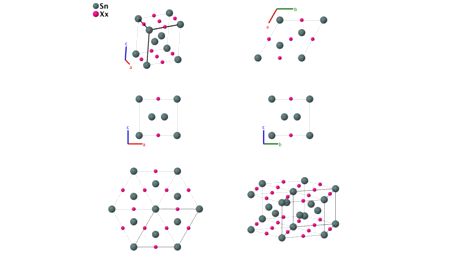}}
\end{figure}
\vspace*{-0.5cm}
\renewcommand{\arraystretch}{1.5}
\begin{equation*}
  \begin{array}{>{$\hspace{-0.15cm}}l<{$}>{$}p{0.5cm}<{$}>{$}p{18.5cm}<{$}}
    \mbox{\large \textbf{Prototype}} &\colon & \ce{CrFe3NiSn5} \\
    \mbox{\large \textbf{\AFLOW\ prototype label}} &\colon & \mbox{AB\_hP6\_183\_c\_ab} \\
    \mbox{\large \textbf{\textit{Strukturbericht} designation}} &\colon & \mbox{None} \\
    \mbox{\large \textbf{Pearson symbol}} &\colon & \mbox{hP6} \\
    \mbox{\large \textbf{Space group number}} &\colon & 183 \\
    \mbox{\large \textbf{Space group symbol}} &\colon & P6mm \\
    \mbox{\large \textbf{\AFLOW\ prototype command}} &\colon &  \texttt{aflow} \,  \, \texttt{-{}-proto=AB\_hP6\_183\_c\_ab } \, \newline \texttt{-{}-params=}{a,c/a,z_{1},z_{2},z_{3} }
  \end{array}
\end{equation*}
\renewcommand{\arraystretch}{1.0}

\vspace*{-0.25cm}
\noindent \hrulefill
\begin{itemize}
  \item{Here, the M sites are mixed occupation 0.6Fe+0.2Cr+0.2Ni
The Jmol image does not distinguish between the different M labels and is represented
by the "Xx" atoms.
}
\end{itemize}

\noindent \parbox{1 \linewidth}{
\noindent \hrulefill
\\
\textbf{Hexagonal primitive vectors:} \\
\vspace*{-0.25cm}
\begin{tabular}{cc}
  \begin{tabular}{c}
    \parbox{0.6 \linewidth}{
      \renewcommand{\arraystretch}{1.5}
      \begin{equation*}
        \centering
        \begin{array}{ccc}
              \mathbf{a}_1 & = & \frac12 \, a \, \mathbf{\hat{x}} - \frac{\sqrt3}2 \, a \, \mathbf{\hat{y}} \\
    \mathbf{a}_2 & = & \frac12 \, a \, \mathbf{\hat{x}} + \frac{\sqrt3}2 \, a \, \mathbf{\hat{y}} \\
    \mathbf{a}_3 & = & c \, \mathbf{\hat{z}} \\

        \end{array}
      \end{equation*}
    }
    \renewcommand{\arraystretch}{1.0}
  \end{tabular}
  \begin{tabular}{c}
    \includegraphics[width=0.3\linewidth]{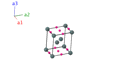} \\
  \end{tabular}
\end{tabular}

}
\vspace*{-0.25cm}

\noindent \hrulefill
\\
\textbf{Basis vectors:}
\vspace*{-0.25cm}
\renewcommand{\arraystretch}{1.5}
\begin{longtabu} to \textwidth{>{\centering $}X[-1,c,c]<{$}>{\centering $}X[-1,c,c]<{$}>{\centering $}X[-1,c,c]<{$}>{\centering $}X[-1,c,c]<{$}>{\centering $}X[-1,c,c]<{$}>{\centering $}X[-1,c,c]<{$}>{\centering $}X[-1,c,c]<{$}}
  & & \mbox{Lattice Coordinates} & & \mbox{Cartesian Coordinates} &\mbox{Wyckoff Position} & \mbox{Atom Type} \\  
  \mathbf{B}_{1} & = & z_{1} \, \mathbf{a}_{3} & = & z_{1}c \, \mathbf{\hat{z}} & \left(1a\right) & \mbox{Sn I} \\ 
\mathbf{B}_{2} & = & \frac{1}{3} \, \mathbf{a}_{1} + \frac{2}{3} \, \mathbf{a}_{2} + z_{2} \, \mathbf{a}_{3} & = & \frac{1}{2}a \, \mathbf{\hat{x}} + \frac{1}{2\sqrt{3}}a \, \mathbf{\hat{y}} + z_{2}c \, \mathbf{\hat{z}} & \left(2b\right) & \mbox{Sn II} \\ 
\mathbf{B}_{3} & = & \frac{2}{3} \, \mathbf{a}_{1} + \frac{1}{3} \, \mathbf{a}_{2} + z_{2} \, \mathbf{a}_{3} & = & \frac{1}{2}a \, \mathbf{\hat{x}}- \frac{1}{2\sqrt{3}}a  \, \mathbf{\hat{y}} + z_{2}c \, \mathbf{\hat{z}} & \left(2b\right) & \mbox{Sn II} \\ 
\mathbf{B}_{4} & = & \frac{1}{2} \, \mathbf{a}_{1} + z_{3} \, \mathbf{a}_{3} & = & \frac{1}{4}a \, \mathbf{\hat{x}}- \frac{\sqrt{3}}{4}a  \, \mathbf{\hat{y}} + z_{3}c \, \mathbf{\hat{z}} & \left(3c\right) & \mbox{M} \\ 
\mathbf{B}_{5} & = & \frac{1}{2} \, \mathbf{a}_{2} + z_{3} \, \mathbf{a}_{3} & = & \frac{1}{4}a \, \mathbf{\hat{x}} + \frac{\sqrt{3}}{4}a \, \mathbf{\hat{y}} + z_{3}c \, \mathbf{\hat{z}} & \left(3c\right) & \mbox{M} \\ 
\mathbf{B}_{6} & = & \frac{1}{2} \, \mathbf{a}_{1} + \frac{1}{2} \, \mathbf{a}_{2} + z_{3} \, \mathbf{a}_{3} & = & \frac{1}{2}a \, \mathbf{\hat{x}} + z_{3}c \, \mathbf{\hat{z}} & \left(3c\right) & \mbox{M} \\ 
\end{longtabu}
\renewcommand{\arraystretch}{1.0}
\noindent \hrulefill
\\
\textbf{References:}
\vspace*{-0.25cm}
\begin{flushleft}
  - \bibentry{Huang_CrFe3NiSn5_PwdrDiff_2004}. \\
\end{flushleft}
\textbf{Found in:}
\vspace*{-0.25cm}
\begin{flushleft}
  - \bibentry{Villars_PearsonsCrystalData_2013}. \\
\end{flushleft}
\noindent \hrulefill
\\
\textbf{Geometry files:}
\\
\noindent  - CIF: pp. {\hyperref[AB_hP6_183_c_ab_cif]{\pageref{AB_hP6_183_c_ab_cif}}} \\
\noindent  - POSCAR: pp. {\hyperref[AB_hP6_183_c_ab_poscar]{\pageref{AB_hP6_183_c_ab_poscar}}} \\
\onecolumn
{\phantomsection\label{AB4C_hP72_184_d_4d_d}}
\subsection*{\huge \textbf{{\normalfont \begin{raggedleft}Al[PO$_{4}$] (Framework type AFI) Structure: \end{raggedleft} \\ AB4C\_hP72\_184\_d\_4d\_d}}}
\noindent \hrulefill
\vspace*{0.25cm}
\begin{figure}[htp]
  \centering
  \vspace{-1em}
  {\includegraphics[width=1\textwidth]{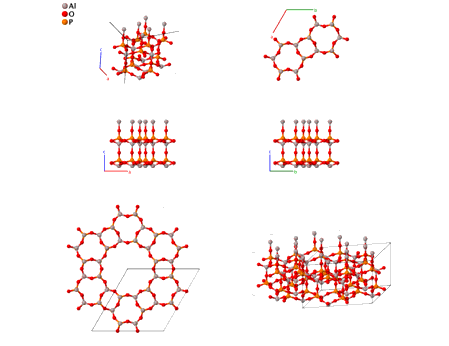}}
\end{figure}
\vspace*{-0.5cm}
\renewcommand{\arraystretch}{1.5}
\begin{equation*}
  \begin{array}{>{$\hspace{-0.15cm}}l<{$}>{$}p{0.5cm}<{$}>{$}p{18.5cm}<{$}}
    \mbox{\large \textbf{Prototype}} &\colon & \ce{Al[PO4]} \\
    \mbox{\large \textbf{\AFLOW\ prototype label}} &\colon & \mbox{AB4C\_hP72\_184\_d\_4d\_d} \\
    \mbox{\large \textbf{\textit{Strukturbericht} designation}} &\colon & \mbox{None} \\
    \mbox{\large \textbf{Pearson symbol}} &\colon & \mbox{hP72} \\
    \mbox{\large \textbf{Space group number}} &\colon & 184 \\
    \mbox{\large \textbf{Space group symbol}} &\colon & P6cc \\
    \mbox{\large \textbf{\AFLOW\ prototype command}} &\colon &  \texttt{aflow} \,  \, \texttt{-{}-proto=AB4C\_hP72\_184\_d\_4d\_d } \, \newline \texttt{-{}-params=}{a,c/a,x_{1},y_{1},z_{1},x_{2},y_{2},z_{2},x_{3},y_{3},z_{3},x_{4},y_{4},z_{4},x_{5},y_{5},z_{5},x_{6},y_{6},z_{6} }
  \end{array}
\end{equation*}
\renewcommand{\arraystretch}{1.0}

\vspace*{-0.25cm}
\noindent \hrulefill
\begin{itemize}
  \item{This is the structure of AlPO$_{4}$-5, which has the zeolite framework designation AFI. 
}
\end{itemize}

\noindent \parbox{1 \linewidth}{
\noindent \hrulefill
\\
\textbf{Hexagonal primitive vectors:} \\
\vspace*{-0.25cm}
\begin{tabular}{cc}
  \begin{tabular}{c}
    \parbox{0.6 \linewidth}{
      \renewcommand{\arraystretch}{1.5}
      \begin{equation*}
        \centering
        \begin{array}{ccc}
              \mathbf{a}_1 & = & \frac12 \, a \, \mathbf{\hat{x}} - \frac{\sqrt3}2 \, a \, \mathbf{\hat{y}} \\
    \mathbf{a}_2 & = & \frac12 \, a \, \mathbf{\hat{x}} + \frac{\sqrt3}2 \, a \, \mathbf{\hat{y}} \\
    \mathbf{a}_3 & = & c \, \mathbf{\hat{z}} \\

        \end{array}
      \end{equation*}
    }
    \renewcommand{\arraystretch}{1.0}
  \end{tabular}
  \begin{tabular}{c}
    \includegraphics[width=0.3\linewidth]{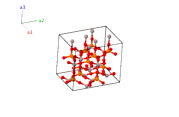} \\
  \end{tabular}
\end{tabular}

}
\vspace*{-0.25cm}

\noindent \hrulefill
\\
\textbf{Basis vectors:}
\vspace*{-0.25cm}
\renewcommand{\arraystretch}{1.5}
\begin{longtabu} to \textwidth{>{\centering $}X[-1,c,c]<{$}>{\centering $}X[-1,c,c]<{$}>{\centering $}X[-1,c,c]<{$}>{\centering $}X[-1,c,c]<{$}>{\centering $}X[-1,c,c]<{$}>{\centering $}X[-1,c,c]<{$}>{\centering $}X[-1,c,c]<{$}}
  & & \mbox{Lattice Coordinates} & & \mbox{Cartesian Coordinates} &\mbox{Wyckoff Position} & \mbox{Atom Type} \\  
  \mathbf{B}_{1} & = & x_{1} \, \mathbf{a}_{1} + y_{1} \, \mathbf{a}_{2} + z_{1} \, \mathbf{a}_{3} & = & \frac{1}{2}\left(x_{1}+y_{1}\right)a \, \mathbf{\hat{x}} + \frac{\sqrt{3}}{2}\left(-x_{1}+y_{1}\right)a \, \mathbf{\hat{y}} + z_{1}c \, \mathbf{\hat{z}} & \left(12d\right) & \mbox{Al} \\ 
\mathbf{B}_{2} & = & -y_{1} \, \mathbf{a}_{1} + \left(x_{1}-y_{1}\right) \, \mathbf{a}_{2} + z_{1} \, \mathbf{a}_{3} & = & \left(\frac{1}{2}x_{1}-y_{1}\right)a \, \mathbf{\hat{x}} + \frac{\sqrt{3}}{2}x_{1}a \, \mathbf{\hat{y}} + z_{1}c \, \mathbf{\hat{z}} & \left(12d\right) & \mbox{Al} \\ 
\mathbf{B}_{3} & = & \left(-x_{1}+y_{1}\right) \, \mathbf{a}_{1}-x_{1} \, \mathbf{a}_{2} + z_{1} \, \mathbf{a}_{3} & = & \left(-x_{1}+\frac{1}{2}y_{1}\right)a \, \mathbf{\hat{x}}-\frac{\sqrt{3}}{2}y_{1}a \, \mathbf{\hat{y}} + z_{1}c \, \mathbf{\hat{z}} & \left(12d\right) & \mbox{Al} \\ 
\mathbf{B}_{4} & = & -x_{1} \, \mathbf{a}_{1}-y_{1} \, \mathbf{a}_{2} + z_{1} \, \mathbf{a}_{3} & = & -\frac{1}{2}\left(x_{1}+y_{1}\right)a \, \mathbf{\hat{x}} + \frac{\sqrt{3}}{2}\left(x_{1}-y_{1}\right)a \, \mathbf{\hat{y}} + z_{1}c \, \mathbf{\hat{z}} & \left(12d\right) & \mbox{Al} \\ 
\mathbf{B}_{5} & = & y_{1} \, \mathbf{a}_{1} + \left(-x_{1}+y_{1}\right) \, \mathbf{a}_{2} + z_{1} \, \mathbf{a}_{3} & = & \left(-\frac{1}{2}x_{1}+y_{1}\right)a \, \mathbf{\hat{x}}-\frac{\sqrt{3}}{2}x_{1}a \, \mathbf{\hat{y}} + z_{1}c \, \mathbf{\hat{z}} & \left(12d\right) & \mbox{Al} \\ 
\mathbf{B}_{6} & = & \left(x_{1}-y_{1}\right) \, \mathbf{a}_{1} + x_{1} \, \mathbf{a}_{2} + z_{1} \, \mathbf{a}_{3} & = & \left(x_{1}-\frac{1}{2}y_{1}\right)a \, \mathbf{\hat{x}} + \frac{\sqrt{3}}{2}y_{1}a \, \mathbf{\hat{y}} + z_{1}c \, \mathbf{\hat{z}} & \left(12d\right) & \mbox{Al} \\ 
\mathbf{B}_{7} & = & -y_{1} \, \mathbf{a}_{1}-x_{1} \, \mathbf{a}_{2} + \left(\frac{1}{2} +z_{1}\right) \, \mathbf{a}_{3} & = & -\frac{1}{2}\left(x_{1}+y_{1}\right)a \, \mathbf{\hat{x}} + \frac{\sqrt{3}}{2}\left(-x_{1}+y_{1}\right)a \, \mathbf{\hat{y}} + \left(\frac{1}{2} +z_{1}\right)c \, \mathbf{\hat{z}} & \left(12d\right) & \mbox{Al} \\ 
\mathbf{B}_{8} & = & \left(-x_{1}+y_{1}\right) \, \mathbf{a}_{1} + y_{1} \, \mathbf{a}_{2} + \left(\frac{1}{2} +z_{1}\right) \, \mathbf{a}_{3} & = & \left(-\frac{1}{2}x_{1}+y_{1}\right)a \, \mathbf{\hat{x}} + \frac{\sqrt{3}}{2}x_{1}a \, \mathbf{\hat{y}} + \left(\frac{1}{2} +z_{1}\right)c \, \mathbf{\hat{z}} & \left(12d\right) & \mbox{Al} \\ 
\mathbf{B}_{9} & = & x_{1} \, \mathbf{a}_{1} + \left(x_{1}-y_{1}\right) \, \mathbf{a}_{2} + \left(\frac{1}{2} +z_{1}\right) \, \mathbf{a}_{3} & = & \left(x_{1}-\frac{1}{2}y_{1}\right)a \, \mathbf{\hat{x}}-\frac{\sqrt{3}}{2}y_{1}a \, \mathbf{\hat{y}} + \left(\frac{1}{2} +z_{1}\right)c \, \mathbf{\hat{z}} & \left(12d\right) & \mbox{Al} \\ 
\mathbf{B}_{10} & = & y_{1} \, \mathbf{a}_{1} + x_{1} \, \mathbf{a}_{2} + \left(\frac{1}{2} +z_{1}\right) \, \mathbf{a}_{3} & = & \frac{1}{2}\left(x_{1}+y_{1}\right)a \, \mathbf{\hat{x}} + \frac{\sqrt{3}}{2}\left(x_{1}-y_{1}\right)a \, \mathbf{\hat{y}} + \left(\frac{1}{2} +z_{1}\right)c \, \mathbf{\hat{z}} & \left(12d\right) & \mbox{Al} \\ 
\mathbf{B}_{11} & = & \left(x_{1}-y_{1}\right) \, \mathbf{a}_{1}-y_{1} \, \mathbf{a}_{2} + \left(\frac{1}{2} +z_{1}\right) \, \mathbf{a}_{3} & = & \left(\frac{1}{2}x_{1}-y_{1}\right)a \, \mathbf{\hat{x}}-\frac{\sqrt{3}}{2}x_{1}a \, \mathbf{\hat{y}} + \left(\frac{1}{2} +z_{1}\right)c \, \mathbf{\hat{z}} & \left(12d\right) & \mbox{Al} \\ 
\mathbf{B}_{12} & = & -x_{1} \, \mathbf{a}_{1} + \left(-x_{1}+y_{1}\right) \, \mathbf{a}_{2} + \left(\frac{1}{2} +z_{1}\right) \, \mathbf{a}_{3} & = & \left(-x_{1}+\frac{1}{2}y_{1}\right)a \, \mathbf{\hat{x}} + \frac{\sqrt{3}}{2}y_{1}a \, \mathbf{\hat{y}} + \left(\frac{1}{2} +z_{1}\right)c \, \mathbf{\hat{z}} & \left(12d\right) & \mbox{Al} \\ 
\mathbf{B}_{13} & = & x_{2} \, \mathbf{a}_{1} + y_{2} \, \mathbf{a}_{2} + z_{2} \, \mathbf{a}_{3} & = & \frac{1}{2}\left(x_{2}+y_{2}\right)a \, \mathbf{\hat{x}} + \frac{\sqrt{3}}{2}\left(-x_{2}+y_{2}\right)a \, \mathbf{\hat{y}} + z_{2}c \, \mathbf{\hat{z}} & \left(12d\right) & \mbox{O I} \\ 
\mathbf{B}_{14} & = & -y_{2} \, \mathbf{a}_{1} + \left(x_{2}-y_{2}\right) \, \mathbf{a}_{2} + z_{2} \, \mathbf{a}_{3} & = & \left(\frac{1}{2}x_{2}-y_{2}\right)a \, \mathbf{\hat{x}} + \frac{\sqrt{3}}{2}x_{2}a \, \mathbf{\hat{y}} + z_{2}c \, \mathbf{\hat{z}} & \left(12d\right) & \mbox{O I} \\ 
\mathbf{B}_{15} & = & \left(-x_{2}+y_{2}\right) \, \mathbf{a}_{1}-x_{2} \, \mathbf{a}_{2} + z_{2} \, \mathbf{a}_{3} & = & \left(-x_{2}+\frac{1}{2}y_{2}\right)a \, \mathbf{\hat{x}}-\frac{\sqrt{3}}{2}y_{2}a \, \mathbf{\hat{y}} + z_{2}c \, \mathbf{\hat{z}} & \left(12d\right) & \mbox{O I} \\ 
\mathbf{B}_{16} & = & -x_{2} \, \mathbf{a}_{1}-y_{2} \, \mathbf{a}_{2} + z_{2} \, \mathbf{a}_{3} & = & -\frac{1}{2}\left(x_{2}+y_{2}\right)a \, \mathbf{\hat{x}} + \frac{\sqrt{3}}{2}\left(x_{2}-y_{2}\right)a \, \mathbf{\hat{y}} + z_{2}c \, \mathbf{\hat{z}} & \left(12d\right) & \mbox{O I} \\ 
\mathbf{B}_{17} & = & y_{2} \, \mathbf{a}_{1} + \left(-x_{2}+y_{2}\right) \, \mathbf{a}_{2} + z_{2} \, \mathbf{a}_{3} & = & \left(-\frac{1}{2}x_{2}+y_{2}\right)a \, \mathbf{\hat{x}}-\frac{\sqrt{3}}{2}x_{2}a \, \mathbf{\hat{y}} + z_{2}c \, \mathbf{\hat{z}} & \left(12d\right) & \mbox{O I} \\ 
\mathbf{B}_{18} & = & \left(x_{2}-y_{2}\right) \, \mathbf{a}_{1} + x_{2} \, \mathbf{a}_{2} + z_{2} \, \mathbf{a}_{3} & = & \left(x_{2}-\frac{1}{2}y_{2}\right)a \, \mathbf{\hat{x}} + \frac{\sqrt{3}}{2}y_{2}a \, \mathbf{\hat{y}} + z_{2}c \, \mathbf{\hat{z}} & \left(12d\right) & \mbox{O I} \\ 
\mathbf{B}_{19} & = & -y_{2} \, \mathbf{a}_{1}-x_{2} \, \mathbf{a}_{2} + \left(\frac{1}{2} +z_{2}\right) \, \mathbf{a}_{3} & = & -\frac{1}{2}\left(x_{2}+y_{2}\right)a \, \mathbf{\hat{x}} + \frac{\sqrt{3}}{2}\left(-x_{2}+y_{2}\right)a \, \mathbf{\hat{y}} + \left(\frac{1}{2} +z_{2}\right)c \, \mathbf{\hat{z}} & \left(12d\right) & \mbox{O I} \\ 
\mathbf{B}_{20} & = & \left(-x_{2}+y_{2}\right) \, \mathbf{a}_{1} + y_{2} \, \mathbf{a}_{2} + \left(\frac{1}{2} +z_{2}\right) \, \mathbf{a}_{3} & = & \left(-\frac{1}{2}x_{2}+y_{2}\right)a \, \mathbf{\hat{x}} + \frac{\sqrt{3}}{2}x_{2}a \, \mathbf{\hat{y}} + \left(\frac{1}{2} +z_{2}\right)c \, \mathbf{\hat{z}} & \left(12d\right) & \mbox{O I} \\ 
\mathbf{B}_{21} & = & x_{2} \, \mathbf{a}_{1} + \left(x_{2}-y_{2}\right) \, \mathbf{a}_{2} + \left(\frac{1}{2} +z_{2}\right) \, \mathbf{a}_{3} & = & \left(x_{2}-\frac{1}{2}y_{2}\right)a \, \mathbf{\hat{x}}-\frac{\sqrt{3}}{2}y_{2}a \, \mathbf{\hat{y}} + \left(\frac{1}{2} +z_{2}\right)c \, \mathbf{\hat{z}} & \left(12d\right) & \mbox{O I} \\ 
\mathbf{B}_{22} & = & y_{2} \, \mathbf{a}_{1} + x_{2} \, \mathbf{a}_{2} + \left(\frac{1}{2} +z_{2}\right) \, \mathbf{a}_{3} & = & \frac{1}{2}\left(x_{2}+y_{2}\right)a \, \mathbf{\hat{x}} + \frac{\sqrt{3}}{2}\left(x_{2}-y_{2}\right)a \, \mathbf{\hat{y}} + \left(\frac{1}{2} +z_{2}\right)c \, \mathbf{\hat{z}} & \left(12d\right) & \mbox{O I} \\ 
\mathbf{B}_{23} & = & \left(x_{2}-y_{2}\right) \, \mathbf{a}_{1}-y_{2} \, \mathbf{a}_{2} + \left(\frac{1}{2} +z_{2}\right) \, \mathbf{a}_{3} & = & \left(\frac{1}{2}x_{2}-y_{2}\right)a \, \mathbf{\hat{x}}-\frac{\sqrt{3}}{2}x_{2}a \, \mathbf{\hat{y}} + \left(\frac{1}{2} +z_{2}\right)c \, \mathbf{\hat{z}} & \left(12d\right) & \mbox{O I} \\ 
\mathbf{B}_{24} & = & -x_{2} \, \mathbf{a}_{1} + \left(-x_{2}+y_{2}\right) \, \mathbf{a}_{2} + \left(\frac{1}{2} +z_{2}\right) \, \mathbf{a}_{3} & = & \left(-x_{2}+\frac{1}{2}y_{2}\right)a \, \mathbf{\hat{x}} + \frac{\sqrt{3}}{2}y_{2}a \, \mathbf{\hat{y}} + \left(\frac{1}{2} +z_{2}\right)c \, \mathbf{\hat{z}} & \left(12d\right) & \mbox{O I} \\ 
\mathbf{B}_{25} & = & x_{3} \, \mathbf{a}_{1} + y_{3} \, \mathbf{a}_{2} + z_{3} \, \mathbf{a}_{3} & = & \frac{1}{2}\left(x_{3}+y_{3}\right)a \, \mathbf{\hat{x}} + \frac{\sqrt{3}}{2}\left(-x_{3}+y_{3}\right)a \, \mathbf{\hat{y}} + z_{3}c \, \mathbf{\hat{z}} & \left(12d\right) & \mbox{O II} \\ 
\mathbf{B}_{26} & = & -y_{3} \, \mathbf{a}_{1} + \left(x_{3}-y_{3}\right) \, \mathbf{a}_{2} + z_{3} \, \mathbf{a}_{3} & = & \left(\frac{1}{2}x_{3}-y_{3}\right)a \, \mathbf{\hat{x}} + \frac{\sqrt{3}}{2}x_{3}a \, \mathbf{\hat{y}} + z_{3}c \, \mathbf{\hat{z}} & \left(12d\right) & \mbox{O II} \\ 
\mathbf{B}_{27} & = & \left(-x_{3}+y_{3}\right) \, \mathbf{a}_{1}-x_{3} \, \mathbf{a}_{2} + z_{3} \, \mathbf{a}_{3} & = & \left(-x_{3}+\frac{1}{2}y_{3}\right)a \, \mathbf{\hat{x}}-\frac{\sqrt{3}}{2}y_{3}a \, \mathbf{\hat{y}} + z_{3}c \, \mathbf{\hat{z}} & \left(12d\right) & \mbox{O II} \\ 
\mathbf{B}_{28} & = & -x_{3} \, \mathbf{a}_{1}-y_{3} \, \mathbf{a}_{2} + z_{3} \, \mathbf{a}_{3} & = & -\frac{1}{2}\left(x_{3}+y_{3}\right)a \, \mathbf{\hat{x}} + \frac{\sqrt{3}}{2}\left(x_{3}-y_{3}\right)a \, \mathbf{\hat{y}} + z_{3}c \, \mathbf{\hat{z}} & \left(12d\right) & \mbox{O II} \\ 
\mathbf{B}_{29} & = & y_{3} \, \mathbf{a}_{1} + \left(-x_{3}+y_{3}\right) \, \mathbf{a}_{2} + z_{3} \, \mathbf{a}_{3} & = & \left(-\frac{1}{2}x_{3}+y_{3}\right)a \, \mathbf{\hat{x}}-\frac{\sqrt{3}}{2}x_{3}a \, \mathbf{\hat{y}} + z_{3}c \, \mathbf{\hat{z}} & \left(12d\right) & \mbox{O II} \\ 
\mathbf{B}_{30} & = & \left(x_{3}-y_{3}\right) \, \mathbf{a}_{1} + x_{3} \, \mathbf{a}_{2} + z_{3} \, \mathbf{a}_{3} & = & \left(x_{3}-\frac{1}{2}y_{3}\right)a \, \mathbf{\hat{x}} + \frac{\sqrt{3}}{2}y_{3}a \, \mathbf{\hat{y}} + z_{3}c \, \mathbf{\hat{z}} & \left(12d\right) & \mbox{O II} \\ 
\mathbf{B}_{31} & = & -y_{3} \, \mathbf{a}_{1}-x_{3} \, \mathbf{a}_{2} + \left(\frac{1}{2} +z_{3}\right) \, \mathbf{a}_{3} & = & -\frac{1}{2}\left(x_{3}+y_{3}\right)a \, \mathbf{\hat{x}} + \frac{\sqrt{3}}{2}\left(-x_{3}+y_{3}\right)a \, \mathbf{\hat{y}} + \left(\frac{1}{2} +z_{3}\right)c \, \mathbf{\hat{z}} & \left(12d\right) & \mbox{O II} \\ 
\mathbf{B}_{32} & = & \left(-x_{3}+y_{3}\right) \, \mathbf{a}_{1} + y_{3} \, \mathbf{a}_{2} + \left(\frac{1}{2} +z_{3}\right) \, \mathbf{a}_{3} & = & \left(-\frac{1}{2}x_{3}+y_{3}\right)a \, \mathbf{\hat{x}} + \frac{\sqrt{3}}{2}x_{3}a \, \mathbf{\hat{y}} + \left(\frac{1}{2} +z_{3}\right)c \, \mathbf{\hat{z}} & \left(12d\right) & \mbox{O II} \\ 
\mathbf{B}_{33} & = & x_{3} \, \mathbf{a}_{1} + \left(x_{3}-y_{3}\right) \, \mathbf{a}_{2} + \left(\frac{1}{2} +z_{3}\right) \, \mathbf{a}_{3} & = & \left(x_{3}-\frac{1}{2}y_{3}\right)a \, \mathbf{\hat{x}}-\frac{\sqrt{3}}{2}y_{3}a \, \mathbf{\hat{y}} + \left(\frac{1}{2} +z_{3}\right)c \, \mathbf{\hat{z}} & \left(12d\right) & \mbox{O II} \\ 
\mathbf{B}_{34} & = & y_{3} \, \mathbf{a}_{1} + x_{3} \, \mathbf{a}_{2} + \left(\frac{1}{2} +z_{3}\right) \, \mathbf{a}_{3} & = & \frac{1}{2}\left(x_{3}+y_{3}\right)a \, \mathbf{\hat{x}} + \frac{\sqrt{3}}{2}\left(x_{3}-y_{3}\right)a \, \mathbf{\hat{y}} + \left(\frac{1}{2} +z_{3}\right)c \, \mathbf{\hat{z}} & \left(12d\right) & \mbox{O II} \\ 
\mathbf{B}_{35} & = & \left(x_{3}-y_{3}\right) \, \mathbf{a}_{1}-y_{3} \, \mathbf{a}_{2} + \left(\frac{1}{2} +z_{3}\right) \, \mathbf{a}_{3} & = & \left(\frac{1}{2}x_{3}-y_{3}\right)a \, \mathbf{\hat{x}}-\frac{\sqrt{3}}{2}x_{3}a \, \mathbf{\hat{y}} + \left(\frac{1}{2} +z_{3}\right)c \, \mathbf{\hat{z}} & \left(12d\right) & \mbox{O II} \\ 
\mathbf{B}_{36} & = & -x_{3} \, \mathbf{a}_{1} + \left(-x_{3}+y_{3}\right) \, \mathbf{a}_{2} + \left(\frac{1}{2} +z_{3}\right) \, \mathbf{a}_{3} & = & \left(-x_{3}+\frac{1}{2}y_{3}\right)a \, \mathbf{\hat{x}} + \frac{\sqrt{3}}{2}y_{3}a \, \mathbf{\hat{y}} + \left(\frac{1}{2} +z_{3}\right)c \, \mathbf{\hat{z}} & \left(12d\right) & \mbox{O II} \\ 
\mathbf{B}_{37} & = & x_{4} \, \mathbf{a}_{1} + y_{4} \, \mathbf{a}_{2} + z_{4} \, \mathbf{a}_{3} & = & \frac{1}{2}\left(x_{4}+y_{4}\right)a \, \mathbf{\hat{x}} + \frac{\sqrt{3}}{2}\left(-x_{4}+y_{4}\right)a \, \mathbf{\hat{y}} + z_{4}c \, \mathbf{\hat{z}} & \left(12d\right) & \mbox{O III} \\ 
\mathbf{B}_{38} & = & -y_{4} \, \mathbf{a}_{1} + \left(x_{4}-y_{4}\right) \, \mathbf{a}_{2} + z_{4} \, \mathbf{a}_{3} & = & \left(\frac{1}{2}x_{4}-y_{4}\right)a \, \mathbf{\hat{x}} + \frac{\sqrt{3}}{2}x_{4}a \, \mathbf{\hat{y}} + z_{4}c \, \mathbf{\hat{z}} & \left(12d\right) & \mbox{O III} \\ 
\mathbf{B}_{39} & = & \left(-x_{4}+y_{4}\right) \, \mathbf{a}_{1}-x_{4} \, \mathbf{a}_{2} + z_{4} \, \mathbf{a}_{3} & = & \left(-x_{4}+\frac{1}{2}y_{4}\right)a \, \mathbf{\hat{x}}-\frac{\sqrt{3}}{2}y_{4}a \, \mathbf{\hat{y}} + z_{4}c \, \mathbf{\hat{z}} & \left(12d\right) & \mbox{O III} \\ 
\mathbf{B}_{40} & = & -x_{4} \, \mathbf{a}_{1}-y_{4} \, \mathbf{a}_{2} + z_{4} \, \mathbf{a}_{3} & = & -\frac{1}{2}\left(x_{4}+y_{4}\right)a \, \mathbf{\hat{x}} + \frac{\sqrt{3}}{2}\left(x_{4}-y_{4}\right)a \, \mathbf{\hat{y}} + z_{4}c \, \mathbf{\hat{z}} & \left(12d\right) & \mbox{O III} \\ 
\mathbf{B}_{41} & = & y_{4} \, \mathbf{a}_{1} + \left(-x_{4}+y_{4}\right) \, \mathbf{a}_{2} + z_{4} \, \mathbf{a}_{3} & = & \left(-\frac{1}{2}x_{4}+y_{4}\right)a \, \mathbf{\hat{x}}-\frac{\sqrt{3}}{2}x_{4}a \, \mathbf{\hat{y}} + z_{4}c \, \mathbf{\hat{z}} & \left(12d\right) & \mbox{O III} \\ 
\mathbf{B}_{42} & = & \left(x_{4}-y_{4}\right) \, \mathbf{a}_{1} + x_{4} \, \mathbf{a}_{2} + z_{4} \, \mathbf{a}_{3} & = & \left(x_{4}-\frac{1}{2}y_{4}\right)a \, \mathbf{\hat{x}} + \frac{\sqrt{3}}{2}y_{4}a \, \mathbf{\hat{y}} + z_{4}c \, \mathbf{\hat{z}} & \left(12d\right) & \mbox{O III} \\ 
\mathbf{B}_{43} & = & -y_{4} \, \mathbf{a}_{1}-x_{4} \, \mathbf{a}_{2} + \left(\frac{1}{2} +z_{4}\right) \, \mathbf{a}_{3} & = & -\frac{1}{2}\left(x_{4}+y_{4}\right)a \, \mathbf{\hat{x}} + \frac{\sqrt{3}}{2}\left(-x_{4}+y_{4}\right)a \, \mathbf{\hat{y}} + \left(\frac{1}{2} +z_{4}\right)c \, \mathbf{\hat{z}} & \left(12d\right) & \mbox{O III} \\ 
\mathbf{B}_{44} & = & \left(-x_{4}+y_{4}\right) \, \mathbf{a}_{1} + y_{4} \, \mathbf{a}_{2} + \left(\frac{1}{2} +z_{4}\right) \, \mathbf{a}_{3} & = & \left(-\frac{1}{2}x_{4}+y_{4}\right)a \, \mathbf{\hat{x}} + \frac{\sqrt{3}}{2}x_{4}a \, \mathbf{\hat{y}} + \left(\frac{1}{2} +z_{4}\right)c \, \mathbf{\hat{z}} & \left(12d\right) & \mbox{O III} \\ 
\mathbf{B}_{45} & = & x_{4} \, \mathbf{a}_{1} + \left(x_{4}-y_{4}\right) \, \mathbf{a}_{2} + \left(\frac{1}{2} +z_{4}\right) \, \mathbf{a}_{3} & = & \left(x_{4}-\frac{1}{2}y_{4}\right)a \, \mathbf{\hat{x}}-\frac{\sqrt{3}}{2}y_{4}a \, \mathbf{\hat{y}} + \left(\frac{1}{2} +z_{4}\right)c \, \mathbf{\hat{z}} & \left(12d\right) & \mbox{O III} \\ 
\mathbf{B}_{46} & = & y_{4} \, \mathbf{a}_{1} + x_{4} \, \mathbf{a}_{2} + \left(\frac{1}{2} +z_{4}\right) \, \mathbf{a}_{3} & = & \frac{1}{2}\left(x_{4}+y_{4}\right)a \, \mathbf{\hat{x}} + \frac{\sqrt{3}}{2}\left(x_{4}-y_{4}\right)a \, \mathbf{\hat{y}} + \left(\frac{1}{2} +z_{4}\right)c \, \mathbf{\hat{z}} & \left(12d\right) & \mbox{O III} \\ 
\mathbf{B}_{47} & = & \left(x_{4}-y_{4}\right) \, \mathbf{a}_{1}-y_{4} \, \mathbf{a}_{2} + \left(\frac{1}{2} +z_{4}\right) \, \mathbf{a}_{3} & = & \left(\frac{1}{2}x_{4}-y_{4}\right)a \, \mathbf{\hat{x}}-\frac{\sqrt{3}}{2}x_{4}a \, \mathbf{\hat{y}} + \left(\frac{1}{2} +z_{4}\right)c \, \mathbf{\hat{z}} & \left(12d\right) & \mbox{O III} \\ 
\mathbf{B}_{48} & = & -x_{4} \, \mathbf{a}_{1} + \left(-x_{4}+y_{4}\right) \, \mathbf{a}_{2} + \left(\frac{1}{2} +z_{4}\right) \, \mathbf{a}_{3} & = & \left(-x_{4}+\frac{1}{2}y_{4}\right)a \, \mathbf{\hat{x}} + \frac{\sqrt{3}}{2}y_{4}a \, \mathbf{\hat{y}} + \left(\frac{1}{2} +z_{4}\right)c \, \mathbf{\hat{z}} & \left(12d\right) & \mbox{O III} \\ 
\mathbf{B}_{49} & = & x_{5} \, \mathbf{a}_{1} + y_{5} \, \mathbf{a}_{2} + z_{5} \, \mathbf{a}_{3} & = & \frac{1}{2}\left(x_{5}+y_{5}\right)a \, \mathbf{\hat{x}} + \frac{\sqrt{3}}{2}\left(-x_{5}+y_{5}\right)a \, \mathbf{\hat{y}} + z_{5}c \, \mathbf{\hat{z}} & \left(12d\right) & \mbox{O IV} \\ 
\mathbf{B}_{50} & = & -y_{5} \, \mathbf{a}_{1} + \left(x_{5}-y_{5}\right) \, \mathbf{a}_{2} + z_{5} \, \mathbf{a}_{3} & = & \left(\frac{1}{2}x_{5}-y_{5}\right)a \, \mathbf{\hat{x}} + \frac{\sqrt{3}}{2}x_{5}a \, \mathbf{\hat{y}} + z_{5}c \, \mathbf{\hat{z}} & \left(12d\right) & \mbox{O IV} \\ 
\mathbf{B}_{51} & = & \left(-x_{5}+y_{5}\right) \, \mathbf{a}_{1}-x_{5} \, \mathbf{a}_{2} + z_{5} \, \mathbf{a}_{3} & = & \left(-x_{5}+\frac{1}{2}y_{5}\right)a \, \mathbf{\hat{x}}-\frac{\sqrt{3}}{2}y_{5}a \, \mathbf{\hat{y}} + z_{5}c \, \mathbf{\hat{z}} & \left(12d\right) & \mbox{O IV} \\ 
\mathbf{B}_{52} & = & -x_{5} \, \mathbf{a}_{1}-y_{5} \, \mathbf{a}_{2} + z_{5} \, \mathbf{a}_{3} & = & -\frac{1}{2}\left(x_{5}+y_{5}\right)a \, \mathbf{\hat{x}} + \frac{\sqrt{3}}{2}\left(x_{5}-y_{5}\right)a \, \mathbf{\hat{y}} + z_{5}c \, \mathbf{\hat{z}} & \left(12d\right) & \mbox{O IV} \\ 
\mathbf{B}_{53} & = & y_{5} \, \mathbf{a}_{1} + \left(-x_{5}+y_{5}\right) \, \mathbf{a}_{2} + z_{5} \, \mathbf{a}_{3} & = & \left(-\frac{1}{2}x_{5}+y_{5}\right)a \, \mathbf{\hat{x}}-\frac{\sqrt{3}}{2}x_{5}a \, \mathbf{\hat{y}} + z_{5}c \, \mathbf{\hat{z}} & \left(12d\right) & \mbox{O IV} \\ 
\mathbf{B}_{54} & = & \left(x_{5}-y_{5}\right) \, \mathbf{a}_{1} + x_{5} \, \mathbf{a}_{2} + z_{5} \, \mathbf{a}_{3} & = & \left(x_{5}-\frac{1}{2}y_{5}\right)a \, \mathbf{\hat{x}} + \frac{\sqrt{3}}{2}y_{5}a \, \mathbf{\hat{y}} + z_{5}c \, \mathbf{\hat{z}} & \left(12d\right) & \mbox{O IV} \\ 
\mathbf{B}_{55} & = & -y_{5} \, \mathbf{a}_{1}-x_{5} \, \mathbf{a}_{2} + \left(\frac{1}{2} +z_{5}\right) \, \mathbf{a}_{3} & = & -\frac{1}{2}\left(x_{5}+y_{5}\right)a \, \mathbf{\hat{x}} + \frac{\sqrt{3}}{2}\left(-x_{5}+y_{5}\right)a \, \mathbf{\hat{y}} + \left(\frac{1}{2} +z_{5}\right)c \, \mathbf{\hat{z}} & \left(12d\right) & \mbox{O IV} \\ 
\mathbf{B}_{56} & = & \left(-x_{5}+y_{5}\right) \, \mathbf{a}_{1} + y_{5} \, \mathbf{a}_{2} + \left(\frac{1}{2} +z_{5}\right) \, \mathbf{a}_{3} & = & \left(-\frac{1}{2}x_{5}+y_{5}\right)a \, \mathbf{\hat{x}} + \frac{\sqrt{3}}{2}x_{5}a \, \mathbf{\hat{y}} + \left(\frac{1}{2} +z_{5}\right)c \, \mathbf{\hat{z}} & \left(12d\right) & \mbox{O IV} \\ 
\mathbf{B}_{57} & = & x_{5} \, \mathbf{a}_{1} + \left(x_{5}-y_{5}\right) \, \mathbf{a}_{2} + \left(\frac{1}{2} +z_{5}\right) \, \mathbf{a}_{3} & = & \left(x_{5}-\frac{1}{2}y_{5}\right)a \, \mathbf{\hat{x}}-\frac{\sqrt{3}}{2}y_{5}a \, \mathbf{\hat{y}} + \left(\frac{1}{2} +z_{5}\right)c \, \mathbf{\hat{z}} & \left(12d\right) & \mbox{O IV} \\ 
\mathbf{B}_{58} & = & y_{5} \, \mathbf{a}_{1} + x_{5} \, \mathbf{a}_{2} + \left(\frac{1}{2} +z_{5}\right) \, \mathbf{a}_{3} & = & \frac{1}{2}\left(x_{5}+y_{5}\right)a \, \mathbf{\hat{x}} + \frac{\sqrt{3}}{2}\left(x_{5}-y_{5}\right)a \, \mathbf{\hat{y}} + \left(\frac{1}{2} +z_{5}\right)c \, \mathbf{\hat{z}} & \left(12d\right) & \mbox{O IV} \\ 
\mathbf{B}_{59} & = & \left(x_{5}-y_{5}\right) \, \mathbf{a}_{1}-y_{5} \, \mathbf{a}_{2} + \left(\frac{1}{2} +z_{5}\right) \, \mathbf{a}_{3} & = & \left(\frac{1}{2}x_{5}-y_{5}\right)a \, \mathbf{\hat{x}}-\frac{\sqrt{3}}{2}x_{5}a \, \mathbf{\hat{y}} + \left(\frac{1}{2} +z_{5}\right)c \, \mathbf{\hat{z}} & \left(12d\right) & \mbox{O IV} \\ 
\mathbf{B}_{60} & = & -x_{5} \, \mathbf{a}_{1} + \left(-x_{5}+y_{5}\right) \, \mathbf{a}_{2} + \left(\frac{1}{2} +z_{5}\right) \, \mathbf{a}_{3} & = & \left(-x_{5}+\frac{1}{2}y_{5}\right)a \, \mathbf{\hat{x}} + \frac{\sqrt{3}}{2}y_{5}a \, \mathbf{\hat{y}} + \left(\frac{1}{2} +z_{5}\right)c \, \mathbf{\hat{z}} & \left(12d\right) & \mbox{O IV} \\ 
\mathbf{B}_{61} & = & x_{6} \, \mathbf{a}_{1} + y_{6} \, \mathbf{a}_{2} + z_{6} \, \mathbf{a}_{3} & = & \frac{1}{2}\left(x_{6}+y_{6}\right)a \, \mathbf{\hat{x}} + \frac{\sqrt{3}}{2}\left(-x_{6}+y_{6}\right)a \, \mathbf{\hat{y}} + z_{6}c \, \mathbf{\hat{z}} & \left(12d\right) & \mbox{P} \\ 
\mathbf{B}_{62} & = & -y_{6} \, \mathbf{a}_{1} + \left(x_{6}-y_{6}\right) \, \mathbf{a}_{2} + z_{6} \, \mathbf{a}_{3} & = & \left(\frac{1}{2}x_{6}-y_{6}\right)a \, \mathbf{\hat{x}} + \frac{\sqrt{3}}{2}x_{6}a \, \mathbf{\hat{y}} + z_{6}c \, \mathbf{\hat{z}} & \left(12d\right) & \mbox{P} \\ 
\mathbf{B}_{63} & = & \left(-x_{6}+y_{6}\right) \, \mathbf{a}_{1}-x_{6} \, \mathbf{a}_{2} + z_{6} \, \mathbf{a}_{3} & = & \left(-x_{6}+\frac{1}{2}y_{6}\right)a \, \mathbf{\hat{x}}-\frac{\sqrt{3}}{2}y_{6}a \, \mathbf{\hat{y}} + z_{6}c \, \mathbf{\hat{z}} & \left(12d\right) & \mbox{P} \\ 
\mathbf{B}_{64} & = & -x_{6} \, \mathbf{a}_{1}-y_{6} \, \mathbf{a}_{2} + z_{6} \, \mathbf{a}_{3} & = & -\frac{1}{2}\left(x_{6}+y_{6}\right)a \, \mathbf{\hat{x}} + \frac{\sqrt{3}}{2}\left(x_{6}-y_{6}\right)a \, \mathbf{\hat{y}} + z_{6}c \, \mathbf{\hat{z}} & \left(12d\right) & \mbox{P} \\ 
\mathbf{B}_{65} & = & y_{6} \, \mathbf{a}_{1} + \left(-x_{6}+y_{6}\right) \, \mathbf{a}_{2} + z_{6} \, \mathbf{a}_{3} & = & \left(-\frac{1}{2}x_{6}+y_{6}\right)a \, \mathbf{\hat{x}}-\frac{\sqrt{3}}{2}x_{6}a \, \mathbf{\hat{y}} + z_{6}c \, \mathbf{\hat{z}} & \left(12d\right) & \mbox{P} \\ 
\mathbf{B}_{66} & = & \left(x_{6}-y_{6}\right) \, \mathbf{a}_{1} + x_{6} \, \mathbf{a}_{2} + z_{6} \, \mathbf{a}_{3} & = & \left(x_{6}-\frac{1}{2}y_{6}\right)a \, \mathbf{\hat{x}} + \frac{\sqrt{3}}{2}y_{6}a \, \mathbf{\hat{y}} + z_{6}c \, \mathbf{\hat{z}} & \left(12d\right) & \mbox{P} \\ 
\mathbf{B}_{67} & = & -y_{6} \, \mathbf{a}_{1}-x_{6} \, \mathbf{a}_{2} + \left(\frac{1}{2} +z_{6}\right) \, \mathbf{a}_{3} & = & -\frac{1}{2}\left(x_{6}+y_{6}\right)a \, \mathbf{\hat{x}} + \frac{\sqrt{3}}{2}\left(-x_{6}+y_{6}\right)a \, \mathbf{\hat{y}} + \left(\frac{1}{2} +z_{6}\right)c \, \mathbf{\hat{z}} & \left(12d\right) & \mbox{P} \\ 
\mathbf{B}_{68} & = & \left(-x_{6}+y_{6}\right) \, \mathbf{a}_{1} + y_{6} \, \mathbf{a}_{2} + \left(\frac{1}{2} +z_{6}\right) \, \mathbf{a}_{3} & = & \left(-\frac{1}{2}x_{6}+y_{6}\right)a \, \mathbf{\hat{x}} + \frac{\sqrt{3}}{2}x_{6}a \, \mathbf{\hat{y}} + \left(\frac{1}{2} +z_{6}\right)c \, \mathbf{\hat{z}} & \left(12d\right) & \mbox{P} \\ 
\mathbf{B}_{69} & = & x_{6} \, \mathbf{a}_{1} + \left(x_{6}-y_{6}\right) \, \mathbf{a}_{2} + \left(\frac{1}{2} +z_{6}\right) \, \mathbf{a}_{3} & = & \left(x_{6}-\frac{1}{2}y_{6}\right)a \, \mathbf{\hat{x}}-\frac{\sqrt{3}}{2}y_{6}a \, \mathbf{\hat{y}} + \left(\frac{1}{2} +z_{6}\right)c \, \mathbf{\hat{z}} & \left(12d\right) & \mbox{P} \\ 
\mathbf{B}_{70} & = & y_{6} \, \mathbf{a}_{1} + x_{6} \, \mathbf{a}_{2} + \left(\frac{1}{2} +z_{6}\right) \, \mathbf{a}_{3} & = & \frac{1}{2}\left(x_{6}+y_{6}\right)a \, \mathbf{\hat{x}} + \frac{\sqrt{3}}{2}\left(x_{6}-y_{6}\right)a \, \mathbf{\hat{y}} + \left(\frac{1}{2} +z_{6}\right)c \, \mathbf{\hat{z}} & \left(12d\right) & \mbox{P} \\ 
\mathbf{B}_{71} & = & \left(x_{6}-y_{6}\right) \, \mathbf{a}_{1}-y_{6} \, \mathbf{a}_{2} + \left(\frac{1}{2} +z_{6}\right) \, \mathbf{a}_{3} & = & \left(\frac{1}{2}x_{6}-y_{6}\right)a \, \mathbf{\hat{x}}-\frac{\sqrt{3}}{2}x_{6}a \, \mathbf{\hat{y}} + \left(\frac{1}{2} +z_{6}\right)c \, \mathbf{\hat{z}} & \left(12d\right) & \mbox{P} \\ 
\mathbf{B}_{72} & = & -x_{6} \, \mathbf{a}_{1} + \left(-x_{6}+y_{6}\right) \, \mathbf{a}_{2} + \left(\frac{1}{2} +z_{6}\right) \, \mathbf{a}_{3} & = & \left(-x_{6}+\frac{1}{2}y_{6}\right)a \, \mathbf{\hat{x}} + \frac{\sqrt{3}}{2}y_{6}a \, \mathbf{\hat{y}} + \left(\frac{1}{2} +z_{6}\right)c \, \mathbf{\hat{z}} & \left(12d\right) & \mbox{P} \\ 
\end{longtabu}
\renewcommand{\arraystretch}{1.0}
\noindent \hrulefill
\\
\textbf{References:}
\vspace*{-0.25cm}
\begin{flushleft}
  - \bibentry{Klap_AlPO4_MicroMesoMater_2000}. \\
\end{flushleft}
\textbf{Found in:}
\vspace*{-0.25cm}
\begin{flushleft}
  - \bibentry{Villars_PearsonsCrystalData_2013}. \\
\end{flushleft}
\noindent \hrulefill
\\
\textbf{Geometry files:}
\\
\noindent  - CIF: pp. {\hyperref[AB4C_hP72_184_d_4d_d_cif]{\pageref{AB4C_hP72_184_d_4d_d_cif}}} \\
\noindent  - POSCAR: pp. {\hyperref[AB4C_hP72_184_d_4d_d_poscar]{\pageref{AB4C_hP72_184_d_4d_d_poscar}}} \\
\onecolumn
{\phantomsection\label{A3BC_hP30_185_cd_c_ab}}
\subsection*{\huge \textbf{{\normalfont \begin{raggedleft}KNiCl$_{3}$ (Room-temperature) Structure: \end{raggedleft} \\ A3BC\_hP30\_185\_cd\_c\_ab}}}
\noindent \hrulefill
\vspace*{0.25cm}
\begin{figure}[htp]
  \centering
  \vspace{-1em}
  {\includegraphics[width=1\textwidth]{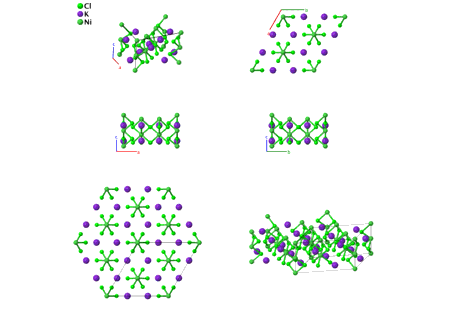}}
\end{figure}
\vspace*{-0.5cm}
\renewcommand{\arraystretch}{1.5}
\begin{equation*}
  \begin{array}{>{$\hspace{-0.15cm}}l<{$}>{$}p{0.5cm}<{$}>{$}p{18.5cm}<{$}}
    \mbox{\large \textbf{Prototype}} &\colon & \ce{KNiCl3} \\
    \mbox{\large \textbf{\AFLOW\ prototype label}} &\colon & \mbox{A3BC\_hP30\_185\_cd\_c\_ab} \\
    \mbox{\large \textbf{\textit{Strukturbericht} designation}} &\colon & \mbox{None} \\
    \mbox{\large \textbf{Pearson symbol}} &\colon & \mbox{hP30} \\
    \mbox{\large \textbf{Space group number}} &\colon & 185 \\
    \mbox{\large \textbf{Space group symbol}} &\colon & P6_{3}cm \\
    \mbox{\large \textbf{\AFLOW\ prototype command}} &\colon &  \texttt{aflow} \,  \, \texttt{-{}-proto=A3BC\_hP30\_185\_cd\_c\_ab } \, \newline \texttt{-{}-params=}{a,c/a,z_{1},z_{2},x_{3},z_{3},x_{4},z_{4},x_{5},y_{5},z_{5} }
  \end{array}
\end{equation*}
\renewcommand{\arraystretch}{1.0}

\noindent \parbox{1 \linewidth}{
\noindent \hrulefill
\\
\textbf{Hexagonal primitive vectors:} \\
\vspace*{-0.25cm}
\begin{tabular}{cc}
  \begin{tabular}{c}
    \parbox{0.6 \linewidth}{
      \renewcommand{\arraystretch}{1.5}
      \begin{equation*}
        \centering
        \begin{array}{ccc}
              \mathbf{a}_1 & = & \frac12 \, a \, \mathbf{\hat{x}} - \frac{\sqrt3}2 \, a \, \mathbf{\hat{y}} \\
    \mathbf{a}_2 & = & \frac12 \, a \, \mathbf{\hat{x}} + \frac{\sqrt3}2 \, a \, \mathbf{\hat{y}} \\
    \mathbf{a}_3 & = & c \, \mathbf{\hat{z}} \\

        \end{array}
      \end{equation*}
    }
    \renewcommand{\arraystretch}{1.0}
  \end{tabular}
  \begin{tabular}{c}
    \includegraphics[width=0.3\linewidth]{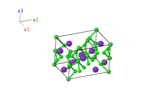} \\
  \end{tabular}
\end{tabular}

}
\vspace*{-0.25cm}

\noindent \hrulefill
\\
\textbf{Basis vectors:}
\vspace*{-0.25cm}
\renewcommand{\arraystretch}{1.5}
\begin{longtabu} to \textwidth{>{\centering $}X[-1,c,c]<{$}>{\centering $}X[-1,c,c]<{$}>{\centering $}X[-1,c,c]<{$}>{\centering $}X[-1,c,c]<{$}>{\centering $}X[-1,c,c]<{$}>{\centering $}X[-1,c,c]<{$}>{\centering $}X[-1,c,c]<{$}}
  & & \mbox{Lattice Coordinates} & & \mbox{Cartesian Coordinates} &\mbox{Wyckoff Position} & \mbox{Atom Type} \\  
  \mathbf{B}_{1} & = & z_{1} \, \mathbf{a}_{3} & = & z_{1}c \, \mathbf{\hat{z}} & \left(2a\right) & \mbox{Ni I} \\ 
\mathbf{B}_{2} & = & \left(\frac{1}{2} +z_{1}\right) \, \mathbf{a}_{3} & = & \left(\frac{1}{2} +z_{1}\right)c \, \mathbf{\hat{z}} & \left(2a\right) & \mbox{Ni I} \\ 
\mathbf{B}_{3} & = & \frac{1}{3} \, \mathbf{a}_{1} + \frac{2}{3} \, \mathbf{a}_{2} + z_{2} \, \mathbf{a}_{3} & = & \frac{1}{2}a \, \mathbf{\hat{x}} + \frac{1}{2\sqrt{3}}a \, \mathbf{\hat{y}} + z_{2}c \, \mathbf{\hat{z}} & \left(4b\right) & \mbox{Ni II} \\ 
\mathbf{B}_{4} & = & \frac{2}{3} \, \mathbf{a}_{1} + \frac{1}{3} \, \mathbf{a}_{2} + \left(\frac{1}{2} +z_{2}\right) \, \mathbf{a}_{3} & = & \frac{1}{2}a \, \mathbf{\hat{x}}- \frac{1}{2\sqrt{3}}a  \, \mathbf{\hat{y}} + \left(\frac{1}{2} +z_{2}\right)c \, \mathbf{\hat{z}} & \left(4b\right) & \mbox{Ni II} \\ 
\mathbf{B}_{5} & = & \frac{1}{3} \, \mathbf{a}_{1} + \frac{2}{3} \, \mathbf{a}_{2} + \left(\frac{1}{2} +z_{2}\right) \, \mathbf{a}_{3} & = & \frac{1}{2}a \, \mathbf{\hat{x}} + \frac{1}{2\sqrt{3}}a \, \mathbf{\hat{y}} + \left(\frac{1}{2} +z_{2}\right)c \, \mathbf{\hat{z}} & \left(4b\right) & \mbox{Ni II} \\ 
\mathbf{B}_{6} & = & \frac{2}{3} \, \mathbf{a}_{1} + \frac{1}{3} \, \mathbf{a}_{2} + z_{2} \, \mathbf{a}_{3} & = & \frac{1}{2}a \, \mathbf{\hat{x}}- \frac{1}{2\sqrt{3}}a  \, \mathbf{\hat{y}} + z_{2}c \, \mathbf{\hat{z}} & \left(4b\right) & \mbox{Ni II} \\ 
\mathbf{B}_{7} & = & x_{3} \, \mathbf{a}_{1} + z_{3} \, \mathbf{a}_{3} & = & \frac{1}{2}x_{3}a \, \mathbf{\hat{x}}-\frac{\sqrt{3}}{2}x_{3}a \, \mathbf{\hat{y}} + z_{3}c \, \mathbf{\hat{z}} & \left(6c\right) & \mbox{Cl I} \\ 
\mathbf{B}_{8} & = & x_{3} \, \mathbf{a}_{2} + z_{3} \, \mathbf{a}_{3} & = & \frac{1}{2}x_{3}a \, \mathbf{\hat{x}} + \frac{\sqrt{3}}{2}x_{3}a \, \mathbf{\hat{y}} + z_{3}c \, \mathbf{\hat{z}} & \left(6c\right) & \mbox{Cl I} \\ 
\mathbf{B}_{9} & = & -x_{3} \, \mathbf{a}_{1}-x_{3} \, \mathbf{a}_{2} + z_{3} \, \mathbf{a}_{3} & = & -x_{3}a \, \mathbf{\hat{x}} + z_{3}c \, \mathbf{\hat{z}} & \left(6c\right) & \mbox{Cl I} \\ 
\mathbf{B}_{10} & = & -x_{3} \, \mathbf{a}_{1} + \left(\frac{1}{2} +z_{3}\right) \, \mathbf{a}_{3} & = & -\frac{1}{2}x_{3}a \, \mathbf{\hat{x}} + \frac{\sqrt{3}}{2}x_{3}a \, \mathbf{\hat{y}} + \left(\frac{1}{2} +z_{3}\right)c \, \mathbf{\hat{z}} & \left(6c\right) & \mbox{Cl I} \\ 
\mathbf{B}_{11} & = & -x_{3} \, \mathbf{a}_{2} + \left(\frac{1}{2} +z_{3}\right) \, \mathbf{a}_{3} & = & -\frac{1}{2}x_{3}a \, \mathbf{\hat{x}}-\frac{\sqrt{3}}{2}x_{3}a \, \mathbf{\hat{y}} + \left(\frac{1}{2} +z_{3}\right)c \, \mathbf{\hat{z}} & \left(6c\right) & \mbox{Cl I} \\ 
\mathbf{B}_{12} & = & x_{3} \, \mathbf{a}_{1} + x_{3} \, \mathbf{a}_{2} + \left(\frac{1}{2} +z_{3}\right) \, \mathbf{a}_{3} & = & x_{3}a \, \mathbf{\hat{x}} + \left(\frac{1}{2} +z_{3}\right)c \, \mathbf{\hat{z}} & \left(6c\right) & \mbox{Cl I} \\ 
\mathbf{B}_{13} & = & x_{4} \, \mathbf{a}_{1} + z_{4} \, \mathbf{a}_{3} & = & \frac{1}{2}x_{4}a \, \mathbf{\hat{x}}-\frac{\sqrt{3}}{2}x_{4}a \, \mathbf{\hat{y}} + z_{4}c \, \mathbf{\hat{z}} & \left(6c\right) & \mbox{K} \\ 
\mathbf{B}_{14} & = & x_{4} \, \mathbf{a}_{2} + z_{4} \, \mathbf{a}_{3} & = & \frac{1}{2}x_{4}a \, \mathbf{\hat{x}} + \frac{\sqrt{3}}{2}x_{4}a \, \mathbf{\hat{y}} + z_{4}c \, \mathbf{\hat{z}} & \left(6c\right) & \mbox{K} \\ 
\mathbf{B}_{15} & = & -x_{4} \, \mathbf{a}_{1}-x_{4} \, \mathbf{a}_{2} + z_{4} \, \mathbf{a}_{3} & = & -x_{4}a \, \mathbf{\hat{x}} + z_{4}c \, \mathbf{\hat{z}} & \left(6c\right) & \mbox{K} \\ 
\mathbf{B}_{16} & = & -x_{4} \, \mathbf{a}_{1} + \left(\frac{1}{2} +z_{4}\right) \, \mathbf{a}_{3} & = & -\frac{1}{2}x_{4}a \, \mathbf{\hat{x}} + \frac{\sqrt{3}}{2}x_{4}a \, \mathbf{\hat{y}} + \left(\frac{1}{2} +z_{4}\right)c \, \mathbf{\hat{z}} & \left(6c\right) & \mbox{K} \\ 
\mathbf{B}_{17} & = & -x_{4} \, \mathbf{a}_{2} + \left(\frac{1}{2} +z_{4}\right) \, \mathbf{a}_{3} & = & -\frac{1}{2}x_{4}a \, \mathbf{\hat{x}}-\frac{\sqrt{3}}{2}x_{4}a \, \mathbf{\hat{y}} + \left(\frac{1}{2} +z_{4}\right)c \, \mathbf{\hat{z}} & \left(6c\right) & \mbox{K} \\ 
\mathbf{B}_{18} & = & x_{4} \, \mathbf{a}_{1} + x_{4} \, \mathbf{a}_{2} + \left(\frac{1}{2} +z_{4}\right) \, \mathbf{a}_{3} & = & x_{4}a \, \mathbf{\hat{x}} + \left(\frac{1}{2} +z_{4}\right)c \, \mathbf{\hat{z}} & \left(6c\right) & \mbox{K} \\ 
\mathbf{B}_{19} & = & x_{5} \, \mathbf{a}_{1} + y_{5} \, \mathbf{a}_{2} + z_{5} \, \mathbf{a}_{3} & = & \frac{1}{2}\left(x_{5}+y_{5}\right)a \, \mathbf{\hat{x}} + \frac{\sqrt{3}}{2}\left(-x_{5}+y_{5}\right)a \, \mathbf{\hat{y}} + z_{5}c \, \mathbf{\hat{z}} & \left(12d\right) & \mbox{Cl II} \\ 
\mathbf{B}_{20} & = & -y_{5} \, \mathbf{a}_{1} + \left(x_{5}-y_{5}\right) \, \mathbf{a}_{2} + z_{5} \, \mathbf{a}_{3} & = & \left(\frac{1}{2}x_{5}-y_{5}\right)a \, \mathbf{\hat{x}} + \frac{\sqrt{3}}{2}x_{5}a \, \mathbf{\hat{y}} + z_{5}c \, \mathbf{\hat{z}} & \left(12d\right) & \mbox{Cl II} \\ 
\mathbf{B}_{21} & = & \left(-x_{5}+y_{5}\right) \, \mathbf{a}_{1}-x_{5} \, \mathbf{a}_{2} + z_{5} \, \mathbf{a}_{3} & = & \left(-x_{5}+\frac{1}{2}y_{5}\right)a \, \mathbf{\hat{x}}-\frac{\sqrt{3}}{2}y_{5}a \, \mathbf{\hat{y}} + z_{5}c \, \mathbf{\hat{z}} & \left(12d\right) & \mbox{Cl II} \\ 
\mathbf{B}_{22} & = & -x_{5} \, \mathbf{a}_{1}-y_{5} \, \mathbf{a}_{2} + \left(\frac{1}{2} +z_{5}\right) \, \mathbf{a}_{3} & = & -\frac{1}{2}\left(x_{5}+y_{5}\right)a \, \mathbf{\hat{x}} + \frac{\sqrt{3}}{2}\left(x_{5}-y_{5}\right)a \, \mathbf{\hat{y}} + \left(\frac{1}{2} +z_{5}\right)c \, \mathbf{\hat{z}} & \left(12d\right) & \mbox{Cl II} \\ 
\mathbf{B}_{23} & = & y_{5} \, \mathbf{a}_{1} + \left(-x_{5}+y_{5}\right) \, \mathbf{a}_{2} + \left(\frac{1}{2} +z_{5}\right) \, \mathbf{a}_{3} & = & \left(-\frac{1}{2}x_{5}+y_{5}\right)a \, \mathbf{\hat{x}}-\frac{\sqrt{3}}{2}x_{5}a \, \mathbf{\hat{y}} + \left(\frac{1}{2} +z_{5}\right)c \, \mathbf{\hat{z}} & \left(12d\right) & \mbox{Cl II} \\ 
\mathbf{B}_{24} & = & \left(x_{5}-y_{5}\right) \, \mathbf{a}_{1} + x_{5} \, \mathbf{a}_{2} + \left(\frac{1}{2} +z_{5}\right) \, \mathbf{a}_{3} & = & \left(x_{5}-\frac{1}{2}y_{5}\right)a \, \mathbf{\hat{x}} + \frac{\sqrt{3}}{2}y_{5}a \, \mathbf{\hat{y}} + \left(\frac{1}{2} +z_{5}\right)c \, \mathbf{\hat{z}} & \left(12d\right) & \mbox{Cl II} \\ 
\mathbf{B}_{25} & = & -y_{5} \, \mathbf{a}_{1}-x_{5} \, \mathbf{a}_{2} + \left(\frac{1}{2} +z_{5}\right) \, \mathbf{a}_{3} & = & -\frac{1}{2}\left(x_{5}+y_{5}\right)a \, \mathbf{\hat{x}} + \frac{\sqrt{3}}{2}\left(-x_{5}+y_{5}\right)a \, \mathbf{\hat{y}} + \left(\frac{1}{2} +z_{5}\right)c \, \mathbf{\hat{z}} & \left(12d\right) & \mbox{Cl II} \\ 
\mathbf{B}_{26} & = & \left(-x_{5}+y_{5}\right) \, \mathbf{a}_{1} + y_{5} \, \mathbf{a}_{2} + \left(\frac{1}{2} +z_{5}\right) \, \mathbf{a}_{3} & = & \left(-\frac{1}{2}x_{5}+y_{5}\right)a \, \mathbf{\hat{x}} + \frac{\sqrt{3}}{2}x_{5}a \, \mathbf{\hat{y}} + \left(\frac{1}{2} +z_{5}\right)c \, \mathbf{\hat{z}} & \left(12d\right) & \mbox{Cl II} \\ 
\mathbf{B}_{27} & = & x_{5} \, \mathbf{a}_{1} + \left(x_{5}-y_{5}\right) \, \mathbf{a}_{2} + \left(\frac{1}{2} +z_{5}\right) \, \mathbf{a}_{3} & = & \left(x_{5}-\frac{1}{2}y_{5}\right)a \, \mathbf{\hat{x}}-\frac{\sqrt{3}}{2}y_{5}a \, \mathbf{\hat{y}} + \left(\frac{1}{2} +z_{5}\right)c \, \mathbf{\hat{z}} & \left(12d\right) & \mbox{Cl II} \\ 
\mathbf{B}_{28} & = & y_{5} \, \mathbf{a}_{1} + x_{5} \, \mathbf{a}_{2} + z_{5} \, \mathbf{a}_{3} & = & \frac{1}{2}\left(x_{5}+y_{5}\right)a \, \mathbf{\hat{x}} + \frac{\sqrt{3}}{2}\left(x_{5}-y_{5}\right)a \, \mathbf{\hat{y}} + z_{5}c \, \mathbf{\hat{z}} & \left(12d\right) & \mbox{Cl II} \\ 
\mathbf{B}_{29} & = & \left(x_{5}-y_{5}\right) \, \mathbf{a}_{1}-y_{5} \, \mathbf{a}_{2} + z_{5} \, \mathbf{a}_{3} & = & \left(\frac{1}{2}x_{5}-y_{5}\right)a \, \mathbf{\hat{x}}-\frac{\sqrt{3}}{2}x_{5}a \, \mathbf{\hat{y}} + z_{5}c \, \mathbf{\hat{z}} & \left(12d\right) & \mbox{Cl II} \\ 
\mathbf{B}_{30} & = & -x_{5} \, \mathbf{a}_{1} + \left(-x_{5}+y_{5}\right) \, \mathbf{a}_{2} + z_{5} \, \mathbf{a}_{3} & = & \left(-x_{5}+\frac{1}{2}y_{5}\right)a \, \mathbf{\hat{x}} + \frac{\sqrt{3}}{2}y_{5}a \, \mathbf{\hat{y}} + z_{5}c \, \mathbf{\hat{z}} & \left(12d\right) & \mbox{Cl II} \\ 
\end{longtabu}
\renewcommand{\arraystretch}{1.0}
\noindent \hrulefill
\\
\textbf{References:}
\vspace*{-0.25cm}
\begin{flushleft}
  - \bibentry{Visser_KNiCl3_ActCrystallogSecB_1980}. \\
\end{flushleft}
\textbf{Found in:}
\vspace*{-0.25cm}
\begin{flushleft}
  - \bibentry{Villars_PearsonsCrystalData_2013}. \\
\end{flushleft}
\noindent \hrulefill
\\
\textbf{Geometry files:}
\\
\noindent  - CIF: pp. {\hyperref[A3BC_hP30_185_cd_c_ab_cif]{\pageref{A3BC_hP30_185_cd_c_ab_cif}}} \\
\noindent  - POSCAR: pp. {\hyperref[A3BC_hP30_185_cd_c_ab_poscar]{\pageref{A3BC_hP30_185_cd_c_ab_poscar}}} \\
\onecolumn
{\phantomsection\label{A3B_hP24_185_ab2c_c}}
\subsection*{\huge \textbf{{\normalfont Cu$_{3}$P Structure: A3B\_hP24\_185\_ab2c\_c}}}
\noindent \hrulefill
\vspace*{0.25cm}
\begin{figure}[htp]
  \centering
  \vspace{-1em}
  {\includegraphics[width=1\textwidth]{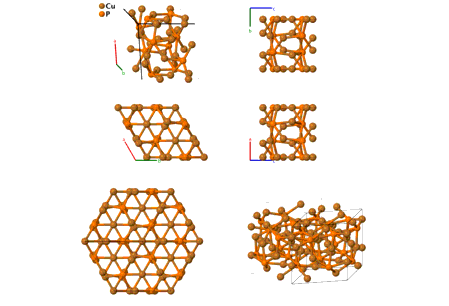}}
\end{figure}
\vspace*{-0.5cm}
\renewcommand{\arraystretch}{1.5}
\begin{equation*}
  \begin{array}{>{$\hspace{-0.15cm}}l<{$}>{$}p{0.5cm}<{$}>{$}p{18.5cm}<{$}}
    \mbox{\large \textbf{Prototype}} &\colon & \ce{Cu$_{3}$P} \\
    \mbox{\large \textbf{\AFLOW\ prototype label}} &\colon & \mbox{A3B\_hP24\_185\_ab2c\_c} \\
    \mbox{\large \textbf{\textit{Strukturbericht} designation}} &\colon & \mbox{None} \\
    \mbox{\large \textbf{Pearson symbol}} &\colon & \mbox{hP24} \\
    \mbox{\large \textbf{Space group number}} &\colon & 185 \\
    \mbox{\large \textbf{Space group symbol}} &\colon & P6_{3}cm \\
    \mbox{\large \textbf{\AFLOW\ prototype command}} &\colon &  \texttt{aflow} \,  \, \texttt{-{}-proto=A3B\_hP24\_185\_ab2c\_c } \, \newline \texttt{-{}-params=}{a,c/a,z_{1},z_{2},x_{3},z_{3},x_{4},z_{4},x_{5},z_{5} }
  \end{array}
\end{equation*}
\renewcommand{\arraystretch}{1.0}

\vspace*{-0.25cm}
\noindent \hrulefill
\begin{itemize}
  \item{Olofsson (Olofsson, 1972) argues that this is the correct structure
for Cu$_{3}$P rather than the structure which was originally
identified as {\em Strukturbericht} $D0_{21}$, space group
$P\bar{3}c1$, \href{http://aflow.org/CrystalDatabase/AB3_hP8_194_c_bf.html}{A3B\_hP24\_165\_bdg\_f}.}
  \item{Hafner and Range (Hafner, 1994) state that this is also the correct
structure for AsNa$_{3}$, rather than {\em Strukturbericht} $D0_{18}$,
space group $P6_3/mmc$,
\href{http://aflow.org/CrystalDatabase/AB3_hP8_194_c_bf.html}{AB3\_hP8\_194\_c\_bf}.}
  \item{Finally, AuMg$_{3}$, IrMg$_{3}$ and Mg$_{3}$Pt, which have been
described by both the $D0_{18}$ and $D0_{21}$ structures, are also
claimed to actually be in this structure (Range, 1993).
}
\end{itemize}

\noindent \parbox{1 \linewidth}{
\noindent \hrulefill
\\
\textbf{Hexagonal primitive vectors:} \\
\vspace*{-0.25cm}
\begin{tabular}{cc}
  \begin{tabular}{c}
    \parbox{0.6 \linewidth}{
      \renewcommand{\arraystretch}{1.5}
      \begin{equation*}
        \centering
        \begin{array}{ccc}
              \mathbf{a}_1 & = & \frac12 \, a \, \mathbf{\hat{x}} - \frac{\sqrt3}2 \, a \, \mathbf{\hat{y}} \\
    \mathbf{a}_2 & = & \frac12 \, a \, \mathbf{\hat{x}} + \frac{\sqrt3}2 \, a \, \mathbf{\hat{y}} \\
    \mathbf{a}_3 & = & c \, \mathbf{\hat{z}} \\

        \end{array}
      \end{equation*}
    }
    \renewcommand{\arraystretch}{1.0}
  \end{tabular}
  \begin{tabular}{c}
    \includegraphics[width=0.3\linewidth]{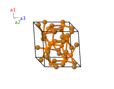} \\
  \end{tabular}
\end{tabular}

}
\vspace*{-0.25cm}

\noindent \hrulefill
\\
\textbf{Basis vectors:}
\vspace*{-0.25cm}
\renewcommand{\arraystretch}{1.5}
\begin{longtabu} to \textwidth{>{\centering $}X[-1,c,c]<{$}>{\centering $}X[-1,c,c]<{$}>{\centering $}X[-1,c,c]<{$}>{\centering $}X[-1,c,c]<{$}>{\centering $}X[-1,c,c]<{$}>{\centering $}X[-1,c,c]<{$}>{\centering $}X[-1,c,c]<{$}}
  & & \mbox{Lattice Coordinates} & & \mbox{Cartesian Coordinates} &\mbox{Wyckoff Position} & \mbox{Atom Type} \\  
  \mathbf{B}_{1} & = & z_{1} \, \mathbf{a}_{3} & = & z_{1}c \, \mathbf{\hat{z}} & \left(2a\right) & \mbox{Cu I} \\ 
\mathbf{B}_{2} & = & \left(\frac{1}{2} +z_{1}\right) \, \mathbf{a}_{3} & = & \left(\frac{1}{2} +z_{1}\right)c \, \mathbf{\hat{z}} & \left(2a\right) & \mbox{Cu I} \\ 
\mathbf{B}_{3} & = & \frac{1}{3} \, \mathbf{a}_{1} + \frac{2}{3} \, \mathbf{a}_{2} + z_{2} \, \mathbf{a}_{3} & = & \frac{1}{2}a \, \mathbf{\hat{x}} + \frac{1}{2\sqrt{3}}a \, \mathbf{\hat{y}} + z_{2}c \, \mathbf{\hat{z}} & \left(4b\right) & \mbox{Cu II} \\ 
\mathbf{B}_{4} & = & \frac{2}{3} \, \mathbf{a}_{1} + \frac{1}{3} \, \mathbf{a}_{2} + \left(\frac{1}{2} +z_{2}\right) \, \mathbf{a}_{3} & = & \frac{1}{2}a \, \mathbf{\hat{x}}- \frac{1}{2\sqrt{3}}a  \, \mathbf{\hat{y}} + \left(\frac{1}{2} +z_{2}\right)c \, \mathbf{\hat{z}} & \left(4b\right) & \mbox{Cu II} \\ 
\mathbf{B}_{5} & = & \frac{1}{3} \, \mathbf{a}_{1} + \frac{2}{3} \, \mathbf{a}_{2} + \left(\frac{1}{2} +z_{2}\right) \, \mathbf{a}_{3} & = & \frac{1}{2}a \, \mathbf{\hat{x}} + \frac{1}{2\sqrt{3}}a \, \mathbf{\hat{y}} + \left(\frac{1}{2} +z_{2}\right)c \, \mathbf{\hat{z}} & \left(4b\right) & \mbox{Cu II} \\ 
\mathbf{B}_{6} & = & \frac{2}{3} \, \mathbf{a}_{1} + \frac{1}{3} \, \mathbf{a}_{2} + z_{2} \, \mathbf{a}_{3} & = & \frac{1}{2}a \, \mathbf{\hat{x}}- \frac{1}{2\sqrt{3}}a  \, \mathbf{\hat{y}} + z_{2}c \, \mathbf{\hat{z}} & \left(4b\right) & \mbox{Cu II} \\ 
\mathbf{B}_{7} & = & x_{3} \, \mathbf{a}_{1} + z_{3} \, \mathbf{a}_{3} & = & \frac{1}{2}x_{3}a \, \mathbf{\hat{x}}-\frac{\sqrt{3}}{2}x_{3}a \, \mathbf{\hat{y}} + z_{3}c \, \mathbf{\hat{z}} & \left(6c\right) & \mbox{Cu III} \\ 
\mathbf{B}_{8} & = & x_{3} \, \mathbf{a}_{2} + z_{3} \, \mathbf{a}_{3} & = & \frac{1}{2}x_{3}a \, \mathbf{\hat{x}} + \frac{\sqrt{3}}{2}x_{3}a \, \mathbf{\hat{y}} + z_{3}c \, \mathbf{\hat{z}} & \left(6c\right) & \mbox{Cu III} \\ 
\mathbf{B}_{9} & = & -x_{3} \, \mathbf{a}_{1}-x_{3} \, \mathbf{a}_{2} + z_{3} \, \mathbf{a}_{3} & = & -x_{3}a \, \mathbf{\hat{x}} + z_{3}c \, \mathbf{\hat{z}} & \left(6c\right) & \mbox{Cu III} \\ 
\mathbf{B}_{10} & = & -x_{3} \, \mathbf{a}_{1} + \left(\frac{1}{2} +z_{3}\right) \, \mathbf{a}_{3} & = & -\frac{1}{2}x_{3}a \, \mathbf{\hat{x}} + \frac{\sqrt{3}}{2}x_{3}a \, \mathbf{\hat{y}} + \left(\frac{1}{2} +z_{3}\right)c \, \mathbf{\hat{z}} & \left(6c\right) & \mbox{Cu III} \\ 
\mathbf{B}_{11} & = & -x_{3} \, \mathbf{a}_{2} + \left(\frac{1}{2} +z_{3}\right) \, \mathbf{a}_{3} & = & -\frac{1}{2}x_{3}a \, \mathbf{\hat{x}}-\frac{\sqrt{3}}{2}x_{3}a \, \mathbf{\hat{y}} + \left(\frac{1}{2} +z_{3}\right)c \, \mathbf{\hat{z}} & \left(6c\right) & \mbox{Cu III} \\ 
\mathbf{B}_{12} & = & x_{3} \, \mathbf{a}_{1} + x_{3} \, \mathbf{a}_{2} + \left(\frac{1}{2} +z_{3}\right) \, \mathbf{a}_{3} & = & x_{3}a \, \mathbf{\hat{x}} + \left(\frac{1}{2} +z_{3}\right)c \, \mathbf{\hat{z}} & \left(6c\right) & \mbox{Cu III} \\ 
\mathbf{B}_{13} & = & x_{4} \, \mathbf{a}_{1} + z_{4} \, \mathbf{a}_{3} & = & \frac{1}{2}x_{4}a \, \mathbf{\hat{x}}-\frac{\sqrt{3}}{2}x_{4}a \, \mathbf{\hat{y}} + z_{4}c \, \mathbf{\hat{z}} & \left(6c\right) & \mbox{Cu IV} \\ 
\mathbf{B}_{14} & = & x_{4} \, \mathbf{a}_{2} + z_{4} \, \mathbf{a}_{3} & = & \frac{1}{2}x_{4}a \, \mathbf{\hat{x}} + \frac{\sqrt{3}}{2}x_{4}a \, \mathbf{\hat{y}} + z_{4}c \, \mathbf{\hat{z}} & \left(6c\right) & \mbox{Cu IV} \\ 
\mathbf{B}_{15} & = & -x_{4} \, \mathbf{a}_{1}-x_{4} \, \mathbf{a}_{2} + z_{4} \, \mathbf{a}_{3} & = & -x_{4}a \, \mathbf{\hat{x}} + z_{4}c \, \mathbf{\hat{z}} & \left(6c\right) & \mbox{Cu IV} \\ 
\mathbf{B}_{16} & = & -x_{4} \, \mathbf{a}_{1} + \left(\frac{1}{2} +z_{4}\right) \, \mathbf{a}_{3} & = & -\frac{1}{2}x_{4}a \, \mathbf{\hat{x}} + \frac{\sqrt{3}}{2}x_{4}a \, \mathbf{\hat{y}} + \left(\frac{1}{2} +z_{4}\right)c \, \mathbf{\hat{z}} & \left(6c\right) & \mbox{Cu IV} \\ 
\mathbf{B}_{17} & = & -x_{4} \, \mathbf{a}_{2} + \left(\frac{1}{2} +z_{4}\right) \, \mathbf{a}_{3} & = & -\frac{1}{2}x_{4}a \, \mathbf{\hat{x}}-\frac{\sqrt{3}}{2}x_{4}a \, \mathbf{\hat{y}} + \left(\frac{1}{2} +z_{4}\right)c \, \mathbf{\hat{z}} & \left(6c\right) & \mbox{Cu IV} \\ 
\mathbf{B}_{18} & = & x_{4} \, \mathbf{a}_{1} + x_{4} \, \mathbf{a}_{2} + \left(\frac{1}{2} +z_{4}\right) \, \mathbf{a}_{3} & = & x_{4}a \, \mathbf{\hat{x}} + \left(\frac{1}{2} +z_{4}\right)c \, \mathbf{\hat{z}} & \left(6c\right) & \mbox{Cu IV} \\ 
\mathbf{B}_{19} & = & x_{5} \, \mathbf{a}_{1} + z_{5} \, \mathbf{a}_{3} & = & \frac{1}{2}x_{5}a \, \mathbf{\hat{x}}-\frac{\sqrt{3}}{2}x_{5}a \, \mathbf{\hat{y}} + z_{5}c \, \mathbf{\hat{z}} & \left(6c\right) & \mbox{P} \\ 
\mathbf{B}_{20} & = & x_{5} \, \mathbf{a}_{2} + z_{5} \, \mathbf{a}_{3} & = & \frac{1}{2}x_{5}a \, \mathbf{\hat{x}} + \frac{\sqrt{3}}{2}x_{5}a \, \mathbf{\hat{y}} + z_{5}c \, \mathbf{\hat{z}} & \left(6c\right) & \mbox{P} \\ 
\mathbf{B}_{21} & = & -x_{5} \, \mathbf{a}_{1}-x_{5} \, \mathbf{a}_{2} + z_{5} \, \mathbf{a}_{3} & = & -x_{5}a \, \mathbf{\hat{x}} + z_{5}c \, \mathbf{\hat{z}} & \left(6c\right) & \mbox{P} \\ 
\mathbf{B}_{22} & = & -x_{5} \, \mathbf{a}_{1} + \left(\frac{1}{2} +z_{5}\right) \, \mathbf{a}_{3} & = & -\frac{1}{2}x_{5}a \, \mathbf{\hat{x}} + \frac{\sqrt{3}}{2}x_{5}a \, \mathbf{\hat{y}} + \left(\frac{1}{2} +z_{5}\right)c \, \mathbf{\hat{z}} & \left(6c\right) & \mbox{P} \\ 
\mathbf{B}_{23} & = & -x_{5} \, \mathbf{a}_{2} + \left(\frac{1}{2} +z_{5}\right) \, \mathbf{a}_{3} & = & -\frac{1}{2}x_{5}a \, \mathbf{\hat{x}}-\frac{\sqrt{3}}{2}x_{5}a \, \mathbf{\hat{y}} + \left(\frac{1}{2} +z_{5}\right)c \, \mathbf{\hat{z}} & \left(6c\right) & \mbox{P} \\ 
\mathbf{B}_{24} & = & x_{5} \, \mathbf{a}_{1} + x_{5} \, \mathbf{a}_{2} + \left(\frac{1}{2} +z_{5}\right) \, \mathbf{a}_{3} & = & x_{5}a \, \mathbf{\hat{x}} + \left(\frac{1}{2} +z_{5}\right)c \, \mathbf{\hat{z}} & \left(6c\right) & \mbox{P} \\ 
\end{longtabu}
\renewcommand{\arraystretch}{1.0}
\noindent \hrulefill
\\
\textbf{References:}
\vspace*{-0.25cm}
\begin{flushleft}
  - \bibentry{Olofsson_Acta_Chem_Scand_26_1972}. \\
  - \bibentry{Hafner_JAC_216_1994}. \\
  - \bibentry{Range_JACS_191_1993}. \\
\end{flushleft}
\noindent \hrulefill
\\
\textbf{Geometry files:}
\\
\noindent  - CIF: pp. {\hyperref[A3B_hP24_185_ab2c_c_cif]{\pageref{A3B_hP24_185_ab2c_c_cif}}} \\
\noindent  - POSCAR: pp. {\hyperref[A3B_hP24_185_ab2c_c_poscar]{\pageref{A3B_hP24_185_ab2c_c_poscar}}} \\
\onecolumn
{\phantomsection\label{A3B_hP8_185_c_a}}
\subsection*{\huge \textbf{{\normalfont $\beta$-RuCl$_{3}$ Structure: A3B\_hP8\_185\_c\_a}}}
\noindent \hrulefill
\vspace*{0.25cm}
\begin{figure}[htp]
  \centering
  \vspace{-1em}
  {\includegraphics[width=1\textwidth]{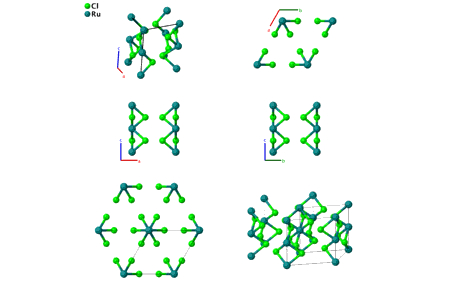}}
\end{figure}
\vspace*{-0.5cm}
\renewcommand{\arraystretch}{1.5}
\begin{equation*}
  \begin{array}{>{$\hspace{-0.15cm}}l<{$}>{$}p{0.5cm}<{$}>{$}p{18.5cm}<{$}}
    \mbox{\large \textbf{Prototype}} &\colon & \ce{$\beta$-RuCl3} \\
    \mbox{\large \textbf{\AFLOW\ prototype label}} &\colon & \mbox{A3B\_hP8\_185\_c\_a} \\
    \mbox{\large \textbf{\textit{Strukturbericht} designation}} &\colon & \mbox{None} \\
    \mbox{\large \textbf{Pearson symbol}} &\colon & \mbox{hP8} \\
    \mbox{\large \textbf{Space group number}} &\colon & 185 \\
    \mbox{\large \textbf{Space group symbol}} &\colon & P6_{3}cm \\
    \mbox{\large \textbf{\AFLOW\ prototype command}} &\colon &  \texttt{aflow} \,  \, \texttt{-{}-proto=A3B\_hP8\_185\_c\_a } \, \newline \texttt{-{}-params=}{a,c/a,z_{1},x_{2},z_{2} }
  \end{array}
\end{equation*}
\renewcommand{\arraystretch}{1.0}

\vspace*{-0.25cm}
\noindent \hrulefill
\begin{itemize}
  \item{Pearson comments that space groups \#158, \#188, \#193, could not be rejected, but this structure is consistent with space group \#185. 
We also provide the structure with space group \#158: \hyperref[A3B_hP8_158_d_a]{$\beta$-RuCl$_{3}$ (A3B\_hP8\_158\_d\_a) structure}.
}
\end{itemize}

\noindent \parbox{1 \linewidth}{
\noindent \hrulefill
\\
\textbf{Hexagonal primitive vectors:} \\
\vspace*{-0.25cm}
\begin{tabular}{cc}
  \begin{tabular}{c}
    \parbox{0.6 \linewidth}{
      \renewcommand{\arraystretch}{1.5}
      \begin{equation*}
        \centering
        \begin{array}{ccc}
              \mathbf{a}_1 & = & \frac12 \, a \, \mathbf{\hat{x}} - \frac{\sqrt3}2 \, a \, \mathbf{\hat{y}} \\
    \mathbf{a}_2 & = & \frac12 \, a \, \mathbf{\hat{x}} + \frac{\sqrt3}2 \, a \, \mathbf{\hat{y}} \\
    \mathbf{a}_3 & = & c \, \mathbf{\hat{z}} \\

        \end{array}
      \end{equation*}
    }
    \renewcommand{\arraystretch}{1.0}
  \end{tabular}
  \begin{tabular}{c}
    \includegraphics[width=0.3\linewidth]{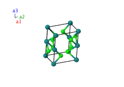} \\
  \end{tabular}
\end{tabular}

}
\vspace*{-0.25cm}

\noindent \hrulefill
\\
\textbf{Basis vectors:}
\vspace*{-0.25cm}
\renewcommand{\arraystretch}{1.5}
\begin{longtabu} to \textwidth{>{\centering $}X[-1,c,c]<{$}>{\centering $}X[-1,c,c]<{$}>{\centering $}X[-1,c,c]<{$}>{\centering $}X[-1,c,c]<{$}>{\centering $}X[-1,c,c]<{$}>{\centering $}X[-1,c,c]<{$}>{\centering $}X[-1,c,c]<{$}}
  & & \mbox{Lattice Coordinates} & & \mbox{Cartesian Coordinates} &\mbox{Wyckoff Position} & \mbox{Atom Type} \\  
  \mathbf{B}_{1} & = & z_{1} \, \mathbf{a}_{3} & = & z_{1}c \, \mathbf{\hat{z}} & \left(2a\right) & \mbox{Ru} \\ 
\mathbf{B}_{2} & = & \left(\frac{1}{2} +z_{1}\right) \, \mathbf{a}_{3} & = & \left(\frac{1}{2} +z_{1}\right)c \, \mathbf{\hat{z}} & \left(2a\right) & \mbox{Ru} \\ 
\mathbf{B}_{3} & = & x_{2} \, \mathbf{a}_{1} + z_{2} \, \mathbf{a}_{3} & = & \frac{1}{2}x_{2}a \, \mathbf{\hat{x}}-\frac{\sqrt{3}}{2}x_{2}a \, \mathbf{\hat{y}} + z_{2}c \, \mathbf{\hat{z}} & \left(6c\right) & \mbox{Cl} \\ 
\mathbf{B}_{4} & = & x_{2} \, \mathbf{a}_{2} + z_{2} \, \mathbf{a}_{3} & = & \frac{1}{2}x_{2}a \, \mathbf{\hat{x}} + \frac{\sqrt{3}}{2}x_{2}a \, \mathbf{\hat{y}} + z_{2}c \, \mathbf{\hat{z}} & \left(6c\right) & \mbox{Cl} \\ 
\mathbf{B}_{5} & = & -x_{2} \, \mathbf{a}_{1}-x_{2} \, \mathbf{a}_{2} + z_{2} \, \mathbf{a}_{3} & = & -x_{2}a \, \mathbf{\hat{x}} + z_{2}c \, \mathbf{\hat{z}} & \left(6c\right) & \mbox{Cl} \\ 
\mathbf{B}_{6} & = & -x_{2} \, \mathbf{a}_{1} + \left(\frac{1}{2} +z_{2}\right) \, \mathbf{a}_{3} & = & -\frac{1}{2}x_{2}a \, \mathbf{\hat{x}} + \frac{\sqrt{3}}{2}x_{2}a \, \mathbf{\hat{y}} + \left(\frac{1}{2} +z_{2}\right)c \, \mathbf{\hat{z}} & \left(6c\right) & \mbox{Cl} \\ 
\mathbf{B}_{7} & = & -x_{2} \, \mathbf{a}_{2} + \left(\frac{1}{2} +z_{2}\right) \, \mathbf{a}_{3} & = & -\frac{1}{2}x_{2}a \, \mathbf{\hat{x}}-\frac{\sqrt{3}}{2}x_{2}a \, \mathbf{\hat{y}} + \left(\frac{1}{2} +z_{2}\right)c \, \mathbf{\hat{z}} & \left(6c\right) & \mbox{Cl} \\ 
\mathbf{B}_{8} & = & x_{2} \, \mathbf{a}_{1} + x_{2} \, \mathbf{a}_{2} + \left(\frac{1}{2} +z_{2}\right) \, \mathbf{a}_{3} & = & x_{2}a \, \mathbf{\hat{x}} + \left(\frac{1}{2} +z_{2}\right)c \, \mathbf{\hat{z}} & \left(6c\right) & \mbox{Cl} \\ 
\end{longtabu}
\renewcommand{\arraystretch}{1.0}
\noindent \hrulefill
\\
\textbf{References:}
\vspace*{-0.25cm}
\begin{flushleft}
  - \bibentry{Fletcher_Cl3Ru_JChemSocA_1967}. \\
\end{flushleft}
\textbf{Found in:}
\vspace*{-0.25cm}
\begin{flushleft}
  - \bibentry{Villars_PearsonsCrystalData_2013}. \\
\end{flushleft}
\noindent \hrulefill
\\
\textbf{Geometry files:}
\\
\noindent  - CIF: pp. {\hyperref[A3B_hP8_185_c_a_cif]{\pageref{A3B_hP8_185_c_a_cif}}} \\
\noindent  - POSCAR: pp. {\hyperref[A3B_hP8_185_c_a_poscar]{\pageref{A3B_hP8_185_c_a_poscar}}} \\
\onecolumn
{\phantomsection\label{AB3_hP24_185_c_ab2c}}
\subsection*{\huge \textbf{{\normalfont Na$_{3}$As Structure: AB3\_hP24\_185\_c\_ab2c}}}
\noindent \hrulefill
\vspace*{0.25cm}
\begin{figure}[htp]
  \centering
  \vspace{-1em}
  {\includegraphics[width=1\textwidth]{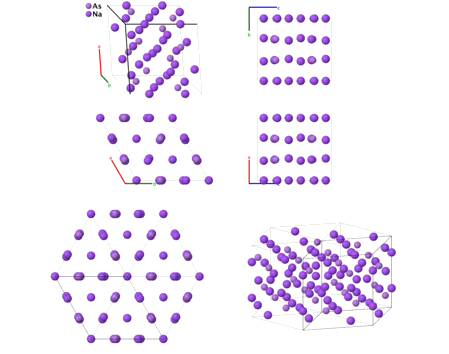}}
\end{figure}
\vspace*{-0.5cm}
\renewcommand{\arraystretch}{1.5}
\begin{equation*}
  \begin{array}{>{$\hspace{-0.15cm}}l<{$}>{$}p{0.5cm}<{$}>{$}p{18.5cm}<{$}}
    \mbox{\large \textbf{Prototype}} &\colon & \ce{Na3As} \\
    \mbox{\large \textbf{\AFLOW\ prototype label}} &\colon & \mbox{AB3\_hP24\_185\_c\_ab2c} \\
    \mbox{\large \textbf{\textit{Strukturbericht} designation}} &\colon & \mbox{None} \\
    \mbox{\large \textbf{Pearson symbol}} &\colon & \mbox{hP24} \\
    \mbox{\large \textbf{Space group number}} &\colon & 185 \\
    \mbox{\large \textbf{Space group symbol}} &\colon & P6_{3}cm \\
    \mbox{\large \textbf{\AFLOW\ prototype command}} &\colon &  \texttt{aflow} \,  \, \texttt{-{}-proto=AB3\_hP24\_185\_c\_ab2c } \, \newline \texttt{-{}-params=}{a,c/a,z_{1},z_{2},x_{3},z_{3},x_{4},z_{4},x_{5},z_{5} }
  \end{array}
\end{equation*}
\renewcommand{\arraystretch}{1.0}

\vspace*{-0.25cm}
\noindent \hrulefill
\\
\textbf{ Other compounds with the structure:}
\begin{itemize}
   \item{ Cu$_{3}$P  }
\end{itemize}
\vspace*{-0.25cm}
\noindent \hrulefill
\begin{itemize}
  \item{The authors state that this is a correction of the \href{http://aflowlib.org/CrystalDatabase/AB3_hP8_194_c_bf.html}{$D0_{18}$}
Na$_{3}$As structure.
Cu$_{3}$P (pp. {\hyperref[A3B_hP24_185_ab2c_c]{\pageref{A3B_hP24_185_ab2c_c}}}) and
Na$_{3}$As (pp. {\hyperref[AB3_hP24_185_c_ab2c]{\pageref{AB3_hP24_185_c_ab2c}}})
have similar \AFLOW\ prototype labels ({\it{i.e.}}, same symmetry and set of
Wyckoff positions with different stoichiometry labels due to alphabetic ordering of atomic species).
They are generated by the same symmetry operations with different sets of parameters
(\texttt{-{}-params}) specified in their corresponding \CIF\ files.
}
\end{itemize}

\noindent \parbox{1 \linewidth}{
\noindent \hrulefill
\\
\textbf{Hexagonal primitive vectors:} \\
\vspace*{-0.25cm}
\begin{tabular}{cc}
  \begin{tabular}{c}
    \parbox{0.6 \linewidth}{
      \renewcommand{\arraystretch}{1.5}
      \begin{equation*}
        \centering
        \begin{array}{ccc}
              \mathbf{a}_1 & = & \frac12 \, a \, \mathbf{\hat{x}} - \frac{\sqrt3}2 \, a \, \mathbf{\hat{y}} \\
    \mathbf{a}_2 & = & \frac12 \, a \, \mathbf{\hat{x}} + \frac{\sqrt3}2 \, a \, \mathbf{\hat{y}} \\
    \mathbf{a}_3 & = & c \, \mathbf{\hat{z}} \\

        \end{array}
      \end{equation*}
    }
    \renewcommand{\arraystretch}{1.0}
  \end{tabular}
  \begin{tabular}{c}
    \includegraphics[width=0.3\linewidth]{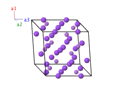} \\
  \end{tabular}
\end{tabular}

}
\vspace*{-0.25cm}

\noindent \hrulefill
\\
\textbf{Basis vectors:}
\vspace*{-0.25cm}
\renewcommand{\arraystretch}{1.5}
\begin{longtabu} to \textwidth{>{\centering $}X[-1,c,c]<{$}>{\centering $}X[-1,c,c]<{$}>{\centering $}X[-1,c,c]<{$}>{\centering $}X[-1,c,c]<{$}>{\centering $}X[-1,c,c]<{$}>{\centering $}X[-1,c,c]<{$}>{\centering $}X[-1,c,c]<{$}}
  & & \mbox{Lattice Coordinates} & & \mbox{Cartesian Coordinates} &\mbox{Wyckoff Position} & \mbox{Atom Type} \\  
  \mathbf{B}_{1} & = & z_{1} \, \mathbf{a}_{3} & = & z_{1}c \, \mathbf{\hat{z}} & \left(2a\right) & \mbox{Na I} \\ 
\mathbf{B}_{2} & = & \left(\frac{1}{2} +z_{1}\right) \, \mathbf{a}_{3} & = & \left(\frac{1}{2} +z_{1}\right)c \, \mathbf{\hat{z}} & \left(2a\right) & \mbox{Na I} \\ 
\mathbf{B}_{3} & = & \frac{1}{3} \, \mathbf{a}_{1} + \frac{2}{3} \, \mathbf{a}_{2} + z_{2} \, \mathbf{a}_{3} & = & \frac{1}{2}a \, \mathbf{\hat{x}} + \frac{1}{2\sqrt{3}}a \, \mathbf{\hat{y}} + z_{2}c \, \mathbf{\hat{z}} & \left(4b\right) & \mbox{Na II} \\ 
\mathbf{B}_{4} & = & \frac{2}{3} \, \mathbf{a}_{1} + \frac{1}{3} \, \mathbf{a}_{2} + \left(\frac{1}{2} +z_{2}\right) \, \mathbf{a}_{3} & = & \frac{1}{2}a \, \mathbf{\hat{x}}- \frac{1}{2\sqrt{3}}a  \, \mathbf{\hat{y}} + \left(\frac{1}{2} +z_{2}\right)c \, \mathbf{\hat{z}} & \left(4b\right) & \mbox{Na II} \\ 
\mathbf{B}_{5} & = & \frac{1}{3} \, \mathbf{a}_{1} + \frac{2}{3} \, \mathbf{a}_{2} + \left(\frac{1}{2} +z_{2}\right) \, \mathbf{a}_{3} & = & \frac{1}{2}a \, \mathbf{\hat{x}} + \frac{1}{2\sqrt{3}}a \, \mathbf{\hat{y}} + \left(\frac{1}{2} +z_{2}\right)c \, \mathbf{\hat{z}} & \left(4b\right) & \mbox{Na II} \\ 
\mathbf{B}_{6} & = & \frac{2}{3} \, \mathbf{a}_{1} + \frac{1}{3} \, \mathbf{a}_{2} + z_{2} \, \mathbf{a}_{3} & = & \frac{1}{2}a \, \mathbf{\hat{x}}- \frac{1}{2\sqrt{3}}a  \, \mathbf{\hat{y}} + z_{2}c \, \mathbf{\hat{z}} & \left(4b\right) & \mbox{Na II} \\ 
\mathbf{B}_{7} & = & x_{3} \, \mathbf{a}_{1} + z_{3} \, \mathbf{a}_{3} & = & \frac{1}{2}x_{3}a \, \mathbf{\hat{x}}-\frac{\sqrt{3}}{2}x_{3}a \, \mathbf{\hat{y}} + z_{3}c \, \mathbf{\hat{z}} & \left(6c\right) & \mbox{As} \\ 
\mathbf{B}_{8} & = & x_{3} \, \mathbf{a}_{2} + z_{3} \, \mathbf{a}_{3} & = & \frac{1}{2}x_{3}a \, \mathbf{\hat{x}} + \frac{\sqrt{3}}{2}x_{3}a \, \mathbf{\hat{y}} + z_{3}c \, \mathbf{\hat{z}} & \left(6c\right) & \mbox{As} \\ 
\mathbf{B}_{9} & = & -x_{3} \, \mathbf{a}_{1}-x_{3} \, \mathbf{a}_{2} + z_{3} \, \mathbf{a}_{3} & = & -x_{3}a \, \mathbf{\hat{x}} + z_{3}c \, \mathbf{\hat{z}} & \left(6c\right) & \mbox{As} \\ 
\mathbf{B}_{10} & = & -x_{3} \, \mathbf{a}_{1} + \left(\frac{1}{2} +z_{3}\right) \, \mathbf{a}_{3} & = & -\frac{1}{2}x_{3}a \, \mathbf{\hat{x}} + \frac{\sqrt{3}}{2}x_{3}a \, \mathbf{\hat{y}} + \left(\frac{1}{2} +z_{3}\right)c \, \mathbf{\hat{z}} & \left(6c\right) & \mbox{As} \\ 
\mathbf{B}_{11} & = & -x_{3} \, \mathbf{a}_{2} + \left(\frac{1}{2} +z_{3}\right) \, \mathbf{a}_{3} & = & -\frac{1}{2}x_{3}a \, \mathbf{\hat{x}}-\frac{\sqrt{3}}{2}x_{3}a \, \mathbf{\hat{y}} + \left(\frac{1}{2} +z_{3}\right)c \, \mathbf{\hat{z}} & \left(6c\right) & \mbox{As} \\ 
\mathbf{B}_{12} & = & x_{3} \, \mathbf{a}_{1} + x_{3} \, \mathbf{a}_{2} + \left(\frac{1}{2} +z_{3}\right) \, \mathbf{a}_{3} & = & x_{3}a \, \mathbf{\hat{x}} + \left(\frac{1}{2} +z_{3}\right)c \, \mathbf{\hat{z}} & \left(6c\right) & \mbox{As} \\ 
\mathbf{B}_{13} & = & x_{4} \, \mathbf{a}_{1} + z_{4} \, \mathbf{a}_{3} & = & \frac{1}{2}x_{4}a \, \mathbf{\hat{x}}-\frac{\sqrt{3}}{2}x_{4}a \, \mathbf{\hat{y}} + z_{4}c \, \mathbf{\hat{z}} & \left(6c\right) & \mbox{Na III} \\ 
\mathbf{B}_{14} & = & x_{4} \, \mathbf{a}_{2} + z_{4} \, \mathbf{a}_{3} & = & \frac{1}{2}x_{4}a \, \mathbf{\hat{x}} + \frac{\sqrt{3}}{2}x_{4}a \, \mathbf{\hat{y}} + z_{4}c \, \mathbf{\hat{z}} & \left(6c\right) & \mbox{Na III} \\ 
\mathbf{B}_{15} & = & -x_{4} \, \mathbf{a}_{1}-x_{4} \, \mathbf{a}_{2} + z_{4} \, \mathbf{a}_{3} & = & -x_{4}a \, \mathbf{\hat{x}} + z_{4}c \, \mathbf{\hat{z}} & \left(6c\right) & \mbox{Na III} \\ 
\mathbf{B}_{16} & = & -x_{4} \, \mathbf{a}_{1} + \left(\frac{1}{2} +z_{4}\right) \, \mathbf{a}_{3} & = & -\frac{1}{2}x_{4}a \, \mathbf{\hat{x}} + \frac{\sqrt{3}}{2}x_{4}a \, \mathbf{\hat{y}} + \left(\frac{1}{2} +z_{4}\right)c \, \mathbf{\hat{z}} & \left(6c\right) & \mbox{Na III} \\ 
\mathbf{B}_{17} & = & -x_{4} \, \mathbf{a}_{2} + \left(\frac{1}{2} +z_{4}\right) \, \mathbf{a}_{3} & = & -\frac{1}{2}x_{4}a \, \mathbf{\hat{x}}-\frac{\sqrt{3}}{2}x_{4}a \, \mathbf{\hat{y}} + \left(\frac{1}{2} +z_{4}\right)c \, \mathbf{\hat{z}} & \left(6c\right) & \mbox{Na III} \\ 
\mathbf{B}_{18} & = & x_{4} \, \mathbf{a}_{1} + x_{4} \, \mathbf{a}_{2} + \left(\frac{1}{2} +z_{4}\right) \, \mathbf{a}_{3} & = & x_{4}a \, \mathbf{\hat{x}} + \left(\frac{1}{2} +z_{4}\right)c \, \mathbf{\hat{z}} & \left(6c\right) & \mbox{Na III} \\ 
\mathbf{B}_{19} & = & x_{5} \, \mathbf{a}_{1} + z_{5} \, \mathbf{a}_{3} & = & \frac{1}{2}x_{5}a \, \mathbf{\hat{x}}-\frac{\sqrt{3}}{2}x_{5}a \, \mathbf{\hat{y}} + z_{5}c \, \mathbf{\hat{z}} & \left(6c\right) & \mbox{Na IV} \\ 
\mathbf{B}_{20} & = & x_{5} \, \mathbf{a}_{2} + z_{5} \, \mathbf{a}_{3} & = & \frac{1}{2}x_{5}a \, \mathbf{\hat{x}} + \frac{\sqrt{3}}{2}x_{5}a \, \mathbf{\hat{y}} + z_{5}c \, \mathbf{\hat{z}} & \left(6c\right) & \mbox{Na IV} \\ 
\mathbf{B}_{21} & = & -x_{5} \, \mathbf{a}_{1}-x_{5} \, \mathbf{a}_{2} + z_{5} \, \mathbf{a}_{3} & = & -x_{5}a \, \mathbf{\hat{x}} + z_{5}c \, \mathbf{\hat{z}} & \left(6c\right) & \mbox{Na IV} \\ 
\mathbf{B}_{22} & = & -x_{5} \, \mathbf{a}_{1} + \left(\frac{1}{2} +z_{5}\right) \, \mathbf{a}_{3} & = & -\frac{1}{2}x_{5}a \, \mathbf{\hat{x}} + \frac{\sqrt{3}}{2}x_{5}a \, \mathbf{\hat{y}} + \left(\frac{1}{2} +z_{5}\right)c \, \mathbf{\hat{z}} & \left(6c\right) & \mbox{Na IV} \\ 
\mathbf{B}_{23} & = & -x_{5} \, \mathbf{a}_{2} + \left(\frac{1}{2} +z_{5}\right) \, \mathbf{a}_{3} & = & -\frac{1}{2}x_{5}a \, \mathbf{\hat{x}}-\frac{\sqrt{3}}{2}x_{5}a \, \mathbf{\hat{y}} + \left(\frac{1}{2} +z_{5}\right)c \, \mathbf{\hat{z}} & \left(6c\right) & \mbox{Na IV} \\ 
\mathbf{B}_{24} & = & x_{5} \, \mathbf{a}_{1} + x_{5} \, \mathbf{a}_{2} + \left(\frac{1}{2} +z_{5}\right) \, \mathbf{a}_{3} & = & x_{5}a \, \mathbf{\hat{x}} + \left(\frac{1}{2} +z_{5}\right)c \, \mathbf{\hat{z}} & \left(6c\right) & \mbox{Na IV} \\ 
\end{longtabu}
\renewcommand{\arraystretch}{1.0}
\noindent \hrulefill
\\
\textbf{References:}
\vspace*{-0.25cm}
\begin{flushleft}
  - \bibentry{Hafner_JAC_216_1994}. \\
\end{flushleft}
\noindent \hrulefill
\\
\textbf{Geometry files:}
\\
\noindent  - CIF: pp. {\hyperref[AB3_hP24_185_c_ab2c_cif]{\pageref{AB3_hP24_185_c_ab2c_cif}}} \\
\noindent  - POSCAR: pp. {\hyperref[AB3_hP24_185_c_ab2c_poscar]{\pageref{AB3_hP24_185_c_ab2c_poscar}}} \\
\onecolumn
{\phantomsection\label{A3B7_hP20_186_c_b2c}}
\subsection*{\huge \textbf{{\normalfont Fe$_{3}$Th$_{7}$ ($D10_{2}$) Structure: A3B7\_hP20\_186\_c\_b2c}}}
\noindent \hrulefill
\vspace*{0.25cm}
\begin{figure}[htp]
  \centering
  \vspace{-1em}
  {\includegraphics[width=1\textwidth]{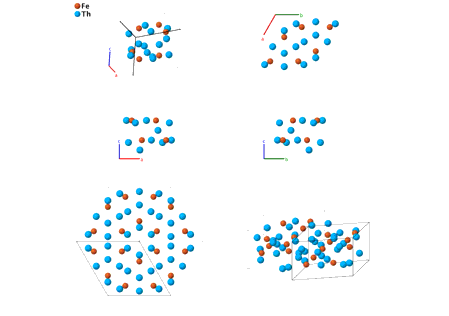}}
\end{figure}
\vspace*{-0.5cm}
\renewcommand{\arraystretch}{1.5}
\begin{equation*}
  \begin{array}{>{$\hspace{-0.15cm}}l<{$}>{$}p{0.5cm}<{$}>{$}p{18.5cm}<{$}}
    \mbox{\large \textbf{Prototype}} &\colon & \ce{Fe$_{3}$Th$_{7}$} \\
    \mbox{\large \textbf{\AFLOW\ prototype label}} &\colon & \mbox{A3B7\_hP20\_186\_c\_b2c} \\
    \mbox{\large \textbf{\textit{Strukturbericht} designation}} &\colon & \mbox{$D10_{2}$} \\
    \mbox{\large \textbf{Pearson symbol}} &\colon & \mbox{hP20} \\
    \mbox{\large \textbf{Space group number}} &\colon & 186 \\
    \mbox{\large \textbf{Space group symbol}} &\colon & P6_{3}mc \\
    \mbox{\large \textbf{\AFLOW\ prototype command}} &\colon &  \texttt{aflow} \,  \, \texttt{-{}-proto=A3B7\_hP20\_186\_c\_b2c } \, \newline \texttt{-{}-params=}{a,c/a,z_{1},x_{2},z_{2},x_{3},z_{3},x_{4},z_{4} }
  \end{array}
\end{equation*}
\renewcommand{\arraystretch}{1.0}

\vspace*{-0.25cm}
\noindent \hrulefill
\\
\textbf{ Other compounds with this structure:}
\begin{itemize}
   \item{ Th$_{7}$Co$_{3}$, Th$_{7}$Ni$_{3}$  }
\end{itemize}
\noindent \parbox{1 \linewidth}{
\noindent \hrulefill
\\
\textbf{Hexagonal primitive vectors:} \\
\vspace*{-0.25cm}
\begin{tabular}{cc}
  \begin{tabular}{c}
    \parbox{0.6 \linewidth}{
      \renewcommand{\arraystretch}{1.5}
      \begin{equation*}
        \centering
        \begin{array}{ccc}
              \mathbf{a}_1 & = & \frac12 \, a \, \mathbf{\hat{x}} - \frac{\sqrt3}2 \, a \, \mathbf{\hat{y}} \\
    \mathbf{a}_2 & = & \frac12 \, a \, \mathbf{\hat{x}} + \frac{\sqrt3}2 \, a \, \mathbf{\hat{y}} \\
    \mathbf{a}_3 & = & c \, \mathbf{\hat{z}} \\

        \end{array}
      \end{equation*}
    }
    \renewcommand{\arraystretch}{1.0}
  \end{tabular}
  \begin{tabular}{c}
    \includegraphics[width=0.3\linewidth]{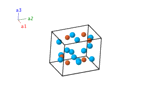} \\
  \end{tabular}
\end{tabular}

}
\vspace*{-0.25cm}

\noindent \hrulefill
\\
\textbf{Basis vectors:}
\vspace*{-0.25cm}
\renewcommand{\arraystretch}{1.5}
\begin{longtabu} to \textwidth{>{\centering $}X[-1,c,c]<{$}>{\centering $}X[-1,c,c]<{$}>{\centering $}X[-1,c,c]<{$}>{\centering $}X[-1,c,c]<{$}>{\centering $}X[-1,c,c]<{$}>{\centering $}X[-1,c,c]<{$}>{\centering $}X[-1,c,c]<{$}}
  & & \mbox{Lattice Coordinates} & & \mbox{Cartesian Coordinates} &\mbox{Wyckoff Position} & \mbox{Atom Type} \\  
  \mathbf{B}_{1} & = & \frac{1}{3} \, \mathbf{a}_{1} + \frac{2}{3} \, \mathbf{a}_{2} + z_{1} \, \mathbf{a}_{3} & = & \frac{1}{2}a \, \mathbf{\hat{x}} + \frac{1}{2\sqrt{3}}a \, \mathbf{\hat{y}} + z_{1}c \, \mathbf{\hat{z}} & \left(2b\right) & \mbox{Th I} \\ 
\mathbf{B}_{2} & = & \frac{2}{3} \, \mathbf{a}_{1} + \frac{1}{3} \, \mathbf{a}_{2} + \left(\frac{1}{2} +z_{1}\right) \, \mathbf{a}_{3} & = & \frac{1}{2}a \, \mathbf{\hat{x}}- \frac{1}{2\sqrt{3}}a  \, \mathbf{\hat{y}} + \left(\frac{1}{2} +z_{1}\right)c \, \mathbf{\hat{z}} & \left(2b\right) & \mbox{Th I} \\ 
\mathbf{B}_{3} & = & x_{2} \, \mathbf{a}_{1}-x_{2} \, \mathbf{a}_{2} + z_{2} \, \mathbf{a}_{3} & = & -\sqrt{3}x_{2}a \, \mathbf{\hat{y}} + z_{2}c \, \mathbf{\hat{z}} & \left(6c\right) & \mbox{Fe} \\ 
\mathbf{B}_{4} & = & x_{2} \, \mathbf{a}_{1} + 2x_{2} \, \mathbf{a}_{2} + z_{2} \, \mathbf{a}_{3} & = & \frac{3}{2}x_{2}a \, \mathbf{\hat{x}} + \frac{\sqrt{3}}{2}x_{2}a \, \mathbf{\hat{y}} + z_{2}c \, \mathbf{\hat{z}} & \left(6c\right) & \mbox{Fe} \\ 
\mathbf{B}_{5} & = & -2x_{2} \, \mathbf{a}_{1}-x_{2} \, \mathbf{a}_{2} + z_{2} \, \mathbf{a}_{3} & = & -\frac{3}{2}x_{2}a \, \mathbf{\hat{x}} + \frac{\sqrt{3}}{2}x_{2}a \, \mathbf{\hat{y}} + z_{2}c \, \mathbf{\hat{z}} & \left(6c\right) & \mbox{Fe} \\ 
\mathbf{B}_{6} & = & -x_{2} \, \mathbf{a}_{1} + x_{2} \, \mathbf{a}_{2} + \left(\frac{1}{2} +z_{2}\right) \, \mathbf{a}_{3} & = & \sqrt{3}x_{2}a \, \mathbf{\hat{y}} + \left(\frac{1}{2} +z_{2}\right)c \, \mathbf{\hat{z}} & \left(6c\right) & \mbox{Fe} \\ 
\mathbf{B}_{7} & = & -x_{2} \, \mathbf{a}_{1}-2x_{2} \, \mathbf{a}_{2} + \left(\frac{1}{2} +z_{2}\right) \, \mathbf{a}_{3} & = & -\frac{3}{2}x_{2}a \, \mathbf{\hat{x}}-\frac{\sqrt{3}}{2}x_{2}a \, \mathbf{\hat{y}} + \left(\frac{1}{2} +z_{2}\right)c \, \mathbf{\hat{z}} & \left(6c\right) & \mbox{Fe} \\ 
\mathbf{B}_{8} & = & 2x_{2} \, \mathbf{a}_{1} + x_{2} \, \mathbf{a}_{2} + \left(\frac{1}{2} +z_{2}\right) \, \mathbf{a}_{3} & = & \frac{3}{2}x_{2}a \, \mathbf{\hat{x}}-\frac{\sqrt{3}}{2}x_{2}a \, \mathbf{\hat{y}} + \left(\frac{1}{2} +z_{2}\right)c \, \mathbf{\hat{z}} & \left(6c\right) & \mbox{Fe} \\ 
\mathbf{B}_{9} & = & x_{3} \, \mathbf{a}_{1}-x_{3} \, \mathbf{a}_{2} + z_{3} \, \mathbf{a}_{3} & = & -\sqrt{3}x_{3}a \, \mathbf{\hat{y}} + z_{3}c \, \mathbf{\hat{z}} & \left(6c\right) & \mbox{Th II} \\ 
\mathbf{B}_{10} & = & x_{3} \, \mathbf{a}_{1} + 2x_{3} \, \mathbf{a}_{2} + z_{3} \, \mathbf{a}_{3} & = & \frac{3}{2}x_{3}a \, \mathbf{\hat{x}} + \frac{\sqrt{3}}{2}x_{3}a \, \mathbf{\hat{y}} + z_{3}c \, \mathbf{\hat{z}} & \left(6c\right) & \mbox{Th II} \\ 
\mathbf{B}_{11} & = & -2x_{3} \, \mathbf{a}_{1}-x_{3} \, \mathbf{a}_{2} + z_{3} \, \mathbf{a}_{3} & = & -\frac{3}{2}x_{3}a \, \mathbf{\hat{x}} + \frac{\sqrt{3}}{2}x_{3}a \, \mathbf{\hat{y}} + z_{3}c \, \mathbf{\hat{z}} & \left(6c\right) & \mbox{Th II} \\ 
\mathbf{B}_{12} & = & -x_{3} \, \mathbf{a}_{1} + x_{3} \, \mathbf{a}_{2} + \left(\frac{1}{2} +z_{3}\right) \, \mathbf{a}_{3} & = & \sqrt{3}x_{3}a \, \mathbf{\hat{y}} + \left(\frac{1}{2} +z_{3}\right)c \, \mathbf{\hat{z}} & \left(6c\right) & \mbox{Th II} \\ 
\mathbf{B}_{13} & = & -x_{3} \, \mathbf{a}_{1}-2x_{3} \, \mathbf{a}_{2} + \left(\frac{1}{2} +z_{3}\right) \, \mathbf{a}_{3} & = & -\frac{3}{2}x_{3}a \, \mathbf{\hat{x}}-\frac{\sqrt{3}}{2}x_{3}a \, \mathbf{\hat{y}} + \left(\frac{1}{2} +z_{3}\right)c \, \mathbf{\hat{z}} & \left(6c\right) & \mbox{Th II} \\ 
\mathbf{B}_{14} & = & 2x_{3} \, \mathbf{a}_{1} + x_{3} \, \mathbf{a}_{2} + \left(\frac{1}{2} +z_{3}\right) \, \mathbf{a}_{3} & = & \frac{3}{2}x_{3}a \, \mathbf{\hat{x}}-\frac{\sqrt{3}}{2}x_{3}a \, \mathbf{\hat{y}} + \left(\frac{1}{2} +z_{3}\right)c \, \mathbf{\hat{z}} & \left(6c\right) & \mbox{Th II} \\ 
\mathbf{B}_{15} & = & x_{4} \, \mathbf{a}_{1}-x_{4} \, \mathbf{a}_{2} + z_{4} \, \mathbf{a}_{3} & = & -\sqrt{3}x_{4}a \, \mathbf{\hat{y}} + z_{4}c \, \mathbf{\hat{z}} & \left(6c\right) & \mbox{Th III} \\ 
\mathbf{B}_{16} & = & x_{4} \, \mathbf{a}_{1} + 2x_{4} \, \mathbf{a}_{2} + z_{4} \, \mathbf{a}_{3} & = & \frac{3}{2}x_{4}a \, \mathbf{\hat{x}} + \frac{\sqrt{3}}{2}x_{4}a \, \mathbf{\hat{y}} + z_{4}c \, \mathbf{\hat{z}} & \left(6c\right) & \mbox{Th III} \\ 
\mathbf{B}_{17} & = & -2x_{4} \, \mathbf{a}_{1}-x_{4} \, \mathbf{a}_{2} + z_{4} \, \mathbf{a}_{3} & = & -\frac{3}{2}x_{4}a \, \mathbf{\hat{x}} + \frac{\sqrt{3}}{2}x_{4}a \, \mathbf{\hat{y}} + z_{4}c \, \mathbf{\hat{z}} & \left(6c\right) & \mbox{Th III} \\ 
\mathbf{B}_{18} & = & -x_{4} \, \mathbf{a}_{1} + x_{4} \, \mathbf{a}_{2} + \left(\frac{1}{2} +z_{4}\right) \, \mathbf{a}_{3} & = & \sqrt{3}x_{4}a \, \mathbf{\hat{y}} + \left(\frac{1}{2} +z_{4}\right)c \, \mathbf{\hat{z}} & \left(6c\right) & \mbox{Th III} \\ 
\mathbf{B}_{19} & = & -x_{4} \, \mathbf{a}_{1}-2x_{4} \, \mathbf{a}_{2} + \left(\frac{1}{2} +z_{4}\right) \, \mathbf{a}_{3} & = & -\frac{3}{2}x_{4}a \, \mathbf{\hat{x}}-\frac{\sqrt{3}}{2}x_{4}a \, \mathbf{\hat{y}} + \left(\frac{1}{2} +z_{4}\right)c \, \mathbf{\hat{z}} & \left(6c\right) & \mbox{Th III} \\ 
\mathbf{B}_{20} & = & 2x_{4} \, \mathbf{a}_{1} + x_{4} \, \mathbf{a}_{2} + \left(\frac{1}{2} +z_{4}\right) \, \mathbf{a}_{3} & = & \frac{3}{2}x_{4}a \, \mathbf{\hat{x}}-\frac{\sqrt{3}}{2}x_{4}a \, \mathbf{\hat{y}} + \left(\frac{1}{2} +z_{4}\right)c \, \mathbf{\hat{z}} & \left(6c\right) & \mbox{Th III} \\ 
\end{longtabu}
\renewcommand{\arraystretch}{1.0}
\noindent \hrulefill
\\
\textbf{References:}
\vspace*{-0.25cm}
\begin{flushleft}
  - \bibentry{Florio_Acta_Cryst_9_1956}. \\
\end{flushleft}
\textbf{Found in:}
\vspace*{-0.25cm}
\begin{flushleft}
  - \bibentry{icsd:ICSD_401657}. \\
\end{flushleft}
\noindent \hrulefill
\\
\textbf{Geometry files:}
\\
\noindent  - CIF: pp. {\hyperref[A3B7_hP20_186_c_b2c_cif]{\pageref{A3B7_hP20_186_c_b2c_cif}}} \\
\noindent  - POSCAR: pp. {\hyperref[A3B7_hP20_186_c_b2c_poscar]{\pageref{A3B7_hP20_186_c_b2c_poscar}}} \\
\onecolumn
{\phantomsection\label{AB3_hP4_187_e_fh}}
\subsection*{\huge \textbf{{\normalfont Re$_{3}$N Structure: AB3\_hP4\_187\_e\_fh}}}
\noindent \hrulefill
\vspace*{0.25cm}
\begin{figure}[htp]
  \centering
  \vspace{-1em}
  {\includegraphics[width=1\textwidth]{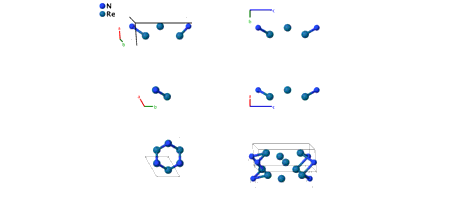}}
\end{figure}
\vspace*{-0.5cm}
\renewcommand{\arraystretch}{1.5}
\begin{equation*}
  \begin{array}{>{$\hspace{-0.15cm}}l<{$}>{$}p{0.5cm}<{$}>{$}p{18.5cm}<{$}}
    \mbox{\large \textbf{Prototype}} &\colon & \ce{Re3N} \\
    \mbox{\large \textbf{\AFLOW\ prototype label}} &\colon & \mbox{AB3\_hP4\_187\_e\_fh} \\
    \mbox{\large \textbf{\textit{Strukturbericht} designation}} &\colon & \mbox{None} \\
    \mbox{\large \textbf{Pearson symbol}} &\colon & \mbox{hP4} \\
    \mbox{\large \textbf{Space group number}} &\colon & 187 \\
    \mbox{\large \textbf{Space group symbol}} &\colon & P\bar{6}m2 \\
    \mbox{\large \textbf{\AFLOW\ prototype command}} &\colon &  \texttt{aflow} \,  \, \texttt{-{}-proto=AB3\_hP4\_187\_e\_fh } \, \newline \texttt{-{}-params=}{a,c/a,z_{3} }
  \end{array}
\end{equation*}
\renewcommand{\arraystretch}{1.0}

\vspace*{-0.25cm}
\noindent \hrulefill
\begin{itemize}
  \item{The reference presents both experimental findings and the results of Density Functional Theory calculations.  
We obtain our data from the density functional theory calculations at equilibrium (P = 0), which are consistent with the lattice constants found experimentally.
}
\end{itemize}

\noindent \parbox{1 \linewidth}{
\noindent \hrulefill
\\
\textbf{Hexagonal primitive vectors:} \\
\vspace*{-0.25cm}
\begin{tabular}{cc}
  \begin{tabular}{c}
    \parbox{0.6 \linewidth}{
      \renewcommand{\arraystretch}{1.5}
      \begin{equation*}
        \centering
        \begin{array}{ccc}
              \mathbf{a}_1 & = & \frac12 \, a \, \mathbf{\hat{x}} - \frac{\sqrt3}2 \, a \, \mathbf{\hat{y}} \\
    \mathbf{a}_2 & = & \frac12 \, a \, \mathbf{\hat{x}} + \frac{\sqrt3}2 \, a \, \mathbf{\hat{y}} \\
    \mathbf{a}_3 & = & c \, \mathbf{\hat{z}} \\

        \end{array}
      \end{equation*}
    }
    \renewcommand{\arraystretch}{1.0}
  \end{tabular}
  \begin{tabular}{c}
    \includegraphics[width=0.3\linewidth]{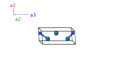} \\
  \end{tabular}
\end{tabular}

}
\vspace*{-0.25cm}

\noindent \hrulefill
\\
\textbf{Basis vectors:}
\vspace*{-0.25cm}
\renewcommand{\arraystretch}{1.5}
\begin{longtabu} to \textwidth{>{\centering $}X[-1,c,c]<{$}>{\centering $}X[-1,c,c]<{$}>{\centering $}X[-1,c,c]<{$}>{\centering $}X[-1,c,c]<{$}>{\centering $}X[-1,c,c]<{$}>{\centering $}X[-1,c,c]<{$}>{\centering $}X[-1,c,c]<{$}}
  & & \mbox{Lattice Coordinates} & & \mbox{Cartesian Coordinates} &\mbox{Wyckoff Position} & \mbox{Atom Type} \\  
  \mathbf{B}_{1} & = & \frac{2}{3} \, \mathbf{a}_{1} + \frac{1}{3} \, \mathbf{a}_{2} & = & \frac{1}{2}a \, \mathbf{\hat{x}}- \frac{1}{2\sqrt{3}}a  \, \mathbf{\hat{y}} & \left(1e\right) & \mbox{N} \\ 
\mathbf{B}_{2} & = & \frac{2}{3} \, \mathbf{a}_{1} + \frac{1}{3} \, \mathbf{a}_{2} + \frac{1}{2} \, \mathbf{a}_{3} & = & \frac{1}{2}a \, \mathbf{\hat{x}}- \frac{1}{2\sqrt{3}}a  \, \mathbf{\hat{y}} + \frac{1}{2}c \, \mathbf{\hat{z}} & \left(1f\right) & \mbox{Re I} \\ 
\mathbf{B}_{3} & = & \frac{1}{3} \, \mathbf{a}_{1} + \frac{2}{3} \, \mathbf{a}_{2} + z_{3} \, \mathbf{a}_{3} & = & \frac{1}{2}a \, \mathbf{\hat{x}} + \frac{1}{2\sqrt{3}}a \, \mathbf{\hat{y}} + z_{3}c \, \mathbf{\hat{z}} & \left(2h\right) & \mbox{Re II} \\ 
\mathbf{B}_{4} & = & \frac{1}{3} \, \mathbf{a}_{1} + \frac{2}{3} \, \mathbf{a}_{2}-z_{3} \, \mathbf{a}_{3} & = & \frac{1}{2}a \, \mathbf{\hat{x}} + \frac{1}{2\sqrt{3}}a \, \mathbf{\hat{y}}-z_{3}c \, \mathbf{\hat{z}} & \left(2h\right) & \mbox{Re II} \\ 
\end{longtabu}
\renewcommand{\arraystretch}{1.0}
\noindent \hrulefill
\\
\textbf{References:}
\vspace*{-0.25cm}
\begin{flushleft}
  - \bibentry{Friedrich_PRL_105_2010}. \\
\end{flushleft}
\noindent \hrulefill
\\
\textbf{Geometry files:}
\\
\noindent  - CIF: pp. {\hyperref[AB3_hP4_187_e_fh_cif]{\pageref{AB3_hP4_187_e_fh_cif}}} \\
\noindent  - POSCAR: pp. {\hyperref[AB3_hP4_187_e_fh_poscar]{\pageref{AB3_hP4_187_e_fh_poscar}}} \\
\onecolumn
{\phantomsection\label{A3BC_hP10_188_k_a_e}}
\subsection*{\huge \textbf{{\normalfont LiScI$_{3}$ Structure: A3BC\_hP10\_188\_k\_a\_e}}}
\noindent \hrulefill
\vspace*{0.25cm}
\begin{figure}[htp]
  \centering
  \vspace{-1em}
  {\includegraphics[width=1\textwidth]{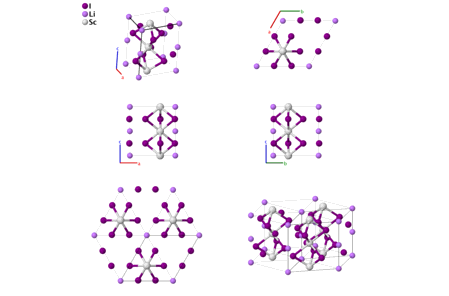}}
\end{figure}
\vspace*{-0.5cm}
\renewcommand{\arraystretch}{1.5}
\begin{equation*}
  \begin{array}{>{$\hspace{-0.15cm}}l<{$}>{$}p{0.5cm}<{$}>{$}p{18.5cm}<{$}}
    \mbox{\large \textbf{Prototype}} &\colon & \ce{LiScI3} \\
    \mbox{\large \textbf{\AFLOW\ prototype label}} &\colon & \mbox{A3BC\_hP10\_188\_k\_a\_e} \\
    \mbox{\large \textbf{\textit{Strukturbericht} designation}} &\colon & \mbox{None} \\
    \mbox{\large \textbf{Pearson symbol}} &\colon & \mbox{hP10} \\
    \mbox{\large \textbf{Space group number}} &\colon & 188 \\
    \mbox{\large \textbf{Space group symbol}} &\colon & P\bar{6}c2 \\
    \mbox{\large \textbf{\AFLOW\ prototype command}} &\colon &  \texttt{aflow} \,  \, \texttt{-{}-proto=A3BC\_hP10\_188\_k\_a\_e } \, \newline \texttt{-{}-params=}{a,c/a,x_{3},y_{3} }
  \end{array}
\end{equation*}
\renewcommand{\arraystretch}{1.0}

\noindent \parbox{1 \linewidth}{
\noindent \hrulefill
\\
\textbf{Hexagonal primitive vectors:} \\
\vspace*{-0.25cm}
\begin{tabular}{cc}
  \begin{tabular}{c}
    \parbox{0.6 \linewidth}{
      \renewcommand{\arraystretch}{1.5}
      \begin{equation*}
        \centering
        \begin{array}{ccc}
              \mathbf{a}_1 & = & \frac12 \, a \, \mathbf{\hat{x}} - \frac{\sqrt3}2 \, a \, \mathbf{\hat{y}} \\
    \mathbf{a}_2 & = & \frac12 \, a \, \mathbf{\hat{x}} + \frac{\sqrt3}2 \, a \, \mathbf{\hat{y}} \\
    \mathbf{a}_3 & = & c \, \mathbf{\hat{z}} \\

        \end{array}
      \end{equation*}
    }
    \renewcommand{\arraystretch}{1.0}
  \end{tabular}
  \begin{tabular}{c}
    \includegraphics[width=0.3\linewidth]{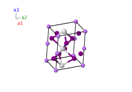} \\
  \end{tabular}
\end{tabular}

}
\vspace*{-0.25cm}

\noindent \hrulefill
\\
\textbf{Basis vectors:}
\vspace*{-0.25cm}
\renewcommand{\arraystretch}{1.5}
\begin{longtabu} to \textwidth{>{\centering $}X[-1,c,c]<{$}>{\centering $}X[-1,c,c]<{$}>{\centering $}X[-1,c,c]<{$}>{\centering $}X[-1,c,c]<{$}>{\centering $}X[-1,c,c]<{$}>{\centering $}X[-1,c,c]<{$}>{\centering $}X[-1,c,c]<{$}}
  & & \mbox{Lattice Coordinates} & & \mbox{Cartesian Coordinates} &\mbox{Wyckoff Position} & \mbox{Atom Type} \\  
  \mathbf{B}_{1} & = & 0 \, \mathbf{a}_{1} + 0 \, \mathbf{a}_{2} + 0 \, \mathbf{a}_{3} & = & 0 \, \mathbf{\hat{x}} + 0 \, \mathbf{\hat{y}} + 0 \, \mathbf{\hat{z}} & \left(2a\right) & \mbox{Li} \\ 
\mathbf{B}_{2} & = & \frac{1}{2} \, \mathbf{a}_{3} & = & \frac{1}{2}c \, \mathbf{\hat{z}} & \left(2a\right) & \mbox{Li} \\ 
\mathbf{B}_{3} & = & \frac{2}{3} \, \mathbf{a}_{1} + \frac{1}{3} \, \mathbf{a}_{2} & = & \frac{1}{2}a \, \mathbf{\hat{x}}- \frac{1}{2\sqrt{3}}a  \, \mathbf{\hat{y}} & \left(2e\right) & \mbox{Sc} \\ 
\mathbf{B}_{4} & = & \frac{2}{3} \, \mathbf{a}_{1} + \frac{1}{3} \, \mathbf{a}_{2} + \frac{1}{2} \, \mathbf{a}_{3} & = & \frac{1}{2}a \, \mathbf{\hat{x}}- \frac{1}{2\sqrt{3}}a  \, \mathbf{\hat{y}} + \frac{1}{2}c \, \mathbf{\hat{z}} & \left(2e\right) & \mbox{Sc} \\ 
\mathbf{B}_{5} & = & x_{3} \, \mathbf{a}_{1} + y_{3} \, \mathbf{a}_{2} + \frac{1}{4} \, \mathbf{a}_{3} & = & \frac{1}{2}\left(x_{3}+y_{3}\right)a \, \mathbf{\hat{x}} + \frac{\sqrt{3}}{2}\left(-x_{3}+y_{3}\right)a \, \mathbf{\hat{y}} + \frac{1}{4}c \, \mathbf{\hat{z}} & \left(6k\right) & \mbox{I} \\ 
\mathbf{B}_{6} & = & -y_{3} \, \mathbf{a}_{1} + \left(x_{3}-y_{3}\right) \, \mathbf{a}_{2} + \frac{1}{4} \, \mathbf{a}_{3} & = & \left(\frac{1}{2}x_{3}-y_{3}\right)a \, \mathbf{\hat{x}} + \frac{\sqrt{3}}{2}x_{3}a \, \mathbf{\hat{y}} + \frac{1}{4}c \, \mathbf{\hat{z}} & \left(6k\right) & \mbox{I} \\ 
\mathbf{B}_{7} & = & \left(-x_{3}+y_{3}\right) \, \mathbf{a}_{1}-x_{3} \, \mathbf{a}_{2} + \frac{1}{4} \, \mathbf{a}_{3} & = & \left(-x_{3}+\frac{1}{2}y_{3}\right)a \, \mathbf{\hat{x}}-\frac{\sqrt{3}}{2}y_{3}a \, \mathbf{\hat{y}} + \frac{1}{4}c \, \mathbf{\hat{z}} & \left(6k\right) & \mbox{I} \\ 
\mathbf{B}_{8} & = & -y_{3} \, \mathbf{a}_{1}-x_{3} \, \mathbf{a}_{2} + \frac{3}{4} \, \mathbf{a}_{3} & = & -\frac{1}{2}\left(x_{3}+y_{3}\right)a \, \mathbf{\hat{x}} + \frac{\sqrt{3}}{2}\left(-x_{3}+y_{3}\right)a \, \mathbf{\hat{y}} + \frac{3}{4}c \, \mathbf{\hat{z}} & \left(6k\right) & \mbox{I} \\ 
\mathbf{B}_{9} & = & \left(-x_{3}+y_{3}\right) \, \mathbf{a}_{1} + y_{3} \, \mathbf{a}_{2} + \frac{3}{4} \, \mathbf{a}_{3} & = & \left(-\frac{1}{2}x_{3}+y_{3}\right)a \, \mathbf{\hat{x}} + \frac{\sqrt{3}}{2}x_{3}a \, \mathbf{\hat{y}} + \frac{3}{4}c \, \mathbf{\hat{z}} & \left(6k\right) & \mbox{I} \\ 
\mathbf{B}_{10} & = & x_{3} \, \mathbf{a}_{1} + \left(x_{3}-y_{3}\right) \, \mathbf{a}_{2} + \frac{3}{4} \, \mathbf{a}_{3} & = & \left(x_{3}-\frac{1}{2}y_{3}\right)a \, \mathbf{\hat{x}}-\frac{\sqrt{3}}{2}y_{3}a \, \mathbf{\hat{y}} + \frac{3}{4}c \, \mathbf{\hat{z}} & \left(6k\right) & \mbox{I} \\ 
\end{longtabu}
\renewcommand{\arraystretch}{1.0}
\noindent \hrulefill
\\
\textbf{References:}
\vspace*{-0.25cm}
\begin{flushleft}
  - \bibentry{Lachgar_LiScI3_InorgChem_1991}. \\
\end{flushleft}
\textbf{Found in:}
\vspace*{-0.25cm}
\begin{flushleft}
  - \bibentry{Villars_PearsonsCrystalData_2013}. \\
\end{flushleft}
\noindent \hrulefill
\\
\textbf{Geometry files:}
\\
\noindent  - CIF: pp. {\hyperref[A3BC_hP10_188_k_a_e_cif]{\pageref{A3BC_hP10_188_k_a_e_cif}}} \\
\noindent  - POSCAR: pp. {\hyperref[A3BC_hP10_188_k_a_e_poscar]{\pageref{A3BC_hP10_188_k_a_e_poscar}}} \\
\onecolumn
{\phantomsection\label{AB9C4_hP28_188_e_kl_ak}}
\subsection*{\huge \textbf{{\normalfont BaSi$_{4}$O$_{9}$ Structure: AB9C4\_hP28\_188\_e\_kl\_ak}}}
\noindent \hrulefill
\vspace*{0.25cm}
\begin{figure}[htp]
  \centering
  \vspace{-1em}
  {\includegraphics[width=1\textwidth]{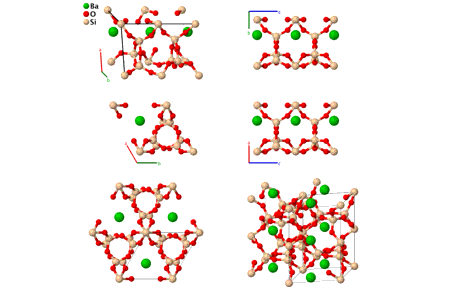}}
\end{figure}
\vspace*{-0.5cm}
\renewcommand{\arraystretch}{1.5}
\begin{equation*}
  \begin{array}{>{$\hspace{-0.15cm}}l<{$}>{$}p{0.5cm}<{$}>{$}p{18.5cm}<{$}}
    \mbox{\large \textbf{Prototype}} &\colon & \ce{BaSi4O9} \\
    \mbox{\large \textbf{\AFLOW\ prototype label}} &\colon & \mbox{AB9C4\_hP28\_188\_e\_kl\_ak} \\
    \mbox{\large \textbf{\textit{Strukturbericht} designation}} &\colon & \mbox{None} \\
    \mbox{\large \textbf{Pearson symbol}} &\colon & \mbox{hP28} \\
    \mbox{\large \textbf{Space group number}} &\colon & 188 \\
    \mbox{\large \textbf{Space group symbol}} &\colon & P\bar{6}c2 \\
    \mbox{\large \textbf{\AFLOW\ prototype command}} &\colon &  \texttt{aflow} \,  \, \texttt{-{}-proto=AB9C4\_hP28\_188\_e\_kl\_ak } \, \newline \texttt{-{}-params=}{a,c/a,x_{3},y_{3},x_{4},y_{4},x_{5},y_{5},z_{5} }
  \end{array}
\end{equation*}
\renewcommand{\arraystretch}{1.0}

\noindent \parbox{1 \linewidth}{
\noindent \hrulefill
\\
\textbf{Hexagonal primitive vectors:} \\
\vspace*{-0.25cm}
\begin{tabular}{cc}
  \begin{tabular}{c}
    \parbox{0.6 \linewidth}{
      \renewcommand{\arraystretch}{1.5}
      \begin{equation*}
        \centering
        \begin{array}{ccc}
              \mathbf{a}_1 & = & \frac12 \, a \, \mathbf{\hat{x}} - \frac{\sqrt3}2 \, a \, \mathbf{\hat{y}} \\
    \mathbf{a}_2 & = & \frac12 \, a \, \mathbf{\hat{x}} + \frac{\sqrt3}2 \, a \, \mathbf{\hat{y}} \\
    \mathbf{a}_3 & = & c \, \mathbf{\hat{z}} \\

        \end{array}
      \end{equation*}
    }
    \renewcommand{\arraystretch}{1.0}
  \end{tabular}
  \begin{tabular}{c}
    \includegraphics[width=0.3\linewidth]{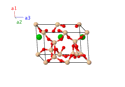} \\
  \end{tabular}
\end{tabular}

}
\vspace*{-0.25cm}

\noindent \hrulefill
\\
\textbf{Basis vectors:}
\vspace*{-0.25cm}
\renewcommand{\arraystretch}{1.5}
\begin{longtabu} to \textwidth{>{\centering $}X[-1,c,c]<{$}>{\centering $}X[-1,c,c]<{$}>{\centering $}X[-1,c,c]<{$}>{\centering $}X[-1,c,c]<{$}>{\centering $}X[-1,c,c]<{$}>{\centering $}X[-1,c,c]<{$}>{\centering $}X[-1,c,c]<{$}}
  & & \mbox{Lattice Coordinates} & & \mbox{Cartesian Coordinates} &\mbox{Wyckoff Position} & \mbox{Atom Type} \\  
  \mathbf{B}_{1} & = & 0 \, \mathbf{a}_{1} + 0 \, \mathbf{a}_{2} + 0 \, \mathbf{a}_{3} & = & 0 \, \mathbf{\hat{x}} + 0 \, \mathbf{\hat{y}} + 0 \, \mathbf{\hat{z}} & \left(2a\right) & \mbox{Si I} \\ 
\mathbf{B}_{2} & = & \frac{1}{2} \, \mathbf{a}_{3} & = & \frac{1}{2}c \, \mathbf{\hat{z}} & \left(2a\right) & \mbox{Si I} \\ 
\mathbf{B}_{3} & = & \frac{2}{3} \, \mathbf{a}_{1} + \frac{1}{3} \, \mathbf{a}_{2} & = & \frac{1}{2}a \, \mathbf{\hat{x}}- \frac{1}{2\sqrt{3}}a  \, \mathbf{\hat{y}} & \left(2e\right) & \mbox{Ba} \\ 
\mathbf{B}_{4} & = & \frac{2}{3} \, \mathbf{a}_{1} + \frac{1}{3} \, \mathbf{a}_{2} + \frac{1}{2} \, \mathbf{a}_{3} & = & \frac{1}{2}a \, \mathbf{\hat{x}}- \frac{1}{2\sqrt{3}}a  \, \mathbf{\hat{y}} + \frac{1}{2}c \, \mathbf{\hat{z}} & \left(2e\right) & \mbox{Ba} \\ 
\mathbf{B}_{5} & = & x_{3} \, \mathbf{a}_{1} + y_{3} \, \mathbf{a}_{2} + \frac{1}{4} \, \mathbf{a}_{3} & = & \frac{1}{2}\left(x_{3}+y_{3}\right)a \, \mathbf{\hat{x}} + \frac{\sqrt{3}}{2}\left(-x_{3}+y_{3}\right)a \, \mathbf{\hat{y}} + \frac{1}{4}c \, \mathbf{\hat{z}} & \left(6k\right) & \mbox{O I} \\ 
\mathbf{B}_{6} & = & -y_{3} \, \mathbf{a}_{1} + \left(x_{3}-y_{3}\right) \, \mathbf{a}_{2} + \frac{1}{4} \, \mathbf{a}_{3} & = & \left(\frac{1}{2}x_{3}-y_{3}\right)a \, \mathbf{\hat{x}} + \frac{\sqrt{3}}{2}x_{3}a \, \mathbf{\hat{y}} + \frac{1}{4}c \, \mathbf{\hat{z}} & \left(6k\right) & \mbox{O I} \\ 
\mathbf{B}_{7} & = & \left(-x_{3}+y_{3}\right) \, \mathbf{a}_{1}-x_{3} \, \mathbf{a}_{2} + \frac{1}{4} \, \mathbf{a}_{3} & = & \left(-x_{3}+\frac{1}{2}y_{3}\right)a \, \mathbf{\hat{x}}-\frac{\sqrt{3}}{2}y_{3}a \, \mathbf{\hat{y}} + \frac{1}{4}c \, \mathbf{\hat{z}} & \left(6k\right) & \mbox{O I} \\ 
\mathbf{B}_{8} & = & -y_{3} \, \mathbf{a}_{1}-x_{3} \, \mathbf{a}_{2} + \frac{3}{4} \, \mathbf{a}_{3} & = & -\frac{1}{2}\left(x_{3}+y_{3}\right)a \, \mathbf{\hat{x}} + \frac{\sqrt{3}}{2}\left(-x_{3}+y_{3}\right)a \, \mathbf{\hat{y}} + \frac{3}{4}c \, \mathbf{\hat{z}} & \left(6k\right) & \mbox{O I} \\ 
\mathbf{B}_{9} & = & \left(-x_{3}+y_{3}\right) \, \mathbf{a}_{1} + y_{3} \, \mathbf{a}_{2} + \frac{3}{4} \, \mathbf{a}_{3} & = & \left(-\frac{1}{2}x_{3}+y_{3}\right)a \, \mathbf{\hat{x}} + \frac{\sqrt{3}}{2}x_{3}a \, \mathbf{\hat{y}} + \frac{3}{4}c \, \mathbf{\hat{z}} & \left(6k\right) & \mbox{O I} \\ 
\mathbf{B}_{10} & = & x_{3} \, \mathbf{a}_{1} + \left(x_{3}-y_{3}\right) \, \mathbf{a}_{2} + \frac{3}{4} \, \mathbf{a}_{3} & = & \left(x_{3}-\frac{1}{2}y_{3}\right)a \, \mathbf{\hat{x}}-\frac{\sqrt{3}}{2}y_{3}a \, \mathbf{\hat{y}} + \frac{3}{4}c \, \mathbf{\hat{z}} & \left(6k\right) & \mbox{O I} \\ 
\mathbf{B}_{11} & = & x_{4} \, \mathbf{a}_{1} + y_{4} \, \mathbf{a}_{2} + \frac{1}{4} \, \mathbf{a}_{3} & = & \frac{1}{2}\left(x_{4}+y_{4}\right)a \, \mathbf{\hat{x}} + \frac{\sqrt{3}}{2}\left(-x_{4}+y_{4}\right)a \, \mathbf{\hat{y}} + \frac{1}{4}c \, \mathbf{\hat{z}} & \left(6k\right) & \mbox{Si II} \\ 
\mathbf{B}_{12} & = & -y_{4} \, \mathbf{a}_{1} + \left(x_{4}-y_{4}\right) \, \mathbf{a}_{2} + \frac{1}{4} \, \mathbf{a}_{3} & = & \left(\frac{1}{2}x_{4}-y_{4}\right)a \, \mathbf{\hat{x}} + \frac{\sqrt{3}}{2}x_{4}a \, \mathbf{\hat{y}} + \frac{1}{4}c \, \mathbf{\hat{z}} & \left(6k\right) & \mbox{Si II} \\ 
\mathbf{B}_{13} & = & \left(-x_{4}+y_{4}\right) \, \mathbf{a}_{1}-x_{4} \, \mathbf{a}_{2} + \frac{1}{4} \, \mathbf{a}_{3} & = & \left(-x_{4}+\frac{1}{2}y_{4}\right)a \, \mathbf{\hat{x}}-\frac{\sqrt{3}}{2}y_{4}a \, \mathbf{\hat{y}} + \frac{1}{4}c \, \mathbf{\hat{z}} & \left(6k\right) & \mbox{Si II} \\ 
\mathbf{B}_{14} & = & -y_{4} \, \mathbf{a}_{1}-x_{4} \, \mathbf{a}_{2} + \frac{3}{4} \, \mathbf{a}_{3} & = & -\frac{1}{2}\left(x_{4}+y_{4}\right)a \, \mathbf{\hat{x}} + \frac{\sqrt{3}}{2}\left(-x_{4}+y_{4}\right)a \, \mathbf{\hat{y}} + \frac{3}{4}c \, \mathbf{\hat{z}} & \left(6k\right) & \mbox{Si II} \\ 
\mathbf{B}_{15} & = & \left(-x_{4}+y_{4}\right) \, \mathbf{a}_{1} + y_{4} \, \mathbf{a}_{2} + \frac{3}{4} \, \mathbf{a}_{3} & = & \left(-\frac{1}{2}x_{4}+y_{4}\right)a \, \mathbf{\hat{x}} + \frac{\sqrt{3}}{2}x_{4}a \, \mathbf{\hat{y}} + \frac{3}{4}c \, \mathbf{\hat{z}} & \left(6k\right) & \mbox{Si II} \\ 
\mathbf{B}_{16} & = & x_{4} \, \mathbf{a}_{1} + \left(x_{4}-y_{4}\right) \, \mathbf{a}_{2} + \frac{3}{4} \, \mathbf{a}_{3} & = & \left(x_{4}-\frac{1}{2}y_{4}\right)a \, \mathbf{\hat{x}}-\frac{\sqrt{3}}{2}y_{4}a \, \mathbf{\hat{y}} + \frac{3}{4}c \, \mathbf{\hat{z}} & \left(6k\right) & \mbox{Si II} \\ 
\mathbf{B}_{17} & = & x_{5} \, \mathbf{a}_{1} + y_{5} \, \mathbf{a}_{2} + z_{5} \, \mathbf{a}_{3} & = & \frac{1}{2}\left(x_{5}+y_{5}\right)a \, \mathbf{\hat{x}} + \frac{\sqrt{3}}{2}\left(-x_{5}+y_{5}\right)a \, \mathbf{\hat{y}} + z_{5}c \, \mathbf{\hat{z}} & \left(12l\right) & \mbox{O II} \\ 
\mathbf{B}_{18} & = & -y_{5} \, \mathbf{a}_{1} + \left(x_{5}-y_{5}\right) \, \mathbf{a}_{2} + z_{5} \, \mathbf{a}_{3} & = & \left(\frac{1}{2}x_{5}-y_{5}\right)a \, \mathbf{\hat{x}} + \frac{\sqrt{3}}{2}x_{5}a \, \mathbf{\hat{y}} + z_{5}c \, \mathbf{\hat{z}} & \left(12l\right) & \mbox{O II} \\ 
\mathbf{B}_{19} & = & \left(-x_{5}+y_{5}\right) \, \mathbf{a}_{1}-x_{5} \, \mathbf{a}_{2} + z_{5} \, \mathbf{a}_{3} & = & \left(-x_{5}+\frac{1}{2}y_{5}\right)a \, \mathbf{\hat{x}}-\frac{\sqrt{3}}{2}y_{5}a \, \mathbf{\hat{y}} + z_{5}c \, \mathbf{\hat{z}} & \left(12l\right) & \mbox{O II} \\ 
\mathbf{B}_{20} & = & x_{5} \, \mathbf{a}_{1} + y_{5} \, \mathbf{a}_{2} + \left(\frac{1}{2} - z_{5}\right) \, \mathbf{a}_{3} & = & \frac{1}{2}\left(x_{5}+y_{5}\right)a \, \mathbf{\hat{x}} + \frac{\sqrt{3}}{2}\left(-x_{5}+y_{5}\right)a \, \mathbf{\hat{y}} + \left(\frac{1}{2} - z_{5}\right)c \, \mathbf{\hat{z}} & \left(12l\right) & \mbox{O II} \\ 
\mathbf{B}_{21} & = & -y_{5} \, \mathbf{a}_{1} + \left(x_{5}-y_{5}\right) \, \mathbf{a}_{2} + \left(\frac{1}{2} - z_{5}\right) \, \mathbf{a}_{3} & = & \left(\frac{1}{2}x_{5}-y_{5}\right)a \, \mathbf{\hat{x}} + \frac{\sqrt{3}}{2}x_{5}a \, \mathbf{\hat{y}} + \left(\frac{1}{2} - z_{5}\right)c \, \mathbf{\hat{z}} & \left(12l\right) & \mbox{O II} \\ 
\mathbf{B}_{22} & = & \left(-x_{5}+y_{5}\right) \, \mathbf{a}_{1}-x_{5} \, \mathbf{a}_{2} + \left(\frac{1}{2} - z_{5}\right) \, \mathbf{a}_{3} & = & \left(-x_{5}+\frac{1}{2}y_{5}\right)a \, \mathbf{\hat{x}}-\frac{\sqrt{3}}{2}y_{5}a \, \mathbf{\hat{y}} + \left(\frac{1}{2} - z_{5}\right)c \, \mathbf{\hat{z}} & \left(12l\right) & \mbox{O II} \\ 
\mathbf{B}_{23} & = & -y_{5} \, \mathbf{a}_{1}-x_{5} \, \mathbf{a}_{2} + \left(\frac{1}{2} +z_{5}\right) \, \mathbf{a}_{3} & = & -\frac{1}{2}\left(x_{5}+y_{5}\right)a \, \mathbf{\hat{x}} + \frac{\sqrt{3}}{2}\left(-x_{5}+y_{5}\right)a \, \mathbf{\hat{y}} + \left(\frac{1}{2} +z_{5}\right)c \, \mathbf{\hat{z}} & \left(12l\right) & \mbox{O II} \\ 
\mathbf{B}_{24} & = & \left(-x_{5}+y_{5}\right) \, \mathbf{a}_{1} + y_{5} \, \mathbf{a}_{2} + \left(\frac{1}{2} +z_{5}\right) \, \mathbf{a}_{3} & = & \left(-\frac{1}{2}x_{5}+y_{5}\right)a \, \mathbf{\hat{x}} + \frac{\sqrt{3}}{2}x_{5}a \, \mathbf{\hat{y}} + \left(\frac{1}{2} +z_{5}\right)c \, \mathbf{\hat{z}} & \left(12l\right) & \mbox{O II} \\ 
\mathbf{B}_{25} & = & x_{5} \, \mathbf{a}_{1} + \left(x_{5}-y_{5}\right) \, \mathbf{a}_{2} + \left(\frac{1}{2} +z_{5}\right) \, \mathbf{a}_{3} & = & \left(x_{5}-\frac{1}{2}y_{5}\right)a \, \mathbf{\hat{x}}-\frac{\sqrt{3}}{2}y_{5}a \, \mathbf{\hat{y}} + \left(\frac{1}{2} +z_{5}\right)c \, \mathbf{\hat{z}} & \left(12l\right) & \mbox{O II} \\ 
\mathbf{B}_{26} & = & -y_{5} \, \mathbf{a}_{1}-x_{5} \, \mathbf{a}_{2}-z_{5} \, \mathbf{a}_{3} & = & -\frac{1}{2}\left(x_{5}+y_{5}\right)a \, \mathbf{\hat{x}} + \frac{\sqrt{3}}{2}\left(-x_{5}+y_{5}\right)a \, \mathbf{\hat{y}}-z_{5}c \, \mathbf{\hat{z}} & \left(12l\right) & \mbox{O II} \\ 
\mathbf{B}_{27} & = & \left(-x_{5}+y_{5}\right) \, \mathbf{a}_{1} + y_{5} \, \mathbf{a}_{2}-z_{5} \, \mathbf{a}_{3} & = & \left(-\frac{1}{2}x_{5}+y_{5}\right)a \, \mathbf{\hat{x}} + \frac{\sqrt{3}}{2}x_{5}a \, \mathbf{\hat{y}}-z_{5}c \, \mathbf{\hat{z}} & \left(12l\right) & \mbox{O II} \\ 
\mathbf{B}_{28} & = & x_{5} \, \mathbf{a}_{1} + \left(x_{5}-y_{5}\right) \, \mathbf{a}_{2}-z_{5} \, \mathbf{a}_{3} & = & \left(x_{5}-\frac{1}{2}y_{5}\right)a \, \mathbf{\hat{x}}-\frac{\sqrt{3}}{2}y_{5}a \, \mathbf{\hat{y}}-z_{5}c \, \mathbf{\hat{z}} & \left(12l\right) & \mbox{O II} \\ 
\end{longtabu}
\renewcommand{\arraystretch}{1.0}
\noindent \hrulefill
\\
\textbf{References:}
\vspace*{-0.25cm}
\begin{flushleft}
  - \bibentry{Finger_BaSi4O9_JPhysChemSol_1995}. \\
\end{flushleft}
\textbf{Found in:}
\vspace*{-0.25cm}
\begin{flushleft}
  - \bibentry{Villars_PearsonsCrystalData_2013}. \\
\end{flushleft}
\noindent \hrulefill
\\
\textbf{Geometry files:}
\\
\noindent  - CIF: pp. {\hyperref[AB9C4_hP28_188_e_kl_ak_cif]{\pageref{AB9C4_hP28_188_e_kl_ak_cif}}} \\
\noindent  - POSCAR: pp. {\hyperref[AB9C4_hP28_188_e_kl_ak_poscar]{\pageref{AB9C4_hP28_188_e_kl_ak_poscar}}} \\
\onecolumn
{\phantomsection\label{A8BC3D6_hP18_189_bfh_a_g_i}}
\subsection*{\huge \textbf{{\normalfont \begin{raggedleft}$\pi$-FeMg$_{3}$Al$_{8}$Si$_{6}$ ($E9_{b}$) Structure: \end{raggedleft} \\ A8BC3D6\_hP18\_189\_bfh\_a\_g\_i}}}
\noindent \hrulefill
\vspace*{0.25cm}
\begin{figure}[htp]
  \centering
  \vspace{-1em}
  {\includegraphics[width=1\textwidth]{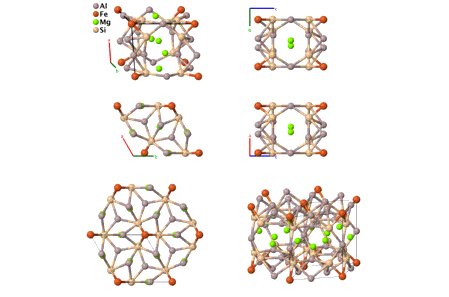}}
\end{figure}
\vspace*{-0.5cm}
\renewcommand{\arraystretch}{1.5}
\begin{equation*}
  \begin{array}{>{$\hspace{-0.15cm}}l<{$}>{$}p{0.5cm}<{$}>{$}p{18.5cm}<{$}}
    \mbox{\large \textbf{Prototype}} &\colon & \ce{$\pi$-FeMg3Al8Si6} \\
    \mbox{\large \textbf{\AFLOW\ prototype label}} &\colon & \mbox{A8BC3D6\_hP18\_189\_bfh\_a\_g\_i} \\
    \mbox{\large \textbf{\textit{Strukturbericht} designation}} &\colon & \mbox{$E9_{b}$} \\
    \mbox{\large \textbf{Pearson symbol}} &\colon & \mbox{hP18} \\
    \mbox{\large \textbf{Space group number}} &\colon & 189 \\
    \mbox{\large \textbf{Space group symbol}} &\colon & P\bar{6}2m \\
    \mbox{\large \textbf{\AFLOW\ prototype command}} &\colon &  \texttt{aflow} \,  \, \texttt{-{}-proto=A8BC3D6\_hP18\_189\_bfh\_a\_g\_i } \, \newline \texttt{-{}-params=}{a,c/a,x_{3},x_{4},z_{5},x_{6},z_{6} }
  \end{array}
\end{equation*}
\renewcommand{\arraystretch}{1.0}

\vspace*{-0.25cm}
\noindent \hrulefill
\begin{itemize}
  \item{We have been unable to obtain a copy of (Perlitz, 1942), 
and use the data provided by (Brandes, 1992) and (Foss, 2003).
Foss {\em et al.} argue that the actual compositition of this phase
should be FeMg$_{3}$Al$_{9}$Si$_{5}$.  This requires a reordering of
the atomic positions, as described in
\hyperref[A9BC3D5_hP18_189_fi_a_g_bh]{A9BC3D5\_hP18\_189\_fi\_a\_g\_bh}.
}
\end{itemize}

\noindent \parbox{1 \linewidth}{
\noindent \hrulefill
\\
\textbf{Hexagonal primitive vectors:} \\
\vspace*{-0.25cm}
\begin{tabular}{cc}
  \begin{tabular}{c}
    \parbox{0.6 \linewidth}{
      \renewcommand{\arraystretch}{1.5}
      \begin{equation*}
        \centering
        \begin{array}{ccc}
              \mathbf{a}_1 & = & \frac12 \, a \, \mathbf{\hat{x}} - \frac{\sqrt3}2 \, a \, \mathbf{\hat{y}} \\
    \mathbf{a}_2 & = & \frac12 \, a \, \mathbf{\hat{x}} + \frac{\sqrt3}2 \, a \, \mathbf{\hat{y}} \\
    \mathbf{a}_3 & = & c \, \mathbf{\hat{z}} \\

        \end{array}
      \end{equation*}
    }
    \renewcommand{\arraystretch}{1.0}
  \end{tabular}
  \begin{tabular}{c}
    \includegraphics[width=0.3\linewidth]{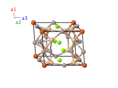} \\
  \end{tabular}
\end{tabular}

}
\vspace*{-0.25cm}

\noindent \hrulefill
\\
\textbf{Basis vectors:}
\vspace*{-0.25cm}
\renewcommand{\arraystretch}{1.5}
\begin{longtabu} to \textwidth{>{\centering $}X[-1,c,c]<{$}>{\centering $}X[-1,c,c]<{$}>{\centering $}X[-1,c,c]<{$}>{\centering $}X[-1,c,c]<{$}>{\centering $}X[-1,c,c]<{$}>{\centering $}X[-1,c,c]<{$}>{\centering $}X[-1,c,c]<{$}}
  & & \mbox{Lattice Coordinates} & & \mbox{Cartesian Coordinates} &\mbox{Wyckoff Position} & \mbox{Atom Type} \\  
  \mathbf{B}_{1} & = & 0 \, \mathbf{a}_{1} + 0 \, \mathbf{a}_{2} + 0 \, \mathbf{a}_{3} & = & 0 \, \mathbf{\hat{x}} + 0 \, \mathbf{\hat{y}} + 0 \, \mathbf{\hat{z}} & \left(1a\right) & \mbox{Fe} \\ 
\mathbf{B}_{2} & = & \frac{1}{2} \, \mathbf{a}_{3} & = & \frac{1}{2}c \, \mathbf{\hat{z}} & \left(1b\right) & \mbox{Al I} \\ 
\mathbf{B}_{3} & = & x_{3} \, \mathbf{a}_{1} & = & \frac{1}{2}x_{3}a \, \mathbf{\hat{x}}-\frac{\sqrt{3}}{2}x_{3}a \, \mathbf{\hat{y}} & \left(3f\right) & \mbox{Al II} \\ 
\mathbf{B}_{4} & = & x_{3} \, \mathbf{a}_{2} & = & \frac{1}{2}x_{3}a \, \mathbf{\hat{x}} + \frac{\sqrt{3}}{2}x_{3}a \, \mathbf{\hat{y}} & \left(3f\right) & \mbox{Al II} \\ 
\mathbf{B}_{5} & = & -x_{3} \, \mathbf{a}_{1}-x_{3} \, \mathbf{a}_{2} & = & -x_{3}a \, \mathbf{\hat{x}} & \left(3f\right) & \mbox{Al II} \\ 
\mathbf{B}_{6} & = & x_{4} \, \mathbf{a}_{1} + \frac{1}{2} \, \mathbf{a}_{3} & = & \frac{1}{2}x_{4}a \, \mathbf{\hat{x}}-\frac{\sqrt{3}}{2}x_{4}a \, \mathbf{\hat{y}} + \frac{1}{2}c \, \mathbf{\hat{z}} & \left(3g\right) & \mbox{Mg} \\ 
\mathbf{B}_{7} & = & x_{4} \, \mathbf{a}_{2} + \frac{1}{2} \, \mathbf{a}_{3} & = & \frac{1}{2}x_{4}a \, \mathbf{\hat{x}} + \frac{\sqrt{3}}{2}x_{4}a \, \mathbf{\hat{y}} + \frac{1}{2}c \, \mathbf{\hat{z}} & \left(3g\right) & \mbox{Mg} \\ 
\mathbf{B}_{8} & = & -x_{4} \, \mathbf{a}_{1}-x_{4} \, \mathbf{a}_{2} + \frac{1}{2} \, \mathbf{a}_{3} & = & -x_{4}a \, \mathbf{\hat{x}} + \frac{1}{2}c \, \mathbf{\hat{z}} & \left(3g\right) & \mbox{Mg} \\ 
\mathbf{B}_{9} & = & \frac{1}{3} \, \mathbf{a}_{1} + \frac{2}{3} \, \mathbf{a}_{2} + z_{5} \, \mathbf{a}_{3} & = & \frac{1}{2}a \, \mathbf{\hat{x}} + \frac{1}{2\sqrt{3}}a \, \mathbf{\hat{y}} + z_{5}c \, \mathbf{\hat{z}} & \left(4h\right) & \mbox{Al III} \\ 
\mathbf{B}_{10} & = & \frac{1}{3} \, \mathbf{a}_{1} + \frac{2}{3} \, \mathbf{a}_{2}-z_{5} \, \mathbf{a}_{3} & = & \frac{1}{2}a \, \mathbf{\hat{x}} + \frac{1}{2\sqrt{3}}a \, \mathbf{\hat{y}}-z_{5}c \, \mathbf{\hat{z}} & \left(4h\right) & \mbox{Al III} \\ 
\mathbf{B}_{11} & = & \frac{2}{3} \, \mathbf{a}_{1} + \frac{1}{3} \, \mathbf{a}_{2}-z_{5} \, \mathbf{a}_{3} & = & \frac{1}{2}a \, \mathbf{\hat{x}}- \frac{1}{2\sqrt{3}}a  \, \mathbf{\hat{y}}-z_{5}c \, \mathbf{\hat{z}} & \left(4h\right) & \mbox{Al III} \\ 
\mathbf{B}_{12} & = & \frac{2}{3} \, \mathbf{a}_{1} + \frac{1}{3} \, \mathbf{a}_{2} + z_{5} \, \mathbf{a}_{3} & = & \frac{1}{2}a \, \mathbf{\hat{x}}- \frac{1}{2\sqrt{3}}a  \, \mathbf{\hat{y}} + z_{5}c \, \mathbf{\hat{z}} & \left(4h\right) & \mbox{Al III} \\ 
\mathbf{B}_{13} & = & x_{6} \, \mathbf{a}_{1} + z_{6} \, \mathbf{a}_{3} & = & \frac{1}{2}x_{6}a \, \mathbf{\hat{x}}-\frac{\sqrt{3}}{2}x_{6}a \, \mathbf{\hat{y}} + z_{6}c \, \mathbf{\hat{z}} & \left(6i\right) & \mbox{Si} \\ 
\mathbf{B}_{14} & = & x_{6} \, \mathbf{a}_{2} + z_{6} \, \mathbf{a}_{3} & = & \frac{1}{2}x_{6}a \, \mathbf{\hat{x}} + \frac{\sqrt{3}}{2}x_{6}a \, \mathbf{\hat{y}} + z_{6}c \, \mathbf{\hat{z}} & \left(6i\right) & \mbox{Si} \\ 
\mathbf{B}_{15} & = & -x_{6} \, \mathbf{a}_{1}-x_{6} \, \mathbf{a}_{2} + z_{6} \, \mathbf{a}_{3} & = & -x_{6}a \, \mathbf{\hat{x}} + z_{6}c \, \mathbf{\hat{z}} & \left(6i\right) & \mbox{Si} \\ 
\mathbf{B}_{16} & = & x_{6} \, \mathbf{a}_{1} + -z_{6} \, \mathbf{a}_{3} & = & \frac{1}{2}x_{6}a \, \mathbf{\hat{x}}-\frac{\sqrt{3}}{2}x_{6}a \, \mathbf{\hat{y}}-z_{6}c \, \mathbf{\hat{z}} & \left(6i\right) & \mbox{Si} \\ 
\mathbf{B}_{17} & = & x_{6} \, \mathbf{a}_{2}-z_{6} \, \mathbf{a}_{3} & = & \frac{1}{2}x_{6}a \, \mathbf{\hat{x}} + \frac{\sqrt{3}}{2}x_{6}a \, \mathbf{\hat{y}}-z_{6}c \, \mathbf{\hat{z}} & \left(6i\right) & \mbox{Si} \\ 
\mathbf{B}_{18} & = & -x_{6} \, \mathbf{a}_{1}-x_{6} \, \mathbf{a}_{2}-z_{6} \, \mathbf{a}_{3} & = & -x_{6}a \, \mathbf{\hat{x}} + -z_{6}c \, \mathbf{\hat{z}} & \left(6i\right) & \mbox{Si} \\ 
\end{longtabu}
\renewcommand{\arraystretch}{1.0}
\noindent \hrulefill
\\
\textbf{References:}
\vspace*{-0.25cm}
\begin{flushleft}
  - \bibentry{Perlitz_ArkKemiMinGeo_15B_1_1942}. \\
  - \bibentry{Brandes_Smithells_1992}. \\
\end{flushleft}
\textbf{Found in:}
\vspace*{-0.25cm}
\begin{flushleft}
  - \bibentry{Foss_ActaCrystB_59_36_2003}. \\
\end{flushleft}
\noindent \hrulefill
\\
\textbf{Geometry files:}
\\
\noindent  - CIF: pp. {\hyperref[A8BC3D6_hP18_189_bfh_a_g_i_cif]{\pageref{A8BC3D6_hP18_189_bfh_a_g_i_cif}}} \\
\noindent  - POSCAR: pp. {\hyperref[A8BC3D6_hP18_189_bfh_a_g_i_poscar]{\pageref{A8BC3D6_hP18_189_bfh_a_g_i_poscar}}} \\
\onecolumn
{\phantomsection\label{A9BC3D5_hP18_189_fi_a_g_bh}}
\subsection*{\huge \textbf{{\normalfont $\pi$-FeMg$_{3}$Al$_{9}$Si$_{5}$ Structure: A9BC3D5\_hP18\_189\_fi\_a\_g\_bh}}}
\noindent \hrulefill
\vspace*{0.25cm}
\begin{figure}[htp]
  \centering
  \vspace{-1em}
  {\includegraphics[width=1\textwidth]{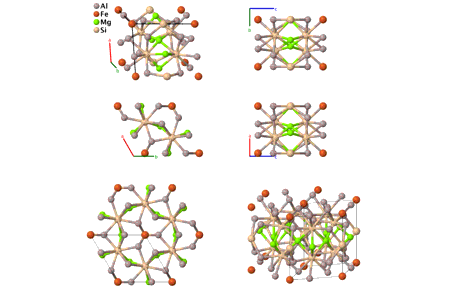}}
\end{figure}
\vspace*{-0.5cm}
\renewcommand{\arraystretch}{1.5}
\begin{equation*}
  \begin{array}{>{$\hspace{-0.15cm}}l<{$}>{$}p{0.5cm}<{$}>{$}p{18.5cm}<{$}}
    \mbox{\large \textbf{Prototype}} &\colon & \ce{$\pi$-FeMg3Al9Si5} \\
    \mbox{\large \textbf{\AFLOW\ prototype label}} &\colon & \mbox{A9BC3D5\_hP18\_189\_fi\_a\_g\_bh} \\
    \mbox{\large \textbf{\textit{Strukturbericht} designation}} &\colon & \mbox{None} \\
    \mbox{\large \textbf{Pearson symbol}} &\colon & \mbox{hP18} \\
    \mbox{\large \textbf{Space group number}} &\colon & 189 \\
    \mbox{\large \textbf{Space group symbol}} &\colon & P\bar{6}2m \\
    \mbox{\large \textbf{\AFLOW\ prototype command}} &\colon &  \texttt{aflow} \,  \, \texttt{-{}-proto=A9BC3D5\_hP18\_189\_fi\_a\_g\_bh } \, \newline \texttt{-{}-params=}{a,c/a,x_{3},x_{4},z_{5},x_{6},z_{6} }
  \end{array}
\end{equation*}
\renewcommand{\arraystretch}{1.0}

\vspace*{-0.25cm}
\noindent \hrulefill
\begin{itemize}
  \item{This is a reanalysis of the
\hyperref[A8BC3D6_hP18_189_bfh_a_g_i]{$\pi$-FeMg$_{3}$Al$_{8}$Si$_{6}$ ($E9_{b}$)
  structure}.  The space group and occupied Wyckoff positions are
unchanged, but the ordering, stoichiometry, and atomic positions are
different.
}
\end{itemize}

\noindent \parbox{1 \linewidth}{
\noindent \hrulefill
\\
\textbf{Hexagonal primitive vectors:} \\
\vspace*{-0.25cm}
\begin{tabular}{cc}
  \begin{tabular}{c}
    \parbox{0.6 \linewidth}{
      \renewcommand{\arraystretch}{1.5}
      \begin{equation*}
        \centering
        \begin{array}{ccc}
              \mathbf{a}_1 & = & \frac12 \, a \, \mathbf{\hat{x}} - \frac{\sqrt3}2 \, a \, \mathbf{\hat{y}} \\
    \mathbf{a}_2 & = & \frac12 \, a \, \mathbf{\hat{x}} + \frac{\sqrt3}2 \, a \, \mathbf{\hat{y}} \\
    \mathbf{a}_3 & = & c \, \mathbf{\hat{z}} \\

        \end{array}
      \end{equation*}
    }
    \renewcommand{\arraystretch}{1.0}
  \end{tabular}
  \begin{tabular}{c}
    \includegraphics[width=0.3\linewidth]{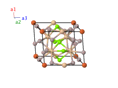} \\
  \end{tabular}
\end{tabular}

}
\vspace*{-0.25cm}

\noindent \hrulefill
\\
\textbf{Basis vectors:}
\vspace*{-0.25cm}
\renewcommand{\arraystretch}{1.5}
\begin{longtabu} to \textwidth{>{\centering $}X[-1,c,c]<{$}>{\centering $}X[-1,c,c]<{$}>{\centering $}X[-1,c,c]<{$}>{\centering $}X[-1,c,c]<{$}>{\centering $}X[-1,c,c]<{$}>{\centering $}X[-1,c,c]<{$}>{\centering $}X[-1,c,c]<{$}}
  & & \mbox{Lattice Coordinates} & & \mbox{Cartesian Coordinates} &\mbox{Wyckoff Position} & \mbox{Atom Type} \\  
  \mathbf{B}_{1} & = & 0 \, \mathbf{a}_{1} + 0 \, \mathbf{a}_{2} + 0 \, \mathbf{a}_{3} & = & 0 \, \mathbf{\hat{x}} + 0 \, \mathbf{\hat{y}} + 0 \, \mathbf{\hat{z}} & \left(1a\right) & \mbox{Fe} \\ 
\mathbf{B}_{2} & = & \frac{1}{2} \, \mathbf{a}_{3} & = & \frac{1}{2}c \, \mathbf{\hat{z}} & \left(1b\right) & \mbox{Si I} \\ 
\mathbf{B}_{3} & = & x_{3} \, \mathbf{a}_{1} & = & \frac{1}{2}x_{3}a \, \mathbf{\hat{x}}-\frac{\sqrt{3}}{2}x_{3}a \, \mathbf{\hat{y}} & \left(3f\right) & \mbox{Al I} \\ 
\mathbf{B}_{4} & = & x_{3} \, \mathbf{a}_{2} & = & \frac{1}{2}x_{3}a \, \mathbf{\hat{x}} + \frac{\sqrt{3}}{2}x_{3}a \, \mathbf{\hat{y}} & \left(3f\right) & \mbox{Al I} \\ 
\mathbf{B}_{5} & = & -x_{3} \, \mathbf{a}_{1}-x_{3} \, \mathbf{a}_{2} & = & -x_{3}a \, \mathbf{\hat{x}} & \left(3f\right) & \mbox{Al I} \\ 
\mathbf{B}_{6} & = & x_{4} \, \mathbf{a}_{1} + \frac{1}{2} \, \mathbf{a}_{3} & = & \frac{1}{2}x_{4}a \, \mathbf{\hat{x}}-\frac{\sqrt{3}}{2}x_{4}a \, \mathbf{\hat{y}} + \frac{1}{2}c \, \mathbf{\hat{z}} & \left(3g\right) & \mbox{Mg} \\ 
\mathbf{B}_{7} & = & x_{4} \, \mathbf{a}_{2} + \frac{1}{2} \, \mathbf{a}_{3} & = & \frac{1}{2}x_{4}a \, \mathbf{\hat{x}} + \frac{\sqrt{3}}{2}x_{4}a \, \mathbf{\hat{y}} + \frac{1}{2}c \, \mathbf{\hat{z}} & \left(3g\right) & \mbox{Mg} \\ 
\mathbf{B}_{8} & = & -x_{4} \, \mathbf{a}_{1}-x_{4} \, \mathbf{a}_{2} + \frac{1}{2} \, \mathbf{a}_{3} & = & -x_{4}a \, \mathbf{\hat{x}} + \frac{1}{2}c \, \mathbf{\hat{z}} & \left(3g\right) & \mbox{Mg} \\ 
\mathbf{B}_{9} & = & \frac{1}{3} \, \mathbf{a}_{1} + \frac{2}{3} \, \mathbf{a}_{2} + z_{5} \, \mathbf{a}_{3} & = & \frac{1}{2}a \, \mathbf{\hat{x}} + \frac{1}{2\sqrt{3}}a \, \mathbf{\hat{y}} + z_{5}c \, \mathbf{\hat{z}} & \left(4h\right) & \mbox{Si II} \\ 
\mathbf{B}_{10} & = & \frac{1}{3} \, \mathbf{a}_{1} + \frac{2}{3} \, \mathbf{a}_{2}-z_{5} \, \mathbf{a}_{3} & = & \frac{1}{2}a \, \mathbf{\hat{x}} + \frac{1}{2\sqrt{3}}a \, \mathbf{\hat{y}}-z_{5}c \, \mathbf{\hat{z}} & \left(4h\right) & \mbox{Si II} \\ 
\mathbf{B}_{11} & = & \frac{2}{3} \, \mathbf{a}_{1} + \frac{1}{3} \, \mathbf{a}_{2}-z_{5} \, \mathbf{a}_{3} & = & \frac{1}{2}a \, \mathbf{\hat{x}}- \frac{1}{2\sqrt{3}}a  \, \mathbf{\hat{y}}-z_{5}c \, \mathbf{\hat{z}} & \left(4h\right) & \mbox{Si II} \\ 
\mathbf{B}_{12} & = & \frac{2}{3} \, \mathbf{a}_{1} + \frac{1}{3} \, \mathbf{a}_{2} + z_{5} \, \mathbf{a}_{3} & = & \frac{1}{2}a \, \mathbf{\hat{x}}- \frac{1}{2\sqrt{3}}a  \, \mathbf{\hat{y}} + z_{5}c \, \mathbf{\hat{z}} & \left(4h\right) & \mbox{Si II} \\ 
\mathbf{B}_{13} & = & x_{6} \, \mathbf{a}_{1} + z_{6} \, \mathbf{a}_{3} & = & \frac{1}{2}x_{6}a \, \mathbf{\hat{x}}-\frac{\sqrt{3}}{2}x_{6}a \, \mathbf{\hat{y}} + z_{6}c \, \mathbf{\hat{z}} & \left(6i\right) & \mbox{Al II} \\ 
\mathbf{B}_{14} & = & x_{6} \, \mathbf{a}_{2} + z_{6} \, \mathbf{a}_{3} & = & \frac{1}{2}x_{6}a \, \mathbf{\hat{x}} + \frac{\sqrt{3}}{2}x_{6}a \, \mathbf{\hat{y}} + z_{6}c \, \mathbf{\hat{z}} & \left(6i\right) & \mbox{Al II} \\ 
\mathbf{B}_{15} & = & -x_{6} \, \mathbf{a}_{1}-x_{6} \, \mathbf{a}_{2} + z_{6} \, \mathbf{a}_{3} & = & -x_{6}a \, \mathbf{\hat{x}} + z_{6}c \, \mathbf{\hat{z}} & \left(6i\right) & \mbox{Al II} \\ 
\mathbf{B}_{16} & = & x_{6} \, \mathbf{a}_{1} + -z_{6} \, \mathbf{a}_{3} & = & \frac{1}{2}x_{6}a \, \mathbf{\hat{x}}-\frac{\sqrt{3}}{2}x_{6}a \, \mathbf{\hat{y}}-z_{6}c \, \mathbf{\hat{z}} & \left(6i\right) & \mbox{Al II} \\ 
\mathbf{B}_{17} & = & x_{6} \, \mathbf{a}_{2}-z_{6} \, \mathbf{a}_{3} & = & \frac{1}{2}x_{6}a \, \mathbf{\hat{x}} + \frac{\sqrt{3}}{2}x_{6}a \, \mathbf{\hat{y}}-z_{6}c \, \mathbf{\hat{z}} & \left(6i\right) & \mbox{Al II} \\ 
\mathbf{B}_{18} & = & -x_{6} \, \mathbf{a}_{1}-x_{6} \, \mathbf{a}_{2}-z_{6} \, \mathbf{a}_{3} & = & -x_{6}a \, \mathbf{\hat{x}} + -z_{6}c \, \mathbf{\hat{z}} & \left(6i\right) & \mbox{Al II} \\ 
\end{longtabu}
\renewcommand{\arraystretch}{1.0}
\noindent \hrulefill
\\
\textbf{References:}
\vspace*{-0.25cm}
\begin{flushleft}
  - \bibentry{Foss_Acta_Cryst_B_59_36_2003}. \\
\end{flushleft}
\textbf{Found in:}
\vspace*{-0.25cm}
\begin{flushleft}
  - \bibentry{MaterialsProject_mp_7062}. \\
  - \bibentry{icsd:ICSD_96905}. \\
\end{flushleft}
\noindent \hrulefill
\\
\textbf{Geometry files:}
\\
\noindent  - CIF: pp. {\hyperref[A9BC3D5_hP18_189_fi_a_g_bh_cif]{\pageref{A9BC3D5_hP18_189_fi_a_g_bh_cif}}} \\
\noindent  - POSCAR: pp. {\hyperref[A9BC3D5_hP18_189_fi_a_g_bh_poscar]{\pageref{A9BC3D5_hP18_189_fi_a_g_bh_poscar}}} \\
\onecolumn
{\phantomsection\label{A2B_hP18_190_gh_bf}}
\subsection*{\huge \textbf{{\normalfont Li$_{2}$Sb Structure: A2B\_hP18\_190\_gh\_bf}}}
\noindent \hrulefill
\vspace*{0.25cm}
\begin{figure}[htp]
  \centering
  \vspace{-1em}
  {\includegraphics[width=1\textwidth]{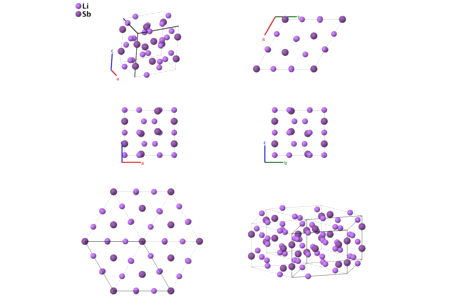}}
\end{figure}
\vspace*{-0.5cm}
\renewcommand{\arraystretch}{1.5}
\begin{equation*}
  \begin{array}{>{$\hspace{-0.15cm}}l<{$}>{$}p{0.5cm}<{$}>{$}p{18.5cm}<{$}}
    \mbox{\large \textbf{Prototype}} &\colon & \ce{Li$_{2}$Sb} \\
    \mbox{\large \textbf{\AFLOW\ prototype label}} &\colon & \mbox{A2B\_hP18\_190\_gh\_bf} \\
    \mbox{\large \textbf{\textit{Strukturbericht} designation}} &\colon & \mbox{None} \\
    \mbox{\large \textbf{Pearson symbol}} &\colon & \mbox{hP18} \\
    \mbox{\large \textbf{Space group number}} &\colon & 190 \\
    \mbox{\large \textbf{Space group symbol}} &\colon & P\bar{6}2c \\
    \mbox{\large \textbf{\AFLOW\ prototype command}} &\colon &  \texttt{aflow} \,  \, \texttt{-{}-proto=A2B\_hP18\_190\_gh\_bf } \, \newline \texttt{-{}-params=}{a,c/a,z_{2},x_{3},x_{4},y_{4} }
  \end{array}
\end{equation*}
\renewcommand{\arraystretch}{1.0}

\noindent \parbox{1 \linewidth}{
\noindent \hrulefill
\\
\textbf{Hexagonal primitive vectors:} \\
\vspace*{-0.25cm}
\begin{tabular}{cc}
  \begin{tabular}{c}
    \parbox{0.6 \linewidth}{
      \renewcommand{\arraystretch}{1.5}
      \begin{equation*}
        \centering
        \begin{array}{ccc}
              \mathbf{a}_1 & = & \frac12 \, a \, \mathbf{\hat{x}} - \frac{\sqrt3}2 \, a \, \mathbf{\hat{y}} \\
    \mathbf{a}_2 & = & \frac12 \, a \, \mathbf{\hat{x}} + \frac{\sqrt3}2 \, a \, \mathbf{\hat{y}} \\
    \mathbf{a}_3 & = & c \, \mathbf{\hat{z}} \\

        \end{array}
      \end{equation*}
    }
    \renewcommand{\arraystretch}{1.0}
  \end{tabular}
  \begin{tabular}{c}
    \includegraphics[width=0.3\linewidth]{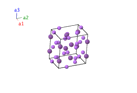} \\
  \end{tabular}
\end{tabular}

}
\vspace*{-0.25cm}

\noindent \hrulefill
\\
\textbf{Basis vectors:}
\vspace*{-0.25cm}
\renewcommand{\arraystretch}{1.5}
\begin{longtabu} to \textwidth{>{\centering $}X[-1,c,c]<{$}>{\centering $}X[-1,c,c]<{$}>{\centering $}X[-1,c,c]<{$}>{\centering $}X[-1,c,c]<{$}>{\centering $}X[-1,c,c]<{$}>{\centering $}X[-1,c,c]<{$}>{\centering $}X[-1,c,c]<{$}}
  & & \mbox{Lattice Coordinates} & & \mbox{Cartesian Coordinates} &\mbox{Wyckoff Position} & \mbox{Atom Type} \\  
  \mathbf{B}_{1} & = & \frac{1}{4} \, \mathbf{a}_{3} & = & \frac{1}{4}c \, \mathbf{\hat{z}} & \left(2b\right) & \mbox{Sb I} \\ 
\mathbf{B}_{2} & = & \frac{3}{4} \, \mathbf{a}_{3} & = & \frac{3}{4}c \, \mathbf{\hat{z}} & \left(2b\right) & \mbox{Sb I} \\ 
\mathbf{B}_{3} & = & \frac{1}{3} \, \mathbf{a}_{1} + \frac{2}{3} \, \mathbf{a}_{2} + z_{2} \, \mathbf{a}_{3} & = & \frac{1}{2}a \, \mathbf{\hat{x}} + \frac{1}{2\sqrt{3}}a \, \mathbf{\hat{y}} + z_{2}c \, \mathbf{\hat{z}} & \left(4f\right) & \mbox{Sb II} \\ 
\mathbf{B}_{4} & = & \frac{1}{3} \, \mathbf{a}_{1} + \frac{2}{3} \, \mathbf{a}_{2} + \left(\frac{1}{2} - z_{2}\right) \, \mathbf{a}_{3} & = & \frac{1}{2}a \, \mathbf{\hat{x}} + \frac{1}{2\sqrt{3}}a \, \mathbf{\hat{y}} + \left(\frac{1}{2} - z_{2}\right)c \, \mathbf{\hat{z}} & \left(4f\right) & \mbox{Sb II} \\ 
\mathbf{B}_{5} & = & \frac{2}{3} \, \mathbf{a}_{1} + \frac{1}{3} \, \mathbf{a}_{2}-z_{2} \, \mathbf{a}_{3} & = & \frac{1}{2}a \, \mathbf{\hat{x}}- \frac{1}{2\sqrt{3}}a  \, \mathbf{\hat{y}}-z_{2}c \, \mathbf{\hat{z}} & \left(4f\right) & \mbox{Sb II} \\ 
\mathbf{B}_{6} & = & \frac{2}{3} \, \mathbf{a}_{1} + \frac{1}{3} \, \mathbf{a}_{2} + \left(\frac{1}{2} +z_{2}\right) \, \mathbf{a}_{3} & = & \frac{1}{2}a \, \mathbf{\hat{x}}- \frac{1}{2\sqrt{3}}a  \, \mathbf{\hat{y}} + \left(\frac{1}{2} +z_{2}\right)c \, \mathbf{\hat{z}} & \left(4f\right) & \mbox{Sb II} \\ 
\mathbf{B}_{7} & = & x_{3} \, \mathbf{a}_{1} & = & \frac{1}{2}x_{3}a \, \mathbf{\hat{x}}-\frac{\sqrt{3}}{2}x_{3}a \, \mathbf{\hat{y}} & \left(6g\right) & \mbox{Li I} \\ 
\mathbf{B}_{8} & = & x_{3} \, \mathbf{a}_{2} & = & \frac{1}{2}x_{3}a \, \mathbf{\hat{x}} + \frac{\sqrt{3}}{2}x_{3}a \, \mathbf{\hat{y}} & \left(6g\right) & \mbox{Li I} \\ 
\mathbf{B}_{9} & = & -x_{3} \, \mathbf{a}_{1}-x_{3} \, \mathbf{a}_{2} & = & -x_{3}a \, \mathbf{\hat{x}} & \left(6g\right) & \mbox{Li I} \\ 
\mathbf{B}_{10} & = & x_{3} \, \mathbf{a}_{1} + \frac{1}{2} \, \mathbf{a}_{3} & = & \frac{1}{2}x_{3}a \, \mathbf{\hat{x}}-\frac{\sqrt{3}}{2}x_{3}a \, \mathbf{\hat{y}} + \frac{1}{2}c \, \mathbf{\hat{z}} & \left(6g\right) & \mbox{Li I} \\ 
\mathbf{B}_{11} & = & x_{3} \, \mathbf{a}_{2} + \frac{1}{2} \, \mathbf{a}_{3} & = & \frac{1}{2}x_{3}a \, \mathbf{\hat{x}} + \frac{\sqrt{3}}{2}x_{3}a \, \mathbf{\hat{y}} + \frac{1}{2}c \, \mathbf{\hat{z}} & \left(6g\right) & \mbox{Li I} \\ 
\mathbf{B}_{12} & = & -x_{3} \, \mathbf{a}_{1}-x_{3} \, \mathbf{a}_{2} + \frac{1}{2} \, \mathbf{a}_{3} & = & -x_{3}a \, \mathbf{\hat{x}} + \frac{1}{2}c \, \mathbf{\hat{z}} & \left(6g\right) & \mbox{Li I} \\ 
\mathbf{B}_{13} & = & x_{4} \, \mathbf{a}_{1} + y_{4} \, \mathbf{a}_{2} + \frac{1}{4} \, \mathbf{a}_{3} & = & \frac{1}{2}\left(x_{4}+y_{4}\right)a \, \mathbf{\hat{x}} + \frac{\sqrt{3}}{2}\left(-x_{4}+y_{4}\right)a \, \mathbf{\hat{y}} + \frac{1}{4}c \, \mathbf{\hat{z}} & \left(6h\right) & \mbox{Li II} \\ 
\mathbf{B}_{14} & = & -y_{4} \, \mathbf{a}_{1} + \left(x_{4}-y_{4}\right) \, \mathbf{a}_{2} + \frac{1}{4} \, \mathbf{a}_{3} & = & \left(\frac{1}{2}x_{4}-y_{4}\right)a \, \mathbf{\hat{x}} + \frac{\sqrt{3}}{2}x_{4}a \, \mathbf{\hat{y}} + \frac{1}{4}c \, \mathbf{\hat{z}} & \left(6h\right) & \mbox{Li II} \\ 
\mathbf{B}_{15} & = & \left(-x_{4}+y_{4}\right) \, \mathbf{a}_{1}-x_{4} \, \mathbf{a}_{2} + \frac{1}{4} \, \mathbf{a}_{3} & = & \left(-x_{4}+\frac{1}{2}y_{4}\right)a \, \mathbf{\hat{x}}-\frac{\sqrt{3}}{2}y_{4}a \, \mathbf{\hat{y}} + \frac{1}{4}c \, \mathbf{\hat{z}} & \left(6h\right) & \mbox{Li II} \\ 
\mathbf{B}_{16} & = & y_{4} \, \mathbf{a}_{1} + x_{4} \, \mathbf{a}_{2} + \frac{3}{4} \, \mathbf{a}_{3} & = & \frac{1}{2}\left(x_{4}+y_{4}\right)a \, \mathbf{\hat{x}} + \frac{\sqrt{3}}{2}\left(x_{4}-y_{4}\right)a \, \mathbf{\hat{y}} + \frac{3}{4}c \, \mathbf{\hat{z}} & \left(6h\right) & \mbox{Li II} \\ 
\mathbf{B}_{17} & = & \left(x_{4}-y_{4}\right) \, \mathbf{a}_{1}-y_{4} \, \mathbf{a}_{2} + \frac{3}{4} \, \mathbf{a}_{3} & = & \left(\frac{1}{2}x_{4}-y_{4}\right)a \, \mathbf{\hat{x}}-\frac{\sqrt{3}}{2}x_{4}a \, \mathbf{\hat{y}} + \frac{3}{4}c \, \mathbf{\hat{z}} & \left(6h\right) & \mbox{Li II} \\ 
\mathbf{B}_{18} & = & -x_{4} \, \mathbf{a}_{1} + \left(-x_{4}+y_{4}\right) \, \mathbf{a}_{2} + \frac{3}{4} \, \mathbf{a}_{3} & = & \left(-x_{4}+\frac{1}{2}y_{4}\right)a \, \mathbf{\hat{x}} + \frac{\sqrt{3}}{2}y_{4}a \, \mathbf{\hat{y}} + \frac{3}{4}c \, \mathbf{\hat{z}} & \left(6h\right) & \mbox{Li II} \\ 
\end{longtabu}
\renewcommand{\arraystretch}{1.0}
\noindent \hrulefill
\\
\textbf{References:}
\vspace*{-0.25cm}
\begin{flushleft}
  - \bibentry{Muller_ZFNB_32_1977}. \\
\end{flushleft}
\noindent \hrulefill
\\
\textbf{Geometry files:}
\\
\noindent  - CIF: pp. {\hyperref[A2B_hP18_190_gh_bf_cif]{\pageref{A2B_hP18_190_gh_bf_cif}}} \\
\noindent  - POSCAR: pp. {\hyperref[A2B_hP18_190_gh_bf_poscar]{\pageref{A2B_hP18_190_gh_bf_poscar}}} \\
\onecolumn
{\phantomsection\label{A5B3_hP16_190_bdh_g}}
\subsection*{\huge \textbf{{\normalfont \begin{raggedleft}$\alpha$-Sm$_{3}$Ge$_{5}$ (High-temperature) Structure: \end{raggedleft} \\ A5B3\_hP16\_190\_bdh\_g}}}
\noindent \hrulefill
\vspace*{0.25cm}
\begin{figure}[htp]
  \centering
  \vspace{-1em}
  {\includegraphics[width=1\textwidth]{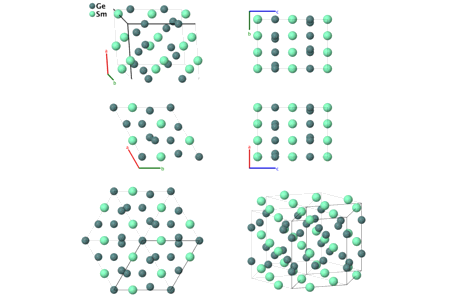}}
\end{figure}
\vspace*{-0.5cm}
\renewcommand{\arraystretch}{1.5}
\begin{equation*}
  \begin{array}{>{$\hspace{-0.15cm}}l<{$}>{$}p{0.5cm}<{$}>{$}p{18.5cm}<{$}}
    \mbox{\large \textbf{Prototype}} &\colon & \ce{$\alpha$-Sm3Ge5} \\
    \mbox{\large \textbf{\AFLOW\ prototype label}} &\colon & \mbox{A5B3\_hP16\_190\_bdh\_g} \\
    \mbox{\large \textbf{\textit{Strukturbericht} designation}} &\colon & \mbox{None} \\
    \mbox{\large \textbf{Pearson symbol}} &\colon & \mbox{hP16} \\
    \mbox{\large \textbf{Space group number}} &\colon & 190 \\
    \mbox{\large \textbf{Space group symbol}} &\colon & P\bar{6}2c \\
    \mbox{\large \textbf{\AFLOW\ prototype command}} &\colon &  \texttt{aflow} \,  \, \texttt{-{}-proto=A5B3\_hP16\_190\_bdh\_g } \, \newline \texttt{-{}-params=}{a,c/a,x_{3},x_{4},y_{4} }
  \end{array}
\end{equation*}
\renewcommand{\arraystretch}{1.0}

\noindent \parbox{1 \linewidth}{
\noindent \hrulefill
\\
\textbf{Hexagonal primitive vectors:} \\
\vspace*{-0.25cm}
\begin{tabular}{cc}
  \begin{tabular}{c}
    \parbox{0.6 \linewidth}{
      \renewcommand{\arraystretch}{1.5}
      \begin{equation*}
        \centering
        \begin{array}{ccc}
              \mathbf{a}_1 & = & \frac12 \, a \, \mathbf{\hat{x}} - \frac{\sqrt3}2 \, a \, \mathbf{\hat{y}} \\
    \mathbf{a}_2 & = & \frac12 \, a \, \mathbf{\hat{x}} + \frac{\sqrt3}2 \, a \, \mathbf{\hat{y}} \\
    \mathbf{a}_3 & = & c \, \mathbf{\hat{z}} \\

        \end{array}
      \end{equation*}
    }
    \renewcommand{\arraystretch}{1.0}
  \end{tabular}
  \begin{tabular}{c}
    \includegraphics[width=0.3\linewidth]{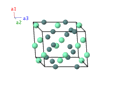} \\
  \end{tabular}
\end{tabular}

}
\vspace*{-0.25cm}

\noindent \hrulefill
\\
\textbf{Basis vectors:}
\vspace*{-0.25cm}
\renewcommand{\arraystretch}{1.5}
\begin{longtabu} to \textwidth{>{\centering $}X[-1,c,c]<{$}>{\centering $}X[-1,c,c]<{$}>{\centering $}X[-1,c,c]<{$}>{\centering $}X[-1,c,c]<{$}>{\centering $}X[-1,c,c]<{$}>{\centering $}X[-1,c,c]<{$}>{\centering $}X[-1,c,c]<{$}}
  & & \mbox{Lattice Coordinates} & & \mbox{Cartesian Coordinates} &\mbox{Wyckoff Position} & \mbox{Atom Type} \\  
  \mathbf{B}_{1} & = & \frac{1}{4} \, \mathbf{a}_{3} & = & \frac{1}{4}c \, \mathbf{\hat{z}} & \left(2b\right) & \mbox{Ge I} \\ 
\mathbf{B}_{2} & = & \frac{3}{4} \, \mathbf{a}_{3} & = & \frac{3}{4}c \, \mathbf{\hat{z}} & \left(2b\right) & \mbox{Ge I} \\ 
\mathbf{B}_{3} & = & \frac{2}{3} \, \mathbf{a}_{1} + \frac{1}{3} \, \mathbf{a}_{2} + \frac{1}{4} \, \mathbf{a}_{3} & = & \frac{1}{2}a \, \mathbf{\hat{x}}- \frac{1}{2\sqrt{3}}a  \, \mathbf{\hat{y}} + \frac{1}{4}c \, \mathbf{\hat{z}} & \left(2d\right) & \mbox{Ge II} \\ 
\mathbf{B}_{4} & = & \frac{1}{3} \, \mathbf{a}_{1} + \frac{2}{3} \, \mathbf{a}_{2} + \frac{3}{4} \, \mathbf{a}_{3} & = & \frac{1}{2}a \, \mathbf{\hat{x}} + \frac{1}{2\sqrt{3}}a \, \mathbf{\hat{y}} + \frac{3}{4}c \, \mathbf{\hat{z}} & \left(2d\right) & \mbox{Ge II} \\ 
\mathbf{B}_{5} & = & x_{3} \, \mathbf{a}_{1} & = & \frac{1}{2}x_{3}a \, \mathbf{\hat{x}}-\frac{\sqrt{3}}{2}x_{3}a \, \mathbf{\hat{y}} & \left(6g\right) & \mbox{Sm} \\ 
\mathbf{B}_{6} & = & x_{3} \, \mathbf{a}_{2} & = & \frac{1}{2}x_{3}a \, \mathbf{\hat{x}} + \frac{\sqrt{3}}{2}x_{3}a \, \mathbf{\hat{y}} & \left(6g\right) & \mbox{Sm} \\ 
\mathbf{B}_{7} & = & -x_{3} \, \mathbf{a}_{1}-x_{3} \, \mathbf{a}_{2} & = & -x_{3}a \, \mathbf{\hat{x}} & \left(6g\right) & \mbox{Sm} \\ 
\mathbf{B}_{8} & = & x_{3} \, \mathbf{a}_{1} + \frac{1}{2} \, \mathbf{a}_{3} & = & \frac{1}{2}x_{3}a \, \mathbf{\hat{x}}-\frac{\sqrt{3}}{2}x_{3}a \, \mathbf{\hat{y}} + \frac{1}{2}c \, \mathbf{\hat{z}} & \left(6g\right) & \mbox{Sm} \\ 
\mathbf{B}_{9} & = & x_{3} \, \mathbf{a}_{2} + \frac{1}{2} \, \mathbf{a}_{3} & = & \frac{1}{2}x_{3}a \, \mathbf{\hat{x}} + \frac{\sqrt{3}}{2}x_{3}a \, \mathbf{\hat{y}} + \frac{1}{2}c \, \mathbf{\hat{z}} & \left(6g\right) & \mbox{Sm} \\ 
\mathbf{B}_{10} & = & -x_{3} \, \mathbf{a}_{1}-x_{3} \, \mathbf{a}_{2} + \frac{1}{2} \, \mathbf{a}_{3} & = & -x_{3}a \, \mathbf{\hat{x}} + \frac{1}{2}c \, \mathbf{\hat{z}} & \left(6g\right) & \mbox{Sm} \\ 
\mathbf{B}_{11} & = & x_{4} \, \mathbf{a}_{1} + y_{4} \, \mathbf{a}_{2} + \frac{1}{4} \, \mathbf{a}_{3} & = & \frac{1}{2}\left(x_{4}+y_{4}\right)a \, \mathbf{\hat{x}} + \frac{\sqrt{3}}{2}\left(-x_{4}+y_{4}\right)a \, \mathbf{\hat{y}} + \frac{1}{4}c \, \mathbf{\hat{z}} & \left(6h\right) & \mbox{Ge III} \\ 
\mathbf{B}_{12} & = & -y_{4} \, \mathbf{a}_{1} + \left(x_{4}-y_{4}\right) \, \mathbf{a}_{2} + \frac{1}{4} \, \mathbf{a}_{3} & = & \left(\frac{1}{2}x_{4}-y_{4}\right)a \, \mathbf{\hat{x}} + \frac{\sqrt{3}}{2}x_{4}a \, \mathbf{\hat{y}} + \frac{1}{4}c \, \mathbf{\hat{z}} & \left(6h\right) & \mbox{Ge III} \\ 
\mathbf{B}_{13} & = & \left(-x_{4}+y_{4}\right) \, \mathbf{a}_{1}-x_{4} \, \mathbf{a}_{2} + \frac{1}{4} \, \mathbf{a}_{3} & = & \left(-x_{4}+\frac{1}{2}y_{4}\right)a \, \mathbf{\hat{x}}-\frac{\sqrt{3}}{2}y_{4}a \, \mathbf{\hat{y}} + \frac{1}{4}c \, \mathbf{\hat{z}} & \left(6h\right) & \mbox{Ge III} \\ 
\mathbf{B}_{14} & = & y_{4} \, \mathbf{a}_{1} + x_{4} \, \mathbf{a}_{2} + \frac{3}{4} \, \mathbf{a}_{3} & = & \frac{1}{2}\left(x_{4}+y_{4}\right)a \, \mathbf{\hat{x}} + \frac{\sqrt{3}}{2}\left(x_{4}-y_{4}\right)a \, \mathbf{\hat{y}} + \frac{3}{4}c \, \mathbf{\hat{z}} & \left(6h\right) & \mbox{Ge III} \\ 
\mathbf{B}_{15} & = & \left(x_{4}-y_{4}\right) \, \mathbf{a}_{1}-y_{4} \, \mathbf{a}_{2} + \frac{3}{4} \, \mathbf{a}_{3} & = & \left(\frac{1}{2}x_{4}-y_{4}\right)a \, \mathbf{\hat{x}}-\frac{\sqrt{3}}{2}x_{4}a \, \mathbf{\hat{y}} + \frac{3}{4}c \, \mathbf{\hat{z}} & \left(6h\right) & \mbox{Ge III} \\ 
\mathbf{B}_{16} & = & -x_{4} \, \mathbf{a}_{1} + \left(-x_{4}+y_{4}\right) \, \mathbf{a}_{2} + \frac{3}{4} \, \mathbf{a}_{3} & = & \left(-x_{4}+\frac{1}{2}y_{4}\right)a \, \mathbf{\hat{x}} + \frac{\sqrt{3}}{2}y_{4}a \, \mathbf{\hat{y}} + \frac{3}{4}c \, \mathbf{\hat{z}} & \left(6h\right) & \mbox{Ge III} \\ 
\end{longtabu}
\renewcommand{\arraystretch}{1.0}
\noindent \hrulefill
\\
\textbf{References:}
\vspace*{-0.25cm}
\begin{flushleft}
  - \bibentry{Tobash_Sm3Ge5_InorgChem_2006}. \\
\end{flushleft}
\textbf{Found in:}
\vspace*{-0.25cm}
\begin{flushleft}
  - \bibentry{Villars_PearsonsCrystalData_2013}. \\
\end{flushleft}
\noindent \hrulefill
\\
\textbf{Geometry files:}
\\
\noindent  - CIF: pp. {\hyperref[A5B3_hP16_190_bdh_g_cif]{\pageref{A5B3_hP16_190_bdh_g_cif}}} \\
\noindent  - POSCAR: pp. {\hyperref[A5B3_hP16_190_bdh_g_poscar]{\pageref{A5B3_hP16_190_bdh_g_poscar}}} \\
\onecolumn
{\phantomsection\label{AB_hP24_190_i_afh}}
\subsection*{\huge \textbf{{\normalfont Troilite (FeS) Structure: AB\_hP24\_190\_i\_afh}}}
\noindent \hrulefill
\vspace*{0.25cm}
\begin{figure}[htp]
  \centering
  \vspace{-1em}
  {\includegraphics[width=1\textwidth]{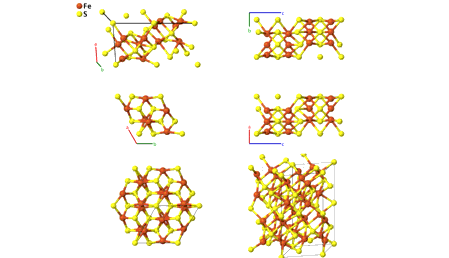}}
\end{figure}
\vspace*{-0.5cm}
\renewcommand{\arraystretch}{1.5}
\begin{equation*}
  \begin{array}{>{$\hspace{-0.15cm}}l<{$}>{$}p{0.5cm}<{$}>{$}p{18.5cm}<{$}}
    \mbox{\large \textbf{Prototype}} &\colon & \ce{FeS} \\
    \mbox{\large \textbf{\AFLOW\ prototype label}} &\colon & \mbox{AB\_hP24\_190\_i\_afh} \\
    \mbox{\large \textbf{\textit{Strukturbericht} designation}} &\colon & \mbox{None} \\
    \mbox{\large \textbf{Pearson symbol}} &\colon & \mbox{hP24} \\
    \mbox{\large \textbf{Space group number}} &\colon & 190 \\
    \mbox{\large \textbf{Space group symbol}} &\colon & P\bar{6}2c \\
    \mbox{\large \textbf{\AFLOW\ prototype command}} &\colon &  \texttt{aflow} \,  \, \texttt{-{}-proto=AB\_hP24\_190\_i\_afh } \, \newline \texttt{-{}-params=}{a,c/a,z_{2},x_{3},y_{3},x_{4},y_{4},z_{4} }
  \end{array}
\end{equation*}
\renewcommand{\arraystretch}{1.0}

\noindent \parbox{1 \linewidth}{
\noindent \hrulefill
\\
\textbf{Hexagonal primitive vectors:} \\
\vspace*{-0.25cm}
\begin{tabular}{cc}
  \begin{tabular}{c}
    \parbox{0.6 \linewidth}{
      \renewcommand{\arraystretch}{1.5}
      \begin{equation*}
        \centering
        \begin{array}{ccc}
              \mathbf{a}_1 & = & \frac12 \, a \, \mathbf{\hat{x}} - \frac{\sqrt3}2 \, a \, \mathbf{\hat{y}} \\
    \mathbf{a}_2 & = & \frac12 \, a \, \mathbf{\hat{x}} + \frac{\sqrt3}2 \, a \, \mathbf{\hat{y}} \\
    \mathbf{a}_3 & = & c \, \mathbf{\hat{z}} \\

        \end{array}
      \end{equation*}
    }
    \renewcommand{\arraystretch}{1.0}
  \end{tabular}
  \begin{tabular}{c}
    \includegraphics[width=0.3\linewidth]{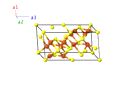} \\
  \end{tabular}
\end{tabular}

}
\vspace*{-0.25cm}

\noindent \hrulefill
\\
\textbf{Basis vectors:}
\vspace*{-0.25cm}
\renewcommand{\arraystretch}{1.5}
\begin{longtabu} to \textwidth{>{\centering $}X[-1,c,c]<{$}>{\centering $}X[-1,c,c]<{$}>{\centering $}X[-1,c,c]<{$}>{\centering $}X[-1,c,c]<{$}>{\centering $}X[-1,c,c]<{$}>{\centering $}X[-1,c,c]<{$}>{\centering $}X[-1,c,c]<{$}}
  & & \mbox{Lattice Coordinates} & & \mbox{Cartesian Coordinates} &\mbox{Wyckoff Position} & \mbox{Atom Type} \\  
  \mathbf{B}_{1} & = & 0 \, \mathbf{a}_{1} + 0 \, \mathbf{a}_{2} + 0 \, \mathbf{a}_{3} & = & 0 \, \mathbf{\hat{x}} + 0 \, \mathbf{\hat{y}} + 0 \, \mathbf{\hat{z}} & \left(2a\right) & \mbox{S I} \\ 
\mathbf{B}_{2} & = & \frac{1}{2} \, \mathbf{a}_{3} & = & \frac{1}{2}c \, \mathbf{\hat{z}} & \left(2a\right) & \mbox{S I} \\ 
\mathbf{B}_{3} & = & \frac{1}{3} \, \mathbf{a}_{1} + \frac{2}{3} \, \mathbf{a}_{2} + z_{2} \, \mathbf{a}_{3} & = & \frac{1}{2}a \, \mathbf{\hat{x}} + \frac{1}{2\sqrt{3}}a \, \mathbf{\hat{y}} + z_{2}c \, \mathbf{\hat{z}} & \left(4f\right) & \mbox{S II} \\ 
\mathbf{B}_{4} & = & \frac{1}{3} \, \mathbf{a}_{1} + \frac{2}{3} \, \mathbf{a}_{2} + \left(\frac{1}{2} - z_{2}\right) \, \mathbf{a}_{3} & = & \frac{1}{2}a \, \mathbf{\hat{x}} + \frac{1}{2\sqrt{3}}a \, \mathbf{\hat{y}} + \left(\frac{1}{2} - z_{2}\right)c \, \mathbf{\hat{z}} & \left(4f\right) & \mbox{S II} \\ 
\mathbf{B}_{5} & = & \frac{2}{3} \, \mathbf{a}_{1} + \frac{1}{3} \, \mathbf{a}_{2}-z_{2} \, \mathbf{a}_{3} & = & \frac{1}{2}a \, \mathbf{\hat{x}}- \frac{1}{2\sqrt{3}}a  \, \mathbf{\hat{y}}-z_{2}c \, \mathbf{\hat{z}} & \left(4f\right) & \mbox{S II} \\ 
\mathbf{B}_{6} & = & \frac{2}{3} \, \mathbf{a}_{1} + \frac{1}{3} \, \mathbf{a}_{2} + \left(\frac{1}{2} +z_{2}\right) \, \mathbf{a}_{3} & = & \frac{1}{2}a \, \mathbf{\hat{x}}- \frac{1}{2\sqrt{3}}a  \, \mathbf{\hat{y}} + \left(\frac{1}{2} +z_{2}\right)c \, \mathbf{\hat{z}} & \left(4f\right) & \mbox{S II} \\ 
\mathbf{B}_{7} & = & x_{3} \, \mathbf{a}_{1} + y_{3} \, \mathbf{a}_{2} + \frac{1}{4} \, \mathbf{a}_{3} & = & \frac{1}{2}\left(x_{3}+y_{3}\right)a \, \mathbf{\hat{x}} + \frac{\sqrt{3}}{2}\left(-x_{3}+y_{3}\right)a \, \mathbf{\hat{y}} + \frac{1}{4}c \, \mathbf{\hat{z}} & \left(6h\right) & \mbox{S III} \\ 
\mathbf{B}_{8} & = & -y_{3} \, \mathbf{a}_{1} + \left(x_{3}-y_{3}\right) \, \mathbf{a}_{2} + \frac{1}{4} \, \mathbf{a}_{3} & = & \left(\frac{1}{2}x_{3}-y_{3}\right)a \, \mathbf{\hat{x}} + \frac{\sqrt{3}}{2}x_{3}a \, \mathbf{\hat{y}} + \frac{1}{4}c \, \mathbf{\hat{z}} & \left(6h\right) & \mbox{S III} \\ 
\mathbf{B}_{9} & = & \left(-x_{3}+y_{3}\right) \, \mathbf{a}_{1}-x_{3} \, \mathbf{a}_{2} + \frac{1}{4} \, \mathbf{a}_{3} & = & \left(-x_{3}+\frac{1}{2}y_{3}\right)a \, \mathbf{\hat{x}}-\frac{\sqrt{3}}{2}y_{3}a \, \mathbf{\hat{y}} + \frac{1}{4}c \, \mathbf{\hat{z}} & \left(6h\right) & \mbox{S III} \\ 
\mathbf{B}_{10} & = & y_{3} \, \mathbf{a}_{1} + x_{3} \, \mathbf{a}_{2} + \frac{3}{4} \, \mathbf{a}_{3} & = & \frac{1}{2}\left(x_{3}+y_{3}\right)a \, \mathbf{\hat{x}} + \frac{\sqrt{3}}{2}\left(x_{3}-y_{3}\right)a \, \mathbf{\hat{y}} + \frac{3}{4}c \, \mathbf{\hat{z}} & \left(6h\right) & \mbox{S III} \\ 
\mathbf{B}_{11} & = & \left(x_{3}-y_{3}\right) \, \mathbf{a}_{1}-y_{3} \, \mathbf{a}_{2} + \frac{3}{4} \, \mathbf{a}_{3} & = & \left(\frac{1}{2}x_{3}-y_{3}\right)a \, \mathbf{\hat{x}}-\frac{\sqrt{3}}{2}x_{3}a \, \mathbf{\hat{y}} + \frac{3}{4}c \, \mathbf{\hat{z}} & \left(6h\right) & \mbox{S III} \\ 
\mathbf{B}_{12} & = & -x_{3} \, \mathbf{a}_{1} + \left(-x_{3}+y_{3}\right) \, \mathbf{a}_{2} + \frac{3}{4} \, \mathbf{a}_{3} & = & \left(-x_{3}+\frac{1}{2}y_{3}\right)a \, \mathbf{\hat{x}} + \frac{\sqrt{3}}{2}y_{3}a \, \mathbf{\hat{y}} + \frac{3}{4}c \, \mathbf{\hat{z}} & \left(6h\right) & \mbox{S III} \\ 
\mathbf{B}_{13} & = & x_{4} \, \mathbf{a}_{1} + y_{4} \, \mathbf{a}_{2} + z_{4} \, \mathbf{a}_{3} & = & \frac{1}{2}\left(x_{4}+y_{4}\right)a \, \mathbf{\hat{x}} + \frac{\sqrt{3}}{2}\left(-x_{4}+y_{4}\right)a \, \mathbf{\hat{y}} + z_{4}c \, \mathbf{\hat{z}} & \left(12i\right) & \mbox{Fe} \\ 
\mathbf{B}_{14} & = & -y_{4} \, \mathbf{a}_{1} + \left(x_{4}-y_{4}\right) \, \mathbf{a}_{2} + z_{4} \, \mathbf{a}_{3} & = & \left(\frac{1}{2}x_{4}-y_{4}\right)a \, \mathbf{\hat{x}} + \frac{\sqrt{3}}{2}x_{4}a \, \mathbf{\hat{y}} + z_{4}c \, \mathbf{\hat{z}} & \left(12i\right) & \mbox{Fe} \\ 
\mathbf{B}_{15} & = & \left(-x_{4}+y_{4}\right) \, \mathbf{a}_{1}-x_{4} \, \mathbf{a}_{2} + z_{4} \, \mathbf{a}_{3} & = & \left(-x_{4}+\frac{1}{2}y_{4}\right)a \, \mathbf{\hat{x}}-\frac{\sqrt{3}}{2}y_{4}a \, \mathbf{\hat{y}} + z_{4}c \, \mathbf{\hat{z}} & \left(12i\right) & \mbox{Fe} \\ 
\mathbf{B}_{16} & = & x_{4} \, \mathbf{a}_{1} + y_{4} \, \mathbf{a}_{2} + \left(\frac{1}{2} - z_{4}\right) \, \mathbf{a}_{3} & = & \frac{1}{2}\left(x_{4}+y_{4}\right)a \, \mathbf{\hat{x}} + \frac{\sqrt{3}}{2}\left(-x_{4}+y_{4}\right)a \, \mathbf{\hat{y}} + \left(\frac{1}{2} - z_{4}\right)c \, \mathbf{\hat{z}} & \left(12i\right) & \mbox{Fe} \\ 
\mathbf{B}_{17} & = & -y_{4} \, \mathbf{a}_{1} + \left(x_{4}-y_{4}\right) \, \mathbf{a}_{2} + \left(\frac{1}{2} - z_{4}\right) \, \mathbf{a}_{3} & = & \left(\frac{1}{2}x_{4}-y_{4}\right)a \, \mathbf{\hat{x}} + \frac{\sqrt{3}}{2}x_{4}a \, \mathbf{\hat{y}} + \left(\frac{1}{2} - z_{4}\right)c \, \mathbf{\hat{z}} & \left(12i\right) & \mbox{Fe} \\ 
\mathbf{B}_{18} & = & \left(-x_{4}+y_{4}\right) \, \mathbf{a}_{1}-x_{4} \, \mathbf{a}_{2} + \left(\frac{1}{2} - z_{4}\right) \, \mathbf{a}_{3} & = & \left(-x_{4}+\frac{1}{2}y_{4}\right)a \, \mathbf{\hat{x}}-\frac{\sqrt{3}}{2}y_{4}a \, \mathbf{\hat{y}} + \left(\frac{1}{2} - z_{4}\right)c \, \mathbf{\hat{z}} & \left(12i\right) & \mbox{Fe} \\ 
\mathbf{B}_{19} & = & y_{4} \, \mathbf{a}_{1} + x_{4} \, \mathbf{a}_{2}-z_{4} \, \mathbf{a}_{3} & = & \frac{1}{2}\left(x_{4}+y_{4}\right)a \, \mathbf{\hat{x}} + \frac{\sqrt{3}}{2}\left(x_{4}-y_{4}\right)a \, \mathbf{\hat{y}}-z_{4}c \, \mathbf{\hat{z}} & \left(12i\right) & \mbox{Fe} \\ 
\mathbf{B}_{20} & = & \left(x_{4}-y_{4}\right) \, \mathbf{a}_{1}-y_{4} \, \mathbf{a}_{2}-z_{4} \, \mathbf{a}_{3} & = & \left(\frac{1}{2}x_{4}-y_{4}\right)a \, \mathbf{\hat{x}}-\frac{\sqrt{3}}{2}x_{4}a \, \mathbf{\hat{y}}-z_{4}c \, \mathbf{\hat{z}} & \left(12i\right) & \mbox{Fe} \\ 
\mathbf{B}_{21} & = & -x_{4} \, \mathbf{a}_{1} + \left(-x_{4}+y_{4}\right) \, \mathbf{a}_{2}-z_{4} \, \mathbf{a}_{3} & = & \left(-x_{4}+\frac{1}{2}y_{4}\right)a \, \mathbf{\hat{x}} + \frac{\sqrt{3}}{2}y_{4}a \, \mathbf{\hat{y}}-z_{4}c \, \mathbf{\hat{z}} & \left(12i\right) & \mbox{Fe} \\ 
\mathbf{B}_{22} & = & y_{4} \, \mathbf{a}_{1} + x_{4} \, \mathbf{a}_{2} + \left(\frac{1}{2} +z_{4}\right) \, \mathbf{a}_{3} & = & \frac{1}{2}\left(x_{4}+y_{4}\right)a \, \mathbf{\hat{x}} + \frac{\sqrt{3}}{2}\left(x_{4}-y_{4}\right)a \, \mathbf{\hat{y}} + \left(\frac{1}{2} +z_{4}\right)c \, \mathbf{\hat{z}} & \left(12i\right) & \mbox{Fe} \\ 
\mathbf{B}_{23} & = & \left(x_{4}-y_{4}\right) \, \mathbf{a}_{1}-y_{4} \, \mathbf{a}_{2} + \left(\frac{1}{2} +z_{4}\right) \, \mathbf{a}_{3} & = & \left(\frac{1}{2}x_{4}-y_{4}\right)a \, \mathbf{\hat{x}}-\frac{\sqrt{3}}{2}x_{4}a \, \mathbf{\hat{y}} + \left(\frac{1}{2} +z_{4}\right)c \, \mathbf{\hat{z}} & \left(12i\right) & \mbox{Fe} \\ 
\mathbf{B}_{24} & = & -x_{4} \, \mathbf{a}_{1} + \left(-x_{4}+y_{4}\right) \, \mathbf{a}_{2} + \left(\frac{1}{2} +z_{4}\right) \, \mathbf{a}_{3} & = & \left(-x_{4}+\frac{1}{2}y_{4}\right)a \, \mathbf{\hat{x}} + \frac{\sqrt{3}}{2}y_{4}a \, \mathbf{\hat{y}} + \left(\frac{1}{2} +z_{4}\right)c \, \mathbf{\hat{z}} & \left(12i\right) & \mbox{Fe} \\ 
\end{longtabu}
\renewcommand{\arraystretch}{1.0}
\noindent \hrulefill
\\
\textbf{References:}
\vspace*{-0.25cm}
\begin{flushleft}
  - \bibentry{Morimoto_FeS_Science_1970}. \\
\end{flushleft}
\textbf{Found in:}
\vspace*{-0.25cm}
\begin{flushleft}
  - \bibentry{Villars_PearsonsCrystalData_2013}. \\
\end{flushleft}
\noindent \hrulefill
\\
\textbf{Geometry files:}
\\
\noindent  - CIF: pp. {\hyperref[AB_hP24_190_i_afh_cif]{\pageref{AB_hP24_190_i_afh_cif}}} \\
\noindent  - POSCAR: pp. {\hyperref[AB_hP24_190_i_afh_poscar]{\pageref{AB_hP24_190_i_afh_poscar}}} \\
\onecolumn
{\phantomsection\label{A2B3C18D6_hP58_192_c_f_lm_l}}
\subsection*{\huge \textbf{{\normalfont \begin{raggedleft}Beryl (Be$_{3}$Al$_{2}$Si$_{6}$O$_{18}$, $G3_{1}$) Structure: \end{raggedleft} \\ A2B3C18D6\_hP58\_192\_c\_f\_lm\_l}}}
\noindent \hrulefill
\vspace*{0.25cm}
\begin{figure}[htp]
  \centering
  \vspace{-1em}
  {\includegraphics[width=1\textwidth]{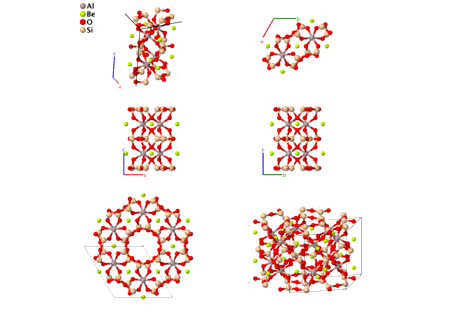}}
\end{figure}
\vspace*{-0.5cm}
\renewcommand{\arraystretch}{1.5}
\begin{equation*}
  \begin{array}{>{$\hspace{-0.15cm}}l<{$}>{$}p{0.5cm}<{$}>{$}p{18.5cm}<{$}}
    \mbox{\large \textbf{Prototype}} &\colon & \ce{Be3Al2Si6O18} \\
    \mbox{\large \textbf{\AFLOW\ prototype label}} &\colon & \mbox{A2B3C18D6\_hP58\_192\_c\_f\_lm\_l} \\
    \mbox{\large \textbf{\textit{Strukturbericht} designation}} &\colon & \mbox{$G3_{1}$} \\
    \mbox{\large \textbf{Pearson symbol}} &\colon & \mbox{hP58} \\
    \mbox{\large \textbf{Space group number}} &\colon & 192 \\
    \mbox{\large \textbf{Space group symbol}} &\colon & P6/mcc \\
    \mbox{\large \textbf{\AFLOW\ prototype command}} &\colon &  \texttt{aflow} \,  \, \texttt{-{}-proto=A2B3C18D6\_hP58\_192\_c\_f\_lm\_l } \, \newline \texttt{-{}-params=}{a,c/a,x_{3},y_{3},x_{4},y_{4},x_{5},y_{5},z_{5} }
  \end{array}
\end{equation*}
\renewcommand{\arraystretch}{1.0}

\vspace*{-0.25cm}
\noindent \hrulefill
\begin{itemize}
  \item{(Morosin, 1972) places oxygen atoms on the (2a) Wyckoff site (lattice
coordinates $(0,0,\pm1/4)$, with an occupation of (0.0991).  We
follow (Hazen, 1986) and ignore this small contribution to the
structure.
}
\end{itemize}

\noindent \parbox{1 \linewidth}{
\noindent \hrulefill
\\
\textbf{Hexagonal primitive vectors:} \\
\vspace*{-0.25cm}
\begin{tabular}{cc}
  \begin{tabular}{c}
    \parbox{0.6 \linewidth}{
      \renewcommand{\arraystretch}{1.5}
      \begin{equation*}
        \centering
        \begin{array}{ccc}
              \mathbf{a}_1 & = & \frac12 \, a \, \mathbf{\hat{x}} - \frac{\sqrt3}2 \, a \, \mathbf{\hat{y}} \\
    \mathbf{a}_2 & = & \frac12 \, a \, \mathbf{\hat{x}} + \frac{\sqrt3}2 \, a \, \mathbf{\hat{y}} \\
    \mathbf{a}_3 & = & c \, \mathbf{\hat{z}} \\

        \end{array}
      \end{equation*}
    }
    \renewcommand{\arraystretch}{1.0}
  \end{tabular}
  \begin{tabular}{c}
    \includegraphics[width=0.3\linewidth]{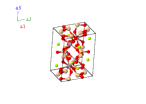} \\
  \end{tabular}
\end{tabular}

}
\vspace*{-0.25cm}

\noindent \hrulefill
\\
\textbf{Basis vectors:}
\vspace*{-0.25cm}
\renewcommand{\arraystretch}{1.5}
\begin{longtabu} to \textwidth{>{\centering $}X[-1,c,c]<{$}>{\centering $}X[-1,c,c]<{$}>{\centering $}X[-1,c,c]<{$}>{\centering $}X[-1,c,c]<{$}>{\centering $}X[-1,c,c]<{$}>{\centering $}X[-1,c,c]<{$}>{\centering $}X[-1,c,c]<{$}}
  & & \mbox{Lattice Coordinates} & & \mbox{Cartesian Coordinates} &\mbox{Wyckoff Position} & \mbox{Atom Type} \\  
  \mathbf{B}_{1} & = & \frac{1}{3} \, \mathbf{a}_{1} + \frac{2}{3} \, \mathbf{a}_{2} + \frac{1}{4} \, \mathbf{a}_{3} & = & \frac{1}{2}a \, \mathbf{\hat{x}} + \frac{1}{2\sqrt{3}}a \, \mathbf{\hat{y}} + \frac{1}{4}c \, \mathbf{\hat{z}} & \left(4c\right) & \mbox{Al} \\ 
\mathbf{B}_{2} & = & \frac{2}{3} \, \mathbf{a}_{1} + \frac{1}{3} \, \mathbf{a}_{2} + \frac{1}{4} \, \mathbf{a}_{3} & = & \frac{1}{2}a \, \mathbf{\hat{x}}- \frac{1}{2\sqrt{3}}a  \, \mathbf{\hat{y}} + \frac{1}{4}c \, \mathbf{\hat{z}} & \left(4c\right) & \mbox{Al} \\ 
\mathbf{B}_{3} & = & \frac{2}{3} \, \mathbf{a}_{1} + \frac{1}{3} \, \mathbf{a}_{2} + \frac{3}{4} \, \mathbf{a}_{3} & = & \frac{1}{2}a \, \mathbf{\hat{x}}- \frac{1}{2\sqrt{3}}a  \, \mathbf{\hat{y}} + \frac{3}{4}c \, \mathbf{\hat{z}} & \left(4c\right) & \mbox{Al} \\ 
\mathbf{B}_{4} & = & \frac{1}{3} \, \mathbf{a}_{1} + \frac{2}{3} \, \mathbf{a}_{2} + \frac{3}{4} \, \mathbf{a}_{3} & = & \frac{1}{2}a \, \mathbf{\hat{x}} + \frac{1}{2\sqrt{3}}a \, \mathbf{\hat{y}} + \frac{3}{4}c \, \mathbf{\hat{z}} & \left(4c\right) & \mbox{Al} \\ 
\mathbf{B}_{5} & = & \frac{1}{2} \, \mathbf{a}_{1} + \frac{1}{4} \, \mathbf{a}_{3} & = & \frac{1}{4}a \, \mathbf{\hat{x}}- \frac{\sqrt{3}}{4}a  \, \mathbf{\hat{y}} + \frac{1}{4}c \, \mathbf{\hat{z}} & \left(6f\right) & \mbox{Be} \\ 
\mathbf{B}_{6} & = & \frac{1}{2} \, \mathbf{a}_{2} + \frac{1}{4} \, \mathbf{a}_{3} & = & \frac{1}{4}a \, \mathbf{\hat{x}} + \frac{\sqrt{3}}{4}a \, \mathbf{\hat{y}} + \frac{1}{4}c \, \mathbf{\hat{z}} & \left(6f\right) & \mbox{Be} \\ 
\mathbf{B}_{7} & = & \frac{1}{2} \, \mathbf{a}_{1} + \frac{1}{2} \, \mathbf{a}_{2} + \frac{1}{4} \, \mathbf{a}_{3} & = & \frac{1}{2}a \, \mathbf{\hat{x}} + \frac{1}{4}c \, \mathbf{\hat{z}} & \left(6f\right) & \mbox{Be} \\ 
\mathbf{B}_{8} & = & \frac{1}{2} \, \mathbf{a}_{1} + \frac{3}{4} \, \mathbf{a}_{3} & = & \frac{1}{4}a \, \mathbf{\hat{x}}- \frac{\sqrt{3}}{4}a  \, \mathbf{\hat{y}} + \frac{3}{4}c \, \mathbf{\hat{z}} & \left(6f\right) & \mbox{Be} \\ 
\mathbf{B}_{9} & = & \frac{1}{2} \, \mathbf{a}_{2} + \frac{3}{4} \, \mathbf{a}_{3} & = & \frac{1}{4}a \, \mathbf{\hat{x}} + \frac{\sqrt{3}}{4}a \, \mathbf{\hat{y}} + \frac{3}{4}c \, \mathbf{\hat{z}} & \left(6f\right) & \mbox{Be} \\ 
\mathbf{B}_{10} & = & \frac{1}{2} \, \mathbf{a}_{1} + \frac{1}{2} \, \mathbf{a}_{2} + \frac{3}{4} \, \mathbf{a}_{3} & = & \frac{1}{2}a \, \mathbf{\hat{x}} + \frac{3}{4}c \, \mathbf{\hat{z}} & \left(6f\right) & \mbox{Be} \\ 
\mathbf{B}_{11} & = & x_{3} \, \mathbf{a}_{1} + y_{3} \, \mathbf{a}_{2} & = & \frac{1}{2}\left(x_{3}+y_{3}\right)a \, \mathbf{\hat{x}} + \frac{\sqrt{3}}{2}\left(-x_{3}+y_{3}\right)a \, \mathbf{\hat{y}} & \left(12l\right) & \mbox{O I} \\ 
\mathbf{B}_{12} & = & -y_{3} \, \mathbf{a}_{1} + \left(x_{3}-y_{3}\right) \, \mathbf{a}_{2} & = & \left(\frac{1}{2}x_{3}-y_{3}\right)a \, \mathbf{\hat{x}} + \frac{\sqrt{3}}{2}x_{3}a \, \mathbf{\hat{y}} & \left(12l\right) & \mbox{O I} \\ 
\mathbf{B}_{13} & = & \left(-x_{3}+y_{3}\right) \, \mathbf{a}_{1}-x_{3} \, \mathbf{a}_{2} & = & \left(-x_{3}+\frac{1}{2}y_{3}\right)a \, \mathbf{\hat{x}}-\frac{\sqrt{3}}{2}y_{3}a \, \mathbf{\hat{y}} & \left(12l\right) & \mbox{O I} \\ 
\mathbf{B}_{14} & = & -x_{3} \, \mathbf{a}_{1}-y_{3} \, \mathbf{a}_{2} & = & -\frac{1}{2}\left(x_{3}+y_{3}\right)a \, \mathbf{\hat{x}} + \frac{\sqrt{3}}{2}\left(x_{3}-y_{3}\right)a \, \mathbf{\hat{y}} & \left(12l\right) & \mbox{O I} \\ 
\mathbf{B}_{15} & = & y_{3} \, \mathbf{a}_{1} + \left(-x_{3}+y_{3}\right) \, \mathbf{a}_{2} & = & \left(-\frac{1}{2}x_{3}+y_{3}\right)a \, \mathbf{\hat{x}}-\frac{\sqrt{3}}{2}x_{3}a \, \mathbf{\hat{y}} & \left(12l\right) & \mbox{O I} \\ 
\mathbf{B}_{16} & = & \left(x_{3}-y_{3}\right) \, \mathbf{a}_{1} + x_{3} \, \mathbf{a}_{2} & = & \left(x_{3}-\frac{1}{2}y_{3}\right)a \, \mathbf{\hat{x}} + \frac{\sqrt{3}}{2}y_{3}a \, \mathbf{\hat{y}} & \left(12l\right) & \mbox{O I} \\ 
\mathbf{B}_{17} & = & y_{3} \, \mathbf{a}_{1} + x_{3} \, \mathbf{a}_{2} + \frac{1}{2} \, \mathbf{a}_{3} & = & \frac{1}{2}\left(x_{3}+y_{3}\right)a \, \mathbf{\hat{x}} + \frac{\sqrt{3}}{2}\left(x_{3}-y_{3}\right)a \, \mathbf{\hat{y}} + \frac{1}{2}c \, \mathbf{\hat{z}} & \left(12l\right) & \mbox{O I} \\ 
\mathbf{B}_{18} & = & \left(x_{3}-y_{3}\right) \, \mathbf{a}_{1}-y_{3} \, \mathbf{a}_{2} + \frac{1}{2} \, \mathbf{a}_{3} & = & \left(\frac{1}{2}x_{3}-y_{3}\right)a \, \mathbf{\hat{x}}-\frac{\sqrt{3}}{2}x_{3}a \, \mathbf{\hat{y}} + \frac{1}{2}c \, \mathbf{\hat{z}} & \left(12l\right) & \mbox{O I} \\ 
\mathbf{B}_{19} & = & -x_{3} \, \mathbf{a}_{1} + \left(-x_{3}+y_{3}\right) \, \mathbf{a}_{2} + \frac{1}{2} \, \mathbf{a}_{3} & = & \left(-x_{3}+\frac{1}{2}y_{3}\right)a \, \mathbf{\hat{x}} + \frac{\sqrt{3}}{2}y_{3}a \, \mathbf{\hat{y}} + \frac{1}{2}c \, \mathbf{\hat{z}} & \left(12l\right) & \mbox{O I} \\ 
\mathbf{B}_{20} & = & -y_{3} \, \mathbf{a}_{1}-x_{3} \, \mathbf{a}_{2} + \frac{1}{2} \, \mathbf{a}_{3} & = & -\frac{1}{2}\left(x_{3}+y_{3}\right)a \, \mathbf{\hat{x}} + \frac{\sqrt{3}}{2}\left(-x_{3}+y_{3}\right)a \, \mathbf{\hat{y}} + \frac{1}{2}c \, \mathbf{\hat{z}} & \left(12l\right) & \mbox{O I} \\ 
\mathbf{B}_{21} & = & \left(-x_{3}+y_{3}\right) \, \mathbf{a}_{1} + y_{3} \, \mathbf{a}_{2} + \frac{1}{2} \, \mathbf{a}_{3} & = & \left(-\frac{1}{2}x_{3}+y_{3}\right)a \, \mathbf{\hat{x}} + \frac{\sqrt{3}}{2}x_{3}a \, \mathbf{\hat{y}} + \frac{1}{2}c \, \mathbf{\hat{z}} & \left(12l\right) & \mbox{O I} \\ 
\mathbf{B}_{22} & = & x_{3} \, \mathbf{a}_{1} + \left(x_{3}-y_{3}\right) \, \mathbf{a}_{2} + \frac{1}{2} \, \mathbf{a}_{3} & = & \left(x_{3}-\frac{1}{2}y_{3}\right)a \, \mathbf{\hat{x}}-\frac{\sqrt{3}}{2}y_{3}a \, \mathbf{\hat{y}} + \frac{1}{2}c \, \mathbf{\hat{z}} & \left(12l\right) & \mbox{O I} \\ 
\mathbf{B}_{23} & = & x_{4} \, \mathbf{a}_{1} + y_{4} \, \mathbf{a}_{2} & = & \frac{1}{2}\left(x_{4}+y_{4}\right)a \, \mathbf{\hat{x}} + \frac{\sqrt{3}}{2}\left(-x_{4}+y_{4}\right)a \, \mathbf{\hat{y}} & \left(12l\right) & \mbox{Si} \\ 
\mathbf{B}_{24} & = & -y_{4} \, \mathbf{a}_{1} + \left(x_{4}-y_{4}\right) \, \mathbf{a}_{2} & = & \left(\frac{1}{2}x_{4}-y_{4}\right)a \, \mathbf{\hat{x}} + \frac{\sqrt{3}}{2}x_{4}a \, \mathbf{\hat{y}} & \left(12l\right) & \mbox{Si} \\ 
\mathbf{B}_{25} & = & \left(-x_{4}+y_{4}\right) \, \mathbf{a}_{1}-x_{4} \, \mathbf{a}_{2} & = & \left(-x_{4}+\frac{1}{2}y_{4}\right)a \, \mathbf{\hat{x}}-\frac{\sqrt{3}}{2}y_{4}a \, \mathbf{\hat{y}} & \left(12l\right) & \mbox{Si} \\ 
\mathbf{B}_{26} & = & -x_{4} \, \mathbf{a}_{1}-y_{4} \, \mathbf{a}_{2} & = & -\frac{1}{2}\left(x_{4}+y_{4}\right)a \, \mathbf{\hat{x}} + \frac{\sqrt{3}}{2}\left(x_{4}-y_{4}\right)a \, \mathbf{\hat{y}} & \left(12l\right) & \mbox{Si} \\ 
\mathbf{B}_{27} & = & y_{4} \, \mathbf{a}_{1} + \left(-x_{4}+y_{4}\right) \, \mathbf{a}_{2} & = & \left(-\frac{1}{2}x_{4}+y_{4}\right)a \, \mathbf{\hat{x}}-\frac{\sqrt{3}}{2}x_{4}a \, \mathbf{\hat{y}} & \left(12l\right) & \mbox{Si} \\ 
\mathbf{B}_{28} & = & \left(x_{4}-y_{4}\right) \, \mathbf{a}_{1} + x_{4} \, \mathbf{a}_{2} & = & \left(x_{4}-\frac{1}{2}y_{4}\right)a \, \mathbf{\hat{x}} + \frac{\sqrt{3}}{2}y_{4}a \, \mathbf{\hat{y}} & \left(12l\right) & \mbox{Si} \\ 
\mathbf{B}_{29} & = & y_{4} \, \mathbf{a}_{1} + x_{4} \, \mathbf{a}_{2} + \frac{1}{2} \, \mathbf{a}_{3} & = & \frac{1}{2}\left(x_{4}+y_{4}\right)a \, \mathbf{\hat{x}} + \frac{\sqrt{3}}{2}\left(x_{4}-y_{4}\right)a \, \mathbf{\hat{y}} + \frac{1}{2}c \, \mathbf{\hat{z}} & \left(12l\right) & \mbox{Si} \\ 
\mathbf{B}_{30} & = & \left(x_{4}-y_{4}\right) \, \mathbf{a}_{1}-y_{4} \, \mathbf{a}_{2} + \frac{1}{2} \, \mathbf{a}_{3} & = & \left(\frac{1}{2}x_{4}-y_{4}\right)a \, \mathbf{\hat{x}}-\frac{\sqrt{3}}{2}x_{4}a \, \mathbf{\hat{y}} + \frac{1}{2}c \, \mathbf{\hat{z}} & \left(12l\right) & \mbox{Si} \\ 
\mathbf{B}_{31} & = & -x_{4} \, \mathbf{a}_{1} + \left(-x_{4}+y_{4}\right) \, \mathbf{a}_{2} + \frac{1}{2} \, \mathbf{a}_{3} & = & \left(-x_{4}+\frac{1}{2}y_{4}\right)a \, \mathbf{\hat{x}} + \frac{\sqrt{3}}{2}y_{4}a \, \mathbf{\hat{y}} + \frac{1}{2}c \, \mathbf{\hat{z}} & \left(12l\right) & \mbox{Si} \\ 
\mathbf{B}_{32} & = & -y_{4} \, \mathbf{a}_{1}-x_{4} \, \mathbf{a}_{2} + \frac{1}{2} \, \mathbf{a}_{3} & = & -\frac{1}{2}\left(x_{4}+y_{4}\right)a \, \mathbf{\hat{x}} + \frac{\sqrt{3}}{2}\left(-x_{4}+y_{4}\right)a \, \mathbf{\hat{y}} + \frac{1}{2}c \, \mathbf{\hat{z}} & \left(12l\right) & \mbox{Si} \\ 
\mathbf{B}_{33} & = & \left(-x_{4}+y_{4}\right) \, \mathbf{a}_{1} + y_{4} \, \mathbf{a}_{2} + \frac{1}{2} \, \mathbf{a}_{3} & = & \left(-\frac{1}{2}x_{4}+y_{4}\right)a \, \mathbf{\hat{x}} + \frac{\sqrt{3}}{2}x_{4}a \, \mathbf{\hat{y}} + \frac{1}{2}c \, \mathbf{\hat{z}} & \left(12l\right) & \mbox{Si} \\ 
\mathbf{B}_{34} & = & x_{4} \, \mathbf{a}_{1} + \left(x_{4}-y_{4}\right) \, \mathbf{a}_{2} + \frac{1}{2} \, \mathbf{a}_{3} & = & \left(x_{4}-\frac{1}{2}y_{4}\right)a \, \mathbf{\hat{x}}-\frac{\sqrt{3}}{2}y_{4}a \, \mathbf{\hat{y}} + \frac{1}{2}c \, \mathbf{\hat{z}} & \left(12l\right) & \mbox{Si} \\ 
\mathbf{B}_{35} & = & x_{5} \, \mathbf{a}_{1} + y_{5} \, \mathbf{a}_{2} + z_{5} \, \mathbf{a}_{3} & = & \frac{1}{2}\left(x_{5}+y_{5}\right)a \, \mathbf{\hat{x}} + \frac{\sqrt{3}}{2}\left(-x_{5}+y_{5}\right)a \, \mathbf{\hat{y}} + z_{5}c \, \mathbf{\hat{z}} & \left(24m\right) & \mbox{O II} \\ 
\mathbf{B}_{36} & = & -y_{5} \, \mathbf{a}_{1} + \left(x_{5}-y_{5}\right) \, \mathbf{a}_{2} + z_{5} \, \mathbf{a}_{3} & = & \left(\frac{1}{2}x_{5}-y_{5}\right)a \, \mathbf{\hat{x}} + \frac{\sqrt{3}}{2}x_{5}a \, \mathbf{\hat{y}} + z_{5}c \, \mathbf{\hat{z}} & \left(24m\right) & \mbox{O II} \\ 
\mathbf{B}_{37} & = & \left(-x_{5}+y_{5}\right) \, \mathbf{a}_{1}-x_{5} \, \mathbf{a}_{2} + z_{5} \, \mathbf{a}_{3} & = & \left(-x_{5}+\frac{1}{2}y_{5}\right)a \, \mathbf{\hat{x}}-\frac{\sqrt{3}}{2}y_{5}a \, \mathbf{\hat{y}} + z_{5}c \, \mathbf{\hat{z}} & \left(24m\right) & \mbox{O II} \\ 
\mathbf{B}_{38} & = & -x_{5} \, \mathbf{a}_{1}-y_{5} \, \mathbf{a}_{2} + z_{5} \, \mathbf{a}_{3} & = & -\frac{1}{2}\left(x_{5}+y_{5}\right)a \, \mathbf{\hat{x}} + \frac{\sqrt{3}}{2}\left(x_{5}-y_{5}\right)a \, \mathbf{\hat{y}} + z_{5}c \, \mathbf{\hat{z}} & \left(24m\right) & \mbox{O II} \\ 
\mathbf{B}_{39} & = & y_{5} \, \mathbf{a}_{1} + \left(-x_{5}+y_{5}\right) \, \mathbf{a}_{2} + z_{5} \, \mathbf{a}_{3} & = & \left(-\frac{1}{2}x_{5}+y_{5}\right)a \, \mathbf{\hat{x}}-\frac{\sqrt{3}}{2}x_{5}a \, \mathbf{\hat{y}} + z_{5}c \, \mathbf{\hat{z}} & \left(24m\right) & \mbox{O II} \\ 
\mathbf{B}_{40} & = & \left(x_{5}-y_{5}\right) \, \mathbf{a}_{1} + x_{5} \, \mathbf{a}_{2} + z_{5} \, \mathbf{a}_{3} & = & \left(x_{5}-\frac{1}{2}y_{5}\right)a \, \mathbf{\hat{x}} + \frac{\sqrt{3}}{2}y_{5}a \, \mathbf{\hat{y}} + z_{5}c \, \mathbf{\hat{z}} & \left(24m\right) & \mbox{O II} \\ 
\mathbf{B}_{41} & = & y_{5} \, \mathbf{a}_{1} + x_{5} \, \mathbf{a}_{2} + \left(\frac{1}{2} - z_{5}\right) \, \mathbf{a}_{3} & = & \frac{1}{2}\left(x_{5}+y_{5}\right)a \, \mathbf{\hat{x}} + \frac{\sqrt{3}}{2}\left(x_{5}-y_{5}\right)a \, \mathbf{\hat{y}} + \left(\frac{1}{2} - z_{5}\right)c \, \mathbf{\hat{z}} & \left(24m\right) & \mbox{O II} \\ 
\mathbf{B}_{42} & = & \left(x_{5}-y_{5}\right) \, \mathbf{a}_{1}-y_{5} \, \mathbf{a}_{2} + \left(\frac{1}{2} - z_{5}\right) \, \mathbf{a}_{3} & = & \left(\frac{1}{2}x_{5}-y_{5}\right)a \, \mathbf{\hat{x}}-\frac{\sqrt{3}}{2}x_{5}a \, \mathbf{\hat{y}} + \left(\frac{1}{2} - z_{5}\right)c \, \mathbf{\hat{z}} & \left(24m\right) & \mbox{O II} \\ 
\mathbf{B}_{43} & = & -x_{5} \, \mathbf{a}_{1} + \left(-x_{5}+y_{5}\right) \, \mathbf{a}_{2} + \left(\frac{1}{2} - z_{5}\right) \, \mathbf{a}_{3} & = & \left(-x_{5}+\frac{1}{2}y_{5}\right)a \, \mathbf{\hat{x}} + \frac{\sqrt{3}}{2}y_{5}a \, \mathbf{\hat{y}} + \left(\frac{1}{2} - z_{5}\right)c \, \mathbf{\hat{z}} & \left(24m\right) & \mbox{O II} \\ 
\mathbf{B}_{44} & = & -y_{5} \, \mathbf{a}_{1}-x_{5} \, \mathbf{a}_{2} + \left(\frac{1}{2} - z_{5}\right) \, \mathbf{a}_{3} & = & -\frac{1}{2}\left(x_{5}+y_{5}\right)a \, \mathbf{\hat{x}} + \frac{\sqrt{3}}{2}\left(-x_{5}+y_{5}\right)a \, \mathbf{\hat{y}} + \left(\frac{1}{2} - z_{5}\right)c \, \mathbf{\hat{z}} & \left(24m\right) & \mbox{O II} \\ 
\mathbf{B}_{45} & = & \left(-x_{5}+y_{5}\right) \, \mathbf{a}_{1} + y_{5} \, \mathbf{a}_{2} + \left(\frac{1}{2} - z_{5}\right) \, \mathbf{a}_{3} & = & \left(-\frac{1}{2}x_{5}+y_{5}\right)a \, \mathbf{\hat{x}} + \frac{\sqrt{3}}{2}x_{5}a \, \mathbf{\hat{y}} + \left(\frac{1}{2} - z_{5}\right)c \, \mathbf{\hat{z}} & \left(24m\right) & \mbox{O II} \\ 
\mathbf{B}_{46} & = & x_{5} \, \mathbf{a}_{1} + \left(x_{5}-y_{5}\right) \, \mathbf{a}_{2} + \left(\frac{1}{2} - z_{5}\right) \, \mathbf{a}_{3} & = & \left(x_{5}-\frac{1}{2}y_{5}\right)a \, \mathbf{\hat{x}}-\frac{\sqrt{3}}{2}y_{5}a \, \mathbf{\hat{y}} + \left(\frac{1}{2} - z_{5}\right)c \, \mathbf{\hat{z}} & \left(24m\right) & \mbox{O II} \\ 
\mathbf{B}_{47} & = & -x_{5} \, \mathbf{a}_{1}-y_{5} \, \mathbf{a}_{2}-z_{5} \, \mathbf{a}_{3} & = & -\frac{1}{2}\left(x_{5}+y_{5}\right)a \, \mathbf{\hat{x}} + \frac{\sqrt{3}}{2}\left(x_{5}-y_{5}\right)a \, \mathbf{\hat{y}}-z_{5}c \, \mathbf{\hat{z}} & \left(24m\right) & \mbox{O II} \\ 
\mathbf{B}_{48} & = & y_{5} \, \mathbf{a}_{1} + \left(-x_{5}+y_{5}\right) \, \mathbf{a}_{2}-z_{5} \, \mathbf{a}_{3} & = & \left(-\frac{1}{2}x_{5}+y_{5}\right)a \, \mathbf{\hat{x}}-\frac{\sqrt{3}}{2}x_{5}a \, \mathbf{\hat{y}}-z_{5}c \, \mathbf{\hat{z}} & \left(24m\right) & \mbox{O II} \\ 
\mathbf{B}_{49} & = & \left(x_{5}-y_{5}\right) \, \mathbf{a}_{1} + x_{5} \, \mathbf{a}_{2}-z_{5} \, \mathbf{a}_{3} & = & \left(x_{5}-\frac{1}{2}y_{5}\right)a \, \mathbf{\hat{x}} + \frac{\sqrt{3}}{2}y_{5}a \, \mathbf{\hat{y}}-z_{5}c \, \mathbf{\hat{z}} & \left(24m\right) & \mbox{O II} \\ 
\mathbf{B}_{50} & = & x_{5} \, \mathbf{a}_{1} + y_{5} \, \mathbf{a}_{2}-z_{5} \, \mathbf{a}_{3} & = & \frac{1}{2}\left(x_{5}+y_{5}\right)a \, \mathbf{\hat{x}} + \frac{\sqrt{3}}{2}\left(-x_{5}+y_{5}\right)a \, \mathbf{\hat{y}}-z_{5}c \, \mathbf{\hat{z}} & \left(24m\right) & \mbox{O II} \\ 
\mathbf{B}_{51} & = & -y_{5} \, \mathbf{a}_{1} + \left(x_{5}-y_{5}\right) \, \mathbf{a}_{2}-z_{5} \, \mathbf{a}_{3} & = & \left(\frac{1}{2}x_{5}-y_{5}\right)a \, \mathbf{\hat{x}} + \frac{\sqrt{3}}{2}x_{5}a \, \mathbf{\hat{y}}-z_{5}c \, \mathbf{\hat{z}} & \left(24m\right) & \mbox{O II} \\ 
\mathbf{B}_{52} & = & \left(-x_{5}+y_{5}\right) \, \mathbf{a}_{1}-x_{5} \, \mathbf{a}_{2}-z_{5} \, \mathbf{a}_{3} & = & \left(-x_{5}+\frac{1}{2}y_{5}\right)a \, \mathbf{\hat{x}}-\frac{\sqrt{3}}{2}y_{5}a \, \mathbf{\hat{y}}-z_{5}c \, \mathbf{\hat{z}} & \left(24m\right) & \mbox{O II} \\ 
\mathbf{B}_{53} & = & -y_{5} \, \mathbf{a}_{1}-x_{5} \, \mathbf{a}_{2} + \left(\frac{1}{2} +z_{5}\right) \, \mathbf{a}_{3} & = & -\frac{1}{2}\left(x_{5}+y_{5}\right)a \, \mathbf{\hat{x}} + \frac{\sqrt{3}}{2}\left(-x_{5}+y_{5}\right)a \, \mathbf{\hat{y}} + \left(\frac{1}{2} +z_{5}\right)c \, \mathbf{\hat{z}} & \left(24m\right) & \mbox{O II} \\ 
\mathbf{B}_{54} & = & \left(-x_{5}+y_{5}\right) \, \mathbf{a}_{1} + y_{5} \, \mathbf{a}_{2} + \left(\frac{1}{2} +z_{5}\right) \, \mathbf{a}_{3} & = & \left(-\frac{1}{2}x_{5}+y_{5}\right)a \, \mathbf{\hat{x}} + \frac{\sqrt{3}}{2}x_{5}a \, \mathbf{\hat{y}} + \left(\frac{1}{2} +z_{5}\right)c \, \mathbf{\hat{z}} & \left(24m\right) & \mbox{O II} \\ 
\mathbf{B}_{55} & = & x_{5} \, \mathbf{a}_{1} + \left(x_{5}-y_{5}\right) \, \mathbf{a}_{2} + \left(\frac{1}{2} +z_{5}\right) \, \mathbf{a}_{3} & = & \left(x_{5}-\frac{1}{2}y_{5}\right)a \, \mathbf{\hat{x}}-\frac{\sqrt{3}}{2}y_{5}a \, \mathbf{\hat{y}} + \left(\frac{1}{2} +z_{5}\right)c \, \mathbf{\hat{z}} & \left(24m\right) & \mbox{O II} \\ 
\mathbf{B}_{56} & = & y_{5} \, \mathbf{a}_{1} + x_{5} \, \mathbf{a}_{2} + \left(\frac{1}{2} +z_{5}\right) \, \mathbf{a}_{3} & = & \frac{1}{2}\left(x_{5}+y_{5}\right)a \, \mathbf{\hat{x}} + \frac{\sqrt{3}}{2}\left(x_{5}-y_{5}\right)a \, \mathbf{\hat{y}} + \left(\frac{1}{2} +z_{5}\right)c \, \mathbf{\hat{z}} & \left(24m\right) & \mbox{O II} \\ 
\mathbf{B}_{57} & = & \left(x_{5}-y_{5}\right) \, \mathbf{a}_{1}-y_{5} \, \mathbf{a}_{2} + \left(\frac{1}{2} +z_{5}\right) \, \mathbf{a}_{3} & = & \left(\frac{1}{2}x_{5}-y_{5}\right)a \, \mathbf{\hat{x}}-\frac{\sqrt{3}}{2}x_{5}a \, \mathbf{\hat{y}} + \left(\frac{1}{2} +z_{5}\right)c \, \mathbf{\hat{z}} & \left(24m\right) & \mbox{O II} \\ 
\mathbf{B}_{58} & = & -x_{5} \, \mathbf{a}_{1} + \left(-x_{5}+y_{5}\right) \, \mathbf{a}_{2} + \left(\frac{1}{2} +z_{5}\right) \, \mathbf{a}_{3} & = & \left(-x_{5}+\frac{1}{2}y_{5}\right)a \, \mathbf{\hat{x}} + \frac{\sqrt{3}}{2}y_{5}a \, \mathbf{\hat{y}} + \left(\frac{1}{2} +z_{5}\right)c \, \mathbf{\hat{z}} & \left(24m\right) & \mbox{O II} \\ 
\end{longtabu}
\renewcommand{\arraystretch}{1.0}
\noindent \hrulefill
\\
\textbf{References:}
\vspace*{-0.25cm}
\begin{flushleft}
  - \bibentry{Hazen_Am_Min_71_1986}. \\
  - \bibentry{Morosin_Acta_Cryst_B_28_1927}. \\
\end{flushleft}
\textbf{Found in:}
\vspace*{-0.25cm}
\begin{flushleft}
  - \bibentry{Downs_Am_Min_88_2003}. \\
\end{flushleft}
\noindent \hrulefill
\\
\textbf{Geometry files:}
\\
\noindent  - CIF: pp. {\hyperref[A2B3C18D6_hP58_192_c_f_lm_l_cif]{\pageref{A2B3C18D6_hP58_192_c_f_lm_l_cif}}} \\
\noindent  - POSCAR: pp. {\hyperref[A2B3C18D6_hP58_192_c_f_lm_l_poscar]{\pageref{A2B3C18D6_hP58_192_c_f_lm_l_poscar}}} \\
\onecolumn
{\phantomsection\label{AB2_hP72_192_m_j2kl}}
\subsection*{\huge \textbf{{\normalfont AlPO$_{4}$ Structure: AB2\_hP72\_192\_m\_j2kl}}}
\noindent \hrulefill
\vspace*{0.25cm}
\begin{figure}[htp]
  \centering
  \vspace{-1em}
  {\includegraphics[width=1\textwidth]{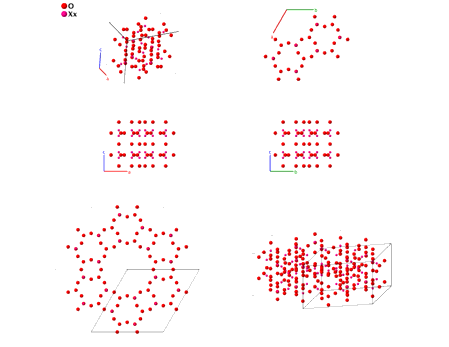}}
\end{figure}
\vspace*{-0.5cm}
\renewcommand{\arraystretch}{1.5}
\begin{equation*}
  \begin{array}{>{$\hspace{-0.15cm}}l<{$}>{$}p{0.5cm}<{$}>{$}p{18.5cm}<{$}}
    \mbox{\large \textbf{Prototype}} &\colon & \ce{AlPO4} \\
    \mbox{\large \textbf{\AFLOW\ prototype label}} &\colon & \mbox{AB2\_hP72\_192\_m\_j2kl} \\
    \mbox{\large \textbf{\textit{Strukturbericht} designation}} &\colon & \mbox{None} \\
    \mbox{\large \textbf{Pearson symbol}} &\colon & \mbox{hP72} \\
    \mbox{\large \textbf{Space group number}} &\colon & 192 \\
    \mbox{\large \textbf{Space group symbol}} &\colon & P6/mcc \\
    \mbox{\large \textbf{\AFLOW\ prototype command}} &\colon &  \texttt{aflow} \,  \, \texttt{-{}-proto=AB2\_hP72\_192\_m\_j2kl } \, \newline \texttt{-{}-params=}{a,c/a,x_{1},x_{2},x_{3},x_{4},y_{4},x_{5},y_{5},z_{5} }
  \end{array}
\end{equation*}
\renewcommand{\arraystretch}{1.0}

\vspace*{-0.25cm}
\noindent \hrulefill
\begin{itemize}
  \item{Here, the M sites are partially occupied with 0.5Al+0.5P.
The Jmol image does not distinguish between the different M labels and is represented
by the "Xx" atoms.
Polytypes of this compound also appear in a space groups \#168 and \#184.
}
\end{itemize}

\noindent \parbox{1 \linewidth}{
\noindent \hrulefill
\\
\textbf{Hexagonal primitive vectors:} \\
\vspace*{-0.25cm}
\begin{tabular}{cc}
  \begin{tabular}{c}
    \parbox{0.6 \linewidth}{
      \renewcommand{\arraystretch}{1.5}
      \begin{equation*}
        \centering
        \begin{array}{ccc}
              \mathbf{a}_1 & = & \frac12 \, a \, \mathbf{\hat{x}} - \frac{\sqrt3}2 \, a \, \mathbf{\hat{y}} \\
    \mathbf{a}_2 & = & \frac12 \, a \, \mathbf{\hat{x}} + \frac{\sqrt3}2 \, a \, \mathbf{\hat{y}} \\
    \mathbf{a}_3 & = & c \, \mathbf{\hat{z}} \\

        \end{array}
      \end{equation*}
    }
    \renewcommand{\arraystretch}{1.0}
  \end{tabular}
  \begin{tabular}{c}
    \includegraphics[width=0.3\linewidth]{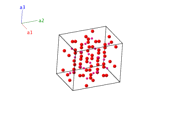} \\
  \end{tabular}
\end{tabular}

}
\vspace*{-0.25cm}

\noindent \hrulefill
\\
\textbf{Basis vectors:}
\vspace*{-0.25cm}
\renewcommand{\arraystretch}{1.5}
\begin{longtabu} to \textwidth{>{\centering $}X[-1,c,c]<{$}>{\centering $}X[-1,c,c]<{$}>{\centering $}X[-1,c,c]<{$}>{\centering $}X[-1,c,c]<{$}>{\centering $}X[-1,c,c]<{$}>{\centering $}X[-1,c,c]<{$}>{\centering $}X[-1,c,c]<{$}}
  & & \mbox{Lattice Coordinates} & & \mbox{Cartesian Coordinates} &\mbox{Wyckoff Position} & \mbox{Atom Type} \\  
  \mathbf{B}_{1} & = & x_{1} \, \mathbf{a}_{1} + \frac{1}{4} \, \mathbf{a}_{3} & = & \frac{1}{2}x_{1}a \, \mathbf{\hat{x}}-\frac{\sqrt{3}}{2}x_{1}a \, \mathbf{\hat{y}} + \frac{1}{4}c \, \mathbf{\hat{z}} & \left(12j\right) & \mbox{O I} \\ 
\mathbf{B}_{2} & = & x_{1} \, \mathbf{a}_{2} + \frac{1}{4} \, \mathbf{a}_{3} & = & \frac{1}{2}x_{1}a \, \mathbf{\hat{x}} + \frac{\sqrt{3}}{2}x_{1}a \, \mathbf{\hat{y}} + \frac{1}{4}c \, \mathbf{\hat{z}} & \left(12j\right) & \mbox{O I} \\ 
\mathbf{B}_{3} & = & -x_{1} \, \mathbf{a}_{1}-x_{1} \, \mathbf{a}_{2} + \frac{1}{4} \, \mathbf{a}_{3} & = & -x_{1}a \, \mathbf{\hat{x}} + \frac{1}{4}c \, \mathbf{\hat{z}} & \left(12j\right) & \mbox{O I} \\ 
\mathbf{B}_{4} & = & -x_{1} \, \mathbf{a}_{1} + \frac{1}{4} \, \mathbf{a}_{3} & = & -\frac{1}{2}x_{1}a \, \mathbf{\hat{x}} + \frac{\sqrt{3}}{2}x_{1}a \, \mathbf{\hat{y}} + \frac{1}{4}c \, \mathbf{\hat{z}} & \left(12j\right) & \mbox{O I} \\ 
\mathbf{B}_{5} & = & -x_{1} \, \mathbf{a}_{2} + \frac{1}{4} \, \mathbf{a}_{3} & = & -\frac{1}{2}x_{1}a \, \mathbf{\hat{x}}-\frac{\sqrt{3}}{2}x_{1}a \, \mathbf{\hat{y}} + \frac{1}{4}c \, \mathbf{\hat{z}} & \left(12j\right) & \mbox{O I} \\ 
\mathbf{B}_{6} & = & x_{1} \, \mathbf{a}_{1} + x_{1} \, \mathbf{a}_{2} + \frac{1}{4} \, \mathbf{a}_{3} & = & x_{1}a \, \mathbf{\hat{x}} + \frac{1}{4}c \, \mathbf{\hat{z}} & \left(12j\right) & \mbox{O I} \\ 
\mathbf{B}_{7} & = & -x_{1} \, \mathbf{a}_{1} + \frac{3}{4} \, \mathbf{a}_{3} & = & -\frac{1}{2}x_{1}a \, \mathbf{\hat{x}} + \frac{\sqrt{3}}{2}x_{1}a \, \mathbf{\hat{y}} + \frac{3}{4}c \, \mathbf{\hat{z}} & \left(12j\right) & \mbox{O I} \\ 
\mathbf{B}_{8} & = & -x_{1} \, \mathbf{a}_{2} + \frac{3}{4} \, \mathbf{a}_{3} & = & -\frac{1}{2}x_{1}a \, \mathbf{\hat{x}}-\frac{\sqrt{3}}{2}x_{1}a \, \mathbf{\hat{y}} + \frac{3}{4}c \, \mathbf{\hat{z}} & \left(12j\right) & \mbox{O I} \\ 
\mathbf{B}_{9} & = & x_{1} \, \mathbf{a}_{1} + x_{1} \, \mathbf{a}_{2} + \frac{3}{4} \, \mathbf{a}_{3} & = & x_{1}a \, \mathbf{\hat{x}} + \frac{3}{4}c \, \mathbf{\hat{z}} & \left(12j\right) & \mbox{O I} \\ 
\mathbf{B}_{10} & = & x_{1} \, \mathbf{a}_{1} + \frac{3}{4} \, \mathbf{a}_{3} & = & \frac{1}{2}x_{1}a \, \mathbf{\hat{x}}-\frac{\sqrt{3}}{2}x_{1}a \, \mathbf{\hat{y}} + \frac{3}{4}c \, \mathbf{\hat{z}} & \left(12j\right) & \mbox{O I} \\ 
\mathbf{B}_{11} & = & x_{1} \, \mathbf{a}_{2} + \frac{3}{4} \, \mathbf{a}_{3} & = & \frac{1}{2}x_{1}a \, \mathbf{\hat{x}} + \frac{\sqrt{3}}{2}x_{1}a \, \mathbf{\hat{y}} + \frac{3}{4}c \, \mathbf{\hat{z}} & \left(12j\right) & \mbox{O I} \\ 
\mathbf{B}_{12} & = & -x_{1} \, \mathbf{a}_{1}-x_{1} \, \mathbf{a}_{2} + \frac{3}{4} \, \mathbf{a}_{3} & = & -x_{1}a \, \mathbf{\hat{x}} + \frac{3}{4}c \, \mathbf{\hat{z}} & \left(12j\right) & \mbox{O I} \\ 
\mathbf{B}_{13} & = & x_{2} \, \mathbf{a}_{1} + 2x_{2} \, \mathbf{a}_{2} + \frac{1}{4} \, \mathbf{a}_{3} & = & \frac{3}{2}x_{2}a \, \mathbf{\hat{x}} + \frac{\sqrt{3}}{2}x_{2}a \, \mathbf{\hat{y}} + \frac{1}{4}c \, \mathbf{\hat{z}} & \left(12k\right) & \mbox{O II} \\ 
\mathbf{B}_{14} & = & -2x_{2} \, \mathbf{a}_{1}-x_{2} \, \mathbf{a}_{2} + \frac{1}{4} \, \mathbf{a}_{3} & = & -\frac{3}{2}x_{2}a \, \mathbf{\hat{x}} + \frac{\sqrt{3}}{2}x_{2}a \, \mathbf{\hat{y}} + \frac{1}{4}c \, \mathbf{\hat{z}} & \left(12k\right) & \mbox{O II} \\ 
\mathbf{B}_{15} & = & x_{2} \, \mathbf{a}_{1}-x_{2} \, \mathbf{a}_{2} + \frac{1}{4} \, \mathbf{a}_{3} & = & -\sqrt{3}x_{2}a \, \mathbf{\hat{y}} + \frac{1}{4}c \, \mathbf{\hat{z}} & \left(12k\right) & \mbox{O II} \\ 
\mathbf{B}_{16} & = & -x_{2} \, \mathbf{a}_{1}-2x_{2} \, \mathbf{a}_{2} + \frac{1}{4} \, \mathbf{a}_{3} & = & -\frac{3}{2}x_{2}a \, \mathbf{\hat{x}}-\frac{\sqrt{3}}{2}x_{2}a \, \mathbf{\hat{y}} + \frac{1}{4}c \, \mathbf{\hat{z}} & \left(12k\right) & \mbox{O II} \\ 
\mathbf{B}_{17} & = & 2x_{2} \, \mathbf{a}_{1} + x_{2} \, \mathbf{a}_{2} + \frac{1}{4} \, \mathbf{a}_{3} & = & \frac{3}{2}x_{2}a \, \mathbf{\hat{x}}-\frac{\sqrt{3}}{2}x_{2}a \, \mathbf{\hat{y}} + \frac{1}{4}c \, \mathbf{\hat{z}} & \left(12k\right) & \mbox{O II} \\ 
\mathbf{B}_{18} & = & -x_{2} \, \mathbf{a}_{1} + x_{2} \, \mathbf{a}_{2} + \frac{1}{4} \, \mathbf{a}_{3} & = & \sqrt{3}x_{2}a \, \mathbf{\hat{y}} + \frac{1}{4}c \, \mathbf{\hat{z}} & \left(12k\right) & \mbox{O II} \\ 
\mathbf{B}_{19} & = & -x_{2} \, \mathbf{a}_{1}-2x_{2} \, \mathbf{a}_{2} + \frac{3}{4} \, \mathbf{a}_{3} & = & -\frac{3}{2}x_{2}a \, \mathbf{\hat{x}}-\frac{\sqrt{3}}{2}x_{2}a \, \mathbf{\hat{y}} + \frac{3}{4}c \, \mathbf{\hat{z}} & \left(12k\right) & \mbox{O II} \\ 
\mathbf{B}_{20} & = & 2x_{2} \, \mathbf{a}_{1} + x_{2} \, \mathbf{a}_{2} + \frac{3}{4} \, \mathbf{a}_{3} & = & \frac{3}{2}x_{2}a \, \mathbf{\hat{x}}-\frac{\sqrt{3}}{2}x_{2}a \, \mathbf{\hat{y}} + \frac{3}{4}c \, \mathbf{\hat{z}} & \left(12k\right) & \mbox{O II} \\ 
\mathbf{B}_{21} & = & -x_{2} \, \mathbf{a}_{1} + x_{2} \, \mathbf{a}_{2} + \frac{3}{4} \, \mathbf{a}_{3} & = & \sqrt{3}x_{2}a \, \mathbf{\hat{y}} + \frac{3}{4}c \, \mathbf{\hat{z}} & \left(12k\right) & \mbox{O II} \\ 
\mathbf{B}_{22} & = & x_{2} \, \mathbf{a}_{1} + 2x_{2} \, \mathbf{a}_{2} + \frac{3}{4} \, \mathbf{a}_{3} & = & \frac{3}{2}x_{2}a \, \mathbf{\hat{x}} + \frac{\sqrt{3}}{2}x_{2}a \, \mathbf{\hat{y}} + \frac{3}{4}c \, \mathbf{\hat{z}} & \left(12k\right) & \mbox{O II} \\ 
\mathbf{B}_{23} & = & -2x_{2} \, \mathbf{a}_{1}-x_{2} \, \mathbf{a}_{2} + \frac{3}{4} \, \mathbf{a}_{3} & = & -\frac{3}{2}x_{2}a \, \mathbf{\hat{x}} + \frac{\sqrt{3}}{2}x_{2}a \, \mathbf{\hat{y}} + \frac{3}{4}c \, \mathbf{\hat{z}} & \left(12k\right) & \mbox{O II} \\ 
\mathbf{B}_{24} & = & x_{2} \, \mathbf{a}_{1}-x_{2} \, \mathbf{a}_{2} + \frac{3}{4} \, \mathbf{a}_{3} & = & -\sqrt{3}x_{2}a \, \mathbf{\hat{y}} + \frac{3}{4}c \, \mathbf{\hat{z}} & \left(12k\right) & \mbox{O II} \\ 
\mathbf{B}_{25} & = & x_{3} \, \mathbf{a}_{1} + 2x_{3} \, \mathbf{a}_{2} + \frac{1}{4} \, \mathbf{a}_{3} & = & \frac{3}{2}x_{3}a \, \mathbf{\hat{x}} + \frac{\sqrt{3}}{2}x_{3}a \, \mathbf{\hat{y}} + \frac{1}{4}c \, \mathbf{\hat{z}} & \left(12k\right) & \mbox{O III} \\ 
\mathbf{B}_{26} & = & -2x_{3} \, \mathbf{a}_{1}-x_{3} \, \mathbf{a}_{2} + \frac{1}{4} \, \mathbf{a}_{3} & = & -\frac{3}{2}x_{3}a \, \mathbf{\hat{x}} + \frac{\sqrt{3}}{2}x_{3}a \, \mathbf{\hat{y}} + \frac{1}{4}c \, \mathbf{\hat{z}} & \left(12k\right) & \mbox{O III} \\ 
\mathbf{B}_{27} & = & x_{3} \, \mathbf{a}_{1}-x_{3} \, \mathbf{a}_{2} + \frac{1}{4} \, \mathbf{a}_{3} & = & -\sqrt{3}x_{3}a \, \mathbf{\hat{y}} + \frac{1}{4}c \, \mathbf{\hat{z}} & \left(12k\right) & \mbox{O III} \\ 
\mathbf{B}_{28} & = & -x_{3} \, \mathbf{a}_{1}-2x_{3} \, \mathbf{a}_{2} + \frac{1}{4} \, \mathbf{a}_{3} & = & -\frac{3}{2}x_{3}a \, \mathbf{\hat{x}}-\frac{\sqrt{3}}{2}x_{3}a \, \mathbf{\hat{y}} + \frac{1}{4}c \, \mathbf{\hat{z}} & \left(12k\right) & \mbox{O III} \\ 
\mathbf{B}_{29} & = & 2x_{3} \, \mathbf{a}_{1} + x_{3} \, \mathbf{a}_{2} + \frac{1}{4} \, \mathbf{a}_{3} & = & \frac{3}{2}x_{3}a \, \mathbf{\hat{x}}-\frac{\sqrt{3}}{2}x_{3}a \, \mathbf{\hat{y}} + \frac{1}{4}c \, \mathbf{\hat{z}} & \left(12k\right) & \mbox{O III} \\ 
\mathbf{B}_{30} & = & -x_{3} \, \mathbf{a}_{1} + x_{3} \, \mathbf{a}_{2} + \frac{1}{4} \, \mathbf{a}_{3} & = & \sqrt{3}x_{3}a \, \mathbf{\hat{y}} + \frac{1}{4}c \, \mathbf{\hat{z}} & \left(12k\right) & \mbox{O III} \\ 
\mathbf{B}_{31} & = & -x_{3} \, \mathbf{a}_{1}-2x_{3} \, \mathbf{a}_{2} + \frac{3}{4} \, \mathbf{a}_{3} & = & -\frac{3}{2}x_{3}a \, \mathbf{\hat{x}}-\frac{\sqrt{3}}{2}x_{3}a \, \mathbf{\hat{y}} + \frac{3}{4}c \, \mathbf{\hat{z}} & \left(12k\right) & \mbox{O III} \\ 
\mathbf{B}_{32} & = & 2x_{3} \, \mathbf{a}_{1} + x_{3} \, \mathbf{a}_{2} + \frac{3}{4} \, \mathbf{a}_{3} & = & \frac{3}{2}x_{3}a \, \mathbf{\hat{x}}-\frac{\sqrt{3}}{2}x_{3}a \, \mathbf{\hat{y}} + \frac{3}{4}c \, \mathbf{\hat{z}} & \left(12k\right) & \mbox{O III} \\ 
\mathbf{B}_{33} & = & -x_{3} \, \mathbf{a}_{1} + x_{3} \, \mathbf{a}_{2} + \frac{3}{4} \, \mathbf{a}_{3} & = & \sqrt{3}x_{3}a \, \mathbf{\hat{y}} + \frac{3}{4}c \, \mathbf{\hat{z}} & \left(12k\right) & \mbox{O III} \\ 
\mathbf{B}_{34} & = & x_{3} \, \mathbf{a}_{1} + 2x_{3} \, \mathbf{a}_{2} + \frac{3}{4} \, \mathbf{a}_{3} & = & \frac{3}{2}x_{3}a \, \mathbf{\hat{x}} + \frac{\sqrt{3}}{2}x_{3}a \, \mathbf{\hat{y}} + \frac{3}{4}c \, \mathbf{\hat{z}} & \left(12k\right) & \mbox{O III} \\ 
\mathbf{B}_{35} & = & -2x_{3} \, \mathbf{a}_{1}-x_{3} \, \mathbf{a}_{2} + \frac{3}{4} \, \mathbf{a}_{3} & = & -\frac{3}{2}x_{3}a \, \mathbf{\hat{x}} + \frac{\sqrt{3}}{2}x_{3}a \, \mathbf{\hat{y}} + \frac{3}{4}c \, \mathbf{\hat{z}} & \left(12k\right) & \mbox{O III} \\ 
\mathbf{B}_{36} & = & x_{3} \, \mathbf{a}_{1}-x_{3} \, \mathbf{a}_{2} + \frac{3}{4} \, \mathbf{a}_{3} & = & -\sqrt{3}x_{3}a \, \mathbf{\hat{y}} + \frac{3}{4}c \, \mathbf{\hat{z}} & \left(12k\right) & \mbox{O III} \\ 
\mathbf{B}_{37} & = & x_{4} \, \mathbf{a}_{1} + y_{4} \, \mathbf{a}_{2} & = & \frac{1}{2}\left(x_{4}+y_{4}\right)a \, \mathbf{\hat{x}} + \frac{\sqrt{3}}{2}\left(-x_{4}+y_{4}\right)a \, \mathbf{\hat{y}} & \left(12l\right) & \mbox{O IV} \\ 
\mathbf{B}_{38} & = & -y_{4} \, \mathbf{a}_{1} + \left(x_{4}-y_{4}\right) \, \mathbf{a}_{2} & = & \left(\frac{1}{2}x_{4}-y_{4}\right)a \, \mathbf{\hat{x}} + \frac{\sqrt{3}}{2}x_{4}a \, \mathbf{\hat{y}} & \left(12l\right) & \mbox{O IV} \\ 
\mathbf{B}_{39} & = & \left(-x_{4}+y_{4}\right) \, \mathbf{a}_{1}-x_{4} \, \mathbf{a}_{2} & = & \left(-x_{4}+\frac{1}{2}y_{4}\right)a \, \mathbf{\hat{x}}-\frac{\sqrt{3}}{2}y_{4}a \, \mathbf{\hat{y}} & \left(12l\right) & \mbox{O IV} \\ 
\mathbf{B}_{40} & = & -x_{4} \, \mathbf{a}_{1}-y_{4} \, \mathbf{a}_{2} & = & -\frac{1}{2}\left(x_{4}+y_{4}\right)a \, \mathbf{\hat{x}} + \frac{\sqrt{3}}{2}\left(x_{4}-y_{4}\right)a \, \mathbf{\hat{y}} & \left(12l\right) & \mbox{O IV} \\ 
\mathbf{B}_{41} & = & y_{4} \, \mathbf{a}_{1} + \left(-x_{4}+y_{4}\right) \, \mathbf{a}_{2} & = & \left(-\frac{1}{2}x_{4}+y_{4}\right)a \, \mathbf{\hat{x}}-\frac{\sqrt{3}}{2}x_{4}a \, \mathbf{\hat{y}} & \left(12l\right) & \mbox{O IV} \\ 
\mathbf{B}_{42} & = & \left(x_{4}-y_{4}\right) \, \mathbf{a}_{1} + x_{4} \, \mathbf{a}_{2} & = & \left(x_{4}-\frac{1}{2}y_{4}\right)a \, \mathbf{\hat{x}} + \frac{\sqrt{3}}{2}y_{4}a \, \mathbf{\hat{y}} & \left(12l\right) & \mbox{O IV} \\ 
\mathbf{B}_{43} & = & y_{4} \, \mathbf{a}_{1} + x_{4} \, \mathbf{a}_{2} + \frac{1}{2} \, \mathbf{a}_{3} & = & \frac{1}{2}\left(x_{4}+y_{4}\right)a \, \mathbf{\hat{x}} + \frac{\sqrt{3}}{2}\left(x_{4}-y_{4}\right)a \, \mathbf{\hat{y}} + \frac{1}{2}c \, \mathbf{\hat{z}} & \left(12l\right) & \mbox{O IV} \\ 
\mathbf{B}_{44} & = & \left(x_{4}-y_{4}\right) \, \mathbf{a}_{1}-y_{4} \, \mathbf{a}_{2} + \frac{1}{2} \, \mathbf{a}_{3} & = & \left(\frac{1}{2}x_{4}-y_{4}\right)a \, \mathbf{\hat{x}}-\frac{\sqrt{3}}{2}x_{4}a \, \mathbf{\hat{y}} + \frac{1}{2}c \, \mathbf{\hat{z}} & \left(12l\right) & \mbox{O IV} \\ 
\mathbf{B}_{45} & = & -x_{4} \, \mathbf{a}_{1} + \left(-x_{4}+y_{4}\right) \, \mathbf{a}_{2} + \frac{1}{2} \, \mathbf{a}_{3} & = & \left(-x_{4}+\frac{1}{2}y_{4}\right)a \, \mathbf{\hat{x}} + \frac{\sqrt{3}}{2}y_{4}a \, \mathbf{\hat{y}} + \frac{1}{2}c \, \mathbf{\hat{z}} & \left(12l\right) & \mbox{O IV} \\ 
\mathbf{B}_{46} & = & -y_{4} \, \mathbf{a}_{1}-x_{4} \, \mathbf{a}_{2} + \frac{1}{2} \, \mathbf{a}_{3} & = & -\frac{1}{2}\left(x_{4}+y_{4}\right)a \, \mathbf{\hat{x}} + \frac{\sqrt{3}}{2}\left(-x_{4}+y_{4}\right)a \, \mathbf{\hat{y}} + \frac{1}{2}c \, \mathbf{\hat{z}} & \left(12l\right) & \mbox{O IV} \\ 
\mathbf{B}_{47} & = & \left(-x_{4}+y_{4}\right) \, \mathbf{a}_{1} + y_{4} \, \mathbf{a}_{2} + \frac{1}{2} \, \mathbf{a}_{3} & = & \left(-\frac{1}{2}x_{4}+y_{4}\right)a \, \mathbf{\hat{x}} + \frac{\sqrt{3}}{2}x_{4}a \, \mathbf{\hat{y}} + \frac{1}{2}c \, \mathbf{\hat{z}} & \left(12l\right) & \mbox{O IV} \\ 
\mathbf{B}_{48} & = & x_{4} \, \mathbf{a}_{1} + \left(x_{4}-y_{4}\right) \, \mathbf{a}_{2} + \frac{1}{2} \, \mathbf{a}_{3} & = & \left(x_{4}-\frac{1}{2}y_{4}\right)a \, \mathbf{\hat{x}}-\frac{\sqrt{3}}{2}y_{4}a \, \mathbf{\hat{y}} + \frac{1}{2}c \, \mathbf{\hat{z}} & \left(12l\right) & \mbox{O IV} \\ 
\mathbf{B}_{49} & = & x_{5} \, \mathbf{a}_{1} + y_{5} \, \mathbf{a}_{2} + z_{5} \, \mathbf{a}_{3} & = & \frac{1}{2}\left(x_{5}+y_{5}\right)a \, \mathbf{\hat{x}} + \frac{\sqrt{3}}{2}\left(-x_{5}+y_{5}\right)a \, \mathbf{\hat{y}} + z_{5}c \, \mathbf{\hat{z}} & \left(24m\right) & \mbox{M} \\ 
\mathbf{B}_{50} & = & -y_{5} \, \mathbf{a}_{1} + \left(x_{5}-y_{5}\right) \, \mathbf{a}_{2} + z_{5} \, \mathbf{a}_{3} & = & \left(\frac{1}{2}x_{5}-y_{5}\right)a \, \mathbf{\hat{x}} + \frac{\sqrt{3}}{2}x_{5}a \, \mathbf{\hat{y}} + z_{5}c \, \mathbf{\hat{z}} & \left(24m\right) & \mbox{M} \\ 
\mathbf{B}_{51} & = & \left(-x_{5}+y_{5}\right) \, \mathbf{a}_{1}-x_{5} \, \mathbf{a}_{2} + z_{5} \, \mathbf{a}_{3} & = & \left(-x_{5}+\frac{1}{2}y_{5}\right)a \, \mathbf{\hat{x}}-\frac{\sqrt{3}}{2}y_{5}a \, \mathbf{\hat{y}} + z_{5}c \, \mathbf{\hat{z}} & \left(24m\right) & \mbox{M} \\ 
\mathbf{B}_{52} & = & -x_{5} \, \mathbf{a}_{1}-y_{5} \, \mathbf{a}_{2} + z_{5} \, \mathbf{a}_{3} & = & -\frac{1}{2}\left(x_{5}+y_{5}\right)a \, \mathbf{\hat{x}} + \frac{\sqrt{3}}{2}\left(x_{5}-y_{5}\right)a \, \mathbf{\hat{y}} + z_{5}c \, \mathbf{\hat{z}} & \left(24m\right) & \mbox{M} \\ 
\mathbf{B}_{53} & = & y_{5} \, \mathbf{a}_{1} + \left(-x_{5}+y_{5}\right) \, \mathbf{a}_{2} + z_{5} \, \mathbf{a}_{3} & = & \left(-\frac{1}{2}x_{5}+y_{5}\right)a \, \mathbf{\hat{x}}-\frac{\sqrt{3}}{2}x_{5}a \, \mathbf{\hat{y}} + z_{5}c \, \mathbf{\hat{z}} & \left(24m\right) & \mbox{M} \\ 
\mathbf{B}_{54} & = & \left(x_{5}-y_{5}\right) \, \mathbf{a}_{1} + x_{5} \, \mathbf{a}_{2} + z_{5} \, \mathbf{a}_{3} & = & \left(x_{5}-\frac{1}{2}y_{5}\right)a \, \mathbf{\hat{x}} + \frac{\sqrt{3}}{2}y_{5}a \, \mathbf{\hat{y}} + z_{5}c \, \mathbf{\hat{z}} & \left(24m\right) & \mbox{M} \\ 
\mathbf{B}_{55} & = & y_{5} \, \mathbf{a}_{1} + x_{5} \, \mathbf{a}_{2} + \left(\frac{1}{2} - z_{5}\right) \, \mathbf{a}_{3} & = & \frac{1}{2}\left(x_{5}+y_{5}\right)a \, \mathbf{\hat{x}} + \frac{\sqrt{3}}{2}\left(x_{5}-y_{5}\right)a \, \mathbf{\hat{y}} + \left(\frac{1}{2} - z_{5}\right)c \, \mathbf{\hat{z}} & \left(24m\right) & \mbox{M} \\ 
\mathbf{B}_{56} & = & \left(x_{5}-y_{5}\right) \, \mathbf{a}_{1}-y_{5} \, \mathbf{a}_{2} + \left(\frac{1}{2} - z_{5}\right) \, \mathbf{a}_{3} & = & \left(\frac{1}{2}x_{5}-y_{5}\right)a \, \mathbf{\hat{x}}-\frac{\sqrt{3}}{2}x_{5}a \, \mathbf{\hat{y}} + \left(\frac{1}{2} - z_{5}\right)c \, \mathbf{\hat{z}} & \left(24m\right) & \mbox{M} \\ 
\mathbf{B}_{57} & = & -x_{5} \, \mathbf{a}_{1} + \left(-x_{5}+y_{5}\right) \, \mathbf{a}_{2} + \left(\frac{1}{2} - z_{5}\right) \, \mathbf{a}_{3} & = & \left(-x_{5}+\frac{1}{2}y_{5}\right)a \, \mathbf{\hat{x}} + \frac{\sqrt{3}}{2}y_{5}a \, \mathbf{\hat{y}} + \left(\frac{1}{2} - z_{5}\right)c \, \mathbf{\hat{z}} & \left(24m\right) & \mbox{M} \\ 
\mathbf{B}_{58} & = & -y_{5} \, \mathbf{a}_{1}-x_{5} \, \mathbf{a}_{2} + \left(\frac{1}{2} - z_{5}\right) \, \mathbf{a}_{3} & = & -\frac{1}{2}\left(x_{5}+y_{5}\right)a \, \mathbf{\hat{x}} + \frac{\sqrt{3}}{2}\left(-x_{5}+y_{5}\right)a \, \mathbf{\hat{y}} + \left(\frac{1}{2} - z_{5}\right)c \, \mathbf{\hat{z}} & \left(24m\right) & \mbox{M} \\ 
\mathbf{B}_{59} & = & \left(-x_{5}+y_{5}\right) \, \mathbf{a}_{1} + y_{5} \, \mathbf{a}_{2} + \left(\frac{1}{2} - z_{5}\right) \, \mathbf{a}_{3} & = & \left(-\frac{1}{2}x_{5}+y_{5}\right)a \, \mathbf{\hat{x}} + \frac{\sqrt{3}}{2}x_{5}a \, \mathbf{\hat{y}} + \left(\frac{1}{2} - z_{5}\right)c \, \mathbf{\hat{z}} & \left(24m\right) & \mbox{M} \\ 
\mathbf{B}_{60} & = & x_{5} \, \mathbf{a}_{1} + \left(x_{5}-y_{5}\right) \, \mathbf{a}_{2} + \left(\frac{1}{2} - z_{5}\right) \, \mathbf{a}_{3} & = & \left(x_{5}-\frac{1}{2}y_{5}\right)a \, \mathbf{\hat{x}}-\frac{\sqrt{3}}{2}y_{5}a \, \mathbf{\hat{y}} + \left(\frac{1}{2} - z_{5}\right)c \, \mathbf{\hat{z}} & \left(24m\right) & \mbox{M} \\ 
\mathbf{B}_{61} & = & -x_{5} \, \mathbf{a}_{1}-y_{5} \, \mathbf{a}_{2}-z_{5} \, \mathbf{a}_{3} & = & -\frac{1}{2}\left(x_{5}+y_{5}\right)a \, \mathbf{\hat{x}} + \frac{\sqrt{3}}{2}\left(x_{5}-y_{5}\right)a \, \mathbf{\hat{y}}-z_{5}c \, \mathbf{\hat{z}} & \left(24m\right) & \mbox{M} \\ 
\mathbf{B}_{62} & = & y_{5} \, \mathbf{a}_{1} + \left(-x_{5}+y_{5}\right) \, \mathbf{a}_{2}-z_{5} \, \mathbf{a}_{3} & = & \left(-\frac{1}{2}x_{5}+y_{5}\right)a \, \mathbf{\hat{x}}-\frac{\sqrt{3}}{2}x_{5}a \, \mathbf{\hat{y}}-z_{5}c \, \mathbf{\hat{z}} & \left(24m\right) & \mbox{M} \\ 
\mathbf{B}_{63} & = & \left(x_{5}-y_{5}\right) \, \mathbf{a}_{1} + x_{5} \, \mathbf{a}_{2}-z_{5} \, \mathbf{a}_{3} & = & \left(x_{5}-\frac{1}{2}y_{5}\right)a \, \mathbf{\hat{x}} + \frac{\sqrt{3}}{2}y_{5}a \, \mathbf{\hat{y}}-z_{5}c \, \mathbf{\hat{z}} & \left(24m\right) & \mbox{M} \\ 
\mathbf{B}_{64} & = & x_{5} \, \mathbf{a}_{1} + y_{5} \, \mathbf{a}_{2}-z_{5} \, \mathbf{a}_{3} & = & \frac{1}{2}\left(x_{5}+y_{5}\right)a \, \mathbf{\hat{x}} + \frac{\sqrt{3}}{2}\left(-x_{5}+y_{5}\right)a \, \mathbf{\hat{y}}-z_{5}c \, \mathbf{\hat{z}} & \left(24m\right) & \mbox{M} \\ 
\mathbf{B}_{65} & = & -y_{5} \, \mathbf{a}_{1} + \left(x_{5}-y_{5}\right) \, \mathbf{a}_{2}-z_{5} \, \mathbf{a}_{3} & = & \left(\frac{1}{2}x_{5}-y_{5}\right)a \, \mathbf{\hat{x}} + \frac{\sqrt{3}}{2}x_{5}a \, \mathbf{\hat{y}}-z_{5}c \, \mathbf{\hat{z}} & \left(24m\right) & \mbox{M} \\ 
\mathbf{B}_{66} & = & \left(-x_{5}+y_{5}\right) \, \mathbf{a}_{1}-x_{5} \, \mathbf{a}_{2}-z_{5} \, \mathbf{a}_{3} & = & \left(-x_{5}+\frac{1}{2}y_{5}\right)a \, \mathbf{\hat{x}}-\frac{\sqrt{3}}{2}y_{5}a \, \mathbf{\hat{y}}-z_{5}c \, \mathbf{\hat{z}} & \left(24m\right) & \mbox{M} \\ 
\mathbf{B}_{67} & = & -y_{5} \, \mathbf{a}_{1}-x_{5} \, \mathbf{a}_{2} + \left(\frac{1}{2} +z_{5}\right) \, \mathbf{a}_{3} & = & -\frac{1}{2}\left(x_{5}+y_{5}\right)a \, \mathbf{\hat{x}} + \frac{\sqrt{3}}{2}\left(-x_{5}+y_{5}\right)a \, \mathbf{\hat{y}} + \left(\frac{1}{2} +z_{5}\right)c \, \mathbf{\hat{z}} & \left(24m\right) & \mbox{M} \\ 
\mathbf{B}_{68} & = & \left(-x_{5}+y_{5}\right) \, \mathbf{a}_{1} + y_{5} \, \mathbf{a}_{2} + \left(\frac{1}{2} +z_{5}\right) \, \mathbf{a}_{3} & = & \left(-\frac{1}{2}x_{5}+y_{5}\right)a \, \mathbf{\hat{x}} + \frac{\sqrt{3}}{2}x_{5}a \, \mathbf{\hat{y}} + \left(\frac{1}{2} +z_{5}\right)c \, \mathbf{\hat{z}} & \left(24m\right) & \mbox{M} \\ 
\mathbf{B}_{69} & = & x_{5} \, \mathbf{a}_{1} + \left(x_{5}-y_{5}\right) \, \mathbf{a}_{2} + \left(\frac{1}{2} +z_{5}\right) \, \mathbf{a}_{3} & = & \left(x_{5}-\frac{1}{2}y_{5}\right)a \, \mathbf{\hat{x}}-\frac{\sqrt{3}}{2}y_{5}a \, \mathbf{\hat{y}} + \left(\frac{1}{2} +z_{5}\right)c \, \mathbf{\hat{z}} & \left(24m\right) & \mbox{M} \\ 
\mathbf{B}_{70} & = & y_{5} \, \mathbf{a}_{1} + x_{5} \, \mathbf{a}_{2} + \left(\frac{1}{2} +z_{5}\right) \, \mathbf{a}_{3} & = & \frac{1}{2}\left(x_{5}+y_{5}\right)a \, \mathbf{\hat{x}} + \frac{\sqrt{3}}{2}\left(x_{5}-y_{5}\right)a \, \mathbf{\hat{y}} + \left(\frac{1}{2} +z_{5}\right)c \, \mathbf{\hat{z}} & \left(24m\right) & \mbox{M} \\ 
\mathbf{B}_{71} & = & \left(x_{5}-y_{5}\right) \, \mathbf{a}_{1}-y_{5} \, \mathbf{a}_{2} + \left(\frac{1}{2} +z_{5}\right) \, \mathbf{a}_{3} & = & \left(\frac{1}{2}x_{5}-y_{5}\right)a \, \mathbf{\hat{x}}-\frac{\sqrt{3}}{2}x_{5}a \, \mathbf{\hat{y}} + \left(\frac{1}{2} +z_{5}\right)c \, \mathbf{\hat{z}} & \left(24m\right) & \mbox{M} \\ 
\mathbf{B}_{72} & = & -x_{5} \, \mathbf{a}_{1} + \left(-x_{5}+y_{5}\right) \, \mathbf{a}_{2} + \left(\frac{1}{2} +z_{5}\right) \, \mathbf{a}_{3} & = & \left(-x_{5}+\frac{1}{2}y_{5}\right)a \, \mathbf{\hat{x}} + \frac{\sqrt{3}}{2}y_{5}a \, \mathbf{\hat{y}} + \left(\frac{1}{2} +z_{5}\right)c \, \mathbf{\hat{z}} & \left(24m\right) & \mbox{M} \\ 
\end{longtabu}
\renewcommand{\arraystretch}{1.0}
\noindent \hrulefill
\\
\textbf{References:}
\vspace*{-0.25cm}
\begin{flushleft}
  - \bibentry{Richardson_AlPO4_ActaCrystallogrSecC_1987}. \\
\end{flushleft}
\textbf{Found in:}
\vspace*{-0.25cm}
\begin{flushleft}
  - \bibentry{Villars_PearsonsCrystalData_2013}. \\
\end{flushleft}
\noindent \hrulefill
\\
\textbf{Geometry files:}
\\
\noindent  - CIF: pp. {\hyperref[AB2_hP72_192_m_j2kl_cif]{\pageref{AB2_hP72_192_m_j2kl_cif}}} \\
\noindent  - POSCAR: pp. {\hyperref[AB2_hP72_192_m_j2kl_poscar]{\pageref{AB2_hP72_192_m_j2kl_poscar}}} \\
\onecolumn
{\phantomsection\label{A5B3_hP16_193_dg_g}}
\subsection*{\huge \textbf{{\normalfont Mavlyanovite (Mn$_{5}$Si$_{3}$) Structure: A5B3\_hP16\_193\_dg\_g}}}
\noindent \hrulefill
\vspace*{0.25cm}
\begin{figure}[htp]
  \centering
  \vspace{-1em}
  {\includegraphics[width=1\textwidth]{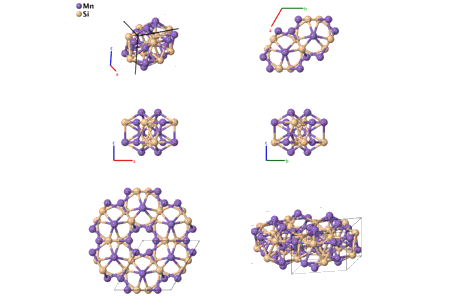}}
\end{figure}
\vspace*{-0.5cm}
\renewcommand{\arraystretch}{1.5}
\begin{equation*}
  \begin{array}{>{$\hspace{-0.15cm}}l<{$}>{$}p{0.5cm}<{$}>{$}p{18.5cm}<{$}}
    \mbox{\large \textbf{Prototype}} &\colon & \ce{Mn5Si3} \\
    \mbox{\large \textbf{\AFLOW\ prototype label}} &\colon & \mbox{A5B3\_hP16\_193\_dg\_g} \\
    \mbox{\large \textbf{\textit{Strukturbericht} designation}} &\colon & \mbox{None} \\
    \mbox{\large \textbf{Pearson symbol}} &\colon & \mbox{hP16} \\
    \mbox{\large \textbf{Space group number}} &\colon & 193 \\
    \mbox{\large \textbf{Space group symbol}} &\colon & P6_{3}/mcm \\
    \mbox{\large \textbf{\AFLOW\ prototype command}} &\colon &  \texttt{aflow} \,  \, \texttt{-{}-proto=A5B3\_hP16\_193\_dg\_g } \, \newline \texttt{-{}-params=}{a,c/a,x_{2},x_{3} }
  \end{array}
\end{equation*}
\renewcommand{\arraystretch}{1.0}

\noindent \parbox{1 \linewidth}{
\noindent \hrulefill
\\
\textbf{Hexagonal primitive vectors:} \\
\vspace*{-0.25cm}
\begin{tabular}{cc}
  \begin{tabular}{c}
    \parbox{0.6 \linewidth}{
      \renewcommand{\arraystretch}{1.5}
      \begin{equation*}
        \centering
        \begin{array}{ccc}
              \mathbf{a}_1 & = & \frac12 \, a \, \mathbf{\hat{x}} - \frac{\sqrt3}2 \, a \, \mathbf{\hat{y}} \\
    \mathbf{a}_2 & = & \frac12 \, a \, \mathbf{\hat{x}} + \frac{\sqrt3}2 \, a \, \mathbf{\hat{y}} \\
    \mathbf{a}_3 & = & c \, \mathbf{\hat{z}} \\

        \end{array}
      \end{equation*}
    }
    \renewcommand{\arraystretch}{1.0}
  \end{tabular}
  \begin{tabular}{c}
    \includegraphics[width=0.3\linewidth]{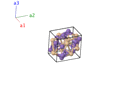} \\
  \end{tabular}
\end{tabular}

}
\vspace*{-0.25cm}

\noindent \hrulefill
\\
\textbf{Basis vectors:}
\vspace*{-0.25cm}
\renewcommand{\arraystretch}{1.5}
\begin{longtabu} to \textwidth{>{\centering $}X[-1,c,c]<{$}>{\centering $}X[-1,c,c]<{$}>{\centering $}X[-1,c,c]<{$}>{\centering $}X[-1,c,c]<{$}>{\centering $}X[-1,c,c]<{$}>{\centering $}X[-1,c,c]<{$}>{\centering $}X[-1,c,c]<{$}}
  & & \mbox{Lattice Coordinates} & & \mbox{Cartesian Coordinates} &\mbox{Wyckoff Position} & \mbox{Atom Type} \\  
  \mathbf{B}_{1} & = & \frac{1}{3} \, \mathbf{a}_{1} + \frac{2}{3} \, \mathbf{a}_{2} & = & \frac{1}{2}a \, \mathbf{\hat{x}} + \frac{1}{2\sqrt{3}}a \, \mathbf{\hat{y}} & \left(4d\right) & \mbox{Mn I} \\ 
\mathbf{B}_{2} & = & \frac{2}{3} \, \mathbf{a}_{1} + \frac{1}{3} \, \mathbf{a}_{2} + \frac{1}{2} \, \mathbf{a}_{3} & = & \frac{1}{2}a \, \mathbf{\hat{x}}- \frac{1}{2\sqrt{3}}a  \, \mathbf{\hat{y}} + \frac{1}{2}c \, \mathbf{\hat{z}} & \left(4d\right) & \mbox{Mn I} \\ 
\mathbf{B}_{3} & = & \frac{2}{3} \, \mathbf{a}_{1} + \frac{1}{3} \, \mathbf{a}_{2} & = & \frac{1}{2}a \, \mathbf{\hat{x}}- \frac{1}{2\sqrt{3}}a  \, \mathbf{\hat{y}} & \left(4d\right) & \mbox{Mn I} \\ 
\mathbf{B}_{4} & = & \frac{1}{3} \, \mathbf{a}_{1} + \frac{2}{3} \, \mathbf{a}_{2} + \frac{1}{2} \, \mathbf{a}_{3} & = & \frac{1}{2}a \, \mathbf{\hat{x}} + \frac{1}{2\sqrt{3}}a \, \mathbf{\hat{y}} + \frac{1}{2}c \, \mathbf{\hat{z}} & \left(4d\right) & \mbox{Mn I} \\ 
\mathbf{B}_{5} & = & x_{2} \, \mathbf{a}_{1} + \frac{1}{4} \, \mathbf{a}_{3} & = & \frac{1}{2}x_{2}a \, \mathbf{\hat{x}}-\frac{\sqrt{3}}{2}x_{2}a \, \mathbf{\hat{y}} + \frac{1}{4}c \, \mathbf{\hat{z}} & \left(6g\right) & \mbox{Mn II} \\ 
\mathbf{B}_{6} & = & x_{2} \, \mathbf{a}_{2} + \frac{1}{4} \, \mathbf{a}_{3} & = & \frac{1}{2}x_{2}a \, \mathbf{\hat{x}} + \frac{\sqrt{3}}{2}x_{2}a \, \mathbf{\hat{y}} + \frac{1}{4}c \, \mathbf{\hat{z}} & \left(6g\right) & \mbox{Mn II} \\ 
\mathbf{B}_{7} & = & -x_{2} \, \mathbf{a}_{1}-x_{2} \, \mathbf{a}_{2} + \frac{1}{4} \, \mathbf{a}_{3} & = & -x_{2}a \, \mathbf{\hat{x}} + \frac{1}{4}c \, \mathbf{\hat{z}} & \left(6g\right) & \mbox{Mn II} \\ 
\mathbf{B}_{8} & = & -x_{2} \, \mathbf{a}_{1} + \frac{3}{4} \, \mathbf{a}_{3} & = & -\frac{1}{2}x_{2}a \, \mathbf{\hat{x}} + \frac{\sqrt{3}}{2}x_{2}a \, \mathbf{\hat{y}} + \frac{3}{4}c \, \mathbf{\hat{z}} & \left(6g\right) & \mbox{Mn II} \\ 
\mathbf{B}_{9} & = & -x_{2} \, \mathbf{a}_{2} + \frac{3}{4} \, \mathbf{a}_{3} & = & -\frac{1}{2}x_{2}a \, \mathbf{\hat{x}}-\frac{\sqrt{3}}{2}x_{2}a \, \mathbf{\hat{y}} + \frac{3}{4}c \, \mathbf{\hat{z}} & \left(6g\right) & \mbox{Mn II} \\ 
\mathbf{B}_{10} & = & x_{2} \, \mathbf{a}_{1} + x_{2} \, \mathbf{a}_{2} + \frac{3}{4} \, \mathbf{a}_{3} & = & x_{2}a \, \mathbf{\hat{x}} + \frac{3}{4}c \, \mathbf{\hat{z}} & \left(6g\right) & \mbox{Mn II} \\ 
\mathbf{B}_{11} & = & x_{3} \, \mathbf{a}_{1} + \frac{1}{4} \, \mathbf{a}_{3} & = & \frac{1}{2}x_{3}a \, \mathbf{\hat{x}}-\frac{\sqrt{3}}{2}x_{3}a \, \mathbf{\hat{y}} + \frac{1}{4}c \, \mathbf{\hat{z}} & \left(6g\right) & \mbox{Si} \\ 
\mathbf{B}_{12} & = & x_{3} \, \mathbf{a}_{2} + \frac{1}{4} \, \mathbf{a}_{3} & = & \frac{1}{2}x_{3}a \, \mathbf{\hat{x}} + \frac{\sqrt{3}}{2}x_{3}a \, \mathbf{\hat{y}} + \frac{1}{4}c \, \mathbf{\hat{z}} & \left(6g\right) & \mbox{Si} \\ 
\mathbf{B}_{13} & = & -x_{3} \, \mathbf{a}_{1}-x_{3} \, \mathbf{a}_{2} + \frac{1}{4} \, \mathbf{a}_{3} & = & -x_{3}a \, \mathbf{\hat{x}} + \frac{1}{4}c \, \mathbf{\hat{z}} & \left(6g\right) & \mbox{Si} \\ 
\mathbf{B}_{14} & = & -x_{3} \, \mathbf{a}_{1} + \frac{3}{4} \, \mathbf{a}_{3} & = & -\frac{1}{2}x_{3}a \, \mathbf{\hat{x}} + \frac{\sqrt{3}}{2}x_{3}a \, \mathbf{\hat{y}} + \frac{3}{4}c \, \mathbf{\hat{z}} & \left(6g\right) & \mbox{Si} \\ 
\mathbf{B}_{15} & = & -x_{3} \, \mathbf{a}_{2} + \frac{3}{4} \, \mathbf{a}_{3} & = & -\frac{1}{2}x_{3}a \, \mathbf{\hat{x}}-\frac{\sqrt{3}}{2}x_{3}a \, \mathbf{\hat{y}} + \frac{3}{4}c \, \mathbf{\hat{z}} & \left(6g\right) & \mbox{Si} \\ 
\mathbf{B}_{16} & = & x_{3} \, \mathbf{a}_{1} + x_{3} \, \mathbf{a}_{2} + \frac{3}{4} \, \mathbf{a}_{3} & = & x_{3}a \, \mathbf{\hat{x}} + \frac{3}{4}c \, \mathbf{\hat{z}} & \left(6g\right) & \mbox{Si} \\ 
\end{longtabu}
\renewcommand{\arraystretch}{1.0}
\noindent \hrulefill
\\
\textbf{References:}
\vspace*{-0.25cm}
\begin{flushleft}
  - \bibentry{Aronsson_Mn5Si3_ActChemScand_1960}. \\
\end{flushleft}
\textbf{Found in:}
\vspace*{-0.25cm}
\begin{flushleft}
  - \bibentry{Villars_PearsonsCrystalData_2013}. \\
\end{flushleft}
\noindent \hrulefill
\\
\textbf{Geometry files:}
\\
\noindent  - CIF: pp. {\hyperref[A5B3_hP16_193_dg_g_cif]{\pageref{A5B3_hP16_193_dg_g_cif}}} \\
\noindent  - POSCAR: pp. {\hyperref[A5B3_hP16_193_dg_g_poscar]{\pageref{A5B3_hP16_193_dg_g_poscar}}} \\
\onecolumn
{\phantomsection\label{A3B_hP16_194_gh_ac}}
\subsection*{\huge \textbf{{\normalfont Ni$_{3}$Ti ($D0_{24}$) Structure: A3B\_hP16\_194\_gh\_ac}}}
\noindent \hrulefill
\vspace*{0.25cm}
\begin{figure}[htp]
  \centering
  \vspace{-1em}
  {\includegraphics[width=1\textwidth]{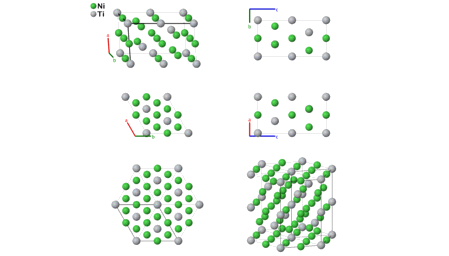}}
\end{figure}
\vspace*{-0.5cm}
\renewcommand{\arraystretch}{1.5}
\begin{equation*}
  \begin{array}{>{$\hspace{-0.15cm}}l<{$}>{$}p{0.5cm}<{$}>{$}p{18.5cm}<{$}}
    \mbox{\large \textbf{Prototype}} &\colon & \ce{Ni$_{3}$Ti} \\
    \mbox{\large \textbf{\AFLOW\ prototype label}} &\colon & \mbox{A3B\_hP16\_194\_gh\_ac} \\
    \mbox{\large \textbf{\textit{Strukturbericht} designation}} &\colon & \mbox{$D0_{24}$} \\
    \mbox{\large \textbf{Pearson symbol}} &\colon & \mbox{hP16} \\
    \mbox{\large \textbf{Space group number}} &\colon & 194 \\
    \mbox{\large \textbf{Space group symbol}} &\colon & P6_{3}/mmc \\
    \mbox{\large \textbf{\AFLOW\ prototype command}} &\colon &  \texttt{aflow} \,  \, \texttt{-{}-proto=A3B\_hP16\_194\_gh\_ac } \, \newline \texttt{-{}-params=}{a,c/a,x_{4} }
  \end{array}
\end{equation*}
\renewcommand{\arraystretch}{1.0}

\vspace*{-0.25cm}
\noindent \hrulefill
\\
\textbf{ Other compounds with this structure:}
\begin{itemize}
   \item{ NpPd$_{3}$, HfPd$_{3}$, TiPd$_{3}$, ZrPd$_{3}$, HfPt$_{3}$, ZrPt$_{3}$  }
\end{itemize}
\vspace*{-0.25cm}
\noindent \hrulefill
\begin{itemize}
  \item{The internal coordinate $x_{4}$ was not determined by any reference we
could find.  We follow (Villars, 2016) and set $x_{4} = -1/6$, which
places the Nb atoms in line with the Ti atoms in the $z = 1/4$ and $z =
3/4$ planes. This is not required by symmetry, and it is likely that the
actual value of $x_{4}$ will be close, but not equal to $-1/6$.
}
\end{itemize}

\noindent \parbox{1 \linewidth}{
\noindent \hrulefill
\\
\textbf{Hexagonal primitive vectors:} \\
\vspace*{-0.25cm}
\begin{tabular}{cc}
  \begin{tabular}{c}
    \parbox{0.6 \linewidth}{
      \renewcommand{\arraystretch}{1.5}
      \begin{equation*}
        \centering
        \begin{array}{ccc}
              \mathbf{a}_1 & = & \frac12 \, a \, \mathbf{\hat{x}} - \frac{\sqrt3}2 \, a \, \mathbf{\hat{y}} \\
    \mathbf{a}_2 & = & \frac12 \, a \, \mathbf{\hat{x}} + \frac{\sqrt3}2 \, a \, \mathbf{\hat{y}} \\
    \mathbf{a}_3 & = & c \, \mathbf{\hat{z}} \\

        \end{array}
      \end{equation*}
    }
    \renewcommand{\arraystretch}{1.0}
  \end{tabular}
  \begin{tabular}{c}
    \includegraphics[width=0.3\linewidth]{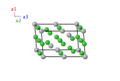} \\
  \end{tabular}
\end{tabular}

}
\vspace*{-0.25cm}

\noindent \hrulefill
\\
\textbf{Basis vectors:}
\vspace*{-0.25cm}
\renewcommand{\arraystretch}{1.5}
\begin{longtabu} to \textwidth{>{\centering $}X[-1,c,c]<{$}>{\centering $}X[-1,c,c]<{$}>{\centering $}X[-1,c,c]<{$}>{\centering $}X[-1,c,c]<{$}>{\centering $}X[-1,c,c]<{$}>{\centering $}X[-1,c,c]<{$}>{\centering $}X[-1,c,c]<{$}}
  & & \mbox{Lattice Coordinates} & & \mbox{Cartesian Coordinates} &\mbox{Wyckoff Position} & \mbox{Atom Type} \\  
  \mathbf{B}_{1} & = & 0 \, \mathbf{a}_{1} + 0 \, \mathbf{a}_{2} + 0 \, \mathbf{a}_{3} & = & 0 \, \mathbf{\hat{x}} + 0 \, \mathbf{\hat{y}} + 0 \, \mathbf{\hat{z}} & \left(2a\right) & \mbox{Ti I} \\ 
\mathbf{B}_{2} & = & \frac{1}{2} \, \mathbf{a}_{3} & = & \frac{1}{2}c \, \mathbf{\hat{z}} & \left(2a\right) & \mbox{Ti I} \\ 
\mathbf{B}_{3} & = & \frac{1}{3} \, \mathbf{a}_{1} + \frac{2}{3} \, \mathbf{a}_{2} + \frac{1}{4} \, \mathbf{a}_{3} & = & \frac{1}{2}a \, \mathbf{\hat{x}} + \frac{1}{2\sqrt{3}}a \, \mathbf{\hat{y}} + \frac{1}{4}c \, \mathbf{\hat{z}} & \left(2c\right) & \mbox{Ti II} \\ 
\mathbf{B}_{4} & = & \frac{2}{3} \, \mathbf{a}_{1} + \frac{1}{3} \, \mathbf{a}_{2} + \frac{3}{4} \, \mathbf{a}_{3} & = & \frac{1}{2}a \, \mathbf{\hat{x}}- \frac{1}{2\sqrt{3}}a  \, \mathbf{\hat{y}} + \frac{3}{4}c \, \mathbf{\hat{z}} & \left(2c\right) & \mbox{Ti II} \\ 
\mathbf{B}_{5} & = & \frac{1}{2} \, \mathbf{a}_{1} & = & \frac{1}{4}a \, \mathbf{\hat{x}}- \frac{\sqrt{3}}{4}a  \, \mathbf{\hat{y}} & \left(6g\right) & \mbox{Ni I} \\ 
\mathbf{B}_{6} & = & \frac{1}{2} \, \mathbf{a}_{2} & = & \frac{1}{4}a \, \mathbf{\hat{x}} + \frac{\sqrt{3}}{4}a \, \mathbf{\hat{y}} & \left(6g\right) & \mbox{Ni I} \\ 
\mathbf{B}_{7} & = & \frac{1}{2} \, \mathbf{a}_{1} + \frac{1}{2} \, \mathbf{a}_{2} & = & \frac{1}{2}a \, \mathbf{\hat{x}} & \left(6g\right) & \mbox{Ni I} \\ 
\mathbf{B}_{8} & = & \frac{1}{2} \, \mathbf{a}_{1} + \frac{1}{2} \, \mathbf{a}_{3} & = & \frac{1}{4}a \, \mathbf{\hat{x}}- \frac{\sqrt{3}}{4}a  \, \mathbf{\hat{y}} + \frac{1}{2}c \, \mathbf{\hat{z}} & \left(6g\right) & \mbox{Ni I} \\ 
\mathbf{B}_{9} & = & \frac{1}{2} \, \mathbf{a}_{2} + \frac{1}{2} \, \mathbf{a}_{3} & = & \frac{1}{4}a \, \mathbf{\hat{x}} + \frac{\sqrt{3}}{4}a \, \mathbf{\hat{y}} + \frac{1}{2}c \, \mathbf{\hat{z}} & \left(6g\right) & \mbox{Ni I} \\ 
\mathbf{B}_{10} & = & \frac{1}{2} \, \mathbf{a}_{1} + \frac{1}{2} \, \mathbf{a}_{2} + \frac{1}{2} \, \mathbf{a}_{3} & = & \frac{1}{2}a \, \mathbf{\hat{x}} + \frac{1}{2}c \, \mathbf{\hat{z}} & \left(6g\right) & \mbox{Ni I} \\ 
\mathbf{B}_{11} & = & x_{4} \, \mathbf{a}_{1} + 2x_{4} \, \mathbf{a}_{2} + \frac{1}{4} \, \mathbf{a}_{3} & = & \frac{3}{2}x_{4}a \, \mathbf{\hat{x}} + \frac{\sqrt{3}}{2}x_{4}a \, \mathbf{\hat{y}} + \frac{1}{4}c \, \mathbf{\hat{z}} & \left(6h\right) & \mbox{Ni II} \\ 
\mathbf{B}_{12} & = & -2x_{4} \, \mathbf{a}_{1}-x_{4} \, \mathbf{a}_{2} + \frac{1}{4} \, \mathbf{a}_{3} & = & -\frac{3}{2}x_{4}a \, \mathbf{\hat{x}} + \frac{\sqrt{3}}{2}x_{4}a \, \mathbf{\hat{y}} + \frac{1}{4}c \, \mathbf{\hat{z}} & \left(6h\right) & \mbox{Ni II} \\ 
\mathbf{B}_{13} & = & x_{4} \, \mathbf{a}_{1}-x_{4} \, \mathbf{a}_{2} + \frac{1}{4} \, \mathbf{a}_{3} & = & -\sqrt{3}x_{4}a \, \mathbf{\hat{y}} + \frac{1}{4}c \, \mathbf{\hat{z}} & \left(6h\right) & \mbox{Ni II} \\ 
\mathbf{B}_{14} & = & -x_{4} \, \mathbf{a}_{1}-2x_{4} \, \mathbf{a}_{2} + \frac{3}{4} \, \mathbf{a}_{3} & = & -\frac{3}{2}x_{4}a \, \mathbf{\hat{x}}-\frac{\sqrt{3}}{2}x_{4}a \, \mathbf{\hat{y}} + \frac{3}{4}c \, \mathbf{\hat{z}} & \left(6h\right) & \mbox{Ni II} \\ 
\mathbf{B}_{15} & = & 2x_{4} \, \mathbf{a}_{1} + x_{4} \, \mathbf{a}_{2} + \frac{3}{4} \, \mathbf{a}_{3} & = & \frac{3}{2}x_{4}a \, \mathbf{\hat{x}}-\frac{\sqrt{3}}{2}x_{4}a \, \mathbf{\hat{y}} + \frac{3}{4}c \, \mathbf{\hat{z}} & \left(6h\right) & \mbox{Ni II} \\ 
\mathbf{B}_{16} & = & -x_{4} \, \mathbf{a}_{1} + x_{4} \, \mathbf{a}_{2} + \frac{3}{4} \, \mathbf{a}_{3} & = & \sqrt{3}x_{4}a \, \mathbf{\hat{y}} + \frac{3}{4}c \, \mathbf{\hat{z}} & \left(6h\right) & \mbox{Ni II} \\ 
\end{longtabu}
\renewcommand{\arraystretch}{1.0}
\noindent \hrulefill
\\
\textbf{References:}
\vspace*{-0.25cm}
\begin{flushleft}
  - \bibentry{Laves_ZKrist_101_1939}. \\
  - \bibentry{Villars_Pauling_2016}. \\
\end{flushleft}
\noindent \hrulefill
\\
\textbf{Geometry files:}
\\
\noindent  - CIF: pp. {\hyperref[A3B_hP16_194_gh_ac_cif]{\pageref{A3B_hP16_194_gh_ac_cif}}} \\
\noindent  - POSCAR: pp. {\hyperref[A3B_hP16_194_gh_ac_poscar]{\pageref{A3B_hP16_194_gh_ac_poscar}}} \\
\onecolumn
{\phantomsection\label{A5B2_hP28_194_ahk_ch}}
\subsection*{\huge \textbf{{\normalfont Co$_{2}$Al$_{5}$ ($D8_{11}$) Structure: A5B2\_hP28\_194\_ahk\_ch}}}
\noindent \hrulefill
\vspace*{0.25cm}
\begin{figure}[htp]
  \centering
  \vspace{-1em}
  {\includegraphics[width=1\textwidth]{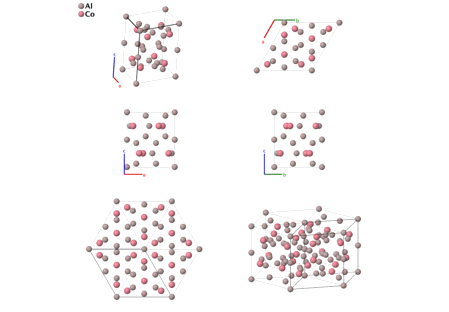}}
\end{figure}
\vspace*{-0.5cm}
\renewcommand{\arraystretch}{1.5}
\begin{equation*}
  \begin{array}{>{$\hspace{-0.15cm}}l<{$}>{$}p{0.5cm}<{$}>{$}p{18.5cm}<{$}}
    \mbox{\large \textbf{Prototype}} &\colon & \ce{Co2Al5} \\
    \mbox{\large \textbf{\AFLOW\ prototype label}} &\colon & \mbox{A5B2\_hP28\_194\_ahk\_ch} \\
    \mbox{\large \textbf{\textit{Strukturbericht} designation}} &\colon & \mbox{$D8_{11}$} \\
    \mbox{\large \textbf{Pearson symbol}} &\colon & \mbox{hP28} \\
    \mbox{\large \textbf{Space group number}} &\colon & 194 \\
    \mbox{\large \textbf{Space group symbol}} &\colon & P6_{3}/mmc \\
    \mbox{\large \textbf{\AFLOW\ prototype command}} &\colon &  \texttt{aflow} \,  \, \texttt{-{}-proto=A5B2\_hP28\_194\_ahk\_ch } \, \newline \texttt{-{}-params=}{a,c/a,x_{3},x_{4},x_{5},z_{5} }
  \end{array}
\end{equation*}
\renewcommand{\arraystretch}{1.0}

\vspace*{-0.25cm}
\noindent \hrulefill
\\
\textbf{ Other compounds with this structure:}
\begin{itemize}
   \item{ Rh$_{2}$Mg$_{5}$, Pd$_{2}$Mg$_{5}$  }
\end{itemize}
\vspace*{-0.25cm}
\noindent \hrulefill
\begin{itemize}
  \item{(Newkirk, 1961) puts the Co I atoms at the (2d) Wyckoff sites.  We
have shifted the origin by $1/2 c \mathbf{\hat{z}}$, which shifts
the Co atoms to the (2c) sites.
}
\end{itemize}

\noindent \parbox{1 \linewidth}{
\noindent \hrulefill
\\
\textbf{Hexagonal primitive vectors:} \\
\vspace*{-0.25cm}
\begin{tabular}{cc}
  \begin{tabular}{c}
    \parbox{0.6 \linewidth}{
      \renewcommand{\arraystretch}{1.5}
      \begin{equation*}
        \centering
        \begin{array}{ccc}
              \mathbf{a}_1 & = & \frac12 \, a \, \mathbf{\hat{x}} - \frac{\sqrt3}2 \, a \, \mathbf{\hat{y}} \\
    \mathbf{a}_2 & = & \frac12 \, a \, \mathbf{\hat{x}} + \frac{\sqrt3}2 \, a \, \mathbf{\hat{y}} \\
    \mathbf{a}_3 & = & c \, \mathbf{\hat{z}} \\

        \end{array}
      \end{equation*}
    }
    \renewcommand{\arraystretch}{1.0}
  \end{tabular}
  \begin{tabular}{c}
    \includegraphics[width=0.3\linewidth]{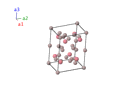} \\
  \end{tabular}
\end{tabular}

}
\vspace*{-0.25cm}

\noindent \hrulefill
\\
\textbf{Basis vectors:}
\vspace*{-0.25cm}
\renewcommand{\arraystretch}{1.5}
\begin{longtabu} to \textwidth{>{\centering $}X[-1,c,c]<{$}>{\centering $}X[-1,c,c]<{$}>{\centering $}X[-1,c,c]<{$}>{\centering $}X[-1,c,c]<{$}>{\centering $}X[-1,c,c]<{$}>{\centering $}X[-1,c,c]<{$}>{\centering $}X[-1,c,c]<{$}}
  & & \mbox{Lattice Coordinates} & & \mbox{Cartesian Coordinates} &\mbox{Wyckoff Position} & \mbox{Atom Type} \\  
  \mathbf{B}_{1} & = & 0 \, \mathbf{a}_{1} + 0 \, \mathbf{a}_{2} + 0 \, \mathbf{a}_{3} & = & 0 \, \mathbf{\hat{x}} + 0 \, \mathbf{\hat{y}} + 0 \, \mathbf{\hat{z}} & \left(2a\right) & \mbox{Al I} \\ 
\mathbf{B}_{2} & = & \frac{1}{2} \, \mathbf{a}_{3} & = & \frac{1}{2}c \, \mathbf{\hat{z}} & \left(2a\right) & \mbox{Al I} \\ 
\mathbf{B}_{3} & = & \frac{1}{3} \, \mathbf{a}_{1} + \frac{2}{3} \, \mathbf{a}_{2} + \frac{1}{4} \, \mathbf{a}_{3} & = & \frac{1}{2}a \, \mathbf{\hat{x}} + \frac{1}{2\sqrt{3}}a \, \mathbf{\hat{y}} + \frac{1}{4}c \, \mathbf{\hat{z}} & \left(2c\right) & \mbox{Co I} \\ 
\mathbf{B}_{4} & = & \frac{2}{3} \, \mathbf{a}_{1} + \frac{1}{3} \, \mathbf{a}_{2} + \frac{3}{4} \, \mathbf{a}_{3} & = & \frac{1}{2}a \, \mathbf{\hat{x}}- \frac{1}{2\sqrt{3}}a  \, \mathbf{\hat{y}} + \frac{3}{4}c \, \mathbf{\hat{z}} & \left(2c\right) & \mbox{Co I} \\ 
\mathbf{B}_{5} & = & x_{3} \, \mathbf{a}_{1} + 2x_{3} \, \mathbf{a}_{2} + \frac{1}{4} \, \mathbf{a}_{3} & = & \frac{3}{2}x_{3}a \, \mathbf{\hat{x}} + \frac{\sqrt{3}}{2}x_{3}a \, \mathbf{\hat{y}} + \frac{1}{4}c \, \mathbf{\hat{z}} & \left(6h\right) & \mbox{Al II} \\ 
\mathbf{B}_{6} & = & -2x_{3} \, \mathbf{a}_{1}-x_{3} \, \mathbf{a}_{2} + \frac{1}{4} \, \mathbf{a}_{3} & = & -\frac{3}{2}x_{3}a \, \mathbf{\hat{x}} + \frac{\sqrt{3}}{2}x_{3}a \, \mathbf{\hat{y}} + \frac{1}{4}c \, \mathbf{\hat{z}} & \left(6h\right) & \mbox{Al II} \\ 
\mathbf{B}_{7} & = & x_{3} \, \mathbf{a}_{1}-x_{3} \, \mathbf{a}_{2} + \frac{1}{4} \, \mathbf{a}_{3} & = & -\sqrt{3}x_{3}a \, \mathbf{\hat{y}} + \frac{1}{4}c \, \mathbf{\hat{z}} & \left(6h\right) & \mbox{Al II} \\ 
\mathbf{B}_{8} & = & -x_{3} \, \mathbf{a}_{1}-2x_{3} \, \mathbf{a}_{2} + \frac{3}{4} \, \mathbf{a}_{3} & = & -\frac{3}{2}x_{3}a \, \mathbf{\hat{x}}-\frac{\sqrt{3}}{2}x_{3}a \, \mathbf{\hat{y}} + \frac{3}{4}c \, \mathbf{\hat{z}} & \left(6h\right) & \mbox{Al II} \\ 
\mathbf{B}_{9} & = & 2x_{3} \, \mathbf{a}_{1} + x_{3} \, \mathbf{a}_{2} + \frac{3}{4} \, \mathbf{a}_{3} & = & \frac{3}{2}x_{3}a \, \mathbf{\hat{x}}-\frac{\sqrt{3}}{2}x_{3}a \, \mathbf{\hat{y}} + \frac{3}{4}c \, \mathbf{\hat{z}} & \left(6h\right) & \mbox{Al II} \\ 
\mathbf{B}_{10} & = & -x_{3} \, \mathbf{a}_{1} + x_{3} \, \mathbf{a}_{2} + \frac{3}{4} \, \mathbf{a}_{3} & = & \sqrt{3}x_{3}a \, \mathbf{\hat{y}} + \frac{3}{4}c \, \mathbf{\hat{z}} & \left(6h\right) & \mbox{Al II} \\ 
\mathbf{B}_{11} & = & x_{4} \, \mathbf{a}_{1} + 2x_{4} \, \mathbf{a}_{2} + \frac{1}{4} \, \mathbf{a}_{3} & = & \frac{3}{2}x_{4}a \, \mathbf{\hat{x}} + \frac{\sqrt{3}}{2}x_{4}a \, \mathbf{\hat{y}} + \frac{1}{4}c \, \mathbf{\hat{z}} & \left(6h\right) & \mbox{Co II} \\ 
\mathbf{B}_{12} & = & -2x_{4} \, \mathbf{a}_{1}-x_{4} \, \mathbf{a}_{2} + \frac{1}{4} \, \mathbf{a}_{3} & = & -\frac{3}{2}x_{4}a \, \mathbf{\hat{x}} + \frac{\sqrt{3}}{2}x_{4}a \, \mathbf{\hat{y}} + \frac{1}{4}c \, \mathbf{\hat{z}} & \left(6h\right) & \mbox{Co II} \\ 
\mathbf{B}_{13} & = & x_{4} \, \mathbf{a}_{1}-x_{4} \, \mathbf{a}_{2} + \frac{1}{4} \, \mathbf{a}_{3} & = & -\sqrt{3}x_{4}a \, \mathbf{\hat{y}} + \frac{1}{4}c \, \mathbf{\hat{z}} & \left(6h\right) & \mbox{Co II} \\ 
\mathbf{B}_{14} & = & -x_{4} \, \mathbf{a}_{1}-2x_{4} \, \mathbf{a}_{2} + \frac{3}{4} \, \mathbf{a}_{3} & = & -\frac{3}{2}x_{4}a \, \mathbf{\hat{x}}-\frac{\sqrt{3}}{2}x_{4}a \, \mathbf{\hat{y}} + \frac{3}{4}c \, \mathbf{\hat{z}} & \left(6h\right) & \mbox{Co II} \\ 
\mathbf{B}_{15} & = & 2x_{4} \, \mathbf{a}_{1} + x_{4} \, \mathbf{a}_{2} + \frac{3}{4} \, \mathbf{a}_{3} & = & \frac{3}{2}x_{4}a \, \mathbf{\hat{x}}-\frac{\sqrt{3}}{2}x_{4}a \, \mathbf{\hat{y}} + \frac{3}{4}c \, \mathbf{\hat{z}} & \left(6h\right) & \mbox{Co II} \\ 
\mathbf{B}_{16} & = & -x_{4} \, \mathbf{a}_{1} + x_{4} \, \mathbf{a}_{2} + \frac{3}{4} \, \mathbf{a}_{3} & = & \sqrt{3}x_{4}a \, \mathbf{\hat{y}} + \frac{3}{4}c \, \mathbf{\hat{z}} & \left(6h\right) & \mbox{Co II} \\ 
\mathbf{B}_{17} & = & x_{5} \, \mathbf{a}_{1} + 2x_{5} \, \mathbf{a}_{2} + z_{5} \, \mathbf{a}_{3} & = & \frac{3}{2}x_{5}a \, \mathbf{\hat{x}} + \frac{\sqrt{3}}{2}x_{5}a \, \mathbf{\hat{y}} + z_{5}c \, \mathbf{\hat{z}} & \left(12k\right) & \mbox{Al III} \\ 
\mathbf{B}_{18} & = & -2x_{5} \, \mathbf{a}_{1}-x_{5} \, \mathbf{a}_{2} + z_{5} \, \mathbf{a}_{3} & = & -\frac{3}{2}x_{5}a \, \mathbf{\hat{x}} + \frac{\sqrt{3}}{2}x_{5}a \, \mathbf{\hat{y}} + z_{5}c \, \mathbf{\hat{z}} & \left(12k\right) & \mbox{Al III} \\ 
\mathbf{B}_{19} & = & x_{5} \, \mathbf{a}_{1}-x_{5} \, \mathbf{a}_{2} + z_{5} \, \mathbf{a}_{3} & = & -\sqrt{3}x_{5}a \, \mathbf{\hat{y}} + z_{5}c \, \mathbf{\hat{z}} & \left(12k\right) & \mbox{Al III} \\ 
\mathbf{B}_{20} & = & -x_{5} \, \mathbf{a}_{1}-2x_{5} \, \mathbf{a}_{2} + \left(\frac{1}{2} +z_{5}\right) \, \mathbf{a}_{3} & = & -\frac{3}{2}x_{5}a \, \mathbf{\hat{x}}-\frac{\sqrt{3}}{2}x_{5}a \, \mathbf{\hat{y}} + \left(\frac{1}{2} +z_{5}\right)c \, \mathbf{\hat{z}} & \left(12k\right) & \mbox{Al III} \\ 
\mathbf{B}_{21} & = & 2x_{5} \, \mathbf{a}_{1} + x_{5} \, \mathbf{a}_{2} + \left(\frac{1}{2} +z_{5}\right) \, \mathbf{a}_{3} & = & \frac{3}{2}x_{5}a \, \mathbf{\hat{x}}-\frac{\sqrt{3}}{2}x_{5}a \, \mathbf{\hat{y}} + \left(\frac{1}{2} +z_{5}\right)c \, \mathbf{\hat{z}} & \left(12k\right) & \mbox{Al III} \\ 
\mathbf{B}_{22} & = & -x_{5} \, \mathbf{a}_{1} + x_{5} \, \mathbf{a}_{2} + \left(\frac{1}{2} +z_{5}\right) \, \mathbf{a}_{3} & = & \sqrt{3}x_{5}a \, \mathbf{\hat{y}} + \left(\frac{1}{2} +z_{5}\right)c \, \mathbf{\hat{z}} & \left(12k\right) & \mbox{Al III} \\ 
\mathbf{B}_{23} & = & 2x_{5} \, \mathbf{a}_{1} + x_{5} \, \mathbf{a}_{2}-z_{5} \, \mathbf{a}_{3} & = & \frac{3}{2}x_{5}a \, \mathbf{\hat{x}}-\frac{\sqrt{3}}{2}x_{5}a \, \mathbf{\hat{y}}-z_{5}c \, \mathbf{\hat{z}} & \left(12k\right) & \mbox{Al III} \\ 
\mathbf{B}_{24} & = & -x_{5} \, \mathbf{a}_{1}-2x_{5} \, \mathbf{a}_{2}-z_{5} \, \mathbf{a}_{3} & = & -\frac{3}{2}x_{5}a \, \mathbf{\hat{x}}-\frac{\sqrt{3}}{2}x_{5}a \, \mathbf{\hat{y}}-z_{5}c \, \mathbf{\hat{z}} & \left(12k\right) & \mbox{Al III} \\ 
\mathbf{B}_{25} & = & -x_{5} \, \mathbf{a}_{1} + x_{5} \, \mathbf{a}_{2}-z_{5} \, \mathbf{a}_{3} & = & \sqrt{3}x_{5}a \, \mathbf{\hat{y}}-z_{5}c \, \mathbf{\hat{z}} & \left(12k\right) & \mbox{Al III} \\ 
\mathbf{B}_{26} & = & -2x_{5} \, \mathbf{a}_{1}-x_{5} \, \mathbf{a}_{2} + \left(\frac{1}{2} - z_{5}\right) \, \mathbf{a}_{3} & = & -\frac{3}{2}x_{5}a \, \mathbf{\hat{x}} + \frac{\sqrt{3}}{2}x_{5}a \, \mathbf{\hat{y}} + \left(\frac{1}{2} - z_{5}\right)c \, \mathbf{\hat{z}} & \left(12k\right) & \mbox{Al III} \\ 
\mathbf{B}_{27} & = & x_{5} \, \mathbf{a}_{1} + 2x_{5} \, \mathbf{a}_{2} + \left(\frac{1}{2} - z_{5}\right) \, \mathbf{a}_{3} & = & \frac{3}{2}x_{5}a \, \mathbf{\hat{x}} + \frac{\sqrt{3}}{2}x_{5}a \, \mathbf{\hat{y}} + \left(\frac{1}{2} - z_{5}\right)c \, \mathbf{\hat{z}} & \left(12k\right) & \mbox{Al III} \\ 
\mathbf{B}_{28} & = & x_{5} \, \mathbf{a}_{1}-x_{5} \, \mathbf{a}_{2} + \left(\frac{1}{2} - z_{5}\right) \, \mathbf{a}_{3} & = & -\sqrt{3}x_{5}a \, \mathbf{\hat{y}} + \left(\frac{1}{2} - z_{5}\right)c \, \mathbf{\hat{z}} & \left(12k\right) & \mbox{Al III} \\ 
\end{longtabu}
\renewcommand{\arraystretch}{1.0}
\noindent \hrulefill
\\
\textbf{References:}
\vspace*{-0.25cm}
\begin{flushleft}
  - \bibentry{Newkirk_Acta_Cryst_14_1961}. \\
\end{flushleft}
\textbf{Found in:}
\vspace*{-0.25cm}
\begin{flushleft}
  - \bibentry{Westin_Acta_Chem_Scand_22_1968}. \\
\end{flushleft}
\noindent \hrulefill
\\
\textbf{Geometry files:}
\\
\noindent  - CIF: pp. {\hyperref[A5B2_hP28_194_ahk_ch_cif]{\pageref{A5B2_hP28_194_ahk_ch_cif}}} \\
\noindent  - POSCAR: pp. {\hyperref[A5B2_hP28_194_ahk_ch_poscar]{\pageref{A5B2_hP28_194_ahk_ch_poscar}}} \\
\onecolumn
{\phantomsection\label{A9B3C_hP26_194_hk_h_a}}
\subsection*{\huge \textbf{{\normalfont Al$_{9}$Mn$_{3}$Si ($E9_{c}$) Structure: A9B3C\_hP26\_194\_hk\_h\_a}}}
\noindent \hrulefill
\vspace*{0.25cm}
\begin{figure}[htp]
  \centering
  \vspace{-1em}
  {\includegraphics[width=1\textwidth]{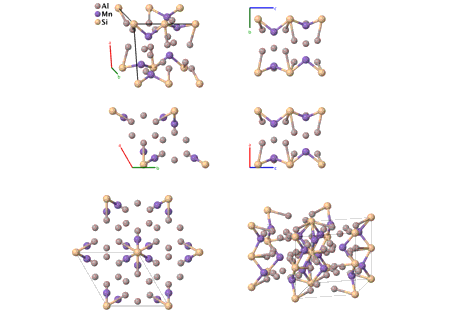}}
\end{figure}
\vspace*{-0.5cm}
\renewcommand{\arraystretch}{1.5}
\begin{equation*}
  \begin{array}{>{$\hspace{-0.15cm}}l<{$}>{$}p{0.5cm}<{$}>{$}p{18.5cm}<{$}}
    \mbox{\large \textbf{Prototype}} &\colon & \ce{Al9Mn3Si} \\
    \mbox{\large \textbf{\AFLOW\ prototype label}} &\colon & \mbox{A9B3C\_hP26\_194\_hk\_h\_a} \\
    \mbox{\large \textbf{\textit{Strukturbericht} designation}} &\colon & \mbox{$E9_{c}$} \\
    \mbox{\large \textbf{Pearson symbol}} &\colon & \mbox{hP26} \\
    \mbox{\large \textbf{Space group number}} &\colon & 194 \\
    \mbox{\large \textbf{Space group symbol}} &\colon & P6_{3}/mmc \\
    \mbox{\large \textbf{\AFLOW\ prototype command}} &\colon &  \texttt{aflow} \,  \, \texttt{-{}-proto=A9B3C\_hP26\_194\_hk\_h\_a } \, \newline \texttt{-{}-params=}{a,c/a,x_{2},x_{3},x_{4},z_{4} }
  \end{array}
\end{equation*}
\renewcommand{\arraystretch}{1.0}

\vspace*{-0.25cm}
\noindent \hrulefill
\begin{itemize}
  \item{(Pearson, 1958) quotes (Pratt, 1951) for this structure, 
but this paper does not contain useful structural information.
Pearson also cites (Robinson, 1952) for this structure, but that reference actually discusses
Ni$_{4}$Mn$_{11}$Al$_{60}$.  We use the lattice constants given by
Pearson, who states that the structure is stabilized by vacancies.
Pearson calls this structure $\beta$-Al-Mn-Si.  The atomic postitions
are taken from (Brandes, 1992), who give no reference.
}
\end{itemize}

\noindent \parbox{1 \linewidth}{
\noindent \hrulefill
\\
\textbf{Hexagonal primitive vectors:} \\
\vspace*{-0.25cm}
\begin{tabular}{cc}
  \begin{tabular}{c}
    \parbox{0.6 \linewidth}{
      \renewcommand{\arraystretch}{1.5}
      \begin{equation*}
        \centering
        \begin{array}{ccc}
              \mathbf{a}_1 & = & \frac12 \, a \, \mathbf{\hat{x}} - \frac{\sqrt3}2 \, a \, \mathbf{\hat{y}} \\
    \mathbf{a}_2 & = & \frac12 \, a \, \mathbf{\hat{x}} + \frac{\sqrt3}2 \, a \, \mathbf{\hat{y}} \\
    \mathbf{a}_3 & = & c \, \mathbf{\hat{z}} \\

        \end{array}
      \end{equation*}
    }
    \renewcommand{\arraystretch}{1.0}
  \end{tabular}
  \begin{tabular}{c}
    \includegraphics[width=0.3\linewidth]{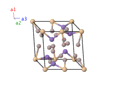} \\
  \end{tabular}
\end{tabular}

}
\vspace*{-0.25cm}

\noindent \hrulefill
\\
\textbf{Basis vectors:}
\vspace*{-0.25cm}
\renewcommand{\arraystretch}{1.5}
\begin{longtabu} to \textwidth{>{\centering $}X[-1,c,c]<{$}>{\centering $}X[-1,c,c]<{$}>{\centering $}X[-1,c,c]<{$}>{\centering $}X[-1,c,c]<{$}>{\centering $}X[-1,c,c]<{$}>{\centering $}X[-1,c,c]<{$}>{\centering $}X[-1,c,c]<{$}}
  & & \mbox{Lattice Coordinates} & & \mbox{Cartesian Coordinates} &\mbox{Wyckoff Position} & \mbox{Atom Type} \\  
  \mathbf{B}_{1} & = & 0 \, \mathbf{a}_{1} + 0 \, \mathbf{a}_{2} + 0 \, \mathbf{a}_{3} & = & 0 \, \mathbf{\hat{x}} + 0 \, \mathbf{\hat{y}} + 0 \, \mathbf{\hat{z}} & \left(2a\right) & \mbox{Si} \\ 
\mathbf{B}_{2} & = & \frac{1}{2} \, \mathbf{a}_{3} & = & \frac{1}{2}c \, \mathbf{\hat{z}} & \left(2a\right) & \mbox{Si} \\ 
\mathbf{B}_{3} & = & x_{2} \, \mathbf{a}_{1} + 2x_{2} \, \mathbf{a}_{2} + \frac{1}{4} \, \mathbf{a}_{3} & = & \frac{3}{2}x_{2}a \, \mathbf{\hat{x}} + \frac{\sqrt{3}}{2}x_{2}a \, \mathbf{\hat{y}} + \frac{1}{4}c \, \mathbf{\hat{z}} & \left(6h\right) & \mbox{Al I} \\ 
\mathbf{B}_{4} & = & -2x_{2} \, \mathbf{a}_{1}-x_{2} \, \mathbf{a}_{2} + \frac{1}{4} \, \mathbf{a}_{3} & = & -\frac{3}{2}x_{2}a \, \mathbf{\hat{x}} + \frac{\sqrt{3}}{2}x_{2}a \, \mathbf{\hat{y}} + \frac{1}{4}c \, \mathbf{\hat{z}} & \left(6h\right) & \mbox{Al I} \\ 
\mathbf{B}_{5} & = & x_{2} \, \mathbf{a}_{1}-x_{2} \, \mathbf{a}_{2} + \frac{1}{4} \, \mathbf{a}_{3} & = & -\sqrt{3}x_{2}a \, \mathbf{\hat{y}} + \frac{1}{4}c \, \mathbf{\hat{z}} & \left(6h\right) & \mbox{Al I} \\ 
\mathbf{B}_{6} & = & -x_{2} \, \mathbf{a}_{1}-2x_{2} \, \mathbf{a}_{2} + \frac{3}{4} \, \mathbf{a}_{3} & = & -\frac{3}{2}x_{2}a \, \mathbf{\hat{x}}-\frac{\sqrt{3}}{2}x_{2}a \, \mathbf{\hat{y}} + \frac{3}{4}c \, \mathbf{\hat{z}} & \left(6h\right) & \mbox{Al I} \\ 
\mathbf{B}_{7} & = & 2x_{2} \, \mathbf{a}_{1} + x_{2} \, \mathbf{a}_{2} + \frac{3}{4} \, \mathbf{a}_{3} & = & \frac{3}{2}x_{2}a \, \mathbf{\hat{x}}-\frac{\sqrt{3}}{2}x_{2}a \, \mathbf{\hat{y}} + \frac{3}{4}c \, \mathbf{\hat{z}} & \left(6h\right) & \mbox{Al I} \\ 
\mathbf{B}_{8} & = & -x_{2} \, \mathbf{a}_{1} + x_{2} \, \mathbf{a}_{2} + \frac{3}{4} \, \mathbf{a}_{3} & = & \sqrt{3}x_{2}a \, \mathbf{\hat{y}} + \frac{3}{4}c \, \mathbf{\hat{z}} & \left(6h\right) & \mbox{Al I} \\ 
\mathbf{B}_{9} & = & x_{3} \, \mathbf{a}_{1} + 2x_{3} \, \mathbf{a}_{2} + \frac{1}{4} \, \mathbf{a}_{3} & = & \frac{3}{2}x_{3}a \, \mathbf{\hat{x}} + \frac{\sqrt{3}}{2}x_{3}a \, \mathbf{\hat{y}} + \frac{1}{4}c \, \mathbf{\hat{z}} & \left(6h\right) & \mbox{Mn} \\ 
\mathbf{B}_{10} & = & -2x_{3} \, \mathbf{a}_{1}-x_{3} \, \mathbf{a}_{2} + \frac{1}{4} \, \mathbf{a}_{3} & = & -\frac{3}{2}x_{3}a \, \mathbf{\hat{x}} + \frac{\sqrt{3}}{2}x_{3}a \, \mathbf{\hat{y}} + \frac{1}{4}c \, \mathbf{\hat{z}} & \left(6h\right) & \mbox{Mn} \\ 
\mathbf{B}_{11} & = & x_{3} \, \mathbf{a}_{1}-x_{3} \, \mathbf{a}_{2} + \frac{1}{4} \, \mathbf{a}_{3} & = & -\sqrt{3}x_{3}a \, \mathbf{\hat{y}} + \frac{1}{4}c \, \mathbf{\hat{z}} & \left(6h\right) & \mbox{Mn} \\ 
\mathbf{B}_{12} & = & -x_{3} \, \mathbf{a}_{1}-2x_{3} \, \mathbf{a}_{2} + \frac{3}{4} \, \mathbf{a}_{3} & = & -\frac{3}{2}x_{3}a \, \mathbf{\hat{x}}-\frac{\sqrt{3}}{2}x_{3}a \, \mathbf{\hat{y}} + \frac{3}{4}c \, \mathbf{\hat{z}} & \left(6h\right) & \mbox{Mn} \\ 
\mathbf{B}_{13} & = & 2x_{3} \, \mathbf{a}_{1} + x_{3} \, \mathbf{a}_{2} + \frac{3}{4} \, \mathbf{a}_{3} & = & \frac{3}{2}x_{3}a \, \mathbf{\hat{x}}-\frac{\sqrt{3}}{2}x_{3}a \, \mathbf{\hat{y}} + \frac{3}{4}c \, \mathbf{\hat{z}} & \left(6h\right) & \mbox{Mn} \\ 
\mathbf{B}_{14} & = & -x_{3} \, \mathbf{a}_{1} + x_{3} \, \mathbf{a}_{2} + \frac{3}{4} \, \mathbf{a}_{3} & = & \sqrt{3}x_{3}a \, \mathbf{\hat{y}} + \frac{3}{4}c \, \mathbf{\hat{z}} & \left(6h\right) & \mbox{Mn} \\ 
\mathbf{B}_{15} & = & x_{4} \, \mathbf{a}_{1} + 2x_{4} \, \mathbf{a}_{2} + z_{4} \, \mathbf{a}_{3} & = & \frac{3}{2}x_{4}a \, \mathbf{\hat{x}} + \frac{\sqrt{3}}{2}x_{4}a \, \mathbf{\hat{y}} + z_{4}c \, \mathbf{\hat{z}} & \left(12k\right) & \mbox{Al II} \\ 
\mathbf{B}_{16} & = & -2x_{4} \, \mathbf{a}_{1}-x_{4} \, \mathbf{a}_{2} + z_{4} \, \mathbf{a}_{3} & = & -\frac{3}{2}x_{4}a \, \mathbf{\hat{x}} + \frac{\sqrt{3}}{2}x_{4}a \, \mathbf{\hat{y}} + z_{4}c \, \mathbf{\hat{z}} & \left(12k\right) & \mbox{Al II} \\ 
\mathbf{B}_{17} & = & x_{4} \, \mathbf{a}_{1}-x_{4} \, \mathbf{a}_{2} + z_{4} \, \mathbf{a}_{3} & = & -\sqrt{3}x_{4}a \, \mathbf{\hat{y}} + z_{4}c \, \mathbf{\hat{z}} & \left(12k\right) & \mbox{Al II} \\ 
\mathbf{B}_{18} & = & -x_{4} \, \mathbf{a}_{1}-2x_{4} \, \mathbf{a}_{2} + \left(\frac{1}{2} +z_{4}\right) \, \mathbf{a}_{3} & = & -\frac{3}{2}x_{4}a \, \mathbf{\hat{x}}-\frac{\sqrt{3}}{2}x_{4}a \, \mathbf{\hat{y}} + \left(\frac{1}{2} +z_{4}\right)c \, \mathbf{\hat{z}} & \left(12k\right) & \mbox{Al II} \\ 
\mathbf{B}_{19} & = & 2x_{4} \, \mathbf{a}_{1} + x_{4} \, \mathbf{a}_{2} + \left(\frac{1}{2} +z_{4}\right) \, \mathbf{a}_{3} & = & \frac{3}{2}x_{4}a \, \mathbf{\hat{x}}-\frac{\sqrt{3}}{2}x_{4}a \, \mathbf{\hat{y}} + \left(\frac{1}{2} +z_{4}\right)c \, \mathbf{\hat{z}} & \left(12k\right) & \mbox{Al II} \\ 
\mathbf{B}_{20} & = & -x_{4} \, \mathbf{a}_{1} + x_{4} \, \mathbf{a}_{2} + \left(\frac{1}{2} +z_{4}\right) \, \mathbf{a}_{3} & = & \sqrt{3}x_{4}a \, \mathbf{\hat{y}} + \left(\frac{1}{2} +z_{4}\right)c \, \mathbf{\hat{z}} & \left(12k\right) & \mbox{Al II} \\ 
\mathbf{B}_{21} & = & 2x_{4} \, \mathbf{a}_{1} + x_{4} \, \mathbf{a}_{2}-z_{4} \, \mathbf{a}_{3} & = & \frac{3}{2}x_{4}a \, \mathbf{\hat{x}}-\frac{\sqrt{3}}{2}x_{4}a \, \mathbf{\hat{y}}-z_{4}c \, \mathbf{\hat{z}} & \left(12k\right) & \mbox{Al II} \\ 
\mathbf{B}_{22} & = & -x_{4} \, \mathbf{a}_{1}-2x_{4} \, \mathbf{a}_{2}-z_{4} \, \mathbf{a}_{3} & = & -\frac{3}{2}x_{4}a \, \mathbf{\hat{x}}-\frac{\sqrt{3}}{2}x_{4}a \, \mathbf{\hat{y}}-z_{4}c \, \mathbf{\hat{z}} & \left(12k\right) & \mbox{Al II} \\ 
\mathbf{B}_{23} & = & -x_{4} \, \mathbf{a}_{1} + x_{4} \, \mathbf{a}_{2}-z_{4} \, \mathbf{a}_{3} & = & \sqrt{3}x_{4}a \, \mathbf{\hat{y}}-z_{4}c \, \mathbf{\hat{z}} & \left(12k\right) & \mbox{Al II} \\ 
\mathbf{B}_{24} & = & -2x_{4} \, \mathbf{a}_{1}-x_{4} \, \mathbf{a}_{2} + \left(\frac{1}{2} - z_{4}\right) \, \mathbf{a}_{3} & = & -\frac{3}{2}x_{4}a \, \mathbf{\hat{x}} + \frac{\sqrt{3}}{2}x_{4}a \, \mathbf{\hat{y}} + \left(\frac{1}{2} - z_{4}\right)c \, \mathbf{\hat{z}} & \left(12k\right) & \mbox{Al II} \\ 
\mathbf{B}_{25} & = & x_{4} \, \mathbf{a}_{1} + 2x_{4} \, \mathbf{a}_{2} + \left(\frac{1}{2} - z_{4}\right) \, \mathbf{a}_{3} & = & \frac{3}{2}x_{4}a \, \mathbf{\hat{x}} + \frac{\sqrt{3}}{2}x_{4}a \, \mathbf{\hat{y}} + \left(\frac{1}{2} - z_{4}\right)c \, \mathbf{\hat{z}} & \left(12k\right) & \mbox{Al II} \\ 
\mathbf{B}_{26} & = & x_{4} \, \mathbf{a}_{1}-x_{4} \, \mathbf{a}_{2} + \left(\frac{1}{2} - z_{4}\right) \, \mathbf{a}_{3} & = & -\sqrt{3}x_{4}a \, \mathbf{\hat{y}} + \left(\frac{1}{2} - z_{4}\right)c \, \mathbf{\hat{z}} & \left(12k\right) & \mbox{Al II} \\ 
\end{longtabu}
\renewcommand{\arraystretch}{1.0}
\noindent \hrulefill
\\
\textbf{References:}
\vspace*{-0.25cm}
\begin{flushleft}
  - \bibentry{Robinson_PhilMag_43_775_1952}. \\
  - \bibentry{Pratt_JIM_79_211_1951}. \\
\end{flushleft}
\textbf{Found in:}
\vspace*{-0.25cm}
\begin{flushleft}
  - \bibentry{Pearson_NRC_1958}. \\
  - \bibentry{Brandes_Smithells_1992}. \\
\end{flushleft}
\noindent \hrulefill
\\
\textbf{Geometry files:}
\\
\noindent  - CIF: pp. {\hyperref[A9B3C_hP26_194_hk_h_a_cif]{\pageref{A9B3C_hP26_194_hk_h_a_cif}}} \\
\noindent  - POSCAR: pp. {\hyperref[A9B3C_hP26_194_hk_h_a_poscar]{\pageref{A9B3C_hP26_194_hk_h_a_poscar}}} \\
\onecolumn
{\phantomsection\label{A12BC4_cP34_195_2j_ab_2e}}
\subsection*{\huge \textbf{{\normalfont PrRu$_{4}$P$_{12}$ Structure: A12BC4\_cP34\_195\_2j\_ab\_2e}}}
\noindent \hrulefill
\vspace*{0.25cm}
\begin{figure}[htp]
  \centering
  \vspace{-1em}
  {\includegraphics[width=1\textwidth]{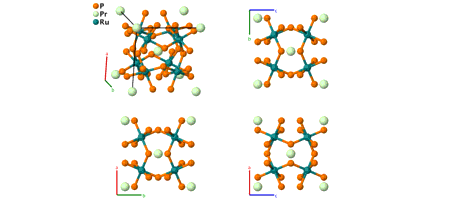}}
\end{figure}
\vspace*{-0.5cm}
\renewcommand{\arraystretch}{1.5}
\begin{equation*}
  \begin{array}{>{$\hspace{-0.15cm}}l<{$}>{$}p{0.5cm}<{$}>{$}p{18.5cm}<{$}}
    \mbox{\large \textbf{Prototype}} &\colon & \ce{PrRu4P12} \\
    \mbox{\large \textbf{\AFLOW\ prototype label}} &\colon & \mbox{A12BC4\_cP34\_195\_2j\_ab\_2e} \\
    \mbox{\large \textbf{\textit{Strukturbericht} designation}} &\colon & \mbox{None} \\
    \mbox{\large \textbf{Pearson symbol}} &\colon & \mbox{cP34} \\
    \mbox{\large \textbf{Space group number}} &\colon & 195 \\
    \mbox{\large \textbf{Space group symbol}} &\colon & P23 \\
    \mbox{\large \textbf{\AFLOW\ prototype command}} &\colon &  \texttt{aflow} \,  \, \texttt{-{}-proto=A12BC4\_cP34\_195\_2j\_ab\_2e } \, \newline \texttt{-{}-params=}{a,x_{3},x_{4},x_{5},y_{5},z_{5},x_{6},y_{6},z_{6} }
  \end{array}
\end{equation*}
\renewcommand{\arraystretch}{1.0}

\noindent \parbox{1 \linewidth}{
\noindent \hrulefill
\\
\textbf{Simple Cubic primitive vectors:} \\
\vspace*{-0.25cm}
\begin{tabular}{cc}
  \begin{tabular}{c}
    \parbox{0.6 \linewidth}{
      \renewcommand{\arraystretch}{1.5}
      \begin{equation*}
        \centering
        \begin{array}{ccc}
              \mathbf{a}_1 & = & a \, \mathbf{\hat{x}} \\
    \mathbf{a}_2 & = & a \, \mathbf{\hat{y}} \\
    \mathbf{a}_3 & = & a \, \mathbf{\hat{z}} \\

        \end{array}
      \end{equation*}
    }
    \renewcommand{\arraystretch}{1.0}
  \end{tabular}
  \begin{tabular}{c}
    \includegraphics[width=0.3\linewidth]{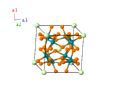} \\
  \end{tabular}
\end{tabular}

}
\vspace*{-0.25cm}

\noindent \hrulefill
\\
\textbf{Basis vectors:}
\vspace*{-0.25cm}
\renewcommand{\arraystretch}{1.5}
\begin{longtabu} to \textwidth{>{\centering $}X[-1,c,c]<{$}>{\centering $}X[-1,c,c]<{$}>{\centering $}X[-1,c,c]<{$}>{\centering $}X[-1,c,c]<{$}>{\centering $}X[-1,c,c]<{$}>{\centering $}X[-1,c,c]<{$}>{\centering $}X[-1,c,c]<{$}}
  & & \mbox{Lattice Coordinates} & & \mbox{Cartesian Coordinates} &\mbox{Wyckoff Position} & \mbox{Atom Type} \\  
  \mathbf{B}_{1} & = & 0 \, \mathbf{a}_{1} + 0 \, \mathbf{a}_{2} + 0 \, \mathbf{a}_{3} & = & 0 \, \mathbf{\hat{x}} + 0 \, \mathbf{\hat{y}} + 0 \, \mathbf{\hat{z}} & \left(1a\right) & \mbox{Pr I} \\ 
\mathbf{B}_{2} & = & \frac{1}{2} \, \mathbf{a}_{1} + \frac{1}{2} \, \mathbf{a}_{2} + \frac{1}{2} \, \mathbf{a}_{3} & = & \frac{1}{2}a \, \mathbf{\hat{x}} + \frac{1}{2}a \, \mathbf{\hat{y}} + \frac{1}{2}a \, \mathbf{\hat{z}} & \left(1b\right) & \mbox{Pr II} \\ 
\mathbf{B}_{3} & = & x_{3} \, \mathbf{a}_{1} + x_{3} \, \mathbf{a}_{2} + x_{3} \, \mathbf{a}_{3} & = & x_{3}a \, \mathbf{\hat{x}} + x_{3}a \, \mathbf{\hat{y}} + x_{3}a \, \mathbf{\hat{z}} & \left(4e\right) & \mbox{Ru I} \\ 
\mathbf{B}_{4} & = & -x_{3} \, \mathbf{a}_{1}-x_{3} \, \mathbf{a}_{2} + x_{3} \, \mathbf{a}_{3} & = & -x_{3}a \, \mathbf{\hat{x}}-x_{3}a \, \mathbf{\hat{y}} + x_{3}a \, \mathbf{\hat{z}} & \left(4e\right) & \mbox{Ru I} \\ 
\mathbf{B}_{5} & = & -x_{3} \, \mathbf{a}_{1} + x_{3} \, \mathbf{a}_{2}-x_{3} \, \mathbf{a}_{3} & = & -x_{3}a \, \mathbf{\hat{x}} + x_{3}a \, \mathbf{\hat{y}}-x_{3}a \, \mathbf{\hat{z}} & \left(4e\right) & \mbox{Ru I} \\ 
\mathbf{B}_{6} & = & x_{3} \, \mathbf{a}_{1}-x_{3} \, \mathbf{a}_{2}-x_{3} \, \mathbf{a}_{3} & = & x_{3}a \, \mathbf{\hat{x}}-x_{3}a \, \mathbf{\hat{y}}-x_{3}a \, \mathbf{\hat{z}} & \left(4e\right) & \mbox{Ru I} \\ 
\mathbf{B}_{7} & = & x_{4} \, \mathbf{a}_{1} + x_{4} \, \mathbf{a}_{2} + x_{4} \, \mathbf{a}_{3} & = & x_{4}a \, \mathbf{\hat{x}} + x_{4}a \, \mathbf{\hat{y}} + x_{4}a \, \mathbf{\hat{z}} & \left(4e\right) & \mbox{Ru II} \\ 
\mathbf{B}_{8} & = & -x_{4} \, \mathbf{a}_{1}-x_{4} \, \mathbf{a}_{2} + x_{4} \, \mathbf{a}_{3} & = & -x_{4}a \, \mathbf{\hat{x}}-x_{4}a \, \mathbf{\hat{y}} + x_{4}a \, \mathbf{\hat{z}} & \left(4e\right) & \mbox{Ru II} \\ 
\mathbf{B}_{9} & = & -x_{4} \, \mathbf{a}_{1} + x_{4} \, \mathbf{a}_{2}-x_{4} \, \mathbf{a}_{3} & = & -x_{4}a \, \mathbf{\hat{x}} + x_{4}a \, \mathbf{\hat{y}}-x_{4}a \, \mathbf{\hat{z}} & \left(4e\right) & \mbox{Ru II} \\ 
\mathbf{B}_{10} & = & x_{4} \, \mathbf{a}_{1}-x_{4} \, \mathbf{a}_{2}-x_{4} \, \mathbf{a}_{3} & = & x_{4}a \, \mathbf{\hat{x}}-x_{4}a \, \mathbf{\hat{y}}-x_{4}a \, \mathbf{\hat{z}} & \left(4e\right) & \mbox{Ru II} \\ 
\mathbf{B}_{11} & = & x_{5} \, \mathbf{a}_{1} + y_{5} \, \mathbf{a}_{2} + z_{5} \, \mathbf{a}_{3} & = & x_{5}a \, \mathbf{\hat{x}} + y_{5}a \, \mathbf{\hat{y}} + z_{5}a \, \mathbf{\hat{z}} & \left(12j\right) & \mbox{P I} \\ 
\mathbf{B}_{12} & = & -x_{5} \, \mathbf{a}_{1}-y_{5} \, \mathbf{a}_{2} + z_{5} \, \mathbf{a}_{3} & = & -x_{5}a \, \mathbf{\hat{x}}-y_{5}a \, \mathbf{\hat{y}} + z_{5}a \, \mathbf{\hat{z}} & \left(12j\right) & \mbox{P I} \\ 
\mathbf{B}_{13} & = & -x_{5} \, \mathbf{a}_{1} + y_{5} \, \mathbf{a}_{2}-z_{5} \, \mathbf{a}_{3} & = & -x_{5}a \, \mathbf{\hat{x}} + y_{5}a \, \mathbf{\hat{y}}-z_{5}a \, \mathbf{\hat{z}} & \left(12j\right) & \mbox{P I} \\ 
\mathbf{B}_{14} & = & x_{5} \, \mathbf{a}_{1}-y_{5} \, \mathbf{a}_{2}-z_{5} \, \mathbf{a}_{3} & = & x_{5}a \, \mathbf{\hat{x}}-y_{5}a \, \mathbf{\hat{y}}-z_{5}a \, \mathbf{\hat{z}} & \left(12j\right) & \mbox{P I} \\ 
\mathbf{B}_{15} & = & z_{5} \, \mathbf{a}_{1} + x_{5} \, \mathbf{a}_{2} + y_{5} \, \mathbf{a}_{3} & = & z_{5}a \, \mathbf{\hat{x}} + x_{5}a \, \mathbf{\hat{y}} + y_{5}a \, \mathbf{\hat{z}} & \left(12j\right) & \mbox{P I} \\ 
\mathbf{B}_{16} & = & z_{5} \, \mathbf{a}_{1}-x_{5} \, \mathbf{a}_{2}-y_{5} \, \mathbf{a}_{3} & = & z_{5}a \, \mathbf{\hat{x}}-x_{5}a \, \mathbf{\hat{y}}-y_{5}a \, \mathbf{\hat{z}} & \left(12j\right) & \mbox{P I} \\ 
\mathbf{B}_{17} & = & -z_{5} \, \mathbf{a}_{1}-x_{5} \, \mathbf{a}_{2} + y_{5} \, \mathbf{a}_{3} & = & -z_{5}a \, \mathbf{\hat{x}}-x_{5}a \, \mathbf{\hat{y}} + y_{5}a \, \mathbf{\hat{z}} & \left(12j\right) & \mbox{P I} \\ 
\mathbf{B}_{18} & = & -z_{5} \, \mathbf{a}_{1} + x_{5} \, \mathbf{a}_{2}-y_{5} \, \mathbf{a}_{3} & = & -z_{5}a \, \mathbf{\hat{x}} + x_{5}a \, \mathbf{\hat{y}}-y_{5}a \, \mathbf{\hat{z}} & \left(12j\right) & \mbox{P I} \\ 
\mathbf{B}_{19} & = & y_{5} \, \mathbf{a}_{1} + z_{5} \, \mathbf{a}_{2} + x_{5} \, \mathbf{a}_{3} & = & y_{5}a \, \mathbf{\hat{x}} + z_{5}a \, \mathbf{\hat{y}} + x_{5}a \, \mathbf{\hat{z}} & \left(12j\right) & \mbox{P I} \\ 
\mathbf{B}_{20} & = & -y_{5} \, \mathbf{a}_{1} + z_{5} \, \mathbf{a}_{2}-x_{5} \, \mathbf{a}_{3} & = & -y_{5}a \, \mathbf{\hat{x}} + z_{5}a \, \mathbf{\hat{y}}-x_{5}a \, \mathbf{\hat{z}} & \left(12j\right) & \mbox{P I} \\ 
\mathbf{B}_{21} & = & y_{5} \, \mathbf{a}_{1}-z_{5} \, \mathbf{a}_{2}-x_{5} \, \mathbf{a}_{3} & = & y_{5}a \, \mathbf{\hat{x}}-z_{5}a \, \mathbf{\hat{y}}-x_{5}a \, \mathbf{\hat{z}} & \left(12j\right) & \mbox{P I} \\ 
\mathbf{B}_{22} & = & -y_{5} \, \mathbf{a}_{1}-z_{5} \, \mathbf{a}_{2} + x_{5} \, \mathbf{a}_{3} & = & -y_{5}a \, \mathbf{\hat{x}}-z_{5}a \, \mathbf{\hat{y}} + x_{5}a \, \mathbf{\hat{z}} & \left(12j\right) & \mbox{P I} \\ 
\mathbf{B}_{23} & = & x_{6} \, \mathbf{a}_{1} + y_{6} \, \mathbf{a}_{2} + z_{6} \, \mathbf{a}_{3} & = & x_{6}a \, \mathbf{\hat{x}} + y_{6}a \, \mathbf{\hat{y}} + z_{6}a \, \mathbf{\hat{z}} & \left(12j\right) & \mbox{P II} \\ 
\mathbf{B}_{24} & = & -x_{6} \, \mathbf{a}_{1}-y_{6} \, \mathbf{a}_{2} + z_{6} \, \mathbf{a}_{3} & = & -x_{6}a \, \mathbf{\hat{x}}-y_{6}a \, \mathbf{\hat{y}} + z_{6}a \, \mathbf{\hat{z}} & \left(12j\right) & \mbox{P II} \\ 
\mathbf{B}_{25} & = & -x_{6} \, \mathbf{a}_{1} + y_{6} \, \mathbf{a}_{2}-z_{6} \, \mathbf{a}_{3} & = & -x_{6}a \, \mathbf{\hat{x}} + y_{6}a \, \mathbf{\hat{y}}-z_{6}a \, \mathbf{\hat{z}} & \left(12j\right) & \mbox{P II} \\ 
\mathbf{B}_{26} & = & x_{6} \, \mathbf{a}_{1}-y_{6} \, \mathbf{a}_{2}-z_{6} \, \mathbf{a}_{3} & = & x_{6}a \, \mathbf{\hat{x}}-y_{6}a \, \mathbf{\hat{y}}-z_{6}a \, \mathbf{\hat{z}} & \left(12j\right) & \mbox{P II} \\ 
\mathbf{B}_{27} & = & z_{6} \, \mathbf{a}_{1} + x_{6} \, \mathbf{a}_{2} + y_{6} \, \mathbf{a}_{3} & = & z_{6}a \, \mathbf{\hat{x}} + x_{6}a \, \mathbf{\hat{y}} + y_{6}a \, \mathbf{\hat{z}} & \left(12j\right) & \mbox{P II} \\ 
\mathbf{B}_{28} & = & z_{6} \, \mathbf{a}_{1}-x_{6} \, \mathbf{a}_{2}-y_{6} \, \mathbf{a}_{3} & = & z_{6}a \, \mathbf{\hat{x}}-x_{6}a \, \mathbf{\hat{y}}-y_{6}a \, \mathbf{\hat{z}} & \left(12j\right) & \mbox{P II} \\ 
\mathbf{B}_{29} & = & -z_{6} \, \mathbf{a}_{1}-x_{6} \, \mathbf{a}_{2} + y_{6} \, \mathbf{a}_{3} & = & -z_{6}a \, \mathbf{\hat{x}}-x_{6}a \, \mathbf{\hat{y}} + y_{6}a \, \mathbf{\hat{z}} & \left(12j\right) & \mbox{P II} \\ 
\mathbf{B}_{30} & = & -z_{6} \, \mathbf{a}_{1} + x_{6} \, \mathbf{a}_{2}-y_{6} \, \mathbf{a}_{3} & = & -z_{6}a \, \mathbf{\hat{x}} + x_{6}a \, \mathbf{\hat{y}}-y_{6}a \, \mathbf{\hat{z}} & \left(12j\right) & \mbox{P II} \\ 
\mathbf{B}_{31} & = & y_{6} \, \mathbf{a}_{1} + z_{6} \, \mathbf{a}_{2} + x_{6} \, \mathbf{a}_{3} & = & y_{6}a \, \mathbf{\hat{x}} + z_{6}a \, \mathbf{\hat{y}} + x_{6}a \, \mathbf{\hat{z}} & \left(12j\right) & \mbox{P II} \\ 
\mathbf{B}_{32} & = & -y_{6} \, \mathbf{a}_{1} + z_{6} \, \mathbf{a}_{2}-x_{6} \, \mathbf{a}_{3} & = & -y_{6}a \, \mathbf{\hat{x}} + z_{6}a \, \mathbf{\hat{y}}-x_{6}a \, \mathbf{\hat{z}} & \left(12j\right) & \mbox{P II} \\ 
\mathbf{B}_{33} & = & y_{6} \, \mathbf{a}_{1}-z_{6} \, \mathbf{a}_{2}-x_{6} \, \mathbf{a}_{3} & = & y_{6}a \, \mathbf{\hat{x}}-z_{6}a \, \mathbf{\hat{y}}-x_{6}a \, \mathbf{\hat{z}} & \left(12j\right) & \mbox{P II} \\ 
\mathbf{B}_{34} & = & -y_{6} \, \mathbf{a}_{1}-z_{6} \, \mathbf{a}_{2} + x_{6} \, \mathbf{a}_{3} & = & -y_{6}a \, \mathbf{\hat{x}}-z_{6}a \, \mathbf{\hat{y}} + x_{6}a \, \mathbf{\hat{z}} & \left(12j\right) & \mbox{P II} \\ 
\end{longtabu}
\renewcommand{\arraystretch}{1.0}
\noindent \hrulefill
\\
\textbf{References:}
\vspace*{-0.25cm}
\begin{flushleft}
  - \bibentry{Lee_PrRu4P12_JMagsmMagMat_2004}. \\
  - \bibentry{stokes_findsym}. \\
  - \bibentry{aflowsym_2018}. \\
  - \bibentry{platon_2003}. \\
\end{flushleft}
\textbf{Found in:}
\vspace*{-0.25cm}
\begin{flushleft}
  - \bibentry{Villars_PearsonsCrystalData_2013}. \\
\end{flushleft}
\noindent \hrulefill
\\
\textbf{Geometry files:}
\\
\noindent  - CIF: pp. {\hyperref[A12BC4_cP34_195_2j_ab_2e_cif]{\pageref{A12BC4_cP34_195_2j_ab_2e_cif}}} \\
\noindent  - POSCAR: pp. {\hyperref[A12BC4_cP34_195_2j_ab_2e_poscar]{\pageref{A12BC4_cP34_195_2j_ab_2e_poscar}}} \\
\onecolumn
{\phantomsection\label{A12B2C_cF60_196_h_bc_a}}
\subsection*{\huge \textbf{{\normalfont Cu$_{2}$Fe[CN]$_{6}$ Structure: A12B2C\_cF60\_196\_h\_bc\_a}}}
\noindent \hrulefill
\vspace*{0.25cm}
\begin{figure}[htp]
  \centering
  \vspace{-1em}
  {\includegraphics[width=1\textwidth]{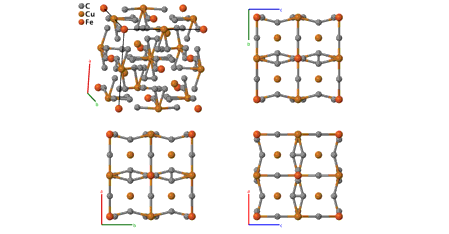}}
\end{figure}
\vspace*{-0.5cm}
\renewcommand{\arraystretch}{1.5}
\begin{equation*}
  \begin{array}{>{$\hspace{-0.15cm}}l<{$}>{$}p{0.5cm}<{$}>{$}p{18.5cm}<{$}}
    \mbox{\large \textbf{Prototype}} &\colon & \ce{Cu2Fe[CN]6} \\
    \mbox{\large \textbf{\AFLOW\ prototype label}} &\colon & \mbox{A12B2C\_cF60\_196\_h\_bc\_a} \\
    \mbox{\large \textbf{\textit{Strukturbericht} designation}} &\colon & \mbox{None} \\
    \mbox{\large \textbf{Pearson symbol}} &\colon & \mbox{cF60} \\
    \mbox{\large \textbf{Space group number}} &\colon & 196 \\
    \mbox{\large \textbf{Space group symbol}} &\colon & F23 \\
    \mbox{\large \textbf{\AFLOW\ prototype command}} &\colon &  \texttt{aflow} \,  \, \texttt{-{}-proto=A12B2C\_cF60\_196\_h\_bc\_a } \, \newline \texttt{-{}-params=}{a,x_{4},y_{4},z_{4} }
  \end{array}
\end{equation*}
\renewcommand{\arraystretch}{1.0}

\vspace*{-0.25cm}
\noindent \hrulefill
\begin{itemize}
  \item{The C sites are partially occupied with 0.5C + 0.5N.
}
\end{itemize}

\noindent \parbox{1 \linewidth}{
\noindent \hrulefill
\\
\textbf{Face-centered Cubic primitive vectors:} \\
\vspace*{-0.25cm}
\begin{tabular}{cc}
  \begin{tabular}{c}
    \parbox{0.6 \linewidth}{
      \renewcommand{\arraystretch}{1.5}
      \begin{equation*}
        \centering
        \begin{array}{ccc}
              \mathbf{a}_1 & = & \frac12 \, a \, \mathbf{\hat{y}} + \frac12 \, a \, \mathbf{\hat{z}} \\
    \mathbf{a}_2 & = & \frac12 \, a \, \mathbf{\hat{x}} + \frac12 \, a \, \mathbf{\hat{z}} \\
    \mathbf{a}_3 & = & \frac12 \, a \, \mathbf{\hat{x}} + \frac12 \, a \, \mathbf{\hat{y}} \\

        \end{array}
      \end{equation*}
    }
    \renewcommand{\arraystretch}{1.0}
  \end{tabular}
  \begin{tabular}{c}
    \includegraphics[width=0.3\linewidth]{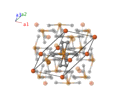} \\
  \end{tabular}
\end{tabular}

}
\vspace*{-0.25cm}

\noindent \hrulefill
\\
\textbf{Basis vectors:}
\vspace*{-0.25cm}
\renewcommand{\arraystretch}{1.5}
\begin{longtabu} to \textwidth{>{\centering $}X[-1,c,c]<{$}>{\centering $}X[-1,c,c]<{$}>{\centering $}X[-1,c,c]<{$}>{\centering $}X[-1,c,c]<{$}>{\centering $}X[-1,c,c]<{$}>{\centering $}X[-1,c,c]<{$}>{\centering $}X[-1,c,c]<{$}}
  & & \mbox{Lattice Coordinates} & & \mbox{Cartesian Coordinates} &\mbox{Wyckoff Position} & \mbox{Atom Type} \\  
  \mathbf{B}_{1} & = & 0 \, \mathbf{a}_{1} + 0 \, \mathbf{a}_{2} + 0 \, \mathbf{a}_{3} & = & 0 \, \mathbf{\hat{x}} + 0 \, \mathbf{\hat{y}} + 0 \, \mathbf{\hat{z}} & \left(4a\right) & \mbox{Fe} \\ 
\mathbf{B}_{2} & = & \frac{1}{2} \, \mathbf{a}_{1} + \frac{1}{2} \, \mathbf{a}_{2} + \frac{1}{2} \, \mathbf{a}_{3} & = & \frac{1}{2}a \, \mathbf{\hat{x}} + \frac{1}{2}a \, \mathbf{\hat{y}} + \frac{1}{2}a \, \mathbf{\hat{z}} & \left(4b\right) & \mbox{Cu I} \\ 
\mathbf{B}_{3} & = & \frac{1}{4} \, \mathbf{a}_{1} + \frac{1}{4} \, \mathbf{a}_{2} + \frac{1}{4} \, \mathbf{a}_{3} & = & \frac{1}{4}a \, \mathbf{\hat{x}} + \frac{1}{4}a \, \mathbf{\hat{y}} + \frac{1}{4}a \, \mathbf{\hat{z}} & \left(4c\right) & \mbox{Cu II} \\ 
\mathbf{B}_{4} & = & \left(-x_{4}+y_{4}+z_{4}\right) \, \mathbf{a}_{1} + \left(x_{4}-y_{4}+z_{4}\right) \, \mathbf{a}_{2} + \left(x_{4}+y_{4}-z_{4}\right) \, \mathbf{a}_{3} & = & x_{4}a \, \mathbf{\hat{x}} + y_{4}a \, \mathbf{\hat{y}} + z_{4}a \, \mathbf{\hat{z}} & \left(48h\right) & \mbox{C} \\ 
\mathbf{B}_{5} & = & \left(x_{4}-y_{4}+z_{4}\right) \, \mathbf{a}_{1} + \left(-x_{4}+y_{4}+z_{4}\right) \, \mathbf{a}_{2} + \left(-x_{4}-y_{4}-z_{4}\right) \, \mathbf{a}_{3} & = & -x_{4}a \, \mathbf{\hat{x}}-y_{4}a \, \mathbf{\hat{y}} + z_{4}a \, \mathbf{\hat{z}} & \left(48h\right) & \mbox{C} \\ 
\mathbf{B}_{6} & = & \left(x_{4}+y_{4}-z_{4}\right) \, \mathbf{a}_{1} + \left(-x_{4}-y_{4}-z_{4}\right) \, \mathbf{a}_{2} + \left(-x_{4}+y_{4}+z_{4}\right) \, \mathbf{a}_{3} & = & -x_{4}a \, \mathbf{\hat{x}} + y_{4}a \, \mathbf{\hat{y}}-z_{4}a \, \mathbf{\hat{z}} & \left(48h\right) & \mbox{C} \\ 
\mathbf{B}_{7} & = & \left(-x_{4}-y_{4}-z_{4}\right) \, \mathbf{a}_{1} + \left(x_{4}+y_{4}-z_{4}\right) \, \mathbf{a}_{2} + \left(x_{4}-y_{4}+z_{4}\right) \, \mathbf{a}_{3} & = & x_{4}a \, \mathbf{\hat{x}}-y_{4}a \, \mathbf{\hat{y}}-z_{4}a \, \mathbf{\hat{z}} & \left(48h\right) & \mbox{C} \\ 
\mathbf{B}_{8} & = & \left(x_{4}+y_{4}-z_{4}\right) \, \mathbf{a}_{1} + \left(-x_{4}+y_{4}+z_{4}\right) \, \mathbf{a}_{2} + \left(x_{4}-y_{4}+z_{4}\right) \, \mathbf{a}_{3} & = & z_{4}a \, \mathbf{\hat{x}} + x_{4}a \, \mathbf{\hat{y}} + y_{4}a \, \mathbf{\hat{z}} & \left(48h\right) & \mbox{C} \\ 
\mathbf{B}_{9} & = & \left(-x_{4}-y_{4}-z_{4}\right) \, \mathbf{a}_{1} + \left(x_{4}-y_{4}+z_{4}\right) \, \mathbf{a}_{2} + \left(-x_{4}+y_{4}+z_{4}\right) \, \mathbf{a}_{3} & = & z_{4}a \, \mathbf{\hat{x}}-x_{4}a \, \mathbf{\hat{y}}-y_{4}a \, \mathbf{\hat{z}} & \left(48h\right) & \mbox{C} \\ 
\mathbf{B}_{10} & = & \left(-x_{4}+y_{4}+z_{4}\right) \, \mathbf{a}_{1} + \left(x_{4}+y_{4}-z_{4}\right) \, \mathbf{a}_{2} + \left(-x_{4}-y_{4}-z_{4}\right) \, \mathbf{a}_{3} & = & -z_{4}a \, \mathbf{\hat{x}}-x_{4}a \, \mathbf{\hat{y}} + y_{4}a \, \mathbf{\hat{z}} & \left(48h\right) & \mbox{C} \\ 
\mathbf{B}_{11} & = & \left(x_{4}-y_{4}+z_{4}\right) \, \mathbf{a}_{1} + \left(-x_{4}-y_{4}-z_{4}\right) \, \mathbf{a}_{2} + \left(x_{4}+y_{4}-z_{4}\right) \, \mathbf{a}_{3} & = & -z_{4}a \, \mathbf{\hat{x}} + x_{4}a \, \mathbf{\hat{y}}-y_{4}a \, \mathbf{\hat{z}} & \left(48h\right) & \mbox{C} \\ 
\mathbf{B}_{12} & = & \left(x_{4}-y_{4}+z_{4}\right) \, \mathbf{a}_{1} + \left(x_{4}+y_{4}-z_{4}\right) \, \mathbf{a}_{2} + \left(-x_{4}+y_{4}+z_{4}\right) \, \mathbf{a}_{3} & = & y_{4}a \, \mathbf{\hat{x}} + z_{4}a \, \mathbf{\hat{y}} + x_{4}a \, \mathbf{\hat{z}} & \left(48h\right) & \mbox{C} \\ 
\mathbf{B}_{13} & = & \left(-x_{4}+y_{4}+z_{4}\right) \, \mathbf{a}_{1} + \left(-x_{4}-y_{4}-z_{4}\right) \, \mathbf{a}_{2} + \left(x_{4}-y_{4}+z_{4}\right) \, \mathbf{a}_{3} & = & -y_{4}a \, \mathbf{\hat{x}} + z_{4}a \, \mathbf{\hat{y}}-x_{4}a \, \mathbf{\hat{z}} & \left(48h\right) & \mbox{C} \\ 
\mathbf{B}_{14} & = & \left(-x_{4}-y_{4}-z_{4}\right) \, \mathbf{a}_{1} + \left(-x_{4}+y_{4}+z_{4}\right) \, \mathbf{a}_{2} + \left(x_{4}+y_{4}-z_{4}\right) \, \mathbf{a}_{3} & = & y_{4}a \, \mathbf{\hat{x}}-z_{4}a \, \mathbf{\hat{y}}-x_{4}a \, \mathbf{\hat{z}} & \left(48h\right) & \mbox{C} \\ 
\mathbf{B}_{15} & = & \left(x_{4}+y_{4}-z_{4}\right) \, \mathbf{a}_{1} + \left(x_{4}-y_{4}+z_{4}\right) \, \mathbf{a}_{2} + \left(-x_{4}-y_{4}-z_{4}\right) \, \mathbf{a}_{3} & = & -y_{4}a \, \mathbf{\hat{x}}-z_{4}a \, \mathbf{\hat{y}} + x_{4}a \, \mathbf{\hat{z}} & \left(48h\right) & \mbox{C} \\ 
\end{longtabu}
\renewcommand{\arraystretch}{1.0}
\noindent \hrulefill
\\
\textbf{References:}
\vspace*{-0.25cm}
\begin{flushleft}
  - \bibentry{Rigamonti_Cu2FeCN6_GazzChimItal_1937}. \\
\end{flushleft}
\textbf{Found in:}
\vspace*{-0.25cm}
\begin{flushleft}
  - \bibentry{Villars_PearsonsCrystalData_2013}. \\
\end{flushleft}
\noindent \hrulefill
\\
\textbf{Geometry files:}
\\
\noindent  - CIF: pp. {\hyperref[A12B2C_cF60_196_h_bc_a_cif]{\pageref{A12B2C_cF60_196_h_bc_a_cif}}} \\
\noindent  - POSCAR: pp. {\hyperref[A12B2C_cF60_196_h_bc_a_poscar]{\pageref{A12B2C_cF60_196_h_bc_a_poscar}}} \\
\onecolumn
{\phantomsection\label{A12B36CD12_cF488_196_2h_6h_ac_fgh}}
\subsection*{\huge \textbf{{\normalfont \begin{raggedleft}MgB$_{12}$H$_{12}$[H$_{2}$O]$_{12}$ Structure: \end{raggedleft} \\ A12B36CD12\_cF488\_196\_2h\_6h\_ac\_fgh}}}
\noindent \hrulefill
\vspace*{0.25cm}
\begin{figure}[htp]
  \centering
  \vspace{-1em}
  {\includegraphics[width=1\textwidth]{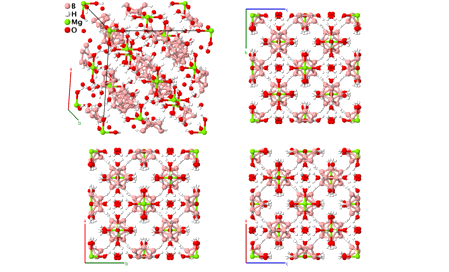}}
\end{figure}
\vspace*{-0.5cm}
\renewcommand{\arraystretch}{1.5}
\begin{equation*}
  \begin{array}{>{$\hspace{-0.15cm}}l<{$}>{$}p{0.5cm}<{$}>{$}p{18.5cm}<{$}}
    \mbox{\large \textbf{Prototype}} &\colon & \ce{MgB12H12[H2O]12} \\
    \mbox{\large \textbf{\AFLOW\ prototype label}} &\colon & \mbox{A12B36CD12\_cF488\_196\_2h\_6h\_ac\_fgh} \\
    \mbox{\large \textbf{\textit{Strukturbericht} designation}} &\colon & \mbox{None} \\
    \mbox{\large \textbf{Pearson symbol}} &\colon & \mbox{cF488} \\
    \mbox{\large \textbf{Space group number}} &\colon & 196 \\
    \mbox{\large \textbf{Space group symbol}} &\colon & F23 \\
    \mbox{\large \textbf{\AFLOW\ prototype command}} &\colon &  \texttt{aflow} \,  \, \texttt{-{}-proto=A12B36CD12\_cF488\_196\_2h\_6h\_ac\_fgh } \, \newline \texttt{-{}-params=}{a,x_{3},x_{4},x_{5},y_{5},z_{5},x_{6},y_{6},z_{6},x_{7},y_{7},z_{7},x_{8},y_{8},z_{8},x_{9},y_{9},z_{9},x_{10},y_{10},z_{10},x_{11},} \newline {y_{11},z_{11},x_{12},y_{12},z_{12},x_{13},y_{13},z_{13} }
  \end{array}
\end{equation*}
\renewcommand{\arraystretch}{1.0}

\noindent \parbox{1 \linewidth}{
\noindent \hrulefill
\\
\textbf{Face-centered Cubic primitive vectors:} \\
\vspace*{-0.25cm}
\begin{tabular}{cc}
  \begin{tabular}{c}
    \parbox{0.6 \linewidth}{
      \renewcommand{\arraystretch}{1.5}
      \begin{equation*}
        \centering
        \begin{array}{ccc}
              \mathbf{a}_1 & = & \frac12 \, a \, \mathbf{\hat{y}} + \frac12 \, a \, \mathbf{\hat{z}} \\
    \mathbf{a}_2 & = & \frac12 \, a \, \mathbf{\hat{x}} + \frac12 \, a \, \mathbf{\hat{z}} \\
    \mathbf{a}_3 & = & \frac12 \, a \, \mathbf{\hat{x}} + \frac12 \, a \, \mathbf{\hat{y}} \\

        \end{array}
      \end{equation*}
    }
    \renewcommand{\arraystretch}{1.0}
  \end{tabular}
  \begin{tabular}{c}
    \includegraphics[width=0.3\linewidth]{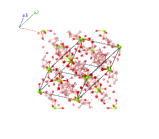} \\
  \end{tabular}
\end{tabular}

}
\vspace*{-0.25cm}

\noindent \hrulefill
\\
\textbf{Basis vectors:}
\vspace*{-0.25cm}
\renewcommand{\arraystretch}{1.5}
\begin{longtabu} to \textwidth{>{\centering $}X[-1,c,c]<{$}>{\centering $}X[-1,c,c]<{$}>{\centering $}X[-1,c,c]<{$}>{\centering $}X[-1,c,c]<{$}>{\centering $}X[-1,c,c]<{$}>{\centering $}X[-1,c,c]<{$}>{\centering $}X[-1,c,c]<{$}}
  & & \mbox{Lattice Coordinates} & & \mbox{Cartesian Coordinates} &\mbox{Wyckoff Position} & \mbox{Atom Type} \\  
  \mathbf{B}_{1} & = & 0 \, \mathbf{a}_{1} + 0 \, \mathbf{a}_{2} + 0 \, \mathbf{a}_{3} & = & 0 \, \mathbf{\hat{x}} + 0 \, \mathbf{\hat{y}} + 0 \, \mathbf{\hat{z}} & \left(4a\right) & \mbox{Mg I} \\ 
\mathbf{B}_{2} & = & \frac{1}{4} \, \mathbf{a}_{1} + \frac{1}{4} \, \mathbf{a}_{2} + \frac{1}{4} \, \mathbf{a}_{3} & = & \frac{1}{4}a \, \mathbf{\hat{x}} + \frac{1}{4}a \, \mathbf{\hat{y}} + \frac{1}{4}a \, \mathbf{\hat{z}} & \left(4c\right) & \mbox{Mg II} \\ 
\mathbf{B}_{3} & = & -x_{3} \, \mathbf{a}_{1} + x_{3} \, \mathbf{a}_{2} + x_{3} \, \mathbf{a}_{3} & = & x_{3}a \, \mathbf{\hat{x}} & \left(24f\right) & \mbox{O I} \\ 
\mathbf{B}_{4} & = & x_{3} \, \mathbf{a}_{1}-x_{3} \, \mathbf{a}_{2}-x_{3} \, \mathbf{a}_{3} & = & -x_{3}a \, \mathbf{\hat{x}} & \left(24f\right) & \mbox{O I} \\ 
\mathbf{B}_{5} & = & x_{3} \, \mathbf{a}_{1}-x_{3} \, \mathbf{a}_{2} + x_{3} \, \mathbf{a}_{3} & = & x_{3}a \, \mathbf{\hat{y}} & \left(24f\right) & \mbox{O I} \\ 
\mathbf{B}_{6} & = & -x_{3} \, \mathbf{a}_{1} + x_{3} \, \mathbf{a}_{2}-x_{3} \, \mathbf{a}_{3} & = & -x_{3}a \, \mathbf{\hat{y}} & \left(24f\right) & \mbox{O I} \\ 
\mathbf{B}_{7} & = & x_{3} \, \mathbf{a}_{1} + x_{3} \, \mathbf{a}_{2}-x_{3} \, \mathbf{a}_{3} & = & x_{3}a \, \mathbf{\hat{z}} & \left(24f\right) & \mbox{O I} \\ 
\mathbf{B}_{8} & = & -x_{3} \, \mathbf{a}_{1}-x_{3} \, \mathbf{a}_{2} + x_{3} \, \mathbf{a}_{3} & = & -x_{3}a \, \mathbf{\hat{z}} & \left(24f\right) & \mbox{O I} \\ 
\mathbf{B}_{9} & = & \left(\frac{1}{2} - x_{4}\right) \, \mathbf{a}_{1} + x_{4} \, \mathbf{a}_{2} + x_{4} \, \mathbf{a}_{3} & = & x_{4}a \, \mathbf{\hat{x}} + \frac{1}{4}a \, \mathbf{\hat{y}} + \frac{1}{4}a \, \mathbf{\hat{z}} & \left(24g\right) & \mbox{O II} \\ 
\mathbf{B}_{10} & = & x_{4} \, \mathbf{a}_{1} + \left(\frac{1}{2} - x_{4}\right) \, \mathbf{a}_{2} + \left(\frac{1}{2} - x_{4}\right) \, \mathbf{a}_{3} & = & \left(\frac{1}{2} - x_{4}\right)a \, \mathbf{\hat{x}} + \frac{1}{4}a \, \mathbf{\hat{y}} + \frac{1}{4}a \, \mathbf{\hat{z}} & \left(24g\right) & \mbox{O II} \\ 
\mathbf{B}_{11} & = & x_{4} \, \mathbf{a}_{1} + \left(\frac{1}{2} - x_{4}\right) \, \mathbf{a}_{2} + x_{4} \, \mathbf{a}_{3} & = & \frac{1}{4}a \, \mathbf{\hat{x}} + x_{4}a \, \mathbf{\hat{y}} + \frac{1}{4}a \, \mathbf{\hat{z}} & \left(24g\right) & \mbox{O II} \\ 
\mathbf{B}_{12} & = & \left(\frac{1}{2} - x_{4}\right) \, \mathbf{a}_{1} + x_{4} \, \mathbf{a}_{2} + \left(\frac{1}{2} - x_{4}\right) \, \mathbf{a}_{3} & = & \frac{1}{4}a \, \mathbf{\hat{x}} + \left(\frac{1}{2} - x_{4}\right)a \, \mathbf{\hat{y}} + \frac{1}{4}a \, \mathbf{\hat{z}} & \left(24g\right) & \mbox{O II} \\ 
\mathbf{B}_{13} & = & x_{4} \, \mathbf{a}_{1} + x_{4} \, \mathbf{a}_{2} + \left(\frac{1}{2} - x_{4}\right) \, \mathbf{a}_{3} & = & \frac{1}{4}a \, \mathbf{\hat{x}} + \frac{1}{4}a \, \mathbf{\hat{y}} + x_{4}a \, \mathbf{\hat{z}} & \left(24g\right) & \mbox{O II} \\ 
\mathbf{B}_{14} & = & \left(\frac{1}{2} - x_{4}\right) \, \mathbf{a}_{1} + \left(\frac{1}{2} - x_{4}\right) \, \mathbf{a}_{2} + x_{4} \, \mathbf{a}_{3} & = & \frac{1}{4}a \, \mathbf{\hat{x}} + \frac{1}{4}a \, \mathbf{\hat{y}} + \left(\frac{1}{2} - x_{4}\right)a \, \mathbf{\hat{z}} & \left(24g\right) & \mbox{O II} \\ 
\mathbf{B}_{15} & = & \left(-x_{5}+y_{5}+z_{5}\right) \, \mathbf{a}_{1} + \left(x_{5}-y_{5}+z_{5}\right) \, \mathbf{a}_{2} + \left(x_{5}+y_{5}-z_{5}\right) \, \mathbf{a}_{3} & = & x_{5}a \, \mathbf{\hat{x}} + y_{5}a \, \mathbf{\hat{y}} + z_{5}a \, \mathbf{\hat{z}} & \left(48h\right) & \mbox{B I} \\ 
\mathbf{B}_{16} & = & \left(x_{5}-y_{5}+z_{5}\right) \, \mathbf{a}_{1} + \left(-x_{5}+y_{5}+z_{5}\right) \, \mathbf{a}_{2} + \left(-x_{5}-y_{5}-z_{5}\right) \, \mathbf{a}_{3} & = & -x_{5}a \, \mathbf{\hat{x}}-y_{5}a \, \mathbf{\hat{y}} + z_{5}a \, \mathbf{\hat{z}} & \left(48h\right) & \mbox{B I} \\ 
\mathbf{B}_{17} & = & \left(x_{5}+y_{5}-z_{5}\right) \, \mathbf{a}_{1} + \left(-x_{5}-y_{5}-z_{5}\right) \, \mathbf{a}_{2} + \left(-x_{5}+y_{5}+z_{5}\right) \, \mathbf{a}_{3} & = & -x_{5}a \, \mathbf{\hat{x}} + y_{5}a \, \mathbf{\hat{y}}-z_{5}a \, \mathbf{\hat{z}} & \left(48h\right) & \mbox{B I} \\ 
\mathbf{B}_{18} & = & \left(-x_{5}-y_{5}-z_{5}\right) \, \mathbf{a}_{1} + \left(x_{5}+y_{5}-z_{5}\right) \, \mathbf{a}_{2} + \left(x_{5}-y_{5}+z_{5}\right) \, \mathbf{a}_{3} & = & x_{5}a \, \mathbf{\hat{x}}-y_{5}a \, \mathbf{\hat{y}}-z_{5}a \, \mathbf{\hat{z}} & \left(48h\right) & \mbox{B I} \\ 
\mathbf{B}_{19} & = & \left(x_{5}+y_{5}-z_{5}\right) \, \mathbf{a}_{1} + \left(-x_{5}+y_{5}+z_{5}\right) \, \mathbf{a}_{2} + \left(x_{5}-y_{5}+z_{5}\right) \, \mathbf{a}_{3} & = & z_{5}a \, \mathbf{\hat{x}} + x_{5}a \, \mathbf{\hat{y}} + y_{5}a \, \mathbf{\hat{z}} & \left(48h\right) & \mbox{B I} \\ 
\mathbf{B}_{20} & = & \left(-x_{5}-y_{5}-z_{5}\right) \, \mathbf{a}_{1} + \left(x_{5}-y_{5}+z_{5}\right) \, \mathbf{a}_{2} + \left(-x_{5}+y_{5}+z_{5}\right) \, \mathbf{a}_{3} & = & z_{5}a \, \mathbf{\hat{x}}-x_{5}a \, \mathbf{\hat{y}}-y_{5}a \, \mathbf{\hat{z}} & \left(48h\right) & \mbox{B I} \\ 
\mathbf{B}_{21} & = & \left(-x_{5}+y_{5}+z_{5}\right) \, \mathbf{a}_{1} + \left(x_{5}+y_{5}-z_{5}\right) \, \mathbf{a}_{2} + \left(-x_{5}-y_{5}-z_{5}\right) \, \mathbf{a}_{3} & = & -z_{5}a \, \mathbf{\hat{x}}-x_{5}a \, \mathbf{\hat{y}} + y_{5}a \, \mathbf{\hat{z}} & \left(48h\right) & \mbox{B I} \\ 
\mathbf{B}_{22} & = & \left(x_{5}-y_{5}+z_{5}\right) \, \mathbf{a}_{1} + \left(-x_{5}-y_{5}-z_{5}\right) \, \mathbf{a}_{2} + \left(x_{5}+y_{5}-z_{5}\right) \, \mathbf{a}_{3} & = & -z_{5}a \, \mathbf{\hat{x}} + x_{5}a \, \mathbf{\hat{y}}-y_{5}a \, \mathbf{\hat{z}} & \left(48h\right) & \mbox{B I} \\ 
\mathbf{B}_{23} & = & \left(x_{5}-y_{5}+z_{5}\right) \, \mathbf{a}_{1} + \left(x_{5}+y_{5}-z_{5}\right) \, \mathbf{a}_{2} + \left(-x_{5}+y_{5}+z_{5}\right) \, \mathbf{a}_{3} & = & y_{5}a \, \mathbf{\hat{x}} + z_{5}a \, \mathbf{\hat{y}} + x_{5}a \, \mathbf{\hat{z}} & \left(48h\right) & \mbox{B I} \\ 
\mathbf{B}_{24} & = & \left(-x_{5}+y_{5}+z_{5}\right) \, \mathbf{a}_{1} + \left(-x_{5}-y_{5}-z_{5}\right) \, \mathbf{a}_{2} + \left(x_{5}-y_{5}+z_{5}\right) \, \mathbf{a}_{3} & = & -y_{5}a \, \mathbf{\hat{x}} + z_{5}a \, \mathbf{\hat{y}}-x_{5}a \, \mathbf{\hat{z}} & \left(48h\right) & \mbox{B I} \\ 
\mathbf{B}_{25} & = & \left(-x_{5}-y_{5}-z_{5}\right) \, \mathbf{a}_{1} + \left(-x_{5}+y_{5}+z_{5}\right) \, \mathbf{a}_{2} + \left(x_{5}+y_{5}-z_{5}\right) \, \mathbf{a}_{3} & = & y_{5}a \, \mathbf{\hat{x}}-z_{5}a \, \mathbf{\hat{y}}-x_{5}a \, \mathbf{\hat{z}} & \left(48h\right) & \mbox{B I} \\ 
\mathbf{B}_{26} & = & \left(x_{5}+y_{5}-z_{5}\right) \, \mathbf{a}_{1} + \left(x_{5}-y_{5}+z_{5}\right) \, \mathbf{a}_{2} + \left(-x_{5}-y_{5}-z_{5}\right) \, \mathbf{a}_{3} & = & -y_{5}a \, \mathbf{\hat{x}}-z_{5}a \, \mathbf{\hat{y}} + x_{5}a \, \mathbf{\hat{z}} & \left(48h\right) & \mbox{B I} \\ 
\mathbf{B}_{27} & = & \left(-x_{6}+y_{6}+z_{6}\right) \, \mathbf{a}_{1} + \left(x_{6}-y_{6}+z_{6}\right) \, \mathbf{a}_{2} + \left(x_{6}+y_{6}-z_{6}\right) \, \mathbf{a}_{3} & = & x_{6}a \, \mathbf{\hat{x}} + y_{6}a \, \mathbf{\hat{y}} + z_{6}a \, \mathbf{\hat{z}} & \left(48h\right) & \mbox{B II} \\ 
\mathbf{B}_{28} & = & \left(x_{6}-y_{6}+z_{6}\right) \, \mathbf{a}_{1} + \left(-x_{6}+y_{6}+z_{6}\right) \, \mathbf{a}_{2} + \left(-x_{6}-y_{6}-z_{6}\right) \, \mathbf{a}_{3} & = & -x_{6}a \, \mathbf{\hat{x}}-y_{6}a \, \mathbf{\hat{y}} + z_{6}a \, \mathbf{\hat{z}} & \left(48h\right) & \mbox{B II} \\ 
\mathbf{B}_{29} & = & \left(x_{6}+y_{6}-z_{6}\right) \, \mathbf{a}_{1} + \left(-x_{6}-y_{6}-z_{6}\right) \, \mathbf{a}_{2} + \left(-x_{6}+y_{6}+z_{6}\right) \, \mathbf{a}_{3} & = & -x_{6}a \, \mathbf{\hat{x}} + y_{6}a \, \mathbf{\hat{y}}-z_{6}a \, \mathbf{\hat{z}} & \left(48h\right) & \mbox{B II} \\ 
\mathbf{B}_{30} & = & \left(-x_{6}-y_{6}-z_{6}\right) \, \mathbf{a}_{1} + \left(x_{6}+y_{6}-z_{6}\right) \, \mathbf{a}_{2} + \left(x_{6}-y_{6}+z_{6}\right) \, \mathbf{a}_{3} & = & x_{6}a \, \mathbf{\hat{x}}-y_{6}a \, \mathbf{\hat{y}}-z_{6}a \, \mathbf{\hat{z}} & \left(48h\right) & \mbox{B II} \\ 
\mathbf{B}_{31} & = & \left(x_{6}+y_{6}-z_{6}\right) \, \mathbf{a}_{1} + \left(-x_{6}+y_{6}+z_{6}\right) \, \mathbf{a}_{2} + \left(x_{6}-y_{6}+z_{6}\right) \, \mathbf{a}_{3} & = & z_{6}a \, \mathbf{\hat{x}} + x_{6}a \, \mathbf{\hat{y}} + y_{6}a \, \mathbf{\hat{z}} & \left(48h\right) & \mbox{B II} \\ 
\mathbf{B}_{32} & = & \left(-x_{6}-y_{6}-z_{6}\right) \, \mathbf{a}_{1} + \left(x_{6}-y_{6}+z_{6}\right) \, \mathbf{a}_{2} + \left(-x_{6}+y_{6}+z_{6}\right) \, \mathbf{a}_{3} & = & z_{6}a \, \mathbf{\hat{x}}-x_{6}a \, \mathbf{\hat{y}}-y_{6}a \, \mathbf{\hat{z}} & \left(48h\right) & \mbox{B II} \\ 
\mathbf{B}_{33} & = & \left(-x_{6}+y_{6}+z_{6}\right) \, \mathbf{a}_{1} + \left(x_{6}+y_{6}-z_{6}\right) \, \mathbf{a}_{2} + \left(-x_{6}-y_{6}-z_{6}\right) \, \mathbf{a}_{3} & = & -z_{6}a \, \mathbf{\hat{x}}-x_{6}a \, \mathbf{\hat{y}} + y_{6}a \, \mathbf{\hat{z}} & \left(48h\right) & \mbox{B II} \\ 
\mathbf{B}_{34} & = & \left(x_{6}-y_{6}+z_{6}\right) \, \mathbf{a}_{1} + \left(-x_{6}-y_{6}-z_{6}\right) \, \mathbf{a}_{2} + \left(x_{6}+y_{6}-z_{6}\right) \, \mathbf{a}_{3} & = & -z_{6}a \, \mathbf{\hat{x}} + x_{6}a \, \mathbf{\hat{y}}-y_{6}a \, \mathbf{\hat{z}} & \left(48h\right) & \mbox{B II} \\ 
\mathbf{B}_{35} & = & \left(x_{6}-y_{6}+z_{6}\right) \, \mathbf{a}_{1} + \left(x_{6}+y_{6}-z_{6}\right) \, \mathbf{a}_{2} + \left(-x_{6}+y_{6}+z_{6}\right) \, \mathbf{a}_{3} & = & y_{6}a \, \mathbf{\hat{x}} + z_{6}a \, \mathbf{\hat{y}} + x_{6}a \, \mathbf{\hat{z}} & \left(48h\right) & \mbox{B II} \\ 
\mathbf{B}_{36} & = & \left(-x_{6}+y_{6}+z_{6}\right) \, \mathbf{a}_{1} + \left(-x_{6}-y_{6}-z_{6}\right) \, \mathbf{a}_{2} + \left(x_{6}-y_{6}+z_{6}\right) \, \mathbf{a}_{3} & = & -y_{6}a \, \mathbf{\hat{x}} + z_{6}a \, \mathbf{\hat{y}}-x_{6}a \, \mathbf{\hat{z}} & \left(48h\right) & \mbox{B II} \\ 
\mathbf{B}_{37} & = & \left(-x_{6}-y_{6}-z_{6}\right) \, \mathbf{a}_{1} + \left(-x_{6}+y_{6}+z_{6}\right) \, \mathbf{a}_{2} + \left(x_{6}+y_{6}-z_{6}\right) \, \mathbf{a}_{3} & = & y_{6}a \, \mathbf{\hat{x}}-z_{6}a \, \mathbf{\hat{y}}-x_{6}a \, \mathbf{\hat{z}} & \left(48h\right) & \mbox{B II} \\ 
\mathbf{B}_{38} & = & \left(x_{6}+y_{6}-z_{6}\right) \, \mathbf{a}_{1} + \left(x_{6}-y_{6}+z_{6}\right) \, \mathbf{a}_{2} + \left(-x_{6}-y_{6}-z_{6}\right) \, \mathbf{a}_{3} & = & -y_{6}a \, \mathbf{\hat{x}}-z_{6}a \, \mathbf{\hat{y}} + x_{6}a \, \mathbf{\hat{z}} & \left(48h\right) & \mbox{B II} \\ 
\mathbf{B}_{39} & = & \left(-x_{7}+y_{7}+z_{7}\right) \, \mathbf{a}_{1} + \left(x_{7}-y_{7}+z_{7}\right) \, \mathbf{a}_{2} + \left(x_{7}+y_{7}-z_{7}\right) \, \mathbf{a}_{3} & = & x_{7}a \, \mathbf{\hat{x}} + y_{7}a \, \mathbf{\hat{y}} + z_{7}a \, \mathbf{\hat{z}} & \left(48h\right) & \mbox{H I} \\ 
\mathbf{B}_{40} & = & \left(x_{7}-y_{7}+z_{7}\right) \, \mathbf{a}_{1} + \left(-x_{7}+y_{7}+z_{7}\right) \, \mathbf{a}_{2} + \left(-x_{7}-y_{7}-z_{7}\right) \, \mathbf{a}_{3} & = & -x_{7}a \, \mathbf{\hat{x}}-y_{7}a \, \mathbf{\hat{y}} + z_{7}a \, \mathbf{\hat{z}} & \left(48h\right) & \mbox{H I} \\ 
\mathbf{B}_{41} & = & \left(x_{7}+y_{7}-z_{7}\right) \, \mathbf{a}_{1} + \left(-x_{7}-y_{7}-z_{7}\right) \, \mathbf{a}_{2} + \left(-x_{7}+y_{7}+z_{7}\right) \, \mathbf{a}_{3} & = & -x_{7}a \, \mathbf{\hat{x}} + y_{7}a \, \mathbf{\hat{y}}-z_{7}a \, \mathbf{\hat{z}} & \left(48h\right) & \mbox{H I} \\ 
\mathbf{B}_{42} & = & \left(-x_{7}-y_{7}-z_{7}\right) \, \mathbf{a}_{1} + \left(x_{7}+y_{7}-z_{7}\right) \, \mathbf{a}_{2} + \left(x_{7}-y_{7}+z_{7}\right) \, \mathbf{a}_{3} & = & x_{7}a \, \mathbf{\hat{x}}-y_{7}a \, \mathbf{\hat{y}}-z_{7}a \, \mathbf{\hat{z}} & \left(48h\right) & \mbox{H I} \\ 
\mathbf{B}_{43} & = & \left(x_{7}+y_{7}-z_{7}\right) \, \mathbf{a}_{1} + \left(-x_{7}+y_{7}+z_{7}\right) \, \mathbf{a}_{2} + \left(x_{7}-y_{7}+z_{7}\right) \, \mathbf{a}_{3} & = & z_{7}a \, \mathbf{\hat{x}} + x_{7}a \, \mathbf{\hat{y}} + y_{7}a \, \mathbf{\hat{z}} & \left(48h\right) & \mbox{H I} \\ 
\mathbf{B}_{44} & = & \left(-x_{7}-y_{7}-z_{7}\right) \, \mathbf{a}_{1} + \left(x_{7}-y_{7}+z_{7}\right) \, \mathbf{a}_{2} + \left(-x_{7}+y_{7}+z_{7}\right) \, \mathbf{a}_{3} & = & z_{7}a \, \mathbf{\hat{x}}-x_{7}a \, \mathbf{\hat{y}}-y_{7}a \, \mathbf{\hat{z}} & \left(48h\right) & \mbox{H I} \\ 
\mathbf{B}_{45} & = & \left(-x_{7}+y_{7}+z_{7}\right) \, \mathbf{a}_{1} + \left(x_{7}+y_{7}-z_{7}\right) \, \mathbf{a}_{2} + \left(-x_{7}-y_{7}-z_{7}\right) \, \mathbf{a}_{3} & = & -z_{7}a \, \mathbf{\hat{x}}-x_{7}a \, \mathbf{\hat{y}} + y_{7}a \, \mathbf{\hat{z}} & \left(48h\right) & \mbox{H I} \\ 
\mathbf{B}_{46} & = & \left(x_{7}-y_{7}+z_{7}\right) \, \mathbf{a}_{1} + \left(-x_{7}-y_{7}-z_{7}\right) \, \mathbf{a}_{2} + \left(x_{7}+y_{7}-z_{7}\right) \, \mathbf{a}_{3} & = & -z_{7}a \, \mathbf{\hat{x}} + x_{7}a \, \mathbf{\hat{y}}-y_{7}a \, \mathbf{\hat{z}} & \left(48h\right) & \mbox{H I} \\ 
\mathbf{B}_{47} & = & \left(x_{7}-y_{7}+z_{7}\right) \, \mathbf{a}_{1} + \left(x_{7}+y_{7}-z_{7}\right) \, \mathbf{a}_{2} + \left(-x_{7}+y_{7}+z_{7}\right) \, \mathbf{a}_{3} & = & y_{7}a \, \mathbf{\hat{x}} + z_{7}a \, \mathbf{\hat{y}} + x_{7}a \, \mathbf{\hat{z}} & \left(48h\right) & \mbox{H I} \\ 
\mathbf{B}_{48} & = & \left(-x_{7}+y_{7}+z_{7}\right) \, \mathbf{a}_{1} + \left(-x_{7}-y_{7}-z_{7}\right) \, \mathbf{a}_{2} + \left(x_{7}-y_{7}+z_{7}\right) \, \mathbf{a}_{3} & = & -y_{7}a \, \mathbf{\hat{x}} + z_{7}a \, \mathbf{\hat{y}}-x_{7}a \, \mathbf{\hat{z}} & \left(48h\right) & \mbox{H I} \\ 
\mathbf{B}_{49} & = & \left(-x_{7}-y_{7}-z_{7}\right) \, \mathbf{a}_{1} + \left(-x_{7}+y_{7}+z_{7}\right) \, \mathbf{a}_{2} + \left(x_{7}+y_{7}-z_{7}\right) \, \mathbf{a}_{3} & = & y_{7}a \, \mathbf{\hat{x}}-z_{7}a \, \mathbf{\hat{y}}-x_{7}a \, \mathbf{\hat{z}} & \left(48h\right) & \mbox{H I} \\ 
\mathbf{B}_{50} & = & \left(x_{7}+y_{7}-z_{7}\right) \, \mathbf{a}_{1} + \left(x_{7}-y_{7}+z_{7}\right) \, \mathbf{a}_{2} + \left(-x_{7}-y_{7}-z_{7}\right) \, \mathbf{a}_{3} & = & -y_{7}a \, \mathbf{\hat{x}}-z_{7}a \, \mathbf{\hat{y}} + x_{7}a \, \mathbf{\hat{z}} & \left(48h\right) & \mbox{H I} \\ 
\mathbf{B}_{51} & = & \left(-x_{8}+y_{8}+z_{8}\right) \, \mathbf{a}_{1} + \left(x_{8}-y_{8}+z_{8}\right) \, \mathbf{a}_{2} + \left(x_{8}+y_{8}-z_{8}\right) \, \mathbf{a}_{3} & = & x_{8}a \, \mathbf{\hat{x}} + y_{8}a \, \mathbf{\hat{y}} + z_{8}a \, \mathbf{\hat{z}} & \left(48h\right) & \mbox{H II} \\ 
\mathbf{B}_{52} & = & \left(x_{8}-y_{8}+z_{8}\right) \, \mathbf{a}_{1} + \left(-x_{8}+y_{8}+z_{8}\right) \, \mathbf{a}_{2} + \left(-x_{8}-y_{8}-z_{8}\right) \, \mathbf{a}_{3} & = & -x_{8}a \, \mathbf{\hat{x}}-y_{8}a \, \mathbf{\hat{y}} + z_{8}a \, \mathbf{\hat{z}} & \left(48h\right) & \mbox{H II} \\ 
\mathbf{B}_{53} & = & \left(x_{8}+y_{8}-z_{8}\right) \, \mathbf{a}_{1} + \left(-x_{8}-y_{8}-z_{8}\right) \, \mathbf{a}_{2} + \left(-x_{8}+y_{8}+z_{8}\right) \, \mathbf{a}_{3} & = & -x_{8}a \, \mathbf{\hat{x}} + y_{8}a \, \mathbf{\hat{y}}-z_{8}a \, \mathbf{\hat{z}} & \left(48h\right) & \mbox{H II} \\ 
\mathbf{B}_{54} & = & \left(-x_{8}-y_{8}-z_{8}\right) \, \mathbf{a}_{1} + \left(x_{8}+y_{8}-z_{8}\right) \, \mathbf{a}_{2} + \left(x_{8}-y_{8}+z_{8}\right) \, \mathbf{a}_{3} & = & x_{8}a \, \mathbf{\hat{x}}-y_{8}a \, \mathbf{\hat{y}}-z_{8}a \, \mathbf{\hat{z}} & \left(48h\right) & \mbox{H II} \\ 
\mathbf{B}_{55} & = & \left(x_{8}+y_{8}-z_{8}\right) \, \mathbf{a}_{1} + \left(-x_{8}+y_{8}+z_{8}\right) \, \mathbf{a}_{2} + \left(x_{8}-y_{8}+z_{8}\right) \, \mathbf{a}_{3} & = & z_{8}a \, \mathbf{\hat{x}} + x_{8}a \, \mathbf{\hat{y}} + y_{8}a \, \mathbf{\hat{z}} & \left(48h\right) & \mbox{H II} \\ 
\mathbf{B}_{56} & = & \left(-x_{8}-y_{8}-z_{8}\right) \, \mathbf{a}_{1} + \left(x_{8}-y_{8}+z_{8}\right) \, \mathbf{a}_{2} + \left(-x_{8}+y_{8}+z_{8}\right) \, \mathbf{a}_{3} & = & z_{8}a \, \mathbf{\hat{x}}-x_{8}a \, \mathbf{\hat{y}}-y_{8}a \, \mathbf{\hat{z}} & \left(48h\right) & \mbox{H II} \\ 
\mathbf{B}_{57} & = & \left(-x_{8}+y_{8}+z_{8}\right) \, \mathbf{a}_{1} + \left(x_{8}+y_{8}-z_{8}\right) \, \mathbf{a}_{2} + \left(-x_{8}-y_{8}-z_{8}\right) \, \mathbf{a}_{3} & = & -z_{8}a \, \mathbf{\hat{x}}-x_{8}a \, \mathbf{\hat{y}} + y_{8}a \, \mathbf{\hat{z}} & \left(48h\right) & \mbox{H II} \\ 
\mathbf{B}_{58} & = & \left(x_{8}-y_{8}+z_{8}\right) \, \mathbf{a}_{1} + \left(-x_{8}-y_{8}-z_{8}\right) \, \mathbf{a}_{2} + \left(x_{8}+y_{8}-z_{8}\right) \, \mathbf{a}_{3} & = & -z_{8}a \, \mathbf{\hat{x}} + x_{8}a \, \mathbf{\hat{y}}-y_{8}a \, \mathbf{\hat{z}} & \left(48h\right) & \mbox{H II} \\ 
\mathbf{B}_{59} & = & \left(x_{8}-y_{8}+z_{8}\right) \, \mathbf{a}_{1} + \left(x_{8}+y_{8}-z_{8}\right) \, \mathbf{a}_{2} + \left(-x_{8}+y_{8}+z_{8}\right) \, \mathbf{a}_{3} & = & y_{8}a \, \mathbf{\hat{x}} + z_{8}a \, \mathbf{\hat{y}} + x_{8}a \, \mathbf{\hat{z}} & \left(48h\right) & \mbox{H II} \\ 
\mathbf{B}_{60} & = & \left(-x_{8}+y_{8}+z_{8}\right) \, \mathbf{a}_{1} + \left(-x_{8}-y_{8}-z_{8}\right) \, \mathbf{a}_{2} + \left(x_{8}-y_{8}+z_{8}\right) \, \mathbf{a}_{3} & = & -y_{8}a \, \mathbf{\hat{x}} + z_{8}a \, \mathbf{\hat{y}}-x_{8}a \, \mathbf{\hat{z}} & \left(48h\right) & \mbox{H II} \\ 
\mathbf{B}_{61} & = & \left(-x_{8}-y_{8}-z_{8}\right) \, \mathbf{a}_{1} + \left(-x_{8}+y_{8}+z_{8}\right) \, \mathbf{a}_{2} + \left(x_{8}+y_{8}-z_{8}\right) \, \mathbf{a}_{3} & = & y_{8}a \, \mathbf{\hat{x}}-z_{8}a \, \mathbf{\hat{y}}-x_{8}a \, \mathbf{\hat{z}} & \left(48h\right) & \mbox{H II} \\ 
\mathbf{B}_{62} & = & \left(x_{8}+y_{8}-z_{8}\right) \, \mathbf{a}_{1} + \left(x_{8}-y_{8}+z_{8}\right) \, \mathbf{a}_{2} + \left(-x_{8}-y_{8}-z_{8}\right) \, \mathbf{a}_{3} & = & -y_{8}a \, \mathbf{\hat{x}}-z_{8}a \, \mathbf{\hat{y}} + x_{8}a \, \mathbf{\hat{z}} & \left(48h\right) & \mbox{H II} \\ 
\mathbf{B}_{63} & = & \left(-x_{9}+y_{9}+z_{9}\right) \, \mathbf{a}_{1} + \left(x_{9}-y_{9}+z_{9}\right) \, \mathbf{a}_{2} + \left(x_{9}+y_{9}-z_{9}\right) \, \mathbf{a}_{3} & = & x_{9}a \, \mathbf{\hat{x}} + y_{9}a \, \mathbf{\hat{y}} + z_{9}a \, \mathbf{\hat{z}} & \left(48h\right) & \mbox{H III} \\ 
\mathbf{B}_{64} & = & \left(x_{9}-y_{9}+z_{9}\right) \, \mathbf{a}_{1} + \left(-x_{9}+y_{9}+z_{9}\right) \, \mathbf{a}_{2} + \left(-x_{9}-y_{9}-z_{9}\right) \, \mathbf{a}_{3} & = & -x_{9}a \, \mathbf{\hat{x}}-y_{9}a \, \mathbf{\hat{y}} + z_{9}a \, \mathbf{\hat{z}} & \left(48h\right) & \mbox{H III} \\ 
\mathbf{B}_{65} & = & \left(x_{9}+y_{9}-z_{9}\right) \, \mathbf{a}_{1} + \left(-x_{9}-y_{9}-z_{9}\right) \, \mathbf{a}_{2} + \left(-x_{9}+y_{9}+z_{9}\right) \, \mathbf{a}_{3} & = & -x_{9}a \, \mathbf{\hat{x}} + y_{9}a \, \mathbf{\hat{y}}-z_{9}a \, \mathbf{\hat{z}} & \left(48h\right) & \mbox{H III} \\ 
\mathbf{B}_{66} & = & \left(-x_{9}-y_{9}-z_{9}\right) \, \mathbf{a}_{1} + \left(x_{9}+y_{9}-z_{9}\right) \, \mathbf{a}_{2} + \left(x_{9}-y_{9}+z_{9}\right) \, \mathbf{a}_{3} & = & x_{9}a \, \mathbf{\hat{x}}-y_{9}a \, \mathbf{\hat{y}}-z_{9}a \, \mathbf{\hat{z}} & \left(48h\right) & \mbox{H III} \\ 
\mathbf{B}_{67} & = & \left(x_{9}+y_{9}-z_{9}\right) \, \mathbf{a}_{1} + \left(-x_{9}+y_{9}+z_{9}\right) \, \mathbf{a}_{2} + \left(x_{9}-y_{9}+z_{9}\right) \, \mathbf{a}_{3} & = & z_{9}a \, \mathbf{\hat{x}} + x_{9}a \, \mathbf{\hat{y}} + y_{9}a \, \mathbf{\hat{z}} & \left(48h\right) & \mbox{H III} \\ 
\mathbf{B}_{68} & = & \left(-x_{9}-y_{9}-z_{9}\right) \, \mathbf{a}_{1} + \left(x_{9}-y_{9}+z_{9}\right) \, \mathbf{a}_{2} + \left(-x_{9}+y_{9}+z_{9}\right) \, \mathbf{a}_{3} & = & z_{9}a \, \mathbf{\hat{x}}-x_{9}a \, \mathbf{\hat{y}}-y_{9}a \, \mathbf{\hat{z}} & \left(48h\right) & \mbox{H III} \\ 
\mathbf{B}_{69} & = & \left(-x_{9}+y_{9}+z_{9}\right) \, \mathbf{a}_{1} + \left(x_{9}+y_{9}-z_{9}\right) \, \mathbf{a}_{2} + \left(-x_{9}-y_{9}-z_{9}\right) \, \mathbf{a}_{3} & = & -z_{9}a \, \mathbf{\hat{x}}-x_{9}a \, \mathbf{\hat{y}} + y_{9}a \, \mathbf{\hat{z}} & \left(48h\right) & \mbox{H III} \\ 
\mathbf{B}_{70} & = & \left(x_{9}-y_{9}+z_{9}\right) \, \mathbf{a}_{1} + \left(-x_{9}-y_{9}-z_{9}\right) \, \mathbf{a}_{2} + \left(x_{9}+y_{9}-z_{9}\right) \, \mathbf{a}_{3} & = & -z_{9}a \, \mathbf{\hat{x}} + x_{9}a \, \mathbf{\hat{y}}-y_{9}a \, \mathbf{\hat{z}} & \left(48h\right) & \mbox{H III} \\ 
\mathbf{B}_{71} & = & \left(x_{9}-y_{9}+z_{9}\right) \, \mathbf{a}_{1} + \left(x_{9}+y_{9}-z_{9}\right) \, \mathbf{a}_{2} + \left(-x_{9}+y_{9}+z_{9}\right) \, \mathbf{a}_{3} & = & y_{9}a \, \mathbf{\hat{x}} + z_{9}a \, \mathbf{\hat{y}} + x_{9}a \, \mathbf{\hat{z}} & \left(48h\right) & \mbox{H III} \\ 
\mathbf{B}_{72} & = & \left(-x_{9}+y_{9}+z_{9}\right) \, \mathbf{a}_{1} + \left(-x_{9}-y_{9}-z_{9}\right) \, \mathbf{a}_{2} + \left(x_{9}-y_{9}+z_{9}\right) \, \mathbf{a}_{3} & = & -y_{9}a \, \mathbf{\hat{x}} + z_{9}a \, \mathbf{\hat{y}}-x_{9}a \, \mathbf{\hat{z}} & \left(48h\right) & \mbox{H III} \\ 
\mathbf{B}_{73} & = & \left(-x_{9}-y_{9}-z_{9}\right) \, \mathbf{a}_{1} + \left(-x_{9}+y_{9}+z_{9}\right) \, \mathbf{a}_{2} + \left(x_{9}+y_{9}-z_{9}\right) \, \mathbf{a}_{3} & = & y_{9}a \, \mathbf{\hat{x}}-z_{9}a \, \mathbf{\hat{y}}-x_{9}a \, \mathbf{\hat{z}} & \left(48h\right) & \mbox{H III} \\ 
\mathbf{B}_{74} & = & \left(x_{9}+y_{9}-z_{9}\right) \, \mathbf{a}_{1} + \left(x_{9}-y_{9}+z_{9}\right) \, \mathbf{a}_{2} + \left(-x_{9}-y_{9}-z_{9}\right) \, \mathbf{a}_{3} & = & -y_{9}a \, \mathbf{\hat{x}}-z_{9}a \, \mathbf{\hat{y}} + x_{9}a \, \mathbf{\hat{z}} & \left(48h\right) & \mbox{H III} \\ 
\mathbf{B}_{75} & = & \left(-x_{10}+y_{10}+z_{10}\right) \, \mathbf{a}_{1} + \left(x_{10}-y_{10}+z_{10}\right) \, \mathbf{a}_{2} + \left(x_{10}+y_{10}-z_{10}\right) \, \mathbf{a}_{3} & = & x_{10}a \, \mathbf{\hat{x}} + y_{10}a \, \mathbf{\hat{y}} + z_{10}a \, \mathbf{\hat{z}} & \left(48h\right) & \mbox{H IV} \\ 
\mathbf{B}_{76} & = & \left(x_{10}-y_{10}+z_{10}\right) \, \mathbf{a}_{1} + \left(-x_{10}+y_{10}+z_{10}\right) \, \mathbf{a}_{2} + \left(-x_{10}-y_{10}-z_{10}\right) \, \mathbf{a}_{3} & = & -x_{10}a \, \mathbf{\hat{x}}-y_{10}a \, \mathbf{\hat{y}} + z_{10}a \, \mathbf{\hat{z}} & \left(48h\right) & \mbox{H IV} \\ 
\mathbf{B}_{77} & = & \left(x_{10}+y_{10}-z_{10}\right) \, \mathbf{a}_{1} + \left(-x_{10}-y_{10}-z_{10}\right) \, \mathbf{a}_{2} + \left(-x_{10}+y_{10}+z_{10}\right) \, \mathbf{a}_{3} & = & -x_{10}a \, \mathbf{\hat{x}} + y_{10}a \, \mathbf{\hat{y}}-z_{10}a \, \mathbf{\hat{z}} & \left(48h\right) & \mbox{H IV} \\ 
\mathbf{B}_{78} & = & \left(-x_{10}-y_{10}-z_{10}\right) \, \mathbf{a}_{1} + \left(x_{10}+y_{10}-z_{10}\right) \, \mathbf{a}_{2} + \left(x_{10}-y_{10}+z_{10}\right) \, \mathbf{a}_{3} & = & x_{10}a \, \mathbf{\hat{x}}-y_{10}a \, \mathbf{\hat{y}}-z_{10}a \, \mathbf{\hat{z}} & \left(48h\right) & \mbox{H IV} \\ 
\mathbf{B}_{79} & = & \left(x_{10}+y_{10}-z_{10}\right) \, \mathbf{a}_{1} + \left(-x_{10}+y_{10}+z_{10}\right) \, \mathbf{a}_{2} + \left(x_{10}-y_{10}+z_{10}\right) \, \mathbf{a}_{3} & = & z_{10}a \, \mathbf{\hat{x}} + x_{10}a \, \mathbf{\hat{y}} + y_{10}a \, \mathbf{\hat{z}} & \left(48h\right) & \mbox{H IV} \\ 
\mathbf{B}_{80} & = & \left(-x_{10}-y_{10}-z_{10}\right) \, \mathbf{a}_{1} + \left(x_{10}-y_{10}+z_{10}\right) \, \mathbf{a}_{2} + \left(-x_{10}+y_{10}+z_{10}\right) \, \mathbf{a}_{3} & = & z_{10}a \, \mathbf{\hat{x}}-x_{10}a \, \mathbf{\hat{y}}-y_{10}a \, \mathbf{\hat{z}} & \left(48h\right) & \mbox{H IV} \\ 
\mathbf{B}_{81} & = & \left(-x_{10}+y_{10}+z_{10}\right) \, \mathbf{a}_{1} + \left(x_{10}+y_{10}-z_{10}\right) \, \mathbf{a}_{2} + \left(-x_{10}-y_{10}-z_{10}\right) \, \mathbf{a}_{3} & = & -z_{10}a \, \mathbf{\hat{x}}-x_{10}a \, \mathbf{\hat{y}} + y_{10}a \, \mathbf{\hat{z}} & \left(48h\right) & \mbox{H IV} \\ 
\mathbf{B}_{82} & = & \left(x_{10}-y_{10}+z_{10}\right) \, \mathbf{a}_{1} + \left(-x_{10}-y_{10}-z_{10}\right) \, \mathbf{a}_{2} + \left(x_{10}+y_{10}-z_{10}\right) \, \mathbf{a}_{3} & = & -z_{10}a \, \mathbf{\hat{x}} + x_{10}a \, \mathbf{\hat{y}}-y_{10}a \, \mathbf{\hat{z}} & \left(48h\right) & \mbox{H IV} \\ 
\mathbf{B}_{83} & = & \left(x_{10}-y_{10}+z_{10}\right) \, \mathbf{a}_{1} + \left(x_{10}+y_{10}-z_{10}\right) \, \mathbf{a}_{2} + \left(-x_{10}+y_{10}+z_{10}\right) \, \mathbf{a}_{3} & = & y_{10}a \, \mathbf{\hat{x}} + z_{10}a \, \mathbf{\hat{y}} + x_{10}a \, \mathbf{\hat{z}} & \left(48h\right) & \mbox{H IV} \\ 
\mathbf{B}_{84} & = & \left(-x_{10}+y_{10}+z_{10}\right) \, \mathbf{a}_{1} + \left(-x_{10}-y_{10}-z_{10}\right) \, \mathbf{a}_{2} + \left(x_{10}-y_{10}+z_{10}\right) \, \mathbf{a}_{3} & = & -y_{10}a \, \mathbf{\hat{x}} + z_{10}a \, \mathbf{\hat{y}}-x_{10}a \, \mathbf{\hat{z}} & \left(48h\right) & \mbox{H IV} \\ 
\mathbf{B}_{85} & = & \left(-x_{10}-y_{10}-z_{10}\right) \, \mathbf{a}_{1} + \left(-x_{10}+y_{10}+z_{10}\right) \, \mathbf{a}_{2} + \left(x_{10}+y_{10}-z_{10}\right) \, \mathbf{a}_{3} & = & y_{10}a \, \mathbf{\hat{x}}-z_{10}a \, \mathbf{\hat{y}}-x_{10}a \, \mathbf{\hat{z}} & \left(48h\right) & \mbox{H IV} \\ 
\mathbf{B}_{86} & = & \left(x_{10}+y_{10}-z_{10}\right) \, \mathbf{a}_{1} + \left(x_{10}-y_{10}+z_{10}\right) \, \mathbf{a}_{2} + \left(-x_{10}-y_{10}-z_{10}\right) \, \mathbf{a}_{3} & = & -y_{10}a \, \mathbf{\hat{x}}-z_{10}a \, \mathbf{\hat{y}} + x_{10}a \, \mathbf{\hat{z}} & \left(48h\right) & \mbox{H IV} \\ 
\mathbf{B}_{87} & = & \left(-x_{11}+y_{11}+z_{11}\right) \, \mathbf{a}_{1} + \left(x_{11}-y_{11}+z_{11}\right) \, \mathbf{a}_{2} + \left(x_{11}+y_{11}-z_{11}\right) \, \mathbf{a}_{3} & = & x_{11}a \, \mathbf{\hat{x}} + y_{11}a \, \mathbf{\hat{y}} + z_{11}a \, \mathbf{\hat{z}} & \left(48h\right) & \mbox{H V} \\ 
\mathbf{B}_{88} & = & \left(x_{11}-y_{11}+z_{11}\right) \, \mathbf{a}_{1} + \left(-x_{11}+y_{11}+z_{11}\right) \, \mathbf{a}_{2} + \left(-x_{11}-y_{11}-z_{11}\right) \, \mathbf{a}_{3} & = & -x_{11}a \, \mathbf{\hat{x}}-y_{11}a \, \mathbf{\hat{y}} + z_{11}a \, \mathbf{\hat{z}} & \left(48h\right) & \mbox{H V} \\ 
\mathbf{B}_{89} & = & \left(x_{11}+y_{11}-z_{11}\right) \, \mathbf{a}_{1} + \left(-x_{11}-y_{11}-z_{11}\right) \, \mathbf{a}_{2} + \left(-x_{11}+y_{11}+z_{11}\right) \, \mathbf{a}_{3} & = & -x_{11}a \, \mathbf{\hat{x}} + y_{11}a \, \mathbf{\hat{y}}-z_{11}a \, \mathbf{\hat{z}} & \left(48h\right) & \mbox{H V} \\ 
\mathbf{B}_{90} & = & \left(-x_{11}-y_{11}-z_{11}\right) \, \mathbf{a}_{1} + \left(x_{11}+y_{11}-z_{11}\right) \, \mathbf{a}_{2} + \left(x_{11}-y_{11}+z_{11}\right) \, \mathbf{a}_{3} & = & x_{11}a \, \mathbf{\hat{x}}-y_{11}a \, \mathbf{\hat{y}}-z_{11}a \, \mathbf{\hat{z}} & \left(48h\right) & \mbox{H V} \\ 
\mathbf{B}_{91} & = & \left(x_{11}+y_{11}-z_{11}\right) \, \mathbf{a}_{1} + \left(-x_{11}+y_{11}+z_{11}\right) \, \mathbf{a}_{2} + \left(x_{11}-y_{11}+z_{11}\right) \, \mathbf{a}_{3} & = & z_{11}a \, \mathbf{\hat{x}} + x_{11}a \, \mathbf{\hat{y}} + y_{11}a \, \mathbf{\hat{z}} & \left(48h\right) & \mbox{H V} \\ 
\mathbf{B}_{92} & = & \left(-x_{11}-y_{11}-z_{11}\right) \, \mathbf{a}_{1} + \left(x_{11}-y_{11}+z_{11}\right) \, \mathbf{a}_{2} + \left(-x_{11}+y_{11}+z_{11}\right) \, \mathbf{a}_{3} & = & z_{11}a \, \mathbf{\hat{x}}-x_{11}a \, \mathbf{\hat{y}}-y_{11}a \, \mathbf{\hat{z}} & \left(48h\right) & \mbox{H V} \\ 
\mathbf{B}_{93} & = & \left(-x_{11}+y_{11}+z_{11}\right) \, \mathbf{a}_{1} + \left(x_{11}+y_{11}-z_{11}\right) \, \mathbf{a}_{2} + \left(-x_{11}-y_{11}-z_{11}\right) \, \mathbf{a}_{3} & = & -z_{11}a \, \mathbf{\hat{x}}-x_{11}a \, \mathbf{\hat{y}} + y_{11}a \, \mathbf{\hat{z}} & \left(48h\right) & \mbox{H V} \\ 
\mathbf{B}_{94} & = & \left(x_{11}-y_{11}+z_{11}\right) \, \mathbf{a}_{1} + \left(-x_{11}-y_{11}-z_{11}\right) \, \mathbf{a}_{2} + \left(x_{11}+y_{11}-z_{11}\right) \, \mathbf{a}_{3} & = & -z_{11}a \, \mathbf{\hat{x}} + x_{11}a \, \mathbf{\hat{y}}-y_{11}a \, \mathbf{\hat{z}} & \left(48h\right) & \mbox{H V} \\ 
\mathbf{B}_{95} & = & \left(x_{11}-y_{11}+z_{11}\right) \, \mathbf{a}_{1} + \left(x_{11}+y_{11}-z_{11}\right) \, \mathbf{a}_{2} + \left(-x_{11}+y_{11}+z_{11}\right) \, \mathbf{a}_{3} & = & y_{11}a \, \mathbf{\hat{x}} + z_{11}a \, \mathbf{\hat{y}} + x_{11}a \, \mathbf{\hat{z}} & \left(48h\right) & \mbox{H V} \\ 
\mathbf{B}_{96} & = & \left(-x_{11}+y_{11}+z_{11}\right) \, \mathbf{a}_{1} + \left(-x_{11}-y_{11}-z_{11}\right) \, \mathbf{a}_{2} + \left(x_{11}-y_{11}+z_{11}\right) \, \mathbf{a}_{3} & = & -y_{11}a \, \mathbf{\hat{x}} + z_{11}a \, \mathbf{\hat{y}}-x_{11}a \, \mathbf{\hat{z}} & \left(48h\right) & \mbox{H V} \\ 
\mathbf{B}_{97} & = & \left(-x_{11}-y_{11}-z_{11}\right) \, \mathbf{a}_{1} + \left(-x_{11}+y_{11}+z_{11}\right) \, \mathbf{a}_{2} + \left(x_{11}+y_{11}-z_{11}\right) \, \mathbf{a}_{3} & = & y_{11}a \, \mathbf{\hat{x}}-z_{11}a \, \mathbf{\hat{y}}-x_{11}a \, \mathbf{\hat{z}} & \left(48h\right) & \mbox{H V} \\ 
\mathbf{B}_{98} & = & \left(x_{11}+y_{11}-z_{11}\right) \, \mathbf{a}_{1} + \left(x_{11}-y_{11}+z_{11}\right) \, \mathbf{a}_{2} + \left(-x_{11}-y_{11}-z_{11}\right) \, \mathbf{a}_{3} & = & -y_{11}a \, \mathbf{\hat{x}}-z_{11}a \, \mathbf{\hat{y}} + x_{11}a \, \mathbf{\hat{z}} & \left(48h\right) & \mbox{H V} \\ 
\mathbf{B}_{99} & = & \left(-x_{12}+y_{12}+z_{12}\right) \, \mathbf{a}_{1} + \left(x_{12}-y_{12}+z_{12}\right) \, \mathbf{a}_{2} + \left(x_{12}+y_{12}-z_{12}\right) \, \mathbf{a}_{3} & = & x_{12}a \, \mathbf{\hat{x}} + y_{12}a \, \mathbf{\hat{y}} + z_{12}a \, \mathbf{\hat{z}} & \left(48h\right) & \mbox{H VI} \\ 
\mathbf{B}_{100} & = & \left(x_{12}-y_{12}+z_{12}\right) \, \mathbf{a}_{1} + \left(-x_{12}+y_{12}+z_{12}\right) \, \mathbf{a}_{2} + \left(-x_{12}-y_{12}-z_{12}\right) \, \mathbf{a}_{3} & = & -x_{12}a \, \mathbf{\hat{x}}-y_{12}a \, \mathbf{\hat{y}} + z_{12}a \, \mathbf{\hat{z}} & \left(48h\right) & \mbox{H VI} \\ 
\mathbf{B}_{101} & = & \left(x_{12}+y_{12}-z_{12}\right) \, \mathbf{a}_{1} + \left(-x_{12}-y_{12}-z_{12}\right) \, \mathbf{a}_{2} + \left(-x_{12}+y_{12}+z_{12}\right) \, \mathbf{a}_{3} & = & -x_{12}a \, \mathbf{\hat{x}} + y_{12}a \, \mathbf{\hat{y}}-z_{12}a \, \mathbf{\hat{z}} & \left(48h\right) & \mbox{H VI} \\ 
\mathbf{B}_{102} & = & \left(-x_{12}-y_{12}-z_{12}\right) \, \mathbf{a}_{1} + \left(x_{12}+y_{12}-z_{12}\right) \, \mathbf{a}_{2} + \left(x_{12}-y_{12}+z_{12}\right) \, \mathbf{a}_{3} & = & x_{12}a \, \mathbf{\hat{x}}-y_{12}a \, \mathbf{\hat{y}}-z_{12}a \, \mathbf{\hat{z}} & \left(48h\right) & \mbox{H VI} \\ 
\mathbf{B}_{103} & = & \left(x_{12}+y_{12}-z_{12}\right) \, \mathbf{a}_{1} + \left(-x_{12}+y_{12}+z_{12}\right) \, \mathbf{a}_{2} + \left(x_{12}-y_{12}+z_{12}\right) \, \mathbf{a}_{3} & = & z_{12}a \, \mathbf{\hat{x}} + x_{12}a \, \mathbf{\hat{y}} + y_{12}a \, \mathbf{\hat{z}} & \left(48h\right) & \mbox{H VI} \\ 
\mathbf{B}_{104} & = & \left(-x_{12}-y_{12}-z_{12}\right) \, \mathbf{a}_{1} + \left(x_{12}-y_{12}+z_{12}\right) \, \mathbf{a}_{2} + \left(-x_{12}+y_{12}+z_{12}\right) \, \mathbf{a}_{3} & = & z_{12}a \, \mathbf{\hat{x}}-x_{12}a \, \mathbf{\hat{y}}-y_{12}a \, \mathbf{\hat{z}} & \left(48h\right) & \mbox{H VI} \\ 
\mathbf{B}_{105} & = & \left(-x_{12}+y_{12}+z_{12}\right) \, \mathbf{a}_{1} + \left(x_{12}+y_{12}-z_{12}\right) \, \mathbf{a}_{2} + \left(-x_{12}-y_{12}-z_{12}\right) \, \mathbf{a}_{3} & = & -z_{12}a \, \mathbf{\hat{x}}-x_{12}a \, \mathbf{\hat{y}} + y_{12}a \, \mathbf{\hat{z}} & \left(48h\right) & \mbox{H VI} \\ 
\mathbf{B}_{106} & = & \left(x_{12}-y_{12}+z_{12}\right) \, \mathbf{a}_{1} + \left(-x_{12}-y_{12}-z_{12}\right) \, \mathbf{a}_{2} + \left(x_{12}+y_{12}-z_{12}\right) \, \mathbf{a}_{3} & = & -z_{12}a \, \mathbf{\hat{x}} + x_{12}a \, \mathbf{\hat{y}}-y_{12}a \, \mathbf{\hat{z}} & \left(48h\right) & \mbox{H VI} \\ 
\mathbf{B}_{107} & = & \left(x_{12}-y_{12}+z_{12}\right) \, \mathbf{a}_{1} + \left(x_{12}+y_{12}-z_{12}\right) \, \mathbf{a}_{2} + \left(-x_{12}+y_{12}+z_{12}\right) \, \mathbf{a}_{3} & = & y_{12}a \, \mathbf{\hat{x}} + z_{12}a \, \mathbf{\hat{y}} + x_{12}a \, \mathbf{\hat{z}} & \left(48h\right) & \mbox{H VI} \\ 
\mathbf{B}_{108} & = & \left(-x_{12}+y_{12}+z_{12}\right) \, \mathbf{a}_{1} + \left(-x_{12}-y_{12}-z_{12}\right) \, \mathbf{a}_{2} + \left(x_{12}-y_{12}+z_{12}\right) \, \mathbf{a}_{3} & = & -y_{12}a \, \mathbf{\hat{x}} + z_{12}a \, \mathbf{\hat{y}}-x_{12}a \, \mathbf{\hat{z}} & \left(48h\right) & \mbox{H VI} \\ 
\mathbf{B}_{109} & = & \left(-x_{12}-y_{12}-z_{12}\right) \, \mathbf{a}_{1} + \left(-x_{12}+y_{12}+z_{12}\right) \, \mathbf{a}_{2} + \left(x_{12}+y_{12}-z_{12}\right) \, \mathbf{a}_{3} & = & y_{12}a \, \mathbf{\hat{x}}-z_{12}a \, \mathbf{\hat{y}}-x_{12}a \, \mathbf{\hat{z}} & \left(48h\right) & \mbox{H VI} \\ 
\mathbf{B}_{110} & = & \left(x_{12}+y_{12}-z_{12}\right) \, \mathbf{a}_{1} + \left(x_{12}-y_{12}+z_{12}\right) \, \mathbf{a}_{2} + \left(-x_{12}-y_{12}-z_{12}\right) \, \mathbf{a}_{3} & = & -y_{12}a \, \mathbf{\hat{x}}-z_{12}a \, \mathbf{\hat{y}} + x_{12}a \, \mathbf{\hat{z}} & \left(48h\right) & \mbox{H VI} \\ 
\mathbf{B}_{111} & = & \left(-x_{13}+y_{13}+z_{13}\right) \, \mathbf{a}_{1} + \left(x_{13}-y_{13}+z_{13}\right) \, \mathbf{a}_{2} + \left(x_{13}+y_{13}-z_{13}\right) \, \mathbf{a}_{3} & = & x_{13}a \, \mathbf{\hat{x}} + y_{13}a \, \mathbf{\hat{y}} + z_{13}a \, \mathbf{\hat{z}} & \left(48h\right) & \mbox{O III} \\ 
\mathbf{B}_{112} & = & \left(x_{13}-y_{13}+z_{13}\right) \, \mathbf{a}_{1} + \left(-x_{13}+y_{13}+z_{13}\right) \, \mathbf{a}_{2} + \left(-x_{13}-y_{13}-z_{13}\right) \, \mathbf{a}_{3} & = & -x_{13}a \, \mathbf{\hat{x}}-y_{13}a \, \mathbf{\hat{y}} + z_{13}a \, \mathbf{\hat{z}} & \left(48h\right) & \mbox{O III} \\ 
\mathbf{B}_{113} & = & \left(x_{13}+y_{13}-z_{13}\right) \, \mathbf{a}_{1} + \left(-x_{13}-y_{13}-z_{13}\right) \, \mathbf{a}_{2} + \left(-x_{13}+y_{13}+z_{13}\right) \, \mathbf{a}_{3} & = & -x_{13}a \, \mathbf{\hat{x}} + y_{13}a \, \mathbf{\hat{y}}-z_{13}a \, \mathbf{\hat{z}} & \left(48h\right) & \mbox{O III} \\ 
\mathbf{B}_{114} & = & \left(-x_{13}-y_{13}-z_{13}\right) \, \mathbf{a}_{1} + \left(x_{13}+y_{13}-z_{13}\right) \, \mathbf{a}_{2} + \left(x_{13}-y_{13}+z_{13}\right) \, \mathbf{a}_{3} & = & x_{13}a \, \mathbf{\hat{x}}-y_{13}a \, \mathbf{\hat{y}}-z_{13}a \, \mathbf{\hat{z}} & \left(48h\right) & \mbox{O III} \\ 
\mathbf{B}_{115} & = & \left(x_{13}+y_{13}-z_{13}\right) \, \mathbf{a}_{1} + \left(-x_{13}+y_{13}+z_{13}\right) \, \mathbf{a}_{2} + \left(x_{13}-y_{13}+z_{13}\right) \, \mathbf{a}_{3} & = & z_{13}a \, \mathbf{\hat{x}} + x_{13}a \, \mathbf{\hat{y}} + y_{13}a \, \mathbf{\hat{z}} & \left(48h\right) & \mbox{O III} \\ 
\mathbf{B}_{116} & = & \left(-x_{13}-y_{13}-z_{13}\right) \, \mathbf{a}_{1} + \left(x_{13}-y_{13}+z_{13}\right) \, \mathbf{a}_{2} + \left(-x_{13}+y_{13}+z_{13}\right) \, \mathbf{a}_{3} & = & z_{13}a \, \mathbf{\hat{x}}-x_{13}a \, \mathbf{\hat{y}}-y_{13}a \, \mathbf{\hat{z}} & \left(48h\right) & \mbox{O III} \\ 
\mathbf{B}_{117} & = & \left(-x_{13}+y_{13}+z_{13}\right) \, \mathbf{a}_{1} + \left(x_{13}+y_{13}-z_{13}\right) \, \mathbf{a}_{2} + \left(-x_{13}-y_{13}-z_{13}\right) \, \mathbf{a}_{3} & = & -z_{13}a \, \mathbf{\hat{x}}-x_{13}a \, \mathbf{\hat{y}} + y_{13}a \, \mathbf{\hat{z}} & \left(48h\right) & \mbox{O III} \\ 
\mathbf{B}_{118} & = & \left(x_{13}-y_{13}+z_{13}\right) \, \mathbf{a}_{1} + \left(-x_{13}-y_{13}-z_{13}\right) \, \mathbf{a}_{2} + \left(x_{13}+y_{13}-z_{13}\right) \, \mathbf{a}_{3} & = & -z_{13}a \, \mathbf{\hat{x}} + x_{13}a \, \mathbf{\hat{y}}-y_{13}a \, \mathbf{\hat{z}} & \left(48h\right) & \mbox{O III} \\ 
\mathbf{B}_{119} & = & \left(x_{13}-y_{13}+z_{13}\right) \, \mathbf{a}_{1} + \left(x_{13}+y_{13}-z_{13}\right) \, \mathbf{a}_{2} + \left(-x_{13}+y_{13}+z_{13}\right) \, \mathbf{a}_{3} & = & y_{13}a \, \mathbf{\hat{x}} + z_{13}a \, \mathbf{\hat{y}} + x_{13}a \, \mathbf{\hat{z}} & \left(48h\right) & \mbox{O III} \\ 
\mathbf{B}_{120} & = & \left(-x_{13}+y_{13}+z_{13}\right) \, \mathbf{a}_{1} + \left(-x_{13}-y_{13}-z_{13}\right) \, \mathbf{a}_{2} + \left(x_{13}-y_{13}+z_{13}\right) \, \mathbf{a}_{3} & = & -y_{13}a \, \mathbf{\hat{x}} + z_{13}a \, \mathbf{\hat{y}}-x_{13}a \, \mathbf{\hat{z}} & \left(48h\right) & \mbox{O III} \\ 
\mathbf{B}_{121} & = & \left(-x_{13}-y_{13}-z_{13}\right) \, \mathbf{a}_{1} + \left(-x_{13}+y_{13}+z_{13}\right) \, \mathbf{a}_{2} + \left(x_{13}+y_{13}-z_{13}\right) \, \mathbf{a}_{3} & = & y_{13}a \, \mathbf{\hat{x}}-z_{13}a \, \mathbf{\hat{y}}-x_{13}a \, \mathbf{\hat{z}} & \left(48h\right) & \mbox{O III} \\ 
\mathbf{B}_{122} & = & \left(x_{13}+y_{13}-z_{13}\right) \, \mathbf{a}_{1} + \left(x_{13}-y_{13}+z_{13}\right) \, \mathbf{a}_{2} + \left(-x_{13}-y_{13}-z_{13}\right) \, \mathbf{a}_{3} & = & -y_{13}a \, \mathbf{\hat{x}}-z_{13}a \, \mathbf{\hat{y}} + x_{13}a \, \mathbf{\hat{z}} & \left(48h\right) & \mbox{O III} \\ 
\end{longtabu}
\renewcommand{\arraystretch}{1.0}
\noindent \hrulefill
\\
\textbf{References:}
\vspace*{-0.25cm}
\begin{flushleft}
  - \bibentry{tiritiris_MgB12H12H2O12_ChemInfo_2004}. \\
\end{flushleft}
\textbf{Found in:}
\vspace*{-0.25cm}
\begin{flushleft}
  - \bibentry{Villars_PearsonsCrystalData_2013}. \\
\end{flushleft}
\noindent \hrulefill
\\
\textbf{Geometry files:}
\\
\noindent  - CIF: pp. {\hyperref[A12B36CD12_cF488_196_2h_6h_ac_fgh_cif]{\pageref{A12B36CD12_cF488_196_2h_6h_ac_fgh_cif}}} \\
\noindent  - POSCAR: pp. {\hyperref[A12B36CD12_cF488_196_2h_6h_ac_fgh_poscar]{\pageref{A12B36CD12_cF488_196_2h_6h_ac_fgh_poscar}}} \\
\onecolumn
{\phantomsection\label{ABC3_cP20_198_a_a_b}}
\subsection*{\huge \textbf{{\normalfont \begin{raggedleft}Sodium Chlorate (NaClO$_{3}$, $G3$) Structure: \end{raggedleft} \\ ABC3\_cP20\_198\_a\_a\_b}}}
\noindent \hrulefill
\vspace*{0.25cm}
\begin{figure}[htp]
  \centering
  \vspace{-1em}
  {\includegraphics[width=1\textwidth]{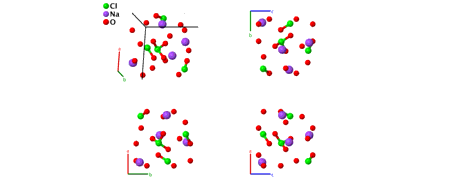}}
\end{figure}
\vspace*{-0.5cm}
\renewcommand{\arraystretch}{1.5}
\begin{equation*}
  \begin{array}{>{$\hspace{-0.15cm}}l<{$}>{$}p{0.5cm}<{$}>{$}p{18.5cm}<{$}}
    \mbox{\large \textbf{Prototype}} &\colon & \ce{NaClO3} \\
    \mbox{\large \textbf{\AFLOW\ prototype label}} &\colon & \mbox{ABC3\_cP20\_198\_a\_a\_b} \\
    \mbox{\large \textbf{\textit{Strukturbericht} designation}} &\colon & \mbox{$G3$} \\
    \mbox{\large \textbf{Pearson symbol}} &\colon & \mbox{cP20} \\
    \mbox{\large \textbf{Space group number}} &\colon & 198 \\
    \mbox{\large \textbf{Space group symbol}} &\colon & P2_{1}3 \\
    \mbox{\large \textbf{\AFLOW\ prototype command}} &\colon &  \texttt{aflow} \,  \, \texttt{-{}-proto=ABC3\_cP20\_198\_a\_a\_b } \, \newline \texttt{-{}-params=}{a,x_{1},x_{2},x_{3},y_{3},z_{3} }
  \end{array}
\end{equation*}
\renewcommand{\arraystretch}{1.0}

\vspace*{-0.25cm}
\noindent \hrulefill
\\
\textbf{ Other compounds with this structure:}
\begin{itemize}
   \item{ NaBrO$_{3}$   }
\end{itemize}
\noindent \parbox{1 \linewidth}{
\noindent \hrulefill
\\
\textbf{Simple Cubic primitive vectors:} \\
\vspace*{-0.25cm}
\begin{tabular}{cc}
  \begin{tabular}{c}
    \parbox{0.6 \linewidth}{
      \renewcommand{\arraystretch}{1.5}
      \begin{equation*}
        \centering
        \begin{array}{ccc}
              \mathbf{a}_1 & = & a \, \mathbf{\hat{x}} \\
    \mathbf{a}_2 & = & a \, \mathbf{\hat{y}} \\
    \mathbf{a}_3 & = & a \, \mathbf{\hat{z}} \\

        \end{array}
      \end{equation*}
    }
    \renewcommand{\arraystretch}{1.0}
  \end{tabular}
  \begin{tabular}{c}
    \includegraphics[width=0.3\linewidth]{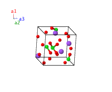} \\
  \end{tabular}
\end{tabular}

}
\vspace*{-0.25cm}

\noindent \hrulefill
\\
\textbf{Basis vectors:}
\vspace*{-0.25cm}
\renewcommand{\arraystretch}{1.5}
\begin{longtabu} to \textwidth{>{\centering $}X[-1,c,c]<{$}>{\centering $}X[-1,c,c]<{$}>{\centering $}X[-1,c,c]<{$}>{\centering $}X[-1,c,c]<{$}>{\centering $}X[-1,c,c]<{$}>{\centering $}X[-1,c,c]<{$}>{\centering $}X[-1,c,c]<{$}}
  & & \mbox{Lattice Coordinates} & & \mbox{Cartesian Coordinates} &\mbox{Wyckoff Position} & \mbox{Atom Type} \\  
  \mathbf{B}_{1} & = & x_{1} \, \mathbf{a}_{1} + x_{1} \, \mathbf{a}_{2} + x_{1} \, \mathbf{a}_{3} & = & x_{1}a \, \mathbf{\hat{x}} + x_{1}a \, \mathbf{\hat{y}} + x_{1}a \, \mathbf{\hat{z}} & \left(4a\right) & \mbox{Cl} \\ 
\mathbf{B}_{2} & = & \left(\frac{1}{2} - x_{1}\right) \, \mathbf{a}_{1}-x_{1} \, \mathbf{a}_{2} + \left(\frac{1}{2} +x_{1}\right) \, \mathbf{a}_{3} & = & \left(\frac{1}{2} - x_{1}\right)a \, \mathbf{\hat{x}}-x_{1}a \, \mathbf{\hat{y}} + \left(\frac{1}{2} +x_{1}\right)a \, \mathbf{\hat{z}} & \left(4a\right) & \mbox{Cl} \\ 
\mathbf{B}_{3} & = & -x_{1} \, \mathbf{a}_{1} + \left(\frac{1}{2} +x_{1}\right) \, \mathbf{a}_{2} + \left(\frac{1}{2} - x_{1}\right) \, \mathbf{a}_{3} & = & -x_{1}a \, \mathbf{\hat{x}} + \left(\frac{1}{2} +x_{1}\right)a \, \mathbf{\hat{y}} + \left(\frac{1}{2} - x_{1}\right)a \, \mathbf{\hat{z}} & \left(4a\right) & \mbox{Cl} \\ 
\mathbf{B}_{4} & = & \left(\frac{1}{2} +x_{1}\right) \, \mathbf{a}_{1} + \left(\frac{1}{2} - x_{1}\right) \, \mathbf{a}_{2}-x_{1} \, \mathbf{a}_{3} & = & \left(\frac{1}{2} +x_{1}\right)a \, \mathbf{\hat{x}} + \left(\frac{1}{2} - x_{1}\right)a \, \mathbf{\hat{y}}-x_{1}a \, \mathbf{\hat{z}} & \left(4a\right) & \mbox{Cl} \\ 
\mathbf{B}_{5} & = & x_{2} \, \mathbf{a}_{1} + x_{2} \, \mathbf{a}_{2} + x_{2} \, \mathbf{a}_{3} & = & x_{2}a \, \mathbf{\hat{x}} + x_{2}a \, \mathbf{\hat{y}} + x_{2}a \, \mathbf{\hat{z}} & \left(4a\right) & \mbox{Na} \\ 
\mathbf{B}_{6} & = & \left(\frac{1}{2} - x_{2}\right) \, \mathbf{a}_{1}-x_{2} \, \mathbf{a}_{2} + \left(\frac{1}{2} +x_{2}\right) \, \mathbf{a}_{3} & = & \left(\frac{1}{2} - x_{2}\right)a \, \mathbf{\hat{x}}-x_{2}a \, \mathbf{\hat{y}} + \left(\frac{1}{2} +x_{2}\right)a \, \mathbf{\hat{z}} & \left(4a\right) & \mbox{Na} \\ 
\mathbf{B}_{7} & = & -x_{2} \, \mathbf{a}_{1} + \left(\frac{1}{2} +x_{2}\right) \, \mathbf{a}_{2} + \left(\frac{1}{2} - x_{2}\right) \, \mathbf{a}_{3} & = & -x_{2}a \, \mathbf{\hat{x}} + \left(\frac{1}{2} +x_{2}\right)a \, \mathbf{\hat{y}} + \left(\frac{1}{2} - x_{2}\right)a \, \mathbf{\hat{z}} & \left(4a\right) & \mbox{Na} \\ 
\mathbf{B}_{8} & = & \left(\frac{1}{2} +x_{2}\right) \, \mathbf{a}_{1} + \left(\frac{1}{2} - x_{2}\right) \, \mathbf{a}_{2}-x_{2} \, \mathbf{a}_{3} & = & \left(\frac{1}{2} +x_{2}\right)a \, \mathbf{\hat{x}} + \left(\frac{1}{2} - x_{2}\right)a \, \mathbf{\hat{y}}-x_{2}a \, \mathbf{\hat{z}} & \left(4a\right) & \mbox{Na} \\ 
\mathbf{B}_{9} & = & x_{3} \, \mathbf{a}_{1} + y_{3} \, \mathbf{a}_{2} + z_{3} \, \mathbf{a}_{3} & = & x_{3}a \, \mathbf{\hat{x}} + y_{3}a \, \mathbf{\hat{y}} + z_{3}a \, \mathbf{\hat{z}} & \left(12b\right) & \mbox{O} \\ 
\mathbf{B}_{10} & = & \left(\frac{1}{2} - x_{3}\right) \, \mathbf{a}_{1}-y_{3} \, \mathbf{a}_{2} + \left(\frac{1}{2} +z_{3}\right) \, \mathbf{a}_{3} & = & \left(\frac{1}{2} - x_{3}\right)a \, \mathbf{\hat{x}}-y_{3}a \, \mathbf{\hat{y}} + \left(\frac{1}{2} +z_{3}\right)a \, \mathbf{\hat{z}} & \left(12b\right) & \mbox{O} \\ 
\mathbf{B}_{11} & = & -x_{3} \, \mathbf{a}_{1} + \left(\frac{1}{2} +y_{3}\right) \, \mathbf{a}_{2} + \left(\frac{1}{2} - z_{3}\right) \, \mathbf{a}_{3} & = & -x_{3}a \, \mathbf{\hat{x}} + \left(\frac{1}{2} +y_{3}\right)a \, \mathbf{\hat{y}} + \left(\frac{1}{2} - z_{3}\right)a \, \mathbf{\hat{z}} & \left(12b\right) & \mbox{O} \\ 
\mathbf{B}_{12} & = & \left(\frac{1}{2} +x_{3}\right) \, \mathbf{a}_{1} + \left(\frac{1}{2} - y_{3}\right) \, \mathbf{a}_{2}-z_{3} \, \mathbf{a}_{3} & = & \left(\frac{1}{2} +x_{3}\right)a \, \mathbf{\hat{x}} + \left(\frac{1}{2} - y_{3}\right)a \, \mathbf{\hat{y}}-z_{3}a \, \mathbf{\hat{z}} & \left(12b\right) & \mbox{O} \\ 
\mathbf{B}_{13} & = & z_{3} \, \mathbf{a}_{1} + x_{3} \, \mathbf{a}_{2} + y_{3} \, \mathbf{a}_{3} & = & z_{3}a \, \mathbf{\hat{x}} + x_{3}a \, \mathbf{\hat{y}} + y_{3}a \, \mathbf{\hat{z}} & \left(12b\right) & \mbox{O} \\ 
\mathbf{B}_{14} & = & \left(\frac{1}{2} +z_{3}\right) \, \mathbf{a}_{1} + \left(\frac{1}{2} - x_{3}\right) \, \mathbf{a}_{2}-y_{3} \, \mathbf{a}_{3} & = & \left(\frac{1}{2} +z_{3}\right)a \, \mathbf{\hat{x}} + \left(\frac{1}{2} - x_{3}\right)a \, \mathbf{\hat{y}}-y_{3}a \, \mathbf{\hat{z}} & \left(12b\right) & \mbox{O} \\ 
\mathbf{B}_{15} & = & \left(\frac{1}{2} - z_{3}\right) \, \mathbf{a}_{1}-x_{3} \, \mathbf{a}_{2} + \left(\frac{1}{2} +y_{3}\right) \, \mathbf{a}_{3} & = & \left(\frac{1}{2} - z_{3}\right)a \, \mathbf{\hat{x}}-x_{3}a \, \mathbf{\hat{y}} + \left(\frac{1}{2} +y_{3}\right)a \, \mathbf{\hat{z}} & \left(12b\right) & \mbox{O} \\ 
\mathbf{B}_{16} & = & -z_{3} \, \mathbf{a}_{1} + \left(\frac{1}{2} +x_{3}\right) \, \mathbf{a}_{2} + \left(\frac{1}{2} - y_{3}\right) \, \mathbf{a}_{3} & = & -z_{3}a \, \mathbf{\hat{x}} + \left(\frac{1}{2} +x_{3}\right)a \, \mathbf{\hat{y}} + \left(\frac{1}{2} - y_{3}\right)a \, \mathbf{\hat{z}} & \left(12b\right) & \mbox{O} \\ 
\mathbf{B}_{17} & = & y_{3} \, \mathbf{a}_{1} + z_{3} \, \mathbf{a}_{2} + x_{3} \, \mathbf{a}_{3} & = & y_{3}a \, \mathbf{\hat{x}} + z_{3}a \, \mathbf{\hat{y}} + x_{3}a \, \mathbf{\hat{z}} & \left(12b\right) & \mbox{O} \\ 
\mathbf{B}_{18} & = & -y_{3} \, \mathbf{a}_{1} + \left(\frac{1}{2} +z_{3}\right) \, \mathbf{a}_{2} + \left(\frac{1}{2} - x_{3}\right) \, \mathbf{a}_{3} & = & -y_{3}a \, \mathbf{\hat{x}} + \left(\frac{1}{2} +z_{3}\right)a \, \mathbf{\hat{y}} + \left(\frac{1}{2} - x_{3}\right)a \, \mathbf{\hat{z}} & \left(12b\right) & \mbox{O} \\ 
\mathbf{B}_{19} & = & \left(\frac{1}{2} +y_{3}\right) \, \mathbf{a}_{1} + \left(\frac{1}{2} - z_{3}\right) \, \mathbf{a}_{2}-x_{3} \, \mathbf{a}_{3} & = & \left(\frac{1}{2} +y_{3}\right)a \, \mathbf{\hat{x}} + \left(\frac{1}{2} - z_{3}\right)a \, \mathbf{\hat{y}}-x_{3}a \, \mathbf{\hat{z}} & \left(12b\right) & \mbox{O} \\ 
\mathbf{B}_{20} & = & \left(\frac{1}{2} - y_{3}\right) \, \mathbf{a}_{1}-z_{3} \, \mathbf{a}_{2} + \left(\frac{1}{2} +x_{3}\right) \, \mathbf{a}_{3} & = & \left(\frac{1}{2} - y_{3}\right)a \, \mathbf{\hat{x}}-z_{3}a \, \mathbf{\hat{y}} + \left(\frac{1}{2} +x_{3}\right)a \, \mathbf{\hat{z}} & \left(12b\right) & \mbox{O} \\ 
\end{longtabu}
\renewcommand{\arraystretch}{1.0}
\noindent \hrulefill
\\
\textbf{References:}
\vspace*{-0.25cm}
\begin{flushleft}
  - \bibentry{Ramachandran_Acta_Cryst_10_1957}. \\
\end{flushleft}
\textbf{Found in:}
\vspace*{-0.25cm}
\begin{flushleft}
  - \bibentry{Kaminskii_App_Phys_B_67_1998}. \\
\end{flushleft}
\noindent \hrulefill
\\
\textbf{Geometry files:}
\\
\noindent  - CIF: pp. {\hyperref[ABC3_cP20_198_a_a_b_cif]{\pageref{ABC3_cP20_198_a_a_b_cif}}} \\
\noindent  - POSCAR: pp. {\hyperref[ABC3_cP20_198_a_a_b_poscar]{\pageref{ABC3_cP20_198_a_a_b_poscar}}} \\
\onecolumn
{\phantomsection\label{A2B11_cP39_200_f_aghij}}
\subsection*{\huge \textbf{{\normalfont Mg$_{2}$Zn$_{11}$ Structure: A2B11\_cP39\_200\_f\_aghij}}}
\noindent \hrulefill
\vspace*{0.25cm}
\begin{figure}[htp]
  \centering
  \vspace{-1em}
  {\includegraphics[width=1\textwidth]{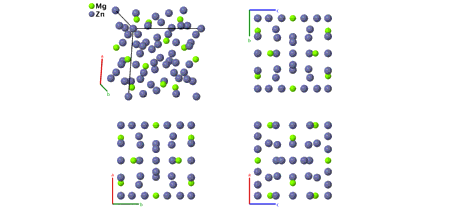}}
\end{figure}
\vspace*{-0.5cm}
\renewcommand{\arraystretch}{1.5}
\begin{equation*}
  \begin{array}{>{$\hspace{-0.15cm}}l<{$}>{$}p{0.5cm}<{$}>{$}p{18.5cm}<{$}}
    \mbox{\large \textbf{Prototype}} &\colon & \ce{Mg2Zn11} \\
    \mbox{\large \textbf{\AFLOW\ prototype label}} &\colon & \mbox{A2B11\_cP39\_200\_f\_aghij} \\
    \mbox{\large \textbf{\textit{Strukturbericht} designation}} &\colon & \mbox{None} \\
    \mbox{\large \textbf{Pearson symbol}} &\colon & \mbox{cP39} \\
    \mbox{\large \textbf{Space group number}} &\colon & 200 \\
    \mbox{\large \textbf{Space group symbol}} &\colon & Pm\bar{3} \\
    \mbox{\large \textbf{\AFLOW\ prototype command}} &\colon &  \texttt{aflow} \,  \, \texttt{-{}-proto=A2B11\_cP39\_200\_f\_aghij } \, \newline \texttt{-{}-params=}{a,x_{2},x_{3},x_{4},x_{5},y_{6},z_{6} }
  \end{array}
\end{equation*}
\renewcommand{\arraystretch}{1.0}

\noindent \parbox{1 \linewidth}{
\noindent \hrulefill
\\
\textbf{Simple Cubic primitive vectors:} \\
\vspace*{-0.25cm}
\begin{tabular}{cc}
  \begin{tabular}{c}
    \parbox{0.6 \linewidth}{
      \renewcommand{\arraystretch}{1.5}
      \begin{equation*}
        \centering
        \begin{array}{ccc}
              \mathbf{a}_1 & = & a \, \mathbf{\hat{x}} \\
    \mathbf{a}_2 & = & a \, \mathbf{\hat{y}} \\
    \mathbf{a}_3 & = & a \, \mathbf{\hat{z}} \\

        \end{array}
      \end{equation*}
    }
    \renewcommand{\arraystretch}{1.0}
  \end{tabular}
  \begin{tabular}{c}
    \includegraphics[width=0.3\linewidth]{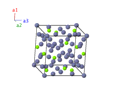} \\
  \end{tabular}
\end{tabular}

}
\vspace*{-0.25cm}

\noindent \hrulefill
\\
\textbf{Basis vectors:}
\vspace*{-0.25cm}
\renewcommand{\arraystretch}{1.5}
\begin{longtabu} to \textwidth{>{\centering $}X[-1,c,c]<{$}>{\centering $}X[-1,c,c]<{$}>{\centering $}X[-1,c,c]<{$}>{\centering $}X[-1,c,c]<{$}>{\centering $}X[-1,c,c]<{$}>{\centering $}X[-1,c,c]<{$}>{\centering $}X[-1,c,c]<{$}}
  & & \mbox{Lattice Coordinates} & & \mbox{Cartesian Coordinates} &\mbox{Wyckoff Position} & \mbox{Atom Type} \\  
  \mathbf{B}_{1} & = & 0 \, \mathbf{a}_{1} + 0 \, \mathbf{a}_{2} + 0 \, \mathbf{a}_{3} & = & 0 \, \mathbf{\hat{x}} + 0 \, \mathbf{\hat{y}} + 0 \, \mathbf{\hat{z}} & \left(1a\right) & \mbox{Zn I} \\ 
\mathbf{B}_{2} & = & x_{2} \, \mathbf{a}_{1} + \frac{1}{2} \, \mathbf{a}_{3} & = & x_{2}a \, \mathbf{\hat{x}} + \frac{1}{2}a \, \mathbf{\hat{z}} & \left(6f\right) & \mbox{Mg} \\ 
\mathbf{B}_{3} & = & -x_{2} \, \mathbf{a}_{1} + \frac{1}{2} \, \mathbf{a}_{3} & = & -x_{2}a \, \mathbf{\hat{x}} + \frac{1}{2}a \, \mathbf{\hat{z}} & \left(6f\right) & \mbox{Mg} \\ 
\mathbf{B}_{4} & = & \frac{1}{2} \, \mathbf{a}_{1} + x_{2} \, \mathbf{a}_{2} & = & \frac{1}{2}a \, \mathbf{\hat{x}} + x_{2}a \, \mathbf{\hat{y}} & \left(6f\right) & \mbox{Mg} \\ 
\mathbf{B}_{5} & = & \frac{1}{2} \, \mathbf{a}_{1}-x_{2} \, \mathbf{a}_{2} & = & \frac{1}{2}a \, \mathbf{\hat{x}}-x_{2}a \, \mathbf{\hat{y}} & \left(6f\right) & \mbox{Mg} \\ 
\mathbf{B}_{6} & = & \frac{1}{2} \, \mathbf{a}_{2} + x_{2} \, \mathbf{a}_{3} & = & \frac{1}{2}a \, \mathbf{\hat{y}} + x_{2}a \, \mathbf{\hat{z}} & \left(6f\right) & \mbox{Mg} \\ 
\mathbf{B}_{7} & = & \frac{1}{2} \, \mathbf{a}_{2}-x_{2} \, \mathbf{a}_{3} & = & \frac{1}{2}a \, \mathbf{\hat{y}}-x_{2}a \, \mathbf{\hat{z}} & \left(6f\right) & \mbox{Mg} \\ 
\mathbf{B}_{8} & = & x_{3} \, \mathbf{a}_{1} + \frac{1}{2} \, \mathbf{a}_{2} & = & x_{3}a \, \mathbf{\hat{x}} + \frac{1}{2}a \, \mathbf{\hat{y}} & \left(6g\right) & \mbox{Zn II} \\ 
\mathbf{B}_{9} & = & -x_{3} \, \mathbf{a}_{1} + \frac{1}{2} \, \mathbf{a}_{2} & = & -x_{3}a \, \mathbf{\hat{x}} + \frac{1}{2}a \, \mathbf{\hat{y}} & \left(6g\right) & \mbox{Zn II} \\ 
\mathbf{B}_{10} & = & x_{3} \, \mathbf{a}_{2} + \frac{1}{2} \, \mathbf{a}_{3} & = & x_{3}a \, \mathbf{\hat{y}} + \frac{1}{2}a \, \mathbf{\hat{z}} & \left(6g\right) & \mbox{Zn II} \\ 
\mathbf{B}_{11} & = & -x_{3} \, \mathbf{a}_{2} + \frac{1}{2} \, \mathbf{a}_{3} & = & -x_{3}a \, \mathbf{\hat{y}} + \frac{1}{2}a \, \mathbf{\hat{z}} & \left(6g\right) & \mbox{Zn II} \\ 
\mathbf{B}_{12} & = & \frac{1}{2} \, \mathbf{a}_{1} + x_{3} \, \mathbf{a}_{3} & = & \frac{1}{2}a \, \mathbf{\hat{x}} + x_{3}a \, \mathbf{\hat{z}} & \left(6g\right) & \mbox{Zn II} \\ 
\mathbf{B}_{13} & = & \frac{1}{2} \, \mathbf{a}_{1} + -x_{3} \, \mathbf{a}_{3} & = & \frac{1}{2}a \, \mathbf{\hat{x}} + -x_{3}a \, \mathbf{\hat{z}} & \left(6g\right) & \mbox{Zn II} \\ 
\mathbf{B}_{14} & = & x_{4} \, \mathbf{a}_{1} + \frac{1}{2} \, \mathbf{a}_{2} + \frac{1}{2} \, \mathbf{a}_{3} & = & x_{4}a \, \mathbf{\hat{x}} + \frac{1}{2}a \, \mathbf{\hat{y}} + \frac{1}{2}a \, \mathbf{\hat{z}} & \left(6h\right) & \mbox{Zn III} \\ 
\mathbf{B}_{15} & = & -x_{4} \, \mathbf{a}_{1} + \frac{1}{2} \, \mathbf{a}_{2} + \frac{1}{2} \, \mathbf{a}_{3} & = & -x_{4}a \, \mathbf{\hat{x}} + \frac{1}{2}a \, \mathbf{\hat{y}} + \frac{1}{2}a \, \mathbf{\hat{z}} & \left(6h\right) & \mbox{Zn III} \\ 
\mathbf{B}_{16} & = & \frac{1}{2} \, \mathbf{a}_{1} + x_{4} \, \mathbf{a}_{2} + \frac{1}{2} \, \mathbf{a}_{3} & = & \frac{1}{2}a \, \mathbf{\hat{x}} + x_{4}a \, \mathbf{\hat{y}} + \frac{1}{2}a \, \mathbf{\hat{z}} & \left(6h\right) & \mbox{Zn III} \\ 
\mathbf{B}_{17} & = & \frac{1}{2} \, \mathbf{a}_{1}-x_{4} \, \mathbf{a}_{2} + \frac{1}{2} \, \mathbf{a}_{3} & = & \frac{1}{2}a \, \mathbf{\hat{x}}-x_{4}a \, \mathbf{\hat{y}} + \frac{1}{2}a \, \mathbf{\hat{z}} & \left(6h\right) & \mbox{Zn III} \\ 
\mathbf{B}_{18} & = & \frac{1}{2} \, \mathbf{a}_{1} + \frac{1}{2} \, \mathbf{a}_{2} + x_{4} \, \mathbf{a}_{3} & = & \frac{1}{2}a \, \mathbf{\hat{x}} + \frac{1}{2}a \, \mathbf{\hat{y}} + x_{4}a \, \mathbf{\hat{z}} & \left(6h\right) & \mbox{Zn III} \\ 
\mathbf{B}_{19} & = & \frac{1}{2} \, \mathbf{a}_{1} + \frac{1}{2} \, \mathbf{a}_{2}-x_{4} \, \mathbf{a}_{3} & = & \frac{1}{2}a \, \mathbf{\hat{x}} + \frac{1}{2}a \, \mathbf{\hat{y}}-x_{4}a \, \mathbf{\hat{z}} & \left(6h\right) & \mbox{Zn III} \\ 
\mathbf{B}_{20} & = & x_{5} \, \mathbf{a}_{1} + x_{5} \, \mathbf{a}_{2} + x_{5} \, \mathbf{a}_{3} & = & x_{5}a \, \mathbf{\hat{x}} + x_{5}a \, \mathbf{\hat{y}} + x_{5}a \, \mathbf{\hat{z}} & \left(8i\right) & \mbox{Zn IV} \\ 
\mathbf{B}_{21} & = & -x_{5} \, \mathbf{a}_{1}-x_{5} \, \mathbf{a}_{2} + x_{5} \, \mathbf{a}_{3} & = & -x_{5}a \, \mathbf{\hat{x}}-x_{5}a \, \mathbf{\hat{y}} + x_{5}a \, \mathbf{\hat{z}} & \left(8i\right) & \mbox{Zn IV} \\ 
\mathbf{B}_{22} & = & -x_{5} \, \mathbf{a}_{1} + x_{5} \, \mathbf{a}_{2}-x_{5} \, \mathbf{a}_{3} & = & -x_{5}a \, \mathbf{\hat{x}} + x_{5}a \, \mathbf{\hat{y}}-x_{5}a \, \mathbf{\hat{z}} & \left(8i\right) & \mbox{Zn IV} \\ 
\mathbf{B}_{23} & = & x_{5} \, \mathbf{a}_{1}-x_{5} \, \mathbf{a}_{2}-x_{5} \, \mathbf{a}_{3} & = & x_{5}a \, \mathbf{\hat{x}}-x_{5}a \, \mathbf{\hat{y}}-x_{5}a \, \mathbf{\hat{z}} & \left(8i\right) & \mbox{Zn IV} \\ 
\mathbf{B}_{24} & = & -x_{5} \, \mathbf{a}_{1}-x_{5} \, \mathbf{a}_{2}-x_{5} \, \mathbf{a}_{3} & = & -x_{5}a \, \mathbf{\hat{x}}-x_{5}a \, \mathbf{\hat{y}}-x_{5}a \, \mathbf{\hat{z}} & \left(8i\right) & \mbox{Zn IV} \\ 
\mathbf{B}_{25} & = & x_{5} \, \mathbf{a}_{1} + x_{5} \, \mathbf{a}_{2}-x_{5} \, \mathbf{a}_{3} & = & x_{5}a \, \mathbf{\hat{x}} + x_{5}a \, \mathbf{\hat{y}}-x_{5}a \, \mathbf{\hat{z}} & \left(8i\right) & \mbox{Zn IV} \\ 
\mathbf{B}_{26} & = & x_{5} \, \mathbf{a}_{1}-x_{5} \, \mathbf{a}_{2} + x_{5} \, \mathbf{a}_{3} & = & x_{5}a \, \mathbf{\hat{x}}-x_{5}a \, \mathbf{\hat{y}} + x_{5}a \, \mathbf{\hat{z}} & \left(8i\right) & \mbox{Zn IV} \\ 
\mathbf{B}_{27} & = & -x_{5} \, \mathbf{a}_{1} + x_{5} \, \mathbf{a}_{2} + x_{5} \, \mathbf{a}_{3} & = & -x_{5}a \, \mathbf{\hat{x}} + x_{5}a \, \mathbf{\hat{y}} + x_{5}a \, \mathbf{\hat{z}} & \left(8i\right) & \mbox{Zn IV} \\ 
\mathbf{B}_{28} & = & y_{6} \, \mathbf{a}_{2} + z_{6} \, \mathbf{a}_{3} & = & y_{6}a \, \mathbf{\hat{y}} + z_{6}a \, \mathbf{\hat{z}} & \left(12j\right) & \mbox{Zn V} \\ 
\mathbf{B}_{29} & = & -y_{6} \, \mathbf{a}_{2} + z_{6} \, \mathbf{a}_{3} & = & -y_{6}a \, \mathbf{\hat{y}} + z_{6}a \, \mathbf{\hat{z}} & \left(12j\right) & \mbox{Zn V} \\ 
\mathbf{B}_{30} & = & y_{6} \, \mathbf{a}_{2}-z_{6} \, \mathbf{a}_{3} & = & y_{6}a \, \mathbf{\hat{y}}-z_{6}a \, \mathbf{\hat{z}} & \left(12j\right) & \mbox{Zn V} \\ 
\mathbf{B}_{31} & = & -y_{6} \, \mathbf{a}_{2}-z_{6} \, \mathbf{a}_{3} & = & -y_{6}a \, \mathbf{\hat{y}}-z_{6}a \, \mathbf{\hat{z}} & \left(12j\right) & \mbox{Zn V} \\ 
\mathbf{B}_{32} & = & z_{6} \, \mathbf{a}_{1} + y_{6} \, \mathbf{a}_{3} & = & z_{6}a \, \mathbf{\hat{x}} + y_{6}a \, \mathbf{\hat{z}} & \left(12j\right) & \mbox{Zn V} \\ 
\mathbf{B}_{33} & = & z_{6} \, \mathbf{a}_{1} + -y_{6} \, \mathbf{a}_{3} & = & z_{6}a \, \mathbf{\hat{x}} + -y_{6}a \, \mathbf{\hat{z}} & \left(12j\right) & \mbox{Zn V} \\ 
\mathbf{B}_{34} & = & -z_{6} \, \mathbf{a}_{1} + y_{6} \, \mathbf{a}_{3} & = & -z_{6}a \, \mathbf{\hat{x}} + y_{6}a \, \mathbf{\hat{z}} & \left(12j\right) & \mbox{Zn V} \\ 
\mathbf{B}_{35} & = & -z_{6} \, \mathbf{a}_{1} + -y_{6} \, \mathbf{a}_{3} & = & -z_{6}a \, \mathbf{\hat{x}} + -y_{6}a \, \mathbf{\hat{z}} & \left(12j\right) & \mbox{Zn V} \\ 
\mathbf{B}_{36} & = & y_{6} \, \mathbf{a}_{1} + z_{6} \, \mathbf{a}_{2} & = & y_{6}a \, \mathbf{\hat{x}} + z_{6}a \, \mathbf{\hat{y}} & \left(12j\right) & \mbox{Zn V} \\ 
\mathbf{B}_{37} & = & -y_{6} \, \mathbf{a}_{1} + z_{6} \, \mathbf{a}_{2} & = & -y_{6}a \, \mathbf{\hat{x}} + z_{6}a \, \mathbf{\hat{y}} & \left(12j\right) & \mbox{Zn V} \\ 
\mathbf{B}_{38} & = & y_{6} \, \mathbf{a}_{1}-z_{6} \, \mathbf{a}_{2} & = & y_{6}a \, \mathbf{\hat{x}}-z_{6}a \, \mathbf{\hat{y}} & \left(12j\right) & \mbox{Zn V} \\ 
\mathbf{B}_{39} & = & -y_{6} \, \mathbf{a}_{1}-z_{6} \, \mathbf{a}_{2} & = & -y_{6}a \, \mathbf{\hat{x}}-z_{6}a \, \mathbf{\hat{y}} & \left(12j\right) & \mbox{Zn V} \\ 
\end{longtabu}
\renewcommand{\arraystretch}{1.0}
\noindent \hrulefill
\\
\textbf{References:}
\vspace*{-0.25cm}
\begin{flushleft}
  - \bibentry{Samson_Mg2Zn11_ActChemScand_1949}. \\
\end{flushleft}
\textbf{Found in:}
\vspace*{-0.25cm}
\begin{flushleft}
  - \bibentry{Villars_PearsonsCrystalData_2013}. \\
\end{flushleft}
\noindent \hrulefill
\\
\textbf{Geometry files:}
\\
\noindent  - CIF: pp. {\hyperref[A2B11_cP39_200_f_aghij_cif]{\pageref{A2B11_cP39_200_f_aghij_cif}}} \\
\noindent  - POSCAR: pp. {\hyperref[A2B11_cP39_200_f_aghij_poscar]{\pageref{A2B11_cP39_200_f_aghij_poscar}}} \\
\onecolumn
{\phantomsection\label{AB3C_cP60_201_ce_fh_g}}
\subsection*{\huge \textbf{{\normalfont \begin{raggedleft}KSbO$_{3}$ (High-temperature) Structure: \end{raggedleft} \\ AB3C\_cP60\_201\_ce\_fh\_g}}}
\noindent \hrulefill
\vspace*{0.25cm}
\begin{figure}[htp]
  \centering
  \vspace{-1em}
  {\includegraphics[width=1\textwidth]{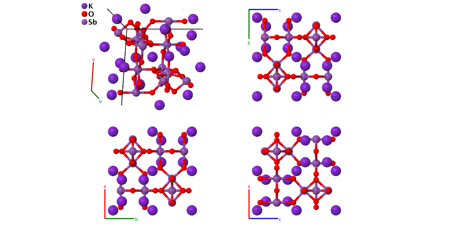}}
\end{figure}
\vspace*{-0.5cm}
\renewcommand{\arraystretch}{1.5}
\begin{equation*}
  \begin{array}{>{$\hspace{-0.15cm}}l<{$}>{$}p{0.5cm}<{$}>{$}p{18.5cm}<{$}}
    \mbox{\large \textbf{Prototype}} &\colon & \ce{KSbO3} \\
    \mbox{\large \textbf{\AFLOW\ prototype label}} &\colon & \mbox{AB3C\_cP60\_201\_ce\_fh\_g} \\
    \mbox{\large \textbf{\textit{Strukturbericht} designation}} &\colon & \mbox{None} \\
    \mbox{\large \textbf{Pearson symbol}} &\colon & \mbox{cP60} \\
    \mbox{\large \textbf{Space group number}} &\colon & 201 \\
    \mbox{\large \textbf{Space group symbol}} &\colon & Pn\bar{3} \\
    \mbox{\large \textbf{\AFLOW\ prototype command}} &\colon &  \texttt{aflow} \,  \, \texttt{-{}-proto=AB3C\_cP60\_201\_ce\_fh\_g } \, \newline \texttt{-{}-params=}{a,x_{2},x_{3},x_{4},x_{5},y_{5},z_{5} }
  \end{array}
\end{equation*}
\renewcommand{\arraystretch}{1.0}

\noindent \parbox{1 \linewidth}{
\noindent \hrulefill
\\
\textbf{Simple Cubic primitive vectors:} \\
\vspace*{-0.25cm}
\begin{tabular}{cc}
  \begin{tabular}{c}
    \parbox{0.6 \linewidth}{
      \renewcommand{\arraystretch}{1.5}
      \begin{equation*}
        \centering
        \begin{array}{ccc}
              \mathbf{a}_1 & = & a \, \mathbf{\hat{x}} \\
    \mathbf{a}_2 & = & a \, \mathbf{\hat{y}} \\
    \mathbf{a}_3 & = & a \, \mathbf{\hat{z}} \\

        \end{array}
      \end{equation*}
    }
    \renewcommand{\arraystretch}{1.0}
  \end{tabular}
  \begin{tabular}{c}
    \includegraphics[width=0.3\linewidth]{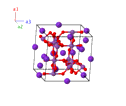} \\
  \end{tabular}
\end{tabular}

}
\vspace*{-0.25cm}

\noindent \hrulefill
\\
\textbf{Basis vectors:}
\vspace*{-0.25cm}
\renewcommand{\arraystretch}{1.5}
\begin{longtabu} to \textwidth{>{\centering $}X[-1,c,c]<{$}>{\centering $}X[-1,c,c]<{$}>{\centering $}X[-1,c,c]<{$}>{\centering $}X[-1,c,c]<{$}>{\centering $}X[-1,c,c]<{$}>{\centering $}X[-1,c,c]<{$}>{\centering $}X[-1,c,c]<{$}}
  & & \mbox{Lattice Coordinates} & & \mbox{Cartesian Coordinates} &\mbox{Wyckoff Position} & \mbox{Atom Type} \\  
  \mathbf{B}_{1} & = & \frac{1}{2} \, \mathbf{a}_{1} + \frac{1}{2} \, \mathbf{a}_{2} + \frac{1}{2} \, \mathbf{a}_{3} & = & \frac{1}{2}a \, \mathbf{\hat{x}} + \frac{1}{2}a \, \mathbf{\hat{y}} + \frac{1}{2}a \, \mathbf{\hat{z}} & \left(4c\right) & \mbox{K I} \\ 
\mathbf{B}_{2} & = & \frac{1}{2} \, \mathbf{a}_{3} & = & \frac{1}{2}a \, \mathbf{\hat{z}} & \left(4c\right) & \mbox{K I} \\ 
\mathbf{B}_{3} & = & \frac{1}{2} \, \mathbf{a}_{2} & = & \frac{1}{2}a \, \mathbf{\hat{y}} & \left(4c\right) & \mbox{K I} \\ 
\mathbf{B}_{4} & = & \frac{1}{2} \, \mathbf{a}_{1} & = & \frac{1}{2}a \, \mathbf{\hat{x}} & \left(4c\right) & \mbox{K I} \\ 
\mathbf{B}_{5} & = & x_{2} \, \mathbf{a}_{1} + x_{2} \, \mathbf{a}_{2} + x_{2} \, \mathbf{a}_{3} & = & x_{2}a \, \mathbf{\hat{x}} + x_{2}a \, \mathbf{\hat{y}} + x_{2}a \, \mathbf{\hat{z}} & \left(8e\right) & \mbox{K II} \\ 
\mathbf{B}_{6} & = & \left(\frac{1}{2} - x_{2}\right) \, \mathbf{a}_{1} + \left(\frac{1}{2} - x_{2}\right) \, \mathbf{a}_{2} + x_{2} \, \mathbf{a}_{3} & = & \left(\frac{1}{2} - x_{2}\right)a \, \mathbf{\hat{x}} + \left(\frac{1}{2} - x_{2}\right)a \, \mathbf{\hat{y}} + x_{2}a \, \mathbf{\hat{z}} & \left(8e\right) & \mbox{K II} \\ 
\mathbf{B}_{7} & = & \left(\frac{1}{2} - x_{2}\right) \, \mathbf{a}_{1} + x_{2} \, \mathbf{a}_{2} + \left(\frac{1}{2} - x_{2}\right) \, \mathbf{a}_{3} & = & \left(\frac{1}{2} - x_{2}\right)a \, \mathbf{\hat{x}} + x_{2}a \, \mathbf{\hat{y}} + \left(\frac{1}{2} - x_{2}\right)a \, \mathbf{\hat{z}} & \left(8e\right) & \mbox{K II} \\ 
\mathbf{B}_{8} & = & x_{2} \, \mathbf{a}_{1} + \left(\frac{1}{2} - x_{2}\right) \, \mathbf{a}_{2} + \left(\frac{1}{2} - x_{2}\right) \, \mathbf{a}_{3} & = & x_{2}a \, \mathbf{\hat{x}} + \left(\frac{1}{2} - x_{2}\right)a \, \mathbf{\hat{y}} + \left(\frac{1}{2} - x_{2}\right)a \, \mathbf{\hat{z}} & \left(8e\right) & \mbox{K II} \\ 
\mathbf{B}_{9} & = & -x_{2} \, \mathbf{a}_{1}-x_{2} \, \mathbf{a}_{2}-x_{2} \, \mathbf{a}_{3} & = & -x_{2}a \, \mathbf{\hat{x}}-x_{2}a \, \mathbf{\hat{y}}-x_{2}a \, \mathbf{\hat{z}} & \left(8e\right) & \mbox{K II} \\ 
\mathbf{B}_{10} & = & \left(\frac{1}{2} +x_{2}\right) \, \mathbf{a}_{1} + \left(\frac{1}{2} +x_{2}\right) \, \mathbf{a}_{2}-x_{2} \, \mathbf{a}_{3} & = & \left(\frac{1}{2} +x_{2}\right)a \, \mathbf{\hat{x}} + \left(\frac{1}{2} +x_{2}\right)a \, \mathbf{\hat{y}}-x_{2}a \, \mathbf{\hat{z}} & \left(8e\right) & \mbox{K II} \\ 
\mathbf{B}_{11} & = & \left(\frac{1}{2} +x_{2}\right) \, \mathbf{a}_{1}-x_{2} \, \mathbf{a}_{2} + \left(\frac{1}{2} +x_{2}\right) \, \mathbf{a}_{3} & = & \left(\frac{1}{2} +x_{2}\right)a \, \mathbf{\hat{x}}-x_{2}a \, \mathbf{\hat{y}} + \left(\frac{1}{2} +x_{2}\right)a \, \mathbf{\hat{z}} & \left(8e\right) & \mbox{K II} \\ 
\mathbf{B}_{12} & = & -x_{2} \, \mathbf{a}_{1} + \left(\frac{1}{2} +x_{2}\right) \, \mathbf{a}_{2} + \left(\frac{1}{2} +x_{2}\right) \, \mathbf{a}_{3} & = & -x_{2}a \, \mathbf{\hat{x}} + \left(\frac{1}{2} +x_{2}\right)a \, \mathbf{\hat{y}} + \left(\frac{1}{2} +x_{2}\right)a \, \mathbf{\hat{z}} & \left(8e\right) & \mbox{K II} \\ 
\mathbf{B}_{13} & = & x_{3} \, \mathbf{a}_{1} + \frac{1}{4} \, \mathbf{a}_{2} + \frac{1}{4} \, \mathbf{a}_{3} & = & x_{3}a \, \mathbf{\hat{x}} + \frac{1}{4}a \, \mathbf{\hat{y}} + \frac{1}{4}a \, \mathbf{\hat{z}} & \left(12f\right) & \mbox{O I} \\ 
\mathbf{B}_{14} & = & \left(\frac{1}{2} - x_{3}\right) \, \mathbf{a}_{1} + \frac{1}{4} \, \mathbf{a}_{2} + \frac{1}{4} \, \mathbf{a}_{3} & = & \left(\frac{1}{2} - x_{3}\right)a \, \mathbf{\hat{x}} + \frac{1}{4}a \, \mathbf{\hat{y}} + \frac{1}{4}a \, \mathbf{\hat{z}} & \left(12f\right) & \mbox{O I} \\ 
\mathbf{B}_{15} & = & \frac{1}{4} \, \mathbf{a}_{1} + x_{3} \, \mathbf{a}_{2} + \frac{1}{4} \, \mathbf{a}_{3} & = & \frac{1}{4}a \, \mathbf{\hat{x}} + x_{3}a \, \mathbf{\hat{y}} + \frac{1}{4}a \, \mathbf{\hat{z}} & \left(12f\right) & \mbox{O I} \\ 
\mathbf{B}_{16} & = & \frac{1}{4} \, \mathbf{a}_{1} + \left(\frac{1}{2} - x_{3}\right) \, \mathbf{a}_{2} + \frac{1}{4} \, \mathbf{a}_{3} & = & \frac{1}{4}a \, \mathbf{\hat{x}} + \left(\frac{1}{2} - x_{3}\right)a \, \mathbf{\hat{y}} + \frac{1}{4}a \, \mathbf{\hat{z}} & \left(12f\right) & \mbox{O I} \\ 
\mathbf{B}_{17} & = & \frac{1}{4} \, \mathbf{a}_{1} + \frac{1}{4} \, \mathbf{a}_{2} + x_{3} \, \mathbf{a}_{3} & = & \frac{1}{4}a \, \mathbf{\hat{x}} + \frac{1}{4}a \, \mathbf{\hat{y}} + x_{3}a \, \mathbf{\hat{z}} & \left(12f\right) & \mbox{O I} \\ 
\mathbf{B}_{18} & = & \frac{1}{4} \, \mathbf{a}_{1} + \frac{1}{4} \, \mathbf{a}_{2} + \left(\frac{1}{2} - x_{3}\right) \, \mathbf{a}_{3} & = & \frac{1}{4}a \, \mathbf{\hat{x}} + \frac{1}{4}a \, \mathbf{\hat{y}} + \left(\frac{1}{2} - x_{3}\right)a \, \mathbf{\hat{z}} & \left(12f\right) & \mbox{O I} \\ 
\mathbf{B}_{19} & = & -x_{3} \, \mathbf{a}_{1} + \frac{3}{4} \, \mathbf{a}_{2} + \frac{3}{4} \, \mathbf{a}_{3} & = & -x_{3}a \, \mathbf{\hat{x}} + \frac{3}{4}a \, \mathbf{\hat{y}} + \frac{3}{4}a \, \mathbf{\hat{z}} & \left(12f\right) & \mbox{O I} \\ 
\mathbf{B}_{20} & = & \left(\frac{1}{2} +x_{3}\right) \, \mathbf{a}_{1} + \frac{3}{4} \, \mathbf{a}_{2} + \frac{3}{4} \, \mathbf{a}_{3} & = & \left(\frac{1}{2} +x_{3}\right)a \, \mathbf{\hat{x}} + \frac{3}{4}a \, \mathbf{\hat{y}} + \frac{3}{4}a \, \mathbf{\hat{z}} & \left(12f\right) & \mbox{O I} \\ 
\mathbf{B}_{21} & = & \frac{3}{4} \, \mathbf{a}_{1}-x_{3} \, \mathbf{a}_{2} + \frac{3}{4} \, \mathbf{a}_{3} & = & \frac{3}{4}a \, \mathbf{\hat{x}}-x_{3}a \, \mathbf{\hat{y}} + \frac{3}{4}a \, \mathbf{\hat{z}} & \left(12f\right) & \mbox{O I} \\ 
\mathbf{B}_{22} & = & \frac{3}{4} \, \mathbf{a}_{1} + \left(\frac{1}{2} +x_{3}\right) \, \mathbf{a}_{2} + \frac{3}{4} \, \mathbf{a}_{3} & = & \frac{3}{4}a \, \mathbf{\hat{x}} + \left(\frac{1}{2} +x_{3}\right)a \, \mathbf{\hat{y}} + \frac{3}{4}a \, \mathbf{\hat{z}} & \left(12f\right) & \mbox{O I} \\ 
\mathbf{B}_{23} & = & \frac{3}{4} \, \mathbf{a}_{1} + \frac{3}{4} \, \mathbf{a}_{2}-x_{3} \, \mathbf{a}_{3} & = & \frac{3}{4}a \, \mathbf{\hat{x}} + \frac{3}{4}a \, \mathbf{\hat{y}}-x_{3}a \, \mathbf{\hat{z}} & \left(12f\right) & \mbox{O I} \\ 
\mathbf{B}_{24} & = & \frac{3}{4} \, \mathbf{a}_{1} + \frac{3}{4} \, \mathbf{a}_{2} + \left(\frac{1}{2} +x_{3}\right) \, \mathbf{a}_{3} & = & \frac{3}{4}a \, \mathbf{\hat{x}} + \frac{3}{4}a \, \mathbf{\hat{y}} + \left(\frac{1}{2} +x_{3}\right)a \, \mathbf{\hat{z}} & \left(12f\right) & \mbox{O I} \\ 
\mathbf{B}_{25} & = & x_{4} \, \mathbf{a}_{1} + \frac{3}{4} \, \mathbf{a}_{2} + \frac{1}{4} \, \mathbf{a}_{3} & = & x_{4}a \, \mathbf{\hat{x}} + \frac{3}{4}a \, \mathbf{\hat{y}} + \frac{1}{4}a \, \mathbf{\hat{z}} & \left(12g\right) & \mbox{Sb} \\ 
\mathbf{B}_{26} & = & \left(\frac{1}{2} - x_{4}\right) \, \mathbf{a}_{1} + \frac{3}{4} \, \mathbf{a}_{2} + \frac{1}{4} \, \mathbf{a}_{3} & = & \left(\frac{1}{2} - x_{4}\right)a \, \mathbf{\hat{x}} + \frac{3}{4}a \, \mathbf{\hat{y}} + \frac{1}{4}a \, \mathbf{\hat{z}} & \left(12g\right) & \mbox{Sb} \\ 
\mathbf{B}_{27} & = & \frac{1}{4} \, \mathbf{a}_{1} + x_{4} \, \mathbf{a}_{2} + \frac{3}{4} \, \mathbf{a}_{3} & = & \frac{1}{4}a \, \mathbf{\hat{x}} + x_{4}a \, \mathbf{\hat{y}} + \frac{3}{4}a \, \mathbf{\hat{z}} & \left(12g\right) & \mbox{Sb} \\ 
\mathbf{B}_{28} & = & \frac{1}{4} \, \mathbf{a}_{1} + \left(\frac{1}{2} - x_{4}\right) \, \mathbf{a}_{2} + \frac{3}{4} \, \mathbf{a}_{3} & = & \frac{1}{4}a \, \mathbf{\hat{x}} + \left(\frac{1}{2} - x_{4}\right)a \, \mathbf{\hat{y}} + \frac{3}{4}a \, \mathbf{\hat{z}} & \left(12g\right) & \mbox{Sb} \\ 
\mathbf{B}_{29} & = & \frac{3}{4} \, \mathbf{a}_{1} + \frac{1}{4} \, \mathbf{a}_{2} + x_{4} \, \mathbf{a}_{3} & = & \frac{3}{4}a \, \mathbf{\hat{x}} + \frac{1}{4}a \, \mathbf{\hat{y}} + x_{4}a \, \mathbf{\hat{z}} & \left(12g\right) & \mbox{Sb} \\ 
\mathbf{B}_{30} & = & \frac{3}{4} \, \mathbf{a}_{1} + \frac{1}{4} \, \mathbf{a}_{2} + \left(\frac{1}{2} - x_{4}\right) \, \mathbf{a}_{3} & = & \frac{3}{4}a \, \mathbf{\hat{x}} + \frac{1}{4}a \, \mathbf{\hat{y}} + \left(\frac{1}{2} - x_{4}\right)a \, \mathbf{\hat{z}} & \left(12g\right) & \mbox{Sb} \\ 
\mathbf{B}_{31} & = & -x_{4} \, \mathbf{a}_{1} + \frac{1}{4} \, \mathbf{a}_{2} + \frac{3}{4} \, \mathbf{a}_{3} & = & -x_{4}a \, \mathbf{\hat{x}} + \frac{1}{4}a \, \mathbf{\hat{y}} + \frac{3}{4}a \, \mathbf{\hat{z}} & \left(12g\right) & \mbox{Sb} \\ 
\mathbf{B}_{32} & = & \left(\frac{1}{2} +x_{4}\right) \, \mathbf{a}_{1} + \frac{1}{4} \, \mathbf{a}_{2} + \frac{3}{4} \, \mathbf{a}_{3} & = & \left(\frac{1}{2} +x_{4}\right)a \, \mathbf{\hat{x}} + \frac{1}{4}a \, \mathbf{\hat{y}} + \frac{3}{4}a \, \mathbf{\hat{z}} & \left(12g\right) & \mbox{Sb} \\ 
\mathbf{B}_{33} & = & \frac{3}{4} \, \mathbf{a}_{1}-x_{4} \, \mathbf{a}_{2} + \frac{1}{4} \, \mathbf{a}_{3} & = & \frac{3}{4}a \, \mathbf{\hat{x}}-x_{4}a \, \mathbf{\hat{y}} + \frac{1}{4}a \, \mathbf{\hat{z}} & \left(12g\right) & \mbox{Sb} \\ 
\mathbf{B}_{34} & = & \frac{3}{4} \, \mathbf{a}_{1} + \left(\frac{1}{2} +x_{4}\right) \, \mathbf{a}_{2} + \frac{1}{4} \, \mathbf{a}_{3} & = & \frac{3}{4}a \, \mathbf{\hat{x}} + \left(\frac{1}{2} +x_{4}\right)a \, \mathbf{\hat{y}} + \frac{1}{4}a \, \mathbf{\hat{z}} & \left(12g\right) & \mbox{Sb} \\ 
\mathbf{B}_{35} & = & \frac{1}{4} \, \mathbf{a}_{1} + \frac{3}{4} \, \mathbf{a}_{2}-x_{4} \, \mathbf{a}_{3} & = & \frac{1}{4}a \, \mathbf{\hat{x}} + \frac{3}{4}a \, \mathbf{\hat{y}}-x_{4}a \, \mathbf{\hat{z}} & \left(12g\right) & \mbox{Sb} \\ 
\mathbf{B}_{36} & = & \frac{1}{4} \, \mathbf{a}_{1} + \frac{3}{4} \, \mathbf{a}_{2} + \left(\frac{1}{2} +x_{4}\right) \, \mathbf{a}_{3} & = & \frac{1}{4}a \, \mathbf{\hat{x}} + \frac{3}{4}a \, \mathbf{\hat{y}} + \left(\frac{1}{2} +x_{4}\right)a \, \mathbf{\hat{z}} & \left(12g\right) & \mbox{Sb} \\ 
\mathbf{B}_{37} & = & x_{5} \, \mathbf{a}_{1} + y_{5} \, \mathbf{a}_{2} + z_{5} \, \mathbf{a}_{3} & = & x_{5}a \, \mathbf{\hat{x}} + y_{5}a \, \mathbf{\hat{y}} + z_{5}a \, \mathbf{\hat{z}} & \left(24h\right) & \mbox{O II} \\ 
\mathbf{B}_{38} & = & \left(\frac{1}{2} - x_{5}\right) \, \mathbf{a}_{1} + \left(\frac{1}{2} - y_{5}\right) \, \mathbf{a}_{2} + z_{5} \, \mathbf{a}_{3} & = & \left(\frac{1}{2} - x_{5}\right)a \, \mathbf{\hat{x}} + \left(\frac{1}{2} - y_{5}\right)a \, \mathbf{\hat{y}} + z_{5}a \, \mathbf{\hat{z}} & \left(24h\right) & \mbox{O II} \\ 
\mathbf{B}_{39} & = & \left(\frac{1}{2} - x_{5}\right) \, \mathbf{a}_{1} + y_{5} \, \mathbf{a}_{2} + \left(\frac{1}{2} - z_{5}\right) \, \mathbf{a}_{3} & = & \left(\frac{1}{2} - x_{5}\right)a \, \mathbf{\hat{x}} + y_{5}a \, \mathbf{\hat{y}} + \left(\frac{1}{2} - z_{5}\right)a \, \mathbf{\hat{z}} & \left(24h\right) & \mbox{O II} \\ 
\mathbf{B}_{40} & = & x_{5} \, \mathbf{a}_{1} + \left(\frac{1}{2} - y_{5}\right) \, \mathbf{a}_{2} + \left(\frac{1}{2} - z_{5}\right) \, \mathbf{a}_{3} & = & x_{5}a \, \mathbf{\hat{x}} + \left(\frac{1}{2} - y_{5}\right)a \, \mathbf{\hat{y}} + \left(\frac{1}{2} - z_{5}\right)a \, \mathbf{\hat{z}} & \left(24h\right) & \mbox{O II} \\ 
\mathbf{B}_{41} & = & z_{5} \, \mathbf{a}_{1} + x_{5} \, \mathbf{a}_{2} + y_{5} \, \mathbf{a}_{3} & = & z_{5}a \, \mathbf{\hat{x}} + x_{5}a \, \mathbf{\hat{y}} + y_{5}a \, \mathbf{\hat{z}} & \left(24h\right) & \mbox{O II} \\ 
\mathbf{B}_{42} & = & z_{5} \, \mathbf{a}_{1} + \left(\frac{1}{2} - x_{5}\right) \, \mathbf{a}_{2} + \left(\frac{1}{2} - y_{5}\right) \, \mathbf{a}_{3} & = & z_{5}a \, \mathbf{\hat{x}} + \left(\frac{1}{2} - x_{5}\right)a \, \mathbf{\hat{y}} + \left(\frac{1}{2} - y_{5}\right)a \, \mathbf{\hat{z}} & \left(24h\right) & \mbox{O II} \\ 
\mathbf{B}_{43} & = & \left(\frac{1}{2} - z_{5}\right) \, \mathbf{a}_{1} + \left(\frac{1}{2} - x_{5}\right) \, \mathbf{a}_{2} + y_{5} \, \mathbf{a}_{3} & = & \left(\frac{1}{2} - z_{5}\right)a \, \mathbf{\hat{x}} + \left(\frac{1}{2} - x_{5}\right)a \, \mathbf{\hat{y}} + y_{5}a \, \mathbf{\hat{z}} & \left(24h\right) & \mbox{O II} \\ 
\mathbf{B}_{44} & = & \left(\frac{1}{2} - z_{5}\right) \, \mathbf{a}_{1} + x_{5} \, \mathbf{a}_{2} + \left(\frac{1}{2} - y_{5}\right) \, \mathbf{a}_{3} & = & \left(\frac{1}{2} - z_{5}\right)a \, \mathbf{\hat{x}} + x_{5}a \, \mathbf{\hat{y}} + \left(\frac{1}{2} - y_{5}\right)a \, \mathbf{\hat{z}} & \left(24h\right) & \mbox{O II} \\ 
\mathbf{B}_{45} & = & y_{5} \, \mathbf{a}_{1} + z_{5} \, \mathbf{a}_{2} + x_{5} \, \mathbf{a}_{3} & = & y_{5}a \, \mathbf{\hat{x}} + z_{5}a \, \mathbf{\hat{y}} + x_{5}a \, \mathbf{\hat{z}} & \left(24h\right) & \mbox{O II} \\ 
\mathbf{B}_{46} & = & \left(\frac{1}{2} - y_{5}\right) \, \mathbf{a}_{1} + z_{5} \, \mathbf{a}_{2} + \left(\frac{1}{2} - x_{5}\right) \, \mathbf{a}_{3} & = & \left(\frac{1}{2} - y_{5}\right)a \, \mathbf{\hat{x}} + z_{5}a \, \mathbf{\hat{y}} + \left(\frac{1}{2} - x_{5}\right)a \, \mathbf{\hat{z}} & \left(24h\right) & \mbox{O II} \\ 
\mathbf{B}_{47} & = & y_{5} \, \mathbf{a}_{1} + \left(\frac{1}{2} - z_{5}\right) \, \mathbf{a}_{2} + \left(\frac{1}{2} - x_{5}\right) \, \mathbf{a}_{3} & = & y_{5}a \, \mathbf{\hat{x}} + \left(\frac{1}{2} - z_{5}\right)a \, \mathbf{\hat{y}} + \left(\frac{1}{2} - x_{5}\right)a \, \mathbf{\hat{z}} & \left(24h\right) & \mbox{O II} \\ 
\mathbf{B}_{48} & = & \left(\frac{1}{2} - y_{5}\right) \, \mathbf{a}_{1} + \left(\frac{1}{2} - z_{5}\right) \, \mathbf{a}_{2} + x_{5} \, \mathbf{a}_{3} & = & \left(\frac{1}{2} - y_{5}\right)a \, \mathbf{\hat{x}} + \left(\frac{1}{2} - z_{5}\right)a \, \mathbf{\hat{y}} + x_{5}a \, \mathbf{\hat{z}} & \left(24h\right) & \mbox{O II} \\ 
\mathbf{B}_{49} & = & -x_{5} \, \mathbf{a}_{1}-y_{5} \, \mathbf{a}_{2}-z_{5} \, \mathbf{a}_{3} & = & -x_{5}a \, \mathbf{\hat{x}}-y_{5}a \, \mathbf{\hat{y}}-z_{5}a \, \mathbf{\hat{z}} & \left(24h\right) & \mbox{O II} \\ 
\mathbf{B}_{50} & = & \left(\frac{1}{2} +x_{5}\right) \, \mathbf{a}_{1} + \left(\frac{1}{2} +y_{5}\right) \, \mathbf{a}_{2}-z_{5} \, \mathbf{a}_{3} & = & \left(\frac{1}{2} +x_{5}\right)a \, \mathbf{\hat{x}} + \left(\frac{1}{2} +y_{5}\right)a \, \mathbf{\hat{y}}-z_{5}a \, \mathbf{\hat{z}} & \left(24h\right) & \mbox{O II} \\ 
\mathbf{B}_{51} & = & \left(\frac{1}{2} +x_{5}\right) \, \mathbf{a}_{1}-y_{5} \, \mathbf{a}_{2} + \left(\frac{1}{2} +z_{5}\right) \, \mathbf{a}_{3} & = & \left(\frac{1}{2} +x_{5}\right)a \, \mathbf{\hat{x}}-y_{5}a \, \mathbf{\hat{y}} + \left(\frac{1}{2} +z_{5}\right)a \, \mathbf{\hat{z}} & \left(24h\right) & \mbox{O II} \\ 
\mathbf{B}_{52} & = & -x_{5} \, \mathbf{a}_{1} + \left(\frac{1}{2} +y_{5}\right) \, \mathbf{a}_{2} + \left(\frac{1}{2} +z_{5}\right) \, \mathbf{a}_{3} & = & -x_{5}a \, \mathbf{\hat{x}} + \left(\frac{1}{2} +y_{5}\right)a \, \mathbf{\hat{y}} + \left(\frac{1}{2} +z_{5}\right)a \, \mathbf{\hat{z}} & \left(24h\right) & \mbox{O II} \\ 
\mathbf{B}_{53} & = & -z_{5} \, \mathbf{a}_{1}-x_{5} \, \mathbf{a}_{2}-y_{5} \, \mathbf{a}_{3} & = & -z_{5}a \, \mathbf{\hat{x}}-x_{5}a \, \mathbf{\hat{y}}-y_{5}a \, \mathbf{\hat{z}} & \left(24h\right) & \mbox{O II} \\ 
\mathbf{B}_{54} & = & -z_{5} \, \mathbf{a}_{1} + \left(\frac{1}{2} +x_{5}\right) \, \mathbf{a}_{2} + \left(\frac{1}{2} +y_{5}\right) \, \mathbf{a}_{3} & = & -z_{5}a \, \mathbf{\hat{x}} + \left(\frac{1}{2} +x_{5}\right)a \, \mathbf{\hat{y}} + \left(\frac{1}{2} +y_{5}\right)a \, \mathbf{\hat{z}} & \left(24h\right) & \mbox{O II} \\ 
\mathbf{B}_{55} & = & \left(\frac{1}{2} +z_{5}\right) \, \mathbf{a}_{1} + \left(\frac{1}{2} +x_{5}\right) \, \mathbf{a}_{2}-y_{5} \, \mathbf{a}_{3} & = & \left(\frac{1}{2} +z_{5}\right)a \, \mathbf{\hat{x}} + \left(\frac{1}{2} +x_{5}\right)a \, \mathbf{\hat{y}}-y_{5}a \, \mathbf{\hat{z}} & \left(24h\right) & \mbox{O II} \\ 
\mathbf{B}_{56} & = & \left(\frac{1}{2} +z_{5}\right) \, \mathbf{a}_{1}-x_{5} \, \mathbf{a}_{2} + \left(\frac{1}{2} +y_{5}\right) \, \mathbf{a}_{3} & = & \left(\frac{1}{2} +z_{5}\right)a \, \mathbf{\hat{x}}-x_{5}a \, \mathbf{\hat{y}} + \left(\frac{1}{2} +y_{5}\right)a \, \mathbf{\hat{z}} & \left(24h\right) & \mbox{O II} \\ 
\mathbf{B}_{57} & = & -y_{5} \, \mathbf{a}_{1}-z_{5} \, \mathbf{a}_{2}-x_{5} \, \mathbf{a}_{3} & = & -y_{5}a \, \mathbf{\hat{x}}-z_{5}a \, \mathbf{\hat{y}}-x_{5}a \, \mathbf{\hat{z}} & \left(24h\right) & \mbox{O II} \\ 
\mathbf{B}_{58} & = & \left(\frac{1}{2} +y_{5}\right) \, \mathbf{a}_{1}-z_{5} \, \mathbf{a}_{2} + \left(\frac{1}{2} +x_{5}\right) \, \mathbf{a}_{3} & = & \left(\frac{1}{2} +y_{5}\right)a \, \mathbf{\hat{x}}-z_{5}a \, \mathbf{\hat{y}} + \left(\frac{1}{2} +x_{5}\right)a \, \mathbf{\hat{z}} & \left(24h\right) & \mbox{O II} \\ 
\mathbf{B}_{59} & = & -y_{5} \, \mathbf{a}_{1} + \left(\frac{1}{2} +z_{5}\right) \, \mathbf{a}_{2} + \left(\frac{1}{2} +x_{5}\right) \, \mathbf{a}_{3} & = & -y_{5}a \, \mathbf{\hat{x}} + \left(\frac{1}{2} +z_{5}\right)a \, \mathbf{\hat{y}} + \left(\frac{1}{2} +x_{5}\right)a \, \mathbf{\hat{z}} & \left(24h\right) & \mbox{O II} \\ 
\mathbf{B}_{60} & = & \left(\frac{1}{2} +y_{5}\right) \, \mathbf{a}_{1} + \left(\frac{1}{2} +z_{5}\right) \, \mathbf{a}_{2}-x_{5} \, \mathbf{a}_{3} & = & \left(\frac{1}{2} +y_{5}\right)a \, \mathbf{\hat{x}} + \left(\frac{1}{2} +z_{5}\right)a \, \mathbf{\hat{y}}-x_{5}a \, \mathbf{\hat{z}} & \left(24h\right) & \mbox{O II} \\ 
\end{longtabu}
\renewcommand{\arraystretch}{1.0}
\noindent \hrulefill
\\
\textbf{References:}
\vspace*{-0.25cm}
\begin{flushleft}
  - \bibentry{Spiegelberg_KSbO3_ArkKemiMon_1940}. \\
\end{flushleft}
\textbf{Found in:}
\vspace*{-0.25cm}
\begin{flushleft}
  - \bibentry{Villars_PearsonsCrystalData_2013}. \\
\end{flushleft}
\noindent \hrulefill
\\
\textbf{Geometry files:}
\\
\noindent  - CIF: pp. {\hyperref[AB3C_cP60_201_ce_fh_g_cif]{\pageref{AB3C_cP60_201_ce_fh_g_cif}}} \\
\noindent  - POSCAR: pp. {\hyperref[AB3C_cP60_201_ce_fh_g_poscar]{\pageref{AB3C_cP60_201_ce_fh_g_poscar}}} \\
\onecolumn
{\phantomsection\label{A6B6C_cF104_202_h_h_c}}
\subsection*{\huge \textbf{{\normalfont KB$_{6}$H$_{6}$ Structure: A6B6C\_cF104\_202\_h\_h\_c}}}
\noindent \hrulefill
\vspace*{0.25cm}
\begin{figure}[htp]
  \centering
  \vspace{-1em}
  {\includegraphics[width=1\textwidth]{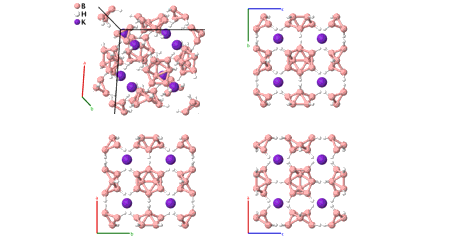}}
\end{figure}
\vspace*{-0.5cm}
\renewcommand{\arraystretch}{1.5}
\begin{equation*}
  \begin{array}{>{$\hspace{-0.15cm}}l<{$}>{$}p{0.5cm}<{$}>{$}p{18.5cm}<{$}}
    \mbox{\large \textbf{Prototype}} &\colon & \ce{KB6H6} \\
    \mbox{\large \textbf{\AFLOW\ prototype label}} &\colon & \mbox{A6B6C\_cF104\_202\_h\_h\_c} \\
    \mbox{\large \textbf{\textit{Strukturbericht} designation}} &\colon & \mbox{None} \\
    \mbox{\large \textbf{Pearson symbol}} &\colon & \mbox{cF104} \\
    \mbox{\large \textbf{Space group number}} &\colon & 202 \\
    \mbox{\large \textbf{Space group symbol}} &\colon & Fm\bar{3} \\
    \mbox{\large \textbf{\AFLOW\ prototype command}} &\colon &  \texttt{aflow} \,  \, \texttt{-{}-proto=A6B6C\_cF104\_202\_h\_h\_c } \, \newline \texttt{-{}-params=}{a,y_{2},z_{2},y_{3},z_{3} }
  \end{array}
\end{equation*}
\renewcommand{\arraystretch}{1.0}

\noindent \parbox{1 \linewidth}{
\noindent \hrulefill
\\
\textbf{Face-centered Cubic primitive vectors:} \\
\vspace*{-0.25cm}
\begin{tabular}{cc}
  \begin{tabular}{c}
    \parbox{0.6 \linewidth}{
      \renewcommand{\arraystretch}{1.5}
      \begin{equation*}
        \centering
        \begin{array}{ccc}
              \mathbf{a}_1 & = & \frac12 \, a \, \mathbf{\hat{y}} + \frac12 \, a \, \mathbf{\hat{z}} \\
    \mathbf{a}_2 & = & \frac12 \, a \, \mathbf{\hat{x}} + \frac12 \, a \, \mathbf{\hat{z}} \\
    \mathbf{a}_3 & = & \frac12 \, a \, \mathbf{\hat{x}} + \frac12 \, a \, \mathbf{\hat{y}} \\

        \end{array}
      \end{equation*}
    }
    \renewcommand{\arraystretch}{1.0}
  \end{tabular}
  \begin{tabular}{c}
    \includegraphics[width=0.3\linewidth]{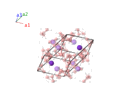} \\
  \end{tabular}
\end{tabular}

}
\vspace*{-0.25cm}

\noindent \hrulefill
\\
\textbf{Basis vectors:}
\vspace*{-0.25cm}
\renewcommand{\arraystretch}{1.5}
\begin{longtabu} to \textwidth{>{\centering $}X[-1,c,c]<{$}>{\centering $}X[-1,c,c]<{$}>{\centering $}X[-1,c,c]<{$}>{\centering $}X[-1,c,c]<{$}>{\centering $}X[-1,c,c]<{$}>{\centering $}X[-1,c,c]<{$}>{\centering $}X[-1,c,c]<{$}}
  & & \mbox{Lattice Coordinates} & & \mbox{Cartesian Coordinates} &\mbox{Wyckoff Position} & \mbox{Atom Type} \\  
  \mathbf{B}_{1} & = & \frac{1}{4} \, \mathbf{a}_{1} + \frac{1}{4} \, \mathbf{a}_{2} + \frac{1}{4} \, \mathbf{a}_{3} & = & \frac{1}{4}a \, \mathbf{\hat{x}} + \frac{1}{4}a \, \mathbf{\hat{y}} + \frac{1}{4}a \, \mathbf{\hat{z}} & \left(8c\right) & \mbox{K} \\ 
\mathbf{B}_{2} & = & \frac{3}{4} \, \mathbf{a}_{1} + \frac{3}{4} \, \mathbf{a}_{2} + \frac{3}{4} \, \mathbf{a}_{3} & = & \frac{3}{4}a \, \mathbf{\hat{x}} + \frac{3}{4}a \, \mathbf{\hat{y}} + \frac{3}{4}a \, \mathbf{\hat{z}} & \left(8c\right) & \mbox{K} \\ 
\mathbf{B}_{3} & = & \left(y_{2}+z_{2}\right) \, \mathbf{a}_{1} + \left(-y_{2}+z_{2}\right) \, \mathbf{a}_{2} + \left(y_{2}-z_{2}\right) \, \mathbf{a}_{3} & = & y_{2}a \, \mathbf{\hat{y}} + z_{2}a \, \mathbf{\hat{z}} & \left(48h\right) & \mbox{B} \\ 
\mathbf{B}_{4} & = & \left(-y_{2}+z_{2}\right) \, \mathbf{a}_{1} + \left(y_{2}+z_{2}\right) \, \mathbf{a}_{2} + \left(-y_{2}-z_{2}\right) \, \mathbf{a}_{3} & = & -y_{2}a \, \mathbf{\hat{y}} + z_{2}a \, \mathbf{\hat{z}} & \left(48h\right) & \mbox{B} \\ 
\mathbf{B}_{5} & = & \left(y_{2}-z_{2}\right) \, \mathbf{a}_{1} + \left(-y_{2}-z_{2}\right) \, \mathbf{a}_{2} + \left(y_{2}+z_{2}\right) \, \mathbf{a}_{3} & = & y_{2}a \, \mathbf{\hat{y}}-z_{2}a \, \mathbf{\hat{z}} & \left(48h\right) & \mbox{B} \\ 
\mathbf{B}_{6} & = & \left(-y_{2}-z_{2}\right) \, \mathbf{a}_{1} + \left(y_{2}-z_{2}\right) \, \mathbf{a}_{2} + \left(-y_{2}+z_{2}\right) \, \mathbf{a}_{3} & = & -y_{2}a \, \mathbf{\hat{y}}-z_{2}a \, \mathbf{\hat{z}} & \left(48h\right) & \mbox{B} \\ 
\mathbf{B}_{7} & = & \left(y_{2}-z_{2}\right) \, \mathbf{a}_{1} + \left(y_{2}+z_{2}\right) \, \mathbf{a}_{2} + \left(-y_{2}+z_{2}\right) \, \mathbf{a}_{3} & = & z_{2}a \, \mathbf{\hat{x}} + y_{2}a \, \mathbf{\hat{z}} & \left(48h\right) & \mbox{B} \\ 
\mathbf{B}_{8} & = & \left(-y_{2}-z_{2}\right) \, \mathbf{a}_{1} + \left(-y_{2}+z_{2}\right) \, \mathbf{a}_{2} + \left(y_{2}+z_{2}\right) \, \mathbf{a}_{3} & = & z_{2}a \, \mathbf{\hat{x}} + -y_{2}a \, \mathbf{\hat{z}} & \left(48h\right) & \mbox{B} \\ 
\mathbf{B}_{9} & = & \left(y_{2}+z_{2}\right) \, \mathbf{a}_{1} + \left(y_{2}-z_{2}\right) \, \mathbf{a}_{2} + \left(-y_{2}-z_{2}\right) \, \mathbf{a}_{3} & = & -z_{2}a \, \mathbf{\hat{x}} + y_{2}a \, \mathbf{\hat{z}} & \left(48h\right) & \mbox{B} \\ 
\mathbf{B}_{10} & = & \left(-y_{2}+z_{2}\right) \, \mathbf{a}_{1} + \left(-y_{2}-z_{2}\right) \, \mathbf{a}_{2} + \left(y_{2}-z_{2}\right) \, \mathbf{a}_{3} & = & -z_{2}a \, \mathbf{\hat{x}} + -y_{2}a \, \mathbf{\hat{z}} & \left(48h\right) & \mbox{B} \\ 
\mathbf{B}_{11} & = & \left(-y_{2}+z_{2}\right) \, \mathbf{a}_{1} + \left(y_{2}-z_{2}\right) \, \mathbf{a}_{2} + \left(y_{2}+z_{2}\right) \, \mathbf{a}_{3} & = & y_{2}a \, \mathbf{\hat{x}} + z_{2}a \, \mathbf{\hat{y}} & \left(48h\right) & \mbox{B} \\ 
\mathbf{B}_{12} & = & \left(y_{2}+z_{2}\right) \, \mathbf{a}_{1} + \left(-y_{2}-z_{2}\right) \, \mathbf{a}_{2} + \left(-y_{2}+z_{2}\right) \, \mathbf{a}_{3} & = & -y_{2}a \, \mathbf{\hat{x}} + z_{2}a \, \mathbf{\hat{y}} & \left(48h\right) & \mbox{B} \\ 
\mathbf{B}_{13} & = & \left(-y_{2}-z_{2}\right) \, \mathbf{a}_{1} + \left(y_{2}+z_{2}\right) \, \mathbf{a}_{2} + \left(y_{2}-z_{2}\right) \, \mathbf{a}_{3} & = & y_{2}a \, \mathbf{\hat{x}}-z_{2}a \, \mathbf{\hat{y}} & \left(48h\right) & \mbox{B} \\ 
\mathbf{B}_{14} & = & \left(y_{2}-z_{2}\right) \, \mathbf{a}_{1} + \left(-y_{2}+z_{2}\right) \, \mathbf{a}_{2} + \left(-y_{2}-z_{2}\right) \, \mathbf{a}_{3} & = & -y_{2}a \, \mathbf{\hat{x}}-z_{2}a \, \mathbf{\hat{y}} & \left(48h\right) & \mbox{B} \\ 
\mathbf{B}_{15} & = & \left(y_{3}+z_{3}\right) \, \mathbf{a}_{1} + \left(-y_{3}+z_{3}\right) \, \mathbf{a}_{2} + \left(y_{3}-z_{3}\right) \, \mathbf{a}_{3} & = & y_{3}a \, \mathbf{\hat{y}} + z_{3}a \, \mathbf{\hat{z}} & \left(48h\right) & \mbox{H} \\ 
\mathbf{B}_{16} & = & \left(-y_{3}+z_{3}\right) \, \mathbf{a}_{1} + \left(y_{3}+z_{3}\right) \, \mathbf{a}_{2} + \left(-y_{3}-z_{3}\right) \, \mathbf{a}_{3} & = & -y_{3}a \, \mathbf{\hat{y}} + z_{3}a \, \mathbf{\hat{z}} & \left(48h\right) & \mbox{H} \\ 
\mathbf{B}_{17} & = & \left(y_{3}-z_{3}\right) \, \mathbf{a}_{1} + \left(-y_{3}-z_{3}\right) \, \mathbf{a}_{2} + \left(y_{3}+z_{3}\right) \, \mathbf{a}_{3} & = & y_{3}a \, \mathbf{\hat{y}}-z_{3}a \, \mathbf{\hat{z}} & \left(48h\right) & \mbox{H} \\ 
\mathbf{B}_{18} & = & \left(-y_{3}-z_{3}\right) \, \mathbf{a}_{1} + \left(y_{3}-z_{3}\right) \, \mathbf{a}_{2} + \left(-y_{3}+z_{3}\right) \, \mathbf{a}_{3} & = & -y_{3}a \, \mathbf{\hat{y}}-z_{3}a \, \mathbf{\hat{z}} & \left(48h\right) & \mbox{H} \\ 
\mathbf{B}_{19} & = & \left(y_{3}-z_{3}\right) \, \mathbf{a}_{1} + \left(y_{3}+z_{3}\right) \, \mathbf{a}_{2} + \left(-y_{3}+z_{3}\right) \, \mathbf{a}_{3} & = & z_{3}a \, \mathbf{\hat{x}} + y_{3}a \, \mathbf{\hat{z}} & \left(48h\right) & \mbox{H} \\ 
\mathbf{B}_{20} & = & \left(-y_{3}-z_{3}\right) \, \mathbf{a}_{1} + \left(-y_{3}+z_{3}\right) \, \mathbf{a}_{2} + \left(y_{3}+z_{3}\right) \, \mathbf{a}_{3} & = & z_{3}a \, \mathbf{\hat{x}} + -y_{3}a \, \mathbf{\hat{z}} & \left(48h\right) & \mbox{H} \\ 
\mathbf{B}_{21} & = & \left(y_{3}+z_{3}\right) \, \mathbf{a}_{1} + \left(y_{3}-z_{3}\right) \, \mathbf{a}_{2} + \left(-y_{3}-z_{3}\right) \, \mathbf{a}_{3} & = & -z_{3}a \, \mathbf{\hat{x}} + y_{3}a \, \mathbf{\hat{z}} & \left(48h\right) & \mbox{H} \\ 
\mathbf{B}_{22} & = & \left(-y_{3}+z_{3}\right) \, \mathbf{a}_{1} + \left(-y_{3}-z_{3}\right) \, \mathbf{a}_{2} + \left(y_{3}-z_{3}\right) \, \mathbf{a}_{3} & = & -z_{3}a \, \mathbf{\hat{x}} + -y_{3}a \, \mathbf{\hat{z}} & \left(48h\right) & \mbox{H} \\ 
\mathbf{B}_{23} & = & \left(-y_{3}+z_{3}\right) \, \mathbf{a}_{1} + \left(y_{3}-z_{3}\right) \, \mathbf{a}_{2} + \left(y_{3}+z_{3}\right) \, \mathbf{a}_{3} & = & y_{3}a \, \mathbf{\hat{x}} + z_{3}a \, \mathbf{\hat{y}} & \left(48h\right) & \mbox{H} \\ 
\mathbf{B}_{24} & = & \left(y_{3}+z_{3}\right) \, \mathbf{a}_{1} + \left(-y_{3}-z_{3}\right) \, \mathbf{a}_{2} + \left(-y_{3}+z_{3}\right) \, \mathbf{a}_{3} & = & -y_{3}a \, \mathbf{\hat{x}} + z_{3}a \, \mathbf{\hat{y}} & \left(48h\right) & \mbox{H} \\ 
\mathbf{B}_{25} & = & \left(-y_{3}-z_{3}\right) \, \mathbf{a}_{1} + \left(y_{3}+z_{3}\right) \, \mathbf{a}_{2} + \left(y_{3}-z_{3}\right) \, \mathbf{a}_{3} & = & y_{3}a \, \mathbf{\hat{x}}-z_{3}a \, \mathbf{\hat{y}} & \left(48h\right) & \mbox{H} \\ 
\mathbf{B}_{26} & = & \left(y_{3}-z_{3}\right) \, \mathbf{a}_{1} + \left(-y_{3}+z_{3}\right) \, \mathbf{a}_{2} + \left(-y_{3}-z_{3}\right) \, \mathbf{a}_{3} & = & -y_{3}a \, \mathbf{\hat{x}}-z_{3}a \, \mathbf{\hat{y}} & \left(48h\right) & \mbox{H} \\ 
\end{longtabu}
\renewcommand{\arraystretch}{1.0}
\noindent \hrulefill
\\
\textbf{References:}
\vspace*{-0.25cm}
\begin{flushleft}
  - \bibentry{Wunderlich_B6H6K_JAmerChemSoc_1960}. \\
\end{flushleft}
\textbf{Found in:}
\vspace*{-0.25cm}
\begin{flushleft}
  - \bibentry{Villars_PearsonsCrystalData_2013}. \\
\end{flushleft}
\noindent \hrulefill
\\
\textbf{Geometry files:}
\\
\noindent  - CIF: pp. {\hyperref[A6B6C_cF104_202_h_h_c_cif]{\pageref{A6B6C_cF104_202_h_h_c_cif}}} \\
\noindent  - POSCAR: pp. {\hyperref[A6B6C_cF104_202_h_h_c_poscar]{\pageref{A6B6C_cF104_202_h_h_c_poscar}}} \\
\onecolumn
{\phantomsection\label{A_cF240_202_h2i}}
\subsection*{\huge \textbf{{\normalfont \begin{raggedleft}FCC C$_{60}$ Buckminsterfullerine Structure: \end{raggedleft} \\ A\_cF240\_202\_h2i}}}
\noindent \hrulefill
\vspace*{0.25cm}
\begin{figure}[htp]
  \centering
  \vspace{-1em}
  {\includegraphics[width=1\textwidth]{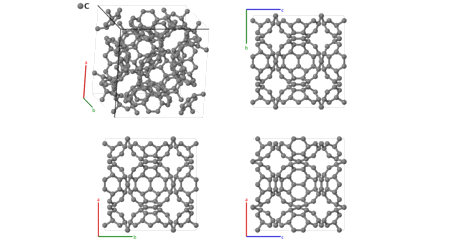}}
\end{figure}
\vspace*{-0.5cm}
\renewcommand{\arraystretch}{1.5}
\begin{equation*}
  \begin{array}{>{$\hspace{-0.15cm}}l<{$}>{$}p{0.5cm}<{$}>{$}p{18.5cm}<{$}}
    \mbox{\large \textbf{Prototype}} &\colon & \ce{C} \\
    \mbox{\large \textbf{\AFLOW\ prototype label}} &\colon & \mbox{A\_cF240\_202\_h2i} \\
    \mbox{\large \textbf{\textit{Strukturbericht} designation}} &\colon & \mbox{None} \\
    \mbox{\large \textbf{Pearson symbol}} &\colon & \mbox{cF240} \\
    \mbox{\large \textbf{Space group number}} &\colon & 202 \\
    \mbox{\large \textbf{Space group symbol}} &\colon & Fm\bar{3} \\
    \mbox{\large \textbf{\AFLOW\ prototype command}} &\colon &  \texttt{aflow} \,  \, \texttt{-{}-proto=A\_cF240\_202\_h2i } \, \newline \texttt{-{}-params=}{a,y_{1},z_{1},x_{2},y_{2},z_{2},x_{3},y_{3},z_{3} }
  \end{array}
\end{equation*}
\renewcommand{\arraystretch}{1.0}

\vspace*{-0.25cm}
\noindent \hrulefill
\begin{itemize}
  \item{This is an {\em approximate} representation of the structure of
C$_{60}$ buckminsterfullerene. As noted by the authors, ``a careful
analysis of the intensity data reveals that the molecules must pack in
an uncorrelated array, in full agreement with the results from most
previous diffraction and spectroscopic determinations.''
The C$_{60}$ molecules are on the sites of an fcc lattice.  Below 249~K
there is a transition to a \hyperref[A_cP240_205_10d]{simple cubic
phase of C$_{60}$}.
}
\end{itemize}

\noindent \parbox{1 \linewidth}{
\noindent \hrulefill
\\
\textbf{Face-centered Cubic primitive vectors:} \\
\vspace*{-0.25cm}
\begin{tabular}{cc}
  \begin{tabular}{c}
    \parbox{0.6 \linewidth}{
      \renewcommand{\arraystretch}{1.5}
      \begin{equation*}
        \centering
        \begin{array}{ccc}
              \mathbf{a}_1 & = & \frac12 \, a \, \mathbf{\hat{y}} + \frac12 \, a \, \mathbf{\hat{z}} \\
    \mathbf{a}_2 & = & \frac12 \, a \, \mathbf{\hat{x}} + \frac12 \, a \, \mathbf{\hat{z}} \\
    \mathbf{a}_3 & = & \frac12 \, a \, \mathbf{\hat{x}} + \frac12 \, a \, \mathbf{\hat{y}} \\

        \end{array}
      \end{equation*}
    }
    \renewcommand{\arraystretch}{1.0}
  \end{tabular}
  \begin{tabular}{c}
    \includegraphics[width=0.3\linewidth]{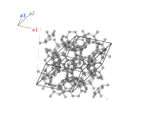} \\
  \end{tabular}
\end{tabular}

}
\vspace*{-0.25cm}

\noindent \hrulefill
\\
\textbf{Basis vectors:}
\vspace*{-0.25cm}
\renewcommand{\arraystretch}{1.5}
\begin{longtabu} to \textwidth{>{\centering $}X[-1,c,c]<{$}>{\centering $}X[-1,c,c]<{$}>{\centering $}X[-1,c,c]<{$}>{\centering $}X[-1,c,c]<{$}>{\centering $}X[-1,c,c]<{$}>{\centering $}X[-1,c,c]<{$}>{\centering $}X[-1,c,c]<{$}}
  & & \mbox{Lattice Coordinates} & & \mbox{Cartesian Coordinates} &\mbox{Wyckoff Position} & \mbox{Atom Type} \\  
  \mathbf{B}_{1} & = & \left(y_{1}+z_{1}\right) \, \mathbf{a}_{1} + \left(-y_{1}+z_{1}\right) \, \mathbf{a}_{2} + \left(y_{1}-z_{1}\right) \, \mathbf{a}_{3} & = & y_{1}a \, \mathbf{\hat{y}} + z_{1}a \, \mathbf{\hat{z}} & \left(48h\right) & \mbox{C I} \\ 
\mathbf{B}_{2} & = & \left(-y_{1}+z_{1}\right) \, \mathbf{a}_{1} + \left(y_{1}+z_{1}\right) \, \mathbf{a}_{2} + \left(-y_{1}-z_{1}\right) \, \mathbf{a}_{3} & = & -y_{1}a \, \mathbf{\hat{y}} + z_{1}a \, \mathbf{\hat{z}} & \left(48h\right) & \mbox{C I} \\ 
\mathbf{B}_{3} & = & \left(y_{1}-z_{1}\right) \, \mathbf{a}_{1} + \left(-y_{1}-z_{1}\right) \, \mathbf{a}_{2} + \left(y_{1}+z_{1}\right) \, \mathbf{a}_{3} & = & y_{1}a \, \mathbf{\hat{y}}-z_{1}a \, \mathbf{\hat{z}} & \left(48h\right) & \mbox{C I} \\ 
\mathbf{B}_{4} & = & \left(-y_{1}-z_{1}\right) \, \mathbf{a}_{1} + \left(y_{1}-z_{1}\right) \, \mathbf{a}_{2} + \left(-y_{1}+z_{1}\right) \, \mathbf{a}_{3} & = & -y_{1}a \, \mathbf{\hat{y}}-z_{1}a \, \mathbf{\hat{z}} & \left(48h\right) & \mbox{C I} \\ 
\mathbf{B}_{5} & = & \left(y_{1}-z_{1}\right) \, \mathbf{a}_{1} + \left(y_{1}+z_{1}\right) \, \mathbf{a}_{2} + \left(-y_{1}+z_{1}\right) \, \mathbf{a}_{3} & = & z_{1}a \, \mathbf{\hat{x}} + y_{1}a \, \mathbf{\hat{z}} & \left(48h\right) & \mbox{C I} \\ 
\mathbf{B}_{6} & = & \left(-y_{1}-z_{1}\right) \, \mathbf{a}_{1} + \left(-y_{1}+z_{1}\right) \, \mathbf{a}_{2} + \left(y_{1}+z_{1}\right) \, \mathbf{a}_{3} & = & z_{1}a \, \mathbf{\hat{x}} + -y_{1}a \, \mathbf{\hat{z}} & \left(48h\right) & \mbox{C I} \\ 
\mathbf{B}_{7} & = & \left(y_{1}+z_{1}\right) \, \mathbf{a}_{1} + \left(y_{1}-z_{1}\right) \, \mathbf{a}_{2} + \left(-y_{1}-z_{1}\right) \, \mathbf{a}_{3} & = & -z_{1}a \, \mathbf{\hat{x}} + y_{1}a \, \mathbf{\hat{z}} & \left(48h\right) & \mbox{C I} \\ 
\mathbf{B}_{8} & = & \left(-y_{1}+z_{1}\right) \, \mathbf{a}_{1} + \left(-y_{1}-z_{1}\right) \, \mathbf{a}_{2} + \left(y_{1}-z_{1}\right) \, \mathbf{a}_{3} & = & -z_{1}a \, \mathbf{\hat{x}} + -y_{1}a \, \mathbf{\hat{z}} & \left(48h\right) & \mbox{C I} \\ 
\mathbf{B}_{9} & = & \left(-y_{1}+z_{1}\right) \, \mathbf{a}_{1} + \left(y_{1}-z_{1}\right) \, \mathbf{a}_{2} + \left(y_{1}+z_{1}\right) \, \mathbf{a}_{3} & = & y_{1}a \, \mathbf{\hat{x}} + z_{1}a \, \mathbf{\hat{y}} & \left(48h\right) & \mbox{C I} \\ 
\mathbf{B}_{10} & = & \left(y_{1}+z_{1}\right) \, \mathbf{a}_{1} + \left(-y_{1}-z_{1}\right) \, \mathbf{a}_{2} + \left(-y_{1}+z_{1}\right) \, \mathbf{a}_{3} & = & -y_{1}a \, \mathbf{\hat{x}} + z_{1}a \, \mathbf{\hat{y}} & \left(48h\right) & \mbox{C I} \\ 
\mathbf{B}_{11} & = & \left(-y_{1}-z_{1}\right) \, \mathbf{a}_{1} + \left(y_{1}+z_{1}\right) \, \mathbf{a}_{2} + \left(y_{1}-z_{1}\right) \, \mathbf{a}_{3} & = & y_{1}a \, \mathbf{\hat{x}}-z_{1}a \, \mathbf{\hat{y}} & \left(48h\right) & \mbox{C I} \\ 
\mathbf{B}_{12} & = & \left(y_{1}-z_{1}\right) \, \mathbf{a}_{1} + \left(-y_{1}+z_{1}\right) \, \mathbf{a}_{2} + \left(-y_{1}-z_{1}\right) \, \mathbf{a}_{3} & = & -y_{1}a \, \mathbf{\hat{x}}-z_{1}a \, \mathbf{\hat{y}} & \left(48h\right) & \mbox{C I} \\ 
\mathbf{B}_{13} & = & \left(-x_{2}+y_{2}+z_{2}\right) \, \mathbf{a}_{1} + \left(x_{2}-y_{2}+z_{2}\right) \, \mathbf{a}_{2} + \left(x_{2}+y_{2}-z_{2}\right) \, \mathbf{a}_{3} & = & x_{2}a \, \mathbf{\hat{x}} + y_{2}a \, \mathbf{\hat{y}} + z_{2}a \, \mathbf{\hat{z}} & \left(96i\right) & \mbox{C II} \\ 
\mathbf{B}_{14} & = & \left(x_{2}-y_{2}+z_{2}\right) \, \mathbf{a}_{1} + \left(-x_{2}+y_{2}+z_{2}\right) \, \mathbf{a}_{2} + \left(-x_{2}-y_{2}-z_{2}\right) \, \mathbf{a}_{3} & = & -x_{2}a \, \mathbf{\hat{x}}-y_{2}a \, \mathbf{\hat{y}} + z_{2}a \, \mathbf{\hat{z}} & \left(96i\right) & \mbox{C II} \\ 
\mathbf{B}_{15} & = & \left(x_{2}+y_{2}-z_{2}\right) \, \mathbf{a}_{1} + \left(-x_{2}-y_{2}-z_{2}\right) \, \mathbf{a}_{2} + \left(-x_{2}+y_{2}+z_{2}\right) \, \mathbf{a}_{3} & = & -x_{2}a \, \mathbf{\hat{x}} + y_{2}a \, \mathbf{\hat{y}}-z_{2}a \, \mathbf{\hat{z}} & \left(96i\right) & \mbox{C II} \\ 
\mathbf{B}_{16} & = & \left(-x_{2}-y_{2}-z_{2}\right) \, \mathbf{a}_{1} + \left(x_{2}+y_{2}-z_{2}\right) \, \mathbf{a}_{2} + \left(x_{2}-y_{2}+z_{2}\right) \, \mathbf{a}_{3} & = & x_{2}a \, \mathbf{\hat{x}}-y_{2}a \, \mathbf{\hat{y}}-z_{2}a \, \mathbf{\hat{z}} & \left(96i\right) & \mbox{C II} \\ 
\mathbf{B}_{17} & = & \left(x_{2}+y_{2}-z_{2}\right) \, \mathbf{a}_{1} + \left(-x_{2}+y_{2}+z_{2}\right) \, \mathbf{a}_{2} + \left(x_{2}-y_{2}+z_{2}\right) \, \mathbf{a}_{3} & = & z_{2}a \, \mathbf{\hat{x}} + x_{2}a \, \mathbf{\hat{y}} + y_{2}a \, \mathbf{\hat{z}} & \left(96i\right) & \mbox{C II} \\ 
\mathbf{B}_{18} & = & \left(-x_{2}-y_{2}-z_{2}\right) \, \mathbf{a}_{1} + \left(x_{2}-y_{2}+z_{2}\right) \, \mathbf{a}_{2} + \left(-x_{2}+y_{2}+z_{2}\right) \, \mathbf{a}_{3} & = & z_{2}a \, \mathbf{\hat{x}}-x_{2}a \, \mathbf{\hat{y}}-y_{2}a \, \mathbf{\hat{z}} & \left(96i\right) & \mbox{C II} \\ 
\mathbf{B}_{19} & = & \left(-x_{2}+y_{2}+z_{2}\right) \, \mathbf{a}_{1} + \left(x_{2}+y_{2}-z_{2}\right) \, \mathbf{a}_{2} + \left(-x_{2}-y_{2}-z_{2}\right) \, \mathbf{a}_{3} & = & -z_{2}a \, \mathbf{\hat{x}}-x_{2}a \, \mathbf{\hat{y}} + y_{2}a \, \mathbf{\hat{z}} & \left(96i\right) & \mbox{C II} \\ 
\mathbf{B}_{20} & = & \left(x_{2}-y_{2}+z_{2}\right) \, \mathbf{a}_{1} + \left(-x_{2}-y_{2}-z_{2}\right) \, \mathbf{a}_{2} + \left(x_{2}+y_{2}-z_{2}\right) \, \mathbf{a}_{3} & = & -z_{2}a \, \mathbf{\hat{x}} + x_{2}a \, \mathbf{\hat{y}}-y_{2}a \, \mathbf{\hat{z}} & \left(96i\right) & \mbox{C II} \\ 
\mathbf{B}_{21} & = & \left(x_{2}-y_{2}+z_{2}\right) \, \mathbf{a}_{1} + \left(x_{2}+y_{2}-z_{2}\right) \, \mathbf{a}_{2} + \left(-x_{2}+y_{2}+z_{2}\right) \, \mathbf{a}_{3} & = & y_{2}a \, \mathbf{\hat{x}} + z_{2}a \, \mathbf{\hat{y}} + x_{2}a \, \mathbf{\hat{z}} & \left(96i\right) & \mbox{C II} \\ 
\mathbf{B}_{22} & = & \left(-x_{2}+y_{2}+z_{2}\right) \, \mathbf{a}_{1} + \left(-x_{2}-y_{2}-z_{2}\right) \, \mathbf{a}_{2} + \left(x_{2}-y_{2}+z_{2}\right) \, \mathbf{a}_{3} & = & -y_{2}a \, \mathbf{\hat{x}} + z_{2}a \, \mathbf{\hat{y}}-x_{2}a \, \mathbf{\hat{z}} & \left(96i\right) & \mbox{C II} \\ 
\mathbf{B}_{23} & = & \left(-x_{2}-y_{2}-z_{2}\right) \, \mathbf{a}_{1} + \left(-x_{2}+y_{2}+z_{2}\right) \, \mathbf{a}_{2} + \left(x_{2}+y_{2}-z_{2}\right) \, \mathbf{a}_{3} & = & y_{2}a \, \mathbf{\hat{x}}-z_{2}a \, \mathbf{\hat{y}}-x_{2}a \, \mathbf{\hat{z}} & \left(96i\right) & \mbox{C II} \\ 
\mathbf{B}_{24} & = & \left(x_{2}+y_{2}-z_{2}\right) \, \mathbf{a}_{1} + \left(x_{2}-y_{2}+z_{2}\right) \, \mathbf{a}_{2} + \left(-x_{2}-y_{2}-z_{2}\right) \, \mathbf{a}_{3} & = & -y_{2}a \, \mathbf{\hat{x}}-z_{2}a \, \mathbf{\hat{y}} + x_{2}a \, \mathbf{\hat{z}} & \left(96i\right) & \mbox{C II} \\ 
\mathbf{B}_{25} & = & \left(x_{2}-y_{2}-z_{2}\right) \, \mathbf{a}_{1} + \left(-x_{2}+y_{2}-z_{2}\right) \, \mathbf{a}_{2} + \left(-x_{2}-y_{2}+z_{2}\right) \, \mathbf{a}_{3} & = & -x_{2}a \, \mathbf{\hat{x}}-y_{2}a \, \mathbf{\hat{y}}-z_{2}a \, \mathbf{\hat{z}} & \left(96i\right) & \mbox{C II} \\ 
\mathbf{B}_{26} & = & \left(-x_{2}+y_{2}-z_{2}\right) \, \mathbf{a}_{1} + \left(x_{2}-y_{2}-z_{2}\right) \, \mathbf{a}_{2} + \left(x_{2}+y_{2}+z_{2}\right) \, \mathbf{a}_{3} & = & x_{2}a \, \mathbf{\hat{x}} + y_{2}a \, \mathbf{\hat{y}}-z_{2}a \, \mathbf{\hat{z}} & \left(96i\right) & \mbox{C II} \\ 
\mathbf{B}_{27} & = & \left(-x_{2}-y_{2}+z_{2}\right) \, \mathbf{a}_{1} + \left(x_{2}+y_{2}+z_{2}\right) \, \mathbf{a}_{2} + \left(x_{2}-y_{2}-z_{2}\right) \, \mathbf{a}_{3} & = & x_{2}a \, \mathbf{\hat{x}}-y_{2}a \, \mathbf{\hat{y}} + z_{2}a \, \mathbf{\hat{z}} & \left(96i\right) & \mbox{C II} \\ 
\mathbf{B}_{28} & = & \left(x_{2}+y_{2}+z_{2}\right) \, \mathbf{a}_{1} + \left(-x_{2}-y_{2}+z_{2}\right) \, \mathbf{a}_{2} + \left(-x_{2}+y_{2}-z_{2}\right) \, \mathbf{a}_{3} & = & -x_{2}a \, \mathbf{\hat{x}} + y_{2}a \, \mathbf{\hat{y}} + z_{2}a \, \mathbf{\hat{z}} & \left(96i\right) & \mbox{C II} \\ 
\mathbf{B}_{29} & = & \left(-x_{2}-y_{2}+z_{2}\right) \, \mathbf{a}_{1} + \left(x_{2}-y_{2}-z_{2}\right) \, \mathbf{a}_{2} + \left(-x_{2}+y_{2}-z_{2}\right) \, \mathbf{a}_{3} & = & -z_{2}a \, \mathbf{\hat{x}}-x_{2}a \, \mathbf{\hat{y}}-y_{2}a \, \mathbf{\hat{z}} & \left(96i\right) & \mbox{C II} \\ 
\mathbf{B}_{30} & = & \left(x_{2}+y_{2}+z_{2}\right) \, \mathbf{a}_{1} + \left(-x_{2}+y_{2}-z_{2}\right) \, \mathbf{a}_{2} + \left(x_{2}-y_{2}-z_{2}\right) \, \mathbf{a}_{3} & = & -z_{2}a \, \mathbf{\hat{x}} + x_{2}a \, \mathbf{\hat{y}} + y_{2}a \, \mathbf{\hat{z}} & \left(96i\right) & \mbox{C II} \\ 
\mathbf{B}_{31} & = & \left(x_{2}-y_{2}-z_{2}\right) \, \mathbf{a}_{1} + \left(-x_{2}-y_{2}+z_{2}\right) \, \mathbf{a}_{2} + \left(x_{2}+y_{2}+z_{2}\right) \, \mathbf{a}_{3} & = & z_{2}a \, \mathbf{\hat{x}} + x_{2}a \, \mathbf{\hat{y}}-y_{2}a \, \mathbf{\hat{z}} & \left(96i\right) & \mbox{C II} \\ 
\mathbf{B}_{32} & = & \left(-x_{2}+y_{2}-z_{2}\right) \, \mathbf{a}_{1} + \left(x_{2}+y_{2}+z_{2}\right) \, \mathbf{a}_{2} + \left(-x_{2}-y_{2}+z_{2}\right) \, \mathbf{a}_{3} & = & z_{2}a \, \mathbf{\hat{x}}-x_{2}a \, \mathbf{\hat{y}} + y_{2}a \, \mathbf{\hat{z}} & \left(96i\right) & \mbox{C II} \\ 
\mathbf{B}_{33} & = & \left(-x_{2}+y_{2}-z_{2}\right) \, \mathbf{a}_{1} + \left(-x_{2}-y_{2}+z_{2}\right) \, \mathbf{a}_{2} + \left(x_{2}-y_{2}-z_{2}\right) \, \mathbf{a}_{3} & = & -y_{2}a \, \mathbf{\hat{x}}-z_{2}a \, \mathbf{\hat{y}}-x_{2}a \, \mathbf{\hat{z}} & \left(96i\right) & \mbox{C II} \\ 
\mathbf{B}_{34} & = & \left(x_{2}-y_{2}-z_{2}\right) \, \mathbf{a}_{1} + \left(x_{2}+y_{2}+z_{2}\right) \, \mathbf{a}_{2} + \left(-x_{2}+y_{2}-z_{2}\right) \, \mathbf{a}_{3} & = & y_{2}a \, \mathbf{\hat{x}}-z_{2}a \, \mathbf{\hat{y}} + x_{2}a \, \mathbf{\hat{z}} & \left(96i\right) & \mbox{C II} \\ 
\mathbf{B}_{35} & = & \left(x_{2}+y_{2}+z_{2}\right) \, \mathbf{a}_{1} + \left(x_{2}-y_{2}-z_{2}\right) \, \mathbf{a}_{2} + \left(-x_{2}-y_{2}+z_{2}\right) \, \mathbf{a}_{3} & = & -y_{2}a \, \mathbf{\hat{x}} + z_{2}a \, \mathbf{\hat{y}} + x_{2}a \, \mathbf{\hat{z}} & \left(96i\right) & \mbox{C II} \\ 
\mathbf{B}_{36} & = & \left(-x_{2}-y_{2}+z_{2}\right) \, \mathbf{a}_{1} + \left(-x_{2}+y_{2}-z_{2}\right) \, \mathbf{a}_{2} + \left(x_{2}+y_{2}+z_{2}\right) \, \mathbf{a}_{3} & = & y_{2}a \, \mathbf{\hat{x}} + z_{2}a \, \mathbf{\hat{y}}-x_{2}a \, \mathbf{\hat{z}} & \left(96i\right) & \mbox{C II} \\ 
\mathbf{B}_{37} & = & \left(-x_{3}+y_{3}+z_{3}\right) \, \mathbf{a}_{1} + \left(x_{3}-y_{3}+z_{3}\right) \, \mathbf{a}_{2} + \left(x_{3}+y_{3}-z_{3}\right) \, \mathbf{a}_{3} & = & x_{3}a \, \mathbf{\hat{x}} + y_{3}a \, \mathbf{\hat{y}} + z_{3}a \, \mathbf{\hat{z}} & \left(96i\right) & \mbox{C III} \\ 
\mathbf{B}_{38} & = & \left(x_{3}-y_{3}+z_{3}\right) \, \mathbf{a}_{1} + \left(-x_{3}+y_{3}+z_{3}\right) \, \mathbf{a}_{2} + \left(-x_{3}-y_{3}-z_{3}\right) \, \mathbf{a}_{3} & = & -x_{3}a \, \mathbf{\hat{x}}-y_{3}a \, \mathbf{\hat{y}} + z_{3}a \, \mathbf{\hat{z}} & \left(96i\right) & \mbox{C III} \\ 
\mathbf{B}_{39} & = & \left(x_{3}+y_{3}-z_{3}\right) \, \mathbf{a}_{1} + \left(-x_{3}-y_{3}-z_{3}\right) \, \mathbf{a}_{2} + \left(-x_{3}+y_{3}+z_{3}\right) \, \mathbf{a}_{3} & = & -x_{3}a \, \mathbf{\hat{x}} + y_{3}a \, \mathbf{\hat{y}}-z_{3}a \, \mathbf{\hat{z}} & \left(96i\right) & \mbox{C III} \\ 
\mathbf{B}_{40} & = & \left(-x_{3}-y_{3}-z_{3}\right) \, \mathbf{a}_{1} + \left(x_{3}+y_{3}-z_{3}\right) \, \mathbf{a}_{2} + \left(x_{3}-y_{3}+z_{3}\right) \, \mathbf{a}_{3} & = & x_{3}a \, \mathbf{\hat{x}}-y_{3}a \, \mathbf{\hat{y}}-z_{3}a \, \mathbf{\hat{z}} & \left(96i\right) & \mbox{C III} \\ 
\mathbf{B}_{41} & = & \left(x_{3}+y_{3}-z_{3}\right) \, \mathbf{a}_{1} + \left(-x_{3}+y_{3}+z_{3}\right) \, \mathbf{a}_{2} + \left(x_{3}-y_{3}+z_{3}\right) \, \mathbf{a}_{3} & = & z_{3}a \, \mathbf{\hat{x}} + x_{3}a \, \mathbf{\hat{y}} + y_{3}a \, \mathbf{\hat{z}} & \left(96i\right) & \mbox{C III} \\ 
\mathbf{B}_{42} & = & \left(-x_{3}-y_{3}-z_{3}\right) \, \mathbf{a}_{1} + \left(x_{3}-y_{3}+z_{3}\right) \, \mathbf{a}_{2} + \left(-x_{3}+y_{3}+z_{3}\right) \, \mathbf{a}_{3} & = & z_{3}a \, \mathbf{\hat{x}}-x_{3}a \, \mathbf{\hat{y}}-y_{3}a \, \mathbf{\hat{z}} & \left(96i\right) & \mbox{C III} \\ 
\mathbf{B}_{43} & = & \left(-x_{3}+y_{3}+z_{3}\right) \, \mathbf{a}_{1} + \left(x_{3}+y_{3}-z_{3}\right) \, \mathbf{a}_{2} + \left(-x_{3}-y_{3}-z_{3}\right) \, \mathbf{a}_{3} & = & -z_{3}a \, \mathbf{\hat{x}}-x_{3}a \, \mathbf{\hat{y}} + y_{3}a \, \mathbf{\hat{z}} & \left(96i\right) & \mbox{C III} \\ 
\mathbf{B}_{44} & = & \left(x_{3}-y_{3}+z_{3}\right) \, \mathbf{a}_{1} + \left(-x_{3}-y_{3}-z_{3}\right) \, \mathbf{a}_{2} + \left(x_{3}+y_{3}-z_{3}\right) \, \mathbf{a}_{3} & = & -z_{3}a \, \mathbf{\hat{x}} + x_{3}a \, \mathbf{\hat{y}}-y_{3}a \, \mathbf{\hat{z}} & \left(96i\right) & \mbox{C III} \\ 
\mathbf{B}_{45} & = & \left(x_{3}-y_{3}+z_{3}\right) \, \mathbf{a}_{1} + \left(x_{3}+y_{3}-z_{3}\right) \, \mathbf{a}_{2} + \left(-x_{3}+y_{3}+z_{3}\right) \, \mathbf{a}_{3} & = & y_{3}a \, \mathbf{\hat{x}} + z_{3}a \, \mathbf{\hat{y}} + x_{3}a \, \mathbf{\hat{z}} & \left(96i\right) & \mbox{C III} \\ 
\mathbf{B}_{46} & = & \left(-x_{3}+y_{3}+z_{3}\right) \, \mathbf{a}_{1} + \left(-x_{3}-y_{3}-z_{3}\right) \, \mathbf{a}_{2} + \left(x_{3}-y_{3}+z_{3}\right) \, \mathbf{a}_{3} & = & -y_{3}a \, \mathbf{\hat{x}} + z_{3}a \, \mathbf{\hat{y}}-x_{3}a \, \mathbf{\hat{z}} & \left(96i\right) & \mbox{C III} \\ 
\mathbf{B}_{47} & = & \left(-x_{3}-y_{3}-z_{3}\right) \, \mathbf{a}_{1} + \left(-x_{3}+y_{3}+z_{3}\right) \, \mathbf{a}_{2} + \left(x_{3}+y_{3}-z_{3}\right) \, \mathbf{a}_{3} & = & y_{3}a \, \mathbf{\hat{x}}-z_{3}a \, \mathbf{\hat{y}}-x_{3}a \, \mathbf{\hat{z}} & \left(96i\right) & \mbox{C III} \\ 
\mathbf{B}_{48} & = & \left(x_{3}+y_{3}-z_{3}\right) \, \mathbf{a}_{1} + \left(x_{3}-y_{3}+z_{3}\right) \, \mathbf{a}_{2} + \left(-x_{3}-y_{3}-z_{3}\right) \, \mathbf{a}_{3} & = & -y_{3}a \, \mathbf{\hat{x}}-z_{3}a \, \mathbf{\hat{y}} + x_{3}a \, \mathbf{\hat{z}} & \left(96i\right) & \mbox{C III} \\ 
\mathbf{B}_{49} & = & \left(x_{3}-y_{3}-z_{3}\right) \, \mathbf{a}_{1} + \left(-x_{3}+y_{3}-z_{3}\right) \, \mathbf{a}_{2} + \left(-x_{3}-y_{3}+z_{3}\right) \, \mathbf{a}_{3} & = & -x_{3}a \, \mathbf{\hat{x}}-y_{3}a \, \mathbf{\hat{y}}-z_{3}a \, \mathbf{\hat{z}} & \left(96i\right) & \mbox{C III} \\ 
\mathbf{B}_{50} & = & \left(-x_{3}+y_{3}-z_{3}\right) \, \mathbf{a}_{1} + \left(x_{3}-y_{3}-z_{3}\right) \, \mathbf{a}_{2} + \left(x_{3}+y_{3}+z_{3}\right) \, \mathbf{a}_{3} & = & x_{3}a \, \mathbf{\hat{x}} + y_{3}a \, \mathbf{\hat{y}}-z_{3}a \, \mathbf{\hat{z}} & \left(96i\right) & \mbox{C III} \\ 
\mathbf{B}_{51} & = & \left(-x_{3}-y_{3}+z_{3}\right) \, \mathbf{a}_{1} + \left(x_{3}+y_{3}+z_{3}\right) \, \mathbf{a}_{2} + \left(x_{3}-y_{3}-z_{3}\right) \, \mathbf{a}_{3} & = & x_{3}a \, \mathbf{\hat{x}}-y_{3}a \, \mathbf{\hat{y}} + z_{3}a \, \mathbf{\hat{z}} & \left(96i\right) & \mbox{C III} \\ 
\mathbf{B}_{52} & = & \left(x_{3}+y_{3}+z_{3}\right) \, \mathbf{a}_{1} + \left(-x_{3}-y_{3}+z_{3}\right) \, \mathbf{a}_{2} + \left(-x_{3}+y_{3}-z_{3}\right) \, \mathbf{a}_{3} & = & -x_{3}a \, \mathbf{\hat{x}} + y_{3}a \, \mathbf{\hat{y}} + z_{3}a \, \mathbf{\hat{z}} & \left(96i\right) & \mbox{C III} \\ 
\mathbf{B}_{53} & = & \left(-x_{3}-y_{3}+z_{3}\right) \, \mathbf{a}_{1} + \left(x_{3}-y_{3}-z_{3}\right) \, \mathbf{a}_{2} + \left(-x_{3}+y_{3}-z_{3}\right) \, \mathbf{a}_{3} & = & -z_{3}a \, \mathbf{\hat{x}}-x_{3}a \, \mathbf{\hat{y}}-y_{3}a \, \mathbf{\hat{z}} & \left(96i\right) & \mbox{C III} \\ 
\mathbf{B}_{54} & = & \left(x_{3}+y_{3}+z_{3}\right) \, \mathbf{a}_{1} + \left(-x_{3}+y_{3}-z_{3}\right) \, \mathbf{a}_{2} + \left(x_{3}-y_{3}-z_{3}\right) \, \mathbf{a}_{3} & = & -z_{3}a \, \mathbf{\hat{x}} + x_{3}a \, \mathbf{\hat{y}} + y_{3}a \, \mathbf{\hat{z}} & \left(96i\right) & \mbox{C III} \\ 
\mathbf{B}_{55} & = & \left(x_{3}-y_{3}-z_{3}\right) \, \mathbf{a}_{1} + \left(-x_{3}-y_{3}+z_{3}\right) \, \mathbf{a}_{2} + \left(x_{3}+y_{3}+z_{3}\right) \, \mathbf{a}_{3} & = & z_{3}a \, \mathbf{\hat{x}} + x_{3}a \, \mathbf{\hat{y}}-y_{3}a \, \mathbf{\hat{z}} & \left(96i\right) & \mbox{C III} \\ 
\mathbf{B}_{56} & = & \left(-x_{3}+y_{3}-z_{3}\right) \, \mathbf{a}_{1} + \left(x_{3}+y_{3}+z_{3}\right) \, \mathbf{a}_{2} + \left(-x_{3}-y_{3}+z_{3}\right) \, \mathbf{a}_{3} & = & z_{3}a \, \mathbf{\hat{x}}-x_{3}a \, \mathbf{\hat{y}} + y_{3}a \, \mathbf{\hat{z}} & \left(96i\right) & \mbox{C III} \\ 
\mathbf{B}_{57} & = & \left(-x_{3}+y_{3}-z_{3}\right) \, \mathbf{a}_{1} + \left(-x_{3}-y_{3}+z_{3}\right) \, \mathbf{a}_{2} + \left(x_{3}-y_{3}-z_{3}\right) \, \mathbf{a}_{3} & = & -y_{3}a \, \mathbf{\hat{x}}-z_{3}a \, \mathbf{\hat{y}}-x_{3}a \, \mathbf{\hat{z}} & \left(96i\right) & \mbox{C III} \\ 
\mathbf{B}_{58} & = & \left(x_{3}-y_{3}-z_{3}\right) \, \mathbf{a}_{1} + \left(x_{3}+y_{3}+z_{3}\right) \, \mathbf{a}_{2} + \left(-x_{3}+y_{3}-z_{3}\right) \, \mathbf{a}_{3} & = & y_{3}a \, \mathbf{\hat{x}}-z_{3}a \, \mathbf{\hat{y}} + x_{3}a \, \mathbf{\hat{z}} & \left(96i\right) & \mbox{C III} \\ 
\mathbf{B}_{59} & = & \left(x_{3}+y_{3}+z_{3}\right) \, \mathbf{a}_{1} + \left(x_{3}-y_{3}-z_{3}\right) \, \mathbf{a}_{2} + \left(-x_{3}-y_{3}+z_{3}\right) \, \mathbf{a}_{3} & = & -y_{3}a \, \mathbf{\hat{x}} + z_{3}a \, \mathbf{\hat{y}} + x_{3}a \, \mathbf{\hat{z}} & \left(96i\right) & \mbox{C III} \\ 
\mathbf{B}_{60} & = & \left(-x_{3}-y_{3}+z_{3}\right) \, \mathbf{a}_{1} + \left(-x_{3}+y_{3}-z_{3}\right) \, \mathbf{a}_{2} + \left(x_{3}+y_{3}+z_{3}\right) \, \mathbf{a}_{3} & = & y_{3}a \, \mathbf{\hat{x}} + z_{3}a \, \mathbf{\hat{y}}-x_{3}a \, \mathbf{\hat{z}} & \left(96i\right) & \mbox{C III} \\ 
\end{longtabu}
\renewcommand{\arraystretch}{1.0}
\noindent \hrulefill
\\
\textbf{References:}
\vspace*{-0.25cm}
\begin{flushleft}
  - \bibentry{Dorset_ACristA_50_94}. \\
\end{flushleft}
\noindent \hrulefill
\\
\textbf{Geometry files:}
\\
\noindent  - CIF: pp. {\hyperref[A_cF240_202_h2i_cif]{\pageref{A_cF240_202_h2i_cif}}} \\
\noindent  - POSCAR: pp. {\hyperref[A_cF240_202_h2i_poscar]{\pageref{A_cF240_202_h2i_poscar}}} \\
\onecolumn
{\phantomsection\label{A2BCD3E6_cF208_203_e_c_d_f_g}}
\subsection*{\huge \textbf{{\normalfont \begin{raggedleft}Pyrochlore (Na$_3$Co(CO$_3$)$_2$Cl) Structure: \end{raggedleft} \\ A2BCD3E6\_cF208\_203\_e\_c\_d\_f\_g}}}
\noindent \hrulefill
\vspace*{0.25cm}
\begin{figure}[htp]
  \centering
  \vspace{-1em}
  {\includegraphics[width=1\textwidth]{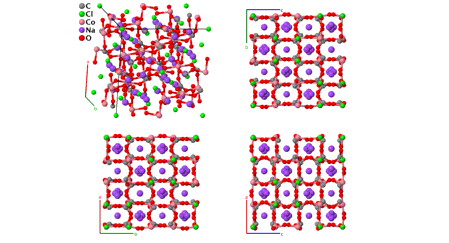}}
\end{figure}
\vspace*{-0.5cm}
\renewcommand{\arraystretch}{1.5}
\begin{equation*}
  \begin{array}{>{$\hspace{-0.15cm}}l<{$}>{$}p{0.5cm}<{$}>{$}p{18.5cm}<{$}}
    \mbox{\large \textbf{Prototype}} &\colon & \ce{Na3Co(CO3)2Cl} \\
    \mbox{\large \textbf{\AFLOW\ prototype label}} &\colon & \mbox{A2BCD3E6\_cF208\_203\_e\_c\_d\_f\_g} \\
    \mbox{\large \textbf{\textit{Strukturbericht} designation}} &\colon & \mbox{None} \\
    \mbox{\large \textbf{Pearson symbol}} &\colon & \mbox{cF208} \\
    \mbox{\large \textbf{Space group number}} &\colon & 203 \\
    \mbox{\large \textbf{Space group symbol}} &\colon & Fd\bar{3} \\
    \mbox{\large \textbf{\AFLOW\ prototype command}} &\colon &  \texttt{aflow} \,  \, \texttt{-{}-proto=A2BCD3E6\_cF208\_203\_e\_c\_d\_f\_g } \, \newline \texttt{-{}-params=}{a,x_{3},x_{4},x_{5},y_{5},z_{5} }
  \end{array}
\end{equation*}
\renewcommand{\arraystretch}{1.0}

\vspace*{-0.25cm}
\noindent \hrulefill
\begin{itemize}
  \item{This structure was suggested to us by Prof. Joel Helton, United States
Naval Academy.
This is a pyrochlore-like antiferromagnet, which we loosely define as
a structure with magnetic ions on the corners of corner-sharing
tetrahedra.
}
\end{itemize}

\noindent \parbox{1 \linewidth}{
\noindent \hrulefill
\\
\textbf{Face-centered Cubic primitive vectors:} \\
\vspace*{-0.25cm}
\begin{tabular}{cc}
  \begin{tabular}{c}
    \parbox{0.6 \linewidth}{
      \renewcommand{\arraystretch}{1.5}
      \begin{equation*}
        \centering
        \begin{array}{ccc}
              \mathbf{a}_1 & = & \frac12 \, a \, \mathbf{\hat{y}} + \frac12 \, a \, \mathbf{\hat{z}} \\
    \mathbf{a}_2 & = & \frac12 \, a \, \mathbf{\hat{x}} + \frac12 \, a \, \mathbf{\hat{z}} \\
    \mathbf{a}_3 & = & \frac12 \, a \, \mathbf{\hat{x}} + \frac12 \, a \, \mathbf{\hat{y}} \\

        \end{array}
      \end{equation*}
    }
    \renewcommand{\arraystretch}{1.0}
  \end{tabular}
  \begin{tabular}{c}
    \includegraphics[width=0.3\linewidth]{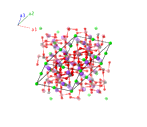} \\
  \end{tabular}
\end{tabular}

}
\vspace*{-0.25cm}

\noindent \hrulefill
\\
\textbf{Basis vectors:}
\vspace*{-0.25cm}
\renewcommand{\arraystretch}{1.5}
\begin{longtabu} to \textwidth{>{\centering $}X[-1,c,c]<{$}>{\centering $}X[-1,c,c]<{$}>{\centering $}X[-1,c,c]<{$}>{\centering $}X[-1,c,c]<{$}>{\centering $}X[-1,c,c]<{$}>{\centering $}X[-1,c,c]<{$}>{\centering $}X[-1,c,c]<{$}}
  & & \mbox{Lattice Coordinates} & & \mbox{Cartesian Coordinates} &\mbox{Wyckoff Position} & \mbox{Atom Type} \\  
  \mathbf{B}_{1} & = & 0 \, \mathbf{a}_{1} + 0 \, \mathbf{a}_{2} + 0 \, \mathbf{a}_{3} & = & 0 \, \mathbf{\hat{x}} + 0 \, \mathbf{\hat{y}} + 0 \, \mathbf{\hat{z}} & \left(16c\right) & \mbox{Cl} \\ 
\mathbf{B}_{2} & = & \frac{1}{2} \, \mathbf{a}_{3} & = & \frac{1}{4}a \, \mathbf{\hat{x}} + \frac{1}{4}a \, \mathbf{\hat{y}} & \left(16c\right) & \mbox{Cl} \\ 
\mathbf{B}_{3} & = & \frac{1}{2} \, \mathbf{a}_{2} & = & \frac{1}{4}a \, \mathbf{\hat{x}} + \frac{1}{4}a \, \mathbf{\hat{z}} & \left(16c\right) & \mbox{Cl} \\ 
\mathbf{B}_{4} & = & \frac{1}{2} \, \mathbf{a}_{1} & = & \frac{1}{4}a \, \mathbf{\hat{y}} + \frac{1}{4}a \, \mathbf{\hat{z}} & \left(16c\right) & \mbox{Cl} \\ 
\mathbf{B}_{5} & = & \frac{1}{2} \, \mathbf{a}_{1} + \frac{1}{2} \, \mathbf{a}_{2} + \frac{1}{2} \, \mathbf{a}_{3} & = & \frac{1}{2}a \, \mathbf{\hat{x}} + \frac{1}{2}a \, \mathbf{\hat{y}} + \frac{1}{2}a \, \mathbf{\hat{z}} & \left(16d\right) & \mbox{Co} \\ 
\mathbf{B}_{6} & = & \frac{1}{2} \, \mathbf{a}_{1} + \frac{1}{2} \, \mathbf{a}_{2} & = & \frac{1}{4}a \, \mathbf{\hat{x}} + \frac{1}{4}a \, \mathbf{\hat{y}} + \frac{1}{2}a \, \mathbf{\hat{z}} & \left(16d\right) & \mbox{Co} \\ 
\mathbf{B}_{7} & = & \frac{1}{2} \, \mathbf{a}_{1} + \frac{1}{2} \, \mathbf{a}_{3} & = & \frac{1}{4}a \, \mathbf{\hat{x}} + \frac{1}{2}a \, \mathbf{\hat{y}} + \frac{1}{4}a \, \mathbf{\hat{z}} & \left(16d\right) & \mbox{Co} \\ 
\mathbf{B}_{8} & = & \frac{1}{2} \, \mathbf{a}_{2} + \frac{1}{2} \, \mathbf{a}_{3} & = & \frac{1}{2}a \, \mathbf{\hat{x}} + \frac{1}{4}a \, \mathbf{\hat{y}} + \frac{1}{4}a \, \mathbf{\hat{z}} & \left(16d\right) & \mbox{Co} \\ 
\mathbf{B}_{9} & = & x_{3} \, \mathbf{a}_{1} + x_{3} \, \mathbf{a}_{2} + x_{3} \, \mathbf{a}_{3} & = & x_{3}a \, \mathbf{\hat{x}} + x_{3}a \, \mathbf{\hat{y}} + x_{3}a \, \mathbf{\hat{z}} & \left(32e\right) & \mbox{C} \\ 
\mathbf{B}_{10} & = & x_{3} \, \mathbf{a}_{1} + x_{3} \, \mathbf{a}_{2} + \left(\frac{1}{2} - 3x_{3}\right) \, \mathbf{a}_{3} & = & \left(\frac{1}{4} - x_{3}\right)a \, \mathbf{\hat{x}} + \left(\frac{1}{4} - x_{3}\right)a \, \mathbf{\hat{y}} + x_{3}a \, \mathbf{\hat{z}} & \left(32e\right) & \mbox{C} \\ 
\mathbf{B}_{11} & = & x_{3} \, \mathbf{a}_{1} + \left(\frac{1}{2} - 3x_{3}\right) \, \mathbf{a}_{2} + x_{3} \, \mathbf{a}_{3} & = & \left(\frac{1}{4} - x_{3}\right)a \, \mathbf{\hat{x}} + x_{3}a \, \mathbf{\hat{y}} + \left(\frac{1}{4} - x_{3}\right)a \, \mathbf{\hat{z}} & \left(32e\right) & \mbox{C} \\ 
\mathbf{B}_{12} & = & \left(\frac{1}{2} - 3x_{3}\right) \, \mathbf{a}_{1} + x_{3} \, \mathbf{a}_{2} + x_{3} \, \mathbf{a}_{3} & = & x_{3}a \, \mathbf{\hat{x}} + \left(\frac{1}{4} - x_{3}\right)a \, \mathbf{\hat{y}} + \left(\frac{1}{4} - x_{3}\right)a \, \mathbf{\hat{z}} & \left(32e\right) & \mbox{C} \\ 
\mathbf{B}_{13} & = & -x_{3} \, \mathbf{a}_{1}-x_{3} \, \mathbf{a}_{2}-x_{3} \, \mathbf{a}_{3} & = & -x_{3}a \, \mathbf{\hat{x}}-x_{3}a \, \mathbf{\hat{y}}-x_{3}a \, \mathbf{\hat{z}} & \left(32e\right) & \mbox{C} \\ 
\mathbf{B}_{14} & = & -x_{3} \, \mathbf{a}_{1}-x_{3} \, \mathbf{a}_{2} + \left(\frac{1}{2} +3x_{3}\right) \, \mathbf{a}_{3} & = & \left(\frac{1}{4} +x_{3}\right)a \, \mathbf{\hat{x}} + \left(\frac{1}{4} +x_{3}\right)a \, \mathbf{\hat{y}}-x_{3}a \, \mathbf{\hat{z}} & \left(32e\right) & \mbox{C} \\ 
\mathbf{B}_{15} & = & -x_{3} \, \mathbf{a}_{1} + \left(\frac{1}{2} +3x_{3}\right) \, \mathbf{a}_{2}-x_{3} \, \mathbf{a}_{3} & = & \left(\frac{1}{4} +x_{3}\right)a \, \mathbf{\hat{x}}-x_{3}a \, \mathbf{\hat{y}} + \left(\frac{1}{4} +x_{3}\right)a \, \mathbf{\hat{z}} & \left(32e\right) & \mbox{C} \\ 
\mathbf{B}_{16} & = & \left(\frac{1}{2} +3x_{3}\right) \, \mathbf{a}_{1}-x_{3} \, \mathbf{a}_{2}-x_{3} \, \mathbf{a}_{3} & = & -x_{3}a \, \mathbf{\hat{x}} + \left(\frac{1}{4} +x_{3}\right)a \, \mathbf{\hat{y}} + \left(\frac{1}{4} +x_{3}\right)a \, \mathbf{\hat{z}} & \left(32e\right) & \mbox{C} \\ 
\mathbf{B}_{17} & = & \left(\frac{1}{4} - x_{4}\right) \, \mathbf{a}_{1} + x_{4} \, \mathbf{a}_{2} + x_{4} \, \mathbf{a}_{3} & = & x_{4}a \, \mathbf{\hat{x}} + \frac{1}{8}a \, \mathbf{\hat{y}} + \frac{1}{8}a \, \mathbf{\hat{z}} & \left(48f\right) & \mbox{Na} \\ 
\mathbf{B}_{18} & = & x_{4} \, \mathbf{a}_{1} + \left(\frac{1}{4} - x_{4}\right) \, \mathbf{a}_{2} + \left(\frac{1}{4} - x_{4}\right) \, \mathbf{a}_{3} & = & \left(\frac{1}{4} - x_{4}\right)a \, \mathbf{\hat{x}} + \frac{1}{8}a \, \mathbf{\hat{y}} + \frac{1}{8}a \, \mathbf{\hat{z}} & \left(48f\right) & \mbox{Na} \\ 
\mathbf{B}_{19} & = & x_{4} \, \mathbf{a}_{1} + \left(\frac{1}{4} - x_{4}\right) \, \mathbf{a}_{2} + x_{4} \, \mathbf{a}_{3} & = & \frac{1}{8}a \, \mathbf{\hat{x}} + x_{4}a \, \mathbf{\hat{y}} + \frac{1}{8}a \, \mathbf{\hat{z}} & \left(48f\right) & \mbox{Na} \\ 
\mathbf{B}_{20} & = & \left(\frac{1}{4} - x_{4}\right) \, \mathbf{a}_{1} + x_{4} \, \mathbf{a}_{2} + \left(\frac{1}{4} - x_{4}\right) \, \mathbf{a}_{3} & = & \frac{1}{8}a \, \mathbf{\hat{x}} + \left(\frac{1}{4} - x_{4}\right)a \, \mathbf{\hat{y}} + \frac{1}{8}a \, \mathbf{\hat{z}} & \left(48f\right) & \mbox{Na} \\ 
\mathbf{B}_{21} & = & x_{4} \, \mathbf{a}_{1} + x_{4} \, \mathbf{a}_{2} + \left(\frac{1}{4} - x_{4}\right) \, \mathbf{a}_{3} & = & \frac{1}{8}a \, \mathbf{\hat{x}} + \frac{1}{8}a \, \mathbf{\hat{y}} + x_{4}a \, \mathbf{\hat{z}} & \left(48f\right) & \mbox{Na} \\ 
\mathbf{B}_{22} & = & \left(\frac{1}{4} - x_{4}\right) \, \mathbf{a}_{1} + \left(\frac{1}{4} - x_{4}\right) \, \mathbf{a}_{2} + x_{4} \, \mathbf{a}_{3} & = & \frac{1}{8}a \, \mathbf{\hat{x}} + \frac{1}{8}a \, \mathbf{\hat{y}} + \left(\frac{1}{4} - x_{4}\right)a \, \mathbf{\hat{z}} & \left(48f\right) & \mbox{Na} \\ 
\mathbf{B}_{23} & = & \left(\frac{3}{4} +x_{4}\right) \, \mathbf{a}_{1}-x_{4} \, \mathbf{a}_{2}-x_{4} \, \mathbf{a}_{3} & = & -x_{4}a \, \mathbf{\hat{x}} + \frac{3}{8}a \, \mathbf{\hat{y}} + \frac{3}{8}a \, \mathbf{\hat{z}} & \left(48f\right) & \mbox{Na} \\ 
\mathbf{B}_{24} & = & -x_{4} \, \mathbf{a}_{1} + \left(\frac{3}{4} +x_{4}\right) \, \mathbf{a}_{2} + \left(\frac{3}{4} +x_{4}\right) \, \mathbf{a}_{3} & = & \left(\frac{3}{4} +x_{4}\right)a \, \mathbf{\hat{x}} + \frac{3}{8}a \, \mathbf{\hat{y}} + \frac{3}{8}a \, \mathbf{\hat{z}} & \left(48f\right) & \mbox{Na} \\ 
\mathbf{B}_{25} & = & -x_{4} \, \mathbf{a}_{1} + \left(\frac{3}{4} +x_{4}\right) \, \mathbf{a}_{2}-x_{4} \, \mathbf{a}_{3} & = & \frac{3}{8}a \, \mathbf{\hat{x}}-x_{4}a \, \mathbf{\hat{y}} + \frac{3}{8}a \, \mathbf{\hat{z}} & \left(48f\right) & \mbox{Na} \\ 
\mathbf{B}_{26} & = & \left(\frac{3}{4} +x_{4}\right) \, \mathbf{a}_{1}-x_{4} \, \mathbf{a}_{2} + \left(\frac{3}{4} +x_{4}\right) \, \mathbf{a}_{3} & = & \frac{3}{8}a \, \mathbf{\hat{x}} + \left(\frac{3}{4} +x_{4}\right)a \, \mathbf{\hat{y}} + \frac{3}{8}a \, \mathbf{\hat{z}} & \left(48f\right) & \mbox{Na} \\ 
\mathbf{B}_{27} & = & -x_{4} \, \mathbf{a}_{1}-x_{4} \, \mathbf{a}_{2} + \left(\frac{3}{4} +x_{4}\right) \, \mathbf{a}_{3} & = & \frac{3}{8}a \, \mathbf{\hat{x}} + \frac{3}{8}a \, \mathbf{\hat{y}}-x_{4}a \, \mathbf{\hat{z}} & \left(48f\right) & \mbox{Na} \\ 
\mathbf{B}_{28} & = & \left(\frac{3}{4} +x_{4}\right) \, \mathbf{a}_{1} + \left(\frac{3}{4} +x_{4}\right) \, \mathbf{a}_{2}-x_{4} \, \mathbf{a}_{3} & = & \frac{3}{8}a \, \mathbf{\hat{x}} + \frac{3}{8}a \, \mathbf{\hat{y}} + \left(\frac{3}{4} +x_{4}\right)a \, \mathbf{\hat{z}} & \left(48f\right) & \mbox{Na} \\ 
\mathbf{B}_{29} & = & \left(-x_{5}+y_{5}+z_{5}\right) \, \mathbf{a}_{1} + \left(x_{5}-y_{5}+z_{5}\right) \, \mathbf{a}_{2} + \left(x_{5}+y_{5}-z_{5}\right) \, \mathbf{a}_{3} & = & x_{5}a \, \mathbf{\hat{x}} + y_{5}a \, \mathbf{\hat{y}} + z_{5}a \, \mathbf{\hat{z}} & \left(96g\right) & \mbox{O} \\ 
\mathbf{B}_{30} & = & \left(x_{5}-y_{5}+z_{5}\right) \, \mathbf{a}_{1} + \left(-x_{5}+y_{5}+z_{5}\right) \, \mathbf{a}_{2} + \left(\frac{1}{2} - x_{5} - y_{5} - z_{5}\right) \, \mathbf{a}_{3} & = & \left(\frac{1}{4} - x_{5}\right)a \, \mathbf{\hat{x}} + \left(\frac{1}{4} - y_{5}\right)a \, \mathbf{\hat{y}} + z_{5}a \, \mathbf{\hat{z}} & \left(96g\right) & \mbox{O} \\ 
\mathbf{B}_{31} & = & \left(x_{5}+y_{5}-z_{5}\right) \, \mathbf{a}_{1} + \left(\frac{1}{2} - x_{5} - y_{5} - z_{5}\right) \, \mathbf{a}_{2} + \left(-x_{5}+y_{5}+z_{5}\right) \, \mathbf{a}_{3} & = & \left(\frac{1}{4} - x_{5}\right)a \, \mathbf{\hat{x}} + y_{5}a \, \mathbf{\hat{y}} + \left(\frac{1}{4} - z_{5}\right)a \, \mathbf{\hat{z}} & \left(96g\right) & \mbox{O} \\ 
\mathbf{B}_{32} & = & \left(\frac{1}{2} - x_{5} - y_{5} - z_{5}\right) \, \mathbf{a}_{1} + \left(x_{5}+y_{5}-z_{5}\right) \, \mathbf{a}_{2} + \left(x_{5}-y_{5}+z_{5}\right) \, \mathbf{a}_{3} & = & x_{5}a \, \mathbf{\hat{x}} + \left(\frac{1}{4} - y_{5}\right)a \, \mathbf{\hat{y}} + \left(\frac{1}{4} - z_{5}\right)a \, \mathbf{\hat{z}} & \left(96g\right) & \mbox{O} \\ 
\mathbf{B}_{33} & = & \left(x_{5}+y_{5}-z_{5}\right) \, \mathbf{a}_{1} + \left(-x_{5}+y_{5}+z_{5}\right) \, \mathbf{a}_{2} + \left(x_{5}-y_{5}+z_{5}\right) \, \mathbf{a}_{3} & = & z_{5}a \, \mathbf{\hat{x}} + x_{5}a \, \mathbf{\hat{y}} + y_{5}a \, \mathbf{\hat{z}} & \left(96g\right) & \mbox{O} \\ 
\mathbf{B}_{34} & = & \left(\frac{1}{2} - x_{5} - y_{5} - z_{5}\right) \, \mathbf{a}_{1} + \left(x_{5}-y_{5}+z_{5}\right) \, \mathbf{a}_{2} + \left(-x_{5}+y_{5}+z_{5}\right) \, \mathbf{a}_{3} & = & z_{5}a \, \mathbf{\hat{x}} + \left(\frac{1}{4} - x_{5}\right)a \, \mathbf{\hat{y}} + \left(\frac{1}{4} - y_{5}\right)a \, \mathbf{\hat{z}} & \left(96g\right) & \mbox{O} \\ 
\mathbf{B}_{35} & = & \left(-x_{5}+y_{5}+z_{5}\right) \, \mathbf{a}_{1} + \left(x_{5}+y_{5}-z_{5}\right) \, \mathbf{a}_{2} + \left(\frac{1}{2} - x_{5} - y_{5} - z_{5}\right) \, \mathbf{a}_{3} & = & \left(\frac{1}{4} - z_{5}\right)a \, \mathbf{\hat{x}} + \left(\frac{1}{4} - x_{5}\right)a \, \mathbf{\hat{y}} + y_{5}a \, \mathbf{\hat{z}} & \left(96g\right) & \mbox{O} \\ 
\mathbf{B}_{36} & = & \left(x_{5}-y_{5}+z_{5}\right) \, \mathbf{a}_{1} + \left(\frac{1}{2} - x_{5} - y_{5} - z_{5}\right) \, \mathbf{a}_{2} + \left(x_{5}+y_{5}-z_{5}\right) \, \mathbf{a}_{3} & = & \left(\frac{1}{4} - z_{5}\right)a \, \mathbf{\hat{x}} + x_{5}a \, \mathbf{\hat{y}} + \left(\frac{1}{4} - y_{5}\right)a \, \mathbf{\hat{z}} & \left(96g\right) & \mbox{O} \\ 
\mathbf{B}_{37} & = & \left(x_{5}-y_{5}+z_{5}\right) \, \mathbf{a}_{1} + \left(x_{5}+y_{5}-z_{5}\right) \, \mathbf{a}_{2} + \left(-x_{5}+y_{5}+z_{5}\right) \, \mathbf{a}_{3} & = & y_{5}a \, \mathbf{\hat{x}} + z_{5}a \, \mathbf{\hat{y}} + x_{5}a \, \mathbf{\hat{z}} & \left(96g\right) & \mbox{O} \\ 
\mathbf{B}_{38} & = & \left(-x_{5}+y_{5}+z_{5}\right) \, \mathbf{a}_{1} + \left(\frac{1}{2} - x_{5} - y_{5} - z_{5}\right) \, \mathbf{a}_{2} + \left(x_{5}-y_{5}+z_{5}\right) \, \mathbf{a}_{3} & = & \left(\frac{1}{4} - y_{5}\right)a \, \mathbf{\hat{x}} + z_{5}a \, \mathbf{\hat{y}} + \left(\frac{1}{4} - x_{5}\right)a \, \mathbf{\hat{z}} & \left(96g\right) & \mbox{O} \\ 
\mathbf{B}_{39} & = & \left(\frac{1}{2} - x_{5} - y_{5} - z_{5}\right) \, \mathbf{a}_{1} + \left(-x_{5}+y_{5}+z_{5}\right) \, \mathbf{a}_{2} + \left(x_{5}+y_{5}-z_{5}\right) \, \mathbf{a}_{3} & = & y_{5}a \, \mathbf{\hat{x}} + \left(\frac{1}{4} - z_{5}\right)a \, \mathbf{\hat{y}} + \left(\frac{1}{4} - x_{5}\right)a \, \mathbf{\hat{z}} & \left(96g\right) & \mbox{O} \\ 
\mathbf{B}_{40} & = & \left(x_{5}+y_{5}-z_{5}\right) \, \mathbf{a}_{1} + \left(x_{5}-y_{5}+z_{5}\right) \, \mathbf{a}_{2} + \left(\frac{1}{2} - x_{5} - y_{5} - z_{5}\right) \, \mathbf{a}_{3} & = & \left(\frac{1}{4} - y_{5}\right)a \, \mathbf{\hat{x}} + \left(\frac{1}{4} - z_{5}\right)a \, \mathbf{\hat{y}} + x_{5}a \, \mathbf{\hat{z}} & \left(96g\right) & \mbox{O} \\ 
\mathbf{B}_{41} & = & \left(x_{5}-y_{5}-z_{5}\right) \, \mathbf{a}_{1} + \left(-x_{5}+y_{5}-z_{5}\right) \, \mathbf{a}_{2} + \left(-x_{5}-y_{5}+z_{5}\right) \, \mathbf{a}_{3} & = & -x_{5}a \, \mathbf{\hat{x}}-y_{5}a \, \mathbf{\hat{y}}-z_{5}a \, \mathbf{\hat{z}} & \left(96g\right) & \mbox{O} \\ 
\mathbf{B}_{42} & = & \left(-x_{5}+y_{5}-z_{5}\right) \, \mathbf{a}_{1} + \left(x_{5}-y_{5}-z_{5}\right) \, \mathbf{a}_{2} + \left(\frac{1}{2} +x_{5} + y_{5} + z_{5}\right) \, \mathbf{a}_{3} & = & \left(\frac{1}{4} +x_{5}\right)a \, \mathbf{\hat{x}} + \left(\frac{1}{4} +y_{5}\right)a \, \mathbf{\hat{y}}-z_{5}a \, \mathbf{\hat{z}} & \left(96g\right) & \mbox{O} \\ 
\mathbf{B}_{43} & = & \left(-x_{5}-y_{5}+z_{5}\right) \, \mathbf{a}_{1} + \left(\frac{1}{2} +x_{5} + y_{5} + z_{5}\right) \, \mathbf{a}_{2} + \left(x_{5}-y_{5}-z_{5}\right) \, \mathbf{a}_{3} & = & \left(\frac{1}{4} +x_{5}\right)a \, \mathbf{\hat{x}}-y_{5}a \, \mathbf{\hat{y}} + \left(\frac{1}{4} +z_{5}\right)a \, \mathbf{\hat{z}} & \left(96g\right) & \mbox{O} \\ 
\mathbf{B}_{44} & = & \left(\frac{1}{2} +x_{5} + y_{5} + z_{5}\right) \, \mathbf{a}_{1} + \left(-x_{5}-y_{5}+z_{5}\right) \, \mathbf{a}_{2} + \left(-x_{5}+y_{5}-z_{5}\right) \, \mathbf{a}_{3} & = & -x_{5}a \, \mathbf{\hat{x}} + \left(\frac{1}{4} +y_{5}\right)a \, \mathbf{\hat{y}} + \left(\frac{1}{4} +z_{5}\right)a \, \mathbf{\hat{z}} & \left(96g\right) & \mbox{O} \\ 
\mathbf{B}_{45} & = & \left(-x_{5}-y_{5}+z_{5}\right) \, \mathbf{a}_{1} + \left(x_{5}-y_{5}-z_{5}\right) \, \mathbf{a}_{2} + \left(-x_{5}+y_{5}-z_{5}\right) \, \mathbf{a}_{3} & = & -z_{5}a \, \mathbf{\hat{x}}-x_{5}a \, \mathbf{\hat{y}}-y_{5}a \, \mathbf{\hat{z}} & \left(96g\right) & \mbox{O} \\ 
\mathbf{B}_{46} & = & \left(\frac{1}{2} +x_{5} + y_{5} + z_{5}\right) \, \mathbf{a}_{1} + \left(-x_{5}+y_{5}-z_{5}\right) \, \mathbf{a}_{2} + \left(x_{5}-y_{5}-z_{5}\right) \, \mathbf{a}_{3} & = & -z_{5}a \, \mathbf{\hat{x}} + \left(\frac{1}{4} +x_{5}\right)a \, \mathbf{\hat{y}} + \left(\frac{1}{4} +y_{5}\right)a \, \mathbf{\hat{z}} & \left(96g\right) & \mbox{O} \\ 
\mathbf{B}_{47} & = & \left(x_{5}-y_{5}-z_{5}\right) \, \mathbf{a}_{1} + \left(-x_{5}-y_{5}+z_{5}\right) \, \mathbf{a}_{2} + \left(\frac{1}{2} +x_{5} + y_{5} + z_{5}\right) \, \mathbf{a}_{3} & = & \left(\frac{1}{4} +z_{5}\right)a \, \mathbf{\hat{x}} + \left(\frac{1}{4} +x_{5}\right)a \, \mathbf{\hat{y}}-y_{5}a \, \mathbf{\hat{z}} & \left(96g\right) & \mbox{O} \\ 
\mathbf{B}_{48} & = & \left(-x_{5}+y_{5}-z_{5}\right) \, \mathbf{a}_{1} + \left(\frac{1}{2} +x_{5} + y_{5} + z_{5}\right) \, \mathbf{a}_{2} + \left(-x_{5}-y_{5}+z_{5}\right) \, \mathbf{a}_{3} & = & \left(\frac{1}{4} +z_{5}\right)a \, \mathbf{\hat{x}}-x_{5}a \, \mathbf{\hat{y}} + \left(\frac{1}{4} +y_{5}\right)a \, \mathbf{\hat{z}} & \left(96g\right) & \mbox{O} \\ 
\mathbf{B}_{49} & = & \left(-x_{5}+y_{5}-z_{5}\right) \, \mathbf{a}_{1} + \left(-x_{5}-y_{5}+z_{5}\right) \, \mathbf{a}_{2} + \left(x_{5}-y_{5}-z_{5}\right) \, \mathbf{a}_{3} & = & -y_{5}a \, \mathbf{\hat{x}}-z_{5}a \, \mathbf{\hat{y}}-x_{5}a \, \mathbf{\hat{z}} & \left(96g\right) & \mbox{O} \\ 
\mathbf{B}_{50} & = & \left(x_{5}-y_{5}-z_{5}\right) \, \mathbf{a}_{1} + \left(\frac{1}{2} +x_{5} + y_{5} + z_{5}\right) \, \mathbf{a}_{2} + \left(-x_{5}+y_{5}-z_{5}\right) \, \mathbf{a}_{3} & = & \left(\frac{1}{4} +y_{5}\right)a \, \mathbf{\hat{x}}-z_{5}a \, \mathbf{\hat{y}} + \left(\frac{1}{4} +x_{5}\right)a \, \mathbf{\hat{z}} & \left(96g\right) & \mbox{O} \\ 
\mathbf{B}_{51} & = & \left(\frac{1}{2} +x_{5} + y_{5} + z_{5}\right) \, \mathbf{a}_{1} + \left(x_{5}-y_{5}-z_{5}\right) \, \mathbf{a}_{2} + \left(-x_{5}-y_{5}+z_{5}\right) \, \mathbf{a}_{3} & = & -y_{5}a \, \mathbf{\hat{x}} + \left(\frac{1}{4} +z_{5}\right)a \, \mathbf{\hat{y}} + \left(\frac{1}{4} +x_{5}\right)a \, \mathbf{\hat{z}} & \left(96g\right) & \mbox{O} \\ 
\mathbf{B}_{52} & = & \left(-x_{5}-y_{5}+z_{5}\right) \, \mathbf{a}_{1} + \left(-x_{5}+y_{5}-z_{5}\right) \, \mathbf{a}_{2} + \left(\frac{1}{2} +x_{5} + y_{5} + z_{5}\right) \, \mathbf{a}_{3} & = & \left(\frac{1}{4} +y_{5}\right)a \, \mathbf{\hat{x}} + \left(\frac{1}{4} +z_{5}\right)a \, \mathbf{\hat{y}}-x_{5}a \, \mathbf{\hat{z}} & \left(96g\right) & \mbox{O} \\ 
\end{longtabu}
\renewcommand{\arraystretch}{1.0}
\noindent \hrulefill
\\
\textbf{References:}
\vspace*{-0.25cm}
\begin{flushleft}
  - \bibentry{Fu_2013}. \\
\end{flushleft}
\noindent \hrulefill
\\
\textbf{Geometry files:}
\\
\noindent  - CIF: pp. {\hyperref[A2BCD3E6_cF208_203_e_c_d_f_g_cif]{\pageref{A2BCD3E6_cF208_203_e_c_d_f_g_cif}}} \\
\noindent  - POSCAR: pp. {\hyperref[A2BCD3E6_cF208_203_e_c_d_f_g_poscar]{\pageref{A2BCD3E6_cF208_203_e_c_d_f_g_poscar}}} \\
\onecolumn
{\phantomsection\label{A4B2C6D16E_cF232_203_e_d_f_eg_a}}
\subsection*{\huge \textbf{{\normalfont \begin{raggedleft}Tychite (Na$_{6}$Mg$_{2}$(SO$_{4}$)(CO$_{3}$)$_{4}$) Structure: \end{raggedleft} \\ A4B2C6D16E\_cF232\_203\_e\_d\_f\_eg\_a}}}
\noindent \hrulefill
\vspace*{0.25cm}
\begin{figure}[htp]
  \centering
  \vspace{-1em}
  {\includegraphics[width=1\textwidth]{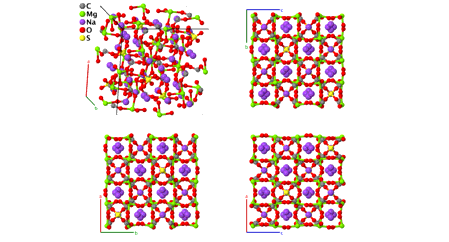}}
\end{figure}
\vspace*{-0.5cm}
\renewcommand{\arraystretch}{1.5}
\begin{equation*}
  \begin{array}{>{$\hspace{-0.15cm}}l<{$}>{$}p{0.5cm}<{$}>{$}p{18.5cm}<{$}}
    \mbox{\large \textbf{Prototype}} &\colon & \ce{Na6Mg2(SO4)(CO3)4} \\
    \mbox{\large \textbf{\AFLOW\ prototype label}} &\colon & \mbox{A4B2C6D16E\_cF232\_203\_e\_d\_f\_eg\_a} \\
    \mbox{\large \textbf{\textit{Strukturbericht} designation}} &\colon & \mbox{None} \\
    \mbox{\large \textbf{Pearson symbol}} &\colon & \mbox{cF232} \\
    \mbox{\large \textbf{Space group number}} &\colon & 203 \\
    \mbox{\large \textbf{Space group symbol}} &\colon & Fd\bar{3} \\
    \mbox{\large \textbf{\AFLOW\ prototype command}} &\colon &  \texttt{aflow} \,  \, \texttt{-{}-proto=A4B2C6D16E\_cF232\_203\_e\_d\_f\_eg\_a } \, \newline \texttt{-{}-params=}{a,x_{3},x_{4},x_{5},x_{6},y_{6},z_{6} }
  \end{array}
\end{equation*}
\renewcommand{\arraystretch}{1.0}

\vspace*{-0.25cm}
\noindent \hrulefill
\begin{itemize}
  \item{The data was obtained from an X-ray diffraction study of a single crystal at room temperature (298~K).
}
\end{itemize}

\noindent \parbox{1 \linewidth}{
\noindent \hrulefill
\\
\textbf{Face-centered Cubic primitive vectors:} \\
\vspace*{-0.25cm}
\begin{tabular}{cc}
  \begin{tabular}{c}
    \parbox{0.6 \linewidth}{
      \renewcommand{\arraystretch}{1.5}
      \begin{equation*}
        \centering
        \begin{array}{ccc}
              \mathbf{a}_1 & = & \frac12 \, a \, \mathbf{\hat{y}} + \frac12 \, a \, \mathbf{\hat{z}} \\
    \mathbf{a}_2 & = & \frac12 \, a \, \mathbf{\hat{x}} + \frac12 \, a \, \mathbf{\hat{z}} \\
    \mathbf{a}_3 & = & \frac12 \, a \, \mathbf{\hat{x}} + \frac12 \, a \, \mathbf{\hat{y}} \\

        \end{array}
      \end{equation*}
    }
    \renewcommand{\arraystretch}{1.0}
  \end{tabular}
  \begin{tabular}{c}
    \includegraphics[width=0.3\linewidth]{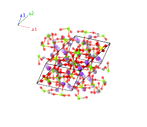} \\
  \end{tabular}
\end{tabular}

}
\vspace*{-0.25cm}

\noindent \hrulefill
\\
\textbf{Basis vectors:}
\vspace*{-0.25cm}
\renewcommand{\arraystretch}{1.5}
\begin{longtabu} to \textwidth{>{\centering $}X[-1,c,c]<{$}>{\centering $}X[-1,c,c]<{$}>{\centering $}X[-1,c,c]<{$}>{\centering $}X[-1,c,c]<{$}>{\centering $}X[-1,c,c]<{$}>{\centering $}X[-1,c,c]<{$}>{\centering $}X[-1,c,c]<{$}}
  & & \mbox{Lattice Coordinates} & & \mbox{Cartesian Coordinates} &\mbox{Wyckoff Position} & \mbox{Atom Type} \\  
  \mathbf{B}_{1} & = & \frac{1}{8} \, \mathbf{a}_{1} + \frac{1}{8} \, \mathbf{a}_{2} + \frac{1}{8} \, \mathbf{a}_{3} & = & \frac{1}{8}a \, \mathbf{\hat{x}} + \frac{1}{8}a \, \mathbf{\hat{y}} + \frac{1}{8}a \, \mathbf{\hat{z}} & \left(8a\right) & \mbox{S} \\ 
\mathbf{B}_{2} & = & \frac{7}{8} \, \mathbf{a}_{1} + \frac{7}{8} \, \mathbf{a}_{2} + \frac{7}{8} \, \mathbf{a}_{3} & = & \frac{7}{8}a \, \mathbf{\hat{x}} + \frac{7}{8}a \, \mathbf{\hat{y}} + \frac{7}{8}a \, \mathbf{\hat{z}} & \left(8a\right) & \mbox{S} \\ 
\mathbf{B}_{3} & = & \frac{1}{2} \, \mathbf{a}_{1} + \frac{1}{2} \, \mathbf{a}_{2} + \frac{1}{2} \, \mathbf{a}_{3} & = & \frac{1}{2}a \, \mathbf{\hat{x}} + \frac{1}{2}a \, \mathbf{\hat{y}} + \frac{1}{2}a \, \mathbf{\hat{z}} & \left(16d\right) & \mbox{Mg} \\ 
\mathbf{B}_{4} & = & \frac{1}{2} \, \mathbf{a}_{1} + \frac{1}{2} \, \mathbf{a}_{2} & = & \frac{1}{4}a \, \mathbf{\hat{x}} + \frac{1}{4}a \, \mathbf{\hat{y}} + \frac{1}{2}a \, \mathbf{\hat{z}} & \left(16d\right) & \mbox{Mg} \\ 
\mathbf{B}_{5} & = & \frac{1}{2} \, \mathbf{a}_{1} + \frac{1}{2} \, \mathbf{a}_{3} & = & \frac{1}{4}a \, \mathbf{\hat{x}} + \frac{1}{2}a \, \mathbf{\hat{y}} + \frac{1}{4}a \, \mathbf{\hat{z}} & \left(16d\right) & \mbox{Mg} \\ 
\mathbf{B}_{6} & = & \frac{1}{2} \, \mathbf{a}_{2} + \frac{1}{2} \, \mathbf{a}_{3} & = & \frac{1}{2}a \, \mathbf{\hat{x}} + \frac{1}{4}a \, \mathbf{\hat{y}} + \frac{1}{4}a \, \mathbf{\hat{z}} & \left(16d\right) & \mbox{Mg} \\ 
\mathbf{B}_{7} & = & x_{3} \, \mathbf{a}_{1} + x_{3} \, \mathbf{a}_{2} + x_{3} \, \mathbf{a}_{3} & = & x_{3}a \, \mathbf{\hat{x}} + x_{3}a \, \mathbf{\hat{y}} + x_{3}a \, \mathbf{\hat{z}} & \left(32e\right) & \mbox{C} \\ 
\mathbf{B}_{8} & = & x_{3} \, \mathbf{a}_{1} + x_{3} \, \mathbf{a}_{2} + \left(\frac{1}{2} - 3x_{3}\right) \, \mathbf{a}_{3} & = & \left(\frac{1}{4} - x_{3}\right)a \, \mathbf{\hat{x}} + \left(\frac{1}{4} - x_{3}\right)a \, \mathbf{\hat{y}} + x_{3}a \, \mathbf{\hat{z}} & \left(32e\right) & \mbox{C} \\ 
\mathbf{B}_{9} & = & x_{3} \, \mathbf{a}_{1} + \left(\frac{1}{2} - 3x_{3}\right) \, \mathbf{a}_{2} + x_{3} \, \mathbf{a}_{3} & = & \left(\frac{1}{4} - x_{3}\right)a \, \mathbf{\hat{x}} + x_{3}a \, \mathbf{\hat{y}} + \left(\frac{1}{4} - x_{3}\right)a \, \mathbf{\hat{z}} & \left(32e\right) & \mbox{C} \\ 
\mathbf{B}_{10} & = & \left(\frac{1}{2} - 3x_{3}\right) \, \mathbf{a}_{1} + x_{3} \, \mathbf{a}_{2} + x_{3} \, \mathbf{a}_{3} & = & x_{3}a \, \mathbf{\hat{x}} + \left(\frac{1}{4} - x_{3}\right)a \, \mathbf{\hat{y}} + \left(\frac{1}{4} - x_{3}\right)a \, \mathbf{\hat{z}} & \left(32e\right) & \mbox{C} \\ 
\mathbf{B}_{11} & = & -x_{3} \, \mathbf{a}_{1}-x_{3} \, \mathbf{a}_{2}-x_{3} \, \mathbf{a}_{3} & = & -x_{3}a \, \mathbf{\hat{x}}-x_{3}a \, \mathbf{\hat{y}}-x_{3}a \, \mathbf{\hat{z}} & \left(32e\right) & \mbox{C} \\ 
\mathbf{B}_{12} & = & -x_{3} \, \mathbf{a}_{1}-x_{3} \, \mathbf{a}_{2} + \left(\frac{1}{2} +3x_{3}\right) \, \mathbf{a}_{3} & = & \left(\frac{1}{4} +x_{3}\right)a \, \mathbf{\hat{x}} + \left(\frac{1}{4} +x_{3}\right)a \, \mathbf{\hat{y}}-x_{3}a \, \mathbf{\hat{z}} & \left(32e\right) & \mbox{C} \\ 
\mathbf{B}_{13} & = & -x_{3} \, \mathbf{a}_{1} + \left(\frac{1}{2} +3x_{3}\right) \, \mathbf{a}_{2}-x_{3} \, \mathbf{a}_{3} & = & \left(\frac{1}{4} +x_{3}\right)a \, \mathbf{\hat{x}}-x_{3}a \, \mathbf{\hat{y}} + \left(\frac{1}{4} +x_{3}\right)a \, \mathbf{\hat{z}} & \left(32e\right) & \mbox{C} \\ 
\mathbf{B}_{14} & = & \left(\frac{1}{2} +3x_{3}\right) \, \mathbf{a}_{1}-x_{3} \, \mathbf{a}_{2}-x_{3} \, \mathbf{a}_{3} & = & -x_{3}a \, \mathbf{\hat{x}} + \left(\frac{1}{4} +x_{3}\right)a \, \mathbf{\hat{y}} + \left(\frac{1}{4} +x_{3}\right)a \, \mathbf{\hat{z}} & \left(32e\right) & \mbox{C} \\ 
\mathbf{B}_{15} & = & x_{4} \, \mathbf{a}_{1} + x_{4} \, \mathbf{a}_{2} + x_{4} \, \mathbf{a}_{3} & = & x_{4}a \, \mathbf{\hat{x}} + x_{4}a \, \mathbf{\hat{y}} + x_{4}a \, \mathbf{\hat{z}} & \left(32e\right) & \mbox{O I} \\ 
\mathbf{B}_{16} & = & x_{4} \, \mathbf{a}_{1} + x_{4} \, \mathbf{a}_{2} + \left(\frac{1}{2} - 3x_{4}\right) \, \mathbf{a}_{3} & = & \left(\frac{1}{4} - x_{4}\right)a \, \mathbf{\hat{x}} + \left(\frac{1}{4} - x_{4}\right)a \, \mathbf{\hat{y}} + x_{4}a \, \mathbf{\hat{z}} & \left(32e\right) & \mbox{O I} \\ 
\mathbf{B}_{17} & = & x_{4} \, \mathbf{a}_{1} + \left(\frac{1}{2} - 3x_{4}\right) \, \mathbf{a}_{2} + x_{4} \, \mathbf{a}_{3} & = & \left(\frac{1}{4} - x_{4}\right)a \, \mathbf{\hat{x}} + x_{4}a \, \mathbf{\hat{y}} + \left(\frac{1}{4} - x_{4}\right)a \, \mathbf{\hat{z}} & \left(32e\right) & \mbox{O I} \\ 
\mathbf{B}_{18} & = & \left(\frac{1}{2} - 3x_{4}\right) \, \mathbf{a}_{1} + x_{4} \, \mathbf{a}_{2} + x_{4} \, \mathbf{a}_{3} & = & x_{4}a \, \mathbf{\hat{x}} + \left(\frac{1}{4} - x_{4}\right)a \, \mathbf{\hat{y}} + \left(\frac{1}{4} - x_{4}\right)a \, \mathbf{\hat{z}} & \left(32e\right) & \mbox{O I} \\ 
\mathbf{B}_{19} & = & -x_{4} \, \mathbf{a}_{1}-x_{4} \, \mathbf{a}_{2}-x_{4} \, \mathbf{a}_{3} & = & -x_{4}a \, \mathbf{\hat{x}}-x_{4}a \, \mathbf{\hat{y}}-x_{4}a \, \mathbf{\hat{z}} & \left(32e\right) & \mbox{O I} \\ 
\mathbf{B}_{20} & = & -x_{4} \, \mathbf{a}_{1}-x_{4} \, \mathbf{a}_{2} + \left(\frac{1}{2} +3x_{4}\right) \, \mathbf{a}_{3} & = & \left(\frac{1}{4} +x_{4}\right)a \, \mathbf{\hat{x}} + \left(\frac{1}{4} +x_{4}\right)a \, \mathbf{\hat{y}}-x_{4}a \, \mathbf{\hat{z}} & \left(32e\right) & \mbox{O I} \\ 
\mathbf{B}_{21} & = & -x_{4} \, \mathbf{a}_{1} + \left(\frac{1}{2} +3x_{4}\right) \, \mathbf{a}_{2}-x_{4} \, \mathbf{a}_{3} & = & \left(\frac{1}{4} +x_{4}\right)a \, \mathbf{\hat{x}}-x_{4}a \, \mathbf{\hat{y}} + \left(\frac{1}{4} +x_{4}\right)a \, \mathbf{\hat{z}} & \left(32e\right) & \mbox{O I} \\ 
\mathbf{B}_{22} & = & \left(\frac{1}{2} +3x_{4}\right) \, \mathbf{a}_{1}-x_{4} \, \mathbf{a}_{2}-x_{4} \, \mathbf{a}_{3} & = & -x_{4}a \, \mathbf{\hat{x}} + \left(\frac{1}{4} +x_{4}\right)a \, \mathbf{\hat{y}} + \left(\frac{1}{4} +x_{4}\right)a \, \mathbf{\hat{z}} & \left(32e\right) & \mbox{O I} \\ 
\mathbf{B}_{23} & = & \left(\frac{1}{4} - x_{5}\right) \, \mathbf{a}_{1} + x_{5} \, \mathbf{a}_{2} + x_{5} \, \mathbf{a}_{3} & = & x_{5}a \, \mathbf{\hat{x}} + \frac{1}{8}a \, \mathbf{\hat{y}} + \frac{1}{8}a \, \mathbf{\hat{z}} & \left(48f\right) & \mbox{Na} \\ 
\mathbf{B}_{24} & = & x_{5} \, \mathbf{a}_{1} + \left(\frac{1}{4} - x_{5}\right) \, \mathbf{a}_{2} + \left(\frac{1}{4} - x_{5}\right) \, \mathbf{a}_{3} & = & \left(\frac{1}{4} - x_{5}\right)a \, \mathbf{\hat{x}} + \frac{1}{8}a \, \mathbf{\hat{y}} + \frac{1}{8}a \, \mathbf{\hat{z}} & \left(48f\right) & \mbox{Na} \\ 
\mathbf{B}_{25} & = & x_{5} \, \mathbf{a}_{1} + \left(\frac{1}{4} - x_{5}\right) \, \mathbf{a}_{2} + x_{5} \, \mathbf{a}_{3} & = & \frac{1}{8}a \, \mathbf{\hat{x}} + x_{5}a \, \mathbf{\hat{y}} + \frac{1}{8}a \, \mathbf{\hat{z}} & \left(48f\right) & \mbox{Na} \\ 
\mathbf{B}_{26} & = & \left(\frac{1}{4} - x_{5}\right) \, \mathbf{a}_{1} + x_{5} \, \mathbf{a}_{2} + \left(\frac{1}{4} - x_{5}\right) \, \mathbf{a}_{3} & = & \frac{1}{8}a \, \mathbf{\hat{x}} + \left(\frac{1}{4} - x_{5}\right)a \, \mathbf{\hat{y}} + \frac{1}{8}a \, \mathbf{\hat{z}} & \left(48f\right) & \mbox{Na} \\ 
\mathbf{B}_{27} & = & x_{5} \, \mathbf{a}_{1} + x_{5} \, \mathbf{a}_{2} + \left(\frac{1}{4} - x_{5}\right) \, \mathbf{a}_{3} & = & \frac{1}{8}a \, \mathbf{\hat{x}} + \frac{1}{8}a \, \mathbf{\hat{y}} + x_{5}a \, \mathbf{\hat{z}} & \left(48f\right) & \mbox{Na} \\ 
\mathbf{B}_{28} & = & \left(\frac{1}{4} - x_{5}\right) \, \mathbf{a}_{1} + \left(\frac{1}{4} - x_{5}\right) \, \mathbf{a}_{2} + x_{5} \, \mathbf{a}_{3} & = & \frac{1}{8}a \, \mathbf{\hat{x}} + \frac{1}{8}a \, \mathbf{\hat{y}} + \left(\frac{1}{4} - x_{5}\right)a \, \mathbf{\hat{z}} & \left(48f\right) & \mbox{Na} \\ 
\mathbf{B}_{29} & = & \left(\frac{3}{4} +x_{5}\right) \, \mathbf{a}_{1}-x_{5} \, \mathbf{a}_{2}-x_{5} \, \mathbf{a}_{3} & = & -x_{5}a \, \mathbf{\hat{x}} + \frac{3}{8}a \, \mathbf{\hat{y}} + \frac{3}{8}a \, \mathbf{\hat{z}} & \left(48f\right) & \mbox{Na} \\ 
\mathbf{B}_{30} & = & -x_{5} \, \mathbf{a}_{1} + \left(\frac{3}{4} +x_{5}\right) \, \mathbf{a}_{2} + \left(\frac{3}{4} +x_{5}\right) \, \mathbf{a}_{3} & = & \left(\frac{3}{4} +x_{5}\right)a \, \mathbf{\hat{x}} + \frac{3}{8}a \, \mathbf{\hat{y}} + \frac{3}{8}a \, \mathbf{\hat{z}} & \left(48f\right) & \mbox{Na} \\ 
\mathbf{B}_{31} & = & -x_{5} \, \mathbf{a}_{1} + \left(\frac{3}{4} +x_{5}\right) \, \mathbf{a}_{2}-x_{5} \, \mathbf{a}_{3} & = & \frac{3}{8}a \, \mathbf{\hat{x}}-x_{5}a \, \mathbf{\hat{y}} + \frac{3}{8}a \, \mathbf{\hat{z}} & \left(48f\right) & \mbox{Na} \\ 
\mathbf{B}_{32} & = & \left(\frac{3}{4} +x_{5}\right) \, \mathbf{a}_{1}-x_{5} \, \mathbf{a}_{2} + \left(\frac{3}{4} +x_{5}\right) \, \mathbf{a}_{3} & = & \frac{3}{8}a \, \mathbf{\hat{x}} + \left(\frac{3}{4} +x_{5}\right)a \, \mathbf{\hat{y}} + \frac{3}{8}a \, \mathbf{\hat{z}} & \left(48f\right) & \mbox{Na} \\ 
\mathbf{B}_{33} & = & -x_{5} \, \mathbf{a}_{1}-x_{5} \, \mathbf{a}_{2} + \left(\frac{3}{4} +x_{5}\right) \, \mathbf{a}_{3} & = & \frac{3}{8}a \, \mathbf{\hat{x}} + \frac{3}{8}a \, \mathbf{\hat{y}}-x_{5}a \, \mathbf{\hat{z}} & \left(48f\right) & \mbox{Na} \\ 
\mathbf{B}_{34} & = & \left(\frac{3}{4} +x_{5}\right) \, \mathbf{a}_{1} + \left(\frac{3}{4} +x_{5}\right) \, \mathbf{a}_{2}-x_{5} \, \mathbf{a}_{3} & = & \frac{3}{8}a \, \mathbf{\hat{x}} + \frac{3}{8}a \, \mathbf{\hat{y}} + \left(\frac{3}{4} +x_{5}\right)a \, \mathbf{\hat{z}} & \left(48f\right) & \mbox{Na} \\ 
\mathbf{B}_{35} & = & \left(-x_{6}+y_{6}+z_{6}\right) \, \mathbf{a}_{1} + \left(x_{6}-y_{6}+z_{6}\right) \, \mathbf{a}_{2} + \left(x_{6}+y_{6}-z_{6}\right) \, \mathbf{a}_{3} & = & x_{6}a \, \mathbf{\hat{x}} + y_{6}a \, \mathbf{\hat{y}} + z_{6}a \, \mathbf{\hat{z}} & \left(96g\right) & \mbox{O II} \\ 
\mathbf{B}_{36} & = & \left(x_{6}-y_{6}+z_{6}\right) \, \mathbf{a}_{1} + \left(-x_{6}+y_{6}+z_{6}\right) \, \mathbf{a}_{2} + \left(\frac{1}{2} - x_{6} - y_{6} - z_{6}\right) \, \mathbf{a}_{3} & = & \left(\frac{1}{4} - x_{6}\right)a \, \mathbf{\hat{x}} + \left(\frac{1}{4} - y_{6}\right)a \, \mathbf{\hat{y}} + z_{6}a \, \mathbf{\hat{z}} & \left(96g\right) & \mbox{O II} \\ 
\mathbf{B}_{37} & = & \left(x_{6}+y_{6}-z_{6}\right) \, \mathbf{a}_{1} + \left(\frac{1}{2} - x_{6} - y_{6} - z_{6}\right) \, \mathbf{a}_{2} + \left(-x_{6}+y_{6}+z_{6}\right) \, \mathbf{a}_{3} & = & \left(\frac{1}{4} - x_{6}\right)a \, \mathbf{\hat{x}} + y_{6}a \, \mathbf{\hat{y}} + \left(\frac{1}{4} - z_{6}\right)a \, \mathbf{\hat{z}} & \left(96g\right) & \mbox{O II} \\ 
\mathbf{B}_{38} & = & \left(\frac{1}{2} - x_{6} - y_{6} - z_{6}\right) \, \mathbf{a}_{1} + \left(x_{6}+y_{6}-z_{6}\right) \, \mathbf{a}_{2} + \left(x_{6}-y_{6}+z_{6}\right) \, \mathbf{a}_{3} & = & x_{6}a \, \mathbf{\hat{x}} + \left(\frac{1}{4} - y_{6}\right)a \, \mathbf{\hat{y}} + \left(\frac{1}{4} - z_{6}\right)a \, \mathbf{\hat{z}} & \left(96g\right) & \mbox{O II} \\ 
\mathbf{B}_{39} & = & \left(x_{6}+y_{6}-z_{6}\right) \, \mathbf{a}_{1} + \left(-x_{6}+y_{6}+z_{6}\right) \, \mathbf{a}_{2} + \left(x_{6}-y_{6}+z_{6}\right) \, \mathbf{a}_{3} & = & z_{6}a \, \mathbf{\hat{x}} + x_{6}a \, \mathbf{\hat{y}} + y_{6}a \, \mathbf{\hat{z}} & \left(96g\right) & \mbox{O II} \\ 
\mathbf{B}_{40} & = & \left(\frac{1}{2} - x_{6} - y_{6} - z_{6}\right) \, \mathbf{a}_{1} + \left(x_{6}-y_{6}+z_{6}\right) \, \mathbf{a}_{2} + \left(-x_{6}+y_{6}+z_{6}\right) \, \mathbf{a}_{3} & = & z_{6}a \, \mathbf{\hat{x}} + \left(\frac{1}{4} - x_{6}\right)a \, \mathbf{\hat{y}} + \left(\frac{1}{4} - y_{6}\right)a \, \mathbf{\hat{z}} & \left(96g\right) & \mbox{O II} \\ 
\mathbf{B}_{41} & = & \left(-x_{6}+y_{6}+z_{6}\right) \, \mathbf{a}_{1} + \left(x_{6}+y_{6}-z_{6}\right) \, \mathbf{a}_{2} + \left(\frac{1}{2} - x_{6} - y_{6} - z_{6}\right) \, \mathbf{a}_{3} & = & \left(\frac{1}{4} - z_{6}\right)a \, \mathbf{\hat{x}} + \left(\frac{1}{4} - x_{6}\right)a \, \mathbf{\hat{y}} + y_{6}a \, \mathbf{\hat{z}} & \left(96g\right) & \mbox{O II} \\ 
\mathbf{B}_{42} & = & \left(x_{6}-y_{6}+z_{6}\right) \, \mathbf{a}_{1} + \left(\frac{1}{2} - x_{6} - y_{6} - z_{6}\right) \, \mathbf{a}_{2} + \left(x_{6}+y_{6}-z_{6}\right) \, \mathbf{a}_{3} & = & \left(\frac{1}{4} - z_{6}\right)a \, \mathbf{\hat{x}} + x_{6}a \, \mathbf{\hat{y}} + \left(\frac{1}{4} - y_{6}\right)a \, \mathbf{\hat{z}} & \left(96g\right) & \mbox{O II} \\ 
\mathbf{B}_{43} & = & \left(x_{6}-y_{6}+z_{6}\right) \, \mathbf{a}_{1} + \left(x_{6}+y_{6}-z_{6}\right) \, \mathbf{a}_{2} + \left(-x_{6}+y_{6}+z_{6}\right) \, \mathbf{a}_{3} & = & y_{6}a \, \mathbf{\hat{x}} + z_{6}a \, \mathbf{\hat{y}} + x_{6}a \, \mathbf{\hat{z}} & \left(96g\right) & \mbox{O II} \\ 
\mathbf{B}_{44} & = & \left(-x_{6}+y_{6}+z_{6}\right) \, \mathbf{a}_{1} + \left(\frac{1}{2} - x_{6} - y_{6} - z_{6}\right) \, \mathbf{a}_{2} + \left(x_{6}-y_{6}+z_{6}\right) \, \mathbf{a}_{3} & = & \left(\frac{1}{4} - y_{6}\right)a \, \mathbf{\hat{x}} + z_{6}a \, \mathbf{\hat{y}} + \left(\frac{1}{4} - x_{6}\right)a \, \mathbf{\hat{z}} & \left(96g\right) & \mbox{O II} \\ 
\mathbf{B}_{45} & = & \left(\frac{1}{2} - x_{6} - y_{6} - z_{6}\right) \, \mathbf{a}_{1} + \left(-x_{6}+y_{6}+z_{6}\right) \, \mathbf{a}_{2} + \left(x_{6}+y_{6}-z_{6}\right) \, \mathbf{a}_{3} & = & y_{6}a \, \mathbf{\hat{x}} + \left(\frac{1}{4} - z_{6}\right)a \, \mathbf{\hat{y}} + \left(\frac{1}{4} - x_{6}\right)a \, \mathbf{\hat{z}} & \left(96g\right) & \mbox{O II} \\ 
\mathbf{B}_{46} & = & \left(x_{6}+y_{6}-z_{6}\right) \, \mathbf{a}_{1} + \left(x_{6}-y_{6}+z_{6}\right) \, \mathbf{a}_{2} + \left(\frac{1}{2} - x_{6} - y_{6} - z_{6}\right) \, \mathbf{a}_{3} & = & \left(\frac{1}{4} - y_{6}\right)a \, \mathbf{\hat{x}} + \left(\frac{1}{4} - z_{6}\right)a \, \mathbf{\hat{y}} + x_{6}a \, \mathbf{\hat{z}} & \left(96g\right) & \mbox{O II} \\ 
\mathbf{B}_{47} & = & \left(x_{6}-y_{6}-z_{6}\right) \, \mathbf{a}_{1} + \left(-x_{6}+y_{6}-z_{6}\right) \, \mathbf{a}_{2} + \left(-x_{6}-y_{6}+z_{6}\right) \, \mathbf{a}_{3} & = & -x_{6}a \, \mathbf{\hat{x}}-y_{6}a \, \mathbf{\hat{y}}-z_{6}a \, \mathbf{\hat{z}} & \left(96g\right) & \mbox{O II} \\ 
\mathbf{B}_{48} & = & \left(-x_{6}+y_{6}-z_{6}\right) \, \mathbf{a}_{1} + \left(x_{6}-y_{6}-z_{6}\right) \, \mathbf{a}_{2} + \left(\frac{1}{2} +x_{6} + y_{6} + z_{6}\right) \, \mathbf{a}_{3} & = & \left(\frac{1}{4} +x_{6}\right)a \, \mathbf{\hat{x}} + \left(\frac{1}{4} +y_{6}\right)a \, \mathbf{\hat{y}}-z_{6}a \, \mathbf{\hat{z}} & \left(96g\right) & \mbox{O II} \\ 
\mathbf{B}_{49} & = & \left(-x_{6}-y_{6}+z_{6}\right) \, \mathbf{a}_{1} + \left(\frac{1}{2} +x_{6} + y_{6} + z_{6}\right) \, \mathbf{a}_{2} + \left(x_{6}-y_{6}-z_{6}\right) \, \mathbf{a}_{3} & = & \left(\frac{1}{4} +x_{6}\right)a \, \mathbf{\hat{x}}-y_{6}a \, \mathbf{\hat{y}} + \left(\frac{1}{4} +z_{6}\right)a \, \mathbf{\hat{z}} & \left(96g\right) & \mbox{O II} \\ 
\mathbf{B}_{50} & = & \left(\frac{1}{2} +x_{6} + y_{6} + z_{6}\right) \, \mathbf{a}_{1} + \left(-x_{6}-y_{6}+z_{6}\right) \, \mathbf{a}_{2} + \left(-x_{6}+y_{6}-z_{6}\right) \, \mathbf{a}_{3} & = & -x_{6}a \, \mathbf{\hat{x}} + \left(\frac{1}{4} +y_{6}\right)a \, \mathbf{\hat{y}} + \left(\frac{1}{4} +z_{6}\right)a \, \mathbf{\hat{z}} & \left(96g\right) & \mbox{O II} \\ 
\mathbf{B}_{51} & = & \left(-x_{6}-y_{6}+z_{6}\right) \, \mathbf{a}_{1} + \left(x_{6}-y_{6}-z_{6}\right) \, \mathbf{a}_{2} + \left(-x_{6}+y_{6}-z_{6}\right) \, \mathbf{a}_{3} & = & -z_{6}a \, \mathbf{\hat{x}}-x_{6}a \, \mathbf{\hat{y}}-y_{6}a \, \mathbf{\hat{z}} & \left(96g\right) & \mbox{O II} \\ 
\mathbf{B}_{52} & = & \left(\frac{1}{2} +x_{6} + y_{6} + z_{6}\right) \, \mathbf{a}_{1} + \left(-x_{6}+y_{6}-z_{6}\right) \, \mathbf{a}_{2} + \left(x_{6}-y_{6}-z_{6}\right) \, \mathbf{a}_{3} & = & -z_{6}a \, \mathbf{\hat{x}} + \left(\frac{1}{4} +x_{6}\right)a \, \mathbf{\hat{y}} + \left(\frac{1}{4} +y_{6}\right)a \, \mathbf{\hat{z}} & \left(96g\right) & \mbox{O II} \\ 
\mathbf{B}_{53} & = & \left(x_{6}-y_{6}-z_{6}\right) \, \mathbf{a}_{1} + \left(-x_{6}-y_{6}+z_{6}\right) \, \mathbf{a}_{2} + \left(\frac{1}{2} +x_{6} + y_{6} + z_{6}\right) \, \mathbf{a}_{3} & = & \left(\frac{1}{4} +z_{6}\right)a \, \mathbf{\hat{x}} + \left(\frac{1}{4} +x_{6}\right)a \, \mathbf{\hat{y}}-y_{6}a \, \mathbf{\hat{z}} & \left(96g\right) & \mbox{O II} \\ 
\mathbf{B}_{54} & = & \left(-x_{6}+y_{6}-z_{6}\right) \, \mathbf{a}_{1} + \left(\frac{1}{2} +x_{6} + y_{6} + z_{6}\right) \, \mathbf{a}_{2} + \left(-x_{6}-y_{6}+z_{6}\right) \, \mathbf{a}_{3} & = & \left(\frac{1}{4} +z_{6}\right)a \, \mathbf{\hat{x}}-x_{6}a \, \mathbf{\hat{y}} + \left(\frac{1}{4} +y_{6}\right)a \, \mathbf{\hat{z}} & \left(96g\right) & \mbox{O II} \\ 
\mathbf{B}_{55} & = & \left(-x_{6}+y_{6}-z_{6}\right) \, \mathbf{a}_{1} + \left(-x_{6}-y_{6}+z_{6}\right) \, \mathbf{a}_{2} + \left(x_{6}-y_{6}-z_{6}\right) \, \mathbf{a}_{3} & = & -y_{6}a \, \mathbf{\hat{x}}-z_{6}a \, \mathbf{\hat{y}}-x_{6}a \, \mathbf{\hat{z}} & \left(96g\right) & \mbox{O II} \\ 
\mathbf{B}_{56} & = & \left(x_{6}-y_{6}-z_{6}\right) \, \mathbf{a}_{1} + \left(\frac{1}{2} +x_{6} + y_{6} + z_{6}\right) \, \mathbf{a}_{2} + \left(-x_{6}+y_{6}-z_{6}\right) \, \mathbf{a}_{3} & = & \left(\frac{1}{4} +y_{6}\right)a \, \mathbf{\hat{x}}-z_{6}a \, \mathbf{\hat{y}} + \left(\frac{1}{4} +x_{6}\right)a \, \mathbf{\hat{z}} & \left(96g\right) & \mbox{O II} \\ 
\mathbf{B}_{57} & = & \left(\frac{1}{2} +x_{6} + y_{6} + z_{6}\right) \, \mathbf{a}_{1} + \left(x_{6}-y_{6}-z_{6}\right) \, \mathbf{a}_{2} + \left(-x_{6}-y_{6}+z_{6}\right) \, \mathbf{a}_{3} & = & -y_{6}a \, \mathbf{\hat{x}} + \left(\frac{1}{4} +z_{6}\right)a \, \mathbf{\hat{y}} + \left(\frac{1}{4} +x_{6}\right)a \, \mathbf{\hat{z}} & \left(96g\right) & \mbox{O II} \\ 
\mathbf{B}_{58} & = & \left(-x_{6}-y_{6}+z_{6}\right) \, \mathbf{a}_{1} + \left(-x_{6}+y_{6}-z_{6}\right) \, \mathbf{a}_{2} + \left(\frac{1}{2} +x_{6} + y_{6} + z_{6}\right) \, \mathbf{a}_{3} & = & \left(\frac{1}{4} +y_{6}\right)a \, \mathbf{\hat{x}} + \left(\frac{1}{4} +z_{6}\right)a \, \mathbf{\hat{y}}-x_{6}a \, \mathbf{\hat{z}} & \left(96g\right) & \mbox{O II} \\ 
\end{longtabu}
\renewcommand{\arraystretch}{1.0}
\noindent \hrulefill
\\
\textbf{References:}
\vspace*{-0.25cm}
\begin{flushleft}
  - \bibentry{Schmidt_2006}. \\
\end{flushleft}
\textbf{Found in:}
\vspace*{-0.25cm}
\begin{flushleft}
  - \bibentry{Downs_2003}. \\
\end{flushleft}
\noindent \hrulefill
\\
\textbf{Geometry files:}
\\
\noindent  - CIF: pp. {\hyperref[A4B2C6D16E_cF232_203_e_d_f_eg_a_cif]{\pageref{A4B2C6D16E_cF232_203_e_d_f_eg_a_cif}}} \\
\noindent  - POSCAR: pp. {\hyperref[A4B2C6D16E_cF232_203_e_d_f_eg_a_poscar]{\pageref{A4B2C6D16E_cF232_203_e_d_f_eg_a_poscar}}} \\
\onecolumn
{\phantomsection\label{AB3C16_cF160_203_b_ad_eg}}
\subsection*{\huge \textbf{{\normalfont Rb$_{3}$AsSe$_{16}$ Structure: AB3C16\_cF160\_203\_b\_ad\_eg}}}
\noindent \hrulefill
\vspace*{0.25cm}
\begin{figure}[htp]
  \centering
  \vspace{-1em}
  {\includegraphics[width=1\textwidth]{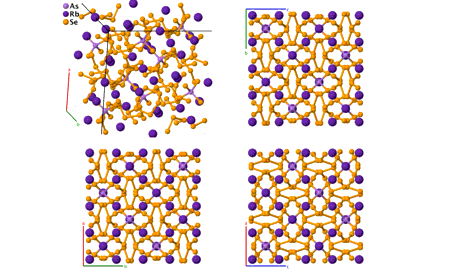}}
\end{figure}
\vspace*{-0.5cm}
\renewcommand{\arraystretch}{1.5}
\begin{equation*}
  \begin{array}{>{$\hspace{-0.15cm}}l<{$}>{$}p{0.5cm}<{$}>{$}p{18.5cm}<{$}}
    \mbox{\large \textbf{Prototype}} &\colon & \ce{Rb3AsSe16} \\
    \mbox{\large \textbf{\AFLOW\ prototype label}} &\colon & \mbox{AB3C16\_cF160\_203\_b\_ad\_eg} \\
    \mbox{\large \textbf{\textit{Strukturbericht} designation}} &\colon & \mbox{None} \\
    \mbox{\large \textbf{Pearson symbol}} &\colon & \mbox{cF160} \\
    \mbox{\large \textbf{Space group number}} &\colon & 203 \\
    \mbox{\large \textbf{Space group symbol}} &\colon & Fd\bar{3} \\
    \mbox{\large \textbf{\AFLOW\ prototype command}} &\colon &  \texttt{aflow} \,  \, \texttt{-{}-proto=AB3C16\_cF160\_203\_b\_ad\_eg } \, \newline \texttt{-{}-params=}{a,x_{4},x_{5},y_{5},z_{5} }
  \end{array}
\end{equation*}
\renewcommand{\arraystretch}{1.0}

\noindent \parbox{1 \linewidth}{
\noindent \hrulefill
\\
\textbf{Face-centered Cubic primitive vectors:} \\
\vspace*{-0.25cm}
\begin{tabular}{cc}
  \begin{tabular}{c}
    \parbox{0.6 \linewidth}{
      \renewcommand{\arraystretch}{1.5}
      \begin{equation*}
        \centering
        \begin{array}{ccc}
              \mathbf{a}_1 & = & \frac12 \, a \, \mathbf{\hat{y}} + \frac12 \, a \, \mathbf{\hat{z}} \\
    \mathbf{a}_2 & = & \frac12 \, a \, \mathbf{\hat{x}} + \frac12 \, a \, \mathbf{\hat{z}} \\
    \mathbf{a}_3 & = & \frac12 \, a \, \mathbf{\hat{x}} + \frac12 \, a \, \mathbf{\hat{y}} \\

        \end{array}
      \end{equation*}
    }
    \renewcommand{\arraystretch}{1.0}
  \end{tabular}
  \begin{tabular}{c}
    \includegraphics[width=0.3\linewidth]{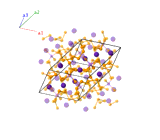} \\
  \end{tabular}
\end{tabular}

}
\vspace*{-0.25cm}

\noindent \hrulefill
\\
\textbf{Basis vectors:}
\vspace*{-0.25cm}
\renewcommand{\arraystretch}{1.5}
\begin{longtabu} to \textwidth{>{\centering $}X[-1,c,c]<{$}>{\centering $}X[-1,c,c]<{$}>{\centering $}X[-1,c,c]<{$}>{\centering $}X[-1,c,c]<{$}>{\centering $}X[-1,c,c]<{$}>{\centering $}X[-1,c,c]<{$}>{\centering $}X[-1,c,c]<{$}}
  & & \mbox{Lattice Coordinates} & & \mbox{Cartesian Coordinates} &\mbox{Wyckoff Position} & \mbox{Atom Type} \\  
  \mathbf{B}_{1} & = & \frac{1}{8} \, \mathbf{a}_{1} + \frac{1}{8} \, \mathbf{a}_{2} + \frac{1}{8} \, \mathbf{a}_{3} & = & \frac{1}{8}a \, \mathbf{\hat{x}} + \frac{1}{8}a \, \mathbf{\hat{y}} + \frac{1}{8}a \, \mathbf{\hat{z}} & \left(8a\right) & \mbox{Rb I} \\ 
\mathbf{B}_{2} & = & \frac{7}{8} \, \mathbf{a}_{1} + \frac{7}{8} \, \mathbf{a}_{2} + \frac{7}{8} \, \mathbf{a}_{3} & = & \frac{7}{8}a \, \mathbf{\hat{x}} + \frac{7}{8}a \, \mathbf{\hat{y}} + \frac{7}{8}a \, \mathbf{\hat{z}} & \left(8a\right) & \mbox{Rb I} \\ 
\mathbf{B}_{3} & = & \frac{5}{8} \, \mathbf{a}_{1} + \frac{5}{8} \, \mathbf{a}_{2} + \frac{5}{8} \, \mathbf{a}_{3} & = & \frac{5}{8}a \, \mathbf{\hat{x}} + \frac{5}{8}a \, \mathbf{\hat{y}} + \frac{5}{8}a \, \mathbf{\hat{z}} & \left(8b\right) & \mbox{As} \\ 
\mathbf{B}_{4} & = & \frac{3}{8} \, \mathbf{a}_{1} + \frac{3}{8} \, \mathbf{a}_{2} + \frac{3}{8} \, \mathbf{a}_{3} & = & \frac{3}{8}a \, \mathbf{\hat{x}} + \frac{3}{8}a \, \mathbf{\hat{y}} + \frac{3}{8}a \, \mathbf{\hat{z}} & \left(8b\right) & \mbox{As} \\ 
\mathbf{B}_{5} & = & \frac{1}{2} \, \mathbf{a}_{1} + \frac{1}{2} \, \mathbf{a}_{2} + \frac{1}{2} \, \mathbf{a}_{3} & = & \frac{1}{2}a \, \mathbf{\hat{x}} + \frac{1}{2}a \, \mathbf{\hat{y}} + \frac{1}{2}a \, \mathbf{\hat{z}} & \left(16d\right) & \mbox{Rb II} \\ 
\mathbf{B}_{6} & = & \frac{1}{2} \, \mathbf{a}_{1} + \frac{1}{2} \, \mathbf{a}_{2} & = & \frac{1}{4}a \, \mathbf{\hat{x}} + \frac{1}{4}a \, \mathbf{\hat{y}} + \frac{1}{2}a \, \mathbf{\hat{z}} & \left(16d\right) & \mbox{Rb II} \\ 
\mathbf{B}_{7} & = & \frac{1}{2} \, \mathbf{a}_{1} + \frac{1}{2} \, \mathbf{a}_{3} & = & \frac{1}{4}a \, \mathbf{\hat{x}} + \frac{1}{2}a \, \mathbf{\hat{y}} + \frac{1}{4}a \, \mathbf{\hat{z}} & \left(16d\right) & \mbox{Rb II} \\ 
\mathbf{B}_{8} & = & \frac{1}{2} \, \mathbf{a}_{2} + \frac{1}{2} \, \mathbf{a}_{3} & = & \frac{1}{2}a \, \mathbf{\hat{x}} + \frac{1}{4}a \, \mathbf{\hat{y}} + \frac{1}{4}a \, \mathbf{\hat{z}} & \left(16d\right) & \mbox{Rb II} \\ 
\mathbf{B}_{9} & = & x_{4} \, \mathbf{a}_{1} + x_{4} \, \mathbf{a}_{2} + x_{4} \, \mathbf{a}_{3} & = & x_{4}a \, \mathbf{\hat{x}} + x_{4}a \, \mathbf{\hat{y}} + x_{4}a \, \mathbf{\hat{z}} & \left(32e\right) & \mbox{Se I} \\ 
\mathbf{B}_{10} & = & x_{4} \, \mathbf{a}_{1} + x_{4} \, \mathbf{a}_{2} + \left(\frac{1}{2} - 3x_{4}\right) \, \mathbf{a}_{3} & = & \left(\frac{1}{4} - x_{4}\right)a \, \mathbf{\hat{x}} + \left(\frac{1}{4} - x_{4}\right)a \, \mathbf{\hat{y}} + x_{4}a \, \mathbf{\hat{z}} & \left(32e\right) & \mbox{Se I} \\ 
\mathbf{B}_{11} & = & x_{4} \, \mathbf{a}_{1} + \left(\frac{1}{2} - 3x_{4}\right) \, \mathbf{a}_{2} + x_{4} \, \mathbf{a}_{3} & = & \left(\frac{1}{4} - x_{4}\right)a \, \mathbf{\hat{x}} + x_{4}a \, \mathbf{\hat{y}} + \left(\frac{1}{4} - x_{4}\right)a \, \mathbf{\hat{z}} & \left(32e\right) & \mbox{Se I} \\ 
\mathbf{B}_{12} & = & \left(\frac{1}{2} - 3x_{4}\right) \, \mathbf{a}_{1} + x_{4} \, \mathbf{a}_{2} + x_{4} \, \mathbf{a}_{3} & = & x_{4}a \, \mathbf{\hat{x}} + \left(\frac{1}{4} - x_{4}\right)a \, \mathbf{\hat{y}} + \left(\frac{1}{4} - x_{4}\right)a \, \mathbf{\hat{z}} & \left(32e\right) & \mbox{Se I} \\ 
\mathbf{B}_{13} & = & -x_{4} \, \mathbf{a}_{1}-x_{4} \, \mathbf{a}_{2}-x_{4} \, \mathbf{a}_{3} & = & -x_{4}a \, \mathbf{\hat{x}}-x_{4}a \, \mathbf{\hat{y}}-x_{4}a \, \mathbf{\hat{z}} & \left(32e\right) & \mbox{Se I} \\ 
\mathbf{B}_{14} & = & -x_{4} \, \mathbf{a}_{1}-x_{4} \, \mathbf{a}_{2} + \left(\frac{1}{2} +3x_{4}\right) \, \mathbf{a}_{3} & = & \left(\frac{1}{4} +x_{4}\right)a \, \mathbf{\hat{x}} + \left(\frac{1}{4} +x_{4}\right)a \, \mathbf{\hat{y}}-x_{4}a \, \mathbf{\hat{z}} & \left(32e\right) & \mbox{Se I} \\ 
\mathbf{B}_{15} & = & -x_{4} \, \mathbf{a}_{1} + \left(\frac{1}{2} +3x_{4}\right) \, \mathbf{a}_{2}-x_{4} \, \mathbf{a}_{3} & = & \left(\frac{1}{4} +x_{4}\right)a \, \mathbf{\hat{x}}-x_{4}a \, \mathbf{\hat{y}} + \left(\frac{1}{4} +x_{4}\right)a \, \mathbf{\hat{z}} & \left(32e\right) & \mbox{Se I} \\ 
\mathbf{B}_{16} & = & \left(\frac{1}{2} +3x_{4}\right) \, \mathbf{a}_{1}-x_{4} \, \mathbf{a}_{2}-x_{4} \, \mathbf{a}_{3} & = & -x_{4}a \, \mathbf{\hat{x}} + \left(\frac{1}{4} +x_{4}\right)a \, \mathbf{\hat{y}} + \left(\frac{1}{4} +x_{4}\right)a \, \mathbf{\hat{z}} & \left(32e\right) & \mbox{Se I} \\ 
\mathbf{B}_{17} & = & \left(-x_{5}+y_{5}+z_{5}\right) \, \mathbf{a}_{1} + \left(x_{5}-y_{5}+z_{5}\right) \, \mathbf{a}_{2} + \left(x_{5}+y_{5}-z_{5}\right) \, \mathbf{a}_{3} & = & x_{5}a \, \mathbf{\hat{x}} + y_{5}a \, \mathbf{\hat{y}} + z_{5}a \, \mathbf{\hat{z}} & \left(96g\right) & \mbox{Se II} \\ 
\mathbf{B}_{18} & = & \left(x_{5}-y_{5}+z_{5}\right) \, \mathbf{a}_{1} + \left(-x_{5}+y_{5}+z_{5}\right) \, \mathbf{a}_{2} + \left(\frac{1}{2} - x_{5} - y_{5} - z_{5}\right) \, \mathbf{a}_{3} & = & \left(\frac{1}{4} - x_{5}\right)a \, \mathbf{\hat{x}} + \left(\frac{1}{4} - y_{5}\right)a \, \mathbf{\hat{y}} + z_{5}a \, \mathbf{\hat{z}} & \left(96g\right) & \mbox{Se II} \\ 
\mathbf{B}_{19} & = & \left(x_{5}+y_{5}-z_{5}\right) \, \mathbf{a}_{1} + \left(\frac{1}{2} - x_{5} - y_{5} - z_{5}\right) \, \mathbf{a}_{2} + \left(-x_{5}+y_{5}+z_{5}\right) \, \mathbf{a}_{3} & = & \left(\frac{1}{4} - x_{5}\right)a \, \mathbf{\hat{x}} + y_{5}a \, \mathbf{\hat{y}} + \left(\frac{1}{4} - z_{5}\right)a \, \mathbf{\hat{z}} & \left(96g\right) & \mbox{Se II} \\ 
\mathbf{B}_{20} & = & \left(\frac{1}{2} - x_{5} - y_{5} - z_{5}\right) \, \mathbf{a}_{1} + \left(x_{5}+y_{5}-z_{5}\right) \, \mathbf{a}_{2} + \left(x_{5}-y_{5}+z_{5}\right) \, \mathbf{a}_{3} & = & x_{5}a \, \mathbf{\hat{x}} + \left(\frac{1}{4} - y_{5}\right)a \, \mathbf{\hat{y}} + \left(\frac{1}{4} - z_{5}\right)a \, \mathbf{\hat{z}} & \left(96g\right) & \mbox{Se II} \\ 
\mathbf{B}_{21} & = & \left(x_{5}+y_{5}-z_{5}\right) \, \mathbf{a}_{1} + \left(-x_{5}+y_{5}+z_{5}\right) \, \mathbf{a}_{2} + \left(x_{5}-y_{5}+z_{5}\right) \, \mathbf{a}_{3} & = & z_{5}a \, \mathbf{\hat{x}} + x_{5}a \, \mathbf{\hat{y}} + y_{5}a \, \mathbf{\hat{z}} & \left(96g\right) & \mbox{Se II} \\ 
\mathbf{B}_{22} & = & \left(\frac{1}{2} - x_{5} - y_{5} - z_{5}\right) \, \mathbf{a}_{1} + \left(x_{5}-y_{5}+z_{5}\right) \, \mathbf{a}_{2} + \left(-x_{5}+y_{5}+z_{5}\right) \, \mathbf{a}_{3} & = & z_{5}a \, \mathbf{\hat{x}} + \left(\frac{1}{4} - x_{5}\right)a \, \mathbf{\hat{y}} + \left(\frac{1}{4} - y_{5}\right)a \, \mathbf{\hat{z}} & \left(96g\right) & \mbox{Se II} \\ 
\mathbf{B}_{23} & = & \left(-x_{5}+y_{5}+z_{5}\right) \, \mathbf{a}_{1} + \left(x_{5}+y_{5}-z_{5}\right) \, \mathbf{a}_{2} + \left(\frac{1}{2} - x_{5} - y_{5} - z_{5}\right) \, \mathbf{a}_{3} & = & \left(\frac{1}{4} - z_{5}\right)a \, \mathbf{\hat{x}} + \left(\frac{1}{4} - x_{5}\right)a \, \mathbf{\hat{y}} + y_{5}a \, \mathbf{\hat{z}} & \left(96g\right) & \mbox{Se II} \\ 
\mathbf{B}_{24} & = & \left(x_{5}-y_{5}+z_{5}\right) \, \mathbf{a}_{1} + \left(\frac{1}{2} - x_{5} - y_{5} - z_{5}\right) \, \mathbf{a}_{2} + \left(x_{5}+y_{5}-z_{5}\right) \, \mathbf{a}_{3} & = & \left(\frac{1}{4} - z_{5}\right)a \, \mathbf{\hat{x}} + x_{5}a \, \mathbf{\hat{y}} + \left(\frac{1}{4} - y_{5}\right)a \, \mathbf{\hat{z}} & \left(96g\right) & \mbox{Se II} \\ 
\mathbf{B}_{25} & = & \left(x_{5}-y_{5}+z_{5}\right) \, \mathbf{a}_{1} + \left(x_{5}+y_{5}-z_{5}\right) \, \mathbf{a}_{2} + \left(-x_{5}+y_{5}+z_{5}\right) \, \mathbf{a}_{3} & = & y_{5}a \, \mathbf{\hat{x}} + z_{5}a \, \mathbf{\hat{y}} + x_{5}a \, \mathbf{\hat{z}} & \left(96g\right) & \mbox{Se II} \\ 
\mathbf{B}_{26} & = & \left(-x_{5}+y_{5}+z_{5}\right) \, \mathbf{a}_{1} + \left(\frac{1}{2} - x_{5} - y_{5} - z_{5}\right) \, \mathbf{a}_{2} + \left(x_{5}-y_{5}+z_{5}\right) \, \mathbf{a}_{3} & = & \left(\frac{1}{4} - y_{5}\right)a \, \mathbf{\hat{x}} + z_{5}a \, \mathbf{\hat{y}} + \left(\frac{1}{4} - x_{5}\right)a \, \mathbf{\hat{z}} & \left(96g\right) & \mbox{Se II} \\ 
\mathbf{B}_{27} & = & \left(\frac{1}{2} - x_{5} - y_{5} - z_{5}\right) \, \mathbf{a}_{1} + \left(-x_{5}+y_{5}+z_{5}\right) \, \mathbf{a}_{2} + \left(x_{5}+y_{5}-z_{5}\right) \, \mathbf{a}_{3} & = & y_{5}a \, \mathbf{\hat{x}} + \left(\frac{1}{4} - z_{5}\right)a \, \mathbf{\hat{y}} + \left(\frac{1}{4} - x_{5}\right)a \, \mathbf{\hat{z}} & \left(96g\right) & \mbox{Se II} \\ 
\mathbf{B}_{28} & = & \left(x_{5}+y_{5}-z_{5}\right) \, \mathbf{a}_{1} + \left(x_{5}-y_{5}+z_{5}\right) \, \mathbf{a}_{2} + \left(\frac{1}{2} - x_{5} - y_{5} - z_{5}\right) \, \mathbf{a}_{3} & = & \left(\frac{1}{4} - y_{5}\right)a \, \mathbf{\hat{x}} + \left(\frac{1}{4} - z_{5}\right)a \, \mathbf{\hat{y}} + x_{5}a \, \mathbf{\hat{z}} & \left(96g\right) & \mbox{Se II} \\ 
\mathbf{B}_{29} & = & \left(x_{5}-y_{5}-z_{5}\right) \, \mathbf{a}_{1} + \left(-x_{5}+y_{5}-z_{5}\right) \, \mathbf{a}_{2} + \left(-x_{5}-y_{5}+z_{5}\right) \, \mathbf{a}_{3} & = & -x_{5}a \, \mathbf{\hat{x}}-y_{5}a \, \mathbf{\hat{y}}-z_{5}a \, \mathbf{\hat{z}} & \left(96g\right) & \mbox{Se II} \\ 
\mathbf{B}_{30} & = & \left(-x_{5}+y_{5}-z_{5}\right) \, \mathbf{a}_{1} + \left(x_{5}-y_{5}-z_{5}\right) \, \mathbf{a}_{2} + \left(\frac{1}{2} +x_{5} + y_{5} + z_{5}\right) \, \mathbf{a}_{3} & = & \left(\frac{1}{4} +x_{5}\right)a \, \mathbf{\hat{x}} + \left(\frac{1}{4} +y_{5}\right)a \, \mathbf{\hat{y}}-z_{5}a \, \mathbf{\hat{z}} & \left(96g\right) & \mbox{Se II} \\ 
\mathbf{B}_{31} & = & \left(-x_{5}-y_{5}+z_{5}\right) \, \mathbf{a}_{1} + \left(\frac{1}{2} +x_{5} + y_{5} + z_{5}\right) \, \mathbf{a}_{2} + \left(x_{5}-y_{5}-z_{5}\right) \, \mathbf{a}_{3} & = & \left(\frac{1}{4} +x_{5}\right)a \, \mathbf{\hat{x}}-y_{5}a \, \mathbf{\hat{y}} + \left(\frac{1}{4} +z_{5}\right)a \, \mathbf{\hat{z}} & \left(96g\right) & \mbox{Se II} \\ 
\mathbf{B}_{32} & = & \left(\frac{1}{2} +x_{5} + y_{5} + z_{5}\right) \, \mathbf{a}_{1} + \left(-x_{5}-y_{5}+z_{5}\right) \, \mathbf{a}_{2} + \left(-x_{5}+y_{5}-z_{5}\right) \, \mathbf{a}_{3} & = & -x_{5}a \, \mathbf{\hat{x}} + \left(\frac{1}{4} +y_{5}\right)a \, \mathbf{\hat{y}} + \left(\frac{1}{4} +z_{5}\right)a \, \mathbf{\hat{z}} & \left(96g\right) & \mbox{Se II} \\ 
\mathbf{B}_{33} & = & \left(-x_{5}-y_{5}+z_{5}\right) \, \mathbf{a}_{1} + \left(x_{5}-y_{5}-z_{5}\right) \, \mathbf{a}_{2} + \left(-x_{5}+y_{5}-z_{5}\right) \, \mathbf{a}_{3} & = & -z_{5}a \, \mathbf{\hat{x}}-x_{5}a \, \mathbf{\hat{y}}-y_{5}a \, \mathbf{\hat{z}} & \left(96g\right) & \mbox{Se II} \\ 
\mathbf{B}_{34} & = & \left(\frac{1}{2} +x_{5} + y_{5} + z_{5}\right) \, \mathbf{a}_{1} + \left(-x_{5}+y_{5}-z_{5}\right) \, \mathbf{a}_{2} + \left(x_{5}-y_{5}-z_{5}\right) \, \mathbf{a}_{3} & = & -z_{5}a \, \mathbf{\hat{x}} + \left(\frac{1}{4} +x_{5}\right)a \, \mathbf{\hat{y}} + \left(\frac{1}{4} +y_{5}\right)a \, \mathbf{\hat{z}} & \left(96g\right) & \mbox{Se II} \\ 
\mathbf{B}_{35} & = & \left(x_{5}-y_{5}-z_{5}\right) \, \mathbf{a}_{1} + \left(-x_{5}-y_{5}+z_{5}\right) \, \mathbf{a}_{2} + \left(\frac{1}{2} +x_{5} + y_{5} + z_{5}\right) \, \mathbf{a}_{3} & = & \left(\frac{1}{4} +z_{5}\right)a \, \mathbf{\hat{x}} + \left(\frac{1}{4} +x_{5}\right)a \, \mathbf{\hat{y}}-y_{5}a \, \mathbf{\hat{z}} & \left(96g\right) & \mbox{Se II} \\ 
\mathbf{B}_{36} & = & \left(-x_{5}+y_{5}-z_{5}\right) \, \mathbf{a}_{1} + \left(\frac{1}{2} +x_{5} + y_{5} + z_{5}\right) \, \mathbf{a}_{2} + \left(-x_{5}-y_{5}+z_{5}\right) \, \mathbf{a}_{3} & = & \left(\frac{1}{4} +z_{5}\right)a \, \mathbf{\hat{x}}-x_{5}a \, \mathbf{\hat{y}} + \left(\frac{1}{4} +y_{5}\right)a \, \mathbf{\hat{z}} & \left(96g\right) & \mbox{Se II} \\ 
\mathbf{B}_{37} & = & \left(-x_{5}+y_{5}-z_{5}\right) \, \mathbf{a}_{1} + \left(-x_{5}-y_{5}+z_{5}\right) \, \mathbf{a}_{2} + \left(x_{5}-y_{5}-z_{5}\right) \, \mathbf{a}_{3} & = & -y_{5}a \, \mathbf{\hat{x}}-z_{5}a \, \mathbf{\hat{y}}-x_{5}a \, \mathbf{\hat{z}} & \left(96g\right) & \mbox{Se II} \\ 
\mathbf{B}_{38} & = & \left(x_{5}-y_{5}-z_{5}\right) \, \mathbf{a}_{1} + \left(\frac{1}{2} +x_{5} + y_{5} + z_{5}\right) \, \mathbf{a}_{2} + \left(-x_{5}+y_{5}-z_{5}\right) \, \mathbf{a}_{3} & = & \left(\frac{1}{4} +y_{5}\right)a \, \mathbf{\hat{x}}-z_{5}a \, \mathbf{\hat{y}} + \left(\frac{1}{4} +x_{5}\right)a \, \mathbf{\hat{z}} & \left(96g\right) & \mbox{Se II} \\ 
\mathbf{B}_{39} & = & \left(\frac{1}{2} +x_{5} + y_{5} + z_{5}\right) \, \mathbf{a}_{1} + \left(x_{5}-y_{5}-z_{5}\right) \, \mathbf{a}_{2} + \left(-x_{5}-y_{5}+z_{5}\right) \, \mathbf{a}_{3} & = & -y_{5}a \, \mathbf{\hat{x}} + \left(\frac{1}{4} +z_{5}\right)a \, \mathbf{\hat{y}} + \left(\frac{1}{4} +x_{5}\right)a \, \mathbf{\hat{z}} & \left(96g\right) & \mbox{Se II} \\ 
\mathbf{B}_{40} & = & \left(-x_{5}-y_{5}+z_{5}\right) \, \mathbf{a}_{1} + \left(-x_{5}+y_{5}-z_{5}\right) \, \mathbf{a}_{2} + \left(\frac{1}{2} +x_{5} + y_{5} + z_{5}\right) \, \mathbf{a}_{3} & = & \left(\frac{1}{4} +y_{5}\right)a \, \mathbf{\hat{x}} + \left(\frac{1}{4} +z_{5}\right)a \, \mathbf{\hat{y}}-x_{5}a \, \mathbf{\hat{z}} & \left(96g\right) & \mbox{Se II} \\ 
\end{longtabu}
\renewcommand{\arraystretch}{1.0}
\noindent \hrulefill
\\
\textbf{References:}
\vspace*{-0.25cm}
\begin{flushleft}
  - \bibentry{Wachhold_AsRb3Se16_ZNaturB_1997}. \\
\end{flushleft}
\textbf{Found in:}
\vspace*{-0.25cm}
\begin{flushleft}
  - \bibentry{Villars_PearsonsCrystalData_2013}. \\
\end{flushleft}
\noindent \hrulefill
\\
\textbf{Geometry files:}
\\
\noindent  - CIF: pp. {\hyperref[AB3C16_cF160_203_b_ad_eg_cif]{\pageref{AB3C16_cF160_203_b_ad_eg_cif}}} \\
\noindent  - POSCAR: pp. {\hyperref[AB3C16_cF160_203_b_ad_eg_poscar]{\pageref{AB3C16_cF160_203_b_ad_eg_poscar}}} \\
\onecolumn
{\phantomsection\label{A2B3C6_cP264_205_2d_ab2c2d_6d}}
\subsection*{\huge \textbf{{\normalfont Ca$_{3}$Al$_{2}$O$_{6}$ Structure: A2B3C6\_cP264\_205\_2d\_ab2c2d\_6d}}}
\noindent \hrulefill
\vspace*{0.25cm}
\begin{figure}[htp]
  \centering
  \vspace{-1em}
  {\includegraphics[width=1\textwidth]{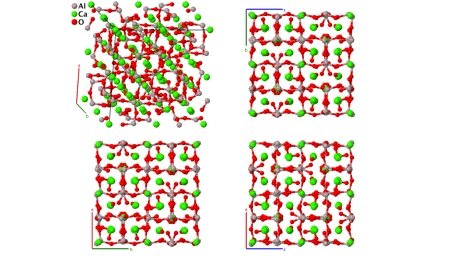}}
\end{figure}
\vspace*{-0.5cm}
\renewcommand{\arraystretch}{1.5}
\begin{equation*}
  \begin{array}{>{$\hspace{-0.15cm}}l<{$}>{$}p{0.5cm}<{$}>{$}p{18.5cm}<{$}}
    \mbox{\large \textbf{Prototype}} &\colon & \ce{Ca3Al2O6} \\
    \mbox{\large \textbf{\AFLOW\ prototype label}} &\colon & \mbox{A2B3C6\_cP264\_205\_2d\_ab2c2d\_6d} \\
    \mbox{\large \textbf{\textit{Strukturbericht} designation}} &\colon & \mbox{None} \\
    \mbox{\large \textbf{Pearson symbol}} &\colon & \mbox{cP264} \\
    \mbox{\large \textbf{Space group number}} &\colon & 205 \\
    \mbox{\large \textbf{Space group symbol}} &\colon & Pa\bar{3} \\
    \mbox{\large \textbf{\AFLOW\ prototype command}} &\colon &  \texttt{aflow} \,  \, \texttt{-{}-proto=A2B3C6\_cP264\_205\_2d\_ab2c2d\_6d } \, \newline \texttt{-{}-params=}{a,x_{3},x_{4},x_{5},y_{5},z_{5},x_{6},y_{6},z_{6},x_{7},y_{7},z_{7},x_{8},y_{8},z_{8},x_{9},y_{9},z_{9},x_{10},y_{10},z_{10},x_{11},} \newline {y_{11},z_{11},x_{12},y_{12},z_{12},x_{13},y_{13},z_{13},x_{14},y_{14},z_{14} }
  \end{array}
\end{equation*}
\renewcommand{\arraystretch}{1.0}

\vspace*{-0.25cm}
\noindent \hrulefill
\begin{itemize}
  \item{This is a redetermination of the
\hyperref[A2B3C6_cP33_221_cd_ad_fh]{$E9_{1}$ (Ca$_{3}$Al$_{2}$O$_{6}$) structure}.  
The lattice constant of the new unit cell is twice the original, giving a volume eight times larger.
}
\end{itemize}

\noindent \parbox{1 \linewidth}{
\noindent \hrulefill
\\
\textbf{Simple Cubic primitive vectors:} \\
\vspace*{-0.25cm}
\begin{tabular}{cc}
  \begin{tabular}{c}
    \parbox{0.6 \linewidth}{
      \renewcommand{\arraystretch}{1.5}
      \begin{equation*}
        \centering
        \begin{array}{ccc}
              \mathbf{a}_1 & = & a \, \mathbf{\hat{x}} \\
    \mathbf{a}_2 & = & a \, \mathbf{\hat{y}} \\
    \mathbf{a}_3 & = & a \, \mathbf{\hat{z}} \\

        \end{array}
      \end{equation*}
    }
    \renewcommand{\arraystretch}{1.0}
  \end{tabular}
  \begin{tabular}{c}
    \includegraphics[width=0.3\linewidth]{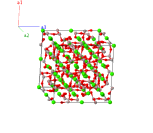} \\
  \end{tabular}
\end{tabular}

}
\vspace*{-0.25cm}

\noindent \hrulefill
\\
\textbf{Basis vectors:}
\vspace*{-0.25cm}
\renewcommand{\arraystretch}{1.5}
\begin{longtabu} to \textwidth{>{\centering $}X[-1,c,c]<{$}>{\centering $}X[-1,c,c]<{$}>{\centering $}X[-1,c,c]<{$}>{\centering $}X[-1,c,c]<{$}>{\centering $}X[-1,c,c]<{$}>{\centering $}X[-1,c,c]<{$}>{\centering $}X[-1,c,c]<{$}}
  & & \mbox{Lattice Coordinates} & & \mbox{Cartesian Coordinates} &\mbox{Wyckoff Position} & \mbox{Atom Type} \\  
  \mathbf{B}_{1} & = & 0 \, \mathbf{a}_{1} + 0 \, \mathbf{a}_{2} + 0 \, \mathbf{a}_{3} & = & 0 \, \mathbf{\hat{x}} + 0 \, \mathbf{\hat{y}} + 0 \, \mathbf{\hat{z}} & \left(4a\right) & \mbox{Ca I} \\ 
\mathbf{B}_{2} & = & \frac{1}{2} \, \mathbf{a}_{1} + \frac{1}{2} \, \mathbf{a}_{3} & = & \frac{1}{2}a \, \mathbf{\hat{x}} + \frac{1}{2}a \, \mathbf{\hat{z}} & \left(4a\right) & \mbox{Ca I} \\ 
\mathbf{B}_{3} & = & \frac{1}{2} \, \mathbf{a}_{2} + \frac{1}{2} \, \mathbf{a}_{3} & = & \frac{1}{2}a \, \mathbf{\hat{y}} + \frac{1}{2}a \, \mathbf{\hat{z}} & \left(4a\right) & \mbox{Ca I} \\ 
\mathbf{B}_{4} & = & \frac{1}{2} \, \mathbf{a}_{1} + \frac{1}{2} \, \mathbf{a}_{2} & = & \frac{1}{2}a \, \mathbf{\hat{x}} + \frac{1}{2}a \, \mathbf{\hat{y}} & \left(4a\right) & \mbox{Ca I} \\ 
\mathbf{B}_{5} & = & \frac{1}{2} \, \mathbf{a}_{1} + \frac{1}{2} \, \mathbf{a}_{2} + \frac{1}{2} \, \mathbf{a}_{3} & = & \frac{1}{2}a \, \mathbf{\hat{x}} + \frac{1}{2}a \, \mathbf{\hat{y}} + \frac{1}{2}a \, \mathbf{\hat{z}} & \left(4b\right) & \mbox{Ca II} \\ 
\mathbf{B}_{6} & = & \frac{1}{2} \, \mathbf{a}_{2} & = & \frac{1}{2}a \, \mathbf{\hat{y}} & \left(4b\right) & \mbox{Ca II} \\ 
\mathbf{B}_{7} & = & \frac{1}{2} \, \mathbf{a}_{1} & = & \frac{1}{2}a \, \mathbf{\hat{x}} & \left(4b\right) & \mbox{Ca II} \\ 
\mathbf{B}_{8} & = & \frac{1}{2} \, \mathbf{a}_{3} & = & \frac{1}{2}a \, \mathbf{\hat{z}} & \left(4b\right) & \mbox{Ca II} \\ 
\mathbf{B}_{9} & = & x_{3} \, \mathbf{a}_{1} + x_{3} \, \mathbf{a}_{2} + x_{3} \, \mathbf{a}_{3} & = & x_{3}a \, \mathbf{\hat{x}} + x_{3}a \, \mathbf{\hat{y}} + x_{3}a \, \mathbf{\hat{z}} & \left(8c\right) & \mbox{Ca III} \\ 
\mathbf{B}_{10} & = & \left(\frac{1}{2} - x_{3}\right) \, \mathbf{a}_{1}-x_{3} \, \mathbf{a}_{2} + \left(\frac{1}{2} +x_{3}\right) \, \mathbf{a}_{3} & = & \left(\frac{1}{2} - x_{3}\right)a \, \mathbf{\hat{x}}-x_{3}a \, \mathbf{\hat{y}} + \left(\frac{1}{2} +x_{3}\right)a \, \mathbf{\hat{z}} & \left(8c\right) & \mbox{Ca III} \\ 
\mathbf{B}_{11} & = & -x_{3} \, \mathbf{a}_{1} + \left(\frac{1}{2} +x_{3}\right) \, \mathbf{a}_{2} + \left(\frac{1}{2} - x_{3}\right) \, \mathbf{a}_{3} & = & -x_{3}a \, \mathbf{\hat{x}} + \left(\frac{1}{2} +x_{3}\right)a \, \mathbf{\hat{y}} + \left(\frac{1}{2} - x_{3}\right)a \, \mathbf{\hat{z}} & \left(8c\right) & \mbox{Ca III} \\ 
\mathbf{B}_{12} & = & \left(\frac{1}{2} +x_{3}\right) \, \mathbf{a}_{1} + \left(\frac{1}{2} - x_{3}\right) \, \mathbf{a}_{2}-x_{3} \, \mathbf{a}_{3} & = & \left(\frac{1}{2} +x_{3}\right)a \, \mathbf{\hat{x}} + \left(\frac{1}{2} - x_{3}\right)a \, \mathbf{\hat{y}}-x_{3}a \, \mathbf{\hat{z}} & \left(8c\right) & \mbox{Ca III} \\ 
\mathbf{B}_{13} & = & -x_{3} \, \mathbf{a}_{1}-x_{3} \, \mathbf{a}_{2}-x_{3} \, \mathbf{a}_{3} & = & -x_{3}a \, \mathbf{\hat{x}}-x_{3}a \, \mathbf{\hat{y}}-x_{3}a \, \mathbf{\hat{z}} & \left(8c\right) & \mbox{Ca III} \\ 
\mathbf{B}_{14} & = & \left(\frac{1}{2} +x_{3}\right) \, \mathbf{a}_{1} + x_{3} \, \mathbf{a}_{2} + \left(\frac{1}{2} - x_{3}\right) \, \mathbf{a}_{3} & = & \left(\frac{1}{2} +x_{3}\right)a \, \mathbf{\hat{x}} + x_{3}a \, \mathbf{\hat{y}} + \left(\frac{1}{2} - x_{3}\right)a \, \mathbf{\hat{z}} & \left(8c\right) & \mbox{Ca III} \\ 
\mathbf{B}_{15} & = & x_{3} \, \mathbf{a}_{1} + \left(\frac{1}{2} - x_{3}\right) \, \mathbf{a}_{2} + \left(\frac{1}{2} +x_{3}\right) \, \mathbf{a}_{3} & = & x_{3}a \, \mathbf{\hat{x}} + \left(\frac{1}{2} - x_{3}\right)a \, \mathbf{\hat{y}} + \left(\frac{1}{2} +x_{3}\right)a \, \mathbf{\hat{z}} & \left(8c\right) & \mbox{Ca III} \\ 
\mathbf{B}_{16} & = & \left(\frac{1}{2} - x_{3}\right) \, \mathbf{a}_{1} + \left(\frac{1}{2} +x_{3}\right) \, \mathbf{a}_{2} + x_{3} \, \mathbf{a}_{3} & = & \left(\frac{1}{2} - x_{3}\right)a \, \mathbf{\hat{x}} + \left(\frac{1}{2} +x_{3}\right)a \, \mathbf{\hat{y}} + x_{3}a \, \mathbf{\hat{z}} & \left(8c\right) & \mbox{Ca III} \\ 
\mathbf{B}_{17} & = & x_{4} \, \mathbf{a}_{1} + x_{4} \, \mathbf{a}_{2} + x_{4} \, \mathbf{a}_{3} & = & x_{4}a \, \mathbf{\hat{x}} + x_{4}a \, \mathbf{\hat{y}} + x_{4}a \, \mathbf{\hat{z}} & \left(8c\right) & \mbox{Ca IV} \\ 
\mathbf{B}_{18} & = & \left(\frac{1}{2} - x_{4}\right) \, \mathbf{a}_{1}-x_{4} \, \mathbf{a}_{2} + \left(\frac{1}{2} +x_{4}\right) \, \mathbf{a}_{3} & = & \left(\frac{1}{2} - x_{4}\right)a \, \mathbf{\hat{x}}-x_{4}a \, \mathbf{\hat{y}} + \left(\frac{1}{2} +x_{4}\right)a \, \mathbf{\hat{z}} & \left(8c\right) & \mbox{Ca IV} \\ 
\mathbf{B}_{19} & = & -x_{4} \, \mathbf{a}_{1} + \left(\frac{1}{2} +x_{4}\right) \, \mathbf{a}_{2} + \left(\frac{1}{2} - x_{4}\right) \, \mathbf{a}_{3} & = & -x_{4}a \, \mathbf{\hat{x}} + \left(\frac{1}{2} +x_{4}\right)a \, \mathbf{\hat{y}} + \left(\frac{1}{2} - x_{4}\right)a \, \mathbf{\hat{z}} & \left(8c\right) & \mbox{Ca IV} \\ 
\mathbf{B}_{20} & = & \left(\frac{1}{2} +x_{4}\right) \, \mathbf{a}_{1} + \left(\frac{1}{2} - x_{4}\right) \, \mathbf{a}_{2}-x_{4} \, \mathbf{a}_{3} & = & \left(\frac{1}{2} +x_{4}\right)a \, \mathbf{\hat{x}} + \left(\frac{1}{2} - x_{4}\right)a \, \mathbf{\hat{y}}-x_{4}a \, \mathbf{\hat{z}} & \left(8c\right) & \mbox{Ca IV} \\ 
\mathbf{B}_{21} & = & -x_{4} \, \mathbf{a}_{1}-x_{4} \, \mathbf{a}_{2}-x_{4} \, \mathbf{a}_{3} & = & -x_{4}a \, \mathbf{\hat{x}}-x_{4}a \, \mathbf{\hat{y}}-x_{4}a \, \mathbf{\hat{z}} & \left(8c\right) & \mbox{Ca IV} \\ 
\mathbf{B}_{22} & = & \left(\frac{1}{2} +x_{4}\right) \, \mathbf{a}_{1} + x_{4} \, \mathbf{a}_{2} + \left(\frac{1}{2} - x_{4}\right) \, \mathbf{a}_{3} & = & \left(\frac{1}{2} +x_{4}\right)a \, \mathbf{\hat{x}} + x_{4}a \, \mathbf{\hat{y}} + \left(\frac{1}{2} - x_{4}\right)a \, \mathbf{\hat{z}} & \left(8c\right) & \mbox{Ca IV} \\ 
\mathbf{B}_{23} & = & x_{4} \, \mathbf{a}_{1} + \left(\frac{1}{2} - x_{4}\right) \, \mathbf{a}_{2} + \left(\frac{1}{2} +x_{4}\right) \, \mathbf{a}_{3} & = & x_{4}a \, \mathbf{\hat{x}} + \left(\frac{1}{2} - x_{4}\right)a \, \mathbf{\hat{y}} + \left(\frac{1}{2} +x_{4}\right)a \, \mathbf{\hat{z}} & \left(8c\right) & \mbox{Ca IV} \\ 
\mathbf{B}_{24} & = & \left(\frac{1}{2} - x_{4}\right) \, \mathbf{a}_{1} + \left(\frac{1}{2} +x_{4}\right) \, \mathbf{a}_{2} + x_{4} \, \mathbf{a}_{3} & = & \left(\frac{1}{2} - x_{4}\right)a \, \mathbf{\hat{x}} + \left(\frac{1}{2} +x_{4}\right)a \, \mathbf{\hat{y}} + x_{4}a \, \mathbf{\hat{z}} & \left(8c\right) & \mbox{Ca IV} \\ 
\mathbf{B}_{25} & = & x_{5} \, \mathbf{a}_{1} + y_{5} \, \mathbf{a}_{2} + z_{5} \, \mathbf{a}_{3} & = & x_{5}a \, \mathbf{\hat{x}} + y_{5}a \, \mathbf{\hat{y}} + z_{5}a \, \mathbf{\hat{z}} & \left(24d\right) & \mbox{Al I} \\ 
\mathbf{B}_{26} & = & \left(\frac{1}{2} - x_{5}\right) \, \mathbf{a}_{1}-y_{5} \, \mathbf{a}_{2} + \left(\frac{1}{2} +z_{5}\right) \, \mathbf{a}_{3} & = & \left(\frac{1}{2} - x_{5}\right)a \, \mathbf{\hat{x}}-y_{5}a \, \mathbf{\hat{y}} + \left(\frac{1}{2} +z_{5}\right)a \, \mathbf{\hat{z}} & \left(24d\right) & \mbox{Al I} \\ 
\mathbf{B}_{27} & = & -x_{5} \, \mathbf{a}_{1} + \left(\frac{1}{2} +y_{5}\right) \, \mathbf{a}_{2} + \left(\frac{1}{2} - z_{5}\right) \, \mathbf{a}_{3} & = & -x_{5}a \, \mathbf{\hat{x}} + \left(\frac{1}{2} +y_{5}\right)a \, \mathbf{\hat{y}} + \left(\frac{1}{2} - z_{5}\right)a \, \mathbf{\hat{z}} & \left(24d\right) & \mbox{Al I} \\ 
\mathbf{B}_{28} & = & \left(\frac{1}{2} +x_{5}\right) \, \mathbf{a}_{1} + \left(\frac{1}{2} - y_{5}\right) \, \mathbf{a}_{2}-z_{5} \, \mathbf{a}_{3} & = & \left(\frac{1}{2} +x_{5}\right)a \, \mathbf{\hat{x}} + \left(\frac{1}{2} - y_{5}\right)a \, \mathbf{\hat{y}}-z_{5}a \, \mathbf{\hat{z}} & \left(24d\right) & \mbox{Al I} \\ 
\mathbf{B}_{29} & = & z_{5} \, \mathbf{a}_{1} + x_{5} \, \mathbf{a}_{2} + y_{5} \, \mathbf{a}_{3} & = & z_{5}a \, \mathbf{\hat{x}} + x_{5}a \, \mathbf{\hat{y}} + y_{5}a \, \mathbf{\hat{z}} & \left(24d\right) & \mbox{Al I} \\ 
\mathbf{B}_{30} & = & \left(\frac{1}{2} +z_{5}\right) \, \mathbf{a}_{1} + \left(\frac{1}{2} - x_{5}\right) \, \mathbf{a}_{2}-y_{5} \, \mathbf{a}_{3} & = & \left(\frac{1}{2} +z_{5}\right)a \, \mathbf{\hat{x}} + \left(\frac{1}{2} - x_{5}\right)a \, \mathbf{\hat{y}}-y_{5}a \, \mathbf{\hat{z}} & \left(24d\right) & \mbox{Al I} \\ 
\mathbf{B}_{31} & = & \left(\frac{1}{2} - z_{5}\right) \, \mathbf{a}_{1}-x_{5} \, \mathbf{a}_{2} + \left(\frac{1}{2} +y_{5}\right) \, \mathbf{a}_{3} & = & \left(\frac{1}{2} - z_{5}\right)a \, \mathbf{\hat{x}}-x_{5}a \, \mathbf{\hat{y}} + \left(\frac{1}{2} +y_{5}\right)a \, \mathbf{\hat{z}} & \left(24d\right) & \mbox{Al I} \\ 
\mathbf{B}_{32} & = & -z_{5} \, \mathbf{a}_{1} + \left(\frac{1}{2} +x_{5}\right) \, \mathbf{a}_{2} + \left(\frac{1}{2} - y_{5}\right) \, \mathbf{a}_{3} & = & -z_{5}a \, \mathbf{\hat{x}} + \left(\frac{1}{2} +x_{5}\right)a \, \mathbf{\hat{y}} + \left(\frac{1}{2} - y_{5}\right)a \, \mathbf{\hat{z}} & \left(24d\right) & \mbox{Al I} \\ 
\mathbf{B}_{33} & = & y_{5} \, \mathbf{a}_{1} + z_{5} \, \mathbf{a}_{2} + x_{5} \, \mathbf{a}_{3} & = & y_{5}a \, \mathbf{\hat{x}} + z_{5}a \, \mathbf{\hat{y}} + x_{5}a \, \mathbf{\hat{z}} & \left(24d\right) & \mbox{Al I} \\ 
\mathbf{B}_{34} & = & -y_{5} \, \mathbf{a}_{1} + \left(\frac{1}{2} +z_{5}\right) \, \mathbf{a}_{2} + \left(\frac{1}{2} - x_{5}\right) \, \mathbf{a}_{3} & = & -y_{5}a \, \mathbf{\hat{x}} + \left(\frac{1}{2} +z_{5}\right)a \, \mathbf{\hat{y}} + \left(\frac{1}{2} - x_{5}\right)a \, \mathbf{\hat{z}} & \left(24d\right) & \mbox{Al I} \\ 
\mathbf{B}_{35} & = & \left(\frac{1}{2} +y_{5}\right) \, \mathbf{a}_{1} + \left(\frac{1}{2} - z_{5}\right) \, \mathbf{a}_{2}-x_{5} \, \mathbf{a}_{3} & = & \left(\frac{1}{2} +y_{5}\right)a \, \mathbf{\hat{x}} + \left(\frac{1}{2} - z_{5}\right)a \, \mathbf{\hat{y}}-x_{5}a \, \mathbf{\hat{z}} & \left(24d\right) & \mbox{Al I} \\ 
\mathbf{B}_{36} & = & \left(\frac{1}{2} - y_{5}\right) \, \mathbf{a}_{1}-z_{5} \, \mathbf{a}_{2} + \left(\frac{1}{2} +x_{5}\right) \, \mathbf{a}_{3} & = & \left(\frac{1}{2} - y_{5}\right)a \, \mathbf{\hat{x}}-z_{5}a \, \mathbf{\hat{y}} + \left(\frac{1}{2} +x_{5}\right)a \, \mathbf{\hat{z}} & \left(24d\right) & \mbox{Al I} \\ 
\mathbf{B}_{37} & = & -x_{5} \, \mathbf{a}_{1}-y_{5} \, \mathbf{a}_{2}-z_{5} \, \mathbf{a}_{3} & = & -x_{5}a \, \mathbf{\hat{x}}-y_{5}a \, \mathbf{\hat{y}}-z_{5}a \, \mathbf{\hat{z}} & \left(24d\right) & \mbox{Al I} \\ 
\mathbf{B}_{38} & = & \left(\frac{1}{2} +x_{5}\right) \, \mathbf{a}_{1} + y_{5} \, \mathbf{a}_{2} + \left(\frac{1}{2} - z_{5}\right) \, \mathbf{a}_{3} & = & \left(\frac{1}{2} +x_{5}\right)a \, \mathbf{\hat{x}} + y_{5}a \, \mathbf{\hat{y}} + \left(\frac{1}{2} - z_{5}\right)a \, \mathbf{\hat{z}} & \left(24d\right) & \mbox{Al I} \\ 
\mathbf{B}_{39} & = & x_{5} \, \mathbf{a}_{1} + \left(\frac{1}{2} - y_{5}\right) \, \mathbf{a}_{2} + \left(\frac{1}{2} +z_{5}\right) \, \mathbf{a}_{3} & = & x_{5}a \, \mathbf{\hat{x}} + \left(\frac{1}{2} - y_{5}\right)a \, \mathbf{\hat{y}} + \left(\frac{1}{2} +z_{5}\right)a \, \mathbf{\hat{z}} & \left(24d\right) & \mbox{Al I} \\ 
\mathbf{B}_{40} & = & \left(\frac{1}{2} - x_{5}\right) \, \mathbf{a}_{1} + \left(\frac{1}{2} +y_{5}\right) \, \mathbf{a}_{2} + z_{5} \, \mathbf{a}_{3} & = & \left(\frac{1}{2} - x_{5}\right)a \, \mathbf{\hat{x}} + \left(\frac{1}{2} +y_{5}\right)a \, \mathbf{\hat{y}} + z_{5}a \, \mathbf{\hat{z}} & \left(24d\right) & \mbox{Al I} \\ 
\mathbf{B}_{41} & = & -z_{5} \, \mathbf{a}_{1}-x_{5} \, \mathbf{a}_{2}-y_{5} \, \mathbf{a}_{3} & = & -z_{5}a \, \mathbf{\hat{x}}-x_{5}a \, \mathbf{\hat{y}}-y_{5}a \, \mathbf{\hat{z}} & \left(24d\right) & \mbox{Al I} \\ 
\mathbf{B}_{42} & = & \left(\frac{1}{2} - z_{5}\right) \, \mathbf{a}_{1} + \left(\frac{1}{2} +x_{5}\right) \, \mathbf{a}_{2} + y_{5} \, \mathbf{a}_{3} & = & \left(\frac{1}{2} - z_{5}\right)a \, \mathbf{\hat{x}} + \left(\frac{1}{2} +x_{5}\right)a \, \mathbf{\hat{y}} + y_{5}a \, \mathbf{\hat{z}} & \left(24d\right) & \mbox{Al I} \\ 
\mathbf{B}_{43} & = & \left(\frac{1}{2} +z_{5}\right) \, \mathbf{a}_{1} + x_{5} \, \mathbf{a}_{2} + \left(\frac{1}{2} - y_{5}\right) \, \mathbf{a}_{3} & = & \left(\frac{1}{2} +z_{5}\right)a \, \mathbf{\hat{x}} + x_{5}a \, \mathbf{\hat{y}} + \left(\frac{1}{2} - y_{5}\right)a \, \mathbf{\hat{z}} & \left(24d\right) & \mbox{Al I} \\ 
\mathbf{B}_{44} & = & z_{5} \, \mathbf{a}_{1} + \left(\frac{1}{2} - x_{5}\right) \, \mathbf{a}_{2} + \left(\frac{1}{2} +y_{5}\right) \, \mathbf{a}_{3} & = & z_{5}a \, \mathbf{\hat{x}} + \left(\frac{1}{2} - x_{5}\right)a \, \mathbf{\hat{y}} + \left(\frac{1}{2} +y_{5}\right)a \, \mathbf{\hat{z}} & \left(24d\right) & \mbox{Al I} \\ 
\mathbf{B}_{45} & = & -y_{5} \, \mathbf{a}_{1}-z_{5} \, \mathbf{a}_{2}-x_{5} \, \mathbf{a}_{3} & = & -y_{5}a \, \mathbf{\hat{x}}-z_{5}a \, \mathbf{\hat{y}}-x_{5}a \, \mathbf{\hat{z}} & \left(24d\right) & \mbox{Al I} \\ 
\mathbf{B}_{46} & = & y_{5} \, \mathbf{a}_{1} + \left(\frac{1}{2} - z_{5}\right) \, \mathbf{a}_{2} + \left(\frac{1}{2} +x_{5}\right) \, \mathbf{a}_{3} & = & y_{5}a \, \mathbf{\hat{x}} + \left(\frac{1}{2} - z_{5}\right)a \, \mathbf{\hat{y}} + \left(\frac{1}{2} +x_{5}\right)a \, \mathbf{\hat{z}} & \left(24d\right) & \mbox{Al I} \\ 
\mathbf{B}_{47} & = & \left(\frac{1}{2} - y_{5}\right) \, \mathbf{a}_{1} + \left(\frac{1}{2} +z_{5}\right) \, \mathbf{a}_{2} + x_{5} \, \mathbf{a}_{3} & = & \left(\frac{1}{2} - y_{5}\right)a \, \mathbf{\hat{x}} + \left(\frac{1}{2} +z_{5}\right)a \, \mathbf{\hat{y}} + x_{5}a \, \mathbf{\hat{z}} & \left(24d\right) & \mbox{Al I} \\ 
\mathbf{B}_{48} & = & \left(\frac{1}{2} +y_{5}\right) \, \mathbf{a}_{1} + z_{5} \, \mathbf{a}_{2} + \left(\frac{1}{2} - x_{5}\right) \, \mathbf{a}_{3} & = & \left(\frac{1}{2} +y_{5}\right)a \, \mathbf{\hat{x}} + z_{5}a \, \mathbf{\hat{y}} + \left(\frac{1}{2} - x_{5}\right)a \, \mathbf{\hat{z}} & \left(24d\right) & \mbox{Al I} \\ 
\mathbf{B}_{49} & = & x_{6} \, \mathbf{a}_{1} + y_{6} \, \mathbf{a}_{2} + z_{6} \, \mathbf{a}_{3} & = & x_{6}a \, \mathbf{\hat{x}} + y_{6}a \, \mathbf{\hat{y}} + z_{6}a \, \mathbf{\hat{z}} & \left(24d\right) & \mbox{Al II} \\ 
\mathbf{B}_{50} & = & \left(\frac{1}{2} - x_{6}\right) \, \mathbf{a}_{1}-y_{6} \, \mathbf{a}_{2} + \left(\frac{1}{2} +z_{6}\right) \, \mathbf{a}_{3} & = & \left(\frac{1}{2} - x_{6}\right)a \, \mathbf{\hat{x}}-y_{6}a \, \mathbf{\hat{y}} + \left(\frac{1}{2} +z_{6}\right)a \, \mathbf{\hat{z}} & \left(24d\right) & \mbox{Al II} \\ 
\mathbf{B}_{51} & = & -x_{6} \, \mathbf{a}_{1} + \left(\frac{1}{2} +y_{6}\right) \, \mathbf{a}_{2} + \left(\frac{1}{2} - z_{6}\right) \, \mathbf{a}_{3} & = & -x_{6}a \, \mathbf{\hat{x}} + \left(\frac{1}{2} +y_{6}\right)a \, \mathbf{\hat{y}} + \left(\frac{1}{2} - z_{6}\right)a \, \mathbf{\hat{z}} & \left(24d\right) & \mbox{Al II} \\ 
\mathbf{B}_{52} & = & \left(\frac{1}{2} +x_{6}\right) \, \mathbf{a}_{1} + \left(\frac{1}{2} - y_{6}\right) \, \mathbf{a}_{2}-z_{6} \, \mathbf{a}_{3} & = & \left(\frac{1}{2} +x_{6}\right)a \, \mathbf{\hat{x}} + \left(\frac{1}{2} - y_{6}\right)a \, \mathbf{\hat{y}}-z_{6}a \, \mathbf{\hat{z}} & \left(24d\right) & \mbox{Al II} \\ 
\mathbf{B}_{53} & = & z_{6} \, \mathbf{a}_{1} + x_{6} \, \mathbf{a}_{2} + y_{6} \, \mathbf{a}_{3} & = & z_{6}a \, \mathbf{\hat{x}} + x_{6}a \, \mathbf{\hat{y}} + y_{6}a \, \mathbf{\hat{z}} & \left(24d\right) & \mbox{Al II} \\ 
\mathbf{B}_{54} & = & \left(\frac{1}{2} +z_{6}\right) \, \mathbf{a}_{1} + \left(\frac{1}{2} - x_{6}\right) \, \mathbf{a}_{2}-y_{6} \, \mathbf{a}_{3} & = & \left(\frac{1}{2} +z_{6}\right)a \, \mathbf{\hat{x}} + \left(\frac{1}{2} - x_{6}\right)a \, \mathbf{\hat{y}}-y_{6}a \, \mathbf{\hat{z}} & \left(24d\right) & \mbox{Al II} \\ 
\mathbf{B}_{55} & = & \left(\frac{1}{2} - z_{6}\right) \, \mathbf{a}_{1}-x_{6} \, \mathbf{a}_{2} + \left(\frac{1}{2} +y_{6}\right) \, \mathbf{a}_{3} & = & \left(\frac{1}{2} - z_{6}\right)a \, \mathbf{\hat{x}}-x_{6}a \, \mathbf{\hat{y}} + \left(\frac{1}{2} +y_{6}\right)a \, \mathbf{\hat{z}} & \left(24d\right) & \mbox{Al II} \\ 
\mathbf{B}_{56} & = & -z_{6} \, \mathbf{a}_{1} + \left(\frac{1}{2} +x_{6}\right) \, \mathbf{a}_{2} + \left(\frac{1}{2} - y_{6}\right) \, \mathbf{a}_{3} & = & -z_{6}a \, \mathbf{\hat{x}} + \left(\frac{1}{2} +x_{6}\right)a \, \mathbf{\hat{y}} + \left(\frac{1}{2} - y_{6}\right)a \, \mathbf{\hat{z}} & \left(24d\right) & \mbox{Al II} \\ 
\mathbf{B}_{57} & = & y_{6} \, \mathbf{a}_{1} + z_{6} \, \mathbf{a}_{2} + x_{6} \, \mathbf{a}_{3} & = & y_{6}a \, \mathbf{\hat{x}} + z_{6}a \, \mathbf{\hat{y}} + x_{6}a \, \mathbf{\hat{z}} & \left(24d\right) & \mbox{Al II} \\ 
\mathbf{B}_{58} & = & -y_{6} \, \mathbf{a}_{1} + \left(\frac{1}{2} +z_{6}\right) \, \mathbf{a}_{2} + \left(\frac{1}{2} - x_{6}\right) \, \mathbf{a}_{3} & = & -y_{6}a \, \mathbf{\hat{x}} + \left(\frac{1}{2} +z_{6}\right)a \, \mathbf{\hat{y}} + \left(\frac{1}{2} - x_{6}\right)a \, \mathbf{\hat{z}} & \left(24d\right) & \mbox{Al II} \\ 
\mathbf{B}_{59} & = & \left(\frac{1}{2} +y_{6}\right) \, \mathbf{a}_{1} + \left(\frac{1}{2} - z_{6}\right) \, \mathbf{a}_{2}-x_{6} \, \mathbf{a}_{3} & = & \left(\frac{1}{2} +y_{6}\right)a \, \mathbf{\hat{x}} + \left(\frac{1}{2} - z_{6}\right)a \, \mathbf{\hat{y}}-x_{6}a \, \mathbf{\hat{z}} & \left(24d\right) & \mbox{Al II} \\ 
\mathbf{B}_{60} & = & \left(\frac{1}{2} - y_{6}\right) \, \mathbf{a}_{1}-z_{6} \, \mathbf{a}_{2} + \left(\frac{1}{2} +x_{6}\right) \, \mathbf{a}_{3} & = & \left(\frac{1}{2} - y_{6}\right)a \, \mathbf{\hat{x}}-z_{6}a \, \mathbf{\hat{y}} + \left(\frac{1}{2} +x_{6}\right)a \, \mathbf{\hat{z}} & \left(24d\right) & \mbox{Al II} \\ 
\mathbf{B}_{61} & = & -x_{6} \, \mathbf{a}_{1}-y_{6} \, \mathbf{a}_{2}-z_{6} \, \mathbf{a}_{3} & = & -x_{6}a \, \mathbf{\hat{x}}-y_{6}a \, \mathbf{\hat{y}}-z_{6}a \, \mathbf{\hat{z}} & \left(24d\right) & \mbox{Al II} \\ 
\mathbf{B}_{62} & = & \left(\frac{1}{2} +x_{6}\right) \, \mathbf{a}_{1} + y_{6} \, \mathbf{a}_{2} + \left(\frac{1}{2} - z_{6}\right) \, \mathbf{a}_{3} & = & \left(\frac{1}{2} +x_{6}\right)a \, \mathbf{\hat{x}} + y_{6}a \, \mathbf{\hat{y}} + \left(\frac{1}{2} - z_{6}\right)a \, \mathbf{\hat{z}} & \left(24d\right) & \mbox{Al II} \\ 
\mathbf{B}_{63} & = & x_{6} \, \mathbf{a}_{1} + \left(\frac{1}{2} - y_{6}\right) \, \mathbf{a}_{2} + \left(\frac{1}{2} +z_{6}\right) \, \mathbf{a}_{3} & = & x_{6}a \, \mathbf{\hat{x}} + \left(\frac{1}{2} - y_{6}\right)a \, \mathbf{\hat{y}} + \left(\frac{1}{2} +z_{6}\right)a \, \mathbf{\hat{z}} & \left(24d\right) & \mbox{Al II} \\ 
\mathbf{B}_{64} & = & \left(\frac{1}{2} - x_{6}\right) \, \mathbf{a}_{1} + \left(\frac{1}{2} +y_{6}\right) \, \mathbf{a}_{2} + z_{6} \, \mathbf{a}_{3} & = & \left(\frac{1}{2} - x_{6}\right)a \, \mathbf{\hat{x}} + \left(\frac{1}{2} +y_{6}\right)a \, \mathbf{\hat{y}} + z_{6}a \, \mathbf{\hat{z}} & \left(24d\right) & \mbox{Al II} \\ 
\mathbf{B}_{65} & = & -z_{6} \, \mathbf{a}_{1}-x_{6} \, \mathbf{a}_{2}-y_{6} \, \mathbf{a}_{3} & = & -z_{6}a \, \mathbf{\hat{x}}-x_{6}a \, \mathbf{\hat{y}}-y_{6}a \, \mathbf{\hat{z}} & \left(24d\right) & \mbox{Al II} \\ 
\mathbf{B}_{66} & = & \left(\frac{1}{2} - z_{6}\right) \, \mathbf{a}_{1} + \left(\frac{1}{2} +x_{6}\right) \, \mathbf{a}_{2} + y_{6} \, \mathbf{a}_{3} & = & \left(\frac{1}{2} - z_{6}\right)a \, \mathbf{\hat{x}} + \left(\frac{1}{2} +x_{6}\right)a \, \mathbf{\hat{y}} + y_{6}a \, \mathbf{\hat{z}} & \left(24d\right) & \mbox{Al II} \\ 
\mathbf{B}_{67} & = & \left(\frac{1}{2} +z_{6}\right) \, \mathbf{a}_{1} + x_{6} \, \mathbf{a}_{2} + \left(\frac{1}{2} - y_{6}\right) \, \mathbf{a}_{3} & = & \left(\frac{1}{2} +z_{6}\right)a \, \mathbf{\hat{x}} + x_{6}a \, \mathbf{\hat{y}} + \left(\frac{1}{2} - y_{6}\right)a \, \mathbf{\hat{z}} & \left(24d\right) & \mbox{Al II} \\ 
\mathbf{B}_{68} & = & z_{6} \, \mathbf{a}_{1} + \left(\frac{1}{2} - x_{6}\right) \, \mathbf{a}_{2} + \left(\frac{1}{2} +y_{6}\right) \, \mathbf{a}_{3} & = & z_{6}a \, \mathbf{\hat{x}} + \left(\frac{1}{2} - x_{6}\right)a \, \mathbf{\hat{y}} + \left(\frac{1}{2} +y_{6}\right)a \, \mathbf{\hat{z}} & \left(24d\right) & \mbox{Al II} \\ 
\mathbf{B}_{69} & = & -y_{6} \, \mathbf{a}_{1}-z_{6} \, \mathbf{a}_{2}-x_{6} \, \mathbf{a}_{3} & = & -y_{6}a \, \mathbf{\hat{x}}-z_{6}a \, \mathbf{\hat{y}}-x_{6}a \, \mathbf{\hat{z}} & \left(24d\right) & \mbox{Al II} \\ 
\mathbf{B}_{70} & = & y_{6} \, \mathbf{a}_{1} + \left(\frac{1}{2} - z_{6}\right) \, \mathbf{a}_{2} + \left(\frac{1}{2} +x_{6}\right) \, \mathbf{a}_{3} & = & y_{6}a \, \mathbf{\hat{x}} + \left(\frac{1}{2} - z_{6}\right)a \, \mathbf{\hat{y}} + \left(\frac{1}{2} +x_{6}\right)a \, \mathbf{\hat{z}} & \left(24d\right) & \mbox{Al II} \\ 
\mathbf{B}_{71} & = & \left(\frac{1}{2} - y_{6}\right) \, \mathbf{a}_{1} + \left(\frac{1}{2} +z_{6}\right) \, \mathbf{a}_{2} + x_{6} \, \mathbf{a}_{3} & = & \left(\frac{1}{2} - y_{6}\right)a \, \mathbf{\hat{x}} + \left(\frac{1}{2} +z_{6}\right)a \, \mathbf{\hat{y}} + x_{6}a \, \mathbf{\hat{z}} & \left(24d\right) & \mbox{Al II} \\ 
\mathbf{B}_{72} & = & \left(\frac{1}{2} +y_{6}\right) \, \mathbf{a}_{1} + z_{6} \, \mathbf{a}_{2} + \left(\frac{1}{2} - x_{6}\right) \, \mathbf{a}_{3} & = & \left(\frac{1}{2} +y_{6}\right)a \, \mathbf{\hat{x}} + z_{6}a \, \mathbf{\hat{y}} + \left(\frac{1}{2} - x_{6}\right)a \, \mathbf{\hat{z}} & \left(24d\right) & \mbox{Al II} \\ 
\mathbf{B}_{73} & = & x_{7} \, \mathbf{a}_{1} + y_{7} \, \mathbf{a}_{2} + z_{7} \, \mathbf{a}_{3} & = & x_{7}a \, \mathbf{\hat{x}} + y_{7}a \, \mathbf{\hat{y}} + z_{7}a \, \mathbf{\hat{z}} & \left(24d\right) & \mbox{Ca V} \\ 
\mathbf{B}_{74} & = & \left(\frac{1}{2} - x_{7}\right) \, \mathbf{a}_{1}-y_{7} \, \mathbf{a}_{2} + \left(\frac{1}{2} +z_{7}\right) \, \mathbf{a}_{3} & = & \left(\frac{1}{2} - x_{7}\right)a \, \mathbf{\hat{x}}-y_{7}a \, \mathbf{\hat{y}} + \left(\frac{1}{2} +z_{7}\right)a \, \mathbf{\hat{z}} & \left(24d\right) & \mbox{Ca V} \\ 
\mathbf{B}_{75} & = & -x_{7} \, \mathbf{a}_{1} + \left(\frac{1}{2} +y_{7}\right) \, \mathbf{a}_{2} + \left(\frac{1}{2} - z_{7}\right) \, \mathbf{a}_{3} & = & -x_{7}a \, \mathbf{\hat{x}} + \left(\frac{1}{2} +y_{7}\right)a \, \mathbf{\hat{y}} + \left(\frac{1}{2} - z_{7}\right)a \, \mathbf{\hat{z}} & \left(24d\right) & \mbox{Ca V} \\ 
\mathbf{B}_{76} & = & \left(\frac{1}{2} +x_{7}\right) \, \mathbf{a}_{1} + \left(\frac{1}{2} - y_{7}\right) \, \mathbf{a}_{2}-z_{7} \, \mathbf{a}_{3} & = & \left(\frac{1}{2} +x_{7}\right)a \, \mathbf{\hat{x}} + \left(\frac{1}{2} - y_{7}\right)a \, \mathbf{\hat{y}}-z_{7}a \, \mathbf{\hat{z}} & \left(24d\right) & \mbox{Ca V} \\ 
\mathbf{B}_{77} & = & z_{7} \, \mathbf{a}_{1} + x_{7} \, \mathbf{a}_{2} + y_{7} \, \mathbf{a}_{3} & = & z_{7}a \, \mathbf{\hat{x}} + x_{7}a \, \mathbf{\hat{y}} + y_{7}a \, \mathbf{\hat{z}} & \left(24d\right) & \mbox{Ca V} \\ 
\mathbf{B}_{78} & = & \left(\frac{1}{2} +z_{7}\right) \, \mathbf{a}_{1} + \left(\frac{1}{2} - x_{7}\right) \, \mathbf{a}_{2}-y_{7} \, \mathbf{a}_{3} & = & \left(\frac{1}{2} +z_{7}\right)a \, \mathbf{\hat{x}} + \left(\frac{1}{2} - x_{7}\right)a \, \mathbf{\hat{y}}-y_{7}a \, \mathbf{\hat{z}} & \left(24d\right) & \mbox{Ca V} \\ 
\mathbf{B}_{79} & = & \left(\frac{1}{2} - z_{7}\right) \, \mathbf{a}_{1}-x_{7} \, \mathbf{a}_{2} + \left(\frac{1}{2} +y_{7}\right) \, \mathbf{a}_{3} & = & \left(\frac{1}{2} - z_{7}\right)a \, \mathbf{\hat{x}}-x_{7}a \, \mathbf{\hat{y}} + \left(\frac{1}{2} +y_{7}\right)a \, \mathbf{\hat{z}} & \left(24d\right) & \mbox{Ca V} \\ 
\mathbf{B}_{80} & = & -z_{7} \, \mathbf{a}_{1} + \left(\frac{1}{2} +x_{7}\right) \, \mathbf{a}_{2} + \left(\frac{1}{2} - y_{7}\right) \, \mathbf{a}_{3} & = & -z_{7}a \, \mathbf{\hat{x}} + \left(\frac{1}{2} +x_{7}\right)a \, \mathbf{\hat{y}} + \left(\frac{1}{2} - y_{7}\right)a \, \mathbf{\hat{z}} & \left(24d\right) & \mbox{Ca V} \\ 
\mathbf{B}_{81} & = & y_{7} \, \mathbf{a}_{1} + z_{7} \, \mathbf{a}_{2} + x_{7} \, \mathbf{a}_{3} & = & y_{7}a \, \mathbf{\hat{x}} + z_{7}a \, \mathbf{\hat{y}} + x_{7}a \, \mathbf{\hat{z}} & \left(24d\right) & \mbox{Ca V} \\ 
\mathbf{B}_{82} & = & -y_{7} \, \mathbf{a}_{1} + \left(\frac{1}{2} +z_{7}\right) \, \mathbf{a}_{2} + \left(\frac{1}{2} - x_{7}\right) \, \mathbf{a}_{3} & = & -y_{7}a \, \mathbf{\hat{x}} + \left(\frac{1}{2} +z_{7}\right)a \, \mathbf{\hat{y}} + \left(\frac{1}{2} - x_{7}\right)a \, \mathbf{\hat{z}} & \left(24d\right) & \mbox{Ca V} \\ 
\mathbf{B}_{83} & = & \left(\frac{1}{2} +y_{7}\right) \, \mathbf{a}_{1} + \left(\frac{1}{2} - z_{7}\right) \, \mathbf{a}_{2}-x_{7} \, \mathbf{a}_{3} & = & \left(\frac{1}{2} +y_{7}\right)a \, \mathbf{\hat{x}} + \left(\frac{1}{2} - z_{7}\right)a \, \mathbf{\hat{y}}-x_{7}a \, \mathbf{\hat{z}} & \left(24d\right) & \mbox{Ca V} \\ 
\mathbf{B}_{84} & = & \left(\frac{1}{2} - y_{7}\right) \, \mathbf{a}_{1}-z_{7} \, \mathbf{a}_{2} + \left(\frac{1}{2} +x_{7}\right) \, \mathbf{a}_{3} & = & \left(\frac{1}{2} - y_{7}\right)a \, \mathbf{\hat{x}}-z_{7}a \, \mathbf{\hat{y}} + \left(\frac{1}{2} +x_{7}\right)a \, \mathbf{\hat{z}} & \left(24d\right) & \mbox{Ca V} \\ 
\mathbf{B}_{85} & = & -x_{7} \, \mathbf{a}_{1}-y_{7} \, \mathbf{a}_{2}-z_{7} \, \mathbf{a}_{3} & = & -x_{7}a \, \mathbf{\hat{x}}-y_{7}a \, \mathbf{\hat{y}}-z_{7}a \, \mathbf{\hat{z}} & \left(24d\right) & \mbox{Ca V} \\ 
\mathbf{B}_{86} & = & \left(\frac{1}{2} +x_{7}\right) \, \mathbf{a}_{1} + y_{7} \, \mathbf{a}_{2} + \left(\frac{1}{2} - z_{7}\right) \, \mathbf{a}_{3} & = & \left(\frac{1}{2} +x_{7}\right)a \, \mathbf{\hat{x}} + y_{7}a \, \mathbf{\hat{y}} + \left(\frac{1}{2} - z_{7}\right)a \, \mathbf{\hat{z}} & \left(24d\right) & \mbox{Ca V} \\ 
\mathbf{B}_{87} & = & x_{7} \, \mathbf{a}_{1} + \left(\frac{1}{2} - y_{7}\right) \, \mathbf{a}_{2} + \left(\frac{1}{2} +z_{7}\right) \, \mathbf{a}_{3} & = & x_{7}a \, \mathbf{\hat{x}} + \left(\frac{1}{2} - y_{7}\right)a \, \mathbf{\hat{y}} + \left(\frac{1}{2} +z_{7}\right)a \, \mathbf{\hat{z}} & \left(24d\right) & \mbox{Ca V} \\ 
\mathbf{B}_{88} & = & \left(\frac{1}{2} - x_{7}\right) \, \mathbf{a}_{1} + \left(\frac{1}{2} +y_{7}\right) \, \mathbf{a}_{2} + z_{7} \, \mathbf{a}_{3} & = & \left(\frac{1}{2} - x_{7}\right)a \, \mathbf{\hat{x}} + \left(\frac{1}{2} +y_{7}\right)a \, \mathbf{\hat{y}} + z_{7}a \, \mathbf{\hat{z}} & \left(24d\right) & \mbox{Ca V} \\ 
\mathbf{B}_{89} & = & -z_{7} \, \mathbf{a}_{1}-x_{7} \, \mathbf{a}_{2}-y_{7} \, \mathbf{a}_{3} & = & -z_{7}a \, \mathbf{\hat{x}}-x_{7}a \, \mathbf{\hat{y}}-y_{7}a \, \mathbf{\hat{z}} & \left(24d\right) & \mbox{Ca V} \\ 
\mathbf{B}_{90} & = & \left(\frac{1}{2} - z_{7}\right) \, \mathbf{a}_{1} + \left(\frac{1}{2} +x_{7}\right) \, \mathbf{a}_{2} + y_{7} \, \mathbf{a}_{3} & = & \left(\frac{1}{2} - z_{7}\right)a \, \mathbf{\hat{x}} + \left(\frac{1}{2} +x_{7}\right)a \, \mathbf{\hat{y}} + y_{7}a \, \mathbf{\hat{z}} & \left(24d\right) & \mbox{Ca V} \\ 
\mathbf{B}_{91} & = & \left(\frac{1}{2} +z_{7}\right) \, \mathbf{a}_{1} + x_{7} \, \mathbf{a}_{2} + \left(\frac{1}{2} - y_{7}\right) \, \mathbf{a}_{3} & = & \left(\frac{1}{2} +z_{7}\right)a \, \mathbf{\hat{x}} + x_{7}a \, \mathbf{\hat{y}} + \left(\frac{1}{2} - y_{7}\right)a \, \mathbf{\hat{z}} & \left(24d\right) & \mbox{Ca V} \\ 
\mathbf{B}_{92} & = & z_{7} \, \mathbf{a}_{1} + \left(\frac{1}{2} - x_{7}\right) \, \mathbf{a}_{2} + \left(\frac{1}{2} +y_{7}\right) \, \mathbf{a}_{3} & = & z_{7}a \, \mathbf{\hat{x}} + \left(\frac{1}{2} - x_{7}\right)a \, \mathbf{\hat{y}} + \left(\frac{1}{2} +y_{7}\right)a \, \mathbf{\hat{z}} & \left(24d\right) & \mbox{Ca V} \\ 
\mathbf{B}_{93} & = & -y_{7} \, \mathbf{a}_{1}-z_{7} \, \mathbf{a}_{2}-x_{7} \, \mathbf{a}_{3} & = & -y_{7}a \, \mathbf{\hat{x}}-z_{7}a \, \mathbf{\hat{y}}-x_{7}a \, \mathbf{\hat{z}} & \left(24d\right) & \mbox{Ca V} \\ 
\mathbf{B}_{94} & = & y_{7} \, \mathbf{a}_{1} + \left(\frac{1}{2} - z_{7}\right) \, \mathbf{a}_{2} + \left(\frac{1}{2} +x_{7}\right) \, \mathbf{a}_{3} & = & y_{7}a \, \mathbf{\hat{x}} + \left(\frac{1}{2} - z_{7}\right)a \, \mathbf{\hat{y}} + \left(\frac{1}{2} +x_{7}\right)a \, \mathbf{\hat{z}} & \left(24d\right) & \mbox{Ca V} \\ 
\mathbf{B}_{95} & = & \left(\frac{1}{2} - y_{7}\right) \, \mathbf{a}_{1} + \left(\frac{1}{2} +z_{7}\right) \, \mathbf{a}_{2} + x_{7} \, \mathbf{a}_{3} & = & \left(\frac{1}{2} - y_{7}\right)a \, \mathbf{\hat{x}} + \left(\frac{1}{2} +z_{7}\right)a \, \mathbf{\hat{y}} + x_{7}a \, \mathbf{\hat{z}} & \left(24d\right) & \mbox{Ca V} \\ 
\mathbf{B}_{96} & = & \left(\frac{1}{2} +y_{7}\right) \, \mathbf{a}_{1} + z_{7} \, \mathbf{a}_{2} + \left(\frac{1}{2} - x_{7}\right) \, \mathbf{a}_{3} & = & \left(\frac{1}{2} +y_{7}\right)a \, \mathbf{\hat{x}} + z_{7}a \, \mathbf{\hat{y}} + \left(\frac{1}{2} - x_{7}\right)a \, \mathbf{\hat{z}} & \left(24d\right) & \mbox{Ca V} \\ 
\mathbf{B}_{97} & = & x_{8} \, \mathbf{a}_{1} + y_{8} \, \mathbf{a}_{2} + z_{8} \, \mathbf{a}_{3} & = & x_{8}a \, \mathbf{\hat{x}} + y_{8}a \, \mathbf{\hat{y}} + z_{8}a \, \mathbf{\hat{z}} & \left(24d\right) & \mbox{Ca VI} \\ 
\mathbf{B}_{98} & = & \left(\frac{1}{2} - x_{8}\right) \, \mathbf{a}_{1}-y_{8} \, \mathbf{a}_{2} + \left(\frac{1}{2} +z_{8}\right) \, \mathbf{a}_{3} & = & \left(\frac{1}{2} - x_{8}\right)a \, \mathbf{\hat{x}}-y_{8}a \, \mathbf{\hat{y}} + \left(\frac{1}{2} +z_{8}\right)a \, \mathbf{\hat{z}} & \left(24d\right) & \mbox{Ca VI} \\ 
\mathbf{B}_{99} & = & -x_{8} \, \mathbf{a}_{1} + \left(\frac{1}{2} +y_{8}\right) \, \mathbf{a}_{2} + \left(\frac{1}{2} - z_{8}\right) \, \mathbf{a}_{3} & = & -x_{8}a \, \mathbf{\hat{x}} + \left(\frac{1}{2} +y_{8}\right)a \, \mathbf{\hat{y}} + \left(\frac{1}{2} - z_{8}\right)a \, \mathbf{\hat{z}} & \left(24d\right) & \mbox{Ca VI} \\ 
\mathbf{B}_{100} & = & \left(\frac{1}{2} +x_{8}\right) \, \mathbf{a}_{1} + \left(\frac{1}{2} - y_{8}\right) \, \mathbf{a}_{2}-z_{8} \, \mathbf{a}_{3} & = & \left(\frac{1}{2} +x_{8}\right)a \, \mathbf{\hat{x}} + \left(\frac{1}{2} - y_{8}\right)a \, \mathbf{\hat{y}}-z_{8}a \, \mathbf{\hat{z}} & \left(24d\right) & \mbox{Ca VI} \\ 
\mathbf{B}_{101} & = & z_{8} \, \mathbf{a}_{1} + x_{8} \, \mathbf{a}_{2} + y_{8} \, \mathbf{a}_{3} & = & z_{8}a \, \mathbf{\hat{x}} + x_{8}a \, \mathbf{\hat{y}} + y_{8}a \, \mathbf{\hat{z}} & \left(24d\right) & \mbox{Ca VI} \\ 
\mathbf{B}_{102} & = & \left(\frac{1}{2} +z_{8}\right) \, \mathbf{a}_{1} + \left(\frac{1}{2} - x_{8}\right) \, \mathbf{a}_{2}-y_{8} \, \mathbf{a}_{3} & = & \left(\frac{1}{2} +z_{8}\right)a \, \mathbf{\hat{x}} + \left(\frac{1}{2} - x_{8}\right)a \, \mathbf{\hat{y}}-y_{8}a \, \mathbf{\hat{z}} & \left(24d\right) & \mbox{Ca VI} \\ 
\mathbf{B}_{103} & = & \left(\frac{1}{2} - z_{8}\right) \, \mathbf{a}_{1}-x_{8} \, \mathbf{a}_{2} + \left(\frac{1}{2} +y_{8}\right) \, \mathbf{a}_{3} & = & \left(\frac{1}{2} - z_{8}\right)a \, \mathbf{\hat{x}}-x_{8}a \, \mathbf{\hat{y}} + \left(\frac{1}{2} +y_{8}\right)a \, \mathbf{\hat{z}} & \left(24d\right) & \mbox{Ca VI} \\ 
\mathbf{B}_{104} & = & -z_{8} \, \mathbf{a}_{1} + \left(\frac{1}{2} +x_{8}\right) \, \mathbf{a}_{2} + \left(\frac{1}{2} - y_{8}\right) \, \mathbf{a}_{3} & = & -z_{8}a \, \mathbf{\hat{x}} + \left(\frac{1}{2} +x_{8}\right)a \, \mathbf{\hat{y}} + \left(\frac{1}{2} - y_{8}\right)a \, \mathbf{\hat{z}} & \left(24d\right) & \mbox{Ca VI} \\ 
\mathbf{B}_{105} & = & y_{8} \, \mathbf{a}_{1} + z_{8} \, \mathbf{a}_{2} + x_{8} \, \mathbf{a}_{3} & = & y_{8}a \, \mathbf{\hat{x}} + z_{8}a \, \mathbf{\hat{y}} + x_{8}a \, \mathbf{\hat{z}} & \left(24d\right) & \mbox{Ca VI} \\ 
\mathbf{B}_{106} & = & -y_{8} \, \mathbf{a}_{1} + \left(\frac{1}{2} +z_{8}\right) \, \mathbf{a}_{2} + \left(\frac{1}{2} - x_{8}\right) \, \mathbf{a}_{3} & = & -y_{8}a \, \mathbf{\hat{x}} + \left(\frac{1}{2} +z_{8}\right)a \, \mathbf{\hat{y}} + \left(\frac{1}{2} - x_{8}\right)a \, \mathbf{\hat{z}} & \left(24d\right) & \mbox{Ca VI} \\ 
\mathbf{B}_{107} & = & \left(\frac{1}{2} +y_{8}\right) \, \mathbf{a}_{1} + \left(\frac{1}{2} - z_{8}\right) \, \mathbf{a}_{2}-x_{8} \, \mathbf{a}_{3} & = & \left(\frac{1}{2} +y_{8}\right)a \, \mathbf{\hat{x}} + \left(\frac{1}{2} - z_{8}\right)a \, \mathbf{\hat{y}}-x_{8}a \, \mathbf{\hat{z}} & \left(24d\right) & \mbox{Ca VI} \\ 
\mathbf{B}_{108} & = & \left(\frac{1}{2} - y_{8}\right) \, \mathbf{a}_{1}-z_{8} \, \mathbf{a}_{2} + \left(\frac{1}{2} +x_{8}\right) \, \mathbf{a}_{3} & = & \left(\frac{1}{2} - y_{8}\right)a \, \mathbf{\hat{x}}-z_{8}a \, \mathbf{\hat{y}} + \left(\frac{1}{2} +x_{8}\right)a \, \mathbf{\hat{z}} & \left(24d\right) & \mbox{Ca VI} \\ 
\mathbf{B}_{109} & = & -x_{8} \, \mathbf{a}_{1}-y_{8} \, \mathbf{a}_{2}-z_{8} \, \mathbf{a}_{3} & = & -x_{8}a \, \mathbf{\hat{x}}-y_{8}a \, \mathbf{\hat{y}}-z_{8}a \, \mathbf{\hat{z}} & \left(24d\right) & \mbox{Ca VI} \\ 
\mathbf{B}_{110} & = & \left(\frac{1}{2} +x_{8}\right) \, \mathbf{a}_{1} + y_{8} \, \mathbf{a}_{2} + \left(\frac{1}{2} - z_{8}\right) \, \mathbf{a}_{3} & = & \left(\frac{1}{2} +x_{8}\right)a \, \mathbf{\hat{x}} + y_{8}a \, \mathbf{\hat{y}} + \left(\frac{1}{2} - z_{8}\right)a \, \mathbf{\hat{z}} & \left(24d\right) & \mbox{Ca VI} \\ 
\mathbf{B}_{111} & = & x_{8} \, \mathbf{a}_{1} + \left(\frac{1}{2} - y_{8}\right) \, \mathbf{a}_{2} + \left(\frac{1}{2} +z_{8}\right) \, \mathbf{a}_{3} & = & x_{8}a \, \mathbf{\hat{x}} + \left(\frac{1}{2} - y_{8}\right)a \, \mathbf{\hat{y}} + \left(\frac{1}{2} +z_{8}\right)a \, \mathbf{\hat{z}} & \left(24d\right) & \mbox{Ca VI} \\ 
\mathbf{B}_{112} & = & \left(\frac{1}{2} - x_{8}\right) \, \mathbf{a}_{1} + \left(\frac{1}{2} +y_{8}\right) \, \mathbf{a}_{2} + z_{8} \, \mathbf{a}_{3} & = & \left(\frac{1}{2} - x_{8}\right)a \, \mathbf{\hat{x}} + \left(\frac{1}{2} +y_{8}\right)a \, \mathbf{\hat{y}} + z_{8}a \, \mathbf{\hat{z}} & \left(24d\right) & \mbox{Ca VI} \\ 
\mathbf{B}_{113} & = & -z_{8} \, \mathbf{a}_{1}-x_{8} \, \mathbf{a}_{2}-y_{8} \, \mathbf{a}_{3} & = & -z_{8}a \, \mathbf{\hat{x}}-x_{8}a \, \mathbf{\hat{y}}-y_{8}a \, \mathbf{\hat{z}} & \left(24d\right) & \mbox{Ca VI} \\ 
\mathbf{B}_{114} & = & \left(\frac{1}{2} - z_{8}\right) \, \mathbf{a}_{1} + \left(\frac{1}{2} +x_{8}\right) \, \mathbf{a}_{2} + y_{8} \, \mathbf{a}_{3} & = & \left(\frac{1}{2} - z_{8}\right)a \, \mathbf{\hat{x}} + \left(\frac{1}{2} +x_{8}\right)a \, \mathbf{\hat{y}} + y_{8}a \, \mathbf{\hat{z}} & \left(24d\right) & \mbox{Ca VI} \\ 
\mathbf{B}_{115} & = & \left(\frac{1}{2} +z_{8}\right) \, \mathbf{a}_{1} + x_{8} \, \mathbf{a}_{2} + \left(\frac{1}{2} - y_{8}\right) \, \mathbf{a}_{3} & = & \left(\frac{1}{2} +z_{8}\right)a \, \mathbf{\hat{x}} + x_{8}a \, \mathbf{\hat{y}} + \left(\frac{1}{2} - y_{8}\right)a \, \mathbf{\hat{z}} & \left(24d\right) & \mbox{Ca VI} \\ 
\mathbf{B}_{116} & = & z_{8} \, \mathbf{a}_{1} + \left(\frac{1}{2} - x_{8}\right) \, \mathbf{a}_{2} + \left(\frac{1}{2} +y_{8}\right) \, \mathbf{a}_{3} & = & z_{8}a \, \mathbf{\hat{x}} + \left(\frac{1}{2} - x_{8}\right)a \, \mathbf{\hat{y}} + \left(\frac{1}{2} +y_{8}\right)a \, \mathbf{\hat{z}} & \left(24d\right) & \mbox{Ca VI} \\ 
\mathbf{B}_{117} & = & -y_{8} \, \mathbf{a}_{1}-z_{8} \, \mathbf{a}_{2}-x_{8} \, \mathbf{a}_{3} & = & -y_{8}a \, \mathbf{\hat{x}}-z_{8}a \, \mathbf{\hat{y}}-x_{8}a \, \mathbf{\hat{z}} & \left(24d\right) & \mbox{Ca VI} \\ 
\mathbf{B}_{118} & = & y_{8} \, \mathbf{a}_{1} + \left(\frac{1}{2} - z_{8}\right) \, \mathbf{a}_{2} + \left(\frac{1}{2} +x_{8}\right) \, \mathbf{a}_{3} & = & y_{8}a \, \mathbf{\hat{x}} + \left(\frac{1}{2} - z_{8}\right)a \, \mathbf{\hat{y}} + \left(\frac{1}{2} +x_{8}\right)a \, \mathbf{\hat{z}} & \left(24d\right) & \mbox{Ca VI} \\ 
\mathbf{B}_{119} & = & \left(\frac{1}{2} - y_{8}\right) \, \mathbf{a}_{1} + \left(\frac{1}{2} +z_{8}\right) \, \mathbf{a}_{2} + x_{8} \, \mathbf{a}_{3} & = & \left(\frac{1}{2} - y_{8}\right)a \, \mathbf{\hat{x}} + \left(\frac{1}{2} +z_{8}\right)a \, \mathbf{\hat{y}} + x_{8}a \, \mathbf{\hat{z}} & \left(24d\right) & \mbox{Ca VI} \\ 
\mathbf{B}_{120} & = & \left(\frac{1}{2} +y_{8}\right) \, \mathbf{a}_{1} + z_{8} \, \mathbf{a}_{2} + \left(\frac{1}{2} - x_{8}\right) \, \mathbf{a}_{3} & = & \left(\frac{1}{2} +y_{8}\right)a \, \mathbf{\hat{x}} + z_{8}a \, \mathbf{\hat{y}} + \left(\frac{1}{2} - x_{8}\right)a \, \mathbf{\hat{z}} & \left(24d\right) & \mbox{Ca VI} \\ 
\mathbf{B}_{121} & = & x_{9} \, \mathbf{a}_{1} + y_{9} \, \mathbf{a}_{2} + z_{9} \, \mathbf{a}_{3} & = & x_{9}a \, \mathbf{\hat{x}} + y_{9}a \, \mathbf{\hat{y}} + z_{9}a \, \mathbf{\hat{z}} & \left(24d\right) & \mbox{O I} \\ 
\mathbf{B}_{122} & = & \left(\frac{1}{2} - x_{9}\right) \, \mathbf{a}_{1}-y_{9} \, \mathbf{a}_{2} + \left(\frac{1}{2} +z_{9}\right) \, \mathbf{a}_{3} & = & \left(\frac{1}{2} - x_{9}\right)a \, \mathbf{\hat{x}}-y_{9}a \, \mathbf{\hat{y}} + \left(\frac{1}{2} +z_{9}\right)a \, \mathbf{\hat{z}} & \left(24d\right) & \mbox{O I} \\ 
\mathbf{B}_{123} & = & -x_{9} \, \mathbf{a}_{1} + \left(\frac{1}{2} +y_{9}\right) \, \mathbf{a}_{2} + \left(\frac{1}{2} - z_{9}\right) \, \mathbf{a}_{3} & = & -x_{9}a \, \mathbf{\hat{x}} + \left(\frac{1}{2} +y_{9}\right)a \, \mathbf{\hat{y}} + \left(\frac{1}{2} - z_{9}\right)a \, \mathbf{\hat{z}} & \left(24d\right) & \mbox{O I} \\ 
\mathbf{B}_{124} & = & \left(\frac{1}{2} +x_{9}\right) \, \mathbf{a}_{1} + \left(\frac{1}{2} - y_{9}\right) \, \mathbf{a}_{2}-z_{9} \, \mathbf{a}_{3} & = & \left(\frac{1}{2} +x_{9}\right)a \, \mathbf{\hat{x}} + \left(\frac{1}{2} - y_{9}\right)a \, \mathbf{\hat{y}}-z_{9}a \, \mathbf{\hat{z}} & \left(24d\right) & \mbox{O I} \\ 
\mathbf{B}_{125} & = & z_{9} \, \mathbf{a}_{1} + x_{9} \, \mathbf{a}_{2} + y_{9} \, \mathbf{a}_{3} & = & z_{9}a \, \mathbf{\hat{x}} + x_{9}a \, \mathbf{\hat{y}} + y_{9}a \, \mathbf{\hat{z}} & \left(24d\right) & \mbox{O I} \\ 
\mathbf{B}_{126} & = & \left(\frac{1}{2} +z_{9}\right) \, \mathbf{a}_{1} + \left(\frac{1}{2} - x_{9}\right) \, \mathbf{a}_{2}-y_{9} \, \mathbf{a}_{3} & = & \left(\frac{1}{2} +z_{9}\right)a \, \mathbf{\hat{x}} + \left(\frac{1}{2} - x_{9}\right)a \, \mathbf{\hat{y}}-y_{9}a \, \mathbf{\hat{z}} & \left(24d\right) & \mbox{O I} \\ 
\mathbf{B}_{127} & = & \left(\frac{1}{2} - z_{9}\right) \, \mathbf{a}_{1}-x_{9} \, \mathbf{a}_{2} + \left(\frac{1}{2} +y_{9}\right) \, \mathbf{a}_{3} & = & \left(\frac{1}{2} - z_{9}\right)a \, \mathbf{\hat{x}}-x_{9}a \, \mathbf{\hat{y}} + \left(\frac{1}{2} +y_{9}\right)a \, \mathbf{\hat{z}} & \left(24d\right) & \mbox{O I} \\ 
\mathbf{B}_{128} & = & -z_{9} \, \mathbf{a}_{1} + \left(\frac{1}{2} +x_{9}\right) \, \mathbf{a}_{2} + \left(\frac{1}{2} - y_{9}\right) \, \mathbf{a}_{3} & = & -z_{9}a \, \mathbf{\hat{x}} + \left(\frac{1}{2} +x_{9}\right)a \, \mathbf{\hat{y}} + \left(\frac{1}{2} - y_{9}\right)a \, \mathbf{\hat{z}} & \left(24d\right) & \mbox{O I} \\ 
\mathbf{B}_{129} & = & y_{9} \, \mathbf{a}_{1} + z_{9} \, \mathbf{a}_{2} + x_{9} \, \mathbf{a}_{3} & = & y_{9}a \, \mathbf{\hat{x}} + z_{9}a \, \mathbf{\hat{y}} + x_{9}a \, \mathbf{\hat{z}} & \left(24d\right) & \mbox{O I} \\ 
\mathbf{B}_{130} & = & -y_{9} \, \mathbf{a}_{1} + \left(\frac{1}{2} +z_{9}\right) \, \mathbf{a}_{2} + \left(\frac{1}{2} - x_{9}\right) \, \mathbf{a}_{3} & = & -y_{9}a \, \mathbf{\hat{x}} + \left(\frac{1}{2} +z_{9}\right)a \, \mathbf{\hat{y}} + \left(\frac{1}{2} - x_{9}\right)a \, \mathbf{\hat{z}} & \left(24d\right) & \mbox{O I} \\ 
\mathbf{B}_{131} & = & \left(\frac{1}{2} +y_{9}\right) \, \mathbf{a}_{1} + \left(\frac{1}{2} - z_{9}\right) \, \mathbf{a}_{2}-x_{9} \, \mathbf{a}_{3} & = & \left(\frac{1}{2} +y_{9}\right)a \, \mathbf{\hat{x}} + \left(\frac{1}{2} - z_{9}\right)a \, \mathbf{\hat{y}}-x_{9}a \, \mathbf{\hat{z}} & \left(24d\right) & \mbox{O I} \\ 
\mathbf{B}_{132} & = & \left(\frac{1}{2} - y_{9}\right) \, \mathbf{a}_{1}-z_{9} \, \mathbf{a}_{2} + \left(\frac{1}{2} +x_{9}\right) \, \mathbf{a}_{3} & = & \left(\frac{1}{2} - y_{9}\right)a \, \mathbf{\hat{x}}-z_{9}a \, \mathbf{\hat{y}} + \left(\frac{1}{2} +x_{9}\right)a \, \mathbf{\hat{z}} & \left(24d\right) & \mbox{O I} \\ 
\mathbf{B}_{133} & = & -x_{9} \, \mathbf{a}_{1}-y_{9} \, \mathbf{a}_{2}-z_{9} \, \mathbf{a}_{3} & = & -x_{9}a \, \mathbf{\hat{x}}-y_{9}a \, \mathbf{\hat{y}}-z_{9}a \, \mathbf{\hat{z}} & \left(24d\right) & \mbox{O I} \\ 
\mathbf{B}_{134} & = & \left(\frac{1}{2} +x_{9}\right) \, \mathbf{a}_{1} + y_{9} \, \mathbf{a}_{2} + \left(\frac{1}{2} - z_{9}\right) \, \mathbf{a}_{3} & = & \left(\frac{1}{2} +x_{9}\right)a \, \mathbf{\hat{x}} + y_{9}a \, \mathbf{\hat{y}} + \left(\frac{1}{2} - z_{9}\right)a \, \mathbf{\hat{z}} & \left(24d\right) & \mbox{O I} \\ 
\mathbf{B}_{135} & = & x_{9} \, \mathbf{a}_{1} + \left(\frac{1}{2} - y_{9}\right) \, \mathbf{a}_{2} + \left(\frac{1}{2} +z_{9}\right) \, \mathbf{a}_{3} & = & x_{9}a \, \mathbf{\hat{x}} + \left(\frac{1}{2} - y_{9}\right)a \, \mathbf{\hat{y}} + \left(\frac{1}{2} +z_{9}\right)a \, \mathbf{\hat{z}} & \left(24d\right) & \mbox{O I} \\ 
\mathbf{B}_{136} & = & \left(\frac{1}{2} - x_{9}\right) \, \mathbf{a}_{1} + \left(\frac{1}{2} +y_{9}\right) \, \mathbf{a}_{2} + z_{9} \, \mathbf{a}_{3} & = & \left(\frac{1}{2} - x_{9}\right)a \, \mathbf{\hat{x}} + \left(\frac{1}{2} +y_{9}\right)a \, \mathbf{\hat{y}} + z_{9}a \, \mathbf{\hat{z}} & \left(24d\right) & \mbox{O I} \\ 
\mathbf{B}_{137} & = & -z_{9} \, \mathbf{a}_{1}-x_{9} \, \mathbf{a}_{2}-y_{9} \, \mathbf{a}_{3} & = & -z_{9}a \, \mathbf{\hat{x}}-x_{9}a \, \mathbf{\hat{y}}-y_{9}a \, \mathbf{\hat{z}} & \left(24d\right) & \mbox{O I} \\ 
\mathbf{B}_{138} & = & \left(\frac{1}{2} - z_{9}\right) \, \mathbf{a}_{1} + \left(\frac{1}{2} +x_{9}\right) \, \mathbf{a}_{2} + y_{9} \, \mathbf{a}_{3} & = & \left(\frac{1}{2} - z_{9}\right)a \, \mathbf{\hat{x}} + \left(\frac{1}{2} +x_{9}\right)a \, \mathbf{\hat{y}} + y_{9}a \, \mathbf{\hat{z}} & \left(24d\right) & \mbox{O I} \\ 
\mathbf{B}_{139} & = & \left(\frac{1}{2} +z_{9}\right) \, \mathbf{a}_{1} + x_{9} \, \mathbf{a}_{2} + \left(\frac{1}{2} - y_{9}\right) \, \mathbf{a}_{3} & = & \left(\frac{1}{2} +z_{9}\right)a \, \mathbf{\hat{x}} + x_{9}a \, \mathbf{\hat{y}} + \left(\frac{1}{2} - y_{9}\right)a \, \mathbf{\hat{z}} & \left(24d\right) & \mbox{O I} \\ 
\mathbf{B}_{140} & = & z_{9} \, \mathbf{a}_{1} + \left(\frac{1}{2} - x_{9}\right) \, \mathbf{a}_{2} + \left(\frac{1}{2} +y_{9}\right) \, \mathbf{a}_{3} & = & z_{9}a \, \mathbf{\hat{x}} + \left(\frac{1}{2} - x_{9}\right)a \, \mathbf{\hat{y}} + \left(\frac{1}{2} +y_{9}\right)a \, \mathbf{\hat{z}} & \left(24d\right) & \mbox{O I} \\ 
\mathbf{B}_{141} & = & -y_{9} \, \mathbf{a}_{1}-z_{9} \, \mathbf{a}_{2}-x_{9} \, \mathbf{a}_{3} & = & -y_{9}a \, \mathbf{\hat{x}}-z_{9}a \, \mathbf{\hat{y}}-x_{9}a \, \mathbf{\hat{z}} & \left(24d\right) & \mbox{O I} \\ 
\mathbf{B}_{142} & = & y_{9} \, \mathbf{a}_{1} + \left(\frac{1}{2} - z_{9}\right) \, \mathbf{a}_{2} + \left(\frac{1}{2} +x_{9}\right) \, \mathbf{a}_{3} & = & y_{9}a \, \mathbf{\hat{x}} + \left(\frac{1}{2} - z_{9}\right)a \, \mathbf{\hat{y}} + \left(\frac{1}{2} +x_{9}\right)a \, \mathbf{\hat{z}} & \left(24d\right) & \mbox{O I} \\ 
\mathbf{B}_{143} & = & \left(\frac{1}{2} - y_{9}\right) \, \mathbf{a}_{1} + \left(\frac{1}{2} +z_{9}\right) \, \mathbf{a}_{2} + x_{9} \, \mathbf{a}_{3} & = & \left(\frac{1}{2} - y_{9}\right)a \, \mathbf{\hat{x}} + \left(\frac{1}{2} +z_{9}\right)a \, \mathbf{\hat{y}} + x_{9}a \, \mathbf{\hat{z}} & \left(24d\right) & \mbox{O I} \\ 
\mathbf{B}_{144} & = & \left(\frac{1}{2} +y_{9}\right) \, \mathbf{a}_{1} + z_{9} \, \mathbf{a}_{2} + \left(\frac{1}{2} - x_{9}\right) \, \mathbf{a}_{3} & = & \left(\frac{1}{2} +y_{9}\right)a \, \mathbf{\hat{x}} + z_{9}a \, \mathbf{\hat{y}} + \left(\frac{1}{2} - x_{9}\right)a \, \mathbf{\hat{z}} & \left(24d\right) & \mbox{O I} \\ 
\mathbf{B}_{145} & = & x_{10} \, \mathbf{a}_{1} + y_{10} \, \mathbf{a}_{2} + z_{10} \, \mathbf{a}_{3} & = & x_{10}a \, \mathbf{\hat{x}} + y_{10}a \, \mathbf{\hat{y}} + z_{10}a \, \mathbf{\hat{z}} & \left(24d\right) & \mbox{O II} \\ 
\mathbf{B}_{146} & = & \left(\frac{1}{2} - x_{10}\right) \, \mathbf{a}_{1}-y_{10} \, \mathbf{a}_{2} + \left(\frac{1}{2} +z_{10}\right) \, \mathbf{a}_{3} & = & \left(\frac{1}{2} - x_{10}\right)a \, \mathbf{\hat{x}}-y_{10}a \, \mathbf{\hat{y}} + \left(\frac{1}{2} +z_{10}\right)a \, \mathbf{\hat{z}} & \left(24d\right) & \mbox{O II} \\ 
\mathbf{B}_{147} & = & -x_{10} \, \mathbf{a}_{1} + \left(\frac{1}{2} +y_{10}\right) \, \mathbf{a}_{2} + \left(\frac{1}{2} - z_{10}\right) \, \mathbf{a}_{3} & = & -x_{10}a \, \mathbf{\hat{x}} + \left(\frac{1}{2} +y_{10}\right)a \, \mathbf{\hat{y}} + \left(\frac{1}{2} - z_{10}\right)a \, \mathbf{\hat{z}} & \left(24d\right) & \mbox{O II} \\ 
\mathbf{B}_{148} & = & \left(\frac{1}{2} +x_{10}\right) \, \mathbf{a}_{1} + \left(\frac{1}{2} - y_{10}\right) \, \mathbf{a}_{2}-z_{10} \, \mathbf{a}_{3} & = & \left(\frac{1}{2} +x_{10}\right)a \, \mathbf{\hat{x}} + \left(\frac{1}{2} - y_{10}\right)a \, \mathbf{\hat{y}}-z_{10}a \, \mathbf{\hat{z}} & \left(24d\right) & \mbox{O II} \\ 
\mathbf{B}_{149} & = & z_{10} \, \mathbf{a}_{1} + x_{10} \, \mathbf{a}_{2} + y_{10} \, \mathbf{a}_{3} & = & z_{10}a \, \mathbf{\hat{x}} + x_{10}a \, \mathbf{\hat{y}} + y_{10}a \, \mathbf{\hat{z}} & \left(24d\right) & \mbox{O II} \\ 
\mathbf{B}_{150} & = & \left(\frac{1}{2} +z_{10}\right) \, \mathbf{a}_{1} + \left(\frac{1}{2} - x_{10}\right) \, \mathbf{a}_{2}-y_{10} \, \mathbf{a}_{3} & = & \left(\frac{1}{2} +z_{10}\right)a \, \mathbf{\hat{x}} + \left(\frac{1}{2} - x_{10}\right)a \, \mathbf{\hat{y}}-y_{10}a \, \mathbf{\hat{z}} & \left(24d\right) & \mbox{O II} \\ 
\mathbf{B}_{151} & = & \left(\frac{1}{2} - z_{10}\right) \, \mathbf{a}_{1}-x_{10} \, \mathbf{a}_{2} + \left(\frac{1}{2} +y_{10}\right) \, \mathbf{a}_{3} & = & \left(\frac{1}{2} - z_{10}\right)a \, \mathbf{\hat{x}}-x_{10}a \, \mathbf{\hat{y}} + \left(\frac{1}{2} +y_{10}\right)a \, \mathbf{\hat{z}} & \left(24d\right) & \mbox{O II} \\ 
\mathbf{B}_{152} & = & -z_{10} \, \mathbf{a}_{1} + \left(\frac{1}{2} +x_{10}\right) \, \mathbf{a}_{2} + \left(\frac{1}{2} - y_{10}\right) \, \mathbf{a}_{3} & = & -z_{10}a \, \mathbf{\hat{x}} + \left(\frac{1}{2} +x_{10}\right)a \, \mathbf{\hat{y}} + \left(\frac{1}{2} - y_{10}\right)a \, \mathbf{\hat{z}} & \left(24d\right) & \mbox{O II} \\ 
\mathbf{B}_{153} & = & y_{10} \, \mathbf{a}_{1} + z_{10} \, \mathbf{a}_{2} + x_{10} \, \mathbf{a}_{3} & = & y_{10}a \, \mathbf{\hat{x}} + z_{10}a \, \mathbf{\hat{y}} + x_{10}a \, \mathbf{\hat{z}} & \left(24d\right) & \mbox{O II} \\ 
\mathbf{B}_{154} & = & -y_{10} \, \mathbf{a}_{1} + \left(\frac{1}{2} +z_{10}\right) \, \mathbf{a}_{2} + \left(\frac{1}{2} - x_{10}\right) \, \mathbf{a}_{3} & = & -y_{10}a \, \mathbf{\hat{x}} + \left(\frac{1}{2} +z_{10}\right)a \, \mathbf{\hat{y}} + \left(\frac{1}{2} - x_{10}\right)a \, \mathbf{\hat{z}} & \left(24d\right) & \mbox{O II} \\ 
\mathbf{B}_{155} & = & \left(\frac{1}{2} +y_{10}\right) \, \mathbf{a}_{1} + \left(\frac{1}{2} - z_{10}\right) \, \mathbf{a}_{2}-x_{10} \, \mathbf{a}_{3} & = & \left(\frac{1}{2} +y_{10}\right)a \, \mathbf{\hat{x}} + \left(\frac{1}{2} - z_{10}\right)a \, \mathbf{\hat{y}}-x_{10}a \, \mathbf{\hat{z}} & \left(24d\right) & \mbox{O II} \\ 
\mathbf{B}_{156} & = & \left(\frac{1}{2} - y_{10}\right) \, \mathbf{a}_{1}-z_{10} \, \mathbf{a}_{2} + \left(\frac{1}{2} +x_{10}\right) \, \mathbf{a}_{3} & = & \left(\frac{1}{2} - y_{10}\right)a \, \mathbf{\hat{x}}-z_{10}a \, \mathbf{\hat{y}} + \left(\frac{1}{2} +x_{10}\right)a \, \mathbf{\hat{z}} & \left(24d\right) & \mbox{O II} \\ 
\mathbf{B}_{157} & = & -x_{10} \, \mathbf{a}_{1}-y_{10} \, \mathbf{a}_{2}-z_{10} \, \mathbf{a}_{3} & = & -x_{10}a \, \mathbf{\hat{x}}-y_{10}a \, \mathbf{\hat{y}}-z_{10}a \, \mathbf{\hat{z}} & \left(24d\right) & \mbox{O II} \\ 
\mathbf{B}_{158} & = & \left(\frac{1}{2} +x_{10}\right) \, \mathbf{a}_{1} + y_{10} \, \mathbf{a}_{2} + \left(\frac{1}{2} - z_{10}\right) \, \mathbf{a}_{3} & = & \left(\frac{1}{2} +x_{10}\right)a \, \mathbf{\hat{x}} + y_{10}a \, \mathbf{\hat{y}} + \left(\frac{1}{2} - z_{10}\right)a \, \mathbf{\hat{z}} & \left(24d\right) & \mbox{O II} \\ 
\mathbf{B}_{159} & = & x_{10} \, \mathbf{a}_{1} + \left(\frac{1}{2} - y_{10}\right) \, \mathbf{a}_{2} + \left(\frac{1}{2} +z_{10}\right) \, \mathbf{a}_{3} & = & x_{10}a \, \mathbf{\hat{x}} + \left(\frac{1}{2} - y_{10}\right)a \, \mathbf{\hat{y}} + \left(\frac{1}{2} +z_{10}\right)a \, \mathbf{\hat{z}} & \left(24d\right) & \mbox{O II} \\ 
\mathbf{B}_{160} & = & \left(\frac{1}{2} - x_{10}\right) \, \mathbf{a}_{1} + \left(\frac{1}{2} +y_{10}\right) \, \mathbf{a}_{2} + z_{10} \, \mathbf{a}_{3} & = & \left(\frac{1}{2} - x_{10}\right)a \, \mathbf{\hat{x}} + \left(\frac{1}{2} +y_{10}\right)a \, \mathbf{\hat{y}} + z_{10}a \, \mathbf{\hat{z}} & \left(24d\right) & \mbox{O II} \\ 
\mathbf{B}_{161} & = & -z_{10} \, \mathbf{a}_{1}-x_{10} \, \mathbf{a}_{2}-y_{10} \, \mathbf{a}_{3} & = & -z_{10}a \, \mathbf{\hat{x}}-x_{10}a \, \mathbf{\hat{y}}-y_{10}a \, \mathbf{\hat{z}} & \left(24d\right) & \mbox{O II} \\ 
\mathbf{B}_{162} & = & \left(\frac{1}{2} - z_{10}\right) \, \mathbf{a}_{1} + \left(\frac{1}{2} +x_{10}\right) \, \mathbf{a}_{2} + y_{10} \, \mathbf{a}_{3} & = & \left(\frac{1}{2} - z_{10}\right)a \, \mathbf{\hat{x}} + \left(\frac{1}{2} +x_{10}\right)a \, \mathbf{\hat{y}} + y_{10}a \, \mathbf{\hat{z}} & \left(24d\right) & \mbox{O II} \\ 
\mathbf{B}_{163} & = & \left(\frac{1}{2} +z_{10}\right) \, \mathbf{a}_{1} + x_{10} \, \mathbf{a}_{2} + \left(\frac{1}{2} - y_{10}\right) \, \mathbf{a}_{3} & = & \left(\frac{1}{2} +z_{10}\right)a \, \mathbf{\hat{x}} + x_{10}a \, \mathbf{\hat{y}} + \left(\frac{1}{2} - y_{10}\right)a \, \mathbf{\hat{z}} & \left(24d\right) & \mbox{O II} \\ 
\mathbf{B}_{164} & = & z_{10} \, \mathbf{a}_{1} + \left(\frac{1}{2} - x_{10}\right) \, \mathbf{a}_{2} + \left(\frac{1}{2} +y_{10}\right) \, \mathbf{a}_{3} & = & z_{10}a \, \mathbf{\hat{x}} + \left(\frac{1}{2} - x_{10}\right)a \, \mathbf{\hat{y}} + \left(\frac{1}{2} +y_{10}\right)a \, \mathbf{\hat{z}} & \left(24d\right) & \mbox{O II} \\ 
\mathbf{B}_{165} & = & -y_{10} \, \mathbf{a}_{1}-z_{10} \, \mathbf{a}_{2}-x_{10} \, \mathbf{a}_{3} & = & -y_{10}a \, \mathbf{\hat{x}}-z_{10}a \, \mathbf{\hat{y}}-x_{10}a \, \mathbf{\hat{z}} & \left(24d\right) & \mbox{O II} \\ 
\mathbf{B}_{166} & = & y_{10} \, \mathbf{a}_{1} + \left(\frac{1}{2} - z_{10}\right) \, \mathbf{a}_{2} + \left(\frac{1}{2} +x_{10}\right) \, \mathbf{a}_{3} & = & y_{10}a \, \mathbf{\hat{x}} + \left(\frac{1}{2} - z_{10}\right)a \, \mathbf{\hat{y}} + \left(\frac{1}{2} +x_{10}\right)a \, \mathbf{\hat{z}} & \left(24d\right) & \mbox{O II} \\ 
\mathbf{B}_{167} & = & \left(\frac{1}{2} - y_{10}\right) \, \mathbf{a}_{1} + \left(\frac{1}{2} +z_{10}\right) \, \mathbf{a}_{2} + x_{10} \, \mathbf{a}_{3} & = & \left(\frac{1}{2} - y_{10}\right)a \, \mathbf{\hat{x}} + \left(\frac{1}{2} +z_{10}\right)a \, \mathbf{\hat{y}} + x_{10}a \, \mathbf{\hat{z}} & \left(24d\right) & \mbox{O II} \\ 
\mathbf{B}_{168} & = & \left(\frac{1}{2} +y_{10}\right) \, \mathbf{a}_{1} + z_{10} \, \mathbf{a}_{2} + \left(\frac{1}{2} - x_{10}\right) \, \mathbf{a}_{3} & = & \left(\frac{1}{2} +y_{10}\right)a \, \mathbf{\hat{x}} + z_{10}a \, \mathbf{\hat{y}} + \left(\frac{1}{2} - x_{10}\right)a \, \mathbf{\hat{z}} & \left(24d\right) & \mbox{O II} \\ 
\mathbf{B}_{169} & = & x_{11} \, \mathbf{a}_{1} + y_{11} \, \mathbf{a}_{2} + z_{11} \, \mathbf{a}_{3} & = & x_{11}a \, \mathbf{\hat{x}} + y_{11}a \, \mathbf{\hat{y}} + z_{11}a \, \mathbf{\hat{z}} & \left(24d\right) & \mbox{O III} \\ 
\mathbf{B}_{170} & = & \left(\frac{1}{2} - x_{11}\right) \, \mathbf{a}_{1}-y_{11} \, \mathbf{a}_{2} + \left(\frac{1}{2} +z_{11}\right) \, \mathbf{a}_{3} & = & \left(\frac{1}{2} - x_{11}\right)a \, \mathbf{\hat{x}}-y_{11}a \, \mathbf{\hat{y}} + \left(\frac{1}{2} +z_{11}\right)a \, \mathbf{\hat{z}} & \left(24d\right) & \mbox{O III} \\ 
\mathbf{B}_{171} & = & -x_{11} \, \mathbf{a}_{1} + \left(\frac{1}{2} +y_{11}\right) \, \mathbf{a}_{2} + \left(\frac{1}{2} - z_{11}\right) \, \mathbf{a}_{3} & = & -x_{11}a \, \mathbf{\hat{x}} + \left(\frac{1}{2} +y_{11}\right)a \, \mathbf{\hat{y}} + \left(\frac{1}{2} - z_{11}\right)a \, \mathbf{\hat{z}} & \left(24d\right) & \mbox{O III} \\ 
\mathbf{B}_{172} & = & \left(\frac{1}{2} +x_{11}\right) \, \mathbf{a}_{1} + \left(\frac{1}{2} - y_{11}\right) \, \mathbf{a}_{2}-z_{11} \, \mathbf{a}_{3} & = & \left(\frac{1}{2} +x_{11}\right)a \, \mathbf{\hat{x}} + \left(\frac{1}{2} - y_{11}\right)a \, \mathbf{\hat{y}}-z_{11}a \, \mathbf{\hat{z}} & \left(24d\right) & \mbox{O III} \\ 
\mathbf{B}_{173} & = & z_{11} \, \mathbf{a}_{1} + x_{11} \, \mathbf{a}_{2} + y_{11} \, \mathbf{a}_{3} & = & z_{11}a \, \mathbf{\hat{x}} + x_{11}a \, \mathbf{\hat{y}} + y_{11}a \, \mathbf{\hat{z}} & \left(24d\right) & \mbox{O III} \\ 
\mathbf{B}_{174} & = & \left(\frac{1}{2} +z_{11}\right) \, \mathbf{a}_{1} + \left(\frac{1}{2} - x_{11}\right) \, \mathbf{a}_{2}-y_{11} \, \mathbf{a}_{3} & = & \left(\frac{1}{2} +z_{11}\right)a \, \mathbf{\hat{x}} + \left(\frac{1}{2} - x_{11}\right)a \, \mathbf{\hat{y}}-y_{11}a \, \mathbf{\hat{z}} & \left(24d\right) & \mbox{O III} \\ 
\mathbf{B}_{175} & = & \left(\frac{1}{2} - z_{11}\right) \, \mathbf{a}_{1}-x_{11} \, \mathbf{a}_{2} + \left(\frac{1}{2} +y_{11}\right) \, \mathbf{a}_{3} & = & \left(\frac{1}{2} - z_{11}\right)a \, \mathbf{\hat{x}}-x_{11}a \, \mathbf{\hat{y}} + \left(\frac{1}{2} +y_{11}\right)a \, \mathbf{\hat{z}} & \left(24d\right) & \mbox{O III} \\ 
\mathbf{B}_{176} & = & -z_{11} \, \mathbf{a}_{1} + \left(\frac{1}{2} +x_{11}\right) \, \mathbf{a}_{2} + \left(\frac{1}{2} - y_{11}\right) \, \mathbf{a}_{3} & = & -z_{11}a \, \mathbf{\hat{x}} + \left(\frac{1}{2} +x_{11}\right)a \, \mathbf{\hat{y}} + \left(\frac{1}{2} - y_{11}\right)a \, \mathbf{\hat{z}} & \left(24d\right) & \mbox{O III} \\ 
\mathbf{B}_{177} & = & y_{11} \, \mathbf{a}_{1} + z_{11} \, \mathbf{a}_{2} + x_{11} \, \mathbf{a}_{3} & = & y_{11}a \, \mathbf{\hat{x}} + z_{11}a \, \mathbf{\hat{y}} + x_{11}a \, \mathbf{\hat{z}} & \left(24d\right) & \mbox{O III} \\ 
\mathbf{B}_{178} & = & -y_{11} \, \mathbf{a}_{1} + \left(\frac{1}{2} +z_{11}\right) \, \mathbf{a}_{2} + \left(\frac{1}{2} - x_{11}\right) \, \mathbf{a}_{3} & = & -y_{11}a \, \mathbf{\hat{x}} + \left(\frac{1}{2} +z_{11}\right)a \, \mathbf{\hat{y}} + \left(\frac{1}{2} - x_{11}\right)a \, \mathbf{\hat{z}} & \left(24d\right) & \mbox{O III} \\ 
\mathbf{B}_{179} & = & \left(\frac{1}{2} +y_{11}\right) \, \mathbf{a}_{1} + \left(\frac{1}{2} - z_{11}\right) \, \mathbf{a}_{2}-x_{11} \, \mathbf{a}_{3} & = & \left(\frac{1}{2} +y_{11}\right)a \, \mathbf{\hat{x}} + \left(\frac{1}{2} - z_{11}\right)a \, \mathbf{\hat{y}}-x_{11}a \, \mathbf{\hat{z}} & \left(24d\right) & \mbox{O III} \\ 
\mathbf{B}_{180} & = & \left(\frac{1}{2} - y_{11}\right) \, \mathbf{a}_{1}-z_{11} \, \mathbf{a}_{2} + \left(\frac{1}{2} +x_{11}\right) \, \mathbf{a}_{3} & = & \left(\frac{1}{2} - y_{11}\right)a \, \mathbf{\hat{x}}-z_{11}a \, \mathbf{\hat{y}} + \left(\frac{1}{2} +x_{11}\right)a \, \mathbf{\hat{z}} & \left(24d\right) & \mbox{O III} \\ 
\mathbf{B}_{181} & = & -x_{11} \, \mathbf{a}_{1}-y_{11} \, \mathbf{a}_{2}-z_{11} \, \mathbf{a}_{3} & = & -x_{11}a \, \mathbf{\hat{x}}-y_{11}a \, \mathbf{\hat{y}}-z_{11}a \, \mathbf{\hat{z}} & \left(24d\right) & \mbox{O III} \\ 
\mathbf{B}_{182} & = & \left(\frac{1}{2} +x_{11}\right) \, \mathbf{a}_{1} + y_{11} \, \mathbf{a}_{2} + \left(\frac{1}{2} - z_{11}\right) \, \mathbf{a}_{3} & = & \left(\frac{1}{2} +x_{11}\right)a \, \mathbf{\hat{x}} + y_{11}a \, \mathbf{\hat{y}} + \left(\frac{1}{2} - z_{11}\right)a \, \mathbf{\hat{z}} & \left(24d\right) & \mbox{O III} \\ 
\mathbf{B}_{183} & = & x_{11} \, \mathbf{a}_{1} + \left(\frac{1}{2} - y_{11}\right) \, \mathbf{a}_{2} + \left(\frac{1}{2} +z_{11}\right) \, \mathbf{a}_{3} & = & x_{11}a \, \mathbf{\hat{x}} + \left(\frac{1}{2} - y_{11}\right)a \, \mathbf{\hat{y}} + \left(\frac{1}{2} +z_{11}\right)a \, \mathbf{\hat{z}} & \left(24d\right) & \mbox{O III} \\ 
\mathbf{B}_{184} & = & \left(\frac{1}{2} - x_{11}\right) \, \mathbf{a}_{1} + \left(\frac{1}{2} +y_{11}\right) \, \mathbf{a}_{2} + z_{11} \, \mathbf{a}_{3} & = & \left(\frac{1}{2} - x_{11}\right)a \, \mathbf{\hat{x}} + \left(\frac{1}{2} +y_{11}\right)a \, \mathbf{\hat{y}} + z_{11}a \, \mathbf{\hat{z}} & \left(24d\right) & \mbox{O III} \\ 
\mathbf{B}_{185} & = & -z_{11} \, \mathbf{a}_{1}-x_{11} \, \mathbf{a}_{2}-y_{11} \, \mathbf{a}_{3} & = & -z_{11}a \, \mathbf{\hat{x}}-x_{11}a \, \mathbf{\hat{y}}-y_{11}a \, \mathbf{\hat{z}} & \left(24d\right) & \mbox{O III} \\ 
\mathbf{B}_{186} & = & \left(\frac{1}{2} - z_{11}\right) \, \mathbf{a}_{1} + \left(\frac{1}{2} +x_{11}\right) \, \mathbf{a}_{2} + y_{11} \, \mathbf{a}_{3} & = & \left(\frac{1}{2} - z_{11}\right)a \, \mathbf{\hat{x}} + \left(\frac{1}{2} +x_{11}\right)a \, \mathbf{\hat{y}} + y_{11}a \, \mathbf{\hat{z}} & \left(24d\right) & \mbox{O III} \\ 
\mathbf{B}_{187} & = & \left(\frac{1}{2} +z_{11}\right) \, \mathbf{a}_{1} + x_{11} \, \mathbf{a}_{2} + \left(\frac{1}{2} - y_{11}\right) \, \mathbf{a}_{3} & = & \left(\frac{1}{2} +z_{11}\right)a \, \mathbf{\hat{x}} + x_{11}a \, \mathbf{\hat{y}} + \left(\frac{1}{2} - y_{11}\right)a \, \mathbf{\hat{z}} & \left(24d\right) & \mbox{O III} \\ 
\mathbf{B}_{188} & = & z_{11} \, \mathbf{a}_{1} + \left(\frac{1}{2} - x_{11}\right) \, \mathbf{a}_{2} + \left(\frac{1}{2} +y_{11}\right) \, \mathbf{a}_{3} & = & z_{11}a \, \mathbf{\hat{x}} + \left(\frac{1}{2} - x_{11}\right)a \, \mathbf{\hat{y}} + \left(\frac{1}{2} +y_{11}\right)a \, \mathbf{\hat{z}} & \left(24d\right) & \mbox{O III} \\ 
\mathbf{B}_{189} & = & -y_{11} \, \mathbf{a}_{1}-z_{11} \, \mathbf{a}_{2}-x_{11} \, \mathbf{a}_{3} & = & -y_{11}a \, \mathbf{\hat{x}}-z_{11}a \, \mathbf{\hat{y}}-x_{11}a \, \mathbf{\hat{z}} & \left(24d\right) & \mbox{O III} \\ 
\mathbf{B}_{190} & = & y_{11} \, \mathbf{a}_{1} + \left(\frac{1}{2} - z_{11}\right) \, \mathbf{a}_{2} + \left(\frac{1}{2} +x_{11}\right) \, \mathbf{a}_{3} & = & y_{11}a \, \mathbf{\hat{x}} + \left(\frac{1}{2} - z_{11}\right)a \, \mathbf{\hat{y}} + \left(\frac{1}{2} +x_{11}\right)a \, \mathbf{\hat{z}} & \left(24d\right) & \mbox{O III} \\ 
\mathbf{B}_{191} & = & \left(\frac{1}{2} - y_{11}\right) \, \mathbf{a}_{1} + \left(\frac{1}{2} +z_{11}\right) \, \mathbf{a}_{2} + x_{11} \, \mathbf{a}_{3} & = & \left(\frac{1}{2} - y_{11}\right)a \, \mathbf{\hat{x}} + \left(\frac{1}{2} +z_{11}\right)a \, \mathbf{\hat{y}} + x_{11}a \, \mathbf{\hat{z}} & \left(24d\right) & \mbox{O III} \\ 
\mathbf{B}_{192} & = & \left(\frac{1}{2} +y_{11}\right) \, \mathbf{a}_{1} + z_{11} \, \mathbf{a}_{2} + \left(\frac{1}{2} - x_{11}\right) \, \mathbf{a}_{3} & = & \left(\frac{1}{2} +y_{11}\right)a \, \mathbf{\hat{x}} + z_{11}a \, \mathbf{\hat{y}} + \left(\frac{1}{2} - x_{11}\right)a \, \mathbf{\hat{z}} & \left(24d\right) & \mbox{O III} \\ 
\mathbf{B}_{193} & = & x_{12} \, \mathbf{a}_{1} + y_{12} \, \mathbf{a}_{2} + z_{12} \, \mathbf{a}_{3} & = & x_{12}a \, \mathbf{\hat{x}} + y_{12}a \, \mathbf{\hat{y}} + z_{12}a \, \mathbf{\hat{z}} & \left(24d\right) & \mbox{O IV} \\ 
\mathbf{B}_{194} & = & \left(\frac{1}{2} - x_{12}\right) \, \mathbf{a}_{1}-y_{12} \, \mathbf{a}_{2} + \left(\frac{1}{2} +z_{12}\right) \, \mathbf{a}_{3} & = & \left(\frac{1}{2} - x_{12}\right)a \, \mathbf{\hat{x}}-y_{12}a \, \mathbf{\hat{y}} + \left(\frac{1}{2} +z_{12}\right)a \, \mathbf{\hat{z}} & \left(24d\right) & \mbox{O IV} \\ 
\mathbf{B}_{195} & = & -x_{12} \, \mathbf{a}_{1} + \left(\frac{1}{2} +y_{12}\right) \, \mathbf{a}_{2} + \left(\frac{1}{2} - z_{12}\right) \, \mathbf{a}_{3} & = & -x_{12}a \, \mathbf{\hat{x}} + \left(\frac{1}{2} +y_{12}\right)a \, \mathbf{\hat{y}} + \left(\frac{1}{2} - z_{12}\right)a \, \mathbf{\hat{z}} & \left(24d\right) & \mbox{O IV} \\ 
\mathbf{B}_{196} & = & \left(\frac{1}{2} +x_{12}\right) \, \mathbf{a}_{1} + \left(\frac{1}{2} - y_{12}\right) \, \mathbf{a}_{2}-z_{12} \, \mathbf{a}_{3} & = & \left(\frac{1}{2} +x_{12}\right)a \, \mathbf{\hat{x}} + \left(\frac{1}{2} - y_{12}\right)a \, \mathbf{\hat{y}}-z_{12}a \, \mathbf{\hat{z}} & \left(24d\right) & \mbox{O IV} \\ 
\mathbf{B}_{197} & = & z_{12} \, \mathbf{a}_{1} + x_{12} \, \mathbf{a}_{2} + y_{12} \, \mathbf{a}_{3} & = & z_{12}a \, \mathbf{\hat{x}} + x_{12}a \, \mathbf{\hat{y}} + y_{12}a \, \mathbf{\hat{z}} & \left(24d\right) & \mbox{O IV} \\ 
\mathbf{B}_{198} & = & \left(\frac{1}{2} +z_{12}\right) \, \mathbf{a}_{1} + \left(\frac{1}{2} - x_{12}\right) \, \mathbf{a}_{2}-y_{12} \, \mathbf{a}_{3} & = & \left(\frac{1}{2} +z_{12}\right)a \, \mathbf{\hat{x}} + \left(\frac{1}{2} - x_{12}\right)a \, \mathbf{\hat{y}}-y_{12}a \, \mathbf{\hat{z}} & \left(24d\right) & \mbox{O IV} \\ 
\mathbf{B}_{199} & = & \left(\frac{1}{2} - z_{12}\right) \, \mathbf{a}_{1}-x_{12} \, \mathbf{a}_{2} + \left(\frac{1}{2} +y_{12}\right) \, \mathbf{a}_{3} & = & \left(\frac{1}{2} - z_{12}\right)a \, \mathbf{\hat{x}}-x_{12}a \, \mathbf{\hat{y}} + \left(\frac{1}{2} +y_{12}\right)a \, \mathbf{\hat{z}} & \left(24d\right) & \mbox{O IV} \\ 
\mathbf{B}_{200} & = & -z_{12} \, \mathbf{a}_{1} + \left(\frac{1}{2} +x_{12}\right) \, \mathbf{a}_{2} + \left(\frac{1}{2} - y_{12}\right) \, \mathbf{a}_{3} & = & -z_{12}a \, \mathbf{\hat{x}} + \left(\frac{1}{2} +x_{12}\right)a \, \mathbf{\hat{y}} + \left(\frac{1}{2} - y_{12}\right)a \, \mathbf{\hat{z}} & \left(24d\right) & \mbox{O IV} \\ 
\mathbf{B}_{201} & = & y_{12} \, \mathbf{a}_{1} + z_{12} \, \mathbf{a}_{2} + x_{12} \, \mathbf{a}_{3} & = & y_{12}a \, \mathbf{\hat{x}} + z_{12}a \, \mathbf{\hat{y}} + x_{12}a \, \mathbf{\hat{z}} & \left(24d\right) & \mbox{O IV} \\ 
\mathbf{B}_{202} & = & -y_{12} \, \mathbf{a}_{1} + \left(\frac{1}{2} +z_{12}\right) \, \mathbf{a}_{2} + \left(\frac{1}{2} - x_{12}\right) \, \mathbf{a}_{3} & = & -y_{12}a \, \mathbf{\hat{x}} + \left(\frac{1}{2} +z_{12}\right)a \, \mathbf{\hat{y}} + \left(\frac{1}{2} - x_{12}\right)a \, \mathbf{\hat{z}} & \left(24d\right) & \mbox{O IV} \\ 
\mathbf{B}_{203} & = & \left(\frac{1}{2} +y_{12}\right) \, \mathbf{a}_{1} + \left(\frac{1}{2} - z_{12}\right) \, \mathbf{a}_{2}-x_{12} \, \mathbf{a}_{3} & = & \left(\frac{1}{2} +y_{12}\right)a \, \mathbf{\hat{x}} + \left(\frac{1}{2} - z_{12}\right)a \, \mathbf{\hat{y}}-x_{12}a \, \mathbf{\hat{z}} & \left(24d\right) & \mbox{O IV} \\ 
\mathbf{B}_{204} & = & \left(\frac{1}{2} - y_{12}\right) \, \mathbf{a}_{1}-z_{12} \, \mathbf{a}_{2} + \left(\frac{1}{2} +x_{12}\right) \, \mathbf{a}_{3} & = & \left(\frac{1}{2} - y_{12}\right)a \, \mathbf{\hat{x}}-z_{12}a \, \mathbf{\hat{y}} + \left(\frac{1}{2} +x_{12}\right)a \, \mathbf{\hat{z}} & \left(24d\right) & \mbox{O IV} \\ 
\mathbf{B}_{205} & = & -x_{12} \, \mathbf{a}_{1}-y_{12} \, \mathbf{a}_{2}-z_{12} \, \mathbf{a}_{3} & = & -x_{12}a \, \mathbf{\hat{x}}-y_{12}a \, \mathbf{\hat{y}}-z_{12}a \, \mathbf{\hat{z}} & \left(24d\right) & \mbox{O IV} \\ 
\mathbf{B}_{206} & = & \left(\frac{1}{2} +x_{12}\right) \, \mathbf{a}_{1} + y_{12} \, \mathbf{a}_{2} + \left(\frac{1}{2} - z_{12}\right) \, \mathbf{a}_{3} & = & \left(\frac{1}{2} +x_{12}\right)a \, \mathbf{\hat{x}} + y_{12}a \, \mathbf{\hat{y}} + \left(\frac{1}{2} - z_{12}\right)a \, \mathbf{\hat{z}} & \left(24d\right) & \mbox{O IV} \\ 
\mathbf{B}_{207} & = & x_{12} \, \mathbf{a}_{1} + \left(\frac{1}{2} - y_{12}\right) \, \mathbf{a}_{2} + \left(\frac{1}{2} +z_{12}\right) \, \mathbf{a}_{3} & = & x_{12}a \, \mathbf{\hat{x}} + \left(\frac{1}{2} - y_{12}\right)a \, \mathbf{\hat{y}} + \left(\frac{1}{2} +z_{12}\right)a \, \mathbf{\hat{z}} & \left(24d\right) & \mbox{O IV} \\ 
\mathbf{B}_{208} & = & \left(\frac{1}{2} - x_{12}\right) \, \mathbf{a}_{1} + \left(\frac{1}{2} +y_{12}\right) \, \mathbf{a}_{2} + z_{12} \, \mathbf{a}_{3} & = & \left(\frac{1}{2} - x_{12}\right)a \, \mathbf{\hat{x}} + \left(\frac{1}{2} +y_{12}\right)a \, \mathbf{\hat{y}} + z_{12}a \, \mathbf{\hat{z}} & \left(24d\right) & \mbox{O IV} \\ 
\mathbf{B}_{209} & = & -z_{12} \, \mathbf{a}_{1}-x_{12} \, \mathbf{a}_{2}-y_{12} \, \mathbf{a}_{3} & = & -z_{12}a \, \mathbf{\hat{x}}-x_{12}a \, \mathbf{\hat{y}}-y_{12}a \, \mathbf{\hat{z}} & \left(24d\right) & \mbox{O IV} \\ 
\mathbf{B}_{210} & = & \left(\frac{1}{2} - z_{12}\right) \, \mathbf{a}_{1} + \left(\frac{1}{2} +x_{12}\right) \, \mathbf{a}_{2} + y_{12} \, \mathbf{a}_{3} & = & \left(\frac{1}{2} - z_{12}\right)a \, \mathbf{\hat{x}} + \left(\frac{1}{2} +x_{12}\right)a \, \mathbf{\hat{y}} + y_{12}a \, \mathbf{\hat{z}} & \left(24d\right) & \mbox{O IV} \\ 
\mathbf{B}_{211} & = & \left(\frac{1}{2} +z_{12}\right) \, \mathbf{a}_{1} + x_{12} \, \mathbf{a}_{2} + \left(\frac{1}{2} - y_{12}\right) \, \mathbf{a}_{3} & = & \left(\frac{1}{2} +z_{12}\right)a \, \mathbf{\hat{x}} + x_{12}a \, \mathbf{\hat{y}} + \left(\frac{1}{2} - y_{12}\right)a \, \mathbf{\hat{z}} & \left(24d\right) & \mbox{O IV} \\ 
\mathbf{B}_{212} & = & z_{12} \, \mathbf{a}_{1} + \left(\frac{1}{2} - x_{12}\right) \, \mathbf{a}_{2} + \left(\frac{1}{2} +y_{12}\right) \, \mathbf{a}_{3} & = & z_{12}a \, \mathbf{\hat{x}} + \left(\frac{1}{2} - x_{12}\right)a \, \mathbf{\hat{y}} + \left(\frac{1}{2} +y_{12}\right)a \, \mathbf{\hat{z}} & \left(24d\right) & \mbox{O IV} \\ 
\mathbf{B}_{213} & = & -y_{12} \, \mathbf{a}_{1}-z_{12} \, \mathbf{a}_{2}-x_{12} \, \mathbf{a}_{3} & = & -y_{12}a \, \mathbf{\hat{x}}-z_{12}a \, \mathbf{\hat{y}}-x_{12}a \, \mathbf{\hat{z}} & \left(24d\right) & \mbox{O IV} \\ 
\mathbf{B}_{214} & = & y_{12} \, \mathbf{a}_{1} + \left(\frac{1}{2} - z_{12}\right) \, \mathbf{a}_{2} + \left(\frac{1}{2} +x_{12}\right) \, \mathbf{a}_{3} & = & y_{12}a \, \mathbf{\hat{x}} + \left(\frac{1}{2} - z_{12}\right)a \, \mathbf{\hat{y}} + \left(\frac{1}{2} +x_{12}\right)a \, \mathbf{\hat{z}} & \left(24d\right) & \mbox{O IV} \\ 
\mathbf{B}_{215} & = & \left(\frac{1}{2} - y_{12}\right) \, \mathbf{a}_{1} + \left(\frac{1}{2} +z_{12}\right) \, \mathbf{a}_{2} + x_{12} \, \mathbf{a}_{3} & = & \left(\frac{1}{2} - y_{12}\right)a \, \mathbf{\hat{x}} + \left(\frac{1}{2} +z_{12}\right)a \, \mathbf{\hat{y}} + x_{12}a \, \mathbf{\hat{z}} & \left(24d\right) & \mbox{O IV} \\ 
\mathbf{B}_{216} & = & \left(\frac{1}{2} +y_{12}\right) \, \mathbf{a}_{1} + z_{12} \, \mathbf{a}_{2} + \left(\frac{1}{2} - x_{12}\right) \, \mathbf{a}_{3} & = & \left(\frac{1}{2} +y_{12}\right)a \, \mathbf{\hat{x}} + z_{12}a \, \mathbf{\hat{y}} + \left(\frac{1}{2} - x_{12}\right)a \, \mathbf{\hat{z}} & \left(24d\right) & \mbox{O IV} \\ 
\mathbf{B}_{217} & = & x_{13} \, \mathbf{a}_{1} + y_{13} \, \mathbf{a}_{2} + z_{13} \, \mathbf{a}_{3} & = & x_{13}a \, \mathbf{\hat{x}} + y_{13}a \, \mathbf{\hat{y}} + z_{13}a \, \mathbf{\hat{z}} & \left(24d\right) & \mbox{O V} \\ 
\mathbf{B}_{218} & = & \left(\frac{1}{2} - x_{13}\right) \, \mathbf{a}_{1}-y_{13} \, \mathbf{a}_{2} + \left(\frac{1}{2} +z_{13}\right) \, \mathbf{a}_{3} & = & \left(\frac{1}{2} - x_{13}\right)a \, \mathbf{\hat{x}}-y_{13}a \, \mathbf{\hat{y}} + \left(\frac{1}{2} +z_{13}\right)a \, \mathbf{\hat{z}} & \left(24d\right) & \mbox{O V} \\ 
\mathbf{B}_{219} & = & -x_{13} \, \mathbf{a}_{1} + \left(\frac{1}{2} +y_{13}\right) \, \mathbf{a}_{2} + \left(\frac{1}{2} - z_{13}\right) \, \mathbf{a}_{3} & = & -x_{13}a \, \mathbf{\hat{x}} + \left(\frac{1}{2} +y_{13}\right)a \, \mathbf{\hat{y}} + \left(\frac{1}{2} - z_{13}\right)a \, \mathbf{\hat{z}} & \left(24d\right) & \mbox{O V} \\ 
\mathbf{B}_{220} & = & \left(\frac{1}{2} +x_{13}\right) \, \mathbf{a}_{1} + \left(\frac{1}{2} - y_{13}\right) \, \mathbf{a}_{2}-z_{13} \, \mathbf{a}_{3} & = & \left(\frac{1}{2} +x_{13}\right)a \, \mathbf{\hat{x}} + \left(\frac{1}{2} - y_{13}\right)a \, \mathbf{\hat{y}}-z_{13}a \, \mathbf{\hat{z}} & \left(24d\right) & \mbox{O V} \\ 
\mathbf{B}_{221} & = & z_{13} \, \mathbf{a}_{1} + x_{13} \, \mathbf{a}_{2} + y_{13} \, \mathbf{a}_{3} & = & z_{13}a \, \mathbf{\hat{x}} + x_{13}a \, \mathbf{\hat{y}} + y_{13}a \, \mathbf{\hat{z}} & \left(24d\right) & \mbox{O V} \\ 
\mathbf{B}_{222} & = & \left(\frac{1}{2} +z_{13}\right) \, \mathbf{a}_{1} + \left(\frac{1}{2} - x_{13}\right) \, \mathbf{a}_{2}-y_{13} \, \mathbf{a}_{3} & = & \left(\frac{1}{2} +z_{13}\right)a \, \mathbf{\hat{x}} + \left(\frac{1}{2} - x_{13}\right)a \, \mathbf{\hat{y}}-y_{13}a \, \mathbf{\hat{z}} & \left(24d\right) & \mbox{O V} \\ 
\mathbf{B}_{223} & = & \left(\frac{1}{2} - z_{13}\right) \, \mathbf{a}_{1}-x_{13} \, \mathbf{a}_{2} + \left(\frac{1}{2} +y_{13}\right) \, \mathbf{a}_{3} & = & \left(\frac{1}{2} - z_{13}\right)a \, \mathbf{\hat{x}}-x_{13}a \, \mathbf{\hat{y}} + \left(\frac{1}{2} +y_{13}\right)a \, \mathbf{\hat{z}} & \left(24d\right) & \mbox{O V} \\ 
\mathbf{B}_{224} & = & -z_{13} \, \mathbf{a}_{1} + \left(\frac{1}{2} +x_{13}\right) \, \mathbf{a}_{2} + \left(\frac{1}{2} - y_{13}\right) \, \mathbf{a}_{3} & = & -z_{13}a \, \mathbf{\hat{x}} + \left(\frac{1}{2} +x_{13}\right)a \, \mathbf{\hat{y}} + \left(\frac{1}{2} - y_{13}\right)a \, \mathbf{\hat{z}} & \left(24d\right) & \mbox{O V} \\ 
\mathbf{B}_{225} & = & y_{13} \, \mathbf{a}_{1} + z_{13} \, \mathbf{a}_{2} + x_{13} \, \mathbf{a}_{3} & = & y_{13}a \, \mathbf{\hat{x}} + z_{13}a \, \mathbf{\hat{y}} + x_{13}a \, \mathbf{\hat{z}} & \left(24d\right) & \mbox{O V} \\ 
\mathbf{B}_{226} & = & -y_{13} \, \mathbf{a}_{1} + \left(\frac{1}{2} +z_{13}\right) \, \mathbf{a}_{2} + \left(\frac{1}{2} - x_{13}\right) \, \mathbf{a}_{3} & = & -y_{13}a \, \mathbf{\hat{x}} + \left(\frac{1}{2} +z_{13}\right)a \, \mathbf{\hat{y}} + \left(\frac{1}{2} - x_{13}\right)a \, \mathbf{\hat{z}} & \left(24d\right) & \mbox{O V} \\ 
\mathbf{B}_{227} & = & \left(\frac{1}{2} +y_{13}\right) \, \mathbf{a}_{1} + \left(\frac{1}{2} - z_{13}\right) \, \mathbf{a}_{2}-x_{13} \, \mathbf{a}_{3} & = & \left(\frac{1}{2} +y_{13}\right)a \, \mathbf{\hat{x}} + \left(\frac{1}{2} - z_{13}\right)a \, \mathbf{\hat{y}}-x_{13}a \, \mathbf{\hat{z}} & \left(24d\right) & \mbox{O V} \\ 
\mathbf{B}_{228} & = & \left(\frac{1}{2} - y_{13}\right) \, \mathbf{a}_{1}-z_{13} \, \mathbf{a}_{2} + \left(\frac{1}{2} +x_{13}\right) \, \mathbf{a}_{3} & = & \left(\frac{1}{2} - y_{13}\right)a \, \mathbf{\hat{x}}-z_{13}a \, \mathbf{\hat{y}} + \left(\frac{1}{2} +x_{13}\right)a \, \mathbf{\hat{z}} & \left(24d\right) & \mbox{O V} \\ 
\mathbf{B}_{229} & = & -x_{13} \, \mathbf{a}_{1}-y_{13} \, \mathbf{a}_{2}-z_{13} \, \mathbf{a}_{3} & = & -x_{13}a \, \mathbf{\hat{x}}-y_{13}a \, \mathbf{\hat{y}}-z_{13}a \, \mathbf{\hat{z}} & \left(24d\right) & \mbox{O V} \\ 
\mathbf{B}_{230} & = & \left(\frac{1}{2} +x_{13}\right) \, \mathbf{a}_{1} + y_{13} \, \mathbf{a}_{2} + \left(\frac{1}{2} - z_{13}\right) \, \mathbf{a}_{3} & = & \left(\frac{1}{2} +x_{13}\right)a \, \mathbf{\hat{x}} + y_{13}a \, \mathbf{\hat{y}} + \left(\frac{1}{2} - z_{13}\right)a \, \mathbf{\hat{z}} & \left(24d\right) & \mbox{O V} \\ 
\mathbf{B}_{231} & = & x_{13} \, \mathbf{a}_{1} + \left(\frac{1}{2} - y_{13}\right) \, \mathbf{a}_{2} + \left(\frac{1}{2} +z_{13}\right) \, \mathbf{a}_{3} & = & x_{13}a \, \mathbf{\hat{x}} + \left(\frac{1}{2} - y_{13}\right)a \, \mathbf{\hat{y}} + \left(\frac{1}{2} +z_{13}\right)a \, \mathbf{\hat{z}} & \left(24d\right) & \mbox{O V} \\ 
\mathbf{B}_{232} & = & \left(\frac{1}{2} - x_{13}\right) \, \mathbf{a}_{1} + \left(\frac{1}{2} +y_{13}\right) \, \mathbf{a}_{2} + z_{13} \, \mathbf{a}_{3} & = & \left(\frac{1}{2} - x_{13}\right)a \, \mathbf{\hat{x}} + \left(\frac{1}{2} +y_{13}\right)a \, \mathbf{\hat{y}} + z_{13}a \, \mathbf{\hat{z}} & \left(24d\right) & \mbox{O V} \\ 
\mathbf{B}_{233} & = & -z_{13} \, \mathbf{a}_{1}-x_{13} \, \mathbf{a}_{2}-y_{13} \, \mathbf{a}_{3} & = & -z_{13}a \, \mathbf{\hat{x}}-x_{13}a \, \mathbf{\hat{y}}-y_{13}a \, \mathbf{\hat{z}} & \left(24d\right) & \mbox{O V} \\ 
\mathbf{B}_{234} & = & \left(\frac{1}{2} - z_{13}\right) \, \mathbf{a}_{1} + \left(\frac{1}{2} +x_{13}\right) \, \mathbf{a}_{2} + y_{13} \, \mathbf{a}_{3} & = & \left(\frac{1}{2} - z_{13}\right)a \, \mathbf{\hat{x}} + \left(\frac{1}{2} +x_{13}\right)a \, \mathbf{\hat{y}} + y_{13}a \, \mathbf{\hat{z}} & \left(24d\right) & \mbox{O V} \\ 
\mathbf{B}_{235} & = & \left(\frac{1}{2} +z_{13}\right) \, \mathbf{a}_{1} + x_{13} \, \mathbf{a}_{2} + \left(\frac{1}{2} - y_{13}\right) \, \mathbf{a}_{3} & = & \left(\frac{1}{2} +z_{13}\right)a \, \mathbf{\hat{x}} + x_{13}a \, \mathbf{\hat{y}} + \left(\frac{1}{2} - y_{13}\right)a \, \mathbf{\hat{z}} & \left(24d\right) & \mbox{O V} \\ 
\mathbf{B}_{236} & = & z_{13} \, \mathbf{a}_{1} + \left(\frac{1}{2} - x_{13}\right) \, \mathbf{a}_{2} + \left(\frac{1}{2} +y_{13}\right) \, \mathbf{a}_{3} & = & z_{13}a \, \mathbf{\hat{x}} + \left(\frac{1}{2} - x_{13}\right)a \, \mathbf{\hat{y}} + \left(\frac{1}{2} +y_{13}\right)a \, \mathbf{\hat{z}} & \left(24d\right) & \mbox{O V} \\ 
\mathbf{B}_{237} & = & -y_{13} \, \mathbf{a}_{1}-z_{13} \, \mathbf{a}_{2}-x_{13} \, \mathbf{a}_{3} & = & -y_{13}a \, \mathbf{\hat{x}}-z_{13}a \, \mathbf{\hat{y}}-x_{13}a \, \mathbf{\hat{z}} & \left(24d\right) & \mbox{O V} \\ 
\mathbf{B}_{238} & = & y_{13} \, \mathbf{a}_{1} + \left(\frac{1}{2} - z_{13}\right) \, \mathbf{a}_{2} + \left(\frac{1}{2} +x_{13}\right) \, \mathbf{a}_{3} & = & y_{13}a \, \mathbf{\hat{x}} + \left(\frac{1}{2} - z_{13}\right)a \, \mathbf{\hat{y}} + \left(\frac{1}{2} +x_{13}\right)a \, \mathbf{\hat{z}} & \left(24d\right) & \mbox{O V} \\ 
\mathbf{B}_{239} & = & \left(\frac{1}{2} - y_{13}\right) \, \mathbf{a}_{1} + \left(\frac{1}{2} +z_{13}\right) \, \mathbf{a}_{2} + x_{13} \, \mathbf{a}_{3} & = & \left(\frac{1}{2} - y_{13}\right)a \, \mathbf{\hat{x}} + \left(\frac{1}{2} +z_{13}\right)a \, \mathbf{\hat{y}} + x_{13}a \, \mathbf{\hat{z}} & \left(24d\right) & \mbox{O V} \\ 
\mathbf{B}_{240} & = & \left(\frac{1}{2} +y_{13}\right) \, \mathbf{a}_{1} + z_{13} \, \mathbf{a}_{2} + \left(\frac{1}{2} - x_{13}\right) \, \mathbf{a}_{3} & = & \left(\frac{1}{2} +y_{13}\right)a \, \mathbf{\hat{x}} + z_{13}a \, \mathbf{\hat{y}} + \left(\frac{1}{2} - x_{13}\right)a \, \mathbf{\hat{z}} & \left(24d\right) & \mbox{O V} \\ 
\mathbf{B}_{241} & = & x_{14} \, \mathbf{a}_{1} + y_{14} \, \mathbf{a}_{2} + z_{14} \, \mathbf{a}_{3} & = & x_{14}a \, \mathbf{\hat{x}} + y_{14}a \, \mathbf{\hat{y}} + z_{14}a \, \mathbf{\hat{z}} & \left(24d\right) & \mbox{O VI} \\ 
\mathbf{B}_{242} & = & \left(\frac{1}{2} - x_{14}\right) \, \mathbf{a}_{1}-y_{14} \, \mathbf{a}_{2} + \left(\frac{1}{2} +z_{14}\right) \, \mathbf{a}_{3} & = & \left(\frac{1}{2} - x_{14}\right)a \, \mathbf{\hat{x}}-y_{14}a \, \mathbf{\hat{y}} + \left(\frac{1}{2} +z_{14}\right)a \, \mathbf{\hat{z}} & \left(24d\right) & \mbox{O VI} \\ 
\mathbf{B}_{243} & = & -x_{14} \, \mathbf{a}_{1} + \left(\frac{1}{2} +y_{14}\right) \, \mathbf{a}_{2} + \left(\frac{1}{2} - z_{14}\right) \, \mathbf{a}_{3} & = & -x_{14}a \, \mathbf{\hat{x}} + \left(\frac{1}{2} +y_{14}\right)a \, \mathbf{\hat{y}} + \left(\frac{1}{2} - z_{14}\right)a \, \mathbf{\hat{z}} & \left(24d\right) & \mbox{O VI} \\ 
\mathbf{B}_{244} & = & \left(\frac{1}{2} +x_{14}\right) \, \mathbf{a}_{1} + \left(\frac{1}{2} - y_{14}\right) \, \mathbf{a}_{2}-z_{14} \, \mathbf{a}_{3} & = & \left(\frac{1}{2} +x_{14}\right)a \, \mathbf{\hat{x}} + \left(\frac{1}{2} - y_{14}\right)a \, \mathbf{\hat{y}}-z_{14}a \, \mathbf{\hat{z}} & \left(24d\right) & \mbox{O VI} \\ 
\mathbf{B}_{245} & = & z_{14} \, \mathbf{a}_{1} + x_{14} \, \mathbf{a}_{2} + y_{14} \, \mathbf{a}_{3} & = & z_{14}a \, \mathbf{\hat{x}} + x_{14}a \, \mathbf{\hat{y}} + y_{14}a \, \mathbf{\hat{z}} & \left(24d\right) & \mbox{O VI} \\ 
\mathbf{B}_{246} & = & \left(\frac{1}{2} +z_{14}\right) \, \mathbf{a}_{1} + \left(\frac{1}{2} - x_{14}\right) \, \mathbf{a}_{2}-y_{14} \, \mathbf{a}_{3} & = & \left(\frac{1}{2} +z_{14}\right)a \, \mathbf{\hat{x}} + \left(\frac{1}{2} - x_{14}\right)a \, \mathbf{\hat{y}}-y_{14}a \, \mathbf{\hat{z}} & \left(24d\right) & \mbox{O VI} \\ 
\mathbf{B}_{247} & = & \left(\frac{1}{2} - z_{14}\right) \, \mathbf{a}_{1}-x_{14} \, \mathbf{a}_{2} + \left(\frac{1}{2} +y_{14}\right) \, \mathbf{a}_{3} & = & \left(\frac{1}{2} - z_{14}\right)a \, \mathbf{\hat{x}}-x_{14}a \, \mathbf{\hat{y}} + \left(\frac{1}{2} +y_{14}\right)a \, \mathbf{\hat{z}} & \left(24d\right) & \mbox{O VI} \\ 
\mathbf{B}_{248} & = & -z_{14} \, \mathbf{a}_{1} + \left(\frac{1}{2} +x_{14}\right) \, \mathbf{a}_{2} + \left(\frac{1}{2} - y_{14}\right) \, \mathbf{a}_{3} & = & -z_{14}a \, \mathbf{\hat{x}} + \left(\frac{1}{2} +x_{14}\right)a \, \mathbf{\hat{y}} + \left(\frac{1}{2} - y_{14}\right)a \, \mathbf{\hat{z}} & \left(24d\right) & \mbox{O VI} \\ 
\mathbf{B}_{249} & = & y_{14} \, \mathbf{a}_{1} + z_{14} \, \mathbf{a}_{2} + x_{14} \, \mathbf{a}_{3} & = & y_{14}a \, \mathbf{\hat{x}} + z_{14}a \, \mathbf{\hat{y}} + x_{14}a \, \mathbf{\hat{z}} & \left(24d\right) & \mbox{O VI} \\ 
\mathbf{B}_{250} & = & -y_{14} \, \mathbf{a}_{1} + \left(\frac{1}{2} +z_{14}\right) \, \mathbf{a}_{2} + \left(\frac{1}{2} - x_{14}\right) \, \mathbf{a}_{3} & = & -y_{14}a \, \mathbf{\hat{x}} + \left(\frac{1}{2} +z_{14}\right)a \, \mathbf{\hat{y}} + \left(\frac{1}{2} - x_{14}\right)a \, \mathbf{\hat{z}} & \left(24d\right) & \mbox{O VI} \\ 
\mathbf{B}_{251} & = & \left(\frac{1}{2} +y_{14}\right) \, \mathbf{a}_{1} + \left(\frac{1}{2} - z_{14}\right) \, \mathbf{a}_{2}-x_{14} \, \mathbf{a}_{3} & = & \left(\frac{1}{2} +y_{14}\right)a \, \mathbf{\hat{x}} + \left(\frac{1}{2} - z_{14}\right)a \, \mathbf{\hat{y}}-x_{14}a \, \mathbf{\hat{z}} & \left(24d\right) & \mbox{O VI} \\ 
\mathbf{B}_{252} & = & \left(\frac{1}{2} - y_{14}\right) \, \mathbf{a}_{1}-z_{14} \, \mathbf{a}_{2} + \left(\frac{1}{2} +x_{14}\right) \, \mathbf{a}_{3} & = & \left(\frac{1}{2} - y_{14}\right)a \, \mathbf{\hat{x}}-z_{14}a \, \mathbf{\hat{y}} + \left(\frac{1}{2} +x_{14}\right)a \, \mathbf{\hat{z}} & \left(24d\right) & \mbox{O VI} \\ 
\mathbf{B}_{253} & = & -x_{14} \, \mathbf{a}_{1}-y_{14} \, \mathbf{a}_{2}-z_{14} \, \mathbf{a}_{3} & = & -x_{14}a \, \mathbf{\hat{x}}-y_{14}a \, \mathbf{\hat{y}}-z_{14}a \, \mathbf{\hat{z}} & \left(24d\right) & \mbox{O VI} \\ 
\mathbf{B}_{254} & = & \left(\frac{1}{2} +x_{14}\right) \, \mathbf{a}_{1} + y_{14} \, \mathbf{a}_{2} + \left(\frac{1}{2} - z_{14}\right) \, \mathbf{a}_{3} & = & \left(\frac{1}{2} +x_{14}\right)a \, \mathbf{\hat{x}} + y_{14}a \, \mathbf{\hat{y}} + \left(\frac{1}{2} - z_{14}\right)a \, \mathbf{\hat{z}} & \left(24d\right) & \mbox{O VI} \\ 
\mathbf{B}_{255} & = & x_{14} \, \mathbf{a}_{1} + \left(\frac{1}{2} - y_{14}\right) \, \mathbf{a}_{2} + \left(\frac{1}{2} +z_{14}\right) \, \mathbf{a}_{3} & = & x_{14}a \, \mathbf{\hat{x}} + \left(\frac{1}{2} - y_{14}\right)a \, \mathbf{\hat{y}} + \left(\frac{1}{2} +z_{14}\right)a \, \mathbf{\hat{z}} & \left(24d\right) & \mbox{O VI} \\ 
\mathbf{B}_{256} & = & \left(\frac{1}{2} - x_{14}\right) \, \mathbf{a}_{1} + \left(\frac{1}{2} +y_{14}\right) \, \mathbf{a}_{2} + z_{14} \, \mathbf{a}_{3} & = & \left(\frac{1}{2} - x_{14}\right)a \, \mathbf{\hat{x}} + \left(\frac{1}{2} +y_{14}\right)a \, \mathbf{\hat{y}} + z_{14}a \, \mathbf{\hat{z}} & \left(24d\right) & \mbox{O VI} \\ 
\mathbf{B}_{257} & = & -z_{14} \, \mathbf{a}_{1}-x_{14} \, \mathbf{a}_{2}-y_{14} \, \mathbf{a}_{3} & = & -z_{14}a \, \mathbf{\hat{x}}-x_{14}a \, \mathbf{\hat{y}}-y_{14}a \, \mathbf{\hat{z}} & \left(24d\right) & \mbox{O VI} \\ 
\mathbf{B}_{258} & = & \left(\frac{1}{2} - z_{14}\right) \, \mathbf{a}_{1} + \left(\frac{1}{2} +x_{14}\right) \, \mathbf{a}_{2} + y_{14} \, \mathbf{a}_{3} & = & \left(\frac{1}{2} - z_{14}\right)a \, \mathbf{\hat{x}} + \left(\frac{1}{2} +x_{14}\right)a \, \mathbf{\hat{y}} + y_{14}a \, \mathbf{\hat{z}} & \left(24d\right) & \mbox{O VI} \\ 
\mathbf{B}_{259} & = & \left(\frac{1}{2} +z_{14}\right) \, \mathbf{a}_{1} + x_{14} \, \mathbf{a}_{2} + \left(\frac{1}{2} - y_{14}\right) \, \mathbf{a}_{3} & = & \left(\frac{1}{2} +z_{14}\right)a \, \mathbf{\hat{x}} + x_{14}a \, \mathbf{\hat{y}} + \left(\frac{1}{2} - y_{14}\right)a \, \mathbf{\hat{z}} & \left(24d\right) & \mbox{O VI} \\ 
\mathbf{B}_{260} & = & z_{14} \, \mathbf{a}_{1} + \left(\frac{1}{2} - x_{14}\right) \, \mathbf{a}_{2} + \left(\frac{1}{2} +y_{14}\right) \, \mathbf{a}_{3} & = & z_{14}a \, \mathbf{\hat{x}} + \left(\frac{1}{2} - x_{14}\right)a \, \mathbf{\hat{y}} + \left(\frac{1}{2} +y_{14}\right)a \, \mathbf{\hat{z}} & \left(24d\right) & \mbox{O VI} \\ 
\mathbf{B}_{261} & = & -y_{14} \, \mathbf{a}_{1}-z_{14} \, \mathbf{a}_{2}-x_{14} \, \mathbf{a}_{3} & = & -y_{14}a \, \mathbf{\hat{x}}-z_{14}a \, \mathbf{\hat{y}}-x_{14}a \, \mathbf{\hat{z}} & \left(24d\right) & \mbox{O VI} \\ 
\mathbf{B}_{262} & = & y_{14} \, \mathbf{a}_{1} + \left(\frac{1}{2} - z_{14}\right) \, \mathbf{a}_{2} + \left(\frac{1}{2} +x_{14}\right) \, \mathbf{a}_{3} & = & y_{14}a \, \mathbf{\hat{x}} + \left(\frac{1}{2} - z_{14}\right)a \, \mathbf{\hat{y}} + \left(\frac{1}{2} +x_{14}\right)a \, \mathbf{\hat{z}} & \left(24d\right) & \mbox{O VI} \\ 
\mathbf{B}_{263} & = & \left(\frac{1}{2} - y_{14}\right) \, \mathbf{a}_{1} + \left(\frac{1}{2} +z_{14}\right) \, \mathbf{a}_{2} + x_{14} \, \mathbf{a}_{3} & = & \left(\frac{1}{2} - y_{14}\right)a \, \mathbf{\hat{x}} + \left(\frac{1}{2} +z_{14}\right)a \, \mathbf{\hat{y}} + x_{14}a \, \mathbf{\hat{z}} & \left(24d\right) & \mbox{O VI} \\ 
\mathbf{B}_{264} & = & \left(\frac{1}{2} +y_{14}\right) \, \mathbf{a}_{1} + z_{14} \, \mathbf{a}_{2} + \left(\frac{1}{2} - x_{14}\right) \, \mathbf{a}_{3} & = & \left(\frac{1}{2} +y_{14}\right)a \, \mathbf{\hat{x}} + z_{14}a \, \mathbf{\hat{y}} + \left(\frac{1}{2} - x_{14}\right)a \, \mathbf{\hat{z}} & \left(24d\right) & \mbox{O VI} \\ 
\end{longtabu}
\renewcommand{\arraystretch}{1.0}
\noindent \hrulefill
\\
\textbf{References:}
\vspace*{-0.25cm}
\begin{flushleft}
  - \bibentry{Mondal_Acta_Cryst_B_1975}. \\
\end{flushleft}
\noindent \hrulefill
\\
\textbf{Geometry files:}
\\
\noindent  - CIF: pp. {\hyperref[A2B3C6_cP264_205_2d_ab2c2d_6d_cif]{\pageref{A2B3C6_cP264_205_2d_ab2c2d_6d_cif}}} \\
\noindent  - POSCAR: pp. {\hyperref[A2B3C6_cP264_205_2d_ab2c2d_6d_poscar]{\pageref{A2B3C6_cP264_205_2d_ab2c2d_6d_poscar}}} \\
\onecolumn
{\phantomsection\label{A_cP240_205_10d}}
\subsection*{\huge \textbf{{\normalfont \begin{raggedleft}Simple Cubic C$_{60}$ Buckminsterfullerine Structure: \end{raggedleft} \\ A\_cP240\_205\_10d}}}
\noindent \hrulefill
\vspace*{0.25cm}
\begin{figure}[htp]
  \centering
  \vspace{-1em}
  {\includegraphics[width=1\textwidth]{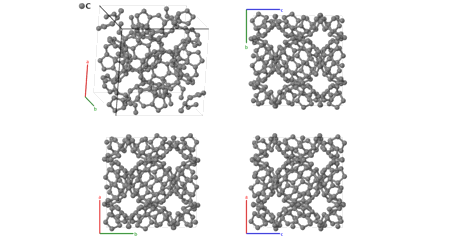}}
\end{figure}
\vspace*{-0.5cm}
\renewcommand{\arraystretch}{1.5}
\begin{equation*}
  \begin{array}{>{$\hspace{-0.15cm}}l<{$}>{$}p{0.5cm}<{$}>{$}p{18.5cm}<{$}}
    \mbox{\large \textbf{Prototype}} &\colon & \ce{C} \\
    \mbox{\large \textbf{\AFLOW\ prototype label}} &\colon & \mbox{A\_cP240\_205\_10d} \\
    \mbox{\large \textbf{\textit{Strukturbericht} designation}} &\colon & \mbox{None} \\
    \mbox{\large \textbf{Pearson symbol}} &\colon & \mbox{cP240} \\
    \mbox{\large \textbf{Space group number}} &\colon & 205 \\
    \mbox{\large \textbf{Space group symbol}} &\colon & Pa\bar{3} \\
    \mbox{\large \textbf{\AFLOW\ prototype command}} &\colon &  \texttt{aflow} \,  \, \texttt{-{}-proto=A\_cP240\_205\_10d } \, \newline \texttt{-{}-params=}{a,x_{1},y_{1},z_{1},x_{2},y_{2},z_{2},x_{3},y_{3},z_{3},x_{4},y_{4},z_{4},x_{5},y_{5},z_{5},x_{6},y_{6},z_{6},x_{7},y_{7},z_{7},} \newline {x_{8},y_{8},z_{8},x_{9},y_{9},z_{9},x_{10},y_{10},z_{10} }
  \end{array}
\end{equation*}
\renewcommand{\arraystretch}{1.0}

\vspace*{-0.25cm}
\noindent \hrulefill
\begin{itemize}
  \item{This is the experimentally determined structure of C$_{60}$
buckminsterfullerne ({\em a.k.a} ``buckyballs'') below 249~K. Above
that temperature the C$_{60}$ molecules are orientationally disordered
and set on a face-centered cubic lattice.  For computational purposes
that structure is approximated by the
\hyperref[A_cF240_202_h2i]{FCC C$_{60}$ buckminsterfullerine
  structure}.
  
}
\end{itemize}

\noindent \parbox{1 \linewidth}{
\noindent \hrulefill
\\
\textbf{Simple Cubic primitive vectors:} \\
\vspace*{-0.25cm}
\begin{tabular}{cc}
  \begin{tabular}{c}
    \parbox{0.6 \linewidth}{
      \renewcommand{\arraystretch}{1.5}
      \begin{equation*}
        \centering
        \begin{array}{ccc}
              \mathbf{a}_1 & = & a \, \mathbf{\hat{x}} \\
    \mathbf{a}_2 & = & a \, \mathbf{\hat{y}} \\
    \mathbf{a}_3 & = & a \, \mathbf{\hat{z}} \\

        \end{array}
      \end{equation*}
    }
    \renewcommand{\arraystretch}{1.0}
  \end{tabular}
  \begin{tabular}{c}
    \includegraphics[width=0.3\linewidth]{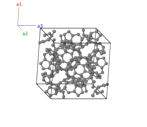} \\
  \end{tabular}
\end{tabular}

}
\vspace*{-0.25cm}

\noindent \hrulefill
\\
\textbf{Basis vectors:}
\vspace*{-0.25cm}
\renewcommand{\arraystretch}{1.5}
\begin{longtabu} to \textwidth{>{\centering $}X[-1,c,c]<{$}>{\centering $}X[-1,c,c]<{$}>{\centering $}X[-1,c,c]<{$}>{\centering $}X[-1,c,c]<{$}>{\centering $}X[-1,c,c]<{$}>{\centering $}X[-1,c,c]<{$}>{\centering $}X[-1,c,c]<{$}}
  & & \mbox{Lattice Coordinates} & & \mbox{Cartesian Coordinates} &\mbox{Wyckoff Position} & \mbox{Atom Type} \\  
  \mathbf{B}_{1} & = & x_{1} \, \mathbf{a}_{1} + y_{1} \, \mathbf{a}_{2} + z_{1} \, \mathbf{a}_{3} & = & x_{1}a \, \mathbf{\hat{x}} + y_{1}a \, \mathbf{\hat{y}} + z_{1}a \, \mathbf{\hat{z}} & \left(24d\right) & \mbox{C I} \\ 
\mathbf{B}_{2} & = & \left(\frac{1}{2} - x_{1}\right) \, \mathbf{a}_{1}-y_{1} \, \mathbf{a}_{2} + \left(\frac{1}{2} +z_{1}\right) \, \mathbf{a}_{3} & = & \left(\frac{1}{2} - x_{1}\right)a \, \mathbf{\hat{x}}-y_{1}a \, \mathbf{\hat{y}} + \left(\frac{1}{2} +z_{1}\right)a \, \mathbf{\hat{z}} & \left(24d\right) & \mbox{C I} \\ 
\mathbf{B}_{3} & = & -x_{1} \, \mathbf{a}_{1} + \left(\frac{1}{2} +y_{1}\right) \, \mathbf{a}_{2} + \left(\frac{1}{2} - z_{1}\right) \, \mathbf{a}_{3} & = & -x_{1}a \, \mathbf{\hat{x}} + \left(\frac{1}{2} +y_{1}\right)a \, \mathbf{\hat{y}} + \left(\frac{1}{2} - z_{1}\right)a \, \mathbf{\hat{z}} & \left(24d\right) & \mbox{C I} \\ 
\mathbf{B}_{4} & = & \left(\frac{1}{2} +x_{1}\right) \, \mathbf{a}_{1} + \left(\frac{1}{2} - y_{1}\right) \, \mathbf{a}_{2}-z_{1} \, \mathbf{a}_{3} & = & \left(\frac{1}{2} +x_{1}\right)a \, \mathbf{\hat{x}} + \left(\frac{1}{2} - y_{1}\right)a \, \mathbf{\hat{y}}-z_{1}a \, \mathbf{\hat{z}} & \left(24d\right) & \mbox{C I} \\ 
\mathbf{B}_{5} & = & z_{1} \, \mathbf{a}_{1} + x_{1} \, \mathbf{a}_{2} + y_{1} \, \mathbf{a}_{3} & = & z_{1}a \, \mathbf{\hat{x}} + x_{1}a \, \mathbf{\hat{y}} + y_{1}a \, \mathbf{\hat{z}} & \left(24d\right) & \mbox{C I} \\ 
\mathbf{B}_{6} & = & \left(\frac{1}{2} +z_{1}\right) \, \mathbf{a}_{1} + \left(\frac{1}{2} - x_{1}\right) \, \mathbf{a}_{2}-y_{1} \, \mathbf{a}_{3} & = & \left(\frac{1}{2} +z_{1}\right)a \, \mathbf{\hat{x}} + \left(\frac{1}{2} - x_{1}\right)a \, \mathbf{\hat{y}}-y_{1}a \, \mathbf{\hat{z}} & \left(24d\right) & \mbox{C I} \\ 
\mathbf{B}_{7} & = & \left(\frac{1}{2} - z_{1}\right) \, \mathbf{a}_{1}-x_{1} \, \mathbf{a}_{2} + \left(\frac{1}{2} +y_{1}\right) \, \mathbf{a}_{3} & = & \left(\frac{1}{2} - z_{1}\right)a \, \mathbf{\hat{x}}-x_{1}a \, \mathbf{\hat{y}} + \left(\frac{1}{2} +y_{1}\right)a \, \mathbf{\hat{z}} & \left(24d\right) & \mbox{C I} \\ 
\mathbf{B}_{8} & = & -z_{1} \, \mathbf{a}_{1} + \left(\frac{1}{2} +x_{1}\right) \, \mathbf{a}_{2} + \left(\frac{1}{2} - y_{1}\right) \, \mathbf{a}_{3} & = & -z_{1}a \, \mathbf{\hat{x}} + \left(\frac{1}{2} +x_{1}\right)a \, \mathbf{\hat{y}} + \left(\frac{1}{2} - y_{1}\right)a \, \mathbf{\hat{z}} & \left(24d\right) & \mbox{C I} \\ 
\mathbf{B}_{9} & = & y_{1} \, \mathbf{a}_{1} + z_{1} \, \mathbf{a}_{2} + x_{1} \, \mathbf{a}_{3} & = & y_{1}a \, \mathbf{\hat{x}} + z_{1}a \, \mathbf{\hat{y}} + x_{1}a \, \mathbf{\hat{z}} & \left(24d\right) & \mbox{C I} \\ 
\mathbf{B}_{10} & = & -y_{1} \, \mathbf{a}_{1} + \left(\frac{1}{2} +z_{1}\right) \, \mathbf{a}_{2} + \left(\frac{1}{2} - x_{1}\right) \, \mathbf{a}_{3} & = & -y_{1}a \, \mathbf{\hat{x}} + \left(\frac{1}{2} +z_{1}\right)a \, \mathbf{\hat{y}} + \left(\frac{1}{2} - x_{1}\right)a \, \mathbf{\hat{z}} & \left(24d\right) & \mbox{C I} \\ 
\mathbf{B}_{11} & = & \left(\frac{1}{2} +y_{1}\right) \, \mathbf{a}_{1} + \left(\frac{1}{2} - z_{1}\right) \, \mathbf{a}_{2}-x_{1} \, \mathbf{a}_{3} & = & \left(\frac{1}{2} +y_{1}\right)a \, \mathbf{\hat{x}} + \left(\frac{1}{2} - z_{1}\right)a \, \mathbf{\hat{y}}-x_{1}a \, \mathbf{\hat{z}} & \left(24d\right) & \mbox{C I} \\ 
\mathbf{B}_{12} & = & \left(\frac{1}{2} - y_{1}\right) \, \mathbf{a}_{1}-z_{1} \, \mathbf{a}_{2} + \left(\frac{1}{2} +x_{1}\right) \, \mathbf{a}_{3} & = & \left(\frac{1}{2} - y_{1}\right)a \, \mathbf{\hat{x}}-z_{1}a \, \mathbf{\hat{y}} + \left(\frac{1}{2} +x_{1}\right)a \, \mathbf{\hat{z}} & \left(24d\right) & \mbox{C I} \\ 
\mathbf{B}_{13} & = & -x_{1} \, \mathbf{a}_{1}-y_{1} \, \mathbf{a}_{2}-z_{1} \, \mathbf{a}_{3} & = & -x_{1}a \, \mathbf{\hat{x}}-y_{1}a \, \mathbf{\hat{y}}-z_{1}a \, \mathbf{\hat{z}} & \left(24d\right) & \mbox{C I} \\ 
\mathbf{B}_{14} & = & \left(\frac{1}{2} +x_{1}\right) \, \mathbf{a}_{1} + y_{1} \, \mathbf{a}_{2} + \left(\frac{1}{2} - z_{1}\right) \, \mathbf{a}_{3} & = & \left(\frac{1}{2} +x_{1}\right)a \, \mathbf{\hat{x}} + y_{1}a \, \mathbf{\hat{y}} + \left(\frac{1}{2} - z_{1}\right)a \, \mathbf{\hat{z}} & \left(24d\right) & \mbox{C I} \\ 
\mathbf{B}_{15} & = & x_{1} \, \mathbf{a}_{1} + \left(\frac{1}{2} - y_{1}\right) \, \mathbf{a}_{2} + \left(\frac{1}{2} +z_{1}\right) \, \mathbf{a}_{3} & = & x_{1}a \, \mathbf{\hat{x}} + \left(\frac{1}{2} - y_{1}\right)a \, \mathbf{\hat{y}} + \left(\frac{1}{2} +z_{1}\right)a \, \mathbf{\hat{z}} & \left(24d\right) & \mbox{C I} \\ 
\mathbf{B}_{16} & = & \left(\frac{1}{2} - x_{1}\right) \, \mathbf{a}_{1} + \left(\frac{1}{2} +y_{1}\right) \, \mathbf{a}_{2} + z_{1} \, \mathbf{a}_{3} & = & \left(\frac{1}{2} - x_{1}\right)a \, \mathbf{\hat{x}} + \left(\frac{1}{2} +y_{1}\right)a \, \mathbf{\hat{y}} + z_{1}a \, \mathbf{\hat{z}} & \left(24d\right) & \mbox{C I} \\ 
\mathbf{B}_{17} & = & -z_{1} \, \mathbf{a}_{1}-x_{1} \, \mathbf{a}_{2}-y_{1} \, \mathbf{a}_{3} & = & -z_{1}a \, \mathbf{\hat{x}}-x_{1}a \, \mathbf{\hat{y}}-y_{1}a \, \mathbf{\hat{z}} & \left(24d\right) & \mbox{C I} \\ 
\mathbf{B}_{18} & = & \left(\frac{1}{2} - z_{1}\right) \, \mathbf{a}_{1} + \left(\frac{1}{2} +x_{1}\right) \, \mathbf{a}_{2} + y_{1} \, \mathbf{a}_{3} & = & \left(\frac{1}{2} - z_{1}\right)a \, \mathbf{\hat{x}} + \left(\frac{1}{2} +x_{1}\right)a \, \mathbf{\hat{y}} + y_{1}a \, \mathbf{\hat{z}} & \left(24d\right) & \mbox{C I} \\ 
\mathbf{B}_{19} & = & \left(\frac{1}{2} +z_{1}\right) \, \mathbf{a}_{1} + x_{1} \, \mathbf{a}_{2} + \left(\frac{1}{2} - y_{1}\right) \, \mathbf{a}_{3} & = & \left(\frac{1}{2} +z_{1}\right)a \, \mathbf{\hat{x}} + x_{1}a \, \mathbf{\hat{y}} + \left(\frac{1}{2} - y_{1}\right)a \, \mathbf{\hat{z}} & \left(24d\right) & \mbox{C I} \\ 
\mathbf{B}_{20} & = & z_{1} \, \mathbf{a}_{1} + \left(\frac{1}{2} - x_{1}\right) \, \mathbf{a}_{2} + \left(\frac{1}{2} +y_{1}\right) \, \mathbf{a}_{3} & = & z_{1}a \, \mathbf{\hat{x}} + \left(\frac{1}{2} - x_{1}\right)a \, \mathbf{\hat{y}} + \left(\frac{1}{2} +y_{1}\right)a \, \mathbf{\hat{z}} & \left(24d\right) & \mbox{C I} \\ 
\mathbf{B}_{21} & = & -y_{1} \, \mathbf{a}_{1}-z_{1} \, \mathbf{a}_{2}-x_{1} \, \mathbf{a}_{3} & = & -y_{1}a \, \mathbf{\hat{x}}-z_{1}a \, \mathbf{\hat{y}}-x_{1}a \, \mathbf{\hat{z}} & \left(24d\right) & \mbox{C I} \\ 
\mathbf{B}_{22} & = & y_{1} \, \mathbf{a}_{1} + \left(\frac{1}{2} - z_{1}\right) \, \mathbf{a}_{2} + \left(\frac{1}{2} +x_{1}\right) \, \mathbf{a}_{3} & = & y_{1}a \, \mathbf{\hat{x}} + \left(\frac{1}{2} - z_{1}\right)a \, \mathbf{\hat{y}} + \left(\frac{1}{2} +x_{1}\right)a \, \mathbf{\hat{z}} & \left(24d\right) & \mbox{C I} \\ 
\mathbf{B}_{23} & = & \left(\frac{1}{2} - y_{1}\right) \, \mathbf{a}_{1} + \left(\frac{1}{2} +z_{1}\right) \, \mathbf{a}_{2} + x_{1} \, \mathbf{a}_{3} & = & \left(\frac{1}{2} - y_{1}\right)a \, \mathbf{\hat{x}} + \left(\frac{1}{2} +z_{1}\right)a \, \mathbf{\hat{y}} + x_{1}a \, \mathbf{\hat{z}} & \left(24d\right) & \mbox{C I} \\ 
\mathbf{B}_{24} & = & \left(\frac{1}{2} +y_{1}\right) \, \mathbf{a}_{1} + z_{1} \, \mathbf{a}_{2} + \left(\frac{1}{2} - x_{1}\right) \, \mathbf{a}_{3} & = & \left(\frac{1}{2} +y_{1}\right)a \, \mathbf{\hat{x}} + z_{1}a \, \mathbf{\hat{y}} + \left(\frac{1}{2} - x_{1}\right)a \, \mathbf{\hat{z}} & \left(24d\right) & \mbox{C I} \\ 
\mathbf{B}_{25} & = & x_{2} \, \mathbf{a}_{1} + y_{2} \, \mathbf{a}_{2} + z_{2} \, \mathbf{a}_{3} & = & x_{2}a \, \mathbf{\hat{x}} + y_{2}a \, \mathbf{\hat{y}} + z_{2}a \, \mathbf{\hat{z}} & \left(24d\right) & \mbox{C II} \\ 
\mathbf{B}_{26} & = & \left(\frac{1}{2} - x_{2}\right) \, \mathbf{a}_{1}-y_{2} \, \mathbf{a}_{2} + \left(\frac{1}{2} +z_{2}\right) \, \mathbf{a}_{3} & = & \left(\frac{1}{2} - x_{2}\right)a \, \mathbf{\hat{x}}-y_{2}a \, \mathbf{\hat{y}} + \left(\frac{1}{2} +z_{2}\right)a \, \mathbf{\hat{z}} & \left(24d\right) & \mbox{C II} \\ 
\mathbf{B}_{27} & = & -x_{2} \, \mathbf{a}_{1} + \left(\frac{1}{2} +y_{2}\right) \, \mathbf{a}_{2} + \left(\frac{1}{2} - z_{2}\right) \, \mathbf{a}_{3} & = & -x_{2}a \, \mathbf{\hat{x}} + \left(\frac{1}{2} +y_{2}\right)a \, \mathbf{\hat{y}} + \left(\frac{1}{2} - z_{2}\right)a \, \mathbf{\hat{z}} & \left(24d\right) & \mbox{C II} \\ 
\mathbf{B}_{28} & = & \left(\frac{1}{2} +x_{2}\right) \, \mathbf{a}_{1} + \left(\frac{1}{2} - y_{2}\right) \, \mathbf{a}_{2}-z_{2} \, \mathbf{a}_{3} & = & \left(\frac{1}{2} +x_{2}\right)a \, \mathbf{\hat{x}} + \left(\frac{1}{2} - y_{2}\right)a \, \mathbf{\hat{y}}-z_{2}a \, \mathbf{\hat{z}} & \left(24d\right) & \mbox{C II} \\ 
\mathbf{B}_{29} & = & z_{2} \, \mathbf{a}_{1} + x_{2} \, \mathbf{a}_{2} + y_{2} \, \mathbf{a}_{3} & = & z_{2}a \, \mathbf{\hat{x}} + x_{2}a \, \mathbf{\hat{y}} + y_{2}a \, \mathbf{\hat{z}} & \left(24d\right) & \mbox{C II} \\ 
\mathbf{B}_{30} & = & \left(\frac{1}{2} +z_{2}\right) \, \mathbf{a}_{1} + \left(\frac{1}{2} - x_{2}\right) \, \mathbf{a}_{2}-y_{2} \, \mathbf{a}_{3} & = & \left(\frac{1}{2} +z_{2}\right)a \, \mathbf{\hat{x}} + \left(\frac{1}{2} - x_{2}\right)a \, \mathbf{\hat{y}}-y_{2}a \, \mathbf{\hat{z}} & \left(24d\right) & \mbox{C II} \\ 
\mathbf{B}_{31} & = & \left(\frac{1}{2} - z_{2}\right) \, \mathbf{a}_{1}-x_{2} \, \mathbf{a}_{2} + \left(\frac{1}{2} +y_{2}\right) \, \mathbf{a}_{3} & = & \left(\frac{1}{2} - z_{2}\right)a \, \mathbf{\hat{x}}-x_{2}a \, \mathbf{\hat{y}} + \left(\frac{1}{2} +y_{2}\right)a \, \mathbf{\hat{z}} & \left(24d\right) & \mbox{C II} \\ 
\mathbf{B}_{32} & = & -z_{2} \, \mathbf{a}_{1} + \left(\frac{1}{2} +x_{2}\right) \, \mathbf{a}_{2} + \left(\frac{1}{2} - y_{2}\right) \, \mathbf{a}_{3} & = & -z_{2}a \, \mathbf{\hat{x}} + \left(\frac{1}{2} +x_{2}\right)a \, \mathbf{\hat{y}} + \left(\frac{1}{2} - y_{2}\right)a \, \mathbf{\hat{z}} & \left(24d\right) & \mbox{C II} \\ 
\mathbf{B}_{33} & = & y_{2} \, \mathbf{a}_{1} + z_{2} \, \mathbf{a}_{2} + x_{2} \, \mathbf{a}_{3} & = & y_{2}a \, \mathbf{\hat{x}} + z_{2}a \, \mathbf{\hat{y}} + x_{2}a \, \mathbf{\hat{z}} & \left(24d\right) & \mbox{C II} \\ 
\mathbf{B}_{34} & = & -y_{2} \, \mathbf{a}_{1} + \left(\frac{1}{2} +z_{2}\right) \, \mathbf{a}_{2} + \left(\frac{1}{2} - x_{2}\right) \, \mathbf{a}_{3} & = & -y_{2}a \, \mathbf{\hat{x}} + \left(\frac{1}{2} +z_{2}\right)a \, \mathbf{\hat{y}} + \left(\frac{1}{2} - x_{2}\right)a \, \mathbf{\hat{z}} & \left(24d\right) & \mbox{C II} \\ 
\mathbf{B}_{35} & = & \left(\frac{1}{2} +y_{2}\right) \, \mathbf{a}_{1} + \left(\frac{1}{2} - z_{2}\right) \, \mathbf{a}_{2}-x_{2} \, \mathbf{a}_{3} & = & \left(\frac{1}{2} +y_{2}\right)a \, \mathbf{\hat{x}} + \left(\frac{1}{2} - z_{2}\right)a \, \mathbf{\hat{y}}-x_{2}a \, \mathbf{\hat{z}} & \left(24d\right) & \mbox{C II} \\ 
\mathbf{B}_{36} & = & \left(\frac{1}{2} - y_{2}\right) \, \mathbf{a}_{1}-z_{2} \, \mathbf{a}_{2} + \left(\frac{1}{2} +x_{2}\right) \, \mathbf{a}_{3} & = & \left(\frac{1}{2} - y_{2}\right)a \, \mathbf{\hat{x}}-z_{2}a \, \mathbf{\hat{y}} + \left(\frac{1}{2} +x_{2}\right)a \, \mathbf{\hat{z}} & \left(24d\right) & \mbox{C II} \\ 
\mathbf{B}_{37} & = & -x_{2} \, \mathbf{a}_{1}-y_{2} \, \mathbf{a}_{2}-z_{2} \, \mathbf{a}_{3} & = & -x_{2}a \, \mathbf{\hat{x}}-y_{2}a \, \mathbf{\hat{y}}-z_{2}a \, \mathbf{\hat{z}} & \left(24d\right) & \mbox{C II} \\ 
\mathbf{B}_{38} & = & \left(\frac{1}{2} +x_{2}\right) \, \mathbf{a}_{1} + y_{2} \, \mathbf{a}_{2} + \left(\frac{1}{2} - z_{2}\right) \, \mathbf{a}_{3} & = & \left(\frac{1}{2} +x_{2}\right)a \, \mathbf{\hat{x}} + y_{2}a \, \mathbf{\hat{y}} + \left(\frac{1}{2} - z_{2}\right)a \, \mathbf{\hat{z}} & \left(24d\right) & \mbox{C II} \\ 
\mathbf{B}_{39} & = & x_{2} \, \mathbf{a}_{1} + \left(\frac{1}{2} - y_{2}\right) \, \mathbf{a}_{2} + \left(\frac{1}{2} +z_{2}\right) \, \mathbf{a}_{3} & = & x_{2}a \, \mathbf{\hat{x}} + \left(\frac{1}{2} - y_{2}\right)a \, \mathbf{\hat{y}} + \left(\frac{1}{2} +z_{2}\right)a \, \mathbf{\hat{z}} & \left(24d\right) & \mbox{C II} \\ 
\mathbf{B}_{40} & = & \left(\frac{1}{2} - x_{2}\right) \, \mathbf{a}_{1} + \left(\frac{1}{2} +y_{2}\right) \, \mathbf{a}_{2} + z_{2} \, \mathbf{a}_{3} & = & \left(\frac{1}{2} - x_{2}\right)a \, \mathbf{\hat{x}} + \left(\frac{1}{2} +y_{2}\right)a \, \mathbf{\hat{y}} + z_{2}a \, \mathbf{\hat{z}} & \left(24d\right) & \mbox{C II} \\ 
\mathbf{B}_{41} & = & -z_{2} \, \mathbf{a}_{1}-x_{2} \, \mathbf{a}_{2}-y_{2} \, \mathbf{a}_{3} & = & -z_{2}a \, \mathbf{\hat{x}}-x_{2}a \, \mathbf{\hat{y}}-y_{2}a \, \mathbf{\hat{z}} & \left(24d\right) & \mbox{C II} \\ 
\mathbf{B}_{42} & = & \left(\frac{1}{2} - z_{2}\right) \, \mathbf{a}_{1} + \left(\frac{1}{2} +x_{2}\right) \, \mathbf{a}_{2} + y_{2} \, \mathbf{a}_{3} & = & \left(\frac{1}{2} - z_{2}\right)a \, \mathbf{\hat{x}} + \left(\frac{1}{2} +x_{2}\right)a \, \mathbf{\hat{y}} + y_{2}a \, \mathbf{\hat{z}} & \left(24d\right) & \mbox{C II} \\ 
\mathbf{B}_{43} & = & \left(\frac{1}{2} +z_{2}\right) \, \mathbf{a}_{1} + x_{2} \, \mathbf{a}_{2} + \left(\frac{1}{2} - y_{2}\right) \, \mathbf{a}_{3} & = & \left(\frac{1}{2} +z_{2}\right)a \, \mathbf{\hat{x}} + x_{2}a \, \mathbf{\hat{y}} + \left(\frac{1}{2} - y_{2}\right)a \, \mathbf{\hat{z}} & \left(24d\right) & \mbox{C II} \\ 
\mathbf{B}_{44} & = & z_{2} \, \mathbf{a}_{1} + \left(\frac{1}{2} - x_{2}\right) \, \mathbf{a}_{2} + \left(\frac{1}{2} +y_{2}\right) \, \mathbf{a}_{3} & = & z_{2}a \, \mathbf{\hat{x}} + \left(\frac{1}{2} - x_{2}\right)a \, \mathbf{\hat{y}} + \left(\frac{1}{2} +y_{2}\right)a \, \mathbf{\hat{z}} & \left(24d\right) & \mbox{C II} \\ 
\mathbf{B}_{45} & = & -y_{2} \, \mathbf{a}_{1}-z_{2} \, \mathbf{a}_{2}-x_{2} \, \mathbf{a}_{3} & = & -y_{2}a \, \mathbf{\hat{x}}-z_{2}a \, \mathbf{\hat{y}}-x_{2}a \, \mathbf{\hat{z}} & \left(24d\right) & \mbox{C II} \\ 
\mathbf{B}_{46} & = & y_{2} \, \mathbf{a}_{1} + \left(\frac{1}{2} - z_{2}\right) \, \mathbf{a}_{2} + \left(\frac{1}{2} +x_{2}\right) \, \mathbf{a}_{3} & = & y_{2}a \, \mathbf{\hat{x}} + \left(\frac{1}{2} - z_{2}\right)a \, \mathbf{\hat{y}} + \left(\frac{1}{2} +x_{2}\right)a \, \mathbf{\hat{z}} & \left(24d\right) & \mbox{C II} \\ 
\mathbf{B}_{47} & = & \left(\frac{1}{2} - y_{2}\right) \, \mathbf{a}_{1} + \left(\frac{1}{2} +z_{2}\right) \, \mathbf{a}_{2} + x_{2} \, \mathbf{a}_{3} & = & \left(\frac{1}{2} - y_{2}\right)a \, \mathbf{\hat{x}} + \left(\frac{1}{2} +z_{2}\right)a \, \mathbf{\hat{y}} + x_{2}a \, \mathbf{\hat{z}} & \left(24d\right) & \mbox{C II} \\ 
\mathbf{B}_{48} & = & \left(\frac{1}{2} +y_{2}\right) \, \mathbf{a}_{1} + z_{2} \, \mathbf{a}_{2} + \left(\frac{1}{2} - x_{2}\right) \, \mathbf{a}_{3} & = & \left(\frac{1}{2} +y_{2}\right)a \, \mathbf{\hat{x}} + z_{2}a \, \mathbf{\hat{y}} + \left(\frac{1}{2} - x_{2}\right)a \, \mathbf{\hat{z}} & \left(24d\right) & \mbox{C II} \\ 
\mathbf{B}_{49} & = & x_{3} \, \mathbf{a}_{1} + y_{3} \, \mathbf{a}_{2} + z_{3} \, \mathbf{a}_{3} & = & x_{3}a \, \mathbf{\hat{x}} + y_{3}a \, \mathbf{\hat{y}} + z_{3}a \, \mathbf{\hat{z}} & \left(24d\right) & \mbox{C III} \\ 
\mathbf{B}_{50} & = & \left(\frac{1}{2} - x_{3}\right) \, \mathbf{a}_{1}-y_{3} \, \mathbf{a}_{2} + \left(\frac{1}{2} +z_{3}\right) \, \mathbf{a}_{3} & = & \left(\frac{1}{2} - x_{3}\right)a \, \mathbf{\hat{x}}-y_{3}a \, \mathbf{\hat{y}} + \left(\frac{1}{2} +z_{3}\right)a \, \mathbf{\hat{z}} & \left(24d\right) & \mbox{C III} \\ 
\mathbf{B}_{51} & = & -x_{3} \, \mathbf{a}_{1} + \left(\frac{1}{2} +y_{3}\right) \, \mathbf{a}_{2} + \left(\frac{1}{2} - z_{3}\right) \, \mathbf{a}_{3} & = & -x_{3}a \, \mathbf{\hat{x}} + \left(\frac{1}{2} +y_{3}\right)a \, \mathbf{\hat{y}} + \left(\frac{1}{2} - z_{3}\right)a \, \mathbf{\hat{z}} & \left(24d\right) & \mbox{C III} \\ 
\mathbf{B}_{52} & = & \left(\frac{1}{2} +x_{3}\right) \, \mathbf{a}_{1} + \left(\frac{1}{2} - y_{3}\right) \, \mathbf{a}_{2}-z_{3} \, \mathbf{a}_{3} & = & \left(\frac{1}{2} +x_{3}\right)a \, \mathbf{\hat{x}} + \left(\frac{1}{2} - y_{3}\right)a \, \mathbf{\hat{y}}-z_{3}a \, \mathbf{\hat{z}} & \left(24d\right) & \mbox{C III} \\ 
\mathbf{B}_{53} & = & z_{3} \, \mathbf{a}_{1} + x_{3} \, \mathbf{a}_{2} + y_{3} \, \mathbf{a}_{3} & = & z_{3}a \, \mathbf{\hat{x}} + x_{3}a \, \mathbf{\hat{y}} + y_{3}a \, \mathbf{\hat{z}} & \left(24d\right) & \mbox{C III} \\ 
\mathbf{B}_{54} & = & \left(\frac{1}{2} +z_{3}\right) \, \mathbf{a}_{1} + \left(\frac{1}{2} - x_{3}\right) \, \mathbf{a}_{2}-y_{3} \, \mathbf{a}_{3} & = & \left(\frac{1}{2} +z_{3}\right)a \, \mathbf{\hat{x}} + \left(\frac{1}{2} - x_{3}\right)a \, \mathbf{\hat{y}}-y_{3}a \, \mathbf{\hat{z}} & \left(24d\right) & \mbox{C III} \\ 
\mathbf{B}_{55} & = & \left(\frac{1}{2} - z_{3}\right) \, \mathbf{a}_{1}-x_{3} \, \mathbf{a}_{2} + \left(\frac{1}{2} +y_{3}\right) \, \mathbf{a}_{3} & = & \left(\frac{1}{2} - z_{3}\right)a \, \mathbf{\hat{x}}-x_{3}a \, \mathbf{\hat{y}} + \left(\frac{1}{2} +y_{3}\right)a \, \mathbf{\hat{z}} & \left(24d\right) & \mbox{C III} \\ 
\mathbf{B}_{56} & = & -z_{3} \, \mathbf{a}_{1} + \left(\frac{1}{2} +x_{3}\right) \, \mathbf{a}_{2} + \left(\frac{1}{2} - y_{3}\right) \, \mathbf{a}_{3} & = & -z_{3}a \, \mathbf{\hat{x}} + \left(\frac{1}{2} +x_{3}\right)a \, \mathbf{\hat{y}} + \left(\frac{1}{2} - y_{3}\right)a \, \mathbf{\hat{z}} & \left(24d\right) & \mbox{C III} \\ 
\mathbf{B}_{57} & = & y_{3} \, \mathbf{a}_{1} + z_{3} \, \mathbf{a}_{2} + x_{3} \, \mathbf{a}_{3} & = & y_{3}a \, \mathbf{\hat{x}} + z_{3}a \, \mathbf{\hat{y}} + x_{3}a \, \mathbf{\hat{z}} & \left(24d\right) & \mbox{C III} \\ 
\mathbf{B}_{58} & = & -y_{3} \, \mathbf{a}_{1} + \left(\frac{1}{2} +z_{3}\right) \, \mathbf{a}_{2} + \left(\frac{1}{2} - x_{3}\right) \, \mathbf{a}_{3} & = & -y_{3}a \, \mathbf{\hat{x}} + \left(\frac{1}{2} +z_{3}\right)a \, \mathbf{\hat{y}} + \left(\frac{1}{2} - x_{3}\right)a \, \mathbf{\hat{z}} & \left(24d\right) & \mbox{C III} \\ 
\mathbf{B}_{59} & = & \left(\frac{1}{2} +y_{3}\right) \, \mathbf{a}_{1} + \left(\frac{1}{2} - z_{3}\right) \, \mathbf{a}_{2}-x_{3} \, \mathbf{a}_{3} & = & \left(\frac{1}{2} +y_{3}\right)a \, \mathbf{\hat{x}} + \left(\frac{1}{2} - z_{3}\right)a \, \mathbf{\hat{y}}-x_{3}a \, \mathbf{\hat{z}} & \left(24d\right) & \mbox{C III} \\ 
\mathbf{B}_{60} & = & \left(\frac{1}{2} - y_{3}\right) \, \mathbf{a}_{1}-z_{3} \, \mathbf{a}_{2} + \left(\frac{1}{2} +x_{3}\right) \, \mathbf{a}_{3} & = & \left(\frac{1}{2} - y_{3}\right)a \, \mathbf{\hat{x}}-z_{3}a \, \mathbf{\hat{y}} + \left(\frac{1}{2} +x_{3}\right)a \, \mathbf{\hat{z}} & \left(24d\right) & \mbox{C III} \\ 
\mathbf{B}_{61} & = & -x_{3} \, \mathbf{a}_{1}-y_{3} \, \mathbf{a}_{2}-z_{3} \, \mathbf{a}_{3} & = & -x_{3}a \, \mathbf{\hat{x}}-y_{3}a \, \mathbf{\hat{y}}-z_{3}a \, \mathbf{\hat{z}} & \left(24d\right) & \mbox{C III} \\ 
\mathbf{B}_{62} & = & \left(\frac{1}{2} +x_{3}\right) \, \mathbf{a}_{1} + y_{3} \, \mathbf{a}_{2} + \left(\frac{1}{2} - z_{3}\right) \, \mathbf{a}_{3} & = & \left(\frac{1}{2} +x_{3}\right)a \, \mathbf{\hat{x}} + y_{3}a \, \mathbf{\hat{y}} + \left(\frac{1}{2} - z_{3}\right)a \, \mathbf{\hat{z}} & \left(24d\right) & \mbox{C III} \\ 
\mathbf{B}_{63} & = & x_{3} \, \mathbf{a}_{1} + \left(\frac{1}{2} - y_{3}\right) \, \mathbf{a}_{2} + \left(\frac{1}{2} +z_{3}\right) \, \mathbf{a}_{3} & = & x_{3}a \, \mathbf{\hat{x}} + \left(\frac{1}{2} - y_{3}\right)a \, \mathbf{\hat{y}} + \left(\frac{1}{2} +z_{3}\right)a \, \mathbf{\hat{z}} & \left(24d\right) & \mbox{C III} \\ 
\mathbf{B}_{64} & = & \left(\frac{1}{2} - x_{3}\right) \, \mathbf{a}_{1} + \left(\frac{1}{2} +y_{3}\right) \, \mathbf{a}_{2} + z_{3} \, \mathbf{a}_{3} & = & \left(\frac{1}{2} - x_{3}\right)a \, \mathbf{\hat{x}} + \left(\frac{1}{2} +y_{3}\right)a \, \mathbf{\hat{y}} + z_{3}a \, \mathbf{\hat{z}} & \left(24d\right) & \mbox{C III} \\ 
\mathbf{B}_{65} & = & -z_{3} \, \mathbf{a}_{1}-x_{3} \, \mathbf{a}_{2}-y_{3} \, \mathbf{a}_{3} & = & -z_{3}a \, \mathbf{\hat{x}}-x_{3}a \, \mathbf{\hat{y}}-y_{3}a \, \mathbf{\hat{z}} & \left(24d\right) & \mbox{C III} \\ 
\mathbf{B}_{66} & = & \left(\frac{1}{2} - z_{3}\right) \, \mathbf{a}_{1} + \left(\frac{1}{2} +x_{3}\right) \, \mathbf{a}_{2} + y_{3} \, \mathbf{a}_{3} & = & \left(\frac{1}{2} - z_{3}\right)a \, \mathbf{\hat{x}} + \left(\frac{1}{2} +x_{3}\right)a \, \mathbf{\hat{y}} + y_{3}a \, \mathbf{\hat{z}} & \left(24d\right) & \mbox{C III} \\ 
\mathbf{B}_{67} & = & \left(\frac{1}{2} +z_{3}\right) \, \mathbf{a}_{1} + x_{3} \, \mathbf{a}_{2} + \left(\frac{1}{2} - y_{3}\right) \, \mathbf{a}_{3} & = & \left(\frac{1}{2} +z_{3}\right)a \, \mathbf{\hat{x}} + x_{3}a \, \mathbf{\hat{y}} + \left(\frac{1}{2} - y_{3}\right)a \, \mathbf{\hat{z}} & \left(24d\right) & \mbox{C III} \\ 
\mathbf{B}_{68} & = & z_{3} \, \mathbf{a}_{1} + \left(\frac{1}{2} - x_{3}\right) \, \mathbf{a}_{2} + \left(\frac{1}{2} +y_{3}\right) \, \mathbf{a}_{3} & = & z_{3}a \, \mathbf{\hat{x}} + \left(\frac{1}{2} - x_{3}\right)a \, \mathbf{\hat{y}} + \left(\frac{1}{2} +y_{3}\right)a \, \mathbf{\hat{z}} & \left(24d\right) & \mbox{C III} \\ 
\mathbf{B}_{69} & = & -y_{3} \, \mathbf{a}_{1}-z_{3} \, \mathbf{a}_{2}-x_{3} \, \mathbf{a}_{3} & = & -y_{3}a \, \mathbf{\hat{x}}-z_{3}a \, \mathbf{\hat{y}}-x_{3}a \, \mathbf{\hat{z}} & \left(24d\right) & \mbox{C III} \\ 
\mathbf{B}_{70} & = & y_{3} \, \mathbf{a}_{1} + \left(\frac{1}{2} - z_{3}\right) \, \mathbf{a}_{2} + \left(\frac{1}{2} +x_{3}\right) \, \mathbf{a}_{3} & = & y_{3}a \, \mathbf{\hat{x}} + \left(\frac{1}{2} - z_{3}\right)a \, \mathbf{\hat{y}} + \left(\frac{1}{2} +x_{3}\right)a \, \mathbf{\hat{z}} & \left(24d\right) & \mbox{C III} \\ 
\mathbf{B}_{71} & = & \left(\frac{1}{2} - y_{3}\right) \, \mathbf{a}_{1} + \left(\frac{1}{2} +z_{3}\right) \, \mathbf{a}_{2} + x_{3} \, \mathbf{a}_{3} & = & \left(\frac{1}{2} - y_{3}\right)a \, \mathbf{\hat{x}} + \left(\frac{1}{2} +z_{3}\right)a \, \mathbf{\hat{y}} + x_{3}a \, \mathbf{\hat{z}} & \left(24d\right) & \mbox{C III} \\ 
\mathbf{B}_{72} & = & \left(\frac{1}{2} +y_{3}\right) \, \mathbf{a}_{1} + z_{3} \, \mathbf{a}_{2} + \left(\frac{1}{2} - x_{3}\right) \, \mathbf{a}_{3} & = & \left(\frac{1}{2} +y_{3}\right)a \, \mathbf{\hat{x}} + z_{3}a \, \mathbf{\hat{y}} + \left(\frac{1}{2} - x_{3}\right)a \, \mathbf{\hat{z}} & \left(24d\right) & \mbox{C III} \\ 
\mathbf{B}_{73} & = & x_{4} \, \mathbf{a}_{1} + y_{4} \, \mathbf{a}_{2} + z_{4} \, \mathbf{a}_{3} & = & x_{4}a \, \mathbf{\hat{x}} + y_{4}a \, \mathbf{\hat{y}} + z_{4}a \, \mathbf{\hat{z}} & \left(24d\right) & \mbox{C IV} \\ 
\mathbf{B}_{74} & = & \left(\frac{1}{2} - x_{4}\right) \, \mathbf{a}_{1}-y_{4} \, \mathbf{a}_{2} + \left(\frac{1}{2} +z_{4}\right) \, \mathbf{a}_{3} & = & \left(\frac{1}{2} - x_{4}\right)a \, \mathbf{\hat{x}}-y_{4}a \, \mathbf{\hat{y}} + \left(\frac{1}{2} +z_{4}\right)a \, \mathbf{\hat{z}} & \left(24d\right) & \mbox{C IV} \\ 
\mathbf{B}_{75} & = & -x_{4} \, \mathbf{a}_{1} + \left(\frac{1}{2} +y_{4}\right) \, \mathbf{a}_{2} + \left(\frac{1}{2} - z_{4}\right) \, \mathbf{a}_{3} & = & -x_{4}a \, \mathbf{\hat{x}} + \left(\frac{1}{2} +y_{4}\right)a \, \mathbf{\hat{y}} + \left(\frac{1}{2} - z_{4}\right)a \, \mathbf{\hat{z}} & \left(24d\right) & \mbox{C IV} \\ 
\mathbf{B}_{76} & = & \left(\frac{1}{2} +x_{4}\right) \, \mathbf{a}_{1} + \left(\frac{1}{2} - y_{4}\right) \, \mathbf{a}_{2}-z_{4} \, \mathbf{a}_{3} & = & \left(\frac{1}{2} +x_{4}\right)a \, \mathbf{\hat{x}} + \left(\frac{1}{2} - y_{4}\right)a \, \mathbf{\hat{y}}-z_{4}a \, \mathbf{\hat{z}} & \left(24d\right) & \mbox{C IV} \\ 
\mathbf{B}_{77} & = & z_{4} \, \mathbf{a}_{1} + x_{4} \, \mathbf{a}_{2} + y_{4} \, \mathbf{a}_{3} & = & z_{4}a \, \mathbf{\hat{x}} + x_{4}a \, \mathbf{\hat{y}} + y_{4}a \, \mathbf{\hat{z}} & \left(24d\right) & \mbox{C IV} \\ 
\mathbf{B}_{78} & = & \left(\frac{1}{2} +z_{4}\right) \, \mathbf{a}_{1} + \left(\frac{1}{2} - x_{4}\right) \, \mathbf{a}_{2}-y_{4} \, \mathbf{a}_{3} & = & \left(\frac{1}{2} +z_{4}\right)a \, \mathbf{\hat{x}} + \left(\frac{1}{2} - x_{4}\right)a \, \mathbf{\hat{y}}-y_{4}a \, \mathbf{\hat{z}} & \left(24d\right) & \mbox{C IV} \\ 
\mathbf{B}_{79} & = & \left(\frac{1}{2} - z_{4}\right) \, \mathbf{a}_{1}-x_{4} \, \mathbf{a}_{2} + \left(\frac{1}{2} +y_{4}\right) \, \mathbf{a}_{3} & = & \left(\frac{1}{2} - z_{4}\right)a \, \mathbf{\hat{x}}-x_{4}a \, \mathbf{\hat{y}} + \left(\frac{1}{2} +y_{4}\right)a \, \mathbf{\hat{z}} & \left(24d\right) & \mbox{C IV} \\ 
\mathbf{B}_{80} & = & -z_{4} \, \mathbf{a}_{1} + \left(\frac{1}{2} +x_{4}\right) \, \mathbf{a}_{2} + \left(\frac{1}{2} - y_{4}\right) \, \mathbf{a}_{3} & = & -z_{4}a \, \mathbf{\hat{x}} + \left(\frac{1}{2} +x_{4}\right)a \, \mathbf{\hat{y}} + \left(\frac{1}{2} - y_{4}\right)a \, \mathbf{\hat{z}} & \left(24d\right) & \mbox{C IV} \\ 
\mathbf{B}_{81} & = & y_{4} \, \mathbf{a}_{1} + z_{4} \, \mathbf{a}_{2} + x_{4} \, \mathbf{a}_{3} & = & y_{4}a \, \mathbf{\hat{x}} + z_{4}a \, \mathbf{\hat{y}} + x_{4}a \, \mathbf{\hat{z}} & \left(24d\right) & \mbox{C IV} \\ 
\mathbf{B}_{82} & = & -y_{4} \, \mathbf{a}_{1} + \left(\frac{1}{2} +z_{4}\right) \, \mathbf{a}_{2} + \left(\frac{1}{2} - x_{4}\right) \, \mathbf{a}_{3} & = & -y_{4}a \, \mathbf{\hat{x}} + \left(\frac{1}{2} +z_{4}\right)a \, \mathbf{\hat{y}} + \left(\frac{1}{2} - x_{4}\right)a \, \mathbf{\hat{z}} & \left(24d\right) & \mbox{C IV} \\ 
\mathbf{B}_{83} & = & \left(\frac{1}{2} +y_{4}\right) \, \mathbf{a}_{1} + \left(\frac{1}{2} - z_{4}\right) \, \mathbf{a}_{2}-x_{4} \, \mathbf{a}_{3} & = & \left(\frac{1}{2} +y_{4}\right)a \, \mathbf{\hat{x}} + \left(\frac{1}{2} - z_{4}\right)a \, \mathbf{\hat{y}}-x_{4}a \, \mathbf{\hat{z}} & \left(24d\right) & \mbox{C IV} \\ 
\mathbf{B}_{84} & = & \left(\frac{1}{2} - y_{4}\right) \, \mathbf{a}_{1}-z_{4} \, \mathbf{a}_{2} + \left(\frac{1}{2} +x_{4}\right) \, \mathbf{a}_{3} & = & \left(\frac{1}{2} - y_{4}\right)a \, \mathbf{\hat{x}}-z_{4}a \, \mathbf{\hat{y}} + \left(\frac{1}{2} +x_{4}\right)a \, \mathbf{\hat{z}} & \left(24d\right) & \mbox{C IV} \\ 
\mathbf{B}_{85} & = & -x_{4} \, \mathbf{a}_{1}-y_{4} \, \mathbf{a}_{2}-z_{4} \, \mathbf{a}_{3} & = & -x_{4}a \, \mathbf{\hat{x}}-y_{4}a \, \mathbf{\hat{y}}-z_{4}a \, \mathbf{\hat{z}} & \left(24d\right) & \mbox{C IV} \\ 
\mathbf{B}_{86} & = & \left(\frac{1}{2} +x_{4}\right) \, \mathbf{a}_{1} + y_{4} \, \mathbf{a}_{2} + \left(\frac{1}{2} - z_{4}\right) \, \mathbf{a}_{3} & = & \left(\frac{1}{2} +x_{4}\right)a \, \mathbf{\hat{x}} + y_{4}a \, \mathbf{\hat{y}} + \left(\frac{1}{2} - z_{4}\right)a \, \mathbf{\hat{z}} & \left(24d\right) & \mbox{C IV} \\ 
\mathbf{B}_{87} & = & x_{4} \, \mathbf{a}_{1} + \left(\frac{1}{2} - y_{4}\right) \, \mathbf{a}_{2} + \left(\frac{1}{2} +z_{4}\right) \, \mathbf{a}_{3} & = & x_{4}a \, \mathbf{\hat{x}} + \left(\frac{1}{2} - y_{4}\right)a \, \mathbf{\hat{y}} + \left(\frac{1}{2} +z_{4}\right)a \, \mathbf{\hat{z}} & \left(24d\right) & \mbox{C IV} \\ 
\mathbf{B}_{88} & = & \left(\frac{1}{2} - x_{4}\right) \, \mathbf{a}_{1} + \left(\frac{1}{2} +y_{4}\right) \, \mathbf{a}_{2} + z_{4} \, \mathbf{a}_{3} & = & \left(\frac{1}{2} - x_{4}\right)a \, \mathbf{\hat{x}} + \left(\frac{1}{2} +y_{4}\right)a \, \mathbf{\hat{y}} + z_{4}a \, \mathbf{\hat{z}} & \left(24d\right) & \mbox{C IV} \\ 
\mathbf{B}_{89} & = & -z_{4} \, \mathbf{a}_{1}-x_{4} \, \mathbf{a}_{2}-y_{4} \, \mathbf{a}_{3} & = & -z_{4}a \, \mathbf{\hat{x}}-x_{4}a \, \mathbf{\hat{y}}-y_{4}a \, \mathbf{\hat{z}} & \left(24d\right) & \mbox{C IV} \\ 
\mathbf{B}_{90} & = & \left(\frac{1}{2} - z_{4}\right) \, \mathbf{a}_{1} + \left(\frac{1}{2} +x_{4}\right) \, \mathbf{a}_{2} + y_{4} \, \mathbf{a}_{3} & = & \left(\frac{1}{2} - z_{4}\right)a \, \mathbf{\hat{x}} + \left(\frac{1}{2} +x_{4}\right)a \, \mathbf{\hat{y}} + y_{4}a \, \mathbf{\hat{z}} & \left(24d\right) & \mbox{C IV} \\ 
\mathbf{B}_{91} & = & \left(\frac{1}{2} +z_{4}\right) \, \mathbf{a}_{1} + x_{4} \, \mathbf{a}_{2} + \left(\frac{1}{2} - y_{4}\right) \, \mathbf{a}_{3} & = & \left(\frac{1}{2} +z_{4}\right)a \, \mathbf{\hat{x}} + x_{4}a \, \mathbf{\hat{y}} + \left(\frac{1}{2} - y_{4}\right)a \, \mathbf{\hat{z}} & \left(24d\right) & \mbox{C IV} \\ 
\mathbf{B}_{92} & = & z_{4} \, \mathbf{a}_{1} + \left(\frac{1}{2} - x_{4}\right) \, \mathbf{a}_{2} + \left(\frac{1}{2} +y_{4}\right) \, \mathbf{a}_{3} & = & z_{4}a \, \mathbf{\hat{x}} + \left(\frac{1}{2} - x_{4}\right)a \, \mathbf{\hat{y}} + \left(\frac{1}{2} +y_{4}\right)a \, \mathbf{\hat{z}} & \left(24d\right) & \mbox{C IV} \\ 
\mathbf{B}_{93} & = & -y_{4} \, \mathbf{a}_{1}-z_{4} \, \mathbf{a}_{2}-x_{4} \, \mathbf{a}_{3} & = & -y_{4}a \, \mathbf{\hat{x}}-z_{4}a \, \mathbf{\hat{y}}-x_{4}a \, \mathbf{\hat{z}} & \left(24d\right) & \mbox{C IV} \\ 
\mathbf{B}_{94} & = & y_{4} \, \mathbf{a}_{1} + \left(\frac{1}{2} - z_{4}\right) \, \mathbf{a}_{2} + \left(\frac{1}{2} +x_{4}\right) \, \mathbf{a}_{3} & = & y_{4}a \, \mathbf{\hat{x}} + \left(\frac{1}{2} - z_{4}\right)a \, \mathbf{\hat{y}} + \left(\frac{1}{2} +x_{4}\right)a \, \mathbf{\hat{z}} & \left(24d\right) & \mbox{C IV} \\ 
\mathbf{B}_{95} & = & \left(\frac{1}{2} - y_{4}\right) \, \mathbf{a}_{1} + \left(\frac{1}{2} +z_{4}\right) \, \mathbf{a}_{2} + x_{4} \, \mathbf{a}_{3} & = & \left(\frac{1}{2} - y_{4}\right)a \, \mathbf{\hat{x}} + \left(\frac{1}{2} +z_{4}\right)a \, \mathbf{\hat{y}} + x_{4}a \, \mathbf{\hat{z}} & \left(24d\right) & \mbox{C IV} \\ 
\mathbf{B}_{96} & = & \left(\frac{1}{2} +y_{4}\right) \, \mathbf{a}_{1} + z_{4} \, \mathbf{a}_{2} + \left(\frac{1}{2} - x_{4}\right) \, \mathbf{a}_{3} & = & \left(\frac{1}{2} +y_{4}\right)a \, \mathbf{\hat{x}} + z_{4}a \, \mathbf{\hat{y}} + \left(\frac{1}{2} - x_{4}\right)a \, \mathbf{\hat{z}} & \left(24d\right) & \mbox{C IV} \\ 
\mathbf{B}_{97} & = & x_{5} \, \mathbf{a}_{1} + y_{5} \, \mathbf{a}_{2} + z_{5} \, \mathbf{a}_{3} & = & x_{5}a \, \mathbf{\hat{x}} + y_{5}a \, \mathbf{\hat{y}} + z_{5}a \, \mathbf{\hat{z}} & \left(24d\right) & \mbox{C V} \\ 
\mathbf{B}_{98} & = & \left(\frac{1}{2} - x_{5}\right) \, \mathbf{a}_{1}-y_{5} \, \mathbf{a}_{2} + \left(\frac{1}{2} +z_{5}\right) \, \mathbf{a}_{3} & = & \left(\frac{1}{2} - x_{5}\right)a \, \mathbf{\hat{x}}-y_{5}a \, \mathbf{\hat{y}} + \left(\frac{1}{2} +z_{5}\right)a \, \mathbf{\hat{z}} & \left(24d\right) & \mbox{C V} \\ 
\mathbf{B}_{99} & = & -x_{5} \, \mathbf{a}_{1} + \left(\frac{1}{2} +y_{5}\right) \, \mathbf{a}_{2} + \left(\frac{1}{2} - z_{5}\right) \, \mathbf{a}_{3} & = & -x_{5}a \, \mathbf{\hat{x}} + \left(\frac{1}{2} +y_{5}\right)a \, \mathbf{\hat{y}} + \left(\frac{1}{2} - z_{5}\right)a \, \mathbf{\hat{z}} & \left(24d\right) & \mbox{C V} \\ 
\mathbf{B}_{100} & = & \left(\frac{1}{2} +x_{5}\right) \, \mathbf{a}_{1} + \left(\frac{1}{2} - y_{5}\right) \, \mathbf{a}_{2}-z_{5} \, \mathbf{a}_{3} & = & \left(\frac{1}{2} +x_{5}\right)a \, \mathbf{\hat{x}} + \left(\frac{1}{2} - y_{5}\right)a \, \mathbf{\hat{y}}-z_{5}a \, \mathbf{\hat{z}} & \left(24d\right) & \mbox{C V} \\ 
\mathbf{B}_{101} & = & z_{5} \, \mathbf{a}_{1} + x_{5} \, \mathbf{a}_{2} + y_{5} \, \mathbf{a}_{3} & = & z_{5}a \, \mathbf{\hat{x}} + x_{5}a \, \mathbf{\hat{y}} + y_{5}a \, \mathbf{\hat{z}} & \left(24d\right) & \mbox{C V} \\ 
\mathbf{B}_{102} & = & \left(\frac{1}{2} +z_{5}\right) \, \mathbf{a}_{1} + \left(\frac{1}{2} - x_{5}\right) \, \mathbf{a}_{2}-y_{5} \, \mathbf{a}_{3} & = & \left(\frac{1}{2} +z_{5}\right)a \, \mathbf{\hat{x}} + \left(\frac{1}{2} - x_{5}\right)a \, \mathbf{\hat{y}}-y_{5}a \, \mathbf{\hat{z}} & \left(24d\right) & \mbox{C V} \\ 
\mathbf{B}_{103} & = & \left(\frac{1}{2} - z_{5}\right) \, \mathbf{a}_{1}-x_{5} \, \mathbf{a}_{2} + \left(\frac{1}{2} +y_{5}\right) \, \mathbf{a}_{3} & = & \left(\frac{1}{2} - z_{5}\right)a \, \mathbf{\hat{x}}-x_{5}a \, \mathbf{\hat{y}} + \left(\frac{1}{2} +y_{5}\right)a \, \mathbf{\hat{z}} & \left(24d\right) & \mbox{C V} \\ 
\mathbf{B}_{104} & = & -z_{5} \, \mathbf{a}_{1} + \left(\frac{1}{2} +x_{5}\right) \, \mathbf{a}_{2} + \left(\frac{1}{2} - y_{5}\right) \, \mathbf{a}_{3} & = & -z_{5}a \, \mathbf{\hat{x}} + \left(\frac{1}{2} +x_{5}\right)a \, \mathbf{\hat{y}} + \left(\frac{1}{2} - y_{5}\right)a \, \mathbf{\hat{z}} & \left(24d\right) & \mbox{C V} \\ 
\mathbf{B}_{105} & = & y_{5} \, \mathbf{a}_{1} + z_{5} \, \mathbf{a}_{2} + x_{5} \, \mathbf{a}_{3} & = & y_{5}a \, \mathbf{\hat{x}} + z_{5}a \, \mathbf{\hat{y}} + x_{5}a \, \mathbf{\hat{z}} & \left(24d\right) & \mbox{C V} \\ 
\mathbf{B}_{106} & = & -y_{5} \, \mathbf{a}_{1} + \left(\frac{1}{2} +z_{5}\right) \, \mathbf{a}_{2} + \left(\frac{1}{2} - x_{5}\right) \, \mathbf{a}_{3} & = & -y_{5}a \, \mathbf{\hat{x}} + \left(\frac{1}{2} +z_{5}\right)a \, \mathbf{\hat{y}} + \left(\frac{1}{2} - x_{5}\right)a \, \mathbf{\hat{z}} & \left(24d\right) & \mbox{C V} \\ 
\mathbf{B}_{107} & = & \left(\frac{1}{2} +y_{5}\right) \, \mathbf{a}_{1} + \left(\frac{1}{2} - z_{5}\right) \, \mathbf{a}_{2}-x_{5} \, \mathbf{a}_{3} & = & \left(\frac{1}{2} +y_{5}\right)a \, \mathbf{\hat{x}} + \left(\frac{1}{2} - z_{5}\right)a \, \mathbf{\hat{y}}-x_{5}a \, \mathbf{\hat{z}} & \left(24d\right) & \mbox{C V} \\ 
\mathbf{B}_{108} & = & \left(\frac{1}{2} - y_{5}\right) \, \mathbf{a}_{1}-z_{5} \, \mathbf{a}_{2} + \left(\frac{1}{2} +x_{5}\right) \, \mathbf{a}_{3} & = & \left(\frac{1}{2} - y_{5}\right)a \, \mathbf{\hat{x}}-z_{5}a \, \mathbf{\hat{y}} + \left(\frac{1}{2} +x_{5}\right)a \, \mathbf{\hat{z}} & \left(24d\right) & \mbox{C V} \\ 
\mathbf{B}_{109} & = & -x_{5} \, \mathbf{a}_{1}-y_{5} \, \mathbf{a}_{2}-z_{5} \, \mathbf{a}_{3} & = & -x_{5}a \, \mathbf{\hat{x}}-y_{5}a \, \mathbf{\hat{y}}-z_{5}a \, \mathbf{\hat{z}} & \left(24d\right) & \mbox{C V} \\ 
\mathbf{B}_{110} & = & \left(\frac{1}{2} +x_{5}\right) \, \mathbf{a}_{1} + y_{5} \, \mathbf{a}_{2} + \left(\frac{1}{2} - z_{5}\right) \, \mathbf{a}_{3} & = & \left(\frac{1}{2} +x_{5}\right)a \, \mathbf{\hat{x}} + y_{5}a \, \mathbf{\hat{y}} + \left(\frac{1}{2} - z_{5}\right)a \, \mathbf{\hat{z}} & \left(24d\right) & \mbox{C V} \\ 
\mathbf{B}_{111} & = & x_{5} \, \mathbf{a}_{1} + \left(\frac{1}{2} - y_{5}\right) \, \mathbf{a}_{2} + \left(\frac{1}{2} +z_{5}\right) \, \mathbf{a}_{3} & = & x_{5}a \, \mathbf{\hat{x}} + \left(\frac{1}{2} - y_{5}\right)a \, \mathbf{\hat{y}} + \left(\frac{1}{2} +z_{5}\right)a \, \mathbf{\hat{z}} & \left(24d\right) & \mbox{C V} \\ 
\mathbf{B}_{112} & = & \left(\frac{1}{2} - x_{5}\right) \, \mathbf{a}_{1} + \left(\frac{1}{2} +y_{5}\right) \, \mathbf{a}_{2} + z_{5} \, \mathbf{a}_{3} & = & \left(\frac{1}{2} - x_{5}\right)a \, \mathbf{\hat{x}} + \left(\frac{1}{2} +y_{5}\right)a \, \mathbf{\hat{y}} + z_{5}a \, \mathbf{\hat{z}} & \left(24d\right) & \mbox{C V} \\ 
\mathbf{B}_{113} & = & -z_{5} \, \mathbf{a}_{1}-x_{5} \, \mathbf{a}_{2}-y_{5} \, \mathbf{a}_{3} & = & -z_{5}a \, \mathbf{\hat{x}}-x_{5}a \, \mathbf{\hat{y}}-y_{5}a \, \mathbf{\hat{z}} & \left(24d\right) & \mbox{C V} \\ 
\mathbf{B}_{114} & = & \left(\frac{1}{2} - z_{5}\right) \, \mathbf{a}_{1} + \left(\frac{1}{2} +x_{5}\right) \, \mathbf{a}_{2} + y_{5} \, \mathbf{a}_{3} & = & \left(\frac{1}{2} - z_{5}\right)a \, \mathbf{\hat{x}} + \left(\frac{1}{2} +x_{5}\right)a \, \mathbf{\hat{y}} + y_{5}a \, \mathbf{\hat{z}} & \left(24d\right) & \mbox{C V} \\ 
\mathbf{B}_{115} & = & \left(\frac{1}{2} +z_{5}\right) \, \mathbf{a}_{1} + x_{5} \, \mathbf{a}_{2} + \left(\frac{1}{2} - y_{5}\right) \, \mathbf{a}_{3} & = & \left(\frac{1}{2} +z_{5}\right)a \, \mathbf{\hat{x}} + x_{5}a \, \mathbf{\hat{y}} + \left(\frac{1}{2} - y_{5}\right)a \, \mathbf{\hat{z}} & \left(24d\right) & \mbox{C V} \\ 
\mathbf{B}_{116} & = & z_{5} \, \mathbf{a}_{1} + \left(\frac{1}{2} - x_{5}\right) \, \mathbf{a}_{2} + \left(\frac{1}{2} +y_{5}\right) \, \mathbf{a}_{3} & = & z_{5}a \, \mathbf{\hat{x}} + \left(\frac{1}{2} - x_{5}\right)a \, \mathbf{\hat{y}} + \left(\frac{1}{2} +y_{5}\right)a \, \mathbf{\hat{z}} & \left(24d\right) & \mbox{C V} \\ 
\mathbf{B}_{117} & = & -y_{5} \, \mathbf{a}_{1}-z_{5} \, \mathbf{a}_{2}-x_{5} \, \mathbf{a}_{3} & = & -y_{5}a \, \mathbf{\hat{x}}-z_{5}a \, \mathbf{\hat{y}}-x_{5}a \, \mathbf{\hat{z}} & \left(24d\right) & \mbox{C V} \\ 
\mathbf{B}_{118} & = & y_{5} \, \mathbf{a}_{1} + \left(\frac{1}{2} - z_{5}\right) \, \mathbf{a}_{2} + \left(\frac{1}{2} +x_{5}\right) \, \mathbf{a}_{3} & = & y_{5}a \, \mathbf{\hat{x}} + \left(\frac{1}{2} - z_{5}\right)a \, \mathbf{\hat{y}} + \left(\frac{1}{2} +x_{5}\right)a \, \mathbf{\hat{z}} & \left(24d\right) & \mbox{C V} \\ 
\mathbf{B}_{119} & = & \left(\frac{1}{2} - y_{5}\right) \, \mathbf{a}_{1} + \left(\frac{1}{2} +z_{5}\right) \, \mathbf{a}_{2} + x_{5} \, \mathbf{a}_{3} & = & \left(\frac{1}{2} - y_{5}\right)a \, \mathbf{\hat{x}} + \left(\frac{1}{2} +z_{5}\right)a \, \mathbf{\hat{y}} + x_{5}a \, \mathbf{\hat{z}} & \left(24d\right) & \mbox{C V} \\ 
\mathbf{B}_{120} & = & \left(\frac{1}{2} +y_{5}\right) \, \mathbf{a}_{1} + z_{5} \, \mathbf{a}_{2} + \left(\frac{1}{2} - x_{5}\right) \, \mathbf{a}_{3} & = & \left(\frac{1}{2} +y_{5}\right)a \, \mathbf{\hat{x}} + z_{5}a \, \mathbf{\hat{y}} + \left(\frac{1}{2} - x_{5}\right)a \, \mathbf{\hat{z}} & \left(24d\right) & \mbox{C V} \\ 
\mathbf{B}_{121} & = & x_{6} \, \mathbf{a}_{1} + y_{6} \, \mathbf{a}_{2} + z_{6} \, \mathbf{a}_{3} & = & x_{6}a \, \mathbf{\hat{x}} + y_{6}a \, \mathbf{\hat{y}} + z_{6}a \, \mathbf{\hat{z}} & \left(24d\right) & \mbox{C VI} \\ 
\mathbf{B}_{122} & = & \left(\frac{1}{2} - x_{6}\right) \, \mathbf{a}_{1}-y_{6} \, \mathbf{a}_{2} + \left(\frac{1}{2} +z_{6}\right) \, \mathbf{a}_{3} & = & \left(\frac{1}{2} - x_{6}\right)a \, \mathbf{\hat{x}}-y_{6}a \, \mathbf{\hat{y}} + \left(\frac{1}{2} +z_{6}\right)a \, \mathbf{\hat{z}} & \left(24d\right) & \mbox{C VI} \\ 
\mathbf{B}_{123} & = & -x_{6} \, \mathbf{a}_{1} + \left(\frac{1}{2} +y_{6}\right) \, \mathbf{a}_{2} + \left(\frac{1}{2} - z_{6}\right) \, \mathbf{a}_{3} & = & -x_{6}a \, \mathbf{\hat{x}} + \left(\frac{1}{2} +y_{6}\right)a \, \mathbf{\hat{y}} + \left(\frac{1}{2} - z_{6}\right)a \, \mathbf{\hat{z}} & \left(24d\right) & \mbox{C VI} \\ 
\mathbf{B}_{124} & = & \left(\frac{1}{2} +x_{6}\right) \, \mathbf{a}_{1} + \left(\frac{1}{2} - y_{6}\right) \, \mathbf{a}_{2}-z_{6} \, \mathbf{a}_{3} & = & \left(\frac{1}{2} +x_{6}\right)a \, \mathbf{\hat{x}} + \left(\frac{1}{2} - y_{6}\right)a \, \mathbf{\hat{y}}-z_{6}a \, \mathbf{\hat{z}} & \left(24d\right) & \mbox{C VI} \\ 
\mathbf{B}_{125} & = & z_{6} \, \mathbf{a}_{1} + x_{6} \, \mathbf{a}_{2} + y_{6} \, \mathbf{a}_{3} & = & z_{6}a \, \mathbf{\hat{x}} + x_{6}a \, \mathbf{\hat{y}} + y_{6}a \, \mathbf{\hat{z}} & \left(24d\right) & \mbox{C VI} \\ 
\mathbf{B}_{126} & = & \left(\frac{1}{2} +z_{6}\right) \, \mathbf{a}_{1} + \left(\frac{1}{2} - x_{6}\right) \, \mathbf{a}_{2}-y_{6} \, \mathbf{a}_{3} & = & \left(\frac{1}{2} +z_{6}\right)a \, \mathbf{\hat{x}} + \left(\frac{1}{2} - x_{6}\right)a \, \mathbf{\hat{y}}-y_{6}a \, \mathbf{\hat{z}} & \left(24d\right) & \mbox{C VI} \\ 
\mathbf{B}_{127} & = & \left(\frac{1}{2} - z_{6}\right) \, \mathbf{a}_{1}-x_{6} \, \mathbf{a}_{2} + \left(\frac{1}{2} +y_{6}\right) \, \mathbf{a}_{3} & = & \left(\frac{1}{2} - z_{6}\right)a \, \mathbf{\hat{x}}-x_{6}a \, \mathbf{\hat{y}} + \left(\frac{1}{2} +y_{6}\right)a \, \mathbf{\hat{z}} & \left(24d\right) & \mbox{C VI} \\ 
\mathbf{B}_{128} & = & -z_{6} \, \mathbf{a}_{1} + \left(\frac{1}{2} +x_{6}\right) \, \mathbf{a}_{2} + \left(\frac{1}{2} - y_{6}\right) \, \mathbf{a}_{3} & = & -z_{6}a \, \mathbf{\hat{x}} + \left(\frac{1}{2} +x_{6}\right)a \, \mathbf{\hat{y}} + \left(\frac{1}{2} - y_{6}\right)a \, \mathbf{\hat{z}} & \left(24d\right) & \mbox{C VI} \\ 
\mathbf{B}_{129} & = & y_{6} \, \mathbf{a}_{1} + z_{6} \, \mathbf{a}_{2} + x_{6} \, \mathbf{a}_{3} & = & y_{6}a \, \mathbf{\hat{x}} + z_{6}a \, \mathbf{\hat{y}} + x_{6}a \, \mathbf{\hat{z}} & \left(24d\right) & \mbox{C VI} \\ 
\mathbf{B}_{130} & = & -y_{6} \, \mathbf{a}_{1} + \left(\frac{1}{2} +z_{6}\right) \, \mathbf{a}_{2} + \left(\frac{1}{2} - x_{6}\right) \, \mathbf{a}_{3} & = & -y_{6}a \, \mathbf{\hat{x}} + \left(\frac{1}{2} +z_{6}\right)a \, \mathbf{\hat{y}} + \left(\frac{1}{2} - x_{6}\right)a \, \mathbf{\hat{z}} & \left(24d\right) & \mbox{C VI} \\ 
\mathbf{B}_{131} & = & \left(\frac{1}{2} +y_{6}\right) \, \mathbf{a}_{1} + \left(\frac{1}{2} - z_{6}\right) \, \mathbf{a}_{2}-x_{6} \, \mathbf{a}_{3} & = & \left(\frac{1}{2} +y_{6}\right)a \, \mathbf{\hat{x}} + \left(\frac{1}{2} - z_{6}\right)a \, \mathbf{\hat{y}}-x_{6}a \, \mathbf{\hat{z}} & \left(24d\right) & \mbox{C VI} \\ 
\mathbf{B}_{132} & = & \left(\frac{1}{2} - y_{6}\right) \, \mathbf{a}_{1}-z_{6} \, \mathbf{a}_{2} + \left(\frac{1}{2} +x_{6}\right) \, \mathbf{a}_{3} & = & \left(\frac{1}{2} - y_{6}\right)a \, \mathbf{\hat{x}}-z_{6}a \, \mathbf{\hat{y}} + \left(\frac{1}{2} +x_{6}\right)a \, \mathbf{\hat{z}} & \left(24d\right) & \mbox{C VI} \\ 
\mathbf{B}_{133} & = & -x_{6} \, \mathbf{a}_{1}-y_{6} \, \mathbf{a}_{2}-z_{6} \, \mathbf{a}_{3} & = & -x_{6}a \, \mathbf{\hat{x}}-y_{6}a \, \mathbf{\hat{y}}-z_{6}a \, \mathbf{\hat{z}} & \left(24d\right) & \mbox{C VI} \\ 
\mathbf{B}_{134} & = & \left(\frac{1}{2} +x_{6}\right) \, \mathbf{a}_{1} + y_{6} \, \mathbf{a}_{2} + \left(\frac{1}{2} - z_{6}\right) \, \mathbf{a}_{3} & = & \left(\frac{1}{2} +x_{6}\right)a \, \mathbf{\hat{x}} + y_{6}a \, \mathbf{\hat{y}} + \left(\frac{1}{2} - z_{6}\right)a \, \mathbf{\hat{z}} & \left(24d\right) & \mbox{C VI} \\ 
\mathbf{B}_{135} & = & x_{6} \, \mathbf{a}_{1} + \left(\frac{1}{2} - y_{6}\right) \, \mathbf{a}_{2} + \left(\frac{1}{2} +z_{6}\right) \, \mathbf{a}_{3} & = & x_{6}a \, \mathbf{\hat{x}} + \left(\frac{1}{2} - y_{6}\right)a \, \mathbf{\hat{y}} + \left(\frac{1}{2} +z_{6}\right)a \, \mathbf{\hat{z}} & \left(24d\right) & \mbox{C VI} \\ 
\mathbf{B}_{136} & = & \left(\frac{1}{2} - x_{6}\right) \, \mathbf{a}_{1} + \left(\frac{1}{2} +y_{6}\right) \, \mathbf{a}_{2} + z_{6} \, \mathbf{a}_{3} & = & \left(\frac{1}{2} - x_{6}\right)a \, \mathbf{\hat{x}} + \left(\frac{1}{2} +y_{6}\right)a \, \mathbf{\hat{y}} + z_{6}a \, \mathbf{\hat{z}} & \left(24d\right) & \mbox{C VI} \\ 
\mathbf{B}_{137} & = & -z_{6} \, \mathbf{a}_{1}-x_{6} \, \mathbf{a}_{2}-y_{6} \, \mathbf{a}_{3} & = & -z_{6}a \, \mathbf{\hat{x}}-x_{6}a \, \mathbf{\hat{y}}-y_{6}a \, \mathbf{\hat{z}} & \left(24d\right) & \mbox{C VI} \\ 
\mathbf{B}_{138} & = & \left(\frac{1}{2} - z_{6}\right) \, \mathbf{a}_{1} + \left(\frac{1}{2} +x_{6}\right) \, \mathbf{a}_{2} + y_{6} \, \mathbf{a}_{3} & = & \left(\frac{1}{2} - z_{6}\right)a \, \mathbf{\hat{x}} + \left(\frac{1}{2} +x_{6}\right)a \, \mathbf{\hat{y}} + y_{6}a \, \mathbf{\hat{z}} & \left(24d\right) & \mbox{C VI} \\ 
\mathbf{B}_{139} & = & \left(\frac{1}{2} +z_{6}\right) \, \mathbf{a}_{1} + x_{6} \, \mathbf{a}_{2} + \left(\frac{1}{2} - y_{6}\right) \, \mathbf{a}_{3} & = & \left(\frac{1}{2} +z_{6}\right)a \, \mathbf{\hat{x}} + x_{6}a \, \mathbf{\hat{y}} + \left(\frac{1}{2} - y_{6}\right)a \, \mathbf{\hat{z}} & \left(24d\right) & \mbox{C VI} \\ 
\mathbf{B}_{140} & = & z_{6} \, \mathbf{a}_{1} + \left(\frac{1}{2} - x_{6}\right) \, \mathbf{a}_{2} + \left(\frac{1}{2} +y_{6}\right) \, \mathbf{a}_{3} & = & z_{6}a \, \mathbf{\hat{x}} + \left(\frac{1}{2} - x_{6}\right)a \, \mathbf{\hat{y}} + \left(\frac{1}{2} +y_{6}\right)a \, \mathbf{\hat{z}} & \left(24d\right) & \mbox{C VI} \\ 
\mathbf{B}_{141} & = & -y_{6} \, \mathbf{a}_{1}-z_{6} \, \mathbf{a}_{2}-x_{6} \, \mathbf{a}_{3} & = & -y_{6}a \, \mathbf{\hat{x}}-z_{6}a \, \mathbf{\hat{y}}-x_{6}a \, \mathbf{\hat{z}} & \left(24d\right) & \mbox{C VI} \\ 
\mathbf{B}_{142} & = & y_{6} \, \mathbf{a}_{1} + \left(\frac{1}{2} - z_{6}\right) \, \mathbf{a}_{2} + \left(\frac{1}{2} +x_{6}\right) \, \mathbf{a}_{3} & = & y_{6}a \, \mathbf{\hat{x}} + \left(\frac{1}{2} - z_{6}\right)a \, \mathbf{\hat{y}} + \left(\frac{1}{2} +x_{6}\right)a \, \mathbf{\hat{z}} & \left(24d\right) & \mbox{C VI} \\ 
\mathbf{B}_{143} & = & \left(\frac{1}{2} - y_{6}\right) \, \mathbf{a}_{1} + \left(\frac{1}{2} +z_{6}\right) \, \mathbf{a}_{2} + x_{6} \, \mathbf{a}_{3} & = & \left(\frac{1}{2} - y_{6}\right)a \, \mathbf{\hat{x}} + \left(\frac{1}{2} +z_{6}\right)a \, \mathbf{\hat{y}} + x_{6}a \, \mathbf{\hat{z}} & \left(24d\right) & \mbox{C VI} \\ 
\mathbf{B}_{144} & = & \left(\frac{1}{2} +y_{6}\right) \, \mathbf{a}_{1} + z_{6} \, \mathbf{a}_{2} + \left(\frac{1}{2} - x_{6}\right) \, \mathbf{a}_{3} & = & \left(\frac{1}{2} +y_{6}\right)a \, \mathbf{\hat{x}} + z_{6}a \, \mathbf{\hat{y}} + \left(\frac{1}{2} - x_{6}\right)a \, \mathbf{\hat{z}} & \left(24d\right) & \mbox{C VI} \\ 
\mathbf{B}_{145} & = & x_{7} \, \mathbf{a}_{1} + y_{7} \, \mathbf{a}_{2} + z_{7} \, \mathbf{a}_{3} & = & x_{7}a \, \mathbf{\hat{x}} + y_{7}a \, \mathbf{\hat{y}} + z_{7}a \, \mathbf{\hat{z}} & \left(24d\right) & \mbox{C VII} \\ 
\mathbf{B}_{146} & = & \left(\frac{1}{2} - x_{7}\right) \, \mathbf{a}_{1}-y_{7} \, \mathbf{a}_{2} + \left(\frac{1}{2} +z_{7}\right) \, \mathbf{a}_{3} & = & \left(\frac{1}{2} - x_{7}\right)a \, \mathbf{\hat{x}}-y_{7}a \, \mathbf{\hat{y}} + \left(\frac{1}{2} +z_{7}\right)a \, \mathbf{\hat{z}} & \left(24d\right) & \mbox{C VII} \\ 
\mathbf{B}_{147} & = & -x_{7} \, \mathbf{a}_{1} + \left(\frac{1}{2} +y_{7}\right) \, \mathbf{a}_{2} + \left(\frac{1}{2} - z_{7}\right) \, \mathbf{a}_{3} & = & -x_{7}a \, \mathbf{\hat{x}} + \left(\frac{1}{2} +y_{7}\right)a \, \mathbf{\hat{y}} + \left(\frac{1}{2} - z_{7}\right)a \, \mathbf{\hat{z}} & \left(24d\right) & \mbox{C VII} \\ 
\mathbf{B}_{148} & = & \left(\frac{1}{2} +x_{7}\right) \, \mathbf{a}_{1} + \left(\frac{1}{2} - y_{7}\right) \, \mathbf{a}_{2}-z_{7} \, \mathbf{a}_{3} & = & \left(\frac{1}{2} +x_{7}\right)a \, \mathbf{\hat{x}} + \left(\frac{1}{2} - y_{7}\right)a \, \mathbf{\hat{y}}-z_{7}a \, \mathbf{\hat{z}} & \left(24d\right) & \mbox{C VII} \\ 
\mathbf{B}_{149} & = & z_{7} \, \mathbf{a}_{1} + x_{7} \, \mathbf{a}_{2} + y_{7} \, \mathbf{a}_{3} & = & z_{7}a \, \mathbf{\hat{x}} + x_{7}a \, \mathbf{\hat{y}} + y_{7}a \, \mathbf{\hat{z}} & \left(24d\right) & \mbox{C VII} \\ 
\mathbf{B}_{150} & = & \left(\frac{1}{2} +z_{7}\right) \, \mathbf{a}_{1} + \left(\frac{1}{2} - x_{7}\right) \, \mathbf{a}_{2}-y_{7} \, \mathbf{a}_{3} & = & \left(\frac{1}{2} +z_{7}\right)a \, \mathbf{\hat{x}} + \left(\frac{1}{2} - x_{7}\right)a \, \mathbf{\hat{y}}-y_{7}a \, \mathbf{\hat{z}} & \left(24d\right) & \mbox{C VII} \\ 
\mathbf{B}_{151} & = & \left(\frac{1}{2} - z_{7}\right) \, \mathbf{a}_{1}-x_{7} \, \mathbf{a}_{2} + \left(\frac{1}{2} +y_{7}\right) \, \mathbf{a}_{3} & = & \left(\frac{1}{2} - z_{7}\right)a \, \mathbf{\hat{x}}-x_{7}a \, \mathbf{\hat{y}} + \left(\frac{1}{2} +y_{7}\right)a \, \mathbf{\hat{z}} & \left(24d\right) & \mbox{C VII} \\ 
\mathbf{B}_{152} & = & -z_{7} \, \mathbf{a}_{1} + \left(\frac{1}{2} +x_{7}\right) \, \mathbf{a}_{2} + \left(\frac{1}{2} - y_{7}\right) \, \mathbf{a}_{3} & = & -z_{7}a \, \mathbf{\hat{x}} + \left(\frac{1}{2} +x_{7}\right)a \, \mathbf{\hat{y}} + \left(\frac{1}{2} - y_{7}\right)a \, \mathbf{\hat{z}} & \left(24d\right) & \mbox{C VII} \\ 
\mathbf{B}_{153} & = & y_{7} \, \mathbf{a}_{1} + z_{7} \, \mathbf{a}_{2} + x_{7} \, \mathbf{a}_{3} & = & y_{7}a \, \mathbf{\hat{x}} + z_{7}a \, \mathbf{\hat{y}} + x_{7}a \, \mathbf{\hat{z}} & \left(24d\right) & \mbox{C VII} \\ 
\mathbf{B}_{154} & = & -y_{7} \, \mathbf{a}_{1} + \left(\frac{1}{2} +z_{7}\right) \, \mathbf{a}_{2} + \left(\frac{1}{2} - x_{7}\right) \, \mathbf{a}_{3} & = & -y_{7}a \, \mathbf{\hat{x}} + \left(\frac{1}{2} +z_{7}\right)a \, \mathbf{\hat{y}} + \left(\frac{1}{2} - x_{7}\right)a \, \mathbf{\hat{z}} & \left(24d\right) & \mbox{C VII} \\ 
\mathbf{B}_{155} & = & \left(\frac{1}{2} +y_{7}\right) \, \mathbf{a}_{1} + \left(\frac{1}{2} - z_{7}\right) \, \mathbf{a}_{2}-x_{7} \, \mathbf{a}_{3} & = & \left(\frac{1}{2} +y_{7}\right)a \, \mathbf{\hat{x}} + \left(\frac{1}{2} - z_{7}\right)a \, \mathbf{\hat{y}}-x_{7}a \, \mathbf{\hat{z}} & \left(24d\right) & \mbox{C VII} \\ 
\mathbf{B}_{156} & = & \left(\frac{1}{2} - y_{7}\right) \, \mathbf{a}_{1}-z_{7} \, \mathbf{a}_{2} + \left(\frac{1}{2} +x_{7}\right) \, \mathbf{a}_{3} & = & \left(\frac{1}{2} - y_{7}\right)a \, \mathbf{\hat{x}}-z_{7}a \, \mathbf{\hat{y}} + \left(\frac{1}{2} +x_{7}\right)a \, \mathbf{\hat{z}} & \left(24d\right) & \mbox{C VII} \\ 
\mathbf{B}_{157} & = & -x_{7} \, \mathbf{a}_{1}-y_{7} \, \mathbf{a}_{2}-z_{7} \, \mathbf{a}_{3} & = & -x_{7}a \, \mathbf{\hat{x}}-y_{7}a \, \mathbf{\hat{y}}-z_{7}a \, \mathbf{\hat{z}} & \left(24d\right) & \mbox{C VII} \\ 
\mathbf{B}_{158} & = & \left(\frac{1}{2} +x_{7}\right) \, \mathbf{a}_{1} + y_{7} \, \mathbf{a}_{2} + \left(\frac{1}{2} - z_{7}\right) \, \mathbf{a}_{3} & = & \left(\frac{1}{2} +x_{7}\right)a \, \mathbf{\hat{x}} + y_{7}a \, \mathbf{\hat{y}} + \left(\frac{1}{2} - z_{7}\right)a \, \mathbf{\hat{z}} & \left(24d\right) & \mbox{C VII} \\ 
\mathbf{B}_{159} & = & x_{7} \, \mathbf{a}_{1} + \left(\frac{1}{2} - y_{7}\right) \, \mathbf{a}_{2} + \left(\frac{1}{2} +z_{7}\right) \, \mathbf{a}_{3} & = & x_{7}a \, \mathbf{\hat{x}} + \left(\frac{1}{2} - y_{7}\right)a \, \mathbf{\hat{y}} + \left(\frac{1}{2} +z_{7}\right)a \, \mathbf{\hat{z}} & \left(24d\right) & \mbox{C VII} \\ 
\mathbf{B}_{160} & = & \left(\frac{1}{2} - x_{7}\right) \, \mathbf{a}_{1} + \left(\frac{1}{2} +y_{7}\right) \, \mathbf{a}_{2} + z_{7} \, \mathbf{a}_{3} & = & \left(\frac{1}{2} - x_{7}\right)a \, \mathbf{\hat{x}} + \left(\frac{1}{2} +y_{7}\right)a \, \mathbf{\hat{y}} + z_{7}a \, \mathbf{\hat{z}} & \left(24d\right) & \mbox{C VII} \\ 
\mathbf{B}_{161} & = & -z_{7} \, \mathbf{a}_{1}-x_{7} \, \mathbf{a}_{2}-y_{7} \, \mathbf{a}_{3} & = & -z_{7}a \, \mathbf{\hat{x}}-x_{7}a \, \mathbf{\hat{y}}-y_{7}a \, \mathbf{\hat{z}} & \left(24d\right) & \mbox{C VII} \\ 
\mathbf{B}_{162} & = & \left(\frac{1}{2} - z_{7}\right) \, \mathbf{a}_{1} + \left(\frac{1}{2} +x_{7}\right) \, \mathbf{a}_{2} + y_{7} \, \mathbf{a}_{3} & = & \left(\frac{1}{2} - z_{7}\right)a \, \mathbf{\hat{x}} + \left(\frac{1}{2} +x_{7}\right)a \, \mathbf{\hat{y}} + y_{7}a \, \mathbf{\hat{z}} & \left(24d\right) & \mbox{C VII} \\ 
\mathbf{B}_{163} & = & \left(\frac{1}{2} +z_{7}\right) \, \mathbf{a}_{1} + x_{7} \, \mathbf{a}_{2} + \left(\frac{1}{2} - y_{7}\right) \, \mathbf{a}_{3} & = & \left(\frac{1}{2} +z_{7}\right)a \, \mathbf{\hat{x}} + x_{7}a \, \mathbf{\hat{y}} + \left(\frac{1}{2} - y_{7}\right)a \, \mathbf{\hat{z}} & \left(24d\right) & \mbox{C VII} \\ 
\mathbf{B}_{164} & = & z_{7} \, \mathbf{a}_{1} + \left(\frac{1}{2} - x_{7}\right) \, \mathbf{a}_{2} + \left(\frac{1}{2} +y_{7}\right) \, \mathbf{a}_{3} & = & z_{7}a \, \mathbf{\hat{x}} + \left(\frac{1}{2} - x_{7}\right)a \, \mathbf{\hat{y}} + \left(\frac{1}{2} +y_{7}\right)a \, \mathbf{\hat{z}} & \left(24d\right) & \mbox{C VII} \\ 
\mathbf{B}_{165} & = & -y_{7} \, \mathbf{a}_{1}-z_{7} \, \mathbf{a}_{2}-x_{7} \, \mathbf{a}_{3} & = & -y_{7}a \, \mathbf{\hat{x}}-z_{7}a \, \mathbf{\hat{y}}-x_{7}a \, \mathbf{\hat{z}} & \left(24d\right) & \mbox{C VII} \\ 
\mathbf{B}_{166} & = & y_{7} \, \mathbf{a}_{1} + \left(\frac{1}{2} - z_{7}\right) \, \mathbf{a}_{2} + \left(\frac{1}{2} +x_{7}\right) \, \mathbf{a}_{3} & = & y_{7}a \, \mathbf{\hat{x}} + \left(\frac{1}{2} - z_{7}\right)a \, \mathbf{\hat{y}} + \left(\frac{1}{2} +x_{7}\right)a \, \mathbf{\hat{z}} & \left(24d\right) & \mbox{C VII} \\ 
\mathbf{B}_{167} & = & \left(\frac{1}{2} - y_{7}\right) \, \mathbf{a}_{1} + \left(\frac{1}{2} +z_{7}\right) \, \mathbf{a}_{2} + x_{7} \, \mathbf{a}_{3} & = & \left(\frac{1}{2} - y_{7}\right)a \, \mathbf{\hat{x}} + \left(\frac{1}{2} +z_{7}\right)a \, \mathbf{\hat{y}} + x_{7}a \, \mathbf{\hat{z}} & \left(24d\right) & \mbox{C VII} \\ 
\mathbf{B}_{168} & = & \left(\frac{1}{2} +y_{7}\right) \, \mathbf{a}_{1} + z_{7} \, \mathbf{a}_{2} + \left(\frac{1}{2} - x_{7}\right) \, \mathbf{a}_{3} & = & \left(\frac{1}{2} +y_{7}\right)a \, \mathbf{\hat{x}} + z_{7}a \, \mathbf{\hat{y}} + \left(\frac{1}{2} - x_{7}\right)a \, \mathbf{\hat{z}} & \left(24d\right) & \mbox{C VII} \\ 
\mathbf{B}_{169} & = & x_{8} \, \mathbf{a}_{1} + y_{8} \, \mathbf{a}_{2} + z_{8} \, \mathbf{a}_{3} & = & x_{8}a \, \mathbf{\hat{x}} + y_{8}a \, \mathbf{\hat{y}} + z_{8}a \, \mathbf{\hat{z}} & \left(24d\right) & \mbox{C VIII} \\ 
\mathbf{B}_{170} & = & \left(\frac{1}{2} - x_{8}\right) \, \mathbf{a}_{1}-y_{8} \, \mathbf{a}_{2} + \left(\frac{1}{2} +z_{8}\right) \, \mathbf{a}_{3} & = & \left(\frac{1}{2} - x_{8}\right)a \, \mathbf{\hat{x}}-y_{8}a \, \mathbf{\hat{y}} + \left(\frac{1}{2} +z_{8}\right)a \, \mathbf{\hat{z}} & \left(24d\right) & \mbox{C VIII} \\ 
\mathbf{B}_{171} & = & -x_{8} \, \mathbf{a}_{1} + \left(\frac{1}{2} +y_{8}\right) \, \mathbf{a}_{2} + \left(\frac{1}{2} - z_{8}\right) \, \mathbf{a}_{3} & = & -x_{8}a \, \mathbf{\hat{x}} + \left(\frac{1}{2} +y_{8}\right)a \, \mathbf{\hat{y}} + \left(\frac{1}{2} - z_{8}\right)a \, \mathbf{\hat{z}} & \left(24d\right) & \mbox{C VIII} \\ 
\mathbf{B}_{172} & = & \left(\frac{1}{2} +x_{8}\right) \, \mathbf{a}_{1} + \left(\frac{1}{2} - y_{8}\right) \, \mathbf{a}_{2}-z_{8} \, \mathbf{a}_{3} & = & \left(\frac{1}{2} +x_{8}\right)a \, \mathbf{\hat{x}} + \left(\frac{1}{2} - y_{8}\right)a \, \mathbf{\hat{y}}-z_{8}a \, \mathbf{\hat{z}} & \left(24d\right) & \mbox{C VIII} \\ 
\mathbf{B}_{173} & = & z_{8} \, \mathbf{a}_{1} + x_{8} \, \mathbf{a}_{2} + y_{8} \, \mathbf{a}_{3} & = & z_{8}a \, \mathbf{\hat{x}} + x_{8}a \, \mathbf{\hat{y}} + y_{8}a \, \mathbf{\hat{z}} & \left(24d\right) & \mbox{C VIII} \\ 
\mathbf{B}_{174} & = & \left(\frac{1}{2} +z_{8}\right) \, \mathbf{a}_{1} + \left(\frac{1}{2} - x_{8}\right) \, \mathbf{a}_{2}-y_{8} \, \mathbf{a}_{3} & = & \left(\frac{1}{2} +z_{8}\right)a \, \mathbf{\hat{x}} + \left(\frac{1}{2} - x_{8}\right)a \, \mathbf{\hat{y}}-y_{8}a \, \mathbf{\hat{z}} & \left(24d\right) & \mbox{C VIII} \\ 
\mathbf{B}_{175} & = & \left(\frac{1}{2} - z_{8}\right) \, \mathbf{a}_{1}-x_{8} \, \mathbf{a}_{2} + \left(\frac{1}{2} +y_{8}\right) \, \mathbf{a}_{3} & = & \left(\frac{1}{2} - z_{8}\right)a \, \mathbf{\hat{x}}-x_{8}a \, \mathbf{\hat{y}} + \left(\frac{1}{2} +y_{8}\right)a \, \mathbf{\hat{z}} & \left(24d\right) & \mbox{C VIII} \\ 
\mathbf{B}_{176} & = & -z_{8} \, \mathbf{a}_{1} + \left(\frac{1}{2} +x_{8}\right) \, \mathbf{a}_{2} + \left(\frac{1}{2} - y_{8}\right) \, \mathbf{a}_{3} & = & -z_{8}a \, \mathbf{\hat{x}} + \left(\frac{1}{2} +x_{8}\right)a \, \mathbf{\hat{y}} + \left(\frac{1}{2} - y_{8}\right)a \, \mathbf{\hat{z}} & \left(24d\right) & \mbox{C VIII} \\ 
\mathbf{B}_{177} & = & y_{8} \, \mathbf{a}_{1} + z_{8} \, \mathbf{a}_{2} + x_{8} \, \mathbf{a}_{3} & = & y_{8}a \, \mathbf{\hat{x}} + z_{8}a \, \mathbf{\hat{y}} + x_{8}a \, \mathbf{\hat{z}} & \left(24d\right) & \mbox{C VIII} \\ 
\mathbf{B}_{178} & = & -y_{8} \, \mathbf{a}_{1} + \left(\frac{1}{2} +z_{8}\right) \, \mathbf{a}_{2} + \left(\frac{1}{2} - x_{8}\right) \, \mathbf{a}_{3} & = & -y_{8}a \, \mathbf{\hat{x}} + \left(\frac{1}{2} +z_{8}\right)a \, \mathbf{\hat{y}} + \left(\frac{1}{2} - x_{8}\right)a \, \mathbf{\hat{z}} & \left(24d\right) & \mbox{C VIII} \\ 
\mathbf{B}_{179} & = & \left(\frac{1}{2} +y_{8}\right) \, \mathbf{a}_{1} + \left(\frac{1}{2} - z_{8}\right) \, \mathbf{a}_{2}-x_{8} \, \mathbf{a}_{3} & = & \left(\frac{1}{2} +y_{8}\right)a \, \mathbf{\hat{x}} + \left(\frac{1}{2} - z_{8}\right)a \, \mathbf{\hat{y}}-x_{8}a \, \mathbf{\hat{z}} & \left(24d\right) & \mbox{C VIII} \\ 
\mathbf{B}_{180} & = & \left(\frac{1}{2} - y_{8}\right) \, \mathbf{a}_{1}-z_{8} \, \mathbf{a}_{2} + \left(\frac{1}{2} +x_{8}\right) \, \mathbf{a}_{3} & = & \left(\frac{1}{2} - y_{8}\right)a \, \mathbf{\hat{x}}-z_{8}a \, \mathbf{\hat{y}} + \left(\frac{1}{2} +x_{8}\right)a \, \mathbf{\hat{z}} & \left(24d\right) & \mbox{C VIII} \\ 
\mathbf{B}_{181} & = & -x_{8} \, \mathbf{a}_{1}-y_{8} \, \mathbf{a}_{2}-z_{8} \, \mathbf{a}_{3} & = & -x_{8}a \, \mathbf{\hat{x}}-y_{8}a \, \mathbf{\hat{y}}-z_{8}a \, \mathbf{\hat{z}} & \left(24d\right) & \mbox{C VIII} \\ 
\mathbf{B}_{182} & = & \left(\frac{1}{2} +x_{8}\right) \, \mathbf{a}_{1} + y_{8} \, \mathbf{a}_{2} + \left(\frac{1}{2} - z_{8}\right) \, \mathbf{a}_{3} & = & \left(\frac{1}{2} +x_{8}\right)a \, \mathbf{\hat{x}} + y_{8}a \, \mathbf{\hat{y}} + \left(\frac{1}{2} - z_{8}\right)a \, \mathbf{\hat{z}} & \left(24d\right) & \mbox{C VIII} \\ 
\mathbf{B}_{183} & = & x_{8} \, \mathbf{a}_{1} + \left(\frac{1}{2} - y_{8}\right) \, \mathbf{a}_{2} + \left(\frac{1}{2} +z_{8}\right) \, \mathbf{a}_{3} & = & x_{8}a \, \mathbf{\hat{x}} + \left(\frac{1}{2} - y_{8}\right)a \, \mathbf{\hat{y}} + \left(\frac{1}{2} +z_{8}\right)a \, \mathbf{\hat{z}} & \left(24d\right) & \mbox{C VIII} \\ 
\mathbf{B}_{184} & = & \left(\frac{1}{2} - x_{8}\right) \, \mathbf{a}_{1} + \left(\frac{1}{2} +y_{8}\right) \, \mathbf{a}_{2} + z_{8} \, \mathbf{a}_{3} & = & \left(\frac{1}{2} - x_{8}\right)a \, \mathbf{\hat{x}} + \left(\frac{1}{2} +y_{8}\right)a \, \mathbf{\hat{y}} + z_{8}a \, \mathbf{\hat{z}} & \left(24d\right) & \mbox{C VIII} \\ 
\mathbf{B}_{185} & = & -z_{8} \, \mathbf{a}_{1}-x_{8} \, \mathbf{a}_{2}-y_{8} \, \mathbf{a}_{3} & = & -z_{8}a \, \mathbf{\hat{x}}-x_{8}a \, \mathbf{\hat{y}}-y_{8}a \, \mathbf{\hat{z}} & \left(24d\right) & \mbox{C VIII} \\ 
\mathbf{B}_{186} & = & \left(\frac{1}{2} - z_{8}\right) \, \mathbf{a}_{1} + \left(\frac{1}{2} +x_{8}\right) \, \mathbf{a}_{2} + y_{8} \, \mathbf{a}_{3} & = & \left(\frac{1}{2} - z_{8}\right)a \, \mathbf{\hat{x}} + \left(\frac{1}{2} +x_{8}\right)a \, \mathbf{\hat{y}} + y_{8}a \, \mathbf{\hat{z}} & \left(24d\right) & \mbox{C VIII} \\ 
\mathbf{B}_{187} & = & \left(\frac{1}{2} +z_{8}\right) \, \mathbf{a}_{1} + x_{8} \, \mathbf{a}_{2} + \left(\frac{1}{2} - y_{8}\right) \, \mathbf{a}_{3} & = & \left(\frac{1}{2} +z_{8}\right)a \, \mathbf{\hat{x}} + x_{8}a \, \mathbf{\hat{y}} + \left(\frac{1}{2} - y_{8}\right)a \, \mathbf{\hat{z}} & \left(24d\right) & \mbox{C VIII} \\ 
\mathbf{B}_{188} & = & z_{8} \, \mathbf{a}_{1} + \left(\frac{1}{2} - x_{8}\right) \, \mathbf{a}_{2} + \left(\frac{1}{2} +y_{8}\right) \, \mathbf{a}_{3} & = & z_{8}a \, \mathbf{\hat{x}} + \left(\frac{1}{2} - x_{8}\right)a \, \mathbf{\hat{y}} + \left(\frac{1}{2} +y_{8}\right)a \, \mathbf{\hat{z}} & \left(24d\right) & \mbox{C VIII} \\ 
\mathbf{B}_{189} & = & -y_{8} \, \mathbf{a}_{1}-z_{8} \, \mathbf{a}_{2}-x_{8} \, \mathbf{a}_{3} & = & -y_{8}a \, \mathbf{\hat{x}}-z_{8}a \, \mathbf{\hat{y}}-x_{8}a \, \mathbf{\hat{z}} & \left(24d\right) & \mbox{C VIII} \\ 
\mathbf{B}_{190} & = & y_{8} \, \mathbf{a}_{1} + \left(\frac{1}{2} - z_{8}\right) \, \mathbf{a}_{2} + \left(\frac{1}{2} +x_{8}\right) \, \mathbf{a}_{3} & = & y_{8}a \, \mathbf{\hat{x}} + \left(\frac{1}{2} - z_{8}\right)a \, \mathbf{\hat{y}} + \left(\frac{1}{2} +x_{8}\right)a \, \mathbf{\hat{z}} & \left(24d\right) & \mbox{C VIII} \\ 
\mathbf{B}_{191} & = & \left(\frac{1}{2} - y_{8}\right) \, \mathbf{a}_{1} + \left(\frac{1}{2} +z_{8}\right) \, \mathbf{a}_{2} + x_{8} \, \mathbf{a}_{3} & = & \left(\frac{1}{2} - y_{8}\right)a \, \mathbf{\hat{x}} + \left(\frac{1}{2} +z_{8}\right)a \, \mathbf{\hat{y}} + x_{8}a \, \mathbf{\hat{z}} & \left(24d\right) & \mbox{C VIII} \\ 
\mathbf{B}_{192} & = & \left(\frac{1}{2} +y_{8}\right) \, \mathbf{a}_{1} + z_{8} \, \mathbf{a}_{2} + \left(\frac{1}{2} - x_{8}\right) \, \mathbf{a}_{3} & = & \left(\frac{1}{2} +y_{8}\right)a \, \mathbf{\hat{x}} + z_{8}a \, \mathbf{\hat{y}} + \left(\frac{1}{2} - x_{8}\right)a \, \mathbf{\hat{z}} & \left(24d\right) & \mbox{C VIII} \\ 
\mathbf{B}_{193} & = & x_{9} \, \mathbf{a}_{1} + y_{9} \, \mathbf{a}_{2} + z_{9} \, \mathbf{a}_{3} & = & x_{9}a \, \mathbf{\hat{x}} + y_{9}a \, \mathbf{\hat{y}} + z_{9}a \, \mathbf{\hat{z}} & \left(24d\right) & \mbox{C IX} \\ 
\mathbf{B}_{194} & = & \left(\frac{1}{2} - x_{9}\right) \, \mathbf{a}_{1}-y_{9} \, \mathbf{a}_{2} + \left(\frac{1}{2} +z_{9}\right) \, \mathbf{a}_{3} & = & \left(\frac{1}{2} - x_{9}\right)a \, \mathbf{\hat{x}}-y_{9}a \, \mathbf{\hat{y}} + \left(\frac{1}{2} +z_{9}\right)a \, \mathbf{\hat{z}} & \left(24d\right) & \mbox{C IX} \\ 
\mathbf{B}_{195} & = & -x_{9} \, \mathbf{a}_{1} + \left(\frac{1}{2} +y_{9}\right) \, \mathbf{a}_{2} + \left(\frac{1}{2} - z_{9}\right) \, \mathbf{a}_{3} & = & -x_{9}a \, \mathbf{\hat{x}} + \left(\frac{1}{2} +y_{9}\right)a \, \mathbf{\hat{y}} + \left(\frac{1}{2} - z_{9}\right)a \, \mathbf{\hat{z}} & \left(24d\right) & \mbox{C IX} \\ 
\mathbf{B}_{196} & = & \left(\frac{1}{2} +x_{9}\right) \, \mathbf{a}_{1} + \left(\frac{1}{2} - y_{9}\right) \, \mathbf{a}_{2}-z_{9} \, \mathbf{a}_{3} & = & \left(\frac{1}{2} +x_{9}\right)a \, \mathbf{\hat{x}} + \left(\frac{1}{2} - y_{9}\right)a \, \mathbf{\hat{y}}-z_{9}a \, \mathbf{\hat{z}} & \left(24d\right) & \mbox{C IX} \\ 
\mathbf{B}_{197} & = & z_{9} \, \mathbf{a}_{1} + x_{9} \, \mathbf{a}_{2} + y_{9} \, \mathbf{a}_{3} & = & z_{9}a \, \mathbf{\hat{x}} + x_{9}a \, \mathbf{\hat{y}} + y_{9}a \, \mathbf{\hat{z}} & \left(24d\right) & \mbox{C IX} \\ 
\mathbf{B}_{198} & = & \left(\frac{1}{2} +z_{9}\right) \, \mathbf{a}_{1} + \left(\frac{1}{2} - x_{9}\right) \, \mathbf{a}_{2}-y_{9} \, \mathbf{a}_{3} & = & \left(\frac{1}{2} +z_{9}\right)a \, \mathbf{\hat{x}} + \left(\frac{1}{2} - x_{9}\right)a \, \mathbf{\hat{y}}-y_{9}a \, \mathbf{\hat{z}} & \left(24d\right) & \mbox{C IX} \\ 
\mathbf{B}_{199} & = & \left(\frac{1}{2} - z_{9}\right) \, \mathbf{a}_{1}-x_{9} \, \mathbf{a}_{2} + \left(\frac{1}{2} +y_{9}\right) \, \mathbf{a}_{3} & = & \left(\frac{1}{2} - z_{9}\right)a \, \mathbf{\hat{x}}-x_{9}a \, \mathbf{\hat{y}} + \left(\frac{1}{2} +y_{9}\right)a \, \mathbf{\hat{z}} & \left(24d\right) & \mbox{C IX} \\ 
\mathbf{B}_{200} & = & -z_{9} \, \mathbf{a}_{1} + \left(\frac{1}{2} +x_{9}\right) \, \mathbf{a}_{2} + \left(\frac{1}{2} - y_{9}\right) \, \mathbf{a}_{3} & = & -z_{9}a \, \mathbf{\hat{x}} + \left(\frac{1}{2} +x_{9}\right)a \, \mathbf{\hat{y}} + \left(\frac{1}{2} - y_{9}\right)a \, \mathbf{\hat{z}} & \left(24d\right) & \mbox{C IX} \\ 
\mathbf{B}_{201} & = & y_{9} \, \mathbf{a}_{1} + z_{9} \, \mathbf{a}_{2} + x_{9} \, \mathbf{a}_{3} & = & y_{9}a \, \mathbf{\hat{x}} + z_{9}a \, \mathbf{\hat{y}} + x_{9}a \, \mathbf{\hat{z}} & \left(24d\right) & \mbox{C IX} \\ 
\mathbf{B}_{202} & = & -y_{9} \, \mathbf{a}_{1} + \left(\frac{1}{2} +z_{9}\right) \, \mathbf{a}_{2} + \left(\frac{1}{2} - x_{9}\right) \, \mathbf{a}_{3} & = & -y_{9}a \, \mathbf{\hat{x}} + \left(\frac{1}{2} +z_{9}\right)a \, \mathbf{\hat{y}} + \left(\frac{1}{2} - x_{9}\right)a \, \mathbf{\hat{z}} & \left(24d\right) & \mbox{C IX} \\ 
\mathbf{B}_{203} & = & \left(\frac{1}{2} +y_{9}\right) \, \mathbf{a}_{1} + \left(\frac{1}{2} - z_{9}\right) \, \mathbf{a}_{2}-x_{9} \, \mathbf{a}_{3} & = & \left(\frac{1}{2} +y_{9}\right)a \, \mathbf{\hat{x}} + \left(\frac{1}{2} - z_{9}\right)a \, \mathbf{\hat{y}}-x_{9}a \, \mathbf{\hat{z}} & \left(24d\right) & \mbox{C IX} \\ 
\mathbf{B}_{204} & = & \left(\frac{1}{2} - y_{9}\right) \, \mathbf{a}_{1}-z_{9} \, \mathbf{a}_{2} + \left(\frac{1}{2} +x_{9}\right) \, \mathbf{a}_{3} & = & \left(\frac{1}{2} - y_{9}\right)a \, \mathbf{\hat{x}}-z_{9}a \, \mathbf{\hat{y}} + \left(\frac{1}{2} +x_{9}\right)a \, \mathbf{\hat{z}} & \left(24d\right) & \mbox{C IX} \\ 
\mathbf{B}_{205} & = & -x_{9} \, \mathbf{a}_{1}-y_{9} \, \mathbf{a}_{2}-z_{9} \, \mathbf{a}_{3} & = & -x_{9}a \, \mathbf{\hat{x}}-y_{9}a \, \mathbf{\hat{y}}-z_{9}a \, \mathbf{\hat{z}} & \left(24d\right) & \mbox{C IX} \\ 
\mathbf{B}_{206} & = & \left(\frac{1}{2} +x_{9}\right) \, \mathbf{a}_{1} + y_{9} \, \mathbf{a}_{2} + \left(\frac{1}{2} - z_{9}\right) \, \mathbf{a}_{3} & = & \left(\frac{1}{2} +x_{9}\right)a \, \mathbf{\hat{x}} + y_{9}a \, \mathbf{\hat{y}} + \left(\frac{1}{2} - z_{9}\right)a \, \mathbf{\hat{z}} & \left(24d\right) & \mbox{C IX} \\ 
\mathbf{B}_{207} & = & x_{9} \, \mathbf{a}_{1} + \left(\frac{1}{2} - y_{9}\right) \, \mathbf{a}_{2} + \left(\frac{1}{2} +z_{9}\right) \, \mathbf{a}_{3} & = & x_{9}a \, \mathbf{\hat{x}} + \left(\frac{1}{2} - y_{9}\right)a \, \mathbf{\hat{y}} + \left(\frac{1}{2} +z_{9}\right)a \, \mathbf{\hat{z}} & \left(24d\right) & \mbox{C IX} \\ 
\mathbf{B}_{208} & = & \left(\frac{1}{2} - x_{9}\right) \, \mathbf{a}_{1} + \left(\frac{1}{2} +y_{9}\right) \, \mathbf{a}_{2} + z_{9} \, \mathbf{a}_{3} & = & \left(\frac{1}{2} - x_{9}\right)a \, \mathbf{\hat{x}} + \left(\frac{1}{2} +y_{9}\right)a \, \mathbf{\hat{y}} + z_{9}a \, \mathbf{\hat{z}} & \left(24d\right) & \mbox{C IX} \\ 
\mathbf{B}_{209} & = & -z_{9} \, \mathbf{a}_{1}-x_{9} \, \mathbf{a}_{2}-y_{9} \, \mathbf{a}_{3} & = & -z_{9}a \, \mathbf{\hat{x}}-x_{9}a \, \mathbf{\hat{y}}-y_{9}a \, \mathbf{\hat{z}} & \left(24d\right) & \mbox{C IX} \\ 
\mathbf{B}_{210} & = & \left(\frac{1}{2} - z_{9}\right) \, \mathbf{a}_{1} + \left(\frac{1}{2} +x_{9}\right) \, \mathbf{a}_{2} + y_{9} \, \mathbf{a}_{3} & = & \left(\frac{1}{2} - z_{9}\right)a \, \mathbf{\hat{x}} + \left(\frac{1}{2} +x_{9}\right)a \, \mathbf{\hat{y}} + y_{9}a \, \mathbf{\hat{z}} & \left(24d\right) & \mbox{C IX} \\ 
\mathbf{B}_{211} & = & \left(\frac{1}{2} +z_{9}\right) \, \mathbf{a}_{1} + x_{9} \, \mathbf{a}_{2} + \left(\frac{1}{2} - y_{9}\right) \, \mathbf{a}_{3} & = & \left(\frac{1}{2} +z_{9}\right)a \, \mathbf{\hat{x}} + x_{9}a \, \mathbf{\hat{y}} + \left(\frac{1}{2} - y_{9}\right)a \, \mathbf{\hat{z}} & \left(24d\right) & \mbox{C IX} \\ 
\mathbf{B}_{212} & = & z_{9} \, \mathbf{a}_{1} + \left(\frac{1}{2} - x_{9}\right) \, \mathbf{a}_{2} + \left(\frac{1}{2} +y_{9}\right) \, \mathbf{a}_{3} & = & z_{9}a \, \mathbf{\hat{x}} + \left(\frac{1}{2} - x_{9}\right)a \, \mathbf{\hat{y}} + \left(\frac{1}{2} +y_{9}\right)a \, \mathbf{\hat{z}} & \left(24d\right) & \mbox{C IX} \\ 
\mathbf{B}_{213} & = & -y_{9} \, \mathbf{a}_{1}-z_{9} \, \mathbf{a}_{2}-x_{9} \, \mathbf{a}_{3} & = & -y_{9}a \, \mathbf{\hat{x}}-z_{9}a \, \mathbf{\hat{y}}-x_{9}a \, \mathbf{\hat{z}} & \left(24d\right) & \mbox{C IX} \\ 
\mathbf{B}_{214} & = & y_{9} \, \mathbf{a}_{1} + \left(\frac{1}{2} - z_{9}\right) \, \mathbf{a}_{2} + \left(\frac{1}{2} +x_{9}\right) \, \mathbf{a}_{3} & = & y_{9}a \, \mathbf{\hat{x}} + \left(\frac{1}{2} - z_{9}\right)a \, \mathbf{\hat{y}} + \left(\frac{1}{2} +x_{9}\right)a \, \mathbf{\hat{z}} & \left(24d\right) & \mbox{C IX} \\ 
\mathbf{B}_{215} & = & \left(\frac{1}{2} - y_{9}\right) \, \mathbf{a}_{1} + \left(\frac{1}{2} +z_{9}\right) \, \mathbf{a}_{2} + x_{9} \, \mathbf{a}_{3} & = & \left(\frac{1}{2} - y_{9}\right)a \, \mathbf{\hat{x}} + \left(\frac{1}{2} +z_{9}\right)a \, \mathbf{\hat{y}} + x_{9}a \, \mathbf{\hat{z}} & \left(24d\right) & \mbox{C IX} \\ 
\mathbf{B}_{216} & = & \left(\frac{1}{2} +y_{9}\right) \, \mathbf{a}_{1} + z_{9} \, \mathbf{a}_{2} + \left(\frac{1}{2} - x_{9}\right) \, \mathbf{a}_{3} & = & \left(\frac{1}{2} +y_{9}\right)a \, \mathbf{\hat{x}} + z_{9}a \, \mathbf{\hat{y}} + \left(\frac{1}{2} - x_{9}\right)a \, \mathbf{\hat{z}} & \left(24d\right) & \mbox{C IX} \\ 
\mathbf{B}_{217} & = & x_{10} \, \mathbf{a}_{1} + y_{10} \, \mathbf{a}_{2} + z_{10} \, \mathbf{a}_{3} & = & x_{10}a \, \mathbf{\hat{x}} + y_{10}a \, \mathbf{\hat{y}} + z_{10}a \, \mathbf{\hat{z}} & \left(24d\right) & \mbox{C X} \\ 
\mathbf{B}_{218} & = & \left(\frac{1}{2} - x_{10}\right) \, \mathbf{a}_{1}-y_{10} \, \mathbf{a}_{2} + \left(\frac{1}{2} +z_{10}\right) \, \mathbf{a}_{3} & = & \left(\frac{1}{2} - x_{10}\right)a \, \mathbf{\hat{x}}-y_{10}a \, \mathbf{\hat{y}} + \left(\frac{1}{2} +z_{10}\right)a \, \mathbf{\hat{z}} & \left(24d\right) & \mbox{C X} \\ 
\mathbf{B}_{219} & = & -x_{10} \, \mathbf{a}_{1} + \left(\frac{1}{2} +y_{10}\right) \, \mathbf{a}_{2} + \left(\frac{1}{2} - z_{10}\right) \, \mathbf{a}_{3} & = & -x_{10}a \, \mathbf{\hat{x}} + \left(\frac{1}{2} +y_{10}\right)a \, \mathbf{\hat{y}} + \left(\frac{1}{2} - z_{10}\right)a \, \mathbf{\hat{z}} & \left(24d\right) & \mbox{C X} \\ 
\mathbf{B}_{220} & = & \left(\frac{1}{2} +x_{10}\right) \, \mathbf{a}_{1} + \left(\frac{1}{2} - y_{10}\right) \, \mathbf{a}_{2}-z_{10} \, \mathbf{a}_{3} & = & \left(\frac{1}{2} +x_{10}\right)a \, \mathbf{\hat{x}} + \left(\frac{1}{2} - y_{10}\right)a \, \mathbf{\hat{y}}-z_{10}a \, \mathbf{\hat{z}} & \left(24d\right) & \mbox{C X} \\ 
\mathbf{B}_{221} & = & z_{10} \, \mathbf{a}_{1} + x_{10} \, \mathbf{a}_{2} + y_{10} \, \mathbf{a}_{3} & = & z_{10}a \, \mathbf{\hat{x}} + x_{10}a \, \mathbf{\hat{y}} + y_{10}a \, \mathbf{\hat{z}} & \left(24d\right) & \mbox{C X} \\ 
\mathbf{B}_{222} & = & \left(\frac{1}{2} +z_{10}\right) \, \mathbf{a}_{1} + \left(\frac{1}{2} - x_{10}\right) \, \mathbf{a}_{2}-y_{10} \, \mathbf{a}_{3} & = & \left(\frac{1}{2} +z_{10}\right)a \, \mathbf{\hat{x}} + \left(\frac{1}{2} - x_{10}\right)a \, \mathbf{\hat{y}}-y_{10}a \, \mathbf{\hat{z}} & \left(24d\right) & \mbox{C X} \\ 
\mathbf{B}_{223} & = & \left(\frac{1}{2} - z_{10}\right) \, \mathbf{a}_{1}-x_{10} \, \mathbf{a}_{2} + \left(\frac{1}{2} +y_{10}\right) \, \mathbf{a}_{3} & = & \left(\frac{1}{2} - z_{10}\right)a \, \mathbf{\hat{x}}-x_{10}a \, \mathbf{\hat{y}} + \left(\frac{1}{2} +y_{10}\right)a \, \mathbf{\hat{z}} & \left(24d\right) & \mbox{C X} \\ 
\mathbf{B}_{224} & = & -z_{10} \, \mathbf{a}_{1} + \left(\frac{1}{2} +x_{10}\right) \, \mathbf{a}_{2} + \left(\frac{1}{2} - y_{10}\right) \, \mathbf{a}_{3} & = & -z_{10}a \, \mathbf{\hat{x}} + \left(\frac{1}{2} +x_{10}\right)a \, \mathbf{\hat{y}} + \left(\frac{1}{2} - y_{10}\right)a \, \mathbf{\hat{z}} & \left(24d\right) & \mbox{C X} \\ 
\mathbf{B}_{225} & = & y_{10} \, \mathbf{a}_{1} + z_{10} \, \mathbf{a}_{2} + x_{10} \, \mathbf{a}_{3} & = & y_{10}a \, \mathbf{\hat{x}} + z_{10}a \, \mathbf{\hat{y}} + x_{10}a \, \mathbf{\hat{z}} & \left(24d\right) & \mbox{C X} \\ 
\mathbf{B}_{226} & = & -y_{10} \, \mathbf{a}_{1} + \left(\frac{1}{2} +z_{10}\right) \, \mathbf{a}_{2} + \left(\frac{1}{2} - x_{10}\right) \, \mathbf{a}_{3} & = & -y_{10}a \, \mathbf{\hat{x}} + \left(\frac{1}{2} +z_{10}\right)a \, \mathbf{\hat{y}} + \left(\frac{1}{2} - x_{10}\right)a \, \mathbf{\hat{z}} & \left(24d\right) & \mbox{C X} \\ 
\mathbf{B}_{227} & = & \left(\frac{1}{2} +y_{10}\right) \, \mathbf{a}_{1} + \left(\frac{1}{2} - z_{10}\right) \, \mathbf{a}_{2}-x_{10} \, \mathbf{a}_{3} & = & \left(\frac{1}{2} +y_{10}\right)a \, \mathbf{\hat{x}} + \left(\frac{1}{2} - z_{10}\right)a \, \mathbf{\hat{y}}-x_{10}a \, \mathbf{\hat{z}} & \left(24d\right) & \mbox{C X} \\ 
\mathbf{B}_{228} & = & \left(\frac{1}{2} - y_{10}\right) \, \mathbf{a}_{1}-z_{10} \, \mathbf{a}_{2} + \left(\frac{1}{2} +x_{10}\right) \, \mathbf{a}_{3} & = & \left(\frac{1}{2} - y_{10}\right)a \, \mathbf{\hat{x}}-z_{10}a \, \mathbf{\hat{y}} + \left(\frac{1}{2} +x_{10}\right)a \, \mathbf{\hat{z}} & \left(24d\right) & \mbox{C X} \\ 
\mathbf{B}_{229} & = & -x_{10} \, \mathbf{a}_{1}-y_{10} \, \mathbf{a}_{2}-z_{10} \, \mathbf{a}_{3} & = & -x_{10}a \, \mathbf{\hat{x}}-y_{10}a \, \mathbf{\hat{y}}-z_{10}a \, \mathbf{\hat{z}} & \left(24d\right) & \mbox{C X} \\ 
\mathbf{B}_{230} & = & \left(\frac{1}{2} +x_{10}\right) \, \mathbf{a}_{1} + y_{10} \, \mathbf{a}_{2} + \left(\frac{1}{2} - z_{10}\right) \, \mathbf{a}_{3} & = & \left(\frac{1}{2} +x_{10}\right)a \, \mathbf{\hat{x}} + y_{10}a \, \mathbf{\hat{y}} + \left(\frac{1}{2} - z_{10}\right)a \, \mathbf{\hat{z}} & \left(24d\right) & \mbox{C X} \\ 
\mathbf{B}_{231} & = & x_{10} \, \mathbf{a}_{1} + \left(\frac{1}{2} - y_{10}\right) \, \mathbf{a}_{2} + \left(\frac{1}{2} +z_{10}\right) \, \mathbf{a}_{3} & = & x_{10}a \, \mathbf{\hat{x}} + \left(\frac{1}{2} - y_{10}\right)a \, \mathbf{\hat{y}} + \left(\frac{1}{2} +z_{10}\right)a \, \mathbf{\hat{z}} & \left(24d\right) & \mbox{C X} \\ 
\mathbf{B}_{232} & = & \left(\frac{1}{2} - x_{10}\right) \, \mathbf{a}_{1} + \left(\frac{1}{2} +y_{10}\right) \, \mathbf{a}_{2} + z_{10} \, \mathbf{a}_{3} & = & \left(\frac{1}{2} - x_{10}\right)a \, \mathbf{\hat{x}} + \left(\frac{1}{2} +y_{10}\right)a \, \mathbf{\hat{y}} + z_{10}a \, \mathbf{\hat{z}} & \left(24d\right) & \mbox{C X} \\ 
\mathbf{B}_{233} & = & -z_{10} \, \mathbf{a}_{1}-x_{10} \, \mathbf{a}_{2}-y_{10} \, \mathbf{a}_{3} & = & -z_{10}a \, \mathbf{\hat{x}}-x_{10}a \, \mathbf{\hat{y}}-y_{10}a \, \mathbf{\hat{z}} & \left(24d\right) & \mbox{C X} \\ 
\mathbf{B}_{234} & = & \left(\frac{1}{2} - z_{10}\right) \, \mathbf{a}_{1} + \left(\frac{1}{2} +x_{10}\right) \, \mathbf{a}_{2} + y_{10} \, \mathbf{a}_{3} & = & \left(\frac{1}{2} - z_{10}\right)a \, \mathbf{\hat{x}} + \left(\frac{1}{2} +x_{10}\right)a \, \mathbf{\hat{y}} + y_{10}a \, \mathbf{\hat{z}} & \left(24d\right) & \mbox{C X} \\ 
\mathbf{B}_{235} & = & \left(\frac{1}{2} +z_{10}\right) \, \mathbf{a}_{1} + x_{10} \, \mathbf{a}_{2} + \left(\frac{1}{2} - y_{10}\right) \, \mathbf{a}_{3} & = & \left(\frac{1}{2} +z_{10}\right)a \, \mathbf{\hat{x}} + x_{10}a \, \mathbf{\hat{y}} + \left(\frac{1}{2} - y_{10}\right)a \, \mathbf{\hat{z}} & \left(24d\right) & \mbox{C X} \\ 
\mathbf{B}_{236} & = & z_{10} \, \mathbf{a}_{1} + \left(\frac{1}{2} - x_{10}\right) \, \mathbf{a}_{2} + \left(\frac{1}{2} +y_{10}\right) \, \mathbf{a}_{3} & = & z_{10}a \, \mathbf{\hat{x}} + \left(\frac{1}{2} - x_{10}\right)a \, \mathbf{\hat{y}} + \left(\frac{1}{2} +y_{10}\right)a \, \mathbf{\hat{z}} & \left(24d\right) & \mbox{C X} \\ 
\mathbf{B}_{237} & = & -y_{10} \, \mathbf{a}_{1}-z_{10} \, \mathbf{a}_{2}-x_{10} \, \mathbf{a}_{3} & = & -y_{10}a \, \mathbf{\hat{x}}-z_{10}a \, \mathbf{\hat{y}}-x_{10}a \, \mathbf{\hat{z}} & \left(24d\right) & \mbox{C X} \\ 
\mathbf{B}_{238} & = & y_{10} \, \mathbf{a}_{1} + \left(\frac{1}{2} - z_{10}\right) \, \mathbf{a}_{2} + \left(\frac{1}{2} +x_{10}\right) \, \mathbf{a}_{3} & = & y_{10}a \, \mathbf{\hat{x}} + \left(\frac{1}{2} - z_{10}\right)a \, \mathbf{\hat{y}} + \left(\frac{1}{2} +x_{10}\right)a \, \mathbf{\hat{z}} & \left(24d\right) & \mbox{C X} \\ 
\mathbf{B}_{239} & = & \left(\frac{1}{2} - y_{10}\right) \, \mathbf{a}_{1} + \left(\frac{1}{2} +z_{10}\right) \, \mathbf{a}_{2} + x_{10} \, \mathbf{a}_{3} & = & \left(\frac{1}{2} - y_{10}\right)a \, \mathbf{\hat{x}} + \left(\frac{1}{2} +z_{10}\right)a \, \mathbf{\hat{y}} + x_{10}a \, \mathbf{\hat{z}} & \left(24d\right) & \mbox{C X} \\ 
\mathbf{B}_{240} & = & \left(\frac{1}{2} +y_{10}\right) \, \mathbf{a}_{1} + z_{10} \, \mathbf{a}_{2} + \left(\frac{1}{2} - x_{10}\right) \, \mathbf{a}_{3} & = & \left(\frac{1}{2} +y_{10}\right)a \, \mathbf{\hat{x}} + z_{10}a \, \mathbf{\hat{y}} + \left(\frac{1}{2} - x_{10}\right)a \, \mathbf{\hat{z}} & \left(24d\right) & \mbox{C X} \\ 
\end{longtabu}
\renewcommand{\arraystretch}{1.0}
\noindent \hrulefill
\\
\textbf{References:}
\vspace*{-0.25cm}
\begin{flushleft}
  - \bibentry{David_Nature_353_1991}. \\
\end{flushleft}
\noindent \hrulefill
\\
\textbf{Geometry files:}
\\
\noindent  - CIF: pp. {\hyperref[A_cP240_205_10d_cif]{\pageref{A_cP240_205_10d_cif}}} \\
\noindent  - POSCAR: pp. {\hyperref[A_cP240_205_10d_poscar]{\pageref{A_cP240_205_10d_poscar}}} \\
\onecolumn
{\phantomsection\label{AB3C2_cI96_206_c_e_ad}}
\subsection*{\huge \textbf{{\normalfont AlLi$_{3}$N$_{2}$ ($E9_{d}$) Structure: AB3C2\_cI96\_206\_c\_e\_ad}}}
\noindent \hrulefill
\vspace*{0.25cm}
\begin{figure}[htp]
  \centering
  \vspace{-1em}
  {\includegraphics[width=1\textwidth]{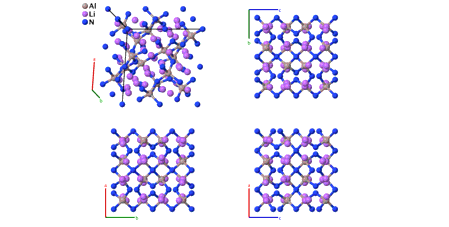}}
\end{figure}
\vspace*{-0.5cm}
\renewcommand{\arraystretch}{1.5}
\begin{equation*}
  \begin{array}{>{$\hspace{-0.15cm}}l<{$}>{$}p{0.5cm}<{$}>{$}p{18.5cm}<{$}}
    \mbox{\large \textbf{Prototype}} &\colon & \ce{AlLi$_{3}$N$_{2}$} \\
    \mbox{\large \textbf{\AFLOW\ prototype label}} &\colon & \mbox{AB3C2\_cI96\_206\_c\_e\_ad} \\
    \mbox{\large \textbf{\textit{Strukturbericht} designation}} &\colon & \mbox{$E9_{d}$} \\
    \mbox{\large \textbf{Pearson symbol}} &\colon & \mbox{cI96} \\
    \mbox{\large \textbf{Space group number}} &\colon & 206 \\
    \mbox{\large \textbf{Space group symbol}} &\colon & Ia\bar{3} \\
    \mbox{\large \textbf{\AFLOW\ prototype command}} &\colon &  \texttt{aflow} \,  \, \texttt{-{}-proto=AB3C2\_cI96\_206\_c\_e\_ad } \, \newline \texttt{-{}-params=}{a,x_{2},x_{3},x_{4},y_{4},z_{4} }
  \end{array}
\end{equation*}
\renewcommand{\arraystretch}{1.0}

\vspace*{-0.25cm}
\noindent \hrulefill
\\
\textbf{ Other compounds with this structure:}
\begin{itemize}
   \item{ GaLi$_{3}$N$_{2}$, ScLi$_{3}$N$_{2}$, TiLi$_{3}$N$_{2}$, ZnLi$_{3}$N$_{2}$, SiLi$_{3}$N$_{2}$, GeLi$_{3}$N$_{2}$  }
\end{itemize}
\noindent \parbox{1 \linewidth}{
\noindent \hrulefill
\\
\textbf{Body-centered Cubic primitive vectors:} \\
\vspace*{-0.25cm}
\begin{tabular}{cc}
  \begin{tabular}{c}
    \parbox{0.6 \linewidth}{
      \renewcommand{\arraystretch}{1.5}
      \begin{equation*}
        \centering
        \begin{array}{ccc}
              \mathbf{a}_1 & = & - \frac12 \, a \, \mathbf{\hat{x}} + \frac12 \, a \, \mathbf{\hat{y}} + \frac12 \, a \, \mathbf{\hat{z}} \\
    \mathbf{a}_2 & = & ~ \frac12 \, a \, \mathbf{\hat{x}} - \frac12 \, a \, \mathbf{\hat{y}} + \frac12 \, a \, \mathbf{\hat{z}} \\
    \mathbf{a}_3 & = & ~ \frac12 \, a \, \mathbf{\hat{x}} + \frac12 \, a \, \mathbf{\hat{y}} - \frac12 \, a \, \mathbf{\hat{z}} \\

        \end{array}
      \end{equation*}
    }
    \renewcommand{\arraystretch}{1.0}
  \end{tabular}
  \begin{tabular}{c}
    \includegraphics[width=0.3\linewidth]{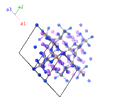} \\
  \end{tabular}
\end{tabular}

}
\vspace*{-0.25cm}

\noindent \hrulefill
\\
\textbf{Basis vectors:}
\vspace*{-0.25cm}
\renewcommand{\arraystretch}{1.5}
\begin{longtabu} to \textwidth{>{\centering $}X[-1,c,c]<{$}>{\centering $}X[-1,c,c]<{$}>{\centering $}X[-1,c,c]<{$}>{\centering $}X[-1,c,c]<{$}>{\centering $}X[-1,c,c]<{$}>{\centering $}X[-1,c,c]<{$}>{\centering $}X[-1,c,c]<{$}}
  & & \mbox{Lattice Coordinates} & & \mbox{Cartesian Coordinates} &\mbox{Wyckoff Position} & \mbox{Atom Type} \\  
  \mathbf{B}_{1} & = & 0 \, \mathbf{a}_{1} + 0 \, \mathbf{a}_{2} + 0 \, \mathbf{a}_{3} & = & 0 \, \mathbf{\hat{x}} + 0 \, \mathbf{\hat{y}} + 0 \, \mathbf{\hat{z}} & \left(8a\right) & \mbox{N I} \\ 
\mathbf{B}_{2} & = & \frac{1}{2} \, \mathbf{a}_{1} + \frac{1}{2} \, \mathbf{a}_{3} & = & \frac{1}{2}a \, \mathbf{\hat{y}} & \left(8a\right) & \mbox{N I} \\ 
\mathbf{B}_{3} & = & \frac{1}{2} \, \mathbf{a}_{2} + \frac{1}{2} \, \mathbf{a}_{3} & = & \frac{1}{2}a \, \mathbf{\hat{x}} & \left(8a\right) & \mbox{N I} \\ 
\mathbf{B}_{4} & = & \frac{1}{2} \, \mathbf{a}_{1} + \frac{1}{2} \, \mathbf{a}_{2} & = & \frac{1}{2}a \, \mathbf{\hat{z}} & \left(8a\right) & \mbox{N I} \\ 
\mathbf{B}_{5} & = & 2x_{2} \, \mathbf{a}_{1} + 2x_{2} \, \mathbf{a}_{2} + 2x_{2} \, \mathbf{a}_{3} & = & x_{2}a \, \mathbf{\hat{x}} + x_{2}a \, \mathbf{\hat{y}} + x_{2}a \, \mathbf{\hat{z}} & \left(16c\right) & \mbox{Al} \\ 
\mathbf{B}_{6} & = & \frac{1}{2} \, \mathbf{a}_{1} + \left(\frac{1}{2} - 2x_{2}\right) \, \mathbf{a}_{3} & = & -x_{2}a \, \mathbf{\hat{x}} + \left(\frac{1}{2} - x_{2}\right)a \, \mathbf{\hat{y}} + x_{2}a \, \mathbf{\hat{z}} & \left(16c\right) & \mbox{Al} \\ 
\mathbf{B}_{7} & = & \left(\frac{1}{2} - 2x_{2}\right) \, \mathbf{a}_{2} + \frac{1}{2} \, \mathbf{a}_{3} & = & \left(\frac{1}{2} - x_{2}\right)a \, \mathbf{\hat{x}} + x_{2}a \, \mathbf{\hat{y}}-x_{2}a \, \mathbf{\hat{z}} & \left(16c\right) & \mbox{Al} \\ 
\mathbf{B}_{8} & = & \left(\frac{1}{2} - 2x_{2}\right) \, \mathbf{a}_{1} + \frac{1}{2} \, \mathbf{a}_{2} & = & x_{2}a \, \mathbf{\hat{x}}-x_{2}a \, \mathbf{\hat{y}} + \left(\frac{1}{2} - x_{2}\right)a \, \mathbf{\hat{z}} & \left(16c\right) & \mbox{Al} \\ 
\mathbf{B}_{9} & = & -2x_{2} \, \mathbf{a}_{1}-2x_{2} \, \mathbf{a}_{2}-2x_{2} \, \mathbf{a}_{3} & = & -x_{2}a \, \mathbf{\hat{x}}-x_{2}a \, \mathbf{\hat{y}}-x_{2}a \, \mathbf{\hat{z}} & \left(16c\right) & \mbox{Al} \\ 
\mathbf{B}_{10} & = & \frac{1}{2} \, \mathbf{a}_{1} + \left(\frac{1}{2} +2x_{2}\right) \, \mathbf{a}_{3} & = & x_{2}a \, \mathbf{\hat{x}} + \left(\frac{1}{2} +x_{2}\right)a \, \mathbf{\hat{y}}-x_{2}a \, \mathbf{\hat{z}} & \left(16c\right) & \mbox{Al} \\ 
\mathbf{B}_{11} & = & \left(\frac{1}{2} +2x_{2}\right) \, \mathbf{a}_{2} + \frac{1}{2} \, \mathbf{a}_{3} & = & \left(\frac{1}{2} +x_{2}\right)a \, \mathbf{\hat{x}}-x_{2}a \, \mathbf{\hat{y}} + x_{2}a \, \mathbf{\hat{z}} & \left(16c\right) & \mbox{Al} \\ 
\mathbf{B}_{12} & = & \left(\frac{1}{2} +2x_{2}\right) \, \mathbf{a}_{1} + \frac{1}{2} \, \mathbf{a}_{2} & = & -x_{2}a \, \mathbf{\hat{x}} + x_{2}a \, \mathbf{\hat{y}} + \left(\frac{1}{2} +x_{2}\right)a \, \mathbf{\hat{z}} & \left(16c\right) & \mbox{Al} \\ 
\mathbf{B}_{13} & = & \frac{1}{4} \, \mathbf{a}_{1} + \left(\frac{1}{4} +x_{3}\right) \, \mathbf{a}_{2} + x_{3} \, \mathbf{a}_{3} & = & x_{3}a \, \mathbf{\hat{x}} + \frac{1}{4}a \, \mathbf{\hat{z}} & \left(24d\right) & \mbox{N II} \\ 
\mathbf{B}_{14} & = & \frac{3}{4} \, \mathbf{a}_{1} + \left(\frac{1}{4} - x_{3}\right) \, \mathbf{a}_{2} + \left(\frac{1}{2} - x_{3}\right) \, \mathbf{a}_{3} & = & -x_{3}a \, \mathbf{\hat{x}} + \frac{1}{2}a \, \mathbf{\hat{y}} + \frac{1}{4}a \, \mathbf{\hat{z}} & \left(24d\right) & \mbox{N II} \\ 
\mathbf{B}_{15} & = & x_{3} \, \mathbf{a}_{1} + \frac{1}{4} \, \mathbf{a}_{2} + \left(\frac{1}{4} +x_{3}\right) \, \mathbf{a}_{3} & = & \frac{1}{4}a \, \mathbf{\hat{x}} + x_{3}a \, \mathbf{\hat{y}} & \left(24d\right) & \mbox{N II} \\ 
\mathbf{B}_{16} & = & \left(\frac{1}{2} - x_{3}\right) \, \mathbf{a}_{1} + \frac{3}{4} \, \mathbf{a}_{2} + \left(\frac{1}{4} - x_{3}\right) \, \mathbf{a}_{3} & = & \frac{1}{4}a \, \mathbf{\hat{x}}-x_{3}a \, \mathbf{\hat{y}} + \frac{1}{2}a \, \mathbf{\hat{z}} & \left(24d\right) & \mbox{N II} \\ 
\mathbf{B}_{17} & = & \left(\frac{1}{4} +x_{3}\right) \, \mathbf{a}_{1} + x_{3} \, \mathbf{a}_{2} + \frac{1}{4} \, \mathbf{a}_{3} & = & \frac{1}{4}a \, \mathbf{\hat{y}} + x_{3}a \, \mathbf{\hat{z}} & \left(24d\right) & \mbox{N II} \\ 
\mathbf{B}_{18} & = & \left(\frac{1}{4} - x_{3}\right) \, \mathbf{a}_{1} + \left(\frac{1}{2} - x_{3}\right) \, \mathbf{a}_{2} + \frac{3}{4} \, \mathbf{a}_{3} & = & \frac{1}{2}a \, \mathbf{\hat{x}} + \frac{1}{4}a \, \mathbf{\hat{y}}-x_{3}a \, \mathbf{\hat{z}} & \left(24d\right) & \mbox{N II} \\ 
\mathbf{B}_{19} & = & \frac{3}{4} \, \mathbf{a}_{1} + \left(\frac{3}{4} - x_{3}\right) \, \mathbf{a}_{2}-x_{3} \, \mathbf{a}_{3} & = & -x_{3}a \, \mathbf{\hat{x}} + \frac{3}{4}a \, \mathbf{\hat{z}} & \left(24d\right) & \mbox{N II} \\ 
\mathbf{B}_{20} & = & \frac{1}{4} \, \mathbf{a}_{1} + \left(\frac{3}{4} +x_{3}\right) \, \mathbf{a}_{2} + \left(\frac{1}{2} +x_{3}\right) \, \mathbf{a}_{3} & = & \left(\frac{1}{2} +x_{3}\right)a \, \mathbf{\hat{x}} + \frac{1}{4}a \, \mathbf{\hat{z}} & \left(24d\right) & \mbox{N II} \\ 
\mathbf{B}_{21} & = & -x_{3} \, \mathbf{a}_{1} + \frac{3}{4} \, \mathbf{a}_{2} + \left(\frac{3}{4} - x_{3}\right) \, \mathbf{a}_{3} & = & \frac{3}{4}a \, \mathbf{\hat{x}}-x_{3}a \, \mathbf{\hat{y}} & \left(24d\right) & \mbox{N II} \\ 
\mathbf{B}_{22} & = & \left(\frac{1}{2} +x_{3}\right) \, \mathbf{a}_{1} + \frac{1}{4} \, \mathbf{a}_{2} + \left(\frac{3}{4} +x_{3}\right) \, \mathbf{a}_{3} & = & \frac{1}{4}a \, \mathbf{\hat{x}} + \left(\frac{1}{2} +x_{3}\right)a \, \mathbf{\hat{y}} & \left(24d\right) & \mbox{N II} \\ 
\mathbf{B}_{23} & = & \left(\frac{3}{4} - x_{3}\right) \, \mathbf{a}_{1}-x_{3} \, \mathbf{a}_{2} + \frac{3}{4} \, \mathbf{a}_{3} & = & \frac{3}{4}a \, \mathbf{\hat{y}}-x_{3}a \, \mathbf{\hat{z}} & \left(24d\right) & \mbox{N II} \\ 
\mathbf{B}_{24} & = & \left(\frac{3}{4} +x_{3}\right) \, \mathbf{a}_{1} + \left(\frac{1}{2} +x_{3}\right) \, \mathbf{a}_{2} + \frac{1}{4} \, \mathbf{a}_{3} & = & \frac{1}{4}a \, \mathbf{\hat{y}} + \left(\frac{1}{2} +x_{3}\right)a \, \mathbf{\hat{z}} & \left(24d\right) & \mbox{N II} \\ 
\mathbf{B}_{25} & = & \left(y_{4}+z_{4}\right) \, \mathbf{a}_{1} + \left(x_{4}+z_{4}\right) \, \mathbf{a}_{2} + \left(x_{4}+y_{4}\right) \, \mathbf{a}_{3} & = & x_{4}a \, \mathbf{\hat{x}} + y_{4}a \, \mathbf{\hat{y}} + z_{4}a \, \mathbf{\hat{z}} & \left(48e\right) & \mbox{Li} \\ 
\mathbf{B}_{26} & = & \left(\frac{1}{2} - y_{4} + z_{4}\right) \, \mathbf{a}_{1} + \left(-x_{4}+z_{4}\right) \, \mathbf{a}_{2} + \left(\frac{1}{2} - x_{4} - y_{4}\right) \, \mathbf{a}_{3} & = & -x_{4}a \, \mathbf{\hat{x}} + \left(\frac{1}{2} - y_{4}\right)a \, \mathbf{\hat{y}} + z_{4}a \, \mathbf{\hat{z}} & \left(48e\right) & \mbox{Li} \\ 
\mathbf{B}_{27} & = & \left(y_{4}-z_{4}\right) \, \mathbf{a}_{1} + \left(\frac{1}{2} - x_{4} - z_{4}\right) \, \mathbf{a}_{2} + \left(\frac{1}{2} - x_{4} + y_{4}\right) \, \mathbf{a}_{3} & = & \left(\frac{1}{2} - x_{4}\right)a \, \mathbf{\hat{x}} + y_{4}a \, \mathbf{\hat{y}}-z_{4}a \, \mathbf{\hat{z}} & \left(48e\right) & \mbox{Li} \\ 
\mathbf{B}_{28} & = & \left(\frac{1}{2} - y_{4} - z_{4}\right) \, \mathbf{a}_{1} + \left(\frac{1}{2} +x_{4} - z_{4}\right) \, \mathbf{a}_{2} + \left(x_{4}-y_{4}\right) \, \mathbf{a}_{3} & = & x_{4}a \, \mathbf{\hat{x}}-y_{4}a \, \mathbf{\hat{y}} + \left(\frac{1}{2} - z_{4}\right)a \, \mathbf{\hat{z}} & \left(48e\right) & \mbox{Li} \\ 
\mathbf{B}_{29} & = & \left(x_{4}+y_{4}\right) \, \mathbf{a}_{1} + \left(y_{4}+z_{4}\right) \, \mathbf{a}_{2} + \left(x_{4}+z_{4}\right) \, \mathbf{a}_{3} & = & z_{4}a \, \mathbf{\hat{x}} + x_{4}a \, \mathbf{\hat{y}} + y_{4}a \, \mathbf{\hat{z}} & \left(48e\right) & \mbox{Li} \\ 
\mathbf{B}_{30} & = & \left(\frac{1}{2} - x_{4} - y_{4}\right) \, \mathbf{a}_{1} + \left(\frac{1}{2} - y_{4} + z_{4}\right) \, \mathbf{a}_{2} + \left(-x_{4}+z_{4}\right) \, \mathbf{a}_{3} & = & z_{4}a \, \mathbf{\hat{x}}-x_{4}a \, \mathbf{\hat{y}} + \left(\frac{1}{2} - y_{4}\right)a \, \mathbf{\hat{z}} & \left(48e\right) & \mbox{Li} \\ 
\mathbf{B}_{31} & = & \left(\frac{1}{2} - x_{4} + y_{4}\right) \, \mathbf{a}_{1} + \left(y_{4}-z_{4}\right) \, \mathbf{a}_{2} + \left(\frac{1}{2} - x_{4} - z_{4}\right) \, \mathbf{a}_{3} & = & -z_{4}a \, \mathbf{\hat{x}} + \left(\frac{1}{2} - x_{4}\right)a \, \mathbf{\hat{y}} + y_{4}a \, \mathbf{\hat{z}} & \left(48e\right) & \mbox{Li} \\ 
\mathbf{B}_{32} & = & \left(x_{4}-y_{4}\right) \, \mathbf{a}_{1} + \left(\frac{1}{2} - y_{4} - z_{4}\right) \, \mathbf{a}_{2} + \left(\frac{1}{2} +x_{4} - z_{4}\right) \, \mathbf{a}_{3} & = & \left(\frac{1}{2} - z_{4}\right)a \, \mathbf{\hat{x}} + x_{4}a \, \mathbf{\hat{y}}-y_{4}a \, \mathbf{\hat{z}} & \left(48e\right) & \mbox{Li} \\ 
\mathbf{B}_{33} & = & \left(x_{4}+z_{4}\right) \, \mathbf{a}_{1} + \left(x_{4}+y_{4}\right) \, \mathbf{a}_{2} + \left(y_{4}+z_{4}\right) \, \mathbf{a}_{3} & = & y_{4}a \, \mathbf{\hat{x}} + z_{4}a \, \mathbf{\hat{y}} + x_{4}a \, \mathbf{\hat{z}} & \left(48e\right) & \mbox{Li} \\ 
\mathbf{B}_{34} & = & \left(-x_{4}+z_{4}\right) \, \mathbf{a}_{1} + \left(\frac{1}{2} - x_{4} - y_{4}\right) \, \mathbf{a}_{2} + \left(\frac{1}{2} - y_{4} + z_{4}\right) \, \mathbf{a}_{3} & = & \left(\frac{1}{2} - y_{4}\right)a \, \mathbf{\hat{x}} + z_{4}a \, \mathbf{\hat{y}}-x_{4}a \, \mathbf{\hat{z}} & \left(48e\right) & \mbox{Li} \\ 
\mathbf{B}_{35} & = & \left(\frac{1}{2} - x_{4} - z_{4}\right) \, \mathbf{a}_{1} + \left(\frac{1}{2} - x_{4} + y_{4}\right) \, \mathbf{a}_{2} + \left(y_{4}-z_{4}\right) \, \mathbf{a}_{3} & = & y_{4}a \, \mathbf{\hat{x}}-z_{4}a \, \mathbf{\hat{y}} + \left(\frac{1}{2} - x_{4}\right)a \, \mathbf{\hat{z}} & \left(48e\right) & \mbox{Li} \\ 
\mathbf{B}_{36} & = & \left(\frac{1}{2} +x_{4} - z_{4}\right) \, \mathbf{a}_{1} + \left(x_{4}-y_{4}\right) \, \mathbf{a}_{2} + \left(\frac{1}{2} - y_{4} - z_{4}\right) \, \mathbf{a}_{3} & = & -y_{4}a \, \mathbf{\hat{x}} + \left(\frac{1}{2} - z_{4}\right)a \, \mathbf{\hat{y}} + x_{4}a \, \mathbf{\hat{z}} & \left(48e\right) & \mbox{Li} \\ 
\mathbf{B}_{37} & = & \left(-y_{4}-z_{4}\right) \, \mathbf{a}_{1} + \left(-x_{4}-z_{4}\right) \, \mathbf{a}_{2} + \left(-x_{4}-y_{4}\right) \, \mathbf{a}_{3} & = & -x_{4}a \, \mathbf{\hat{x}}-y_{4}a \, \mathbf{\hat{y}}-z_{4}a \, \mathbf{\hat{z}} & \left(48e\right) & \mbox{Li} \\ 
\mathbf{B}_{38} & = & \left(\frac{1}{2} +y_{4} - z_{4}\right) \, \mathbf{a}_{1} + \left(x_{4}-z_{4}\right) \, \mathbf{a}_{2} + \left(\frac{1}{2} +x_{4} + y_{4}\right) \, \mathbf{a}_{3} & = & x_{4}a \, \mathbf{\hat{x}} + \left(\frac{1}{2} +y_{4}\right)a \, \mathbf{\hat{y}}-z_{4}a \, \mathbf{\hat{z}} & \left(48e\right) & \mbox{Li} \\ 
\mathbf{B}_{39} & = & \left(-y_{4}+z_{4}\right) \, \mathbf{a}_{1} + \left(\frac{1}{2} +x_{4} + z_{4}\right) \, \mathbf{a}_{2} + \left(\frac{1}{2} +x_{4} - y_{4}\right) \, \mathbf{a}_{3} & = & \left(\frac{1}{2} +x_{4}\right)a \, \mathbf{\hat{x}}-y_{4}a \, \mathbf{\hat{y}} + z_{4}a \, \mathbf{\hat{z}} & \left(48e\right) & \mbox{Li} \\ 
\mathbf{B}_{40} & = & \left(\frac{1}{2} +y_{4} + z_{4}\right) \, \mathbf{a}_{1} + \left(\frac{1}{2} - x_{4} + z_{4}\right) \, \mathbf{a}_{2} + \left(-x_{4}+y_{4}\right) \, \mathbf{a}_{3} & = & -x_{4}a \, \mathbf{\hat{x}} + y_{4}a \, \mathbf{\hat{y}} + \left(\frac{1}{2} +z_{4}\right)a \, \mathbf{\hat{z}} & \left(48e\right) & \mbox{Li} \\ 
\mathbf{B}_{41} & = & \left(-x_{4}-y_{4}\right) \, \mathbf{a}_{1} + \left(-y_{4}-z_{4}\right) \, \mathbf{a}_{2} + \left(-x_{4}-z_{4}\right) \, \mathbf{a}_{3} & = & -z_{4}a \, \mathbf{\hat{x}}-x_{4}a \, \mathbf{\hat{y}}-y_{4}a \, \mathbf{\hat{z}} & \left(48e\right) & \mbox{Li} \\ 
\mathbf{B}_{42} & = & \left(\frac{1}{2} +x_{4} + y_{4}\right) \, \mathbf{a}_{1} + \left(\frac{1}{2} +y_{4} - z_{4}\right) \, \mathbf{a}_{2} + \left(x_{4}-z_{4}\right) \, \mathbf{a}_{3} & = & -z_{4}a \, \mathbf{\hat{x}} + x_{4}a \, \mathbf{\hat{y}} + \left(\frac{1}{2} +y_{4}\right)a \, \mathbf{\hat{z}} & \left(48e\right) & \mbox{Li} \\ 
\mathbf{B}_{43} & = & \left(\frac{1}{2} +x_{4} - y_{4}\right) \, \mathbf{a}_{1} + \left(-y_{4}+z_{4}\right) \, \mathbf{a}_{2} + \left(\frac{1}{2} +x_{4} + z_{4}\right) \, \mathbf{a}_{3} & = & z_{4}a \, \mathbf{\hat{x}} + \left(\frac{1}{2} +x_{4}\right)a \, \mathbf{\hat{y}}-y_{4}a \, \mathbf{\hat{z}} & \left(48e\right) & \mbox{Li} \\ 
\mathbf{B}_{44} & = & \left(-x_{4}+y_{4}\right) \, \mathbf{a}_{1} + \left(\frac{1}{2} +y_{4} + z_{4}\right) \, \mathbf{a}_{2} + \left(\frac{1}{2} - x_{4} + z_{4}\right) \, \mathbf{a}_{3} & = & \left(\frac{1}{2} +z_{4}\right)a \, \mathbf{\hat{x}}-x_{4}a \, \mathbf{\hat{y}} + y_{4}a \, \mathbf{\hat{z}} & \left(48e\right) & \mbox{Li} \\ 
\mathbf{B}_{45} & = & \left(-x_{4}-z_{4}\right) \, \mathbf{a}_{1} + \left(-x_{4}-y_{4}\right) \, \mathbf{a}_{2} + \left(-y_{4}-z_{4}\right) \, \mathbf{a}_{3} & = & -y_{4}a \, \mathbf{\hat{x}}-z_{4}a \, \mathbf{\hat{y}}-x_{4}a \, \mathbf{\hat{z}} & \left(48e\right) & \mbox{Li} \\ 
\mathbf{B}_{46} & = & \left(x_{4}-z_{4}\right) \, \mathbf{a}_{1} + \left(\frac{1}{2} +x_{4} + y_{4}\right) \, \mathbf{a}_{2} + \left(\frac{1}{2} +y_{4} - z_{4}\right) \, \mathbf{a}_{3} & = & \left(\frac{1}{2} +y_{4}\right)a \, \mathbf{\hat{x}}-z_{4}a \, \mathbf{\hat{y}} + x_{4}a \, \mathbf{\hat{z}} & \left(48e\right) & \mbox{Li} \\ 
\mathbf{B}_{47} & = & \left(\frac{1}{2} +x_{4} + z_{4}\right) \, \mathbf{a}_{1} + \left(\frac{1}{2} +x_{4} - y_{4}\right) \, \mathbf{a}_{2} + \left(-y_{4}+z_{4}\right) \, \mathbf{a}_{3} & = & -y_{4}a \, \mathbf{\hat{x}} + z_{4}a \, \mathbf{\hat{y}} + \left(\frac{1}{2} +x_{4}\right)a \, \mathbf{\hat{z}} & \left(48e\right) & \mbox{Li} \\ 
\mathbf{B}_{48} & = & \left(\frac{1}{2} - x_{4} + z_{4}\right) \, \mathbf{a}_{1} + \left(-x_{4}+y_{4}\right) \, \mathbf{a}_{2} + \left(\frac{1}{2} +y_{4} + z_{4}\right) \, \mathbf{a}_{3} & = & y_{4}a \, \mathbf{\hat{x}} + \left(\frac{1}{2} +z_{4}\right)a \, \mathbf{\hat{y}}-x_{4}a \, \mathbf{\hat{z}} & \left(48e\right) & \mbox{Li} \\ 
\end{longtabu}
\renewcommand{\arraystretch}{1.0}
\noindent \hrulefill
\\
\textbf{References:}
\vspace*{-0.25cm}
\begin{flushleft}
  - \bibentry{Juza_ZAAC_257_13_1948}. \\
\end{flushleft}
\textbf{Found in:}
\vspace*{-0.25cm}
\begin{flushleft}
  - \bibentry{herbst12:Herbst_PRB_85_195137_2012}. \\
\end{flushleft}
\noindent \hrulefill
\\
\textbf{Geometry files:}
\\
\noindent  - CIF: pp. {\hyperref[AB3C2_cI96_206_c_e_ad_cif]{\pageref{AB3C2_cI96_206_c_e_ad_cif}}} \\
\noindent  - POSCAR: pp. {\hyperref[AB3C2_cI96_206_c_e_ad_poscar]{\pageref{AB3C2_cI96_206_c_e_ad_poscar}}} \\
\onecolumn
{\phantomsection\label{A17B15_cP64_207_acfk_eij}}
\subsection*{\huge \textbf{{\normalfont Pd$_{17}$Se$_{15}$ Structure: A17B15\_cP64\_207\_acfk\_eij}}}
\noindent \hrulefill
\vspace*{0.25cm}
\begin{figure}[htp]
  \centering
  \vspace{-1em}
  {\includegraphics[width=1\textwidth]{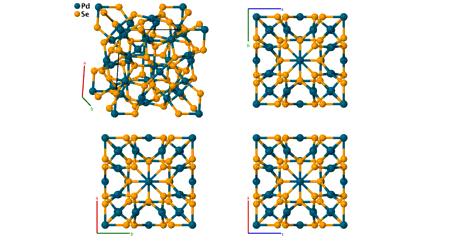}}
\end{figure}
\vspace*{-0.5cm}
\renewcommand{\arraystretch}{1.5}
\begin{equation*}
  \begin{array}{>{$\hspace{-0.15cm}}l<{$}>{$}p{0.5cm}<{$}>{$}p{18.5cm}<{$}}
    \mbox{\large \textbf{Prototype}} &\colon & \ce{Pd17Se15} \\
    \mbox{\large \textbf{\AFLOW\ prototype label}} &\colon & \mbox{A17B15\_cP64\_207\_acfk\_eij} \\
    \mbox{\large \textbf{\textit{Strukturbericht} designation}} &\colon & \mbox{None} \\
    \mbox{\large \textbf{Pearson symbol}} &\colon & \mbox{cP64} \\
    \mbox{\large \textbf{Space group number}} &\colon & 207 \\
    \mbox{\large \textbf{Space group symbol}} &\colon & P432 \\
    \mbox{\large \textbf{\AFLOW\ prototype command}} &\colon &  \texttt{aflow} \,  \, \texttt{-{}-proto=A17B15\_cP64\_207\_acfk\_eij } \, \newline \texttt{-{}-params=}{a,x_{3},x_{4},y_{5},y_{6},x_{7},y_{7},z_{7} }
  \end{array}
\end{equation*}
\renewcommand{\arraystretch}{1.0}

\noindent \parbox{1 \linewidth}{
\noindent \hrulefill
\\
\textbf{Simple Cubic primitive vectors:} \\
\vspace*{-0.25cm}
\begin{tabular}{cc}
  \begin{tabular}{c}
    \parbox{0.6 \linewidth}{
      \renewcommand{\arraystretch}{1.5}
      \begin{equation*}
        \centering
        \begin{array}{ccc}
              \mathbf{a}_1 & = & a \, \mathbf{\hat{x}} \\
    \mathbf{a}_2 & = & a \, \mathbf{\hat{y}} \\
    \mathbf{a}_3 & = & a \, \mathbf{\hat{z}} \\

        \end{array}
      \end{equation*}
    }
    \renewcommand{\arraystretch}{1.0}
  \end{tabular}
  \begin{tabular}{c}
    \includegraphics[width=0.3\linewidth]{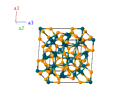} \\
  \end{tabular}
\end{tabular}

}
\vspace*{-0.25cm}

\noindent \hrulefill
\\
\textbf{Basis vectors:}
\vspace*{-0.25cm}
\renewcommand{\arraystretch}{1.5}
\begin{longtabu} to \textwidth{>{\centering $}X[-1,c,c]<{$}>{\centering $}X[-1,c,c]<{$}>{\centering $}X[-1,c,c]<{$}>{\centering $}X[-1,c,c]<{$}>{\centering $}X[-1,c,c]<{$}>{\centering $}X[-1,c,c]<{$}>{\centering $}X[-1,c,c]<{$}}
  & & \mbox{Lattice Coordinates} & & \mbox{Cartesian Coordinates} &\mbox{Wyckoff Position} & \mbox{Atom Type} \\  
  \mathbf{B}_{1} & = & 0 \, \mathbf{a}_{1} + 0 \, \mathbf{a}_{2} + 0 \, \mathbf{a}_{3} & = & 0 \, \mathbf{\hat{x}} + 0 \, \mathbf{\hat{y}} + 0 \, \mathbf{\hat{z}} & \left(1a\right) & \mbox{Pd I} \\ 
\mathbf{B}_{2} & = & \frac{1}{2} \, \mathbf{a}_{2} + \frac{1}{2} \, \mathbf{a}_{3} & = & \frac{1}{2}a \, \mathbf{\hat{y}} + \frac{1}{2}a \, \mathbf{\hat{z}} & \left(3c\right) & \mbox{Pd II} \\ 
\mathbf{B}_{3} & = & \frac{1}{2} \, \mathbf{a}_{1} + \frac{1}{2} \, \mathbf{a}_{3} & = & \frac{1}{2}a \, \mathbf{\hat{x}} + \frac{1}{2}a \, \mathbf{\hat{z}} & \left(3c\right) & \mbox{Pd II} \\ 
\mathbf{B}_{4} & = & \frac{1}{2} \, \mathbf{a}_{1} + \frac{1}{2} \, \mathbf{a}_{2} & = & \frac{1}{2}a \, \mathbf{\hat{x}} + \frac{1}{2}a \, \mathbf{\hat{y}} & \left(3c\right) & \mbox{Pd II} \\ 
\mathbf{B}_{5} & = & x_{3} \, \mathbf{a}_{1} & = & x_{3}a \, \mathbf{\hat{x}} & \left(6e\right) & \mbox{Se I} \\ 
\mathbf{B}_{6} & = & -x_{3} \, \mathbf{a}_{1} & = & -x_{3}a \, \mathbf{\hat{x}} & \left(6e\right) & \mbox{Se I} \\ 
\mathbf{B}_{7} & = & x_{3} \, \mathbf{a}_{2} & = & x_{3}a \, \mathbf{\hat{y}} & \left(6e\right) & \mbox{Se I} \\ 
\mathbf{B}_{8} & = & -x_{3} \, \mathbf{a}_{2} & = & -x_{3}a \, \mathbf{\hat{y}} & \left(6e\right) & \mbox{Se I} \\ 
\mathbf{B}_{9} & = & x_{3} \, \mathbf{a}_{3} & = & x_{3}a \, \mathbf{\hat{z}} & \left(6e\right) & \mbox{Se I} \\ 
\mathbf{B}_{10} & = & -x_{3} \, \mathbf{a}_{3} & = & -x_{3}a \, \mathbf{\hat{z}} & \left(6e\right) & \mbox{Se I} \\ 
\mathbf{B}_{11} & = & x_{4} \, \mathbf{a}_{1} + \frac{1}{2} \, \mathbf{a}_{2} + \frac{1}{2} \, \mathbf{a}_{3} & = & x_{4}a \, \mathbf{\hat{x}} + \frac{1}{2}a \, \mathbf{\hat{y}} + \frac{1}{2}a \, \mathbf{\hat{z}} & \left(6f\right) & \mbox{Pd III} \\ 
\mathbf{B}_{12} & = & -x_{4} \, \mathbf{a}_{1} + \frac{1}{2} \, \mathbf{a}_{2} + \frac{1}{2} \, \mathbf{a}_{3} & = & -x_{4}a \, \mathbf{\hat{x}} + \frac{1}{2}a \, \mathbf{\hat{y}} + \frac{1}{2}a \, \mathbf{\hat{z}} & \left(6f\right) & \mbox{Pd III} \\ 
\mathbf{B}_{13} & = & \frac{1}{2} \, \mathbf{a}_{1} + x_{4} \, \mathbf{a}_{2} + \frac{1}{2} \, \mathbf{a}_{3} & = & \frac{1}{2}a \, \mathbf{\hat{x}} + x_{4}a \, \mathbf{\hat{y}} + \frac{1}{2}a \, \mathbf{\hat{z}} & \left(6f\right) & \mbox{Pd III} \\ 
\mathbf{B}_{14} & = & \frac{1}{2} \, \mathbf{a}_{1}-x_{4} \, \mathbf{a}_{2} + \frac{1}{2} \, \mathbf{a}_{3} & = & \frac{1}{2}a \, \mathbf{\hat{x}}-x_{4}a \, \mathbf{\hat{y}} + \frac{1}{2}a \, \mathbf{\hat{z}} & \left(6f\right) & \mbox{Pd III} \\ 
\mathbf{B}_{15} & = & \frac{1}{2} \, \mathbf{a}_{1} + \frac{1}{2} \, \mathbf{a}_{2} + x_{4} \, \mathbf{a}_{3} & = & \frac{1}{2}a \, \mathbf{\hat{x}} + \frac{1}{2}a \, \mathbf{\hat{y}} + x_{4}a \, \mathbf{\hat{z}} & \left(6f\right) & \mbox{Pd III} \\ 
\mathbf{B}_{16} & = & \frac{1}{2} \, \mathbf{a}_{1} + \frac{1}{2} \, \mathbf{a}_{2}-x_{4} \, \mathbf{a}_{3} & = & \frac{1}{2}a \, \mathbf{\hat{x}} + \frac{1}{2}a \, \mathbf{\hat{y}}-x_{4}a \, \mathbf{\hat{z}} & \left(6f\right) & \mbox{Pd III} \\ 
\mathbf{B}_{17} & = & y_{5} \, \mathbf{a}_{2} + y_{5} \, \mathbf{a}_{3} & = & y_{5}a \, \mathbf{\hat{y}} + y_{5}a \, \mathbf{\hat{z}} & \left(12i\right) & \mbox{Se II} \\ 
\mathbf{B}_{18} & = & -y_{5} \, \mathbf{a}_{2} + y_{5} \, \mathbf{a}_{3} & = & -y_{5}a \, \mathbf{\hat{y}} + y_{5}a \, \mathbf{\hat{z}} & \left(12i\right) & \mbox{Se II} \\ 
\mathbf{B}_{19} & = & y_{5} \, \mathbf{a}_{2}-y_{5} \, \mathbf{a}_{3} & = & y_{5}a \, \mathbf{\hat{y}}-y_{5}a \, \mathbf{\hat{z}} & \left(12i\right) & \mbox{Se II} \\ 
\mathbf{B}_{20} & = & -y_{5} \, \mathbf{a}_{2}-y_{5} \, \mathbf{a}_{3} & = & -y_{5}a \, \mathbf{\hat{y}}-y_{5}a \, \mathbf{\hat{z}} & \left(12i\right) & \mbox{Se II} \\ 
\mathbf{B}_{21} & = & y_{5} \, \mathbf{a}_{1} + y_{5} \, \mathbf{a}_{3} & = & y_{5}a \, \mathbf{\hat{x}} + y_{5}a \, \mathbf{\hat{z}} & \left(12i\right) & \mbox{Se II} \\ 
\mathbf{B}_{22} & = & y_{5} \, \mathbf{a}_{1} + -y_{5} \, \mathbf{a}_{3} & = & y_{5}a \, \mathbf{\hat{x}} + -y_{5}a \, \mathbf{\hat{z}} & \left(12i\right) & \mbox{Se II} \\ 
\mathbf{B}_{23} & = & -y_{5} \, \mathbf{a}_{1} + y_{5} \, \mathbf{a}_{3} & = & -y_{5}a \, \mathbf{\hat{x}} + y_{5}a \, \mathbf{\hat{z}} & \left(12i\right) & \mbox{Se II} \\ 
\mathbf{B}_{24} & = & -y_{5} \, \mathbf{a}_{1} + -y_{5} \, \mathbf{a}_{3} & = & -y_{5}a \, \mathbf{\hat{x}} + -y_{5}a \, \mathbf{\hat{z}} & \left(12i\right) & \mbox{Se II} \\ 
\mathbf{B}_{25} & = & y_{5} \, \mathbf{a}_{1} + y_{5} \, \mathbf{a}_{2} & = & y_{5}a \, \mathbf{\hat{x}} + y_{5}a \, \mathbf{\hat{y}} & \left(12i\right) & \mbox{Se II} \\ 
\mathbf{B}_{26} & = & -y_{5} \, \mathbf{a}_{1} + y_{5} \, \mathbf{a}_{2} & = & -y_{5}a \, \mathbf{\hat{x}} + y_{5}a \, \mathbf{\hat{y}} & \left(12i\right) & \mbox{Se II} \\ 
\mathbf{B}_{27} & = & y_{5} \, \mathbf{a}_{1}-y_{5} \, \mathbf{a}_{2} & = & y_{5}a \, \mathbf{\hat{x}}-y_{5}a \, \mathbf{\hat{y}} & \left(12i\right) & \mbox{Se II} \\ 
\mathbf{B}_{28} & = & -y_{5} \, \mathbf{a}_{1}-y_{5} \, \mathbf{a}_{2} & = & -y_{5}a \, \mathbf{\hat{x}}-y_{5}a \, \mathbf{\hat{y}} & \left(12i\right) & \mbox{Se II} \\ 
\mathbf{B}_{29} & = & \frac{1}{2} \, \mathbf{a}_{1} + y_{6} \, \mathbf{a}_{2} + y_{6} \, \mathbf{a}_{3} & = & \frac{1}{2}a \, \mathbf{\hat{x}} + y_{6}a \, \mathbf{\hat{y}} + y_{6}a \, \mathbf{\hat{z}} & \left(12j\right) & \mbox{Se III} \\ 
\mathbf{B}_{30} & = & \frac{1}{2} \, \mathbf{a}_{1}-y_{6} \, \mathbf{a}_{2} + y_{6} \, \mathbf{a}_{3} & = & \frac{1}{2}a \, \mathbf{\hat{x}}-y_{6}a \, \mathbf{\hat{y}} + y_{6}a \, \mathbf{\hat{z}} & \left(12j\right) & \mbox{Se III} \\ 
\mathbf{B}_{31} & = & \frac{1}{2} \, \mathbf{a}_{1} + y_{6} \, \mathbf{a}_{2}-y_{6} \, \mathbf{a}_{3} & = & \frac{1}{2}a \, \mathbf{\hat{x}} + y_{6}a \, \mathbf{\hat{y}}-y_{6}a \, \mathbf{\hat{z}} & \left(12j\right) & \mbox{Se III} \\ 
\mathbf{B}_{32} & = & \frac{1}{2} \, \mathbf{a}_{1}-y_{6} \, \mathbf{a}_{2}-y_{6} \, \mathbf{a}_{3} & = & \frac{1}{2}a \, \mathbf{\hat{x}}-y_{6}a \, \mathbf{\hat{y}}-y_{6}a \, \mathbf{\hat{z}} & \left(12j\right) & \mbox{Se III} \\ 
\mathbf{B}_{33} & = & y_{6} \, \mathbf{a}_{1} + \frac{1}{2} \, \mathbf{a}_{2} + y_{6} \, \mathbf{a}_{3} & = & y_{6}a \, \mathbf{\hat{x}} + \frac{1}{2}a \, \mathbf{\hat{y}} + y_{6}a \, \mathbf{\hat{z}} & \left(12j\right) & \mbox{Se III} \\ 
\mathbf{B}_{34} & = & y_{6} \, \mathbf{a}_{1} + \frac{1}{2} \, \mathbf{a}_{2}-y_{6} \, \mathbf{a}_{3} & = & y_{6}a \, \mathbf{\hat{x}} + \frac{1}{2}a \, \mathbf{\hat{y}}-y_{6}a \, \mathbf{\hat{z}} & \left(12j\right) & \mbox{Se III} \\ 
\mathbf{B}_{35} & = & -y_{6} \, \mathbf{a}_{1} + \frac{1}{2} \, \mathbf{a}_{2} + y_{6} \, \mathbf{a}_{3} & = & -y_{6}a \, \mathbf{\hat{x}} + \frac{1}{2}a \, \mathbf{\hat{y}} + y_{6}a \, \mathbf{\hat{z}} & \left(12j\right) & \mbox{Se III} \\ 
\mathbf{B}_{36} & = & -y_{6} \, \mathbf{a}_{1} + \frac{1}{2} \, \mathbf{a}_{2}-y_{6} \, \mathbf{a}_{3} & = & -y_{6}a \, \mathbf{\hat{x}} + \frac{1}{2}a \, \mathbf{\hat{y}}-y_{6}a \, \mathbf{\hat{z}} & \left(12j\right) & \mbox{Se III} \\ 
\mathbf{B}_{37} & = & y_{6} \, \mathbf{a}_{1} + y_{6} \, \mathbf{a}_{2} + \frac{1}{2} \, \mathbf{a}_{3} & = & y_{6}a \, \mathbf{\hat{x}} + y_{6}a \, \mathbf{\hat{y}} + \frac{1}{2}a \, \mathbf{\hat{z}} & \left(12j\right) & \mbox{Se III} \\ 
\mathbf{B}_{38} & = & -y_{6} \, \mathbf{a}_{1} + y_{6} \, \mathbf{a}_{2} + \frac{1}{2} \, \mathbf{a}_{3} & = & -y_{6}a \, \mathbf{\hat{x}} + y_{6}a \, \mathbf{\hat{y}} + \frac{1}{2}a \, \mathbf{\hat{z}} & \left(12j\right) & \mbox{Se III} \\ 
\mathbf{B}_{39} & = & y_{6} \, \mathbf{a}_{1}-y_{6} \, \mathbf{a}_{2} + \frac{1}{2} \, \mathbf{a}_{3} & = & y_{6}a \, \mathbf{\hat{x}}-y_{6}a \, \mathbf{\hat{y}} + \frac{1}{2}a \, \mathbf{\hat{z}} & \left(12j\right) & \mbox{Se III} \\ 
\mathbf{B}_{40} & = & -y_{6} \, \mathbf{a}_{1}-y_{6} \, \mathbf{a}_{2} + \frac{1}{2} \, \mathbf{a}_{3} & = & -y_{6}a \, \mathbf{\hat{x}}-y_{6}a \, \mathbf{\hat{y}} + \frac{1}{2}a \, \mathbf{\hat{z}} & \left(12j\right) & \mbox{Se III} \\ 
\mathbf{B}_{41} & = & x_{7} \, \mathbf{a}_{1} + y_{7} \, \mathbf{a}_{2} + z_{7} \, \mathbf{a}_{3} & = & x_{7}a \, \mathbf{\hat{x}} + y_{7}a \, \mathbf{\hat{y}} + z_{7}a \, \mathbf{\hat{z}} & \left(24k\right) & \mbox{Pd IV} \\ 
\mathbf{B}_{42} & = & -x_{7} \, \mathbf{a}_{1}-y_{7} \, \mathbf{a}_{2} + z_{7} \, \mathbf{a}_{3} & = & -x_{7}a \, \mathbf{\hat{x}}-y_{7}a \, \mathbf{\hat{y}} + z_{7}a \, \mathbf{\hat{z}} & \left(24k\right) & \mbox{Pd IV} \\ 
\mathbf{B}_{43} & = & -x_{7} \, \mathbf{a}_{1} + y_{7} \, \mathbf{a}_{2}-z_{7} \, \mathbf{a}_{3} & = & -x_{7}a \, \mathbf{\hat{x}} + y_{7}a \, \mathbf{\hat{y}}-z_{7}a \, \mathbf{\hat{z}} & \left(24k\right) & \mbox{Pd IV} \\ 
\mathbf{B}_{44} & = & x_{7} \, \mathbf{a}_{1}-y_{7} \, \mathbf{a}_{2}-z_{7} \, \mathbf{a}_{3} & = & x_{7}a \, \mathbf{\hat{x}}-y_{7}a \, \mathbf{\hat{y}}-z_{7}a \, \mathbf{\hat{z}} & \left(24k\right) & \mbox{Pd IV} \\ 
\mathbf{B}_{45} & = & z_{7} \, \mathbf{a}_{1} + x_{7} \, \mathbf{a}_{2} + y_{7} \, \mathbf{a}_{3} & = & z_{7}a \, \mathbf{\hat{x}} + x_{7}a \, \mathbf{\hat{y}} + y_{7}a \, \mathbf{\hat{z}} & \left(24k\right) & \mbox{Pd IV} \\ 
\mathbf{B}_{46} & = & z_{7} \, \mathbf{a}_{1}-x_{7} \, \mathbf{a}_{2}-y_{7} \, \mathbf{a}_{3} & = & z_{7}a \, \mathbf{\hat{x}}-x_{7}a \, \mathbf{\hat{y}}-y_{7}a \, \mathbf{\hat{z}} & \left(24k\right) & \mbox{Pd IV} \\ 
\mathbf{B}_{47} & = & -z_{7} \, \mathbf{a}_{1}-x_{7} \, \mathbf{a}_{2} + y_{7} \, \mathbf{a}_{3} & = & -z_{7}a \, \mathbf{\hat{x}}-x_{7}a \, \mathbf{\hat{y}} + y_{7}a \, \mathbf{\hat{z}} & \left(24k\right) & \mbox{Pd IV} \\ 
\mathbf{B}_{48} & = & -z_{7} \, \mathbf{a}_{1} + x_{7} \, \mathbf{a}_{2}-y_{7} \, \mathbf{a}_{3} & = & -z_{7}a \, \mathbf{\hat{x}} + x_{7}a \, \mathbf{\hat{y}}-y_{7}a \, \mathbf{\hat{z}} & \left(24k\right) & \mbox{Pd IV} \\ 
\mathbf{B}_{49} & = & y_{7} \, \mathbf{a}_{1} + z_{7} \, \mathbf{a}_{2} + x_{7} \, \mathbf{a}_{3} & = & y_{7}a \, \mathbf{\hat{x}} + z_{7}a \, \mathbf{\hat{y}} + x_{7}a \, \mathbf{\hat{z}} & \left(24k\right) & \mbox{Pd IV} \\ 
\mathbf{B}_{50} & = & -y_{7} \, \mathbf{a}_{1} + z_{7} \, \mathbf{a}_{2}-x_{7} \, \mathbf{a}_{3} & = & -y_{7}a \, \mathbf{\hat{x}} + z_{7}a \, \mathbf{\hat{y}}-x_{7}a \, \mathbf{\hat{z}} & \left(24k\right) & \mbox{Pd IV} \\ 
\mathbf{B}_{51} & = & y_{7} \, \mathbf{a}_{1}-z_{7} \, \mathbf{a}_{2}-x_{7} \, \mathbf{a}_{3} & = & y_{7}a \, \mathbf{\hat{x}}-z_{7}a \, \mathbf{\hat{y}}-x_{7}a \, \mathbf{\hat{z}} & \left(24k\right) & \mbox{Pd IV} \\ 
\mathbf{B}_{52} & = & -y_{7} \, \mathbf{a}_{1}-z_{7} \, \mathbf{a}_{2} + x_{7} \, \mathbf{a}_{3} & = & -y_{7}a \, \mathbf{\hat{x}}-z_{7}a \, \mathbf{\hat{y}} + x_{7}a \, \mathbf{\hat{z}} & \left(24k\right) & \mbox{Pd IV} \\ 
\mathbf{B}_{53} & = & y_{7} \, \mathbf{a}_{1} + x_{7} \, \mathbf{a}_{2}-z_{7} \, \mathbf{a}_{3} & = & y_{7}a \, \mathbf{\hat{x}} + x_{7}a \, \mathbf{\hat{y}}-z_{7}a \, \mathbf{\hat{z}} & \left(24k\right) & \mbox{Pd IV} \\ 
\mathbf{B}_{54} & = & -y_{7} \, \mathbf{a}_{1}-x_{7} \, \mathbf{a}_{2}-z_{7} \, \mathbf{a}_{3} & = & -y_{7}a \, \mathbf{\hat{x}}-x_{7}a \, \mathbf{\hat{y}}-z_{7}a \, \mathbf{\hat{z}} & \left(24k\right) & \mbox{Pd IV} \\ 
\mathbf{B}_{55} & = & y_{7} \, \mathbf{a}_{1}-x_{7} \, \mathbf{a}_{2} + z_{7} \, \mathbf{a}_{3} & = & y_{7}a \, \mathbf{\hat{x}}-x_{7}a \, \mathbf{\hat{y}} + z_{7}a \, \mathbf{\hat{z}} & \left(24k\right) & \mbox{Pd IV} \\ 
\mathbf{B}_{56} & = & -y_{7} \, \mathbf{a}_{1} + x_{7} \, \mathbf{a}_{2} + z_{7} \, \mathbf{a}_{3} & = & -y_{7}a \, \mathbf{\hat{x}} + x_{7}a \, \mathbf{\hat{y}} + z_{7}a \, \mathbf{\hat{z}} & \left(24k\right) & \mbox{Pd IV} \\ 
\mathbf{B}_{57} & = & x_{7} \, \mathbf{a}_{1} + z_{7} \, \mathbf{a}_{2}-y_{7} \, \mathbf{a}_{3} & = & x_{7}a \, \mathbf{\hat{x}} + z_{7}a \, \mathbf{\hat{y}}-y_{7}a \, \mathbf{\hat{z}} & \left(24k\right) & \mbox{Pd IV} \\ 
\mathbf{B}_{58} & = & -x_{7} \, \mathbf{a}_{1} + z_{7} \, \mathbf{a}_{2} + y_{7} \, \mathbf{a}_{3} & = & -x_{7}a \, \mathbf{\hat{x}} + z_{7}a \, \mathbf{\hat{y}} + y_{7}a \, \mathbf{\hat{z}} & \left(24k\right) & \mbox{Pd IV} \\ 
\mathbf{B}_{59} & = & -x_{7} \, \mathbf{a}_{1}-z_{7} \, \mathbf{a}_{2}-y_{7} \, \mathbf{a}_{3} & = & -x_{7}a \, \mathbf{\hat{x}}-z_{7}a \, \mathbf{\hat{y}}-y_{7}a \, \mathbf{\hat{z}} & \left(24k\right) & \mbox{Pd IV} \\ 
\mathbf{B}_{60} & = & x_{7} \, \mathbf{a}_{1}-z_{7} \, \mathbf{a}_{2} + y_{7} \, \mathbf{a}_{3} & = & x_{7}a \, \mathbf{\hat{x}}-z_{7}a \, \mathbf{\hat{y}} + y_{7}a \, \mathbf{\hat{z}} & \left(24k\right) & \mbox{Pd IV} \\ 
\mathbf{B}_{61} & = & z_{7} \, \mathbf{a}_{1} + y_{7} \, \mathbf{a}_{2}-x_{7} \, \mathbf{a}_{3} & = & z_{7}a \, \mathbf{\hat{x}} + y_{7}a \, \mathbf{\hat{y}}-x_{7}a \, \mathbf{\hat{z}} & \left(24k\right) & \mbox{Pd IV} \\ 
\mathbf{B}_{62} & = & z_{7} \, \mathbf{a}_{1}-y_{7} \, \mathbf{a}_{2} + x_{7} \, \mathbf{a}_{3} & = & z_{7}a \, \mathbf{\hat{x}}-y_{7}a \, \mathbf{\hat{y}} + x_{7}a \, \mathbf{\hat{z}} & \left(24k\right) & \mbox{Pd IV} \\ 
\mathbf{B}_{63} & = & -z_{7} \, \mathbf{a}_{1} + y_{7} \, \mathbf{a}_{2} + x_{7} \, \mathbf{a}_{3} & = & -z_{7}a \, \mathbf{\hat{x}} + y_{7}a \, \mathbf{\hat{y}} + x_{7}a \, \mathbf{\hat{z}} & \left(24k\right) & \mbox{Pd IV} \\ 
\mathbf{B}_{64} & = & -z_{7} \, \mathbf{a}_{1}-y_{7} \, \mathbf{a}_{2}-x_{7} \, \mathbf{a}_{3} & = & -z_{7}a \, \mathbf{\hat{x}}-y_{7}a \, \mathbf{\hat{y}}-x_{7}a \, \mathbf{\hat{z}} & \left(24k\right) & \mbox{Pd IV} \\ 
\end{longtabu}
\renewcommand{\arraystretch}{1.0}
\noindent \hrulefill
\\
\textbf{References:}
\vspace*{-0.25cm}
\begin{flushleft}
  - \bibentry{Geller_Pd17Se15_ActCrystallogr_1962}. \\
\end{flushleft}
\textbf{Found in:}
\vspace*{-0.25cm}
\begin{flushleft}
  - \bibentry{Villars_PearsonsCrystalData_2013}. \\
\end{flushleft}
\noindent \hrulefill
\\
\textbf{Geometry files:}
\\
\noindent  - CIF: pp. {\hyperref[A17B15_cP64_207_acfk_eij_cif]{\pageref{A17B15_cP64_207_acfk_eij_cif}}} \\
\noindent  - POSCAR: pp. {\hyperref[A17B15_cP64_207_acfk_eij_poscar]{\pageref{A17B15_cP64_207_acfk_eij_poscar}}} \\
\onecolumn
{\phantomsection\label{A3B_cP16_208_j_b}}
\subsection*{\huge \textbf{{\normalfont PH$_{3}$ Structure: A3B\_cP16\_208\_j\_b}}}
\noindent \hrulefill
\vspace*{0.25cm}
\begin{figure}[htp]
  \centering
  \vspace{-1em}
  {\includegraphics[width=1\textwidth]{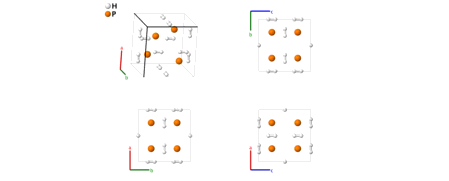}}
\end{figure}
\vspace*{-0.5cm}
\renewcommand{\arraystretch}{1.5}
\begin{equation*}
  \begin{array}{>{$\hspace{-0.15cm}}l<{$}>{$}p{0.5cm}<{$}>{$}p{18.5cm}<{$}}
    \mbox{\large \textbf{Prototype}} &\colon & \ce{PH3} \\
    \mbox{\large \textbf{\AFLOW\ prototype label}} &\colon & \mbox{A3B\_cP16\_208\_j\_b} \\
    \mbox{\large \textbf{\textit{Strukturbericht} designation}} &\colon & \mbox{None} \\
    \mbox{\large \textbf{Pearson symbol}} &\colon & \mbox{cP16} \\
    \mbox{\large \textbf{Space group number}} &\colon & 208 \\
    \mbox{\large \textbf{Space group symbol}} &\colon & P4_{2}32 \\
    \mbox{\large \textbf{\AFLOW\ prototype command}} &\colon &  \texttt{aflow} \,  \, \texttt{-{}-proto=A3B\_cP16\_208\_j\_b } \, \newline \texttt{-{}-params=}{a,x_{2} }
  \end{array}
\end{equation*}
\renewcommand{\arraystretch}{1.0}

\noindent \parbox{1 \linewidth}{
\noindent \hrulefill
\\
\textbf{Simple Cubic primitive vectors:} \\
\vspace*{-0.25cm}
\begin{tabular}{cc}
  \begin{tabular}{c}
    \parbox{0.6 \linewidth}{
      \renewcommand{\arraystretch}{1.5}
      \begin{equation*}
        \centering
        \begin{array}{ccc}
              \mathbf{a}_1 & = & a \, \mathbf{\hat{x}} \\
    \mathbf{a}_2 & = & a \, \mathbf{\hat{y}} \\
    \mathbf{a}_3 & = & a \, \mathbf{\hat{z}} \\

        \end{array}
      \end{equation*}
    }
    \renewcommand{\arraystretch}{1.0}
  \end{tabular}
  \begin{tabular}{c}
    \includegraphics[width=0.3\linewidth]{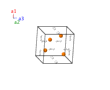} \\
  \end{tabular}
\end{tabular}

}
\vspace*{-0.25cm}

\noindent \hrulefill
\\
\textbf{Basis vectors:}
\vspace*{-0.25cm}
\renewcommand{\arraystretch}{1.5}
\begin{longtabu} to \textwidth{>{\centering $}X[-1,c,c]<{$}>{\centering $}X[-1,c,c]<{$}>{\centering $}X[-1,c,c]<{$}>{\centering $}X[-1,c,c]<{$}>{\centering $}X[-1,c,c]<{$}>{\centering $}X[-1,c,c]<{$}>{\centering $}X[-1,c,c]<{$}}
  & & \mbox{Lattice Coordinates} & & \mbox{Cartesian Coordinates} &\mbox{Wyckoff Position} & \mbox{Atom Type} \\  
  \mathbf{B}_{1} & = & \frac{1}{4} \, \mathbf{a}_{1} + \frac{1}{4} \, \mathbf{a}_{2} + \frac{1}{4} \, \mathbf{a}_{3} & = & \frac{1}{4}a \, \mathbf{\hat{x}} + \frac{1}{4}a \, \mathbf{\hat{y}} + \frac{1}{4}a \, \mathbf{\hat{z}} & \left(4b\right) & \mbox{P} \\ 
\mathbf{B}_{2} & = & \frac{3}{4} \, \mathbf{a}_{1} + \frac{3}{4} \, \mathbf{a}_{2} + \frac{1}{4} \, \mathbf{a}_{3} & = & \frac{3}{4}a \, \mathbf{\hat{x}} + \frac{3}{4}a \, \mathbf{\hat{y}} + \frac{1}{4}a \, \mathbf{\hat{z}} & \left(4b\right) & \mbox{P} \\ 
\mathbf{B}_{3} & = & \frac{3}{4} \, \mathbf{a}_{1} + \frac{1}{4} \, \mathbf{a}_{2} + \frac{3}{4} \, \mathbf{a}_{3} & = & \frac{3}{4}a \, \mathbf{\hat{x}} + \frac{1}{4}a \, \mathbf{\hat{y}} + \frac{3}{4}a \, \mathbf{\hat{z}} & \left(4b\right) & \mbox{P} \\ 
\mathbf{B}_{4} & = & \frac{1}{4} \, \mathbf{a}_{1} + \frac{3}{4} \, \mathbf{a}_{2} + \frac{3}{4} \, \mathbf{a}_{3} & = & \frac{1}{4}a \, \mathbf{\hat{x}} + \frac{3}{4}a \, \mathbf{\hat{y}} + \frac{3}{4}a \, \mathbf{\hat{z}} & \left(4b\right) & \mbox{P} \\ 
\mathbf{B}_{5} & = & x_{2} \, \mathbf{a}_{1} + \frac{1}{2} \, \mathbf{a}_{2} & = & x_{2}a \, \mathbf{\hat{x}} + \frac{1}{2}a \, \mathbf{\hat{y}} & \left(12j\right) & \mbox{H} \\ 
\mathbf{B}_{6} & = & -x_{2} \, \mathbf{a}_{1} + \frac{1}{2} \, \mathbf{a}_{2} & = & -x_{2}a \, \mathbf{\hat{x}} + \frac{1}{2}a \, \mathbf{\hat{y}} & \left(12j\right) & \mbox{H} \\ 
\mathbf{B}_{7} & = & x_{2} \, \mathbf{a}_{2} + \frac{1}{2} \, \mathbf{a}_{3} & = & x_{2}a \, \mathbf{\hat{y}} + \frac{1}{2}a \, \mathbf{\hat{z}} & \left(12j\right) & \mbox{H} \\ 
\mathbf{B}_{8} & = & -x_{2} \, \mathbf{a}_{2} + \frac{1}{2} \, \mathbf{a}_{3} & = & -x_{2}a \, \mathbf{\hat{y}} + \frac{1}{2}a \, \mathbf{\hat{z}} & \left(12j\right) & \mbox{H} \\ 
\mathbf{B}_{9} & = & \frac{1}{2} \, \mathbf{a}_{1} + x_{2} \, \mathbf{a}_{3} & = & \frac{1}{2}a \, \mathbf{\hat{x}} + x_{2}a \, \mathbf{\hat{z}} & \left(12j\right) & \mbox{H} \\ 
\mathbf{B}_{10} & = & \frac{1}{2} \, \mathbf{a}_{1} + -x_{2} \, \mathbf{a}_{3} & = & \frac{1}{2}a \, \mathbf{\hat{x}} + -x_{2}a \, \mathbf{\hat{z}} & \left(12j\right) & \mbox{H} \\ 
\mathbf{B}_{11} & = & \left(\frac{1}{2} +x_{2}\right) \, \mathbf{a}_{2} + \frac{1}{2} \, \mathbf{a}_{3} & = & \left(\frac{1}{2} +x_{2}\right)a \, \mathbf{\hat{y}} + \frac{1}{2}a \, \mathbf{\hat{z}} & \left(12j\right) & \mbox{H} \\ 
\mathbf{B}_{12} & = & \left(\frac{1}{2} - x_{2}\right) \, \mathbf{a}_{2} + \frac{1}{2} \, \mathbf{a}_{3} & = & \left(\frac{1}{2} - x_{2}\right)a \, \mathbf{\hat{y}} + \frac{1}{2}a \, \mathbf{\hat{z}} & \left(12j\right) & \mbox{H} \\ 
\mathbf{B}_{13} & = & \left(\frac{1}{2} +x_{2}\right) \, \mathbf{a}_{1} + \frac{1}{2} \, \mathbf{a}_{2} & = & \left(\frac{1}{2} +x_{2}\right)a \, \mathbf{\hat{x}} + \frac{1}{2}a \, \mathbf{\hat{y}} & \left(12j\right) & \mbox{H} \\ 
\mathbf{B}_{14} & = & \left(\frac{1}{2} - x_{2}\right) \, \mathbf{a}_{1} + \frac{1}{2} \, \mathbf{a}_{2} & = & \left(\frac{1}{2} - x_{2}\right)a \, \mathbf{\hat{x}} + \frac{1}{2}a \, \mathbf{\hat{y}} & \left(12j\right) & \mbox{H} \\ 
\mathbf{B}_{15} & = & \frac{1}{2} \, \mathbf{a}_{1} + \left(\frac{1}{2} - x_{2}\right) \, \mathbf{a}_{3} & = & \frac{1}{2}a \, \mathbf{\hat{x}} + \left(\frac{1}{2} - x_{2}\right)a \, \mathbf{\hat{z}} & \left(12j\right) & \mbox{H} \\ 
\mathbf{B}_{16} & = & \frac{1}{2} \, \mathbf{a}_{1} + \left(\frac{1}{2} +x_{2}\right) \, \mathbf{a}_{3} & = & \frac{1}{2}a \, \mathbf{\hat{x}} + \left(\frac{1}{2} +x_{2}\right)a \, \mathbf{\hat{z}} & \left(12j\right) & \mbox{H} \\ 
\end{longtabu}
\renewcommand{\arraystretch}{1.0}
\noindent \hrulefill
\\
\textbf{References:}
\vspace*{-0.25cm}
\begin{flushleft}
  - \bibentry{Natta_Gazzetta_Chimica_Italiana_60_1930}. \\
\end{flushleft}
\textbf{Found in:}
\vspace*{-0.25cm}
\begin{flushleft}
  - \bibentry{Downs_AM_88_2003}. \\
\end{flushleft}
\noindent \hrulefill
\\
\textbf{Geometry files:}
\\
\noindent  - CIF: pp. {\hyperref[A3B_cP16_208_j_b_cif]{\pageref{A3B_cP16_208_j_b_cif}}} \\
\noindent  - POSCAR: pp. {\hyperref[A3B_cP16_208_j_b_poscar]{\pageref{A3B_cP16_208_j_b_poscar}}} \\
\onecolumn
{\phantomsection\label{A6B2CD6E_cP64_208_m_ad_b_m_c}}
\subsection*{\huge \textbf{{\normalfont \begin{raggedleft}Cs$_{2}$ZnFe[CN]$_{6}$ Structure: \end{raggedleft} \\ A6B2CD6E\_cP64\_208\_m\_ad\_b\_m\_c}}}
\noindent \hrulefill
\vspace*{0.25cm}
\begin{figure}[htp]
  \centering
  \vspace{-1em}
  {\includegraphics[width=1\textwidth]{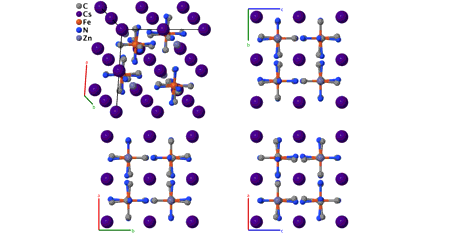}}
\end{figure}
\vspace*{-0.5cm}
\renewcommand{\arraystretch}{1.5}
\begin{equation*}
  \begin{array}{>{$\hspace{-0.15cm}}l<{$}>{$}p{0.5cm}<{$}>{$}p{18.5cm}<{$}}
    \mbox{\large \textbf{Prototype}} &\colon & \ce{Cs2ZnFe[CN]6} \\
    \mbox{\large \textbf{\AFLOW\ prototype label}} &\colon & \mbox{A6B2CD6E\_cP64\_208\_m\_ad\_b\_m\_c} \\
    \mbox{\large \textbf{\textit{Strukturbericht} designation}} &\colon & \mbox{None} \\
    \mbox{\large \textbf{Pearson symbol}} &\colon & \mbox{cP64} \\
    \mbox{\large \textbf{Space group number}} &\colon & 208 \\
    \mbox{\large \textbf{Space group symbol}} &\colon & P4_{2}32 \\
    \mbox{\large \textbf{\AFLOW\ prototype command}} &\colon &  \texttt{aflow} \,  \, \texttt{-{}-proto=A6B2CD6E\_cP64\_208\_m\_ad\_b\_m\_c } \, \newline \texttt{-{}-params=}{a,x_{5},y_{5},z_{5},x_{6},y_{6},z_{6} }
  \end{array}
\end{equation*}
\renewcommand{\arraystretch}{1.0}

\noindent \parbox{1 \linewidth}{
\noindent \hrulefill
\\
\textbf{Simple Cubic primitive vectors:} \\
\vspace*{-0.25cm}
\begin{tabular}{cc}
  \begin{tabular}{c}
    \parbox{0.6 \linewidth}{
      \renewcommand{\arraystretch}{1.5}
      \begin{equation*}
        \centering
        \begin{array}{ccc}
              \mathbf{a}_1 & = & a \, \mathbf{\hat{x}} \\
    \mathbf{a}_2 & = & a \, \mathbf{\hat{y}} \\
    \mathbf{a}_3 & = & a \, \mathbf{\hat{z}} \\

        \end{array}
      \end{equation*}
    }
    \renewcommand{\arraystretch}{1.0}
  \end{tabular}
  \begin{tabular}{c}
    \includegraphics[width=0.3\linewidth]{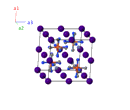} \\
  \end{tabular}
\end{tabular}

}
\vspace*{-0.25cm}

\noindent \hrulefill
\\
\textbf{Basis vectors:}
\vspace*{-0.25cm}
\renewcommand{\arraystretch}{1.5}
\begin{longtabu} to \textwidth{>{\centering $}X[-1,c,c]<{$}>{\centering $}X[-1,c,c]<{$}>{\centering $}X[-1,c,c]<{$}>{\centering $}X[-1,c,c]<{$}>{\centering $}X[-1,c,c]<{$}>{\centering $}X[-1,c,c]<{$}>{\centering $}X[-1,c,c]<{$}}
  & & \mbox{Lattice Coordinates} & & \mbox{Cartesian Coordinates} &\mbox{Wyckoff Position} & \mbox{Atom Type} \\  
  \mathbf{B}_{1} & = & 0 \, \mathbf{a}_{1} + 0 \, \mathbf{a}_{2} + 0 \, \mathbf{a}_{3} & = & 0 \, \mathbf{\hat{x}} + 0 \, \mathbf{\hat{y}} + 0 \, \mathbf{\hat{z}} & \left(2a\right) & \mbox{Cs I} \\ 
\mathbf{B}_{2} & = & \frac{1}{2} \, \mathbf{a}_{1} + \frac{1}{2} \, \mathbf{a}_{2} + \frac{1}{2} \, \mathbf{a}_{3} & = & \frac{1}{2}a \, \mathbf{\hat{x}} + \frac{1}{2}a \, \mathbf{\hat{y}} + \frac{1}{2}a \, \mathbf{\hat{z}} & \left(2a\right) & \mbox{Cs I} \\ 
\mathbf{B}_{3} & = & \frac{1}{4} \, \mathbf{a}_{1} + \frac{1}{4} \, \mathbf{a}_{2} + \frac{1}{4} \, \mathbf{a}_{3} & = & \frac{1}{4}a \, \mathbf{\hat{x}} + \frac{1}{4}a \, \mathbf{\hat{y}} + \frac{1}{4}a \, \mathbf{\hat{z}} & \left(4b\right) & \mbox{Fe} \\ 
\mathbf{B}_{4} & = & \frac{3}{4} \, \mathbf{a}_{1} + \frac{3}{4} \, \mathbf{a}_{2} + \frac{1}{4} \, \mathbf{a}_{3} & = & \frac{3}{4}a \, \mathbf{\hat{x}} + \frac{3}{4}a \, \mathbf{\hat{y}} + \frac{1}{4}a \, \mathbf{\hat{z}} & \left(4b\right) & \mbox{Fe} \\ 
\mathbf{B}_{5} & = & \frac{3}{4} \, \mathbf{a}_{1} + \frac{1}{4} \, \mathbf{a}_{2} + \frac{3}{4} \, \mathbf{a}_{3} & = & \frac{3}{4}a \, \mathbf{\hat{x}} + \frac{1}{4}a \, \mathbf{\hat{y}} + \frac{3}{4}a \, \mathbf{\hat{z}} & \left(4b\right) & \mbox{Fe} \\ 
\mathbf{B}_{6} & = & \frac{1}{4} \, \mathbf{a}_{1} + \frac{3}{4} \, \mathbf{a}_{2} + \frac{3}{4} \, \mathbf{a}_{3} & = & \frac{1}{4}a \, \mathbf{\hat{x}} + \frac{3}{4}a \, \mathbf{\hat{y}} + \frac{3}{4}a \, \mathbf{\hat{z}} & \left(4b\right) & \mbox{Fe} \\ 
\mathbf{B}_{7} & = & \frac{3}{4} \, \mathbf{a}_{1} + \frac{3}{4} \, \mathbf{a}_{2} + \frac{3}{4} \, \mathbf{a}_{3} & = & \frac{3}{4}a \, \mathbf{\hat{x}} + \frac{3}{4}a \, \mathbf{\hat{y}} + \frac{3}{4}a \, \mathbf{\hat{z}} & \left(4c\right) & \mbox{Zn} \\ 
\mathbf{B}_{8} & = & \frac{1}{4} \, \mathbf{a}_{1} + \frac{1}{4} \, \mathbf{a}_{2} + \frac{3}{4} \, \mathbf{a}_{3} & = & \frac{1}{4}a \, \mathbf{\hat{x}} + \frac{1}{4}a \, \mathbf{\hat{y}} + \frac{3}{4}a \, \mathbf{\hat{z}} & \left(4c\right) & \mbox{Zn} \\ 
\mathbf{B}_{9} & = & \frac{1}{4} \, \mathbf{a}_{1} + \frac{3}{4} \, \mathbf{a}_{2} + \frac{1}{4} \, \mathbf{a}_{3} & = & \frac{1}{4}a \, \mathbf{\hat{x}} + \frac{3}{4}a \, \mathbf{\hat{y}} + \frac{1}{4}a \, \mathbf{\hat{z}} & \left(4c\right) & \mbox{Zn} \\ 
\mathbf{B}_{10} & = & \frac{3}{4} \, \mathbf{a}_{1} + \frac{1}{4} \, \mathbf{a}_{2} + \frac{1}{4} \, \mathbf{a}_{3} & = & \frac{3}{4}a \, \mathbf{\hat{x}} + \frac{1}{4}a \, \mathbf{\hat{y}} + \frac{1}{4}a \, \mathbf{\hat{z}} & \left(4c\right) & \mbox{Zn} \\ 
\mathbf{B}_{11} & = & \frac{1}{2} \, \mathbf{a}_{2} + \frac{1}{2} \, \mathbf{a}_{3} & = & \frac{1}{2}a \, \mathbf{\hat{y}} + \frac{1}{2}a \, \mathbf{\hat{z}} & \left(6d\right) & \mbox{Cs II} \\ 
\mathbf{B}_{12} & = & \frac{1}{2} \, \mathbf{a}_{1} + \frac{1}{2} \, \mathbf{a}_{3} & = & \frac{1}{2}a \, \mathbf{\hat{x}} + \frac{1}{2}a \, \mathbf{\hat{z}} & \left(6d\right) & \mbox{Cs II} \\ 
\mathbf{B}_{13} & = & \frac{1}{2} \, \mathbf{a}_{1} + \frac{1}{2} \, \mathbf{a}_{2} & = & \frac{1}{2}a \, \mathbf{\hat{x}} + \frac{1}{2}a \, \mathbf{\hat{y}} & \left(6d\right) & \mbox{Cs II} \\ 
\mathbf{B}_{14} & = & \frac{1}{2} \, \mathbf{a}_{2} & = & \frac{1}{2}a \, \mathbf{\hat{y}} & \left(6d\right) & \mbox{Cs II} \\ 
\mathbf{B}_{15} & = & \frac{1}{2} \, \mathbf{a}_{1} & = & \frac{1}{2}a \, \mathbf{\hat{x}} & \left(6d\right) & \mbox{Cs II} \\ 
\mathbf{B}_{16} & = & \frac{1}{2} \, \mathbf{a}_{3} & = & \frac{1}{2}a \, \mathbf{\hat{z}} & \left(6d\right) & \mbox{Cs II} \\ 
\mathbf{B}_{17} & = & x_{5} \, \mathbf{a}_{1} + y_{5} \, \mathbf{a}_{2} + z_{5} \, \mathbf{a}_{3} & = & x_{5}a \, \mathbf{\hat{x}} + y_{5}a \, \mathbf{\hat{y}} + z_{5}a \, \mathbf{\hat{z}} & \left(24m\right) & \mbox{C} \\ 
\mathbf{B}_{18} & = & -x_{5} \, \mathbf{a}_{1}-y_{5} \, \mathbf{a}_{2} + z_{5} \, \mathbf{a}_{3} & = & -x_{5}a \, \mathbf{\hat{x}}-y_{5}a \, \mathbf{\hat{y}} + z_{5}a \, \mathbf{\hat{z}} & \left(24m\right) & \mbox{C} \\ 
\mathbf{B}_{19} & = & -x_{5} \, \mathbf{a}_{1} + y_{5} \, \mathbf{a}_{2}-z_{5} \, \mathbf{a}_{3} & = & -x_{5}a \, \mathbf{\hat{x}} + y_{5}a \, \mathbf{\hat{y}}-z_{5}a \, \mathbf{\hat{z}} & \left(24m\right) & \mbox{C} \\ 
\mathbf{B}_{20} & = & x_{5} \, \mathbf{a}_{1}-y_{5} \, \mathbf{a}_{2}-z_{5} \, \mathbf{a}_{3} & = & x_{5}a \, \mathbf{\hat{x}}-y_{5}a \, \mathbf{\hat{y}}-z_{5}a \, \mathbf{\hat{z}} & \left(24m\right) & \mbox{C} \\ 
\mathbf{B}_{21} & = & z_{5} \, \mathbf{a}_{1} + x_{5} \, \mathbf{a}_{2} + y_{5} \, \mathbf{a}_{3} & = & z_{5}a \, \mathbf{\hat{x}} + x_{5}a \, \mathbf{\hat{y}} + y_{5}a \, \mathbf{\hat{z}} & \left(24m\right) & \mbox{C} \\ 
\mathbf{B}_{22} & = & z_{5} \, \mathbf{a}_{1}-x_{5} \, \mathbf{a}_{2}-y_{5} \, \mathbf{a}_{3} & = & z_{5}a \, \mathbf{\hat{x}}-x_{5}a \, \mathbf{\hat{y}}-y_{5}a \, \mathbf{\hat{z}} & \left(24m\right) & \mbox{C} \\ 
\mathbf{B}_{23} & = & -z_{5} \, \mathbf{a}_{1}-x_{5} \, \mathbf{a}_{2} + y_{5} \, \mathbf{a}_{3} & = & -z_{5}a \, \mathbf{\hat{x}}-x_{5}a \, \mathbf{\hat{y}} + y_{5}a \, \mathbf{\hat{z}} & \left(24m\right) & \mbox{C} \\ 
\mathbf{B}_{24} & = & -z_{5} \, \mathbf{a}_{1} + x_{5} \, \mathbf{a}_{2}-y_{5} \, \mathbf{a}_{3} & = & -z_{5}a \, \mathbf{\hat{x}} + x_{5}a \, \mathbf{\hat{y}}-y_{5}a \, \mathbf{\hat{z}} & \left(24m\right) & \mbox{C} \\ 
\mathbf{B}_{25} & = & y_{5} \, \mathbf{a}_{1} + z_{5} \, \mathbf{a}_{2} + x_{5} \, \mathbf{a}_{3} & = & y_{5}a \, \mathbf{\hat{x}} + z_{5}a \, \mathbf{\hat{y}} + x_{5}a \, \mathbf{\hat{z}} & \left(24m\right) & \mbox{C} \\ 
\mathbf{B}_{26} & = & -y_{5} \, \mathbf{a}_{1} + z_{5} \, \mathbf{a}_{2}-x_{5} \, \mathbf{a}_{3} & = & -y_{5}a \, \mathbf{\hat{x}} + z_{5}a \, \mathbf{\hat{y}}-x_{5}a \, \mathbf{\hat{z}} & \left(24m\right) & \mbox{C} \\ 
\mathbf{B}_{27} & = & y_{5} \, \mathbf{a}_{1}-z_{5} \, \mathbf{a}_{2}-x_{5} \, \mathbf{a}_{3} & = & y_{5}a \, \mathbf{\hat{x}}-z_{5}a \, \mathbf{\hat{y}}-x_{5}a \, \mathbf{\hat{z}} & \left(24m\right) & \mbox{C} \\ 
\mathbf{B}_{28} & = & -y_{5} \, \mathbf{a}_{1}-z_{5} \, \mathbf{a}_{2} + x_{5} \, \mathbf{a}_{3} & = & -y_{5}a \, \mathbf{\hat{x}}-z_{5}a \, \mathbf{\hat{y}} + x_{5}a \, \mathbf{\hat{z}} & \left(24m\right) & \mbox{C} \\ 
\mathbf{B}_{29} & = & \left(\frac{1}{2} +y_{5}\right) \, \mathbf{a}_{1} + \left(\frac{1}{2} +x_{5}\right) \, \mathbf{a}_{2} + \left(\frac{1}{2} - z_{5}\right) \, \mathbf{a}_{3} & = & \left(\frac{1}{2} +y_{5}\right)a \, \mathbf{\hat{x}} + \left(\frac{1}{2} +x_{5}\right)a \, \mathbf{\hat{y}} + \left(\frac{1}{2} - z_{5}\right)a \, \mathbf{\hat{z}} & \left(24m\right) & \mbox{C} \\ 
\mathbf{B}_{30} & = & \left(\frac{1}{2} - y_{5}\right) \, \mathbf{a}_{1} + \left(\frac{1}{2} - x_{5}\right) \, \mathbf{a}_{2} + \left(\frac{1}{2} - z_{5}\right) \, \mathbf{a}_{3} & = & \left(\frac{1}{2} - y_{5}\right)a \, \mathbf{\hat{x}} + \left(\frac{1}{2} - x_{5}\right)a \, \mathbf{\hat{y}} + \left(\frac{1}{2} - z_{5}\right)a \, \mathbf{\hat{z}} & \left(24m\right) & \mbox{C} \\ 
\mathbf{B}_{31} & = & \left(\frac{1}{2} +y_{5}\right) \, \mathbf{a}_{1} + \left(\frac{1}{2} - x_{5}\right) \, \mathbf{a}_{2} + \left(\frac{1}{2} +z_{5}\right) \, \mathbf{a}_{3} & = & \left(\frac{1}{2} +y_{5}\right)a \, \mathbf{\hat{x}} + \left(\frac{1}{2} - x_{5}\right)a \, \mathbf{\hat{y}} + \left(\frac{1}{2} +z_{5}\right)a \, \mathbf{\hat{z}} & \left(24m\right) & \mbox{C} \\ 
\mathbf{B}_{32} & = & \left(\frac{1}{2} - y_{5}\right) \, \mathbf{a}_{1} + \left(\frac{1}{2} +x_{5}\right) \, \mathbf{a}_{2} + \left(\frac{1}{2} +z_{5}\right) \, \mathbf{a}_{3} & = & \left(\frac{1}{2} - y_{5}\right)a \, \mathbf{\hat{x}} + \left(\frac{1}{2} +x_{5}\right)a \, \mathbf{\hat{y}} + \left(\frac{1}{2} +z_{5}\right)a \, \mathbf{\hat{z}} & \left(24m\right) & \mbox{C} \\ 
\mathbf{B}_{33} & = & \left(\frac{1}{2} +x_{5}\right) \, \mathbf{a}_{1} + \left(\frac{1}{2} +z_{5}\right) \, \mathbf{a}_{2} + \left(\frac{1}{2} - y_{5}\right) \, \mathbf{a}_{3} & = & \left(\frac{1}{2} +x_{5}\right)a \, \mathbf{\hat{x}} + \left(\frac{1}{2} +z_{5}\right)a \, \mathbf{\hat{y}} + \left(\frac{1}{2} - y_{5}\right)a \, \mathbf{\hat{z}} & \left(24m\right) & \mbox{C} \\ 
\mathbf{B}_{34} & = & \left(\frac{1}{2} - x_{5}\right) \, \mathbf{a}_{1} + \left(\frac{1}{2} +z_{5}\right) \, \mathbf{a}_{2} + \left(\frac{1}{2} +y_{5}\right) \, \mathbf{a}_{3} & = & \left(\frac{1}{2} - x_{5}\right)a \, \mathbf{\hat{x}} + \left(\frac{1}{2} +z_{5}\right)a \, \mathbf{\hat{y}} + \left(\frac{1}{2} +y_{5}\right)a \, \mathbf{\hat{z}} & \left(24m\right) & \mbox{C} \\ 
\mathbf{B}_{35} & = & \left(\frac{1}{2} - x_{5}\right) \, \mathbf{a}_{1} + \left(\frac{1}{2} - z_{5}\right) \, \mathbf{a}_{2} + \left(\frac{1}{2} - y_{5}\right) \, \mathbf{a}_{3} & = & \left(\frac{1}{2} - x_{5}\right)a \, \mathbf{\hat{x}} + \left(\frac{1}{2} - z_{5}\right)a \, \mathbf{\hat{y}} + \left(\frac{1}{2} - y_{5}\right)a \, \mathbf{\hat{z}} & \left(24m\right) & \mbox{C} \\ 
\mathbf{B}_{36} & = & \left(\frac{1}{2} +x_{5}\right) \, \mathbf{a}_{1} + \left(\frac{1}{2} - z_{5}\right) \, \mathbf{a}_{2} + \left(\frac{1}{2} +y_{5}\right) \, \mathbf{a}_{3} & = & \left(\frac{1}{2} +x_{5}\right)a \, \mathbf{\hat{x}} + \left(\frac{1}{2} - z_{5}\right)a \, \mathbf{\hat{y}} + \left(\frac{1}{2} +y_{5}\right)a \, \mathbf{\hat{z}} & \left(24m\right) & \mbox{C} \\ 
\mathbf{B}_{37} & = & \left(\frac{1}{2} +z_{5}\right) \, \mathbf{a}_{1} + \left(\frac{1}{2} +y_{5}\right) \, \mathbf{a}_{2} + \left(\frac{1}{2} - x_{5}\right) \, \mathbf{a}_{3} & = & \left(\frac{1}{2} +z_{5}\right)a \, \mathbf{\hat{x}} + \left(\frac{1}{2} +y_{5}\right)a \, \mathbf{\hat{y}} + \left(\frac{1}{2} - x_{5}\right)a \, \mathbf{\hat{z}} & \left(24m\right) & \mbox{C} \\ 
\mathbf{B}_{38} & = & \left(\frac{1}{2} +z_{5}\right) \, \mathbf{a}_{1} + \left(\frac{1}{2} - y_{5}\right) \, \mathbf{a}_{2} + \left(\frac{1}{2} +x_{5}\right) \, \mathbf{a}_{3} & = & \left(\frac{1}{2} +z_{5}\right)a \, \mathbf{\hat{x}} + \left(\frac{1}{2} - y_{5}\right)a \, \mathbf{\hat{y}} + \left(\frac{1}{2} +x_{5}\right)a \, \mathbf{\hat{z}} & \left(24m\right) & \mbox{C} \\ 
\mathbf{B}_{39} & = & \left(\frac{1}{2} - z_{5}\right) \, \mathbf{a}_{1} + \left(\frac{1}{2} +y_{5}\right) \, \mathbf{a}_{2} + \left(\frac{1}{2} +x_{5}\right) \, \mathbf{a}_{3} & = & \left(\frac{1}{2} - z_{5}\right)a \, \mathbf{\hat{x}} + \left(\frac{1}{2} +y_{5}\right)a \, \mathbf{\hat{y}} + \left(\frac{1}{2} +x_{5}\right)a \, \mathbf{\hat{z}} & \left(24m\right) & \mbox{C} \\ 
\mathbf{B}_{40} & = & \left(\frac{1}{2} - z_{5}\right) \, \mathbf{a}_{1} + \left(\frac{1}{2} - y_{5}\right) \, \mathbf{a}_{2} + \left(\frac{1}{2} - x_{5}\right) \, \mathbf{a}_{3} & = & \left(\frac{1}{2} - z_{5}\right)a \, \mathbf{\hat{x}} + \left(\frac{1}{2} - y_{5}\right)a \, \mathbf{\hat{y}} + \left(\frac{1}{2} - x_{5}\right)a \, \mathbf{\hat{z}} & \left(24m\right) & \mbox{C} \\ 
\mathbf{B}_{41} & = & x_{6} \, \mathbf{a}_{1} + y_{6} \, \mathbf{a}_{2} + z_{6} \, \mathbf{a}_{3} & = & x_{6}a \, \mathbf{\hat{x}} + y_{6}a \, \mathbf{\hat{y}} + z_{6}a \, \mathbf{\hat{z}} & \left(24m\right) & \mbox{N} \\ 
\mathbf{B}_{42} & = & -x_{6} \, \mathbf{a}_{1}-y_{6} \, \mathbf{a}_{2} + z_{6} \, \mathbf{a}_{3} & = & -x_{6}a \, \mathbf{\hat{x}}-y_{6}a \, \mathbf{\hat{y}} + z_{6}a \, \mathbf{\hat{z}} & \left(24m\right) & \mbox{N} \\ 
\mathbf{B}_{43} & = & -x_{6} \, \mathbf{a}_{1} + y_{6} \, \mathbf{a}_{2}-z_{6} \, \mathbf{a}_{3} & = & -x_{6}a \, \mathbf{\hat{x}} + y_{6}a \, \mathbf{\hat{y}}-z_{6}a \, \mathbf{\hat{z}} & \left(24m\right) & \mbox{N} \\ 
\mathbf{B}_{44} & = & x_{6} \, \mathbf{a}_{1}-y_{6} \, \mathbf{a}_{2}-z_{6} \, \mathbf{a}_{3} & = & x_{6}a \, \mathbf{\hat{x}}-y_{6}a \, \mathbf{\hat{y}}-z_{6}a \, \mathbf{\hat{z}} & \left(24m\right) & \mbox{N} \\ 
\mathbf{B}_{45} & = & z_{6} \, \mathbf{a}_{1} + x_{6} \, \mathbf{a}_{2} + y_{6} \, \mathbf{a}_{3} & = & z_{6}a \, \mathbf{\hat{x}} + x_{6}a \, \mathbf{\hat{y}} + y_{6}a \, \mathbf{\hat{z}} & \left(24m\right) & \mbox{N} \\ 
\mathbf{B}_{46} & = & z_{6} \, \mathbf{a}_{1}-x_{6} \, \mathbf{a}_{2}-y_{6} \, \mathbf{a}_{3} & = & z_{6}a \, \mathbf{\hat{x}}-x_{6}a \, \mathbf{\hat{y}}-y_{6}a \, \mathbf{\hat{z}} & \left(24m\right) & \mbox{N} \\ 
\mathbf{B}_{47} & = & -z_{6} \, \mathbf{a}_{1}-x_{6} \, \mathbf{a}_{2} + y_{6} \, \mathbf{a}_{3} & = & -z_{6}a \, \mathbf{\hat{x}}-x_{6}a \, \mathbf{\hat{y}} + y_{6}a \, \mathbf{\hat{z}} & \left(24m\right) & \mbox{N} \\ 
\mathbf{B}_{48} & = & -z_{6} \, \mathbf{a}_{1} + x_{6} \, \mathbf{a}_{2}-y_{6} \, \mathbf{a}_{3} & = & -z_{6}a \, \mathbf{\hat{x}} + x_{6}a \, \mathbf{\hat{y}}-y_{6}a \, \mathbf{\hat{z}} & \left(24m\right) & \mbox{N} \\ 
\mathbf{B}_{49} & = & y_{6} \, \mathbf{a}_{1} + z_{6} \, \mathbf{a}_{2} + x_{6} \, \mathbf{a}_{3} & = & y_{6}a \, \mathbf{\hat{x}} + z_{6}a \, \mathbf{\hat{y}} + x_{6}a \, \mathbf{\hat{z}} & \left(24m\right) & \mbox{N} \\ 
\mathbf{B}_{50} & = & -y_{6} \, \mathbf{a}_{1} + z_{6} \, \mathbf{a}_{2}-x_{6} \, \mathbf{a}_{3} & = & -y_{6}a \, \mathbf{\hat{x}} + z_{6}a \, \mathbf{\hat{y}}-x_{6}a \, \mathbf{\hat{z}} & \left(24m\right) & \mbox{N} \\ 
\mathbf{B}_{51} & = & y_{6} \, \mathbf{a}_{1}-z_{6} \, \mathbf{a}_{2}-x_{6} \, \mathbf{a}_{3} & = & y_{6}a \, \mathbf{\hat{x}}-z_{6}a \, \mathbf{\hat{y}}-x_{6}a \, \mathbf{\hat{z}} & \left(24m\right) & \mbox{N} \\ 
\mathbf{B}_{52} & = & -y_{6} \, \mathbf{a}_{1}-z_{6} \, \mathbf{a}_{2} + x_{6} \, \mathbf{a}_{3} & = & -y_{6}a \, \mathbf{\hat{x}}-z_{6}a \, \mathbf{\hat{y}} + x_{6}a \, \mathbf{\hat{z}} & \left(24m\right) & \mbox{N} \\ 
\mathbf{B}_{53} & = & \left(\frac{1}{2} +y_{6}\right) \, \mathbf{a}_{1} + \left(\frac{1}{2} +x_{6}\right) \, \mathbf{a}_{2} + \left(\frac{1}{2} - z_{6}\right) \, \mathbf{a}_{3} & = & \left(\frac{1}{2} +y_{6}\right)a \, \mathbf{\hat{x}} + \left(\frac{1}{2} +x_{6}\right)a \, \mathbf{\hat{y}} + \left(\frac{1}{2} - z_{6}\right)a \, \mathbf{\hat{z}} & \left(24m\right) & \mbox{N} \\ 
\mathbf{B}_{54} & = & \left(\frac{1}{2} - y_{6}\right) \, \mathbf{a}_{1} + \left(\frac{1}{2} - x_{6}\right) \, \mathbf{a}_{2} + \left(\frac{1}{2} - z_{6}\right) \, \mathbf{a}_{3} & = & \left(\frac{1}{2} - y_{6}\right)a \, \mathbf{\hat{x}} + \left(\frac{1}{2} - x_{6}\right)a \, \mathbf{\hat{y}} + \left(\frac{1}{2} - z_{6}\right)a \, \mathbf{\hat{z}} & \left(24m\right) & \mbox{N} \\ 
\mathbf{B}_{55} & = & \left(\frac{1}{2} +y_{6}\right) \, \mathbf{a}_{1} + \left(\frac{1}{2} - x_{6}\right) \, \mathbf{a}_{2} + \left(\frac{1}{2} +z_{6}\right) \, \mathbf{a}_{3} & = & \left(\frac{1}{2} +y_{6}\right)a \, \mathbf{\hat{x}} + \left(\frac{1}{2} - x_{6}\right)a \, \mathbf{\hat{y}} + \left(\frac{1}{2} +z_{6}\right)a \, \mathbf{\hat{z}} & \left(24m\right) & \mbox{N} \\ 
\mathbf{B}_{56} & = & \left(\frac{1}{2} - y_{6}\right) \, \mathbf{a}_{1} + \left(\frac{1}{2} +x_{6}\right) \, \mathbf{a}_{2} + \left(\frac{1}{2} +z_{6}\right) \, \mathbf{a}_{3} & = & \left(\frac{1}{2} - y_{6}\right)a \, \mathbf{\hat{x}} + \left(\frac{1}{2} +x_{6}\right)a \, \mathbf{\hat{y}} + \left(\frac{1}{2} +z_{6}\right)a \, \mathbf{\hat{z}} & \left(24m\right) & \mbox{N} \\ 
\mathbf{B}_{57} & = & \left(\frac{1}{2} +x_{6}\right) \, \mathbf{a}_{1} + \left(\frac{1}{2} +z_{6}\right) \, \mathbf{a}_{2} + \left(\frac{1}{2} - y_{6}\right) \, \mathbf{a}_{3} & = & \left(\frac{1}{2} +x_{6}\right)a \, \mathbf{\hat{x}} + \left(\frac{1}{2} +z_{6}\right)a \, \mathbf{\hat{y}} + \left(\frac{1}{2} - y_{6}\right)a \, \mathbf{\hat{z}} & \left(24m\right) & \mbox{N} \\ 
\mathbf{B}_{58} & = & \left(\frac{1}{2} - x_{6}\right) \, \mathbf{a}_{1} + \left(\frac{1}{2} +z_{6}\right) \, \mathbf{a}_{2} + \left(\frac{1}{2} +y_{6}\right) \, \mathbf{a}_{3} & = & \left(\frac{1}{2} - x_{6}\right)a \, \mathbf{\hat{x}} + \left(\frac{1}{2} +z_{6}\right)a \, \mathbf{\hat{y}} + \left(\frac{1}{2} +y_{6}\right)a \, \mathbf{\hat{z}} & \left(24m\right) & \mbox{N} \\ 
\mathbf{B}_{59} & = & \left(\frac{1}{2} - x_{6}\right) \, \mathbf{a}_{1} + \left(\frac{1}{2} - z_{6}\right) \, \mathbf{a}_{2} + \left(\frac{1}{2} - y_{6}\right) \, \mathbf{a}_{3} & = & \left(\frac{1}{2} - x_{6}\right)a \, \mathbf{\hat{x}} + \left(\frac{1}{2} - z_{6}\right)a \, \mathbf{\hat{y}} + \left(\frac{1}{2} - y_{6}\right)a \, \mathbf{\hat{z}} & \left(24m\right) & \mbox{N} \\ 
\mathbf{B}_{60} & = & \left(\frac{1}{2} +x_{6}\right) \, \mathbf{a}_{1} + \left(\frac{1}{2} - z_{6}\right) \, \mathbf{a}_{2} + \left(\frac{1}{2} +y_{6}\right) \, \mathbf{a}_{3} & = & \left(\frac{1}{2} +x_{6}\right)a \, \mathbf{\hat{x}} + \left(\frac{1}{2} - z_{6}\right)a \, \mathbf{\hat{y}} + \left(\frac{1}{2} +y_{6}\right)a \, \mathbf{\hat{z}} & \left(24m\right) & \mbox{N} \\ 
\mathbf{B}_{61} & = & \left(\frac{1}{2} +z_{6}\right) \, \mathbf{a}_{1} + \left(\frac{1}{2} +y_{6}\right) \, \mathbf{a}_{2} + \left(\frac{1}{2} - x_{6}\right) \, \mathbf{a}_{3} & = & \left(\frac{1}{2} +z_{6}\right)a \, \mathbf{\hat{x}} + \left(\frac{1}{2} +y_{6}\right)a \, \mathbf{\hat{y}} + \left(\frac{1}{2} - x_{6}\right)a \, \mathbf{\hat{z}} & \left(24m\right) & \mbox{N} \\ 
\mathbf{B}_{62} & = & \left(\frac{1}{2} +z_{6}\right) \, \mathbf{a}_{1} + \left(\frac{1}{2} - y_{6}\right) \, \mathbf{a}_{2} + \left(\frac{1}{2} +x_{6}\right) \, \mathbf{a}_{3} & = & \left(\frac{1}{2} +z_{6}\right)a \, \mathbf{\hat{x}} + \left(\frac{1}{2} - y_{6}\right)a \, \mathbf{\hat{y}} + \left(\frac{1}{2} +x_{6}\right)a \, \mathbf{\hat{z}} & \left(24m\right) & \mbox{N} \\ 
\mathbf{B}_{63} & = & \left(\frac{1}{2} - z_{6}\right) \, \mathbf{a}_{1} + \left(\frac{1}{2} +y_{6}\right) \, \mathbf{a}_{2} + \left(\frac{1}{2} +x_{6}\right) \, \mathbf{a}_{3} & = & \left(\frac{1}{2} - z_{6}\right)a \, \mathbf{\hat{x}} + \left(\frac{1}{2} +y_{6}\right)a \, \mathbf{\hat{y}} + \left(\frac{1}{2} +x_{6}\right)a \, \mathbf{\hat{z}} & \left(24m\right) & \mbox{N} \\ 
\mathbf{B}_{64} & = & \left(\frac{1}{2} - z_{6}\right) \, \mathbf{a}_{1} + \left(\frac{1}{2} - y_{6}\right) \, \mathbf{a}_{2} + \left(\frac{1}{2} - x_{6}\right) \, \mathbf{a}_{3} & = & \left(\frac{1}{2} - z_{6}\right)a \, \mathbf{\hat{x}} + \left(\frac{1}{2} - y_{6}\right)a \, \mathbf{\hat{y}} + \left(\frac{1}{2} - x_{6}\right)a \, \mathbf{\hat{z}} & \left(24m\right) & \mbox{N} \\ 
\end{longtabu}
\renewcommand{\arraystretch}{1.0}
\noindent \hrulefill
\\
\textbf{References:}
\vspace*{-0.25cm}
\begin{flushleft}
  - \bibentry{Kuznetsov_Cs2ZnFeCN6_RussJInorgChem_1970}. \\
\end{flushleft}
\textbf{Found in:}
\vspace*{-0.25cm}
\begin{flushleft}
  - \bibentry{Villars_PearsonsCrystalData_2013}. \\
\end{flushleft}
\noindent \hrulefill
\\
\textbf{Geometry files:}
\\
\noindent  - CIF: pp. {\hyperref[A6B2CD6E_cP64_208_m_ad_b_m_c_cif]{\pageref{A6B2CD6E_cP64_208_m_ad_b_m_c_cif}}} \\
\noindent  - POSCAR: pp. {\hyperref[A6B2CD6E_cP64_208_m_ad_b_m_c_poscar]{\pageref{A6B2CD6E_cP64_208_m_ad_b_m_c_poscar}}} \\
\onecolumn
{\phantomsection\label{A24BC_cF104_209_j_a_b}}
\subsection*{\huge \textbf{{\normalfont F$_{6}$KP Structure: A24BC\_cF104\_209\_j\_a\_b}}}
\noindent \hrulefill
\vspace*{0.25cm}
\begin{figure}[htp]
  \centering
  \vspace{-1em}
  {\includegraphics[width=1\textwidth]{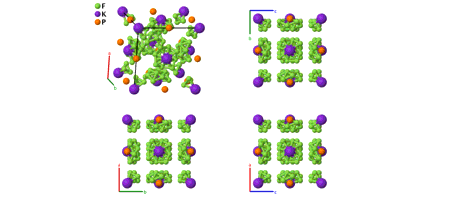}}
\end{figure}
\vspace*{-0.5cm}
\renewcommand{\arraystretch}{1.5}
\begin{equation*}
  \begin{array}{>{$\hspace{-0.15cm}}l<{$}>{$}p{0.5cm}<{$}>{$}p{18.5cm}<{$}}
    \mbox{\large \textbf{Prototype}} &\colon & \ce{F6KP} \\
    \mbox{\large \textbf{\AFLOW\ prototype label}} &\colon & \mbox{A24BC\_cF104\_209\_j\_a\_b} \\
    \mbox{\large \textbf{\textit{Strukturbericht} designation}} &\colon & \mbox{None} \\
    \mbox{\large \textbf{Pearson symbol}} &\colon & \mbox{cF104} \\
    \mbox{\large \textbf{Space group number}} &\colon & 209 \\
    \mbox{\large \textbf{Space group symbol}} &\colon & F432 \\
    \mbox{\large \textbf{\AFLOW\ prototype command}} &\colon &  \texttt{aflow} \,  \, \texttt{-{}-proto=A24BC\_cF104\_209\_j\_a\_b } \, \newline \texttt{-{}-params=}{a,x_{3},y_{3},z_{3} }
  \end{array}
\end{equation*}
\renewcommand{\arraystretch}{1.0}

\vspace*{-0.25cm}
\noindent \hrulefill
\begin{itemize}
  \item{The (96j) Wyckoff positions are decorated by F atoms with a site occupation of 0.25.  
Hence, the prototype material is F$_{6}$KP as opposed to F$_{24}$KP.
}
\end{itemize}

\noindent \parbox{1 \linewidth}{
\noindent \hrulefill
\\
\textbf{Face-centered Cubic primitive vectors:} \\
\vspace*{-0.25cm}
\begin{tabular}{cc}
  \begin{tabular}{c}
    \parbox{0.6 \linewidth}{
      \renewcommand{\arraystretch}{1.5}
      \begin{equation*}
        \centering
        \begin{array}{ccc}
              \mathbf{a}_1 & = & \frac12 \, a \, \mathbf{\hat{y}} + \frac12 \, a \, \mathbf{\hat{z}} \\
    \mathbf{a}_2 & = & \frac12 \, a \, \mathbf{\hat{x}} + \frac12 \, a \, \mathbf{\hat{z}} \\
    \mathbf{a}_3 & = & \frac12 \, a \, \mathbf{\hat{x}} + \frac12 \, a \, \mathbf{\hat{y}} \\

        \end{array}
      \end{equation*}
    }
    \renewcommand{\arraystretch}{1.0}
  \end{tabular}
  \begin{tabular}{c}
    \includegraphics[width=0.3\linewidth]{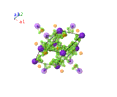} \\
  \end{tabular}
\end{tabular}

}
\vspace*{-0.25cm}

\noindent \hrulefill
\\
\textbf{Basis vectors:}
\vspace*{-0.25cm}
\renewcommand{\arraystretch}{1.5}
\begin{longtabu} to \textwidth{>{\centering $}X[-1,c,c]<{$}>{\centering $}X[-1,c,c]<{$}>{\centering $}X[-1,c,c]<{$}>{\centering $}X[-1,c,c]<{$}>{\centering $}X[-1,c,c]<{$}>{\centering $}X[-1,c,c]<{$}>{\centering $}X[-1,c,c]<{$}}
  & & \mbox{Lattice Coordinates} & & \mbox{Cartesian Coordinates} &\mbox{Wyckoff Position} & \mbox{Atom Type} \\  
  \mathbf{B}_{1} & = & 0 \, \mathbf{a}_{1} + 0 \, \mathbf{a}_{2} + 0 \, \mathbf{a}_{3} & = & 0 \, \mathbf{\hat{x}} + 0 \, \mathbf{\hat{y}} + 0 \, \mathbf{\hat{z}} & \left(4a\right) & \mbox{K} \\ 
\mathbf{B}_{2} & = & \frac{1}{2} \, \mathbf{a}_{1} + \frac{1}{2} \, \mathbf{a}_{2} + \frac{1}{2} \, \mathbf{a}_{3} & = & \frac{1}{2}a \, \mathbf{\hat{x}} + \frac{1}{2}a \, \mathbf{\hat{y}} + \frac{1}{2}a \, \mathbf{\hat{z}} & \left(4b\right) & \mbox{P} \\ 
\mathbf{B}_{3} & = & \left(-x_{3}+y_{3}+z_{3}\right) \, \mathbf{a}_{1} + \left(x_{3}-y_{3}+z_{3}\right) \, \mathbf{a}_{2} + \left(x_{3}+y_{3}-z_{3}\right) \, \mathbf{a}_{3} & = & x_{3}a \, \mathbf{\hat{x}} + y_{3}a \, \mathbf{\hat{y}} + z_{3}a \, \mathbf{\hat{z}} & \left(96j\right) & \mbox{F} \\ 
\mathbf{B}_{4} & = & \left(x_{3}-y_{3}+z_{3}\right) \, \mathbf{a}_{1} + \left(-x_{3}+y_{3}+z_{3}\right) \, \mathbf{a}_{2} + \left(-x_{3}-y_{3}-z_{3}\right) \, \mathbf{a}_{3} & = & -x_{3}a \, \mathbf{\hat{x}}-y_{3}a \, \mathbf{\hat{y}} + z_{3}a \, \mathbf{\hat{z}} & \left(96j\right) & \mbox{F} \\ 
\mathbf{B}_{5} & = & \left(x_{3}+y_{3}-z_{3}\right) \, \mathbf{a}_{1} + \left(-x_{3}-y_{3}-z_{3}\right) \, \mathbf{a}_{2} + \left(-x_{3}+y_{3}+z_{3}\right) \, \mathbf{a}_{3} & = & -x_{3}a \, \mathbf{\hat{x}} + y_{3}a \, \mathbf{\hat{y}}-z_{3}a \, \mathbf{\hat{z}} & \left(96j\right) & \mbox{F} \\ 
\mathbf{B}_{6} & = & \left(-x_{3}-y_{3}-z_{3}\right) \, \mathbf{a}_{1} + \left(x_{3}+y_{3}-z_{3}\right) \, \mathbf{a}_{2} + \left(x_{3}-y_{3}+z_{3}\right) \, \mathbf{a}_{3} & = & x_{3}a \, \mathbf{\hat{x}}-y_{3}a \, \mathbf{\hat{y}}-z_{3}a \, \mathbf{\hat{z}} & \left(96j\right) & \mbox{F} \\ 
\mathbf{B}_{7} & = & \left(x_{3}+y_{3}-z_{3}\right) \, \mathbf{a}_{1} + \left(-x_{3}+y_{3}+z_{3}\right) \, \mathbf{a}_{2} + \left(x_{3}-y_{3}+z_{3}\right) \, \mathbf{a}_{3} & = & z_{3}a \, \mathbf{\hat{x}} + x_{3}a \, \mathbf{\hat{y}} + y_{3}a \, \mathbf{\hat{z}} & \left(96j\right) & \mbox{F} \\ 
\mathbf{B}_{8} & = & \left(-x_{3}-y_{3}-z_{3}\right) \, \mathbf{a}_{1} + \left(x_{3}-y_{3}+z_{3}\right) \, \mathbf{a}_{2} + \left(-x_{3}+y_{3}+z_{3}\right) \, \mathbf{a}_{3} & = & z_{3}a \, \mathbf{\hat{x}}-x_{3}a \, \mathbf{\hat{y}}-y_{3}a \, \mathbf{\hat{z}} & \left(96j\right) & \mbox{F} \\ 
\mathbf{B}_{9} & = & \left(-x_{3}+y_{3}+z_{3}\right) \, \mathbf{a}_{1} + \left(x_{3}+y_{3}-z_{3}\right) \, \mathbf{a}_{2} + \left(-x_{3}-y_{3}-z_{3}\right) \, \mathbf{a}_{3} & = & -z_{3}a \, \mathbf{\hat{x}}-x_{3}a \, \mathbf{\hat{y}} + y_{3}a \, \mathbf{\hat{z}} & \left(96j\right) & \mbox{F} \\ 
\mathbf{B}_{10} & = & \left(x_{3}-y_{3}+z_{3}\right) \, \mathbf{a}_{1} + \left(-x_{3}-y_{3}-z_{3}\right) \, \mathbf{a}_{2} + \left(x_{3}+y_{3}-z_{3}\right) \, \mathbf{a}_{3} & = & -z_{3}a \, \mathbf{\hat{x}} + x_{3}a \, \mathbf{\hat{y}}-y_{3}a \, \mathbf{\hat{z}} & \left(96j\right) & \mbox{F} \\ 
\mathbf{B}_{11} & = & \left(x_{3}-y_{3}+z_{3}\right) \, \mathbf{a}_{1} + \left(x_{3}+y_{3}-z_{3}\right) \, \mathbf{a}_{2} + \left(-x_{3}+y_{3}+z_{3}\right) \, \mathbf{a}_{3} & = & y_{3}a \, \mathbf{\hat{x}} + z_{3}a \, \mathbf{\hat{y}} + x_{3}a \, \mathbf{\hat{z}} & \left(96j\right) & \mbox{F} \\ 
\mathbf{B}_{12} & = & \left(-x_{3}+y_{3}+z_{3}\right) \, \mathbf{a}_{1} + \left(-x_{3}-y_{3}-z_{3}\right) \, \mathbf{a}_{2} + \left(x_{3}-y_{3}+z_{3}\right) \, \mathbf{a}_{3} & = & -y_{3}a \, \mathbf{\hat{x}} + z_{3}a \, \mathbf{\hat{y}}-x_{3}a \, \mathbf{\hat{z}} & \left(96j\right) & \mbox{F} \\ 
\mathbf{B}_{13} & = & \left(-x_{3}-y_{3}-z_{3}\right) \, \mathbf{a}_{1} + \left(-x_{3}+y_{3}+z_{3}\right) \, \mathbf{a}_{2} + \left(x_{3}+y_{3}-z_{3}\right) \, \mathbf{a}_{3} & = & y_{3}a \, \mathbf{\hat{x}}-z_{3}a \, \mathbf{\hat{y}}-x_{3}a \, \mathbf{\hat{z}} & \left(96j\right) & \mbox{F} \\ 
\mathbf{B}_{14} & = & \left(x_{3}+y_{3}-z_{3}\right) \, \mathbf{a}_{1} + \left(x_{3}-y_{3}+z_{3}\right) \, \mathbf{a}_{2} + \left(-x_{3}-y_{3}-z_{3}\right) \, \mathbf{a}_{3} & = & -y_{3}a \, \mathbf{\hat{x}}-z_{3}a \, \mathbf{\hat{y}} + x_{3}a \, \mathbf{\hat{z}} & \left(96j\right) & \mbox{F} \\ 
\mathbf{B}_{15} & = & \left(x_{3}-y_{3}-z_{3}\right) \, \mathbf{a}_{1} + \left(-x_{3}+y_{3}-z_{3}\right) \, \mathbf{a}_{2} + \left(x_{3}+y_{3}+z_{3}\right) \, \mathbf{a}_{3} & = & y_{3}a \, \mathbf{\hat{x}} + x_{3}a \, \mathbf{\hat{y}}-z_{3}a \, \mathbf{\hat{z}} & \left(96j\right) & \mbox{F} \\ 
\mathbf{B}_{16} & = & \left(-x_{3}+y_{3}-z_{3}\right) \, \mathbf{a}_{1} + \left(x_{3}-y_{3}-z_{3}\right) \, \mathbf{a}_{2} + \left(-x_{3}-y_{3}+z_{3}\right) \, \mathbf{a}_{3} & = & -y_{3}a \, \mathbf{\hat{x}}-x_{3}a \, \mathbf{\hat{y}}-z_{3}a \, \mathbf{\hat{z}} & \left(96j\right) & \mbox{F} \\ 
\mathbf{B}_{17} & = & \left(-x_{3}-y_{3}+z_{3}\right) \, \mathbf{a}_{1} + \left(x_{3}+y_{3}+z_{3}\right) \, \mathbf{a}_{2} + \left(-x_{3}+y_{3}-z_{3}\right) \, \mathbf{a}_{3} & = & y_{3}a \, \mathbf{\hat{x}}-x_{3}a \, \mathbf{\hat{y}} + z_{3}a \, \mathbf{\hat{z}} & \left(96j\right) & \mbox{F} \\ 
\mathbf{B}_{18} & = & \left(x_{3}+y_{3}+z_{3}\right) \, \mathbf{a}_{1} + \left(-x_{3}-y_{3}+z_{3}\right) \, \mathbf{a}_{2} + \left(x_{3}-y_{3}-z_{3}\right) \, \mathbf{a}_{3} & = & -y_{3}a \, \mathbf{\hat{x}} + x_{3}a \, \mathbf{\hat{y}} + z_{3}a \, \mathbf{\hat{z}} & \left(96j\right) & \mbox{F} \\ 
\mathbf{B}_{19} & = & \left(-x_{3}-y_{3}+z_{3}\right) \, \mathbf{a}_{1} + \left(x_{3}-y_{3}-z_{3}\right) \, \mathbf{a}_{2} + \left(x_{3}+y_{3}+z_{3}\right) \, \mathbf{a}_{3} & = & x_{3}a \, \mathbf{\hat{x}} + z_{3}a \, \mathbf{\hat{y}}-y_{3}a \, \mathbf{\hat{z}} & \left(96j\right) & \mbox{F} \\ 
\mathbf{B}_{20} & = & \left(x_{3}+y_{3}+z_{3}\right) \, \mathbf{a}_{1} + \left(-x_{3}+y_{3}-z_{3}\right) \, \mathbf{a}_{2} + \left(-x_{3}-y_{3}+z_{3}\right) \, \mathbf{a}_{3} & = & -x_{3}a \, \mathbf{\hat{x}} + z_{3}a \, \mathbf{\hat{y}} + y_{3}a \, \mathbf{\hat{z}} & \left(96j\right) & \mbox{F} \\ 
\mathbf{B}_{21} & = & \left(x_{3}-y_{3}-z_{3}\right) \, \mathbf{a}_{1} + \left(-x_{3}-y_{3}+z_{3}\right) \, \mathbf{a}_{2} + \left(-x_{3}+y_{3}-z_{3}\right) \, \mathbf{a}_{3} & = & -x_{3}a \, \mathbf{\hat{x}}-z_{3}a \, \mathbf{\hat{y}}-y_{3}a \, \mathbf{\hat{z}} & \left(96j\right) & \mbox{F} \\ 
\mathbf{B}_{22} & = & \left(-x_{3}+y_{3}-z_{3}\right) \, \mathbf{a}_{1} + \left(x_{3}+y_{3}+z_{3}\right) \, \mathbf{a}_{2} + \left(x_{3}-y_{3}-z_{3}\right) \, \mathbf{a}_{3} & = & x_{3}a \, \mathbf{\hat{x}}-z_{3}a \, \mathbf{\hat{y}} + y_{3}a \, \mathbf{\hat{z}} & \left(96j\right) & \mbox{F} \\ 
\mathbf{B}_{23} & = & \left(-x_{3}+y_{3}-z_{3}\right) \, \mathbf{a}_{1} + \left(-x_{3}-y_{3}+z_{3}\right) \, \mathbf{a}_{2} + \left(x_{3}+y_{3}+z_{3}\right) \, \mathbf{a}_{3} & = & z_{3}a \, \mathbf{\hat{x}} + y_{3}a \, \mathbf{\hat{y}}-x_{3}a \, \mathbf{\hat{z}} & \left(96j\right) & \mbox{F} \\ 
\mathbf{B}_{24} & = & \left(x_{3}-y_{3}-z_{3}\right) \, \mathbf{a}_{1} + \left(x_{3}+y_{3}+z_{3}\right) \, \mathbf{a}_{2} + \left(-x_{3}-y_{3}+z_{3}\right) \, \mathbf{a}_{3} & = & z_{3}a \, \mathbf{\hat{x}}-y_{3}a \, \mathbf{\hat{y}} + x_{3}a \, \mathbf{\hat{z}} & \left(96j\right) & \mbox{F} \\ 
\mathbf{B}_{25} & = & \left(x_{3}+y_{3}+z_{3}\right) \, \mathbf{a}_{1} + \left(x_{3}-y_{3}-z_{3}\right) \, \mathbf{a}_{2} + \left(-x_{3}+y_{3}-z_{3}\right) \, \mathbf{a}_{3} & = & -z_{3}a \, \mathbf{\hat{x}} + y_{3}a \, \mathbf{\hat{y}} + x_{3}a \, \mathbf{\hat{z}} & \left(96j\right) & \mbox{F} \\ 
\mathbf{B}_{26} & = & \left(-x_{3}-y_{3}+z_{3}\right) \, \mathbf{a}_{1} + \left(-x_{3}+y_{3}-z_{3}\right) \, \mathbf{a}_{2} + \left(x_{3}-y_{3}-z_{3}\right) \, \mathbf{a}_{3} & = & -z_{3}a \, \mathbf{\hat{x}}-y_{3}a \, \mathbf{\hat{y}}-x_{3}a \, \mathbf{\hat{z}} & \left(96j\right) & \mbox{F} \\ 
\end{longtabu}
\renewcommand{\arraystretch}{1.0}
\noindent \hrulefill
\\
\textbf{References:}
\vspace*{-0.25cm}
\begin{flushleft}
  - \bibentry{Mascarenhas_KF24P_actacristA_1981}. \\
\end{flushleft}
\textbf{Found in:}
\vspace*{-0.25cm}
\begin{flushleft}
  - \bibentry{Villars_PearsonsCrystalData_2013}. \\
\end{flushleft}
\noindent \hrulefill
\\
\textbf{Geometry files:}
\\
\noindent  - CIF: pp. {\hyperref[A24BC_cF104_209_j_a_b_cif]{\pageref{A24BC_cF104_209_j_a_b_cif}}} \\
\noindent  - POSCAR: pp. {\hyperref[A24BC_cF104_209_j_a_b_poscar]{\pageref{A24BC_cF104_209_j_a_b_poscar}}} \\
\onecolumn
{\phantomsection\label{A12B6C_cF608_210_4h_2h_e}}
\subsection*{\huge \textbf{{\normalfont Te[OH]$_{6}$ Structure: A12B6C\_cF608\_210\_4h\_2h\_e}}}
\noindent \hrulefill
\vspace*{0.25cm}
\begin{figure}[htp]
  \centering
  \vspace{-1em}
  {\includegraphics[width=1\textwidth]{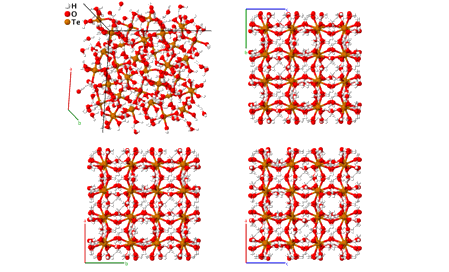}}
\end{figure}
\vspace*{-0.5cm}
\renewcommand{\arraystretch}{1.5}
\begin{equation*}
  \begin{array}{>{$\hspace{-0.15cm}}l<{$}>{$}p{0.5cm}<{$}>{$}p{18.5cm}<{$}}
    \mbox{\large \textbf{Prototype}} &\colon & \ce{Te[OH]6} \\
    \mbox{\large \textbf{\AFLOW\ prototype label}} &\colon & \mbox{A12B6C\_cF608\_210\_4h\_2h\_e} \\
    \mbox{\large \textbf{\textit{Strukturbericht} designation}} &\colon & \mbox{None} \\
    \mbox{\large \textbf{Pearson symbol}} &\colon & \mbox{cF608} \\
    \mbox{\large \textbf{Space group number}} &\colon & 210 \\
    \mbox{\large \textbf{Space group symbol}} &\colon & F4_{1}32 \\
    \mbox{\large \textbf{\AFLOW\ prototype command}} &\colon &  \texttt{aflow} \,  \, \texttt{-{}-proto=A12B6C\_cF608\_210\_4h\_2h\_e } \, \newline \texttt{-{}-params=}{a,x_{1},x_{2},y_{2},z_{2},x_{3},y_{3},z_{3},x_{4},y_{4},z_{4},x_{5},y_{5},z_{5},x_{6},y_{6},z_{6},x_{7},y_{7},z_{7} }
  \end{array}
\end{equation*}
\renewcommand{\arraystretch}{1.0}

\vspace*{-0.25cm}
\noindent \hrulefill
\begin{itemize}
  \item{All H sites are half occupation.
Polytypes appear in space groups \#14, \#228, and \#225.
}
\end{itemize}

\noindent \parbox{1 \linewidth}{
\noindent \hrulefill
\\
\textbf{Face-centered Cubic primitive vectors:} \\
\vspace*{-0.25cm}
\begin{tabular}{cc}
  \begin{tabular}{c}
    \parbox{0.6 \linewidth}{
      \renewcommand{\arraystretch}{1.5}
      \begin{equation*}
        \centering
        \begin{array}{ccc}
              \mathbf{a}_1 & = & \frac12 \, a \, \mathbf{\hat{y}} + \frac12 \, a \, \mathbf{\hat{z}} \\
    \mathbf{a}_2 & = & \frac12 \, a \, \mathbf{\hat{x}} + \frac12 \, a \, \mathbf{\hat{z}} \\
    \mathbf{a}_3 & = & \frac12 \, a \, \mathbf{\hat{x}} + \frac12 \, a \, \mathbf{\hat{y}} \\

        \end{array}
      \end{equation*}
    }
    \renewcommand{\arraystretch}{1.0}
  \end{tabular}
  \begin{tabular}{c}
    \includegraphics[width=0.3\linewidth]{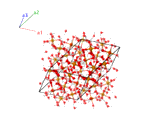} \\
  \end{tabular}
\end{tabular}

}
\vspace*{-0.25cm}

\noindent \hrulefill
\\
\textbf{Basis vectors:}
\vspace*{-0.25cm}
\renewcommand{\arraystretch}{1.5}
\begin{longtabu} to \textwidth{>{\centering $}X[-1,c,c]<{$}>{\centering $}X[-1,c,c]<{$}>{\centering $}X[-1,c,c]<{$}>{\centering $}X[-1,c,c]<{$}>{\centering $}X[-1,c,c]<{$}>{\centering $}X[-1,c,c]<{$}>{\centering $}X[-1,c,c]<{$}}
  & & \mbox{Lattice Coordinates} & & \mbox{Cartesian Coordinates} &\mbox{Wyckoff Position} & \mbox{Atom Type} \\  
  \mathbf{B}_{1} & = & x_{1} \, \mathbf{a}_{1} + x_{1} \, \mathbf{a}_{2} + x_{1} \, \mathbf{a}_{3} & = & x_{1}a \, \mathbf{\hat{x}} + x_{1}a \, \mathbf{\hat{y}} + x_{1}a \, \mathbf{\hat{z}} & \left(32e\right) & \mbox{Te} \\ 
\mathbf{B}_{2} & = & x_{1} \, \mathbf{a}_{1} + x_{1} \, \mathbf{a}_{2}-3x_{1} \, \mathbf{a}_{3} & = & -x_{1}a \, \mathbf{\hat{x}}-x_{1}a \, \mathbf{\hat{y}} + x_{1}a \, \mathbf{\hat{z}} & \left(32e\right) & \mbox{Te} \\ 
\mathbf{B}_{3} & = & x_{1} \, \mathbf{a}_{1}-3x_{1} \, \mathbf{a}_{2} + x_{1} \, \mathbf{a}_{3} & = & -x_{1}a \, \mathbf{\hat{x}} + x_{1}a \, \mathbf{\hat{y}}-x_{1}a \, \mathbf{\hat{z}} & \left(32e\right) & \mbox{Te} \\ 
\mathbf{B}_{4} & = & -3x_{1} \, \mathbf{a}_{1} + x_{1} \, \mathbf{a}_{2} + x_{1} \, \mathbf{a}_{3} & = & x_{1}a \, \mathbf{\hat{x}}-x_{1}a \, \mathbf{\hat{y}}-x_{1}a \, \mathbf{\hat{z}} & \left(32e\right) & \mbox{Te} \\ 
\mathbf{B}_{5} & = & \left(\frac{1}{4} - x_{1}\right) \, \mathbf{a}_{1} + \left(\frac{1}{4} - x_{1}\right) \, \mathbf{a}_{2} + \left(\frac{1}{4} +3x_{1}\right) \, \mathbf{a}_{3} & = & \left(\frac{1}{4} +x_{1}\right)a \, \mathbf{\hat{x}} + \left(\frac{1}{4} +x_{1}\right)a \, \mathbf{\hat{y}} + \left(\frac{1}{4} - x_{1}\right)a \, \mathbf{\hat{z}} & \left(32e\right) & \mbox{Te} \\ 
\mathbf{B}_{6} & = & \left(\frac{1}{4} - x_{1}\right) \, \mathbf{a}_{1} + \left(\frac{1}{4} - x_{1}\right) \, \mathbf{a}_{2} + \left(\frac{1}{4} - x_{1}\right) \, \mathbf{a}_{3} & = & \left(\frac{1}{4} - x_{1}\right)a \, \mathbf{\hat{x}} + \left(\frac{1}{4} - x_{1}\right)a \, \mathbf{\hat{y}} + \left(\frac{1}{4} - x_{1}\right)a \, \mathbf{\hat{z}} & \left(32e\right) & \mbox{Te} \\ 
\mathbf{B}_{7} & = & \left(\frac{1}{4} - x_{1}\right) \, \mathbf{a}_{1} + \left(\frac{1}{4} +3x_{1}\right) \, \mathbf{a}_{2} + \left(\frac{1}{4} - x_{1}\right) \, \mathbf{a}_{3} & = & \left(\frac{1}{4} +x_{1}\right)a \, \mathbf{\hat{x}} + \left(\frac{1}{4} - x_{1}\right)a \, \mathbf{\hat{y}} + \left(\frac{1}{4} +x_{1}\right)a \, \mathbf{\hat{z}} & \left(32e\right) & \mbox{Te} \\ 
\mathbf{B}_{8} & = & \left(\frac{1}{4} +3x_{1}\right) \, \mathbf{a}_{1} + \left(\frac{1}{4} - x_{1}\right) \, \mathbf{a}_{2} + \left(\frac{1}{4} - x_{1}\right) \, \mathbf{a}_{3} & = & \left(\frac{1}{4} - x_{1}\right)a \, \mathbf{\hat{x}} + \left(\frac{1}{4} +x_{1}\right)a \, \mathbf{\hat{y}} + \left(\frac{1}{4} +x_{1}\right)a \, \mathbf{\hat{z}} & \left(32e\right) & \mbox{Te} \\ 
\mathbf{B}_{9} & = & \left(-x_{2}+y_{2}+z_{2}\right) \, \mathbf{a}_{1} + \left(x_{2}-y_{2}+z_{2}\right) \, \mathbf{a}_{2} + \left(x_{2}+y_{2}-z_{2}\right) \, \mathbf{a}_{3} & = & x_{2}a \, \mathbf{\hat{x}} + y_{2}a \, \mathbf{\hat{y}} + z_{2}a \, \mathbf{\hat{z}} & \left(96h\right) & \mbox{H I} \\ 
\mathbf{B}_{10} & = & \left(x_{2}-y_{2}+z_{2}\right) \, \mathbf{a}_{1} + \left(-x_{2}+y_{2}+z_{2}\right) \, \mathbf{a}_{2} + \left(-x_{2}-y_{2}-z_{2}\right) \, \mathbf{a}_{3} & = & -x_{2}a \, \mathbf{\hat{x}}-y_{2}a \, \mathbf{\hat{y}} + z_{2}a \, \mathbf{\hat{z}} & \left(96h\right) & \mbox{H I} \\ 
\mathbf{B}_{11} & = & \left(x_{2}+y_{2}-z_{2}\right) \, \mathbf{a}_{1} + \left(-x_{2}-y_{2}-z_{2}\right) \, \mathbf{a}_{2} + \left(-x_{2}+y_{2}+z_{2}\right) \, \mathbf{a}_{3} & = & -x_{2}a \, \mathbf{\hat{x}} + y_{2}a \, \mathbf{\hat{y}}-z_{2}a \, \mathbf{\hat{z}} & \left(96h\right) & \mbox{H I} \\ 
\mathbf{B}_{12} & = & \left(-x_{2}-y_{2}-z_{2}\right) \, \mathbf{a}_{1} + \left(x_{2}+y_{2}-z_{2}\right) \, \mathbf{a}_{2} + \left(x_{2}-y_{2}+z_{2}\right) \, \mathbf{a}_{3} & = & x_{2}a \, \mathbf{\hat{x}}-y_{2}a \, \mathbf{\hat{y}}-z_{2}a \, \mathbf{\hat{z}} & \left(96h\right) & \mbox{H I} \\ 
\mathbf{B}_{13} & = & \left(x_{2}+y_{2}-z_{2}\right) \, \mathbf{a}_{1} + \left(-x_{2}+y_{2}+z_{2}\right) \, \mathbf{a}_{2} + \left(x_{2}-y_{2}+z_{2}\right) \, \mathbf{a}_{3} & = & z_{2}a \, \mathbf{\hat{x}} + x_{2}a \, \mathbf{\hat{y}} + y_{2}a \, \mathbf{\hat{z}} & \left(96h\right) & \mbox{H I} \\ 
\mathbf{B}_{14} & = & \left(-x_{2}-y_{2}-z_{2}\right) \, \mathbf{a}_{1} + \left(x_{2}-y_{2}+z_{2}\right) \, \mathbf{a}_{2} + \left(-x_{2}+y_{2}+z_{2}\right) \, \mathbf{a}_{3} & = & z_{2}a \, \mathbf{\hat{x}}-x_{2}a \, \mathbf{\hat{y}}-y_{2}a \, \mathbf{\hat{z}} & \left(96h\right) & \mbox{H I} \\ 
\mathbf{B}_{15} & = & \left(-x_{2}+y_{2}+z_{2}\right) \, \mathbf{a}_{1} + \left(x_{2}+y_{2}-z_{2}\right) \, \mathbf{a}_{2} + \left(-x_{2}-y_{2}-z_{2}\right) \, \mathbf{a}_{3} & = & -z_{2}a \, \mathbf{\hat{x}}-x_{2}a \, \mathbf{\hat{y}} + y_{2}a \, \mathbf{\hat{z}} & \left(96h\right) & \mbox{H I} \\ 
\mathbf{B}_{16} & = & \left(x_{2}-y_{2}+z_{2}\right) \, \mathbf{a}_{1} + \left(-x_{2}-y_{2}-z_{2}\right) \, \mathbf{a}_{2} + \left(x_{2}+y_{2}-z_{2}\right) \, \mathbf{a}_{3} & = & -z_{2}a \, \mathbf{\hat{x}} + x_{2}a \, \mathbf{\hat{y}}-y_{2}a \, \mathbf{\hat{z}} & \left(96h\right) & \mbox{H I} \\ 
\mathbf{B}_{17} & = & \left(x_{2}-y_{2}+z_{2}\right) \, \mathbf{a}_{1} + \left(x_{2}+y_{2}-z_{2}\right) \, \mathbf{a}_{2} + \left(-x_{2}+y_{2}+z_{2}\right) \, \mathbf{a}_{3} & = & y_{2}a \, \mathbf{\hat{x}} + z_{2}a \, \mathbf{\hat{y}} + x_{2}a \, \mathbf{\hat{z}} & \left(96h\right) & \mbox{H I} \\ 
\mathbf{B}_{18} & = & \left(-x_{2}+y_{2}+z_{2}\right) \, \mathbf{a}_{1} + \left(-x_{2}-y_{2}-z_{2}\right) \, \mathbf{a}_{2} + \left(x_{2}-y_{2}+z_{2}\right) \, \mathbf{a}_{3} & = & -y_{2}a \, \mathbf{\hat{x}} + z_{2}a \, \mathbf{\hat{y}}-x_{2}a \, \mathbf{\hat{z}} & \left(96h\right) & \mbox{H I} \\ 
\mathbf{B}_{19} & = & \left(-x_{2}-y_{2}-z_{2}\right) \, \mathbf{a}_{1} + \left(-x_{2}+y_{2}+z_{2}\right) \, \mathbf{a}_{2} + \left(x_{2}+y_{2}-z_{2}\right) \, \mathbf{a}_{3} & = & y_{2}a \, \mathbf{\hat{x}}-z_{2}a \, \mathbf{\hat{y}}-x_{2}a \, \mathbf{\hat{z}} & \left(96h\right) & \mbox{H I} \\ 
\mathbf{B}_{20} & = & \left(x_{2}+y_{2}-z_{2}\right) \, \mathbf{a}_{1} + \left(x_{2}-y_{2}+z_{2}\right) \, \mathbf{a}_{2} + \left(-x_{2}-y_{2}-z_{2}\right) \, \mathbf{a}_{3} & = & -y_{2}a \, \mathbf{\hat{x}}-z_{2}a \, \mathbf{\hat{y}} + x_{2}a \, \mathbf{\hat{z}} & \left(96h\right) & \mbox{H I} \\ 
\mathbf{B}_{21} & = & \left(\frac{1}{4} +x_{2} - y_{2} - z_{2}\right) \, \mathbf{a}_{1} + \left(\frac{1}{4} - x_{2} + y_{2} - z_{2}\right) \, \mathbf{a}_{2} + \left(\frac{1}{4} +x_{2} + y_{2} + z_{2}\right) \, \mathbf{a}_{3} & = & \left(\frac{1}{4} +y_{2}\right)a \, \mathbf{\hat{x}} + \left(\frac{1}{4} +x_{2}\right)a \, \mathbf{\hat{y}} + \left(\frac{1}{4} - z_{2}\right)a \, \mathbf{\hat{z}} & \left(96h\right) & \mbox{H I} \\ 
\mathbf{B}_{22} & = & \left(\frac{1}{4} - x_{2} + y_{2} - z_{2}\right) \, \mathbf{a}_{1} + \left(\frac{1}{4} +x_{2} - y_{2} - z_{2}\right) \, \mathbf{a}_{2} + \left(\frac{1}{4} - x_{2} - y_{2} + z_{2}\right) \, \mathbf{a}_{3} & = & \left(\frac{1}{4} - y_{2}\right)a \, \mathbf{\hat{x}} + \left(\frac{1}{4} - x_{2}\right)a \, \mathbf{\hat{y}} + \left(\frac{1}{4} - z_{2}\right)a \, \mathbf{\hat{z}} & \left(96h\right) & \mbox{H I} \\ 
\mathbf{B}_{23} & = & \left(\frac{1}{4} - x_{2} - y_{2} + z_{2}\right) \, \mathbf{a}_{1} + \left(\frac{1}{4} +x_{2} + y_{2} + z_{2}\right) \, \mathbf{a}_{2} + \left(\frac{1}{4} - x_{2} + y_{2} - z_{2}\right) \, \mathbf{a}_{3} & = & \left(\frac{1}{4} +y_{2}\right)a \, \mathbf{\hat{x}} + \left(\frac{1}{4} - x_{2}\right)a \, \mathbf{\hat{y}} + \left(\frac{1}{4} +z_{2}\right)a \, \mathbf{\hat{z}} & \left(96h\right) & \mbox{H I} \\ 
\mathbf{B}_{24} & = & \left(\frac{1}{4} +x_{2} + y_{2} + z_{2}\right) \, \mathbf{a}_{1} + \left(\frac{1}{4} - x_{2} - y_{2} + z_{2}\right) \, \mathbf{a}_{2} + \left(\frac{1}{4} +x_{2} - y_{2} - z_{2}\right) \, \mathbf{a}_{3} & = & \left(\frac{1}{4} - y_{2}\right)a \, \mathbf{\hat{x}} + \left(\frac{1}{4} +x_{2}\right)a \, \mathbf{\hat{y}} + \left(\frac{1}{4} +z_{2}\right)a \, \mathbf{\hat{z}} & \left(96h\right) & \mbox{H I} \\ 
\mathbf{B}_{25} & = & \left(\frac{1}{4} - x_{2} - y_{2} + z_{2}\right) \, \mathbf{a}_{1} + \left(\frac{1}{4} +x_{2} - y_{2} - z_{2}\right) \, \mathbf{a}_{2} + \left(\frac{1}{4} +x_{2} + y_{2} + z_{2}\right) \, \mathbf{a}_{3} & = & \left(\frac{1}{4} +x_{2}\right)a \, \mathbf{\hat{x}} + \left(\frac{1}{4} +z_{2}\right)a \, \mathbf{\hat{y}} + \left(\frac{1}{4} - y_{2}\right)a \, \mathbf{\hat{z}} & \left(96h\right) & \mbox{H I} \\ 
\mathbf{B}_{26} & = & \left(\frac{1}{4} +x_{2} + y_{2} + z_{2}\right) \, \mathbf{a}_{1} + \left(\frac{1}{4} - x_{2} + y_{2} - z_{2}\right) \, \mathbf{a}_{2} + \left(\frac{1}{4} - x_{2} - y_{2} + z_{2}\right) \, \mathbf{a}_{3} & = & \left(\frac{1}{4} - x_{2}\right)a \, \mathbf{\hat{x}} + \left(\frac{1}{4} +z_{2}\right)a \, \mathbf{\hat{y}} + \left(\frac{1}{4} +y_{2}\right)a \, \mathbf{\hat{z}} & \left(96h\right) & \mbox{H I} \\ 
\mathbf{B}_{27} & = & \left(\frac{1}{4} +x_{2} - y_{2} - z_{2}\right) \, \mathbf{a}_{1} + \left(\frac{1}{4} - x_{2} - y_{2} + z_{2}\right) \, \mathbf{a}_{2} + \left(\frac{1}{4} - x_{2} + y_{2} - z_{2}\right) \, \mathbf{a}_{3} & = & \left(\frac{1}{4} - x_{2}\right)a \, \mathbf{\hat{x}} + \left(\frac{1}{4} - z_{2}\right)a \, \mathbf{\hat{y}} + \left(\frac{1}{4} - y_{2}\right)a \, \mathbf{\hat{z}} & \left(96h\right) & \mbox{H I} \\ 
\mathbf{B}_{28} & = & \left(\frac{1}{4} - x_{2} + y_{2} - z_{2}\right) \, \mathbf{a}_{1} + \left(\frac{1}{4} +x_{2} + y_{2} + z_{2}\right) \, \mathbf{a}_{2} + \left(\frac{1}{4} +x_{2} - y_{2} - z_{2}\right) \, \mathbf{a}_{3} & = & \left(\frac{1}{4} +x_{2}\right)a \, \mathbf{\hat{x}} + \left(\frac{1}{4} - z_{2}\right)a \, \mathbf{\hat{y}} + \left(\frac{1}{4} +y_{2}\right)a \, \mathbf{\hat{z}} & \left(96h\right) & \mbox{H I} \\ 
\mathbf{B}_{29} & = & \left(\frac{1}{4} - x_{2} + y_{2} - z_{2}\right) \, \mathbf{a}_{1} + \left(\frac{1}{4} - x_{2} - y_{2} + z_{2}\right) \, \mathbf{a}_{2} + \left(\frac{1}{4} +x_{2} + y_{2} + z_{2}\right) \, \mathbf{a}_{3} & = & \left(\frac{1}{4} +z_{2}\right)a \, \mathbf{\hat{x}} + \left(\frac{1}{4} +y_{2}\right)a \, \mathbf{\hat{y}} + \left(\frac{1}{4} - x_{2}\right)a \, \mathbf{\hat{z}} & \left(96h\right) & \mbox{H I} \\ 
\mathbf{B}_{30} & = & \left(\frac{1}{4} +x_{2} - y_{2} - z_{2}\right) \, \mathbf{a}_{1} + \left(\frac{1}{4} +x_{2} + y_{2} + z_{2}\right) \, \mathbf{a}_{2} + \left(\frac{1}{4} - x_{2} - y_{2} + z_{2}\right) \, \mathbf{a}_{3} & = & \left(\frac{1}{4} +z_{2}\right)a \, \mathbf{\hat{x}} + \left(\frac{1}{4} - y_{2}\right)a \, \mathbf{\hat{y}} + \left(\frac{1}{4} +x_{2}\right)a \, \mathbf{\hat{z}} & \left(96h\right) & \mbox{H I} \\ 
\mathbf{B}_{31} & = & \left(\frac{1}{4} +x_{2} + y_{2} + z_{2}\right) \, \mathbf{a}_{1} + \left(\frac{1}{4} +x_{2} - y_{2} - z_{2}\right) \, \mathbf{a}_{2} + \left(\frac{1}{4} - x_{2} + y_{2} - z_{2}\right) \, \mathbf{a}_{3} & = & \left(\frac{1}{4} - z_{2}\right)a \, \mathbf{\hat{x}} + \left(\frac{1}{4} +y_{2}\right)a \, \mathbf{\hat{y}} + \left(\frac{1}{4} +x_{2}\right)a \, \mathbf{\hat{z}} & \left(96h\right) & \mbox{H I} \\ 
\mathbf{B}_{32} & = & \left(\frac{1}{4} - x_{2} - y_{2} + z_{2}\right) \, \mathbf{a}_{1} + \left(\frac{1}{4} - x_{2} + y_{2} - z_{2}\right) \, \mathbf{a}_{2} + \left(\frac{1}{4} +x_{2} - y_{2} - z_{2}\right) \, \mathbf{a}_{3} & = & \left(\frac{1}{4} - z_{2}\right)a \, \mathbf{\hat{x}} + \left(\frac{1}{4} - y_{2}\right)a \, \mathbf{\hat{y}} + \left(\frac{1}{4} - x_{2}\right)a \, \mathbf{\hat{z}} & \left(96h\right) & \mbox{H I} \\ 
\mathbf{B}_{33} & = & \left(-x_{3}+y_{3}+z_{3}\right) \, \mathbf{a}_{1} + \left(x_{3}-y_{3}+z_{3}\right) \, \mathbf{a}_{2} + \left(x_{3}+y_{3}-z_{3}\right) \, \mathbf{a}_{3} & = & x_{3}a \, \mathbf{\hat{x}} + y_{3}a \, \mathbf{\hat{y}} + z_{3}a \, \mathbf{\hat{z}} & \left(96h\right) & \mbox{H II} \\ 
\mathbf{B}_{34} & = & \left(x_{3}-y_{3}+z_{3}\right) \, \mathbf{a}_{1} + \left(-x_{3}+y_{3}+z_{3}\right) \, \mathbf{a}_{2} + \left(-x_{3}-y_{3}-z_{3}\right) \, \mathbf{a}_{3} & = & -x_{3}a \, \mathbf{\hat{x}}-y_{3}a \, \mathbf{\hat{y}} + z_{3}a \, \mathbf{\hat{z}} & \left(96h\right) & \mbox{H II} \\ 
\mathbf{B}_{35} & = & \left(x_{3}+y_{3}-z_{3}\right) \, \mathbf{a}_{1} + \left(-x_{3}-y_{3}-z_{3}\right) \, \mathbf{a}_{2} + \left(-x_{3}+y_{3}+z_{3}\right) \, \mathbf{a}_{3} & = & -x_{3}a \, \mathbf{\hat{x}} + y_{3}a \, \mathbf{\hat{y}}-z_{3}a \, \mathbf{\hat{z}} & \left(96h\right) & \mbox{H II} \\ 
\mathbf{B}_{36} & = & \left(-x_{3}-y_{3}-z_{3}\right) \, \mathbf{a}_{1} + \left(x_{3}+y_{3}-z_{3}\right) \, \mathbf{a}_{2} + \left(x_{3}-y_{3}+z_{3}\right) \, \mathbf{a}_{3} & = & x_{3}a \, \mathbf{\hat{x}}-y_{3}a \, \mathbf{\hat{y}}-z_{3}a \, \mathbf{\hat{z}} & \left(96h\right) & \mbox{H II} \\ 
\mathbf{B}_{37} & = & \left(x_{3}+y_{3}-z_{3}\right) \, \mathbf{a}_{1} + \left(-x_{3}+y_{3}+z_{3}\right) \, \mathbf{a}_{2} + \left(x_{3}-y_{3}+z_{3}\right) \, \mathbf{a}_{3} & = & z_{3}a \, \mathbf{\hat{x}} + x_{3}a \, \mathbf{\hat{y}} + y_{3}a \, \mathbf{\hat{z}} & \left(96h\right) & \mbox{H II} \\ 
\mathbf{B}_{38} & = & \left(-x_{3}-y_{3}-z_{3}\right) \, \mathbf{a}_{1} + \left(x_{3}-y_{3}+z_{3}\right) \, \mathbf{a}_{2} + \left(-x_{3}+y_{3}+z_{3}\right) \, \mathbf{a}_{3} & = & z_{3}a \, \mathbf{\hat{x}}-x_{3}a \, \mathbf{\hat{y}}-y_{3}a \, \mathbf{\hat{z}} & \left(96h\right) & \mbox{H II} \\ 
\mathbf{B}_{39} & = & \left(-x_{3}+y_{3}+z_{3}\right) \, \mathbf{a}_{1} + \left(x_{3}+y_{3}-z_{3}\right) \, \mathbf{a}_{2} + \left(-x_{3}-y_{3}-z_{3}\right) \, \mathbf{a}_{3} & = & -z_{3}a \, \mathbf{\hat{x}}-x_{3}a \, \mathbf{\hat{y}} + y_{3}a \, \mathbf{\hat{z}} & \left(96h\right) & \mbox{H II} \\ 
\mathbf{B}_{40} & = & \left(x_{3}-y_{3}+z_{3}\right) \, \mathbf{a}_{1} + \left(-x_{3}-y_{3}-z_{3}\right) \, \mathbf{a}_{2} + \left(x_{3}+y_{3}-z_{3}\right) \, \mathbf{a}_{3} & = & -z_{3}a \, \mathbf{\hat{x}} + x_{3}a \, \mathbf{\hat{y}}-y_{3}a \, \mathbf{\hat{z}} & \left(96h\right) & \mbox{H II} \\ 
\mathbf{B}_{41} & = & \left(x_{3}-y_{3}+z_{3}\right) \, \mathbf{a}_{1} + \left(x_{3}+y_{3}-z_{3}\right) \, \mathbf{a}_{2} + \left(-x_{3}+y_{3}+z_{3}\right) \, \mathbf{a}_{3} & = & y_{3}a \, \mathbf{\hat{x}} + z_{3}a \, \mathbf{\hat{y}} + x_{3}a \, \mathbf{\hat{z}} & \left(96h\right) & \mbox{H II} \\ 
\mathbf{B}_{42} & = & \left(-x_{3}+y_{3}+z_{3}\right) \, \mathbf{a}_{1} + \left(-x_{3}-y_{3}-z_{3}\right) \, \mathbf{a}_{2} + \left(x_{3}-y_{3}+z_{3}\right) \, \mathbf{a}_{3} & = & -y_{3}a \, \mathbf{\hat{x}} + z_{3}a \, \mathbf{\hat{y}}-x_{3}a \, \mathbf{\hat{z}} & \left(96h\right) & \mbox{H II} \\ 
\mathbf{B}_{43} & = & \left(-x_{3}-y_{3}-z_{3}\right) \, \mathbf{a}_{1} + \left(-x_{3}+y_{3}+z_{3}\right) \, \mathbf{a}_{2} + \left(x_{3}+y_{3}-z_{3}\right) \, \mathbf{a}_{3} & = & y_{3}a \, \mathbf{\hat{x}}-z_{3}a \, \mathbf{\hat{y}}-x_{3}a \, \mathbf{\hat{z}} & \left(96h\right) & \mbox{H II} \\ 
\mathbf{B}_{44} & = & \left(x_{3}+y_{3}-z_{3}\right) \, \mathbf{a}_{1} + \left(x_{3}-y_{3}+z_{3}\right) \, \mathbf{a}_{2} + \left(-x_{3}-y_{3}-z_{3}\right) \, \mathbf{a}_{3} & = & -y_{3}a \, \mathbf{\hat{x}}-z_{3}a \, \mathbf{\hat{y}} + x_{3}a \, \mathbf{\hat{z}} & \left(96h\right) & \mbox{H II} \\ 
\mathbf{B}_{45} & = & \left(\frac{1}{4} +x_{3} - y_{3} - z_{3}\right) \, \mathbf{a}_{1} + \left(\frac{1}{4} - x_{3} + y_{3} - z_{3}\right) \, \mathbf{a}_{2} + \left(\frac{1}{4} +x_{3} + y_{3} + z_{3}\right) \, \mathbf{a}_{3} & = & \left(\frac{1}{4} +y_{3}\right)a \, \mathbf{\hat{x}} + \left(\frac{1}{4} +x_{3}\right)a \, \mathbf{\hat{y}} + \left(\frac{1}{4} - z_{3}\right)a \, \mathbf{\hat{z}} & \left(96h\right) & \mbox{H II} \\ 
\mathbf{B}_{46} & = & \left(\frac{1}{4} - x_{3} + y_{3} - z_{3}\right) \, \mathbf{a}_{1} + \left(\frac{1}{4} +x_{3} - y_{3} - z_{3}\right) \, \mathbf{a}_{2} + \left(\frac{1}{4} - x_{3} - y_{3} + z_{3}\right) \, \mathbf{a}_{3} & = & \left(\frac{1}{4} - y_{3}\right)a \, \mathbf{\hat{x}} + \left(\frac{1}{4} - x_{3}\right)a \, \mathbf{\hat{y}} + \left(\frac{1}{4} - z_{3}\right)a \, \mathbf{\hat{z}} & \left(96h\right) & \mbox{H II} \\ 
\mathbf{B}_{47} & = & \left(\frac{1}{4} - x_{3} - y_{3} + z_{3}\right) \, \mathbf{a}_{1} + \left(\frac{1}{4} +x_{3} + y_{3} + z_{3}\right) \, \mathbf{a}_{2} + \left(\frac{1}{4} - x_{3} + y_{3} - z_{3}\right) \, \mathbf{a}_{3} & = & \left(\frac{1}{4} +y_{3}\right)a \, \mathbf{\hat{x}} + \left(\frac{1}{4} - x_{3}\right)a \, \mathbf{\hat{y}} + \left(\frac{1}{4} +z_{3}\right)a \, \mathbf{\hat{z}} & \left(96h\right) & \mbox{H II} \\ 
\mathbf{B}_{48} & = & \left(\frac{1}{4} +x_{3} + y_{3} + z_{3}\right) \, \mathbf{a}_{1} + \left(\frac{1}{4} - x_{3} - y_{3} + z_{3}\right) \, \mathbf{a}_{2} + \left(\frac{1}{4} +x_{3} - y_{3} - z_{3}\right) \, \mathbf{a}_{3} & = & \left(\frac{1}{4} - y_{3}\right)a \, \mathbf{\hat{x}} + \left(\frac{1}{4} +x_{3}\right)a \, \mathbf{\hat{y}} + \left(\frac{1}{4} +z_{3}\right)a \, \mathbf{\hat{z}} & \left(96h\right) & \mbox{H II} \\ 
\mathbf{B}_{49} & = & \left(\frac{1}{4} - x_{3} - y_{3} + z_{3}\right) \, \mathbf{a}_{1} + \left(\frac{1}{4} +x_{3} - y_{3} - z_{3}\right) \, \mathbf{a}_{2} + \left(\frac{1}{4} +x_{3} + y_{3} + z_{3}\right) \, \mathbf{a}_{3} & = & \left(\frac{1}{4} +x_{3}\right)a \, \mathbf{\hat{x}} + \left(\frac{1}{4} +z_{3}\right)a \, \mathbf{\hat{y}} + \left(\frac{1}{4} - y_{3}\right)a \, \mathbf{\hat{z}} & \left(96h\right) & \mbox{H II} \\ 
\mathbf{B}_{50} & = & \left(\frac{1}{4} +x_{3} + y_{3} + z_{3}\right) \, \mathbf{a}_{1} + \left(\frac{1}{4} - x_{3} + y_{3} - z_{3}\right) \, \mathbf{a}_{2} + \left(\frac{1}{4} - x_{3} - y_{3} + z_{3}\right) \, \mathbf{a}_{3} & = & \left(\frac{1}{4} - x_{3}\right)a \, \mathbf{\hat{x}} + \left(\frac{1}{4} +z_{3}\right)a \, \mathbf{\hat{y}} + \left(\frac{1}{4} +y_{3}\right)a \, \mathbf{\hat{z}} & \left(96h\right) & \mbox{H II} \\ 
\mathbf{B}_{51} & = & \left(\frac{1}{4} +x_{3} - y_{3} - z_{3}\right) \, \mathbf{a}_{1} + \left(\frac{1}{4} - x_{3} - y_{3} + z_{3}\right) \, \mathbf{a}_{2} + \left(\frac{1}{4} - x_{3} + y_{3} - z_{3}\right) \, \mathbf{a}_{3} & = & \left(\frac{1}{4} - x_{3}\right)a \, \mathbf{\hat{x}} + \left(\frac{1}{4} - z_{3}\right)a \, \mathbf{\hat{y}} + \left(\frac{1}{4} - y_{3}\right)a \, \mathbf{\hat{z}} & \left(96h\right) & \mbox{H II} \\ 
\mathbf{B}_{52} & = & \left(\frac{1}{4} - x_{3} + y_{3} - z_{3}\right) \, \mathbf{a}_{1} + \left(\frac{1}{4} +x_{3} + y_{3} + z_{3}\right) \, \mathbf{a}_{2} + \left(\frac{1}{4} +x_{3} - y_{3} - z_{3}\right) \, \mathbf{a}_{3} & = & \left(\frac{1}{4} +x_{3}\right)a \, \mathbf{\hat{x}} + \left(\frac{1}{4} - z_{3}\right)a \, \mathbf{\hat{y}} + \left(\frac{1}{4} +y_{3}\right)a \, \mathbf{\hat{z}} & \left(96h\right) & \mbox{H II} \\ 
\mathbf{B}_{53} & = & \left(\frac{1}{4} - x_{3} + y_{3} - z_{3}\right) \, \mathbf{a}_{1} + \left(\frac{1}{4} - x_{3} - y_{3} + z_{3}\right) \, \mathbf{a}_{2} + \left(\frac{1}{4} +x_{3} + y_{3} + z_{3}\right) \, \mathbf{a}_{3} & = & \left(\frac{1}{4} +z_{3}\right)a \, \mathbf{\hat{x}} + \left(\frac{1}{4} +y_{3}\right)a \, \mathbf{\hat{y}} + \left(\frac{1}{4} - x_{3}\right)a \, \mathbf{\hat{z}} & \left(96h\right) & \mbox{H II} \\ 
\mathbf{B}_{54} & = & \left(\frac{1}{4} +x_{3} - y_{3} - z_{3}\right) \, \mathbf{a}_{1} + \left(\frac{1}{4} +x_{3} + y_{3} + z_{3}\right) \, \mathbf{a}_{2} + \left(\frac{1}{4} - x_{3} - y_{3} + z_{3}\right) \, \mathbf{a}_{3} & = & \left(\frac{1}{4} +z_{3}\right)a \, \mathbf{\hat{x}} + \left(\frac{1}{4} - y_{3}\right)a \, \mathbf{\hat{y}} + \left(\frac{1}{4} +x_{3}\right)a \, \mathbf{\hat{z}} & \left(96h\right) & \mbox{H II} \\ 
\mathbf{B}_{55} & = & \left(\frac{1}{4} +x_{3} + y_{3} + z_{3}\right) \, \mathbf{a}_{1} + \left(\frac{1}{4} +x_{3} - y_{3} - z_{3}\right) \, \mathbf{a}_{2} + \left(\frac{1}{4} - x_{3} + y_{3} - z_{3}\right) \, \mathbf{a}_{3} & = & \left(\frac{1}{4} - z_{3}\right)a \, \mathbf{\hat{x}} + \left(\frac{1}{4} +y_{3}\right)a \, \mathbf{\hat{y}} + \left(\frac{1}{4} +x_{3}\right)a \, \mathbf{\hat{z}} & \left(96h\right) & \mbox{H II} \\ 
\mathbf{B}_{56} & = & \left(\frac{1}{4} - x_{3} - y_{3} + z_{3}\right) \, \mathbf{a}_{1} + \left(\frac{1}{4} - x_{3} + y_{3} - z_{3}\right) \, \mathbf{a}_{2} + \left(\frac{1}{4} +x_{3} - y_{3} - z_{3}\right) \, \mathbf{a}_{3} & = & \left(\frac{1}{4} - z_{3}\right)a \, \mathbf{\hat{x}} + \left(\frac{1}{4} - y_{3}\right)a \, \mathbf{\hat{y}} + \left(\frac{1}{4} - x_{3}\right)a \, \mathbf{\hat{z}} & \left(96h\right) & \mbox{H II} \\ 
\mathbf{B}_{57} & = & \left(-x_{4}+y_{4}+z_{4}\right) \, \mathbf{a}_{1} + \left(x_{4}-y_{4}+z_{4}\right) \, \mathbf{a}_{2} + \left(x_{4}+y_{4}-z_{4}\right) \, \mathbf{a}_{3} & = & x_{4}a \, \mathbf{\hat{x}} + y_{4}a \, \mathbf{\hat{y}} + z_{4}a \, \mathbf{\hat{z}} & \left(96h\right) & \mbox{H III} \\ 
\mathbf{B}_{58} & = & \left(x_{4}-y_{4}+z_{4}\right) \, \mathbf{a}_{1} + \left(-x_{4}+y_{4}+z_{4}\right) \, \mathbf{a}_{2} + \left(-x_{4}-y_{4}-z_{4}\right) \, \mathbf{a}_{3} & = & -x_{4}a \, \mathbf{\hat{x}}-y_{4}a \, \mathbf{\hat{y}} + z_{4}a \, \mathbf{\hat{z}} & \left(96h\right) & \mbox{H III} \\ 
\mathbf{B}_{59} & = & \left(x_{4}+y_{4}-z_{4}\right) \, \mathbf{a}_{1} + \left(-x_{4}-y_{4}-z_{4}\right) \, \mathbf{a}_{2} + \left(-x_{4}+y_{4}+z_{4}\right) \, \mathbf{a}_{3} & = & -x_{4}a \, \mathbf{\hat{x}} + y_{4}a \, \mathbf{\hat{y}}-z_{4}a \, \mathbf{\hat{z}} & \left(96h\right) & \mbox{H III} \\ 
\mathbf{B}_{60} & = & \left(-x_{4}-y_{4}-z_{4}\right) \, \mathbf{a}_{1} + \left(x_{4}+y_{4}-z_{4}\right) \, \mathbf{a}_{2} + \left(x_{4}-y_{4}+z_{4}\right) \, \mathbf{a}_{3} & = & x_{4}a \, \mathbf{\hat{x}}-y_{4}a \, \mathbf{\hat{y}}-z_{4}a \, \mathbf{\hat{z}} & \left(96h\right) & \mbox{H III} \\ 
\mathbf{B}_{61} & = & \left(x_{4}+y_{4}-z_{4}\right) \, \mathbf{a}_{1} + \left(-x_{4}+y_{4}+z_{4}\right) \, \mathbf{a}_{2} + \left(x_{4}-y_{4}+z_{4}\right) \, \mathbf{a}_{3} & = & z_{4}a \, \mathbf{\hat{x}} + x_{4}a \, \mathbf{\hat{y}} + y_{4}a \, \mathbf{\hat{z}} & \left(96h\right) & \mbox{H III} \\ 
\mathbf{B}_{62} & = & \left(-x_{4}-y_{4}-z_{4}\right) \, \mathbf{a}_{1} + \left(x_{4}-y_{4}+z_{4}\right) \, \mathbf{a}_{2} + \left(-x_{4}+y_{4}+z_{4}\right) \, \mathbf{a}_{3} & = & z_{4}a \, \mathbf{\hat{x}}-x_{4}a \, \mathbf{\hat{y}}-y_{4}a \, \mathbf{\hat{z}} & \left(96h\right) & \mbox{H III} \\ 
\mathbf{B}_{63} & = & \left(-x_{4}+y_{4}+z_{4}\right) \, \mathbf{a}_{1} + \left(x_{4}+y_{4}-z_{4}\right) \, \mathbf{a}_{2} + \left(-x_{4}-y_{4}-z_{4}\right) \, \mathbf{a}_{3} & = & -z_{4}a \, \mathbf{\hat{x}}-x_{4}a \, \mathbf{\hat{y}} + y_{4}a \, \mathbf{\hat{z}} & \left(96h\right) & \mbox{H III} \\ 
\mathbf{B}_{64} & = & \left(x_{4}-y_{4}+z_{4}\right) \, \mathbf{a}_{1} + \left(-x_{4}-y_{4}-z_{4}\right) \, \mathbf{a}_{2} + \left(x_{4}+y_{4}-z_{4}\right) \, \mathbf{a}_{3} & = & -z_{4}a \, \mathbf{\hat{x}} + x_{4}a \, \mathbf{\hat{y}}-y_{4}a \, \mathbf{\hat{z}} & \left(96h\right) & \mbox{H III} \\ 
\mathbf{B}_{65} & = & \left(x_{4}-y_{4}+z_{4}\right) \, \mathbf{a}_{1} + \left(x_{4}+y_{4}-z_{4}\right) \, \mathbf{a}_{2} + \left(-x_{4}+y_{4}+z_{4}\right) \, \mathbf{a}_{3} & = & y_{4}a \, \mathbf{\hat{x}} + z_{4}a \, \mathbf{\hat{y}} + x_{4}a \, \mathbf{\hat{z}} & \left(96h\right) & \mbox{H III} \\ 
\mathbf{B}_{66} & = & \left(-x_{4}+y_{4}+z_{4}\right) \, \mathbf{a}_{1} + \left(-x_{4}-y_{4}-z_{4}\right) \, \mathbf{a}_{2} + \left(x_{4}-y_{4}+z_{4}\right) \, \mathbf{a}_{3} & = & -y_{4}a \, \mathbf{\hat{x}} + z_{4}a \, \mathbf{\hat{y}}-x_{4}a \, \mathbf{\hat{z}} & \left(96h\right) & \mbox{H III} \\ 
\mathbf{B}_{67} & = & \left(-x_{4}-y_{4}-z_{4}\right) \, \mathbf{a}_{1} + \left(-x_{4}+y_{4}+z_{4}\right) \, \mathbf{a}_{2} + \left(x_{4}+y_{4}-z_{4}\right) \, \mathbf{a}_{3} & = & y_{4}a \, \mathbf{\hat{x}}-z_{4}a \, \mathbf{\hat{y}}-x_{4}a \, \mathbf{\hat{z}} & \left(96h\right) & \mbox{H III} \\ 
\mathbf{B}_{68} & = & \left(x_{4}+y_{4}-z_{4}\right) \, \mathbf{a}_{1} + \left(x_{4}-y_{4}+z_{4}\right) \, \mathbf{a}_{2} + \left(-x_{4}-y_{4}-z_{4}\right) \, \mathbf{a}_{3} & = & -y_{4}a \, \mathbf{\hat{x}}-z_{4}a \, \mathbf{\hat{y}} + x_{4}a \, \mathbf{\hat{z}} & \left(96h\right) & \mbox{H III} \\ 
\mathbf{B}_{69} & = & \left(\frac{1}{4} +x_{4} - y_{4} - z_{4}\right) \, \mathbf{a}_{1} + \left(\frac{1}{4} - x_{4} + y_{4} - z_{4}\right) \, \mathbf{a}_{2} + \left(\frac{1}{4} +x_{4} + y_{4} + z_{4}\right) \, \mathbf{a}_{3} & = & \left(\frac{1}{4} +y_{4}\right)a \, \mathbf{\hat{x}} + \left(\frac{1}{4} +x_{4}\right)a \, \mathbf{\hat{y}} + \left(\frac{1}{4} - z_{4}\right)a \, \mathbf{\hat{z}} & \left(96h\right) & \mbox{H III} \\ 
\mathbf{B}_{70} & = & \left(\frac{1}{4} - x_{4} + y_{4} - z_{4}\right) \, \mathbf{a}_{1} + \left(\frac{1}{4} +x_{4} - y_{4} - z_{4}\right) \, \mathbf{a}_{2} + \left(\frac{1}{4} - x_{4} - y_{4} + z_{4}\right) \, \mathbf{a}_{3} & = & \left(\frac{1}{4} - y_{4}\right)a \, \mathbf{\hat{x}} + \left(\frac{1}{4} - x_{4}\right)a \, \mathbf{\hat{y}} + \left(\frac{1}{4} - z_{4}\right)a \, \mathbf{\hat{z}} & \left(96h\right) & \mbox{H III} \\ 
\mathbf{B}_{71} & = & \left(\frac{1}{4} - x_{4} - y_{4} + z_{4}\right) \, \mathbf{a}_{1} + \left(\frac{1}{4} +x_{4} + y_{4} + z_{4}\right) \, \mathbf{a}_{2} + \left(\frac{1}{4} - x_{4} + y_{4} - z_{4}\right) \, \mathbf{a}_{3} & = & \left(\frac{1}{4} +y_{4}\right)a \, \mathbf{\hat{x}} + \left(\frac{1}{4} - x_{4}\right)a \, \mathbf{\hat{y}} + \left(\frac{1}{4} +z_{4}\right)a \, \mathbf{\hat{z}} & \left(96h\right) & \mbox{H III} \\ 
\mathbf{B}_{72} & = & \left(\frac{1}{4} +x_{4} + y_{4} + z_{4}\right) \, \mathbf{a}_{1} + \left(\frac{1}{4} - x_{4} - y_{4} + z_{4}\right) \, \mathbf{a}_{2} + \left(\frac{1}{4} +x_{4} - y_{4} - z_{4}\right) \, \mathbf{a}_{3} & = & \left(\frac{1}{4} - y_{4}\right)a \, \mathbf{\hat{x}} + \left(\frac{1}{4} +x_{4}\right)a \, \mathbf{\hat{y}} + \left(\frac{1}{4} +z_{4}\right)a \, \mathbf{\hat{z}} & \left(96h\right) & \mbox{H III} \\ 
\mathbf{B}_{73} & = & \left(\frac{1}{4} - x_{4} - y_{4} + z_{4}\right) \, \mathbf{a}_{1} + \left(\frac{1}{4} +x_{4} - y_{4} - z_{4}\right) \, \mathbf{a}_{2} + \left(\frac{1}{4} +x_{4} + y_{4} + z_{4}\right) \, \mathbf{a}_{3} & = & \left(\frac{1}{4} +x_{4}\right)a \, \mathbf{\hat{x}} + \left(\frac{1}{4} +z_{4}\right)a \, \mathbf{\hat{y}} + \left(\frac{1}{4} - y_{4}\right)a \, \mathbf{\hat{z}} & \left(96h\right) & \mbox{H III} \\ 
\mathbf{B}_{74} & = & \left(\frac{1}{4} +x_{4} + y_{4} + z_{4}\right) \, \mathbf{a}_{1} + \left(\frac{1}{4} - x_{4} + y_{4} - z_{4}\right) \, \mathbf{a}_{2} + \left(\frac{1}{4} - x_{4} - y_{4} + z_{4}\right) \, \mathbf{a}_{3} & = & \left(\frac{1}{4} - x_{4}\right)a \, \mathbf{\hat{x}} + \left(\frac{1}{4} +z_{4}\right)a \, \mathbf{\hat{y}} + \left(\frac{1}{4} +y_{4}\right)a \, \mathbf{\hat{z}} & \left(96h\right) & \mbox{H III} \\ 
\mathbf{B}_{75} & = & \left(\frac{1}{4} +x_{4} - y_{4} - z_{4}\right) \, \mathbf{a}_{1} + \left(\frac{1}{4} - x_{4} - y_{4} + z_{4}\right) \, \mathbf{a}_{2} + \left(\frac{1}{4} - x_{4} + y_{4} - z_{4}\right) \, \mathbf{a}_{3} & = & \left(\frac{1}{4} - x_{4}\right)a \, \mathbf{\hat{x}} + \left(\frac{1}{4} - z_{4}\right)a \, \mathbf{\hat{y}} + \left(\frac{1}{4} - y_{4}\right)a \, \mathbf{\hat{z}} & \left(96h\right) & \mbox{H III} \\ 
\mathbf{B}_{76} & = & \left(\frac{1}{4} - x_{4} + y_{4} - z_{4}\right) \, \mathbf{a}_{1} + \left(\frac{1}{4} +x_{4} + y_{4} + z_{4}\right) \, \mathbf{a}_{2} + \left(\frac{1}{4} +x_{4} - y_{4} - z_{4}\right) \, \mathbf{a}_{3} & = & \left(\frac{1}{4} +x_{4}\right)a \, \mathbf{\hat{x}} + \left(\frac{1}{4} - z_{4}\right)a \, \mathbf{\hat{y}} + \left(\frac{1}{4} +y_{4}\right)a \, \mathbf{\hat{z}} & \left(96h\right) & \mbox{H III} \\ 
\mathbf{B}_{77} & = & \left(\frac{1}{4} - x_{4} + y_{4} - z_{4}\right) \, \mathbf{a}_{1} + \left(\frac{1}{4} - x_{4} - y_{4} + z_{4}\right) \, \mathbf{a}_{2} + \left(\frac{1}{4} +x_{4} + y_{4} + z_{4}\right) \, \mathbf{a}_{3} & = & \left(\frac{1}{4} +z_{4}\right)a \, \mathbf{\hat{x}} + \left(\frac{1}{4} +y_{4}\right)a \, \mathbf{\hat{y}} + \left(\frac{1}{4} - x_{4}\right)a \, \mathbf{\hat{z}} & \left(96h\right) & \mbox{H III} \\ 
\mathbf{B}_{78} & = & \left(\frac{1}{4} +x_{4} - y_{4} - z_{4}\right) \, \mathbf{a}_{1} + \left(\frac{1}{4} +x_{4} + y_{4} + z_{4}\right) \, \mathbf{a}_{2} + \left(\frac{1}{4} - x_{4} - y_{4} + z_{4}\right) \, \mathbf{a}_{3} & = & \left(\frac{1}{4} +z_{4}\right)a \, \mathbf{\hat{x}} + \left(\frac{1}{4} - y_{4}\right)a \, \mathbf{\hat{y}} + \left(\frac{1}{4} +x_{4}\right)a \, \mathbf{\hat{z}} & \left(96h\right) & \mbox{H III} \\ 
\mathbf{B}_{79} & = & \left(\frac{1}{4} +x_{4} + y_{4} + z_{4}\right) \, \mathbf{a}_{1} + \left(\frac{1}{4} +x_{4} - y_{4} - z_{4}\right) \, \mathbf{a}_{2} + \left(\frac{1}{4} - x_{4} + y_{4} - z_{4}\right) \, \mathbf{a}_{3} & = & \left(\frac{1}{4} - z_{4}\right)a \, \mathbf{\hat{x}} + \left(\frac{1}{4} +y_{4}\right)a \, \mathbf{\hat{y}} + \left(\frac{1}{4} +x_{4}\right)a \, \mathbf{\hat{z}} & \left(96h\right) & \mbox{H III} \\ 
\mathbf{B}_{80} & = & \left(\frac{1}{4} - x_{4} - y_{4} + z_{4}\right) \, \mathbf{a}_{1} + \left(\frac{1}{4} - x_{4} + y_{4} - z_{4}\right) \, \mathbf{a}_{2} + \left(\frac{1}{4} +x_{4} - y_{4} - z_{4}\right) \, \mathbf{a}_{3} & = & \left(\frac{1}{4} - z_{4}\right)a \, \mathbf{\hat{x}} + \left(\frac{1}{4} - y_{4}\right)a \, \mathbf{\hat{y}} + \left(\frac{1}{4} - x_{4}\right)a \, \mathbf{\hat{z}} & \left(96h\right) & \mbox{H III} \\ 
\mathbf{B}_{81} & = & \left(-x_{5}+y_{5}+z_{5}\right) \, \mathbf{a}_{1} + \left(x_{5}-y_{5}+z_{5}\right) \, \mathbf{a}_{2} + \left(x_{5}+y_{5}-z_{5}\right) \, \mathbf{a}_{3} & = & x_{5}a \, \mathbf{\hat{x}} + y_{5}a \, \mathbf{\hat{y}} + z_{5}a \, \mathbf{\hat{z}} & \left(96h\right) & \mbox{H IV} \\ 
\mathbf{B}_{82} & = & \left(x_{5}-y_{5}+z_{5}\right) \, \mathbf{a}_{1} + \left(-x_{5}+y_{5}+z_{5}\right) \, \mathbf{a}_{2} + \left(-x_{5}-y_{5}-z_{5}\right) \, \mathbf{a}_{3} & = & -x_{5}a \, \mathbf{\hat{x}}-y_{5}a \, \mathbf{\hat{y}} + z_{5}a \, \mathbf{\hat{z}} & \left(96h\right) & \mbox{H IV} \\ 
\mathbf{B}_{83} & = & \left(x_{5}+y_{5}-z_{5}\right) \, \mathbf{a}_{1} + \left(-x_{5}-y_{5}-z_{5}\right) \, \mathbf{a}_{2} + \left(-x_{5}+y_{5}+z_{5}\right) \, \mathbf{a}_{3} & = & -x_{5}a \, \mathbf{\hat{x}} + y_{5}a \, \mathbf{\hat{y}}-z_{5}a \, \mathbf{\hat{z}} & \left(96h\right) & \mbox{H IV} \\ 
\mathbf{B}_{84} & = & \left(-x_{5}-y_{5}-z_{5}\right) \, \mathbf{a}_{1} + \left(x_{5}+y_{5}-z_{5}\right) \, \mathbf{a}_{2} + \left(x_{5}-y_{5}+z_{5}\right) \, \mathbf{a}_{3} & = & x_{5}a \, \mathbf{\hat{x}}-y_{5}a \, \mathbf{\hat{y}}-z_{5}a \, \mathbf{\hat{z}} & \left(96h\right) & \mbox{H IV} \\ 
\mathbf{B}_{85} & = & \left(x_{5}+y_{5}-z_{5}\right) \, \mathbf{a}_{1} + \left(-x_{5}+y_{5}+z_{5}\right) \, \mathbf{a}_{2} + \left(x_{5}-y_{5}+z_{5}\right) \, \mathbf{a}_{3} & = & z_{5}a \, \mathbf{\hat{x}} + x_{5}a \, \mathbf{\hat{y}} + y_{5}a \, \mathbf{\hat{z}} & \left(96h\right) & \mbox{H IV} \\ 
\mathbf{B}_{86} & = & \left(-x_{5}-y_{5}-z_{5}\right) \, \mathbf{a}_{1} + \left(x_{5}-y_{5}+z_{5}\right) \, \mathbf{a}_{2} + \left(-x_{5}+y_{5}+z_{5}\right) \, \mathbf{a}_{3} & = & z_{5}a \, \mathbf{\hat{x}}-x_{5}a \, \mathbf{\hat{y}}-y_{5}a \, \mathbf{\hat{z}} & \left(96h\right) & \mbox{H IV} \\ 
\mathbf{B}_{87} & = & \left(-x_{5}+y_{5}+z_{5}\right) \, \mathbf{a}_{1} + \left(x_{5}+y_{5}-z_{5}\right) \, \mathbf{a}_{2} + \left(-x_{5}-y_{5}-z_{5}\right) \, \mathbf{a}_{3} & = & -z_{5}a \, \mathbf{\hat{x}}-x_{5}a \, \mathbf{\hat{y}} + y_{5}a \, \mathbf{\hat{z}} & \left(96h\right) & \mbox{H IV} \\ 
\mathbf{B}_{88} & = & \left(x_{5}-y_{5}+z_{5}\right) \, \mathbf{a}_{1} + \left(-x_{5}-y_{5}-z_{5}\right) \, \mathbf{a}_{2} + \left(x_{5}+y_{5}-z_{5}\right) \, \mathbf{a}_{3} & = & -z_{5}a \, \mathbf{\hat{x}} + x_{5}a \, \mathbf{\hat{y}}-y_{5}a \, \mathbf{\hat{z}} & \left(96h\right) & \mbox{H IV} \\ 
\mathbf{B}_{89} & = & \left(x_{5}-y_{5}+z_{5}\right) \, \mathbf{a}_{1} + \left(x_{5}+y_{5}-z_{5}\right) \, \mathbf{a}_{2} + \left(-x_{5}+y_{5}+z_{5}\right) \, \mathbf{a}_{3} & = & y_{5}a \, \mathbf{\hat{x}} + z_{5}a \, \mathbf{\hat{y}} + x_{5}a \, \mathbf{\hat{z}} & \left(96h\right) & \mbox{H IV} \\ 
\mathbf{B}_{90} & = & \left(-x_{5}+y_{5}+z_{5}\right) \, \mathbf{a}_{1} + \left(-x_{5}-y_{5}-z_{5}\right) \, \mathbf{a}_{2} + \left(x_{5}-y_{5}+z_{5}\right) \, \mathbf{a}_{3} & = & -y_{5}a \, \mathbf{\hat{x}} + z_{5}a \, \mathbf{\hat{y}}-x_{5}a \, \mathbf{\hat{z}} & \left(96h\right) & \mbox{H IV} \\ 
\mathbf{B}_{91} & = & \left(-x_{5}-y_{5}-z_{5}\right) \, \mathbf{a}_{1} + \left(-x_{5}+y_{5}+z_{5}\right) \, \mathbf{a}_{2} + \left(x_{5}+y_{5}-z_{5}\right) \, \mathbf{a}_{3} & = & y_{5}a \, \mathbf{\hat{x}}-z_{5}a \, \mathbf{\hat{y}}-x_{5}a \, \mathbf{\hat{z}} & \left(96h\right) & \mbox{H IV} \\ 
\mathbf{B}_{92} & = & \left(x_{5}+y_{5}-z_{5}\right) \, \mathbf{a}_{1} + \left(x_{5}-y_{5}+z_{5}\right) \, \mathbf{a}_{2} + \left(-x_{5}-y_{5}-z_{5}\right) \, \mathbf{a}_{3} & = & -y_{5}a \, \mathbf{\hat{x}}-z_{5}a \, \mathbf{\hat{y}} + x_{5}a \, \mathbf{\hat{z}} & \left(96h\right) & \mbox{H IV} \\ 
\mathbf{B}_{93} & = & \left(\frac{1}{4} +x_{5} - y_{5} - z_{5}\right) \, \mathbf{a}_{1} + \left(\frac{1}{4} - x_{5} + y_{5} - z_{5}\right) \, \mathbf{a}_{2} + \left(\frac{1}{4} +x_{5} + y_{5} + z_{5}\right) \, \mathbf{a}_{3} & = & \left(\frac{1}{4} +y_{5}\right)a \, \mathbf{\hat{x}} + \left(\frac{1}{4} +x_{5}\right)a \, \mathbf{\hat{y}} + \left(\frac{1}{4} - z_{5}\right)a \, \mathbf{\hat{z}} & \left(96h\right) & \mbox{H IV} \\ 
\mathbf{B}_{94} & = & \left(\frac{1}{4} - x_{5} + y_{5} - z_{5}\right) \, \mathbf{a}_{1} + \left(\frac{1}{4} +x_{5} - y_{5} - z_{5}\right) \, \mathbf{a}_{2} + \left(\frac{1}{4} - x_{5} - y_{5} + z_{5}\right) \, \mathbf{a}_{3} & = & \left(\frac{1}{4} - y_{5}\right)a \, \mathbf{\hat{x}} + \left(\frac{1}{4} - x_{5}\right)a \, \mathbf{\hat{y}} + \left(\frac{1}{4} - z_{5}\right)a \, \mathbf{\hat{z}} & \left(96h\right) & \mbox{H IV} \\ 
\mathbf{B}_{95} & = & \left(\frac{1}{4} - x_{5} - y_{5} + z_{5}\right) \, \mathbf{a}_{1} + \left(\frac{1}{4} +x_{5} + y_{5} + z_{5}\right) \, \mathbf{a}_{2} + \left(\frac{1}{4} - x_{5} + y_{5} - z_{5}\right) \, \mathbf{a}_{3} & = & \left(\frac{1}{4} +y_{5}\right)a \, \mathbf{\hat{x}} + \left(\frac{1}{4} - x_{5}\right)a \, \mathbf{\hat{y}} + \left(\frac{1}{4} +z_{5}\right)a \, \mathbf{\hat{z}} & \left(96h\right) & \mbox{H IV} \\ 
\mathbf{B}_{96} & = & \left(\frac{1}{4} +x_{5} + y_{5} + z_{5}\right) \, \mathbf{a}_{1} + \left(\frac{1}{4} - x_{5} - y_{5} + z_{5}\right) \, \mathbf{a}_{2} + \left(\frac{1}{4} +x_{5} - y_{5} - z_{5}\right) \, \mathbf{a}_{3} & = & \left(\frac{1}{4} - y_{5}\right)a \, \mathbf{\hat{x}} + \left(\frac{1}{4} +x_{5}\right)a \, \mathbf{\hat{y}} + \left(\frac{1}{4} +z_{5}\right)a \, \mathbf{\hat{z}} & \left(96h\right) & \mbox{H IV} \\ 
\mathbf{B}_{97} & = & \left(\frac{1}{4} - x_{5} - y_{5} + z_{5}\right) \, \mathbf{a}_{1} + \left(\frac{1}{4} +x_{5} - y_{5} - z_{5}\right) \, \mathbf{a}_{2} + \left(\frac{1}{4} +x_{5} + y_{5} + z_{5}\right) \, \mathbf{a}_{3} & = & \left(\frac{1}{4} +x_{5}\right)a \, \mathbf{\hat{x}} + \left(\frac{1}{4} +z_{5}\right)a \, \mathbf{\hat{y}} + \left(\frac{1}{4} - y_{5}\right)a \, \mathbf{\hat{z}} & \left(96h\right) & \mbox{H IV} \\ 
\mathbf{B}_{98} & = & \left(\frac{1}{4} +x_{5} + y_{5} + z_{5}\right) \, \mathbf{a}_{1} + \left(\frac{1}{4} - x_{5} + y_{5} - z_{5}\right) \, \mathbf{a}_{2} + \left(\frac{1}{4} - x_{5} - y_{5} + z_{5}\right) \, \mathbf{a}_{3} & = & \left(\frac{1}{4} - x_{5}\right)a \, \mathbf{\hat{x}} + \left(\frac{1}{4} +z_{5}\right)a \, \mathbf{\hat{y}} + \left(\frac{1}{4} +y_{5}\right)a \, \mathbf{\hat{z}} & \left(96h\right) & \mbox{H IV} \\ 
\mathbf{B}_{99} & = & \left(\frac{1}{4} +x_{5} - y_{5} - z_{5}\right) \, \mathbf{a}_{1} + \left(\frac{1}{4} - x_{5} - y_{5} + z_{5}\right) \, \mathbf{a}_{2} + \left(\frac{1}{4} - x_{5} + y_{5} - z_{5}\right) \, \mathbf{a}_{3} & = & \left(\frac{1}{4} - x_{5}\right)a \, \mathbf{\hat{x}} + \left(\frac{1}{4} - z_{5}\right)a \, \mathbf{\hat{y}} + \left(\frac{1}{4} - y_{5}\right)a \, \mathbf{\hat{z}} & \left(96h\right) & \mbox{H IV} \\ 
\mathbf{B}_{100} & = & \left(\frac{1}{4} - x_{5} + y_{5} - z_{5}\right) \, \mathbf{a}_{1} + \left(\frac{1}{4} +x_{5} + y_{5} + z_{5}\right) \, \mathbf{a}_{2} + \left(\frac{1}{4} +x_{5} - y_{5} - z_{5}\right) \, \mathbf{a}_{3} & = & \left(\frac{1}{4} +x_{5}\right)a \, \mathbf{\hat{x}} + \left(\frac{1}{4} - z_{5}\right)a \, \mathbf{\hat{y}} + \left(\frac{1}{4} +y_{5}\right)a \, \mathbf{\hat{z}} & \left(96h\right) & \mbox{H IV} \\ 
\mathbf{B}_{101} & = & \left(\frac{1}{4} - x_{5} + y_{5} - z_{5}\right) \, \mathbf{a}_{1} + \left(\frac{1}{4} - x_{5} - y_{5} + z_{5}\right) \, \mathbf{a}_{2} + \left(\frac{1}{4} +x_{5} + y_{5} + z_{5}\right) \, \mathbf{a}_{3} & = & \left(\frac{1}{4} +z_{5}\right)a \, \mathbf{\hat{x}} + \left(\frac{1}{4} +y_{5}\right)a \, \mathbf{\hat{y}} + \left(\frac{1}{4} - x_{5}\right)a \, \mathbf{\hat{z}} & \left(96h\right) & \mbox{H IV} \\ 
\mathbf{B}_{102} & = & \left(\frac{1}{4} +x_{5} - y_{5} - z_{5}\right) \, \mathbf{a}_{1} + \left(\frac{1}{4} +x_{5} + y_{5} + z_{5}\right) \, \mathbf{a}_{2} + \left(\frac{1}{4} - x_{5} - y_{5} + z_{5}\right) \, \mathbf{a}_{3} & = & \left(\frac{1}{4} +z_{5}\right)a \, \mathbf{\hat{x}} + \left(\frac{1}{4} - y_{5}\right)a \, \mathbf{\hat{y}} + \left(\frac{1}{4} +x_{5}\right)a \, \mathbf{\hat{z}} & \left(96h\right) & \mbox{H IV} \\ 
\mathbf{B}_{103} & = & \left(\frac{1}{4} +x_{5} + y_{5} + z_{5}\right) \, \mathbf{a}_{1} + \left(\frac{1}{4} +x_{5} - y_{5} - z_{5}\right) \, \mathbf{a}_{2} + \left(\frac{1}{4} - x_{5} + y_{5} - z_{5}\right) \, \mathbf{a}_{3} & = & \left(\frac{1}{4} - z_{5}\right)a \, \mathbf{\hat{x}} + \left(\frac{1}{4} +y_{5}\right)a \, \mathbf{\hat{y}} + \left(\frac{1}{4} +x_{5}\right)a \, \mathbf{\hat{z}} & \left(96h\right) & \mbox{H IV} \\ 
\mathbf{B}_{104} & = & \left(\frac{1}{4} - x_{5} - y_{5} + z_{5}\right) \, \mathbf{a}_{1} + \left(\frac{1}{4} - x_{5} + y_{5} - z_{5}\right) \, \mathbf{a}_{2} + \left(\frac{1}{4} +x_{5} - y_{5} - z_{5}\right) \, \mathbf{a}_{3} & = & \left(\frac{1}{4} - z_{5}\right)a \, \mathbf{\hat{x}} + \left(\frac{1}{4} - y_{5}\right)a \, \mathbf{\hat{y}} + \left(\frac{1}{4} - x_{5}\right)a \, \mathbf{\hat{z}} & \left(96h\right) & \mbox{H IV} \\ 
\mathbf{B}_{105} & = & \left(-x_{6}+y_{6}+z_{6}\right) \, \mathbf{a}_{1} + \left(x_{6}-y_{6}+z_{6}\right) \, \mathbf{a}_{2} + \left(x_{6}+y_{6}-z_{6}\right) \, \mathbf{a}_{3} & = & x_{6}a \, \mathbf{\hat{x}} + y_{6}a \, \mathbf{\hat{y}} + z_{6}a \, \mathbf{\hat{z}} & \left(96h\right) & \mbox{O I} \\ 
\mathbf{B}_{106} & = & \left(x_{6}-y_{6}+z_{6}\right) \, \mathbf{a}_{1} + \left(-x_{6}+y_{6}+z_{6}\right) \, \mathbf{a}_{2} + \left(-x_{6}-y_{6}-z_{6}\right) \, \mathbf{a}_{3} & = & -x_{6}a \, \mathbf{\hat{x}}-y_{6}a \, \mathbf{\hat{y}} + z_{6}a \, \mathbf{\hat{z}} & \left(96h\right) & \mbox{O I} \\ 
\mathbf{B}_{107} & = & \left(x_{6}+y_{6}-z_{6}\right) \, \mathbf{a}_{1} + \left(-x_{6}-y_{6}-z_{6}\right) \, \mathbf{a}_{2} + \left(-x_{6}+y_{6}+z_{6}\right) \, \mathbf{a}_{3} & = & -x_{6}a \, \mathbf{\hat{x}} + y_{6}a \, \mathbf{\hat{y}}-z_{6}a \, \mathbf{\hat{z}} & \left(96h\right) & \mbox{O I} \\ 
\mathbf{B}_{108} & = & \left(-x_{6}-y_{6}-z_{6}\right) \, \mathbf{a}_{1} + \left(x_{6}+y_{6}-z_{6}\right) \, \mathbf{a}_{2} + \left(x_{6}-y_{6}+z_{6}\right) \, \mathbf{a}_{3} & = & x_{6}a \, \mathbf{\hat{x}}-y_{6}a \, \mathbf{\hat{y}}-z_{6}a \, \mathbf{\hat{z}} & \left(96h\right) & \mbox{O I} \\ 
\mathbf{B}_{109} & = & \left(x_{6}+y_{6}-z_{6}\right) \, \mathbf{a}_{1} + \left(-x_{6}+y_{6}+z_{6}\right) \, \mathbf{a}_{2} + \left(x_{6}-y_{6}+z_{6}\right) \, \mathbf{a}_{3} & = & z_{6}a \, \mathbf{\hat{x}} + x_{6}a \, \mathbf{\hat{y}} + y_{6}a \, \mathbf{\hat{z}} & \left(96h\right) & \mbox{O I} \\ 
\mathbf{B}_{110} & = & \left(-x_{6}-y_{6}-z_{6}\right) \, \mathbf{a}_{1} + \left(x_{6}-y_{6}+z_{6}\right) \, \mathbf{a}_{2} + \left(-x_{6}+y_{6}+z_{6}\right) \, \mathbf{a}_{3} & = & z_{6}a \, \mathbf{\hat{x}}-x_{6}a \, \mathbf{\hat{y}}-y_{6}a \, \mathbf{\hat{z}} & \left(96h\right) & \mbox{O I} \\ 
\mathbf{B}_{111} & = & \left(-x_{6}+y_{6}+z_{6}\right) \, \mathbf{a}_{1} + \left(x_{6}+y_{6}-z_{6}\right) \, \mathbf{a}_{2} + \left(-x_{6}-y_{6}-z_{6}\right) \, \mathbf{a}_{3} & = & -z_{6}a \, \mathbf{\hat{x}}-x_{6}a \, \mathbf{\hat{y}} + y_{6}a \, \mathbf{\hat{z}} & \left(96h\right) & \mbox{O I} \\ 
\mathbf{B}_{112} & = & \left(x_{6}-y_{6}+z_{6}\right) \, \mathbf{a}_{1} + \left(-x_{6}-y_{6}-z_{6}\right) \, \mathbf{a}_{2} + \left(x_{6}+y_{6}-z_{6}\right) \, \mathbf{a}_{3} & = & -z_{6}a \, \mathbf{\hat{x}} + x_{6}a \, \mathbf{\hat{y}}-y_{6}a \, \mathbf{\hat{z}} & \left(96h\right) & \mbox{O I} \\ 
\mathbf{B}_{113} & = & \left(x_{6}-y_{6}+z_{6}\right) \, \mathbf{a}_{1} + \left(x_{6}+y_{6}-z_{6}\right) \, \mathbf{a}_{2} + \left(-x_{6}+y_{6}+z_{6}\right) \, \mathbf{a}_{3} & = & y_{6}a \, \mathbf{\hat{x}} + z_{6}a \, \mathbf{\hat{y}} + x_{6}a \, \mathbf{\hat{z}} & \left(96h\right) & \mbox{O I} \\ 
\mathbf{B}_{114} & = & \left(-x_{6}+y_{6}+z_{6}\right) \, \mathbf{a}_{1} + \left(-x_{6}-y_{6}-z_{6}\right) \, \mathbf{a}_{2} + \left(x_{6}-y_{6}+z_{6}\right) \, \mathbf{a}_{3} & = & -y_{6}a \, \mathbf{\hat{x}} + z_{6}a \, \mathbf{\hat{y}}-x_{6}a \, \mathbf{\hat{z}} & \left(96h\right) & \mbox{O I} \\ 
\mathbf{B}_{115} & = & \left(-x_{6}-y_{6}-z_{6}\right) \, \mathbf{a}_{1} + \left(-x_{6}+y_{6}+z_{6}\right) \, \mathbf{a}_{2} + \left(x_{6}+y_{6}-z_{6}\right) \, \mathbf{a}_{3} & = & y_{6}a \, \mathbf{\hat{x}}-z_{6}a \, \mathbf{\hat{y}}-x_{6}a \, \mathbf{\hat{z}} & \left(96h\right) & \mbox{O I} \\ 
\mathbf{B}_{116} & = & \left(x_{6}+y_{6}-z_{6}\right) \, \mathbf{a}_{1} + \left(x_{6}-y_{6}+z_{6}\right) \, \mathbf{a}_{2} + \left(-x_{6}-y_{6}-z_{6}\right) \, \mathbf{a}_{3} & = & -y_{6}a \, \mathbf{\hat{x}}-z_{6}a \, \mathbf{\hat{y}} + x_{6}a \, \mathbf{\hat{z}} & \left(96h\right) & \mbox{O I} \\ 
\mathbf{B}_{117} & = & \left(\frac{1}{4} +x_{6} - y_{6} - z_{6}\right) \, \mathbf{a}_{1} + \left(\frac{1}{4} - x_{6} + y_{6} - z_{6}\right) \, \mathbf{a}_{2} + \left(\frac{1}{4} +x_{6} + y_{6} + z_{6}\right) \, \mathbf{a}_{3} & = & \left(\frac{1}{4} +y_{6}\right)a \, \mathbf{\hat{x}} + \left(\frac{1}{4} +x_{6}\right)a \, \mathbf{\hat{y}} + \left(\frac{1}{4} - z_{6}\right)a \, \mathbf{\hat{z}} & \left(96h\right) & \mbox{O I} \\ 
\mathbf{B}_{118} & = & \left(\frac{1}{4} - x_{6} + y_{6} - z_{6}\right) \, \mathbf{a}_{1} + \left(\frac{1}{4} +x_{6} - y_{6} - z_{6}\right) \, \mathbf{a}_{2} + \left(\frac{1}{4} - x_{6} - y_{6} + z_{6}\right) \, \mathbf{a}_{3} & = & \left(\frac{1}{4} - y_{6}\right)a \, \mathbf{\hat{x}} + \left(\frac{1}{4} - x_{6}\right)a \, \mathbf{\hat{y}} + \left(\frac{1}{4} - z_{6}\right)a \, \mathbf{\hat{z}} & \left(96h\right) & \mbox{O I} \\ 
\mathbf{B}_{119} & = & \left(\frac{1}{4} - x_{6} - y_{6} + z_{6}\right) \, \mathbf{a}_{1} + \left(\frac{1}{4} +x_{6} + y_{6} + z_{6}\right) \, \mathbf{a}_{2} + \left(\frac{1}{4} - x_{6} + y_{6} - z_{6}\right) \, \mathbf{a}_{3} & = & \left(\frac{1}{4} +y_{6}\right)a \, \mathbf{\hat{x}} + \left(\frac{1}{4} - x_{6}\right)a \, \mathbf{\hat{y}} + \left(\frac{1}{4} +z_{6}\right)a \, \mathbf{\hat{z}} & \left(96h\right) & \mbox{O I} \\ 
\mathbf{B}_{120} & = & \left(\frac{1}{4} +x_{6} + y_{6} + z_{6}\right) \, \mathbf{a}_{1} + \left(\frac{1}{4} - x_{6} - y_{6} + z_{6}\right) \, \mathbf{a}_{2} + \left(\frac{1}{4} +x_{6} - y_{6} - z_{6}\right) \, \mathbf{a}_{3} & = & \left(\frac{1}{4} - y_{6}\right)a \, \mathbf{\hat{x}} + \left(\frac{1}{4} +x_{6}\right)a \, \mathbf{\hat{y}} + \left(\frac{1}{4} +z_{6}\right)a \, \mathbf{\hat{z}} & \left(96h\right) & \mbox{O I} \\ 
\mathbf{B}_{121} & = & \left(\frac{1}{4} - x_{6} - y_{6} + z_{6}\right) \, \mathbf{a}_{1} + \left(\frac{1}{4} +x_{6} - y_{6} - z_{6}\right) \, \mathbf{a}_{2} + \left(\frac{1}{4} +x_{6} + y_{6} + z_{6}\right) \, \mathbf{a}_{3} & = & \left(\frac{1}{4} +x_{6}\right)a \, \mathbf{\hat{x}} + \left(\frac{1}{4} +z_{6}\right)a \, \mathbf{\hat{y}} + \left(\frac{1}{4} - y_{6}\right)a \, \mathbf{\hat{z}} & \left(96h\right) & \mbox{O I} \\ 
\mathbf{B}_{122} & = & \left(\frac{1}{4} +x_{6} + y_{6} + z_{6}\right) \, \mathbf{a}_{1} + \left(\frac{1}{4} - x_{6} + y_{6} - z_{6}\right) \, \mathbf{a}_{2} + \left(\frac{1}{4} - x_{6} - y_{6} + z_{6}\right) \, \mathbf{a}_{3} & = & \left(\frac{1}{4} - x_{6}\right)a \, \mathbf{\hat{x}} + \left(\frac{1}{4} +z_{6}\right)a \, \mathbf{\hat{y}} + \left(\frac{1}{4} +y_{6}\right)a \, \mathbf{\hat{z}} & \left(96h\right) & \mbox{O I} \\ 
\mathbf{B}_{123} & = & \left(\frac{1}{4} +x_{6} - y_{6} - z_{6}\right) \, \mathbf{a}_{1} + \left(\frac{1}{4} - x_{6} - y_{6} + z_{6}\right) \, \mathbf{a}_{2} + \left(\frac{1}{4} - x_{6} + y_{6} - z_{6}\right) \, \mathbf{a}_{3} & = & \left(\frac{1}{4} - x_{6}\right)a \, \mathbf{\hat{x}} + \left(\frac{1}{4} - z_{6}\right)a \, \mathbf{\hat{y}} + \left(\frac{1}{4} - y_{6}\right)a \, \mathbf{\hat{z}} & \left(96h\right) & \mbox{O I} \\ 
\mathbf{B}_{124} & = & \left(\frac{1}{4} - x_{6} + y_{6} - z_{6}\right) \, \mathbf{a}_{1} + \left(\frac{1}{4} +x_{6} + y_{6} + z_{6}\right) \, \mathbf{a}_{2} + \left(\frac{1}{4} +x_{6} - y_{6} - z_{6}\right) \, \mathbf{a}_{3} & = & \left(\frac{1}{4} +x_{6}\right)a \, \mathbf{\hat{x}} + \left(\frac{1}{4} - z_{6}\right)a \, \mathbf{\hat{y}} + \left(\frac{1}{4} +y_{6}\right)a \, \mathbf{\hat{z}} & \left(96h\right) & \mbox{O I} \\ 
\mathbf{B}_{125} & = & \left(\frac{1}{4} - x_{6} + y_{6} - z_{6}\right) \, \mathbf{a}_{1} + \left(\frac{1}{4} - x_{6} - y_{6} + z_{6}\right) \, \mathbf{a}_{2} + \left(\frac{1}{4} +x_{6} + y_{6} + z_{6}\right) \, \mathbf{a}_{3} & = & \left(\frac{1}{4} +z_{6}\right)a \, \mathbf{\hat{x}} + \left(\frac{1}{4} +y_{6}\right)a \, \mathbf{\hat{y}} + \left(\frac{1}{4} - x_{6}\right)a \, \mathbf{\hat{z}} & \left(96h\right) & \mbox{O I} \\ 
\mathbf{B}_{126} & = & \left(\frac{1}{4} +x_{6} - y_{6} - z_{6}\right) \, \mathbf{a}_{1} + \left(\frac{1}{4} +x_{6} + y_{6} + z_{6}\right) \, \mathbf{a}_{2} + \left(\frac{1}{4} - x_{6} - y_{6} + z_{6}\right) \, \mathbf{a}_{3} & = & \left(\frac{1}{4} +z_{6}\right)a \, \mathbf{\hat{x}} + \left(\frac{1}{4} - y_{6}\right)a \, \mathbf{\hat{y}} + \left(\frac{1}{4} +x_{6}\right)a \, \mathbf{\hat{z}} & \left(96h\right) & \mbox{O I} \\ 
\mathbf{B}_{127} & = & \left(\frac{1}{4} +x_{6} + y_{6} + z_{6}\right) \, \mathbf{a}_{1} + \left(\frac{1}{4} +x_{6} - y_{6} - z_{6}\right) \, \mathbf{a}_{2} + \left(\frac{1}{4} - x_{6} + y_{6} - z_{6}\right) \, \mathbf{a}_{3} & = & \left(\frac{1}{4} - z_{6}\right)a \, \mathbf{\hat{x}} + \left(\frac{1}{4} +y_{6}\right)a \, \mathbf{\hat{y}} + \left(\frac{1}{4} +x_{6}\right)a \, \mathbf{\hat{z}} & \left(96h\right) & \mbox{O I} \\ 
\mathbf{B}_{128} & = & \left(\frac{1}{4} - x_{6} - y_{6} + z_{6}\right) \, \mathbf{a}_{1} + \left(\frac{1}{4} - x_{6} + y_{6} - z_{6}\right) \, \mathbf{a}_{2} + \left(\frac{1}{4} +x_{6} - y_{6} - z_{6}\right) \, \mathbf{a}_{3} & = & \left(\frac{1}{4} - z_{6}\right)a \, \mathbf{\hat{x}} + \left(\frac{1}{4} - y_{6}\right)a \, \mathbf{\hat{y}} + \left(\frac{1}{4} - x_{6}\right)a \, \mathbf{\hat{z}} & \left(96h\right) & \mbox{O I} \\ 
\mathbf{B}_{129} & = & \left(-x_{7}+y_{7}+z_{7}\right) \, \mathbf{a}_{1} + \left(x_{7}-y_{7}+z_{7}\right) \, \mathbf{a}_{2} + \left(x_{7}+y_{7}-z_{7}\right) \, \mathbf{a}_{3} & = & x_{7}a \, \mathbf{\hat{x}} + y_{7}a \, \mathbf{\hat{y}} + z_{7}a \, \mathbf{\hat{z}} & \left(96h\right) & \mbox{O II} \\ 
\mathbf{B}_{130} & = & \left(x_{7}-y_{7}+z_{7}\right) \, \mathbf{a}_{1} + \left(-x_{7}+y_{7}+z_{7}\right) \, \mathbf{a}_{2} + \left(-x_{7}-y_{7}-z_{7}\right) \, \mathbf{a}_{3} & = & -x_{7}a \, \mathbf{\hat{x}}-y_{7}a \, \mathbf{\hat{y}} + z_{7}a \, \mathbf{\hat{z}} & \left(96h\right) & \mbox{O II} \\ 
\mathbf{B}_{131} & = & \left(x_{7}+y_{7}-z_{7}\right) \, \mathbf{a}_{1} + \left(-x_{7}-y_{7}-z_{7}\right) \, \mathbf{a}_{2} + \left(-x_{7}+y_{7}+z_{7}\right) \, \mathbf{a}_{3} & = & -x_{7}a \, \mathbf{\hat{x}} + y_{7}a \, \mathbf{\hat{y}}-z_{7}a \, \mathbf{\hat{z}} & \left(96h\right) & \mbox{O II} \\ 
\mathbf{B}_{132} & = & \left(-x_{7}-y_{7}-z_{7}\right) \, \mathbf{a}_{1} + \left(x_{7}+y_{7}-z_{7}\right) \, \mathbf{a}_{2} + \left(x_{7}-y_{7}+z_{7}\right) \, \mathbf{a}_{3} & = & x_{7}a \, \mathbf{\hat{x}}-y_{7}a \, \mathbf{\hat{y}}-z_{7}a \, \mathbf{\hat{z}} & \left(96h\right) & \mbox{O II} \\ 
\mathbf{B}_{133} & = & \left(x_{7}+y_{7}-z_{7}\right) \, \mathbf{a}_{1} + \left(-x_{7}+y_{7}+z_{7}\right) \, \mathbf{a}_{2} + \left(x_{7}-y_{7}+z_{7}\right) \, \mathbf{a}_{3} & = & z_{7}a \, \mathbf{\hat{x}} + x_{7}a \, \mathbf{\hat{y}} + y_{7}a \, \mathbf{\hat{z}} & \left(96h\right) & \mbox{O II} \\ 
\mathbf{B}_{134} & = & \left(-x_{7}-y_{7}-z_{7}\right) \, \mathbf{a}_{1} + \left(x_{7}-y_{7}+z_{7}\right) \, \mathbf{a}_{2} + \left(-x_{7}+y_{7}+z_{7}\right) \, \mathbf{a}_{3} & = & z_{7}a \, \mathbf{\hat{x}}-x_{7}a \, \mathbf{\hat{y}}-y_{7}a \, \mathbf{\hat{z}} & \left(96h\right) & \mbox{O II} \\ 
\mathbf{B}_{135} & = & \left(-x_{7}+y_{7}+z_{7}\right) \, \mathbf{a}_{1} + \left(x_{7}+y_{7}-z_{7}\right) \, \mathbf{a}_{2} + \left(-x_{7}-y_{7}-z_{7}\right) \, \mathbf{a}_{3} & = & -z_{7}a \, \mathbf{\hat{x}}-x_{7}a \, \mathbf{\hat{y}} + y_{7}a \, \mathbf{\hat{z}} & \left(96h\right) & \mbox{O II} \\ 
\mathbf{B}_{136} & = & \left(x_{7}-y_{7}+z_{7}\right) \, \mathbf{a}_{1} + \left(-x_{7}-y_{7}-z_{7}\right) \, \mathbf{a}_{2} + \left(x_{7}+y_{7}-z_{7}\right) \, \mathbf{a}_{3} & = & -z_{7}a \, \mathbf{\hat{x}} + x_{7}a \, \mathbf{\hat{y}}-y_{7}a \, \mathbf{\hat{z}} & \left(96h\right) & \mbox{O II} \\ 
\mathbf{B}_{137} & = & \left(x_{7}-y_{7}+z_{7}\right) \, \mathbf{a}_{1} + \left(x_{7}+y_{7}-z_{7}\right) \, \mathbf{a}_{2} + \left(-x_{7}+y_{7}+z_{7}\right) \, \mathbf{a}_{3} & = & y_{7}a \, \mathbf{\hat{x}} + z_{7}a \, \mathbf{\hat{y}} + x_{7}a \, \mathbf{\hat{z}} & \left(96h\right) & \mbox{O II} \\ 
\mathbf{B}_{138} & = & \left(-x_{7}+y_{7}+z_{7}\right) \, \mathbf{a}_{1} + \left(-x_{7}-y_{7}-z_{7}\right) \, \mathbf{a}_{2} + \left(x_{7}-y_{7}+z_{7}\right) \, \mathbf{a}_{3} & = & -y_{7}a \, \mathbf{\hat{x}} + z_{7}a \, \mathbf{\hat{y}}-x_{7}a \, \mathbf{\hat{z}} & \left(96h\right) & \mbox{O II} \\ 
\mathbf{B}_{139} & = & \left(-x_{7}-y_{7}-z_{7}\right) \, \mathbf{a}_{1} + \left(-x_{7}+y_{7}+z_{7}\right) \, \mathbf{a}_{2} + \left(x_{7}+y_{7}-z_{7}\right) \, \mathbf{a}_{3} & = & y_{7}a \, \mathbf{\hat{x}}-z_{7}a \, \mathbf{\hat{y}}-x_{7}a \, \mathbf{\hat{z}} & \left(96h\right) & \mbox{O II} \\ 
\mathbf{B}_{140} & = & \left(x_{7}+y_{7}-z_{7}\right) \, \mathbf{a}_{1} + \left(x_{7}-y_{7}+z_{7}\right) \, \mathbf{a}_{2} + \left(-x_{7}-y_{7}-z_{7}\right) \, \mathbf{a}_{3} & = & -y_{7}a \, \mathbf{\hat{x}}-z_{7}a \, \mathbf{\hat{y}} + x_{7}a \, \mathbf{\hat{z}} & \left(96h\right) & \mbox{O II} \\ 
\mathbf{B}_{141} & = & \left(\frac{1}{4} +x_{7} - y_{7} - z_{7}\right) \, \mathbf{a}_{1} + \left(\frac{1}{4} - x_{7} + y_{7} - z_{7}\right) \, \mathbf{a}_{2} + \left(\frac{1}{4} +x_{7} + y_{7} + z_{7}\right) \, \mathbf{a}_{3} & = & \left(\frac{1}{4} +y_{7}\right)a \, \mathbf{\hat{x}} + \left(\frac{1}{4} +x_{7}\right)a \, \mathbf{\hat{y}} + \left(\frac{1}{4} - z_{7}\right)a \, \mathbf{\hat{z}} & \left(96h\right) & \mbox{O II} \\ 
\mathbf{B}_{142} & = & \left(\frac{1}{4} - x_{7} + y_{7} - z_{7}\right) \, \mathbf{a}_{1} + \left(\frac{1}{4} +x_{7} - y_{7} - z_{7}\right) \, \mathbf{a}_{2} + \left(\frac{1}{4} - x_{7} - y_{7} + z_{7}\right) \, \mathbf{a}_{3} & = & \left(\frac{1}{4} - y_{7}\right)a \, \mathbf{\hat{x}} + \left(\frac{1}{4} - x_{7}\right)a \, \mathbf{\hat{y}} + \left(\frac{1}{4} - z_{7}\right)a \, \mathbf{\hat{z}} & \left(96h\right) & \mbox{O II} \\ 
\mathbf{B}_{143} & = & \left(\frac{1}{4} - x_{7} - y_{7} + z_{7}\right) \, \mathbf{a}_{1} + \left(\frac{1}{4} +x_{7} + y_{7} + z_{7}\right) \, \mathbf{a}_{2} + \left(\frac{1}{4} - x_{7} + y_{7} - z_{7}\right) \, \mathbf{a}_{3} & = & \left(\frac{1}{4} +y_{7}\right)a \, \mathbf{\hat{x}} + \left(\frac{1}{4} - x_{7}\right)a \, \mathbf{\hat{y}} + \left(\frac{1}{4} +z_{7}\right)a \, \mathbf{\hat{z}} & \left(96h\right) & \mbox{O II} \\ 
\mathbf{B}_{144} & = & \left(\frac{1}{4} +x_{7} + y_{7} + z_{7}\right) \, \mathbf{a}_{1} + \left(\frac{1}{4} - x_{7} - y_{7} + z_{7}\right) \, \mathbf{a}_{2} + \left(\frac{1}{4} +x_{7} - y_{7} - z_{7}\right) \, \mathbf{a}_{3} & = & \left(\frac{1}{4} - y_{7}\right)a \, \mathbf{\hat{x}} + \left(\frac{1}{4} +x_{7}\right)a \, \mathbf{\hat{y}} + \left(\frac{1}{4} +z_{7}\right)a \, \mathbf{\hat{z}} & \left(96h\right) & \mbox{O II} \\ 
\mathbf{B}_{145} & = & \left(\frac{1}{4} - x_{7} - y_{7} + z_{7}\right) \, \mathbf{a}_{1} + \left(\frac{1}{4} +x_{7} - y_{7} - z_{7}\right) \, \mathbf{a}_{2} + \left(\frac{1}{4} +x_{7} + y_{7} + z_{7}\right) \, \mathbf{a}_{3} & = & \left(\frac{1}{4} +x_{7}\right)a \, \mathbf{\hat{x}} + \left(\frac{1}{4} +z_{7}\right)a \, \mathbf{\hat{y}} + \left(\frac{1}{4} - y_{7}\right)a \, \mathbf{\hat{z}} & \left(96h\right) & \mbox{O II} \\ 
\mathbf{B}_{146} & = & \left(\frac{1}{4} +x_{7} + y_{7} + z_{7}\right) \, \mathbf{a}_{1} + \left(\frac{1}{4} - x_{7} + y_{7} - z_{7}\right) \, \mathbf{a}_{2} + \left(\frac{1}{4} - x_{7} - y_{7} + z_{7}\right) \, \mathbf{a}_{3} & = & \left(\frac{1}{4} - x_{7}\right)a \, \mathbf{\hat{x}} + \left(\frac{1}{4} +z_{7}\right)a \, \mathbf{\hat{y}} + \left(\frac{1}{4} +y_{7}\right)a \, \mathbf{\hat{z}} & \left(96h\right) & \mbox{O II} \\ 
\mathbf{B}_{147} & = & \left(\frac{1}{4} +x_{7} - y_{7} - z_{7}\right) \, \mathbf{a}_{1} + \left(\frac{1}{4} - x_{7} - y_{7} + z_{7}\right) \, \mathbf{a}_{2} + \left(\frac{1}{4} - x_{7} + y_{7} - z_{7}\right) \, \mathbf{a}_{3} & = & \left(\frac{1}{4} - x_{7}\right)a \, \mathbf{\hat{x}} + \left(\frac{1}{4} - z_{7}\right)a \, \mathbf{\hat{y}} + \left(\frac{1}{4} - y_{7}\right)a \, \mathbf{\hat{z}} & \left(96h\right) & \mbox{O II} \\ 
\mathbf{B}_{148} & = & \left(\frac{1}{4} - x_{7} + y_{7} - z_{7}\right) \, \mathbf{a}_{1} + \left(\frac{1}{4} +x_{7} + y_{7} + z_{7}\right) \, \mathbf{a}_{2} + \left(\frac{1}{4} +x_{7} - y_{7} - z_{7}\right) \, \mathbf{a}_{3} & = & \left(\frac{1}{4} +x_{7}\right)a \, \mathbf{\hat{x}} + \left(\frac{1}{4} - z_{7}\right)a \, \mathbf{\hat{y}} + \left(\frac{1}{4} +y_{7}\right)a \, \mathbf{\hat{z}} & \left(96h\right) & \mbox{O II} \\ 
\mathbf{B}_{149} & = & \left(\frac{1}{4} - x_{7} + y_{7} - z_{7}\right) \, \mathbf{a}_{1} + \left(\frac{1}{4} - x_{7} - y_{7} + z_{7}\right) \, \mathbf{a}_{2} + \left(\frac{1}{4} +x_{7} + y_{7} + z_{7}\right) \, \mathbf{a}_{3} & = & \left(\frac{1}{4} +z_{7}\right)a \, \mathbf{\hat{x}} + \left(\frac{1}{4} +y_{7}\right)a \, \mathbf{\hat{y}} + \left(\frac{1}{4} - x_{7}\right)a \, \mathbf{\hat{z}} & \left(96h\right) & \mbox{O II} \\ 
\mathbf{B}_{150} & = & \left(\frac{1}{4} +x_{7} - y_{7} - z_{7}\right) \, \mathbf{a}_{1} + \left(\frac{1}{4} +x_{7} + y_{7} + z_{7}\right) \, \mathbf{a}_{2} + \left(\frac{1}{4} - x_{7} - y_{7} + z_{7}\right) \, \mathbf{a}_{3} & = & \left(\frac{1}{4} +z_{7}\right)a \, \mathbf{\hat{x}} + \left(\frac{1}{4} - y_{7}\right)a \, \mathbf{\hat{y}} + \left(\frac{1}{4} +x_{7}\right)a \, \mathbf{\hat{z}} & \left(96h\right) & \mbox{O II} \\ 
\mathbf{B}_{151} & = & \left(\frac{1}{4} +x_{7} + y_{7} + z_{7}\right) \, \mathbf{a}_{1} + \left(\frac{1}{4} +x_{7} - y_{7} - z_{7}\right) \, \mathbf{a}_{2} + \left(\frac{1}{4} - x_{7} + y_{7} - z_{7}\right) \, \mathbf{a}_{3} & = & \left(\frac{1}{4} - z_{7}\right)a \, \mathbf{\hat{x}} + \left(\frac{1}{4} +y_{7}\right)a \, \mathbf{\hat{y}} + \left(\frac{1}{4} +x_{7}\right)a \, \mathbf{\hat{z}} & \left(96h\right) & \mbox{O II} \\ 
\mathbf{B}_{152} & = & \left(\frac{1}{4} - x_{7} - y_{7} + z_{7}\right) \, \mathbf{a}_{1} + \left(\frac{1}{4} - x_{7} + y_{7} - z_{7}\right) \, \mathbf{a}_{2} + \left(\frac{1}{4} +x_{7} - y_{7} - z_{7}\right) \, \mathbf{a}_{3} & = & \left(\frac{1}{4} - z_{7}\right)a \, \mathbf{\hat{x}} + \left(\frac{1}{4} - y_{7}\right)a \, \mathbf{\hat{y}} + \left(\frac{1}{4} - x_{7}\right)a \, \mathbf{\hat{z}} & \left(96h\right) & \mbox{O II} \\ 
\end{longtabu}
\renewcommand{\arraystretch}{1.0}
\noindent \hrulefill
\\
\textbf{References:}
\vspace*{-0.25cm}
\begin{flushleft}
  - \bibentry{Mullica_H6TeO6_actacristB_1980}. \\
\end{flushleft}
\textbf{Found in:}
\vspace*{-0.25cm}
\begin{flushleft}
  - \bibentry{Villars_PearsonsCrystalData_2013}. \\
\end{flushleft}
\noindent \hrulefill
\\
\textbf{Geometry files:}
\\
\noindent  - CIF: pp. {\hyperref[A12B6C_cF608_210_4h_2h_e_cif]{\pageref{A12B6C_cF608_210_4h_2h_e_cif}}} \\
\noindent  - POSCAR: pp. {\hyperref[A12B6C_cF608_210_4h_2h_e_poscar]{\pageref{A12B6C_cF608_210_4h_2h_e_poscar}}} \\
\onecolumn
{\phantomsection\label{A2B_cI72_211_hi_i}}
\subsection*{\huge \textbf{{\normalfont SiO$_{2}$ Structure: A2B\_cI72\_211\_hi\_i}}}
\noindent \hrulefill
\vspace*{0.25cm}
\begin{figure}[htp]
  \centering
  \vspace{-1em}
  {\includegraphics[width=1\textwidth]{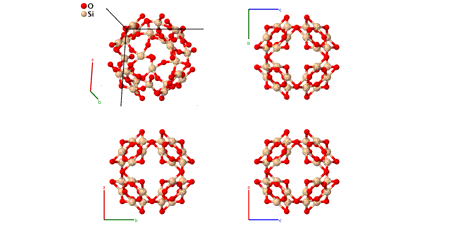}}
\end{figure}
\vspace*{-0.5cm}
\renewcommand{\arraystretch}{1.5}
\begin{equation*}
  \begin{array}{>{$\hspace{-0.15cm}}l<{$}>{$}p{0.5cm}<{$}>{$}p{18.5cm}<{$}}
    \mbox{\large \textbf{Prototype}} &\colon & \ce{SiO2} \\
    \mbox{\large \textbf{\AFLOW\ prototype label}} &\colon & \mbox{A2B\_cI72\_211\_hi\_i} \\
    \mbox{\large \textbf{\textit{Strukturbericht} designation}} &\colon & \mbox{None} \\
    \mbox{\large \textbf{Pearson symbol}} &\colon & \mbox{cI72} \\
    \mbox{\large \textbf{Space group number}} &\colon & 211 \\
    \mbox{\large \textbf{Space group symbol}} &\colon & I432 \\
    \mbox{\large \textbf{\AFLOW\ prototype command}} &\colon &  \texttt{aflow} \,  \, \texttt{-{}-proto=A2B\_cI72\_211\_hi\_i } \, \newline \texttt{-{}-params=}{a,y_{1},y_{2},y_{3} }
  \end{array}
\end{equation*}
\renewcommand{\arraystretch}{1.0}

\noindent \parbox{1 \linewidth}{
\noindent \hrulefill
\\
\textbf{Body-centered Cubic primitive vectors:} \\
\vspace*{-0.25cm}
\begin{tabular}{cc}
  \begin{tabular}{c}
    \parbox{0.6 \linewidth}{
      \renewcommand{\arraystretch}{1.5}
      \begin{equation*}
        \centering
        \begin{array}{ccc}
              \mathbf{a}_1 & = & - \frac12 \, a \, \mathbf{\hat{x}} + \frac12 \, a \, \mathbf{\hat{y}} + \frac12 \, a \, \mathbf{\hat{z}} \\
    \mathbf{a}_2 & = & ~ \frac12 \, a \, \mathbf{\hat{x}} - \frac12 \, a \, \mathbf{\hat{y}} + \frac12 \, a \, \mathbf{\hat{z}} \\
    \mathbf{a}_3 & = & ~ \frac12 \, a \, \mathbf{\hat{x}} + \frac12 \, a \, \mathbf{\hat{y}} - \frac12 \, a \, \mathbf{\hat{z}} \\

        \end{array}
      \end{equation*}
    }
    \renewcommand{\arraystretch}{1.0}
  \end{tabular}
  \begin{tabular}{c}
    \includegraphics[width=0.3\linewidth]{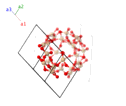} \\
  \end{tabular}
\end{tabular}

}
\vspace*{-0.25cm}

\noindent \hrulefill
\\
\textbf{Basis vectors:}
\vspace*{-0.25cm}
\renewcommand{\arraystretch}{1.5}
\begin{longtabu} to \textwidth{>{\centering $}X[-1,c,c]<{$}>{\centering $}X[-1,c,c]<{$}>{\centering $}X[-1,c,c]<{$}>{\centering $}X[-1,c,c]<{$}>{\centering $}X[-1,c,c]<{$}>{\centering $}X[-1,c,c]<{$}>{\centering $}X[-1,c,c]<{$}}
  & & \mbox{Lattice Coordinates} & & \mbox{Cartesian Coordinates} &\mbox{Wyckoff Position} & \mbox{Atom Type} \\  
  \mathbf{B}_{1} & = & 2y_{1} \, \mathbf{a}_{1} + y_{1} \, \mathbf{a}_{2} + y_{1} \, \mathbf{a}_{3} & = & y_{1}a \, \mathbf{\hat{y}} + y_{1}a \, \mathbf{\hat{z}} & \left(24h\right) & \mbox{O I} \\ 
\mathbf{B}_{2} & = & y_{1} \, \mathbf{a}_{2}-y_{1} \, \mathbf{a}_{3} & = & -y_{1}a \, \mathbf{\hat{y}} + y_{1}a \, \mathbf{\hat{z}} & \left(24h\right) & \mbox{O I} \\ 
\mathbf{B}_{3} & = & -y_{1} \, \mathbf{a}_{2} + y_{1} \, \mathbf{a}_{3} & = & y_{1}a \, \mathbf{\hat{y}}-y_{1}a \, \mathbf{\hat{z}} & \left(24h\right) & \mbox{O I} \\ 
\mathbf{B}_{4} & = & -2y_{1} \, \mathbf{a}_{1}-y_{1} \, \mathbf{a}_{2}-y_{1} \, \mathbf{a}_{3} & = & -y_{1}a \, \mathbf{\hat{y}}-y_{1}a \, \mathbf{\hat{z}} & \left(24h\right) & \mbox{O I} \\ 
\mathbf{B}_{5} & = & y_{1} \, \mathbf{a}_{1} + 2y_{1} \, \mathbf{a}_{2} + y_{1} \, \mathbf{a}_{3} & = & y_{1}a \, \mathbf{\hat{x}} + y_{1}a \, \mathbf{\hat{z}} & \left(24h\right) & \mbox{O I} \\ 
\mathbf{B}_{6} & = & -y_{1} \, \mathbf{a}_{1} + y_{1} \, \mathbf{a}_{3} & = & y_{1}a \, \mathbf{\hat{x}} + -y_{1}a \, \mathbf{\hat{z}} & \left(24h\right) & \mbox{O I} \\ 
\mathbf{B}_{7} & = & y_{1} \, \mathbf{a}_{1} + -y_{1} \, \mathbf{a}_{3} & = & -y_{1}a \, \mathbf{\hat{x}} + y_{1}a \, \mathbf{\hat{z}} & \left(24h\right) & \mbox{O I} \\ 
\mathbf{B}_{8} & = & -y_{1} \, \mathbf{a}_{1}-2y_{1} \, \mathbf{a}_{2}-y_{1} \, \mathbf{a}_{3} & = & -y_{1}a \, \mathbf{\hat{x}} + -y_{1}a \, \mathbf{\hat{z}} & \left(24h\right) & \mbox{O I} \\ 
\mathbf{B}_{9} & = & y_{1} \, \mathbf{a}_{1} + y_{1} \, \mathbf{a}_{2} + 2y_{1} \, \mathbf{a}_{3} & = & y_{1}a \, \mathbf{\hat{x}} + y_{1}a \, \mathbf{\hat{y}} & \left(24h\right) & \mbox{O I} \\ 
\mathbf{B}_{10} & = & y_{1} \, \mathbf{a}_{1}-y_{1} \, \mathbf{a}_{2} & = & -y_{1}a \, \mathbf{\hat{x}} + y_{1}a \, \mathbf{\hat{y}} & \left(24h\right) & \mbox{O I} \\ 
\mathbf{B}_{11} & = & -y_{1} \, \mathbf{a}_{1} + y_{1} \, \mathbf{a}_{2} & = & y_{1}a \, \mathbf{\hat{x}}-y_{1}a \, \mathbf{\hat{y}} & \left(24h\right) & \mbox{O I} \\ 
\mathbf{B}_{12} & = & -y_{1} \, \mathbf{a}_{1}-y_{1} \, \mathbf{a}_{2}-2y_{1} \, \mathbf{a}_{3} & = & -y_{1}a \, \mathbf{\hat{x}}-y_{1}a \, \mathbf{\hat{y}} & \left(24h\right) & \mbox{O I} \\ 
\mathbf{B}_{13} & = & \frac{1}{2} \, \mathbf{a}_{1} + \left(\frac{3}{4} - y_{2}\right) \, \mathbf{a}_{2} + \left(\frac{1}{4} +y_{2}\right) \, \mathbf{a}_{3} & = & \frac{1}{4}a \, \mathbf{\hat{x}} + y_{2}a \, \mathbf{\hat{y}} + \left(\frac{1}{2} - y_{2}\right)a \, \mathbf{\hat{z}} & \left(24i\right) & \mbox{O II} \\ 
\mathbf{B}_{14} & = & \left(\frac{1}{2} - 2y_{2}\right) \, \mathbf{a}_{1} + \left(\frac{1}{4} - y_{2}\right) \, \mathbf{a}_{2} + \left(\frac{3}{4} - y_{2}\right) \, \mathbf{a}_{3} & = & \frac{1}{4}a \, \mathbf{\hat{x}} + \left(\frac{1}{2} - y_{2}\right)a \, \mathbf{\hat{y}}-y_{2}a \, \mathbf{\hat{z}} & \left(24i\right) & \mbox{O II} \\ 
\mathbf{B}_{15} & = & \left(\frac{1}{2} +2y_{2}\right) \, \mathbf{a}_{1} + \left(\frac{1}{4} +y_{2}\right) \, \mathbf{a}_{2} + \left(\frac{3}{4} +y_{2}\right) \, \mathbf{a}_{3} & = & \frac{1}{4}a \, \mathbf{\hat{x}} + \left(\frac{1}{2} +y_{2}\right)a \, \mathbf{\hat{y}} + y_{2}a \, \mathbf{\hat{z}} & \left(24i\right) & \mbox{O II} \\ 
\mathbf{B}_{16} & = & \frac{1}{2} \, \mathbf{a}_{1} + \left(\frac{3}{4} +y_{2}\right) \, \mathbf{a}_{2} + \left(\frac{1}{4} - y_{2}\right) \, \mathbf{a}_{3} & = & \frac{1}{4}a \, \mathbf{\hat{x}}-y_{2}a \, \mathbf{\hat{y}} + \left(\frac{1}{2} +y_{2}\right)a \, \mathbf{\hat{z}} & \left(24i\right) & \mbox{O II} \\ 
\mathbf{B}_{17} & = & \left(\frac{1}{4} +y_{2}\right) \, \mathbf{a}_{1} + \frac{1}{2} \, \mathbf{a}_{2} + \left(\frac{3}{4} - y_{2}\right) \, \mathbf{a}_{3} & = & \left(\frac{1}{2} - y_{2}\right)a \, \mathbf{\hat{x}} + \frac{1}{4}a \, \mathbf{\hat{y}} + y_{2}a \, \mathbf{\hat{z}} & \left(24i\right) & \mbox{O II} \\ 
\mathbf{B}_{18} & = & \left(\frac{3}{4} - y_{2}\right) \, \mathbf{a}_{1} + \left(\frac{1}{2} - 2y_{2}\right) \, \mathbf{a}_{2} + \left(\frac{1}{4} - y_{2}\right) \, \mathbf{a}_{3} & = & -y_{2}a \, \mathbf{\hat{x}} + \frac{1}{4}a \, \mathbf{\hat{y}} + \left(\frac{1}{2} - y_{2}\right)a \, \mathbf{\hat{z}} & \left(24i\right) & \mbox{O II} \\ 
\mathbf{B}_{19} & = & \left(\frac{3}{4} +y_{2}\right) \, \mathbf{a}_{1} + \left(\frac{1}{2} +2y_{2}\right) \, \mathbf{a}_{2} + \left(\frac{1}{4} +y_{2}\right) \, \mathbf{a}_{3} & = & y_{2}a \, \mathbf{\hat{x}} + \frac{1}{4}a \, \mathbf{\hat{y}} + \left(\frac{1}{2} +y_{2}\right)a \, \mathbf{\hat{z}} & \left(24i\right) & \mbox{O II} \\ 
\mathbf{B}_{20} & = & \left(\frac{1}{4} - y_{2}\right) \, \mathbf{a}_{1} + \frac{1}{2} \, \mathbf{a}_{2} + \left(\frac{3}{4} +y_{2}\right) \, \mathbf{a}_{3} & = & \left(\frac{1}{2} +y_{2}\right)a \, \mathbf{\hat{x}} + \frac{1}{4}a \, \mathbf{\hat{y}}-y_{2}a \, \mathbf{\hat{z}} & \left(24i\right) & \mbox{O II} \\ 
\mathbf{B}_{21} & = & \left(\frac{3}{4} - y_{2}\right) \, \mathbf{a}_{1} + \left(\frac{1}{4} +y_{2}\right) \, \mathbf{a}_{2} + \frac{1}{2} \, \mathbf{a}_{3} & = & y_{2}a \, \mathbf{\hat{x}} + \left(\frac{1}{2} - y_{2}\right)a \, \mathbf{\hat{y}} + \frac{1}{4}a \, \mathbf{\hat{z}} & \left(24i\right) & \mbox{O II} \\ 
\mathbf{B}_{22} & = & \left(\frac{1}{4} - y_{2}\right) \, \mathbf{a}_{1} + \left(\frac{3}{4} - y_{2}\right) \, \mathbf{a}_{2} + \left(\frac{1}{2} - 2y_{2}\right) \, \mathbf{a}_{3} & = & \left(\frac{1}{2} - y_{2}\right)a \, \mathbf{\hat{x}}-y_{2}a \, \mathbf{\hat{y}} + \frac{1}{4}a \, \mathbf{\hat{z}} & \left(24i\right) & \mbox{O II} \\ 
\mathbf{B}_{23} & = & \left(\frac{1}{4} +y_{2}\right) \, \mathbf{a}_{1} + \left(\frac{3}{4} +y_{2}\right) \, \mathbf{a}_{2} + \left(\frac{1}{2} +2y_{2}\right) \, \mathbf{a}_{3} & = & \left(\frac{1}{2} +y_{2}\right)a \, \mathbf{\hat{x}} + y_{2}a \, \mathbf{\hat{y}} + \frac{1}{4}a \, \mathbf{\hat{z}} & \left(24i\right) & \mbox{O II} \\ 
\mathbf{B}_{24} & = & \left(\frac{3}{4} +y_{2}\right) \, \mathbf{a}_{1} + \left(\frac{1}{4} - y_{2}\right) \, \mathbf{a}_{2} + \frac{1}{2} \, \mathbf{a}_{3} & = & -y_{2}a \, \mathbf{\hat{x}} + \left(\frac{1}{2} +y_{2}\right)a \, \mathbf{\hat{y}} + \frac{1}{4}a \, \mathbf{\hat{z}} & \left(24i\right) & \mbox{O II} \\ 
\mathbf{B}_{25} & = & \frac{1}{2} \, \mathbf{a}_{1} + \left(\frac{3}{4} - y_{3}\right) \, \mathbf{a}_{2} + \left(\frac{1}{4} +y_{3}\right) \, \mathbf{a}_{3} & = & \frac{1}{4}a \, \mathbf{\hat{x}} + y_{3}a \, \mathbf{\hat{y}} + \left(\frac{1}{2} - y_{3}\right)a \, \mathbf{\hat{z}} & \left(24i\right) & \mbox{Si} \\ 
\mathbf{B}_{26} & = & \left(\frac{1}{2} - 2y_{3}\right) \, \mathbf{a}_{1} + \left(\frac{1}{4} - y_{3}\right) \, \mathbf{a}_{2} + \left(\frac{3}{4} - y_{3}\right) \, \mathbf{a}_{3} & = & \frac{1}{4}a \, \mathbf{\hat{x}} + \left(\frac{1}{2} - y_{3}\right)a \, \mathbf{\hat{y}}-y_{3}a \, \mathbf{\hat{z}} & \left(24i\right) & \mbox{Si} \\ 
\mathbf{B}_{27} & = & \left(\frac{1}{2} +2y_{3}\right) \, \mathbf{a}_{1} + \left(\frac{1}{4} +y_{3}\right) \, \mathbf{a}_{2} + \left(\frac{3}{4} +y_{3}\right) \, \mathbf{a}_{3} & = & \frac{1}{4}a \, \mathbf{\hat{x}} + \left(\frac{1}{2} +y_{3}\right)a \, \mathbf{\hat{y}} + y_{3}a \, \mathbf{\hat{z}} & \left(24i\right) & \mbox{Si} \\ 
\mathbf{B}_{28} & = & \frac{1}{2} \, \mathbf{a}_{1} + \left(\frac{3}{4} +y_{3}\right) \, \mathbf{a}_{2} + \left(\frac{1}{4} - y_{3}\right) \, \mathbf{a}_{3} & = & \frac{1}{4}a \, \mathbf{\hat{x}}-y_{3}a \, \mathbf{\hat{y}} + \left(\frac{1}{2} +y_{3}\right)a \, \mathbf{\hat{z}} & \left(24i\right) & \mbox{Si} \\ 
\mathbf{B}_{29} & = & \left(\frac{1}{4} +y_{3}\right) \, \mathbf{a}_{1} + \frac{1}{2} \, \mathbf{a}_{2} + \left(\frac{3}{4} - y_{3}\right) \, \mathbf{a}_{3} & = & \left(\frac{1}{2} - y_{3}\right)a \, \mathbf{\hat{x}} + \frac{1}{4}a \, \mathbf{\hat{y}} + y_{3}a \, \mathbf{\hat{z}} & \left(24i\right) & \mbox{Si} \\ 
\mathbf{B}_{30} & = & \left(\frac{3}{4} - y_{3}\right) \, \mathbf{a}_{1} + \left(\frac{1}{2} - 2y_{3}\right) \, \mathbf{a}_{2} + \left(\frac{1}{4} - y_{3}\right) \, \mathbf{a}_{3} & = & -y_{3}a \, \mathbf{\hat{x}} + \frac{1}{4}a \, \mathbf{\hat{y}} + \left(\frac{1}{2} - y_{3}\right)a \, \mathbf{\hat{z}} & \left(24i\right) & \mbox{Si} \\ 
\mathbf{B}_{31} & = & \left(\frac{3}{4} +y_{3}\right) \, \mathbf{a}_{1} + \left(\frac{1}{2} +2y_{3}\right) \, \mathbf{a}_{2} + \left(\frac{1}{4} +y_{3}\right) \, \mathbf{a}_{3} & = & y_{3}a \, \mathbf{\hat{x}} + \frac{1}{4}a \, \mathbf{\hat{y}} + \left(\frac{1}{2} +y_{3}\right)a \, \mathbf{\hat{z}} & \left(24i\right) & \mbox{Si} \\ 
\mathbf{B}_{32} & = & \left(\frac{1}{4} - y_{3}\right) \, \mathbf{a}_{1} + \frac{1}{2} \, \mathbf{a}_{2} + \left(\frac{3}{4} +y_{3}\right) \, \mathbf{a}_{3} & = & \left(\frac{1}{2} +y_{3}\right)a \, \mathbf{\hat{x}} + \frac{1}{4}a \, \mathbf{\hat{y}}-y_{3}a \, \mathbf{\hat{z}} & \left(24i\right) & \mbox{Si} \\ 
\mathbf{B}_{33} & = & \left(\frac{3}{4} - y_{3}\right) \, \mathbf{a}_{1} + \left(\frac{1}{4} +y_{3}\right) \, \mathbf{a}_{2} + \frac{1}{2} \, \mathbf{a}_{3} & = & y_{3}a \, \mathbf{\hat{x}} + \left(\frac{1}{2} - y_{3}\right)a \, \mathbf{\hat{y}} + \frac{1}{4}a \, \mathbf{\hat{z}} & \left(24i\right) & \mbox{Si} \\ 
\mathbf{B}_{34} & = & \left(\frac{1}{4} - y_{3}\right) \, \mathbf{a}_{1} + \left(\frac{3}{4} - y_{3}\right) \, \mathbf{a}_{2} + \left(\frac{1}{2} - 2y_{3}\right) \, \mathbf{a}_{3} & = & \left(\frac{1}{2} - y_{3}\right)a \, \mathbf{\hat{x}}-y_{3}a \, \mathbf{\hat{y}} + \frac{1}{4}a \, \mathbf{\hat{z}} & \left(24i\right) & \mbox{Si} \\ 
\mathbf{B}_{35} & = & \left(\frac{1}{4} +y_{3}\right) \, \mathbf{a}_{1} + \left(\frac{3}{4} +y_{3}\right) \, \mathbf{a}_{2} + \left(\frac{1}{2} +2y_{3}\right) \, \mathbf{a}_{3} & = & \left(\frac{1}{2} +y_{3}\right)a \, \mathbf{\hat{x}} + y_{3}a \, \mathbf{\hat{y}} + \frac{1}{4}a \, \mathbf{\hat{z}} & \left(24i\right) & \mbox{Si} \\ 
\mathbf{B}_{36} & = & \left(\frac{3}{4} +y_{3}\right) \, \mathbf{a}_{1} + \left(\frac{1}{4} - y_{3}\right) \, \mathbf{a}_{2} + \frac{1}{2} \, \mathbf{a}_{3} & = & -y_{3}a \, \mathbf{\hat{x}} + \left(\frac{1}{2} +y_{3}\right)a \, \mathbf{\hat{y}} + \frac{1}{4}a \, \mathbf{\hat{z}} & \left(24i\right) & \mbox{Si} \\ 
\end{longtabu}
\renewcommand{\arraystretch}{1.0}
\noindent \hrulefill
\\
\textbf{References:}
\vspace*{-0.25cm}
\begin{flushleft}
  - \bibentry{Foster_SiO2_jacs_126_2004}. \\
\end{flushleft}
\textbf{Found in:}
\vspace*{-0.25cm}
\begin{flushleft}
  - \bibentry{icsd:ICSD_170506}. \\
\end{flushleft}
\noindent \hrulefill
\\
\textbf{Geometry files:}
\\
\noindent  - CIF: pp. {\hyperref[A2B_cI72_211_hi_i_cif]{\pageref{A2B_cI72_211_hi_i_cif}}} \\
\noindent  - POSCAR: pp. {\hyperref[A2B_cI72_211_hi_i_poscar]{\pageref{A2B_cI72_211_hi_i_poscar}}} \\
\onecolumn
{\phantomsection\label{A2B_cP12_212_c_a}}
\subsection*{\huge \textbf{{\normalfont SrSi$_{2}$ Structure: A2B\_cP12\_212\_c\_a}}}
\noindent \hrulefill
\vspace*{0.25cm}
\begin{figure}[htp]
  \centering
  \vspace{-1em}
  {\includegraphics[width=1\textwidth]{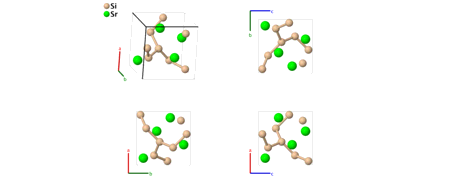}}
\end{figure}
\vspace*{-0.5cm}
\renewcommand{\arraystretch}{1.5}
\begin{equation*}
  \begin{array}{>{$\hspace{-0.15cm}}l<{$}>{$}p{0.5cm}<{$}>{$}p{18.5cm}<{$}}
    \mbox{\large \textbf{Prototype}} &\colon & \ce{SrSi2} \\
    \mbox{\large \textbf{\AFLOW\ prototype label}} &\colon & \mbox{A2B\_cP12\_212\_c\_a} \\
    \mbox{\large \textbf{\textit{Strukturbericht} designation}} &\colon & \mbox{None} \\
    \mbox{\large \textbf{Pearson symbol}} &\colon & \mbox{cP12} \\
    \mbox{\large \textbf{Space group number}} &\colon & 212 \\
    \mbox{\large \textbf{Space group symbol}} &\colon & P4_{3}32 \\
    \mbox{\large \textbf{\AFLOW\ prototype command}} &\colon &  \texttt{aflow} \,  \, \texttt{-{}-proto=A2B\_cP12\_212\_c\_a } \, \newline \texttt{-{}-params=}{a,x_{2} }
  \end{array}
\end{equation*}
\renewcommand{\arraystretch}{1.0}

\vspace*{-0.25cm}
\noindent \hrulefill
\\
\textbf{ Other compounds with this structure:}
\begin{itemize}
   \item{ BaSi$_{2}$, BaSi$_{4}$Sr  }
\end{itemize}
\noindent \parbox{1 \linewidth}{
\noindent \hrulefill
\\
\textbf{Simple Cubic primitive vectors:} \\
\vspace*{-0.25cm}
\begin{tabular}{cc}
  \begin{tabular}{c}
    \parbox{0.6 \linewidth}{
      \renewcommand{\arraystretch}{1.5}
      \begin{equation*}
        \centering
        \begin{array}{ccc}
              \mathbf{a}_1 & = & a \, \mathbf{\hat{x}} \\
    \mathbf{a}_2 & = & a \, \mathbf{\hat{y}} \\
    \mathbf{a}_3 & = & a \, \mathbf{\hat{z}} \\

        \end{array}
      \end{equation*}
    }
    \renewcommand{\arraystretch}{1.0}
  \end{tabular}
  \begin{tabular}{c}
    \includegraphics[width=0.3\linewidth]{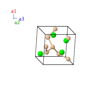} \\
  \end{tabular}
\end{tabular}

}
\vspace*{-0.25cm}

\noindent \hrulefill
\\
\textbf{Basis vectors:}
\vspace*{-0.25cm}
\renewcommand{\arraystretch}{1.5}
\begin{longtabu} to \textwidth{>{\centering $}X[-1,c,c]<{$}>{\centering $}X[-1,c,c]<{$}>{\centering $}X[-1,c,c]<{$}>{\centering $}X[-1,c,c]<{$}>{\centering $}X[-1,c,c]<{$}>{\centering $}X[-1,c,c]<{$}>{\centering $}X[-1,c,c]<{$}}
  & & \mbox{Lattice Coordinates} & & \mbox{Cartesian Coordinates} &\mbox{Wyckoff Position} & \mbox{Atom Type} \\  
  \mathbf{B}_{1} & = & \frac{1}{8} \, \mathbf{a}_{1} + \frac{1}{8} \, \mathbf{a}_{2} + \frac{1}{8} \, \mathbf{a}_{3} & = & \frac{1}{8}a \, \mathbf{\hat{x}} + \frac{1}{8}a \, \mathbf{\hat{y}} + \frac{1}{8}a \, \mathbf{\hat{z}} & \left(4a\right) & \mbox{Sr} \\ 
\mathbf{B}_{2} & = & \frac{3}{8} \, \mathbf{a}_{1} + \frac{7}{8} \, \mathbf{a}_{2} + \frac{5}{8} \, \mathbf{a}_{3} & = & \frac{3}{8}a \, \mathbf{\hat{x}} + \frac{7}{8}a \, \mathbf{\hat{y}} + \frac{5}{8}a \, \mathbf{\hat{z}} & \left(4a\right) & \mbox{Sr} \\ 
\mathbf{B}_{3} & = & \frac{7}{8} \, \mathbf{a}_{1} + \frac{5}{8} \, \mathbf{a}_{2} + \frac{3}{8} \, \mathbf{a}_{3} & = & \frac{7}{8}a \, \mathbf{\hat{x}} + \frac{5}{8}a \, \mathbf{\hat{y}} + \frac{3}{8}a \, \mathbf{\hat{z}} & \left(4a\right) & \mbox{Sr} \\ 
\mathbf{B}_{4} & = & \frac{5}{8} \, \mathbf{a}_{1} + \frac{3}{8} \, \mathbf{a}_{2} + \frac{7}{8} \, \mathbf{a}_{3} & = & \frac{5}{8}a \, \mathbf{\hat{x}} + \frac{3}{8}a \, \mathbf{\hat{y}} + \frac{7}{8}a \, \mathbf{\hat{z}} & \left(4a\right) & \mbox{Sr} \\ 
\mathbf{B}_{5} & = & x_{2} \, \mathbf{a}_{1} + x_{2} \, \mathbf{a}_{2} + x_{2} \, \mathbf{a}_{3} & = & x_{2}a \, \mathbf{\hat{x}} + x_{2}a \, \mathbf{\hat{y}} + x_{2}a \, \mathbf{\hat{z}} & \left(8c\right) & \mbox{Si} \\ 
\mathbf{B}_{6} & = & \left(\frac{1}{2} - x_{2}\right) \, \mathbf{a}_{1}-x_{2} \, \mathbf{a}_{2} + \left(\frac{1}{2} +x_{2}\right) \, \mathbf{a}_{3} & = & \left(\frac{1}{2} - x_{2}\right)a \, \mathbf{\hat{x}}-x_{2}a \, \mathbf{\hat{y}} + \left(\frac{1}{2} +x_{2}\right)a \, \mathbf{\hat{z}} & \left(8c\right) & \mbox{Si} \\ 
\mathbf{B}_{7} & = & -x_{2} \, \mathbf{a}_{1} + \left(\frac{1}{2} +x_{2}\right) \, \mathbf{a}_{2} + \left(\frac{1}{2} - x_{2}\right) \, \mathbf{a}_{3} & = & -x_{2}a \, \mathbf{\hat{x}} + \left(\frac{1}{2} +x_{2}\right)a \, \mathbf{\hat{y}} + \left(\frac{1}{2} - x_{2}\right)a \, \mathbf{\hat{z}} & \left(8c\right) & \mbox{Si} \\ 
\mathbf{B}_{8} & = & \left(\frac{1}{2} +x_{2}\right) \, \mathbf{a}_{1} + \left(\frac{1}{2} - x_{2}\right) \, \mathbf{a}_{2}-x_{2} \, \mathbf{a}_{3} & = & \left(\frac{1}{2} +x_{2}\right)a \, \mathbf{\hat{x}} + \left(\frac{1}{2} - x_{2}\right)a \, \mathbf{\hat{y}}-x_{2}a \, \mathbf{\hat{z}} & \left(8c\right) & \mbox{Si} \\ 
\mathbf{B}_{9} & = & \left(\frac{1}{4} +x_{2}\right) \, \mathbf{a}_{1} + \left(\frac{3}{4} +x_{2}\right) \, \mathbf{a}_{2} + \left(\frac{3}{4} - x_{2}\right) \, \mathbf{a}_{3} & = & \left(\frac{1}{4} +x_{2}\right)a \, \mathbf{\hat{x}} + \left(\frac{3}{4} +x_{2}\right)a \, \mathbf{\hat{y}} + \left(\frac{3}{4} - x_{2}\right)a \, \mathbf{\hat{z}} & \left(8c\right) & \mbox{Si} \\ 
\mathbf{B}_{10} & = & \left(\frac{1}{4} - x_{2}\right) \, \mathbf{a}_{1} + \left(\frac{1}{4} - x_{2}\right) \, \mathbf{a}_{2} + \left(\frac{1}{4} - x_{2}\right) \, \mathbf{a}_{3} & = & \left(\frac{1}{4} - x_{2}\right)a \, \mathbf{\hat{x}} + \left(\frac{1}{4} - x_{2}\right)a \, \mathbf{\hat{y}} + \left(\frac{1}{4} - x_{2}\right)a \, \mathbf{\hat{z}} & \left(8c\right) & \mbox{Si} \\ 
\mathbf{B}_{11} & = & \left(\frac{3}{4} +x_{2}\right) \, \mathbf{a}_{1} + \left(\frac{3}{4} - x_{2}\right) \, \mathbf{a}_{2} + \left(\frac{1}{4} +x_{2}\right) \, \mathbf{a}_{3} & = & \left(\frac{3}{4} +x_{2}\right)a \, \mathbf{\hat{x}} + \left(\frac{3}{4} - x_{2}\right)a \, \mathbf{\hat{y}} + \left(\frac{1}{4} +x_{2}\right)a \, \mathbf{\hat{z}} & \left(8c\right) & \mbox{Si} \\ 
\mathbf{B}_{12} & = & \left(\frac{3}{4} - x_{2}\right) \, \mathbf{a}_{1} + \left(\frac{1}{4} +x_{2}\right) \, \mathbf{a}_{2} + \left(\frac{3}{4} +x_{2}\right) \, \mathbf{a}_{3} & = & \left(\frac{3}{4} - x_{2}\right)a \, \mathbf{\hat{x}} + \left(\frac{1}{4} +x_{2}\right)a \, \mathbf{\hat{y}} + \left(\frac{3}{4} +x_{2}\right)a \, \mathbf{\hat{z}} & \left(8c\right) & \mbox{Si} \\ 
\end{longtabu}
\renewcommand{\arraystretch}{1.0}
\noindent \hrulefill
\\
\textbf{References:}
\vspace*{-0.25cm}
\begin{flushleft}
  - \bibentry{Janzon_SrSi2_angchemint_77_1965}. \\
\end{flushleft}
\textbf{Found in:}
\vspace*{-0.25cm}
\begin{flushleft}
  - \bibentry{villars91:pearson_Si2Sr}. \\
\end{flushleft}
\noindent \hrulefill
\\
\textbf{Geometry files:}
\\
\noindent  - CIF: pp. {\hyperref[A2B_cP12_212_c_a_cif]{\pageref{A2B_cP12_212_c_a_cif}}} \\
\noindent  - POSCAR: pp. {\hyperref[A2B_cP12_212_c_a_poscar]{\pageref{A2B_cP12_212_c_a_poscar}}} \\
\onecolumn
{\phantomsection\label{A3B3C_cI56_214_g_h_a}}
\subsection*{\huge \textbf{{\normalfont Ca$_{3}$PI$_{3}$ Structure: A3B3C\_cI56\_214\_g\_h\_a}}}
\noindent \hrulefill
\vspace*{0.25cm}
\begin{figure}[htp]
  \centering
  \vspace{-1em}
  {\includegraphics[width=1\textwidth]{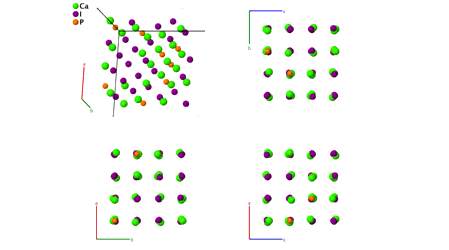}}
\end{figure}
\vspace*{-0.5cm}
\renewcommand{\arraystretch}{1.5}
\begin{equation*}
  \begin{array}{>{$\hspace{-0.15cm}}l<{$}>{$}p{0.5cm}<{$}>{$}p{18.5cm}<{$}}
    \mbox{\large \textbf{Prototype}} &\colon & \ce{Ca3PI3} \\
    \mbox{\large \textbf{\AFLOW\ prototype label}} &\colon & \mbox{A3B3C\_cI56\_214\_g\_h\_a} \\
    \mbox{\large \textbf{\textit{Strukturbericht} designation}} &\colon & \mbox{None} \\
    \mbox{\large \textbf{Pearson symbol}} &\colon & \mbox{cI56} \\
    \mbox{\large \textbf{Space group number}} &\colon & 214 \\
    \mbox{\large \textbf{Space group symbol}} &\colon & I4_{1}32 \\
    \mbox{\large \textbf{\AFLOW\ prototype command}} &\colon &  \texttt{aflow} \,  \, \texttt{-{}-proto=A3B3C\_cI56\_214\_g\_h\_a } \, \newline \texttt{-{}-params=}{a,y_{2},y_{3} }
  \end{array}
\end{equation*}
\renewcommand{\arraystretch}{1.0}

\noindent \parbox{1 \linewidth}{
\noindent \hrulefill
\\
\textbf{Body-centered Cubic primitive vectors:} \\
\vspace*{-0.25cm}
\begin{tabular}{cc}
  \begin{tabular}{c}
    \parbox{0.6 \linewidth}{
      \renewcommand{\arraystretch}{1.5}
      \begin{equation*}
        \centering
        \begin{array}{ccc}
              \mathbf{a}_1 & = & - \frac12 \, a \, \mathbf{\hat{x}} + \frac12 \, a \, \mathbf{\hat{y}} + \frac12 \, a \, \mathbf{\hat{z}} \\
    \mathbf{a}_2 & = & ~ \frac12 \, a \, \mathbf{\hat{x}} - \frac12 \, a \, \mathbf{\hat{y}} + \frac12 \, a \, \mathbf{\hat{z}} \\
    \mathbf{a}_3 & = & ~ \frac12 \, a \, \mathbf{\hat{x}} + \frac12 \, a \, \mathbf{\hat{y}} - \frac12 \, a \, \mathbf{\hat{z}} \\

        \end{array}
      \end{equation*}
    }
    \renewcommand{\arraystretch}{1.0}
  \end{tabular}
  \begin{tabular}{c}
    \includegraphics[width=0.3\linewidth]{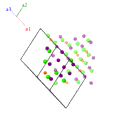} \\
  \end{tabular}
\end{tabular}

}
\vspace*{-0.25cm}

\noindent \hrulefill
\\
\textbf{Basis vectors:}
\vspace*{-0.25cm}
\renewcommand{\arraystretch}{1.5}
\begin{longtabu} to \textwidth{>{\centering $}X[-1,c,c]<{$}>{\centering $}X[-1,c,c]<{$}>{\centering $}X[-1,c,c]<{$}>{\centering $}X[-1,c,c]<{$}>{\centering $}X[-1,c,c]<{$}>{\centering $}X[-1,c,c]<{$}>{\centering $}X[-1,c,c]<{$}}
  & & \mbox{Lattice Coordinates} & & \mbox{Cartesian Coordinates} &\mbox{Wyckoff Position} & \mbox{Atom Type} \\  
  \mathbf{B}_{1} & = & \frac{1}{4} \, \mathbf{a}_{1} + \frac{1}{4} \, \mathbf{a}_{2} + \frac{1}{4} \, \mathbf{a}_{3} & = & \frac{1}{8}a \, \mathbf{\hat{x}} + \frac{1}{8}a \, \mathbf{\hat{y}} + \frac{1}{8}a \, \mathbf{\hat{z}} & \left(8a\right) & \mbox{P} \\ 
\mathbf{B}_{2} & = & \frac{1}{2} \, \mathbf{a}_{1} + \frac{1}{4} \, \mathbf{a}_{3} & = & - \frac{1}{8}a  \, \mathbf{\hat{x}} + \frac{3}{8}a \, \mathbf{\hat{y}} + \frac{1}{8}a \, \mathbf{\hat{z}} & \left(8a\right) & \mbox{P} \\ 
\mathbf{B}_{3} & = & \frac{1}{4} \, \mathbf{a}_{2} + \frac{1}{2} \, \mathbf{a}_{3} & = & \frac{3}{8}a \, \mathbf{\hat{x}} + \frac{1}{8}a \, \mathbf{\hat{y}}- \frac{1}{8}a  \, \mathbf{\hat{z}} & \left(8a\right) & \mbox{P} \\ 
\mathbf{B}_{4} & = & \frac{1}{4} \, \mathbf{a}_{1} + \frac{1}{2} \, \mathbf{a}_{2} & = & \frac{1}{8}a \, \mathbf{\hat{x}}- \frac{1}{8}a  \, \mathbf{\hat{y}} + \frac{3}{8}a \, \mathbf{\hat{z}} & \left(8a\right) & \mbox{P} \\ 
\mathbf{B}_{5} & = & \left(\frac{1}{4} +2y_{2}\right) \, \mathbf{a}_{1} + \left(\frac{3}{8} +y_{2}\right) \, \mathbf{a}_{2} + \left(\frac{1}{8} +y_{2}\right) \, \mathbf{a}_{3} & = & \frac{1}{8}a \, \mathbf{\hat{x}} + y_{2}a \, \mathbf{\hat{y}} + \left(\frac{1}{4} +y_{2}\right)a \, \mathbf{\hat{z}} & \left(24g\right) & \mbox{Ca} \\ 
\mathbf{B}_{6} & = & \frac{3}{4} \, \mathbf{a}_{1} + \left(\frac{1}{8} +y_{2}\right) \, \mathbf{a}_{2} + \left(\frac{3}{8} - y_{2}\right) \, \mathbf{a}_{3} & = & - \frac{1}{8}a  \, \mathbf{\hat{x}} + \left(\frac{1}{2} - y_{2}\right)a \, \mathbf{\hat{y}} + \left(\frac{1}{4} +y_{2}\right)a \, \mathbf{\hat{z}} & \left(24g\right) & \mbox{Ca} \\ 
\mathbf{B}_{7} & = & \frac{3}{4} \, \mathbf{a}_{1} + \left(\frac{1}{8} - y_{2}\right) \, \mathbf{a}_{2} + \left(\frac{3}{8} +y_{2}\right) \, \mathbf{a}_{3} & = & \frac{7}{8}a \, \mathbf{\hat{x}} + \left(\frac{1}{2} +y_{2}\right)a \, \mathbf{\hat{y}} + \left(\frac{1}{4} - y_{2}\right)a \, \mathbf{\hat{z}} & \left(24g\right) & \mbox{Ca} \\ 
\mathbf{B}_{8} & = & \left(\frac{1}{4} - 2y_{2}\right) \, \mathbf{a}_{1} + \left(\frac{3}{8} - y_{2}\right) \, \mathbf{a}_{2} + \left(\frac{1}{8} - y_{2}\right) \, \mathbf{a}_{3} & = & \frac{1}{8}a \, \mathbf{\hat{x}}-y_{2}a \, \mathbf{\hat{y}} + \left(\frac{1}{4} - y_{2}\right)a \, \mathbf{\hat{z}} & \left(24g\right) & \mbox{Ca} \\ 
\mathbf{B}_{9} & = & \left(\frac{1}{8} +y_{2}\right) \, \mathbf{a}_{1} + \left(\frac{1}{4} +2y_{2}\right) \, \mathbf{a}_{2} + \left(\frac{3}{8} +y_{2}\right) \, \mathbf{a}_{3} & = & \left(\frac{1}{4} +y_{2}\right)a \, \mathbf{\hat{x}} + \frac{1}{8}a \, \mathbf{\hat{y}} + y_{2}a \, \mathbf{\hat{z}} & \left(24g\right) & \mbox{Ca} \\ 
\mathbf{B}_{10} & = & \left(\frac{3}{8} - y_{2}\right) \, \mathbf{a}_{1} + \frac{3}{4} \, \mathbf{a}_{2} + \left(\frac{1}{8} +y_{2}\right) \, \mathbf{a}_{3} & = & \left(\frac{1}{4} +y_{2}\right)a \, \mathbf{\hat{x}}- \frac{1}{8}a  \, \mathbf{\hat{y}} + \left(\frac{1}{2} - y_{2}\right)a \, \mathbf{\hat{z}} & \left(24g\right) & \mbox{Ca} \\ 
\mathbf{B}_{11} & = & \left(\frac{3}{8} +y_{2}\right) \, \mathbf{a}_{1} + \frac{3}{4} \, \mathbf{a}_{2} + \left(\frac{1}{8} - y_{2}\right) \, \mathbf{a}_{3} & = & \left(\frac{1}{4} - y_{2}\right)a \, \mathbf{\hat{x}} + \frac{7}{8}a \, \mathbf{\hat{y}} + \left(\frac{1}{2} +y_{2}\right)a \, \mathbf{\hat{z}} & \left(24g\right) & \mbox{Ca} \\ 
\mathbf{B}_{12} & = & \left(\frac{1}{8} - y_{2}\right) \, \mathbf{a}_{1} + \left(\frac{1}{4} - 2y_{2}\right) \, \mathbf{a}_{2} + \left(\frac{3}{8} - y_{2}\right) \, \mathbf{a}_{3} & = & \left(\frac{1}{4} - y_{2}\right)a \, \mathbf{\hat{x}} + \frac{1}{8}a \, \mathbf{\hat{y}}-y_{2}a \, \mathbf{\hat{z}} & \left(24g\right) & \mbox{Ca} \\ 
\mathbf{B}_{13} & = & \left(\frac{3}{8} +y_{2}\right) \, \mathbf{a}_{1} + \left(\frac{1}{8} +y_{2}\right) \, \mathbf{a}_{2} + \left(\frac{1}{4} +2y_{2}\right) \, \mathbf{a}_{3} & = & y_{2}a \, \mathbf{\hat{x}} + \left(\frac{1}{4} +y_{2}\right)a \, \mathbf{\hat{y}} + \frac{1}{8}a \, \mathbf{\hat{z}} & \left(24g\right) & \mbox{Ca} \\ 
\mathbf{B}_{14} & = & \left(\frac{1}{8} +y_{2}\right) \, \mathbf{a}_{1} + \left(\frac{3}{8} - y_{2}\right) \, \mathbf{a}_{2} + \frac{3}{4} \, \mathbf{a}_{3} & = & \left(\frac{1}{2} - y_{2}\right)a \, \mathbf{\hat{x}} + \left(\frac{1}{4} +y_{2}\right)a \, \mathbf{\hat{y}}- \frac{1}{8}a  \, \mathbf{\hat{z}} & \left(24g\right) & \mbox{Ca} \\ 
\mathbf{B}_{15} & = & \left(\frac{1}{8} - y_{2}\right) \, \mathbf{a}_{1} + \left(\frac{3}{8} +y_{2}\right) \, \mathbf{a}_{2} + \frac{3}{4} \, \mathbf{a}_{3} & = & \left(\frac{1}{2} +y_{2}\right)a \, \mathbf{\hat{x}} + \left(\frac{1}{4} - y_{2}\right)a \, \mathbf{\hat{y}} + \frac{7}{8}a \, \mathbf{\hat{z}} & \left(24g\right) & \mbox{Ca} \\ 
\mathbf{B}_{16} & = & \left(\frac{3}{8} - y_{2}\right) \, \mathbf{a}_{1} + \left(\frac{1}{8} - y_{2}\right) \, \mathbf{a}_{2} + \left(\frac{1}{4} - 2y_{2}\right) \, \mathbf{a}_{3} & = & -y_{2}a \, \mathbf{\hat{x}} + \left(\frac{1}{4} - y_{2}\right)a \, \mathbf{\hat{y}} + \frac{1}{8}a \, \mathbf{\hat{z}} & \left(24g\right) & \mbox{Ca} \\ 
\mathbf{B}_{17} & = & \frac{1}{4} \, \mathbf{a}_{1} + \left(\frac{3}{8} - y_{3}\right) \, \mathbf{a}_{2} + \left(\frac{1}{8} +y_{3}\right) \, \mathbf{a}_{3} & = & \frac{1}{8}a \, \mathbf{\hat{x}} + y_{3}a \, \mathbf{\hat{y}} + \left(\frac{1}{4} - y_{3}\right)a \, \mathbf{\hat{z}} & \left(24h\right) & \mbox{I} \\ 
\mathbf{B}_{18} & = & \left(\frac{3}{4} - 2y_{3}\right) \, \mathbf{a}_{1} + \left(\frac{1}{8} - y_{3}\right) \, \mathbf{a}_{2} + \left(\frac{3}{8} - y_{3}\right) \, \mathbf{a}_{3} & = & - \frac{1}{8}a  \, \mathbf{\hat{x}} + \left(\frac{1}{2} - y_{3}\right)a \, \mathbf{\hat{y}} + \left(\frac{1}{4} - y_{3}\right)a \, \mathbf{\hat{z}} & \left(24h\right) & \mbox{I} \\ 
\mathbf{B}_{19} & = & \left(\frac{3}{4} +2y_{3}\right) \, \mathbf{a}_{1} + \left(\frac{1}{8} +y_{3}\right) \, \mathbf{a}_{2} + \left(\frac{3}{8} +y_{3}\right) \, \mathbf{a}_{3} & = & \frac{7}{8}a \, \mathbf{\hat{x}} + \left(\frac{1}{2} +y_{3}\right)a \, \mathbf{\hat{y}} + \left(\frac{1}{4} +y_{3}\right)a \, \mathbf{\hat{z}} & \left(24h\right) & \mbox{I} \\ 
\mathbf{B}_{20} & = & \frac{1}{4} \, \mathbf{a}_{1} + \left(\frac{3}{8} +y_{3}\right) \, \mathbf{a}_{2} + \left(\frac{1}{8} - y_{3}\right) \, \mathbf{a}_{3} & = & \frac{1}{8}a \, \mathbf{\hat{x}}-y_{3}a \, \mathbf{\hat{y}} + \left(\frac{1}{4} +y_{3}\right)a \, \mathbf{\hat{z}} & \left(24h\right) & \mbox{I} \\ 
\mathbf{B}_{21} & = & \left(\frac{1}{8} +y_{3}\right) \, \mathbf{a}_{1} + \frac{1}{4} \, \mathbf{a}_{2} + \left(\frac{3}{8} - y_{3}\right) \, \mathbf{a}_{3} & = & \left(\frac{1}{4} - y_{3}\right)a \, \mathbf{\hat{x}} + \frac{1}{8}a \, \mathbf{\hat{y}} + y_{3}a \, \mathbf{\hat{z}} & \left(24h\right) & \mbox{I} \\ 
\mathbf{B}_{22} & = & \left(\frac{3}{8} - y_{3}\right) \, \mathbf{a}_{1} + \left(\frac{3}{4} - 2y_{3}\right) \, \mathbf{a}_{2} + \left(\frac{1}{8} - y_{3}\right) \, \mathbf{a}_{3} & = & \left(\frac{1}{4} - y_{3}\right)a \, \mathbf{\hat{x}}- \frac{1}{8}a  \, \mathbf{\hat{y}} + \left(\frac{1}{2} - y_{3}\right)a \, \mathbf{\hat{z}} & \left(24h\right) & \mbox{I} \\ 
\mathbf{B}_{23} & = & \left(\frac{3}{8} +y_{3}\right) \, \mathbf{a}_{1} + \left(\frac{3}{4} +2y_{3}\right) \, \mathbf{a}_{2} + \left(\frac{1}{8} +y_{3}\right) \, \mathbf{a}_{3} & = & \left(\frac{1}{4} +y_{3}\right)a \, \mathbf{\hat{x}} + \frac{7}{8}a \, \mathbf{\hat{y}} + \left(\frac{1}{2} +y_{3}\right)a \, \mathbf{\hat{z}} & \left(24h\right) & \mbox{I} \\ 
\mathbf{B}_{24} & = & \left(\frac{1}{8} - y_{3}\right) \, \mathbf{a}_{1} + \frac{1}{4} \, \mathbf{a}_{2} + \left(\frac{3}{8} +y_{3}\right) \, \mathbf{a}_{3} & = & \left(\frac{1}{4} +y_{3}\right)a \, \mathbf{\hat{x}} + \frac{1}{8}a \, \mathbf{\hat{y}}-y_{3}a \, \mathbf{\hat{z}} & \left(24h\right) & \mbox{I} \\ 
\mathbf{B}_{25} & = & \left(\frac{3}{8} - y_{3}\right) \, \mathbf{a}_{1} + \left(\frac{1}{8} +y_{3}\right) \, \mathbf{a}_{2} + \frac{1}{4} \, \mathbf{a}_{3} & = & y_{3}a \, \mathbf{\hat{x}} + \left(\frac{1}{4} - y_{3}\right)a \, \mathbf{\hat{y}} + \frac{1}{8}a \, \mathbf{\hat{z}} & \left(24h\right) & \mbox{I} \\ 
\mathbf{B}_{26} & = & \left(\frac{1}{8} - y_{3}\right) \, \mathbf{a}_{1} + \left(\frac{3}{8} - y_{3}\right) \, \mathbf{a}_{2} + \left(\frac{3}{4} - 2y_{3}\right) \, \mathbf{a}_{3} & = & \left(\frac{1}{2} - y_{3}\right)a \, \mathbf{\hat{x}} + \left(\frac{1}{4} - y_{3}\right)a \, \mathbf{\hat{y}}- \frac{1}{8}a  \, \mathbf{\hat{z}} & \left(24h\right) & \mbox{I} \\ 
\mathbf{B}_{27} & = & \left(\frac{1}{8} +y_{3}\right) \, \mathbf{a}_{1} + \left(\frac{3}{8} +y_{3}\right) \, \mathbf{a}_{2} + \left(\frac{3}{4} +2y_{3}\right) \, \mathbf{a}_{3} & = & \left(\frac{1}{2} +y_{3}\right)a \, \mathbf{\hat{x}} + \left(\frac{1}{4} +y_{3}\right)a \, \mathbf{\hat{y}} + \frac{7}{8}a \, \mathbf{\hat{z}} & \left(24h\right) & \mbox{I} \\ 
\mathbf{B}_{28} & = & \left(\frac{3}{8} +y_{3}\right) \, \mathbf{a}_{1} + \left(\frac{1}{8} - y_{3}\right) \, \mathbf{a}_{2} + \frac{1}{4} \, \mathbf{a}_{3} & = & -y_{3}a \, \mathbf{\hat{x}} + \left(\frac{1}{4} +y_{3}\right)a \, \mathbf{\hat{y}} + \frac{1}{8}a \, \mathbf{\hat{z}} & \left(24h\right) & \mbox{I} \\ 
\end{longtabu}
\renewcommand{\arraystretch}{1.0}
\noindent \hrulefill
\\
\textbf{References:}
\vspace*{-0.25cm}
\begin{flushleft}
  - \bibentry{Hamon_Ca3PI3_BullSocFrMinCristallogr_1974}. \\
\end{flushleft}
\textbf{Found in:}
\vspace*{-0.25cm}
\begin{flushleft}
  - \bibentry{Villars_PearsonsCrystalData_2013}. \\
\end{flushleft}
\noindent \hrulefill
\\
\textbf{Geometry files:}
\\
\noindent  - CIF: pp. {\hyperref[A3B3C_cI56_214_g_h_a_cif]{\pageref{A3B3C_cI56_214_g_h_a_cif}}} \\
\noindent  - POSCAR: pp. {\hyperref[A3B3C_cI56_214_g_h_a_poscar]{\pageref{A3B3C_cI56_214_g_h_a_poscar}}} \\
\onecolumn
{\phantomsection\label{A3BC2_cI48_214_f_a_e}}
\subsection*{\huge \textbf{{\normalfont Petzite (Ag$_{3}$AuTe$_{2}$) Structure: A3BC2\_cI48\_214\_f\_a\_e}}}
\noindent \hrulefill
\vspace*{0.25cm}
\begin{figure}[htp]
  \centering
  \vspace{-1em}
  {\includegraphics[width=1\textwidth]{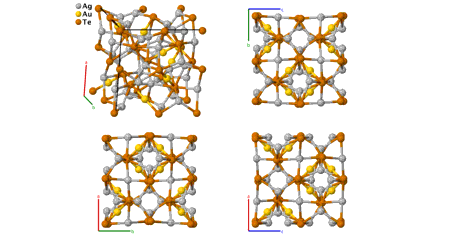}}
\end{figure}
\vspace*{-0.5cm}
\renewcommand{\arraystretch}{1.5}
\begin{equation*}
  \begin{array}{>{$\hspace{-0.15cm}}l<{$}>{$}p{0.5cm}<{$}>{$}p{18.5cm}<{$}}
    \mbox{\large \textbf{Prototype}} &\colon & \ce{Ag3AuTe2} \\
    \mbox{\large \textbf{\AFLOW\ prototype label}} &\colon & \mbox{A3BC2\_cI48\_214\_f\_a\_e} \\
    \mbox{\large \textbf{\textit{Strukturbericht} designation}} &\colon & \mbox{None} \\
    \mbox{\large \textbf{Pearson symbol}} &\colon & \mbox{cI48} \\
    \mbox{\large \textbf{Space group number}} &\colon & 214 \\
    \mbox{\large \textbf{Space group symbol}} &\colon & I4_{1}32 \\
    \mbox{\large \textbf{\AFLOW\ prototype command}} &\colon &  \texttt{aflow} \,  \, \texttt{-{}-proto=A3BC2\_cI48\_214\_f\_a\_e } \, \newline \texttt{-{}-params=}{a,x_{2},x_{3} }
  \end{array}
\end{equation*}
\renewcommand{\arraystretch}{1.0}

\vspace*{-0.25cm}
\noindent \hrulefill
\begin{itemize}
  \item{The Wyckoff positions given in the {\em International Tables} have
been shifted by one or more body-centered cubic primitive lattice
vectors to provide a more compact set of coordinates in both Cartesian
and lattice space.
}
\end{itemize}

\noindent \parbox{1 \linewidth}{
\noindent \hrulefill
\\
\textbf{Body-centered Cubic primitive vectors:} \\
\vspace*{-0.25cm}
\begin{tabular}{cc}
  \begin{tabular}{c}
    \parbox{0.6 \linewidth}{
      \renewcommand{\arraystretch}{1.5}
      \begin{equation*}
        \centering
        \begin{array}{ccc}
              \mathbf{a}_1 & = & - \frac12 \, a \, \mathbf{\hat{x}} + \frac12 \, a \, \mathbf{\hat{y}} + \frac12 \, a \, \mathbf{\hat{z}} \\
    \mathbf{a}_2 & = & ~ \frac12 \, a \, \mathbf{\hat{x}} - \frac12 \, a \, \mathbf{\hat{y}} + \frac12 \, a \, \mathbf{\hat{z}} \\
    \mathbf{a}_3 & = & ~ \frac12 \, a \, \mathbf{\hat{x}} + \frac12 \, a \, \mathbf{\hat{y}} - \frac12 \, a \, \mathbf{\hat{z}} \\

        \end{array}
      \end{equation*}
    }
    \renewcommand{\arraystretch}{1.0}
  \end{tabular}
  \begin{tabular}{c}
    \includegraphics[width=0.3\linewidth]{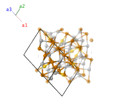} \\
  \end{tabular}
\end{tabular}

}
\vspace*{-0.25cm}

\noindent \hrulefill
\\
\textbf{Basis vectors:}
\vspace*{-0.25cm}
\renewcommand{\arraystretch}{1.5}
\begin{longtabu} to \textwidth{>{\centering $}X[-1,c,c]<{$}>{\centering $}X[-1,c,c]<{$}>{\centering $}X[-1,c,c]<{$}>{\centering $}X[-1,c,c]<{$}>{\centering $}X[-1,c,c]<{$}>{\centering $}X[-1,c,c]<{$}>{\centering $}X[-1,c,c]<{$}}
  & & \mbox{Lattice Coordinates} & & \mbox{Cartesian Coordinates} &\mbox{Wyckoff Position} & \mbox{Atom Type} \\  
  \mathbf{B}_{1} & = & \frac{1}{4} \, \mathbf{a}_{1} + \frac{1}{4} \, \mathbf{a}_{2} + \frac{1}{4} \, \mathbf{a}_{3} & = & \frac{1}{8}a \, \mathbf{\hat{x}} + \frac{1}{8}a \, \mathbf{\hat{y}} + \frac{1}{8}a \, \mathbf{\hat{z}} & \left(8a\right) & \mbox{Au} \\ 
\mathbf{B}_{2} & = & \frac{1}{2} \, \mathbf{a}_{1} + \frac{1}{4} \, \mathbf{a}_{3} & = & - \frac{1}{8}a  \, \mathbf{\hat{x}} + \frac{3}{8}a \, \mathbf{\hat{y}} + \frac{1}{8}a \, \mathbf{\hat{z}} & \left(8a\right) & \mbox{Au} \\ 
\mathbf{B}_{3} & = & \frac{1}{4} \, \mathbf{a}_{2} + \frac{1}{2} \, \mathbf{a}_{3} & = & \frac{3}{8}a \, \mathbf{\hat{x}} + \frac{1}{8}a \, \mathbf{\hat{y}}- \frac{1}{8}a  \, \mathbf{\hat{z}} & \left(8a\right) & \mbox{Au} \\ 
\mathbf{B}_{4} & = & \frac{1}{4} \, \mathbf{a}_{1} + \frac{1}{2} \, \mathbf{a}_{2} & = & \frac{1}{8}a \, \mathbf{\hat{x}}- \frac{1}{8}a  \, \mathbf{\hat{y}} + \frac{3}{8}a \, \mathbf{\hat{z}} & \left(8a\right) & \mbox{Au} \\ 
\mathbf{B}_{5} & = & 2x_{2} \, \mathbf{a}_{1} + 2x_{2} \, \mathbf{a}_{2} + 2x_{2} \, \mathbf{a}_{3} & = & x_{2}a \, \mathbf{\hat{x}} + x_{2}a \, \mathbf{\hat{y}} + x_{2}a \, \mathbf{\hat{z}} & \left(16e\right) & \mbox{Te} \\ 
\mathbf{B}_{6} & = & \frac{1}{2} \, \mathbf{a}_{1} + \left(\frac{1}{2} - 2x_{2}\right) \, \mathbf{a}_{3} & = & -x_{2}a \, \mathbf{\hat{x}} + \left(\frac{1}{2} - x_{2}\right)a \, \mathbf{\hat{y}} + x_{2}a \, \mathbf{\hat{z}} & \left(16e\right) & \mbox{Te} \\ 
\mathbf{B}_{7} & = & \left(\frac{1}{2} - 2x_{2}\right) \, \mathbf{a}_{2} + \frac{1}{2} \, \mathbf{a}_{3} & = & \left(\frac{1}{2} - x_{2}\right)a \, \mathbf{\hat{x}} + x_{2}a \, \mathbf{\hat{y}}-x_{2}a \, \mathbf{\hat{z}} & \left(16e\right) & \mbox{Te} \\ 
\mathbf{B}_{8} & = & \left(\frac{1}{2} - 2x_{2}\right) \, \mathbf{a}_{1} + \frac{1}{2} \, \mathbf{a}_{2} & = & x_{2}a \, \mathbf{\hat{x}}-x_{2}a \, \mathbf{\hat{y}} + \left(\frac{1}{2} - x_{2}\right)a \, \mathbf{\hat{z}} & \left(16e\right) & \mbox{Te} \\ 
\mathbf{B}_{9} & = & \frac{1}{2} \, \mathbf{a}_{1} + 2x_{2} \, \mathbf{a}_{3} & = & \left(\frac{3}{4} +x_{2}\right)a \, \mathbf{\hat{x}} + \left(\frac{1}{4} +x_{2}\right)a \, \mathbf{\hat{y}} + \left(\frac{1}{4} - x_{2}\right)a \, \mathbf{\hat{z}} & \left(16e\right) & \mbox{Te} \\ 
\mathbf{B}_{10} & = & \left(\frac{1}{2} - 2x_{2}\right) \, \mathbf{a}_{1} + \left(\frac{1}{2} - 2x_{2}\right) \, \mathbf{a}_{2} + \left(\frac{1}{2} - 2x_{2}\right) \, \mathbf{a}_{3} & = & \left(\frac{1}{4} - x_{2}\right)a \, \mathbf{\hat{x}} + \left(\frac{1}{4} - x_{2}\right)a \, \mathbf{\hat{y}} + \left(\frac{1}{4} - x_{2}\right)a \, \mathbf{\hat{z}} & \left(16e\right) & \mbox{Te} \\ 
\mathbf{B}_{11} & = & 2x_{2} \, \mathbf{a}_{2} + \frac{1}{2} \, \mathbf{a}_{3} & = & \left(\frac{1}{4} +x_{2}\right)a \, \mathbf{\hat{x}} + \left(\frac{1}{4} - x_{2}\right)a \, \mathbf{\hat{y}} + \left(\frac{3}{4} +x_{2}\right)a \, \mathbf{\hat{z}} & \left(16e\right) & \mbox{Te} \\ 
\mathbf{B}_{12} & = & 2x_{2} \, \mathbf{a}_{1} + \frac{1}{2} \, \mathbf{a}_{2} & = & \left(\frac{1}{4} - x_{2}\right)a \, \mathbf{\hat{x}} + \left(\frac{3}{4} +x_{2}\right)a \, \mathbf{\hat{y}} + \left(\frac{1}{4} +x_{2}\right)a \, \mathbf{\hat{z}} & \left(16e\right) & \mbox{Te} \\ 
\mathbf{B}_{13} & = & \frac{1}{4} \, \mathbf{a}_{1} + \left(\frac{1}{4} +x_{3}\right) \, \mathbf{a}_{2} + x_{3} \, \mathbf{a}_{3} & = & x_{3}a \, \mathbf{\hat{x}} + \frac{1}{4}a \, \mathbf{\hat{z}} & \left(24f\right) & \mbox{Ag} \\ 
\mathbf{B}_{14} & = & \frac{3}{4} \, \mathbf{a}_{1} + \left(\frac{1}{4} - x_{3}\right) \, \mathbf{a}_{2} + \left(\frac{1}{2} - x_{3}\right) \, \mathbf{a}_{3} & = & -x_{3}a \, \mathbf{\hat{x}} + \frac{1}{2}a \, \mathbf{\hat{y}} + \frac{1}{4}a \, \mathbf{\hat{z}} & \left(24f\right) & \mbox{Ag} \\ 
\mathbf{B}_{15} & = & x_{3} \, \mathbf{a}_{1} + \frac{1}{4} \, \mathbf{a}_{2} + \left(\frac{1}{4} +x_{3}\right) \, \mathbf{a}_{3} & = & \frac{1}{4}a \, \mathbf{\hat{x}} + x_{3}a \, \mathbf{\hat{y}} & \left(24f\right) & \mbox{Ag} \\ 
\mathbf{B}_{16} & = & \left(\frac{1}{2} - x_{3}\right) \, \mathbf{a}_{1} + \frac{3}{4} \, \mathbf{a}_{2} + \left(\frac{1}{4} - x_{3}\right) \, \mathbf{a}_{3} & = & \frac{1}{4}a \, \mathbf{\hat{x}}-x_{3}a \, \mathbf{\hat{y}} + \frac{1}{2}a \, \mathbf{\hat{z}} & \left(24f\right) & \mbox{Ag} \\ 
\mathbf{B}_{17} & = & \left(\frac{1}{4} +x_{3}\right) \, \mathbf{a}_{1} + x_{3} \, \mathbf{a}_{2} + \frac{1}{4} \, \mathbf{a}_{3} & = & \frac{1}{4}a \, \mathbf{\hat{y}} + x_{3}a \, \mathbf{\hat{z}} & \left(24f\right) & \mbox{Ag} \\ 
\mathbf{B}_{18} & = & \left(\frac{1}{4} - x_{3}\right) \, \mathbf{a}_{1} + \left(\frac{1}{2} - x_{3}\right) \, \mathbf{a}_{2} + \frac{3}{4} \, \mathbf{a}_{3} & = & \frac{1}{2}a \, \mathbf{\hat{x}} + \frac{1}{4}a \, \mathbf{\hat{y}}-x_{3}a \, \mathbf{\hat{z}} & \left(24f\right) & \mbox{Ag} \\ 
\mathbf{B}_{19} & = & \left(\frac{1}{4} +x_{3}\right) \, \mathbf{a}_{1} + \frac{3}{4} \, \mathbf{a}_{2} + x_{3} \, \mathbf{a}_{3} & = & \frac{1}{4}a \, \mathbf{\hat{x}} + \left(- \frac{1}{4} +x_{3}\right)a \, \mathbf{\hat{y}} + \frac{1}{2}a \, \mathbf{\hat{z}} & \left(24f\right) & \mbox{Ag} \\ 
\mathbf{B}_{20} & = & \left(\frac{1}{4} - x_{3}\right) \, \mathbf{a}_{1} + \frac{1}{4} \, \mathbf{a}_{2} + \left(\frac{1}{2} - x_{3}\right) \, \mathbf{a}_{3} & = & \frac{1}{4}a \, \mathbf{\hat{x}} + \left(\frac{1}{4} - x_{3}\right)a \, \mathbf{\hat{y}} & \left(24f\right) & \mbox{Ag} \\ 
\mathbf{B}_{21} & = & \frac{3}{4} \, \mathbf{a}_{1} + x_{3} \, \mathbf{a}_{2} + \left(\frac{1}{4} +x_{3}\right) \, \mathbf{a}_{3} & = & \left(\frac{3}{4} +x_{3}\right)a \, \mathbf{\hat{x}} + \frac{1}{2}a \, \mathbf{\hat{y}} + \frac{1}{4}a \, \mathbf{\hat{z}} & \left(24f\right) & \mbox{Ag} \\ 
\mathbf{B}_{22} & = & \frac{1}{4} \, \mathbf{a}_{1} + \left(\frac{1}{2} - x_{3}\right) \, \mathbf{a}_{2} + \left(\frac{1}{4} - x_{3}\right) \, \mathbf{a}_{3} & = & \left(\frac{1}{4} - x_{3}\right)a \, \mathbf{\hat{x}} + \frac{1}{4}a \, \mathbf{\hat{z}} & \left(24f\right) & \mbox{Ag} \\ 
\mathbf{B}_{23} & = & \left(\frac{1}{2} - x_{3}\right) \, \mathbf{a}_{1} + \left(\frac{1}{4} - x_{3}\right) \, \mathbf{a}_{2} + \frac{1}{4} \, \mathbf{a}_{3} & = & \frac{1}{4}a \, \mathbf{\hat{y}} + \left(\frac{1}{4} - x_{3}\right)a \, \mathbf{\hat{z}} & \left(24f\right) & \mbox{Ag} \\ 
\mathbf{B}_{24} & = & x_{3} \, \mathbf{a}_{1} + \left(\frac{1}{4} +x_{3}\right) \, \mathbf{a}_{2} + \frac{3}{4} \, \mathbf{a}_{3} & = & \frac{1}{2}a \, \mathbf{\hat{x}} + \frac{1}{4}a \, \mathbf{\hat{y}} + \left(\frac{3}{4} +x_{3}\right)a \, \mathbf{\hat{z}} & \left(24f\right) & \mbox{Ag} \\ 
\end{longtabu}
\renewcommand{\arraystretch}{1.0}
\noindent \hrulefill
\\
\textbf{References:}
\vspace*{-0.25cm}
\begin{flushleft}
  - \bibentry{Frueh_Am_Min_44_1959}. \\
\end{flushleft}
\textbf{Found in:}
\vspace*{-0.25cm}
\begin{flushleft}
  - \bibentry{Downs_AM_88_2003}. \\
\end{flushleft}
\noindent \hrulefill
\\
\textbf{Geometry files:}
\\
\noindent  - CIF: pp. {\hyperref[A3BC2_cI48_214_f_a_e_cif]{\pageref{A3BC2_cI48_214_f_a_e_cif}}} \\
\noindent  - POSCAR: pp. {\hyperref[A3BC2_cI48_214_f_a_e_poscar]{\pageref{A3BC2_cI48_214_f_a_e_poscar}}} \\
\onecolumn
{\phantomsection\label{A4B9_cP52_215_ei_3efgi}}
\subsection*{\huge \textbf{{\normalfont \begin{raggedleft}$\gamma$-brass (Cu$_{9}$Al$_{4}$, $D8_{3}$) Structure: \end{raggedleft} \\ A4B9\_cP52\_215\_ei\_3efgi}}}
\noindent \hrulefill
\vspace*{0.25cm}
\begin{figure}[htp]
  \centering
  \vspace{-1em}
  {\includegraphics[width=1\textwidth]{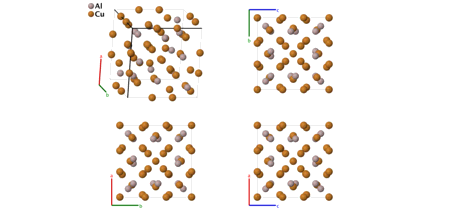}}
\end{figure}
\vspace*{-0.5cm}
\renewcommand{\arraystretch}{1.5}
\begin{equation*}
  \begin{array}{>{$\hspace{-0.15cm}}l<{$}>{$}p{0.5cm}<{$}>{$}p{18.5cm}<{$}}
    \mbox{\large \textbf{Prototype}} &\colon & \ce{$\gamma$-Cu9Al4} \\
    \mbox{\large \textbf{\AFLOW\ prototype label}} &\colon & \mbox{A4B9\_cP52\_215\_ei\_3efgi} \\
    \mbox{\large \textbf{\textit{Strukturbericht} designation}} &\colon & \mbox{$D8_{3}$} \\
    \mbox{\large \textbf{Pearson symbol}} &\colon & \mbox{cP52} \\
    \mbox{\large \textbf{Space group number}} &\colon & 215 \\
    \mbox{\large \textbf{Space group symbol}} &\colon & P\bar{4}3m \\
    \mbox{\large \textbf{\AFLOW\ prototype command}} &\colon &  \texttt{aflow} \,  \, \texttt{-{}-proto=A4B9\_cP52\_215\_ei\_3efgi } \, \newline \texttt{-{}-params=}{a,x_{1},x_{2},x_{3},x_{4},x_{5},x_{6},x_{7},z_{7},x_{8},z_{8} }
  \end{array}
\end{equation*}
\renewcommand{\arraystretch}{1.0}

\vspace*{-0.25cm}
\noindent \hrulefill
\\
\textbf{ Other compounds with this structure:}
\begin{itemize}
   \item{ Cu$_{9}$Ga$_{4}$. (Pearson, 1958), pp.~252, gives a list of compounds which can take on the $D8_{1}$, $D8_{2}$, or $D8_{3}$ structure, depending on the exact composition.  }
\end{itemize}
\vspace*{-0.25cm}
\noindent \hrulefill
\begin{itemize}
  \item{(Arnberg, 1978) give the Wyckoff positions of the Cu IV and Cu V atoms
as (6g) $(x , 1/2 , 1/2)$, but give the coordinates in the form
$(x , 0 , 0)$ corresponding to the (6f) site.  
(Stokhuyzen, 1974) used
(6f) for both types of atoms in the isostructural system
Ga$_{9}$Al$_{4}$.  (Pearson, 1958) places the Cu IV atoms on a (6f)
site and Cu V on (6g), but does not give explicit coordinates.}
  \item{Placing the Cu V atoms on (6f) sites yields an interatomic distance of
1.8\AA.  
This contradicts (Arnberg, 1978), who say that the minimum
interatomic distance is 2.48~\AA\ between the Cu IV and Cu V atoms.
Placing the Cu V atoms on (6g) sites gives this distance, in agreement
with (Pearson, 1958), so we make this choice for the crystal
structure.
This is a variety of $\gamma$-brass comparable to
the \href{http://aflow.org/CrystalDatabase/A5B8_cI52_217_ce_cg.html}{$D8_{2}$ structure}.  
In fact, if we
\begin{itemize}
\item Replace the Al and Cu III atoms by Zn, while setting $x_{4} = x_{1} +
1/2$ ,
\item Replace the Al II and Cu VI atoms by Zn, with $x_{8} = x_{7} +
1/2$ and $z_{8} = z_{7} + 1/2$,
\item Set $x_{3} = x_{2} + 1/2$ and
\item Set $x_{6} = x_{5} + 1/2$,
\end{itemize}
then this structure is identical to $D8_{2}$ $\gamma$-brass.
}
\end{itemize}

\noindent \parbox{1 \linewidth}{
\noindent \hrulefill
\\
\textbf{Simple Cubic primitive vectors:} \\
\vspace*{-0.25cm}
\begin{tabular}{cc}
  \begin{tabular}{c}
    \parbox{0.6 \linewidth}{
      \renewcommand{\arraystretch}{1.5}
      \begin{equation*}
        \centering
        \begin{array}{ccc}
              \mathbf{a}_1 & = & a \, \mathbf{\hat{x}} \\
    \mathbf{a}_2 & = & a \, \mathbf{\hat{y}} \\
    \mathbf{a}_3 & = & a \, \mathbf{\hat{z}} \\

        \end{array}
      \end{equation*}
    }
    \renewcommand{\arraystretch}{1.0}
  \end{tabular}
  \begin{tabular}{c}
    \includegraphics[width=0.3\linewidth]{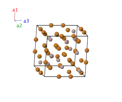} \\
  \end{tabular}
\end{tabular}

}
\vspace*{-0.25cm}

\noindent \hrulefill
\\
\textbf{Basis vectors:}
\vspace*{-0.25cm}
\renewcommand{\arraystretch}{1.5}
\begin{longtabu} to \textwidth{>{\centering $}X[-1,c,c]<{$}>{\centering $}X[-1,c,c]<{$}>{\centering $}X[-1,c,c]<{$}>{\centering $}X[-1,c,c]<{$}>{\centering $}X[-1,c,c]<{$}>{\centering $}X[-1,c,c]<{$}>{\centering $}X[-1,c,c]<{$}}
  & & \mbox{Lattice Coordinates} & & \mbox{Cartesian Coordinates} &\mbox{Wyckoff Position} & \mbox{Atom Type} \\  
  \mathbf{B}_{1} & = & x_{1} \, \mathbf{a}_{1} + x_{1} \, \mathbf{a}_{2} + x_{1} \, \mathbf{a}_{3} & = & x_{1}a \, \mathbf{\hat{x}} + x_{1}a \, \mathbf{\hat{y}} + x_{1}a \, \mathbf{\hat{z}} & \left(4e\right) & \mbox{Al I} \\ 
\mathbf{B}_{2} & = & -x_{1} \, \mathbf{a}_{1}-x_{1} \, \mathbf{a}_{2} + x_{1} \, \mathbf{a}_{3} & = & -x_{1}a \, \mathbf{\hat{x}}-x_{1}a \, \mathbf{\hat{y}} + x_{1}a \, \mathbf{\hat{z}} & \left(4e\right) & \mbox{Al I} \\ 
\mathbf{B}_{3} & = & -x_{1} \, \mathbf{a}_{1} + x_{1} \, \mathbf{a}_{2}-x_{1} \, \mathbf{a}_{3} & = & -x_{1}a \, \mathbf{\hat{x}} + x_{1}a \, \mathbf{\hat{y}}-x_{1}a \, \mathbf{\hat{z}} & \left(4e\right) & \mbox{Al I} \\ 
\mathbf{B}_{4} & = & x_{1} \, \mathbf{a}_{1}-x_{1} \, \mathbf{a}_{2}-x_{1} \, \mathbf{a}_{3} & = & x_{1}a \, \mathbf{\hat{x}}-x_{1}a \, \mathbf{\hat{y}}-x_{1}a \, \mathbf{\hat{z}} & \left(4e\right) & \mbox{Al I} \\ 
\mathbf{B}_{5} & = & x_{2} \, \mathbf{a}_{1} + x_{2} \, \mathbf{a}_{2} + x_{2} \, \mathbf{a}_{3} & = & x_{2}a \, \mathbf{\hat{x}} + x_{2}a \, \mathbf{\hat{y}} + x_{2}a \, \mathbf{\hat{z}} & \left(4e\right) & \mbox{Cu I} \\ 
\mathbf{B}_{6} & = & -x_{2} \, \mathbf{a}_{1}-x_{2} \, \mathbf{a}_{2} + x_{2} \, \mathbf{a}_{3} & = & -x_{2}a \, \mathbf{\hat{x}}-x_{2}a \, \mathbf{\hat{y}} + x_{2}a \, \mathbf{\hat{z}} & \left(4e\right) & \mbox{Cu I} \\ 
\mathbf{B}_{7} & = & -x_{2} \, \mathbf{a}_{1} + x_{2} \, \mathbf{a}_{2}-x_{2} \, \mathbf{a}_{3} & = & -x_{2}a \, \mathbf{\hat{x}} + x_{2}a \, \mathbf{\hat{y}}-x_{2}a \, \mathbf{\hat{z}} & \left(4e\right) & \mbox{Cu I} \\ 
\mathbf{B}_{8} & = & x_{2} \, \mathbf{a}_{1}-x_{2} \, \mathbf{a}_{2}-x_{2} \, \mathbf{a}_{3} & = & x_{2}a \, \mathbf{\hat{x}}-x_{2}a \, \mathbf{\hat{y}}-x_{2}a \, \mathbf{\hat{z}} & \left(4e\right) & \mbox{Cu I} \\ 
\mathbf{B}_{9} & = & x_{3} \, \mathbf{a}_{1} + x_{3} \, \mathbf{a}_{2} + x_{3} \, \mathbf{a}_{3} & = & x_{3}a \, \mathbf{\hat{x}} + x_{3}a \, \mathbf{\hat{y}} + x_{3}a \, \mathbf{\hat{z}} & \left(4e\right) & \mbox{Cu II} \\ 
\mathbf{B}_{10} & = & -x_{3} \, \mathbf{a}_{1}-x_{3} \, \mathbf{a}_{2} + x_{3} \, \mathbf{a}_{3} & = & -x_{3}a \, \mathbf{\hat{x}}-x_{3}a \, \mathbf{\hat{y}} + x_{3}a \, \mathbf{\hat{z}} & \left(4e\right) & \mbox{Cu II} \\ 
\mathbf{B}_{11} & = & -x_{3} \, \mathbf{a}_{1} + x_{3} \, \mathbf{a}_{2}-x_{3} \, \mathbf{a}_{3} & = & -x_{3}a \, \mathbf{\hat{x}} + x_{3}a \, \mathbf{\hat{y}}-x_{3}a \, \mathbf{\hat{z}} & \left(4e\right) & \mbox{Cu II} \\ 
\mathbf{B}_{12} & = & x_{3} \, \mathbf{a}_{1}-x_{3} \, \mathbf{a}_{2}-x_{3} \, \mathbf{a}_{3} & = & x_{3}a \, \mathbf{\hat{x}}-x_{3}a \, \mathbf{\hat{y}}-x_{3}a \, \mathbf{\hat{z}} & \left(4e\right) & \mbox{Cu II} \\ 
\mathbf{B}_{13} & = & x_{4} \, \mathbf{a}_{1} + x_{4} \, \mathbf{a}_{2} + x_{4} \, \mathbf{a}_{3} & = & x_{4}a \, \mathbf{\hat{x}} + x_{4}a \, \mathbf{\hat{y}} + x_{4}a \, \mathbf{\hat{z}} & \left(4e\right) & \mbox{Cu III} \\ 
\mathbf{B}_{14} & = & -x_{4} \, \mathbf{a}_{1}-x_{4} \, \mathbf{a}_{2} + x_{4} \, \mathbf{a}_{3} & = & -x_{4}a \, \mathbf{\hat{x}}-x_{4}a \, \mathbf{\hat{y}} + x_{4}a \, \mathbf{\hat{z}} & \left(4e\right) & \mbox{Cu III} \\ 
\mathbf{B}_{15} & = & -x_{4} \, \mathbf{a}_{1} + x_{4} \, \mathbf{a}_{2}-x_{4} \, \mathbf{a}_{3} & = & -x_{4}a \, \mathbf{\hat{x}} + x_{4}a \, \mathbf{\hat{y}}-x_{4}a \, \mathbf{\hat{z}} & \left(4e\right) & \mbox{Cu III} \\ 
\mathbf{B}_{16} & = & x_{4} \, \mathbf{a}_{1}-x_{4} \, \mathbf{a}_{2}-x_{4} \, \mathbf{a}_{3} & = & x_{4}a \, \mathbf{\hat{x}}-x_{4}a \, \mathbf{\hat{y}}-x_{4}a \, \mathbf{\hat{z}} & \left(4e\right) & \mbox{Cu III} \\ 
\mathbf{B}_{17} & = & x_{5} \, \mathbf{a}_{1} & = & x_{5}a \, \mathbf{\hat{x}} & \left(6f\right) & \mbox{Cu IV} \\ 
\mathbf{B}_{18} & = & -x_{5} \, \mathbf{a}_{1} & = & -x_{5}a \, \mathbf{\hat{x}} & \left(6f\right) & \mbox{Cu IV} \\ 
\mathbf{B}_{19} & = & x_{5} \, \mathbf{a}_{2} & = & x_{5}a \, \mathbf{\hat{y}} & \left(6f\right) & \mbox{Cu IV} \\ 
\mathbf{B}_{20} & = & -x_{5} \, \mathbf{a}_{2} & = & -x_{5}a \, \mathbf{\hat{y}} & \left(6f\right) & \mbox{Cu IV} \\ 
\mathbf{B}_{21} & = & x_{5} \, \mathbf{a}_{3} & = & x_{5}a \, \mathbf{\hat{z}} & \left(6f\right) & \mbox{Cu IV} \\ 
\mathbf{B}_{22} & = & -x_{5} \, \mathbf{a}_{3} & = & -x_{5}a \, \mathbf{\hat{z}} & \left(6f\right) & \mbox{Cu IV} \\ 
\mathbf{B}_{23} & = & x_{6} \, \mathbf{a}_{1} + \frac{1}{2} \, \mathbf{a}_{2} + \frac{1}{2} \, \mathbf{a}_{3} & = & x_{6}a \, \mathbf{\hat{x}} + \frac{1}{2}a \, \mathbf{\hat{y}} + \frac{1}{2}a \, \mathbf{\hat{z}} & \left(6g\right) & \mbox{Cu V} \\ 
\mathbf{B}_{24} & = & -x_{6} \, \mathbf{a}_{1} + \frac{1}{2} \, \mathbf{a}_{2} + \frac{1}{2} \, \mathbf{a}_{3} & = & -x_{6}a \, \mathbf{\hat{x}} + \frac{1}{2}a \, \mathbf{\hat{y}} + \frac{1}{2}a \, \mathbf{\hat{z}} & \left(6g\right) & \mbox{Cu V} \\ 
\mathbf{B}_{25} & = & \frac{1}{2} \, \mathbf{a}_{1} + x_{6} \, \mathbf{a}_{2} + \frac{1}{2} \, \mathbf{a}_{3} & = & \frac{1}{2}a \, \mathbf{\hat{x}} + x_{6}a \, \mathbf{\hat{y}} + \frac{1}{2}a \, \mathbf{\hat{z}} & \left(6g\right) & \mbox{Cu V} \\ 
\mathbf{B}_{26} & = & \frac{1}{2} \, \mathbf{a}_{1}-x_{6} \, \mathbf{a}_{2} + \frac{1}{2} \, \mathbf{a}_{3} & = & \frac{1}{2}a \, \mathbf{\hat{x}}-x_{6}a \, \mathbf{\hat{y}} + \frac{1}{2}a \, \mathbf{\hat{z}} & \left(6g\right) & \mbox{Cu V} \\ 
\mathbf{B}_{27} & = & \frac{1}{2} \, \mathbf{a}_{1} + \frac{1}{2} \, \mathbf{a}_{2} + x_{6} \, \mathbf{a}_{3} & = & \frac{1}{2}a \, \mathbf{\hat{x}} + \frac{1}{2}a \, \mathbf{\hat{y}} + x_{6}a \, \mathbf{\hat{z}} & \left(6g\right) & \mbox{Cu V} \\ 
\mathbf{B}_{28} & = & \frac{1}{2} \, \mathbf{a}_{1} + \frac{1}{2} \, \mathbf{a}_{2}-x_{6} \, \mathbf{a}_{3} & = & \frac{1}{2}a \, \mathbf{\hat{x}} + \frac{1}{2}a \, \mathbf{\hat{y}}-x_{6}a \, \mathbf{\hat{z}} & \left(6g\right) & \mbox{Cu V} \\ 
\mathbf{B}_{29} & = & x_{7} \, \mathbf{a}_{1} + x_{7} \, \mathbf{a}_{2} + z_{7} \, \mathbf{a}_{3} & = & x_{7}a \, \mathbf{\hat{x}} + x_{7}a \, \mathbf{\hat{y}} + z_{7}a \, \mathbf{\hat{z}} & \left(12i\right) & \mbox{Al II} \\ 
\mathbf{B}_{30} & = & -x_{7} \, \mathbf{a}_{1}-x_{7} \, \mathbf{a}_{2} + z_{7} \, \mathbf{a}_{3} & = & -x_{7}a \, \mathbf{\hat{x}}-x_{7}a \, \mathbf{\hat{y}} + z_{7}a \, \mathbf{\hat{z}} & \left(12i\right) & \mbox{Al II} \\ 
\mathbf{B}_{31} & = & -x_{7} \, \mathbf{a}_{1} + x_{7} \, \mathbf{a}_{2}-z_{7} \, \mathbf{a}_{3} & = & -x_{7}a \, \mathbf{\hat{x}} + x_{7}a \, \mathbf{\hat{y}}-z_{7}a \, \mathbf{\hat{z}} & \left(12i\right) & \mbox{Al II} \\ 
\mathbf{B}_{32} & = & x_{7} \, \mathbf{a}_{1}-x_{7} \, \mathbf{a}_{2}-z_{7} \, \mathbf{a}_{3} & = & x_{7}a \, \mathbf{\hat{x}}-x_{7}a \, \mathbf{\hat{y}}-z_{7}a \, \mathbf{\hat{z}} & \left(12i\right) & \mbox{Al II} \\ 
\mathbf{B}_{33} & = & z_{7} \, \mathbf{a}_{1} + x_{7} \, \mathbf{a}_{2} + x_{7} \, \mathbf{a}_{3} & = & z_{7}a \, \mathbf{\hat{x}} + x_{7}a \, \mathbf{\hat{y}} + x_{7}a \, \mathbf{\hat{z}} & \left(12i\right) & \mbox{Al II} \\ 
\mathbf{B}_{34} & = & z_{7} \, \mathbf{a}_{1}-x_{7} \, \mathbf{a}_{2}-x_{7} \, \mathbf{a}_{3} & = & z_{7}a \, \mathbf{\hat{x}}-x_{7}a \, \mathbf{\hat{y}}-x_{7}a \, \mathbf{\hat{z}} & \left(12i\right) & \mbox{Al II} \\ 
\mathbf{B}_{35} & = & -z_{7} \, \mathbf{a}_{1}-x_{7} \, \mathbf{a}_{2} + x_{7} \, \mathbf{a}_{3} & = & -z_{7}a \, \mathbf{\hat{x}}-x_{7}a \, \mathbf{\hat{y}} + x_{7}a \, \mathbf{\hat{z}} & \left(12i\right) & \mbox{Al II} \\ 
\mathbf{B}_{36} & = & -z_{7} \, \mathbf{a}_{1} + x_{7} \, \mathbf{a}_{2}-x_{7} \, \mathbf{a}_{3} & = & -z_{7}a \, \mathbf{\hat{x}} + x_{7}a \, \mathbf{\hat{y}}-x_{7}a \, \mathbf{\hat{z}} & \left(12i\right) & \mbox{Al II} \\ 
\mathbf{B}_{37} & = & x_{7} \, \mathbf{a}_{1} + z_{7} \, \mathbf{a}_{2} + x_{7} \, \mathbf{a}_{3} & = & x_{7}a \, \mathbf{\hat{x}} + z_{7}a \, \mathbf{\hat{y}} + x_{7}a \, \mathbf{\hat{z}} & \left(12i\right) & \mbox{Al II} \\ 
\mathbf{B}_{38} & = & -x_{7} \, \mathbf{a}_{1} + z_{7} \, \mathbf{a}_{2}-x_{7} \, \mathbf{a}_{3} & = & -x_{7}a \, \mathbf{\hat{x}} + z_{7}a \, \mathbf{\hat{y}}-x_{7}a \, \mathbf{\hat{z}} & \left(12i\right) & \mbox{Al II} \\ 
\mathbf{B}_{39} & = & x_{7} \, \mathbf{a}_{1}-z_{7} \, \mathbf{a}_{2}-x_{7} \, \mathbf{a}_{3} & = & x_{7}a \, \mathbf{\hat{x}}-z_{7}a \, \mathbf{\hat{y}}-x_{7}a \, \mathbf{\hat{z}} & \left(12i\right) & \mbox{Al II} \\ 
\mathbf{B}_{40} & = & -x_{7} \, \mathbf{a}_{1}-z_{7} \, \mathbf{a}_{2} + x_{7} \, \mathbf{a}_{3} & = & -x_{7}a \, \mathbf{\hat{x}}-z_{7}a \, \mathbf{\hat{y}} + x_{7}a \, \mathbf{\hat{z}} & \left(12i\right) & \mbox{Al II} \\ 
\mathbf{B}_{41} & = & x_{8} \, \mathbf{a}_{1} + x_{8} \, \mathbf{a}_{2} + z_{8} \, \mathbf{a}_{3} & = & x_{8}a \, \mathbf{\hat{x}} + x_{8}a \, \mathbf{\hat{y}} + z_{8}a \, \mathbf{\hat{z}} & \left(12i\right) & \mbox{Cu VI} \\ 
\mathbf{B}_{42} & = & -x_{8} \, \mathbf{a}_{1}-x_{8} \, \mathbf{a}_{2} + z_{8} \, \mathbf{a}_{3} & = & -x_{8}a \, \mathbf{\hat{x}}-x_{8}a \, \mathbf{\hat{y}} + z_{8}a \, \mathbf{\hat{z}} & \left(12i\right) & \mbox{Cu VI} \\ 
\mathbf{B}_{43} & = & -x_{8} \, \mathbf{a}_{1} + x_{8} \, \mathbf{a}_{2}-z_{8} \, \mathbf{a}_{3} & = & -x_{8}a \, \mathbf{\hat{x}} + x_{8}a \, \mathbf{\hat{y}}-z_{8}a \, \mathbf{\hat{z}} & \left(12i\right) & \mbox{Cu VI} \\ 
\mathbf{B}_{44} & = & x_{8} \, \mathbf{a}_{1}-x_{8} \, \mathbf{a}_{2}-z_{8} \, \mathbf{a}_{3} & = & x_{8}a \, \mathbf{\hat{x}}-x_{8}a \, \mathbf{\hat{y}}-z_{8}a \, \mathbf{\hat{z}} & \left(12i\right) & \mbox{Cu VI} \\ 
\mathbf{B}_{45} & = & z_{8} \, \mathbf{a}_{1} + x_{8} \, \mathbf{a}_{2} + x_{8} \, \mathbf{a}_{3} & = & z_{8}a \, \mathbf{\hat{x}} + x_{8}a \, \mathbf{\hat{y}} + x_{8}a \, \mathbf{\hat{z}} & \left(12i\right) & \mbox{Cu VI} \\ 
\mathbf{B}_{46} & = & z_{8} \, \mathbf{a}_{1}-x_{8} \, \mathbf{a}_{2}-x_{8} \, \mathbf{a}_{3} & = & z_{8}a \, \mathbf{\hat{x}}-x_{8}a \, \mathbf{\hat{y}}-x_{8}a \, \mathbf{\hat{z}} & \left(12i\right) & \mbox{Cu VI} \\ 
\mathbf{B}_{47} & = & -z_{8} \, \mathbf{a}_{1}-x_{8} \, \mathbf{a}_{2} + x_{8} \, \mathbf{a}_{3} & = & -z_{8}a \, \mathbf{\hat{x}}-x_{8}a \, \mathbf{\hat{y}} + x_{8}a \, \mathbf{\hat{z}} & \left(12i\right) & \mbox{Cu VI} \\ 
\mathbf{B}_{48} & = & -z_{8} \, \mathbf{a}_{1} + x_{8} \, \mathbf{a}_{2}-x_{8} \, \mathbf{a}_{3} & = & -z_{8}a \, \mathbf{\hat{x}} + x_{8}a \, \mathbf{\hat{y}}-x_{8}a \, \mathbf{\hat{z}} & \left(12i\right) & \mbox{Cu VI} \\ 
\mathbf{B}_{49} & = & x_{8} \, \mathbf{a}_{1} + z_{8} \, \mathbf{a}_{2} + x_{8} \, \mathbf{a}_{3} & = & x_{8}a \, \mathbf{\hat{x}} + z_{8}a \, \mathbf{\hat{y}} + x_{8}a \, \mathbf{\hat{z}} & \left(12i\right) & \mbox{Cu VI} \\ 
\mathbf{B}_{50} & = & -x_{8} \, \mathbf{a}_{1} + z_{8} \, \mathbf{a}_{2}-x_{8} \, \mathbf{a}_{3} & = & -x_{8}a \, \mathbf{\hat{x}} + z_{8}a \, \mathbf{\hat{y}}-x_{8}a \, \mathbf{\hat{z}} & \left(12i\right) & \mbox{Cu VI} \\ 
\mathbf{B}_{51} & = & x_{8} \, \mathbf{a}_{1}-z_{8} \, \mathbf{a}_{2}-x_{8} \, \mathbf{a}_{3} & = & x_{8}a \, \mathbf{\hat{x}}-z_{8}a \, \mathbf{\hat{y}}-x_{8}a \, \mathbf{\hat{z}} & \left(12i\right) & \mbox{Cu VI} \\ 
\mathbf{B}_{52} & = & -x_{8} \, \mathbf{a}_{1}-z_{8} \, \mathbf{a}_{2} + x_{8} \, \mathbf{a}_{3} & = & -x_{8}a \, \mathbf{\hat{x}}-z_{8}a \, \mathbf{\hat{y}} + x_{8}a \, \mathbf{\hat{z}} & \left(12i\right) & \mbox{Cu VI} \\ 
\end{longtabu}
\renewcommand{\arraystretch}{1.0}
\noindent \hrulefill
\\
\textbf{References:}
\vspace*{-0.25cm}
\begin{flushleft}
  - \bibentry{Arnberg_Acta_Cryst_A_34_1978}. \\
  - \bibentry{Stokhuyzen_Acta_Cryst_B_30_1974}. \\
  - \bibentry{Pearson_NRC_1958}. \\
\end{flushleft}
\textbf{Found in:}
\vspace*{-0.25cm}
\begin{flushleft}
  - \bibentry{Villars_LB_2005}. \\
\end{flushleft}
\noindent \hrulefill
\\
\textbf{Geometry files:}
\\
\noindent  - CIF: pp. {\hyperref[A4B9_cP52_215_ei_3efgi_cif]{\pageref{A4B9_cP52_215_ei_3efgi_cif}}} \\
\noindent  - POSCAR: pp. {\hyperref[A4B9_cP52_215_ei_3efgi_poscar]{\pageref{A4B9_cP52_215_ei_3efgi_poscar}}} \\
\onecolumn
{\phantomsection\label{ABCD_cF16_216_c_d_b_a}}
\subsection*{\huge \textbf{{\normalfont \begin{raggedleft}Quartenary Heusler (LiMgAuSn) Structure: \end{raggedleft} \\ ABCD\_cF16\_216\_c\_d\_b\_a}}}
\noindent \hrulefill
\vspace*{0.25cm}
\begin{figure}[htp]
  \centering
  \vspace{-1em}
  {\includegraphics[width=1\textwidth]{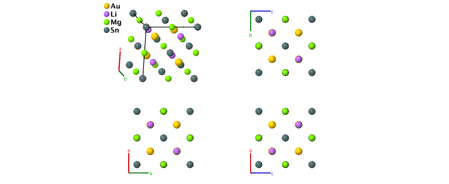}}
\end{figure}
\vspace*{-0.5cm}
\renewcommand{\arraystretch}{1.5}
\begin{equation*}
  \begin{array}{>{$\hspace{-0.15cm}}l<{$}>{$}p{0.5cm}<{$}>{$}p{18.5cm}<{$}}
    \mbox{\large \textbf{Prototype}} &\colon & \ce{LiMgAuSn} \\
    \mbox{\large \textbf{\AFLOW\ prototype label}} &\colon & \mbox{ABCD\_cF16\_216\_c\_d\_b\_a} \\
    \mbox{\large \textbf{\textit{Strukturbericht} designation}} &\colon & \mbox{None} \\
    \mbox{\large \textbf{Pearson symbol}} &\colon & \mbox{cF16} \\
    \mbox{\large \textbf{Space group number}} &\colon & 216 \\
    \mbox{\large \textbf{Space group symbol}} &\colon & F\bar{4}3m \\
    \mbox{\large \textbf{\AFLOW\ prototype command}} &\colon &  \texttt{aflow} \,  \, \texttt{-{}-proto=ABCD\_cF16\_216\_c\_d\_b\_a } \, \newline \texttt{-{}-params=}{a }
  \end{array}
\end{equation*}
\renewcommand{\arraystretch}{1.0}

\vspace*{-0.25cm}
\noindent \hrulefill
\\
\textbf{ Other compounds with this structure:}
\begin{itemize}
   \item{ AuLiMgSn, CuMg$_{2}$Ti, AuBiLi$_{2}$, AgLi$_{2}$Sn, AuLi$_{2}$Sn, CuHfHg$_{2}$, MnPd$_{2}$Sn, and many more.  See (Eberz, 1980).  }
\end{itemize}
\vspace*{-0.25cm}
\noindent \hrulefill
\begin{itemize}
  \item{This ``quaternary-Heusler'' structure can be considered as the parent
of a wide variety of structures, depending on the occupancy of the
(4a), (4b), (4c), and (4d) Wyckoff positions.  Consider atoms of type
A, B, C, D, distributed in this structure.  By placing these atoms on
the appropriate Wyckoff positions we find the following structures:
\begin{center}
\begin{tabular}{|c|c|c|c|c|c|c|}
  \hline
  Structure & {\em Strukturbericht} & \AFLOW\ label & (4a) & (4b) & (4c) & (4d) \\
  \hline
  \hline
  \href{http://aflow.org/CrystalDatabase/A_cP1_221_a.html}{simple cubic} & $A_{h}$ & A\_cP1\_221\_a & A & A & - & - \\
  \hline
  \href{http://aflow.org/CrystalDatabase/A_cF4_225_a.html}{fcc} & $A1$ & A\_cF4\_225\_a & A & - & - & - \\
  \hline
  \href{http://aflow.org/CrystalDatabase/A_cI2_229_a.html}{bcc} & $A2$ & A\_cI2\_229\_a & A & A & - & - \\
  \hline
  \href{http://aflow.org/CrystalDatabase/A_cF8_227_a.html}{diamond} & $A4$ & A\_cF8\_227\_a & A & - & A & - \\
  \hline
  \href{http://aflow.org/CrystalDatabase/AB_cF8_225_a_b.html}{NaCl} & $B1$ & AB\_cF8\_225\_a\_b & A & B & - & - \\
  \hline
  \href{http://aflow.org/CrystalDatabase/AB_cP2_221_a_b.html}{CsCl} & $B2$ & AB\_cP2\_221\_a\_b & A & B & - & - \\
  \hline
  \href{http://aflow.org/CrystalDatabase/AB_cF8_216_c_a.html}{ZnS (zincblende)} & $B3$ & AB\_cF8\_216\_c\_a & B & - & A & - \\
  \hline
  \href{http://aflow.org/CrystalDatabase/ABC_cF12_216_b_c_a.html}{half-Heusler} & $C1_{b}$ & ABC\_cF12\_216\_b\_c\_a & C & A & B & - \\
  \hline
  \href{http://aflow.org/CrystalDatabase/AB2C_cF16_225_a_c_b.html}{Heusler} & $L2_{1}$ & AB2C\_cF16\_225\_a\_c\_b & A & C & B & B  \\
  \hline
\end{tabular}
\end{center}
The ordering of this structure is somewhat arbitary.  So long as Sn
and Mg are on either the (4a)/(4b) or (4c)/(4d) sites, with Au and Li
on the opposite sites, we will get the same structure.
}
\end{itemize}

\noindent \parbox{1 \linewidth}{
\noindent \hrulefill
\\
\textbf{Face-centered Cubic primitive vectors:} \\
\vspace*{-0.25cm}
\begin{tabular}{cc}
  \begin{tabular}{c}
    \parbox{0.6 \linewidth}{
      \renewcommand{\arraystretch}{1.5}
      \begin{equation*}
        \centering
        \begin{array}{ccc}
              \mathbf{a}_1 & = & \frac12 \, a \, \mathbf{\hat{y}} + \frac12 \, a \, \mathbf{\hat{z}} \\
    \mathbf{a}_2 & = & \frac12 \, a \, \mathbf{\hat{x}} + \frac12 \, a \, \mathbf{\hat{z}} \\
    \mathbf{a}_3 & = & \frac12 \, a \, \mathbf{\hat{x}} + \frac12 \, a \, \mathbf{\hat{y}} \\

        \end{array}
      \end{equation*}
    }
    \renewcommand{\arraystretch}{1.0}
  \end{tabular}
  \begin{tabular}{c}
    \includegraphics[width=0.3\linewidth]{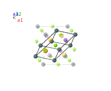} \\
  \end{tabular}
\end{tabular}

}
\vspace*{-0.25cm}

\noindent \hrulefill
\\
\textbf{Basis vectors:}
\vspace*{-0.25cm}
\renewcommand{\arraystretch}{1.5}
\begin{longtabu} to \textwidth{>{\centering $}X[-1,c,c]<{$}>{\centering $}X[-1,c,c]<{$}>{\centering $}X[-1,c,c]<{$}>{\centering $}X[-1,c,c]<{$}>{\centering $}X[-1,c,c]<{$}>{\centering $}X[-1,c,c]<{$}>{\centering $}X[-1,c,c]<{$}}
  & & \mbox{Lattice Coordinates} & & \mbox{Cartesian Coordinates} &\mbox{Wyckoff Position} & \mbox{Atom Type} \\  
  \mathbf{B}_{1} & = & 0 \, \mathbf{a}_{1} + 0 \, \mathbf{a}_{2} + 0 \, \mathbf{a}_{3} & = & 0 \, \mathbf{\hat{x}} + 0 \, \mathbf{\hat{y}} + 0 \, \mathbf{\hat{z}} & \left(4a\right) & \mbox{Sn} \\ 
\mathbf{B}_{2} & = & \frac{1}{2} \, \mathbf{a}_{1} + \frac{1}{2} \, \mathbf{a}_{2} + \frac{1}{2} \, \mathbf{a}_{3} & = & \frac{1}{2}a \, \mathbf{\hat{x}} + \frac{1}{2}a \, \mathbf{\hat{y}} + \frac{1}{2}a \, \mathbf{\hat{z}} & \left(4b\right) & \mbox{Mg} \\ 
\mathbf{B}_{3} & = & \frac{1}{4} \, \mathbf{a}_{1} + \frac{1}{4} \, \mathbf{a}_{2} + \frac{1}{4} \, \mathbf{a}_{3} & = & \frac{1}{4}a \, \mathbf{\hat{x}} + \frac{1}{4}a \, \mathbf{\hat{y}} + \frac{1}{4}a \, \mathbf{\hat{z}} & \left(4c\right) & \mbox{Au} \\ 
\mathbf{B}_{4} & = & \frac{3}{4} \, \mathbf{a}_{1} + \frac{3}{4} \, \mathbf{a}_{2} + \frac{3}{4} \, \mathbf{a}_{3} & = & \frac{3}{4}a \, \mathbf{\hat{x}} + \frac{3}{4}a \, \mathbf{\hat{y}} + \frac{3}{4}a \, \mathbf{\hat{z}} & \left(4d\right) & \mbox{Li} \\ 
\end{longtabu}
\renewcommand{\arraystretch}{1.0}
\noindent \hrulefill
\\
\textbf{References:}
\vspace*{-0.25cm}
\begin{flushleft}
  - \bibentry{Eberz_ZfNaturfB_35_1341_1980}. \\
\end{flushleft}
\textbf{Found in:}
\vspace*{-0.25cm}
\begin{flushleft}
  - \bibentry{Villars_Pauling_File_LiMgPdSn_2016}. \\
\end{flushleft}
\noindent \hrulefill
\\
\textbf{Geometry files:}
\\
\noindent  - CIF: pp. {\hyperref[ABCD_cF16_216_c_d_b_a_cif]{\pageref{ABCD_cF16_216_c_d_b_a_cif}}} \\
\noindent  - POSCAR: pp. {\hyperref[ABCD_cF16_216_c_d_b_a_poscar]{\pageref{ABCD_cF16_216_c_d_b_a_poscar}}} \\
\onecolumn
{\phantomsection\label{A3B4C_cP16_218_c_e_a}}
\subsection*{\huge \textbf{{\normalfont Ag$_{3}$[PO$_{4}$] Structure: A3B4C\_cP16\_218\_c\_e\_a}}}
\noindent \hrulefill
\vspace*{0.25cm}
\begin{figure}[htp]
  \centering
  \vspace{-1em}
  {\includegraphics[width=1\textwidth]{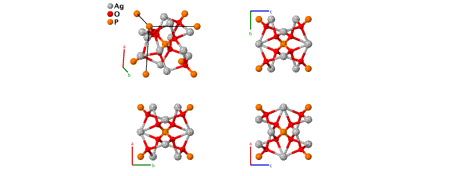}}
\end{figure}
\vspace*{-0.5cm}
\renewcommand{\arraystretch}{1.5}
\begin{equation*}
  \begin{array}{>{$\hspace{-0.15cm}}l<{$}>{$}p{0.5cm}<{$}>{$}p{18.5cm}<{$}}
    \mbox{\large \textbf{Prototype}} &\colon & \ce{Ag3[PO4]} \\
    \mbox{\large \textbf{\AFLOW\ prototype label}} &\colon & \mbox{A3B4C\_cP16\_218\_c\_e\_a} \\
    \mbox{\large \textbf{\textit{Strukturbericht} designation}} &\colon & \mbox{None} \\
    \mbox{\large \textbf{Pearson symbol}} &\colon & \mbox{cP16} \\
    \mbox{\large \textbf{Space group number}} &\colon & 218 \\
    \mbox{\large \textbf{Space group symbol}} &\colon & P\bar{4}3n \\
    \mbox{\large \textbf{\AFLOW\ prototype command}} &\colon &  \texttt{aflow} \,  \, \texttt{-{}-proto=A3B4C\_cP16\_218\_c\_e\_a } \, \newline \texttt{-{}-params=}{a,x_{3} }
  \end{array}
\end{equation*}
\renewcommand{\arraystretch}{1.0}

\noindent \parbox{1 \linewidth}{
\noindent \hrulefill
\\
\textbf{Simple Cubic primitive vectors:} \\
\vspace*{-0.25cm}
\begin{tabular}{cc}
  \begin{tabular}{c}
    \parbox{0.6 \linewidth}{
      \renewcommand{\arraystretch}{1.5}
      \begin{equation*}
        \centering
        \begin{array}{ccc}
              \mathbf{a}_1 & = & a \, \mathbf{\hat{x}} \\
    \mathbf{a}_2 & = & a \, \mathbf{\hat{y}} \\
    \mathbf{a}_3 & = & a \, \mathbf{\hat{z}} \\

        \end{array}
      \end{equation*}
    }
    \renewcommand{\arraystretch}{1.0}
  \end{tabular}
  \begin{tabular}{c}
    \includegraphics[width=0.3\linewidth]{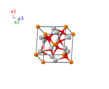} \\
  \end{tabular}
\end{tabular}

}
\vspace*{-0.25cm}

\noindent \hrulefill
\\
\textbf{Basis vectors:}
\vspace*{-0.25cm}
\renewcommand{\arraystretch}{1.5}
\begin{longtabu} to \textwidth{>{\centering $}X[-1,c,c]<{$}>{\centering $}X[-1,c,c]<{$}>{\centering $}X[-1,c,c]<{$}>{\centering $}X[-1,c,c]<{$}>{\centering $}X[-1,c,c]<{$}>{\centering $}X[-1,c,c]<{$}>{\centering $}X[-1,c,c]<{$}}
  & & \mbox{Lattice Coordinates} & & \mbox{Cartesian Coordinates} &\mbox{Wyckoff Position} & \mbox{Atom Type} \\  
  \mathbf{B}_{1} & = & 0 \, \mathbf{a}_{1} + 0 \, \mathbf{a}_{2} + 0 \, \mathbf{a}_{3} & = & 0 \, \mathbf{\hat{x}} + 0 \, \mathbf{\hat{y}} + 0 \, \mathbf{\hat{z}} & \left(2a\right) & \mbox{P} \\ 
\mathbf{B}_{2} & = & \frac{1}{2} \, \mathbf{a}_{1} + \frac{1}{2} \, \mathbf{a}_{2} + \frac{1}{2} \, \mathbf{a}_{3} & = & \frac{1}{2}a \, \mathbf{\hat{x}} + \frac{1}{2}a \, \mathbf{\hat{y}} + \frac{1}{2}a \, \mathbf{\hat{z}} & \left(2a\right) & \mbox{P} \\ 
\mathbf{B}_{3} & = & \frac{1}{4} \, \mathbf{a}_{1} + \frac{1}{2} \, \mathbf{a}_{2} & = & \frac{1}{4}a \, \mathbf{\hat{x}} + \frac{1}{2}a \, \mathbf{\hat{y}} & \left(6c\right) & \mbox{Ag} \\ 
\mathbf{B}_{4} & = & \frac{3}{4} \, \mathbf{a}_{1} + \frac{1}{2} \, \mathbf{a}_{2} & = & \frac{3}{4}a \, \mathbf{\hat{x}} + \frac{1}{2}a \, \mathbf{\hat{y}} & \left(6c\right) & \mbox{Ag} \\ 
\mathbf{B}_{5} & = & \frac{1}{4} \, \mathbf{a}_{2} + \frac{1}{2} \, \mathbf{a}_{3} & = & \frac{1}{4}a \, \mathbf{\hat{y}} + \frac{1}{2}a \, \mathbf{\hat{z}} & \left(6c\right) & \mbox{Ag} \\ 
\mathbf{B}_{6} & = & \frac{3}{4} \, \mathbf{a}_{2} + \frac{1}{2} \, \mathbf{a}_{3} & = & \frac{3}{4}a \, \mathbf{\hat{y}} + \frac{1}{2}a \, \mathbf{\hat{z}} & \left(6c\right) & \mbox{Ag} \\ 
\mathbf{B}_{7} & = & \frac{1}{2} \, \mathbf{a}_{1} + \frac{1}{4} \, \mathbf{a}_{3} & = & \frac{1}{2}a \, \mathbf{\hat{x}} + \frac{1}{4}a \, \mathbf{\hat{z}} & \left(6c\right) & \mbox{Ag} \\ 
\mathbf{B}_{8} & = & \frac{1}{2} \, \mathbf{a}_{1} + \frac{3}{4} \, \mathbf{a}_{3} & = & \frac{1}{2}a \, \mathbf{\hat{x}} + \frac{3}{4}a \, \mathbf{\hat{z}} & \left(6c\right) & \mbox{Ag} \\ 
\mathbf{B}_{9} & = & x_{3} \, \mathbf{a}_{1} + x_{3} \, \mathbf{a}_{2} + x_{3} \, \mathbf{a}_{3} & = & x_{3}a \, \mathbf{\hat{x}} + x_{3}a \, \mathbf{\hat{y}} + x_{3}a \, \mathbf{\hat{z}} & \left(8e\right) & \mbox{O} \\ 
\mathbf{B}_{10} & = & -x_{3} \, \mathbf{a}_{1}-x_{3} \, \mathbf{a}_{2} + x_{3} \, \mathbf{a}_{3} & = & -x_{3}a \, \mathbf{\hat{x}}-x_{3}a \, \mathbf{\hat{y}} + x_{3}a \, \mathbf{\hat{z}} & \left(8e\right) & \mbox{O} \\ 
\mathbf{B}_{11} & = & -x_{3} \, \mathbf{a}_{1} + x_{3} \, \mathbf{a}_{2}-x_{3} \, \mathbf{a}_{3} & = & -x_{3}a \, \mathbf{\hat{x}} + x_{3}a \, \mathbf{\hat{y}}-x_{3}a \, \mathbf{\hat{z}} & \left(8e\right) & \mbox{O} \\ 
\mathbf{B}_{12} & = & x_{3} \, \mathbf{a}_{1}-x_{3} \, \mathbf{a}_{2}-x_{3} \, \mathbf{a}_{3} & = & x_{3}a \, \mathbf{\hat{x}}-x_{3}a \, \mathbf{\hat{y}}-x_{3}a \, \mathbf{\hat{z}} & \left(8e\right) & \mbox{O} \\ 
\mathbf{B}_{13} & = & \left(\frac{1}{2} +x_{3}\right) \, \mathbf{a}_{1} + \left(\frac{1}{2} +x_{3}\right) \, \mathbf{a}_{2} + \left(\frac{1}{2} +x_{3}\right) \, \mathbf{a}_{3} & = & \left(\frac{1}{2} +x_{3}\right)a \, \mathbf{\hat{x}} + \left(\frac{1}{2} +x_{3}\right)a \, \mathbf{\hat{y}} + \left(\frac{1}{2} +x_{3}\right)a \, \mathbf{\hat{z}} & \left(8e\right) & \mbox{O} \\ 
\mathbf{B}_{14} & = & \left(\frac{1}{2} - x_{3}\right) \, \mathbf{a}_{1} + \left(\frac{1}{2} - x_{3}\right) \, \mathbf{a}_{2} + \left(\frac{1}{2} +x_{3}\right) \, \mathbf{a}_{3} & = & \left(\frac{1}{2} - x_{3}\right)a \, \mathbf{\hat{x}} + \left(\frac{1}{2} - x_{3}\right)a \, \mathbf{\hat{y}} + \left(\frac{1}{2} +x_{3}\right)a \, \mathbf{\hat{z}} & \left(8e\right) & \mbox{O} \\ 
\mathbf{B}_{15} & = & \left(\frac{1}{2} +x_{3}\right) \, \mathbf{a}_{1} + \left(\frac{1}{2} - x_{3}\right) \, \mathbf{a}_{2} + \left(\frac{1}{2} - x_{3}\right) \, \mathbf{a}_{3} & = & \left(\frac{1}{2} +x_{3}\right)a \, \mathbf{\hat{x}} + \left(\frac{1}{2} - x_{3}\right)a \, \mathbf{\hat{y}} + \left(\frac{1}{2} - x_{3}\right)a \, \mathbf{\hat{z}} & \left(8e\right) & \mbox{O} \\ 
\mathbf{B}_{16} & = & \left(\frac{1}{2} - x_{3}\right) \, \mathbf{a}_{1} + \left(\frac{1}{2} +x_{3}\right) \, \mathbf{a}_{2} + \left(\frac{1}{2} - x_{3}\right) \, \mathbf{a}_{3} & = & \left(\frac{1}{2} - x_{3}\right)a \, \mathbf{\hat{x}} + \left(\frac{1}{2} +x_{3}\right)a \, \mathbf{\hat{y}} + \left(\frac{1}{2} - x_{3}\right)a \, \mathbf{\hat{z}} & \left(8e\right) & \mbox{O} \\ 
\end{longtabu}
\renewcommand{\arraystretch}{1.0}
\noindent \hrulefill
\\
\textbf{References:}
\vspace*{-0.25cm}
\begin{flushleft}
  - \bibentry{Masse_Ag3PO4_ZKristallogrMat_1976}. \\
\end{flushleft}
\textbf{Found in:}
\vspace*{-0.25cm}
\begin{flushleft}
  - \bibentry{Villars_PearsonsCrystalData_2013}. \\
\end{flushleft}
\noindent \hrulefill
\\
\textbf{Geometry files:}
\\
\noindent  - CIF: pp. {\hyperref[A3B4C_cP16_218_c_e_a_cif]{\pageref{A3B4C_cP16_218_c_e_a_cif}}} \\
\noindent  - POSCAR: pp. {\hyperref[A3B4C_cP16_218_c_e_a_poscar]{\pageref{A3B4C_cP16_218_c_e_a_poscar}}} \\
\onecolumn
{\phantomsection\label{A7BC3D13_cF192_219_de_b_c_ah}}
\subsection*{\huge \textbf{{\normalfont \begin{raggedleft}Boracite (Mg$_{3}$B$_{7}$ClO$_{13}$) Structure: \end{raggedleft} \\ A7BC3D13\_cF192\_219\_de\_b\_c\_ah}}}
\noindent \hrulefill
\vspace*{0.25cm}
\begin{figure}[htp]
  \centering
  \vspace{-1em}
  {\includegraphics[width=1\textwidth]{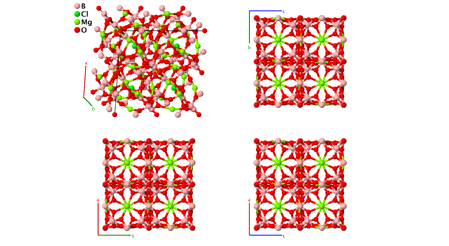}}
\end{figure}
\vspace*{-0.5cm}
\renewcommand{\arraystretch}{1.5}
\begin{equation*}
  \begin{array}{>{$\hspace{-0.15cm}}l<{$}>{$}p{0.5cm}<{$}>{$}p{18.5cm}<{$}}
    \mbox{\large \textbf{Prototype}} &\colon & \ce{Mg3B7ClO13} \\
    \mbox{\large \textbf{\AFLOW\ prototype label}} &\colon & \mbox{A7BC3D13\_cF192\_219\_de\_b\_c\_ah} \\
    \mbox{\large \textbf{\textit{Strukturbericht} designation}} &\colon & \mbox{None} \\
    \mbox{\large \textbf{Pearson symbol}} &\colon & \mbox{cF192} \\
    \mbox{\large \textbf{Space group number}} &\colon & 219 \\
    \mbox{\large \textbf{Space group symbol}} &\colon & F\bar{4}3c \\
    \mbox{\large \textbf{\AFLOW\ prototype command}} &\colon &  \texttt{aflow} \,  \, \texttt{-{}-proto=A7BC3D13\_cF192\_219\_de\_b\_c\_ah } \, \newline \texttt{-{}-params=}{a,x_{5},x_{6},y_{6},z_{6} }
  \end{array}
\end{equation*}
\renewcommand{\arraystretch}{1.0}

\vspace*{-0.25cm}
\noindent \hrulefill
\\
\textbf{ Other compounds with this structure:}
\begin{itemize}
   \item{ $M_{3}$B$_{7}$O$_{13}X$; $M$ = Mg, Cr, Mn, Fe, Co; $X$ = Cl, Br, I  }
\end{itemize}
\vspace*{-0.25cm}
\noindent \hrulefill
\begin{itemize}
  \item{Experimental data was obtained at 400$^\circ$~C.
}
\end{itemize}

\noindent \parbox{1 \linewidth}{
\noindent \hrulefill
\\
\textbf{Face-centered Cubic primitive vectors:} \\
\vspace*{-0.25cm}
\begin{tabular}{cc}
  \begin{tabular}{c}
    \parbox{0.6 \linewidth}{
      \renewcommand{\arraystretch}{1.5}
      \begin{equation*}
        \centering
        \begin{array}{ccc}
              \mathbf{a}_1 & = & \frac12 \, a \, \mathbf{\hat{y}} + \frac12 \, a \, \mathbf{\hat{z}} \\
    \mathbf{a}_2 & = & \frac12 \, a \, \mathbf{\hat{x}} + \frac12 \, a \, \mathbf{\hat{z}} \\
    \mathbf{a}_3 & = & \frac12 \, a \, \mathbf{\hat{x}} + \frac12 \, a \, \mathbf{\hat{y}} \\

        \end{array}
      \end{equation*}
    }
    \renewcommand{\arraystretch}{1.0}
  \end{tabular}
  \begin{tabular}{c}
    \includegraphics[width=0.3\linewidth]{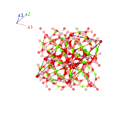} \\
  \end{tabular}
\end{tabular}

}
\vspace*{-0.25cm}

\noindent \hrulefill
\\
\textbf{Basis vectors:}
\vspace*{-0.25cm}
\renewcommand{\arraystretch}{1.5}
\begin{longtabu} to \textwidth{>{\centering $}X[-1,c,c]<{$}>{\centering $}X[-1,c,c]<{$}>{\centering $}X[-1,c,c]<{$}>{\centering $}X[-1,c,c]<{$}>{\centering $}X[-1,c,c]<{$}>{\centering $}X[-1,c,c]<{$}>{\centering $}X[-1,c,c]<{$}}
  & & \mbox{Lattice Coordinates} & & \mbox{Cartesian Coordinates} &\mbox{Wyckoff Position} & \mbox{Atom Type} \\  
  \mathbf{B}_{1} & = & 0 \, \mathbf{a}_{1} + 0 \, \mathbf{a}_{2} + 0 \, \mathbf{a}_{3} & = & 0 \, \mathbf{\hat{x}} + 0 \, \mathbf{\hat{y}} + 0 \, \mathbf{\hat{z}} & \left(8a\right) & \mbox{O I} \\ 
\mathbf{B}_{2} & = & \frac{1}{2} \, \mathbf{a}_{1} + \frac{1}{2} \, \mathbf{a}_{2} + \frac{1}{2} \, \mathbf{a}_{3} & = & \frac{1}{2}a \, \mathbf{\hat{x}} + \frac{1}{2}a \, \mathbf{\hat{y}} + \frac{1}{2}a \, \mathbf{\hat{z}} & \left(8a\right) & \mbox{O I} \\ 
\mathbf{B}_{3} & = & \frac{1}{4} \, \mathbf{a}_{1} + \frac{1}{4} \, \mathbf{a}_{2} + \frac{1}{4} \, \mathbf{a}_{3} & = & \frac{1}{4}a \, \mathbf{\hat{x}} + \frac{1}{4}a \, \mathbf{\hat{y}} + \frac{1}{4}a \, \mathbf{\hat{z}} & \left(8b\right) & \mbox{Cl} \\ 
\mathbf{B}_{4} & = & \frac{3}{4} \, \mathbf{a}_{1} + \frac{3}{4} \, \mathbf{a}_{2} + \frac{3}{4} \, \mathbf{a}_{3} & = & \frac{3}{4}a \, \mathbf{\hat{x}} + \frac{3}{4}a \, \mathbf{\hat{y}} + \frac{3}{4}a \, \mathbf{\hat{z}} & \left(8b\right) & \mbox{Cl} \\ 
\mathbf{B}_{5} & = & \frac{1}{2} \, \mathbf{a}_{1} & = & \frac{1}{4}a \, \mathbf{\hat{y}} + \frac{1}{4}a \, \mathbf{\hat{z}} & \left(24c\right) & \mbox{Mg} \\ 
\mathbf{B}_{6} & = & \frac{1}{2} \, \mathbf{a}_{2} + \frac{1}{2} \, \mathbf{a}_{3} & = & \frac{1}{2}a \, \mathbf{\hat{x}} + \frac{1}{4}a \, \mathbf{\hat{y}} + \frac{1}{4}a \, \mathbf{\hat{z}} & \left(24c\right) & \mbox{Mg} \\ 
\mathbf{B}_{7} & = & \frac{1}{2} \, \mathbf{a}_{2} & = & \frac{1}{4}a \, \mathbf{\hat{x}} + \frac{1}{4}a \, \mathbf{\hat{z}} & \left(24c\right) & \mbox{Mg} \\ 
\mathbf{B}_{8} & = & \frac{1}{2} \, \mathbf{a}_{1} + \frac{1}{2} \, \mathbf{a}_{3} & = & \frac{1}{4}a \, \mathbf{\hat{x}} + \frac{1}{2}a \, \mathbf{\hat{y}} + \frac{1}{4}a \, \mathbf{\hat{z}} & \left(24c\right) & \mbox{Mg} \\ 
\mathbf{B}_{9} & = & \frac{1}{2} \, \mathbf{a}_{3} & = & \frac{1}{4}a \, \mathbf{\hat{x}} + \frac{1}{4}a \, \mathbf{\hat{y}} & \left(24c\right) & \mbox{Mg} \\ 
\mathbf{B}_{10} & = & \frac{1}{2} \, \mathbf{a}_{1} + \frac{1}{2} \, \mathbf{a}_{2} & = & \frac{1}{4}a \, \mathbf{\hat{x}} + \frac{1}{4}a \, \mathbf{\hat{y}} + \frac{1}{2}a \, \mathbf{\hat{z}} & \left(24c\right) & \mbox{Mg} \\ 
\mathbf{B}_{11} & = & \frac{3}{4} \, \mathbf{a}_{1} + \frac{1}{4} \, \mathbf{a}_{2} + \frac{1}{4} \, \mathbf{a}_{3} & = & \frac{1}{4}a \, \mathbf{\hat{x}} + \frac{1}{2}a \, \mathbf{\hat{y}} + \frac{1}{2}a \, \mathbf{\hat{z}} & \left(24d\right) & \mbox{B I} \\ 
\mathbf{B}_{12} & = & \frac{1}{4} \, \mathbf{a}_{1} + \frac{3}{4} \, \mathbf{a}_{2} + \frac{3}{4} \, \mathbf{a}_{3} & = & \frac{3}{4}a \, \mathbf{\hat{x}} + \frac{1}{2}a \, \mathbf{\hat{y}} + \frac{1}{2}a \, \mathbf{\hat{z}} & \left(24d\right) & \mbox{B I} \\ 
\mathbf{B}_{13} & = & \frac{1}{4} \, \mathbf{a}_{1} + \frac{3}{4} \, \mathbf{a}_{2} + \frac{1}{4} \, \mathbf{a}_{3} & = & \frac{1}{2}a \, \mathbf{\hat{x}} + \frac{1}{4}a \, \mathbf{\hat{y}} + \frac{1}{2}a \, \mathbf{\hat{z}} & \left(24d\right) & \mbox{B I} \\ 
\mathbf{B}_{14} & = & \frac{3}{4} \, \mathbf{a}_{1} + \frac{1}{4} \, \mathbf{a}_{2} + \frac{3}{4} \, \mathbf{a}_{3} & = & \frac{1}{2}a \, \mathbf{\hat{x}} + \frac{3}{4}a \, \mathbf{\hat{y}} + \frac{1}{2}a \, \mathbf{\hat{z}} & \left(24d\right) & \mbox{B I} \\ 
\mathbf{B}_{15} & = & \frac{1}{4} \, \mathbf{a}_{1} + \frac{1}{4} \, \mathbf{a}_{2} + \frac{3}{4} \, \mathbf{a}_{3} & = & \frac{1}{2}a \, \mathbf{\hat{x}} + \frac{1}{2}a \, \mathbf{\hat{y}} + \frac{1}{4}a \, \mathbf{\hat{z}} & \left(24d\right) & \mbox{B I} \\ 
\mathbf{B}_{16} & = & \frac{3}{4} \, \mathbf{a}_{1} + \frac{3}{4} \, \mathbf{a}_{2} + \frac{1}{4} \, \mathbf{a}_{3} & = & \frac{1}{2}a \, \mathbf{\hat{x}} + \frac{1}{2}a \, \mathbf{\hat{y}} + \frac{3}{4}a \, \mathbf{\hat{z}} & \left(24d\right) & \mbox{B I} \\ 
\mathbf{B}_{17} & = & x_{5} \, \mathbf{a}_{1} + x_{5} \, \mathbf{a}_{2} + x_{5} \, \mathbf{a}_{3} & = & x_{5}a \, \mathbf{\hat{x}} + x_{5}a \, \mathbf{\hat{y}} + x_{5}a \, \mathbf{\hat{z}} & \left(32e\right) & \mbox{B II} \\ 
\mathbf{B}_{18} & = & x_{5} \, \mathbf{a}_{1} + x_{5} \, \mathbf{a}_{2}-3x_{5} \, \mathbf{a}_{3} & = & -x_{5}a \, \mathbf{\hat{x}}-x_{5}a \, \mathbf{\hat{y}} + x_{5}a \, \mathbf{\hat{z}} & \left(32e\right) & \mbox{B II} \\ 
\mathbf{B}_{19} & = & x_{5} \, \mathbf{a}_{1}-3x_{5} \, \mathbf{a}_{2} + x_{5} \, \mathbf{a}_{3} & = & -x_{5}a \, \mathbf{\hat{x}} + x_{5}a \, \mathbf{\hat{y}}-x_{5}a \, \mathbf{\hat{z}} & \left(32e\right) & \mbox{B II} \\ 
\mathbf{B}_{20} & = & -3x_{5} \, \mathbf{a}_{1} + x_{5} \, \mathbf{a}_{2} + x_{5} \, \mathbf{a}_{3} & = & x_{5}a \, \mathbf{\hat{x}}-x_{5}a \, \mathbf{\hat{y}}-x_{5}a \, \mathbf{\hat{z}} & \left(32e\right) & \mbox{B II} \\ 
\mathbf{B}_{21} & = & \left(\frac{1}{2} +x_{5}\right) \, \mathbf{a}_{1} + \left(\frac{1}{2} +x_{5}\right) \, \mathbf{a}_{2} + \left(\frac{1}{2} +x_{5}\right) \, \mathbf{a}_{3} & = & \left(\frac{1}{2} +x_{5}\right)a \, \mathbf{\hat{x}} + \left(\frac{1}{2} +x_{5}\right)a \, \mathbf{\hat{y}} + \left(\frac{1}{2} +x_{5}\right)a \, \mathbf{\hat{z}} & \left(32e\right) & \mbox{B II} \\ 
\mathbf{B}_{22} & = & \left(\frac{1}{2} +x_{5}\right) \, \mathbf{a}_{1} + \left(\frac{1}{2} +x_{5}\right) \, \mathbf{a}_{2} + \left(\frac{1}{2} - 3x_{5}\right) \, \mathbf{a}_{3} & = & \left(\frac{1}{2} - x_{5}\right)a \, \mathbf{\hat{x}} + \left(\frac{1}{2} - x_{5}\right)a \, \mathbf{\hat{y}} + \left(\frac{1}{2} +x_{5}\right)a \, \mathbf{\hat{z}} & \left(32e\right) & \mbox{B II} \\ 
\mathbf{B}_{23} & = & \left(\frac{1}{2} - 3x_{5}\right) \, \mathbf{a}_{1} + \left(\frac{1}{2} +x_{5}\right) \, \mathbf{a}_{2} + \left(\frac{1}{2} +x_{5}\right) \, \mathbf{a}_{3} & = & \left(\frac{1}{2} +x_{5}\right)a \, \mathbf{\hat{x}} + \left(\frac{1}{2} - x_{5}\right)a \, \mathbf{\hat{y}} + \left(\frac{1}{2} - x_{5}\right)a \, \mathbf{\hat{z}} & \left(32e\right) & \mbox{B II} \\ 
\mathbf{B}_{24} & = & \left(\frac{1}{2} +x_{5}\right) \, \mathbf{a}_{1} + \left(\frac{1}{2} - 3x_{5}\right) \, \mathbf{a}_{2} + \left(\frac{1}{2} +x_{5}\right) \, \mathbf{a}_{3} & = & \left(\frac{1}{2} - x_{5}\right)a \, \mathbf{\hat{x}} + \left(\frac{1}{2} +x_{5}\right)a \, \mathbf{\hat{y}} + \left(\frac{1}{2} - x_{5}\right)a \, \mathbf{\hat{z}} & \left(32e\right) & \mbox{B II} \\ 
\mathbf{B}_{25} & = & \left(-x_{6}+y_{6}+z_{6}\right) \, \mathbf{a}_{1} + \left(x_{6}-y_{6}+z_{6}\right) \, \mathbf{a}_{2} + \left(x_{6}+y_{6}-z_{6}\right) \, \mathbf{a}_{3} & = & x_{6}a \, \mathbf{\hat{x}} + y_{6}a \, \mathbf{\hat{y}} + z_{6}a \, \mathbf{\hat{z}} & \left(96h\right) & \mbox{O II} \\ 
\mathbf{B}_{26} & = & \left(x_{6}-y_{6}+z_{6}\right) \, \mathbf{a}_{1} + \left(-x_{6}+y_{6}+z_{6}\right) \, \mathbf{a}_{2} + \left(-x_{6}-y_{6}-z_{6}\right) \, \mathbf{a}_{3} & = & -x_{6}a \, \mathbf{\hat{x}}-y_{6}a \, \mathbf{\hat{y}} + z_{6}a \, \mathbf{\hat{z}} & \left(96h\right) & \mbox{O II} \\ 
\mathbf{B}_{27} & = & \left(x_{6}+y_{6}-z_{6}\right) \, \mathbf{a}_{1} + \left(-x_{6}-y_{6}-z_{6}\right) \, \mathbf{a}_{2} + \left(-x_{6}+y_{6}+z_{6}\right) \, \mathbf{a}_{3} & = & -x_{6}a \, \mathbf{\hat{x}} + y_{6}a \, \mathbf{\hat{y}}-z_{6}a \, \mathbf{\hat{z}} & \left(96h\right) & \mbox{O II} \\ 
\mathbf{B}_{28} & = & \left(-x_{6}-y_{6}-z_{6}\right) \, \mathbf{a}_{1} + \left(x_{6}+y_{6}-z_{6}\right) \, \mathbf{a}_{2} + \left(x_{6}-y_{6}+z_{6}\right) \, \mathbf{a}_{3} & = & x_{6}a \, \mathbf{\hat{x}}-y_{6}a \, \mathbf{\hat{y}}-z_{6}a \, \mathbf{\hat{z}} & \left(96h\right) & \mbox{O II} \\ 
\mathbf{B}_{29} & = & \left(x_{6}+y_{6}-z_{6}\right) \, \mathbf{a}_{1} + \left(-x_{6}+y_{6}+z_{6}\right) \, \mathbf{a}_{2} + \left(x_{6}-y_{6}+z_{6}\right) \, \mathbf{a}_{3} & = & z_{6}a \, \mathbf{\hat{x}} + x_{6}a \, \mathbf{\hat{y}} + y_{6}a \, \mathbf{\hat{z}} & \left(96h\right) & \mbox{O II} \\ 
\mathbf{B}_{30} & = & \left(-x_{6}-y_{6}-z_{6}\right) \, \mathbf{a}_{1} + \left(x_{6}-y_{6}+z_{6}\right) \, \mathbf{a}_{2} + \left(-x_{6}+y_{6}+z_{6}\right) \, \mathbf{a}_{3} & = & z_{6}a \, \mathbf{\hat{x}}-x_{6}a \, \mathbf{\hat{y}}-y_{6}a \, \mathbf{\hat{z}} & \left(96h\right) & \mbox{O II} \\ 
\mathbf{B}_{31} & = & \left(-x_{6}+y_{6}+z_{6}\right) \, \mathbf{a}_{1} + \left(x_{6}+y_{6}-z_{6}\right) \, \mathbf{a}_{2} + \left(-x_{6}-y_{6}-z_{6}\right) \, \mathbf{a}_{3} & = & -z_{6}a \, \mathbf{\hat{x}}-x_{6}a \, \mathbf{\hat{y}} + y_{6}a \, \mathbf{\hat{z}} & \left(96h\right) & \mbox{O II} \\ 
\mathbf{B}_{32} & = & \left(x_{6}-y_{6}+z_{6}\right) \, \mathbf{a}_{1} + \left(-x_{6}-y_{6}-z_{6}\right) \, \mathbf{a}_{2} + \left(x_{6}+y_{6}-z_{6}\right) \, \mathbf{a}_{3} & = & -z_{6}a \, \mathbf{\hat{x}} + x_{6}a \, \mathbf{\hat{y}}-y_{6}a \, \mathbf{\hat{z}} & \left(96h\right) & \mbox{O II} \\ 
\mathbf{B}_{33} & = & \left(x_{6}-y_{6}+z_{6}\right) \, \mathbf{a}_{1} + \left(x_{6}+y_{6}-z_{6}\right) \, \mathbf{a}_{2} + \left(-x_{6}+y_{6}+z_{6}\right) \, \mathbf{a}_{3} & = & y_{6}a \, \mathbf{\hat{x}} + z_{6}a \, \mathbf{\hat{y}} + x_{6}a \, \mathbf{\hat{z}} & \left(96h\right) & \mbox{O II} \\ 
\mathbf{B}_{34} & = & \left(-x_{6}+y_{6}+z_{6}\right) \, \mathbf{a}_{1} + \left(-x_{6}-y_{6}-z_{6}\right) \, \mathbf{a}_{2} + \left(x_{6}-y_{6}+z_{6}\right) \, \mathbf{a}_{3} & = & -y_{6}a \, \mathbf{\hat{x}} + z_{6}a \, \mathbf{\hat{y}}-x_{6}a \, \mathbf{\hat{z}} & \left(96h\right) & \mbox{O II} \\ 
\mathbf{B}_{35} & = & \left(-x_{6}-y_{6}-z_{6}\right) \, \mathbf{a}_{1} + \left(-x_{6}+y_{6}+z_{6}\right) \, \mathbf{a}_{2} + \left(x_{6}+y_{6}-z_{6}\right) \, \mathbf{a}_{3} & = & y_{6}a \, \mathbf{\hat{x}}-z_{6}a \, \mathbf{\hat{y}}-x_{6}a \, \mathbf{\hat{z}} & \left(96h\right) & \mbox{O II} \\ 
\mathbf{B}_{36} & = & \left(x_{6}+y_{6}-z_{6}\right) \, \mathbf{a}_{1} + \left(x_{6}-y_{6}+z_{6}\right) \, \mathbf{a}_{2} + \left(-x_{6}-y_{6}-z_{6}\right) \, \mathbf{a}_{3} & = & -y_{6}a \, \mathbf{\hat{x}}-z_{6}a \, \mathbf{\hat{y}} + x_{6}a \, \mathbf{\hat{z}} & \left(96h\right) & \mbox{O II} \\ 
\mathbf{B}_{37} & = & \left(\frac{1}{2} +x_{6} - y_{6} + z_{6}\right) \, \mathbf{a}_{1} + \left(\frac{1}{2} - x_{6} + y_{6} + z_{6}\right) \, \mathbf{a}_{2} + \left(\frac{1}{2} +x_{6} + y_{6} - z_{6}\right) \, \mathbf{a}_{3} & = & \left(\frac{1}{2} +y_{6}\right)a \, \mathbf{\hat{x}} + \left(\frac{1}{2} +x_{6}\right)a \, \mathbf{\hat{y}} + \left(\frac{1}{2} +z_{6}\right)a \, \mathbf{\hat{z}} & \left(96h\right) & \mbox{O II} \\ 
\mathbf{B}_{38} & = & \left(\frac{1}{2} - x_{6} + y_{6} + z_{6}\right) \, \mathbf{a}_{1} + \left(\frac{1}{2} +x_{6} - y_{6} + z_{6}\right) \, \mathbf{a}_{2} + \left(\frac{1}{2} - x_{6} - y_{6} - z_{6}\right) \, \mathbf{a}_{3} & = & \left(\frac{1}{2} - y_{6}\right)a \, \mathbf{\hat{x}} + \left(\frac{1}{2} - x_{6}\right)a \, \mathbf{\hat{y}} + \left(\frac{1}{2} +z_{6}\right)a \, \mathbf{\hat{z}} & \left(96h\right) & \mbox{O II} \\ 
\mathbf{B}_{39} & = & \left(\frac{1}{2} - x_{6} - y_{6} - z_{6}\right) \, \mathbf{a}_{1} + \left(\frac{1}{2} +x_{6} + y_{6} - z_{6}\right) \, \mathbf{a}_{2} + \left(\frac{1}{2} - x_{6} + y_{6} + z_{6}\right) \, \mathbf{a}_{3} & = & \left(\frac{1}{2} +y_{6}\right)a \, \mathbf{\hat{x}} + \left(\frac{1}{2} - x_{6}\right)a \, \mathbf{\hat{y}} + \left(\frac{1}{2} - z_{6}\right)a \, \mathbf{\hat{z}} & \left(96h\right) & \mbox{O II} \\ 
\mathbf{B}_{40} & = & \left(\frac{1}{2} +x_{6} + y_{6} - z_{6}\right) \, \mathbf{a}_{1} + \left(\frac{1}{2} - x_{6} - y_{6} - z_{6}\right) \, \mathbf{a}_{2} + \left(\frac{1}{2} +x_{6} - y_{6} + z_{6}\right) \, \mathbf{a}_{3} & = & \left(\frac{1}{2} - y_{6}\right)a \, \mathbf{\hat{x}} + \left(\frac{1}{2} +x_{6}\right)a \, \mathbf{\hat{y}} + \left(\frac{1}{2} - z_{6}\right)a \, \mathbf{\hat{z}} & \left(96h\right) & \mbox{O II} \\ 
\mathbf{B}_{41} & = & \left(\frac{1}{2} - x_{6} + y_{6} + z_{6}\right) \, \mathbf{a}_{1} + \left(\frac{1}{2} +x_{6} + y_{6} - z_{6}\right) \, \mathbf{a}_{2} + \left(\frac{1}{2} +x_{6} - y_{6} + z_{6}\right) \, \mathbf{a}_{3} & = & \left(\frac{1}{2} +x_{6}\right)a \, \mathbf{\hat{x}} + \left(\frac{1}{2} +z_{6}\right)a \, \mathbf{\hat{y}} + \left(\frac{1}{2} +y_{6}\right)a \, \mathbf{\hat{z}} & \left(96h\right) & \mbox{O II} \\ 
\mathbf{B}_{42} & = & \left(\frac{1}{2} +x_{6} - y_{6} + z_{6}\right) \, \mathbf{a}_{1} + \left(\frac{1}{2} - x_{6} - y_{6} - z_{6}\right) \, \mathbf{a}_{2} + \left(\frac{1}{2} - x_{6} + y_{6} + z_{6}\right) \, \mathbf{a}_{3} & = & \left(\frac{1}{2} - x_{6}\right)a \, \mathbf{\hat{x}} + \left(\frac{1}{2} +z_{6}\right)a \, \mathbf{\hat{y}} + \left(\frac{1}{2} - y_{6}\right)a \, \mathbf{\hat{z}} & \left(96h\right) & \mbox{O II} \\ 
\mathbf{B}_{43} & = & \left(\frac{1}{2} +x_{6} + y_{6} - z_{6}\right) \, \mathbf{a}_{1} + \left(\frac{1}{2} - x_{6} + y_{6} + z_{6}\right) \, \mathbf{a}_{2} + \left(\frac{1}{2} - x_{6} - y_{6} - z_{6}\right) \, \mathbf{a}_{3} & = & \left(\frac{1}{2} - x_{6}\right)a \, \mathbf{\hat{x}} + \left(\frac{1}{2} - z_{6}\right)a \, \mathbf{\hat{y}} + \left(\frac{1}{2} +y_{6}\right)a \, \mathbf{\hat{z}} & \left(96h\right) & \mbox{O II} \\ 
\mathbf{B}_{44} & = & \left(\frac{1}{2} - x_{6} - y_{6} - z_{6}\right) \, \mathbf{a}_{1} + \left(\frac{1}{2} +x_{6} - y_{6} + z_{6}\right) \, \mathbf{a}_{2} + \left(\frac{1}{2} +x_{6} + y_{6} - z_{6}\right) \, \mathbf{a}_{3} & = & \left(\frac{1}{2} +x_{6}\right)a \, \mathbf{\hat{x}} + \left(\frac{1}{2} - z_{6}\right)a \, \mathbf{\hat{y}} + \left(\frac{1}{2} - y_{6}\right)a \, \mathbf{\hat{z}} & \left(96h\right) & \mbox{O II} \\ 
\mathbf{B}_{45} & = & \left(\frac{1}{2} +x_{6} + y_{6} - z_{6}\right) \, \mathbf{a}_{1} + \left(\frac{1}{2} +x_{6} - y_{6} + z_{6}\right) \, \mathbf{a}_{2} + \left(\frac{1}{2} - x_{6} + y_{6} + z_{6}\right) \, \mathbf{a}_{3} & = & \left(\frac{1}{2} +z_{6}\right)a \, \mathbf{\hat{x}} + \left(\frac{1}{2} +y_{6}\right)a \, \mathbf{\hat{y}} + \left(\frac{1}{2} +x_{6}\right)a \, \mathbf{\hat{z}} & \left(96h\right) & \mbox{O II} \\ 
\mathbf{B}_{46} & = & \left(\frac{1}{2} - x_{6} - y_{6} - z_{6}\right) \, \mathbf{a}_{1} + \left(\frac{1}{2} - x_{6} + y_{6} + z_{6}\right) \, \mathbf{a}_{2} + \left(\frac{1}{2} +x_{6} - y_{6} + z_{6}\right) \, \mathbf{a}_{3} & = & \left(\frac{1}{2} +z_{6}\right)a \, \mathbf{\hat{x}} + \left(\frac{1}{2} - y_{6}\right)a \, \mathbf{\hat{y}} + \left(\frac{1}{2} - x_{6}\right)a \, \mathbf{\hat{z}} & \left(96h\right) & \mbox{O II} \\ 
\mathbf{B}_{47} & = & \left(\frac{1}{2} - x_{6} + y_{6} + z_{6}\right) \, \mathbf{a}_{1} + \left(\frac{1}{2} - x_{6} - y_{6} - z_{6}\right) \, \mathbf{a}_{2} + \left(\frac{1}{2} +x_{6} + y_{6} - z_{6}\right) \, \mathbf{a}_{3} & = & \left(\frac{1}{2} - z_{6}\right)a \, \mathbf{\hat{x}} + \left(\frac{1}{2} +y_{6}\right)a \, \mathbf{\hat{y}} + \left(\frac{1}{2} - x_{6}\right)a \, \mathbf{\hat{z}} & \left(96h\right) & \mbox{O II} \\ 
\mathbf{B}_{48} & = & \left(\frac{1}{2} +x_{6} - y_{6} + z_{6}\right) \, \mathbf{a}_{1} + \left(\frac{1}{2} +x_{6} + y_{6} - z_{6}\right) \, \mathbf{a}_{2} + \left(\frac{1}{2} - x_{6} - y_{6} - z_{6}\right) \, \mathbf{a}_{3} & = & \left(\frac{1}{2} - z_{6}\right)a \, \mathbf{\hat{x}} + \left(\frac{1}{2} - y_{6}\right)a \, \mathbf{\hat{y}} + \left(\frac{1}{2} +x_{6}\right)a \, \mathbf{\hat{z}} & \left(96h\right) & \mbox{O II} \\ 
\end{longtabu}
\renewcommand{\arraystretch}{1.0}
\noindent \hrulefill
\\
\textbf{References:}
\vspace*{-0.25cm}
\begin{flushleft}
  - \bibentry{Sueng_Am_Min_1973}. \\
\end{flushleft}
\noindent \hrulefill
\\
\textbf{Geometry files:}
\\
\noindent  - CIF: pp. {\hyperref[A7BC3D13_cF192_219_de_b_c_ah_cif]{\pageref{A7BC3D13_cF192_219_de_b_c_ah_cif}}} \\
\noindent  - POSCAR: pp. {\hyperref[A7BC3D13_cF192_219_de_b_c_ah_poscar]{\pageref{A7BC3D13_cF192_219_de_b_c_ah_poscar}}} \\
\onecolumn
{\phantomsection\label{A15B4_cI76_220_ae_c}}
\subsection*{\huge \textbf{{\normalfont Cu$_{15}$Si$_{4}$ ($D8_{6}$) Structure: A15B4\_cI76\_220\_ae\_c}}}
\noindent \hrulefill
\vspace*{0.25cm}
\begin{figure}[htp]
  \centering
  \vspace{-1em}
  {\includegraphics[width=1\textwidth]{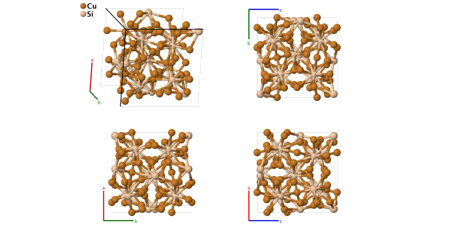}}
\end{figure}
\vspace*{-0.5cm}
\renewcommand{\arraystretch}{1.5}
\begin{equation*}
  \begin{array}{>{$\hspace{-0.15cm}}l<{$}>{$}p{0.5cm}<{$}>{$}p{18.5cm}<{$}}
    \mbox{\large \textbf{Prototype}} &\colon & \ce{Cu$_{15}$Si$_{4}$} \\
    \mbox{\large \textbf{\AFLOW\ prototype label}} &\colon & \mbox{A15B4\_cI76\_220\_ae\_c} \\
    \mbox{\large \textbf{\textit{Strukturbericht} designation}} &\colon & \mbox{$D8_{6}$} \\
    \mbox{\large \textbf{Pearson symbol}} &\colon & \mbox{cI76} \\
    \mbox{\large \textbf{Space group number}} &\colon & 220 \\
    \mbox{\large \textbf{Space group symbol}} &\colon & I\bar{4}3d \\
    \mbox{\large \textbf{\AFLOW\ prototype command}} &\colon &  \texttt{aflow} \,  \, \texttt{-{}-proto=A15B4\_cI76\_220\_ae\_c } \, \newline \texttt{-{}-params=}{a,x_{2},x_{3},y_{3},z_{3} }
  \end{array}
\end{equation*}
\renewcommand{\arraystretch}{1.0}

\vspace*{-0.25cm}
\noindent \hrulefill
\\
\textbf{ Other compounds with this structure:}
\begin{itemize}
   \item{ Cu$_{15}$As$_{4}$, Li$_{15}$Si$_{4}$, Na$_{15}$Pb$_{4}$   }
\end{itemize}
\noindent \parbox{1 \linewidth}{
\noindent \hrulefill
\\
\textbf{Body-centered Cubic primitive vectors:} \\
\vspace*{-0.25cm}
\begin{tabular}{cc}
  \begin{tabular}{c}
    \parbox{0.6 \linewidth}{
      \renewcommand{\arraystretch}{1.5}
      \begin{equation*}
        \centering
        \begin{array}{ccc}
              \mathbf{a}_1 & = & - \frac12 \, a \, \mathbf{\hat{x}} + \frac12 \, a \, \mathbf{\hat{y}} + \frac12 \, a \, \mathbf{\hat{z}} \\
    \mathbf{a}_2 & = & ~ \frac12 \, a \, \mathbf{\hat{x}} - \frac12 \, a \, \mathbf{\hat{y}} + \frac12 \, a \, \mathbf{\hat{z}} \\
    \mathbf{a}_3 & = & ~ \frac12 \, a \, \mathbf{\hat{x}} + \frac12 \, a \, \mathbf{\hat{y}} - \frac12 \, a \, \mathbf{\hat{z}} \\

        \end{array}
      \end{equation*}
    }
    \renewcommand{\arraystretch}{1.0}
  \end{tabular}
  \begin{tabular}{c}
    \includegraphics[width=0.3\linewidth]{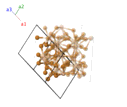} \\
  \end{tabular}
\end{tabular}

}
\vspace*{-0.25cm}

\noindent \hrulefill
\\
\textbf{Basis vectors:}
\vspace*{-0.25cm}
\renewcommand{\arraystretch}{1.5}
\begin{longtabu} to \textwidth{>{\centering $}X[-1,c,c]<{$}>{\centering $}X[-1,c,c]<{$}>{\centering $}X[-1,c,c]<{$}>{\centering $}X[-1,c,c]<{$}>{\centering $}X[-1,c,c]<{$}>{\centering $}X[-1,c,c]<{$}>{\centering $}X[-1,c,c]<{$}}
  & & \mbox{Lattice Coordinates} & & \mbox{Cartesian Coordinates} &\mbox{Wyckoff Position} & \mbox{Atom Type} \\  
  \mathbf{B}_{1} & = & \frac{1}{4} \, \mathbf{a}_{1} + \frac{5}{8} \, \mathbf{a}_{2} + \frac{3}{8} \, \mathbf{a}_{3} & = & \frac{3}{8}a \, \mathbf{\hat{x}} + \frac{1}{4}a \, \mathbf{\hat{z}} & \left(12a\right) & \mbox{Cu I} \\ 
\mathbf{B}_{2} & = & \frac{3}{4} \, \mathbf{a}_{1} + \frac{7}{8} \, \mathbf{a}_{2} + \frac{1}{8} \, \mathbf{a}_{3} & = & \frac{1}{8}a \, \mathbf{\hat{x}} + \frac{3}{4}a \, \mathbf{\hat{z}} & \left(12a\right) & \mbox{Cu I} \\ 
\mathbf{B}_{3} & = & \frac{3}{8} \, \mathbf{a}_{1} + \frac{1}{4} \, \mathbf{a}_{2} + \frac{5}{8} \, \mathbf{a}_{3} & = & \frac{1}{4}a \, \mathbf{\hat{x}} + \frac{3}{8}a \, \mathbf{\hat{y}} & \left(12a\right) & \mbox{Cu I} \\ 
\mathbf{B}_{4} & = & \frac{1}{8} \, \mathbf{a}_{1} + \frac{3}{4} \, \mathbf{a}_{2} + \frac{7}{8} \, \mathbf{a}_{3} & = & \frac{3}{4}a \, \mathbf{\hat{x}} + \frac{1}{8}a \, \mathbf{\hat{y}} & \left(12a\right) & \mbox{Cu I} \\ 
\mathbf{B}_{5} & = & \frac{5}{8} \, \mathbf{a}_{1} + \frac{3}{8} \, \mathbf{a}_{2} + \frac{1}{4} \, \mathbf{a}_{3} & = & \frac{1}{4}a \, \mathbf{\hat{y}} + \frac{3}{8}a \, \mathbf{\hat{z}} & \left(12a\right) & \mbox{Cu I} \\ 
\mathbf{B}_{6} & = & \frac{7}{8} \, \mathbf{a}_{1} + \frac{1}{8} \, \mathbf{a}_{2} + \frac{3}{4} \, \mathbf{a}_{3} & = & \frac{3}{4}a \, \mathbf{\hat{y}} + \frac{1}{8}a \, \mathbf{\hat{z}} & \left(12a\right) & \mbox{Cu I} \\ 
\mathbf{B}_{7} & = & 2x_{2} \, \mathbf{a}_{1} + 2x_{2} \, \mathbf{a}_{2} + 2x_{2} \, \mathbf{a}_{3} & = & x_{2}a \, \mathbf{\hat{x}} + x_{2}a \, \mathbf{\hat{y}} + x_{2}a \, \mathbf{\hat{z}} & \left(16c\right) & \mbox{Si} \\ 
\mathbf{B}_{8} & = & \frac{1}{2} \, \mathbf{a}_{1} + \left(\frac{1}{2} - 2x_{2}\right) \, \mathbf{a}_{3} & = & -x_{2}a \, \mathbf{\hat{x}} + \left(\frac{1}{2} - x_{2}\right)a \, \mathbf{\hat{y}} + x_{2}a \, \mathbf{\hat{z}} & \left(16c\right) & \mbox{Si} \\ 
\mathbf{B}_{9} & = & \left(\frac{1}{2} - 2x_{2}\right) \, \mathbf{a}_{2} + \frac{1}{2} \, \mathbf{a}_{3} & = & \left(\frac{1}{2} - x_{2}\right)a \, \mathbf{\hat{x}} + x_{2}a \, \mathbf{\hat{y}}-x_{2}a \, \mathbf{\hat{z}} & \left(16c\right) & \mbox{Si} \\ 
\mathbf{B}_{10} & = & \left(\frac{1}{2} - 2x_{2}\right) \, \mathbf{a}_{1} + \frac{1}{2} \, \mathbf{a}_{2} & = & x_{2}a \, \mathbf{\hat{x}}-x_{2}a \, \mathbf{\hat{y}} + \left(\frac{1}{2} - x_{2}\right)a \, \mathbf{\hat{z}} & \left(16c\right) & \mbox{Si} \\ 
\mathbf{B}_{11} & = & \left(\frac{1}{2} +2x_{2}\right) \, \mathbf{a}_{1} + \left(\frac{1}{2} +2x_{2}\right) \, \mathbf{a}_{2} + \left(\frac{1}{2} +2x_{2}\right) \, \mathbf{a}_{3} & = & \left(\frac{1}{4} +x_{2}\right)a \, \mathbf{\hat{x}} + \left(\frac{1}{4} +x_{2}\right)a \, \mathbf{\hat{y}} + \left(\frac{1}{4} +x_{2}\right)a \, \mathbf{\hat{z}} & \left(16c\right) & \mbox{Si} \\ 
\mathbf{B}_{12} & = & \frac{1}{2} \, \mathbf{a}_{1} + -2x_{2} \, \mathbf{a}_{3} & = & -a\left(x_{2}+\frac{1}{4}\right) \, \mathbf{\hat{x}} + \left(\frac{1}{4} - x_{2}\right)a \, \mathbf{\hat{y}} + \left(\frac{1}{4} +x_{2}\right)a \, \mathbf{\hat{z}} & \left(16c\right) & \mbox{Si} \\ 
\mathbf{B}_{13} & = & -2x_{2} \, \mathbf{a}_{1} + \frac{1}{2} \, \mathbf{a}_{2} & = & \left(\frac{1}{4} +x_{2}\right)a \, \mathbf{\hat{x}}-a\left(x_{2}+\frac{1}{4}\right) \, \mathbf{\hat{y}} + \left(\frac{1}{4} - x_{2}\right)a \, \mathbf{\hat{z}} & \left(16c\right) & \mbox{Si} \\ 
\mathbf{B}_{14} & = & -2x_{2} \, \mathbf{a}_{2} + \frac{1}{2} \, \mathbf{a}_{3} & = & \left(\frac{1}{4} - x_{2}\right)a \, \mathbf{\hat{x}} + \left(\frac{1}{4} +x_{2}\right)a \, \mathbf{\hat{y}}-a\left(x_{2}+\frac{1}{4}\right) \, \mathbf{\hat{z}} & \left(16c\right) & \mbox{Si} \\ 
\mathbf{B}_{15} & = & \left(y_{3}+z_{3}\right) \, \mathbf{a}_{1} + \left(x_{3}+z_{3}\right) \, \mathbf{a}_{2} + \left(x_{3}+y_{3}\right) \, \mathbf{a}_{3} & = & x_{3}a \, \mathbf{\hat{x}} + y_{3}a \, \mathbf{\hat{y}} + z_{3}a \, \mathbf{\hat{z}} & \left(48e\right) & \mbox{Cu II} \\ 
\mathbf{B}_{16} & = & \left(\frac{1}{2} - y_{3} + z_{3}\right) \, \mathbf{a}_{1} + \left(-x_{3}+z_{3}\right) \, \mathbf{a}_{2} + \left(\frac{1}{2} - x_{3} - y_{3}\right) \, \mathbf{a}_{3} & = & -x_{3}a \, \mathbf{\hat{x}} + \left(\frac{1}{2} - y_{3}\right)a \, \mathbf{\hat{y}} + z_{3}a \, \mathbf{\hat{z}} & \left(48e\right) & \mbox{Cu II} \\ 
\mathbf{B}_{17} & = & \left(y_{3}-z_{3}\right) \, \mathbf{a}_{1} + \left(\frac{1}{2} - x_{3} - z_{3}\right) \, \mathbf{a}_{2} + \left(\frac{1}{2} - x_{3} + y_{3}\right) \, \mathbf{a}_{3} & = & \left(\frac{1}{2} - x_{3}\right)a \, \mathbf{\hat{x}} + y_{3}a \, \mathbf{\hat{y}}-z_{3}a \, \mathbf{\hat{z}} & \left(48e\right) & \mbox{Cu II} \\ 
\mathbf{B}_{18} & = & \left(\frac{1}{2} - y_{3} - z_{3}\right) \, \mathbf{a}_{1} + \left(\frac{1}{2} +x_{3} - z_{3}\right) \, \mathbf{a}_{2} + \left(x_{3}-y_{3}\right) \, \mathbf{a}_{3} & = & x_{3}a \, \mathbf{\hat{x}}-y_{3}a \, \mathbf{\hat{y}} + \left(\frac{1}{2} - z_{3}\right)a \, \mathbf{\hat{z}} & \left(48e\right) & \mbox{Cu II} \\ 
\mathbf{B}_{19} & = & \left(x_{3}+y_{3}\right) \, \mathbf{a}_{1} + \left(y_{3}+z_{3}\right) \, \mathbf{a}_{2} + \left(x_{3}+z_{3}\right) \, \mathbf{a}_{3} & = & z_{3}a \, \mathbf{\hat{x}} + x_{3}a \, \mathbf{\hat{y}} + y_{3}a \, \mathbf{\hat{z}} & \left(48e\right) & \mbox{Cu II} \\ 
\mathbf{B}_{20} & = & \left(\frac{1}{2} - x_{3} - y_{3}\right) \, \mathbf{a}_{1} + \left(\frac{1}{2} - y_{3} + z_{3}\right) \, \mathbf{a}_{2} + \left(-x_{3}+z_{3}\right) \, \mathbf{a}_{3} & = & z_{3}a \, \mathbf{\hat{x}}-x_{3}a \, \mathbf{\hat{y}} + \left(\frac{1}{2} - y_{3}\right)a \, \mathbf{\hat{z}} & \left(48e\right) & \mbox{Cu II} \\ 
\mathbf{B}_{21} & = & \left(\frac{1}{2} - x_{3} + y_{3}\right) \, \mathbf{a}_{1} + \left(y_{3}-z_{3}\right) \, \mathbf{a}_{2} + \left(\frac{1}{2} - x_{3} - z_{3}\right) \, \mathbf{a}_{3} & = & -z_{3}a \, \mathbf{\hat{x}} + \left(\frac{1}{2} - x_{3}\right)a \, \mathbf{\hat{y}} + y_{3}a \, \mathbf{\hat{z}} & \left(48e\right) & \mbox{Cu II} \\ 
\mathbf{B}_{22} & = & \left(x_{3}-y_{3}\right) \, \mathbf{a}_{1} + \left(\frac{1}{2} - y_{3} - z_{3}\right) \, \mathbf{a}_{2} + \left(\frac{1}{2} +x_{3} - z_{3}\right) \, \mathbf{a}_{3} & = & \left(\frac{1}{2} - z_{3}\right)a \, \mathbf{\hat{x}} + x_{3}a \, \mathbf{\hat{y}}-y_{3}a \, \mathbf{\hat{z}} & \left(48e\right) & \mbox{Cu II} \\ 
\mathbf{B}_{23} & = & \left(x_{3}+z_{3}\right) \, \mathbf{a}_{1} + \left(x_{3}+y_{3}\right) \, \mathbf{a}_{2} + \left(y_{3}+z_{3}\right) \, \mathbf{a}_{3} & = & y_{3}a \, \mathbf{\hat{x}} + z_{3}a \, \mathbf{\hat{y}} + x_{3}a \, \mathbf{\hat{z}} & \left(48e\right) & \mbox{Cu II} \\ 
\mathbf{B}_{24} & = & \left(-x_{3}+z_{3}\right) \, \mathbf{a}_{1} + \left(\frac{1}{2} - x_{3} - y_{3}\right) \, \mathbf{a}_{2} + \left(\frac{1}{2} - y_{3} + z_{3}\right) \, \mathbf{a}_{3} & = & \left(\frac{1}{2} - y_{3}\right)a \, \mathbf{\hat{x}} + z_{3}a \, \mathbf{\hat{y}}-x_{3}a \, \mathbf{\hat{z}} & \left(48e\right) & \mbox{Cu II} \\ 
\mathbf{B}_{25} & = & \left(\frac{1}{2} - x_{3} - z_{3}\right) \, \mathbf{a}_{1} + \left(\frac{1}{2} - x_{3} + y_{3}\right) \, \mathbf{a}_{2} + \left(y_{3}-z_{3}\right) \, \mathbf{a}_{3} & = & y_{3}a \, \mathbf{\hat{x}}-z_{3}a \, \mathbf{\hat{y}} + \left(\frac{1}{2} - x_{3}\right)a \, \mathbf{\hat{z}} & \left(48e\right) & \mbox{Cu II} \\ 
\mathbf{B}_{26} & = & \left(\frac{1}{2} +x_{3} - z_{3}\right) \, \mathbf{a}_{1} + \left(x_{3}-y_{3}\right) \, \mathbf{a}_{2} + \left(\frac{1}{2} - y_{3} - z_{3}\right) \, \mathbf{a}_{3} & = & -y_{3}a \, \mathbf{\hat{x}} + \left(\frac{1}{2} - z_{3}\right)a \, \mathbf{\hat{y}} + x_{3}a \, \mathbf{\hat{z}} & \left(48e\right) & \mbox{Cu II} \\ 
\mathbf{B}_{27} & = & \left(\frac{1}{2} +x_{3} + z_{3}\right) \, \mathbf{a}_{1} + \left(\frac{1}{2} +y_{3} + z_{3}\right) \, \mathbf{a}_{2} + \left(\frac{1}{2} +x_{3} + y_{3}\right) \, \mathbf{a}_{3} & = & \left(\frac{1}{4} +y_{3}\right)a \, \mathbf{\hat{x}} + \left(\frac{1}{4} +x_{3}\right)a \, \mathbf{\hat{y}} + \left(\frac{1}{4} +z_{3}\right)a \, \mathbf{\hat{z}} & \left(48e\right) & \mbox{Cu II} \\ 
\mathbf{B}_{28} & = & \left(\frac{1}{2} - x_{3} + z_{3}\right) \, \mathbf{a}_{1} + \left(-y_{3}+z_{3}\right) \, \mathbf{a}_{2} + \left(-x_{3}-y_{3}\right) \, \mathbf{a}_{3} & = & -a\left(y_{3}+\frac{1}{4}\right) \, \mathbf{\hat{x}} + \left(\frac{1}{4} - x_{3}\right)a \, \mathbf{\hat{y}} + \left(\frac{1}{4} +z_{3}\right)a \, \mathbf{\hat{z}} & \left(48e\right) & \mbox{Cu II} \\ 
\mathbf{B}_{29} & = & \left(-x_{3}-z_{3}\right) \, \mathbf{a}_{1} + \left(\frac{1}{2} +y_{3} - z_{3}\right) \, \mathbf{a}_{2} + \left(-x_{3}+y_{3}\right) \, \mathbf{a}_{3} & = & \left(\frac{1}{4} +y_{3}\right)a \, \mathbf{\hat{x}}-a\left(x_{3}+\frac{1}{4}\right) \, \mathbf{\hat{y}} + \left(\frac{1}{4} - z_{3}\right)a \, \mathbf{\hat{z}} & \left(48e\right) & \mbox{Cu II} \\ 
\mathbf{B}_{30} & = & \left(x_{3}-z_{3}\right) \, \mathbf{a}_{1} + \left(-y_{3}-z_{3}\right) \, \mathbf{a}_{2} + \left(\frac{1}{2} +x_{3} - y_{3}\right) \, \mathbf{a}_{3} & = & \left(\frac{1}{4} - y_{3}\right)a \, \mathbf{\hat{x}} + \left(\frac{1}{4} +x_{3}\right)a \, \mathbf{\hat{y}}-a\left(z_{3}+\frac{1}{4}\right) \, \mathbf{\hat{z}} & \left(48e\right) & \mbox{Cu II} \\ 
\mathbf{B}_{31} & = & \left(\frac{1}{2} +y_{3} + z_{3}\right) \, \mathbf{a}_{1} + \left(\frac{1}{2} +x_{3} + y_{3}\right) \, \mathbf{a}_{2} + \left(\frac{1}{2} +x_{3} + z_{3}\right) \, \mathbf{a}_{3} & = & \left(\frac{1}{4} +x_{3}\right)a \, \mathbf{\hat{x}} + \left(\frac{1}{4} +z_{3}\right)a \, \mathbf{\hat{y}} + \left(\frac{1}{4} +y_{3}\right)a \, \mathbf{\hat{z}} & \left(48e\right) & \mbox{Cu II} \\ 
\mathbf{B}_{32} & = & \left(-y_{3}+z_{3}\right) \, \mathbf{a}_{1} + \left(-x_{3}-y_{3}\right) \, \mathbf{a}_{2} + \left(\frac{1}{2} - x_{3} + z_{3}\right) \, \mathbf{a}_{3} & = & \left(\frac{1}{4} - x_{3}\right)a \, \mathbf{\hat{x}} + \left(\frac{1}{4} +z_{3}\right)a \, \mathbf{\hat{y}}-a\left(y_{3}+\frac{1}{4}\right) \, \mathbf{\hat{z}} & \left(48e\right) & \mbox{Cu II} \\ 
\mathbf{B}_{33} & = & \left(\frac{1}{2} +y_{3} - z_{3}\right) \, \mathbf{a}_{1} + \left(-x_{3}+y_{3}\right) \, \mathbf{a}_{2} + \left(-x_{3}-z_{3}\right) \, \mathbf{a}_{3} & = & -a\left(x_{3}+\frac{1}{4}\right) \, \mathbf{\hat{x}} + \left(\frac{1}{4} - z_{3}\right)a \, \mathbf{\hat{y}} + \left(\frac{1}{4} +y_{3}\right)a \, \mathbf{\hat{z}} & \left(48e\right) & \mbox{Cu II} \\ 
\mathbf{B}_{34} & = & \left(-y_{3}-z_{3}\right) \, \mathbf{a}_{1} + \left(\frac{1}{2} +x_{3} - y_{3}\right) \, \mathbf{a}_{2} + \left(x_{3}-z_{3}\right) \, \mathbf{a}_{3} & = & \left(\frac{1}{4} +x_{3}\right)a \, \mathbf{\hat{x}}-a\left(z_{3}+\frac{1}{4}\right) \, \mathbf{\hat{y}} + \left(\frac{1}{4} - y_{3}\right)a \, \mathbf{\hat{z}} & \left(48e\right) & \mbox{Cu II} \\ 
\mathbf{B}_{35} & = & \left(\frac{1}{2} +x_{3} + y_{3}\right) \, \mathbf{a}_{1} + \left(\frac{1}{2} +x_{3} + z_{3}\right) \, \mathbf{a}_{2} + \left(\frac{1}{2} +y_{3} + z_{3}\right) \, \mathbf{a}_{3} & = & \left(\frac{1}{4} +z_{3}\right)a \, \mathbf{\hat{x}} + \left(\frac{1}{4} +y_{3}\right)a \, \mathbf{\hat{y}} + \left(\frac{1}{4} +x_{3}\right)a \, \mathbf{\hat{z}} & \left(48e\right) & \mbox{Cu II} \\ 
\mathbf{B}_{36} & = & \left(-x_{3}-y_{3}\right) \, \mathbf{a}_{1} + \left(\frac{1}{2} - x_{3} + z_{3}\right) \, \mathbf{a}_{2} + \left(-y_{3}+z_{3}\right) \, \mathbf{a}_{3} & = & \left(\frac{1}{4} +z_{3}\right)a \, \mathbf{\hat{x}}-a\left(y_{3}+\frac{1}{4}\right) \, \mathbf{\hat{y}} + \left(\frac{1}{4} - x_{3}\right)a \, \mathbf{\hat{z}} & \left(48e\right) & \mbox{Cu II} \\ 
\mathbf{B}_{37} & = & \left(-x_{3}+y_{3}\right) \, \mathbf{a}_{1} + \left(-x_{3}-z_{3}\right) \, \mathbf{a}_{2} + \left(\frac{1}{2} +y_{3} - z_{3}\right) \, \mathbf{a}_{3} & = & \left(\frac{1}{4} - z_{3}\right)a \, \mathbf{\hat{x}} + \left(\frac{1}{4} +y_{3}\right)a \, \mathbf{\hat{y}}-a\left(x_{3}+\frac{1}{4}\right) \, \mathbf{\hat{z}} & \left(48e\right) & \mbox{Cu II} \\ 
\mathbf{B}_{38} & = & \left(\frac{1}{2} +x_{3} - y_{3}\right) \, \mathbf{a}_{1} + \left(x_{3}-z_{3}\right) \, \mathbf{a}_{2} + \left(-y_{3}-z_{3}\right) \, \mathbf{a}_{3} & = & -a\left(z_{3}+\frac{1}{4}\right) \, \mathbf{\hat{x}} + \left(\frac{1}{4} - y_{3}\right)a \, \mathbf{\hat{y}} + \left(\frac{1}{4} +x_{3}\right)a \, \mathbf{\hat{z}} & \left(48e\right) & \mbox{Cu II} \\ 
\end{longtabu}
\renewcommand{\arraystretch}{1.0}
\noindent \hrulefill
\\
\textbf{References:}
\vspace*{-0.25cm}
\begin{flushleft}
  - \bibentry{Mattern_JAC_249_2007}. \\
\end{flushleft}
\textbf{Found in:}
\vspace*{-0.25cm}
\begin{flushleft}
  - \bibentry{Sufryd_Intermetallics_19_2011}. \\
\end{flushleft}
\noindent \hrulefill
\\
\textbf{Geometry files:}
\\
\noindent  - CIF: pp. {\hyperref[A15B4_cI76_220_ae_c_cif]{\pageref{A15B4_cI76_220_ae_c_cif}}} \\
\noindent  - POSCAR: pp. {\hyperref[A15B4_cI76_220_ae_c_poscar]{\pageref{A15B4_cI76_220_ae_c_poscar}}} \\
\onecolumn
{\phantomsection\label{A4B3_cI28_220_c_a}}
\subsection*{\huge \textbf{{\normalfont Th$_{3}$P$_{4}$ ($D7_{3}$) Structure: A4B3\_cI28\_220\_c\_a}}}
\noindent \hrulefill
\vspace*{0.25cm}
\begin{figure}[htp]
  \centering
  \vspace{-1em}
  {\includegraphics[width=1\textwidth]{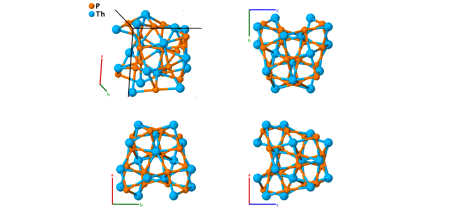}}
\end{figure}
\vspace*{-0.5cm}
\renewcommand{\arraystretch}{1.5}
\begin{equation*}
  \begin{array}{>{$\hspace{-0.15cm}}l<{$}>{$}p{0.5cm}<{$}>{$}p{18.5cm}<{$}}
    \mbox{\large \textbf{Prototype}} &\colon & \ce{Th3P4} \\
    \mbox{\large \textbf{\AFLOW\ prototype label}} &\colon & \mbox{A4B3\_cI28\_220\_c\_a} \\
    \mbox{\large \textbf{\textit{Strukturbericht} designation}} &\colon & \mbox{$D7_{3}$} \\
    \mbox{\large \textbf{Pearson symbol}} &\colon & \mbox{cI28} \\
    \mbox{\large \textbf{Space group number}} &\colon & 220 \\
    \mbox{\large \textbf{Space group symbol}} &\colon & I\bar{4}3d \\
    \mbox{\large \textbf{\AFLOW\ prototype command}} &\colon &  \texttt{aflow} \,  \, \texttt{-{}-proto=A4B3\_cI28\_220\_c\_a } \, \newline \texttt{-{}-params=}{a,x_{2} }
  \end{array}
\end{equation*}
\renewcommand{\arraystretch}{1.0}

\vspace*{-0.25cm}
\noindent \hrulefill
\\
\textbf{ Other compounds with this structure:}
\begin{itemize}
   \item{ Th$_{3}$As$_{4}$, U$_{3}$AS$_{4}$, U$_{3}$Bi$_{4}$, N$_{3}$P$_{4}$, Th$_{3}$P$_{4}$, U$_{3}$P$_{4}$, Th$_{3}$Sb$_{4}$, U$_{3}$Sb$_{4}$, U$_{3}$Te$_{4}$  }
\end{itemize}
\noindent \parbox{1 \linewidth}{
\noindent \hrulefill
\\
\textbf{Body-centered Cubic primitive vectors:} \\
\vspace*{-0.25cm}
\begin{tabular}{cc}
  \begin{tabular}{c}
    \parbox{0.6 \linewidth}{
      \renewcommand{\arraystretch}{1.5}
      \begin{equation*}
        \centering
        \begin{array}{ccc}
              \mathbf{a}_1 & = & - \frac12 \, a \, \mathbf{\hat{x}} + \frac12 \, a \, \mathbf{\hat{y}} + \frac12 \, a \, \mathbf{\hat{z}} \\
    \mathbf{a}_2 & = & ~ \frac12 \, a \, \mathbf{\hat{x}} - \frac12 \, a \, \mathbf{\hat{y}} + \frac12 \, a \, \mathbf{\hat{z}} \\
    \mathbf{a}_3 & = & ~ \frac12 \, a \, \mathbf{\hat{x}} + \frac12 \, a \, \mathbf{\hat{y}} - \frac12 \, a \, \mathbf{\hat{z}} \\

        \end{array}
      \end{equation*}
    }
    \renewcommand{\arraystretch}{1.0}
  \end{tabular}
  \begin{tabular}{c}
    \includegraphics[width=0.3\linewidth]{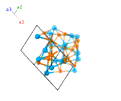} \\
  \end{tabular}
\end{tabular}

}
\vspace*{-0.25cm}

\noindent \hrulefill
\\
\textbf{Basis vectors:}
\vspace*{-0.25cm}
\renewcommand{\arraystretch}{1.5}
\begin{longtabu} to \textwidth{>{\centering $}X[-1,c,c]<{$}>{\centering $}X[-1,c,c]<{$}>{\centering $}X[-1,c,c]<{$}>{\centering $}X[-1,c,c]<{$}>{\centering $}X[-1,c,c]<{$}>{\centering $}X[-1,c,c]<{$}>{\centering $}X[-1,c,c]<{$}}
  & & \mbox{Lattice Coordinates} & & \mbox{Cartesian Coordinates} &\mbox{Wyckoff Position} & \mbox{Atom Type} \\  
  \mathbf{B}_{1} & = & \frac{1}{4} \, \mathbf{a}_{1} + \frac{5}{8} \, \mathbf{a}_{2} + \frac{3}{8} \, \mathbf{a}_{3} & = & \frac{3}{8}a \, \mathbf{\hat{x}} + \frac{1}{4}a \, \mathbf{\hat{z}} & \left(12a\right) & \mbox{Th} \\ 
\mathbf{B}_{2} & = & \frac{3}{4} \, \mathbf{a}_{1} + \frac{7}{8} \, \mathbf{a}_{2} + \frac{1}{8} \, \mathbf{a}_{3} & = & \frac{1}{8}a \, \mathbf{\hat{x}} + \frac{3}{4}a \, \mathbf{\hat{z}} & \left(12a\right) & \mbox{Th} \\ 
\mathbf{B}_{3} & = & \frac{3}{8} \, \mathbf{a}_{1} + \frac{1}{4} \, \mathbf{a}_{2} + \frac{5}{8} \, \mathbf{a}_{3} & = & \frac{1}{4}a \, \mathbf{\hat{x}} + \frac{3}{8}a \, \mathbf{\hat{y}} & \left(12a\right) & \mbox{Th} \\ 
\mathbf{B}_{4} & = & \frac{1}{8} \, \mathbf{a}_{1} + \frac{3}{4} \, \mathbf{a}_{2} + \frac{7}{8} \, \mathbf{a}_{3} & = & \frac{3}{4}a \, \mathbf{\hat{x}} + \frac{1}{8}a \, \mathbf{\hat{y}} & \left(12a\right) & \mbox{Th} \\ 
\mathbf{B}_{5} & = & \frac{5}{8} \, \mathbf{a}_{1} + \frac{3}{8} \, \mathbf{a}_{2} + \frac{1}{4} \, \mathbf{a}_{3} & = & \frac{1}{4}a \, \mathbf{\hat{y}} + \frac{3}{8}a \, \mathbf{\hat{z}} & \left(12a\right) & \mbox{Th} \\ 
\mathbf{B}_{6} & = & \frac{7}{8} \, \mathbf{a}_{1} + \frac{1}{8} \, \mathbf{a}_{2} + \frac{3}{4} \, \mathbf{a}_{3} & = & \frac{3}{4}a \, \mathbf{\hat{y}} + \frac{1}{8}a \, \mathbf{\hat{z}} & \left(12a\right) & \mbox{Th} \\ 
\mathbf{B}_{7} & = & 2x_{2} \, \mathbf{a}_{1} + 2x_{2} \, \mathbf{a}_{2} + 2x_{2} \, \mathbf{a}_{3} & = & x_{2}a \, \mathbf{\hat{x}} + x_{2}a \, \mathbf{\hat{y}} + x_{2}a \, \mathbf{\hat{z}} & \left(16c\right) & \mbox{P} \\ 
\mathbf{B}_{8} & = & \frac{1}{2} \, \mathbf{a}_{1} + \left(\frac{1}{2} - 2x_{2}\right) \, \mathbf{a}_{3} & = & -x_{2}a \, \mathbf{\hat{x}} + \left(\frac{1}{2} - x_{2}\right)a \, \mathbf{\hat{y}} + x_{2}a \, \mathbf{\hat{z}} & \left(16c\right) & \mbox{P} \\ 
\mathbf{B}_{9} & = & \left(\frac{1}{2} - 2x_{2}\right) \, \mathbf{a}_{2} + \frac{1}{2} \, \mathbf{a}_{3} & = & \left(\frac{1}{2} - x_{2}\right)a \, \mathbf{\hat{x}} + x_{2}a \, \mathbf{\hat{y}}-x_{2}a \, \mathbf{\hat{z}} & \left(16c\right) & \mbox{P} \\ 
\mathbf{B}_{10} & = & \left(\frac{1}{2} - 2x_{2}\right) \, \mathbf{a}_{1} + \frac{1}{2} \, \mathbf{a}_{2} & = & x_{2}a \, \mathbf{\hat{x}}-x_{2}a \, \mathbf{\hat{y}} + \left(\frac{1}{2} - x_{2}\right)a \, \mathbf{\hat{z}} & \left(16c\right) & \mbox{P} \\ 
\mathbf{B}_{11} & = & \left(\frac{1}{2} +2x_{2}\right) \, \mathbf{a}_{1} + \left(\frac{1}{2} +2x_{2}\right) \, \mathbf{a}_{2} + \left(\frac{1}{2} +2x_{2}\right) \, \mathbf{a}_{3} & = & \left(\frac{1}{4} +x_{2}\right)a \, \mathbf{\hat{x}} + \left(\frac{1}{4} +x_{2}\right)a \, \mathbf{\hat{y}} + \left(\frac{1}{4} +x_{2}\right)a \, \mathbf{\hat{z}} & \left(16c\right) & \mbox{P} \\ 
\mathbf{B}_{12} & = & \frac{1}{2} \, \mathbf{a}_{1} + -2x_{2} \, \mathbf{a}_{3} & = & -a\left(x_{2}+\frac{1}{4}\right) \, \mathbf{\hat{x}} + \left(\frac{1}{4} - x_{2}\right)a \, \mathbf{\hat{y}} + \left(\frac{1}{4} +x_{2}\right)a \, \mathbf{\hat{z}} & \left(16c\right) & \mbox{P} \\ 
\mathbf{B}_{13} & = & -2x_{2} \, \mathbf{a}_{1} + \frac{1}{2} \, \mathbf{a}_{2} & = & \left(\frac{1}{4} +x_{2}\right)a \, \mathbf{\hat{x}}-a\left(x_{2}+\frac{1}{4}\right) \, \mathbf{\hat{y}} + \left(\frac{1}{4} - x_{2}\right)a \, \mathbf{\hat{z}} & \left(16c\right) & \mbox{P} \\ 
\mathbf{B}_{14} & = & -2x_{2} \, \mathbf{a}_{2} + \frac{1}{2} \, \mathbf{a}_{3} & = & \left(\frac{1}{4} - x_{2}\right)a \, \mathbf{\hat{x}} + \left(\frac{1}{4} +x_{2}\right)a \, \mathbf{\hat{y}}-a\left(x_{2}+\frac{1}{4}\right) \, \mathbf{\hat{z}} & \left(16c\right) & \mbox{P} \\ 
\end{longtabu}
\renewcommand{\arraystretch}{1.0}
\noindent \hrulefill
\\
\textbf{References:}
\vspace*{-0.25cm}
\begin{flushleft}
  - \bibentry{Meisel_ZAAC_240_1939}. \\
\end{flushleft}
\noindent \hrulefill
\\
\textbf{Geometry files:}
\\
\noindent  - CIF: pp. {\hyperref[A4B3_cI28_220_c_a_cif]{\pageref{A4B3_cI28_220_c_a_cif}}} \\
\noindent  - POSCAR: pp. {\hyperref[A4B3_cI28_220_c_a_poscar]{\pageref{A4B3_cI28_220_c_a_poscar}}} \\
\onecolumn
{\phantomsection\label{A2B3C6_cP33_221_cd_ag_fh}}
\subsection*{\huge \textbf{{\normalfont \begin{raggedleft}Ca$_{3}$Al$_{2}$O$_{6}$ ($E9_{1}$) Structure: \end{raggedleft} \\ A2B3C6\_cP33\_221\_cd\_ag\_fh}}}
\noindent \hrulefill
\vspace*{0.25cm}
\begin{figure}[htp]
  \centering
  \vspace{-1em}
  {\includegraphics[width=1\textwidth]{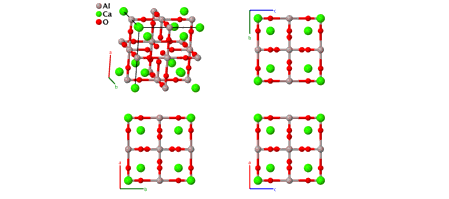}}
\end{figure}
\vspace*{-0.5cm}
\renewcommand{\arraystretch}{1.5}
\begin{equation*}
  \begin{array}{>{$\hspace{-0.15cm}}l<{$}>{$}p{0.5cm}<{$}>{$}p{18.5cm}<{$}}
    \mbox{\large \textbf{Prototype}} &\colon & \ce{Ca3Al2O6} \\
    \mbox{\large \textbf{\AFLOW\ prototype label}} &\colon & \mbox{A2B3C6\_cP33\_221\_cd\_ag\_fh} \\
    \mbox{\large \textbf{\textit{Strukturbericht} designation}} &\colon & \mbox{$E9_{1}$} \\
    \mbox{\large \textbf{Pearson symbol}} &\colon & \mbox{cP33} \\
    \mbox{\large \textbf{Space group number}} &\colon & 221 \\
    \mbox{\large \textbf{Space group symbol}} &\colon & Pm\bar{3}m \\
    \mbox{\large \textbf{\AFLOW\ prototype command}} &\colon &  \texttt{aflow} \,  \, \texttt{-{}-proto=A2B3C6\_cP33\_221\_cd\_ag\_fh } \, \newline \texttt{-{}-params=}{a,x_{4},x_{5},x_{6} }
  \end{array}
\end{equation*}
\renewcommand{\arraystretch}{1.0}

\vspace*{-0.25cm}
\noindent \hrulefill
\begin{itemize}
  \item{(Steele, 1929) do not use the standard Wyckoff position notation to
describe the atomic positions, so we use the parameters found in
(Herman, 1937).
An alternative description of the structure places the O I atoms on
the (6e) $(\pm x , 0 , 0) \dots$ site rather than the (6f) site.}
  \item{(Mondal, 1975) reanalyzed this structure and concluded that the true
structure was one where the lattice constant was doubled and contained
264 atoms. See \hyperref[A2B3C6_cP264_205_2d_ab2c2d_6d]{the
  Ca$_{3}$Al$_{2}$O$_{6}$ structure page}.
}
\end{itemize}

\noindent \parbox{1 \linewidth}{
\noindent \hrulefill
\\
\textbf{Simple Cubic primitive vectors:} \\
\vspace*{-0.25cm}
\begin{tabular}{cc}
  \begin{tabular}{c}
    \parbox{0.6 \linewidth}{
      \renewcommand{\arraystretch}{1.5}
      \begin{equation*}
        \centering
        \begin{array}{ccc}
              \mathbf{a}_1 & = & a \, \mathbf{\hat{x}} \\
    \mathbf{a}_2 & = & a \, \mathbf{\hat{y}} \\
    \mathbf{a}_3 & = & a \, \mathbf{\hat{z}} \\

        \end{array}
      \end{equation*}
    }
    \renewcommand{\arraystretch}{1.0}
  \end{tabular}
  \begin{tabular}{c}
    \includegraphics[width=0.3\linewidth]{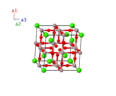} \\
  \end{tabular}
\end{tabular}

}
\vspace*{-0.25cm}

\noindent \hrulefill
\\
\textbf{Basis vectors:}
\vspace*{-0.25cm}
\renewcommand{\arraystretch}{1.5}
\begin{longtabu} to \textwidth{>{\centering $}X[-1,c,c]<{$}>{\centering $}X[-1,c,c]<{$}>{\centering $}X[-1,c,c]<{$}>{\centering $}X[-1,c,c]<{$}>{\centering $}X[-1,c,c]<{$}>{\centering $}X[-1,c,c]<{$}>{\centering $}X[-1,c,c]<{$}}
  & & \mbox{Lattice Coordinates} & & \mbox{Cartesian Coordinates} &\mbox{Wyckoff Position} & \mbox{Atom Type} \\  
  \mathbf{B}_{1} & = & 0 \, \mathbf{a}_{1} + 0 \, \mathbf{a}_{2} + 0 \, \mathbf{a}_{3} & = & 0 \, \mathbf{\hat{x}} + 0 \, \mathbf{\hat{y}} + 0 \, \mathbf{\hat{z}} & \left(1a\right) & \mbox{Ca I} \\ 
\mathbf{B}_{2} & = & \frac{1}{2} \, \mathbf{a}_{2} + \frac{1}{2} \, \mathbf{a}_{3} & = & \frac{1}{2}a \, \mathbf{\hat{y}} + \frac{1}{2}a \, \mathbf{\hat{z}} & \left(3c\right) & \mbox{Al I} \\ 
\mathbf{B}_{3} & = & \frac{1}{2} \, \mathbf{a}_{1} + \frac{1}{2} \, \mathbf{a}_{3} & = & \frac{1}{2}a \, \mathbf{\hat{x}} + \frac{1}{2}a \, \mathbf{\hat{z}} & \left(3c\right) & \mbox{Al I} \\ 
\mathbf{B}_{4} & = & \frac{1}{2} \, \mathbf{a}_{1} + \frac{1}{2} \, \mathbf{a}_{2} & = & \frac{1}{2}a \, \mathbf{\hat{x}} + \frac{1}{2}a \, \mathbf{\hat{y}} & \left(3c\right) & \mbox{Al I} \\ 
\mathbf{B}_{5} & = & \frac{1}{2} \, \mathbf{a}_{1} & = & \frac{1}{2}a \, \mathbf{\hat{x}} & \left(3d\right) & \mbox{Al II} \\ 
\mathbf{B}_{6} & = & \frac{1}{2} \, \mathbf{a}_{2} & = & \frac{1}{2}a \, \mathbf{\hat{y}} & \left(3d\right) & \mbox{Al II} \\ 
\mathbf{B}_{7} & = & \frac{1}{2} \, \mathbf{a}_{3} & = & \frac{1}{2}a \, \mathbf{\hat{z}} & \left(3d\right) & \mbox{Al II} \\ 
\mathbf{B}_{8} & = & x_{4} \, \mathbf{a}_{1} + \frac{1}{2} \, \mathbf{a}_{2} + \frac{1}{2} \, \mathbf{a}_{3} & = & x_{4}a \, \mathbf{\hat{x}} + \frac{1}{2}a \, \mathbf{\hat{y}} + \frac{1}{2}a \, \mathbf{\hat{z}} & \left(6f\right) & \mbox{O I} \\ 
\mathbf{B}_{9} & = & -x_{4} \, \mathbf{a}_{1} + \frac{1}{2} \, \mathbf{a}_{2} + \frac{1}{2} \, \mathbf{a}_{3} & = & -x_{4}a \, \mathbf{\hat{x}} + \frac{1}{2}a \, \mathbf{\hat{y}} + \frac{1}{2}a \, \mathbf{\hat{z}} & \left(6f\right) & \mbox{O I} \\ 
\mathbf{B}_{10} & = & \frac{1}{2} \, \mathbf{a}_{1} + x_{4} \, \mathbf{a}_{2} + \frac{1}{2} \, \mathbf{a}_{3} & = & \frac{1}{2}a \, \mathbf{\hat{x}} + x_{4}a \, \mathbf{\hat{y}} + \frac{1}{2}a \, \mathbf{\hat{z}} & \left(6f\right) & \mbox{O I} \\ 
\mathbf{B}_{11} & = & \frac{1}{2} \, \mathbf{a}_{1}-x_{4} \, \mathbf{a}_{2} + \frac{1}{2} \, \mathbf{a}_{3} & = & \frac{1}{2}a \, \mathbf{\hat{x}}-x_{4}a \, \mathbf{\hat{y}} + \frac{1}{2}a \, \mathbf{\hat{z}} & \left(6f\right) & \mbox{O I} \\ 
\mathbf{B}_{12} & = & \frac{1}{2} \, \mathbf{a}_{1} + \frac{1}{2} \, \mathbf{a}_{2} + x_{4} \, \mathbf{a}_{3} & = & \frac{1}{2}a \, \mathbf{\hat{x}} + \frac{1}{2}a \, \mathbf{\hat{y}} + x_{4}a \, \mathbf{\hat{z}} & \left(6f\right) & \mbox{O I} \\ 
\mathbf{B}_{13} & = & \frac{1}{2} \, \mathbf{a}_{1} + \frac{1}{2} \, \mathbf{a}_{2}-x_{4} \, \mathbf{a}_{3} & = & \frac{1}{2}a \, \mathbf{\hat{x}} + \frac{1}{2}a \, \mathbf{\hat{y}}-x_{4}a \, \mathbf{\hat{z}} & \left(6f\right) & \mbox{O I} \\ 
\mathbf{B}_{14} & = & x_{5} \, \mathbf{a}_{1} + x_{5} \, \mathbf{a}_{2} + x_{5} \, \mathbf{a}_{3} & = & x_{5}a \, \mathbf{\hat{x}} + x_{5}a \, \mathbf{\hat{y}} + x_{5}a \, \mathbf{\hat{z}} & \left(8g\right) & \mbox{Ca II} \\ 
\mathbf{B}_{15} & = & -x_{5} \, \mathbf{a}_{1}-x_{5} \, \mathbf{a}_{2} + x_{5} \, \mathbf{a}_{3} & = & -x_{5}a \, \mathbf{\hat{x}}-x_{5}a \, \mathbf{\hat{y}} + x_{5}a \, \mathbf{\hat{z}} & \left(8g\right) & \mbox{Ca II} \\ 
\mathbf{B}_{16} & = & -x_{5} \, \mathbf{a}_{1} + x_{5} \, \mathbf{a}_{2}-x_{5} \, \mathbf{a}_{3} & = & -x_{5}a \, \mathbf{\hat{x}} + x_{5}a \, \mathbf{\hat{y}}-x_{5}a \, \mathbf{\hat{z}} & \left(8g\right) & \mbox{Ca II} \\ 
\mathbf{B}_{17} & = & x_{5} \, \mathbf{a}_{1}-x_{5} \, \mathbf{a}_{2}-x_{5} \, \mathbf{a}_{3} & = & x_{5}a \, \mathbf{\hat{x}}-x_{5}a \, \mathbf{\hat{y}}-x_{5}a \, \mathbf{\hat{z}} & \left(8g\right) & \mbox{Ca II} \\ 
\mathbf{B}_{18} & = & x_{5} \, \mathbf{a}_{1} + x_{5} \, \mathbf{a}_{2}-x_{5} \, \mathbf{a}_{3} & = & x_{5}a \, \mathbf{\hat{x}} + x_{5}a \, \mathbf{\hat{y}}-x_{5}a \, \mathbf{\hat{z}} & \left(8g\right) & \mbox{Ca II} \\ 
\mathbf{B}_{19} & = & -x_{5} \, \mathbf{a}_{1}-x_{5} \, \mathbf{a}_{2}-x_{5} \, \mathbf{a}_{3} & = & -x_{5}a \, \mathbf{\hat{x}}-x_{5}a \, \mathbf{\hat{y}}-x_{5}a \, \mathbf{\hat{z}} & \left(8g\right) & \mbox{Ca II} \\ 
\mathbf{B}_{20} & = & x_{5} \, \mathbf{a}_{1}-x_{5} \, \mathbf{a}_{2} + x_{5} \, \mathbf{a}_{3} & = & x_{5}a \, \mathbf{\hat{x}}-x_{5}a \, \mathbf{\hat{y}} + x_{5}a \, \mathbf{\hat{z}} & \left(8g\right) & \mbox{Ca II} \\ 
\mathbf{B}_{21} & = & -x_{5} \, \mathbf{a}_{1} + x_{5} \, \mathbf{a}_{2} + x_{5} \, \mathbf{a}_{3} & = & -x_{5}a \, \mathbf{\hat{x}} + x_{5}a \, \mathbf{\hat{y}} + x_{5}a \, \mathbf{\hat{z}} & \left(8g\right) & \mbox{Ca II} \\ 
\mathbf{B}_{22} & = & x_{6} \, \mathbf{a}_{1} + \frac{1}{2} \, \mathbf{a}_{2} & = & x_{6}a \, \mathbf{\hat{x}} + \frac{1}{2}a \, \mathbf{\hat{y}} & \left(12h\right) & \mbox{O II} \\ 
\mathbf{B}_{23} & = & -x_{6} \, \mathbf{a}_{1} + \frac{1}{2} \, \mathbf{a}_{2} & = & -x_{6}a \, \mathbf{\hat{x}} + \frac{1}{2}a \, \mathbf{\hat{y}} & \left(12h\right) & \mbox{O II} \\ 
\mathbf{B}_{24} & = & x_{6} \, \mathbf{a}_{2} + \frac{1}{2} \, \mathbf{a}_{3} & = & x_{6}a \, \mathbf{\hat{y}} + \frac{1}{2}a \, \mathbf{\hat{z}} & \left(12h\right) & \mbox{O II} \\ 
\mathbf{B}_{25} & = & -x_{6} \, \mathbf{a}_{2} + \frac{1}{2} \, \mathbf{a}_{3} & = & -x_{6}a \, \mathbf{\hat{y}} + \frac{1}{2}a \, \mathbf{\hat{z}} & \left(12h\right) & \mbox{O II} \\ 
\mathbf{B}_{26} & = & \frac{1}{2} \, \mathbf{a}_{1} + x_{6} \, \mathbf{a}_{3} & = & \frac{1}{2}a \, \mathbf{\hat{x}} + x_{6}a \, \mathbf{\hat{z}} & \left(12h\right) & \mbox{O II} \\ 
\mathbf{B}_{27} & = & \frac{1}{2} \, \mathbf{a}_{1} + -x_{6} \, \mathbf{a}_{3} & = & \frac{1}{2}a \, \mathbf{\hat{x}} + -x_{6}a \, \mathbf{\hat{z}} & \left(12h\right) & \mbox{O II} \\ 
\mathbf{B}_{28} & = & \frac{1}{2} \, \mathbf{a}_{1} + x_{6} \, \mathbf{a}_{2} & = & \frac{1}{2}a \, \mathbf{\hat{x}} + x_{6}a \, \mathbf{\hat{y}} & \left(12h\right) & \mbox{O II} \\ 
\mathbf{B}_{29} & = & \frac{1}{2} \, \mathbf{a}_{1}-x_{6} \, \mathbf{a}_{2} & = & \frac{1}{2}a \, \mathbf{\hat{x}}-x_{6}a \, \mathbf{\hat{y}} & \left(12h\right) & \mbox{O II} \\ 
\mathbf{B}_{30} & = & x_{6} \, \mathbf{a}_{1} + \frac{1}{2} \, \mathbf{a}_{3} & = & x_{6}a \, \mathbf{\hat{x}} + \frac{1}{2}a \, \mathbf{\hat{z}} & \left(12h\right) & \mbox{O II} \\ 
\mathbf{B}_{31} & = & -x_{6} \, \mathbf{a}_{1} + \frac{1}{2} \, \mathbf{a}_{3} & = & -x_{6}a \, \mathbf{\hat{x}} + \frac{1}{2}a \, \mathbf{\hat{z}} & \left(12h\right) & \mbox{O II} \\ 
\mathbf{B}_{32} & = & \frac{1}{2} \, \mathbf{a}_{2}-x_{6} \, \mathbf{a}_{3} & = & \frac{1}{2}a \, \mathbf{\hat{y}}-x_{6}a \, \mathbf{\hat{z}} & \left(12h\right) & \mbox{O II} \\ 
\mathbf{B}_{33} & = & \frac{1}{2} \, \mathbf{a}_{2} + x_{6} \, \mathbf{a}_{3} & = & \frac{1}{2}a \, \mathbf{\hat{y}} + x_{6}a \, \mathbf{\hat{z}} & \left(12h\right) & \mbox{O II} \\ 
\end{longtabu}
\renewcommand{\arraystretch}{1.0}
\noindent \hrulefill
\\
\textbf{References:}
\vspace*{-0.25cm}
\begin{flushleft}
  - \bibentry{Steele_JACS_1929}. \\
  - \bibentry{Hermann_StrukII_1937}. \\
\end{flushleft}
\textbf{Found in:}
\vspace*{-0.25cm}
\begin{flushleft}
  - \bibentry{mondal75:Ca3Al2O6}. \\
\end{flushleft}
\noindent \hrulefill
\\
\textbf{Geometry files:}
\\
\noindent  - CIF: pp. {\hyperref[A2B3C6_cP33_221_cd_ag_fh_cif]{\pageref{A2B3C6_cP33_221_cd_ag_fh_cif}}} \\
\noindent  - POSCAR: pp. {\hyperref[A2B3C6_cP33_221_cd_ag_fh_poscar]{\pageref{A2B3C6_cP33_221_cd_ag_fh_poscar}}} \\
\onecolumn
{\phantomsection\label{A5B3C16_cP96_222_ce_d_fi}}
\subsection*{\huge \textbf{{\normalfont Ce$_{5}$Mo$_{3}$O$_{16}$ Structure: A5B3C16\_cP96\_222\_ce\_d\_fi}}}
\noindent \hrulefill
\vspace*{0.25cm}
\begin{figure}[htp]
  \centering
  \vspace{-1em}
  {\includegraphics[width=1\textwidth]{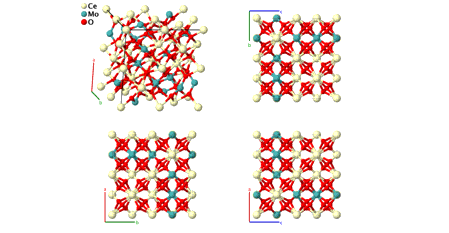}}
\end{figure}
\vspace*{-0.5cm}
\renewcommand{\arraystretch}{1.5}
\begin{equation*}
  \begin{array}{>{$\hspace{-0.15cm}}l<{$}>{$}p{0.5cm}<{$}>{$}p{18.5cm}<{$}}
    \mbox{\large \textbf{Prototype}} &\colon & \ce{Ce5Mo3O16} \\
    \mbox{\large \textbf{\AFLOW\ prototype label}} &\colon & \mbox{A5B3C16\_cP96\_222\_ce\_d\_fi} \\
    \mbox{\large \textbf{\textit{Strukturbericht} designation}} &\colon & \mbox{None} \\
    \mbox{\large \textbf{Pearson symbol}} &\colon & \mbox{cP96} \\
    \mbox{\large \textbf{Space group number}} &\colon & 222 \\
    \mbox{\large \textbf{Space group symbol}} &\colon & Pn\bar{3}n \\
    \mbox{\large \textbf{\AFLOW\ prototype command}} &\colon &  \texttt{aflow} \,  \, \texttt{-{}-proto=A5B3C16\_cP96\_222\_ce\_d\_fi } \, \newline \texttt{-{}-params=}{a,x_{3},x_{4},x_{5},y_{5},z_{5} }
  \end{array}
\end{equation*}
\renewcommand{\arraystretch}{1.0}

\noindent \parbox{1 \linewidth}{
\noindent \hrulefill
\\
\textbf{Simple Cubic primitive vectors:} \\
\vspace*{-0.25cm}
\begin{tabular}{cc}
  \begin{tabular}{c}
    \parbox{0.6 \linewidth}{
      \renewcommand{\arraystretch}{1.5}
      \begin{equation*}
        \centering
        \begin{array}{ccc}
              \mathbf{a}_1 & = & a \, \mathbf{\hat{x}} \\
    \mathbf{a}_2 & = & a \, \mathbf{\hat{y}} \\
    \mathbf{a}_3 & = & a \, \mathbf{\hat{z}} \\

        \end{array}
      \end{equation*}
    }
    \renewcommand{\arraystretch}{1.0}
  \end{tabular}
  \begin{tabular}{c}
    \includegraphics[width=0.3\linewidth]{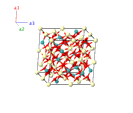} \\
  \end{tabular}
\end{tabular}

}
\vspace*{-0.25cm}

\noindent \hrulefill
\\
\textbf{Basis vectors:}
\vspace*{-0.25cm}
\renewcommand{\arraystretch}{1.5}
\begin{longtabu} to \textwidth{>{\centering $}X[-1,c,c]<{$}>{\centering $}X[-1,c,c]<{$}>{\centering $}X[-1,c,c]<{$}>{\centering $}X[-1,c,c]<{$}>{\centering $}X[-1,c,c]<{$}>{\centering $}X[-1,c,c]<{$}>{\centering $}X[-1,c,c]<{$}}
  & & \mbox{Lattice Coordinates} & & \mbox{Cartesian Coordinates} &\mbox{Wyckoff Position} & \mbox{Atom Type} \\  
  \mathbf{B}_{1} & = & 0 \, \mathbf{a}_{1} + 0 \, \mathbf{a}_{2} + 0 \, \mathbf{a}_{3} & = & 0 \, \mathbf{\hat{x}} + 0 \, \mathbf{\hat{y}} + 0 \, \mathbf{\hat{z}} & \left(8c\right) & \mbox{Ce I} \\ 
\mathbf{B}_{2} & = & \frac{1}{2} \, \mathbf{a}_{1} + \frac{1}{2} \, \mathbf{a}_{2} & = & \frac{1}{2}a \, \mathbf{\hat{x}} + \frac{1}{2}a \, \mathbf{\hat{y}} & \left(8c\right) & \mbox{Ce I} \\ 
\mathbf{B}_{3} & = & \frac{1}{2} \, \mathbf{a}_{1} + \frac{1}{2} \, \mathbf{a}_{3} & = & \frac{1}{2}a \, \mathbf{\hat{x}} + \frac{1}{2}a \, \mathbf{\hat{z}} & \left(8c\right) & \mbox{Ce I} \\ 
\mathbf{B}_{4} & = & \frac{1}{2} \, \mathbf{a}_{2} + \frac{1}{2} \, \mathbf{a}_{3} & = & \frac{1}{2}a \, \mathbf{\hat{y}} + \frac{1}{2}a \, \mathbf{\hat{z}} & \left(8c\right) & \mbox{Ce I} \\ 
\mathbf{B}_{5} & = & \frac{1}{2} \, \mathbf{a}_{3} & = & \frac{1}{2}a \, \mathbf{\hat{z}} & \left(8c\right) & \mbox{Ce I} \\ 
\mathbf{B}_{6} & = & \frac{1}{2} \, \mathbf{a}_{1} + \frac{1}{2} \, \mathbf{a}_{2} + \frac{1}{2} \, \mathbf{a}_{3} & = & \frac{1}{2}a \, \mathbf{\hat{x}} + \frac{1}{2}a \, \mathbf{\hat{y}} + \frac{1}{2}a \, \mathbf{\hat{z}} & \left(8c\right) & \mbox{Ce I} \\ 
\mathbf{B}_{7} & = & \frac{1}{2} \, \mathbf{a}_{2} & = & \frac{1}{2}a \, \mathbf{\hat{y}} & \left(8c\right) & \mbox{Ce I} \\ 
\mathbf{B}_{8} & = & \frac{1}{2} \, \mathbf{a}_{1} & = & \frac{1}{2}a \, \mathbf{\hat{x}} & \left(8c\right) & \mbox{Ce I} \\ 
\mathbf{B}_{9} & = & \frac{3}{4} \, \mathbf{a}_{2} + \frac{1}{4} \, \mathbf{a}_{3} & = & \frac{3}{4}a \, \mathbf{\hat{y}} + \frac{1}{4}a \, \mathbf{\hat{z}} & \left(12d\right) & \mbox{Mo} \\ 
\mathbf{B}_{10} & = & \frac{1}{2} \, \mathbf{a}_{1} + \frac{3}{4} \, \mathbf{a}_{2} + \frac{1}{4} \, \mathbf{a}_{3} & = & \frac{1}{2}a \, \mathbf{\hat{x}} + \frac{3}{4}a \, \mathbf{\hat{y}} + \frac{1}{4}a \, \mathbf{\hat{z}} & \left(12d\right) & \mbox{Mo} \\ 
\mathbf{B}_{11} & = & \frac{1}{4} \, \mathbf{a}_{1} + \frac{3}{4} \, \mathbf{a}_{3} & = & \frac{1}{4}a \, \mathbf{\hat{x}} + \frac{3}{4}a \, \mathbf{\hat{z}} & \left(12d\right) & \mbox{Mo} \\ 
\mathbf{B}_{12} & = & \frac{1}{4} \, \mathbf{a}_{1} + \frac{1}{2} \, \mathbf{a}_{2} + \frac{3}{4} \, \mathbf{a}_{3} & = & \frac{1}{4}a \, \mathbf{\hat{x}} + \frac{1}{2}a \, \mathbf{\hat{y}} + \frac{3}{4}a \, \mathbf{\hat{z}} & \left(12d\right) & \mbox{Mo} \\ 
\mathbf{B}_{13} & = & \frac{3}{4} \, \mathbf{a}_{1} + \frac{1}{4} \, \mathbf{a}_{2} & = & \frac{3}{4}a \, \mathbf{\hat{x}} + \frac{1}{4}a \, \mathbf{\hat{y}} & \left(12d\right) & \mbox{Mo} \\ 
\mathbf{B}_{14} & = & \frac{3}{4} \, \mathbf{a}_{1} + \frac{1}{4} \, \mathbf{a}_{2} + \frac{1}{2} \, \mathbf{a}_{3} & = & \frac{3}{4}a \, \mathbf{\hat{x}} + \frac{1}{4}a \, \mathbf{\hat{y}} + \frac{1}{2}a \, \mathbf{\hat{z}} & \left(12d\right) & \mbox{Mo} \\ 
\mathbf{B}_{15} & = & \frac{3}{4} \, \mathbf{a}_{1} + \frac{1}{4} \, \mathbf{a}_{3} & = & \frac{3}{4}a \, \mathbf{\hat{x}} + \frac{1}{4}a \, \mathbf{\hat{z}} & \left(12d\right) & \mbox{Mo} \\ 
\mathbf{B}_{16} & = & \frac{3}{4} \, \mathbf{a}_{1} + \frac{1}{2} \, \mathbf{a}_{2} + \frac{1}{4} \, \mathbf{a}_{3} & = & \frac{3}{4}a \, \mathbf{\hat{x}} + \frac{1}{2}a \, \mathbf{\hat{y}} + \frac{1}{4}a \, \mathbf{\hat{z}} & \left(12d\right) & \mbox{Mo} \\ 
\mathbf{B}_{17} & = & \frac{1}{4} \, \mathbf{a}_{2} + \frac{3}{4} \, \mathbf{a}_{3} & = & \frac{1}{4}a \, \mathbf{\hat{y}} + \frac{3}{4}a \, \mathbf{\hat{z}} & \left(12d\right) & \mbox{Mo} \\ 
\mathbf{B}_{18} & = & \frac{1}{2} \, \mathbf{a}_{1} + \frac{1}{4} \, \mathbf{a}_{2} + \frac{3}{4} \, \mathbf{a}_{3} & = & \frac{1}{2}a \, \mathbf{\hat{x}} + \frac{1}{4}a \, \mathbf{\hat{y}} + \frac{3}{4}a \, \mathbf{\hat{z}} & \left(12d\right) & \mbox{Mo} \\ 
\mathbf{B}_{19} & = & \frac{1}{4} \, \mathbf{a}_{1} + \frac{3}{4} \, \mathbf{a}_{2} + \frac{1}{2} \, \mathbf{a}_{3} & = & \frac{1}{4}a \, \mathbf{\hat{x}} + \frac{3}{4}a \, \mathbf{\hat{y}} + \frac{1}{2}a \, \mathbf{\hat{z}} & \left(12d\right) & \mbox{Mo} \\ 
\mathbf{B}_{20} & = & \frac{1}{4} \, \mathbf{a}_{1} + \frac{3}{4} \, \mathbf{a}_{2} & = & \frac{1}{4}a \, \mathbf{\hat{x}} + \frac{3}{4}a \, \mathbf{\hat{y}} & \left(12d\right) & \mbox{Mo} \\ 
\mathbf{B}_{21} & = & x_{3} \, \mathbf{a}_{1} + \frac{1}{4} \, \mathbf{a}_{2} + \frac{1}{4} \, \mathbf{a}_{3} & = & x_{3}a \, \mathbf{\hat{x}} + \frac{1}{4}a \, \mathbf{\hat{y}} + \frac{1}{4}a \, \mathbf{\hat{z}} & \left(12e\right) & \mbox{Ce II} \\ 
\mathbf{B}_{22} & = & \left(\frac{1}{2} - x_{3}\right) \, \mathbf{a}_{1} + \frac{1}{4} \, \mathbf{a}_{2} + \frac{1}{4} \, \mathbf{a}_{3} & = & \left(\frac{1}{2} - x_{3}\right)a \, \mathbf{\hat{x}} + \frac{1}{4}a \, \mathbf{\hat{y}} + \frac{1}{4}a \, \mathbf{\hat{z}} & \left(12e\right) & \mbox{Ce II} \\ 
\mathbf{B}_{23} & = & \frac{1}{4} \, \mathbf{a}_{1} + x_{3} \, \mathbf{a}_{2} + \frac{1}{4} \, \mathbf{a}_{3} & = & \frac{1}{4}a \, \mathbf{\hat{x}} + x_{3}a \, \mathbf{\hat{y}} + \frac{1}{4}a \, \mathbf{\hat{z}} & \left(12e\right) & \mbox{Ce II} \\ 
\mathbf{B}_{24} & = & \frac{1}{4} \, \mathbf{a}_{1} + \left(\frac{1}{2} - x_{3}\right) \, \mathbf{a}_{2} + \frac{1}{4} \, \mathbf{a}_{3} & = & \frac{1}{4}a \, \mathbf{\hat{x}} + \left(\frac{1}{2} - x_{3}\right)a \, \mathbf{\hat{y}} + \frac{1}{4}a \, \mathbf{\hat{z}} & \left(12e\right) & \mbox{Ce II} \\ 
\mathbf{B}_{25} & = & \frac{1}{4} \, \mathbf{a}_{1} + \frac{1}{4} \, \mathbf{a}_{2} + x_{3} \, \mathbf{a}_{3} & = & \frac{1}{4}a \, \mathbf{\hat{x}} + \frac{1}{4}a \, \mathbf{\hat{y}} + x_{3}a \, \mathbf{\hat{z}} & \left(12e\right) & \mbox{Ce II} \\ 
\mathbf{B}_{26} & = & \frac{1}{4} \, \mathbf{a}_{1} + \frac{1}{4} \, \mathbf{a}_{2} + \left(\frac{1}{2} - x_{3}\right) \, \mathbf{a}_{3} & = & \frac{1}{4}a \, \mathbf{\hat{x}} + \frac{1}{4}a \, \mathbf{\hat{y}} + \left(\frac{1}{2} - x_{3}\right)a \, \mathbf{\hat{z}} & \left(12e\right) & \mbox{Ce II} \\ 
\mathbf{B}_{27} & = & -x_{3} \, \mathbf{a}_{1} + \frac{3}{4} \, \mathbf{a}_{2} + \frac{3}{4} \, \mathbf{a}_{3} & = & -x_{3}a \, \mathbf{\hat{x}} + \frac{3}{4}a \, \mathbf{\hat{y}} + \frac{3}{4}a \, \mathbf{\hat{z}} & \left(12e\right) & \mbox{Ce II} \\ 
\mathbf{B}_{28} & = & \left(\frac{1}{2} +x_{3}\right) \, \mathbf{a}_{1} + \frac{3}{4} \, \mathbf{a}_{2} + \frac{3}{4} \, \mathbf{a}_{3} & = & \left(\frac{1}{2} +x_{3}\right)a \, \mathbf{\hat{x}} + \frac{3}{4}a \, \mathbf{\hat{y}} + \frac{3}{4}a \, \mathbf{\hat{z}} & \left(12e\right) & \mbox{Ce II} \\ 
\mathbf{B}_{29} & = & \frac{3}{4} \, \mathbf{a}_{1}-x_{3} \, \mathbf{a}_{2} + \frac{3}{4} \, \mathbf{a}_{3} & = & \frac{3}{4}a \, \mathbf{\hat{x}}-x_{3}a \, \mathbf{\hat{y}} + \frac{3}{4}a \, \mathbf{\hat{z}} & \left(12e\right) & \mbox{Ce II} \\ 
\mathbf{B}_{30} & = & \frac{3}{4} \, \mathbf{a}_{1} + \left(\frac{1}{2} +x_{3}\right) \, \mathbf{a}_{2} + \frac{3}{4} \, \mathbf{a}_{3} & = & \frac{3}{4}a \, \mathbf{\hat{x}} + \left(\frac{1}{2} +x_{3}\right)a \, \mathbf{\hat{y}} + \frac{3}{4}a \, \mathbf{\hat{z}} & \left(12e\right) & \mbox{Ce II} \\ 
\mathbf{B}_{31} & = & \frac{3}{4} \, \mathbf{a}_{1} + \frac{3}{4} \, \mathbf{a}_{2}-x_{3} \, \mathbf{a}_{3} & = & \frac{3}{4}a \, \mathbf{\hat{x}} + \frac{3}{4}a \, \mathbf{\hat{y}}-x_{3}a \, \mathbf{\hat{z}} & \left(12e\right) & \mbox{Ce II} \\ 
\mathbf{B}_{32} & = & \frac{3}{4} \, \mathbf{a}_{1} + \frac{3}{4} \, \mathbf{a}_{2} + \left(\frac{1}{2} +x_{3}\right) \, \mathbf{a}_{3} & = & \frac{3}{4}a \, \mathbf{\hat{x}} + \frac{3}{4}a \, \mathbf{\hat{y}} + \left(\frac{1}{2} +x_{3}\right)a \, \mathbf{\hat{z}} & \left(12e\right) & \mbox{Ce II} \\ 
\mathbf{B}_{33} & = & x_{4} \, \mathbf{a}_{1} + x_{4} \, \mathbf{a}_{2} + x_{4} \, \mathbf{a}_{3} & = & x_{4}a \, \mathbf{\hat{x}} + x_{4}a \, \mathbf{\hat{y}} + x_{4}a \, \mathbf{\hat{z}} & \left(16f\right) & \mbox{O I} \\ 
\mathbf{B}_{34} & = & \left(\frac{1}{2} - x_{4}\right) \, \mathbf{a}_{1} + \left(\frac{1}{2} - x_{4}\right) \, \mathbf{a}_{2} + x_{4} \, \mathbf{a}_{3} & = & \left(\frac{1}{2} - x_{4}\right)a \, \mathbf{\hat{x}} + \left(\frac{1}{2} - x_{4}\right)a \, \mathbf{\hat{y}} + x_{4}a \, \mathbf{\hat{z}} & \left(16f\right) & \mbox{O I} \\ 
\mathbf{B}_{35} & = & \left(\frac{1}{2} - x_{4}\right) \, \mathbf{a}_{1} + x_{4} \, \mathbf{a}_{2} + \left(\frac{1}{2} - x_{4}\right) \, \mathbf{a}_{3} & = & \left(\frac{1}{2} - x_{4}\right)a \, \mathbf{\hat{x}} + x_{4}a \, \mathbf{\hat{y}} + \left(\frac{1}{2} - x_{4}\right)a \, \mathbf{\hat{z}} & \left(16f\right) & \mbox{O I} \\ 
\mathbf{B}_{36} & = & x_{4} \, \mathbf{a}_{1} + \left(\frac{1}{2} - x_{4}\right) \, \mathbf{a}_{2} + \left(\frac{1}{2} - x_{4}\right) \, \mathbf{a}_{3} & = & x_{4}a \, \mathbf{\hat{x}} + \left(\frac{1}{2} - x_{4}\right)a \, \mathbf{\hat{y}} + \left(\frac{1}{2} - x_{4}\right)a \, \mathbf{\hat{z}} & \left(16f\right) & \mbox{O I} \\ 
\mathbf{B}_{37} & = & x_{4} \, \mathbf{a}_{1} + x_{4} \, \mathbf{a}_{2} + \left(\frac{1}{2} - x_{4}\right) \, \mathbf{a}_{3} & = & x_{4}a \, \mathbf{\hat{x}} + x_{4}a \, \mathbf{\hat{y}} + \left(\frac{1}{2} - x_{4}\right)a \, \mathbf{\hat{z}} & \left(16f\right) & \mbox{O I} \\ 
\mathbf{B}_{38} & = & \left(\frac{1}{2} - x_{4}\right) \, \mathbf{a}_{1} + \left(\frac{1}{2} - x_{4}\right) \, \mathbf{a}_{2} + \left(\frac{1}{2} - x_{4}\right) \, \mathbf{a}_{3} & = & \left(\frac{1}{2} - x_{4}\right)a \, \mathbf{\hat{x}} + \left(\frac{1}{2} - x_{4}\right)a \, \mathbf{\hat{y}} + \left(\frac{1}{2} - x_{4}\right)a \, \mathbf{\hat{z}} & \left(16f\right) & \mbox{O I} \\ 
\mathbf{B}_{39} & = & x_{4} \, \mathbf{a}_{1} + \left(\frac{1}{2} - x_{4}\right) \, \mathbf{a}_{2} + x_{4} \, \mathbf{a}_{3} & = & x_{4}a \, \mathbf{\hat{x}} + \left(\frac{1}{2} - x_{4}\right)a \, \mathbf{\hat{y}} + x_{4}a \, \mathbf{\hat{z}} & \left(16f\right) & \mbox{O I} \\ 
\mathbf{B}_{40} & = & \left(\frac{1}{2} - x_{4}\right) \, \mathbf{a}_{1} + x_{4} \, \mathbf{a}_{2} + x_{4} \, \mathbf{a}_{3} & = & \left(\frac{1}{2} - x_{4}\right)a \, \mathbf{\hat{x}} + x_{4}a \, \mathbf{\hat{y}} + x_{4}a \, \mathbf{\hat{z}} & \left(16f\right) & \mbox{O I} \\ 
\mathbf{B}_{41} & = & -x_{4} \, \mathbf{a}_{1}-x_{4} \, \mathbf{a}_{2}-x_{4} \, \mathbf{a}_{3} & = & -x_{4}a \, \mathbf{\hat{x}}-x_{4}a \, \mathbf{\hat{y}}-x_{4}a \, \mathbf{\hat{z}} & \left(16f\right) & \mbox{O I} \\ 
\mathbf{B}_{42} & = & \left(\frac{1}{2} +x_{4}\right) \, \mathbf{a}_{1} + \left(\frac{1}{2} +x_{4}\right) \, \mathbf{a}_{2}-x_{4} \, \mathbf{a}_{3} & = & \left(\frac{1}{2} +x_{4}\right)a \, \mathbf{\hat{x}} + \left(\frac{1}{2} +x_{4}\right)a \, \mathbf{\hat{y}}-x_{4}a \, \mathbf{\hat{z}} & \left(16f\right) & \mbox{O I} \\ 
\mathbf{B}_{43} & = & \left(\frac{1}{2} +x_{4}\right) \, \mathbf{a}_{1}-x_{4} \, \mathbf{a}_{2} + \left(\frac{1}{2} +x_{4}\right) \, \mathbf{a}_{3} & = & \left(\frac{1}{2} +x_{4}\right)a \, \mathbf{\hat{x}}-x_{4}a \, \mathbf{\hat{y}} + \left(\frac{1}{2} +x_{4}\right)a \, \mathbf{\hat{z}} & \left(16f\right) & \mbox{O I} \\ 
\mathbf{B}_{44} & = & -x_{4} \, \mathbf{a}_{1} + \left(\frac{1}{2} +x_{4}\right) \, \mathbf{a}_{2} + \left(\frac{1}{2} +x_{4}\right) \, \mathbf{a}_{3} & = & -x_{4}a \, \mathbf{\hat{x}} + \left(\frac{1}{2} +x_{4}\right)a \, \mathbf{\hat{y}} + \left(\frac{1}{2} +x_{4}\right)a \, \mathbf{\hat{z}} & \left(16f\right) & \mbox{O I} \\ 
\mathbf{B}_{45} & = & -x_{4} \, \mathbf{a}_{1}-x_{4} \, \mathbf{a}_{2} + \left(\frac{1}{2} +x_{4}\right) \, \mathbf{a}_{3} & = & -x_{4}a \, \mathbf{\hat{x}}-x_{4}a \, \mathbf{\hat{y}} + \left(\frac{1}{2} +x_{4}\right)a \, \mathbf{\hat{z}} & \left(16f\right) & \mbox{O I} \\ 
\mathbf{B}_{46} & = & \left(\frac{1}{2} +x_{4}\right) \, \mathbf{a}_{1} + \left(\frac{1}{2} +x_{4}\right) \, \mathbf{a}_{2} + \left(\frac{1}{2} +x_{4}\right) \, \mathbf{a}_{3} & = & \left(\frac{1}{2} +x_{4}\right)a \, \mathbf{\hat{x}} + \left(\frac{1}{2} +x_{4}\right)a \, \mathbf{\hat{y}} + \left(\frac{1}{2} +x_{4}\right)a \, \mathbf{\hat{z}} & \left(16f\right) & \mbox{O I} \\ 
\mathbf{B}_{47} & = & -x_{4} \, \mathbf{a}_{1} + \left(\frac{1}{2} +x_{4}\right) \, \mathbf{a}_{2}-x_{4} \, \mathbf{a}_{3} & = & -x_{4}a \, \mathbf{\hat{x}} + \left(\frac{1}{2} +x_{4}\right)a \, \mathbf{\hat{y}}-x_{4}a \, \mathbf{\hat{z}} & \left(16f\right) & \mbox{O I} \\ 
\mathbf{B}_{48} & = & \left(\frac{1}{2} +x_{4}\right) \, \mathbf{a}_{1}-x_{4} \, \mathbf{a}_{2}-x_{4} \, \mathbf{a}_{3} & = & \left(\frac{1}{2} +x_{4}\right)a \, \mathbf{\hat{x}}-x_{4}a \, \mathbf{\hat{y}}-x_{4}a \, \mathbf{\hat{z}} & \left(16f\right) & \mbox{O I} \\ 
\mathbf{B}_{49} & = & x_{5} \, \mathbf{a}_{1} + y_{5} \, \mathbf{a}_{2} + z_{5} \, \mathbf{a}_{3} & = & x_{5}a \, \mathbf{\hat{x}} + y_{5}a \, \mathbf{\hat{y}} + z_{5}a \, \mathbf{\hat{z}} & \left(48i\right) & \mbox{O II} \\ 
\mathbf{B}_{50} & = & \left(\frac{1}{2} - x_{5}\right) \, \mathbf{a}_{1} + \left(\frac{1}{2} - y_{5}\right) \, \mathbf{a}_{2} + z_{5} \, \mathbf{a}_{3} & = & \left(\frac{1}{2} - x_{5}\right)a \, \mathbf{\hat{x}} + \left(\frac{1}{2} - y_{5}\right)a \, \mathbf{\hat{y}} + z_{5}a \, \mathbf{\hat{z}} & \left(48i\right) & \mbox{O II} \\ 
\mathbf{B}_{51} & = & \left(\frac{1}{2} - x_{5}\right) \, \mathbf{a}_{1} + y_{5} \, \mathbf{a}_{2} + \left(\frac{1}{2} - z_{5}\right) \, \mathbf{a}_{3} & = & \left(\frac{1}{2} - x_{5}\right)a \, \mathbf{\hat{x}} + y_{5}a \, \mathbf{\hat{y}} + \left(\frac{1}{2} - z_{5}\right)a \, \mathbf{\hat{z}} & \left(48i\right) & \mbox{O II} \\ 
\mathbf{B}_{52} & = & x_{5} \, \mathbf{a}_{1} + \left(\frac{1}{2} - y_{5}\right) \, \mathbf{a}_{2} + \left(\frac{1}{2} - z_{5}\right) \, \mathbf{a}_{3} & = & x_{5}a \, \mathbf{\hat{x}} + \left(\frac{1}{2} - y_{5}\right)a \, \mathbf{\hat{y}} + \left(\frac{1}{2} - z_{5}\right)a \, \mathbf{\hat{z}} & \left(48i\right) & \mbox{O II} \\ 
\mathbf{B}_{53} & = & z_{5} \, \mathbf{a}_{1} + x_{5} \, \mathbf{a}_{2} + y_{5} \, \mathbf{a}_{3} & = & z_{5}a \, \mathbf{\hat{x}} + x_{5}a \, \mathbf{\hat{y}} + y_{5}a \, \mathbf{\hat{z}} & \left(48i\right) & \mbox{O II} \\ 
\mathbf{B}_{54} & = & z_{5} \, \mathbf{a}_{1} + \left(\frac{1}{2} - x_{5}\right) \, \mathbf{a}_{2} + \left(\frac{1}{2} - y_{5}\right) \, \mathbf{a}_{3} & = & z_{5}a \, \mathbf{\hat{x}} + \left(\frac{1}{2} - x_{5}\right)a \, \mathbf{\hat{y}} + \left(\frac{1}{2} - y_{5}\right)a \, \mathbf{\hat{z}} & \left(48i\right) & \mbox{O II} \\ 
\mathbf{B}_{55} & = & \left(\frac{1}{2} - z_{5}\right) \, \mathbf{a}_{1} + \left(\frac{1}{2} - x_{5}\right) \, \mathbf{a}_{2} + y_{5} \, \mathbf{a}_{3} & = & \left(\frac{1}{2} - z_{5}\right)a \, \mathbf{\hat{x}} + \left(\frac{1}{2} - x_{5}\right)a \, \mathbf{\hat{y}} + y_{5}a \, \mathbf{\hat{z}} & \left(48i\right) & \mbox{O II} \\ 
\mathbf{B}_{56} & = & \left(\frac{1}{2} - z_{5}\right) \, \mathbf{a}_{1} + x_{5} \, \mathbf{a}_{2} + \left(\frac{1}{2} - y_{5}\right) \, \mathbf{a}_{3} & = & \left(\frac{1}{2} - z_{5}\right)a \, \mathbf{\hat{x}} + x_{5}a \, \mathbf{\hat{y}} + \left(\frac{1}{2} - y_{5}\right)a \, \mathbf{\hat{z}} & \left(48i\right) & \mbox{O II} \\ 
\mathbf{B}_{57} & = & y_{5} \, \mathbf{a}_{1} + z_{5} \, \mathbf{a}_{2} + x_{5} \, \mathbf{a}_{3} & = & y_{5}a \, \mathbf{\hat{x}} + z_{5}a \, \mathbf{\hat{y}} + x_{5}a \, \mathbf{\hat{z}} & \left(48i\right) & \mbox{O II} \\ 
\mathbf{B}_{58} & = & \left(\frac{1}{2} - y_{5}\right) \, \mathbf{a}_{1} + z_{5} \, \mathbf{a}_{2} + \left(\frac{1}{2} - x_{5}\right) \, \mathbf{a}_{3} & = & \left(\frac{1}{2} - y_{5}\right)a \, \mathbf{\hat{x}} + z_{5}a \, \mathbf{\hat{y}} + \left(\frac{1}{2} - x_{5}\right)a \, \mathbf{\hat{z}} & \left(48i\right) & \mbox{O II} \\ 
\mathbf{B}_{59} & = & y_{5} \, \mathbf{a}_{1} + \left(\frac{1}{2} - z_{5}\right) \, \mathbf{a}_{2} + \left(\frac{1}{2} - x_{5}\right) \, \mathbf{a}_{3} & = & y_{5}a \, \mathbf{\hat{x}} + \left(\frac{1}{2} - z_{5}\right)a \, \mathbf{\hat{y}} + \left(\frac{1}{2} - x_{5}\right)a \, \mathbf{\hat{z}} & \left(48i\right) & \mbox{O II} \\ 
\mathbf{B}_{60} & = & \left(\frac{1}{2} - y_{5}\right) \, \mathbf{a}_{1} + \left(\frac{1}{2} - z_{5}\right) \, \mathbf{a}_{2} + x_{5} \, \mathbf{a}_{3} & = & \left(\frac{1}{2} - y_{5}\right)a \, \mathbf{\hat{x}} + \left(\frac{1}{2} - z_{5}\right)a \, \mathbf{\hat{y}} + x_{5}a \, \mathbf{\hat{z}} & \left(48i\right) & \mbox{O II} \\ 
\mathbf{B}_{61} & = & y_{5} \, \mathbf{a}_{1} + x_{5} \, \mathbf{a}_{2} + \left(\frac{1}{2} - z_{5}\right) \, \mathbf{a}_{3} & = & y_{5}a \, \mathbf{\hat{x}} + x_{5}a \, \mathbf{\hat{y}} + \left(\frac{1}{2} - z_{5}\right)a \, \mathbf{\hat{z}} & \left(48i\right) & \mbox{O II} \\ 
\mathbf{B}_{62} & = & \left(\frac{1}{2} - y_{5}\right) \, \mathbf{a}_{1} + \left(\frac{1}{2} - x_{5}\right) \, \mathbf{a}_{2} + \left(\frac{1}{2} - z_{5}\right) \, \mathbf{a}_{3} & = & \left(\frac{1}{2} - y_{5}\right)a \, \mathbf{\hat{x}} + \left(\frac{1}{2} - x_{5}\right)a \, \mathbf{\hat{y}} + \left(\frac{1}{2} - z_{5}\right)a \, \mathbf{\hat{z}} & \left(48i\right) & \mbox{O II} \\ 
\mathbf{B}_{63} & = & y_{5} \, \mathbf{a}_{1} + \left(\frac{1}{2} - x_{5}\right) \, \mathbf{a}_{2} + z_{5} \, \mathbf{a}_{3} & = & y_{5}a \, \mathbf{\hat{x}} + \left(\frac{1}{2} - x_{5}\right)a \, \mathbf{\hat{y}} + z_{5}a \, \mathbf{\hat{z}} & \left(48i\right) & \mbox{O II} \\ 
\mathbf{B}_{64} & = & \left(\frac{1}{2} - y_{5}\right) \, \mathbf{a}_{1} + x_{5} \, \mathbf{a}_{2} + z_{5} \, \mathbf{a}_{3} & = & \left(\frac{1}{2} - y_{5}\right)a \, \mathbf{\hat{x}} + x_{5}a \, \mathbf{\hat{y}} + z_{5}a \, \mathbf{\hat{z}} & \left(48i\right) & \mbox{O II} \\ 
\mathbf{B}_{65} & = & x_{5} \, \mathbf{a}_{1} + z_{5} \, \mathbf{a}_{2} + \left(\frac{1}{2} - y_{5}\right) \, \mathbf{a}_{3} & = & x_{5}a \, \mathbf{\hat{x}} + z_{5}a \, \mathbf{\hat{y}} + \left(\frac{1}{2} - y_{5}\right)a \, \mathbf{\hat{z}} & \left(48i\right) & \mbox{O II} \\ 
\mathbf{B}_{66} & = & \left(\frac{1}{2} - x_{5}\right) \, \mathbf{a}_{1} + z_{5} \, \mathbf{a}_{2} + y_{5} \, \mathbf{a}_{3} & = & \left(\frac{1}{2} - x_{5}\right)a \, \mathbf{\hat{x}} + z_{5}a \, \mathbf{\hat{y}} + y_{5}a \, \mathbf{\hat{z}} & \left(48i\right) & \mbox{O II} \\ 
\mathbf{B}_{67} & = & \left(\frac{1}{2} - x_{5}\right) \, \mathbf{a}_{1} + \left(\frac{1}{2} - z_{5}\right) \, \mathbf{a}_{2} + \left(\frac{1}{2} - y_{5}\right) \, \mathbf{a}_{3} & = & \left(\frac{1}{2} - x_{5}\right)a \, \mathbf{\hat{x}} + \left(\frac{1}{2} - z_{5}\right)a \, \mathbf{\hat{y}} + \left(\frac{1}{2} - y_{5}\right)a \, \mathbf{\hat{z}} & \left(48i\right) & \mbox{O II} \\ 
\mathbf{B}_{68} & = & x_{5} \, \mathbf{a}_{1} + \left(\frac{1}{2} - z_{5}\right) \, \mathbf{a}_{2} + y_{5} \, \mathbf{a}_{3} & = & x_{5}a \, \mathbf{\hat{x}} + \left(\frac{1}{2} - z_{5}\right)a \, \mathbf{\hat{y}} + y_{5}a \, \mathbf{\hat{z}} & \left(48i\right) & \mbox{O II} \\ 
\mathbf{B}_{69} & = & z_{5} \, \mathbf{a}_{1} + y_{5} \, \mathbf{a}_{2} + \left(\frac{1}{2} - x_{5}\right) \, \mathbf{a}_{3} & = & z_{5}a \, \mathbf{\hat{x}} + y_{5}a \, \mathbf{\hat{y}} + \left(\frac{1}{2} - x_{5}\right)a \, \mathbf{\hat{z}} & \left(48i\right) & \mbox{O II} \\ 
\mathbf{B}_{70} & = & z_{5} \, \mathbf{a}_{1} + \left(\frac{1}{2} - y_{5}\right) \, \mathbf{a}_{2} + x_{5} \, \mathbf{a}_{3} & = & z_{5}a \, \mathbf{\hat{x}} + \left(\frac{1}{2} - y_{5}\right)a \, \mathbf{\hat{y}} + x_{5}a \, \mathbf{\hat{z}} & \left(48i\right) & \mbox{O II} \\ 
\mathbf{B}_{71} & = & \left(\frac{1}{2} - z_{5}\right) \, \mathbf{a}_{1} + y_{5} \, \mathbf{a}_{2} + x_{5} \, \mathbf{a}_{3} & = & \left(\frac{1}{2} - z_{5}\right)a \, \mathbf{\hat{x}} + y_{5}a \, \mathbf{\hat{y}} + x_{5}a \, \mathbf{\hat{z}} & \left(48i\right) & \mbox{O II} \\ 
\mathbf{B}_{72} & = & \left(\frac{1}{2} - z_{5}\right) \, \mathbf{a}_{1} + \left(\frac{1}{2} - y_{5}\right) \, \mathbf{a}_{2} + \left(\frac{1}{2} - x_{5}\right) \, \mathbf{a}_{3} & = & \left(\frac{1}{2} - z_{5}\right)a \, \mathbf{\hat{x}} + \left(\frac{1}{2} - y_{5}\right)a \, \mathbf{\hat{y}} + \left(\frac{1}{2} - x_{5}\right)a \, \mathbf{\hat{z}} & \left(48i\right) & \mbox{O II} \\ 
\mathbf{B}_{73} & = & -x_{5} \, \mathbf{a}_{1}-y_{5} \, \mathbf{a}_{2}-z_{5} \, \mathbf{a}_{3} & = & -x_{5}a \, \mathbf{\hat{x}}-y_{5}a \, \mathbf{\hat{y}}-z_{5}a \, \mathbf{\hat{z}} & \left(48i\right) & \mbox{O II} \\ 
\mathbf{B}_{74} & = & \left(\frac{1}{2} +x_{5}\right) \, \mathbf{a}_{1} + \left(\frac{1}{2} +y_{5}\right) \, \mathbf{a}_{2}-z_{5} \, \mathbf{a}_{3} & = & \left(\frac{1}{2} +x_{5}\right)a \, \mathbf{\hat{x}} + \left(\frac{1}{2} +y_{5}\right)a \, \mathbf{\hat{y}}-z_{5}a \, \mathbf{\hat{z}} & \left(48i\right) & \mbox{O II} \\ 
\mathbf{B}_{75} & = & \left(\frac{1}{2} +x_{5}\right) \, \mathbf{a}_{1}-y_{5} \, \mathbf{a}_{2} + \left(\frac{1}{2} +z_{5}\right) \, \mathbf{a}_{3} & = & \left(\frac{1}{2} +x_{5}\right)a \, \mathbf{\hat{x}}-y_{5}a \, \mathbf{\hat{y}} + \left(\frac{1}{2} +z_{5}\right)a \, \mathbf{\hat{z}} & \left(48i\right) & \mbox{O II} \\ 
\mathbf{B}_{76} & = & -x_{5} \, \mathbf{a}_{1} + \left(\frac{1}{2} +y_{5}\right) \, \mathbf{a}_{2} + \left(\frac{1}{2} +z_{5}\right) \, \mathbf{a}_{3} & = & -x_{5}a \, \mathbf{\hat{x}} + \left(\frac{1}{2} +y_{5}\right)a \, \mathbf{\hat{y}} + \left(\frac{1}{2} +z_{5}\right)a \, \mathbf{\hat{z}} & \left(48i\right) & \mbox{O II} \\ 
\mathbf{B}_{77} & = & -z_{5} \, \mathbf{a}_{1}-x_{5} \, \mathbf{a}_{2}-y_{5} \, \mathbf{a}_{3} & = & -z_{5}a \, \mathbf{\hat{x}}-x_{5}a \, \mathbf{\hat{y}}-y_{5}a \, \mathbf{\hat{z}} & \left(48i\right) & \mbox{O II} \\ 
\mathbf{B}_{78} & = & -z_{5} \, \mathbf{a}_{1} + \left(\frac{1}{2} +x_{5}\right) \, \mathbf{a}_{2} + \left(\frac{1}{2} +y_{5}\right) \, \mathbf{a}_{3} & = & -z_{5}a \, \mathbf{\hat{x}} + \left(\frac{1}{2} +x_{5}\right)a \, \mathbf{\hat{y}} + \left(\frac{1}{2} +y_{5}\right)a \, \mathbf{\hat{z}} & \left(48i\right) & \mbox{O II} \\ 
\mathbf{B}_{79} & = & \left(\frac{1}{2} +z_{5}\right) \, \mathbf{a}_{1} + \left(\frac{1}{2} +x_{5}\right) \, \mathbf{a}_{2}-y_{5} \, \mathbf{a}_{3} & = & \left(\frac{1}{2} +z_{5}\right)a \, \mathbf{\hat{x}} + \left(\frac{1}{2} +x_{5}\right)a \, \mathbf{\hat{y}}-y_{5}a \, \mathbf{\hat{z}} & \left(48i\right) & \mbox{O II} \\ 
\mathbf{B}_{80} & = & \left(\frac{1}{2} +z_{5}\right) \, \mathbf{a}_{1}-x_{5} \, \mathbf{a}_{2} + \left(\frac{1}{2} +y_{5}\right) \, \mathbf{a}_{3} & = & \left(\frac{1}{2} +z_{5}\right)a \, \mathbf{\hat{x}}-x_{5}a \, \mathbf{\hat{y}} + \left(\frac{1}{2} +y_{5}\right)a \, \mathbf{\hat{z}} & \left(48i\right) & \mbox{O II} \\ 
\mathbf{B}_{81} & = & -y_{5} \, \mathbf{a}_{1}-z_{5} \, \mathbf{a}_{2}-x_{5} \, \mathbf{a}_{3} & = & -y_{5}a \, \mathbf{\hat{x}}-z_{5}a \, \mathbf{\hat{y}}-x_{5}a \, \mathbf{\hat{z}} & \left(48i\right) & \mbox{O II} \\ 
\mathbf{B}_{82} & = & \left(\frac{1}{2} +y_{5}\right) \, \mathbf{a}_{1}-z_{5} \, \mathbf{a}_{2} + \left(\frac{1}{2} +x_{5}\right) \, \mathbf{a}_{3} & = & \left(\frac{1}{2} +y_{5}\right)a \, \mathbf{\hat{x}}-z_{5}a \, \mathbf{\hat{y}} + \left(\frac{1}{2} +x_{5}\right)a \, \mathbf{\hat{z}} & \left(48i\right) & \mbox{O II} \\ 
\mathbf{B}_{83} & = & -y_{5} \, \mathbf{a}_{1} + \left(\frac{1}{2} +z_{5}\right) \, \mathbf{a}_{2} + \left(\frac{1}{2} +x_{5}\right) \, \mathbf{a}_{3} & = & -y_{5}a \, \mathbf{\hat{x}} + \left(\frac{1}{2} +z_{5}\right)a \, \mathbf{\hat{y}} + \left(\frac{1}{2} +x_{5}\right)a \, \mathbf{\hat{z}} & \left(48i\right) & \mbox{O II} \\ 
\mathbf{B}_{84} & = & \left(\frac{1}{2} +y_{5}\right) \, \mathbf{a}_{1} + \left(\frac{1}{2} +z_{5}\right) \, \mathbf{a}_{2}-x_{5} \, \mathbf{a}_{3} & = & \left(\frac{1}{2} +y_{5}\right)a \, \mathbf{\hat{x}} + \left(\frac{1}{2} +z_{5}\right)a \, \mathbf{\hat{y}}-x_{5}a \, \mathbf{\hat{z}} & \left(48i\right) & \mbox{O II} \\ 
\mathbf{B}_{85} & = & -y_{5} \, \mathbf{a}_{1}-x_{5} \, \mathbf{a}_{2} + \left(\frac{1}{2} +z_{5}\right) \, \mathbf{a}_{3} & = & -y_{5}a \, \mathbf{\hat{x}}-x_{5}a \, \mathbf{\hat{y}} + \left(\frac{1}{2} +z_{5}\right)a \, \mathbf{\hat{z}} & \left(48i\right) & \mbox{O II} \\ 
\mathbf{B}_{86} & = & \left(\frac{1}{2} +y_{5}\right) \, \mathbf{a}_{1} + \left(\frac{1}{2} +x_{5}\right) \, \mathbf{a}_{2} + \left(\frac{1}{2} +z_{5}\right) \, \mathbf{a}_{3} & = & \left(\frac{1}{2} +y_{5}\right)a \, \mathbf{\hat{x}} + \left(\frac{1}{2} +x_{5}\right)a \, \mathbf{\hat{y}} + \left(\frac{1}{2} +z_{5}\right)a \, \mathbf{\hat{z}} & \left(48i\right) & \mbox{O II} \\ 
\mathbf{B}_{87} & = & -y_{5} \, \mathbf{a}_{1} + \left(\frac{1}{2} +x_{5}\right) \, \mathbf{a}_{2}-z_{5} \, \mathbf{a}_{3} & = & -y_{5}a \, \mathbf{\hat{x}} + \left(\frac{1}{2} +x_{5}\right)a \, \mathbf{\hat{y}}-z_{5}a \, \mathbf{\hat{z}} & \left(48i\right) & \mbox{O II} \\ 
\mathbf{B}_{88} & = & \left(\frac{1}{2} +y_{5}\right) \, \mathbf{a}_{1}-x_{5} \, \mathbf{a}_{2}-z_{5} \, \mathbf{a}_{3} & = & \left(\frac{1}{2} +y_{5}\right)a \, \mathbf{\hat{x}}-x_{5}a \, \mathbf{\hat{y}}-z_{5}a \, \mathbf{\hat{z}} & \left(48i\right) & \mbox{O II} \\ 
\mathbf{B}_{89} & = & -x_{5} \, \mathbf{a}_{1}-z_{5} \, \mathbf{a}_{2} + \left(\frac{1}{2} +y_{5}\right) \, \mathbf{a}_{3} & = & -x_{5}a \, \mathbf{\hat{x}}-z_{5}a \, \mathbf{\hat{y}} + \left(\frac{1}{2} +y_{5}\right)a \, \mathbf{\hat{z}} & \left(48i\right) & \mbox{O II} \\ 
\mathbf{B}_{90} & = & \left(\frac{1}{2} +x_{5}\right) \, \mathbf{a}_{1}-z_{5} \, \mathbf{a}_{2}-y_{5} \, \mathbf{a}_{3} & = & \left(\frac{1}{2} +x_{5}\right)a \, \mathbf{\hat{x}}-z_{5}a \, \mathbf{\hat{y}}-y_{5}a \, \mathbf{\hat{z}} & \left(48i\right) & \mbox{O II} \\ 
\mathbf{B}_{91} & = & \left(\frac{1}{2} +x_{5}\right) \, \mathbf{a}_{1} + \left(\frac{1}{2} +z_{5}\right) \, \mathbf{a}_{2} + \left(\frac{1}{2} +y_{5}\right) \, \mathbf{a}_{3} & = & \left(\frac{1}{2} +x_{5}\right)a \, \mathbf{\hat{x}} + \left(\frac{1}{2} +z_{5}\right)a \, \mathbf{\hat{y}} + \left(\frac{1}{2} +y_{5}\right)a \, \mathbf{\hat{z}} & \left(48i\right) & \mbox{O II} \\ 
\mathbf{B}_{92} & = & -x_{5} \, \mathbf{a}_{1} + \left(\frac{1}{2} +z_{5}\right) \, \mathbf{a}_{2}-y_{5} \, \mathbf{a}_{3} & = & -x_{5}a \, \mathbf{\hat{x}} + \left(\frac{1}{2} +z_{5}\right)a \, \mathbf{\hat{y}}-y_{5}a \, \mathbf{\hat{z}} & \left(48i\right) & \mbox{O II} \\ 
\mathbf{B}_{93} & = & -z_{5} \, \mathbf{a}_{1}-y_{5} \, \mathbf{a}_{2} + \left(\frac{1}{2} +x_{5}\right) \, \mathbf{a}_{3} & = & -z_{5}a \, \mathbf{\hat{x}}-y_{5}a \, \mathbf{\hat{y}} + \left(\frac{1}{2} +x_{5}\right)a \, \mathbf{\hat{z}} & \left(48i\right) & \mbox{O II} \\ 
\mathbf{B}_{94} & = & -z_{5} \, \mathbf{a}_{1} + \left(\frac{1}{2} +y_{5}\right) \, \mathbf{a}_{2}-x_{5} \, \mathbf{a}_{3} & = & -z_{5}a \, \mathbf{\hat{x}} + \left(\frac{1}{2} +y_{5}\right)a \, \mathbf{\hat{y}}-x_{5}a \, \mathbf{\hat{z}} & \left(48i\right) & \mbox{O II} \\ 
\mathbf{B}_{95} & = & \left(\frac{1}{2} +z_{5}\right) \, \mathbf{a}_{1}-y_{5} \, \mathbf{a}_{2}-x_{5} \, \mathbf{a}_{3} & = & \left(\frac{1}{2} +z_{5}\right)a \, \mathbf{\hat{x}}-y_{5}a \, \mathbf{\hat{y}}-x_{5}a \, \mathbf{\hat{z}} & \left(48i\right) & \mbox{O II} \\ 
\mathbf{B}_{96} & = & \left(\frac{1}{2} +z_{5}\right) \, \mathbf{a}_{1} + \left(\frac{1}{2} +y_{5}\right) \, \mathbf{a}_{2} + \left(\frac{1}{2} +x_{5}\right) \, \mathbf{a}_{3} & = & \left(\frac{1}{2} +z_{5}\right)a \, \mathbf{\hat{x}} + \left(\frac{1}{2} +y_{5}\right)a \, \mathbf{\hat{y}} + \left(\frac{1}{2} +x_{5}\right)a \, \mathbf{\hat{z}} & \left(48i\right) & \mbox{O II} \\ 
\end{longtabu}
\renewcommand{\arraystretch}{1.0}
\noindent \hrulefill
\\
\textbf{References:}
\vspace*{-0.25cm}
\begin{flushleft}
  - \bibentry{Hubert_Ce5Mo3O16_BullSocChimFr_1974}. \\
\end{flushleft}
\textbf{Found in:}
\vspace*{-0.25cm}
\begin{flushleft}
  - \bibentry{Villars_PearsonsCrystalData_2013}. \\
\end{flushleft}
\noindent \hrulefill
\\
\textbf{Geometry files:}
\\
\noindent  - CIF: pp. {\hyperref[A5B3C16_cP96_222_ce_d_fi_cif]{\pageref{A5B3C16_cP96_222_ce_d_fi_cif}}} \\
\noindent  - POSCAR: pp. {\hyperref[A5B3C16_cP96_222_ce_d_fi_poscar]{\pageref{A5B3C16_cP96_222_ce_d_fi_poscar}}} \\
\onecolumn
{\phantomsection\label{A23B6_cF116_225_bd2f_e}}
\subsection*{\huge \textbf{{\normalfont Th$_{6}$Mn$_{23}$ ($D8_{a}$) Structure: A23B6\_cF116\_225\_bd2f\_e}}}
\noindent \hrulefill
\vspace*{0.25cm}
\begin{figure}[htp]
  \centering
  \vspace{-1em}
  {\includegraphics[width=1\textwidth]{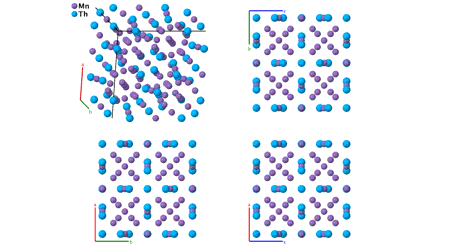}}
\end{figure}
\vspace*{-0.5cm}
\renewcommand{\arraystretch}{1.5}
\begin{equation*}
  \begin{array}{>{$\hspace{-0.15cm}}l<{$}>{$}p{0.5cm}<{$}>{$}p{18.5cm}<{$}}
    \mbox{\large \textbf{Prototype}} &\colon & \ce{Th6Mn23} \\
    \mbox{\large \textbf{\AFLOW\ prototype label}} &\colon & \mbox{A23B6\_cF116\_225\_bd2f\_e} \\
    \mbox{\large \textbf{\textit{Strukturbericht} designation}} &\colon & \mbox{$D8_{a}$} \\
    \mbox{\large \textbf{Pearson symbol}} &\colon & \mbox{cF116} \\
    \mbox{\large \textbf{Space group number}} &\colon & 225 \\
    \mbox{\large \textbf{Space group symbol}} &\colon & Fm\bar{3}m \\
    \mbox{\large \textbf{\AFLOW\ prototype command}} &\colon &  \texttt{aflow} \,  \, \texttt{-{}-proto=A23B6\_cF116\_225\_bd2f\_e } \, \newline \texttt{-{}-params=}{a,x_{3},x_{4},x_{5} }
  \end{array}
\end{equation*}
\renewcommand{\arraystretch}{1.0}

\vspace*{-0.25cm}
\noindent \hrulefill
\\
\textbf{ Other compounds with this structure:}
\begin{itemize}
   \item{ Ba$_{6}$Mg$_{23}$, Cu$_{16}$Mg$_{4}$Si$_{7}$, Er${6}$Fe${23}$, Fe$_{10}$Ge$_{13}$Ti${6}$, Fe${3}$Zn, Ho${6}$Fe${23}$, Mn$_{6}$Ni$_{16}$Si$_{7}$, Sm${6}$Fe${23}$, Sr$_{6}$Li$_{23}$, Tb${6}$Fe${23}$, Th$_{6-x}$Y$_{x}$Mn$_{23}$, Th$_{6}$Mn$_{23}$, Y$_{6}$Mn$_{23}$  }
\end{itemize}
\noindent \parbox{1 \linewidth}{
\noindent \hrulefill
\\
\textbf{Face-centered Cubic primitive vectors:} \\
\vspace*{-0.25cm}
\begin{tabular}{cc}
  \begin{tabular}{c}
    \parbox{0.6 \linewidth}{
      \renewcommand{\arraystretch}{1.5}
      \begin{equation*}
        \centering
        \begin{array}{ccc}
              \mathbf{a}_1 & = & \frac12 \, a \, \mathbf{\hat{y}} + \frac12 \, a \, \mathbf{\hat{z}} \\
    \mathbf{a}_2 & = & \frac12 \, a \, \mathbf{\hat{x}} + \frac12 \, a \, \mathbf{\hat{z}} \\
    \mathbf{a}_3 & = & \frac12 \, a \, \mathbf{\hat{x}} + \frac12 \, a \, \mathbf{\hat{y}} \\

        \end{array}
      \end{equation*}
    }
    \renewcommand{\arraystretch}{1.0}
  \end{tabular}
  \begin{tabular}{c}
    \includegraphics[width=0.3\linewidth]{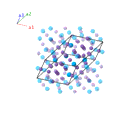} \\
  \end{tabular}
\end{tabular}

}
\vspace*{-0.25cm}

\noindent \hrulefill
\\
\textbf{Basis vectors:}
\vspace*{-0.25cm}
\renewcommand{\arraystretch}{1.5}
\begin{longtabu} to \textwidth{>{\centering $}X[-1,c,c]<{$}>{\centering $}X[-1,c,c]<{$}>{\centering $}X[-1,c,c]<{$}>{\centering $}X[-1,c,c]<{$}>{\centering $}X[-1,c,c]<{$}>{\centering $}X[-1,c,c]<{$}>{\centering $}X[-1,c,c]<{$}}
  & & \mbox{Lattice Coordinates} & & \mbox{Cartesian Coordinates} &\mbox{Wyckoff Position} & \mbox{Atom Type} \\  
  \mathbf{B}_{1} & = & \frac{1}{2} \, \mathbf{a}_{1} + \frac{1}{2} \, \mathbf{a}_{2} + \frac{1}{2} \, \mathbf{a}_{3} & = & \frac{1}{2}a \, \mathbf{\hat{x}} + \frac{1}{2}a \, \mathbf{\hat{y}} + \frac{1}{2}a \, \mathbf{\hat{z}} & \left(4b\right) & \mbox{Mn I} \\ 
\mathbf{B}_{2} & = & \frac{1}{2} \, \mathbf{a}_{1} & = & \frac{1}{4}a \, \mathbf{\hat{y}} + \frac{1}{4}a \, \mathbf{\hat{z}} & \left(24d\right) & \mbox{Mn II} \\ 
\mathbf{B}_{3} & = & \frac{1}{2} \, \mathbf{a}_{2} + \frac{1}{2} \, \mathbf{a}_{3} & = & \frac{1}{2}a \, \mathbf{\hat{x}} + \frac{1}{4}a \, \mathbf{\hat{y}} + \frac{1}{4}a \, \mathbf{\hat{z}} & \left(24d\right) & \mbox{Mn II} \\ 
\mathbf{B}_{4} & = & \frac{1}{2} \, \mathbf{a}_{2} & = & \frac{1}{4}a \, \mathbf{\hat{x}} + \frac{1}{4}a \, \mathbf{\hat{z}} & \left(24d\right) & \mbox{Mn II} \\ 
\mathbf{B}_{5} & = & \frac{1}{2} \, \mathbf{a}_{1} + \frac{1}{2} \, \mathbf{a}_{3} & = & \frac{1}{4}a \, \mathbf{\hat{x}} + \frac{1}{2}a \, \mathbf{\hat{y}} + \frac{1}{4}a \, \mathbf{\hat{z}} & \left(24d\right) & \mbox{Mn II} \\ 
\mathbf{B}_{6} & = & \frac{1}{2} \, \mathbf{a}_{3} & = & \frac{1}{4}a \, \mathbf{\hat{x}} + \frac{1}{4}a \, \mathbf{\hat{y}} & \left(24d\right) & \mbox{Mn II} \\ 
\mathbf{B}_{7} & = & \frac{1}{2} \, \mathbf{a}_{1} + \frac{1}{2} \, \mathbf{a}_{2} & = & \frac{1}{4}a \, \mathbf{\hat{x}} + \frac{1}{4}a \, \mathbf{\hat{y}} + \frac{1}{2}a \, \mathbf{\hat{z}} & \left(24d\right) & \mbox{Mn II} \\ 
\mathbf{B}_{8} & = & -x_{3} \, \mathbf{a}_{1} + x_{3} \, \mathbf{a}_{2} + x_{3} \, \mathbf{a}_{3} & = & x_{3}a \, \mathbf{\hat{x}} & \left(24e\right) & \mbox{Th} \\ 
\mathbf{B}_{9} & = & x_{3} \, \mathbf{a}_{1}-x_{3} \, \mathbf{a}_{2}-x_{3} \, \mathbf{a}_{3} & = & -x_{3}a \, \mathbf{\hat{x}} & \left(24e\right) & \mbox{Th} \\ 
\mathbf{B}_{10} & = & x_{3} \, \mathbf{a}_{1}-x_{3} \, \mathbf{a}_{2} + x_{3} \, \mathbf{a}_{3} & = & x_{3}a \, \mathbf{\hat{y}} & \left(24e\right) & \mbox{Th} \\ 
\mathbf{B}_{11} & = & -x_{3} \, \mathbf{a}_{1} + x_{3} \, \mathbf{a}_{2}-x_{3} \, \mathbf{a}_{3} & = & -x_{3}a \, \mathbf{\hat{y}} & \left(24e\right) & \mbox{Th} \\ 
\mathbf{B}_{12} & = & x_{3} \, \mathbf{a}_{1} + x_{3} \, \mathbf{a}_{2}-x_{3} \, \mathbf{a}_{3} & = & x_{3}a \, \mathbf{\hat{z}} & \left(24e\right) & \mbox{Th} \\ 
\mathbf{B}_{13} & = & -x_{3} \, \mathbf{a}_{1}-x_{3} \, \mathbf{a}_{2} + x_{3} \, \mathbf{a}_{3} & = & -x_{3}a \, \mathbf{\hat{z}} & \left(24e\right) & \mbox{Th} \\ 
\mathbf{B}_{14} & = & x_{4} \, \mathbf{a}_{1} + x_{4} \, \mathbf{a}_{2} + x_{4} \, \mathbf{a}_{3} & = & x_{4}a \, \mathbf{\hat{x}} + x_{4}a \, \mathbf{\hat{y}} + x_{4}a \, \mathbf{\hat{z}} & \left(32f\right) & \mbox{Mn III} \\ 
\mathbf{B}_{15} & = & x_{4} \, \mathbf{a}_{1} + x_{4} \, \mathbf{a}_{2}-3x_{4} \, \mathbf{a}_{3} & = & -x_{4}a \, \mathbf{\hat{x}}-x_{4}a \, \mathbf{\hat{y}} + x_{4}a \, \mathbf{\hat{z}} & \left(32f\right) & \mbox{Mn III} \\ 
\mathbf{B}_{16} & = & x_{4} \, \mathbf{a}_{1}-3x_{4} \, \mathbf{a}_{2} + x_{4} \, \mathbf{a}_{3} & = & -x_{4}a \, \mathbf{\hat{x}} + x_{4}a \, \mathbf{\hat{y}}-x_{4}a \, \mathbf{\hat{z}} & \left(32f\right) & \mbox{Mn III} \\ 
\mathbf{B}_{17} & = & -3x_{4} \, \mathbf{a}_{1} + x_{4} \, \mathbf{a}_{2} + x_{4} \, \mathbf{a}_{3} & = & x_{4}a \, \mathbf{\hat{x}}-x_{4}a \, \mathbf{\hat{y}}-x_{4}a \, \mathbf{\hat{z}} & \left(32f\right) & \mbox{Mn III} \\ 
\mathbf{B}_{18} & = & -x_{4} \, \mathbf{a}_{1}-x_{4} \, \mathbf{a}_{2} + 3x_{4} \, \mathbf{a}_{3} & = & x_{4}a \, \mathbf{\hat{x}} + x_{4}a \, \mathbf{\hat{y}}-x_{4}a \, \mathbf{\hat{z}} & \left(32f\right) & \mbox{Mn III} \\ 
\mathbf{B}_{19} & = & -x_{4} \, \mathbf{a}_{1}-x_{4} \, \mathbf{a}_{2}-x_{4} \, \mathbf{a}_{3} & = & -x_{4}a \, \mathbf{\hat{x}}-x_{4}a \, \mathbf{\hat{y}}-x_{4}a \, \mathbf{\hat{z}} & \left(32f\right) & \mbox{Mn III} \\ 
\mathbf{B}_{20} & = & -x_{4} \, \mathbf{a}_{1} + 3x_{4} \, \mathbf{a}_{2}-x_{4} \, \mathbf{a}_{3} & = & x_{4}a \, \mathbf{\hat{x}}-x_{4}a \, \mathbf{\hat{y}} + x_{4}a \, \mathbf{\hat{z}} & \left(32f\right) & \mbox{Mn III} \\ 
\mathbf{B}_{21} & = & 3x_{4} \, \mathbf{a}_{1}-x_{4} \, \mathbf{a}_{2}-x_{4} \, \mathbf{a}_{3} & = & -x_{4}a \, \mathbf{\hat{x}} + x_{4}a \, \mathbf{\hat{y}} + x_{4}a \, \mathbf{\hat{z}} & \left(32f\right) & \mbox{Mn III} \\ 
\mathbf{B}_{22} & = & x_{5} \, \mathbf{a}_{1} + x_{5} \, \mathbf{a}_{2} + x_{5} \, \mathbf{a}_{3} & = & x_{5}a \, \mathbf{\hat{x}} + x_{5}a \, \mathbf{\hat{y}} + x_{5}a \, \mathbf{\hat{z}} & \left(32f\right) & \mbox{Mn IV} \\ 
\mathbf{B}_{23} & = & x_{5} \, \mathbf{a}_{1} + x_{5} \, \mathbf{a}_{2}-3x_{5} \, \mathbf{a}_{3} & = & -x_{5}a \, \mathbf{\hat{x}}-x_{5}a \, \mathbf{\hat{y}} + x_{5}a \, \mathbf{\hat{z}} & \left(32f\right) & \mbox{Mn IV} \\ 
\mathbf{B}_{24} & = & x_{5} \, \mathbf{a}_{1}-3x_{5} \, \mathbf{a}_{2} + x_{5} \, \mathbf{a}_{3} & = & -x_{5}a \, \mathbf{\hat{x}} + x_{5}a \, \mathbf{\hat{y}}-x_{5}a \, \mathbf{\hat{z}} & \left(32f\right) & \mbox{Mn IV} \\ 
\mathbf{B}_{25} & = & -3x_{5} \, \mathbf{a}_{1} + x_{5} \, \mathbf{a}_{2} + x_{5} \, \mathbf{a}_{3} & = & x_{5}a \, \mathbf{\hat{x}}-x_{5}a \, \mathbf{\hat{y}}-x_{5}a \, \mathbf{\hat{z}} & \left(32f\right) & \mbox{Mn IV} \\ 
\mathbf{B}_{26} & = & -x_{5} \, \mathbf{a}_{1}-x_{5} \, \mathbf{a}_{2} + 3x_{5} \, \mathbf{a}_{3} & = & x_{5}a \, \mathbf{\hat{x}} + x_{5}a \, \mathbf{\hat{y}}-x_{5}a \, \mathbf{\hat{z}} & \left(32f\right) & \mbox{Mn IV} \\ 
\mathbf{B}_{27} & = & -x_{5} \, \mathbf{a}_{1}-x_{5} \, \mathbf{a}_{2}-x_{5} \, \mathbf{a}_{3} & = & -x_{5}a \, \mathbf{\hat{x}}-x_{5}a \, \mathbf{\hat{y}}-x_{5}a \, \mathbf{\hat{z}} & \left(32f\right) & \mbox{Mn IV} \\ 
\mathbf{B}_{28} & = & -x_{5} \, \mathbf{a}_{1} + 3x_{5} \, \mathbf{a}_{2}-x_{5} \, \mathbf{a}_{3} & = & x_{5}a \, \mathbf{\hat{x}}-x_{5}a \, \mathbf{\hat{y}} + x_{5}a \, \mathbf{\hat{z}} & \left(32f\right) & \mbox{Mn IV} \\ 
\mathbf{B}_{29} & = & 3x_{5} \, \mathbf{a}_{1}-x_{5} \, \mathbf{a}_{2}-x_{5} \, \mathbf{a}_{3} & = & -x_{5}a \, \mathbf{\hat{x}} + x_{5}a \, \mathbf{\hat{y}} + x_{5}a \, \mathbf{\hat{z}} & \left(32f\right) & \mbox{Mn IV} \\ 
\end{longtabu}
\renewcommand{\arraystretch}{1.0}
\noindent \hrulefill
\\
\textbf{References:}
\vspace*{-0.25cm}
\begin{flushleft}
  - \bibentry{Florio_Acta_Cryst_5_1952}. \\
\end{flushleft}
\textbf{Found in:}
\vspace*{-0.25cm}
\begin{flushleft}
  - \bibentry{Pearson_NRC_1958}. \\
\end{flushleft}
\noindent \hrulefill
\\
\textbf{Geometry files:}
\\
\noindent  - CIF: pp. {\hyperref[A23B6_cF116_225_bd2f_e_cif]{\pageref{A23B6_cF116_225_bd2f_e_cif}}} \\
\noindent  - POSCAR: pp. {\hyperref[A23B6_cF116_225_bd2f_e_poscar]{\pageref{A23B6_cF116_225_bd2f_e_poscar}}} \\
\onecolumn
{\phantomsection\label{A6B2C_cF36_225_e_c_a}}
\subsection*{\huge \textbf{{\normalfont K$_{2}$PtCl$_{6}$ ($J1_{1}$) Structure: A6B2C\_cF36\_225\_e\_c\_a}}}
\noindent \hrulefill
\vspace*{0.25cm}
\begin{figure}[htp]
  \centering
  \vspace{-1em}
  {\includegraphics[width=1\textwidth]{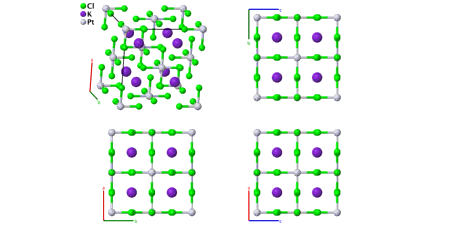}}
\end{figure}
\vspace*{-0.5cm}
\renewcommand{\arraystretch}{1.5}
\begin{equation*}
  \begin{array}{>{$\hspace{-0.15cm}}l<{$}>{$}p{0.5cm}<{$}>{$}p{18.5cm}<{$}}
    \mbox{\large \textbf{Prototype}} &\colon & \ce{K2PtCl6} \\
    \mbox{\large \textbf{\AFLOW\ prototype label}} &\colon & \mbox{A6B2C\_cF36\_225\_e\_c\_a} \\
    \mbox{\large \textbf{\textit{Strukturbericht} designation}} &\colon & \mbox{$J1_{1}$} \\
    \mbox{\large \textbf{Pearson symbol}} &\colon & \mbox{cF36} \\
    \mbox{\large \textbf{Space group number}} &\colon & 225 \\
    \mbox{\large \textbf{Space group symbol}} &\colon & Fm\bar{3}m \\
    \mbox{\large \textbf{\AFLOW\ prototype command}} &\colon &  \texttt{aflow} \,  \, \texttt{-{}-proto=A6B2C\_cF36\_225\_e\_c\_a } \, \newline \texttt{-{}-params=}{a,x_{3} }
  \end{array}
\end{equation*}
\renewcommand{\arraystretch}{1.0}

\vspace*{-0.25cm}
\noindent \hrulefill
\\
\textbf{ Other compounds with this structure:}
\begin{itemize}
   \item{ Gd$_{2}$MnGa$_{6}$, K$_{2}$TeBr$_{6}$, Sr$_{2}$RuH$_{6}$   }
\end{itemize}
\vspace*{-0.25cm}
\noindent \hrulefill
\begin{itemize}
  \item{In (Douglas, 2006), Table 6.6 provides an extensive list of compounds
with this structure.  Most have the formula $A_{2}MX_{6}$, where $A$
is an alkali metal, $M$ is a metal, and $X$ is a halide.  An
ammonium ion (NH$_{4}^+$) can also substitute for the alkali.
}
\end{itemize}

\noindent \parbox{1 \linewidth}{
\noindent \hrulefill
\\
\textbf{Face-centered Cubic primitive vectors:} \\
\vspace*{-0.25cm}
\begin{tabular}{cc}
  \begin{tabular}{c}
    \parbox{0.6 \linewidth}{
      \renewcommand{\arraystretch}{1.5}
      \begin{equation*}
        \centering
        \begin{array}{ccc}
              \mathbf{a}_1 & = & \frac12 \, a \, \mathbf{\hat{y}} + \frac12 \, a \, \mathbf{\hat{z}} \\
    \mathbf{a}_2 & = & \frac12 \, a \, \mathbf{\hat{x}} + \frac12 \, a \, \mathbf{\hat{z}} \\
    \mathbf{a}_3 & = & \frac12 \, a \, \mathbf{\hat{x}} + \frac12 \, a \, \mathbf{\hat{y}} \\

        \end{array}
      \end{equation*}
    }
    \renewcommand{\arraystretch}{1.0}
  \end{tabular}
  \begin{tabular}{c}
    \includegraphics[width=0.3\linewidth]{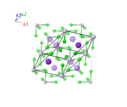} \\
  \end{tabular}
\end{tabular}

}
\vspace*{-0.25cm}

\noindent \hrulefill
\\
\textbf{Basis vectors:}
\vspace*{-0.25cm}
\renewcommand{\arraystretch}{1.5}
\begin{longtabu} to \textwidth{>{\centering $}X[-1,c,c]<{$}>{\centering $}X[-1,c,c]<{$}>{\centering $}X[-1,c,c]<{$}>{\centering $}X[-1,c,c]<{$}>{\centering $}X[-1,c,c]<{$}>{\centering $}X[-1,c,c]<{$}>{\centering $}X[-1,c,c]<{$}}
  & & \mbox{Lattice Coordinates} & & \mbox{Cartesian Coordinates} &\mbox{Wyckoff Position} & \mbox{Atom Type} \\  
  \mathbf{B}_{1} & = & 0 \, \mathbf{a}_{1} + 0 \, \mathbf{a}_{2} + 0 \, \mathbf{a}_{3} & = & 0 \, \mathbf{\hat{x}} + 0 \, \mathbf{\hat{y}} + 0 \, \mathbf{\hat{z}} & \left(4a\right) & \mbox{Pt} \\ 
\mathbf{B}_{2} & = & \frac{1}{4} \, \mathbf{a}_{1} + \frac{1}{4} \, \mathbf{a}_{2} + \frac{1}{4} \, \mathbf{a}_{3} & = & \frac{1}{4}a \, \mathbf{\hat{x}} + \frac{1}{4}a \, \mathbf{\hat{y}} + \frac{1}{4}a \, \mathbf{\hat{z}} & \left(8c\right) & \mbox{K} \\ 
\mathbf{B}_{3} & = & \frac{3}{4} \, \mathbf{a}_{1} + \frac{3}{4} \, \mathbf{a}_{2} + \frac{3}{4} \, \mathbf{a}_{3} & = & \frac{3}{4}a \, \mathbf{\hat{x}} + \frac{3}{4}a \, \mathbf{\hat{y}} + \frac{3}{4}a \, \mathbf{\hat{z}} & \left(8c\right) & \mbox{K} \\ 
\mathbf{B}_{4} & = & -x_{3} \, \mathbf{a}_{1} + x_{3} \, \mathbf{a}_{2} + x_{3} \, \mathbf{a}_{3} & = & x_{3}a \, \mathbf{\hat{x}} & \left(24e\right) & \mbox{Cl} \\ 
\mathbf{B}_{5} & = & x_{3} \, \mathbf{a}_{1}-x_{3} \, \mathbf{a}_{2}-x_{3} \, \mathbf{a}_{3} & = & -x_{3}a \, \mathbf{\hat{x}} & \left(24e\right) & \mbox{Cl} \\ 
\mathbf{B}_{6} & = & x_{3} \, \mathbf{a}_{1}-x_{3} \, \mathbf{a}_{2} + x_{3} \, \mathbf{a}_{3} & = & x_{3}a \, \mathbf{\hat{y}} & \left(24e\right) & \mbox{Cl} \\ 
\mathbf{B}_{7} & = & -x_{3} \, \mathbf{a}_{1} + x_{3} \, \mathbf{a}_{2}-x_{3} \, \mathbf{a}_{3} & = & -x_{3}a \, \mathbf{\hat{y}} & \left(24e\right) & \mbox{Cl} \\ 
\mathbf{B}_{8} & = & x_{3} \, \mathbf{a}_{1} + x_{3} \, \mathbf{a}_{2}-x_{3} \, \mathbf{a}_{3} & = & x_{3}a \, \mathbf{\hat{z}} & \left(24e\right) & \mbox{Cl} \\ 
\mathbf{B}_{9} & = & -x_{3} \, \mathbf{a}_{1}-x_{3} \, \mathbf{a}_{2} + x_{3} \, \mathbf{a}_{3} & = & -x_{3}a \, \mathbf{\hat{z}} & \left(24e\right) & \mbox{Cl} \\ 
\end{longtabu}
\renewcommand{\arraystretch}{1.0}
\noindent \hrulefill
\\
\textbf{References:}
\vspace*{-0.25cm}
\begin{flushleft}
  - \bibentry{Engel_Z_Krist_90_1935}. \\
  - \bibentry{Douglas_2006}. \\
\end{flushleft}
\textbf{Found in:}
\vspace*{-0.25cm}
\begin{flushleft}
  - \bibentry{Downs_Am_Min_88_2003}. \\
\end{flushleft}
\noindent \hrulefill
\\
\textbf{Geometry files:}
\\
\noindent  - CIF: pp. {\hyperref[A6B2C_cF36_225_e_c_a_cif]{\pageref{A6B2C_cF36_225_e_c_a_cif}}} \\
\noindent  - POSCAR: pp. {\hyperref[A6B2C_cF36_225_e_c_a_poscar]{\pageref{A6B2C_cF36_225_e_c_a_poscar}}} \\
\onecolumn
{\phantomsection\label{AB13_cF112_226_a_bi}}
\subsection*{\huge \textbf{{\normalfont NaZn$_{13}$ ($D2_{3}$) Structure: AB13\_cF112\_226\_a\_bi}}}
\noindent \hrulefill
\vspace*{0.25cm}
\begin{figure}[htp]
  \centering
  \vspace{-1em}
  {\includegraphics[width=1\textwidth]{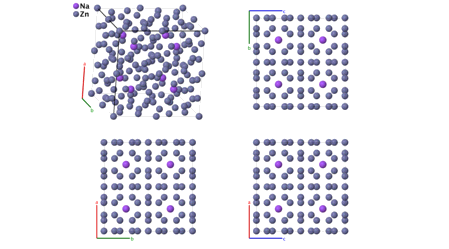}}
\end{figure}
\vspace*{-0.5cm}
\renewcommand{\arraystretch}{1.5}
\begin{equation*}
  \begin{array}{>{$\hspace{-0.15cm}}l<{$}>{$}p{0.5cm}<{$}>{$}p{18.5cm}<{$}}
    \mbox{\large \textbf{Prototype}} &\colon & \ce{NaZn$_{13}$} \\
    \mbox{\large \textbf{\AFLOW\ prototype label}} &\colon & \mbox{AB13\_cF112\_226\_a\_bi} \\
    \mbox{\large \textbf{\textit{Strukturbericht} designation}} &\colon & \mbox{$D2_{3}$} \\
    \mbox{\large \textbf{Pearson symbol}} &\colon & \mbox{cF112} \\
    \mbox{\large \textbf{Space group number}} &\colon & 226 \\
    \mbox{\large \textbf{Space group symbol}} &\colon & Fm\bar{3}c \\
    \mbox{\large \textbf{\AFLOW\ prototype command}} &\colon &  \texttt{aflow} \,  \, \texttt{-{}-proto=AB13\_cF112\_226\_a\_bi } \, \newline \texttt{-{}-params=}{a,y_{3},z_{3} }
  \end{array}
\end{equation*}
\renewcommand{\arraystretch}{1.0}

\vspace*{-0.25cm}
\noindent \hrulefill
\\
\textbf{ Other compounds with this structure:}
\begin{itemize}
   \item{ AmBe$_{13}$, BaZn$_{13}$, CaBe$_{13}$, CaZn$_{13}$, CdZn$_{13}$, CeBe$_{13}$, CsCd$_{13}$, KCd$_{13}$, KZn$_{13}$, MgBe$_{13}$, NbBe$_{13}$, RbCd$_{13}$, SrZn$_{13}$, ThBe$_{13}$, UBe$_{13}$, VBe$_{13}$, ZrBe$_{13}$, CeNi$_{8.5}$Si$_{4.5}$, LaFe$_{13-x-y}$Co$_{y}$Al$_{x}$, LaFe$_{13-x-y}$Co$_{y}$Si$_{x}$, NdFe$_{13-x-y}$Co$_{y}$Si$_{x}$  }
\end{itemize}
\noindent \parbox{1 \linewidth}{
\noindent \hrulefill
\\
\textbf{Face-centered Cubic primitive vectors:} \\
\vspace*{-0.25cm}
\begin{tabular}{cc}
  \begin{tabular}{c}
    \parbox{0.6 \linewidth}{
      \renewcommand{\arraystretch}{1.5}
      \begin{equation*}
        \centering
        \begin{array}{ccc}
              \mathbf{a}_1 & = & \frac12 \, a \, \mathbf{\hat{y}} + \frac12 \, a \, \mathbf{\hat{z}} \\
    \mathbf{a}_2 & = & \frac12 \, a \, \mathbf{\hat{x}} + \frac12 \, a \, \mathbf{\hat{z}} \\
    \mathbf{a}_3 & = & \frac12 \, a \, \mathbf{\hat{x}} + \frac12 \, a \, \mathbf{\hat{y}} \\

        \end{array}
      \end{equation*}
    }
    \renewcommand{\arraystretch}{1.0}
  \end{tabular}
  \begin{tabular}{c}
    \includegraphics[width=0.3\linewidth]{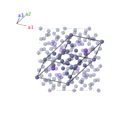} \\
  \end{tabular}
\end{tabular}

}
\vspace*{-0.25cm}

\noindent \hrulefill
\\
\textbf{Basis vectors:}
\vspace*{-0.25cm}
\renewcommand{\arraystretch}{1.5}
\begin{longtabu} to \textwidth{>{\centering $}X[-1,c,c]<{$}>{\centering $}X[-1,c,c]<{$}>{\centering $}X[-1,c,c]<{$}>{\centering $}X[-1,c,c]<{$}>{\centering $}X[-1,c,c]<{$}>{\centering $}X[-1,c,c]<{$}>{\centering $}X[-1,c,c]<{$}}
  & & \mbox{Lattice Coordinates} & & \mbox{Cartesian Coordinates} &\mbox{Wyckoff Position} & \mbox{Atom Type} \\  
  \mathbf{B}_{1} & = & \frac{1}{4} \, \mathbf{a}_{1} + \frac{1}{4} \, \mathbf{a}_{2} + \frac{1}{4} \, \mathbf{a}_{3} & = & \frac{1}{4}a \, \mathbf{\hat{x}} + \frac{1}{4}a \, \mathbf{\hat{y}} + \frac{1}{4}a \, \mathbf{\hat{z}} & \left(8a\right) & \mbox{Na} \\ 
\mathbf{B}_{2} & = & \frac{3}{4} \, \mathbf{a}_{1} + \frac{3}{4} \, \mathbf{a}_{2} + \frac{3}{4} \, \mathbf{a}_{3} & = & \frac{3}{4}a \, \mathbf{\hat{x}} + \frac{3}{4}a \, \mathbf{\hat{y}} + \frac{3}{4}a \, \mathbf{\hat{z}} & \left(8a\right) & \mbox{Na} \\ 
\mathbf{B}_{3} & = & 0 \, \mathbf{a}_{1} + 0 \, \mathbf{a}_{2} + 0 \, \mathbf{a}_{3} & = & 0 \, \mathbf{\hat{x}} + 0 \, \mathbf{\hat{y}} + 0 \, \mathbf{\hat{z}} & \left(8b\right) & \mbox{Zn I} \\ 
\mathbf{B}_{4} & = & \frac{1}{2} \, \mathbf{a}_{1} + \frac{1}{2} \, \mathbf{a}_{2} + \frac{1}{2} \, \mathbf{a}_{3} & = & \frac{1}{2}a \, \mathbf{\hat{x}} + \frac{1}{2}a \, \mathbf{\hat{y}} + \frac{1}{2}a \, \mathbf{\hat{z}} & \left(8b\right) & \mbox{Zn I} \\ 
\mathbf{B}_{5} & = & \left(y_{3}+z_{3}\right) \, \mathbf{a}_{1} + \left(-y_{3}+z_{3}\right) \, \mathbf{a}_{2} + \left(y_{3}-z_{3}\right) \, \mathbf{a}_{3} & = & y_{3}a \, \mathbf{\hat{y}} + z_{3}a \, \mathbf{\hat{z}} & \left(96i\right) & \mbox{Zn II} \\ 
\mathbf{B}_{6} & = & \left(-y_{3}+z_{3}\right) \, \mathbf{a}_{1} + \left(y_{3}+z_{3}\right) \, \mathbf{a}_{2} + \left(-y_{3}-z_{3}\right) \, \mathbf{a}_{3} & = & -y_{3}a \, \mathbf{\hat{y}} + z_{3}a \, \mathbf{\hat{z}} & \left(96i\right) & \mbox{Zn II} \\ 
\mathbf{B}_{7} & = & \left(y_{3}-z_{3}\right) \, \mathbf{a}_{1} + \left(-y_{3}-z_{3}\right) \, \mathbf{a}_{2} + \left(y_{3}+z_{3}\right) \, \mathbf{a}_{3} & = & y_{3}a \, \mathbf{\hat{y}}-z_{3}a \, \mathbf{\hat{z}} & \left(96i\right) & \mbox{Zn II} \\ 
\mathbf{B}_{8} & = & \left(-y_{3}-z_{3}\right) \, \mathbf{a}_{1} + \left(y_{3}-z_{3}\right) \, \mathbf{a}_{2} + \left(-y_{3}+z_{3}\right) \, \mathbf{a}_{3} & = & -y_{3}a \, \mathbf{\hat{y}}-z_{3}a \, \mathbf{\hat{z}} & \left(96i\right) & \mbox{Zn II} \\ 
\mathbf{B}_{9} & = & \left(y_{3}-z_{3}\right) \, \mathbf{a}_{1} + \left(y_{3}+z_{3}\right) \, \mathbf{a}_{2} + \left(-y_{3}+z_{3}\right) \, \mathbf{a}_{3} & = & z_{3}a \, \mathbf{\hat{x}} + y_{3}a \, \mathbf{\hat{z}} & \left(96i\right) & \mbox{Zn II} \\ 
\mathbf{B}_{10} & = & \left(-y_{3}-z_{3}\right) \, \mathbf{a}_{1} + \left(-y_{3}+z_{3}\right) \, \mathbf{a}_{2} + \left(y_{3}+z_{3}\right) \, \mathbf{a}_{3} & = & z_{3}a \, \mathbf{\hat{x}} + -y_{3}a \, \mathbf{\hat{z}} & \left(96i\right) & \mbox{Zn II} \\ 
\mathbf{B}_{11} & = & \left(y_{3}+z_{3}\right) \, \mathbf{a}_{1} + \left(y_{3}-z_{3}\right) \, \mathbf{a}_{2} + \left(-y_{3}-z_{3}\right) \, \mathbf{a}_{3} & = & -z_{3}a \, \mathbf{\hat{x}} + y_{3}a \, \mathbf{\hat{z}} & \left(96i\right) & \mbox{Zn II} \\ 
\mathbf{B}_{12} & = & \left(-y_{3}+z_{3}\right) \, \mathbf{a}_{1} + \left(-y_{3}-z_{3}\right) \, \mathbf{a}_{2} + \left(y_{3}-z_{3}\right) \, \mathbf{a}_{3} & = & -z_{3}a \, \mathbf{\hat{x}} + -y_{3}a \, \mathbf{\hat{z}} & \left(96i\right) & \mbox{Zn II} \\ 
\mathbf{B}_{13} & = & \left(-y_{3}+z_{3}\right) \, \mathbf{a}_{1} + \left(y_{3}-z_{3}\right) \, \mathbf{a}_{2} + \left(y_{3}+z_{3}\right) \, \mathbf{a}_{3} & = & y_{3}a \, \mathbf{\hat{x}} + z_{3}a \, \mathbf{\hat{y}} & \left(96i\right) & \mbox{Zn II} \\ 
\mathbf{B}_{14} & = & \left(y_{3}+z_{3}\right) \, \mathbf{a}_{1} + \left(-y_{3}-z_{3}\right) \, \mathbf{a}_{2} + \left(-y_{3}+z_{3}\right) \, \mathbf{a}_{3} & = & -y_{3}a \, \mathbf{\hat{x}} + z_{3}a \, \mathbf{\hat{y}} & \left(96i\right) & \mbox{Zn II} \\ 
\mathbf{B}_{15} & = & \left(-y_{3}-z_{3}\right) \, \mathbf{a}_{1} + \left(y_{3}+z_{3}\right) \, \mathbf{a}_{2} + \left(y_{3}-z_{3}\right) \, \mathbf{a}_{3} & = & y_{3}a \, \mathbf{\hat{x}}-z_{3}a \, \mathbf{\hat{y}} & \left(96i\right) & \mbox{Zn II} \\ 
\mathbf{B}_{16} & = & \left(y_{3}-z_{3}\right) \, \mathbf{a}_{1} + \left(-y_{3}+z_{3}\right) \, \mathbf{a}_{2} + \left(-y_{3}-z_{3}\right) \, \mathbf{a}_{3} & = & -y_{3}a \, \mathbf{\hat{x}}-z_{3}a \, \mathbf{\hat{y}} & \left(96i\right) & \mbox{Zn II} \\ 
\mathbf{B}_{17} & = & \left(\frac{1}{2} - y_{3} - z_{3}\right) \, \mathbf{a}_{1} + \left(\frac{1}{2} +y_{3} - z_{3}\right) \, \mathbf{a}_{2} + \left(\frac{1}{2} +y_{3} + z_{3}\right) \, \mathbf{a}_{3} & = & \left(\frac{1}{2} +y_{3}\right)a \, \mathbf{\hat{x}} + \frac{1}{2}a \, \mathbf{\hat{y}} + \left(\frac{1}{2} - z_{3}\right)a \, \mathbf{\hat{z}} & \left(96i\right) & \mbox{Zn II} \\ 
\mathbf{B}_{18} & = & \left(\frac{1}{2} +y_{3} - z_{3}\right) \, \mathbf{a}_{1} + \left(\frac{1}{2} - y_{3} - z_{3}\right) \, \mathbf{a}_{2} + \left(\frac{1}{2} - y_{3} + z_{3}\right) \, \mathbf{a}_{3} & = & \left(\frac{1}{2} - y_{3}\right)a \, \mathbf{\hat{x}} + \frac{1}{2}a \, \mathbf{\hat{y}} + \left(\frac{1}{2} - z_{3}\right)a \, \mathbf{\hat{z}} & \left(96i\right) & \mbox{Zn II} \\ 
\mathbf{B}_{19} & = & \left(\frac{1}{2} - y_{3} + z_{3}\right) \, \mathbf{a}_{1} + \left(\frac{1}{2} +y_{3} + z_{3}\right) \, \mathbf{a}_{2} + \left(\frac{1}{2} +y_{3} - z_{3}\right) \, \mathbf{a}_{3} & = & \left(\frac{1}{2} +y_{3}\right)a \, \mathbf{\hat{x}} + \frac{1}{2}a \, \mathbf{\hat{y}} + \left(\frac{1}{2} +z_{3}\right)a \, \mathbf{\hat{z}} & \left(96i\right) & \mbox{Zn II} \\ 
\mathbf{B}_{20} & = & \left(\frac{1}{2} +y_{3} + z_{3}\right) \, \mathbf{a}_{1} + \left(\frac{1}{2} - y_{3} + z_{3}\right) \, \mathbf{a}_{2} + \left(\frac{1}{2} - y_{3} - z_{3}\right) \, \mathbf{a}_{3} & = & \left(\frac{1}{2} - y_{3}\right)a \, \mathbf{\hat{x}} + \frac{1}{2}a \, \mathbf{\hat{y}} + \left(\frac{1}{2} +z_{3}\right)a \, \mathbf{\hat{z}} & \left(96i\right) & \mbox{Zn II} \\ 
\mathbf{B}_{21} & = & \left(\frac{1}{2} - y_{3} + z_{3}\right) \, \mathbf{a}_{1} + \left(\frac{1}{2} - y_{3} - z_{3}\right) \, \mathbf{a}_{2} + \left(\frac{1}{2} +y_{3} + z_{3}\right) \, \mathbf{a}_{3} & = & \frac{1}{2}a \, \mathbf{\hat{x}} + \left(\frac{1}{2} +z_{3}\right)a \, \mathbf{\hat{y}} + \left(\frac{1}{2} - y_{3}\right)a \, \mathbf{\hat{z}} & \left(96i\right) & \mbox{Zn II} \\ 
\mathbf{B}_{22} & = & \left(\frac{1}{2} +y_{3} + z_{3}\right) \, \mathbf{a}_{1} + \left(\frac{1}{2} +y_{3} - z_{3}\right) \, \mathbf{a}_{2} + \left(\frac{1}{2} - y_{3} + z_{3}\right) \, \mathbf{a}_{3} & = & \frac{1}{2}a \, \mathbf{\hat{x}} + \left(\frac{1}{2} +z_{3}\right)a \, \mathbf{\hat{y}} + \left(\frac{1}{2} +y_{3}\right)a \, \mathbf{\hat{z}} & \left(96i\right) & \mbox{Zn II} \\ 
\mathbf{B}_{23} & = & \left(\frac{1}{2} - y_{3} - z_{3}\right) \, \mathbf{a}_{1} + \left(\frac{1}{2} - y_{3} + z_{3}\right) \, \mathbf{a}_{2} + \left(\frac{1}{2} +y_{3} - z_{3}\right) \, \mathbf{a}_{3} & = & \frac{1}{2}a \, \mathbf{\hat{x}} + \left(\frac{1}{2} - z_{3}\right)a \, \mathbf{\hat{y}} + \left(\frac{1}{2} - y_{3}\right)a \, \mathbf{\hat{z}} & \left(96i\right) & \mbox{Zn II} \\ 
\mathbf{B}_{24} & = & \left(\frac{1}{2} +y_{3} - z_{3}\right) \, \mathbf{a}_{1} + \left(\frac{1}{2} +y_{3} + z_{3}\right) \, \mathbf{a}_{2} + \left(\frac{1}{2} - y_{3} - z_{3}\right) \, \mathbf{a}_{3} & = & \frac{1}{2}a \, \mathbf{\hat{x}} + \left(\frac{1}{2} - z_{3}\right)a \, \mathbf{\hat{y}} + \left(\frac{1}{2} +y_{3}\right)a \, \mathbf{\hat{z}} & \left(96i\right) & \mbox{Zn II} \\ 
\mathbf{B}_{25} & = & \left(\frac{1}{2} +y_{3} - z_{3}\right) \, \mathbf{a}_{1} + \left(\frac{1}{2} - y_{3} + z_{3}\right) \, \mathbf{a}_{2} + \left(\frac{1}{2} +y_{3} + z_{3}\right) \, \mathbf{a}_{3} & = & \left(\frac{1}{2} +z_{3}\right)a \, \mathbf{\hat{x}} + \left(\frac{1}{2} +y_{3}\right)a \, \mathbf{\hat{y}} + \frac{1}{2}a \, \mathbf{\hat{z}} & \left(96i\right) & \mbox{Zn II} \\ 
\mathbf{B}_{26} & = & \left(\frac{1}{2} - y_{3} - z_{3}\right) \, \mathbf{a}_{1} + \left(\frac{1}{2} +y_{3} + z_{3}\right) \, \mathbf{a}_{2} + \left(\frac{1}{2} - y_{3} + z_{3}\right) \, \mathbf{a}_{3} & = & \left(\frac{1}{2} +z_{3}\right)a \, \mathbf{\hat{x}} + \left(\frac{1}{2} - y_{3}\right)a \, \mathbf{\hat{y}} + \frac{1}{2}a \, \mathbf{\hat{z}} & \left(96i\right) & \mbox{Zn II} \\ 
\mathbf{B}_{27} & = & \left(\frac{1}{2} +y_{3} + z_{3}\right) \, \mathbf{a}_{1} + \left(\frac{1}{2} - y_{3} - z_{3}\right) \, \mathbf{a}_{2} + \left(\frac{1}{2} +y_{3} - z_{3}\right) \, \mathbf{a}_{3} & = & \left(\frac{1}{2} - z_{3}\right)a \, \mathbf{\hat{x}} + \left(\frac{1}{2} +y_{3}\right)a \, \mathbf{\hat{y}} + \frac{1}{2}a \, \mathbf{\hat{z}} & \left(96i\right) & \mbox{Zn II} \\ 
\mathbf{B}_{28} & = & \left(\frac{1}{2} - y_{3} + z_{3}\right) \, \mathbf{a}_{1} + \left(\frac{1}{2} +y_{3} - z_{3}\right) \, \mathbf{a}_{2} + \left(\frac{1}{2} - y_{3} - z_{3}\right) \, \mathbf{a}_{3} & = & \left(\frac{1}{2} - z_{3}\right)a \, \mathbf{\hat{x}} + \left(\frac{1}{2} - y_{3}\right)a \, \mathbf{\hat{y}} + \frac{1}{2}a \, \mathbf{\hat{z}} & \left(96i\right) & \mbox{Zn II} \\ 
\end{longtabu}
\renewcommand{\arraystretch}{1.0}
\noindent \hrulefill
\\
\textbf{References:}
\vspace*{-0.25cm}
\begin{flushleft}
  - \bibentry{Shoemaker_1952}. \\
\end{flushleft}
\noindent \hrulefill
\\
\textbf{Geometry files:}
\\
\noindent  - CIF: pp. {\hyperref[AB13_cF112_226_a_bi_cif]{\pageref{AB13_cF112_226_a_bi_cif}}} \\
\noindent  - POSCAR: pp. {\hyperref[AB13_cF112_226_a_bi_poscar]{\pageref{AB13_cF112_226_a_bi_poscar}}} \\
\onecolumn
{\phantomsection\label{A2B2C7_cF88_227_c_d_af}}
\subsection*{\huge \textbf{{\normalfont \begin{raggedleft}Pyrochlore Iridate (Eu$_{2}$Ir$_{2}$O$_{7}$) Structure: \end{raggedleft} \\ A2B2C7\_cF88\_227\_c\_d\_af}}}
\noindent \hrulefill
\vspace*{0.25cm}
\begin{figure}[htp]
  \centering
  \vspace{-1em}
  {\includegraphics[width=1\textwidth]{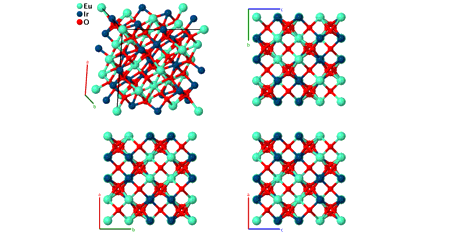}}
\end{figure}
\vspace*{-0.5cm}
\renewcommand{\arraystretch}{1.5}
\begin{equation*}
  \begin{array}{>{$\hspace{-0.15cm}}l<{$}>{$}p{0.5cm}<{$}>{$}p{18.5cm}<{$}}
    \mbox{\large \textbf{Prototype}} &\colon & \ce{Eu2Ir2O7} \\
    \mbox{\large \textbf{\AFLOW\ prototype label}} &\colon & \mbox{A2B2C7\_cF88\_227\_c\_d\_af} \\
    \mbox{\large \textbf{\textit{Strukturbericht} designation}} &\colon & \mbox{None} \\
    \mbox{\large \textbf{Pearson symbol}} &\colon & \mbox{cF88} \\
    \mbox{\large \textbf{Space group number}} &\colon & 227 \\
    \mbox{\large \textbf{Space group symbol}} &\colon & Fd\bar{3}m \\
    \mbox{\large \textbf{\AFLOW\ prototype command}} &\colon &  \texttt{aflow} \,  \, \texttt{-{}-proto=A2B2C7\_cF88\_227\_c\_d\_af } \, \newline \texttt{-{}-params=}{a,x_{4} }
  \end{array}
\end{equation*}
\renewcommand{\arraystretch}{1.0}

\noindent \parbox{1 \linewidth}{
\noindent \hrulefill
\\
\textbf{Face-centered Cubic primitive vectors:} \\
\vspace*{-0.25cm}
\begin{tabular}{cc}
  \begin{tabular}{c}
    \parbox{0.6 \linewidth}{
      \renewcommand{\arraystretch}{1.5}
      \begin{equation*}
        \centering
        \begin{array}{ccc}
              \mathbf{a}_1 & = & \frac12 \, a \, \mathbf{\hat{y}} + \frac12 \, a \, \mathbf{\hat{z}} \\
    \mathbf{a}_2 & = & \frac12 \, a \, \mathbf{\hat{x}} + \frac12 \, a \, \mathbf{\hat{z}} \\
    \mathbf{a}_3 & = & \frac12 \, a \, \mathbf{\hat{x}} + \frac12 \, a \, \mathbf{\hat{y}} \\

        \end{array}
      \end{equation*}
    }
    \renewcommand{\arraystretch}{1.0}
  \end{tabular}
  \begin{tabular}{c}
    \includegraphics[width=0.3\linewidth]{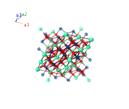} \\
  \end{tabular}
\end{tabular}

}
\vspace*{-0.25cm}

\noindent \hrulefill
\\
\textbf{Basis vectors:}
\vspace*{-0.25cm}
\renewcommand{\arraystretch}{1.5}
\begin{longtabu} to \textwidth{>{\centering $}X[-1,c,c]<{$}>{\centering $}X[-1,c,c]<{$}>{\centering $}X[-1,c,c]<{$}>{\centering $}X[-1,c,c]<{$}>{\centering $}X[-1,c,c]<{$}>{\centering $}X[-1,c,c]<{$}>{\centering $}X[-1,c,c]<{$}}
  & & \mbox{Lattice Coordinates} & & \mbox{Cartesian Coordinates} &\mbox{Wyckoff Position} & \mbox{Atom Type} \\  
  \mathbf{B}_{1} & = & \frac{1}{8} \, \mathbf{a}_{1} + \frac{1}{8} \, \mathbf{a}_{2} + \frac{1}{8} \, \mathbf{a}_{3} & = & \frac{1}{8}a \, \mathbf{\hat{x}} + \frac{1}{8}a \, \mathbf{\hat{y}} + \frac{1}{8}a \, \mathbf{\hat{z}} & \left(8a\right) & \mbox{O I} \\ 
\mathbf{B}_{2} & = & \frac{7}{8} \, \mathbf{a}_{1} + \frac{7}{8} \, \mathbf{a}_{2} + \frac{7}{8} \, \mathbf{a}_{3} & = & \frac{7}{8}a \, \mathbf{\hat{x}} + \frac{7}{8}a \, \mathbf{\hat{y}} + \frac{7}{8}a \, \mathbf{\hat{z}} & \left(8a\right) & \mbox{O I} \\ 
\mathbf{B}_{3} & = & 0 \, \mathbf{a}_{1} + 0 \, \mathbf{a}_{2} + 0 \, \mathbf{a}_{3} & = & 0 \, \mathbf{\hat{x}} + 0 \, \mathbf{\hat{y}} + 0 \, \mathbf{\hat{z}} & \left(16c\right) & \mbox{Eu} \\ 
\mathbf{B}_{4} & = & \frac{1}{2} \, \mathbf{a}_{3} & = & \frac{1}{4}a \, \mathbf{\hat{x}} + \frac{1}{4}a \, \mathbf{\hat{y}} & \left(16c\right) & \mbox{Eu} \\ 
\mathbf{B}_{5} & = & \frac{1}{2} \, \mathbf{a}_{2} & = & \frac{1}{4}a \, \mathbf{\hat{x}} + \frac{1}{4}a \, \mathbf{\hat{z}} & \left(16c\right) & \mbox{Eu} \\ 
\mathbf{B}_{6} & = & \frac{1}{2} \, \mathbf{a}_{1} & = & \frac{1}{4}a \, \mathbf{\hat{y}} + \frac{1}{4}a \, \mathbf{\hat{z}} & \left(16c\right) & \mbox{Eu} \\ 
\mathbf{B}_{7} & = & \frac{1}{2} \, \mathbf{a}_{1} + \frac{1}{2} \, \mathbf{a}_{2} + \frac{1}{2} \, \mathbf{a}_{3} & = & \frac{1}{2}a \, \mathbf{\hat{x}} + \frac{1}{2}a \, \mathbf{\hat{y}} + \frac{1}{2}a \, \mathbf{\hat{z}} & \left(16d\right) & \mbox{Ir} \\ 
\mathbf{B}_{8} & = & \frac{1}{2} \, \mathbf{a}_{1} + \frac{1}{2} \, \mathbf{a}_{2} & = & \frac{1}{4}a \, \mathbf{\hat{x}} + \frac{1}{4}a \, \mathbf{\hat{y}} + \frac{1}{2}a \, \mathbf{\hat{z}} & \left(16d\right) & \mbox{Ir} \\ 
\mathbf{B}_{9} & = & \frac{1}{2} \, \mathbf{a}_{1} + \frac{1}{2} \, \mathbf{a}_{3} & = & \frac{1}{4}a \, \mathbf{\hat{x}} + \frac{1}{2}a \, \mathbf{\hat{y}} + \frac{1}{4}a \, \mathbf{\hat{z}} & \left(16d\right) & \mbox{Ir} \\ 
\mathbf{B}_{10} & = & \frac{1}{2} \, \mathbf{a}_{2} + \frac{1}{2} \, \mathbf{a}_{3} & = & \frac{1}{2}a \, \mathbf{\hat{x}} + \frac{1}{4}a \, \mathbf{\hat{y}} + \frac{1}{4}a \, \mathbf{\hat{z}} & \left(16d\right) & \mbox{Ir} \\ 
\mathbf{B}_{11} & = & \left(\frac{1}{4} - x_{4}\right) \, \mathbf{a}_{1} + x_{4} \, \mathbf{a}_{2} + x_{4} \, \mathbf{a}_{3} & = & x_{4}a \, \mathbf{\hat{x}} + \frac{1}{8}a \, \mathbf{\hat{y}} + \frac{1}{8}a \, \mathbf{\hat{z}} & \left(48f\right) & \mbox{O II} \\ 
\mathbf{B}_{12} & = & x_{4} \, \mathbf{a}_{1} + \left(\frac{1}{4} - x_{4}\right) \, \mathbf{a}_{2} + \left(\frac{1}{4} - x_{4}\right) \, \mathbf{a}_{3} & = & \left(\frac{1}{4} - x_{4}\right)a \, \mathbf{\hat{x}} + \frac{1}{8}a \, \mathbf{\hat{y}} + \frac{1}{8}a \, \mathbf{\hat{z}} & \left(48f\right) & \mbox{O II} \\ 
\mathbf{B}_{13} & = & x_{4} \, \mathbf{a}_{1} + \left(\frac{1}{4} - x_{4}\right) \, \mathbf{a}_{2} + x_{4} \, \mathbf{a}_{3} & = & \frac{1}{8}a \, \mathbf{\hat{x}} + x_{4}a \, \mathbf{\hat{y}} + \frac{1}{8}a \, \mathbf{\hat{z}} & \left(48f\right) & \mbox{O II} \\ 
\mathbf{B}_{14} & = & \left(\frac{1}{4} - x_{4}\right) \, \mathbf{a}_{1} + x_{4} \, \mathbf{a}_{2} + \left(\frac{1}{4} - x_{4}\right) \, \mathbf{a}_{3} & = & \frac{1}{8}a \, \mathbf{\hat{x}} + \left(\frac{1}{4} - x_{4}\right)a \, \mathbf{\hat{y}} + \frac{1}{8}a \, \mathbf{\hat{z}} & \left(48f\right) & \mbox{O II} \\ 
\mathbf{B}_{15} & = & x_{4} \, \mathbf{a}_{1} + x_{4} \, \mathbf{a}_{2} + \left(\frac{1}{4} - x_{4}\right) \, \mathbf{a}_{3} & = & \frac{1}{8}a \, \mathbf{\hat{x}} + \frac{1}{8}a \, \mathbf{\hat{y}} + x_{4}a \, \mathbf{\hat{z}} & \left(48f\right) & \mbox{O II} \\ 
\mathbf{B}_{16} & = & \left(\frac{1}{4} - x_{4}\right) \, \mathbf{a}_{1} + \left(\frac{1}{4} - x_{4}\right) \, \mathbf{a}_{2} + x_{4} \, \mathbf{a}_{3} & = & \frac{1}{8}a \, \mathbf{\hat{x}} + \frac{1}{8}a \, \mathbf{\hat{y}} + \left(\frac{1}{4} - x_{4}\right)a \, \mathbf{\hat{z}} & \left(48f\right) & \mbox{O II} \\ 
\mathbf{B}_{17} & = & \left(\frac{3}{4} +x_{4}\right) \, \mathbf{a}_{1}-x_{4} \, \mathbf{a}_{2} + \left(\frac{3}{4} +x_{4}\right) \, \mathbf{a}_{3} & = & \frac{3}{8}a \, \mathbf{\hat{x}} + \left(\frac{3}{4} +x_{4}\right)a \, \mathbf{\hat{y}} + \frac{3}{8}a \, \mathbf{\hat{z}} & \left(48f\right) & \mbox{O II} \\ 
\mathbf{B}_{18} & = & -x_{4} \, \mathbf{a}_{1} + \left(\frac{3}{4} +x_{4}\right) \, \mathbf{a}_{2}-x_{4} \, \mathbf{a}_{3} & = & \frac{3}{8}a \, \mathbf{\hat{x}}-x_{4}a \, \mathbf{\hat{y}} + \frac{3}{8}a \, \mathbf{\hat{z}} & \left(48f\right) & \mbox{O II} \\ 
\mathbf{B}_{19} & = & -x_{4} \, \mathbf{a}_{1} + \left(\frac{3}{4} +x_{4}\right) \, \mathbf{a}_{2} + \left(\frac{3}{4} +x_{4}\right) \, \mathbf{a}_{3} & = & \left(\frac{3}{4} +x_{4}\right)a \, \mathbf{\hat{x}} + \frac{3}{8}a \, \mathbf{\hat{y}} + \frac{3}{8}a \, \mathbf{\hat{z}} & \left(48f\right) & \mbox{O II} \\ 
\mathbf{B}_{20} & = & \left(\frac{3}{4} +x_{4}\right) \, \mathbf{a}_{1}-x_{4} \, \mathbf{a}_{2}-x_{4} \, \mathbf{a}_{3} & = & -x_{4}a \, \mathbf{\hat{x}} + \frac{3}{8}a \, \mathbf{\hat{y}} + \frac{3}{8}a \, \mathbf{\hat{z}} & \left(48f\right) & \mbox{O II} \\ 
\mathbf{B}_{21} & = & -x_{4} \, \mathbf{a}_{1}-x_{4} \, \mathbf{a}_{2} + \left(\frac{3}{4} +x_{4}\right) \, \mathbf{a}_{3} & = & \frac{3}{8}a \, \mathbf{\hat{x}} + \frac{3}{8}a \, \mathbf{\hat{y}}-x_{4}a \, \mathbf{\hat{z}} & \left(48f\right) & \mbox{O II} \\ 
\mathbf{B}_{22} & = & \left(\frac{3}{4} +x_{4}\right) \, \mathbf{a}_{1} + \left(\frac{3}{4} +x_{4}\right) \, \mathbf{a}_{2}-x_{4} \, \mathbf{a}_{3} & = & \frac{3}{8}a \, \mathbf{\hat{x}} + \frac{3}{8}a \, \mathbf{\hat{y}} + \left(\frac{3}{4} +x_{4}\right)a \, \mathbf{\hat{z}} & \left(48f\right) & \mbox{O II} \\ 
\end{longtabu}
\renewcommand{\arraystretch}{1.0}
\noindent \hrulefill
\\
\textbf{References:}
\vspace*{-0.25cm}
\begin{flushleft}
  - \bibentry{Sagayama_PRB_87_100403_2013}. \\
\end{flushleft}
\textbf{Found in:}
\vspace*{-0.25cm}
\begin{flushleft}
  - \bibentry{Chun_PRL_120_177203_2018}. \\
\end{flushleft}
\noindent \hrulefill
\\
\textbf{Geometry files:}
\\
\noindent  - CIF: pp. {\hyperref[A2B2C7_cF88_227_c_d_af_cif]{\pageref{A2B2C7_cF88_227_c_d_af_cif}}} \\
\noindent  - POSCAR: pp. {\hyperref[A2B2C7_cF88_227_c_d_af_poscar]{\pageref{A2B2C7_cF88_227_c_d_af_poscar}}} \\
\onecolumn
{\phantomsection\label{A3B4_cF56_227_ad_e}}
\subsection*{\huge \textbf{{\normalfont \begin{raggedleft}Spinel (Co$_{3}$O$_{4}$, $D7_{2}$) Structure: \end{raggedleft} \\ A3B4\_cF56\_227\_ad\_e}}}
\noindent \hrulefill
\vspace*{0.25cm}
\begin{figure}[htp]
  \centering
  \vspace{-1em}
  {\includegraphics[width=1\textwidth]{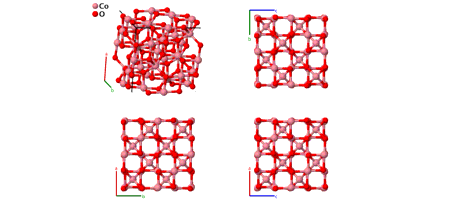}}
\end{figure}
\vspace*{-0.5cm}
\renewcommand{\arraystretch}{1.5}
\begin{equation*}
  \begin{array}{>{$\hspace{-0.15cm}}l<{$}>{$}p{0.5cm}<{$}>{$}p{18.5cm}<{$}}
    \mbox{\large \textbf{Prototype}} &\colon & \ce{Co3O4} \\
    \mbox{\large \textbf{\AFLOW\ prototype label}} &\colon & \mbox{A3B4\_cF56\_227\_ad\_e} \\
    \mbox{\large \textbf{\textit{Strukturbericht} designation}} &\colon & \mbox{$D7_{2}$} \\
    \mbox{\large \textbf{Pearson symbol}} &\colon & \mbox{cF56} \\
    \mbox{\large \textbf{Space group number}} &\colon & 227 \\
    \mbox{\large \textbf{Space group symbol}} &\colon & Fd\bar{3}m \\
    \mbox{\large \textbf{\AFLOW\ prototype command}} &\colon &  \texttt{aflow} \,  \, \texttt{-{}-proto=A3B4\_cF56\_227\_ad\_e } \, \newline \texttt{-{}-params=}{a,x_{3} }
  \end{array}
\end{equation*}
\renewcommand{\arraystretch}{1.0}

\vspace*{-0.25cm}
\noindent \hrulefill
\\
\textbf{ Other compounds with this structure: }
\begin{itemize}
   \item{ NiCo$_{2}$O$_{4}$, Co$_{3}$S$_{4}$, NiCo$_{2}$S$_{4}$, FeNi$_{2}$S$_{4}$  }
\end{itemize}
\vspace*{-0.25cm}
\noindent \hrulefill
\begin{itemize}
  \item{The \hyperref[A3B4_cF56_227_ad_e]{$D7_{2}$} and
\href{http://aflow.org/CrystalDatabase/A2BC4_cF56_227_d_a_e.html}{$H1_{1}$} Spinel structures are
for all intents and purposes identical. We could use $D7_{3}$ for the
binary spinels and $H1_{1}$ for the ternaries, but historically this
has not been the case.  We dual-list this structure only to keep the
historical record intact.
(Hahn, 1955) has an extensive list of ternary spinels and inverse
spinels.
}
\end{itemize}

\noindent \parbox{1 \linewidth}{
\noindent \hrulefill
\\
\textbf{Face-centered Cubic primitive vectors:} \\
\vspace*{-0.25cm}
\begin{tabular}{cc}
  \begin{tabular}{c}
    \parbox{0.6 \linewidth}{
      \renewcommand{\arraystretch}{1.5}
      \begin{equation*}
        \centering
        \begin{array}{ccc}
              \mathbf{a}_1 & = & \frac12 \, a \, \mathbf{\hat{y}} + \frac12 \, a \, \mathbf{\hat{z}} \\
    \mathbf{a}_2 & = & \frac12 \, a \, \mathbf{\hat{x}} + \frac12 \, a \, \mathbf{\hat{z}} \\
    \mathbf{a}_3 & = & \frac12 \, a \, \mathbf{\hat{x}} + \frac12 \, a \, \mathbf{\hat{y}} \\

        \end{array}
      \end{equation*}
    }
    \renewcommand{\arraystretch}{1.0}
  \end{tabular}
  \begin{tabular}{c}
    \includegraphics[width=0.3\linewidth]{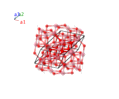} \\
  \end{tabular}
\end{tabular}

}
\vspace*{-0.25cm}

\noindent \hrulefill
\\
\textbf{Basis vectors:}
\vspace*{-0.25cm}
\renewcommand{\arraystretch}{1.5}
\begin{longtabu} to \textwidth{>{\centering $}X[-1,c,c]<{$}>{\centering $}X[-1,c,c]<{$}>{\centering $}X[-1,c,c]<{$}>{\centering $}X[-1,c,c]<{$}>{\centering $}X[-1,c,c]<{$}>{\centering $}X[-1,c,c]<{$}>{\centering $}X[-1,c,c]<{$}}
  & & \mbox{Lattice Coordinates} & & \mbox{Cartesian Coordinates} &\mbox{Wyckoff Position} & \mbox{Atom Type} \\  
  \mathbf{B}_{1} & = & \frac{1}{8} \, \mathbf{a}_{1} + \frac{1}{8} \, \mathbf{a}_{2} + \frac{1}{8} \, \mathbf{a}_{3} & = & \frac{1}{8}a \, \mathbf{\hat{x}} + \frac{1}{8}a \, \mathbf{\hat{y}} + \frac{1}{8}a \, \mathbf{\hat{z}} & \left(8a\right) & \mbox{Co I} \\ 
\mathbf{B}_{2} & = & \frac{7}{8} \, \mathbf{a}_{1} + \frac{7}{8} \, \mathbf{a}_{2} + \frac{7}{8} \, \mathbf{a}_{3} & = & \frac{7}{8}a \, \mathbf{\hat{x}} + \frac{7}{8}a \, \mathbf{\hat{y}} + \frac{7}{8}a \, \mathbf{\hat{z}} & \left(8a\right) & \mbox{Co I} \\ 
\mathbf{B}_{3} & = & \frac{1}{2} \, \mathbf{a}_{1} + \frac{1}{2} \, \mathbf{a}_{2} + \frac{1}{2} \, \mathbf{a}_{3} & = & \frac{1}{2}a \, \mathbf{\hat{x}} + \frac{1}{2}a \, \mathbf{\hat{y}} + \frac{1}{2}a \, \mathbf{\hat{z}} & \left(16d\right) & \mbox{Co II} \\ 
\mathbf{B}_{4} & = & \frac{1}{2} \, \mathbf{a}_{1} + \frac{1}{2} \, \mathbf{a}_{2} & = & \frac{1}{4}a \, \mathbf{\hat{x}} + \frac{1}{4}a \, \mathbf{\hat{y}} + \frac{1}{2}a \, \mathbf{\hat{z}} & \left(16d\right) & \mbox{Co II} \\ 
\mathbf{B}_{5} & = & \frac{1}{2} \, \mathbf{a}_{1} + \frac{1}{2} \, \mathbf{a}_{3} & = & \frac{1}{4}a \, \mathbf{\hat{x}} + \frac{1}{2}a \, \mathbf{\hat{y}} + \frac{1}{4}a \, \mathbf{\hat{z}} & \left(16d\right) & \mbox{Co II} \\ 
\mathbf{B}_{6} & = & \frac{1}{2} \, \mathbf{a}_{2} + \frac{1}{2} \, \mathbf{a}_{3} & = & \frac{1}{2}a \, \mathbf{\hat{x}} + \frac{1}{4}a \, \mathbf{\hat{y}} + \frac{1}{4}a \, \mathbf{\hat{z}} & \left(16d\right) & \mbox{Co II} \\ 
\mathbf{B}_{7} & = & x_{3} \, \mathbf{a}_{1} + x_{3} \, \mathbf{a}_{2} + x_{3} \, \mathbf{a}_{3} & = & x_{3}a \, \mathbf{\hat{x}} + x_{3}a \, \mathbf{\hat{y}} + x_{3}a \, \mathbf{\hat{z}} & \left(32e\right) & \mbox{O} \\ 
\mathbf{B}_{8} & = & x_{3} \, \mathbf{a}_{1} + x_{3} \, \mathbf{a}_{2} + \left(\frac{1}{2} - 3x_{3}\right) \, \mathbf{a}_{3} & = & \left(\frac{1}{4} - x_{3}\right)a \, \mathbf{\hat{x}} + \left(\frac{1}{4} - x_{3}\right)a \, \mathbf{\hat{y}} + x_{3}a \, \mathbf{\hat{z}} & \left(32e\right) & \mbox{O} \\ 
\mathbf{B}_{9} & = & x_{3} \, \mathbf{a}_{1} + \left(\frac{1}{2} - 3x_{3}\right) \, \mathbf{a}_{2} + x_{3} \, \mathbf{a}_{3} & = & \left(\frac{1}{4} - x_{3}\right)a \, \mathbf{\hat{x}} + x_{3}a \, \mathbf{\hat{y}} + \left(\frac{1}{4} - x_{3}\right)a \, \mathbf{\hat{z}} & \left(32e\right) & \mbox{O} \\ 
\mathbf{B}_{10} & = & \left(\frac{1}{2} - 3x_{3}\right) \, \mathbf{a}_{1} + x_{3} \, \mathbf{a}_{2} + x_{3} \, \mathbf{a}_{3} & = & x_{3}a \, \mathbf{\hat{x}} + \left(\frac{1}{4} - x_{3}\right)a \, \mathbf{\hat{y}} + \left(\frac{1}{4} - x_{3}\right)a \, \mathbf{\hat{z}} & \left(32e\right) & \mbox{O} \\ 
\mathbf{B}_{11} & = & -x_{3} \, \mathbf{a}_{1}-x_{3} \, \mathbf{a}_{2} + \left(\frac{1}{2} +3x_{3}\right) \, \mathbf{a}_{3} & = & \left(\frac{1}{4} +x_{3}\right)a \, \mathbf{\hat{x}} + \left(\frac{1}{4} +x_{3}\right)a \, \mathbf{\hat{y}}-x_{3}a \, \mathbf{\hat{z}} & \left(32e\right) & \mbox{O} \\ 
\mathbf{B}_{12} & = & -x_{3} \, \mathbf{a}_{1}-x_{3} \, \mathbf{a}_{2}-x_{3} \, \mathbf{a}_{3} & = & -x_{3}a \, \mathbf{\hat{x}}-x_{3}a \, \mathbf{\hat{y}}-x_{3}a \, \mathbf{\hat{z}} & \left(32e\right) & \mbox{O} \\ 
\mathbf{B}_{13} & = & -x_{3} \, \mathbf{a}_{1} + \left(\frac{1}{2} +3x_{3}\right) \, \mathbf{a}_{2}-x_{3} \, \mathbf{a}_{3} & = & \left(\frac{1}{4} +x_{3}\right)a \, \mathbf{\hat{x}}-x_{3}a \, \mathbf{\hat{y}} + \left(\frac{1}{4} +x_{3}\right)a \, \mathbf{\hat{z}} & \left(32e\right) & \mbox{O} \\ 
\mathbf{B}_{14} & = & \left(\frac{1}{2} +3x_{3}\right) \, \mathbf{a}_{1}-x_{3} \, \mathbf{a}_{2}-x_{3} \, \mathbf{a}_{3} & = & -x_{3}a \, \mathbf{\hat{x}} + \left(\frac{1}{4} +x_{3}\right)a \, \mathbf{\hat{y}} + \left(\frac{1}{4} +x_{3}\right)a \, \mathbf{\hat{z}} & \left(32e\right) & \mbox{O} \\ 
\end{longtabu}
\renewcommand{\arraystretch}{1.0}
\noindent \hrulefill
\\
\textbf{References:}
\vspace*{-0.25cm}
\begin{flushleft}
  - \bibentry{Knop_Can_J_Chem_46_1968}. \\
  - \bibentry{Hahn_ZAAC_279_1955}. \\
\end{flushleft}
\noindent \hrulefill
\\
\textbf{Geometry files:}
\\
\noindent  - CIF: pp. {\hyperref[A3B4_cF56_227_ad_e_cif]{\pageref{A3B4_cF56_227_ad_e_cif}}} \\
\noindent  - POSCAR: pp. {\hyperref[A3B4_cF56_227_ad_e_poscar]{\pageref{A3B4_cF56_227_ad_e_poscar}}} \\
\onecolumn
{\phantomsection\label{A5BCD6_cF416_228_eg_c_b_h}}
\subsection*{\huge \textbf{{\normalfont CuCrCl$_{5}$[NH$_{3}$]$_{6}$ Structure: A5BCD6\_cF416\_228\_eg\_c\_b\_h}}}
\noindent \hrulefill
\vspace*{0.25cm}
\begin{figure}[htp]
  \centering
  \vspace{-1em}
  {\includegraphics[width=1\textwidth]{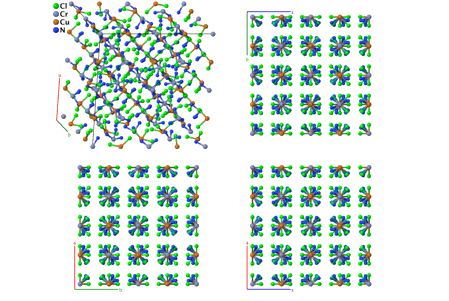}}
\end{figure}
\vspace*{-0.5cm}
\renewcommand{\arraystretch}{1.5}
\begin{equation*}
  \begin{array}{>{$\hspace{-0.15cm}}l<{$}>{$}p{0.5cm}<{$}>{$}p{18.5cm}<{$}}
    \mbox{\large \textbf{Prototype}} &\colon & \ce{CuCrCl5[NH3]6} \\
    \mbox{\large \textbf{\AFLOW\ prototype label}} &\colon & \mbox{A5BCD6\_cF416\_228\_eg\_c\_b\_h} \\
    \mbox{\large \textbf{\textit{Strukturbericht} designation}} &\colon & \mbox{None} \\
    \mbox{\large \textbf{Pearson symbol}} &\colon & \mbox{cF416} \\
    \mbox{\large \textbf{Space group number}} &\colon & 228 \\
    \mbox{\large \textbf{Space group symbol}} &\colon & Fd\bar{3}c \\
    \mbox{\large \textbf{\AFLOW\ prototype command}} &\colon &  \texttt{aflow} \,  \, \texttt{-{}-proto=A5BCD6\_cF416\_228\_eg\_c\_b\_h } \, \newline \texttt{-{}-params=}{a,x_{3},y_{4},x_{5},y_{5},z_{5} }
  \end{array}
\end{equation*}
\renewcommand{\arraystretch}{1.0}

\vspace*{-0.25cm}
\noindent \hrulefill
\begin{itemize}
  \item{The N atoms correspond to NH$_{3}$ units centered on the (192h) Wyckoff positions.
}
\end{itemize}

\noindent \parbox{1 \linewidth}{
\noindent \hrulefill
\\
\textbf{Face-centered Cubic primitive vectors:} \\
\vspace*{-0.25cm}
\begin{tabular}{cc}
  \begin{tabular}{c}
    \parbox{0.6 \linewidth}{
      \renewcommand{\arraystretch}{1.5}
      \begin{equation*}
        \centering
        \begin{array}{ccc}
              \mathbf{a}_1 & = & \frac12 \, a \, \mathbf{\hat{y}} + \frac12 \, a \, \mathbf{\hat{z}} \\
    \mathbf{a}_2 & = & \frac12 \, a \, \mathbf{\hat{x}} + \frac12 \, a \, \mathbf{\hat{z}} \\
    \mathbf{a}_3 & = & \frac12 \, a \, \mathbf{\hat{x}} + \frac12 \, a \, \mathbf{\hat{y}} \\

        \end{array}
      \end{equation*}
    }
    \renewcommand{\arraystretch}{1.0}
  \end{tabular}
  \begin{tabular}{c}
    \includegraphics[width=0.3\linewidth]{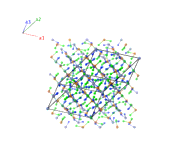} \\
  \end{tabular}
\end{tabular}

}
\vspace*{-0.25cm}

\noindent \hrulefill
\\
\textbf{Basis vectors:}
\vspace*{-0.25cm}
\renewcommand{\arraystretch}{1.5}
\begin{longtabu} to \textwidth{>{\centering $}X[-1,c,c]<{$}>{\centering $}X[-1,c,c]<{$}>{\centering $}X[-1,c,c]<{$}>{\centering $}X[-1,c,c]<{$}>{\centering $}X[-1,c,c]<{$}>{\centering $}X[-1,c,c]<{$}>{\centering $}X[-1,c,c]<{$}}
  & & \mbox{Lattice Coordinates} & & \mbox{Cartesian Coordinates} &\mbox{Wyckoff Position} & \mbox{Atom Type} \\  
  \mathbf{B}_{1} & = & \frac{1}{4} \, \mathbf{a}_{1} + \frac{1}{4} \, \mathbf{a}_{2} + \frac{1}{4} \, \mathbf{a}_{3} & = & \frac{1}{4}a \, \mathbf{\hat{x}} + \frac{1}{4}a \, \mathbf{\hat{y}} + \frac{1}{4}a \, \mathbf{\hat{z}} & \left(32b\right) & \mbox{Cu} \\ 
\mathbf{B}_{2} & = & \frac{1}{4} \, \mathbf{a}_{1} + \frac{1}{4} \, \mathbf{a}_{2} + \frac{3}{4} \, \mathbf{a}_{3} & = & \frac{1}{2}a \, \mathbf{\hat{x}} + \frac{1}{2}a \, \mathbf{\hat{y}} + \frac{1}{4}a \, \mathbf{\hat{z}} & \left(32b\right) & \mbox{Cu} \\ 
\mathbf{B}_{3} & = & \frac{1}{4} \, \mathbf{a}_{1} + \frac{3}{4} \, \mathbf{a}_{2} + \frac{1}{4} \, \mathbf{a}_{3} & = & \frac{1}{2}a \, \mathbf{\hat{x}} + \frac{1}{4}a \, \mathbf{\hat{y}} + \frac{1}{2}a \, \mathbf{\hat{z}} & \left(32b\right) & \mbox{Cu} \\ 
\mathbf{B}_{4} & = & \frac{3}{4} \, \mathbf{a}_{1} + \frac{1}{4} \, \mathbf{a}_{2} + \frac{1}{4} \, \mathbf{a}_{3} & = & \frac{1}{4}a \, \mathbf{\hat{x}} + \frac{1}{2}a \, \mathbf{\hat{y}} + \frac{1}{2}a \, \mathbf{\hat{z}} & \left(32b\right) & \mbox{Cu} \\ 
\mathbf{B}_{5} & = & \frac{3}{4} \, \mathbf{a}_{1} + \frac{3}{4} \, \mathbf{a}_{2} + \frac{3}{4} \, \mathbf{a}_{3} & = & \frac{3}{4}a \, \mathbf{\hat{x}} + \frac{3}{4}a \, \mathbf{\hat{y}} + \frac{3}{4}a \, \mathbf{\hat{z}} & \left(32b\right) & \mbox{Cu} \\ 
\mathbf{B}_{6} & = & \frac{3}{4} \, \mathbf{a}_{1} + \frac{3}{4} \, \mathbf{a}_{2} + \frac{1}{4} \, \mathbf{a}_{3} & = & \frac{1}{2}a \, \mathbf{\hat{x}} + \frac{1}{2}a \, \mathbf{\hat{y}} + \frac{3}{4}a \, \mathbf{\hat{z}} & \left(32b\right) & \mbox{Cu} \\ 
\mathbf{B}_{7} & = & \frac{3}{4} \, \mathbf{a}_{1} + \frac{1}{4} \, \mathbf{a}_{2} + \frac{3}{4} \, \mathbf{a}_{3} & = & \frac{1}{2}a \, \mathbf{\hat{x}} + \frac{3}{4}a \, \mathbf{\hat{y}} + \frac{1}{2}a \, \mathbf{\hat{z}} & \left(32b\right) & \mbox{Cu} \\ 
\mathbf{B}_{8} & = & \frac{1}{4} \, \mathbf{a}_{1} + \frac{3}{4} \, \mathbf{a}_{2} + \frac{3}{4} \, \mathbf{a}_{3} & = & \frac{3}{4}a \, \mathbf{\hat{x}} + \frac{1}{2}a \, \mathbf{\hat{y}} + \frac{1}{2}a \, \mathbf{\hat{z}} & \left(32b\right) & \mbox{Cu} \\ 
\mathbf{B}_{9} & = & 0 \, \mathbf{a}_{1} + 0 \, \mathbf{a}_{2} + 0 \, \mathbf{a}_{3} & = & 0 \, \mathbf{\hat{x}} + 0 \, \mathbf{\hat{y}} + 0 \, \mathbf{\hat{z}} & \left(32c\right) & \mbox{Cr} \\ 
\mathbf{B}_{10} & = & \frac{1}{2} \, \mathbf{a}_{3} & = & \frac{1}{4}a \, \mathbf{\hat{x}} + \frac{1}{4}a \, \mathbf{\hat{y}} & \left(32c\right) & \mbox{Cr} \\ 
\mathbf{B}_{11} & = & \frac{1}{2} \, \mathbf{a}_{2} & = & \frac{1}{4}a \, \mathbf{\hat{x}} + \frac{1}{4}a \, \mathbf{\hat{z}} & \left(32c\right) & \mbox{Cr} \\ 
\mathbf{B}_{12} & = & \frac{1}{2} \, \mathbf{a}_{1} & = & \frac{1}{4}a \, \mathbf{\hat{y}} + \frac{1}{4}a \, \mathbf{\hat{z}} & \left(32c\right) & \mbox{Cr} \\ 
\mathbf{B}_{13} & = & \frac{1}{2} \, \mathbf{a}_{1} + \frac{1}{2} \, \mathbf{a}_{2} & = & \frac{1}{4}a \, \mathbf{\hat{x}} + \frac{1}{4}a \, \mathbf{\hat{y}} + \frac{1}{2}a \, \mathbf{\hat{z}} & \left(32c\right) & \mbox{Cr} \\ 
\mathbf{B}_{14} & = & \frac{1}{2} \, \mathbf{a}_{1} + \frac{1}{2} \, \mathbf{a}_{2} + \frac{1}{2} \, \mathbf{a}_{3} & = & \frac{1}{2}a \, \mathbf{\hat{x}} + \frac{1}{2}a \, \mathbf{\hat{y}} + \frac{1}{2}a \, \mathbf{\hat{z}} & \left(32c\right) & \mbox{Cr} \\ 
\mathbf{B}_{15} & = & \frac{1}{2} \, \mathbf{a}_{1} + \frac{1}{2} \, \mathbf{a}_{3} & = & \frac{1}{4}a \, \mathbf{\hat{x}} + \frac{1}{2}a \, \mathbf{\hat{y}} + \frac{1}{4}a \, \mathbf{\hat{z}} & \left(32c\right) & \mbox{Cr} \\ 
\mathbf{B}_{16} & = & \frac{1}{2} \, \mathbf{a}_{2} + \frac{1}{2} \, \mathbf{a}_{3} & = & \frac{1}{2}a \, \mathbf{\hat{x}} + \frac{1}{4}a \, \mathbf{\hat{y}} + \frac{1}{4}a \, \mathbf{\hat{z}} & \left(32c\right) & \mbox{Cr} \\ 
\mathbf{B}_{17} & = & x_{3} \, \mathbf{a}_{1} + x_{3} \, \mathbf{a}_{2} + x_{3} \, \mathbf{a}_{3} & = & x_{3}a \, \mathbf{\hat{x}} + x_{3}a \, \mathbf{\hat{y}} + x_{3}a \, \mathbf{\hat{z}} & \left(64e\right) & \mbox{Cl I} \\ 
\mathbf{B}_{18} & = & x_{3} \, \mathbf{a}_{1} + x_{3} \, \mathbf{a}_{2} + \left(\frac{1}{2} - 3x_{3}\right) \, \mathbf{a}_{3} & = & \left(\frac{1}{4} - x_{3}\right)a \, \mathbf{\hat{x}} + \left(\frac{1}{4} - x_{3}\right)a \, \mathbf{\hat{y}} + x_{3}a \, \mathbf{\hat{z}} & \left(64e\right) & \mbox{Cl I} \\ 
\mathbf{B}_{19} & = & x_{3} \, \mathbf{a}_{1} + \left(\frac{1}{2} - 3x_{3}\right) \, \mathbf{a}_{2} + x_{3} \, \mathbf{a}_{3} & = & \left(\frac{1}{4} - x_{3}\right)a \, \mathbf{\hat{x}} + x_{3}a \, \mathbf{\hat{y}} + \left(\frac{1}{4} - x_{3}\right)a \, \mathbf{\hat{z}} & \left(64e\right) & \mbox{Cl I} \\ 
\mathbf{B}_{20} & = & \left(\frac{1}{2} - 3x_{3}\right) \, \mathbf{a}_{1} + x_{3} \, \mathbf{a}_{2} + x_{3} \, \mathbf{a}_{3} & = & x_{3}a \, \mathbf{\hat{x}} + \left(\frac{1}{4} - x_{3}\right)a \, \mathbf{\hat{y}} + \left(\frac{1}{4} - x_{3}\right)a \, \mathbf{\hat{z}} & \left(64e\right) & \mbox{Cl I} \\ 
\mathbf{B}_{21} & = & \left(\frac{1}{2} - x_{3}\right) \, \mathbf{a}_{1} + \left(\frac{1}{2} - x_{3}\right) \, \mathbf{a}_{2} + 3x_{3} \, \mathbf{a}_{3} & = & \left(\frac{1}{4} +x_{3}\right)a \, \mathbf{\hat{x}} + \left(\frac{1}{4} +x_{3}\right)a \, \mathbf{\hat{y}} + \left(\frac{1}{2} - x_{3}\right)a \, \mathbf{\hat{z}} & \left(64e\right) & \mbox{Cl I} \\ 
\mathbf{B}_{22} & = & \left(\frac{1}{2} - x_{3}\right) \, \mathbf{a}_{1} + \left(\frac{1}{2} - x_{3}\right) \, \mathbf{a}_{2} + \left(\frac{1}{2} - x_{3}\right) \, \mathbf{a}_{3} & = & \left(\frac{1}{2} - x_{3}\right)a \, \mathbf{\hat{x}} + \left(\frac{1}{2} - x_{3}\right)a \, \mathbf{\hat{y}} + \left(\frac{1}{2} - x_{3}\right)a \, \mathbf{\hat{z}} & \left(64e\right) & \mbox{Cl I} \\ 
\mathbf{B}_{23} & = & \left(\frac{1}{2} - x_{3}\right) \, \mathbf{a}_{1} + 3x_{3} \, \mathbf{a}_{2} + \left(\frac{1}{2} - x_{3}\right) \, \mathbf{a}_{3} & = & \left(\frac{1}{4} +x_{3}\right)a \, \mathbf{\hat{x}} + \left(\frac{1}{2} - x_{3}\right)a \, \mathbf{\hat{y}} + \left(\frac{1}{4} +x_{3}\right)a \, \mathbf{\hat{z}} & \left(64e\right) & \mbox{Cl I} \\ 
\mathbf{B}_{24} & = & 3x_{3} \, \mathbf{a}_{1} + \left(\frac{1}{2} - x_{3}\right) \, \mathbf{a}_{2} + \left(\frac{1}{2} - x_{3}\right) \, \mathbf{a}_{3} & = & \left(\frac{1}{2} - x_{3}\right)a \, \mathbf{\hat{x}} + \left(\frac{1}{4} +x_{3}\right)a \, \mathbf{\hat{y}} + \left(\frac{1}{4} +x_{3}\right)a \, \mathbf{\hat{z}} & \left(64e\right) & \mbox{Cl I} \\ 
\mathbf{B}_{25} & = & -x_{3} \, \mathbf{a}_{1}-x_{3} \, \mathbf{a}_{2}-x_{3} \, \mathbf{a}_{3} & = & -x_{3}a \, \mathbf{\hat{x}}-x_{3}a \, \mathbf{\hat{y}}-x_{3}a \, \mathbf{\hat{z}} & \left(64e\right) & \mbox{Cl I} \\ 
\mathbf{B}_{26} & = & -x_{3} \, \mathbf{a}_{1}-x_{3} \, \mathbf{a}_{2} + \left(\frac{1}{2} +3x_{3}\right) \, \mathbf{a}_{3} & = & \left(\frac{1}{4} +x_{3}\right)a \, \mathbf{\hat{x}} + \left(\frac{1}{4} +x_{3}\right)a \, \mathbf{\hat{y}}-x_{3}a \, \mathbf{\hat{z}} & \left(64e\right) & \mbox{Cl I} \\ 
\mathbf{B}_{27} & = & -x_{3} \, \mathbf{a}_{1} + \left(\frac{1}{2} +3x_{3}\right) \, \mathbf{a}_{2}-x_{3} \, \mathbf{a}_{3} & = & \left(\frac{1}{4} +x_{3}\right)a \, \mathbf{\hat{x}}-x_{3}a \, \mathbf{\hat{y}} + \left(\frac{1}{4} +x_{3}\right)a \, \mathbf{\hat{z}} & \left(64e\right) & \mbox{Cl I} \\ 
\mathbf{B}_{28} & = & \left(\frac{1}{2} +3x_{3}\right) \, \mathbf{a}_{1}-x_{3} \, \mathbf{a}_{2}-x_{3} \, \mathbf{a}_{3} & = & -x_{3}a \, \mathbf{\hat{x}} + \left(\frac{1}{4} +x_{3}\right)a \, \mathbf{\hat{y}} + \left(\frac{1}{4} +x_{3}\right)a \, \mathbf{\hat{z}} & \left(64e\right) & \mbox{Cl I} \\ 
\mathbf{B}_{29} & = & \left(\frac{1}{2} +x_{3}\right) \, \mathbf{a}_{1} + \left(\frac{1}{2} +x_{3}\right) \, \mathbf{a}_{2}-3x_{3} \, \mathbf{a}_{3} & = & \left(\frac{1}{4} - x_{3}\right)a \, \mathbf{\hat{x}} + \left(\frac{1}{4} - x_{3}\right)a \, \mathbf{\hat{y}} + \left(\frac{1}{2} +x_{3}\right)a \, \mathbf{\hat{z}} & \left(64e\right) & \mbox{Cl I} \\ 
\mathbf{B}_{30} & = & \left(\frac{1}{2} +x_{3}\right) \, \mathbf{a}_{1} + \left(\frac{1}{2} +x_{3}\right) \, \mathbf{a}_{2} + \left(\frac{1}{2} +x_{3}\right) \, \mathbf{a}_{3} & = & \left(\frac{1}{2} +x_{3}\right)a \, \mathbf{\hat{x}} + \left(\frac{1}{2} +x_{3}\right)a \, \mathbf{\hat{y}} + \left(\frac{1}{2} +x_{3}\right)a \, \mathbf{\hat{z}} & \left(64e\right) & \mbox{Cl I} \\ 
\mathbf{B}_{31} & = & \left(\frac{1}{2} +x_{3}\right) \, \mathbf{a}_{1}-3x_{3} \, \mathbf{a}_{2} + \left(\frac{1}{2} +x_{3}\right) \, \mathbf{a}_{3} & = & \left(\frac{1}{4} - x_{3}\right)a \, \mathbf{\hat{x}} + \left(\frac{1}{2} +x_{3}\right)a \, \mathbf{\hat{y}} + \left(\frac{1}{4} - x_{3}\right)a \, \mathbf{\hat{z}} & \left(64e\right) & \mbox{Cl I} \\ 
\mathbf{B}_{32} & = & -3x_{3} \, \mathbf{a}_{1} + \left(\frac{1}{2} +x_{3}\right) \, \mathbf{a}_{2} + \left(\frac{1}{2} +x_{3}\right) \, \mathbf{a}_{3} & = & \left(\frac{1}{2} +x_{3}\right)a \, \mathbf{\hat{x}} + \left(\frac{1}{4} - x_{3}\right)a \, \mathbf{\hat{y}} + \left(\frac{1}{4} - x_{3}\right)a \, \mathbf{\hat{z}} & \left(64e\right) & \mbox{Cl I} \\ 
\mathbf{B}_{33} & = & \left(\frac{1}{2} - 2y_{4}\right) \, \mathbf{a}_{2} + 2y_{4} \, \mathbf{a}_{3} & = & \frac{1}{4}a \, \mathbf{\hat{x}} + y_{4}a \, \mathbf{\hat{y}} + \left(\frac{1}{4} - y_{4}\right)a \, \mathbf{\hat{z}} & \left(96g\right) & \mbox{Cl II} \\ 
\mathbf{B}_{34} & = & \left(\frac{1}{4} - 2y_{4}\right) \, \mathbf{a}_{1} + \frac{3}{4} \, \mathbf{a}_{2} + \frac{1}{4} \, \mathbf{a}_{3} & = & \frac{1}{2}a \, \mathbf{\hat{x}} + \left(\frac{1}{4} - y_{4}\right)a \, \mathbf{\hat{y}} + \left(\frac{1}{2} - y_{4}\right)a \, \mathbf{\hat{z}} & \left(96g\right) & \mbox{Cl II} \\ 
\mathbf{B}_{35} & = & \left(\frac{1}{4} +2y_{4}\right) \, \mathbf{a}_{1} + \frac{1}{4} \, \mathbf{a}_{2} + \frac{3}{4} \, \mathbf{a}_{3} & = & \frac{1}{2}a \, \mathbf{\hat{x}} + \left(\frac{1}{2} +y_{4}\right)a \, \mathbf{\hat{y}} + \left(\frac{1}{4} +y_{4}\right)a \, \mathbf{\hat{z}} & \left(96g\right) & \mbox{Cl II} \\ 
\mathbf{B}_{36} & = & \frac{1}{4} \, \mathbf{a}_{1} + \left(\frac{1}{4} +2y_{4}\right) \, \mathbf{a}_{2} + \left(\frac{1}{4} - 2y_{4}\right) \, \mathbf{a}_{3} & = & \frac{1}{4}a \, \mathbf{\hat{x}} + \left(\frac{1}{4} - y_{4}\right)a \, \mathbf{\hat{y}} + \left(\frac{1}{4} +y_{4}\right)a \, \mathbf{\hat{z}} & \left(96g\right) & \mbox{Cl II} \\ 
\mathbf{B}_{37} & = & \left(\frac{1}{4} +2y_{4}\right) \, \mathbf{a}_{1} + \frac{3}{4} \, \mathbf{a}_{2} + \left(\frac{1}{4} - 2y_{4}\right) \, \mathbf{a}_{3} & = & \left(\frac{1}{2} - y_{4}\right)a \, \mathbf{\hat{x}} + \frac{1}{4}a \, \mathbf{\hat{y}} + \left(\frac{1}{2} +y_{4}\right)a \, \mathbf{\hat{z}} & \left(96g\right) & \mbox{Cl II} \\ 
\mathbf{B}_{38} & = & \frac{1}{4} \, \mathbf{a}_{1} + \left(\frac{1}{4} - 2y_{4}\right) \, \mathbf{a}_{2} + \frac{3}{4} \, \mathbf{a}_{3} & = & \left(\frac{1}{2} - y_{4}\right)a \, \mathbf{\hat{x}} + \frac{1}{2}a \, \mathbf{\hat{y}} + \left(\frac{1}{4} - y_{4}\right)a \, \mathbf{\hat{z}} & \left(96g\right) & \mbox{Cl II} \\ 
\mathbf{B}_{39} & = & \frac{3}{4} \, \mathbf{a}_{1} + \left(\frac{1}{4} +2y_{4}\right) \, \mathbf{a}_{2} + \frac{1}{4} \, \mathbf{a}_{3} & = & \left(\frac{1}{4} +y_{4}\right)a \, \mathbf{\hat{x}} + \frac{1}{2}a \, \mathbf{\hat{y}} + \left(\frac{1}{2} +y_{4}\right)a \, \mathbf{\hat{z}} & \left(96g\right) & \mbox{Cl II} \\ 
\mathbf{B}_{40} & = & \left(\frac{1}{4} - 2y_{4}\right) \, \mathbf{a}_{1} + \frac{1}{4} \, \mathbf{a}_{2} + \left(\frac{1}{4} +2y_{4}\right) \, \mathbf{a}_{3} & = & \left(\frac{1}{4} +y_{4}\right)a \, \mathbf{\hat{x}} + \frac{1}{4}a \, \mathbf{\hat{y}} + \left(\frac{1}{4} - y_{4}\right)a \, \mathbf{\hat{z}} & \left(96g\right) & \mbox{Cl II} \\ 
\mathbf{B}_{41} & = & \left(\frac{1}{4} - 2y_{4}\right) \, \mathbf{a}_{1} + \left(\frac{1}{4} +2y_{4}\right) \, \mathbf{a}_{2} + \frac{3}{4} \, \mathbf{a}_{3} & = & \left(\frac{1}{2} +y_{4}\right)a \, \mathbf{\hat{x}} + \left(\frac{1}{2} - y_{4}\right)a \, \mathbf{\hat{y}} + \frac{1}{4}a \, \mathbf{\hat{z}} & \left(96g\right) & \mbox{Cl II} \\ 
\mathbf{B}_{42} & = & \frac{3}{4} \, \mathbf{a}_{1} + \frac{1}{4} \, \mathbf{a}_{2} + \left(\frac{1}{4} - 2y_{4}\right) \, \mathbf{a}_{3} & = & \left(\frac{1}{4} - y_{4}\right)a \, \mathbf{\hat{x}} + \left(\frac{1}{2} - y_{4}\right)a \, \mathbf{\hat{y}} + \frac{1}{2}a \, \mathbf{\hat{z}} & \left(96g\right) & \mbox{Cl II} \\ 
\mathbf{B}_{43} & = & \frac{1}{4} \, \mathbf{a}_{1} + \frac{3}{4} \, \mathbf{a}_{2} + \left(\frac{1}{4} +2y_{4}\right) \, \mathbf{a}_{3} & = & \left(\frac{1}{2} +y_{4}\right)a \, \mathbf{\hat{x}} + \left(\frac{1}{4} +y_{4}\right)a \, \mathbf{\hat{y}} + \frac{1}{2}a \, \mathbf{\hat{z}} & \left(96g\right) & \mbox{Cl II} \\ 
\mathbf{B}_{44} & = & \left(\frac{1}{4} +2y_{4}\right) \, \mathbf{a}_{1} + \left(\frac{1}{4} - 2y_{4}\right) \, \mathbf{a}_{2} + \frac{1}{4} \, \mathbf{a}_{3} & = & \left(\frac{1}{4} - y_{4}\right)a \, \mathbf{\hat{x}} + \left(\frac{1}{4} +y_{4}\right)a \, \mathbf{\hat{y}} + \frac{1}{4}a \, \mathbf{\hat{z}} & \left(96g\right) & \mbox{Cl II} \\ 
\mathbf{B}_{45} & = & \frac{1}{4} \, \mathbf{a}_{1} + \left(\frac{3}{4} +2y_{4}\right) \, \mathbf{a}_{2} + \left(\frac{3}{4} - 2y_{4}\right) \, \mathbf{a}_{3} & = & \frac{3}{4}a \, \mathbf{\hat{x}} + \left(\frac{1}{2} - y_{4}\right)a \, \mathbf{\hat{y}} + \left(\frac{1}{2} +y_{4}\right)a \, \mathbf{\hat{z}} & \left(96g\right) & \mbox{Cl II} \\ 
\mathbf{B}_{46} & = & \left(\frac{3}{4} +2y_{4}\right) \, \mathbf{a}_{1} + \frac{1}{4} \, \mathbf{a}_{2} + \frac{3}{4} \, \mathbf{a}_{3} & = & \frac{1}{2}a \, \mathbf{\hat{x}} + \left(\frac{3}{4} +y_{4}\right)a \, \mathbf{\hat{y}} + \left(\frac{1}{2} +y_{4}\right)a \, \mathbf{\hat{z}} & \left(96g\right) & \mbox{Cl II} \\ 
\mathbf{B}_{47} & = & \left(\frac{3}{4} - 2y_{4}\right) \, \mathbf{a}_{1} + \frac{3}{4} \, \mathbf{a}_{2} + \frac{1}{4} \, \mathbf{a}_{3} & = & \frac{1}{2}a \, \mathbf{\hat{x}} + \left(\frac{1}{2} - y_{4}\right)a \, \mathbf{\hat{y}} + \left(\frac{3}{4} - y_{4}\right)a \, \mathbf{\hat{z}} & \left(96g\right) & \mbox{Cl II} \\ 
\mathbf{B}_{48} & = & \frac{3}{4} \, \mathbf{a}_{1} + \left(\frac{3}{4} - 2y_{4}\right) \, \mathbf{a}_{2} + \left(\frac{3}{4} +2y_{4}\right) \, \mathbf{a}_{3} & = & \frac{3}{4}a \, \mathbf{\hat{x}} + \left(\frac{3}{4} +y_{4}\right)a \, \mathbf{\hat{y}} + \left(\frac{3}{4} - y_{4}\right)a \, \mathbf{\hat{z}} & \left(96g\right) & \mbox{Cl II} \\ 
\mathbf{B}_{49} & = & \left(\frac{3}{4} - 2y_{4}\right) \, \mathbf{a}_{1} + \frac{1}{4} \, \mathbf{a}_{2} + \left(\frac{3}{4} +2y_{4}\right) \, \mathbf{a}_{3} & = & \left(\frac{1}{2} +y_{4}\right)a \, \mathbf{\hat{x}} + \frac{3}{4}a \, \mathbf{\hat{y}} + \left(\frac{1}{2} - y_{4}\right)a \, \mathbf{\hat{z}} & \left(96g\right) & \mbox{Cl II} \\ 
\mathbf{B}_{50} & = & \frac{3}{4} \, \mathbf{a}_{1} + \left(\frac{3}{4} +2y_{4}\right) \, \mathbf{a}_{2} + \frac{1}{4} \, \mathbf{a}_{3} & = & \left(\frac{1}{2} +y_{4}\right)a \, \mathbf{\hat{x}} + \frac{1}{2}a \, \mathbf{\hat{y}} + \left(\frac{3}{4} +y_{4}\right)a \, \mathbf{\hat{z}} & \left(96g\right) & \mbox{Cl II} \\ 
\mathbf{B}_{51} & = & \frac{1}{4} \, \mathbf{a}_{1} + \left(\frac{3}{4} - 2y_{4}\right) \, \mathbf{a}_{2} + \frac{3}{4} \, \mathbf{a}_{3} & = & \left(\frac{3}{4} - y_{4}\right)a \, \mathbf{\hat{x}} + \frac{1}{2}a \, \mathbf{\hat{y}} + \left(\frac{1}{2} - y_{4}\right)a \, \mathbf{\hat{z}} & \left(96g\right) & \mbox{Cl II} \\ 
\mathbf{B}_{52} & = & \left(\frac{3}{4} +2y_{4}\right) \, \mathbf{a}_{1} + \frac{3}{4} \, \mathbf{a}_{2} + \left(\frac{3}{4} - 2y_{4}\right) \, \mathbf{a}_{3} & = & \left(\frac{3}{4} - y_{4}\right)a \, \mathbf{\hat{x}} + \frac{3}{4}a \, \mathbf{\hat{y}} + \left(\frac{3}{4} +y_{4}\right)a \, \mathbf{\hat{z}} & \left(96g\right) & \mbox{Cl II} \\ 
\mathbf{B}_{53} & = & \left(\frac{3}{4} +2y_{4}\right) \, \mathbf{a}_{1} + \left(\frac{3}{4} - 2y_{4}\right) \, \mathbf{a}_{2} + \frac{1}{4} \, \mathbf{a}_{3} & = & \left(\frac{1}{2} - y_{4}\right)a \, \mathbf{\hat{x}} + \left(\frac{1}{2} +y_{4}\right)a \, \mathbf{\hat{y}} + \frac{3}{4}a \, \mathbf{\hat{z}} & \left(96g\right) & \mbox{Cl II} \\ 
\mathbf{B}_{54} & = & \frac{1}{4} \, \mathbf{a}_{1} + \frac{3}{4} \, \mathbf{a}_{2} + \left(\frac{3}{4} +2y_{4}\right) \, \mathbf{a}_{3} & = & \left(\frac{3}{4} +y_{4}\right)a \, \mathbf{\hat{x}} + \left(\frac{1}{2} +y_{4}\right)a \, \mathbf{\hat{y}} + \frac{1}{2}a \, \mathbf{\hat{z}} & \left(96g\right) & \mbox{Cl II} \\ 
\mathbf{B}_{55} & = & \frac{3}{4} \, \mathbf{a}_{1} + \frac{1}{4} \, \mathbf{a}_{2} + \left(\frac{3}{4} - 2y_{4}\right) \, \mathbf{a}_{3} & = & \left(\frac{1}{2} - y_{4}\right)a \, \mathbf{\hat{x}} + \left(\frac{3}{4} - y_{4}\right)a \, \mathbf{\hat{y}} + \frac{1}{2}a \, \mathbf{\hat{z}} & \left(96g\right) & \mbox{Cl II} \\ 
\mathbf{B}_{56} & = & \left(\frac{3}{4} - 2y_{4}\right) \, \mathbf{a}_{1} + \left(\frac{3}{4} +2y_{4}\right) \, \mathbf{a}_{2} + \frac{3}{4} \, \mathbf{a}_{3} & = & \left(\frac{3}{4} +y_{4}\right)a \, \mathbf{\hat{x}} + \left(\frac{3}{4} - y_{4}\right)a \, \mathbf{\hat{y}} + \frac{3}{4}a \, \mathbf{\hat{z}} & \left(96g\right) & \mbox{Cl II} \\ 
\mathbf{B}_{57} & = & \left(-x_{5}+y_{5}+z_{5}\right) \, \mathbf{a}_{1} + \left(x_{5}-y_{5}+z_{5}\right) \, \mathbf{a}_{2} + \left(x_{5}+y_{5}-z_{5}\right) \, \mathbf{a}_{3} & = & x_{5}a \, \mathbf{\hat{x}} + y_{5}a \, \mathbf{\hat{y}} + z_{5}a \, \mathbf{\hat{z}} & \left(192h\right) & \mbox{N} \\ 
\mathbf{B}_{58} & = & \left(x_{5}-y_{5}+z_{5}\right) \, \mathbf{a}_{1} + \left(-x_{5}+y_{5}+z_{5}\right) \, \mathbf{a}_{2} + \left(\frac{1}{2} - x_{5} - y_{5} - z_{5}\right) \, \mathbf{a}_{3} & = & \left(\frac{1}{4} - x_{5}\right)a \, \mathbf{\hat{x}} + \left(\frac{1}{4} - y_{5}\right)a \, \mathbf{\hat{y}} + z_{5}a \, \mathbf{\hat{z}} & \left(192h\right) & \mbox{N} \\ 
\mathbf{B}_{59} & = & \left(x_{5}+y_{5}-z_{5}\right) \, \mathbf{a}_{1} + \left(\frac{1}{2} - x_{5} - y_{5} - z_{5}\right) \, \mathbf{a}_{2} + \left(-x_{5}+y_{5}+z_{5}\right) \, \mathbf{a}_{3} & = & \left(\frac{1}{4} - x_{5}\right)a \, \mathbf{\hat{x}} + y_{5}a \, \mathbf{\hat{y}} + \left(\frac{1}{4} - z_{5}\right)a \, \mathbf{\hat{z}} & \left(192h\right) & \mbox{N} \\ 
\mathbf{B}_{60} & = & \left(\frac{1}{2} - x_{5} - y_{5} - z_{5}\right) \, \mathbf{a}_{1} + \left(x_{5}+y_{5}-z_{5}\right) \, \mathbf{a}_{2} + \left(x_{5}-y_{5}+z_{5}\right) \, \mathbf{a}_{3} & = & x_{5}a \, \mathbf{\hat{x}} + \left(\frac{1}{4} - y_{5}\right)a \, \mathbf{\hat{y}} + \left(\frac{1}{4} - z_{5}\right)a \, \mathbf{\hat{z}} & \left(192h\right) & \mbox{N} \\ 
\mathbf{B}_{61} & = & \left(x_{5}+y_{5}-z_{5}\right) \, \mathbf{a}_{1} + \left(-x_{5}+y_{5}+z_{5}\right) \, \mathbf{a}_{2} + \left(x_{5}-y_{5}+z_{5}\right) \, \mathbf{a}_{3} & = & z_{5}a \, \mathbf{\hat{x}} + x_{5}a \, \mathbf{\hat{y}} + y_{5}a \, \mathbf{\hat{z}} & \left(192h\right) & \mbox{N} \\ 
\mathbf{B}_{62} & = & \left(\frac{1}{2} - x_{5} - y_{5} - z_{5}\right) \, \mathbf{a}_{1} + \left(x_{5}-y_{5}+z_{5}\right) \, \mathbf{a}_{2} + \left(-x_{5}+y_{5}+z_{5}\right) \, \mathbf{a}_{3} & = & z_{5}a \, \mathbf{\hat{x}} + \left(\frac{1}{4} - x_{5}\right)a \, \mathbf{\hat{y}} + \left(\frac{1}{4} - y_{5}\right)a \, \mathbf{\hat{z}} & \left(192h\right) & \mbox{N} \\ 
\mathbf{B}_{63} & = & \left(-x_{5}+y_{5}+z_{5}\right) \, \mathbf{a}_{1} + \left(x_{5}+y_{5}-z_{5}\right) \, \mathbf{a}_{2} + \left(\frac{1}{2} - x_{5} - y_{5} - z_{5}\right) \, \mathbf{a}_{3} & = & \left(\frac{1}{4} - z_{5}\right)a \, \mathbf{\hat{x}} + \left(\frac{1}{4} - x_{5}\right)a \, \mathbf{\hat{y}} + y_{5}a \, \mathbf{\hat{z}} & \left(192h\right) & \mbox{N} \\ 
\mathbf{B}_{64} & = & \left(x_{5}-y_{5}+z_{5}\right) \, \mathbf{a}_{1} + \left(\frac{1}{2} - x_{5} - y_{5} - z_{5}\right) \, \mathbf{a}_{2} + \left(x_{5}+y_{5}-z_{5}\right) \, \mathbf{a}_{3} & = & \left(\frac{1}{4} - z_{5}\right)a \, \mathbf{\hat{x}} + x_{5}a \, \mathbf{\hat{y}} + \left(\frac{1}{4} - y_{5}\right)a \, \mathbf{\hat{z}} & \left(192h\right) & \mbox{N} \\ 
\mathbf{B}_{65} & = & \left(x_{5}-y_{5}+z_{5}\right) \, \mathbf{a}_{1} + \left(x_{5}+y_{5}-z_{5}\right) \, \mathbf{a}_{2} + \left(-x_{5}+y_{5}+z_{5}\right) \, \mathbf{a}_{3} & = & y_{5}a \, \mathbf{\hat{x}} + z_{5}a \, \mathbf{\hat{y}} + x_{5}a \, \mathbf{\hat{z}} & \left(192h\right) & \mbox{N} \\ 
\mathbf{B}_{66} & = & \left(-x_{5}+y_{5}+z_{5}\right) \, \mathbf{a}_{1} + \left(\frac{1}{2} - x_{5} - y_{5} - z_{5}\right) \, \mathbf{a}_{2} + \left(x_{5}-y_{5}+z_{5}\right) \, \mathbf{a}_{3} & = & \left(\frac{1}{4} - y_{5}\right)a \, \mathbf{\hat{x}} + z_{5}a \, \mathbf{\hat{y}} + \left(\frac{1}{4} - x_{5}\right)a \, \mathbf{\hat{z}} & \left(192h\right) & \mbox{N} \\ 
\mathbf{B}_{67} & = & \left(\frac{1}{2} - x_{5} - y_{5} - z_{5}\right) \, \mathbf{a}_{1} + \left(-x_{5}+y_{5}+z_{5}\right) \, \mathbf{a}_{2} + \left(x_{5}+y_{5}-z_{5}\right) \, \mathbf{a}_{3} & = & y_{5}a \, \mathbf{\hat{x}} + \left(\frac{1}{4} - z_{5}\right)a \, \mathbf{\hat{y}} + \left(\frac{1}{4} - x_{5}\right)a \, \mathbf{\hat{z}} & \left(192h\right) & \mbox{N} \\ 
\mathbf{B}_{68} & = & \left(x_{5}+y_{5}-z_{5}\right) \, \mathbf{a}_{1} + \left(x_{5}-y_{5}+z_{5}\right) \, \mathbf{a}_{2} + \left(\frac{1}{2} - x_{5} - y_{5} - z_{5}\right) \, \mathbf{a}_{3} & = & \left(\frac{1}{4} - y_{5}\right)a \, \mathbf{\hat{x}} + \left(\frac{1}{4} - z_{5}\right)a \, \mathbf{\hat{y}} + x_{5}a \, \mathbf{\hat{z}} & \left(192h\right) & \mbox{N} \\ 
\mathbf{B}_{69} & = & \left(\frac{1}{2} +x_{5} - y_{5} - z_{5}\right) \, \mathbf{a}_{1} + \left(\frac{1}{2} - x_{5} + y_{5} - z_{5}\right) \, \mathbf{a}_{2} + \left(x_{5}+y_{5}+z_{5}\right) \, \mathbf{a}_{3} & = & \left(\frac{1}{4} +y_{5}\right)a \, \mathbf{\hat{x}} + \left(\frac{1}{4} +x_{5}\right)a \, \mathbf{\hat{y}} + \left(\frac{1}{2} - z_{5}\right)a \, \mathbf{\hat{z}} & \left(192h\right) & \mbox{N} \\ 
\mathbf{B}_{70} & = & \left(\frac{1}{2} - x_{5} + y_{5} - z_{5}\right) \, \mathbf{a}_{1} + \left(\frac{1}{2} +x_{5} - y_{5} - z_{5}\right) \, \mathbf{a}_{2} + \left(\frac{1}{2} - x_{5} - y_{5} + z_{5}\right) \, \mathbf{a}_{3} & = & \left(\frac{1}{2} - y_{5}\right)a \, \mathbf{\hat{x}} + \left(\frac{1}{2} - x_{5}\right)a \, \mathbf{\hat{y}} + \left(\frac{1}{2} - z_{5}\right)a \, \mathbf{\hat{z}} & \left(192h\right) & \mbox{N} \\ 
\mathbf{B}_{71} & = & \left(\frac{1}{2} - x_{5} - y_{5} + z_{5}\right) \, \mathbf{a}_{1} + \left(x_{5}+y_{5}+z_{5}\right) \, \mathbf{a}_{2} + \left(\frac{1}{2} - x_{5} + y_{5} - z_{5}\right) \, \mathbf{a}_{3} & = & \left(\frac{1}{4} +y_{5}\right)a \, \mathbf{\hat{x}} + \left(\frac{1}{2} - x_{5}\right)a \, \mathbf{\hat{y}} + \left(\frac{1}{4} +z_{5}\right)a \, \mathbf{\hat{z}} & \left(192h\right) & \mbox{N} \\ 
\mathbf{B}_{72} & = & \left(x_{5}+y_{5}+z_{5}\right) \, \mathbf{a}_{1} + \left(\frac{1}{2} - x_{5} - y_{5} + z_{5}\right) \, \mathbf{a}_{2} + \left(\frac{1}{2} +x_{5} - y_{5} - z_{5}\right) \, \mathbf{a}_{3} & = & \left(\frac{1}{2} - y_{5}\right)a \, \mathbf{\hat{x}} + \left(\frac{1}{4} +x_{5}\right)a \, \mathbf{\hat{y}} + \left(\frac{1}{4} +z_{5}\right)a \, \mathbf{\hat{z}} & \left(192h\right) & \mbox{N} \\ 
\mathbf{B}_{73} & = & \left(\frac{1}{2} - x_{5} - y_{5} + z_{5}\right) \, \mathbf{a}_{1} + \left(\frac{1}{2} +x_{5} - y_{5} - z_{5}\right) \, \mathbf{a}_{2} + \left(x_{5}+y_{5}+z_{5}\right) \, \mathbf{a}_{3} & = & \left(\frac{1}{4} +x_{5}\right)a \, \mathbf{\hat{x}} + \left(\frac{1}{4} +z_{5}\right)a \, \mathbf{\hat{y}} + \left(\frac{1}{2} - y_{5}\right)a \, \mathbf{\hat{z}} & \left(192h\right) & \mbox{N} \\ 
\mathbf{B}_{74} & = & \left(x_{5}+y_{5}+z_{5}\right) \, \mathbf{a}_{1} + \left(\frac{1}{2} - x_{5} + y_{5} - z_{5}\right) \, \mathbf{a}_{2} + \left(\frac{1}{2} - x_{5} - y_{5} + z_{5}\right) \, \mathbf{a}_{3} & = & \left(\frac{1}{2} - x_{5}\right)a \, \mathbf{\hat{x}} + \left(\frac{1}{4} +z_{5}\right)a \, \mathbf{\hat{y}} + \left(\frac{1}{4} +y_{5}\right)a \, \mathbf{\hat{z}} & \left(192h\right) & \mbox{N} \\ 
\mathbf{B}_{75} & = & \left(\frac{1}{2} +x_{5} - y_{5} - z_{5}\right) \, \mathbf{a}_{1} + \left(\frac{1}{2} - x_{5} - y_{5} + z_{5}\right) \, \mathbf{a}_{2} + \left(\frac{1}{2} - x_{5} + y_{5} - z_{5}\right) \, \mathbf{a}_{3} & = & \left(\frac{1}{2} - x_{5}\right)a \, \mathbf{\hat{x}} + \left(\frac{1}{2} - z_{5}\right)a \, \mathbf{\hat{y}} + \left(\frac{1}{2} - y_{5}\right)a \, \mathbf{\hat{z}} & \left(192h\right) & \mbox{N} \\ 
\mathbf{B}_{76} & = & \left(\frac{1}{2} - x_{5} + y_{5} - z_{5}\right) \, \mathbf{a}_{1} + \left(x_{5}+y_{5}+z_{5}\right) \, \mathbf{a}_{2} + \left(\frac{1}{2} +x_{5} - y_{5} - z_{5}\right) \, \mathbf{a}_{3} & = & \left(\frac{1}{4} +x_{5}\right)a \, \mathbf{\hat{x}} + \left(\frac{1}{2} - z_{5}\right)a \, \mathbf{\hat{y}} + \left(\frac{1}{4} +y_{5}\right)a \, \mathbf{\hat{z}} & \left(192h\right) & \mbox{N} \\ 
\mathbf{B}_{77} & = & \left(\frac{1}{2} - x_{5} + y_{5} - z_{5}\right) \, \mathbf{a}_{1} + \left(\frac{1}{2} - x_{5} - y_{5} + z_{5}\right) \, \mathbf{a}_{2} + \left(x_{5}+y_{5}+z_{5}\right) \, \mathbf{a}_{3} & = & \left(\frac{1}{4} +z_{5}\right)a \, \mathbf{\hat{x}} + \left(\frac{1}{4} +y_{5}\right)a \, \mathbf{\hat{y}} + \left(\frac{1}{2} - x_{5}\right)a \, \mathbf{\hat{z}} & \left(192h\right) & \mbox{N} \\ 
\mathbf{B}_{78} & = & \left(\frac{1}{2} +x_{5} - y_{5} - z_{5}\right) \, \mathbf{a}_{1} + \left(x_{5}+y_{5}+z_{5}\right) \, \mathbf{a}_{2} + \left(\frac{1}{2} - x_{5} - y_{5} + z_{5}\right) \, \mathbf{a}_{3} & = & \left(\frac{1}{4} +z_{5}\right)a \, \mathbf{\hat{x}} + \left(\frac{1}{2} - y_{5}\right)a \, \mathbf{\hat{y}} + \left(\frac{1}{4} +x_{5}\right)a \, \mathbf{\hat{z}} & \left(192h\right) & \mbox{N} \\ 
\mathbf{B}_{79} & = & \left(x_{5}+y_{5}+z_{5}\right) \, \mathbf{a}_{1} + \left(\frac{1}{2} +x_{5} - y_{5} - z_{5}\right) \, \mathbf{a}_{2} + \left(\frac{1}{2} - x_{5} + y_{5} - z_{5}\right) \, \mathbf{a}_{3} & = & \left(\frac{1}{2} - z_{5}\right)a \, \mathbf{\hat{x}} + \left(\frac{1}{4} +y_{5}\right)a \, \mathbf{\hat{y}} + \left(\frac{1}{4} +x_{5}\right)a \, \mathbf{\hat{z}} & \left(192h\right) & \mbox{N} \\ 
\mathbf{B}_{80} & = & \left(\frac{1}{2} - x_{5} - y_{5} + z_{5}\right) \, \mathbf{a}_{1} + \left(\frac{1}{2} - x_{5} + y_{5} - z_{5}\right) \, \mathbf{a}_{2} + \left(\frac{1}{2} +x_{5} - y_{5} - z_{5}\right) \, \mathbf{a}_{3} & = & \left(\frac{1}{2} - z_{5}\right)a \, \mathbf{\hat{x}} + \left(\frac{1}{2} - y_{5}\right)a \, \mathbf{\hat{y}} + \left(\frac{1}{2} - x_{5}\right)a \, \mathbf{\hat{z}} & \left(192h\right) & \mbox{N} \\ 
\mathbf{B}_{81} & = & \left(x_{5}-y_{5}-z_{5}\right) \, \mathbf{a}_{1} + \left(-x_{5}+y_{5}-z_{5}\right) \, \mathbf{a}_{2} + \left(-x_{5}-y_{5}+z_{5}\right) \, \mathbf{a}_{3} & = & -x_{5}a \, \mathbf{\hat{x}}-y_{5}a \, \mathbf{\hat{y}}-z_{5}a \, \mathbf{\hat{z}} & \left(192h\right) & \mbox{N} \\ 
\mathbf{B}_{82} & = & \left(-x_{5}+y_{5}-z_{5}\right) \, \mathbf{a}_{1} + \left(x_{5}-y_{5}-z_{5}\right) \, \mathbf{a}_{2} + \left(\frac{1}{2} +x_{5} + y_{5} + z_{5}\right) \, \mathbf{a}_{3} & = & \left(\frac{1}{4} +x_{5}\right)a \, \mathbf{\hat{x}} + \left(\frac{1}{4} +y_{5}\right)a \, \mathbf{\hat{y}}-z_{5}a \, \mathbf{\hat{z}} & \left(192h\right) & \mbox{N} \\ 
\mathbf{B}_{83} & = & \left(-x_{5}-y_{5}+z_{5}\right) \, \mathbf{a}_{1} + \left(\frac{1}{2} +x_{5} + y_{5} + z_{5}\right) \, \mathbf{a}_{2} + \left(x_{5}-y_{5}-z_{5}\right) \, \mathbf{a}_{3} & = & \left(\frac{1}{4} +x_{5}\right)a \, \mathbf{\hat{x}}-y_{5}a \, \mathbf{\hat{y}} + \left(\frac{1}{4} +z_{5}\right)a \, \mathbf{\hat{z}} & \left(192h\right) & \mbox{N} \\ 
\mathbf{B}_{84} & = & \left(\frac{1}{2} +x_{5} + y_{5} + z_{5}\right) \, \mathbf{a}_{1} + \left(-x_{5}-y_{5}+z_{5}\right) \, \mathbf{a}_{2} + \left(-x_{5}+y_{5}-z_{5}\right) \, \mathbf{a}_{3} & = & -x_{5}a \, \mathbf{\hat{x}} + \left(\frac{1}{4} +y_{5}\right)a \, \mathbf{\hat{y}} + \left(\frac{1}{4} +z_{5}\right)a \, \mathbf{\hat{z}} & \left(192h\right) & \mbox{N} \\ 
\mathbf{B}_{85} & = & \left(-x_{5}-y_{5}+z_{5}\right) \, \mathbf{a}_{1} + \left(x_{5}-y_{5}-z_{5}\right) \, \mathbf{a}_{2} + \left(-x_{5}+y_{5}-z_{5}\right) \, \mathbf{a}_{3} & = & -z_{5}a \, \mathbf{\hat{x}}-x_{5}a \, \mathbf{\hat{y}}-y_{5}a \, \mathbf{\hat{z}} & \left(192h\right) & \mbox{N} \\ 
\mathbf{B}_{86} & = & \left(\frac{1}{2} +x_{5} + y_{5} + z_{5}\right) \, \mathbf{a}_{1} + \left(-x_{5}+y_{5}-z_{5}\right) \, \mathbf{a}_{2} + \left(x_{5}-y_{5}-z_{5}\right) \, \mathbf{a}_{3} & = & -z_{5}a \, \mathbf{\hat{x}} + \left(\frac{1}{4} +x_{5}\right)a \, \mathbf{\hat{y}} + \left(\frac{1}{4} +y_{5}\right)a \, \mathbf{\hat{z}} & \left(192h\right) & \mbox{N} \\ 
\mathbf{B}_{87} & = & \left(x_{5}-y_{5}-z_{5}\right) \, \mathbf{a}_{1} + \left(-x_{5}-y_{5}+z_{5}\right) \, \mathbf{a}_{2} + \left(\frac{1}{2} +x_{5} + y_{5} + z_{5}\right) \, \mathbf{a}_{3} & = & \left(\frac{1}{4} +z_{5}\right)a \, \mathbf{\hat{x}} + \left(\frac{1}{4} +x_{5}\right)a \, \mathbf{\hat{y}}-y_{5}a \, \mathbf{\hat{z}} & \left(192h\right) & \mbox{N} \\ 
\mathbf{B}_{88} & = & \left(-x_{5}+y_{5}-z_{5}\right) \, \mathbf{a}_{1} + \left(\frac{1}{2} +x_{5} + y_{5} + z_{5}\right) \, \mathbf{a}_{2} + \left(-x_{5}-y_{5}+z_{5}\right) \, \mathbf{a}_{3} & = & \left(\frac{1}{4} +z_{5}\right)a \, \mathbf{\hat{x}}-x_{5}a \, \mathbf{\hat{y}} + \left(\frac{1}{4} +y_{5}\right)a \, \mathbf{\hat{z}} & \left(192h\right) & \mbox{N} \\ 
\mathbf{B}_{89} & = & \left(-x_{5}+y_{5}-z_{5}\right) \, \mathbf{a}_{1} + \left(-x_{5}-y_{5}+z_{5}\right) \, \mathbf{a}_{2} + \left(x_{5}-y_{5}-z_{5}\right) \, \mathbf{a}_{3} & = & -y_{5}a \, \mathbf{\hat{x}}-z_{5}a \, \mathbf{\hat{y}}-x_{5}a \, \mathbf{\hat{z}} & \left(192h\right) & \mbox{N} \\ 
\mathbf{B}_{90} & = & \left(x_{5}-y_{5}-z_{5}\right) \, \mathbf{a}_{1} + \left(\frac{1}{2} +x_{5} + y_{5} + z_{5}\right) \, \mathbf{a}_{2} + \left(-x_{5}+y_{5}-z_{5}\right) \, \mathbf{a}_{3} & = & \left(\frac{1}{4} +y_{5}\right)a \, \mathbf{\hat{x}}-z_{5}a \, \mathbf{\hat{y}} + \left(\frac{1}{4} +x_{5}\right)a \, \mathbf{\hat{z}} & \left(192h\right) & \mbox{N} \\ 
\mathbf{B}_{91} & = & \left(\frac{1}{2} +x_{5} + y_{5} + z_{5}\right) \, \mathbf{a}_{1} + \left(x_{5}-y_{5}-z_{5}\right) \, \mathbf{a}_{2} + \left(-x_{5}-y_{5}+z_{5}\right) \, \mathbf{a}_{3} & = & -y_{5}a \, \mathbf{\hat{x}} + \left(\frac{1}{4} +z_{5}\right)a \, \mathbf{\hat{y}} + \left(\frac{1}{4} +x_{5}\right)a \, \mathbf{\hat{z}} & \left(192h\right) & \mbox{N} \\ 
\mathbf{B}_{92} & = & \left(-x_{5}-y_{5}+z_{5}\right) \, \mathbf{a}_{1} + \left(-x_{5}+y_{5}-z_{5}\right) \, \mathbf{a}_{2} + \left(\frac{1}{2} +x_{5} + y_{5} + z_{5}\right) \, \mathbf{a}_{3} & = & \left(\frac{1}{4} +y_{5}\right)a \, \mathbf{\hat{x}} + \left(\frac{1}{4} +z_{5}\right)a \, \mathbf{\hat{y}}-x_{5}a \, \mathbf{\hat{z}} & \left(192h\right) & \mbox{N} \\ 
\mathbf{B}_{93} & = & \left(\frac{1}{2} - x_{5} + y_{5} + z_{5}\right) \, \mathbf{a}_{1} + \left(\frac{1}{2} +x_{5} - y_{5} + z_{5}\right) \, \mathbf{a}_{2} + \left(-x_{5}-y_{5}-z_{5}\right) \, \mathbf{a}_{3} & = & \left(\frac{1}{4} - y_{5}\right)a \, \mathbf{\hat{x}} + \left(\frac{1}{4} - x_{5}\right)a \, \mathbf{\hat{y}} + \left(\frac{1}{2} +z_{5}\right)a \, \mathbf{\hat{z}} & \left(192h\right) & \mbox{N} \\ 
\mathbf{B}_{94} & = & \left(\frac{1}{2} +x_{5} - y_{5} + z_{5}\right) \, \mathbf{a}_{1} + \left(\frac{1}{2} - x_{5} + y_{5} + z_{5}\right) \, \mathbf{a}_{2} + \left(\frac{1}{2} +x_{5} + y_{5} - z_{5}\right) \, \mathbf{a}_{3} & = & \left(\frac{1}{2} +y_{5}\right)a \, \mathbf{\hat{x}} + \left(\frac{1}{2} +x_{5}\right)a \, \mathbf{\hat{y}} + \left(\frac{1}{2} +z_{5}\right)a \, \mathbf{\hat{z}} & \left(192h\right) & \mbox{N} \\ 
\mathbf{B}_{95} & = & \left(\frac{1}{2} +x_{5} + y_{5} - z_{5}\right) \, \mathbf{a}_{1} + \left(-x_{5}-y_{5}-z_{5}\right) \, \mathbf{a}_{2} + \left(\frac{1}{2} +x_{5} - y_{5} + z_{5}\right) \, \mathbf{a}_{3} & = & \left(\frac{1}{4} - y_{5}\right)a \, \mathbf{\hat{x}} + \left(\frac{1}{2} +x_{5}\right)a \, \mathbf{\hat{y}} + \left(\frac{1}{4} - z_{5}\right)a \, \mathbf{\hat{z}} & \left(192h\right) & \mbox{N} \\ 
\mathbf{B}_{96} & = & \left(-x_{5}-y_{5}-z_{5}\right) \, \mathbf{a}_{1} + \left(\frac{1}{2} +x_{5} + y_{5} - z_{5}\right) \, \mathbf{a}_{2} + \left(\frac{1}{2} - x_{5} + y_{5} + z_{5}\right) \, \mathbf{a}_{3} & = & \left(\frac{1}{2} +y_{5}\right)a \, \mathbf{\hat{x}} + \left(\frac{1}{4} - x_{5}\right)a \, \mathbf{\hat{y}} + \left(\frac{1}{4} - z_{5}\right)a \, \mathbf{\hat{z}} & \left(192h\right) & \mbox{N} \\ 
\mathbf{B}_{97} & = & \left(\frac{1}{2} +x_{5} + y_{5} - z_{5}\right) \, \mathbf{a}_{1} + \left(\frac{1}{2} - x_{5} + y_{5} + z_{5}\right) \, \mathbf{a}_{2} + \left(-x_{5}-y_{5}-z_{5}\right) \, \mathbf{a}_{3} & = & \left(\frac{1}{4} - x_{5}\right)a \, \mathbf{\hat{x}} + \left(\frac{1}{4} - z_{5}\right)a \, \mathbf{\hat{y}} + \left(\frac{1}{2} +y_{5}\right)a \, \mathbf{\hat{z}} & \left(192h\right) & \mbox{N} \\ 
\mathbf{B}_{98} & = & \left(-x_{5}-y_{5}-z_{5}\right) \, \mathbf{a}_{1} + \left(\frac{1}{2} +x_{5} - y_{5} + z_{5}\right) \, \mathbf{a}_{2} + \left(\frac{1}{2} +x_{5} + y_{5} - z_{5}\right) \, \mathbf{a}_{3} & = & \left(\frac{1}{2} +x_{5}\right)a \, \mathbf{\hat{x}} + \left(\frac{1}{4} - z_{5}\right)a \, \mathbf{\hat{y}} + \left(\frac{1}{4} - y_{5}\right)a \, \mathbf{\hat{z}} & \left(192h\right) & \mbox{N} \\ 
\mathbf{B}_{99} & = & \left(\frac{1}{2} - x_{5} + y_{5} + z_{5}\right) \, \mathbf{a}_{1} + \left(\frac{1}{2} +x_{5} + y_{5} - z_{5}\right) \, \mathbf{a}_{2} + \left(\frac{1}{2} +x_{5} - y_{5} + z_{5}\right) \, \mathbf{a}_{3} & = & \left(\frac{1}{2} +x_{5}\right)a \, \mathbf{\hat{x}} + \left(\frac{1}{2} +z_{5}\right)a \, \mathbf{\hat{y}} + \left(\frac{1}{2} +y_{5}\right)a \, \mathbf{\hat{z}} & \left(192h\right) & \mbox{N} \\ 
\mathbf{B}_{100} & = & \left(\frac{1}{2} +x_{5} - y_{5} + z_{5}\right) \, \mathbf{a}_{1} + \left(-x_{5}-y_{5}-z_{5}\right) \, \mathbf{a}_{2} + \left(\frac{1}{2} - x_{5} + y_{5} + z_{5}\right) \, \mathbf{a}_{3} & = & \left(\frac{1}{4} - x_{5}\right)a \, \mathbf{\hat{x}} + \left(\frac{1}{2} +z_{5}\right)a \, \mathbf{\hat{y}} + \left(\frac{1}{4} - y_{5}\right)a \, \mathbf{\hat{z}} & \left(192h\right) & \mbox{N} \\ 
\mathbf{B}_{101} & = & \left(\frac{1}{2} +x_{5} - y_{5} + z_{5}\right) \, \mathbf{a}_{1} + \left(\frac{1}{2} +x_{5} + y_{5} - z_{5}\right) \, \mathbf{a}_{2} + \left(-x_{5}-y_{5}-z_{5}\right) \, \mathbf{a}_{3} & = & \left(\frac{1}{4} - z_{5}\right)a \, \mathbf{\hat{x}} + \left(\frac{1}{4} - y_{5}\right)a \, \mathbf{\hat{y}} + \left(\frac{1}{2} +x_{5}\right)a \, \mathbf{\hat{z}} & \left(192h\right) & \mbox{N} \\ 
\mathbf{B}_{102} & = & \left(\frac{1}{2} - x_{5} + y_{5} + z_{5}\right) \, \mathbf{a}_{1} + \left(-x_{5}-y_{5}-z_{5}\right) \, \mathbf{a}_{2} + \left(\frac{1}{2} +x_{5} + y_{5} - z_{5}\right) \, \mathbf{a}_{3} & = & \left(\frac{1}{4} - z_{5}\right)a \, \mathbf{\hat{x}} + \left(\frac{1}{2} +y_{5}\right)a \, \mathbf{\hat{y}} + \left(\frac{1}{4} - x_{5}\right)a \, \mathbf{\hat{z}} & \left(192h\right) & \mbox{N} \\ 
\mathbf{B}_{103} & = & \left(-x_{5}-y_{5}-z_{5}\right) \, \mathbf{a}_{1} + \left(\frac{1}{2} - x_{5} + y_{5} + z_{5}\right) \, \mathbf{a}_{2} + \left(\frac{1}{2} +x_{5} - y_{5} + z_{5}\right) \, \mathbf{a}_{3} & = & \left(\frac{1}{2} +z_{5}\right)a \, \mathbf{\hat{x}} + \left(\frac{1}{4} - y_{5}\right)a \, \mathbf{\hat{y}} + \left(\frac{1}{4} - x_{5}\right)a \, \mathbf{\hat{z}} & \left(192h\right) & \mbox{N} \\ 
\mathbf{B}_{104} & = & \left(\frac{1}{2} +x_{5} + y_{5} - z_{5}\right) \, \mathbf{a}_{1} + \left(\frac{1}{2} +x_{5} - y_{5} + z_{5}\right) \, \mathbf{a}_{2} + \left(\frac{1}{2} - x_{5} + y_{5} + z_{5}\right) \, \mathbf{a}_{3} & = & \left(\frac{1}{2} +z_{5}\right)a \, \mathbf{\hat{x}} + \left(\frac{1}{2} +y_{5}\right)a \, \mathbf{\hat{y}} + \left(\frac{1}{2} +x_{5}\right)a \, \mathbf{\hat{z}} & \left(192h\right) & \mbox{N} \\ 
\end{longtabu}
\renewcommand{\arraystretch}{1.0}
\noindent \hrulefill
\\
\textbf{References:}
\vspace*{-0.25cm}
\begin{flushleft}
  - \bibentry{Masayasu_Cr6-NH3-6CuCl5_bcsj_34_1961}. \\
\end{flushleft}
\textbf{Found in:}
\vspace*{-0.25cm}
\begin{flushleft}
  - \bibentry{Villars_PearsonsCrystalData_2013}. \\
\end{flushleft}
\noindent \hrulefill
\\
\textbf{Geometry files:}
\\
\noindent  - CIF: pp. {\hyperref[A5BCD6_cF416_228_eg_c_b_h_cif]{\pageref{A5BCD6_cF416_228_eg_c_b_h_cif}}} \\
\noindent  - POSCAR: pp. {\hyperref[A5BCD6_cF416_228_eg_c_b_h_poscar]{\pageref{A5BCD6_cF416_228_eg_c_b_h_poscar}}} \\
\onecolumn
{\phantomsection\label{A6B_cF224_228_h_c}}
\subsection*{\huge \textbf{{\normalfont TeO$_{6}$H$_{6}$ Structure: A6B\_cF224\_228\_h\_c}}}
\noindent \hrulefill
\vspace*{0.25cm}
\begin{figure}[htp]
  \centering
  \vspace{-1em}
  {\includegraphics[width=1\textwidth]{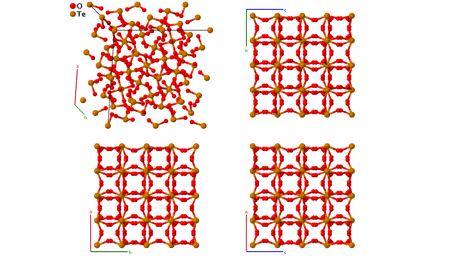}}
\end{figure}
\vspace*{-0.5cm}
\renewcommand{\arraystretch}{1.5}
\begin{equation*}
  \begin{array}{>{$\hspace{-0.15cm}}l<{$}>{$}p{0.5cm}<{$}>{$}p{18.5cm}<{$}}
    \mbox{\large \textbf{Prototype}} &\colon & \ce{TeO6H6} \\
    \mbox{\large \textbf{\AFLOW\ prototype label}} &\colon & \mbox{A6B\_cF224\_228\_h\_c} \\
    \mbox{\large \textbf{\textit{Strukturbericht} designation}} &\colon & \mbox{None} \\
    \mbox{\large \textbf{Pearson symbol}} &\colon & \mbox{cF224} \\
    \mbox{\large \textbf{Space group number}} &\colon & 228 \\
    \mbox{\large \textbf{Space group symbol}} &\colon & Fd\bar{3}c \\
    \mbox{\large \textbf{\AFLOW\ prototype command}} &\colon &  \texttt{aflow} \,  \, \texttt{-{}-proto=A6B\_cF224\_228\_h\_c } \, \newline \texttt{-{}-params=}{a,x_{2},y_{2},z_{2} }
  \end{array}
\end{equation*}
\renewcommand{\arraystretch}{1.0}

\vspace*{-0.25cm}
\noindent \hrulefill
\begin{itemize}
  \item{Polytypes appear in space groups \#14, \#210 and \#225. 
Only the non-hydrogen atoms are listed.
}
\end{itemize}

\noindent \parbox{1 \linewidth}{
\noindent \hrulefill
\\
\textbf{Face-centered Cubic primitive vectors:} \\
\vspace*{-0.25cm}
\begin{tabular}{cc}
  \begin{tabular}{c}
    \parbox{0.6 \linewidth}{
      \renewcommand{\arraystretch}{1.5}
      \begin{equation*}
        \centering
        \begin{array}{ccc}
              \mathbf{a}_1 & = & \frac12 \, a \, \mathbf{\hat{y}} + \frac12 \, a \, \mathbf{\hat{z}} \\
    \mathbf{a}_2 & = & \frac12 \, a \, \mathbf{\hat{x}} + \frac12 \, a \, \mathbf{\hat{z}} \\
    \mathbf{a}_3 & = & \frac12 \, a \, \mathbf{\hat{x}} + \frac12 \, a \, \mathbf{\hat{y}} \\

        \end{array}
      \end{equation*}
    }
    \renewcommand{\arraystretch}{1.0}
  \end{tabular}
  \begin{tabular}{c}
    \includegraphics[width=0.3\linewidth]{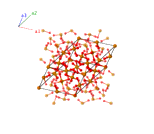} \\
  \end{tabular}
\end{tabular}

}
\vspace*{-0.25cm}

\noindent \hrulefill
\\
\textbf{Basis vectors:}
\vspace*{-0.25cm}
\renewcommand{\arraystretch}{1.5}
\begin{longtabu} to \textwidth{>{\centering $}X[-1,c,c]<{$}>{\centering $}X[-1,c,c]<{$}>{\centering $}X[-1,c,c]<{$}>{\centering $}X[-1,c,c]<{$}>{\centering $}X[-1,c,c]<{$}>{\centering $}X[-1,c,c]<{$}>{\centering $}X[-1,c,c]<{$}}
  & & \mbox{Lattice Coordinates} & & \mbox{Cartesian Coordinates} &\mbox{Wyckoff Position} & \mbox{Atom Type} \\  
  \mathbf{B}_{1} & = & 0 \, \mathbf{a}_{1} + 0 \, \mathbf{a}_{2} + 0 \, \mathbf{a}_{3} & = & 0 \, \mathbf{\hat{x}} + 0 \, \mathbf{\hat{y}} + 0 \, \mathbf{\hat{z}} & \left(32c\right) & \mbox{Te} \\ 
\mathbf{B}_{2} & = & \frac{1}{2} \, \mathbf{a}_{3} & = & \frac{1}{4}a \, \mathbf{\hat{x}} + \frac{1}{4}a \, \mathbf{\hat{y}} & \left(32c\right) & \mbox{Te} \\ 
\mathbf{B}_{3} & = & \frac{1}{2} \, \mathbf{a}_{2} & = & \frac{1}{4}a \, \mathbf{\hat{x}} + \frac{1}{4}a \, \mathbf{\hat{z}} & \left(32c\right) & \mbox{Te} \\ 
\mathbf{B}_{4} & = & \frac{1}{2} \, \mathbf{a}_{1} & = & \frac{1}{4}a \, \mathbf{\hat{y}} + \frac{1}{4}a \, \mathbf{\hat{z}} & \left(32c\right) & \mbox{Te} \\ 
\mathbf{B}_{5} & = & \frac{1}{2} \, \mathbf{a}_{1} + \frac{1}{2} \, \mathbf{a}_{2} & = & \frac{1}{4}a \, \mathbf{\hat{x}} + \frac{1}{4}a \, \mathbf{\hat{y}} + \frac{1}{2}a \, \mathbf{\hat{z}} & \left(32c\right) & \mbox{Te} \\ 
\mathbf{B}_{6} & = & \frac{1}{2} \, \mathbf{a}_{1} + \frac{1}{2} \, \mathbf{a}_{2} + \frac{1}{2} \, \mathbf{a}_{3} & = & \frac{1}{2}a \, \mathbf{\hat{x}} + \frac{1}{2}a \, \mathbf{\hat{y}} + \frac{1}{2}a \, \mathbf{\hat{z}} & \left(32c\right) & \mbox{Te} \\ 
\mathbf{B}_{7} & = & \frac{1}{2} \, \mathbf{a}_{1} + \frac{1}{2} \, \mathbf{a}_{3} & = & \frac{1}{4}a \, \mathbf{\hat{x}} + \frac{1}{2}a \, \mathbf{\hat{y}} + \frac{1}{4}a \, \mathbf{\hat{z}} & \left(32c\right) & \mbox{Te} \\ 
\mathbf{B}_{8} & = & \frac{1}{2} \, \mathbf{a}_{2} + \frac{1}{2} \, \mathbf{a}_{3} & = & \frac{1}{2}a \, \mathbf{\hat{x}} + \frac{1}{4}a \, \mathbf{\hat{y}} + \frac{1}{4}a \, \mathbf{\hat{z}} & \left(32c\right) & \mbox{Te} \\ 
\mathbf{B}_{9} & = & \left(-x_{2}+y_{2}+z_{2}\right) \, \mathbf{a}_{1} + \left(x_{2}-y_{2}+z_{2}\right) \, \mathbf{a}_{2} + \left(x_{2}+y_{2}-z_{2}\right) \, \mathbf{a}_{3} & = & x_{2}a \, \mathbf{\hat{x}} + y_{2}a \, \mathbf{\hat{y}} + z_{2}a \, \mathbf{\hat{z}} & \left(192h\right) & \mbox{O} \\ 
\mathbf{B}_{10} & = & \left(x_{2}-y_{2}+z_{2}\right) \, \mathbf{a}_{1} + \left(-x_{2}+y_{2}+z_{2}\right) \, \mathbf{a}_{2} + \left(\frac{1}{2} - x_{2} - y_{2} - z_{2}\right) \, \mathbf{a}_{3} & = & \left(\frac{1}{4} - x_{2}\right)a \, \mathbf{\hat{x}} + \left(\frac{1}{4} - y_{2}\right)a \, \mathbf{\hat{y}} + z_{2}a \, \mathbf{\hat{z}} & \left(192h\right) & \mbox{O} \\ 
\mathbf{B}_{11} & = & \left(x_{2}+y_{2}-z_{2}\right) \, \mathbf{a}_{1} + \left(\frac{1}{2} - x_{2} - y_{2} - z_{2}\right) \, \mathbf{a}_{2} + \left(-x_{2}+y_{2}+z_{2}\right) \, \mathbf{a}_{3} & = & \left(\frac{1}{4} - x_{2}\right)a \, \mathbf{\hat{x}} + y_{2}a \, \mathbf{\hat{y}} + \left(\frac{1}{4} - z_{2}\right)a \, \mathbf{\hat{z}} & \left(192h\right) & \mbox{O} \\ 
\mathbf{B}_{12} & = & \left(\frac{1}{2} - x_{2} - y_{2} - z_{2}\right) \, \mathbf{a}_{1} + \left(x_{2}+y_{2}-z_{2}\right) \, \mathbf{a}_{2} + \left(x_{2}-y_{2}+z_{2}\right) \, \mathbf{a}_{3} & = & x_{2}a \, \mathbf{\hat{x}} + \left(\frac{1}{4} - y_{2}\right)a \, \mathbf{\hat{y}} + \left(\frac{1}{4} - z_{2}\right)a \, \mathbf{\hat{z}} & \left(192h\right) & \mbox{O} \\ 
\mathbf{B}_{13} & = & \left(x_{2}+y_{2}-z_{2}\right) \, \mathbf{a}_{1} + \left(-x_{2}+y_{2}+z_{2}\right) \, \mathbf{a}_{2} + \left(x_{2}-y_{2}+z_{2}\right) \, \mathbf{a}_{3} & = & z_{2}a \, \mathbf{\hat{x}} + x_{2}a \, \mathbf{\hat{y}} + y_{2}a \, \mathbf{\hat{z}} & \left(192h\right) & \mbox{O} \\ 
\mathbf{B}_{14} & = & \left(\frac{1}{2} - x_{2} - y_{2} - z_{2}\right) \, \mathbf{a}_{1} + \left(x_{2}-y_{2}+z_{2}\right) \, \mathbf{a}_{2} + \left(-x_{2}+y_{2}+z_{2}\right) \, \mathbf{a}_{3} & = & z_{2}a \, \mathbf{\hat{x}} + \left(\frac{1}{4} - x_{2}\right)a \, \mathbf{\hat{y}} + \left(\frac{1}{4} - y_{2}\right)a \, \mathbf{\hat{z}} & \left(192h\right) & \mbox{O} \\ 
\mathbf{B}_{15} & = & \left(-x_{2}+y_{2}+z_{2}\right) \, \mathbf{a}_{1} + \left(x_{2}+y_{2}-z_{2}\right) \, \mathbf{a}_{2} + \left(\frac{1}{2} - x_{2} - y_{2} - z_{2}\right) \, \mathbf{a}_{3} & = & \left(\frac{1}{4} - z_{2}\right)a \, \mathbf{\hat{x}} + \left(\frac{1}{4} - x_{2}\right)a \, \mathbf{\hat{y}} + y_{2}a \, \mathbf{\hat{z}} & \left(192h\right) & \mbox{O} \\ 
\mathbf{B}_{16} & = & \left(x_{2}-y_{2}+z_{2}\right) \, \mathbf{a}_{1} + \left(\frac{1}{2} - x_{2} - y_{2} - z_{2}\right) \, \mathbf{a}_{2} + \left(x_{2}+y_{2}-z_{2}\right) \, \mathbf{a}_{3} & = & \left(\frac{1}{4} - z_{2}\right)a \, \mathbf{\hat{x}} + x_{2}a \, \mathbf{\hat{y}} + \left(\frac{1}{4} - y_{2}\right)a \, \mathbf{\hat{z}} & \left(192h\right) & \mbox{O} \\ 
\mathbf{B}_{17} & = & \left(x_{2}-y_{2}+z_{2}\right) \, \mathbf{a}_{1} + \left(x_{2}+y_{2}-z_{2}\right) \, \mathbf{a}_{2} + \left(-x_{2}+y_{2}+z_{2}\right) \, \mathbf{a}_{3} & = & y_{2}a \, \mathbf{\hat{x}} + z_{2}a \, \mathbf{\hat{y}} + x_{2}a \, \mathbf{\hat{z}} & \left(192h\right) & \mbox{O} \\ 
\mathbf{B}_{18} & = & \left(-x_{2}+y_{2}+z_{2}\right) \, \mathbf{a}_{1} + \left(\frac{1}{2} - x_{2} - y_{2} - z_{2}\right) \, \mathbf{a}_{2} + \left(x_{2}-y_{2}+z_{2}\right) \, \mathbf{a}_{3} & = & \left(\frac{1}{4} - y_{2}\right)a \, \mathbf{\hat{x}} + z_{2}a \, \mathbf{\hat{y}} + \left(\frac{1}{4} - x_{2}\right)a \, \mathbf{\hat{z}} & \left(192h\right) & \mbox{O} \\ 
\mathbf{B}_{19} & = & \left(\frac{1}{2} - x_{2} - y_{2} - z_{2}\right) \, \mathbf{a}_{1} + \left(-x_{2}+y_{2}+z_{2}\right) \, \mathbf{a}_{2} + \left(x_{2}+y_{2}-z_{2}\right) \, \mathbf{a}_{3} & = & y_{2}a \, \mathbf{\hat{x}} + \left(\frac{1}{4} - z_{2}\right)a \, \mathbf{\hat{y}} + \left(\frac{1}{4} - x_{2}\right)a \, \mathbf{\hat{z}} & \left(192h\right) & \mbox{O} \\ 
\mathbf{B}_{20} & = & \left(x_{2}+y_{2}-z_{2}\right) \, \mathbf{a}_{1} + \left(x_{2}-y_{2}+z_{2}\right) \, \mathbf{a}_{2} + \left(\frac{1}{2} - x_{2} - y_{2} - z_{2}\right) \, \mathbf{a}_{3} & = & \left(\frac{1}{4} - y_{2}\right)a \, \mathbf{\hat{x}} + \left(\frac{1}{4} - z_{2}\right)a \, \mathbf{\hat{y}} + x_{2}a \, \mathbf{\hat{z}} & \left(192h\right) & \mbox{O} \\ 
\mathbf{B}_{21} & = & \left(\frac{1}{2} +x_{2} - y_{2} - z_{2}\right) \, \mathbf{a}_{1} + \left(\frac{1}{2} - x_{2} + y_{2} - z_{2}\right) \, \mathbf{a}_{2} + \left(x_{2}+y_{2}+z_{2}\right) \, \mathbf{a}_{3} & = & \left(\frac{1}{4} +y_{2}\right)a \, \mathbf{\hat{x}} + \left(\frac{1}{4} +x_{2}\right)a \, \mathbf{\hat{y}} + \left(\frac{1}{2} - z_{2}\right)a \, \mathbf{\hat{z}} & \left(192h\right) & \mbox{O} \\ 
\mathbf{B}_{22} & = & \left(\frac{1}{2} - x_{2} + y_{2} - z_{2}\right) \, \mathbf{a}_{1} + \left(\frac{1}{2} +x_{2} - y_{2} - z_{2}\right) \, \mathbf{a}_{2} + \left(\frac{1}{2} - x_{2} - y_{2} + z_{2}\right) \, \mathbf{a}_{3} & = & \left(\frac{1}{2} - y_{2}\right)a \, \mathbf{\hat{x}} + \left(\frac{1}{2} - x_{2}\right)a \, \mathbf{\hat{y}} + \left(\frac{1}{2} - z_{2}\right)a \, \mathbf{\hat{z}} & \left(192h\right) & \mbox{O} \\ 
\mathbf{B}_{23} & = & \left(\frac{1}{2} - x_{2} - y_{2} + z_{2}\right) \, \mathbf{a}_{1} + \left(x_{2}+y_{2}+z_{2}\right) \, \mathbf{a}_{2} + \left(\frac{1}{2} - x_{2} + y_{2} - z_{2}\right) \, \mathbf{a}_{3} & = & \left(\frac{1}{4} +y_{2}\right)a \, \mathbf{\hat{x}} + \left(\frac{1}{2} - x_{2}\right)a \, \mathbf{\hat{y}} + \left(\frac{1}{4} +z_{2}\right)a \, \mathbf{\hat{z}} & \left(192h\right) & \mbox{O} \\ 
\mathbf{B}_{24} & = & \left(x_{2}+y_{2}+z_{2}\right) \, \mathbf{a}_{1} + \left(\frac{1}{2} - x_{2} - y_{2} + z_{2}\right) \, \mathbf{a}_{2} + \left(\frac{1}{2} +x_{2} - y_{2} - z_{2}\right) \, \mathbf{a}_{3} & = & \left(\frac{1}{2} - y_{2}\right)a \, \mathbf{\hat{x}} + \left(\frac{1}{4} +x_{2}\right)a \, \mathbf{\hat{y}} + \left(\frac{1}{4} +z_{2}\right)a \, \mathbf{\hat{z}} & \left(192h\right) & \mbox{O} \\ 
\mathbf{B}_{25} & = & \left(\frac{1}{2} - x_{2} - y_{2} + z_{2}\right) \, \mathbf{a}_{1} + \left(\frac{1}{2} +x_{2} - y_{2} - z_{2}\right) \, \mathbf{a}_{2} + \left(x_{2}+y_{2}+z_{2}\right) \, \mathbf{a}_{3} & = & \left(\frac{1}{4} +x_{2}\right)a \, \mathbf{\hat{x}} + \left(\frac{1}{4} +z_{2}\right)a \, \mathbf{\hat{y}} + \left(\frac{1}{2} - y_{2}\right)a \, \mathbf{\hat{z}} & \left(192h\right) & \mbox{O} \\ 
\mathbf{B}_{26} & = & \left(x_{2}+y_{2}+z_{2}\right) \, \mathbf{a}_{1} + \left(\frac{1}{2} - x_{2} + y_{2} - z_{2}\right) \, \mathbf{a}_{2} + \left(\frac{1}{2} - x_{2} - y_{2} + z_{2}\right) \, \mathbf{a}_{3} & = & \left(\frac{1}{2} - x_{2}\right)a \, \mathbf{\hat{x}} + \left(\frac{1}{4} +z_{2}\right)a \, \mathbf{\hat{y}} + \left(\frac{1}{4} +y_{2}\right)a \, \mathbf{\hat{z}} & \left(192h\right) & \mbox{O} \\ 
\mathbf{B}_{27} & = & \left(\frac{1}{2} +x_{2} - y_{2} - z_{2}\right) \, \mathbf{a}_{1} + \left(\frac{1}{2} - x_{2} - y_{2} + z_{2}\right) \, \mathbf{a}_{2} + \left(\frac{1}{2} - x_{2} + y_{2} - z_{2}\right) \, \mathbf{a}_{3} & = & \left(\frac{1}{2} - x_{2}\right)a \, \mathbf{\hat{x}} + \left(\frac{1}{2} - z_{2}\right)a \, \mathbf{\hat{y}} + \left(\frac{1}{2} - y_{2}\right)a \, \mathbf{\hat{z}} & \left(192h\right) & \mbox{O} \\ 
\mathbf{B}_{28} & = & \left(\frac{1}{2} - x_{2} + y_{2} - z_{2}\right) \, \mathbf{a}_{1} + \left(x_{2}+y_{2}+z_{2}\right) \, \mathbf{a}_{2} + \left(\frac{1}{2} +x_{2} - y_{2} - z_{2}\right) \, \mathbf{a}_{3} & = & \left(\frac{1}{4} +x_{2}\right)a \, \mathbf{\hat{x}} + \left(\frac{1}{2} - z_{2}\right)a \, \mathbf{\hat{y}} + \left(\frac{1}{4} +y_{2}\right)a \, \mathbf{\hat{z}} & \left(192h\right) & \mbox{O} \\ 
\mathbf{B}_{29} & = & \left(\frac{1}{2} - x_{2} + y_{2} - z_{2}\right) \, \mathbf{a}_{1} + \left(\frac{1}{2} - x_{2} - y_{2} + z_{2}\right) \, \mathbf{a}_{2} + \left(x_{2}+y_{2}+z_{2}\right) \, \mathbf{a}_{3} & = & \left(\frac{1}{4} +z_{2}\right)a \, \mathbf{\hat{x}} + \left(\frac{1}{4} +y_{2}\right)a \, \mathbf{\hat{y}} + \left(\frac{1}{2} - x_{2}\right)a \, \mathbf{\hat{z}} & \left(192h\right) & \mbox{O} \\ 
\mathbf{B}_{30} & = & \left(\frac{1}{2} +x_{2} - y_{2} - z_{2}\right) \, \mathbf{a}_{1} + \left(x_{2}+y_{2}+z_{2}\right) \, \mathbf{a}_{2} + \left(\frac{1}{2} - x_{2} - y_{2} + z_{2}\right) \, \mathbf{a}_{3} & = & \left(\frac{1}{4} +z_{2}\right)a \, \mathbf{\hat{x}} + \left(\frac{1}{2} - y_{2}\right)a \, \mathbf{\hat{y}} + \left(\frac{1}{4} +x_{2}\right)a \, \mathbf{\hat{z}} & \left(192h\right) & \mbox{O} \\ 
\mathbf{B}_{31} & = & \left(x_{2}+y_{2}+z_{2}\right) \, \mathbf{a}_{1} + \left(\frac{1}{2} +x_{2} - y_{2} - z_{2}\right) \, \mathbf{a}_{2} + \left(\frac{1}{2} - x_{2} + y_{2} - z_{2}\right) \, \mathbf{a}_{3} & = & \left(\frac{1}{2} - z_{2}\right)a \, \mathbf{\hat{x}} + \left(\frac{1}{4} +y_{2}\right)a \, \mathbf{\hat{y}} + \left(\frac{1}{4} +x_{2}\right)a \, \mathbf{\hat{z}} & \left(192h\right) & \mbox{O} \\ 
\mathbf{B}_{32} & = & \left(\frac{1}{2} - x_{2} - y_{2} + z_{2}\right) \, \mathbf{a}_{1} + \left(\frac{1}{2} - x_{2} + y_{2} - z_{2}\right) \, \mathbf{a}_{2} + \left(\frac{1}{2} +x_{2} - y_{2} - z_{2}\right) \, \mathbf{a}_{3} & = & \left(\frac{1}{2} - z_{2}\right)a \, \mathbf{\hat{x}} + \left(\frac{1}{2} - y_{2}\right)a \, \mathbf{\hat{y}} + \left(\frac{1}{2} - x_{2}\right)a \, \mathbf{\hat{z}} & \left(192h\right) & \mbox{O} \\ 
\mathbf{B}_{33} & = & \left(x_{2}-y_{2}-z_{2}\right) \, \mathbf{a}_{1} + \left(-x_{2}+y_{2}-z_{2}\right) \, \mathbf{a}_{2} + \left(-x_{2}-y_{2}+z_{2}\right) \, \mathbf{a}_{3} & = & -x_{2}a \, \mathbf{\hat{x}}-y_{2}a \, \mathbf{\hat{y}}-z_{2}a \, \mathbf{\hat{z}} & \left(192h\right) & \mbox{O} \\ 
\mathbf{B}_{34} & = & \left(-x_{2}+y_{2}-z_{2}\right) \, \mathbf{a}_{1} + \left(x_{2}-y_{2}-z_{2}\right) \, \mathbf{a}_{2} + \left(\frac{1}{2} +x_{2} + y_{2} + z_{2}\right) \, \mathbf{a}_{3} & = & \left(\frac{1}{4} +x_{2}\right)a \, \mathbf{\hat{x}} + \left(\frac{1}{4} +y_{2}\right)a \, \mathbf{\hat{y}}-z_{2}a \, \mathbf{\hat{z}} & \left(192h\right) & \mbox{O} \\ 
\mathbf{B}_{35} & = & \left(-x_{2}-y_{2}+z_{2}\right) \, \mathbf{a}_{1} + \left(\frac{1}{2} +x_{2} + y_{2} + z_{2}\right) \, \mathbf{a}_{2} + \left(x_{2}-y_{2}-z_{2}\right) \, \mathbf{a}_{3} & = & \left(\frac{1}{4} +x_{2}\right)a \, \mathbf{\hat{x}}-y_{2}a \, \mathbf{\hat{y}} + \left(\frac{1}{4} +z_{2}\right)a \, \mathbf{\hat{z}} & \left(192h\right) & \mbox{O} \\ 
\mathbf{B}_{36} & = & \left(\frac{1}{2} +x_{2} + y_{2} + z_{2}\right) \, \mathbf{a}_{1} + \left(-x_{2}-y_{2}+z_{2}\right) \, \mathbf{a}_{2} + \left(-x_{2}+y_{2}-z_{2}\right) \, \mathbf{a}_{3} & = & -x_{2}a \, \mathbf{\hat{x}} + \left(\frac{1}{4} +y_{2}\right)a \, \mathbf{\hat{y}} + \left(\frac{1}{4} +z_{2}\right)a \, \mathbf{\hat{z}} & \left(192h\right) & \mbox{O} \\ 
\mathbf{B}_{37} & = & \left(-x_{2}-y_{2}+z_{2}\right) \, \mathbf{a}_{1} + \left(x_{2}-y_{2}-z_{2}\right) \, \mathbf{a}_{2} + \left(-x_{2}+y_{2}-z_{2}\right) \, \mathbf{a}_{3} & = & -z_{2}a \, \mathbf{\hat{x}}-x_{2}a \, \mathbf{\hat{y}}-y_{2}a \, \mathbf{\hat{z}} & \left(192h\right) & \mbox{O} \\ 
\mathbf{B}_{38} & = & \left(\frac{1}{2} +x_{2} + y_{2} + z_{2}\right) \, \mathbf{a}_{1} + \left(-x_{2}+y_{2}-z_{2}\right) \, \mathbf{a}_{2} + \left(x_{2}-y_{2}-z_{2}\right) \, \mathbf{a}_{3} & = & -z_{2}a \, \mathbf{\hat{x}} + \left(\frac{1}{4} +x_{2}\right)a \, \mathbf{\hat{y}} + \left(\frac{1}{4} +y_{2}\right)a \, \mathbf{\hat{z}} & \left(192h\right) & \mbox{O} \\ 
\mathbf{B}_{39} & = & \left(x_{2}-y_{2}-z_{2}\right) \, \mathbf{a}_{1} + \left(-x_{2}-y_{2}+z_{2}\right) \, \mathbf{a}_{2} + \left(\frac{1}{2} +x_{2} + y_{2} + z_{2}\right) \, \mathbf{a}_{3} & = & \left(\frac{1}{4} +z_{2}\right)a \, \mathbf{\hat{x}} + \left(\frac{1}{4} +x_{2}\right)a \, \mathbf{\hat{y}}-y_{2}a \, \mathbf{\hat{z}} & \left(192h\right) & \mbox{O} \\ 
\mathbf{B}_{40} & = & \left(-x_{2}+y_{2}-z_{2}\right) \, \mathbf{a}_{1} + \left(\frac{1}{2} +x_{2} + y_{2} + z_{2}\right) \, \mathbf{a}_{2} + \left(-x_{2}-y_{2}+z_{2}\right) \, \mathbf{a}_{3} & = & \left(\frac{1}{4} +z_{2}\right)a \, \mathbf{\hat{x}}-x_{2}a \, \mathbf{\hat{y}} + \left(\frac{1}{4} +y_{2}\right)a \, \mathbf{\hat{z}} & \left(192h\right) & \mbox{O} \\ 
\mathbf{B}_{41} & = & \left(-x_{2}+y_{2}-z_{2}\right) \, \mathbf{a}_{1} + \left(-x_{2}-y_{2}+z_{2}\right) \, \mathbf{a}_{2} + \left(x_{2}-y_{2}-z_{2}\right) \, \mathbf{a}_{3} & = & -y_{2}a \, \mathbf{\hat{x}}-z_{2}a \, \mathbf{\hat{y}}-x_{2}a \, \mathbf{\hat{z}} & \left(192h\right) & \mbox{O} \\ 
\mathbf{B}_{42} & = & \left(x_{2}-y_{2}-z_{2}\right) \, \mathbf{a}_{1} + \left(\frac{1}{2} +x_{2} + y_{2} + z_{2}\right) \, \mathbf{a}_{2} + \left(-x_{2}+y_{2}-z_{2}\right) \, \mathbf{a}_{3} & = & \left(\frac{1}{4} +y_{2}\right)a \, \mathbf{\hat{x}}-z_{2}a \, \mathbf{\hat{y}} + \left(\frac{1}{4} +x_{2}\right)a \, \mathbf{\hat{z}} & \left(192h\right) & \mbox{O} \\ 
\mathbf{B}_{43} & = & \left(\frac{1}{2} +x_{2} + y_{2} + z_{2}\right) \, \mathbf{a}_{1} + \left(x_{2}-y_{2}-z_{2}\right) \, \mathbf{a}_{2} + \left(-x_{2}-y_{2}+z_{2}\right) \, \mathbf{a}_{3} & = & -y_{2}a \, \mathbf{\hat{x}} + \left(\frac{1}{4} +z_{2}\right)a \, \mathbf{\hat{y}} + \left(\frac{1}{4} +x_{2}\right)a \, \mathbf{\hat{z}} & \left(192h\right) & \mbox{O} \\ 
\mathbf{B}_{44} & = & \left(-x_{2}-y_{2}+z_{2}\right) \, \mathbf{a}_{1} + \left(-x_{2}+y_{2}-z_{2}\right) \, \mathbf{a}_{2} + \left(\frac{1}{2} +x_{2} + y_{2} + z_{2}\right) \, \mathbf{a}_{3} & = & \left(\frac{1}{4} +y_{2}\right)a \, \mathbf{\hat{x}} + \left(\frac{1}{4} +z_{2}\right)a \, \mathbf{\hat{y}}-x_{2}a \, \mathbf{\hat{z}} & \left(192h\right) & \mbox{O} \\ 
\mathbf{B}_{45} & = & \left(\frac{1}{2} - x_{2} + y_{2} + z_{2}\right) \, \mathbf{a}_{1} + \left(\frac{1}{2} +x_{2} - y_{2} + z_{2}\right) \, \mathbf{a}_{2} + \left(-x_{2}-y_{2}-z_{2}\right) \, \mathbf{a}_{3} & = & \left(\frac{1}{4} - y_{2}\right)a \, \mathbf{\hat{x}} + \left(\frac{1}{4} - x_{2}\right)a \, \mathbf{\hat{y}} + \left(\frac{1}{2} +z_{2}\right)a \, \mathbf{\hat{z}} & \left(192h\right) & \mbox{O} \\ 
\mathbf{B}_{46} & = & \left(\frac{1}{2} +x_{2} - y_{2} + z_{2}\right) \, \mathbf{a}_{1} + \left(\frac{1}{2} - x_{2} + y_{2} + z_{2}\right) \, \mathbf{a}_{2} + \left(\frac{1}{2} +x_{2} + y_{2} - z_{2}\right) \, \mathbf{a}_{3} & = & \left(\frac{1}{2} +y_{2}\right)a \, \mathbf{\hat{x}} + \left(\frac{1}{2} +x_{2}\right)a \, \mathbf{\hat{y}} + \left(\frac{1}{2} +z_{2}\right)a \, \mathbf{\hat{z}} & \left(192h\right) & \mbox{O} \\ 
\mathbf{B}_{47} & = & \left(\frac{1}{2} +x_{2} + y_{2} - z_{2}\right) \, \mathbf{a}_{1} + \left(-x_{2}-y_{2}-z_{2}\right) \, \mathbf{a}_{2} + \left(\frac{1}{2} +x_{2} - y_{2} + z_{2}\right) \, \mathbf{a}_{3} & = & \left(\frac{1}{4} - y_{2}\right)a \, \mathbf{\hat{x}} + \left(\frac{1}{2} +x_{2}\right)a \, \mathbf{\hat{y}} + \left(\frac{1}{4} - z_{2}\right)a \, \mathbf{\hat{z}} & \left(192h\right) & \mbox{O} \\ 
\mathbf{B}_{48} & = & \left(-x_{2}-y_{2}-z_{2}\right) \, \mathbf{a}_{1} + \left(\frac{1}{2} +x_{2} + y_{2} - z_{2}\right) \, \mathbf{a}_{2} + \left(\frac{1}{2} - x_{2} + y_{2} + z_{2}\right) \, \mathbf{a}_{3} & = & \left(\frac{1}{2} +y_{2}\right)a \, \mathbf{\hat{x}} + \left(\frac{1}{4} - x_{2}\right)a \, \mathbf{\hat{y}} + \left(\frac{1}{4} - z_{2}\right)a \, \mathbf{\hat{z}} & \left(192h\right) & \mbox{O} \\ 
\mathbf{B}_{49} & = & \left(\frac{1}{2} +x_{2} + y_{2} - z_{2}\right) \, \mathbf{a}_{1} + \left(\frac{1}{2} - x_{2} + y_{2} + z_{2}\right) \, \mathbf{a}_{2} + \left(-x_{2}-y_{2}-z_{2}\right) \, \mathbf{a}_{3} & = & \left(\frac{1}{4} - x_{2}\right)a \, \mathbf{\hat{x}} + \left(\frac{1}{4} - z_{2}\right)a \, \mathbf{\hat{y}} + \left(\frac{1}{2} +y_{2}\right)a \, \mathbf{\hat{z}} & \left(192h\right) & \mbox{O} \\ 
\mathbf{B}_{50} & = & \left(-x_{2}-y_{2}-z_{2}\right) \, \mathbf{a}_{1} + \left(\frac{1}{2} +x_{2} - y_{2} + z_{2}\right) \, \mathbf{a}_{2} + \left(\frac{1}{2} +x_{2} + y_{2} - z_{2}\right) \, \mathbf{a}_{3} & = & \left(\frac{1}{2} +x_{2}\right)a \, \mathbf{\hat{x}} + \left(\frac{1}{4} - z_{2}\right)a \, \mathbf{\hat{y}} + \left(\frac{1}{4} - y_{2}\right)a \, \mathbf{\hat{z}} & \left(192h\right) & \mbox{O} \\ 
\mathbf{B}_{51} & = & \left(\frac{1}{2} - x_{2} + y_{2} + z_{2}\right) \, \mathbf{a}_{1} + \left(\frac{1}{2} +x_{2} + y_{2} - z_{2}\right) \, \mathbf{a}_{2} + \left(\frac{1}{2} +x_{2} - y_{2} + z_{2}\right) \, \mathbf{a}_{3} & = & \left(\frac{1}{2} +x_{2}\right)a \, \mathbf{\hat{x}} + \left(\frac{1}{2} +z_{2}\right)a \, \mathbf{\hat{y}} + \left(\frac{1}{2} +y_{2}\right)a \, \mathbf{\hat{z}} & \left(192h\right) & \mbox{O} \\ 
\mathbf{B}_{52} & = & \left(\frac{1}{2} +x_{2} - y_{2} + z_{2}\right) \, \mathbf{a}_{1} + \left(-x_{2}-y_{2}-z_{2}\right) \, \mathbf{a}_{2} + \left(\frac{1}{2} - x_{2} + y_{2} + z_{2}\right) \, \mathbf{a}_{3} & = & \left(\frac{1}{4} - x_{2}\right)a \, \mathbf{\hat{x}} + \left(\frac{1}{2} +z_{2}\right)a \, \mathbf{\hat{y}} + \left(\frac{1}{4} - y_{2}\right)a \, \mathbf{\hat{z}} & \left(192h\right) & \mbox{O} \\ 
\mathbf{B}_{53} & = & \left(\frac{1}{2} +x_{2} - y_{2} + z_{2}\right) \, \mathbf{a}_{1} + \left(\frac{1}{2} +x_{2} + y_{2} - z_{2}\right) \, \mathbf{a}_{2} + \left(-x_{2}-y_{2}-z_{2}\right) \, \mathbf{a}_{3} & = & \left(\frac{1}{4} - z_{2}\right)a \, \mathbf{\hat{x}} + \left(\frac{1}{4} - y_{2}\right)a \, \mathbf{\hat{y}} + \left(\frac{1}{2} +x_{2}\right)a \, \mathbf{\hat{z}} & \left(192h\right) & \mbox{O} \\ 
\mathbf{B}_{54} & = & \left(\frac{1}{2} - x_{2} + y_{2} + z_{2}\right) \, \mathbf{a}_{1} + \left(-x_{2}-y_{2}-z_{2}\right) \, \mathbf{a}_{2} + \left(\frac{1}{2} +x_{2} + y_{2} - z_{2}\right) \, \mathbf{a}_{3} & = & \left(\frac{1}{4} - z_{2}\right)a \, \mathbf{\hat{x}} + \left(\frac{1}{2} +y_{2}\right)a \, \mathbf{\hat{y}} + \left(\frac{1}{4} - x_{2}\right)a \, \mathbf{\hat{z}} & \left(192h\right) & \mbox{O} \\ 
\mathbf{B}_{55} & = & \left(-x_{2}-y_{2}-z_{2}\right) \, \mathbf{a}_{1} + \left(\frac{1}{2} - x_{2} + y_{2} + z_{2}\right) \, \mathbf{a}_{2} + \left(\frac{1}{2} +x_{2} - y_{2} + z_{2}\right) \, \mathbf{a}_{3} & = & \left(\frac{1}{2} +z_{2}\right)a \, \mathbf{\hat{x}} + \left(\frac{1}{4} - y_{2}\right)a \, \mathbf{\hat{y}} + \left(\frac{1}{4} - x_{2}\right)a \, \mathbf{\hat{z}} & \left(192h\right) & \mbox{O} \\ 
\mathbf{B}_{56} & = & \left(\frac{1}{2} +x_{2} + y_{2} - z_{2}\right) \, \mathbf{a}_{1} + \left(\frac{1}{2} +x_{2} - y_{2} + z_{2}\right) \, \mathbf{a}_{2} + \left(\frac{1}{2} - x_{2} + y_{2} + z_{2}\right) \, \mathbf{a}_{3} & = & \left(\frac{1}{2} +z_{2}\right)a \, \mathbf{\hat{x}} + \left(\frac{1}{2} +y_{2}\right)a \, \mathbf{\hat{y}} + \left(\frac{1}{2} +x_{2}\right)a \, \mathbf{\hat{z}} & \left(192h\right) & \mbox{O} \\ 
\end{longtabu}
\renewcommand{\arraystretch}{1.0}
\noindent \hrulefill
\\
\textbf{References:}
\vspace*{-0.25cm}
\begin{flushleft}
  - \bibentry{Kirkpatrick_TeO6H6_ZKristallogCrystMat_1926}. \\
\end{flushleft}
\textbf{Found in:}
\vspace*{-0.25cm}
\begin{flushleft}
  - \bibentry{Villars_PearsonsCrystalData_2013}. \\
\end{flushleft}
\noindent \hrulefill
\\
\textbf{Geometry files:}
\\
\noindent  - CIF: pp. {\hyperref[A6B_cF224_228_h_c_cif]{\pageref{A6B_cF224_228_h_c_cif}}} \\
\noindent  - POSCAR: pp. {\hyperref[A6B_cF224_228_h_c_poscar]{\pageref{A6B_cF224_228_h_c_poscar}}} \\
\onecolumn
{\phantomsection\label{A3B10_cI52_229_e_fh}}
\subsection*{\huge \textbf{{\normalfont \begin{raggedleft}$\gamma$-brass (Fe$_{3}$Zn$_{10}$, $D8_{1}$) Structure: \end{raggedleft} \\ A3B10\_cI52\_229\_e\_fh}}}
\noindent \hrulefill
\vspace*{0.25cm}
\begin{figure}[htp]
  \centering
  \vspace{-1em}
  {\includegraphics[width=1\textwidth]{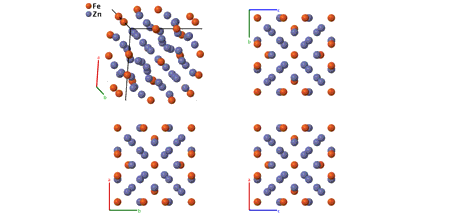}}
\end{figure}
\vspace*{-0.5cm}
\renewcommand{\arraystretch}{1.5}
\begin{equation*}
  \begin{array}{>{$\hspace{-0.15cm}}l<{$}>{$}p{0.5cm}<{$}>{$}p{18.5cm}<{$}}
    \mbox{\large \textbf{Prototype}} &\colon & \ce{$\gamma$-Fe3Zn10} \\
    \mbox{\large \textbf{\AFLOW\ prototype label}} &\colon & \mbox{A3B10\_cI52\_229\_e\_fh} \\
    \mbox{\large \textbf{\textit{Strukturbericht} designation}} &\colon & \mbox{$D8_{1}$} \\
    \mbox{\large \textbf{Pearson symbol}} &\colon & \mbox{cI52} \\
    \mbox{\large \textbf{Space group number}} &\colon & 229 \\
    \mbox{\large \textbf{Space group symbol}} &\colon & Im\bar{3}m \\
    \mbox{\large \textbf{\AFLOW\ prototype command}} &\colon &  \texttt{aflow} \,  \, \texttt{-{}-proto=A3B10\_cI52\_229\_e\_fh } \, \newline \texttt{-{}-params=}{a,x_{1},x_{2},y_{3} }
  \end{array}
\end{equation*}
\renewcommand{\arraystretch}{1.0}

\vspace*{-0.25cm}
\noindent \hrulefill
\\
\textbf{ Other compounds with this structure:}
\begin{itemize}
   \item{ (Pearson, 1958), p. 252, gives a list of compounds which can take on the $D8_{1}$, $D8_{2}$, or $D8_{3}$ structure, depending on the exact composition.  }
\end{itemize}
\vspace*{-0.25cm}
\noindent \hrulefill
\begin{itemize}
  \item{Adding another atom at the origin changes this to the
\href{http://aflow.org/CrystalDatabase/A2B7_cI54_229_e_afh.html}{$L2_{2}$ structure}.
This structure is defined in (Pearson, 1958) quoting (Schramm,
1938). More recent investigations such as (Johannsson, 1968),
(Brandon, 1974) and (Yu, 2005) find that $\gamma$-Fe$_{3}$Zn$_{10}$
forms in the \href{http://aflow.org/CrystalDatabase/A5B8_cI52_217_ce_cg.html}{$D8_{2}$ structure},
with Fe atoms on one (8c) site, Zn atoms on the other (8e) site and
the (24g) sites, and a 50-50 alloy of Fe and Zn on the other (8e)
site.  We use Brandon's data, mapping (12g) $\rightarrow$ (12e), (24g)
$\rightarrow$ (24h), and averaging the two (8e) sites to produce the
(12e) coordinate here.
}
\end{itemize}

\noindent \parbox{1 \linewidth}{
\noindent \hrulefill
\\
\textbf{Body-centered Cubic primitive vectors:} \\
\vspace*{-0.25cm}
\begin{tabular}{cc}
  \begin{tabular}{c}
    \parbox{0.6 \linewidth}{
      \renewcommand{\arraystretch}{1.5}
      \begin{equation*}
        \centering
        \begin{array}{ccc}
              \mathbf{a}_1 & = & - \frac12 \, a \, \mathbf{\hat{x}} + \frac12 \, a \, \mathbf{\hat{y}} + \frac12 \, a \, \mathbf{\hat{z}} \\
    \mathbf{a}_2 & = & ~ \frac12 \, a \, \mathbf{\hat{x}} - \frac12 \, a \, \mathbf{\hat{y}} + \frac12 \, a \, \mathbf{\hat{z}} \\
    \mathbf{a}_3 & = & ~ \frac12 \, a \, \mathbf{\hat{x}} + \frac12 \, a \, \mathbf{\hat{y}} - \frac12 \, a \, \mathbf{\hat{z}} \\

        \end{array}
      \end{equation*}
    }
    \renewcommand{\arraystretch}{1.0}
  \end{tabular}
  \begin{tabular}{c}
    \includegraphics[width=0.3\linewidth]{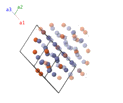} \\
  \end{tabular}
\end{tabular}

}
\vspace*{-0.25cm}

\noindent \hrulefill
\\
\textbf{Basis vectors:}
\vspace*{-0.25cm}
\renewcommand{\arraystretch}{1.5}
\begin{longtabu} to \textwidth{>{\centering $}X[-1,c,c]<{$}>{\centering $}X[-1,c,c]<{$}>{\centering $}X[-1,c,c]<{$}>{\centering $}X[-1,c,c]<{$}>{\centering $}X[-1,c,c]<{$}>{\centering $}X[-1,c,c]<{$}>{\centering $}X[-1,c,c]<{$}}
  & & \mbox{Lattice Coordinates} & & \mbox{Cartesian Coordinates} &\mbox{Wyckoff Position} & \mbox{Atom Type} \\  
  \mathbf{B}_{1} & = & x_{1} \, \mathbf{a}_{2} + x_{1} \, \mathbf{a}_{3} & = & x_{1}a \, \mathbf{\hat{x}} & \left(12e\right) & \mbox{Fe} \\ 
\mathbf{B}_{2} & = & -x_{1} \, \mathbf{a}_{2}-x_{1} \, \mathbf{a}_{3} & = & -x_{1}a \, \mathbf{\hat{x}} & \left(12e\right) & \mbox{Fe} \\ 
\mathbf{B}_{3} & = & x_{1} \, \mathbf{a}_{1} + x_{1} \, \mathbf{a}_{3} & = & x_{1}a \, \mathbf{\hat{y}} & \left(12e\right) & \mbox{Fe} \\ 
\mathbf{B}_{4} & = & -x_{1} \, \mathbf{a}_{1} + -x_{1} \, \mathbf{a}_{3} & = & -x_{1}a \, \mathbf{\hat{y}} & \left(12e\right) & \mbox{Fe} \\ 
\mathbf{B}_{5} & = & x_{1} \, \mathbf{a}_{1} + x_{1} \, \mathbf{a}_{2} & = & x_{1}a \, \mathbf{\hat{z}} & \left(12e\right) & \mbox{Fe} \\ 
\mathbf{B}_{6} & = & -x_{1} \, \mathbf{a}_{1}-x_{1} \, \mathbf{a}_{2} & = & -x_{1}a \, \mathbf{\hat{z}} & \left(12e\right) & \mbox{Fe} \\ 
\mathbf{B}_{7} & = & 2x_{2} \, \mathbf{a}_{1} + 2x_{2} \, \mathbf{a}_{2} + 2x_{2} \, \mathbf{a}_{3} & = & x_{2}a \, \mathbf{\hat{x}} + x_{2}a \, \mathbf{\hat{y}} + x_{2}a \, \mathbf{\hat{z}} & \left(16f\right) & \mbox{Zn I} \\ 
\mathbf{B}_{8} & = & -2x_{2} \, \mathbf{a}_{3} & = & -x_{2}a \, \mathbf{\hat{x}}-x_{2}a \, \mathbf{\hat{y}} + x_{2}a \, \mathbf{\hat{z}} & \left(16f\right) & \mbox{Zn I} \\ 
\mathbf{B}_{9} & = & -2x_{2} \, \mathbf{a}_{2} & = & -x_{2}a \, \mathbf{\hat{x}} + x_{2}a \, \mathbf{\hat{y}}-x_{2}a \, \mathbf{\hat{z}} & \left(16f\right) & \mbox{Zn I} \\ 
\mathbf{B}_{10} & = & -2x_{2} \, \mathbf{a}_{1} & = & x_{2}a \, \mathbf{\hat{x}}-x_{2}a \, \mathbf{\hat{y}}-x_{2}a \, \mathbf{\hat{z}} & \left(16f\right) & \mbox{Zn I} \\ 
\mathbf{B}_{11} & = & 2x_{2} \, \mathbf{a}_{3} & = & x_{2}a \, \mathbf{\hat{x}} + x_{2}a \, \mathbf{\hat{y}}-x_{2}a \, \mathbf{\hat{z}} & \left(16f\right) & \mbox{Zn I} \\ 
\mathbf{B}_{12} & = & -2x_{2} \, \mathbf{a}_{1}-2x_{2} \, \mathbf{a}_{2}-2x_{2} \, \mathbf{a}_{3} & = & -x_{2}a \, \mathbf{\hat{x}}-x_{2}a \, \mathbf{\hat{y}}-x_{2}a \, \mathbf{\hat{z}} & \left(16f\right) & \mbox{Zn I} \\ 
\mathbf{B}_{13} & = & 2x_{2} \, \mathbf{a}_{2} & = & x_{2}a \, \mathbf{\hat{x}}-x_{2}a \, \mathbf{\hat{y}} + x_{2}a \, \mathbf{\hat{z}} & \left(16f\right) & \mbox{Zn I} \\ 
\mathbf{B}_{14} & = & 2x_{2} \, \mathbf{a}_{1} & = & -x_{2}a \, \mathbf{\hat{x}} + x_{2}a \, \mathbf{\hat{y}} + x_{2}a \, \mathbf{\hat{z}} & \left(16f\right) & \mbox{Zn I} \\ 
\mathbf{B}_{15} & = & 2y_{3} \, \mathbf{a}_{1} + y_{3} \, \mathbf{a}_{2} + y_{3} \, \mathbf{a}_{3} & = & y_{3}a \, \mathbf{\hat{y}} + y_{3}a \, \mathbf{\hat{z}} & \left(24h\right) & \mbox{Zn II} \\ 
\mathbf{B}_{16} & = & y_{3} \, \mathbf{a}_{2}-y_{3} \, \mathbf{a}_{3} & = & -y_{3}a \, \mathbf{\hat{y}} + y_{3}a \, \mathbf{\hat{z}} & \left(24h\right) & \mbox{Zn II} \\ 
\mathbf{B}_{17} & = & -y_{3} \, \mathbf{a}_{2} + y_{3} \, \mathbf{a}_{3} & = & y_{3}a \, \mathbf{\hat{y}}-y_{3}a \, \mathbf{\hat{z}} & \left(24h\right) & \mbox{Zn II} \\ 
\mathbf{B}_{18} & = & -2y_{3} \, \mathbf{a}_{1}-y_{3} \, \mathbf{a}_{2}-y_{3} \, \mathbf{a}_{3} & = & -y_{3}a \, \mathbf{\hat{y}}-y_{3}a \, \mathbf{\hat{z}} & \left(24h\right) & \mbox{Zn II} \\ 
\mathbf{B}_{19} & = & y_{3} \, \mathbf{a}_{1} + 2y_{3} \, \mathbf{a}_{2} + y_{3} \, \mathbf{a}_{3} & = & y_{3}a \, \mathbf{\hat{x}} + y_{3}a \, \mathbf{\hat{z}} & \left(24h\right) & \mbox{Zn II} \\ 
\mathbf{B}_{20} & = & -y_{3} \, \mathbf{a}_{1} + y_{3} \, \mathbf{a}_{3} & = & y_{3}a \, \mathbf{\hat{x}} + -y_{3}a \, \mathbf{\hat{z}} & \left(24h\right) & \mbox{Zn II} \\ 
\mathbf{B}_{21} & = & y_{3} \, \mathbf{a}_{1} + -y_{3} \, \mathbf{a}_{3} & = & -y_{3}a \, \mathbf{\hat{x}} + y_{3}a \, \mathbf{\hat{z}} & \left(24h\right) & \mbox{Zn II} \\ 
\mathbf{B}_{22} & = & -y_{3} \, \mathbf{a}_{1}-2y_{3} \, \mathbf{a}_{2}-y_{3} \, \mathbf{a}_{3} & = & -y_{3}a \, \mathbf{\hat{x}} + -y_{3}a \, \mathbf{\hat{z}} & \left(24h\right) & \mbox{Zn II} \\ 
\mathbf{B}_{23} & = & y_{3} \, \mathbf{a}_{1} + y_{3} \, \mathbf{a}_{2} + 2y_{3} \, \mathbf{a}_{3} & = & y_{3}a \, \mathbf{\hat{x}} + y_{3}a \, \mathbf{\hat{y}} & \left(24h\right) & \mbox{Zn II} \\ 
\mathbf{B}_{24} & = & y_{3} \, \mathbf{a}_{1}-y_{3} \, \mathbf{a}_{2} & = & -y_{3}a \, \mathbf{\hat{x}} + y_{3}a \, \mathbf{\hat{y}} & \left(24h\right) & \mbox{Zn II} \\ 
\mathbf{B}_{25} & = & -y_{3} \, \mathbf{a}_{1} + y_{3} \, \mathbf{a}_{2} & = & y_{3}a \, \mathbf{\hat{x}}-y_{3}a \, \mathbf{\hat{y}} & \left(24h\right) & \mbox{Zn II} \\ 
\mathbf{B}_{26} & = & -y_{3} \, \mathbf{a}_{1}-y_{3} \, \mathbf{a}_{2}-2y_{3} \, \mathbf{a}_{3} & = & -y_{3}a \, \mathbf{\hat{x}}-y_{3}a \, \mathbf{\hat{y}} & \left(24h\right) & \mbox{Zn II} \\ 
\end{longtabu}
\renewcommand{\arraystretch}{1.0}
\noindent \hrulefill
\\
\textbf{References:}
\vspace*{-0.25cm}
\begin{flushleft}
  - \bibentry{Schramm_Z_Metallkd_30_1938}. \\
  - \bibentry{Johansson_Acta_Chem_Scand_22_1968}. \\
  - \bibentry{Brandon_Acta_Cryst_B_30_1974}. \\
  - \bibentry{Yu_Mat_Trans_46_2005}. \\
\end{flushleft}
\textbf{Found in:}
\vspace*{-0.25cm}
\begin{flushleft}
  - \bibentry{Pearson_NRC_1958}. \\
\end{flushleft}
\noindent \hrulefill
\\
\textbf{Geometry files:}
\\
\noindent  - CIF: pp. {\hyperref[A3B10_cI52_229_e_fh_cif]{\pageref{A3B10_cI52_229_e_fh_cif}}} \\
\noindent  - POSCAR: pp. {\hyperref[A3B10_cI52_229_e_fh_poscar]{\pageref{A3B10_cI52_229_e_fh_poscar}}} \\
\onecolumn
{\phantomsection\label{A4B_cI10_229_c_a}}
\subsection*{\huge \textbf{{\normalfont $\beta$-Hg$_{4}$Pt Structure: A4B\_cI10\_229\_c\_a}}}
\noindent \hrulefill
\vspace*{0.25cm}
\begin{figure}[htp]
  \centering
  \vspace{-1em}
  {\includegraphics[width=1\textwidth]{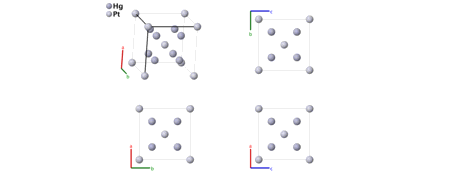}}
\end{figure}
\vspace*{-0.5cm}
\renewcommand{\arraystretch}{1.5}
\begin{equation*}
  \begin{array}{>{$\hspace{-0.15cm}}l<{$}>{$}p{0.5cm}<{$}>{$}p{18.5cm}<{$}}
    \mbox{\large \textbf{Prototype}} &\colon & \ce{$\beta$-Hg4Pt} \\
    \mbox{\large \textbf{\AFLOW\ prototype label}} &\colon & \mbox{A4B\_cI10\_229\_c\_a} \\
    \mbox{\large \textbf{\textit{Strukturbericht} designation}} &\colon & \mbox{None} \\
    \mbox{\large \textbf{Pearson symbol}} &\colon & \mbox{cI10} \\
    \mbox{\large \textbf{Space group number}} &\colon & 229 \\
    \mbox{\large \textbf{Space group symbol}} &\colon & Im\bar{3}m \\
    \mbox{\large \textbf{\AFLOW\ prototype command}} &\colon &  \texttt{aflow} \,  \, \texttt{-{}-proto=A4B\_cI10\_229\_c\_a } \, \newline \texttt{-{}-params=}{a }
  \end{array}
\end{equation*}
\renewcommand{\arraystretch}{1.0}

\vspace*{-0.25cm}
\noindent \hrulefill
\\
\textbf{ Other compounds with this structure:}
\begin{itemize}
   \item{ Hg$_{4}$Pd, Hg$_{4}$U  }
\end{itemize}
\noindent \parbox{1 \linewidth}{
\noindent \hrulefill
\\
\textbf{Body-centered Cubic primitive vectors:} \\
\vspace*{-0.25cm}
\begin{tabular}{cc}
  \begin{tabular}{c}
    \parbox{0.6 \linewidth}{
      \renewcommand{\arraystretch}{1.5}
      \begin{equation*}
        \centering
        \begin{array}{ccc}
              \mathbf{a}_1 & = & - \frac12 \, a \, \mathbf{\hat{x}} + \frac12 \, a \, \mathbf{\hat{y}} + \frac12 \, a \, \mathbf{\hat{z}} \\
    \mathbf{a}_2 & = & ~ \frac12 \, a \, \mathbf{\hat{x}} - \frac12 \, a \, \mathbf{\hat{y}} + \frac12 \, a \, \mathbf{\hat{z}} \\
    \mathbf{a}_3 & = & ~ \frac12 \, a \, \mathbf{\hat{x}} + \frac12 \, a \, \mathbf{\hat{y}} - \frac12 \, a \, \mathbf{\hat{z}} \\

        \end{array}
      \end{equation*}
    }
    \renewcommand{\arraystretch}{1.0}
  \end{tabular}
  \begin{tabular}{c}
    \includegraphics[width=0.3\linewidth]{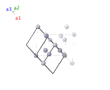} \\
  \end{tabular}
\end{tabular}

}
\vspace*{-0.25cm}

\noindent \hrulefill
\\
\textbf{Basis vectors:}
\vspace*{-0.25cm}
\renewcommand{\arraystretch}{1.5}
\begin{longtabu} to \textwidth{>{\centering $}X[-1,c,c]<{$}>{\centering $}X[-1,c,c]<{$}>{\centering $}X[-1,c,c]<{$}>{\centering $}X[-1,c,c]<{$}>{\centering $}X[-1,c,c]<{$}>{\centering $}X[-1,c,c]<{$}>{\centering $}X[-1,c,c]<{$}}
  & & \mbox{Lattice Coordinates} & & \mbox{Cartesian Coordinates} &\mbox{Wyckoff Position} & \mbox{Atom Type} \\  
  \mathbf{B}_{1} & = & 0 \, \mathbf{a}_{1} + 0 \, \mathbf{a}_{2} + 0 \, \mathbf{a}_{3} & = & 0 \, \mathbf{\hat{x}} + 0 \, \mathbf{\hat{y}} + 0 \, \mathbf{\hat{z}} & \left(2a\right) & \mbox{Pt} \\ 
\mathbf{B}_{2} & = & \frac{1}{2} \, \mathbf{a}_{1} + \frac{1}{2} \, \mathbf{a}_{2} + \frac{1}{2} \, \mathbf{a}_{3} & = & \frac{1}{4}a \, \mathbf{\hat{x}} + \frac{1}{4}a \, \mathbf{\hat{y}} + \frac{1}{4}a \, \mathbf{\hat{z}} & \left(8c\right) & \mbox{Hg} \\ 
\mathbf{B}_{3} & = & \frac{1}{2} \, \mathbf{a}_{3} & = & \frac{1}{4}a \, \mathbf{\hat{x}} + \frac{1}{4}a \, \mathbf{\hat{y}}- \frac{1}{4}a  \, \mathbf{\hat{z}} & \left(8c\right) & \mbox{Hg} \\ 
\mathbf{B}_{4} & = & \frac{1}{2} \, \mathbf{a}_{2} & = & \frac{1}{4}a \, \mathbf{\hat{x}}- \frac{1}{4}a  \, \mathbf{\hat{y}} + \frac{1}{4}a \, \mathbf{\hat{z}} & \left(8c\right) & \mbox{Hg} \\ 
\mathbf{B}_{5} & = & \frac{1}{2} \, \mathbf{a}_{1} & = & - \frac{1}{4}a  \, \mathbf{\hat{x}} + \frac{1}{4}a \, \mathbf{\hat{y}} + \frac{1}{4}a \, \mathbf{\hat{z}} & \left(8c\right) & \mbox{Hg} \\ 
\end{longtabu}
\renewcommand{\arraystretch}{1.0}
\noindent \hrulefill
\\
\textbf{References:}
\vspace*{-0.25cm}
\begin{flushleft}
  - \bibentry{Bauer_MonatshChem_84_1953}. \\
\end{flushleft}
\noindent \hrulefill
\\
\textbf{Geometry files:}
\\
\noindent  - CIF: pp. {\hyperref[A4B_cI10_229_c_a_cif]{\pageref{A4B_cI10_229_c_a_cif}}} \\
\noindent  - POSCAR: pp. {\hyperref[A4B_cI10_229_c_a_poscar]{\pageref{A4B_cI10_229_c_a_poscar}}} \\
\onecolumn
{\phantomsection\label{A7B3_cI40_229_df_e}}
\subsection*{\huge \textbf{{\normalfont Ir$_{3}$Ge$_{7}$ ($D8_{f}$) Structure: A7B3\_cI40\_229\_df\_e}}}
\noindent \hrulefill
\vspace*{0.25cm}
\begin{figure}[htp]
  \centering
  \vspace{-1em}
  {\includegraphics[width=1\textwidth]{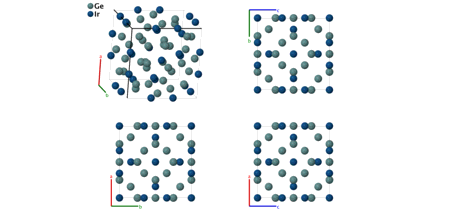}}
\end{figure}
\vspace*{-0.5cm}
\renewcommand{\arraystretch}{1.5}
\begin{equation*}
  \begin{array}{>{$\hspace{-0.15cm}}l<{$}>{$}p{0.5cm}<{$}>{$}p{18.5cm}<{$}}
    \mbox{\large \textbf{Prototype}} &\colon & \ce{Ir3Ge7} \\
    \mbox{\large \textbf{\AFLOW\ prototype label}} &\colon & \mbox{A7B3\_cI40\_229\_df\_e} \\
    \mbox{\large \textbf{\textit{Strukturbericht} designation}} &\colon & \mbox{$D8_{f}$} \\
    \mbox{\large \textbf{Pearson symbol}} &\colon & \mbox{cI40} \\
    \mbox{\large \textbf{Space group number}} &\colon & 229 \\
    \mbox{\large \textbf{Space group symbol}} &\colon & Im\bar{3}m \\
    \mbox{\large \textbf{\AFLOW\ prototype command}} &\colon &  \texttt{aflow} \,  \, \texttt{-{}-proto=A7B3\_cI40\_229\_df\_e } \, \newline \texttt{-{}-params=}{a,x_{2},x_{3} }
  \end{array}
\end{equation*}
\renewcommand{\arraystretch}{1.0}

\vspace*{-0.25cm}
\noindent \hrulefill
\\
\textbf{ Other compounds with this structure}
\begin{itemize}
   \item{ Ga$_{7}$Ni$_{3}$, In$_{7}$Pd$_{3}$, Sb$_{7}$Mo$_{3}$, As$_{7}$Re$_{3}$, Sn$_{7}$Ru$_{3}$  }
\end{itemize}
\noindent \parbox{1 \linewidth}{
\noindent \hrulefill
\\
\textbf{Body-centered Cubic primitive vectors:} \\
\vspace*{-0.25cm}
\begin{tabular}{cc}
  \begin{tabular}{c}
    \parbox{0.6 \linewidth}{
      \renewcommand{\arraystretch}{1.5}
      \begin{equation*}
        \centering
        \begin{array}{ccc}
              \mathbf{a}_1 & = & - \frac12 \, a \, \mathbf{\hat{x}} + \frac12 \, a \, \mathbf{\hat{y}} + \frac12 \, a \, \mathbf{\hat{z}} \\
    \mathbf{a}_2 & = & ~ \frac12 \, a \, \mathbf{\hat{x}} - \frac12 \, a \, \mathbf{\hat{y}} + \frac12 \, a \, \mathbf{\hat{z}} \\
    \mathbf{a}_3 & = & ~ \frac12 \, a \, \mathbf{\hat{x}} + \frac12 \, a \, \mathbf{\hat{y}} - \frac12 \, a \, \mathbf{\hat{z}} \\

        \end{array}
      \end{equation*}
    }
    \renewcommand{\arraystretch}{1.0}
  \end{tabular}
  \begin{tabular}{c}
    \includegraphics[width=0.3\linewidth]{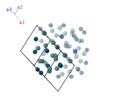} \\
  \end{tabular}
\end{tabular}

}
\vspace*{-0.25cm}

\noindent \hrulefill
\\
\textbf{Basis vectors:}
\vspace*{-0.25cm}
\renewcommand{\arraystretch}{1.5}
\begin{longtabu} to \textwidth{>{\centering $}X[-1,c,c]<{$}>{\centering $}X[-1,c,c]<{$}>{\centering $}X[-1,c,c]<{$}>{\centering $}X[-1,c,c]<{$}>{\centering $}X[-1,c,c]<{$}>{\centering $}X[-1,c,c]<{$}>{\centering $}X[-1,c,c]<{$}}
  & & \mbox{Lattice Coordinates} & & \mbox{Cartesian Coordinates} &\mbox{Wyckoff Position} & \mbox{Atom Type} \\  
  \mathbf{B}_{1} & = & \frac{1}{2} \, \mathbf{a}_{1} + \frac{3}{4} \, \mathbf{a}_{2} + \frac{1}{4} \, \mathbf{a}_{3} & = & \frac{1}{4}a \, \mathbf{\hat{x}} + \frac{1}{2}a \, \mathbf{\hat{z}} & \left(12d\right) & \mbox{Ge I} \\ 
\mathbf{B}_{2} & = & \frac{1}{2} \, \mathbf{a}_{1} + \frac{1}{4} \, \mathbf{a}_{2} + \frac{3}{4} \, \mathbf{a}_{3} & = & \frac{1}{4}a \, \mathbf{\hat{x}} + \frac{1}{2}a \, \mathbf{\hat{y}} & \left(12d\right) & \mbox{Ge I} \\ 
\mathbf{B}_{3} & = & \frac{1}{4} \, \mathbf{a}_{1} + \frac{1}{2} \, \mathbf{a}_{2} + \frac{3}{4} \, \mathbf{a}_{3} & = & \frac{1}{2}a \, \mathbf{\hat{x}} + \frac{1}{4}a \, \mathbf{\hat{y}} & \left(12d\right) & \mbox{Ge I} \\ 
\mathbf{B}_{4} & = & \frac{3}{4} \, \mathbf{a}_{1} + \frac{1}{2} \, \mathbf{a}_{2} + \frac{1}{4} \, \mathbf{a}_{3} & = & \frac{1}{4}a \, \mathbf{\hat{y}} + \frac{1}{2}a \, \mathbf{\hat{z}} & \left(12d\right) & \mbox{Ge I} \\ 
\mathbf{B}_{5} & = & \frac{3}{4} \, \mathbf{a}_{1} + \frac{1}{4} \, \mathbf{a}_{2} + \frac{1}{2} \, \mathbf{a}_{3} & = & \frac{1}{2}a \, \mathbf{\hat{y}} + \frac{1}{4}a \, \mathbf{\hat{z}} & \left(12d\right) & \mbox{Ge I} \\ 
\mathbf{B}_{6} & = & \frac{1}{4} \, \mathbf{a}_{1} + \frac{3}{4} \, \mathbf{a}_{2} + \frac{1}{2} \, \mathbf{a}_{3} & = & \frac{1}{2}a \, \mathbf{\hat{x}} + \frac{1}{4}a \, \mathbf{\hat{z}} & \left(12d\right) & \mbox{Ge I} \\ 
\mathbf{B}_{7} & = & x_{2} \, \mathbf{a}_{2} + x_{2} \, \mathbf{a}_{3} & = & x_{2}a \, \mathbf{\hat{x}} & \left(12e\right) & \mbox{Ir} \\ 
\mathbf{B}_{8} & = & -x_{2} \, \mathbf{a}_{2}-x_{2} \, \mathbf{a}_{3} & = & -x_{2}a \, \mathbf{\hat{x}} & \left(12e\right) & \mbox{Ir} \\ 
\mathbf{B}_{9} & = & x_{2} \, \mathbf{a}_{1} + x_{2} \, \mathbf{a}_{3} & = & x_{2}a \, \mathbf{\hat{y}} & \left(12e\right) & \mbox{Ir} \\ 
\mathbf{B}_{10} & = & -x_{2} \, \mathbf{a}_{1} + -x_{2} \, \mathbf{a}_{3} & = & -x_{2}a \, \mathbf{\hat{y}} & \left(12e\right) & \mbox{Ir} \\ 
\mathbf{B}_{11} & = & x_{2} \, \mathbf{a}_{1} + x_{2} \, \mathbf{a}_{2} & = & x_{2}a \, \mathbf{\hat{z}} & \left(12e\right) & \mbox{Ir} \\ 
\mathbf{B}_{12} & = & -x_{2} \, \mathbf{a}_{1}-x_{2} \, \mathbf{a}_{2} & = & -x_{2}a \, \mathbf{\hat{z}} & \left(12e\right) & \mbox{Ir} \\ 
\mathbf{B}_{13} & = & 2x_{3} \, \mathbf{a}_{1} + 2x_{3} \, \mathbf{a}_{2} + 2x_{3} \, \mathbf{a}_{3} & = & x_{3}a \, \mathbf{\hat{x}} + x_{3}a \, \mathbf{\hat{y}} + x_{3}a \, \mathbf{\hat{z}} & \left(16f\right) & \mbox{Ge II} \\ 
\mathbf{B}_{14} & = & -2x_{3} \, \mathbf{a}_{3} & = & -x_{3}a \, \mathbf{\hat{x}}-x_{3}a \, \mathbf{\hat{y}} + x_{3}a \, \mathbf{\hat{z}} & \left(16f\right) & \mbox{Ge II} \\ 
\mathbf{B}_{15} & = & -2x_{3} \, \mathbf{a}_{2} & = & -x_{3}a \, \mathbf{\hat{x}} + x_{3}a \, \mathbf{\hat{y}}-x_{3}a \, \mathbf{\hat{z}} & \left(16f\right) & \mbox{Ge II} \\ 
\mathbf{B}_{16} & = & -2x_{3} \, \mathbf{a}_{1} & = & x_{3}a \, \mathbf{\hat{x}}-x_{3}a \, \mathbf{\hat{y}}-x_{3}a \, \mathbf{\hat{z}} & \left(16f\right) & \mbox{Ge II} \\ 
\mathbf{B}_{17} & = & 2x_{3} \, \mathbf{a}_{3} & = & x_{3}a \, \mathbf{\hat{x}} + x_{3}a \, \mathbf{\hat{y}}-x_{3}a \, \mathbf{\hat{z}} & \left(16f\right) & \mbox{Ge II} \\ 
\mathbf{B}_{18} & = & -2x_{3} \, \mathbf{a}_{1}-2x_{3} \, \mathbf{a}_{2}-2x_{3} \, \mathbf{a}_{3} & = & -x_{3}a \, \mathbf{\hat{x}}-x_{3}a \, \mathbf{\hat{y}}-x_{3}a \, \mathbf{\hat{z}} & \left(16f\right) & \mbox{Ge II} \\ 
\mathbf{B}_{19} & = & 2x_{3} \, \mathbf{a}_{2} & = & x_{3}a \, \mathbf{\hat{x}}-x_{3}a \, \mathbf{\hat{y}} + x_{3}a \, \mathbf{\hat{z}} & \left(16f\right) & \mbox{Ge II} \\ 
\mathbf{B}_{20} & = & 2x_{3} \, \mathbf{a}_{1} & = & -x_{3}a \, \mathbf{\hat{x}} + x_{3}a \, \mathbf{\hat{y}} + x_{3}a \, \mathbf{\hat{z}} & \left(16f\right) & \mbox{Ge II} \\ 
\end{longtabu}
\renewcommand{\arraystretch}{1.0}
\noindent \hrulefill
\\
\textbf{References:}
\vspace*{-0.25cm}
\begin{flushleft}
  - \bibentry{Haussermann_Chem_Euro_J_4_1998}. \\
\end{flushleft}
\textbf{Found in:}
\vspace*{-0.25cm}
\begin{flushleft}
  - \bibentry{Selim_Hyperf_Int_178_2007}. \\
\end{flushleft}
\noindent \hrulefill
\\
\textbf{Geometry files:}
\\
\noindent  - CIF: pp. {\hyperref[A7B3_cI40_229_df_e_cif]{\pageref{A7B3_cI40_229_df_e_cif}}} \\
\noindent  - POSCAR: pp. {\hyperref[A7B3_cI40_229_df_e_poscar]{\pageref{A7B3_cI40_229_df_e_poscar}}} \\
\onecolumn
{\phantomsection\label{A2B3C12D3_cI160_230_a_c_h_d}}
\subsection*{\huge \textbf{{\normalfont \begin{raggedleft}Garnet (Co$_3$Al$_2$Si$_3$O$_{12}$, $S1_{4}$) Structure: \end{raggedleft} \\ A2B3C12D3\_cI160\_230\_a\_c\_h\_d}}}
\noindent \hrulefill
\vspace*{0.25cm}
\begin{figure}[htp]
  \centering
  \vspace{-1em}
  {\includegraphics[width=1\textwidth]{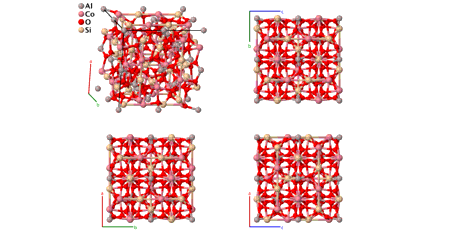}}
\end{figure}
\vspace*{-0.5cm}
\renewcommand{\arraystretch}{1.5}
\begin{equation*}
  \begin{array}{>{$\hspace{-0.15cm}}l<{$}>{$}p{0.5cm}<{$}>{$}p{18.5cm}<{$}}
    \mbox{\large \textbf{Prototype}} &\colon & \ce{Co3Al2Si3O12} \\
    \mbox{\large \textbf{\AFLOW\ prototype label}} &\colon & \mbox{A2B3C12D3\_cI160\_230\_a\_c\_h\_d} \\
    \mbox{\large \textbf{\textit{Strukturbericht} designation}} &\colon & \mbox{$S1_{4}$} \\
    \mbox{\large \textbf{Pearson symbol}} &\colon & \mbox{cI160} \\
    \mbox{\large \textbf{Space group number}} &\colon & 230 \\
    \mbox{\large \textbf{Space group symbol}} &\colon & Ia\bar{3}d \\
    \mbox{\large \textbf{\AFLOW\ prototype command}} &\colon &  \texttt{aflow} \,  \, \texttt{-{}-proto=A2B3C12D3\_cI160\_230\_a\_c\_h\_d } \, \newline \texttt{-{}-params=}{a,x_{4},y_{4},z_{4} }
  \end{array}
\end{equation*}
\renewcommand{\arraystretch}{1.0}

\vspace*{-0.25cm}
\noindent \hrulefill
\\
\textbf{ Other compounds with this structure:}
\begin{itemize}
   \item{ Al$_{2}$(Mg,Ni)$_{3}$Si$_{3}$O$_{12}$, Al$_{2}$Ca$_{3}$Si$_{3}$O$_{12}$, Al$_{2}$Co$_{3}$Si$_{3}$O$_{12}$, Al$_{2}$Mg$_{3}$Si$_{3}$O$_{12}$,  Al$_{2}$Mn$_{3}$Si$_{3}$O$_{12}$, Cr$_{2}$Ca$_{3}$Si$_{3}$O$_{12}$, Fe$_{2}$Ca$_{3}$Si$_{3}$O$_{12}$, Fe$_{2}$Mn$_{3}$Ge$_{3}$O$_{12}$, Mn$_{5}$Si$_{3}$O$_{12}$, Sc$_{2}$Ca$_{3}$Si$_{3}$O$_{12}$  }
\end{itemize}
\vspace*{-0.25cm}
\noindent \hrulefill
\begin{itemize}
  \item{(Ross, 1996) does not explicitly give the positions of the Al and Si
atoms, which we take from (Downs, 2003).
}
\end{itemize}

\noindent \parbox{1 \linewidth}{
\noindent \hrulefill
\\
\textbf{Body-centered Cubic primitive vectors:} \\
\vspace*{-0.25cm}
\begin{tabular}{cc}
  \begin{tabular}{c}
    \parbox{0.6 \linewidth}{
      \renewcommand{\arraystretch}{1.5}
      \begin{equation*}
        \centering
        \begin{array}{ccc}
              \mathbf{a}_1 & = & - \frac12 \, a \, \mathbf{\hat{x}} + \frac12 \, a \, \mathbf{\hat{y}} + \frac12 \, a \, \mathbf{\hat{z}} \\
    \mathbf{a}_2 & = & ~ \frac12 \, a \, \mathbf{\hat{x}} - \frac12 \, a \, \mathbf{\hat{y}} + \frac12 \, a \, \mathbf{\hat{z}} \\
    \mathbf{a}_3 & = & ~ \frac12 \, a \, \mathbf{\hat{x}} + \frac12 \, a \, \mathbf{\hat{y}} - \frac12 \, a \, \mathbf{\hat{z}} \\

        \end{array}
      \end{equation*}
    }
    \renewcommand{\arraystretch}{1.0}
  \end{tabular}
  \begin{tabular}{c}
    \includegraphics[width=0.3\linewidth]{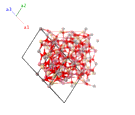} \\
  \end{tabular}
\end{tabular}

}
\vspace*{-0.25cm}

\noindent \hrulefill
\\
\textbf{Basis vectors:}
\vspace*{-0.25cm}
\renewcommand{\arraystretch}{1.5}
\begin{longtabu} to \textwidth{>{\centering $}X[-1,c,c]<{$}>{\centering $}X[-1,c,c]<{$}>{\centering $}X[-1,c,c]<{$}>{\centering $}X[-1,c,c]<{$}>{\centering $}X[-1,c,c]<{$}>{\centering $}X[-1,c,c]<{$}>{\centering $}X[-1,c,c]<{$}}
  & & \mbox{Lattice Coordinates} & & \mbox{Cartesian Coordinates} &\mbox{Wyckoff Position} & \mbox{Atom Type} \\  
  \mathbf{B}_{1} & = & 0 \, \mathbf{a}_{1} + 0 \, \mathbf{a}_{2} + 0 \, \mathbf{a}_{3} & = & 0 \, \mathbf{\hat{x}} + 0 \, \mathbf{\hat{y}} + 0 \, \mathbf{\hat{z}} & \left(16a\right) & \mbox{Al} \\ 
\mathbf{B}_{2} & = & \frac{1}{2} \, \mathbf{a}_{1} + \frac{1}{2} \, \mathbf{a}_{3} & = & \frac{1}{2}a \, \mathbf{\hat{y}} & \left(16a\right) & \mbox{Al} \\ 
\mathbf{B}_{3} & = & \frac{1}{2} \, \mathbf{a}_{2} + \frac{1}{2} \, \mathbf{a}_{3} & = & \frac{1}{2}a \, \mathbf{\hat{x}} & \left(16a\right) & \mbox{Al} \\ 
\mathbf{B}_{4} & = & \frac{1}{2} \, \mathbf{a}_{1} + \frac{1}{2} \, \mathbf{a}_{2} & = & \frac{1}{2}a \, \mathbf{\hat{z}} & \left(16a\right) & \mbox{Al} \\ 
\mathbf{B}_{5} & = & \frac{1}{2} \, \mathbf{a}_{1} & = & \frac{3}{4}a \, \mathbf{\hat{x}} + \frac{1}{4}a \, \mathbf{\hat{y}} + \frac{1}{4}a \, \mathbf{\hat{z}} & \left(16a\right) & \mbox{Al} \\ 
\mathbf{B}_{6} & = & \frac{1}{2} \, \mathbf{a}_{1} + \frac{1}{2} \, \mathbf{a}_{2} + \frac{1}{2} \, \mathbf{a}_{3} & = & \frac{1}{4}a \, \mathbf{\hat{x}} + \frac{1}{4}a \, \mathbf{\hat{y}} + \frac{1}{4}a \, \mathbf{\hat{z}} & \left(16a\right) & \mbox{Al} \\ 
\mathbf{B}_{7} & = & \frac{1}{2} \, \mathbf{a}_{3} & = & \frac{1}{4}a \, \mathbf{\hat{x}} + \frac{1}{4}a \, \mathbf{\hat{y}} + \frac{3}{4}a \, \mathbf{\hat{z}} & \left(16a\right) & \mbox{Al} \\ 
\mathbf{B}_{8} & = & \frac{1}{2} \, \mathbf{a}_{2} & = & \frac{1}{4}a \, \mathbf{\hat{x}} + \frac{3}{4}a \, \mathbf{\hat{y}} + \frac{1}{4}a \, \mathbf{\hat{z}} & \left(16a\right) & \mbox{Al} \\ 
\mathbf{B}_{9} & = & \frac{1}{4} \, \mathbf{a}_{1} + \frac{3}{8} \, \mathbf{a}_{2} + \frac{1}{8} \, \mathbf{a}_{3} & = & \frac{1}{8}a \, \mathbf{\hat{x}} + \frac{1}{4}a \, \mathbf{\hat{z}} & \left(24c\right) & \mbox{Co} \\ 
\mathbf{B}_{10} & = & \frac{3}{4} \, \mathbf{a}_{1} + \frac{1}{8} \, \mathbf{a}_{2} + \frac{3}{8} \, \mathbf{a}_{3} & = & - \frac{1}{8}a  \, \mathbf{\hat{x}} + \frac{1}{2}a \, \mathbf{\hat{y}} + \frac{1}{4}a \, \mathbf{\hat{z}} & \left(24c\right) & \mbox{Co} \\ 
\mathbf{B}_{11} & = & \frac{1}{8} \, \mathbf{a}_{1} + \frac{1}{4} \, \mathbf{a}_{2} + \frac{3}{8} \, \mathbf{a}_{3} & = & \frac{1}{4}a \, \mathbf{\hat{x}} + \frac{1}{8}a \, \mathbf{\hat{y}} & \left(24c\right) & \mbox{Co} \\ 
\mathbf{B}_{12} & = & \frac{3}{8} \, \mathbf{a}_{1} + \frac{3}{4} \, \mathbf{a}_{2} + \frac{1}{8} \, \mathbf{a}_{3} & = & \frac{1}{4}a \, \mathbf{\hat{x}}- \frac{1}{8}a  \, \mathbf{\hat{y}} + \frac{1}{2}a \, \mathbf{\hat{z}} & \left(24c\right) & \mbox{Co} \\ 
\mathbf{B}_{13} & = & \frac{3}{8} \, \mathbf{a}_{1} + \frac{1}{8} \, \mathbf{a}_{2} + \frac{1}{4} \, \mathbf{a}_{3} & = & \frac{1}{4}a \, \mathbf{\hat{y}} + \frac{1}{8}a \, \mathbf{\hat{z}} & \left(24c\right) & \mbox{Co} \\ 
\mathbf{B}_{14} & = & \frac{1}{8} \, \mathbf{a}_{1} + \frac{3}{8} \, \mathbf{a}_{2} + \frac{3}{4} \, \mathbf{a}_{3} & = & \frac{1}{2}a \, \mathbf{\hat{x}} + \frac{1}{4}a \, \mathbf{\hat{y}}- \frac{1}{8}a  \, \mathbf{\hat{z}} & \left(24c\right) & \mbox{Co} \\ 
\mathbf{B}_{15} & = & \frac{3}{4} \, \mathbf{a}_{1} + \frac{5}{8} \, \mathbf{a}_{2} + \frac{7}{8} \, \mathbf{a}_{3} & = & \frac{3}{8}a \, \mathbf{\hat{x}} + \frac{1}{2}a \, \mathbf{\hat{y}} + \frac{1}{4}a \, \mathbf{\hat{z}} & \left(24c\right) & \mbox{Co} \\ 
\mathbf{B}_{16} & = & \frac{1}{4} \, \mathbf{a}_{1} + \frac{7}{8} \, \mathbf{a}_{2} + \frac{5}{8} \, \mathbf{a}_{3} & = & \frac{5}{8}a \, \mathbf{\hat{x}} + \frac{1}{4}a \, \mathbf{\hat{z}} & \left(24c\right) & \mbox{Co} \\ 
\mathbf{B}_{17} & = & \frac{7}{8} \, \mathbf{a}_{1} + \frac{3}{4} \, \mathbf{a}_{2} + \frac{5}{8} \, \mathbf{a}_{3} & = & \frac{1}{4}a \, \mathbf{\hat{x}} + \frac{3}{8}a \, \mathbf{\hat{y}} + \frac{1}{2}a \, \mathbf{\hat{z}} & \left(24c\right) & \mbox{Co} \\ 
\mathbf{B}_{18} & = & \frac{5}{8} \, \mathbf{a}_{1} + \frac{1}{4} \, \mathbf{a}_{2} + \frac{7}{8} \, \mathbf{a}_{3} & = & \frac{1}{4}a \, \mathbf{\hat{x}} + \frac{5}{8}a \, \mathbf{\hat{y}} & \left(24c\right) & \mbox{Co} \\ 
\mathbf{B}_{19} & = & \frac{5}{8} \, \mathbf{a}_{1} + \frac{7}{8} \, \mathbf{a}_{2} + \frac{3}{4} \, \mathbf{a}_{3} & = & \frac{1}{2}a \, \mathbf{\hat{x}} + \frac{1}{4}a \, \mathbf{\hat{y}} + \frac{3}{8}a \, \mathbf{\hat{z}} & \left(24c\right) & \mbox{Co} \\ 
\mathbf{B}_{20} & = & \frac{7}{8} \, \mathbf{a}_{1} + \frac{5}{8} \, \mathbf{a}_{2} + \frac{1}{4} \, \mathbf{a}_{3} & = & \frac{1}{4}a \, \mathbf{\hat{y}} + \frac{5}{8}a \, \mathbf{\hat{z}} & \left(24c\right) & \mbox{Co} \\ 
\mathbf{B}_{21} & = & \frac{1}{4} \, \mathbf{a}_{1} + \frac{5}{8} \, \mathbf{a}_{2} + \frac{3}{8} \, \mathbf{a}_{3} & = & \frac{3}{8}a \, \mathbf{\hat{x}} + \frac{1}{4}a \, \mathbf{\hat{z}} & \left(24d\right) & \mbox{Si} \\ 
\mathbf{B}_{22} & = & \frac{3}{4} \, \mathbf{a}_{1} + \frac{7}{8} \, \mathbf{a}_{2} + \frac{1}{8} \, \mathbf{a}_{3} & = & \frac{1}{8}a \, \mathbf{\hat{x}} + \frac{3}{4}a \, \mathbf{\hat{z}} & \left(24d\right) & \mbox{Si} \\ 
\mathbf{B}_{23} & = & \frac{3}{8} \, \mathbf{a}_{1} + \frac{1}{4} \, \mathbf{a}_{2} + \frac{5}{8} \, \mathbf{a}_{3} & = & \frac{1}{4}a \, \mathbf{\hat{x}} + \frac{3}{8}a \, \mathbf{\hat{y}} & \left(24d\right) & \mbox{Si} \\ 
\mathbf{B}_{24} & = & \frac{1}{8} \, \mathbf{a}_{1} + \frac{3}{4} \, \mathbf{a}_{2} + \frac{7}{8} \, \mathbf{a}_{3} & = & \frac{3}{4}a \, \mathbf{\hat{x}} + \frac{1}{8}a \, \mathbf{\hat{y}} & \left(24d\right) & \mbox{Si} \\ 
\mathbf{B}_{25} & = & \frac{5}{8} \, \mathbf{a}_{1} + \frac{3}{8} \, \mathbf{a}_{2} + \frac{1}{4} \, \mathbf{a}_{3} & = & \frac{1}{4}a \, \mathbf{\hat{y}} + \frac{3}{8}a \, \mathbf{\hat{z}} & \left(24d\right) & \mbox{Si} \\ 
\mathbf{B}_{26} & = & \frac{7}{8} \, \mathbf{a}_{1} + \frac{1}{8} \, \mathbf{a}_{2} + \frac{3}{4} \, \mathbf{a}_{3} & = & \frac{3}{4}a \, \mathbf{\hat{y}} + \frac{1}{8}a \, \mathbf{\hat{z}} & \left(24d\right) & \mbox{Si} \\ 
\mathbf{B}_{27} & = & \frac{5}{8} \, \mathbf{a}_{1} + \frac{3}{4} \, \mathbf{a}_{2} + \frac{3}{8} \, \mathbf{a}_{3} & = & \frac{1}{4}a \, \mathbf{\hat{x}} + \frac{1}{8}a \, \mathbf{\hat{y}} + \frac{1}{2}a \, \mathbf{\hat{z}} & \left(24d\right) & \mbox{Si} \\ 
\mathbf{B}_{28} & = & \frac{7}{8} \, \mathbf{a}_{1} + \frac{1}{4} \, \mathbf{a}_{2} + \frac{1}{8} \, \mathbf{a}_{3} & = & \frac{3}{4}a \, \mathbf{\hat{x}} + \frac{3}{8}a \, \mathbf{\hat{y}} + \frac{1}{2}a \, \mathbf{\hat{z}} & \left(24d\right) & \mbox{Si} \\ 
\mathbf{B}_{29} & = & \frac{3}{4} \, \mathbf{a}_{1} + \frac{3}{8} \, \mathbf{a}_{2} + \frac{5}{8} \, \mathbf{a}_{3} & = & \frac{1}{8}a \, \mathbf{\hat{x}} + \frac{1}{2}a \, \mathbf{\hat{y}} + \frac{1}{4}a \, \mathbf{\hat{z}} & \left(24d\right) & \mbox{Si} \\ 
\mathbf{B}_{30} & = & \frac{1}{4} \, \mathbf{a}_{1} + \frac{1}{8} \, \mathbf{a}_{2} + \frac{7}{8} \, \mathbf{a}_{3} & = & \frac{3}{8}a \, \mathbf{\hat{x}} + \frac{1}{2}a \, \mathbf{\hat{y}}- \frac{1}{4}a  \, \mathbf{\hat{z}} & \left(24d\right) & \mbox{Si} \\ 
\mathbf{B}_{31} & = & \frac{1}{8} \, \mathbf{a}_{1} + \frac{7}{8} \, \mathbf{a}_{2} + \frac{1}{4} \, \mathbf{a}_{3} & = & \frac{1}{2}a \, \mathbf{\hat{x}}- \frac{1}{4}a  \, \mathbf{\hat{y}} + \frac{3}{8}a \, \mathbf{\hat{z}} & \left(24d\right) & \mbox{Si} \\ 
\mathbf{B}_{32} & = & \frac{3}{8} \, \mathbf{a}_{1} + \frac{5}{8} \, \mathbf{a}_{2} + \frac{3}{4} \, \mathbf{a}_{3} & = & \frac{1}{2}a \, \mathbf{\hat{x}} + \frac{1}{4}a \, \mathbf{\hat{y}} + \frac{1}{8}a \, \mathbf{\hat{z}} & \left(24d\right) & \mbox{Si} \\ 
\mathbf{B}_{33} & = & \left(y_{4}+z_{4}\right) \, \mathbf{a}_{1} + \left(x_{4}+z_{4}\right) \, \mathbf{a}_{2} + \left(x_{4}+y_{4}\right) \, \mathbf{a}_{3} & = & x_{4}a \, \mathbf{\hat{x}} + y_{4}a \, \mathbf{\hat{y}} + z_{4}a \, \mathbf{\hat{z}} & \left(96h\right) & \mbox{O} \\ 
\mathbf{B}_{34} & = & \left(\frac{1}{2} - y_{4} + z_{4}\right) \, \mathbf{a}_{1} + \left(-x_{4}+z_{4}\right) \, \mathbf{a}_{2} + \left(\frac{1}{2} - x_{4} - y_{4}\right) \, \mathbf{a}_{3} & = & -x_{4}a \, \mathbf{\hat{x}} + \left(\frac{1}{2} - y_{4}\right)a \, \mathbf{\hat{y}} + z_{4}a \, \mathbf{\hat{z}} & \left(96h\right) & \mbox{O} \\ 
\mathbf{B}_{35} & = & \left(y_{4}-z_{4}\right) \, \mathbf{a}_{1} + \left(\frac{1}{2} - x_{4} - z_{4}\right) \, \mathbf{a}_{2} + \left(\frac{1}{2} - x_{4} + y_{4}\right) \, \mathbf{a}_{3} & = & \left(\frac{1}{2} - x_{4}\right)a \, \mathbf{\hat{x}} + y_{4}a \, \mathbf{\hat{y}}-z_{4}a \, \mathbf{\hat{z}} & \left(96h\right) & \mbox{O} \\ 
\mathbf{B}_{36} & = & \left(\frac{1}{2} - y_{4} - z_{4}\right) \, \mathbf{a}_{1} + \left(\frac{1}{2} +x_{4} - z_{4}\right) \, \mathbf{a}_{2} + \left(x_{4}-y_{4}\right) \, \mathbf{a}_{3} & = & x_{4}a \, \mathbf{\hat{x}}-y_{4}a \, \mathbf{\hat{y}} + \left(\frac{1}{2} - z_{4}\right)a \, \mathbf{\hat{z}} & \left(96h\right) & \mbox{O} \\ 
\mathbf{B}_{37} & = & \left(x_{4}+y_{4}\right) \, \mathbf{a}_{1} + \left(y_{4}+z_{4}\right) \, \mathbf{a}_{2} + \left(x_{4}+z_{4}\right) \, \mathbf{a}_{3} & = & z_{4}a \, \mathbf{\hat{x}} + x_{4}a \, \mathbf{\hat{y}} + y_{4}a \, \mathbf{\hat{z}} & \left(96h\right) & \mbox{O} \\ 
\mathbf{B}_{38} & = & \left(\frac{1}{2} - x_{4} - y_{4}\right) \, \mathbf{a}_{1} + \left(\frac{1}{2} - y_{4} + z_{4}\right) \, \mathbf{a}_{2} + \left(-x_{4}+z_{4}\right) \, \mathbf{a}_{3} & = & z_{4}a \, \mathbf{\hat{x}}-x_{4}a \, \mathbf{\hat{y}} + \left(\frac{1}{2} - y_{4}\right)a \, \mathbf{\hat{z}} & \left(96h\right) & \mbox{O} \\ 
\mathbf{B}_{39} & = & \left(\frac{1}{2} - x_{4} + y_{4}\right) \, \mathbf{a}_{1} + \left(y_{4}-z_{4}\right) \, \mathbf{a}_{2} + \left(\frac{1}{2} - x_{4} - z_{4}\right) \, \mathbf{a}_{3} & = & -z_{4}a \, \mathbf{\hat{x}} + \left(\frac{1}{2} - x_{4}\right)a \, \mathbf{\hat{y}} + y_{4}a \, \mathbf{\hat{z}} & \left(96h\right) & \mbox{O} \\ 
\mathbf{B}_{40} & = & \left(x_{4}-y_{4}\right) \, \mathbf{a}_{1} + \left(\frac{1}{2} - y_{4} - z_{4}\right) \, \mathbf{a}_{2} + \left(\frac{1}{2} +x_{4} - z_{4}\right) \, \mathbf{a}_{3} & = & \left(\frac{1}{2} - z_{4}\right)a \, \mathbf{\hat{x}} + x_{4}a \, \mathbf{\hat{y}}-y_{4}a \, \mathbf{\hat{z}} & \left(96h\right) & \mbox{O} \\ 
\mathbf{B}_{41} & = & \left(x_{4}+z_{4}\right) \, \mathbf{a}_{1} + \left(x_{4}+y_{4}\right) \, \mathbf{a}_{2} + \left(y_{4}+z_{4}\right) \, \mathbf{a}_{3} & = & y_{4}a \, \mathbf{\hat{x}} + z_{4}a \, \mathbf{\hat{y}} + x_{4}a \, \mathbf{\hat{z}} & \left(96h\right) & \mbox{O} \\ 
\mathbf{B}_{42} & = & \left(-x_{4}+z_{4}\right) \, \mathbf{a}_{1} + \left(\frac{1}{2} - x_{4} - y_{4}\right) \, \mathbf{a}_{2} + \left(\frac{1}{2} - y_{4} + z_{4}\right) \, \mathbf{a}_{3} & = & \left(\frac{1}{2} - y_{4}\right)a \, \mathbf{\hat{x}} + z_{4}a \, \mathbf{\hat{y}}-x_{4}a \, \mathbf{\hat{z}} & \left(96h\right) & \mbox{O} \\ 
\mathbf{B}_{43} & = & \left(\frac{1}{2} - x_{4} - z_{4}\right) \, \mathbf{a}_{1} + \left(\frac{1}{2} - x_{4} + y_{4}\right) \, \mathbf{a}_{2} + \left(y_{4}-z_{4}\right) \, \mathbf{a}_{3} & = & y_{4}a \, \mathbf{\hat{x}}-z_{4}a \, \mathbf{\hat{y}} + \left(\frac{1}{2} - x_{4}\right)a \, \mathbf{\hat{z}} & \left(96h\right) & \mbox{O} \\ 
\mathbf{B}_{44} & = & \left(\frac{1}{2} +x_{4} - z_{4}\right) \, \mathbf{a}_{1} + \left(x_{4}-y_{4}\right) \, \mathbf{a}_{2} + \left(\frac{1}{2} - y_{4} - z_{4}\right) \, \mathbf{a}_{3} & = & -y_{4}a \, \mathbf{\hat{x}} + \left(\frac{1}{2} - z_{4}\right)a \, \mathbf{\hat{y}} + x_{4}a \, \mathbf{\hat{z}} & \left(96h\right) & \mbox{O} \\ 
\mathbf{B}_{45} & = & \left(\frac{1}{2} +x_{4} - z_{4}\right) \, \mathbf{a}_{1} + \left(y_{4}-z_{4}\right) \, \mathbf{a}_{2} + \left(x_{4}+y_{4}\right) \, \mathbf{a}_{3} & = & \left(\frac{3}{4} +y_{4}\right)a \, \mathbf{\hat{x}} + \left(\frac{1}{4} +x_{4}\right)a \, \mathbf{\hat{y}} + \left(\frac{1}{4} - z_{4}\right)a \, \mathbf{\hat{z}} & \left(96h\right) & \mbox{O} \\ 
\mathbf{B}_{46} & = & \left(\frac{1}{2} - x_{4} - z_{4}\right) \, \mathbf{a}_{1} + \left(\frac{1}{2} - y_{4} - z_{4}\right) \, \mathbf{a}_{2} + \left(\frac{1}{2} - x_{4} - y_{4}\right) \, \mathbf{a}_{3} & = & \left(\frac{1}{4} - y_{4}\right)a \, \mathbf{\hat{x}} + \left(\frac{1}{4} - x_{4}\right)a \, \mathbf{\hat{y}} + \left(\frac{1}{4} - z_{4}\right)a \, \mathbf{\hat{z}} & \left(96h\right) & \mbox{O} \\ 
\mathbf{B}_{47} & = & \left(-x_{4}+z_{4}\right) \, \mathbf{a}_{1} + \left(y_{4}+z_{4}\right) \, \mathbf{a}_{2} + \left(\frac{1}{2} - x_{4} + y_{4}\right) \, \mathbf{a}_{3} & = & \left(\frac{1}{4} +y_{4}\right)a \, \mathbf{\hat{x}} + \left(\frac{1}{4} - x_{4}\right)a \, \mathbf{\hat{y}} + \left(\frac{3}{4} +z_{4}\right)a \, \mathbf{\hat{z}} & \left(96h\right) & \mbox{O} \\ 
\mathbf{B}_{48} & = & \left(x_{4}+z_{4}\right) \, \mathbf{a}_{1} + \left(\frac{1}{2} - y_{4} + z_{4}\right) \, \mathbf{a}_{2} + \left(x_{4}-y_{4}\right) \, \mathbf{a}_{3} & = & \left(\frac{1}{4} - y_{4}\right)a \, \mathbf{\hat{x}} + \left(\frac{3}{4} +x_{4}\right)a \, \mathbf{\hat{y}} + \left(\frac{1}{4} +z_{4}\right)a \, \mathbf{\hat{z}} & \left(96h\right) & \mbox{O} \\ 
\mathbf{B}_{49} & = & \left(\frac{1}{2} - y_{4} + z_{4}\right) \, \mathbf{a}_{1} + \left(x_{4}-y_{4}\right) \, \mathbf{a}_{2} + \left(x_{4}+z_{4}\right) \, \mathbf{a}_{3} & = & \left(\frac{3}{4} +x_{4}\right)a \, \mathbf{\hat{x}} + \left(\frac{1}{4} +z_{4}\right)a \, \mathbf{\hat{y}} + \left(\frac{1}{4} - y_{4}\right)a \, \mathbf{\hat{z}} & \left(96h\right) & \mbox{O} \\ 
\mathbf{B}_{50} & = & \left(y_{4}+z_{4}\right) \, \mathbf{a}_{1} + \left(\frac{1}{2} - x_{4} + y_{4}\right) \, \mathbf{a}_{2} + \left(-x_{4}+z_{4}\right) \, \mathbf{a}_{3} & = & \left(\frac{1}{4} - x_{4}\right)a \, \mathbf{\hat{x}} + \left(\frac{3}{4} +z_{4}\right)a \, \mathbf{\hat{y}} + \left(\frac{1}{4} +y_{4}\right)a \, \mathbf{\hat{z}} & \left(96h\right) & \mbox{O} \\ 
\mathbf{B}_{51} & = & \left(\frac{1}{2} - y_{4} - z_{4}\right) \, \mathbf{a}_{1} + \left(\frac{1}{2} - x_{4} - y_{4}\right) \, \mathbf{a}_{2} + \left(\frac{1}{2} - x_{4} - z_{4}\right) \, \mathbf{a}_{3} & = & \left(\frac{1}{4} - x_{4}\right)a \, \mathbf{\hat{x}} + \left(\frac{1}{4} - z_{4}\right)a \, \mathbf{\hat{y}} + \left(\frac{1}{4} - y_{4}\right)a \, \mathbf{\hat{z}} & \left(96h\right) & \mbox{O} \\ 
\mathbf{B}_{52} & = & \left(y_{4}-z_{4}\right) \, \mathbf{a}_{1} + \left(x_{4}+y_{4}\right) \, \mathbf{a}_{2} + \left(\frac{1}{2} +x_{4} - z_{4}\right) \, \mathbf{a}_{3} & = & \left(\frac{1}{4} +x_{4}\right)a \, \mathbf{\hat{x}} + \left(\frac{1}{4} - z_{4}\right)a \, \mathbf{\hat{y}} + \left(\frac{3}{4} +y_{4}\right)a \, \mathbf{\hat{z}} & \left(96h\right) & \mbox{O} \\ 
\mathbf{B}_{53} & = & \left(\frac{1}{2} - x_{4} + y_{4}\right) \, \mathbf{a}_{1} + \left(-x_{4}+z_{4}\right) \, \mathbf{a}_{2} + \left(y_{4}+z_{4}\right) \, \mathbf{a}_{3} & = & \left(\frac{3}{4} +z_{4}\right)a \, \mathbf{\hat{x}} + \left(\frac{1}{4} +y_{4}\right)a \, \mathbf{\hat{y}} + \left(\frac{1}{4} - x_{4}\right)a \, \mathbf{\hat{z}} & \left(96h\right) & \mbox{O} \\ 
\mathbf{B}_{54} & = & \left(x_{4}-y_{4}\right) \, \mathbf{a}_{1} + \left(x_{4}+z_{4}\right) \, \mathbf{a}_{2} + \left(\frac{1}{2} - y_{4} + z_{4}\right) \, \mathbf{a}_{3} & = & \left(\frac{1}{4} +z_{4}\right)a \, \mathbf{\hat{x}} + \left(\frac{1}{4} - y_{4}\right)a \, \mathbf{\hat{y}} + \left(\frac{3}{4} +x_{4}\right)a \, \mathbf{\hat{z}} & \left(96h\right) & \mbox{O} \\ 
\mathbf{B}_{55} & = & \left(x_{4}+y_{4}\right) \, \mathbf{a}_{1} + \left(\frac{1}{2} +x_{4} - z_{4}\right) \, \mathbf{a}_{2} + \left(y_{4}-z_{4}\right) \, \mathbf{a}_{3} & = & \left(\frac{1}{4} - z_{4}\right)a \, \mathbf{\hat{x}} + \left(\frac{3}{4} +y_{4}\right)a \, \mathbf{\hat{y}} + \left(\frac{1}{4} +x_{4}\right)a \, \mathbf{\hat{z}} & \left(96h\right) & \mbox{O} \\ 
\mathbf{B}_{56} & = & \left(\frac{1}{2} - x_{4} - y_{4}\right) \, \mathbf{a}_{1} + \left(\frac{1}{2} - x_{4} - z_{4}\right) \, \mathbf{a}_{2} + \left(\frac{1}{2} - y_{4} - z_{4}\right) \, \mathbf{a}_{3} & = & \left(\frac{1}{4} - z_{4}\right)a \, \mathbf{\hat{x}} + \left(\frac{1}{4} - y_{4}\right)a \, \mathbf{\hat{y}} + \left(\frac{1}{4} - x_{4}\right)a \, \mathbf{\hat{z}} & \left(96h\right) & \mbox{O} \\ 
\mathbf{B}_{57} & = & \left(-y_{4}-z_{4}\right) \, \mathbf{a}_{1} + \left(-x_{4}-z_{4}\right) \, \mathbf{a}_{2} + \left(-x_{4}-y_{4}\right) \, \mathbf{a}_{3} & = & -x_{4}a \, \mathbf{\hat{x}}-y_{4}a \, \mathbf{\hat{y}}-z_{4}a \, \mathbf{\hat{z}} & \left(96h\right) & \mbox{O} \\ 
\mathbf{B}_{58} & = & \left(\frac{1}{2} +y_{4} - z_{4}\right) \, \mathbf{a}_{1} + \left(x_{4}-z_{4}\right) \, \mathbf{a}_{2} + \left(\frac{1}{2} +x_{4} + y_{4}\right) \, \mathbf{a}_{3} & = & x_{4}a \, \mathbf{\hat{x}} + \left(\frac{1}{2} +y_{4}\right)a \, \mathbf{\hat{y}}-z_{4}a \, \mathbf{\hat{z}} & \left(96h\right) & \mbox{O} \\ 
\mathbf{B}_{59} & = & \left(-y_{4}+z_{4}\right) \, \mathbf{a}_{1} + \left(\frac{1}{2} +x_{4} + z_{4}\right) \, \mathbf{a}_{2} + \left(\frac{1}{2} +x_{4} - y_{4}\right) \, \mathbf{a}_{3} & = & \left(\frac{1}{2} +x_{4}\right)a \, \mathbf{\hat{x}}-y_{4}a \, \mathbf{\hat{y}} + z_{4}a \, \mathbf{\hat{z}} & \left(96h\right) & \mbox{O} \\ 
\mathbf{B}_{60} & = & \left(\frac{1}{2} +y_{4} + z_{4}\right) \, \mathbf{a}_{1} + \left(\frac{1}{2} - x_{4} + z_{4}\right) \, \mathbf{a}_{2} + \left(-x_{4}+y_{4}\right) \, \mathbf{a}_{3} & = & -x_{4}a \, \mathbf{\hat{x}} + y_{4}a \, \mathbf{\hat{y}} + \left(\frac{1}{2} +z_{4}\right)a \, \mathbf{\hat{z}} & \left(96h\right) & \mbox{O} \\ 
\mathbf{B}_{61} & = & \left(-x_{4}-y_{4}\right) \, \mathbf{a}_{1} + \left(-y_{4}-z_{4}\right) \, \mathbf{a}_{2} + \left(-x_{4}-z_{4}\right) \, \mathbf{a}_{3} & = & -z_{4}a \, \mathbf{\hat{x}}-x_{4}a \, \mathbf{\hat{y}}-y_{4}a \, \mathbf{\hat{z}} & \left(96h\right) & \mbox{O} \\ 
\mathbf{B}_{62} & = & \left(\frac{1}{2} +x_{4} + y_{4}\right) \, \mathbf{a}_{1} + \left(\frac{1}{2} +y_{4} - z_{4}\right) \, \mathbf{a}_{2} + \left(x_{4}-z_{4}\right) \, \mathbf{a}_{3} & = & -z_{4}a \, \mathbf{\hat{x}} + x_{4}a \, \mathbf{\hat{y}} + \left(\frac{1}{2} +y_{4}\right)a \, \mathbf{\hat{z}} & \left(96h\right) & \mbox{O} \\ 
\mathbf{B}_{63} & = & \left(\frac{1}{2} +x_{4} - y_{4}\right) \, \mathbf{a}_{1} + \left(-y_{4}+z_{4}\right) \, \mathbf{a}_{2} + \left(\frac{1}{2} +x_{4} + z_{4}\right) \, \mathbf{a}_{3} & = & z_{4}a \, \mathbf{\hat{x}} + \left(\frac{1}{2} +x_{4}\right)a \, \mathbf{\hat{y}}-y_{4}a \, \mathbf{\hat{z}} & \left(96h\right) & \mbox{O} \\ 
\mathbf{B}_{64} & = & \left(-x_{4}+y_{4}\right) \, \mathbf{a}_{1} + \left(\frac{1}{2} +y_{4} + z_{4}\right) \, \mathbf{a}_{2} + \left(\frac{1}{2} - x_{4} + z_{4}\right) \, \mathbf{a}_{3} & = & \left(\frac{1}{2} +z_{4}\right)a \, \mathbf{\hat{x}}-x_{4}a \, \mathbf{\hat{y}} + y_{4}a \, \mathbf{\hat{z}} & \left(96h\right) & \mbox{O} \\ 
\mathbf{B}_{65} & = & \left(-x_{4}-z_{4}\right) \, \mathbf{a}_{1} + \left(-x_{4}-y_{4}\right) \, \mathbf{a}_{2} + \left(-y_{4}-z_{4}\right) \, \mathbf{a}_{3} & = & -y_{4}a \, \mathbf{\hat{x}}-z_{4}a \, \mathbf{\hat{y}}-x_{4}a \, \mathbf{\hat{z}} & \left(96h\right) & \mbox{O} \\ 
\mathbf{B}_{66} & = & \left(x_{4}-z_{4}\right) \, \mathbf{a}_{1} + \left(\frac{1}{2} +x_{4} + y_{4}\right) \, \mathbf{a}_{2} + \left(\frac{1}{2} +y_{4} - z_{4}\right) \, \mathbf{a}_{3} & = & \left(\frac{1}{2} +y_{4}\right)a \, \mathbf{\hat{x}}-z_{4}a \, \mathbf{\hat{y}} + x_{4}a \, \mathbf{\hat{z}} & \left(96h\right) & \mbox{O} \\ 
\mathbf{B}_{67} & = & \left(\frac{1}{2} +x_{4} + z_{4}\right) \, \mathbf{a}_{1} + \left(\frac{1}{2} +x_{4} - y_{4}\right) \, \mathbf{a}_{2} + \left(-y_{4}+z_{4}\right) \, \mathbf{a}_{3} & = & -y_{4}a \, \mathbf{\hat{x}} + z_{4}a \, \mathbf{\hat{y}} + \left(\frac{1}{2} +x_{4}\right)a \, \mathbf{\hat{z}} & \left(96h\right) & \mbox{O} \\ 
\mathbf{B}_{68} & = & \left(\frac{1}{2} - x_{4} + z_{4}\right) \, \mathbf{a}_{1} + \left(-x_{4}+y_{4}\right) \, \mathbf{a}_{2} + \left(\frac{1}{2} +y_{4} + z_{4}\right) \, \mathbf{a}_{3} & = & y_{4}a \, \mathbf{\hat{x}} + \left(\frac{1}{2} +z_{4}\right)a \, \mathbf{\hat{y}}-x_{4}a \, \mathbf{\hat{z}} & \left(96h\right) & \mbox{O} \\ 
\mathbf{B}_{69} & = & \left(\frac{1}{2} - x_{4} + z_{4}\right) \, \mathbf{a}_{1} + \left(-y_{4}+z_{4}\right) \, \mathbf{a}_{2} + \left(-x_{4}-y_{4}\right) \, \mathbf{a}_{3} & = & -a\left(y_{4}+\frac{1}{4}\right) \, \mathbf{\hat{x}} + \left(\frac{1}{4} - x_{4}\right)a \, \mathbf{\hat{y}} + \left(\frac{1}{4} +z_{4}\right)a \, \mathbf{\hat{z}} & \left(96h\right) & \mbox{O} \\ 
\mathbf{B}_{70} & = & \left(\frac{1}{2} +x_{4} + z_{4}\right) \, \mathbf{a}_{1} + \left(\frac{1}{2} +y_{4} + z_{4}\right) \, \mathbf{a}_{2} + \left(\frac{1}{2} +x_{4} + y_{4}\right) \, \mathbf{a}_{3} & = & \left(\frac{1}{4} +y_{4}\right)a \, \mathbf{\hat{x}} + \left(\frac{1}{4} +x_{4}\right)a \, \mathbf{\hat{y}} + \left(\frac{1}{4} +z_{4}\right)a \, \mathbf{\hat{z}} & \left(96h\right) & \mbox{O} \\ 
\mathbf{B}_{71} & = & \left(x_{4}-z_{4}\right) \, \mathbf{a}_{1} + \left(-y_{4}-z_{4}\right) \, \mathbf{a}_{2} + \left(\frac{1}{2} +x_{4} - y_{4}\right) \, \mathbf{a}_{3} & = & \left(\frac{1}{4} - y_{4}\right)a \, \mathbf{\hat{x}} + \left(\frac{1}{4} +x_{4}\right)a \, \mathbf{\hat{y}}-a\left(z_{4}+\frac{1}{4}\right) \, \mathbf{\hat{z}} & \left(96h\right) & \mbox{O} \\ 
\mathbf{B}_{72} & = & \left(-x_{4}-z_{4}\right) \, \mathbf{a}_{1} + \left(\frac{1}{2} +y_{4} - z_{4}\right) \, \mathbf{a}_{2} + \left(-x_{4}+y_{4}\right) \, \mathbf{a}_{3} & = & \left(\frac{1}{4} +y_{4}\right)a \, \mathbf{\hat{x}}-a\left(x_{4}+\frac{1}{4}\right) \, \mathbf{\hat{y}} + \left(\frac{1}{4} - z_{4}\right)a \, \mathbf{\hat{z}} & \left(96h\right) & \mbox{O} \\ 
\mathbf{B}_{73} & = & \left(\frac{1}{2} +y_{4} - z_{4}\right) \, \mathbf{a}_{1} + \left(-x_{4}+y_{4}\right) \, \mathbf{a}_{2} + \left(-x_{4}-z_{4}\right) \, \mathbf{a}_{3} & = & -a\left(x_{4}+\frac{1}{4}\right) \, \mathbf{\hat{x}} + \left(\frac{1}{4} - z_{4}\right)a \, \mathbf{\hat{y}} + \left(\frac{1}{4} +y_{4}\right)a \, \mathbf{\hat{z}} & \left(96h\right) & \mbox{O} \\ 
\mathbf{B}_{74} & = & \left(-y_{4}-z_{4}\right) \, \mathbf{a}_{1} + \left(\frac{1}{2} +x_{4} - y_{4}\right) \, \mathbf{a}_{2} + \left(x_{4}-z_{4}\right) \, \mathbf{a}_{3} & = & \left(\frac{1}{4} +x_{4}\right)a \, \mathbf{\hat{x}}-a\left(z_{4}+\frac{1}{4}\right) \, \mathbf{\hat{y}} + \left(\frac{1}{4} - y_{4}\right)a \, \mathbf{\hat{z}} & \left(96h\right) & \mbox{O} \\ 
\mathbf{B}_{75} & = & \left(\frac{1}{2} +y_{4} + z_{4}\right) \, \mathbf{a}_{1} + \left(\frac{1}{2} +x_{4} + y_{4}\right) \, \mathbf{a}_{2} + \left(\frac{1}{2} +x_{4} + z_{4}\right) \, \mathbf{a}_{3} & = & \left(\frac{1}{4} +x_{4}\right)a \, \mathbf{\hat{x}} + \left(\frac{1}{4} +z_{4}\right)a \, \mathbf{\hat{y}} + \left(\frac{1}{4} +y_{4}\right)a \, \mathbf{\hat{z}} & \left(96h\right) & \mbox{O} \\ 
\mathbf{B}_{76} & = & \left(-y_{4}+z_{4}\right) \, \mathbf{a}_{1} + \left(-x_{4}-y_{4}\right) \, \mathbf{a}_{2} + \left(\frac{1}{2} - x_{4} + z_{4}\right) \, \mathbf{a}_{3} & = & \left(\frac{1}{4} - x_{4}\right)a \, \mathbf{\hat{x}} + \left(\frac{1}{4} +z_{4}\right)a \, \mathbf{\hat{y}}-a\left(y_{4}+\frac{1}{4}\right) \, \mathbf{\hat{z}} & \left(96h\right) & \mbox{O} \\ 
\mathbf{B}_{77} & = & \left(\frac{1}{2} +x_{4} - y_{4}\right) \, \mathbf{a}_{1} + \left(x_{4}-z_{4}\right) \, \mathbf{a}_{2} + \left(-y_{4}-z_{4}\right) \, \mathbf{a}_{3} & = & -a\left(z_{4}+\frac{1}{4}\right) \, \mathbf{\hat{x}} + \left(\frac{1}{4} - y_{4}\right)a \, \mathbf{\hat{y}} + \left(\frac{1}{4} +x_{4}\right)a \, \mathbf{\hat{z}} & \left(96h\right) & \mbox{O} \\ 
\mathbf{B}_{78} & = & \left(-x_{4}+y_{4}\right) \, \mathbf{a}_{1} + \left(-x_{4}-z_{4}\right) \, \mathbf{a}_{2} + \left(\frac{1}{2} +y_{4} - z_{4}\right) \, \mathbf{a}_{3} & = & \left(\frac{1}{4} - z_{4}\right)a \, \mathbf{\hat{x}} + \left(\frac{1}{4} +y_{4}\right)a \, \mathbf{\hat{y}}-a\left(x_{4}+\frac{1}{4}\right) \, \mathbf{\hat{z}} & \left(96h\right) & \mbox{O} \\ 
\mathbf{B}_{79} & = & \left(-x_{4}-y_{4}\right) \, \mathbf{a}_{1} + \left(\frac{1}{2} - x_{4} + z_{4}\right) \, \mathbf{a}_{2} + \left(-y_{4}+z_{4}\right) \, \mathbf{a}_{3} & = & \left(\frac{1}{4} +z_{4}\right)a \, \mathbf{\hat{x}}-a\left(y_{4}+\frac{1}{4}\right) \, \mathbf{\hat{y}} + \left(\frac{1}{4} - x_{4}\right)a \, \mathbf{\hat{z}} & \left(96h\right) & \mbox{O} \\ 
\mathbf{B}_{80} & = & \left(\frac{1}{2} +x_{4} + y_{4}\right) \, \mathbf{a}_{1} + \left(\frac{1}{2} +x_{4} + z_{4}\right) \, \mathbf{a}_{2} + \left(\frac{1}{2} +y_{4} + z_{4}\right) \, \mathbf{a}_{3} & = & \left(\frac{1}{4} +z_{4}\right)a \, \mathbf{\hat{x}} + \left(\frac{1}{4} +y_{4}\right)a \, \mathbf{\hat{y}} + \left(\frac{1}{4} +x_{4}\right)a \, \mathbf{\hat{z}} & \left(96h\right) & \mbox{O} \\ 
\end{longtabu}
\renewcommand{\arraystretch}{1.0}
\noindent \hrulefill
\\
\textbf{References:}
\vspace*{-0.25cm}
\begin{flushleft}
  - \bibentry{Ross_Am_Min_81_1996}. \\
\end{flushleft}
\textbf{Found in:}
\vspace*{-0.25cm}
\begin{flushleft}
  - \bibentry{Downs_Am_Min_88_2003}. \\
\end{flushleft}
\noindent \hrulefill
\\
\textbf{Geometry files:}
\\
\noindent  - CIF: pp. {\hyperref[A2B3C12D3_cI160_230_a_c_h_d_cif]{\pageref{A2B3C12D3_cI160_230_a_c_h_d_cif}}} \\
\noindent  - POSCAR: pp. {\hyperref[A2B3C12D3_cI160_230_a_c_h_d_poscar]{\pageref{A2B3C12D3_cI160_230_a_c_h_d_poscar}}} \\

\newpage
\twocolumn
\section*{\CIF\ and \POSCAR\ Files}
{\tiny
\noindent{\phantomsection\label{A2B_aP6_2_aei_i_cif}}
{\hyperref[A2B_aP6_2_aei_i]{H$_{2}$S (90~GPa): A2B\_aP6\_2\_aei\_i}} - CIF
\begin{lstlisting}[numbers=none,language={mylang}]
# CIF file 
data_findsym-output
_audit_creation_method FINDSYM

_chemical_name_mineral 'H2S'
_chemical_formula_sum 'H2 S'

loop_
_publ_author_name
 'Y. Li'
 'J. Hao'
 'H. Liu'
 'Y. Li'
 'Y. Ma'
_journal_name_full_name
;
 Journal of Chemical Physics
;
_journal_volume 140
_journal_year 2014
_journal_page_first 174712
_journal_page_last 174712
_publ_Section_title
;
 The metallization and superconductivity of dense hydrogen sulfide
;

_aflow_title 'H$_{2}$S (90~GPa) Structure'
_aflow_proto 'A2B_aP6_2_aei_i'
_aflow_params 'a,b/a,c/a,\alpha,\beta,\gamma,x_{3},y_{3},z_{3},x_{4},y_{4},z_{4}'
_aflow_params_values '2.7804,1.00438785786,1.53100273342,78.28,76.53,70.42,0.259,0.415,0.663,0.192,0.183,0.255'
_aflow_Strukturbericht 'None'
_aflow_Pearson 'aP6'

_symmetry_space_group_name_H-M "P -1"
_symmetry_Int_Tables_number 2
 
_cell_length_a    2.78040
_cell_length_b    2.79260
_cell_length_c    4.25680
_cell_angle_alpha 78.28000
_cell_angle_beta  76.53000
_cell_angle_gamma 70.42000
 
loop_
_space_group_symop_id
_space_group_symop_operation_xyz
1 x,y,z
2 -x,-y,-z
 
loop_
_atom_site_label
_atom_site_type_symbol
_atom_site_symmetry_multiplicity
_atom_site_Wyckoff_label
_atom_site_fract_x
_atom_site_fract_y
_atom_site_fract_z
_atom_site_occupancy
H1 H   1 a 0.00000 0.00000 0.00000 1.00000
H2 H   1 e 0.50000 0.50000 0.00000 1.00000
H3 H   2 i 0.25900 0.41500 0.66300 1.00000
S1 S   2 i 0.19200 0.18300 0.25500 1.00000
\end{lstlisting}
{\phantomsection\label{A2B_aP6_2_aei_i_poscar}}
{\hyperref[A2B_aP6_2_aei_i]{H$_{2}$S (90~GPa): A2B\_aP6\_2\_aei\_i}} - POSCAR
\begin{lstlisting}[numbers=none,language={mylang}]
A2B_aP6_2_aei_i & a,b/a,c/a,alpha,beta,gamma,x3,y3,z3,x4,y4,z4 --params=2.7804,1.00438785786,1.53100273342,78.28,76.53,70.42,0.259,0.415,0.663,0.192,0.183,0.255 & P-1 C_{i}^{1} #2 (aei^2) & aP6 & None & H2S & H2S & Y. Li et al., J. Chem. Phys. 140, 174712(2014)
   1.00000000000000
   2.78040000000000   0.00000000000000   0.00000000000000
   0.93586367780735   2.63111648099450   0.00000000000000
   0.99156281692297   0.56505957748759   4.10095807025448
     H     S
     4     2
Direct
   0.00000000000000   0.00000000000000   0.00000000000000    H   (1a)
   0.50000000000000   0.50000000000000   0.00000000000000    H   (1e)
   0.25900000000000   0.41500000000000   0.66300000000000    H   (2i)
  -0.25900000000000  -0.41500000000000  -0.66300000000000    H   (2i)
   0.19200000000000   0.18300000000000   0.25500000000000    S   (2i)
  -0.19200000000000  -0.18300000000000  -0.25500000000000    S   (2i)
\end{lstlisting}
{\phantomsection\label{A8B5_mP13_6_a7b_3a2b_cif}}
{\hyperref[A8B5_mP13_6_a7b_3a2b]{Mo$_{8}$P$_{5}$ (High-temperature): A8B5\_mP13\_6\_a7b\_3a2b}} - CIF

{\phantomsection\label{A8B5_mP13_6_a7b_3a2b_poscar}}
{\hyperref[A8B5_mP13_6_a7b_3a2b]{Mo$_{8}$P$_{5}$ (High-temperature): A8B5\_mP13\_6\_a7b\_3a2b}} - POSCAR
\begin{lstlisting}[numbers=none,language={mylang}]
A8B5_mP13_6_a7b_3a2b & a,b/a,c/a,beta,x1,z1,x2,z2,x3,z3,x4,z4,x5,z5,x6,z6,x7,z7,x8,z8,x9,z9,x10,z10,x11,z11,x12,z12,x13,z13 --params=6.5369054321,0.490874879533,1.43786810261,109.592,0.5119,0.7172,0.4546,0.4285,0.044,0.5911,0.37,0.0053,0.3887,0.2195,0.6277,0.0001,0.1818,0.4663,0.7456,0.5256,0.1826,0.7901,0.7824,0.2724,0.0,0.0,0.8188,0.8026,0.0123,0.2051 & Pm C_{s}^{1} #6 (a^4b^9) & mP13 & None & Mo8P5 &  & T. Johnsson, Acta Chem. Scand. 26, 365-382 (1972)
   1.00000000000000
   6.53690543210000   0.00000000000000   0.00000000000000
   0.00000000000000   3.20880266650000   0.00000000000000
  -3.15174264922271   0.00000000000000   8.85503392087883
    Mo     P
     8     5
Direct
   0.51190000000000   0.00000000000000   0.71720000000000   Mo   (1a)
   0.38870000000000   0.50000000000000   0.21950000000000   Mo   (1b)
   0.62770000000000   0.50000000000000   0.00010000000000   Mo   (1b)
   0.18180000000000   0.50000000000000   0.46630000000000   Mo   (1b)
   0.74560000000000   0.50000000000000   0.52560000000000   Mo   (1b)
   0.18260000000000   0.50000000000000   0.79010000000000   Mo   (1b)
   0.78240000000000   0.50000000000000   0.27240000000000   Mo   (1b)
   0.00000000000000   0.50000000000000   0.00000000000000   Mo   (1b)
   0.45460000000000   0.00000000000000   0.42850000000000    P   (1a)
   0.04400000000000   0.00000000000000   0.59110000000000    P   (1a)
   0.37000000000000   0.00000000000000   0.00530000000000    P   (1a)
   0.81880000000000   0.50000000000000   0.80260000000000    P   (1b)
   0.01230000000000   0.50000000000000   0.20510000000000    P   (1b)
\end{lstlisting}
{\phantomsection\label{AB_mP4_6_2b_2a_cif}}
{\hyperref[AB_mP4_6_2b_2a]{FeNi: AB\_mP4\_6\_2b\_2a}} - CIF
\begin{lstlisting}[numbers=none,language={mylang}]
# CIF file
data_findsym-output
_audit_creation_method FINDSYM

_chemical_name_mineral 'FeNi'
_chemical_formula_sum 'Fe Ni'

loop_
_publ_author_name
 'T. Tagai'
 'H. Takeda'
 'T. Fukuda'
_journal_name_full_name
;
 Zeitschrift f{\"u}r Kristallographie - Crystalline Materials
;
_journal_volume 210
_journal_year 1995
_journal_page_first 14
_journal_page_last 18
_publ_Section_title
;
 Superstructure of tetrataenite from the Saint Severin meteorite
;

# Found in Pearson's Crystal Data - Crystal Structure Database for Inorganic Compounds, 2013

_aflow_title 'FeNi Structure'
_aflow_proto 'AB_mP4_6_2b_2a'
_aflow_params 'a,b/a,c/a,\beta,x_{1},z_{1},x_{2},z_{2},x_{3},z_{3},x_{4},z_{4}'
_aflow_params_values '3.580975428,1.00027925162,1.00167550965,90.04,0.0,0.0,0.518,0.507,0.026,0.501,0.529,0.027'
_aflow_Strukturbericht 'None'
_aflow_Pearson 'mP4'

_cell_length_a    3.5809754280
_cell_length_b    3.5819754212
_cell_length_c    3.5869753869
_cell_angle_alpha 90.0000000000
_cell_angle_beta  90.0400000000
_cell_angle_gamma 90.0000000000
 
_symmetry_space_group_name_H-M "P 1 m 1"
_symmetry_Int_Tables_number 6
 
loop_
_space_group_symop_id
_space_group_symop_operation_xyz
1 x,y,z
2 x,-y,z
 
loop_
_atom_site_label
_atom_site_type_symbol
_atom_site_symmetry_multiplicity
_atom_site_Wyckoff_label
_atom_site_fract_x
_atom_site_fract_y
_atom_site_fract_z
_atom_site_occupancy
Ni1 Ni   1 a 0.00000 0.00000 0.00000 1.00000
Ni2 Ni   1 a 0.51800 0.00000 0.50700 1.00000
Fe1 Fe   1 b 0.02600 0.50000 0.50100 1.00000
Fe2 Fe   1 b 0.52900 0.50000 0.02700 1.00000
\end{lstlisting}
{\phantomsection\label{AB_mP4_6_2b_2a_poscar}}
{\hyperref[AB_mP4_6_2b_2a]{FeNi: AB\_mP4\_6\_2b\_2a}} - POSCAR
\begin{lstlisting}[numbers=none,language={mylang}]
AB_mP4_6_2b_2a & a,b/a,c/a,beta,x1,z1,x2,z2,x3,z3,x4,z4 --params=3.580975428,1.00027925162,1.00167550965,90.04,0.0,0.0,0.518,0.507,0.026,0.501,0.529,0.027 & Pm C_{s}^{1} #6 (a^2b^2) & mP4 & None & FeNi &  & T. Tagai and H. Takeda and T. Fukuda, Zeitschrift f"{u}r Kristallographie - Crystalline Materials 210, 14-18 (1995)
   1.00000000000000
   3.58097542800000   0.00000000000000   0.00000000000000
   0.00000000000000   3.58197542120000   0.00000000000000
  -0.00250418102416   0.00000000000000   3.58697451277589
    Fe    Ni
     2     2
Direct
   0.02600000000000   0.50000000000000   0.50100000000000   Fe   (1b)
   0.52900000000000   0.50000000000000   0.02700000000000   Fe   (1b)
   0.00000000000000   0.00000000000000   0.00000000000000   Ni   (1a)
   0.51800000000000   0.00000000000000   0.50700000000000   Ni   (1a)
\end{lstlisting}
{\phantomsection\label{A2B_mP12_7_4a_2a_cif}}
{\hyperref[A2B_mP12_7_4a_2a]{H$_{2}$S IV: A2B\_mP12\_7\_4a\_2a}} - CIF
\begin{lstlisting}[numbers=none,language={mylang}]
# CIF file 
data_findsym-output
_audit_creation_method FINDSYM

_chemical_name_mineral 'H2S IV'
_chemical_formula_sum 'H2 S'

loop_
_publ_author_name
 'Y. Li'
 'J. Hao'
 'H. Liu'
 'Y. Li'
 'Y. Ma'
_journal_name_full_name
;
 Journal of Chemical Physics
;
_journal_volume 140
_journal_year 2014
_journal_page_first 174712
_journal_page_last 174712
_publ_Section_title
;
 The metallization and superconductivity of dense hydrogen sulfide
;

_aflow_title 'H$_{2}$S IV Structure'
_aflow_proto 'A2B_mP12_7_4a_2a'
_aflow_params 'a,b/a,c/a,\beta,x_{1},y_{1},z_{1},x_{2},y_{2},z_{2},x_{3},y_{3},z_{3},x_{4},y_{4},z_{4},x_{5},y_{5},z_{5},x_{6},y_{6},z_{6}'
_aflow_params_values '5.0942,0.627360527659,1.04603274312,90.38,0.498,-0.062,0.472,-0.023,0.574,0.143,0.777,-0.052,0.799,0.271,0.151,0.261,-0.001,0.808,0.36,0.494,0.649,0.658'
_aflow_Strukturbericht 'None'
_aflow_Pearson 'mP12'

_symmetry_space_group_name_H-M "P 1 c 1"
_symmetry_Int_Tables_number 7
 
_cell_length_a    5.09420
_cell_length_b    3.19590
_cell_length_c    5.32870
_cell_angle_alpha 90.00000
_cell_angle_beta  90.38000
_cell_angle_gamma 90.00000
 
loop_
_space_group_symop_id
_space_group_symop_operation_xyz
1 x,y,z
2 x,-y,z+1/2
 
loop_
_atom_site_label
_atom_site_type_symbol
_atom_site_symmetry_multiplicity
_atom_site_Wyckoff_label
_atom_site_fract_x
_atom_site_fract_y
_atom_site_fract_z
_atom_site_occupancy
H1 H   2 a 0.49800  -0.06200 0.47200 1.00000
H2 H   2 a -0.02300 0.57400  0.14300 1.00000
H3 H   2 a 0.77700  -0.05200 0.79900 1.00000
H4 H   2 a 0.27100  0.15100  0.26100 1.00000
S1 S   2 a -0.00100 0.80800  0.36000 1.00000
S2 S   2 a 0.49400  0.64900  0.65800 1.00000
\end{lstlisting}
{\phantomsection\label{A2B_mP12_7_4a_2a_poscar}}
{\hyperref[A2B_mP12_7_4a_2a]{H$_{2}$S IV: A2B\_mP12\_7\_4a\_2a}} - POSCAR
\begin{lstlisting}[numbers=none,language={mylang}]
A2B_mP12_7_4a_2a & a,b/a,c/a,beta,x1,y1,z1,x2,y2,z2,x3,y3,z3,x4,y4,z4,x5,y5,z5,x6,y6,z6 --params=5.0942,0.627360527659,1.04603274312,90.38,0.498,-0.062,0.472,-0.023,0.574,0.143,0.777,-0.052,0.799,0.271,0.151,0.261,-0.001,0.808,0.36,0.494,0.649,0.658 & Pc C_{s}^{2} #7 (a^6) & mP12 & None & H2S & H2S IV & Y. Li et al., J. Chem. Phys. 140, 174712(2014)
   1.00000000000000
   5.09420000000000   0.00000000000000   0.00000000000000
   0.00000000000000   3.19590000000000   0.00000000000000
  -0.03534101765261   0.00000000000000   5.32858280431779
     H     S
     8     4
Direct
   0.49800000000000  -0.06200000000000   0.47200000000000    H   (2a)
   0.49800000000000   0.06200000000000   0.97200000000000    H   (2a)
  -0.02300000000000   0.57400000000000   0.14300000000000    H   (2a)
  -0.02300000000000  -0.57400000000000   0.64300000000000    H   (2a)
   0.77700000000000  -0.05200000000000   0.79900000000000    H   (2a)
   0.77700000000000   0.05200000000000   1.29900000000000    H   (2a)
   0.27100000000000   0.15100000000000   0.26100000000000    H   (2a)
   0.27100000000000  -0.15100000000000   0.76100000000000    H   (2a)
  -0.00100000000000   0.80800000000000   0.36000000000000    S   (2a)
  -0.00100000000000  -0.80800000000000   0.86000000000000    S   (2a)
   0.49400000000000   0.64900000000000   0.65800000000000    S   (2a)
   0.49400000000000  -0.64900000000000   1.15800000000000    S   (2a)
\end{lstlisting}
{\phantomsection\label{A2B_mP18_7_6a_3a_cif}}
{\hyperref[A2B_mP18_7_6a_3a]{As$_{2}$Ba: A2B\_mP18\_7\_6a\_3a}} - CIF

{\phantomsection\label{A2B_mP18_7_6a_3a_poscar}}
{\hyperref[A2B_mP18_7_6a_3a]{As$_{2}$Ba: A2B\_mP18\_7\_6a\_3a}} - POSCAR

{\phantomsection\label{A3B_mP16_7_6a_2a_cif}}
{\hyperref[A3B_mP16_7_6a_2a]{$\epsilon$-WO$_{3}$ (Low-temperature): A3B\_mP16\_7\_6a\_2a}} - CIF
\begin{lstlisting}[numbers=none,language={mylang}]
# CIF file
data_findsym-output
_audit_creation_method FINDSYM

_chemical_name_mineral 'epsilon-WO3'
_chemical_formula_sum 'O3 W'

loop_
_publ_author_name
 'P. M. Woodward'
 'A. W. Sleight'
 'T. Vogt'
_journal_name_full_name
;
 Journal of Solid State Chemistry
;
_journal_volume 131
_journal_year 1997
_journal_page_first 9
_journal_page_last 17
_publ_Section_title
;
 Ferroelectric tungsten trioxide
;

# Found in Pearson's Crystal Data - Crystal Structure Database for Inorganic Compounds, 2013

_aflow_title '$\epsilon$-WO$_{3}$ (Low-temperature) Structure'
_aflow_proto 'A3B_mP16_7_6a_2a'
_aflow_params 'a,b/a,c/a,\beta,x_{1},y_{1},z_{1},x_{2},y_{2},z_{2},x_{3},y_{3},z_{3},x_{4},y_{4},z_{4},x_{5},y_{5},z_{5},x_{6},y_{6},z_{6},x_{7},y_{7},z_{7},x_{8},y_{8},z_{8}'
_aflow_params_values '5.2778002048,0.97690325515,1.45210125431,91.762,0.5044,0.292,0.01,0.5764,0.215,0.586,0.0,0.209,0.0,0.0864,0.29,0.58,0.2874,0.0717,0.287,0.7924,0.4201,0.301,0.2874,0.014,0.0012,0.7994,0.528,0.078'
_aflow_Strukturbericht 'None'
_aflow_Pearson 'mP16'

_cell_length_a    5.2778002048
_cell_length_b    5.1559002001
_cell_length_c    7.6639002974
_cell_angle_alpha 90.0000000000
_cell_angle_beta  91.7620000000
_cell_angle_gamma 90.0000000000
 
_symmetry_space_group_name_H-M "P 1 c 1"
_symmetry_Int_Tables_number 7
 
loop_
_space_group_symop_id
_space_group_symop_operation_xyz
1 x,y,z
2 x,-y,z+1/2
 
loop_
_atom_site_label
_atom_site_type_symbol
_atom_site_symmetry_multiplicity
_atom_site_Wyckoff_label
_atom_site_fract_x
_atom_site_fract_y
_atom_site_fract_z
_atom_site_occupancy
O1 O   2 a 0.50440 0.29200 0.01000 1.00000
O2 O   2 a 0.57640 0.21500 0.58600 1.00000
O3 O   2 a 0.00000 0.20900 0.00000 1.00000
O4 O   2 a 0.08640 0.29000 0.58000 1.00000
O5 O   2 a 0.28740 0.07170 0.28700 1.00000
O6 O   2 a 0.79240 0.42010 0.30100 1.00000
W1 W   2 a 0.28740 0.01400 0.00120 1.00000
W2 W   2 a 0.79940 0.52800 0.07800 1.00000
\end{lstlisting}
{\phantomsection\label{A3B_mP16_7_6a_2a_poscar}}
{\hyperref[A3B_mP16_7_6a_2a]{$\epsilon$-WO$_{3}$ (Low-temperature): A3B\_mP16\_7\_6a\_2a}} - POSCAR
\begin{lstlisting}[numbers=none,language={mylang}]
A3B_mP16_7_6a_2a & a,b/a,c/a,beta,x1,y1,z1,x2,y2,z2,x3,y3,z3,x4,y4,z4,x5,y5,z5,x6,y6,z6,x7,y7,z7,x8,y8,z8 --params=5.2778002048,0.97690325515,1.45210125431,91.762,0.5044,0.292,0.01,0.5764,0.215,0.586,0.0,0.209,0.0,0.0864,0.29,0.58,0.2874,0.0717,0.287,0.7924,0.4201,0.301,0.2874,0.014,0.0012,0.7994,0.528,0.078 & Pc C_{s}^{2} #7 (a^8) & mP16 & None & WO3 & epsilon & P. M. Woodward and A. W. Sleight and T. Vogt, J. Solid State Chem. 131, 9-17 (1997)
   1.00000000000000
   5.27780020480000   0.00000000000000   0.00000000000000
   0.00000000000000   5.15590020010000   0.00000000000000
  -0.23564849020651   0.00000000000000   7.66027659797942
     O     W
    12     4
Direct
   0.50440000000000   0.29200000000000   0.01000000000000    O   (2a)
   0.50440000000000  -0.29200000000000   0.51000000000000    O   (2a)
   0.57640000000000   0.21500000000000   0.58600000000000    O   (2a)
   0.57640000000000  -0.21500000000000   1.08600000000000    O   (2a)
   0.00000000000000   0.20900000000000   0.00000000000000    O   (2a)
   0.00000000000000  -0.20900000000000   0.50000000000000    O   (2a)
   0.08640000000000   0.29000000000000   0.58000000000000    O   (2a)
   0.08640000000000  -0.29000000000000   1.08000000000000    O   (2a)
   0.28740000000000   0.07170000000000   0.28700000000000    O   (2a)
   0.28740000000000  -0.07170000000000   0.78700000000000    O   (2a)
   0.79240000000000   0.42010000000000   0.30100000000000    O   (2a)
   0.79240000000000  -0.42010000000000   0.80100000000000    O   (2a)
   0.28740000000000   0.01400000000000   0.00120000000000    W   (2a)
   0.28740000000000  -0.01400000000000   0.50120000000000    W   (2a)
   0.79940000000000   0.52800000000000   0.07800000000000    W   (2a)
   0.79940000000000  -0.52800000000000   0.57800000000000    W   (2a)
\end{lstlisting}
{\phantomsection\label{A9B2_mP22_7_9a_2a_cif}}
{\hyperref[A9B2_mP22_7_9a_2a]{Rh$_{2}$Ga$_{9}$: A9B2\_mP22\_7\_9a\_2a}} - CIF

{\phantomsection\label{A9B2_mP22_7_9a_2a_poscar}}
{\hyperref[A9B2_mP22_7_9a_2a]{Rh$_{2}$Ga$_{9}$: A9B2\_mP22\_7\_9a\_2a}} - POSCAR

{\phantomsection\label{A5B3_mC32_9_5a_3a_cif}}
{\hyperref[A5B3_mC32_9_5a_3a]{$\alpha$-P$_3$N$_5$: A5B3\_mC32\_9\_5a\_3a}} - CIF
\begin{lstlisting}[numbers=none,language={mylang}]
# CIF file
data_findsym-output
_audit_creation_method FINDSYM

_chemical_name_mineral '$\alpha$-P3N5'
_chemical_formula_sum 'N5 P3'

loop_
_publ_author_name
 'S. Horstmann'
 'E. Irran'
 'W. Schnick'
_journal_name_full_name
;
 Angewandte Chemie (International ed.)
;
_journal_volume 36
_journal_year 1997
_journal_page_first 1873
_journal_page_last 1875
_publ_Section_title
;
 Synthesis and Crystal Structure of Phosphorus(V) Nitride $\alpha$-P$_{3}$N$_{5}$
;

_aflow_title '$\alpha$-P$_3$N$_5$ Structure'
_aflow_proto 'A5B3_mC32_9_5a_3a'
_aflow_params 'a,b/a,c/a,\beta,x_{1},y_{1},z_{1},x_{2},y_{2},z_{2},x_{3},y_{3},z_{3},x_{4},y_{4},z_{4},x_{5},y_{5},z_{5},x_{6},y_{6},z_{6},x_{7},y_{7},z_{7},x_{8},y_{8},z_{8}'
_aflow_params_values '8.12077,0.718445418353,1.12797801194,115.809,0.009,-0.003,0.269,0.129,0.341,0.45,0.37,0.119,0.066,0.142,0.351,0.147,0.356,0.135,0.348,0.0,0.5182,0.0,0.136,0.2,0.309,0.365,0.2924,0.196'
_aflow_Strukturbericht 'None'
_aflow_Pearson 'mC32'

_symmetry_space_group_name_H-M "C 1 c 1"
_symmetry_Int_Tables_number 9
 
_cell_length_a    8.12077
_cell_length_b    5.83433
_cell_length_c    9.16005
_cell_angle_alpha 90.00000
_cell_angle_beta  115.80900
_cell_angle_gamma 90.00000
 
loop_
_space_group_symop_id
_space_group_symop_operation_xyz
1 x,y,z
2 x,-y,z+1/2
3 x+1/2,y+1/2,z
4 x+1/2,-y+1/2,z+1/2
 
loop_
_atom_site_label
_atom_site_type_symbol
_atom_site_symmetry_multiplicity
_atom_site_Wyckoff_label
_atom_site_fract_x
_atom_site_fract_y
_atom_site_fract_z
_atom_site_occupancy
N1 N   4 a 0.00900 -0.00300 0.26900 1.00000
N2 N   4 a 0.12900 0.34100 0.45000 1.00000
N3 N   4 a 0.37000 0.11900 0.06600 1.00000
N4 N   4 a 0.14200 0.35100 0.14700 1.00000
N5 N   4 a 0.35600 0.13500 0.34800 1.00000
P1 P   4 a 0.00000 0.51820 0.00000 1.00000
P2 P   4 a 0.13600 0.20000 0.30900 1.00000
P3 P   4 a 0.36500 0.29240 0.19600 1.00000
\end{lstlisting}
{\phantomsection\label{A5B3_mC32_9_5a_3a_poscar}}
{\hyperref[A5B3_mC32_9_5a_3a]{$\alpha$-P$_3$N$_5$: A5B3\_mC32\_9\_5a\_3a}} - POSCAR
\begin{lstlisting}[numbers=none,language={mylang}]
A5B3_mC32_9_5a_3a & a,b/a,c/a,beta,x1,y1,z1,x2,y2,z2,x3,y3,z3,x4,y4,z4,x5,y5,z5,x6,y6,z6,x7,y7,z7,x8,y8,z8 --params=8.12077,0.718445418353,1.12797801194,115.809,0.009,-0.003,0.269,0.129,0.341,0.45,0.37,0.119,0.066,0.142,0.351,0.147,0.356,0.135,0.348,0.0,0.5182,0.0,0.136,0.2,0.309,0.365,0.2924,0.196 & Cc C_{s}^{4} #9 (a^8) & mC32 & None & P3N5 & $\alpha$-P3N5 & S. Horstmann and E. Irran and W. Schnick, Angew. Chem. Int. Ed. 36, 1873-1875 (1997)
   1.00000000000000
   4.06038500000000  -2.91716500000000   0.00000000000000
   4.06038500000000   2.91716500000000   0.00000000000000
  -3.98803401284269   0.00000000000000   8.24633862480252
     N     P
    10     6
Direct
   0.01200000000000   0.00600000000000   0.26900000000000    N   (4a)
   0.00600000000000   0.01200000000000   0.76900000000000    N   (4a)
  -0.21200000000000   0.47000000000000   0.45000000000000    N   (4a)
   0.47000000000000  -0.21200000000000   0.95000000000000    N   (4a)
   0.25100000000000   0.48900000000000   0.06600000000000    N   (4a)
   0.48900000000000   0.25100000000000   0.56600000000000    N   (4a)
  -0.20900000000000   0.49300000000000   0.14700000000000    N   (4a)
   0.49300000000000  -0.20900000000000   0.64700000000000    N   (4a)
   0.22100000000000   0.49100000000000   0.34800000000000    N   (4a)
   0.49100000000000   0.22100000000000   0.84800000000000    N   (4a)
  -0.51820000000000   0.51820000000000   0.00000000000000    P   (4a)
   0.51820000000000  -0.51820000000000   0.50000000000000    P   (4a)
  -0.06400000000000   0.33600000000000   0.30900000000000    P   (4a)
   0.33600000000000  -0.06400000000000   0.80900000000000    P   (4a)
   0.07260000000000   0.65740000000000   0.19600000000000    P   (4a)
   0.65740000000000   0.07260000000000   0.69600000000000    P   (4a)
\end{lstlisting}
{\phantomsection\label{AB3_mC16_9_a_3a_cif}}
{\hyperref[AB3_mC16_9_a_3a]{H$_{3}$Cl (20~GPa): AB3\_mC16\_9\_a\_3a}} - CIF
\begin{lstlisting}[numbers=none,language={mylang}]
# CIF file 
data_findsym-output
_audit_creation_method FINDSYM

_chemical_name_mineral 'H3Cl'
_chemical_formula_sum 'Cl H3'

loop_
_publ_author_name
 'D. Duan'
 'X. Huang'
 'F. Tian'
 'Y. Liu'
 'Da Li'
 'H. Yu'
 'B. Liu'
 'W. Tian'
 'T. Cui'
_journal_name_full_name
;
 Journal of Physical Chemistry A
;
_journal_volume 119
_journal_year 2015
_journal_page_first 11059
_journal_page_last 11065
_publ_Section_title
;
 Predicted Formation of H$_{3}^{+}$ in Solid Halogen Polyhydrides at High Pressures
;

_aflow_title 'H$_{3}$Cl (20~GPa) Structure'
_aflow_proto 'AB3_mC16_9_a_3a'
_aflow_params 'a,b/a,c/a,\beta,x_{1},y_{1},z_{1},x_{2},y_{2},z_{2},x_{3},y_{3},z_{3},x_{4},y_{4},z_{4}'
_aflow_params_values '3.5367,2.67777872028,0.986823875364,93.018,0.50691,0.35551,0.49997,0.28659,0.26502,0.27913,0.57076,0.06256,0.41996,0.42173,0.07037,0.56036'
_aflow_Strukturbericht 'None'
_aflow_Pearson 'mC16'

_symmetry_space_group_name_H-M "C 1 c 1"
_symmetry_Int_Tables_number 9
 
_cell_length_a    3.53670
_cell_length_b    9.47050
_cell_length_c    3.49010
_cell_angle_alpha 90.00000
_cell_angle_beta  93.01800
_cell_angle_gamma 90.00000
 
loop_
_space_group_symop_id
_space_group_symop_operation_xyz
1 x,y,z
2 x,-y,z+1/2
3 x+1/2,y+1/2,z
4 x+1/2,-y+1/2,z+1/2
 
loop_
_atom_site_label
_atom_site_type_symbol
_atom_site_symmetry_multiplicity
_atom_site_Wyckoff_label
_atom_site_fract_x
_atom_site_fract_y
_atom_site_fract_z
_atom_site_occupancy
Cl1 Cl   4 a 0.50691 0.35551 0.49997 1.00000
H1  H    4 a 0.28659 0.26502 0.27913 1.00000
H2  H    4 a 0.57076 0.06256 0.41996 1.00000
H3  H    4 a 0.42173 0.07037 0.56036 1.00000
\end{lstlisting}
{\phantomsection\label{AB3_mC16_9_a_3a_poscar}}
{\hyperref[AB3_mC16_9_a_3a]{H$_{3}$Cl (20~GPa): AB3\_mC16\_9\_a\_3a}} - POSCAR
\begin{lstlisting}[numbers=none,language={mylang}]
AB3_mC16_9_a_3a & a,b/a,c/a,beta,x1,y1,z1,x2,y2,z2,x3,y3,z3,x4,y4,z4 --params=3.5367,2.67777872028,0.986823875364,93.018,0.50691,0.35551,0.49997,0.28659,0.26502,0.27913,0.57076,0.06256,0.41996,0.42173,0.07037,0.56036 & Cc C_{s}^{4} #9 (a^4) & mC16 & None & H3Cl & H3Cl & D. Duan et al., J. Phys. Chem. A 119, 11059-11065 (2015)
   1.00000000000000
   1.76835000000000  -4.73525000000000   0.00000000000000
   1.76835000000000   4.73525000000000   0.00000000000000
  -0.18375265646087   0.00000000000000   3.48525938363898
    Cl     H
     2     6
Direct
   0.15140000000000   0.86242000000000   0.49997000000000   Cl   (4a)
   0.86242000000000   0.15140000000000   0.99997000000000   Cl   (4a)
   0.02157000000000   0.55161000000000   0.27913000000000    H   (4a)
   0.55161000000000   0.02157000000000   0.77913000000000    H   (4a)
   0.50820000000000   0.63332000000000   0.41996000000000    H   (4a)
   0.63332000000000   0.50820000000000   0.91996000000000    H   (4a)
   0.35136000000000   0.49210000000000   0.56036000000000    H   (4a)
   0.49210000000000   0.35136000000000   1.06036000000000    H   (4a)
\end{lstlisting}
{\phantomsection\label{A2B_mP6_10_mn_bg_cif}}
{\hyperref[A2B_mP6_10_mn_bg]{$\delta$-PdCl$_{2}$: A2B\_mP6\_10\_mn\_bg}} - CIF
\begin{lstlisting}[numbers=none,language={mylang}]
# CIF file 
data_findsym-output
_audit_creation_method FINDSYM

_chemical_name_mineral '$\delta$-PdCl2'
_chemical_formula_sum 'Cl2 Pd'

loop_
_publ_author_name
 'J. Evers'
 'W. Beck'
 'M. G\"{o}bel'
 'S. Jakob'
 'P. Mayer'
 'G. Oehlinger'
 'M. Rotter'
 'T. M. Klap\"{o}tke'
_journal_name_full_name
;
 Angewandte Chemie (International ed.)
;
_journal_volume 49
_journal_year 2010
_journal_page_first 5677
_journal_page_last 5682
_publ_Section_title
;
 The Structures of $\delta$-PdCl$_{2}$ and $\gamma$-PdCl$_{2}$: Phases with Negative Thermal Expansion in One Direction
;

_aflow_title '$\delta$-PdCl$_{2}$ Structure'
_aflow_proto 'A2B_mP6_10_mn_bg'
_aflow_params 'a,b/a,c/a,\beta,x_{3},z_{3},x_{4},z_{4}'
_aflow_params_values '4.012,0.819541375872,2.93668993021,97.03,0.843,0.126,0.558,0.644'
_aflow_Strukturbericht 'None'
_aflow_Pearson 'mP6'

_symmetry_space_group_name_H-M "P 1 2/m 1"
_symmetry_Int_Tables_number 10
 
_cell_length_a    4.01200
_cell_length_b    3.28800
_cell_length_c    11.78200
_cell_angle_alpha 90.00000
_cell_angle_beta  97.03000
_cell_angle_gamma 90.00000
 
loop_
_space_group_symop_id
_space_group_symop_operation_xyz
1 x,y,z
2 -x,y,-z
3 -x,-y,-z
4 x,-y,z
 
loop_
_atom_site_label
_atom_site_type_symbol
_atom_site_symmetry_multiplicity
_atom_site_Wyckoff_label
_atom_site_fract_x
_atom_site_fract_y
_atom_site_fract_z
_atom_site_occupancy
Pd1 Pd   1 b 0.00000 0.50000 0.00000 1.00000
Pd2 Pd   1 g 0.50000 0.00000 0.50000 1.00000
Cl1 Cl   2 m 0.84300 0.00000 0.12600 1.00000
Cl2 Cl   2 n 0.55800 0.50000 0.64400 1.00000
\end{lstlisting}
{\phantomsection\label{A2B_mP6_10_mn_bg_poscar}}
{\hyperref[A2B_mP6_10_mn_bg]{$\delta$-PdCl$_{2}$: A2B\_mP6\_10\_mn\_bg}} - POSCAR
\begin{lstlisting}[numbers=none,language={mylang}]
A2B_mP6_10_mn_bg & a,b/a,c/a,beta,x3,z3,x4,z4 --params=4.012,0.819541375872,2.93668993021,97.03,0.843,0.126,0.558,0.644 & P2/m C_{2h}^{1} #10 (bgmn) & mP6 & None & PdCl2 & $\delta$-PdCl2 & J. Evers et al., Angew. Chem. Int. Ed. 49, 5677-5682 (2010)
   1.00000000000000
   4.01200000000000   0.00000000000000   0.00000000000000
   0.00000000000000   3.28800000000000   0.00000000000000
  -1.44198746457411   0.00000000000000  11.69342533871110
    Cl    Pd
     4     2
Direct
   0.84300000000000   0.00000000000000   0.12600000000000   Cl   (2m)
  -0.84300000000000   0.00000000000000  -0.12600000000000   Cl   (2m)
   0.55800000000000   0.50000000000000   0.64400000000000   Cl   (2n)
  -0.55800000000000   0.50000000000000  -0.64400000000000   Cl   (2n)
   0.00000000000000   0.50000000000000   0.00000000000000   Pd   (1b)
   0.50000000000000   0.00000000000000   0.50000000000000   Pd   (1g)
\end{lstlisting}
{\phantomsection\label{AB3_mP16_10_mn_3m3n_cif}}
{\hyperref[AB3_mP16_10_mn_3m3n]{H$_{3}$Cl (400~GPa): AB3\_mP16\_10\_mn\_3m3n}} - CIF
\begin{lstlisting}[numbers=none,language={mylang}]
# CIF file 
data_findsym-output
_audit_creation_method FINDSYM

_chemical_name_mineral 'H3Cl'
_chemical_formula_sum 'Cl H3'

loop_
_publ_author_name
 'Q. Zeng'
 'S. Yu'
 'D. Li'
 'A. R. Oganov'
 'G. Frapper'
_journal_name_full_name
;
 Physical Chemistry Chemical Physics
;
_journal_volume 19
_journal_year 2017
_journal_page_first 8236
_journal_page_last 8242
_publ_Section_title
;
 Emergence of novel hydrogen chlorides under high pressure
;

_aflow_title 'H$_{3}$Cl (400~GPa) Structure'
_aflow_proto 'AB3_mP16_10_mn_3m3n'
_aflow_params 'a,b/a,c/a,\beta,x_{1},z_{1},x_{2},z_{2},x_{3},z_{3},x_{4},z_{4},x_{5},z_{5},x_{6},z_{6},x_{7},z_{7},x_{8},z_{8}'
_aflow_params_values '3.678,0.71098423056,1.23817292007,90.0,0.263,0.339,0.08,0.059,0.795,0.341,0.438,-0.069,0.763,0.161,0.705,0.841,0.58,0.441,0.062,0.431'
_aflow_Strukturbericht 'None'
_aflow_Pearson 'mP16'

_symmetry_space_group_name_H-M "P 1 2/m 1"
_symmetry_Int_Tables_number 10
 
_cell_length_a    3.67800
_cell_length_b    2.61500
_cell_length_c    4.55400
_cell_angle_alpha 90.00000
_cell_angle_beta  90.00000
_cell_angle_gamma 90.00000
 
loop_
_space_group_symop_id
_space_group_symop_operation_xyz
1 x,y,z
2 -x,y,-z
3 -x,-y,-z
4 x,-y,z
 
loop_
_atom_site_label
_atom_site_type_symbol
_atom_site_symmetry_multiplicity
_atom_site_Wyckoff_label
_atom_site_fract_x
_atom_site_fract_y
_atom_site_fract_z
_atom_site_occupancy
Cl1 Cl   2 m 0.26300 0.00000 0.33900  1.00000
H1  H    2 m 0.08000 0.00000 0.05900  1.00000
H2  H    2 m 0.79500 0.00000 0.34100  1.00000
H3  H    2 m 0.43800 0.00000 -0.06900 1.00000
Cl2 Cl   2 n 0.76300 0.50000 0.16100  1.00000
H4  H    2 n 0.70500 0.50000 0.84100  1.00000
H5  H    2 n 0.58000 0.50000 0.44100  1.00000
H6  H    2 n 0.06200 0.50000 0.43100  1.00000
\end{lstlisting}
{\phantomsection\label{AB3_mP16_10_mn_3m3n_poscar}}
{\hyperref[AB3_mP16_10_mn_3m3n]{H$_{3}$Cl (400~GPa): AB3\_mP16\_10\_mn\_3m3n}} - POSCAR
\begin{lstlisting}[numbers=none,language={mylang}]
AB3_mP16_10_mn_3m3n & a,b/a,c/a,beta,x1,z1,x2,z2,x3,z3,x4,z4,x5,z5,x6,z6,x7,z7,x8,z8 --params=3.678,0.71098423056,1.23817292007,90.0,0.263,0.339,0.08,0.059,0.795,0.341,0.438,-0.069,0.763,0.161,0.705,0.841,0.58,0.441,0.062,0.431 & P2/m C_{2h}^{1} #10 (m^4n^4) & mP16 & None & H3Cl & H3Cl & Q. Zeng et al., Phys. Chem. Chem. Phys. 19, 8236-8242 (2017)
   1.00000000000000
   3.67800000000000   0.00000000000000   0.00000000000000
   0.00000000000000   2.61500000000000   0.00000000000000
   0.00000000000000   0.00000000000000   4.55400000000000
    Cl     H
     4    12
Direct
   0.26300000000000   0.00000000000000   0.33900000000000   Cl   (2m)
  -0.26300000000000   0.00000000000000  -0.33900000000000   Cl   (2m)
   0.76300000000000   0.50000000000000   0.16100000000000   Cl   (2n)
  -0.76300000000000   0.50000000000000  -0.16100000000000   Cl   (2n)
   0.08000000000000   0.00000000000000   0.05900000000000    H   (2m)
  -0.08000000000000   0.00000000000000  -0.05900000000000    H   (2m)
   0.79500000000000   0.00000000000000   0.34100000000000    H   (2m)
  -0.79500000000000   0.00000000000000  -0.34100000000000    H   (2m)
   0.43800000000000   0.00000000000000  -0.06900000000000    H   (2m)
  -0.43800000000000   0.00000000000000   0.06900000000000    H   (2m)
   0.70500000000000   0.50000000000000   0.84100000000000    H   (2n)
  -0.70500000000000   0.50000000000000  -0.84100000000000    H   (2n)
   0.58000000000000   0.50000000000000   0.44100000000000    H   (2n)
  -0.58000000000000   0.50000000000000  -0.44100000000000    H   (2n)
   0.06200000000000   0.50000000000000   0.43100000000000    H   (2n)
  -0.06200000000000   0.50000000000000  -0.43100000000000    H   (2n)
\end{lstlisting}
{\phantomsection\label{ABC2_mP8_10_ac_eh_mn_cif}}
{\hyperref[ABC2_mP8_10_ac_eh_mn]{Muthmannite (AuAgTe$_{2}$): ABC2\_mP8\_10\_ac\_eh\_mn}} - CIF
\begin{lstlisting}[numbers=none,language={mylang}]
# CIF file
data_findsym-output
_audit_creation_method FINDSYM

_chemical_name_mineral 'AuAgTe2'
_chemical_formula_sum 'Ag Au Te2'

loop_
_publ_author_name
 'L. Bindi'
_journal_name_full_name
;
 Philosophical Magazine Letters
;
_journal_volume 88
_journal_year 2008
_journal_page_first 533
_journal_page_last 541
_publ_Section_title
;
 Commensurate-incommensurate phase transition in muthmannite, AuAgTe$_2$: first evidence of a modulated structure at low temperature
;

# Found in Pearson's Crystal Data - Crystal Structure Database for Inorganic Compounds, 2013

_aflow_title 'Muthmannite (AuAgTe$_{2}$) Structure'
_aflow_proto 'ABC2_mP8_10_ac_eh_mn'
_aflow_params 'a,b/a,c/a,\beta,x_{5},z_{5},x_{6},z_{6}'
_aflow_params_values '5.1177434753,0.86107854631,1.45123094959,90.021,0.6089,0.24179,0.1277,0.24913'
_aflow_Strukturbericht 'None'
_aflow_Pearson 'mP8'

_cell_length_a    5.1177434753
_cell_length_b    4.4067791121
_cell_length_c    7.4270277234
_cell_angle_alpha 90.0000000000
_cell_angle_beta  90.0210000000
_cell_angle_gamma 90.0000000000
 
_symmetry_space_group_name_H-M "P 1 2/m 1"
_symmetry_Int_Tables_number 10
 
loop_
_space_group_symop_id
_space_group_symop_operation_xyz
1 x,y,z
2 -x,y,-z
3 -x,-y,-z
4 x,-y,z
 
loop_
_atom_site_label
_atom_site_type_symbol
_atom_site_symmetry_multiplicity
_atom_site_Wyckoff_label
_atom_site_fract_x
_atom_site_fract_y
_atom_site_fract_z
_atom_site_occupancy
Ag1 Ag   1 a 0.00000 0.00000 0.00000 1.00000
Ag2 Ag   1 c 0.00000 0.00000 0.50000 1.00000
Au1 Au   1 e 0.50000 0.50000 0.00000 1.00000
Au2 Au   1 h 0.50000 0.50000 0.50000 1.00000
Te1 Te   2 m 0.60890 0.00000 0.24179 1.00000
Te2 Te   2 n 0.12770 0.50000 0.24913 1.00000
\end{lstlisting}
{\phantomsection\label{ABC2_mP8_10_ac_eh_mn_poscar}}
{\hyperref[ABC2_mP8_10_ac_eh_mn]{Muthmannite (AuAgTe$_{2}$): ABC2\_mP8\_10\_ac\_eh\_mn}} - POSCAR
\begin{lstlisting}[numbers=none,language={mylang}]
ABC2_mP8_10_ac_eh_mn & a,b/a,c/a,beta,x5,z5,x6,z6 --params=5.1177434753,0.86107854631,1.45123094959,90.021,0.6089,0.24179,0.1277,0.24913 & P2/m C_{2h}^{1} #10 (acehmn) & mP8 & None & AuAgTe2 &  & L. Bindi, Philos. Mag. Lett. 88, 533-541 (2008)
   1.00000000000000
   5.11774347530000   0.00000000000000   0.00000000000000
   0.00000000000000   4.40677911210000   0.00000000000000
  -0.00272214777467   0.00000000000000   7.42702722454036
    Ag    Au    Te
     2     2     4
Direct
   0.00000000000000   0.00000000000000   0.00000000000000   Ag   (1a)
   0.00000000000000   0.00000000000000   0.50000000000000   Ag   (1c)
   0.50000000000000   0.50000000000000   0.00000000000000   Au   (1e)
   0.50000000000000   0.50000000000000   0.50000000000000   Au   (1h)
   0.60890000000000   0.00000000000000   0.24179000000000   Te   (2m)
  -0.60890000000000   0.00000000000000  -0.24179000000000   Te   (2m)
   0.12770000000000   0.50000000000000   0.24913000000000   Te   (2n)
  -0.12770000000000   0.50000000000000  -0.24913000000000   Te   (2n)
\end{lstlisting}
{\phantomsection\label{AB_mP6_10_en_am_cif}}
{\hyperref[AB_mP6_10_en_am]{LiSn: AB\_mP6\_10\_en\_am}} - CIF
\begin{lstlisting}[numbers=none,language={mylang}]
# CIF file
data_findsym-output
_audit_creation_method FINDSYM

_chemical_name_mineral 'LiSn'
_chemical_formula_sum 'Li Sn'

loop_
_publ_author_name
 'W. M{\"u}ller'
 'H. Sch{\"a}fer'
_journal_name_full_name
;
 Zeitschrift f{\"u}r Naturforschung B
;
_journal_volume 28
_journal_year 1973
_journal_page_first 246
_journal_page_last 248
_publ_Section_title
;
 Die Kristallstruktur der Phase LiSn
;

# Found in Pearson's Crystal Data - Crystal Structure Database for Inorganic Compounds, 2013

_aflow_title 'LiSn Structure'
_aflow_proto 'AB_mP6_10_en_am'
_aflow_params 'a,b/a,c/a,\beta,x_{3},z_{3},x_{4},z_{4}'
_aflow_params_values '5.1700416367,0.61508704062,1.49709864605,104.5,0.234,0.66,0.263,0.336'
_aflow_Strukturbericht 'None'
_aflow_Pearson 'mP6'

_cell_length_a    5.1700416367
_cell_length_b    3.1800256102
_cell_length_c    7.7400623343
_cell_angle_alpha 90.0000000000
_cell_angle_beta  104.5000000000
_cell_angle_gamma 90.0000000000
 
_symmetry_space_group_name_H-M "P 1 2/m 1"
_symmetry_Int_Tables_number 10
 
loop_
_space_group_symop_id
_space_group_symop_operation_xyz
1 x,y,z
2 -x,y,-z
3 -x,-y,-z
4 x,-y,z
 
loop_
_atom_site_label
_atom_site_type_symbol
_atom_site_symmetry_multiplicity
_atom_site_Wyckoff_label
_atom_site_fract_x
_atom_site_fract_y
_atom_site_fract_z
_atom_site_occupancy
Sn1 Sn   1 a 0.00000 0.00000 0.00000 1.00000
Li1 Li   1 e 0.50000 0.50000 0.00000 1.00000
Sn2 Sn   2 m 0.23400 0.00000 0.66000 1.00000
Li2 Li   2 n 0.26300 0.50000 0.33600 1.00000
\end{lstlisting}
{\phantomsection\label{AB_mP6_10_en_am_poscar}}
{\hyperref[AB_mP6_10_en_am]{LiSn: AB\_mP6\_10\_en\_am}} - POSCAR
\begin{lstlisting}[numbers=none,language={mylang}]
AB_mP6_10_en_am & a,b/a,c/a,beta,x3,z3,x4,z4 --params=5.1700416367,0.61508704062,1.49709864605,104.5,0.234,0.66,0.263,0.336 & P2/m C_{2h}^{1} #10 (aemn) & mP6 & None & LiSn &  & W. M{\"u}ller and H. Sch{\"a}fer, Z. Naturforsch. B 28, 246-248 (1973)
   1.00000000000000
   5.17004163670000   0.00000000000000   0.00000000000000
   0.00000000000000   3.18002561020000   0.00000000000000
  -1.93795683864366   0.00000000000000   7.49352308533201
    Li    Sn
     3     3
Direct
   0.50000000000000   0.50000000000000   0.00000000000000   Li   (1e)
   0.26300000000000   0.50000000000000   0.33600000000000   Li   (2n)
  -0.26300000000000   0.50000000000000  -0.33600000000000   Li   (2n)
   0.00000000000000   0.00000000000000   0.00000000000000   Sn   (1a)
   0.23400000000000   0.00000000000000   0.66000000000000   Sn   (2m)
  -0.23400000000000   0.00000000000000  -0.66000000000000   Sn   (2m)
\end{lstlisting}
{\phantomsection\label{A_mP8_10_2m2n_cif}}
{\hyperref[A_mP8_10_2m2n]{S-carbon: A\_mP8\_10\_2m2n}} - CIF
\begin{lstlisting}[numbers=none,language={mylang}]
# CIF file 
data_findsym-output
_audit_creation_method FINDSYM

_chemical_name_mineral 'S-carbon'
_chemical_formula_sum 'C'

loop_
_publ_author_name
 'H. Niu'
 'X.-Q. Chen'
 'S. Wang'
 'D. Li'
 'W. L. Mao'
 'Y. Li'
_journal_name_full_name
;
 Physical Review Letters
;
_journal_volume 108
_journal_year 2012
_journal_page_first 135501
_journal_page_last 135501
_publ_Section_title
;
 Families of Superhard Crystalline Carbon Allotropes Constructed via Cold Compression of Graphite and Nanotubes
;

_aflow_title 'S-carbon Structure'
_aflow_proto 'A_mP8_10_2m2n'
_aflow_params 'a,b/a,c/a,\beta,x_{1},z_{1},x_{2},z_{2},x_{3},z_{3},x_{4},z_{4}'
_aflow_params_values '4.7302,0.527461840937,0.86332501797,106.1,0.1175,0.6746,0.5344,0.3333,0.1131,0.8977,0.4209,0.1319'
_aflow_Strukturbericht 'None'
_aflow_Pearson 'mP8'

_symmetry_space_group_name_H-M "P 1 2/m 1"
_symmetry_Int_Tables_number 10
 
_cell_length_a    4.73020
_cell_length_b    2.49500
_cell_length_c    4.08370
_cell_angle_alpha 90.00000
_cell_angle_beta  106.10000
_cell_angle_gamma 90.00000
 
loop_
_space_group_symop_id
_space_group_symop_operation_xyz
1 x,y,z
2 -x,y,-z
3 -x,-y,-z
4 x,-y,z
 
loop_
_atom_site_label
_atom_site_type_symbol
_atom_site_symmetry_multiplicity
_atom_site_Wyckoff_label
_atom_site_fract_x
_atom_site_fract_y
_atom_site_fract_z
_atom_site_occupancy
C1 C   2 m 0.11750 0.00000 0.67460 1.00000
C2 C   2 m 0.53440 0.00000 0.33330 1.00000
C3 C   2 n 0.11310 0.50000 0.89770 1.00000
C4 C   2 n 0.42090 0.50000 0.13190 1.00000
\end{lstlisting}
{\phantomsection\label{A_mP8_10_2m2n_poscar}}
{\hyperref[A_mP8_10_2m2n]{S-carbon: A\_mP8\_10\_2m2n}} - POSCAR
\begin{lstlisting}[numbers=none,language={mylang}]
A_mP8_10_2m2n & a,b/a,c/a,beta,x1,z1,x2,z2,x3,z3,x4,z4 --params=4.7302,0.527461840937,0.86332501797,106.1,0.1175,0.6746,0.5344,0.3333,0.1131,0.8977,0.4209,0.1319 & P2/m C_{2h}^{1} #10 (m^2n^2) & mP8 & None & C & S-carbon & H. Niu et al., Phys. Rev. Lett. 108, 135501(2012)
   1.00000000000000
   4.73020000000000   0.00000000000000   0.00000000000000
   0.00000000000000   2.49500000000000   0.00000000000000
  -1.13246984969092   0.00000000000000   3.92353383183337
     C
     8
Direct
   0.11750000000000   0.00000000000000   0.67460000000000    C   (2m)
  -0.11750000000000   0.00000000000000  -0.67460000000000    C   (2m)
   0.53440000000000   0.00000000000000   0.33330000000000    C   (2m)
  -0.53440000000000   0.00000000000000  -0.33330000000000    C   (2m)
   0.11310000000000   0.50000000000000   0.89770000000000    C   (2n)
  -0.11310000000000   0.50000000000000  -0.89770000000000    C   (2n)
   0.42090000000000   0.50000000000000   0.13190000000000    C   (2n)
  -0.42090000000000   0.50000000000000  -0.13190000000000    C   (2n)
\end{lstlisting}
{\phantomsection\label{A7B2C2_mC22_12_aij_h_i_cif}}
{\hyperref[A7B2C2_mC22_12_aij_h_i]{Thortveitite ([Sc,Y]$_2$Si$_2$O$_7$, $S2_{1}$): A7B2C2\_mC22\_12\_aij\_h\_i}} - CIF
\begin{lstlisting}[numbers=none,language={mylang}]
# CIF file 
data_findsym-output
_audit_creation_method FINDSYM

_chemical_name_mineral 'Thortveitite'
_chemical_formula_sum 'O7 Sc2 Si2'

loop_
_publ_author_name
 'R. Bianchi'
 'T. Pilati'
 'V. Diella'
 'C. M. Gramaccioli'
 'G. Mannucci'
_journal_name_full_name
;
 American Mineralogist
;
_journal_volume 73
_journal_year 1988
_journal_page_first 601
_journal_page_last 607
_publ_Section_title
;
 A re-examination of thortveitite
;

# Found in The American Mineralogist Crystal Structure Database, 2003

_aflow_title 'Thortveitite ([Sc,Y]$_2$Si$_2$O$_7$, $S2_{1}$) Structure'
_aflow_proto 'A7B2C2_mC22_12_aij_h_i'
_aflow_params 'a,b/a,c/a,\beta,y_{2},x_{3},z_{3},x_{4},z_{4},x_{5},y_{5},z_{5}'
_aflow_params_values '6.65,1.29563909774,0.704661654135,102.2,0.30503,0.38654,0.22171,0.22108,-0.08762,0.23655,0.15499,0.71826'
_aflow_Strukturbericht '$S2_{1}$'
_aflow_Pearson 'mC22'

_symmetry_space_group_name_H-M "C 1 2/m 1"
_symmetry_Int_Tables_number 12
 
_cell_length_a    6.65000
_cell_length_b    8.61600
_cell_length_c    4.68600
_cell_angle_alpha 90.00000
_cell_angle_beta  102.20000
_cell_angle_gamma 90.00000
 
loop_
_space_group_symop_id
_space_group_symop_operation_xyz
1 x,y,z
2 -x,y,-z
3 -x,-y,-z
4 x,-y,z
5 x+1/2,y+1/2,z
6 -x+1/2,y+1/2,-z
7 -x+1/2,-y+1/2,-z
8 x+1/2,-y+1/2,z
 
loop_
_atom_site_label
_atom_site_type_symbol
_atom_site_symmetry_multiplicity
_atom_site_Wyckoff_label
_atom_site_fract_x
_atom_site_fract_y
_atom_site_fract_z
_atom_site_occupancy
O1  O    2 a 0.00000 0.00000 0.00000  1.00000
Sc1 Sc   4 h 0.00000 0.30503 0.50000  1.00000
O2  O    4 i 0.38654 0.00000 0.22171  1.00000
Si1 Si   4 i 0.22108 0.00000 -0.08762 1.00000
O3  O    8 j 0.23655 0.15499 0.71826  1.00000
\end{lstlisting}
{\phantomsection\label{A7B2C2_mC22_12_aij_h_i_poscar}}
{\hyperref[A7B2C2_mC22_12_aij_h_i]{Thortveitite ([Sc,Y]$_2$Si$_2$O$_7$, $S2_{1}$): A7B2C2\_mC22\_12\_aij\_h\_i}} - POSCAR
\begin{lstlisting}[numbers=none,language={mylang}]
A7B2C2_mC22_12_aij_h_i & a,b/a,c/a,beta,y2,x3,z3,x4,z4,x5,y5,z5 --params=6.65,1.29563909774,0.704661654135,102.2,0.30503,0.38654,0.22171,0.22108,-0.08762,0.23655,0.15499,0.71826 & C2/m C_{2h}^{3} #12 (ahi^2j) & mC22 & $S2_{1}$ & [Sc,Y]2Si2O7 & Thortveitite & R. Bianchi et al., Am. Mineral. 73, 601-607 (1988)
   1.00000000000000
   3.32500000000000  -4.30800000000000   0.00000000000000
   3.32500000000000   4.30800000000000   0.00000000000000
  -0.99026799618995   0.00000000000000   4.58017088062465
     O    Sc    Si
     7     2     2
Direct
   0.00000000000000   0.00000000000000   0.00000000000000    O   (2a)
   0.38654000000000   0.38654000000000   0.22171000000000    O   (4i)
  -0.38654000000000  -0.38654000000000  -0.22171000000000    O   (4i)
   0.08156000000000   0.39154000000000   0.71826000000000    O   (8j)
  -0.39154000000000  -0.08156000000000  -0.71826000000000    O   (8j)
  -0.08156000000000  -0.39154000000000  -0.71826000000000    O   (8j)
   0.39154000000000   0.08156000000000   0.71826000000000    O   (8j)
  -0.30503000000000   0.30503000000000   0.50000000000000   Sc   (4h)
   0.30503000000000  -0.30503000000000   0.50000000000000   Sc   (4h)
   0.22108000000000   0.22108000000000  -0.08762000000000   Si   (4i)
  -0.22108000000000  -0.22108000000000   0.08762000000000   Si   (4i)
\end{lstlisting}
{\phantomsection\label{A_mC16_12_4i_cif}}
{\hyperref[A_mC16_12_4i]{M-carbon: A\_mC16\_12\_4i}} - CIF
\begin{lstlisting}[numbers=none,language={mylang}]
# CIF file 
data_findsym-output
_audit_creation_method FINDSYM

_chemical_name_mineral 'M-carbon'
_chemical_formula_sum 'C'

loop_
_publ_author_name
 'Q. Li'
 'Y. Ma'
 'A. R. Oganov'
 'H. Wang'
 'H. Wang'
 'Y. Xu'
 'T. Cui'
 'H.-K. Mao'
 'G. Zou'
_journal_name_full_name
;
 Physical Review Letters
;
_journal_volume 102
_journal_year 2009
_journal_page_first 175506
_journal_page_last 175506
_publ_Section_title
;
 Superhard Monoclinic Polymorph of Carbon
;

_aflow_title 'M-carbon Structure'
_aflow_proto 'A_mC16_12_4i'
_aflow_params 'a,b/a,c/a,\beta,x_{1},z_{1},x_{2},z_{2},x_{3},z_{3},x_{4},z_{4}'
_aflow_params_values '9.089,0.274617669711,0.451534822313,96.96,-0.0572,0.1206,0.4419,0.3467,0.7858,-0.0594,0.2715,0.4149'
_aflow_Strukturbericht 'None'
_aflow_Pearson 'mC16'

_symmetry_space_group_name_H-M "C 1 2/m 1"
_symmetry_Int_Tables_number 12
 
_cell_length_a    9.08900
_cell_length_b    2.49600
_cell_length_c    4.10400
_cell_angle_alpha 90.00000
_cell_angle_beta  96.96000
_cell_angle_gamma 90.00000
 
loop_
_space_group_symop_id
_space_group_symop_operation_xyz
1 x,y,z
2 -x,y,-z
3 -x,-y,-z
4 x,-y,z
5 x+1/2,y+1/2,z
6 -x+1/2,y+1/2,-z
7 -x+1/2,-y+1/2,-z
8 x+1/2,-y+1/2,z
 
loop_
_atom_site_label
_atom_site_type_symbol
_atom_site_symmetry_multiplicity
_atom_site_Wyckoff_label
_atom_site_fract_x
_atom_site_fract_y
_atom_site_fract_z
_atom_site_occupancy
C1 C   4 i -0.05720 0.00000 0.12060  1.00000
C2 C   4 i 0.44190  0.00000 0.34670  1.00000
C3 C   4 i 0.78580  0.00000 -0.05940 1.00000
C4 C   4 i 0.27150  0.00000 0.41490  1.00000
\end{lstlisting}
{\phantomsection\label{A_mC16_12_4i_poscar}}
{\hyperref[A_mC16_12_4i]{M-carbon: A\_mC16\_12\_4i}} - POSCAR
\begin{lstlisting}[numbers=none,language={mylang}]
A_mC16_12_4i & a,b/a,c/a,beta,x1,z1,x2,z2,x3,z3,x4,z4 --params=9.089,0.274617669711,0.451534822313,96.96,-0.0572,0.1206,0.4419,0.3467,0.7858,-0.0594,0.2715,0.4149 & C2/m C_{2h}^{3} #12 (i^4) & mC16 & None & C & M-carbon & Q. Li et al., Phys. Rev. Lett. 102, 175506(2009)
   1.00000000000000
   4.54450000000000  -1.24800000000000   0.00000000000000
   4.54450000000000   1.24800000000000   0.00000000000000
  -0.49730788744489   0.00000000000000   4.07375758545904
     C
     8
Direct
  -0.05720000000000  -0.05720000000000   0.12060000000000    C   (4i)
   0.05720000000000   0.05720000000000  -0.12060000000000    C   (4i)
   0.44190000000000   0.44190000000000   0.34670000000000    C   (4i)
  -0.44190000000000  -0.44190000000000  -0.34670000000000    C   (4i)
   0.78580000000000   0.78580000000000  -0.05940000000000    C   (4i)
  -0.78580000000000  -0.78580000000000   0.05940000000000    C   (4i)
   0.27150000000000   0.27150000000000   0.41490000000000    C   (4i)
  -0.27150000000000  -0.27150000000000  -0.41490000000000    C   (4i)
\end{lstlisting}
{\phantomsection\label{A2B_mP12_13_2g_ef_cif}}
{\hyperref[A2B_mP12_13_2g_ef]{H$_{2}$S (15~GPa): A2B\_mP12\_13\_2g\_ef}} - CIF
\begin{lstlisting}[numbers=none,language={mylang}]
# CIF file 
data_findsym-output
_audit_creation_method FINDSYM

_chemical_name_mineral 'H2S'
_chemical_formula_sum 'H2 S'

loop_
_publ_author_name
 'Y. Li'
 'J. Hao'
 'H. Liu'
 'Y. Li'
 'Y. Ma'
_journal_name_full_name
;
 Journal of Chemical Physics
;
_journal_volume 140
_journal_year 2014
_journal_page_first 174712
_journal_page_last 174712
_publ_Section_title
;
 The metallization and superconductivity of dense hydrogen sulfide
;

_aflow_title 'H$_{2}$S (15~GPa) Structure'
_aflow_proto 'A2B_mP12_13_2g_ef'
_aflow_params 'a,b/a,c/a,\beta,y_{1},y_{2},x_{3},y_{3},z_{3},x_{4},y_{4},z_{4}'
_aflow_params_values '5.6255,0.61198115723,1.23780997245,127.44,0.1808,-0.004,0.155,0.346,0.225,0.345,0.273,0.573'
_aflow_Strukturbericht 'None'
_aflow_Pearson 'mP12'

_symmetry_space_group_name_H-M "P 1 2/c 1"
_symmetry_Int_Tables_number 13
 
_cell_length_a    5.62550
_cell_length_b    3.44270
_cell_length_c    6.96330
_cell_angle_alpha 90.00000
_cell_angle_beta  127.44000
_cell_angle_gamma 90.00000
 
loop_
_space_group_symop_id
_space_group_symop_operation_xyz
1 x,y,z
2 -x,y,-z+1/2
3 -x,-y,-z
4 x,-y,z+1/2
 
loop_
_atom_site_label
_atom_site_type_symbol
_atom_site_symmetry_multiplicity
_atom_site_Wyckoff_label
_atom_site_fract_x
_atom_site_fract_y
_atom_site_fract_z
_atom_site_occupancy
S1 S   2 e 0.00000 0.18080  0.25000 1.00000
S2 S   2 f 0.50000 -0.00400 0.25000 1.00000
H1 H   4 g 0.15500 0.34600  0.22500 1.00000
H2 H   4 g 0.34500 0.27300  0.57300 1.00000
\end{lstlisting}
{\phantomsection\label{A2B_mP12_13_2g_ef_poscar}}
{\hyperref[A2B_mP12_13_2g_ef]{H$_{2}$S (15~GPa): A2B\_mP12\_13\_2g\_ef}} - POSCAR
\begin{lstlisting}[numbers=none,language={mylang}]
A2B_mP12_13_2g_ef & a,b/a,c/a,beta,y1,y2,x3,y3,z3,x4,y4,z4 --params=5.6255,0.61198115723,1.23780997245,127.44,0.1808,-0.004,0.155,0.346,0.225,0.345,0.273,0.573 & P2/c C_{2h}^{4} #13 (efg^2) & mP12 & None & H2S & H2S & Y. Li et al., J. Chem. Phys. 140, 174712(2014)
   1.00000000000000
   5.62550000000000   0.00000000000000   0.00000000000000
   0.00000000000000   3.44270000000000   0.00000000000000
  -4.23320104193685   0.00000000000000   5.52879334290447
     H     S
     8     4
Direct
   0.15500000000000   0.34600000000000   0.22500000000000    H   (4g)
  -0.15500000000000   0.34600000000000   0.27500000000000    H   (4g)
  -0.15500000000000  -0.34600000000000  -0.22500000000000    H   (4g)
   0.15500000000000  -0.34600000000000   0.72500000000000    H   (4g)
   0.34500000000000   0.27300000000000   0.57300000000000    H   (4g)
  -0.34500000000000   0.27300000000000  -0.07300000000000    H   (4g)
  -0.34500000000000  -0.27300000000000  -0.57300000000000    H   (4g)
   0.34500000000000  -0.27300000000000   1.07300000000000    H   (4g)
   0.00000000000000   0.18080000000000   0.25000000000000    S   (2e)
   0.00000000000000  -0.18080000000000   0.75000000000000    S   (2e)
   0.50000000000000  -0.00400000000000   0.25000000000000    S   (2f)
   0.50000000000000   0.00400000000000   0.75000000000000    S   (2f)
\end{lstlisting}
{\phantomsection\label{A2B_mP6_14_e_a_cif}}
{\hyperref[A2B_mP6_14_e_a]{$\gamma$-PdCl$_{2}$: A2B\_mP6\_14\_e\_a}} - CIF
\begin{lstlisting}[numbers=none,language={mylang}]
# CIF file 
data_findsym-output
_audit_creation_method FINDSYM

_chemical_name_mineral '$\gamma$-PdCl2'
_chemical_formula_sum 'Cl2 Pd'

loop_
_publ_author_name
 'J. Evers'
 'W. Beck'
 'M. G\"{o}bel'
 'S. Jakob'
 'P. Mayer'
 'G. Oehlinger'
 'M. Rotter'
 'T. M. Klap\"{o}tke'
_journal_name_full_name
;
 Angewandte Chemie (International ed.)
;
_journal_volume 49
_journal_year 2010
_journal_page_first 5677
_journal_page_last 5682
_publ_Section_title
;
 The Structures of $\delta$-PdCl$_{2}$ and $\gamma$-PdCl$_{2}$: Phases with Negative Thermal Expansion in One Direction
;

_aflow_title '$\gamma$-PdCl$_{2}$ Structure'
_aflow_proto 'A2B_mP6_14_e_a'
_aflow_params 'a,b/a,c/a,\beta,x_{2},y_{2},z_{2}'
_aflow_params_values '5.5496,0.695689779444,1.15512829753,107.151,0.255,0.2573,0.3141'
_aflow_Strukturbericht 'None'
_aflow_Pearson 'mP6'

_symmetry_space_group_name_H-M "P 1 21/c 1"
_symmetry_Int_Tables_number 14
 
_cell_length_a    5.54960
_cell_length_b    3.86080
_cell_length_c    6.41050
_cell_angle_alpha 90.00000
_cell_angle_beta  107.15100
_cell_angle_gamma 90.00000
 
loop_
_space_group_symop_id
_space_group_symop_operation_xyz
1 x,y,z
2 -x,y+1/2,-z+1/2
3 -x,-y,-z
4 x,-y+1/2,z+1/2
 
loop_
_atom_site_label
_atom_site_type_symbol
_atom_site_symmetry_multiplicity
_atom_site_Wyckoff_label
_atom_site_fract_x
_atom_site_fract_y
_atom_site_fract_z
_atom_site_occupancy
Pd1 Pd   2 a 0.00000 0.00000 0.00000 1.00000
Cl1 Cl   4 e 0.25500 0.25730 0.31410 1.00000
\end{lstlisting}
{\phantomsection\label{A2B_mP6_14_e_a_poscar}}
{\hyperref[A2B_mP6_14_e_a]{$\gamma$-PdCl$_{2}$: A2B\_mP6\_14\_e\_a}} - POSCAR
\begin{lstlisting}[numbers=none,language={mylang}]
A2B_mP6_14_e_a & a,b/a,c/a,beta,x2,y2,z2 --params=5.5496,0.695689779444,1.15512829753,107.151,0.255,0.2573,0.3141 & P2_{1}/c C_{2h}^{5} #14 (ae) & mP6 & None & PdCl2 & $\gamma$-PdCl2 & J. Evers et al., Angew. Chem. Int. Ed. 49, 5677-5682 (2010)
   1.00000000000000
   5.54960000000000   0.00000000000000   0.00000000000000
   0.00000000000000   3.86080000000000   0.00000000000000
  -1.89039860884891   0.00000000000000   6.12543087053165
    Cl    Pd
     4     2
Direct
   0.25500000000000   0.25730000000000   0.31410000000000   Cl   (4e)
  -0.25500000000000   0.75730000000000   0.18590000000000   Cl   (4e)
  -0.25500000000000  -0.25730000000000  -0.31410000000000   Cl   (4e)
   0.25500000000000   0.24270000000000   0.81410000000000   Cl   (4e)
   0.00000000000000   0.00000000000000   0.00000000000000   Pd   (2a)
   0.00000000000000   0.50000000000000   0.50000000000000   Pd   (2a)
\end{lstlisting}
{\phantomsection\label{A7B8_mP120_14_14e_16e_cif}}
{\hyperref[A7B8_mP120_14_14e_16e]{$\alpha$-Toluene: A7B8\_mP120\_14\_14e\_16e}} - CIF

{\phantomsection\label{A7B8_mP120_14_14e_16e_poscar}}
{\hyperref[A7B8_mP120_14_14e_16e]{$\alpha$-Toluene: A7B8\_mP120\_14\_14e\_16e}} - POSCAR

{\phantomsection\label{AB3_mC16_15_e_cf_cif}}
{\hyperref[AB3_mC16_15_e_cf]{H$_{3}$Cl (50~GPa): AB3\_mC16\_15\_e\_cf}} - CIF
\begin{lstlisting}[numbers=none,language={mylang}]
# CIF file 
data_findsym-output
_audit_creation_method FINDSYM

_chemical_name_mineral 'H3Cl'
_chemical_formula_sum 'Cl H3'

loop_
_publ_author_name
 'D. Duan'
 'X. Huang'
 'F. Tian'
 'Y. Liu'
 'Da Li'
 'H. Yu'
 'B. Liu'
 'W. Tian'
 'T. Cui'
_journal_name_full_name
;
 Journal of Physical Chemistry A
;
_journal_volume 119
_journal_year 2015
_journal_page_first 11059
_journal_page_last 11065
_publ_Section_title
;
 Predicted Formation of H$_{3}^{+}$ in Solid Halogen Polyhydrides at High Pressures
;

_aflow_title 'H$_{3}$Cl (50~GPa) Structure'
_aflow_proto 'AB3_mC16_15_e_cf'
_aflow_params 'a,b/a,c/a,\beta,y_{2},x_{3},y_{3},z_{3}'
_aflow_params_values '3.282,2.64137720902,0.970749542962,91.9,0.86,0.577,0.068,0.169'
_aflow_Strukturbericht 'None'
_aflow_Pearson 'mC16'

_symmetry_space_group_name_H-M "C 1 2/c 1"
_symmetry_Int_Tables_number 15
 
_cell_length_a    3.28200
_cell_length_b    8.66900
_cell_length_c    3.18600
_cell_angle_alpha 90.00000
_cell_angle_beta  91.90000
_cell_angle_gamma 90.00000
 
loop_
_space_group_symop_id
_space_group_symop_operation_xyz
1 x,y,z
2 -x,y,-z+1/2
3 -x,-y,-z
4 x,-y,z+1/2
5 x+1/2,y+1/2,z
6 -x+1/2,y+1/2,-z+1/2
7 -x+1/2,-y+1/2,-z
8 x+1/2,-y+1/2,z+1/2
 
loop_
_atom_site_label
_atom_site_type_symbol
_atom_site_symmetry_multiplicity
_atom_site_Wyckoff_label
_atom_site_fract_x
_atom_site_fract_y
_atom_site_fract_z
_atom_site_occupancy
H1  H    4 c 0.25000 0.25000 0.00000 1.00000
Cl1 Cl   4 e 0.00000 0.86000 0.25000 1.00000
H2  H    8 f 0.57700 0.06800 0.16900 1.00000
\end{lstlisting}
{\phantomsection\label{AB3_mC16_15_e_cf_poscar}}
{\hyperref[AB3_mC16_15_e_cf]{H$_{3}$Cl (50~GPa): AB3\_mC16\_15\_e\_cf}} - POSCAR
\begin{lstlisting}[numbers=none,language={mylang}]
AB3_mC16_15_e_cf & a,b/a,c/a,beta,y2,x3,y3,z3 --params=3.282,2.64137720902,0.970749542962,91.9,0.86,0.577,0.068,0.169 & C2/c C_{2h}^{6} #15 (cef) & mC16 & None & H3Cl & H3Cl & D. Duan et al., J. Phys. Chem. A 119, 11059-11065 (2015)
   1.00000000000000
   1.64100000000000  -4.33450000000000   0.00000000000000
   1.64100000000000   4.33450000000000   0.00000000000000
  -0.10563239834584   0.00000000000000   3.18424838799044
    Cl     H
     2     6
Direct
  -0.86000000000000   0.86000000000000   0.25000000000000   Cl   (4e)
   0.86000000000000  -0.86000000000000   0.75000000000000   Cl   (4e)
   0.00000000000000   0.50000000000000   0.00000000000000    H   (4c)
   0.50000000000000   0.00000000000000   0.50000000000000    H   (4c)
   0.50900000000000   0.64500000000000   0.16900000000000    H   (8f)
  -0.64500000000000  -0.50900000000000   0.33100000000000    H   (8f)
  -0.50900000000000  -0.64500000000000  -0.16900000000000    H   (8f)
   0.64500000000000   0.50900000000000   0.66900000000000    H   (8f)
\end{lstlisting}
{\phantomsection\label{A_mC24_15_2e2f_cif}}
{\hyperref[A_mC24_15_2e2f]{H-III (300~GPa): A\_mC24\_15\_2e2f}} - CIF
\begin{lstlisting}[numbers=none,language={mylang}]
# CIF file 
data_findsym-output
_audit_creation_method FINDSYM

_chemical_name_mineral 'H'
_chemical_formula_sum 'H'

loop_
_publ_author_name
 'C. J. Pickard'
 'R. J. Needs'
_journal_name_full_name
;
 Nature Physics
;
_journal_volume 3
_journal_year 2007
_journal_page_first 473
_journal_page_last 476
_publ_Section_title
;
 Structure of phase III of solid hydrogen
;

_aflow_title 'H-III (300~GPa) Structure'
_aflow_proto 'A_mC24_15_2e2f'
_aflow_params 'a,b/a,c/a,\beta,y_{1},y_{2},x_{3},y_{3},z_{3},x_{4},y_{4},z_{4}'
_aflow_params_values '4.939,0.569143551326,0.838023891476,142.47,0.1012,0.3684,0.226,0.0672,0.2464,0.3443,0.1958,0.2227'
_aflow_Strukturbericht 'None'
_aflow_Pearson 'mC24'

_symmetry_space_group_name_H-M "C 1 2/c 1"
_symmetry_Int_Tables_number 15
 
_cell_length_a    4.93900
_cell_length_b    2.81100
_cell_length_c    4.13900
_cell_angle_alpha 90.00000
_cell_angle_beta  142.47000
_cell_angle_gamma 90.00000
 
loop_
_space_group_symop_id
_space_group_symop_operation_xyz
1 x,y,z
2 -x,y,-z+1/2
3 -x,-y,-z
4 x,-y,z+1/2
5 x+1/2,y+1/2,z
6 -x+1/2,y+1/2,-z+1/2
7 -x+1/2,-y+1/2,-z
8 x+1/2,-y+1/2,z+1/2
 
loop_
_atom_site_label
_atom_site_type_symbol
_atom_site_symmetry_multiplicity
_atom_site_Wyckoff_label
_atom_site_fract_x
_atom_site_fract_y
_atom_site_fract_z
_atom_site_occupancy
H1 H   4 e 0.00000 0.10120 0.25000 1.00000
H2 H   4 e 0.00000 0.36840 0.25000 1.00000
H3 H   8 f 0.22600 0.06720 0.24640 1.00000
H4 H   8 f 0.34430 0.19580 0.22270 1.00000
\end{lstlisting}
{\phantomsection\label{A_mC24_15_2e2f_poscar}}
{\hyperref[A_mC24_15_2e2f]{H-III (300~GPa): A\_mC24\_15\_2e2f}} - POSCAR
\begin{lstlisting}[numbers=none,language={mylang}]
A_mC24_15_2e2f & a,b/a,c/a,beta,y1,y2,x3,y3,z3,x4,y4,z4 --params=4.939,0.569143551326,0.838023891476,142.47,0.1012,0.3684,0.226,0.0672,0.2464,0.3443,0.1958,0.2227 & C2/c C_{2h}^{6} #15 (e^2f^2) & mC24 & None & H & H & C. J. Pickard and R. J. Needs, Nat. Phys. 3, 473-476 (2007)
   1.00000000000000
   2.46950000000000  -1.40550000000000   0.00000000000000
   2.46950000000000   1.40550000000000   0.00000000000000
  -3.28236973265251   0.00000000000000   2.52138254498731
     H
    12
Direct
  -0.10120000000000   0.10120000000000   0.25000000000000    H   (4e)
   0.10120000000000  -0.10120000000000   0.75000000000000    H   (4e)
  -0.36840000000000   0.36840000000000   0.25000000000000    H   (4e)
   0.36840000000000  -0.36840000000000   0.75000000000000    H   (4e)
   0.15880000000000   0.29320000000000   0.24640000000000    H   (8f)
  -0.29320000000000  -0.15880000000000   0.25360000000000    H   (8f)
  -0.15880000000000  -0.29320000000000  -0.24640000000000    H   (8f)
   0.29320000000000   0.15880000000000   0.74640000000000    H   (8f)
   0.14850000000000   0.54010000000000   0.22270000000000    H   (8f)
  -0.54010000000000  -0.14850000000000   0.27730000000000    H   (8f)
  -0.14850000000000  -0.54010000000000  -0.22270000000000    H   (8f)
   0.54010000000000   0.14850000000000   0.72270000000000    H   (8f)
\end{lstlisting}
{\phantomsection\label{A2B_oP12_17_abe_e_cif}}
{\hyperref[A2B_oP12_17_abe_e]{$\alpha$-Naumannite (Ag$_{2}$Se): A2B\_oP12\_17\_abe\_e}} - CIF
\begin{lstlisting}[numbers=none,language={mylang}]
# CIF file
data_findsym-output
_audit_creation_method FINDSYM

_chemical_name_mineral 'alpha-Ag2Se'
_chemical_formula_sum 'Ag2 Se'

loop_
_publ_author_name
 'Z. G. Pinsker'
 'C. {Ching-liang}'
 'R. M. Imamov'
 'E. L. Lapidus'
_journal_name_full_name
;
 Soviet Physics Crystallography
;
_journal_volume 10
_journal_year 1965
_journal_page_first 225
_journal_page_last 231
_publ_Section_title
;
 Determination of the crystal structure of the low-temperature phase $\alpha$-Ag$_{2}$Se
;

# Found in Pearson's Crystal Data - Crystal Structure Database for Inorganic Compounds, 2013

_aflow_title '$\alpha$-Naumannite (Ag$_{2}$Se) Structure'
_aflow_proto 'A2B_oP12_17_abe_e'
_aflow_params 'a,b/a,c/a,x_{1},x_{2},x_{3},y_{3},z_{3},x_{4},y_{4},z_{4}'
_aflow_params_values '7.0499703347,1.11347517732,0.614184397158,0.893,0.878,0.379,0.225,0.522,0.202,0.275,0.022'
_aflow_Strukturbericht 'None'
_aflow_Pearson 'oP12'

_cell_length_a    7.0499703347
_cell_length_b    7.8499669685
_cell_length_c    4.3299817800
_cell_angle_alpha 90.0000000000
_cell_angle_beta  90.0000000000
_cell_angle_gamma 90.0000000000
 
_symmetry_space_group_name_H-M "P 2 2 21"
_symmetry_Int_Tables_number 17
 
loop_
_space_group_symop_id
_space_group_symop_operation_xyz
1 x,y,z
2 x,-y,-z
3 -x,y,-z+1/2
4 -x,-y,z+1/2
 
loop_
_atom_site_label
_atom_site_type_symbol
_atom_site_symmetry_multiplicity
_atom_site_Wyckoff_label
_atom_site_fract_x
_atom_site_fract_y
_atom_site_fract_z
_atom_site_occupancy
Ag1 Ag   2 a 0.89300 0.00000 0.00000 1.00000
Ag2 Ag   2 b 0.87800 0.50000 0.00000 1.00000
Ag3 Ag   4 e 0.37900 0.22500 0.52200 1.00000
Se1 Se   4 e 0.20200 0.27500 0.02200 1.00000
\end{lstlisting}
{\phantomsection\label{A2B_oP12_17_abe_e_poscar}}
{\hyperref[A2B_oP12_17_abe_e]{$\alpha$-Naumannite (Ag$_{2}$Se): A2B\_oP12\_17\_abe\_e}} - POSCAR
\begin{lstlisting}[numbers=none,language={mylang}]
A2B_oP12_17_abe_e & a,b/a,c/a,x1,x2,x3,y3,z3,x4,y4,z4 --params=7.0499703347,1.11347517732,0.614184397158,0.893,0.878,0.379,0.225,0.522,0.202,0.275,0.022 & P222_{1} D_{2}^{2} #17 (abe^2) & oP12 & None & Ag2Se & alpha & Z. G. Pinsker et al., Sov. Phys. Crystallogr. 10, 225-231 (1965)
   1.00000000000000
   7.04997033470000   0.00000000000000   0.00000000000000
   0.00000000000000   7.84996696850000   0.00000000000000
   0.00000000000000   0.00000000000000   4.32998178000000
    Ag    Se
     8     4
Direct
   0.89300000000000   0.00000000000000   0.00000000000000   Ag   (2a)
  -0.89300000000000   0.00000000000000   0.50000000000000   Ag   (2a)
   0.87800000000000   0.50000000000000   0.00000000000000   Ag   (2b)
  -0.87800000000000   0.50000000000000   0.50000000000000   Ag   (2b)
   0.37900000000000   0.22500000000000   0.52200000000000   Ag   (4e)
  -0.37900000000000  -0.22500000000000   1.02200000000000   Ag   (4e)
  -0.37900000000000   0.22500000000000  -0.02200000000000   Ag   (4e)
   0.37900000000000  -0.22500000000000  -0.52200000000000   Ag   (4e)
   0.20200000000000   0.27500000000000   0.02200000000000   Se   (4e)
  -0.20200000000000  -0.27500000000000   0.52200000000000   Se   (4e)
  -0.20200000000000   0.27500000000000   0.47800000000000   Se   (4e)
   0.20200000000000  -0.27500000000000  -0.02200000000000   Se   (4e)
\end{lstlisting}
{\phantomsection\label{AB3_oP16_19_a_3a_cif}}
{\hyperref[AB3_oP16_19_a_3a]{H$_{3}$Cl (100~GPa): AB3\_oP16\_19\_a\_3a}} - CIF
\begin{lstlisting}[numbers=none,language={mylang}]
# CIF file 
data_findsym-output
_audit_creation_method FINDSYM

_chemical_name_mineral 'H3Cl'
_chemical_formula_sum 'Cl H3'

loop_
_publ_author_name
 'D. Duan'
 'X. Huang'
 'F. Tian'
 'Y. Liu'
 'Da Li'
 'H. Yu'
 'B. Liu'
 'W. Tian'
 'T. Cui'
_journal_name_full_name
;
 Journal of Physical Chemistry A
;
_journal_volume 119
_journal_year 2015
_journal_page_first 11059
_journal_page_last 11065
_publ_Section_title
;
 Predicted Formation of H$_{3}^{+}$ in Solid Halogen Polyhydrides at High Pressures
;

_aflow_title 'H$_{3}$Cl (100~GPa) Structure'
_aflow_proto 'AB3_oP16_19_a_3a'
_aflow_params 'a,b/a,c/a,x_{1},y_{1},z_{1},x_{2},y_{2},z_{2},x_{3},y_{3},z_{3},x_{4},y_{4},z_{4}'
_aflow_params_values '5.668,0.758997882851,0.528757939308,0.584,0.123,0.027,0.31,0.159,0.417,0.257,0.073,0.603,0.983,0.124,0.227'
_aflow_Strukturbericht 'None'
_aflow_Pearson 'oP16'

_symmetry_space_group_name_H-M "P 21 21 21"
_symmetry_Int_Tables_number 19
 
_cell_length_a    5.66800
_cell_length_b    4.30200
_cell_length_c    2.99700
_cell_angle_alpha 90.00000
_cell_angle_beta  90.00000
_cell_angle_gamma 90.00000
 
loop_
_space_group_symop_id
_space_group_symop_operation_xyz
1 x,y,z
2 x+1/2,-y+1/2,-z
3 -x,y+1/2,-z+1/2
4 -x+1/2,-y,z+1/2
 
loop_
_atom_site_label
_atom_site_type_symbol
_atom_site_symmetry_multiplicity
_atom_site_Wyckoff_label
_atom_site_fract_x
_atom_site_fract_y
_atom_site_fract_z
_atom_site_occupancy
Cl1 Cl   4 a 0.58400  0.12300 0.02700  1.00000
H1  H    4 a 0.31000  0.15900 0.41700  1.00000
H2  H    4 a 0.25700  0.07300 0.60300  1.00000
H3  H    4 a 0.98300  0.12400 0.22700  1.00000
\end{lstlisting}
{\phantomsection\label{AB3_oP16_19_a_3a_poscar}}
{\hyperref[AB3_oP16_19_a_3a]{H$_{3}$Cl (100~GPa): AB3\_oP16\_19\_a\_3a}} - POSCAR
\begin{lstlisting}[numbers=none,language={mylang}]
AB3_oP16_19_a_3a & a,b/a,c/a,x1,y1,z1,x2,y2,z2,x3,y3,z3,x4,y4,z4 --params=5.668,0.758997882851,0.528757939308,0.584,0.123,0.027,0.31,0.159,0.417,0.257,0.073,0.603,0.983,0.124,0.227 & P2_{1}2_{1}2_{1} D_{2}^{4} #19 (a^4) & oP16 & None & H3Cl & H3Cl & D. Duan et al., J. Phys. Chem. A 119, 11059-11065 (2015)
   1.00000000000000
   5.66800000000000   0.00000000000000   0.00000000000000
   0.00000000000000   4.30200000000000   0.00000000000000
   0.00000000000000   0.00000000000000   2.99700000000000
    Cl     H
     4    12
Direct
   0.58400000000000   0.12300000000000   0.02700000000000   Cl   (4a)
  -0.08400000000000  -0.12300000000000   0.52700000000000   Cl   (4a)
  -0.58400000000000   0.62300000000000   0.47300000000000   Cl   (4a)
   1.08400000000000   0.37700000000000  -0.02700000000000   Cl   (4a)
   0.31000000000000   0.15900000000000   0.41700000000000    H   (4a)
   0.19000000000000  -0.15900000000000   0.91700000000000    H   (4a)
  -0.31000000000000   0.65900000000000   0.08300000000000    H   (4a)
   0.81000000000000   0.34100000000000  -0.41700000000000    H   (4a)
   0.25700000000000   0.07300000000000   0.60300000000000    H   (4a)
   0.24300000000000  -0.07300000000000   1.10300000000000    H   (4a)
  -0.25700000000000   0.57300000000000  -0.10300000000000    H   (4a)
   0.75700000000000   0.42700000000000  -0.60300000000000    H   (4a)
   0.98300000000000   0.12400000000000   0.22700000000000    H   (4a)
  -0.48300000000000  -0.12400000000000   0.72700000000000    H   (4a)
  -0.98300000000000   0.62400000000000   0.27300000000000    H   (4a)
   1.48300000000000   0.37600000000000  -0.22700000000000    H   (4a)
\end{lstlisting}
{\phantomsection\label{AB2_oC6_21_a_k_cif}}
{\hyperref[AB2_oC6_21_a_k]{Ta$_{2}$H: AB2\_oC6\_21\_a\_k}} - CIF
\begin{lstlisting}[numbers=none,language={mylang}]
# CIF file
data_findsym-output
_audit_creation_method FINDSYM

_chemical_name_mineral 'Ta2H'
_chemical_formula_sum 'H Ta2'

loop_
_publ_author_name
 'H. Asano'
 'Y. Ishikawa'
 'M. Hirabayashi'
_journal_name_full_name
;
 Journal of Applied Crystallography
;
_journal_volume 11
_journal_year 1978
_journal_page_first 681
_journal_page_last 683
_publ_Section_title
;
 Single-crystal X-ray diffraction study on the hydrogen ordering in Ta$_{2}$H
;

# Found in Pearson's Crystal Data - Crystal Structure Database for Inorganic Compounds, 2013

_aflow_title 'Ta$_{2}$H Structure'
_aflow_proto 'AB2_oC6_21_a_k'
_aflow_params 'a,b/a,c/a,z_{2}'
_aflow_params_values '3.3982513,1.3943496174,1.40170688639,0.268'
_aflow_Strukturbericht 'None'
_aflow_Pearson 'oC6'

_cell_length_a    3.3982513000
_cell_length_b    4.7383504000
_cell_length_c    4.7633522489
_cell_angle_alpha 90.0000000000
_cell_angle_beta  90.0000000000
_cell_angle_gamma 90.0000000000
 
_symmetry_space_group_name_H-M "C 2 2 2"
_symmetry_Int_Tables_number 21
 
loop_
_space_group_symop_id
_space_group_symop_operation_xyz
1 x,y,z
2 x,-y,-z
3 -x,y,-z
4 -x,-y,z
5 x+1/2,y+1/2,z
6 x+1/2,-y+1/2,-z
7 -x+1/2,y+1/2,-z
8 -x+1/2,-y+1/2,z
 
loop_
_atom_site_label
_atom_site_type_symbol
_atom_site_symmetry_multiplicity
_atom_site_Wyckoff_label
_atom_site_fract_x
_atom_site_fract_y
_atom_site_fract_z
_atom_site_occupancy
H1  H    2 a 0.00000 0.00000 0.00000 1.00000
Ta1 Ta   4 k 0.25000 0.25000 0.26800 1.00000
\end{lstlisting}
{\phantomsection\label{AB2_oC6_21_a_k_poscar}}
{\hyperref[AB2_oC6_21_a_k]{Ta$_{2}$H: AB2\_oC6\_21\_a\_k}} - POSCAR
\begin{lstlisting}[numbers=none,language={mylang}]
AB2_oC6_21_a_k & a,b/a,c/a,z2 --params=3.3982513,1.3943496174,1.40170688639,0.268 & C222 D_{2}^{6} #21 (ak) & oC6 & None & Ta2H &  & H. Asano and Y. Ishikawa and M. Hirabayashi, J. Appl. Crystallogr. 11, 681-683 (1978)
   1.00000000000000
   1.69912565000000  -2.36917520000000   0.00000000000000
   1.69912565000000   2.36917520000000   0.00000000000000
   0.00000000000000   0.00000000000000   4.76335224890000
     H    Ta
     1     2
Direct
   0.00000000000000   0.00000000000000   0.00000000000000    H   (2a)
   0.00000000000000   0.50000000000000   0.26800000000000   Ta   (4k)
   0.50000000000000   0.00000000000000  -0.26800000000000   Ta   (4k)
\end{lstlisting}
{\phantomsection\label{A2BC2_oF40_22_fi_ad_gh_cif}}
{\hyperref[A2BC2_oF40_22_fi_ad_gh]{CeRu$_{2}$B$_{2}$: A2BC2\_oF40\_22\_fi\_ad\_gh}} - CIF
\begin{lstlisting}[numbers=none,language={mylang}]
# CIF file
data_findsym-output
_audit_creation_method FINDSYM

_chemical_name_mineral 'CeRu2B2'
_chemical_formula_sum 'B2 Ce Ru2'

loop_
_publ_author_name
 'P. Rogl'
_journal_name_full_name
;
 Journal of the Less-Common Metals
;
_journal_volume 110
_journal_year 1985
_journal_page_first 283
_journal_page_last 294
_publ_Section_title
;
 Structural chemistry and phase equilibria of ternary rare earth-platinum metal borides
;

# Found in Pearson's Crystal Data - Crystal Structure Database for Inorganic Compounds, 2013

_aflow_title 'CeRu$_{2}$B$_{2}$ Structure'
_aflow_proto 'A2BC2_oF40_22_fi_ad_gh'
_aflow_params 'a,b/a,c/a,y_{3},z_{4},z_{5},y_{6}'
_aflow_params_values '6.4793068924,1.3977127159,1.54737394473,0.3134,0.3627,0.1144,0.0719'
_aflow_Strukturbericht 'None'
_aflow_Pearson 'oF40'

_cell_length_a    6.4793068924
_cell_length_b    9.0562096337
_cell_length_c    10.0259106652
_cell_angle_alpha 90.0000000000
_cell_angle_beta  90.0000000000
_cell_angle_gamma 90.0000000000
 
_symmetry_space_group_name_H-M "F 2 2 2"
_symmetry_Int_Tables_number 22
 
loop_
_space_group_symop_id
_space_group_symop_operation_xyz
1 x,y,z
2 x,-y,-z
3 -x,y,-z
4 -x,-y,z
5 x,y+1/2,z+1/2
6 x,-y+1/2,-z+1/2
7 -x,y+1/2,-z+1/2
8 -x,-y+1/2,z+1/2
9 x+1/2,y,z+1/2
10 x+1/2,-y,-z+1/2
11 -x+1/2,y,-z+1/2
12 -x+1/2,-y,z+1/2
13 x+1/2,y+1/2,z
14 x+1/2,-y+1/2,-z
15 -x+1/2,y+1/2,-z
16 -x+1/2,-y+1/2,z
 
loop_
_atom_site_label
_atom_site_type_symbol
_atom_site_symmetry_multiplicity
_atom_site_Wyckoff_label
_atom_site_fract_x
_atom_site_fract_y
_atom_site_fract_z
_atom_site_occupancy
Ce1 Ce   4 a 0.00000 0.00000 0.00000 1.00000
Ce2 Ce   4 d 0.25000 0.25000 0.75000 1.00000
B1  B    8 f 0.00000 0.31340 0.00000 1.00000
Ru1 Ru   8 g 0.00000 0.00000 0.36270 1.00000
Ru2 Ru   8 h 0.25000 0.25000 0.11440 1.00000
B2  B    8 i 0.25000 0.07190 0.25000 1.00000
\end{lstlisting}
{\phantomsection\label{A2BC2_oF40_22_fi_ad_gh_poscar}}
{\hyperref[A2BC2_oF40_22_fi_ad_gh]{CeRu$_{2}$B$_{2}$: A2BC2\_oF40\_22\_fi\_ad\_gh}} - POSCAR
\begin{lstlisting}[numbers=none,language={mylang}]
A2BC2_oF40_22_fi_ad_gh & a,b/a,c/a,y3,z4,z5,y6 --params=6.4793068924,1.3977127159,1.54737394473,0.3134,0.3627,0.1144,0.0719 & F222 D_{2}^{7} #22 (adfghi) & oF40 & None & CeRu2B2 &  & P. Rogl, J. Less-Common Met. 110, 283-294 (1985)
   1.00000000000000
   0.00000000000000   4.52810481685000   5.01295533260000
   3.23965344620000   0.00000000000000   5.01295533260000
   3.23965344620000   4.52810481685000   0.00000000000000
     B    Ce    Ru
     4     2     4
Direct
   0.31340000000000  -0.31340000000000   0.31340000000000    B   (8f)
  -0.31340000000000   0.31340000000000  -0.31340000000000    B   (8f)
   0.07190000000000   0.42810000000000   0.07190000000000    B   (8i)
   0.42810000000000   0.07190000000000   0.42810000000000    B   (8i)
   0.00000000000000   0.00000000000000   0.00000000000000   Ce   (4a)
   0.75000000000000   0.75000000000000   0.75000000000000   Ce   (4d)
   0.36270000000000   0.36270000000000  -0.36270000000000   Ru   (8g)
  -0.36270000000000  -0.36270000000000   0.36270000000000   Ru   (8g)
   0.11440000000000   0.11440000000000   0.38560000000000   Ru   (8h)
   0.38560000000000   0.38560000000000   0.11440000000000   Ru   (8h)
\end{lstlisting}
{\phantomsection\label{AB_oF8_22_a_c_cif}}
{\hyperref[AB_oF8_22_a_c]{FeS (Low-temperature): AB\_oF8\_22\_a\_c}} - CIF
\begin{lstlisting}[numbers=none,language={mylang}]
# CIF file
data_findsym-output
_audit_creation_method FINDSYM

_chemical_name_mineral 'FeS'
_chemical_formula_sum 'Fe S'

loop_
_publ_author_name
 'M. Wintenberger'
 'B. Srour'
 'C. Meyer'
 'F. Hartmannboutron'
 'Y. Gros'
 'J. L. Buevoz'
_journal_name_full_name
;
 Acta Crystallographica Section A: Foundations and Advances
;
_journal_volume 34
_journal_year 1978
_journal_page_first S318
_journal_page_last S318
_publ_Section_title
;
 First order transitions and magnetic-structure of zincblende-type iron sulfide
;

# Found in Pearson's Crystal Data - Crystal Structure Database for Inorganic Compounds, 2013

_aflow_title 'FeS (Low-temperature) Structure'
_aflow_proto 'AB_oF8_22_a_c'
_aflow_params 'a,b/a,c/a'
_aflow_params_values '5.5400291632,0.990433212996,0.937725631773'
_aflow_Strukturbericht 'None'
_aflow_Pearson 'oF8'

_cell_length_a    5.5400291632
_cell_length_b    5.4870288842
_cell_length_c    5.1950273471
_cell_angle_alpha 90.0000000000
_cell_angle_beta  90.0000000000
_cell_angle_gamma 90.0000000000
 
_symmetry_space_group_name_H-M "F 2 2 2"
_symmetry_Int_Tables_number 22
 
loop_
_space_group_symop_id
_space_group_symop_operation_xyz
1 x,y,z
2 x,-y,-z
3 -x,y,-z
4 -x,-y,z
5 x,y+1/2,z+1/2
6 x,-y+1/2,-z+1/2
7 -x,y+1/2,-z+1/2
8 -x,-y+1/2,z+1/2
9 x+1/2,y,z+1/2
10 x+1/2,-y,-z+1/2
11 -x+1/2,y,-z+1/2
12 -x+1/2,-y,z+1/2
13 x+1/2,y+1/2,z
14 x+1/2,-y+1/2,-z
15 -x+1/2,y+1/2,-z
16 -x+1/2,-y+1/2,z
 
loop_
_atom_site_label
_atom_site_type_symbol
_atom_site_symmetry_multiplicity
_atom_site_Wyckoff_label
_atom_site_fract_x
_atom_site_fract_y
_atom_site_fract_z
_atom_site_occupancy
Fe1 Fe   4 a 0.00000 0.00000 0.00000 1.00000
S1  S    4 c 0.25000 0.25000 0.25000 1.00000
\end{lstlisting}
{\phantomsection\label{AB_oF8_22_a_c_poscar}}
{\hyperref[AB_oF8_22_a_c]{FeS (Low-temperature): AB\_oF8\_22\_a\_c}} - POSCAR
\begin{lstlisting}[numbers=none,language={mylang}]
AB_oF8_22_a_c & a,b/a,c/a --params=5.5400291632,0.990433212996,0.937725631773 & F222 D_{2}^{7} #22 (ac) & oF8 & None & FeS &  & M. Wintenberger et al., Acta Crystallogr. Sect. A 34, S318-S318 (1978)
   1.00000000000000
   0.00000000000000   2.74351444210000   2.59751367355000
   2.77001458160000   0.00000000000000   2.59751367355000
   2.77001458160000   2.74351444210000   0.00000000000000
    Fe     S
     1     1
Direct
   0.00000000000000   0.00000000000000   0.00000000000000   Fe   (4a)
   0.25000000000000   0.25000000000000   0.25000000000000    S   (4c)
\end{lstlisting}
{\phantomsection\label{A3B_oI32_23_ij2k_k_cif}}
{\hyperref[A3B_oI32_23_ij2k_k]{H$_{3}$S (5~GPa): A3B\_oI32\_23\_ij2k\_k}} - CIF
\begin{lstlisting}[numbers=none,language={mylang}]
# CIF file 
data_findsym-output
_audit_creation_method FINDSYM

_chemical_name_mineral 'H3S'
_chemical_formula_sum 'H3 S'

loop_
_publ_author_name
 'T. A. Strobel'
 'P. Ganesh'
 'M. Somayazulu'
 'P. R. C. Kent'
 'R. J. Hemley'
_journal_name_full_name
;
 Physical Review Letters
;
_journal_volume 107
_journal_year 2011
_journal_page_first 255503
_journal_page_last 255503
_publ_Section_title
;
 Novel Cooperative Interactions and Structural Ordering in H$_{2}$S-H$_{2}$
;

_aflow_title 'H$_{3}$S (5~GPa) Structure'
_aflow_proto 'A3B_oI32_23_ij2k_k'
_aflow_params 'a,b/a,c/a,z_{1},z_{2},x_{3},y_{3},z_{3},x_{4},y_{4},z_{4},x_{5},y_{5},z_{5}'
_aflow_params_values '5.82463,1.24369101557,1.32254065924,0.04851,0.45153,0.75475,0.49255,0.20405,0.4421,0.23223,0.28616,0.76005,0.8216,0.36488'
_aflow_Strukturbericht 'None'
_aflow_Pearson 'oI32'

_symmetry_space_group_name_H-M "I 2 2 2"
_symmetry_Int_Tables_number 23
 
_cell_length_a    5.82463
_cell_length_b    7.24404
_cell_length_c    7.70331
_cell_angle_alpha 90.00000
_cell_angle_beta  90.00000
_cell_angle_gamma 90.00000
 
loop_
_space_group_symop_id
_space_group_symop_operation_xyz
1 x,y,z
2 x,-y,-z
3 -x,y,-z
4 -x,-y,z
5 x+1/2,y+1/2,z+1/2
6 x+1/2,-y+1/2,-z+1/2
7 -x+1/2,y+1/2,-z+1/2
8 -x+1/2,-y+1/2,z+1/2
 
loop_
_atom_site_label
_atom_site_type_symbol
_atom_site_symmetry_multiplicity
_atom_site_Wyckoff_label
_atom_site_fract_x
_atom_site_fract_y
_atom_site_fract_z
_atom_site_occupancy
H1 H   4 i 0.00000 0.00000 0.04851 1.00000
H2 H   4 j 0.00000 0.50000 0.45153 1.00000
H3 H   8 k 0.75475 0.49255 0.20405 1.00000
H4 H   8 k 0.44210 0.23223 0.28616 1.00000
S1 S   8 k 0.76005 0.82160 0.36488 1.00000
\end{lstlisting}
{\phantomsection\label{A3B_oI32_23_ij2k_k_poscar}}
{\hyperref[A3B_oI32_23_ij2k_k]{H$_{3}$S (5~GPa): A3B\_oI32\_23\_ij2k\_k}} - POSCAR
\begin{lstlisting}[numbers=none,language={mylang}]
A3B_oI32_23_ij2k_k & a,b/a,c/a,z1,z2,x3,y3,z3,x4,y4,z4,x5,y5,z5 --params=5.82463,1.24369101557,1.32254065924,0.04851,0.45153,0.75475,0.49255,0.20405,0.4421,0.23223,0.28616,0.76005,0.8216,0.36488 & I222 D_{2}^{8} #23 (ijk^3) & oI32 & None & H3S & H3S & T. A. Strobel et al., Phys. Rev. Lett. 107, 255503(2011)
   1.00000000000000
  -2.91231500000000   3.62202000000000   3.85165500000000
   2.91231500000000  -3.62202000000000   3.85165500000000
   2.91231500000000   3.62202000000000  -3.85165500000000
     H     S
    12     4
Direct
   0.04851000000000   0.04851000000000   0.00000000000000    H   (4i)
  -0.04851000000000  -0.04851000000000   0.00000000000000    H   (4i)
   0.95153000000000   0.45153000000000   0.50000000000000    H   (4j)
   0.04847000000000  -0.45153000000000   0.50000000000000    H   (4j)
   0.69660000000000   0.95880000000000   1.24730000000000    H   (8k)
  -0.28850000000000  -0.55070000000000  -1.24730000000000    H   (8k)
   0.28850000000000  -0.95880000000000  -0.26220000000000    H   (8k)
  -0.69660000000000   0.55070000000000   0.26220000000000    H   (8k)
   0.51839000000000   0.72826000000000   0.67433000000000    H   (8k)
   0.05393000000000  -0.15594000000000  -0.67433000000000    H   (8k)
  -0.05393000000000  -0.72826000000000  -0.20987000000000    H   (8k)
  -0.51839000000000   0.15594000000000   0.20987000000000    H   (8k)
   1.18648000000000   1.12493000000000   1.58165000000000    S   (8k)
  -0.45672000000000  -0.39517000000000  -1.58165000000000    S   (8k)
   0.45672000000000  -1.12493000000000   0.06155000000000    S   (8k)
  -1.18648000000000   0.39517000000000  -0.06155000000000    S   (8k)
\end{lstlisting}
{\phantomsection\label{A8B2C12D2E_oI50_23_bcfk_i_3k_j_a_cif}}
{\hyperref[A8B2C12D2E_oI50_23_bcfk_i_3k_j_a]{Stannoidite (Cu$_{8}$(Fe,Zn)$_{3}$Sn$_{2}$S$_{12}$): A8B2C12D2E\_oI50\_23\_bcfk\_i\_3k\_j\_a}} - CIF

{\phantomsection\label{A8B2C12D2E_oI50_23_bcfk_i_3k_j_a_poscar}}
{\hyperref[A8B2C12D2E_oI50_23_bcfk_i_3k_j_a]{Stannoidite (Cu$_{8}$(Fe,Zn)$_{3}$Sn$_{2}$S$_{12}$): A8B2C12D2E\_oI50\_23\_bcfk\_i\_3k\_j\_a}} - POSCAR

{\phantomsection\label{ABC2_oI16_23_ab_i_k_cif}}
{\hyperref[ABC2_oI16_23_ab_i_k]{NaFeS$_{2}$: ABC2\_oI16\_23\_ab\_i\_k}} - CIF
\begin{lstlisting}[numbers=none,language={mylang}]
# CIF file
data_findsym-output
_audit_creation_method FINDSYM

_chemical_name_mineral 'NaFeS2'
_chemical_formula_sum 'Fe Na S2'

loop_
_publ_author_name
 'H. Boller'
 'H. Blaha'
_journal_name_full_name
;
 Monatshefte f{\"u}r Chemie - Chemical Monthly
;
_journal_volume 114
_journal_year 1983
_journal_page_first 145
_journal_page_last 154
_publ_Section_title
;
 Zur Kenntnis des Natriumthioferrates (III)
;

# Found in Pearson's Crystal Data - Crystal Structure Database for Inorganic Compounds, 2013

_aflow_title 'NaFeS$_{2}$ Structure'
_aflow_proto 'ABC2_oI16_23_ab_i_k'
_aflow_params 'a,b/a,c/a,z_{3},x_{4},y_{4},z_{4}'
_aflow_params_values '5.3999384419,1.15740740741,2.00555555555,0.28,0.25,0.2,0.115'
_aflow_Strukturbericht 'None'
_aflow_Pearson 'oI16'

_cell_length_a    5.3999384419
_cell_length_b    6.2499287522
_cell_length_c    10.8298765418
_cell_angle_alpha 90.0000000000
_cell_angle_beta  90.0000000000
_cell_angle_gamma 90.0000000000
 
_symmetry_space_group_name_H-M "I 2 2 2"
_symmetry_Int_Tables_number 23
 
loop_
_space_group_symop_id
_space_group_symop_operation_xyz
1 x,y,z
2 x,-y,-z
3 -x,y,-z
4 -x,-y,z
5 x+1/2,y+1/2,z+1/2
6 x+1/2,-y+1/2,-z+1/2
7 -x+1/2,y+1/2,-z+1/2
8 -x+1/2,-y+1/2,z+1/2
 
loop_
_atom_site_label
_atom_site_type_symbol
_atom_site_symmetry_multiplicity
_atom_site_Wyckoff_label
_atom_site_fract_x
_atom_site_fract_y
_atom_site_fract_z
_atom_site_occupancy
Fe1 Fe   2 a 0.00000 0.00000 0.00000 1.00000
Fe2 Fe   2 b 0.50000 0.00000 0.00000 1.00000
Na1 Na   4 i 0.00000 0.00000 0.28000 1.00000
S1  S    8 k 0.25000 0.20000 0.11500 1.00000
\end{lstlisting}
{\phantomsection\label{ABC2_oI16_23_ab_i_k_poscar}}
{\hyperref[ABC2_oI16_23_ab_i_k]{NaFeS$_{2}$: ABC2\_oI16\_23\_ab\_i\_k}} - POSCAR
\begin{lstlisting}[numbers=none,language={mylang}]
ABC2_oI16_23_ab_i_k & a,b/a,c/a,z3,x4,y4,z4 --params=5.3999384419,1.15740740741,2.00555555555,0.28,0.25,0.2,0.115 & I222 D_{2}^{8} #23 (abik) & oI16 & None & NaFeS2 &  & H. Boller and H. Blaha, Monatsh. Chem. 114, 145-154 (1983)
   1.00000000000000
  -2.69996922095000   3.12496437610000   5.41493827090000
   2.69996922095000  -3.12496437610000   5.41493827090000
   2.69996922095000   3.12496437610000  -5.41493827090000
    Fe    Na     S
     2     2     4
Direct
   0.00000000000000   0.00000000000000   0.00000000000000   Fe   (2a)
   0.00000000000000   0.50000000000000   0.50000000000000   Fe   (2b)
   0.28000000000000   0.28000000000000   0.00000000000000   Na   (4i)
  -0.28000000000000  -0.28000000000000   0.00000000000000   Na   (4i)
   0.31500000000000   0.36500000000000   0.45000000000000    S   (8k)
  -0.08500000000000  -0.13500000000000  -0.45000000000000    S   (8k)
   0.08500000000000  -0.36500000000000  -0.05000000000000    S   (8k)
  -0.31500000000000   0.13500000000000   0.05000000000000    S   (8k)
\end{lstlisting}
{\phantomsection\label{ABC4_oI12_23_a_b_k_cif}}
{\hyperref[ABC4_oI12_23_a_b_k]{BPS$_{4}$: ABC4\_oI12\_23\_a\_b\_k}} - CIF
\begin{lstlisting}[numbers=none,language={mylang}]
# CIF file
data_findsym-output
_audit_creation_method FINDSYM

_chemical_name_mineral 'BPS4'
_chemical_formula_sum 'B P S4'

loop_
_publ_author_name
 'A. Weiss'
 'H. Sch{\"a}fer'
_journal_name_full_name
;
 Zeitschrift f{\"u}r Naturforschung B
;
_journal_volume 18
_journal_year 1963
_journal_page_first 81
_journal_page_last 82
_publ_Section_title
;
 Zur Kenntnis von Bortetrathiophosphat BPS$_{4}$
;

# Found in Pearson's Crystal Data - Crystal Structure Database for Inorganic Compounds, 2013

_aflow_title 'BPS$_{4}$ Structure'
_aflow_proto 'ABC4_oI12_23_a_b_k'
_aflow_params 'a,b/a,c/a,x_{3},y_{3},z_{3}'
_aflow_params_values '5.2501580231,1.06666666665,1.7219047619,0.21,0.2,0.115'
_aflow_Strukturbericht 'None'
_aflow_Pearson 'oI12'

_cell_length_a    5.2501580231
_cell_length_b    5.6001685579
_cell_length_c    9.0402721007
_cell_angle_alpha 90.0000000000
_cell_angle_beta  90.0000000000
_cell_angle_gamma 90.0000000000
 
_symmetry_space_group_name_H-M "I 2 2 2"
_symmetry_Int_Tables_number 23
 
loop_
_space_group_symop_id
_space_group_symop_operation_xyz
1 x,y,z
2 x,-y,-z
3 -x,y,-z
4 -x,-y,z
5 x+1/2,y+1/2,z+1/2
6 x+1/2,-y+1/2,-z+1/2
7 -x+1/2,y+1/2,-z+1/2
8 -x+1/2,-y+1/2,z+1/2
 
loop_
_atom_site_label
_atom_site_type_symbol
_atom_site_symmetry_multiplicity
_atom_site_Wyckoff_label
_atom_site_fract_x
_atom_site_fract_y
_atom_site_fract_z
_atom_site_occupancy
B1 B   2 a 0.00000 0.00000 0.00000 1.00000
P1 P   2 b 0.50000 0.00000 0.00000 1.00000
S1 S   8 k 0.21000 0.20000 0.11500 1.00000
\end{lstlisting}
{\phantomsection\label{ABC4_oI12_23_a_b_k_poscar}}
{\hyperref[ABC4_oI12_23_a_b_k]{BPS$_{4}$: ABC4\_oI12\_23\_a\_b\_k}} - POSCAR
\begin{lstlisting}[numbers=none,language={mylang}]
ABC4_oI12_23_a_b_k & a,b/a,c/a,x3,y3,z3 --params=5.2501580231,1.06666666665,1.7219047619,0.21,0.2,0.115 & I222 D_{2}^{8} #23 (abk) & oI12 & None & BPS4 &  & A. Weiss and H. Sch{\"a}fer, Z. Naturforsch. B 18, 81-82 (1963)
   1.00000000000000
  -2.62507901155000   2.80008427895000   4.52013605035000
   2.62507901155000  -2.80008427895000   4.52013605035000
   2.62507901155000   2.80008427895000  -4.52013605035000
     B     P     S
     1     1     4
Direct
   0.00000000000000   0.00000000000000   0.00000000000000    B   (2a)
   0.00000000000000   0.50000000000000   0.50000000000000    P   (2b)
   0.31500000000000   0.32500000000000   0.41000000000000    S   (8k)
  -0.08500000000000  -0.09500000000000  -0.41000000000000    S   (8k)
   0.08500000000000  -0.32500000000000  -0.01000000000000    S   (8k)
  -0.31500000000000   0.09500000000000   0.01000000000000    S   (8k)
\end{lstlisting}
{\phantomsection\label{AB7CD2_oI44_24_a_b3d_c_ac_cif}}
{\hyperref[AB7CD2_oI44_24_a_b3d_c_ac]{Weberite (Na$_{2}$MgAlF$_{7}$): AB7CD2\_oI44\_24\_a\_b3d\_c\_ac}} - CIF

{\phantomsection\label{AB7CD2_oI44_24_a_b3d_c_ac_poscar}}
{\hyperref[AB7CD2_oI44_24_a_b3d_c_ac]{Weberite (Na$_{2}$MgAlF$_{7}$): AB7CD2\_oI44\_24\_a\_b3d\_c\_ac}} - POSCAR

{\phantomsection\label{A2B_oP12_26_abc_ab-H2S_cif}}
{\hyperref[A2B_oP12_26_abc_ab-H2S]{H$_{2}$S (70~GPa): A2B\_oP12\_26\_abc\_ab}} - CIF
\begin{lstlisting}[numbers=none,language={mylang}]
# CIF file 
data_findsym-output
_audit_creation_method FINDSYM

_chemical_name_mineral 'H2S'
_chemical_formula_sum 'H2 S'

loop_
_publ_author_name
 'Y. Li'
 'J. Hao'
 'H. Liu'
 'Y. Li'
 'Y. Ma'
_journal_name_full_name
;
 Journal of Chemical Physics
;
_journal_volume 140
_journal_year 2014
_journal_page_first 174712
_journal_page_last 174712
_publ_Section_title
;
 The metallization and superconductivity of dense hydrogen sulfide
;

_aflow_title 'H$_{2}$S (70~GPa) Structure'
_aflow_proto 'A2B_oP12_26_abc_ab'
_aflow_params 'a,b/a,c/a,y_{1},z_{1},y_{2},z_{2},y_{3},z_{3},y_{4},z_{4},x_{5},y_{5},z_{5}'
_aflow_params_values '4.6806,0.627034995513,1.05710806307,0.455,0.858,0.179,0.623,0.048,0.545,0.375,0.355,0.751,0.119,0.213'
_aflow_Strukturbericht 'None'
_aflow_Pearson 'oP12'

_symmetry_space_group_name_H-M "P m c 21"
_symmetry_Int_Tables_number 26
 
_cell_length_a    4.68060
_cell_length_b    2.93490
_cell_length_c    4.94790
_cell_angle_alpha 90.00000
_cell_angle_beta  90.00000
_cell_angle_gamma 90.00000
 
loop_
_space_group_symop_id
_space_group_symop_operation_xyz
1 x,y,z
2 -x,-y,z+1/2
3 -x,y,z
4 x,-y,z+1/2
 
loop_
_atom_site_label
_atom_site_type_symbol
_atom_site_symmetry_multiplicity
_atom_site_Wyckoff_label
_atom_site_fract_x
_atom_site_fract_y
_atom_site_fract_z
_atom_site_occupancy
H1 H   2 a 0.00000 0.45500 0.85800 1.00000
S1 S   2 a 0.00000 0.17900 0.62300 1.00000
H2 H   2 b 0.50000 0.04800 0.54500 1.00000
S2 S   2 b 0.50000 0.37500 0.35500 1.00000
H3 H   4 c 0.75100 0.11900 0.21300 1.00000
\end{lstlisting}
{\phantomsection\label{A2B_oP12_26_abc_ab-H2S_poscar}}
{\hyperref[A2B_oP12_26_abc_ab-H2S]{H$_{2}$S (70~GPa): A2B\_oP12\_26\_abc\_ab}} - POSCAR
\begin{lstlisting}[numbers=none,language={mylang}]
A2B_oP12_26_abc_ab & a,b/a,c/a,y1,z1,y2,z2,y3,z3,y4,z4,x5,y5,z5 --params=4.6806,0.627034995513,1.05710806307,0.455,0.858,0.179,0.623,0.048,0.545,0.375,0.355,0.751,0.119,0.213 & Pmc2_{1} C_{2v}^{2} #26 (a^2b^2c) & oP12 & None & H2S & H2S & Y. Li et al., J. Chem. Phys. 140, 174712(2014)
   1.00000000000000
   4.68060000000000   0.00000000000000   0.00000000000000
   0.00000000000000   2.93490000000000   0.00000000000000
   0.00000000000000   0.00000000000000   4.94790000000000
     H     S
     8     4
Direct
   0.00000000000000   0.45500000000000   0.85800000000000    H   (2a)
   0.00000000000000  -0.45500000000000   1.35800000000000    H   (2a)
   0.50000000000000   0.04800000000000   0.54500000000000    H   (2b)
   0.50000000000000  -0.04800000000000   1.04500000000000    H   (2b)
   0.75100000000000   0.11900000000000   0.21300000000000    H   (4c)
  -0.75100000000000  -0.11900000000000   0.71300000000000    H   (4c)
   0.75100000000000  -0.11900000000000   0.71300000000000    H   (4c)
  -0.75100000000000   0.11900000000000   0.21300000000000    H   (4c)
   0.00000000000000   0.17900000000000   0.62300000000000    S   (2a)
   0.00000000000000  -0.17900000000000   1.12300000000000    S   (2a)
   0.50000000000000   0.37500000000000   0.35500000000000    S   (2b)
   0.50000000000000  -0.37500000000000   0.85500000000000    S   (2b)
\end{lstlisting}
{\phantomsection\label{A2B_oP12_26_abc_ab-SeO2_cif}}
{\hyperref[A2B_oP12_26_abc_ab-SeO2]{$\beta$-SeO$_{2}$: A2B\_oP12\_26\_abc\_ab}} - CIF
\begin{lstlisting}[numbers=none,language={mylang}]
# CIF file
data_findsym-output
_audit_creation_method FINDSYM

_chemical_name_mineral 'beta-SeO2'
_chemical_formula_sum 'O2 Se'

loop_
_publ_author_name
 'D. Orosel'
 'O. Leynaud'
 'P. Balog'
 'M. Jansen'
_journal_name_full_name
;
 Journal of Solid State Chemistry
;
_journal_volume 177
_journal_year 2004
_journal_page_first 1631
_journal_page_last 1638
_publ_Section_title
;
 Pressure-temperature phase diagram of SeO$_{2}$. Characterization of new phases
;

# Found in Pearson's Crystal Data - Crystal Structure Database for Inorganic Compounds, 2013

_aflow_title '$\beta$-SeO$_{2}$ Structure'
_aflow_proto 'A2B_oP12_26_abc_ab'
_aflow_params 'a,b/a,c/a,y_{1},z_{1},y_{2},z_{2},y_{3},z_{3},y_{4},z_{4},x_{5},y_{5},z_{5}'
_aflow_params_values '5.0725824349,0.881353258932,1.48474034935,0.12,0.0,0.2484,0.461,0.246,0.289,0.3781,0.0862,0.253,0.652,0.12'
_aflow_Strukturbericht 'None'
_aflow_Pearson 'oP12'

_cell_length_a    5.0725824349
_cell_length_b    4.4707370602
_cell_length_c    7.5314678165
_cell_angle_alpha 90.0000000000
_cell_angle_beta  90.0000000000
_cell_angle_gamma 90.0000000000
 
_symmetry_space_group_name_H-M "P m c 21"
_symmetry_Int_Tables_number 26
 
loop_
_space_group_symop_id
_space_group_symop_operation_xyz
1 x,y,z
2 -x,-y,z+1/2
3 -x,y,z
4 x,-y,z+1/2
 
loop_
_atom_site_label
_atom_site_type_symbol
_atom_site_symmetry_multiplicity
_atom_site_Wyckoff_label
_atom_site_fract_x
_atom_site_fract_y
_atom_site_fract_z
_atom_site_occupancy
O1 O   2 a 0.00000 0.12000 0.00000 1.00000
Se1 Se   2 a 0.00000 0.24840 0.46100 1.00000
O2 O   2 b 0.50000 0.24600 0.28900 1.00000
Se2 Se   2 b 0.50000 0.37810 0.08620 1.00000
O3 O   4 c 0.25300 0.65200 0.12000 1.00000
\end{lstlisting}
{\phantomsection\label{A2B_oP12_26_abc_ab-SeO2_poscar}}
{\hyperref[A2B_oP12_26_abc_ab-SeO2]{$\beta$-SeO$_{2}$: A2B\_oP12\_26\_abc\_ab}} - POSCAR
\begin{lstlisting}[numbers=none,language={mylang}]
A2B_oP12_26_abc_ab & a,b/a,c/a,y1,z1,y2,z2,y3,z3,y4,z4,x5,y5,z5 --params=5.0725824349,0.881353258932,1.48474034935,0.12,0.0,0.2484,0.461,0.246,0.289,0.3781,0.0862,0.253,0.652,0.12 & Pmc2_{1} C_{2v}^{2} #26 (a^2b^2c) & oP12 & None & SeO2 & beta & D. Orosel et al., J. Solid State Chem. 177, 1631-1638 (2004)
   1.00000000000000
   5.07258243490000   0.00000000000000   0.00000000000000
   0.00000000000000   4.47073706020000   0.00000000000000
   0.00000000000000   0.00000000000000   7.53146781650000
     O    Se
     8     4
Direct
   0.00000000000000   0.12000000000000   0.00000000000000    O   (2a)
   0.00000000000000  -0.12000000000000   0.50000000000000    O   (2a)
   0.50000000000000   0.24600000000000   0.28900000000000    O   (2b)
   0.50000000000000  -0.24600000000000   0.78900000000000    O   (2b)
   0.25300000000000   0.65200000000000   0.12000000000000    O   (4c)
  -0.25300000000000  -0.65200000000000   0.62000000000000    O   (4c)
   0.25300000000000  -0.65200000000000   0.62000000000000    O   (4c)
  -0.25300000000000   0.65200000000000   0.12000000000000    O   (4c)
   0.00000000000000   0.24840000000000   0.46100000000000   Se   (2a)
   0.00000000000000  -0.24840000000000   0.96100000000000   Se   (2a)
   0.50000000000000   0.37810000000000   0.08620000000000   Se   (2b)
   0.50000000000000  -0.37810000000000   0.58620000000000   Se   (2b)
\end{lstlisting}
{\phantomsection\label{A5B_oP24_26_3a3b2c_ab_cif}}
{\hyperref[A5B_oP24_26_3a3b2c_ab]{TlP$_{5}$: A5B\_oP24\_26\_3a3b2c\_ab}} - CIF

{\phantomsection\label{A5B_oP24_26_3a3b2c_ab_poscar}}
{\hyperref[A5B_oP24_26_3a3b2c_ab]{TlP$_{5}$: A5B\_oP24\_26\_3a3b2c\_ab}} - POSCAR

{\phantomsection\label{A6B4C16D_oP108_27_abcd4e_4e_16e_e_cif}}
{\hyperref[A6B4C16D_oP108_27_abcd4e_4e_16e_e]{Ca$_{4}$Al$_{6}$O$_{16}$S: A6B4C16D\_oP108\_27\_abcd4e\_4e\_16e\_e}} - CIF

{\phantomsection\label{A6B4C16D_oP108_27_abcd4e_4e_16e_e_poscar}}
{\hyperref[A6B4C16D_oP108_27_abcd4e_4e_16e_e]{Ca$_{4}$Al$_{6}$O$_{16}$S: A6B4C16D\_oP108\_27\_abcd4e\_4e\_16e\_e}} - POSCAR

{\phantomsection\label{A2B_oP12_29_2a_a_cif}}
{\hyperref[A2B_oP12_29_2a_a]{ZrO$_{2}$: A2B\_oP12\_29\_2a\_a}} - CIF
\begin{lstlisting}[numbers=none,language={mylang}]
# CIF file
data_findsym-output
_audit_creation_method FINDSYM

_chemical_name_mineral 'ZrO2'
_chemical_formula_sum 'O2 Zr'

loop_
_publ_author_name
 'J. Grins'
 'P.-O. K{\"a}ll'
 'G. Svensson'
_journal_name_full_name
;
 Journal of Materials Chemistry
;
_journal_volume 4
_journal_year 1994
_journal_page_first 1293
_journal_page_last 1301
_publ_Section_title
;
 Phases in the Zr$_{x}$Ta$_{1-x}$(O,N)$_{y}$ system, formed by ammonolysis of Zr-Ta gels: Preparation of a baddeleyite-type solid solution phase Zr$_{x}$Ta$_{1-x}$O$_{1+x}$N$_{1-x}$, $0 \le x \le 1$
;

# Found in Pearson's Crystal Data - Crystal Structure Database for Inorganic Compounds, 2013

_aflow_title 'ZrO$_{2}$ Structure'
_aflow_proto 'A2B_oP12_29_2a_a'
_aflow_params 'a,b/a,c/a,x_{1},y_{1},z_{1},x_{2},y_{2},z_{2},x_{3},y_{3},z_{3}'
_aflow_params_values '5.2594682584,0.963498098863,0.965209125484,0.639,0.068,0.0,0.771,0.537,0.106,0.53,0.267,0.356'
_aflow_Strukturbericht 'None'
_aflow_Pearson 'oP12'

_cell_length_a    5.2594682584
_cell_length_b    5.0674876680
_cell_length_c    5.0764867582
_cell_angle_alpha 90.0000000000
_cell_angle_beta  90.0000000000
_cell_angle_gamma 90.0000000000
 
_symmetry_space_group_name_H-M "P c a 21"
_symmetry_Int_Tables_number 29
 
loop_
_space_group_symop_id
_space_group_symop_operation_xyz
1 x,y,z
2 -x,-y,z+1/2
3 -x+1/2,y,z+1/2
4 x+1/2,-y,z
 
loop_
_atom_site_label
_atom_site_type_symbol
_atom_site_symmetry_multiplicity
_atom_site_Wyckoff_label
_atom_site_fract_x
_atom_site_fract_y
_atom_site_fract_z
_atom_site_occupancy
O1  O    4 a 0.63900 0.06800 0.00000 1.00000
O2  O    4 a 0.77100 0.53700 0.10600 1.00000
Zr1 Zr   4 a 0.53000 0.26700 0.35600 1.00000
\end{lstlisting}
{\phantomsection\label{A2B_oP12_29_2a_a_poscar}}
{\hyperref[A2B_oP12_29_2a_a]{ZrO$_{2}$: A2B\_oP12\_29\_2a\_a}} - POSCAR
\begin{lstlisting}[numbers=none,language={mylang}]
A2B_oP12_29_2a_a & a,b/a,c/a,x1,y1,z1,x2,y2,z2,x3,y3,z3 --params=5.2594682584,0.963498098863,0.965209125484,0.639,0.068,0.0,0.771,0.537,0.106,0.53,0.267,0.356 & Pca2_{1} C_{2v}^{5} #29 (a^3) & oP12 & None & ZrO2 &  & J. Grins and P.-O. K{\"a}ll and G. Svensson, J. Mater. Chem. 4, 1293-1301 (1994)
   1.00000000000000
   5.25946825840000   0.00000000000000   0.00000000000000
   0.00000000000000   5.06748766800000   0.00000000000000
   0.00000000000000   0.00000000000000   5.07648675820000
     O    Zr
     8     4
Direct
   0.63900000000000   0.06800000000000   0.00000000000000    O   (4a)
  -0.63900000000000  -0.06800000000000   0.50000000000000    O   (4a)
   1.13900000000000  -0.06800000000000   0.00000000000000    O   (4a)
  -0.13900000000000   0.06800000000000   0.50000000000000    O   (4a)
   0.77100000000000   0.53700000000000   0.10600000000000    O   (4a)
  -0.77100000000000  -0.53700000000000   0.60600000000000    O   (4a)
   1.27100000000000  -0.53700000000000   0.10600000000000    O   (4a)
  -0.27100000000000   0.53700000000000   0.60600000000000    O   (4a)
   0.53000000000000   0.26700000000000   0.35600000000000   Zr   (4a)
  -0.53000000000000  -0.26700000000000   0.85600000000000   Zr   (4a)
   1.03000000000000  -0.26700000000000   0.35600000000000   Zr   (4a)
  -0.03000000000000   0.26700000000000   0.85600000000000   Zr   (4a)
\end{lstlisting}
{\phantomsection\label{AB2_oP12_29_a_2a_cif}}
{\hyperref[AB2_oP12_29_a_2a]{Pyrite (FeS$_{2}$, Low-temperature): AB2\_oP12\_29\_a\_2a}} - CIF
\begin{lstlisting}[numbers=none,language={mylang}]
# CIF file
data_findsym-output
_audit_creation_method FINDSYM

_chemical_name_mineral 'FeS2'
_chemical_formula_sum 'Fe S2'

loop_
_publ_author_name
 'P. Bayliss'
_journal_name_full_name
;
 American Mineralogist
;
_journal_volume 74
_journal_year 1989
_journal_page_first 1168
_journal_page_last 1176
_publ_Section_title
;
 Crystal chemistry and crystallography of some minerals within the pyrite group
;

# Found in Pearson's Crystal Data - Crystal Structure Database for Inorganic Compounds, 2013

_aflow_title 'Pyrite (FeS$_{2}$, Low-temperature) Structure'
_aflow_proto 'AB2_oP12_29_a_2a'
_aflow_params 'a,b/a,c/a,x_{1},y_{1},z_{1},x_{2},y_{2},z_{2},x_{3},y_{3},z_{3}'
_aflow_params_values '5.4179557747,1.0,1.0,0.5049,0.2419,0.0,0.615,0.135,0.3834,0.615,0.635,0.1134'
_aflow_Strukturbericht 'None'
_aflow_Pearson 'oP12'

_cell_length_a    5.4179557747
_cell_length_b    5.4179557747
_cell_length_c    5.4179557747
_cell_angle_alpha 90.0000000000
_cell_angle_beta  90.0000000000
_cell_angle_gamma 90.0000000000
 
_symmetry_space_group_name_H-M "P c a 21"
_symmetry_Int_Tables_number 29
 
loop_
_space_group_symop_id
_space_group_symop_operation_xyz
1 x,y,z
2 -x,-y,z+1/2
3 -x+1/2,y,z+1/2
4 x+1/2,-y,z
 
loop_
_atom_site_label
_atom_site_type_symbol
_atom_site_symmetry_multiplicity
_atom_site_Wyckoff_label
_atom_site_fract_x
_atom_site_fract_y
_atom_site_fract_z
_atom_site_occupancy
Fe1 Fe   4 a 0.50490 0.24190 0.00000 1.00000
S1  S    4 a 0.61500 0.13500 0.38340 1.00000
S2  S    4 a 0.61500 0.63500 0.11340 1.00000
\end{lstlisting}
{\phantomsection\label{AB2_oP12_29_a_2a_poscar}}
{\hyperref[AB2_oP12_29_a_2a]{Pyrite (FeS$_{2}$, Low-temperature): AB2\_oP12\_29\_a\_2a}} - POSCAR
\begin{lstlisting}[numbers=none,language={mylang}]
AB2_oP12_29_a_2a & a,b/a,c/a,x1,y1,z1,x2,y2,z2,x3,y3,z3 --params=5.4179557747,1.0,1.0,0.5049,0.2419,0.0,0.615,0.135,0.3834,0.615,0.635,0.1134 & Pca2_{1} C_{2v}^{5} #29 (a^3) & oP12 & None & FeS2 &  & P. Bayliss, Am. Mineral. 74, 1168-1176 (1989)
   1.00000000000000
   5.41795577470000   0.00000000000000   0.00000000000000
   0.00000000000000   5.41795577470000   0.00000000000000
   0.00000000000000   0.00000000000000   5.41795577470000
    Fe     S
     4     8
Direct
   0.50490000000000   0.24190000000000   0.00000000000000   Fe   (4a)
  -0.50490000000000  -0.24190000000000   0.50000000000000   Fe   (4a)
   1.00490000000000  -0.24190000000000   0.00000000000000   Fe   (4a)
  -0.00490000000000   0.24190000000000   0.50000000000000   Fe   (4a)
   0.61500000000000   0.13500000000000   0.38340000000000    S   (4a)
  -0.61500000000000  -0.13500000000000   0.88340000000000    S   (4a)
   1.11500000000000  -0.13500000000000   0.38340000000000    S   (4a)
  -0.11500000000000   0.13500000000000   0.88340000000000    S   (4a)
   0.61500000000000   0.63500000000000   0.11340000000000    S   (4a)
  -0.61500000000000  -0.63500000000000   0.61340000000000    S   (4a)
   1.11500000000000  -0.63500000000000   0.11340000000000    S   (4a)
  -0.11500000000000   0.63500000000000   0.61340000000000    S   (4a)
\end{lstlisting}
{\phantomsection\label{ABC_oP12_29_a_a_a_cif}}
{\hyperref[ABC_oP12_29_a_a_a]{Cobaltite (CoAsS): ABC\_oP12\_29\_a\_a\_a}} - CIF
\begin{lstlisting}[numbers=none,language={mylang}]
# CIF file
data_findsym-output
_audit_creation_method FINDSYM

_chemical_name_mineral 'CoAsS'
_chemical_formula_sum 'As Co S'

loop_
_publ_author_name
 'M. E. Fleet'
 'P. C. Burns'
_journal_name_full_name
;
 Canadian Mineralogist
;
_journal_volume 28
_journal_year 1990
_journal_page_first 719
_journal_page_last 723
_publ_Section_title
;
 Structure and twinning of cobaltite
;

# Found in Pearson's Crystal Data - Crystal Structure Database for Inorganic Compounds, 2013

_aflow_title 'Cobaltite (CoAsS) Structure'
_aflow_proto 'ABC_oP12_29_a_a_a'
_aflow_params 'a,b/a,c/a,x_{1},y_{1},z_{1},x_{2},y_{2},z_{2},x_{3},y_{3},z_{3}'
_aflow_params_values '5.2594682584,0.963498098863,0.965209125484,0.61885,0.63065,0.11668,0.50496,0.24091,0.0,0.61734,0.13129,0.37996'
_aflow_Strukturbericht 'None'
_aflow_Pearson 'oP12'

_cell_length_a    5.2594682584
_cell_length_b    5.0674876680
_cell_length_c    5.0764867582
_cell_angle_alpha 90.0000000000
_cell_angle_beta  90.0000000000
_cell_angle_gamma 90.0000000000
 
_symmetry_space_group_name_H-M "P c a 21"
_symmetry_Int_Tables_number 29
 
loop_
_space_group_symop_id
_space_group_symop_operation_xyz
1 x,y,z
2 -x,-y,z+1/2
3 -x+1/2,y,z+1/2
4 x+1/2,-y,z
 
loop_
_atom_site_label
_atom_site_type_symbol
_atom_site_symmetry_multiplicity
_atom_site_Wyckoff_label
_atom_site_fract_x
_atom_site_fract_y
_atom_site_fract_z
_atom_site_occupancy
As1 As   4 a 0.61885 0.63065 0.11668 1.00000
Co1 Co   4 a 0.50496 0.24091 0.00000 1.00000
S1  S    4 a 0.61734 0.13129 0.37996 1.00000
\end{lstlisting}
{\phantomsection\label{ABC_oP12_29_a_a_a_poscar}}
{\hyperref[ABC_oP12_29_a_a_a]{Cobaltite (CoAsS): ABC\_oP12\_29\_a\_a\_a}} - POSCAR
\begin{lstlisting}[numbers=none,language={mylang}]
ABC_oP12_29_a_a_a & a,b/a,c/a,x1,y1,z1,x2,y2,z2,x3,y3,z3 --params=5.2594682584,0.963498098863,0.965209125484,0.61885,0.63065,0.11668,0.50496,0.24091,0.0,0.61734,0.13129,0.37996 & Pca2_{1} C_{2v}^{5} #29 (a^3) & oP12 & None & CoAsS &  & M. E. Fleet and P. C. Burns, Can. Mineral. 28, 719-723 (1990)
   1.00000000000000
   5.25946825840000   0.00000000000000   0.00000000000000
   0.00000000000000   5.06748766800000   0.00000000000000
   0.00000000000000   0.00000000000000   5.07648675820000
    As    Co     S
     4     4     4
Direct
   0.61885000000000   0.63065000000000   0.11668000000000   As   (4a)
  -0.61885000000000  -0.63065000000000   0.61668000000000   As   (4a)
   1.11885000000000  -0.63065000000000   0.11668000000000   As   (4a)
  -0.11885000000000   0.63065000000000   0.61668000000000   As   (4a)
   0.50496000000000   0.24091000000000   0.00000000000000   Co   (4a)
  -0.50496000000000  -0.24091000000000   0.50000000000000   Co   (4a)
   1.00496000000000  -0.24091000000000   0.00000000000000   Co   (4a)
  -0.00496000000000   0.24091000000000   0.50000000000000   Co   (4a)
   0.61734000000000   0.13129000000000   0.37996000000000    S   (4a)
  -0.61734000000000  -0.13129000000000   0.87996000000000    S   (4a)
   1.11734000000000  -0.13129000000000   0.37996000000000    S   (4a)
  -0.11734000000000   0.13129000000000   0.87996000000000    S   (4a)
\end{lstlisting}
{\phantomsection\label{A5B3C15_oP46_30_a2c_bc_a7c_cif}}
{\hyperref[A5B3C15_oP46_30_a2c_bc_a7c]{Bi$_{5}$Nb$_{3}$O$_{15}$: A5B3C15\_oP46\_30\_a2c\_bc\_a7c}} - CIF

{\phantomsection\label{A5B3C15_oP46_30_a2c_bc_a7c_poscar}}
{\hyperref[A5B3C15_oP46_30_a2c_bc_a7c]{Bi$_{5}$Nb$_{3}$O$_{15}$: A5B3C15\_oP46\_30\_a2c\_bc\_a7c}} - POSCAR

{\phantomsection\label{ABC3_oP20_30_2a_c_3c_cif}}
{\hyperref[ABC3_oP20_30_2a_c_3c]{CuBrSe$_{3}$: ABC3\_oP20\_30\_2a\_c\_3c}} - CIF
\begin{lstlisting}[numbers=none,language={mylang}]
# CIF file
data_findsym-output
_audit_creation_method FINDSYM

_chemical_name_mineral 'CuBrSe3'
_chemical_formula_sum 'Br Cu Se3'

loop_
_publ_author_name
 'T. Sakuma'
 'T. Kaneko'
 'T. Kurita'
 'H. Takahashi'
_journal_name_full_name
;
 Journal of the Physical Society of Japan
;
_journal_volume 60
_journal_year 1991
_journal_page_first 1608
_journal_page_last 1611
_publ_Section_title
;
 Crystal structure of CuBrSe$_{3}$
;

# Found in Pearson's Crystal Data - Crystal Structure Database for Inorganic Compounds, 2013

_aflow_title 'CuBrSe$_{3}$ Structure'
_aflow_proto 'ABC3_oP20_30_2a_c_3c'
_aflow_params 'a,b/a,c/a,z_{1},z_{2},x_{3},y_{3},z_{3},x_{4},y_{4},z_{4},x_{5},y_{5},z_{5},x_{6},y_{6},z_{6}'
_aflow_params_values '4.480982758,1.71211783083,3.19772372237,0.8543,0.5,0.3011,0.721,0.4096,0.3607,0.7305,0.1733,0.5847,0.8629,0.0496,0.6302,0.8523,0.2991'
_aflow_Strukturbericht 'None'
_aflow_Pearson 'oP20'

_cell_length_a    4.4809827580
_cell_length_b    7.6719704796
_cell_length_c    14.3289448648
_cell_angle_alpha 90.0000000000
_cell_angle_beta  90.0000000000
_cell_angle_gamma 90.0000000000
 
_symmetry_space_group_name_H-M "P n c 2"
_symmetry_Int_Tables_number 30
 
loop_
_space_group_symop_id
_space_group_symop_operation_xyz
1 x,y,z
2 -x,-y,z
3 -x,y+1/2,z+1/2
4 x,-y+1/2,z+1/2
 
loop_
_atom_site_label
_atom_site_type_symbol
_atom_site_symmetry_multiplicity
_atom_site_Wyckoff_label
_atom_site_fract_x
_atom_site_fract_y
_atom_site_fract_z
_atom_site_occupancy
Br1 Br   2 a 0.00000 0.00000 0.85430 1.00000
Br2 Br   2 a 0.00000 0.00000 0.50000 1.00000
Cu1 Cu   4 c 0.30110 0.72100 0.40960 1.00000
Se1 Se   4 c 0.36070 0.73050 0.17330 1.00000
Se2 Se   4 c 0.58470 0.86290 0.04960 1.00000
Se3 Se   4 c 0.63020 0.85230 0.29910 1.00000
\end{lstlisting}
{\phantomsection\label{ABC3_oP20_30_2a_c_3c_poscar}}
{\hyperref[ABC3_oP20_30_2a_c_3c]{CuBrSe$_{3}$: ABC3\_oP20\_30\_2a\_c\_3c}} - POSCAR

{\phantomsection\label{A13B2C2_oP34_32_a6c_c_c_cif}}
{\hyperref[A13B2C2_oP34_32_a6c_c_c]{Re$_{2}$O$_{5}$[SO$_{4}$]$_{2}$: A13B2C2\_oP34\_32\_a6c\_c\_c}} - CIF

{\phantomsection\label{A13B2C2_oP34_32_a6c_c_c_poscar}}
{\hyperref[A13B2C2_oP34_32_a6c_c_c]{Re$_{2}$O$_{5}$[SO$_{4}$]$_{2}$: A13B2C2\_oP34\_32\_a6c\_c\_c}} - POSCAR

{\phantomsection\label{A2B3_oP40_33_4a_6a_cif}}
{\hyperref[A2B3_oP40_33_4a_6a]{$\kappa$-alumina (Al$_{2}$O$_{3}$): A2B3\_oP40\_33\_4a\_6a}} - CIF

{\phantomsection\label{A2B3_oP40_33_4a_6a_poscar}}
{\hyperref[A2B3_oP40_33_4a_6a]{$\kappa$-alumina (Al$_{2}$O$_{3}$): A2B3\_oP40\_33\_4a\_6a}} - POSCAR

{\phantomsection\label{A2B8C_oP22_34_c_4c_a_cif}}
{\hyperref[A2B8C_oP22_34_c_4c_a]{TiAl$_{2}$Br$_{8}$: A2B8C\_oP22\_34\_c\_4c\_a}} - CIF
\begin{lstlisting}[numbers=none,language={mylang}]
# CIF file
data_findsym-output
_audit_creation_method FINDSYM

_chemical_name_mineral 'TiAl2Br8'
_chemical_formula_sum 'Al2 Br8 Ti'

loop_
_publ_author_name
 'S. I. Troyanov'
 'V. B. Rybakov'
 'V. M. Ionov'
_journal_name_full_name
;
 Russian Journal of Inorganic Chemistry
;
_journal_volume 35
_journal_year 1990
_journal_page_first 882
_journal_page_last 887
_publ_Section_title
;
 Synthesis and Crystal Structure of TiBr$_{4}$, TiBr$_{3}$ and Ti(AlBr$_{4}$)$_{2}$
;

# Found in Pearson's Crystal Data - Crystal Structure Database for Inorganic Compounds, 2013

_aflow_title 'TiAl$_{2}$Br$_{8}$ Structure'
_aflow_proto 'A2B8C_oP22_34_c_4c_a'
_aflow_params 'a,b/a,c/a,z_{1},x_{2},y_{2},z_{2},x_{3},y_{3},z_{3},x_{4},y_{4},z_{4},x_{5},y_{5},z_{5},x_{6},y_{6},z_{6}'
_aflow_params_values '6.2740008991,2.04032515143,1.38284985656,0.5,0.605,0.8135,0.499,0.7349,0.6796,0.009,0.743,-0.0925,0.3064,0.2388,0.5935,0.2196,0.7131,0.6459,0.513'
_aflow_Strukturbericht 'None'
_aflow_Pearson 'oP22'

_cell_length_a    6.2740008991
_cell_length_b    12.8010018345
_cell_length_c    8.6760012434
_cell_angle_alpha 90.0000000000
_cell_angle_beta  90.0000000000
_cell_angle_gamma 90.0000000000
 
_symmetry_space_group_name_H-M "P n n 2"
_symmetry_Int_Tables_number 34
 
loop_
_space_group_symop_id
_space_group_symop_operation_xyz
1 x,y,z
2 -x,-y,z
3 -x+1/2,y+1/2,z+1/2
4 x+1/2,-y+1/2,z+1/2
 
loop_
_atom_site_label
_atom_site_type_symbol
_atom_site_symmetry_multiplicity
_atom_site_Wyckoff_label
_atom_site_fract_x
_atom_site_fract_y
_atom_site_fract_z
_atom_site_occupancy
Ti1 Ti   2 a 0.00000 0.00000  0.50000 1.00000
Al1 Al   4 c 0.60500 0.81350  0.49900 1.00000
Br1 Br   4 c 0.73490 0.67960  0.00900 1.00000
Br2 Br   4 c 0.74300 -0.09250 0.30640 1.00000
Br3 Br   4 c 0.23880 0.59350  0.21960 1.00000
Br4 Br   4 c 0.71310 0.64590  0.51300 1.00000
\end{lstlisting}
{\phantomsection\label{A2B8C_oP22_34_c_4c_a_poscar}}
{\hyperref[A2B8C_oP22_34_c_4c_a]{TiAl$_{2}$Br$_{8}$: A2B8C\_oP22\_34\_c\_4c\_a}} - POSCAR

{\phantomsection\label{AB2_oP6_34_a_c_cif}}
{\hyperref[AB2_oP6_34_a_c]{FeSb$_{2}$: AB2\_oP6\_34\_a\_c}} - CIF
\begin{lstlisting}[numbers=none,language={mylang}]
# CIF file
data_findsym-output
_audit_creation_method FINDSYM

_chemical_name_mineral 'FeSb2'
_chemical_formula_sum 'Fe Sb2'

loop_
_publ_author_name
 'H. Holseth'
 'A. Kjekshus'
_journal_name_full_name
;
 Acta Chemica Scandinavica
;
_journal_volume 23
_journal_year 1969
_journal_page_first 3043
_journal_page_last 3050
_publ_Section_title
;
 Compounds with the Marcasite Type Crystal Structure. IV. The Crystal Structure of FeSb$_{2}$
;

# Found in Pearson's Crystal Data - Crystal Structure Database for Inorganic Compounds, 2013

_aflow_title 'FeSb$_{2}$ Structure'
_aflow_proto 'AB2_oP6_34_a_c'
_aflow_params 'a,b/a,c/a,z_{1},x_{2},y_{2},z_{2}'
_aflow_params_values '5.8327827022,1.12083390481,0.548158688784,0.5,0.6881,0.8565,0.0097'
_aflow_Strukturbericht 'None'
_aflow_Pearson 'oP6'

_cell_length_a    5.8327827022
_cell_length_b    6.5375806120
_cell_length_c    3.1972905180
_cell_angle_alpha 90.0000000000
_cell_angle_beta  90.0000000000
_cell_angle_gamma 90.0000000000
 
_symmetry_space_group_name_H-M "P n n 2"
_symmetry_Int_Tables_number 34
 
loop_
_space_group_symop_id
_space_group_symop_operation_xyz
1 x,y,z
2 -x,-y,z
3 -x+1/2,y+1/2,z+1/2
4 x+1/2,-y+1/2,z+1/2
 
loop_
_atom_site_label
_atom_site_type_symbol
_atom_site_symmetry_multiplicity
_atom_site_Wyckoff_label
_atom_site_fract_x
_atom_site_fract_y
_atom_site_fract_z
_atom_site_occupancy
Fe1 Fe   2 a 0.00000 0.00000 0.50000 1.00000
Sb1 Sb   4 c 0.68810 0.85650 0.00970 1.00000
\end{lstlisting}
{\phantomsection\label{AB2_oP6_34_a_c_poscar}}
{\hyperref[AB2_oP6_34_a_c]{FeSb$_{2}$: AB2\_oP6\_34\_a\_c}} - POSCAR
\begin{lstlisting}[numbers=none,language={mylang}]
AB2_oP6_34_a_c & a,b/a,c/a,z1,x2,y2,z2 --params=5.8327827022,1.12083390481,0.548158688784,0.5,0.6881,0.8565,0.0097 & Pnn2 C_{2v}^{10} #34 (ac) & oP6 & None & FeSb2 &  & H. Holseth and A. Kjekshus, Acta Chem. Scand. 23, 3043-3050 (1969)
   1.00000000000000
   5.83278270220000   0.00000000000000   0.00000000000000
   0.00000000000000   6.53758061200000   0.00000000000000
   0.00000000000000   0.00000000000000   3.19729051800000
    Fe    Sb
     2     4
Direct
   0.00000000000000   0.00000000000000   0.50000000000000   Fe   (2a)
   0.50000000000000   0.50000000000000   1.00000000000000   Fe   (2a)
   0.68810000000000   0.85650000000000   0.00970000000000   Sb   (4c)
  -0.68810000000000  -0.85650000000000   0.00970000000000   Sb   (4c)
   1.18810000000000  -0.35650000000000   0.50970000000000   Sb   (4c)
  -0.18810000000000   1.35650000000000   0.50970000000000   Sb   (4c)
\end{lstlisting}
{\phantomsection\label{AB8C2_oC22_35_a_ab3e_e_cif}}
{\hyperref[AB8C2_oC22_35_a_ab3e_e]{V$_{2}$MoO$_{8}$: AB8C2\_oC22\_35\_a\_ab3e\_e}} - CIF
\begin{lstlisting}[numbers=none,language={mylang}]
# CIF file
data_findsym-output
_audit_creation_method FINDSYM

_chemical_name_mineral 'V2MoO8'
_chemical_formula_sum 'Mo O8 V2'

loop_
_publ_author_name
 'P. Mah{\\'e}-Pailleret'
_journal_year 1970
_publ_Section_title
;
 Contribution {\`a} l\'{\\'e}tude chimique et structurale des compos{\\'e}s AB$_{2}$O rencontr{\\'e}s dans les syst{\`e}mes Mo-VO, UVO et U-Mo-O
;

# Found in Pearson's Crystal Data - Crystal Structure Database for Inorganic Compounds, 2013

_aflow_title 'V$_{2}$MoO$_{8}$ Structure'
_aflow_proto 'AB8C2_oC22_35_a_ab3e_e'
_aflow_params 'a,b/a,c/a,z_{1},z_{2},z_{3},y_{4},z_{4},y_{5},z_{5},y_{6},z_{6},y_{7},z_{7}'
_aflow_params_values '4.534758023,1.1408839779,2.72580110498,0.0,0.5838,-0.042,0.189,0.5461,0.0961,0.122,0.2982,0.1399,0.1866,0.0038'
_aflow_Strukturbericht 'None'
_aflow_Pearson 'oC22'

_cell_length_a    4.5347580230
_cell_length_b    5.1736327721
_cell_length_c    12.3608484299
_cell_angle_alpha 90.0000000000
_cell_angle_beta  90.0000000000
_cell_angle_gamma 90.0000000000
 
_symmetry_space_group_name_H-M "C m m 2"
_symmetry_Int_Tables_number 35
 
loop_
_space_group_symop_id
_space_group_symop_operation_xyz
1 x,y,z
2 -x,-y,z
3 -x,y,z
4 x,-y,z
5 x+1/2,y+1/2,z
6 -x+1/2,-y+1/2,z
7 -x+1/2,y+1/2,z
8 x+1/2,-y+1/2,z
 
loop_
_atom_site_label
_atom_site_type_symbol
_atom_site_symmetry_multiplicity
_atom_site_Wyckoff_label
_atom_site_fract_x
_atom_site_fract_y
_atom_site_fract_z
_atom_site_occupancy
Mo1 Mo   2 a 0.00000 0.00000 0.00000  1.00000
O1  O    2 a 0.00000 0.00000 0.58380  1.00000
O2  O    2 b 0.00000 0.50000 -0.04200 1.00000
O3  O    4 e 0.00000 0.18900 0.54610  1.00000
O4  O    4 e 0.00000 0.09610 0.12200  1.00000
O5  O    4 e 0.00000 0.29820 0.13990  1.00000
V1  V    4 e 0.00000 0.18660 0.00380  1.00000
\end{lstlisting}
{\phantomsection\label{AB8C2_oC22_35_a_ab3e_e_poscar}}
{\hyperref[AB8C2_oC22_35_a_ab3e_e]{V$_{2}$MoO$_{8}$: AB8C2\_oC22\_35\_a\_ab3e\_e}} - POSCAR
\begin{lstlisting}[numbers=none,language={mylang}]
AB8C2_oC22_35_a_ab3e_e & a,b/a,c/a,z1,z2,z3,y4,z4,y5,z5,y6,z6,y7,z7 --params=4.534758023,1.1408839779,2.72580110498,0.0,0.5838,-0.042,0.189,0.5461,0.0961,0.122,0.2982,0.1399,0.1866,0.0038 & Cmm2 C_{2v}^{11} #35 (a^2be^4) & oC22 & None & V2MoO8 &  & P. Mah{\'e}-Pailleret, (1970)
   1.00000000000000
   2.26737901150000  -2.58681638605000   0.00000000000000
   2.26737901150000   2.58681638605000   0.00000000000000
   0.00000000000000   0.00000000000000  12.36084842990000
    Mo     O     V
     1     8     2
Direct
   0.00000000000000   0.00000000000000   0.00000000000000   Mo   (2a)
   0.00000000000000   0.00000000000000   0.58380000000000    O   (2a)
   0.50000000000000   0.50000000000000  -0.04200000000000    O   (2b)
  -0.18900000000000   0.18900000000000   0.54610000000000    O   (4e)
   0.18900000000000  -0.18900000000000   0.54610000000000    O   (4e)
  -0.09610000000000   0.09610000000000   0.12200000000000    O   (4e)
   0.09610000000000  -0.09610000000000   0.12200000000000    O   (4e)
  -0.29820000000000   0.29820000000000   0.13990000000000    O   (4e)
   0.29820000000000  -0.29820000000000   0.13990000000000    O   (4e)
  -0.18660000000000   0.18660000000000   0.00380000000000    V   (4e)
   0.18660000000000  -0.18660000000000   0.00380000000000    V   (4e)
\end{lstlisting}
{\phantomsection\label{AB_oC8_36_a_a_cif}}
{\hyperref[AB_oC8_36_a_a]{HCl: AB\_oC8\_36\_a\_a}} - CIF
\begin{lstlisting}[numbers=none,language={mylang}]
# CIF file 
data_findsym-output
_audit_creation_method FINDSYM

_chemical_name_mineral 'HCl'
_chemical_formula_sum 'Cl H'

loop_
_publ_author_name
 'E. S\\'{a}ndor'
 'R. F. C. Farrow'
_journal_name_full_name
;
 Nature
;
_journal_volume 213
_journal_year 1967
_journal_page_first 171
_journal_page_last 172
_publ_Section_title
;
 Crystal Structure of Solid Hydrogen Chloride and Deuterium Chloride
;

_aflow_title 'HCl Structure'
_aflow_proto 'AB_oC8_36_a_a'
_aflow_params 'a,b/a,c/a,y_{1},z_{1},y_{2},z_{2}'
_aflow_params_values '5.825,0.945115879828,0.922403433476,0.25,0.0,0.081,0.83'
_aflow_Strukturbericht 'None'
_aflow_Pearson 'oC8'

_symmetry_space_group_name_H-M "C m c 21"
_symmetry_Int_Tables_number 36
 
_cell_length_a    5.82500
_cell_length_b    5.50530
_cell_length_c    5.37300
_cell_angle_alpha 90.00000
_cell_angle_beta  90.00000
_cell_angle_gamma 90.00000
 
loop_
_space_group_symop_id
_space_group_symop_operation_xyz
1 x,y,z
2 -x,-y,z+1/2
3 -x,y,z
4 x,-y,z+1/2
5 x+1/2,y+1/2,z
6 -x+1/2,-y+1/2,z+1/2
7 -x+1/2,y+1/2,z
8 x+1/2,-y+1/2,z+1/2
 
loop_
_atom_site_label
_atom_site_type_symbol
_atom_site_symmetry_multiplicity
_atom_site_Wyckoff_label
_atom_site_fract_x
_atom_site_fract_y
_atom_site_fract_z
_atom_site_occupancy
Cl1 Cl   4 a 0.00000 0.25000 0.00000 1.00000
H1  H    4 a 0.00000 0.08100 0.83000 1.00000
\end{lstlisting}
{\phantomsection\label{AB_oC8_36_a_a_poscar}}
{\hyperref[AB_oC8_36_a_a]{HCl: AB\_oC8\_36\_a\_a}} - POSCAR
\begin{lstlisting}[numbers=none,language={mylang}]
AB_oC8_36_a_a & a,b/a,c/a,y1,z1,y2,z2 --params=5.825,0.945115879828,0.922403433476,0.25,0.0,0.081,0.83 & Cmc2_{1} C_{2v}^{12} #36 (a^2) & oC8 & None & HCl & HCl & E. S\'{a}ndor and R. F. C. Farrow, Nature 213, 171-172 (1967)
   1.00000000000000
   2.91250000000000  -2.75265000000000   0.00000000000000
   2.91250000000000   2.75265000000000   0.00000000000000
   0.00000000000000   0.00000000000000   5.37300000000000
    Cl     H
     2     2
Direct
  -0.25000000000000   0.25000000000000   0.00000000000000   Cl   (4a)
   0.25000000000000  -0.25000000000000   0.50000000000000   Cl   (4a)
  -0.08100000000000   0.08100000000000   0.83000000000000    H   (4a)
   0.08100000000000  -0.08100000000000   1.33000000000000    H   (4a)
\end{lstlisting}
{\phantomsection\label{A2B5C2_oC36_37_d_c2d_d_cif}}
{\hyperref[A2B5C2_oC36_37_d_c2d_d]{Li$_{2}$Si$_{2}$O$_{5}$: A2B5C2\_oC36\_37\_d\_c2d\_d}} - CIF
\begin{lstlisting}[numbers=none,language={mylang}]
# CIF file
data_findsym-output
_audit_creation_method FINDSYM

_chemical_name_mineral 'Li2Si2O5'
_chemical_formula_sum 'Li2 O5 Si2'

loop_
_publ_author_name
 'B. H. W. S {de Jong}'
 'P. G. G. Slaats'
 'H. T. J. Sup{\`e}r'
 'N. Veldman'
 'A. L. Spek'
_journal_name_full_name
;
 Journal of Non-Crystalline Solids
;
_journal_volume 176
_journal_year 1994
_journal_page_first 164
_journal_page_last 171
_publ_Section_title
;
 Extended structures in crystalline phyllosilicates: silica ring systems in lithium, rubidium, cesium, and cesium/lithium phyllosilicate
;

# Found in Pearson's Crystal Data - Crystal Structure Database for Inorganic Compounds, 2013

_aflow_title 'Li$_{2}$Si$_{2}$O$_{5}$ Structure'
_aflow_proto 'A2B5C2_oC36_37_d_c2d_d'
_aflow_params 'a,b/a,c/a,z_{1},x_{2},y_{2},z_{2},x_{3},y_{3},z_{3},x_{4},y_{4},z_{4},x_{5},y_{5},z_{5}'
_aflow_params_values '5.8071602545,2.51110728429,0.821939039086,0.0,0.654,0.0584,0.0469,0.6705,0.0718,0.471,0.0932,0.1377,0.4004,0.1552,0.14836,0.0571'
_aflow_Strukturbericht 'None'
_aflow_Pearson 'oC36'

_cell_length_a    5.8071602545
_cell_length_b    14.5824024161
_cell_length_c    4.7731317194
_cell_angle_alpha 90.0000000000
_cell_angle_beta  90.0000000000
_cell_angle_gamma 90.0000000000
 
_symmetry_space_group_name_H-M "C c c 2"
_symmetry_Int_Tables_number 37
 
loop_
_space_group_symop_id
_space_group_symop_operation_xyz
1 x,y,z
2 -x,-y,z
3 -x,y,z+1/2
4 x,-y,z+1/2
5 x+1/2,y+1/2,z
6 -x+1/2,-y+1/2,z
7 -x+1/2,y+1/2,z+1/2
8 x+1/2,-y+1/2,z+1/2
 
loop_
_atom_site_label
_atom_site_type_symbol
_atom_site_symmetry_multiplicity
_atom_site_Wyckoff_label
_atom_site_fract_x
_atom_site_fract_y
_atom_site_fract_z
_atom_site_occupancy
O1  O    4 c 0.25000 0.25000 0.00000 1.00000
Li1 Li   8 d 0.65400 0.05840 0.04690 1.00000
O2  O    8 d 0.67050 0.07180 0.47100 1.00000
O3  O    8 d 0.09320 0.13770 0.40040 1.00000
Si1 Si   8 d 0.15520 0.14836 0.05710 1.00000
\end{lstlisting}
{\phantomsection\label{A2B5C2_oC36_37_d_c2d_d_poscar}}
{\hyperref[A2B5C2_oC36_37_d_c2d_d]{Li$_{2}$Si$_{2}$O$_{5}$: A2B5C2\_oC36\_37\_d\_c2d\_d}} - POSCAR
\begin{lstlisting}[numbers=none,language={mylang}]
A2B5C2_oC36_37_d_c2d_d & a,b/a,c/a,z1,x2,y2,z2,x3,y3,z3,x4,y4,z4,x5,y5,z5 --params=5.8071602545,2.51110728429,0.821939039086,0.0,0.654,0.0584,0.0469,0.6705,0.0718,0.471,0.0932,0.1377,0.4004,0.1552,0.14836,0.0571 & Ccc2 C_{2v}^{13} #37 (cd^4) & oC36 & None & Li2Si2O5 &  & B. H. W. S {de Jong} et al., J. Non Cryst. Solids 176, 164-171 (1994)
   1.00000000000000
   2.90358012725000  -7.29120120805000   0.00000000000000
   2.90358012725000   7.29120120805000   0.00000000000000
   0.00000000000000   0.00000000000000   4.77313171940000
    Li     O    Si
     4    10     4
Direct
   0.59560000000000   0.71240000000000   0.04690000000000   Li   (8d)
  -0.59560000000000  -0.71240000000000   0.04690000000000   Li   (8d)
   0.71240000000000   0.59560000000000   0.54690000000000   Li   (8d)
  -0.71240000000000  -0.59560000000000   0.54690000000000   Li   (8d)
   0.00000000000000   0.50000000000000   0.00000000000000    O   (4c)
   0.50000000000000   0.00000000000000   0.50000000000000    O   (4c)
   0.59870000000000   0.74230000000000   0.47100000000000    O   (8d)
  -0.59870000000000  -0.74230000000000   0.47100000000000    O   (8d)
   0.74230000000000   0.59870000000000   0.97100000000000    O   (8d)
  -0.74230000000000  -0.59870000000000   0.97100000000000    O   (8d)
  -0.04450000000000   0.23090000000000   0.40040000000000    O   (8d)
   0.04450000000000  -0.23090000000000   0.40040000000000    O   (8d)
   0.23090000000000  -0.04450000000000   0.90040000000000    O   (8d)
  -0.23090000000000   0.04450000000000   0.90040000000000    O   (8d)
   0.00684000000000   0.30356000000000   0.05710000000000   Si   (8d)
  -0.00684000000000  -0.30356000000000   0.05710000000000   Si   (8d)
   0.30356000000000   0.00684000000000   0.55710000000000   Si   (8d)
  -0.30356000000000  -0.00684000000000   0.55710000000000   Si   (8d)
\end{lstlisting}
{\phantomsection\label{A2B3_oC40_39_2d_2c2d_cif}}
{\hyperref[A2B3_oC40_39_2d_2c2d]{Ta$_{3}$S$_{2}$: A2B3\_oC40\_39\_2d\_2c2d}} - CIF

{\phantomsection\label{A2B3_oC40_39_2d_2c2d_poscar}}
{\hyperref[A2B3_oC40_39_2d_2c2d]{Ta$_{3}$S$_{2}$: A2B3\_oC40\_39\_2d\_2c2d}} - POSCAR

{\phantomsection\label{A9BC_oC44_39_3c3d_a_c_cif}}
{\hyperref[A9BC_oC44_39_3c3d_a_c]{VPCl$_{9}$: A9BC\_oC44\_39\_3c3d\_a\_c}} - CIF

{\phantomsection\label{A9BC_oC44_39_3c3d_a_c_poscar}}
{\hyperref[A9BC_oC44_39_3c3d_a_c]{VPCl$_{9}$: A9BC\_oC44\_39\_3c3d\_a\_c}} - POSCAR

{\phantomsection\label{AB2C_oC16_40_a_2b_b_cif}}
{\hyperref[AB2C_oC16_40_a_2b_b]{K$_{2}$CdPb: AB2C\_oC16\_40\_a\_2b\_b}} - CIF
\begin{lstlisting}[numbers=none,language={mylang}]
# CIF file
data_findsym-output
_audit_creation_method FINDSYM

_chemical_name_mineral 'K2CdPb'
_chemical_formula_sum 'Cd K2 Pb'

loop_
_publ_author_name
 'R. Matthes'
 'H.-U. Schuster'
_journal_name_full_name
;
 Zeitschrift f{\"u}r Naturforschung B
;
_journal_volume 34
_journal_year 1979
_journal_page_first 541
_journal_page_last 543
_publ_Section_title
;
 Synthese und Struktur der Phasen K$_{2}$CdSn und K$_{2}$CdPb/ Synthesis and Structure of the Phase K$_{2}$CdSn and K$_{2}$CdPb
;

# Found in Pearson's Crystal Data - Crystal Structure Database for Inorganic Compounds, 2013

_aflow_title 'K$_{2}$CdPb Structure'
_aflow_proto 'AB2C_oC16_40_a_2b_b'
_aflow_params 'a,b/a,c/a,z_{1},y_{2},z_{2},y_{3},z_{3},y_{4},z_{4}'
_aflow_params_values '6.4580033647,1.33199132859,1.69634561784,0.0,0.3467,0.1773,0.8661,0.3301,0.7281,0.0093'
_aflow_Strukturbericht 'None'
_aflow_Pearson 'oC16'

_cell_length_a    6.4580033647
_cell_length_b    8.6020044818
_cell_length_c    10.9550057077
_cell_angle_alpha 90.0000000000
_cell_angle_beta  90.0000000000
_cell_angle_gamma 90.0000000000
 
_symmetry_space_group_name_H-M "A m a 2"
_symmetry_Int_Tables_number 40
 
loop_
_space_group_symop_id
_space_group_symop_operation_xyz
1 x,y,z
2 -x,-y,z
3 x+1/2,-y,z
4 -x+1/2,y,z
5 x,y+1/2,z+1/2
6 -x,-y+1/2,z+1/2
7 x+1/2,-y+1/2,z+1/2
8 -x+1/2,y+1/2,z+1/2
 
loop_
_atom_site_label
_atom_site_type_symbol
_atom_site_symmetry_multiplicity
_atom_site_Wyckoff_label
_atom_site_fract_x
_atom_site_fract_y
_atom_site_fract_z
_atom_site_occupancy
Cd1 Cd   4 a 0.00000 0.00000 0.00000 1.00000
K1  K    4 b 0.25000 0.34670 0.17730 1.00000
K2  K    4 b 0.25000 0.86610 0.33010 1.00000
Pb1 Pb   4 b 0.25000 0.72810 0.00930 1.00000
\end{lstlisting}
{\phantomsection\label{AB2C_oC16_40_a_2b_b_poscar}}
{\hyperref[AB2C_oC16_40_a_2b_b]{K$_{2}$CdPb: AB2C\_oC16\_40\_a\_2b\_b}} - POSCAR
\begin{lstlisting}[numbers=none,language={mylang}]
AB2C_oC16_40_a_2b_b & a,b/a,c/a,z1,y2,z2,y3,z3,y4,z4 --params=6.4580033647,1.33199132859,1.69634561784,0.0,0.3467,0.1773,0.8661,0.3301,0.7281,0.0093 & Ama2 C_{2v}^{16} #40 (ab^3) & oC16 & None & K2CdPb &  & R. Matthes and H.-U. Schuster, Z. Naturforsch. B 34, 541-543 (1979)
   1.00000000000000
   6.45800336470000   0.00000000000000   0.00000000000000
   0.00000000000000   4.30100224090000  -5.47750285385000
   0.00000000000000   4.30100224090000   5.47750285385000
    Cd     K    Pb
     2     4     2
Direct
   0.00000000000000   0.00000000000000   0.00000000000000   Cd   (4a)
   0.50000000000000   0.00000000000000   0.00000000000000   Cd   (4a)
   0.25000000000000   0.16940000000000   0.52400000000000    K   (4b)
   0.75000000000000  -0.52400000000000  -0.16940000000000    K   (4b)
   0.25000000000000   0.53600000000000   1.19620000000000    K   (4b)
   0.75000000000000  -1.19620000000000  -0.53600000000000    K   (4b)
   0.25000000000000   0.71880000000000   0.73740000000000   Pb   (4b)
   0.75000000000000  -0.73740000000000  -0.71880000000000   Pb   (4b)
\end{lstlisting}
{\phantomsection\label{AB3_oC16_40_b_3b_cif}}
{\hyperref[AB3_oC16_40_b_3b]{CeTe$_{3}$: AB3\_oC16\_40\_b\_3b}} - CIF
\begin{lstlisting}[numbers=none,language={mylang}]
# CIF file
data_findsym-output
_audit_creation_method FINDSYM

_chemical_name_mineral 'CeTe3'
_chemical_formula_sum 'Ce Te3'

loop_
_publ_author_name
 'C. Malliakas'
 'S. J. L. Billinge'
 'H. J. Kim'
 'M. G. Kanatzidis'
_journal_name_full_name
;
 Journal of the American Chemical Society
;
_journal_volume 127
_journal_year 2005
_journal_page_first 6510
_journal_page_last 6511
_publ_Section_title
;
 Square Nets of Tellurium: Rare-Earth Dependent Variation in the Charge-Density Wave of $RE$Te$_{3}$ ($RE$ = Rare-Earth Element)
;

# Found in Pearson's Crystal Data - Crystal Structure Database for Inorganic Compounds, 2013

_aflow_title 'CeTe$_{3}$ Structure'
_aflow_proto 'AB3_oC16_40_b_3b'
_aflow_params 'a,b/a,c/a,y_{1},z_{1},y_{2},z_{2},y_{3},z_{3},y_{4},z_{4}'
_aflow_params_values '4.3850022361,5.92335058951,0.997331752147,0.83109,0.0,0.70417,0.0021,0.56971,0.4966,-0.07002,0.4978'
_aflow_Strukturbericht 'None'
_aflow_Pearson 'oC16'

_cell_length_a    4.3850022361
_cell_length_b    25.9739055802
_cell_length_c    4.3733019633
_cell_angle_alpha 90.0000000000
_cell_angle_beta  90.0000000000
_cell_angle_gamma 90.0000000000
 
_symmetry_space_group_name_H-M "A m a 2"
_symmetry_Int_Tables_number 40
 
loop_
_space_group_symop_id
_space_group_symop_operation_xyz
1 x,y,z
2 -x,-y,z
3 x+1/2,-y,z
4 -x+1/2,y,z
5 x,y+1/2,z+1/2
6 -x,-y+1/2,z+1/2
7 x+1/2,-y+1/2,z+1/2
8 -x+1/2,y+1/2,z+1/2
 
loop_
_atom_site_label
_atom_site_type_symbol
_atom_site_symmetry_multiplicity
_atom_site_Wyckoff_label
_atom_site_fract_x
_atom_site_fract_y
_atom_site_fract_z
_atom_site_occupancy
Ce1 Ce   4 b 0.25000 0.83109  0.00000 1.00000
Te1 Te   4 b 0.25000 0.70417  0.00210 1.00000
Te2 Te   4 b 0.25000 0.56971  0.49660 1.00000
Te3 Te   4 b 0.25000 -0.07002 0.49780 1.00000
\end{lstlisting}
{\phantomsection\label{AB3_oC16_40_b_3b_poscar}}
{\hyperref[AB3_oC16_40_b_3b]{CeTe$_{3}$: AB3\_oC16\_40\_b\_3b}} - POSCAR
\begin{lstlisting}[numbers=none,language={mylang}]
AB3_oC16_40_b_3b & a,b/a,c/a,y1,z1,y2,z2,y3,z3,y4,z4 --params=4.3850022361,5.92335058951,0.997331752147,0.83109,0.0,0.70417,0.0021,0.56971,0.4966,-0.07002,0.4978 & Ama2 C_{2v}^{16} #40 (b^4) & oC16 & None & CeTe3 &  & C. Malliakas et al., J. Am. Chem. Soc. 127, 6510-6511 (2005)
   1.00000000000000
   4.38500223610000   0.00000000000000   0.00000000000000
   0.00000000000000  12.98695279010000  -2.18665098165000
   0.00000000000000  12.98695279010000   2.18665098165000
    Ce    Te
     2     6
Direct
   0.25000000000000   0.83109000000000   0.83109000000000   Ce   (4b)
   0.75000000000000  -0.83109000000000  -0.83109000000000   Ce   (4b)
   0.25000000000000   0.70207000000000   0.70627000000000   Te   (4b)
   0.75000000000000  -0.70627000000000  -0.70207000000000   Te   (4b)
   0.25000000000000   0.07311000000000   1.06631000000000   Te   (4b)
   0.75000000000000  -1.06631000000000  -0.07311000000000   Te   (4b)
   0.25000000000000  -0.56782000000000   0.42778000000000   Te   (4b)
   0.75000000000000  -0.42778000000000   0.56782000000000   Te   (4b)
\end{lstlisting}
{\phantomsection\label{A10B3_oF52_42_2abce_ab_cif}}
{\hyperref[A10B3_oF52_42_2abce_ab]{W$_{3}$O$_{10}$: A10B3\_oF52\_42\_2abce\_ab}} - CIF

{\phantomsection\label{A10B3_oF52_42_2abce_ab_poscar}}
{\hyperref[A10B3_oF52_42_2abce_ab]{W$_{3}$O$_{10}$: A10B3\_oF52\_42\_2abce\_ab}} - POSCAR
\begin{lstlisting}[numbers=none,language={mylang}]
A10B3_oF52_42_2abce_ab & a,b/a,c/a,z1,z2,z3,z4,z5,y6,z6,x7,y7,z7 --params=7.4494846573,1.70036689767,1.04688136975,0.76,0.27,0.0,0.31,0.06,0.79,0.6,0.17,0.11,0.07 & Fmm2 C_{2v}^{18} #42 (a^3b^2ce) & oF52 & None & W3O10 &  & B. Gerand and G. Nowogrocki and M. Figlarz, J. Solid State Chem. 38, 312-320 (1981)
   1.00000000000000
   0.00000000000000   6.33342855800000   3.89936335100000
   3.72474232865000   0.00000000000000   3.89936335100000
   3.72474232865000   6.33342855800000   0.00000000000000
     O     W
    10     3
Direct
   0.76000000000000   0.76000000000000  -0.76000000000000    O   (4a)
   0.27000000000000   0.27000000000000  -0.27000000000000    O   (4a)
   0.31000000000000   0.31000000000000   0.19000000000000    O   (8b)
   0.81000000000000   0.81000000000000  -0.31000000000000    O   (8b)
   1.39000000000000  -0.19000000000000   0.19000000000000    O   (8c)
  -0.19000000000000   1.39000000000000  -1.39000000000000    O   (8c)
   0.01000000000000   0.13000000000000   0.21000000000000    O  (16e)
   0.13000000000000   0.01000000000000  -0.35000000000000    O  (16e)
  -0.21000000000000   0.35000000000000  -0.01000000000000    O  (16e)
   0.35000000000000  -0.21000000000000  -0.13000000000000    O  (16e)
   0.00000000000000   0.00000000000000   0.00000000000000    W   (4a)
   0.06000000000000   0.06000000000000   0.44000000000000    W   (8b)
   0.56000000000000   0.56000000000000  -0.06000000000000    W   (8b)
\end{lstlisting}
{\phantomsection\label{AB_oF8_42_a_a_cif}}
{\hyperref[AB_oF8_42_a_a]{BN (High-pressure, high-temperature): AB\_oF8\_42\_a\_a}} - CIF
\begin{lstlisting}[numbers=none,language={mylang}]
# CIF file
data_findsym-output
_audit_creation_method FINDSYM

_chemical_name_mineral 'BN'
_chemical_formula_sum 'B N'

loop_
_publ_author_name
 'A. V. Kurdyumov'
 'G. S. Olejnik'
_journal_name_full_name
;
 Kristallografiya, English title: Crystallography Reports
;
_journal_volume 29
_journal_year 1984
_journal_page_first 792
_journal_page_last 793
_publ_Section_title
;
 On metastable structures of graphite-like boron nitride
;

# Found in Pearson's Crystal Data - Crystal Structure Database for Inorganic Compounds, 2013

_aflow_title 'BN (High-pressure, high-temperature) Structure'
_aflow_proto 'AB_oF8_42_a_a'
_aflow_params 'a,b/a,c/a,z_{1},z_{2}'
_aflow_params_values '2.5000573158,1.33999999997,1.73599999999,0.0,0.333'
_aflow_Strukturbericht 'None'
_aflow_Pearson 'oF8'

_cell_length_a    2.5000573158
_cell_length_b    3.3500768031
_cell_length_c    4.3400995002
_cell_angle_alpha 90.0000000000
_cell_angle_beta  90.0000000000
_cell_angle_gamma 90.0000000000
 
_symmetry_space_group_name_H-M "F m m 2"
_symmetry_Int_Tables_number 42
 
loop_
_space_group_symop_id
_space_group_symop_operation_xyz
1 x,y,z
2 -x,-y,z
3 -x,y,z
4 x,-y,z
5 x,y+1/2,z+1/2
6 -x,-y+1/2,z+1/2
7 -x,y+1/2,z+1/2
8 x,-y+1/2,z+1/2
9 x+1/2,y,z+1/2
10 -x+1/2,-y,z+1/2
11 -x+1/2,y,z+1/2
12 x+1/2,-y,z+1/2
13 x+1/2,y+1/2,z
14 -x+1/2,-y+1/2,z
15 -x+1/2,y+1/2,z
16 x+1/2,-y+1/2,z
 
loop_
_atom_site_label
_atom_site_type_symbol
_atom_site_symmetry_multiplicity
_atom_site_Wyckoff_label
_atom_site_fract_x
_atom_site_fract_y
_atom_site_fract_z
_atom_site_occupancy
B1 B   4 a 0.00000 0.00000 0.00000 1.00000
N1 N   4 a 0.00000 0.00000 0.33300 1.00000
\end{lstlisting}
{\phantomsection\label{AB_oF8_42_a_a_poscar}}
{\hyperref[AB_oF8_42_a_a]{BN (High-pressure, high-temperature): AB\_oF8\_42\_a\_a}} - POSCAR
\begin{lstlisting}[numbers=none,language={mylang}]
AB_oF8_42_a_a & a,b/a,c/a,z1,z2 --params=2.5000573158,1.33999999997,1.73599999999,0.0,0.333 & Fmm2 C_{2v}^{18} #42 (a^2) & oF8 & None & BN &  & A. V. Kurdyumov and G. S. Olejnik, Kristallografiya 29, 792-793 (1984)
   1.00000000000000
   0.00000000000000   1.67503840155000   2.17004975010000
   1.25002865790000   0.00000000000000   2.17004975010000
   1.25002865790000   1.67503840155000   0.00000000000000
     B     N
     1     1
Direct
   0.00000000000000   0.00000000000000   0.00000000000000    B   (4a)
   0.33300000000000   0.33300000000000  -0.33300000000000    N   (4a)
\end{lstlisting}
{\phantomsection\label{A2BC2_oI20_45_c_b_c_cif}}
{\hyperref[A2BC2_oI20_45_c_b_c]{MnGa$_{2}$Sb$_{2}$: A2BC2\_oI20\_45\_c\_b\_c}} - CIF
\begin{lstlisting}[numbers=none,language={mylang}]
# CIF file
data_findsym-output
_audit_creation_method FINDSYM

_chemical_name_mineral 'MnGa2Sb2'
_chemical_formula_sum 'Ga2 Mn Sb2'

loop_
_publ_author_name
 'W. Sakakibara'
 'Y. Hayashi'
 'H. Takizawa'
_journal_name_full_name
;
 Journal of The Ceramic Society of Japan
;
_journal_volume 117
_journal_year 2009
_journal_page_first 72
_journal_page_last 75
_publ_Section_title
;
 MnGa$_{2}$Sb$_{2}$, a new ferromagnetic compound synthesized under high pressure
;

# Found in Pearson's Crystal Data - Crystal Structure Database for Inorganic Compounds, 2013

_aflow_title 'MnGa$_{2}$Sb$_{2}$ Structure'
_aflow_proto 'A2BC2_oI20_45_c_b_c'
_aflow_params 'a,b/a,c/a,z_{1},x_{2},y_{2},z_{2},x_{3},y_{3},z_{3}'
_aflow_params_values '5.9678340183,1.97721179624,0.981568364609,0.25,0.2729,0.629,0.493,0.2296,0.8629,0.474'
_aflow_Strukturbericht 'None'
_aflow_Pearson 'oI20'

_cell_length_a    5.9678340183
_cell_length_b    11.7996718190
_cell_length_c    5.8578370776
_cell_angle_alpha 90.0000000000
_cell_angle_beta  90.0000000000
_cell_angle_gamma 90.0000000000
 
_symmetry_space_group_name_H-M "I b a 2"
_symmetry_Int_Tables_number 45
 
loop_
_space_group_symop_id
_space_group_symop_operation_xyz
1 x,y,z
2 -x,-y,z
3 -x,y,z+1/2
4 x,-y,z+1/2
5 x+1/2,y+1/2,z+1/2
6 -x+1/2,-y+1/2,z+1/2
7 -x+1/2,y+1/2,z
8 x+1/2,-y+1/2,z
 
loop_
_atom_site_label
_atom_site_type_symbol
_atom_site_symmetry_multiplicity
_atom_site_Wyckoff_label
_atom_site_fract_x
_atom_site_fract_y
_atom_site_fract_z
_atom_site_occupancy
Mn1 Mn   4 b 0.00000 0.50000 0.25000 1.00000
Ga1 Ga   8 c 0.27290 0.62900 0.49300 1.00000
Sb1 Sb   8 c 0.22960 0.86290 0.47400 1.00000
\end{lstlisting}
{\phantomsection\label{A2BC2_oI20_45_c_b_c_poscar}}
{\hyperref[A2BC2_oI20_45_c_b_c]{MnGa$_{2}$Sb$_{2}$: A2BC2\_oI20\_45\_c\_b\_c}} - POSCAR
\begin{lstlisting}[numbers=none,language={mylang}]
A2BC2_oI20_45_c_b_c & a,b/a,c/a,z1,x2,y2,z2,x3,y3,z3 --params=5.9678340183,1.97721179624,0.981568364609,0.25,0.2729,0.629,0.493,0.2296,0.8629,0.474 & Iba2 C_{2v}^{21} #45 (bc^2) & oI20 & None & MnGa2Sb2 &  & W. Sakakibara and Y. Hayashi and H. Takizawa, J. Ceram. Soc. Jpn. 117, 72-75 (2009)
   1.00000000000000
  -2.98391700915000   5.89983590950000   2.92891853880000
   2.98391700915000  -5.89983590950000   2.92891853880000
   2.98391700915000   5.89983590950000  -2.92891853880000
    Ga    Mn    Sb
     4     2     4
Direct
   1.12200000000000   0.76590000000000   0.90190000000000   Ga   (8c)
  -0.13600000000000   0.22010000000000  -0.90190000000000   Ga   (8c)
   0.36400000000000   1.26590000000000  -0.35610000000000   Ga   (8c)
   1.62200000000000   0.72010000000000   0.35610000000000   Ga   (8c)
   0.75000000000000   0.25000000000000   0.50000000000000   Mn   (4b)
   0.25000000000000   0.75000000000000   0.50000000000000   Mn   (4b)
   1.33690000000000   0.70360000000000   1.09250000000000   Sb   (8c)
  -0.38890000000000   0.24440000000000  -1.09250000000000   Sb   (8c)
   0.11110000000000   1.20360000000000  -0.63330000000000   Sb   (8c)
   1.83690000000000   0.74440000000000   0.63330000000000   Sb   (8c)
\end{lstlisting}
{\phantomsection\label{ABC_oI36_46_ac_bc_3b_cif}}
{\hyperref[ABC_oI36_46_ac_bc_3b]{TiFeSi: ABC\_oI36\_46\_ac\_bc\_3b}} - CIF
\begin{lstlisting}[numbers=none,language={mylang}]
# CIF file
data_findsym-output
_audit_creation_method FINDSYM

_chemical_name_mineral 'TiFeSi'
_chemical_formula_sum 'Fe Si Ti'

loop_
_publ_author_name
 'W. Jeitschko'
_journal_name_full_name
;
 Acta Crystallographica Section B: Structural Science
;
_journal_volume 26
_journal_year 1970
_journal_page_first 815
_journal_page_last 822
_publ_Section_title
;
 The crystal structure of TiFeSi and related compounds
;

# Found in Pearson's Crystal Data - Crystal Structure Database for Inorganic Compounds, 2013

_aflow_title 'TiFeSi Structure'
_aflow_proto 'ABC_oI36_46_ac_bc_3b'
_aflow_params 'a,b/a,c/a,z_{1},y_{2},z_{2},y_{3},z_{3},y_{4},z_{4},y_{5},z_{5},x_{6},y_{6},z_{6},x_{7},y_{7},z_{7}'
_aflow_params_values '6.9969353511,1.54780620265,0.898527940538,0.25,0.5253,0.0054,0.7207,0.7706,-0.0021,-0.0823,0.2996,0.7963,0.5295,0.6236,0.1199,0.006,0.3325,0.4952'
_aflow_Strukturbericht 'None'
_aflow_Pearson 'oI36'

_cell_length_a    6.9969353511
_cell_length_b    10.8298999360
_cell_length_c    6.2869419111
_cell_angle_alpha 90.0000000000
_cell_angle_beta  90.0000000000
_cell_angle_gamma 90.0000000000
 
_symmetry_space_group_name_H-M "I m a 2"
_symmetry_Int_Tables_number 46
 
loop_
_space_group_symop_id
_space_group_symop_operation_xyz
1 x,y,z
2 -x,-y,z
3 -x+1/2,y,z
4 x+1/2,-y,z
5 x+1/2,y+1/2,z+1/2
6 -x+1/2,-y+1/2,z+1/2
7 -x,y+1/2,z+1/2
8 x,-y+1/2,z+1/2
 
loop_
_atom_site_label
_atom_site_type_symbol
_atom_site_symmetry_multiplicity
_atom_site_Wyckoff_label
_atom_site_fract_x
_atom_site_fract_y
_atom_site_fract_z
_atom_site_occupancy
Fe1 Fe   4 a 0.00000 0.00000  0.25000  1.00000
Si1 Si   4 b 0.25000 0.52530  0.00540  1.00000
Ti1 Ti   4 b 0.25000 0.72070  0.77060  1.00000
Ti2 Ti   4 b 0.25000 -0.00210 -0.08230 1.00000
Ti3 Ti   4 b 0.25000 0.29960  0.79630  1.00000
Fe2 Fe   8 c 0.52950 0.62360  0.11990  1.00000
Si2 Si   8 c 0.00600 0.33250  0.49520  1.00000
\end{lstlisting}
{\phantomsection\label{ABC_oI36_46_ac_bc_3b_poscar}}
{\hyperref[ABC_oI36_46_ac_bc_3b]{TiFeSi: ABC\_oI36\_46\_ac\_bc\_3b}} - POSCAR
\begin{lstlisting}[numbers=none,language={mylang}]
ABC_oI36_46_ac_bc_3b & a,b/a,c/a,z1,y2,z2,y3,z3,y4,z4,y5,z5,x6,y6,z6,x7,y7,z7 --params=6.9969353511,1.54780620265,0.898527940538,0.25,0.5253,0.0054,0.7207,0.7706,-0.0021,-0.0823,0.2996,0.7963,0.5295,0.6236,0.1199,0.006,0.3325,0.4952 & Ima2 C_{2v}^{22} #46 (ab^4c^2) & oI36 & None & TiFeSi &  & W. Jeitschko, Acta Crystallogr. Sect. B Struct. Sci. 26, 815-822 (1970)
   1.00000000000000
  -3.49846767555000   5.41494996800000   3.14347095555000
   3.49846767555000  -5.41494996800000   3.14347095555000
   3.49846767555000   5.41494996800000  -3.14347095555000
    Fe    Si    Ti
     6     6     6
Direct
   0.25000000000000   0.25000000000000   0.00000000000000   Fe   (4a)
   0.25000000000000   0.75000000000000   0.50000000000000   Fe   (4a)
   0.74350000000000   0.64940000000000   1.15310000000000   Fe   (8c)
  -0.50370000000000  -0.40960000000000  -1.15310000000000   Fe   (8c)
  -0.50370000000000   1.14940000000000   0.40590000000000   Fe   (8c)
   0.74350000000000   0.09040000000000   0.59410000000000   Fe   (8c)
   0.53070000000000   0.25540000000000   0.77530000000000   Si   (4b)
  -0.51990000000000   0.75540000000000   0.22470000000000   Si   (4b)
   0.82770000000000   0.50120000000000   0.33850000000000   Si   (8c)
   0.16270000000000   0.48920000000000  -0.33850000000000   Si   (8c)
   0.16270000000000   1.00120000000000   0.17350000000000   Si   (8c)
   0.82770000000000   0.98920000000000   0.82650000000000   Si   (8c)
   1.49130000000000   1.02060000000000   0.97070000000000   Ti   (4b)
   0.04990000000000   1.52060000000000   0.02930000000000   Ti   (4b)
  -0.08440000000000   0.16770000000000   0.24790000000000   Ti   (4b)
  -0.08020000000000   0.66770000000000   0.75210000000000   Ti   (4b)
   1.09590000000000   1.04630000000000   0.54960000000000   Ti   (4b)
   0.49670000000000   1.54630000000000   0.45040000000000   Ti   (4b)
\end{lstlisting}
{\phantomsection\label{A2B8CD_oP24_48_k_2m_d_b_cif}}
{\hyperref[A2B8CD_oP24_48_k_2m_d_b]{$\alpha$-RbPr[MoO$_{4}$]$_{2}$: A2B8CD\_oP24\_48\_k\_2m\_d\_b}} - CIF
\begin{lstlisting}[numbers=none,language={mylang}]
# CIF file
data_findsym-output
_audit_creation_method FINDSYM

_chemical_name_mineral 'alpha-RbPr[MoO4]2'
_chemical_formula_sum 'Mo2 O8 Pr Rb'

loop_
_publ_author_name
 'R. F. Klevtsova'
 'P. V. Klevtsov'
_journal_name_full_name
;
 Kristallografiya, English title: Crystallography Reports
;
_journal_volume 15
_journal_year 1970
_journal_page_first 466
_journal_page_last 470
_publ_Section_title
;
 Polymorphism of rubidium-praseodymium molybdate, RbPr(MoO$_{4}$)$_{2}$
;

# Found in Pearson's Crystal Data - Crystal Structure Database for Inorganic Compounds, 2013

_aflow_title '$\alpha$-RbPr[MoO$_{4}$]$_{2}$ Structure'
_aflow_proto 'A2B8CD_oP24_48_k_2m_d_b'
_aflow_params 'a,b/a,c/a,z_{3},x_{4},y_{4},z_{4},x_{5},y_{5},z_{5}'
_aflow_params_values '6.3302356554,1.0,1.50710900473,0.4978,0.56,0.629,0.114,0.642,0.558,0.601'
_aflow_Strukturbericht 'None'
_aflow_Pearson 'oP24'

_cell_length_a    6.3302356554
_cell_length_b    6.3302356554
_cell_length_c    9.5403551583
_cell_angle_alpha 90.0000000000
_cell_angle_beta  90.0000000000
_cell_angle_gamma 90.0000000000
 
_symmetry_space_group_name_H-M "P 2/n 2/n 2/n (origin choice 2)"
_symmetry_Int_Tables_number 48
 
loop_
_space_group_symop_id
_space_group_symop_operation_xyz
1 x,y,z
2 x,-y+1/2,-z+1/2
3 -x+1/2,y,-z+1/2
4 -x+1/2,-y+1/2,z
5 -x,-y,-z
6 -x,y+1/2,z+1/2
7 x+1/2,-y,z+1/2
8 x+1/2,y+1/2,-z
 
loop_
_atom_site_label
_atom_site_type_symbol
_atom_site_symmetry_multiplicity
_atom_site_Wyckoff_label
_atom_site_fract_x
_atom_site_fract_y
_atom_site_fract_z
_atom_site_occupancy
Rb1 Rb   2 b 0.75000 0.25000 0.25000 1.00000
Pr1 Pr   2 d 0.25000 0.75000 0.25000 1.00000
Mo1 Mo   4 k 0.25000 0.25000 0.49780 1.00000
O1  O    8 m 0.56000 0.62900 0.11400 1.00000
O2  O    8 m 0.64200 0.55800 0.60100 1.00000
\end{lstlisting}
{\phantomsection\label{A2B8CD_oP24_48_k_2m_d_b_poscar}}
{\hyperref[A2B8CD_oP24_48_k_2m_d_b]{$\alpha$-RbPr[MoO$_{4}$]$_{2}$: A2B8CD\_oP24\_48\_k\_2m\_d\_b}} - POSCAR

{\phantomsection\label{A5B2_oP14_49_dehq_ab_cif}}
{\hyperref[A5B2_oP14_49_dehq_ab]{$\beta$-Ta$_{2}$O$_{5}$: A5B2\_oP14\_49\_dehq\_ab}} - CIF
\begin{lstlisting}[numbers=none,language={mylang}]
# CIF file
data_findsym-output
_audit_creation_method FINDSYM

_chemical_name_mineral 'beta-Ta2O5'
_chemical_formula_sum 'O5 Ta2'

loop_
_publ_author_name
 'L. A. Aleshina'
 'S. V. Loginova'
_journal_name_full_name
;
 Crystallography Reports
;
_journal_volume 47
_journal_year 2002
_journal_page_first 415
_journal_page_last 419
_publ_Section_title
;
 Rietveld analysis of X-ray diffraction pattern from $\beta$-Ta$_{2}$O$_{5}$ oxide
;

# Found in Pearson's Crystal Data - Crystal Structure Database for Inorganic Compounds, 2013

_aflow_title '$\beta$-Ta$_{2}$O$_{5}$ Structure'
_aflow_proto 'A5B2_oP14_49_dehq_ab'
_aflow_params 'a,b/a,c/a,x_{6},y_{6}'
_aflow_params_values '3.6705354001,1.69539132804,2.12544314151,0.002,0.681'
_aflow_Strukturbericht 'None'
_aflow_Pearson 'oP14'

_cell_length_a    3.6705354001
_cell_length_b    6.2229938866
_cell_length_c    7.8015142918
_cell_angle_alpha 90.0000000000
_cell_angle_beta  90.0000000000
_cell_angle_gamma 90.0000000000
 
_symmetry_space_group_name_H-M "P 2/c 2/c 2/m"
_symmetry_Int_Tables_number 49
 
loop_
_space_group_symop_id
_space_group_symop_operation_xyz
1 x,y,z
2 x,-y,-z+1/2
3 -x,y,-z+1/2
4 -x,-y,z
5 -x,-y,-z
6 -x,y,z+1/2
7 x,-y,z+1/2
8 x,y,-z
 
loop_
_atom_site_label
_atom_site_type_symbol
_atom_site_symmetry_multiplicity
_atom_site_Wyckoff_label
_atom_site_fract_x
_atom_site_fract_y
_atom_site_fract_z
_atom_site_occupancy
Ta1 Ta   2 a 0.00000 0.00000 0.00000 1.00000
Ta2 Ta   2 b 0.50000 0.50000 0.00000 1.00000
O1  O    2 d 0.50000 0.00000 0.00000 1.00000
O2  O    2 e 0.00000 0.00000 0.25000 1.00000
O3  O    2 h 0.50000 0.50000 0.25000 1.00000
O4  O    4 q 0.00200 0.68100 0.00000 1.00000
\end{lstlisting}
{\phantomsection\label{A5B2_oP14_49_dehq_ab_poscar}}
{\hyperref[A5B2_oP14_49_dehq_ab]{$\beta$-Ta$_{2}$O$_{5}$: A5B2\_oP14\_49\_dehq\_ab}} - POSCAR
\begin{lstlisting}[numbers=none,language={mylang}]
A5B2_oP14_49_dehq_ab & a,b/a,c/a,x6,y6 --params=3.6705354001,1.69539132804,2.12544314151,0.002,0.681 & Pccm D_{2h}^{3} #49 (abdehq) & oP14 & None & Ta2O5 & beta & L. A. Aleshina and S. V. Loginova, Crystallogr. Rep. 47, 415-419 (2002)
   1.00000000000000
   3.67053540010000   0.00000000000000   0.00000000000000
   0.00000000000000   6.22299388660000   0.00000000000000
   0.00000000000000   0.00000000000000   7.80151429180000
     O    Ta
    10     4
Direct
   0.50000000000000   0.00000000000000   0.00000000000000    O   (2d)
   0.50000000000000   0.00000000000000   0.50000000000000    O   (2d)
   0.00000000000000   0.00000000000000   0.25000000000000    O   (2e)
   0.00000000000000   0.00000000000000   0.75000000000000    O   (2e)
   0.50000000000000   0.50000000000000   0.25000000000000    O   (2h)
   0.50000000000000   0.50000000000000   0.75000000000000    O   (2h)
   0.00200000000000   0.68100000000000   0.00000000000000    O   (4q)
  -0.00200000000000  -0.68100000000000   0.00000000000000    O   (4q)
  -0.00200000000000   0.68100000000000   0.50000000000000    O   (4q)
   0.00200000000000  -0.68100000000000   0.50000000000000    O   (4q)
   0.00000000000000   0.00000000000000   0.00000000000000   Ta   (2a)
   0.00000000000000   0.00000000000000   0.50000000000000   Ta   (2a)
   0.50000000000000   0.50000000000000   0.00000000000000   Ta   (2b)
   0.50000000000000   0.50000000000000   0.50000000000000   Ta   (2b)
\end{lstlisting}
{\phantomsection\label{AB2C8D_oP24_49_g_q_2qr_e_cif}}
{\hyperref[AB2C8D_oP24_49_g_q_2qr_e]{CsPr[MoO$_{4}$]$_{2}$: AB2C8D\_oP24\_49\_g\_q\_2qr\_e}} - CIF
\begin{lstlisting}[numbers=none,language={mylang}]
# CIF file
data_findsym-output
_audit_creation_method FINDSYM

_chemical_name_mineral 'CsPr[MoO4]2'
_chemical_formula_sum 'Cs Mo2 O8 Pr'

loop_
_publ_author_name
 'V. A. Vinokurov'
 'P. V. Klevtsov'
_journal_name_full_name
;
 Soviet Physics Crystallography
;
_journal_volume 17
_journal_year 1972
_journal_page_first 102
_journal_page_last 106
_publ_Section_title
;
 Polymorphism and crystallization of binary cesium-rare earth molybdates CsLn(MoO$_{4}$)$_{2}$
;

# Found in Pearson's Crystal Data - Crystal Structure Database for Inorganic Compounds, 2013

_aflow_title 'CsPr[MoO$_{4}$]$_{2}$ Structure'
_aflow_proto 'AB2C8D_oP24_49_g_q_2qr_e'
_aflow_params 'a,b/a,c/a,x_{3},y_{3},x_{4},y_{4},x_{5},y_{5},x_{6},y_{6},z_{6}'
_aflow_params_values '6.3302356554,1.0,1.50710900473,0.5184,0.7996,0.258,-0.0697,0.3652,0.6307,0.2617,0.1943,0.8286'
_aflow_Strukturbericht 'None'
_aflow_Pearson 'oP24'

_cell_length_a    6.3302356554
_cell_length_b    6.3302356554
_cell_length_c    9.5403551583
_cell_angle_alpha 90.0000000000
_cell_angle_beta  90.0000000000
_cell_angle_gamma 90.0000000000
 
_symmetry_space_group_name_H-M "P 2/c 2/c 2/m"
_symmetry_Int_Tables_number 49
 
loop_
_space_group_symop_id
_space_group_symop_operation_xyz
1 x,y,z
2 x,-y,-z+1/2
3 -x,y,-z+1/2
4 -x,-y,z
5 -x,-y,-z
6 -x,y,z+1/2
7 x,-y,z+1/2
8 x,y,-z
 
loop_
_atom_site_label
_atom_site_type_symbol
_atom_site_symmetry_multiplicity
_atom_site_Wyckoff_label
_atom_site_fract_x
_atom_site_fract_y
_atom_site_fract_z
_atom_site_occupancy
Pr1 Pr   2 e 0.00000 0.00000  0.25000 1.00000
Cs1 Cs   2 g 0.00000 0.50000  0.25000 1.00000
Mo1 Mo   4 q 0.51840 0.79960  0.00000 1.00000
O1  O    4 q 0.25800 -0.06970 0.00000 1.00000
O2  O    4 q 0.36520 0.63070  0.00000 1.00000
O3  O    8 r 0.26170 0.19430  0.82860 1.00000
\end{lstlisting}
{\phantomsection\label{AB2C8D_oP24_49_g_q_2qr_e_poscar}}
{\hyperref[AB2C8D_oP24_49_g_q_2qr_e]{CsPr[MoO$_{4}$]$_{2}$: AB2C8D\_oP24\_49\_g\_q\_2qr\_e}} - POSCAR

{\phantomsection\label{A2BC4_oP28_50_ij_ac_ijm_cif}}
{\hyperref[A2BC4_oP28_50_ij_ac_ijm]{La$_{2}$NiO$_{4}$: A2BC4\_oP28\_50\_ij\_ac\_ijm}} - CIF
\begin{lstlisting}[numbers=none,language={mylang}]
# CIF file
data_findsym-output
_audit_creation_method FINDSYM

_chemical_name_mineral 'La2NiO4'
_chemical_formula_sum 'La2 Ni O4'

loop_
_publ_author_name
 'P. Odier'
 'M. Leblanc'
 'J. Choisnet'
_journal_name_full_name
;
 Materials Research Bulletin
;
_journal_volume 21
_journal_year 1986
_journal_page_first 787
_journal_page_last 796
_publ_Section_title
;
 Structural characterization of an orthorhombic form of La$_{2}$NiO$_{4}$
;

# Found in Pearson's Crystal Data - Crystal Structure Database for Inorganic Compounds, 2013

_aflow_title 'La$_{2}$NiO$_{4}$ Structure'
_aflow_proto 'A2BC4_oP28_50_ij_ac_ijm'
_aflow_params 'a,b/a,c/a,y_{3},y_{4},y_{5},y_{6},x_{7},y_{7},z_{7}'
_aflow_params_values '5.5348069961,2.26684733515,0.987895212291,0.8726,0.445,0.3867,-0.076,0.5,0.743,0.226'
_aflow_Strukturbericht 'None'
_aflow_Pearson 'oP28'

_cell_length_a    5.5348069961
_cell_length_b    12.5465624897
_cell_length_c    5.4678093324
_cell_angle_alpha 90.0000000000
_cell_angle_beta  90.0000000000
_cell_angle_gamma 90.0000000000
 
_symmetry_space_group_name_H-M "P 2/b 2/a 2/n (origin choice 2)"
_symmetry_Int_Tables_number 50
 
loop_
_space_group_symop_id
_space_group_symop_operation_xyz
1 x,y,z
2 x,-y+1/2,-z
3 -x+1/2,y,-z
4 -x+1/2,-y+1/2,z
5 -x,-y,-z
6 -x,y+1/2,z
7 x+1/2,-y,z
8 x+1/2,y+1/2,-z
 
loop_
_atom_site_label
_atom_site_type_symbol
_atom_site_symmetry_multiplicity
_atom_site_Wyckoff_label
_atom_site_fract_x
_atom_site_fract_y
_atom_site_fract_z
_atom_site_occupancy
Ni1 Ni   2 a 0.25000 0.25000  0.00000 1.00000
Ni2 Ni   2 c 0.75000 0.25000  0.50000 1.00000
La1 La   4 i 0.25000 0.87260  0.00000 1.00000
O1  O    4 i 0.25000 0.44500  0.00000 1.00000
La2 La   4 j 0.25000 0.38670  0.50000 1.00000
O2  O    4 j 0.25000 -0.07600 0.50000 1.00000
O3  O    8 m 0.50000 0.74300  0.22600 1.00000
\end{lstlisting}
{\phantomsection\label{A2BC4_oP28_50_ij_ac_ijm_poscar}}
{\hyperref[A2BC4_oP28_50_ij_ac_ijm]{La$_{2}$NiO$_{4}$: A2BC4\_oP28\_50\_ij\_ac\_ijm}} - POSCAR

{\phantomsection\label{A3BC2_oP48_50_3m_m_2m_cif}}
{\hyperref[A3BC2_oP48_50_3m_m_2m]{$\alpha$-Tl$_{2}$TeO$_{3}$: A3BC2\_oP48\_50\_3m\_m\_2m}} - CIF
\begin{lstlisting}[numbers=none,language={mylang}]
# CIF file
data_findsym-output
_audit_creation_method FINDSYM

_chemical_name_mineral 'alpha-Tl2TeO3'
_chemical_formula_sum 'O3 Te Tl2'

loop_
_publ_author_name
 'F. Rieger'
 'A.-V. Mudring'
_journal_name_full_name
;
 Inorganic Chemistry
;
_journal_volume 46
_journal_year 2007
_journal_page_first 446
_journal_page_last 452
_publ_Section_title
;
 Phase transition in Tl$_{2}$TeO$_{3}$: Influence and origin of the thallium lone pair distortion
;

# Found in Pearson's Crystal Data - Crystal Structure Database for Inorganic Compounds, 2013

_aflow_title '$\alpha$-Tl$_{2}$TeO$_{3}$ Structure'
_aflow_proto 'A3BC2_oP48_50_3m_m_2m'
_aflow_params 'a,b/a,c/a,x_{1},y_{1},z_{1},x_{2},y_{2},z_{2},x_{3},y_{3},z_{3},x_{4},y_{4},z_{4},x_{5},y_{5},z_{5},x_{6},y_{6},z_{6}'
_aflow_params_values '11.0940438033,1.50045069408,0.472480620154,0.0515,0.8152,0.238,0.602,0.5567,0.315,0.604,0.8447,0.116,0.61286,0.66266,0.2344,0.11964,0.66839,0.2479,0.63183,0.00412,0.2439'
_aflow_Strukturbericht 'None'
_aflow_Pearson 'oP48'

_cell_length_a    11.0940438033
_cell_length_b    16.6460657248
_cell_length_c    5.2417206962
_cell_angle_alpha 90.0000000000
_cell_angle_beta  90.0000000000
_cell_angle_gamma 90.0000000000
 
_symmetry_space_group_name_H-M "P 2/b 2/a 2/n (origin choice 2)"
_symmetry_Int_Tables_number 50
 
loop_
_space_group_symop_id
_space_group_symop_operation_xyz
1 x,y,z
2 x,-y+1/2,-z
3 -x+1/2,y,-z
4 -x+1/2,-y+1/2,z
5 -x,-y,-z
6 -x,y+1/2,z
7 x+1/2,-y,z
8 x+1/2,y+1/2,-z
 
loop_
_atom_site_label
_atom_site_type_symbol
_atom_site_symmetry_multiplicity
_atom_site_Wyckoff_label
_atom_site_fract_x
_atom_site_fract_y
_atom_site_fract_z
_atom_site_occupancy
O1  O    8 m 0.05150 0.81520 0.23800 1.00000
O2  O    8 m 0.60200 0.55670 0.31500 1.00000
O3  O    8 m 0.60400 0.84470 0.11600 1.00000
Te1 Te   8 m 0.61286 0.66266 0.23440 1.00000
Tl1 Tl   8 m 0.11964 0.66839 0.24790 1.00000
Tl2 Tl   8 m 0.63183 0.00412 0.24390 1.00000
\end{lstlisting}
{\phantomsection\label{A3BC2_oP48_50_3m_m_2m_poscar}}
{\hyperref[A3BC2_oP48_50_3m_m_2m]{$\alpha$-Tl$_{2}$TeO$_{3}$: A3BC2\_oP48\_50\_3m\_m\_2m}} - POSCAR

{\phantomsection\label{A2B_oP24_52_2e_cd_cif}}
{\hyperref[A2B_oP24_52_2e_cd]{GaCl$_{2}$ (High-temperature): A2B\_oP24\_52\_2e\_cd}} - CIF
\begin{lstlisting}[numbers=none,language={mylang}]
# CIF file
data_findsym-output
_audit_creation_method FINDSYM

_chemical_name_mineral 'GaCl2'
_chemical_formula_sum 'Cl2 Ga'

loop_
_publ_author_name
 'A. P. Wilkinson'
 'A. K. Cheetham'
 'D. E. Cox'
_journal_name_full_name
;
 Acta Crystallographica Section B: Structural Science
;
_journal_volume 47
_journal_year 1991
_journal_page_first 155
_journal_page_last 161
_publ_Section_title
;
 Study of oxidation-state contrast in gallium dichloride by synchrotron X-ray anomalous scattering
;

# Found in Pearson's Crystal Data - Crystal Structure Database for Inorganic Compounds, 2013

_aflow_title 'GaCl$_{2}$ (High-temperature) Structure'
_aflow_proto 'A2B_oP24_52_2e_cd'
_aflow_params 'a,b/a,c/a,z_{1},x_{2},x_{3},y_{3},z_{3},x_{4},y_{4},z_{4}'
_aflow_params_values '7.2235008877,1.34576036547,1.32098013428,0.3175,0.6759,0.8271,0.1762,0.5576,0.5093,0.0419,0.8142'
_aflow_Strukturbericht 'None'
_aflow_Pearson 'oP24'

_cell_length_a    7.2235008877
_cell_length_b    9.7211011946
_cell_length_c    9.5421011726
_cell_angle_alpha 90.0000000000
_cell_angle_beta  90.0000000000
_cell_angle_gamma 90.0000000000
 
_symmetry_space_group_name_H-M "P 2/n 21/n 2/a"
_symmetry_Int_Tables_number 52
 
loop_
_space_group_symop_id
_space_group_symop_operation_xyz
1 x,y,z
2 x,-y+1/2,-z+1/2
3 -x+1/2,y+1/2,-z+1/2
4 -x+1/2,-y,z
5 -x,-y,-z
6 -x,y+1/2,z+1/2
7 x+1/2,-y+1/2,z+1/2
8 x+1/2,y,-z
 
loop_
_atom_site_label
_atom_site_type_symbol
_atom_site_symmetry_multiplicity
_atom_site_Wyckoff_label
_atom_site_fract_x
_atom_site_fract_y
_atom_site_fract_z
_atom_site_occupancy
Ga1 Ga   4 c 0.25000 0.00000 0.31750 1.00000
Ga2 Ga   4 d 0.67590 0.25000 0.25000 1.00000
Cl1 Cl   8 e 0.82710 0.17620 0.55760 1.00000
Cl2 Cl   8 e 0.50930 0.04190 0.81420 1.00000
\end{lstlisting}
{\phantomsection\label{A2B_oP24_52_2e_cd_poscar}}
{\hyperref[A2B_oP24_52_2e_cd]{GaCl$_{2}$ (High-temperature): A2B\_oP24\_52\_2e\_cd}} - POSCAR

{\phantomsection\label{A3B2_oP20_52_de_cd_cif}}
{\hyperref[A3B2_oP20_52_de_cd]{Sr$_{2}$Bi$_{3}$: A3B2\_oP20\_52\_de\_cd}} - CIF
\begin{lstlisting}[numbers=none,language={mylang}]
# CIF file
data_findsym-output
_audit_creation_method FINDSYM

_chemical_name_mineral 'Sr2Bi3'
_chemical_formula_sum 'Bi3 Sr2'

loop_
_publ_author_name
 'F. Merlo'
 'M. L. Fornasini'
_journal_name_full_name
;
 Materials Research Bulletin
;
_journal_volume 29
_journal_year 1994
_journal_page_first 149
_journal_page_last 154
_publ_Section_title
;
 Crystal structure of some phases and alloying behaviour in alkaline earths, europium and ytterbium pnictides
;

# Found in Pearson's Crystal Data - Crystal Structure Database for Inorganic Compounds, 2013

_aflow_title 'Sr$_{2}$Bi$_{3}$ Structure'
_aflow_proto 'A3B2_oP20_52_de_cd'
_aflow_params 'a,b/a,c/a,z_{1},x_{2},x_{3},x_{4},y_{4},z_{4}'
_aflow_params_values '15.5832022616,0.434585119311,0.426069989123,0.443,0.4294,0.0001,0.6539,0.064,0.0788'
_aflow_Strukturbericht 'None'
_aflow_Pearson 'oP20'

_cell_length_a    15.5832022616
_cell_length_b    6.7722278141
_cell_length_c    6.6395348181
_cell_angle_alpha 90.0000000000
_cell_angle_beta  90.0000000000
_cell_angle_gamma 90.0000000000
 
_symmetry_space_group_name_H-M "P 2/n 21/n 2/a"
_symmetry_Int_Tables_number 52
 
loop_
_space_group_symop_id
_space_group_symop_operation_xyz
1 x,y,z
2 x,-y+1/2,-z+1/2
3 -x+1/2,y+1/2,-z+1/2
4 -x+1/2,-y,z
5 -x,-y,-z
6 -x,y+1/2,z+1/2
7 x+1/2,-y+1/2,z+1/2
8 x+1/2,y,-z
 
loop_
_atom_site_label
_atom_site_type_symbol
_atom_site_symmetry_multiplicity
_atom_site_Wyckoff_label
_atom_site_fract_x
_atom_site_fract_y
_atom_site_fract_z
_atom_site_occupancy
Sr1 Sr   4 c 0.25000 0.00000 0.44300 1.00000
Bi1 Bi   4 d 0.42940 0.25000 0.25000 1.00000
Sr2 Sr   4 d 0.00010 0.25000 0.25000 1.00000
Bi2 Bi   8 e 0.65390 0.06400 0.07880 1.00000
\end{lstlisting}
{\phantomsection\label{A3B2_oP20_52_de_cd_poscar}}
{\hyperref[A3B2_oP20_52_de_cd]{Sr$_{2}$Bi$_{3}$: A3B2\_oP20\_52\_de\_cd}} - POSCAR
\begin{lstlisting}[numbers=none,language={mylang}]
A3B2_oP20_52_de_cd & a,b/a,c/a,z1,x2,x3,x4,y4,z4 --params=15.5832022616,0.434585119311,0.426069989123,0.443,0.4294,0.0001,0.6539,0.064,0.0788 & Pnna D_{2h}^{6} #52 (cd^2e) & oP20 & None & Sr2Bi3 &  & F. Merlo and M. L. Fornasini, Mater. Res. Bull. 29, 149-154 (1994)
   1.00000000000000
  15.58320226160000   0.00000000000000   0.00000000000000
   0.00000000000000   6.77222781410000   0.00000000000000
   0.00000000000000   0.00000000000000   6.63953481810000
    Bi    Sr
    12     8
Direct
   0.42940000000000   0.25000000000000   0.25000000000000   Bi   (4d)
   0.07060000000000   0.75000000000000   0.25000000000000   Bi   (4d)
  -0.42940000000000   0.75000000000000   0.75000000000000   Bi   (4d)
   0.92940000000000   0.25000000000000   0.75000000000000   Bi   (4d)
   0.65390000000000   0.06400000000000   0.07880000000000   Bi   (8e)
  -0.15390000000000  -0.06400000000000   0.07880000000000   Bi   (8e)
  -0.15390000000000   0.56400000000000   0.42120000000000   Bi   (8e)
   0.65390000000000   0.43600000000000   0.42120000000000   Bi   (8e)
  -0.65390000000000  -0.06400000000000  -0.07880000000000   Bi   (8e)
   1.15390000000000   0.06400000000000  -0.07880000000000   Bi   (8e)
   1.15390000000000   0.43600000000000   0.57880000000000   Bi   (8e)
  -0.65390000000000   0.56400000000000   0.57880000000000   Bi   (8e)
   0.25000000000000   0.00000000000000   0.44300000000000   Sr   (4c)
   0.25000000000000   0.50000000000000   0.05700000000000   Sr   (4c)
   0.75000000000000   0.00000000000000  -0.44300000000000   Sr   (4c)
   0.75000000000000   0.50000000000000   0.94300000000000   Sr   (4c)
   0.00010000000000   0.25000000000000   0.25000000000000   Sr   (4d)
   0.49990000000000   0.75000000000000   0.25000000000000   Sr   (4d)
  -0.00010000000000   0.75000000000000   0.75000000000000   Sr   (4d)
   0.50010000000000   0.25000000000000   0.75000000000000   Sr   (4d)
\end{lstlisting}
{\phantomsection\label{ABC2_oP16_53_h_e_gh_cif}}
{\hyperref[ABC2_oP16_53_h_e_gh]{TaNiTe$_{2}$: ABC2\_oP16\_53\_h\_e\_gh}} - CIF
\begin{lstlisting}[numbers=none,language={mylang}]
# CIF file
data_findsym-output
_audit_creation_method FINDSYM

_chemical_name_mineral 'TaNiTe2'
_chemical_formula_sum 'Ni Ta Te2'

loop_
_publ_author_name
 'W. Tremel'
_journal_name_full_name
;
 Angewandte Chemie (International ed.)
;
_journal_volume 30
_journal_year 1991
_journal_page_first 840
_journal_page_last 843
_publ_Section_title
;
 Isolated and Condensed Ta$_{2}$Ni$_{2}$ Clusters in the Layered Tellurides Ta$_{2}$Ni$_{2}$Te$_{4}$ and Ta$_{2}$Ni$_{3}$Te$_{5}$
;

# Found in Pearson's Crystal Data - Crystal Structure Database for Inorganic Compounds, 2013

_aflow_title 'TaNiTe$_{2}$ Structure'
_aflow_proto 'ABC2_oP16_53_h_e_gh'
_aflow_params 'a,b/a,c/a,x_{1},y_{2},y_{3},z_{3},y_{4},z_{4}'
_aflow_params_values '11.6904618594,0.282132455361,0.78890717994,0.79514,0.3193,0.1198,0.3549,0.224,0.7493'
_aflow_Strukturbericht 'None'
_aflow_Pearson 'oP16'

_cell_length_a    11.6904618594
_cell_length_b    3.2982587087
_cell_length_c    9.2226892977
_cell_angle_alpha 90.0000000000
_cell_angle_beta  90.0000000000
_cell_angle_gamma 90.0000000000
 
_symmetry_space_group_name_H-M "P 2/m 2/n 21/a"
_symmetry_Int_Tables_number 53
 
loop_
_space_group_symop_id
_space_group_symop_operation_xyz
1 x,y,z
2 x,-y,-z
3 -x+1/2,y,-z+1/2
4 -x+1/2,-y,z+1/2
5 -x,-y,-z
6 -x,y,z
7 x+1/2,-y,z+1/2
8 x+1/2,y,-z+1/2
 
loop_
_atom_site_label
_atom_site_type_symbol
_atom_site_symmetry_multiplicity
_atom_site_Wyckoff_label
_atom_site_fract_x
_atom_site_fract_y
_atom_site_fract_z
_atom_site_occupancy
Ta1 Ta   4 e 0.79514 0.00000 0.00000 1.00000
Te1 Te   4 g 0.25000 0.31930 0.25000 1.00000
Ni1 Ni   4 h 0.00000 0.11980 0.35490 1.00000
Te2 Te   4 h 0.00000 0.22400 0.74930 1.00000
\end{lstlisting}
{\phantomsection\label{ABC2_oP16_53_h_e_gh_poscar}}
{\hyperref[ABC2_oP16_53_h_e_gh]{TaNiTe$_{2}$: ABC2\_oP16\_53\_h\_e\_gh}} - POSCAR
\begin{lstlisting}[numbers=none,language={mylang}]
ABC2_oP16_53_h_e_gh & a,b/a,c/a,x1,y2,y3,z3,y4,z4 --params=11.6904618594,0.282132455361,0.78890717994,0.79514,0.3193,0.1198,0.3549,0.224,0.7493 & Pmna D_{2h}^{7} #53 (egh^2) & oP16 & None & TaNiTe2 &  & W. Tremel, Angew. Chem. Int. Ed. 30, 840-843 (1991)
   1.00000000000000
  11.69046185940000   0.00000000000000   0.00000000000000
   0.00000000000000   3.29825870870000   0.00000000000000
   0.00000000000000   0.00000000000000   9.22268929770000
    Ni    Ta    Te
     4     4     8
Direct
   0.00000000000000   0.11980000000000   0.35490000000000   Ni   (4h)
   0.50000000000000  -0.11980000000000   0.85490000000000   Ni   (4h)
   0.50000000000000   0.11980000000000   0.14510000000000   Ni   (4h)
   0.00000000000000  -0.11980000000000  -0.35490000000000   Ni   (4h)
   0.79514000000000   0.00000000000000   0.00000000000000   Ta   (4e)
  -0.29514000000000   0.00000000000000   0.50000000000000   Ta   (4e)
  -0.79514000000000   0.00000000000000   0.00000000000000   Ta   (4e)
   1.29514000000000   0.00000000000000   0.50000000000000   Ta   (4e)
   0.25000000000000   0.31930000000000   0.25000000000000   Te   (4g)
   0.25000000000000  -0.31930000000000   0.75000000000000   Te   (4g)
   0.75000000000000  -0.31930000000000   0.75000000000000   Te   (4g)
   0.75000000000000   0.31930000000000   0.25000000000000   Te   (4g)
   0.00000000000000   0.22400000000000   0.74930000000000   Te   (4h)
   0.50000000000000  -0.22400000000000   1.24930000000000   Te   (4h)
   0.50000000000000   0.22400000000000  -0.24930000000000   Te   (4h)
   0.00000000000000  -0.22400000000000  -0.74930000000000   Te   (4h)
\end{lstlisting}
{\phantomsection\label{ABC3_oP20_53_e_g_hi_cif}}
{\hyperref[ABC3_oP20_53_e_g_hi]{CuBrSe$_{3}$: ABC3\_oP20\_53\_e\_g\_hi}} - CIF
\begin{lstlisting}[numbers=none,language={mylang}]
# CIF file
data_findsym-output
_audit_creation_method FINDSYM

_chemical_name_mineral 'CuBrSe3'
_chemical_formula_sum 'Br Cu Se3'

loop_
_publ_author_name
 'H. M. Haendler'
 'P. M. Carkner'
_journal_name_full_name
;
 Journal of Solid State Chemistry
;
_journal_volume 29
_journal_year 1979
_journal_page_first 35
_journal_page_last 39
_publ_Section_title
;
 The crystal structure of copper bromide triselenide, CuBrSe$_{3}$
;

# Found in Pearson's Crystal Data - Crystal Structure Database for Inorganic Compounds, 2013

_aflow_title 'CuBrSe$_{3}$ Structure'
_aflow_proto 'ABC3_oP20_53_e_g_hi'
_aflow_params 'a,b/a,c/a,x_{1},y_{2},y_{3},z_{3},x_{4},y_{4},z_{4}'
_aflow_params_values '14.3630679002,0.312469539787,0.535821207264,0.6826,0.2856,0.3458,0.2708,0.6247,0.6057,0.3575'
_aflow_Strukturbericht 'None'
_aflow_Pearson 'oP20'

_cell_length_a    14.3630679002
_cell_length_b    4.4880212167
_cell_length_c    7.6960363823
_cell_angle_alpha 90.0000000000
_cell_angle_beta  90.0000000000
_cell_angle_gamma 90.0000000000
 
_symmetry_space_group_name_H-M "P 2/m 2/n 21/a"
_symmetry_Int_Tables_number 53
 
loop_
_space_group_symop_id
_space_group_symop_operation_xyz
1 x,y,z
2 x,-y,-z
3 -x+1/2,y,-z+1/2
4 -x+1/2,-y,z+1/2
5 -x,-y,-z
6 -x,y,z
7 x+1/2,-y,z+1/2
8 x+1/2,y,-z+1/2
 
loop_
_atom_site_label
_atom_site_type_symbol
_atom_site_symmetry_multiplicity
_atom_site_Wyckoff_label
_atom_site_fract_x
_atom_site_fract_y
_atom_site_fract_z
_atom_site_occupancy
Br1 Br   4 e 0.68260 0.00000 0.00000 1.00000
Cu1 Cu   4 g 0.25000 0.28560 0.25000 1.00000
Se1 Se   4 h 0.00000 0.34580 0.27080 1.00000
Se2 Se   8 i 0.62470 0.60570 0.35750 1.00000
\end{lstlisting}
{\phantomsection\label{ABC3_oP20_53_e_g_hi_poscar}}
{\hyperref[ABC3_oP20_53_e_g_hi]{CuBrSe$_{3}$: ABC3\_oP20\_53\_e\_g\_hi}} - POSCAR
\begin{lstlisting}[numbers=none,language={mylang}]
ABC3_oP20_53_e_g_hi & a,b/a,c/a,x1,y2,y3,z3,x4,y4,z4 --params=14.3630679002,0.312469539787,0.535821207264,0.6826,0.2856,0.3458,0.2708,0.6247,0.6057,0.3575 & Pmna D_{2h}^{7} #53 (eghi) & oP20 & None & CuBrSe3 &  & H. M. Haendler and P. M. Carkner, J. Solid State Chem. 29, 35-39 (1979)
   1.00000000000000
  14.36306790020000   0.00000000000000   0.00000000000000
   0.00000000000000   4.48802121670000   0.00000000000000
   0.00000000000000   0.00000000000000   7.69603638230000
    Br    Cu    Se
     4     4    12
Direct
   0.68260000000000   0.00000000000000   0.00000000000000   Br   (4e)
  -0.18260000000000   0.00000000000000   0.50000000000000   Br   (4e)
  -0.68260000000000   0.00000000000000   0.00000000000000   Br   (4e)
   1.18260000000000   0.00000000000000   0.50000000000000   Br   (4e)
   0.25000000000000   0.28560000000000   0.25000000000000   Cu   (4g)
   0.25000000000000  -0.28560000000000   0.75000000000000   Cu   (4g)
   0.75000000000000  -0.28560000000000   0.75000000000000   Cu   (4g)
   0.75000000000000   0.28560000000000   0.25000000000000   Cu   (4g)
   0.00000000000000   0.34580000000000   0.27080000000000   Se   (4h)
   0.50000000000000  -0.34580000000000   0.77080000000000   Se   (4h)
   0.50000000000000   0.34580000000000   0.22920000000000   Se   (4h)
   0.00000000000000  -0.34580000000000  -0.27080000000000   Se   (4h)
   0.62470000000000   0.60570000000000   0.35750000000000   Se   (8i)
  -0.12470000000000  -0.60570000000000   0.85750000000000   Se   (8i)
  -0.12470000000000   0.60570000000000   0.14250000000000   Se   (8i)
   0.62470000000000  -0.60570000000000  -0.35750000000000   Se   (8i)
  -0.62470000000000  -0.60570000000000  -0.35750000000000   Se   (8i)
   1.12470000000000   0.60570000000000   0.14250000000000   Se   (8i)
   1.12470000000000  -0.60570000000000   0.85750000000000   Se   (8i)
  -0.62470000000000   0.60570000000000   0.35750000000000   Se   (8i)
\end{lstlisting}
{\phantomsection\label{ABC3_oP20_54_e_d_cf_cif}}
{\hyperref[ABC3_oP20_54_e_d_cf]{BiGaO$_{3}$: ABC3\_oP20\_54\_e\_d\_cf}} - CIF
\begin{lstlisting}[numbers=none,language={mylang}]
# CIF file
data_findsym-output
_audit_creation_method FINDSYM

_chemical_name_mineral 'BiGaO3'
_chemical_formula_sum 'Bi Ga O3'

loop_
_publ_author_name
 'H. Yusa'
 'A. A. Belik'
 'E. {Takayama-Muromachi}'
 'N. Hirao'
 'Y. Ohishi'
_journal_name_full_name
;
 Physical Review B
;
_journal_volume 80
_journal_year 2009
_journal_page_first 214103
_journal_page_last 214103
_publ_Section_title
;
 High-pressure phase transitions in Bi$M$O$_{3}$ ($M$ = Al, Ga, and In): {\it In situ} x-ray diffraction and Raman scattering experiments
;

# Found in Pearson's Crystal Data - Crystal Structure Database for Inorganic Compounds, 2013

_aflow_title 'BiGaO$_{3}$ Structure'
_aflow_proto 'ABC3_oP20_54_e_d_cf'
_aflow_params 'a,b/a,c/a,y_{1},z_{2},z_{3},x_{4},y_{4},z_{4}'
_aflow_params_values '5.3467489374,0.956627452436,1.81710213776,0.8667,0.6417,0.8902,0.4055,0.2686,0.5503'
_aflow_Strukturbericht 'None'
_aflow_Pearson 'oP20'

_cell_length_a    5.3467489374
_cell_length_b    5.1148468148
_cell_length_c    9.7155889242
_cell_angle_alpha 90.0000000000
_cell_angle_beta  90.0000000000
_cell_angle_gamma 90.0000000000
 
_symmetry_space_group_name_H-M "P 21/c 2/c 2/a"
_symmetry_Int_Tables_number 54
 
loop_
_space_group_symop_id
_space_group_symop_operation_xyz
1 x,y,z
2 x+1/2,-y,-z+1/2
3 -x,y,-z+1/2
4 -x+1/2,-y,z
5 -x,-y,-z
6 -x+1/2,y,z+1/2
7 x,-y,z+1/2
8 x+1/2,y,-z
 
loop_
_atom_site_label
_atom_site_type_symbol
_atom_site_symmetry_multiplicity
_atom_site_Wyckoff_label
_atom_site_fract_x
_atom_site_fract_y
_atom_site_fract_z
_atom_site_occupancy
O1  O    4 c 0.00000 0.86670 0.25000 1.00000
Ga1 Ga   4 d 0.25000 0.00000 0.64170 1.00000
Bi1 Bi   4 e 0.25000 0.50000 0.89020 1.00000
O2  O    8 f 0.40550 0.26860 0.55030 1.00000
\end{lstlisting}
{\phantomsection\label{ABC3_oP20_54_e_d_cf_poscar}}
{\hyperref[ABC3_oP20_54_e_d_cf]{BiGaO$_{3}$: ABC3\_oP20\_54\_e\_d\_cf}} - POSCAR
\begin{lstlisting}[numbers=none,language={mylang}]
ABC3_oP20_54_e_d_cf & a,b/a,c/a,y1,z2,z3,x4,y4,z4 --params=5.3467489374,0.956627452436,1.81710213776,0.8667,0.6417,0.8902,0.4055,0.2686,0.5503 & Pcca D_{2h}^{8} #54 (cdef) & oP20 & None & BiGaO3 &  & H. Yusa et al., Phys. Rev. B 80, 214103(2009)
   1.00000000000000
   5.34674893740000   0.00000000000000   0.00000000000000
   0.00000000000000   5.11484681480000   0.00000000000000
   0.00000000000000   0.00000000000000   9.71558892420000
    Bi    Ga     O
     4     4    12
Direct
   0.25000000000000   0.50000000000000   0.89020000000000   Bi   (4e)
   0.75000000000000   0.50000000000000  -0.39020000000000   Bi   (4e)
   0.75000000000000   0.50000000000000  -0.89020000000000   Bi   (4e)
   0.25000000000000   0.50000000000000   1.39020000000000   Bi   (4e)
   0.25000000000000   0.00000000000000   0.64170000000000   Ga   (4d)
   0.75000000000000   0.00000000000000  -0.14170000000000   Ga   (4d)
   0.75000000000000   0.00000000000000  -0.64170000000000   Ga   (4d)
   0.25000000000000   0.00000000000000   1.14170000000000   Ga   (4d)
   0.00000000000000   0.86670000000000   0.25000000000000    O   (4c)
   0.50000000000000  -0.86670000000000   0.25000000000000    O   (4c)
   0.00000000000000  -0.86670000000000   0.75000000000000    O   (4c)
   0.50000000000000   0.86670000000000   0.75000000000000    O   (4c)
   0.40550000000000   0.26860000000000   0.55030000000000    O   (8f)
   0.09450000000000  -0.26860000000000   0.55030000000000    O   (8f)
  -0.40550000000000   0.26860000000000  -0.05030000000000    O   (8f)
   0.90550000000000  -0.26860000000000  -0.05030000000000    O   (8f)
  -0.40550000000000  -0.26860000000000  -0.55030000000000    O   (8f)
   0.90550000000000   0.26860000000000  -0.55030000000000    O   (8f)
   0.40550000000000  -0.26860000000000   1.05030000000000    O   (8f)
   0.09450000000000   0.26860000000000   1.05030000000000    O   (8f)
\end{lstlisting}
{\phantomsection\label{A2B_oP24_55_2g2h_gh_cif}}
{\hyperref[A2B_oP24_55_2g2h_gh]{GeAs$_{2}$: A2B\_oP24\_55\_2g2h\_gh}} - CIF
\begin{lstlisting}[numbers=none,language={mylang}]
# CIF file
data_findsym-output
_audit_creation_method FINDSYM

_chemical_name_mineral 'GeAs2'
_chemical_formula_sum 'As2 Ge'

loop_
_publ_author_name
 'T. Wadsten'
_journal_name_full_name
;
 Acta Chemica Scandinavica
;
_journal_volume 21
_journal_year 1967
_journal_page_first 593
_journal_page_last 594
_publ_Section_title
;
 Crystal structures of SiP$_{2}$, SiAs$_{2}$, and GeP
;

# Found in Pearson's Crystal Data - Crystal Structure Database for Inorganic Compounds, 2013

_aflow_title 'GeAs$_{2}$ Structure'
_aflow_proto 'A2B_oP24_55_2g2h_gh'
_aflow_params 'a,b/a,c/a,x_{1},y_{1},x_{2},y_{2},x_{3},y_{3},x_{4},y_{4},x_{5},y_{5},x_{6},y_{6}'
_aflow_params_values '10.1600191136,1.45275590551,0.366929133855,0.0628,0.4022,0.1014,0.1118,0.2024,0.2667,0.226,0.0384,0.3532,0.2953,0.4192,0.1378'
_aflow_Strukturbericht 'None'
_aflow_Pearson 'oP24'

_cell_length_a    10.1600191136
_cell_length_b    14.7600277674
_cell_length_c    3.7280070133
_cell_angle_alpha 90.0000000000
_cell_angle_beta  90.0000000000
_cell_angle_gamma 90.0000000000
 
_symmetry_space_group_name_H-M "P 21/b 21/a 2/m"
_symmetry_Int_Tables_number 55
 
loop_
_space_group_symop_id
_space_group_symop_operation_xyz
1 x,y,z
2 x+1/2,-y+1/2,-z
3 -x+1/2,y+1/2,-z
4 -x,-y,z
5 -x,-y,-z
6 -x+1/2,y+1/2,z
7 x+1/2,-y+1/2,z
8 x,y,-z
 
loop_
_atom_site_label
_atom_site_type_symbol
_atom_site_symmetry_multiplicity
_atom_site_Wyckoff_label
_atom_site_fract_x
_atom_site_fract_y
_atom_site_fract_z
_atom_site_occupancy
As1 As   4 g 0.06280 0.40220 0.00000 1.00000
As2 As   4 g 0.10140 0.11180 0.00000 1.00000
Ge1 Ge   4 g 0.20240 0.26670 0.00000 1.00000
As3 As   4 h 0.22600 0.03840 0.50000 1.00000
As4 As   4 h 0.35320 0.29530 0.50000 1.00000
Ge2 Ge   4 h 0.41920 0.13780 0.50000 1.00000
\end{lstlisting}
{\phantomsection\label{A2B_oP24_55_2g2h_gh_poscar}}
{\hyperref[A2B_oP24_55_2g2h_gh]{GeAs$_{2}$: A2B\_oP24\_55\_2g2h\_gh}} - POSCAR

{\phantomsection\label{A3B5_oP16_55_ch_agh_cif}}
{\hyperref[A3B5_oP16_55_ch_agh]{Rh$_{5}$Ge$_{3}$: A3B5\_oP16\_55\_ch\_agh}} - CIF
\begin{lstlisting}[numbers=none,language={mylang}]
# CIF file
data_findsym-output
_audit_creation_method FINDSYM

_chemical_name_mineral 'Rh5Ge3'
_chemical_formula_sum 'Ge3 Rh5'

loop_
_publ_author_name
 'S. Geller'
_journal_name_full_name
;
 Acta Cristallographica
;
_journal_volume 8
_journal_year 1955
_journal_page_first 15
_journal_page_last 21
_publ_Section_title
;
 The rhodium--germanium system. I. The crystal structures of Rh$_{2}$Ge, Rh$_{5}$Ge$_{3}$ and RhGe
;

# Found in Pearson's Crystal Data - Crystal Structure Database for Inorganic Compounds, 2013

_aflow_title 'Rh$_{5}$Ge$_{3}$ Structure'
_aflow_proto 'A3B5_oP16_55_ch_agh'
_aflow_params 'a,b/a,c/a,x_{3},y_{3},x_{4},y_{4},x_{5},y_{5}'
_aflow_params_values '5.4199981729,1.90405904061,0.730627306278,0.348,0.22,0.112,0.152,0.17,0.393'
_aflow_Strukturbericht 'None'
_aflow_Pearson 'oP16'

_cell_length_a    5.4199981729
_cell_length_b    10.3199965212
_cell_length_c    3.9599986651
_cell_angle_alpha 90.0000000000
_cell_angle_beta  90.0000000000
_cell_angle_gamma 90.0000000000
 
_symmetry_space_group_name_H-M "P 21/b 21/a 2/m"
_symmetry_Int_Tables_number 55
 
loop_
_space_group_symop_id
_space_group_symop_operation_xyz
1 x,y,z
2 x+1/2,-y+1/2,-z
3 -x+1/2,y+1/2,-z
4 -x,-y,z
5 -x,-y,-z
6 -x+1/2,y+1/2,z
7 x+1/2,-y+1/2,z
8 x,y,-z
 
loop_
_atom_site_label
_atom_site_type_symbol
_atom_site_symmetry_multiplicity
_atom_site_Wyckoff_label
_atom_site_fract_x
_atom_site_fract_y
_atom_site_fract_z
_atom_site_occupancy
Rh1 Rh   2 a 0.00000 0.00000 0.00000 1.00000
Ge1 Ge   2 c 0.00000 0.50000 0.00000 1.00000
Rh2 Rh   4 g 0.34800 0.22000 0.00000 1.00000
Ge2 Ge   4 h 0.11200 0.15200 0.50000 1.00000
Rh3 Rh   4 h 0.17000 0.39300 0.50000 1.00000
\end{lstlisting}
{\phantomsection\label{A3B5_oP16_55_ch_agh_poscar}}
{\hyperref[A3B5_oP16_55_ch_agh]{Rh$_{5}$Ge$_{3}$: A3B5\_oP16\_55\_ch\_agh}} - POSCAR
\begin{lstlisting}[numbers=none,language={mylang}]
A3B5_oP16_55_ch_agh & a,b/a,c/a,x3,y3,x4,y4,x5,y5 --params=5.4199981729,1.90405904061,0.730627306278,0.348,0.22,0.112,0.152,0.17,0.393 & Pbam D_{2h}^{9} #55 (acgh^2) & oP16 & None & Rh5Ge3 &  & S. Geller, Acta Cryst. 8, 15-21 (1955)
   1.00000000000000
   5.41999817290000   0.00000000000000   0.00000000000000
   0.00000000000000  10.31999652120000   0.00000000000000
   0.00000000000000   0.00000000000000   3.95999866510000
    Ge    Rh
     6    10
Direct
   0.00000000000000   0.50000000000000   0.00000000000000   Ge   (2c)
   0.50000000000000   0.00000000000000   0.00000000000000   Ge   (2c)
   0.11200000000000   0.15200000000000   0.50000000000000   Ge   (4h)
  -0.11200000000000  -0.15200000000000   0.50000000000000   Ge   (4h)
   0.38800000000000   0.65200000000000   0.50000000000000   Ge   (4h)
   0.61200000000000   0.34800000000000   0.50000000000000   Ge   (4h)
   0.00000000000000   0.00000000000000   0.00000000000000   Rh   (2a)
   0.50000000000000   0.50000000000000   0.00000000000000   Rh   (2a)
   0.34800000000000   0.22000000000000   0.00000000000000   Rh   (4g)
  -0.34800000000000  -0.22000000000000   0.00000000000000   Rh   (4g)
   0.15200000000000   0.72000000000000   0.00000000000000   Rh   (4g)
   0.84800000000000   0.28000000000000   0.00000000000000   Rh   (4g)
   0.17000000000000   0.39300000000000   0.50000000000000   Rh   (4h)
  -0.17000000000000  -0.39300000000000   0.50000000000000   Rh   (4h)
   0.33000000000000   0.89300000000000   0.50000000000000   Rh   (4h)
   0.67000000000000   0.10700000000000   0.50000000000000   Rh   (4h)
\end{lstlisting}
{\phantomsection\label{A_oP16_55_2g2h_cif}}
{\hyperref[A_oP16_55_2g2h]{R-carbon: A\_oP16\_55\_2g2h}} - CIF
\begin{lstlisting}[numbers=none,language={mylang}]
# CIF file 
data_findsym-output
_audit_creation_method FINDSYM

_chemical_name_mineral 'R-carbon'
_chemical_formula_sum 'C'

loop_
_publ_author_name
 'H. Niu'
 'X.-Q. Chen'
 'S. Wang'
 'D. Li'
 'W. L. Mao'
 'Y. Li'
_journal_name_full_name
;
 Physical Review Letters
;
_journal_volume 108
_journal_year 2012
_journal_page_first 135501
_journal_page_last 135501
_publ_Section_title
;
 Families of Superhard Crystalline Carbon Allotropes Constructed via Cold Compression of Graphite and Nanotubes
;

_aflow_title 'R-carbon Structure'
_aflow_proto 'A_oP16_55_2g2h'
_aflow_params 'a,b/a,c/a,x_{1},y_{1},x_{2},y_{2},x_{3},y_{3},x_{4},y_{4}'
_aflow_params_values '7.7886,0.613101199189,0.320442698303,0.6731,-0.037,0.8435,0.8087,-0.0454,0.8613,0.5704,0.8926'
_aflow_Strukturbericht 'None'
_aflow_Pearson 'oP16'

_symmetry_space_group_name_H-M "P 21/b 21/a 2/m"
_symmetry_Int_Tables_number 55
 
_cell_length_a    7.78860
_cell_length_b    4.77520
_cell_length_c    2.49580
_cell_angle_alpha 90.00000
_cell_angle_beta  90.00000
_cell_angle_gamma 90.00000
 
loop_
_space_group_symop_id
_space_group_symop_operation_xyz
1 x,y,z
2 x+1/2,-y+1/2,-z
3 -x+1/2,y+1/2,-z
4 -x,-y,z
5 -x,-y,-z
6 -x+1/2,y+1/2,z
7 x+1/2,-y+1/2,z
8 x,y,-z
 
loop_
_atom_site_label
_atom_site_type_symbol
_atom_site_symmetry_multiplicity
_atom_site_Wyckoff_label
_atom_site_fract_x
_atom_site_fract_y
_atom_site_fract_z
_atom_site_occupancy
C1 C   4 g 0.67310  -0.03700 0.00000 1.00000
C2 C   4 g 0.84350  0.80870  0.00000 1.00000
C3 C   4 h -0.04540 0.86130  0.50000 1.00000
C4 C   4 h 0.57040  0.89260  0.50000 1.00000
\end{lstlisting}
{\phantomsection\label{A_oP16_55_2g2h_poscar}}
{\hyperref[A_oP16_55_2g2h]{R-carbon: A\_oP16\_55\_2g2h}} - POSCAR
\begin{lstlisting}[numbers=none,language={mylang}]
A_oP16_55_2g2h & a,b/a,c/a,x1,y1,x2,y2,x3,y3,x4,y4 --params=7.7886,0.613101199189,0.320442698303,0.6731,-0.037,0.8435,0.8087,-0.0454,0.8613,0.5704,0.8926 & Pbam D_{2h}^{9} #55 (g^2h^2) & oP16 & None & C & R-carbon & H. Niu et al., Phys. Rev. Lett. 108, 135501(2012)
   1.00000000000000
   7.78860000000000   0.00000000000000   0.00000000000000
   0.00000000000000   4.77520000000000   0.00000000000000
   0.00000000000000   0.00000000000000   2.49580000000000
     C
    16
Direct
   0.67310000000000  -0.03700000000000   0.00000000000000    C   (4g)
  -0.67310000000000   0.03700000000000   0.00000000000000    C   (4g)
  -0.17310000000000   0.46300000000000   0.00000000000000    C   (4g)
   1.17310000000000   0.53700000000000   0.00000000000000    C   (4g)
   0.84350000000000   0.80870000000000   0.00000000000000    C   (4g)
  -0.84350000000000  -0.80870000000000   0.00000000000000    C   (4g)
  -0.34350000000000   1.30870000000000   0.00000000000000    C   (4g)
   1.34350000000000  -0.30870000000000   0.00000000000000    C   (4g)
  -0.04540000000000   0.86130000000000   0.50000000000000    C   (4h)
   0.04540000000000  -0.86130000000000   0.50000000000000    C   (4h)
   0.54540000000000   1.36130000000000   0.50000000000000    C   (4h)
   0.45460000000000  -0.36130000000000   0.50000000000000    C   (4h)
   0.57040000000000   0.89260000000000   0.50000000000000    C   (4h)
  -0.57040000000000  -0.89260000000000   0.50000000000000    C   (4h)
  -0.07040000000000   1.39260000000000   0.50000000000000    C   (4h)
   1.07040000000000  -0.39260000000000   0.50000000000000    C   (4h)
\end{lstlisting}
{\phantomsection\label{A2B_oP6_58_g_a_cif}}
{\hyperref[A2B_oP6_58_g_a]{$\alpha$-PdCl$_{2}$ ($C50$): A2B\_oP6\_58\_g\_a}} - CIF
\begin{lstlisting}[numbers=none,language={mylang}]
# CIF file 
data_findsym-output
_audit_creation_method FINDSYM

_chemical_name_mineral '$\alpha$-PdCl2'
_chemical_formula_sum 'Cl2 Pd'

loop_
_publ_author_name
 'J. Evers'
 'W. Beck'
 'M. G\"{o}bel'
 'S. Jakob'
 'P. Mayer'
 'G. Oehlinger'
 'M. Rotter'
 'T. M. Klap\"{o}tke'
_journal_name_full_name
;
 Angewandte Chemie (International ed.)
;
_journal_volume 49
_journal_year 2010
_journal_page_first 5677
_journal_page_last 5682
_publ_Section_title
;
 The Structures of $\delta$-PdCl$_{2}$ and $\gamma$-PdCl$_{2}$: Phases with Negative Thermal Expansion in One Direction
;

_aflow_title '$\alpha$-PdCl$_{2}$ ($C50$) Structure'
_aflow_proto 'A2B_oP6_58_g_a'
_aflow_params 'a,b/a,c/a,x_{2},y_{2}'
_aflow_params_values '3.7572,2.89952624295,0.89063664431,0.3326,0.63309'
_aflow_Strukturbericht '$C50$'
_aflow_Pearson 'oP6'

_symmetry_space_group_name_H-M "P 21/n 21/n 2/m"
_symmetry_Int_Tables_number 58
 
_cell_length_a    3.75720
_cell_length_b    10.89410
_cell_length_c    3.34630
_cell_angle_alpha 90.00000
_cell_angle_beta  90.00000
_cell_angle_gamma 90.00000
 
loop_
_space_group_symop_id
_space_group_symop_operation_xyz
1 x,y,z
2 x+1/2,-y+1/2,-z+1/2
3 -x+1/2,y+1/2,-z+1/2
4 -x,-y,z
5 -x,-y,-z
6 -x+1/2,y+1/2,z+1/2
7 x+1/2,-y+1/2,z+1/2
8 x,y,-z
 
loop_
_atom_site_label
_atom_site_type_symbol
_atom_site_symmetry_multiplicity
_atom_site_Wyckoff_label
_atom_site_fract_x
_atom_site_fract_y
_atom_site_fract_z
_atom_site_occupancy
Pd1 Pd   2 a 0.00000 0.00000 0.00000 1.00000
Cl1 Cl   4 g 0.33260 0.63309 0.00000 1.00000
\end{lstlisting}
{\phantomsection\label{A2B_oP6_58_g_a_poscar}}
{\hyperref[A2B_oP6_58_g_a]{$\alpha$-PdCl$_{2}$ ($C50$): A2B\_oP6\_58\_g\_a}} - POSCAR
\begin{lstlisting}[numbers=none,language={mylang}]
A2B_oP6_58_g_a & a,b/a,c/a,x2,y2 --params=3.7572,2.89952624295,0.89063664431,0.3326,0.63309 & Pnnm D_{2h}^{12} #58 (ag) & oP6 & $C50$ & PdCl2 & $\alpha$-PdCl2 & J. Evers et al., Angew. Chem. Int. Ed. 49, 5677-5682 (2010)
   1.00000000000000
   3.75720000000000   0.00000000000000   0.00000000000000
   0.00000000000000  10.89410000000000   0.00000000000000
   0.00000000000000   0.00000000000000   3.34630000000000
    Cl    Pd
     4     2
Direct
   0.33260000000000   0.63309000000000   0.00000000000000   Cl   (4g)
  -0.33260000000000  -0.63309000000000   0.00000000000000   Cl   (4g)
   0.16740000000000   1.13309000000000   0.50000000000000   Cl   (4g)
   0.83260000000000  -0.13309000000000   0.50000000000000   Cl   (4g)
   0.00000000000000   0.00000000000000   0.00000000000000   Pd   (2a)
   0.50000000000000   0.50000000000000   0.50000000000000   Pd   (2a)
\end{lstlisting}
{\phantomsection\label{ABC_oP6_59_a_b_a_cif}}
{\hyperref[ABC_oP6_59_a_b_a]{FeOCl: ABC\_oP6\_59\_a\_b\_a}} - CIF
\begin{lstlisting}[numbers=none,language={mylang}]
# CIF file 
data_findsym-output
_audit_creation_method FINDSYM

_chemical_name_mineral 'FeOCl'
_chemical_formula_sum 'Cl Fe O'

loop_
_publ_author_name
 'S. M. Kauzlarich'
 'J. L. Stanton'
 'J. Faber'
 'B. A. Averill'
_journal_name_full_name
;
 Journal of the American Chemical Society
;
_journal_volume 108
_journal_year 1986
_journal_page_first 7946
_journal_page_last 7951
_publ_Section_title
;
 Neutron profile refinement of the structure of FeOCl and FeOCl(TTF)$_{1/8.5}$
;

_aflow_title 'FeOCl Structure'
_aflow_proto 'ABC_oP6_59_a_b_a'
_aflow_params 'a,b/a,c/a,z_{1},z_{2},z_{3}'
_aflow_params_values '3.301,1.14298697364,2.39612238716,0.32961,-0.04795,0.89243'
_aflow_Strukturbericht 'None'
_aflow_Pearson 'oP6'

_symmetry_space_group_name_H-M "P m m n:2"
_symmetry_Int_Tables_number 59
 
_cell_length_a    3.30100
_cell_length_b    3.77300
_cell_length_c    7.90960
_cell_angle_alpha 90.00000
_cell_angle_beta  90.00000
_cell_angle_gamma 90.00000
 
loop_
_space_group_symop_id
_space_group_symop_operation_xyz
1 x,y,z
2 x+1/2,-y,-z
3 -x,y+1/2,-z
4 -x+1/2,-y+1/2,z
5 -x,-y,-z
6 -x+1/2,y,z
7 x,-y+1/2,z
8 x+1/2,y+1/2,-z
 
loop_
_atom_site_label
_atom_site_type_symbol
_atom_site_symmetry_multiplicity
_atom_site_Wyckoff_label
_atom_site_fract_x
_atom_site_fract_y
_atom_site_fract_z
_atom_site_occupancy
Cl1 Cl   2 a 0.25000 0.25000 0.32961  1.00000
O1  O    2 a 0.25000 0.25000 -0.04795 1.00000
Fe1 Fe   2 b 0.25000 0.75000 0.89243  1.00000
\end{lstlisting}
{\phantomsection\label{ABC_oP6_59_a_b_a_poscar}}
{\hyperref[ABC_oP6_59_a_b_a]{FeOCl: ABC\_oP6\_59\_a\_b\_a}} - POSCAR
\begin{lstlisting}[numbers=none,language={mylang}]
ABC_oP6_59_a_b_a & a,b/a,c/a,z1,z2,z3 --params=3.301,1.14298697364,2.39612238716,0.32961,-0.04795,0.89243 & Pmmn D_{2h}^{13} #59 (a^2b) & oP6 & None & FeOCl & FeOCl & S. M. Kauzlarich et al., J. Am. Chem. Soc. 108, 7946-7951 (1986)
   1.00000000000000
   3.30100000000000   0.00000000000000   0.00000000000000
   0.00000000000000   3.77300000000000   0.00000000000000
   0.00000000000000   0.00000000000000   7.90960000000000
    Cl    Fe     O
     2     2     2
Direct
   0.25000000000000   0.25000000000000   0.32961000000000   Cl   (2a)
   0.75000000000000   0.75000000000000  -0.32961000000000   Cl   (2a)
   0.25000000000000   0.75000000000000   0.89243000000000   Fe   (2b)
   0.75000000000000   0.25000000000000  -0.89243000000000   Fe   (2b)
   0.25000000000000   0.25000000000000  -0.04795000000000    O   (2a)
   0.75000000000000   0.75000000000000   0.04795000000000    O   (2a)
\end{lstlisting}
{\phantomsection\label{A2B3_oP20_60_d_cd_cif}}
{\hyperref[A2B3_oP20_60_d_cd]{Rh$_{2}$S$_{3}$: A2B3\_oP20\_60\_d\_cd}} - CIF
\begin{lstlisting}[numbers=none,language={mylang}]
# CIF file
data_findsym-output
_audit_creation_method FINDSYM

_chemical_name_mineral 'Rh2S3'
_chemical_formula_sum 'Rh2 S3'

loop_
_publ_author_name
 'E. Parth{\\'e} F.'
 'Hulliger'
_journal_year 1966
_publ_Section_title
;
 The crystal structure of Rh$_{2}$S$_{3}$
;

# Found in Pearson's Crystal Data - Crystal Structure Database for Inorganic Compounds, 2013

_aflow_title 'Rh$_{2}$S$_{3}$ Structure'
_aflow_proto 'A2B3_oP20_60_d_cd'
_aflow_params 'a,b/a,c/a,y_{1},x_{2},y_{2},z_{2},x_{3},y_{3},z_{3}'
_aflow_params_values '8.4598035458,0.706855791961,0.724586288418,0.547,0.394,0.75,0.033,0.348,0.611,0.396'
_aflow_Strukturbericht 'None'
_aflow_Pearson 'oP20'

_cell_length_a    8.4598035458
_cell_length_b    5.9798611352
_cell_length_c    6.1298576520
_cell_angle_alpha 90.0000000000
_cell_angle_beta  90.0000000000
_cell_angle_gamma 90.0000000000
 
_symmetry_space_group_name_H-M "P 21/b 2/c 21/n"
_symmetry_Int_Tables_number 60
 
loop_
_space_group_symop_id
_space_group_symop_operation_xyz
1 x,y,z
2 x+1/2,-y+1/2,-z
3 -x,y,-z+1/2
4 -x+1/2,-y+1/2,z+1/2
5 -x,-y,-z
6 -x+1/2,y+1/2,z
7 x,-y,z+1/2
8 x+1/2,y+1/2,-z+1/2
 
loop_
_atom_site_label
_atom_site_type_symbol
_atom_site_symmetry_multiplicity
_atom_site_Wyckoff_label
_atom_site_fract_x
_atom_site_fract_y
_atom_site_fract_z
_atom_site_occupancy
S1  S    4 c 0.00000 0.54700 0.25000 1.00000
Rh1 Rh   8 d 0.39400 0.75000 0.03300 1.00000
S2  S    8 d 0.34800 0.61100 0.39600 1.00000
\end{lstlisting}
{\phantomsection\label{A2B3_oP20_60_d_cd_poscar}}
{\hyperref[A2B3_oP20_60_d_cd]{Rh$_{2}$S$_{3}$: A2B3\_oP20\_60\_d\_cd}} - POSCAR
\begin{lstlisting}[numbers=none,language={mylang}]
A2B3_oP20_60_d_cd & a,b/a,c/a,y1,x2,y2,z2,x3,y3,z3 --params=8.4598035458,0.706855791961,0.724586288418,0.547,0.394,0.75,0.033,0.348,0.611,0.396 & Pbcn D_{2h}^{14} #60 (cd^2) & oP20 & None & Rh2S3 &  & E. Parth{\'e} F. and Hulliger, (1966)
   1.00000000000000
   8.45980354580000   0.00000000000000   0.00000000000000
   0.00000000000000   5.97986113520000   0.00000000000000
   0.00000000000000   0.00000000000000   6.12985765200000
    Rh     S
     8    12
Direct
   0.39400000000000   0.75000000000000   0.03300000000000   Rh   (8d)
   0.10600000000000  -0.25000000000000   0.53300000000000   Rh   (8d)
  -0.39400000000000   0.75000000000000   0.46700000000000   Rh   (8d)
   0.89400000000000  -0.25000000000000  -0.03300000000000   Rh   (8d)
  -0.39400000000000  -0.75000000000000  -0.03300000000000   Rh   (8d)
   0.89400000000000   1.25000000000000   0.46700000000000   Rh   (8d)
   0.39400000000000  -0.75000000000000   0.53300000000000   Rh   (8d)
   0.10600000000000   1.25000000000000   0.03300000000000   Rh   (8d)
   0.00000000000000   0.54700000000000   0.25000000000000    S   (4c)
   0.50000000000000  -0.04700000000000   0.75000000000000    S   (4c)
   0.00000000000000  -0.54700000000000   0.75000000000000    S   (4c)
   0.50000000000000   1.04700000000000   0.25000000000000    S   (4c)
   0.34800000000000   0.61100000000000   0.39600000000000    S   (8d)
   0.15200000000000  -0.11100000000000   0.89600000000000    S   (8d)
  -0.34800000000000   0.61100000000000   0.10400000000000    S   (8d)
   0.84800000000000  -0.11100000000000  -0.39600000000000    S   (8d)
  -0.34800000000000  -0.61100000000000  -0.39600000000000    S   (8d)
   0.84800000000000   1.11100000000000   0.10400000000000    S   (8d)
   0.34800000000000  -0.61100000000000   0.89600000000000    S   (8d)
   0.15200000000000   1.11100000000000   0.39600000000000    S   (8d)
\end{lstlisting}
{\phantomsection\label{A3B_oP32_60_3d_d_cif}}
{\hyperref[A3B_oP32_60_3d_d]{WO$_{3}$: A3B\_oP32\_60\_3d\_d}} - CIF
\begin{lstlisting}[numbers=none,language={mylang}]
# CIF file
data_findsym-output
_audit_creation_method FINDSYM

_chemical_name_mineral 'WO3'
_chemical_formula_sum 'O3 W'

loop_
_publ_author_name
 'T. Vogt'
 'P. M. Woodward'
 'B. A. Hunter'
_journal_name_full_name
;
 Journal of Solid State Chemistry
;
_journal_volume 144
_journal_year 1999
_journal_page_first 209
_journal_page_last 215
_publ_Section_title
;
 The high-temperature phases of WO$_{3}$
;

# Found in Pearson's Crystal Data - Crystal Structure Database for Inorganic Compounds, 2013

_aflow_title 'WO$_{3}$ Structure'
_aflow_proto 'A3B_oP32_60_3d_d'
_aflow_params 'a,b/a,c/a,x_{1},y_{1},z_{1},x_{2},y_{2},z_{2},x_{3},y_{3},z_{3},x_{4},y_{4},z_{4}'
_aflow_params_values '7.3397836195,1.05524748967,1.03197678378,0.5016,0.7205,0.0322,0.2167,0.7591,0.2582,0.2197,0.5016,0.013,0.248,0.783,0.0291'
_aflow_Strukturbericht 'None'
_aflow_Pearson 'oP32'

_cell_length_a    7.3397836195
_cell_length_b    7.7452882392
_cell_length_c    7.5744862933
_cell_angle_alpha 90.0000000000
_cell_angle_beta  90.0000000000
_cell_angle_gamma 90.0000000000
 
_symmetry_space_group_name_H-M "P 21/b 2/c 21/n"
_symmetry_Int_Tables_number 60
 
loop_
_space_group_symop_id
_space_group_symop_operation_xyz
1 x,y,z
2 x+1/2,-y+1/2,-z
3 -x,y,-z+1/2
4 -x+1/2,-y+1/2,z+1/2
5 -x,-y,-z
6 -x+1/2,y+1/2,z
7 x,-y,z+1/2
8 x+1/2,y+1/2,-z+1/2
 
loop_
_atom_site_label
_atom_site_type_symbol
_atom_site_symmetry_multiplicity
_atom_site_Wyckoff_label
_atom_site_fract_x
_atom_site_fract_y
_atom_site_fract_z
_atom_site_occupancy
O1 O   8 d 0.50160 0.72050 0.03220 1.00000
O2 O   8 d 0.21670 0.75910 0.25820 1.00000
O3 O   8 d 0.21970 0.50160 0.01300 1.00000
W1 W   8 d 0.24800 0.78300 0.02910 1.00000
\end{lstlisting}
{\phantomsection\label{A3B_oP32_60_3d_d_poscar}}
{\hyperref[A3B_oP32_60_3d_d]{WO$_{3}$: A3B\_oP32\_60\_3d\_d}} - POSCAR

{\phantomsection\label{A7B8_oP120_60_7d_8d_cif}}
{\hyperref[A7B8_oP120_60_7d_8d]{$\beta$-Toluene: A7B8\_oP120\_60\_7d\_8d}} - CIF

{\phantomsection\label{A7B8_oP120_60_7d_8d_poscar}}
{\hyperref[A7B8_oP120_60_7d_8d]{$\beta$-Toluene: A7B8\_oP120\_60\_7d\_8d}} - POSCAR

{\phantomsection\label{AB_oP48_61_3c_3c_cif}}
{\hyperref[AB_oP48_61_3c_3c]{Benzene: AB\_oP48\_61\_3c\_3c}} - CIF
\begin{lstlisting}[numbers=none,language={mylang}]
# CIF file 
data_findsym-output
_audit_creation_method FINDSYM

_chemical_name_mineral 'Benzene'
_chemical_formula_sum 'C H'

loop_
_publ_author_name
 'S. K Nayak'
 'R. Sathishkumar'
 'T. N. {Guru Row}'
_journal_name_full_name
;
 CrystEngComm
;
_journal_volume 12
_journal_year 2010
_journal_page_first 3112
_journal_page_last 3118
_publ_Section_title
;
 Directing role of functional groups in selective generation of C-H-$\pi$ interactions: In situ cryo-crystallographic studies on benzyl derivatives
;

_aflow_title 'Benzene Structure'
_aflow_proto 'AB_oP48_61_3c_3c'
_aflow_params 'a,b/a,c/a,x_{1},y_{1},z_{1},x_{2},y_{2},z_{2},x_{3},y_{3},z_{3},x_{4},y_{4},z_{4},x_{5},y_{5},z_{5},x_{6},y_{6},z_{6}'
_aflow_params_values '6.914,1.08128435059,1.38313566676,0.1297,0.5762,0.40803,0.1235,0.6328,0.54518,0.0057,0.4432,0.36289,0.2172,0.6275,0.346,0.2068,0.7225,0.5756,0.0095,0.4051,0.2704'
_aflow_Strukturbericht 'None'
_aflow_Pearson 'oP48'

_symmetry_space_group_name_H-M "P 21/b 21/c 21/a"
_symmetry_Int_Tables_number 61
 
_cell_length_a    6.91400
_cell_length_b    7.47600
_cell_length_c    9.56300
_cell_angle_alpha 90.00000
_cell_angle_beta  90.00000
_cell_angle_gamma 90.00000
 
loop_
_space_group_symop_id
_space_group_symop_operation_xyz
1 x,y,z
2 x+1/2,-y+1/2,-z
3 -x,y+1/2,-z+1/2
4 -x+1/2,-y,z+1/2
5 -x,-y,-z
6 -x+1/2,y+1/2,z
7 x,-y+1/2,z+1/2
8 x+1/2,y,-z+1/2
 
loop_
_atom_site_label
_atom_site_type_symbol
_atom_site_symmetry_multiplicity
_atom_site_Wyckoff_label
_atom_site_fract_x
_atom_site_fract_y
_atom_site_fract_z
_atom_site_occupancy
C1 C   8 c 0.12970 0.57620 0.40803  1.00000
C2 C   8 c 0.12350 0.63280 0.54518  1.00000
C3 C   8 c 0.00570 0.44320 0.36289  1.00000
H1 H   8 c 0.21720 0.62750 0.34600  1.00000
H2 H   8 c 0.20680 0.72250 0.57560  1.00000
H3 H   8 c 0.00950 0.40510 0.27040  1.00000
\end{lstlisting}
{\phantomsection\label{AB_oP48_61_3c_3c_poscar}}
{\hyperref[AB_oP48_61_3c_3c]{Benzene: AB\_oP48\_61\_3c\_3c}} - POSCAR

{\phantomsection\label{A2B3_oP20_62_2c_3c_cif}}
{\hyperref[A2B3_oP20_62_2c_3c]{Tongbaite (Cr$_{3}$C$_{2}$, $D5_{10}$): A2B3\_oP20\_62\_2c\_3c}} - CIF
\begin{lstlisting}[numbers=none,language={mylang}]
# CIF file 
data_findsym-output
_audit_creation_method FINDSYM

_chemical_name_mineral 'Tongbaite'
_chemical_formula_sum 'C2 Cr3'

loop_
_publ_author_name
 'S. Rundqvist'
 'G. Runnsj{\"o}'
_journal_name_full_name
;
 Acta Chemica Scandinavica
;
_journal_volume 23
_journal_year 1969
_journal_page_first 1191
_journal_page_last 1199
_publ_Section_title
;
 Crystal Structure Refinement of Cr$_{3}$C$_{2}$
;

_aflow_title 'Tongbaite (Cr$_{3}$C$_{2}$, $D5_{10}$) Structure'
_aflow_proto 'A2B3_oP20_62_2c_3c'
_aflow_params 'a,b/a,c/a,x_{1},z_{1},x_{2},z_{2},x_{3},z_{3},x_{4},z_{4},x_{5},z_{5}'
_aflow_params_values '5.5329,0.511305102207,2.07339731425,0.1008,0.2055,0.2432,-0.0464,0.0157,0.4015,0.1808,0.7737,0.8691,-0.0688'
_aflow_Strukturbericht '$D5_{10}$'
_aflow_Pearson 'oP20'

_symmetry_space_group_name_H-M "P 21/n 21/m 21/a"
_symmetry_Int_Tables_number 62
 
_cell_length_a    5.53290
_cell_length_b    2.82900
_cell_length_c    11.47190
_cell_angle_alpha 90.00000
_cell_angle_beta  90.00000
_cell_angle_gamma 90.00000
 
loop_
_space_group_symop_id
_space_group_symop_operation_xyz
1 x,y,z
2 x+1/2,-y+1/2,-z+1/2
3 -x,y+1/2,-z
4 -x+1/2,-y,z+1/2
5 -x,-y,-z
6 -x+1/2,y+1/2,z+1/2
7 x,-y+1/2,z
8 x+1/2,y,-z+1/2
 
loop_
_atom_site_label
_atom_site_type_symbol
_atom_site_symmetry_multiplicity
_atom_site_Wyckoff_label
_atom_site_fract_x
_atom_site_fract_y
_atom_site_fract_z
_atom_site_occupancy
C1  C    4 c 0.10080 0.25000 0.20550 1.00000
C2  C    4 c 0.24320 0.25000 -0.04640 1.00000
Cr1 Cr   4 c 0.01570 0.25000 0.40150 1.00000
Cr2 Cr   4 c 0.18080 0.25000 0.77370 1.00000
Cr3 Cr   4 c 0.86910 0.25000 -0.06880 1.00000
\end{lstlisting}
{\phantomsection\label{A2B3_oP20_62_2c_3c_poscar}}
{\hyperref[A2B3_oP20_62_2c_3c]{Tongbaite (Cr$_{3}$C$_{2}$, $D5_{10}$): A2B3\_oP20\_62\_2c\_3c}} - POSCAR

{\phantomsection\label{A2B4C_oP28_62_ac_2cd_c_cif}}
{\hyperref[A2B4C_oP28_62_ac_2cd_c]{Forsterite (Mg$_{2}$SiO$_{4}$, $S1_{2}$): A2B4C\_oP28\_62\_ac\_2cd\_c}} - CIF
\begin{lstlisting}[numbers=none,language={mylang}]
# CIF file 
data_findsym-output
_audit_creation_method FINDSYM

_chemical_name_mineral 'Forsterite'
_chemical_formula_sum 'Mg2 O4 Si'

loop_
_publ_author_name
 'R. M. Hazen'
_journal_name_full_name
;
 American Mineralogist
;
_journal_volume 61
_journal_year 1976
_journal_page_first 1280
_journal_page_last 1293
_publ_Section_title
;
 Effects of temperature and pressure on the crystal structure of forsterite
;

_aflow_title 'Forsterite (Mg$_{2}$SiO$_{4}$, $S1_{2}$) Structure'
_aflow_proto 'A2B4C_oP28_62_ac_2cd_c'
_aflow_params 'a,b/a,c/a,x_{2},z_{2},x_{3},z_{3},x_{4},z_{4},x_{5},z_{5},x_{6},y_{6},z_{6}'
_aflow_params_values '10.193,0.586382811734,0.466202295693,0.2774,-0.0085,0.0913,0.7657,0.4474,0.2215,0.094,0.4262,0.1628,0.0331,0.2777'
_aflow_Strukturbericht '$S1_{2}$'
_aflow_Pearson 'oP28'

_symmetry_space_group_name_H-M "P 21/n 21/m 21/a"
_symmetry_Int_Tables_number 62
 
_cell_length_a    10.19300
_cell_length_b    5.97700
_cell_length_c    4.75200
_cell_angle_alpha 90.00000
_cell_angle_beta  90.00000
_cell_angle_gamma 90.00000
 
loop_
_space_group_symop_id
_space_group_symop_operation_xyz
1 x,y,z
2 x+1/2,-y+1/2,-z+1/2
3 -x,y+1/2,-z
4 -x+1/2,-y,z+1/2
5 -x,-y,-z
6 -x+1/2,y+1/2,z+1/2
7 x,-y+1/2,z
8 x+1/2,y,-z+1/2
 
loop_
_atom_site_label
_atom_site_type_symbol
_atom_site_symmetry_multiplicity
_atom_site_Wyckoff_label
_atom_site_fract_x
_atom_site_fract_y
_atom_site_fract_z
_atom_site_occupancy
Mg1 Mg   4 a 0.00000 0.00000 0.00000 1.00000
Mg2 Mg   4 c 0.27740 0.25000 -0.00850 1.00000
O1  O    4 c 0.09130 0.25000 0.76570 1.00000
O2  O    4 c 0.44740 0.25000 0.22150 1.00000
Si1 Si   4 c 0.09400 0.25000 0.42620 1.00000
O3  O    8 d 0.16280 0.03310 0.27770 1.00000
\end{lstlisting}
{\phantomsection\label{A2B4C_oP28_62_ac_2cd_c_poscar}}
{\hyperref[A2B4C_oP28_62_ac_2cd_c]{Forsterite (Mg$_{2}$SiO$_{4}$, $S1_{2}$): A2B4C\_oP28\_62\_ac\_2cd\_c}} - POSCAR

{\phantomsection\label{A2B_oP12_62_2c_c_cif}}
{\hyperref[A2B_oP12_62_2c_c]{SrH$_{2}$ ($C29$): A2B\_oP12\_62\_2c\_c}} - CIF
\begin{lstlisting}[numbers=none,language={mylang}]
# CIF file
data_findsym-output
_audit_creation_method FINDSYM

_chemical_name_mineral 'SrH2'
_chemical_formula_sum 'H2 Sr'

loop_
_publ_author_name
 'R. C. Ropp'
_journal_year 2013
_publ_Section_title
;
 Encyclopedia of the Alkaline Earth Compounds
;

_aflow_title 'SrH$_{2}$ ($C29$) Structure'
_aflow_proto 'A2B_oP12_62_2c_c'
_aflow_params 'a,b/a,c/a,x_{1},z_{1},x_{2},z_{2},x_{3},z_{3}'
_aflow_params_values '3.875,1.64232258065,1.89496774194,0.004,0.758,0.24,0.07,0.24,0.39'
_aflow_Strukturbericht '$C29$'
_aflow_Pearson 'oP12'

_symmetry_space_group_name_H-M "P 21/n 21/m 21/a"
_symmetry_Int_Tables_number 62
 
_cell_length_a    3.87500
_cell_length_b    6.36400
_cell_length_c    7.34300
_cell_angle_alpha 90.00000
_cell_angle_beta  90.00000
_cell_angle_gamma 90.00000
 
loop_
_space_group_symop_id
_space_group_symop_operation_xyz
1 x,y,z
2 x+1/2,-y+1/2,-z+1/2
3 -x,y+1/2,-z
4 -x+1/2,-y,z+1/2
5 -x,-y,-z
6 -x+1/2,y+1/2,z+1/2
7 x,-y+1/2,z
8 x+1/2,y,-z+1/2
 
loop_
_atom_site_label
_atom_site_type_symbol
_atom_site_symmetry_multiplicity
_atom_site_Wyckoff_label
_atom_site_fract_x
_atom_site_fract_y
_atom_site_fract_z
_atom_site_occupancy
H1  H    4 c 0.00400 0.25000 0.75800 1.00000
H2  H    4 c 0.24000 0.25000 0.07000 1.00000
Sr1 Sr   4 c 0.24000 0.25000 0.39000 1.00000
\end{lstlisting}
{\phantomsection\label{A2B_oP12_62_2c_c_poscar}}
{\hyperref[A2B_oP12_62_2c_c]{SrH$_{2}$ ($C29$): A2B\_oP12\_62\_2c\_c}} - POSCAR
\begin{lstlisting}[numbers=none,language={mylang}]
A2B_oP12_62_2c_c & a,b/a,c/a,x1,z1,x2,z2,x3,z3 --params=3.875,1.64232258065,1.89496774194,0.004,0.758,0.24,0.07,0.24,0.39 & Pnma D_{2h}^{16} #62 (c^3) & oP12 & $C29$ & SrH2 & SrH2 & R. C. Ropp, (2013)
   1.00000000000000
   3.87500000000000   0.00000000000000   0.00000000000000
   0.00000000000000   6.36400000000000   0.00000000000000
   0.00000000000000   0.00000000000000   7.34300000000000
     H    Sr
     8     4
Direct
   0.00400000000000   0.25000000000000   0.75800000000000    H   (4c)
   0.49600000000000   0.75000000000000   1.25800000000000    H   (4c)
  -0.00400000000000   0.75000000000000  -0.75800000000000    H   (4c)
   0.50400000000000   0.25000000000000  -0.25800000000000    H   (4c)
   0.24000000000000   0.25000000000000   0.07000000000000    H   (4c)
   0.26000000000000   0.75000000000000   0.57000000000000    H   (4c)
  -0.24000000000000   0.75000000000000  -0.07000000000000    H   (4c)
   0.74000000000000   0.25000000000000   0.43000000000000    H   (4c)
   0.24000000000000   0.25000000000000   0.39000000000000   Sr   (4c)
   0.26000000000000   0.75000000000000   0.89000000000000   Sr   (4c)
  -0.24000000000000   0.75000000000000  -0.39000000000000   Sr   (4c)
   0.74000000000000   0.25000000000000   0.11000000000000   Sr   (4c)
\end{lstlisting}
{\phantomsection\label{A3B_oP16_62_cd_c_cif}}
{\hyperref[A3B_oP16_62_cd_c]{$\epsilon$-NiAl$_{3}$ ($D0_{20}$): A3B\_oP16\_62\_cd\_c}} - CIF
\begin{lstlisting}[numbers=none,language={mylang}]
# CIF file
data_findsym-output
_audit_creation_method FINDSYM

_chemical_name_mineral '$\epsilon$-NiAl$_{3}$'
_chemical_formula_sum 'Al3 Ni'

loop_
_publ_author_name
 'A. J. Bradley'
 'A. Taylor'
_journal_name_full_name
;
 Philosophical Magazine
;
_journal_volume 23
_journal_year 1937
_journal_page_first 1049
_journal_page_last 1067
_publ_Section_title
;
 The crystal structures of Ni$_{2}$Al$_{3}$ and NiAl$_{3}$
;

# Found in A Handbook of Lattice Spacings and Structures of Metals and Alloys, 1958

_aflow_title '$\epsilon$-NiAl$_{3}$ ($D0_{20}$) Structure'
_aflow_proto 'A3B_oP16_62_cd_c'
_aflow_params 'a,b/a,c/a,x_{1},z_{1},x_{2},z_{2},x_{3},y_{3},z_{3}'
_aflow_params_values '6.5982,1.11416750023,0.727789397108,0.011,0.415,0.369,0.555,0.174,0.053,0.856'
_aflow_Strukturbericht '$D0_{20}$'
_aflow_Pearson 'oP16'

_symmetry_space_group_name_H-M "P 21/n 21/m 21/a"
_symmetry_Int_Tables_number 62
 
_cell_length_a    6.59820
_cell_length_b    7.35150
_cell_length_c    4.80210
_cell_angle_alpha 90.00000
_cell_angle_beta  90.00000
_cell_angle_gamma 90.00000
 
loop_
_space_group_symop_id
_space_group_symop_operation_xyz
1 x,y,z
2 x+1/2,-y+1/2,-z+1/2
3 -x,y+1/2,-z
4 -x+1/2,-y,z+1/2
5 -x,-y,-z
6 -x+1/2,y+1/2,z+1/2
7 x,-y+1/2,z
8 x+1/2,y,-z+1/2
 
loop_
_atom_site_label
_atom_site_type_symbol
_atom_site_symmetry_multiplicity
_atom_site_Wyckoff_label
_atom_site_fract_x
_atom_site_fract_y
_atom_site_fract_z
_atom_site_occupancy
Al1 Al   4 c 0.01100 0.25000 0.41500 1.00000
Ni1 Ni   4 c 0.36900 0.25000 0.55500 1.00000
Al2 Al   8 d 0.17400 0.05300 0.85600 1.00000
\end{lstlisting}
{\phantomsection\label{A3B_oP16_62_cd_c_poscar}}
{\hyperref[A3B_oP16_62_cd_c]{$\epsilon$-NiAl$_{3}$ ($D0_{20}$): A3B\_oP16\_62\_cd\_c}} - POSCAR
\begin{lstlisting}[numbers=none,language={mylang}]
A3B_oP16_62_cd_c & a,b/a,c/a,x1,z1,x2,z2,x3,y3,z3 --params=6.5982,1.11416750023,0.727789397108,0.011,0.415,0.369,0.555,0.174,0.053,0.856 & Pnma D_{2h}^{16} #62 (c^2d) & oP16 & $D0_{20}$ & NiAl3 & $\epsilon$-NiAl$_{3}$ & A. J. Bradley and A. Taylor, Philos. Mag. 23, 1049-1067 (1937)
   1.00000000000000
   6.59820000000000   0.00000000000000   0.00000000000000
   0.00000000000000   7.35150000000000   0.00000000000000
   0.00000000000000   0.00000000000000   4.80210000000000
    Al    Ni
    12     4
Direct
   0.01100000000000   0.25000000000000   0.41500000000000   Al   (4c)
   0.48900000000000   0.75000000000000   0.91500000000000   Al   (4c)
  -0.01100000000000   0.75000000000000  -0.41500000000000   Al   (4c)
   0.51100000000000   0.25000000000000   0.08500000000000   Al   (4c)
   0.17400000000000   0.05300000000000   0.85600000000000   Al   (8d)
   0.32600000000000  -0.05300000000000   1.35600000000000   Al   (8d)
  -0.17400000000000   0.55300000000000  -0.85600000000000   Al   (8d)
   0.67400000000000   0.44700000000000  -0.35600000000000   Al   (8d)
  -0.17400000000000  -0.05300000000000  -0.85600000000000   Al   (8d)
   0.67400000000000   0.05300000000000  -0.35600000000000   Al   (8d)
   0.17400000000000   0.44700000000000   0.85600000000000   Al   (8d)
   0.32600000000000   0.55300000000000   1.35600000000000   Al   (8d)
   0.36900000000000   0.25000000000000   0.55500000000000   Ni   (4c)
   0.13100000000000   0.75000000000000   1.05500000000000   Ni   (4c)
  -0.36900000000000   0.75000000000000  -0.55500000000000   Ni   (4c)
   0.86900000000000   0.25000000000000  -0.05500000000000   Ni   (4c)
\end{lstlisting}
{\phantomsection\label{AB2C3_oP24_62_c_d_cd_cif}}
{\hyperref[AB2C3_oP24_62_c_d_cd]{Cubanite (CuFe$_{2}$S$_{3}$, $E9_{e}$): AB2C3\_oP24\_62\_c\_d\_cd}} - CIF
\begin{lstlisting}[numbers=none,language={mylang}]
# CIF file
data_findsym-output
_audit_creation_method FINDSYM

_chemical_name_mineral 'Cubanite'
_chemical_formula_sum 'Cu Fe2 S3'

loop_
_publ_author_name
 'T. Szyma{\\'n}ski'
_journal_name_full_name
;
 Zeitschrift f{\"u}r Kristallographie - Crystalline Materials
;
_journal_volume 140
_journal_year 1974
_journal_page_first 218
_journal_page_last 239
_publ_Section_title
;
 A refinement of the structure of cubanite, CuFe$_{2}$S$_{3}$
;

_aflow_title 'Cubanite (CuFe$_{2}$S$_{3}$, $E9_{e}$) Structure'
_aflow_proto 'AB2C3_oP24_62_c_d_cd'
_aflow_params 'a,b/a,c/a,x_{1},z_{1},x_{2},z_{2},x_{3},y_{3},z_{3},x_{4},y_{4},z_{4}'
_aflow_params_values '6.231,1.78414379714,1.03787514043,0.123,0.0823,0.2579,0.4127,0.1366,0.087,0.5853,0.267,0.0846,-0.088'
_aflow_Strukturbericht '$E9_{e}$'
_aflow_Pearson 'oP24'

_symmetry_space_group_name_H-M "P 21/n 21/m 21/a"
_symmetry_Int_Tables_number 62
 
_cell_length_a    6.23100
_cell_length_b    11.11700
_cell_length_c    6.46700
_cell_angle_alpha 90.00000
_cell_angle_beta  90.00000
_cell_angle_gamma 90.00000
 
loop_
_space_group_symop_id
_space_group_symop_operation_xyz
1 x,y,z
2 x+1/2,-y+1/2,-z+1/2
3 -x,y+1/2,-z
4 -x+1/2,-y,z+1/2
5 -x,-y,-z
6 -x+1/2,y+1/2,z+1/2
7 x,-y+1/2,z
8 x+1/2,y,-z+1/2
 
loop_
_atom_site_label
_atom_site_type_symbol
_atom_site_symmetry_multiplicity
_atom_site_Wyckoff_label
_atom_site_fract_x
_atom_site_fract_y
_atom_site_fract_z
_atom_site_occupancy
Cu1 Cu   4 c 0.12300 0.25000 0.08230  1.00000
S1  S    4 c 0.25790 0.25000 0.41270  1.00000
Fe1 Fe   8 d 0.13660 0.08700 0.58530  1.00000
S2  S    8 d 0.26700 0.08460 -0.08800 1.00000
\end{lstlisting}
{\phantomsection\label{AB2C3_oP24_62_c_d_cd_poscar}}
{\hyperref[AB2C3_oP24_62_c_d_cd]{Cubanite (CuFe$_{2}$S$_{3}$, $E9_{e}$): AB2C3\_oP24\_62\_c\_d\_cd}} - POSCAR

{\phantomsection\label{AB3_oP16_62_c_3c_cif}}
{\hyperref[AB3_oP16_62_c_3c]{Molybdite (MoO$_{3}$, $D0_{8}$): AB3\_oP16\_62\_c\_3c}} - CIF
\begin{lstlisting}[numbers=none,language={mylang}]
# CIF file 
data_findsym-output
_audit_creation_method FINDSYM

_chemical_name_mineral 'Molybdite'
_chemical_formula_sum 'Mo O3'

loop_
_publ_author_name
 'H. Sitepu'
 'B. H. {O\'Connor}'
 'D. Li'
_journal_name_full_name
;
 Journal of Applied Crystallography
;
_journal_volume 38
_journal_year 2005
_journal_page_first 158
_journal_page_last 167
_publ_Section_title
;
 Comparative evaluation of the March and generalized spherical harmonic preferred orientation models using X-ray diffraction data for molybdite and calcite powders
;

_aflow_title 'Molybdite (MoO$_{3}$, $D0_{8}$) Structure'
_aflow_proto 'AB3_oP16_62_c_3c'
_aflow_params 'a,b/a,c/a,x_{1},z_{1},x_{2},z_{2},x_{3},z_{3},x_{4},z_{4}'
_aflow_params_values '13.855,0.266791771923,0.28601948755,0.398,0.425,0.074,-0.026,0.414,-0.066,0.276,0.49'
_aflow_Strukturbericht '$D0_{8}$'
_aflow_Pearson 'oP16'

_symmetry_space_group_name_H-M "P n m a"
_symmetry_Int_Tables_number 62
 
_cell_length_a    13.85500
_cell_length_b    3.69640
_cell_length_c    3.96280
_cell_angle_alpha 90.00000
_cell_angle_beta  90.00000
_cell_angle_gamma 90.00000
 
loop_
_space_group_symop_id
_space_group_symop_operation_xyz
1 x,y,z
2 x+1/2,-y+1/2,-z+1/2
3 -x,y+1/2,-z
4 -x+1/2,-y,z+1/2
5 -x,-y,-z
6 -x+1/2,y+1/2,z+1/2
7 x,-y+1/2,z
8 x+1/2,y,-z+1/2
 
loop_
_atom_site_label
_atom_site_type_symbol
_atom_site_symmetry_multiplicity
_atom_site_Wyckoff_label
_atom_site_fract_x
_atom_site_fract_y
_atom_site_fract_z
_atom_site_occupancy
Mo1 Mo   4 c 0.39800  0.25000 0.42500 1.00000
O1  O    4 c 0.07400  0.25000 -0.02600 1.00000
O2  O    4 c 0.41400  0.25000 -0.06600 1.00000
O3  O    4 c 0.27600  0.25000 0.49000 1.00000
\end{lstlisting}
{\phantomsection\label{AB3_oP16_62_c_3c_poscar}}
{\hyperref[AB3_oP16_62_c_3c]{Molybdite (MoO$_{3}$, $D0_{8}$): AB3\_oP16\_62\_c\_3c}} - POSCAR
\begin{lstlisting}[numbers=none,language={mylang}]
AB3_oP16_62_c_3c & a,b/a,c/a,x1,z1,x2,z2,x3,z3,x4,z4 --params=13.855,0.266791771923,0.28601948755,0.398,0.425,0.074,-0.026,0.414,-0.066,0.276,0.49 & Pnma D_{2h}^{16} #62 (c^4) & oP16 & $D0_{8}$ & MoO3 & Molybdite & H. Sitepu and B. H. {O'Connor} and D. Li, J. Appl. Crystallogr. 38, 158-167 (2005)
   1.00000000000000
  13.85500000000000   0.00000000000000   0.00000000000000
   0.00000000000000   3.69640000000000   0.00000000000000
   0.00000000000000   0.00000000000000   3.96280000000000
    Mo     O
     4    12
Direct
   0.39800000000000   0.25000000000000   0.42500000000000   Mo   (4c)
   0.10200000000000   0.75000000000000   0.92500000000000   Mo   (4c)
  -0.39800000000000   0.75000000000000  -0.42500000000000   Mo   (4c)
   0.89800000000000   0.25000000000000   0.07500000000000   Mo   (4c)
   0.07400000000000   0.25000000000000  -0.02600000000000    O   (4c)
   0.42600000000000   0.75000000000000   0.47400000000000    O   (4c)
  -0.07400000000000   0.75000000000000   0.02600000000000    O   (4c)
   0.57400000000000   0.25000000000000   0.52600000000000    O   (4c)
   0.41400000000000   0.25000000000000  -0.06600000000000    O   (4c)
   0.08600000000000   0.75000000000000   0.43400000000000    O   (4c)
  -0.41400000000000   0.75000000000000   0.06600000000000    O   (4c)
   0.91400000000000   0.25000000000000   0.56600000000000    O   (4c)
   0.27600000000000   0.25000000000000   0.49000000000000    O   (4c)
   0.22400000000000   0.75000000000000   0.99000000000000    O   (4c)
  -0.27600000000000   0.75000000000000  -0.49000000000000    O   (4c)
   0.77600000000000   0.25000000000000   0.01000000000000    O   (4c)
\end{lstlisting}
{\phantomsection\label{AB4C_oP24_62_c_2cd_c_cif}}
{\hyperref[AB4C_oP24_62_c_2cd_c]{Barite (BaSO$_{4}$, $H0_{2}$): AB4C\_oP24\_62\_c\_2cd\_c}} - CIF
\begin{lstlisting}[numbers=none,language={mylang}]
# CIF file 
data_findsym-output
_audit_creation_method FINDSYM

_chemical_name_mineral 'Barite'
_chemical_formula_sum 'Ba O4 S'

loop_
_publ_author_name
 'A. A. Colville'
 'K. Staudhammer'
_journal_name_full_name
;
 American Mineralogist
;
_journal_volume 52
_journal_year 1967
_journal_page_first 1877
_journal_page_last 1880
_publ_Section_title
;
 A refinement of the structure of barite
;

# Found in Mineralogy Database, 2012 Found in Mineralogy Database, {Barite},

_aflow_title 'Barite (BaSO$_{4}$, $H0_{2}$) Structure'
_aflow_proto 'AB4C_oP24_62_c_2cd_c'
_aflow_params 'a,b/a,c/a,x_{1},z_{1},x_{2},z_{2},x_{3},z_{3},x_{4},z_{4},x_{5},y_{5},z_{5}'
_aflow_params_values '8.884,0.614362899595,0.805155335434,0.8154,0.3419,0.5878,0.6062,0.3192,0.5515,0.437,0.6914,0.4186,0.4702,0.819'
_aflow_Strukturbericht '$H0_{2}$'
_aflow_Pearson 'oP24'

_symmetry_space_group_name_H-M "P 21/n 21/m 21/a"
_symmetry_Int_Tables_number 62
 
_cell_length_a    8.88400
_cell_length_b    5.45800
_cell_length_c    7.15300
_cell_angle_alpha 90.00000
_cell_angle_beta  90.00000
_cell_angle_gamma 90.00000
 
loop_
_space_group_symop_id
_space_group_symop_operation_xyz
1 x,y,z
2 x+1/2,-y+1/2,-z+1/2
3 -x,y+1/2,-z
4 -x+1/2,-y,z+1/2
5 -x,-y,-z
6 -x+1/2,y+1/2,z+1/2
7 x,-y+1/2,z
8 x+1/2,y,-z+1/2
 
loop_
_atom_site_label
_atom_site_type_symbol
_atom_site_symmetry_multiplicity
_atom_site_Wyckoff_label
_atom_site_fract_x
_atom_site_fract_y
_atom_site_fract_z
_atom_site_occupancy
Ba1 Ba   4 c 0.81540 0.25000 0.34190 1.00000
O1  O    4 c 0.58780 0.25000 0.60620 1.00000
O2  O    4 c 0.31920 0.25000 0.55150 1.00000
S1  S    4 c 0.43700 0.25000 0.69140 1.00000
O3  O    8 d 0.41860 0.47020 0.81900 1.00000
\end{lstlisting}
{\phantomsection\label{AB4C_oP24_62_c_2cd_c_poscar}}
{\hyperref[AB4C_oP24_62_c_2cd_c]{Barite (BaSO$_{4}$, $H0_{2}$): AB4C\_oP24\_62\_c\_2cd\_c}} - POSCAR

{\phantomsection\label{AB_oP8_62_c_c_cif}}
{\hyperref[AB_oP8_62_c_c]{Westerveldite (FeAs, $B14$): AB\_oP8\_62\_c\_c}} - CIF
\begin{lstlisting}[numbers=none,language={mylang}]
# CIF file 
data_findsym-output
_audit_creation_method FINDSYM

_chemical_name_mineral 'Westerveldite'
_chemical_formula_sum 'As Fe'

loop_
_publ_author_name
 'K. Selte'
 'A. Kjekshus'
 'A. F. Andresen'
_journal_name_full_name
;
 Acta Chemica Scandinavica
;
_journal_volume 26
_journal_year 1972
_journal_page_first 3101
_journal_page_last 3113
_publ_Section_title
;
 Magnetic Structure and Properties of FeAs
;

# Found in Interplay between magnetism, structure, and strong electron-phonon coupling in binary FeAs under pressure, 2011

_aflow_title 'Westerveldite (FeAs, $B14$) Structure'
_aflow_proto 'AB_oP8_62_c_c'
_aflow_params 'a,b/a,c/a,x_{1},z_{1},x_{2},z_{2}'
_aflow_params_values '5.454,0.609644297763,1.10542720939,0.2005,0.5741,0.0058,0.1993'
_aflow_Strukturbericht '$B14$'
_aflow_Pearson 'oP8'

_symmetry_space_group_name_H-M "P 21/n 21/m 21/a"
_symmetry_Int_Tables_number 62
 
_cell_length_a    5.45400
_cell_length_b    3.32500
_cell_length_c    6.02900
_cell_angle_alpha 90.00000
_cell_angle_beta  90.00000
_cell_angle_gamma 90.00000
 
loop_
_space_group_symop_id
_space_group_symop_operation_xyz
1 x,y,z
2 x+1/2,-y+1/2,-z+1/2
3 -x,y+1/2,-z
4 -x+1/2,-y,z+1/2
5 -x,-y,-z
6 -x+1/2,y+1/2,z+1/2
7 x,-y+1/2,z
8 x+1/2,y,-z+1/2
 
loop_
_atom_site_label
_atom_site_type_symbol
_atom_site_symmetry_multiplicity
_atom_site_Wyckoff_label
_atom_site_fract_x
_atom_site_fract_y
_atom_site_fract_z
_atom_site_occupancy
As1 As   4 c 0.20050 0.25000 0.57410 1.00000
Fe1 Fe   4 c 0.00580 0.25000 0.19930 1.00000
\end{lstlisting}
{\phantomsection\label{AB_oP8_62_c_c_poscar}}
{\hyperref[AB_oP8_62_c_c]{Westerveldite (FeAs, $B14$): AB\_oP8\_62\_c\_c}} - POSCAR
\begin{lstlisting}[numbers=none,language={mylang}]
AB_oP8_62_c_c & a,b/a,c/a,x1,z1,x2,z2 --params=5.454,0.609644297763,1.10542720939,0.2005,0.5741,0.0058,0.1993 & Pnma D_{2h}^{16} #62 (c^2) & oP8 & $B14$ & FeAs & Westerveldite & K. Selte and A. Kjekshus and A. F. Andresen, Acta Chem. Scand. 26, 3101-3113 (1972)
   1.00000000000000
   5.45400000000000   0.00000000000000   0.00000000000000
   0.00000000000000   3.32500000000000   0.00000000000000
   0.00000000000000   0.00000000000000   6.02900000000000
    As    Fe
     4     4
Direct
   0.20050000000000   0.25000000000000   0.57410000000000   As   (4c)
   0.29950000000000   0.75000000000000   1.07410000000000   As   (4c)
  -0.20050000000000   0.75000000000000  -0.57410000000000   As   (4c)
   0.70050000000000   0.25000000000000  -0.07410000000000   As   (4c)
   0.00580000000000   0.25000000000000   0.19930000000000   Fe   (4c)
   0.49420000000000   0.75000000000000   0.69930000000000   Fe   (4c)
  -0.00580000000000   0.75000000000000  -0.19930000000000   Fe   (4c)
   0.50580000000000   0.25000000000000   0.30070000000000   Fe   (4c)
\end{lstlisting}
{\phantomsection\label{A2BC3_oC24_63_e_c_cg_cif}}
{\hyperref[A2BC3_oC24_63_e_c_cg]{Rasvumite (KFe$_{2}$S$_{3}$): A2BC3\_oC24\_63\_e\_c\_cg}} - CIF
\begin{lstlisting}[numbers=none,language={mylang}]
# CIF file 
data_findsym-output
_audit_creation_method FINDSYM

_chemical_name_mineral 'Rasvumite'
_chemical_formula_sum 'Fe2 K S3'

loop_
_publ_author_name
 'J. R. Clark'
 'G. E. {Brown, Jr.}'
_journal_name_full_name
;
 American Mineralogist
;
_journal_volume 65
_journal_year 1980
_journal_page_first 477
_journal_page_last 482
_publ_Section_title
;
 Crystal structure of rasvumite, KFe$_{2}$S$_{3}$
;

# Found in The American Mineralogist Crystal Structure Database, 2003

_aflow_title 'Rasvumite (KFe$_{2}$S$_{3}$) Structure'
_aflow_proto 'A2BC3_oC24_63_e_c_cg'
_aflow_params 'a,b/a,c/a,y_{1},y_{2},x_{3},x_{4},y_{4}'
_aflow_params_values '9.049,1.21770361366,0.600176815118,0.6699,0.1191,0.8502,0.2174,0.3859'
_aflow_Strukturbericht 'None'
_aflow_Pearson 'oC24'

_symmetry_space_group_name_H-M "C 2/m 2/c 21/m"
_symmetry_Int_Tables_number 63
 
_cell_length_a    9.04900
_cell_length_b    11.01900
_cell_length_c    5.43100
_cell_angle_alpha 90.00000
_cell_angle_beta  90.00000
_cell_angle_gamma 90.00000
 
loop_
_space_group_symop_id
_space_group_symop_operation_xyz
1 x,y,z
2 x,-y,-z
3 -x,y,-z+1/2
4 -x,-y,z+1/2
5 -x,-y,-z
6 -x,y,z
7 x,-y,z+1/2
8 x,y,-z+1/2
9 x+1/2,y+1/2,z
10 x+1/2,-y+1/2,-z
11 -x+1/2,y+1/2,-z+1/2
12 -x+1/2,-y+1/2,z+1/2
13 -x+1/2,-y+1/2,-z
14 -x+1/2,y+1/2,z
15 x+1/2,-y+1/2,z+1/2
16 x+1/2,y+1/2,-z+1/2
 
loop_
_atom_site_label
_atom_site_type_symbol
_atom_site_symmetry_multiplicity
_atom_site_Wyckoff_label
_atom_site_fract_x
_atom_site_fract_y
_atom_site_fract_z
_atom_site_occupancy
K1  K    4 c 0.00000 0.66990 0.25000 1.00000
S1  S    4 c 0.00000 0.11910 0.25000 1.00000
Fe1 Fe   8 e 0.85020 0.00000 0.00000 1.00000
S2  S    8 g 0.21740 0.38590 0.25000 1.00000
\end{lstlisting}
{\phantomsection\label{A2BC3_oC24_63_e_c_cg_poscar}}
{\hyperref[A2BC3_oC24_63_e_c_cg]{Rasvumite (KFe$_{2}$S$_{3}$): A2BC3\_oC24\_63\_e\_c\_cg}} - POSCAR
\begin{lstlisting}[numbers=none,language={mylang}]
A2BC3_oC24_63_e_c_cg & a,b/a,c/a,y1,y2,x3,x4,y4 --params=9.049,1.21770361366,0.600176815118,0.6699,0.1191,0.8502,0.2174,0.3859 & Cmcm D_{2h}^{17} #63 (c^2eg) & oC24 & None & KFe2S3 & Rasvumite & J. R. Clark and G. E. {Brown, Jr.}, Am. Mineral. 65, 477-482 (1980)
   1.00000000000000
   4.52450000000000  -5.50950000000000   0.00000000000000
   4.52450000000000   5.50950000000000   0.00000000000000
   0.00000000000000   0.00000000000000   5.43100000000000
    Fe     K     S
     4     2     6
Direct
   0.85020000000000   0.85020000000000   0.00000000000000   Fe   (8e)
  -0.85020000000000  -0.85020000000000   0.50000000000000   Fe   (8e)
  -0.85020000000000  -0.85020000000000   0.00000000000000   Fe   (8e)
   0.85020000000000   0.85020000000000   0.50000000000000   Fe   (8e)
  -0.66990000000000   0.66990000000000   0.25000000000000    K   (4c)
   0.66990000000000  -0.66990000000000   0.75000000000000    K   (4c)
  -0.11910000000000   0.11910000000000   0.25000000000000    S   (4c)
   0.11910000000000  -0.11910000000000   0.75000000000000    S   (4c)
  -0.16850000000000   0.60330000000000   0.25000000000000    S   (8g)
   0.16850000000000  -0.60330000000000   0.75000000000000    S   (8g)
  -0.60330000000000   0.16850000000000   0.25000000000000    S   (8g)
   0.60330000000000  -0.16850000000000   0.75000000000000    S   (8g)
\end{lstlisting}
{\phantomsection\label{A43B5C17_oC260_63_c8fg6h_cfg_ce3f2h_cif}}
{\hyperref[A43B5C17_oC260_63_c8fg6h_cfg_ce3f2h]{La$_{43}$Ni$_{17}$Mg$_{5}$: A43B5C17\_oC260\_63\_c8fg6h\_cfg\_ce3f2h}} - CIF

{\phantomsection\label{A43B5C17_oC260_63_c8fg6h_cfg_ce3f2h_poscar}}
{\hyperref[A43B5C17_oC260_63_c8fg6h_cfg_ce3f2h]{La$_{43}$Ni$_{17}$Mg$_{5}$: A43B5C17\_oC260\_63\_c8fg6h\_cfg\_ce3f2h}} - POSCAR

{\phantomsection\label{A6B_oC28_63_efg_c_cif}}
{\hyperref[A6B_oC28_63_efg_c]{MnAl$_{6}$ ($D2_{h}$): A6B\_oC28\_63\_efg\_c}} - CIF
\begin{lstlisting}[numbers=none,language={mylang}]
# CIF file 
data_findsym-output
_audit_creation_method FINDSYM

_chemical_name_mineral 'MnAl6'
_chemical_formula_sum 'Al6 Mn'

loop_
_publ_author_name
 'A. Kontio'
 'P. Coppens'
_journal_name_full_name
;
 Acta Crystallographica Section B: Structural Science
;
_journal_volume 37
_journal_year 1981
_journal_page_first 433
_journal_page_last 435
_publ_Section_title
;
 New study of the structure of MnAl$_{6}$
;

_aflow_title 'MnAl$_{6}$ ($D2_{h}$) Structure'
_aflow_proto 'A6B_oC28_63_efg_c'
_aflow_params 'a,b/a,c/a,y_{1},x_{2},y_{3},z_{3},x_{4},y_{4}'
_aflow_params_values '7.5551,0.860266574896,1.17435904224,0.45686,0.32602,0.13917,0.10039,0.31768,0.28622'
_aflow_Strukturbericht '$D2_{h}$'
_aflow_Pearson 'oC28'

_symmetry_space_group_name_H-M "C 2/m 2/c 21/m"
_symmetry_Int_Tables_number 63
 
_cell_length_a    7.55510
_cell_length_b    6.49940
_cell_length_c    8.87240
_cell_angle_alpha 90.00000
_cell_angle_beta  90.00000
_cell_angle_gamma 90.00000
 
loop_
_space_group_symop_id
_space_group_symop_operation_xyz
1 x,y,z
2 x,-y,-z
3 -x,y,-z+1/2
4 -x,-y,z+1/2
5 -x,-y,-z
6 -x,y,z
7 x,-y,z+1/2
8 x,y,-z+1/2
9 x+1/2,y+1/2,z
10 x+1/2,-y+1/2,-z
11 -x+1/2,y+1/2,-z+1/2
12 -x+1/2,-y+1/2,z+1/2
13 -x+1/2,-y+1/2,-z
14 -x+1/2,y+1/2,z
15 x+1/2,-y+1/2,z+1/2
16 x+1/2,y+1/2,-z+1/2
 
loop_
_atom_site_label
_atom_site_type_symbol
_atom_site_symmetry_multiplicity
_atom_site_Wyckoff_label
_atom_site_fract_x
_atom_site_fract_y
_atom_site_fract_z
_atom_site_occupancy
Mn1 Mn   4 c 0.00000 0.45686 0.25000 1.00000
Al1 Al   8 e 0.32602 0.00000 0.00000 1.00000
Al2 Al   8 f 0.00000 0.13917 0.10039 1.00000
Al3 Al   8 g 0.31768 0.28622 0.25000 1.00000
\end{lstlisting}
{\phantomsection\label{A6B_oC28_63_efg_c_poscar}}
{\hyperref[A6B_oC28_63_efg_c]{MnAl$_{6}$ ($D2_{h}$): A6B\_oC28\_63\_efg\_c}} - POSCAR
\begin{lstlisting}[numbers=none,language={mylang}]
A6B_oC28_63_efg_c & a,b/a,c/a,y1,x2,y3,z3,x4,y4 --params=7.5551,0.860266574896,1.17435904224,0.45686,0.32602,0.13917,0.10039,0.31768,0.28622 & Cmcm D_{2h}^{17} #63 (cefg) & oC28 & $D2_{h}$ & MnAl6 & MnAl6 & A. Kontio and P. Coppens, Acta Crystallogr. Sect. B Struct. Sci. 37, 433-435 (1981)
   1.00000000000000
   3.77755000000000  -3.24970000000000   0.00000000000000
   3.77755000000000   3.24970000000000   0.00000000000000
   0.00000000000000   0.00000000000000   8.87240000000000
    Al    Mn
    12     2
Direct
   0.32602000000000   0.32602000000000   0.00000000000000   Al   (8e)
  -0.32602000000000  -0.32602000000000   0.50000000000000   Al   (8e)
  -0.32602000000000  -0.32602000000000   0.00000000000000   Al   (8e)
   0.32602000000000   0.32602000000000   0.50000000000000   Al   (8e)
  -0.13917000000000   0.13917000000000   0.10039000000000   Al   (8f)
   0.13917000000000  -0.13917000000000   0.60039000000000   Al   (8f)
  -0.13917000000000   0.13917000000000   0.39961000000000   Al   (8f)
   0.13917000000000  -0.13917000000000  -0.10039000000000   Al   (8f)
   0.03146000000000   0.60390000000000   0.25000000000000   Al   (8g)
  -0.03146000000000  -0.60390000000000   0.75000000000000   Al   (8g)
  -0.60390000000000  -0.03146000000000   0.25000000000000   Al   (8g)
   0.60390000000000   0.03146000000000   0.75000000000000   Al   (8g)
  -0.45686000000000   0.45686000000000   0.25000000000000   Mn   (4c)
   0.45686000000000  -0.45686000000000   0.75000000000000   Mn   (4c)
\end{lstlisting}
{\phantomsection\label{AB3C_oC20_63_a_cf_c_cif}}
{\hyperref[AB3C_oC20_63_a_cf_c]{Post-perovskite (MgSiO$_{3}$): AB3C\_oC20\_63\_a\_cf\_c}} - CIF
\begin{lstlisting}[numbers=none,language={mylang}]
# CIF file 
data_findsym-output
_audit_creation_method FINDSYM

_chemical_name_mineral 'Post-perovskite MgSiO$_{3}$'
_chemical_formula_sum 'Mg O3 Si'

loop_
_publ_author_name
 'M. Murakami'
 'K. Hirose'
 'K. Kawamura'
 'N. Sata'
 'Y. Ohishi'
_journal_name_full_name
;
 Science
;
_journal_volume 304
_journal_year 2004
_journal_page_first 855
_journal_page_last 858
_publ_Section_title
;
 Post-Perovskite Phase Transition in MgSiO$_{3}$
;

_aflow_title 'Post-perovskite (MgSiO$_{3}$) Structure'
_aflow_proto 'AB3C_oC20_63_a_cf_c'
_aflow_params 'a,b/a,c/a,y_{2},y_{3},y_{4},z_{4}'
_aflow_params_values '2.456,3.27442996743,2.48086319218,0.077,0.747,0.631,-0.064'
_aflow_Strukturbericht 'None'
_aflow_Pearson 'oC20'

_symmetry_space_group_name_H-M "C m c m"
_symmetry_Int_Tables_number 63
 
_cell_length_a    2.45600
_cell_length_b    8.04200
_cell_length_c    6.09300
_cell_angle_alpha 90.00000
_cell_angle_beta  90.00000
_cell_angle_gamma 90.00000
 
loop_
_space_group_symop_id
_space_group_symop_operation_xyz
1 x,y,z
2 x,-y,-z
3 -x,y,-z+1/2
4 -x,-y,z+1/2
5 -x,-y,-z
6 -x,y,z
7 x,-y,z+1/2
8 x,y,-z+1/2
9 x+1/2,y+1/2,z
10 x+1/2,-y+1/2,-z
11 -x+1/2,y+1/2,-z+1/2
12 -x+1/2,-y+1/2,z+1/2
13 -x+1/2,-y+1/2,-z
14 -x+1/2,y+1/2,z
15 x+1/2,-y+1/2,z+1/2
16 x+1/2,y+1/2,-z+1/2
 
loop_
_atom_site_label
_atom_site_type_symbol
_atom_site_symmetry_multiplicity
_atom_site_Wyckoff_label
_atom_site_fract_x
_atom_site_fract_y
_atom_site_fract_z
_atom_site_occupancy
Mg1 Mg   4 a 0.00000 0.00000 0.00000  1.00000
O1  O    4 c 0.00000 0.07700 0.25000  1.00000
Si1 Si   4 c 0.00000 0.74700 0.25000  1.00000
O2  O    8 f 0.00000 0.63100 -0.06400 1.00000
\end{lstlisting}
{\phantomsection\label{AB3C_oC20_63_a_cf_c_poscar}}
{\hyperref[AB3C_oC20_63_a_cf_c]{Post-perovskite (MgSiO$_{3}$): AB3C\_oC20\_63\_a\_cf\_c}} - POSCAR
\begin{lstlisting}[numbers=none,language={mylang}]
AB3C_oC20_63_a_cf_c & a,b/a,c/a,y2,y3,y4,z4 --params=2.456,3.27442996743,2.48086319218,0.077,0.747,0.631,-0.064 & Cmcm D_{2h}^{17} #63 (ac^2f) & oC20 & None & MgSiO3 & Post-perovskite MgSiO$_{3}$ & M. Murakami et al., Science 304, 855-858 (2004)
   1.00000000000000
   1.22800000000000  -4.02100000000000   0.00000000000000
   1.22800000000000   4.02100000000000   0.00000000000000
   0.00000000000000   0.00000000000000   6.09300000000000
    Mg     O    Si
     2     6     2
Direct
   0.00000000000000   0.00000000000000   0.00000000000000   Mg   (4a)
   0.00000000000000   0.00000000000000   0.50000000000000   Mg   (4a)
  -0.07700000000000   0.07700000000000   0.25000000000000    O   (4c)
   0.07700000000000  -0.07700000000000   0.75000000000000    O   (4c)
  -0.63100000000000   0.63100000000000  -0.06400000000000    O   (8f)
   0.63100000000000  -0.63100000000000   0.43600000000000    O   (8f)
  -0.63100000000000   0.63100000000000   0.56400000000000    O   (8f)
   0.63100000000000  -0.63100000000000   0.06400000000000    O   (8f)
  -0.74700000000000   0.74700000000000   0.25000000000000   Si   (4c)
   0.74700000000000  -0.74700000000000   0.75000000000000   Si   (4c)
\end{lstlisting}
{\phantomsection\label{AB4C_oC24_63_a_fg_c_cif}}
{\hyperref[AB4C_oC24_63_a_fg_c]{MgSO$_{4}$: AB4C\_oC24\_63\_a\_fg\_c}} - CIF
\begin{lstlisting}[numbers=none,language={mylang}]
# CIF file 
data_findsym-output
_audit_creation_method FINDSYM

_chemical_name_mineral ''
_chemical_formula_sum 'Mg O4 S'

loop_
_publ_author_name
 'P. J. Rentzeperis'
 'C. T. Soldatos'
_journal_name_full_name
;
 Acta Cristallographica
;
_journal_volume 11
_journal_year 1958
_journal_page_first 686
_journal_page_last 688
_publ_Section_title
;
 The crystal structure of the anhydrous magnesium sulphate
;

# Found in The American Mineralogist Crystal Structure Database, 2003

_aflow_title 'MgSO$_{4}$ Structure'
_aflow_proto 'AB4C_oC24_63_a_fg_c'
_aflow_params 'a,b/a,c/a,y_{2},y_{3},z_{3},x_{4},y_{4}'
_aflow_params_values '5.182,1.52315708221,1.25549980702,0.37,0.25,0.06,0.25,0.47'
_aflow_Strukturbericht 'None'
_aflow_Pearson 'oC24'

_symmetry_space_group_name_H-M "C 2/m 2/c 21/m"
_symmetry_Int_Tables_number 63
 
_cell_length_a    5.18200
_cell_length_b    7.89300
_cell_length_c    6.50600
_cell_angle_alpha 90.00000
_cell_angle_beta  90.00000
_cell_angle_gamma 90.00000
 
loop_
_space_group_symop_id
_space_group_symop_operation_xyz
1 x,y,z
2 x,-y,-z
3 -x,y,-z+1/2
4 -x,-y,z+1/2
5 -x,-y,-z
6 -x,y,z
7 x,-y,z+1/2
8 x,y,-z+1/2
9 x+1/2,y+1/2,z
10 x+1/2,-y+1/2,-z
11 -x+1/2,y+1/2,-z+1/2
12 -x+1/2,-y+1/2,z+1/2
13 -x+1/2,-y+1/2,-z
14 -x+1/2,y+1/2,z
15 x+1/2,-y+1/2,z+1/2
16 x+1/2,y+1/2,-z+1/2
 
loop_
_atom_site_label
_atom_site_type_symbol
_atom_site_symmetry_multiplicity
_atom_site_Wyckoff_label
_atom_site_fract_x
_atom_site_fract_y
_atom_site_fract_z
_atom_site_occupancy
Mg1 Mg   4 a 0.00000 0.00000 0.00000 1.00000
S1  S    4 c 0.00000 0.37000 0.25000 1.00000
O1  O    8 f 0.00000 0.25000 0.06000 1.00000
O2  O    8 g 0.25000 0.47000 0.25000 1.00000
\end{lstlisting}
{\phantomsection\label{AB4C_oC24_63_a_fg_c_poscar}}
{\hyperref[AB4C_oC24_63_a_fg_c]{MgSO$_{4}$: AB4C\_oC24\_63\_a\_fg\_c}} - POSCAR
\begin{lstlisting}[numbers=none,language={mylang}]
AB4C_oC24_63_a_fg_c & a,b/a,c/a,y2,y3,z3,x4,y4 --params=5.182,1.52315708221,1.25549980702,0.37,0.25,0.06,0.25,0.47 & Cmcm D_{2h}^{17} #63 (acfg) & oC24 & None & MgO4S &  & P. J. Rentzeperis and C. T. Soldatos, Acta Cryst. 11, 686-688 (1958)
   1.00000000000000
   2.59100000000000  -3.94650000000000   0.00000000000000
   2.59100000000000   3.94650000000000   0.00000000000000
   0.00000000000000   0.00000000000000   6.50600000000000
    Mg     O     S
     2     8     2
Direct
   0.00000000000000   0.00000000000000   0.00000000000000   Mg   (4a)
   0.00000000000000   0.00000000000000   0.50000000000000   Mg   (4a)
  -0.25000000000000   0.25000000000000   0.06000000000000    O   (8f)
   0.25000000000000  -0.25000000000000   0.56000000000000    O   (8f)
  -0.25000000000000   0.25000000000000   0.44000000000000    O   (8f)
   0.25000000000000  -0.25000000000000  -0.06000000000000    O   (8f)
  -0.22000000000000   0.72000000000000   0.25000000000000    O   (8g)
   0.22000000000000  -0.72000000000000   0.75000000000000    O   (8g)
  -0.72000000000000   0.22000000000000   0.25000000000000    O   (8g)
   0.72000000000000  -0.22000000000000   0.75000000000000    O   (8g)
  -0.37000000000000   0.37000000000000   0.25000000000000    S   (4c)
   0.37000000000000  -0.37000000000000   0.75000000000000    S   (4c)
\end{lstlisting}
{\phantomsection\label{AB4C_oC24_63_c_fg_c_cif}}
{\hyperref[AB4C_oC24_63_c_fg_c]{Anhydrite (CaSO$_{4}$, $H0_{1}$): AB4C\_oC24\_63\_c\_fg\_c}} - CIF
\begin{lstlisting}[numbers=none,language={mylang}]
# CIF file 
data_findsym-output
_audit_creation_method FINDSYM

_chemical_name_mineral 'Anhydrite'
_chemical_formula_sum 'Ca O4 S'

loop_
_publ_author_name
 'F. C. Hawthorne'
 'R. B. Ferguson'
_journal_name_full_name
;
 Canadian Mineralogist
;
_journal_volume 13
_journal_year 1975
_journal_page_first 289
_journal_page_last 292
_publ_Section_title
;
 Anhydrous sulphates. II. Refinement of the crystal structure of anhydrite
;

_aflow_title 'Anhydrite (CaSO$_{4}$, $H0_{1}$) Structure'
_aflow_proto 'AB4C_oC24_63_c_fg_c'
_aflow_params 'a,b/a,c/a,y_{1},y_{2},y_{3},z_{3},x_{4},y_{4}'
_aflow_params_values '6.995,0.892780557541,0.999714081487,0.6524,0.15556,0.7025,-0.0819,0.1699,0.0162'
_aflow_Strukturbericht '$H0_{1}$'
_aflow_Pearson 'oC24'

_symmetry_space_group_name_H-M "C 2/m 2/c 21/m"
_symmetry_Int_Tables_number 63
 
_cell_length_a    6.99500
_cell_length_b    6.24500
_cell_length_c    6.99300
_cell_angle_alpha 90.00000
_cell_angle_beta  90.00000
_cell_angle_gamma 90.00000
 
loop_
_space_group_symop_id
_space_group_symop_operation_xyz
1 x,y,z
2 x,-y,-z
3 -x,y,-z+1/2
4 -x,-y,z+1/2
5 -x,-y,-z
6 -x,y,z
7 x,-y,z+1/2
8 x,y,-z+1/2
9 x+1/2,y+1/2,z
10 x+1/2,-y+1/2,-z
11 -x+1/2,y+1/2,-z+1/2
12 -x+1/2,-y+1/2,z+1/2
13 -x+1/2,-y+1/2,-z
14 -x+1/2,y+1/2,z
15 x+1/2,-y+1/2,z+1/2
16 x+1/2,y+1/2,-z+1/2
 
loop_
_atom_site_label
_atom_site_type_symbol
_atom_site_symmetry_multiplicity
_atom_site_Wyckoff_label
_atom_site_fract_x
_atom_site_fract_y
_atom_site_fract_z
_atom_site_occupancy
Ca1 Ca   4 c 0.00000 0.65240 0.25000 1.00000
S1  S    4 c 0.00000 0.15556 0.25000 1.00000
O1  O    8 f 0.00000 0.70250 -0.08190 1.00000
O2  O    8 g 0.16990 0.01620 0.25000 1.00000
\end{lstlisting}
{\phantomsection\label{AB4C_oC24_63_c_fg_c_poscar}}
{\hyperref[AB4C_oC24_63_c_fg_c]{Anhydrite (CaSO$_{4}$, $H0_{1}$): AB4C\_oC24\_63\_c\_fg\_c}} - POSCAR
\begin{lstlisting}[numbers=none,language={mylang}]
AB4C_oC24_63_c_fg_c & a,b/a,c/a,y1,y2,y3,z3,x4,y4 --params=6.995,0.892780557541,0.999714081487,0.6524,0.15556,0.7025,-0.0819,0.1699,0.0162 & Cmcm D_{2h}^{17} #63 (c^2fg) & oC24 & $H0_{1}$ & CaSO4 & Anhydrite & F. C. Hawthorne and R. B. Ferguson, Can. Mineral. 13, 289-292 (1975)
   1.00000000000000
   3.49750000000000  -3.12250000000000   0.00000000000000
   3.49750000000000   3.12250000000000   0.00000000000000
   0.00000000000000   0.00000000000000   6.99300000000000
    Ca     O     S
     2     8     2
Direct
  -0.65240000000000   0.65240000000000   0.25000000000000   Ca   (4c)
   0.65240000000000  -0.65240000000000   0.75000000000000   Ca   (4c)
  -0.70250000000000   0.70250000000000  -0.08190000000000    O   (8f)
   0.70250000000000  -0.70250000000000   0.41810000000000    O   (8f)
  -0.70250000000000   0.70250000000000   0.58190000000000    O   (8f)
   0.70250000000000  -0.70250000000000   0.08190000000000    O   (8f)
   0.15370000000000   0.18610000000000   0.25000000000000    O   (8g)
  -0.15370000000000  -0.18610000000000   0.75000000000000    O   (8g)
  -0.18610000000000  -0.15370000000000   0.25000000000000    O   (8g)
   0.18610000000000   0.15370000000000   0.75000000000000    O   (8g)
  -0.15556000000000   0.15556000000000   0.25000000000000    S   (4c)
   0.15556000000000  -0.15556000000000   0.75000000000000    S   (4c)
\end{lstlisting}
{\phantomsection\label{A2B_oC24_64_2f_f_cif}}
{\hyperref[A2B_oC24_64_2f_f]{H$_{2}$S (170~GPa): A2B\_oC24\_64\_2f\_f}} - CIF
\begin{lstlisting}[numbers=none,language={mylang}]
# CIF file 
data_findsym-output
_audit_creation_method FINDSYM

_chemical_name_mineral 'H2S'
_chemical_formula_sum 'H2 S'

loop_
_publ_author_name
 'Y. Li'
 'J. Hao'
 'H. Liu'
 'Y. Li'
 'Y. Ma'
_journal_name_full_name
;
 Journal of Chemical Physics
;
_journal_volume 140
_journal_year 2014
_journal_page_first 174712
_journal_page_last 174712
_publ_Section_title
;
 The metallization and superconductivity of dense hydrogen sulfide
;

_aflow_title 'H$_{2}$S (170~GPa) Structure'
_aflow_proto 'A2B_oC24_64_2f_f'
_aflow_params 'a,b/a,c/a,y_{1},z_{1},y_{2},z_{2},y_{3},z_{3}'
_aflow_params_values '2.9986,1.44314013206,2.534682852,0.372,0.27,0.6,0.092,0.371,0.615'
_aflow_Strukturbericht 'None'
_aflow_Pearson 'oC24'

_symmetry_space_group_name_H-M "C 2/m 2/c 21/a"
_symmetry_Int_Tables_number 64
 
_cell_length_a    2.99860
_cell_length_b    4.32740
_cell_length_c    7.60050
_cell_angle_alpha 90.00000
_cell_angle_beta  90.00000
_cell_angle_gamma 90.00000
 
loop_
_space_group_symop_id
_space_group_symop_operation_xyz
1 x,y,z
2 x,-y,-z
3 -x+1/2,y,-z+1/2
4 -x+1/2,-y,z+1/2
5 -x,-y,-z
6 -x,y,z
7 x+1/2,-y,z+1/2
8 x+1/2,y,-z+1/2
9 x+1/2,y+1/2,z
10 x+1/2,-y+1/2,-z
11 -x,y+1/2,-z+1/2
12 -x,-y+1/2,z+1/2
13 -x+1/2,-y+1/2,-z
14 -x+1/2,y+1/2,z
15 x,-y+1/2,z+1/2
16 x,y+1/2,-z+1/2
 
loop_
_atom_site_label
_atom_site_type_symbol
_atom_site_symmetry_multiplicity
_atom_site_Wyckoff_label
_atom_site_fract_x
_atom_site_fract_y
_atom_site_fract_z
_atom_site_occupancy
H1 H   8 f 0.00000 0.37200 0.27000 1.00000
H2 H   8 f 0.00000 0.60000 0.09200 1.00000
S1 S   8 f 0.00000 0.37100 0.61500 1.00000
\end{lstlisting}
{\phantomsection\label{A2B_oC24_64_2f_f_poscar}}
{\hyperref[A2B_oC24_64_2f_f]{H$_{2}$S (170~GPa): A2B\_oC24\_64\_2f\_f}} - POSCAR
\begin{lstlisting}[numbers=none,language={mylang}]
A2B_oC24_64_2f_f & a,b/a,c/a,y1,z1,y2,z2,y3,z3 --params=2.9986,1.44314013206,2.534682852,0.372,0.27,0.6,0.092,0.371,0.615 & Cmca D_{2h}^{18} #64 (f^3) & oC24 & None & H2S & H2S & Y. Li et al., J. Chem. Phys. 140, 174712(2014)
   1.00000000000000
   1.49930000000000  -2.16370000000000   0.00000000000000
   1.49930000000000   2.16370000000000   0.00000000000000
   0.00000000000000   0.00000000000000   7.60050000000000
     H     S
     8     4
Direct
  -0.37200000000000   0.37200000000000   0.27000000000000    H   (8f)
   0.87200000000000   0.12800000000000   0.77000000000000    H   (8f)
   0.12800000000000   0.87200000000000   0.23000000000000    H   (8f)
   0.37200000000000  -0.37200000000000  -0.27000000000000    H   (8f)
  -0.60000000000000   0.60000000000000   0.09200000000000    H   (8f)
   1.10000000000000  -0.10000000000000   0.59200000000000    H   (8f)
  -0.10000000000000   1.10000000000000   0.40800000000000    H   (8f)
   0.60000000000000  -0.60000000000000  -0.09200000000000    H   (8f)
  -0.37100000000000   0.37100000000000   0.61500000000000    S   (8f)
   0.87100000000000   0.12900000000000   1.11500000000000    S   (8f)
   0.12900000000000   0.87100000000000  -0.11500000000000    S   (8f)
   0.37100000000000  -0.37100000000000  -0.61500000000000    S   (8f)
\end{lstlisting}
{\phantomsection\label{A2B4C_oC28_66_l_kl_a_cif}}
{\hyperref[A2B4C_oC28_66_l_kl_a]{SrAl$_{2}$Se$_{4}$: A2B4C\_oC28\_66\_l\_kl\_a}} - CIF
\begin{lstlisting}[numbers=none,language={mylang}]
# CIF file
data_findsym-output
_audit_creation_method FINDSYM

_chemical_name_mineral 'SrAl2Se4'
_chemical_formula_sum 'Al2 Se4 Sr'

loop_
_publ_author_name
 'W. Klee'
 'H. Sch{\"a}fer'
_journal_name_full_name
;
 Zeitschrift f{\"u}r Naturforschung B
;
_journal_volume 33
_journal_year 1978
_journal_page_first 829
_journal_page_last 833
_publ_Section_title
;
 CaAl$_{2}$Se$_{4}$ und SrAl$_{2}$Se$_{4}$-Strukturvarianten des TlSe-Typs / CaAl$_{2}$Se$_{4}$ and SrAl$_{2}$Se$_{4}$-Variants of the TlSe-Structure
;

# Found in Pearson's Crystal Data - Crystal Structure Database for Inorganic Compounds, 2013

_aflow_title 'SrAl$_{2}$Se$_{4}$ Structure'
_aflow_proto 'A2B4C_oC28_66_l_kl_a'
_aflow_params 'a,b/a,c/a,z_{2},x_{3},y_{3},x_{4},y_{4}'
_aflow_params_values '6.2700590867,1.72567783094,1.73046251993,0.833,0.005,0.268,0.737,0.42'
_aflow_Strukturbericht 'None'
_aflow_Pearson 'oC28'

_cell_length_a    6.2700590867
_cell_length_b    10.8201019646
_cell_length_c    10.8501022473
_cell_angle_alpha 90.0000000000
_cell_angle_beta  90.0000000000
_cell_angle_gamma 90.0000000000
 
_symmetry_space_group_name_H-M "C 2/c 2/c 2/m"
_symmetry_Int_Tables_number 66
 
loop_
_space_group_symop_id
_space_group_symop_operation_xyz
1 x,y,z
2 x,-y,-z+1/2
3 -x,y,-z+1/2
4 -x,-y,z
5 -x,-y,-z
6 -x,y,z+1/2
7 x,-y,z+1/2
8 x,y,-z
9 x+1/2,y+1/2,z
10 x+1/2,-y+1/2,-z+1/2
11 -x+1/2,y+1/2,-z+1/2
12 -x+1/2,-y+1/2,z
13 -x+1/2,-y+1/2,-z
14 -x+1/2,y+1/2,z+1/2
15 x+1/2,-y+1/2,z+1/2
16 x+1/2,y+1/2,-z
 
loop_
_atom_site_label
_atom_site_type_symbol
_atom_site_symmetry_multiplicity
_atom_site_Wyckoff_label
_atom_site_fract_x
_atom_site_fract_y
_atom_site_fract_z
_atom_site_occupancy
Sr1 Sr   4 a 0.00000 0.00000 0.25000 1.00000
Se1 Se   8 k 0.25000 0.25000 0.83300 1.00000
Al1 Al   8 l 0.00500 0.26800 0.00000 1.00000
Se2 Se   8 l 0.73700 0.42000 0.00000 1.00000
\end{lstlisting}
{\phantomsection\label{A2B4C_oC28_66_l_kl_a_poscar}}
{\hyperref[A2B4C_oC28_66_l_kl_a]{SrAl$_{2}$Se$_{4}$: A2B4C\_oC28\_66\_l\_kl\_a}} - POSCAR
\begin{lstlisting}[numbers=none,language={mylang}]
A2B4C_oC28_66_l_kl_a & a,b/a,c/a,z2,x3,y3,x4,y4 --params=6.2700590867,1.72567783094,1.73046251993,0.833,0.005,0.268,0.737,0.42 & Cccm D_{2h}^{20} #66 (akl^2) & oC28 & None & SrAl2Se4 &  & W. Klee and H. Sch{\"a}fer, Z. Naturforsch. B 33, 829-833 (1978)
   1.00000000000000
   3.13502954335000  -5.41005098230000   0.00000000000000
   3.13502954335000   5.41005098230000   0.00000000000000
   0.00000000000000   0.00000000000000  10.85010224730000
    Al    Se    Sr
     4     8     2
Direct
  -0.26300000000000   0.27300000000000   0.00000000000000   Al   (8l)
   0.26300000000000  -0.27300000000000   0.00000000000000   Al   (8l)
  -0.27300000000000   0.26300000000000   0.50000000000000   Al   (8l)
   0.27300000000000  -0.26300000000000   0.50000000000000   Al   (8l)
   0.00000000000000   0.50000000000000   0.83300000000000   Se   (8k)
   0.50000000000000   0.00000000000000  -0.33300000000000   Se   (8k)
   0.00000000000000   0.50000000000000  -0.83300000000000   Se   (8k)
   0.50000000000000   0.00000000000000   1.33300000000000   Se   (8k)
   0.31700000000000   1.15700000000000   0.00000000000000   Se   (8l)
  -0.31700000000000  -1.15700000000000   0.00000000000000   Se   (8l)
  -1.15700000000000  -0.31700000000000   0.50000000000000   Se   (8l)
   1.15700000000000   0.31700000000000   0.50000000000000   Se   (8l)
   0.00000000000000   0.00000000000000   0.25000000000000   Sr   (4a)
   0.00000000000000   0.00000000000000   0.75000000000000   Sr   (4a)
\end{lstlisting}
{\phantomsection\label{A3B_oC64_66_gi2lm_2l_cif}}
{\hyperref[A3B_oC64_66_gi2lm_2l]{H$_{3}$S (60~GPa): A3B\_oC64\_66\_gi2lm\_2l}} - CIF
\begin{lstlisting}[numbers=none,language={mylang}]
# CIF file 
data_findsym-output
_audit_creation_method FINDSYM

_chemical_name_mineral 'H3S'
_chemical_formula_sum 'H3 S'

_aflow_title 'H$_{3}$S (60~GPa) Structure'
_aflow_proto 'A3B_oC64_66_gi2lm_2l'
_aflow_params 'a,b/a,c/a,x_{1},z_{2},x_{3},y_{3},x_{4},y_{4},x_{5},y_{5},x_{6},y_{6},x_{7},y_{7},z_{7}'
_aflow_params_values '8.157,1.00294225818,0.59212945936,0.54552,0.3287,0.39296,0.16145,0.33837,0.89039,0.24136,0.07877,0.42337,0.74171,0.33285,0.66798,0.24948'
_aflow_Strukturbericht 'None'
_aflow_Pearson 'oC64'

_symmetry_space_group_name_H-M "C 2/c 2/c 2/m"
_symmetry_Int_Tables_number 66
 
_cell_length_a    8.15700
_cell_length_b    8.18100
_cell_length_c    4.83000
_cell_angle_alpha 90.00000
_cell_angle_beta  90.00000
_cell_angle_gamma 90.00000
 
loop_
_space_group_symop_id
_space_group_symop_operation_xyz
1 x,y,z
2 x,-y,-z+1/2
3 -x,y,-z+1/2
4 -x,-y,z
5 -x,-y,-z
6 -x,y,z+1/2
7 x,-y,z+1/2
8 x,y,-z
9 x+1/2,y+1/2,z
10 x+1/2,-y+1/2,-z+1/2
11 -x+1/2,y+1/2,-z+1/2
12 -x+1/2,-y+1/2,z
13 -x+1/2,-y+1/2,-z
14 -x+1/2,y+1/2,z+1/2
15 x+1/2,-y+1/2,z+1/2
16 x+1/2,y+1/2,-z
 
loop_
_atom_site_label
_atom_site_type_symbol
_atom_site_symmetry_multiplicity
_atom_site_Wyckoff_label
_atom_site_fract_x
_atom_site_fract_y
_atom_site_fract_z
_atom_site_occupancy
H1 H   8 g 0.54552  0.00000  0.25000 1.00000
H2 H   8 i 0.00000  0.00000  0.32870 1.00000
H3 H   8 l 0.39296  0.16145  0.00000 1.00000
H4 H   8 l 0.33837  0.89039  0.00000 1.00000
S1 S   8 l 0.24136  0.07877  0.00000 1.00000
S2 S   8 l 0.42337  0.74171  0.00000 1.00000
H5 H  16 m 0.33285  0.66798  0.24948 1.00000
\end{lstlisting}
{\phantomsection\label{A3B_oC64_66_gi2lm_2l_poscar}}
{\hyperref[A3B_oC64_66_gi2lm_2l]{H$_{3}$S (60~GPa): A3B\_oC64\_66\_gi2lm\_2l}} - POSCAR

{\phantomsection\label{A3B_oC64_66_kl2m_bdl_cif}}
{\hyperref[A3B_oC64_66_kl2m_bdl]{$\beta$-ThI$_{3}$: A3B\_oC64\_66\_kl2m\_bdl}} - CIF
\begin{lstlisting}[numbers=none,language={mylang}]
# CIF file 
data_findsym-output
_audit_creation_method FINDSYM

_chemical_name_mineral '$\beta$-ThI$_{3}$'
_chemical_formula_sum 'I3 Th'

loop_
_publ_author_name
 'H. P. Beck'
 'C. Strobel'
_journal_name_full_name
;
 Angewandte Chemie
;
_journal_volume 94
_journal_year 1982
_journal_page_first 558
_journal_page_last 559
_publ_Section_title
;
 ThI$_{3}$, ein Janus unter den Verbindungen mit Metall-Metall-Wechselwirkungen
;

# Found in The American Mineralogist Crystal Structure Database, 2003

_aflow_title '$\beta$-ThI$_{3}$ Structure'
_aflow_proto 'A3B_oC64_66_kl2m_bdl'
_aflow_params 'a,b/a,c/a,z_{3},x_{4},y_{4},x_{5},y_{5},x_{6},y_{6},z_{6},x_{7},y_{7},z_{7}'
_aflow_params_values '8.735,2.32364052662,1.67842014883,0.1826,-0.0318,0.1994,0.327,0.1716,0.2894,0.451,0.1302,0.1133,0.3773,0.3708'
_aflow_Strukturbericht 'None'
_aflow_Pearson 'oC64'

_symmetry_space_group_name_H-M "C 2/c 2/c 2/m"
_symmetry_Int_Tables_number 66
 
_cell_length_a    8.73500
_cell_length_b    20.29700
_cell_length_c    14.66100
_cell_angle_alpha 90.00000
_cell_angle_beta  90.00000
_cell_angle_gamma 90.00000
 
loop_
_space_group_symop_id
_space_group_symop_operation_xyz
1 x,y,z
2 x,-y,-z+1/2
3 -x,y,-z+1/2
4 -x,-y,z
5 -x,-y,-z
6 -x,y,z+1/2
7 x,-y,z+1/2
8 x,y,-z
9 x+1/2,y+1/2,z
10 x+1/2,-y+1/2,-z+1/2
11 -x+1/2,y+1/2,-z+1/2
12 -x+1/2,-y+1/2,z
13 -x+1/2,-y+1/2,-z
14 -x+1/2,y+1/2,z+1/2
15 x+1/2,-y+1/2,z+1/2
16 x+1/2,y+1/2,-z
 
loop_
_atom_site_label
_atom_site_type_symbol
_atom_site_symmetry_multiplicity
_atom_site_Wyckoff_label
_atom_site_fract_x
_atom_site_fract_y
_atom_site_fract_z
_atom_site_occupancy
Th1 Th   4 b 0.00000 0.50000 0.25000 1.00000
Th2 Th   4 d 0.00000 0.50000 0.00000 1.00000
I1  I    8 k 0.25000 0.25000 0.18260 1.00000
I2  I    8 l -0.03180  0.19940 0.00000 1.00000
Th3 Th   8 l 0.32700 0.17160 0.00000 1.00000
I3  I   16 m 0.28940 0.45100 0.13020 1.00000
I4  I   16 m 0.11330 0.37730 0.37080 1.00000
\end{lstlisting}
{\phantomsection\label{A3B_oC64_66_kl2m_bdl_poscar}}
{\hyperref[A3B_oC64_66_kl2m_bdl]{$\beta$-ThI$_{3}$: A3B\_oC64\_66\_kl2m\_bdl}} - POSCAR

{\phantomsection\label{A2BC_oC16_67_ag_b_g_cif}}
{\hyperref[A2BC_oC16_67_ag_b_g]{Al$_{2}$CuIr: A2BC\_oC16\_67\_ag\_b\_g}} - CIF
\begin{lstlisting}[numbers=none,language={mylang}]
# CIF file
data_findsym-output
_audit_creation_method FINDSYM

_chemical_name_mineral 'Al2CuIr'
_chemical_formula_sum 'Al2 Cu Ir'

loop_
_publ_author_name
 'L. Meshi'
 'V. Ezersky'
 'D. Kapush'
 'B. Grushko'
_journal_name_full_name
;
 Journal of Alloys and Compounds
;
_journal_volume 496
_journal_year 2010
_journal_page_first 208
_journal_page_last 211
_publ_Section_title
;
 Crystal structure of the Al$_{2}$CuIr phase
;

# Found in Pearson's Crystal Data - Crystal Structure Database for Inorganic Compounds, 2013

_aflow_title 'Al$_{2}$CuIr Structure'
_aflow_proto 'A2BC_oC16_67_ag_b_g'
_aflow_params 'a,b/a,c/a,z_{3},z_{4}'
_aflow_params_values '5.0644067238,1.60320657112,1.02379259962,0.3305,0.8198'
_aflow_Strukturbericht 'None'
_aflow_Pearson 'oC16'

_cell_length_a    5.0644067238
_cell_length_b    8.1192901384
_cell_length_c    5.1849021253
_cell_angle_alpha 90.0000000000
_cell_angle_beta  90.0000000000
_cell_angle_gamma 90.0000000000
 
_symmetry_space_group_name_H-M "C 2/m 2/m 2/a"
_symmetry_Int_Tables_number 67
 
loop_
_space_group_symop_id
_space_group_symop_operation_xyz
1 x,y,z
2 x,-y,-z
3 -x+1/2,y,-z
4 -x+1/2,-y,z
5 -x,-y,-z
6 -x,y,z
7 x+1/2,-y,z
8 x+1/2,y,-z
9 x+1/2,y+1/2,z
10 x+1/2,-y+1/2,-z
11 -x,y+1/2,-z
12 -x,-y+1/2,z
13 -x+1/2,-y+1/2,-z
14 -x+1/2,y+1/2,z
15 x,-y+1/2,z
16 x,y+1/2,-z
 
loop_
_atom_site_label
_atom_site_type_symbol
_atom_site_symmetry_multiplicity
_atom_site_Wyckoff_label
_atom_site_fract_x
_atom_site_fract_y
_atom_site_fract_z
_atom_site_occupancy
Al1 Al   4 a 0.25000 0.00000 0.00000 1.00000
Cu1 Cu   4 b 0.25000 0.00000 0.50000 1.00000
Al2 Al   4 g 0.00000 0.25000 0.33050 1.00000
Ir1 Ir   4 g 0.00000 0.25000 0.81980 1.00000
\end{lstlisting}
{\phantomsection\label{A2BC_oC16_67_ag_b_g_poscar}}
{\hyperref[A2BC_oC16_67_ag_b_g]{Al$_{2}$CuIr: A2BC\_oC16\_67\_ag\_b\_g}} - POSCAR
\begin{lstlisting}[numbers=none,language={mylang}]
A2BC_oC16_67_ag_b_g & a,b/a,c/a,z3,z4 --params=5.0644067238,1.60320657112,1.02379259962,0.3305,0.8198 & Cmma D_{2h}^{21} #67 (abg^2) & oC16 & None & Al2CuIr &  & L. Meshi et al., J. Alloys Compd. 496, 208-211 (2010)
   1.00000000000000
   2.53220336190000  -4.05964506920000   0.00000000000000
   2.53220336190000   4.05964506920000   0.00000000000000
   0.00000000000000   0.00000000000000   5.18490212530000
    Al    Cu    Ir
     4     2     2
Direct
   0.25000000000000   0.25000000000000   0.00000000000000   Al   (4a)
   0.75000000000000   0.75000000000000   0.00000000000000   Al   (4a)
   0.75000000000000   0.25000000000000   0.33050000000000   Al   (4g)
   0.25000000000000   0.75000000000000  -0.33050000000000   Al   (4g)
   0.25000000000000   0.25000000000000   0.50000000000000   Cu   (4b)
   0.75000000000000   0.75000000000000   0.50000000000000   Cu   (4b)
   0.75000000000000   0.25000000000000   0.81980000000000   Ir   (4g)
   0.25000000000000   0.75000000000000  -0.81980000000000   Ir   (4g)
\end{lstlisting}
{\phantomsection\label{ABC2_oC16_67_b_g_ag_cif}}
{\hyperref[ABC2_oC16_67_b_g_ag]{HoCuP$_{2}$: ABC2\_oC16\_67\_b\_g\_ag}} - CIF
\begin{lstlisting}[numbers=none,language={mylang}]
# CIF file
data_findsym-output
_audit_creation_method FINDSYM

_chemical_name_mineral 'HoCuP2'
_chemical_formula_sum 'Cu Ho P2'

loop_
_publ_author_name
 'Y. Mozharivsky'
 'D. Kaczorowski'
 'H. F. Franzen'
_journal_name_full_name
;
 Zeitschrift fur Anorganische und Allgemeine Chemie
;
_journal_volume 627
_journal_year 2001
_journal_page_first 2163
_journal_page_last 2172
_publ_Section_title
;
 Symmetry-Breaking Transitions in HoCuAs$_{2-x}$P$_{x}$ and ErCuAs$_{2-x}$P$_{x}$ ($x=0-2$): Crystal Structure, Application of Landau Theory, Magnetic and Electrical Properties
;

# Found in Pearson's Crystal Data - Crystal Structure Database for Inorganic Compounds, 2013

_aflow_title 'HoCuP$_{2}$ Structure'
_aflow_proto 'ABC2_oC16_67_b_g_ag'
_aflow_params 'a,b/a,c/a,z_{3},z_{4}'
_aflow_params_values '5.2729860526,1.00606865161,1.82912952778,0.765,0.3403'
_aflow_Strukturbericht 'None'
_aflow_Pearson 'oC16'

_cell_length_a    5.2729860526
_cell_length_b    5.3049859679
_cell_length_c    9.6449744884
_cell_angle_alpha 90.0000000000
_cell_angle_beta  90.0000000000
_cell_angle_gamma 90.0000000000
 
_symmetry_space_group_name_H-M "C 2/m 2/m 2/a"
_symmetry_Int_Tables_number 67
 
loop_
_space_group_symop_id
_space_group_symop_operation_xyz
1 x,y,z
2 x,-y,-z
3 -x+1/2,y,-z
4 -x+1/2,-y,z
5 -x,-y,-z
6 -x,y,z
7 x+1/2,-y,z
8 x+1/2,y,-z
9 x+1/2,y+1/2,z
10 x+1/2,-y+1/2,-z
11 -x,y+1/2,-z
12 -x,-y+1/2,z
13 -x+1/2,-y+1/2,-z
14 -x+1/2,y+1/2,z
15 x,-y+1/2,z
16 x,y+1/2,-z
 
loop_
_atom_site_label
_atom_site_type_symbol
_atom_site_symmetry_multiplicity
_atom_site_Wyckoff_label
_atom_site_fract_x
_atom_site_fract_y
_atom_site_fract_z
_atom_site_occupancy
P1  P    4 a 0.25000 0.00000 0.00000 1.00000
Cu1 Cu   4 b 0.25000 0.00000 0.50000 1.00000
Ho1 Ho   4 g 0.00000 0.25000 0.76500 1.00000
P2  P    4 g 0.00000 0.25000 0.34030 1.00000
\end{lstlisting}
{\phantomsection\label{ABC2_oC16_67_b_g_ag_poscar}}
{\hyperref[ABC2_oC16_67_b_g_ag]{HoCuP$_{2}$: ABC2\_oC16\_67\_b\_g\_ag}} - POSCAR
\begin{lstlisting}[numbers=none,language={mylang}]
ABC2_oC16_67_b_g_ag & a,b/a,c/a,z3,z4 --params=5.2729860526,1.00606865161,1.82912952778,0.765,0.3403 & Cmma D_{2h}^{21} #67 (abg^2) & oC16 & None & HoCuP2 &  & Y. Mozharivsky and D. Kaczorowski and H. F. Franzen, Z. Anorg. Allg. Chem. 627, 2163-2172 (2001)
   1.00000000000000
   2.63649302630000  -2.65249298395000   0.00000000000000
   2.63649302630000   2.65249298395000   0.00000000000000
   0.00000000000000   0.00000000000000   9.64497448840000
    Cu    Ho     P
     2     2     4
Direct
   0.25000000000000   0.25000000000000   0.50000000000000   Cu   (4b)
   0.75000000000000   0.75000000000000   0.50000000000000   Cu   (4b)
   0.75000000000000   0.25000000000000   0.76500000000000   Ho   (4g)
   0.25000000000000   0.75000000000000  -0.76500000000000   Ho   (4g)
   0.25000000000000   0.25000000000000   0.00000000000000    P   (4a)
   0.75000000000000   0.75000000000000   0.00000000000000    P   (4a)
   0.75000000000000   0.25000000000000   0.34030000000000    P   (4g)
   0.25000000000000   0.75000000000000  -0.34030000000000    P   (4g)
\end{lstlisting}
{\phantomsection\label{AB_oC8_67_a_g-FeSe_cif}}
{\hyperref[AB_oC8_67_a_g-FeSe]{$\alpha$-FeSe: AB\_oC8\_67\_a\_g}} - CIF
\begin{lstlisting}[numbers=none,language={mylang}]
# CIF file 
data_findsym-output
_audit_creation_method FINDSYM

_chemical_name_mineral '$\alpha$-FeSe'
_chemical_formula_sum 'Fe Se'

loop_
_publ_author_name
 'D. Louca'
 'K. Horigane'
 'A. Llobet'
 'R. Arita'
 'S. Ji'
 'N. Katayama'
 'S. Konbu'
 'K. Nakamura'
 'T.-Y. Koo'
 'P. Tong'
 'K. Yamada'
_journal_name_full_name
;
 Physical Review B
;
_journal_volume 81
_journal_year 2010
_journal_page_first 134524
_journal_page_last 134524
_publ_Section_title
;
 Local atomic structure of superconducting FeSe$_{1-x}$Te$_{x}$
;

_aflow_title '$\alpha$-FeSe Structure'
_aflow_proto 'AB_oC8_67_a_g'
_aflow_params 'a,b/a,c/a,z_{2}'
_aflow_params_values '5.32495,0.997010300566,1.02896740814,0.26686'
_aflow_Strukturbericht 'None'
_aflow_Pearson 'oC8'

_symmetry_space_group_name_H-M "C 2/m 2/m 2/a"
_symmetry_Int_Tables_number 67
 
_cell_length_a    5.32495
_cell_length_b    5.30903
_cell_length_c    5.47920
_cell_angle_alpha 90.00000
_cell_angle_beta  90.00000
_cell_angle_gamma 90.00000
 
loop_
_space_group_symop_id
_space_group_symop_operation_xyz
1 x,y,z
2 x,-y,-z
3 -x+1/2,y,-z
4 -x+1/2,-y,z
5 -x,-y,-z
6 -x,y,z
7 x+1/2,-y,z
8 x+1/2,y,-z
9 x+1/2,y+1/2,z
10 x+1/2,-y+1/2,-z
11 -x,y+1/2,-z
12 -x,-y+1/2,z
13 -x+1/2,-y+1/2,-z
14 -x+1/2,y+1/2,z
15 x,-y+1/2,z
16 x,y+1/2,-z
 
loop_
_atom_site_label
_atom_site_type_symbol
_atom_site_symmetry_multiplicity
_atom_site_Wyckoff_label
_atom_site_fract_x
_atom_site_fract_y
_atom_site_fract_z
_atom_site_occupancy
Fe1 Fe   4 a 0.25000 0.00000 0.00000 1.00000
Se1 Se   4 g 0.00000 0.25000 0.26686 1.00000
\end{lstlisting}
{\phantomsection\label{AB_oC8_67_a_g-FeSe_poscar}}
{\hyperref[AB_oC8_67_a_g-FeSe]{$\alpha$-FeSe: AB\_oC8\_67\_a\_g}} - POSCAR
\begin{lstlisting}[numbers=none,language={mylang}]
AB_oC8_67_a_g & a,b/a,c/a,z2 --params=5.32495,0.997010300566,1.02896740814,0.26686 & Cmma D_{2h}^{21} #67 (ag) & oC8 & None & FeSe & $\alpha$-FeSe & D. Louca et al., Phys. Rev. B 81, 134524(2010)
   1.00000000000000
   2.66247500000000  -2.65451500000000   0.00000000000000
   2.66247500000000   2.65451500000000   0.00000000000000
   0.00000000000000   0.00000000000000   5.47920000000000
    Fe    Se
     2     2
Direct
   0.25000000000000   0.25000000000000   0.00000000000000   Fe   (4a)
   0.75000000000000   0.75000000000000   0.00000000000000   Fe   (4a)
   0.75000000000000   0.25000000000000   0.26686000000000   Se   (4g)
   0.25000000000000   0.75000000000000  -0.26686000000000   Se   (4g)
\end{lstlisting}
{\phantomsection\label{AB_oC8_67_a_g-PbO_cif}}
{\hyperref[AB_oC8_67_a_g-PbO]{$\alpha$-PbO: AB\_oC8\_67\_a\_g}} - CIF
\begin{lstlisting}[numbers=none,language={mylang}]
# CIF file
data_findsym-output
_audit_creation_method FINDSYM

_chemical_name_mineral 'alpha-PbO'
_chemical_formula_sum 'O Pb'

loop_
_publ_author_name
 'P. Boher'
 'P. Garnier'
 'J. R. Gavarri'
 'A. W. Hewat'
_journal_name_full_name
;
 Journal of Solid State Chemistry
;
_journal_volume 57
_journal_year 1985
_journal_page_first 343
_journal_page_last 350
_publ_Section_title
;
 Monoxyde quadratique PbO$\alpha$ (I): Description de la transition structurale ferro{\\'e}lastique
;

# Found in Pearson's Crystal Data - Crystal Structure Database for Inorganic Compounds, 2013

_aflow_title '$\alpha$-PbO Structure'
_aflow_proto 'AB_oC8_67_a_g'
_aflow_params 'a,b/a,c/a,z_{2}'
_aflow_params_values '5.6124272786,0.999376380873,0.8895303257,0.7642'
_aflow_Strukturbericht 'None'
_aflow_Pearson 'oC8'

_cell_length_a    5.6124272786
_cell_length_b    5.6089272616
_cell_length_c    4.9924242651
_cell_angle_alpha 90.0000000000
_cell_angle_beta  90.0000000000
_cell_angle_gamma 90.0000000000
 
_symmetry_space_group_name_H-M "C 2/m 2/m 2/a"
_symmetry_Int_Tables_number 67
 
loop_
_space_group_symop_id
_space_group_symop_operation_xyz
1 x,y,z
2 x,-y,-z
3 -x+1/2,y,-z
4 -x+1/2,-y,z
5 -x,-y,-z
6 -x,y,z
7 x+1/2,-y,z
8 x+1/2,y,-z
9 x+1/2,y+1/2,z
10 x+1/2,-y+1/2,-z
11 -x,y+1/2,-z
12 -x,-y+1/2,z
13 -x+1/2,-y+1/2,-z
14 -x+1/2,y+1/2,z
15 x,-y+1/2,z
16 x,y+1/2,-z
 
loop_
_atom_site_label
_atom_site_type_symbol
_atom_site_symmetry_multiplicity
_atom_site_Wyckoff_label
_atom_site_fract_x
_atom_site_fract_y
_atom_site_fract_z
_atom_site_occupancy
O1  O    4 a 0.25000 0.00000 0.00000 1.00000
Pb1 Pb   4 g 0.00000 0.25000 0.76420 1.00000
\end{lstlisting}
{\phantomsection\label{AB_oC8_67_a_g-PbO_poscar}}
{\hyperref[AB_oC8_67_a_g-PbO]{$\alpha$-PbO: AB\_oC8\_67\_a\_g}} - POSCAR
\begin{lstlisting}[numbers=none,language={mylang}]
AB_oC8_67_a_g & a,b/a,c/a,z2 --params=5.6124272786,0.999376380873,0.8895303257,0.7642 & Cmma D_{2h}^{21} #67 (ag) & oC8 & None & PbO & alpha & P. Boher et al., J. Solid State Chem. 57, 343-350 (1985)
   1.00000000000000
   2.80621363930000  -2.80446363080000   0.00000000000000
   2.80621363930000   2.80446363080000   0.00000000000000
   0.00000000000000   0.00000000000000   4.99242426510000
     O    Pb
     2     2
Direct
   0.25000000000000   0.25000000000000   0.00000000000000    O   (4a)
   0.75000000000000   0.75000000000000   0.00000000000000    O   (4a)
   0.75000000000000   0.25000000000000   0.76420000000000   Pb   (4g)
   0.25000000000000   0.75000000000000  -0.76420000000000   Pb   (4g)
\end{lstlisting}
{\phantomsection\label{AB4_oC20_68_a_i_cif}}
{\hyperref[AB4_oC20_68_a_i]{PdSn$_{4}$: AB4\_oC20\_68\_a\_i}} - CIF
\begin{lstlisting}[numbers=none,language={mylang}]
# CIF file# This file was generated by FINDSYM
# Harold T. Stokes, Branton J. Campbell, Dorian M. Hatch
# Brigham Young University, Provo, Utah, USA

data_findsym-output
_audit_creation_method FINDSYM

_chemical_name_mineral 'PdSn4'
_chemical_formula_sum 'Pd Sn4'

loop_
_publ_author_name
 'J. Nyl{\\'e}n'
 'F. J. {Garc{\\'i}a Garc{\\'i}a}'
 'B. D. Mosel'
 'R. P{\"o}ttgen'
 'U. H{\"a}ussermann'
_journal_name_full_name
;
 Solid State Sciences
;
_journal_volume 6
_journal_year 2004
_journal_page_first 147
_journal_page_last 155
_publ_Section_title
;
 Structural relationships, phase stability and bonding of compounds PdSn$_{n}$ ($n$ = 2, 3, 4)
;

# Found in Pearson's Crystal Data - Crystal Structure Database for Inorganic Compounds, 2013

_aflow_title 'PdSn$_{4}$ Structure'
_aflow_proto 'AB4_oC20_68_a_i'
_aflow_params 'a,b/a,c/a,x_{2},y_{2},z_{2}'
_aflow_params_values '6.44222,1.77662048176,0.991771470083,0.3313,0.1233,0.0801'
_aflow_Strukturbericht 'None'
_aflow_Pearson 'oC20'

_symmetry_space_group_name_H-M "C 2/c 2/c 2/a (origin choice 2)"
_symmetry_Int_Tables_number 68

_cell_length_a    6.44222
_cell_length_b    11.44538
_cell_length_c    6.38921
_cell_angle_alpha 90.00000
_cell_angle_beta  90.00000
_cell_angle_gamma 90.00000

loop_
_space_group_symop_id
_space_group_symop_operation_xyz
1 x,y,z
2 x,-y+1/2,-z+1/2
3 -x,y,-z+1/2
4 -x,-y+1/2,z
5 -x+1/2,-y+1/2,-z
6 -x+1/2,y,z+1/2
7 x+1/2,-y+1/2,z+1/2
8 x+1/2,y,-z
9 x+1/2,y+1/2,z
10 x+1/2,-y,-z+1/2
11 -x+1/2,y+1/2,-z+1/2
12 -x+1/2,-y,z
13 -x,-y,-z
14 -x,y+1/2,z+1/2
15 x,-y,z+1/2
16 x,y+1/2,-z

loop_
_atom_site_label
_atom_site_type_symbol
_atom_site_symmetry_multiplicity
_atom_site_Wyckoff_label
_atom_site_fract_x
_atom_site_fract_y
_atom_site_fract_z
_atom_site_occupancy
Pd1 Pd   4 a 0.00000 0.25000 0.25000 1.00000
Sn1 Sn  16 i 0.33130 0.12330 0.08010 1.00000
\end{lstlisting}
{\phantomsection\label{AB4_oC20_68_a_i_poscar}}
{\hyperref[AB4_oC20_68_a_i]{PdSn$_{4}$: AB4\_oC20\_68\_a\_i}} - POSCAR
\begin{lstlisting}[numbers=none,language={mylang}]
AB4_oC20_68_a_i & a,b/a,c/a,x2,y2,z2 --params=6.44222,1.77662048176,0.991771470083,0.3313,0.1233,0.0801 & Ccca D_{2h}^{22} #68 (ai) & oC20 & None & PdSn4 &  & J. Nyl{\'e}n et al., Solid State Sci. 6, 147-155 (2004)
   1.00000000000000
   3.22111000000000  -5.72269000000000   0.00000000000000
   3.22111000000000   5.72269000000000   0.00000000000000
   0.00000000000000   0.00000000000000   6.38921000000000
    Pd    Sn
     2     8
Direct
   0.75000000000000   0.25000000000000   0.25000000000000   Pd   (4a)
   0.25000000000000   0.75000000000000   0.75000000000000   Pd   (4a)
   0.20800000000000   0.45460000000000   0.08010000000000   Sn  (16i)
   0.29200000000000   0.04540000000000   0.08010000000000   Sn  (16i)
  -0.45460000000000  -0.20800000000000   0.41990000000000   Sn  (16i)
   0.95460000000000   0.70800000000000   0.41990000000000   Sn  (16i)
  -0.20800000000000  -0.45460000000000  -0.08010000000000   Sn  (16i)
   0.70800000000000   0.95460000000000  -0.08010000000000   Sn  (16i)
   0.45460000000000   0.20800000000000   0.58010000000000   Sn  (16i)
   0.04540000000000   0.29200000000000   0.58010000000000   Sn  (16i)
\end{lstlisting}
{\phantomsection\label{AB2_oF48_70_f_fg_cif}}
{\hyperref[AB2_oF48_70_f_fg]{Mn$_{2}$B ($D1_{f}$): AB2\_oF48\_70\_f\_fg}} - CIF

{\phantomsection\label{AB2_oF48_70_f_fg_poscar}}
{\hyperref[AB2_oF48_70_f_fg]{Mn$_{2}$B ($D1_{f}$): AB2\_oF48\_70\_f\_fg}} - POSCAR
\begin{lstlisting}[numbers=none,language={mylang}]
AB2_oF48_70_f_fg & a,b/a,c/a,y1,y2,z3 --params=4.2082,3.45504015969,1.7326647973,0.2495,0.54337,0.29445 & Fddd D_{2h}^{24} #70 (f^2g) & oF48 & $D1_{f}$ & Mn2B & Mn2B & L.-E. Tergenius, J. Less-Common Met. 82, 335-340 (1981)
   1.00000000000000
   0.00000000000000   7.26975000000000   3.64570000000000
   2.10410000000000   0.00000000000000   3.64570000000000
   2.10410000000000   7.26975000000000   0.00000000000000
     B    Mn
     4     8
Direct
   0.24950000000000   0.00050000000000   0.24950000000000    B  (16f)
   0.00050000000000   0.24950000000000   0.00050000000000    B  (16f)
  -0.24950000000000   0.99950000000000  -0.24950000000000    B  (16f)
   0.99950000000000  -0.24950000000000   0.99950000000000    B  (16f)
   0.54337000000000  -0.29337000000000   0.54337000000000   Mn  (16f)
  -0.29337000000000   0.54337000000000  -0.29337000000000   Mn  (16f)
  -0.54337000000000   1.29337000000000  -0.54337000000000   Mn  (16f)
   1.29337000000000  -0.54337000000000   1.29337000000000   Mn  (16f)
   0.29445000000000   0.29445000000000  -0.04445000000000   Mn  (16g)
  -0.04445000000000  -0.04445000000000   0.29445000000000   Mn  (16g)
  -0.29445000000000  -0.29445000000000   1.04445000000000   Mn  (16g)
   1.04445000000000   1.04445000000000  -0.29445000000000   Mn  (16g)
\end{lstlisting}
{\phantomsection\label{A4B3_oI14_71_gh_cg_cif}}
{\hyperref[A4B3_oI14_71_gh_cg]{Ta$_{3}$B$_{4}$ ($D7_{b}$): A4B3\_oI14\_71\_gh\_cg}} - CIF
\begin{lstlisting}[numbers=none,language={mylang}]
# CIF file
data_findsym-output
_audit_creation_method FINDSYM

_chemical_name_mineral 'Ta3B4'
_chemical_formula_sum 'B4 Ta3'

loop_
_publ_author_name
 'R. Kiessling'
_journal_name_full_name
;
 Acta Chemica Scandinavica
;
_journal_volume 3
_journal_year 1949
_journal_page_first 603
_journal_page_last 615
_publ_Section_title
;
 The Borides of Tantalum
;

# Found in Transition-metal borides with the tantalum boride (Ta$_{3}$B$_{4}$) crystal structure: their electronic and bonding properties, 1991

_aflow_title 'Ta$_{3}$B$_{4}$ ($D7_{b}$) Structure'
_aflow_proto 'A4B3_oI14_71_gh_cg'
_aflow_params 'a,b/a,c/a,y_{2},y_{3},y_{4}'
_aflow_params_values '3.29,4.25531914894,0.951367781155,0.375,0.18,0.444'
_aflow_Strukturbericht '$D7_{b}$'
_aflow_Pearson 'oI14'

_symmetry_space_group_name_H-M "I 2/m 2/m 2/m"
_symmetry_Int_Tables_number 71
 
_cell_length_a    3.29000
_cell_length_b    14.00000
_cell_length_c    3.13000
_cell_angle_alpha 90.00000
_cell_angle_beta  90.00000
_cell_angle_gamma 90.00000
 
loop_
_space_group_symop_id
_space_group_symop_operation_xyz
1 x,y,z
2 x,-y,-z
3 -x,y,-z
4 -x,-y,z
5 -x,-y,-z
6 -x,y,z
7 x,-y,z
8 x,y,-z
9 x+1/2,y+1/2,z+1/2
10 x+1/2,-y+1/2,-z+1/2
11 -x+1/2,y+1/2,-z+1/2
12 -x+1/2,-y+1/2,z+1/2
13 -x+1/2,-y+1/2,-z+1/2
14 -x+1/2,y+1/2,z+1/2
15 x+1/2,-y+1/2,z+1/2
16 x+1/2,y+1/2,-z+1/2
 
loop_
_atom_site_label
_atom_site_type_symbol
_atom_site_symmetry_multiplicity
_atom_site_Wyckoff_label
_atom_site_fract_x
_atom_site_fract_y
_atom_site_fract_z
_atom_site_occupancy
Ta1 Ta   2 c 0.50000 0.50000 0.00000  1.00000
B1  B    4 g 0.00000 0.37500 0.00000  1.00000
Ta2 Ta   4 g 0.00000 0.18000 0.00000  1.00000
B2  B    4 h 0.00000 0.44400 0.50000  1.00000
\end{lstlisting}
{\phantomsection\label{A4B3_oI14_71_gh_cg_poscar}}
{\hyperref[A4B3_oI14_71_gh_cg]{Ta$_{3}$B$_{4}$ ($D7_{b}$): A4B3\_oI14\_71\_gh\_cg}} - POSCAR
\begin{lstlisting}[numbers=none,language={mylang}]
A4B3_oI14_71_gh_cg & a,b/a,c/a,y2,y3,y4 --params=3.29,4.25531914894,0.951367781155,0.375,0.18,0.444 & Immm D_{2h}^{25} #71 (cg^2h) & oI14 & $D7_{b}$ & Ta3B4 & Ta3B4 & R. Kiessling, Acta Chem. Scand. 3, 603-615 (1949)
   1.00000000000000
  -1.64500000000000   7.00000000000000   1.56500000000000
   1.64500000000000  -7.00000000000000   1.56500000000000
   1.64500000000000   7.00000000000000  -1.56500000000000
     B    Ta
     4     3
Direct
   0.37500000000000   0.00000000000000   0.37500000000000    B   (4g)
  -0.37500000000000   0.00000000000000  -0.37500000000000    B   (4g)
   0.94400000000000   0.50000000000000   0.44400000000000    B   (4h)
   0.05600000000000   0.50000000000000  -0.44400000000000    B   (4h)
   0.50000000000000   0.50000000000000   0.00000000000000   Ta   (2c)
   0.18000000000000   0.00000000000000   0.18000000000000   Ta   (4g)
  -0.18000000000000   0.00000000000000  -0.18000000000000   Ta   (4g)
\end{lstlisting}
{\phantomsection\label{ABC_oI12_71_h_j_g_cif}}
{\hyperref[ABC_oI12_71_h_j_g]{NbPS: ABC\_oI12\_71\_h\_j\_g}} - CIF
\begin{lstlisting}[numbers=none,language={mylang}]
# CIF file
data_findsym-output
_audit_creation_method FINDSYM

_chemical_name_mineral ''
_chemical_formula_sum 'Nb P S'

loop_
_publ_author_name
 'P. C. Donohue'
 'P. E. Bierstedt'
_journal_name_full_name
;
 Inorganic Chemistry
;
_journal_volume 8
_journal_year 1969
_journal_page_first 2690
_journal_page_last 2694
_publ_Section_title
;
 Synthesis, crystal structure, and superconducting properties of niobium phosphorus sulfide, niobium phosphorus selenide and tantalum phosphorus sulfide
;

_aflow_title 'NbPS Structure'
_aflow_proto 'ABC_oI12_71_h_j_g'
_aflow_params 'a,b/a,c/a,y_{1},y_{2},z_{3}'
_aflow_params_values '3.438,3.4554973822,1.37434554974,0.212,0.1232,0.235'
_aflow_Strukturbericht 'None'
_aflow_Pearson 'oI12'

_symmetry_space_group_name_H-M "I 2/m 2/m 2/m"
_symmetry_Int_Tables_number 71
 
_cell_length_a    3.43800
_cell_length_b    11.88000
_cell_length_c    4.72500
_cell_angle_alpha 90.00000
_cell_angle_beta  90.00000
_cell_angle_gamma 90.00000
 
loop_
_space_group_symop_id
_space_group_symop_operation_xyz
1 x,y,z
2 x,-y,-z
3 -x,y,-z
4 -x,-y,z
5 -x,-y,-z
6 -x,y,z
7 x,-y,z
8 x,y,-z
9 x+1/2,y+1/2,z+1/2
10 x+1/2,-y+1/2,-z+1/2
11 -x+1/2,y+1/2,-z+1/2
12 -x+1/2,-y+1/2,z+1/2
13 -x+1/2,-y+1/2,-z+1/2
14 -x+1/2,y+1/2,z+1/2
15 x+1/2,-y+1/2,z+1/2
16 x+1/2,y+1/2,-z+1/2
 
loop_
_atom_site_label
_atom_site_type_symbol
_atom_site_symmetry_multiplicity
_atom_site_Wyckoff_label
_atom_site_fract_x
_atom_site_fract_y
_atom_site_fract_z
_atom_site_occupancy
S1  S    4 g 0.00000 0.21200 0.00000 1.00000
Nb1 Nb   4 h 0.00000 0.12320 0.50000 1.00000
P1  P    4 j 0.50000 0.00000 0.23500 1.00000
\end{lstlisting}
{\phantomsection\label{ABC_oI12_71_h_j_g_poscar}}
{\hyperref[ABC_oI12_71_h_j_g]{NbPS: ABC\_oI12\_71\_h\_j\_g}} - POSCAR
\begin{lstlisting}[numbers=none,language={mylang}]
ABC_oI12_71_h_j_g & a,b/a,c/a,y1,y2,z3 --params=3.438,3.4554973822,1.37434554974,0.212,0.1232,0.235 & Immm D_{2h}^{25} #71 (ghj) & oI12 & None & NbPS &  & P. C. Donohue and P. E. Bierstedt, Inorg. Chem. 8, 2690-2694 (1969)
   1.00000000000000
  -1.71900000000000   5.94000000000000   2.36250000000000
   1.71900000000000  -5.94000000000000   2.36250000000000
   1.71900000000000   5.94000000000000  -2.36250000000000
    Nb     P     S
     2     2     2
Direct
   0.62320000000000   0.50000000000000   0.12320000000000   Nb   (4h)
   0.37680000000000   0.50000000000000  -0.12320000000000   Nb   (4h)
   0.23500000000000   0.73500000000000   0.50000000000000    P   (4j)
  -0.23500000000000   0.26500000000000   0.50000000000000    P   (4j)
   0.21200000000000   0.00000000000000   0.21200000000000    S   (4g)
  -0.21200000000000   0.00000000000000  -0.21200000000000    S   (4g)
\end{lstlisting}
{\phantomsection\label{ABCD3_oI48_73_d_e_e_ef_cif}}
{\hyperref[ABCD3_oI48_73_d_e_e_ef]{KAg[CO$_{3}$]: ABCD3\_oI48\_73\_d\_e\_e\_ef}} - CIF
\begin{lstlisting}[numbers=none,language={mylang}]
# CIF file
data_findsym-output
_audit_creation_method FINDSYM

_chemical_name_mineral 'KAg[CO3]'
_chemical_formula_sum 'Ag C K O3'

loop_
_publ_author_name
 'Y.-Q. Zheng'
 'L.-X. Zhou'
 'J.-L. Lin'
 'S.-W. Zhang'
_journal_name_full_name
;
 Zeitschrift f{\"u}r Kristallographie - New Crystal Structures
;
_journal_volume 215
_journal_year 2000
_journal_page_first 467
_journal_page_last 468
_publ_Section_title
;
 Refinement of the crystal structure of potassium {\it catena}--carbonato--argentate(I), K[Ag(CO$_{3}$)]
;

# Found in Pearson's Crystal Data - Crystal Structure Database for Inorganic Compounds, 2013

_aflow_title 'KAg[CO$_{3}$] Structure'
_aflow_proto 'ABCD3_oI48_73_d_e_e_ef'
_aflow_params 'a,b/a,c/a,y_{1},z_{2},z_{3},z_{4},x_{5},y_{5},z_{5}'
_aflow_params_values '5.7750411261,1.02926406926,3.53056277057,0.63427,0.3734,0.18032,0.311,0.1349,0.6124,0.0967'
_aflow_Strukturbericht 'None'
_aflow_Pearson 'oI48'

_cell_length_a    5.7750411261
_cell_length_b    5.9440423296
_cell_length_c    20.3891451983
_cell_angle_alpha 90.0000000000
_cell_angle_beta  90.0000000000
_cell_angle_gamma 90.0000000000
 
_symmetry_space_group_name_H-M "I 21/b 21/c 21/a"
_symmetry_Int_Tables_number 73
 
loop_
_space_group_symop_id
_space_group_symop_operation_xyz
1 x,y,z
2 x,-y,-z+1/2
3 -x+1/2,y,-z
4 -x,-y+1/2,z
5 -x,-y,-z
6 -x,y,z+1/2
7 x+1/2,-y,z
8 x,y+1/2,-z
9 x+1/2,y+1/2,z+1/2
10 x+1/2,-y+1/2,-z
11 -x,y+1/2,-z+1/2
12 -x+1/2,-y,z+1/2
13 -x+1/2,-y+1/2,-z+1/2
14 -x+1/2,y+1/2,z
15 x,-y+1/2,z+1/2
16 x+1/2,y,-z+1/2
 
loop_
_atom_site_label
_atom_site_type_symbol
_atom_site_symmetry_multiplicity
_atom_site_Wyckoff_label
_atom_site_fract_x
_atom_site_fract_y
_atom_site_fract_z
_atom_site_occupancy
Ag1 Ag   8 d 0.25000 0.63427 0.00000 1.00000
C1  C    8 e 0.00000 0.25000 0.37340 1.00000
K1  K    8 e 0.00000 0.25000 0.18032 1.00000
O1  O    8 e 0.00000 0.25000 0.31100 1.00000
O2  O   16 f 0.13490 0.61240 0.09670 1.00000
\end{lstlisting}
{\phantomsection\label{ABCD3_oI48_73_d_e_e_ef_poscar}}
{\hyperref[ABCD3_oI48_73_d_e_e_ef]{KAg[CO$_{3}$]: ABCD3\_oI48\_73\_d\_e\_e\_ef}} - POSCAR

{\phantomsection\label{A2B_oI12_74_h_e_cif}}
{\hyperref[A2B_oI12_74_h_e]{KHg$_{2}$: A2B\_oI12\_74\_h\_e}} - CIF
\begin{lstlisting}[numbers=none,language={mylang}]
# CIF file
data_findsym-output
_audit_creation_method FINDSYM

_chemical_name_mineral 'KHg2'
_chemical_formula_sum 'Hg2 K'

loop_
_publ_author_name
 'E. J. Duwell'
 'N. C. Baenziger'
_journal_name_full_name
;
 Acta Cristallographica
;
_journal_volume 8
_journal_year 1955
_journal_page_first 705
_journal_page_last 710
_publ_Section_title
;
 The crystal structures of KHg and KHg$_{2}$
;

# Found in Pearson's Crystal Data - Crystal Structure Database for Inorganic Compounds, 2013

_aflow_title 'KHg$_{2}$ Structure'
_aflow_proto 'A2B_oI12_74_h_e'
_aflow_params 'a,b/a,c/a,z_{1},y_{2},z_{2}'
_aflow_params_values '5.158858099,1.56976744188,1.70096899227,-0.047,0.56,0.663'
_aflow_Strukturbericht 'None'
_aflow_Pearson 'oI12'

_cell_length_a    5.1588580990
_cell_length_b    8.0982074811
_cell_length_c    8.7750576619
_cell_angle_alpha 90.0000000000
_cell_angle_beta  90.0000000000
_cell_angle_gamma 90.0000000000
 
_symmetry_space_group_name_H-M "I 21/m 21/m 21/a"
_symmetry_Int_Tables_number 74
 
loop_
_space_group_symop_id
_space_group_symop_operation_xyz
1 x,y,z
2 x,-y,-z
3 -x,y+1/2,-z
4 -x,-y+1/2,z
5 -x,-y,-z
6 -x,y,z
7 x,-y+1/2,z
8 x,y+1/2,-z
9 x+1/2,y+1/2,z+1/2
10 x+1/2,-y+1/2,-z+1/2
11 -x+1/2,y,-z+1/2
12 -x+1/2,-y,z+1/2
13 -x+1/2,-y+1/2,-z+1/2
14 -x+1/2,y+1/2,z+1/2
15 x+1/2,-y,z+1/2
16 x+1/2,y,-z+1/2
 
loop_
_atom_site_label
_atom_site_type_symbol
_atom_site_symmetry_multiplicity
_atom_site_Wyckoff_label
_atom_site_fract_x
_atom_site_fract_y
_atom_site_fract_z
_atom_site_occupancy
K1  K    4 e 0.00000 0.25000 -0.04700 1.00000
Hg1 Hg   8 h 0.00000 0.56000 0.66300  1.00000
\end{lstlisting}
{\phantomsection\label{A2B_oI12_74_h_e_poscar}}
{\hyperref[A2B_oI12_74_h_e]{KHg$_{2}$: A2B\_oI12\_74\_h\_e}} - POSCAR
\begin{lstlisting}[numbers=none,language={mylang}]
A2B_oI12_74_h_e & a,b/a,c/a,z1,y2,z2 --params=5.158858099,1.56976744188,1.70096899227,-0.047,0.56,0.663 & Imma D_{2h}^{28} #74 (eh) & oI12 & None & KHg2 &  & E. J. Duwell and N. C. Baenziger, Acta Cryst. 8, 705-710 (1955)
   1.00000000000000
  -2.57942904950000   4.04910374055000   4.38752883095000
   2.57942904950000  -4.04910374055000   4.38752883095000
   2.57942904950000   4.04910374055000  -4.38752883095000
    Hg     K
     4     2
Direct
   1.22300000000000   0.66300000000000   0.56000000000000   Hg   (8h)
   0.60300000000000   0.66300000000000  -0.06000000000000   Hg   (8h)
   0.39700000000000  -0.66300000000000   1.06000000000000   Hg   (8h)
  -1.22300000000000  -0.66300000000000  -0.56000000000000   Hg   (8h)
   0.20300000000000  -0.04700000000000   0.25000000000000    K   (4e)
   0.79700000000000   0.04700000000000   0.75000000000000    K   (4e)
\end{lstlisting}
{\phantomsection\label{A4B_oI20_74_beh_e_cif}}
{\hyperref[A4B_oI20_74_beh_e]{Al$_{4}$U ($D1_{b}$): A4B\_oI20\_74\_beh\_e}} - CIF
\begin{lstlisting}[numbers=none,language={mylang}]
# CIF file
data_findsym-output
_audit_creation_method FINDSYM

_chemical_name_mineral ''
_chemical_formula_sum 'Al4 U'

loop_
_publ_author_name
 'H. U. Borgstedt'
 'H. Wedemeyer'
_journal_year 1989
_publ_Section_title
;
 Gmelin Handbook of Inorganic Chemistry
;

_aflow_title 'Al$_{4}$U ($D1_{b}$) Structure'
_aflow_proto 'A4B_oI20_74_beh_e'
_aflow_params 'a,b/a,c/a,z_{2},z_{3},y_{4},z_{4}'
_aflow_params_values '4.39,1.42369020501,3.12528473804,-0.111,0.111,-0.033,0.314'
_aflow_Strukturbericht '$D1_{b}$'
_aflow_Pearson 'oI20'

_symmetry_space_group_name_H-M "I 21/m 21/m 21/a"
_symmetry_Int_Tables_number 74
 
_cell_length_a    4.39000
_cell_length_b    6.25000
_cell_length_c    13.72000
_cell_angle_alpha 90.00000
_cell_angle_beta  90.00000
_cell_angle_gamma 90.00000
 
loop_
_space_group_symop_id
_space_group_symop_operation_xyz
1 x,y,z
2 x,-y,-z
3 -x,y+1/2,-z
4 -x,-y+1/2,z
5 -x,-y,-z
6 -x,y,z
7 x,-y+1/2,z
8 x,y+1/2,-z
9 x+1/2,y+1/2,z+1/2
10 x+1/2,-y+1/2,-z+1/2
11 -x+1/2,y,-z+1/2
12 -x+1/2,-y,z+1/2
13 -x+1/2,-y+1/2,-z+1/2
14 -x+1/2,y+1/2,z+1/2
15 x+1/2,-y,z+1/2
16 x+1/2,y,-z+1/2
 
loop_
_atom_site_label
_atom_site_type_symbol
_atom_site_symmetry_multiplicity
_atom_site_Wyckoff_label
_atom_site_fract_x
_atom_site_fract_y
_atom_site_fract_z
_atom_site_occupancy
Al1 Al   4 b 0.00000 0.00000 0.50000 1.00000
Al2 Al   4 e 0.00000 0.25000 -0.11100 1.00000
U1  U    4 e 0.00000 0.25000 0.11100 1.00000
Al3 Al   8 h 0.00000 -0.03300 0.31400 1.00000
\end{lstlisting}
{\phantomsection\label{A4B_oI20_74_beh_e_poscar}}
{\hyperref[A4B_oI20_74_beh_e]{Al$_{4}$U ($D1_{b}$): A4B\_oI20\_74\_beh\_e}} - POSCAR
\begin{lstlisting}[numbers=none,language={mylang}]
A4B_oI20_74_beh_e & a,b/a,c/a,z2,z3,y4,z4 --params=4.39,1.42369020501,3.12528473804,-0.111,0.111,-0.033,0.314 & Imma D_{2h}^{28} #74 (be^2h) & oI20 & $D1_{b}$ & Al4U &  & H. U. Borgstedt and H. Wedemeyer, (1989)
   1.00000000000000
  -2.19500000000000   3.12500000000000   6.86000000000000
   2.19500000000000  -3.12500000000000   6.86000000000000
   2.19500000000000   3.12500000000000  -6.86000000000000
    Al     U
     8     2
Direct
   0.50000000000000   0.50000000000000   0.00000000000000   Al   (4b)
   0.00000000000000   0.50000000000000   0.50000000000000   Al   (4b)
   0.13900000000000  -0.11100000000000   0.25000000000000   Al   (4e)
   0.86100000000000   0.11100000000000   0.75000000000000   Al   (4e)
   0.28100000000000   0.31400000000000  -0.03300000000000   Al   (8h)
   0.84700000000000   0.31400000000000   0.53300000000000   Al   (8h)
   0.15300000000000  -0.31400000000000   0.46700000000000   Al   (8h)
  -0.28100000000000  -0.31400000000000   0.03300000000000   Al   (8h)
   0.36100000000000   0.11100000000000   0.25000000000000    U   (4e)
   0.63900000000000  -0.11100000000000   0.75000000000000    U   (4e)
\end{lstlisting}
{\phantomsection\label{AB2C12D4_tP76_75_2a2b_2d_12d_4d_cif}}
{\hyperref[AB2C12D4_tP76_75_2a2b_2d_12d_4d]{BaCr$_{2}$Ru$_{4}$O$_{12}$: AB2C12D4\_tP76\_75\_2a2b\_2d\_12d\_4d}} - CIF

{\phantomsection\label{AB2C12D4_tP76_75_2a2b_2d_12d_4d_poscar}}
{\hyperref[AB2C12D4_tP76_75_2a2b_2d_12d_4d]{BaCr$_{2}$Ru$_{4}$O$_{12}$: AB2C12D4\_tP76\_75\_2a2b\_2d\_12d\_4d}} - POSCAR

{\phantomsection\label{A2BC_tP16_76_2a_a_a_cif}}
{\hyperref[A2BC_tP16_76_2a_a_a]{LaRhC$_{2}$: A2BC\_tP16\_76\_2a\_a\_a}} - CIF
\begin{lstlisting}[numbers=none,language={mylang}]
# CIF file
data_findsym-output
_audit_creation_method FINDSYM

_chemical_name_mineral 'LaRhC2'
_chemical_formula_sum 'C2 La Rh'

loop_
_publ_author_name
 'A. O. Tsokol\''
 'O. I. Bodak'
 'E. P. Marusin'
 'V. E. Zavodnik'
_journal_name_full_name
;
 Kristallografiya, English title: Crystallography Reports
;
_journal_volume 33
_journal_year 1988
_journal_page_first 345
_journal_page_last 348
_publ_Section_title
;
 X-ray diffraction studies of ternary $R$RhC$_{2}$ ($R$ = La, Ce, Pr, Nd, Sm) compounds
;

# Found in Pearson's Crystal Data - Crystal Structure Database for Inorganic Compounds, 2013

_aflow_title 'LaRhC$_{2}$ Structure'
_aflow_proto 'A2BC_tP16_76_2a_a_a'
_aflow_params 'a,c/a,x_{1},y_{1},z_{1},x_{2},y_{2},z_{2},x_{3},y_{3},z_{3},x_{4},y_{4},z_{4}'
_aflow_params_values '3.9810644174,3.85606631503,0.143,0.173,0.1347,0.353,0.141,0.2027,0.3461,0.3475,0.5937,0.1519,0.1578,0.0'
_aflow_Strukturbericht 'None'
_aflow_Pearson 'tP16'

_cell_length_a    3.9810644174
_cell_length_b    3.9810644174
_cell_length_c    15.3512483979
_cell_angle_alpha 90.0000000000
_cell_angle_beta  90.0000000000
_cell_angle_gamma 90.0000000000
 
_symmetry_space_group_name_H-M "P 41"
_symmetry_Int_Tables_number 76
 
loop_
_space_group_symop_id
_space_group_symop_operation_xyz
1 x,y,z
2 -x,-y,z+1/2
3 -y,x,z+1/4
4 y,-x,z+3/4
 
loop_
_atom_site_label
_atom_site_type_symbol
_atom_site_symmetry_multiplicity
_atom_site_Wyckoff_label
_atom_site_fract_x
_atom_site_fract_y
_atom_site_fract_z
_atom_site_occupancy
C1  C    4 a 0.14300 0.17300 0.13470 1.00000
C2  C    4 a 0.35300 0.14100 0.20270 1.00000
La1 La   4 a 0.34610 0.34750 0.59370 1.00000
Rh1 Rh   4 a 0.15190 0.15780 0.00000 1.00000
\end{lstlisting}
{\phantomsection\label{A2BC_tP16_76_2a_a_a_poscar}}
{\hyperref[A2BC_tP16_76_2a_a_a]{LaRhC$_{2}$: A2BC\_tP16\_76\_2a\_a\_a}} - POSCAR
\begin{lstlisting}[numbers=none,language={mylang}]
A2BC_tP16_76_2a_a_a & a,c/a,x1,y1,z1,x2,y2,z2,x3,y3,z3,x4,y4,z4 --params=3.9810644174,3.85606631503,0.143,0.173,0.1347,0.353,0.141,0.2027,0.3461,0.3475,0.5937,0.1519,0.1578,0.0 & P4_{1} C_{4}^{2} #76 (a^4) & tP16 & None & LaRhC2 &  & A. O. Tsokol' et al., Kristallografiya 33, 345-348 (1988)
   1.00000000000000
   3.98106441740000   0.00000000000000   0.00000000000000
   0.00000000000000   3.98106441740000   0.00000000000000
   0.00000000000000   0.00000000000000  15.35124839790000
     C    La    Rh
     8     4     4
Direct
   0.14300000000000   0.17300000000000   0.13470000000000    C   (4a)
  -0.14300000000000  -0.17300000000000   0.63470000000000    C   (4a)
  -0.17300000000000   0.14300000000000   0.38470000000000    C   (4a)
   0.17300000000000  -0.14300000000000   0.88470000000000    C   (4a)
   0.35300000000000   0.14100000000000   0.20270000000000    C   (4a)
  -0.35300000000000  -0.14100000000000   0.70270000000000    C   (4a)
  -0.14100000000000   0.35300000000000   0.45270000000000    C   (4a)
   0.14100000000000  -0.35300000000000   0.95270000000000    C   (4a)
   0.34610000000000   0.34750000000000   0.59370000000000   La   (4a)
  -0.34610000000000  -0.34750000000000   1.09370000000000   La   (4a)
  -0.34750000000000   0.34610000000000   0.84370000000000   La   (4a)
   0.34750000000000  -0.34610000000000   1.34370000000000   La   (4a)
   0.15190000000000   0.15780000000000   0.00000000000000   Rh   (4a)
  -0.15190000000000  -0.15780000000000   0.50000000000000   Rh   (4a)
  -0.15780000000000   0.15190000000000   0.25000000000000   Rh   (4a)
   0.15780000000000  -0.15190000000000   0.75000000000000   Rh   (4a)
\end{lstlisting}
{\phantomsection\label{A3B7_tP40_76_3a_7a_cif}}
{\hyperref[A3B7_tP40_76_3a_7a]{Cs$_{3}$P$_{7}$: A3B7\_tP40\_76\_3a\_7a}} - CIF

{\phantomsection\label{A3B7_tP40_76_3a_7a_poscar}}
{\hyperref[A3B7_tP40_76_3a_7a]{Cs$_{3}$P$_{7}$: A3B7\_tP40\_76\_3a\_7a}} - POSCAR

{\phantomsection\label{A2B6CD7_tP64_77_2d_6d_d_ab6d_cif}}
{\hyperref[A2B6CD7_tP64_77_2d_6d_d_ab6d]{Pinnoite (MgB$_{2}$O(OH)$_{6}$): A2B6CD7\_tP64\_77\_2d\_6d\_d\_ab6d}} - CIF

{\phantomsection\label{A2B6CD7_tP64_77_2d_6d_d_ab6d_poscar}}
{\hyperref[A2B6CD7_tP64_77_2d_6d_d_ab6d]{Pinnoite (MgB$_{2}$O(OH)$_{6}$): A2B6CD7\_tP64\_77\_2d\_6d\_d\_ab6d}} - POSCAR

{\phantomsection\label{A2B_tP48_77_8d_4d_cif}}
{\hyperref[A2B_tP48_77_8d_4d]{H$_{2}$S III: A2B\_tP48\_77\_8d\_4d}} - CIF

{\phantomsection\label{A2B_tP48_77_8d_4d_poscar}}
{\hyperref[A2B_tP48_77_8d_4d]{H$_{2}$S III: A2B\_tP48\_77\_8d\_4d}} - POSCAR

{\phantomsection\label{A2B7C2_tP88_78_4a_14a_4a_cif}}
{\hyperref[A2B7C2_tP88_78_4a_14a_4a]{Sr$_{2}$As$_{2}$O$_{7}$: A2B7C2\_tP88\_78\_4a\_14a\_4a}} - CIF

{\phantomsection\label{A2B7C2_tP88_78_4a_14a_4a_poscar}}
{\hyperref[A2B7C2_tP88_78_4a_14a_4a]{Sr$_{2}$As$_{2}$O$_{7}$: A2B7C2\_tP88\_78\_4a\_14a\_4a}} - POSCAR

{\phantomsection\label{A2BC2_tI20_79_c_2a_c_cif}}
{\hyperref[A2BC2_tI20_79_c_2a_c]{TlZn$_{2}$Sb$_{2}$: A2BC2\_tI20\_79\_c\_2a\_c}} - CIF
\begin{lstlisting}[numbers=none,language={mylang}]
# CIF file
data_findsym-output
_audit_creation_method FINDSYM

_chemical_name_mineral 'TlZn2Sb2'
_chemical_formula_sum 'Sb2 Tl Zn2'

loop_
_publ_author_name
 'A. Czybulka'
 'B. Krenkel'
 'H.-U. Schuster'
_journal_name_full_name
;
 Journal of the Less-Common Metals
;
_journal_volume 137
_journal_year 1988
_journal_page_first 311
_journal_page_last 322
_publ_Section_title
;
 Tern{\"a}re zintl-Verbindungen mit thallium als elektronendonator
;

# Found in Pearson's Crystal Data - Crystal Structure Database for Inorganic Compounds, 2013

_aflow_title 'TlZn$_{2}$Sb$_{2}$ Structure'
_aflow_proto 'A2BC2_tI20_79_c_2a_c'
_aflow_params 'a,c/a,z_{1},z_{2},x_{3},y_{3},z_{3},x_{4},y_{4},z_{4}'
_aflow_params_values '8.6489709127,0.842525147414,0.0,0.4896,0.337,0.164,0.2196,0.1519,0.1578,0.0'
_aflow_Strukturbericht 'None'
_aflow_Pearson 'tI20'

_cell_length_a    8.6489709127
_cell_length_b    8.6489709127
_cell_length_c    7.2869754932
_cell_angle_alpha 90.0000000000
_cell_angle_beta  90.0000000000
_cell_angle_gamma 90.0000000000
 
_symmetry_space_group_name_H-M "I 4"
_symmetry_Int_Tables_number 79
 
loop_
_space_group_symop_id
_space_group_symop_operation_xyz
1 x,y,z
2 -x,-y,z
3 -y,x,z
4 y,-x,z
5 x+1/2,y+1/2,z+1/2
6 -x+1/2,-y+1/2,z+1/2
7 -y+1/2,x+1/2,z+1/2
8 y+1/2,-x+1/2,z+1/2
 
loop_
_atom_site_label
_atom_site_type_symbol
_atom_site_symmetry_multiplicity
_atom_site_Wyckoff_label
_atom_site_fract_x
_atom_site_fract_y
_atom_site_fract_z
_atom_site_occupancy
Tl1 Tl   2 a 0.00000 0.00000 0.00000 1.00000
Tl2 Tl   2 a 0.00000 0.00000 0.48960 1.00000
Sb1 Sb   8 c 0.33700 0.16400 0.21960 1.00000
Zn1 Zn   8 c 0.15190 0.15780 0.00000 1.00000
\end{lstlisting}
{\phantomsection\label{A2BC2_tI20_79_c_2a_c_poscar}}
{\hyperref[A2BC2_tI20_79_c_2a_c]{TlZn$_{2}$Sb$_{2}$: A2BC2\_tI20\_79\_c\_2a\_c}} - POSCAR
\begin{lstlisting}[numbers=none,language={mylang}]
A2BC2_tI20_79_c_2a_c & a,c/a,z1,z2,x3,y3,z3,x4,y4,z4 --params=8.6489709127,0.842525147414,0.0,0.4896,0.337,0.164,0.2196,0.1519,0.1578,0.0 & I4 C_{4}^{5} #79 (a^2c^2) & tI20 & None & TlZn2Sb2 &  & A. Czybulka and B. Krenkel and H.-U. Schuster, J. Less-Common Met. 137, 311-322 (1988)
   1.00000000000000
  -4.32448545635000   4.32448545635000   3.64348774660000
   4.32448545635000  -4.32448545635000   3.64348774660000
   4.32448545635000   4.32448545635000  -3.64348774660000
    Sb    Tl    Zn
     4     2     4
Direct
   0.38360000000000   0.55660000000000   0.50100000000000   Sb   (8c)
   0.05560000000000  -0.11740000000000  -0.50100000000000   Sb   (8c)
   0.55660000000000   0.05560000000000   0.17300000000000   Sb   (8c)
  -0.11740000000000   0.38360000000000  -0.17300000000000   Sb   (8c)
   0.00000000000000   0.00000000000000   0.00000000000000   Tl   (2a)
   0.48960000000000   0.48960000000000   0.00000000000000   Tl   (2a)
   0.15780000000000   0.15190000000000   0.30970000000000   Zn   (8c)
  -0.15780000000000  -0.15190000000000  -0.30970000000000   Zn   (8c)
   0.15190000000000  -0.15780000000000  -0.00590000000000   Zn   (8c)
  -0.15190000000000   0.15780000000000   0.00590000000000   Zn   (8c)
\end{lstlisting}
{\phantomsection\label{AB2_tI48_80_2b_4b_cif}}
{\hyperref[AB2_tI48_80_2b_4b]{$\beta$-NbO$_{2}$: AB2\_tI48\_80\_2b\_4b}} - CIF
\begin{lstlisting}[numbers=none,language={mylang}]
# CIF file 
data_findsym-output
_audit_creation_method FINDSYM

_chemical_name_mineral '$\beta$-NbO$_{2}$'
_chemical_formula_sum 'Nb O2'

loop_
_publ_author_name
 'H.-J. Schweizer'
 'R. Gruehn'
_journal_name_full_name
;
 Zeitschrift f{\"u}r Naturforschung B
;
_journal_volume 37
_journal_year 1982
_journal_page_first 1361
_journal_page_last 1368
_publ_Section_title
;
 Zur Darstellung und Kristallstruktur von $\beta$-NbO$_{2}$ / Synthesis and Crystal Structure of $\beta$-NbO$_{2}$
;

# Found in Pearson's Handbook of Crystallographic Data, 1991

_aflow_title '$\beta$-NbO$_{2}$ Structure'
_aflow_proto 'AB2_tI48_80_2b_4b'
_aflow_params 'a,c/a,x_{1},y_{1},z_{1},x_{2},y_{2},z_{2},x_{3},y_{3},z_{3},x_{4},y_{4},z_{4},x_{5},y_{5},z_{5},x_{6},y_{6},z_{6}'
_aflow_params_values '9.693,0.617455896007,0.2621,0.5076,0.0299,0.2455,0.4909,0.4804,0.3974,0.1497,0.0077,0.1102,0.3642,-0.0098,0.6086,0.3609,0.5064,0.65,0.1038,0.2484'
_aflow_Strukturbericht 'None'
_aflow_Pearson 'tI48'

_symmetry_space_group_name_H-M "I 41"
_symmetry_Int_Tables_number 80
 
_cell_length_a    9.69300
_cell_length_b    9.69300
_cell_length_c    5.98500
_cell_angle_alpha 90.00000
_cell_angle_beta  90.00000
_cell_angle_gamma 90.00000
 
loop_
_space_group_symop_id
_space_group_symop_operation_xyz
1 x,y,z
2 -x,-y,z
3 -y,x+1/2,z+1/4
4 y,-x+1/2,z+1/4
5 x+1/2,y+1/2,z+1/2
6 -x+1/2,-y+1/2,z+1/2
7 -y+1/2,x,z+3/4
8 y+1/2,-x,z+3/4
 
loop_
_atom_site_label
_atom_site_type_symbol
_atom_site_symmetry_multiplicity
_atom_site_Wyckoff_label
_atom_site_fract_x
_atom_site_fract_y
_atom_site_fract_z
_atom_site_occupancy
Nb1 Nb   8 b 0.26210 0.50760 0.02990 1.00000
Nb2 Nb   8 b 0.24550 0.49090 0.48040 1.00000
O1  O    8 b 0.39740 0.14970 0.00770 1.00000
O2  O    8 b 0.11020 0.36420 -0.00980 1.00000
O3  O    8 b 0.60860 0.36090 0.50640 1.00000
O4  O    8 b 0.65000 0.10380 0.24840 1.00000
\end{lstlisting}
{\phantomsection\label{AB2_tI48_80_2b_4b_poscar}}
{\hyperref[AB2_tI48_80_2b_4b]{$\beta$-NbO$_{2}$: AB2\_tI48\_80\_2b\_4b}} - POSCAR

{\phantomsection\label{AB2_tP12_81_adg_2h_cif}}
{\hyperref[AB2_tP12_81_adg_2h]{GeSe$_{2}$ (High-pressure): AB2\_tP12\_81\_adg\_2h}} - CIF
\begin{lstlisting}[numbers=none,language={mylang}]
# CIF file
data_findsym-output
_audit_creation_method FINDSYM

_chemical_name_mineral 'GeSe2'
_chemical_formula_sum 'Ge Se2'

loop_
_publ_author_name
 'A. Grzechnik'
 'S. St{\o}len'
 'E. Bakken'
 'T. Grande'
 'M. Mezouar'
_journal_name_full_name
;
 Journal of Solid State Chemistry
;
_journal_volume 150
_journal_year 2000
_journal_page_first 121
_journal_page_last 127
_publ_Section_title
;
 Structural transformations in three-dimensional crystalline GeSe$_{2}$ at high pressures and high temperatures
;

# Found in Pearson's Crystal Data - Crystal Structure Database for Inorganic Compounds, 2013

_aflow_title 'GeSe$_{2}$ (High-pressure) Structure'
_aflow_proto 'AB2_tP12_81_adg_2h'
_aflow_params 'a,c/a,z_{3},x_{4},y_{4},z_{4},x_{5},y_{5},z_{5}'
_aflow_params_values '5.3391020438,1.87980670176,0.25,0.2739,0.234,0.128,0.2289,0.23,0.6373'
_aflow_Strukturbericht 'None'
_aflow_Pearson 'tP12'

_cell_length_a    5.3391020438
_cell_length_b    5.3391020438
_cell_length_c    10.0364798033
_cell_angle_alpha 90.0000000000
_cell_angle_beta  90.0000000000
_cell_angle_gamma 90.0000000000
 
_symmetry_space_group_name_H-M "P -4"
_symmetry_Int_Tables_number 81
 
loop_
_space_group_symop_id
_space_group_symop_operation_xyz
1 x,y,z
2 -x,-y,z
3 y,-x,-z
4 -y,x,-z
 
loop_
_atom_site_label
_atom_site_type_symbol
_atom_site_symmetry_multiplicity
_atom_site_Wyckoff_label
_atom_site_fract_x
_atom_site_fract_y
_atom_site_fract_z
_atom_site_occupancy
Ge1 Ge   1 a 0.00000 0.00000 0.00000 1.00000
Ge2 Ge   1 d 0.50000 0.50000 0.50000 1.00000
Ge3 Ge   2 g 0.00000 0.50000 0.25000 1.00000
Se1 Se   4 h 0.27390 0.23400 0.12800 1.00000
Se2 Se   4 h 0.22890 0.23000 0.63730 1.00000
\end{lstlisting}
{\phantomsection\label{AB2_tP12_81_adg_2h_poscar}}
{\hyperref[AB2_tP12_81_adg_2h]{GeSe$_{2}$ (High-pressure): AB2\_tP12\_81\_adg\_2h}} - POSCAR
\begin{lstlisting}[numbers=none,language={mylang}]
AB2_tP12_81_adg_2h & a,c/a,z3,x4,y4,z4,x5,y5,z5 --params=5.3391020438,1.87980670176,0.25,0.2739,0.234,0.128,0.2289,0.23,0.6373 & P-4 S_{4}^{1} #81 (adgh^2) & tP12 & None & GeSe2 &  & A. Grzechnik et al., J. Solid State Chem. 150, 121-127 (2000)
   1.00000000000000
   5.33910204380000   0.00000000000000   0.00000000000000
   0.00000000000000   5.33910204380000   0.00000000000000
   0.00000000000000   0.00000000000000  10.03647980330000
    Ge    Se
     4     8
Direct
   0.00000000000000   0.00000000000000   0.00000000000000   Ge   (1a)
   0.50000000000000   0.50000000000000   0.50000000000000   Ge   (1d)
   0.00000000000000   0.50000000000000   0.25000000000000   Ge   (2g)
   0.50000000000000   0.00000000000000  -0.25000000000000   Ge   (2g)
   0.27390000000000   0.23400000000000   0.12800000000000   Se   (4h)
  -0.27390000000000  -0.23400000000000   0.12800000000000   Se   (4h)
   0.23400000000000  -0.27390000000000  -0.12800000000000   Se   (4h)
  -0.23400000000000   0.27390000000000  -0.12800000000000   Se   (4h)
   0.22890000000000   0.23000000000000   0.63730000000000   Se   (4h)
  -0.22890000000000  -0.23000000000000   0.63730000000000   Se   (4h)
   0.23000000000000  -0.22890000000000  -0.63730000000000   Se   (4h)
  -0.23000000000000   0.22890000000000  -0.63730000000000   Se   (4h)
\end{lstlisting}
{\phantomsection\label{A3B_tI32_82_3g_g_cif}}
{\hyperref[A3B_tI32_82_3g_g]{Ni$_{3}$P ($D0_{e}$): A3B\_tI32\_82\_3g\_g}} - CIF
\begin{lstlisting}[numbers=none,language={mylang}]
# CIF file 
data_findsym-output
_audit_creation_method FINDSYM

_chemical_name_mineral ''
_chemical_formula_sum 'Ni3 P'

loop_
_publ_author_name
 'S. Rundqvist'
 'E. Hassler'
 'L. Lundvik'
_journal_name_full_name
;
 Acta Chemica Scandinavica
;
_journal_volume 16
_journal_year 1962
_journal_page_first 242
_journal_page_last 243
_publ_Section_title
;
 Refinement of the Ni$_{3}$P Structure
;

_aflow_title 'Ni$_{3}$P ($D0_{e}$) Structure'
_aflow_proto 'A3B_tI32_82_3g_g'
_aflow_params 'a,c/a,x_{1},y_{1},z_{1},x_{2},y_{2},z_{2},x_{3},y_{3},z_{3},x_{4},y_{4},z_{4}'
_aflow_params_values '8.954,0.495711413893,0.0775,0.1117,0.2391,0.3649,0.0321,0.9765,0.1689,0.22,0.7524,0.2862,0.0487,0.4807'
_aflow_Strukturbericht '$D0_{e}$'
_aflow_Pearson 'tI32'

_symmetry_space_group_name_H-M "I-4"
_symmetry_Int_Tables_number 82
 
_cell_length_a    8.95400
_cell_length_b    8.95400
_cell_length_c    4.43860
_cell_angle_alpha 90.00000
_cell_angle_beta  90.00000
_cell_angle_gamma 90.00000
 
loop_
_space_group_symop_id
_space_group_symop_operation_xyz
1 x,y,z
2 -x,-y,z
3 y,-x,-z
4 -y,x,-z
5 x+1/2,y+1/2,z+1/2
6 -x+1/2,-y+1/2,z+1/2
7 y+1/2,-x+1/2,-z+1/2
8 -y+1/2,x+1/2,-z+1/2
 
loop_
_atom_site_label
_atom_site_type_symbol
_atom_site_symmetry_multiplicity
_atom_site_Wyckoff_label
_atom_site_fract_x
_atom_site_fract_y
_atom_site_fract_z
_atom_site_occupancy
Ni1 Ni   8 g  0.07750 0.11170 0.23910 1.00000
Ni2 Ni   8 g  0.36490 0.03210 0.97650 1.00000
Ni3 Ni   8 g  0.16890 0.22000 0.75240 1.00000
P1  P    8 g  0.28620 0.04870 0.48070 1.00000
\end{lstlisting}
{\phantomsection\label{A3B_tI32_82_3g_g_poscar}}
{\hyperref[A3B_tI32_82_3g_g]{Ni$_{3}$P ($D0_{e}$): A3B\_tI32\_82\_3g\_g}} - POSCAR
\begin{lstlisting}[numbers=none,language={mylang}]
A3B_tI32_82_3g_g & a,c/a,x1,y1,z1,x2,y2,z2,x3,y3,z3,x4,y4,z4 --params=8.954,0.495711413893,0.0775,0.1117,0.2391,0.3649,0.0321,0.9765,0.1689,0.22,0.7524,0.2862,0.0487,0.4807 & I-4 S_{4}^{2} #82 (g^4) & tI32 & $D0_{e}$ & Ni3P &  & S. Rundqvist and E. Hassler and L. Lundvik, Acta Chem. Scand. 16, 242-243 (1962)
   1.00000000000000
  -4.47700000000000   4.47700000000000   2.21930000000000
   4.47700000000000  -4.47700000000000   2.21930000000000
   4.47700000000000   4.47700000000000  -2.21930000000000
    Ni     P
    12     4
Direct
   0.35080000000000   0.31660000000000   0.18920000000000   Ni   (8g)
   0.12740000000000   0.16160000000000  -0.18920000000000   Ni   (8g)
  -0.31660000000000  -0.12740000000000   0.03420000000000   Ni   (8g)
  -0.16160000000000  -0.35080000000000  -0.03420000000000   Ni   (8g)
   1.00860000000000   1.34140000000000   0.39700000000000   Ni   (8g)
   0.94440000000000   0.61160000000000  -0.39700000000000   Ni   (8g)
  -1.34140000000000  -0.94440000000000  -0.33280000000000   Ni   (8g)
  -0.61160000000000  -1.00860000000000   0.33280000000000   Ni   (8g)
   0.97240000000000   0.92130000000000   0.38890000000000   Ni   (8g)
   0.53240000000000   0.58350000000000  -0.38890000000000   Ni   (8g)
  -0.92130000000000  -0.53240000000000   0.05110000000000   Ni   (8g)
  -0.58350000000000  -0.97240000000000  -0.05110000000000   Ni   (8g)
   0.52940000000000   0.76690000000000   0.33490000000000    P   (8g)
   0.43200000000000   0.19450000000000  -0.33490000000000    P   (8g)
  -0.76690000000000  -0.43200000000000  -0.23750000000000    P   (8g)
  -0.19450000000000  -0.52940000000000   0.23750000000000    P   (8g)
\end{lstlisting}
{\phantomsection\label{A3B2_tP10_83_adk_j_cif}}
{\hyperref[A3B2_tP10_83_adk_j]{Ti$_{2}$Ge$_{3}$: A3B2\_tP10\_83\_adk\_j}} - CIF
\begin{lstlisting}[numbers=none,language={mylang}]
# CIF file
data_findsym-output
_audit_creation_method FINDSYM

_chemical_name_mineral 'Ti2Ge3'
_chemical_formula_sum 'Ge3 Ti2'

loop_
_publ_author_name
 'K. Schubert'
 'H. G. Meissner'
 'M. P{\"o}tzschke'
 'W. Rossteutscher'
 'E. Stolz'
_journal_name_full_name
;
 Naturwissenschaften
;
_journal_volume 49
_journal_year 1962
_journal_page_first 57
_journal_page_last 57
_publ_Section_title
;
 Einige Strukturdaten metallischer Phasen (7)
;

# Found in Pearson's Crystal Data - Crystal Structure Database for Inorganic Compounds, 2013

_aflow_title 'Ti$_{2}$Ge$_{3}$ Structure'
_aflow_proto 'A3B2_tP10_83_adk_j'
_aflow_params 'a,c/a,x_{3},y_{3},x_{4},y_{4}'
_aflow_params_values '6.2840064744,0.638128580522,0.375,0.191,0.109,0.314'
_aflow_Strukturbericht 'None'
_aflow_Pearson 'tP10'

_cell_length_a    6.2840064744
_cell_length_b    6.2840064744
_cell_length_c    4.0100041315
_cell_angle_alpha 90.0000000000
_cell_angle_beta  90.0000000000
_cell_angle_gamma 90.0000000000
 
_symmetry_space_group_name_H-M "P 4/m"
_symmetry_Int_Tables_number 83
 
loop_
_space_group_symop_id
_space_group_symop_operation_xyz
1 x,y,z
2 -x,-y,z
3 -y,x,z
4 y,-x,z
5 -x,-y,-z
6 x,y,-z
7 y,-x,-z
8 -y,x,-z
 
loop_
_atom_site_label
_atom_site_type_symbol
_atom_site_symmetry_multiplicity
_atom_site_Wyckoff_label
_atom_site_fract_x
_atom_site_fract_y
_atom_site_fract_z
_atom_site_occupancy
Ge1 Ge   1 a 0.00000 0.00000 0.00000 1.00000
Ge2 Ge   1 d 0.50000 0.50000 0.50000 1.00000
Ti1 Ti   4 j 0.37500 0.19100 0.00000 1.00000
Ge3 Ge   4 k 0.10900 0.31400 0.50000 1.00000
\end{lstlisting}
{\phantomsection\label{A3B2_tP10_83_adk_j_poscar}}
{\hyperref[A3B2_tP10_83_adk_j]{Ti$_{2}$Ge$_{3}$: A3B2\_tP10\_83\_adk\_j}} - POSCAR
\begin{lstlisting}[numbers=none,language={mylang}]
A3B2_tP10_83_adk_j & a,c/a,x3,y3,x4,y4 --params=6.2840064744,0.638128580522,0.375,0.191,0.109,0.314 & P4/m C_{4h}^{1} #83 (adjk) & tP10 & None & Ti2Ge3 &  & K. Schubert et al., {Naturwissenschaften 49, 57-57 (1962)
   1.00000000000000
   6.28400647440000   0.00000000000000   0.00000000000000
   0.00000000000000   6.28400647440000   0.00000000000000
   0.00000000000000   0.00000000000000   4.01000413150000
    Ge    Ti
     6     4
Direct
   0.00000000000000   0.00000000000000   0.00000000000000   Ge   (1a)
   0.50000000000000   0.50000000000000   0.50000000000000   Ge   (1d)
   0.10900000000000   0.31400000000000   0.50000000000000   Ge   (4k)
  -0.10900000000000  -0.31400000000000   0.50000000000000   Ge   (4k)
  -0.31400000000000   0.10900000000000   0.50000000000000   Ge   (4k)
   0.31400000000000  -0.10900000000000   0.50000000000000   Ge   (4k)
   0.37500000000000   0.19100000000000   0.00000000000000   Ti   (4j)
  -0.37500000000000  -0.19100000000000   0.00000000000000   Ti   (4j)
  -0.19100000000000   0.37500000000000   0.00000000000000   Ti   (4j)
   0.19100000000000  -0.37500000000000   0.00000000000000   Ti   (4j)
\end{lstlisting}
{\phantomsection\label{A2B_tP30_85_ab2g_cg_cif}}
{\hyperref[A2B_tP30_85_ab2g_cg]{SrBr$_{2}$: A2B\_tP30\_85\_ab2g\_cg}} - CIF
\begin{lstlisting}[numbers=none,language={mylang}]
# CIF file
data_findsym-output
_audit_creation_method FINDSYM

_chemical_name_mineral 'SrBr2'
_chemical_formula_sum 'Br2 Sr'

loop_
_publ_author_name
 'B. Frit'
 'M. M. Chbany'
_journal_name_full_name
;
 Journal of Inorganic and Nuclear Chemistry
;
_journal_volume 31
_journal_year 1969
_journal_page_first 2685
_journal_page_last 2693
_publ_Section_title
;
 Les halogeno-carbonates de strontium
;

# Found in Pearson's Crystal Data - Crystal Structure Database for Inorganic Compounds, 2013

_aflow_title 'SrBr$_{2}$ Structure'
_aflow_proto 'A2B_tP30_85_ab2g_cg'
_aflow_params 'a,c/a,z_{3},x_{4},y_{4},z_{4},x_{5},y_{5},z_{5},x_{6},y_{6},z_{6}'
_aflow_params_values '11.6179902668,0.614391461525,0.6517,0.5428,0.6612,0.5963,0.6531,0.541,0.1258,0.5856,0.1045,0.2524'
_aflow_Strukturbericht 'None'
_aflow_Pearson 'tP30'

_cell_length_a    11.6179902668
_cell_length_b    11.6179902668
_cell_length_c    7.1379940200
_cell_angle_alpha 90.0000000000
_cell_angle_beta  90.0000000000
_cell_angle_gamma 90.0000000000
 
_symmetry_space_group_name_H-M "P 4/n (origin choice 2)"
_symmetry_Int_Tables_number 85
 
loop_
_space_group_symop_id
_space_group_symop_operation_xyz
1 x,y,z
2 -x+1/2,-y+1/2,z
3 -y+1/2,x,z
4 y,-x+1/2,z
5 -x,-y,-z
6 x+1/2,y+1/2,-z
7 y+1/2,-x,-z
8 -y,x+1/2,-z
 
loop_
_atom_site_label
_atom_site_type_symbol
_atom_site_symmetry_multiplicity
_atom_site_Wyckoff_label
_atom_site_fract_x
_atom_site_fract_y
_atom_site_fract_z
_atom_site_occupancy
Br1 Br   2 a 0.25000 0.75000 0.00000 1.00000
Br2 Br   2 b 0.25000 0.75000 0.50000 1.00000
Sr1 Sr   2 c 0.25000 0.25000 0.65170 1.00000
Br3 Br   8 g 0.54280 0.66120 0.59630 1.00000
Br4 Br   8 g 0.65310 0.54100 0.12580 1.00000
Sr2 Sr   8 g 0.58560 0.10450 0.25240 1.00000
\end{lstlisting}
{\phantomsection\label{A2B_tP30_85_ab2g_cg_poscar}}
{\hyperref[A2B_tP30_85_ab2g_cg]{SrBr$_{2}$: A2B\_tP30\_85\_ab2g\_cg}} - POSCAR

{\phantomsection\label{AB3_tP32_86_g_3g_cif}}
{\hyperref[AB3_tP32_86_g_3g]{Ti$_{3}$P: AB3\_tP32\_86\_g\_3g}} - CIF
\begin{lstlisting}[numbers=none,language={mylang}]
# CIF file
data_findsym-output
_audit_creation_method FINDSYM

_chemical_name_mineral 'Ti3P'
_chemical_formula_sum 'P Ti3'

loop_
_publ_author_name
 'V. N. Eremenko'
 'V. E. Listovnichii'
_journal_year 1965
_publ_Section_title
;
 State diagram of the Ti-P system
;

# Found in Pearson's Crystal Data - Crystal Structure Database for Inorganic Compounds, 2013

_aflow_title 'Ti$_{3}$P Structure'
_aflow_proto 'AB3_tP32_86_g_3g'
_aflow_params 'a,c/a,x_{1},y_{1},z_{1},x_{2},y_{2},z_{2},x_{3},y_{3},z_{3},x_{4},y_{4},z_{4}'
_aflow_params_values '9.997999333,0.499899979999,0.0439,0.20812,0.5354,0.11009,0.22151,0.0295,0.14275,0.66613,0.7153,0.53342,0.06957,0.7593'
_aflow_Strukturbericht 'None'
_aflow_Pearson 'tP32'

_cell_length_a    9.9979993330
_cell_length_b    9.9979993330
_cell_length_c    4.9979996666
_cell_angle_alpha 90.0000000000
_cell_angle_beta  90.0000000000
_cell_angle_gamma 90.0000000000
 
_symmetry_space_group_name_H-M "P 42/n (origin choice 2)"
_symmetry_Int_Tables_number 86
 
loop_
_space_group_symop_id
_space_group_symop_operation_xyz
1 x,y,z
2 -x+1/2,-y+1/2,z
3 -y,x+1/2,z+1/2
4 y+1/2,-x,z+1/2
5 -x,-y,-z
6 x+1/2,y+1/2,-z
7 y,-x+1/2,-z+1/2
8 -y+1/2,x,-z+1/2
 
loop_
_atom_site_label
_atom_site_type_symbol
_atom_site_symmetry_multiplicity
_atom_site_Wyckoff_label
_atom_site_fract_x
_atom_site_fract_y
_atom_site_fract_z
_atom_site_occupancy
P1  P    8 g 0.04390 0.20812 0.53540 1.00000
Ti1 Ti   8 g 0.11009 0.22151 0.02950 1.00000
Ti2 Ti   8 g 0.14275 0.66613 0.71530 1.00000
Ti3 Ti   8 g 0.53342 0.06957 0.75930 1.00000
\end{lstlisting}
{\phantomsection\label{AB3_tP32_86_g_3g_poscar}}
{\hyperref[AB3_tP32_86_g_3g]{Ti$_{3}$P: AB3\_tP32\_86\_g\_3g}} - POSCAR

{\phantomsection\label{A4B_tI20_88_f_a_cif}}
{\hyperref[A4B_tI20_88_f_a]{ThCl$_{4}$: A4B\_tI20\_88\_f\_a}} - CIF
\begin{lstlisting}[numbers=none,language={mylang}]
# CIF file
data_findsym-output
_audit_creation_method FINDSYM

_chemical_name_mineral 'ThCl4'
_chemical_formula_sum 'Cl4 Th'

loop_
_publ_author_name
 'J. T. Mason'
 'M. C. Jha'
 'P. Chiotti'
_journal_name_full_name
;
 Journal of the Less-Common Metals
;
_journal_volume 34
_journal_year 1974
_journal_page_first 143
_journal_page_last 151
_publ_Section_title
;
 Crystal structures of ThCl$_{4}$ polymorphs
;

# Found in Pearson's Crystal Data - Crystal Structure Database for Inorganic Compounds, 2013

_aflow_title 'ThCl$_{4}$ Structure'
_aflow_proto 'A4B_tI20_88_f_a'
_aflow_params 'a,c/a,x_{2},y_{2},z_{2}'
_aflow_params_values '6.407994829,2.01685393259,0.147,0.017,0.298'
_aflow_Strukturbericht 'None'
_aflow_Pearson 'tI20'

_cell_length_a    6.4079948290
_cell_length_b    6.4079948290
_cell_length_c    12.9239895709
_cell_angle_alpha 90.0000000000
_cell_angle_beta  90.0000000000
_cell_angle_gamma 90.0000000000
 
_symmetry_space_group_name_H-M "I 41/a (origin choice 2)"
_symmetry_Int_Tables_number 88
 
loop_
_space_group_symop_id
_space_group_symop_operation_xyz
1 x,y,z
2 -x,-y+1/2,z
3 -y+3/4,x+1/4,z+1/4
4 y+1/4,-x+1/4,z+1/4
5 -x,-y,-z
6 x,y+1/2,-z
7 y+1/4,-x+3/4,-z+3/4
8 -y+3/4,x+3/4,-z+3/4
9 x+1/2,y+1/2,z+1/2
10 -x+1/2,-y,z+1/2
11 -y+1/4,x+3/4,z+3/4
12 y+3/4,-x+3/4,z+3/4
13 -x+1/2,-y+1/2,-z+1/2
14 x+1/2,y,-z+1/2
15 y+3/4,-x+1/4,-z+1/4
16 -y+1/4,x+1/4,-z+1/4
 
loop_
_atom_site_label
_atom_site_type_symbol
_atom_site_symmetry_multiplicity
_atom_site_Wyckoff_label
_atom_site_fract_x
_atom_site_fract_y
_atom_site_fract_z
_atom_site_occupancy
Th1 Th   4 a 0.00000 0.25000 0.12500 1.00000
Cl1 Cl  16 f 0.14700 0.01700 0.29800 1.00000
\end{lstlisting}
{\phantomsection\label{A4B_tI20_88_f_a_poscar}}
{\hyperref[A4B_tI20_88_f_a]{ThCl$_{4}$: A4B\_tI20\_88\_f\_a}} - POSCAR
\begin{lstlisting}[numbers=none,language={mylang}]
A4B_tI20_88_f_a & a,c/a,x2,y2,z2 --params=6.407994829,2.01685393259,0.147,0.017,0.298 & I4_{1}/a C_{4h}^{6} #88 (af) & tI20 & None & ThCl4 &  & J. T. Mason and M. C. Jha and P. Chiotti, J. Less-Common Met. 34, 143-151 (1974)
   1.00000000000000
  -3.20399741450000   3.20399741450000   6.46199478545000
   3.20399741450000  -3.20399741450000   6.46199478545000
   3.20399741450000   3.20399741450000  -6.46199478545000
    Cl    Th
     8     2
Direct
   0.31500000000000   0.44500000000000   0.16400000000000   Cl  (16f)
   0.78100000000000   0.15100000000000   0.33600000000000   Cl  (16f)
   0.94500000000000   0.28100000000000   0.13000000000000   Cl  (16f)
   0.65100000000000   0.81500000000000   0.37000000000000   Cl  (16f)
  -0.31500000000000  -0.44500000000000  -0.16400000000000   Cl  (16f)
   0.21900000000000  -0.15100000000000   0.66400000000000   Cl  (16f)
   0.05500000000000  -0.28100000000000  -0.13000000000000   Cl  (16f)
   0.34900000000000   0.18500000000000   0.63000000000000   Cl  (16f)
   0.37500000000000   0.12500000000000   0.25000000000000   Th   (4a)
   0.62500000000000   0.87500000000000   0.75000000000000   Th   (4a)
\end{lstlisting}
{\phantomsection\label{AB2_tI96_88_2f_4f_cif}}
{\hyperref[AB2_tI96_88_2f_4f]{$\alpha$-NbO$_{2}$: AB2\_tI96\_88\_2f\_4f}} - CIF

{\phantomsection\label{AB2_tI96_88_2f_4f_poscar}}
{\hyperref[AB2_tI96_88_2f_4f]{$\alpha$-NbO$_{2}$: AB2\_tI96\_88\_2f\_4f}} - POSCAR

{\phantomsection\label{A17BC4D_tP184_89_17p_p_4p_io_cif}}
{\hyperref[A17BC4D_tP184_89_17p_p_4p_io]{C$_{17}$FeO$_{4}$Pt: A17BC4D\_tP184\_89\_17p\_p\_4p\_io}} - CIF

{\phantomsection\label{A17BC4D_tP184_89_17p_p_4p_io_poscar}}
{\hyperref[A17BC4D_tP184_89_17p_p_4p_io]{C$_{17}$FeO$_{4}$Pt: A17BC4D\_tP184\_89\_17p\_p\_4p\_io}} - POSCAR

{\phantomsection\label{A4B2C13D_tP40_90_g_d_cef2g_c_cif}}
{\hyperref[A4B2C13D_tP40_90_g_d_cef2g_c]{Na$_{4}$Ti$_{2}$Si$_{8}$O$_{22}$[H$_{2}$O]$_{4}$: A4B2C13D\_tP40\_90\_g\_d\_cef2g\_c}} - CIF

{\phantomsection\label{A4B2C13D_tP40_90_g_d_cef2g_c_poscar}}
{\hyperref[A4B2C13D_tP40_90_g_d_cef2g_c]{Na$_{4}$Ti$_{2}$Si$_{8}$O$_{22}$[H$_{2}$O]$_{4}$: A4B2C13D\_tP40\_90\_g\_d\_cef2g\_c}} - POSCAR

{\phantomsection\label{AB4C17D4E_tP54_90_a_g_c4g_g_c_cif}}
{\hyperref[AB4C17D4E_tP54_90_a_g_c4g_g_c]{BaCu$_{4}$[VO][PO$_{4}$]$_{4}$: AB4C17D4E\_tP54\_90\_a\_g\_c4g\_g\_c}} - CIF

{\phantomsection\label{AB4C17D4E_tP54_90_a_g_c4g_g_c_poscar}}
{\hyperref[AB4C17D4E_tP54_90_a_g_c4g_g_c]{BaCu$_{4}$[VO][PO$_{4}$]$_{4}$: AB4C17D4E\_tP54\_90\_a\_g\_c4g\_g\_c}} - POSCAR

{\phantomsection\label{ABC_tP24_91_d_d_d_cif}}
{\hyperref[ABC_tP24_91_d_d_d]{ThBC: ABC\_tP24\_91\_d\_d\_d}} - CIF
\begin{lstlisting}[numbers=none,language={mylang}]
# CIF file
data_findsym-output
_audit_creation_method FINDSYM

_chemical_name_mineral 'ThBC'
_chemical_formula_sum 'B C Th'

loop_
_publ_author_name
 'P. Rogl'
_journal_name_full_name
;
 Journal of Nuclear Materials
;
_journal_volume 73
_journal_year 1978
_journal_page_first 198
_journal_page_last 203
_publ_Section_title
;
 The crystal structure of ThBC
;

# Found in Pearson's Crystal Data - Crystal Structure Database for Inorganic Compounds, 2013

_aflow_title 'ThBC Structure'
_aflow_proto 'ABC_tP24_91_d_d_d'
_aflow_params 'a,c/a,x_{1},y_{1},z_{1},x_{2},y_{2},z_{2},x_{3},y_{3},z_{3}'
_aflow_params_values '3.7620082462,6.71079213192,0.303,0.202,0.019,0.296,0.189,0.08,0.2975,0.1983,0.1795'
_aflow_Strukturbericht 'None'
_aflow_Pearson 'tP24'

_cell_length_a    3.7620082462
_cell_length_b    3.7620082462
_cell_length_c    25.2460553388
_cell_angle_alpha 90.0000000000
_cell_angle_beta  90.0000000000
_cell_angle_gamma 90.0000000000
 
_symmetry_space_group_name_H-M "P 41 2 2"
_symmetry_Int_Tables_number 91
 
loop_
_space_group_symop_id
_space_group_symop_operation_xyz
1 x,y,z
2 x,-y,-z+1/2
3 -x,y,-z
4 -x,-y,z+1/2
5 -y,-x,-z+1/4
6 -y,x,z+1/4
7 y,-x,z+3/4
8 y,x,-z+3/4
 
loop_
_atom_site_label
_atom_site_type_symbol
_atom_site_symmetry_multiplicity
_atom_site_Wyckoff_label
_atom_site_fract_x
_atom_site_fract_y
_atom_site_fract_z
_atom_site_occupancy
B1  B    8 d 0.30300 0.20200 0.01900 1.00000
C1  C    8 d 0.29600 0.18900 0.08000 1.00000
Th1 Th   8 d 0.29750 0.19830 0.17950 1.00000
\end{lstlisting}
{\phantomsection\label{ABC_tP24_91_d_d_d_poscar}}
{\hyperref[ABC_tP24_91_d_d_d]{ThBC: ABC\_tP24\_91\_d\_d\_d}} - POSCAR

{\phantomsection\label{AB32CD4E8_tP184_93_i_16p_af_2p_4p_cif}}
{\hyperref[AB32CD4E8_tP184_93_i_16p_af_2p_4p]{AsPh$_{4}$CeS$_{8}$P$_{4}$Me$_{8}$: AB32CD4E8\_tP184\_93\_i\_16p\_af\_2p\_4p}} - CIF

{\phantomsection\label{AB32CD4E8_tP184_93_i_16p_af_2p_4p_poscar}}
{\hyperref[AB32CD4E8_tP184_93_i_16p_af_2p_4p]{AsPh$_{4}$CeS$_{8}$P$_{4}$Me$_{8}$: AB32CD4E8\_tP184\_93\_i\_16p\_af\_2p\_4p}} - POSCAR

{\phantomsection\label{A14B3C5_tP44_94_c3g_ad_bg_cif}}
{\hyperref[A14B3C5_tP44_94_c3g_ad_bg]{Na$_{5}$Fe$_{3}$F$_{14}$ (High-temperature): A14B3C5\_tP44\_94\_c3g\_ad\_bg}} - CIF

{\phantomsection\label{A14B3C5_tP44_94_c3g_ad_bg_poscar}}
{\hyperref[A14B3C5_tP44_94_c3g_ad_bg]{Na$_{5}$Fe$_{3}$F$_{14}$ (High-temperature): A14B3C5\_tP44\_94\_c3g\_ad\_bg}} - POSCAR

{\phantomsection\label{A6B2C_tP18_94_eg_c_a_cif}}
{\hyperref[A6B2C_tP18_94_eg_c_a]{Li$_{2}$MoF$_{6}$: A6B2C\_tP18\_94\_eg\_c\_a}} - CIF
\begin{lstlisting}[numbers=none,language={mylang}]
# CIF file
data_findsym-output
_audit_creation_method FINDSYM

_chemical_name_mineral 'Li2MoF6'
_chemical_formula_sum 'F6 Li2 Mo'

loop_
_publ_author_name
 'G. Brunton'
_journal_name_full_name
;
 Materials Research Bulletin
;
_journal_volume 6
_journal_year 1971
_journal_page_first 555
_journal_page_last 560
_publ_Section_title
;
 The crystal structure of Li$_{2}$MoF$_{6}$
;

# Found in Pearson's Crystal Data - Crystal Structure Database for Inorganic Compounds, 2013

_aflow_title 'Li$_{2}$MoF$_{6}$ Structure'
_aflow_proto 'A6B2C_tP18_94_eg_c_a'
_aflow_params 'a,c/a,z_{2},x_{3},x_{4},y_{4},z_{4}'
_aflow_params_values '4.6863209101,1.96124874637,0.6623,0.7093,0.684,0.707,0.6579'
_aflow_Strukturbericht 'None'
_aflow_Pearson 'tP18'

_cell_length_a    4.6863209101
_cell_length_b    4.6863209101
_cell_length_c    9.1910410100
_cell_angle_alpha 90.0000000000
_cell_angle_beta  90.0000000000
_cell_angle_gamma 90.0000000000
 
_symmetry_space_group_name_H-M "P 42 21 2"
_symmetry_Int_Tables_number 94
 
loop_
_space_group_symop_id
_space_group_symop_operation_xyz
1 x,y,z
2 x+1/2,-y+1/2,-z+1/2
3 -x+1/2,y+1/2,-z+1/2
4 -x,-y,z
5 -y,-x,-z
6 -y+1/2,x+1/2,z+1/2
7 y+1/2,-x+1/2,z+1/2
8 y,x,-z
 
loop_
_atom_site_label
_atom_site_type_symbol
_atom_site_symmetry_multiplicity
_atom_site_Wyckoff_label
_atom_site_fract_x
_atom_site_fract_y
_atom_site_fract_z
_atom_site_occupancy
Mo1 Mo   2 a 0.00000 0.00000 0.00000 1.00000
Li1 Li   4 c 0.00000 0.00000 0.66230 1.00000
F1  F    4 e 0.70930 0.70930 0.00000 1.00000
F2  F    8 g 0.68400 0.70700 0.65790 1.00000
\end{lstlisting}
{\phantomsection\label{A6B2C_tP18_94_eg_c_a_poscar}}
{\hyperref[A6B2C_tP18_94_eg_c_a]{Li$_{2}$MoF$_{6}$: A6B2C\_tP18\_94\_eg\_c\_a}} - POSCAR
\begin{lstlisting}[numbers=none,language={mylang}]
A6B2C_tP18_94_eg_c_a & a,c/a,z2,x3,x4,y4,z4 --params=4.6863209101,1.96124874637,0.6623,0.7093,0.684,0.707,0.6579 & P4_{2}2_{1}2 D_{4}^{6} #94 (aceg) & tP18 & None & Li2MoF6 &  & G. Brunton, Mater. Res. Bull. 6, 555-560 (1971)
   1.00000000000000
   4.68632091010000   0.00000000000000   0.00000000000000
   0.00000000000000   4.68632091010000   0.00000000000000
   0.00000000000000   0.00000000000000   9.19104101000000
     F    Li    Mo
    12     4     2
Direct
   0.70930000000000   0.70930000000000   0.00000000000000    F   (4e)
  -0.70930000000000  -0.70930000000000   0.00000000000000    F   (4e)
  -0.20930000000000   1.20930000000000   0.50000000000000    F   (4e)
   1.20930000000000  -0.20930000000000   0.50000000000000    F   (4e)
   0.68400000000000   0.70700000000000   0.65790000000000    F   (8g)
  -0.68400000000000  -0.70700000000000   0.65790000000000    F   (8g)
  -0.20700000000000   1.18400000000000   1.15790000000000    F   (8g)
   1.20700000000000  -0.18400000000000   1.15790000000000    F   (8g)
  -0.18400000000000   1.20700000000000  -0.15790000000000    F   (8g)
   1.18400000000000  -0.20700000000000  -0.15790000000000    F   (8g)
   0.70700000000000   0.68400000000000  -0.65790000000000    F   (8g)
  -0.70700000000000  -0.68400000000000  -0.65790000000000    F   (8g)
   0.00000000000000   0.00000000000000   0.66230000000000   Li   (4c)
   0.50000000000000   0.50000000000000   1.16230000000000   Li   (4c)
   0.50000000000000   0.50000000000000  -0.16230000000000   Li   (4c)
   0.00000000000000   0.00000000000000  -0.66230000000000   Li   (4c)
   0.00000000000000   0.00000000000000   0.00000000000000   Mo   (2a)
   0.50000000000000   0.50000000000000   0.50000000000000   Mo   (2a)
\end{lstlisting}
{\phantomsection\label{ABC_tP24_95_d_d_d_cif}}
{\hyperref[ABC_tP24_95_d_d_d]{ThBC: ABC\_tP24\_95\_d\_d\_d}} - CIF
\begin{lstlisting}[numbers=none,language={mylang}]
# CIF file
data_findsym-output
_audit_creation_method FINDSYM

_chemical_name_mineral 'ThBC'
_chemical_formula_sum 'B C Th'

_aflow_title 'ThBC Structure'
_aflow_proto 'ABC_tP24_95_d_d_d'
_aflow_params 'a,c/a,x_{1},y_{1},z_{1},x_{2},y_{2},z_{2},x_{3},y_{3},z_{3}'
_aflow_params_values '3.7620082462,6.71079213192,0.303,0.202,-0.019,0.296,0.189,-0.08,0.2975,0.1983,0.8205'
_aflow_Strukturbericht 'None'
_aflow_Pearson 'tP24'

_cell_length_a    3.7620082462
_cell_length_b    3.7620082462
_cell_length_c    25.2460553388
_cell_angle_alpha 90.0000000000
_cell_angle_beta  90.0000000000
_cell_angle_gamma 90.0000000000
 
_symmetry_space_group_name_H-M "P 43 2 2"
_symmetry_Int_Tables_number 95
 
loop_
_space_group_symop_id
_space_group_symop_operation_xyz
1 x,y,z
2 x,-y,-z+1/2
3 -x,y,-z
4 -x,-y,z+1/2
5 -y,-x,-z+3/4
6 -y,x,z+3/4
7 y,-x,z+1/4
8 y,x,-z+1/4
 
loop_
_atom_site_label
_atom_site_type_symbol
_atom_site_symmetry_multiplicity
_atom_site_Wyckoff_label
_atom_site_fract_x
_atom_site_fract_y
_atom_site_fract_z
_atom_site_occupancy
B1  B    8 d 0.30300 0.20200 -0.01900 1.00000
C1  C    8 d 0.29600 0.18900 -0.08000 1.00000
Th1 Th   8 d 0.29750 0.19830 0.82050  1.00000
\end{lstlisting}
{\phantomsection\label{ABC_tP24_95_d_d_d_poscar}}
{\hyperref[ABC_tP24_95_d_d_d]{ThBC: ABC\_tP24\_95\_d\_d\_d}} - POSCAR

{\phantomsection\label{A2B8CD_tI24_97_d_k_a_b_cif}}
{\hyperref[A2B8CD_tI24_97_d_k_a_b]{NaGdCu$_{2}$F$_{8}$: A2B8CD\_tI24\_97\_d\_k\_a\_b}} - CIF
\begin{lstlisting}[numbers=none,language={mylang}]
# CIF file
data_findsym-output
_audit_creation_method FINDSYM

_chemical_name_mineral 'NaGdCu2F8'
_chemical_formula_sum 'Cu2 F8 Gd Na'

loop_
_publ_author_name
 'C. {De Nada\"i}'
 'A. Demourgues'
 'L. Lozano'
 'P. Gravereau'
 'J. Grannec'
_journal_name_full_name
;
 Journal of Materials Chemistry
;
_journal_volume 8
_journal_year 1998
_journal_page_first 2487
_journal_page_last 2491
_publ_Section_title
;
 Structural investigations of new copper fluorides Na$RE$Cu$_{2}$F$_{8}$ ($RE^{3+}$ = Sm$^{3+}$, Eu$^{3+}$, Gd$^{3+}$, Y$^{3+}$, Er$^{3+}$, Yb$^{3+}$)
;

# Found in Pearson's Crystal Data - Crystal Structure Database for Inorganic Compounds, 2013

_aflow_title 'NaGdCu$_{2}$F$_{8}$ Structure'
_aflow_proto 'A2B8CD_tI24_97_d_k_a_b'
_aflow_params 'a,c/a,x_{4},y_{4},z_{4}'
_aflow_params_values '5.4068544677,1.92010356944,0.1697,0.3128,0.1237'
_aflow_Strukturbericht 'None'
_aflow_Pearson 'tI24'

_cell_length_a    5.4068544677
_cell_length_b    5.4068544677
_cell_length_c    10.3817205629
_cell_angle_alpha 90.0000000000
_cell_angle_beta  90.0000000000
_cell_angle_gamma 90.0000000000
 
_symmetry_space_group_name_H-M "I 4 2 2"
_symmetry_Int_Tables_number 97
 
loop_
_space_group_symop_id
_space_group_symop_operation_xyz
1 x,y,z
2 x,-y,-z
3 -x,y,-z
4 -x,-y,z
5 -y,-x,-z
6 -y,x,z
7 y,-x,z
8 y,x,-z
9 x+1/2,y+1/2,z+1/2
10 x+1/2,-y+1/2,-z+1/2
11 -x+1/2,y+1/2,-z+1/2
12 -x+1/2,-y+1/2,z+1/2
13 -y+1/2,-x+1/2,-z+1/2
14 -y+1/2,x+1/2,z+1/2
15 y+1/2,-x+1/2,z+1/2
16 y+1/2,x+1/2,-z+1/2
 
loop_
_atom_site_label
_atom_site_type_symbol
_atom_site_symmetry_multiplicity
_atom_site_Wyckoff_label
_atom_site_fract_x
_atom_site_fract_y
_atom_site_fract_z
_atom_site_occupancy
Gd1 Gd   2 a 0.00000 0.00000 0.00000 1.00000
Na1 Na   2 b 0.00000 0.00000 0.50000 1.00000
Cu1 Cu   4 d 0.00000 0.50000 0.25000 1.00000
F1  F   16 k 0.16970 0.31280 0.12370 1.00000
\end{lstlisting}
{\phantomsection\label{A2B8CD_tI24_97_d_k_a_b_poscar}}
{\hyperref[A2B8CD_tI24_97_d_k_a_b]{NaGdCu$_{2}$F$_{8}$: A2B8CD\_tI24\_97\_d\_k\_a\_b}} - POSCAR
\begin{lstlisting}[numbers=none,language={mylang}]
A2B8CD_tI24_97_d_k_a_b & a,c/a,x4,y4,z4 --params=5.4068544677,1.92010356944,0.1697,0.3128,0.1237 & I422 D_{4}^{9} #97 (abdk) & tI24 & None & NaGdCu2F8 &  & C. {De Nada\"i} et al., J. Mater. Chem. 8, 2487-2491 (1998)
   1.00000000000000
  -2.70342723385000   2.70342723385000   5.19086028145000
   2.70342723385000  -2.70342723385000   5.19086028145000
   2.70342723385000   2.70342723385000  -5.19086028145000
    Cu     F    Gd    Na
     2     8     1     1
Direct
   0.75000000000000   0.25000000000000   0.50000000000000   Cu   (4d)
   0.25000000000000   0.75000000000000   0.50000000000000   Cu   (4d)
   0.43650000000000   0.29340000000000   0.48250000000000    F  (16k)
  -0.18910000000000  -0.04600000000000  -0.48250000000000    F  (16k)
   0.29340000000000  -0.18910000000000  -0.14310000000000    F  (16k)
  -0.04600000000000   0.43650000000000   0.14310000000000    F  (16k)
   0.18910000000000  -0.29340000000000   0.14310000000000    F  (16k)
  -0.43650000000000   0.04600000000000  -0.14310000000000    F  (16k)
   0.04600000000000   0.18910000000000   0.48250000000000    F  (16k)
  -0.29340000000000  -0.43650000000000  -0.48250000000000    F  (16k)
   0.00000000000000   0.00000000000000   0.00000000000000   Gd   (2a)
   0.50000000000000   0.50000000000000   0.00000000000000   Na   (2b)
\end{lstlisting}
{\phantomsection\label{AB8C2_tI44_97_e_2k_cd_cif}}
{\hyperref[AB8C2_tI44_97_e_2k_cd]{Ta$_{2}$Se$_{8}$I: AB8C2\_tI44\_97\_e\_2k\_cd}} - CIF
\begin{lstlisting}[numbers=none,language={mylang}]
# CIF file
data_findsym-output
_audit_creation_method FINDSYM

_chemical_name_mineral 'Ta2Se8I'
_chemical_formula_sum 'I Se8 Ta2'

loop_
_publ_author_name
 'P. Gressier'
 'A. Meerschaut'
 'L. Guemas'
 'J. Rouxel'
 'P. Monceau'
_journal_name_full_name
;
 Journal of Solid State Chemistry
;
_journal_volume 51
_journal_year 1984
_journal_page_first 141
_journal_page_last 151
_publ_Section_title
;
 Characterization of the new series of quasi one-dimensional compounds ($MX_{4}$)$_{n}Y$ ($M$ = Nb, Ta; $X$ = S, Se; $Y$ = Br, I)
;

# Found in Pearson's Crystal Data - Crystal Structure Database for Inorganic Compounds, 2013

_aflow_title 'Ta$_{2}$Se$_{8}$I Structure'
_aflow_proto 'AB8C2_tI44_97_e_2k_cd'
_aflow_params 'a,c/a,z_{3},x_{4},y_{4},z_{4},x_{5},y_{5},z_{5}'
_aflow_params_values '9.5317026755,1.33879580768,0.1553,0.0449,0.284,0.3693,0.312,0.1212,0.1191'
_aflow_Strukturbericht 'None'
_aflow_Pearson 'tI44'

_cell_length_a    9.5317026755
_cell_length_b    9.5317026755
_cell_length_c    12.7610035820
_cell_angle_alpha 90.0000000000
_cell_angle_beta  90.0000000000
_cell_angle_gamma 90.0000000000
 
_symmetry_space_group_name_H-M "I 4 2 2"
_symmetry_Int_Tables_number 97
 
loop_
_space_group_symop_id
_space_group_symop_operation_xyz
1 x,y,z
2 x,-y,-z
3 -x,y,-z
4 -x,-y,z
5 -y,-x,-z
6 -y,x,z
7 y,-x,z
8 y,x,-z
9 x+1/2,y+1/2,z+1/2
10 x+1/2,-y+1/2,-z+1/2
11 -x+1/2,y+1/2,-z+1/2
12 -x+1/2,-y+1/2,z+1/2
13 -y+1/2,-x+1/2,-z+1/2
14 -y+1/2,x+1/2,z+1/2
15 y+1/2,-x+1/2,z+1/2
16 y+1/2,x+1/2,-z+1/2
 
loop_
_atom_site_label
_atom_site_type_symbol
_atom_site_symmetry_multiplicity
_atom_site_Wyckoff_label
_atom_site_fract_x
_atom_site_fract_y
_atom_site_fract_z
_atom_site_occupancy
Ta1 Ta   4 c 0.00000 0.50000 0.00000 1.00000
Ta2 Ta   4 d 0.00000 0.50000 0.25000 1.00000
I1  I    4 e 0.00000 0.00000 0.15530 1.00000
Se1 Se  16 k 0.04490 0.28400 0.36930 1.00000
Se2 Se  16 k 0.31200 0.12120 0.11910 1.00000
\end{lstlisting}
{\phantomsection\label{AB8C2_tI44_97_e_2k_cd_poscar}}
{\hyperref[AB8C2_tI44_97_e_2k_cd]{Ta$_{2}$Se$_{8}$I: AB8C2\_tI44\_97\_e\_2k\_cd}} - POSCAR

{\phantomsection\label{A2B_tI12_98_f_a_cif}}
{\hyperref[A2B_tI12_98_f_a]{CdAs$_{2}$: A2B\_tI12\_98\_f\_a}} - CIF
\begin{lstlisting}[numbers=none,language={mylang}]
# CIF file
data_findsym-output
_audit_creation_method FINDSYM

_chemical_name_mineral 'CdAs2'
_chemical_formula_sum 'As2 Cd'

loop_
_publ_author_name
 'V. N. Yakimovich'
 'V. A. Rubtsov'
 'V. M. Trukhan'
_journal_name_full_name
;
 Inorganic Materials
;
_journal_volume 32
_journal_year 1996
_journal_page_first 579
_journal_page_last 582
_publ_Section_title
;
 Phase Relationships in the CdP$_{4}$-ZnP$_{2}$-CdAs$_{2}$-ZnAs$_{2}$ System
;

# Found in Pearson's Crystal Data - Crystal Structure Database for Inorganic Compounds, 2013

_aflow_title 'CdAs$_{2}$ Structure'
_aflow_proto 'A2B_tI12_98_f_a'
_aflow_params 'a,c/a,x_{2}'
_aflow_params_values '7.953376649,0.587954231111,0.44'
_aflow_Strukturbericht 'None'
_aflow_Pearson 'tI12'

_cell_length_a    7.9533766490
_cell_length_b    7.9533766490
_cell_length_c    4.6762214524
_cell_angle_alpha 90.0000000000
_cell_angle_beta  90.0000000000
_cell_angle_gamma 90.0000000000
 
_symmetry_space_group_name_H-M "I 41 2 2"
_symmetry_Int_Tables_number 98
 
loop_
_space_group_symop_id
_space_group_symop_operation_xyz
1 x,y,z
2 x,-y+1/2,-z+1/4
3 -x,y+1/2,-z+1/4
4 -x,-y,z
5 -y,-x,-z
6 -y,x+1/2,z+1/4
7 y,-x+1/2,z+1/4
8 y,x,-z
9 x+1/2,y+1/2,z+1/2
10 x+1/2,-y,-z+3/4
11 -x+1/2,y,-z+3/4
12 -x+1/2,-y+1/2,z+1/2
13 -y+1/2,-x+1/2,-z+1/2
14 -y+1/2,x,z+3/4
15 y+1/2,-x,z+3/4
16 y+1/2,x+1/2,-z+1/2
 
loop_
_atom_site_label
_atom_site_type_symbol
_atom_site_symmetry_multiplicity
_atom_site_Wyckoff_label
_atom_site_fract_x
_atom_site_fract_y
_atom_site_fract_z
_atom_site_occupancy
Cd1 Cd   4 a 0.00000 0.00000 0.00000 1.00000
As1 As   8 f 0.44000 0.25000 0.12500 1.00000
\end{lstlisting}
{\phantomsection\label{A2B_tI12_98_f_a_poscar}}
{\hyperref[A2B_tI12_98_f_a]{CdAs$_{2}$: A2B\_tI12\_98\_f\_a}} - POSCAR
\begin{lstlisting}[numbers=none,language={mylang}]
A2B_tI12_98_f_a & a,c/a,x2 --params=7.953376649,0.587954231111,0.44 & I4_{1}22 D_{4}^{10} #98 (af) & tI12 & None & CdAs2 &  & V. N. Yakimovich and V. A. Rubtsov and V. M. Trukhan, Inorg. Mat. 32, 579-582 (1996)
   1.00000000000000
  -3.97668832450000   3.97668832450000   2.33811072620000
   3.97668832450000  -3.97668832450000   2.33811072620000
   3.97668832450000   3.97668832450000  -2.33811072620000
    As    Cd
     4     2
Direct
   0.37500000000000   0.56500000000000   0.69000000000000   As   (8f)
   0.87500000000000  -0.31500000000000   0.31000000000000   As   (8f)
   1.31500000000000   0.12500000000000   0.69000000000000   As   (8f)
   0.43500000000000   0.62500000000000   0.31000000000000   As   (8f)
   0.00000000000000   0.00000000000000   0.00000000000000   Cd   (4a)
   0.75000000000000   0.25000000000000   0.50000000000000   Cd   (4a)
\end{lstlisting}
{\phantomsection\label{A2B8C2D_tP26_100_c_abcd_c_a_cif}}
{\hyperref[A2B8C2D_tP26_100_c_abcd_c_a]{Fresnoite (Ba$_{2}$TiSi$_{2}$O$_{8}$): A2B8C2D\_tP26\_100\_c\_abcd\_c\_a}} - CIF
\begin{lstlisting}[numbers=none,language={mylang}]
# CIF file 
data_findsym-output
_audit_creation_method FINDSYM

_chemical_name_mineral 'Fresnoite'
_chemical_formula_sum 'Ba2 O8 Si2 Ti'

loop_
_publ_author_name
 'S. A. Markgraf'
 'A. Halliya'
 'A. S. Bhalla'
 'R. E. Newnham'
 'C. T. Prewitt'
_journal_name_full_name
;
 Ferroelectrics
;
_journal_volume 62
_journal_year 1985
_journal_page_first 17
_journal_page_last 26
_publ_Section_title
;
 X-ray structure refinement and pyroelectric investigation of fresnoite, Ba$_{2}$TiSi$_{2}$O$_{8}$
;

_aflow_title 'Fresnoite (Ba$_{2}$TiSi$_{2}$O$_{8}$) Structure'
_aflow_proto 'A2B8C2D_tP26_100_c_abcd_c_a'
_aflow_params 'a,c/a,z_{1},z_{2},z_{3},x_{4},z_{4},x_{5},z_{5},x_{6},z_{6},x_{7},y_{7},z_{7}'
_aflow_params_values '8.527,0.611047261639,0.7904,0.4646,0.3707,0.32701,0.0,0.1259,0.7949,0.1282,0.4871,0.2924,0.5772,0.3571'
_aflow_Strukturbericht 'None'
_aflow_Pearson 'tP26'

_symmetry_space_group_name_H-M "P 4 b m"
_symmetry_Int_Tables_number 100
 
_cell_length_a    8.52700
_cell_length_b    8.52700
_cell_length_c    5.21040
_cell_angle_alpha 90.00000
_cell_angle_beta  90.00000
_cell_angle_gamma 90.00000
 
loop_
_space_group_symop_id
_space_group_symop_operation_xyz
1 x,y,z
2 -x,-y,z
3 -y,x,z
4 y,-x,z
5 -x+1/2,y+1/2,z
6 x+1/2,-y+1/2,z
7 y+1/2,x+1/2,z
8 -y+1/2,-x+1/2,z
 
loop_
_atom_site_label
_atom_site_type_symbol
_atom_site_symmetry_multiplicity
_atom_site_Wyckoff_label
_atom_site_fract_x
_atom_site_fract_y
_atom_site_fract_z
_atom_site_occupancy
O1  O    2 a 0.00000 0.00000 0.79040 1.00000
Ti1 Ti   2 a 0.00000 0.00000 0.46460 1.00000
O2  O    2 b 0.50000 0.00000 0.37070 1.00000
Ba1 Ba   4 c 0.32701 0.82701 0.00000 1.00000
O3  O    4 c 0.12590 0.62590 0.79490 1.00000
Si1 Si   4 c 0.12820 0.62820 0.48710 1.00000
O4  O    8 d 0.29240 0.57720 0.35710 1.00000
\end{lstlisting}
{\phantomsection\label{A2B8C2D_tP26_100_c_abcd_c_a_poscar}}
{\hyperref[A2B8C2D_tP26_100_c_abcd_c_a]{Fresnoite (Ba$_{2}$TiSi$_{2}$O$_{8}$): A2B8C2D\_tP26\_100\_c\_abcd\_c\_a}} - POSCAR

{\phantomsection\label{A3B11C6_tP40_100_ac_bc2d_cd_cif}}
{\hyperref[A3B11C6_tP40_100_ac_bc2d_cd]{Ce$_{3}$Si$_{6}$N$_{11}$: A3B11C6\_tP40\_100\_ac\_bc2d\_cd}} - CIF

{\phantomsection\label{A3B11C6_tP40_100_ac_bc2d_cd_poscar}}
{\hyperref[A3B11C6_tP40_100_ac_bc2d_cd]{Ce$_{3}$Si$_{6}$N$_{11}$: A3B11C6\_tP40\_100\_ac\_bc2d\_cd}} - POSCAR

{\phantomsection\label{A7B7C2_tP32_101_bde_ade_d_cif}}
{\hyperref[A7B7C2_tP32_101_bde_ade_d]{$\gamma$-MgNiSn: A7B7C2\_tP32\_101\_bde\_ade\_d}} - CIF
\begin{lstlisting}[numbers=none,language={mylang}]
# CIF file
data_findsym-output
_audit_creation_method FINDSYM

_chemical_name_mineral 'gamma-MgNiSn'
_chemical_formula_sum 'M7 Mg7 Ni2'

loop_
_publ_author_name
 'M. Boudard'
 'P. Bordet'
 'H. Vincent'
 'F. Audebert'
_journal_name_full_name
;
 Journal of Alloys and Compounds
;
_journal_volume 372
_journal_year 2004
_journal_page_first 121
_journal_page_last 128
_publ_Section_title
;
 The structure of the Y-phase in the Mg--Ni--Sn system
;

# Found in Pearson's Crystal Data - Crystal Structure Database for Inorganic Compounds, 2013

_aflow_title '$\gamma$-MgNiSn Structure'
_aflow_proto 'A7B7C2_tP32_101_bde_ade_d'
_aflow_params 'a,c/a,z_{1},z_{2},x_{3},z_{3},x_{4},z_{4},x_{5},z_{5},x_{6},y_{6},z_{6},x_{7},y_{7},z_{7}'
_aflow_params_values '9.8510697809,0.697188102729,0.0,0.4692,0.17136,0.73945,0.2254,0.3402,0.30926,0.0086,0.2384,0.5244,0.2259,0.0352,0.3449,0.0281'
_aflow_Strukturbericht 'None'
_aflow_Pearson 'tP32'

_cell_length_a    9.8510697809
_cell_length_b    9.8510697809
_cell_length_c    6.8680486504
_cell_angle_alpha 90.0000000000
_cell_angle_beta  90.0000000000
_cell_angle_gamma 90.0000000000
 
_symmetry_space_group_name_H-M "P 42 c m"
_symmetry_Int_Tables_number 101
 
loop_
_space_group_symop_id
_space_group_symop_operation_xyz
1 x,y,z
2 -x,-y,z
3 -y,x,z+1/2
4 y,-x,z+1/2
5 -x,y,z+1/2
6 x,-y,z+1/2
7 y,x,z
8 -y,-x,z
 
loop_
_atom_site_label
_atom_site_type_symbol
_atom_site_symmetry_multiplicity
_atom_site_Wyckoff_label
_atom_site_fract_x
_atom_site_fract_y
_atom_site_fract_z
_atom_site_occupancy
Mg1 Mg   2 a 0.00000 0.00000 0.00000 1.00000
M1 M    2 b 0.50000 0.50000 0.46920 1.00000
M2 M    4 d 0.17136 0.17136 0.73945 1.00000
Mg2 Mg   4 d 0.22540 0.22540 0.34020 1.00000
Ni1 Ni   4 d 0.30926 0.30926 0.00860 1.00000
M3 M    8 e 0.23840 0.52440 0.22590 1.00000
Mg3 Mg   8 e 0.03520 0.34490 0.02810 1.00000
\end{lstlisting}
{\phantomsection\label{A7B7C2_tP32_101_bde_ade_d_poscar}}
{\hyperref[A7B7C2_tP32_101_bde_ade_d]{$\gamma$-MgNiSn: A7B7C2\_tP32\_101\_bde\_ade\_d}} - POSCAR

{\phantomsection\label{A2B3_tP20_102_2c_b2c_cif}}
{\hyperref[A2B3_tP20_102_2c_b2c]{Gd$_{3}$Al$_{2}$: A2B3\_tP20\_102\_2c\_b2c}} - CIF
\begin{lstlisting}[numbers=none,language={mylang}]
# CIF file
data_findsym-output
_audit_creation_method FINDSYM

_chemical_name_mineral 'Gd3Al2'
_chemical_formula_sum 'Al2 Gd3'

loop_
_publ_author_name
 'K. H. J. Buschow'
_journal_name_full_name
;
 Journal of the Less-Common Metals
;
_journal_volume 8
_journal_year 1965
_journal_page_first 209
_journal_page_last 212
_publ_Section_title
;
 Rare earth-aluminium intermetallic compounds of the form $R$Al and $R_{3}$Al$_{2}$
;

# Found in Pearson's Crystal Data - Crystal Structure Database for Inorganic Compounds, 2013

_aflow_title 'Gd$_{3}$Al$_{2}$ Structure'
_aflow_proto 'A2B3_tP20_102_2c_b2c'
_aflow_params 'a,c/a,z_{1},x_{2},z_{2},x_{3},z_{3},x_{4},z_{4},x_{5},z_{5}'
_aflow_params_values '8.3289849893,0.909833113223,0.5,0.604,0.439,0.623,0.031,0.795,0.725,0.848,0.251'
_aflow_Strukturbericht 'None'
_aflow_Pearson 'tP20'

_cell_length_a    8.3289849893
_cell_length_b    8.3289849893
_cell_length_c    7.5779863428
_cell_angle_alpha 90.0000000000
_cell_angle_beta  90.0000000000
_cell_angle_gamma 90.0000000000
 
_symmetry_space_group_name_H-M "P 42 n m"
_symmetry_Int_Tables_number 102
 
loop_
_space_group_symop_id
_space_group_symop_operation_xyz
1 x,y,z
2 -x,-y,z
3 -y+1/2,x+1/2,z+1/2
4 y+1/2,-x+1/2,z+1/2
5 -x+1/2,y+1/2,z+1/2
6 x+1/2,-y+1/2,z+1/2
7 y,x,z
8 -y,-x,z
 
loop_
_atom_site_label
_atom_site_type_symbol
_atom_site_symmetry_multiplicity
_atom_site_Wyckoff_label
_atom_site_fract_x
_atom_site_fract_y
_atom_site_fract_z
_atom_site_occupancy
Gd1 Gd   4 b 0.00000 0.50000 0.50000 1.00000
Al1 Al   4 c 0.60400 0.60400 0.43900 1.00000
Al2 Al   4 c 0.62300 0.62300 0.03100 1.00000
Gd2 Gd   4 c 0.79500 0.79500 0.72500 1.00000
Gd3 Gd   4 c 0.84800 0.84800 0.25100 1.00000
\end{lstlisting}
{\phantomsection\label{A2B3_tP20_102_2c_b2c_poscar}}
{\hyperref[A2B3_tP20_102_2c_b2c]{Gd$_{3}$Al$_{2}$: A2B3\_tP20\_102\_2c\_b2c}} - POSCAR
\begin{lstlisting}[numbers=none,language={mylang}]
A2B3_tP20_102_2c_b2c & a,c/a,z1,x2,z2,x3,z3,x4,z4,x5,z5 --params=8.3289849893,0.909833113223,0.5,0.604,0.439,0.623,0.031,0.795,0.725,0.848,0.251 & P4_{2}nm C_{4v}^{4} #102 (bc^4) & tP20 & None & Gd3Al2 &  & K. H. J. Buschow, J. Less-Common Met. 8, 209-212 (1965)
   1.00000000000000
   8.32898498930000   0.00000000000000   0.00000000000000
   0.00000000000000   8.32898498930000   0.00000000000000
   0.00000000000000   0.00000000000000   7.57798634280000
    Al    Gd
     8    12
Direct
   0.60400000000000   0.60400000000000   0.43900000000000   Al   (4c)
  -0.60400000000000  -0.60400000000000   0.43900000000000   Al   (4c)
  -0.10400000000000   1.10400000000000   0.93900000000000   Al   (4c)
   1.10400000000000  -0.10400000000000   0.93900000000000   Al   (4c)
   0.62300000000000   0.62300000000000   0.03100000000000   Al   (4c)
  -0.62300000000000  -0.62300000000000   0.03100000000000   Al   (4c)
  -0.12300000000000   1.12300000000000   0.53100000000000   Al   (4c)
   1.12300000000000  -0.12300000000000   0.53100000000000   Al   (4c)
   0.00000000000000   0.50000000000000   0.50000000000000   Gd   (4b)
   0.00000000000000   0.50000000000000   1.00000000000000   Gd   (4b)
   0.50000000000000   0.00000000000000   1.00000000000000   Gd   (4b)
   0.50000000000000   0.00000000000000   0.50000000000000   Gd   (4b)
   0.79500000000000   0.79500000000000   0.72500000000000   Gd   (4c)
  -0.79500000000000  -0.79500000000000   0.72500000000000   Gd   (4c)
  -0.29500000000000   1.29500000000000   1.22500000000000   Gd   (4c)
   1.29500000000000  -0.29500000000000   1.22500000000000   Gd   (4c)
   0.84800000000000   0.84800000000000   0.25100000000000   Gd   (4c)
  -0.84800000000000  -0.84800000000000   0.25100000000000   Gd   (4c)
  -0.34800000000000   1.34800000000000   0.75100000000000   Gd   (4c)
   1.34800000000000  -0.34800000000000   0.75100000000000   Gd   (4c)
\end{lstlisting}
{\phantomsection\label{AB4_tP10_103_a_d_cif}}
{\hyperref[AB4_tP10_103_a_d]{NbTe$_{4}$: AB4\_tP10\_103\_a\_d}} - CIF
\begin{lstlisting}[numbers=none,language={mylang}]
# CIF file
data_findsym-output
_audit_creation_method FINDSYM

_chemical_name_mineral 'NbTe4'
_chemical_formula_sum 'Nb Te4'

loop_
_publ_author_name
 'H. B{\"o}hm'
_journal_name_full_name
;
 Zeitschrift f{\"u}r Kristallographie - Crystalline Materials
;
_journal_volume 180
_journal_year 1987
_journal_page_first 113
_journal_page_last 122
_publ_Section_title
;
 The high temperature modification of niobium tetratelluride NbTe$_{4}$
;

# Found in Pearson's Crystal Data - Crystal Structure Database for Inorganic Compounds, 2013

_aflow_title 'NbTe$_{4}$ Structure'
_aflow_proto 'AB4_tP10_103_a_d'
_aflow_params 'a,c/a,z_{1},x_{2},y_{2},z_{2}'
_aflow_params_values '6.5509768136,1.04518394138,0.0,0.144,0.3276,0.242'
_aflow_Strukturbericht 'None'
_aflow_Pearson 'tP10'

_cell_length_a    6.5509768136
_cell_length_b    6.5509768136
_cell_length_c    6.8469757659
_cell_angle_alpha 90.0000000000
_cell_angle_beta  90.0000000000
_cell_angle_gamma 90.0000000000
 
_symmetry_space_group_name_H-M "P 4 c c"
_symmetry_Int_Tables_number 103
 
loop_
_space_group_symop_id
_space_group_symop_operation_xyz
1 x,y,z
2 -x,-y,z
3 -y,x,z
4 y,-x,z
5 -x,y,z+1/2
6 x,-y,z+1/2
7 y,x,z+1/2
8 -y,-x,z+1/2
 
loop_
_atom_site_label
_atom_site_type_symbol
_atom_site_symmetry_multiplicity
_atom_site_Wyckoff_label
_atom_site_fract_x
_atom_site_fract_y
_atom_site_fract_z
_atom_site_occupancy
Nb1 Nb   2 a 0.00000 0.00000 0.00000 1.00000
Te1 Te   8 d 0.14400 0.32760 0.24200 1.00000
\end{lstlisting}
{\phantomsection\label{AB4_tP10_103_a_d_poscar}}
{\hyperref[AB4_tP10_103_a_d]{NbTe$_{4}$: AB4\_tP10\_103\_a\_d}} - POSCAR
\begin{lstlisting}[numbers=none,language={mylang}]
AB4_tP10_103_a_d & a,c/a,z1,x2,y2,z2 --params=6.5509768136,1.04518394138,0.0,0.144,0.3276,0.242 & P4cc C_{4v}^{5} #103 (ad) & tP10 & None & NbTe4 &  & H. B{\"o}hm, Zeitschrift f"{u}r Kristallographie - Crystalline Materials 180, 113-122 (1987)
   1.00000000000000
   6.55097681360000   0.00000000000000   0.00000000000000
   0.00000000000000   6.55097681360000   0.00000000000000
   0.00000000000000   0.00000000000000   6.84697576590000
    Nb    Te
     2     8
Direct
   0.00000000000000   0.00000000000000   0.00000000000000   Nb   (2a)
   0.00000000000000   0.00000000000000   0.50000000000000   Nb   (2a)
   0.14400000000000   0.32760000000000   0.24200000000000   Te   (8d)
  -0.14400000000000  -0.32760000000000   0.24200000000000   Te   (8d)
  -0.32760000000000   0.14400000000000   0.24200000000000   Te   (8d)
   0.32760000000000  -0.14400000000000   0.24200000000000   Te   (8d)
   0.14400000000000  -0.32760000000000   0.74200000000000   Te   (8d)
  -0.14400000000000   0.32760000000000   0.74200000000000   Te   (8d)
  -0.32760000000000  -0.14400000000000   0.74200000000000   Te   (8d)
   0.32760000000000   0.14400000000000   0.74200000000000   Te   (8d)
\end{lstlisting}
{\phantomsection\label{A5B5C4_tP28_104_ac_ac_c_cif}}
{\hyperref[A5B5C4_tP28_104_ac_ac_c]{Ba$_{5}$In$_{4}$Bi$_{5}$: A5B5C4\_tP28\_104\_ac\_ac\_c}} - CIF
\begin{lstlisting}[numbers=none,language={mylang}]
# CIF file
data_findsym-output
_audit_creation_method FINDSYM

_chemical_name_mineral 'Ba5In4Bi5'
_chemical_formula_sum 'Ba5 Bi5 In4'

loop_
_publ_author_name
 'S. Ponou'
 'T. F. F{\"a}ssler'
 'G. Tob{\\'i}as'
 'E. Canadell'
 'A. Cho'
 'S. C. Sevov'
_journal_name_full_name
;
 Chemistry - A European Journal
;
_journal_volume 10
_journal_year 2004
_journal_page_first 3615
_journal_page_last 3621
_publ_Section_title
;
 Synthesis, Characterization, and Electronic Structure of Ba$_{5}$In$_{4}$Bi$_{5}$: An Acentric and One-Electron Deficient Phase
;

# Found in Pearson's Crystal Data - Crystal Structure Database for Inorganic Compounds, 2013

_aflow_title 'Ba$_{5}$In$_{4}$Bi$_{5}$ Structure'
_aflow_proto 'A5B5C4_tP28_104_ac_ac_c'
_aflow_params 'a,c/a,z_{1},z_{2},x_{3},y_{3},z_{3},x_{4},y_{4},z_{4},x_{5},y_{5},z_{5}'
_aflow_params_values '10.6225961282,0.84830508475,0.5,0.8821,0.8116,0.6057,0.3261,0.60942,0.80921,0.00978,0.8116,-0.072,0.1681'
_aflow_Strukturbericht 'None'
_aflow_Pearson 'tP28'

_cell_length_a    10.6225961282
_cell_length_b    10.6225961282
_cell_length_c    9.0112023088
_cell_angle_alpha 90.0000000000
_cell_angle_beta  90.0000000000
_cell_angle_gamma 90.0000000000
 
_symmetry_space_group_name_H-M "P 4 n c"
_symmetry_Int_Tables_number 104
 
loop_
_space_group_symop_id
_space_group_symop_operation_xyz
1 x,y,z
2 -x,-y,z
3 -y,x,z
4 y,-x,z
5 -x+1/2,y+1/2,z+1/2
6 x+1/2,-y+1/2,z+1/2
7 y+1/2,x+1/2,z+1/2
8 -y+1/2,-x+1/2,z+1/2
 
loop_
_atom_site_label
_atom_site_type_symbol
_atom_site_symmetry_multiplicity
_atom_site_Wyckoff_label
_atom_site_fract_x
_atom_site_fract_y
_atom_site_fract_z
_atom_site_occupancy
Ba1 Ba   2 a 0.00000 0.00000  0.50000 1.00000
Bi1 Bi   2 a 0.00000 0.00000  0.88210 1.00000
Ba2 Ba   8 c 0.81160 0.60570  0.32610 1.00000
Bi2 Bi   8 c 0.60942 0.80921  0.00978 1.00000
In1 In   8 c 0.81160 -0.07200 0.16810 1.00000
\end{lstlisting}
{\phantomsection\label{A5B5C4_tP28_104_ac_ac_c_poscar}}
{\hyperref[A5B5C4_tP28_104_ac_ac_c]{Ba$_{5}$In$_{4}$Bi$_{5}$: A5B5C4\_tP28\_104\_ac\_ac\_c}} - POSCAR

{\phantomsection\label{AB6C4_tP22_104_a_2ac_c_cif}}
{\hyperref[AB6C4_tP22_104_a_2ac_c]{Tl$_{4}$HgI$_{6}$: AB6C4\_tP22\_104\_a\_2ac\_c}} - CIF
\begin{lstlisting}[numbers=none,language={mylang}]
# CIF file
data_findsym-output
_audit_creation_method FINDSYM

_chemical_name_mineral 'Tl4HgI6'
_chemical_formula_sum 'Hg I6 Tl4'

loop_
_publ_author_name
 'D. V. Badikov'
 'V. V. Badikov'
 'G. M. {Kuz\'micheva}'
 'V. L. Panyutin'
 'V. B. Rybakov'
 'V. I. Chizhikov'
 'G. S. Shevyrdyaeva'
 'E. S. Shcherbakova'
_journal_name_full_name
;
 Inorganic Materials
;
_journal_volume 40
_journal_year 2004
_journal_page_first 314
_journal_page_last 320
_publ_Section_title
;
 Growth and X-ray diffraction study of Tl$_{4}$HgI$_{6}$ crystals
;

# Found in Pearson's Crystal Data - Crystal Structure Database for Inorganic Compounds, 2013

_aflow_title 'Tl$_{4}$HgI$_{6}$ Structure'
_aflow_proto 'AB6C4_tP22_104_a_2ac_c'
_aflow_params 'a,c/a,z_{1},z_{2},z_{3},x_{4},y_{4},z_{4},x_{5},y_{5},z_{5}'
_aflow_params_values '9.3940153509,0.981690440703,0.786,0.5,0.0649,0.8297,0.6458,0.286,0.6491,0.8588,0.036'
_aflow_Strukturbericht 'None'
_aflow_Pearson 'tP22'

_cell_length_a    9.3940153509
_cell_length_b    9.3940153509
_cell_length_c    9.2220150698
_cell_angle_alpha 90.0000000000
_cell_angle_beta  90.0000000000
_cell_angle_gamma 90.0000000000
 
_symmetry_space_group_name_H-M "P 4 n c"
_symmetry_Int_Tables_number 104
 
loop_
_space_group_symop_id
_space_group_symop_operation_xyz
1 x,y,z
2 -x,-y,z
3 -y,x,z
4 y,-x,z
5 -x+1/2,y+1/2,z+1/2
6 x+1/2,-y+1/2,z+1/2
7 y+1/2,x+1/2,z+1/2
8 -y+1/2,-x+1/2,z+1/2
 
loop_
_atom_site_label
_atom_site_type_symbol
_atom_site_symmetry_multiplicity
_atom_site_Wyckoff_label
_atom_site_fract_x
_atom_site_fract_y
_atom_site_fract_z
_atom_site_occupancy
Hg1 Hg   2 a 0.00000 0.00000 0.78600 1.00000
I1  I    2 a 0.00000 0.00000 0.50000 1.00000
I2  I    2 a 0.00000 0.00000 0.06490 1.00000
I3  I    8 c 0.82970 0.64580 0.28600 1.00000
Tl1 Tl   8 c 0.64910 0.85880 0.03600 1.00000
\end{lstlisting}
{\phantomsection\label{AB6C4_tP22_104_a_2ac_c_poscar}}
{\hyperref[AB6C4_tP22_104_a_2ac_c]{Tl$_{4}$HgI$_{6}$: AB6C4\_tP22\_104\_a\_2ac\_c}} - POSCAR

{\phantomsection\label{A2BC2_tP20_105_f_ac_2e_cif}}
{\hyperref[A2BC2_tP20_105_f_ac_2e]{BaGe$_{2}$As$_{2}$: A2BC2\_tP20\_105\_f\_ac\_2e}} - CIF
\begin{lstlisting}[numbers=none,language={mylang}]
# CIF file
data_findsym-output
_audit_creation_method FINDSYM

_chemical_name_mineral 'BaGe2As2'
_chemical_formula_sum 'As2 Ba Ge2'

loop_
_publ_author_name
 'B. Eisenmann'
 'H. Sch{\"a}fer'
_journal_name_full_name
;
 Zeitschrift f{\"u}r Naturforschung B
;
_journal_volume 36
_journal_year 1981
_journal_page_first 415
_journal_page_last 419
_publ_Section_title
;
 Zintlphasen mit bin{\"a}ren Anionen: Zur Kenntnis von BaGe$_{2}$P$_{2}$ und BaGe$_{2}$As$_{2}$ / Zintl Phases with Binary Anions: BaGe$_{2}$P$_{2}$ and BaGe$_{2}$As$_{2}$
;

# Found in Pearson's Crystal Data - Crystal Structure Database for Inorganic Compounds, 2013

_aflow_title 'BaGe$_{2}$As$_{2}$ Structure'
_aflow_proto 'A2BC2_tP20_105_f_ac_2e'
_aflow_params 'a,c/a,z_{1},z_{2},x_{3},z_{3},x_{4},z_{4},x_{5},y_{5},z_{5}'
_aflow_params_values '7.7858653925,1.11276650398,0.0,0.0241,0.3351,0.3664,0.1632,0.6068,0.3453,0.229,0.2558'
_aflow_Strukturbericht 'None'
_aflow_Pearson 'tP20'

_cell_length_a    7.7858653925
_cell_length_b    7.7858653925
_cell_length_c    8.6638502133
_cell_angle_alpha 90.0000000000
_cell_angle_beta  90.0000000000
_cell_angle_gamma 90.0000000000
 
_symmetry_space_group_name_H-M "P 42 m c"
_symmetry_Int_Tables_number 105
 
loop_
_space_group_symop_id
_space_group_symop_operation_xyz
1 x,y,z
2 -x,-y,z
3 -y,x,z+1/2
4 y,-x,z+1/2
5 -x,y,z
6 x,-y,z
7 y,x,z+1/2
8 -y,-x,z+1/2
 
loop_
_atom_site_label
_atom_site_type_symbol
_atom_site_symmetry_multiplicity
_atom_site_Wyckoff_label
_atom_site_fract_x
_atom_site_fract_y
_atom_site_fract_z
_atom_site_occupancy
Ba1 Ba   2 a 0.00000 0.00000 0.00000 1.00000
Ba2 Ba   2 c 0.00000 0.50000 0.02410 1.00000
Ge1 Ge   4 e 0.33510 0.50000 0.36640 1.00000
Ge2 Ge   4 e 0.16320 0.50000 0.60680 1.00000
As1 As   8 f 0.34530 0.22900 0.25580 1.00000
\end{lstlisting}
{\phantomsection\label{A2BC2_tP20_105_f_ac_2e_poscar}}
{\hyperref[A2BC2_tP20_105_f_ac_2e]{BaGe$_{2}$As$_{2}$: A2BC2\_tP20\_105\_f\_ac\_2e}} - POSCAR
\begin{lstlisting}[numbers=none,language={mylang}]
A2BC2_tP20_105_f_ac_2e & a,c/a,z1,z2,x3,z3,x4,z4,x5,y5,z5 --params=7.7858653925,1.11276650398,0.0,0.0241,0.3351,0.3664,0.1632,0.6068,0.3453,0.229,0.2558 & P4_{2}mc C_{4v}^{7} #105 (ace^2f) & tP20 & None & BaGe2As2 &  & B. Eisenmann and H. Sch{\"a}fer, Z. Naturforsch. B 36, 415-419 (1981)
   1.00000000000000
   7.78586539250000   0.00000000000000   0.00000000000000
   0.00000000000000   7.78586539250000   0.00000000000000
   0.00000000000000   0.00000000000000   8.66385021330000
    As    Ba    Ge
     8     4     8
Direct
   0.34530000000000   0.22900000000000   0.25580000000000   As   (8f)
  -0.34530000000000  -0.22900000000000   0.25580000000000   As   (8f)
  -0.22900000000000   0.34530000000000   0.75580000000000   As   (8f)
   0.22900000000000  -0.34530000000000   0.75580000000000   As   (8f)
   0.34530000000000  -0.22900000000000   0.25580000000000   As   (8f)
  -0.34530000000000   0.22900000000000   0.25580000000000   As   (8f)
  -0.22900000000000  -0.34530000000000   0.75580000000000   As   (8f)
   0.22900000000000   0.34530000000000   0.75580000000000   As   (8f)
   0.00000000000000   0.00000000000000   0.00000000000000   Ba   (2a)
   0.00000000000000   0.00000000000000   0.50000000000000   Ba   (2a)
   0.00000000000000   0.50000000000000   0.02410000000000   Ba   (2c)
   0.50000000000000   0.00000000000000   0.52410000000000   Ba   (2c)
   0.33510000000000   0.50000000000000   0.36640000000000   Ge   (4e)
  -0.33510000000000   0.50000000000000   0.36640000000000   Ge   (4e)
   0.50000000000000   0.33510000000000   0.86640000000000   Ge   (4e)
   0.50000000000000  -0.33510000000000   0.86640000000000   Ge   (4e)
   0.16320000000000   0.50000000000000   0.60680000000000   Ge   (4e)
  -0.16320000000000   0.50000000000000   0.60680000000000   Ge   (4e)
   0.50000000000000   0.16320000000000   1.10680000000000   Ge   (4e)
   0.50000000000000  -0.16320000000000   1.10680000000000   Ge   (4e)
\end{lstlisting}
{\phantomsection\label{A3BC3D_tP64_106_3c_c_3c_c_cif}}
{\hyperref[A3BC3D_tP64_106_3c_c_3c_c]{NaZn[OH]$_{3}$: A3BC3D\_tP64\_106\_3c\_c\_3c\_c}} - CIF

{\phantomsection\label{A3BC3D_tP64_106_3c_c_3c_c_poscar}}
{\hyperref[A3BC3D_tP64_106_3c_c_3c_c]{NaZn[OH]$_{3}$: A3BC3D\_tP64\_106\_3c\_c\_3c\_c}} - POSCAR

{\phantomsection\label{A5B7_tI24_107_ac_abd_cif}}
{\hyperref[A5B7_tI24_107_ac_abd]{Co$_{5}$Ge$_{7}$: A5B7\_tI24\_107\_ac\_abd}} - CIF
\begin{lstlisting}[numbers=none,language={mylang}]
# CIF file
data_findsym-output
_audit_creation_method FINDSYM

_chemical_name_mineral 'Co5Ge7'
_chemical_formula_sum 'Co5 Ge7'

loop_
_publ_author_name
 'K. Schubert'
 'T. R. Anantharaman'
 'H. O. K. Ata'
 'H. G. Meissner'
 'M. P{\"o}tzschke'
 'W. Rossteutscher'
 'E. Stolz'
_journal_name_full_name
;
 Naturwissenschaften
;
_journal_volume 47
_journal_year 1960
_journal_page_first 512
_journal_page_last 512
_publ_Section_title
;
 Einige strukturelle Ergebnisse an metallischen Phasen (6)
;

# Found in Pearson's Crystal Data - Crystal Structure Database for Inorganic Compounds, 2013

_aflow_title 'Co$_{5}$Ge$_{7}$ Structure'
_aflow_proto 'A5B7_tI24_107_ac_abd'
_aflow_params 'a,c/a,z_{1},z_{2},z_{3},x_{4},z_{4},x_{5},z_{5}'
_aflow_params_values '7.6400197048,0.760471204184,0.0,0.056,0.04,0.22,0.0,0.243,0.29'
_aflow_Strukturbericht 'None'
_aflow_Pearson 'tI24'

_cell_length_a    7.6400197048
_cell_length_b    7.6400197048
_cell_length_c    5.8100149849
_cell_angle_alpha 90.0000000000
_cell_angle_beta  90.0000000000
_cell_angle_gamma 90.0000000000
 
_symmetry_space_group_name_H-M "I 4 m m"
_symmetry_Int_Tables_number 107
 
loop_
_space_group_symop_id
_space_group_symop_operation_xyz
1 x,y,z
2 -x,-y,z
3 -y,x,z
4 y,-x,z
5 -x,y,z
6 x,-y,z
7 y,x,z
8 -y,-x,z
9 x+1/2,y+1/2,z+1/2
10 -x+1/2,-y+1/2,z+1/2
11 -y+1/2,x+1/2,z+1/2
12 y+1/2,-x+1/2,z+1/2
13 -x+1/2,y+1/2,z+1/2
14 x+1/2,-y+1/2,z+1/2
15 y+1/2,x+1/2,z+1/2
16 -y+1/2,-x+1/2,z+1/2
 
loop_
_atom_site_label
_atom_site_type_symbol
_atom_site_symmetry_multiplicity
_atom_site_Wyckoff_label
_atom_site_fract_x
_atom_site_fract_y
_atom_site_fract_z
_atom_site_occupancy
Co1 Co   2 a 0.00000 0.00000 0.00000 1.00000
Ge1 Ge   2 a 0.00000 0.00000 0.05600 1.00000
Ge2 Ge   4 b 0.00000 0.50000 0.04000 1.00000
Co2 Co   8 c 0.22000 0.22000 0.00000 1.00000
Ge3 Ge   8 d 0.24300 0.00000 0.29000 1.00000
\end{lstlisting}
{\phantomsection\label{A5B7_tI24_107_ac_abd_poscar}}
{\hyperref[A5B7_tI24_107_ac_abd]{Co$_{5}$Ge$_{7}$: A5B7\_tI24\_107\_ac\_abd}} - POSCAR
\begin{lstlisting}[numbers=none,language={mylang}]
A5B7_tI24_107_ac_abd & a,c/a,z1,z2,z3,x4,z4,x5,z5 --params=7.6400197048,0.760471204184,0.0,0.056,0.04,0.22,0.0,0.243,0.29 & I4mm C_{4v}^{9} #107 (a^2bcd) & tI24 & None & Co5Ge7 &  & K. Schubert et al., {Naturwissenschaften 47, 512(1960)
   1.00000000000000
  -3.82000985240000   3.82000985240000   2.90500749245000
   3.82000985240000  -3.82000985240000   2.90500749245000
   3.82000985240000   3.82000985240000  -2.90500749245000
    Co    Ge
     5     7
Direct
   0.00000000000000   0.00000000000000   0.00000000000000   Co   (2a)
   0.22000000000000   0.22000000000000   0.44000000000000   Co   (8c)
  -0.22000000000000  -0.22000000000000  -0.44000000000000   Co   (8c)
   0.22000000000000  -0.22000000000000   0.00000000000000   Co   (8c)
  -0.22000000000000   0.22000000000000   0.00000000000000   Co   (8c)
   0.05600000000000   0.05600000000000   0.00000000000000   Ge   (2a)
   0.54000000000000   0.04000000000000   0.50000000000000   Ge   (4b)
   0.04000000000000   0.54000000000000   0.50000000000000   Ge   (4b)
   0.29000000000000   0.53300000000000   0.24300000000000   Ge   (8d)
   0.29000000000000   0.04700000000000  -0.24300000000000   Ge   (8d)
   0.53300000000000   0.29000000000000   0.24300000000000   Ge   (8d)
   0.04700000000000   0.29000000000000  -0.24300000000000   Ge   (8d)
\end{lstlisting}
{\phantomsection\label{AB_tI4_107_a_a_cif}}
{\hyperref[AB_tI4_107_a_a]{GeP (High-pressure, superconducting): AB\_tI4\_107\_a\_a}} - CIF
\begin{lstlisting}[numbers=none,language={mylang}]
# CIF file
data_findsym-output
_audit_creation_method FINDSYM

_chemical_name_mineral 'GeP'
_chemical_formula_sum 'Ge P'

loop_
_publ_author_name
 'P. C. Donohue'
 'H. S. Young'
_journal_name_full_name
;
 Journal of Solid State Chemistry
;
_journal_volume 1
_journal_year 1970
_journal_page_first 143
_journal_page_last 149
_publ_Section_title
;
 Synthesis, structure, and superconductivity of new high pressure phases in the systems Ge-P and Ge-As
;

# Found in Pearson's Crystal Data - Crystal Structure Database for Inorganic Compounds, 2013

_aflow_title 'GeP (High-pressure, superconducting) Structure'
_aflow_proto 'AB_tI4_107_a_a'
_aflow_params 'a,c/a,z_{1},z_{2}'
_aflow_params_values '3.5440505103,1.57477426639,0.0,0.427'
_aflow_Strukturbericht 'None'
_aflow_Pearson 'tI4'

_cell_length_a    3.5440505103
_cell_length_b    3.5440505103
_cell_length_c    5.5810795424
_cell_angle_alpha 90.0000000000
_cell_angle_beta  90.0000000000
_cell_angle_gamma 90.0000000000
 
_symmetry_space_group_name_H-M "I 4 m m"
_symmetry_Int_Tables_number 107
 
loop_
_space_group_symop_id
_space_group_symop_operation_xyz
1 x,y,z
2 -x,-y,z
3 -y,x,z
4 y,-x,z
5 -x,y,z
6 x,-y,z
7 y,x,z
8 -y,-x,z
9 x+1/2,y+1/2,z+1/2
10 -x+1/2,-y+1/2,z+1/2
11 -y+1/2,x+1/2,z+1/2
12 y+1/2,-x+1/2,z+1/2
13 -x+1/2,y+1/2,z+1/2
14 x+1/2,-y+1/2,z+1/2
15 y+1/2,x+1/2,z+1/2
16 -y+1/2,-x+1/2,z+1/2
 
loop_
_atom_site_label
_atom_site_type_symbol
_atom_site_symmetry_multiplicity
_atom_site_Wyckoff_label
_atom_site_fract_x
_atom_site_fract_y
_atom_site_fract_z
_atom_site_occupancy
Ge1 Ge   2 a 0.00000 0.00000 0.00000 1.00000
P1  P    2 a 0.00000 0.00000 0.42700 1.00000
\end{lstlisting}
{\phantomsection\label{AB_tI4_107_a_a_poscar}}
{\hyperref[AB_tI4_107_a_a]{GeP (High-pressure, superconducting): AB\_tI4\_107\_a\_a}} - POSCAR
\begin{lstlisting}[numbers=none,language={mylang}]
AB_tI4_107_a_a & a,c/a,z1,z2 --params=3.5440505103,1.57477426639,0.0,0.427 & I4mm C_{4v}^{9} #107 (a^2) & tI4 & None & GeP &  & P. C. Donohue and H. S. Young, J. Solid State Chem. 1, 143-149 (1970)
   1.00000000000000
  -1.77202525515000   1.77202525515000   2.79053977120000
   1.77202525515000  -1.77202525515000   2.79053977120000
   1.77202525515000   1.77202525515000  -2.79053977120000
    Ge     P
     1     1
Direct
   0.00000000000000   0.00000000000000   0.00000000000000   Ge   (2a)
   0.42700000000000   0.42700000000000   0.00000000000000    P   (2a)
\end{lstlisting}
{\phantomsection\label{A3B5_tI32_108_ac_a2c_cif}}
{\hyperref[A3B5_tI32_108_ac_a2c]{Sr$_{5}$Si$_{3}$: A3B5\_tI32\_108\_ac\_a2c}} - CIF
\begin{lstlisting}[numbers=none,language={mylang}]
# CIF file
data_findsym-output
_audit_creation_method FINDSYM

_chemical_name_mineral 'Sr5Si3'
_chemical_formula_sum 'Si3 Sr5'

loop_
_publ_author_name
 'G. Nagorsen'
 'G. Rockt{\"a}schel'
 'H. Sch{\"a}fer'
 'A. Weiss'
_journal_name_full_name
;
 Zeitschrift f{\"u}r Naturforschung B
;
_journal_volume 22
_journal_year 1967
_journal_page_first 101
_journal_page_last 102
_publ_Section_title
;
 Die Kristallstruktur der Phase Sr$_{5}$Si$_{3}$
;

# Found in Pearson's Crystal Data - Crystal Structure Database for Inorganic Compounds, 2013

_aflow_title 'Sr$_{5}$Si$_{3}$ Structure'
_aflow_proto 'A3B5_tI32_108_ac_a2c'
_aflow_params 'a,c/a,z_{1},z_{2},x_{3},z_{3},x_{4},z_{4},x_{5},z_{5}'
_aflow_params_values '8.0549870847,1.94761018001,0.007,0.75,0.109,0.257,0.676,0.114,0.676,0.4'
_aflow_Strukturbericht 'None'
_aflow_Pearson 'tI32'

_cell_length_a    8.0549870847
_cell_length_b    8.0549870847
_cell_length_c    15.6879748460
_cell_angle_alpha 90.0000000000
_cell_angle_beta  90.0000000000
_cell_angle_gamma 90.0000000000
 
_symmetry_space_group_name_H-M "I 4 c m"
_symmetry_Int_Tables_number 108
 
loop_
_space_group_symop_id
_space_group_symop_operation_xyz
1 x,y,z
2 -x,-y,z
3 -y,x,z
4 y,-x,z
5 -x,y,z+1/2
6 x,-y,z+1/2
7 y,x,z+1/2
8 -y,-x,z+1/2
9 x+1/2,y+1/2,z+1/2
10 -x+1/2,-y+1/2,z+1/2
11 -y+1/2,x+1/2,z+1/2
12 y+1/2,-x+1/2,z+1/2
13 -x+1/2,y+1/2,z
14 x+1/2,-y+1/2,z
15 y+1/2,x+1/2,z
16 -y+1/2,-x+1/2,z
 
loop_
_atom_site_label
_atom_site_type_symbol
_atom_site_symmetry_multiplicity
_atom_site_Wyckoff_label
_atom_site_fract_x
_atom_site_fract_y
_atom_site_fract_z
_atom_site_occupancy
Si1 Si   4 a 0.00000 0.00000 0.00700 1.00000
Sr1 Sr   4 a 0.00000 0.00000 0.75000 1.00000
Si2 Si   8 c 0.10900 0.60900 0.25700 1.00000
Sr2 Sr   8 c 0.67600 0.17600 0.11400 1.00000
Sr3 Sr   8 c 0.67600 0.17600 0.40000 1.00000
\end{lstlisting}
{\phantomsection\label{A3B5_tI32_108_ac_a2c_poscar}}
{\hyperref[A3B5_tI32_108_ac_a2c]{Sr$_{5}$Si$_{3}$: A3B5\_tI32\_108\_ac\_a2c}} - POSCAR
\begin{lstlisting}[numbers=none,language={mylang}]
A3B5_tI32_108_ac_a2c & a,c/a,z1,z2,x3,z3,x4,z4,x5,z5 --params=8.0549870847,1.94761018001,0.007,0.75,0.109,0.257,0.676,0.114,0.676,0.4 & I4cm C_{4v}^{10} #108 (a^2c^3) & tI32 & None & Sr5Si3 &  & G. Nagorsen et al., Z. Naturforsch. B 22, 101-102 (1967)
   1.00000000000000
  -4.02749354235000   4.02749354235000   7.84398742300000
   4.02749354235000  -4.02749354235000   7.84398742300000
   4.02749354235000   4.02749354235000  -7.84398742300000
    Si    Sr
     6    10
Direct
   0.00700000000000   0.00700000000000   0.00000000000000   Si   (4a)
   0.50700000000000   0.50700000000000   0.00000000000000   Si   (4a)
   0.86600000000000   0.36600000000000   0.71800000000000   Si   (8c)
   0.64800000000000   0.14800000000000   0.28200000000000   Si   (8c)
   0.36600000000000   0.64800000000000   0.50000000000000   Si   (8c)
   0.14800000000000   0.86600000000000   0.50000000000000   Si   (8c)
   0.75000000000000   0.75000000000000   0.00000000000000   Sr   (4a)
   1.25000000000000   1.25000000000000   0.00000000000000   Sr   (4a)
   1.29000000000000   0.79000000000000   1.85200000000000   Sr   (8c)
  -0.06200000000000  -0.56200000000000  -0.85200000000000   Sr   (8c)
   0.79000000000000  -0.06200000000000   0.50000000000000   Sr   (8c)
  -0.56200000000000   1.29000000000000   0.50000000000000   Sr   (8c)
   1.57600000000000   1.07600000000000   1.85200000000000   Sr   (8c)
   0.22400000000000  -0.27600000000000  -0.85200000000000   Sr   (8c)
   1.07600000000000   0.22400000000000   0.50000000000000   Sr   (8c)
  -0.27600000000000   1.57600000000000   0.50000000000000   Sr   (8c)
\end{lstlisting}
{\phantomsection\label{ABC_tI12_109_a_a_a_cif}}
{\hyperref[ABC_tI12_109_a_a_a]{LaPtSi: ABC\_tI12\_109\_a\_a\_a}} - CIF
\begin{lstlisting}[numbers=none,language={mylang}]
# CIF file
data_findsym-output
_audit_creation_method FINDSYM

_chemical_name_mineral 'LaPtSi'
_chemical_formula_sum 'La Pt Si'

loop_
_publ_author_name
 'K. Klepp'
 'E. Parth{\\'e}'
_journal_name_full_name
;
 Acta Crystallographica Section B: Structural Science
;
_journal_volume 38
_journal_year 1982
_journal_page_first 1105
_journal_page_last 1108
_publ_Section_title
;
 $R$PtSi phases ($R$ = La, Ce, Pr, Nd, Sm and Gd) with an ordered ThSi$_{2}$ derivative structure
;

# Found in Pearson's Crystal Data - Crystal Structure Database for Inorganic Compounds, 2013

_aflow_title 'LaPtSi Structure'
_aflow_proto 'ABC_tI12_109_a_a_a'
_aflow_params 'a,c/a,z_{1},z_{2},z_{3}'
_aflow_params_values '4.2490694941,3.42174629325,0.081,0.666,0.5'
_aflow_Strukturbericht 'None'
_aflow_Pearson 'tI12'

_cell_length_a    4.2490694941
_cell_length_b    4.2490694941
_cell_length_c    14.5392377912
_cell_angle_alpha 90.0000000000
_cell_angle_beta  90.0000000000
_cell_angle_gamma 90.0000000000
 
_symmetry_space_group_name_H-M "I 41 m d"
_symmetry_Int_Tables_number 109
 
loop_
_space_group_symop_id
_space_group_symop_operation_xyz
1 x,y,z
2 -x,-y,z
3 -y,x+1/2,z+1/4
4 y,-x+1/2,z+1/4
5 -x,y,z
6 x,-y,z
7 y,x+1/2,z+1/4
8 -y,-x+1/2,z+1/4
9 x+1/2,y+1/2,z+1/2
10 -x+1/2,-y+1/2,z+1/2
11 -y+1/2,x,z+3/4
12 y+1/2,-x,z+3/4
13 -x+1/2,y+1/2,z+1/2
14 x+1/2,-y+1/2,z+1/2
15 y+1/2,x,z+3/4
16 -y+1/2,-x,z+3/4
 
loop_
_atom_site_label
_atom_site_type_symbol
_atom_site_symmetry_multiplicity
_atom_site_Wyckoff_label
_atom_site_fract_x
_atom_site_fract_y
_atom_site_fract_z
_atom_site_occupancy
La1 La   4 a 0.00000 0.00000 0.08100 1.00000
Pt1 Pt   4 a 0.00000 0.00000 0.66600 1.00000
Si1 Si   4 a 0.00000 0.00000 0.50000 1.00000
\end{lstlisting}
{\phantomsection\label{ABC_tI12_109_a_a_a_poscar}}
{\hyperref[ABC_tI12_109_a_a_a]{LaPtSi: ABC\_tI12\_109\_a\_a\_a}} - POSCAR
\begin{lstlisting}[numbers=none,language={mylang}]
ABC_tI12_109_a_a_a & a,c/a,z1,z2,z3 --params=4.2490694941,3.42174629325,0.081,0.666,0.5 & I4_{1}md C_{4v}^{11} #109 (a^3) & tI12 & None & LaPtSi &  & K. Klepp and E. Parth{\'e}, Acta Crystallogr. Sect. B Struct. Sci. 38, 1105-1108 (1982)
   1.00000000000000
  -2.12453474705000   2.12453474705000   7.26961889560000
   2.12453474705000  -2.12453474705000   7.26961889560000
   2.12453474705000   2.12453474705000  -7.26961889560000
    La    Pt    Si
     2     2     2
Direct
   0.08100000000000   0.08100000000000   0.00000000000000   La   (4a)
   0.83100000000000   0.33100000000000   0.50000000000000   La   (4a)
   0.66600000000000   0.66600000000000   0.00000000000000   Pt   (4a)
   1.41600000000000   0.91600000000000   0.50000000000000   Pt   (4a)
   0.50000000000000   0.50000000000000   0.00000000000000   Si   (4a)
   1.25000000000000   0.75000000000000   0.50000000000000   Si   (4a)
\end{lstlisting}
{\phantomsection\label{AB_tI8_109_a_a_cif}}
{\hyperref[AB_tI8_109_a_a]{NbAs: AB\_tI8\_109\_a\_a}} - CIF
\begin{lstlisting}[numbers=none,language={mylang}]
# CIF file
data_findsym-output
_audit_creation_method FINDSYM

_chemical_name_mineral 'NbAs'
_chemical_formula_sum 'As Nb'

loop_
_publ_author_name
 'S. Furuseth'
 'A. Kjekshus'
_journal_name_full_name
;
 Acta Chemica Scandinavica
;
_journal_volume 18
_journal_year 1964
_journal_page_first 1180
_journal_page_last 1195
_publ_Section_title
;
 On the Arsenides and Antimonides of Niobium
;

# Found in Pearson's Crystal Data - Crystal Structure Database for Inorganic Compounds, 2013

_aflow_title 'NbAs Structure'
_aflow_proto 'AB_tI8_109_a_a'
_aflow_params 'a,c/a,z_{1},z_{2}'
_aflow_params_values '3.4517145504,3.38383984705,0.5416,0.5'
_aflow_Strukturbericht 'None'
_aflow_Pearson 'tI8'

_cell_length_a    3.4517145504
_cell_length_b    3.4517145504
_cell_length_c    11.6800492363
_cell_angle_alpha 90.0000000000
_cell_angle_beta  90.0000000000
_cell_angle_gamma 90.0000000000
 
_symmetry_space_group_name_H-M "I 41 m d"
_symmetry_Int_Tables_number 109
 
loop_
_space_group_symop_id
_space_group_symop_operation_xyz
1 x,y,z
2 -x,-y,z
3 -y,x+1/2,z+1/4
4 y,-x+1/2,z+1/4
5 -x,y,z
6 x,-y,z
7 y,x+1/2,z+1/4
8 -y,-x+1/2,z+1/4
9 x+1/2,y+1/2,z+1/2
10 -x+1/2,-y+1/2,z+1/2
11 -y+1/2,x,z+3/4
12 y+1/2,-x,z+3/4
13 -x+1/2,y+1/2,z+1/2
14 x+1/2,-y+1/2,z+1/2
15 y+1/2,x,z+3/4
16 -y+1/2,-x,z+3/4
 
loop_
_atom_site_label
_atom_site_type_symbol
_atom_site_symmetry_multiplicity
_atom_site_Wyckoff_label
_atom_site_fract_x
_atom_site_fract_y
_atom_site_fract_z
_atom_site_occupancy
As1 As   4 a 0.00000 0.00000 0.54160 1.00000
Nb1 Nb   4 a 0.00000 0.00000 0.50000 1.00000
\end{lstlisting}
{\phantomsection\label{AB_tI8_109_a_a_poscar}}
{\hyperref[AB_tI8_109_a_a]{NbAs: AB\_tI8\_109\_a\_a}} - POSCAR
\begin{lstlisting}[numbers=none,language={mylang}]
AB_tI8_109_a_a & a,c/a,z1,z2 --params=3.4517145504,3.38383984705,0.5416,0.5 & I4_{1}md C_{4v}^{11} #109 (a^2) & tI8 & None & NbAs &  & S. Furuseth and A. Kjekshus, Acta Chem. Scand. 18, 1180-1195 (1964)
   1.00000000000000
  -1.72585727520000   1.72585727520000   5.84002461815000
   1.72585727520000  -1.72585727520000   5.84002461815000
   1.72585727520000   1.72585727520000  -5.84002461815000
    As    Nb
     2     2
Direct
   0.54160000000000   0.54160000000000   0.00000000000000   As   (4a)
   1.29160000000000   0.79160000000000   0.50000000000000   As   (4a)
   0.50000000000000   0.50000000000000   0.00000000000000   Nb   (4a)
   1.25000000000000   0.75000000000000   0.50000000000000   Nb   (4a)
\end{lstlisting}
{\phantomsection\label{A2BC8_tI176_110_2b_b_8b_cif}}
{\hyperref[A2BC8_tI176_110_2b_b_8b]{Be[BH$_{4}$]$_{2}$: A2BC8\_tI176\_110\_2b\_b\_8b}} - CIF

{\phantomsection\label{A2BC8_tI176_110_2b_b_8b_poscar}}
{\hyperref[A2BC8_tI176_110_2b_b_8b]{Be[BH$_{4}$]$_{2}$: A2BC8\_tI176\_110\_2b\_b\_8b}} - POSCAR

{\phantomsection\label{A2B_tP12_111_2n_adf_cif}}
{\hyperref[A2B_tP12_111_2n_adf]{MnF$_{2}$: A2B\_tP12\_111\_2n\_adf}} - CIF
\begin{lstlisting}[numbers=none,language={mylang}]
# CIF file
data_findsym-output
_audit_creation_method FINDSYM

_chemical_name_mineral 'MnF2'
_chemical_formula_sum 'F2 Mn'

loop_
_publ_author_name
 'T. Yagi'
 'J. C. Jamieson'
 'P. B. Moore'
_journal_name_full_name
;
 Journal of Geophysical Research
;
_journal_volume 84
_journal_year 1979
_journal_page_first 1113
_journal_page_last 1115
_publ_Section_title
;
 Polymorphism in MnF$_{2}$ (rutile type) at high pressures
;

# Found in Pearson's Crystal Data - Crystal Structure Database for Inorganic Compounds, 2013

_aflow_title 'MnF$_{2}$ Structure'
_aflow_proto 'A2B_tP12_111_2n_adf'
_aflow_params 'a,c/a,x_{4},z_{4},x_{5},z_{5}'
_aflow_params_values '5.1219931862,1.02616165562,0.205,0.28,0.301,0.622'
_aflow_Strukturbericht 'None'
_aflow_Pearson 'tP12'

_cell_length_a    5.1219931862
_cell_length_b    5.1219931862
_cell_length_c    5.2559930080
_cell_angle_alpha 90.0000000000
_cell_angle_beta  90.0000000000
_cell_angle_gamma 90.0000000000
 
_symmetry_space_group_name_H-M "P -4 2 m"
_symmetry_Int_Tables_number 111
 
loop_
_space_group_symop_id
_space_group_symop_operation_xyz
1 x,y,z
2 x,-y,-z
3 -x,y,-z
4 -x,-y,z
5 y,x,z
6 y,-x,-z
7 -y,x,-z
8 -y,-x,z
 
loop_
_atom_site_label
_atom_site_type_symbol
_atom_site_symmetry_multiplicity
_atom_site_Wyckoff_label
_atom_site_fract_x
_atom_site_fract_y
_atom_site_fract_z
_atom_site_occupancy
Mn1 Mn   1 a 0.00000 0.00000 0.00000 1.00000
Mn2 Mn   1 d 0.50000 0.50000 0.00000 1.00000
Mn3 Mn   2 f 0.50000 0.00000 0.50000 1.00000
F1  F    4 n 0.20500 0.20500 0.28000 1.00000
F2  F    4 n 0.30100 0.30100 0.62200 1.00000
\end{lstlisting}
{\phantomsection\label{A2B_tP12_111_2n_adf_poscar}}
{\hyperref[A2B_tP12_111_2n_adf]{MnF$_{2}$: A2B\_tP12\_111\_2n\_adf}} - POSCAR
\begin{lstlisting}[numbers=none,language={mylang}]
A2B_tP12_111_2n_adf & a,c/a,x4,z4,x5,z5 --params=5.1219931862,1.02616165562,0.205,0.28,0.301,0.622 & P-42m D_{2d}^{1} #111 (adfn^2) & tP12 & None & MnF2 &  & T. Yagi and J. C. Jamieson and P. B. Moore, J. Geophys. Res. 84, 1113-1115 (1979)
   1.00000000000000
   5.12199318620000   0.00000000000000   0.00000000000000
   0.00000000000000   5.12199318620000   0.00000000000000
   0.00000000000000   0.00000000000000   5.25599300800000
     F    Mn
     8     4
Direct
   0.20500000000000   0.20500000000000   0.28000000000000    F   (4n)
  -0.20500000000000  -0.20500000000000   0.28000000000000    F   (4n)
   0.20500000000000  -0.20500000000000  -0.28000000000000    F   (4n)
  -0.20500000000000   0.20500000000000  -0.28000000000000    F   (4n)
   0.30100000000000   0.30100000000000   0.62200000000000    F   (4n)
  -0.30100000000000  -0.30100000000000   0.62200000000000    F   (4n)
   0.30100000000000  -0.30100000000000  -0.62200000000000    F   (4n)
  -0.30100000000000   0.30100000000000  -0.62200000000000    F   (4n)
   0.00000000000000   0.00000000000000   0.00000000000000   Mn   (1a)
   0.50000000000000   0.50000000000000   0.00000000000000   Mn   (1d)
   0.50000000000000   0.00000000000000   0.50000000000000   Mn   (2f)
   0.00000000000000   0.50000000000000   0.50000000000000   Mn   (2f)
\end{lstlisting}
{\phantomsection\label{AB_tP8_111_n_n_cif}}
{\hyperref[AB_tP8_111_n_n]{NV (Low-temperature): AB\_tP8\_111\_n\_n}} - CIF
\begin{lstlisting}[numbers=none,language={mylang}]
# CIF file
data_findsym-output
_audit_creation_method FINDSYM

_chemical_name_mineral 'VN'
_chemical_formula_sum 'N V'

loop_
_publ_author_name
 'F. Kubel'
 'W. Lengauer'
 'K. Yvon'
 'K. Knorr'
 'A. Junod'
_journal_name_full_name
;
 Physical Review B
;
_journal_volume 38
_journal_year 1988
_journal_page_first 12908
_journal_page_last 12908
_publ_Section_title
;
 Structural phase transition at 205 K in stoichiometric vanadium nitride
;

# Found in Pearson's Crystal Data - Crystal Structure Database for Inorganic Compounds, 2013

_aflow_title 'NV (Low-temperature) Structure'
_aflow_proto 'AB_tP8_111_n_n'
_aflow_params 'a,c/a,x_{1},z_{1},x_{2},z_{2}'
_aflow_params_values '4.1305540686,0.998959140196,0.2522,0.7473,0.24415,0.24404'
_aflow_Strukturbericht 'None'
_aflow_Pearson 'tP8'

_cell_length_a    4.1305540686
_cell_length_b    4.1305540686
_cell_length_c    4.1262547409
_cell_angle_alpha 90.0000000000
_cell_angle_beta  90.0000000000
_cell_angle_gamma 90.0000000000
 
_symmetry_space_group_name_H-M "P -4 2 m"
_symmetry_Int_Tables_number 111
 
loop_
_space_group_symop_id
_space_group_symop_operation_xyz
1 x,y,z
2 x,-y,-z
3 -x,y,-z
4 -x,-y,z
5 y,x,z
6 y,-x,-z
7 -y,x,-z
8 -y,-x,z
 
loop_
_atom_site_label
_atom_site_type_symbol
_atom_site_symmetry_multiplicity
_atom_site_Wyckoff_label
_atom_site_fract_x
_atom_site_fract_y
_atom_site_fract_z
_atom_site_occupancy
N1 N   4 n 0.25220 0.25220 0.74730 1.00000
V1 V   4 n 0.24415 0.24415 0.24404 1.00000
\end{lstlisting}
{\phantomsection\label{AB_tP8_111_n_n_poscar}}
{\hyperref[AB_tP8_111_n_n]{NV (Low-temperature): AB\_tP8\_111\_n\_n}} - POSCAR
\begin{lstlisting}[numbers=none,language={mylang}]
AB_tP8_111_n_n & a,c/a,x1,z1,x2,z2 --params=4.1305540686,0.998959140196,0.2522,0.7473,0.24415,0.24404 & P-42m D_{2d}^{1} #111 (n^2) & tP8 & None & VN &  & F. Kubel et al., Phys. Rev. B 38, 12908(1988)
   1.00000000000000
   4.13055406860000   0.00000000000000   0.00000000000000
   0.00000000000000   4.13055406860000   0.00000000000000
   0.00000000000000   0.00000000000000   4.12625474090000
     N     V
     4     4
Direct
   0.25220000000000   0.25220000000000   0.74730000000000    N   (4n)
  -0.25220000000000  -0.25220000000000   0.74730000000000    N   (4n)
   0.25220000000000  -0.25220000000000  -0.74730000000000    N   (4n)
  -0.25220000000000   0.25220000000000  -0.74730000000000    N   (4n)
   0.24415000000000   0.24415000000000   0.24404000000000    V   (4n)
  -0.24415000000000  -0.24415000000000   0.24404000000000    V   (4n)
   0.24415000000000  -0.24415000000000  -0.24404000000000    V   (4n)
  -0.24415000000000   0.24415000000000  -0.24404000000000    V   (4n)
\end{lstlisting}
{\phantomsection\label{AB4C_tP12_112_b_n_e_cif}}
{\hyperref[AB4C_tP12_112_b_n_e]{$\alpha$-CuAlCl$_{4}$: AB4C\_tP12\_112\_b\_n\_e}} - CIF
\begin{lstlisting}[numbers=none,language={mylang}]
# CIF file
data_findsym-output
_audit_creation_method FINDSYM

_chemical_name_mineral 'alpha-CuAlCl4'
_chemical_formula_sum 'Al Cl4 Cu'

loop_
_publ_author_name
 'J. D. Martin'
 'B. R. Leafblad'
 'R. M. Sullivan'
 'P. D. Boyle'
_journal_name_full_name
;
 Inorganic Chemistry
;
_journal_volume 37
_journal_year 1998
_journal_page_first 1341
_journal_page_last 1346
_publ_Section_title
;
 $\alpha$-and $\beta$-CuAlCl$_{4}$: Framework Construction Using Corner-Shared Tetrahedral Metal-Halide Building Blocks
;

# Found in Pearson's Crystal Data - Crystal Structure Database for Inorganic Compounds, 2013

_aflow_title '$\alpha$-CuAlCl$_{4}$ Structure'
_aflow_proto 'AB4C_tP12_112_b_n_e'
_aflow_params 'a,c/a,x_{3},y_{3},z_{3}'
_aflow_params_values '5.4410982776,1.85862633019,0.2334,0.2761,0.6295'
_aflow_Strukturbericht 'None'
_aflow_Pearson 'tP12'

_cell_length_a    5.4410982776
_cell_length_b    5.4410982776
_cell_length_c    10.1129685239
_cell_angle_alpha 90.0000000000
_cell_angle_beta  90.0000000000
_cell_angle_gamma 90.0000000000
 
_symmetry_space_group_name_H-M "P -4 2 c"
_symmetry_Int_Tables_number 112
 
loop_
_space_group_symop_id
_space_group_symop_operation_xyz
1 x,y,z
2 x,-y,-z+1/2
3 -x,y,-z+1/2
4 -x,-y,z
5 y,x,z+1/2
6 y,-x,-z
7 -y,x,-z
8 -y,-x,z+1/2
 
loop_
_atom_site_label
_atom_site_type_symbol
_atom_site_symmetry_multiplicity
_atom_site_Wyckoff_label
_atom_site_fract_x
_atom_site_fract_y
_atom_site_fract_z
_atom_site_occupancy
Al1 Al   2 b 0.50000 0.00000 0.25000 1.00000
Cu1 Cu   2 e 0.00000 0.00000 0.00000 1.00000
Cl1 Cl   8 n 0.23340 0.27610 0.62950 1.00000
\end{lstlisting}
{\phantomsection\label{AB4C_tP12_112_b_n_e_poscar}}
{\hyperref[AB4C_tP12_112_b_n_e]{$\alpha$-CuAlCl$_{4}$: AB4C\_tP12\_112\_b\_n\_e}} - POSCAR
\begin{lstlisting}[numbers=none,language={mylang}]
AB4C_tP12_112_b_n_e & a,c/a,x3,y3,z3 --params=5.4410982776,1.85862633019,0.2334,0.2761,0.6295 & P-42c D_{2d}^{2} #112 (ben) & tP12 & None & CuAlCl4 & alpha & J. D. Martin et al., Inorg. Chem. 37, 1341-1346 (1998)
   1.00000000000000
   5.44109827760000   0.00000000000000   0.00000000000000
   0.00000000000000   5.44109827760000   0.00000000000000
   0.00000000000000   0.00000000000000  10.11296852390000
    Al    Cl    Cu
     2     8     2
Direct
   0.50000000000000   0.00000000000000   0.25000000000000   Al   (2b)
   0.00000000000000   0.50000000000000   0.75000000000000   Al   (2b)
   0.23340000000000   0.27610000000000   0.62950000000000   Cl   (8n)
  -0.23340000000000  -0.27610000000000   0.62950000000000   Cl   (8n)
   0.27610000000000  -0.23340000000000  -0.62950000000000   Cl   (8n)
  -0.27610000000000   0.23340000000000  -0.62950000000000   Cl   (8n)
  -0.23340000000000   0.27610000000000  -0.12950000000000   Cl   (8n)
   0.23340000000000  -0.27610000000000  -0.12950000000000   Cl   (8n)
  -0.27610000000000  -0.23340000000000   1.12950000000000   Cl   (8n)
   0.27610000000000   0.23340000000000   1.12950000000000   Cl   (8n)
   0.00000000000000   0.00000000000000   0.00000000000000   Cu   (2e)
   0.00000000000000   0.00000000000000   0.50000000000000   Cu   (2e)
\end{lstlisting}
{\phantomsection\label{A2BC7D2_tP24_113_e_a_cef_e_cif}}
{\hyperref[A2BC7D2_tP24_113_e_a_cef_e]{Akermanite (Ca$_2$MgSi$_2$O$_7$, $S5_{3}$): A2BC7D2\_tP24\_113\_e\_a\_cef\_e}} - CIF
\begin{lstlisting}[numbers=none,language={mylang}]
# CIF file 
data_findsym-output
_audit_creation_method FINDSYM

_chemical_name_mineral 'Akermanite'
_chemical_formula_sum 'Ca2 Mg O7 Si2'

loop_
_publ_author_name
 'H. Yang'
 'R. M. Hazen'
 'R. T. Downs'
 'L. W. Finger'
_journal_name_full_name
;
 Physics and Chemistry of Minerals
;
_journal_volume 24
_journal_year 1997
_journal_page_first 510
_journal_page_last 519
_publ_Section_title
;
 Structural change associated with the incommensurate-normal phase transition in akermanite, Ca$_{2}$MgSi$_{2}$O$_{7}$, at high pressure
;

_aflow_title 'Akermanite (Ca$_2$MgSi$_2$O$_7$, $S5_{3}$) Structure'
_aflow_proto 'A2BC7D2_tP24_113_e_a_cef_e'
_aflow_params 'a,c/a,z_{2},x_{3},z_{3},x_{4},z_{4},x_{5},z_{5},x_{6},y_{6},z_{6}'
_aflow_params_values '7.8338,0.639306594501,0.8201,0.8324,0.4935,0.6407,0.7471,0.6396,0.0642,0.0798,0.1862,0.7856'
_aflow_Strukturbericht '$S5_{3}$'
_aflow_Pearson 'tP24'

_symmetry_space_group_name_H-M "P -4 21 m"
_symmetry_Int_Tables_number 113
 
_cell_length_a    7.83380
_cell_length_b    7.83380
_cell_length_c    5.00820
_cell_angle_alpha 90.00000
_cell_angle_beta  90.00000
_cell_angle_gamma 90.00000
 
loop_
_space_group_symop_id
_space_group_symop_operation_xyz
1 x,y,z
2 x+1/2,-y+1/2,-z
3 -x+1/2,y+1/2,-z
4 -x,-y,z
5 y+1/2,x+1/2,z
6 y,-x,-z
7 -y,x,-z
8 -y+1/2,-x+1/2,z
 
loop_
_atom_site_label
_atom_site_type_symbol
_atom_site_symmetry_multiplicity
_atom_site_Wyckoff_label
_atom_site_fract_x
_atom_site_fract_y
_atom_site_fract_z
_atom_site_occupancy
Mg1 Mg   2 a 0.00000 0.00000 0.00000 1.00000
O1  O    2 c 0.00000 0.50000 0.82010 1.00000
Ca1 Ca   4 e 0.83240 0.33240 0.49350 1.00000
O2  O    4 e 0.64070 0.14070 0.74710 1.00000
Si1 Si   4 e 0.63960 0.13960 0.06420 1.00000
O3  O    8 f 0.07980 0.18620 0.78560 1.00000
\end{lstlisting}
{\phantomsection\label{A2BC7D2_tP24_113_e_a_cef_e_poscar}}
{\hyperref[A2BC7D2_tP24_113_e_a_cef_e]{Akermanite (Ca$_2$MgSi$_2$O$_7$, $S5_{3}$): A2BC7D2\_tP24\_113\_e\_a\_cef\_e}} - POSCAR

{\phantomsection\label{A3B_tP32_114_3e_e_cif}}
{\hyperref[A3B_tP32_114_3e_e]{SeO$_{3}$: A3B\_tP32\_114\_3e\_e}} - CIF
\begin{lstlisting}[numbers=none,language={mylang}]
# CIF file
data_findsym-output
_audit_creation_method FINDSYM

_chemical_name_mineral 'SeO3'
_chemical_formula_sum 'O3 Se'

loop_
_publ_author_name
 'F. C. Mijlhoff'
 'C. H. {MacGillavry}'
_journal_name_full_name
;
 Acta Cristallographica
;
_journal_volume 15
_journal_year 1962
_journal_page_first 620
_journal_page_last 620
_publ_Section_title
;
 Symmetry and unit-cell dimensions of selenium trioxide
;

# Found in Pearson's Crystal Data - Crystal Structure Database for Inorganic Compounds, 2013

_aflow_title 'SeO$_{3}$ Structure'
_aflow_proto 'A3B_tP32_114_3e_e'
_aflow_params 'a,c/a,x_{1},y_{1},z_{1},x_{2},y_{2},z_{2},x_{3},y_{3},z_{3},x_{4},y_{4},z_{4}'
_aflow_params_values '9.6362543341,0.547945205485,0.6,0.743,0.809,0.836,0.555,0.604,0.881,0.899,0.844,0.5125,0.7273,0.563'
_aflow_Strukturbericht 'None'
_aflow_Pearson 'tP32'

_cell_length_a    9.6362543341
_cell_length_b    9.6362543341
_cell_length_c    5.2801393612
_cell_angle_alpha 90.0000000000
_cell_angle_beta  90.0000000000
_cell_angle_gamma 90.0000000000
 
_symmetry_space_group_name_H-M "P -4 21 c"
_symmetry_Int_Tables_number 114
 
loop_
_space_group_symop_id
_space_group_symop_operation_xyz
1 x,y,z
2 x+1/2,-y+1/2,-z+1/2
3 -x+1/2,y+1/2,-z+1/2
4 -x,-y,z
5 y+1/2,x+1/2,z+1/2
6 y,-x,-z
7 -y,x,-z
8 -y+1/2,-x+1/2,z+1/2
 
loop_
_atom_site_label
_atom_site_type_symbol
_atom_site_symmetry_multiplicity
_atom_site_Wyckoff_label
_atom_site_fract_x
_atom_site_fract_y
_atom_site_fract_z
_atom_site_occupancy
O1  O    8 e 0.60000 0.74300 0.80900 1.00000
O2  O    8 e 0.83600 0.55500 0.60400 1.00000
O3  O    8 e 0.88100 0.89900 0.84400 1.00000
Se1 Se   8 e 0.51250 0.72730 0.56300 1.00000
\end{lstlisting}
{\phantomsection\label{A3B_tP32_114_3e_e_poscar}}
{\hyperref[A3B_tP32_114_3e_e]{SeO$_{3}$: A3B\_tP32\_114\_3e\_e}} - POSCAR

{\phantomsection\label{A4B_tP10_114_e_a_cif}}
{\hyperref[A4B_tP10_114_e_a]{Pd$_{4}$Se: A4B\_tP10\_114\_e\_a}} - CIF
\begin{lstlisting}[numbers=none,language={mylang}]
# CIF file
data_findsym-output
_audit_creation_method FINDSYM

_chemical_name_mineral 'Pd4Se'
_chemical_formula_sum 'Pd4 Se'

loop_
_publ_author_name
 'F. Gr{\o}nvold'
 'E. R{\o}st'
_journal_name_full_name
;
 Acta Chemica Scandinavica
;
_journal_volume 10
_journal_year 1956
_journal_page_first 1620
_journal_page_last 1634
_publ_Section_title
;
 On the sulfides, selenides and tellurides of palladium
;

# Found in Pearson's Crystal Data - Crystal Structure Database for Inorganic Compounds, 2013

_aflow_title 'Pd$_{4}$Se Structure'
_aflow_proto 'A4B_tP10_114_e_a'
_aflow_params 'a,c/a,x_{2},y_{2},z_{2}'
_aflow_params_values '5.2323591487,1.07923706139,0.626,0.768,0.846'
_aflow_Strukturbericht 'None'
_aflow_Pearson 'tP10'

_cell_length_a    5.2323591487
_cell_length_b    5.2323591487
_cell_length_c    5.6469559118
_cell_angle_alpha 90.0000000000
_cell_angle_beta  90.0000000000
_cell_angle_gamma 90.0000000000
 
_symmetry_space_group_name_H-M "P -4 21 c"
_symmetry_Int_Tables_number 114
 
loop_
_space_group_symop_id
_space_group_symop_operation_xyz
1 x,y,z
2 x+1/2,-y+1/2,-z+1/2
3 -x+1/2,y+1/2,-z+1/2
4 -x,-y,z
5 y+1/2,x+1/2,z+1/2
6 y,-x,-z
7 -y,x,-z
8 -y+1/2,-x+1/2,z+1/2
 
loop_
_atom_site_label
_atom_site_type_symbol
_atom_site_symmetry_multiplicity
_atom_site_Wyckoff_label
_atom_site_fract_x
_atom_site_fract_y
_atom_site_fract_z
_atom_site_occupancy
Se1 Se   2 a 0.00000 0.00000 0.00000 1.00000
Pd1 Pd   8 e 0.62600 0.76800 0.84600 1.00000
\end{lstlisting}
{\phantomsection\label{A4B_tP10_114_e_a_poscar}}
{\hyperref[A4B_tP10_114_e_a]{Pd$_{4}$Se: A4B\_tP10\_114\_e\_a}} - POSCAR
\begin{lstlisting}[numbers=none,language={mylang}]
A4B_tP10_114_e_a & a,c/a,x2,y2,z2 --params=5.2323591487,1.07923706139,0.626,0.768,0.846 & P-42_{1}c D_{2d}^{4} #114 (ae) & tP10 & None & Pd4Se &  & F. Gr{\o}nvold and E. R{\o}st, Acta Chem. Scand. 10, 1620-1634 (1956)
   1.00000000000000
   5.23235914870000   0.00000000000000   0.00000000000000
   0.00000000000000   5.23235914870000   0.00000000000000
   0.00000000000000   0.00000000000000   5.64695591180000
    Pd    Se
     8     2
Direct
   0.62600000000000   0.76800000000000   0.84600000000000   Pd   (8e)
  -0.62600000000000  -0.76800000000000   0.84600000000000   Pd   (8e)
   0.76800000000000  -0.62600000000000  -0.84600000000000   Pd   (8e)
  -0.76800000000000   0.62600000000000  -0.84600000000000   Pd   (8e)
  -0.12600000000000   1.26800000000000  -0.34600000000000   Pd   (8e)
   1.12600000000000  -0.26800000000000  -0.34600000000000   Pd   (8e)
  -0.26800000000000  -0.12600000000000   1.34600000000000   Pd   (8e)
   1.26800000000000   1.12600000000000   1.34600000000000   Pd   (8e)
   0.00000000000000   0.00000000000000   0.00000000000000   Se   (2a)
   0.50000000000000   0.50000000000000   0.50000000000000   Se   (2a)
\end{lstlisting}
{\phantomsection\label{A2B3_tP5_115_g_ag_cif}}
{\hyperref[A2B3_tP5_115_g_ag]{Rh$_{3}$P$_{2}$: A2B3\_tP5\_115\_g\_ag}} - CIF
\begin{lstlisting}[numbers=none,language={mylang}]
# CIF file
data_findsym-output
_audit_creation_method FINDSYM

_chemical_name_mineral 'Rh3P2'
_chemical_formula_sum 'P2 Rh3'

loop_
_publ_author_name
 'E. H. {El Ghadraoui}'
 'R. Guerin'
 'M. Sergent'
_journal_name_full_name
;
 Acta Crystallographica Section C: Structural Chemistry
;
_journal_volume 39
_journal_year 1983
_journal_page_first 1493
_journal_page_last 1494
_publ_Section_title
;
 Diphosphure de trirhodium, Rh$_{3}$P$_{2}$: premier exemple d\'une structure lacunaire ordonn{\\'e}e de type {\it anti}-PbFCl
;

# Found in Pearson's Crystal Data - Crystal Structure Database for Inorganic Compounds, 2013

_aflow_title 'Rh$_{3}$P$_{2}$ Structure'
_aflow_proto 'A2B3_tP5_115_g_ag'
_aflow_params 'a,c/a,z_{2},z_{3}'
_aflow_params_values '3.3269188443,1.84881274424,0.253,0.6308'
_aflow_Strukturbericht 'None'
_aflow_Pearson 'tP5'

_cell_length_a    3.3269188443
_cell_length_b    3.3269188443
_cell_length_c    6.1508499584
_cell_angle_alpha 90.0000000000
_cell_angle_beta  90.0000000000
_cell_angle_gamma 90.0000000000
 
_symmetry_space_group_name_H-M "P -4 m 2"
_symmetry_Int_Tables_number 115
 
loop_
_space_group_symop_id
_space_group_symop_operation_xyz
1 x,y,z
2 -x,-y,z
3 -y,-x,-z
4 y,x,-z
5 -x,y,z
6 x,-y,z
7 y,-x,-z
8 -y,x,-z
 
loop_
_atom_site_label
_atom_site_type_symbol
_atom_site_symmetry_multiplicity
_atom_site_Wyckoff_label
_atom_site_fract_x
_atom_site_fract_y
_atom_site_fract_z
_atom_site_occupancy
Rh1 Rh   1 a 0.00000 0.00000 0.00000 1.00000
P1  P    2 g 0.00000 0.50000 0.25300 1.00000
Rh2 Rh   2 g 0.00000 0.50000 0.63080 1.00000
\end{lstlisting}
{\phantomsection\label{A2B3_tP5_115_g_ag_poscar}}
{\hyperref[A2B3_tP5_115_g_ag]{Rh$_{3}$P$_{2}$: A2B3\_tP5\_115\_g\_ag}} - POSCAR
\begin{lstlisting}[numbers=none,language={mylang}]
A2B3_tP5_115_g_ag & a,c/a,z2,z3 --params=3.3269188443,1.84881274424,0.253,0.6308 & P-4m2 D_{2d}^{5} #115 (ag^2) & tP5 & None & Rh3P2 &  & E. H. {El Ghadraoui} and R. Guerin and M. Sergent, Acta Crystallogr. C 39, 1493-1494 (1983)
   1.00000000000000
   3.32691884430000   0.00000000000000   0.00000000000000
   0.00000000000000   3.32691884430000   0.00000000000000
   0.00000000000000   0.00000000000000   6.15084995840000
     P    Rh
     2     3
Direct
   0.00000000000000   0.50000000000000   0.25300000000000    P   (2g)
   0.50000000000000   0.00000000000000  -0.25300000000000    P   (2g)
   0.00000000000000   0.00000000000000   0.00000000000000   Rh   (1a)
   0.00000000000000   0.50000000000000   0.63080000000000   Rh   (2g)
   0.50000000000000   0.00000000000000  -0.63080000000000   Rh   (2g)
\end{lstlisting}
{\phantomsection\label{AB2_tP12_115_j_egi_cif}}
{\hyperref[AB2_tP12_115_j_egi]{HgI$_{2}$: AB2\_tP12\_115\_j\_egi}} - CIF
\begin{lstlisting}[numbers=none,language={mylang}]
# CIF file
data_findsym-output
_audit_creation_method FINDSYM

_chemical_name_mineral 'HgI2'
_chemical_formula_sum 'Hg I2'

loop_
_publ_author_name
 'M. Hostettler'
 'H. Birkedal'
 'D. Schwarzenbach'
_journal_name_full_name
;
 Acta Crystallographica Section B: Structural Science
;
_journal_volume 58
_journal_year 2002
_journal_page_first 903
_journal_page_last 913
_publ_Section_title
;
 The structure of orange HgI$_{2}$. I. Polytypic layer structure
;

# Found in Pearson's Crystal Data - Crystal Structure Database for Inorganic Compounds, 2013

_aflow_title 'HgI$_{2}$ Structure'
_aflow_proto 'AB2_tP12_115_j_egi'
_aflow_params 'a,c/a,z_{1},z_{2},x_{3},x_{4},z_{4}'
_aflow_params_values '8.7863400452,0.701854022742,0.0288,0.0315,0.26398,0.24918,0.24945'
_aflow_Strukturbericht 'None'
_aflow_Pearson 'tP12'

_cell_length_a    8.7863400452
_cell_length_b    8.7863400452
_cell_length_c    6.1667281059
_cell_angle_alpha 90.0000000000
_cell_angle_beta  90.0000000000
_cell_angle_gamma 90.0000000000
 
_symmetry_space_group_name_H-M "P -4 m 2"
_symmetry_Int_Tables_number 115
 
loop_
_space_group_symop_id
_space_group_symop_operation_xyz
1 x,y,z
2 -x,-y,z
3 -y,-x,-z
4 y,x,-z
5 -x,y,z
6 x,-y,z
7 y,-x,-z
8 -y,x,-z
 
loop_
_atom_site_label
_atom_site_type_symbol
_atom_site_symmetry_multiplicity
_atom_site_Wyckoff_label
_atom_site_fract_x
_atom_site_fract_y
_atom_site_fract_z
_atom_site_occupancy
I1  I    2 e 0.00000 0.00000 0.02880 1.00000
I2  I    2 g 0.00000 0.50000 0.03150 1.00000
I3  I    4 i 0.26398 0.26398 0.50000 1.00000
Hg1 Hg   4 j 0.24918 0.00000 0.24945 1.00000
\end{lstlisting}
{\phantomsection\label{AB2_tP12_115_j_egi_poscar}}
{\hyperref[AB2_tP12_115_j_egi]{HgI$_{2}$: AB2\_tP12\_115\_j\_egi}} - POSCAR
\begin{lstlisting}[numbers=none,language={mylang}]
AB2_tP12_115_j_egi & a,c/a,z1,z2,x3,x4,z4 --params=8.7863400452,0.701854022742,0.0288,0.0315,0.26398,0.24918,0.24945 & P-4m2 D_{2d}^{5} #115 (egij) & tP12 & None & HgI2 &  & M. Hostettler and H. Birkedal and D. Schwarzenbach, Acta Crystallogr. Sect. B Struct. Sci. 58, 903-913 (2002)
   1.00000000000000
   8.78634004520000   0.00000000000000   0.00000000000000
   0.00000000000000   8.78634004520000   0.00000000000000
   0.00000000000000   0.00000000000000   6.16672810590000
    Hg     I
     4     8
Direct
   0.24918000000000   0.00000000000000   0.24945000000000   Hg   (4j)
  -0.24918000000000   0.00000000000000   0.24945000000000   Hg   (4j)
   0.00000000000000  -0.24918000000000  -0.24945000000000   Hg   (4j)
   0.00000000000000   0.24918000000000  -0.24945000000000   Hg   (4j)
   0.00000000000000   0.00000000000000   0.02880000000000    I   (2e)
   0.00000000000000   0.00000000000000  -0.02880000000000    I   (2e)
   0.00000000000000   0.50000000000000   0.03150000000000    I   (2g)
   0.50000000000000   0.00000000000000  -0.03150000000000    I   (2g)
   0.26398000000000   0.26398000000000   0.50000000000000    I   (4i)
  -0.26398000000000  -0.26398000000000   0.50000000000000    I   (4i)
   0.26398000000000  -0.26398000000000   0.50000000000000    I   (4i)
  -0.26398000000000   0.26398000000000   0.50000000000000    I   (4i)
\end{lstlisting}
{\phantomsection\label{A2B3_tP20_116_bci_fj_cif}}
{\hyperref[A2B3_tP20_116_bci_fj]{Ru$_{2}$Sn$_{3}$: A2B3\_tP20\_116\_bci\_fj}} - CIF
\begin{lstlisting}[numbers=none,language={mylang}]
# CIF file
data_findsym-output
_audit_creation_method FINDSYM

_chemical_name_mineral 'Ru2Sn3'
_chemical_formula_sum 'Ru2 Sn3'

loop_
_publ_author_name
 'O. Schwomma'
 'H. Nowotny'
 'A. Wittmann'
_journal_name_full_name
;
 Monatshefte f{\"u}r Chemie - Chemical Monthly
;
_journal_volume 95
_journal_year 1964
_journal_page_first 1538
_journal_page_last 1543
_publ_Section_title
;
 Untersuchungen im System: Ru--Sn
;

# Found in Pearson's Crystal Data - Crystal Structure Database for Inorganic Compounds, 2013

_aflow_title 'Ru$_{2}$Sn$_{3}$ Structure'
_aflow_proto 'A2B3_tP20_116_bci_fj'
_aflow_params 'a,c/a,x_{3},z_{4},x_{5},y_{5},z_{5}'
_aflow_params_values '6.1720115185,1.606448477,0.177,0.625,0.655,0.216,0.582'
_aflow_Strukturbericht 'None'
_aflow_Pearson 'tP20'

_cell_length_a    6.1720115185
_cell_length_b    6.1720115185
_cell_length_c    9.9150185039
_cell_angle_alpha 90.0000000000
_cell_angle_beta  90.0000000000
_cell_angle_gamma 90.0000000000
 
_symmetry_space_group_name_H-M "P -4 c 2"
_symmetry_Int_Tables_number 116
 
loop_
_space_group_symop_id
_space_group_symop_operation_xyz
1 x,y,z
2 -x,-y,z
3 -y,-x,-z+1/2
4 y,x,-z+1/2
5 -x,y,z+1/2
6 x,-y,z+1/2
7 y,-x,-z
8 -y,x,-z
 
loop_
_atom_site_label
_atom_site_type_symbol
_atom_site_symmetry_multiplicity
_atom_site_Wyckoff_label
_atom_site_fract_x
_atom_site_fract_y
_atom_site_fract_z
_atom_site_occupancy
Ru1 Ru   2 b 0.50000 0.50000 0.25000 1.00000
Ru2 Ru   2 c 0.00000 0.00000 0.00000 1.00000
Sn1 Sn   4 f 0.17700 0.17700 0.75000 1.00000
Ru3 Ru   4 i 0.00000 0.50000 0.62500 1.00000
Sn2 Sn   8 j 0.65500 0.21600 0.58200 1.00000
\end{lstlisting}
{\phantomsection\label{A2B3_tP20_116_bci_fj_poscar}}
{\hyperref[A2B3_tP20_116_bci_fj]{Ru$_{2}$Sn$_{3}$: A2B3\_tP20\_116\_bci\_fj}} - POSCAR
\begin{lstlisting}[numbers=none,language={mylang}]
A2B3_tP20_116_bci_fj & a,c/a,x3,z4,x5,y5,z5 --params=6.1720115185,1.606448477,0.177,0.625,0.655,0.216,0.582 & P-4c2 D_{2d}^{6} #116 (bcfij) & tP20 & None & Ru2Sn3 &  & O. Schwomma and H. Nowotny and A. Wittmann, Monatsh. Chem. 95, 1538-1543 (1964)
   1.00000000000000
   6.17201151850000   0.00000000000000   0.00000000000000
   0.00000000000000   6.17201151850000   0.00000000000000
   0.00000000000000   0.00000000000000   9.91501850390000
    Ru    Sn
     8    12
Direct
   0.50000000000000   0.50000000000000   0.25000000000000   Ru   (2b)
   0.50000000000000   0.50000000000000   0.75000000000000   Ru   (2b)
   0.00000000000000   0.00000000000000   0.00000000000000   Ru   (2c)
   0.00000000000000   0.00000000000000   0.50000000000000   Ru   (2c)
   0.00000000000000   0.50000000000000   0.62500000000000   Ru   (4i)
   0.50000000000000   0.00000000000000  -0.62500000000000   Ru   (4i)
   0.00000000000000   0.50000000000000   1.12500000000000   Ru   (4i)
   0.50000000000000   0.00000000000000  -0.12500000000000   Ru   (4i)
   0.17700000000000   0.17700000000000   0.75000000000000   Sn   (4f)
  -0.17700000000000  -0.17700000000000   0.75000000000000   Sn   (4f)
   0.17700000000000  -0.17700000000000   0.25000000000000   Sn   (4f)
  -0.17700000000000   0.17700000000000   0.25000000000000   Sn   (4f)
   0.65500000000000   0.21600000000000   0.58200000000000   Sn   (8j)
  -0.65500000000000  -0.21600000000000   0.58200000000000   Sn   (8j)
   0.21600000000000  -0.65500000000000  -0.58200000000000   Sn   (8j)
  -0.21600000000000   0.65500000000000  -0.58200000000000   Sn   (8j)
   0.65500000000000  -0.21600000000000   1.08200000000000   Sn   (8j)
  -0.65500000000000   0.21600000000000   1.08200000000000   Sn   (8j)
   0.21600000000000   0.65500000000000  -0.08200000000000   Sn   (8j)
  -0.21600000000000  -0.65500000000000  -0.08200000000000   Sn   (8j)
\end{lstlisting}
{\phantomsection\label{A2B3_tP20_117_i_adgh_cif}}
{\hyperref[A2B3_tP20_117_i_adgh]{$\beta$-Bi$_{2}$O$_{3}$ (High-temperature): A2B3\_tP20\_117\_i\_adgh}} - CIF
\begin{lstlisting}[numbers=none,language={mylang}]
# CIF file
data_findsym-output
_audit_creation_method FINDSYM

_chemical_name_mineral 'beta-Bi2O3'
_chemical_formula_sum 'Bi2 O3'

loop_
_publ_author_name
 'L. G. Sill{\\'e}n'
_journal_name_full_name
;
 Arkiv f{\"o}r Kemi, Mineralogi och Geologi
;
_journal_volume 12A
_journal_year 1937
_journal_page_first 1
_journal_page_last 15
_publ_Section_title
;
 X-ray studies on bismuth trioxide
;

# Found in Pearson's Crystal Data - Crystal Structure Database for Inorganic Compounds, 2013

_aflow_title '$\beta$-Bi$_{2}$O$_{3}$ (High-temperature) Structure'
_aflow_proto 'A2B3_tP20_117_i_adgh'
_aflow_params 'a,c/a,x_{3},x_{4},x_{5},y_{5},z_{5}'
_aflow_params_values '7.7289660931,0.727131582349,0.73,0.73,0.75,0.52,0.25'
_aflow_Strukturbericht 'None'
_aflow_Pearson 'tP20'

_cell_length_a    7.7289660931
_cell_length_b    7.7289660931
_cell_length_c    5.6199753452
_cell_angle_alpha 90.0000000000
_cell_angle_beta  90.0000000000
_cell_angle_gamma 90.0000000000
 
_symmetry_space_group_name_H-M "P -4 b 2"
_symmetry_Int_Tables_number 117
 
loop_
_space_group_symop_id
_space_group_symop_operation_xyz
1 x,y,z
2 -x,-y,z
3 -y+1/2,-x+1/2,-z
4 y+1/2,x+1/2,-z
5 -x+1/2,y+1/2,z
6 x+1/2,-y+1/2,z
7 y,-x,-z
8 -y,x,-z
 
loop_
_atom_site_label
_atom_site_type_symbol
_atom_site_symmetry_multiplicity
_atom_site_Wyckoff_label
_atom_site_fract_x
_atom_site_fract_y
_atom_site_fract_z
_atom_site_occupancy
O1  O    2 a 0.00000 0.00000 0.00000 1.00000
O2  O    2 d 0.00000 0.50000 0.50000 1.00000
O3  O    4 g 0.73000 0.23000 0.00000 1.00000
O4  O    4 h 0.73000 0.23000 0.50000 1.00000
Bi1 Bi   8 i 0.75000 0.52000 0.25000 1.00000
\end{lstlisting}
{\phantomsection\label{A2B3_tP20_117_i_adgh_poscar}}
{\hyperref[A2B3_tP20_117_i_adgh]{$\beta$-Bi$_{2}$O$_{3}$ (High-temperature): A2B3\_tP20\_117\_i\_adgh}} - POSCAR
\begin{lstlisting}[numbers=none,language={mylang}]
A2B3_tP20_117_i_adgh & a,c/a,x3,x4,x5,y5,z5 --params=7.7289660931,0.727131582349,0.73,0.73,0.75,0.52,0.25 & P-4b2 D_{2d}^{7} #117 (adghi) & tP20 & None & Bi2O3 & beta & L. G. Sill{\'e}n, {Ark. Kem. Mineral. Geol. 12A, 1-15 (1937)
   1.00000000000000
   7.72896609310000   0.00000000000000   0.00000000000000
   0.00000000000000   7.72896609310000   0.00000000000000
   0.00000000000000   0.00000000000000   5.61997534520000
    Bi     O
     8    12
Direct
   0.75000000000000   0.52000000000000   0.25000000000000   Bi   (8i)
  -0.75000000000000  -0.52000000000000   0.25000000000000   Bi   (8i)
   0.52000000000000  -0.75000000000000  -0.25000000000000   Bi   (8i)
  -0.52000000000000   0.75000000000000  -0.25000000000000   Bi   (8i)
   1.25000000000000  -0.02000000000000   0.25000000000000   Bi   (8i)
  -0.25000000000000   1.02000000000000   0.25000000000000   Bi   (8i)
   1.02000000000000   1.25000000000000  -0.25000000000000   Bi   (8i)
  -0.02000000000000  -0.25000000000000  -0.25000000000000   Bi   (8i)
   0.00000000000000   0.00000000000000   0.00000000000000    O   (2a)
   0.50000000000000   0.50000000000000   0.00000000000000    O   (2a)
   0.00000000000000   0.50000000000000   0.50000000000000    O   (2d)
   0.50000000000000   0.00000000000000   0.50000000000000    O   (2d)
   0.73000000000000   1.23000000000000   0.00000000000000    O   (4g)
  -0.73000000000000  -0.23000000000000   0.00000000000000    O   (4g)
   1.23000000000000  -0.73000000000000   0.00000000000000    O   (4g)
  -0.23000000000000   0.73000000000000   0.00000000000000    O   (4g)
   0.73000000000000   1.23000000000000   0.50000000000000    O   (4h)
  -0.73000000000000  -0.23000000000000   0.50000000000000    O   (4h)
   1.23000000000000  -0.73000000000000   0.50000000000000    O   (4h)
  -0.23000000000000   0.73000000000000   0.50000000000000    O   (4h)
\end{lstlisting}
{\phantomsection\label{A3B_tP16_118_ei_f_cif}}
{\hyperref[A3B_tP16_118_ei_f]{RuIn$_{3}$: A3B\_tP16\_118\_ei\_f}} - CIF
\begin{lstlisting}[numbers=none,language={mylang}]
# CIF file
data_findsym-output
_audit_creation_method FINDSYM

_chemical_name_mineral 'RuIn3'
_chemical_formula_sum 'In3 Ru'

loop_
_publ_author_name
 'R. B. Roof'
 'Z. Fisk'
 'J. L. Smith'
_journal_name_full_name
;
 Powder Diffraction
;
_journal_volume 1
_journal_year 1986
_journal_page_first 20
_journal_page_last 21
_publ_Section_title
;
 Crystal data for RuIn$_{3}$
;

# Found in Pearson's Crystal Data - Crystal Structure Database for Inorganic Compounds, 2013

_aflow_title 'RuIn$_{3}$ Structure'
_aflow_proto 'A3B_tP16_118_ei_f'
_aflow_params 'a,c/a,z_{1},x_{2},x_{3},y_{3},z_{3}'
_aflow_params_values '6.9983025398,1.03510852635,0.237,0.15,0.343,0.149,0.509'
_aflow_Strukturbericht 'None'
_aflow_Pearson 'tP16'

_cell_length_a    6.9983025398
_cell_length_b    6.9983025398
_cell_length_c    7.2440026289
_cell_angle_alpha 90.0000000000
_cell_angle_beta  90.0000000000
_cell_angle_gamma 90.0000000000
 
_symmetry_space_group_name_H-M "P -4 n 2"
_symmetry_Int_Tables_number 118
 
loop_
_space_group_symop_id
_space_group_symop_operation_xyz
1 x,y,z
2 -x,-y,z
3 -y+1/2,-x+1/2,-z+1/2
4 y+1/2,x+1/2,-z+1/2
5 -x+1/2,y+1/2,z+1/2
6 x+1/2,-y+1/2,z+1/2
7 y,-x,-z
8 -y,x,-z
 
loop_
_atom_site_label
_atom_site_type_symbol
_atom_site_symmetry_multiplicity
_atom_site_Wyckoff_label
_atom_site_fract_x
_atom_site_fract_y
_atom_site_fract_z
_atom_site_occupancy
In1 In   4 e 0.00000 0.00000 0.23700 1.00000
Ru1 Ru   4 f 0.15000 0.35000 0.25000 1.00000
In2 In   8 i 0.34300 0.14900 0.50900 1.00000
\end{lstlisting}
{\phantomsection\label{A3B_tP16_118_ei_f_poscar}}
{\hyperref[A3B_tP16_118_ei_f]{RuIn$_{3}$: A3B\_tP16\_118\_ei\_f}} - POSCAR
\begin{lstlisting}[numbers=none,language={mylang}]
A3B_tP16_118_ei_f & a,c/a,z1,x2,x3,y3,z3 --params=6.9983025398,1.03510852635,0.237,0.15,0.343,0.149,0.509 & P-4n2 D_{2d}^{8} #118 (efi) & tP16 & None & RuIn3 &  & R. B. Roof and Z. Fisk and J. L. Smith, Powder Diffraction 1, 20-21 (1986)
   1.00000000000000
   6.99830253980000   0.00000000000000   0.00000000000000
   0.00000000000000   6.99830253980000   0.00000000000000
   0.00000000000000   0.00000000000000   7.24400262890000
    In    Ru
    12     4
Direct
   0.00000000000000   0.00000000000000   0.23700000000000   In   (4e)
   0.00000000000000   0.00000000000000  -0.23700000000000   In   (4e)
   0.50000000000000   0.50000000000000   0.73700000000000   In   (4e)
   0.50000000000000   0.50000000000000   0.26300000000000   In   (4e)
   0.34300000000000   0.14900000000000   0.50900000000000   In   (8i)
  -0.34300000000000  -0.14900000000000   0.50900000000000   In   (8i)
   0.14900000000000  -0.34300000000000  -0.50900000000000   In   (8i)
  -0.14900000000000   0.34300000000000  -0.50900000000000   In   (8i)
   0.84300000000000   0.35100000000000   1.00900000000000   In   (8i)
   0.15700000000000   0.64900000000000   1.00900000000000   In   (8i)
   0.64900000000000   0.84300000000000  -0.00900000000000   In   (8i)
   0.35100000000000   0.15700000000000  -0.00900000000000   In   (8i)
   0.15000000000000   0.35000000000000   0.25000000000000   Ru   (4f)
  -0.15000000000000   0.65000000000000   0.25000000000000   Ru   (4f)
   0.35000000000000  -0.15000000000000   0.75000000000000   Ru   (4f)
   0.65000000000000   0.15000000000000   0.75000000000000   Ru   (4f)
\end{lstlisting}
{\phantomsection\label{A5B3_tP32_118_g2i_aceh_cif}}
{\hyperref[A5B3_tP32_118_g2i_aceh]{Ir$_{3}$Ga$_{5}$: A5B3\_tP32\_118\_g2i\_aceh}} - CIF
\begin{lstlisting}[numbers=none,language={mylang}]
# CIF file
data_findsym-output
_audit_creation_method FINDSYM

_chemical_name_mineral 'Ir3Ga5'
_chemical_formula_sum 'Ga5 Ir3'

loop_
_publ_author_name
 'H. V{\"o}llenkle'
 'A. Wittmann'
 'H. Nowotny'
_journal_name_full_name
;
 Monatshefte f{\"u}r Chemie - Chemical Monthly
;
_journal_volume 98
_journal_year 1967
_journal_page_first 176
_journal_page_last 183
_publ_Section_title
;
 Die Kristallstrukturen von Rh$_{10}$Ga$_{17}$ und Ir$_{3}$Ga$_{5}$
;

# Found in Pearson's Crystal Data - Crystal Structure Database for Inorganic Compounds, 2013

_aflow_title 'Ir$_{3}$Ga$_{5}$ Structure'
_aflow_proto 'A5B3_tP32_118_g2i_aceh'
_aflow_params 'a,c/a,z_{3},x_{4},z_{5},x_{6},y_{6},z_{6},x_{7},y_{7},z_{7}'
_aflow_params_values '5.8229835854,2.43860552978,0.6709,0.675,0.5861,0.73,0.85,0.5515,0.84,0.8,0.15'
_aflow_Strukturbericht 'None'
_aflow_Pearson 'tP32'

_cell_length_a    5.8229835854
_cell_length_b    5.8229835854
_cell_length_c    14.1999599712
_cell_angle_alpha 90.0000000000
_cell_angle_beta  90.0000000000
_cell_angle_gamma 90.0000000000
 
_symmetry_space_group_name_H-M "P -4 n 2"
_symmetry_Int_Tables_number 118
 
loop_
_space_group_symop_id
_space_group_symop_operation_xyz
1 x,y,z
2 -x,-y,z
3 -y+1/2,-x+1/2,-z+1/2
4 y+1/2,x+1/2,-z+1/2
5 -x+1/2,y+1/2,z+1/2
6 x+1/2,-y+1/2,z+1/2
7 y,-x,-z
8 -y,x,-z
 
loop_
_atom_site_label
_atom_site_type_symbol
_atom_site_symmetry_multiplicity
_atom_site_Wyckoff_label
_atom_site_fract_x
_atom_site_fract_y
_atom_site_fract_z
_atom_site_occupancy
Ir1 Ir   2 a 0.00000 0.00000 0.00000 1.00000
Ir2 Ir   2 c 0.00000 0.50000 0.25000 1.00000
Ir3 Ir   4 e 0.00000 0.00000 0.67090 1.00000
Ga1 Ga   4 g 0.67500 0.17500 0.25000 1.00000
Ir4 Ir   4 h 0.00000 0.50000 0.58610 1.00000
Ga2 Ga   8 i 0.73000 0.85000 0.55150 1.00000
Ga3 Ga   8 i 0.84000 0.80000 0.15000 1.00000
\end{lstlisting}
{\phantomsection\label{A5B3_tP32_118_g2i_aceh_poscar}}
{\hyperref[A5B3_tP32_118_g2i_aceh]{Ir$_{3}$Ga$_{5}$: A5B3\_tP32\_118\_g2i\_aceh}} - POSCAR

{\phantomsection\label{A3B_tI24_119_b2i_af_cif}}
{\hyperref[A3B_tI24_119_b2i_af]{RbGa$_{3}$: A3B\_tI24\_119\_b2i\_af}} - CIF
\begin{lstlisting}[numbers=none,language={mylang}]
# CIF file 
data_findsym-output
_audit_creation_method FINDSYM

_chemical_name_mineral 'RbGa3'
_chemical_formula_sum 'Ga3 Rb'

loop_
_publ_author_name
 'R. G. Ling'
 'C. Belin'
_journal_name_full_name
;
 Zeitschrift fur Anorganische und Allgemeine Chemie
;
_journal_volume 480
_journal_year 1981
_journal_page_first 181
_journal_page_last 185
_publ_Section_title
;
 Preparation and Crystal Structure Determination of the New Intermetallic Compound RbGa$_{3}$
;

# Found in Pearson's Handbook of Crystallographic Data for Intermetallic Phases, 1991

_aflow_title 'RbGa$_{3}$ Structure'
_aflow_proto 'A3B_tI24_119_b2i_af'
_aflow_params 'a,c/a,z_{3},x_{4},z_{4},x_{5},z_{5}'
_aflow_params_values '6.315,2.37529691211,0.372,0.2068,0.2229,0.3067,0.3917'
_aflow_Strukturbericht 'None'
_aflow_Pearson 'tI24'

_symmetry_space_group_name_H-M "I -4 m 2"
_symmetry_Int_Tables_number 119
 
_cell_length_a    6.31500
_cell_length_b    6.31500
_cell_length_c    15.00000
_cell_angle_alpha 90.00000
_cell_angle_beta  90.00000
_cell_angle_gamma 90.00000
 
loop_
_space_group_symop_id
_space_group_symop_operation_xyz
1 x,y,z
2 -x,-y,z
3 -y,-x,-z
4 y,x,-z
5 -x,y,z
6 x,-y,z
7 y,-x,-z
8 -y,x,-z
9 x+1/2,y+1/2,z+1/2
10 -x+1/2,-y+1/2,z+1/2
11 -y+1/2,-x+1/2,-z+1/2
12 y+1/2,x+1/2,-z+1/2
13 -x+1/2,y+1/2,z+1/2
14 x+1/2,-y+1/2,z+1/2
15 y+1/2,-x+1/2,-z+1/2
16 -y+1/2,x+1/2,-z+1/2
 
loop_
_atom_site_label
_atom_site_type_symbol
_atom_site_symmetry_multiplicity
_atom_site_Wyckoff_label
_atom_site_fract_x
_atom_site_fract_y
_atom_site_fract_z
_atom_site_occupancy
Rb1 Rb   2 a 0.00000 0.00000 0.00000 1.00000
Ga1 Ga   2 b 0.00000 0.00000 0.50000 1.00000
Rb2 Rb   4 f 0.00000 0.50000 0.37200 1.00000
Ga2 Ga   8 i 0.20680 0.00000 0.22290 1.00000
Ga3 Ga   8 i 0.30670 0.00000 0.39170 1.00000
\end{lstlisting}
{\phantomsection\label{A3B_tI24_119_b2i_af_poscar}}
{\hyperref[A3B_tI24_119_b2i_af]{RbGa$_{3}$: A3B\_tI24\_119\_b2i\_af}} - POSCAR
\begin{lstlisting}[numbers=none,language={mylang}]
A3B_tI24_119_b2i_af & a,c/a,z3,x4,z4,x5,z5 --params=6.315,2.37529691211,0.372,0.2068,0.2229,0.3067,0.3917 & I-4m2 D_{2d}^{9} #119 (abfi^2) & tI24 & None & RbGa3 & RbGa3 & R. G. Ling and C. Belin, Z. Anorg. Allg. Chem. 480, 181-185 (1981)
   1.00000000000000
  -3.15750000000000   3.15750000000000   7.50000000000000
   3.15750000000000  -3.15750000000000   7.50000000000000
   3.15750000000000   3.15750000000000  -7.50000000000000
    Ga    Rb
     9     3
Direct
   0.50000000000000   0.50000000000000   0.00000000000000   Ga   (2b)
   0.22290000000000   0.42970000000000   0.20680000000000   Ga   (8i)
   0.22290000000000   0.01610000000000  -0.20680000000000   Ga   (8i)
  -0.42970000000000  -0.22290000000000  -0.20680000000000   Ga   (8i)
  -0.01610000000000  -0.22290000000000   0.20680000000000   Ga   (8i)
   0.39170000000000   0.69840000000000   0.30670000000000   Ga   (8i)
   0.39170000000000   0.08500000000000  -0.30670000000000   Ga   (8i)
  -0.69840000000000  -0.39170000000000  -0.30670000000000   Ga   (8i)
  -0.08500000000000  -0.39170000000000   0.30670000000000   Ga   (8i)
   0.00000000000000   0.00000000000000   0.00000000000000   Rb   (2a)
   0.87200000000000   0.37200000000000   0.50000000000000   Rb   (4f)
  -0.37200000000000   0.12800000000000   0.50000000000000   Rb   (4f)
\end{lstlisting}
{\phantomsection\label{AB_tI4_119_c_a_cif}}
{\hyperref[AB_tI4_119_c_a]{GaSb: AB\_tI4\_119\_c\_a}} - CIF
\begin{lstlisting}[numbers=none,language={mylang}]
# CIF file
data_findsym-output
_audit_creation_method FINDSYM

_chemical_name_mineral 'GaSb'
_chemical_formula_sum 'Ga Sb'

loop_
_publ_author_name
 'T. R. R. McDonald'
 'R. Sard'
 'E. Gregory'
_journal_name_full_name
;
 Journal of Applied Physics
;
_journal_volume 36
_journal_year 1965
_journal_page_first 1498
_journal_page_last 1499
_publ_Section_title
;
 Retention of GaSb (II) at low temperatures and one atmosphere pressure
;

# Found in Pearson's Crystal Data - Crystal Structure Database for Inorganic Compounds, 2013

_aflow_title 'GaSb Structure'
_aflow_proto 'AB_tI4_119_c_a'
_aflow_params 'a,c/a'
_aflow_params_values '5.4790101504,0.558496075934'
_aflow_Strukturbericht 'None'
_aflow_Pearson 'tI4'

_cell_length_a    5.4790101504
_cell_length_b    5.4790101504
_cell_length_c    3.0600056690
_cell_angle_alpha 90.0000000000
_cell_angle_beta  90.0000000000
_cell_angle_gamma 90.0000000000
 
_symmetry_space_group_name_H-M "I -4 m 2"
_symmetry_Int_Tables_number 119
 
loop_
_space_group_symop_id
_space_group_symop_operation_xyz
1 x,y,z
2 -x,-y,z
3 -y,-x,-z
4 y,x,-z
5 -x,y,z
6 x,-y,z
7 y,-x,-z
8 -y,x,-z
9 x+1/2,y+1/2,z+1/2
10 -x+1/2,-y+1/2,z+1/2
11 -y+1/2,-x+1/2,-z+1/2
12 y+1/2,x+1/2,-z+1/2
13 -x+1/2,y+1/2,z+1/2
14 x+1/2,-y+1/2,z+1/2
15 y+1/2,-x+1/2,-z+1/2
16 -y+1/2,x+1/2,-z+1/2
 
loop_
_atom_site_label
_atom_site_type_symbol
_atom_site_symmetry_multiplicity
_atom_site_Wyckoff_label
_atom_site_fract_x
_atom_site_fract_y
_atom_site_fract_z
_atom_site_occupancy
Sb1 Sb   2 a 0.00000 0.00000 0.00000 1.00000
Ga1 Ga   2 c 0.00000 0.50000 0.25000 1.00000
\end{lstlisting}
{\phantomsection\label{AB_tI4_119_c_a_poscar}}
{\hyperref[AB_tI4_119_c_a]{GaSb: AB\_tI4\_119\_c\_a}} - POSCAR
\begin{lstlisting}[numbers=none,language={mylang}]
AB_tI4_119_c_a & a,c/a --params=5.4790101504,0.558496075934 & I-4m2 D_{2d}^{9} #119 (ac) & tI4 & None & GaSb &  & T. R. R. McDonald and R. Sard and E. Gregory, J. Appl. Phys. 36, 1498-1499 (1965)
   1.00000000000000
  -2.73950507520000   2.73950507520000   1.53000283450000
   2.73950507520000  -2.73950507520000   1.53000283450000
   2.73950507520000   2.73950507520000  -1.53000283450000
    Ga    Sb
     1     1
Direct
   0.75000000000000   0.25000000000000   0.50000000000000   Ga   (2c)
   0.00000000000000   0.00000000000000   0.00000000000000   Sb   (2a)
\end{lstlisting}
{\phantomsection\label{A4BC2_tI28_120_i_d_e_cif}}
{\hyperref[A4BC2_tI28_120_i_d_e]{KAu$_{4}$Sn$_{2}$: A4BC2\_tI28\_120\_i\_d\_e}} - CIF
\begin{lstlisting}[numbers=none,language={mylang}]
# CIF file
data_findsym-output
_audit_creation_method FINDSYM

_chemical_name_mineral 'KAu4Sn2'
_chemical_formula_sum 'Au4 K Sn2'

loop_
_publ_author_name
 'H.-D. Sinnen'
 'H.-U. Schuster'
_journal_name_full_name
;
 Zeitschrift f{\"u}r Naturforschung B
;
_journal_volume 33
_journal_year 1978
_journal_page_first 1077
_journal_page_last 1079
_publ_Section_title
;
 Darstellung und Struktur des KAu$_{4}$Sn$_{2}$ / Preparation and Crystal Structure of KAu$_{4}$Sn$_{2}$
;

# Found in Pearson's Crystal Data - Crystal Structure Database for Inorganic Compounds, 2013

_aflow_title 'KAu$_{4}$Sn$_{2}$ Structure'
_aflow_proto 'A4BC2_tI28_120_i_d_e'
_aflow_params 'a,c/a,x_{2},x_{3},y_{3},z_{3}'
_aflow_params_values '8.8470588481,0.924381146154,0.856,0.6452,0.6575,0.0851'
_aflow_Strukturbericht 'None'
_aflow_Pearson 'tI28'

_cell_length_a    8.8470588481
_cell_length_b    8.8470588481
_cell_length_c    8.1780543981
_cell_angle_alpha 90.0000000000
_cell_angle_beta  90.0000000000
_cell_angle_gamma 90.0000000000
 
_symmetry_space_group_name_H-M "I -4 c 2"
_symmetry_Int_Tables_number 120
 
loop_
_space_group_symop_id
_space_group_symop_operation_xyz
1 x,y,z
2 -x,-y,z
3 -y,-x,-z+1/2
4 y,x,-z+1/2
5 -x,y,z+1/2
6 x,-y,z+1/2
7 y,-x,-z
8 -y,x,-z
9 x+1/2,y+1/2,z+1/2
10 -x+1/2,-y+1/2,z+1/2
11 -y+1/2,-x+1/2,-z
12 y+1/2,x+1/2,-z
13 -x+1/2,y+1/2,z
14 x+1/2,-y+1/2,z
15 y+1/2,-x+1/2,-z+1/2
16 -y+1/2,x+1/2,-z+1/2
 
loop_
_atom_site_label
_atom_site_type_symbol
_atom_site_symmetry_multiplicity
_atom_site_Wyckoff_label
_atom_site_fract_x
_atom_site_fract_y
_atom_site_fract_z
_atom_site_occupancy
K1  K    4 d 0.00000 0.50000 0.00000 1.00000
Sn1 Sn   8 e 0.85600 0.85600 0.25000 1.00000
Au1 Au  16 i 0.64520 0.65750 0.08510 1.00000
\end{lstlisting}
{\phantomsection\label{A4BC2_tI28_120_i_d_e_poscar}}
{\hyperref[A4BC2_tI28_120_i_d_e]{KAu$_{4}$Sn$_{2}$: A4BC2\_tI28\_120\_i\_d\_e}} - POSCAR
\begin{lstlisting}[numbers=none,language={mylang}]
A4BC2_tI28_120_i_d_e & a,c/a,x2,x3,y3,z3 --params=8.8470588481,0.924381146154,0.856,0.6452,0.6575,0.0851 & I-4c2 D_{2d}^{10} #120 (dei) & tI28 & None & KAu4Sn2 &  & H.-D. Sinnen and H.-U. Schuster, Z. Naturforsch. B 33, 1077-1079 (1978)
   1.00000000000000
  -4.42352942405000   4.42352942405000   4.08902719905000
   4.42352942405000  -4.42352942405000   4.08902719905000
   4.42352942405000   4.42352942405000  -4.08902719905000
    Au     K    Sn
     8     2     4
Direct
   0.74260000000000   0.73030000000000   1.30270000000000   Au  (16i)
  -0.57240000000000  -0.56010000000000  -1.30270000000000   Au  (16i)
  -0.73030000000000   0.57240000000000   0.01230000000000   Au  (16i)
   0.56010000000000  -0.74260000000000  -0.01230000000000   Au  (16i)
  -0.07240000000000   1.23030000000000  -0.01230000000000   Au  (16i)
   1.24260000000000  -0.06010000000000   0.01230000000000   Au  (16i)
   1.06010000000000   1.07240000000000   1.30270000000000   Au  (16i)
  -0.23030000000000  -0.24260000000000  -1.30270000000000   Au  (16i)
   0.50000000000000   0.00000000000000   0.50000000000000    K   (4d)
   0.00000000000000   0.50000000000000   0.50000000000000    K   (4d)
   1.10600000000000   1.10600000000000   1.71200000000000   Sn   (8e)
  -0.60600000000000  -0.60600000000000  -1.71200000000000   Sn   (8e)
  -0.10600000000000   1.60600000000000   0.00000000000000   Sn   (8e)
   1.60600000000000  -0.10600000000000   0.00000000000000   Sn   (8e)
\end{lstlisting}
{\phantomsection\label{A4BC4D_tP10_123_gh_a_i_d_cif}}
{\hyperref[A4BC4D_tP10_123_gh_a_i_d]{CaRbFe$_{4}$As$_{4}$ (Superconducting): A4BC4D\_tP10\_123\_gh\_a\_i\_d}} - CIF
\begin{lstlisting}[numbers=none,language={mylang}]
# CIF file 
data_findsym-output
_audit_creation_method FINDSYM

_chemical_name_mineral 'CsRbFe4As4'
_chemical_formula_sum 'As4 Cs Fe4 Rb'

loop_
_publ_author_name
 'A. Iyo'
 'K. Kawashima'
 'T. Kinjo'
 'T. Nishio'
 'S. Ishida'
 'H. Fujihisa'
 'Y. Gotoh'
 'K. Kihou'
 'H. Eisaki'
 'Y. Yoshida'
_journal_name_full_name
;
 Journal of the American Chemical Society
;
_journal_volume 138
_journal_year 2016
_journal_page_first 3410
_journal_page_last 3415
_publ_Section_title
;
 New-Structure-Type Fe-Based Superconductors: CaAFe$_{4}$As$_{4}$ (A = K, Rb, Cs) and SrAFe$_{4}$As$_{4}$ (A = Rb, Cs)
;

_aflow_title 'CaRbFe$_{4}$As$_{4}$ (Superconducting) Structure'
_aflow_proto 'A4BC4D_tP10_123_gh_a_i_d'
_aflow_params 'a,c/a,z_{3},z_{4},z_{5}'
_aflow_params_values '3.8757,3.38106664603,0.3336,0.1193,0.2246'
_aflow_Strukturbericht 'None'
_aflow_Pearson 'tP10'

_symmetry_space_group_name_H-M "P 4/m 2/m 2/m"
_symmetry_Int_Tables_number 123
 
_cell_length_a    3.87570
_cell_length_b    3.87570
_cell_length_c    13.10400
_cell_angle_alpha 90.00000
_cell_angle_beta  90.00000
_cell_angle_gamma 90.00000
 
loop_
_space_group_symop_id
_space_group_symop_operation_xyz
1 x,y,z
2 x,-y,-z
3 -x,y,-z
4 -x,-y,z
5 -y,-x,-z
6 -y,x,z
7 y,-x,z
8 y,x,-z
9 -x,-y,-z
10 -x,y,z
11 x,-y,z
12 x,y,-z
13 y,x,z
14 y,-x,-z
15 -y,x,-z
16 -y,-x,z
 
loop_
_atom_site_label
_atom_site_type_symbol
_atom_site_symmetry_multiplicity
_atom_site_Wyckoff_label
_atom_site_fract_x
_atom_site_fract_y
_atom_site_fract_z
_atom_site_occupancy
Cs1 Cs   1 a 0.00000 0.00000 0.00000 1.00000
Rb1 Rb   1 d 0.50000 0.50000 0.50000 1.00000
As1 As   2 g 0.00000 0.00000 0.33360 1.00000
As2 As   2 h 0.50000 0.50000 0.11930 1.00000
Fe1 Fe   4 i 0.00000 0.50000 0.22460 1.00000
\end{lstlisting}
{\phantomsection\label{A4BC4D_tP10_123_gh_a_i_d_poscar}}
{\hyperref[A4BC4D_tP10_123_gh_a_i_d]{CaRbFe$_{4}$As$_{4}$ (Superconducting): A4BC4D\_tP10\_123\_gh\_a\_i\_d}} - POSCAR
\begin{lstlisting}[numbers=none,language={mylang}]
A4BC4D_tP10_123_gh_a_i_d & a,c/a,z3,z4,z5 --params=3.8757,3.38106664603,0.3336,0.1193,0.2246 & P4/mmm D_{4h}^{1} #123 (adghi) & tP10 & None & CsRbFe4As4 & CsRbFe4As4 & A. Iyo et al., J. Am. Chem. Soc. 138, 3410-3415 (2016)
   1.00000000000000
   3.87570000000000   0.00000000000000   0.00000000000000
   0.00000000000000   3.87570000000000   0.00000000000000
   0.00000000000000   0.00000000000000  13.10400000000000
    As    Cs    Fe    Rb
     4     1     4     1
Direct
   0.00000000000000   0.00000000000000   0.33360000000000   As   (2g)
   0.00000000000000   0.00000000000000  -0.33360000000000   As   (2g)
   0.50000000000000   0.50000000000000   0.11930000000000   As   (2h)
   0.50000000000000   0.50000000000000  -0.11930000000000   As   (2h)
   0.00000000000000   0.00000000000000   0.00000000000000   Cs   (1a)
   0.00000000000000   0.50000000000000   0.22460000000000   Fe   (4i)
   0.50000000000000   0.00000000000000   0.22460000000000   Fe   (4i)
   0.00000000000000   0.50000000000000  -0.22460000000000   Fe   (4i)
   0.50000000000000   0.00000000000000  -0.22460000000000   Fe   (4i)
   0.50000000000000   0.50000000000000   0.50000000000000   Rb   (1d)
\end{lstlisting}
{\phantomsection\label{AB4C_tP12_124_a_m_c_cif}}
{\hyperref[AB4C_tP12_124_a_m_c]{Nb$_{4}$CoSi: AB4C\_tP12\_124\_a\_m\_c}} - CIF
\begin{lstlisting}[numbers=none,language={mylang}]
# CIF file
data_findsym-output
_audit_creation_method FINDSYM

_chemical_name_mineral 'Nb4CoSi'
_chemical_formula_sum 'Co Nb4 Si'

loop_
_publ_author_name
 'E. I. Gladyshevskii'
 '{Yu}. B. {Kuz\'ma}'
_journal_name_full_name
;
 Journal of Structural Chemistry
;
_journal_volume 6
_journal_year 1965
_journal_page_first 60
_journal_page_last 63
_publ_Section_title
;
 The compounds Nb$_{4}$FeSi, Nb$_{4}$CoSi, Nb$_{4}$NiSi and their crystal structures
;

# Found in Pearson's Crystal Data - Crystal Structure Database for Inorganic Compounds, 2013

_aflow_title 'Nb$_{4}$CoSi Structure'
_aflow_proto 'AB4C_tP12_124_a_m_c'
_aflow_params 'a,c/a,x_{3},y_{3}'
_aflow_params_values '6.1884808485,0.816448537725,0.162,0.662'
_aflow_Strukturbericht 'None'
_aflow_Pearson 'tP12'

_cell_length_a    6.1884808485
_cell_length_b    6.1884808485
_cell_length_c    5.0525761395
_cell_angle_alpha 90.0000000000
_cell_angle_beta  90.0000000000
_cell_angle_gamma 90.0000000000
 
_symmetry_space_group_name_H-M "P 4/m 2/c 2/c"
_symmetry_Int_Tables_number 124
 
loop_
_space_group_symop_id
_space_group_symop_operation_xyz
1 x,y,z
2 x,-y,-z+1/2
3 -x,y,-z+1/2
4 -x,-y,z
5 -y,-x,-z+1/2
6 -y,x,z
7 y,-x,z
8 y,x,-z+1/2
9 -x,-y,-z
10 -x,y,z+1/2
11 x,-y,z+1/2
12 x,y,-z
13 y,x,z+1/2
14 y,-x,-z
15 -y,x,-z
16 -y,-x,z+1/2
 
loop_
_atom_site_label
_atom_site_type_symbol
_atom_site_symmetry_multiplicity
_atom_site_Wyckoff_label
_atom_site_fract_x
_atom_site_fract_y
_atom_site_fract_z
_atom_site_occupancy
Co1 Co   2 a 0.00000 0.00000 0.25000 1.00000
Si1 Si   2 c 0.50000 0.50000 0.25000 1.00000
Nb1 Nb   8 m 0.16200 0.66200 0.00000 1.00000
\end{lstlisting}
{\phantomsection\label{AB4C_tP12_124_a_m_c_poscar}}
{\hyperref[AB4C_tP12_124_a_m_c]{Nb$_{4}$CoSi: AB4C\_tP12\_124\_a\_m\_c}} - POSCAR
\begin{lstlisting}[numbers=none,language={mylang}]
AB4C_tP12_124_a_m_c & a,c/a,x3,y3 --params=6.1884808485,0.816448537725,0.162,0.662 & P4/mcc D_{4h}^{2} #124 (acm) & tP12 & None & Nb4CoSi &  & E. I. Gladyshevskii and {Yu}. B. {Kuz'ma}, J. Struct. Chem. 6, 60-63 (1965)
   1.00000000000000
   6.18848084850000   0.00000000000000   0.00000000000000
   0.00000000000000   6.18848084850000   0.00000000000000
   0.00000000000000   0.00000000000000   5.05257613950000
    Co    Nb    Si
     2     8     2
Direct
   0.00000000000000   0.00000000000000   0.25000000000000   Co   (2a)
   0.00000000000000   0.00000000000000   0.75000000000000   Co   (2a)
   0.16200000000000   0.66200000000000   0.00000000000000   Nb   (8m)
  -0.16200000000000  -0.66200000000000   0.00000000000000   Nb   (8m)
  -0.66200000000000   0.16200000000000   0.00000000000000   Nb   (8m)
   0.66200000000000  -0.16200000000000   0.00000000000000   Nb   (8m)
  -0.16200000000000   0.66200000000000   0.50000000000000   Nb   (8m)
   0.16200000000000  -0.66200000000000   0.50000000000000   Nb   (8m)
   0.66200000000000   0.16200000000000   0.50000000000000   Nb   (8m)
  -0.66200000000000  -0.16200000000000   0.50000000000000   Nb   (8m)
   0.50000000000000   0.50000000000000   0.25000000000000   Si   (2c)
   0.50000000000000   0.50000000000000   0.75000000000000   Si   (2c)
\end{lstlisting}
{\phantomsection\label{AB4_tP10_124_a_m_cif}}
{\hyperref[AB4_tP10_124_a_m]{NbTe$_{4}$: AB4\_tP10\_124\_a\_m}} - CIF
\begin{lstlisting}[numbers=none,language={mylang}]
# CIF file
data_findsym-output
_audit_creation_method FINDSYM

_chemical_name_mineral 'NbTe4'
_chemical_formula_sum 'Nb Te4'

loop_
_publ_author_name
 'K. Selte'
 'A. Kjekshus'
_journal_name_full_name
;
 Acta Chemica Scandinavica
;
_journal_volume 18
_journal_year 1964
_journal_page_first 690
_journal_page_last 696
_publ_Section_title
;
 On the crystal structure of NbTe$_{4}$
;

# Found in Pearson's Crystal Data - Crystal Structure Database for Inorganic Compounds, 2013

_aflow_title 'NbTe$_{4}$ Structure'
_aflow_proto 'AB4_tP10_124_a_m'
_aflow_params 'a,c/a,x_{2},y_{2}'
_aflow_params_values '6.4989671731,1.05200800123,0.1425,0.3361'
_aflow_Strukturbericht 'None'
_aflow_Pearson 'tP10'

_cell_length_a    6.4989671731
_cell_length_b    6.4989671731
_cell_length_c    6.8369654658
_cell_angle_alpha 90.0000000000
_cell_angle_beta  90.0000000000
_cell_angle_gamma 90.0000000000
 
_symmetry_space_group_name_H-M "P 4/m 2/c 2/c"
_symmetry_Int_Tables_number 124
 
loop_
_space_group_symop_id
_space_group_symop_operation_xyz
1 x,y,z
2 x,-y,-z+1/2
3 -x,y,-z+1/2
4 -x,-y,z
5 -y,-x,-z+1/2
6 -y,x,z
7 y,-x,z
8 y,x,-z+1/2
9 -x,-y,-z
10 -x,y,z+1/2
11 x,-y,z+1/2
12 x,y,-z
13 y,x,z+1/2
14 y,-x,-z
15 -y,x,-z
16 -y,-x,z+1/2
 
loop_
_atom_site_label
_atom_site_type_symbol
_atom_site_symmetry_multiplicity
_atom_site_Wyckoff_label
_atom_site_fract_x
_atom_site_fract_y
_atom_site_fract_z
_atom_site_occupancy
Nb1 Nb   2 a 0.00000 0.00000 0.25000 1.00000
Te1 Te   8 m 0.14250 0.33610 0.00000 1.00000
\end{lstlisting}
{\phantomsection\label{AB4_tP10_124_a_m_poscar}}
{\hyperref[AB4_tP10_124_a_m]{NbTe$_{4}$: AB4\_tP10\_124\_a\_m}} - POSCAR
\begin{lstlisting}[numbers=none,language={mylang}]
AB4_tP10_124_a_m & a,c/a,x2,y2 --params=6.4989671731,1.05200800123,0.1425,0.3361 & P4/mcc D_{4h}^{2} #124 (am) & tP10 & None & NbTe4 &  & K. Selte and A. Kjekshus, Acta Chem. Scand. 18, 690-696 (1964)
   1.00000000000000
   6.49896717310000   0.00000000000000   0.00000000000000
   0.00000000000000   6.49896717310000   0.00000000000000
   0.00000000000000   0.00000000000000   6.83696546580000
    Nb    Te
     2     8
Direct
   0.00000000000000   0.00000000000000   0.25000000000000   Nb   (2a)
   0.00000000000000   0.00000000000000   0.75000000000000   Nb   (2a)
   0.14250000000000   0.33610000000000   0.00000000000000   Te   (8m)
  -0.14250000000000  -0.33610000000000   0.00000000000000   Te   (8m)
  -0.33610000000000   0.14250000000000   0.00000000000000   Te   (8m)
   0.33610000000000  -0.14250000000000   0.00000000000000   Te   (8m)
  -0.14250000000000   0.33610000000000   0.50000000000000   Te   (8m)
   0.14250000000000  -0.33610000000000   0.50000000000000   Te   (8m)
   0.33610000000000   0.14250000000000   0.50000000000000   Te   (8m)
  -0.33610000000000  -0.14250000000000   0.50000000000000   Te   (8m)
\end{lstlisting}
{\phantomsection\label{A4B_tP10_125_m_a_cif}}
{\hyperref[A4B_tP10_125_m_a]{PtPb$_{4}$: A4B\_tP10\_125\_m\_a}} - CIF
\begin{lstlisting}[numbers=none,language={mylang}]
# CIF file
data_findsym-output
_audit_creation_method FINDSYM

_chemical_name_mineral 'PtPb4'
_chemical_formula_sum 'Pb4 Pt'

loop_
_publ_author_name
 'R. Graham'
 'G. C. S. Waghorn'
 'P. T. Davies'
_journal_year 1954
_publ_Section_title
;
 An X-ray investigation of the lead-platinum system
;

# Found in Pearson's Crystal Data - Crystal Structure Database for Inorganic Compounds, 2013

_aflow_title 'PtPb$_{4}$ Structure'
_aflow_proto 'A4B_tP10_125_m_a'
_aflow_params 'a,c/a,x_{2},z_{2}'
_aflow_params_values '6.6398746049,0.899096385539,0.425,0.255'
_aflow_Strukturbericht 'None'
_aflow_Pearson 'tP10'

_cell_length_a    6.6398746049
_cell_length_b    6.6398746049
_cell_length_c    5.9698872577
_cell_angle_alpha 90.0000000000
_cell_angle_beta  90.0000000000
_cell_angle_gamma 90.0000000000
 
_symmetry_space_group_name_H-M "P 4/n 2/b 2/m (origin choice 2)"
_symmetry_Int_Tables_number 125
 
loop_
_space_group_symop_id
_space_group_symop_operation_xyz
1 x,y,z
2 x,-y+1/2,-z
3 -x+1/2,y,-z
4 -x+1/2,-y+1/2,z
5 -y+1/2,-x+1/2,-z
6 -y+1/2,x,z
7 y,-x+1/2,z
8 y,x,-z
9 -x,-y,-z
10 -x,y+1/2,z
11 x+1/2,-y,z
12 x+1/2,y+1/2,-z
13 y+1/2,x+1/2,z
14 y+1/2,-x,-z
15 -y,x+1/2,-z
16 -y,-x,z
 
loop_
_atom_site_label
_atom_site_type_symbol
_atom_site_symmetry_multiplicity
_atom_site_Wyckoff_label
_atom_site_fract_x
_atom_site_fract_y
_atom_site_fract_z
_atom_site_occupancy
Pt1 Pt   2 a 0.25000 0.25000 0.00000 1.00000
Pb1 Pb   8 m 0.42500 0.57500 0.25500 1.00000
\end{lstlisting}
{\phantomsection\label{A4B_tP10_125_m_a_poscar}}
{\hyperref[A4B_tP10_125_m_a]{PtPb$_{4}$: A4B\_tP10\_125\_m\_a}} - POSCAR
\begin{lstlisting}[numbers=none,language={mylang}]
A4B_tP10_125_m_a & a,c/a,x2,z2 --params=6.6398746049,0.899096385539,0.425,0.255 & P4/nbm D_{4h}^{3} #125 (am) & tP10 & None & PtPb4 &  & R. Graham and G. C. S. Waghorn and P. T. Davies, (1954)
   1.00000000000000
   6.63987460490000   0.00000000000000   0.00000000000000
   0.00000000000000   6.63987460490000   0.00000000000000
   0.00000000000000   0.00000000000000   5.96988725770000
    Pb    Pt
     8     2
Direct
   0.42500000000000  -0.42500000000000   0.25500000000000   Pb   (8m)
   0.07500000000000   0.92500000000000   0.25500000000000   Pb   (8m)
   0.92500000000000   0.42500000000000   0.25500000000000   Pb   (8m)
  -0.42500000000000   0.07500000000000   0.25500000000000   Pb   (8m)
   0.07500000000000  -0.42500000000000  -0.25500000000000   Pb   (8m)
   0.42500000000000   0.92500000000000  -0.25500000000000   Pb   (8m)
  -0.42500000000000   0.42500000000000  -0.25500000000000   Pb   (8m)
   0.92500000000000   0.07500000000000  -0.25500000000000   Pb   (8m)
   0.25000000000000   0.25000000000000   0.00000000000000   Pt   (2a)
   0.75000000000000   0.75000000000000   0.00000000000000   Pt   (2a)
\end{lstlisting}
{\phantomsection\label{ABC4_tP12_125_a_b_m_cif}}
{\hyperref[ABC4_tP12_125_a_b_m]{KCeSe$_{4}$: ABC4\_tP12\_125\_a\_b\_m}} - CIF
\begin{lstlisting}[numbers=none,language={mylang}]
# CIF file
data_findsym-output
_audit_creation_method FINDSYM

_chemical_name_mineral 'KCeSe4'
_chemical_formula_sum 'Ce K Se4'

loop_
_publ_author_name
 'A. C. Sutorik'
 'M. G. Kanatzidis'
_journal_name_full_name
;
 Angewandte Chemie (International ed.)
;
_journal_volume 31
_journal_year 1992
_journal_page_first 1594
_journal_page_last 1596
_publ_Section_title
;
 KCeSe$_{4}$: A New Solid-State Lanthanide Polychalcogenide
;

# Found in Pearson's Crystal Data - Crystal Structure Database for Inorganic Compounds, 2013

_aflow_title 'KCeSe$_{4}$ Structure'
_aflow_proto 'ABC4_tP12_125_a_b_m'
_aflow_params 'a,c/a,x_{3},z_{3}'
_aflow_params_values '6.3759876428,1.30630489336,0.3822,0.2163'
_aflow_Strukturbericht 'None'
_aflow_Pearson 'tP12'

_cell_length_a    6.3759876428
_cell_length_b    6.3759876428
_cell_length_c    8.3289838578
_cell_angle_alpha 90.0000000000
_cell_angle_beta  90.0000000000
_cell_angle_gamma 90.0000000000
 
_symmetry_space_group_name_H-M "P 4/n 2/b 2/m (origin choice 2)"
_symmetry_Int_Tables_number 125
 
loop_
_space_group_symop_id
_space_group_symop_operation_xyz
1 x,y,z
2 x,-y+1/2,-z
3 -x+1/2,y,-z
4 -x+1/2,-y+1/2,z
5 -y+1/2,-x+1/2,-z
6 -y+1/2,x,z
7 y,-x+1/2,z
8 y,x,-z
9 -x,-y,-z
10 -x,y+1/2,z
11 x+1/2,-y,z
12 x+1/2,y+1/2,-z
13 y+1/2,x+1/2,z
14 y+1/2,-x,-z
15 -y,x+1/2,-z
16 -y,-x,z
 
loop_
_atom_site_label
_atom_site_type_symbol
_atom_site_symmetry_multiplicity
_atom_site_Wyckoff_label
_atom_site_fract_x
_atom_site_fract_y
_atom_site_fract_z
_atom_site_occupancy
Ce1 Ce   2 a 0.25000 0.25000 0.00000 1.00000
K1  K    2 b 0.25000 0.25000 0.50000 1.00000
Se1 Se   8 m 0.38220 0.61780 0.21630 1.00000
\end{lstlisting}
{\phantomsection\label{ABC4_tP12_125_a_b_m_poscar}}
{\hyperref[ABC4_tP12_125_a_b_m]{KCeSe$_{4}$: ABC4\_tP12\_125\_a\_b\_m}} - POSCAR
\begin{lstlisting}[numbers=none,language={mylang}]
ABC4_tP12_125_a_b_m & a,c/a,x3,z3 --params=6.3759876428,1.30630489336,0.3822,0.2163 & P4/nbm D_{4h}^{3} #125 (abm) & tP12 & None & KCeSe4 &  & A. C. Sutorik and M. G. Kanatzidis, Angew. Chem. Int. Ed. 31, 1594-1596 (1992)
   1.00000000000000
   6.37598764280000   0.00000000000000   0.00000000000000
   0.00000000000000   6.37598764280000   0.00000000000000
   0.00000000000000   0.00000000000000   8.32898385780000
    Ce     K    Se
     2     2     8
Direct
   0.25000000000000   0.25000000000000   0.00000000000000   Ce   (2a)
   0.75000000000000   0.75000000000000   0.00000000000000   Ce   (2a)
   0.25000000000000   0.25000000000000   0.50000000000000    K   (2b)
   0.75000000000000   0.75000000000000   0.50000000000000    K   (2b)
   0.38220000000000  -0.38220000000000   0.21630000000000   Se   (8m)
   0.11780000000000   0.88220000000000   0.21630000000000   Se   (8m)
   0.88220000000000   0.38220000000000   0.21630000000000   Se   (8m)
  -0.38220000000000   0.11780000000000   0.21630000000000   Se   (8m)
   0.11780000000000  -0.38220000000000  -0.21630000000000   Se   (8m)
   0.38220000000000   0.88220000000000  -0.21630000000000   Se   (8m)
  -0.38220000000000   0.38220000000000  -0.21630000000000   Se   (8m)
   0.88220000000000   0.11780000000000  -0.21630000000000   Se   (8m)
\end{lstlisting}
{\phantomsection\label{A2BC4_tP28_126_cd_e_k_cif}}
{\hyperref[A2BC4_tP28_126_cd_e_k]{BiAl$_{2}$S$_{4}$: A2BC4\_tP28\_126\_cd\_e\_k}} - CIF
\begin{lstlisting}[numbers=none,language={mylang}]
# CIF file
data_findsym-output
_audit_creation_method FINDSYM

_chemical_name_mineral 'BiAl2S4'
_chemical_formula_sum 'Al2 Bi S4'

loop_
_publ_author_name
 'H. Kalpen'
 'W. H{\"o}nle'
 'M. Somer'
 'U. Schwarz'
 'K. Peters'
 'H. G. {von Schnering}'
 'R. Blachnik'
_journal_name_full_name
;
 Zeitschrift fur Anorganische und Allgemeine Chemie
;
_journal_volume 624
_journal_year 1998
_journal_page_first 1137
_journal_page_last 1147
_publ_Section_title
;
 Bismut(II)-chalkogenometallate(III) Bi$_{2}M_{4}X_{8}$, Verbindungen mit Bi$_{2}^{4+}$-Hanteln ($M$ = Al, Ga; $X$ = S, Se)
;

# Found in Pearson's Crystal Data - Crystal Structure Database for Inorganic Compounds, 2013

_aflow_title 'BiAl$_{2}$S$_{4}$ Structure'
_aflow_proto 'A2BC4_tP28_126_cd_e_k'
_aflow_params 'a,c/a,z_{3},x_{4},y_{4},z_{4}'
_aflow_params_values '7.4920241479,1.58609183128,0.11808,0.0865,0.5924,0.1254'
_aflow_Strukturbericht 'None'
_aflow_Pearson 'tP28'

_cell_length_a    7.4920241479
_cell_length_b    7.4920241479
_cell_length_c    11.8830383007
_cell_angle_alpha 90.0000000000
_cell_angle_beta  90.0000000000
_cell_angle_gamma 90.0000000000
 
_symmetry_space_group_name_H-M "P 4/n 2/n 2/c (origin choice 2)"
_symmetry_Int_Tables_number 126
 
loop_
_space_group_symop_id
_space_group_symop_operation_xyz
1 x,y,z
2 x,-y+1/2,-z+1/2
3 -x+1/2,y,-z+1/2
4 -x+1/2,-y+1/2,z
5 -y+1/2,-x+1/2,-z+1/2
6 -y+1/2,x,z
7 y,-x+1/2,z
8 y,x,-z+1/2
9 -x,-y,-z
10 -x,y+1/2,z+1/2
11 x+1/2,-y,z+1/2
12 x+1/2,y+1/2,-z
13 y+1/2,x+1/2,z+1/2
14 y+1/2,-x,-z
15 -y,x+1/2,-z
16 -y,-x,z+1/2
 
loop_
_atom_site_label
_atom_site_type_symbol
_atom_site_symmetry_multiplicity
_atom_site_Wyckoff_label
_atom_site_fract_x
_atom_site_fract_y
_atom_site_fract_z
_atom_site_occupancy
Al1 Al   4 c 0.25000 0.75000 0.75000 1.00000
Al2 Al   4 d 0.25000 0.75000 0.00000 1.00000
Bi1 Bi   4 e 0.25000 0.25000 0.11808 1.00000
S1  S   16 k 0.08650 0.59240 0.12540 1.00000
\end{lstlisting}
{\phantomsection\label{A2BC4_tP28_126_cd_e_k_poscar}}
{\hyperref[A2BC4_tP28_126_cd_e_k]{BiAl$_{2}$S$_{4}$: A2BC4\_tP28\_126\_cd\_e\_k}} - POSCAR

{\phantomsection\label{A4B_tP20_127_ehj_g_cif}}
{\hyperref[A4B_tP20_127_ehj_g]{ThB$_{4}$ ($D1_{e}$): A4B\_tP20\_127\_ehj\_g}} - CIF
\begin{lstlisting}[numbers=none,language={mylang}]
# CIF file
data_findsym-output
_audit_creation_method FINDSYM

_chemical_name_mineral 'ThB4'
_chemical_formula_sum 'B4 Th'

loop_
_publ_author_name
 'A. Zalkin'
 'D. H. Templeton'
_journal_name_full_name
;
 Journal of Chemical Physics
;
_journal_volume 18
_journal_year 1950
_journal_page_first 391
_journal_page_last 391
_publ_Section_title
;
 The Crystal Structures of CeB$_{4}$, ThB$_{4}$, and UB$_{4}$
;

_aflow_title 'ThB$_{4}$ ($D1_{e}$) Structure'
_aflow_proto 'A4B_tP20_127_ehj_g'
_aflow_params 'a,c/a,z_{1},x_{2},x_{3},x_{4},y_{4}'
_aflow_params_values '7.256,0.56684123484,0.2,0.31,0.1,0.2,0.04'
_aflow_Strukturbericht '$D1_{e}$'
_aflow_Pearson 'tP20'

_symmetry_space_group_name_H-M "P 4/m 21/b 2/m"
_symmetry_Int_Tables_number 127
 
_cell_length_a    7.25600
_cell_length_b    7.25600
_cell_length_c    4.11300
_cell_angle_alpha 90.00000
_cell_angle_beta  90.00000
_cell_angle_gamma 90.00000
 
loop_
_space_group_symop_id
_space_group_symop_operation_xyz
1 x,y,z
2 x+1/2,-y+1/2,-z
3 -x+1/2,y+1/2,-z
4 -x,-y,z
5 -y+1/2,-x+1/2,-z
6 -y,x,z
7 y,-x,z
8 y+1/2,x+1/2,-z
9 -x,-y,-z
10 -x+1/2,y+1/2,z
11 x+1/2,-y+1/2,z
12 x,y,-z
13 y+1/2,x+1/2,z
14 y,-x,-z
15 -y,x,-z
16 -y+1/2,-x+1/2,z
 
loop_
_atom_site_label
_atom_site_type_symbol
_atom_site_symmetry_multiplicity
_atom_site_Wyckoff_label
_atom_site_fract_x
_atom_site_fract_y
_atom_site_fract_z
_atom_site_occupancy
B1  B    4 e 0.00000 0.00000 0.20000 1.00000
Th1 Th   4 g 0.31000 0.81000 0.00000 1.00000
B2  B    4 h 0.10000 0.60000 0.50000 1.00000
B3  B    8 j 0.20000 0.04000 0.50000 1.00000
\end{lstlisting}
{\phantomsection\label{A4B_tP20_127_ehj_g_poscar}}
{\hyperref[A4B_tP20_127_ehj_g]{ThB$_{4}$ ($D1_{e}$): A4B\_tP20\_127\_ehj\_g}} - POSCAR
\begin{lstlisting}[numbers=none,language={mylang}]
A4B_tP20_127_ehj_g & a,c/a,z1,x2,x3,x4,y4 --params=7.256,0.56684123484,0.2,0.31,0.1,0.2,0.04 & P4/mbm D_{4h}^{5} #127 (eghj) & tP20 & $D1_{e}$ & ThB4 & ThB4 & A. Zalkin and D. H. Templeton, J. Chem. Phys. 18, 391(1950)
   1.00000000000000
   7.25600000000000   0.00000000000000   0.00000000000000
   0.00000000000000   7.25600000000000   0.00000000000000
   0.00000000000000   0.00000000000000   4.11300000000000
     B    Th
    16     4
Direct
   0.00000000000000   0.00000000000000   0.20000000000000    B   (4e)
   0.50000000000000   0.50000000000000  -0.20000000000000    B   (4e)
   0.00000000000000   0.00000000000000  -0.20000000000000    B   (4e)
   0.50000000000000   0.50000000000000   0.20000000000000    B   (4e)
   0.10000000000000   0.60000000000000   0.50000000000000    B   (4h)
  -0.10000000000000   0.40000000000000   0.50000000000000    B   (4h)
   0.40000000000000   0.10000000000000   0.50000000000000    B   (4h)
   0.60000000000000  -0.10000000000000   0.50000000000000    B   (4h)
   0.20000000000000   0.04000000000000   0.50000000000000    B   (8j)
  -0.20000000000000  -0.04000000000000   0.50000000000000    B   (8j)
  -0.04000000000000   0.20000000000000   0.50000000000000    B   (8j)
   0.04000000000000  -0.20000000000000   0.50000000000000    B   (8j)
   0.30000000000000   0.54000000000000   0.50000000000000    B   (8j)
   0.70000000000000   0.46000000000000   0.50000000000000    B   (8j)
   0.54000000000000   0.70000000000000   0.50000000000000    B   (8j)
   0.46000000000000   0.30000000000000   0.50000000000000    B   (8j)
   0.31000000000000   0.81000000000000   0.00000000000000   Th   (4g)
  -0.31000000000000   0.19000000000000   0.00000000000000   Th   (4g)
   0.19000000000000   0.31000000000000   0.00000000000000   Th   (4g)
   0.81000000000000  -0.31000000000000   0.00000000000000   Th   (4g)
\end{lstlisting}
{\phantomsection\label{A6B2C_tP18_128_eh_d_b_cif}}
{\hyperref[A6B2C_tP18_128_eh_d_b]{K$_{2}$SnCl$_{6}$ (Low-temperature): A6B2C\_tP18\_128\_eh\_d\_b}} - CIF
\begin{lstlisting}[numbers=none,language={mylang}]
# CIF file
data_findsym-output
_audit_creation_method FINDSYM

_chemical_name_mineral 'K2SnCl6'
_chemical_formula_sum 'Cl6 K2 Sn'

loop_
_publ_author_name
 'H. Boysen'
 'A. W. Hewat'
_journal_name_full_name
;
 Acta Crystallographica Section B: Structural Science
;
_journal_volume 34
_journal_year 1978
_journal_page_first 1412
_journal_page_last 1418
_publ_Section_title
;
 A neutron powder investigation of the structural changes in K$_{2}$SnCl$_{6}$
;

# Found in Pearson's Crystal Data - Crystal Structure Database for Inorganic Compounds, 2013

_aflow_title 'K$_{2}$SnCl$_{6}$ (Low-temperature) Structure'
_aflow_proto 'A6B2C_tP18_128_eh_d_b'
_aflow_params 'a,c/a,z_{3},x_{4},y_{4}'
_aflow_params_values '7.057532571,1.41383170154,0.7523,0.7217,0.2489'
_aflow_Strukturbericht 'None'
_aflow_Pearson 'tP18'

_cell_length_a    7.0575325710
_cell_length_b    7.0575325710
_cell_length_c    9.9781632835
_cell_angle_alpha 90.0000000000
_cell_angle_beta  90.0000000000
_cell_angle_gamma 90.0000000000
 
_symmetry_space_group_name_H-M "P 4/m 21/n 2/c"
_symmetry_Int_Tables_number 128
 
loop_
_space_group_symop_id
_space_group_symop_operation_xyz
1 x,y,z
2 x+1/2,-y+1/2,-z+1/2
3 -x+1/2,y+1/2,-z+1/2
4 -x,-y,z
5 -y+1/2,-x+1/2,-z+1/2
6 -y,x,z
7 y,-x,z
8 y+1/2,x+1/2,-z+1/2
9 -x,-y,-z
10 -x+1/2,y+1/2,z+1/2
11 x+1/2,-y+1/2,z+1/2
12 x,y,-z
13 y+1/2,x+1/2,z+1/2
14 y,-x,-z
15 -y,x,-z
16 -y+1/2,-x+1/2,z+1/2
 
loop_
_atom_site_label
_atom_site_type_symbol
_atom_site_symmetry_multiplicity
_atom_site_Wyckoff_label
_atom_site_fract_x
_atom_site_fract_y
_atom_site_fract_z
_atom_site_occupancy
Sn1 Sn   2 b 0.00000 0.00000 0.50000 1.00000
K1  K    4 d 0.00000 0.50000 0.25000 1.00000
Cl1 Cl   4 e 0.00000 0.00000 0.75230 1.00000
Cl2 Cl   8 h 0.72170 0.24890 0.00000 1.00000
\end{lstlisting}
{\phantomsection\label{A6B2C_tP18_128_eh_d_b_poscar}}
{\hyperref[A6B2C_tP18_128_eh_d_b]{K$_{2}$SnCl$_{6}$ (Low-temperature): A6B2C\_tP18\_128\_eh\_d\_b}} - POSCAR
\begin{lstlisting}[numbers=none,language={mylang}]
A6B2C_tP18_128_eh_d_b & a,c/a,z3,x4,y4 --params=7.057532571,1.41383170154,0.7523,0.7217,0.2489 & P4/mnc D_{4h}^{6} #128 (bdeh) & tP18 & None & K2SnCl6 &  & H. Boysen and A. W. Hewat, Acta Crystallogr. Sect. B Struct. Sci. 34, 1412-1418 (1978)
   1.00000000000000
   7.05753257100000   0.00000000000000   0.00000000000000
   0.00000000000000   7.05753257100000   0.00000000000000
   0.00000000000000   0.00000000000000   9.97816328350000
    Cl     K    Sn
    12     4     2
Direct
   0.00000000000000   0.00000000000000   0.75230000000000   Cl   (4e)
   0.50000000000000   0.50000000000000  -0.25230000000000   Cl   (4e)
   0.00000000000000   0.00000000000000  -0.75230000000000   Cl   (4e)
   0.50000000000000   0.50000000000000   1.25230000000000   Cl   (4e)
   0.72170000000000   0.24890000000000   0.00000000000000   Cl   (8h)
  -0.72170000000000  -0.24890000000000   0.00000000000000   Cl   (8h)
  -0.24890000000000   0.72170000000000   0.00000000000000   Cl   (8h)
   0.24890000000000  -0.72170000000000   0.00000000000000   Cl   (8h)
  -0.22170000000000   0.74890000000000   0.50000000000000   Cl   (8h)
   1.22170000000000   0.25110000000000   0.50000000000000   Cl   (8h)
   0.74890000000000   1.22170000000000   0.50000000000000   Cl   (8h)
   0.25110000000000  -0.22170000000000   0.50000000000000   Cl   (8h)
   0.00000000000000   0.50000000000000   0.25000000000000    K   (4d)
   0.50000000000000   0.00000000000000   0.25000000000000    K   (4d)
   0.00000000000000   0.50000000000000   0.75000000000000    K   (4d)
   0.50000000000000   0.00000000000000   0.75000000000000    K   (4d)
   0.00000000000000   0.00000000000000   0.50000000000000   Sn   (2b)
   0.50000000000000   0.50000000000000   0.00000000000000   Sn   (2b)
\end{lstlisting}
{\phantomsection\label{A7B2C_tP40_128_egi_h_e_cif}}
{\hyperref[A7B2C_tP40_128_egi_h_e]{FeCu$_{2}$Al$_{7}$ ($E9_{a}$): A7B2C\_tP40\_128\_egi\_h\_e}} - CIF
\begin{lstlisting}[numbers=none,language={mylang}]
# CIF file
data_findsym-output
_audit_creation_method FINDSYM

_chemical_name_mineral 'FeCu2Al7'
_chemical_formula_sum 'Al7 Cu2 Fe'

loop_
_publ_author_name
 'M. G. Bown'
 'P. J. Brown'
_journal_name_full_name
;
 Acta Cristallographica
;
_journal_volume 9
_journal_year 1956
_journal_page_first 911
_journal_page_last 914
_publ_Section_title
;
 The structure of FeCu$_{2}$Al$_{7}$ and T(CoCuAl)
;

_aflow_title 'FeCu$_{2}$Al$_{7}$ ($E9_{a}$) Structure'
_aflow_proto 'A7B2C_tP40_128_egi_h_e'
_aflow_params 'a,c/a,z_{1},z_{2},x_{3},x_{4},y_{4},x_{5},y_{5},z_{5}'
_aflow_params_values '6.336,2.34690656566,0.366,0.2008,0.165,0.278,0.088,0.198,0.42,0.1'
_aflow_Strukturbericht '$E9_{a}$'
_aflow_Pearson 'tP40'

_symmetry_space_group_name_H-M "P 4/m 21/n 2/c"
_symmetry_Int_Tables_number 128
 
_cell_length_a    6.33600
_cell_length_b    6.33600
_cell_length_c    14.87000
_cell_angle_alpha 90.00000
_cell_angle_beta  90.00000
_cell_angle_gamma 90.00000
 
loop_
_space_group_symop_id
_space_group_symop_operation_xyz
1 x,y,z
2 x+1/2,-y+1/2,-z+1/2
3 -x+1/2,y+1/2,-z+1/2
4 -x,-y,z
5 -y+1/2,-x+1/2,-z+1/2
6 -y,x,z
7 y,-x,z
8 y+1/2,x+1/2,-z+1/2
9 -x,-y,-z
10 -x+1/2,y+1/2,z+1/2
11 x+1/2,-y+1/2,z+1/2
12 x,y,-z
13 y+1/2,x+1/2,z+1/2
14 y,-x,-z
15 -y,x,-z
16 -y+1/2,-x+1/2,z+1/2
 
loop_
_atom_site_label
_atom_site_type_symbol
_atom_site_symmetry_multiplicity
_atom_site_Wyckoff_label
_atom_site_fract_x
_atom_site_fract_y
_atom_site_fract_z
_atom_site_occupancy
Al1 Al   4 e 0.00000 0.00000  0.36600 1.00000
Fe1 Fe   4 e 0.00000 0.00000  0.20080 1.00000
Al2 Al   8 g 0.16500 0.66500  0.25000 1.00000
Cu1 Cu   8 h 0.27800 0.08800  0.00000 1.00000
Al3 Al  16 i 0.19800 0.42000  0.10000 1.00000
\end{lstlisting}
{\phantomsection\label{A7B2C_tP40_128_egi_h_e_poscar}}
{\hyperref[A7B2C_tP40_128_egi_h_e]{FeCu$_{2}$Al$_{7}$ ($E9_{a}$): A7B2C\_tP40\_128\_egi\_h\_e}} - POSCAR

{\phantomsection\label{A2BC4_tP28_130_f_c_g_cif}}
{\hyperref[A2BC4_tP28_130_f_c_g]{CuBi$_{2}$O$_{4}$: A2BC4\_tP28\_130\_f\_c\_g}} - CIF
\begin{lstlisting}[numbers=none,language={mylang}]
# CIF file
data_findsym-output
_audit_creation_method FINDSYM

_chemical_name_mineral 'CuBi2O4'
_chemical_formula_sum 'Bi2 Cu O4'

loop_
_publ_author_name
 'J.-C. Boivin'
 'J. Trehoux'
 'D. Thomas'
_journal_name_full_name
;
 Bulletin of Research Laboratory of Precision Machinery and Electronics
;
_journal_volume 99
_journal_year 1976
_journal_page_first 193
_journal_page_last 196
_publ_Section_title
;
 {\\'E}tude structurale de CuBi$_{2}$O$_{4}$
;

# Found in Pearson's Crystal Data - Crystal Structure Database for Inorganic Compounds, 2013

_aflow_title 'CuBi$_{2}$O$_{4}$ Structure'
_aflow_proto 'A2BC4_tP28_130_f_c_g'
_aflow_params 'a,c/a,z_{1},x_{2},x_{3},y_{3},z_{3}'
_aflow_params_values '8.5103337343,0.683196239716,0.58,0.5815,0.045,0.136,0.597'
_aflow_Strukturbericht 'None'
_aflow_Pearson 'tP28'

_cell_length_a    8.5103337343
_cell_length_b    8.5103337343
_cell_length_c    5.8142280060
_cell_angle_alpha 90.0000000000
_cell_angle_beta  90.0000000000
_cell_angle_gamma 90.0000000000
 
_symmetry_space_group_name_H-M "P 4/n 21/c 2/c (origin choice 2)"
_symmetry_Int_Tables_number 130
 
loop_
_space_group_symop_id
_space_group_symop_operation_xyz
1 x,y,z
2 x+1/2,-y,-z+1/2
3 -x,y+1/2,-z+1/2
4 -x+1/2,-y+1/2,z
5 -y,-x,-z+1/2
6 -y+1/2,x,z
7 y,-x+1/2,z
8 y+1/2,x+1/2,-z+1/2
9 -x,-y,-z
10 -x+1/2,y,z+1/2
11 x,-y+1/2,z+1/2
12 x+1/2,y+1/2,-z
13 y,x,z+1/2
14 y+1/2,-x,-z
15 -y,x+1/2,-z
16 -y+1/2,-x+1/2,z+1/2
 
loop_
_atom_site_label
_atom_site_type_symbol
_atom_site_symmetry_multiplicity
_atom_site_Wyckoff_label
_atom_site_fract_x
_atom_site_fract_y
_atom_site_fract_z
_atom_site_occupancy
Cu1 Cu   4 c 0.25000 0.25000 0.58000 1.00000
Bi1 Bi   8 f 0.58150 0.41850 0.25000 1.00000
O1  O   16 g 0.04500 0.13600 0.59700 1.00000
\end{lstlisting}
{\phantomsection\label{A2BC4_tP28_130_f_c_g_poscar}}
{\hyperref[A2BC4_tP28_130_f_c_g]{CuBi$_{2}$O$_{4}$: A2BC4\_tP28\_130\_f\_c\_g}} - POSCAR

{\phantomsection\label{A5B3_tP32_130_cg_cf_cif}}
{\hyperref[A5B3_tP32_130_cg_cf]{Ba$_{5}$Si$_{3}$: A5B3\_tP32\_130\_cg\_cf}} - CIF
\begin{lstlisting}[numbers=none,language={mylang}]
# CIF file 
data_findsym-output
_audit_creation_method FINDSYM

_chemical_name_mineral ''
_chemical_formula_sum 'Ba5 Si3'

loop_
_publ_author_name
 'R. Nesper'
 'F. Z{\"u}rcher'
_journal_name_full_name
;
 Zeitschrift f{\"u}r Kristallografiya B
;
_journal_volume 214
_journal_year 1966
_journal_page_first 20
_journal_page_last 20
_publ_Section_title
;
 Refinement of the crystal structure of pentabarium trisilicide, Ba$_{5}$Si$_{3}$
;

_aflow_title 'Ba$_{5}$Si$_{3}$ Structure'
_aflow_proto 'A5B3_tP32_130_cg_cf'
_aflow_params 'a,c/a,z_{1},z_{2},x_{3},x_{4},y_{4},z_{4}'
_aflow_params_values '8.465,1.94329592439,0.2271,0.0095,0.1482,0.57997,0.07997,0.10688'
_aflow_Strukturbericht 'None'
_aflow_Pearson 'tP32'

_symmetry_space_group_name_H-M "P 4/n c c:2"
_symmetry_Int_Tables_number 130
 
_cell_length_a    8.46500
_cell_length_b    8.46500
_cell_length_c    16.45000
_cell_angle_alpha 90.00000
_cell_angle_beta  90.00000
_cell_angle_gamma 90.00000
 
loop_
_space_group_symop_id
_space_group_symop_operation_xyz
1 x,y,z
2 x+1/2,-y,-z+1/2
3 -x,y+1/2,-z+1/2
4 -x+1/2,-y+1/2,z
5 -y,-x,-z+1/2
6 -y+1/2,x,z
7 y,-x+1/2,z
8 y+1/2,x+1/2,-z+1/2
9 -x,-y,-z
10 -x+1/2,y,z+1/2
11 x,-y+1/2,z+1/2
12 x+1/2,y+1/2,-z
13 y,x,z+1/2
14 y+1/2,-x,-z
15 -y,x+1/2,-z
16 -y+1/2,-x+1/2,z+1/2
 
loop_
_atom_site_label
_atom_site_type_symbol
_atom_site_symmetry_multiplicity
_atom_site_Wyckoff_label
_atom_site_fract_x
_atom_site_fract_y
_atom_site_fract_z
_atom_site_occupancy
Ba1 Ba   4 c 0.25000 0.25000 0.22710 1.00000
Si1 Si   4 c 0.25000 0.25000 0.00950 1.00000
Si2 Si   8 f 0.14820 0.85180 0.25000 1.00000
Ba2 Ba  16 g 0.57997 0.07997 0.10688 1.00000
\end{lstlisting}
{\phantomsection\label{A5B3_tP32_130_cg_cf_poscar}}
{\hyperref[A5B3_tP32_130_cg_cf]{Ba$_{5}$Si$_{3}$: A5B3\_tP32\_130\_cg\_cf}} - POSCAR

{\phantomsection\label{A2B2C4D_tP18_132_e_i_o_d_cif}}
{\hyperref[A2B2C4D_tP18_132_e_i_o_d]{Rb$_{2}$TiCu$_{2}$S$_{4}$: A2B2C4D\_tP18\_132\_e\_i\_o\_d}} - CIF
\begin{lstlisting}[numbers=none,language={mylang}]
# CIF file
data_findsym-output
_audit_creation_method FINDSYM

_chemical_name_mineral 'Rb2TiCu2Se4'
_chemical_formula_sum 'Cu2 Rb2 S4 Ti'

loop_
_publ_author_name
 'F. Q. Huang'
 'J. A. Ibers'
_journal_name_full_name
;
 Inorganic Chemistry
;
_journal_volume 40
_journal_year 2001
_journal_page_first 2602
_journal_page_last 2607
_publ_Section_title
;
 New Layered Materials: Syntheses, Structures, and Optical Properties of K$_{2}$TiCu$_{2}$S$_{4}$, Rb$_{2}$TiCu$_{2}$S$_{4}$, Rb$_{2}$TiAg$_{2}$S$_{4}$, Cs$_{2}$TiAg$_{2}$S$_{4}$, and Cs$_{2}$TiCu$_{2}$Se$_{4}$
;

_aflow_title 'Rb$_{2}$TiCu$_{2}$S$_{4}$ Structure'
_aflow_proto 'A2B2C4D_tP18_132_e_i_o_d'
_aflow_params 'a,c/a,x_{3},x_{4},z_{4}'
_aflow_params_values '5.6046,2.34700067801,0.2369,0.26316,0.34803'
_aflow_Strukturbericht 'None'
_aflow_Pearson 'tP18'

_symmetry_space_group_name_H-M "P 42/m 2/c 2/m"
_symmetry_Int_Tables_number 132
 
_cell_length_a    5.60460
_cell_length_b    5.60460
_cell_length_c    13.15400
_cell_angle_alpha 90.00000
_cell_angle_beta  90.00000
_cell_angle_gamma 90.00000
 
loop_
_space_group_symop_id
_space_group_symop_operation_xyz
1 x,y,z
2 x,-y,-z+1/2
3 -x,y,-z+1/2
4 -x,-y,z
5 -y,-x,-z
6 -y,x,z+1/2
7 y,-x,z+1/2
8 y,x,-z
9 -x,-y,-z
10 -x,y,z+1/2
11 x,-y,z+1/2
12 x,y,-z
13 y,x,z
14 y,-x,-z+1/2
15 -y,x,-z+1/2
16 -y,-x,z
 
loop_
_atom_site_label
_atom_site_type_symbol
_atom_site_symmetry_multiplicity
_atom_site_Wyckoff_label
_atom_site_fract_x
_atom_site_fract_y
_atom_site_fract_z
_atom_site_occupancy
Ti1 Ti   2 d 0.50000 0.50000 0.25000 1.00000
Cu1 Cu   4 e 0.00000 0.50000 0.25000 1.00000
Rb1 Rb   4 i 0.23690 0.23690 0.00000 1.00000
S1  S    8 o 0.26316 0.26316 0.34803 1.00000
\end{lstlisting}
{\phantomsection\label{A2B2C4D_tP18_132_e_i_o_d_poscar}}
{\hyperref[A2B2C4D_tP18_132_e_i_o_d]{Rb$_{2}$TiCu$_{2}$S$_{4}$: A2B2C4D\_tP18\_132\_e\_i\_o\_d}} - POSCAR
\begin{lstlisting}[numbers=none,language={mylang}]
A2B2C4D_tP18_132_e_i_o_d & a,c/a,x3,x4,z4 --params=5.6046,2.34700067801,0.2369,0.26316,0.34803 & P4_{2}/mcm D_{4h}^{10} #132 (deio) & tP18 & None & Rb2TiCu2Se4 & Rb2TiCu2Se4 & F. Q. Huang and J. A. Ibers, Inorg. Chem. 40, 2602-2607 (2001)
   1.00000000000000
   5.60460000000000   0.00000000000000   0.00000000000000
   0.00000000000000   5.60460000000000   0.00000000000000
   0.00000000000000   0.00000000000000  13.15400000000000
    Cu    Rb     S    Ti
     4     4     8     2
Direct
   0.00000000000000   0.50000000000000   0.25000000000000   Cu   (4e)
   0.50000000000000   0.00000000000000   0.75000000000000   Cu   (4e)
   0.00000000000000   0.50000000000000   0.75000000000000   Cu   (4e)
   0.50000000000000   0.00000000000000   0.25000000000000   Cu   (4e)
   0.23690000000000   0.23690000000000   0.00000000000000   Rb   (4i)
  -0.23690000000000  -0.23690000000000   0.00000000000000   Rb   (4i)
  -0.23690000000000   0.23690000000000   0.50000000000000   Rb   (4i)
   0.23690000000000  -0.23690000000000   0.50000000000000   Rb   (4i)
   0.26316000000000   0.26316000000000   0.34803000000000    S   (8o)
  -0.26316000000000  -0.26316000000000   0.34803000000000    S   (8o)
  -0.26316000000000   0.26316000000000   0.84803000000000    S   (8o)
   0.26316000000000  -0.26316000000000   0.84803000000000    S   (8o)
  -0.26316000000000   0.26316000000000   0.15197000000000    S   (8o)
   0.26316000000000  -0.26316000000000   0.15197000000000    S   (8o)
   0.26316000000000   0.26316000000000  -0.34803000000000    S   (8o)
  -0.26316000000000  -0.26316000000000  -0.34803000000000    S   (8o)
   0.50000000000000   0.50000000000000   0.25000000000000   Ti   (2d)
   0.50000000000000   0.50000000000000   0.75000000000000   Ti   (2d)
\end{lstlisting}
{\phantomsection\label{AB6C_tP16_132_d_io_a_cif}}
{\hyperref[AB6C_tP16_132_d_io_a]{AgUF$_{6}$: AB6C\_tP16\_132\_d\_io\_a}} - CIF
\begin{lstlisting}[numbers=none,language={mylang}]
# CIF file
data_findsym-output
_audit_creation_method FINDSYM

_chemical_name_mineral 'AgUF6'
_chemical_formula_sum 'Ag F6 U'

loop_
_publ_author_name
 'P. Charpin'
_journal_name_full_name
;
 Comptes Rendus Hebdomadaires des S{\'eances de l'Acad{\'e}mie des Sciences
;
_journal_volume 260
_journal_year 1965
_journal_page_first 1914
_journal_page_last 1916
_publ_Section_title
;
 Structure cristalline des hexafluorures complexes d\'uranium V et d\'argent de potassium d\'ammonium de rubidium ou de thallium
;

# Found in Pearson's Crystal Data - Crystal Structure Database for Inorganic Compounds, 2013

_aflow_title 'AgUF$_{6}$ Structure'
_aflow_proto 'AB6C_tP16_132_d_io_a'
_aflow_params 'a,c/a,x_{3},x_{4},z_{4}'
_aflow_params_values '5.4229923801,1.46597824081,0.3,0.2,0.333'
_aflow_Strukturbericht 'None'
_aflow_Pearson 'tP16'

_cell_length_a    5.4229923801
_cell_length_b    5.4229923801
_cell_length_c    7.9499888293
_cell_angle_alpha 90.0000000000
_cell_angle_beta  90.0000000000
_cell_angle_gamma 90.0000000000
 
_symmetry_space_group_name_H-M "P 42/m 2/c 2/m"
_symmetry_Int_Tables_number 132
 
loop_
_space_group_symop_id
_space_group_symop_operation_xyz
1 x,y,z
2 x,-y,-z+1/2
3 -x,y,-z+1/2
4 -x,-y,z
5 -y,-x,-z
6 -y,x,z+1/2
7 y,-x,z+1/2
8 y,x,-z
9 -x,-y,-z
10 -x,y,z+1/2
11 x,-y,z+1/2
12 x,y,-z
13 y,x,z
14 y,-x,-z+1/2
15 -y,x,-z+1/2
16 -y,-x,z
 
loop_
_atom_site_label
_atom_site_type_symbol
_atom_site_symmetry_multiplicity
_atom_site_Wyckoff_label
_atom_site_fract_x
_atom_site_fract_y
_atom_site_fract_z
_atom_site_occupancy
U1  U    2 a 0.00000 0.00000 0.00000 1.00000
Ag1 Ag   2 d 0.50000 0.50000 0.25000 1.00000
F1  F    4 i 0.30000 0.30000 0.00000 1.00000
F2  F    8 o 0.20000 0.20000 0.33300 1.00000
\end{lstlisting}
{\phantomsection\label{AB6C_tP16_132_d_io_a_poscar}}
{\hyperref[AB6C_tP16_132_d_io_a]{AgUF$_{6}$: AB6C\_tP16\_132\_d\_io\_a}} - POSCAR
\begin{lstlisting}[numbers=none,language={mylang}]
AB6C_tP16_132_d_io_a & a,c/a,x3,x4,z4 --params=5.4229923801,1.46597824081,0.3,0.2,0.333 & P4_{2}/mcm D_{4h}^{10} #132 (adio) & tP16 & None & AgUF6 &  & P. Charpin, {C. R. Hebd. S{'e}ances Acad. Sci. 260, 1914-1916 (1965)
   1.00000000000000
   5.42299238010000   0.00000000000000   0.00000000000000
   0.00000000000000   5.42299238010000   0.00000000000000
   0.00000000000000   0.00000000000000   7.94998882930000
    Ag     F     U
     2    12     2
Direct
   0.50000000000000   0.50000000000000   0.25000000000000   Ag   (2d)
   0.50000000000000   0.50000000000000   0.75000000000000   Ag   (2d)
   0.30000000000000   0.30000000000000   0.00000000000000    F   (4i)
  -0.30000000000000  -0.30000000000000   0.00000000000000    F   (4i)
  -0.30000000000000   0.30000000000000   0.50000000000000    F   (4i)
   0.30000000000000  -0.30000000000000   0.50000000000000    F   (4i)
   0.20000000000000   0.20000000000000   0.33300000000000    F   (8o)
  -0.20000000000000  -0.20000000000000   0.33300000000000    F   (8o)
  -0.20000000000000   0.20000000000000   0.83300000000000    F   (8o)
   0.20000000000000  -0.20000000000000   0.83300000000000    F   (8o)
  -0.20000000000000   0.20000000000000   0.16700000000000    F   (8o)
   0.20000000000000  -0.20000000000000   0.16700000000000    F   (8o)
   0.20000000000000   0.20000000000000  -0.33300000000000    F   (8o)
  -0.20000000000000  -0.20000000000000  -0.33300000000000    F   (8o)
   0.00000000000000   0.00000000000000   0.00000000000000    U   (2a)
   0.00000000000000   0.00000000000000   0.50000000000000    U   (2a)
\end{lstlisting}
{\phantomsection\label{AB3_tP32_133_h_i2j_cif}}
{\hyperref[AB3_tP32_133_h_i2j]{$\beta$-V$_{3}$S: AB3\_tP32\_133\_h\_i2j}} - CIF
\begin{lstlisting}[numbers=none,language={mylang}]
# CIF file
data_findsym-output
_audit_creation_method FINDSYM

_chemical_name_mineral 'beta-V3S'
_chemical_formula_sum 'S V3'

loop_
_publ_author_name
 'B. Pedersen'
 'F. Gr{\o}nvold'
_journal_name_full_name
;
 Acta Cristallographica
;
_journal_volume 12
_journal_year 1959
_journal_page_first 1022
_journal_page_last 1027
_publ_Section_title
;
 The crystal structures of $\alpha$-V$_{3}$S and $\beta$-V$_{3}$S
;

# Found in Pearson's Crystal Data - Crystal Structure Database for Inorganic Compounds, 2013

_aflow_title '$\beta$-V$_{3}$S Structure'
_aflow_proto 'AB3_tP32_133_h_i2j'
_aflow_params 'a,c/a,x_{1},x_{2},x_{3},x_{4}'
_aflow_params_values '9.3810096033,0.497068542799,-0.0329,0.8986,0.658,0.0472'
_aflow_Strukturbericht 'None'
_aflow_Pearson 'tP32'

_cell_length_a    9.3810096033
_cell_length_b    9.3810096033
_cell_length_c    4.6630047735
_cell_angle_alpha 90.0000000000
_cell_angle_beta  90.0000000000
_cell_angle_gamma 90.0000000000
 
_symmetry_space_group_name_H-M "P 42/n 2/b 2/c (origin choice 2)"
_symmetry_Int_Tables_number 133
 
loop_
_space_group_symop_id
_space_group_symop_operation_xyz
1 x,y,z
2 x,-y+1/2,-z
3 -x+1/2,y,-z
4 -x+1/2,-y+1/2,z
5 -y+1/2,-x+1/2,-z+1/2
6 -y+1/2,x,z+1/2
7 y,-x+1/2,z+1/2
8 y,x,-z+1/2
9 -x,-y,-z
10 -x,y+1/2,z
11 x+1/2,-y,z
12 x+1/2,y+1/2,-z
13 y+1/2,x+1/2,z+1/2
14 y+1/2,-x,-z+1/2
15 -y,x+1/2,-z+1/2
16 -y,-x,z+1/2
 
loop_
_atom_site_label
_atom_site_type_symbol
_atom_site_symmetry_multiplicity
_atom_site_Wyckoff_label
_atom_site_fract_x
_atom_site_fract_y
_atom_site_fract_z
_atom_site_occupancy
S1 S   8 h -0.03290 0.25000 0.00000 1.00000
V1 V   8 i 0.89860  0.25000 0.50000 1.00000
V2 V   8 j 0.65800  0.65800 0.25000 1.00000
V3 V   8 j 0.04720  0.04720 0.25000 1.00000
\end{lstlisting}
{\phantomsection\label{AB3_tP32_133_h_i2j_poscar}}
{\hyperref[AB3_tP32_133_h_i2j]{$\beta$-V$_{3}$S: AB3\_tP32\_133\_h\_i2j}} - POSCAR

{\phantomsection\label{A2B_tP24_135_gh_h_cif}}
{\hyperref[A2B_tP24_135_gh_h]{Downeyite (SeO$_{2}$, $C47$): A2B\_tP24\_135\_gh\_h}} - CIF
\begin{lstlisting}[numbers=none,language={mylang}]
# CIF file
data_findsym-output
_audit_creation_method FINDSYM

_chemical_name_mineral 'Downeyite'
_chemical_formula_sum 'O2 Se'

loop_
_publ_author_name
 'K. St{\aa}hl'
 'J. P. Legros'
 'J. Galy'
_journal_name_full_name
;
 Kristallografiya, English title: Crystallography Reports
;
_journal_volume 202
_journal_year 1992
_journal_page_first 99
_journal_page_last 107
_publ_Section_title
;
 The crystal structure of SeO$_{2}$ at 139 and 286 K
;

# Found in The American Mineralogist Crystal Structure Database, 2003

_aflow_title 'Downeyite (SeO$_{2}$, $C47$) Structure'
_aflow_proto 'A2B_tP24_135_gh_h'
_aflow_params 'a,c/a,x_{1},x_{2},y_{2},x_{3},y_{3}'
_aflow_params_values '8.3218,0.607332548247,0.36248,-0.05789,0.17358,0.13396,0.20929'
_aflow_Strukturbericht '$C47$'
_aflow_Pearson 'tP24'

_symmetry_space_group_name_H-M "P 42/m 21/b 2/c"
_symmetry_Int_Tables_number 135
 
_cell_length_a    8.32180
_cell_length_b    8.32180
_cell_length_c    5.05410
_cell_angle_alpha 90.00000
_cell_angle_beta  90.00000
_cell_angle_gamma 90.00000
 
loop_
_space_group_symop_id
_space_group_symop_operation_xyz
1 x,y,z
2 x+1/2,-y+1/2,-z
3 -x+1/2,y+1/2,-z
4 -x,-y,z
5 -y+1/2,-x+1/2,-z+1/2
6 -y,x,z+1/2
7 y,-x,z+1/2
8 y+1/2,x+1/2,-z+1/2
9 -x,-y,-z
10 -x+1/2,y+1/2,z
11 x+1/2,-y+1/2,z
12 x,y,-z
13 y+1/2,x+1/2,z+1/2
14 y,-x,-z+1/2
15 -y,x,-z+1/2
16 -y+1/2,-x+1/2,z+1/2
 
loop_
_atom_site_label
_atom_site_type_symbol
_atom_site_symmetry_multiplicity
_atom_site_Wyckoff_label
_atom_site_fract_x
_atom_site_fract_y
_atom_site_fract_z
_atom_site_occupancy
O1  O    8 g 0.36248 0.86248 0.25000 1.00000
O2  O    8 h -0.05789 0.17358 0.00000 1.00000
Se1 Se   8 h 0.13396 0.20929 0.00000 1.00000
\end{lstlisting}
{\phantomsection\label{A2B_tP24_135_gh_h_poscar}}
{\hyperref[A2B_tP24_135_gh_h]{Downeyite (SeO$_{2}$, $C47$): A2B\_tP24\_135\_gh\_h}} - POSCAR

{\phantomsection\label{A4B2C_tP28_135_gh_h_d_cif}}
{\hyperref[A4B2C_tP28_135_gh_h_d]{ZnSb$_{2}$O$_{4}$: A4B2C\_tP28\_135\_gh\_h\_d}} - CIF
\begin{lstlisting}[numbers=none,language={mylang}]
# CIF file
data_findsym-output
_audit_creation_method FINDSYM

_chemical_name_mineral 'ZnSb2O4'
_chemical_formula_sum 'O4 Sb2 Zn'

loop_
_publ_author_name
 'S. St{\aa}hl'
_journal_name_full_name
;
 Arkiv f{\"o}r Kemi, Mineralogi och Geologi
;
_journal_volume 17B
_journal_year 1943
_journal_page_first 1
_journal_page_last 7
_publ_Section_title
;
 The crystal structure of ZnSb$_{2}$O$_{4}$ and isomorphous compounds
;

# Found in Pearson's Crystal Data - Crystal Structure Database for Inorganic Compounds, 2013

_aflow_title 'ZnSb$_{2}$O$_{4}$ Structure'
_aflow_proto 'A4B2C_tP28_135_gh_h_d'
_aflow_params 'a,c/a,x_{2},x_{3},y_{3},x_{4},y_{4}'
_aflow_params_values '8.4909023894,0.697208809336,0.169,0.114,0.386,0.167,0.175'
_aflow_Strukturbericht 'None'
_aflow_Pearson 'tP28'

_cell_length_a    8.4909023894
_cell_length_b    8.4909023894
_cell_length_c    5.9199319451
_cell_angle_alpha 90.0000000000
_cell_angle_beta  90.0000000000
_cell_angle_gamma 90.0000000000
 
_symmetry_space_group_name_H-M "P 42/m 21/b 2/c"
_symmetry_Int_Tables_number 135
 
loop_
_space_group_symop_id
_space_group_symop_operation_xyz
1 x,y,z
2 x+1/2,-y+1/2,-z
3 -x+1/2,y+1/2,-z
4 -x,-y,z
5 -y+1/2,-x+1/2,-z+1/2
6 -y,x,z+1/2
7 y,-x,z+1/2
8 y+1/2,x+1/2,-z+1/2
9 -x,-y,-z
10 -x+1/2,y+1/2,z
11 x+1/2,-y+1/2,z
12 x,y,-z
13 y+1/2,x+1/2,z+1/2
14 y,-x,-z+1/2
15 -y,x,-z+1/2
16 -y+1/2,-x+1/2,z+1/2
 
loop_
_atom_site_label
_atom_site_type_symbol
_atom_site_symmetry_multiplicity
_atom_site_Wyckoff_label
_atom_site_fract_x
_atom_site_fract_y
_atom_site_fract_z
_atom_site_occupancy
Zn1 Zn   4 d 0.00000 0.50000 0.25000 1.00000
O1  O    8 g 0.16900 0.66900 0.25000 1.00000
O2  O    8 h 0.11400 0.38600 0.00000 1.00000
Sb1 Sb   8 h 0.16700 0.17500 0.00000 1.00000
\end{lstlisting}
{\phantomsection\label{A4B2C_tP28_135_gh_h_d_poscar}}
{\hyperref[A4B2C_tP28_135_gh_h_d]{ZnSb$_{2}$O$_{4}$: A4B2C\_tP28\_135\_gh\_h\_d}} - POSCAR

{\phantomsection\label{A2B3_tP40_137_cdf_3g_cif}}
{\hyperref[A2B3_tP40_137_cdf_3g]{Zn$_{3}$P$_{2}$ ($D5_{9}$): A2B3\_tP40\_137\_cdf\_3g}} - CIF
\begin{lstlisting}[numbers=none,language={mylang}]
# CIF file 
data_findsym-output
_audit_creation_method FINDSYM

_chemical_name_mineral 'Zn3P2'
_chemical_formula_sum 'P2 Zn3'

loop_
_publ_author_name
 'M. v. Stackelberg'
 'R. Paulu'
_journal_name_full_name
;
 Zeitschrift f\"{u}r Physikalische Chemie B
;
_journal_volume 28
_journal_year 1935
_journal_page_first 427
_journal_page_last 460
_publ_Section_title
;
 Untersuchungen an den Phosphiden und Arseniden des Zinks und Cadmiums. Das Zn$_{3}$P$_{2}$-Gitter
;

# Found in The American Mineralogist Crystal Structure Database, 2003

_aflow_title 'Zn$_{3}$P$_{2}$ ($D5_{9}$) Structure'
_aflow_proto 'A2B3_tP40_137_cdf_3g'
_aflow_params 'a,c/a,z_{1},z_{2},x_{3},y_{4},z_{4},y_{5},z_{5},y_{6},z_{6}'
_aflow_params_values '8.097,1.41410398913,0.0,0.011,0.511,0.533,0.147,0.467,0.864,0.5,0.603'
_aflow_Strukturbericht '$D5_{9}$'
_aflow_Pearson 'tP40'

_symmetry_space_group_name_H-M "P 42/n 21/m 2/c (origin choice 2)"
_symmetry_Int_Tables_number 137
 
_cell_length_a    8.09700
_cell_length_b    8.09700
_cell_length_c    11.45000
_cell_angle_alpha 90.00000
_cell_angle_beta  90.00000
_cell_angle_gamma 90.00000
 
loop_
_space_group_symop_id
_space_group_symop_operation_xyz
1 x,y,z
2 x+1/2,-y,-z
3 -x,y+1/2,-z
4 -x+1/2,-y+1/2,z
5 -y,-x,-z+1/2
6 -y+1/2,x,z+1/2
7 y,-x+1/2,z+1/2
8 y+1/2,x+1/2,-z+1/2
9 -x,-y,-z
10 -x+1/2,y,z
11 x,-y+1/2,z
12 x+1/2,y+1/2,-z
13 y,x,z+1/2
14 y+1/2,-x,-z+1/2
15 -y,x+1/2,-z+1/2
16 -y+1/2,-x+1/2,z+1/2
 
loop_
_atom_site_label
_atom_site_type_symbol
_atom_site_symmetry_multiplicity
_atom_site_Wyckoff_label
_atom_site_fract_x
_atom_site_fract_y
_atom_site_fract_z
_atom_site_occupancy
P1  P    4 c 0.75000 0.25000 0.00000 1.00000
P2  P    4 d 0.25000 0.25000 0.01100 1.00000
P3  P    8 f 0.51100 0.48900 0.25000 1.00000
Zn1 Zn   8 g 0.25000 0.53300 0.14700 1.00000
Zn2 Zn   8 g 0.25000 0.46700 0.86400 1.00000
Zn3 Zn   8 g 0.25000 0.50000 0.60300 1.00000
\end{lstlisting}
{\phantomsection\label{A2B3_tP40_137_cdf_3g_poscar}}
{\hyperref[A2B3_tP40_137_cdf_3g]{Zn$_{3}$P$_{2}$ ($D5_{9}$): A2B3\_tP40\_137\_cdf\_3g}} - POSCAR

{\phantomsection\label{A2B_tP6_137_d_a_cif}}
{\hyperref[A2B_tP6_137_d_a]{ZrO$_{2}$ (High-temperature): A2B\_tP6\_137\_d\_a}} - CIF
\begin{lstlisting}[numbers=none,language={mylang}]
# CIF file
data_findsym-output
_audit_creation_method FINDSYM

_chemical_name_mineral 'ZrO2'
_chemical_formula_sum 'O2 Zr'

loop_
_publ_author_name
 'G. Teufer'
_journal_name_full_name
;
 Acta Cristallographica
;
_journal_volume 15
_journal_year 1962
_journal_page_first 1187
_journal_page_last 1187
_publ_Section_title
;
 The crystal structure of tetragonal ZrO$_{2}$
;

# Found in Pearson's Crystal Data - Crystal Structure Database for Inorganic Compounds, 2013

_aflow_title 'ZrO$_{2}$ (High-temperature) Structure'
_aflow_proto 'A2B_tP6_137_d_a'
_aflow_params 'a,c/a,z_{2}'
_aflow_params_values '3.64008007,1.4478021978,0.565'
_aflow_Strukturbericht 'None'
_aflow_Pearson 'tP6'

_cell_length_a    3.6400800700
_cell_length_b    3.6400800700
_cell_length_c    5.2701159255
_cell_angle_alpha 90.0000000000
_cell_angle_beta  90.0000000000
_cell_angle_gamma 90.0000000000
 
_symmetry_space_group_name_H-M "P 42/n 21/m 2/c (origin choice 2)"
_symmetry_Int_Tables_number 137
 
loop_
_space_group_symop_id
_space_group_symop_operation_xyz
1 x,y,z
2 x+1/2,-y,-z
3 -x,y+1/2,-z
4 -x+1/2,-y+1/2,z
5 -y,-x,-z+1/2
6 -y+1/2,x,z+1/2
7 y,-x+1/2,z+1/2
8 y+1/2,x+1/2,-z+1/2
9 -x,-y,-z
10 -x+1/2,y,z
11 x,-y+1/2,z
12 x+1/2,y+1/2,-z
13 y,x,z+1/2
14 y+1/2,-x,-z+1/2
15 -y,x+1/2,-z+1/2
16 -y+1/2,-x+1/2,z+1/2
 
loop_
_atom_site_label
_atom_site_type_symbol
_atom_site_symmetry_multiplicity
_atom_site_Wyckoff_label
_atom_site_fract_x
_atom_site_fract_y
_atom_site_fract_z
_atom_site_occupancy
Zr1 Zr   2 a 0.75000 0.25000 0.75000 1.00000
O1  O    4 d 0.25000 0.25000 0.56500 1.00000
\end{lstlisting}
{\phantomsection\label{A2B_tP6_137_d_a_poscar}}
{\hyperref[A2B_tP6_137_d_a]{ZrO$_{2}$ (High-temperature): A2B\_tP6\_137\_d\_a}} - POSCAR
\begin{lstlisting}[numbers=none,language={mylang}]
A2B_tP6_137_d_a & a,c/a,z2 --params=3.64008007,1.4478021978,0.565 & P4_{2}/nmc D_{4h}^{15} #137 (ad) & tP6 & None & ZrO2 &  & G. Teufer, Acta Cryst. 15, 1187-1187 (1962)
   1.00000000000000
   3.64008007000000   0.00000000000000   0.00000000000000
   0.00000000000000   3.64008007000000   0.00000000000000
   0.00000000000000   0.00000000000000   5.27011592550000
     O    Zr
     4     2
Direct
   0.25000000000000   0.25000000000000   0.56500000000000    O   (4d)
   0.25000000000000   0.25000000000000   1.06500000000000    O   (4d)
   0.75000000000000   0.75000000000000  -0.56500000000000    O   (4d)
   0.75000000000000   0.75000000000000  -0.06500000000000    O   (4d)
   0.75000000000000   0.25000000000000   0.75000000000000   Zr   (2a)
   0.25000000000000   0.75000000000000   0.25000000000000   Zr   (2a)
\end{lstlisting}
{\phantomsection\label{A4BC4_tP18_137_g_b_g_cif}}
{\hyperref[A4BC4_tP18_137_g_b_g]{CeCo$_{4}$B$_{4}$: A4BC4\_tP18\_137\_g\_b\_g}} - CIF
\begin{lstlisting}[numbers=none,language={mylang}]
# CIF file
data_findsym-output
_audit_creation_method FINDSYM

_chemical_name_mineral 'CeCo4B4'
_chemical_formula_sum 'B4 Ce Co4'

loop_
_publ_author_name
 'Y. B. Kuzma'
 'N. S. Bilonizhko'
_journal_name_full_name
;
 Soviet Physics Crystallography
;
_journal_volume 16
_journal_year 1972
_journal_page_first 897
_journal_page_last 898
_publ_Section_title
;
 Crystal structure of the compounds CeCo$_{4}$B$_{4}$ and its analogs
;

# Found in Pearson's Crystal Data - Crystal Structure Database for Inorganic Compounds, 2013

_aflow_title 'CeCo$_{4}$B$_{4}$ Structure'
_aflow_proto 'A4BC4_tP18_137_g_b_g'
_aflow_params 'a,c/a,y_{2},z_{2},y_{3},z_{3}'
_aflow_params_values '5.0593101041,1.39612571654,0.08,0.1,0.503,0.384'
_aflow_Strukturbericht 'None'
_aflow_Pearson 'tP18'

_cell_length_a    5.0593101041
_cell_length_b    5.0593101041
_cell_length_c    7.0634329443
_cell_angle_alpha 90.0000000000
_cell_angle_beta  90.0000000000
_cell_angle_gamma 90.0000000000
 
_symmetry_space_group_name_H-M "P 42/n 21/m 2/c (origin choice 2)"
_symmetry_Int_Tables_number 137
 
loop_
_space_group_symop_id
_space_group_symop_operation_xyz
1 x,y,z
2 x+1/2,-y,-z
3 -x,y+1/2,-z
4 -x+1/2,-y+1/2,z
5 -y,-x,-z+1/2
6 -y+1/2,x,z+1/2
7 y,-x+1/2,z+1/2
8 y+1/2,x+1/2,-z+1/2
9 -x,-y,-z
10 -x+1/2,y,z
11 x,-y+1/2,z
12 x+1/2,y+1/2,-z
13 y,x,z+1/2
14 y+1/2,-x,-z+1/2
15 -y,x+1/2,-z+1/2
16 -y+1/2,-x+1/2,z+1/2
 
loop_
_atom_site_label
_atom_site_type_symbol
_atom_site_symmetry_multiplicity
_atom_site_Wyckoff_label
_atom_site_fract_x
_atom_site_fract_y
_atom_site_fract_z
_atom_site_occupancy
Ce1 Ce   2 b 0.75000 0.25000 0.25000 1.00000
B1  B    8 g 0.25000 0.08000 0.10000 1.00000
Co1 Co   8 g 0.25000 0.50300 0.38400 1.00000
\end{lstlisting}
{\phantomsection\label{A4BC4_tP18_137_g_b_g_poscar}}
{\hyperref[A4BC4_tP18_137_g_b_g]{CeCo$_{4}$B$_{4}$: A4BC4\_tP18\_137\_g\_b\_g}} - POSCAR
\begin{lstlisting}[numbers=none,language={mylang}]
A4BC4_tP18_137_g_b_g & a,c/a,y2,z2,y3,z3 --params=5.0593101041,1.39612571654,0.08,0.1,0.503,0.384 & P4_{2}/nmc D_{4h}^{15} #137 (bg^2) & tP18 & None & CeCo4B4 &  & Y. B. Kuzma and N. S. Bilonizhko, Sov. Phys. Crystallogr. 16, 897-898 (1972)
   1.00000000000000
   5.05931010410000   0.00000000000000   0.00000000000000
   0.00000000000000   5.05931010410000   0.00000000000000
   0.00000000000000   0.00000000000000   7.06343294430000
     B    Ce    Co
     8     2     8
Direct
   0.25000000000000   0.08000000000000   0.10000000000000    B   (8g)
   0.25000000000000   0.42000000000000   0.10000000000000    B   (8g)
   0.42000000000000   0.25000000000000   0.60000000000000    B   (8g)
   0.08000000000000   0.25000000000000   0.60000000000000    B   (8g)
   0.75000000000000   0.58000000000000  -0.10000000000000    B   (8g)
   0.75000000000000  -0.08000000000000  -0.10000000000000    B   (8g)
   0.58000000000000   0.75000000000000   0.40000000000000    B   (8g)
  -0.08000000000000   0.75000000000000   0.40000000000000    B   (8g)
   0.75000000000000   0.25000000000000   0.25000000000000   Ce   (2b)
   0.25000000000000   0.75000000000000   0.75000000000000   Ce   (2b)
   0.25000000000000   0.50300000000000   0.38400000000000   Co   (8g)
   0.25000000000000  -0.00300000000000   0.38400000000000   Co   (8g)
  -0.00300000000000   0.25000000000000   0.88400000000000   Co   (8g)
   0.50300000000000   0.25000000000000   0.88400000000000   Co   (8g)
   0.75000000000000   1.00300000000000  -0.38400000000000   Co   (8g)
   0.75000000000000  -0.50300000000000  -0.38400000000000   Co   (8g)
   1.00300000000000   0.75000000000000   0.11600000000000   Co   (8g)
  -0.50300000000000   0.75000000000000   0.11600000000000   Co   (8g)
\end{lstlisting}
{\phantomsection\label{AB2_tP6_137_a_d_cif}}
{\hyperref[AB2_tP6_137_a_d]{HgI$_{2}$ ($C13$): AB2\_tP6\_137\_a\_d}} - CIF
\begin{lstlisting}[numbers=none,language={mylang}]
# CIF file 
data_findsym-output
_audit_creation_method FINDSYM

_chemical_name_mineral ''
_chemical_formula_sum 'Hg I2'

loop_
_publ_author_name
 'D. Schwarzenbach'
 'H. Birkedal'
 'M. Hostettler'
 'P. Fischer'
_journal_name_full_name
;
 Acta Crystallographica Section B: Structural Science
;
_journal_volume 63
_journal_year 2007
_journal_page_first 826
_journal_page_last 835
_publ_Section_title
;
 Neutron diffraction investigation of the temperature dependence of crystal structure and thermal motions of red HgI$_{2}$
;

# Found in The American Mineralogist Crystal Structure Database, 2003

_aflow_title 'HgI$_{2}$ ($C13$) Structure'
_aflow_proto 'AB2_tP6_137_a_d'
_aflow_params 'a,c/a,z_{2}'
_aflow_params_values '4.3675,2.8551803091,0.389'
_aflow_Strukturbericht '$C13$'
_aflow_Pearson 'tP6'

_symmetry_space_group_name_H-M "P 42/n 21/m 2/c (origin choice 2)"
_symmetry_Int_Tables_number 137
 
_cell_length_a    4.36750
_cell_length_b    4.36750
_cell_length_c    12.47000
_cell_angle_alpha 90.00000
_cell_angle_beta  90.00000
_cell_angle_gamma 90.00000
 
loop_
_space_group_symop_id
_space_group_symop_operation_xyz
1 x,y,z
2 x+1/2,-y,-z
3 -x,y+1/2,-z
4 -x+1/2,-y+1/2,z
5 -y,-x,-z+1/2
6 -y+1/2,x,z+1/2
7 y,-x+1/2,z+1/2
8 y+1/2,x+1/2,-z+1/2
9 -x,-y,-z
10 -x+1/2,y,z
11 x,-y+1/2,z
12 x+1/2,y+1/2,-z
13 y,x,z+1/2
14 y+1/2,-x,-z+1/2
15 -y,x+1/2,-z+1/2
16 -y+1/2,-x+1/2,z+1/2
 
loop_
_atom_site_label
_atom_site_type_symbol
_atom_site_symmetry_multiplicity
_atom_site_Wyckoff_label
_atom_site_fract_x
_atom_site_fract_y
_atom_site_fract_z
_atom_site_occupancy
Hg1 Hg   2 a 0.75000 0.25000 0.75000 1.00000
I1  I    4 d 0.25000 0.25000 0.38900 1.00000
\end{lstlisting}
{\phantomsection\label{AB2_tP6_137_a_d_poscar}}
{\hyperref[AB2_tP6_137_a_d]{HgI$_{2}$ ($C13$): AB2\_tP6\_137\_a\_d}} - POSCAR
\begin{lstlisting}[numbers=none,language={mylang}]
AB2_tP6_137_a_d & a,c/a,z2 --params=4.3675,2.8551803091,0.389 & P4_{2}/nmc D_{4h}^{15} #137 (ad) & tP6 & $C13$ & HgI2 &  & D. Schwarzenbach et al., Acta Crystallogr. Sect. B Struct. Sci. 63, 826-835 (2007)
   1.00000000000000
   4.36750000000000   0.00000000000000   0.00000000000000
   0.00000000000000   4.36750000000000   0.00000000000000
   0.00000000000000   0.00000000000000  12.47000000000000
    Hg     I
     2     4
Direct
   0.75000000000000   0.25000000000000   0.75000000000000   Hg   (2a)
   0.25000000000000   0.75000000000000   0.25000000000000   Hg   (2a)
   0.25000000000000   0.25000000000000   0.38900000000000    I   (4d)
   0.25000000000000   0.25000000000000   0.88900000000000    I   (4d)
   0.75000000000000   0.75000000000000  -0.38900000000000    I   (4d)
   0.75000000000000   0.75000000000000   0.11100000000000    I   (4d)
\end{lstlisting}
{\phantomsection\label{A_tP12_138_bi_cif}}
{\hyperref[A_tP12_138_bi]{C (T12 Group IV): A\_tP12\_138\_bi}} - CIF
\begin{lstlisting}[numbers=none,language={mylang}]
# CIF file 
data_findsym-output
_audit_creation_method FINDSYM

_chemical_name_mineral 'T12'
_chemical_formula_sum 'C'

loop_
_publ_author_name
 'Z. Zhao'
 'F. Tian'
 'X. Dong'
 'Q. Li'
 'Q. Wang'
 'H. Wang'
 'X. Zhong'
 'B. Xu'
 'D. Yu'
 'J. He'
 'H.-T. Wang'
 'Y. Ma'
 'Y. Tian'
_journal_name_full_name
;
 Journal of the American Chemical Society
;
_journal_volume 134
_journal_year 2012
_journal_page_first 12362
_journal_page_last 12365
_publ_Section_title
;
 Tetragonal Allotrope of Group 14 Elements
;

_aflow_title 'C (T12 Group IV) Structure'
_aflow_proto 'A_tP12_138_bi'
_aflow_params 'a,c/a,x_{2},z_{2}'
_aflow_params_values '3.388,1.77420306966,0.086,0.107'
_aflow_Strukturbericht 'None'
_aflow_Pearson 'tP12'

_symmetry_space_group_name_H-M "P 42/n 21/c 2/m (origin choice 2)"
_symmetry_Int_Tables_number 138
 
_cell_length_a    3.38800
_cell_length_b    3.38800
_cell_length_c    6.01100
_cell_angle_alpha 90.00000
_cell_angle_beta  90.00000
_cell_angle_gamma 90.00000
 
loop_
_space_group_symop_id
_space_group_symop_operation_xyz
1 x,y,z
2 x+1/2,-y,-z+1/2
3 -x,y+1/2,-z+1/2
4 -x+1/2,-y+1/2,z
5 -y,-x,-z
6 -y+1/2,x,z+1/2
7 y,-x+1/2,z+1/2
8 y+1/2,x+1/2,-z
9 -x,-y,-z
10 -x+1/2,y,z+1/2
11 x,-y+1/2,z+1/2
12 x+1/2,y+1/2,-z
13 y,x,z
14 y+1/2,-x,-z+1/2
15 -y,x+1/2,-z+1/2
16 -y+1/2,-x+1/2,z
 
loop_
_atom_site_label
_atom_site_type_symbol
_atom_site_symmetry_multiplicity
_atom_site_Wyckoff_label
_atom_site_fract_x
_atom_site_fract_y
_atom_site_fract_z
_atom_site_occupancy
C1 C   4 b 0.75000 0.25000 0.75000 1.00000
C2 C   8 i 0.08600 0.08600 0.10700 1.00000
\end{lstlisting}
{\phantomsection\label{A_tP12_138_bi_poscar}}
{\hyperref[A_tP12_138_bi]{C (T12 Group IV): A\_tP12\_138\_bi}} - POSCAR
\begin{lstlisting}[numbers=none,language={mylang}]
A_tP12_138_bi & a,c/a,x2,z2 --params=3.388,1.77420306966,0.086,0.107 & P4_{2}/ncm D_{4h}^{16} #138 (bi) & tP12 & None & C & T12 & Z. Zhao et al., J. Am. Chem. Soc. 134, 12362-12365 (2012)
   1.00000000000000
   3.38800000000000   0.00000000000000   0.00000000000000
   0.00000000000000   3.38800000000000   0.00000000000000
   0.00000000000000   0.00000000000000   6.01100000000000
     C
    12
Direct
   0.75000000000000   0.25000000000000   0.75000000000000    C   (4b)
   0.25000000000000   0.75000000000000   0.25000000000000    C   (4b)
   0.25000000000000   0.75000000000000   0.75000000000000    C   (4b)
   0.75000000000000   0.25000000000000   0.25000000000000    C   (4b)
   0.08600000000000   0.08600000000000   0.10700000000000    C   (8i)
   0.41400000000000   0.41400000000000   0.10700000000000    C   (8i)
   0.41400000000000   0.08600000000000   0.60700000000000    C   (8i)
   0.08600000000000   0.41400000000000   0.60700000000000    C   (8i)
  -0.08600000000000   0.58600000000000   0.39300000000000    C   (8i)
   0.58600000000000  -0.08600000000000   0.39300000000000    C   (8i)
   0.58600000000000   0.58600000000000  -0.10700000000000    C   (8i)
  -0.08600000000000  -0.08600000000000  -0.10700000000000    C   (8i)
\end{lstlisting}
{\phantomsection\label{AB_tI8_139_e_e_cif}}
{\hyperref[AB_tI8_139_e_e]{Calomel (Hg$_{2}$Cl$_{2}$, $D3_{1}$): AB\_tI8\_139\_e\_e}} - CIF
\begin{lstlisting}[numbers=none,language={mylang}]
# CIF file
data_findsym-output
_audit_creation_method FINDSYM

_chemical_name_mineral 'Calomel'
_chemical_formula_sum 'Cl Hg'

loop_
_publ_author_name
 'N. J. Calos'
 'C. H. L. Kennard'
 'R. L. Davis'
_journal_name_full_name
;
 Zeitschrift f{\"u}r Kristallografiya
;
_journal_volume 187
_journal_year 1989
_journal_page_first 305
_journal_page_last 307
_publ_Section_title
;
 The structure of calomel, Hg$_{2}$Cl$_{2}$, derived from neutron powder data
;

# Found in The American Mineralogist Crystal Structure Database, 2003

_aflow_title 'Calomel (Hg$_{2}$Cl$_{2}$, $D3_{1}$) Structure'
_aflow_proto 'AB_tI8_139_e_e'
_aflow_params 'a,c/a,z_{1},z_{2}'
_aflow_params_values '4.4795,2.43451278044,0.3356,0.119'
_aflow_Strukturbericht '$D3_{1}$'
_aflow_Pearson 'tI8'

_symmetry_space_group_name_H-M "I 4/m 2/m 2/m"
_symmetry_Int_Tables_number 139
 
_cell_length_a    4.47950
_cell_length_b    4.47950
_cell_length_c    10.90540
_cell_angle_alpha 90.00000
_cell_angle_beta  90.00000
_cell_angle_gamma 90.00000
 
loop_
_space_group_symop_id
_space_group_symop_operation_xyz
1 x,y,z
2 x,-y,-z
3 -x,y,-z
4 -x,-y,z
5 -y,-x,-z
6 -y,x,z
7 y,-x,z
8 y,x,-z
9 -x,-y,-z
10 -x,y,z
11 x,-y,z
12 x,y,-z
13 y,x,z
14 y,-x,-z
15 -y,x,-z
16 -y,-x,z
17 x+1/2,y+1/2,z+1/2
18 x+1/2,-y+1/2,-z+1/2
19 -x+1/2,y+1/2,-z+1/2
20 -x+1/2,-y+1/2,z+1/2
21 -y+1/2,-x+1/2,-z+1/2
22 -y+1/2,x+1/2,z+1/2
23 y+1/2,-x+1/2,z+1/2
24 y+1/2,x+1/2,-z+1/2
25 -x+1/2,-y+1/2,-z+1/2
26 -x+1/2,y+1/2,z+1/2
27 x+1/2,-y+1/2,z+1/2
28 x+1/2,y+1/2,-z+1/2
29 y+1/2,x+1/2,z+1/2
30 y+1/2,-x+1/2,-z+1/2
31 -y+1/2,x+1/2,-z+1/2
32 -y+1/2,-x+1/2,z+1/2
 
loop_
_atom_site_label
_atom_site_type_symbol
_atom_site_symmetry_multiplicity
_atom_site_Wyckoff_label
_atom_site_fract_x
_atom_site_fract_y
_atom_site_fract_z
_atom_site_occupancy
Cl1 Cl   4 e 0.00000 0.00000 0.33560 1.00000
Hg1 Hg   4 e 0.00000 0.00000 0.11900 1.00000
\end{lstlisting}
{\phantomsection\label{AB_tI8_139_e_e_poscar}}
{\hyperref[AB_tI8_139_e_e]{Calomel (Hg$_{2}$Cl$_{2}$, $D3_{1}$): AB\_tI8\_139\_e\_e}} - POSCAR
\begin{lstlisting}[numbers=none,language={mylang}]
AB_tI8_139_e_e & a,c/a,z1,z2 --params=4.4795,2.43451278044,0.3356,0.119 & I4/mmm D_{4h}^{17} #139 (e^2) & tI8 & $D3_{1}$ & Hg2Cl2 & Calomel & N. J. Calos and C. H. L. Kennard and R. L. Davis, Z. Kristallogr. 187, 305-307 (1989)
   1.00000000000000
  -2.23975000000000   2.23975000000000   5.45270000000000
   2.23975000000000  -2.23975000000000   5.45270000000000
   2.23975000000000   2.23975000000000  -5.45270000000000
    Cl    Hg
     2     2
Direct
   0.33560000000000   0.33560000000000   0.00000000000000   Cl   (4e)
  -0.33560000000000  -0.33560000000000   0.00000000000000   Cl   (4e)
   0.11900000000000   0.11900000000000   0.00000000000000   Hg   (4e)
  -0.11900000000000  -0.11900000000000   0.00000000000000   Hg   (4e)
\end{lstlisting}
{\phantomsection\label{A3B5_tI32_140_ah_bk_cif}}
{\hyperref[A3B5_tI32_140_ah_bk]{W$_{5}$Si$_{3}$ ($D8_{m}$): A3B5\_tI32\_140\_ah\_bk}} - CIF
\begin{lstlisting}[numbers=none,language={mylang}]
# CIF file
data_findsym-output
_audit_creation_method FINDSYM

_chemical_name_mineral 'W5Si3'
_chemical_formula_sum 'Si3 W5'

loop_
_publ_author_name
 'B. Aronsson'
_journal_name_full_name
;
 Acta Chemica Scandinavica
;
_journal_volume 9
_journal_year 1955
_journal_page_first 1107
_journal_page_last 1110
_publ_Section_title
;
 The Crystal Structure of Mo$_{5}$Si$_{3}$ and W$_{5}$Si$_{3}$
;

_aflow_title 'W$_{5}$Si$_{3}$ ($D8_{m}$) Structure'
_aflow_proto 'A3B5_tI32_140_ah_bk'
_aflow_params 'a,c/a,x_{3},x_{4},y_{4}'
_aflow_params_values '9.64,0.515560165975,0.17,0.074,0.223'
_aflow_Strukturbericht '$D8_{m}$'
_aflow_Pearson 'tI32'

_symmetry_space_group_name_H-M "I 4/m 2/c 2/m"
_symmetry_Int_Tables_number 140
_cell_length_a    9.64000
_cell_length_b    9.64000
_cell_length_c    4.97000
_cell_angle_alpha 90.00000
_cell_angle_beta  90.00000
_cell_angle_gamma 90.00000
 
loop_
_space_group_symop_id
_space_group_symop_operation_xyz
1 x,y,z
2 x,-y,-z+1/2
3 -x,y,-z+1/2
4 -x,-y,z
5 -y,-x,-z+1/2
6 -y,x,z
7 y,-x,z
8 y,x,-z+1/2
9 -x,-y,-z
10 -x,y,z+1/2
11 x,-y,z+1/2
12 x,y,-z
13 y,x,z+1/2
14 y,-x,-z
15 -y,x,-z
16 -y,-x,z+1/2
17 x+1/2,y+1/2,z+1/2
18 x+1/2,-y+1/2,-z
19 -x+1/2,y+1/2,-z
20 -x+1/2,-y+1/2,z+1/2
21 -y+1/2,-x+1/2,-z
22 -y+1/2,x+1/2,z+1/2
23 y+1/2,-x+1/2,z+1/2
24 y+1/2,x+1/2,-z
25 -x+1/2,-y+1/2,-z+1/2
26 -x+1/2,y+1/2,z
27 x+1/2,-y+1/2,z
28 x+1/2,y+1/2,-z+1/2
29 y+1/2,x+1/2,z
30 y+1/2,-x+1/2,-z+1/2
31 -y+1/2,x+1/2,-z+1/2
32 -y+1/2,-x+1/2,z
 
loop_
_atom_site_label
_atom_site_type_symbol
_atom_site_symmetry_multiplicity
_atom_site_Wyckoff_label
_atom_site_fract_x
_atom_site_fract_y
_atom_site_fract_z
_atom_site_occupancy
Si1 Si   4 a 0.00000 0.00000 0.25000 1.00000
W1  W    4 b 0.00000 0.50000 0.25000 1.00000
Si2 Si   8 h 0.17000 0.67000 0.00000 1.00000
W2  W   16 k 0.07400 0.22300 0.00000 1.00000
\end{lstlisting}
{\phantomsection\label{A3B5_tI32_140_ah_bk_poscar}}
{\hyperref[A3B5_tI32_140_ah_bk]{W$_{5}$Si$_{3}$ ($D8_{m}$): A3B5\_tI32\_140\_ah\_bk}} - POSCAR
\begin{lstlisting}[numbers=none,language={mylang}]
A3B5_tI32_140_ah_bk & a,c/a,x3,x4,y4 --params=9.64,0.515560165975,0.17,0.074,0.223 & I4/mcm D_{4h}^{18} #140 (abhk) & tI32 & $D8_{m}$ & W5Si3 & W5Si3 & B. Aronsson, Acta Chem. Scand. 9, 1107-1110 (1955)
   1.00000000000000
  -4.82000000000000   4.82000000000000   2.48500000000000
   4.82000000000000  -4.82000000000000   2.48500000000000
   4.82000000000000   4.82000000000000  -2.48500000000000
    Si     W
     6    10
Direct
   0.25000000000000   0.25000000000000   0.00000000000000   Si   (4a)
   0.75000000000000   0.75000000000000   0.00000000000000   Si   (4a)
   0.67000000000000   0.17000000000000   0.84000000000000   Si   (8h)
   0.33000000000000  -0.17000000000000   0.16000000000000   Si   (8h)
   0.17000000000000   0.33000000000000   0.50000000000000   Si   (8h)
  -0.17000000000000   0.67000000000000   0.50000000000000   Si   (8h)
   0.75000000000000   0.25000000000000   0.50000000000000    W   (4b)
   0.25000000000000   0.75000000000000   0.50000000000000    W   (4b)
   0.22300000000000   0.07400000000000   0.29700000000000    W  (16k)
  -0.22300000000000  -0.07400000000000  -0.29700000000000    W  (16k)
   0.07400000000000  -0.22300000000000  -0.14900000000000    W  (16k)
  -0.07400000000000   0.22300000000000   0.14900000000000    W  (16k)
   0.72300000000000   0.42600000000000   0.14900000000000    W  (16k)
   0.27700000000000   0.57400000000000  -0.14900000000000    W  (16k)
   0.57400000000000   0.72300000000000   0.29700000000000    W  (16k)
   0.42600000000000   0.27700000000000  -0.29700000000000    W  (16k)
\end{lstlisting}
{\phantomsection\label{A3B5_tI32_140_ah_cl_cif}}
{\hyperref[A3B5_tI32_140_ah_cl]{Cr$_{5}$B$_{3}$ ($D8_{l}$): A3B5\_tI32\_140\_ah\_cl}} - CIF
\begin{lstlisting}[numbers=none,language={mylang}]
# CIF file
data_findsym-output
_audit_creation_method FINDSYM

_chemical_name_mineral 'Cr5B3'
_chemical_formula_sum 'B3 Cr5'

loop_
_publ_author_name
 'F. Bertaut'
 'P. Blum'
_journal_name_full_name
;
 Comptes Rendus Hebdomadaires des S{\'eances de l'Acad{\'e}mie des Sciences
;
_journal_volume 236
_journal_year 1953
_journal_page_first 1055
_journal_page_last 1056
_publ_Section_title
;
 Etude des borures de chrome
;

# Found in The American Mineralogist Crystal Structure Database, 2003

_aflow_title 'Cr$_{5}$B$_{3}$ ($D8_{l}$) Structure'
_aflow_proto 'A3B5_tI32_140_ah_cl'
_aflow_params 'a,c/a,x_{3},x_{4},z_{4}'
_aflow_params_values '5.46,1.91575091575,0.625,0.166,0.15'
_aflow_Strukturbericht '$D8_{l}$'
_aflow_Pearson 'tI32'

_symmetry_space_group_name_H-M "I 4/m 2/c 2/m"
_symmetry_Int_Tables_number 140
 
_cell_length_a    5.46000
_cell_length_b    5.46000
_cell_length_c    10.46000
_cell_angle_alpha 90.00000
_cell_angle_beta  90.00000
_cell_angle_gamma 90.00000
 
loop_
_space_group_symop_id
_space_group_symop_operation_xyz
1 x,y,z
2 x,-y,-z+1/2
3 -x,y,-z+1/2
4 -x,-y,z
5 -y,-x,-z+1/2
6 -y,x,z
7 y,-x,z
8 y,x,-z+1/2
9 -x,-y,-z
10 -x,y,z+1/2
11 x,-y,z+1/2
12 x,y,-z
13 y,x,z+1/2
14 y,-x,-z
15 -y,x,-z
16 -y,-x,z+1/2
17 x+1/2,y+1/2,z+1/2
18 x+1/2,-y+1/2,-z
19 -x+1/2,y+1/2,-z
20 -x+1/2,-y+1/2,z+1/2
21 -y+1/2,-x+1/2,-z
22 -y+1/2,x+1/2,z+1/2
23 y+1/2,-x+1/2,z+1/2
24 y+1/2,x+1/2,-z
25 -x+1/2,-y+1/2,-z+1/2
26 -x+1/2,y+1/2,z
27 x+1/2,-y+1/2,z
28 x+1/2,y+1/2,-z+1/2
29 y+1/2,x+1/2,z
30 y+1/2,-x+1/2,-z+1/2
31 -y+1/2,x+1/2,-z+1/2
32 -y+1/2,-x+1/2,z
 
loop_
_atom_site_label
_atom_site_type_symbol
_atom_site_symmetry_multiplicity
_atom_site_Wyckoff_label
_atom_site_fract_x
_atom_site_fract_y
_atom_site_fract_z
_atom_site_occupancy
B1  B    4 a 0.00000 0.00000 0.25000 1.00000
Cr1 Cr   4 c 0.00000 0.00000 0.00000 1.00000
B2  B    8 h 0.62500 0.12500 0.00000 1.00000
Cr2 Cr  16 l 0.16600 0.66600 0.15000 1.00000
\end{lstlisting}
{\phantomsection\label{A3B5_tI32_140_ah_cl_poscar}}
{\hyperref[A3B5_tI32_140_ah_cl]{Cr$_{5}$B$_{3}$ ($D8_{l}$): A3B5\_tI32\_140\_ah\_cl}} - POSCAR
\begin{lstlisting}[numbers=none,language={mylang}]
A3B5_tI32_140_ah_cl & a,c/a,x3,x4,z4 --params=5.46,1.91575091575,0.625,0.166,0.15 & I4/mcm D_{4h}^{18} #140 (achl) & tI32 & $D8_{l}$ & Cr5B3 & Cr5B3 & F. Bertaut and P. Blum, {C. R. Hebd. S{'e}ances Acad. Sci. 236, 1055-1056 (1953)
   1.00000000000000
  -2.73000000000000   2.73000000000000   5.23000000000000
   2.73000000000000  -2.73000000000000   5.23000000000000
   2.73000000000000   2.73000000000000  -5.23000000000000
     B    Cr
     6    10
Direct
   0.25000000000000   0.25000000000000   0.00000000000000    B   (4a)
   0.75000000000000   0.75000000000000   0.00000000000000    B   (4a)
   1.12500000000000   0.62500000000000   1.75000000000000    B   (8h)
  -0.12500000000000  -0.62500000000000  -0.75000000000000    B   (8h)
   0.62500000000000  -0.12500000000000   0.50000000000000    B   (8h)
  -0.62500000000000   1.12500000000000   0.50000000000000    B   (8h)
   0.00000000000000   0.00000000000000   0.00000000000000   Cr   (4c)
   0.50000000000000   0.50000000000000   0.00000000000000   Cr   (4c)
   0.81600000000000   0.31600000000000   0.83200000000000   Cr  (16l)
   0.48400000000000  -0.01600000000000   0.16800000000000   Cr  (16l)
   0.31600000000000   0.48400000000000   0.50000000000000   Cr  (16l)
  -0.01600000000000   0.81600000000000   0.50000000000000   Cr  (16l)
   0.01600000000000   0.18400000000000   0.50000000000000   Cr  (16l)
  -0.31600000000000   0.51600000000000   0.50000000000000   Cr  (16l)
   0.51600000000000   0.01600000000000   0.83200000000000   Cr  (16l)
   0.18400000000000  -0.31600000000000   0.16800000000000   Cr  (16l)
\end{lstlisting}
{\phantomsection\label{A2B_tI12_141_e_a_cif}}
{\hyperref[A2B_tI12_141_e_a]{$\alpha$-ThSi$_{2}$ ($C_{c}$): A2B\_tI12\_141\_e\_a}} - CIF
\begin{lstlisting}[numbers=none,language={mylang}]
# CIF file
data_findsym-output
_audit_creation_method FINDSYM

_chemical_name_mineral '$\alpha$-ThSi$_{2}$'
_chemical_formula_sum 'Si2 Th'

loop_
_publ_author_name
 'G. Brauer'
 'A. Mitius'
_journal_name_full_name
;
 Zeitschrift fur Anorganische und Allgemeine Chemie
;
_journal_volume 249
_journal_year 1942
_journal_page_first 325
_journal_page_last 339
_publ_Section_title
;
 Die Kristallstruktur des Thoriumsilicids ThSi$_2$
;

# Found in The Crystal Chemistry and Physics of Metals and Alloys, 1972

_aflow_title '$\alpha$-ThSi$_{2}$ ($C_{c}$) Structure'
_aflow_proto 'A2B_tI12_141_e_a'
_aflow_params 'a,c/a,z_{2}'
_aflow_params_values '4.126,3.47697527872,0.2915'
_aflow_Strukturbericht '$C_{c}$'
_aflow_Pearson 'tI12'

_symmetry_space_group_name_H-M "I 41/a 2/m 2/d"
_symmetry_Int_Tables_number 141
 
_cell_length_a    4.12600
_cell_length_b    4.12600
_cell_length_c    14.34600
_cell_angle_alpha 90.00000
_cell_angle_beta  90.00000
_cell_angle_gamma 90.00000
 
loop_
_space_group_symop_id
_space_group_symop_operation_xyz
1 x,y,z
2 x,-y,-z
3 -x,y+1/2,-z
4 -x,-y+1/2,z
5 -y+1/4,-x+1/4,-z+3/4
6 -y+1/4,x+3/4,z+1/4
7 y+3/4,-x+3/4,z+1/4
8 y+3/4,x+1/4,-z+3/4
9 -x,-y,-z
10 -x,y,z
11 x,-y+1/2,z
12 x,y+1/2,-z
13 y+3/4,x+3/4,z+1/4
14 y+3/4,-x+1/4,-z+3/4
15 -y+1/4,x+1/4,-z+3/4
16 -y+1/4,-x+3/4,z+1/4
17 x+1/2,y+1/2,z+1/2
18 x+1/2,-y+1/2,-z+1/2
19 -x+1/2,y,-z+1/2
20 -x+1/2,-y,z+1/2
21 -y+3/4,-x+3/4,-z+1/4
22 -y+3/4,x+1/4,z+3/4
23 y+1/4,-x+1/4,z+3/4
24 y+1/4,x+3/4,-z+1/4
25 -x+1/2,-y+1/2,-z+1/2
26 -x+1/2,y+1/2,z+1/2
27 x+1/2,-y,z+1/2
28 x+1/2,y,-z+1/2
29 y+1/4,x+1/4,z+3/4
30 y+1/4,-x+3/4,-z+1/4
31 -y+3/4,x+3/4,-z+1/4
32 -y+3/4,-x+1/4,z+3/4
 
loop_
_atom_site_label
_atom_site_type_symbol
_atom_site_symmetry_multiplicity
_atom_site_Wyckoff_label
_atom_site_fract_x
_atom_site_fract_y
_atom_site_fract_z
_atom_site_occupancy
Th1 Th   4 a 0.00000 0.75000 0.12500 1.00000
Si1 Si   8 e 0.00000 0.25000 0.29150 1.00000
\end{lstlisting}
{\phantomsection\label{A2B_tI12_141_e_a_poscar}}
{\hyperref[A2B_tI12_141_e_a]{$\alpha$-ThSi$_{2}$ ($C_{c}$): A2B\_tI12\_141\_e\_a}} - POSCAR
\begin{lstlisting}[numbers=none,language={mylang}]
A2B_tI12_141_e_a & a,c/a,z2 --params=4.126,3.47697527872,0.2915 & I4_{1}/amd D_{4h}^{19} #141 (ae) & tI12 & $C_{c}$ & ThSi2 & $\alpha$-ThSi$_{2}$ & G. Brauer and A. Mitius, Z. Anorg. Allg. Chem. 249, 325-339 (1942)
   1.00000000000000
  -2.06300000000000   2.06300000000000   7.17300000000000
   2.06300000000000  -2.06300000000000   7.17300000000000
   2.06300000000000   2.06300000000000  -7.17300000000000
    Si    Th
     4     2
Direct
   0.54150000000000   0.29150000000000   0.25000000000000   Si   (8e)
   0.29150000000000   0.54150000000000   0.75000000000000   Si   (8e)
   0.45850000000000  -0.29150000000000   0.75000000000000   Si   (8e)
  -0.29150000000000   0.45850000000000   0.25000000000000   Si   (8e)
   0.87500000000000   0.12500000000000   0.75000000000000   Th   (4a)
   0.12500000000000   0.87500000000000   0.25000000000000   Th   (4a)
\end{lstlisting}
{\phantomsection\label{A_tI16_142_f_cif}}
{\hyperref[A_tI16_142_f]{S-III: A\_tI16\_142\_f}} - CIF
\begin{lstlisting}[numbers=none,language={mylang}]
# CIF file 
data_findsym-output
_audit_creation_method FINDSYM

_chemical_name_mineral 'S-III'
_chemical_formula_sum 'S'

loop_
_publ_author_name
 'Olga Degtyareva'
 'Eugene Gregoryanz'
 'Maddury Somayazulu'
 'Przemyslaw Dera'
 'Ho-kwang Mao'
 'Russell J. Hemley'
_journal_name_full_name
;
 Nature Materials
;
_journal_volume 4
_journal_year 2005
_journal_page_first 152
_journal_page_last 155
_publ_Section_title
;
 Novel chain structures in group VI elements
;

_aflow_title 'S-III Structure'
_aflow_proto 'A_tI16_142_f'
_aflow_params 'a,c/a,x_{1}'
_aflow_params_values '8.5939,0.420984651904,0.1405'
_aflow_Strukturbericht 'None'
_aflow_Pearson 'tI16'

_symmetry_space_group_name_H-M "I 41/a 2/c 2/d (origin choice 2)"
_symmetry_Int_Tables_number 142
 
_cell_length_a    8.59390
_cell_length_b    8.59390
_cell_length_c    3.61790
_cell_angle_alpha 90.00000
_cell_angle_beta  90.00000
_cell_angle_gamma 90.00000
 
loop_
_space_group_symop_id
_space_group_symop_operation_xyz
1 x,y,z
2 x+1/2,-y+1/2,-z
3 -x+1/2,y,-z
4 -x,-y+1/2,z
5 -y+1/4,-x+1/4,-z+1/4
6 -y+1/4,x+3/4,z+1/4
7 y+3/4,-x+3/4,z+1/4
8 y+3/4,x+1/4,-z+1/4
9 -x,-y,-z
10 -x,y,z+1/2
11 x,-y+1/2,z+1/2
12 x,y+1/2,-z
13 y+1/4,x+1/4,z+1/4
14 y+3/4,-x+1/4,-z+3/4
15 -y+1/4,x+1/4,-z+3/4
16 -y+3/4,-x+1/4,z+1/4
17 x+1/2,y+1/2,z+1/2
18 x,-y,-z+1/2
19 -x,y+1/2,-z+1/2
20 -x+1/2,-y,z+1/2
21 -y+3/4,-x+3/4,-z+3/4
22 -y+3/4,x+1/4,z+3/4
23 y+1/4,-x+1/4,z+3/4
24 y+1/4,x+3/4,-z+3/4
25 -x+1/2,-y+1/2,-z+1/2
26 -x+1/2,y+1/2,z
27 x+1/2,-y,z
28 x+1/2,y,-z+1/2
29 y+3/4,x+3/4,z+3/4
30 y+1/4,-x+3/4,-z+1/4
31 -y+3/4,x+3/4,-z+1/4
32 -y+1/4,-x+3/4,z+3/4
 
loop_
_atom_site_label
_atom_site_type_symbol
_atom_site_symmetry_multiplicity
_atom_site_Wyckoff_label
_atom_site_fract_x
_atom_site_fract_y
_atom_site_fract_z
_atom_site_occupancy
S1 S  16 f 0.14050 0.39050 0.12500 1.00000
\end{lstlisting}
{\phantomsection\label{A_tI16_142_f_poscar}}
{\hyperref[A_tI16_142_f]{S-III: A\_tI16\_142\_f}} - POSCAR
\begin{lstlisting}[numbers=none,language={mylang}]
A_tI16_142_f & a,c/a,x1 --params=8.5939,0.420984651904,0.1405 & I4_{1}/acd D_{4h}^{20} #142 (f) & tI16 & None & S & S-III & Olga Degtyareva et al., Nat. Mater. 4, 152-155 (2005)
   1.00000000000000
  -4.29695000000000   4.29695000000000   1.80895000000000
   4.29695000000000  -4.29695000000000   1.80895000000000
   4.29695000000000   4.29695000000000  -1.80895000000000
     S
     8
Direct
   0.51550000000000   0.26550000000000   0.53100000000000    S  (16f)
   0.23450000000000  -0.01550000000000  -0.03100000000000    S  (16f)
   0.26550000000000   0.23450000000000   0.75000000000000    S  (16f)
  -0.01550000000000   0.51550000000000   0.75000000000000    S  (16f)
   0.48450000000000   0.73450000000000   0.46900000000000    S  (16f)
   0.76550000000000   1.01550000000000   1.03100000000000    S  (16f)
   0.73450000000000   0.76550000000000   0.25000000000000    S  (16f)
   1.01550000000000   0.48450000000000   0.25000000000000    S  (16f)
\end{lstlisting}
{\phantomsection\label{A4B14C3_hP21_143_bd_ac4d_d_cif}}
{\hyperref[A4B14C3_hP21_143_bd_ac4d_d]{Simpsonite (Ta$_{3}$Al$_{4}$O$_{13}$[OH]): A4B14C3\_hP21\_143\_bd\_ac4d\_d}} - CIF

{\phantomsection\label{A4B14C3_hP21_143_bd_ac4d_d_poscar}}
{\hyperref[A4B14C3_hP21_143_bd_ac4d_d]{Simpsonite (Ta$_{3}$Al$_{4}$O$_{13}$[OH]): A4B14C3\_hP21\_143\_bd\_ac4d\_d}} - POSCAR

{\phantomsection\label{A4B6C_hP11_143_bd_2d_a_cif}}
{\hyperref[A4B6C_hP11_143_bd_2d_a]{ScRh$_{6}$P$_{4}$: A4B6C\_hP11\_143\_bd\_2d\_a}} - CIF
\begin{lstlisting}[numbers=none,language={mylang}]
# CIF file
data_findsym-output
_audit_creation_method FINDSYM

_chemical_name_mineral 'ScRh6P4'
_chemical_formula_sum 'P4 Rh6 Sc'

loop_
_publ_author_name
 'U. Pfannenschmidt'
 'U. C. Rodewald'
 'R. P{\"o}ttgen'
_journal_name_full_name
;
 Monatshefte f{\"u}r Chemie - Chemical Monthly
;
_journal_volume 142
_journal_year 2011
_journal_page_first 219
_journal_page_last 224
_publ_Section_title
;
 Bismuth flux crystal growth of $RE$Rh$_{6}$P$_{4}$ ($RE$ = Sc, Yb, Lu): new phosphides with a superstructure of the LiCo$_{6}$P$_{4}$ type
;

# Found in Pearson's Crystal Data - Crystal Structure Database for Inorganic Compounds, 2013

_aflow_title 'ScRh$_{6}$P$_{4}$ Structure'
_aflow_proto 'A4B6C_hP11_143_bd_2d_a'
_aflow_params 'a,c/a,z_{1},z_{2},x_{3},y_{3},z_{3},x_{4},y_{4},z_{4},x_{5},y_{5},z_{5}'
_aflow_params_values '6.9672687469,0.526119402977,0.0004,-0.0007,0.8181,0.1915,0.4998,0.5475,0.48,0.0003,0.1859,0.799,0.5002'
_aflow_Strukturbericht 'None'
_aflow_Pearson 'hP11'

_cell_length_a    6.9672687469
_cell_length_b    6.9672687469
_cell_length_c    3.6656152735
_cell_angle_alpha 90.0000000000
_cell_angle_beta  90.0000000000
_cell_angle_gamma 120.0000000000
 
_symmetry_space_group_name_H-M "P 3"
_symmetry_Int_Tables_number 143
 
loop_
_space_group_symop_id
_space_group_symop_operation_xyz
1 x,y,z
2 -y,x-y,z
3 -x+y,-x,z
 
loop_
_atom_site_label
_atom_site_type_symbol
_atom_site_symmetry_multiplicity
_atom_site_Wyckoff_label
_atom_site_fract_x
_atom_site_fract_y
_atom_site_fract_z
_atom_site_occupancy
Sc1 Sc   1 a 0.00000 0.00000 0.00040  1.00000
P1  P    1 b 0.33333 0.66667 -0.00070 1.00000
P2  P    3 d 0.81810 0.19150 0.49980  1.00000
Rh1 Rh   3 d 0.54750 0.48000 0.00030  1.00000
Rh2 Rh   3 d 0.18590 0.79900 0.50020  1.00000
\end{lstlisting}
{\phantomsection\label{A4B6C_hP11_143_bd_2d_a_poscar}}
{\hyperref[A4B6C_hP11_143_bd_2d_a]{ScRh$_{6}$P$_{4}$: A4B6C\_hP11\_143\_bd\_2d\_a}} - POSCAR
\begin{lstlisting}[numbers=none,language={mylang}]
A4B6C_hP11_143_bd_2d_a & a,c/a,z1,z2,x3,y3,z3,x4,y4,z4,x5,y5,z5 --params=6.9672687469,0.526119402977,0.0004,-0.0007,0.8181,0.1915,0.4998,0.5475,0.48,0.0003,0.1859,0.799,0.5002 & P3 C_{3}^{1} #143 (abd^3) & hP11 & None & ScRh6P4 &  & U. Pfannenschmidt and U. C. Rodewald and R. P{\"o}ttgen, Monatsh. Chem. 142, 219-224 (2011)
   1.00000000000000
   3.48363437345000  -6.03383172980877   0.00000000000000
   3.48363437345000   6.03383172980877   0.00000000000000
   0.00000000000000   0.00000000000000   3.66561527350000
     P    Rh    Sc
     4     6     1
Direct
   0.33333333333333   0.66666666666667  -0.00070000000000    P   (1b)
   0.81810000000000   0.19150000000000   0.49980000000000    P   (3d)
  -0.19150000000000   0.62660000000000   0.49980000000000    P   (3d)
  -0.62660000000000  -0.81810000000000   0.49980000000000    P   (3d)
   0.54750000000000   0.48000000000000   0.00030000000000   Rh   (3d)
  -0.48000000000000   0.06750000000000   0.00030000000000   Rh   (3d)
  -0.06750000000000  -0.54750000000000   0.00030000000000   Rh   (3d)
   0.18590000000000   0.79900000000000   0.50020000000000   Rh   (3d)
  -0.79900000000000  -0.61310000000000   0.50020000000000   Rh   (3d)
   0.61310000000000  -0.18590000000000   0.50020000000000   Rh   (3d)
   0.00000000000000   0.00000000000000   0.00040000000000   Sc   (1a)
\end{lstlisting}
{\phantomsection\label{AB2_hP12_143_cd_ab2d_cif}}
{\hyperref[AB2_hP12_143_cd_ab2d]{MoS$_{2}$: AB2\_hP12\_143\_cd\_ab2d}} - CIF
\begin{lstlisting}[numbers=none,language={mylang}]
# CIF file
data_findsym-output
_audit_creation_method FINDSYM

_chemical_name_mineral 'MoS2'
_chemical_formula_sum 'Mo S2'

loop_
_publ_author_name
 'K. E. Dungey'
 'M. D. Curtis'
 'J. E. {Penner-Hahn}'
_journal_name_full_name
;
 Chemistry of Materials
;
_journal_volume 10
_journal_year 1998
_journal_page_first 2152
_journal_page_last 2161
_publ_Section_title
;
 Structural characterization and thermal stability of MoS$_{2}$ intercalation compounds
;

# Found in Pearson's Crystal Data - Crystal Structure Database for Inorganic Compounds, 2013

_aflow_title 'MoS$_{2}$ Structure'
_aflow_proto 'AB2_hP12_143_cd_ab2d'
_aflow_params 'a,c/a,z_{1},z_{2},z_{3},x_{4},y_{4},z_{4},x_{5},y_{5},z_{5},x_{6},y_{6},z_{6}'
_aflow_params_values '6.5003859369,0.944615384626,0.0,0.5,0.25,0.0548,0.2679,0.25,0.33333,0.16667,0.5,0.0,0.5,0.0'
_aflow_Strukturbericht 'None'
_aflow_Pearson 'hP12'

_cell_length_a    6.5003859369
_cell_length_b    6.5003859369
_cell_length_c    6.1403645620
_cell_angle_alpha 90.0000000000
_cell_angle_beta  90.0000000000
_cell_angle_gamma 120.0000000000
 
_symmetry_space_group_name_H-M "P 3"
_symmetry_Int_Tables_number 143
 
loop_
_space_group_symop_id
_space_group_symop_operation_xyz
1 x,y,z
2 -y,x-y,z
3 -x+y,-x,z
 
loop_
_atom_site_label
_atom_site_type_symbol
_atom_site_symmetry_multiplicity
_atom_site_Wyckoff_label
_atom_site_fract_x
_atom_site_fract_y
_atom_site_fract_z
_atom_site_occupancy
S1  S    1 a 0.00000 0.00000 0.00000 1.00000
S2  S    1 b 0.33333 0.66667 0.50000 1.00000
Mo1 Mo   1 c 0.66667 0.33333 0.25000 1.00000
Mo2 Mo   3 d 0.05480 0.26790 0.25000 1.00000
S3  S    3 d 0.33333 0.16667 0.50000 1.00000
S4  S    3 d 0.00000 0.50000 0.00000 1.00000
\end{lstlisting}
{\phantomsection\label{AB2_hP12_143_cd_ab2d_poscar}}
{\hyperref[AB2_hP12_143_cd_ab2d]{MoS$_{2}$: AB2\_hP12\_143\_cd\_ab2d}} - POSCAR
\begin{lstlisting}[numbers=none,language={mylang}]
AB2_hP12_143_cd_ab2d & a,c/a,z1,z2,z3,x4,y4,z4,x5,y5,z5,x6,y6,z6 --params=6.5003859369,0.944615384626,0.0,0.5,0.25,0.0548,0.2679,0.25,0.33333,0.16667,0.5,0.0,0.5,0.0 & P3 C_{3}^{1} #143 (abcd^3) & hP12 & None & MoS2 &  & K. E. Dungey and M. D. Curtis and J. E. {Penner-Hahn}, Chem. Mater. 10, 2152-2161 (1998)
   1.00000000000000
   3.25019296845000  -5.62949935575851   0.00000000000000
   3.25019296845000   5.62949935575851   0.00000000000000
   0.00000000000000   0.00000000000000   6.14036456200000
    Mo     S
     4     8
Direct
   0.66666666666667   0.33333333333333   0.25000000000000   Mo   (1c)
   0.05480000000000   0.26790000000000   0.25000000000000   Mo   (3d)
  -0.26790000000000  -0.21310000000000   0.25000000000000   Mo   (3d)
   0.21310000000000  -0.05480000000000   0.25000000000000   Mo   (3d)
   0.00000000000000   0.00000000000000   0.00000000000000    S   (1a)
   0.33333333333333   0.66666666666667   0.50000000000000    S   (1b)
   0.33333000000000   0.16667000000000   0.50000000000000    S   (3d)
  -0.16667000000000   0.16666000000000   0.50000000000000    S   (3d)
  -0.16666000000000  -0.33333000000000   0.50000000000000    S   (3d)
   0.00000000000000   0.50000000000000   0.00000000000000    S   (3d)
  -0.50000000000000  -0.50000000000000   0.00000000000000    S   (3d)
   0.50000000000000   0.00000000000000   0.00000000000000    S   (3d)
\end{lstlisting}
{\phantomsection\label{A4B_hP15_144_4a_a_cif}}
{\hyperref[A4B_hP15_144_4a_a]{IrGe$_{4}$: A4B\_hP15\_144\_4a\_a}} - CIF
\begin{lstlisting}[numbers=none,language={mylang}]
# CIF file
data_findsym-output
_audit_creation_method FINDSYM

_chemical_name_mineral 'IrGe4'
_chemical_formula_sum 'Ge4 Ir'

loop_
_publ_author_name
 'K. Schubert'
 'S. Bhan'
 'T. K. Biswas'
 'K. Frank'
 'P. K. Panday'
_journal_name_full_name
;
 Naturwissenschaften
;
_journal_volume 55
_journal_year 1968
_journal_page_first 542
_journal_page_last 543
_publ_Section_title
;
 Einige Strukturdaten metallischer Phasen
;

# Found in Pearson's Crystal Data - Crystal Structure Database for Inorganic Compounds, 2013

_aflow_title 'IrGe$_{4}$ Structure'
_aflow_proto 'A4B_hP15_144_4a_a'
_aflow_params 'a,c/a,x_{1},y_{1},z_{1},x_{2},y_{2},z_{2},x_{3},y_{3},z_{3},x_{4},y_{4},z_{4},x_{5},y_{5},z_{5}'
_aflow_params_values '6.2151204722,1.25245374096,0.4904,0.2194,0.2268,0.2226,0.4873,0.1142,0.0775,0.0012,0.0,0.6097,0.0014,0.0018,0.3178,0.0008,0.5062'
_aflow_Strukturbericht 'None'
_aflow_Pearson 'hP15'

_cell_length_a    6.2151204722
_cell_length_b    6.2151204722
_cell_length_c    7.7841508859
_cell_angle_alpha 90.0000000000
_cell_angle_beta  90.0000000000
_cell_angle_gamma 120.0000000000
 
_symmetry_space_group_name_H-M "P 31"
_symmetry_Int_Tables_number 144
 
loop_
_space_group_symop_id
_space_group_symop_operation_xyz
1 x,y,z
2 -y,x-y,z+1/3
3 -x+y,-x,z+2/3
 
loop_
_atom_site_label
_atom_site_type_symbol
_atom_site_symmetry_multiplicity
_atom_site_Wyckoff_label
_atom_site_fract_x
_atom_site_fract_y
_atom_site_fract_z
_atom_site_occupancy
Ge1 Ge   3 a 0.49040 0.21940 0.22680 1.00000
Ge2 Ge   3 a 0.22260 0.48730 0.11420 1.00000
Ge3 Ge   3 a 0.07750 0.00120 0.00000 1.00000
Ge4 Ge   3 a 0.60970 0.00140 0.00180 1.00000
Ir1 Ir   3 a 0.31780 0.00080 0.50620 1.00000
\end{lstlisting}
{\phantomsection\label{A4B_hP15_144_4a_a_poscar}}
{\hyperref[A4B_hP15_144_4a_a]{IrGe$_{4}$: A4B\_hP15\_144\_4a\_a}} - POSCAR
\begin{lstlisting}[numbers=none,language={mylang}]
A4B_hP15_144_4a_a & a,c/a,x1,y1,z1,x2,y2,z2,x3,y3,z3,x4,y4,z4,x5,y5,z5 --params=6.2151204722,1.25245374096,0.4904,0.2194,0.2268,0.2226,0.4873,0.1142,0.0775,0.0012,0.0,0.6097,0.0014,0.0018,0.3178,0.0008,0.5062 & P3_{1} C_{3}^{2} #144 (a^5) & hP15 & None & IrGe4 &  & K. Schubert et al., {Naturwissenschaften 55, 542-543 (1968)
   1.00000000000000
   3.10756023610000  -5.38245221650594   0.00000000000000
   3.10756023610000   5.38245221650594   0.00000000000000
   0.00000000000000   0.00000000000000   7.78415088590000
    Ge    Ir
    12     3
Direct
   0.49040000000000   0.21940000000000   0.22680000000000   Ge   (3a)
  -0.21940000000000   0.27100000000000   0.56013333333333   Ge   (3a)
  -0.27100000000000  -0.49040000000000   0.89346666666667   Ge   (3a)
   0.22260000000000   0.48730000000000   0.11420000000000   Ge   (3a)
  -0.48730000000000  -0.26470000000000   0.44753333333333   Ge   (3a)
   0.26470000000000  -0.22260000000000   0.78086666666667   Ge   (3a)
   0.07750000000000   0.00120000000000   0.00000000000000   Ge   (3a)
  -0.00120000000000   0.07630000000000   0.33333333333333   Ge   (3a)
  -0.07630000000000  -0.07750000000000   0.66666666666667   Ge   (3a)
   0.60970000000000   0.00140000000000   0.00180000000000   Ge   (3a)
  -0.00140000000000   0.60830000000000   0.33513333333333   Ge   (3a)
  -0.60830000000000  -0.60970000000000   0.66846666666667   Ge   (3a)
   0.31780000000000   0.00080000000000   0.50620000000000   Ir   (3a)
  -0.00080000000000   0.31700000000000   0.83953333333333   Ir   (3a)
  -0.31700000000000  -0.31780000000000   1.17286666666667   Ir   (3a)
\end{lstlisting}
{\phantomsection\label{AB_hP6_144_a_a_cif}}
{\hyperref[AB_hP6_144_a_a]{TeZn (High-pressure): AB\_hP6\_144\_a\_a}} - CIF
\begin{lstlisting}[numbers=none,language={mylang}]
# CIF file
data_findsym-output
_audit_creation_method FINDSYM

_chemical_name_mineral 'ZnTe'
_chemical_formula_sum 'Te Zn'

loop_
_publ_author_name
 'K. Kusaba'
 'D. J. Weidner'
_journal_year 1994
_publ_Section_title
;
 Structure of high pressure phase I in ZnTe
;

# Found in Pearson's Crystal Data - Crystal Structure Database for Inorganic Compounds, 2013

_aflow_title 'TeZn (High-pressure) Structure'
_aflow_proto 'AB_hP6_144_a_a'
_aflow_params 'a,c/a,x_{1},y_{1},z_{1},x_{2},y_{2},z_{2}'
_aflow_params_values '4.0452497415,2.30951792334,0.33,0.16,0.197,0.13,0.32,0.0'
_aflow_Strukturbericht 'None'
_aflow_Pearson 'hP6'

_cell_length_a    4.0452497415
_cell_length_b    4.0452497415
_cell_length_c    9.3425767824
_cell_angle_alpha 90.0000000000
_cell_angle_beta  90.0000000000
_cell_angle_gamma 120.0000000000
 
_symmetry_space_group_name_H-M "P 31"
_symmetry_Int_Tables_number 144
 
loop_
_space_group_symop_id
_space_group_symop_operation_xyz
1 x,y,z
2 -y,x-y,z+1/3
3 -x+y,-x,z+2/3
 
loop_
_atom_site_label
_atom_site_type_symbol
_atom_site_symmetry_multiplicity
_atom_site_Wyckoff_label
_atom_site_fract_x
_atom_site_fract_y
_atom_site_fract_z
_atom_site_occupancy
Te1 Te   3 a 0.33000 0.16000 0.19700 1.00000
Zn1 Zn   3 a 0.13000 0.32000 0.00000 1.00000
\end{lstlisting}
{\phantomsection\label{AB_hP6_144_a_a_poscar}}
{\hyperref[AB_hP6_144_a_a]{TeZn (High-pressure): AB\_hP6\_144\_a\_a}} - POSCAR
\begin{lstlisting}[numbers=none,language={mylang}]
AB_hP6_144_a_a & a,c/a,x1,y1,z1,x2,y2,z2 --params=4.0452497415,2.30951792334,0.33,0.16,0.197,0.13,0.32,0.0 & P3_{1} C_{3}^{2} #144 (a^2) & hP6 & None & ZnTe &  & K. Kusaba and D. J. Weidner, (1994)
   1.00000000000000
   2.02262487075000  -3.50328904079143   0.00000000000000
   2.02262487075000   3.50328904079143   0.00000000000000
   0.00000000000000   0.00000000000000   9.34257678240000
    Te    Zn
     3     3
Direct
   0.33000000000000   0.16000000000000   0.19700000000000   Te   (3a)
  -0.16000000000000   0.17000000000000   0.53033333333333   Te   (3a)
  -0.17000000000000  -0.33000000000000   0.86366666666667   Te   (3a)
   0.13000000000000   0.32000000000000   0.00000000000000   Zn   (3a)
  -0.32000000000000  -0.19000000000000   0.33333333333333   Zn   (3a)
   0.19000000000000  -0.13000000000000   0.66666666666667   Zn   (3a)
\end{lstlisting}
{\phantomsection\label{A2B3C3DE7_hP48_145_2a_3a_3a_a_7a_cif}}
{\hyperref[A2B3C3DE7_hP48_145_2a_3a_3a_a_7a]{Sheldrickite (NaCa$_{3}$[CO$_{3}$]$_{2}$F$_{3}$[H$_{2}$O]): A2B3C3DE7\_hP48\_145\_2a\_3a\_3a\_a\_7a}} - CIF

{\phantomsection\label{A2B3C3DE7_hP48_145_2a_3a_3a_a_7a_poscar}}
{\hyperref[A2B3C3DE7_hP48_145_2a_3a_3a_a_7a]{Sheldrickite (NaCa$_{3}$[CO$_{3}$]$_{2}$F$_{3}$[H$_{2}$O]): A2B3C3DE7\_hP48\_145\_2a\_3a\_3a\_a\_7a}} - POSCAR

{\phantomsection\label{A3BC_hR5_146_b_a_a_cif}}
{\hyperref[A3BC_hR5_146_b_a_a]{$\gamma$-Ag$_{3}$SI (Low-temperature): A3BC\_hR5\_146\_b\_a\_a}} - CIF
\begin{lstlisting}[numbers=none,language={mylang}]
# CIF file
data_findsym-output
_audit_creation_method FINDSYM

_chemical_name_mineral 'gamma-Ag3SI'
_chemical_formula_sum 'Ag3 I S'

loop_
_publ_author_name
 'S. Hoshino'
 'T. Sakuma'
 'Y. Fujii'
_journal_name_full_name
;
 Journal of the Physical Society of Japan
;
_journal_volume 47
_journal_year 1979
_journal_page_first 1252
_journal_page_last 1259
_publ_Section_title
;
 A Structural Phase Transition in Superionic Conductor Ag$_{3}$SI
;

# Found in Pearson's Crystal Data - Crystal Structure Database for Inorganic Compounds, 2013

_aflow_title '$\gamma$-Ag$_{3}$SI (Low-temperature) Structure'
_aflow_proto 'A3BC_hR5_146_b_a_a'
_aflow_params 'a,c/a,x_{1},x_{2},x_{3},y_{3},z_{3}'
_aflow_params_values '6.8858547087,1.22475238897,0.47,0.0,0.49,-0.002,-0.14399'
_aflow_Strukturbericht 'None'
_aflow_Pearson 'hR5'

_cell_length_a    6.8858547087
_cell_length_b    6.8858547087
_cell_length_c    8.4334670046
_cell_angle_alpha 90.0000000000
_cell_angle_beta  90.0000000000
_cell_angle_gamma 120.0000000000
 
_symmetry_space_group_name_H-M "R 3 (hexagonal axes)"
_symmetry_Int_Tables_number 146
 
loop_
_space_group_symop_id
_space_group_symop_operation_xyz
1 x,y,z
2 -y,x-y,z
3 -x+y,-x,z
4 x+1/3,y+2/3,z+2/3
5 -y+1/3,x-y+2/3,z+2/3
6 -x+y+1/3,-x+2/3,z+2/3
7 x+2/3,y+1/3,z+1/3
8 -y+2/3,x-y+1/3,z+1/3
9 -x+y+2/3,-x+1/3,z+1/3
 
loop_
_atom_site_label
_atom_site_type_symbol
_atom_site_symmetry_multiplicity
_atom_site_Wyckoff_label
_atom_site_fract_x
_atom_site_fract_y
_atom_site_fract_z
_atom_site_occupancy
I1  I    3 a 0.00000 0.00000 0.47000 1.00000
S1  S    3 a 0.00000 0.00000 0.00000 1.00000
Ag1 Ag   9 b 0.37533 0.11667 0.11467 1.00000
\end{lstlisting}
{\phantomsection\label{A3BC_hR5_146_b_a_a_poscar}}
{\hyperref[A3BC_hR5_146_b_a_a]{$\gamma$-Ag$_{3}$SI (Low-temperature): A3BC\_hR5\_146\_b\_a\_a}} - POSCAR
\begin{lstlisting}[numbers=none,language={mylang}]
A3BC_hR5_146_b_a_a & a,c/a,x1,x2,x3,y3,z3 --params=6.8858547087,1.22475238897,0.47,0.0,0.49,-0.002,-0.14399 & R3 C_{3}^{4} #146 (a^2b) & hR5 & None & Ag3SI & gamma & S. Hoshino and T. Sakuma and Y. Fujii, J. Phys. Soc. Jpn. 47, 1252-1259 (1979)
   1.00000000000000
   3.44292735435000  -1.98777503483430   2.81115566820000
   0.00000000000000   3.97555006966860   2.81115566820000
  -3.44292735435000  -1.98777503483430   2.81115566820000
    Ag     I     S
     3     1     1
Direct
   0.49000000000000  -0.00200000000000  -0.14399000000000   Ag   (3b)
  -0.14399000000000   0.49000000000000  -0.00200000000000   Ag   (3b)
  -0.00200000000000  -0.14399000000000   0.49000000000000   Ag   (3b)
   0.47000000000000   0.47000000000000   0.47000000000000    I   (1a)
   0.00000000000000   0.00000000000000   0.00000000000000    S   (1a)
\end{lstlisting}
{\phantomsection\label{ABC3_hR10_146_2a_2a_2b_cif}}
{\hyperref[ABC3_hR10_146_2a_2a_2b]{FePSe$_{3}$: ABC3\_hR10\_146\_2a\_2a\_2b}} - CIF
\begin{lstlisting}[numbers=none,language={mylang}]
# CIF file
data_findsym-output
_audit_creation_method FINDSYM

_chemical_name_mineral 'FePSe3'
_chemical_formula_sum 'Fe P Se3'

loop_
_publ_author_name
 'W. Klingen'
 'G. Eulenberger'
 'H. Hahn'
_journal_name_full_name
;
 Zeitschrift fur Anorganische und Allgemeine Chemie
;
_journal_volume 401
_journal_year 1973
_journal_page_first 97
_journal_page_last 112
_publ_Section_title
;
 Uber die Kristallstrukturen von Fe$_{2}$P$_{2}$Se$_{6}$ und Fe$_{2}$P$_{2}$S$_{6}$
;

# Found in Pearson's Crystal Data - Crystal Structure Database for Inorganic Compounds, 2013

_aflow_title 'FePSe$_{3}$ Structure'
_aflow_proto 'ABC3_hR10_146_2a_2a_2b'
_aflow_params 'a,c/a,x_{1},x_{2},x_{3},x_{4},x_{5},y_{5},z_{5},x_{6},y_{6},z_{6}'
_aflow_params_values '6.2648898445,3.16041500397,0.3911,0.7256,0.1132,0.0,0.1454,-0.1886,0.474,0.6151,-0.0128,0.3247'
_aflow_Strukturbericht 'None'
_aflow_Pearson 'hR10'

_cell_length_a    6.2648898445
_cell_length_b    6.2648898445
_cell_length_c    19.7996518628
_cell_angle_alpha 90.0000000000
_cell_angle_beta  90.0000000000
_cell_angle_gamma 120.0000000000
 
_symmetry_space_group_name_H-M "R 3 (hexagonal axes)"
_symmetry_Int_Tables_number 146
 
loop_
_space_group_symop_id
_space_group_symop_operation_xyz
1 x,y,z
2 -y,x-y,z
3 -x+y,-x,z
4 x+1/3,y+2/3,z+2/3
5 -y+1/3,x-y+2/3,z+2/3
6 -x+y+1/3,-x+2/3,z+2/3
7 x+2/3,y+1/3,z+1/3
8 -y+2/3,x-y+1/3,z+1/3
9 -x+y+2/3,-x+1/3,z+1/3
 
loop_
_atom_site_label
_atom_site_type_symbol
_atom_site_symmetry_multiplicity
_atom_site_Wyckoff_label
_atom_site_fract_x
_atom_site_fract_y
_atom_site_fract_z
_atom_site_occupancy
Fe1 Fe   3 a 0.00000 0.00000 0.39110 1.00000
Fe2 Fe   3 a 0.00000 0.00000 0.72560 1.00000
P1  P    3 a 0.00000 0.00000 0.11320 1.00000
P2  P    3 a 0.00000 0.00000 0.00000 1.00000
Se1 Se   9 b 0.00180 0.33220 0.14360 1.00000
Se2 Se   9 b 0.30610 0.32180 0.30900 1.00000
\end{lstlisting}
{\phantomsection\label{ABC3_hR10_146_2a_2a_2b_poscar}}
{\hyperref[ABC3_hR10_146_2a_2a_2b]{FePSe$_{3}$: ABC3\_hR10\_146\_2a\_2a\_2b}} - POSCAR
\begin{lstlisting}[numbers=none,language={mylang}]
ABC3_hR10_146_2a_2a_2b & a,c/a,x1,x2,x3,x4,x5,y5,z5,x6,y6,z6 --params=6.2648898445,3.16041500397,0.3911,0.7256,0.1132,0.0,0.1454,-0.1886,0.474,0.6151,-0.0128,0.3247 & R3 C_{3}^{4} #146 (a^4b^2) & hR10 & None & FePSe3 &  & W. Klingen and G. Eulenberger and H. Hahn, Z. Anorg. Allg. Chem. 401, 97-112 (1973)
   1.00000000000000
   3.13244492225000  -1.80851791908271   6.59988395426667
   0.00000000000000   3.61703583816543   6.59988395426667
  -3.13244492225000  -1.80851791908271   6.59988395426667
    Fe     P    Se
     2     2     6
Direct
   0.39110000000000   0.39110000000000   0.39110000000000   Fe   (1a)
   0.72560000000000   0.72560000000000   0.72560000000000   Fe   (1a)
   0.11320000000000   0.11320000000000   0.11320000000000    P   (1a)
   0.00000000000000   0.00000000000000   0.00000000000000    P   (1a)
   0.14540000000000  -0.18860000000000   0.47400000000000   Se   (3b)
   0.47400000000000   0.14540000000000  -0.18860000000000   Se   (3b)
  -0.18860000000000   0.47400000000000   0.14540000000000   Se   (3b)
   0.61510000000000  -0.01280000000000   0.32470000000000   Se   (3b)
   0.32470000000000   0.61510000000000  -0.01280000000000   Se   (3b)
  -0.01280000000000   0.32470000000000   0.61510000000000   Se   (3b)
\end{lstlisting}
{\phantomsection\label{A2B4C_hR42_148_2f_4f_f_cif}}
{\hyperref[A2B4C_hR42_148_2f_4f_f]{Phenakite (Be$_2$SiO$_4$, $S1_{3}$): A2B4C\_hR42\_148\_2f\_4f\_f}} - CIF

{\phantomsection\label{A2B4C_hR42_148_2f_4f_f_poscar}}
{\hyperref[A2B4C_hR42_148_2f_4f_f]{Phenakite (Be$_2$SiO$_4$, $S1_{3}$): A2B4C\_hR42\_148\_2f\_4f\_f}} - POSCAR

{\phantomsection\label{A2B_hR18_148_2f_f_cif}}
{\hyperref[A2B_hR18_148_2f_f]{$\beta$-PdCl$_2$: A2B\_hR18\_148\_2f\_f}} - CIF
\begin{lstlisting}[numbers=none,language={mylang}]
# CIF file 
data_findsym-output
_audit_creation_method FINDSYM

_chemical_name_mineral '$\beta$-PdCl2'
_chemical_formula_sum 'Cl2 Pd'

loop_
_publ_author_name
 'D. B. {Dell\'Amico}'
 'F. Calderazzo'
 'F. Marchetti'
 'S. Ramello'
_journal_name_full_name
;
 Angewandte Chemie (International ed.)
;
_journal_volume 35
_journal_year 1996
_journal_page_first 1331
_journal_page_last 1333
_publ_Section_title
;
 Molecular Structure of [Pd$_{6}$Cl$_{12}$] in Single Crystals Chemically Grown at Room Temperature
;

_aflow_title '$\beta$-PdCl$_2$ Structure'
_aflow_proto 'A2B_hR18_148_2f_f'
_aflow_params 'a,c/a,x_{1},y_{1},z_{1},x_{2},y_{2},z_{2},x_{3},y_{3},z_{3}'
_aflow_params_values '13.0471,0.659280606418,0.7407,-0.76315,0.0238,1.14609,0.38403,-0.59217,0.08714,0.06386,0.3263'
_aflow_Strukturbericht 'None'
_aflow_Pearson 'hR18'

_symmetry_space_group_name_H-M "R -3 (hexagonal axes)"
_symmetry_Int_Tables_number 148
 
_cell_length_a    13.04710
_cell_length_b    13.04710
_cell_length_c    8.60170
_cell_angle_alpha 90.00000
_cell_angle_beta  90.00000
_cell_angle_gamma 120.00000
 
loop_
_space_group_symop_id
_space_group_symop_operation_xyz
1 x,y,z
2 -y,x-y,z
3 -x+y,-x,z
4 -x,-y,-z
5 y,-x+y,-z
6 x-y,x,-z
7 x+1/3,y+2/3,z+2/3
8 -y+1/3,x-y+2/3,z+2/3
9 -x+y+1/3,-x+2/3,z+2/3
10 -x+1/3,-y+2/3,-z+2/3
11 y+1/3,-x+y+2/3,-z+2/3
12 x-y+1/3,x+2/3,-z+2/3
13 x+2/3,y+1/3,z+1/3
14 -y+2/3,x-y+1/3,z+1/3
15 -x+y+2/3,-x+1/3,z+1/3
16 -x+2/3,-y+1/3,-z+1/3
17 y+2/3,-x+y+1/3,-z+1/3
18 x-y+2/3,x+1/3,-z+1/3
 
loop_
_atom_site_label
_atom_site_type_symbol
_atom_site_symmetry_multiplicity
_atom_site_Wyckoff_label
_atom_site_fract_x
_atom_site_fract_y
_atom_site_fract_z
_atom_site_occupancy
Cl1 Cl  18 f 0.74025  0.76360  0.00045 1.00000
Cl2 Cl  18 f 0.83344  -0.07138 0.31265 1.00000
Pd1 Pd  18 f -0.07196 0.09524  0.15910 1.00000
\end{lstlisting}
{\phantomsection\label{A2B_hR18_148_2f_f_poscar}}
{\hyperref[A2B_hR18_148_2f_f]{$\beta$-PdCl$_2$: A2B\_hR18\_148\_2f\_f}} - POSCAR
\begin{lstlisting}[numbers=none,language={mylang}]
A2B_hR18_148_2f_f & a,c/a,x1,y1,z1,x2,y2,z2,x3,y3,z3 --params=13.0471,0.659280606418,0.7407,-0.76315,0.0238,1.14609,0.38403,-0.59217,0.08714,0.06386,0.3263 & R-3 C_{3i}^{2} #148 (f^3) & hR18 & None & PdCl2 & $\beta$-PdCl2 & D. B. {Dell'Amico} et al., Angew. Chem. Int. Ed. 35, 1331-1333 (1996)
   1.00000000000000
   6.52355000000000  -3.76637334857198   2.86723333333333
   0.00000000000000   7.53274669714397   2.86723333333333
  -6.52355000000000  -3.76637334857198   2.86723333333333
    Cl    Pd
    12     6
Direct
   0.74070000000000  -0.76315000000000   0.02380000000000   Cl   (6f)
   0.02380000000000   0.74070000000000  -0.76315000000000   Cl   (6f)
  -0.76315000000000   0.02380000000000   0.74070000000000   Cl   (6f)
  -0.74070000000000   0.76315000000000  -0.02380000000000   Cl   (6f)
  -0.02380000000000  -0.74070000000000   0.76315000000000   Cl   (6f)
   0.76315000000000  -0.02380000000000  -0.74070000000000   Cl   (6f)
   1.14609000000000   0.38403000000000  -0.59217000000000   Cl   (6f)
  -0.59217000000000   1.14609000000000   0.38403000000000   Cl   (6f)
   0.38403000000000  -0.59217000000000   1.14609000000000   Cl   (6f)
  -1.14609000000000  -0.38403000000000   0.59217000000000   Cl   (6f)
   0.59217000000000  -1.14609000000000  -0.38403000000000   Cl   (6f)
  -0.38403000000000   0.59217000000000  -1.14609000000000   Cl   (6f)
   0.08714000000000   0.06386000000000   0.32630000000000   Pd   (6f)
   0.32630000000000   0.08714000000000   0.06386000000000   Pd   (6f)
   0.06386000000000   0.32630000000000   0.08714000000000   Pd   (6f)
  -0.08714000000000  -0.06386000000000  -0.32630000000000   Pd   (6f)
  -0.32630000000000  -0.08714000000000  -0.06386000000000   Pd   (6f)
  -0.06386000000000  -0.32630000000000  -0.08714000000000   Pd   (6f)
\end{lstlisting}
{\phantomsection\label{AB3_hP24_149_acgi_3l_cif}}
{\hyperref[AB3_hP24_149_acgi_3l]{Ti$_{3}$O (Room-temperature): AB3\_hP24\_149\_acgi\_3l}} - CIF
\begin{lstlisting}[numbers=none,language={mylang}]
# CIF file
data_findsym-output
_audit_creation_method FINDSYM

_chemical_name_mineral 'Ti3O'
_chemical_formula_sum 'O Ti3'

loop_
_publ_author_name
 'A. Jostsons'
 'A. S. Malin'
_journal_name_full_name
;
 Acta Crystallographica Section B: Structural Science
;
_journal_volume 24
_journal_year 1968
_journal_page_first 211
_journal_page_last 213
_publ_Section_title
;
 The ordered structure of Ti$_{3}$O
;

# Found in Pearson's Crystal Data - Crystal Structure Database for Inorganic Compounds, 2013

_aflow_title 'Ti$_{3}$O (Room-temperature) Structure'
_aflow_proto 'AB3_hP24_149_acgi_3l'
_aflow_params 'a,c/a,z_{3},z_{4},x_{5},y_{5},z_{5},x_{6},y_{6},z_{6},x_{7},y_{7},z_{7}'
_aflow_params_values '5.1418156677,2.78268310709,0.33333,0.33333,0.0,0.33333,0.421,0.33333,0.33333,0.246,0.0,0.33333,0.088'
_aflow_Strukturbericht 'None'
_aflow_Pearson 'hP24'

_cell_length_a    5.1418156677
_cell_length_b    5.1418156677
_cell_length_c    14.3080435983
_cell_angle_alpha 90.0000000000
_cell_angle_beta  90.0000000000
_cell_angle_gamma 120.0000000000
 
_symmetry_space_group_name_H-M "P 3 1 2"
_symmetry_Int_Tables_number 149
 
loop_
_space_group_symop_id
_space_group_symop_operation_xyz
1 x,y,z
2 -y,x-y,z
3 -x+y,-x,z
4 x,x-y,-z
5 -x+y,y,-z
6 -y,-x,-z
 
loop_
_atom_site_label
_atom_site_type_symbol
_atom_site_symmetry_multiplicity
_atom_site_Wyckoff_label
_atom_site_fract_x
_atom_site_fract_y
_atom_site_fract_z
_atom_site_occupancy
O1  O    1 a 0.00000 0.00000 0.00000 1.00000
O2  O    1 c 0.33333 0.66667 0.00000 1.00000
O3  O    2 g 0.00000 0.00000 0.33333 1.00000
O4  O    2 i 0.66667 0.33333 0.33333 1.00000
Ti1 Ti   6 l 0.00000 0.33333 0.42100 1.00000
Ti2 Ti   6 l 0.33333 0.33333 0.24600 1.00000
Ti3 Ti   6 l 0.00000 0.33333 0.08800 1.00000
\end{lstlisting}
{\phantomsection\label{AB3_hP24_149_acgi_3l_poscar}}
{\hyperref[AB3_hP24_149_acgi_3l]{Ti$_{3}$O (Room-temperature): AB3\_hP24\_149\_acgi\_3l}} - POSCAR

{\phantomsection\label{A3B_hP24_153_3c_2b_cif}}
{\hyperref[A3B_hP24_153_3c_2b]{CrCl$_{3}$: A3B\_hP24\_153\_3c\_2b}} - CIF
\begin{lstlisting}[numbers=none,language={mylang}]
# CIF file
data_findsym-output
_audit_creation_method FINDSYM

_chemical_name_mineral 'CrCl3'
_chemical_formula_sum 'Cl3 Cr'

_aflow_title 'CrCl$_{3}$ Structure'
_aflow_proto 'A3B_hP24_153_3c_2b'
_aflow_params 'a,c/a,x_{1},x_{2},x_{3},y_{3},z_{3},x_{4},y_{4},z_{4},x_{5},y_{5},z_{5}'
_aflow_params_values '6.017,2.87518697025,0.1111,0.4444,0.1111,0.2222,0.09357,0.4444,0.8888,0.09357,0.77778,0.55558,0.09357'
_aflow_Strukturbericht 'None'
_aflow_Pearson 'hP24'

_cell_length_a    6.0170000000
_cell_length_b    6.0170000000
_cell_length_c    17.3000000000
_cell_angle_alpha 90.0000000000
_cell_angle_beta  90.0000000000
_cell_angle_gamma 120.0000000000
 
_symmetry_space_group_name_H-M "P 32 1 2"
_symmetry_Int_Tables_number 153
 
loop_
_space_group_symop_id
_space_group_symop_operation_xyz
1 x,y,z
2 -y,x-y,z+2/3
3 -x+y,-x,z+1/3
4 x,x-y,-z
5 -x+y,y,-z+2/3
6 -y,-x,-z+1/3
 
loop_
_atom_site_label
_atom_site_type_symbol
_atom_site_symmetry_multiplicity
_atom_site_Wyckoff_label
_atom_site_fract_x
_atom_site_fract_y
_atom_site_fract_z
_atom_site_occupancy
Cr1 Cr   3 b 0.11110 0.88890 0.16667 1.00000
Cr2 Cr   3 b 0.44440 0.55560 0.16667 1.00000
Cl1 Cl   6 c 0.11110 0.22220 0.09357 1.00000
Cl2 Cl   6 c 0.44440 0.88880 0.09357 1.00000
Cl3 Cl   6 c 0.77778 0.55558 0.09357 1.00000
\end{lstlisting}
{\phantomsection\label{A3B_hP24_153_3c_2b_poscar}}
{\hyperref[A3B_hP24_153_3c_2b]{CrCl$_{3}$: A3B\_hP24\_153\_3c\_2b}} - POSCAR

{\phantomsection\label{A_hP9_154_bc_cif}}
{\hyperref[A_hP9_154_bc]{S-II: A\_hP9\_154\_bc}} - CIF
\begin{lstlisting}[numbers=none,language={mylang}]
# CIF file 
data_findsym-output
_audit_creation_method FINDSYM

_chemical_name_mineral 'S-II'
_chemical_formula_sum 'S'

loop_
_publ_author_name
 'O. Degtyareva'
 'E. Gregoryanz'
 'M. Somayazulu'
 'P. Dera'
 'H. Mao'
 'R. J. Hemley'
_journal_name_full_name
;
 Nature Materials
;
_journal_volume 4
_journal_year 2005
_journal_page_first 152
_journal_page_last 155
_publ_Section_title
;
 Novel chain structures in group VI elements
;

_aflow_title 'S-II Structure'
_aflow_proto 'A_hP9_154_bc'
_aflow_params 'a,c/a,x_{1},x_{2},y_{2},z_{2}'
_aflow_params_values '6.9082,0.616557134999,0.876,0.23,0.534,0.051'
_aflow_Strukturbericht 'None'
_aflow_Pearson 'hP9'

_symmetry_space_group_name_H-M "P 32 2 1"
_symmetry_Int_Tables_number 154
 
_cell_length_a    6.90820
_cell_length_b    6.90820
_cell_length_c    4.25930
_cell_angle_alpha 90.00000
_cell_angle_beta  90.00000
_cell_angle_gamma 120.00000
 
loop_
_space_group_symop_id
_space_group_symop_operation_xyz
1 x,y,z
2 -y,x-y,z+2/3
3 -x+y,-x,z+1/3
4 x-y,-y,-z+1/3
5 y,x,-z
6 -x,-x+y,-z+2/3
 
loop_
_atom_site_label
_atom_site_type_symbol
_atom_site_symmetry_multiplicity
_atom_site_Wyckoff_label
_atom_site_fract_x
_atom_site_fract_y
_atom_site_fract_z
_atom_site_occupancy
S1 S   3 b 0.87600 0.00000 0.16667 1.00000
S2 S   6 c 0.23000 0.53400 0.05100 1.00000
\end{lstlisting}
{\phantomsection\label{A_hP9_154_bc_poscar}}
{\hyperref[A_hP9_154_bc]{S-II: A\_hP9\_154\_bc}} - POSCAR
\begin{lstlisting}[numbers=none,language={mylang}]
A_hP9_154_bc & a,c/a,x1,x2,y2,z2 --params=6.9082,0.616557134999,0.876,0.23,0.534,0.051 & P3_{2}21 D_{3}^{6} #154 (bc) & hP9 & None & S & S-II & O. Degtyareva et al., Nat. Mater. 4, 152-155 (2005)
   1.00000000000000
   3.45410000000000  -5.98267669442366   0.00000000000000
   3.45410000000000   5.98267669442366   0.00000000000000
   0.00000000000000   0.00000000000000   4.25930000000000
     S
     9
Direct
   0.87600000000000   0.00000000000000   0.16666666666667    S   (3b)
   0.00000000000000   0.87600000000000   0.83333333333333    S   (3b)
  -0.87600000000000  -0.87600000000000   0.50000000000000    S   (3b)
   0.23000000000000   0.53400000000000   0.05100000000000    S   (6c)
  -0.53400000000000  -0.30400000000000   0.71766666666667    S   (6c)
   0.30400000000000  -0.23000000000000   0.38433333333333    S   (6c)
   0.53400000000000   0.23000000000000  -0.05100000000000    S   (6c)
  -0.30400000000000  -0.53400000000000   0.28233333333333    S   (6c)
  -0.23000000000000   0.30400000000000   0.61566666666667    S   (6c)
\end{lstlisting}
{\phantomsection\label{AB2_hP9_156_b2c_3a2bc_cif}}
{\hyperref[AB2_hP9_156_b2c_3a2bc]{CdI$_{2}$ (Polytype 6H$_{1}$): AB2\_hP9\_156\_b2c\_3a2bc}} - CIF
\begin{lstlisting}[numbers=none,language={mylang}]
# CIF file
data_findsym-output
_audit_creation_method FINDSYM

_chemical_name_mineral 'CdI2'
_chemical_formula_sum 'Cd I2'

loop_
_publ_author_name
 'R. S. Mitchell'
_journal_name_full_name
;
 Zeitschrift f{\"u}r Kristallographie - Crystalline Materials
;
_journal_volume 108
_journal_year 1956
_journal_page_first 296
_journal_page_last 315
_publ_Section_title
;
 Polytypism of Cadmium Iodide and its Relationship to Screw Dislocations: I. Cadmium Iodide Polytypes
;

# Found in Pearson's Crystal Data - Crystal Structure Database for Inorganic Compounds, 2013

_aflow_title 'CdI$_{2}$ (Polytype 6H$_{1}$) Structure'
_aflow_proto 'AB2_hP9_156_b2c_3a2bc'
_aflow_params 'a,c/a,z_{1},z_{2},z_{3},z_{4},z_{5},z_{6},z_{7},z_{8},z_{9}'
_aflow_params_values '4.239807232,4.83608490569,0.0,0.66667,0.33333,0.75,0.16667,0.5,0.08333,0.41667,0.83333'
_aflow_Strukturbericht 'None'
_aflow_Pearson 'hP9'

_cell_length_a    4.2398072320
_cell_length_b    4.2398072320
_cell_length_c    20.5040677577
_cell_angle_alpha 90.0000000000
_cell_angle_beta  90.0000000000
_cell_angle_gamma 120.0000000000
 
_symmetry_space_group_name_H-M "P 3 m 1"
_symmetry_Int_Tables_number 156
 
loop_
_space_group_symop_id
_space_group_symop_operation_xyz
1 x,y,z
2 -y,x-y,z
3 -x+y,-x,z
4 -x+y,y,z
5 -y,-x,z
6 x,x-y,z
 
loop_
_atom_site_label
_atom_site_type_symbol
_atom_site_symmetry_multiplicity
_atom_site_Wyckoff_label
_atom_site_fract_x
_atom_site_fract_y
_atom_site_fract_z
_atom_site_occupancy
I1  I    1 a 0.00000 0.00000 0.00000 1.00000
I2  I    1 a 0.00000 0.00000 0.66667 1.00000
I3  I    1 a 0.00000 0.00000 0.33333 1.00000
Cd1 Cd   1 b 0.33333 0.66667 0.75000 1.00000
I4  I    1 b 0.33333 0.66667 0.16667 1.00000
I5  I    1 b 0.33333 0.66667 0.50000 1.00000
Cd2 Cd   1 c 0.66667 0.33333 0.08333 1.00000
Cd3 Cd   1 c 0.66667 0.33333 0.41667 1.00000
I6  I    1 c 0.66667 0.33333 0.83333 1.00000
\end{lstlisting}
{\phantomsection\label{AB2_hP9_156_b2c_3a2bc_poscar}}
{\hyperref[AB2_hP9_156_b2c_3a2bc]{CdI$_{2}$ (Polytype 6H$_{1}$): AB2\_hP9\_156\_b2c\_3a2bc}} - POSCAR
\begin{lstlisting}[numbers=none,language={mylang}]
AB2_hP9_156_b2c_3a2bc & a,c/a,z1,z2,z3,z4,z5,z6,z7,z8,z9 --params=4.239807232,4.83608490569,0.0,0.66667,0.33333,0.75,0.16667,0.5,0.08333,0.41667,0.83333 & P3m1 C_{3v}^{1} #156 (a^3b^3c^3) & hP9 & None & CdI2 &  & R. S. Mitchell, Zeitschrift f"{u}r Kristallographie - Crystalline Materials 108, 296-315 (1956)
   1.00000000000000
   2.11990361600000  -3.67178077006098   0.00000000000000
   2.11990361600000   3.67178077006098   0.00000000000000
   0.00000000000000   0.00000000000000  20.50406775770000
    Cd     I
     3     6
Direct
   0.33333333333333   0.66666666666667   0.75000000000000   Cd   (1b)
   0.66666666666667   0.33333333333333   0.08333000000000   Cd   (1c)
   0.66666666666667   0.33333333333333   0.41667000000000   Cd   (1c)
   0.00000000000000   0.00000000000000   0.00000000000000    I   (1a)
   0.00000000000000   0.00000000000000   0.66667000000000    I   (1a)
   0.00000000000000   0.00000000000000   0.33333000000000    I   (1a)
   0.33333333333333   0.66666666666667   0.16667000000000    I   (1b)
   0.33333333333333   0.66666666666667   0.50000000000000    I   (1b)
   0.66666666666667   0.33333333333333   0.83333000000000    I   (1c)
\end{lstlisting}
{\phantomsection\label{AB_hP12_156_2ab3c_2ab3c_cif}}
{\hyperref[AB_hP12_156_2ab3c_2ab3c]{CuI: AB\_hP12\_156\_2ab3c\_2ab3c}} - CIF

{\phantomsection\label{AB_hP12_156_2ab3c_2ab3c_poscar}}
{\hyperref[AB_hP12_156_2ab3c_2ab3c]{CuI: AB\_hP12\_156\_2ab3c\_2ab3c}} - POSCAR
\begin{lstlisting}[numbers=none,language={mylang}]
AB_hP12_156_2ab3c_2ab3c & a,c/a,z1,z2,z3,z4,z5,z6,z7,z8,z9,z10,z11,z12 --params=4.2499813346,4.90823529409,0.375,0.70833,0.5,0.83333,0.04167,0.16667,0.45833,0.79167,0.125,0.33333,0.66667,0.0 & P3m1 C_{3v}^{1} #156 (a^4b^2c^6) & hP12 & None & CuI &  & R. N. Kurdyumova and R. V. Baranova, Sov. Phys. Crystallogr. 6, 318-321 (1961)
   1.00000000000000
   2.12499066730000  -3.68059180137329   0.00000000000000
   2.12499066730000   3.68059180137329   0.00000000000000
   0.00000000000000   0.00000000000000  20.85990838570000
    Cu     I
     6     6
Direct
   0.00000000000000   0.00000000000000   0.37500000000000   Cu   (1a)
   0.00000000000000   0.00000000000000   0.70833000000000   Cu   (1a)
   0.33333333333333   0.66666666666667   0.04167000000000   Cu   (1b)
   0.66666666666667   0.33333333333333   0.45833000000000   Cu   (1c)
   0.66666666666667   0.33333333333333   0.79167000000000   Cu   (1c)
   0.66666666666667   0.33333333333333   0.12500000000000   Cu   (1c)
   0.00000000000000   0.00000000000000   0.50000000000000    I   (1a)
   0.00000000000000   0.00000000000000   0.83333000000000    I   (1a)
   0.33333333333333   0.66666666666667   0.16667000000000    I   (1b)
   0.66666666666667   0.33333333333333   0.33333000000000    I   (1c)
   0.66666666666667   0.33333333333333   0.66667000000000    I   (1c)
   0.66666666666667   0.33333333333333   0.00000000000000    I   (1c)
\end{lstlisting}
{\phantomsection\label{AB_hP4_156_ac_ac_cif}}
{\hyperref[AB_hP4_156_ac_ac]{$\beta$-CuI: AB\_hP4\_156\_ac\_ac}} - CIF
\begin{lstlisting}[numbers=none,language={mylang}]
# CIF file
data_findsym-output
_audit_creation_method FINDSYM

_chemical_name_mineral 'beta-CuI'
_chemical_formula_sum 'Cu I'

loop_
_publ_author_name
 'T. Sakuma'
_journal_name_full_name
;
 Journal of the Physical Society of Japan
;
_journal_volume 57
_journal_year 1988
_journal_page_first 565
_journal_page_last 569
_publ_Section_title
;
 Crystal structure of $\beta$-CuI
;

# Found in Pearson's Crystal Data - Crystal Structure Database for Inorganic Compounds, 2013

_aflow_title '$\beta$-CuI Structure'
_aflow_proto 'AB_hP4_156_ac_ac'
_aflow_params 'a,c/a,z_{1},z_{2},z_{3},z_{4}'
_aflow_params_values '4.2794836776,1.67515774714,0.104,0.5,0.364,0.0'
_aflow_Strukturbericht 'None'
_aflow_Pearson 'hP4'

_cell_length_a    4.2794836776
_cell_length_b    4.2794836776
_cell_length_c    7.1688102363
_cell_angle_alpha 90.0000000000
_cell_angle_beta  90.0000000000
_cell_angle_gamma 120.0000000000
 
_symmetry_space_group_name_H-M "P 3 m 1"
_symmetry_Int_Tables_number 156
 
loop_
_space_group_symop_id
_space_group_symop_operation_xyz
1 x,y,z
2 -y,x-y,z
3 -x+y,-x,z
4 -x+y,y,z
5 -y,-x,z
6 x,x-y,z
 
loop_
_atom_site_label
_atom_site_type_symbol
_atom_site_symmetry_multiplicity
_atom_site_Wyckoff_label
_atom_site_fract_x
_atom_site_fract_y
_atom_site_fract_z
_atom_site_occupancy
Cu1 Cu   1 a 0.00000 0.00000 0.10400 1.00000
I1  I    1 a 0.00000 0.00000 0.50000 1.00000
Cu2 Cu   1 c 0.66667 0.33333 0.36400 1.00000
I2  I    1 c 0.66667 0.33333 0.00000 1.00000
\end{lstlisting}
{\phantomsection\label{AB_hP4_156_ac_ac_poscar}}
{\hyperref[AB_hP4_156_ac_ac]{$\beta$-CuI: AB\_hP4\_156\_ac\_ac}} - POSCAR
\begin{lstlisting}[numbers=none,language={mylang}]
AB_hP4_156_ac_ac & a,c/a,z1,z2,z3,z4 --params=4.2794836776,1.67515774714,0.104,0.5,0.364,0.0 & P3m1 C_{3v}^{1} #156 (a^2c^2) & hP4 & None & CuI & beta & T. Sakuma, J. Phys. Soc. Jpn. 57, 565-569 (1988)
   1.00000000000000
   2.13974183880000  -3.70614157988245   0.00000000000000
   2.13974183880000   3.70614157988245   0.00000000000000
   0.00000000000000   0.00000000000000   7.16881023630000
    Cu     I
     2     2
Direct
   0.00000000000000   0.00000000000000   0.10400000000000   Cu   (1a)
   0.66666666666667   0.33333333333333   0.36400000000000   Cu   (1c)
   0.00000000000000   0.00000000000000   0.50000000000000    I   (1a)
   0.66666666666667   0.33333333333333   0.00000000000000    I   (1c)
\end{lstlisting}
{\phantomsection\label{A5B6C2_hP13_157_2ac_2c_b_cif}}
{\hyperref[A5B6C2_hP13_157_2ac_2c_b]{Ag$_{5}$Pb$_{2}$O$_{6}$: A5B6C2\_hP13\_157\_2ac\_2c\_b}} - CIF
\begin{lstlisting}[numbers=none,language={mylang}]
# CIF file
data_findsym-output
_audit_creation_method FINDSYM

_chemical_name_mineral 'Ag5Pb2O6'
_chemical_formula_sum 'Ag5 O6 Pb2'

_aflow_title 'Ag$_{5}$Pb$_{2}$O$_{6}$ Structure'
_aflow_proto 'A5B6C2_hP13_157_2ac_2c_b'
_aflow_params 'a,c/a,z_{1},z_{2},z_{3},x_{4},z_{4},x_{5},z_{5},x_{6},z_{6}'
_aflow_params_values '5.939,1.08233709379,0.264,0.736,0.022,0.5,0.522,0.603,0.215,0.397,0.829'
_aflow_Strukturbericht 'None'
_aflow_Pearson 'hP13'

_cell_length_a    5.9390000000
_cell_length_b    5.9390000000
_cell_length_c    6.4280000000
_cell_angle_alpha 90.0000000000
_cell_angle_beta  90.0000000000
_cell_angle_gamma 120.0000000000
 
_symmetry_space_group_name_H-M "P 3 1 m"
_symmetry_Int_Tables_number 157
 
loop_
_space_group_symop_id
_space_group_symop_operation_xyz
1 x,y,z
2 -y,x-y,z
3 -x+y,-x,z
4 -x,-x+y,z
5 x-y,-y,z
6 y,x,z
 
loop_
_atom_site_label
_atom_site_type_symbol
_atom_site_symmetry_multiplicity
_atom_site_Wyckoff_label
_atom_site_fract_x
_atom_site_fract_y
_atom_site_fract_z
_atom_site_occupancy
Ag1 Ag   1 a 0.00000 0.00000 0.26400 1.00000
Ag2 Ag   1 a 0.00000 0.00000 0.73600 1.00000
Pb1 Pb   2 b 0.33333 0.66667 0.02200 1.00000
Ag3 Ag   3 c 0.50000 0.00000 0.52200 1.00000
O1  O    3 c 0.60300 0.00000 0.21500 1.00000
O2  O    3 c 0.39700 0.00000 0.82900 1.00000
\end{lstlisting}
{\phantomsection\label{A5B6C2_hP13_157_2ac_2c_b_poscar}}
{\hyperref[A5B6C2_hP13_157_2ac_2c_b]{Ag$_{5}$Pb$_{2}$O$_{6}$: A5B6C2\_hP13\_157\_2ac\_2c\_b}} - POSCAR
\begin{lstlisting}[numbers=none,language={mylang}]
A5B6C2_hP13_157_2ac_2c_b & a,c/a,z1,z2,z3,x4,z4,x5,z5,x6,z6 --params=5.939,1.08233709379,0.264,0.736,0.022,0.5,0.522,0.603,0.215,0.397,0.829 & P31m C_{3v}^{2} #157 (a^2bc^3) & hP13 & None & Ag5Pb2O6 &  & 
   1.00000000000000
   2.96950000000000  -5.14332487307578   0.00000000000000
   2.96950000000000   5.14332487307578   0.00000000000000
   0.00000000000000   0.00000000000000   6.42800000000000
    Ag     O    Pb
     5     6     2
Direct
   0.00000000000000   0.00000000000000   0.26400000000000   Ag   (1a)
   0.00000000000000   0.00000000000000   0.73600000000000   Ag   (1a)
   0.50000000000000   0.00000000000000   0.52200000000000   Ag   (3c)
   0.00000000000000   0.50000000000000   0.52200000000000   Ag   (3c)
  -0.50000000000000  -0.50000000000000   0.52200000000000   Ag   (3c)
   0.60300000000000   0.00000000000000   0.21500000000000    O   (3c)
   0.00000000000000   0.60300000000000   0.21500000000000    O   (3c)
  -0.60300000000000  -0.60300000000000   0.21500000000000    O   (3c)
   0.39700000000000   0.00000000000000   0.82900000000000    O   (3c)
   0.00000000000000   0.39700000000000   0.82900000000000    O   (3c)
  -0.39700000000000  -0.39700000000000   0.82900000000000    O   (3c)
   0.33333333333333   0.66666666666667   0.02200000000000   Pb   (2b)
   0.66666666666667   0.33333333333333   0.02200000000000   Pb   (2b)
\end{lstlisting}
{\phantomsection\label{A3B_hP8_158_d_a_cif}}
{\hyperref[A3B_hP8_158_d_a]{$\beta$-RuCl$_{3}$: A3B\_hP8\_158\_d\_a}} - CIF
\begin{lstlisting}[numbers=none,language={mylang}]
# CIF file
data_findsym-output
_audit_creation_method FINDSYM

_chemical_name_mineral 'beta-RuCl3'
_chemical_formula_sum 'Cl3 Ru'

_aflow_title '$\beta$-RuCl$_{3}$ Structure'
_aflow_proto 'A3B_hP8_158_d_a'
_aflow_params 'a,c/a,z_{1},x_{2},y_{2},z_{2}'
_aflow_params_values '6.12,0.924509803922,0.0,0.318,0.027,0.237'
_aflow_Strukturbericht 'None'
_aflow_Pearson 'hP8'

_cell_length_a    6.1200000000
_cell_length_b    6.1200000000
_cell_length_c    5.6580000000
_cell_angle_alpha 90.0000000000
_cell_angle_beta  90.0000000000
_cell_angle_gamma 120.0000000000
 
_symmetry_space_group_name_H-M "P 3 c 1"
_symmetry_Int_Tables_number 158
 
loop_
_space_group_symop_id
_space_group_symop_operation_xyz
1 x,y,z
2 -y,x-y,z
3 -x+y,-x,z
4 -x+y,y,z+1/2
5 -y,-x,z+1/2
6 x,x-y,z+1/2
 
loop_
_atom_site_label
_atom_site_type_symbol
_atom_site_symmetry_multiplicity
_atom_site_Wyckoff_label
_atom_site_fract_x
_atom_site_fract_y
_atom_site_fract_z
_atom_site_occupancy
Ru1 Ru   2 a 0.00000 0.00000 0.00000 1.00000
Cl1 Cl   6 d 0.31800 0.02700 0.23700 1.00000
\end{lstlisting}
{\phantomsection\label{A3B_hP8_158_d_a_poscar}}
{\hyperref[A3B_hP8_158_d_a]{$\beta$-RuCl$_{3}$: A3B\_hP8\_158\_d\_a}} - POSCAR
\begin{lstlisting}[numbers=none,language={mylang}]
A3B_hP8_158_d_a & a,c/a,z1,x2,y2,z2 --params=6.12,0.924509803922,0.0,0.318,0.027,0.237 & P3c1 C_{3v}^{3} #158 (ad) & hP8 & None & RuCl3 & beta & 
   1.00000000000000
   3.06000000000000  -5.30007547116076   0.00000000000000
   3.06000000000000   5.30007547116076   0.00000000000000
   0.00000000000000   0.00000000000000   5.65800000000000
    Cl    Ru
     6     2
Direct
   0.31800000000000   0.02700000000000   0.23700000000000   Cl   (6d)
  -0.02700000000000   0.29100000000000   0.23700000000000   Cl   (6d)
  -0.29100000000000  -0.31800000000000   0.23700000000000   Cl   (6d)
  -0.02700000000000  -0.31800000000000   0.73700000000000   Cl   (6d)
  -0.29100000000000   0.02700000000000   0.73700000000000   Cl   (6d)
   0.31800000000000   0.29100000000000   0.73700000000000   Cl   (6d)
   0.00000000000000   0.00000000000000   0.00000000000000   Ru   (2a)
   0.00000000000000   0.00000000000000   0.50000000000000   Ru   (2a)
\end{lstlisting}
{\phantomsection\label{A2B3_hP20_159_bc_2c_cif}}
{\hyperref[A2B3_hP20_159_bc_2c]{Bi$_{2}$O$_{3}$ (High-pressure): A2B3\_hP20\_159\_bc\_2c}} - CIF
\begin{lstlisting}[numbers=none,language={mylang}]
# CIF file
data_findsym-output
_audit_creation_method FINDSYM

_chemical_name_mineral 'Bi2O3'
_chemical_formula_sum 'Bi2 O3'

loop_
_publ_author_name
 'T. Locherer'
 'D. L. V. K. Prasad'
 'R. Dinnebier'
 'U. Wedig'
 'M. Jansen'
 'G. Garbarino'
 'T. Hansen'
_journal_name_full_name
;
 Physical Review B
;
_journal_volume 83
_journal_year 2011
_journal_page_first 214102
_journal_page_last 214102
_publ_Section_title
;
 High-pressure structural evolution of HP-Bi$_{2}$O$_{3}$
;

# Found in Pearson's Crystal Data - Crystal Structure Database for Inorganic Compounds, 2013

_aflow_title 'Bi$_{2}$O$_{3}$ (High-pressure) Structure'
_aflow_proto 'A2B3_hP20_159_bc_2c'
_aflow_params 'a,c/a,z_{1},x_{2},y_{2},z_{2},x_{3},y_{3},z_{3},x_{4},y_{4},z_{4}'
_aflow_params_values '7.7488577892,0.813266227893,0.0,0.337,0.154,0.021,0.451,0.061,0.287,0.149,0.282,0.083'
_aflow_Strukturbericht 'None'
_aflow_Pearson 'hP20'

_cell_length_a    7.7488577892
_cell_length_b    7.7488577892
_cell_length_c    6.3018843447
_cell_angle_alpha 90.0000000000
_cell_angle_beta  90.0000000000
_cell_angle_gamma 120.0000000000
 
_symmetry_space_group_name_H-M "P 3 1 c"
_symmetry_Int_Tables_number 159
 
loop_
_space_group_symop_id
_space_group_symop_operation_xyz
1 x,y,z
2 -y,x-y,z
3 -x+y,-x,z
4 -x,-x+y,z+1/2
5 x-y,-y,z+1/2
6 y,x,z+1/2
 
loop_
_atom_site_label
_atom_site_type_symbol
_atom_site_symmetry_multiplicity
_atom_site_Wyckoff_label
_atom_site_fract_x
_atom_site_fract_y
_atom_site_fract_z
_atom_site_occupancy
Bi1 Bi   2 b 0.33333 0.66667 0.00000 1.00000
Bi2 Bi   6 c 0.33700 0.15400 0.02100 1.00000
O1  O    6 c 0.45100 0.06100 0.28700 1.00000
O2  O    6 c 0.14900 0.28200 0.08300 1.00000
\end{lstlisting}
{\phantomsection\label{A2B3_hP20_159_bc_2c_poscar}}
{\hyperref[A2B3_hP20_159_bc_2c]{Bi$_{2}$O$_{3}$ (High-pressure): A2B3\_hP20\_159\_bc\_2c}} - POSCAR
\begin{lstlisting}[numbers=none,language={mylang}]
A2B3_hP20_159_bc_2c & a,c/a,z1,x2,y2,z2,x3,y3,z3,x4,y4,z4 --params=7.7488577892,0.813266227893,0.0,0.337,0.154,0.021,0.451,0.061,0.287,0.149,0.282,0.083 & P31c C_{3v}^{4} #159 (bc^3) & hP20 & None & Bi2O3 &  & T. Locherer et al., Phys. Rev. B 83, 214102(2011)
   1.00000000000000
   3.87442889460000  -6.71070769576012   0.00000000000000
   3.87442889460000   6.71070769576012   0.00000000000000
   0.00000000000000   0.00000000000000   6.30188434470000
    Bi     O
     8    12
Direct
   0.33333333333333   0.66666666666667   0.00000000000000   Bi   (2b)
   0.66666666666667   0.33333333333333   0.50000000000000   Bi   (2b)
   0.33700000000000   0.15400000000000   0.02100000000000   Bi   (6c)
  -0.15400000000000   0.18300000000000   0.02100000000000   Bi   (6c)
  -0.18300000000000  -0.33700000000000   0.02100000000000   Bi   (6c)
   0.15400000000000   0.33700000000000   0.52100000000000   Bi   (6c)
   0.18300000000000  -0.15400000000000   0.52100000000000   Bi   (6c)
  -0.33700000000000  -0.18300000000000   0.52100000000000   Bi   (6c)
   0.45100000000000   0.06100000000000   0.28700000000000    O   (6c)
  -0.06100000000000   0.39000000000000   0.28700000000000    O   (6c)
  -0.39000000000000  -0.45100000000000   0.28700000000000    O   (6c)
   0.06100000000000   0.45100000000000   0.78700000000000    O   (6c)
   0.39000000000000  -0.06100000000000   0.78700000000000    O   (6c)
  -0.45100000000000  -0.39000000000000   0.78700000000000    O   (6c)
   0.14900000000000   0.28200000000000   0.08300000000000    O   (6c)
  -0.28200000000000  -0.13300000000000   0.08300000000000    O   (6c)
   0.13300000000000  -0.14900000000000   0.08300000000000    O   (6c)
   0.28200000000000   0.14900000000000   0.58300000000000    O   (6c)
  -0.13300000000000  -0.28200000000000   0.58300000000000    O   (6c)
  -0.14900000000000   0.13300000000000   0.58300000000000    O   (6c)
\end{lstlisting}
{\phantomsection\label{A4B3_hP28_159_ab2c_2c_cif}}
{\hyperref[A4B3_hP28_159_ab2c_2c]{Nierite ($\alpha$-Si$_{3}$N$_{4}$): A4B3\_hP28\_159\_ab2c\_2c}} - CIF
\begin{lstlisting}[numbers=none,language={mylang}]
# CIF file
data_findsym-output
_audit_creation_method FINDSYM

_chemical_name_mineral 'alpha-Si3N4'
_chemical_formula_sum 'N4 Si3'

loop_
_publ_author_name
 'D. Hardie'
 'K. H. Jack'
_journal_name_full_name
;
 Nature
;
_journal_volume 180
_journal_year 1957
_journal_page_first 332
_journal_page_last 333
_publ_Section_title
;
 Crystal structures of silicon nitride
;

# Found in Pearson's Crystal Data - Crystal Structure Database for Inorganic Compounds, 2013

_aflow_title 'Nierite ($\alpha$-Si$_{3}$N$_{4}$) Structure'
_aflow_proto 'A4B3_hP28_159_ab2c_2c'
_aflow_params 'a,c/a,z_{1},z_{2},x_{3},y_{3},z_{3},x_{4},y_{4},z_{4},x_{5},y_{5},z_{5},x_{6},y_{6},z_{6}'
_aflow_params_values '7.7479899913,0.724961280333,0.0,0.3649,0.0424,0.3891,0.0408,0.3169,0.3198,0.2712,0.0821,0.5089,0.3172,0.1712,0.2563,0.0274'
_aflow_Strukturbericht 'None'
_aflow_Pearson 'hP28'

_cell_length_a    7.7479899913
_cell_length_b    7.7479899913
_cell_length_c    5.6169927441
_cell_angle_alpha 90.0000000000
_cell_angle_beta  90.0000000000
_cell_angle_gamma 120.0000000000
 
_symmetry_space_group_name_H-M "P 3 1 c"
_symmetry_Int_Tables_number 159
 
loop_
_space_group_symop_id
_space_group_symop_operation_xyz
1 x,y,z
2 -y,x-y,z
3 -x+y,-x,z
4 -x,-x+y,z+1/2
5 x-y,-y,z+1/2
6 y,x,z+1/2
 
loop_
_atom_site_label
_atom_site_type_symbol
_atom_site_symmetry_multiplicity
_atom_site_Wyckoff_label
_atom_site_fract_x
_atom_site_fract_y
_atom_site_fract_z
_atom_site_occupancy
N1  N    2 a 0.00000 0.00000 0.00000 1.00000
N2  N    2 b 0.33333 0.66667 0.36490 1.00000
N3  N    6 c 0.04240 0.38910 0.04080 1.00000
N4  N    6 c 0.31690 0.31980 0.27120 1.00000
Si1 Si   6 c 0.08210 0.50890 0.31720 1.00000
Si2 Si   6 c 0.17120 0.25630 0.02740 1.00000
\end{lstlisting}
{\phantomsection\label{A4B3_hP28_159_ab2c_2c_poscar}}
{\hyperref[A4B3_hP28_159_ab2c_2c]{Nierite ($\alpha$-Si$_{3}$N$_{4}$): A4B3\_hP28\_159\_ab2c\_2c}} - POSCAR

{\phantomsection\label{AB4C7D_hP26_159_b_ac_a2c_b_cif}}
{\hyperref[AB4C7D_hP26_159_b_ac_a2c_b]{YbBaCo$_{4}$O$_{7}$: AB4C7D\_hP26\_159\_b\_ac\_a2c\_b}} - CIF
\begin{lstlisting}[numbers=none,language={mylang}]
# CIF file
data_findsym-output
_audit_creation_method FINDSYM

_chemical_name_mineral 'YbBaCo4O7'
_chemical_formula_sum 'Ba Co4 O7 Y'

loop_
_publ_author_name
 'A. Huq'
 'J. F. Mitchell'
 'H. Zheng'
 'L. C. Chapon'
 'P. G. Radaelli'
 'K. S. Knight'
 'P. W. Stephens'
_journal_name_full_name
;
 Journal of Solid State Chemistry
;
_journal_volume 179
_journal_year 2006
_journal_page_first 1136
_journal_page_last 1145
_publ_Section_title
;
 Structural and magnetic properties of the Kagom{\\'e} antiferromagnet YbBaCo$_{4}$O$_{7}$
;

# Found in Pearson's Crystal Data - Crystal Structure Database for Inorganic Compounds, 2013

_aflow_title 'YbBaCo$_{4}$O$_{7}$ Structure'
_aflow_proto 'AB4C7D_hP26_159_b_ac_a2c_b'
_aflow_params 'a,c/a,z_{1},z_{2},z_{3},z_{4},x_{5},y_{5},z_{5},x_{6},y_{6},z_{6},x_{7},y_{7},z_{7}'
_aflow_params_values '6.2653109789,1.6324735851,0.0,0.3057,0.0613,0.4379,0.3425,0.1575,0.2472,0.0015,0.5149,0.3102,0.3347,0.1157,0.0618'
_aflow_Strukturbericht 'None'
_aflow_Pearson 'hP26'

_cell_length_a    6.2653109789
_cell_length_b    6.2653109789
_cell_length_c    10.2279546755
_cell_angle_alpha 90.0000000000
_cell_angle_beta  90.0000000000
_cell_angle_gamma 120.0000000000
 
_symmetry_space_group_name_H-M "P 3 1 c"
_symmetry_Int_Tables_number 159
 
loop_
_space_group_symop_id
_space_group_symop_operation_xyz
1 x,y,z
2 -y,x-y,z
3 -x+y,-x,z
4 -x,-x+y,z+1/2
5 x-y,-y,z+1/2
6 y,x,z+1/2
 
loop_
_atom_site_label
_atom_site_type_symbol
_atom_site_symmetry_multiplicity
_atom_site_Wyckoff_label
_atom_site_fract_x
_atom_site_fract_y
_atom_site_fract_z
_atom_site_occupancy
Co1 Co   2 a 0.00000 0.00000 0.00000 1.00000
O1  O    2 a 0.00000 0.00000 0.30570 1.00000
Ba1 Ba   2 b 0.33333 0.66667 0.06130 1.00000
Y1  Y    2 b 0.33333 0.66667 0.43790 1.00000
Co2 Co   6 c 0.34250 0.15750 0.24720 1.00000
O2  O    6 c 0.00150 0.51490 0.31020 1.00000
O3  O    6 c 0.33470 0.11570 0.06180 1.00000
\end{lstlisting}
{\phantomsection\label{AB4C7D_hP26_159_b_ac_a2c_b_poscar}}
{\hyperref[AB4C7D_hP26_159_b_ac_a2c_b]{YbBaCo$_{4}$O$_{7}$: AB4C7D\_hP26\_159\_b\_ac\_a2c\_b}} - POSCAR

{\phantomsection\label{A3B_hR4_160_b_a_cif}}
{\hyperref[A3B_hR4_160_b_a]{H$_{3}$S (130~GPa): A3B\_hR4\_160\_b\_a}} - CIF
\begin{lstlisting}[numbers=none,language={mylang}]
# CIF file 
data_findsym-output
_audit_creation_method FINDSYM

_chemical_name_mineral 'H3S'
_chemical_formula_sum 'H3 S'

loop_
_publ_author_name
 'D. Duan'
 'Y. Liu'
 'F. Tian'
 'D. Li'
 'X. Huang'
 'Z. Zhao'
 'H. Yu'
 'B. Liu'
 'W. Tian'
 'T. Cui'
_journal_name_full_name
;
 Scientific Reports
;
_journal_volume 4
_journal_year 2014
_journal_page_first 6968
_journal_page_last 6968
_publ_Section_title
;
 Pressure-induced metallization of dense (H$_2$S)$_2$H$_2$ with high-T$_c$ superconductivity
;

_aflow_title 'H$_{3}$S (130~GPa) Structure'
_aflow_proto 'A3B_hR4_160_b_a'
_aflow_params 'a,c/a,x_{1},x_{2},z_{2}'
_aflow_params_values '4.405,0.610442678774,0.0,0.52073,-0.02217'
_aflow_Strukturbericht 'None'
_aflow_Pearson 'hR4'

_symmetry_space_group_name_H-M "R 3 m (hexagonal axes)"
_symmetry_Int_Tables_number 160
 
_cell_length_a    4.40500
_cell_length_b    4.40500
_cell_length_c    2.68900
_cell_angle_alpha 90.00000
_cell_angle_beta  90.00000
_cell_angle_gamma 120.00000
 
loop_
_space_group_symop_id
_space_group_symop_operation_xyz
1 x,y,z
2 -y,x-y,z
3 -x+y,-x,z
4 -y,-x,z
5 x,x-y,z
6 -x+y,y,z
7 x+1/3,y+2/3,z+2/3
8 -y+1/3,x-y+2/3,z+2/3
9 -x+y+1/3,-x+2/3,z+2/3
10 -y+1/3,-x+2/3,z+2/3
11 x+1/3,x-y+2/3,z+2/3
12 -x+y+1/3,y+2/3,z+2/3
13 x+2/3,y+1/3,z+1/3
14 -y+2/3,x-y+1/3,z+1/3
15 -x+y+2/3,-x+1/3,z+1/3
16 -y+2/3,-x+1/3,z+1/3
17 x+2/3,x-y+1/3,z+1/3
18 -x+y+2/3,y+1/3,z+1/3
 
loop_
_atom_site_label
_atom_site_type_symbol
_atom_site_symmetry_multiplicity
_atom_site_Wyckoff_label
_atom_site_fract_x
_atom_site_fract_y
_atom_site_fract_z
_atom_site_occupancy
S1 S   3 a 0.00000 0.00000 0.00000 1.00000
H1 H   9 b 0.51430 0.48570 0.00643 1.00000
\end{lstlisting}
{\phantomsection\label{A3B_hR4_160_b_a_poscar}}
{\hyperref[A3B_hR4_160_b_a]{H$_{3}$S (130~GPa): A3B\_hR4\_160\_b\_a}} - POSCAR
\begin{lstlisting}[numbers=none,language={mylang}]
A3B_hR4_160_b_a & a,c/a,x1,x2,z2 --params=4.405,0.610442678774,0.0,0.52073,-0.02217 & R3m C_{3v}^{5} #160 (ab) & hR4 & None & H3S & H3S & D. Duan et al., Sci. Rep. 4, 6968(2014)
   1.00000000000000
   2.20250000000000  -1.27161396789015   0.89633333333333
   0.00000000000000   2.54322793578030   0.89633333333333
  -2.20250000000000  -1.27161396789015   0.89633333333333
     H     S
     3     1
Direct
   0.52073000000000   0.52073000000000  -0.02217000000000    H   (3b)
  -0.02217000000000   0.52073000000000   0.52073000000000    H   (3b)
   0.52073000000000  -0.02217000000000   0.52073000000000    H   (3b)
   0.00000000000000   0.00000000000000   0.00000000000000    S   (1a)
\end{lstlisting}
{\phantomsection\label{A8B5_hR26_160_a3bc_a3b_cif}}
{\hyperref[A8B5_hR26_160_a3bc_a3b]{Al$_{8}$Cr$_{5}$ ($D8_{10}$): A8B5\_hR26\_160\_a3bc\_a3b}} - CIF

{\phantomsection\label{A8B5_hR26_160_a3bc_a3b_poscar}}
{\hyperref[A8B5_hR26_160_a3bc_a3b]{Al$_{8}$Cr$_{5}$ ($D8_{10}$): A8B5\_hR26\_160\_a3bc\_a3b}} - POSCAR

{\phantomsection\label{ABC_hR3_160_a_a_a_cif}}
{\hyperref[ABC_hR3_160_a_a_a]{Carbonyl Sulphide (COS, $F0_{2}$): ABC\_hR3\_160\_a\_a\_a}} - CIF
\begin{lstlisting}[numbers=none,language={mylang}]
# CIF file
data_findsym-output
_audit_creation_method FINDSYM

_chemical_name_mineral 'Carbonyl Sulphide'
_chemical_formula_sum 'C O S'

loop_
_publ_author_name
 'J. S. W. Overell'
 'G. S. Pawley'
 'B. M. Powell'
_journal_name_full_name
;
 Acta Crystallographica Section B: Structural Science
;
_journal_volume 38
_journal_year 1982
_journal_page_first 1121
_journal_page_last 1123
_publ_Section_title
;
 Powder refinement of carbonyl sulphide
;

_aflow_title 'Carbonyl Sulphide (COS, $F0_{2}$) Structure'
_aflow_proto 'ABC_hR3_160_a_a_a'
_aflow_params 'a,c/a,x_{1},x_{2},x_{3}'
_aflow_params_values '6.1703,0.949908432329,0.0,0.79356,0.25763'
_aflow_Strukturbericht '$F0_{2}$'
_aflow_Pearson 'hR3'

_symmetry_space_group_name_H-M "R 3 m (hexagonal axes)"
_symmetry_Int_Tables_number 160
 
_cell_length_a    6.17030
_cell_length_b    6.17030
_cell_length_c    5.86122
_cell_angle_alpha 90.00000
_cell_angle_beta  90.00000
_cell_angle_gamma 120.00000
 
loop_
_space_group_symop_id
_space_group_symop_operation_xyz
1 x,y,z
2 -y,x-y,z
3 -x+y,-x,z
4 -y,-x,z
5 x,x-y,z
6 -x+y,y,z
7 x+1/3,y+2/3,z+2/3
8 -y+1/3,x-y+2/3,z+2/3
9 -x+y+1/3,-x+2/3,z+2/3
10 -y+1/3,-x+2/3,z+2/3
11 x+1/3,x-y+2/3,z+2/3
12 -x+y+1/3,y+2/3,z+2/3
13 x+2/3,y+1/3,z+1/3
14 -y+2/3,x-y+1/3,z+1/3
15 -x+y+2/3,-x+1/3,z+1/3
16 -y+2/3,-x+1/3,z+1/3
17 x+2/3,x-y+1/3,z+1/3
18 -x+y+2/3,y+1/3,z+1/3
 
loop_
_atom_site_label
_atom_site_type_symbol
_atom_site_symmetry_multiplicity
_atom_site_Wyckoff_label
_atom_site_fract_x
_atom_site_fract_y
_atom_site_fract_z
_atom_site_occupancy
C1 C   3 a 0.00000 0.00000 0.00000 1.00000
O1 O   3 a 0.00000 0.00000 0.79356 1.00000
S1 S   3 a 0.00000 0.00000 0.25763 1.00000
\end{lstlisting}
{\phantomsection\label{ABC_hR3_160_a_a_a_poscar}}
{\hyperref[ABC_hR3_160_a_a_a]{Carbonyl Sulphide (COS, $F0_{2}$): ABC\_hR3\_160\_a\_a\_a}} - POSCAR
\begin{lstlisting}[numbers=none,language={mylang}]
ABC_hR3_160_a_a_a & a,c/a,x1,x2,x3 --params=6.1703,0.949908432329,0.0,0.79356,0.25763 & R3m C_{3v}^{5} #160 (a^3) & hR3 & $F0_{2}$ & COS & Carbonyl Sulphide & J. S. W. Overell and G. S. Pawley and B. M. Powell, Acta Crystallogr. Sect. B Struct. Sci. 38, 1121-1123 (1982)
   1.00000000000000
   3.08515000000000  -1.78121218299037   1.95374000000000
   0.00000000000000   3.56242436598075   1.95374000000000
  -3.08515000000000  -1.78121218299037   1.95374000000000
     C     O     S
     1     1     1
Direct
   0.00000000000000   0.00000000000000   0.00000000000000    C   (1a)
   0.79356000000000   0.79356000000000   0.79356000000000    O   (1a)
   0.25763000000000   0.25763000000000   0.25763000000000    S   (1a)
\end{lstlisting}
{\phantomsection\label{AB_hR10_160_5a_5a_cif}}
{\hyperref[AB_hR10_160_5a_5a]{Moissanite-15R (SiC, $B7$): AB\_hR10\_160\_5a\_5a}} - CIF

{\phantomsection\label{AB_hR10_160_5a_5a_poscar}}
{\hyperref[AB_hR10_160_5a_5a]{Moissanite-15R (SiC, $B7$): AB\_hR10\_160\_5a\_5a}} - POSCAR
\begin{lstlisting}[numbers=none,language={mylang}]
AB_hR10_160_5a_5a & a,c/a,x1,x2,x3,x4,x5,x6,x7,x8,x9,x10 --params=3.09,12.2653721683,0.0,0.13333,0.4,0.6,0.86667,0.05,0.18333,0.45,0.65,-0.08333 & R3m C_{3v}^{5} #160 (a^10) & hR10 & $B7$ & SiC & Moissanite-15R & N. W. Thibault, Am. Mineral. 29, 327-362 (1944)
   1.00000000000000
   1.54500000000000  -0.89200616589797  12.63333333333330
   0.00000000000000   1.78401233179594  12.63333333333330
  -1.54500000000000  -0.89200616589797  12.63333333333330
     C    Si
     5     5
Direct
   0.00000000000000   0.00000000000000   0.00000000000000    C   (1a)
   0.13333000000000   0.13333000000000   0.13333000000000    C   (1a)
   0.40000000000000   0.40000000000000   0.40000000000000    C   (1a)
   0.60000000000000   0.60000000000000   0.60000000000000    C   (1a)
   0.86667000000000   0.86667000000000   0.86667000000000    C   (1a)
   0.05000000000000   0.05000000000000   0.05000000000000   Si   (1a)
   0.18333000000000   0.18333000000000   0.18333000000000   Si   (1a)
   0.45000000000000   0.45000000000000   0.45000000000000   Si   (1a)
   0.65000000000000   0.65000000000000   0.65000000000000   Si   (1a)
  -0.08333000000000  -0.08333000000000  -0.08333000000000   Si   (1a)
\end{lstlisting}
{\phantomsection\label{A2B3_hP5_164_d_ad_cif}}
{\hyperref[A2B3_hP5_164_d_ad]{La$_{2}$O$_{3}$ ($D5_{2}$): A2B3\_hP5\_164\_d\_ad}} - CIF
\begin{lstlisting}[numbers=none,language={mylang}]
# CIF file 
data_findsym-output
_audit_creation_method FINDSYM

_chemical_name_mineral ''
_chemical_formula_sum 'La2 O3'

loop_
_publ_author_name
 'P. Aldebert'
 'J. P. Traverse'
_journal_name_full_name
;
 Materials Research Bulletin
;
_journal_volume 14
_journal_year 1979
_journal_page_first 303
_journal_page_last 323
_publ_Section_title
;
 Etude par diffraction neutronique des structures de haute temperature de La$_{2}$O$_{3}$ et Nd$_{2}$O$_{3}$
;

_aflow_title 'La$_{2}$O$_{3}$ ($D5_{2}$) Structure'
_aflow_proto 'A2B3_hP5_164_d_ad'
_aflow_params 'a,c/a,z_{2},z_{3}'
_aflow_params_values '3.9381,1.55813717275,0.2467,0.647'
_aflow_Strukturbericht '$D5_{2}$'
_aflow_Pearson 'hP5'

_symmetry_space_group_name_H-M "P -3 2/m 1"
_symmetry_Int_Tables_number 164
 
_cell_length_a    3.93810
_cell_length_b    3.93810
_cell_length_c    6.13610
_cell_angle_alpha 90.00000
_cell_angle_beta  90.00000
_cell_angle_gamma 120.00000
 
loop_
_space_group_symop_id
_space_group_symop_operation_xyz
1 x,y,z
2 -y,x-y,z
3 -x+y,-x,z
4 x-y,-y,-z
5 y,x,-z
6 -x,-x+y,-z
7 -x,-y,-z
8 y,-x+y,-z
9 x-y,x,-z
10 -x+y,y,z
11 -y,-x,z
12 x,x-y,z
 
loop_
_atom_site_label
_atom_site_type_symbol
_atom_site_symmetry_multiplicity
_atom_site_Wyckoff_label
_atom_site_fract_x
_atom_site_fract_y
_atom_site_fract_z
_atom_site_occupancy
O1  O    1 a 0.00000 0.00000 0.00000 1.00000
La1 La   2 d 0.33333 0.66667 0.24670 1.00000
O2  O    2 d 0.33333 0.66667 0.64700 1.00000
\end{lstlisting}
{\phantomsection\label{A2B3_hP5_164_d_ad_poscar}}
{\hyperref[A2B3_hP5_164_d_ad]{La$_{2}$O$_{3}$ ($D5_{2}$): A2B3\_hP5\_164\_d\_ad}} - POSCAR
\begin{lstlisting}[numbers=none,language={mylang}]
A2B3_hP5_164_d_ad & a,c/a,z2,z3 --params=3.9381,1.55813717275,0.2467,0.647 & P-3m1 D_{3d}^{3} #164 (ad^2) & hP5 & $D5_{2}$ & La2O3 &  & P. Aldebert and J. P. Traverse, Mater. Res. Bull. 14, 303-323 (1979)
   1.00000000000000
   1.96905000000000  -3.41049464264350   0.00000000000000
   1.96905000000000   3.41049464264350   0.00000000000000
   0.00000000000000   0.00000000000000   6.13610000000000
    La     O
     2     3
Direct
   0.33333333333333   0.66666666666667   0.24670000000000   La   (2d)
   0.66666666666667   0.33333333333333  -0.24670000000000   La   (2d)
   0.00000000000000   0.00000000000000   0.00000000000000    O   (1a)
   0.33333333333333   0.66666666666667   0.64700000000000    O   (2d)
   0.66666666666667   0.33333333333333  -0.64700000000000    O   (2d)
\end{lstlisting}
{\phantomsection\label{AB2_hP9_164_bd_c2d_cif}}
{\hyperref[AB2_hP9_164_bd_c2d]{$\delta_{H}^{II}$-NW$_2$: AB2\_hP9\_164\_bd\_c2d}} - CIF
\begin{lstlisting}[numbers=none,language={mylang}]
# CIF file 
data_findsym-output
_audit_creation_method FINDSYM

_chemical_name_mineral '$\delta_{H}^{II}$-NW$_2$'
_chemical_formula_sum 'N W2'

loop_
_publ_author_name
 'V. I. Khitrova'
 'Z. G. Pinkser'
_journal_name_full_name
;
 Soviet Physics Crystallography
;
_journal_volume 6
_journal_year 1962
_journal_page_first 712
_journal_page_last 719
_publ_Section_title
;
 Chemical Crystallography of Tungsten Nitrides and of Some Other Interstitial Phases
;

_aflow_title '$\delta_{H}^{II}$-NW$_2$ Structure'
_aflow_proto 'AB2_hP9_164_bd_c2d'
_aflow_params 'a,c/a,z_{2},z_{3},z_{4},z_{5}'
_aflow_params_values '2.89,7.90657439446,0.0607,0.154,0.27263,0.39403'
_aflow_Strukturbericht 'None'
_aflow_Pearson 'hP9'

_symmetry_space_group_name_H-M "P -3 2/m 1"
_symmetry_Int_Tables_number 164
 
_cell_length_a    2.89000
_cell_length_b    2.89000
_cell_length_c    22.85000
_cell_angle_alpha 90.00000
_cell_angle_beta  90.00000
_cell_angle_gamma 120.00000
 
loop_
_space_group_symop_id
_space_group_symop_operation_xyz
1 x,y,z
2 -y,x-y,z
3 -x+y,-x,z
4 x-y,-y,-z
5 y,x,-z
6 -x,-x+y,-z
7 -x,-y,-z
8 y,-x+y,-z
9 x-y,x,-z
10 -x+y,y,z
11 -y,-x,z
12 x,x-y,z
 
loop_
_atom_site_label
_atom_site_type_symbol
_atom_site_symmetry_multiplicity
_atom_site_Wyckoff_label
_atom_site_fract_x
_atom_site_fract_y
_atom_site_fract_z
_atom_site_occupancy
N1 N   1 b 0.00000 0.00000 0.50000 1.00000
W1 W   2 c 0.00000 0.00000 0.06070 1.00000
N2 N   2 d 0.33333 0.66667 0.15400 1.00000
W2 W   2 d 0.33333 0.66667 0.27263 1.00000
W3 W   2 d 0.33333 0.66667 0.39403 1.00000
\end{lstlisting}
{\phantomsection\label{AB2_hP9_164_bd_c2d_poscar}}
{\hyperref[AB2_hP9_164_bd_c2d]{$\delta_{H}^{II}$-NW$_2$: AB2\_hP9\_164\_bd\_c2d}} - POSCAR
\begin{lstlisting}[numbers=none,language={mylang}]
AB2_hP9_164_bd_c2d & a,c/a,z2,z3,z4,z5 --params=2.89,7.90657439446,0.0607,0.154,0.27263,0.39403 & P-3m1 D_{3d}^{3} #164 (bcd^3) & hP9 & None & NW2 & $\delta_{H}^{II}$-NW$_2$ & V. I. Khitrova and Z. G. Pinkser, Sov. Phys. Crystallogr. 6, 712-719 (1962)
   1.00000000000000
   1.44500000000000  -2.50281341693703   0.00000000000000
   1.44500000000000   2.50281341693703   0.00000000000000
   0.00000000000000   0.00000000000000  22.85000000000000
     N     W
     3     6
Direct
   0.00000000000000   0.00000000000000   0.50000000000000    N   (1b)
   0.33333333333333   0.66666666666667   0.15400000000000    N   (2d)
   0.66666666666667   0.33333333333333  -0.15400000000000    N   (2d)
   0.00000000000000   0.00000000000000   0.06070000000000    W   (2c)
   0.00000000000000   0.00000000000000  -0.06070000000000    W   (2c)
   0.33333333333333   0.66666666666667   0.27263000000000    W   (2d)
   0.66666666666667   0.33333333333333  -0.27263000000000    W   (2d)
   0.33333333333333   0.66666666666667   0.39403000000000    W   (2d)
   0.66666666666667   0.33333333333333  -0.39403000000000    W   (2d)
\end{lstlisting}
{\phantomsection\label{ABC2_hP4_164_a_b_d_cif}}
{\hyperref[ABC2_hP4_164_a_b_d]{CuNiSb$_{2}$: ABC2\_hP4\_164\_a\_b\_d}} - CIF
\begin{lstlisting}[numbers=none,language={mylang}]
# CIF file 
data_findsym-output
_audit_creation_method FINDSYM

_chemical_name_mineral ''
_chemical_formula_sum 'Cu Ni Sb2'

loop_
_publ_author_name
 'R. L. Kift'
_journal_year 2010
_publ_Section_title
;
 Intermetallic Compounds by Reductive Annealing
;

_aflow_title 'CuNiSb$_{2}$ Structure'
_aflow_proto 'ABC2_hP4_164_a_b_d'
_aflow_params 'a,c/a,z_{3}'
_aflow_params_values '4.0482,1.26777333136,0.271'
_aflow_Strukturbericht 'None'
_aflow_Pearson 'hP4'

_symmetry_space_group_name_H-M "P -3 2/m 1"
_symmetry_Int_Tables_number 164
 
_cell_length_a    4.04820
_cell_length_b    4.04820
_cell_length_c    5.13220
_cell_angle_alpha 90.00000
_cell_angle_beta  90.00000
_cell_angle_gamma 120.00000
 
loop_
_space_group_symop_id
_space_group_symop_operation_xyz
1 x,y,z
2 -y,x-y,z
3 -x+y,-x,z
4 x-y,-y,-z
5 y,x,-z
6 -x,-x+y,-z
7 -x,-y,-z
8 y,-x+y,-z
9 x-y,x,-z
10 -x+y,y,z
11 -y,-x,z
12 x,x-y,z
 
loop_
_atom_site_label
_atom_site_type_symbol
_atom_site_symmetry_multiplicity
_atom_site_Wyckoff_label
_atom_site_fract_x
_atom_site_fract_y
_atom_site_fract_z
_atom_site_occupancy
Cu1 Cu   1 a 0.00000 0.00000 0.00000 1.00000
Ni1 Ni   1 b 0.00000 0.00000 0.50000 1.00000
Sb1 Sb   2 d 0.33333 0.66667 0.27100 1.00000
\end{lstlisting}
{\phantomsection\label{ABC2_hP4_164_a_b_d_poscar}}
{\hyperref[ABC2_hP4_164_a_b_d]{CuNiSb$_{2}$: ABC2\_hP4\_164\_a\_b\_d}} - POSCAR
\begin{lstlisting}[numbers=none,language={mylang}]
ABC2_hP4_164_a_b_d & a,c/a,z3 --params=4.0482,1.26777333136,0.271 & P-3m1 D_{3d}^{3} #164 (abd) & hP4 & None & CuNiSb2 &  & R. L. Kift, (2010)
   1.00000000000000
   2.02410000000000  -3.50584403960016   0.00000000000000
   2.02410000000000   3.50584403960016   0.00000000000000
   0.00000000000000   0.00000000000000   5.13220000000000
    Cu    Ni    Sb
     1     1     2
Direct
   0.00000000000000   0.00000000000000   0.00000000000000   Cu   (1a)
   0.00000000000000   0.00000000000000   0.50000000000000   Ni   (1b)
   0.33333333333333   0.66666666666667   0.27100000000000   Sb   (2d)
   0.66666666666667   0.33333333333333  -0.27100000000000   Sb   (2d)
\end{lstlisting}
{\phantomsection\label{A3B_hP24_165_bdg_f_cif}}
{\hyperref[A3B_hP24_165_bdg_f]{Cu$_{3}$P ($D0_{21}$): A3B\_hP24\_165\_bdg\_f}} - CIF
\begin{lstlisting}[numbers=none,language={mylang}]
# CIF file 
data_findsym-output
_audit_creation_method FINDSYM

_chemical_name_mineral ''
_chemical_formula_sum 'Cu3 P'

loop_
_publ_author_name
 'B. Steenberg'
_journal_name_full_name
;
 Arkiv f{\"o}r Kemi, Mineralogi och Geologi
;
_journal_volume A12
_journal_year 1938
_journal_page_first 1
_journal_page_last 15
_publ_Section_title
;
 The Crystal Structure of Cu$_{3}$As and Cu$_{3}$P
;

# Found in A Handbook of Lattice Spacings and Structures of Metals and Alloys, 1958

_aflow_title 'Cu$_{3}$P ($D0_{21}$) Structure'
_aflow_proto 'A3B_hP24_165_bdg_f'
_aflow_params 'a,c/a,z_{2},x_{3},x_{4},y_{4},z_{4}'
_aflow_params_values '7.07,1.00919377652,0.17,0.38,0.69,0.07,0.08'
_aflow_Strukturbericht '$D0_{21}$'
_aflow_Pearson 'hP24'

_symmetry_space_group_name_H-M "P -3 2/c 1"
_symmetry_Int_Tables_number 165
 
_cell_length_a    7.07000
_cell_length_b    7.07000
_cell_length_c    7.13500
_cell_angle_alpha 90.00000
_cell_angle_beta  90.00000
_cell_angle_gamma 120.00000
 
loop_
_space_group_symop_id
_space_group_symop_operation_xyz
1 x,y,z
2 -y,x-y,z
3 -x+y,-x,z
4 x-y,-y,-z+1/2
5 y,x,-z+1/2
6 -x,-x+y,-z+1/2
7 -x,-y,-z
8 y,-x+y,-z
9 x-y,x,-z
10 -x+y,y,z+1/2
11 -y,-x,z+1/2
12 x,x-y,z+1/2
 
loop_
_atom_site_label
_atom_site_type_symbol
_atom_site_symmetry_multiplicity
_atom_site_Wyckoff_label
_atom_site_fract_x
_atom_site_fract_y
_atom_site_fract_z
_atom_site_occupancy
Cu1 Cu   2 b 0.00000 0.00000 0.00000 1.00000
Cu2 Cu   4 d 0.33333 0.66667 0.17000 1.00000
P1  P    6 f 0.38000 0.00000 0.25000 1.00000
Cu3 Cu  12 g 0.69000 0.07000 0.08000 1.00000
\end{lstlisting}
{\phantomsection\label{A3B_hP24_165_bdg_f_poscar}}
{\hyperref[A3B_hP24_165_bdg_f]{Cu$_{3}$P ($D0_{21}$): A3B\_hP24\_165\_bdg\_f}} - POSCAR

{\phantomsection\label{A4B3_hR7_166_2c_ac_cif}}
{\hyperref[A4B3_hR7_166_2c_ac]{Al$_{4}$C$_{3}$ ($D7_{1}$): A4B3\_hR7\_166\_2c\_ac}} - CIF

{\phantomsection\label{A4B3_hR7_166_2c_ac_poscar}}
{\hyperref[A4B3_hR7_166_2c_ac]{Al$_{4}$C$_{3}$ ($D7_{1}$): A4B3\_hR7\_166\_2c\_ac}} - POSCAR
\begin{lstlisting}[numbers=none,language={mylang}]
A4B3_hR7_166_2c_ac & a,c/a,x2,x3,x4 --params=3.335,7.48635682159,0.29422,0.12967,0.2168 & R-3m D_{3d}^{5} #166 (ac^3) & hR7 & $D7_{1}$ & Al4C3 &  & T. M. Gesing and W. Jeitschko, Z. Naturforsch. B 50, 196-200 (1995)
   1.00000000000000
   1.66750000000000  -0.96273157387370   8.32233333333333
   0.00000000000000   1.92546314774740   8.32233333333333
  -1.66750000000000  -0.96273157387370   8.32233333333333
    Al     C
     4     3
Direct
   0.29422000000000   0.29422000000000   0.29422000000000   Al   (2c)
  -0.29422000000000  -0.29422000000000  -0.29422000000000   Al   (2c)
   0.12967000000000   0.12967000000000   0.12967000000000   Al   (2c)
  -0.12967000000000  -0.12967000000000  -0.12967000000000   Al   (2c)
   0.00000000000000   0.00000000000000   0.00000000000000    C   (1a)
   0.21680000000000   0.21680000000000   0.21680000000000    C   (2c)
  -0.21680000000000  -0.21680000000000  -0.21680000000000    C   (2c)
\end{lstlisting}
{\phantomsection\label{ABC_hR6_166_c_c_c_cif}}
{\hyperref[ABC_hR6_166_c_c_c]{SmSI: ABC\_hR6\_166\_c\_c\_c}} - CIF

{\phantomsection\label{ABC_hR6_166_c_c_c_poscar}}
{\hyperref[ABC_hR6_166_c_c_c]{SmSI: ABC\_hR6\_166\_c\_c\_c}} - POSCAR
\begin{lstlisting}[numbers=none,language={mylang}]
ABC_hR6_166_c_c_c & a,c/a,x1,x2,x3 --params=3.8548,7.95397945419,0.1159,0.3017,0.3815 & R-3m D_{3d}^{5} #166 (c^3) & hR6 & None & SmSI & SmSI & H. P. Beck and C. Strobel, Z. Anorg. Allg. Chem. 535, 222-239 (1986)
   1.00000000000000
   1.92740000000000  -1.11278490883608  10.22033333333330
   0.00000000000000   2.22556981767217  10.22033333333330
  -1.92740000000000  -1.11278490883608  10.22033333333330
     I     S    Sm
     2     2     2
Direct
   0.11590000000000   0.11590000000000   0.11590000000000    I   (2c)
  -0.11590000000000  -0.11590000000000  -0.11590000000000    I   (2c)
   0.30170000000000   0.30170000000000   0.30170000000000    S   (2c)
  -0.30170000000000  -0.30170000000000  -0.30170000000000    S   (2c)
   0.38150000000000   0.38150000000000   0.38150000000000   Sm   (2c)
  -0.38150000000000  -0.38150000000000  -0.38150000000000   Sm   (2c)
\end{lstlisting}
{\phantomsection\label{AB3C_hR10_167_b_e_a_cif}}
{\hyperref[AB3C_hR10_167_b_e_a]{PrNiO$_{3}$: AB3C\_hR10\_167\_b\_e\_a}} - CIF

{\phantomsection\label{AB3C_hR10_167_b_e_a_poscar}}
{\hyperref[AB3C_hR10_167_b_e_a]{PrNiO$_{3}$: AB3C\_hR10\_167\_b\_e\_a}} - POSCAR
\begin{lstlisting}[numbers=none,language={mylang}]
AB3C_hR10_167_b_e_a & a,c/a,x3 --params=5.4577,2.40134122433,0.6946 & R-3c D_{3d}^{6} #167 (abe) & hR10 & None & PrNiO3 & PrNiO3 & T. C. Huang et al., Mater. Res. Bull. 25, 1091-1098 (1990)
   1.00000000000000
   2.72885000000000  -1.57550228207811   4.36860000000000
   0.00000000000000   3.15100456415622   4.36860000000000
  -2.72885000000000  -1.57550228207811   4.36860000000000
    Ni     O    Pr
     2     6     2
Direct
   0.00000000000000   0.00000000000000   0.00000000000000   Ni   (2b)
   0.50000000000000   0.50000000000000   0.50000000000000   Ni   (2b)
   0.69460000000000  -0.19460000000000   0.25000000000000    O   (6e)
   0.25000000000000   0.69460000000000  -0.19460000000000    O   (6e)
  -0.19460000000000   0.25000000000000   0.69460000000000    O   (6e)
  -0.69460000000000   1.19460000000000   0.75000000000000    O   (6e)
   0.75000000000000  -0.69460000000000   1.19460000000000    O   (6e)
   1.19460000000000   0.75000000000000  -0.69460000000000    O   (6e)
   0.25000000000000   0.25000000000000   0.25000000000000   Pr   (2a)
   0.75000000000000   0.75000000000000   0.75000000000000   Pr   (2a)
\end{lstlisting}
{\phantomsection\label{ABC2_hR24_167_e_e_2e_cif}}
{\hyperref[ABC2_hR24_167_e_e_2e]{KBO$_{2}$ ($F5_{13}$): ABC2\_hR24\_167\_e\_e\_2e}} - CIF

{\phantomsection\label{ABC2_hR24_167_e_e_2e_poscar}}
{\hyperref[ABC2_hR24_167_e_e_2e]{KBO$_{2}$ ($F5_{13}$): ABC2\_hR24\_167\_e\_e\_2e}} - POSCAR

{\phantomsection\label{A2B13C4_hP57_168_d_c6d_2d_cif}}
{\hyperref[A2B13C4_hP57_168_d_c6d_2d]{K$_{2}$Ta$_{4}$O$_{9}$F$_{4}$: A2B13C4\_hP57\_168\_d\_c6d\_2d}} - CIF

{\phantomsection\label{A2B13C4_hP57_168_d_c6d_2d_poscar}}
{\hyperref[A2B13C4_hP57_168_d_c6d_2d]{K$_{2}$Ta$_{4}$O$_{9}$F$_{4}$: A2B13C4\_hP57\_168\_d\_c6d\_2d}} - POSCAR

{\phantomsection\label{AB4C_hP72_168_2d_8d_2d_cif}}
{\hyperref[AB4C_hP72_168_2d_8d_2d]{Al[PO$_{4}$]: AB4C\_hP72\_168\_2d\_8d\_2d}} - CIF

{\phantomsection\label{AB4C_hP72_168_2d_8d_2d_poscar}}
{\hyperref[AB4C_hP72_168_2d_8d_2d]{Al[PO$_{4}$]: AB4C\_hP72\_168\_2d\_8d\_2d}} - POSCAR

{\phantomsection\label{A2B3_hP30_169_2a_3a_cif}}
{\hyperref[A2B3_hP30_169_2a_3a]{$\alpha$-Al$_{2}$S$_{3}$: A2B3\_hP30\_169\_2a\_3a}} - CIF
\begin{lstlisting}[numbers=none,language={mylang}]
# CIF file
data_findsym-output
_audit_creation_method FINDSYM

_chemical_name_mineral 'alpha-Al2S3'
_chemical_formula_sum 'Al2 S3'

loop_
_publ_author_name
 'B. Eisenmann'
_journal_name_full_name
;
 Zeitschrift f{\"u}r Kristallografiya
;
_journal_volume 198
_journal_year 1992
_journal_page_first 307
_journal_page_last 308
_publ_Section_title
;
 Crystal structure of $\alpha$-dialuminium trisulfide, Al$_{2}$S$_{3}$
;

# Found in Pearson's Crystal Data - Crystal Structure Database for Inorganic Compounds, 2013

_aflow_title '$\alpha$-Al$_{2}$S$_{3}$ Structure'
_aflow_proto 'A2B3_hP30_169_2a_3a'
_aflow_params 'a,c/a,x_{1},y_{1},z_{1},x_{2},y_{2},z_{2},x_{3},y_{3},z_{3},x_{4},y_{4},z_{4},x_{5},y_{5},z_{5}'
_aflow_params_values '6.4300116544,2.78071539658,0.013,0.3579,0.12736,0.3339,0.3226,0.29886,0.3347,0.0,0.004,0.0119,0.3343,0.0,0.338,0.0064,0.33823'
_aflow_Strukturbericht 'None'
_aflow_Pearson 'hP30'

_cell_length_a    6.4300116544
_cell_length_b    6.4300116544
_cell_length_c    17.8800324076
_cell_angle_alpha 90.0000000000
_cell_angle_beta  90.0000000000
_cell_angle_gamma 120.0000000000
 
_symmetry_space_group_name_H-M "P 61"
_symmetry_Int_Tables_number 169
 
loop_
_space_group_symop_id
_space_group_symop_operation_xyz
1 x,y,z
2 x-y,x,z+1/6
3 -y,x-y,z+1/3
4 -x,-y,z+1/2
5 -x+y,-x,z+2/3
6 y,-x+y,z+5/6
 
loop_
_atom_site_label
_atom_site_type_symbol
_atom_site_symmetry_multiplicity
_atom_site_Wyckoff_label
_atom_site_fract_x
_atom_site_fract_y
_atom_site_fract_z
_atom_site_occupancy
Al1 Al   6 a 0.01300 0.35790 0.12736 1.00000
Al2 Al   6 a 0.33390 0.32260 0.29886 1.00000
S1  S    6 a 0.33470 0.00000 0.00400 1.00000
S2  S    6 a 0.01190 0.33430 0.00000 1.00000
S3  S    6 a 0.33800 0.00640 0.33823 1.00000
\end{lstlisting}
{\phantomsection\label{A2B3_hP30_169_2a_3a_poscar}}
{\hyperref[A2B3_hP30_169_2a_3a]{$\alpha$-Al$_{2}$S$_{3}$: A2B3\_hP30\_169\_2a\_3a}} - POSCAR

{\phantomsection\label{A2B3_hP30_170_2a_3a_cif}}
{\hyperref[A2B3_hP30_170_2a_3a]{Al$_{2}$S$_{3}$: A2B3\_hP30\_170\_2a\_3a}} - CIF
\begin{lstlisting}[numbers=none,language={mylang}]
# CIF file
data_findsym-output
_audit_creation_method FINDSYM

_chemical_name_mineral 'Al2S3'
_chemical_formula_sum 'Al2 S3'

_aflow_title 'Al$_{2}$S$_{3}$ Structure'
_aflow_proto 'A2B3_hP30_170_2a_3a'
_aflow_params 'a,c/a,x_{1},y_{1},z_{1},x_{2},y_{2},z_{2},x_{3},y_{3},z_{3},x_{4},y_{4},z_{4},x_{5},y_{5},z_{5}'
_aflow_params_values '6.4300116544,2.78071539658,0.013,0.3579,0.87264,0.3339,0.3226,0.70114,0.3347,0.0,-0.004,0.0119,0.3343,0.0,0.338,0.0064,0.66177'
_aflow_Strukturbericht 'None'
_aflow_Pearson 'hP30'

_cell_length_a    6.4300116544
_cell_length_b    6.4300116544
_cell_length_c    17.8800324076
_cell_angle_alpha 90.0000000000
_cell_angle_beta  90.0000000000
_cell_angle_gamma 120.0000000000
 
_symmetry_space_group_name_H-M "P 65"
_symmetry_Int_Tables_number 170
 
loop_
_space_group_symop_id
_space_group_symop_operation_xyz
1 x,y,z
2 x-y,x,z+5/6
3 -y,x-y,z+2/3
4 -x,-y,z+1/2
5 -x+y,-x,z+1/3
6 y,-x+y,z+1/6
 
loop_
_atom_site_label
_atom_site_type_symbol
_atom_site_symmetry_multiplicity
_atom_site_Wyckoff_label
_atom_site_fract_x
_atom_site_fract_y
_atom_site_fract_z
_atom_site_occupancy
Al1 Al   6 a 0.01300 0.35790 0.87264  1.00000
Al2 Al   6 a 0.33390 0.32260 0.70114  1.00000
S1  S    6 a 0.33470 0.00000 -0.00400 1.00000
S2  S    6 a 0.01190 0.33430 0.00000  1.00000
S3  S    6 a 0.33800 0.00640 0.66177  1.00000
\end{lstlisting}
{\phantomsection\label{A2B3_hP30_170_2a_3a_poscar}}
{\hyperref[A2B3_hP30_170_2a_3a]{Al$_{2}$S$_{3}$: A2B3\_hP30\_170\_2a\_3a}} - POSCAR

{\phantomsection\label{A10B2C_hP39_171_5c_c_a_cif}}
{\hyperref[A10B2C_hP39_171_5c_c_a]{Sr[S$_{2}$O$_{6}$][H$_{2}$O]$_{4}$: A10B2C\_hP39\_171\_5c\_c\_a}} - CIF
\begin{lstlisting}[numbers=none,language={mylang}]
# CIF file
data_findsym-output
_audit_creation_method FINDSYM

_chemical_name_mineral 'Sr[S2O6][H2O]4'
_chemical_formula_sum 'O10 S2 Sr'

loop_
_publ_author_name
 'R. N. Hargreaves'
 'E. Stanley'
_journal_name_full_name
;
 Zeitschrift f{\"u}r Kristallographie - Crystalline Materials
;
_journal_volume 135
_journal_year 1972
_journal_page_first 399
_journal_page_last 407
_publ_Section_title
;
 The structure of strontium dithionate tetrahydrate
;

# Found in Pearson's Crystal Data - Crystal Structure Database for Inorganic Compounds, 2013

_aflow_title 'Sr[S$_{2}$O$_{6}$][H$_{2}$O]$_{4}$ Structure'
_aflow_proto 'A10B2C_hP39_171_5c_c_a'
_aflow_params 'a,c/a,z_{1},x_{2},y_{2},z_{2},x_{3},y_{3},z_{3},x_{4},y_{4},z_{4},x_{5},y_{5},z_{5},x_{6},y_{6},z_{6},x_{7},y_{7},z_{7}'
_aflow_params_values '6.3199906634,3.05221518986,0.0,-0.004,0.253,0.23633,0.015,0.248,0.11034,0.681,0.192,0.16733,0.448,0.188,0.29533,0.449,0.25,0.048,0.373,0.065,0.50467'
_aflow_Strukturbericht 'None'
_aflow_Pearson 'hP39'

_cell_length_a    6.3199906634
_cell_length_b    6.3199906634
_cell_length_c    19.2899715026
_cell_angle_alpha 90.0000000000
_cell_angle_beta  90.0000000000
_cell_angle_gamma 120.0000000000
 
_symmetry_space_group_name_H-M "P 62"
_symmetry_Int_Tables_number 171
 
loop_
_space_group_symop_id
_space_group_symop_operation_xyz
1 x,y,z
2 x-y,x,z+1/3
3 -y,x-y,z+2/3
4 -x,-y,z
5 -x+y,-x,z+1/3
6 y,-x+y,z+2/3
 
loop_
_atom_site_label
_atom_site_type_symbol
_atom_site_symmetry_multiplicity
_atom_site_Wyckoff_label
_atom_site_fract_x
_atom_site_fract_y
_atom_site_fract_z
_atom_site_occupancy
Sr1  Sr   3 a 0.00000  0.00000 0.00000 1.00000
O1   O    6 c -0.00400 0.25300 0.23633 1.00000
O2   O    6 c 0.01500  0.24800 0.11034 1.00000
O3   O    6 c 0.68100  0.19200 0.16733 1.00000
O4 O    6 c 0.44800  0.18800 0.29533 1.00000
O5 O    6 c 0.44900  0.25000 0.04800 1.00000
S1   S    6 c 0.37300  0.06500 0.50467 1.00000
\end{lstlisting}
{\phantomsection\label{A10B2C_hP39_171_5c_c_a_poscar}}
{\hyperref[A10B2C_hP39_171_5c_c_a]{Sr[S$_{2}$O$_{6}$][H$_{2}$O]$_{4}$: A10B2C\_hP39\_171\_5c\_c\_a}} - POSCAR

{\phantomsection\label{A10B2C_hP39_172_5c_c_a_cif}}
{\hyperref[A10B2C_hP39_172_5c_c_a]{Sr[S$_{2}$O$_{6}$][H$_{2}$O]$_{4}$: A10B2C\_hP39\_172\_5c\_c\_a}} - CIF
\begin{lstlisting}[numbers=none,language={mylang}]
# CIF file
data_findsym-output
_audit_creation_method FINDSYM

_chemical_name_mineral 'Sr[S2O6][H2O]4'
_chemical_formula_sum 'O10 S2 Sr'

_aflow_title 'Sr[S$_{2}$O$_{6}$][H$_{2}$O]$_{4}$ Structure'
_aflow_proto 'A10B2C_hP39_172_5c_c_a'
_aflow_params 'a,c/a,z_{1},x_{2},y_{2},z_{2},x_{3},y_{3},z_{3},x_{4},y_{4},z_{4},x_{5},y_{5},z_{5},x_{6},y_{6},z_{6},x_{7},y_{7},z_{7}'
_aflow_params_values '6.3199906634,3.05221518986,0.0,-0.004,0.253,0.76367,0.015,0.248,0.88966,0.681,0.192,0.83267,0.448,0.188,0.70467,0.449,0.25,-0.048,0.373,0.065,0.49533'
_aflow_Strukturbericht 'None'
_aflow_Pearson 'hP39'

_cell_length_a    6.3199906634
_cell_length_b    6.3199906634
_cell_length_c    19.2899715026
_cell_angle_alpha 90.0000000000
_cell_angle_beta  90.0000000000
_cell_angle_gamma 120.0000000000
 
_symmetry_space_group_name_H-M "P 64"
_symmetry_Int_Tables_number 172
 
loop_
_space_group_symop_id
_space_group_symop_operation_xyz
1 x,y,z
2 x-y,x,z+2/3
3 -y,x-y,z+1/3
4 -x,-y,z
5 -x+y,-x,z+2/3
6 y,-x+y,z+1/3
 
loop_
_atom_site_label
_atom_site_type_symbol
_atom_site_symmetry_multiplicity
_atom_site_Wyckoff_label
_atom_site_fract_x
_atom_site_fract_y
_atom_site_fract_z
_atom_site_occupancy
Sr1 Sr   3 a 0.00000  0.00000 0.00000  1.00000
O1  O    6 c -0.00400 0.25300 0.76367  1.00000
O2  O    6 c 0.01500  0.24800 0.88966  1.00000
O3  O    6 c 0.68100  0.19200 0.83267  1.00000
O4  O    6 c 0.44800  0.18800 0.70467  1.00000
O5  O    6 c 0.44900  0.25000 -0.04800 1.00000
S1  S    6 c 0.37300  0.06500 0.49533  1.00000
\end{lstlisting}
{\phantomsection\label{A10B2C_hP39_172_5c_c_a_poscar}}
{\hyperref[A10B2C_hP39_172_5c_c_a]{Sr[S$_{2}$O$_{6}$][H$_{2}$O]$_{4}$: A10B2C\_hP39\_172\_5c\_c\_a}} - POSCAR

{\phantomsection\label{A3B_hP8_173_c_b_cif}}
{\hyperref[A3B_hP8_173_c_b]{PI$_{3}$: A3B\_hP8\_173\_c\_b}} - CIF
\begin{lstlisting}[numbers=none,language={mylang}]
# CIF file
data_findsym-output
_audit_creation_method FINDSYM

_chemical_name_mineral 'PI3'
_chemical_formula_sum 'I3 P'

loop_
_publ_author_name
 'E. T. Lance'
 'J. M. Haschke'
 'D. R. Peacor'
_journal_name_full_name
;
 Inorganic Chemistry
;
_journal_volume 15
_journal_year 1976
_journal_page_first 780
_journal_page_last 781
_publ_Section_title
;
 Crystal and molecular structure of phosphorus triiodide
;

# Found in Pearson's Crystal Data - Crystal Structure Database for Inorganic Compounds, 2013

_aflow_title 'PI$_{3}$ Structure'
_aflow_proto 'A3B_hP8_173_c_b'
_aflow_params 'a,c/a,z_{1},x_{2},y_{2},z_{2}'
_aflow_params_values '7.1329719992,1.03939436422,0.0,0.0337,0.3475,0.146'
_aflow_Strukturbericht 'None'
_aflow_Pearson 'hP8'

_cell_length_a    7.1329719992
_cell_length_b    7.1329719992
_cell_length_c    7.4139708961
_cell_angle_alpha 90.0000000000
_cell_angle_beta  90.0000000000
_cell_angle_gamma 120.0000000000
 
_symmetry_space_group_name_H-M "P 63"
_symmetry_Int_Tables_number 173
 
loop_
_space_group_symop_id
_space_group_symop_operation_xyz
1 x,y,z
2 x-y,x,z+1/2
3 -y,x-y,z
4 -x,-y,z+1/2
5 -x+y,-x,z
6 y,-x+y,z+1/2
 
loop_
_atom_site_label
_atom_site_type_symbol
_atom_site_symmetry_multiplicity
_atom_site_Wyckoff_label
_atom_site_fract_x
_atom_site_fract_y
_atom_site_fract_z
_atom_site_occupancy
P1 P   2 b 0.33333 0.66667 0.00000 1.00000
I1 I   6 c 0.03370 0.34750 0.14600 1.00000
\end{lstlisting}
{\phantomsection\label{A3B_hP8_173_c_b_poscar}}
{\hyperref[A3B_hP8_173_c_b]{PI$_{3}$: A3B\_hP8\_173\_c\_b}} - POSCAR
\begin{lstlisting}[numbers=none,language={mylang}]
A3B_hP8_173_c_b & a,c/a,z1,x2,y2,z2 --params=7.1329719992,1.03939436422,0.0,0.0337,0.3475,0.146 & P6_{3} C_{6}^{6} #173 (bc) & hP8 & None & PI3 &  & E. T. Lance and J. M. Haschke and D. R. Peacor, Inorg. Chem. 15, 780-781 (1976)
   1.00000000000000
   3.56648599960000  -6.17733495579027   0.00000000000000
   3.56648599960000   6.17733495579027   0.00000000000000
   0.00000000000000   0.00000000000000   7.41397089610000
     I     P
     6     2
Direct
   0.03370000000000   0.34750000000000   0.14600000000000    I   (6c)
  -0.34750000000000  -0.31380000000000   0.14600000000000    I   (6c)
   0.31380000000000  -0.03370000000000   0.14600000000000    I   (6c)
  -0.03370000000000  -0.34750000000000   0.64600000000000    I   (6c)
   0.34750000000000   0.31380000000000   0.64600000000000    I   (6c)
  -0.31380000000000   0.03370000000000   0.64600000000000    I   (6c)
   0.33333333333333   0.66666666666667   0.00000000000000    P   (2b)
   0.66666666666667   0.33333333333333   0.50000000000000    P   (2b)
\end{lstlisting}
{\phantomsection\label{A4B3_hP14_173_bc_c_cif}}
{\hyperref[A4B3_hP14_173_bc_c]{$\beta$-Si$_{3}$N$_{4}$: A4B3\_hP14\_173\_bc\_c}} - CIF
\begin{lstlisting}[numbers=none,language={mylang}]
# CIF file
data_findsym-output
_audit_creation_method FINDSYM

_chemical_name_mineral 'beta-Si3N4'
_chemical_formula_sum 'N4 Si3'

loop_
_publ_author_name
 'W. D. Forgeng'
 'B. F. Decker'
_journal_name_full_name
;
 Transactions of the American Institute of Mining and Metallurgical Engineers
;
_journal_volume 212
_journal_year 1958
_journal_page_first 343
_journal_page_last 348
_publ_Section_title
;
 Nitrides of silicon
;

# Found in Pearson's Crystal Data - Crystal Structure Database for Inorganic Compounds, 2013

_aflow_title '$\beta$-Si$_{3}$N$_{4}$ Structure'
_aflow_proto 'A4B3_hP14_173_bc_c'
_aflow_params 'a,c/a,z_{1},x_{2},y_{2},z_{2},x_{3},y_{3},z_{3}'
_aflow_params_values '7.603038022,0.382612126795,0.0,0.3284,0.0313,0.05,0.2314,0.4063,0.013'
_aflow_Strukturbericht 'None'
_aflow_Pearson 'hP14'

_cell_length_a    7.6030380220
_cell_length_b    7.6030380220
_cell_length_c    2.9090145477
_cell_angle_alpha 90.0000000000
_cell_angle_beta  90.0000000000
_cell_angle_gamma 120.0000000000
 
_symmetry_space_group_name_H-M "P 63"
_symmetry_Int_Tables_number 173
 
loop_
_space_group_symop_id
_space_group_symop_operation_xyz
1 x,y,z
2 x-y,x,z+1/2
3 -y,x-y,z
4 -x,-y,z+1/2
5 -x+y,-x,z
6 y,-x+y,z+1/2
 
loop_
_atom_site_label
_atom_site_type_symbol
_atom_site_symmetry_multiplicity
_atom_site_Wyckoff_label
_atom_site_fract_x
_atom_site_fract_y
_atom_site_fract_z
_atom_site_occupancy
N1  N    2 b 0.33333 0.66667 0.00000 1.00000
N2  N    6 c 0.32840 0.03130 0.05000 1.00000
Si1 Si   6 c 0.23140 0.40630 0.01300 1.00000
\end{lstlisting}
{\phantomsection\label{A4B3_hP14_173_bc_c_poscar}}
{\hyperref[A4B3_hP14_173_bc_c]{$\beta$-Si$_{3}$N$_{4}$: A4B3\_hP14\_173\_bc\_c}} - POSCAR
\begin{lstlisting}[numbers=none,language={mylang}]
A4B3_hP14_173_bc_c & a,c/a,z1,x2,y2,z2,x3,y3,z3 --params=7.603038022,0.382612126795,0.0,0.3284,0.0313,0.05,0.2314,0.4063,0.013 & P6_{3} C_{6}^{6} #173 (bc^2) & hP14 & None & Si3N4 & beta & W. D. Forgeng and B. F. Decker, T. Am. I. Min. Met. Eng. 212, 343-348 (1958)
   1.00000000000000
   3.80151901100000  -6.58442407299099   0.00000000000000
   3.80151901100000   6.58442407299099   0.00000000000000
   0.00000000000000   0.00000000000000   2.90901454770000
     N    Si
     8     6
Direct
   0.33333333333333   0.66666666666667   0.00000000000000    N   (2b)
   0.66666666666667   0.33333333333333   0.50000000000000    N   (2b)
   0.32840000000000   0.03130000000000   0.05000000000000    N   (6c)
  -0.03130000000000   0.29710000000000   0.05000000000000    N   (6c)
  -0.29710000000000  -0.32840000000000   0.05000000000000    N   (6c)
  -0.32840000000000  -0.03130000000000   0.55000000000000    N   (6c)
   0.03130000000000  -0.29710000000000   0.55000000000000    N   (6c)
   0.29710000000000   0.32840000000000   0.55000000000000    N   (6c)
   0.23140000000000   0.40630000000000   0.01300000000000   Si   (6c)
  -0.40630000000000  -0.17490000000000   0.01300000000000   Si   (6c)
   0.17490000000000  -0.23140000000000   0.01300000000000   Si   (6c)
  -0.23140000000000  -0.40630000000000   0.51300000000000   Si   (6c)
   0.40630000000000   0.17490000000000   0.51300000000000   Si   (6c)
  -0.17490000000000   0.23140000000000   0.51300000000000   Si   (6c)
\end{lstlisting}
{\phantomsection\label{A12B7C2_hP21_174_2j2k_ajk_cf_cif}}
{\hyperref[A12B7C2_hP21_174_2j2k_ajk_cf]{Fe$_{12}$Zr$_{2}$P$_{7}$: A12B7C2\_hP21\_174\_2j2k\_ajk\_cf}} - CIF
\begin{lstlisting}[numbers=none,language={mylang}]
# CIF file
data_findsym-output
_audit_creation_method FINDSYM

_chemical_name_mineral 'Fe12Zr2P7'
_chemical_formula_sum 'Fe12 P7 Zr2'

loop_
_publ_author_name
 'E. Ganglberger'
_journal_name_full_name
;
 Monatshefte f{\"u}r Chemie - Chemical Monthly
;
_journal_volume 99
_journal_year 1968
_journal_page_first 557
_journal_page_last 565
_publ_Section_title
;
 Die Kristallstruktur von Fe$_{12}$Zr$_{2}$P$_{7}$
;

# Found in Pearson's Crystal Data - Crystal Structure Database for Inorganic Compounds, 2013

_aflow_title 'Fe$_{12}$Zr$_{2}$P$_{7}$ Structure'
_aflow_proto 'A12B7C2_hP21_174_2j2k_ajk_cf'
_aflow_params 'a,c/a,x_{4},y_{4},x_{5},y_{5},x_{6},y_{6},x_{7},y_{7},x_{8},y_{8},x_{9},y_{9}'
_aflow_params_values '9.0004021308,0.399102242177,0.4309,0.3719,0.1189,0.2772,0.4163,0.1204,0.0495,0.4359,0.2232,0.124,0.2889,0.4096'
_aflow_Strukturbericht 'None'
_aflow_Pearson 'hP21'

_cell_length_a    9.0004021308
_cell_length_b    9.0004021308
_cell_length_c    3.5920806709
_cell_angle_alpha 90.0000000000
_cell_angle_beta  90.0000000000
_cell_angle_gamma 120.0000000000
 
_symmetry_space_group_name_H-M "P -6"
_symmetry_Int_Tables_number 174
 
loop_
_space_group_symop_id
_space_group_symop_operation_xyz
1 x,y,z
2 -y,x-y,z
3 -x+y,-x,z
4 -x+y,-x,-z
5 x,y,-z
6 -y,x-y,-z
 
loop_
_atom_site_label
_atom_site_type_symbol
_atom_site_symmetry_multiplicity
_atom_site_Wyckoff_label
_atom_site_fract_x
_atom_site_fract_y
_atom_site_fract_z
_atom_site_occupancy
P1  P    1 a 0.00000 0.00000 0.00000 1.00000
Zr1 Zr   1 c 0.33333 0.66667 0.00000 1.00000
Zr2 Zr   1 f 0.66667 0.33333 0.50000 1.00000
Fe1 Fe   3 j 0.43090 0.37190 0.00000 1.00000
Fe2 Fe   3 j 0.11890 0.27720 0.00000 1.00000
P2  P    3 j 0.41630 0.12040 0.00000 1.00000
Fe3 Fe   3 k 0.04950 0.43590 0.50000 1.00000
Fe4 Fe   3 k 0.22320 0.12400 0.50000 1.00000
P3  P    3 k 0.28890 0.40960 0.50000 1.00000
\end{lstlisting}
{\phantomsection\label{A12B7C2_hP21_174_2j2k_ajk_cf_poscar}}
{\hyperref[A12B7C2_hP21_174_2j2k_ajk_cf]{Fe$_{12}$Zr$_{2}$P$_{7}$: A12B7C2\_hP21\_174\_2j2k\_ajk\_cf}} - POSCAR

{\phantomsection\label{ABC_hP12_174_cj_fk_aj_cif}}
{\hyperref[ABC_hP12_174_cj_fk_aj]{GdSI: ABC\_hP12\_174\_cj\_fk\_aj}} - CIF
\begin{lstlisting}[numbers=none,language={mylang}]
# CIF file
data_findsym-output
_audit_creation_method FINDSYM

_chemical_name_mineral 'GdSI'
_chemical_formula_sum 'Gd I S'

loop_
_publ_author_name
 'C. Dagron'
 'F. Thevet'
_journal_name_full_name
;
 Comptes Rendus Hebdomadaires des S{\'eances de l'Acad{\'e}mie des Sciences S{\'e}rie C - Sciences chimiques
;
_journal_volume 268
_journal_year 1969
_journal_page_first 1867
_journal_page_last 1869
_publ_Section_title
;
 R{\\'e}partition des types cristallins dans la s{\\'e}rie des iodosulfures et fluorosulfures des {\\'e}l{\\'e}ments des terres rares et d\'yttrium
;

# Found in Pearson's Crystal Data - Crystal Structure Database for Inorganic Compounds, 2013

_aflow_title 'GdSI Structure'
_aflow_proto 'ABC_hP12_174_cj_fk_aj'
_aflow_params 'a,c/a,x_{4},y_{4},x_{5},y_{5},x_{6},y_{6}'
_aflow_params_values '10.7303215747,0.395153774459,0.30167,0.15433,0.03467,0.51733,0.14967,0.31433'
_aflow_Strukturbericht 'None'
_aflow_Pearson 'hP12'

_cell_length_a    10.7303215747
_cell_length_b    10.7303215747
_cell_length_c    4.2401270714
_cell_angle_alpha 90.0000000000
_cell_angle_beta  90.0000000000
_cell_angle_gamma 120.0000000000
 
_symmetry_space_group_name_H-M "P -6"
_symmetry_Int_Tables_number 174
 
loop_
_space_group_symop_id
_space_group_symop_operation_xyz
1 x,y,z
2 -y,x-y,z
3 -x+y,-x,z
4 -x+y,-x,-z
5 x,y,-z
6 -y,x-y,-z
 
loop_
_atom_site_label
_atom_site_type_symbol
_atom_site_symmetry_multiplicity
_atom_site_Wyckoff_label
_atom_site_fract_x
_atom_site_fract_y
_atom_site_fract_z
_atom_site_occupancy
S1  S    1 a 0.00000 0.00000 0.00000 1.00000
Gd1 Gd   1 c 0.33333 0.66667 0.00000 1.00000
I1  I    1 f 0.66667 0.33333 0.50000 1.00000
Gd2 Gd   3 j 0.30167 0.15433 0.00000 1.00000
S2  S    3 j 0.03467 0.51733 0.00000 1.00000
I2  I    3 k 0.14967 0.31433 0.50000 1.00000
\end{lstlisting}
{\phantomsection\label{ABC_hP12_174_cj_fk_aj_poscar}}
{\hyperref[ABC_hP12_174_cj_fk_aj]{GdSI: ABC\_hP12\_174\_cj\_fk\_aj}} - POSCAR
\begin{lstlisting}[numbers=none,language={mylang}]
ABC_hP12_174_cj_fk_aj & a,c/a,x4,y4,x5,y5,x6,y6 --params=10.7303215747,0.395153774459,0.30167,0.15433,0.03467,0.51733,0.14967,0.31433 & P-6 C_{3h}^{1} #174 (acfj^2k) & hP12 & None & GdSI &  & C. Dagron and F. Thevet, {C. R. Hebd. S{'e}ances Acad. Sci. C 268, 1867-1869 (1969)
   1.00000000000000
   5.36516078735000  -9.29273107446644   0.00000000000000
   5.36516078735000   9.29273107446644   0.00000000000000
   0.00000000000000   0.00000000000000   4.24012707140000
    Gd     I     S
     4     4     4
Direct
   0.33333333333333   0.66666666666667   0.00000000000000   Gd   (1c)
   0.30167000000000   0.15433000000000   0.00000000000000   Gd   (3j)
  -0.15433000000000   0.14734000000000   0.00000000000000   Gd   (3j)
  -0.14734000000000  -0.30167000000000   0.00000000000000   Gd   (3j)
   0.66666666666667   0.33333333333333   0.50000000000000    I   (1f)
   0.14967000000000   0.31433000000000   0.50000000000000    I   (3k)
  -0.31433000000000  -0.16466000000000   0.50000000000000    I   (3k)
   0.16466000000000  -0.14967000000000   0.50000000000000    I   (3k)
   0.00000000000000   0.00000000000000   0.00000000000000    S   (1a)
   0.03467000000000   0.51733000000000   0.00000000000000    S   (3j)
  -0.51733000000000  -0.48266000000000   0.00000000000000    S   (3j)
   0.48266000000000  -0.03467000000000   0.00000000000000    S   (3j)
\end{lstlisting}
{\phantomsection\label{A8B7C6_hP21_175_ck_aj_k_cif}}
{\hyperref[A8B7C6_hP21_175_ck_aj_k]{Nb$_{7}$Ru$_{6}$B$_{8}$: A8B7C6\_hP21\_175\_ck\_aj\_k}} - CIF
\begin{lstlisting}[numbers=none,language={mylang}]
# CIF file
data_findsym-output
_audit_creation_method FINDSYM

_chemical_name_mineral 'Nb7Ru6B8'
_chemical_formula_sum 'B8 Nb7 Ru6'

loop_
_publ_author_name
 'Q. Zheng'
 'M. Kohout'
 'R. Gumeniuk'
 'N. Abramchuk'
 'H. Borrmann'
 'Y. Prots'
 'U. Burkhardt'
 'W. Schnelle'
 'L. Akselrud'
 'H. Gu'
 'A. {Leithe-Jasper}'
 'Y. Grin'
_journal_name_full_name
;
 Inorganic Chemistry
;
_journal_volume 51
_journal_year 2012
_journal_page_first 7472
_journal_page_last 7483
_publ_Section_title
;
 TM$_{7}$ TM\'$_{6}$B$_{8}$ (TM = Ta, Nb; TM\' = Ru, Rh, Ir): New Compounds with [B$_{6}$] Ring Polyanions
;

# Found in Pearson's Crystal Data - Crystal Structure Database for Inorganic Compounds, 2013

_aflow_title 'Nb$_{7}$Ru$_{6}$B$_{8}$ Structure'
_aflow_proto 'A8B7C6_hP21_175_ck_aj_k'
_aflow_params 'a,c/a,x_{3},y_{3},x_{4},y_{4},x_{5},y_{5}'
_aflow_params_values '9.5057625379,0.329093950194,0.36335,0.08627,0.0661,0.221,0.15405,0.51373'
_aflow_Strukturbericht 'None'
_aflow_Pearson 'hP21'

_cell_length_a    9.5057625379
_cell_length_b    9.5057625379
_cell_length_c    3.1282889432
_cell_angle_alpha 90.0000000000
_cell_angle_beta  90.0000000000
_cell_angle_gamma 120.0000000000
 
_symmetry_space_group_name_H-M "P 6/m"
_symmetry_Int_Tables_number 175
 
loop_
_space_group_symop_id
_space_group_symop_operation_xyz
1 x,y,z
2 x-y,x,z
3 -y,x-y,z
4 -x,-y,z
5 -x+y,-x,z
6 y,-x+y,z
7 -x,-y,-z
8 -x+y,-x,-z
9 y,-x+y,-z
10 x,y,-z
11 x-y,x,-z
12 -y,x-y,-z
 
loop_
_atom_site_label
_atom_site_type_symbol
_atom_site_symmetry_multiplicity
_atom_site_Wyckoff_label
_atom_site_fract_x
_atom_site_fract_y
_atom_site_fract_z
_atom_site_occupancy
Nb1 Nb   1 a 0.00000 0.00000 0.00000 1.00000
B1  B    2 c 0.33333 0.66667 0.00000 1.00000
Nb2 Nb   6 j 0.36335 0.08627 0.00000 1.00000
B2  B    6 k 0.06610 0.22100 0.50000 1.00000
Ru1 Ru   6 k 0.15405 0.51373 0.50000 1.00000
\end{lstlisting}
{\phantomsection\label{A8B7C6_hP21_175_ck_aj_k_poscar}}
{\hyperref[A8B7C6_hP21_175_ck_aj_k]{Nb$_{7}$Ru$_{6}$B$_{8}$: A8B7C6\_hP21\_175\_ck\_aj\_k}} - POSCAR

{\phantomsection\label{ABC_hP36_175_jk_jk_jk_cif}}
{\hyperref[ABC_hP36_175_jk_jk_jk]{Mg[NH]: ABC\_hP36\_175\_jk\_jk\_jk}} - CIF

{\phantomsection\label{ABC_hP36_175_jk_jk_jk_poscar}}
{\hyperref[ABC_hP36_175_jk_jk_jk]{Mg[NH]: ABC\_hP36\_175\_jk\_jk\_jk}} - POSCAR

{\phantomsection\label{A3B2_hP10_176_h_bd_cif}}
{\hyperref[A3B2_hP10_176_h_bd]{Er$_{3}$Ru$_{2}$: A3B2\_hP10\_176\_h\_bd}} - CIF
\begin{lstlisting}[numbers=none,language={mylang}]
# CIF file
data_findsym-output
_audit_creation_method FINDSYM

_chemical_name_mineral 'Er3Ru2'
_chemical_formula_sum 'Er3 Ru2'

loop_
_publ_author_name
 'A. Palenzona'
_journal_name_full_name
;
 Journal of the Less-Common Metals
;
_journal_volume 159
_journal_year 1990
_journal_page_first L21
_journal_page_last L23
_publ_Section_title
;
 The phase diagram of the Er-Ru system
;

# Found in Pearson's Crystal Data - Crystal Structure Database for Inorganic Compounds, 2013

_aflow_title 'Er$_{3}$Ru$_{2}$ Structure'
_aflow_proto 'A3B2_hP10_176_h_bd'
_aflow_params 'a,c/a,x_{3},y_{3}'
_aflow_params_values '7.8700439404,0.50025412961,0.0915,0.7068'
_aflow_Strukturbericht 'None'
_aflow_Pearson 'hP10'

_cell_length_a    7.8700439404
_cell_length_b    7.8700439404
_cell_length_c    3.9370219814
_cell_angle_alpha 90.0000000000
_cell_angle_beta  90.0000000000
_cell_angle_gamma 120.0000000000
 
_symmetry_space_group_name_H-M "P 63/m"
_symmetry_Int_Tables_number 176
 
loop_
_space_group_symop_id
_space_group_symop_operation_xyz
1 x,y,z
2 x-y,x,z+1/2
3 -y,x-y,z
4 -x,-y,z+1/2
5 -x+y,-x,z
6 y,-x+y,z+1/2
7 -x,-y,-z
8 -x+y,-x,-z+1/2
9 y,-x+y,-z
10 x,y,-z+1/2
11 x-y,x,-z
12 -y,x-y,-z+1/2
 
loop_
_atom_site_label
_atom_site_type_symbol
_atom_site_symmetry_multiplicity
_atom_site_Wyckoff_label
_atom_site_fract_x
_atom_site_fract_y
_atom_site_fract_z
_atom_site_occupancy
Ru1 Ru   2 b 0.00000 0.00000 0.00000 1.00000
Ru2 Ru   2 d 0.66667 0.33333 0.25000 1.00000
Er1 Er   6 h 0.09150 0.70680 0.25000 1.00000
\end{lstlisting}
{\phantomsection\label{A3B2_hP10_176_h_bd_poscar}}
{\hyperref[A3B2_hP10_176_h_bd]{Er$_{3}$Ru$_{2}$: A3B2\_hP10\_176\_h\_bd}} - POSCAR
\begin{lstlisting}[numbers=none,language={mylang}]
A3B2_hP10_176_h_bd & a,c/a,x3,y3 --params=7.8700439404,0.50025412961,0.0915,0.7068 & P6_{3}/m C_{6h}^{2} #176 (bdh) & hP10 & None & Er3Ru2 &  & A. Palenzona, J. Less-Common Met. 159, L21-L23 (1990)
   1.00000000000000
   3.93502197020000  -6.81565798128618   0.00000000000000
   3.93502197020000   6.81565798128618   0.00000000000000
   0.00000000000000   0.00000000000000   3.93702198140000
    Er    Ru
     6     4
Direct
   0.09150000000000   0.70680000000000   0.25000000000000   Er   (6h)
  -0.70680000000000  -0.61530000000000   0.25000000000000   Er   (6h)
   0.61530000000000  -0.09150000000000   0.25000000000000   Er   (6h)
  -0.09150000000000  -0.70680000000000   0.75000000000000   Er   (6h)
   0.70680000000000   0.61530000000000   0.75000000000000   Er   (6h)
  -0.61530000000000   0.09150000000000   0.75000000000000   Er   (6h)
   0.00000000000000   0.00000000000000   0.00000000000000   Ru   (2b)
   0.00000000000000   0.00000000000000   0.50000000000000   Ru   (2b)
   0.66666666666667   0.33333333333333   0.25000000000000   Ru   (2d)
   0.33333333333333   0.66666666666667   0.75000000000000   Ru   (2d)
\end{lstlisting}
{\phantomsection\label{A3B3C_hP14_176_h_h_d_cif}}
{\hyperref[A3B3C_hP14_176_h_h_d]{Fe$_{3}$Te$_{3}$Tl: A3B3C\_hP14\_176\_h\_h\_d}} - CIF
\begin{lstlisting}[numbers=none,language={mylang}]
# CIF file
data_findsym-output
_audit_creation_method FINDSYM

_chemical_name_mineral 'Fe3Te3Tl'
_chemical_formula_sum 'Fe3 Te3 Tl'

loop_
_publ_author_name
 'K. Klepp'
 'H. Boller'
_journal_name_full_name
;
 Acta Crystallographica Section A: Foundations and Advances
;
_journal_volume 34
_journal_year 1978
_journal_page_first S160
_journal_page_last S161
_publ_Section_title
;
 Crystal Structures of Thallium-Iron Chalcogenides
;

# Found in Pearson's Crystal Data - Crystal Structure Database for Inorganic Compounds, 2013

_aflow_title 'Fe$_{3}$Te$_{3}$Tl Structure'
_aflow_proto 'A3B3C_hP14_176_h_h_d'
_aflow_params 'a,c/a,x_{2},y_{2},x_{3},y_{3}'
_aflow_params_values '9.3500327107,0.451336898402,0.1701,0.0208,0.0462,0.6892'
_aflow_Strukturbericht 'None'
_aflow_Pearson 'hP14'

_cell_length_a    9.3500327107
_cell_length_b    9.3500327107
_cell_length_c    4.2200147636
_cell_angle_alpha 90.0000000000
_cell_angle_beta  90.0000000000
_cell_angle_gamma 120.0000000000
 
_symmetry_space_group_name_H-M "P 63/m"
_symmetry_Int_Tables_number 176
 
loop_
_space_group_symop_id
_space_group_symop_operation_xyz
1 x,y,z
2 x-y,x,z+1/2
3 -y,x-y,z
4 -x,-y,z+1/2
5 -x+y,-x,z
6 y,-x+y,z+1/2
7 -x,-y,-z
8 -x+y,-x,-z+1/2
9 y,-x+y,-z
10 x,y,-z+1/2
11 x-y,x,-z
12 -y,x-y,-z+1/2
 
loop_
_atom_site_label
_atom_site_type_symbol
_atom_site_symmetry_multiplicity
_atom_site_Wyckoff_label
_atom_site_fract_x
_atom_site_fract_y
_atom_site_fract_z
_atom_site_occupancy
Tl1 Tl   2 d 0.66667 0.33333 0.25000 1.00000
Fe1 Fe   6 h 0.17010 0.02080 0.25000 1.00000
Te1 Te   6 h 0.04620 0.68920 0.25000 1.00000
\end{lstlisting}
{\phantomsection\label{A3B3C_hP14_176_h_h_d_poscar}}
{\hyperref[A3B3C_hP14_176_h_h_d]{Fe$_{3}$Te$_{3}$Tl: A3B3C\_hP14\_176\_h\_h\_d}} - POSCAR
\begin{lstlisting}[numbers=none,language={mylang}]
A3B3C_hP14_176_h_h_d & a,c/a,x2,y2,x3,y3 --params=9.3500327107,0.451336898402,0.1701,0.0208,0.0462,0.6892 & P6_{3}/m C_{6h}^{2} #176 (dh^2) & hP14 & None & Fe3Te3Tl &  & K. Klepp and H. Boller, Acta Crystallogr. Sect. A 34, S160-S161 (1978)
   1.00000000000000
   4.67501635535000  -8.09736585368168   0.00000000000000
   4.67501635535000   8.09736585368168   0.00000000000000
   0.00000000000000   0.00000000000000   4.22001476360000
    Fe    Te    Tl
     6     6     2
Direct
   0.17010000000000   0.02080000000000   0.25000000000000   Fe   (6h)
  -0.02080000000000   0.14930000000000   0.25000000000000   Fe   (6h)
  -0.14930000000000  -0.17010000000000   0.25000000000000   Fe   (6h)
  -0.17010000000000  -0.02080000000000   0.75000000000000   Fe   (6h)
   0.02080000000000  -0.14930000000000   0.75000000000000   Fe   (6h)
   0.14930000000000   0.17010000000000   0.75000000000000   Fe   (6h)
   0.04620000000000   0.68920000000000   0.25000000000000   Te   (6h)
  -0.68920000000000  -0.64300000000000   0.25000000000000   Te   (6h)
   0.64300000000000  -0.04620000000000   0.25000000000000   Te   (6h)
  -0.04620000000000  -0.68920000000000   0.75000000000000   Te   (6h)
   0.68920000000000   0.64300000000000   0.75000000000000   Te   (6h)
  -0.64300000000000   0.04620000000000   0.75000000000000   Te   (6h)
   0.66666666666667   0.33333333333333   0.25000000000000   Tl   (2d)
   0.33333333333333   0.66666666666667   0.75000000000000   Tl   (2d)
\end{lstlisting}
{\phantomsection\label{A3B_hP8_176_h_d_cif}}
{\hyperref[A3B_hP8_176_h_d]{UCl$_{3}$: A3B\_hP8\_176\_h\_d}} - CIF
\begin{lstlisting}[numbers=none,language={mylang}]
# CIF file
data_findsym-output
_audit_creation_method FINDSYM

_chemical_name_mineral 'UCl3'
_chemical_formula_sum 'Cl3 U'

loop_
_publ_author_name
 'W. H. Zachariasen'
_journal_name_full_name
;
 Acta Cristallographica
;
_journal_volume 1
_journal_year 1948
_journal_page_first 265
_journal_page_last 268
_publ_Section_title
;
 Crystal chemical studies of the 5f-series of elements. I. New structure types
;

# Found in Pearson's Crystal Data - Crystal Structure Database for Inorganic Compounds, 2013

_aflow_title 'UCl$_{3}$ Structure'
_aflow_proto 'A3B_hP8_176_h_d'
_aflow_params 'a,c/a,x_{2},y_{2}'
_aflow_params_values '7.4429335392,0.580545478976,0.083,0.708'
_aflow_Strukturbericht 'None'
_aflow_Pearson 'hP8'

_cell_length_a    7.4429335392
_cell_length_b    7.4429335392
_cell_length_c    4.3209614165
_cell_angle_alpha 90.0000000000
_cell_angle_beta  90.0000000000
_cell_angle_gamma 120.0000000000
 
_symmetry_space_group_name_H-M "P 63/m"
_symmetry_Int_Tables_number 176
 
loop_
_space_group_symop_id
_space_group_symop_operation_xyz
1 x,y,z
2 x-y,x,z+1/2
3 -y,x-y,z
4 -x,-y,z+1/2
5 -x+y,-x,z
6 y,-x+y,z+1/2
7 -x,-y,-z
8 -x+y,-x,-z+1/2
9 y,-x+y,-z
10 x,y,-z+1/2
11 x-y,x,-z
12 -y,x-y,-z+1/2
 
loop_
_atom_site_label
_atom_site_type_symbol
_atom_site_symmetry_multiplicity
_atom_site_Wyckoff_label
_atom_site_fract_x
_atom_site_fract_y
_atom_site_fract_z
_atom_site_occupancy
U1  U    2 d 0.66667 0.33333 0.25000 1.00000
Cl1 Cl   6 h 0.08300 0.70800 0.25000 1.00000
\end{lstlisting}
{\phantomsection\label{A3B_hP8_176_h_d_poscar}}
{\hyperref[A3B_hP8_176_h_d]{UCl$_{3}$: A3B\_hP8\_176\_h\_d}} - POSCAR
\begin{lstlisting}[numbers=none,language={mylang}]
A3B_hP8_176_h_d & a,c/a,x2,y2 --params=7.4429335392,0.580545478976,0.083,0.708 & P6_{3}/m C_{6h}^{2} #176 (dh) & hP8 & None & UCl3 &  & W. H. Zachariasen, Acta Cryst. 1, 265-268 (1948)
   1.00000000000000
   3.72146676960000  -6.44576952362642   0.00000000000000
   3.72146676960000   6.44576952362642   0.00000000000000
   0.00000000000000   0.00000000000000   4.32096141650000
    Cl     U
     6     2
Direct
   0.08300000000000   0.70800000000000   0.25000000000000   Cl   (6h)
  -0.70800000000000  -0.62500000000000   0.25000000000000   Cl   (6h)
   0.62500000000000  -0.08300000000000   0.25000000000000   Cl   (6h)
  -0.08300000000000  -0.70800000000000   0.75000000000000   Cl   (6h)
   0.70800000000000   0.62500000000000   0.75000000000000   Cl   (6h)
  -0.62500000000000   0.08300000000000   0.75000000000000   Cl   (6h)
   0.66666666666667   0.33333333333333   0.25000000000000    U   (2d)
   0.33333333333333   0.66666666666667   0.75000000000000    U   (2d)
\end{lstlisting}
{\phantomsection\label{A2B_hP36_177_j2lm_n_cif}}
{\hyperref[A2B_hP36_177_j2lm_n]{SiO$_{2}$: A2B\_hP36\_177\_j2lm\_n}} - CIF
\begin{lstlisting}[numbers=none,language={mylang}]
# CIF file
data_findsym-output
_audit_creation_method FINDSYM

_chemical_name_mineral 'SiO2'
_chemical_formula_sum 'O2 Si'

_aflow_title 'SiO$_{2}$ Structure'
_aflow_proto 'A2B_hP36_177_j2lm_n'
_aflow_params 'a,c/a,x_{1},x_{2},x_{3},x_{4},x_{5},y_{5},z_{5}'
_aflow_params_values '12.7835,0.291064262526,0.61855,0.39242,0.79257,0.44445,0.52169,0.86952,0.16458'
_aflow_Strukturbericht 'None'
_aflow_Pearson 'hP36'

_cell_length_a    12.7835000000
_cell_length_b    12.7835000000
_cell_length_c    3.7208200000
_cell_angle_alpha 90.0000000000
_cell_angle_beta  90.0000000000
_cell_angle_gamma 120.0000000000
 
_symmetry_space_group_name_H-M "P 6 2 2"
_symmetry_Int_Tables_number 177
 
loop_
_space_group_symop_id
_space_group_symop_operation_xyz
1 x,y,z
2 x-y,x,z
3 -y,x-y,z
4 -x,-y,z
5 -x+y,-x,z
6 y,-x+y,z
7 x-y,-y,-z
8 x,x-y,-z
9 y,x,-z
10 -x+y,y,-z
11 -x,-x+y,-z
12 -y,-x,-z
 
loop_
_atom_site_label
_atom_site_type_symbol
_atom_site_symmetry_multiplicity
_atom_site_Wyckoff_label
_atom_site_fract_x
_atom_site_fract_y
_atom_site_fract_z
_atom_site_occupancy
O1  O    6 j 0.61855 0.00000 0.00000 1.00000
O2  O    6 l 0.39242 0.60758 0.00000 1.00000
O3  O    6 l 0.79257 0.20743 0.00000 1.00000
O4  O    6 m 0.44445 0.55555 0.50000 1.00000
Si1 Si  12 n 0.52169 0.86952 0.16458 1.00000
\end{lstlisting}
{\phantomsection\label{A2B_hP36_177_j2lm_n_poscar}}
{\hyperref[A2B_hP36_177_j2lm_n]{SiO$_{2}$: A2B\_hP36\_177\_j2lm\_n}} - POSCAR

{\phantomsection\label{AB3_hP24_178_b_ac_cif}}
{\hyperref[AB3_hP24_178_b_ac]{AuF$_{3}$: AB3\_hP24\_178\_b\_ac}} - CIF
\begin{lstlisting}[numbers=none,language={mylang}]
# CIF file
data_findsym-output
_audit_creation_method FINDSYM

_chemical_name_mineral 'AuF3'
_chemical_formula_sum 'Au F3'

loop_
_publ_author_name
 'L. B. Asprey'
 'F. H. Kruse'
 'K. H. Jack'
 'R. Maitland'
_journal_name_full_name
;
 Inorganic Chemistry
;
_journal_volume 3
_journal_year 1964
_journal_page_first 602
_journal_page_last 604
_publ_Section_title
;
 Preparation and properties of crystalline gold trifluoride
;

# Found in Pearson's Crystal Data - Crystal Structure Database for Inorganic Compounds, 2013

_aflow_title 'AuF$_{3}$ Structure'
_aflow_proto 'AB3_hP24_178_b_ac'
_aflow_params 'a,c/a,x_{1},x_{2},x_{3},y_{3},z_{3}'
_aflow_params_values '5.14898393,3.15789473684,0.8361,0.7601,0.5338,0.3099,-0.0053'
_aflow_Strukturbericht 'None'
_aflow_Pearson 'hP24'

_cell_length_a    5.1489839300
_cell_length_b    5.1489839300
_cell_length_c    16.2599492526
_cell_angle_alpha 90.0000000000
_cell_angle_beta  90.0000000000
_cell_angle_gamma 120.0000000000
 
_symmetry_space_group_name_H-M "P 61 2 2"
_symmetry_Int_Tables_number 178
 
loop_
_space_group_symop_id
_space_group_symop_operation_xyz
1 x,y,z
2 x-y,x,z+1/6
3 -y,x-y,z+1/3
4 -x,-y,z+1/2
5 -x+y,-x,z+2/3
6 y,-x+y,z+5/6
7 x-y,-y,-z
8 x,x-y,-z+1/6
9 y,x,-z+1/3
10 -x+y,y,-z+1/2
11 -x,-x+y,-z+2/3
12 -y,-x,-z+5/6
 
loop_
_atom_site_label
_atom_site_type_symbol
_atom_site_symmetry_multiplicity
_atom_site_Wyckoff_label
_atom_site_fract_x
_atom_site_fract_y
_atom_site_fract_z
_atom_site_occupancy
F1  F    6 a 0.83610 0.00000 0.00000  1.00000
Au1 Au   6 b 0.76010 0.52020 0.25000  1.00000
F2  F   12 c 0.53380 0.30990 -0.00530 1.00000
\end{lstlisting}
{\phantomsection\label{AB3_hP24_178_b_ac_poscar}}
{\hyperref[AB3_hP24_178_b_ac]{AuF$_{3}$: AB3\_hP24\_178\_b\_ac}} - POSCAR

{\phantomsection\label{A_hP6_178_a_cif}}
{\hyperref[A_hP6_178_a]{Sc-V (High-pressure): A\_hP6\_178\_a}} - CIF
\begin{lstlisting}[numbers=none,language={mylang}]
# CIF file 
data_findsym-output
_audit_creation_method FINDSYM

_chemical_name_mineral 'Sc V'
_chemical_formula_sum 'Sc'

loop_
_publ_author_name
 'Y. Akahama'
 'H. Fujihisa'
 'H. Kawamura'
_journal_name_full_name
;
 Physical Review Letters
;
_journal_volume 94
_journal_year 2005
_journal_page_first 195503
_journal_page_last 195503
_publ_Section_title
;
 New Helical Chain Structure for Scandium at 240 GPa
;

_aflow_title 'Sc-V (High-pressure) Structure'
_aflow_proto 'A_hP6_178_a'
_aflow_params 'a,c/a,x_{1}'
_aflow_params_values '2.355,4.43566878981,0.461'
_aflow_Strukturbericht 'None'
_aflow_Pearson 'hP6'

_symmetry_space_group_name_H-M "P 61 2 2"
_symmetry_Int_Tables_number 178
 
_cell_length_a    2.35500
_cell_length_b    2.35500
_cell_length_c    10.44600
_cell_angle_alpha 90.00000
_cell_angle_beta  90.00000
_cell_angle_gamma 120.00000
 
loop_
_space_group_symop_id
_space_group_symop_operation_xyz
1 x,y,z
2 x-y,x,z+1/6
3 -y,x-y,z+1/3
4 -x,-y,z+1/2
5 -x+y,-x,z+2/3
6 y,-x+y,z+5/6
7 x-y,-y,-z
8 x,x-y,-z+1/6
9 y,x,-z+1/3
10 -x+y,y,-z+1/2
11 -x,-x+y,-z+2/3
12 -y,-x,-z+5/6
 
loop_
_atom_site_label
_atom_site_type_symbol
_atom_site_symmetry_multiplicity
_atom_site_Wyckoff_label
_atom_site_fract_x
_atom_site_fract_y
_atom_site_fract_z
_atom_site_occupancy
Sc1 Sc   6 a 0.46100 0.00000 0.00000 1.00000
\end{lstlisting}
{\phantomsection\label{A_hP6_178_a_poscar}}
{\hyperref[A_hP6_178_a]{Sc-V (High-pressure): A\_hP6\_178\_a}} - POSCAR
\begin{lstlisting}[numbers=none,language={mylang}]
A_hP6_178_a & a,c/a,x1 --params=2.355,4.43566878981,0.461 & P6_{1}22 D_{6}^{2} #178 (a) & hP6 & None & Sc & Sc V & Y. Akahama and H. Fujihisa and H. Kawamura, Phys. Rev. Lett. 94, 195503(2005)
   1.00000000000000
   1.17750000000000  -2.03948982591235   0.00000000000000
   1.17750000000000   2.03948982591235   0.00000000000000
   0.00000000000000   0.00000000000000  10.44600000000000
    Sc
     6
Direct
   0.46100000000000   0.00000000000000   0.00000000000000   Sc   (6a)
   0.00000000000000   0.46100000000000   0.33333333333333   Sc   (6a)
  -0.46100000000000  -0.46100000000000   0.66666666666667   Sc   (6a)
  -0.46100000000000   0.00000000000000   0.50000000000000   Sc   (6a)
   0.00000000000000  -0.46100000000000   0.83333333333333   Sc   (6a)
   0.46100000000000   0.46100000000000   0.16666666666667   Sc   (6a)
\end{lstlisting}
{\phantomsection\label{AB3_hP24_179_b_ac_cif}}
{\hyperref[AB3_hP24_179_b_ac]{AuF$_{3}$: AB3\_hP24\_179\_b\_ac}} - CIF
\begin{lstlisting}[numbers=none,language={mylang}]
# CIF file
data_findsym-output
_audit_creation_method FINDSYM

_chemical_name_mineral 'AuF3'
_chemical_formula_sum 'Au F3'

_aflow_title 'AuF$_{3}$ Structure'
_aflow_proto 'AB3_hP24_179_b_ac'
_aflow_params 'a,c/a,x_{1},x_{2},x_{3},y_{3},z_{3}'
_aflow_params_values '5.14898393,3.15789473684,0.8361,0.7601,0.5338,0.3099,0.0053'
_aflow_Strukturbericht 'None'
_aflow_Pearson 'hP24'

_cell_length_a    5.1489839300
_cell_length_b    5.1489839300
_cell_length_c    16.2599492526
_cell_angle_alpha 90.0000000000
_cell_angle_beta  90.0000000000
_cell_angle_gamma 120.0000000000
 
_symmetry_space_group_name_H-M "P 65 2 2"
_symmetry_Int_Tables_number 179
 
loop_
_space_group_symop_id
_space_group_symop_operation_xyz
1 x,y,z
2 x-y,x,z+5/6
3 -y,x-y,z+2/3
4 -x,-y,z+1/2
5 -x+y,-x,z+1/3
6 y,-x+y,z+1/6
7 x-y,-y,-z
8 x,x-y,-z+5/6
9 y,x,-z+2/3
10 -x+y,y,-z+1/2
11 -x,-x+y,-z+1/3
12 -y,-x,-z+1/6
 
loop_
_atom_site_label
_atom_site_type_symbol
_atom_site_symmetry_multiplicity
_atom_site_Wyckoff_label
_atom_site_fract_x
_atom_site_fract_y
_atom_site_fract_z
_atom_site_occupancy
F1  F    6 a 0.83610 0.00000 0.00000 1.00000
Au1 Au   6 b 0.76010 0.52020 0.75000 1.00000
F2  F   12 c 0.53380 0.30990 0.00530 1.00000
\end{lstlisting}
{\phantomsection\label{AB3_hP24_179_b_ac_poscar}}
{\hyperref[AB3_hP24_179_b_ac]{AuF$_{3}$: AB3\_hP24\_179\_b\_ac}} - POSCAR

{\phantomsection\label{A2B_hP9_181_j_c_cif}}
{\hyperref[A2B_hP9_181_j_c]{$\beta$-SiO$_{2}$: A2B\_hP9\_181\_j\_c}} - CIF
\begin{lstlisting}[numbers=none,language={mylang}]
# CIF file
data_findsym-output
_audit_creation_method FINDSYM

_chemical_name_mineral 'beta-SiO2'
_chemical_formula_sum 'O2 Si'

_aflow_title '$\beta$-SiO$_{2}$ Structure'
_aflow_proto 'A2B_hP9_181_j_c'
_aflow_params 'a,c/a,x_{2}'
_aflow_params_values '4.9977,1.09252256038,0.2072'
_aflow_Strukturbericht 'None'
_aflow_Pearson 'hP9'

_cell_length_a    4.9977000000
_cell_length_b    4.9977000000
_cell_length_c    5.4601000000
_cell_angle_alpha 90.0000000000
_cell_angle_beta  90.0000000000
_cell_angle_gamma 120.0000000000
 
_symmetry_space_group_name_H-M "P 64 2 2"
_symmetry_Int_Tables_number 181
 
loop_
_space_group_symop_id
_space_group_symop_operation_xyz
1 x,y,z
2 x-y,x,z+2/3
3 -y,x-y,z+1/3
4 -x,-y,z
5 -x+y,-x,z+2/3
6 y,-x+y,z+1/3
7 x-y,-y,-z
8 x,x-y,-z+2/3
9 y,x,-z+1/3
10 -x+y,y,-z
11 -x,-x+y,-z+2/3
12 -y,-x,-z+1/3
 
loop_
_atom_site_label
_atom_site_type_symbol
_atom_site_symmetry_multiplicity
_atom_site_Wyckoff_label
_atom_site_fract_x
_atom_site_fract_y
_atom_site_fract_z
_atom_site_occupancy
Si1 Si   3 c 0.50000 0.00000 0.00000 1.00000
O1  O    6 j 0.20720 0.41440 0.50000 1.00000
\end{lstlisting}
{\phantomsection\label{A2B_hP9_181_j_c_poscar}}
{\hyperref[A2B_hP9_181_j_c]{$\beta$-SiO$_{2}$: A2B\_hP9\_181\_j\_c}} - POSCAR
\begin{lstlisting}[numbers=none,language={mylang}]
A2B_hP9_181_j_c & a,c/a,x2 --params=4.9977,1.09252256038,0.2072 & P6_{4}22 D_{6}^{5} #181 (cj) & hP9 & None & SiO2 & beta & 
   1.00000000000000
   2.49885000000000  -4.32813516049349   0.00000000000000
   2.49885000000000   4.32813516049349   0.00000000000000
   0.00000000000000   0.00000000000000   5.46010000000000
     O    Si
     6     3
Direct
   0.20720000000000   0.41440000000000   0.50000000000000    O   (6j)
  -0.41440000000000  -0.20720000000000   0.83333333333333    O   (6j)
   0.20720000000000  -0.20720000000000   0.16666666666667    O   (6j)
  -0.20720000000000  -0.41440000000000   0.50000000000000    O   (6j)
   0.41440000000000   0.20720000000000   0.83333333333333    O   (6j)
  -0.20720000000000   0.20720000000000   0.16666666666667    O   (6j)
   0.50000000000000   0.00000000000000   0.00000000000000   Si   (3c)
   0.00000000000000   0.50000000000000   0.33333333333333   Si   (3c)
   0.50000000000000   0.50000000000000   0.66666666666667   Si   (3c)
\end{lstlisting}
{\phantomsection\label{ABC_hP3_183_a_a_a_cif}}
{\hyperref[ABC_hP3_183_a_a_a]{AuCN: ABC\_hP3\_183\_a\_a\_a}} - CIF
\begin{lstlisting}[numbers=none,language={mylang}]
# CIF file
data_findsym-output
_audit_creation_method FINDSYM

_chemical_name_mineral 'AuCN'
_chemical_formula_sum 'Au C N'

loop_
_publ_author_name
 'S. J. Hibble'
 'A. C. Hannon'
 'S. M. Cheyne'
_journal_name_full_name
;
 Inorganic Chemistry
;
_journal_volume 42
_journal_year 2003
_journal_page_first 4724
_journal_page_last 4730
_publ_Section_title
;
 Structure of AuCN determined from total neutron diffraction
;

# Found in Pearson's Crystal Data - Crystal Structure Database for Inorganic Compounds, 2013

_aflow_title 'AuCN Structure'
_aflow_proto 'ABC_hP3_183_a_a_a'
_aflow_params 'a,c/a,z_{1},z_{2},z_{3}'
_aflow_params_values '3.3908401495,1.49568037743,0.608,0.0,0.226'
_aflow_Strukturbericht 'None'
_aflow_Pearson 'hP3'

_cell_length_a    3.3908401495
_cell_length_b    3.3908401495
_cell_length_c    5.0716130746
_cell_angle_alpha 90.0000000000
_cell_angle_beta  90.0000000000
_cell_angle_gamma 120.0000000000
 
_symmetry_space_group_name_H-M "P 6 m m"
_symmetry_Int_Tables_number 183
 
loop_
_space_group_symop_id
_space_group_symop_operation_xyz
1 x,y,z
2 x-y,x,z
3 -y,x-y,z
4 -x,-y,z
5 -x+y,-x,z
6 y,-x+y,z
7 -x+y,y,z
8 -x,-x+y,z
9 -y,-x,z
10 x-y,-y,z
11 x,x-y,z
12 y,x,z
 
loop_
_atom_site_label
_atom_site_type_symbol
_atom_site_symmetry_multiplicity
_atom_site_Wyckoff_label
_atom_site_fract_x
_atom_site_fract_y
_atom_site_fract_z
_atom_site_occupancy
Au1 Au   1 a 0.00000 0.00000 0.60800 1.00000
C1  C    1 a 0.00000 0.00000 0.00000 1.00000
N1  N    1 a 0.00000 0.00000 0.22600 1.00000
\end{lstlisting}
{\phantomsection\label{ABC_hP3_183_a_a_a_poscar}}
{\hyperref[ABC_hP3_183_a_a_a]{AuCN: ABC\_hP3\_183\_a\_a\_a}} - POSCAR
\begin{lstlisting}[numbers=none,language={mylang}]
ABC_hP3_183_a_a_a & a,c/a,z1,z2,z3 --params=3.3908401495,1.49568037743,0.608,0.0,0.226 & P6mm C_{6v}^{1} #183 (a^3) & hP3 & None & AuCN &  & S. J. Hibble and A. C. Hannon and S. M. Cheyne, Inorg. Chem. 42, 4724-4730 (2003)
   1.00000000000000
   1.69542007475000  -2.93655370963922   0.00000000000000
   1.69542007475000   2.93655370963922   0.00000000000000
   0.00000000000000   0.00000000000000   5.07161307460000
    Au     C     N
     1     1     1
Direct
   0.00000000000000   0.00000000000000   0.60800000000000   Au   (1a)
   0.00000000000000   0.00000000000000   0.00000000000000    C   (1a)
   0.00000000000000   0.00000000000000   0.22600000000000    N   (1a)
\end{lstlisting}
{\phantomsection\label{AB_hP6_183_c_ab_cif}}
{\hyperref[AB_hP6_183_c_ab]{CrFe$_{3}$NiSn$_{5}$: AB\_hP6\_183\_c\_ab}} - CIF
\begin{lstlisting}[numbers=none,language={mylang}]
# CIF file
data_findsym-output
_audit_creation_method FINDSYM

_chemical_name_mineral 'CrFe3NiSn5'
_chemical_formula_sum 'M Sn'

loop_
_publ_author_name
 'J. Huang'
 'L. Zeng'
 'Z. Sun'
_journal_name_full_name
;
 Powder Diffraction
;
_journal_volume 19
_journal_year 2004
_journal_page_first 372
_journal_page_last 374
_publ_Section_title
;
 X-ray powder diffraction data and Rietveld refinement of CrFe$_{3}$NiSn$_{5}$
;

# Found in Pearson's Crystal Data - Crystal Structure Database for Inorganic Compounds, 2013

_aflow_title 'CrFe$_{3}$NiSn$_{5}$ Structure'
_aflow_proto 'AB_hP6_183_c_ab'
_aflow_params 'a,c/a,z_{1},z_{2},z_{3}'
_aflow_params_values '5.3175214551,0.83247442072,0.0,0.513,0.01'
_aflow_Strukturbericht 'None'
_aflow_Pearson 'hP6'

_cell_length_a    5.3175214551
_cell_length_b    5.3175214551
_cell_length_c    4.4267005930
_cell_angle_alpha 90.0000000000
_cell_angle_beta  90.0000000000
_cell_angle_gamma 120.0000000000
 
_symmetry_space_group_name_H-M "P 6 m m"
_symmetry_Int_Tables_number 183
 
loop_
_space_group_symop_id
_space_group_symop_operation_xyz
1 x,y,z
2 x-y,x,z
3 -y,x-y,z
4 -x,-y,z
5 -x+y,-x,z
6 y,-x+y,z
7 -x+y,y,z
8 -x,-x+y,z
9 -y,-x,z
10 x-y,-y,z
11 x,x-y,z
12 y,x,z
 
loop_
_atom_site_label
_atom_site_type_symbol
_atom_site_symmetry_multiplicity
_atom_site_Wyckoff_label
_atom_site_fract_x
_atom_site_fract_y
_atom_site_fract_z
_atom_site_occupancy
Sn1 Sn   1 a 0.00000 0.00000 0.00000 1.00000
Sn2 Sn   2 b 0.33333 0.66667 0.51300 1.00000
M1  M    3 c 0.50000 0.00000 0.01000 1.00000
\end{lstlisting}
{\phantomsection\label{AB_hP6_183_c_ab_poscar}}
{\hyperref[AB_hP6_183_c_ab]{CrFe$_{3}$NiSn$_{5}$: AB\_hP6\_183\_c\_ab}} - POSCAR
\begin{lstlisting}[numbers=none,language={mylang}]
AB_hP6_183_c_ab & a,c/a,z1,z2,z3 --params=5.3175214551,0.83247442072,0.0,0.513,0.01 & P6mm C_{6v}^{1} #183 (abc) & hP6 & None & CrFe3NiSn5 &  & J. Huang and L. Zeng and Z. Sun, Powder Diffraction 19, 372-374 (2004)
   1.00000000000000
   2.65876072755000  -4.60510866528539   0.00000000000000
   2.65876072755000   4.60510866528539   0.00000000000000
   0.00000000000000   0.00000000000000   4.42670059300000
     M    Sn
     3     3
Direct
   0.50000000000000   0.00000000000000   0.01000000000000    M   (3c)
   0.00000000000000   0.50000000000000   0.01000000000000    M   (3c)
   0.50000000000000   0.50000000000000   0.01000000000000    M   (3c)
   0.00000000000000   0.00000000000000   0.00000000000000   Sn   (1a)
   0.33333333333333   0.66666666666667   0.51300000000000   Sn   (2b)
   0.66666666666667   0.33333333333333   0.51300000000000   Sn   (2b)
\end{lstlisting}
{\phantomsection\label{AB4C_hP72_184_d_4d_d_cif}}
{\hyperref[AB4C_hP72_184_d_4d_d]{Al[PO$_{4}$] (Framework type AFI): AB4C\_hP72\_184\_d\_4d\_d}} - CIF

{\phantomsection\label{AB4C_hP72_184_d_4d_d_poscar}}
{\hyperref[AB4C_hP72_184_d_4d_d]{Al[PO$_{4}$] (Framework type AFI): AB4C\_hP72\_184\_d\_4d\_d}} - POSCAR

{\phantomsection\label{A3BC_hP30_185_cd_c_ab_cif}}
{\hyperref[A3BC_hP30_185_cd_c_ab]{KNiCl$_{3}$ (Room-temperature): A3BC\_hP30\_185\_cd\_c\_ab}} - CIF
\begin{lstlisting}[numbers=none,language={mylang}]
# CIF file
data_findsym-output
_audit_creation_method FINDSYM

_chemical_name_mineral 'KNiCl3'
_chemical_formula_sum 'Cl3 K Ni'

loop_
_publ_author_name
 'D. Visser'
 'G. C. Verschoor'
 'D. J. W. {IJdo}'
_journal_name_full_name
;
 Acta Crystallographica Section B: Structural Science
;
_journal_volume 36
_journal_year 1980
_journal_page_first 28
_journal_page_last 34
_publ_Section_title
;
 The structure of KNiCl$_{3}$ at room temperature
;

# Found in Pearson's Crystal Data - Crystal Structure Database for Inorganic Compounds, 2013

_aflow_title 'KNiCl$_{3}$ (Room-temperature) Structure'
_aflow_proto 'A3BC_hP30_185_cd_c_ab'
_aflow_params 'a,c/a,z_{1},z_{2},x_{3},z_{3},x_{4},z_{4},x_{5},y_{5},z_{5}'
_aflow_params_values '11.795090915,0.502416278086,0.0,0.377,0.1598,0.2396,0.6647,0.1706,0.1732,0.5056,0.1148'
_aflow_Strukturbericht 'None'
_aflow_Pearson 'hP30'

_cell_length_a    11.7950909150
_cell_length_b    11.7950909150
_cell_length_c    5.9260456772
_cell_angle_alpha 90.0000000000
_cell_angle_beta  90.0000000000
_cell_angle_gamma 120.0000000000
 
_symmetry_space_group_name_H-M "P 63 c m"
_symmetry_Int_Tables_number 185
 
loop_
_space_group_symop_id
_space_group_symop_operation_xyz
1 x,y,z
2 x-y,x,z+1/2
3 -y,x-y,z
4 -x,-y,z+1/2
5 -x+y,-x,z
6 y,-x+y,z+1/2
7 -x+y,y,z+1/2
8 -x,-x+y,z
9 -y,-x,z+1/2
10 x-y,-y,z
11 x,x-y,z+1/2
12 y,x,z
 
loop_
_atom_site_label
_atom_site_type_symbol
_atom_site_symmetry_multiplicity
_atom_site_Wyckoff_label
_atom_site_fract_x
_atom_site_fract_y
_atom_site_fract_z
_atom_site_occupancy
Ni1 Ni   2 a 0.00000 0.00000 0.00000 1.00000
Ni2 Ni   4 b 0.33333 0.66667 0.37700 1.00000
Cl1 Cl   6 c 0.15980 0.00000 0.23960 1.00000
K1  K    6 c 0.66470 0.00000 0.17060 1.00000
Cl2 Cl  12 d 0.17320 0.50560 0.11480 1.00000
\end{lstlisting}
{\phantomsection\label{A3BC_hP30_185_cd_c_ab_poscar}}
{\hyperref[A3BC_hP30_185_cd_c_ab]{KNiCl$_{3}$ (Room-temperature): A3BC\_hP30\_185\_cd\_c\_ab}} - POSCAR

{\phantomsection\label{A3B_hP24_185_ab2c_c_cif}}
{\hyperref[A3B_hP24_185_ab2c_c]{Cu$_{3}$P: A3B\_hP24\_185\_ab2c\_c}} - CIF
\begin{lstlisting}[numbers=none,language={mylang}]
# CIF file 
data_findsym-output
_audit_creation_method FINDSYM

_chemical_name_mineral ''
_chemical_formula_sum 'Cu3 P'

loop_
_publ_author_name
 'O. Olofsson'
_journal_name_full_name
;
 Acta Chemica Scandinavica
;
_journal_volume 26
_journal_year 1972
_journal_page_first 2777
_journal_page_last 2787
_publ_Section_title
;
 The Crystal Structure of Cu$_{3}$P
;

_aflow_title 'Cu$_{3}$P Structure'
_aflow_proto 'A3B_hP24_185_ab2c_c'
_aflow_params 'a,c/a,z_{1},z_{2},x_{3},z_{3},x_{4},z_{4},x_{5},z_{5}'
_aflow_params_values '6.9593,1.02639633296,0.3213,0.1998,0.2806,0.0765,0.3761,0.4246,0.3322,0.75'
_aflow_Strukturbericht 'None'
_aflow_Pearson 'hP24'

_symmetry_space_group_name_H-M "P 63 c m"
_symmetry_Int_Tables_number 185
 
_cell_length_a    6.95930
_cell_length_b    6.95930
_cell_length_c    7.14300
_cell_angle_alpha 90.00000
_cell_angle_beta  90.00000
_cell_angle_gamma 120.00000
 
loop_
_space_group_symop_id
_space_group_symop_operation_xyz
1 x,y,z
2 x-y,x,z+1/2
3 -y,x-y,z
4 -x,-y,z+1/2
5 -x+y,-x,z
6 y,-x+y,z+1/2
7 -x+y,y,z+1/2
8 -x,-x+y,z
9 -y,-x,z+1/2
10 x-y,-y,z
11 x,x-y,z+1/2
12 y,x,z
 
loop_
_atom_site_label
_atom_site_type_symbol
_atom_site_symmetry_multiplicity
_atom_site_Wyckoff_label
_atom_site_fract_x
_atom_site_fract_y
_atom_site_fract_z
_atom_site_occupancy
Cu1 Cu   2 a 0.00000 0.00000 0.32130 1.00000
Cu2 Cu   4 b 0.33333 0.66667 0.19980 1.00000
Cu3 Cu   6 c 0.28060 0.00000 0.07650 1.00000
Cu4 Cu   6 c 0.37610 0.00000 0.42460 1.00000
P1  P    6 c 0.33220 0.00000 0.75000 1.00000
\end{lstlisting}
{\phantomsection\label{A3B_hP24_185_ab2c_c_poscar}}
{\hyperref[A3B_hP24_185_ab2c_c]{Cu$_{3}$P: A3B\_hP24\_185\_ab2c\_c}} - POSCAR

{\phantomsection\label{A3B_hP8_185_c_a_cif}}
{\hyperref[A3B_hP8_185_c_a]{$\beta$-RuCl$_{3}$: A3B\_hP8\_185\_c\_a}} - CIF
\begin{lstlisting}[numbers=none,language={mylang}]
# CIF file
data_findsym-output
_audit_creation_method FINDSYM

_chemical_name_mineral 'beta-RuCl3'
_chemical_formula_sum 'Cl3 Ru'

loop_
_publ_author_name
 'J. M. Fletcher'
 'W. E. Gardner'
 'A. C. Fox'
 'G. Topping'
_journal_name_full_name
;
 Journal of the Chemical Society A
;
_journal_volume ~
_journal_year 1967
_journal_page_first 1038
_journal_page_last 1045
_publ_Section_title
;
 X-Ray, infrared, and magnetic studies of $\alpha$-and $\beta$-ruthenium trichloride
;

# Found in Pearson's Crystal Data - Crystal Structure Database for Inorganic Compounds, 2013

_aflow_title '$\beta$-RuCl$_{3}$ Structure'
_aflow_proto 'A3B_hP8_185_c_a'
_aflow_params 'a,c/a,z_{1},x_{2},z_{2}'
_aflow_params_values '6.1197165237,0.924509803925,0.0,0.305,0.245'
_aflow_Strukturbericht 'None'
_aflow_Pearson 'hP8'

_cell_length_a    6.1197165237
_cell_length_b    6.1197165237
_cell_length_c    5.6577379234
_cell_angle_alpha 90.0000000000
_cell_angle_beta  90.0000000000
_cell_angle_gamma 120.0000000000
 
_symmetry_space_group_name_H-M "P 63 c m"
_symmetry_Int_Tables_number 185
 
loop_
_space_group_symop_id
_space_group_symop_operation_xyz
1 x,y,z
2 x-y,x,z+1/2
3 -y,x-y,z
4 -x,-y,z+1/2
5 -x+y,-x,z
6 y,-x+y,z+1/2
7 -x+y,y,z+1/2
8 -x,-x+y,z
9 -y,-x,z+1/2
10 x-y,-y,z
11 x,x-y,z+1/2
12 y,x,z
 
loop_
_atom_site_label
_atom_site_type_symbol
_atom_site_symmetry_multiplicity
_atom_site_Wyckoff_label
_atom_site_fract_x
_atom_site_fract_y
_atom_site_fract_z
_atom_site_occupancy
Ru1 Ru   2 a 0.00000 0.00000 0.00000 1.00000
Cl1 Cl   6 c 0.30500 0.00000 0.24500 1.00000
\end{lstlisting}
{\phantomsection\label{A3B_hP8_185_c_a_poscar}}
{\hyperref[A3B_hP8_185_c_a]{$\beta$-RuCl$_{3}$: A3B\_hP8\_185\_c\_a}} - POSCAR
\begin{lstlisting}[numbers=none,language={mylang}]
A3B_hP8_185_c_a & a,c/a,z1,x2,z2 --params=6.1197165237,0.924509803925,0.0,0.305,0.245 & P6_{3}cm C_{6v}^{3} #185 (ac) & hP8 & None & RuCl3 & beta & J. M. Fletcher et al., J. Chem. Soc. A, 1038-1045 (1967)
   1.00000000000000
   3.05985826185000  -5.29982997348359   0.00000000000000
   3.05985826185000   5.29982997348359   0.00000000000000
   0.00000000000000   0.00000000000000   5.65773792340000
    Cl    Ru
     6     2
Direct
   0.30500000000000   0.00000000000000   0.24500000000000   Cl   (6c)
   0.00000000000000   0.30500000000000   0.24500000000000   Cl   (6c)
  -0.30500000000000  -0.30500000000000   0.24500000000000   Cl   (6c)
  -0.30500000000000   0.00000000000000   0.74500000000000   Cl   (6c)
   0.00000000000000  -0.30500000000000   0.74500000000000   Cl   (6c)
   0.30500000000000   0.30500000000000   0.74500000000000   Cl   (6c)
   0.00000000000000   0.00000000000000   0.00000000000000   Ru   (2a)
   0.00000000000000   0.00000000000000   0.50000000000000   Ru   (2a)
\end{lstlisting}
{\phantomsection\label{AB3_hP24_185_c_ab2c_cif}}
{\hyperref[AB3_hP24_185_c_ab2c]{Na$_{3}$As: AB3\_hP24\_185\_c\_ab2c}} - CIF
\begin{lstlisting}[numbers=none,language={mylang}]
# CIF file 
data_findsym-output
_audit_creation_method FINDSYM

_chemical_name_mineral 'Na3As'
_chemical_formula_sum 'As Na3'

loop_
_publ_author_name
 'P. Hafner'
 'K.-J. Range'
_journal_name_full_name
;
 Journal of Alloys and Compounds
;
_journal_volume 216
_journal_year 1994
_journal_page_first 7
_journal_page_last 10
_publ_Section_title
;
 Na$_{3}$As revisited: high-pressure synthesis of single crystals and structure refinement
;

_aflow_title 'Na$_{3}$As Structure'
_aflow_proto 'AB3_hP24_185_c_ab2c'
_aflow_params 'a,c/a,z_{1},z_{2},x_{3},z_{3},x_{4},z_{4},x_{5},z_{5}'
_aflow_params_values '8.7838,1.02449964708,0.2684,0.2311,0.3321,0.25,0.3153,0.5863,0.3518,-0.0769'
_aflow_Strukturbericht 'None'
_aflow_Pearson 'hP24'

_symmetry_space_group_name_H-M "P 63 c m"
_symmetry_Int_Tables_number 185
 
_cell_length_a    8.78380
_cell_length_b    8.78380
_cell_length_c    8.99900
_cell_angle_alpha 90.00000
_cell_angle_beta  90.00000
_cell_angle_gamma 120.00000
 
loop_
_space_group_symop_id
_space_group_symop_operation_xyz
1 x,y,z
2 x-y,x,z+1/2
3 -y,x-y,z
4 -x,-y,z+1/2
5 -x+y,-x,z
6 y,-x+y,z+1/2
7 -x+y,y,z+1/2
8 -x,-x+y,z
9 -y,-x,z+1/2
10 x-y,-y,z
11 x,x-y,z+1/2
12 y,x,z
 
loop_
_atom_site_label
_atom_site_type_symbol
_atom_site_symmetry_multiplicity
_atom_site_Wyckoff_label
_atom_site_fract_x
_atom_site_fract_y
_atom_site_fract_z
_atom_site_occupancy
Na1 Na   2 a 0.00000 0.00000 0.26840  1.00000
Na2 Na   4 b 0.33333 0.66667 0.23110  1.00000
As1 As   6 c 0.33210 0.00000 0.25000  1.00000
Na3 Na   6 c 0.31530 0.00000 0.58630  1.00000
Na4 Na   6 c 0.35180 0.00000 -0.07690 1.00000
\end{lstlisting}
{\phantomsection\label{AB3_hP24_185_c_ab2c_poscar}}
{\hyperref[AB3_hP24_185_c_ab2c]{Na$_{3}$As: AB3\_hP24\_185\_c\_ab2c}} - POSCAR

{\phantomsection\label{A3B7_hP20_186_c_b2c_cif}}
{\hyperref[A3B7_hP20_186_c_b2c]{Fe$_{3}$Th$_{7}$ ($D10_{2}$): A3B7\_hP20\_186\_c\_b2c}} - CIF
\begin{lstlisting}[numbers=none,language={mylang}]
# CIF file
data_findsym-output
_audit_creation_method FINDSYM

_chemical_name_mineral ''
_chemical_formula_sum 'Fe3 Th7'

loop_
_publ_author_name
 'J. V. Florio'
 'N. C. Baenziger'
 'R. E. Rundle'
_journal_name_full_name
;
 Acta Cristallographica
;
_journal_volume 9
_journal_year 1956
_journal_page_first 367
_journal_page_last 372
_publ_Section_title
;
 Compounds of thorium with transition metals. II. Systems with iron, cobalt and nickel
;

# Found in Inorganic Crystal Structure Database, {ID 401657},

_aflow_title 'Fe$_{3}$Th$_{7}$ ($D10_{2}$) Structure'
_aflow_proto 'A3B7_hP20_186_c_b2c'
_aflow_params 'a,c/a,z_{1},x_{2},z_{2},x_{3},z_{3},x_{4},z_{4}'
_aflow_params_values '9.85,0.624365482234,0.06,0.815,0.31,0.126,0.25,0.544,0.31'
_aflow_Strukturbericht '$D10_{2}$'
_aflow_Pearson 'hP20'

_symmetry_space_group_name_H-M "P 63 m c"
_symmetry_Int_Tables_number 186
 
_cell_length_a    9.85000
_cell_length_b    9.85000
_cell_length_c    6.15000
_cell_angle_alpha 90.00000
_cell_angle_beta  90.00000
_cell_angle_gamma 120.00000
 
loop_
_space_group_symop_id
_space_group_symop_operation_xyz
1 x,y,z
2 x-y,x,z+1/2
3 -y,x-y,z
4 -x,-y,z+1/2
5 -x+y,-x,z
6 y,-x+y,z+1/2
7 -x+y,y,z
8 -x,-x+y,z+1/2
9 -y,-x,z
10 x-y,-y,z+1/2
11 x,x-y,z
12 y,x,z+1/2
 
loop_
_atom_site_label
_atom_site_type_symbol
_atom_site_symmetry_multiplicity
_atom_site_Wyckoff_label
_atom_site_fract_x
_atom_site_fract_y
_atom_site_fract_z
_atom_site_occupancy
Th1 Th   2 b 0.33333 0.66667 0.06000 1.00000
Fe1 Fe   6 c 0.81500 0.18500 0.31000 1.00000
Th2 Th   6 c 0.12600 0.87400 0.25000 1.00000
Th3 Th   6 c 0.54400 0.45600 0.31000 1.00000
\end{lstlisting}
{\phantomsection\label{A3B7_hP20_186_c_b2c_poscar}}
{\hyperref[A3B7_hP20_186_c_b2c]{Fe$_{3}$Th$_{7}$ ($D10_{2}$): A3B7\_hP20\_186\_c\_b2c}} - POSCAR
\begin{lstlisting}[numbers=none,language={mylang}]
A3B7_hP20_186_c_b2c & a,c/a,z1,x2,z2,x3,z3,x4,z4 --params=9.85,0.624365482234,0.06,0.815,0.31,0.126,0.25,0.544,0.31 & P6_{3}mc C_{6v}^{4} #186 (bc^3) & hP20 & $D10_{2}$ & Fe3Th7 &  & J. V. Florio and N. C. Baenziger and R. E. Rundle, Acta Cryst. 9, 367-372 (1956)
   1.00000000000000
   4.92500000000000  -8.53035022727672   0.00000000000000
   4.92500000000000   8.53035022727672   0.00000000000000
   0.00000000000000   0.00000000000000   6.15000000000000
    Fe    Th
     6    14
Direct
   0.81500000000000  -0.81500000000000   0.31000000000000   Fe   (6c)
   0.81500000000000   1.63000000000000   0.31000000000000   Fe   (6c)
  -1.63000000000000  -0.81500000000000   0.31000000000000   Fe   (6c)
  -0.81500000000000   0.81500000000000   0.81000000000000   Fe   (6c)
  -0.81500000000000  -1.63000000000000   0.81000000000000   Fe   (6c)
   1.63000000000000   0.81500000000000   0.81000000000000   Fe   (6c)
   0.33333333333333   0.66666666666667   0.06000000000000   Th   (2b)
   0.66666666666667   0.33333333333333   0.56000000000000   Th   (2b)
   0.12600000000000  -0.12600000000000   0.25000000000000   Th   (6c)
   0.12600000000000   0.25200000000000   0.25000000000000   Th   (6c)
  -0.25200000000000  -0.12600000000000   0.25000000000000   Th   (6c)
  -0.12600000000000   0.12600000000000   0.75000000000000   Th   (6c)
  -0.12600000000000  -0.25200000000000   0.75000000000000   Th   (6c)
   0.25200000000000   0.12600000000000   0.75000000000000   Th   (6c)
   0.54400000000000  -0.54400000000000   0.31000000000000   Th   (6c)
   0.54400000000000   1.08800000000000   0.31000000000000   Th   (6c)
  -1.08800000000000  -0.54400000000000   0.31000000000000   Th   (6c)
  -0.54400000000000   0.54400000000000   0.81000000000000   Th   (6c)
  -0.54400000000000  -1.08800000000000   0.81000000000000   Th   (6c)
   1.08800000000000   0.54400000000000   0.81000000000000   Th   (6c)
\end{lstlisting}
{\phantomsection\label{AB3_hP4_187_e_fh_cif}}
{\hyperref[AB3_hP4_187_e_fh]{Re$_{3}$N: AB3\_hP4\_187\_e\_fh}} - CIF
\begin{lstlisting}[numbers=none,language={mylang}]
# CIF file 
data_findsym-output
_audit_creation_method FINDSYM

_chemical_name_mineral 'Re3N'
_chemical_formula_sum 'N Re3'

loop_
_publ_author_name
 'A. Friedrich'
 'B. Winkler'
 'L. Bayarjargal'
 'W. Morgenroth'
 'E. A. Juarez-Arellano'
 'V. Milman'
 'K. Refson'
 'M. Kunz'
 'K. Chen'
_journal_name_full_name
;
 Physical Review Letters
;
_journal_volume 105
_journal_year 2010
_journal_page_first 085504
_journal_page_last 085504
_publ_Section_title
;
 Novel Rhenium Nitrides
;

_aflow_title 'Re$_{3}$N Structure'
_aflow_proto 'AB3_hP4_187_e_fh'
_aflow_params 'a,c/a,z_{3}'
_aflow_params_values '2.8065,2.53411722786,0.198'
_aflow_Strukturbericht 'None'
_aflow_Pearson 'hP4'

_symmetry_space_group_name_H-M "P -6 m 2"
_symmetry_Int_Tables_number 187
 
_cell_length_a    2.80650
_cell_length_b    2.80650
_cell_length_c    7.11200
_cell_angle_alpha 90.00000
_cell_angle_beta  90.00000
_cell_angle_gamma 120.00000
 
loop_
_space_group_symop_id
_space_group_symop_operation_xyz
1 x,y,z
2 -y,x-y,z
3 -x+y,-x,z
4 x,x-y,-z
5 -x+y,y,-z
6 -y,-x,-z
7 -x+y,-x,-z
8 x,y,-z
9 -y,x-y,-z
10 -x+y,y,z
11 -y,-x,z
12 x,x-y,z
 
loop_
_atom_site_label
_atom_site_type_symbol
_atom_site_symmetry_multiplicity
_atom_site_Wyckoff_label
_atom_site_fract_x
_atom_site_fract_y
_atom_site_fract_z
_atom_site_occupancy
N1  N    1 e 0.66667 0.33333 0.00000 1.00000
Re1 Re   1 f 0.66667 0.33333 0.50000 1.00000
Re2 Re   2 h 0.33333 0.66667 0.19800 1.00000
\end{lstlisting}
{\phantomsection\label{AB3_hP4_187_e_fh_poscar}}
{\hyperref[AB3_hP4_187_e_fh]{Re$_{3}$N: AB3\_hP4\_187\_e\_fh}} - POSCAR
\begin{lstlisting}[numbers=none,language={mylang}]
AB3_hP4_187_e_fh & a,c/a,z3 --params=2.8065,2.53411722786,0.198 & P-6m2 D_{3h}^{1} #187 (efh) & hP4 & None & Re3N & Re3N & A. Friedrich et al., Phys. Rev. Lett. 105, 085504(2010)
   1.00000000000000
   1.40325000000000  -2.43050029572103   0.00000000000000
   1.40325000000000   2.43050029572103   0.00000000000000
   0.00000000000000   0.00000000000000   7.11200000000000
     N    Re
     1     3
Direct
   0.66666666666667   0.33333333333333   0.00000000000000    N   (1e)
   0.66666666666667   0.33333333333333   0.50000000000000   Re   (1f)
   0.33333333333333   0.66666666666667   0.19800000000000   Re   (2h)
   0.33333333333333   0.66666666666667  -0.19800000000000   Re   (2h)
\end{lstlisting}
{\phantomsection\label{A3BC_hP10_188_k_a_e_cif}}
{\hyperref[A3BC_hP10_188_k_a_e]{LiScI$_{3}$: A3BC\_hP10\_188\_k\_a\_e}} - CIF
\begin{lstlisting}[numbers=none,language={mylang}]
# CIF file
data_findsym-output
_audit_creation_method FINDSYM

_chemical_name_mineral 'LiScI3'
_chemical_formula_sum 'I3 Li Sc'

loop_
_publ_author_name
 'A. Lachgar'
 'D. S. Dudis'
 'P. K. Dorhout'
 'J. D. Corbett'
_journal_name_full_name
;
 Inorganic Chemistry
;
_journal_volume 30
_journal_year 1991
_journal_page_first 3321
_journal_page_last 3326
_publ_Section_title
;
 Synthesis and properties of two novel line phases that contain linear scandium chains, lithium scandium iodide (LiScI$_{3}$) and sodium scandium iodide (Na$_{0.5}$ScI$_{3}$)
;

# Found in Pearson's Crystal Data - Crystal Structure Database for Inorganic Compounds, 2013

_aflow_title 'LiScI$_{3}$ Structure'
_aflow_proto 'A3BC_hP10_188_k_a_e'
_aflow_params 'a,c/a,x_{3},y_{3}'
_aflow_params_values '7.2864263258,0.928891999832,0.67077,0.01058'
_aflow_Strukturbericht 'None'
_aflow_Pearson 'hP10'

_cell_length_a    7.2864263258
_cell_length_b    7.2864263258
_cell_length_c    6.7683031214
_cell_angle_alpha 90.0000000000
_cell_angle_beta  90.0000000000
_cell_angle_gamma 120.0000000000
 
_symmetry_space_group_name_H-M "P -6 c 2"
_symmetry_Int_Tables_number 188
 
loop_
_space_group_symop_id
_space_group_symop_operation_xyz
1 x,y,z
2 -y,x-y,z
3 -x+y,-x,z
4 x,x-y,-z
5 -x+y,y,-z
6 -y,-x,-z
7 -x+y,-x,-z+1/2
8 x,y,-z+1/2
9 -y,x-y,-z+1/2
10 -x+y,y,z+1/2
11 -y,-x,z+1/2
12 x,x-y,z+1/2
 
loop_
_atom_site_label
_atom_site_type_symbol
_atom_site_symmetry_multiplicity
_atom_site_Wyckoff_label
_atom_site_fract_x
_atom_site_fract_y
_atom_site_fract_z
_atom_site_occupancy
Li1 Li   2 a 0.00000 0.00000 0.00000 1.00000
Sc1 Sc   2 e 0.66667 0.33333 0.00000 1.00000
I1  I    6 k 0.67077 0.01058 0.25000 1.00000
\end{lstlisting}
{\phantomsection\label{A3BC_hP10_188_k_a_e_poscar}}
{\hyperref[A3BC_hP10_188_k_a_e]{LiScI$_{3}$: A3BC\_hP10\_188\_k\_a\_e}} - POSCAR
\begin{lstlisting}[numbers=none,language={mylang}]
A3BC_hP10_188_k_a_e & a,c/a,x3,y3 --params=7.2864263258,0.928891999832,0.67077,0.01058 & P-6c2 D_{3h}^{2} #188 (aek) & hP10 & None & LiScI3 &  & A. Lachgar et al., Inorg. Chem. 30, 3321-3326 (1991)
   1.00000000000000
   3.64321316290000  -6.31023030094651   0.00000000000000
   3.64321316290000   6.31023030094651   0.00000000000000
   0.00000000000000   0.00000000000000   6.76830312140000
     I    Li    Sc
     6     2     2
Direct
   0.67077000000000   0.01058000000000   0.25000000000000    I   (6k)
  -0.01058000000000   0.66019000000000   0.25000000000000    I   (6k)
  -0.66019000000000  -0.67077000000000   0.25000000000000    I   (6k)
  -0.01058000000000  -0.67077000000000   0.75000000000000    I   (6k)
  -0.66019000000000   0.01058000000000   0.75000000000000    I   (6k)
   0.67077000000000   0.66019000000000   0.75000000000000    I   (6k)
   0.00000000000000   0.00000000000000   0.00000000000000   Li   (2a)
   0.00000000000000   0.00000000000000   0.50000000000000   Li   (2a)
   0.66666666666667   0.33333333333333   0.00000000000000   Sc   (2e)
   0.66666666666667   0.33333333333333   0.50000000000000   Sc   (2e)
\end{lstlisting}
{\phantomsection\label{AB9C4_hP28_188_e_kl_ak_cif}}
{\hyperref[AB9C4_hP28_188_e_kl_ak]{BaSi$_{4}$O$_{9}$: AB9C4\_hP28\_188\_e\_kl\_ak}} - CIF
\begin{lstlisting}[numbers=none,language={mylang}]
# CIF file
data_findsym-output
_audit_creation_method FINDSYM

_chemical_name_mineral 'BaSi4O9'
_chemical_formula_sum 'Ba O9 Si4'

loop_
_publ_author_name
 'L. W. Finger'
 'R. M. Hazen'
 'B. A. Fursenko'
_journal_name_full_name
;
 Journal of Physics and Chemistry of Solids
;
_journal_volume 56
_journal_year 1995
_journal_page_first 1389
_journal_page_last 1393
_publ_Section_title
;
 Refinement of the crystal structure of BaSi$_{4}$O$_{9}$ in the benitoite form
;

# Found in Pearson's Crystal Data - Crystal Structure Database for Inorganic Compounds, 2013

_aflow_title 'BaSi$_{4}$O$_{9}$ Structure'
_aflow_proto 'AB9C4_hP28_188_e_kl_ak'
_aflow_params 'a,c/a,x_{3},y_{3},x_{4},y_{4},x_{5},y_{5},z_{5}'
_aflow_params_values '6.4953629976,1.43896355826,0.07103,0.48306,0.12023,0.75436,0.22923,0.00127,0.6032'
_aflow_Strukturbericht 'None'
_aflow_Pearson 'hP28'

_cell_length_a    6.4953629976
_cell_length_b    6.4953629976
_cell_length_c    9.3465906512
_cell_angle_alpha 90.0000000000
_cell_angle_beta  90.0000000000
_cell_angle_gamma 120.0000000000
 
_symmetry_space_group_name_H-M "P -6 c 2"
_symmetry_Int_Tables_number 188
 
loop_
_space_group_symop_id
_space_group_symop_operation_xyz
1 x,y,z
2 -y,x-y,z
3 -x+y,-x,z
4 x,x-y,-z
5 -x+y,y,-z
6 -y,-x,-z
7 -x+y,-x,-z+1/2
8 x,y,-z+1/2
9 -y,x-y,-z+1/2
10 -x+y,y,z+1/2
11 -y,-x,z+1/2
12 x,x-y,z+1/2
 
loop_
_atom_site_label
_atom_site_type_symbol
_atom_site_symmetry_multiplicity
_atom_site_Wyckoff_label
_atom_site_fract_x
_atom_site_fract_y
_atom_site_fract_z
_atom_site_occupancy
Si1 Si   2 a 0.00000 0.00000 0.00000 1.00000
Ba1 Ba   2 e 0.66667 0.33333 0.00000 1.00000
O1  O    6 k 0.07103 0.48306 0.25000 1.00000
Si2 Si   6 k 0.12023 0.75436 0.25000 1.00000
O2  O   12 l 0.22923 0.00127 0.60320 1.00000
\end{lstlisting}
{\phantomsection\label{AB9C4_hP28_188_e_kl_ak_poscar}}
{\hyperref[AB9C4_hP28_188_e_kl_ak]{BaSi$_{4}$O$_{9}$: AB9C4\_hP28\_188\_e\_kl\_ak}} - POSCAR

{\phantomsection\label{A8BC3D6_hP18_189_bfh_a_g_i_cif}}
{\hyperref[A8BC3D6_hP18_189_bfh_a_g_i]{$\pi$-FeMg$_{3}$Al$_{8}$Si$_{6}$ ($E9_{b}$): A8BC3D6\_hP18\_189\_bfh\_a\_g\_i}} - CIF
\begin{lstlisting}[numbers=none,language={mylang}]
# CIF file 
data_findsym-output
_audit_creation_method FINDSYM

_chemical_name_mineral '$\pi$-FeMg$_{3}$Al$_{8}$Si$_{6}$'
_chemical_formula_sum 'Al8 Fe Mg3 Si6'

loop_
_publ_author_name
 'H. Perlitz'
 'A. Westgren'
_journal_name_full_name
;
 Arkiv f{\"o}r Kemi, Mineralogi och Geologi
;
_journal_volume 15B
_journal_year 1942
_journal_page_first 1
_journal_page_last 8
_publ_Section_title
;
 The Crystal Structure of Al$_{8}$Si$_{6}$Mg$_{3}$Fe
;

# Found in Determination of the crystal structure of the $\pi$-AlFeMgSi phase using symmetry- and site-sensitive electron microscope techniques, 2003

_aflow_title '$\pi$-FeMg$_{3}$Al$_{8}$Si$_{6}$ ($E9_{b}$) Structure'
_aflow_proto 'A8BC3D6_hP18_189_bfh_a_g_i'
_aflow_params 'a,c/a,x_{3},x_{4},z_{5},x_{6},z_{6}'
_aflow_params_values '6.62,1.20241691843,0.403,0.444,0.231,0.75,0.222'
_aflow_Strukturbericht '$E9_{b}$'
_aflow_Pearson 'hP18'

_symmetry_space_group_name_H-M "P -6 2 m"
_symmetry_Int_Tables_number 189
 
_cell_length_a    6.62000
_cell_length_b    6.62000
_cell_length_c    7.96000
_cell_angle_alpha 90.00000
_cell_angle_beta  90.00000
_cell_angle_gamma 120.00000
 
loop_
_space_group_symop_id
_space_group_symop_operation_xyz
1 x,y,z
2 -y,x-y,z
3 -x+y,-x,z
4 x-y,-y,-z
5 y,x,-z
6 -x,-x+y,-z
7 -x+y,-x,-z
8 x,y,-z
9 -y,x-y,-z
10 -x,-x+y,z
11 x-y,-y,z
12 y,x,z
 
loop_
_atom_site_label
_atom_site_type_symbol
_atom_site_symmetry_multiplicity
_atom_site_Wyckoff_label
_atom_site_fract_x
_atom_site_fract_y
_atom_site_fract_z
_atom_site_occupancy
Fe1 Fe   1 a 0.00000 0.00000 0.00000 1.00000
Al1 Al   1 b 0.00000 0.00000 0.50000 1.00000
Al2 Al   3 f 0.40300 0.00000 0.00000 1.00000
Mg1 Mg   3 g 0.44400 0.00000 0.50000 1.00000
Al3 Al   4 h 0.33333 0.66667 0.23100 1.00000
Si1 Si   6 i 0.75000 0.00000 0.22200 1.00000
\end{lstlisting}
{\phantomsection\label{A8BC3D6_hP18_189_bfh_a_g_i_poscar}}
{\hyperref[A8BC3D6_hP18_189_bfh_a_g_i]{$\pi$-FeMg$_{3}$Al$_{8}$Si$_{6}$ ($E9_{b}$): A8BC3D6\_hP18\_189\_bfh\_a\_g\_i}} - POSCAR
\begin{lstlisting}[numbers=none,language={mylang}]
A8BC3D6_hP18_189_bfh_a_g_i & a,c/a,x3,x4,z5,x6,z6 --params=6.62,1.20241691843,0.403,0.444,0.231,0.75,0.222 & P-62m D_{3h}^{3} #189 (abfghi) & hP18 & $E9_{b}$ & FeMg3Al8Si6 & $\pi$-FeMg$_{3}$Al$_{8}$Si$_{6}$ & H. Perlitz and A. Westgren, {Ark. Kem. Mineral. Geol. 15B, 1-8 (1942)
   1.00000000000000
   3.31000000000000  -5.73308817305298   0.00000000000000
   3.31000000000000   5.73308817305298   0.00000000000000
   0.00000000000000   0.00000000000000   7.96000000000000
    Al    Fe    Mg    Si
     8     1     3     6
Direct
   0.00000000000000   0.00000000000000   0.50000000000000   Al   (1b)
   0.40300000000000   0.00000000000000   0.00000000000000   Al   (3f)
   0.00000000000000   0.40300000000000   0.00000000000000   Al   (3f)
  -0.40300000000000  -0.40300000000000   0.00000000000000   Al   (3f)
   0.33333333333333   0.66666666666667   0.23100000000000   Al   (4h)
   0.33333333333333   0.66666666666667  -0.23100000000000   Al   (4h)
   0.66666666666667   0.33333333333333  -0.23100000000000   Al   (4h)
   0.66666666666667   0.33333333333333   0.23100000000000   Al   (4h)
   0.00000000000000   0.00000000000000   0.00000000000000   Fe   (1a)
   0.44400000000000   0.00000000000000   0.50000000000000   Mg   (3g)
   0.00000000000000   0.44400000000000   0.50000000000000   Mg   (3g)
  -0.44400000000000  -0.44400000000000   0.50000000000000   Mg   (3g)
   0.75000000000000   0.00000000000000   0.22200000000000   Si   (6i)
   0.00000000000000   0.75000000000000   0.22200000000000   Si   (6i)
  -0.75000000000000  -0.75000000000000   0.22200000000000   Si   (6i)
   0.75000000000000   0.00000000000000  -0.22200000000000   Si   (6i)
   0.00000000000000   0.75000000000000  -0.22200000000000   Si   (6i)
  -0.75000000000000  -0.75000000000000  -0.22200000000000   Si   (6i)
\end{lstlisting}
{\phantomsection\label{A9BC3D5_hP18_189_fi_a_g_bh_cif}}
{\hyperref[A9BC3D5_hP18_189_fi_a_g_bh]{$\pi$-FeMg$_{3}$Al$_{9}$Si$_{5}$: A9BC3D5\_hP18\_189\_fi\_a\_g\_bh}} - CIF
\begin{lstlisting}[numbers=none,language={mylang}]
# CIF file 
data_findsym-output
_audit_creation_method FINDSYM

_chemical_name_mineral '$\pi$-FeMg$_{3}$Al$_{9}$Si$_{5}$'
_chemical_formula_sum 'Al9 Fe Mg3 Si5'

loop_
_publ_author_name
 'S. Foss'
 'A. Olsen'
 'C. J. Simensen'
 'J. Taft{\o}'
_journal_name_full_name
;
 Acta Crystallographica Section B: Structural Science
;
_journal_volume 59
_journal_year 2003
_journal_page_first 36
_journal_page_last 42
_publ_Section_title
;
 Determination of the crystal structure of the $\pi$-AlFeMgSi phase using symmetry- and site-sensitive electron microscope techniques
;

# Found in The Materials Project, Mg$_{3}$Al$_{9}$FeSi$_{5}$, {ID mp-7062},

_aflow_title '$\pi$-FeMg$_{3}$Al$_{9}$Si$_{5}$ Structure'
_aflow_proto 'A9BC3D5_hP18_189_fi_a_g_bh'
_aflow_params 'a,c/a,x_{3},x_{4},z_{5},x_{6},z_{6}'
_aflow_params_values '6.6,1.19696969697,0.378,0.43,0.266,0.755,0.236'
_aflow_Strukturbericht 'None'
_aflow_Pearson 'hP18'

_symmetry_space_group_name_H-M "P -6 2 m"
_symmetry_Int_Tables_number 189
 
_cell_length_a    6.60000
_cell_length_b    6.60000
_cell_length_c    7.90000
_cell_angle_alpha 90.00000
_cell_angle_beta  90.00000
_cell_angle_gamma 120.00000
 
loop_
_space_group_symop_id
_space_group_symop_operation_xyz
1 x,y,z
2 -y,x-y,z
3 -x+y,-x,z
4 x-y,-y,-z
5 y,x,-z
6 -x,-x+y,-z
7 -x+y,-x,-z
8 x,y,-z
9 -y,x-y,-z
10 -x,-x+y,z
11 x-y,-y,z
12 y,x,z
 
loop_
_atom_site_label
_atom_site_type_symbol
_atom_site_symmetry_multiplicity
_atom_site_Wyckoff_label
_atom_site_fract_x
_atom_site_fract_y
_atom_site_fract_z
_atom_site_occupancy
Fe1 Fe   1 a 0.00000 0.00000 0.00000 1.00000
Si1 Si   1 b 0.00000 0.00000 0.50000 1.00000
Al1 Al   3 f 0.37800 0.00000 0.00000 1.00000
Mg1 Mg   3 g 0.43000 0.00000 0.50000 1.00000
Si2 Si   4 h 0.33333 0.66667 0.26600 1.00000
Al2 Al   6 i 0.75500 0.00000 0.23600 1.00000
\end{lstlisting}
{\phantomsection\label{A9BC3D5_hP18_189_fi_a_g_bh_poscar}}
{\hyperref[A9BC3D5_hP18_189_fi_a_g_bh]{$\pi$-FeMg$_{3}$Al$_{9}$Si$_{5}$: A9BC3D5\_hP18\_189\_fi\_a\_g\_bh}} - POSCAR
\begin{lstlisting}[numbers=none,language={mylang}]
A9BC3D5_hP18_189_fi_a_g_bh & a,c/a,x3,x4,z5,x6,z6 --params=6.6,1.19696969697,0.378,0.43,0.266,0.755,0.236 & P-62m D_{3h}^{3} #189 (abfghi) & hP18 & None & FeMg3Al9Si5 & $\pi$-FeMg$_{3}$Al$_{9}$Si$_{5}$ & S. Foss et al., Acta Crystallogr. Sect. B Struct. Sci. 59, 36-42 (2003)
   1.00000000000000
   3.30000000000000  -5.71576766497729   0.00000000000000
   3.30000000000000   5.71576766497729   0.00000000000000
   0.00000000000000   0.00000000000000   7.90000000000000
    Al    Fe    Mg    Si
     9     1     3     5
Direct
   0.37800000000000   0.00000000000000   0.00000000000000   Al   (3f)
   0.00000000000000   0.37800000000000   0.00000000000000   Al   (3f)
  -0.37800000000000  -0.37800000000000   0.00000000000000   Al   (3f)
   0.75500000000000   0.00000000000000   0.23600000000000   Al   (6i)
   0.00000000000000   0.75500000000000   0.23600000000000   Al   (6i)
  -0.75500000000000  -0.75500000000000   0.23600000000000   Al   (6i)
   0.75500000000000   0.00000000000000  -0.23600000000000   Al   (6i)
   0.00000000000000   0.75500000000000  -0.23600000000000   Al   (6i)
  -0.75500000000000  -0.75500000000000  -0.23600000000000   Al   (6i)
   0.00000000000000   0.00000000000000   0.00000000000000   Fe   (1a)
   0.43000000000000   0.00000000000000   0.50000000000000   Mg   (3g)
   0.00000000000000   0.43000000000000   0.50000000000000   Mg   (3g)
  -0.43000000000000  -0.43000000000000   0.50000000000000   Mg   (3g)
   0.00000000000000   0.00000000000000   0.50000000000000   Si   (1b)
   0.33333333333333   0.66666666666667   0.26600000000000   Si   (4h)
   0.33333333333333   0.66666666666667  -0.26600000000000   Si   (4h)
   0.66666666666667   0.33333333333333  -0.26600000000000   Si   (4h)
   0.66666666666667   0.33333333333333   0.26600000000000   Si   (4h)
\end{lstlisting}
{\phantomsection\label{A2B_hP18_190_gh_bf_cif}}
{\hyperref[A2B_hP18_190_gh_bf]{Li$_{2}$Sb: A2B\_hP18\_190\_gh\_bf}} - CIF
\begin{lstlisting}[numbers=none,language={mylang}]
# CIF file 
data_findsym-output
_audit_creation_method FINDSYM

_chemical_name_mineral ''
_chemical_formula_sum 'Li2 Sb'

loop_
_publ_author_name
 'Wiking M\"{u}ller'
_journal_name_full_name
;
 Zeitschrift f{\"u}r Naturforschung B
;
_journal_volume 32
_journal_year 1977
_journal_page_first 357
_journal_page_last 359
_publ_Section_title
;
 Darstellung und Struktur der Phase Li$_{2}$Sb
;

_aflow_title 'Li$_{2}$Sb Structure'
_aflow_proto 'A2B_hP18_190_gh_bf'
_aflow_params 'a,c/a,z_{2},x_{3},x_{4},y_{4}'
_aflow_params_values '7.946,0.789076264787,0.0225,0.294,0.612,0.01'
_aflow_Strukturbericht 'None'
_aflow_Pearson 'hP18'

_symmetry_space_group_name_H-M "P -6 2 c"
_symmetry_Int_Tables_number 190
 
_cell_length_a    7.94600
_cell_length_b    7.94600
_cell_length_c    6.27000
_cell_angle_alpha 90.00000
_cell_angle_beta  90.00000
_cell_angle_gamma 120.00000
 
loop_
_space_group_symop_id
_space_group_symop_operation_xyz
1 x,y,z
2 -y,x-y,z
3 -x+y,-x,z
4 x-y,-y,-z
5 y,x,-z
6 -x,-x+y,-z
7 -x+y,-x,-z+1/2
8 x,y,-z+1/2
9 -y,x-y,-z+1/2
10 -x,-x+y,z+1/2
11 x-y,-y,z+1/2
12 y,x,z+1/2
 
loop_
_atom_site_label
_atom_site_type_symbol
_atom_site_symmetry_multiplicity
_atom_site_Wyckoff_label
_atom_site_fract_x
_atom_site_fract_y
_atom_site_fract_z
_atom_site_occupancy
Sb1 Sb   2 b 0.00000 0.00000  0.25000 1.00000
Sb2 Sb   4 f 0.33333 0.66667  0.02250 1.00000
Li1 Li   6 g 0.29400 0.00000  0.00000 1.00000
Li2 Li   6 h 0.61200 0.01000  0.25000 1.00000
\end{lstlisting}
{\phantomsection\label{A2B_hP18_190_gh_bf_poscar}}
{\hyperref[A2B_hP18_190_gh_bf]{Li$_{2}$Sb: A2B\_hP18\_190\_gh\_bf}} - POSCAR
\begin{lstlisting}[numbers=none,language={mylang}]
A2B_hP18_190_gh_bf & a,c/a,z2,x3,x4,y4 --params=7.946,0.789076264787,0.0225,0.294,0.612,0.01 & P-62c D_{3h}^{4} #190 (bfgh) & hP18 & None & Li2Sb &  & Wiking M\"{u}ller, Z. Naturforsch. B 32, 357-359 (1977)
   1.00000000000000
   3.97300000000000  -6.88143785847115   0.00000000000000
   3.97300000000000   6.88143785847115   0.00000000000000
   0.00000000000000   0.00000000000000   6.27000000000000
    Li    Sb
    12     6
Direct
   0.29400000000000   0.00000000000000   0.00000000000000   Li   (6g)
   0.00000000000000   0.29400000000000   0.00000000000000   Li   (6g)
  -0.29400000000000  -0.29400000000000   0.00000000000000   Li   (6g)
   0.29400000000000   0.00000000000000   0.50000000000000   Li   (6g)
   0.00000000000000   0.29400000000000   0.50000000000000   Li   (6g)
  -0.29400000000000  -0.29400000000000   0.50000000000000   Li   (6g)
   0.61200000000000   0.01000000000000   0.25000000000000   Li   (6h)
  -0.01000000000000   0.60200000000000   0.25000000000000   Li   (6h)
  -0.60200000000000  -0.61200000000000   0.25000000000000   Li   (6h)
   0.01000000000000   0.61200000000000   0.75000000000000   Li   (6h)
   0.60200000000000  -0.01000000000000   0.75000000000000   Li   (6h)
  -0.61200000000000  -0.60200000000000   0.75000000000000   Li   (6h)
   0.00000000000000   0.00000000000000   0.25000000000000   Sb   (2b)
   0.00000000000000   0.00000000000000   0.75000000000000   Sb   (2b)
   0.33333333333333   0.66666666666667   0.02250000000000   Sb   (4f)
   0.33333333333333   0.66666666666667   0.47750000000000   Sb   (4f)
   0.66666666666667   0.33333333333333  -0.02250000000000   Sb   (4f)
   0.66666666666667   0.33333333333333   0.52250000000000   Sb   (4f)
\end{lstlisting}
{\phantomsection\label{A5B3_hP16_190_bdh_g_cif}}
{\hyperref[A5B3_hP16_190_bdh_g]{$\alpha$-Sm$_{3}$Ge$_{5}$ (High-temperature): A5B3\_hP16\_190\_bdh\_g}} - CIF
\begin{lstlisting}[numbers=none,language={mylang}]
# CIF file
data_findsym-output
_audit_creation_method FINDSYM

_chemical_name_mineral 'alpha-Sm3Ge5'
_chemical_formula_sum 'Ge5 Sm3'

loop_
_publ_author_name
 'P. H. Tobash'
 'D. Lins'
 'S. Bobev'
 'N. Hur'
 'J. D. Thompson'
 'J. L. Sarrao'
_journal_name_full_name
;
 Inorganic Chemistry
;
_journal_volume 45
_journal_year 2006
_journal_page_first 7286
_journal_page_last 7294
_publ_Section_title
;
 Vacancy ordering in SmGe$_{2-x}$ and GdGe$_{2-x}$ ($x$ = 0.33): Structure and properties of two Sm$_{3}$Ge$_{5}$ polymorphs and of Gd$_{3}$Ge$_{5}$
;

# Found in Pearson's Crystal Data - Crystal Structure Database for Inorganic Compounds, 2013

_aflow_title '$\alpha$-Sm$_{3}$Ge$_{5}$ (High-temperature) Structure'
_aflow_proto 'A5B3_hP16_190_bdh_g'
_aflow_params 'a,c/a,x_{3},x_{4},y_{4}'
_aflow_params_values '6.9236970109,1.22634969237,0.3313,0.0628,0.6682'
_aflow_Strukturbericht 'None'
_aflow_Pearson 'hP16'

_cell_length_a    6.9236970109
_cell_length_b    6.9236970109
_cell_length_c    8.4908736994
_cell_angle_alpha 90.0000000000
_cell_angle_beta  90.0000000000
_cell_angle_gamma 120.0000000000
 
_symmetry_space_group_name_H-M "P -6 2 c"
_symmetry_Int_Tables_number 190
 
loop_
_space_group_symop_id
_space_group_symop_operation_xyz
1 x,y,z
2 -y,x-y,z
3 -x+y,-x,z
4 x-y,-y,-z
5 y,x,-z
6 -x,-x+y,-z
7 -x+y,-x,-z+1/2
8 x,y,-z+1/2
9 -y,x-y,-z+1/2
10 -x,-x+y,z+1/2
11 x-y,-y,z+1/2
12 y,x,z+1/2
 
loop_
_atom_site_label
_atom_site_type_symbol
_atom_site_symmetry_multiplicity
_atom_site_Wyckoff_label
_atom_site_fract_x
_atom_site_fract_y
_atom_site_fract_z
_atom_site_occupancy
Ge1 Ge   2 b 0.00000 0.00000 0.25000 1.00000
Ge2 Ge   2 d 0.66667 0.33333 0.25000 1.00000
Sm1 Sm   6 g 0.33130 0.00000 0.00000 1.00000
Ge3 Ge   6 h 0.06280 0.66820 0.25000 1.00000
\end{lstlisting}
{\phantomsection\label{A5B3_hP16_190_bdh_g_poscar}}
{\hyperref[A5B3_hP16_190_bdh_g]{$\alpha$-Sm$_{3}$Ge$_{5}$ (High-temperature): A5B3\_hP16\_190\_bdh\_g}} - POSCAR
\begin{lstlisting}[numbers=none,language={mylang}]
A5B3_hP16_190_bdh_g & a,c/a,x3,x4,y4 --params=6.9236970109,1.22634969237,0.3313,0.0628,0.6682 & P-62c D_{3h}^{4} #190 (bdgh) & hP16 & None & Sm3Ge5 & alpha & P. H. Tobash et al., Inorg. Chem. 45, 7286-7294 (2006)
   1.00000000000000
   3.46184850545000  -5.99609749954578   0.00000000000000
   3.46184850545000   5.99609749954578   0.00000000000000
   0.00000000000000   0.00000000000000   8.49087369940000
    Ge    Sm
    10     6
Direct
   0.00000000000000   0.00000000000000   0.25000000000000   Ge   (2b)
   0.00000000000000   0.00000000000000   0.75000000000000   Ge   (2b)
   0.66666666666667   0.33333333333333   0.25000000000000   Ge   (2d)
   0.33333333333333   0.66666666666667   0.75000000000000   Ge   (2d)
   0.06280000000000   0.66820000000000   0.25000000000000   Ge   (6h)
  -0.66820000000000  -0.60540000000000   0.25000000000000   Ge   (6h)
   0.60540000000000  -0.06280000000000   0.25000000000000   Ge   (6h)
   0.66820000000000   0.06280000000000   0.75000000000000   Ge   (6h)
  -0.60540000000000  -0.66820000000000   0.75000000000000   Ge   (6h)
  -0.06280000000000   0.60540000000000   0.75000000000000   Ge   (6h)
   0.33130000000000   0.00000000000000   0.00000000000000   Sm   (6g)
   0.00000000000000   0.33130000000000   0.00000000000000   Sm   (6g)
  -0.33130000000000  -0.33130000000000   0.00000000000000   Sm   (6g)
   0.33130000000000   0.00000000000000   0.50000000000000   Sm   (6g)
   0.00000000000000   0.33130000000000   0.50000000000000   Sm   (6g)
  -0.33130000000000  -0.33130000000000   0.50000000000000   Sm   (6g)
\end{lstlisting}
{\phantomsection\label{AB_hP24_190_i_afh_cif}}
{\hyperref[AB_hP24_190_i_afh]{Troilite (FeS): AB\_hP24\_190\_i\_afh}} - CIF
\begin{lstlisting}[numbers=none,language={mylang}]
# CIF file
data_findsym-output
_audit_creation_method FINDSYM

_chemical_name_mineral 'FeS'
_chemical_formula_sum 'Fe S'

loop_
_publ_author_name
 'N. Morimoto'
 'H. Nakazawa'
 'K. Nishigucmi'
 'M. Tokonami'
_journal_name_full_name
;
 Science
;
_journal_volume 168
_journal_year 1970
_journal_page_first 964
_journal_page_last 966
_publ_Section_title
;
 Pyrrhotites: Stoichiometric Compounds with Composition Fe$_{n-1}$S$_{n}$ ($n \ge 8$)
;

# Found in Pearson's Crystal Data - Crystal Structure Database for Inorganic Compounds, 2013

_aflow_title 'Troilite (FeS) Structure'
_aflow_proto 'AB_hP24_190_i_afh'
_aflow_params 'a,c/a,z_{2},x_{3},y_{3},x_{4},y_{4},z_{4}'
_aflow_params_values '5.9699820408,1.96984924623,0.52,0.6683,0.6653,0.3786,0.3233,0.623'
_aflow_Strukturbericht 'None'
_aflow_Pearson 'hP24'

_cell_length_a    5.9699820408
_cell_length_b    5.9699820408
_cell_length_c    11.7599646231
_cell_angle_alpha 90.0000000000
_cell_angle_beta  90.0000000000
_cell_angle_gamma 120.0000000000
 
_symmetry_space_group_name_H-M "P -6 2 c"
_symmetry_Int_Tables_number 190
 
loop_
_space_group_symop_id
_space_group_symop_operation_xyz
1 x,y,z
2 -y,x-y,z
3 -x+y,-x,z
4 x-y,-y,-z
5 y,x,-z
6 -x,-x+y,-z
7 -x+y,-x,-z+1/2
8 x,y,-z+1/2
9 -y,x-y,-z+1/2
10 -x,-x+y,z+1/2
11 x-y,-y,z+1/2
12 y,x,z+1/2
 
loop_
_atom_site_label
_atom_site_type_symbol
_atom_site_symmetry_multiplicity
_atom_site_Wyckoff_label
_atom_site_fract_x
_atom_site_fract_y
_atom_site_fract_z
_atom_site_occupancy
S1  S    2 a 0.00000 0.00000 0.00000 1.00000
S2  S    4 f 0.33333 0.66667 0.52000 1.00000
S3  S    6 h 0.66830 0.66530 0.25000 1.00000
Fe1 Fe  12 i 0.37860 0.32330 0.62300 1.00000
\end{lstlisting}
{\phantomsection\label{AB_hP24_190_i_afh_poscar}}
{\hyperref[AB_hP24_190_i_afh]{Troilite (FeS): AB\_hP24\_190\_i\_afh}} - POSCAR

{\phantomsection\label{A2B3C18D6_hP58_192_c_f_lm_l_cif}}
{\hyperref[A2B3C18D6_hP58_192_c_f_lm_l]{Beryl (Be$_{3}$Al$_{2}$Si$_{6}$O$_{18}$, $G3_{1}$): A2B3C18D6\_hP58\_192\_c\_f\_lm\_l}} - CIF
\begin{lstlisting}[numbers=none,language={mylang}]
# CIF file
data_findsym-output
_audit_creation_method FINDSYM

_chemical_name_mineral 'Beryl'
_chemical_formula_sum 'Al2 Be3 O18 Si6'

loop_
_publ_author_name
 'R. M. Hazen'
 'A. Y. Au'
 'L. W. Finger'
_journal_name_full_name
;
 American Mineralogist
;
_journal_volume 71
_journal_year 1986
_journal_page_first 977
_journal_page_last 984
_publ_Section_title
;
 High-pressure crystal chemistry of beryl (Be$_{3}$Al$_{2}$Si$_{6}$O$_{18}$) and euclase (BeAlSiO$_{4}$OH)
;

# Found in The American Mineralogist Crystal Structure Database, 2003

_aflow_title 'Beryl (Be$_{3}$Al$_{2}$Si$_{6}$O$_{18}$, $G3_{1}$) Structure'
_aflow_proto 'A2B3C18D6_hP58_192_c_f_lm_l'
_aflow_params 'a,c/a,x_{3},y_{3},x_{4},y_{4},x_{5},y_{5},z_{5}'
_aflow_params_values '9.214,0.997829390059,0.3103,0.2369,0.3876,0.1159,0.4985,0.1456,0.1453'
_aflow_Strukturbericht '$G3_{1}$'
_aflow_Pearson 'hP58'

_symmetry_space_group_name_H-M "P 6/m 2/c 2/c"
_symmetry_Int_Tables_number 192
 
_cell_length_a    9.21400
_cell_length_b    9.21400
_cell_length_c    9.19400
_cell_angle_alpha 90.00000
_cell_angle_beta  90.00000
_cell_angle_gamma 120.00000
 
loop_
_space_group_symop_id
_space_group_symop_operation_xyz
1 x,y,z
2 x-y,x,z
3 -y,x-y,z
4 -x,-y,z
5 -x+y,-x,z
6 y,-x+y,z
7 x-y,-y,-z+1/2
8 x,x-y,-z+1/2
9 y,x,-z+1/2
10 -x+y,y,-z+1/2
11 -x,-x+y,-z+1/2
12 -y,-x,-z+1/2
13 -x,-y,-z
14 -x+y,-x,-z
15 y,-x+y,-z
16 x,y,-z
17 x-y,x,-z
18 -y,x-y,-z
19 -x+y,y,z+1/2
20 -x,-x+y,z+1/2
21 -y,-x,z+1/2
22 x-y,-y,z+1/2
23 x,x-y,z+1/2
24 y,x,z+1/2
 
loop_
_atom_site_label
_atom_site_type_symbol
_atom_site_symmetry_multiplicity
_atom_site_Wyckoff_label
_atom_site_fract_x
_atom_site_fract_y
_atom_site_fract_z
_atom_site_occupancy
Al1 Al   4 c 0.33333 0.66667 0.25000 1.00000
Be1 Be   6 f 0.50000 0.00000 0.25000 1.00000
O1  O   12 l 0.31030 0.23690 0.00000 1.00000
Si1 Si  12 l 0.38760 0.11590 0.00000 1.00000
O2  O   24 m 0.49850 0.14560 0.14530 1.00000
\end{lstlisting}
{\phantomsection\label{A2B3C18D6_hP58_192_c_f_lm_l_poscar}}
{\hyperref[A2B3C18D6_hP58_192_c_f_lm_l]{Beryl (Be$_{3}$Al$_{2}$Si$_{6}$O$_{18}$, $G3_{1}$): A2B3C18D6\_hP58\_192\_c\_f\_lm\_l}} - POSCAR

{\phantomsection\label{AB2_hP72_192_m_j2kl_cif}}
{\hyperref[AB2_hP72_192_m_j2kl]{AlPO$_{4}$: AB2\_hP72\_192\_m\_j2kl}} - CIF

{\phantomsection\label{AB2_hP72_192_m_j2kl_poscar}}
{\hyperref[AB2_hP72_192_m_j2kl]{AlPO$_{4}$: AB2\_hP72\_192\_m\_j2kl}} - POSCAR

{\phantomsection\label{A5B3_hP16_193_dg_g_cif}}
{\hyperref[A5B3_hP16_193_dg_g]{Mavlyanovite (Mn$_{5}$Si$_{3}$): A5B3\_hP16\_193\_dg\_g}} - CIF
\begin{lstlisting}[numbers=none,language={mylang}]
# CIF file
data_findsym-output
_audit_creation_method FINDSYM

_chemical_name_mineral 'Mn5Si3'
_chemical_formula_sum 'Mn5 Si3'

loop_
_publ_author_name
 'B. Aronsson'
_journal_name_full_name
;
 Acta Chemica Scandinavica
;
_journal_volume 14
_journal_year 1960
_journal_page_first 1414
_journal_page_last 1418
_publ_Section_title
;
 A note on the compositions and crystal structures of MnB$_{2}$, Mn$_{3}$Si, Mn$_{5}$Si$_{3}$, and FeSi$_{2}$
;

# Found in Pearson's Crystal Data - Crystal Structure Database for Inorganic Compounds, 2013

_aflow_title 'Mavlyanovite (Mn$_{5}$Si$_{3}$) Structure'
_aflow_proto 'A5B3_hP16_193_dg_g'
_aflow_params 'a,c/a,x_{2},x_{3}'
_aflow_params_values '6.9104160691,0.696671490596,0.2358,0.5992'
_aflow_Strukturbericht 'None'
_aflow_Pearson 'hP16'

_cell_length_a    6.9104160691
_cell_length_b    6.9104160691
_cell_length_c    4.8142898635
_cell_angle_alpha 90.0000000000
_cell_angle_beta  90.0000000000
_cell_angle_gamma 120.0000000000
 
_symmetry_space_group_name_H-M "P 63/m 2/c 2/m"
_symmetry_Int_Tables_number 193
 
loop_
_space_group_symop_id
_space_group_symop_operation_xyz
1 x,y,z
2 x-y,x,z+1/2
3 -y,x-y,z
4 -x,-y,z+1/2
5 -x+y,-x,z
6 y,-x+y,z+1/2
7 x-y,-y,-z+1/2
8 x,x-y,-z
9 y,x,-z+1/2
10 -x+y,y,-z
11 -x,-x+y,-z+1/2
12 -y,-x,-z
13 -x,-y,-z
14 -x+y,-x,-z+1/2
15 y,-x+y,-z
16 x,y,-z+1/2
17 x-y,x,-z
18 -y,x-y,-z+1/2
19 -x+y,y,z+1/2
20 -x,-x+y,z
21 -y,-x,z+1/2
22 x-y,-y,z
23 x,x-y,z+1/2
24 y,x,z
 
loop_
_atom_site_label
_atom_site_type_symbol
_atom_site_symmetry_multiplicity
_atom_site_Wyckoff_label
_atom_site_fract_x
_atom_site_fract_y
_atom_site_fract_z
_atom_site_occupancy
Mn1 Mn   4 d 0.33333 0.66667 0.00000 1.00000
Mn2 Mn   6 g 0.23580 0.00000 0.25000 1.00000
Si1 Si   6 g 0.59920 0.00000 0.25000 1.00000
\end{lstlisting}
{\phantomsection\label{A5B3_hP16_193_dg_g_poscar}}
{\hyperref[A5B3_hP16_193_dg_g]{Mavlyanovite (Mn$_{5}$Si$_{3}$): A5B3\_hP16\_193\_dg\_g}} - POSCAR
\begin{lstlisting}[numbers=none,language={mylang}]
A5B3_hP16_193_dg_g & a,c/a,x2,x3 --params=6.9104160691,0.696671490596,0.2358,0.5992 & P6_{3}/mcm D_{6h}^{3} #193 (dg^2) & hP16 & None & Mn5Si3 &  & B. Aronsson, Acta Chem. Scand. 14, 1414-1418 (1960)
   1.00000000000000
   3.45520803455000  -5.98459586656080   0.00000000000000
   3.45520803455000   5.98459586656080   0.00000000000000
   0.00000000000000   0.00000000000000   4.81428986350000
    Mn    Si
    10     6
Direct
   0.33333333333333   0.66666666666667   0.00000000000000   Mn   (4d)
   0.66666666666667   0.33333333333333   0.50000000000000   Mn   (4d)
   0.66666666666667   0.33333333333333   0.00000000000000   Mn   (4d)
   0.33333333333333   0.66666666666667   0.50000000000000   Mn   (4d)
   0.23580000000000   0.00000000000000   0.25000000000000   Mn   (6g)
   0.00000000000000   0.23580000000000   0.25000000000000   Mn   (6g)
  -0.23580000000000  -0.23580000000000   0.25000000000000   Mn   (6g)
  -0.23580000000000   0.00000000000000   0.75000000000000   Mn   (6g)
   0.00000000000000  -0.23580000000000   0.75000000000000   Mn   (6g)
   0.23580000000000   0.23580000000000   0.75000000000000   Mn   (6g)
   0.59920000000000   0.00000000000000   0.25000000000000   Si   (6g)
   0.00000000000000   0.59920000000000   0.25000000000000   Si   (6g)
  -0.59920000000000  -0.59920000000000   0.25000000000000   Si   (6g)
  -0.59920000000000   0.00000000000000   0.75000000000000   Si   (6g)
   0.00000000000000  -0.59920000000000   0.75000000000000   Si   (6g)
   0.59920000000000   0.59920000000000   0.75000000000000   Si   (6g)
\end{lstlisting}
{\phantomsection\label{A3B_hP16_194_gh_ac_cif}}
{\hyperref[A3B_hP16_194_gh_ac]{Ni$_{3}$Ti ($D0_{24}$): A3B\_hP16\_194\_gh\_ac}} - CIF
\begin{lstlisting}[numbers=none,language={mylang}]
# CIF file
data_findsym-output
_audit_creation_method FINDSYM

_chemical_name_mineral ''
_chemical_formula_sum 'Ni3 Ti'

loop_
_publ_author_name
 'F. Laves'
 'H. J. Wallbaum'
_journal_name_full_name
;
 Zeitschrift f{\"u}r Kristallografiya
;
_journal_volume 101
_journal_year 1939
_journal_page_first 78
_journal_page_last 93
_publ_Section_title
;
 Die Kristallstruktur von Ni$_{3}$Ti und Si$_{2}$Ti (Zwei neue Typen.)
;

_aflow_title 'Ni$_{3}$Ti ($D0_{24}$) Structure'
_aflow_proto 'A3B_hP16_194_gh_ac'
_aflow_params 'a,c/a,x_{4}'
_aflow_params_values '5.096,1.6295133438,-0.16667'
_aflow_Strukturbericht '$D0_{24}$'
_aflow_Pearson 'hP16'

_symmetry_space_group_name_H-M "P 63/m 2/m 2/c"
_symmetry_Int_Tables_number 194
 
_cell_length_a    5.09600
_cell_length_b    5.09600
_cell_length_c    8.30400
_cell_angle_alpha 90.00000
_cell_angle_beta  90.00000
_cell_angle_gamma 120.00000
 
loop_
_space_group_symop_id
_space_group_symop_operation_xyz
1 x,y,z
2 x-y,x,z+1/2
3 -y,x-y,z
4 -x,-y,z+1/2
5 -x+y,-x,z
6 y,-x+y,z+1/2
7 x-y,-y,-z
8 x,x-y,-z+1/2
9 y,x,-z
10 -x+y,y,-z+1/2
11 -x,-x+y,-z
12 -y,-x,-z+1/2
13 -x,-y,-z
14 -x+y,-x,-z+1/2
15 y,-x+y,-z
16 x,y,-z+1/2
17 x-y,x,-z
18 -y,x-y,-z+1/2
19 -x+y,y,z
20 -x,-x+y,z+1/2
21 -y,-x,z
22 x-y,-y,z+1/2
23 x,x-y,z
24 y,x,z+1/2
 
loop_
_atom_site_label
_atom_site_type_symbol
_atom_site_symmetry_multiplicity
_atom_site_Wyckoff_label
_atom_site_fract_x
_atom_site_fract_y
_atom_site_fract_z
_atom_site_occupancy
Ti1 Ti   2 a 0.00000 0.00000 0.00000 1.00000
Ti2 Ti   2 c 0.33333 0.66667 0.25000 1.00000
Ni1 Ni   6 g 0.50000 0.00000 0.00000 1.00000
Ni2 Ni   6 h -0.16667 -0.33333 0.25000 1.00000
\end{lstlisting}
{\phantomsection\label{A3B_hP16_194_gh_ac_poscar}}
{\hyperref[A3B_hP16_194_gh_ac]{Ni$_{3}$Ti ($D0_{24}$): A3B\_hP16\_194\_gh\_ac}} - POSCAR
\begin{lstlisting}[numbers=none,language={mylang}]
A3B_hP16_194_gh_ac & a,c/a,x4 --params=5.096,1.6295133438,-0.16667 & P6_{3}/mmc D_{6h}^{4} #194 (acgh) & hP16 & $D0_{24}$ & Ni3Ti &  & F. Laves and H. J. Wallbaum, Z. Kristallogr. 101, 78-93 (1939)
   1.00000000000000
   2.54800000000000  -4.41326545768550   0.00000000000000
   2.54800000000000   4.41326545768550   0.00000000000000
   0.00000000000000   0.00000000000000   8.30400000000000
    Ni    Ti
    12     4
Direct
   0.50000000000000   0.00000000000000   0.00000000000000   Ni   (6g)
   0.00000000000000   0.50000000000000   0.00000000000000   Ni   (6g)
   0.50000000000000   0.50000000000000   0.00000000000000   Ni   (6g)
   0.50000000000000   0.00000000000000   0.50000000000000   Ni   (6g)
   0.00000000000000   0.50000000000000   0.50000000000000   Ni   (6g)
   0.50000000000000   0.50000000000000   0.50000000000000   Ni   (6g)
  -0.16667000000000  -0.33334000000000   0.25000000000000   Ni   (6h)
   0.33334000000000   0.16667000000000   0.25000000000000   Ni   (6h)
  -0.16667000000000   0.16667000000000   0.25000000000000   Ni   (6h)
   0.16667000000000   0.33334000000000   0.75000000000000   Ni   (6h)
  -0.33334000000000  -0.16667000000000   0.75000000000000   Ni   (6h)
   0.16667000000000  -0.16667000000000   0.75000000000000   Ni   (6h)
   0.00000000000000   0.00000000000000   0.00000000000000   Ti   (2a)
   0.00000000000000   0.00000000000000   0.50000000000000   Ti   (2a)
   0.33333333333333   0.66666666666667   0.25000000000000   Ti   (2c)
   0.66666666666667   0.33333333333333   0.75000000000000   Ti   (2c)
\end{lstlisting}
{\phantomsection\label{A5B2_hP28_194_ahk_ch_cif}}
{\hyperref[A5B2_hP28_194_ahk_ch]{Co$_{2}$Al$_{5}$ ($D8_{11}$): A5B2\_hP28\_194\_ahk\_ch}} - CIF
\begin{lstlisting}[numbers=none,language={mylang}]
# CIF file
data_findsym-output
_audit_creation_method FINDSYM

_chemical_name_mineral 'Co2Al5'
_chemical_formula_sum 'Al5 Co2'

loop_
_publ_author_name
 'J. B. Newkirk'
 'P. J. Black'
 'A. Damjanovic'
_journal_name_full_name
;
 Acta Cristallographica
;
_journal_volume 14
_journal_year 1961
_journal_page_first 532
_journal_page_last 533
_publ_Section_title
;
 The refinement of the Co$_{2}$Al$_{5}$ structures
;

# Found in A Palladium-Magnesium Alloy Phase of Co$_{2}$Al$_{5}$ Type, 1968

_aflow_title 'Co$_{2}$Al$_{5}$ ($D8_{11}$) Structure'
_aflow_proto 'A5B2_hP28_194_ahk_ch'
_aflow_params 'a,c/a,x_{3},x_{4},x_{5},z_{5}'
_aflow_params_values '7.656,0.991771159875,0.533,0.872,0.196,0.439'
_aflow_Strukturbericht '$D8_{11}$'
_aflow_Pearson 'hP28'

_symmetry_space_group_name_H-M "P 63/m 2/m 2/c"
_symmetry_Int_Tables_number 194
 
_cell_length_a    7.65600
_cell_length_b    7.65600
_cell_length_c    7.59300
_cell_angle_alpha 90.00000
_cell_angle_beta  90.00000
_cell_angle_gamma 120.00000
 
loop_
_space_group_symop_id
_space_group_symop_operation_xyz
1 x,y,z
2 x-y,x,z+1/2
3 -y,x-y,z
4 -x,-y,z+1/2
5 -x+y,-x,z
6 y,-x+y,z+1/2
7 x-y,-y,-z
8 x,x-y,-z+1/2
9 y,x,-z
10 -x+y,y,-z+1/2
11 -x,-x+y,-z
12 -y,-x,-z+1/2
13 -x,-y,-z
14 -x+y,-x,-z+1/2
15 y,-x+y,-z
16 x,y,-z+1/2
17 x-y,x,-z
18 -y,x-y,-z+1/2
19 -x+y,y,z
20 -x,-x+y,z+1/2
21 -y,-x,z
22 x-y,-y,z+1/2
23 x,x-y,z
24 y,x,z+1/2
 
loop_
_atom_site_label
_atom_site_type_symbol
_atom_site_symmetry_multiplicity
_atom_site_Wyckoff_label
_atom_site_fract_x
_atom_site_fract_y
_atom_site_fract_z
_atom_site_occupancy
Al1 Al   2 a 0.00000 0.00000 0.00000 1.00000
Co1 Co   2 c 0.33333 0.66667 0.25000 1.00000
Al2 Al   6 h 0.53300 0.06600 0.25000 1.00000
Co2 Co   6 h 0.87200 0.74400 0.25000 1.00000
Al3 Al  12 k 0.19600 0.39200 0.43900 1.00000
\end{lstlisting}
{\phantomsection\label{A5B2_hP28_194_ahk_ch_poscar}}
{\hyperref[A5B2_hP28_194_ahk_ch]{Co$_{2}$Al$_{5}$ ($D8_{11}$): A5B2\_hP28\_194\_ahk\_ch}} - POSCAR

{\phantomsection\label{A9B3C_hP26_194_hk_h_a_cif}}
{\hyperref[A9B3C_hP26_194_hk_h_a]{Al$_{9}$Mn$_{3}$Si ($E9_{c}$): A9B3C\_hP26\_194\_hk\_h\_a}} - CIF
\begin{lstlisting}[numbers=none,language={mylang}]
# CIF file
data_findsym-output
_audit_creation_method FINDSYM

_chemical_name_mineral 'Al9Mn3Si'
_chemical_formula_sum 'Al9 Mn3 Si'

loop_
_publ_author_name
 'K. Robinson'
_journal_name_full_name
;
 Philosophical Magazine
;
_journal_volume 43
_journal_year 1952
_journal_page_first 775
_journal_page_last 782
_publ_Section_title
;
 LXXIII. The unit cell and Brillouin Zones of Ni$_{4}$Mn$_{11}$Al$_{60}$ and belated compounds
;

# Found in A Handbook of Lattice Spacings and Structures of Metals and Alloys, 1958

_aflow_title 'Al$_{9}$Mn$_{3}$Si ($E9_{c}$) Structure'
_aflow_proto 'A9B3C_hP26_194_hk_h_a'
_aflow_params 'a,c/a,x_{2},x_{3},x_{4},z_{4}'
_aflow_params_values '7.513,1.03087980833,0.458,0.12,0.201,-0.067'
_aflow_Strukturbericht '$E9_{c}$'
_aflow_Pearson 'hP26'

_symmetry_space_group_name_H-M "P 63/m 2/m 2/c"
_symmetry_Int_Tables_number 194
 
_cell_length_a    7.51300
_cell_length_b    7.51300
_cell_length_c    7.74500
_cell_angle_alpha 90.00000
_cell_angle_beta  90.00000
_cell_angle_gamma 120.00000
 
loop_
_space_group_symop_id
_space_group_symop_operation_xyz
1 x,y,z
2 x-y,x,z+1/2
3 -y,x-y,z
4 -x,-y,z+1/2
5 -x+y,-x,z
6 y,-x+y,z+1/2
7 x-y,-y,-z
8 x,x-y,-z+1/2
9 y,x,-z
10 -x+y,y,-z+1/2
11 -x,-x+y,-z
12 -y,-x,-z+1/2
13 -x,-y,-z
14 -x+y,-x,-z+1/2
15 y,-x+y,-z
16 x,y,-z+1/2
17 x-y,x,-z
18 -y,x-y,-z+1/2
19 -x+y,y,z
20 -x,-x+y,z+1/2
21 -y,-x,z
22 x-y,-y,z+1/2
23 x,x-y,z
24 y,x,z+1/2
 
loop_
_atom_site_label
_atom_site_type_symbol
_atom_site_symmetry_multiplicity
_atom_site_Wyckoff_label
_atom_site_fract_x
_atom_site_fract_y
_atom_site_fract_z
_atom_site_occupancy
Si1 Si   2 a 0.00000 0.00000 0.00000 1.00000
Al1 Al   6 h 0.45800 0.91600 0.25000 1.00000
Mn1 Mn   6 h 0.12000 0.24000 0.25000 1.00000
Al2 Al  12 k 0.20100 0.40200 -0.06700 1.00000
\end{lstlisting}
{\phantomsection\label{A9B3C_hP26_194_hk_h_a_poscar}}
{\hyperref[A9B3C_hP26_194_hk_h_a]{Al$_{9}$Mn$_{3}$Si ($E9_{c}$): A9B3C\_hP26\_194\_hk\_h\_a}} - POSCAR

{\phantomsection\label{A12BC4_cP34_195_2j_ab_2e_cif}}
{\hyperref[A12BC4_cP34_195_2j_ab_2e]{PrRu$_{4}$P$_{12}$: A12BC4\_cP34\_195\_2j\_ab\_2e}} - CIF
\begin{lstlisting}[numbers=none,language={mylang}]
# CIF file
data_findsym-output
_audit_creation_method FINDSYM

_chemical_name_mineral 'PrRu4P12'
_chemical_formula_sum 'P12 Pr Ru4'

loop_
_publ_author_name
 'C. H. Lee'
 'H. Matsuhata'
 'H. Yamaguchi'
 'C. Sekine'
 'K. Kihou'
 'I. Shirotani'
_journal_name_full_name
;
 Journal of Magnetism and Magnetic Materials
;
_journal_volume 272
_journal_year 2004
_journal_page_first 426
_journal_page_last 427
_publ_Section_title
;
 A study of the crystal structure at low temperature in the metal--insulator transition compound PrRu$_{4}$P$_{12}$
;

# Found in Pearson's Crystal Data - Crystal Structure Database for Inorganic Compounds, 2013

_aflow_title 'PrRu$_{4}$P$_{12}$ Structure'
_aflow_proto 'A12BC4_cP34_195_2j_ab_2e'
_aflow_params 'a,x_{3},x_{4},x_{5},y_{5},z_{5},x_{6},y_{6},z_{6}'
_aflow_params_values '8.0357772599,0.2493,0.7507,0.14225,0.5,0.35579,0.0002,0.14286,0.35831'
_aflow_Strukturbericht 'None'
_aflow_Pearson 'cP34'

_cell_length_a    8.0357772599
_cell_length_b    8.0357772599
_cell_length_c    8.0357772599
_cell_angle_alpha 90.0000000000
_cell_angle_beta  90.0000000000
_cell_angle_gamma 90.0000000000
 
_symmetry_space_group_name_H-M "P 2 3"
_symmetry_Int_Tables_number 195
 
loop_
_space_group_symop_id
_space_group_symop_operation_xyz
1 x,y,z
2 x,-y,-z
3 -x,y,-z
4 -x,-y,z
5 y,z,x
6 y,-z,-x
7 -y,z,-x
8 -y,-z,x
9 z,x,y
10 z,-x,-y
11 -z,x,-y
12 -z,-x,y
 
loop_
_atom_site_label
_atom_site_type_symbol
_atom_site_symmetry_multiplicity
_atom_site_Wyckoff_label
_atom_site_fract_x
_atom_site_fract_y
_atom_site_fract_z
_atom_site_occupancy
Pr1 Pr   1 a 0.00000 0.00000 0.00000 1.00000
Pr2 Pr   1 b 0.50000 0.50000 0.50000 1.00000
Ru1 Ru   4 e 0.24930 0.24930 0.24930 1.00000
Ru2 Ru   4 e 0.75070 0.75070 0.75070 1.00000
P1  P   12 j 0.14225 0.50000 0.35579 1.00000
P2  P   12 j 0.00020 0.14286 0.35831 1.00000
\end{lstlisting}
{\phantomsection\label{A12BC4_cP34_195_2j_ab_2e_poscar}}
{\hyperref[A12BC4_cP34_195_2j_ab_2e]{PrRu$_{4}$P$_{12}$: A12BC4\_cP34\_195\_2j\_ab\_2e}} - POSCAR

{\phantomsection\label{A12B2C_cF60_196_h_bc_a_cif}}
{\hyperref[A12B2C_cF60_196_h_bc_a]{Cu$_{2}$Fe[CN]$_{6}$: A12B2C\_cF60\_196\_h\_bc\_a}} - CIF

{\phantomsection\label{A12B2C_cF60_196_h_bc_a_poscar}}
{\hyperref[A12B2C_cF60_196_h_bc_a]{Cu$_{2}$Fe[CN]$_{6}$: A12B2C\_cF60\_196\_h\_bc\_a}} - POSCAR
\begin{lstlisting}[numbers=none,language={mylang}]
A12B2C_cF60_196_h_bc_a & a,x4,y4,z4 --params=9.9799933334,0.0,0.0625,0.25 & F23 T^{2} #196 (abch) & cF60 & None & Cu2Fe[CN]6 &  & R. Rigamonti, Gazz. Chim. Ital. 67, 137-146 (1937)
   1.00000000000000
   0.00000000000000   4.98999666670000   4.98999666670000
   4.98999666670000   0.00000000000000   4.98999666670000
   4.98999666670000   4.98999666670000   0.00000000000000
     C    Cu    Fe
    12     2     1
Direct
   0.31250000000000   0.18750000000000  -0.18750000000000    C  (48h)
   0.18750000000000   0.31250000000000  -0.31250000000000    C  (48h)
  -0.18750000000000  -0.31250000000000   0.31250000000000    C  (48h)
  -0.31250000000000  -0.18750000000000   0.18750000000000    C  (48h)
  -0.18750000000000   0.31250000000000   0.18750000000000    C  (48h)
  -0.31250000000000   0.18750000000000   0.31250000000000    C  (48h)
   0.31250000000000  -0.18750000000000  -0.31250000000000    C  (48h)
   0.18750000000000  -0.31250000000000  -0.18750000000000    C  (48h)
   0.18750000000000  -0.18750000000000   0.31250000000000    C  (48h)
   0.31250000000000  -0.31250000000000   0.18750000000000    C  (48h)
  -0.31250000000000   0.31250000000000  -0.18750000000000    C  (48h)
  -0.18750000000000   0.18750000000000  -0.31250000000000    C  (48h)
   0.50000000000000   0.50000000000000   0.50000000000000   Cu   (4b)
   0.25000000000000   0.25000000000000   0.25000000000000   Cu   (4c)
   0.00000000000000   0.00000000000000   0.00000000000000   Fe   (4a)
\end{lstlisting}
{\phantomsection\label{A12B36CD12_cF488_196_2h_6h_ac_fgh_cif}}
{\hyperref[A12B36CD12_cF488_196_2h_6h_ac_fgh]{MgB$_{12}$H$_{12}$[H$_{2}$O]$_{12}$: A12B36CD12\_cF488\_196\_2h\_6h\_ac\_fgh}} - CIF

{\phantomsection\label{A12B36CD12_cF488_196_2h_6h_ac_fgh_poscar}}
{\hyperref[A12B36CD12_cF488_196_2h_6h_ac_fgh]{MgB$_{12}$H$_{12}$[H$_{2}$O]$_{12}$: A12B36CD12\_cF488\_196\_2h\_6h\_ac\_fgh}} - POSCAR

{\phantomsection\label{ABC3_cP20_198_a_a_b_cif}}
{\hyperref[ABC3_cP20_198_a_a_b]{Sodium Chlorate (NaClO$_{3}$, $G3$): ABC3\_cP20\_198\_a\_a\_b}} - CIF
\begin{lstlisting}[numbers=none,language={mylang}]
# CIF file
data_findsym-output
_audit_creation_method FINDSYM

_chemical_name_mineral 'Sodium chlorate'
_chemical_formula_sum 'Cl Na O3'

loop_
_publ_author_name
 'G. N. Ramachandran'
 'K. S. Chandrasekaran'
_journal_name_full_name
;
 Acta Cristallographica
;
_journal_volume 10
_journal_year 1957
_journal_page_first 671
_journal_page_last 675
_publ_Section_title
;
 The absolute configuration of sodium chlorate
;

# Found in Acentric cubic NaClO$_{3}$--a new crystal for Raman lasers, 1998

_aflow_title 'Sodium Chlorate (NaClO$_{3}$, $G3$) Structure'
_aflow_proto 'ABC3_cP20_198_a_a_b'
_aflow_params 'a,x_{1},x_{2},x_{3},y_{3},z_{3}'
_aflow_params_values '6.57,0.417,0.064,0.303,0.592,0.5'
_aflow_Strukturbericht '$G3$'
_aflow_Pearson 'cP20'

_symmetry_space_group_name_H-M "P 21 3"
_symmetry_Int_Tables_number 198
 
_cell_length_a    6.57000
_cell_length_b    6.57000
_cell_length_c    6.57000
_cell_angle_alpha 90.00000
_cell_angle_beta  90.00000
_cell_angle_gamma 90.00000
 
loop_
_space_group_symop_id
_space_group_symop_operation_xyz
1 x,y,z
2 x+1/2,-y+1/2,-z
3 -x,y+1/2,-z+1/2
4 -x+1/2,-y,z+1/2
5 y,z,x
6 y+1/2,-z+1/2,-x
7 -y,z+1/2,-x+1/2
8 -y+1/2,-z,x+1/2
9 z,x,y
10 z+1/2,-x+1/2,-y
11 -z,x+1/2,-y+1/2
12 -z+1/2,-x,y+1/2
 
loop_
_atom_site_label
_atom_site_type_symbol
_atom_site_symmetry_multiplicity
_atom_site_Wyckoff_label
_atom_site_fract_x
_atom_site_fract_y
_atom_site_fract_z
_atom_site_occupancy
Cl1 Cl   4 a 0.41700 0.41700 0.41700 1.00000
Na1 Na   4 a 0.06400 0.06400 0.06400 1.00000
O1  O   12 b 0.30300 0.59200 0.50000 1.00000
\end{lstlisting}
{\phantomsection\label{ABC3_cP20_198_a_a_b_poscar}}
{\hyperref[ABC3_cP20_198_a_a_b]{Sodium Chlorate (NaClO$_{3}$, $G3$): ABC3\_cP20\_198\_a\_a\_b}} - POSCAR
\begin{lstlisting}[numbers=none,language={mylang}]
ABC3_cP20_198_a_a_b & a,x1,x2,x3,y3,z3 --params=6.57,0.417,0.064,0.303,0.592,0.5 & P2_{1}3 T^{4} #198 (a^2b) & cP20 & $G3$ & NaClO3 & Sodium chlorate & G. N. Ramachandran and K. S. Chandrasekaran, Acta Cryst. 10, 671-675 (1957)
   1.00000000000000
   6.57000000000000   0.00000000000000   0.00000000000000
   0.00000000000000   6.57000000000000   0.00000000000000
   0.00000000000000   0.00000000000000   6.57000000000000
    Cl    Na     O
     4     4    12
Direct
   0.41700000000000   0.41700000000000   0.41700000000000   Cl   (4a)
   0.08300000000000  -0.41700000000000   0.91700000000000   Cl   (4a)
  -0.41700000000000   0.91700000000000   0.08300000000000   Cl   (4a)
   0.91700000000000   0.08300000000000  -0.41700000000000   Cl   (4a)
   0.06400000000000   0.06400000000000   0.06400000000000   Na   (4a)
   0.43600000000000  -0.06400000000000   0.56400000000000   Na   (4a)
  -0.06400000000000   0.56400000000000   0.43600000000000   Na   (4a)
   0.56400000000000   0.43600000000000  -0.06400000000000   Na   (4a)
   0.30300000000000   0.59200000000000   0.50000000000000    O  (12b)
   0.19700000000000  -0.59200000000000   1.00000000000000    O  (12b)
  -0.30300000000000   1.09200000000000   0.00000000000000    O  (12b)
   0.80300000000000  -0.09200000000000  -0.50000000000000    O  (12b)
   0.50000000000000   0.30300000000000   0.59200000000000    O  (12b)
   1.00000000000000   0.19700000000000  -0.59200000000000    O  (12b)
   0.00000000000000  -0.30300000000000   1.09200000000000    O  (12b)
  -0.50000000000000   0.80300000000000  -0.09200000000000    O  (12b)
   0.59200000000000   0.50000000000000   0.30300000000000    O  (12b)
  -0.59200000000000   1.00000000000000   0.19700000000000    O  (12b)
   1.09200000000000   0.00000000000000  -0.30300000000000    O  (12b)
  -0.09200000000000  -0.50000000000000   0.80300000000000    O  (12b)
\end{lstlisting}
{\phantomsection\label{A2B11_cP39_200_f_aghij_cif}}
{\hyperref[A2B11_cP39_200_f_aghij]{Mg$_{2}$Zn$_{11}$: A2B11\_cP39\_200\_f\_aghij}} - CIF
\begin{lstlisting}[numbers=none,language={mylang}]
# CIF file
data_findsym-output
_audit_creation_method FINDSYM

_chemical_name_mineral 'Mg2Zn11'
_chemical_formula_sum 'Mg2 Zn11'

loop_
_publ_author_name
 'S. Samson'
_journal_name_full_name
;
 Acta Chemica Scandinavica
;
_journal_volume 3
_journal_year 1949
_journal_page_first 835
_journal_page_last 843
_publ_Section_title
;
 Die Kristallstruktur von Mg$_{2}$Zn$_{11}$ Isomorphie zwischen Mg$_{2}$Zn$_{11}$ und Mg$_{2}$Cu$_{6}$Al$_{5}$
;

# Found in Pearson's Crystal Data - Crystal Structure Database for Inorganic Compounds, 2013

_aflow_title 'Mg$_{2}$Zn$_{11}$ Structure'
_aflow_proto 'A2B11_cP39_200_f_aghij'
_aflow_params 'a,x_{2},x_{3},x_{4},x_{5},y_{6},z_{6}'
_aflow_params_values '8.5520223662,0.18,0.34,0.265,0.278,0.157,0.257'
_aflow_Strukturbericht 'None'
_aflow_Pearson 'cP39'

_cell_length_a    8.5520223662
_cell_length_b    8.5520223662
_cell_length_c    8.5520223662
_cell_angle_alpha 90.0000000000
_cell_angle_beta  90.0000000000
_cell_angle_gamma 90.0000000000
 
_symmetry_space_group_name_H-M "P 2/m -3"
_symmetry_Int_Tables_number 200
 
loop_
_space_group_symop_id
_space_group_symop_operation_xyz
1 x,y,z
2 x,-y,-z
3 -x,y,-z
4 -x,-y,z
5 y,z,x
6 y,-z,-x
7 -y,z,-x
8 -y,-z,x
9 z,x,y
10 z,-x,-y
11 -z,x,-y
12 -z,-x,y
13 -x,-y,-z
14 -x,y,z
15 x,-y,z
16 x,y,-z
17 -y,-z,-x
18 -y,z,x
19 y,-z,x
20 y,z,-x
21 -z,-x,-y
22 -z,x,y
23 z,-x,y
24 z,x,-y
 
loop_
_atom_site_label
_atom_site_type_symbol
_atom_site_symmetry_multiplicity
_atom_site_Wyckoff_label
_atom_site_fract_x
_atom_site_fract_y
_atom_site_fract_z
_atom_site_occupancy
Zn1 Zn   1 a 0.00000 0.00000 0.00000 1.00000
Mg1 Mg   6 f 0.18000 0.00000 0.50000 1.00000
Zn2 Zn   6 g 0.34000 0.50000 0.00000 1.00000
Zn3 Zn   6 h 0.26500 0.50000 0.50000 1.00000
Zn4 Zn   8 i 0.27800 0.27800 0.27800 1.00000
Zn5 Zn  12 j 0.00000 0.15700 0.25700 1.00000
\end{lstlisting}
{\phantomsection\label{A2B11_cP39_200_f_aghij_poscar}}
{\hyperref[A2B11_cP39_200_f_aghij]{Mg$_{2}$Zn$_{11}$: A2B11\_cP39\_200\_f\_aghij}} - POSCAR

{\phantomsection\label{AB3C_cP60_201_ce_fh_g_cif}}
{\hyperref[AB3C_cP60_201_ce_fh_g]{KSbO$_{3}$ (High-temperature): AB3C\_cP60\_201\_ce\_fh\_g}} - CIF
\begin{lstlisting}[numbers=none,language={mylang}]
# CIF file
data_findsym-output
_audit_creation_method FINDSYM

_chemical_name_mineral 'KSbO3'
_chemical_formula_sum 'K O3 Sb'

loop_
_publ_author_name
 'P. Spiegelberg'
_journal_name_full_name
;
 Arkiv f{\"o}r Kemi, Mineralogi och Geologi
;
_journal_volume 14A
_journal_year 1940
_journal_page_first 1
_journal_page_last 12
_publ_Section_title
;
 X-ray studies on potassium antimonates
;

# Found in Pearson's Crystal Data - Crystal Structure Database for Inorganic Compounds, 2013

_aflow_title 'KSbO$_{3}$ (High-temperature) Structure'
_aflow_proto 'AB3C_cP60_201_ce_fh_g'
_aflow_params 'a,x_{2},x_{3},x_{4},x_{5},y_{5},z_{5}'
_aflow_params_values '9.5599167841,0.8889,0.8889,0.4028,0.5389,0.5972,0.25'
_aflow_Strukturbericht 'None'
_aflow_Pearson 'cP60'

_cell_length_a    9.5599167841
_cell_length_b    9.5599167841
_cell_length_c    9.5599167841
_cell_angle_alpha 90.0000000000
_cell_angle_beta  90.0000000000
_cell_angle_gamma 90.0000000000
 
_symmetry_space_group_name_H-M "P 2/n -3 (origin choice 2)"
_symmetry_Int_Tables_number 201
 
loop_
_space_group_symop_id
_space_group_symop_operation_xyz
1 x,y,z
2 x,-y+1/2,-z+1/2
3 -x+1/2,y,-z+1/2
4 -x+1/2,-y+1/2,z
5 y,z,x
6 y,-z+1/2,-x+1/2
7 -y+1/2,z,-x+1/2
8 -y+1/2,-z+1/2,x
9 z,x,y
10 z,-x+1/2,-y+1/2
11 -z+1/2,x,-y+1/2
12 -z+1/2,-x+1/2,y
13 -x,-y,-z
14 -x,y+1/2,z+1/2
15 x+1/2,-y,z+1/2
16 x+1/2,y+1/2,-z
17 -y,-z,-x
18 -y,z+1/2,x+1/2
19 y+1/2,-z,x+1/2
20 y+1/2,z+1/2,-x
21 -z,-x,-y
22 -z,x+1/2,y+1/2
23 z+1/2,-x,y+1/2
24 z+1/2,x+1/2,-y
 
loop_
_atom_site_label
_atom_site_type_symbol
_atom_site_symmetry_multiplicity
_atom_site_Wyckoff_label
_atom_site_fract_x
_atom_site_fract_y
_atom_site_fract_z
_atom_site_occupancy
K1  K    4 c 0.50000 0.50000 0.50000 1.00000
K2  K    8 e 0.88890 0.88890 0.88890 1.00000
O1  O   12 f 0.88890 0.25000 0.25000 1.00000
Sb1 Sb  12 g 0.40280 0.75000 0.25000 1.00000
O2  O   24 h 0.53890 0.59720 0.25000 1.00000
\end{lstlisting}
{\phantomsection\label{AB3C_cP60_201_ce_fh_g_poscar}}
{\hyperref[AB3C_cP60_201_ce_fh_g]{KSbO$_{3}$ (High-temperature): AB3C\_cP60\_201\_ce\_fh\_g}} - POSCAR

{\phantomsection\label{A6B6C_cF104_202_h_h_c_cif}}
{\hyperref[A6B6C_cF104_202_h_h_c]{KB$_{6}$H$_{6}$: A6B6C\_cF104\_202\_h\_h\_c}} - CIF

{\phantomsection\label{A6B6C_cF104_202_h_h_c_poscar}}
{\hyperref[A6B6C_cF104_202_h_h_c]{KB$_{6}$H$_{6}$: A6B6C\_cF104\_202\_h\_h\_c}} - POSCAR

{\phantomsection\label{A_cF240_202_h2i_cif}}
{\hyperref[A_cF240_202_h2i]{FCC C$_{60}$ Buckminsterfullerine: A\_cF240\_202\_h2i}} - CIF

{\phantomsection\label{A_cF240_202_h2i_poscar}}
{\hyperref[A_cF240_202_h2i]{FCC C$_{60}$ Buckminsterfullerine: A\_cF240\_202\_h2i}} - POSCAR

{\phantomsection\label{A2BCD3E6_cF208_203_e_c_d_f_g_cif}}
{\hyperref[A2BCD3E6_cF208_203_e_c_d_f_g]{Pyrochlore (Na$_3$Co(CO$_3$)$_2$Cl): A2BCD3E6\_cF208\_203\_e\_c\_d\_f\_g}} - CIF

{\phantomsection\label{A2BCD3E6_cF208_203_e_c_d_f_g_poscar}}
{\hyperref[A2BCD3E6_cF208_203_e_c_d_f_g]{Pyrochlore (Na$_3$Co(CO$_3$)$_2$Cl): A2BCD3E6\_cF208\_203\_e\_c\_d\_f\_g}} - POSCAR

{\phantomsection\label{A4B2C6D16E_cF232_203_e_d_f_eg_a_cif}}
{\hyperref[A4B2C6D16E_cF232_203_e_d_f_eg_a]{Tychite (Na$_{6}$Mg$_{2}$(SO$_{4}$)(CO$_{3}$)$_{4}$): A4B2C6D16E\_cF232\_203\_e\_d\_f\_eg\_a}} - CIF

{\phantomsection\label{A4B2C6D16E_cF232_203_e_d_f_eg_a_poscar}}
{\hyperref[A4B2C6D16E_cF232_203_e_d_f_eg_a]{Tychite (Na$_{6}$Mg$_{2}$(SO$_{4}$)(CO$_{3}$)$_{4}$): A4B2C6D16E\_cF232\_203\_e\_d\_f\_eg\_a}} - POSCAR

{\phantomsection\label{AB3C16_cF160_203_b_ad_eg_cif}}
{\hyperref[AB3C16_cF160_203_b_ad_eg]{Rb$_{3}$AsSe$_{16}$: AB3C16\_cF160\_203\_b\_ad\_eg}} - CIF

{\phantomsection\label{AB3C16_cF160_203_b_ad_eg_poscar}}
{\hyperref[AB3C16_cF160_203_b_ad_eg]{Rb$_{3}$AsSe$_{16}$: AB3C16\_cF160\_203\_b\_ad\_eg}} - POSCAR

{\phantomsection\label{A2B3C6_cP264_205_2d_ab2c2d_6d_cif}}
{\hyperref[A2B3C6_cP264_205_2d_ab2c2d_6d]{Ca$_{3}$Al$_{2}$O$_{6}$: A2B3C6\_cP264\_205\_2d\_ab2c2d\_6d}} - CIF

{\phantomsection\label{A2B3C6_cP264_205_2d_ab2c2d_6d_poscar}}
{\hyperref[A2B3C6_cP264_205_2d_ab2c2d_6d]{Ca$_{3}$Al$_{2}$O$_{6}$: A2B3C6\_cP264\_205\_2d\_ab2c2d\_6d}} - POSCAR

{\phantomsection\label{A_cP240_205_10d_cif}}
{\hyperref[A_cP240_205_10d]{Simple Cubic C$_{60}$ Buckminsterfullerine: A\_cP240\_205\_10d}} - CIF

{\phantomsection\label{A_cP240_205_10d_poscar}}
{\hyperref[A_cP240_205_10d]{Simple Cubic C$_{60}$ Buckminsterfullerine: A\_cP240\_205\_10d}} - POSCAR

{\phantomsection\label{AB3C2_cI96_206_c_e_ad_cif}}
{\hyperref[AB3C2_cI96_206_c_e_ad]{AlLi$_{3}$N$_{2}$ ($E9_{d}$): AB3C2\_cI96\_206\_c\_e\_ad}} - CIF

{\phantomsection\label{AB3C2_cI96_206_c_e_ad_poscar}}
{\hyperref[AB3C2_cI96_206_c_e_ad]{AlLi$_{3}$N$_{2}$ ($E9_{d}$): AB3C2\_cI96\_206\_c\_e\_ad}} - POSCAR

{\phantomsection\label{A17B15_cP64_207_acfk_eij_cif}}
{\hyperref[A17B15_cP64_207_acfk_eij]{Pd$_{17}$Se$_{15}$: A17B15\_cP64\_207\_acfk\_eij}} - CIF
\begin{lstlisting}[numbers=none,language={mylang}]
# CIF file
data_findsym-output
_audit_creation_method FINDSYM

_chemical_name_mineral 'Pd17Se15'
_chemical_formula_sum 'Pd17 Se15'

loop_
_publ_author_name
 'S. Geller'
_journal_name_full_name
;
 Acta Cristallographica
;
_journal_volume 15
_journal_year 1962
_journal_page_first 713
_journal_page_last 721
_publ_Section_title
;
 The crystal structure of Pd$_{17}$Se$_{15}$
;

# Found in Pearson's Crystal Data - Crystal Structure Database for Inorganic Compounds, 2013

_aflow_title 'Pd$_{17}$Se$_{15}$ Structure'
_aflow_proto 'A17B15_cP64_207_acfk_eij'
_aflow_params 'a,x_{3},x_{4},y_{5},y_{6},x_{7},y_{7},z_{7}'
_aflow_params_values '10.6058825779,0.2422,0.2622,0.3319,0.2701,0.142,0.1539,0.3498'
_aflow_Strukturbericht 'None'
_aflow_Pearson 'cP64'

_cell_length_a    10.6058825779
_cell_length_b    10.6058825779
_cell_length_c    10.6058825779
_cell_angle_alpha 90.0000000000
_cell_angle_beta  90.0000000000
_cell_angle_gamma 90.0000000000
 
_symmetry_space_group_name_H-M "P 4 3 2"
_symmetry_Int_Tables_number 207
 
loop_
_space_group_symop_id
_space_group_symop_operation_xyz
1 x,y,z
2 x,-y,-z
3 -x,y,-z
4 -x,-y,z
5 y,z,x
6 y,-z,-x
7 -y,z,-x
8 -y,-z,x
9 z,x,y
10 z,-x,-y
11 -z,x,-y
12 -z,-x,y
13 -y,-x,-z
14 -y,x,z
15 y,-x,z
16 y,x,-z
17 -x,-z,-y
18 -x,z,y
19 x,-z,y
20 x,z,-y
21 -z,-y,-x
22 -z,y,x
23 z,-y,x
24 z,y,-x
 
loop_
_atom_site_label
_atom_site_type_symbol
_atom_site_symmetry_multiplicity
_atom_site_Wyckoff_label
_atom_site_fract_x
_atom_site_fract_y
_atom_site_fract_z
_atom_site_occupancy
Pd1 Pd   1 a 0.00000 0.00000 0.00000 1.00000
Pd2 Pd   3 c 0.00000 0.50000 0.50000 1.00000
Se1 Se   6 e 0.24220 0.00000 0.00000 1.00000
Pd3 Pd   6 f 0.26220 0.50000 0.50000 1.00000
Se2 Se  12 i 0.00000 0.33190 0.33190 1.00000
Se3 Se  12 j 0.50000 0.27010 0.27010 1.00000
Pd4 Pd  24 k 0.14200 0.15390 0.34980 1.00000
\end{lstlisting}
{\phantomsection\label{A17B15_cP64_207_acfk_eij_poscar}}
{\hyperref[A17B15_cP64_207_acfk_eij]{Pd$_{17}$Se$_{15}$: A17B15\_cP64\_207\_acfk\_eij}} - POSCAR

{\phantomsection\label{A3B_cP16_208_j_b_cif}}
{\hyperref[A3B_cP16_208_j_b]{PH$_{3}$: A3B\_cP16\_208\_j\_b}} - CIF
\begin{lstlisting}[numbers=none,language={mylang}]
# CIF file 
data_findsym-output
_audit_creation_method FINDSYM

_chemical_name_mineral 'PH3'
_chemical_formula_sum 'H3 P'

loop_
_publ_author_name
 'G. Natta'
 'E. Casazza'
_journal_name_full_name
;
 Gazzetta Chimica Italiana
;
_journal_volume 60
_journal_year 1930
_journal_page_first 851
_journal_page_last 859
_publ_Section_title
;
 La struttura dell\'idrogeno fosforato (PH$_{3}$) e dell\'idrogeno arsenicale (AsH$_{3}$)
;

# Found in The American Mineralogist Crystal Structure Database, 2003

_aflow_title 'PH$_{3}$ Structure'
_aflow_proto 'A3B_cP16_208_j_b'
_aflow_params 'a,x_{2}'
_aflow_params_values '6.31,0.184'
_aflow_Strukturbericht 'None'
_aflow_Pearson 'cP16'

_symmetry_space_group_name_H-M "P 42 3 2"
_symmetry_Int_Tables_number 208
 
_cell_length_a    6.31000
_cell_length_b    6.31000
_cell_length_c    6.31000
_cell_angle_alpha 90.00000
_cell_angle_beta  90.00000
_cell_angle_gamma 90.00000
 
loop_
_space_group_symop_id
_space_group_symop_operation_xyz
1 x,y,z
2 x,-y,-z
3 -x,y,-z
4 -x,-y,z
5 y,z,x
6 y,-z,-x
7 -y,z,-x
8 -y,-z,x
9 z,x,y
10 z,-x,-y
11 -z,x,-y
12 -z,-x,y
13 -y+1/2,-x+1/2,-z+1/2
14 -y+1/2,x+1/2,z+1/2
15 y+1/2,-x+1/2,z+1/2
16 y+1/2,x+1/2,-z+1/2
17 -x+1/2,-z+1/2,-y+1/2
18 -x+1/2,z+1/2,y+1/2
19 x+1/2,-z+1/2,y+1/2
20 x+1/2,z+1/2,-y+1/2
21 -z+1/2,-y+1/2,-x+1/2
22 -z+1/2,y+1/2,x+1/2
23 z+1/2,-y+1/2,x+1/2
24 z+1/2,y+1/2,-x+1/2
 
loop_
_atom_site_label
_atom_site_type_symbol
_atom_site_symmetry_multiplicity
_atom_site_Wyckoff_label
_atom_site_fract_x
_atom_site_fract_y
_atom_site_fract_z
_atom_site_occupancy
P1 P   4 b 0.25000 0.25000 0.25000 1.00000
H1 H  12 j 0.18400 0.50000 0.00000 1.00000
\end{lstlisting}
{\phantomsection\label{A3B_cP16_208_j_b_poscar}}
{\hyperref[A3B_cP16_208_j_b]{PH$_{3}$: A3B\_cP16\_208\_j\_b}} - POSCAR
\begin{lstlisting}[numbers=none,language={mylang}]
A3B_cP16_208_j_b & a,x2 --params=6.31,0.184 & P4_{2}32 O^{2} #208 (bj) & cP16 & None & PH3 & PH3 & G. Natta and E. Casazza, Gazz. Chim. Ital. 60, 851-859 (1930)
   1.00000000000000
   6.31000000000000   0.00000000000000   0.00000000000000
   0.00000000000000   6.31000000000000   0.00000000000000
   0.00000000000000   0.00000000000000   6.31000000000000
     H     P
    12     4
Direct
   0.18400000000000   0.50000000000000   0.00000000000000    H  (12j)
  -0.18400000000000   0.50000000000000   0.00000000000000    H  (12j)
   0.00000000000000   0.18400000000000   0.50000000000000    H  (12j)
   0.00000000000000  -0.18400000000000   0.50000000000000    H  (12j)
   0.50000000000000   0.00000000000000   0.18400000000000    H  (12j)
   0.50000000000000   0.00000000000000  -0.18400000000000    H  (12j)
   0.00000000000000   0.68400000000000   0.50000000000000    H  (12j)
   0.00000000000000   0.31600000000000   0.50000000000000    H  (12j)
   0.68400000000000   0.50000000000000   0.00000000000000    H  (12j)
   0.31600000000000   0.50000000000000   0.00000000000000    H  (12j)
   0.50000000000000   0.00000000000000   0.31600000000000    H  (12j)
   0.50000000000000   0.00000000000000   0.68400000000000    H  (12j)
   0.25000000000000   0.25000000000000   0.25000000000000    P   (4b)
   0.75000000000000   0.75000000000000   0.25000000000000    P   (4b)
   0.75000000000000   0.25000000000000   0.75000000000000    P   (4b)
   0.25000000000000   0.75000000000000   0.75000000000000    P   (4b)
\end{lstlisting}
{\phantomsection\label{A6B2CD6E_cP64_208_m_ad_b_m_c_cif}}
{\hyperref[A6B2CD6E_cP64_208_m_ad_b_m_c]{Cs$_{2}$ZnFe[CN]$_{6}$: A6B2CD6E\_cP64\_208\_m\_ad\_b\_m\_c}} - CIF

{\phantomsection\label{A6B2CD6E_cP64_208_m_ad_b_m_c_poscar}}
{\hyperref[A6B2CD6E_cP64_208_m_ad_b_m_c]{Cs$_{2}$ZnFe[CN]$_{6}$: A6B2CD6E\_cP64\_208\_m\_ad\_b\_m\_c}} - POSCAR

{\phantomsection\label{A24BC_cF104_209_j_a_b_cif}}
{\hyperref[A24BC_cF104_209_j_a_b]{F$_{6}$KP: A24BC\_cF104\_209\_j\_a\_b}} - CIF

{\phantomsection\label{A24BC_cF104_209_j_a_b_poscar}}
{\hyperref[A24BC_cF104_209_j_a_b]{F$_{6}$KP: A24BC\_cF104\_209\_j\_a\_b}} - POSCAR

{\phantomsection\label{A12B6C_cF608_210_4h_2h_e_cif}}
{\hyperref[A12B6C_cF608_210_4h_2h_e]{Te[OH]$_{6}$: A12B6C\_cF608\_210\_4h\_2h\_e}} - CIF

{\phantomsection\label{A12B6C_cF608_210_4h_2h_e_poscar}}
{\hyperref[A12B6C_cF608_210_4h_2h_e]{Te[OH]$_{6}$: A12B6C\_cF608\_210\_4h\_2h\_e}} - POSCAR

{\phantomsection\label{A2B_cI72_211_hi_i_cif}}
{\hyperref[A2B_cI72_211_hi_i]{SiO$_{2}$: A2B\_cI72\_211\_hi\_i}} - CIF
\begin{lstlisting}[numbers=none,language={mylang}]
# CIF file
data_findsym-output
_audit_creation_method FINDSYM

_chemical_name_mineral 'SiO2'
_chemical_formula_sum 'O2 Si'

_aflow_title 'SiO$_{2}$ Structure'
_aflow_proto 'A2B_cI72_211_hi_i'
_aflow_params 'a,y_{1},y_{2},y_{3}'
_aflow_params_values '9.68882,0.37338,0.15866,0.38235'
_aflow_Strukturbericht 'None'
_aflow_Pearson 'cI72'

_cell_length_a    9.6888200000
_cell_length_b    9.6888200000
_cell_length_c    9.6888200000
_cell_angle_alpha 90.0000000000
_cell_angle_beta  90.0000000000
_cell_angle_gamma 90.0000000000
 
_symmetry_space_group_name_H-M "I 4 3 2"
_symmetry_Int_Tables_number 211
 
loop_
_space_group_symop_id
_space_group_symop_operation_xyz
1 x,y,z
2 x,-y,-z
3 -x,y,-z
4 -x,-y,z
5 y,z,x
6 y,-z,-x
7 -y,z,-x
8 -y,-z,x
9 z,x,y
10 z,-x,-y
11 -z,x,-y
12 -z,-x,y
13 -y,-x,-z
14 -y,x,z
15 y,-x,z
16 y,x,-z
17 -x,-z,-y
18 -x,z,y
19 x,-z,y
20 x,z,-y
21 -z,-y,-x
22 -z,y,x
23 z,-y,x
24 z,y,-x
25 x+1/2,y+1/2,z+1/2
26 x+1/2,-y+1/2,-z+1/2
27 -x+1/2,y+1/2,-z+1/2
28 -x+1/2,-y+1/2,z+1/2
29 y+1/2,z+1/2,x+1/2
30 y+1/2,-z+1/2,-x+1/2
31 -y+1/2,z+1/2,-x+1/2
32 -y+1/2,-z+1/2,x+1/2
33 z+1/2,x+1/2,y+1/2
34 z+1/2,-x+1/2,-y+1/2
35 -z+1/2,x+1/2,-y+1/2
36 -z+1/2,-x+1/2,y+1/2
37 -y+1/2,-x+1/2,-z+1/2
38 -y+1/2,x+1/2,z+1/2
39 y+1/2,-x+1/2,z+1/2
40 y+1/2,x+1/2,-z+1/2
41 -x+1/2,-z+1/2,-y+1/2
42 -x+1/2,z+1/2,y+1/2
43 x+1/2,-z+1/2,y+1/2
44 x+1/2,z+1/2,-y+1/2
45 -z+1/2,-y+1/2,-x+1/2
46 -z+1/2,y+1/2,x+1/2
47 z+1/2,-y+1/2,x+1/2
48 z+1/2,y+1/2,-x+1/2
 
loop_
_atom_site_label
_atom_site_type_symbol
_atom_site_symmetry_multiplicity
_atom_site_Wyckoff_label
_atom_site_fract_x
_atom_site_fract_y
_atom_site_fract_z
_atom_site_occupancy
O1  O   24 h 0.00000 0.37338 0.37338 1.00000
O2  O   24 i 0.25000 0.15866 0.34134 1.00000
Si1 Si  24 i 0.25000 0.38235 0.11765 1.00000
\end{lstlisting}
{\phantomsection\label{A2B_cI72_211_hi_i_poscar}}
{\hyperref[A2B_cI72_211_hi_i]{SiO$_{2}$: A2B\_cI72\_211\_hi\_i}} - POSCAR

{\phantomsection\label{A2B_cP12_212_c_a_cif}}
{\hyperref[A2B_cP12_212_c_a]{SrSi$_{2}$: A2B\_cP12\_212\_c\_a}} - CIF
\begin{lstlisting}[numbers=none,language={mylang}]
# CIF file
data_findsym-output
_audit_creation_method FINDSYM

_chemical_name_mineral 'SrSi2'
_chemical_formula_sum 'Si2 Sr'

_aflow_title 'SrSi$_{2}$ Structure'
_aflow_proto 'A2B_cP12_212_c_a'
_aflow_params 'a,x_{2}'
_aflow_params_values '6.54,0.428'
_aflow_Strukturbericht 'None'
_aflow_Pearson 'cP12'

_cell_length_a    6.5400000000
_cell_length_b    6.5400000000
_cell_length_c    6.5400000000
_cell_angle_alpha 90.0000000000
_cell_angle_beta  90.0000000000
_cell_angle_gamma 90.0000000000
 
_symmetry_space_group_name_H-M "P 43 3 2"
_symmetry_Int_Tables_number 212
 
loop_
_space_group_symop_id
_space_group_symop_operation_xyz
1 x,y,z
2 x+1/2,-y+1/2,-z
3 -x,y+1/2,-z+1/2
4 -x+1/2,-y,z+1/2
5 y,z,x
6 y+1/2,-z+1/2,-x
7 -y,z+1/2,-x+1/2
8 -y+1/2,-z,x+1/2
9 z,x,y
10 z+1/2,-x+1/2,-y
11 -z,x+1/2,-y+1/2
12 -z+1/2,-x,y+1/2
13 -y+1/4,-x+1/4,-z+1/4
14 -y+3/4,x+1/4,z+3/4
15 y+3/4,-x+3/4,z+1/4
16 y+1/4,x+3/4,-z+3/4
17 -x+1/4,-z+1/4,-y+1/4
18 -x+3/4,z+1/4,y+3/4
19 x+3/4,-z+3/4,y+1/4
20 x+1/4,z+3/4,-y+3/4
21 -z+1/4,-y+1/4,-x+1/4
22 -z+3/4,y+1/4,x+3/4
23 z+3/4,-y+3/4,x+1/4
24 z+1/4,y+3/4,-x+3/4
 
loop_
_atom_site_label
_atom_site_type_symbol
_atom_site_symmetry_multiplicity
_atom_site_Wyckoff_label
_atom_site_fract_x
_atom_site_fract_y
_atom_site_fract_z
_atom_site_occupancy
Sr1 Sr   4 a 0.12500 0.12500 0.12500 1.00000
Si1 Si   8 c 0.42800 0.42800 0.42800 1.00000
\end{lstlisting}
{\phantomsection\label{A2B_cP12_212_c_a_poscar}}
{\hyperref[A2B_cP12_212_c_a]{SrSi$_{2}$: A2B\_cP12\_212\_c\_a}} - POSCAR
\begin{lstlisting}[numbers=none,language={mylang}]
A2B_cP12_212_c_a & a,x2 --params=6.54,0.428 & P4_{3}32 O^{6} #212 (ac) & cP12 & None & SrSi2 &  & 
   1.00000000000000
   6.54000000000000   0.00000000000000   0.00000000000000
   0.00000000000000   6.54000000000000   0.00000000000000
   0.00000000000000   0.00000000000000   6.54000000000000
    Si    Sr
     8     4
Direct
   0.42800000000000   0.42800000000000   0.42800000000000   Si   (8c)
   0.07200000000000  -0.42800000000000   0.92800000000000   Si   (8c)
  -0.42800000000000   0.92800000000000   0.07200000000000   Si   (8c)
   0.92800000000000   0.07200000000000  -0.42800000000000   Si   (8c)
   0.67800000000000   1.17800000000000   0.32200000000000   Si   (8c)
  -0.17800000000000  -0.17800000000000  -0.17800000000000   Si   (8c)
   1.17800000000000   0.32200000000000   0.67800000000000   Si   (8c)
   0.32200000000000   0.67800000000000   1.17800000000000   Si   (8c)
   0.12500000000000   0.12500000000000   0.12500000000000   Sr   (4a)
   0.37500000000000   0.87500000000000   0.62500000000000   Sr   (4a)
   0.87500000000000   0.62500000000000   0.37500000000000   Sr   (4a)
   0.62500000000000   0.37500000000000   0.87500000000000   Sr   (4a)
\end{lstlisting}
{\phantomsection\label{A3B3C_cI56_214_g_h_a_cif}}
{\hyperref[A3B3C_cI56_214_g_h_a]{Ca$_{3}$PI$_{3}$: A3B3C\_cI56\_214\_g\_h\_a}} - CIF

{\phantomsection\label{A3B3C_cI56_214_g_h_a_poscar}}
{\hyperref[A3B3C_cI56_214_g_h_a]{Ca$_{3}$PI$_{3}$: A3B3C\_cI56\_214\_g\_h\_a}} - POSCAR

{\phantomsection\label{A3BC2_cI48_214_f_a_e_cif}}
{\hyperref[A3BC2_cI48_214_f_a_e]{Petzite (Ag$_{3}$AuTe$_{2}$): A3BC2\_cI48\_214\_f\_a\_e}} - CIF

{\phantomsection\label{A3BC2_cI48_214_f_a_e_poscar}}
{\hyperref[A3BC2_cI48_214_f_a_e]{Petzite (Ag$_{3}$AuTe$_{2}$): A3BC2\_cI48\_214\_f\_a\_e}} - POSCAR

{\phantomsection\label{A4B9_cP52_215_ei_3efgi_cif}}
{\hyperref[A4B9_cP52_215_ei_3efgi]{$\gamma$-brass (Cu$_{9}$Al$_{4}$, $D8_{3}$): A4B9\_cP52\_215\_ei\_3efgi}} - CIF

{\phantomsection\label{A4B9_cP52_215_ei_3efgi_poscar}}
{\hyperref[A4B9_cP52_215_ei_3efgi]{$\gamma$-brass (Cu$_{9}$Al$_{4}$, $D8_{3}$): A4B9\_cP52\_215\_ei\_3efgi}} - POSCAR

{\phantomsection\label{ABCD_cF16_216_c_d_b_a_cif}}
{\hyperref[ABCD_cF16_216_c_d_b_a]{Quartenary Heusler (LiMgAuSn): ABCD\_cF16\_216\_c\_d\_b\_a}} - CIF

{\phantomsection\label{ABCD_cF16_216_c_d_b_a_poscar}}
{\hyperref[ABCD_cF16_216_c_d_b_a]{Quartenary Heusler (LiMgAuSn): ABCD\_cF16\_216\_c\_d\_b\_a}} - POSCAR
\begin{lstlisting}[numbers=none,language={mylang}]
ABCD_cF16_216_c_d_b_a & a --params=6.465 & F-43m T_{d}^{2} #216 (abcd) & cF16 & None & LiMgAuSn & Quartenary Heusler & U. Eberz and W. Seelentag and H.-U. Schuster, Z. Naturforsch. B 35, 1341-1343 (1980)
   1.00000000000000
   0.00000000000000   3.23250000000000   3.23250000000000
   3.23250000000000   0.00000000000000   3.23250000000000
   3.23250000000000   3.23250000000000   0.00000000000000
    Au    Li    Mg    Sn
     1     1     1     1
Direct
   0.25000000000000   0.25000000000000   0.25000000000000   Au   (4c)
   0.75000000000000   0.75000000000000   0.75000000000000   Li   (4d)
   0.50000000000000   0.50000000000000   0.50000000000000   Mg   (4b)
   0.00000000000000   0.00000000000000   0.00000000000000   Sn   (4a)
\end{lstlisting}
{\phantomsection\label{A3B4C_cP16_218_c_e_a_cif}}
{\hyperref[A3B4C_cP16_218_c_e_a]{Ag$_{3}$[PO$_{4}$]: A3B4C\_cP16\_218\_c\_e\_a}} - CIF
\begin{lstlisting}[numbers=none,language={mylang}]
# CIF file
data_findsym-output
_audit_creation_method FINDSYM

_chemical_name_mineral 'Ag3[PO4]'
_chemical_formula_sum 'Ag3 O4 P'

loop_
_publ_author_name
 'R. Masse'
 'I. Tordjman'
 'A. Durif'
_journal_name_full_name
;
 Zeitschrift f{\"u}r Kristallographie - Crystalline Materials
;
_journal_volume 144
_journal_year 1976
_journal_page_first 76
_journal_page_last 81
_publ_Section_title
;
 Affinement de la structure cristalline du monophosphate d\'argent Ag$_{3}$PO$_{4}$. Existence d\'une forme haute temperature
;

# Found in Pearson's Crystal Data - Crystal Structure Database for Inorganic Compounds, 2013

_aflow_title 'Ag$_{3}$[PO$_{4}$] Structure'
_aflow_proto 'A3B4C_cP16_218_c_e_a'
_aflow_params 'a,x_{3}'
_aflow_params_values '6.0258147002,0.6486'
_aflow_Strukturbericht 'None'
_aflow_Pearson 'cP16'

_cell_length_a    6.0258147002
_cell_length_b    6.0258147002
_cell_length_c    6.0258147002
_cell_angle_alpha 90.0000000000
_cell_angle_beta  90.0000000000
_cell_angle_gamma 90.0000000000
 
_symmetry_space_group_name_H-M "P -4 3 n"
_symmetry_Int_Tables_number 218
 
loop_
_space_group_symop_id
_space_group_symop_operation_xyz
1 x,y,z
2 x,-y,-z
3 -x,y,-z
4 -x,-y,z
5 y,z,x
6 y,-z,-x
7 -y,z,-x
8 -y,-z,x
9 z,x,y
10 z,-x,-y
11 -z,x,-y
12 -z,-x,y
13 y+1/2,x+1/2,z+1/2
14 y+1/2,-x+1/2,-z+1/2
15 -y+1/2,x+1/2,-z+1/2
16 -y+1/2,-x+1/2,z+1/2
17 x+1/2,z+1/2,y+1/2
18 x+1/2,-z+1/2,-y+1/2
19 -x+1/2,z+1/2,-y+1/2
20 -x+1/2,-z+1/2,y+1/2
21 z+1/2,y+1/2,x+1/2
22 z+1/2,-y+1/2,-x+1/2
23 -z+1/2,y+1/2,-x+1/2
24 -z+1/2,-y+1/2,x+1/2
 
loop_
_atom_site_label
_atom_site_type_symbol
_atom_site_symmetry_multiplicity
_atom_site_Wyckoff_label
_atom_site_fract_x
_atom_site_fract_y
_atom_site_fract_z
_atom_site_occupancy
P1  P    2 a 0.00000 0.00000 0.00000 1.00000
Ag1 Ag   6 c 0.25000 0.50000 0.00000 1.00000
O1  O    8 e 0.64860 0.64860 0.64860 1.00000
\end{lstlisting}
{\phantomsection\label{A3B4C_cP16_218_c_e_a_poscar}}
{\hyperref[A3B4C_cP16_218_c_e_a]{Ag$_{3}$[PO$_{4}$]: A3B4C\_cP16\_218\_c\_e\_a}} - POSCAR
\begin{lstlisting}[numbers=none,language={mylang}]
A3B4C_cP16_218_c_e_a & a,x3 --params=6.0258147002,0.6486 & P-43n T_{d}^{4} #218 (ace) & cP16 & None & Ag3[PO4] &  & R. Masse and I. Tordjman and A. Durif, Zeitschrift f"{u}r Kristallographie - Crystalline Materials 144, 76-81 (1976)
   1.00000000000000
   6.02581470020000   0.00000000000000   0.00000000000000
   0.00000000000000   6.02581470020000   0.00000000000000
   0.00000000000000   0.00000000000000   6.02581470020000
    Ag     O     P
     6     8     2
Direct
   0.25000000000000   0.50000000000000   0.00000000000000   Ag   (6c)
   0.75000000000000   0.50000000000000   0.00000000000000   Ag   (6c)
   0.00000000000000   0.25000000000000   0.50000000000000   Ag   (6c)
   0.00000000000000   0.75000000000000   0.50000000000000   Ag   (6c)
   0.50000000000000   0.00000000000000   0.25000000000000   Ag   (6c)
   0.50000000000000   0.00000000000000   0.75000000000000   Ag   (6c)
   0.64860000000000   0.64860000000000   0.64860000000000    O   (8e)
  -0.64860000000000  -0.64860000000000   0.64860000000000    O   (8e)
  -0.64860000000000   0.64860000000000  -0.64860000000000    O   (8e)
   0.64860000000000  -0.64860000000000  -0.64860000000000    O   (8e)
   1.14860000000000   1.14860000000000   1.14860000000000    O   (8e)
  -0.14860000000000  -0.14860000000000   1.14860000000000    O   (8e)
   1.14860000000000  -0.14860000000000  -0.14860000000000    O   (8e)
  -0.14860000000000   1.14860000000000  -0.14860000000000    O   (8e)
   0.00000000000000   0.00000000000000   0.00000000000000    P   (2a)
   0.50000000000000   0.50000000000000   0.50000000000000    P   (2a)
\end{lstlisting}
{\phantomsection\label{A7BC3D13_cF192_219_de_b_c_ah_cif}}
{\hyperref[A7BC3D13_cF192_219_de_b_c_ah]{Boracite (Mg$_{3}$B$_{7}$ClO$_{13}$): A7BC3D13\_cF192\_219\_de\_b\_c\_ah}} - CIF

{\phantomsection\label{A7BC3D13_cF192_219_de_b_c_ah_poscar}}
{\hyperref[A7BC3D13_cF192_219_de_b_c_ah]{Boracite (Mg$_{3}$B$_{7}$ClO$_{13}$): A7BC3D13\_cF192\_219\_de\_b\_c\_ah}} - POSCAR

{\phantomsection\label{A15B4_cI76_220_ae_c_cif}}
{\hyperref[A15B4_cI76_220_ae_c]{Cu$_{15}$Si$_{4}$ ($D8_{6}$): A15B4\_cI76\_220\_ae\_c}} - CIF

{\phantomsection\label{A15B4_cI76_220_ae_c_poscar}}
{\hyperref[A15B4_cI76_220_ae_c]{Cu$_{15}$Si$_{4}$ ($D8_{6}$): A15B4\_cI76\_220\_ae\_c}} - POSCAR

{\phantomsection\label{A4B3_cI28_220_c_a_cif}}
{\hyperref[A4B3_cI28_220_c_a]{Th$_{3}$P$_{4}$ ($D7_{3}$): A4B3\_cI28\_220\_c\_a}} - CIF

{\phantomsection\label{A4B3_cI28_220_c_a_poscar}}
{\hyperref[A4B3_cI28_220_c_a]{Th$_{3}$P$_{4}$ ($D7_{3}$): A4B3\_cI28\_220\_c\_a}} - POSCAR
\begin{lstlisting}[numbers=none,language={mylang}]
A4B3_cI28_220_c_a & a,x2 --params=8.6,0.08333 & I-43d T_{d}^{6} #220 (ac) & cI28 & $D7_{3}$ & Th3P4 & Th3P4 & K. Meisel, Z. Anorg. Allg. Chem. 240, 300-312 (1939)
   1.00000000000000
  -4.30000000000000   4.30000000000000   4.30000000000000
   4.30000000000000  -4.30000000000000   4.30000000000000
   4.30000000000000   4.30000000000000  -4.30000000000000
     P    Th
     8     6
Direct
   0.16666000000000   0.16666000000000   0.16666000000000    P  (16c)
   0.50000000000000   0.00000000000000   0.33334000000000    P  (16c)
   0.00000000000000   0.33334000000000   0.50000000000000    P  (16c)
   0.33334000000000   0.50000000000000   0.00000000000000    P  (16c)
   0.66666000000000   0.66666000000000   0.66666000000000    P  (16c)
   0.50000000000000   0.00000000000000  -0.16666000000000    P  (16c)
  -0.16666000000000   0.50000000000000   0.00000000000000    P  (16c)
   0.00000000000000  -0.16666000000000   0.50000000000000    P  (16c)
   0.25000000000000   0.62500000000000   0.37500000000000   Th  (12a)
   0.75000000000000   0.87500000000000   0.12500000000000   Th  (12a)
   0.37500000000000   0.25000000000000   0.62500000000000   Th  (12a)
   0.12500000000000   0.75000000000000   0.87500000000000   Th  (12a)
   0.62500000000000   0.37500000000000   0.25000000000000   Th  (12a)
   0.87500000000000   0.12500000000000   0.75000000000000   Th  (12a)
\end{lstlisting}
{\phantomsection\label{A2B3C6_cP33_221_cd_ag_fh_cif}}
{\hyperref[A2B3C6_cP33_221_cd_ag_fh]{Ca$_{3}$Al$_{2}$O$_{6}$ ($E9_{1}$): A2B3C6\_cP33\_221\_cd\_ag\_fh}} - CIF

{\phantomsection\label{A2B3C6_cP33_221_cd_ag_fh_poscar}}
{\hyperref[A2B3C6_cP33_221_cd_ag_fh]{Ca$_{3}$Al$_{2}$O$_{6}$ ($E9_{1}$): A2B3C6\_cP33\_221\_cd\_ag\_fh}} - POSCAR

{\phantomsection\label{A5B3C16_cP96_222_ce_d_fi_cif}}
{\hyperref[A5B3C16_cP96_222_ce_d_fi]{Ce$_{5}$Mo$_{3}$O$_{16}$: A5B3C16\_cP96\_222\_ce\_d\_fi}} - CIF

{\phantomsection\label{A5B3C16_cP96_222_ce_d_fi_poscar}}
{\hyperref[A5B3C16_cP96_222_ce_d_fi]{Ce$_{5}$Mo$_{3}$O$_{16}$: A5B3C16\_cP96\_222\_ce\_d\_fi}} - POSCAR

{\phantomsection\label{A23B6_cF116_225_bd2f_e_cif}}
{\hyperref[A23B6_cF116_225_bd2f_e]{Th$_{6}$Mn$_{23}$ ($D8_{a}$): A23B6\_cF116\_225\_bd2f\_e}} - CIF

{\phantomsection\label{A23B6_cF116_225_bd2f_e_poscar}}
{\hyperref[A23B6_cF116_225_bd2f_e]{Th$_{6}$Mn$_{23}$ ($D8_{a}$): A23B6\_cF116\_225\_bd2f\_e}} - POSCAR

{\phantomsection\label{A6B2C_cF36_225_e_c_a_cif}}
{\hyperref[A6B2C_cF36_225_e_c_a]{K$_{2}$PtCl$_{6}$ ($J1_{1}$): A6B2C\_cF36\_225\_e\_c\_a}} - CIF

{\phantomsection\label{A6B2C_cF36_225_e_c_a_poscar}}
{\hyperref[A6B2C_cF36_225_e_c_a]{K$_{2}$PtCl$_{6}$ ($J1_{1}$): A6B2C\_cF36\_225\_e\_c\_a}} - POSCAR
\begin{lstlisting}[numbers=none,language={mylang}]
A6B2C_cF36_225_e_c_a & a,x3 --params=9.725,0.24 & Fm-3m O_{h}^{5} #225 (ace) & cF36 & $J1_{1}$ & K2PtCl6 & K2PtCl6 & G. Engel, Z. Kristallogr. 90, 341-373 (1935)
   1.00000000000000
   0.00000000000000   4.86250000000000   4.86250000000000
   4.86250000000000   0.00000000000000   4.86250000000000
   4.86250000000000   4.86250000000000   0.00000000000000
    Cl     K    Pt
     6     2     1
Direct
  -0.24000000000000   0.24000000000000   0.24000000000000   Cl  (24e)
   0.24000000000000  -0.24000000000000  -0.24000000000000   Cl  (24e)
   0.24000000000000  -0.24000000000000   0.24000000000000   Cl  (24e)
  -0.24000000000000   0.24000000000000  -0.24000000000000   Cl  (24e)
   0.24000000000000   0.24000000000000  -0.24000000000000   Cl  (24e)
  -0.24000000000000  -0.24000000000000   0.24000000000000   Cl  (24e)
   0.25000000000000   0.25000000000000   0.25000000000000    K   (8c)
   0.75000000000000   0.75000000000000   0.75000000000000    K   (8c)
   0.00000000000000   0.00000000000000   0.00000000000000   Pt   (4a)
\end{lstlisting}
{\phantomsection\label{AB13_cF112_226_a_bi_cif}}
{\hyperref[AB13_cF112_226_a_bi]{NaZn$_{13}$ ($D2_{3}$): AB13\_cF112\_226\_a\_bi}} - CIF

{\phantomsection\label{AB13_cF112_226_a_bi_poscar}}
{\hyperref[AB13_cF112_226_a_bi]{NaZn$_{13}$ ($D2_{3}$): AB13\_cF112\_226\_a\_bi}} - POSCAR

{\phantomsection\label{A2B2C7_cF88_227_c_d_af_cif}}
{\hyperref[A2B2C7_cF88_227_c_d_af]{Pyrochlore Iridate (Eu$_{2}$Ir$_{2}$O$_{7}$): A2B2C7\_cF88\_227\_c\_d\_af}} - CIF

{\phantomsection\label{A2B2C7_cF88_227_c_d_af_poscar}}
{\hyperref[A2B2C7_cF88_227_c_d_af]{Pyrochlore Iridate (Eu$_{2}$Ir$_{2}$O$_{7}$): A2B2C7\_cF88\_227\_c\_d\_af}} - POSCAR

{\phantomsection\label{A3B4_cF56_227_ad_e_cif}}
{\hyperref[A3B4_cF56_227_ad_e]{Spinel (Co$_{3}$O$_{4}$, $D7_{2}$): A3B4\_cF56\_227\_ad\_e}} - CIF

{\phantomsection\label{A3B4_cF56_227_ad_e_poscar}}
{\hyperref[A3B4_cF56_227_ad_e]{Spinel (Co$_{3}$O$_{4}$, $D7_{2}$): A3B4\_cF56\_227\_ad\_e}} - POSCAR
\begin{lstlisting}[numbers=none,language={mylang}]
A3B4_cF56_227_ad_e & a,x3 --params=8.0835,0.2642 & Fd-3m O_{h}^{7} #227 (ade) & cF56 & $D7_{2}$ & Co3O4 & Spinel & O. Knop et al., Can. J. Chem. 46, 3463-3476 (1968)
   1.00000000000000
   0.00000000000000   4.04175000000000   4.04175000000000
   4.04175000000000   0.00000000000000   4.04175000000000
   4.04175000000000   4.04175000000000   0.00000000000000
    Co     O
     6     8
Direct
   0.12500000000000   0.12500000000000   0.12500000000000   Co   (8a)
   0.87500000000000   0.87500000000000   0.87500000000000   Co   (8a)
   0.50000000000000   0.50000000000000   0.50000000000000   Co  (16d)
   0.50000000000000   0.50000000000000   0.00000000000000   Co  (16d)
   0.50000000000000   0.00000000000000   0.50000000000000   Co  (16d)
   0.00000000000000   0.50000000000000   0.50000000000000   Co  (16d)
   0.26420000000000   0.26420000000000   0.26420000000000    O  (32e)
   0.26420000000000   0.26420000000000  -0.29260000000000    O  (32e)
   0.26420000000000  -0.29260000000000   0.26420000000000    O  (32e)
  -0.29260000000000   0.26420000000000   0.26420000000000    O  (32e)
  -0.26420000000000  -0.26420000000000   1.29260000000000    O  (32e)
  -0.26420000000000  -0.26420000000000  -0.26420000000000    O  (32e)
  -0.26420000000000   1.29260000000000  -0.26420000000000    O  (32e)
   1.29260000000000  -0.26420000000000  -0.26420000000000    O  (32e)
\end{lstlisting}
{\phantomsection\label{A5BCD6_cF416_228_eg_c_b_h_cif}}
{\hyperref[A5BCD6_cF416_228_eg_c_b_h]{CuCrCl$_{5}$[NH$_{3}$]$_{6}$: A5BCD6\_cF416\_228\_eg\_c\_b\_h}} - CIF

{\phantomsection\label{A5BCD6_cF416_228_eg_c_b_h_poscar}}
{\hyperref[A5BCD6_cF416_228_eg_c_b_h]{CuCrCl$_{5}$[NH$_{3}$]$_{6}$: A5BCD6\_cF416\_228\_eg\_c\_b\_h}} - POSCAR

{\phantomsection\label{A6B_cF224_228_h_c_cif}}
{\hyperref[A6B_cF224_228_h_c]{TeO$_{6}$H$_{6}$: A6B\_cF224\_228\_h\_c}} - CIF

{\phantomsection\label{A6B_cF224_228_h_c_poscar}}
{\hyperref[A6B_cF224_228_h_c]{TeO$_{6}$H$_{6}$: A6B\_cF224\_228\_h\_c}} - POSCAR

{\phantomsection\label{A3B10_cI52_229_e_fh_cif}}
{\hyperref[A3B10_cI52_229_e_fh]{$\gamma$-brass (Fe$_{3}$Zn$_{10}$, $D8_{1}$): A3B10\_cI52\_229\_e\_fh}} - CIF

{\phantomsection\label{A3B10_cI52_229_e_fh_poscar}}
{\hyperref[A3B10_cI52_229_e_fh]{$\gamma$-brass (Fe$_{3}$Zn$_{10}$, $D8_{1}$): A3B10\_cI52\_229\_e\_fh}} - POSCAR

{\phantomsection\label{A4B_cI10_229_c_a_cif}}
{\hyperref[A4B_cI10_229_c_a]{$\beta$-Hg$_{4}$Pt: A4B\_cI10\_229\_c\_a}} - CIF

{\phantomsection\label{A4B_cI10_229_c_a_poscar}}
{\hyperref[A4B_cI10_229_c_a]{$\beta$-Hg$_{4}$Pt: A4B\_cI10\_229\_c\_a}} - POSCAR
\begin{lstlisting}[numbers=none,language={mylang}]
A4B_cI10_229_c_a & a --params=6.186 & Im-3m O_{h}^{9} #229 (ac) & cI10 & None & Hg4Pt & $\beta$-Hg$_{4}$Pt & E. Bauer and H. Nowotny and A. Stempfl, Monatsh. Chem. 84, 211-212 (1953)
   1.00000000000000
  -3.09300000000000   3.09300000000000   3.09300000000000
   3.09300000000000  -3.09300000000000   3.09300000000000
   3.09300000000000   3.09300000000000  -3.09300000000000
    Hg    Pt
     4     1
Direct
   0.50000000000000   0.50000000000000   0.50000000000000   Hg   (8c)
   0.00000000000000   0.00000000000000   0.50000000000000   Hg   (8c)
   0.00000000000000   0.50000000000000   0.00000000000000   Hg   (8c)
   0.50000000000000   0.00000000000000   0.00000000000000   Hg   (8c)
   0.00000000000000   0.00000000000000   0.00000000000000   Pt   (2a)
\end{lstlisting}
{\phantomsection\label{A7B3_cI40_229_df_e_cif}}
{\hyperref[A7B3_cI40_229_df_e]{Ir$_{3}$Ge$_{7}$ ($D8_{f}$): A7B3\_cI40\_229\_df\_e}} - CIF

{\phantomsection\label{A7B3_cI40_229_df_e_poscar}}
{\hyperref[A7B3_cI40_229_df_e]{Ir$_{3}$Ge$_{7}$ ($D8_{f}$): A7B3\_cI40\_229\_df\_e}} - POSCAR
\begin{lstlisting}[numbers=none,language={mylang}]
A7B3_cI40_229_df_e & a,x2,x3 --params=8.735,0.342,0.156 & Im-3m O_{h}^{9} #229 (def) & cI40 & $D8_{f}$ & Ir3Ge7 & Ir3Ge7 & U. H\"{a}ussermann et al., Chem. Euro. J. 4, 1007-1015 (1998)
   1.00000000000000
  -4.36750000000000   4.36750000000000   4.36750000000000
   4.36750000000000  -4.36750000000000   4.36750000000000
   4.36750000000000   4.36750000000000  -4.36750000000000
    Ge    Ir
    14     6
Direct
   0.50000000000000   0.75000000000000   0.25000000000000   Ge  (12d)
   0.50000000000000   0.25000000000000   0.75000000000000   Ge  (12d)
   0.25000000000000   0.50000000000000   0.75000000000000   Ge  (12d)
   0.75000000000000   0.50000000000000   0.25000000000000   Ge  (12d)
   0.75000000000000   0.25000000000000   0.50000000000000   Ge  (12d)
   0.25000000000000   0.75000000000000   0.50000000000000   Ge  (12d)
   0.31200000000000   0.31200000000000   0.31200000000000   Ge  (16f)
   0.00000000000000   0.00000000000000  -0.31200000000000   Ge  (16f)
   0.00000000000000  -0.31200000000000   0.00000000000000   Ge  (16f)
  -0.31200000000000   0.00000000000000   0.00000000000000   Ge  (16f)
   0.00000000000000   0.00000000000000   0.31200000000000   Ge  (16f)
  -0.31200000000000  -0.31200000000000  -0.31200000000000   Ge  (16f)
   0.00000000000000   0.31200000000000   0.00000000000000   Ge  (16f)
   0.31200000000000   0.00000000000000   0.00000000000000   Ge  (16f)
   0.00000000000000   0.34200000000000   0.34200000000000   Ir  (12e)
   0.00000000000000  -0.34200000000000  -0.34200000000000   Ir  (12e)
   0.34200000000000   0.00000000000000   0.34200000000000   Ir  (12e)
  -0.34200000000000   0.00000000000000  -0.34200000000000   Ir  (12e)
   0.34200000000000   0.34200000000000   0.00000000000000   Ir  (12e)
  -0.34200000000000  -0.34200000000000   0.00000000000000   Ir  (12e)
\end{lstlisting}
{\phantomsection\label{A2B3C12D3_cI160_230_a_c_h_d_cif}}
{\hyperref[A2B3C12D3_cI160_230_a_c_h_d]{Garnet (Co$_3$Al$_2$Si$_3$O$_{12}$, $S1_{4}$): A2B3C12D3\_cI160\_230\_a\_c\_h\_d}} - CIF

{\phantomsection\label{A2B3C12D3_cI160_230_a_c_h_d_poscar}}
{\hyperref[A2B3C12D3_cI160_230_a_c_h_d]{Garnet (Co$_3$Al$_2$Si$_3$O$_{12}$, $S1_{4}$): A2B3C12D3\_cI160\_230\_a\_c\_h\_d}} - POSCAR

}

\twocolumn
\newgeometry{left=1.0cm,right=1.0cm,top=1.5cm,bottom=2.0cm,footskip=1.5cm}
\renewcommand{\thefootnote}{\fnsymbol{footnote}} 
\section*{\label{sec:protoInd}Prototype Index}
\noindent
\begin{enumerate}
\vspace{-0.75cm} \item $\alpha$-Al$_{2}$S$_{3}$: {\small A2B3\_hP30\_169\_2a\_3a} \dotfill {\hyperref[A2B3_hP30_169_2a_3a]{\pageref{A2B3_hP30_169_2a_3a}}} \\
\vspace{-0.75cm} \item $\alpha$-CuAlCl$_{4}$: {\small AB4C\_tP12\_112\_b\_n\_e} \dotfill {\hyperref[AB4C_tP12_112_b_n_e]{\pageref{AB4C_tP12_112_b_n_e}}} \\
\vspace{-0.75cm} \item $\alpha$-FeSe\footnote[2]{\label{note:AB_oC8_67_a_g-prototype}$\alpha$-FeSe and $\alpha$-PbO have the same \AFLOW\ prototype label. They are generated by the same symmetry operations with different sets of parameters.}: {\small AB\_oC8\_67\_a\_g} \dotfill {\hyperref[AB_oC8_67_a_g-FeSe]{\pageref{AB_oC8_67_a_g-FeSe}}} \\
\vspace{-0.75cm} \item $\alpha$-Naumannite: {\small A2B\_oP12\_17\_abe\_e} \dotfill {\hyperref[A2B_oP12_17_abe_e]{\pageref{A2B_oP12_17_abe_e}}} \\
\vspace{-0.75cm} \item $\alpha$-NbO$_{2}$: {\small AB2\_tI96\_88\_2f\_4f} \dotfill {\hyperref[AB2_tI96_88_2f_4f]{\pageref{AB2_tI96_88_2f_4f}}} \\
\vspace{-0.75cm} \item $\alpha$-P$_3$N$_5$: {\small A5B3\_mC32\_9\_5a\_3a} \dotfill {\hyperref[A5B3_mC32_9_5a_3a]{\pageref{A5B3_mC32_9_5a_3a}}} \\
\vspace{-0.75cm} \item $\alpha$-PbO\footnoteref{note:AB_oC8_67_a_g-prototype}: {\small AB\_oC8\_67\_a\_g} \dotfill {\hyperref[AB_oC8_67_a_g-PbO]{\pageref{AB_oC8_67_a_g-PbO}}} \\
\vspace{-0.75cm} \item $\alpha$-PdCl$_{2}$: {\small A2B\_oP6\_58\_g\_a} \dotfill {\hyperref[A2B_oP6_58_g_a]{\pageref{A2B_oP6_58_g_a}}} \\
\vspace{-0.75cm} \item \begin{raggedleft}$\alpha$-RbPr[MoO$_{4}$]$_{2}$: \end{raggedleft} \\ {\small A2B8CD\_oP24\_48\_k\_2m\_d\_b} \dotfill {\hyperref[A2B8CD_oP24_48_k_2m_d_b]{\pageref{A2B8CD_oP24_48_k_2m_d_b}}} \\
\vspace{-0.75cm} \item $\alpha$-Sm$_{3}$Ge$_{5}$: {\small A5B3\_hP16\_190\_bdh\_g} \dotfill {\hyperref[A5B3_hP16_190_bdh_g]{\pageref{A5B3_hP16_190_bdh_g}}} \\
\vspace{-0.75cm} \item $\alpha$-ThSi$_{2}$: {\small A2B\_tI12\_141\_e\_a} \dotfill {\hyperref[A2B_tI12_141_e_a]{\pageref{A2B_tI12_141_e_a}}} \\
\vspace{-0.75cm} \item $\alpha$-Tl$_{2}$TeO$_{3}$: {\small A3BC2\_oP48\_50\_3m\_m\_2m} \dotfill {\hyperref[A3BC2_oP48_50_3m_m_2m]{\pageref{A3BC2_oP48_50_3m_m_2m}}} \\
\vspace{-0.75cm} \item $\alpha$-Toluene: {\small A7B8\_mP120\_14\_14e\_16e} \dotfill {\hyperref[A7B8_mP120_14_14e_16e]{\pageref{A7B8_mP120_14_14e_16e}}} \\
\vspace{-0.75cm} \item $\beta$-Bi$_{2}$O$_{3}$: {\small A2B3\_tP20\_117\_i\_adgh} \dotfill {\hyperref[A2B3_tP20_117_i_adgh]{\pageref{A2B3_tP20_117_i_adgh}}} \\
\vspace{-0.75cm} \item $\beta$-CuI: {\small AB\_hP4\_156\_ac\_ac} \dotfill {\hyperref[AB_hP4_156_ac_ac]{\pageref{AB_hP4_156_ac_ac}}} \\
\vspace{-0.75cm} \item $\beta$-Hg$_{4}$Pt: {\small A4B\_cI10\_229\_c\_a} \dotfill {\hyperref[A4B_cI10_229_c_a]{\pageref{A4B_cI10_229_c_a}}} \\
\vspace{-0.75cm} \item $\beta$-NbO$_{2}$: {\small AB2\_tI48\_80\_2b\_4b} \dotfill {\hyperref[AB2_tI48_80_2b_4b]{\pageref{AB2_tI48_80_2b_4b}}} \\
\vspace{-0.75cm} \item $\beta$-PdCl$_2$: {\small A2B\_hR18\_148\_2f\_f} \dotfill {\hyperref[A2B_hR18_148_2f_f]{\pageref{A2B_hR18_148_2f_f}}} \\
\vspace{-0.75cm} \item $\beta$-RuCl$_{3}$: {\small A3B\_hP8\_158\_d\_a} \dotfill {\hyperref[A3B_hP8_158_d_a]{\pageref{A3B_hP8_158_d_a}}} \\
\vspace{-0.75cm} \item $\beta$-RuCl$_{3}$: {\small A3B\_hP8\_185\_c\_a} \dotfill {\hyperref[A3B_hP8_185_c_a]{\pageref{A3B_hP8_185_c_a}}} \\
\vspace{-0.75cm} \item $\beta$-SeO$_{2}$\footnote[1]{\label{note:A2B_oP12_26_abc_ab-prototype}H$_{2}$S and $\beta$-SeO$_{2}$ have the same \AFLOW\ prototype label. They are generated by the same symmetry operations with different sets of parameters.}: {\small A2B\_oP12\_26\_abc\_ab} \dotfill {\hyperref[A2B_oP12_26_abc_ab-SeO2]{\pageref{A2B_oP12_26_abc_ab-SeO2}}} \\
\vspace{-0.75cm} \item $\beta$-Si$_{3}$N$_{4}$: {\small A4B3\_hP14\_173\_bc\_c} \dotfill {\hyperref[A4B3_hP14_173_bc_c]{\pageref{A4B3_hP14_173_bc_c}}} \\
\vspace{-0.75cm} \item $\beta$-SiO$_{2}$: {\small A2B\_hP9\_181\_j\_c} \dotfill {\hyperref[A2B_hP9_181_j_c]{\pageref{A2B_hP9_181_j_c}}} \\
\vspace{-0.75cm} \item $\beta$-Ta$_{2}$O$_{5}$: {\small A5B2\_oP14\_49\_dehq\_ab} \dotfill {\hyperref[A5B2_oP14_49_dehq_ab]{\pageref{A5B2_oP14_49_dehq_ab}}} \\
\vspace{-0.75cm} \item $\beta$-ThI$_{3}$: {\small A3B\_oC64\_66\_kl2m\_bdl} \dotfill {\hyperref[A3B_oC64_66_kl2m_bdl]{\pageref{A3B_oC64_66_kl2m_bdl}}} \\
\vspace{-0.75cm} \item $\beta$-Toluene: {\small A7B8\_oP120\_60\_7d\_8d} \dotfill {\hyperref[A7B8_oP120_60_7d_8d]{\pageref{A7B8_oP120_60_7d_8d}}} \\
\vspace{-0.75cm} \item $\beta$-V$_{3}$S: {\small AB3\_tP32\_133\_h\_i2j} \dotfill {\hyperref[AB3_tP32_133_h_i2j]{\pageref{AB3_tP32_133_h_i2j}}} \\
\vspace{-0.75cm} \item $\delta$-PdCl$_{2}$: {\small A2B\_mP6\_10\_mn\_bg} \dotfill {\hyperref[A2B_mP6_10_mn_bg]{\pageref{A2B_mP6_10_mn_bg}}} \\
\vspace{-0.75cm} \item $\delta_{H}^{II}$-NW$_2$: {\small AB2\_hP9\_164\_bd\_c2d} \dotfill {\hyperref[AB2_hP9_164_bd_c2d]{\pageref{AB2_hP9_164_bd_c2d}}} \\
\vspace{-0.75cm} \item $\epsilon$-NiAl$_{3}$: {\small A3B\_oP16\_62\_cd\_c} \dotfill {\hyperref[A3B_oP16_62_cd_c]{\pageref{A3B_oP16_62_cd_c}}} \\
\vspace{-0.75cm} \item $\epsilon$-WO$_{3}$: {\small A3B\_mP16\_7\_6a\_2a} \dotfill {\hyperref[A3B_mP16_7_6a_2a]{\pageref{A3B_mP16_7_6a_2a}}} \\
\vspace{-0.75cm} \item $\gamma$-Ag$_{3}$SI: {\small A3BC\_hR5\_146\_b\_a\_a} \dotfill {\hyperref[A3BC_hR5_146_b_a_a]{\pageref{A3BC_hR5_146_b_a_a}}} \\
\vspace{-0.75cm} \item $\gamma$-MgNiSn: {\small A7B7C2\_tP32\_101\_bde\_ade\_d} \dotfill {\hyperref[A7B7C2_tP32_101_bde_ade_d]{\pageref{A7B7C2_tP32_101_bde_ade_d}}} \\
\vspace{-0.75cm} \item $\gamma$-PdCl$_{2}$: {\small A2B\_mP6\_14\_e\_a} \dotfill {\hyperref[A2B_mP6_14_e_a]{\pageref{A2B_mP6_14_e_a}}} \\
\vspace{-0.75cm} \item $\gamma$-brass: {\small A4B9\_cP52\_215\_ei\_3efgi} \dotfill {\hyperref[A4B9_cP52_215_ei_3efgi]{\pageref{A4B9_cP52_215_ei_3efgi}}} \\
\vspace{-0.75cm} \item $\gamma$-brass: {\small A3B10\_cI52\_229\_e\_fh} \dotfill {\hyperref[A3B10_cI52_229_e_fh]{\pageref{A3B10_cI52_229_e_fh}}} \\
\vspace{-0.75cm} \item $\kappa$-alumina: {\small A2B3\_oP40\_33\_4a\_6a} \dotfill {\hyperref[A2B3_oP40_33_4a_6a]{\pageref{A2B3_oP40_33_4a_6a}}} \\
\vspace{-0.75cm} \item \begin{raggedleft}$\pi$-FeMg$_{3}$Al$_{8}$Si$_{6}$: \end{raggedleft} \\ {\small A8BC3D6\_hP18\_189\_bfh\_a\_g\_i} \dotfill {\hyperref[A8BC3D6_hP18_189_bfh_a_g_i]{\pageref{A8BC3D6_hP18_189_bfh_a_g_i}}} \\
\vspace{-0.75cm} \item \begin{raggedleft}$\pi$-FeMg$_{3}$Al$_{9}$Si$_{5}$: \end{raggedleft} \\ {\small A9BC3D5\_hP18\_189\_fi\_a\_g\_bh} \dotfill {\hyperref[A9BC3D5_hP18_189_fi_a_g_bh]{\pageref{A9BC3D5_hP18_189_fi_a_g_bh}}} \\
\vspace{-0.75cm} \item Ag$_{3}$[PO$_{4}$]: {\small A3B4C\_cP16\_218\_c\_e\_a} \dotfill {\hyperref[A3B4C_cP16_218_c_e_a]{\pageref{A3B4C_cP16_218_c_e_a}}} \\
\vspace{-0.75cm} \item Ag$_{5}$Pb$_{2}$O$_{6}$: {\small A5B6C2\_hP13\_157\_2ac\_2c\_b} \dotfill {\hyperref[A5B6C2_hP13_157_2ac_2c_b]{\pageref{A5B6C2_hP13_157_2ac_2c_b}}} \\
\vspace{-0.75cm} \item AgUF$_{6}$: {\small AB6C\_tP16\_132\_d\_io\_a} \dotfill {\hyperref[AB6C_tP16_132_d_io_a]{\pageref{AB6C_tP16_132_d_io_a}}} \\
\vspace{-0.75cm} \item Akermanite: {\small A2BC7D2\_tP24\_113\_e\_a\_cef\_e} \dotfill {\hyperref[A2BC7D2_tP24_113_e_a_cef_e]{\pageref{A2BC7D2_tP24_113_e_a_cef_e}}} \\
\vspace{-0.75cm} \item Al$_{2}$CuIr\footnote[4]{\label{note:ABC2_oC16_67_b_g_ag-prototype}Al$_{2}$CuIr and HoCuP$_{2}$ have similar \AFLOW\ prototype labels ({\it{i.e.}}, same symmetry and set of Wyckoff positions with different stoichiometry labels due to alphabetic ordering of atomic species). They are generated by the same symmetry operations with different sets of parameters.}: {\small A2BC\_oC16\_67\_ag\_b\_g} \dotfill {\hyperref[A2BC_oC16_67_ag_b_g]{\pageref{A2BC_oC16_67_ag_b_g}}} \\
\vspace{-0.75cm} \item Al$_{2}$S$_{3}$: {\small A2B3\_hP30\_170\_2a\_3a} \dotfill {\hyperref[A2B3_hP30_170_2a_3a]{\pageref{A2B3_hP30_170_2a_3a}}} \\
\vspace{-0.75cm} \item Al$_{4}$C$_{3}$: {\small A4B3\_hR7\_166\_2c\_ac} \dotfill {\hyperref[A4B3_hR7_166_2c_ac]{\pageref{A4B3_hR7_166_2c_ac}}} \\
\vspace{-0.75cm} \item Al$_{4}$U: {\small A4B\_oI20\_74\_beh\_e} \dotfill {\hyperref[A4B_oI20_74_beh_e]{\pageref{A4B_oI20_74_beh_e}}} \\
\vspace{-0.75cm} \item Al$_{8}$Cr$_{5}$: {\small A8B5\_hR26\_160\_a3bc\_a3b} \dotfill {\hyperref[A8B5_hR26_160_a3bc_a3b]{\pageref{A8B5_hR26_160_a3bc_a3b}}} \\
\vspace{-0.75cm} \item Al$_{9}$Mn$_{3}$Si: {\small A9B3C\_hP26\_194\_hk\_h\_a} \dotfill {\hyperref[A9B3C_hP26_194_hk_h_a]{\pageref{A9B3C_hP26_194_hk_h_a}}} \\
\vspace{-0.75cm} \item AlLi$_{3}$N$_{2}$: {\small AB3C2\_cI96\_206\_c\_e\_ad} \dotfill {\hyperref[AB3C2_cI96_206_c_e_ad]{\pageref{AB3C2_cI96_206_c_e_ad}}} \\
\vspace{-0.75cm} \item AlPO$_{4}$: {\small AB2\_hP72\_192\_m\_j2kl} \dotfill {\hyperref[AB2_hP72_192_m_j2kl]{\pageref{AB2_hP72_192_m_j2kl}}} \\
\vspace{-0.75cm} \item Al[PO$_{4}$]: {\small AB4C\_hP72\_168\_2d\_8d\_2d} \dotfill {\hyperref[AB4C_hP72_168_2d_8d_2d]{\pageref{AB4C_hP72_168_2d_8d_2d}}} \\
\vspace{-0.75cm} \item Al[PO$_{4}$]: {\small AB4C\_hP72\_184\_d\_4d\_d} \dotfill {\hyperref[AB4C_hP72_184_d_4d_d]{\pageref{AB4C_hP72_184_d_4d_d}}} \\
\vspace{-0.75cm} \item Anhydrite: {\small AB4C\_oC24\_63\_c\_fg\_c} \dotfill {\hyperref[AB4C_oC24_63_c_fg_c]{\pageref{AB4C_oC24_63_c_fg_c}}} \\
\vspace{-0.75cm} \item As$_{2}$Ba: {\small A2B\_mP18\_7\_6a\_3a} \dotfill {\hyperref[A2B_mP18_7_6a_3a]{\pageref{A2B_mP18_7_6a_3a}}} \\
\vspace{-0.75cm} \item \begin{raggedleft}AsPh$_{4}$CeS$_{8}$P$_{4}$Me$_{8}$: \end{raggedleft} \\ {\small AB32CD4E8\_tP184\_93\_i\_16p\_af\_2p\_4p} \dotfill {\hyperref[AB32CD4E8_tP184_93_i_16p_af_2p_4p]{\pageref{AB32CD4E8_tP184_93_i_16p_af_2p_4p}}} \\
\vspace{-0.75cm} \item AuCN: {\small ABC\_hP3\_183\_a\_a\_a} \dotfill {\hyperref[ABC_hP3_183_a_a_a]{\pageref{ABC_hP3_183_a_a_a}}} \\
\vspace{-0.75cm} \item AuF$_{3}$: {\small AB3\_hP24\_178\_b\_ac} \dotfill {\hyperref[AB3_hP24_178_b_ac]{\pageref{AB3_hP24_178_b_ac}}} \\
\vspace{-0.75cm} \item AuF$_{3}$: {\small AB3\_hP24\_179\_b\_ac} \dotfill {\hyperref[AB3_hP24_179_b_ac]{\pageref{AB3_hP24_179_b_ac}}} \\
\vspace{-0.75cm} \item BN: {\small AB\_oF8\_42\_a\_a} \dotfill {\hyperref[AB_oF8_42_a_a]{\pageref{AB_oF8_42_a_a}}} \\
\vspace{-0.75cm} \item BPS$_{4}$: {\small ABC4\_oI12\_23\_a\_b\_k} \dotfill {\hyperref[ABC4_oI12_23_a_b_k]{\pageref{ABC4_oI12_23_a_b_k}}} \\
\vspace{-0.75cm} \item Ba$_{5}$In$_{4}$Bi$_{5}$: {\small A5B5C4\_tP28\_104\_ac\_ac\_c} \dotfill {\hyperref[A5B5C4_tP28_104_ac_ac_c]{\pageref{A5B5C4_tP28_104_ac_ac_c}}} \\
\vspace{-0.75cm} \item Ba$_{5}$Si$_{3}$: {\small A5B3\_tP32\_130\_cg\_cf} \dotfill {\hyperref[A5B3_tP32_130_cg_cf]{\pageref{A5B3_tP32_130_cg_cf}}} \\
\vspace{-0.75cm} \item \begin{raggedleft}BaCr$_{2}$Ru$_{4}$O$_{12}$: \end{raggedleft} \\ {\small AB2C12D4\_tP76\_75\_2a2b\_2d\_12d\_4d} \dotfill {\hyperref[AB2C12D4_tP76_75_2a2b_2d_12d_4d]{\pageref{AB2C12D4_tP76_75_2a2b_2d_12d_4d}}} \\
\vspace{-0.75cm} \item \begin{raggedleft}BaCu$_{4}$[VO][PO$_{4}$]$_{4}$: \end{raggedleft} \\ {\small AB4C17D4E\_tP54\_90\_a\_g\_c4g\_g\_c} \dotfill {\hyperref[AB4C17D4E_tP54_90_a_g_c4g_g_c]{\pageref{AB4C17D4E_tP54_90_a_g_c4g_g_c}}} \\
\vspace{-0.75cm} \item BaGe$_{2}$As$_{2}$: {\small A2BC2\_tP20\_105\_f\_ac\_2e} \dotfill {\hyperref[A2BC2_tP20_105_f_ac_2e]{\pageref{A2BC2_tP20_105_f_ac_2e}}} \\
\vspace{-0.75cm} \item BaSi$_{4}$O$_{9}$: {\small AB9C4\_hP28\_188\_e\_kl\_ak} \dotfill {\hyperref[AB9C4_hP28_188_e_kl_ak]{\pageref{AB9C4_hP28_188_e_kl_ak}}} \\
\vspace{-0.75cm} \item Barite: {\small AB4C\_oP24\_62\_c\_2cd\_c} \dotfill {\hyperref[AB4C_oP24_62_c_2cd_c]{\pageref{AB4C_oP24_62_c_2cd_c}}} \\
\vspace{-0.75cm} \item Be[BH$_{4}$]$_{2}$: {\small A2BC8\_tI176\_110\_2b\_b\_8b} \dotfill {\hyperref[A2BC8_tI176_110_2b_b_8b]{\pageref{A2BC8_tI176_110_2b_b_8b}}} \\
\vspace{-0.75cm} \item Benzene: {\small AB\_oP48\_61\_3c\_3c} \dotfill {\hyperref[AB_oP48_61_3c_3c]{\pageref{AB_oP48_61_3c_3c}}} \\
\vspace{-0.75cm} \item Beryl: {\small A2B3C18D6\_hP58\_192\_c\_f\_lm\_l} \dotfill {\hyperref[A2B3C18D6_hP58_192_c_f_lm_l]{\pageref{A2B3C18D6_hP58_192_c_f_lm_l}}} \\
\vspace{-0.75cm} \item Bi$_{2}$O$_{3}$: {\small A2B3\_hP20\_159\_bc\_2c} \dotfill {\hyperref[A2B3_hP20_159_bc_2c]{\pageref{A2B3_hP20_159_bc_2c}}} \\
\vspace{-0.75cm} \item Bi$_{5}$Nb$_{3}$O$_{15}$: {\small A5B3C15\_oP46\_30\_a2c\_bc\_a7c} \dotfill {\hyperref[A5B3C15_oP46_30_a2c_bc_a7c]{\pageref{A5B3C15_oP46_30_a2c_bc_a7c}}} \\
\vspace{-0.75cm} \item BiAl$_{2}$S$_{4}$: {\small A2BC4\_tP28\_126\_cd\_e\_k} \dotfill {\hyperref[A2BC4_tP28_126_cd_e_k]{\pageref{A2BC4_tP28_126_cd_e_k}}} \\
\vspace{-0.75cm} \item BiGaO$_{3}$: {\small ABC3\_oP20\_54\_e\_d\_cf} \dotfill {\hyperref[ABC3_oP20_54_e_d_cf]{\pageref{ABC3_oP20_54_e_d_cf}}} \\
\vspace{-0.75cm} \item Boracite: {\small A7BC3D13\_cF192\_219\_de\_b\_c\_ah} \dotfill {\hyperref[A7BC3D13_cF192_219_de_b_c_ah]{\pageref{A7BC3D13_cF192_219_de_b_c_ah}}} \\
\vspace{-0.75cm} \item C: {\small A\_tP12\_138\_bi} \dotfill {\hyperref[A_tP12_138_bi]{\pageref{A_tP12_138_bi}}} \\
\vspace{-0.75cm} \item C$_{17}$FeO$_{4}$Pt: {\small A17BC4D\_tP184\_89\_17p\_p\_4p\_io} \dotfill {\hyperref[A17BC4D_tP184_89_17p_p_4p_io]{\pageref{A17BC4D_tP184_89_17p_p_4p_io}}} \\
\vspace{-0.75cm} \item Ca$_{3}$Al$_{2}$O$_{6}$: {\small A2B3C6\_cP264\_205\_2d\_ab2c2d\_6d} \dotfill {\hyperref[A2B3C6_cP264_205_2d_ab2c2d_6d]{\pageref{A2B3C6_cP264_205_2d_ab2c2d_6d}}} \\
\vspace{-0.75cm} \item Ca$_{3}$Al$_{2}$O$_{6}$: {\small A2B3C6\_cP33\_221\_cd\_ag\_fh} \dotfill {\hyperref[A2B3C6_cP33_221_cd_ag_fh]{\pageref{A2B3C6_cP33_221_cd_ag_fh}}} \\
\vspace{-0.75cm} \item Ca$_{3}$PI$_{3}$: {\small A3B3C\_cI56\_214\_g\_h\_a} \dotfill {\hyperref[A3B3C_cI56_214_g_h_a]{\pageref{A3B3C_cI56_214_g_h_a}}} \\
\vspace{-0.75cm} \item \begin{raggedleft}Ca$_{4}$Al$_{6}$O$_{16}$S: \end{raggedleft} \\ {\small A6B4C16D\_oP108\_27\_abcd4e\_4e\_16e\_e} \dotfill {\hyperref[A6B4C16D_oP108_27_abcd4e_4e_16e_e]{\pageref{A6B4C16D_oP108_27_abcd4e_4e_16e_e}}} \\
\vspace{-0.75cm} \item CaRbFe$_{4}$As$_{4}$: {\small A4BC4D\_tP10\_123\_gh\_a\_i\_d} \dotfill {\hyperref[A4BC4D_tP10_123_gh_a_i_d]{\pageref{A4BC4D_tP10_123_gh_a_i_d}}} \\
\vspace{-0.75cm} \item Calomel: {\small AB\_tI8\_139\_e\_e} \dotfill {\hyperref[AB_tI8_139_e_e]{\pageref{AB_tI8_139_e_e}}} \\
\vspace{-0.75cm} \item Carbonyl Sulphide: {\small ABC\_hR3\_160\_a\_a\_a} \dotfill {\hyperref[ABC_hR3_160_a_a_a]{\pageref{ABC_hR3_160_a_a_a}}} \\
\vspace{-0.75cm} \item CdAs$_{2}$: {\small A2B\_tI12\_98\_f\_a} \dotfill {\hyperref[A2B_tI12_98_f_a]{\pageref{A2B_tI12_98_f_a}}} \\
\vspace{-0.75cm} \item CdI$_{2}$: {\small AB2\_hP9\_156\_b2c\_3a2bc} \dotfill {\hyperref[AB2_hP9_156_b2c_3a2bc]{\pageref{AB2_hP9_156_b2c_3a2bc}}} \\
\vspace{-0.75cm} \item Ce$_{3}$Si$_{6}$N$_{11}$: {\small A3B11C6\_tP40\_100\_ac\_bc2d\_cd} \dotfill {\hyperref[A3B11C6_tP40_100_ac_bc2d_cd]{\pageref{A3B11C6_tP40_100_ac_bc2d_cd}}} \\
\vspace{-0.75cm} \item Ce$_{5}$Mo$_{3}$O$_{16}$: {\small A5B3C16\_cP96\_222\_ce\_d\_fi} \dotfill {\hyperref[A5B3C16_cP96_222_ce_d_fi]{\pageref{A5B3C16_cP96_222_ce_d_fi}}} \\
\vspace{-0.75cm} \item CeCo$_{4}$B$_{4}$: {\small A4BC4\_tP18\_137\_g\_b\_g} \dotfill {\hyperref[A4BC4_tP18_137_g_b_g]{\pageref{A4BC4_tP18_137_g_b_g}}} \\
\vspace{-0.75cm} \item CeRu$_{2}$B$_{2}$: {\small A2BC2\_oF40\_22\_fi\_ad\_gh} \dotfill {\hyperref[A2BC2_oF40_22_fi_ad_gh]{\pageref{A2BC2_oF40_22_fi_ad_gh}}} \\
\vspace{-0.75cm} \item CeTe$_{3}$: {\small AB3\_oC16\_40\_b\_3b} \dotfill {\hyperref[AB3_oC16_40_b_3b]{\pageref{AB3_oC16_40_b_3b}}} \\
\vspace{-0.75cm} \item Co$_{2}$Al$_{5}$: {\small A5B2\_hP28\_194\_ahk\_ch} \dotfill {\hyperref[A5B2_hP28_194_ahk_ch]{\pageref{A5B2_hP28_194_ahk_ch}}} \\
\vspace{-0.75cm} \item Co$_{5}$Ge$_{7}$: {\small A5B7\_tI24\_107\_ac\_abd} \dotfill {\hyperref[A5B7_tI24_107_ac_abd]{\pageref{A5B7_tI24_107_ac_abd}}} \\
\vspace{-0.75cm} \item Cobaltite: {\small ABC\_oP12\_29\_a\_a\_a} \dotfill {\hyperref[ABC_oP12_29_a_a_a]{\pageref{ABC_oP12_29_a_a_a}}} \\
\vspace{-0.75cm} \item Cr$_{5}$B$_{3}$: {\small A3B5\_tI32\_140\_ah\_cl} \dotfill {\hyperref[A3B5_tI32_140_ah_cl]{\pageref{A3B5_tI32_140_ah_cl}}} \\
\vspace{-0.75cm} \item CrCl$_{3}$: {\small A3B\_hP24\_153\_3c\_2b} \dotfill {\hyperref[A3B_hP24_153_3c_2b]{\pageref{A3B_hP24_153_3c_2b}}} \\
\vspace{-0.75cm} \item CrFe$_{3}$NiSn$_{5}$: {\small AB\_hP6\_183\_c\_ab} \dotfill {\hyperref[AB_hP6_183_c_ab]{\pageref{AB_hP6_183_c_ab}}} \\
\vspace{-0.75cm} \item \begin{raggedleft}Cs$_{2}$ZnFe[CN]$_{6}$: \end{raggedleft} \\ {\small A6B2CD6E\_cP64\_208\_m\_ad\_b\_m\_c} \dotfill {\hyperref[A6B2CD6E_cP64_208_m_ad_b_m_c]{\pageref{A6B2CD6E_cP64_208_m_ad_b_m_c}}} \\
\vspace{-0.75cm} \item Cs$_{3}$P$_{7}$: {\small A3B7\_tP40\_76\_3a\_7a} \dotfill {\hyperref[A3B7_tP40_76_3a_7a]{\pageref{A3B7_tP40_76_3a_7a}}} \\
\vspace{-0.75cm} \item CsPr[MoO$_{4}$]$_{2}$: {\small AB2C8D\_oP24\_49\_g\_q\_2qr\_e} \dotfill {\hyperref[AB2C8D_oP24_49_g_q_2qr_e]{\pageref{AB2C8D_oP24_49_g_q_2qr_e}}} \\
\vspace{-0.75cm} \item Cu$_{15}$Si$_{4}$: {\small A15B4\_cI76\_220\_ae\_c} \dotfill {\hyperref[A15B4_cI76_220_ae_c]{\pageref{A15B4_cI76_220_ae_c}}} \\
\vspace{-0.75cm} \item Cu$_{2}$Fe[CN]$_{6}$: {\small A12B2C\_cF60\_196\_h\_bc\_a} \dotfill {\hyperref[A12B2C_cF60_196_h_bc_a]{\pageref{A12B2C_cF60_196_h_bc_a}}} \\
\vspace{-0.75cm} \item Cu$_{3}$P: {\small A3B\_hP24\_165\_bdg\_f} \dotfill {\hyperref[A3B_hP24_165_bdg_f]{\pageref{A3B_hP24_165_bdg_f}}} \\
\vspace{-0.75cm} \item Cu$_{3}$P\footnote[6]{\label{note:AB3_hP24_185_c_ab2c-prototype}Cu$_{3}$P and Na$_{3}$As have similar \AFLOW\ prototype labels ({\it{i.e.}}, same symmetry and set of Wyckoff positions with different stoichiometry labels due to alphabetic ordering of atomic species). They are generated by the same symmetry operations with different sets of parameters.}: {\small A3B\_hP24\_185\_ab2c\_c} \dotfill {\hyperref[A3B_hP24_185_ab2c_c]{\pageref{A3B_hP24_185_ab2c_c}}} \\
\vspace{-0.75cm} \item CuBi$_{2}$O$_{4}$: {\small A2BC4\_tP28\_130\_f\_c\_g} \dotfill {\hyperref[A2BC4_tP28_130_f_c_g]{\pageref{A2BC4_tP28_130_f_c_g}}} \\
\vspace{-0.75cm} \item CuBrSe$_{3}$: {\small ABC3\_oP20\_30\_2a\_c\_3c} \dotfill {\hyperref[ABC3_oP20_30_2a_c_3c]{\pageref{ABC3_oP20_30_2a_c_3c}}} \\
\vspace{-0.75cm} \item CuBrSe$_{3}$: {\small ABC3\_oP20\_53\_e\_g\_hi} \dotfill {\hyperref[ABC3_oP20_53_e_g_hi]{\pageref{ABC3_oP20_53_e_g_hi}}} \\
\vspace{-0.75cm} \item CuCrCl$_{5}$[NH$_{3}$]$_{6}$: {\small A5BCD6\_cF416\_228\_eg\_c\_b\_h} \dotfill {\hyperref[A5BCD6_cF416_228_eg_c_b_h]{\pageref{A5BCD6_cF416_228_eg_c_b_h}}} \\
\vspace{-0.75cm} \item CuI: {\small AB\_hP12\_156\_2ab3c\_2ab3c} \dotfill {\hyperref[AB_hP12_156_2ab3c_2ab3c]{\pageref{AB_hP12_156_2ab3c_2ab3c}}} \\
\vspace{-0.75cm} \item CuNiSb$_{2}$: {\small ABC2\_hP4\_164\_a\_b\_d} \dotfill {\hyperref[ABC2_hP4_164_a_b_d]{\pageref{ABC2_hP4_164_a_b_d}}} \\
\vspace{-0.75cm} \item Cubanite: {\small AB2C3\_oP24\_62\_c\_d\_cd} \dotfill {\hyperref[AB2C3_oP24_62_c_d_cd]{\pageref{AB2C3_oP24_62_c_d_cd}}} \\
\vspace{-0.75cm} \item Downeyite: {\small A2B\_tP24\_135\_gh\_h} \dotfill {\hyperref[A2B_tP24_135_gh_h]{\pageref{A2B_tP24_135_gh_h}}} \\
\vspace{-0.75cm} \item Er$_{3}$Ru$_{2}$: {\small A3B2\_hP10\_176\_h\_bd} \dotfill {\hyperref[A3B2_hP10_176_h_bd]{\pageref{A3B2_hP10_176_h_bd}}} \\
\vspace{-0.75cm} \item F$_{6}$KP: {\small A24BC\_cF104\_209\_j\_a\_b} \dotfill {\hyperref[A24BC_cF104_209_j_a_b]{\pageref{A24BC_cF104_209_j_a_b}}} \\
\vspace{-0.75cm} \item \begin{raggedleft}FCC C$_{60}$ Buckminsterfullerine: \end{raggedleft} \\ {\small A\_cF240\_202\_h2i} \dotfill {\hyperref[A_cF240_202_h2i]{\pageref{A_cF240_202_h2i}}} \\
\vspace{-0.75cm} \item Fe$_{12}$Zr$_{2}$P$_{7}$: {\small A12B7C2\_hP21\_174\_2j2k\_ajk\_cf} \dotfill {\hyperref[A12B7C2_hP21_174_2j2k_ajk_cf]{\pageref{A12B7C2_hP21_174_2j2k_ajk_cf}}} \\
\vspace{-0.75cm} \item Fe$_{3}$Te$_{3}$Tl: {\small A3B3C\_hP14\_176\_h\_h\_d} \dotfill {\hyperref[A3B3C_hP14_176_h_h_d]{\pageref{A3B3C_hP14_176_h_h_d}}} \\
\vspace{-0.75cm} \item Fe$_{3}$Th$_{7}$: {\small A3B7\_hP20\_186\_c\_b2c} \dotfill {\hyperref[A3B7_hP20_186_c_b2c]{\pageref{A3B7_hP20_186_c_b2c}}} \\
\vspace{-0.75cm} \item FeCu$_{2}$Al$_{7}$: {\small A7B2C\_tP40\_128\_egi\_h\_e} \dotfill {\hyperref[A7B2C_tP40_128_egi_h_e]{\pageref{A7B2C_tP40_128_egi_h_e}}} \\
\vspace{-0.75cm} \item FeNi: {\small AB\_mP4\_6\_2b\_2a} \dotfill {\hyperref[AB_mP4_6_2b_2a]{\pageref{AB_mP4_6_2b_2a}}} \\
\vspace{-0.75cm} \item FeOCl: {\small ABC\_oP6\_59\_a\_b\_a} \dotfill {\hyperref[ABC_oP6_59_a_b_a]{\pageref{ABC_oP6_59_a_b_a}}} \\
\vspace{-0.75cm} \item FePSe$_{3}$: {\small ABC3\_hR10\_146\_2a\_2a\_2b} \dotfill {\hyperref[ABC3_hR10_146_2a_2a_2b]{\pageref{ABC3_hR10_146_2a_2a_2b}}} \\
\vspace{-0.75cm} \item FeS: {\small AB\_oF8\_22\_a\_c} \dotfill {\hyperref[AB_oF8_22_a_c]{\pageref{AB_oF8_22_a_c}}} \\
\vspace{-0.75cm} \item FeSb$_{2}$: {\small AB2\_oP6\_34\_a\_c} \dotfill {\hyperref[AB2_oP6_34_a_c]{\pageref{AB2_oP6_34_a_c}}} \\
\vspace{-0.75cm} \item Forsterite: {\small A2B4C\_oP28\_62\_ac\_2cd\_c} \dotfill {\hyperref[A2B4C_oP28_62_ac_2cd_c]{\pageref{A2B4C_oP28_62_ac_2cd_c}}} \\
\vspace{-0.75cm} \item Fresnoite: {\small A2B8C2D\_tP26\_100\_c\_abcd\_c\_a} \dotfill {\hyperref[A2B8C2D_tP26_100_c_abcd_c_a]{\pageref{A2B8C2D_tP26_100_c_abcd_c_a}}} \\
\vspace{-0.75cm} \item GaCl$_{2}$: {\small A2B\_oP24\_52\_2e\_cd} \dotfill {\hyperref[A2B_oP24_52_2e_cd]{\pageref{A2B_oP24_52_2e_cd}}} \\
\vspace{-0.75cm} \item GaSb: {\small AB\_tI4\_119\_c\_a} \dotfill {\hyperref[AB_tI4_119_c_a]{\pageref{AB_tI4_119_c_a}}} \\
\vspace{-0.75cm} \item Garnet: {\small A2B3C12D3\_cI160\_230\_a\_c\_h\_d} \dotfill {\hyperref[A2B3C12D3_cI160_230_a_c_h_d]{\pageref{A2B3C12D3_cI160_230_a_c_h_d}}} \\
\vspace{-0.75cm} \item Gd$_{3}$Al$_{2}$: {\small A2B3\_tP20\_102\_2c\_b2c} \dotfill {\hyperref[A2B3_tP20_102_2c_b2c]{\pageref{A2B3_tP20_102_2c_b2c}}} \\
\vspace{-0.75cm} \item GdSI: {\small ABC\_hP12\_174\_cj\_fk\_aj} \dotfill {\hyperref[ABC_hP12_174_cj_fk_aj]{\pageref{ABC_hP12_174_cj_fk_aj}}} \\
\vspace{-0.75cm} \item GeAs$_{2}$: {\small A2B\_oP24\_55\_2g2h\_gh} \dotfill {\hyperref[A2B_oP24_55_2g2h_gh]{\pageref{A2B_oP24_55_2g2h_gh}}} \\
\vspace{-0.75cm} \item GeP: {\small AB\_tI4\_107\_a\_a} \dotfill {\hyperref[AB_tI4_107_a_a]{\pageref{AB_tI4_107_a_a}}} \\
\vspace{-0.75cm} \item GeSe$_{2}$: {\small AB2\_tP12\_81\_adg\_2h} \dotfill {\hyperref[AB2_tP12_81_adg_2h]{\pageref{AB2_tP12_81_adg_2h}}} \\
\vspace{-0.75cm} \item H$_{2}$S: {\small A2B\_aP6\_2\_aei\_i} \dotfill {\hyperref[A2B_aP6_2_aei_i]{\pageref{A2B_aP6_2_aei_i}}} \\
\vspace{-0.75cm} \item H$_{2}$S: {\small A2B\_mP12\_13\_2g\_ef} \dotfill {\hyperref[A2B_mP12_13_2g_ef]{\pageref{A2B_mP12_13_2g_ef}}} \\
\vspace{-0.75cm} \item H$_{2}$S\footnoteref{note:A2B_oP12_26_abc_ab-prototype}: {\small A2B\_oP12\_26\_abc\_ab} \dotfill {\hyperref[A2B_oP12_26_abc_ab-H2S]{\pageref{A2B_oP12_26_abc_ab-H2S}}} \\
\vspace{-0.75cm} \item H$_{2}$S: {\small A2B\_oC24\_64\_2f\_f} \dotfill {\hyperref[A2B_oC24_64_2f_f]{\pageref{A2B_oC24_64_2f_f}}} \\
\vspace{-0.75cm} \item H$_{2}$S III: {\small A2B\_tP48\_77\_8d\_4d} \dotfill {\hyperref[A2B_tP48_77_8d_4d]{\pageref{A2B_tP48_77_8d_4d}}} \\
\vspace{-0.75cm} \item H$_{2}$S IV: {\small A2B\_mP12\_7\_4a\_2a} \dotfill {\hyperref[A2B_mP12_7_4a_2a]{\pageref{A2B_mP12_7_4a_2a}}} \\
\vspace{-0.75cm} \item H$_{3}$Cl: {\small AB3\_mC16\_9\_a\_3a} \dotfill {\hyperref[AB3_mC16_9_a_3a]{\pageref{AB3_mC16_9_a_3a}}} \\
\vspace{-0.75cm} \item H$_{3}$Cl: {\small AB3\_mP16\_10\_mn\_3m3n} \dotfill {\hyperref[AB3_mP16_10_mn_3m3n]{\pageref{AB3_mP16_10_mn_3m3n}}} \\
\vspace{-0.75cm} \item H$_{3}$Cl: {\small AB3\_mC16\_15\_e\_cf} \dotfill {\hyperref[AB3_mC16_15_e_cf]{\pageref{AB3_mC16_15_e_cf}}} \\
\vspace{-0.75cm} \item H$_{3}$Cl: {\small AB3\_oP16\_19\_a\_3a} \dotfill {\hyperref[AB3_oP16_19_a_3a]{\pageref{AB3_oP16_19_a_3a}}} \\
\vspace{-0.75cm} \item H$_{3}$S: {\small A3B\_oI32\_23\_ij2k\_k} \dotfill {\hyperref[A3B_oI32_23_ij2k_k]{\pageref{A3B_oI32_23_ij2k_k}}} \\
\vspace{-0.75cm} \item H$_{3}$S: {\small A3B\_oC64\_66\_gi2lm\_2l} \dotfill {\hyperref[A3B_oC64_66_gi2lm_2l]{\pageref{A3B_oC64_66_gi2lm_2l}}} \\
\vspace{-0.75cm} \item H$_{3}$S: {\small A3B\_hR4\_160\_b\_a} \dotfill {\hyperref[A3B_hR4_160_b_a]{\pageref{A3B_hR4_160_b_a}}} \\
\vspace{-0.75cm} \item H-III: {\small A\_mC24\_15\_2e2f} \dotfill {\hyperref[A_mC24_15_2e2f]{\pageref{A_mC24_15_2e2f}}} \\
\vspace{-0.75cm} \item HCl: {\small AB\_oC8\_36\_a\_a} \dotfill {\hyperref[AB_oC8_36_a_a]{\pageref{AB_oC8_36_a_a}}} \\
\vspace{-0.75cm} \item HgI$_{2}$: {\small AB2\_tP12\_115\_j\_egi} \dotfill {\hyperref[AB2_tP12_115_j_egi]{\pageref{AB2_tP12_115_j_egi}}} \\
\vspace{-0.75cm} \item HgI$_{2}$\footnote[5]{\label{note:AB2_tP6_137_a_d-prototype}ZrO$_{2}$ and HgI$_{2}$ have similar \AFLOW\ prototype labels ({\it{i.e.}}, same symmetry and set of Wyckoff positions with different stoichiometry labels due to alphabetic ordering of atomic species). They are generated by the same symmetry operations with different sets of parameters.}: {\small AB2\_tP6\_137\_a\_d} \dotfill {\hyperref[AB2_tP6_137_a_d]{\pageref{AB2_tP6_137_a_d}}} \\
\vspace{-0.75cm} \item HoCuP$_{2}$\footnoteref{note:ABC2_oC16_67_b_g_ag-prototype}: {\small ABC2\_oC16\_67\_b\_g\_ag} \dotfill {\hyperref[ABC2_oC16_67_b_g_ag]{\pageref{ABC2_oC16_67_b_g_ag}}} \\
\vspace{-0.75cm} \item Ir$_{3}$Ga$_{5}$: {\small A5B3\_tP32\_118\_g2i\_aceh} \dotfill {\hyperref[A5B3_tP32_118_g2i_aceh]{\pageref{A5B3_tP32_118_g2i_aceh}}} \\
\vspace{-0.75cm} \item Ir$_{3}$Ge$_{7}$: {\small A7B3\_cI40\_229\_df\_e} \dotfill {\hyperref[A7B3_cI40_229_df_e]{\pageref{A7B3_cI40_229_df_e}}} \\
\vspace{-0.75cm} \item IrGe$_{4}$: {\small A4B\_hP15\_144\_4a\_a} \dotfill {\hyperref[A4B_hP15_144_4a_a]{\pageref{A4B_hP15_144_4a_a}}} \\
\vspace{-0.75cm} \item K$_{2}$CdPb: {\small AB2C\_oC16\_40\_a\_2b\_b} \dotfill {\hyperref[AB2C_oC16_40_a_2b_b]{\pageref{AB2C_oC16_40_a_2b_b}}} \\
\vspace{-0.75cm} \item K$_{2}$PtCl$_{6}$: {\small A6B2C\_cF36\_225\_e\_c\_a} \dotfill {\hyperref[A6B2C_cF36_225_e_c_a]{\pageref{A6B2C_cF36_225_e_c_a}}} \\
\vspace{-0.75cm} \item K$_{2}$SnCl$_{6}$: {\small A6B2C\_tP18\_128\_eh\_d\_b} \dotfill {\hyperref[A6B2C_tP18_128_eh_d_b]{\pageref{A6B2C_tP18_128_eh_d_b}}} \\
\vspace{-0.75cm} \item K$_{2}$Ta$_{4}$O$_{9}$F$_{4}$: {\small A2B13C4\_hP57\_168\_d\_c6d\_2d} \dotfill {\hyperref[A2B13C4_hP57_168_d_c6d_2d]{\pageref{A2B13C4_hP57_168_d_c6d_2d}}} \\
\vspace{-0.75cm} \item KAg[CO$_{3}$]: {\small ABCD3\_oI48\_73\_d\_e\_e\_ef} \dotfill {\hyperref[ABCD3_oI48_73_d_e_e_ef]{\pageref{ABCD3_oI48_73_d_e_e_ef}}} \\
\vspace{-0.75cm} \item KAu$_{4}$Sn$_{2}$: {\small A4BC2\_tI28\_120\_i\_d\_e} \dotfill {\hyperref[A4BC2_tI28_120_i_d_e]{\pageref{A4BC2_tI28_120_i_d_e}}} \\
\vspace{-0.75cm} \item KB$_{6}$H$_{6}$: {\small A6B6C\_cF104\_202\_h\_h\_c} \dotfill {\hyperref[A6B6C_cF104_202_h_h_c]{\pageref{A6B6C_cF104_202_h_h_c}}} \\
\vspace{-0.75cm} \item KBO$_{2}$: {\small ABC2\_hR24\_167\_e\_e\_2e} \dotfill {\hyperref[ABC2_hR24_167_e_e_2e]{\pageref{ABC2_hR24_167_e_e_2e}}} \\
\vspace{-0.75cm} \item KCeSe$_{4}$: {\small ABC4\_tP12\_125\_a\_b\_m} \dotfill {\hyperref[ABC4_tP12_125_a_b_m]{\pageref{ABC4_tP12_125_a_b_m}}} \\
\vspace{-0.75cm} \item KHg$_{2}$: {\small A2B\_oI12\_74\_h\_e} \dotfill {\hyperref[A2B_oI12_74_h_e]{\pageref{A2B_oI12_74_h_e}}} \\
\vspace{-0.75cm} \item KNiCl$_{3}$: {\small A3BC\_hP30\_185\_cd\_c\_ab} \dotfill {\hyperref[A3BC_hP30_185_cd_c_ab]{\pageref{A3BC_hP30_185_cd_c_ab}}} \\
\vspace{-0.75cm} \item KSbO$_{3}$: {\small AB3C\_cP60\_201\_ce\_fh\_g} \dotfill {\hyperref[AB3C_cP60_201_ce_fh_g]{\pageref{AB3C_cP60_201_ce_fh_g}}} \\
\vspace{-0.75cm} \item La$_{2}$NiO$_{4}$: {\small A2BC4\_oP28\_50\_ij\_ac\_ijm} \dotfill {\hyperref[A2BC4_oP28_50_ij_ac_ijm]{\pageref{A2BC4_oP28_50_ij_ac_ijm}}} \\
\vspace{-0.75cm} \item La$_{2}$O$_{3}$: {\small A2B3\_hP5\_164\_d\_ad} \dotfill {\hyperref[A2B3_hP5_164_d_ad]{\pageref{A2B3_hP5_164_d_ad}}} \\
\vspace{-0.75cm} \item \begin{raggedleft}La$_{43}$Ni$_{17}$Mg$_{5}$: \end{raggedleft} \\ {\small A43B5C17\_oC260\_63\_c8fg6h\_cfg\_ce3f2h} \dotfill {\hyperref[A43B5C17_oC260_63_c8fg6h_cfg_ce3f2h]{\pageref{A43B5C17_oC260_63_c8fg6h_cfg_ce3f2h}}} \\
\vspace{-0.75cm} \item LaPtSi: {\small ABC\_tI12\_109\_a\_a\_a} \dotfill {\hyperref[ABC_tI12_109_a_a_a]{\pageref{ABC_tI12_109_a_a_a}}} \\
\vspace{-0.75cm} \item LaRhC$_{2}$: {\small A2BC\_tP16\_76\_2a\_a\_a} \dotfill {\hyperref[A2BC_tP16_76_2a_a_a]{\pageref{A2BC_tP16_76_2a_a_a}}} \\
\vspace{-0.75cm} \item Li$_{2}$MoF$_{6}$: {\small A6B2C\_tP18\_94\_eg\_c\_a} \dotfill {\hyperref[A6B2C_tP18_94_eg_c_a]{\pageref{A6B2C_tP18_94_eg_c_a}}} \\
\vspace{-0.75cm} \item Li$_{2}$Sb: {\small A2B\_hP18\_190\_gh\_bf} \dotfill {\hyperref[A2B_hP18_190_gh_bf]{\pageref{A2B_hP18_190_gh_bf}}} \\
\vspace{-0.75cm} \item Li$_{2}$Si$_{2}$O$_{5}$: {\small A2B5C2\_oC36\_37\_d\_c2d\_d} \dotfill {\hyperref[A2B5C2_oC36_37_d_c2d_d]{\pageref{A2B5C2_oC36_37_d_c2d_d}}} \\
\vspace{-0.75cm} \item LiScI$_{3}$: {\small A3BC\_hP10\_188\_k\_a\_e} \dotfill {\hyperref[A3BC_hP10_188_k_a_e]{\pageref{A3BC_hP10_188_k_a_e}}} \\
\vspace{-0.75cm} \item LiSn: {\small AB\_mP6\_10\_en\_am} \dotfill {\hyperref[AB_mP6_10_en_am]{\pageref{AB_mP6_10_en_am}}} \\
\vspace{-0.75cm} \item M-carbon: {\small A\_mC16\_12\_4i} \dotfill {\hyperref[A_mC16_12_4i]{\pageref{A_mC16_12_4i}}} \\
\vspace{-0.75cm} \item Mavlyanovite: {\small A5B3\_hP16\_193\_dg\_g} \dotfill {\hyperref[A5B3_hP16_193_dg_g]{\pageref{A5B3_hP16_193_dg_g}}} \\
\vspace{-0.75cm} \item Mg$_{2}$Zn$_{11}$: {\small A2B11\_cP39\_200\_f\_aghij} \dotfill {\hyperref[A2B11_cP39_200_f_aghij]{\pageref{A2B11_cP39_200_f_aghij}}} \\
\vspace{-0.75cm} \item \begin{raggedleft}MgB$_{12}$H$_{12}$[H$_{2}$O]$_{12}$: \end{raggedleft} \\ {\small A12B36CD12\_cF488\_196\_2h\_6h\_ac\_fgh} \dotfill {\hyperref[A12B36CD12_cF488_196_2h_6h_ac_fgh]{\pageref{A12B36CD12_cF488_196_2h_6h_ac_fgh}}} \\
\vspace{-0.75cm} \item MgSO$_{4}$: {\small AB4C\_oC24\_63\_a\_fg\_c} \dotfill {\hyperref[AB4C_oC24_63_a_fg_c]{\pageref{AB4C_oC24_63_a_fg_c}}} \\
\vspace{-0.75cm} \item Mg[NH]: {\small ABC\_hP36\_175\_jk\_jk\_jk} \dotfill {\hyperref[ABC_hP36_175_jk_jk_jk]{\pageref{ABC_hP36_175_jk_jk_jk}}} \\
\vspace{-0.75cm} \item Mn$_{2}$B: {\small AB2\_oF48\_70\_f\_fg} \dotfill {\hyperref[AB2_oF48_70_f_fg]{\pageref{AB2_oF48_70_f_fg}}} \\
\vspace{-0.75cm} \item MnAl$_{6}$: {\small A6B\_oC28\_63\_efg\_c} \dotfill {\hyperref[A6B_oC28_63_efg_c]{\pageref{A6B_oC28_63_efg_c}}} \\
\vspace{-0.75cm} \item MnF$_{2}$: {\small A2B\_tP12\_111\_2n\_adf} \dotfill {\hyperref[A2B_tP12_111_2n_adf]{\pageref{A2B_tP12_111_2n_adf}}} \\
\vspace{-0.75cm} \item MnGa$_{2}$Sb$_{2}$: {\small A2BC2\_oI20\_45\_c\_b\_c} \dotfill {\hyperref[A2BC2_oI20_45_c_b_c]{\pageref{A2BC2_oI20_45_c_b_c}}} \\
\vspace{-0.75cm} \item Mo$_{8}$P$_{5}$: {\small A8B5\_mP13\_6\_a7b\_3a2b} \dotfill {\hyperref[A8B5_mP13_6_a7b_3a2b]{\pageref{A8B5_mP13_6_a7b_3a2b}}} \\
\vspace{-0.75cm} \item MoS$_{2}$: {\small AB2\_hP12\_143\_cd\_ab2d} \dotfill {\hyperref[AB2_hP12_143_cd_ab2d]{\pageref{AB2_hP12_143_cd_ab2d}}} \\
\vspace{-0.75cm} \item Moissanite-15R: {\small AB\_hR10\_160\_5a\_5a} \dotfill {\hyperref[AB_hR10_160_5a_5a]{\pageref{AB_hR10_160_5a_5a}}} \\
\vspace{-0.75cm} \item Molybdite: {\small AB3\_oP16\_62\_c\_3c} \dotfill {\hyperref[AB3_oP16_62_c_3c]{\pageref{AB3_oP16_62_c_3c}}} \\
\vspace{-0.75cm} \item Muthmannite: {\small ABC2\_mP8\_10\_ac\_eh\_mn} \dotfill {\hyperref[ABC2_mP8_10_ac_eh_mn]{\pageref{ABC2_mP8_10_ac_eh_mn}}} \\
\vspace{-0.75cm} \item NV: {\small AB\_tP8\_111\_n\_n} \dotfill {\hyperref[AB_tP8_111_n_n]{\pageref{AB_tP8_111_n_n}}} \\
\vspace{-0.75cm} \item Na$_{3}$As\footnoteref{note:AB3_hP24_185_c_ab2c-prototype}: {\small AB3\_hP24\_185\_c\_ab2c} \dotfill {\hyperref[AB3_hP24_185_c_ab2c]{\pageref{AB3_hP24_185_c_ab2c}}} \\
\vspace{-0.75cm} \item \begin{raggedleft}Na$_{4}$Ti$_{2}$Si$_{8}$O$_{22}$[H$_{2}$O]$_{4}$: \end{raggedleft} \\ {\small A4B2C13D\_tP40\_90\_g\_d\_cef2g\_c} \dotfill {\hyperref[A4B2C13D_tP40_90_g_d_cef2g_c]{\pageref{A4B2C13D_tP40_90_g_d_cef2g_c}}} \\
\vspace{-0.75cm} \item Na$_{5}$Fe$_{3}$F$_{14}$: {\small A14B3C5\_tP44\_94\_c3g\_ad\_bg} \dotfill {\hyperref[A14B3C5_tP44_94_c3g_ad_bg]{\pageref{A14B3C5_tP44_94_c3g_ad_bg}}} \\
\vspace{-0.75cm} \item NaFeS$_{2}$: {\small ABC2\_oI16\_23\_ab\_i\_k} \dotfill {\hyperref[ABC2_oI16_23_ab_i_k]{\pageref{ABC2_oI16_23_ab_i_k}}} \\
\vspace{-0.75cm} \item NaGdCu$_{2}$F$_{8}$: {\small A2B8CD\_tI24\_97\_d\_k\_a\_b} \dotfill {\hyperref[A2B8CD_tI24_97_d_k_a_b]{\pageref{A2B8CD_tI24_97_d_k_a_b}}} \\
\vspace{-0.75cm} \item NaZn$_{13}$: {\small AB13\_cF112\_226\_a\_bi} \dotfill {\hyperref[AB13_cF112_226_a_bi]{\pageref{AB13_cF112_226_a_bi}}} \\
\vspace{-0.75cm} \item NaZn[OH]$_{3}$: {\small A3BC3D\_tP64\_106\_3c\_c\_3c\_c} \dotfill {\hyperref[A3BC3D_tP64_106_3c_c_3c_c]{\pageref{A3BC3D_tP64_106_3c_c_3c_c}}} \\
\vspace{-0.75cm} \item Nb$_{4}$CoSi: {\small AB4C\_tP12\_124\_a\_m\_c} \dotfill {\hyperref[AB4C_tP12_124_a_m_c]{\pageref{AB4C_tP12_124_a_m_c}}} \\
\vspace{-0.75cm} \item Nb$_{7}$Ru$_{6}$B$_{8}$: {\small A8B7C6\_hP21\_175\_ck\_aj\_k} \dotfill {\hyperref[A8B7C6_hP21_175_ck_aj_k]{\pageref{A8B7C6_hP21_175_ck_aj_k}}} \\
\vspace{-0.75cm} \item NbAs: {\small AB\_tI8\_109\_a\_a} \dotfill {\hyperref[AB_tI8_109_a_a]{\pageref{AB_tI8_109_a_a}}} \\
\vspace{-0.75cm} \item NbPS: {\small ABC\_oI12\_71\_h\_j\_g} \dotfill {\hyperref[ABC_oI12_71_h_j_g]{\pageref{ABC_oI12_71_h_j_g}}} \\
\vspace{-0.75cm} \item NbTe$_{4}$: {\small AB4\_tP10\_103\_a\_d} \dotfill {\hyperref[AB4_tP10_103_a_d]{\pageref{AB4_tP10_103_a_d}}} \\
\vspace{-0.75cm} \item NbTe$_{4}$: {\small AB4\_tP10\_124\_a\_m} \dotfill {\hyperref[AB4_tP10_124_a_m]{\pageref{AB4_tP10_124_a_m}}} \\
\vspace{-0.75cm} \item Ni$_{3}$P: {\small A3B\_tI32\_82\_3g\_g} \dotfill {\hyperref[A3B_tI32_82_3g_g]{\pageref{A3B_tI32_82_3g_g}}} \\
\vspace{-0.75cm} \item Ni$_{3}$Ti: {\small A3B\_hP16\_194\_gh\_ac} \dotfill {\hyperref[A3B_hP16_194_gh_ac]{\pageref{A3B_hP16_194_gh_ac}}} \\
\vspace{-0.75cm} \item Nierite: {\small A4B3\_hP28\_159\_ab2c\_2c} \dotfill {\hyperref[A4B3_hP28_159_ab2c_2c]{\pageref{A4B3_hP28_159_ab2c_2c}}} \\
\vspace{-0.75cm} \item PH$_{3}$: {\small A3B\_cP16\_208\_j\_b} \dotfill {\hyperref[A3B_cP16_208_j_b]{\pageref{A3B_cP16_208_j_b}}} \\
\vspace{-0.75cm} \item PI$_{3}$: {\small A3B\_hP8\_173\_c\_b} \dotfill {\hyperref[A3B_hP8_173_c_b]{\pageref{A3B_hP8_173_c_b}}} \\
\vspace{-0.75cm} \item Pd$_{17}$Se$_{15}$: {\small A17B15\_cP64\_207\_acfk\_eij} \dotfill {\hyperref[A17B15_cP64_207_acfk_eij]{\pageref{A17B15_cP64_207_acfk_eij}}} \\
\vspace{-0.75cm} \item Pd$_{4}$Se: {\small A4B\_tP10\_114\_e\_a} \dotfill {\hyperref[A4B_tP10_114_e_a]{\pageref{A4B_tP10_114_e_a}}} \\
\vspace{-0.75cm} \item PdSn$_{4}$: {\small AB4\_oC20\_68\_a\_i} \dotfill {\hyperref[AB4_oC20_68_a_i]{\pageref{AB4_oC20_68_a_i}}} \\
\vspace{-0.75cm} \item Petzite: {\small A3BC2\_cI48\_214\_f\_a\_e} \dotfill {\hyperref[A3BC2_cI48_214_f_a_e]{\pageref{A3BC2_cI48_214_f_a_e}}} \\
\vspace{-0.75cm} \item Phenakite: {\small A2B4C\_hR42\_148\_2f\_4f\_f} \dotfill {\hyperref[A2B4C_hR42_148_2f_4f_f]{\pageref{A2B4C_hR42_148_2f_4f_f}}} \\
\vspace{-0.75cm} \item Pinnoite: {\small A2B6CD7\_tP64\_77\_2d\_6d\_d\_ab6d} \dotfill {\hyperref[A2B6CD7_tP64_77_2d_6d_d_ab6d]{\pageref{A2B6CD7_tP64_77_2d_6d_d_ab6d}}} \\
\vspace{-0.75cm} \item Post-perovskite: {\small AB3C\_oC20\_63\_a\_cf\_c} \dotfill {\hyperref[AB3C_oC20_63_a_cf_c]{\pageref{AB3C_oC20_63_a_cf_c}}} \\
\vspace{-0.75cm} \item PrNiO$_{3}$: {\small AB3C\_hR10\_167\_b\_e\_a} \dotfill {\hyperref[AB3C_hR10_167_b_e_a]{\pageref{AB3C_hR10_167_b_e_a}}} \\
\vspace{-0.75cm} \item PrRu$_{4}$P$_{12}$: {\small A12BC4\_cP34\_195\_2j\_ab\_2e} \dotfill {\hyperref[A12BC4_cP34_195_2j_ab_2e]{\pageref{A12BC4_cP34_195_2j_ab_2e}}} \\
\vspace{-0.75cm} \item PtPb$_{4}$: {\small A4B\_tP10\_125\_m\_a} \dotfill {\hyperref[A4B_tP10_125_m_a]{\pageref{A4B_tP10_125_m_a}}} \\
\vspace{-0.75cm} \item Pyrite\footnote[3]{\label{note:AB2_oP12_29_a_2a-prototype}ZrO$_{2}$ and Pyrite have similar \AFLOW\ prototype labels ({\it{i.e.}}, same symmetry and set of Wyckoff positions with different stoichiometry labels due to alphabetic ordering of atomic species). They are generated by the same symmetry operations with different sets of parameters.}: {\small AB2\_oP12\_29\_a\_2a} \dotfill {\hyperref[AB2_oP12_29_a_2a]{\pageref{AB2_oP12_29_a_2a}}} \\
\vspace{-0.75cm} \item Pyrochlore: {\small A2BCD3E6\_cF208\_203\_e\_c\_d\_f\_g} \dotfill {\hyperref[A2BCD3E6_cF208_203_e_c_d_f_g]{\pageref{A2BCD3E6_cF208_203_e_c_d_f_g}}} \\
\vspace{-0.75cm} \item \begin{raggedleft}Pyrochlore Iridate: \end{raggedleft} \\ {\small A2B2C7\_cF88\_227\_c\_d\_af} \dotfill {\hyperref[A2B2C7_cF88_227_c_d_af]{\pageref{A2B2C7_cF88_227_c_d_af}}} \\
\vspace{-0.75cm} \item \begin{raggedleft}Quartenary Heusler: \end{raggedleft} \\ {\small ABCD\_cF16\_216\_c\_d\_b\_a} \dotfill {\hyperref[ABCD_cF16_216_c_d_b_a]{\pageref{ABCD_cF16_216_c_d_b_a}}} \\
\vspace{-0.75cm} \item R-carbon: {\small A\_oP16\_55\_2g2h} \dotfill {\hyperref[A_oP16_55_2g2h]{\pageref{A_oP16_55_2g2h}}} \\
\vspace{-0.75cm} \item Rasvumite: {\small A2BC3\_oC24\_63\_e\_c\_cg} \dotfill {\hyperref[A2BC3_oC24_63_e_c_cg]{\pageref{A2BC3_oC24_63_e_c_cg}}} \\
\vspace{-0.75cm} \item Rb$_{2}$TiCu$_{2}$S$_{4}$: {\small A2B2C4D\_tP18\_132\_e\_i\_o\_d} \dotfill {\hyperref[A2B2C4D_tP18_132_e_i_o_d]{\pageref{A2B2C4D_tP18_132_e_i_o_d}}} \\
\vspace{-0.75cm} \item Rb$_{3}$AsSe$_{16}$: {\small AB3C16\_cF160\_203\_b\_ad\_eg} \dotfill {\hyperref[AB3C16_cF160_203_b_ad_eg]{\pageref{AB3C16_cF160_203_b_ad_eg}}} \\
\vspace{-0.75cm} \item RbGa$_{3}$: {\small A3B\_tI24\_119\_b2i\_af} \dotfill {\hyperref[A3B_tI24_119_b2i_af]{\pageref{A3B_tI24_119_b2i_af}}} \\
\vspace{-0.75cm} \item Re$_{2}$O$_{5}$[SO$_{4}$]$_{2}$: {\small A13B2C2\_oP34\_32\_a6c\_c\_c} \dotfill {\hyperref[A13B2C2_oP34_32_a6c_c_c]{\pageref{A13B2C2_oP34_32_a6c_c_c}}} \\
\vspace{-0.75cm} \item Re$_{3}$N: {\small AB3\_hP4\_187\_e\_fh} \dotfill {\hyperref[AB3_hP4_187_e_fh]{\pageref{AB3_hP4_187_e_fh}}} \\
\vspace{-0.75cm} \item Rh$_{2}$Ga$_{9}$: {\small A9B2\_mP22\_7\_9a\_2a} \dotfill {\hyperref[A9B2_mP22_7_9a_2a]{\pageref{A9B2_mP22_7_9a_2a}}} \\
\vspace{-0.75cm} \item Rh$_{2}$S$_{3}$: {\small A2B3\_oP20\_60\_d\_cd} \dotfill {\hyperref[A2B3_oP20_60_d_cd]{\pageref{A2B3_oP20_60_d_cd}}} \\
\vspace{-0.75cm} \item Rh$_{3}$P$_{2}$: {\small A2B3\_tP5\_115\_g\_ag} \dotfill {\hyperref[A2B3_tP5_115_g_ag]{\pageref{A2B3_tP5_115_g_ag}}} \\
\vspace{-0.75cm} \item Rh$_{5}$Ge$_{3}$: {\small A3B5\_oP16\_55\_ch\_agh} \dotfill {\hyperref[A3B5_oP16_55_ch_agh]{\pageref{A3B5_oP16_55_ch_agh}}} \\
\vspace{-0.75cm} \item Ru$_{2}$Sn$_{3}$: {\small A2B3\_tP20\_116\_bci\_fj} \dotfill {\hyperref[A2B3_tP20_116_bci_fj]{\pageref{A2B3_tP20_116_bci_fj}}} \\
\vspace{-0.75cm} \item RuIn$_{3}$: {\small A3B\_tP16\_118\_ei\_f} \dotfill {\hyperref[A3B_tP16_118_ei_f]{\pageref{A3B_tP16_118_ei_f}}} \\
\vspace{-0.75cm} \item S-II: {\small A\_hP9\_154\_bc} \dotfill {\hyperref[A_hP9_154_bc]{\pageref{A_hP9_154_bc}}} \\
\vspace{-0.75cm} \item S-III: {\small A\_tI16\_142\_f} \dotfill {\hyperref[A_tI16_142_f]{\pageref{A_tI16_142_f}}} \\
\vspace{-0.75cm} \item S-carbon: {\small A\_mP8\_10\_2m2n} \dotfill {\hyperref[A_mP8_10_2m2n]{\pageref{A_mP8_10_2m2n}}} \\
\vspace{-0.75cm} \item Sc-V: {\small A\_hP6\_178\_a} \dotfill {\hyperref[A_hP6_178_a]{\pageref{A_hP6_178_a}}} \\
\vspace{-0.75cm} \item ScRh$_{6}$P$_{4}$: {\small A4B6C\_hP11\_143\_bd\_2d\_a} \dotfill {\hyperref[A4B6C_hP11_143_bd_2d_a]{\pageref{A4B6C_hP11_143_bd_2d_a}}} \\
\vspace{-0.75cm} \item SeO$_{3}$: {\small A3B\_tP32\_114\_3e\_e} \dotfill {\hyperref[A3B_tP32_114_3e_e]{\pageref{A3B_tP32_114_3e_e}}} \\
\vspace{-0.75cm} \item \begin{raggedleft}Sheldrickite: \end{raggedleft} \\ {\small A2B3C3DE7\_hP48\_145\_2a\_3a\_3a\_a\_7a} \dotfill {\hyperref[A2B3C3DE7_hP48_145_2a_3a_3a_a_7a]{\pageref{A2B3C3DE7_hP48_145_2a_3a_3a_a_7a}}} \\
\vspace{-0.75cm} \item SiO$_{2}$: {\small A2B\_hP36\_177\_j2lm\_n} \dotfill {\hyperref[A2B_hP36_177_j2lm_n]{\pageref{A2B_hP36_177_j2lm_n}}} \\
\vspace{-0.75cm} \item SiO$_{2}$: {\small A2B\_cI72\_211\_hi\_i} \dotfill {\hyperref[A2B_cI72_211_hi_i]{\pageref{A2B_cI72_211_hi_i}}} \\
\vspace{-0.75cm} \item \begin{raggedleft}Simple Cubic C$_{60}$ Buckminsterfullerine: \end{raggedleft} \\ {\small A\_cP240\_205\_10d} \dotfill {\hyperref[A_cP240_205_10d]{\pageref{A_cP240_205_10d}}} \\
\vspace{-0.75cm} \item Simpsonite: {\small A4B14C3\_hP21\_143\_bd\_ac4d\_d} \dotfill {\hyperref[A4B14C3_hP21_143_bd_ac4d_d]{\pageref{A4B14C3_hP21_143_bd_ac4d_d}}} \\
\vspace{-0.75cm} \item SmSI: {\small ABC\_hR6\_166\_c\_c\_c} \dotfill {\hyperref[ABC_hR6_166_c_c_c]{\pageref{ABC_hR6_166_c_c_c}}} \\
\vspace{-0.75cm} \item Sodium Chlorate: {\small ABC3\_cP20\_198\_a\_a\_b} \dotfill {\hyperref[ABC3_cP20_198_a_a_b]{\pageref{ABC3_cP20_198_a_a_b}}} \\
\vspace{-0.75cm} \item Spinel: {\small A3B4\_cF56\_227\_ad\_e} \dotfill {\hyperref[A3B4_cF56_227_ad_e]{\pageref{A3B4_cF56_227_ad_e}}} \\
\vspace{-0.75cm} \item Sr$_{2}$As$_{2}$O$_{7}$: {\small A2B7C2\_tP88\_78\_4a\_14a\_4a} \dotfill {\hyperref[A2B7C2_tP88_78_4a_14a_4a]{\pageref{A2B7C2_tP88_78_4a_14a_4a}}} \\
\vspace{-0.75cm} \item Sr$_{2}$Bi$_{3}$: {\small A3B2\_oP20\_52\_de\_cd} \dotfill {\hyperref[A3B2_oP20_52_de_cd]{\pageref{A3B2_oP20_52_de_cd}}} \\
\vspace{-0.75cm} \item Sr$_{5}$Si$_{3}$: {\small A3B5\_tI32\_108\_ac\_a2c} \dotfill {\hyperref[A3B5_tI32_108_ac_a2c]{\pageref{A3B5_tI32_108_ac_a2c}}} \\
\vspace{-0.75cm} \item SrAl$_{2}$Se$_{4}$: {\small A2B4C\_oC28\_66\_l\_kl\_a} \dotfill {\hyperref[A2B4C_oC28_66_l_kl_a]{\pageref{A2B4C_oC28_66_l_kl_a}}} \\
\vspace{-0.75cm} \item SrBr$_{2}$: {\small A2B\_tP30\_85\_ab2g\_cg} \dotfill {\hyperref[A2B_tP30_85_ab2g_cg]{\pageref{A2B_tP30_85_ab2g_cg}}} \\
\vspace{-0.75cm} \item SrH$_{2}$: {\small A2B\_oP12\_62\_2c\_c} \dotfill {\hyperref[A2B_oP12_62_2c_c]{\pageref{A2B_oP12_62_2c_c}}} \\
\vspace{-0.75cm} \item SrSi$_{2}$: {\small A2B\_cP12\_212\_c\_a} \dotfill {\hyperref[A2B_cP12_212_c_a]{\pageref{A2B_cP12_212_c_a}}} \\
\vspace{-0.75cm} \item Sr[S$_{2}$O$_{6}$][H$_{2}$O]$_{4}$: {\small A10B2C\_hP39\_171\_5c\_c\_a} \dotfill {\hyperref[A10B2C_hP39_171_5c_c_a]{\pageref{A10B2C_hP39_171_5c_c_a}}} \\
\vspace{-0.75cm} \item Sr[S$_{2}$O$_{6}$][H$_{2}$O]$_{4}$: {\small A10B2C\_hP39\_172\_5c\_c\_a} \dotfill {\hyperref[A10B2C_hP39_172_5c_c_a]{\pageref{A10B2C_hP39_172_5c_c_a}}} \\
\vspace{-0.75cm} \item \begin{raggedleft}Stannoidite: \end{raggedleft} \\ {\small A8B2C12D2E\_oI50\_23\_bcfk\_i\_3k\_j\_a} \dotfill {\hyperref[A8B2C12D2E_oI50_23_bcfk_i_3k_j_a]{\pageref{A8B2C12D2E_oI50_23_bcfk_i_3k_j_a}}} \\
\vspace{-0.75cm} \item Ta$_{2}$H: {\small AB2\_oC6\_21\_a\_k} \dotfill {\hyperref[AB2_oC6_21_a_k]{\pageref{AB2_oC6_21_a_k}}} \\
\vspace{-0.75cm} \item Ta$_{2}$Se$_{8}$I: {\small AB8C2\_tI44\_97\_e\_2k\_cd} \dotfill {\hyperref[AB8C2_tI44_97_e_2k_cd]{\pageref{AB8C2_tI44_97_e_2k_cd}}} \\
\vspace{-0.75cm} \item Ta$_{3}$B$_{4}$: {\small A4B3\_oI14\_71\_gh\_cg} \dotfill {\hyperref[A4B3_oI14_71_gh_cg]{\pageref{A4B3_oI14_71_gh_cg}}} \\
\vspace{-0.75cm} \item Ta$_{3}$S$_{2}$: {\small A2B3\_oC40\_39\_2d\_2c2d} \dotfill {\hyperref[A2B3_oC40_39_2d_2c2d]{\pageref{A2B3_oC40_39_2d_2c2d}}} \\
\vspace{-0.75cm} \item TaNiTe$_{2}$: {\small ABC2\_oP16\_53\_h\_e\_gh} \dotfill {\hyperref[ABC2_oP16_53_h_e_gh]{\pageref{ABC2_oP16_53_h_e_gh}}} \\
\vspace{-0.75cm} \item TeO$_{6}$H$_{6}$: {\small A6B\_cF224\_228\_h\_c} \dotfill {\hyperref[A6B_cF224_228_h_c]{\pageref{A6B_cF224_228_h_c}}} \\
\vspace{-0.75cm} \item TeZn: {\small AB\_hP6\_144\_a\_a} \dotfill {\hyperref[AB_hP6_144_a_a]{\pageref{AB_hP6_144_a_a}}} \\
\vspace{-0.75cm} \item Te[OH]$_{6}$: {\small A12B6C\_cF608\_210\_4h\_2h\_e} \dotfill {\hyperref[A12B6C_cF608_210_4h_2h_e]{\pageref{A12B6C_cF608_210_4h_2h_e}}} \\
\vspace{-0.75cm} \item Th$_{3}$P$_{4}$: {\small A4B3\_cI28\_220\_c\_a} \dotfill {\hyperref[A4B3_cI28_220_c_a]{\pageref{A4B3_cI28_220_c_a}}} \\
\vspace{-0.75cm} \item Th$_{6}$Mn$_{23}$: {\small A23B6\_cF116\_225\_bd2f\_e} \dotfill {\hyperref[A23B6_cF116_225_bd2f_e]{\pageref{A23B6_cF116_225_bd2f_e}}} \\
\vspace{-0.75cm} \item ThB$_{4}$: {\small A4B\_tP20\_127\_ehj\_g} \dotfill {\hyperref[A4B_tP20_127_ehj_g]{\pageref{A4B_tP20_127_ehj_g}}} \\
\vspace{-0.75cm} \item ThBC: {\small ABC\_tP24\_91\_d\_d\_d} \dotfill {\hyperref[ABC_tP24_91_d_d_d]{\pageref{ABC_tP24_91_d_d_d}}} \\
\vspace{-0.75cm} \item ThBC: {\small ABC\_tP24\_95\_d\_d\_d} \dotfill {\hyperref[ABC_tP24_95_d_d_d]{\pageref{ABC_tP24_95_d_d_d}}} \\
\vspace{-0.75cm} \item ThCl$_{4}$: {\small A4B\_tI20\_88\_f\_a} \dotfill {\hyperref[A4B_tI20_88_f_a]{\pageref{A4B_tI20_88_f_a}}} \\
\vspace{-0.75cm} \item Thortveitite: {\small A7B2C2\_mC22\_12\_aij\_h\_i} \dotfill {\hyperref[A7B2C2_mC22_12_aij_h_i]{\pageref{A7B2C2_mC22_12_aij_h_i}}} \\
\vspace{-0.75cm} \item Ti$_{2}$Ge$_{3}$: {\small A3B2\_tP10\_83\_adk\_j} \dotfill {\hyperref[A3B2_tP10_83_adk_j]{\pageref{A3B2_tP10_83_adk_j}}} \\
\vspace{-0.75cm} \item Ti$_{3}$O: {\small AB3\_hP24\_149\_acgi\_3l} \dotfill {\hyperref[AB3_hP24_149_acgi_3l]{\pageref{AB3_hP24_149_acgi_3l}}} \\
\vspace{-0.75cm} \item Ti$_{3}$P: {\small AB3\_tP32\_86\_g\_3g} \dotfill {\hyperref[AB3_tP32_86_g_3g]{\pageref{AB3_tP32_86_g_3g}}} \\
\vspace{-0.75cm} \item TiAl$_{2}$Br$_{8}$: {\small A2B8C\_oP22\_34\_c\_4c\_a} \dotfill {\hyperref[A2B8C_oP22_34_c_4c_a]{\pageref{A2B8C_oP22_34_c_4c_a}}} \\
\vspace{-0.75cm} \item TiFeSi: {\small ABC\_oI36\_46\_ac\_bc\_3b} \dotfill {\hyperref[ABC_oI36_46_ac_bc_3b]{\pageref{ABC_oI36_46_ac_bc_3b}}} \\
\vspace{-0.75cm} \item Tl$_{4}$HgI$_{6}$: {\small AB6C4\_tP22\_104\_a\_2ac\_c} \dotfill {\hyperref[AB6C4_tP22_104_a_2ac_c]{\pageref{AB6C4_tP22_104_a_2ac_c}}} \\
\vspace{-0.75cm} \item TlP$_{5}$: {\small A5B\_oP24\_26\_3a3b2c\_ab} \dotfill {\hyperref[A5B_oP24_26_3a3b2c_ab]{\pageref{A5B_oP24_26_3a3b2c_ab}}} \\
\vspace{-0.75cm} \item TlZn$_{2}$Sb$_{2}$: {\small A2BC2\_tI20\_79\_c\_2a\_c} \dotfill {\hyperref[A2BC2_tI20_79_c_2a_c]{\pageref{A2BC2_tI20_79_c_2a_c}}} \\
\vspace{-0.75cm} \item Tongbaite: {\small A2B3\_oP20\_62\_2c\_3c} \dotfill {\hyperref[A2B3_oP20_62_2c_3c]{\pageref{A2B3_oP20_62_2c_3c}}} \\
\vspace{-0.75cm} \item Troilite: {\small AB\_hP24\_190\_i\_afh} \dotfill {\hyperref[AB_hP24_190_i_afh]{\pageref{AB_hP24_190_i_afh}}} \\
\vspace{-0.75cm} \item Tychite: {\small A4B2C6D16E\_cF232\_203\_e\_d\_f\_eg\_a} \dotfill {\hyperref[A4B2C6D16E_cF232_203_e_d_f_eg_a]{\pageref{A4B2C6D16E_cF232_203_e_d_f_eg_a}}} \\
\vspace{-0.75cm} \item UCl$_{3}$: {\small A3B\_hP8\_176\_h\_d} \dotfill {\hyperref[A3B_hP8_176_h_d]{\pageref{A3B_hP8_176_h_d}}} \\
\vspace{-0.75cm} \item V$_{2}$MoO$_{8}$: {\small AB8C2\_oC22\_35\_a\_ab3e\_e} \dotfill {\hyperref[AB8C2_oC22_35_a_ab3e_e]{\pageref{AB8C2_oC22_35_a_ab3e_e}}} \\
\vspace{-0.75cm} \item VPCl$_{9}$: {\small A9BC\_oC44\_39\_3c3d\_a\_c} \dotfill {\hyperref[A9BC_oC44_39_3c3d_a_c]{\pageref{A9BC_oC44_39_3c3d_a_c}}} \\
\vspace{-0.75cm} \item W$_{3}$O$_{10}$: {\small A10B3\_oF52\_42\_2abce\_ab} \dotfill {\hyperref[A10B3_oF52_42_2abce_ab]{\pageref{A10B3_oF52_42_2abce_ab}}} \\
\vspace{-0.75cm} \item W$_{5}$Si$_{3}$: {\small A3B5\_tI32\_140\_ah\_bk} \dotfill {\hyperref[A3B5_tI32_140_ah_bk]{\pageref{A3B5_tI32_140_ah_bk}}} \\
\vspace{-0.75cm} \item WO$_{3}$: {\small A3B\_oP32\_60\_3d\_d} \dotfill {\hyperref[A3B_oP32_60_3d_d]{\pageref{A3B_oP32_60_3d_d}}} \\
\vspace{-0.75cm} \item Weberite: {\small AB7CD2\_oI44\_24\_a\_b3d\_c\_ac} \dotfill {\hyperref[AB7CD2_oI44_24_a_b3d_c_ac]{\pageref{AB7CD2_oI44_24_a_b3d_c_ac}}} \\
\vspace{-0.75cm} \item Westerveldite: {\small AB\_oP8\_62\_c\_c} \dotfill {\hyperref[AB_oP8_62_c_c]{\pageref{AB_oP8_62_c_c}}} \\
\vspace{-0.75cm} \item YbBaCo$_{4}$O$_{7}$: {\small AB4C7D\_hP26\_159\_b\_ac\_a2c\_b} \dotfill {\hyperref[AB4C7D_hP26_159_b_ac_a2c_b]{\pageref{AB4C7D_hP26_159_b_ac_a2c_b}}} \\
\vspace{-0.75cm} \item Zn$_{3}$P$_{2}$: {\small A2B3\_tP40\_137\_cdf\_3g} \dotfill {\hyperref[A2B3_tP40_137_cdf_3g]{\pageref{A2B3_tP40_137_cdf_3g}}} \\
\vspace{-0.75cm} \item ZnSb$_{2}$O$_{4}$: {\small A4B2C\_tP28\_135\_gh\_h\_d} \dotfill {\hyperref[A4B2C_tP28_135_gh_h_d]{\pageref{A4B2C_tP28_135_gh_h_d}}} \\
\vspace{-0.75cm} \item ZrO$_{2}$\footnoteref{note:AB2_oP12_29_a_2a-prototype}: {\small A2B\_oP12\_29\_2a\_a} \dotfill {\hyperref[A2B_oP12_29_2a_a]{\pageref{A2B_oP12_29_2a_a}}} \\
\vspace{-0.75cm} \item ZrO$_{2}$\footnoteref{note:AB2_tP6_137_a_d-prototype}: {\small A2B\_tP6\_137\_d\_a} \dotfill {\hyperref[A2B_tP6_137_d_a]{\pageref{A2B_tP6_137_d_a}}} \\
\end{enumerate}
\section*{\label{sec:pearsonInd}Pearson Symbol Index}
\noindent
$\mathbf{aP}$ \textbf{\dotfill} \\
\begin{enumerate}
\vspace{-0.85cm} \item \textbf{ aP6} \vspace{-0.15cm} \\
\begin{enumerate}
\vspace{-0.65cm} \item H$_{2}$S: {\small A2B\_aP6\_2\_aei\_i} \dotfill {\hyperref[A2B_aP6_2_aei_i]{\pageref{A2B_aP6_2_aei_i}}} \\
\end{enumerate}
\end{enumerate}
\vspace{-0.75cm}
$\mathbf{cF}$ \textbf{\dotfill} \\
\begin{enumerate}
\vspace{-0.85cm} \item \textbf{ cF16} \vspace{-0.15cm} \\
\begin{enumerate}
\vspace{-0.65cm} \item \begin{raggedleft}Quartenary Heusler: \end{raggedleft} \\ {\small ABCD\_cF16\_216\_c\_d\_b\_a} \dotfill {\hyperref[ABCD_cF16_216_c_d_b_a]{\pageref{ABCD_cF16_216_c_d_b_a}}} \\
\end{enumerate}
\vspace{-0.85cm} \item \textbf{ cF36} \vspace{-0.15cm} \\
\begin{enumerate}
\vspace{-0.65cm} \item K$_{2}$PtCl$_{6}$: {\small A6B2C\_cF36\_225\_e\_c\_a} \dotfill {\hyperref[A6B2C_cF36_225_e_c_a]{\pageref{A6B2C_cF36_225_e_c_a}}} \\
\end{enumerate}
\vspace{-0.85cm} \item \textbf{ cF56} \vspace{-0.15cm} \\
\begin{enumerate}
\vspace{-0.65cm} \item Spinel: {\small A3B4\_cF56\_227\_ad\_e} \dotfill {\hyperref[A3B4_cF56_227_ad_e]{\pageref{A3B4_cF56_227_ad_e}}} \\
\end{enumerate}
\vspace{-0.85cm} \item \textbf{ cF60} \vspace{-0.15cm} \\
\begin{enumerate}
\vspace{-0.65cm} \item Cu$_{2}$Fe[CN]$_{6}$: {\small A12B2C\_cF60\_196\_h\_bc\_a} \dotfill {\hyperref[A12B2C_cF60_196_h_bc_a]{\pageref{A12B2C_cF60_196_h_bc_a}}} \\
\end{enumerate}
\vspace{-0.85cm} \item \textbf{ cF88} \vspace{-0.15cm} \\
\begin{enumerate}
\vspace{-0.65cm} \item \begin{raggedleft}Pyrochlore Iridate: \end{raggedleft} \\ {\small A2B2C7\_cF88\_227\_c\_d\_af} \dotfill {\hyperref[A2B2C7_cF88_227_c_d_af]{\pageref{A2B2C7_cF88_227_c_d_af}}} \\
\end{enumerate}
\vspace{-0.85cm} \item \textbf{ cF104} \vspace{-0.15cm} \\
\begin{enumerate}
\vspace{-0.65cm} \item KB$_{6}$H$_{6}$: {\small A6B6C\_cF104\_202\_h\_h\_c} \dotfill {\hyperref[A6B6C_cF104_202_h_h_c]{\pageref{A6B6C_cF104_202_h_h_c}}} \\
\vspace{-0.65cm} \item F$_{6}$KP: {\small A24BC\_cF104\_209\_j\_a\_b} \dotfill {\hyperref[A24BC_cF104_209_j_a_b]{\pageref{A24BC_cF104_209_j_a_b}}} \\
\end{enumerate}
\vspace{-0.85cm} \item \textbf{ cF112} \vspace{-0.15cm} \\
\begin{enumerate}
\vspace{-0.65cm} \item NaZn$_{13}$: {\small AB13\_cF112\_226\_a\_bi} \dotfill {\hyperref[AB13_cF112_226_a_bi]{\pageref{AB13_cF112_226_a_bi}}} \\
\end{enumerate}
\vspace{-0.85cm} \item \textbf{ cF116} \vspace{-0.15cm} \\
\begin{enumerate}
\vspace{-0.65cm} \item Th$_{6}$Mn$_{23}$: {\small A23B6\_cF116\_225\_bd2f\_e} \dotfill {\hyperref[A23B6_cF116_225_bd2f_e]{\pageref{A23B6_cF116_225_bd2f_e}}} \\
\end{enumerate}
\vspace{-0.85cm} \item \textbf{ cF160} \vspace{-0.15cm} \\
\begin{enumerate}
\vspace{-0.65cm} \item Rb$_{3}$AsSe$_{16}$: {\small AB3C16\_cF160\_203\_b\_ad\_eg} \dotfill {\hyperref[AB3C16_cF160_203_b_ad_eg]{\pageref{AB3C16_cF160_203_b_ad_eg}}} \\
\end{enumerate}
\vspace{-0.85cm} \item \textbf{ cF192} \vspace{-0.15cm} \\
\begin{enumerate}
\vspace{-0.65cm} \item \begin{raggedleft}Boracite: \end{raggedleft} \\ {\small A7BC3D13\_cF192\_219\_de\_b\_c\_ah} \dotfill {\hyperref[A7BC3D13_cF192_219_de_b_c_ah]{\pageref{A7BC3D13_cF192_219_de_b_c_ah}}} \\
\end{enumerate}
\vspace{-0.85cm} \item \textbf{ cF208} \vspace{-0.15cm} \\
\begin{enumerate}
\vspace{-0.65cm} \item \begin{raggedleft}Pyrochlore: \end{raggedleft} \\ {\small A2BCD3E6\_cF208\_203\_e\_c\_d\_f\_g} \dotfill {\hyperref[A2BCD3E6_cF208_203_e_c_d_f_g]{\pageref{A2BCD3E6_cF208_203_e_c_d_f_g}}} \\
\end{enumerate}
\vspace{-0.85cm} \item \textbf{ cF224} \vspace{-0.15cm} \\
\begin{enumerate}
\vspace{-0.65cm} \item TeO$_{6}$H$_{6}$: {\small A6B\_cF224\_228\_h\_c} \dotfill {\hyperref[A6B_cF224_228_h_c]{\pageref{A6B_cF224_228_h_c}}} \\
\end{enumerate}
\vspace{-0.85cm} \item \textbf{ cF232} \vspace{-0.15cm} \\
\begin{enumerate}
\vspace{-0.65cm} \item \begin{raggedleft}Tychite: \end{raggedleft} \\ {\small A4B2C6D16E\_cF232\_203\_e\_d\_f\_eg\_a} \dotfill {\hyperref[A4B2C6D16E_cF232_203_e_d_f_eg_a]{\pageref{A4B2C6D16E_cF232_203_e_d_f_eg_a}}} \\
\end{enumerate}
\vspace{-0.85cm} \item \textbf{ cF240} \vspace{-0.15cm} \\
\begin{enumerate}
\vspace{-0.65cm} \item \begin{raggedleft}FCC C$_{60}$ Buckminsterfullerine: \end{raggedleft} \\ {\small A\_cF240\_202\_h2i} \dotfill {\hyperref[A_cF240_202_h2i]{\pageref{A_cF240_202_h2i}}} \\
\end{enumerate}
\vspace{-0.85cm} \item \textbf{ cF416} \vspace{-0.15cm} \\
\begin{enumerate}
\vspace{-0.65cm} \item \begin{raggedleft}CuCrCl$_{5}$[NH$_{3}$]$_{6}$: \end{raggedleft} \\ {\small A5BCD6\_cF416\_228\_eg\_c\_b\_h} \dotfill {\hyperref[A5BCD6_cF416_228_eg_c_b_h]{\pageref{A5BCD6_cF416_228_eg_c_b_h}}} \\
\end{enumerate}
\vspace{-0.85cm} \item \textbf{ cF488} \vspace{-0.15cm} \\
\begin{enumerate}
\vspace{-0.65cm} \item \begin{raggedleft}MgB$_{12}$H$_{12}$[H$_{2}$O]$_{12}$: \end{raggedleft} \\ {\small A12B36CD12\_cF488\_196\_2h\_6h\_ac\_fgh} \dotfill {\hyperref[A12B36CD12_cF488_196_2h_6h_ac_fgh]{\pageref{A12B36CD12_cF488_196_2h_6h_ac_fgh}}} \\
\end{enumerate}
\vspace{-0.85cm} \item \textbf{ cF608} \vspace{-0.15cm} \\
\begin{enumerate}
\vspace{-0.65cm} \item Te[OH]$_{6}$: {\small A12B6C\_cF608\_210\_4h\_2h\_e} \dotfill {\hyperref[A12B6C_cF608_210_4h_2h_e]{\pageref{A12B6C_cF608_210_4h_2h_e}}} \\
\end{enumerate}
\end{enumerate}
\vspace{-0.75cm}
$\mathbf{cI}$ \textbf{\dotfill} \\
\begin{enumerate}
\vspace{-0.85cm} \item \textbf{ cI10} \vspace{-0.15cm} \\
\begin{enumerate}
\vspace{-0.65cm} \item $\beta$-Hg$_{4}$Pt: {\small A4B\_cI10\_229\_c\_a} \dotfill {\hyperref[A4B_cI10_229_c_a]{\pageref{A4B_cI10_229_c_a}}} \\
\end{enumerate}
\vspace{-0.85cm} \item \textbf{ cI28} \vspace{-0.15cm} \\
\begin{enumerate}
\vspace{-0.65cm} \item Th$_{3}$P$_{4}$: {\small A4B3\_cI28\_220\_c\_a} \dotfill {\hyperref[A4B3_cI28_220_c_a]{\pageref{A4B3_cI28_220_c_a}}} \\
\end{enumerate}
\vspace{-0.85cm} \item \textbf{ cI40} \vspace{-0.15cm} \\
\begin{enumerate}
\vspace{-0.65cm} \item Ir$_{3}$Ge$_{7}$: {\small A7B3\_cI40\_229\_df\_e} \dotfill {\hyperref[A7B3_cI40_229_df_e]{\pageref{A7B3_cI40_229_df_e}}} \\
\end{enumerate}
\vspace{-0.85cm} \item \textbf{ cI48} \vspace{-0.15cm} \\
\begin{enumerate}
\vspace{-0.65cm} \item Petzite: {\small A3BC2\_cI48\_214\_f\_a\_e} \dotfill {\hyperref[A3BC2_cI48_214_f_a_e]{\pageref{A3BC2_cI48_214_f_a_e}}} \\
\end{enumerate}
\vspace{-0.85cm} \item \textbf{ cI52} \vspace{-0.15cm} \\
\begin{enumerate}
\vspace{-0.65cm} \item $\gamma$-brass: {\small A3B10\_cI52\_229\_e\_fh} \dotfill {\hyperref[A3B10_cI52_229_e_fh]{\pageref{A3B10_cI52_229_e_fh}}} \\
\end{enumerate}
\vspace{-0.85cm} \item \textbf{ cI56} \vspace{-0.15cm} \\
\begin{enumerate}
\vspace{-0.65cm} \item Ca$_{3}$PI$_{3}$: {\small A3B3C\_cI56\_214\_g\_h\_a} \dotfill {\hyperref[A3B3C_cI56_214_g_h_a]{\pageref{A3B3C_cI56_214_g_h_a}}} \\
\end{enumerate}
\vspace{-0.85cm} \item \textbf{ cI72} \vspace{-0.15cm} \\
\begin{enumerate}
\vspace{-0.65cm} \item SiO$_{2}$: {\small A2B\_cI72\_211\_hi\_i} \dotfill {\hyperref[A2B_cI72_211_hi_i]{\pageref{A2B_cI72_211_hi_i}}} \\
\end{enumerate}
\vspace{-0.85cm} \item \textbf{ cI76} \vspace{-0.15cm} \\
\begin{enumerate}
\vspace{-0.65cm} \item Cu$_{15}$Si$_{4}$: {\small A15B4\_cI76\_220\_ae\_c} \dotfill {\hyperref[A15B4_cI76_220_ae_c]{\pageref{A15B4_cI76_220_ae_c}}} \\
\end{enumerate}
\vspace{-0.85cm} \item \textbf{ cI96} \vspace{-0.15cm} \\
\begin{enumerate}
\vspace{-0.65cm} \item AlLi$_{3}$N$_{2}$: {\small AB3C2\_cI96\_206\_c\_e\_ad} \dotfill {\hyperref[AB3C2_cI96_206_c_e_ad]{\pageref{AB3C2_cI96_206_c_e_ad}}} \\
\end{enumerate}
\vspace{-0.85cm} \item \textbf{ cI160} \vspace{-0.15cm} \\
\begin{enumerate}
\vspace{-0.65cm} \item Garnet: {\small A2B3C12D3\_cI160\_230\_a\_c\_h\_d} \dotfill {\hyperref[A2B3C12D3_cI160_230_a_c_h_d]{\pageref{A2B3C12D3_cI160_230_a_c_h_d}}} \\
\end{enumerate}
\end{enumerate}
\vspace{-0.75cm}
$\mathbf{cP}$ \textbf{\dotfill} \\
\begin{enumerate}
\vspace{-0.85cm} \item \textbf{ cP12} \vspace{-0.15cm} \\
\begin{enumerate}
\vspace{-0.65cm} \item SrSi$_{2}$: {\small A2B\_cP12\_212\_c\_a} \dotfill {\hyperref[A2B_cP12_212_c_a]{\pageref{A2B_cP12_212_c_a}}} \\
\end{enumerate}
\vspace{-0.85cm} \item \textbf{ cP16} \vspace{-0.15cm} \\
\begin{enumerate}
\vspace{-0.65cm} \item PH$_{3}$: {\small A3B\_cP16\_208\_j\_b} \dotfill {\hyperref[A3B_cP16_208_j_b]{\pageref{A3B_cP16_208_j_b}}} \\
\vspace{-0.65cm} \item Ag$_{3}$[PO$_{4}$]: {\small A3B4C\_cP16\_218\_c\_e\_a} \dotfill {\hyperref[A3B4C_cP16_218_c_e_a]{\pageref{A3B4C_cP16_218_c_e_a}}} \\
\end{enumerate}
\vspace{-0.85cm} \item \textbf{ cP20} \vspace{-0.15cm} \\
\begin{enumerate}
\vspace{-0.65cm} \item Sodium Chlorate: {\small ABC3\_cP20\_198\_a\_a\_b} \dotfill {\hyperref[ABC3_cP20_198_a_a_b]{\pageref{ABC3_cP20_198_a_a_b}}} \\
\end{enumerate}
\vspace{-0.85cm} \item \textbf{ cP33} \vspace{-0.15cm} \\
\begin{enumerate}
\vspace{-0.65cm} \item Ca$_{3}$Al$_{2}$O$_{6}$: {\small A2B3C6\_cP33\_221\_cd\_ag\_fh} \dotfill {\hyperref[A2B3C6_cP33_221_cd_ag_fh]{\pageref{A2B3C6_cP33_221_cd_ag_fh}}} \\
\end{enumerate}
\vspace{-0.85cm} \item \textbf{ cP34} \vspace{-0.15cm} \\
\begin{enumerate}
\vspace{-0.65cm} \item PrRu$_{4}$P$_{12}$: {\small A12BC4\_cP34\_195\_2j\_ab\_2e} \dotfill {\hyperref[A12BC4_cP34_195_2j_ab_2e]{\pageref{A12BC4_cP34_195_2j_ab_2e}}} \\
\end{enumerate}
\vspace{-0.85cm} \item \textbf{ cP39} \vspace{-0.15cm} \\
\begin{enumerate}
\vspace{-0.65cm} \item Mg$_{2}$Zn$_{11}$: {\small A2B11\_cP39\_200\_f\_aghij} \dotfill {\hyperref[A2B11_cP39_200_f_aghij]{\pageref{A2B11_cP39_200_f_aghij}}} \\
\end{enumerate}
\vspace{-0.85cm} \item \textbf{ cP52} \vspace{-0.15cm} \\
\begin{enumerate}
\vspace{-0.65cm} \item $\gamma$-brass: {\small A4B9\_cP52\_215\_ei\_3efgi} \dotfill {\hyperref[A4B9_cP52_215_ei_3efgi]{\pageref{A4B9_cP52_215_ei_3efgi}}} \\
\end{enumerate}
\vspace{-0.85cm} \item \textbf{ cP60} \vspace{-0.15cm} \\
\begin{enumerate}
\vspace{-0.65cm} \item KSbO$_{3}$: {\small AB3C\_cP60\_201\_ce\_fh\_g} \dotfill {\hyperref[AB3C_cP60_201_ce_fh_g]{\pageref{AB3C_cP60_201_ce_fh_g}}} \\
\end{enumerate}
\vspace{-0.85cm} \item \textbf{ cP64} \vspace{-0.15cm} \\
\begin{enumerate}
\vspace{-0.65cm} \item Pd$_{17}$Se$_{15}$: {\small A17B15\_cP64\_207\_acfk\_eij} \dotfill {\hyperref[A17B15_cP64_207_acfk_eij]{\pageref{A17B15_cP64_207_acfk_eij}}} \\
\vspace{-0.65cm} \item \begin{raggedleft}Cs$_{2}$ZnFe[CN]$_{6}$: \end{raggedleft} \\ {\small A6B2CD6E\_cP64\_208\_m\_ad\_b\_m\_c} \dotfill {\hyperref[A6B2CD6E_cP64_208_m_ad_b_m_c]{\pageref{A6B2CD6E_cP64_208_m_ad_b_m_c}}} \\
\end{enumerate}
\vspace{-0.85cm} \item \textbf{ cP96} \vspace{-0.15cm} \\
\begin{enumerate}
\vspace{-0.65cm} \item Ce$_{5}$Mo$_{3}$O$_{16}$: {\small A5B3C16\_cP96\_222\_ce\_d\_fi} \dotfill {\hyperref[A5B3C16_cP96_222_ce_d_fi]{\pageref{A5B3C16_cP96_222_ce_d_fi}}} \\
\end{enumerate}
\vspace{-0.85cm} \item \textbf{ cP240} \vspace{-0.15cm} \\
\begin{enumerate}
\vspace{-0.65cm} \item \begin{raggedleft}Simple Cubic C$_{60}$ Buckminsterfullerine: \end{raggedleft} \\ {\small A\_cP240\_205\_10d} \dotfill {\hyperref[A_cP240_205_10d]{\pageref{A_cP240_205_10d}}} \\
\end{enumerate}
\vspace{-0.85cm} \item \textbf{ cP264} \vspace{-0.15cm} \\
\begin{enumerate}
\vspace{-0.65cm} \item \begin{raggedleft}Ca$_{3}$Al$_{2}$O$_{6}$: \end{raggedleft} \\ {\small A2B3C6\_cP264\_205\_2d\_ab2c2d\_6d} \dotfill {\hyperref[A2B3C6_cP264_205_2d_ab2c2d_6d]{\pageref{A2B3C6_cP264_205_2d_ab2c2d_6d}}} \\
\end{enumerate}
\end{enumerate}
\vspace{-0.75cm}
$\mathbf{hP}$ \textbf{\dotfill} \\
\begin{enumerate}
\vspace{-0.85cm} \item \textbf{ hP3} \vspace{-0.15cm} \\
\begin{enumerate}
\vspace{-0.65cm} \item AuCN: {\small ABC\_hP3\_183\_a\_a\_a} \dotfill {\hyperref[ABC_hP3_183_a_a_a]{\pageref{ABC_hP3_183_a_a_a}}} \\
\end{enumerate}
\vspace{-0.85cm} \item \textbf{ hP4} \vspace{-0.15cm} \\
\begin{enumerate}
\vspace{-0.65cm} \item $\beta$-CuI: {\small AB\_hP4\_156\_ac\_ac} \dotfill {\hyperref[AB_hP4_156_ac_ac]{\pageref{AB_hP4_156_ac_ac}}} \\
\vspace{-0.65cm} \item CuNiSb$_{2}$: {\small ABC2\_hP4\_164\_a\_b\_d} \dotfill {\hyperref[ABC2_hP4_164_a_b_d]{\pageref{ABC2_hP4_164_a_b_d}}} \\
\vspace{-0.65cm} \item Re$_{3}$N: {\small AB3\_hP4\_187\_e\_fh} \dotfill {\hyperref[AB3_hP4_187_e_fh]{\pageref{AB3_hP4_187_e_fh}}} \\
\end{enumerate}
\vspace{-0.85cm} \item \textbf{ hP5} \vspace{-0.15cm} \\
\begin{enumerate}
\vspace{-0.65cm} \item La$_{2}$O$_{3}$: {\small A2B3\_hP5\_164\_d\_ad} \dotfill {\hyperref[A2B3_hP5_164_d_ad]{\pageref{A2B3_hP5_164_d_ad}}} \\
\end{enumerate}
\vspace{-0.85cm} \item \textbf{ hP6} \vspace{-0.15cm} \\
\begin{enumerate}
\vspace{-0.65cm} \item TeZn: {\small AB\_hP6\_144\_a\_a} \dotfill {\hyperref[AB_hP6_144_a_a]{\pageref{AB_hP6_144_a_a}}} \\
\vspace{-0.65cm} \item Sc-V: {\small A\_hP6\_178\_a} \dotfill {\hyperref[A_hP6_178_a]{\pageref{A_hP6_178_a}}} \\
\vspace{-0.65cm} \item CrFe$_{3}$NiSn$_{5}$: {\small AB\_hP6\_183\_c\_ab} \dotfill {\hyperref[AB_hP6_183_c_ab]{\pageref{AB_hP6_183_c_ab}}} \\
\end{enumerate}
\vspace{-0.85cm} \item \textbf{ hP8} \vspace{-0.15cm} \\
\begin{enumerate}
\vspace{-0.65cm} \item $\beta$-RuCl$_{3}$: {\small A3B\_hP8\_158\_d\_a} \dotfill {\hyperref[A3B_hP8_158_d_a]{\pageref{A3B_hP8_158_d_a}}} \\
\vspace{-0.65cm} \item PI$_{3}$: {\small A3B\_hP8\_173\_c\_b} \dotfill {\hyperref[A3B_hP8_173_c_b]{\pageref{A3B_hP8_173_c_b}}} \\
\vspace{-0.65cm} \item UCl$_{3}$: {\small A3B\_hP8\_176\_h\_d} \dotfill {\hyperref[A3B_hP8_176_h_d]{\pageref{A3B_hP8_176_h_d}}} \\
\vspace{-0.65cm} \item $\beta$-RuCl$_{3}$: {\small A3B\_hP8\_185\_c\_a} \dotfill {\hyperref[A3B_hP8_185_c_a]{\pageref{A3B_hP8_185_c_a}}} \\
\end{enumerate}
\vspace{-0.85cm} \item \textbf{ hP9} \vspace{-0.15cm} \\
\begin{enumerate}
\vspace{-0.65cm} \item S-II: {\small A\_hP9\_154\_bc} \dotfill {\hyperref[A_hP9_154_bc]{\pageref{A_hP9_154_bc}}} \\
\vspace{-0.65cm} \item CdI$_{2}$: {\small AB2\_hP9\_156\_b2c\_3a2bc} \dotfill {\hyperref[AB2_hP9_156_b2c_3a2bc]{\pageref{AB2_hP9_156_b2c_3a2bc}}} \\
\vspace{-0.65cm} \item $\delta_{H}^{II}$-NW$_2$: {\small AB2\_hP9\_164\_bd\_c2d} \dotfill {\hyperref[AB2_hP9_164_bd_c2d]{\pageref{AB2_hP9_164_bd_c2d}}} \\
\vspace{-0.65cm} \item $\beta$-SiO$_{2}$: {\small A2B\_hP9\_181\_j\_c} \dotfill {\hyperref[A2B_hP9_181_j_c]{\pageref{A2B_hP9_181_j_c}}} \\
\end{enumerate}
\vspace{-0.85cm} \item \textbf{ hP10} \vspace{-0.15cm} \\
\begin{enumerate}
\vspace{-0.65cm} \item Er$_{3}$Ru$_{2}$: {\small A3B2\_hP10\_176\_h\_bd} \dotfill {\hyperref[A3B2_hP10_176_h_bd]{\pageref{A3B2_hP10_176_h_bd}}} \\
\vspace{-0.65cm} \item LiScI$_{3}$: {\small A3BC\_hP10\_188\_k\_a\_e} \dotfill {\hyperref[A3BC_hP10_188_k_a_e]{\pageref{A3BC_hP10_188_k_a_e}}} \\
\end{enumerate}
\vspace{-0.85cm} \item \textbf{ hP11} \vspace{-0.15cm} \\
\begin{enumerate}
\vspace{-0.65cm} \item ScRh$_{6}$P$_{4}$: {\small A4B6C\_hP11\_143\_bd\_2d\_a} \dotfill {\hyperref[A4B6C_hP11_143_bd_2d_a]{\pageref{A4B6C_hP11_143_bd_2d_a}}} \\
\end{enumerate}
\vspace{-0.85cm} \item \textbf{ hP12} \vspace{-0.15cm} \\
\begin{enumerate}
\vspace{-0.65cm} \item MoS$_{2}$: {\small AB2\_hP12\_143\_cd\_ab2d} \dotfill {\hyperref[AB2_hP12_143_cd_ab2d]{\pageref{AB2_hP12_143_cd_ab2d}}} \\
\vspace{-0.65cm} \item CuI: {\small AB\_hP12\_156\_2ab3c\_2ab3c} \dotfill {\hyperref[AB_hP12_156_2ab3c_2ab3c]{\pageref{AB_hP12_156_2ab3c_2ab3c}}} \\
\vspace{-0.65cm} \item GdSI: {\small ABC\_hP12\_174\_cj\_fk\_aj} \dotfill {\hyperref[ABC_hP12_174_cj_fk_aj]{\pageref{ABC_hP12_174_cj_fk_aj}}} \\
\end{enumerate}
\vspace{-0.85cm} \item \textbf{ hP13} \vspace{-0.15cm} \\
\begin{enumerate}
\vspace{-0.65cm} \item Ag$_{5}$Pb$_{2}$O$_{6}$: {\small A5B6C2\_hP13\_157\_2ac\_2c\_b} \dotfill {\hyperref[A5B6C2_hP13_157_2ac_2c_b]{\pageref{A5B6C2_hP13_157_2ac_2c_b}}} \\
\end{enumerate}
\vspace{-0.85cm} \item \textbf{ hP14} \vspace{-0.15cm} \\
\begin{enumerate}
\vspace{-0.65cm} \item $\beta$-Si$_{3}$N$_{4}$: {\small A4B3\_hP14\_173\_bc\_c} \dotfill {\hyperref[A4B3_hP14_173_bc_c]{\pageref{A4B3_hP14_173_bc_c}}} \\
\vspace{-0.65cm} \item Fe$_{3}$Te$_{3}$Tl: {\small A3B3C\_hP14\_176\_h\_h\_d} \dotfill {\hyperref[A3B3C_hP14_176_h_h_d]{\pageref{A3B3C_hP14_176_h_h_d}}} \\
\end{enumerate}
\vspace{-0.85cm} \item \textbf{ hP15} \vspace{-0.15cm} \\
\begin{enumerate}
\vspace{-0.65cm} \item IrGe$_{4}$: {\small A4B\_hP15\_144\_4a\_a} \dotfill {\hyperref[A4B_hP15_144_4a_a]{\pageref{A4B_hP15_144_4a_a}}} \\
\end{enumerate}
\vspace{-0.85cm} \item \textbf{ hP16} \vspace{-0.15cm} \\
\begin{enumerate}
\vspace{-0.65cm} \item $\alpha$-Sm$_{3}$Ge$_{5}$: {\small A5B3\_hP16\_190\_bdh\_g} \dotfill {\hyperref[A5B3_hP16_190_bdh_g]{\pageref{A5B3_hP16_190_bdh_g}}} \\
\vspace{-0.65cm} \item Mavlyanovite: {\small A5B3\_hP16\_193\_dg\_g} \dotfill {\hyperref[A5B3_hP16_193_dg_g]{\pageref{A5B3_hP16_193_dg_g}}} \\
\vspace{-0.65cm} \item Ni$_{3}$Ti: {\small A3B\_hP16\_194\_gh\_ac} \dotfill {\hyperref[A3B_hP16_194_gh_ac]{\pageref{A3B_hP16_194_gh_ac}}} \\
\end{enumerate}
\vspace{-0.85cm} \item \textbf{ hP18} \vspace{-0.15cm} \\
\begin{enumerate}
\vspace{-0.65cm} \item \begin{raggedleft}$\pi$-FeMg$_{3}$Al$_{8}$Si$_{6}$: \end{raggedleft} \\ {\small A8BC3D6\_hP18\_189\_bfh\_a\_g\_i} \dotfill {\hyperref[A8BC3D6_hP18_189_bfh_a_g_i]{\pageref{A8BC3D6_hP18_189_bfh_a_g_i}}} \\
\vspace{-0.65cm} \item \begin{raggedleft}$\pi$-FeMg$_{3}$Al$_{9}$Si$_{5}$: \end{raggedleft} \\ {\small A9BC3D5\_hP18\_189\_fi\_a\_g\_bh} \dotfill {\hyperref[A9BC3D5_hP18_189_fi_a_g_bh]{\pageref{A9BC3D5_hP18_189_fi_a_g_bh}}} \\
\vspace{-0.65cm} \item Li$_{2}$Sb: {\small A2B\_hP18\_190\_gh\_bf} \dotfill {\hyperref[A2B_hP18_190_gh_bf]{\pageref{A2B_hP18_190_gh_bf}}} \\
\end{enumerate}
\vspace{-0.85cm} \item \textbf{ hP20} \vspace{-0.15cm} \\
\begin{enumerate}
\vspace{-0.65cm} \item Bi$_{2}$O$_{3}$: {\small A2B3\_hP20\_159\_bc\_2c} \dotfill {\hyperref[A2B3_hP20_159_bc_2c]{\pageref{A2B3_hP20_159_bc_2c}}} \\
\vspace{-0.65cm} \item Fe$_{3}$Th$_{7}$: {\small A3B7\_hP20\_186\_c\_b2c} \dotfill {\hyperref[A3B7_hP20_186_c_b2c]{\pageref{A3B7_hP20_186_c_b2c}}} \\
\end{enumerate}
\vspace{-0.85cm} \item \textbf{ hP21} \vspace{-0.15cm} \\
\begin{enumerate}
\vspace{-0.65cm} \item \begin{raggedleft}Simpsonite: \end{raggedleft} \\ {\small A4B14C3\_hP21\_143\_bd\_ac4d\_d} \dotfill {\hyperref[A4B14C3_hP21_143_bd_ac4d_d]{\pageref{A4B14C3_hP21_143_bd_ac4d_d}}} \\
\vspace{-0.65cm} \item \begin{raggedleft}Fe$_{12}$Zr$_{2}$P$_{7}$: \end{raggedleft} \\ {\small A12B7C2\_hP21\_174\_2j2k\_ajk\_cf} \dotfill {\hyperref[A12B7C2_hP21_174_2j2k_ajk_cf]{\pageref{A12B7C2_hP21_174_2j2k_ajk_cf}}} \\
\vspace{-0.65cm} \item Nb$_{7}$Ru$_{6}$B$_{8}$: {\small A8B7C6\_hP21\_175\_ck\_aj\_k} \dotfill {\hyperref[A8B7C6_hP21_175_ck_aj_k]{\pageref{A8B7C6_hP21_175_ck_aj_k}}} \\
\end{enumerate}
\vspace{-0.85cm} \item \textbf{ hP24} \vspace{-0.15cm} \\
\begin{enumerate}
\vspace{-0.65cm} \item Ti$_{3}$O: {\small AB3\_hP24\_149\_acgi\_3l} \dotfill {\hyperref[AB3_hP24_149_acgi_3l]{\pageref{AB3_hP24_149_acgi_3l}}} \\
\vspace{-0.65cm} \item CrCl$_{3}$: {\small A3B\_hP24\_153\_3c\_2b} \dotfill {\hyperref[A3B_hP24_153_3c_2b]{\pageref{A3B_hP24_153_3c_2b}}} \\
\vspace{-0.65cm} \item Cu$_{3}$P: {\small A3B\_hP24\_165\_bdg\_f} \dotfill {\hyperref[A3B_hP24_165_bdg_f]{\pageref{A3B_hP24_165_bdg_f}}} \\
\vspace{-0.65cm} \item AuF$_{3}$: {\small AB3\_hP24\_178\_b\_ac} \dotfill {\hyperref[AB3_hP24_178_b_ac]{\pageref{AB3_hP24_178_b_ac}}} \\
\vspace{-0.65cm} \item AuF$_{3}$: {\small AB3\_hP24\_179\_b\_ac} \dotfill {\hyperref[AB3_hP24_179_b_ac]{\pageref{AB3_hP24_179_b_ac}}} \\
\vspace{-0.65cm} \item Cu$_{3}$P\footnote[6]{\label{note:AB3_hP24_185_c_ab2c-Pearson}Cu$_{3}$P and Na$_{3}$As have similar \AFLOW\ prototype labels ({\it{i.e.}}, same symmetry and set of Wyckoff positions with different stoichiometry labels due to alphabetic ordering of atomic species). They are generated by the same symmetry operations with different sets of parameters.}: {\small A3B\_hP24\_185\_ab2c\_c} \dotfill {\hyperref[A3B_hP24_185_ab2c_c]{\pageref{A3B_hP24_185_ab2c_c}}} \\
\vspace{-0.65cm} \item Na$_{3}$As\footnoteref{note:AB3_hP24_185_c_ab2c-Pearson}: {\small AB3\_hP24\_185\_c\_ab2c} \dotfill {\hyperref[AB3_hP24_185_c_ab2c]{\pageref{AB3_hP24_185_c_ab2c}}} \\
\vspace{-0.65cm} \item Troilite: {\small AB\_hP24\_190\_i\_afh} \dotfill {\hyperref[AB_hP24_190_i_afh]{\pageref{AB_hP24_190_i_afh}}} \\
\end{enumerate}
\vspace{-0.85cm} \item \textbf{ hP26} \vspace{-0.15cm} \\
\begin{enumerate}
\vspace{-0.65cm} \item YbBaCo$_{4}$O$_{7}$: {\small AB4C7D\_hP26\_159\_b\_ac\_a2c\_b} \dotfill {\hyperref[AB4C7D_hP26_159_b_ac_a2c_b]{\pageref{AB4C7D_hP26_159_b_ac_a2c_b}}} \\
\vspace{-0.65cm} \item Al$_{9}$Mn$_{3}$Si: {\small A9B3C\_hP26\_194\_hk\_h\_a} \dotfill {\hyperref[A9B3C_hP26_194_hk_h_a]{\pageref{A9B3C_hP26_194_hk_h_a}}} \\
\end{enumerate}
\vspace{-0.85cm} \item \textbf{ hP28} \vspace{-0.15cm} \\
\begin{enumerate}
\vspace{-0.65cm} \item Nierite: {\small A4B3\_hP28\_159\_ab2c\_2c} \dotfill {\hyperref[A4B3_hP28_159_ab2c_2c]{\pageref{A4B3_hP28_159_ab2c_2c}}} \\
\vspace{-0.65cm} \item BaSi$_{4}$O$_{9}$: {\small AB9C4\_hP28\_188\_e\_kl\_ak} \dotfill {\hyperref[AB9C4_hP28_188_e_kl_ak]{\pageref{AB9C4_hP28_188_e_kl_ak}}} \\
\vspace{-0.65cm} \item Co$_{2}$Al$_{5}$: {\small A5B2\_hP28\_194\_ahk\_ch} \dotfill {\hyperref[A5B2_hP28_194_ahk_ch]{\pageref{A5B2_hP28_194_ahk_ch}}} \\
\end{enumerate}
\vspace{-0.85cm} \item \textbf{ hP30} \vspace{-0.15cm} \\
\begin{enumerate}
\vspace{-0.65cm} \item $\alpha$-Al$_{2}$S$_{3}$: {\small A2B3\_hP30\_169\_2a\_3a} \dotfill {\hyperref[A2B3_hP30_169_2a_3a]{\pageref{A2B3_hP30_169_2a_3a}}} \\
\vspace{-0.65cm} \item Al$_{2}$S$_{3}$: {\small A2B3\_hP30\_170\_2a\_3a} \dotfill {\hyperref[A2B3_hP30_170_2a_3a]{\pageref{A2B3_hP30_170_2a_3a}}} \\
\vspace{-0.65cm} \item KNiCl$_{3}$: {\small A3BC\_hP30\_185\_cd\_c\_ab} \dotfill {\hyperref[A3BC_hP30_185_cd_c_ab]{\pageref{A3BC_hP30_185_cd_c_ab}}} \\
\end{enumerate}
\vspace{-0.85cm} \item \textbf{ hP36} \vspace{-0.15cm} \\
\begin{enumerate}
\vspace{-0.65cm} \item Mg[NH]: {\small ABC\_hP36\_175\_jk\_jk\_jk} \dotfill {\hyperref[ABC_hP36_175_jk_jk_jk]{\pageref{ABC_hP36_175_jk_jk_jk}}} \\
\vspace{-0.65cm} \item SiO$_{2}$: {\small A2B\_hP36\_177\_j2lm\_n} \dotfill {\hyperref[A2B_hP36_177_j2lm_n]{\pageref{A2B_hP36_177_j2lm_n}}} \\
\end{enumerate}
\vspace{-0.85cm} \item \textbf{ hP39} \vspace{-0.15cm} \\
\begin{enumerate}
\vspace{-0.65cm} \item \begin{raggedleft}Sr[S$_{2}$O$_{6}$][H$_{2}$O]$_{4}$: \end{raggedleft} \\ {\small A10B2C\_hP39\_171\_5c\_c\_a} \dotfill {\hyperref[A10B2C_hP39_171_5c_c_a]{\pageref{A10B2C_hP39_171_5c_c_a}}} \\
\vspace{-0.65cm} \item \begin{raggedleft}Sr[S$_{2}$O$_{6}$][H$_{2}$O]$_{4}$: \end{raggedleft} \\ {\small A10B2C\_hP39\_172\_5c\_c\_a} \dotfill {\hyperref[A10B2C_hP39_172_5c_c_a]{\pageref{A10B2C_hP39_172_5c_c_a}}} \\
\end{enumerate}
\vspace{-0.85cm} \item \textbf{ hP48} \vspace{-0.15cm} \\
\begin{enumerate}
\vspace{-0.65cm} \item \begin{raggedleft}Sheldrickite: \end{raggedleft} \\ {\small A2B3C3DE7\_hP48\_145\_2a\_3a\_3a\_a\_7a} \dotfill {\hyperref[A2B3C3DE7_hP48_145_2a_3a_3a_a_7a]{\pageref{A2B3C3DE7_hP48_145_2a_3a_3a_a_7a}}} \\
\end{enumerate}
\vspace{-0.85cm} \item \textbf{ hP57} \vspace{-0.15cm} \\
\begin{enumerate}
\vspace{-0.65cm} \item K$_{2}$Ta$_{4}$O$_{9}$F$_{4}$: {\small A2B13C4\_hP57\_168\_d\_c6d\_2d} \dotfill {\hyperref[A2B13C4_hP57_168_d_c6d_2d]{\pageref{A2B13C4_hP57_168_d_c6d_2d}}} \\
\end{enumerate}
\vspace{-0.85cm} \item \textbf{ hP58} \vspace{-0.15cm} \\
\begin{enumerate}
\vspace{-0.65cm} \item Beryl: {\small A2B3C18D6\_hP58\_192\_c\_f\_lm\_l} \dotfill {\hyperref[A2B3C18D6_hP58_192_c_f_lm_l]{\pageref{A2B3C18D6_hP58_192_c_f_lm_l}}} \\
\end{enumerate}
\vspace{-0.85cm} \item \textbf{ hP72} \vspace{-0.15cm} \\
\begin{enumerate}
\vspace{-0.65cm} \item Al[PO$_{4}$]: {\small AB4C\_hP72\_168\_2d\_8d\_2d} \dotfill {\hyperref[AB4C_hP72_168_2d_8d_2d]{\pageref{AB4C_hP72_168_2d_8d_2d}}} \\
\vspace{-0.65cm} \item Al[PO$_{4}$]: {\small AB4C\_hP72\_184\_d\_4d\_d} \dotfill {\hyperref[AB4C_hP72_184_d_4d_d]{\pageref{AB4C_hP72_184_d_4d_d}}} \\
\vspace{-0.65cm} \item AlPO$_{4}$: {\small AB2\_hP72\_192\_m\_j2kl} \dotfill {\hyperref[AB2_hP72_192_m_j2kl]{\pageref{AB2_hP72_192_m_j2kl}}} \\
\end{enumerate}
\end{enumerate}
\vspace{-0.75cm}
$\mathbf{hR}$ \textbf{\dotfill} \\
\begin{enumerate}
\vspace{-0.85cm} \item \textbf{ hR3} \vspace{-0.15cm} \\
\begin{enumerate}
\vspace{-0.65cm} \item Carbonyl Sulphide: {\small ABC\_hR3\_160\_a\_a\_a} \dotfill {\hyperref[ABC_hR3_160_a_a_a]{\pageref{ABC_hR3_160_a_a_a}}} \\
\end{enumerate}
\vspace{-0.85cm} \item \textbf{ hR4} \vspace{-0.15cm} \\
\begin{enumerate}
\vspace{-0.65cm} \item H$_{3}$S: {\small A3B\_hR4\_160\_b\_a} \dotfill {\hyperref[A3B_hR4_160_b_a]{\pageref{A3B_hR4_160_b_a}}} \\
\end{enumerate}
\vspace{-0.85cm} \item \textbf{ hR5} \vspace{-0.15cm} \\
\begin{enumerate}
\vspace{-0.65cm} \item $\gamma$-Ag$_{3}$SI: {\small A3BC\_hR5\_146\_b\_a\_a} \dotfill {\hyperref[A3BC_hR5_146_b_a_a]{\pageref{A3BC_hR5_146_b_a_a}}} \\
\end{enumerate}
\vspace{-0.85cm} \item \textbf{ hR6} \vspace{-0.15cm} \\
\begin{enumerate}
\vspace{-0.65cm} \item SmSI: {\small ABC\_hR6\_166\_c\_c\_c} \dotfill {\hyperref[ABC_hR6_166_c_c_c]{\pageref{ABC_hR6_166_c_c_c}}} \\
\end{enumerate}
\vspace{-0.85cm} \item \textbf{ hR7} \vspace{-0.15cm} \\
\begin{enumerate}
\vspace{-0.65cm} \item Al$_{4}$C$_{3}$: {\small A4B3\_hR7\_166\_2c\_ac} \dotfill {\hyperref[A4B3_hR7_166_2c_ac]{\pageref{A4B3_hR7_166_2c_ac}}} \\
\end{enumerate}
\vspace{-0.85cm} \item \textbf{ hR10} \vspace{-0.15cm} \\
\begin{enumerate}
\vspace{-0.65cm} \item FePSe$_{3}$: {\small ABC3\_hR10\_146\_2a\_2a\_2b} \dotfill {\hyperref[ABC3_hR10_146_2a_2a_2b]{\pageref{ABC3_hR10_146_2a_2a_2b}}} \\
\vspace{-0.65cm} \item Moissanite-15R: {\small AB\_hR10\_160\_5a\_5a} \dotfill {\hyperref[AB_hR10_160_5a_5a]{\pageref{AB_hR10_160_5a_5a}}} \\
\vspace{-0.65cm} \item PrNiO$_{3}$: {\small AB3C\_hR10\_167\_b\_e\_a} \dotfill {\hyperref[AB3C_hR10_167_b_e_a]{\pageref{AB3C_hR10_167_b_e_a}}} \\
\end{enumerate}
\vspace{-0.85cm} \item \textbf{ hR18} \vspace{-0.15cm} \\
\begin{enumerate}
\vspace{-0.65cm} \item $\beta$-PdCl$_2$: {\small A2B\_hR18\_148\_2f\_f} \dotfill {\hyperref[A2B_hR18_148_2f_f]{\pageref{A2B_hR18_148_2f_f}}} \\
\end{enumerate}
\vspace{-0.85cm} \item \textbf{ hR24} \vspace{-0.15cm} \\
\begin{enumerate}
\vspace{-0.65cm} \item KBO$_{2}$: {\small ABC2\_hR24\_167\_e\_e\_2e} \dotfill {\hyperref[ABC2_hR24_167_e_e_2e]{\pageref{ABC2_hR24_167_e_e_2e}}} \\
\end{enumerate}
\vspace{-0.85cm} \item \textbf{ hR26} \vspace{-0.15cm} \\
\begin{enumerate}
\vspace{-0.65cm} \item Al$_{8}$Cr$_{5}$: {\small A8B5\_hR26\_160\_a3bc\_a3b} \dotfill {\hyperref[A8B5_hR26_160_a3bc_a3b]{\pageref{A8B5_hR26_160_a3bc_a3b}}} \\
\end{enumerate}
\vspace{-0.85cm} \item \textbf{ hR42} \vspace{-0.15cm} \\
\begin{enumerate}
\vspace{-0.65cm} \item Phenakite: {\small A2B4C\_hR42\_148\_2f\_4f\_f} \dotfill {\hyperref[A2B4C_hR42_148_2f_4f_f]{\pageref{A2B4C_hR42_148_2f_4f_f}}} \\
\end{enumerate}
\end{enumerate}
\vspace{-0.75cm}
$\mathbf{mC}$ \textbf{\dotfill} \\
\begin{enumerate}
\vspace{-0.85cm} \item \textbf{ mC16} \vspace{-0.15cm} \\
\begin{enumerate}
\vspace{-0.65cm} \item H$_{3}$Cl: {\small AB3\_mC16\_9\_a\_3a} \dotfill {\hyperref[AB3_mC16_9_a_3a]{\pageref{AB3_mC16_9_a_3a}}} \\
\vspace{-0.65cm} \item M-carbon: {\small A\_mC16\_12\_4i} \dotfill {\hyperref[A_mC16_12_4i]{\pageref{A_mC16_12_4i}}} \\
\vspace{-0.65cm} \item H$_{3}$Cl: {\small AB3\_mC16\_15\_e\_cf} \dotfill {\hyperref[AB3_mC16_15_e_cf]{\pageref{AB3_mC16_15_e_cf}}} \\
\end{enumerate}
\vspace{-0.85cm} \item \textbf{ mC22} \vspace{-0.15cm} \\
\begin{enumerate}
\vspace{-0.65cm} \item Thortveitite: {\small A7B2C2\_mC22\_12\_aij\_h\_i} \dotfill {\hyperref[A7B2C2_mC22_12_aij_h_i]{\pageref{A7B2C2_mC22_12_aij_h_i}}} \\
\end{enumerate}
\vspace{-0.85cm} \item \textbf{ mC24} \vspace{-0.15cm} \\
\begin{enumerate}
\vspace{-0.65cm} \item H-III: {\small A\_mC24\_15\_2e2f} \dotfill {\hyperref[A_mC24_15_2e2f]{\pageref{A_mC24_15_2e2f}}} \\
\end{enumerate}
\vspace{-0.85cm} \item \textbf{ mC32} \vspace{-0.15cm} \\
\begin{enumerate}
\vspace{-0.65cm} \item $\alpha$-P$_3$N$_5$: {\small A5B3\_mC32\_9\_5a\_3a} \dotfill {\hyperref[A5B3_mC32_9_5a_3a]{\pageref{A5B3_mC32_9_5a_3a}}} \\
\end{enumerate}
\end{enumerate}
\vspace{-0.75cm}
$\mathbf{mP}$ \textbf{\dotfill} \\
\begin{enumerate}
\vspace{-0.85cm} \item \textbf{ mP4} \vspace{-0.15cm} \\
\begin{enumerate}
\vspace{-0.65cm} \item FeNi: {\small AB\_mP4\_6\_2b\_2a} \dotfill {\hyperref[AB_mP4_6_2b_2a]{\pageref{AB_mP4_6_2b_2a}}} \\
\end{enumerate}
\vspace{-0.85cm} \item \textbf{ mP6} \vspace{-0.15cm} \\
\begin{enumerate}
\vspace{-0.65cm} \item $\delta$-PdCl$_{2}$: {\small A2B\_mP6\_10\_mn\_bg} \dotfill {\hyperref[A2B_mP6_10_mn_bg]{\pageref{A2B_mP6_10_mn_bg}}} \\
\vspace{-0.65cm} \item LiSn: {\small AB\_mP6\_10\_en\_am} \dotfill {\hyperref[AB_mP6_10_en_am]{\pageref{AB_mP6_10_en_am}}} \\
\vspace{-0.65cm} \item $\gamma$-PdCl$_{2}$: {\small A2B\_mP6\_14\_e\_a} \dotfill {\hyperref[A2B_mP6_14_e_a]{\pageref{A2B_mP6_14_e_a}}} \\
\end{enumerate}
\vspace{-0.85cm} \item \textbf{ mP8} \vspace{-0.15cm} \\
\begin{enumerate}
\vspace{-0.65cm} \item Muthmannite: {\small ABC2\_mP8\_10\_ac\_eh\_mn} \dotfill {\hyperref[ABC2_mP8_10_ac_eh_mn]{\pageref{ABC2_mP8_10_ac_eh_mn}}} \\
\vspace{-0.65cm} \item S-carbon: {\small A\_mP8\_10\_2m2n} \dotfill {\hyperref[A_mP8_10_2m2n]{\pageref{A_mP8_10_2m2n}}} \\
\end{enumerate}
\vspace{-0.85cm} \item \textbf{ mP12} \vspace{-0.15cm} \\
\begin{enumerate}
\vspace{-0.65cm} \item H$_{2}$S IV: {\small A2B\_mP12\_7\_4a\_2a} \dotfill {\hyperref[A2B_mP12_7_4a_2a]{\pageref{A2B_mP12_7_4a_2a}}} \\
\vspace{-0.65cm} \item H$_{2}$S: {\small A2B\_mP12\_13\_2g\_ef} \dotfill {\hyperref[A2B_mP12_13_2g_ef]{\pageref{A2B_mP12_13_2g_ef}}} \\
\end{enumerate}
\vspace{-0.85cm} \item \textbf{ mP13} \vspace{-0.15cm} \\
\begin{enumerate}
\vspace{-0.65cm} \item Mo$_{8}$P$_{5}$: {\small A8B5\_mP13\_6\_a7b\_3a2b} \dotfill {\hyperref[A8B5_mP13_6_a7b_3a2b]{\pageref{A8B5_mP13_6_a7b_3a2b}}} \\
\end{enumerate}
\vspace{-0.85cm} \item \textbf{ mP16} \vspace{-0.15cm} \\
\begin{enumerate}
\vspace{-0.65cm} \item $\epsilon$-WO$_{3}$: {\small A3B\_mP16\_7\_6a\_2a} \dotfill {\hyperref[A3B_mP16_7_6a_2a]{\pageref{A3B_mP16_7_6a_2a}}} \\
\vspace{-0.65cm} \item H$_{3}$Cl: {\small AB3\_mP16\_10\_mn\_3m3n} \dotfill {\hyperref[AB3_mP16_10_mn_3m3n]{\pageref{AB3_mP16_10_mn_3m3n}}} \\
\end{enumerate}
\vspace{-0.85cm} \item \textbf{ mP18} \vspace{-0.15cm} \\
\begin{enumerate}
\vspace{-0.65cm} \item As$_{2}$Ba: {\small A2B\_mP18\_7\_6a\_3a} \dotfill {\hyperref[A2B_mP18_7_6a_3a]{\pageref{A2B_mP18_7_6a_3a}}} \\
\end{enumerate}
\vspace{-0.85cm} \item \textbf{ mP22} \vspace{-0.15cm} \\
\begin{enumerate}
\vspace{-0.65cm} \item Rh$_{2}$Ga$_{9}$: {\small A9B2\_mP22\_7\_9a\_2a} \dotfill {\hyperref[A9B2_mP22_7_9a_2a]{\pageref{A9B2_mP22_7_9a_2a}}} \\
\end{enumerate}
\vspace{-0.85cm} \item \textbf{ mP120} \vspace{-0.15cm} \\
\begin{enumerate}
\vspace{-0.65cm} \item \begin{raggedleft}$\alpha$-Toluene: \end{raggedleft} \\ {\small A7B8\_mP120\_14\_14e\_16e} \dotfill {\hyperref[A7B8_mP120_14_14e_16e]{\pageref{A7B8_mP120_14_14e_16e}}} \\
\end{enumerate}
\end{enumerate}
\vspace{-0.75cm}
$\mathbf{oC}$ \textbf{\dotfill} \\
\begin{enumerate}
\vspace{-0.85cm} \item \textbf{ oC6} \vspace{-0.15cm} \\
\begin{enumerate}
\vspace{-0.65cm} \item Ta$_{2}$H: {\small AB2\_oC6\_21\_a\_k} \dotfill {\hyperref[AB2_oC6_21_a_k]{\pageref{AB2_oC6_21_a_k}}} \\
\end{enumerate}
\vspace{-0.85cm} \item \textbf{ oC8} \vspace{-0.15cm} \\
\begin{enumerate}
\vspace{-0.65cm} \item HCl: {\small AB\_oC8\_36\_a\_a} \dotfill {\hyperref[AB_oC8_36_a_a]{\pageref{AB_oC8_36_a_a}}} \\
\vspace{-0.65cm} \item $\alpha$-FeSe\footnote[2]{\label{note:AB_oC8_67_a_g-Pearson}$\alpha$-FeSe and $\alpha$-PbO have the same \AFLOW\ prototype label. They are generated by the same symmetry operations with different sets of parameters.}: {\small AB\_oC8\_67\_a\_g} \dotfill {\hyperref[AB_oC8_67_a_g-FeSe]{\pageref{AB_oC8_67_a_g-FeSe}}} \\
\vspace{-0.65cm} \item $\alpha$-PbO\footnoteref{note:AB_oC8_67_a_g-Pearson}: {\small AB\_oC8\_67\_a\_g} \dotfill {\hyperref[AB_oC8_67_a_g-PbO]{\pageref{AB_oC8_67_a_g-PbO}}} \\
\end{enumerate}
\vspace{-0.85cm} \item \textbf{ oC16} \vspace{-0.15cm} \\
\begin{enumerate}
\vspace{-0.65cm} \item K$_{2}$CdPb: {\small AB2C\_oC16\_40\_a\_2b\_b} \dotfill {\hyperref[AB2C_oC16_40_a_2b_b]{\pageref{AB2C_oC16_40_a_2b_b}}} \\
\vspace{-0.65cm} \item CeTe$_{3}$: {\small AB3\_oC16\_40\_b\_3b} \dotfill {\hyperref[AB3_oC16_40_b_3b]{\pageref{AB3_oC16_40_b_3b}}} \\
\vspace{-0.65cm} \item Al$_{2}$CuIr\footnote[4]{\label{note:ABC2_oC16_67_b_g_ag-Pearson}Al$_{2}$CuIr and HoCuP$_{2}$ have similar \AFLOW\ prototype labels ({\it{i.e.}}, same symmetry and set of Wyckoff positions with different stoichiometry labels due to alphabetic ordering of atomic species). They are generated by the same symmetry operations with different sets of parameters.}: {\small A2BC\_oC16\_67\_ag\_b\_g} \dotfill {\hyperref[A2BC_oC16_67_ag_b_g]{\pageref{A2BC_oC16_67_ag_b_g}}} \\
\vspace{-0.65cm} \item HoCuP$_{2}$\footnoteref{note:ABC2_oC16_67_b_g_ag-Pearson}: {\small ABC2\_oC16\_67\_b\_g\_ag} \dotfill {\hyperref[ABC2_oC16_67_b_g_ag]{\pageref{ABC2_oC16_67_b_g_ag}}} \\
\end{enumerate}
\vspace{-0.85cm} \item \textbf{ oC20} \vspace{-0.15cm} \\
\begin{enumerate}
\vspace{-0.65cm} \item Post-perovskite: {\small AB3C\_oC20\_63\_a\_cf\_c} \dotfill {\hyperref[AB3C_oC20_63_a_cf_c]{\pageref{AB3C_oC20_63_a_cf_c}}} \\
\vspace{-0.65cm} \item PdSn$_{4}$: {\small AB4\_oC20\_68\_a\_i} \dotfill {\hyperref[AB4_oC20_68_a_i]{\pageref{AB4_oC20_68_a_i}}} \\
\end{enumerate}
\vspace{-0.85cm} \item \textbf{ oC22} \vspace{-0.15cm} \\
\begin{enumerate}
\vspace{-0.65cm} \item V$_{2}$MoO$_{8}$: {\small AB8C2\_oC22\_35\_a\_ab3e\_e} \dotfill {\hyperref[AB8C2_oC22_35_a_ab3e_e]{\pageref{AB8C2_oC22_35_a_ab3e_e}}} \\
\end{enumerate}
\vspace{-0.85cm} \item \textbf{ oC24} \vspace{-0.15cm} \\
\begin{enumerate}
\vspace{-0.65cm} \item Rasvumite: {\small A2BC3\_oC24\_63\_e\_c\_cg} \dotfill {\hyperref[A2BC3_oC24_63_e_c_cg]{\pageref{A2BC3_oC24_63_e_c_cg}}} \\
\vspace{-0.65cm} \item MgSO$_{4}$: {\small AB4C\_oC24\_63\_a\_fg\_c} \dotfill {\hyperref[AB4C_oC24_63_a_fg_c]{\pageref{AB4C_oC24_63_a_fg_c}}} \\
\vspace{-0.65cm} \item Anhydrite: {\small AB4C\_oC24\_63\_c\_fg\_c} \dotfill {\hyperref[AB4C_oC24_63_c_fg_c]{\pageref{AB4C_oC24_63_c_fg_c}}} \\
\vspace{-0.65cm} \item H$_{2}$S: {\small A2B\_oC24\_64\_2f\_f} \dotfill {\hyperref[A2B_oC24_64_2f_f]{\pageref{A2B_oC24_64_2f_f}}} \\
\end{enumerate}
\vspace{-0.85cm} \item \textbf{ oC28} \vspace{-0.15cm} \\
\begin{enumerate}
\vspace{-0.65cm} \item MnAl$_{6}$: {\small A6B\_oC28\_63\_efg\_c} \dotfill {\hyperref[A6B_oC28_63_efg_c]{\pageref{A6B_oC28_63_efg_c}}} \\
\vspace{-0.65cm} \item SrAl$_{2}$Se$_{4}$: {\small A2B4C\_oC28\_66\_l\_kl\_a} \dotfill {\hyperref[A2B4C_oC28_66_l_kl_a]{\pageref{A2B4C_oC28_66_l_kl_a}}} \\
\end{enumerate}
\vspace{-0.85cm} \item \textbf{ oC36} \vspace{-0.15cm} \\
\begin{enumerate}
\vspace{-0.65cm} \item Li$_{2}$Si$_{2}$O$_{5}$: {\small A2B5C2\_oC36\_37\_d\_c2d\_d} \dotfill {\hyperref[A2B5C2_oC36_37_d_c2d_d]{\pageref{A2B5C2_oC36_37_d_c2d_d}}} \\
\end{enumerate}
\vspace{-0.85cm} \item \textbf{ oC40} \vspace{-0.15cm} \\
\begin{enumerate}
\vspace{-0.65cm} \item Ta$_{3}$S$_{2}$: {\small A2B3\_oC40\_39\_2d\_2c2d} \dotfill {\hyperref[A2B3_oC40_39_2d_2c2d]{\pageref{A2B3_oC40_39_2d_2c2d}}} \\
\end{enumerate}
\vspace{-0.85cm} \item \textbf{ oC44} \vspace{-0.15cm} \\
\begin{enumerate}
\vspace{-0.65cm} \item VPCl$_{9}$: {\small A9BC\_oC44\_39\_3c3d\_a\_c} \dotfill {\hyperref[A9BC_oC44_39_3c3d_a_c]{\pageref{A9BC_oC44_39_3c3d_a_c}}} \\
\end{enumerate}
\vspace{-0.85cm} \item \textbf{ oC64} \vspace{-0.15cm} \\
\begin{enumerate}
\vspace{-0.65cm} \item H$_{3}$S: {\small A3B\_oC64\_66\_gi2lm\_2l} \dotfill {\hyperref[A3B_oC64_66_gi2lm_2l]{\pageref{A3B_oC64_66_gi2lm_2l}}} \\
\vspace{-0.65cm} \item $\beta$-ThI$_{3}$: {\small A3B\_oC64\_66\_kl2m\_bdl} \dotfill {\hyperref[A3B_oC64_66_kl2m_bdl]{\pageref{A3B_oC64_66_kl2m_bdl}}} \\
\end{enumerate}
\vspace{-0.85cm} \item \textbf{ oC260} \vspace{-0.15cm} \\
\begin{enumerate}
\vspace{-0.65cm} \item \begin{raggedleft}La$_{43}$Ni$_{17}$Mg$_{5}$: \end{raggedleft} \\ {\small A43B5C17\_oC260\_63\_c8fg6h\_cfg\_ce3f2h} \dotfill {\hyperref[A43B5C17_oC260_63_c8fg6h_cfg_ce3f2h]{\pageref{A43B5C17_oC260_63_c8fg6h_cfg_ce3f2h}}} \\
\end{enumerate}
\end{enumerate}
\vspace{-0.75cm}
$\mathbf{oF}$ \textbf{\dotfill} \\
\begin{enumerate}
\vspace{-0.85cm} \item \textbf{ oF8} \vspace{-0.15cm} \\
\begin{enumerate}
\vspace{-0.65cm} \item FeS: {\small AB\_oF8\_22\_a\_c} \dotfill {\hyperref[AB_oF8_22_a_c]{\pageref{AB_oF8_22_a_c}}} \\
\vspace{-0.65cm} \item BN: {\small AB\_oF8\_42\_a\_a} \dotfill {\hyperref[AB_oF8_42_a_a]{\pageref{AB_oF8_42_a_a}}} \\
\end{enumerate}
\vspace{-0.85cm} \item \textbf{ oF40} \vspace{-0.15cm} \\
\begin{enumerate}
\vspace{-0.65cm} \item CeRu$_{2}$B$_{2}$: {\small A2BC2\_oF40\_22\_fi\_ad\_gh} \dotfill {\hyperref[A2BC2_oF40_22_fi_ad_gh]{\pageref{A2BC2_oF40_22_fi_ad_gh}}} \\
\end{enumerate}
\vspace{-0.85cm} \item \textbf{ oF48} \vspace{-0.15cm} \\
\begin{enumerate}
\vspace{-0.65cm} \item Mn$_{2}$B: {\small AB2\_oF48\_70\_f\_fg} \dotfill {\hyperref[AB2_oF48_70_f_fg]{\pageref{AB2_oF48_70_f_fg}}} \\
\end{enumerate}
\vspace{-0.85cm} \item \textbf{ oF52} \vspace{-0.15cm} \\
\begin{enumerate}
\vspace{-0.65cm} \item W$_{3}$O$_{10}$: {\small A10B3\_oF52\_42\_2abce\_ab} \dotfill {\hyperref[A10B3_oF52_42_2abce_ab]{\pageref{A10B3_oF52_42_2abce_ab}}} \\
\end{enumerate}
\end{enumerate}
\vspace{-0.75cm}
$\mathbf{oI}$ \textbf{\dotfill} \\
\begin{enumerate}
\vspace{-0.85cm} \item \textbf{ oI12} \vspace{-0.15cm} \\
\begin{enumerate}
\vspace{-0.65cm} \item BPS$_{4}$: {\small ABC4\_oI12\_23\_a\_b\_k} \dotfill {\hyperref[ABC4_oI12_23_a_b_k]{\pageref{ABC4_oI12_23_a_b_k}}} \\
\vspace{-0.65cm} \item NbPS: {\small ABC\_oI12\_71\_h\_j\_g} \dotfill {\hyperref[ABC_oI12_71_h_j_g]{\pageref{ABC_oI12_71_h_j_g}}} \\
\vspace{-0.65cm} \item KHg$_{2}$: {\small A2B\_oI12\_74\_h\_e} \dotfill {\hyperref[A2B_oI12_74_h_e]{\pageref{A2B_oI12_74_h_e}}} \\
\end{enumerate}
\vspace{-0.85cm} \item \textbf{ oI14} \vspace{-0.15cm} \\
\begin{enumerate}
\vspace{-0.65cm} \item Ta$_{3}$B$_{4}$: {\small A4B3\_oI14\_71\_gh\_cg} \dotfill {\hyperref[A4B3_oI14_71_gh_cg]{\pageref{A4B3_oI14_71_gh_cg}}} \\
\end{enumerate}
\vspace{-0.85cm} \item \textbf{ oI16} \vspace{-0.15cm} \\
\begin{enumerate}
\vspace{-0.65cm} \item NaFeS$_{2}$: {\small ABC2\_oI16\_23\_ab\_i\_k} \dotfill {\hyperref[ABC2_oI16_23_ab_i_k]{\pageref{ABC2_oI16_23_ab_i_k}}} \\
\end{enumerate}
\vspace{-0.85cm} \item \textbf{ oI20} \vspace{-0.15cm} \\
\begin{enumerate}
\vspace{-0.65cm} \item MnGa$_{2}$Sb$_{2}$: {\small A2BC2\_oI20\_45\_c\_b\_c} \dotfill {\hyperref[A2BC2_oI20_45_c_b_c]{\pageref{A2BC2_oI20_45_c_b_c}}} \\
\vspace{-0.65cm} \item Al$_{4}$U: {\small A4B\_oI20\_74\_beh\_e} \dotfill {\hyperref[A4B_oI20_74_beh_e]{\pageref{A4B_oI20_74_beh_e}}} \\
\end{enumerate}
\vspace{-0.85cm} \item \textbf{ oI32} \vspace{-0.15cm} \\
\begin{enumerate}
\vspace{-0.65cm} \item H$_{3}$S: {\small A3B\_oI32\_23\_ij2k\_k} \dotfill {\hyperref[A3B_oI32_23_ij2k_k]{\pageref{A3B_oI32_23_ij2k_k}}} \\
\end{enumerate}
\vspace{-0.85cm} \item \textbf{ oI36} \vspace{-0.15cm} \\
\begin{enumerate}
\vspace{-0.65cm} \item TiFeSi: {\small ABC\_oI36\_46\_ac\_bc\_3b} \dotfill {\hyperref[ABC_oI36_46_ac_bc_3b]{\pageref{ABC_oI36_46_ac_bc_3b}}} \\
\end{enumerate}
\vspace{-0.85cm} \item \textbf{ oI44} \vspace{-0.15cm} \\
\begin{enumerate}
\vspace{-0.65cm} \item Weberite: {\small AB7CD2\_oI44\_24\_a\_b3d\_c\_ac} \dotfill {\hyperref[AB7CD2_oI44_24_a_b3d_c_ac]{\pageref{AB7CD2_oI44_24_a_b3d_c_ac}}} \\
\end{enumerate}
\vspace{-0.85cm} \item \textbf{ oI48} \vspace{-0.15cm} \\
\begin{enumerate}
\vspace{-0.65cm} \item KAg[CO$_{3}$]: {\small ABCD3\_oI48\_73\_d\_e\_e\_ef} \dotfill {\hyperref[ABCD3_oI48_73_d_e_e_ef]{\pageref{ABCD3_oI48_73_d_e_e_ef}}} \\
\end{enumerate}
\vspace{-0.85cm} \item \textbf{ oI50} \vspace{-0.15cm} \\
\begin{enumerate}
\vspace{-0.65cm} \item \begin{raggedleft}Stannoidite: \end{raggedleft} \\ {\small A8B2C12D2E\_oI50\_23\_bcfk\_i\_3k\_j\_a} \dotfill {\hyperref[A8B2C12D2E_oI50_23_bcfk_i_3k_j_a]{\pageref{A8B2C12D2E_oI50_23_bcfk_i_3k_j_a}}} \\
\end{enumerate}
\end{enumerate}
\vspace{-0.75cm}
$\mathbf{oP}$ \textbf{\dotfill} \\
\begin{enumerate}
\vspace{-0.85cm} \item \textbf{ oP6} \vspace{-0.15cm} \\
\begin{enumerate}
\vspace{-0.65cm} \item FeSb$_{2}$: {\small AB2\_oP6\_34\_a\_c} \dotfill {\hyperref[AB2_oP6_34_a_c]{\pageref{AB2_oP6_34_a_c}}} \\
\vspace{-0.65cm} \item $\alpha$-PdCl$_{2}$: {\small A2B\_oP6\_58\_g\_a} \dotfill {\hyperref[A2B_oP6_58_g_a]{\pageref{A2B_oP6_58_g_a}}} \\
\vspace{-0.65cm} \item FeOCl: {\small ABC\_oP6\_59\_a\_b\_a} \dotfill {\hyperref[ABC_oP6_59_a_b_a]{\pageref{ABC_oP6_59_a_b_a}}} \\
\end{enumerate}
\vspace{-0.85cm} \item \textbf{ oP8} \vspace{-0.15cm} \\
\begin{enumerate}
\vspace{-0.65cm} \item Westerveldite: {\small AB\_oP8\_62\_c\_c} \dotfill {\hyperref[AB_oP8_62_c_c]{\pageref{AB_oP8_62_c_c}}} \\
\end{enumerate}
\vspace{-0.85cm} \item \textbf{ oP12} \vspace{-0.15cm} \\
\begin{enumerate}
\vspace{-0.65cm} \item $\alpha$-Naumannite: {\small A2B\_oP12\_17\_abe\_e} \dotfill {\hyperref[A2B_oP12_17_abe_e]{\pageref{A2B_oP12_17_abe_e}}} \\
\vspace{-0.65cm} \item H$_{2}$S\footnote[1]{\label{note:A2B_oP12_26_abc_ab-Pearson}H$_{2}$S and $\beta$-SeO$_{2}$ have the same \AFLOW\ prototype label. They are generated by the same symmetry operations with different sets of parameters.}: {\small A2B\_oP12\_26\_abc\_ab} \dotfill {\hyperref[A2B_oP12_26_abc_ab-H2S]{\pageref{A2B_oP12_26_abc_ab-H2S}}} \\
\vspace{-0.65cm} \item $\beta$-SeO$_{2}$\footnoteref{note:A2B_oP12_26_abc_ab-Pearson}: {\small A2B\_oP12\_26\_abc\_ab} \dotfill {\hyperref[A2B_oP12_26_abc_ab-SeO2]{\pageref{A2B_oP12_26_abc_ab-SeO2}}} \\
\vspace{-0.65cm} \item ZrO$_{2}$\footnote[3]{\label{note:AB2_oP12_29_a_2a-Pearson}ZrO$_{2}$ and Pyrite have similar \AFLOW\ prototype labels ({\it{i.e.}}, same symmetry and set of Wyckoff positions with different stoichiometry labels due to alphabetic ordering of atomic species). They are generated by the same symmetry operations with different sets of parameters.}: {\small A2B\_oP12\_29\_2a\_a} \dotfill {\hyperref[A2B_oP12_29_2a_a]{\pageref{A2B_oP12_29_2a_a}}} \\
\vspace{-0.65cm} \item Pyrite\footnoteref{note:AB2_oP12_29_a_2a-Pearson}: {\small AB2\_oP12\_29\_a\_2a} \dotfill {\hyperref[AB2_oP12_29_a_2a]{\pageref{AB2_oP12_29_a_2a}}} \\
\vspace{-0.65cm} \item Cobaltite: {\small ABC\_oP12\_29\_a\_a\_a} \dotfill {\hyperref[ABC_oP12_29_a_a_a]{\pageref{ABC_oP12_29_a_a_a}}} \\
\vspace{-0.65cm} \item SrH$_{2}$: {\small A2B\_oP12\_62\_2c\_c} \dotfill {\hyperref[A2B_oP12_62_2c_c]{\pageref{A2B_oP12_62_2c_c}}} \\
\end{enumerate}
\vspace{-0.85cm} \item \textbf{ oP14} \vspace{-0.15cm} \\
\begin{enumerate}
\vspace{-0.65cm} \item $\beta$-Ta$_{2}$O$_{5}$: {\small A5B2\_oP14\_49\_dehq\_ab} \dotfill {\hyperref[A5B2_oP14_49_dehq_ab]{\pageref{A5B2_oP14_49_dehq_ab}}} \\
\end{enumerate}
\vspace{-0.85cm} \item \textbf{ oP16} \vspace{-0.15cm} \\
\begin{enumerate}
\vspace{-0.65cm} \item H$_{3}$Cl: {\small AB3\_oP16\_19\_a\_3a} \dotfill {\hyperref[AB3_oP16_19_a_3a]{\pageref{AB3_oP16_19_a_3a}}} \\
\vspace{-0.65cm} \item TaNiTe$_{2}$: {\small ABC2\_oP16\_53\_h\_e\_gh} \dotfill {\hyperref[ABC2_oP16_53_h_e_gh]{\pageref{ABC2_oP16_53_h_e_gh}}} \\
\vspace{-0.65cm} \item Rh$_{5}$Ge$_{3}$: {\small A3B5\_oP16\_55\_ch\_agh} \dotfill {\hyperref[A3B5_oP16_55_ch_agh]{\pageref{A3B5_oP16_55_ch_agh}}} \\
\vspace{-0.65cm} \item R-carbon: {\small A\_oP16\_55\_2g2h} \dotfill {\hyperref[A_oP16_55_2g2h]{\pageref{A_oP16_55_2g2h}}} \\
\vspace{-0.65cm} \item $\epsilon$-NiAl$_{3}$: {\small A3B\_oP16\_62\_cd\_c} \dotfill {\hyperref[A3B_oP16_62_cd_c]{\pageref{A3B_oP16_62_cd_c}}} \\
\vspace{-0.65cm} \item Molybdite: {\small AB3\_oP16\_62\_c\_3c} \dotfill {\hyperref[AB3_oP16_62_c_3c]{\pageref{AB3_oP16_62_c_3c}}} \\
\end{enumerate}
\vspace{-0.85cm} \item \textbf{ oP20} \vspace{-0.15cm} \\
\begin{enumerate}
\vspace{-0.65cm} \item CuBrSe$_{3}$: {\small ABC3\_oP20\_30\_2a\_c\_3c} \dotfill {\hyperref[ABC3_oP20_30_2a_c_3c]{\pageref{ABC3_oP20_30_2a_c_3c}}} \\
\vspace{-0.65cm} \item Sr$_{2}$Bi$_{3}$: {\small A3B2\_oP20\_52\_de\_cd} \dotfill {\hyperref[A3B2_oP20_52_de_cd]{\pageref{A3B2_oP20_52_de_cd}}} \\
\vspace{-0.65cm} \item CuBrSe$_{3}$: {\small ABC3\_oP20\_53\_e\_g\_hi} \dotfill {\hyperref[ABC3_oP20_53_e_g_hi]{\pageref{ABC3_oP20_53_e_g_hi}}} \\
\vspace{-0.65cm} \item BiGaO$_{3}$: {\small ABC3\_oP20\_54\_e\_d\_cf} \dotfill {\hyperref[ABC3_oP20_54_e_d_cf]{\pageref{ABC3_oP20_54_e_d_cf}}} \\
\vspace{-0.65cm} \item Rh$_{2}$S$_{3}$: {\small A2B3\_oP20\_60\_d\_cd} \dotfill {\hyperref[A2B3_oP20_60_d_cd]{\pageref{A2B3_oP20_60_d_cd}}} \\
\vspace{-0.65cm} \item Tongbaite: {\small A2B3\_oP20\_62\_2c\_3c} \dotfill {\hyperref[A2B3_oP20_62_2c_3c]{\pageref{A2B3_oP20_62_2c_3c}}} \\
\end{enumerate}
\vspace{-0.85cm} \item \textbf{ oP22} \vspace{-0.15cm} \\
\begin{enumerate}
\vspace{-0.65cm} \item TiAl$_{2}$Br$_{8}$: {\small A2B8C\_oP22\_34\_c\_4c\_a} \dotfill {\hyperref[A2B8C_oP22_34_c_4c_a]{\pageref{A2B8C_oP22_34_c_4c_a}}} \\
\end{enumerate}
\vspace{-0.85cm} \item \textbf{ oP24} \vspace{-0.15cm} \\
\begin{enumerate}
\vspace{-0.65cm} \item TlP$_{5}$: {\small A5B\_oP24\_26\_3a3b2c\_ab} \dotfill {\hyperref[A5B_oP24_26_3a3b2c_ab]{\pageref{A5B_oP24_26_3a3b2c_ab}}} \\
\vspace{-0.65cm} \item \begin{raggedleft}$\alpha$-RbPr[MoO$_{4}$]$_{2}$: \end{raggedleft} \\ {\small A2B8CD\_oP24\_48\_k\_2m\_d\_b} \dotfill {\hyperref[A2B8CD_oP24_48_k_2m_d_b]{\pageref{A2B8CD_oP24_48_k_2m_d_b}}} \\
\vspace{-0.65cm} \item CsPr[MoO$_{4}$]$_{2}$: {\small AB2C8D\_oP24\_49\_g\_q\_2qr\_e} \dotfill {\hyperref[AB2C8D_oP24_49_g_q_2qr_e]{\pageref{AB2C8D_oP24_49_g_q_2qr_e}}} \\
\vspace{-0.65cm} \item GaCl$_{2}$: {\small A2B\_oP24\_52\_2e\_cd} \dotfill {\hyperref[A2B_oP24_52_2e_cd]{\pageref{A2B_oP24_52_2e_cd}}} \\
\vspace{-0.65cm} \item GeAs$_{2}$: {\small A2B\_oP24\_55\_2g2h\_gh} \dotfill {\hyperref[A2B_oP24_55_2g2h_gh]{\pageref{A2B_oP24_55_2g2h_gh}}} \\
\vspace{-0.65cm} \item Cubanite: {\small AB2C3\_oP24\_62\_c\_d\_cd} \dotfill {\hyperref[AB2C3_oP24_62_c_d_cd]{\pageref{AB2C3_oP24_62_c_d_cd}}} \\
\vspace{-0.65cm} \item Barite: {\small AB4C\_oP24\_62\_c\_2cd\_c} \dotfill {\hyperref[AB4C_oP24_62_c_2cd_c]{\pageref{AB4C_oP24_62_c_2cd_c}}} \\
\end{enumerate}
\vspace{-0.85cm} \item \textbf{ oP28} \vspace{-0.15cm} \\
\begin{enumerate}
\vspace{-0.65cm} \item La$_{2}$NiO$_{4}$: {\small A2BC4\_oP28\_50\_ij\_ac\_ijm} \dotfill {\hyperref[A2BC4_oP28_50_ij_ac_ijm]{\pageref{A2BC4_oP28_50_ij_ac_ijm}}} \\
\vspace{-0.65cm} \item Forsterite: {\small A2B4C\_oP28\_62\_ac\_2cd\_c} \dotfill {\hyperref[A2B4C_oP28_62_ac_2cd_c]{\pageref{A2B4C_oP28_62_ac_2cd_c}}} \\
\end{enumerate}
\vspace{-0.85cm} \item \textbf{ oP32} \vspace{-0.15cm} \\
\begin{enumerate}
\vspace{-0.65cm} \item WO$_{3}$: {\small A3B\_oP32\_60\_3d\_d} \dotfill {\hyperref[A3B_oP32_60_3d_d]{\pageref{A3B_oP32_60_3d_d}}} \\
\end{enumerate}
\vspace{-0.85cm} \item \textbf{ oP34} \vspace{-0.15cm} \\
\begin{enumerate}
\vspace{-0.65cm} \item Re$_{2}$O$_{5}$[SO$_{4}$]$_{2}$: {\small A13B2C2\_oP34\_32\_a6c\_c\_c} \dotfill {\hyperref[A13B2C2_oP34_32_a6c_c_c]{\pageref{A13B2C2_oP34_32_a6c_c_c}}} \\
\end{enumerate}
\vspace{-0.85cm} \item \textbf{ oP40} \vspace{-0.15cm} \\
\begin{enumerate}
\vspace{-0.65cm} \item $\kappa$-alumina: {\small A2B3\_oP40\_33\_4a\_6a} \dotfill {\hyperref[A2B3_oP40_33_4a_6a]{\pageref{A2B3_oP40_33_4a_6a}}} \\
\end{enumerate}
\vspace{-0.85cm} \item \textbf{ oP46} \vspace{-0.15cm} \\
\begin{enumerate}
\vspace{-0.65cm} \item Bi$_{5}$Nb$_{3}$O$_{15}$: {\small A5B3C15\_oP46\_30\_a2c\_bc\_a7c} \dotfill {\hyperref[A5B3C15_oP46_30_a2c_bc_a7c]{\pageref{A5B3C15_oP46_30_a2c_bc_a7c}}} \\
\end{enumerate}
\vspace{-0.85cm} \item \textbf{ oP48} \vspace{-0.15cm} \\
\begin{enumerate}
\vspace{-0.65cm} \item \begin{raggedleft}$\alpha$-Tl$_{2}$TeO$_{3}$: \end{raggedleft} \\ {\small A3BC2\_oP48\_50\_3m\_m\_2m} \dotfill {\hyperref[A3BC2_oP48_50_3m_m_2m]{\pageref{A3BC2_oP48_50_3m_m_2m}}} \\
\vspace{-0.65cm} \item Benzene: {\small AB\_oP48\_61\_3c\_3c} \dotfill {\hyperref[AB_oP48_61_3c_3c]{\pageref{AB_oP48_61_3c_3c}}} \\
\end{enumerate}
\vspace{-0.85cm} \item \textbf{ oP108} \vspace{-0.15cm} \\
\begin{enumerate}
\vspace{-0.65cm} \item \begin{raggedleft}Ca$_{4}$Al$_{6}$O$_{16}$S: \end{raggedleft} \\ {\small A6B4C16D\_oP108\_27\_abcd4e\_4e\_16e\_e} \dotfill {\hyperref[A6B4C16D_oP108_27_abcd4e_4e_16e_e]{\pageref{A6B4C16D_oP108_27_abcd4e_4e_16e_e}}} \\
\end{enumerate}
\vspace{-0.85cm} \item \textbf{ oP120} \vspace{-0.15cm} \\
\begin{enumerate}
\vspace{-0.65cm} \item $\beta$-Toluene: {\small A7B8\_oP120\_60\_7d\_8d} \dotfill {\hyperref[A7B8_oP120_60_7d_8d]{\pageref{A7B8_oP120_60_7d_8d}}} \\
\end{enumerate}
\end{enumerate}
\vspace{-0.75cm}
$\mathbf{tI}$ \textbf{\dotfill} \\
\begin{enumerate}
\vspace{-0.85cm} \item \textbf{ tI4} \vspace{-0.15cm} \\
\begin{enumerate}
\vspace{-0.65cm} \item GeP: {\small AB\_tI4\_107\_a\_a} \dotfill {\hyperref[AB_tI4_107_a_a]{\pageref{AB_tI4_107_a_a}}} \\
\vspace{-0.65cm} \item GaSb: {\small AB\_tI4\_119\_c\_a} \dotfill {\hyperref[AB_tI4_119_c_a]{\pageref{AB_tI4_119_c_a}}} \\
\end{enumerate}
\vspace{-0.85cm} \item \textbf{ tI8} \vspace{-0.15cm} \\
\begin{enumerate}
\vspace{-0.65cm} \item NbAs: {\small AB\_tI8\_109\_a\_a} \dotfill {\hyperref[AB_tI8_109_a_a]{\pageref{AB_tI8_109_a_a}}} \\
\vspace{-0.65cm} \item Calomel: {\small AB\_tI8\_139\_e\_e} \dotfill {\hyperref[AB_tI8_139_e_e]{\pageref{AB_tI8_139_e_e}}} \\
\end{enumerate}
\vspace{-0.85cm} \item \textbf{ tI12} \vspace{-0.15cm} \\
\begin{enumerate}
\vspace{-0.65cm} \item CdAs$_{2}$: {\small A2B\_tI12\_98\_f\_a} \dotfill {\hyperref[A2B_tI12_98_f_a]{\pageref{A2B_tI12_98_f_a}}} \\
\vspace{-0.65cm} \item LaPtSi: {\small ABC\_tI12\_109\_a\_a\_a} \dotfill {\hyperref[ABC_tI12_109_a_a_a]{\pageref{ABC_tI12_109_a_a_a}}} \\
\vspace{-0.65cm} \item $\alpha$-ThSi$_{2}$: {\small A2B\_tI12\_141\_e\_a} \dotfill {\hyperref[A2B_tI12_141_e_a]{\pageref{A2B_tI12_141_e_a}}} \\
\end{enumerate}
\vspace{-0.85cm} \item \textbf{ tI16} \vspace{-0.15cm} \\
\begin{enumerate}
\vspace{-0.65cm} \item S-III: {\small A\_tI16\_142\_f} \dotfill {\hyperref[A_tI16_142_f]{\pageref{A_tI16_142_f}}} \\
\end{enumerate}
\vspace{-0.85cm} \item \textbf{ tI20} \vspace{-0.15cm} \\
\begin{enumerate}
\vspace{-0.65cm} \item TlZn$_{2}$Sb$_{2}$: {\small A2BC2\_tI20\_79\_c\_2a\_c} \dotfill {\hyperref[A2BC2_tI20_79_c_2a_c]{\pageref{A2BC2_tI20_79_c_2a_c}}} \\
\vspace{-0.65cm} \item ThCl$_{4}$: {\small A4B\_tI20\_88\_f\_a} \dotfill {\hyperref[A4B_tI20_88_f_a]{\pageref{A4B_tI20_88_f_a}}} \\
\end{enumerate}
\vspace{-0.85cm} \item \textbf{ tI24} \vspace{-0.15cm} \\
\begin{enumerate}
\vspace{-0.65cm} \item NaGdCu$_{2}$F$_{8}$: {\small A2B8CD\_tI24\_97\_d\_k\_a\_b} \dotfill {\hyperref[A2B8CD_tI24_97_d_k_a_b]{\pageref{A2B8CD_tI24_97_d_k_a_b}}} \\
\vspace{-0.65cm} \item Co$_{5}$Ge$_{7}$: {\small A5B7\_tI24\_107\_ac\_abd} \dotfill {\hyperref[A5B7_tI24_107_ac_abd]{\pageref{A5B7_tI24_107_ac_abd}}} \\
\vspace{-0.65cm} \item RbGa$_{3}$: {\small A3B\_tI24\_119\_b2i\_af} \dotfill {\hyperref[A3B_tI24_119_b2i_af]{\pageref{A3B_tI24_119_b2i_af}}} \\
\end{enumerate}
\vspace{-0.85cm} \item \textbf{ tI28} \vspace{-0.15cm} \\
\begin{enumerate}
\vspace{-0.65cm} \item KAu$_{4}$Sn$_{2}$: {\small A4BC2\_tI28\_120\_i\_d\_e} \dotfill {\hyperref[A4BC2_tI28_120_i_d_e]{\pageref{A4BC2_tI28_120_i_d_e}}} \\
\end{enumerate}
\vspace{-0.85cm} \item \textbf{ tI32} \vspace{-0.15cm} \\
\begin{enumerate}
\vspace{-0.65cm} \item Ni$_{3}$P: {\small A3B\_tI32\_82\_3g\_g} \dotfill {\hyperref[A3B_tI32_82_3g_g]{\pageref{A3B_tI32_82_3g_g}}} \\
\vspace{-0.65cm} \item Sr$_{5}$Si$_{3}$: {\small A3B5\_tI32\_108\_ac\_a2c} \dotfill {\hyperref[A3B5_tI32_108_ac_a2c]{\pageref{A3B5_tI32_108_ac_a2c}}} \\
\vspace{-0.65cm} \item W$_{5}$Si$_{3}$: {\small A3B5\_tI32\_140\_ah\_bk} \dotfill {\hyperref[A3B5_tI32_140_ah_bk]{\pageref{A3B5_tI32_140_ah_bk}}} \\
\vspace{-0.65cm} \item Cr$_{5}$B$_{3}$: {\small A3B5\_tI32\_140\_ah\_cl} \dotfill {\hyperref[A3B5_tI32_140_ah_cl]{\pageref{A3B5_tI32_140_ah_cl}}} \\
\end{enumerate}
\vspace{-0.85cm} \item \textbf{ tI44} \vspace{-0.15cm} \\
\begin{enumerate}
\vspace{-0.65cm} \item Ta$_{2}$Se$_{8}$I: {\small AB8C2\_tI44\_97\_e\_2k\_cd} \dotfill {\hyperref[AB8C2_tI44_97_e_2k_cd]{\pageref{AB8C2_tI44_97_e_2k_cd}}} \\
\end{enumerate}
\vspace{-0.85cm} \item \textbf{ tI48} \vspace{-0.15cm} \\
\begin{enumerate}
\vspace{-0.65cm} \item $\beta$-NbO$_{2}$: {\small AB2\_tI48\_80\_2b\_4b} \dotfill {\hyperref[AB2_tI48_80_2b_4b]{\pageref{AB2_tI48_80_2b_4b}}} \\
\end{enumerate}
\vspace{-0.85cm} \item \textbf{ tI96} \vspace{-0.15cm} \\
\begin{enumerate}
\vspace{-0.65cm} \item $\alpha$-NbO$_{2}$: {\small AB2\_tI96\_88\_2f\_4f} \dotfill {\hyperref[AB2_tI96_88_2f_4f]{\pageref{AB2_tI96_88_2f_4f}}} \\
\end{enumerate}
\vspace{-0.85cm} \item \textbf{ tI176} \vspace{-0.15cm} \\
\begin{enumerate}
\vspace{-0.65cm} \item Be[BH$_{4}$]$_{2}$: {\small A2BC8\_tI176\_110\_2b\_b\_8b} \dotfill {\hyperref[A2BC8_tI176_110_2b_b_8b]{\pageref{A2BC8_tI176_110_2b_b_8b}}} \\
\end{enumerate}
\end{enumerate}
\vspace{-0.75cm}
$\mathbf{tP}$ \textbf{\dotfill} \\
\begin{enumerate}
\vspace{-0.85cm} \item \textbf{ tP5} \vspace{-0.15cm} \\
\begin{enumerate}
\vspace{-0.65cm} \item Rh$_{3}$P$_{2}$: {\small A2B3\_tP5\_115\_g\_ag} \dotfill {\hyperref[A2B3_tP5_115_g_ag]{\pageref{A2B3_tP5_115_g_ag}}} \\
\end{enumerate}
\vspace{-0.85cm} \item \textbf{ tP6} \vspace{-0.15cm} \\
\begin{enumerate}
\vspace{-0.65cm} \item ZrO$_{2}$\footnote[5]{\label{note:AB2_tP6_137_a_d-Pearson}ZrO$_{2}$ and HgI$_{2}$ have similar \AFLOW\ prototype labels ({\it{i.e.}}, same symmetry and set of Wyckoff positions with different stoichiometry labels due to alphabetic ordering of atomic species). They are generated by the same symmetry operations with different sets of parameters.}: {\small A2B\_tP6\_137\_d\_a} \dotfill {\hyperref[A2B_tP6_137_d_a]{\pageref{A2B_tP6_137_d_a}}} \\
\vspace{-0.65cm} \item HgI$_{2}$\footnoteref{note:AB2_tP6_137_a_d-Pearson}: {\small AB2\_tP6\_137\_a\_d} \dotfill {\hyperref[AB2_tP6_137_a_d]{\pageref{AB2_tP6_137_a_d}}} \\
\end{enumerate}
\vspace{-0.85cm} \item \textbf{ tP8} \vspace{-0.15cm} \\
\begin{enumerate}
\vspace{-0.65cm} \item NV: {\small AB\_tP8\_111\_n\_n} \dotfill {\hyperref[AB_tP8_111_n_n]{\pageref{AB_tP8_111_n_n}}} \\
\end{enumerate}
\vspace{-0.85cm} \item \textbf{ tP10} \vspace{-0.15cm} \\
\begin{enumerate}
\vspace{-0.65cm} \item Ti$_{2}$Ge$_{3}$: {\small A3B2\_tP10\_83\_adk\_j} \dotfill {\hyperref[A3B2_tP10_83_adk_j]{\pageref{A3B2_tP10_83_adk_j}}} \\
\vspace{-0.65cm} \item NbTe$_{4}$: {\small AB4\_tP10\_103\_a\_d} \dotfill {\hyperref[AB4_tP10_103_a_d]{\pageref{AB4_tP10_103_a_d}}} \\
\vspace{-0.65cm} \item Pd$_{4}$Se: {\small A4B\_tP10\_114\_e\_a} \dotfill {\hyperref[A4B_tP10_114_e_a]{\pageref{A4B_tP10_114_e_a}}} \\
\vspace{-0.65cm} \item CaRbFe$_{4}$As$_{4}$: {\small A4BC4D\_tP10\_123\_gh\_a\_i\_d} \dotfill {\hyperref[A4BC4D_tP10_123_gh_a_i_d]{\pageref{A4BC4D_tP10_123_gh_a_i_d}}} \\
\vspace{-0.65cm} \item NbTe$_{4}$: {\small AB4\_tP10\_124\_a\_m} \dotfill {\hyperref[AB4_tP10_124_a_m]{\pageref{AB4_tP10_124_a_m}}} \\
\vspace{-0.65cm} \item PtPb$_{4}$: {\small A4B\_tP10\_125\_m\_a} \dotfill {\hyperref[A4B_tP10_125_m_a]{\pageref{A4B_tP10_125_m_a}}} \\
\end{enumerate}
\vspace{-0.85cm} \item \textbf{ tP12} \vspace{-0.15cm} \\
\begin{enumerate}
\vspace{-0.65cm} \item GeSe$_{2}$: {\small AB2\_tP12\_81\_adg\_2h} \dotfill {\hyperref[AB2_tP12_81_adg_2h]{\pageref{AB2_tP12_81_adg_2h}}} \\
\vspace{-0.65cm} \item MnF$_{2}$: {\small A2B\_tP12\_111\_2n\_adf} \dotfill {\hyperref[A2B_tP12_111_2n_adf]{\pageref{A2B_tP12_111_2n_adf}}} \\
\vspace{-0.65cm} \item $\alpha$-CuAlCl$_{4}$: {\small AB4C\_tP12\_112\_b\_n\_e} \dotfill {\hyperref[AB4C_tP12_112_b_n_e]{\pageref{AB4C_tP12_112_b_n_e}}} \\
\vspace{-0.65cm} \item HgI$_{2}$: {\small AB2\_tP12\_115\_j\_egi} \dotfill {\hyperref[AB2_tP12_115_j_egi]{\pageref{AB2_tP12_115_j_egi}}} \\
\vspace{-0.65cm} \item Nb$_{4}$CoSi: {\small AB4C\_tP12\_124\_a\_m\_c} \dotfill {\hyperref[AB4C_tP12_124_a_m_c]{\pageref{AB4C_tP12_124_a_m_c}}} \\
\vspace{-0.65cm} \item KCeSe$_{4}$: {\small ABC4\_tP12\_125\_a\_b\_m} \dotfill {\hyperref[ABC4_tP12_125_a_b_m]{\pageref{ABC4_tP12_125_a_b_m}}} \\
\vspace{-0.65cm} \item C: {\small A\_tP12\_138\_bi} \dotfill {\hyperref[A_tP12_138_bi]{\pageref{A_tP12_138_bi}}} \\
\end{enumerate}
\vspace{-0.85cm} \item \textbf{ tP16} \vspace{-0.15cm} \\
\begin{enumerate}
\vspace{-0.65cm} \item LaRhC$_{2}$: {\small A2BC\_tP16\_76\_2a\_a\_a} \dotfill {\hyperref[A2BC_tP16_76_2a_a_a]{\pageref{A2BC_tP16_76_2a_a_a}}} \\
\vspace{-0.65cm} \item RuIn$_{3}$: {\small A3B\_tP16\_118\_ei\_f} \dotfill {\hyperref[A3B_tP16_118_ei_f]{\pageref{A3B_tP16_118_ei_f}}} \\
\vspace{-0.65cm} \item AgUF$_{6}$: {\small AB6C\_tP16\_132\_d\_io\_a} \dotfill {\hyperref[AB6C_tP16_132_d_io_a]{\pageref{AB6C_tP16_132_d_io_a}}} \\
\end{enumerate}
\vspace{-0.85cm} \item \textbf{ tP18} \vspace{-0.15cm} \\
\begin{enumerate}
\vspace{-0.65cm} \item Li$_{2}$MoF$_{6}$: {\small A6B2C\_tP18\_94\_eg\_c\_a} \dotfill {\hyperref[A6B2C_tP18_94_eg_c_a]{\pageref{A6B2C_tP18_94_eg_c_a}}} \\
\vspace{-0.65cm} \item K$_{2}$SnCl$_{6}$: {\small A6B2C\_tP18\_128\_eh\_d\_b} \dotfill {\hyperref[A6B2C_tP18_128_eh_d_b]{\pageref{A6B2C_tP18_128_eh_d_b}}} \\
\vspace{-0.65cm} \item Rb$_{2}$TiCu$_{2}$S$_{4}$: {\small A2B2C4D\_tP18\_132\_e\_i\_o\_d} \dotfill {\hyperref[A2B2C4D_tP18_132_e_i_o_d]{\pageref{A2B2C4D_tP18_132_e_i_o_d}}} \\
\vspace{-0.65cm} \item CeCo$_{4}$B$_{4}$: {\small A4BC4\_tP18\_137\_g\_b\_g} \dotfill {\hyperref[A4BC4_tP18_137_g_b_g]{\pageref{A4BC4_tP18_137_g_b_g}}} \\
\end{enumerate}
\vspace{-0.85cm} \item \textbf{ tP20} \vspace{-0.15cm} \\
\begin{enumerate}
\vspace{-0.65cm} \item Gd$_{3}$Al$_{2}$: {\small A2B3\_tP20\_102\_2c\_b2c} \dotfill {\hyperref[A2B3_tP20_102_2c_b2c]{\pageref{A2B3_tP20_102_2c_b2c}}} \\
\vspace{-0.65cm} \item BaGe$_{2}$As$_{2}$: {\small A2BC2\_tP20\_105\_f\_ac\_2e} \dotfill {\hyperref[A2BC2_tP20_105_f_ac_2e]{\pageref{A2BC2_tP20_105_f_ac_2e}}} \\
\vspace{-0.65cm} \item Ru$_{2}$Sn$_{3}$: {\small A2B3\_tP20\_116\_bci\_fj} \dotfill {\hyperref[A2B3_tP20_116_bci_fj]{\pageref{A2B3_tP20_116_bci_fj}}} \\
\vspace{-0.65cm} \item $\beta$-Bi$_{2}$O$_{3}$: {\small A2B3\_tP20\_117\_i\_adgh} \dotfill {\hyperref[A2B3_tP20_117_i_adgh]{\pageref{A2B3_tP20_117_i_adgh}}} \\
\vspace{-0.65cm} \item ThB$_{4}$: {\small A4B\_tP20\_127\_ehj\_g} \dotfill {\hyperref[A4B_tP20_127_ehj_g]{\pageref{A4B_tP20_127_ehj_g}}} \\
\end{enumerate}
\vspace{-0.85cm} \item \textbf{ tP22} \vspace{-0.15cm} \\
\begin{enumerate}
\vspace{-0.65cm} \item Tl$_{4}$HgI$_{6}$: {\small AB6C4\_tP22\_104\_a\_2ac\_c} \dotfill {\hyperref[AB6C4_tP22_104_a_2ac_c]{\pageref{AB6C4_tP22_104_a_2ac_c}}} \\
\end{enumerate}
\vspace{-0.85cm} \item \textbf{ tP24} \vspace{-0.15cm} \\
\begin{enumerate}
\vspace{-0.65cm} \item ThBC: {\small ABC\_tP24\_91\_d\_d\_d} \dotfill {\hyperref[ABC_tP24_91_d_d_d]{\pageref{ABC_tP24_91_d_d_d}}} \\
\vspace{-0.65cm} \item ThBC: {\small ABC\_tP24\_95\_d\_d\_d} \dotfill {\hyperref[ABC_tP24_95_d_d_d]{\pageref{ABC_tP24_95_d_d_d}}} \\
\vspace{-0.65cm} \item \begin{raggedleft}Akermanite: \end{raggedleft} \\ {\small A2BC7D2\_tP24\_113\_e\_a\_cef\_e} \dotfill {\hyperref[A2BC7D2_tP24_113_e_a_cef_e]{\pageref{A2BC7D2_tP24_113_e_a_cef_e}}} \\
\vspace{-0.65cm} \item Downeyite: {\small A2B\_tP24\_135\_gh\_h} \dotfill {\hyperref[A2B_tP24_135_gh_h]{\pageref{A2B_tP24_135_gh_h}}} \\
\end{enumerate}
\vspace{-0.85cm} \item \textbf{ tP26} \vspace{-0.15cm} \\
\begin{enumerate}
\vspace{-0.65cm} \item \begin{raggedleft}Fresnoite: \end{raggedleft} \\ {\small A2B8C2D\_tP26\_100\_c\_abcd\_c\_a} \dotfill {\hyperref[A2B8C2D_tP26_100_c_abcd_c_a]{\pageref{A2B8C2D_tP26_100_c_abcd_c_a}}} \\
\end{enumerate}
\vspace{-0.85cm} \item \textbf{ tP28} \vspace{-0.15cm} \\
\begin{enumerate}
\vspace{-0.65cm} \item Ba$_{5}$In$_{4}$Bi$_{5}$: {\small A5B5C4\_tP28\_104\_ac\_ac\_c} \dotfill {\hyperref[A5B5C4_tP28_104_ac_ac_c]{\pageref{A5B5C4_tP28_104_ac_ac_c}}} \\
\vspace{-0.65cm} \item BiAl$_{2}$S$_{4}$: {\small A2BC4\_tP28\_126\_cd\_e\_k} \dotfill {\hyperref[A2BC4_tP28_126_cd_e_k]{\pageref{A2BC4_tP28_126_cd_e_k}}} \\
\vspace{-0.65cm} \item CuBi$_{2}$O$_{4}$: {\small A2BC4\_tP28\_130\_f\_c\_g} \dotfill {\hyperref[A2BC4_tP28_130_f_c_g]{\pageref{A2BC4_tP28_130_f_c_g}}} \\
\vspace{-0.65cm} \item ZnSb$_{2}$O$_{4}$: {\small A4B2C\_tP28\_135\_gh\_h\_d} \dotfill {\hyperref[A4B2C_tP28_135_gh_h_d]{\pageref{A4B2C_tP28_135_gh_h_d}}} \\
\end{enumerate}
\vspace{-0.85cm} \item \textbf{ tP30} \vspace{-0.15cm} \\
\begin{enumerate}
\vspace{-0.65cm} \item SrBr$_{2}$: {\small A2B\_tP30\_85\_ab2g\_cg} \dotfill {\hyperref[A2B_tP30_85_ab2g_cg]{\pageref{A2B_tP30_85_ab2g_cg}}} \\
\end{enumerate}
\vspace{-0.85cm} \item \textbf{ tP32} \vspace{-0.15cm} \\
\begin{enumerate}
\vspace{-0.65cm} \item Ti$_{3}$P: {\small AB3\_tP32\_86\_g\_3g} \dotfill {\hyperref[AB3_tP32_86_g_3g]{\pageref{AB3_tP32_86_g_3g}}} \\
\vspace{-0.65cm} \item \begin{raggedleft}$\gamma$-MgNiSn: \end{raggedleft} \\ {\small A7B7C2\_tP32\_101\_bde\_ade\_d} \dotfill {\hyperref[A7B7C2_tP32_101_bde_ade_d]{\pageref{A7B7C2_tP32_101_bde_ade_d}}} \\
\vspace{-0.65cm} \item SeO$_{3}$: {\small A3B\_tP32\_114\_3e\_e} \dotfill {\hyperref[A3B_tP32_114_3e_e]{\pageref{A3B_tP32_114_3e_e}}} \\
\vspace{-0.65cm} \item Ir$_{3}$Ga$_{5}$: {\small A5B3\_tP32\_118\_g2i\_aceh} \dotfill {\hyperref[A5B3_tP32_118_g2i_aceh]{\pageref{A5B3_tP32_118_g2i_aceh}}} \\
\vspace{-0.65cm} \item Ba$_{5}$Si$_{3}$: {\small A5B3\_tP32\_130\_cg\_cf} \dotfill {\hyperref[A5B3_tP32_130_cg_cf]{\pageref{A5B3_tP32_130_cg_cf}}} \\
\vspace{-0.65cm} \item $\beta$-V$_{3}$S: {\small AB3\_tP32\_133\_h\_i2j} \dotfill {\hyperref[AB3_tP32_133_h_i2j]{\pageref{AB3_tP32_133_h_i2j}}} \\
\end{enumerate}
\vspace{-0.85cm} \item \textbf{ tP40} \vspace{-0.15cm} \\
\begin{enumerate}
\vspace{-0.65cm} \item Cs$_{3}$P$_{7}$: {\small A3B7\_tP40\_76\_3a\_7a} \dotfill {\hyperref[A3B7_tP40_76_3a_7a]{\pageref{A3B7_tP40_76_3a_7a}}} \\
\vspace{-0.65cm} \item \begin{raggedleft}Na$_{4}$Ti$_{2}$Si$_{8}$O$_{22}$[H$_{2}$O]$_{4}$: \end{raggedleft} \\ {\small A4B2C13D\_tP40\_90\_g\_d\_cef2g\_c} \dotfill {\hyperref[A4B2C13D_tP40_90_g_d_cef2g_c]{\pageref{A4B2C13D_tP40_90_g_d_cef2g_c}}} \\
\vspace{-0.65cm} \item \begin{raggedleft}Ce$_{3}$Si$_{6}$N$_{11}$: \end{raggedleft} \\ {\small A3B11C6\_tP40\_100\_ac\_bc2d\_cd} \dotfill {\hyperref[A3B11C6_tP40_100_ac_bc2d_cd]{\pageref{A3B11C6_tP40_100_ac_bc2d_cd}}} \\
\vspace{-0.65cm} \item FeCu$_{2}$Al$_{7}$: {\small A7B2C\_tP40\_128\_egi\_h\_e} \dotfill {\hyperref[A7B2C_tP40_128_egi_h_e]{\pageref{A7B2C_tP40_128_egi_h_e}}} \\
\vspace{-0.65cm} \item Zn$_{3}$P$_{2}$: {\small A2B3\_tP40\_137\_cdf\_3g} \dotfill {\hyperref[A2B3_tP40_137_cdf_3g]{\pageref{A2B3_tP40_137_cdf_3g}}} \\
\end{enumerate}
\vspace{-0.85cm} \item \textbf{ tP44} \vspace{-0.15cm} \\
\begin{enumerate}
\vspace{-0.65cm} \item Na$_{5}$Fe$_{3}$F$_{14}$: {\small A14B3C5\_tP44\_94\_c3g\_ad\_bg} \dotfill {\hyperref[A14B3C5_tP44_94_c3g_ad_bg]{\pageref{A14B3C5_tP44_94_c3g_ad_bg}}} \\
\end{enumerate}
\vspace{-0.85cm} \item \textbf{ tP48} \vspace{-0.15cm} \\
\begin{enumerate}
\vspace{-0.65cm} \item H$_{2}$S III: {\small A2B\_tP48\_77\_8d\_4d} \dotfill {\hyperref[A2B_tP48_77_8d_4d]{\pageref{A2B_tP48_77_8d_4d}}} \\
\end{enumerate}
\vspace{-0.85cm} \item \textbf{ tP54} \vspace{-0.15cm} \\
\begin{enumerate}
\vspace{-0.65cm} \item \begin{raggedleft}BaCu$_{4}$[VO][PO$_{4}$]$_{4}$: \end{raggedleft} \\ {\small AB4C17D4E\_tP54\_90\_a\_g\_c4g\_g\_c} \dotfill {\hyperref[AB4C17D4E_tP54_90_a_g_c4g_g_c]{\pageref{AB4C17D4E_tP54_90_a_g_c4g_g_c}}} \\
\end{enumerate}
\vspace{-0.85cm} \item \textbf{ tP64} \vspace{-0.15cm} \\
\begin{enumerate}
\vspace{-0.65cm} \item \begin{raggedleft}Pinnoite: \end{raggedleft} \\ {\small A2B6CD7\_tP64\_77\_2d\_6d\_d\_ab6d} \dotfill {\hyperref[A2B6CD7_tP64_77_2d_6d_d_ab6d]{\pageref{A2B6CD7_tP64_77_2d_6d_d_ab6d}}} \\
\vspace{-0.65cm} \item NaZn[OH]$_{3}$: {\small A3BC3D\_tP64\_106\_3c\_c\_3c\_c} \dotfill {\hyperref[A3BC3D_tP64_106_3c_c_3c_c]{\pageref{A3BC3D_tP64_106_3c_c_3c_c}}} \\
\end{enumerate}
\vspace{-0.85cm} \item \textbf{ tP76} \vspace{-0.15cm} \\
\begin{enumerate}
\vspace{-0.65cm} \item \begin{raggedleft}BaCr$_{2}$Ru$_{4}$O$_{12}$: \end{raggedleft} \\ {\small AB2C12D4\_tP76\_75\_2a2b\_2d\_12d\_4d} \dotfill {\hyperref[AB2C12D4_tP76_75_2a2b_2d_12d_4d]{\pageref{AB2C12D4_tP76_75_2a2b_2d_12d_4d}}} \\
\end{enumerate}
\vspace{-0.85cm} \item \textbf{ tP88} \vspace{-0.15cm} \\
\begin{enumerate}
\vspace{-0.65cm} \item Sr$_{2}$As$_{2}$O$_{7}$: {\small A2B7C2\_tP88\_78\_4a\_14a\_4a} \dotfill {\hyperref[A2B7C2_tP88_78_4a_14a_4a]{\pageref{A2B7C2_tP88_78_4a_14a_4a}}} \\
\end{enumerate}
\vspace{-0.85cm} \item \textbf{ tP184} \vspace{-0.15cm} \\
\begin{enumerate}
\vspace{-0.65cm} \item \begin{raggedleft}C$_{17}$FeO$_{4}$Pt: \end{raggedleft} \\ {\small A17BC4D\_tP184\_89\_17p\_p\_4p\_io} \dotfill {\hyperref[A17BC4D_tP184_89_17p_p_4p_io]{\pageref{A17BC4D_tP184_89_17p_p_4p_io}}} \\
\vspace{-0.65cm} \item \begin{raggedleft}AsPh$_{4}$CeS$_{8}$P$_{4}$Me$_{8}$: \end{raggedleft} \\ {\small AB32CD4E8\_tP184\_93\_i\_16p\_af\_2p\_4p} \dotfill {\hyperref[AB32CD4E8_tP184_93_i_16p_af_2p_4p]{\pageref{AB32CD4E8_tP184_93_i_16p_af_2p_4p}}} \\
\end{enumerate}
\end{enumerate}
\section*{\label{sec:strukInd}Strukturbericht Designation Index}
\noindent
\textbf{$\bm{B7}$ \dotfill} \\
\begin{enumerate}
\vspace{-0.75cm} \item Moissanite-15R: {\small AB\_hR10\_160\_5a\_5a} \dotfill {\hyperref[AB_hR10_160_5a_5a]{\pageref{AB_hR10_160_5a_5a}}} \\
\end{enumerate} \vspace{-0.75cm}
\textbf{$\bm{B14}$ \dotfill} \\
\begin{enumerate}
\vspace{-0.75cm} \item Westerveldite: {\small AB\_oP8\_62\_c\_c} \dotfill {\hyperref[AB_oP8_62_c_c]{\pageref{AB_oP8_62_c_c}}} \\
\end{enumerate} \vspace{-0.75cm}
\textbf{$\bm{C13}$ \dotfill} \\
\begin{enumerate}
\vspace{-0.75cm} \item HgI$_{2}$\footnote[5]{\label{note:AB2_tP6_137_a_d-struk}ZrO$_{2}$ and HgI$_{2}$ have similar \AFLOW\ prototype labels ({\it{i.e.}}, same symmetry and set of Wyckoff positions with different stoichiometry labels due to alphabetic ordering of atomic species). They are generated by the same symmetry operations with different sets of parameters.}: {\small AB2\_tP6\_137\_a\_d} \dotfill {\hyperref[AB2_tP6_137_a_d]{\pageref{AB2_tP6_137_a_d}}} \\
\end{enumerate} \vspace{-0.75cm}
\textbf{$\bm{C29}$ \dotfill} \\
\begin{enumerate}
\vspace{-0.75cm} \item SrH$_{2}$: {\small A2B\_oP12\_62\_2c\_c} \dotfill {\hyperref[A2B_oP12_62_2c_c]{\pageref{A2B_oP12_62_2c_c}}} \\
\end{enumerate} \vspace{-0.75cm}
\textbf{$\bm{C47}$ \dotfill} \\
\begin{enumerate}
\vspace{-0.75cm} \item Downeyite: {\small A2B\_tP24\_135\_gh\_h} \dotfill {\hyperref[A2B_tP24_135_gh_h]{\pageref{A2B_tP24_135_gh_h}}} \\
\end{enumerate} \vspace{-0.75cm}
\textbf{$\bm{C50}$ \dotfill} \\
\begin{enumerate}
\vspace{-0.75cm} \item $\alpha$-PdCl$_{2}$: {\small A2B\_oP6\_58\_g\_a} \dotfill {\hyperref[A2B_oP6_58_g_a]{\pageref{A2B_oP6_58_g_a}}} \\
\end{enumerate} \vspace{-0.75cm}
\textbf{$\bm{C_{c}}$ \dotfill} \\
\begin{enumerate}
\vspace{-0.75cm} \item $\alpha$-ThSi$_{2}$: {\small A2B\_tI12\_141\_e\_a} \dotfill {\hyperref[A2B_tI12_141_e_a]{\pageref{A2B_tI12_141_e_a}}} \\
\end{enumerate} \vspace{-0.75cm}
\textbf{$\bm{D0_{8}}$ \dotfill} \\
\begin{enumerate}
\vspace{-0.75cm} \item Molybdite: {\small AB3\_oP16\_62\_c\_3c} \dotfill {\hyperref[AB3_oP16_62_c_3c]{\pageref{AB3_oP16_62_c_3c}}} \\
\end{enumerate} \vspace{-0.75cm}
\textbf{$\bm{D0_{20}}$ \dotfill} \\
\begin{enumerate}
\vspace{-0.75cm} \item $\epsilon$-NiAl$_{3}$: {\small A3B\_oP16\_62\_cd\_c} \dotfill {\hyperref[A3B_oP16_62_cd_c]{\pageref{A3B_oP16_62_cd_c}}} \\
\end{enumerate} \vspace{-0.75cm}
\textbf{$\bm{D0_{21}}$ \dotfill} \\
\begin{enumerate}
\vspace{-0.75cm} \item Cu$_{3}$P: {\small A3B\_hP24\_165\_bdg\_f} \dotfill {\hyperref[A3B_hP24_165_bdg_f]{\pageref{A3B_hP24_165_bdg_f}}} \\
\end{enumerate} \vspace{-0.75cm}
\textbf{$\bm{D0_{24}}$ \dotfill} \\
\begin{enumerate}
\vspace{-0.75cm} \item Ni$_{3}$Ti: {\small A3B\_hP16\_194\_gh\_ac} \dotfill {\hyperref[A3B_hP16_194_gh_ac]{\pageref{A3B_hP16_194_gh_ac}}} \\
\end{enumerate} \vspace{-0.75cm}
\textbf{$\bm{D0_{e}}$ \dotfill} \\
\begin{enumerate}
\vspace{-0.75cm} \item Ni$_{3}$P: {\small A3B\_tI32\_82\_3g\_g} \dotfill {\hyperref[A3B_tI32_82_3g_g]{\pageref{A3B_tI32_82_3g_g}}} \\
\end{enumerate} \vspace{-0.75cm}
\textbf{$\bm{D1_{b}}$ \dotfill} \\
\begin{enumerate}
\vspace{-0.75cm} \item Al$_{4}$U: {\small A4B\_oI20\_74\_beh\_e} \dotfill {\hyperref[A4B_oI20_74_beh_e]{\pageref{A4B_oI20_74_beh_e}}} \\
\end{enumerate} \vspace{-0.75cm}
\textbf{$\bm{D1_{e}}$ \dotfill} \\
\begin{enumerate}
\vspace{-0.75cm} \item ThB$_{4}$: {\small A4B\_tP20\_127\_ehj\_g} \dotfill {\hyperref[A4B_tP20_127_ehj_g]{\pageref{A4B_tP20_127_ehj_g}}} \\
\end{enumerate} \vspace{-0.75cm}
\textbf{$\bm{D1_{f}}$ \dotfill} \\
\begin{enumerate}
\vspace{-0.75cm} \item Mn$_{2}$B: {\small AB2\_oF48\_70\_f\_fg} \dotfill {\hyperref[AB2_oF48_70_f_fg]{\pageref{AB2_oF48_70_f_fg}}} \\
\end{enumerate} \vspace{-0.75cm}
\textbf{$\bm{D2_{3}}$ \dotfill} \\
\begin{enumerate}
\vspace{-0.75cm} \item NaZn$_{13}$: {\small AB13\_cF112\_226\_a\_bi} \dotfill {\hyperref[AB13_cF112_226_a_bi]{\pageref{AB13_cF112_226_a_bi}}} \\
\end{enumerate} \vspace{-0.75cm}
\textbf{$\bm{D2_{h}}$ \dotfill} \\
\begin{enumerate}
\vspace{-0.75cm} \item MnAl$_{6}$: {\small A6B\_oC28\_63\_efg\_c} \dotfill {\hyperref[A6B_oC28_63_efg_c]{\pageref{A6B_oC28_63_efg_c}}} \\
\end{enumerate} \vspace{-0.75cm}
\textbf{$\bm{D3_{1}}$ \dotfill} \\
\begin{enumerate}
\vspace{-0.75cm} \item Calomel: {\small AB\_tI8\_139\_e\_e} \dotfill {\hyperref[AB_tI8_139_e_e]{\pageref{AB_tI8_139_e_e}}} \\
\end{enumerate} \vspace{-0.75cm}
\textbf{$\bm{D5_{2}}$ \dotfill} \\
\begin{enumerate}
\vspace{-0.75cm} \item La$_{2}$O$_{3}$: {\small A2B3\_hP5\_164\_d\_ad} \dotfill {\hyperref[A2B3_hP5_164_d_ad]{\pageref{A2B3_hP5_164_d_ad}}} \\
\end{enumerate} \vspace{-0.75cm}
\textbf{$\bm{D5_{9}}$ \dotfill} \\
\begin{enumerate}
\vspace{-0.75cm} \item Zn$_{3}$P$_{2}$: {\small A2B3\_tP40\_137\_cdf\_3g} \dotfill {\hyperref[A2B3_tP40_137_cdf_3g]{\pageref{A2B3_tP40_137_cdf_3g}}} \\
\end{enumerate} \vspace{-0.75cm}
\textbf{$\bm{D5_{10}}$ \dotfill} \\
\begin{enumerate}
\vspace{-0.75cm} \item Tongbaite: {\small A2B3\_oP20\_62\_2c\_3c} \dotfill {\hyperref[A2B3_oP20_62_2c_3c]{\pageref{A2B3_oP20_62_2c_3c}}} \\
\end{enumerate} \vspace{-0.75cm}
\textbf{$\bm{D7_{1}}$ \dotfill} \\
\begin{enumerate}
\vspace{-0.75cm} \item Al$_{4}$C$_{3}$: {\small A4B3\_hR7\_166\_2c\_ac} \dotfill {\hyperref[A4B3_hR7_166_2c_ac]{\pageref{A4B3_hR7_166_2c_ac}}} \\
\end{enumerate} \vspace{-0.75cm}
\textbf{$\bm{D7_{2}}$ \dotfill} \\
\begin{enumerate}
\vspace{-0.75cm} \item Spinel: {\small A3B4\_cF56\_227\_ad\_e} \dotfill {\hyperref[A3B4_cF56_227_ad_e]{\pageref{A3B4_cF56_227_ad_e}}} \\
\end{enumerate} \vspace{-0.75cm}
\textbf{$\bm{D7_{3}}$ \dotfill} \\
\begin{enumerate}
\vspace{-0.75cm} \item Th$_{3}$P$_{4}$: {\small A4B3\_cI28\_220\_c\_a} \dotfill {\hyperref[A4B3_cI28_220_c_a]{\pageref{A4B3_cI28_220_c_a}}} \\
\end{enumerate} \vspace{-0.75cm}
\textbf{$\bm{D7_{b}}$ \dotfill} \\
\begin{enumerate}
\vspace{-0.75cm} \item Ta$_{3}$B$_{4}$: {\small A4B3\_oI14\_71\_gh\_cg} \dotfill {\hyperref[A4B3_oI14_71_gh_cg]{\pageref{A4B3_oI14_71_gh_cg}}} \\
\end{enumerate} \vspace{-0.75cm}
\textbf{$\bm{D8_{1}}$ \dotfill} \\
\begin{enumerate}
\vspace{-0.75cm} \item $\gamma$-brass: {\small A3B10\_cI52\_229\_e\_fh} \dotfill {\hyperref[A3B10_cI52_229_e_fh]{\pageref{A3B10_cI52_229_e_fh}}} \\
\end{enumerate} \vspace{-0.75cm}
\textbf{$\bm{D8_{3}}$ \dotfill} \\
\begin{enumerate}
\vspace{-0.75cm} \item $\gamma$-brass: {\small A4B9\_cP52\_215\_ei\_3efgi} \dotfill {\hyperref[A4B9_cP52_215_ei_3efgi]{\pageref{A4B9_cP52_215_ei_3efgi}}} \\
\end{enumerate} \vspace{-0.75cm}
\textbf{$\bm{D8_{6}}$ \dotfill} \\
\begin{enumerate}
\vspace{-0.75cm} \item Cu$_{15}$Si$_{4}$: {\small A15B4\_cI76\_220\_ae\_c} \dotfill {\hyperref[A15B4_cI76_220_ae_c]{\pageref{A15B4_cI76_220_ae_c}}} \\
\end{enumerate} \vspace{-0.75cm}
\textbf{$\bm{D8_{10}}$ \dotfill} \\
\begin{enumerate}
\vspace{-0.75cm} \item Al$_{8}$Cr$_{5}$: {\small A8B5\_hR26\_160\_a3bc\_a3b} \dotfill {\hyperref[A8B5_hR26_160_a3bc_a3b]{\pageref{A8B5_hR26_160_a3bc_a3b}}} \\
\end{enumerate} \vspace{-0.75cm}
\textbf{$\bm{D8_{11}}$ \dotfill} \\
\begin{enumerate}
\vspace{-0.75cm} \item Co$_{2}$Al$_{5}$: {\small A5B2\_hP28\_194\_ahk\_ch} \dotfill {\hyperref[A5B2_hP28_194_ahk_ch]{\pageref{A5B2_hP28_194_ahk_ch}}} \\
\end{enumerate} \vspace{-0.75cm}
\textbf{$\bm{D8_{a}}$ \dotfill} \\
\begin{enumerate}
\vspace{-0.75cm} \item Th$_{6}$Mn$_{23}$: {\small A23B6\_cF116\_225\_bd2f\_e} \dotfill {\hyperref[A23B6_cF116_225_bd2f_e]{\pageref{A23B6_cF116_225_bd2f_e}}} \\
\end{enumerate} \vspace{-0.75cm}
\textbf{$\bm{D8_{f}}$ \dotfill} \\
\begin{enumerate}
\vspace{-0.75cm} \item Ir$_{3}$Ge$_{7}$: {\small A7B3\_cI40\_229\_df\_e} \dotfill {\hyperref[A7B3_cI40_229_df_e]{\pageref{A7B3_cI40_229_df_e}}} \\
\end{enumerate} \vspace{-0.75cm}
\textbf{$\bm{D8_{l}}$ \dotfill} \\
\begin{enumerate}
\vspace{-0.75cm} \item Cr$_{5}$B$_{3}$: {\small A3B5\_tI32\_140\_ah\_cl} \dotfill {\hyperref[A3B5_tI32_140_ah_cl]{\pageref{A3B5_tI32_140_ah_cl}}} \\
\end{enumerate} \vspace{-0.75cm}
\textbf{$\bm{D8_{m}}$ \dotfill} \\
\begin{enumerate}
\vspace{-0.75cm} \item W$_{5}$Si$_{3}$: {\small A3B5\_tI32\_140\_ah\_bk} \dotfill {\hyperref[A3B5_tI32_140_ah_bk]{\pageref{A3B5_tI32_140_ah_bk}}} \\
\end{enumerate} \vspace{-0.75cm}
\textbf{$\bm{D10_{2}}$ \dotfill} \\
\begin{enumerate}
\vspace{-0.75cm} \item Fe$_{3}$Th$_{7}$: {\small A3B7\_hP20\_186\_c\_b2c} \dotfill {\hyperref[A3B7_hP20_186_c_b2c]{\pageref{A3B7_hP20_186_c_b2c}}} \\
\end{enumerate} \vspace{-0.75cm}
\textbf{$\bm{E9_{1}}$ \dotfill} \\
\begin{enumerate}
\vspace{-0.75cm} \item Ca$_{3}$Al$_{2}$O$_{6}$: {\small A2B3C6\_cP33\_221\_cd\_ag\_fh} \dotfill {\hyperref[A2B3C6_cP33_221_cd_ag_fh]{\pageref{A2B3C6_cP33_221_cd_ag_fh}}} \\
\end{enumerate} \vspace{-0.75cm}
\textbf{$\bm{E9_{a}}$ \dotfill} \\
\begin{enumerate}
\vspace{-0.75cm} \item FeCu$_{2}$Al$_{7}$: {\small A7B2C\_tP40\_128\_egi\_h\_e} \dotfill {\hyperref[A7B2C_tP40_128_egi_h_e]{\pageref{A7B2C_tP40_128_egi_h_e}}} \\
\end{enumerate} \vspace{-0.75cm}
\textbf{$\bm{E9_{b}}$ \dotfill} \\
\begin{enumerate}
\vspace{-0.75cm} \item \begin{raggedleft}$\pi$-FeMg$_{3}$Al$_{8}$Si$_{6}$: \end{raggedleft} \\ {\small A8BC3D6\_hP18\_189\_bfh\_a\_g\_i} \dotfill {\hyperref[A8BC3D6_hP18_189_bfh_a_g_i]{\pageref{A8BC3D6_hP18_189_bfh_a_g_i}}} \\
\end{enumerate} \vspace{-0.75cm}
\textbf{$\bm{E9_{c}}$ \dotfill} \\
\begin{enumerate}
\vspace{-0.75cm} \item Al$_{9}$Mn$_{3}$Si: {\small A9B3C\_hP26\_194\_hk\_h\_a} \dotfill {\hyperref[A9B3C_hP26_194_hk_h_a]{\pageref{A9B3C_hP26_194_hk_h_a}}} \\
\end{enumerate} \vspace{-0.75cm}
\textbf{$\bm{E9_{d}}$ \dotfill} \\
\begin{enumerate}
\vspace{-0.75cm} \item AlLi$_{3}$N$_{2}$: {\small AB3C2\_cI96\_206\_c\_e\_ad} \dotfill {\hyperref[AB3C2_cI96_206_c_e_ad]{\pageref{AB3C2_cI96_206_c_e_ad}}} \\
\end{enumerate} \vspace{-0.75cm}
\textbf{$\bm{E9_{e}}$ \dotfill} \\
\begin{enumerate}
\vspace{-0.75cm} \item Cubanite: {\small AB2C3\_oP24\_62\_c\_d\_cd} \dotfill {\hyperref[AB2C3_oP24_62_c_d_cd]{\pageref{AB2C3_oP24_62_c_d_cd}}} \\
\end{enumerate} \vspace{-0.75cm}
\textbf{$\bm{F0_{2}}$ \dotfill} \\
\begin{enumerate}
\vspace{-0.75cm} \item Carbonyl Sulphide: {\small ABC\_hR3\_160\_a\_a\_a} \dotfill {\hyperref[ABC_hR3_160_a_a_a]{\pageref{ABC_hR3_160_a_a_a}}} \\
\end{enumerate} \vspace{-0.75cm}
\textbf{$\bm{F5_{13}}$ \dotfill} \\
\begin{enumerate}
\vspace{-0.75cm} \item KBO$_{2}$: {\small ABC2\_hR24\_167\_e\_e\_2e} \dotfill {\hyperref[ABC2_hR24_167_e_e_2e]{\pageref{ABC2_hR24_167_e_e_2e}}} \\
\end{enumerate} \vspace{-0.75cm}
\textbf{$\bm{G3}$ \dotfill} \\
\begin{enumerate}
\vspace{-0.75cm} \item Sodium Chlorate: {\small ABC3\_cP20\_198\_a\_a\_b} \dotfill {\hyperref[ABC3_cP20_198_a_a_b]{\pageref{ABC3_cP20_198_a_a_b}}} \\
\end{enumerate} \vspace{-0.75cm}
\textbf{$\bm{G3_{1}}$ \dotfill} \\
\begin{enumerate}
\vspace{-0.75cm} \item Beryl: {\small A2B3C18D6\_hP58\_192\_c\_f\_lm\_l} \dotfill {\hyperref[A2B3C18D6_hP58_192_c_f_lm_l]{\pageref{A2B3C18D6_hP58_192_c_f_lm_l}}} \\
\end{enumerate} \vspace{-0.75cm}
\textbf{$\bm{H0_{1}}$ \dotfill} \\
\begin{enumerate}
\vspace{-0.75cm} \item Anhydrite: {\small AB4C\_oC24\_63\_c\_fg\_c} \dotfill {\hyperref[AB4C_oC24_63_c_fg_c]{\pageref{AB4C_oC24_63_c_fg_c}}} \\
\end{enumerate} \vspace{-0.75cm}
\textbf{$\bm{H0_{2}}$ \dotfill} \\
\begin{enumerate}
\vspace{-0.75cm} \item Barite: {\small AB4C\_oP24\_62\_c\_2cd\_c} \dotfill {\hyperref[AB4C_oP24_62_c_2cd_c]{\pageref{AB4C_oP24_62_c_2cd_c}}} \\
\end{enumerate} \vspace{-0.75cm}
\textbf{$\bm{J1_{1}}$ \dotfill} \\
\begin{enumerate}
\vspace{-0.75cm} \item K$_{2}$PtCl$_{6}$: {\small A6B2C\_cF36\_225\_e\_c\_a} \dotfill {\hyperref[A6B2C_cF36_225_e_c_a]{\pageref{A6B2C_cF36_225_e_c_a}}} \\
\end{enumerate} \vspace{-0.75cm}
\textbf{$\bm{S1_{2}}$ \dotfill} \\
\begin{enumerate}
\vspace{-0.75cm} \item Forsterite: {\small A2B4C\_oP28\_62\_ac\_2cd\_c} \dotfill {\hyperref[A2B4C_oP28_62_ac_2cd_c]{\pageref{A2B4C_oP28_62_ac_2cd_c}}} \\
\end{enumerate} \vspace{-0.75cm}
\textbf{$\bm{S1_{3}}$ \dotfill} \\
\begin{enumerate}
\vspace{-0.75cm} \item Phenakite: {\small A2B4C\_hR42\_148\_2f\_4f\_f} \dotfill {\hyperref[A2B4C_hR42_148_2f_4f_f]{\pageref{A2B4C_hR42_148_2f_4f_f}}} \\
\end{enumerate} \vspace{-0.75cm}
\textbf{$\bm{S1_{4}}$ \dotfill} \\
\begin{enumerate}
\vspace{-0.75cm} \item Garnet: {\small A2B3C12D3\_cI160\_230\_a\_c\_h\_d} \dotfill {\hyperref[A2B3C12D3_cI160_230_a_c_h_d]{\pageref{A2B3C12D3_cI160_230_a_c_h_d}}} \\
\end{enumerate} \vspace{-0.75cm}
\textbf{$\bm{S2_{1}}$ \dotfill} \\
\begin{enumerate}
\vspace{-0.75cm} \item Thortveitite: {\small A7B2C2\_mC22\_12\_aij\_h\_i} \dotfill {\hyperref[A7B2C2_mC22_12_aij_h_i]{\pageref{A7B2C2_mC22_12_aij_h_i}}} \\
\end{enumerate} \vspace{-0.75cm}
\textbf{$\bm{S5_{3}}$ \dotfill} \\
\begin{enumerate}
\vspace{-0.75cm} \item Akermanite: {\small A2BC7D2\_tP24\_113\_e\_a\_cef\_e} \dotfill {\hyperref[A2BC7D2_tP24_113_e_a_cef_e]{\pageref{A2BC7D2_tP24_113_e_a_cef_e}}} \\
\end{enumerate} \vspace{-0.75cm}
\textbf{None \dotfill} \\
\begin{enumerate}
\vspace{-0.75cm} \item H$_{2}$S: {\small A2B\_aP6\_2\_aei\_i} \dotfill {\hyperref[A2B_aP6_2_aei_i]{\pageref{A2B_aP6_2_aei_i}}} \\
\vspace{-0.75cm} \item Mo$_{8}$P$_{5}$: {\small A8B5\_mP13\_6\_a7b\_3a2b} \dotfill {\hyperref[A8B5_mP13_6_a7b_3a2b]{\pageref{A8B5_mP13_6_a7b_3a2b}}} \\
\vspace{-0.75cm} \item FeNi: {\small AB\_mP4\_6\_2b\_2a} \dotfill {\hyperref[AB_mP4_6_2b_2a]{\pageref{AB_mP4_6_2b_2a}}} \\
\vspace{-0.75cm} \item H$_{2}$S IV: {\small A2B\_mP12\_7\_4a\_2a} \dotfill {\hyperref[A2B_mP12_7_4a_2a]{\pageref{A2B_mP12_7_4a_2a}}} \\
\vspace{-0.75cm} \item As$_{2}$Ba: {\small A2B\_mP18\_7\_6a\_3a} \dotfill {\hyperref[A2B_mP18_7_6a_3a]{\pageref{A2B_mP18_7_6a_3a}}} \\
\vspace{-0.75cm} \item $\epsilon$-WO$_{3}$: {\small A3B\_mP16\_7\_6a\_2a} \dotfill {\hyperref[A3B_mP16_7_6a_2a]{\pageref{A3B_mP16_7_6a_2a}}} \\
\vspace{-0.75cm} \item Rh$_{2}$Ga$_{9}$: {\small A9B2\_mP22\_7\_9a\_2a} \dotfill {\hyperref[A9B2_mP22_7_9a_2a]{\pageref{A9B2_mP22_7_9a_2a}}} \\
\vspace{-0.75cm} \item $\alpha$-P$_3$N$_5$: {\small A5B3\_mC32\_9\_5a\_3a} \dotfill {\hyperref[A5B3_mC32_9_5a_3a]{\pageref{A5B3_mC32_9_5a_3a}}} \\
\vspace{-0.75cm} \item H$_{3}$Cl: {\small AB3\_mC16\_9\_a\_3a} \dotfill {\hyperref[AB3_mC16_9_a_3a]{\pageref{AB3_mC16_9_a_3a}}} \\
\vspace{-0.75cm} \item $\delta$-PdCl$_{2}$: {\small A2B\_mP6\_10\_mn\_bg} \dotfill {\hyperref[A2B_mP6_10_mn_bg]{\pageref{A2B_mP6_10_mn_bg}}} \\
\vspace{-0.75cm} \item H$_{3}$Cl: {\small AB3\_mP16\_10\_mn\_3m3n} \dotfill {\hyperref[AB3_mP16_10_mn_3m3n]{\pageref{AB3_mP16_10_mn_3m3n}}} \\
\vspace{-0.75cm} \item Muthmannite: {\small ABC2\_mP8\_10\_ac\_eh\_mn} \dotfill {\hyperref[ABC2_mP8_10_ac_eh_mn]{\pageref{ABC2_mP8_10_ac_eh_mn}}} \\
\vspace{-0.75cm} \item LiSn: {\small AB\_mP6\_10\_en\_am} \dotfill {\hyperref[AB_mP6_10_en_am]{\pageref{AB_mP6_10_en_am}}} \\
\vspace{-0.75cm} \item S-carbon: {\small A\_mP8\_10\_2m2n} \dotfill {\hyperref[A_mP8_10_2m2n]{\pageref{A_mP8_10_2m2n}}} \\
\vspace{-0.75cm} \item M-carbon: {\small A\_mC16\_12\_4i} \dotfill {\hyperref[A_mC16_12_4i]{\pageref{A_mC16_12_4i}}} \\
\vspace{-0.75cm} \item H$_{2}$S: {\small A2B\_mP12\_13\_2g\_ef} \dotfill {\hyperref[A2B_mP12_13_2g_ef]{\pageref{A2B_mP12_13_2g_ef}}} \\
\vspace{-0.75cm} \item $\gamma$-PdCl$_{2}$: {\small A2B\_mP6\_14\_e\_a} \dotfill {\hyperref[A2B_mP6_14_e_a]{\pageref{A2B_mP6_14_e_a}}} \\
\vspace{-0.75cm} \item $\alpha$-Toluene: {\small A7B8\_mP120\_14\_14e\_16e} \dotfill {\hyperref[A7B8_mP120_14_14e_16e]{\pageref{A7B8_mP120_14_14e_16e}}} \\
\vspace{-0.75cm} \item H$_{3}$Cl: {\small AB3\_mC16\_15\_e\_cf} \dotfill {\hyperref[AB3_mC16_15_e_cf]{\pageref{AB3_mC16_15_e_cf}}} \\
\vspace{-0.75cm} \item H-III: {\small A\_mC24\_15\_2e2f} \dotfill {\hyperref[A_mC24_15_2e2f]{\pageref{A_mC24_15_2e2f}}} \\
\vspace{-0.75cm} \item $\alpha$-Naumannite: {\small A2B\_oP12\_17\_abe\_e} \dotfill {\hyperref[A2B_oP12_17_abe_e]{\pageref{A2B_oP12_17_abe_e}}} \\
\vspace{-0.75cm} \item H$_{3}$Cl: {\small AB3\_oP16\_19\_a\_3a} \dotfill {\hyperref[AB3_oP16_19_a_3a]{\pageref{AB3_oP16_19_a_3a}}} \\
\vspace{-0.75cm} \item Ta$_{2}$H: {\small AB2\_oC6\_21\_a\_k} \dotfill {\hyperref[AB2_oC6_21_a_k]{\pageref{AB2_oC6_21_a_k}}} \\
\vspace{-0.75cm} \item CeRu$_{2}$B$_{2}$: {\small A2BC2\_oF40\_22\_fi\_ad\_gh} \dotfill {\hyperref[A2BC2_oF40_22_fi_ad_gh]{\pageref{A2BC2_oF40_22_fi_ad_gh}}} \\
\vspace{-0.75cm} \item FeS: {\small AB\_oF8\_22\_a\_c} \dotfill {\hyperref[AB_oF8_22_a_c]{\pageref{AB_oF8_22_a_c}}} \\
\vspace{-0.75cm} \item H$_{3}$S: {\small A3B\_oI32\_23\_ij2k\_k} \dotfill {\hyperref[A3B_oI32_23_ij2k_k]{\pageref{A3B_oI32_23_ij2k_k}}} \\
\vspace{-0.75cm} \item \begin{raggedleft}Stannoidite: \end{raggedleft} \\ {\small A8B2C12D2E\_oI50\_23\_bcfk\_i\_3k\_j\_a} \dotfill {\hyperref[A8B2C12D2E_oI50_23_bcfk_i_3k_j_a]{\pageref{A8B2C12D2E_oI50_23_bcfk_i_3k_j_a}}} \\
\vspace{-0.75cm} \item NaFeS$_{2}$: {\small ABC2\_oI16\_23\_ab\_i\_k} \dotfill {\hyperref[ABC2_oI16_23_ab_i_k]{\pageref{ABC2_oI16_23_ab_i_k}}} \\
\vspace{-0.75cm} \item BPS$_{4}$: {\small ABC4\_oI12\_23\_a\_b\_k} \dotfill {\hyperref[ABC4_oI12_23_a_b_k]{\pageref{ABC4_oI12_23_a_b_k}}} \\
\vspace{-0.75cm} \item Weberite: {\small AB7CD2\_oI44\_24\_a\_b3d\_c\_ac} \dotfill {\hyperref[AB7CD2_oI44_24_a_b3d_c_ac]{\pageref{AB7CD2_oI44_24_a_b3d_c_ac}}} \\
\vspace{-0.75cm} \item H$_{2}$S\footnote[1]{\label{note:A2B_oP12_26_abc_ab-struk}H$_{2}$S and $\beta$-SeO$_{2}$ have the same \AFLOW\ prototype label. They are generated by the same symmetry operations with different sets of parameters.}: {\small A2B\_oP12\_26\_abc\_ab} \dotfill {\hyperref[A2B_oP12_26_abc_ab-H2S]{\pageref{A2B_oP12_26_abc_ab-H2S}}} \\
\vspace{-0.75cm} \item $\beta$-SeO$_{2}$\footnoteref{note:A2B_oP12_26_abc_ab-struk}: {\small A2B\_oP12\_26\_abc\_ab} \dotfill {\hyperref[A2B_oP12_26_abc_ab-SeO2]{\pageref{A2B_oP12_26_abc_ab-SeO2}}} \\
\vspace{-0.75cm} \item TlP$_{5}$: {\small A5B\_oP24\_26\_3a3b2c\_ab} \dotfill {\hyperref[A5B_oP24_26_3a3b2c_ab]{\pageref{A5B_oP24_26_3a3b2c_ab}}} \\
\vspace{-0.75cm} \item \begin{raggedleft}Ca$_{4}$Al$_{6}$O$_{16}$S: \end{raggedleft} \\ {\small A6B4C16D\_oP108\_27\_abcd4e\_4e\_16e\_e} \dotfill {\hyperref[A6B4C16D_oP108_27_abcd4e_4e_16e_e]{\pageref{A6B4C16D_oP108_27_abcd4e_4e_16e_e}}} \\
\vspace{-0.75cm} \item ZrO$_{2}$\footnote[3]{\label{note:AB2_oP12_29_a_2a-struk}ZrO$_{2}$ and Pyrite have similar \AFLOW\ prototype labels ({\it{i.e.}}, same symmetry and set of Wyckoff positions with different stoichiometry labels due to alphabetic ordering of atomic species). They are generated by the same symmetry operations with different sets of parameters.}: {\small A2B\_oP12\_29\_2a\_a} \dotfill {\hyperref[A2B_oP12_29_2a_a]{\pageref{A2B_oP12_29_2a_a}}} \\
\vspace{-0.75cm} \item Pyrite\footnoteref{note:AB2_oP12_29_a_2a-struk}: {\small AB2\_oP12\_29\_a\_2a} \dotfill {\hyperref[AB2_oP12_29_a_2a]{\pageref{AB2_oP12_29_a_2a}}} \\
\vspace{-0.75cm} \item Cobaltite: {\small ABC\_oP12\_29\_a\_a\_a} \dotfill {\hyperref[ABC_oP12_29_a_a_a]{\pageref{ABC_oP12_29_a_a_a}}} \\
\vspace{-0.75cm} \item Bi$_{5}$Nb$_{3}$O$_{15}$: {\small A5B3C15\_oP46\_30\_a2c\_bc\_a7c} \dotfill {\hyperref[A5B3C15_oP46_30_a2c_bc_a7c]{\pageref{A5B3C15_oP46_30_a2c_bc_a7c}}} \\
\vspace{-0.75cm} \item CuBrSe$_{3}$: {\small ABC3\_oP20\_30\_2a\_c\_3c} \dotfill {\hyperref[ABC3_oP20_30_2a_c_3c]{\pageref{ABC3_oP20_30_2a_c_3c}}} \\
\vspace{-0.75cm} \item Re$_{2}$O$_{5}$[SO$_{4}$]$_{2}$: {\small A13B2C2\_oP34\_32\_a6c\_c\_c} \dotfill {\hyperref[A13B2C2_oP34_32_a6c_c_c]{\pageref{A13B2C2_oP34_32_a6c_c_c}}} \\
\vspace{-0.75cm} \item $\kappa$-alumina: {\small A2B3\_oP40\_33\_4a\_6a} \dotfill {\hyperref[A2B3_oP40_33_4a_6a]{\pageref{A2B3_oP40_33_4a_6a}}} \\
\vspace{-0.75cm} \item TiAl$_{2}$Br$_{8}$: {\small A2B8C\_oP22\_34\_c\_4c\_a} \dotfill {\hyperref[A2B8C_oP22_34_c_4c_a]{\pageref{A2B8C_oP22_34_c_4c_a}}} \\
\vspace{-0.75cm} \item FeSb$_{2}$: {\small AB2\_oP6\_34\_a\_c} \dotfill {\hyperref[AB2_oP6_34_a_c]{\pageref{AB2_oP6_34_a_c}}} \\
\vspace{-0.75cm} \item V$_{2}$MoO$_{8}$: {\small AB8C2\_oC22\_35\_a\_ab3e\_e} \dotfill {\hyperref[AB8C2_oC22_35_a_ab3e_e]{\pageref{AB8C2_oC22_35_a_ab3e_e}}} \\
\vspace{-0.75cm} \item HCl: {\small AB\_oC8\_36\_a\_a} \dotfill {\hyperref[AB_oC8_36_a_a]{\pageref{AB_oC8_36_a_a}}} \\
\vspace{-0.75cm} \item Li$_{2}$Si$_{2}$O$_{5}$: {\small A2B5C2\_oC36\_37\_d\_c2d\_d} \dotfill {\hyperref[A2B5C2_oC36_37_d_c2d_d]{\pageref{A2B5C2_oC36_37_d_c2d_d}}} \\
\vspace{-0.75cm} \item Ta$_{3}$S$_{2}$: {\small A2B3\_oC40\_39\_2d\_2c2d} \dotfill {\hyperref[A2B3_oC40_39_2d_2c2d]{\pageref{A2B3_oC40_39_2d_2c2d}}} \\
\vspace{-0.75cm} \item VPCl$_{9}$: {\small A9BC\_oC44\_39\_3c3d\_a\_c} \dotfill {\hyperref[A9BC_oC44_39_3c3d_a_c]{\pageref{A9BC_oC44_39_3c3d_a_c}}} \\
\vspace{-0.75cm} \item K$_{2}$CdPb: {\small AB2C\_oC16\_40\_a\_2b\_b} \dotfill {\hyperref[AB2C_oC16_40_a_2b_b]{\pageref{AB2C_oC16_40_a_2b_b}}} \\
\vspace{-0.75cm} \item CeTe$_{3}$: {\small AB3\_oC16\_40\_b\_3b} \dotfill {\hyperref[AB3_oC16_40_b_3b]{\pageref{AB3_oC16_40_b_3b}}} \\
\vspace{-0.75cm} \item W$_{3}$O$_{10}$: {\small A10B3\_oF52\_42\_2abce\_ab} \dotfill {\hyperref[A10B3_oF52_42_2abce_ab]{\pageref{A10B3_oF52_42_2abce_ab}}} \\
\vspace{-0.75cm} \item BN: {\small AB\_oF8\_42\_a\_a} \dotfill {\hyperref[AB_oF8_42_a_a]{\pageref{AB_oF8_42_a_a}}} \\
\vspace{-0.75cm} \item MnGa$_{2}$Sb$_{2}$: {\small A2BC2\_oI20\_45\_c\_b\_c} \dotfill {\hyperref[A2BC2_oI20_45_c_b_c]{\pageref{A2BC2_oI20_45_c_b_c}}} \\
\vspace{-0.75cm} \item TiFeSi: {\small ABC\_oI36\_46\_ac\_bc\_3b} \dotfill {\hyperref[ABC_oI36_46_ac_bc_3b]{\pageref{ABC_oI36_46_ac_bc_3b}}} \\
\vspace{-0.75cm} \item \begin{raggedleft}$\alpha$-RbPr[MoO$_{4}$]$_{2}$: \end{raggedleft} \\ {\small A2B8CD\_oP24\_48\_k\_2m\_d\_b} \dotfill {\hyperref[A2B8CD_oP24_48_k_2m_d_b]{\pageref{A2B8CD_oP24_48_k_2m_d_b}}} \\
\vspace{-0.75cm} \item $\beta$-Ta$_{2}$O$_{5}$: {\small A5B2\_oP14\_49\_dehq\_ab} \dotfill {\hyperref[A5B2_oP14_49_dehq_ab]{\pageref{A5B2_oP14_49_dehq_ab}}} \\
\vspace{-0.75cm} \item CsPr[MoO$_{4}$]$_{2}$: {\small AB2C8D\_oP24\_49\_g\_q\_2qr\_e} \dotfill {\hyperref[AB2C8D_oP24_49_g_q_2qr_e]{\pageref{AB2C8D_oP24_49_g_q_2qr_e}}} \\
\vspace{-0.75cm} \item La$_{2}$NiO$_{4}$: {\small A2BC4\_oP28\_50\_ij\_ac\_ijm} \dotfill {\hyperref[A2BC4_oP28_50_ij_ac_ijm]{\pageref{A2BC4_oP28_50_ij_ac_ijm}}} \\
\vspace{-0.75cm} \item $\alpha$-Tl$_{2}$TeO$_{3}$: {\small A3BC2\_oP48\_50\_3m\_m\_2m} \dotfill {\hyperref[A3BC2_oP48_50_3m_m_2m]{\pageref{A3BC2_oP48_50_3m_m_2m}}} \\
\vspace{-0.75cm} \item GaCl$_{2}$: {\small A2B\_oP24\_52\_2e\_cd} \dotfill {\hyperref[A2B_oP24_52_2e_cd]{\pageref{A2B_oP24_52_2e_cd}}} \\
\vspace{-0.75cm} \item Sr$_{2}$Bi$_{3}$: {\small A3B2\_oP20\_52\_de\_cd} \dotfill {\hyperref[A3B2_oP20_52_de_cd]{\pageref{A3B2_oP20_52_de_cd}}} \\
\vspace{-0.75cm} \item TaNiTe$_{2}$: {\small ABC2\_oP16\_53\_h\_e\_gh} \dotfill {\hyperref[ABC2_oP16_53_h_e_gh]{\pageref{ABC2_oP16_53_h_e_gh}}} \\
\vspace{-0.75cm} \item CuBrSe$_{3}$: {\small ABC3\_oP20\_53\_e\_g\_hi} \dotfill {\hyperref[ABC3_oP20_53_e_g_hi]{\pageref{ABC3_oP20_53_e_g_hi}}} \\
\vspace{-0.75cm} \item BiGaO$_{3}$: {\small ABC3\_oP20\_54\_e\_d\_cf} \dotfill {\hyperref[ABC3_oP20_54_e_d_cf]{\pageref{ABC3_oP20_54_e_d_cf}}} \\
\vspace{-0.75cm} \item GeAs$_{2}$: {\small A2B\_oP24\_55\_2g2h\_gh} \dotfill {\hyperref[A2B_oP24_55_2g2h_gh]{\pageref{A2B_oP24_55_2g2h_gh}}} \\
\vspace{-0.75cm} \item Rh$_{5}$Ge$_{3}$: {\small A3B5\_oP16\_55\_ch\_agh} \dotfill {\hyperref[A3B5_oP16_55_ch_agh]{\pageref{A3B5_oP16_55_ch_agh}}} \\
\vspace{-0.75cm} \item R-carbon: {\small A\_oP16\_55\_2g2h} \dotfill {\hyperref[A_oP16_55_2g2h]{\pageref{A_oP16_55_2g2h}}} \\
\vspace{-0.75cm} \item FeOCl: {\small ABC\_oP6\_59\_a\_b\_a} \dotfill {\hyperref[ABC_oP6_59_a_b_a]{\pageref{ABC_oP6_59_a_b_a}}} \\
\vspace{-0.75cm} \item Rh$_{2}$S$_{3}$: {\small A2B3\_oP20\_60\_d\_cd} \dotfill {\hyperref[A2B3_oP20_60_d_cd]{\pageref{A2B3_oP20_60_d_cd}}} \\
\vspace{-0.75cm} \item WO$_{3}$: {\small A3B\_oP32\_60\_3d\_d} \dotfill {\hyperref[A3B_oP32_60_3d_d]{\pageref{A3B_oP32_60_3d_d}}} \\
\vspace{-0.75cm} \item $\beta$-Toluene: {\small A7B8\_oP120\_60\_7d\_8d} \dotfill {\hyperref[A7B8_oP120_60_7d_8d]{\pageref{A7B8_oP120_60_7d_8d}}} \\
\vspace{-0.75cm} \item Benzene: {\small AB\_oP48\_61\_3c\_3c} \dotfill {\hyperref[AB_oP48_61_3c_3c]{\pageref{AB_oP48_61_3c_3c}}} \\
\vspace{-0.75cm} \item Rasvumite: {\small A2BC3\_oC24\_63\_e\_c\_cg} \dotfill {\hyperref[A2BC3_oC24_63_e_c_cg]{\pageref{A2BC3_oC24_63_e_c_cg}}} \\
\vspace{-0.75cm} \item \begin{raggedleft}La$_{43}$Ni$_{17}$Mg$_{5}$: \end{raggedleft} \\ {\small A43B5C17\_oC260\_63\_c8fg6h\_cfg\_ce3f2h} \dotfill {\hyperref[A43B5C17_oC260_63_c8fg6h_cfg_ce3f2h]{\pageref{A43B5C17_oC260_63_c8fg6h_cfg_ce3f2h}}} \\
\vspace{-0.75cm} \item Post-perovskite: {\small AB3C\_oC20\_63\_a\_cf\_c} \dotfill {\hyperref[AB3C_oC20_63_a_cf_c]{\pageref{AB3C_oC20_63_a_cf_c}}} \\
\vspace{-0.75cm} \item MgSO$_{4}$: {\small AB4C\_oC24\_63\_a\_fg\_c} \dotfill {\hyperref[AB4C_oC24_63_a_fg_c]{\pageref{AB4C_oC24_63_a_fg_c}}} \\
\vspace{-0.75cm} \item H$_{2}$S: {\small A2B\_oC24\_64\_2f\_f} \dotfill {\hyperref[A2B_oC24_64_2f_f]{\pageref{A2B_oC24_64_2f_f}}} \\
\vspace{-0.75cm} \item SrAl$_{2}$Se$_{4}$: {\small A2B4C\_oC28\_66\_l\_kl\_a} \dotfill {\hyperref[A2B4C_oC28_66_l_kl_a]{\pageref{A2B4C_oC28_66_l_kl_a}}} \\
\vspace{-0.75cm} \item H$_{3}$S: {\small A3B\_oC64\_66\_gi2lm\_2l} \dotfill {\hyperref[A3B_oC64_66_gi2lm_2l]{\pageref{A3B_oC64_66_gi2lm_2l}}} \\
\vspace{-0.75cm} \item $\beta$-ThI$_{3}$: {\small A3B\_oC64\_66\_kl2m\_bdl} \dotfill {\hyperref[A3B_oC64_66_kl2m_bdl]{\pageref{A3B_oC64_66_kl2m_bdl}}} \\
\vspace{-0.75cm} \item Al$_{2}$CuIr\footnote[4]{\label{note:ABC2_oC16_67_b_g_ag-struk}Al$_{2}$CuIr and HoCuP$_{2}$ have similar \AFLOW\ prototype labels ({\it{i.e.}}, same symmetry and set of Wyckoff positions with different stoichiometry labels due to alphabetic ordering of atomic species). They are generated by the same symmetry operations with different sets of parameters.}: {\small A2BC\_oC16\_67\_ag\_b\_g} \dotfill {\hyperref[A2BC_oC16_67_ag_b_g]{\pageref{A2BC_oC16_67_ag_b_g}}} \\
\vspace{-0.75cm} \item HoCuP$_{2}$\footnoteref{note:ABC2_oC16_67_b_g_ag-struk}: {\small ABC2\_oC16\_67\_b\_g\_ag} \dotfill {\hyperref[ABC2_oC16_67_b_g_ag]{\pageref{ABC2_oC16_67_b_g_ag}}} \\
\vspace{-0.75cm} \item $\alpha$-FeSe\footnote[2]{\label{note:AB_oC8_67_a_g-struk}$\alpha$-FeSe and $\alpha$-PbO have the same \AFLOW\ prototype label. They are generated by the same symmetry operations with different sets of parameters.}: {\small AB\_oC8\_67\_a\_g} \dotfill {\hyperref[AB_oC8_67_a_g-FeSe]{\pageref{AB_oC8_67_a_g-FeSe}}} \\
\vspace{-0.75cm} \item $\alpha$-PbO\footnoteref{note:AB_oC8_67_a_g-struk}: {\small AB\_oC8\_67\_a\_g} \dotfill {\hyperref[AB_oC8_67_a_g-PbO]{\pageref{AB_oC8_67_a_g-PbO}}} \\
\vspace{-0.75cm} \item PdSn$_{4}$: {\small AB4\_oC20\_68\_a\_i} \dotfill {\hyperref[AB4_oC20_68_a_i]{\pageref{AB4_oC20_68_a_i}}} \\
\vspace{-0.75cm} \item NbPS: {\small ABC\_oI12\_71\_h\_j\_g} \dotfill {\hyperref[ABC_oI12_71_h_j_g]{\pageref{ABC_oI12_71_h_j_g}}} \\
\vspace{-0.75cm} \item KAg[CO$_{3}$]: {\small ABCD3\_oI48\_73\_d\_e\_e\_ef} \dotfill {\hyperref[ABCD3_oI48_73_d_e_e_ef]{\pageref{ABCD3_oI48_73_d_e_e_ef}}} \\
\vspace{-0.75cm} \item KHg$_{2}$: {\small A2B\_oI12\_74\_h\_e} \dotfill {\hyperref[A2B_oI12_74_h_e]{\pageref{A2B_oI12_74_h_e}}} \\
\vspace{-0.75cm} \item \begin{raggedleft}BaCr$_{2}$Ru$_{4}$O$_{12}$: \end{raggedleft} \\ {\small AB2C12D4\_tP76\_75\_2a2b\_2d\_12d\_4d} \dotfill {\hyperref[AB2C12D4_tP76_75_2a2b_2d_12d_4d]{\pageref{AB2C12D4_tP76_75_2a2b_2d_12d_4d}}} \\
\vspace{-0.75cm} \item LaRhC$_{2}$: {\small A2BC\_tP16\_76\_2a\_a\_a} \dotfill {\hyperref[A2BC_tP16_76_2a_a_a]{\pageref{A2BC_tP16_76_2a_a_a}}} \\
\vspace{-0.75cm} \item Cs$_{3}$P$_{7}$: {\small A3B7\_tP40\_76\_3a\_7a} \dotfill {\hyperref[A3B7_tP40_76_3a_7a]{\pageref{A3B7_tP40_76_3a_7a}}} \\
\vspace{-0.75cm} \item Pinnoite: {\small A2B6CD7\_tP64\_77\_2d\_6d\_d\_ab6d} \dotfill {\hyperref[A2B6CD7_tP64_77_2d_6d_d_ab6d]{\pageref{A2B6CD7_tP64_77_2d_6d_d_ab6d}}} \\
\vspace{-0.75cm} \item H$_{2}$S III: {\small A2B\_tP48\_77\_8d\_4d} \dotfill {\hyperref[A2B_tP48_77_8d_4d]{\pageref{A2B_tP48_77_8d_4d}}} \\
\vspace{-0.75cm} \item Sr$_{2}$As$_{2}$O$_{7}$: {\small A2B7C2\_tP88\_78\_4a\_14a\_4a} \dotfill {\hyperref[A2B7C2_tP88_78_4a_14a_4a]{\pageref{A2B7C2_tP88_78_4a_14a_4a}}} \\
\vspace{-0.75cm} \item TlZn$_{2}$Sb$_{2}$: {\small A2BC2\_tI20\_79\_c\_2a\_c} \dotfill {\hyperref[A2BC2_tI20_79_c_2a_c]{\pageref{A2BC2_tI20_79_c_2a_c}}} \\
\vspace{-0.75cm} \item $\beta$-NbO$_{2}$: {\small AB2\_tI48\_80\_2b\_4b} \dotfill {\hyperref[AB2_tI48_80_2b_4b]{\pageref{AB2_tI48_80_2b_4b}}} \\
\vspace{-0.75cm} \item GeSe$_{2}$: {\small AB2\_tP12\_81\_adg\_2h} \dotfill {\hyperref[AB2_tP12_81_adg_2h]{\pageref{AB2_tP12_81_adg_2h}}} \\
\vspace{-0.75cm} \item Ti$_{2}$Ge$_{3}$: {\small A3B2\_tP10\_83\_adk\_j} \dotfill {\hyperref[A3B2_tP10_83_adk_j]{\pageref{A3B2_tP10_83_adk_j}}} \\
\vspace{-0.75cm} \item SrBr$_{2}$: {\small A2B\_tP30\_85\_ab2g\_cg} \dotfill {\hyperref[A2B_tP30_85_ab2g_cg]{\pageref{A2B_tP30_85_ab2g_cg}}} \\
\vspace{-0.75cm} \item Ti$_{3}$P: {\small AB3\_tP32\_86\_g\_3g} \dotfill {\hyperref[AB3_tP32_86_g_3g]{\pageref{AB3_tP32_86_g_3g}}} \\
\vspace{-0.75cm} \item ThCl$_{4}$: {\small A4B\_tI20\_88\_f\_a} \dotfill {\hyperref[A4B_tI20_88_f_a]{\pageref{A4B_tI20_88_f_a}}} \\
\vspace{-0.75cm} \item $\alpha$-NbO$_{2}$: {\small AB2\_tI96\_88\_2f\_4f} \dotfill {\hyperref[AB2_tI96_88_2f_4f]{\pageref{AB2_tI96_88_2f_4f}}} \\
\vspace{-0.75cm} \item C$_{17}$FeO$_{4}$Pt: {\small A17BC4D\_tP184\_89\_17p\_p\_4p\_io} \dotfill {\hyperref[A17BC4D_tP184_89_17p_p_4p_io]{\pageref{A17BC4D_tP184_89_17p_p_4p_io}}} \\
\vspace{-0.75cm} \item \begin{raggedleft}Na$_{4}$Ti$_{2}$Si$_{8}$O$_{22}$[H$_{2}$O]$_{4}$: \end{raggedleft} \\ {\small A4B2C13D\_tP40\_90\_g\_d\_cef2g\_c} \dotfill {\hyperref[A4B2C13D_tP40_90_g_d_cef2g_c]{\pageref{A4B2C13D_tP40_90_g_d_cef2g_c}}} \\
\vspace{-0.75cm} \item \begin{raggedleft}BaCu$_{4}$[VO][PO$_{4}$]$_{4}$: \end{raggedleft} \\ {\small AB4C17D4E\_tP54\_90\_a\_g\_c4g\_g\_c} \dotfill {\hyperref[AB4C17D4E_tP54_90_a_g_c4g_g_c]{\pageref{AB4C17D4E_tP54_90_a_g_c4g_g_c}}} \\
\vspace{-0.75cm} \item ThBC: {\small ABC\_tP24\_91\_d\_d\_d} \dotfill {\hyperref[ABC_tP24_91_d_d_d]{\pageref{ABC_tP24_91_d_d_d}}} \\
\vspace{-0.75cm} \item \begin{raggedleft}AsPh$_{4}$CeS$_{8}$P$_{4}$Me$_{8}$: \end{raggedleft} \\ {\small AB32CD4E8\_tP184\_93\_i\_16p\_af\_2p\_4p} \dotfill {\hyperref[AB32CD4E8_tP184_93_i_16p_af_2p_4p]{\pageref{AB32CD4E8_tP184_93_i_16p_af_2p_4p}}} \\
\vspace{-0.75cm} \item Na$_{5}$Fe$_{3}$F$_{14}$: {\small A14B3C5\_tP44\_94\_c3g\_ad\_bg} \dotfill {\hyperref[A14B3C5_tP44_94_c3g_ad_bg]{\pageref{A14B3C5_tP44_94_c3g_ad_bg}}} \\
\vspace{-0.75cm} \item Li$_{2}$MoF$_{6}$: {\small A6B2C\_tP18\_94\_eg\_c\_a} \dotfill {\hyperref[A6B2C_tP18_94_eg_c_a]{\pageref{A6B2C_tP18_94_eg_c_a}}} \\
\vspace{-0.75cm} \item ThBC: {\small ABC\_tP24\_95\_d\_d\_d} \dotfill {\hyperref[ABC_tP24_95_d_d_d]{\pageref{ABC_tP24_95_d_d_d}}} \\
\vspace{-0.75cm} \item NaGdCu$_{2}$F$_{8}$: {\small A2B8CD\_tI24\_97\_d\_k\_a\_b} \dotfill {\hyperref[A2B8CD_tI24_97_d_k_a_b]{\pageref{A2B8CD_tI24_97_d_k_a_b}}} \\
\vspace{-0.75cm} \item Ta$_{2}$Se$_{8}$I: {\small AB8C2\_tI44\_97\_e\_2k\_cd} \dotfill {\hyperref[AB8C2_tI44_97_e_2k_cd]{\pageref{AB8C2_tI44_97_e_2k_cd}}} \\
\vspace{-0.75cm} \item CdAs$_{2}$: {\small A2B\_tI12\_98\_f\_a} \dotfill {\hyperref[A2B_tI12_98_f_a]{\pageref{A2B_tI12_98_f_a}}} \\
\vspace{-0.75cm} \item Fresnoite: {\small A2B8C2D\_tP26\_100\_c\_abcd\_c\_a} \dotfill {\hyperref[A2B8C2D_tP26_100_c_abcd_c_a]{\pageref{A2B8C2D_tP26_100_c_abcd_c_a}}} \\
\vspace{-0.75cm} \item Ce$_{3}$Si$_{6}$N$_{11}$: {\small A3B11C6\_tP40\_100\_ac\_bc2d\_cd} \dotfill {\hyperref[A3B11C6_tP40_100_ac_bc2d_cd]{\pageref{A3B11C6_tP40_100_ac_bc2d_cd}}} \\
\vspace{-0.75cm} \item $\gamma$-MgNiSn: {\small A7B7C2\_tP32\_101\_bde\_ade\_d} \dotfill {\hyperref[A7B7C2_tP32_101_bde_ade_d]{\pageref{A7B7C2_tP32_101_bde_ade_d}}} \\
\vspace{-0.75cm} \item Gd$_{3}$Al$_{2}$: {\small A2B3\_tP20\_102\_2c\_b2c} \dotfill {\hyperref[A2B3_tP20_102_2c_b2c]{\pageref{A2B3_tP20_102_2c_b2c}}} \\
\vspace{-0.75cm} \item NbTe$_{4}$: {\small AB4\_tP10\_103\_a\_d} \dotfill {\hyperref[AB4_tP10_103_a_d]{\pageref{AB4_tP10_103_a_d}}} \\
\vspace{-0.75cm} \item Ba$_{5}$In$_{4}$Bi$_{5}$: {\small A5B5C4\_tP28\_104\_ac\_ac\_c} \dotfill {\hyperref[A5B5C4_tP28_104_ac_ac_c]{\pageref{A5B5C4_tP28_104_ac_ac_c}}} \\
\vspace{-0.75cm} \item Tl$_{4}$HgI$_{6}$: {\small AB6C4\_tP22\_104\_a\_2ac\_c} \dotfill {\hyperref[AB6C4_tP22_104_a_2ac_c]{\pageref{AB6C4_tP22_104_a_2ac_c}}} \\
\vspace{-0.75cm} \item BaGe$_{2}$As$_{2}$: {\small A2BC2\_tP20\_105\_f\_ac\_2e} \dotfill {\hyperref[A2BC2_tP20_105_f_ac_2e]{\pageref{A2BC2_tP20_105_f_ac_2e}}} \\
\vspace{-0.75cm} \item NaZn[OH]$_{3}$: {\small A3BC3D\_tP64\_106\_3c\_c\_3c\_c} \dotfill {\hyperref[A3BC3D_tP64_106_3c_c_3c_c]{\pageref{A3BC3D_tP64_106_3c_c_3c_c}}} \\
\vspace{-0.75cm} \item Co$_{5}$Ge$_{7}$: {\small A5B7\_tI24\_107\_ac\_abd} \dotfill {\hyperref[A5B7_tI24_107_ac_abd]{\pageref{A5B7_tI24_107_ac_abd}}} \\
\vspace{-0.75cm} \item GeP: {\small AB\_tI4\_107\_a\_a} \dotfill {\hyperref[AB_tI4_107_a_a]{\pageref{AB_tI4_107_a_a}}} \\
\vspace{-0.75cm} \item Sr$_{5}$Si$_{3}$: {\small A3B5\_tI32\_108\_ac\_a2c} \dotfill {\hyperref[A3B5_tI32_108_ac_a2c]{\pageref{A3B5_tI32_108_ac_a2c}}} \\
\vspace{-0.75cm} \item LaPtSi: {\small ABC\_tI12\_109\_a\_a\_a} \dotfill {\hyperref[ABC_tI12_109_a_a_a]{\pageref{ABC_tI12_109_a_a_a}}} \\
\vspace{-0.75cm} \item NbAs: {\small AB\_tI8\_109\_a\_a} \dotfill {\hyperref[AB_tI8_109_a_a]{\pageref{AB_tI8_109_a_a}}} \\
\vspace{-0.75cm} \item Be[BH$_{4}$]$_{2}$: {\small A2BC8\_tI176\_110\_2b\_b\_8b} \dotfill {\hyperref[A2BC8_tI176_110_2b_b_8b]{\pageref{A2BC8_tI176_110_2b_b_8b}}} \\
\vspace{-0.75cm} \item MnF$_{2}$: {\small A2B\_tP12\_111\_2n\_adf} \dotfill {\hyperref[A2B_tP12_111_2n_adf]{\pageref{A2B_tP12_111_2n_adf}}} \\
\vspace{-0.75cm} \item NV: {\small AB\_tP8\_111\_n\_n} \dotfill {\hyperref[AB_tP8_111_n_n]{\pageref{AB_tP8_111_n_n}}} \\
\vspace{-0.75cm} \item $\alpha$-CuAlCl$_{4}$: {\small AB4C\_tP12\_112\_b\_n\_e} \dotfill {\hyperref[AB4C_tP12_112_b_n_e]{\pageref{AB4C_tP12_112_b_n_e}}} \\
\vspace{-0.75cm} \item SeO$_{3}$: {\small A3B\_tP32\_114\_3e\_e} \dotfill {\hyperref[A3B_tP32_114_3e_e]{\pageref{A3B_tP32_114_3e_e}}} \\
\vspace{-0.75cm} \item Pd$_{4}$Se: {\small A4B\_tP10\_114\_e\_a} \dotfill {\hyperref[A4B_tP10_114_e_a]{\pageref{A4B_tP10_114_e_a}}} \\
\vspace{-0.75cm} \item Rh$_{3}$P$_{2}$: {\small A2B3\_tP5\_115\_g\_ag} \dotfill {\hyperref[A2B3_tP5_115_g_ag]{\pageref{A2B3_tP5_115_g_ag}}} \\
\vspace{-0.75cm} \item HgI$_{2}$: {\small AB2\_tP12\_115\_j\_egi} \dotfill {\hyperref[AB2_tP12_115_j_egi]{\pageref{AB2_tP12_115_j_egi}}} \\
\vspace{-0.75cm} \item Ru$_{2}$Sn$_{3}$: {\small A2B3\_tP20\_116\_bci\_fj} \dotfill {\hyperref[A2B3_tP20_116_bci_fj]{\pageref{A2B3_tP20_116_bci_fj}}} \\
\vspace{-0.75cm} \item $\beta$-Bi$_{2}$O$_{3}$: {\small A2B3\_tP20\_117\_i\_adgh} \dotfill {\hyperref[A2B3_tP20_117_i_adgh]{\pageref{A2B3_tP20_117_i_adgh}}} \\
\vspace{-0.75cm} \item RuIn$_{3}$: {\small A3B\_tP16\_118\_ei\_f} \dotfill {\hyperref[A3B_tP16_118_ei_f]{\pageref{A3B_tP16_118_ei_f}}} \\
\vspace{-0.75cm} \item Ir$_{3}$Ga$_{5}$: {\small A5B3\_tP32\_118\_g2i\_aceh} \dotfill {\hyperref[A5B3_tP32_118_g2i_aceh]{\pageref{A5B3_tP32_118_g2i_aceh}}} \\
\vspace{-0.75cm} \item RbGa$_{3}$: {\small A3B\_tI24\_119\_b2i\_af} \dotfill {\hyperref[A3B_tI24_119_b2i_af]{\pageref{A3B_tI24_119_b2i_af}}} \\
\vspace{-0.75cm} \item GaSb: {\small AB\_tI4\_119\_c\_a} \dotfill {\hyperref[AB_tI4_119_c_a]{\pageref{AB_tI4_119_c_a}}} \\
\vspace{-0.75cm} \item KAu$_{4}$Sn$_{2}$: {\small A4BC2\_tI28\_120\_i\_d\_e} \dotfill {\hyperref[A4BC2_tI28_120_i_d_e]{\pageref{A4BC2_tI28_120_i_d_e}}} \\
\vspace{-0.75cm} \item CaRbFe$_{4}$As$_{4}$: {\small A4BC4D\_tP10\_123\_gh\_a\_i\_d} \dotfill {\hyperref[A4BC4D_tP10_123_gh_a_i_d]{\pageref{A4BC4D_tP10_123_gh_a_i_d}}} \\
\vspace{-0.75cm} \item Nb$_{4}$CoSi: {\small AB4C\_tP12\_124\_a\_m\_c} \dotfill {\hyperref[AB4C_tP12_124_a_m_c]{\pageref{AB4C_tP12_124_a_m_c}}} \\
\vspace{-0.75cm} \item NbTe$_{4}$: {\small AB4\_tP10\_124\_a\_m} \dotfill {\hyperref[AB4_tP10_124_a_m]{\pageref{AB4_tP10_124_a_m}}} \\
\vspace{-0.75cm} \item PtPb$_{4}$: {\small A4B\_tP10\_125\_m\_a} \dotfill {\hyperref[A4B_tP10_125_m_a]{\pageref{A4B_tP10_125_m_a}}} \\
\vspace{-0.75cm} \item KCeSe$_{4}$: {\small ABC4\_tP12\_125\_a\_b\_m} \dotfill {\hyperref[ABC4_tP12_125_a_b_m]{\pageref{ABC4_tP12_125_a_b_m}}} \\
\vspace{-0.75cm} \item BiAl$_{2}$S$_{4}$: {\small A2BC4\_tP28\_126\_cd\_e\_k} \dotfill {\hyperref[A2BC4_tP28_126_cd_e_k]{\pageref{A2BC4_tP28_126_cd_e_k}}} \\
\vspace{-0.75cm} \item K$_{2}$SnCl$_{6}$: {\small A6B2C\_tP18\_128\_eh\_d\_b} \dotfill {\hyperref[A6B2C_tP18_128_eh_d_b]{\pageref{A6B2C_tP18_128_eh_d_b}}} \\
\vspace{-0.75cm} \item CuBi$_{2}$O$_{4}$: {\small A2BC4\_tP28\_130\_f\_c\_g} \dotfill {\hyperref[A2BC4_tP28_130_f_c_g]{\pageref{A2BC4_tP28_130_f_c_g}}} \\
\vspace{-0.75cm} \item Ba$_{5}$Si$_{3}$: {\small A5B3\_tP32\_130\_cg\_cf} \dotfill {\hyperref[A5B3_tP32_130_cg_cf]{\pageref{A5B3_tP32_130_cg_cf}}} \\
\vspace{-0.75cm} \item Rb$_{2}$TiCu$_{2}$S$_{4}$: {\small A2B2C4D\_tP18\_132\_e\_i\_o\_d} \dotfill {\hyperref[A2B2C4D_tP18_132_e_i_o_d]{\pageref{A2B2C4D_tP18_132_e_i_o_d}}} \\
\vspace{-0.75cm} \item AgUF$_{6}$: {\small AB6C\_tP16\_132\_d\_io\_a} \dotfill {\hyperref[AB6C_tP16_132_d_io_a]{\pageref{AB6C_tP16_132_d_io_a}}} \\
\vspace{-0.75cm} \item $\beta$-V$_{3}$S: {\small AB3\_tP32\_133\_h\_i2j} \dotfill {\hyperref[AB3_tP32_133_h_i2j]{\pageref{AB3_tP32_133_h_i2j}}} \\
\vspace{-0.75cm} \item ZnSb$_{2}$O$_{4}$: {\small A4B2C\_tP28\_135\_gh\_h\_d} \dotfill {\hyperref[A4B2C_tP28_135_gh_h_d]{\pageref{A4B2C_tP28_135_gh_h_d}}} \\
\vspace{-0.75cm} \item ZrO$_{2}$\footnoteref{note:AB2_tP6_137_a_d-struk}: {\small A2B\_tP6\_137\_d\_a} \dotfill {\hyperref[A2B_tP6_137_d_a]{\pageref{A2B_tP6_137_d_a}}} \\
\vspace{-0.75cm} \item CeCo$_{4}$B$_{4}$: {\small A4BC4\_tP18\_137\_g\_b\_g} \dotfill {\hyperref[A4BC4_tP18_137_g_b_g]{\pageref{A4BC4_tP18_137_g_b_g}}} \\
\vspace{-0.75cm} \item C: {\small A\_tP12\_138\_bi} \dotfill {\hyperref[A_tP12_138_bi]{\pageref{A_tP12_138_bi}}} \\
\vspace{-0.75cm} \item S-III: {\small A\_tI16\_142\_f} \dotfill {\hyperref[A_tI16_142_f]{\pageref{A_tI16_142_f}}} \\
\vspace{-0.75cm} \item Simpsonite: {\small A4B14C3\_hP21\_143\_bd\_ac4d\_d} \dotfill {\hyperref[A4B14C3_hP21_143_bd_ac4d_d]{\pageref{A4B14C3_hP21_143_bd_ac4d_d}}} \\
\vspace{-0.75cm} \item ScRh$_{6}$P$_{4}$: {\small A4B6C\_hP11\_143\_bd\_2d\_a} \dotfill {\hyperref[A4B6C_hP11_143_bd_2d_a]{\pageref{A4B6C_hP11_143_bd_2d_a}}} \\
\vspace{-0.75cm} \item MoS$_{2}$: {\small AB2\_hP12\_143\_cd\_ab2d} \dotfill {\hyperref[AB2_hP12_143_cd_ab2d]{\pageref{AB2_hP12_143_cd_ab2d}}} \\
\vspace{-0.75cm} \item IrGe$_{4}$: {\small A4B\_hP15\_144\_4a\_a} \dotfill {\hyperref[A4B_hP15_144_4a_a]{\pageref{A4B_hP15_144_4a_a}}} \\
\vspace{-0.75cm} \item TeZn: {\small AB\_hP6\_144\_a\_a} \dotfill {\hyperref[AB_hP6_144_a_a]{\pageref{AB_hP6_144_a_a}}} \\
\vspace{-0.75cm} \item \begin{raggedleft}Sheldrickite: \end{raggedleft} \\ {\small A2B3C3DE7\_hP48\_145\_2a\_3a\_3a\_a\_7a} \dotfill {\hyperref[A2B3C3DE7_hP48_145_2a_3a_3a_a_7a]{\pageref{A2B3C3DE7_hP48_145_2a_3a_3a_a_7a}}} \\
\vspace{-0.75cm} \item $\gamma$-Ag$_{3}$SI: {\small A3BC\_hR5\_146\_b\_a\_a} \dotfill {\hyperref[A3BC_hR5_146_b_a_a]{\pageref{A3BC_hR5_146_b_a_a}}} \\
\vspace{-0.75cm} \item FePSe$_{3}$: {\small ABC3\_hR10\_146\_2a\_2a\_2b} \dotfill {\hyperref[ABC3_hR10_146_2a_2a_2b]{\pageref{ABC3_hR10_146_2a_2a_2b}}} \\
\vspace{-0.75cm} \item $\beta$-PdCl$_2$: {\small A2B\_hR18\_148\_2f\_f} \dotfill {\hyperref[A2B_hR18_148_2f_f]{\pageref{A2B_hR18_148_2f_f}}} \\
\vspace{-0.75cm} \item Ti$_{3}$O: {\small AB3\_hP24\_149\_acgi\_3l} \dotfill {\hyperref[AB3_hP24_149_acgi_3l]{\pageref{AB3_hP24_149_acgi_3l}}} \\
\vspace{-0.75cm} \item CrCl$_{3}$: {\small A3B\_hP24\_153\_3c\_2b} \dotfill {\hyperref[A3B_hP24_153_3c_2b]{\pageref{A3B_hP24_153_3c_2b}}} \\
\vspace{-0.75cm} \item S-II: {\small A\_hP9\_154\_bc} \dotfill {\hyperref[A_hP9_154_bc]{\pageref{A_hP9_154_bc}}} \\
\vspace{-0.75cm} \item CdI$_{2}$: {\small AB2\_hP9\_156\_b2c\_3a2bc} \dotfill {\hyperref[AB2_hP9_156_b2c_3a2bc]{\pageref{AB2_hP9_156_b2c_3a2bc}}} \\
\vspace{-0.75cm} \item CuI: {\small AB\_hP12\_156\_2ab3c\_2ab3c} \dotfill {\hyperref[AB_hP12_156_2ab3c_2ab3c]{\pageref{AB_hP12_156_2ab3c_2ab3c}}} \\
\vspace{-0.75cm} \item $\beta$-CuI: {\small AB\_hP4\_156\_ac\_ac} \dotfill {\hyperref[AB_hP4_156_ac_ac]{\pageref{AB_hP4_156_ac_ac}}} \\
\vspace{-0.75cm} \item Ag$_{5}$Pb$_{2}$O$_{6}$: {\small A5B6C2\_hP13\_157\_2ac\_2c\_b} \dotfill {\hyperref[A5B6C2_hP13_157_2ac_2c_b]{\pageref{A5B6C2_hP13_157_2ac_2c_b}}} \\
\vspace{-0.75cm} \item $\beta$-RuCl$_{3}$: {\small A3B\_hP8\_158\_d\_a} \dotfill {\hyperref[A3B_hP8_158_d_a]{\pageref{A3B_hP8_158_d_a}}} \\
\vspace{-0.75cm} \item Bi$_{2}$O$_{3}$: {\small A2B3\_hP20\_159\_bc\_2c} \dotfill {\hyperref[A2B3_hP20_159_bc_2c]{\pageref{A2B3_hP20_159_bc_2c}}} \\
\vspace{-0.75cm} \item Nierite: {\small A4B3\_hP28\_159\_ab2c\_2c} \dotfill {\hyperref[A4B3_hP28_159_ab2c_2c]{\pageref{A4B3_hP28_159_ab2c_2c}}} \\
\vspace{-0.75cm} \item YbBaCo$_{4}$O$_{7}$: {\small AB4C7D\_hP26\_159\_b\_ac\_a2c\_b} \dotfill {\hyperref[AB4C7D_hP26_159_b_ac_a2c_b]{\pageref{AB4C7D_hP26_159_b_ac_a2c_b}}} \\
\vspace{-0.75cm} \item H$_{3}$S: {\small A3B\_hR4\_160\_b\_a} \dotfill {\hyperref[A3B_hR4_160_b_a]{\pageref{A3B_hR4_160_b_a}}} \\
\vspace{-0.75cm} \item $\delta_{H}^{II}$-NW$_2$: {\small AB2\_hP9\_164\_bd\_c2d} \dotfill {\hyperref[AB2_hP9_164_bd_c2d]{\pageref{AB2_hP9_164_bd_c2d}}} \\
\vspace{-0.75cm} \item CuNiSb$_{2}$: {\small ABC2\_hP4\_164\_a\_b\_d} \dotfill {\hyperref[ABC2_hP4_164_a_b_d]{\pageref{ABC2_hP4_164_a_b_d}}} \\
\vspace{-0.75cm} \item SmSI: {\small ABC\_hR6\_166\_c\_c\_c} \dotfill {\hyperref[ABC_hR6_166_c_c_c]{\pageref{ABC_hR6_166_c_c_c}}} \\
\vspace{-0.75cm} \item PrNiO$_{3}$: {\small AB3C\_hR10\_167\_b\_e\_a} \dotfill {\hyperref[AB3C_hR10_167_b_e_a]{\pageref{AB3C_hR10_167_b_e_a}}} \\
\vspace{-0.75cm} \item K$_{2}$Ta$_{4}$O$_{9}$F$_{4}$: {\small A2B13C4\_hP57\_168\_d\_c6d\_2d} \dotfill {\hyperref[A2B13C4_hP57_168_d_c6d_2d]{\pageref{A2B13C4_hP57_168_d_c6d_2d}}} \\
\vspace{-0.75cm} \item Al[PO$_{4}$]: {\small AB4C\_hP72\_168\_2d\_8d\_2d} \dotfill {\hyperref[AB4C_hP72_168_2d_8d_2d]{\pageref{AB4C_hP72_168_2d_8d_2d}}} \\
\vspace{-0.75cm} \item $\alpha$-Al$_{2}$S$_{3}$: {\small A2B3\_hP30\_169\_2a\_3a} \dotfill {\hyperref[A2B3_hP30_169_2a_3a]{\pageref{A2B3_hP30_169_2a_3a}}} \\
\vspace{-0.75cm} \item Al$_{2}$S$_{3}$: {\small A2B3\_hP30\_170\_2a\_3a} \dotfill {\hyperref[A2B3_hP30_170_2a_3a]{\pageref{A2B3_hP30_170_2a_3a}}} \\
\vspace{-0.75cm} \item Sr[S$_{2}$O$_{6}$][H$_{2}$O]$_{4}$: {\small A10B2C\_hP39\_171\_5c\_c\_a} \dotfill {\hyperref[A10B2C_hP39_171_5c_c_a]{\pageref{A10B2C_hP39_171_5c_c_a}}} \\
\vspace{-0.75cm} \item Sr[S$_{2}$O$_{6}$][H$_{2}$O]$_{4}$: {\small A10B2C\_hP39\_172\_5c\_c\_a} \dotfill {\hyperref[A10B2C_hP39_172_5c_c_a]{\pageref{A10B2C_hP39_172_5c_c_a}}} \\
\vspace{-0.75cm} \item PI$_{3}$: {\small A3B\_hP8\_173\_c\_b} \dotfill {\hyperref[A3B_hP8_173_c_b]{\pageref{A3B_hP8_173_c_b}}} \\
\vspace{-0.75cm} \item $\beta$-Si$_{3}$N$_{4}$: {\small A4B3\_hP14\_173\_bc\_c} \dotfill {\hyperref[A4B3_hP14_173_bc_c]{\pageref{A4B3_hP14_173_bc_c}}} \\
\vspace{-0.75cm} \item Fe$_{12}$Zr$_{2}$P$_{7}$: {\small A12B7C2\_hP21\_174\_2j2k\_ajk\_cf} \dotfill {\hyperref[A12B7C2_hP21_174_2j2k_ajk_cf]{\pageref{A12B7C2_hP21_174_2j2k_ajk_cf}}} \\
\vspace{-0.75cm} \item GdSI: {\small ABC\_hP12\_174\_cj\_fk\_aj} \dotfill {\hyperref[ABC_hP12_174_cj_fk_aj]{\pageref{ABC_hP12_174_cj_fk_aj}}} \\
\vspace{-0.75cm} \item Nb$_{7}$Ru$_{6}$B$_{8}$: {\small A8B7C6\_hP21\_175\_ck\_aj\_k} \dotfill {\hyperref[A8B7C6_hP21_175_ck_aj_k]{\pageref{A8B7C6_hP21_175_ck_aj_k}}} \\
\vspace{-0.75cm} \item Mg[NH]: {\small ABC\_hP36\_175\_jk\_jk\_jk} \dotfill {\hyperref[ABC_hP36_175_jk_jk_jk]{\pageref{ABC_hP36_175_jk_jk_jk}}} \\
\vspace{-0.75cm} \item Er$_{3}$Ru$_{2}$: {\small A3B2\_hP10\_176\_h\_bd} \dotfill {\hyperref[A3B2_hP10_176_h_bd]{\pageref{A3B2_hP10_176_h_bd}}} \\
\vspace{-0.75cm} \item Fe$_{3}$Te$_{3}$Tl: {\small A3B3C\_hP14\_176\_h\_h\_d} \dotfill {\hyperref[A3B3C_hP14_176_h_h_d]{\pageref{A3B3C_hP14_176_h_h_d}}} \\
\vspace{-0.75cm} \item UCl$_{3}$: {\small A3B\_hP8\_176\_h\_d} \dotfill {\hyperref[A3B_hP8_176_h_d]{\pageref{A3B_hP8_176_h_d}}} \\
\vspace{-0.75cm} \item SiO$_{2}$: {\small A2B\_hP36\_177\_j2lm\_n} \dotfill {\hyperref[A2B_hP36_177_j2lm_n]{\pageref{A2B_hP36_177_j2lm_n}}} \\
\vspace{-0.75cm} \item AuF$_{3}$: {\small AB3\_hP24\_178\_b\_ac} \dotfill {\hyperref[AB3_hP24_178_b_ac]{\pageref{AB3_hP24_178_b_ac}}} \\
\vspace{-0.75cm} \item Sc-V: {\small A\_hP6\_178\_a} \dotfill {\hyperref[A_hP6_178_a]{\pageref{A_hP6_178_a}}} \\
\vspace{-0.75cm} \item AuF$_{3}$: {\small AB3\_hP24\_179\_b\_ac} \dotfill {\hyperref[AB3_hP24_179_b_ac]{\pageref{AB3_hP24_179_b_ac}}} \\
\vspace{-0.75cm} \item $\beta$-SiO$_{2}$: {\small A2B\_hP9\_181\_j\_c} \dotfill {\hyperref[A2B_hP9_181_j_c]{\pageref{A2B_hP9_181_j_c}}} \\
\vspace{-0.75cm} \item AuCN: {\small ABC\_hP3\_183\_a\_a\_a} \dotfill {\hyperref[ABC_hP3_183_a_a_a]{\pageref{ABC_hP3_183_a_a_a}}} \\
\vspace{-0.75cm} \item CrFe$_{3}$NiSn$_{5}$: {\small AB\_hP6\_183\_c\_ab} \dotfill {\hyperref[AB_hP6_183_c_ab]{\pageref{AB_hP6_183_c_ab}}} \\
\vspace{-0.75cm} \item Al[PO$_{4}$]: {\small AB4C\_hP72\_184\_d\_4d\_d} \dotfill {\hyperref[AB4C_hP72_184_d_4d_d]{\pageref{AB4C_hP72_184_d_4d_d}}} \\
\vspace{-0.75cm} \item KNiCl$_{3}$: {\small A3BC\_hP30\_185\_cd\_c\_ab} \dotfill {\hyperref[A3BC_hP30_185_cd_c_ab]{\pageref{A3BC_hP30_185_cd_c_ab}}} \\
\vspace{-0.75cm} \item Cu$_{3}$P\footnote[6]{\label{note:AB3_hP24_185_c_ab2c-struk}Cu$_{3}$P and Na$_{3}$As have similar \AFLOW\ prototype labels ({\it{i.e.}}, same symmetry and set of Wyckoff positions with different stoichiometry labels due to alphabetic ordering of atomic species). They are generated by the same symmetry operations with different sets of parameters.}: {\small A3B\_hP24\_185\_ab2c\_c} \dotfill {\hyperref[A3B_hP24_185_ab2c_c]{\pageref{A3B_hP24_185_ab2c_c}}} \\
\vspace{-0.75cm} \item $\beta$-RuCl$_{3}$: {\small A3B\_hP8\_185\_c\_a} \dotfill {\hyperref[A3B_hP8_185_c_a]{\pageref{A3B_hP8_185_c_a}}} \\
\vspace{-0.75cm} \item Na$_{3}$As\footnoteref{note:AB3_hP24_185_c_ab2c-struk}: {\small AB3\_hP24\_185\_c\_ab2c} \dotfill {\hyperref[AB3_hP24_185_c_ab2c]{\pageref{AB3_hP24_185_c_ab2c}}} \\
\vspace{-0.75cm} \item Re$_{3}$N: {\small AB3\_hP4\_187\_e\_fh} \dotfill {\hyperref[AB3_hP4_187_e_fh]{\pageref{AB3_hP4_187_e_fh}}} \\
\vspace{-0.75cm} \item LiScI$_{3}$: {\small A3BC\_hP10\_188\_k\_a\_e} \dotfill {\hyperref[A3BC_hP10_188_k_a_e]{\pageref{A3BC_hP10_188_k_a_e}}} \\
\vspace{-0.75cm} \item BaSi$_{4}$O$_{9}$: {\small AB9C4\_hP28\_188\_e\_kl\_ak} \dotfill {\hyperref[AB9C4_hP28_188_e_kl_ak]{\pageref{AB9C4_hP28_188_e_kl_ak}}} \\
\vspace{-0.75cm} \item \begin{raggedleft}$\pi$-FeMg$_{3}$Al$_{9}$Si$_{5}$: \end{raggedleft} \\ {\small A9BC3D5\_hP18\_189\_fi\_a\_g\_bh} \dotfill {\hyperref[A9BC3D5_hP18_189_fi_a_g_bh]{\pageref{A9BC3D5_hP18_189_fi_a_g_bh}}} \\
\vspace{-0.75cm} \item Li$_{2}$Sb: {\small A2B\_hP18\_190\_gh\_bf} \dotfill {\hyperref[A2B_hP18_190_gh_bf]{\pageref{A2B_hP18_190_gh_bf}}} \\
\vspace{-0.75cm} \item $\alpha$-Sm$_{3}$Ge$_{5}$: {\small A5B3\_hP16\_190\_bdh\_g} \dotfill {\hyperref[A5B3_hP16_190_bdh_g]{\pageref{A5B3_hP16_190_bdh_g}}} \\
\vspace{-0.75cm} \item Troilite: {\small AB\_hP24\_190\_i\_afh} \dotfill {\hyperref[AB_hP24_190_i_afh]{\pageref{AB_hP24_190_i_afh}}} \\
\vspace{-0.75cm} \item AlPO$_{4}$: {\small AB2\_hP72\_192\_m\_j2kl} \dotfill {\hyperref[AB2_hP72_192_m_j2kl]{\pageref{AB2_hP72_192_m_j2kl}}} \\
\vspace{-0.75cm} \item Mavlyanovite: {\small A5B3\_hP16\_193\_dg\_g} \dotfill {\hyperref[A5B3_hP16_193_dg_g]{\pageref{A5B3_hP16_193_dg_g}}} \\
\vspace{-0.75cm} \item PrRu$_{4}$P$_{12}$: {\small A12BC4\_cP34\_195\_2j\_ab\_2e} \dotfill {\hyperref[A12BC4_cP34_195_2j_ab_2e]{\pageref{A12BC4_cP34_195_2j_ab_2e}}} \\
\vspace{-0.75cm} \item Cu$_{2}$Fe[CN]$_{6}$: {\small A12B2C\_cF60\_196\_h\_bc\_a} \dotfill {\hyperref[A12B2C_cF60_196_h_bc_a]{\pageref{A12B2C_cF60_196_h_bc_a}}} \\
\vspace{-0.75cm} \item \begin{raggedleft}MgB$_{12}$H$_{12}$[H$_{2}$O]$_{12}$: \end{raggedleft} \\ {\small A12B36CD12\_cF488\_196\_2h\_6h\_ac\_fgh} \dotfill {\hyperref[A12B36CD12_cF488_196_2h_6h_ac_fgh]{\pageref{A12B36CD12_cF488_196_2h_6h_ac_fgh}}} \\
\vspace{-0.75cm} \item Mg$_{2}$Zn$_{11}$: {\small A2B11\_cP39\_200\_f\_aghij} \dotfill {\hyperref[A2B11_cP39_200_f_aghij]{\pageref{A2B11_cP39_200_f_aghij}}} \\
\vspace{-0.75cm} \item KSbO$_{3}$: {\small AB3C\_cP60\_201\_ce\_fh\_g} \dotfill {\hyperref[AB3C_cP60_201_ce_fh_g]{\pageref{AB3C_cP60_201_ce_fh_g}}} \\
\vspace{-0.75cm} \item KB$_{6}$H$_{6}$: {\small A6B6C\_cF104\_202\_h\_h\_c} \dotfill {\hyperref[A6B6C_cF104_202_h_h_c]{\pageref{A6B6C_cF104_202_h_h_c}}} \\
\vspace{-0.75cm} \item \begin{raggedleft}FCC C$_{60}$ Buckminsterfullerine: \end{raggedleft} \\ {\small A\_cF240\_202\_h2i} \dotfill {\hyperref[A_cF240_202_h2i]{\pageref{A_cF240_202_h2i}}} \\
\vspace{-0.75cm} \item Pyrochlore: {\small A2BCD3E6\_cF208\_203\_e\_c\_d\_f\_g} \dotfill {\hyperref[A2BCD3E6_cF208_203_e_c_d_f_g]{\pageref{A2BCD3E6_cF208_203_e_c_d_f_g}}} \\
\vspace{-0.75cm} \item Tychite: {\small A4B2C6D16E\_cF232\_203\_e\_d\_f\_eg\_a} \dotfill {\hyperref[A4B2C6D16E_cF232_203_e_d_f_eg_a]{\pageref{A4B2C6D16E_cF232_203_e_d_f_eg_a}}} \\
\vspace{-0.75cm} \item Rb$_{3}$AsSe$_{16}$: {\small AB3C16\_cF160\_203\_b\_ad\_eg} \dotfill {\hyperref[AB3C16_cF160_203_b_ad_eg]{\pageref{AB3C16_cF160_203_b_ad_eg}}} \\
\vspace{-0.75cm} \item Ca$_{3}$Al$_{2}$O$_{6}$: {\small A2B3C6\_cP264\_205\_2d\_ab2c2d\_6d} \dotfill {\hyperref[A2B3C6_cP264_205_2d_ab2c2d_6d]{\pageref{A2B3C6_cP264_205_2d_ab2c2d_6d}}} \\
\vspace{-0.75cm} \item \begin{raggedleft}Simple Cubic C$_{60}$ Buckminsterfullerine: \end{raggedleft} \\ {\small A\_cP240\_205\_10d} \dotfill {\hyperref[A_cP240_205_10d]{\pageref{A_cP240_205_10d}}} \\
\vspace{-0.75cm} \item Pd$_{17}$Se$_{15}$: {\small A17B15\_cP64\_207\_acfk\_eij} \dotfill {\hyperref[A17B15_cP64_207_acfk_eij]{\pageref{A17B15_cP64_207_acfk_eij}}} \\
\vspace{-0.75cm} \item PH$_{3}$: {\small A3B\_cP16\_208\_j\_b} \dotfill {\hyperref[A3B_cP16_208_j_b]{\pageref{A3B_cP16_208_j_b}}} \\
\vspace{-0.75cm} \item \begin{raggedleft}Cs$_{2}$ZnFe[CN]$_{6}$: \end{raggedleft} \\ {\small A6B2CD6E\_cP64\_208\_m\_ad\_b\_m\_c} \dotfill {\hyperref[A6B2CD6E_cP64_208_m_ad_b_m_c]{\pageref{A6B2CD6E_cP64_208_m_ad_b_m_c}}} \\
\vspace{-0.75cm} \item F$_{6}$KP: {\small A24BC\_cF104\_209\_j\_a\_b} \dotfill {\hyperref[A24BC_cF104_209_j_a_b]{\pageref{A24BC_cF104_209_j_a_b}}} \\
\vspace{-0.75cm} \item Te[OH]$_{6}$: {\small A12B6C\_cF608\_210\_4h\_2h\_e} \dotfill {\hyperref[A12B6C_cF608_210_4h_2h_e]{\pageref{A12B6C_cF608_210_4h_2h_e}}} \\
\vspace{-0.75cm} \item SiO$_{2}$: {\small A2B\_cI72\_211\_hi\_i} \dotfill {\hyperref[A2B_cI72_211_hi_i]{\pageref{A2B_cI72_211_hi_i}}} \\
\vspace{-0.75cm} \item SrSi$_{2}$: {\small A2B\_cP12\_212\_c\_a} \dotfill {\hyperref[A2B_cP12_212_c_a]{\pageref{A2B_cP12_212_c_a}}} \\
\vspace{-0.75cm} \item Ca$_{3}$PI$_{3}$: {\small A3B3C\_cI56\_214\_g\_h\_a} \dotfill {\hyperref[A3B3C_cI56_214_g_h_a]{\pageref{A3B3C_cI56_214_g_h_a}}} \\
\vspace{-0.75cm} \item Petzite: {\small A3BC2\_cI48\_214\_f\_a\_e} \dotfill {\hyperref[A3BC2_cI48_214_f_a_e]{\pageref{A3BC2_cI48_214_f_a_e}}} \\
\vspace{-0.75cm} \item \begin{raggedleft}Quartenary Heusler: \end{raggedleft} \\ {\small ABCD\_cF16\_216\_c\_d\_b\_a} \dotfill {\hyperref[ABCD_cF16_216_c_d_b_a]{\pageref{ABCD_cF16_216_c_d_b_a}}} \\
\vspace{-0.75cm} \item Ag$_{3}$[PO$_{4}$]: {\small A3B4C\_cP16\_218\_c\_e\_a} \dotfill {\hyperref[A3B4C_cP16_218_c_e_a]{\pageref{A3B4C_cP16_218_c_e_a}}} \\
\vspace{-0.75cm} \item Boracite: {\small A7BC3D13\_cF192\_219\_de\_b\_c\_ah} \dotfill {\hyperref[A7BC3D13_cF192_219_de_b_c_ah]{\pageref{A7BC3D13_cF192_219_de_b_c_ah}}} \\
\vspace{-0.75cm} \item Ce$_{5}$Mo$_{3}$O$_{16}$: {\small A5B3C16\_cP96\_222\_ce\_d\_fi} \dotfill {\hyperref[A5B3C16_cP96_222_ce_d_fi]{\pageref{A5B3C16_cP96_222_ce_d_fi}}} \\
\vspace{-0.75cm} \item \begin{raggedleft}Pyrochlore Iridate: \end{raggedleft} \\ {\small A2B2C7\_cF88\_227\_c\_d\_af} \dotfill {\hyperref[A2B2C7_cF88_227_c_d_af]{\pageref{A2B2C7_cF88_227_c_d_af}}} \\
\vspace{-0.75cm} \item CuCrCl$_{5}$[NH$_{3}$]$_{6}$: {\small A5BCD6\_cF416\_228\_eg\_c\_b\_h} \dotfill {\hyperref[A5BCD6_cF416_228_eg_c_b_h]{\pageref{A5BCD6_cF416_228_eg_c_b_h}}} \\
\vspace{-0.75cm} \item TeO$_{6}$H$_{6}$: {\small A6B\_cF224\_228\_h\_c} \dotfill {\hyperref[A6B_cF224_228_h_c]{\pageref{A6B_cF224_228_h_c}}} \\
\vspace{-0.75cm} \item $\beta$-Hg$_{4}$Pt: {\small A4B\_cI10\_229\_c\_a} \dotfill {\hyperref[A4B_cI10_229_c_a]{\pageref{A4B_cI10_229_c_a}}} \\
\end{enumerate}
\section*{\label{sec:dupInd}Duplicate \AFLOW\ Label Index}
\noindent
\textbf{A2B\_oP12\_26\_abc\_ab \dotfill} \\
\begin{enumerate}
\vspace{-0.75cm} \item H$_{2}$S \dotfill {\hyperref[A2B_oP12_26_abc_ab-H2S]{\pageref{A2B_oP12_26_abc_ab-H2S}}} \\
\vspace{-0.75cm} \item $\beta$-SeO$_{2}$ \dotfill {\hyperref[A2B_oP12_26_abc_ab-SeO2]{\pageref{A2B_oP12_26_abc_ab-SeO2}}} \\
\end{enumerate} \vspace{-0.75cm}
\textbf{AB\_oC8\_67\_a\_g \dotfill} \\
\begin{enumerate}
\vspace{-0.75cm} \item $\alpha$-FeSe \dotfill {\hyperref[AB_oC8_67_a_g-FeSe]{\pageref{AB_oC8_67_a_g-FeSe}}} \\
\vspace{-0.75cm} \item $\alpha$-PbO \dotfill {\hyperref[AB_oC8_67_a_g-PbO]{\pageref{AB_oC8_67_a_g-PbO}}} \\
\end{enumerate} \vspace{-0.75cm}
\section*{\label{sec:simInd}Similar \AFLOW\ Label Index}
\noindent
\textbf{A2B\_oP12\_29\_2a\_a \dotfill} \\
\begin{enumerate}
\vspace{-0.75cm} \item ZrO$_{2}$: {\small A2B\_oP12\_29\_2a\_a} \dotfill {\hyperref[A2B_oP12_29_2a_a]{\pageref{A2B_oP12_29_2a_a}}} \\
\vspace{-0.75cm} \item Pyrite: {\small AB2\_oP12\_29\_a\_2a} \dotfill {\hyperref[AB2_oP12_29_a_2a]{\pageref{AB2_oP12_29_a_2a}}} \\
\end{enumerate} \vspace{-0.75cm}
\textbf{A2BC\_oC16\_67\_ag\_b\_g \dotfill} \\
\begin{enumerate}
\vspace{-0.75cm} \item Al$_{2}$CuIr: {\small A2BC\_oC16\_67\_ag\_b\_g} \dotfill {\hyperref[A2BC_oC16_67_ag_b_g]{\pageref{A2BC_oC16_67_ag_b_g}}} \\
\vspace{-0.75cm} \item HoCuP$_{2}$: {\small ABC2\_oC16\_67\_b\_g\_ag} \dotfill {\hyperref[ABC2_oC16_67_b_g_ag]{\pageref{ABC2_oC16_67_b_g_ag}}} \\
\end{enumerate} \vspace{-0.75cm}
\textbf{A2B\_tP6\_137\_d\_a \dotfill} \\
\begin{enumerate}
\vspace{-0.75cm} \item ZrO$_{2}$: {\small A2B\_tP6\_137\_d\_a} \dotfill {\hyperref[A2B_tP6_137_d_a]{\pageref{A2B_tP6_137_d_a}}} \\
\vspace{-0.75cm} \item HgI$_{2}$: {\small AB2\_tP6\_137\_a\_d} \dotfill {\hyperref[AB2_tP6_137_a_d]{\pageref{AB2_tP6_137_a_d}}} \\
\end{enumerate} \vspace{-0.75cm}
\textbf{A3B\_hP24\_185\_ab2c\_c \dotfill} \\
\begin{enumerate}
\vspace{-0.75cm} \item Cu$_{3}$P: {\small A3B\_hP24\_185\_ab2c\_c} \dotfill {\hyperref[A3B_hP24_185_ab2c_c]{\pageref{A3B_hP24_185_ab2c_c}}} \\
\vspace{-0.75cm} \item Na$_{3}$As: {\small AB3\_hP24\_185\_c\_ab2c} \dotfill {\hyperref[AB3_hP24_185_c_ab2c]{\pageref{AB3_hP24_185_c_ab2c}}} \\
\end{enumerate} \vspace{-0.75cm}
\section*{\label{sec:cifInd}CIF Index}
\noindent
\begin{enumerate}
\vspace{-0.75cm} \item $\alpha$-Al$_{2}$S$_{3}$: {\small A2B3\_hP30\_169\_2a\_3a} \dotfill {\hyperref[A2B3_hP30_169_2a_3a_cif]{\pageref{A2B3_hP30_169_2a_3a_cif}}} \\
\vspace{-0.75cm} \item $\alpha$-CuAlCl$_{4}$: {\small AB4C\_tP12\_112\_b\_n\_e} \dotfill {\hyperref[AB4C_tP12_112_b_n_e_cif]{\pageref{AB4C_tP12_112_b_n_e_cif}}} \\
\vspace{-0.75cm} \item $\alpha$-FeSe\footnote[2]{\label{note:AB_oC8_67_a_g-cif}$\alpha$-FeSe and $\alpha$-PbO have the same \AFLOW\ prototype label. They are generated by the same symmetry operations with different sets of parameters.}: {\small AB\_oC8\_67\_a\_g} \dotfill {\hyperref[AB_oC8_67_a_g-FeSe_cif]{\pageref{AB_oC8_67_a_g-FeSe_cif}}} \\
\vspace{-0.75cm} \item $\alpha$-Naumannite: {\small A2B\_oP12\_17\_abe\_e} \dotfill {\hyperref[A2B_oP12_17_abe_e_cif]{\pageref{A2B_oP12_17_abe_e_cif}}} \\
\vspace{-0.75cm} \item $\alpha$-NbO$_{2}$: {\small AB2\_tI96\_88\_2f\_4f} \dotfill {\hyperref[AB2_tI96_88_2f_4f_cif]{\pageref{AB2_tI96_88_2f_4f_cif}}} \\
\vspace{-0.75cm} \item $\alpha$-P$_3$N$_5$: {\small A5B3\_mC32\_9\_5a\_3a} \dotfill {\hyperref[A5B3_mC32_9_5a_3a_cif]{\pageref{A5B3_mC32_9_5a_3a_cif}}} \\
\vspace{-0.75cm} \item $\alpha$-PbO\footnoteref{note:AB_oC8_67_a_g-cif}: {\small AB\_oC8\_67\_a\_g} \dotfill {\hyperref[AB_oC8_67_a_g-PbO_cif]{\pageref{AB_oC8_67_a_g-PbO_cif}}} \\
\vspace{-0.75cm} \item $\alpha$-PdCl$_{2}$: {\small A2B\_oP6\_58\_g\_a} \dotfill {\hyperref[A2B_oP6_58_g_a_cif]{\pageref{A2B_oP6_58_g_a_cif}}} \\
\vspace{-0.75cm} \item \begin{raggedleft}$\alpha$-RbPr[MoO$_{4}$]$_{2}$: \end{raggedleft} \\ {\small A2B8CD\_oP24\_48\_k\_2m\_d\_b} \dotfill {\hyperref[A2B8CD_oP24_48_k_2m_d_b_cif]{\pageref{A2B8CD_oP24_48_k_2m_d_b_cif}}} \\
\vspace{-0.75cm} \item $\alpha$-Sm$_{3}$Ge$_{5}$: {\small A5B3\_hP16\_190\_bdh\_g} \dotfill {\hyperref[A5B3_hP16_190_bdh_g_cif]{\pageref{A5B3_hP16_190_bdh_g_cif}}} \\
\vspace{-0.75cm} \item $\alpha$-ThSi$_{2}$: {\small A2B\_tI12\_141\_e\_a} \dotfill {\hyperref[A2B_tI12_141_e_a_cif]{\pageref{A2B_tI12_141_e_a_cif}}} \\
\vspace{-0.75cm} \item $\alpha$-Tl$_{2}$TeO$_{3}$: {\small A3BC2\_oP48\_50\_3m\_m\_2m} \dotfill {\hyperref[A3BC2_oP48_50_3m_m_2m_cif]{\pageref{A3BC2_oP48_50_3m_m_2m_cif}}} \\
\vspace{-0.75cm} \item $\alpha$-Toluene: {\small A7B8\_mP120\_14\_14e\_16e} \dotfill {\hyperref[A7B8_mP120_14_14e_16e_cif]{\pageref{A7B8_mP120_14_14e_16e_cif}}} \\
\vspace{-0.75cm} \item $\beta$-Bi$_{2}$O$_{3}$: {\small A2B3\_tP20\_117\_i\_adgh} \dotfill {\hyperref[A2B3_tP20_117_i_adgh_cif]{\pageref{A2B3_tP20_117_i_adgh_cif}}} \\
\vspace{-0.75cm} \item $\beta$-CuI: {\small AB\_hP4\_156\_ac\_ac} \dotfill {\hyperref[AB_hP4_156_ac_ac_cif]{\pageref{AB_hP4_156_ac_ac_cif}}} \\
\vspace{-0.75cm} \item $\beta$-Hg$_{4}$Pt: {\small A4B\_cI10\_229\_c\_a} \dotfill {\hyperref[A4B_cI10_229_c_a_cif]{\pageref{A4B_cI10_229_c_a_cif}}} \\
\vspace{-0.75cm} \item $\beta$-NbO$_{2}$: {\small AB2\_tI48\_80\_2b\_4b} \dotfill {\hyperref[AB2_tI48_80_2b_4b_cif]{\pageref{AB2_tI48_80_2b_4b_cif}}} \\
\vspace{-0.75cm} \item $\beta$-PdCl$_2$: {\small A2B\_hR18\_148\_2f\_f} \dotfill {\hyperref[A2B_hR18_148_2f_f_cif]{\pageref{A2B_hR18_148_2f_f_cif}}} \\
\vspace{-0.75cm} \item $\beta$-RuCl$_{3}$: {\small A3B\_hP8\_158\_d\_a} \dotfill {\hyperref[A3B_hP8_158_d_a_cif]{\pageref{A3B_hP8_158_d_a_cif}}} \\
\vspace{-0.75cm} \item $\beta$-RuCl$_{3}$: {\small A3B\_hP8\_185\_c\_a} \dotfill {\hyperref[A3B_hP8_185_c_a_cif]{\pageref{A3B_hP8_185_c_a_cif}}} \\
\vspace{-0.75cm} \item $\beta$-SeO$_{2}$\footnote[1]{\label{note:A2B_oP12_26_abc_ab-cif}H$_{2}$S and $\beta$-SeO$_{2}$ have the same \AFLOW\ prototype label. They are generated by the same symmetry operations with different sets of parameters.}: {\small A2B\_oP12\_26\_abc\_ab} \dotfill {\hyperref[A2B_oP12_26_abc_ab-SeO2_cif]{\pageref{A2B_oP12_26_abc_ab-SeO2_cif}}} \\
\vspace{-0.75cm} \item $\beta$-Si$_{3}$N$_{4}$: {\small A4B3\_hP14\_173\_bc\_c} \dotfill {\hyperref[A4B3_hP14_173_bc_c_cif]{\pageref{A4B3_hP14_173_bc_c_cif}}} \\
\vspace{-0.75cm} \item $\beta$-SiO$_{2}$: {\small A2B\_hP9\_181\_j\_c} \dotfill {\hyperref[A2B_hP9_181_j_c_cif]{\pageref{A2B_hP9_181_j_c_cif}}} \\
\vspace{-0.75cm} \item $\beta$-Ta$_{2}$O$_{5}$: {\small A5B2\_oP14\_49\_dehq\_ab} \dotfill {\hyperref[A5B2_oP14_49_dehq_ab_cif]{\pageref{A5B2_oP14_49_dehq_ab_cif}}} \\
\vspace{-0.75cm} \item $\beta$-ThI$_{3}$: {\small A3B\_oC64\_66\_kl2m\_bdl} \dotfill {\hyperref[A3B_oC64_66_kl2m_bdl_cif]{\pageref{A3B_oC64_66_kl2m_bdl_cif}}} \\
\vspace{-0.75cm} \item $\beta$-Toluene: {\small A7B8\_oP120\_60\_7d\_8d} \dotfill {\hyperref[A7B8_oP120_60_7d_8d_cif]{\pageref{A7B8_oP120_60_7d_8d_cif}}} \\
\vspace{-0.75cm} \item $\beta$-V$_{3}$S: {\small AB3\_tP32\_133\_h\_i2j} \dotfill {\hyperref[AB3_tP32_133_h_i2j_cif]{\pageref{AB3_tP32_133_h_i2j_cif}}} \\
\vspace{-0.75cm} \item $\delta$-PdCl$_{2}$: {\small A2B\_mP6\_10\_mn\_bg} \dotfill {\hyperref[A2B_mP6_10_mn_bg_cif]{\pageref{A2B_mP6_10_mn_bg_cif}}} \\
\vspace{-0.75cm} \item $\delta_{H}^{II}$-NW$_2$: {\small AB2\_hP9\_164\_bd\_c2d} \dotfill {\hyperref[AB2_hP9_164_bd_c2d_cif]{\pageref{AB2_hP9_164_bd_c2d_cif}}} \\
\vspace{-0.75cm} \item $\epsilon$-NiAl$_{3}$: {\small A3B\_oP16\_62\_cd\_c} \dotfill {\hyperref[A3B_oP16_62_cd_c_cif]{\pageref{A3B_oP16_62_cd_c_cif}}} \\
\vspace{-0.75cm} \item $\epsilon$-WO$_{3}$: {\small A3B\_mP16\_7\_6a\_2a} \dotfill {\hyperref[A3B_mP16_7_6a_2a_cif]{\pageref{A3B_mP16_7_6a_2a_cif}}} \\
\vspace{-0.75cm} \item $\gamma$-Ag$_{3}$SI: {\small A3BC\_hR5\_146\_b\_a\_a} \dotfill {\hyperref[A3BC_hR5_146_b_a_a_cif]{\pageref{A3BC_hR5_146_b_a_a_cif}}} \\
\vspace{-0.75cm} \item $\gamma$-MgNiSn: {\small A7B7C2\_tP32\_101\_bde\_ade\_d} \dotfill {\hyperref[A7B7C2_tP32_101_bde_ade_d_cif]{\pageref{A7B7C2_tP32_101_bde_ade_d_cif}}} \\
\vspace{-0.75cm} \item $\gamma$-PdCl$_{2}$: {\small A2B\_mP6\_14\_e\_a} \dotfill {\hyperref[A2B_mP6_14_e_a_cif]{\pageref{A2B_mP6_14_e_a_cif}}} \\
\vspace{-0.75cm} \item $\gamma$-brass: {\small A4B9\_cP52\_215\_ei\_3efgi} \dotfill {\hyperref[A4B9_cP52_215_ei_3efgi_cif]{\pageref{A4B9_cP52_215_ei_3efgi_cif}}} \\
\vspace{-0.75cm} \item $\gamma$-brass: {\small A3B10\_cI52\_229\_e\_fh} \dotfill {\hyperref[A3B10_cI52_229_e_fh_cif]{\pageref{A3B10_cI52_229_e_fh_cif}}} \\
\vspace{-0.75cm} \item $\kappa$-alumina: {\small A2B3\_oP40\_33\_4a\_6a} \dotfill {\hyperref[A2B3_oP40_33_4a_6a_cif]{\pageref{A2B3_oP40_33_4a_6a_cif}}} \\
\vspace{-0.75cm} \item \begin{raggedleft}$\pi$-FeMg$_{3}$Al$_{8}$Si$_{6}$: \end{raggedleft} \\ {\small A8BC3D6\_hP18\_189\_bfh\_a\_g\_i} \dotfill {\hyperref[A8BC3D6_hP18_189_bfh_a_g_i_cif]{\pageref{A8BC3D6_hP18_189_bfh_a_g_i_cif}}} \\
\vspace{-0.75cm} \item \begin{raggedleft}$\pi$-FeMg$_{3}$Al$_{9}$Si$_{5}$: \end{raggedleft} \\ {\small A9BC3D5\_hP18\_189\_fi\_a\_g\_bh} \dotfill {\hyperref[A9BC3D5_hP18_189_fi_a_g_bh_cif]{\pageref{A9BC3D5_hP18_189_fi_a_g_bh_cif}}} \\
\vspace{-0.75cm} \item Ag$_{3}$[PO$_{4}$]: {\small A3B4C\_cP16\_218\_c\_e\_a} \dotfill {\hyperref[A3B4C_cP16_218_c_e_a_cif]{\pageref{A3B4C_cP16_218_c_e_a_cif}}} \\
\vspace{-0.75cm} \item Ag$_{5}$Pb$_{2}$O$_{6}$: {\small A5B6C2\_hP13\_157\_2ac\_2c\_b} \dotfill {\hyperref[A5B6C2_hP13_157_2ac_2c_b_cif]{\pageref{A5B6C2_hP13_157_2ac_2c_b_cif}}} \\
\vspace{-0.75cm} \item AgUF$_{6}$: {\small AB6C\_tP16\_132\_d\_io\_a} \dotfill {\hyperref[AB6C_tP16_132_d_io_a_cif]{\pageref{AB6C_tP16_132_d_io_a_cif}}} \\
\vspace{-0.75cm} \item Akermanite: {\small A2BC7D2\_tP24\_113\_e\_a\_cef\_e} \dotfill {\hyperref[A2BC7D2_tP24_113_e_a_cef_e_cif]{\pageref{A2BC7D2_tP24_113_e_a_cef_e_cif}}} \\
\vspace{-0.75cm} \item Al$_{2}$CuIr\footnote[4]{\label{note:ABC2_oC16_67_b_g_ag-cif}Al$_{2}$CuIr and HoCuP$_{2}$ have similar \AFLOW\ prototype labels ({\it{i.e.}}, same symmetry and set of Wyckoff positions with different stoichiometry labels due to alphabetic ordering of atomic species). They are generated by the same symmetry operations with different sets of parameters.}: {\small A2BC\_oC16\_67\_ag\_b\_g} \dotfill {\hyperref[A2BC_oC16_67_ag_b_g_cif]{\pageref{A2BC_oC16_67_ag_b_g_cif}}} \\
\vspace{-0.75cm} \item Al$_{2}$S$_{3}$: {\small A2B3\_hP30\_170\_2a\_3a} \dotfill {\hyperref[A2B3_hP30_170_2a_3a_cif]{\pageref{A2B3_hP30_170_2a_3a_cif}}} \\
\vspace{-0.75cm} \item Al$_{4}$C$_{3}$: {\small A4B3\_hR7\_166\_2c\_ac} \dotfill {\hyperref[A4B3_hR7_166_2c_ac_cif]{\pageref{A4B3_hR7_166_2c_ac_cif}}} \\
\vspace{-0.75cm} \item Al$_{4}$U: {\small A4B\_oI20\_74\_beh\_e} \dotfill {\hyperref[A4B_oI20_74_beh_e_cif]{\pageref{A4B_oI20_74_beh_e_cif}}} \\
\vspace{-0.75cm} \item Al$_{8}$Cr$_{5}$: {\small A8B5\_hR26\_160\_a3bc\_a3b} \dotfill {\hyperref[A8B5_hR26_160_a3bc_a3b_cif]{\pageref{A8B5_hR26_160_a3bc_a3b_cif}}} \\
\vspace{-0.75cm} \item Al$_{9}$Mn$_{3}$Si: {\small A9B3C\_hP26\_194\_hk\_h\_a} \dotfill {\hyperref[A9B3C_hP26_194_hk_h_a_cif]{\pageref{A9B3C_hP26_194_hk_h_a_cif}}} \\
\vspace{-0.75cm} \item AlLi$_{3}$N$_{2}$: {\small AB3C2\_cI96\_206\_c\_e\_ad} \dotfill {\hyperref[AB3C2_cI96_206_c_e_ad_cif]{\pageref{AB3C2_cI96_206_c_e_ad_cif}}} \\
\vspace{-0.75cm} \item AlPO$_{4}$: {\small AB2\_hP72\_192\_m\_j2kl} \dotfill {\hyperref[AB2_hP72_192_m_j2kl_cif]{\pageref{AB2_hP72_192_m_j2kl_cif}}} \\
\vspace{-0.75cm} \item Al[PO$_{4}$]: {\small AB4C\_hP72\_168\_2d\_8d\_2d} \dotfill {\hyperref[AB4C_hP72_168_2d_8d_2d_cif]{\pageref{AB4C_hP72_168_2d_8d_2d_cif}}} \\
\vspace{-0.75cm} \item Al[PO$_{4}$]: {\small AB4C\_hP72\_184\_d\_4d\_d} \dotfill {\hyperref[AB4C_hP72_184_d_4d_d_cif]{\pageref{AB4C_hP72_184_d_4d_d_cif}}} \\
\vspace{-0.75cm} \item Anhydrite: {\small AB4C\_oC24\_63\_c\_fg\_c} \dotfill {\hyperref[AB4C_oC24_63_c_fg_c_cif]{\pageref{AB4C_oC24_63_c_fg_c_cif}}} \\
\vspace{-0.75cm} \item As$_{2}$Ba: {\small A2B\_mP18\_7\_6a\_3a} \dotfill {\hyperref[A2B_mP18_7_6a_3a_cif]{\pageref{A2B_mP18_7_6a_3a_cif}}} \\
\vspace{-0.75cm} \item \begin{raggedleft}AsPh$_{4}$CeS$_{8}$P$_{4}$Me$_{8}$: \end{raggedleft} \\ {\small AB32CD4E8\_tP184\_93\_i\_16p\_af\_2p\_4p} \dotfill {\hyperref[AB32CD4E8_tP184_93_i_16p_af_2p_4p_cif]{\pageref{AB32CD4E8_tP184_93_i_16p_af_2p_4p_cif}}} \\
\vspace{-0.75cm} \item AuCN: {\small ABC\_hP3\_183\_a\_a\_a} \dotfill {\hyperref[ABC_hP3_183_a_a_a_cif]{\pageref{ABC_hP3_183_a_a_a_cif}}} \\
\vspace{-0.75cm} \item AuF$_{3}$: {\small AB3\_hP24\_178\_b\_ac} \dotfill {\hyperref[AB3_hP24_178_b_ac_cif]{\pageref{AB3_hP24_178_b_ac_cif}}} \\
\vspace{-0.75cm} \item AuF$_{3}$: {\small AB3\_hP24\_179\_b\_ac} \dotfill {\hyperref[AB3_hP24_179_b_ac_cif]{\pageref{AB3_hP24_179_b_ac_cif}}} \\
\vspace{-0.75cm} \item BN: {\small AB\_oF8\_42\_a\_a} \dotfill {\hyperref[AB_oF8_42_a_a_cif]{\pageref{AB_oF8_42_a_a_cif}}} \\
\vspace{-0.75cm} \item BPS$_{4}$: {\small ABC4\_oI12\_23\_a\_b\_k} \dotfill {\hyperref[ABC4_oI12_23_a_b_k_cif]{\pageref{ABC4_oI12_23_a_b_k_cif}}} \\
\vspace{-0.75cm} \item Ba$_{5}$In$_{4}$Bi$_{5}$: {\small A5B5C4\_tP28\_104\_ac\_ac\_c} \dotfill {\hyperref[A5B5C4_tP28_104_ac_ac_c_cif]{\pageref{A5B5C4_tP28_104_ac_ac_c_cif}}} \\
\vspace{-0.75cm} \item Ba$_{5}$Si$_{3}$: {\small A5B3\_tP32\_130\_cg\_cf} \dotfill {\hyperref[A5B3_tP32_130_cg_cf_cif]{\pageref{A5B3_tP32_130_cg_cf_cif}}} \\
\vspace{-0.75cm} \item \begin{raggedleft}BaCr$_{2}$Ru$_{4}$O$_{12}$: \end{raggedleft} \\ {\small AB2C12D4\_tP76\_75\_2a2b\_2d\_12d\_4d} \dotfill {\hyperref[AB2C12D4_tP76_75_2a2b_2d_12d_4d_cif]{\pageref{AB2C12D4_tP76_75_2a2b_2d_12d_4d_cif}}} \\
\vspace{-0.75cm} \item \begin{raggedleft}BaCu$_{4}$[VO][PO$_{4}$]$_{4}$: \end{raggedleft} \\ {\small AB4C17D4E\_tP54\_90\_a\_g\_c4g\_g\_c} \dotfill {\hyperref[AB4C17D4E_tP54_90_a_g_c4g_g_c_cif]{\pageref{AB4C17D4E_tP54_90_a_g_c4g_g_c_cif}}} \\
\vspace{-0.75cm} \item BaGe$_{2}$As$_{2}$: {\small A2BC2\_tP20\_105\_f\_ac\_2e} \dotfill {\hyperref[A2BC2_tP20_105_f_ac_2e_cif]{\pageref{A2BC2_tP20_105_f_ac_2e_cif}}} \\
\vspace{-0.75cm} \item BaSi$_{4}$O$_{9}$: {\small AB9C4\_hP28\_188\_e\_kl\_ak} \dotfill {\hyperref[AB9C4_hP28_188_e_kl_ak_cif]{\pageref{AB9C4_hP28_188_e_kl_ak_cif}}} \\
\vspace{-0.75cm} \item Barite: {\small AB4C\_oP24\_62\_c\_2cd\_c} \dotfill {\hyperref[AB4C_oP24_62_c_2cd_c_cif]{\pageref{AB4C_oP24_62_c_2cd_c_cif}}} \\
\vspace{-0.75cm} \item Be[BH$_{4}$]$_{2}$: {\small A2BC8\_tI176\_110\_2b\_b\_8b} \dotfill {\hyperref[A2BC8_tI176_110_2b_b_8b_cif]{\pageref{A2BC8_tI176_110_2b_b_8b_cif}}} \\
\vspace{-0.75cm} \item Benzene: {\small AB\_oP48\_61\_3c\_3c} \dotfill {\hyperref[AB_oP48_61_3c_3c_cif]{\pageref{AB_oP48_61_3c_3c_cif}}} \\
\vspace{-0.75cm} \item Beryl: {\small A2B3C18D6\_hP58\_192\_c\_f\_lm\_l} \dotfill {\hyperref[A2B3C18D6_hP58_192_c_f_lm_l_cif]{\pageref{A2B3C18D6_hP58_192_c_f_lm_l_cif}}} \\
\vspace{-0.75cm} \item Bi$_{2}$O$_{3}$: {\small A2B3\_hP20\_159\_bc\_2c} \dotfill {\hyperref[A2B3_hP20_159_bc_2c_cif]{\pageref{A2B3_hP20_159_bc_2c_cif}}} \\
\vspace{-0.75cm} \item Bi$_{5}$Nb$_{3}$O$_{15}$: {\small A5B3C15\_oP46\_30\_a2c\_bc\_a7c} \dotfill {\hyperref[A5B3C15_oP46_30_a2c_bc_a7c_cif]{\pageref{A5B3C15_oP46_30_a2c_bc_a7c_cif}}} \\
\vspace{-0.75cm} \item BiAl$_{2}$S$_{4}$: {\small A2BC4\_tP28\_126\_cd\_e\_k} \dotfill {\hyperref[A2BC4_tP28_126_cd_e_k_cif]{\pageref{A2BC4_tP28_126_cd_e_k_cif}}} \\
\vspace{-0.75cm} \item BiGaO$_{3}$: {\small ABC3\_oP20\_54\_e\_d\_cf} \dotfill {\hyperref[ABC3_oP20_54_e_d_cf_cif]{\pageref{ABC3_oP20_54_e_d_cf_cif}}} \\
\vspace{-0.75cm} \item Boracite: {\small A7BC3D13\_cF192\_219\_de\_b\_c\_ah} \dotfill {\hyperref[A7BC3D13_cF192_219_de_b_c_ah_cif]{\pageref{A7BC3D13_cF192_219_de_b_c_ah_cif}}} \\
\vspace{-0.75cm} \item C: {\small A\_tP12\_138\_bi} \dotfill {\hyperref[A_tP12_138_bi_cif]{\pageref{A_tP12_138_bi_cif}}} \\
\vspace{-0.75cm} \item C$_{17}$FeO$_{4}$Pt: {\small A17BC4D\_tP184\_89\_17p\_p\_4p\_io} \dotfill {\hyperref[A17BC4D_tP184_89_17p_p_4p_io_cif]{\pageref{A17BC4D_tP184_89_17p_p_4p_io_cif}}} \\
\vspace{-0.75cm} \item Ca$_{3}$Al$_{2}$O$_{6}$: {\small A2B3C6\_cP264\_205\_2d\_ab2c2d\_6d} \dotfill {\hyperref[A2B3C6_cP264_205_2d_ab2c2d_6d_cif]{\pageref{A2B3C6_cP264_205_2d_ab2c2d_6d_cif}}} \\
\vspace{-0.75cm} \item Ca$_{3}$Al$_{2}$O$_{6}$: {\small A2B3C6\_cP33\_221\_cd\_ag\_fh} \dotfill {\hyperref[A2B3C6_cP33_221_cd_ag_fh_cif]{\pageref{A2B3C6_cP33_221_cd_ag_fh_cif}}} \\
\vspace{-0.75cm} \item Ca$_{3}$PI$_{3}$: {\small A3B3C\_cI56\_214\_g\_h\_a} \dotfill {\hyperref[A3B3C_cI56_214_g_h_a_cif]{\pageref{A3B3C_cI56_214_g_h_a_cif}}} \\
\vspace{-0.75cm} \item \begin{raggedleft}Ca$_{4}$Al$_{6}$O$_{16}$S: \end{raggedleft} \\ {\small A6B4C16D\_oP108\_27\_abcd4e\_4e\_16e\_e} \dotfill {\hyperref[A6B4C16D_oP108_27_abcd4e_4e_16e_e_cif]{\pageref{A6B4C16D_oP108_27_abcd4e_4e_16e_e_cif}}} \\
\vspace{-0.75cm} \item CaRbFe$_{4}$As$_{4}$: {\small A4BC4D\_tP10\_123\_gh\_a\_i\_d} \dotfill {\hyperref[A4BC4D_tP10_123_gh_a_i_d_cif]{\pageref{A4BC4D_tP10_123_gh_a_i_d_cif}}} \\
\vspace{-0.75cm} \item Calomel: {\small AB\_tI8\_139\_e\_e} \dotfill {\hyperref[AB_tI8_139_e_e_cif]{\pageref{AB_tI8_139_e_e_cif}}} \\
\vspace{-0.75cm} \item Carbonyl Sulphide: {\small ABC\_hR3\_160\_a\_a\_a} \dotfill {\hyperref[ABC_hR3_160_a_a_a_cif]{\pageref{ABC_hR3_160_a_a_a_cif}}} \\
\vspace{-0.75cm} \item CdAs$_{2}$: {\small A2B\_tI12\_98\_f\_a} \dotfill {\hyperref[A2B_tI12_98_f_a_cif]{\pageref{A2B_tI12_98_f_a_cif}}} \\
\vspace{-0.75cm} \item CdI$_{2}$: {\small AB2\_hP9\_156\_b2c\_3a2bc} \dotfill {\hyperref[AB2_hP9_156_b2c_3a2bc_cif]{\pageref{AB2_hP9_156_b2c_3a2bc_cif}}} \\
\vspace{-0.75cm} \item Ce$_{3}$Si$_{6}$N$_{11}$: {\small A3B11C6\_tP40\_100\_ac\_bc2d\_cd} \dotfill {\hyperref[A3B11C6_tP40_100_ac_bc2d_cd_cif]{\pageref{A3B11C6_tP40_100_ac_bc2d_cd_cif}}} \\
\vspace{-0.75cm} \item Ce$_{5}$Mo$_{3}$O$_{16}$: {\small A5B3C16\_cP96\_222\_ce\_d\_fi} \dotfill {\hyperref[A5B3C16_cP96_222_ce_d_fi_cif]{\pageref{A5B3C16_cP96_222_ce_d_fi_cif}}} \\
\vspace{-0.75cm} \item CeCo$_{4}$B$_{4}$: {\small A4BC4\_tP18\_137\_g\_b\_g} \dotfill {\hyperref[A4BC4_tP18_137_g_b_g_cif]{\pageref{A4BC4_tP18_137_g_b_g_cif}}} \\
\vspace{-0.75cm} \item CeRu$_{2}$B$_{2}$: {\small A2BC2\_oF40\_22\_fi\_ad\_gh} \dotfill {\hyperref[A2BC2_oF40_22_fi_ad_gh_cif]{\pageref{A2BC2_oF40_22_fi_ad_gh_cif}}} \\
\vspace{-0.75cm} \item CeTe$_{3}$: {\small AB3\_oC16\_40\_b\_3b} \dotfill {\hyperref[AB3_oC16_40_b_3b_cif]{\pageref{AB3_oC16_40_b_3b_cif}}} \\
\vspace{-0.75cm} \item Co$_{2}$Al$_{5}$: {\small A5B2\_hP28\_194\_ahk\_ch} \dotfill {\hyperref[A5B2_hP28_194_ahk_ch_cif]{\pageref{A5B2_hP28_194_ahk_ch_cif}}} \\
\vspace{-0.75cm} \item Co$_{5}$Ge$_{7}$: {\small A5B7\_tI24\_107\_ac\_abd} \dotfill {\hyperref[A5B7_tI24_107_ac_abd_cif]{\pageref{A5B7_tI24_107_ac_abd_cif}}} \\
\vspace{-0.75cm} \item Cobaltite: {\small ABC\_oP12\_29\_a\_a\_a} \dotfill {\hyperref[ABC_oP12_29_a_a_a_cif]{\pageref{ABC_oP12_29_a_a_a_cif}}} \\
\vspace{-0.75cm} \item Cr$_{5}$B$_{3}$: {\small A3B5\_tI32\_140\_ah\_cl} \dotfill {\hyperref[A3B5_tI32_140_ah_cl_cif]{\pageref{A3B5_tI32_140_ah_cl_cif}}} \\
\vspace{-0.75cm} \item CrCl$_{3}$: {\small A3B\_hP24\_153\_3c\_2b} \dotfill {\hyperref[A3B_hP24_153_3c_2b_cif]{\pageref{A3B_hP24_153_3c_2b_cif}}} \\
\vspace{-0.75cm} \item CrFe$_{3}$NiSn$_{5}$: {\small AB\_hP6\_183\_c\_ab} \dotfill {\hyperref[AB_hP6_183_c_ab_cif]{\pageref{AB_hP6_183_c_ab_cif}}} \\
\vspace{-0.75cm} \item \begin{raggedleft}Cs$_{2}$ZnFe[CN]$_{6}$: \end{raggedleft} \\ {\small A6B2CD6E\_cP64\_208\_m\_ad\_b\_m\_c} \dotfill {\hyperref[A6B2CD6E_cP64_208_m_ad_b_m_c_cif]{\pageref{A6B2CD6E_cP64_208_m_ad_b_m_c_cif}}} \\
\vspace{-0.75cm} \item Cs$_{3}$P$_{7}$: {\small A3B7\_tP40\_76\_3a\_7a} \dotfill {\hyperref[A3B7_tP40_76_3a_7a_cif]{\pageref{A3B7_tP40_76_3a_7a_cif}}} \\
\vspace{-0.75cm} \item CsPr[MoO$_{4}$]$_{2}$: {\small AB2C8D\_oP24\_49\_g\_q\_2qr\_e} \dotfill {\hyperref[AB2C8D_oP24_49_g_q_2qr_e_cif]{\pageref{AB2C8D_oP24_49_g_q_2qr_e_cif}}} \\
\vspace{-0.75cm} \item Cu$_{15}$Si$_{4}$: {\small A15B4\_cI76\_220\_ae\_c} \dotfill {\hyperref[A15B4_cI76_220_ae_c_cif]{\pageref{A15B4_cI76_220_ae_c_cif}}} \\
\vspace{-0.75cm} \item Cu$_{2}$Fe[CN]$_{6}$: {\small A12B2C\_cF60\_196\_h\_bc\_a} \dotfill {\hyperref[A12B2C_cF60_196_h_bc_a_cif]{\pageref{A12B2C_cF60_196_h_bc_a_cif}}} \\
\vspace{-0.75cm} \item Cu$_{3}$P: {\small A3B\_hP24\_165\_bdg\_f} \dotfill {\hyperref[A3B_hP24_165_bdg_f_cif]{\pageref{A3B_hP24_165_bdg_f_cif}}} \\
\vspace{-0.75cm} \item Cu$_{3}$P\footnote[6]{\label{note:AB3_hP24_185_c_ab2c-cif}Cu$_{3}$P and Na$_{3}$As have similar \AFLOW\ prototype labels ({\it{i.e.}}, same symmetry and set of Wyckoff positions with different stoichiometry labels due to alphabetic ordering of atomic species). They are generated by the same symmetry operations with different sets of parameters.}: {\small A3B\_hP24\_185\_ab2c\_c} \dotfill {\hyperref[A3B_hP24_185_ab2c_c_cif]{\pageref{A3B_hP24_185_ab2c_c_cif}}} \\
\vspace{-0.75cm} \item CuBi$_{2}$O$_{4}$: {\small A2BC4\_tP28\_130\_f\_c\_g} \dotfill {\hyperref[A2BC4_tP28_130_f_c_g_cif]{\pageref{A2BC4_tP28_130_f_c_g_cif}}} \\
\vspace{-0.75cm} \item CuBrSe$_{3}$: {\small ABC3\_oP20\_30\_2a\_c\_3c} \dotfill {\hyperref[ABC3_oP20_30_2a_c_3c_cif]{\pageref{ABC3_oP20_30_2a_c_3c_cif}}} \\
\vspace{-0.75cm} \item CuBrSe$_{3}$: {\small ABC3\_oP20\_53\_e\_g\_hi} \dotfill {\hyperref[ABC3_oP20_53_e_g_hi_cif]{\pageref{ABC3_oP20_53_e_g_hi_cif}}} \\
\vspace{-0.75cm} \item CuCrCl$_{5}$[NH$_{3}$]$_{6}$: {\small A5BCD6\_cF416\_228\_eg\_c\_b\_h} \dotfill {\hyperref[A5BCD6_cF416_228_eg_c_b_h_cif]{\pageref{A5BCD6_cF416_228_eg_c_b_h_cif}}} \\
\vspace{-0.75cm} \item CuI: {\small AB\_hP12\_156\_2ab3c\_2ab3c} \dotfill {\hyperref[AB_hP12_156_2ab3c_2ab3c_cif]{\pageref{AB_hP12_156_2ab3c_2ab3c_cif}}} \\
\vspace{-0.75cm} \item CuNiSb$_{2}$: {\small ABC2\_hP4\_164\_a\_b\_d} \dotfill {\hyperref[ABC2_hP4_164_a_b_d_cif]{\pageref{ABC2_hP4_164_a_b_d_cif}}} \\
\vspace{-0.75cm} \item Cubanite: {\small AB2C3\_oP24\_62\_c\_d\_cd} \dotfill {\hyperref[AB2C3_oP24_62_c_d_cd_cif]{\pageref{AB2C3_oP24_62_c_d_cd_cif}}} \\
\vspace{-0.75cm} \item Downeyite: {\small A2B\_tP24\_135\_gh\_h} \dotfill {\hyperref[A2B_tP24_135_gh_h_cif]{\pageref{A2B_tP24_135_gh_h_cif}}} \\
\vspace{-0.75cm} \item Er$_{3}$Ru$_{2}$: {\small A3B2\_hP10\_176\_h\_bd} \dotfill {\hyperref[A3B2_hP10_176_h_bd_cif]{\pageref{A3B2_hP10_176_h_bd_cif}}} \\
\vspace{-0.75cm} \item F$_{6}$KP: {\small A24BC\_cF104\_209\_j\_a\_b} \dotfill {\hyperref[A24BC_cF104_209_j_a_b_cif]{\pageref{A24BC_cF104_209_j_a_b_cif}}} \\
\vspace{-0.75cm} \item \begin{raggedleft}FCC C$_{60}$ Buckminsterfullerine: \end{raggedleft} \\ {\small A\_cF240\_202\_h2i} \dotfill {\hyperref[A_cF240_202_h2i_cif]{\pageref{A_cF240_202_h2i_cif}}} \\
\vspace{-0.75cm} \item Fe$_{12}$Zr$_{2}$P$_{7}$: {\small A12B7C2\_hP21\_174\_2j2k\_ajk\_cf} \dotfill {\hyperref[A12B7C2_hP21_174_2j2k_ajk_cf_cif]{\pageref{A12B7C2_hP21_174_2j2k_ajk_cf_cif}}} \\
\vspace{-0.75cm} \item Fe$_{3}$Te$_{3}$Tl: {\small A3B3C\_hP14\_176\_h\_h\_d} \dotfill {\hyperref[A3B3C_hP14_176_h_h_d_cif]{\pageref{A3B3C_hP14_176_h_h_d_cif}}} \\
\vspace{-0.75cm} \item Fe$_{3}$Th$_{7}$: {\small A3B7\_hP20\_186\_c\_b2c} \dotfill {\hyperref[A3B7_hP20_186_c_b2c_cif]{\pageref{A3B7_hP20_186_c_b2c_cif}}} \\
\vspace{-0.75cm} \item FeCu$_{2}$Al$_{7}$: {\small A7B2C\_tP40\_128\_egi\_h\_e} \dotfill {\hyperref[A7B2C_tP40_128_egi_h_e_cif]{\pageref{A7B2C_tP40_128_egi_h_e_cif}}} \\
\vspace{-0.75cm} \item FeNi: {\small AB\_mP4\_6\_2b\_2a} \dotfill {\hyperref[AB_mP4_6_2b_2a_cif]{\pageref{AB_mP4_6_2b_2a_cif}}} \\
\vspace{-0.75cm} \item FeOCl: {\small ABC\_oP6\_59\_a\_b\_a} \dotfill {\hyperref[ABC_oP6_59_a_b_a_cif]{\pageref{ABC_oP6_59_a_b_a_cif}}} \\
\vspace{-0.75cm} \item FePSe$_{3}$: {\small ABC3\_hR10\_146\_2a\_2a\_2b} \dotfill {\hyperref[ABC3_hR10_146_2a_2a_2b_cif]{\pageref{ABC3_hR10_146_2a_2a_2b_cif}}} \\
\vspace{-0.75cm} \item FeS: {\small AB\_oF8\_22\_a\_c} \dotfill {\hyperref[AB_oF8_22_a_c_cif]{\pageref{AB_oF8_22_a_c_cif}}} \\
\vspace{-0.75cm} \item FeSb$_{2}$: {\small AB2\_oP6\_34\_a\_c} \dotfill {\hyperref[AB2_oP6_34_a_c_cif]{\pageref{AB2_oP6_34_a_c_cif}}} \\
\vspace{-0.75cm} \item Forsterite: {\small A2B4C\_oP28\_62\_ac\_2cd\_c} \dotfill {\hyperref[A2B4C_oP28_62_ac_2cd_c_cif]{\pageref{A2B4C_oP28_62_ac_2cd_c_cif}}} \\
\vspace{-0.75cm} \item Fresnoite: {\small A2B8C2D\_tP26\_100\_c\_abcd\_c\_a} \dotfill {\hyperref[A2B8C2D_tP26_100_c_abcd_c_a_cif]{\pageref{A2B8C2D_tP26_100_c_abcd_c_a_cif}}} \\
\vspace{-0.75cm} \item GaCl$_{2}$: {\small A2B\_oP24\_52\_2e\_cd} \dotfill {\hyperref[A2B_oP24_52_2e_cd_cif]{\pageref{A2B_oP24_52_2e_cd_cif}}} \\
\vspace{-0.75cm} \item GaSb: {\small AB\_tI4\_119\_c\_a} \dotfill {\hyperref[AB_tI4_119_c_a_cif]{\pageref{AB_tI4_119_c_a_cif}}} \\
\vspace{-0.75cm} \item Garnet: {\small A2B3C12D3\_cI160\_230\_a\_c\_h\_d} \dotfill {\hyperref[A2B3C12D3_cI160_230_a_c_h_d_cif]{\pageref{A2B3C12D3_cI160_230_a_c_h_d_cif}}} \\
\vspace{-0.75cm} \item Gd$_{3}$Al$_{2}$: {\small A2B3\_tP20\_102\_2c\_b2c} \dotfill {\hyperref[A2B3_tP20_102_2c_b2c_cif]{\pageref{A2B3_tP20_102_2c_b2c_cif}}} \\
\vspace{-0.75cm} \item GdSI: {\small ABC\_hP12\_174\_cj\_fk\_aj} \dotfill {\hyperref[ABC_hP12_174_cj_fk_aj_cif]{\pageref{ABC_hP12_174_cj_fk_aj_cif}}} \\
\vspace{-0.75cm} \item GeAs$_{2}$: {\small A2B\_oP24\_55\_2g2h\_gh} \dotfill {\hyperref[A2B_oP24_55_2g2h_gh_cif]{\pageref{A2B_oP24_55_2g2h_gh_cif}}} \\
\vspace{-0.75cm} \item GeP: {\small AB\_tI4\_107\_a\_a} \dotfill {\hyperref[AB_tI4_107_a_a_cif]{\pageref{AB_tI4_107_a_a_cif}}} \\
\vspace{-0.75cm} \item GeSe$_{2}$: {\small AB2\_tP12\_81\_adg\_2h} \dotfill {\hyperref[AB2_tP12_81_adg_2h_cif]{\pageref{AB2_tP12_81_adg_2h_cif}}} \\
\vspace{-0.75cm} \item H$_{2}$S: {\small A2B\_aP6\_2\_aei\_i} \dotfill {\hyperref[A2B_aP6_2_aei_i_cif]{\pageref{A2B_aP6_2_aei_i_cif}}} \\
\vspace{-0.75cm} \item H$_{2}$S: {\small A2B\_mP12\_13\_2g\_ef} \dotfill {\hyperref[A2B_mP12_13_2g_ef_cif]{\pageref{A2B_mP12_13_2g_ef_cif}}} \\
\vspace{-0.75cm} \item H$_{2}$S\footnoteref{note:A2B_oP12_26_abc_ab-cif}: {\small A2B\_oP12\_26\_abc\_ab} \dotfill {\hyperref[A2B_oP12_26_abc_ab-H2S_cif]{\pageref{A2B_oP12_26_abc_ab-H2S_cif}}} \\
\vspace{-0.75cm} \item H$_{2}$S: {\small A2B\_oC24\_64\_2f\_f} \dotfill {\hyperref[A2B_oC24_64_2f_f_cif]{\pageref{A2B_oC24_64_2f_f_cif}}} \\
\vspace{-0.75cm} \item H$_{2}$S III: {\small A2B\_tP48\_77\_8d\_4d} \dotfill {\hyperref[A2B_tP48_77_8d_4d_cif]{\pageref{A2B_tP48_77_8d_4d_cif}}} \\
\vspace{-0.75cm} \item H$_{2}$S IV: {\small A2B\_mP12\_7\_4a\_2a} \dotfill {\hyperref[A2B_mP12_7_4a_2a_cif]{\pageref{A2B_mP12_7_4a_2a_cif}}} \\
\vspace{-0.75cm} \item H$_{3}$Cl: {\small AB3\_mC16\_9\_a\_3a} \dotfill {\hyperref[AB3_mC16_9_a_3a_cif]{\pageref{AB3_mC16_9_a_3a_cif}}} \\
\vspace{-0.75cm} \item H$_{3}$Cl: {\small AB3\_mP16\_10\_mn\_3m3n} \dotfill {\hyperref[AB3_mP16_10_mn_3m3n_cif]{\pageref{AB3_mP16_10_mn_3m3n_cif}}} \\
\vspace{-0.75cm} \item H$_{3}$Cl: {\small AB3\_mC16\_15\_e\_cf} \dotfill {\hyperref[AB3_mC16_15_e_cf_cif]{\pageref{AB3_mC16_15_e_cf_cif}}} \\
\vspace{-0.75cm} \item H$_{3}$Cl: {\small AB3\_oP16\_19\_a\_3a} \dotfill {\hyperref[AB3_oP16_19_a_3a_cif]{\pageref{AB3_oP16_19_a_3a_cif}}} \\
\vspace{-0.75cm} \item H$_{3}$S: {\small A3B\_oI32\_23\_ij2k\_k} \dotfill {\hyperref[A3B_oI32_23_ij2k_k_cif]{\pageref{A3B_oI32_23_ij2k_k_cif}}} \\
\vspace{-0.75cm} \item H$_{3}$S: {\small A3B\_oC64\_66\_gi2lm\_2l} \dotfill {\hyperref[A3B_oC64_66_gi2lm_2l_cif]{\pageref{A3B_oC64_66_gi2lm_2l_cif}}} \\
\vspace{-0.75cm} \item H$_{3}$S: {\small A3B\_hR4\_160\_b\_a} \dotfill {\hyperref[A3B_hR4_160_b_a_cif]{\pageref{A3B_hR4_160_b_a_cif}}} \\
\vspace{-0.75cm} \item H-III: {\small A\_mC24\_15\_2e2f} \dotfill {\hyperref[A_mC24_15_2e2f_cif]{\pageref{A_mC24_15_2e2f_cif}}} \\
\vspace{-0.75cm} \item HCl: {\small AB\_oC8\_36\_a\_a} \dotfill {\hyperref[AB_oC8_36_a_a_cif]{\pageref{AB_oC8_36_a_a_cif}}} \\
\vspace{-0.75cm} \item HgI$_{2}$: {\small AB2\_tP12\_115\_j\_egi} \dotfill {\hyperref[AB2_tP12_115_j_egi_cif]{\pageref{AB2_tP12_115_j_egi_cif}}} \\
\vspace{-0.75cm} \item HgI$_{2}$\footnote[5]{\label{note:AB2_tP6_137_a_d-cif}ZrO$_{2}$ and HgI$_{2}$ have similar \AFLOW\ prototype labels ({\it{i.e.}}, same symmetry and set of Wyckoff positions with different stoichiometry labels due to alphabetic ordering of atomic species). They are generated by the same symmetry operations with different sets of parameters.}: {\small AB2\_tP6\_137\_a\_d} \dotfill {\hyperref[AB2_tP6_137_a_d_cif]{\pageref{AB2_tP6_137_a_d_cif}}} \\
\vspace{-0.75cm} \item HoCuP$_{2}$\footnoteref{note:ABC2_oC16_67_b_g_ag-cif}: {\small ABC2\_oC16\_67\_b\_g\_ag} \dotfill {\hyperref[ABC2_oC16_67_b_g_ag_cif]{\pageref{ABC2_oC16_67_b_g_ag_cif}}} \\
\vspace{-0.75cm} \item Ir$_{3}$Ga$_{5}$: {\small A5B3\_tP32\_118\_g2i\_aceh} \dotfill {\hyperref[A5B3_tP32_118_g2i_aceh_cif]{\pageref{A5B3_tP32_118_g2i_aceh_cif}}} \\
\vspace{-0.75cm} \item Ir$_{3}$Ge$_{7}$: {\small A7B3\_cI40\_229\_df\_e} \dotfill {\hyperref[A7B3_cI40_229_df_e_cif]{\pageref{A7B3_cI40_229_df_e_cif}}} \\
\vspace{-0.75cm} \item IrGe$_{4}$: {\small A4B\_hP15\_144\_4a\_a} \dotfill {\hyperref[A4B_hP15_144_4a_a_cif]{\pageref{A4B_hP15_144_4a_a_cif}}} \\
\vspace{-0.75cm} \item K$_{2}$CdPb: {\small AB2C\_oC16\_40\_a\_2b\_b} \dotfill {\hyperref[AB2C_oC16_40_a_2b_b_cif]{\pageref{AB2C_oC16_40_a_2b_b_cif}}} \\
\vspace{-0.75cm} \item K$_{2}$PtCl$_{6}$: {\small A6B2C\_cF36\_225\_e\_c\_a} \dotfill {\hyperref[A6B2C_cF36_225_e_c_a_cif]{\pageref{A6B2C_cF36_225_e_c_a_cif}}} \\
\vspace{-0.75cm} \item K$_{2}$SnCl$_{6}$: {\small A6B2C\_tP18\_128\_eh\_d\_b} \dotfill {\hyperref[A6B2C_tP18_128_eh_d_b_cif]{\pageref{A6B2C_tP18_128_eh_d_b_cif}}} \\
\vspace{-0.75cm} \item K$_{2}$Ta$_{4}$O$_{9}$F$_{4}$: {\small A2B13C4\_hP57\_168\_d\_c6d\_2d} \dotfill {\hyperref[A2B13C4_hP57_168_d_c6d_2d_cif]{\pageref{A2B13C4_hP57_168_d_c6d_2d_cif}}} \\
\vspace{-0.75cm} \item KAg[CO$_{3}$]: {\small ABCD3\_oI48\_73\_d\_e\_e\_ef} \dotfill {\hyperref[ABCD3_oI48_73_d_e_e_ef_cif]{\pageref{ABCD3_oI48_73_d_e_e_ef_cif}}} \\
\vspace{-0.75cm} \item KAu$_{4}$Sn$_{2}$: {\small A4BC2\_tI28\_120\_i\_d\_e} \dotfill {\hyperref[A4BC2_tI28_120_i_d_e_cif]{\pageref{A4BC2_tI28_120_i_d_e_cif}}} \\
\vspace{-0.75cm} \item KB$_{6}$H$_{6}$: {\small A6B6C\_cF104\_202\_h\_h\_c} \dotfill {\hyperref[A6B6C_cF104_202_h_h_c_cif]{\pageref{A6B6C_cF104_202_h_h_c_cif}}} \\
\vspace{-0.75cm} \item KBO$_{2}$: {\small ABC2\_hR24\_167\_e\_e\_2e} \dotfill {\hyperref[ABC2_hR24_167_e_e_2e_cif]{\pageref{ABC2_hR24_167_e_e_2e_cif}}} \\
\vspace{-0.75cm} \item KCeSe$_{4}$: {\small ABC4\_tP12\_125\_a\_b\_m} \dotfill {\hyperref[ABC4_tP12_125_a_b_m_cif]{\pageref{ABC4_tP12_125_a_b_m_cif}}} \\
\vspace{-0.75cm} \item KHg$_{2}$: {\small A2B\_oI12\_74\_h\_e} \dotfill {\hyperref[A2B_oI12_74_h_e_cif]{\pageref{A2B_oI12_74_h_e_cif}}} \\
\vspace{-0.75cm} \item KNiCl$_{3}$: {\small A3BC\_hP30\_185\_cd\_c\_ab} \dotfill {\hyperref[A3BC_hP30_185_cd_c_ab_cif]{\pageref{A3BC_hP30_185_cd_c_ab_cif}}} \\
\vspace{-0.75cm} \item KSbO$_{3}$: {\small AB3C\_cP60\_201\_ce\_fh\_g} \dotfill {\hyperref[AB3C_cP60_201_ce_fh_g_cif]{\pageref{AB3C_cP60_201_ce_fh_g_cif}}} \\
\vspace{-0.75cm} \item La$_{2}$NiO$_{4}$: {\small A2BC4\_oP28\_50\_ij\_ac\_ijm} \dotfill {\hyperref[A2BC4_oP28_50_ij_ac_ijm_cif]{\pageref{A2BC4_oP28_50_ij_ac_ijm_cif}}} \\
\vspace{-0.75cm} \item La$_{2}$O$_{3}$: {\small A2B3\_hP5\_164\_d\_ad} \dotfill {\hyperref[A2B3_hP5_164_d_ad_cif]{\pageref{A2B3_hP5_164_d_ad_cif}}} \\
\vspace{-0.75cm} \item \begin{raggedleft}La$_{43}$Ni$_{17}$Mg$_{5}$: \end{raggedleft} \\ {\small A43B5C17\_oC260\_63\_c8fg6h\_cfg\_ce3f2h} \dotfill {\hyperref[A43B5C17_oC260_63_c8fg6h_cfg_ce3f2h_cif]{\pageref{A43B5C17_oC260_63_c8fg6h_cfg_ce3f2h_cif}}} \\
\vspace{-0.75cm} \item LaPtSi: {\small ABC\_tI12\_109\_a\_a\_a} \dotfill {\hyperref[ABC_tI12_109_a_a_a_cif]{\pageref{ABC_tI12_109_a_a_a_cif}}} \\
\vspace{-0.75cm} \item LaRhC$_{2}$: {\small A2BC\_tP16\_76\_2a\_a\_a} \dotfill {\hyperref[A2BC_tP16_76_2a_a_a_cif]{\pageref{A2BC_tP16_76_2a_a_a_cif}}} \\
\vspace{-0.75cm} \item Li$_{2}$MoF$_{6}$: {\small A6B2C\_tP18\_94\_eg\_c\_a} \dotfill {\hyperref[A6B2C_tP18_94_eg_c_a_cif]{\pageref{A6B2C_tP18_94_eg_c_a_cif}}} \\
\vspace{-0.75cm} \item Li$_{2}$Sb: {\small A2B\_hP18\_190\_gh\_bf} \dotfill {\hyperref[A2B_hP18_190_gh_bf_cif]{\pageref{A2B_hP18_190_gh_bf_cif}}} \\
\vspace{-0.75cm} \item Li$_{2}$Si$_{2}$O$_{5}$: {\small A2B5C2\_oC36\_37\_d\_c2d\_d} \dotfill {\hyperref[A2B5C2_oC36_37_d_c2d_d_cif]{\pageref{A2B5C2_oC36_37_d_c2d_d_cif}}} \\
\vspace{-0.75cm} \item LiScI$_{3}$: {\small A3BC\_hP10\_188\_k\_a\_e} \dotfill {\hyperref[A3BC_hP10_188_k_a_e_cif]{\pageref{A3BC_hP10_188_k_a_e_cif}}} \\
\vspace{-0.75cm} \item LiSn: {\small AB\_mP6\_10\_en\_am} \dotfill {\hyperref[AB_mP6_10_en_am_cif]{\pageref{AB_mP6_10_en_am_cif}}} \\
\vspace{-0.75cm} \item M-carbon: {\small A\_mC16\_12\_4i} \dotfill {\hyperref[A_mC16_12_4i_cif]{\pageref{A_mC16_12_4i_cif}}} \\
\vspace{-0.75cm} \item Mavlyanovite: {\small A5B3\_hP16\_193\_dg\_g} \dotfill {\hyperref[A5B3_hP16_193_dg_g_cif]{\pageref{A5B3_hP16_193_dg_g_cif}}} \\
\vspace{-0.75cm} \item Mg$_{2}$Zn$_{11}$: {\small A2B11\_cP39\_200\_f\_aghij} \dotfill {\hyperref[A2B11_cP39_200_f_aghij_cif]{\pageref{A2B11_cP39_200_f_aghij_cif}}} \\
\vspace{-0.75cm} \item \begin{raggedleft}MgB$_{12}$H$_{12}$[H$_{2}$O]$_{12}$: \end{raggedleft} \\ {\small A12B36CD12\_cF488\_196\_2h\_6h\_ac\_fgh} \dotfill {\hyperref[A12B36CD12_cF488_196_2h_6h_ac_fgh_cif]{\pageref{A12B36CD12_cF488_196_2h_6h_ac_fgh_cif}}} \\
\vspace{-0.75cm} \item MgSO$_{4}$: {\small AB4C\_oC24\_63\_a\_fg\_c} \dotfill {\hyperref[AB4C_oC24_63_a_fg_c_cif]{\pageref{AB4C_oC24_63_a_fg_c_cif}}} \\
\vspace{-0.75cm} \item Mg[NH]: {\small ABC\_hP36\_175\_jk\_jk\_jk} \dotfill {\hyperref[ABC_hP36_175_jk_jk_jk_cif]{\pageref{ABC_hP36_175_jk_jk_jk_cif}}} \\
\vspace{-0.75cm} \item Mn$_{2}$B: {\small AB2\_oF48\_70\_f\_fg} \dotfill {\hyperref[AB2_oF48_70_f_fg_cif]{\pageref{AB2_oF48_70_f_fg_cif}}} \\
\vspace{-0.75cm} \item MnAl$_{6}$: {\small A6B\_oC28\_63\_efg\_c} \dotfill {\hyperref[A6B_oC28_63_efg_c_cif]{\pageref{A6B_oC28_63_efg_c_cif}}} \\
\vspace{-0.75cm} \item MnF$_{2}$: {\small A2B\_tP12\_111\_2n\_adf} \dotfill {\hyperref[A2B_tP12_111_2n_adf_cif]{\pageref{A2B_tP12_111_2n_adf_cif}}} \\
\vspace{-0.75cm} \item MnGa$_{2}$Sb$_{2}$: {\small A2BC2\_oI20\_45\_c\_b\_c} \dotfill {\hyperref[A2BC2_oI20_45_c_b_c_cif]{\pageref{A2BC2_oI20_45_c_b_c_cif}}} \\
\vspace{-0.75cm} \item Mo$_{8}$P$_{5}$: {\small A8B5\_mP13\_6\_a7b\_3a2b} \dotfill {\hyperref[A8B5_mP13_6_a7b_3a2b_cif]{\pageref{A8B5_mP13_6_a7b_3a2b_cif}}} \\
\vspace{-0.75cm} \item MoS$_{2}$: {\small AB2\_hP12\_143\_cd\_ab2d} \dotfill {\hyperref[AB2_hP12_143_cd_ab2d_cif]{\pageref{AB2_hP12_143_cd_ab2d_cif}}} \\
\vspace{-0.75cm} \item Moissanite-15R: {\small AB\_hR10\_160\_5a\_5a} \dotfill {\hyperref[AB_hR10_160_5a_5a_cif]{\pageref{AB_hR10_160_5a_5a_cif}}} \\
\vspace{-0.75cm} \item Molybdite: {\small AB3\_oP16\_62\_c\_3c} \dotfill {\hyperref[AB3_oP16_62_c_3c_cif]{\pageref{AB3_oP16_62_c_3c_cif}}} \\
\vspace{-0.75cm} \item Muthmannite: {\small ABC2\_mP8\_10\_ac\_eh\_mn} \dotfill {\hyperref[ABC2_mP8_10_ac_eh_mn_cif]{\pageref{ABC2_mP8_10_ac_eh_mn_cif}}} \\
\vspace{-0.75cm} \item NV: {\small AB\_tP8\_111\_n\_n} \dotfill {\hyperref[AB_tP8_111_n_n_cif]{\pageref{AB_tP8_111_n_n_cif}}} \\
\vspace{-0.75cm} \item Na$_{3}$As\footnoteref{note:AB3_hP24_185_c_ab2c-cif}: {\small AB3\_hP24\_185\_c\_ab2c} \dotfill {\hyperref[AB3_hP24_185_c_ab2c_cif]{\pageref{AB3_hP24_185_c_ab2c_cif}}} \\
\vspace{-0.75cm} \item \begin{raggedleft}Na$_{4}$Ti$_{2}$Si$_{8}$O$_{22}$[H$_{2}$O]$_{4}$: \end{raggedleft} \\ {\small A4B2C13D\_tP40\_90\_g\_d\_cef2g\_c} \dotfill {\hyperref[A4B2C13D_tP40_90_g_d_cef2g_c_cif]{\pageref{A4B2C13D_tP40_90_g_d_cef2g_c_cif}}} \\
\vspace{-0.75cm} \item Na$_{5}$Fe$_{3}$F$_{14}$: {\small A14B3C5\_tP44\_94\_c3g\_ad\_bg} \dotfill {\hyperref[A14B3C5_tP44_94_c3g_ad_bg_cif]{\pageref{A14B3C5_tP44_94_c3g_ad_bg_cif}}} \\
\vspace{-0.75cm} \item NaFeS$_{2}$: {\small ABC2\_oI16\_23\_ab\_i\_k} \dotfill {\hyperref[ABC2_oI16_23_ab_i_k_cif]{\pageref{ABC2_oI16_23_ab_i_k_cif}}} \\
\vspace{-0.75cm} \item NaGdCu$_{2}$F$_{8}$: {\small A2B8CD\_tI24\_97\_d\_k\_a\_b} \dotfill {\hyperref[A2B8CD_tI24_97_d_k_a_b_cif]{\pageref{A2B8CD_tI24_97_d_k_a_b_cif}}} \\
\vspace{-0.75cm} \item NaZn$_{13}$: {\small AB13\_cF112\_226\_a\_bi} \dotfill {\hyperref[AB13_cF112_226_a_bi_cif]{\pageref{AB13_cF112_226_a_bi_cif}}} \\
\vspace{-0.75cm} \item NaZn[OH]$_{3}$: {\small A3BC3D\_tP64\_106\_3c\_c\_3c\_c} \dotfill {\hyperref[A3BC3D_tP64_106_3c_c_3c_c_cif]{\pageref{A3BC3D_tP64_106_3c_c_3c_c_cif}}} \\
\vspace{-0.75cm} \item Nb$_{4}$CoSi: {\small AB4C\_tP12\_124\_a\_m\_c} \dotfill {\hyperref[AB4C_tP12_124_a_m_c_cif]{\pageref{AB4C_tP12_124_a_m_c_cif}}} \\
\vspace{-0.75cm} \item Nb$_{7}$Ru$_{6}$B$_{8}$: {\small A8B7C6\_hP21\_175\_ck\_aj\_k} \dotfill {\hyperref[A8B7C6_hP21_175_ck_aj_k_cif]{\pageref{A8B7C6_hP21_175_ck_aj_k_cif}}} \\
\vspace{-0.75cm} \item NbAs: {\small AB\_tI8\_109\_a\_a} \dotfill {\hyperref[AB_tI8_109_a_a_cif]{\pageref{AB_tI8_109_a_a_cif}}} \\
\vspace{-0.75cm} \item NbPS: {\small ABC\_oI12\_71\_h\_j\_g} \dotfill {\hyperref[ABC_oI12_71_h_j_g_cif]{\pageref{ABC_oI12_71_h_j_g_cif}}} \\
\vspace{-0.75cm} \item NbTe$_{4}$: {\small AB4\_tP10\_103\_a\_d} \dotfill {\hyperref[AB4_tP10_103_a_d_cif]{\pageref{AB4_tP10_103_a_d_cif}}} \\
\vspace{-0.75cm} \item NbTe$_{4}$: {\small AB4\_tP10\_124\_a\_m} \dotfill {\hyperref[AB4_tP10_124_a_m_cif]{\pageref{AB4_tP10_124_a_m_cif}}} \\
\vspace{-0.75cm} \item Ni$_{3}$P: {\small A3B\_tI32\_82\_3g\_g} \dotfill {\hyperref[A3B_tI32_82_3g_g_cif]{\pageref{A3B_tI32_82_3g_g_cif}}} \\
\vspace{-0.75cm} \item Ni$_{3}$Ti: {\small A3B\_hP16\_194\_gh\_ac} \dotfill {\hyperref[A3B_hP16_194_gh_ac_cif]{\pageref{A3B_hP16_194_gh_ac_cif}}} \\
\vspace{-0.75cm} \item Nierite: {\small A4B3\_hP28\_159\_ab2c\_2c} \dotfill {\hyperref[A4B3_hP28_159_ab2c_2c_cif]{\pageref{A4B3_hP28_159_ab2c_2c_cif}}} \\
\vspace{-0.75cm} \item PH$_{3}$: {\small A3B\_cP16\_208\_j\_b} \dotfill {\hyperref[A3B_cP16_208_j_b_cif]{\pageref{A3B_cP16_208_j_b_cif}}} \\
\vspace{-0.75cm} \item PI$_{3}$: {\small A3B\_hP8\_173\_c\_b} \dotfill {\hyperref[A3B_hP8_173_c_b_cif]{\pageref{A3B_hP8_173_c_b_cif}}} \\
\vspace{-0.75cm} \item Pd$_{17}$Se$_{15}$: {\small A17B15\_cP64\_207\_acfk\_eij} \dotfill {\hyperref[A17B15_cP64_207_acfk_eij_cif]{\pageref{A17B15_cP64_207_acfk_eij_cif}}} \\
\vspace{-0.75cm} \item Pd$_{4}$Se: {\small A4B\_tP10\_114\_e\_a} \dotfill {\hyperref[A4B_tP10_114_e_a_cif]{\pageref{A4B_tP10_114_e_a_cif}}} \\
\vspace{-0.75cm} \item PdSn$_{4}$: {\small AB4\_oC20\_68\_a\_i} \dotfill {\hyperref[AB4_oC20_68_a_i_cif]{\pageref{AB4_oC20_68_a_i_cif}}} \\
\vspace{-0.75cm} \item Petzite: {\small A3BC2\_cI48\_214\_f\_a\_e} \dotfill {\hyperref[A3BC2_cI48_214_f_a_e_cif]{\pageref{A3BC2_cI48_214_f_a_e_cif}}} \\
\vspace{-0.75cm} \item Phenakite: {\small A2B4C\_hR42\_148\_2f\_4f\_f} \dotfill {\hyperref[A2B4C_hR42_148_2f_4f_f_cif]{\pageref{A2B4C_hR42_148_2f_4f_f_cif}}} \\
\vspace{-0.75cm} \item Pinnoite: {\small A2B6CD7\_tP64\_77\_2d\_6d\_d\_ab6d} \dotfill {\hyperref[A2B6CD7_tP64_77_2d_6d_d_ab6d_cif]{\pageref{A2B6CD7_tP64_77_2d_6d_d_ab6d_cif}}} \\
\vspace{-0.75cm} \item Post-perovskite: {\small AB3C\_oC20\_63\_a\_cf\_c} \dotfill {\hyperref[AB3C_oC20_63_a_cf_c_cif]{\pageref{AB3C_oC20_63_a_cf_c_cif}}} \\
\vspace{-0.75cm} \item PrNiO$_{3}$: {\small AB3C\_hR10\_167\_b\_e\_a} \dotfill {\hyperref[AB3C_hR10_167_b_e_a_cif]{\pageref{AB3C_hR10_167_b_e_a_cif}}} \\
\vspace{-0.75cm} \item PrRu$_{4}$P$_{12}$: {\small A12BC4\_cP34\_195\_2j\_ab\_2e} \dotfill {\hyperref[A12BC4_cP34_195_2j_ab_2e_cif]{\pageref{A12BC4_cP34_195_2j_ab_2e_cif}}} \\
\vspace{-0.75cm} \item PtPb$_{4}$: {\small A4B\_tP10\_125\_m\_a} \dotfill {\hyperref[A4B_tP10_125_m_a_cif]{\pageref{A4B_tP10_125_m_a_cif}}} \\
\vspace{-0.75cm} \item Pyrite\footnote[3]{\label{note:AB2_oP12_29_a_2a-cif}ZrO$_{2}$ and Pyrite have similar \AFLOW\ prototype labels ({\it{i.e.}}, same symmetry and set of Wyckoff positions with different stoichiometry labels due to alphabetic ordering of atomic species). They are generated by the same symmetry operations with different sets of parameters.}: {\small AB2\_oP12\_29\_a\_2a} \dotfill {\hyperref[AB2_oP12_29_a_2a_cif]{\pageref{AB2_oP12_29_a_2a_cif}}} \\
\vspace{-0.75cm} \item Pyrochlore: {\small A2BCD3E6\_cF208\_203\_e\_c\_d\_f\_g} \dotfill {\hyperref[A2BCD3E6_cF208_203_e_c_d_f_g_cif]{\pageref{A2BCD3E6_cF208_203_e_c_d_f_g_cif}}} \\
\vspace{-0.75cm} \item \begin{raggedleft}Pyrochlore Iridate: \end{raggedleft} \\ {\small A2B2C7\_cF88\_227\_c\_d\_af} \dotfill {\hyperref[A2B2C7_cF88_227_c_d_af_cif]{\pageref{A2B2C7_cF88_227_c_d_af_cif}}} \\
\vspace{-0.75cm} \item \begin{raggedleft}Quartenary Heusler: \end{raggedleft} \\ {\small ABCD\_cF16\_216\_c\_d\_b\_a} \dotfill {\hyperref[ABCD_cF16_216_c_d_b_a_cif]{\pageref{ABCD_cF16_216_c_d_b_a_cif}}} \\
\vspace{-0.75cm} \item R-carbon: {\small A\_oP16\_55\_2g2h} \dotfill {\hyperref[A_oP16_55_2g2h_cif]{\pageref{A_oP16_55_2g2h_cif}}} \\
\vspace{-0.75cm} \item Rasvumite: {\small A2BC3\_oC24\_63\_e\_c\_cg} \dotfill {\hyperref[A2BC3_oC24_63_e_c_cg_cif]{\pageref{A2BC3_oC24_63_e_c_cg_cif}}} \\
\vspace{-0.75cm} \item Rb$_{2}$TiCu$_{2}$S$_{4}$: {\small A2B2C4D\_tP18\_132\_e\_i\_o\_d} \dotfill {\hyperref[A2B2C4D_tP18_132_e_i_o_d_cif]{\pageref{A2B2C4D_tP18_132_e_i_o_d_cif}}} \\
\vspace{-0.75cm} \item Rb$_{3}$AsSe$_{16}$: {\small AB3C16\_cF160\_203\_b\_ad\_eg} \dotfill {\hyperref[AB3C16_cF160_203_b_ad_eg_cif]{\pageref{AB3C16_cF160_203_b_ad_eg_cif}}} \\
\vspace{-0.75cm} \item RbGa$_{3}$: {\small A3B\_tI24\_119\_b2i\_af} \dotfill {\hyperref[A3B_tI24_119_b2i_af_cif]{\pageref{A3B_tI24_119_b2i_af_cif}}} \\
\vspace{-0.75cm} \item Re$_{2}$O$_{5}$[SO$_{4}$]$_{2}$: {\small A13B2C2\_oP34\_32\_a6c\_c\_c} \dotfill {\hyperref[A13B2C2_oP34_32_a6c_c_c_cif]{\pageref{A13B2C2_oP34_32_a6c_c_c_cif}}} \\
\vspace{-0.75cm} \item Re$_{3}$N: {\small AB3\_hP4\_187\_e\_fh} \dotfill {\hyperref[AB3_hP4_187_e_fh_cif]{\pageref{AB3_hP4_187_e_fh_cif}}} \\
\vspace{-0.75cm} \item Rh$_{2}$Ga$_{9}$: {\small A9B2\_mP22\_7\_9a\_2a} \dotfill {\hyperref[A9B2_mP22_7_9a_2a_cif]{\pageref{A9B2_mP22_7_9a_2a_cif}}} \\
\vspace{-0.75cm} \item Rh$_{2}$S$_{3}$: {\small A2B3\_oP20\_60\_d\_cd} \dotfill {\hyperref[A2B3_oP20_60_d_cd_cif]{\pageref{A2B3_oP20_60_d_cd_cif}}} \\
\vspace{-0.75cm} \item Rh$_{3}$P$_{2}$: {\small A2B3\_tP5\_115\_g\_ag} \dotfill {\hyperref[A2B3_tP5_115_g_ag_cif]{\pageref{A2B3_tP5_115_g_ag_cif}}} \\
\vspace{-0.75cm} \item Rh$_{5}$Ge$_{3}$: {\small A3B5\_oP16\_55\_ch\_agh} \dotfill {\hyperref[A3B5_oP16_55_ch_agh_cif]{\pageref{A3B5_oP16_55_ch_agh_cif}}} \\
\vspace{-0.75cm} \item Ru$_{2}$Sn$_{3}$: {\small A2B3\_tP20\_116\_bci\_fj} \dotfill {\hyperref[A2B3_tP20_116_bci_fj_cif]{\pageref{A2B3_tP20_116_bci_fj_cif}}} \\
\vspace{-0.75cm} \item RuIn$_{3}$: {\small A3B\_tP16\_118\_ei\_f} \dotfill {\hyperref[A3B_tP16_118_ei_f_cif]{\pageref{A3B_tP16_118_ei_f_cif}}} \\
\vspace{-0.75cm} \item S-II: {\small A\_hP9\_154\_bc} \dotfill {\hyperref[A_hP9_154_bc_cif]{\pageref{A_hP9_154_bc_cif}}} \\
\vspace{-0.75cm} \item S-III: {\small A\_tI16\_142\_f} \dotfill {\hyperref[A_tI16_142_f_cif]{\pageref{A_tI16_142_f_cif}}} \\
\vspace{-0.75cm} \item S-carbon: {\small A\_mP8\_10\_2m2n} \dotfill {\hyperref[A_mP8_10_2m2n_cif]{\pageref{A_mP8_10_2m2n_cif}}} \\
\vspace{-0.75cm} \item Sc-V: {\small A\_hP6\_178\_a} \dotfill {\hyperref[A_hP6_178_a_cif]{\pageref{A_hP6_178_a_cif}}} \\
\vspace{-0.75cm} \item ScRh$_{6}$P$_{4}$: {\small A4B6C\_hP11\_143\_bd\_2d\_a} \dotfill {\hyperref[A4B6C_hP11_143_bd_2d_a_cif]{\pageref{A4B6C_hP11_143_bd_2d_a_cif}}} \\
\vspace{-0.75cm} \item SeO$_{3}$: {\small A3B\_tP32\_114\_3e\_e} \dotfill {\hyperref[A3B_tP32_114_3e_e_cif]{\pageref{A3B_tP32_114_3e_e_cif}}} \\
\vspace{-0.75cm} \item \begin{raggedleft}Sheldrickite: \end{raggedleft} \\ {\small A2B3C3DE7\_hP48\_145\_2a\_3a\_3a\_a\_7a} \dotfill {\hyperref[A2B3C3DE7_hP48_145_2a_3a_3a_a_7a_cif]{\pageref{A2B3C3DE7_hP48_145_2a_3a_3a_a_7a_cif}}} \\
\vspace{-0.75cm} \item SiO$_{2}$: {\small A2B\_hP36\_177\_j2lm\_n} \dotfill {\hyperref[A2B_hP36_177_j2lm_n_cif]{\pageref{A2B_hP36_177_j2lm_n_cif}}} \\
\vspace{-0.75cm} \item SiO$_{2}$: {\small A2B\_cI72\_211\_hi\_i} \dotfill {\hyperref[A2B_cI72_211_hi_i_cif]{\pageref{A2B_cI72_211_hi_i_cif}}} \\
\vspace{-0.75cm} \item \begin{raggedleft}Simple Cubic C$_{60}$ Buckminsterfullerine: \end{raggedleft} \\ {\small A\_cP240\_205\_10d} \dotfill {\hyperref[A_cP240_205_10d_cif]{\pageref{A_cP240_205_10d_cif}}} \\
\vspace{-0.75cm} \item Simpsonite: {\small A4B14C3\_hP21\_143\_bd\_ac4d\_d} \dotfill {\hyperref[A4B14C3_hP21_143_bd_ac4d_d_cif]{\pageref{A4B14C3_hP21_143_bd_ac4d_d_cif}}} \\
\vspace{-0.75cm} \item SmSI: {\small ABC\_hR6\_166\_c\_c\_c} \dotfill {\hyperref[ABC_hR6_166_c_c_c_cif]{\pageref{ABC_hR6_166_c_c_c_cif}}} \\
\vspace{-0.75cm} \item Sodium Chlorate: {\small ABC3\_cP20\_198\_a\_a\_b} \dotfill {\hyperref[ABC3_cP20_198_a_a_b_cif]{\pageref{ABC3_cP20_198_a_a_b_cif}}} \\
\vspace{-0.75cm} \item Spinel: {\small A3B4\_cF56\_227\_ad\_e} \dotfill {\hyperref[A3B4_cF56_227_ad_e_cif]{\pageref{A3B4_cF56_227_ad_e_cif}}} \\
\vspace{-0.75cm} \item Sr$_{2}$As$_{2}$O$_{7}$: {\small A2B7C2\_tP88\_78\_4a\_14a\_4a} \dotfill {\hyperref[A2B7C2_tP88_78_4a_14a_4a_cif]{\pageref{A2B7C2_tP88_78_4a_14a_4a_cif}}} \\
\vspace{-0.75cm} \item Sr$_{2}$Bi$_{3}$: {\small A3B2\_oP20\_52\_de\_cd} \dotfill {\hyperref[A3B2_oP20_52_de_cd_cif]{\pageref{A3B2_oP20_52_de_cd_cif}}} \\
\vspace{-0.75cm} \item Sr$_{5}$Si$_{3}$: {\small A3B5\_tI32\_108\_ac\_a2c} \dotfill {\hyperref[A3B5_tI32_108_ac_a2c_cif]{\pageref{A3B5_tI32_108_ac_a2c_cif}}} \\
\vspace{-0.75cm} \item SrAl$_{2}$Se$_{4}$: {\small A2B4C\_oC28\_66\_l\_kl\_a} \dotfill {\hyperref[A2B4C_oC28_66_l_kl_a_cif]{\pageref{A2B4C_oC28_66_l_kl_a_cif}}} \\
\vspace{-0.75cm} \item SrBr$_{2}$: {\small A2B\_tP30\_85\_ab2g\_cg} \dotfill {\hyperref[A2B_tP30_85_ab2g_cg_cif]{\pageref{A2B_tP30_85_ab2g_cg_cif}}} \\
\vspace{-0.75cm} \item SrH$_{2}$: {\small A2B\_oP12\_62\_2c\_c} \dotfill {\hyperref[A2B_oP12_62_2c_c_cif]{\pageref{A2B_oP12_62_2c_c_cif}}} \\
\vspace{-0.75cm} \item SrSi$_{2}$: {\small A2B\_cP12\_212\_c\_a} \dotfill {\hyperref[A2B_cP12_212_c_a_cif]{\pageref{A2B_cP12_212_c_a_cif}}} \\
\vspace{-0.75cm} \item Sr[S$_{2}$O$_{6}$][H$_{2}$O]$_{4}$: {\small A10B2C\_hP39\_171\_5c\_c\_a} \dotfill {\hyperref[A10B2C_hP39_171_5c_c_a_cif]{\pageref{A10B2C_hP39_171_5c_c_a_cif}}} \\
\vspace{-0.75cm} \item Sr[S$_{2}$O$_{6}$][H$_{2}$O]$_{4}$: {\small A10B2C\_hP39\_172\_5c\_c\_a} \dotfill {\hyperref[A10B2C_hP39_172_5c_c_a_cif]{\pageref{A10B2C_hP39_172_5c_c_a_cif}}} \\
\vspace{-0.75cm} \item \begin{raggedleft}Stannoidite: \end{raggedleft} \\ {\small A8B2C12D2E\_oI50\_23\_bcfk\_i\_3k\_j\_a} \dotfill {\hyperref[A8B2C12D2E_oI50_23_bcfk_i_3k_j_a_cif]{\pageref{A8B2C12D2E_oI50_23_bcfk_i_3k_j_a_cif}}} \\
\vspace{-0.75cm} \item Ta$_{2}$H: {\small AB2\_oC6\_21\_a\_k} \dotfill {\hyperref[AB2_oC6_21_a_k_cif]{\pageref{AB2_oC6_21_a_k_cif}}} \\
\vspace{-0.75cm} \item Ta$_{2}$Se$_{8}$I: {\small AB8C2\_tI44\_97\_e\_2k\_cd} \dotfill {\hyperref[AB8C2_tI44_97_e_2k_cd_cif]{\pageref{AB8C2_tI44_97_e_2k_cd_cif}}} \\
\vspace{-0.75cm} \item Ta$_{3}$B$_{4}$: {\small A4B3\_oI14\_71\_gh\_cg} \dotfill {\hyperref[A4B3_oI14_71_gh_cg_cif]{\pageref{A4B3_oI14_71_gh_cg_cif}}} \\
\vspace{-0.75cm} \item Ta$_{3}$S$_{2}$: {\small A2B3\_oC40\_39\_2d\_2c2d} \dotfill {\hyperref[A2B3_oC40_39_2d_2c2d_cif]{\pageref{A2B3_oC40_39_2d_2c2d_cif}}} \\
\vspace{-0.75cm} \item TaNiTe$_{2}$: {\small ABC2\_oP16\_53\_h\_e\_gh} \dotfill {\hyperref[ABC2_oP16_53_h_e_gh_cif]{\pageref{ABC2_oP16_53_h_e_gh_cif}}} \\
\vspace{-0.75cm} \item TeO$_{6}$H$_{6}$: {\small A6B\_cF224\_228\_h\_c} \dotfill {\hyperref[A6B_cF224_228_h_c_cif]{\pageref{A6B_cF224_228_h_c_cif}}} \\
\vspace{-0.75cm} \item TeZn: {\small AB\_hP6\_144\_a\_a} \dotfill {\hyperref[AB_hP6_144_a_a_cif]{\pageref{AB_hP6_144_a_a_cif}}} \\
\vspace{-0.75cm} \item Te[OH]$_{6}$: {\small A12B6C\_cF608\_210\_4h\_2h\_e} \dotfill {\hyperref[A12B6C_cF608_210_4h_2h_e_cif]{\pageref{A12B6C_cF608_210_4h_2h_e_cif}}} \\
\vspace{-0.75cm} \item Th$_{3}$P$_{4}$: {\small A4B3\_cI28\_220\_c\_a} \dotfill {\hyperref[A4B3_cI28_220_c_a_cif]{\pageref{A4B3_cI28_220_c_a_cif}}} \\
\vspace{-0.75cm} \item Th$_{6}$Mn$_{23}$: {\small A23B6\_cF116\_225\_bd2f\_e} \dotfill {\hyperref[A23B6_cF116_225_bd2f_e_cif]{\pageref{A23B6_cF116_225_bd2f_e_cif}}} \\
\vspace{-0.75cm} \item ThB$_{4}$: {\small A4B\_tP20\_127\_ehj\_g} \dotfill {\hyperref[A4B_tP20_127_ehj_g_cif]{\pageref{A4B_tP20_127_ehj_g_cif}}} \\
\vspace{-0.75cm} \item ThBC: {\small ABC\_tP24\_91\_d\_d\_d} \dotfill {\hyperref[ABC_tP24_91_d_d_d_cif]{\pageref{ABC_tP24_91_d_d_d_cif}}} \\
\vspace{-0.75cm} \item ThBC: {\small ABC\_tP24\_95\_d\_d\_d} \dotfill {\hyperref[ABC_tP24_95_d_d_d_cif]{\pageref{ABC_tP24_95_d_d_d_cif}}} \\
\vspace{-0.75cm} \item ThCl$_{4}$: {\small A4B\_tI20\_88\_f\_a} \dotfill {\hyperref[A4B_tI20_88_f_a_cif]{\pageref{A4B_tI20_88_f_a_cif}}} \\
\vspace{-0.75cm} \item Thortveitite: {\small A7B2C2\_mC22\_12\_aij\_h\_i} \dotfill {\hyperref[A7B2C2_mC22_12_aij_h_i_cif]{\pageref{A7B2C2_mC22_12_aij_h_i_cif}}} \\
\vspace{-0.75cm} \item Ti$_{2}$Ge$_{3}$: {\small A3B2\_tP10\_83\_adk\_j} \dotfill {\hyperref[A3B2_tP10_83_adk_j_cif]{\pageref{A3B2_tP10_83_adk_j_cif}}} \\
\vspace{-0.75cm} \item Ti$_{3}$O: {\small AB3\_hP24\_149\_acgi\_3l} \dotfill {\hyperref[AB3_hP24_149_acgi_3l_cif]{\pageref{AB3_hP24_149_acgi_3l_cif}}} \\
\vspace{-0.75cm} \item Ti$_{3}$P: {\small AB3\_tP32\_86\_g\_3g} \dotfill {\hyperref[AB3_tP32_86_g_3g_cif]{\pageref{AB3_tP32_86_g_3g_cif}}} \\
\vspace{-0.75cm} \item TiAl$_{2}$Br$_{8}$: {\small A2B8C\_oP22\_34\_c\_4c\_a} \dotfill {\hyperref[A2B8C_oP22_34_c_4c_a_cif]{\pageref{A2B8C_oP22_34_c_4c_a_cif}}} \\
\vspace{-0.75cm} \item TiFeSi: {\small ABC\_oI36\_46\_ac\_bc\_3b} \dotfill {\hyperref[ABC_oI36_46_ac_bc_3b_cif]{\pageref{ABC_oI36_46_ac_bc_3b_cif}}} \\
\vspace{-0.75cm} \item Tl$_{4}$HgI$_{6}$: {\small AB6C4\_tP22\_104\_a\_2ac\_c} \dotfill {\hyperref[AB6C4_tP22_104_a_2ac_c_cif]{\pageref{AB6C4_tP22_104_a_2ac_c_cif}}} \\
\vspace{-0.75cm} \item TlP$_{5}$: {\small A5B\_oP24\_26\_3a3b2c\_ab} \dotfill {\hyperref[A5B_oP24_26_3a3b2c_ab_cif]{\pageref{A5B_oP24_26_3a3b2c_ab_cif}}} \\
\vspace{-0.75cm} \item TlZn$_{2}$Sb$_{2}$: {\small A2BC2\_tI20\_79\_c\_2a\_c} \dotfill {\hyperref[A2BC2_tI20_79_c_2a_c_cif]{\pageref{A2BC2_tI20_79_c_2a_c_cif}}} \\
\vspace{-0.75cm} \item Tongbaite: {\small A2B3\_oP20\_62\_2c\_3c} \dotfill {\hyperref[A2B3_oP20_62_2c_3c_cif]{\pageref{A2B3_oP20_62_2c_3c_cif}}} \\
\vspace{-0.75cm} \item Troilite: {\small AB\_hP24\_190\_i\_afh} \dotfill {\hyperref[AB_hP24_190_i_afh_cif]{\pageref{AB_hP24_190_i_afh_cif}}} \\
\vspace{-0.75cm} \item Tychite: {\small A4B2C6D16E\_cF232\_203\_e\_d\_f\_eg\_a} \dotfill {\hyperref[A4B2C6D16E_cF232_203_e_d_f_eg_a_cif]{\pageref{A4B2C6D16E_cF232_203_e_d_f_eg_a_cif}}} \\
\vspace{-0.75cm} \item UCl$_{3}$: {\small A3B\_hP8\_176\_h\_d} \dotfill {\hyperref[A3B_hP8_176_h_d_cif]{\pageref{A3B_hP8_176_h_d_cif}}} \\
\vspace{-0.75cm} \item V$_{2}$MoO$_{8}$: {\small AB8C2\_oC22\_35\_a\_ab3e\_e} \dotfill {\hyperref[AB8C2_oC22_35_a_ab3e_e_cif]{\pageref{AB8C2_oC22_35_a_ab3e_e_cif}}} \\
\vspace{-0.75cm} \item VPCl$_{9}$: {\small A9BC\_oC44\_39\_3c3d\_a\_c} \dotfill {\hyperref[A9BC_oC44_39_3c3d_a_c_cif]{\pageref{A9BC_oC44_39_3c3d_a_c_cif}}} \\
\vspace{-0.75cm} \item W$_{3}$O$_{10}$: {\small A10B3\_oF52\_42\_2abce\_ab} \dotfill {\hyperref[A10B3_oF52_42_2abce_ab_cif]{\pageref{A10B3_oF52_42_2abce_ab_cif}}} \\
\vspace{-0.75cm} \item W$_{5}$Si$_{3}$: {\small A3B5\_tI32\_140\_ah\_bk} \dotfill {\hyperref[A3B5_tI32_140_ah_bk_cif]{\pageref{A3B5_tI32_140_ah_bk_cif}}} \\
\vspace{-0.75cm} \item WO$_{3}$: {\small A3B\_oP32\_60\_3d\_d} \dotfill {\hyperref[A3B_oP32_60_3d_d_cif]{\pageref{A3B_oP32_60_3d_d_cif}}} \\
\vspace{-0.75cm} \item Weberite: {\small AB7CD2\_oI44\_24\_a\_b3d\_c\_ac} \dotfill {\hyperref[AB7CD2_oI44_24_a_b3d_c_ac_cif]{\pageref{AB7CD2_oI44_24_a_b3d_c_ac_cif}}} \\
\vspace{-0.75cm} \item Westerveldite: {\small AB\_oP8\_62\_c\_c} \dotfill {\hyperref[AB_oP8_62_c_c_cif]{\pageref{AB_oP8_62_c_c_cif}}} \\
\vspace{-0.75cm} \item YbBaCo$_{4}$O$_{7}$: {\small AB4C7D\_hP26\_159\_b\_ac\_a2c\_b} \dotfill {\hyperref[AB4C7D_hP26_159_b_ac_a2c_b_cif]{\pageref{AB4C7D_hP26_159_b_ac_a2c_b_cif}}} \\
\vspace{-0.75cm} \item Zn$_{3}$P$_{2}$: {\small A2B3\_tP40\_137\_cdf\_3g} \dotfill {\hyperref[A2B3_tP40_137_cdf_3g_cif]{\pageref{A2B3_tP40_137_cdf_3g_cif}}} \\
\vspace{-0.75cm} \item ZnSb$_{2}$O$_{4}$: {\small A4B2C\_tP28\_135\_gh\_h\_d} \dotfill {\hyperref[A4B2C_tP28_135_gh_h_d_cif]{\pageref{A4B2C_tP28_135_gh_h_d_cif}}} \\
\vspace{-0.75cm} \item ZrO$_{2}$\footnoteref{note:AB2_oP12_29_a_2a-cif}: {\small A2B\_oP12\_29\_2a\_a} \dotfill {\hyperref[A2B_oP12_29_2a_a_cif]{\pageref{A2B_oP12_29_2a_a_cif}}} \\
\vspace{-0.75cm} \item ZrO$_{2}$\footnoteref{note:AB2_tP6_137_a_d-cif}: {\small A2B\_tP6\_137\_d\_a} \dotfill {\hyperref[A2B_tP6_137_d_a_cif]{\pageref{A2B_tP6_137_d_a_cif}}} \\
\end{enumerate}
\section*{\label{sec:poscarInd}POSCAR Index}
\noindent
\begin{enumerate}
\vspace{-0.75cm} \item $\alpha$-Al$_{2}$S$_{3}$: {\small A2B3\_hP30\_169\_2a\_3a} \dotfill {\hyperref[A2B3_hP30_169_2a_3a_poscar]{\pageref{A2B3_hP30_169_2a_3a_poscar}}} \\
\vspace{-0.75cm} \item $\alpha$-CuAlCl$_{4}$: {\small AB4C\_tP12\_112\_b\_n\_e} \dotfill {\hyperref[AB4C_tP12_112_b_n_e_poscar]{\pageref{AB4C_tP12_112_b_n_e_poscar}}} \\
\vspace{-0.75cm} \item $\alpha$-FeSe\footnote[2]{\label{note:AB_oC8_67_a_g-poscar}$\alpha$-FeSe and $\alpha$-PbO have the same \AFLOW\ prototype label. They are generated by the same symmetry operations with different sets of parameters.}: {\small AB\_oC8\_67\_a\_g} \dotfill {\hyperref[AB_oC8_67_a_g-FeSe_poscar]{\pageref{AB_oC8_67_a_g-FeSe_poscar}}} \\
\vspace{-0.75cm} \item $\alpha$-Naumannite: {\small A2B\_oP12\_17\_abe\_e} \dotfill {\hyperref[A2B_oP12_17_abe_e_poscar]{\pageref{A2B_oP12_17_abe_e_poscar}}} \\
\vspace{-0.75cm} \item $\alpha$-NbO$_{2}$: {\small AB2\_tI96\_88\_2f\_4f} \dotfill {\hyperref[AB2_tI96_88_2f_4f_poscar]{\pageref{AB2_tI96_88_2f_4f_poscar}}} \\
\vspace{-0.75cm} \item $\alpha$-P$_3$N$_5$: {\small A5B3\_mC32\_9\_5a\_3a} \dotfill {\hyperref[A5B3_mC32_9_5a_3a_poscar]{\pageref{A5B3_mC32_9_5a_3a_poscar}}} \\
\vspace{-0.75cm} \item $\alpha$-PbO\footnoteref{note:AB_oC8_67_a_g-poscar}: {\small AB\_oC8\_67\_a\_g} \dotfill {\hyperref[AB_oC8_67_a_g-PbO_poscar]{\pageref{AB_oC8_67_a_g-PbO_poscar}}} \\
\vspace{-0.75cm} \item $\alpha$-PdCl$_{2}$: {\small A2B\_oP6\_58\_g\_a} \dotfill {\hyperref[A2B_oP6_58_g_a_poscar]{\pageref{A2B_oP6_58_g_a_poscar}}} \\
\vspace{-0.75cm} \item \begin{raggedleft}$\alpha$-RbPr[MoO$_{4}$]$_{2}$: \end{raggedleft} \\ {\small A2B8CD\_oP24\_48\_k\_2m\_d\_b} \dotfill {\hyperref[A2B8CD_oP24_48_k_2m_d_b_poscar]{\pageref{A2B8CD_oP24_48_k_2m_d_b_poscar}}} \\
\vspace{-0.75cm} \item $\alpha$-Sm$_{3}$Ge$_{5}$: {\small A5B3\_hP16\_190\_bdh\_g} \dotfill {\hyperref[A5B3_hP16_190_bdh_g_poscar]{\pageref{A5B3_hP16_190_bdh_g_poscar}}} \\
\vspace{-0.75cm} \item $\alpha$-ThSi$_{2}$: {\small A2B\_tI12\_141\_e\_a} \dotfill {\hyperref[A2B_tI12_141_e_a_poscar]{\pageref{A2B_tI12_141_e_a_poscar}}} \\
\vspace{-0.75cm} \item $\alpha$-Tl$_{2}$TeO$_{3}$: {\small A3BC2\_oP48\_50\_3m\_m\_2m} \dotfill {\hyperref[A3BC2_oP48_50_3m_m_2m_poscar]{\pageref{A3BC2_oP48_50_3m_m_2m_poscar}}} \\
\vspace{-0.75cm} \item $\alpha$-Toluene: {\small A7B8\_mP120\_14\_14e\_16e} \dotfill {\hyperref[A7B8_mP120_14_14e_16e_poscar]{\pageref{A7B8_mP120_14_14e_16e_poscar}}} \\
\vspace{-0.75cm} \item $\beta$-Bi$_{2}$O$_{3}$: {\small A2B3\_tP20\_117\_i\_adgh} \dotfill {\hyperref[A2B3_tP20_117_i_adgh_poscar]{\pageref{A2B3_tP20_117_i_adgh_poscar}}} \\
\vspace{-0.75cm} \item $\beta$-CuI: {\small AB\_hP4\_156\_ac\_ac} \dotfill {\hyperref[AB_hP4_156_ac_ac_poscar]{\pageref{AB_hP4_156_ac_ac_poscar}}} \\
\vspace{-0.75cm} \item $\beta$-Hg$_{4}$Pt: {\small A4B\_cI10\_229\_c\_a} \dotfill {\hyperref[A4B_cI10_229_c_a_poscar]{\pageref{A4B_cI10_229_c_a_poscar}}} \\
\vspace{-0.75cm} \item $\beta$-NbO$_{2}$: {\small AB2\_tI48\_80\_2b\_4b} \dotfill {\hyperref[AB2_tI48_80_2b_4b_poscar]{\pageref{AB2_tI48_80_2b_4b_poscar}}} \\
\vspace{-0.75cm} \item $\beta$-PdCl$_2$: {\small A2B\_hR18\_148\_2f\_f} \dotfill {\hyperref[A2B_hR18_148_2f_f_poscar]{\pageref{A2B_hR18_148_2f_f_poscar}}} \\
\vspace{-0.75cm} \item $\beta$-RuCl$_{3}$: {\small A3B\_hP8\_158\_d\_a} \dotfill {\hyperref[A3B_hP8_158_d_a_poscar]{\pageref{A3B_hP8_158_d_a_poscar}}} \\
\vspace{-0.75cm} \item $\beta$-RuCl$_{3}$: {\small A3B\_hP8\_185\_c\_a} \dotfill {\hyperref[A3B_hP8_185_c_a_poscar]{\pageref{A3B_hP8_185_c_a_poscar}}} \\
\vspace{-0.75cm} \item $\beta$-SeO$_{2}$\footnote[1]{\label{note:A2B_oP12_26_abc_ab-poscar}H$_{2}$S and $\beta$-SeO$_{2}$ have the same \AFLOW\ prototype label. They are generated by the same symmetry operations with different sets of parameters.}: {\small A2B\_oP12\_26\_abc\_ab} \dotfill {\hyperref[A2B_oP12_26_abc_ab-SeO2_poscar]{\pageref{A2B_oP12_26_abc_ab-SeO2_poscar}}} \\
\vspace{-0.75cm} \item $\beta$-Si$_{3}$N$_{4}$: {\small A4B3\_hP14\_173\_bc\_c} \dotfill {\hyperref[A4B3_hP14_173_bc_c_poscar]{\pageref{A4B3_hP14_173_bc_c_poscar}}} \\
\vspace{-0.75cm} \item $\beta$-SiO$_{2}$: {\small A2B\_hP9\_181\_j\_c} \dotfill {\hyperref[A2B_hP9_181_j_c_poscar]{\pageref{A2B_hP9_181_j_c_poscar}}} \\
\vspace{-0.75cm} \item $\beta$-Ta$_{2}$O$_{5}$: {\small A5B2\_oP14\_49\_dehq\_ab} \dotfill {\hyperref[A5B2_oP14_49_dehq_ab_poscar]{\pageref{A5B2_oP14_49_dehq_ab_poscar}}} \\
\vspace{-0.75cm} \item $\beta$-ThI$_{3}$: {\small A3B\_oC64\_66\_kl2m\_bdl} \dotfill {\hyperref[A3B_oC64_66_kl2m_bdl_poscar]{\pageref{A3B_oC64_66_kl2m_bdl_poscar}}} \\
\vspace{-0.75cm} \item $\beta$-Toluene: {\small A7B8\_oP120\_60\_7d\_8d} \dotfill {\hyperref[A7B8_oP120_60_7d_8d_poscar]{\pageref{A7B8_oP120_60_7d_8d_poscar}}} \\
\vspace{-0.75cm} \item $\beta$-V$_{3}$S: {\small AB3\_tP32\_133\_h\_i2j} \dotfill {\hyperref[AB3_tP32_133_h_i2j_poscar]{\pageref{AB3_tP32_133_h_i2j_poscar}}} \\
\vspace{-0.75cm} \item $\delta$-PdCl$_{2}$: {\small A2B\_mP6\_10\_mn\_bg} \dotfill {\hyperref[A2B_mP6_10_mn_bg_poscar]{\pageref{A2B_mP6_10_mn_bg_poscar}}} \\
\vspace{-0.75cm} \item $\delta_{H}^{II}$-NW$_2$: {\small AB2\_hP9\_164\_bd\_c2d} \dotfill {\hyperref[AB2_hP9_164_bd_c2d_poscar]{\pageref{AB2_hP9_164_bd_c2d_poscar}}} \\
\vspace{-0.75cm} \item $\epsilon$-NiAl$_{3}$: {\small A3B\_oP16\_62\_cd\_c} \dotfill {\hyperref[A3B_oP16_62_cd_c_poscar]{\pageref{A3B_oP16_62_cd_c_poscar}}} \\
\vspace{-0.75cm} \item $\epsilon$-WO$_{3}$: {\small A3B\_mP16\_7\_6a\_2a} \dotfill {\hyperref[A3B_mP16_7_6a_2a_poscar]{\pageref{A3B_mP16_7_6a_2a_poscar}}} \\
\vspace{-0.75cm} \item $\gamma$-Ag$_{3}$SI: {\small A3BC\_hR5\_146\_b\_a\_a} \dotfill {\hyperref[A3BC_hR5_146_b_a_a_poscar]{\pageref{A3BC_hR5_146_b_a_a_poscar}}} \\
\vspace{-0.75cm} \item $\gamma$-MgNiSn: {\small A7B7C2\_tP32\_101\_bde\_ade\_d} \dotfill {\hyperref[A7B7C2_tP32_101_bde_ade_d_poscar]{\pageref{A7B7C2_tP32_101_bde_ade_d_poscar}}} \\
\vspace{-0.75cm} \item $\gamma$-PdCl$_{2}$: {\small A2B\_mP6\_14\_e\_a} \dotfill {\hyperref[A2B_mP6_14_e_a_poscar]{\pageref{A2B_mP6_14_e_a_poscar}}} \\
\vspace{-0.75cm} \item $\gamma$-brass: {\small A4B9\_cP52\_215\_ei\_3efgi} \dotfill {\hyperref[A4B9_cP52_215_ei_3efgi_poscar]{\pageref{A4B9_cP52_215_ei_3efgi_poscar}}} \\
\vspace{-0.75cm} \item $\gamma$-brass: {\small A3B10\_cI52\_229\_e\_fh} \dotfill {\hyperref[A3B10_cI52_229_e_fh_poscar]{\pageref{A3B10_cI52_229_e_fh_poscar}}} \\
\vspace{-0.75cm} \item $\kappa$-alumina: {\small A2B3\_oP40\_33\_4a\_6a} \dotfill {\hyperref[A2B3_oP40_33_4a_6a_poscar]{\pageref{A2B3_oP40_33_4a_6a_poscar}}} \\
\vspace{-0.75cm} \item \begin{raggedleft}$\pi$-FeMg$_{3}$Al$_{8}$Si$_{6}$: \end{raggedleft} \\ {\small A8BC3D6\_hP18\_189\_bfh\_a\_g\_i} \dotfill {\hyperref[A8BC3D6_hP18_189_bfh_a_g_i_poscar]{\pageref{A8BC3D6_hP18_189_bfh_a_g_i_poscar}}} \\
\vspace{-0.75cm} \item \begin{raggedleft}$\pi$-FeMg$_{3}$Al$_{9}$Si$_{5}$: \end{raggedleft} \\ {\small A9BC3D5\_hP18\_189\_fi\_a\_g\_bh} \dotfill {\hyperref[A9BC3D5_hP18_189_fi_a_g_bh_poscar]{\pageref{A9BC3D5_hP18_189_fi_a_g_bh_poscar}}} \\
\vspace{-0.75cm} \item Ag$_{3}$[PO$_{4}$]: {\small A3B4C\_cP16\_218\_c\_e\_a} \dotfill {\hyperref[A3B4C_cP16_218_c_e_a_poscar]{\pageref{A3B4C_cP16_218_c_e_a_poscar}}} \\
\vspace{-0.75cm} \item Ag$_{5}$Pb$_{2}$O$_{6}$: {\small A5B6C2\_hP13\_157\_2ac\_2c\_b} \dotfill {\hyperref[A5B6C2_hP13_157_2ac_2c_b_poscar]{\pageref{A5B6C2_hP13_157_2ac_2c_b_poscar}}} \\
\vspace{-0.75cm} \item AgUF$_{6}$: {\small AB6C\_tP16\_132\_d\_io\_a} \dotfill {\hyperref[AB6C_tP16_132_d_io_a_poscar]{\pageref{AB6C_tP16_132_d_io_a_poscar}}} \\
\vspace{-0.75cm} \item Akermanite: {\small A2BC7D2\_tP24\_113\_e\_a\_cef\_e} \dotfill {\hyperref[A2BC7D2_tP24_113_e_a_cef_e_poscar]{\pageref{A2BC7D2_tP24_113_e_a_cef_e_poscar}}} \\
\vspace{-0.75cm} \item Al$_{2}$CuIr\footnote[4]{\label{note:ABC2_oC16_67_b_g_ag-poscar}Al$_{2}$CuIr and HoCuP$_{2}$ have similar \AFLOW\ prototype labels ({\it{i.e.}}, same symmetry and set of Wyckoff positions with different stoichiometry labels due to alphabetic ordering of atomic species). They are generated by the same symmetry operations with different sets of parameters.}: {\small A2BC\_oC16\_67\_ag\_b\_g} \dotfill {\hyperref[A2BC_oC16_67_ag_b_g_poscar]{\pageref{A2BC_oC16_67_ag_b_g_poscar}}} \\
\vspace{-0.75cm} \item Al$_{2}$S$_{3}$: {\small A2B3\_hP30\_170\_2a\_3a} \dotfill {\hyperref[A2B3_hP30_170_2a_3a_poscar]{\pageref{A2B3_hP30_170_2a_3a_poscar}}} \\
\vspace{-0.75cm} \item Al$_{4}$C$_{3}$: {\small A4B3\_hR7\_166\_2c\_ac} \dotfill {\hyperref[A4B3_hR7_166_2c_ac_poscar]{\pageref{A4B3_hR7_166_2c_ac_poscar}}} \\
\vspace{-0.75cm} \item Al$_{4}$U: {\small A4B\_oI20\_74\_beh\_e} \dotfill {\hyperref[A4B_oI20_74_beh_e_poscar]{\pageref{A4B_oI20_74_beh_e_poscar}}} \\
\vspace{-0.75cm} \item Al$_{8}$Cr$_{5}$: {\small A8B5\_hR26\_160\_a3bc\_a3b} \dotfill {\hyperref[A8B5_hR26_160_a3bc_a3b_poscar]{\pageref{A8B5_hR26_160_a3bc_a3b_poscar}}} \\
\vspace{-0.75cm} \item Al$_{9}$Mn$_{3}$Si: {\small A9B3C\_hP26\_194\_hk\_h\_a} \dotfill {\hyperref[A9B3C_hP26_194_hk_h_a_poscar]{\pageref{A9B3C_hP26_194_hk_h_a_poscar}}} \\
\vspace{-0.75cm} \item AlLi$_{3}$N$_{2}$: {\small AB3C2\_cI96\_206\_c\_e\_ad} \dotfill {\hyperref[AB3C2_cI96_206_c_e_ad_poscar]{\pageref{AB3C2_cI96_206_c_e_ad_poscar}}} \\
\vspace{-0.75cm} \item AlPO$_{4}$: {\small AB2\_hP72\_192\_m\_j2kl} \dotfill {\hyperref[AB2_hP72_192_m_j2kl_poscar]{\pageref{AB2_hP72_192_m_j2kl_poscar}}} \\
\vspace{-0.75cm} \item Al[PO$_{4}$]: {\small AB4C\_hP72\_168\_2d\_8d\_2d} \dotfill {\hyperref[AB4C_hP72_168_2d_8d_2d_poscar]{\pageref{AB4C_hP72_168_2d_8d_2d_poscar}}} \\
\vspace{-0.75cm} \item Al[PO$_{4}$]: {\small AB4C\_hP72\_184\_d\_4d\_d} \dotfill {\hyperref[AB4C_hP72_184_d_4d_d_poscar]{\pageref{AB4C_hP72_184_d_4d_d_poscar}}} \\
\vspace{-0.75cm} \item Anhydrite: {\small AB4C\_oC24\_63\_c\_fg\_c} \dotfill {\hyperref[AB4C_oC24_63_c_fg_c_poscar]{\pageref{AB4C_oC24_63_c_fg_c_poscar}}} \\
\vspace{-0.75cm} \item As$_{2}$Ba: {\small A2B\_mP18\_7\_6a\_3a} \dotfill {\hyperref[A2B_mP18_7_6a_3a_poscar]{\pageref{A2B_mP18_7_6a_3a_poscar}}} \\
\vspace{-0.75cm} \item \begin{raggedleft}AsPh$_{4}$CeS$_{8}$P$_{4}$Me$_{8}$: \end{raggedleft} \\ {\small AB32CD4E8\_tP184\_93\_i\_16p\_af\_2p\_4p} \dotfill {\hyperref[AB32CD4E8_tP184_93_i_16p_af_2p_4p_poscar]{\pageref{AB32CD4E8_tP184_93_i_16p_af_2p_4p_poscar}}} \\
\vspace{-0.75cm} \item AuCN: {\small ABC\_hP3\_183\_a\_a\_a} \dotfill {\hyperref[ABC_hP3_183_a_a_a_poscar]{\pageref{ABC_hP3_183_a_a_a_poscar}}} \\
\vspace{-0.75cm} \item AuF$_{3}$: {\small AB3\_hP24\_178\_b\_ac} \dotfill {\hyperref[AB3_hP24_178_b_ac_poscar]{\pageref{AB3_hP24_178_b_ac_poscar}}} \\
\vspace{-0.75cm} \item AuF$_{3}$: {\small AB3\_hP24\_179\_b\_ac} \dotfill {\hyperref[AB3_hP24_179_b_ac_poscar]{\pageref{AB3_hP24_179_b_ac_poscar}}} \\
\vspace{-0.75cm} \item BN: {\small AB\_oF8\_42\_a\_a} \dotfill {\hyperref[AB_oF8_42_a_a_poscar]{\pageref{AB_oF8_42_a_a_poscar}}} \\
\vspace{-0.75cm} \item BPS$_{4}$: {\small ABC4\_oI12\_23\_a\_b\_k} \dotfill {\hyperref[ABC4_oI12_23_a_b_k_poscar]{\pageref{ABC4_oI12_23_a_b_k_poscar}}} \\
\vspace{-0.75cm} \item Ba$_{5}$In$_{4}$Bi$_{5}$: {\small A5B5C4\_tP28\_104\_ac\_ac\_c} \dotfill {\hyperref[A5B5C4_tP28_104_ac_ac_c_poscar]{\pageref{A5B5C4_tP28_104_ac_ac_c_poscar}}} \\
\vspace{-0.75cm} \item Ba$_{5}$Si$_{3}$: {\small A5B3\_tP32\_130\_cg\_cf} \dotfill {\hyperref[A5B3_tP32_130_cg_cf_poscar]{\pageref{A5B3_tP32_130_cg_cf_poscar}}} \\
\vspace{-0.75cm} \item \begin{raggedleft}BaCr$_{2}$Ru$_{4}$O$_{12}$: \end{raggedleft} \\ {\small AB2C12D4\_tP76\_75\_2a2b\_2d\_12d\_4d} \dotfill {\hyperref[AB2C12D4_tP76_75_2a2b_2d_12d_4d_poscar]{\pageref{AB2C12D4_tP76_75_2a2b_2d_12d_4d_poscar}}} \\
\vspace{-0.75cm} \item \begin{raggedleft}BaCu$_{4}$[VO][PO$_{4}$]$_{4}$: \end{raggedleft} \\ {\small AB4C17D4E\_tP54\_90\_a\_g\_c4g\_g\_c} \dotfill {\hyperref[AB4C17D4E_tP54_90_a_g_c4g_g_c_poscar]{\pageref{AB4C17D4E_tP54_90_a_g_c4g_g_c_poscar}}} \\
\vspace{-0.75cm} \item BaGe$_{2}$As$_{2}$: {\small A2BC2\_tP20\_105\_f\_ac\_2e} \dotfill {\hyperref[A2BC2_tP20_105_f_ac_2e_poscar]{\pageref{A2BC2_tP20_105_f_ac_2e_poscar}}} \\
\vspace{-0.75cm} \item BaSi$_{4}$O$_{9}$: {\small AB9C4\_hP28\_188\_e\_kl\_ak} \dotfill {\hyperref[AB9C4_hP28_188_e_kl_ak_poscar]{\pageref{AB9C4_hP28_188_e_kl_ak_poscar}}} \\
\vspace{-0.75cm} \item Barite: {\small AB4C\_oP24\_62\_c\_2cd\_c} \dotfill {\hyperref[AB4C_oP24_62_c_2cd_c_poscar]{\pageref{AB4C_oP24_62_c_2cd_c_poscar}}} \\
\vspace{-0.75cm} \item Be[BH$_{4}$]$_{2}$: {\small A2BC8\_tI176\_110\_2b\_b\_8b} \dotfill {\hyperref[A2BC8_tI176_110_2b_b_8b_poscar]{\pageref{A2BC8_tI176_110_2b_b_8b_poscar}}} \\
\vspace{-0.75cm} \item Benzene: {\small AB\_oP48\_61\_3c\_3c} \dotfill {\hyperref[AB_oP48_61_3c_3c_poscar]{\pageref{AB_oP48_61_3c_3c_poscar}}} \\
\vspace{-0.75cm} \item Beryl: {\small A2B3C18D6\_hP58\_192\_c\_f\_lm\_l} \dotfill {\hyperref[A2B3C18D6_hP58_192_c_f_lm_l_poscar]{\pageref{A2B3C18D6_hP58_192_c_f_lm_l_poscar}}} \\
\vspace{-0.75cm} \item Bi$_{2}$O$_{3}$: {\small A2B3\_hP20\_159\_bc\_2c} \dotfill {\hyperref[A2B3_hP20_159_bc_2c_poscar]{\pageref{A2B3_hP20_159_bc_2c_poscar}}} \\
\vspace{-0.75cm} \item Bi$_{5}$Nb$_{3}$O$_{15}$: {\small A5B3C15\_oP46\_30\_a2c\_bc\_a7c} \dotfill {\hyperref[A5B3C15_oP46_30_a2c_bc_a7c_poscar]{\pageref{A5B3C15_oP46_30_a2c_bc_a7c_poscar}}} \\
\vspace{-0.75cm} \item BiAl$_{2}$S$_{4}$: {\small A2BC4\_tP28\_126\_cd\_e\_k} \dotfill {\hyperref[A2BC4_tP28_126_cd_e_k_poscar]{\pageref{A2BC4_tP28_126_cd_e_k_poscar}}} \\
\vspace{-0.75cm} \item BiGaO$_{3}$: {\small ABC3\_oP20\_54\_e\_d\_cf} \dotfill {\hyperref[ABC3_oP20_54_e_d_cf_poscar]{\pageref{ABC3_oP20_54_e_d_cf_poscar}}} \\
\vspace{-0.75cm} \item Boracite: {\small A7BC3D13\_cF192\_219\_de\_b\_c\_ah} \dotfill {\hyperref[A7BC3D13_cF192_219_de_b_c_ah_poscar]{\pageref{A7BC3D13_cF192_219_de_b_c_ah_poscar}}} \\
\vspace{-0.75cm} \item C: {\small A\_tP12\_138\_bi} \dotfill {\hyperref[A_tP12_138_bi_poscar]{\pageref{A_tP12_138_bi_poscar}}} \\
\vspace{-0.75cm} \item C$_{17}$FeO$_{4}$Pt: {\small A17BC4D\_tP184\_89\_17p\_p\_4p\_io} \dotfill {\hyperref[A17BC4D_tP184_89_17p_p_4p_io_poscar]{\pageref{A17BC4D_tP184_89_17p_p_4p_io_poscar}}} \\
\vspace{-0.75cm} \item Ca$_{3}$Al$_{2}$O$_{6}$: {\small A2B3C6\_cP264\_205\_2d\_ab2c2d\_6d} \dotfill {\hyperref[A2B3C6_cP264_205_2d_ab2c2d_6d_poscar]{\pageref{A2B3C6_cP264_205_2d_ab2c2d_6d_poscar}}} \\
\vspace{-0.75cm} \item Ca$_{3}$Al$_{2}$O$_{6}$: {\small A2B3C6\_cP33\_221\_cd\_ag\_fh} \dotfill {\hyperref[A2B3C6_cP33_221_cd_ag_fh_poscar]{\pageref{A2B3C6_cP33_221_cd_ag_fh_poscar}}} \\
\vspace{-0.75cm} \item Ca$_{3}$PI$_{3}$: {\small A3B3C\_cI56\_214\_g\_h\_a} \dotfill {\hyperref[A3B3C_cI56_214_g_h_a_poscar]{\pageref{A3B3C_cI56_214_g_h_a_poscar}}} \\
\vspace{-0.75cm} \item \begin{raggedleft}Ca$_{4}$Al$_{6}$O$_{16}$S: \end{raggedleft} \\ {\small A6B4C16D\_oP108\_27\_abcd4e\_4e\_16e\_e} \dotfill {\hyperref[A6B4C16D_oP108_27_abcd4e_4e_16e_e_poscar]{\pageref{A6B4C16D_oP108_27_abcd4e_4e_16e_e_poscar}}} \\
\vspace{-0.75cm} \item CaRbFe$_{4}$As$_{4}$: {\small A4BC4D\_tP10\_123\_gh\_a\_i\_d} \dotfill {\hyperref[A4BC4D_tP10_123_gh_a_i_d_poscar]{\pageref{A4BC4D_tP10_123_gh_a_i_d_poscar}}} \\
\vspace{-0.75cm} \item Calomel: {\small AB\_tI8\_139\_e\_e} \dotfill {\hyperref[AB_tI8_139_e_e_poscar]{\pageref{AB_tI8_139_e_e_poscar}}} \\
\vspace{-0.75cm} \item Carbonyl Sulphide: {\small ABC\_hR3\_160\_a\_a\_a} \dotfill {\hyperref[ABC_hR3_160_a_a_a_poscar]{\pageref{ABC_hR3_160_a_a_a_poscar}}} \\
\vspace{-0.75cm} \item CdAs$_{2}$: {\small A2B\_tI12\_98\_f\_a} \dotfill {\hyperref[A2B_tI12_98_f_a_poscar]{\pageref{A2B_tI12_98_f_a_poscar}}} \\
\vspace{-0.75cm} \item CdI$_{2}$: {\small AB2\_hP9\_156\_b2c\_3a2bc} \dotfill {\hyperref[AB2_hP9_156_b2c_3a2bc_poscar]{\pageref{AB2_hP9_156_b2c_3a2bc_poscar}}} \\
\vspace{-0.75cm} \item Ce$_{3}$Si$_{6}$N$_{11}$: {\small A3B11C6\_tP40\_100\_ac\_bc2d\_cd} \dotfill {\hyperref[A3B11C6_tP40_100_ac_bc2d_cd_poscar]{\pageref{A3B11C6_tP40_100_ac_bc2d_cd_poscar}}} \\
\vspace{-0.75cm} \item Ce$_{5}$Mo$_{3}$O$_{16}$: {\small A5B3C16\_cP96\_222\_ce\_d\_fi} \dotfill {\hyperref[A5B3C16_cP96_222_ce_d_fi_poscar]{\pageref{A5B3C16_cP96_222_ce_d_fi_poscar}}} \\
\vspace{-0.75cm} \item CeCo$_{4}$B$_{4}$: {\small A4BC4\_tP18\_137\_g\_b\_g} \dotfill {\hyperref[A4BC4_tP18_137_g_b_g_poscar]{\pageref{A4BC4_tP18_137_g_b_g_poscar}}} \\
\vspace{-0.75cm} \item CeRu$_{2}$B$_{2}$: {\small A2BC2\_oF40\_22\_fi\_ad\_gh} \dotfill {\hyperref[A2BC2_oF40_22_fi_ad_gh_poscar]{\pageref{A2BC2_oF40_22_fi_ad_gh_poscar}}} \\
\vspace{-0.75cm} \item CeTe$_{3}$: {\small AB3\_oC16\_40\_b\_3b} \dotfill {\hyperref[AB3_oC16_40_b_3b_poscar]{\pageref{AB3_oC16_40_b_3b_poscar}}} \\
\vspace{-0.75cm} \item Co$_{2}$Al$_{5}$: {\small A5B2\_hP28\_194\_ahk\_ch} \dotfill {\hyperref[A5B2_hP28_194_ahk_ch_poscar]{\pageref{A5B2_hP28_194_ahk_ch_poscar}}} \\
\vspace{-0.75cm} \item Co$_{5}$Ge$_{7}$: {\small A5B7\_tI24\_107\_ac\_abd} \dotfill {\hyperref[A5B7_tI24_107_ac_abd_poscar]{\pageref{A5B7_tI24_107_ac_abd_poscar}}} \\
\vspace{-0.75cm} \item Cobaltite: {\small ABC\_oP12\_29\_a\_a\_a} \dotfill {\hyperref[ABC_oP12_29_a_a_a_poscar]{\pageref{ABC_oP12_29_a_a_a_poscar}}} \\
\vspace{-0.75cm} \item Cr$_{5}$B$_{3}$: {\small A3B5\_tI32\_140\_ah\_cl} \dotfill {\hyperref[A3B5_tI32_140_ah_cl_poscar]{\pageref{A3B5_tI32_140_ah_cl_poscar}}} \\
\vspace{-0.75cm} \item CrCl$_{3}$: {\small A3B\_hP24\_153\_3c\_2b} \dotfill {\hyperref[A3B_hP24_153_3c_2b_poscar]{\pageref{A3B_hP24_153_3c_2b_poscar}}} \\
\vspace{-0.75cm} \item CrFe$_{3}$NiSn$_{5}$: {\small AB\_hP6\_183\_c\_ab} \dotfill {\hyperref[AB_hP6_183_c_ab_poscar]{\pageref{AB_hP6_183_c_ab_poscar}}} \\
\vspace{-0.75cm} \item \begin{raggedleft}Cs$_{2}$ZnFe[CN]$_{6}$: \end{raggedleft} \\ {\small A6B2CD6E\_cP64\_208\_m\_ad\_b\_m\_c} \dotfill {\hyperref[A6B2CD6E_cP64_208_m_ad_b_m_c_poscar]{\pageref{A6B2CD6E_cP64_208_m_ad_b_m_c_poscar}}} \\
\vspace{-0.75cm} \item Cs$_{3}$P$_{7}$: {\small A3B7\_tP40\_76\_3a\_7a} \dotfill {\hyperref[A3B7_tP40_76_3a_7a_poscar]{\pageref{A3B7_tP40_76_3a_7a_poscar}}} \\
\vspace{-0.75cm} \item CsPr[MoO$_{4}$]$_{2}$: {\small AB2C8D\_oP24\_49\_g\_q\_2qr\_e} \dotfill {\hyperref[AB2C8D_oP24_49_g_q_2qr_e_poscar]{\pageref{AB2C8D_oP24_49_g_q_2qr_e_poscar}}} \\
\vspace{-0.75cm} \item Cu$_{15}$Si$_{4}$: {\small A15B4\_cI76\_220\_ae\_c} \dotfill {\hyperref[A15B4_cI76_220_ae_c_poscar]{\pageref{A15B4_cI76_220_ae_c_poscar}}} \\
\vspace{-0.75cm} \item Cu$_{2}$Fe[CN]$_{6}$: {\small A12B2C\_cF60\_196\_h\_bc\_a} \dotfill {\hyperref[A12B2C_cF60_196_h_bc_a_poscar]{\pageref{A12B2C_cF60_196_h_bc_a_poscar}}} \\
\vspace{-0.75cm} \item Cu$_{3}$P: {\small A3B\_hP24\_165\_bdg\_f} \dotfill {\hyperref[A3B_hP24_165_bdg_f_poscar]{\pageref{A3B_hP24_165_bdg_f_poscar}}} \\
\vspace{-0.75cm} \item Cu$_{3}$P\footnote[6]{\label{note:AB3_hP24_185_c_ab2c-poscar}Cu$_{3}$P and Na$_{3}$As have similar \AFLOW\ prototype labels ({\it{i.e.}}, same symmetry and set of Wyckoff positions with different stoichiometry labels due to alphabetic ordering of atomic species). They are generated by the same symmetry operations with different sets of parameters.}: {\small A3B\_hP24\_185\_ab2c\_c} \dotfill {\hyperref[A3B_hP24_185_ab2c_c_poscar]{\pageref{A3B_hP24_185_ab2c_c_poscar}}} \\
\vspace{-0.75cm} \item CuBi$_{2}$O$_{4}$: {\small A2BC4\_tP28\_130\_f\_c\_g} \dotfill {\hyperref[A2BC4_tP28_130_f_c_g_poscar]{\pageref{A2BC4_tP28_130_f_c_g_poscar}}} \\
\vspace{-0.75cm} \item CuBrSe$_{3}$: {\small ABC3\_oP20\_30\_2a\_c\_3c} \dotfill {\hyperref[ABC3_oP20_30_2a_c_3c_poscar]{\pageref{ABC3_oP20_30_2a_c_3c_poscar}}} \\
\vspace{-0.75cm} \item CuBrSe$_{3}$: {\small ABC3\_oP20\_53\_e\_g\_hi} \dotfill {\hyperref[ABC3_oP20_53_e_g_hi_poscar]{\pageref{ABC3_oP20_53_e_g_hi_poscar}}} \\
\vspace{-0.75cm} \item CuCrCl$_{5}$[NH$_{3}$]$_{6}$: {\small A5BCD6\_cF416\_228\_eg\_c\_b\_h} \dotfill {\hyperref[A5BCD6_cF416_228_eg_c_b_h_poscar]{\pageref{A5BCD6_cF416_228_eg_c_b_h_poscar}}} \\
\vspace{-0.75cm} \item CuI: {\small AB\_hP12\_156\_2ab3c\_2ab3c} \dotfill {\hyperref[AB_hP12_156_2ab3c_2ab3c_poscar]{\pageref{AB_hP12_156_2ab3c_2ab3c_poscar}}} \\
\vspace{-0.75cm} \item CuNiSb$_{2}$: {\small ABC2\_hP4\_164\_a\_b\_d} \dotfill {\hyperref[ABC2_hP4_164_a_b_d_poscar]{\pageref{ABC2_hP4_164_a_b_d_poscar}}} \\
\vspace{-0.75cm} \item Cubanite: {\small AB2C3\_oP24\_62\_c\_d\_cd} \dotfill {\hyperref[AB2C3_oP24_62_c_d_cd_poscar]{\pageref{AB2C3_oP24_62_c_d_cd_poscar}}} \\
\vspace{-0.75cm} \item Downeyite: {\small A2B\_tP24\_135\_gh\_h} \dotfill {\hyperref[A2B_tP24_135_gh_h_poscar]{\pageref{A2B_tP24_135_gh_h_poscar}}} \\
\vspace{-0.75cm} \item Er$_{3}$Ru$_{2}$: {\small A3B2\_hP10\_176\_h\_bd} \dotfill {\hyperref[A3B2_hP10_176_h_bd_poscar]{\pageref{A3B2_hP10_176_h_bd_poscar}}} \\
\vspace{-0.75cm} \item F$_{6}$KP: {\small A24BC\_cF104\_209\_j\_a\_b} \dotfill {\hyperref[A24BC_cF104_209_j_a_b_poscar]{\pageref{A24BC_cF104_209_j_a_b_poscar}}} \\
\vspace{-0.75cm} \item \begin{raggedleft}FCC C$_{60}$ Buckminsterfullerine: \end{raggedleft} \\ {\small A\_cF240\_202\_h2i} \dotfill {\hyperref[A_cF240_202_h2i_poscar]{\pageref{A_cF240_202_h2i_poscar}}} \\
\vspace{-0.75cm} \item Fe$_{12}$Zr$_{2}$P$_{7}$: {\small A12B7C2\_hP21\_174\_2j2k\_ajk\_cf} \dotfill {\hyperref[A12B7C2_hP21_174_2j2k_ajk_cf_poscar]{\pageref{A12B7C2_hP21_174_2j2k_ajk_cf_poscar}}} \\
\vspace{-0.75cm} \item Fe$_{3}$Te$_{3}$Tl: {\small A3B3C\_hP14\_176\_h\_h\_d} \dotfill {\hyperref[A3B3C_hP14_176_h_h_d_poscar]{\pageref{A3B3C_hP14_176_h_h_d_poscar}}} \\
\vspace{-0.75cm} \item Fe$_{3}$Th$_{7}$: {\small A3B7\_hP20\_186\_c\_b2c} \dotfill {\hyperref[A3B7_hP20_186_c_b2c_poscar]{\pageref{A3B7_hP20_186_c_b2c_poscar}}} \\
\vspace{-0.75cm} \item FeCu$_{2}$Al$_{7}$: {\small A7B2C\_tP40\_128\_egi\_h\_e} \dotfill {\hyperref[A7B2C_tP40_128_egi_h_e_poscar]{\pageref{A7B2C_tP40_128_egi_h_e_poscar}}} \\
\vspace{-0.75cm} \item FeNi: {\small AB\_mP4\_6\_2b\_2a} \dotfill {\hyperref[AB_mP4_6_2b_2a_poscar]{\pageref{AB_mP4_6_2b_2a_poscar}}} \\
\vspace{-0.75cm} \item FeOCl: {\small ABC\_oP6\_59\_a\_b\_a} \dotfill {\hyperref[ABC_oP6_59_a_b_a_poscar]{\pageref{ABC_oP6_59_a_b_a_poscar}}} \\
\vspace{-0.75cm} \item FePSe$_{3}$: {\small ABC3\_hR10\_146\_2a\_2a\_2b} \dotfill {\hyperref[ABC3_hR10_146_2a_2a_2b_poscar]{\pageref{ABC3_hR10_146_2a_2a_2b_poscar}}} \\
\vspace{-0.75cm} \item FeS: {\small AB\_oF8\_22\_a\_c} \dotfill {\hyperref[AB_oF8_22_a_c_poscar]{\pageref{AB_oF8_22_a_c_poscar}}} \\
\vspace{-0.75cm} \item FeSb$_{2}$: {\small AB2\_oP6\_34\_a\_c} \dotfill {\hyperref[AB2_oP6_34_a_c_poscar]{\pageref{AB2_oP6_34_a_c_poscar}}} \\
\vspace{-0.75cm} \item Forsterite: {\small A2B4C\_oP28\_62\_ac\_2cd\_c} \dotfill {\hyperref[A2B4C_oP28_62_ac_2cd_c_poscar]{\pageref{A2B4C_oP28_62_ac_2cd_c_poscar}}} \\
\vspace{-0.75cm} \item Fresnoite: {\small A2B8C2D\_tP26\_100\_c\_abcd\_c\_a} \dotfill {\hyperref[A2B8C2D_tP26_100_c_abcd_c_a_poscar]{\pageref{A2B8C2D_tP26_100_c_abcd_c_a_poscar}}} \\
\vspace{-0.75cm} \item GaCl$_{2}$: {\small A2B\_oP24\_52\_2e\_cd} \dotfill {\hyperref[A2B_oP24_52_2e_cd_poscar]{\pageref{A2B_oP24_52_2e_cd_poscar}}} \\
\vspace{-0.75cm} \item GaSb: {\small AB\_tI4\_119\_c\_a} \dotfill {\hyperref[AB_tI4_119_c_a_poscar]{\pageref{AB_tI4_119_c_a_poscar}}} \\
\vspace{-0.75cm} \item Garnet: {\small A2B3C12D3\_cI160\_230\_a\_c\_h\_d} \dotfill {\hyperref[A2B3C12D3_cI160_230_a_c_h_d_poscar]{\pageref{A2B3C12D3_cI160_230_a_c_h_d_poscar}}} \\
\vspace{-0.75cm} \item Gd$_{3}$Al$_{2}$: {\small A2B3\_tP20\_102\_2c\_b2c} \dotfill {\hyperref[A2B3_tP20_102_2c_b2c_poscar]{\pageref{A2B3_tP20_102_2c_b2c_poscar}}} \\
\vspace{-0.75cm} \item GdSI: {\small ABC\_hP12\_174\_cj\_fk\_aj} \dotfill {\hyperref[ABC_hP12_174_cj_fk_aj_poscar]{\pageref{ABC_hP12_174_cj_fk_aj_poscar}}} \\
\vspace{-0.75cm} \item GeAs$_{2}$: {\small A2B\_oP24\_55\_2g2h\_gh} \dotfill {\hyperref[A2B_oP24_55_2g2h_gh_poscar]{\pageref{A2B_oP24_55_2g2h_gh_poscar}}} \\
\vspace{-0.75cm} \item GeP: {\small AB\_tI4\_107\_a\_a} \dotfill {\hyperref[AB_tI4_107_a_a_poscar]{\pageref{AB_tI4_107_a_a_poscar}}} \\
\vspace{-0.75cm} \item GeSe$_{2}$: {\small AB2\_tP12\_81\_adg\_2h} \dotfill {\hyperref[AB2_tP12_81_adg_2h_poscar]{\pageref{AB2_tP12_81_adg_2h_poscar}}} \\
\vspace{-0.75cm} \item H$_{2}$S: {\small A2B\_aP6\_2\_aei\_i} \dotfill {\hyperref[A2B_aP6_2_aei_i_poscar]{\pageref{A2B_aP6_2_aei_i_poscar}}} \\
\vspace{-0.75cm} \item H$_{2}$S: {\small A2B\_mP12\_13\_2g\_ef} \dotfill {\hyperref[A2B_mP12_13_2g_ef_poscar]{\pageref{A2B_mP12_13_2g_ef_poscar}}} \\
\vspace{-0.75cm} \item H$_{2}$S\footnoteref{note:A2B_oP12_26_abc_ab-poscar}: {\small A2B\_oP12\_26\_abc\_ab} \dotfill {\hyperref[A2B_oP12_26_abc_ab-H2S_poscar]{\pageref{A2B_oP12_26_abc_ab-H2S_poscar}}} \\
\vspace{-0.75cm} \item H$_{2}$S: {\small A2B\_oC24\_64\_2f\_f} \dotfill {\hyperref[A2B_oC24_64_2f_f_poscar]{\pageref{A2B_oC24_64_2f_f_poscar}}} \\
\vspace{-0.75cm} \item H$_{2}$S III: {\small A2B\_tP48\_77\_8d\_4d} \dotfill {\hyperref[A2B_tP48_77_8d_4d_poscar]{\pageref{A2B_tP48_77_8d_4d_poscar}}} \\
\vspace{-0.75cm} \item H$_{2}$S IV: {\small A2B\_mP12\_7\_4a\_2a} \dotfill {\hyperref[A2B_mP12_7_4a_2a_poscar]{\pageref{A2B_mP12_7_4a_2a_poscar}}} \\
\vspace{-0.75cm} \item H$_{3}$Cl: {\small AB3\_mC16\_9\_a\_3a} \dotfill {\hyperref[AB3_mC16_9_a_3a_poscar]{\pageref{AB3_mC16_9_a_3a_poscar}}} \\
\vspace{-0.75cm} \item H$_{3}$Cl: {\small AB3\_mP16\_10\_mn\_3m3n} \dotfill {\hyperref[AB3_mP16_10_mn_3m3n_poscar]{\pageref{AB3_mP16_10_mn_3m3n_poscar}}} \\
\vspace{-0.75cm} \item H$_{3}$Cl: {\small AB3\_mC16\_15\_e\_cf} \dotfill {\hyperref[AB3_mC16_15_e_cf_poscar]{\pageref{AB3_mC16_15_e_cf_poscar}}} \\
\vspace{-0.75cm} \item H$_{3}$Cl: {\small AB3\_oP16\_19\_a\_3a} \dotfill {\hyperref[AB3_oP16_19_a_3a_poscar]{\pageref{AB3_oP16_19_a_3a_poscar}}} \\
\vspace{-0.75cm} \item H$_{3}$S: {\small A3B\_oI32\_23\_ij2k\_k} \dotfill {\hyperref[A3B_oI32_23_ij2k_k_poscar]{\pageref{A3B_oI32_23_ij2k_k_poscar}}} \\
\vspace{-0.75cm} \item H$_{3}$S: {\small A3B\_oC64\_66\_gi2lm\_2l} \dotfill {\hyperref[A3B_oC64_66_gi2lm_2l_poscar]{\pageref{A3B_oC64_66_gi2lm_2l_poscar}}} \\
\vspace{-0.75cm} \item H$_{3}$S: {\small A3B\_hR4\_160\_b\_a} \dotfill {\hyperref[A3B_hR4_160_b_a_poscar]{\pageref{A3B_hR4_160_b_a_poscar}}} \\
\vspace{-0.75cm} \item H-III: {\small A\_mC24\_15\_2e2f} \dotfill {\hyperref[A_mC24_15_2e2f_poscar]{\pageref{A_mC24_15_2e2f_poscar}}} \\
\vspace{-0.75cm} \item HCl: {\small AB\_oC8\_36\_a\_a} \dotfill {\hyperref[AB_oC8_36_a_a_poscar]{\pageref{AB_oC8_36_a_a_poscar}}} \\
\vspace{-0.75cm} \item HgI$_{2}$: {\small AB2\_tP12\_115\_j\_egi} \dotfill {\hyperref[AB2_tP12_115_j_egi_poscar]{\pageref{AB2_tP12_115_j_egi_poscar}}} \\
\vspace{-0.75cm} \item HgI$_{2}$\footnote[5]{\label{note:AB2_tP6_137_a_d-poscar}ZrO$_{2}$ and HgI$_{2}$ have similar \AFLOW\ prototype labels ({\it{i.e.}}, same symmetry and set of Wyckoff positions with different stoichiometry labels due to alphabetic ordering of atomic species). They are generated by the same symmetry operations with different sets of parameters.}: {\small AB2\_tP6\_137\_a\_d} \dotfill {\hyperref[AB2_tP6_137_a_d_poscar]{\pageref{AB2_tP6_137_a_d_poscar}}} \\
\vspace{-0.75cm} \item HoCuP$_{2}$\footnoteref{note:ABC2_oC16_67_b_g_ag-poscar}: {\small ABC2\_oC16\_67\_b\_g\_ag} \dotfill {\hyperref[ABC2_oC16_67_b_g_ag_poscar]{\pageref{ABC2_oC16_67_b_g_ag_poscar}}} \\
\vspace{-0.75cm} \item Ir$_{3}$Ga$_{5}$: {\small A5B3\_tP32\_118\_g2i\_aceh} \dotfill {\hyperref[A5B3_tP32_118_g2i_aceh_poscar]{\pageref{A5B3_tP32_118_g2i_aceh_poscar}}} \\
\vspace{-0.75cm} \item Ir$_{3}$Ge$_{7}$: {\small A7B3\_cI40\_229\_df\_e} \dotfill {\hyperref[A7B3_cI40_229_df_e_poscar]{\pageref{A7B3_cI40_229_df_e_poscar}}} \\
\vspace{-0.75cm} \item IrGe$_{4}$: {\small A4B\_hP15\_144\_4a\_a} \dotfill {\hyperref[A4B_hP15_144_4a_a_poscar]{\pageref{A4B_hP15_144_4a_a_poscar}}} \\
\vspace{-0.75cm} \item K$_{2}$CdPb: {\small AB2C\_oC16\_40\_a\_2b\_b} \dotfill {\hyperref[AB2C_oC16_40_a_2b_b_poscar]{\pageref{AB2C_oC16_40_a_2b_b_poscar}}} \\
\vspace{-0.75cm} \item K$_{2}$PtCl$_{6}$: {\small A6B2C\_cF36\_225\_e\_c\_a} \dotfill {\hyperref[A6B2C_cF36_225_e_c_a_poscar]{\pageref{A6B2C_cF36_225_e_c_a_poscar}}} \\
\vspace{-0.75cm} \item K$_{2}$SnCl$_{6}$: {\small A6B2C\_tP18\_128\_eh\_d\_b} \dotfill {\hyperref[A6B2C_tP18_128_eh_d_b_poscar]{\pageref{A6B2C_tP18_128_eh_d_b_poscar}}} \\
\vspace{-0.75cm} \item K$_{2}$Ta$_{4}$O$_{9}$F$_{4}$: {\small A2B13C4\_hP57\_168\_d\_c6d\_2d} \dotfill {\hyperref[A2B13C4_hP57_168_d_c6d_2d_poscar]{\pageref{A2B13C4_hP57_168_d_c6d_2d_poscar}}} \\
\vspace{-0.75cm} \item KAg[CO$_{3}$]: {\small ABCD3\_oI48\_73\_d\_e\_e\_ef} \dotfill {\hyperref[ABCD3_oI48_73_d_e_e_ef_poscar]{\pageref{ABCD3_oI48_73_d_e_e_ef_poscar}}} \\
\vspace{-0.75cm} \item KAu$_{4}$Sn$_{2}$: {\small A4BC2\_tI28\_120\_i\_d\_e} \dotfill {\hyperref[A4BC2_tI28_120_i_d_e_poscar]{\pageref{A4BC2_tI28_120_i_d_e_poscar}}} \\
\vspace{-0.75cm} \item KB$_{6}$H$_{6}$: {\small A6B6C\_cF104\_202\_h\_h\_c} \dotfill {\hyperref[A6B6C_cF104_202_h_h_c_poscar]{\pageref{A6B6C_cF104_202_h_h_c_poscar}}} \\
\vspace{-0.75cm} \item KBO$_{2}$: {\small ABC2\_hR24\_167\_e\_e\_2e} \dotfill {\hyperref[ABC2_hR24_167_e_e_2e_poscar]{\pageref{ABC2_hR24_167_e_e_2e_poscar}}} \\
\vspace{-0.75cm} \item KCeSe$_{4}$: {\small ABC4\_tP12\_125\_a\_b\_m} \dotfill {\hyperref[ABC4_tP12_125_a_b_m_poscar]{\pageref{ABC4_tP12_125_a_b_m_poscar}}} \\
\vspace{-0.75cm} \item KHg$_{2}$: {\small A2B\_oI12\_74\_h\_e} \dotfill {\hyperref[A2B_oI12_74_h_e_poscar]{\pageref{A2B_oI12_74_h_e_poscar}}} \\
\vspace{-0.75cm} \item KNiCl$_{3}$: {\small A3BC\_hP30\_185\_cd\_c\_ab} \dotfill {\hyperref[A3BC_hP30_185_cd_c_ab_poscar]{\pageref{A3BC_hP30_185_cd_c_ab_poscar}}} \\
\vspace{-0.75cm} \item KSbO$_{3}$: {\small AB3C\_cP60\_201\_ce\_fh\_g} \dotfill {\hyperref[AB3C_cP60_201_ce_fh_g_poscar]{\pageref{AB3C_cP60_201_ce_fh_g_poscar}}} \\
\vspace{-0.75cm} \item La$_{2}$NiO$_{4}$: {\small A2BC4\_oP28\_50\_ij\_ac\_ijm} \dotfill {\hyperref[A2BC4_oP28_50_ij_ac_ijm_poscar]{\pageref{A2BC4_oP28_50_ij_ac_ijm_poscar}}} \\
\vspace{-0.75cm} \item La$_{2}$O$_{3}$: {\small A2B3\_hP5\_164\_d\_ad} \dotfill {\hyperref[A2B3_hP5_164_d_ad_poscar]{\pageref{A2B3_hP5_164_d_ad_poscar}}} \\
\vspace{-0.75cm} \item \begin{raggedleft}La$_{43}$Ni$_{17}$Mg$_{5}$: \end{raggedleft} \\ {\small A43B5C17\_oC260\_63\_c8fg6h\_cfg\_ce3f2h} \dotfill {\hyperref[A43B5C17_oC260_63_c8fg6h_cfg_ce3f2h_poscar]{\pageref{A43B5C17_oC260_63_c8fg6h_cfg_ce3f2h_poscar}}} \\
\vspace{-0.75cm} \item LaPtSi: {\small ABC\_tI12\_109\_a\_a\_a} \dotfill {\hyperref[ABC_tI12_109_a_a_a_poscar]{\pageref{ABC_tI12_109_a_a_a_poscar}}} \\
\vspace{-0.75cm} \item LaRhC$_{2}$: {\small A2BC\_tP16\_76\_2a\_a\_a} \dotfill {\hyperref[A2BC_tP16_76_2a_a_a_poscar]{\pageref{A2BC_tP16_76_2a_a_a_poscar}}} \\
\vspace{-0.75cm} \item Li$_{2}$MoF$_{6}$: {\small A6B2C\_tP18\_94\_eg\_c\_a} \dotfill {\hyperref[A6B2C_tP18_94_eg_c_a_poscar]{\pageref{A6B2C_tP18_94_eg_c_a_poscar}}} \\
\vspace{-0.75cm} \item Li$_{2}$Sb: {\small A2B\_hP18\_190\_gh\_bf} \dotfill {\hyperref[A2B_hP18_190_gh_bf_poscar]{\pageref{A2B_hP18_190_gh_bf_poscar}}} \\
\vspace{-0.75cm} \item Li$_{2}$Si$_{2}$O$_{5}$: {\small A2B5C2\_oC36\_37\_d\_c2d\_d} \dotfill {\hyperref[A2B5C2_oC36_37_d_c2d_d_poscar]{\pageref{A2B5C2_oC36_37_d_c2d_d_poscar}}} \\
\vspace{-0.75cm} \item LiScI$_{3}$: {\small A3BC\_hP10\_188\_k\_a\_e} \dotfill {\hyperref[A3BC_hP10_188_k_a_e_poscar]{\pageref{A3BC_hP10_188_k_a_e_poscar}}} \\
\vspace{-0.75cm} \item LiSn: {\small AB\_mP6\_10\_en\_am} \dotfill {\hyperref[AB_mP6_10_en_am_poscar]{\pageref{AB_mP6_10_en_am_poscar}}} \\
\vspace{-0.75cm} \item M-carbon: {\small A\_mC16\_12\_4i} \dotfill {\hyperref[A_mC16_12_4i_poscar]{\pageref{A_mC16_12_4i_poscar}}} \\
\vspace{-0.75cm} \item Mavlyanovite: {\small A5B3\_hP16\_193\_dg\_g} \dotfill {\hyperref[A5B3_hP16_193_dg_g_poscar]{\pageref{A5B3_hP16_193_dg_g_poscar}}} \\
\vspace{-0.75cm} \item Mg$_{2}$Zn$_{11}$: {\small A2B11\_cP39\_200\_f\_aghij} \dotfill {\hyperref[A2B11_cP39_200_f_aghij_poscar]{\pageref{A2B11_cP39_200_f_aghij_poscar}}} \\
\vspace{-0.75cm} \item \begin{raggedleft}MgB$_{12}$H$_{12}$[H$_{2}$O]$_{12}$: \end{raggedleft} \\ {\small A12B36CD12\_cF488\_196\_2h\_6h\_ac\_fgh} \dotfill {\hyperref[A12B36CD12_cF488_196_2h_6h_ac_fgh_poscar]{\pageref{A12B36CD12_cF488_196_2h_6h_ac_fgh_poscar}}} \\
\vspace{-0.75cm} \item MgSO$_{4}$: {\small AB4C\_oC24\_63\_a\_fg\_c} \dotfill {\hyperref[AB4C_oC24_63_a_fg_c_poscar]{\pageref{AB4C_oC24_63_a_fg_c_poscar}}} \\
\vspace{-0.75cm} \item Mg[NH]: {\small ABC\_hP36\_175\_jk\_jk\_jk} \dotfill {\hyperref[ABC_hP36_175_jk_jk_jk_poscar]{\pageref{ABC_hP36_175_jk_jk_jk_poscar}}} \\
\vspace{-0.75cm} \item Mn$_{2}$B: {\small AB2\_oF48\_70\_f\_fg} \dotfill {\hyperref[AB2_oF48_70_f_fg_poscar]{\pageref{AB2_oF48_70_f_fg_poscar}}} \\
\vspace{-0.75cm} \item MnAl$_{6}$: {\small A6B\_oC28\_63\_efg\_c} \dotfill {\hyperref[A6B_oC28_63_efg_c_poscar]{\pageref{A6B_oC28_63_efg_c_poscar}}} \\
\vspace{-0.75cm} \item MnF$_{2}$: {\small A2B\_tP12\_111\_2n\_adf} \dotfill {\hyperref[A2B_tP12_111_2n_adf_poscar]{\pageref{A2B_tP12_111_2n_adf_poscar}}} \\
\vspace{-0.75cm} \item MnGa$_{2}$Sb$_{2}$: {\small A2BC2\_oI20\_45\_c\_b\_c} \dotfill {\hyperref[A2BC2_oI20_45_c_b_c_poscar]{\pageref{A2BC2_oI20_45_c_b_c_poscar}}} \\
\vspace{-0.75cm} \item Mo$_{8}$P$_{5}$: {\small A8B5\_mP13\_6\_a7b\_3a2b} \dotfill {\hyperref[A8B5_mP13_6_a7b_3a2b_poscar]{\pageref{A8B5_mP13_6_a7b_3a2b_poscar}}} \\
\vspace{-0.75cm} \item MoS$_{2}$: {\small AB2\_hP12\_143\_cd\_ab2d} \dotfill {\hyperref[AB2_hP12_143_cd_ab2d_poscar]{\pageref{AB2_hP12_143_cd_ab2d_poscar}}} \\
\vspace{-0.75cm} \item Moissanite-15R: {\small AB\_hR10\_160\_5a\_5a} \dotfill {\hyperref[AB_hR10_160_5a_5a_poscar]{\pageref{AB_hR10_160_5a_5a_poscar}}} \\
\vspace{-0.75cm} \item Molybdite: {\small AB3\_oP16\_62\_c\_3c} \dotfill {\hyperref[AB3_oP16_62_c_3c_poscar]{\pageref{AB3_oP16_62_c_3c_poscar}}} \\
\vspace{-0.75cm} \item Muthmannite: {\small ABC2\_mP8\_10\_ac\_eh\_mn} \dotfill {\hyperref[ABC2_mP8_10_ac_eh_mn_poscar]{\pageref{ABC2_mP8_10_ac_eh_mn_poscar}}} \\
\vspace{-0.75cm} \item NV: {\small AB\_tP8\_111\_n\_n} \dotfill {\hyperref[AB_tP8_111_n_n_poscar]{\pageref{AB_tP8_111_n_n_poscar}}} \\
\vspace{-0.75cm} \item Na$_{3}$As\footnoteref{note:AB3_hP24_185_c_ab2c-poscar}: {\small AB3\_hP24\_185\_c\_ab2c} \dotfill {\hyperref[AB3_hP24_185_c_ab2c_poscar]{\pageref{AB3_hP24_185_c_ab2c_poscar}}} \\
\vspace{-0.75cm} \item \begin{raggedleft}Na$_{4}$Ti$_{2}$Si$_{8}$O$_{22}$[H$_{2}$O]$_{4}$: \end{raggedleft} \\ {\small A4B2C13D\_tP40\_90\_g\_d\_cef2g\_c} \dotfill {\hyperref[A4B2C13D_tP40_90_g_d_cef2g_c_poscar]{\pageref{A4B2C13D_tP40_90_g_d_cef2g_c_poscar}}} \\
\vspace{-0.75cm} \item Na$_{5}$Fe$_{3}$F$_{14}$: {\small A14B3C5\_tP44\_94\_c3g\_ad\_bg} \dotfill {\hyperref[A14B3C5_tP44_94_c3g_ad_bg_poscar]{\pageref{A14B3C5_tP44_94_c3g_ad_bg_poscar}}} \\
\vspace{-0.75cm} \item NaFeS$_{2}$: {\small ABC2\_oI16\_23\_ab\_i\_k} \dotfill {\hyperref[ABC2_oI16_23_ab_i_k_poscar]{\pageref{ABC2_oI16_23_ab_i_k_poscar}}} \\
\vspace{-0.75cm} \item NaGdCu$_{2}$F$_{8}$: {\small A2B8CD\_tI24\_97\_d\_k\_a\_b} \dotfill {\hyperref[A2B8CD_tI24_97_d_k_a_b_poscar]{\pageref{A2B8CD_tI24_97_d_k_a_b_poscar}}} \\
\vspace{-0.75cm} \item NaZn$_{13}$: {\small AB13\_cF112\_226\_a\_bi} \dotfill {\hyperref[AB13_cF112_226_a_bi_poscar]{\pageref{AB13_cF112_226_a_bi_poscar}}} \\
\vspace{-0.75cm} \item NaZn[OH]$_{3}$: {\small A3BC3D\_tP64\_106\_3c\_c\_3c\_c} \dotfill {\hyperref[A3BC3D_tP64_106_3c_c_3c_c_poscar]{\pageref{A3BC3D_tP64_106_3c_c_3c_c_poscar}}} \\
\vspace{-0.75cm} \item Nb$_{4}$CoSi: {\small AB4C\_tP12\_124\_a\_m\_c} \dotfill {\hyperref[AB4C_tP12_124_a_m_c_poscar]{\pageref{AB4C_tP12_124_a_m_c_poscar}}} \\
\vspace{-0.75cm} \item Nb$_{7}$Ru$_{6}$B$_{8}$: {\small A8B7C6\_hP21\_175\_ck\_aj\_k} \dotfill {\hyperref[A8B7C6_hP21_175_ck_aj_k_poscar]{\pageref{A8B7C6_hP21_175_ck_aj_k_poscar}}} \\
\vspace{-0.75cm} \item NbAs: {\small AB\_tI8\_109\_a\_a} \dotfill {\hyperref[AB_tI8_109_a_a_poscar]{\pageref{AB_tI8_109_a_a_poscar}}} \\
\vspace{-0.75cm} \item NbPS: {\small ABC\_oI12\_71\_h\_j\_g} \dotfill {\hyperref[ABC_oI12_71_h_j_g_poscar]{\pageref{ABC_oI12_71_h_j_g_poscar}}} \\
\vspace{-0.75cm} \item NbTe$_{4}$: {\small AB4\_tP10\_103\_a\_d} \dotfill {\hyperref[AB4_tP10_103_a_d_poscar]{\pageref{AB4_tP10_103_a_d_poscar}}} \\
\vspace{-0.75cm} \item NbTe$_{4}$: {\small AB4\_tP10\_124\_a\_m} \dotfill {\hyperref[AB4_tP10_124_a_m_poscar]{\pageref{AB4_tP10_124_a_m_poscar}}} \\
\vspace{-0.75cm} \item Ni$_{3}$P: {\small A3B\_tI32\_82\_3g\_g} \dotfill {\hyperref[A3B_tI32_82_3g_g_poscar]{\pageref{A3B_tI32_82_3g_g_poscar}}} \\
\vspace{-0.75cm} \item Ni$_{3}$Ti: {\small A3B\_hP16\_194\_gh\_ac} \dotfill {\hyperref[A3B_hP16_194_gh_ac_poscar]{\pageref{A3B_hP16_194_gh_ac_poscar}}} \\
\vspace{-0.75cm} \item Nierite: {\small A4B3\_hP28\_159\_ab2c\_2c} \dotfill {\hyperref[A4B3_hP28_159_ab2c_2c_poscar]{\pageref{A4B3_hP28_159_ab2c_2c_poscar}}} \\
\vspace{-0.75cm} \item PH$_{3}$: {\small A3B\_cP16\_208\_j\_b} \dotfill {\hyperref[A3B_cP16_208_j_b_poscar]{\pageref{A3B_cP16_208_j_b_poscar}}} \\
\vspace{-0.75cm} \item PI$_{3}$: {\small A3B\_hP8\_173\_c\_b} \dotfill {\hyperref[A3B_hP8_173_c_b_poscar]{\pageref{A3B_hP8_173_c_b_poscar}}} \\
\vspace{-0.75cm} \item Pd$_{17}$Se$_{15}$: {\small A17B15\_cP64\_207\_acfk\_eij} \dotfill {\hyperref[A17B15_cP64_207_acfk_eij_poscar]{\pageref{A17B15_cP64_207_acfk_eij_poscar}}} \\
\vspace{-0.75cm} \item Pd$_{4}$Se: {\small A4B\_tP10\_114\_e\_a} \dotfill {\hyperref[A4B_tP10_114_e_a_poscar]{\pageref{A4B_tP10_114_e_a_poscar}}} \\
\vspace{-0.75cm} \item PdSn$_{4}$: {\small AB4\_oC20\_68\_a\_i} \dotfill {\hyperref[AB4_oC20_68_a_i_poscar]{\pageref{AB4_oC20_68_a_i_poscar}}} \\
\vspace{-0.75cm} \item Petzite: {\small A3BC2\_cI48\_214\_f\_a\_e} \dotfill {\hyperref[A3BC2_cI48_214_f_a_e_poscar]{\pageref{A3BC2_cI48_214_f_a_e_poscar}}} \\
\vspace{-0.75cm} \item Phenakite: {\small A2B4C\_hR42\_148\_2f\_4f\_f} \dotfill {\hyperref[A2B4C_hR42_148_2f_4f_f_poscar]{\pageref{A2B4C_hR42_148_2f_4f_f_poscar}}} \\
\vspace{-0.75cm} \item Pinnoite: {\small A2B6CD7\_tP64\_77\_2d\_6d\_d\_ab6d} \dotfill {\hyperref[A2B6CD7_tP64_77_2d_6d_d_ab6d_poscar]{\pageref{A2B6CD7_tP64_77_2d_6d_d_ab6d_poscar}}} \\
\vspace{-0.75cm} \item Post-perovskite: {\small AB3C\_oC20\_63\_a\_cf\_c} \dotfill {\hyperref[AB3C_oC20_63_a_cf_c_poscar]{\pageref{AB3C_oC20_63_a_cf_c_poscar}}} \\
\vspace{-0.75cm} \item PrNiO$_{3}$: {\small AB3C\_hR10\_167\_b\_e\_a} \dotfill {\hyperref[AB3C_hR10_167_b_e_a_poscar]{\pageref{AB3C_hR10_167_b_e_a_poscar}}} \\
\vspace{-0.75cm} \item PrRu$_{4}$P$_{12}$: {\small A12BC4\_cP34\_195\_2j\_ab\_2e} \dotfill {\hyperref[A12BC4_cP34_195_2j_ab_2e_poscar]{\pageref{A12BC4_cP34_195_2j_ab_2e_poscar}}} \\
\vspace{-0.75cm} \item PtPb$_{4}$: {\small A4B\_tP10\_125\_m\_a} \dotfill {\hyperref[A4B_tP10_125_m_a_poscar]{\pageref{A4B_tP10_125_m_a_poscar}}} \\
\vspace{-0.75cm} \item Pyrite\footnote[3]{\label{note:AB2_oP12_29_a_2a-poscar}ZrO$_{2}$ and Pyrite have similar \AFLOW\ prototype labels ({\it{i.e.}}, same symmetry and set of Wyckoff positions with different stoichiometry labels due to alphabetic ordering of atomic species). They are generated by the same symmetry operations with different sets of parameters.}: {\small AB2\_oP12\_29\_a\_2a} \dotfill {\hyperref[AB2_oP12_29_a_2a_poscar]{\pageref{AB2_oP12_29_a_2a_poscar}}} \\
\vspace{-0.75cm} \item Pyrochlore: {\small A2BCD3E6\_cF208\_203\_e\_c\_d\_f\_g} \dotfill {\hyperref[A2BCD3E6_cF208_203_e_c_d_f_g_poscar]{\pageref{A2BCD3E6_cF208_203_e_c_d_f_g_poscar}}} \\
\vspace{-0.75cm} \item \begin{raggedleft}Pyrochlore Iridate: \end{raggedleft} \\ {\small A2B2C7\_cF88\_227\_c\_d\_af} \dotfill {\hyperref[A2B2C7_cF88_227_c_d_af_poscar]{\pageref{A2B2C7_cF88_227_c_d_af_poscar}}} \\
\vspace{-0.75cm} \item \begin{raggedleft}Quartenary Heusler: \end{raggedleft} \\ {\small ABCD\_cF16\_216\_c\_d\_b\_a} \dotfill {\hyperref[ABCD_cF16_216_c_d_b_a_poscar]{\pageref{ABCD_cF16_216_c_d_b_a_poscar}}} \\
\vspace{-0.75cm} \item R-carbon: {\small A\_oP16\_55\_2g2h} \dotfill {\hyperref[A_oP16_55_2g2h_poscar]{\pageref{A_oP16_55_2g2h_poscar}}} \\
\vspace{-0.75cm} \item Rasvumite: {\small A2BC3\_oC24\_63\_e\_c\_cg} \dotfill {\hyperref[A2BC3_oC24_63_e_c_cg_poscar]{\pageref{A2BC3_oC24_63_e_c_cg_poscar}}} \\
\vspace{-0.75cm} \item Rb$_{2}$TiCu$_{2}$S$_{4}$: {\small A2B2C4D\_tP18\_132\_e\_i\_o\_d} \dotfill {\hyperref[A2B2C4D_tP18_132_e_i_o_d_poscar]{\pageref{A2B2C4D_tP18_132_e_i_o_d_poscar}}} \\
\vspace{-0.75cm} \item Rb$_{3}$AsSe$_{16}$: {\small AB3C16\_cF160\_203\_b\_ad\_eg} \dotfill {\hyperref[AB3C16_cF160_203_b_ad_eg_poscar]{\pageref{AB3C16_cF160_203_b_ad_eg_poscar}}} \\
\vspace{-0.75cm} \item RbGa$_{3}$: {\small A3B\_tI24\_119\_b2i\_af} \dotfill {\hyperref[A3B_tI24_119_b2i_af_poscar]{\pageref{A3B_tI24_119_b2i_af_poscar}}} \\
\vspace{-0.75cm} \item Re$_{2}$O$_{5}$[SO$_{4}$]$_{2}$: {\small A13B2C2\_oP34\_32\_a6c\_c\_c} \dotfill {\hyperref[A13B2C2_oP34_32_a6c_c_c_poscar]{\pageref{A13B2C2_oP34_32_a6c_c_c_poscar}}} \\
\vspace{-0.75cm} \item Re$_{3}$N: {\small AB3\_hP4\_187\_e\_fh} \dotfill {\hyperref[AB3_hP4_187_e_fh_poscar]{\pageref{AB3_hP4_187_e_fh_poscar}}} \\
\vspace{-0.75cm} \item Rh$_{2}$Ga$_{9}$: {\small A9B2\_mP22\_7\_9a\_2a} \dotfill {\hyperref[A9B2_mP22_7_9a_2a_poscar]{\pageref{A9B2_mP22_7_9a_2a_poscar}}} \\
\vspace{-0.75cm} \item Rh$_{2}$S$_{3}$: {\small A2B3\_oP20\_60\_d\_cd} \dotfill {\hyperref[A2B3_oP20_60_d_cd_poscar]{\pageref{A2B3_oP20_60_d_cd_poscar}}} \\
\vspace{-0.75cm} \item Rh$_{3}$P$_{2}$: {\small A2B3\_tP5\_115\_g\_ag} \dotfill {\hyperref[A2B3_tP5_115_g_ag_poscar]{\pageref{A2B3_tP5_115_g_ag_poscar}}} \\
\vspace{-0.75cm} \item Rh$_{5}$Ge$_{3}$: {\small A3B5\_oP16\_55\_ch\_agh} \dotfill {\hyperref[A3B5_oP16_55_ch_agh_poscar]{\pageref{A3B5_oP16_55_ch_agh_poscar}}} \\
\vspace{-0.75cm} \item Ru$_{2}$Sn$_{3}$: {\small A2B3\_tP20\_116\_bci\_fj} \dotfill {\hyperref[A2B3_tP20_116_bci_fj_poscar]{\pageref{A2B3_tP20_116_bci_fj_poscar}}} \\
\vspace{-0.75cm} \item RuIn$_{3}$: {\small A3B\_tP16\_118\_ei\_f} \dotfill {\hyperref[A3B_tP16_118_ei_f_poscar]{\pageref{A3B_tP16_118_ei_f_poscar}}} \\
\vspace{-0.75cm} \item S-II: {\small A\_hP9\_154\_bc} \dotfill {\hyperref[A_hP9_154_bc_poscar]{\pageref{A_hP9_154_bc_poscar}}} \\
\vspace{-0.75cm} \item S-III: {\small A\_tI16\_142\_f} \dotfill {\hyperref[A_tI16_142_f_poscar]{\pageref{A_tI16_142_f_poscar}}} \\
\vspace{-0.75cm} \item S-carbon: {\small A\_mP8\_10\_2m2n} \dotfill {\hyperref[A_mP8_10_2m2n_poscar]{\pageref{A_mP8_10_2m2n_poscar}}} \\
\vspace{-0.75cm} \item Sc-V: {\small A\_hP6\_178\_a} \dotfill {\hyperref[A_hP6_178_a_poscar]{\pageref{A_hP6_178_a_poscar}}} \\
\vspace{-0.75cm} \item ScRh$_{6}$P$_{4}$: {\small A4B6C\_hP11\_143\_bd\_2d\_a} \dotfill {\hyperref[A4B6C_hP11_143_bd_2d_a_poscar]{\pageref{A4B6C_hP11_143_bd_2d_a_poscar}}} \\
\vspace{-0.75cm} \item SeO$_{3}$: {\small A3B\_tP32\_114\_3e\_e} \dotfill {\hyperref[A3B_tP32_114_3e_e_poscar]{\pageref{A3B_tP32_114_3e_e_poscar}}} \\
\vspace{-0.75cm} \item \begin{raggedleft}Sheldrickite: \end{raggedleft} \\ {\small A2B3C3DE7\_hP48\_145\_2a\_3a\_3a\_a\_7a} \dotfill {\hyperref[A2B3C3DE7_hP48_145_2a_3a_3a_a_7a_poscar]{\pageref{A2B3C3DE7_hP48_145_2a_3a_3a_a_7a_poscar}}} \\
\vspace{-0.75cm} \item SiO$_{2}$: {\small A2B\_hP36\_177\_j2lm\_n} \dotfill {\hyperref[A2B_hP36_177_j2lm_n_poscar]{\pageref{A2B_hP36_177_j2lm_n_poscar}}} \\
\vspace{-0.75cm} \item SiO$_{2}$: {\small A2B\_cI72\_211\_hi\_i} \dotfill {\hyperref[A2B_cI72_211_hi_i_poscar]{\pageref{A2B_cI72_211_hi_i_poscar}}} \\
\vspace{-0.75cm} \item \begin{raggedleft}Simple Cubic C$_{60}$ Buckminsterfullerine: \end{raggedleft} \\ {\small A\_cP240\_205\_10d} \dotfill {\hyperref[A_cP240_205_10d_poscar]{\pageref{A_cP240_205_10d_poscar}}} \\
\vspace{-0.75cm} \item Simpsonite: {\small A4B14C3\_hP21\_143\_bd\_ac4d\_d} \dotfill {\hyperref[A4B14C3_hP21_143_bd_ac4d_d_poscar]{\pageref{A4B14C3_hP21_143_bd_ac4d_d_poscar}}} \\
\vspace{-0.75cm} \item SmSI: {\small ABC\_hR6\_166\_c\_c\_c} \dotfill {\hyperref[ABC_hR6_166_c_c_c_poscar]{\pageref{ABC_hR6_166_c_c_c_poscar}}} \\
\vspace{-0.75cm} \item Sodium Chlorate: {\small ABC3\_cP20\_198\_a\_a\_b} \dotfill {\hyperref[ABC3_cP20_198_a_a_b_poscar]{\pageref{ABC3_cP20_198_a_a_b_poscar}}} \\
\vspace{-0.75cm} \item Spinel: {\small A3B4\_cF56\_227\_ad\_e} \dotfill {\hyperref[A3B4_cF56_227_ad_e_poscar]{\pageref{A3B4_cF56_227_ad_e_poscar}}} \\
\vspace{-0.75cm} \item Sr$_{2}$As$_{2}$O$_{7}$: {\small A2B7C2\_tP88\_78\_4a\_14a\_4a} \dotfill {\hyperref[A2B7C2_tP88_78_4a_14a_4a_poscar]{\pageref{A2B7C2_tP88_78_4a_14a_4a_poscar}}} \\
\vspace{-0.75cm} \item Sr$_{2}$Bi$_{3}$: {\small A3B2\_oP20\_52\_de\_cd} \dotfill {\hyperref[A3B2_oP20_52_de_cd_poscar]{\pageref{A3B2_oP20_52_de_cd_poscar}}} \\
\vspace{-0.75cm} \item Sr$_{5}$Si$_{3}$: {\small A3B5\_tI32\_108\_ac\_a2c} \dotfill {\hyperref[A3B5_tI32_108_ac_a2c_poscar]{\pageref{A3B5_tI32_108_ac_a2c_poscar}}} \\
\vspace{-0.75cm} \item SrAl$_{2}$Se$_{4}$: {\small A2B4C\_oC28\_66\_l\_kl\_a} \dotfill {\hyperref[A2B4C_oC28_66_l_kl_a_poscar]{\pageref{A2B4C_oC28_66_l_kl_a_poscar}}} \\
\vspace{-0.75cm} \item SrBr$_{2}$: {\small A2B\_tP30\_85\_ab2g\_cg} \dotfill {\hyperref[A2B_tP30_85_ab2g_cg_poscar]{\pageref{A2B_tP30_85_ab2g_cg_poscar}}} \\
\vspace{-0.75cm} \item SrH$_{2}$: {\small A2B\_oP12\_62\_2c\_c} \dotfill {\hyperref[A2B_oP12_62_2c_c_poscar]{\pageref{A2B_oP12_62_2c_c_poscar}}} \\
\vspace{-0.75cm} \item SrSi$_{2}$: {\small A2B\_cP12\_212\_c\_a} \dotfill {\hyperref[A2B_cP12_212_c_a_poscar]{\pageref{A2B_cP12_212_c_a_poscar}}} \\
\vspace{-0.75cm} \item Sr[S$_{2}$O$_{6}$][H$_{2}$O]$_{4}$: {\small A10B2C\_hP39\_171\_5c\_c\_a} \dotfill {\hyperref[A10B2C_hP39_171_5c_c_a_poscar]{\pageref{A10B2C_hP39_171_5c_c_a_poscar}}} \\
\vspace{-0.75cm} \item Sr[S$_{2}$O$_{6}$][H$_{2}$O]$_{4}$: {\small A10B2C\_hP39\_172\_5c\_c\_a} \dotfill {\hyperref[A10B2C_hP39_172_5c_c_a_poscar]{\pageref{A10B2C_hP39_172_5c_c_a_poscar}}} \\
\vspace{-0.75cm} \item \begin{raggedleft}Stannoidite: \end{raggedleft} \\ {\small A8B2C12D2E\_oI50\_23\_bcfk\_i\_3k\_j\_a} \dotfill {\hyperref[A8B2C12D2E_oI50_23_bcfk_i_3k_j_a_poscar]{\pageref{A8B2C12D2E_oI50_23_bcfk_i_3k_j_a_poscar}}} \\
\vspace{-0.75cm} \item Ta$_{2}$H: {\small AB2\_oC6\_21\_a\_k} \dotfill {\hyperref[AB2_oC6_21_a_k_poscar]{\pageref{AB2_oC6_21_a_k_poscar}}} \\
\vspace{-0.75cm} \item Ta$_{2}$Se$_{8}$I: {\small AB8C2\_tI44\_97\_e\_2k\_cd} \dotfill {\hyperref[AB8C2_tI44_97_e_2k_cd_poscar]{\pageref{AB8C2_tI44_97_e_2k_cd_poscar}}} \\
\vspace{-0.75cm} \item Ta$_{3}$B$_{4}$: {\small A4B3\_oI14\_71\_gh\_cg} \dotfill {\hyperref[A4B3_oI14_71_gh_cg_poscar]{\pageref{A4B3_oI14_71_gh_cg_poscar}}} \\
\vspace{-0.75cm} \item Ta$_{3}$S$_{2}$: {\small A2B3\_oC40\_39\_2d\_2c2d} \dotfill {\hyperref[A2B3_oC40_39_2d_2c2d_poscar]{\pageref{A2B3_oC40_39_2d_2c2d_poscar}}} \\
\vspace{-0.75cm} \item TaNiTe$_{2}$: {\small ABC2\_oP16\_53\_h\_e\_gh} \dotfill {\hyperref[ABC2_oP16_53_h_e_gh_poscar]{\pageref{ABC2_oP16_53_h_e_gh_poscar}}} \\
\vspace{-0.75cm} \item TeO$_{6}$H$_{6}$: {\small A6B\_cF224\_228\_h\_c} \dotfill {\hyperref[A6B_cF224_228_h_c_poscar]{\pageref{A6B_cF224_228_h_c_poscar}}} \\
\vspace{-0.75cm} \item TeZn: {\small AB\_hP6\_144\_a\_a} \dotfill {\hyperref[AB_hP6_144_a_a_poscar]{\pageref{AB_hP6_144_a_a_poscar}}} \\
\vspace{-0.75cm} \item Te[OH]$_{6}$: {\small A12B6C\_cF608\_210\_4h\_2h\_e} \dotfill {\hyperref[A12B6C_cF608_210_4h_2h_e_poscar]{\pageref{A12B6C_cF608_210_4h_2h_e_poscar}}} \\
\vspace{-0.75cm} \item Th$_{3}$P$_{4}$: {\small A4B3\_cI28\_220\_c\_a} \dotfill {\hyperref[A4B3_cI28_220_c_a_poscar]{\pageref{A4B3_cI28_220_c_a_poscar}}} \\
\vspace{-0.75cm} \item Th$_{6}$Mn$_{23}$: {\small A23B6\_cF116\_225\_bd2f\_e} \dotfill {\hyperref[A23B6_cF116_225_bd2f_e_poscar]{\pageref{A23B6_cF116_225_bd2f_e_poscar}}} \\
\vspace{-0.75cm} \item ThB$_{4}$: {\small A4B\_tP20\_127\_ehj\_g} \dotfill {\hyperref[A4B_tP20_127_ehj_g_poscar]{\pageref{A4B_tP20_127_ehj_g_poscar}}} \\
\vspace{-0.75cm} \item ThBC: {\small ABC\_tP24\_91\_d\_d\_d} \dotfill {\hyperref[ABC_tP24_91_d_d_d_poscar]{\pageref{ABC_tP24_91_d_d_d_poscar}}} \\
\vspace{-0.75cm} \item ThBC: {\small ABC\_tP24\_95\_d\_d\_d} \dotfill {\hyperref[ABC_tP24_95_d_d_d_poscar]{\pageref{ABC_tP24_95_d_d_d_poscar}}} \\
\vspace{-0.75cm} \item ThCl$_{4}$: {\small A4B\_tI20\_88\_f\_a} \dotfill {\hyperref[A4B_tI20_88_f_a_poscar]{\pageref{A4B_tI20_88_f_a_poscar}}} \\
\vspace{-0.75cm} \item Thortveitite: {\small A7B2C2\_mC22\_12\_aij\_h\_i} \dotfill {\hyperref[A7B2C2_mC22_12_aij_h_i_poscar]{\pageref{A7B2C2_mC22_12_aij_h_i_poscar}}} \\
\vspace{-0.75cm} \item Ti$_{2}$Ge$_{3}$: {\small A3B2\_tP10\_83\_adk\_j} \dotfill {\hyperref[A3B2_tP10_83_adk_j_poscar]{\pageref{A3B2_tP10_83_adk_j_poscar}}} \\
\vspace{-0.75cm} \item Ti$_{3}$O: {\small AB3\_hP24\_149\_acgi\_3l} \dotfill {\hyperref[AB3_hP24_149_acgi_3l_poscar]{\pageref{AB3_hP24_149_acgi_3l_poscar}}} \\
\vspace{-0.75cm} \item Ti$_{3}$P: {\small AB3\_tP32\_86\_g\_3g} \dotfill {\hyperref[AB3_tP32_86_g_3g_poscar]{\pageref{AB3_tP32_86_g_3g_poscar}}} \\
\vspace{-0.75cm} \item TiAl$_{2}$Br$_{8}$: {\small A2B8C\_oP22\_34\_c\_4c\_a} \dotfill {\hyperref[A2B8C_oP22_34_c_4c_a_poscar]{\pageref{A2B8C_oP22_34_c_4c_a_poscar}}} \\
\vspace{-0.75cm} \item TiFeSi: {\small ABC\_oI36\_46\_ac\_bc\_3b} \dotfill {\hyperref[ABC_oI36_46_ac_bc_3b_poscar]{\pageref{ABC_oI36_46_ac_bc_3b_poscar}}} \\
\vspace{-0.75cm} \item Tl$_{4}$HgI$_{6}$: {\small AB6C4\_tP22\_104\_a\_2ac\_c} \dotfill {\hyperref[AB6C4_tP22_104_a_2ac_c_poscar]{\pageref{AB6C4_tP22_104_a_2ac_c_poscar}}} \\
\vspace{-0.75cm} \item TlP$_{5}$: {\small A5B\_oP24\_26\_3a3b2c\_ab} \dotfill {\hyperref[A5B_oP24_26_3a3b2c_ab_poscar]{\pageref{A5B_oP24_26_3a3b2c_ab_poscar}}} \\
\vspace{-0.75cm} \item TlZn$_{2}$Sb$_{2}$: {\small A2BC2\_tI20\_79\_c\_2a\_c} \dotfill {\hyperref[A2BC2_tI20_79_c_2a_c_poscar]{\pageref{A2BC2_tI20_79_c_2a_c_poscar}}} \\
\vspace{-0.75cm} \item Tongbaite: {\small A2B3\_oP20\_62\_2c\_3c} \dotfill {\hyperref[A2B3_oP20_62_2c_3c_poscar]{\pageref{A2B3_oP20_62_2c_3c_poscar}}} \\
\vspace{-0.75cm} \item Troilite: {\small AB\_hP24\_190\_i\_afh} \dotfill {\hyperref[AB_hP24_190_i_afh_poscar]{\pageref{AB_hP24_190_i_afh_poscar}}} \\
\vspace{-0.75cm} \item Tychite: {\small A4B2C6D16E\_cF232\_203\_e\_d\_f\_eg\_a} \dotfill {\hyperref[A4B2C6D16E_cF232_203_e_d_f_eg_a_poscar]{\pageref{A4B2C6D16E_cF232_203_e_d_f_eg_a_poscar}}} \\
\vspace{-0.75cm} \item UCl$_{3}$: {\small A3B\_hP8\_176\_h\_d} \dotfill {\hyperref[A3B_hP8_176_h_d_poscar]{\pageref{A3B_hP8_176_h_d_poscar}}} \\
\vspace{-0.75cm} \item V$_{2}$MoO$_{8}$: {\small AB8C2\_oC22\_35\_a\_ab3e\_e} \dotfill {\hyperref[AB8C2_oC22_35_a_ab3e_e_poscar]{\pageref{AB8C2_oC22_35_a_ab3e_e_poscar}}} \\
\vspace{-0.75cm} \item VPCl$_{9}$: {\small A9BC\_oC44\_39\_3c3d\_a\_c} \dotfill {\hyperref[A9BC_oC44_39_3c3d_a_c_poscar]{\pageref{A9BC_oC44_39_3c3d_a_c_poscar}}} \\
\vspace{-0.75cm} \item W$_{3}$O$_{10}$: {\small A10B3\_oF52\_42\_2abce\_ab} \dotfill {\hyperref[A10B3_oF52_42_2abce_ab_poscar]{\pageref{A10B3_oF52_42_2abce_ab_poscar}}} \\
\vspace{-0.75cm} \item W$_{5}$Si$_{3}$: {\small A3B5\_tI32\_140\_ah\_bk} \dotfill {\hyperref[A3B5_tI32_140_ah_bk_poscar]{\pageref{A3B5_tI32_140_ah_bk_poscar}}} \\
\vspace{-0.75cm} \item WO$_{3}$: {\small A3B\_oP32\_60\_3d\_d} \dotfill {\hyperref[A3B_oP32_60_3d_d_poscar]{\pageref{A3B_oP32_60_3d_d_poscar}}} \\
\vspace{-0.75cm} \item Weberite: {\small AB7CD2\_oI44\_24\_a\_b3d\_c\_ac} \dotfill {\hyperref[AB7CD2_oI44_24_a_b3d_c_ac_poscar]{\pageref{AB7CD2_oI44_24_a_b3d_c_ac_poscar}}} \\
\vspace{-0.75cm} \item Westerveldite: {\small AB\_oP8\_62\_c\_c} \dotfill {\hyperref[AB_oP8_62_c_c_poscar]{\pageref{AB_oP8_62_c_c_poscar}}} \\
\vspace{-0.75cm} \item YbBaCo$_{4}$O$_{7}$: {\small AB4C7D\_hP26\_159\_b\_ac\_a2c\_b} \dotfill {\hyperref[AB4C7D_hP26_159_b_ac_a2c_b_poscar]{\pageref{AB4C7D_hP26_159_b_ac_a2c_b_poscar}}} \\
\vspace{-0.75cm} \item Zn$_{3}$P$_{2}$: {\small A2B3\_tP40\_137\_cdf\_3g} \dotfill {\hyperref[A2B3_tP40_137_cdf_3g_poscar]{\pageref{A2B3_tP40_137_cdf_3g_poscar}}} \\
\vspace{-0.75cm} \item ZnSb$_{2}$O$_{4}$: {\small A4B2C\_tP28\_135\_gh\_h\_d} \dotfill {\hyperref[A4B2C_tP28_135_gh_h_d_poscar]{\pageref{A4B2C_tP28_135_gh_h_d_poscar}}} \\
\vspace{-0.75cm} \item ZrO$_{2}$\footnoteref{note:AB2_oP12_29_a_2a-poscar}: {\small A2B\_oP12\_29\_2a\_a} \dotfill {\hyperref[A2B_oP12_29_2a_a_poscar]{\pageref{A2B_oP12_29_2a_a_poscar}}} \\
\vspace{-0.75cm} \item ZrO$_{2}$\footnoteref{note:AB2_tP6_137_a_d-poscar}: {\small A2B\_tP6\_137\_d\_a} \dotfill {\hyperref[A2B_tP6_137_d_a_poscar]{\pageref{A2B_tP6_137_d_a_poscar}}} \\
\end{enumerate}
\vfill
\renewcommand{\thefootnote}{\arabic{footnote}}

\bibliographystyle{PhysRevwithTitles_DOI_v1b}

\nobibliography{xextra,xstefano-20180531}

\nobibliography{manuscript.bbl}
\end{document}